\documentclass{eczhandbook}

\flmFinalPreambleSetup

\input{EcZooHandbook.eczaux}

\eczhbkSetGitVersion\eczhbkGitVersionData{2026-06-08 0b2ff250c5aff9cf25d2957b0c7e2a84caa9d4d9}
\eczhbkSetGitVersion\eczhbkGitVersionSitegen{2026-06-09 6bbebab66fb95072b377f61196580e16ce9f3866}
\eczhbkSetGitVersion\eczhbkGitVersionHandbook{2026-06-09 c73b681977bb40918d7eae6601fe8d2d027cbcf0}

\begin{document}
\title{Handbook of Error-Correcting Codes}
\author{Victor V. Albert}
\author{Philippe Faist}
\date{Jun 9, 2026}
\eczmaketitlepage

%
%

\def\fineprintvgap{\vspace{12pt}}

\begin{eczhbkInitialPages}
\begin{eczhbkInitialPage}

The error correction zoo can be navigated online at \url{https://errorcorrectionzoo.org/}.

All data and sources available at
\url{https://github.com/errorcorrectionzoo/}.

\fineprintvgap

{
\eczhbkGitVersionInTabular
\begin{tabular}{l@{\hspace*{8pt}}lll}
  \texttt{\detokenize{eczoo_data}} & \eczhbkGitVersionData\\
  \texttt{\detokenize{eczoo_sitegen}} & \eczhbkGitVersionSitegen\\
  \texttt{\detokenize{eczoo_handbook}} &\eczhbkGitVersionHandbook
\end{tabular}\par}

\fineprintvgap

{
  Certain products, commercial and otherwise, are mentioned in this publication. These mentions are for 
informational purposes only, and do not imply recommendation or endorsement by NIST.\par}

\fineprintvgap

This work is distributed under CC-BY-SA 4.0.  See
\url{https://creativecommons.org/licenses/by-sa/4.0/}

\end{eczhbkInitialPage}
\begin{eczhbkInitialPage}[t]

\large\raggedleft

\vspace*{72pt}

{\Large\bfseries
  Handbook of Error-Correcting Codes}
\vspace{32pt}

Victor V. Albert\\[1ex]
\emph{Joint Center for Quantum Information and Computer Science\\
  NIST/University of Maryland, College Park, MD 20742, USA}

\vspace{24pt}

Philippe Faist\\[1ex]
\emph{Dahlem Center for Complex Quantum Systems\\
  Freie Universit\"at Berlin, 14195 Berlin, Germany}


\end{eczhbkInitialPage}
\begin{eczhbkInitialPage}[cb]

{\centering\itshape  Special Acknowledgments
\par}

\fineprintvgap

V.V.A.\@ thanks Olga Albert, Ryhor Kandratsenia, as well as Tatyana and Thomas Albert for
providing the support necessary to perform this work.

Ph.F.\@ thanks Evita Varela, Corinne Kassapoglou Faist, Jérôme Faist, Anna Vitali, and
Georgios Varelas for providing the support necessary to perform this work.

\end{eczhbkInitialPage}
\end{eczhbkInitialPages}

\part{Welcome to the Error Correction Zoo Handbook}
\twocolumngrid

\begin{quote}
\flmQuoteSetup{quote}%
All animals are equal, but some animals are more equal than others.
\flmQuoteAttributed{George Orwell}
\end{quote}

\section{Why a Zoo?}

You can extract a URL by taking a picture of a checkerboard-like grid called a QR code, even when the picture is blurry. You can look up the price of a grocery item by scanning its universal product barcode. You can ensure that a file was downloaded correctly by verifying it against the file's checksum. These are all examples of error-detecting or error-correcting encodings, or codes for short.

Error-correcting codes detect and correct errors by redundantly encoding a “message” into a larger “alphabet”. A QR code encodes a URL into a binary string arranged in a checkerboard-like grid. A checksum encoding of a file consists of the file together with an error-detecting checksum. The International Radiotelephony Spelling Alphabet encodes each spoken letter into a word: A \(\to\) “Alfa”, B \(\to\) “Bravo”, and so on and so forth.

In the error-correcting repetition code, a one-bit message is encoded into the three-bit alphabet as \(0 \to 000\) and \(1 \to 111\). When these three bits are transmitted through a noisy channel that flips one of the bits, the receiver is able to discern the message by taking the larger of the number of zeroes and ones she received.

There are codes for communicating information over fiber, satellite, broadcast, cellular, and wireless channels. There are codes for storing information on hard drives, compact disks, in datacenters, and on USB drives. Codes also protect classical and quantum information communicated over a quantum link or stored in a quantum computer.

Coding theory~\NoCaseChange{\protect\cite{cite1}} began in the late 1940s and was followed in the 1960s by quantum estimation theory~\NoCaseChange{\protect\cite{cite2}} --- the study of (classical) information transmission over quantum devices. The ability to robustly store and process quantum information was discovered in 1995--1996~\NoCaseChange{\protect\cite{cite3}}. Countless codes have been proposed using insights from computer science, mathematics, physics, engineering, and even biology.

The ability to coherently manipulate individual quantum systems is ushering in a
new era of quantum technologies and quantum computing.  Quantum error correction
plays a central role in this development, given that quantum information is
notoriously fragile and difficult to preserve over time.  While borrowing many
concepts from classical error correction, the field's quantum counterpart faces
challenges, such as the no-cloning theorem, that are not present in the
classical world.

Certain codes excel at packing information densely on the smallest physical
supports.  Some have blazingly fast encoding and decoding procedures.  Others
offer the ability to easily perform logical operations on the encoded data.
Codes can exploit features of the error model to squeeze out more performance.
Other codes have elegant mathematical formulations that facilitate their
analysis or that draw tight connections with abstract mathematical constructs,
condensed matter physics, or even quantum gravity.

There is no one code to rule them all, and codes are typically chosen on a
case-by-case basis for a specific application, with considerations including
physical support, type and magnitude of noise, and desired performance.
Wouldn't it be great if there were a place where you could look up which code to use?


\section{What is the Zoo and who should use it?}

The Error Correction Zoo 
aims to be a living and breathing catalogue of error-correcting codes and related structures: 
sphere packings, lattices, designs, groups, and classical and quantum phases of matter.

A qualitative companion to numerical tables, the EC Zoo contains
\begin{enumerate}\item succinct high-level descriptions 
of code constructions, features, historical context, and practical uses;
    \item pointers to technical details;
    \item a taxonomic hierarchy revealing the relations between different codes;
      and 
\item a structured repository with easily maintainable data files and
      advanced automatic processing of the data files into user-friendly
      formats, including an interactive website and a print handbook.
\end{enumerate}

The EC Zoo was inspired by the
Complexity Zoo~\NoCaseChange{\protect\cite{cite4}} and the
Quantum Algorithm Zoo~\NoCaseChange{\protect\cite{cite5}}.  What sets it
apart is that data is stored in a structured format, the YAML file format~\NoCaseChange{\protect\cite{cite6}},
which is readable by
\textit{both} humans and machines.  
Humans can populate the data by editing these files manually.
Computers can process it into webpages, this PDF handbook, and visual outputs such as graphs illustrating the codes' relations.
Artificially intelligent agents can pick up on this data to provide more accurate prompt responses and, eventually, help
humans contribute to the EC Zoo~\NoCaseChange{\protect\cite{cite7}}.

The EC Zoo organizes the literature on codes in a modular fashion, with links to thousands of academic papers compartmentalized among roughly a thousand code entries.

If you are working on a new code and are worried it is already known, if you are interested in how codes relate to deep concepts in science and mathematics, or if you just want to surf the highly interconnected code graph, the EC Zoo is for you!





















\section{What makes a Zoo entry?}

Code entries constitute the basic objects of the EC Zoo and are added based on
community interest.  Entries range from specific well-known codes, such as
\flmRefsHyperref{code:hamming}{Hamming codes} or \flmRefsHyperref{code:shor_nine}{Shor codes},
to general code constructions, such as
\flmRefsHyperref{code:q-ary_linear}{linear codes} or
\flmRefsHyperref{code:qubit_stabilizer}{stabilizer codes}, to codes with certain properties,
such as \flmRefsHyperref{code:perfect}{perfect codes} or 
\flmRefsHyperref{code:quantum_perfect}{perfect quantum codes}.  In this way, the EC Zoo
grows either as special cases of particular codes are studied in greater depth,
or as codes unifying disparate families are constructed.

Code entries are stored as YAML files in a structured data format on
\flmHref{https://github.com/errorcorrectionzoo/eczoo_data}{Github}.  Data
is organized into fixed field names (or \emph{keys}) and associated field
\emph{values}, which are provided as free-form text.  The text is written in a
custom LaTeX-inspired language and may contain math, citations,
figures, tables, and references to other codes.
The possible data fields for a code entry
are as follows:
\begin{description}\item[{\EscVerb[formatcom=\flmFmtVRB{verba}]{code\_id}}] The all-lower-case unique code database identifier;
\item[{\EscVerb[formatcom=\flmFmtVRB{verba}]{physical}}] The alphabet (for quantum,
physical space) over which the code is defined;
\item[{\EscVerb[formatcom=\flmFmtVRB{verba}]{logical}}] The alphabet (for quantum, logical space) over which the
  message is defined;
\item[{\EscVerb[formatcom=\flmFmtVRB{verba}]{name}}] Primary name for the code;
\item[{\EscVerb[formatcom=\flmFmtVRB{verba}]{alternative\_names}}] Other names used for the code;
\item[{\EscVerb[formatcom=\flmFmtVRB{verba}]{description}}] A short paragraph succinctly describing the
  code, followed by additional details;
\item[{\EscVerb[formatcom=\flmFmtVRB{verba}]{protection}}] Noise model(s) protected by the code, code
  distance(s), etc.;
\item[{\EscVerb[formatcom=\flmFmtVRB{verba}]{rate}}] Density of storage of the code, asymptotic or otherwise;
\item[{\EscVerb[formatcom=\flmFmtVRB{verba}]{encoders}}] List of encoding maps for the code;
\item[{\EscVerb[formatcom=\flmFmtVRB{verba}]{general\_gates}}] List of logical gates that can be performed on the encoded
  information;
\item[{\EscVerb[formatcom=\flmFmtVRB{verba}]{decoders}}] List of recipes for decoding the message after noise;
\item[{\EscVerb[formatcom=\flmFmtVRB{verba}]{fault\_tolerance}}] List of code gadgets (encoders, gates, decoders,
  etc.) that are fault tolerant to imperfections;
\item[{\EscVerb[formatcom=\flmFmtVRB{verba}]{realizations}}] List of realizations of the code in the real world;
\item[{\EscVerb[formatcom=\flmFmtVRB{verba}]{notes}}] List of expository references, code databases,
  implementations, popular summaries, other notable facts, etc.;
\item[{\EscVerb[formatcom=\flmFmtVRB{verba}]{relations\/parents}}] List of \EscVerb[formatcom=\flmFmtVRB{verba}]{code\_id}'s and details of codes
  that include the code as a special case;
\item[{\EscVerb[formatcom=\flmFmtVRB{verba}]{relations\/cousins}}] List of \EscVerb[formatcom=\flmFmtVRB{verba}]{code\_id}'s and details of codes
  that are otherwise related to the code;
\item[{\EscVerb[formatcom=\flmFmtVRB{verba}]{transversal\_gates} *}] Gates that are implemented via tensor-product
  unitary operations or tensor factor permutations;
\item[{\EscVerb[formatcom=\flmFmtVRB{verba}]{magic\_scaling\_exponent} *}] Resource estimates for magic-state
  distillation protocols based on the code;
\item[{\EscVerb[formatcom=\flmFmtVRB{verba}]{code\_capacity\_threshold} *}] Threshold for a noisy channel with
  perfect decoding;
\item[{\EscVerb[formatcom=\flmFmtVRB{verba}]{threshold} *}] Threshold under noisy logical gates with perfect or
  imperfect decoding, along with other types of measurement or entanglement thresholds;
\item[{\EscVerb[formatcom=\flmFmtVRB{verba}]{\_meta}}] Entry meta-information, including names and dates of
  contributors to the entry.
\end{description}
Not all fields are meaningful for every type of code.  Fields that are not
applicable to a particular code entry can be omitted.  The keys marked above
with an asterisk (*) are only meaningful for quantum codes.


The Zoo strikes a balance between fine-grained structured data and high-level
qualitative descriptions.  
Different code properties are separated into
property-specific fields, whose values are qualitative descriptions in
free-form text.  
These qualitative values let the Zoo cover past, present, and future literature 
without restructuring database fields for new features or regimes.
For example, we found that code \emph{distance} was too specific to store as a
dedicated key with a numerical value: codes can have different notions of
distance (e.g., analog displacement or number of bits),
some codes may have no distance at all (\flmRefsCref{code:approximate_qecc},
\flmRefsCref{code:lexicographic}, \flmRefsCref{code:jw},
\flmRefsCref{code:t-designs}, etc.), and the distance of many code families is 
only known in terms of its scaling with block size.










\section{How are Codes Organized?}\label{ref8}

\begin{flmFloat}{figure}{NumCap}\includegraphics[width=438.91200000000003bp,max width=\linewidth]{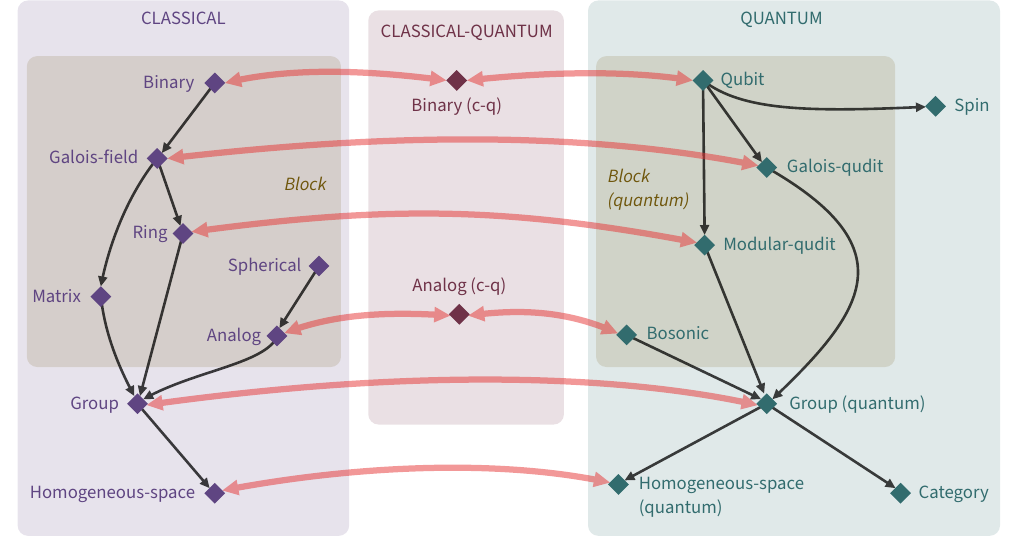}\caption{The EC Zoo consists of three Domains, with each Domain consisting
      of several Kingdoms.  Codes are primarily organized by the alphabet (for
      quantum, physical space) used to define the codes; each alphabet is a
      Kingdom (diamonds).  Horizontal red arrows show correspondences between
      Kingdoms in different Domains; e.g., qubit codes are quantum analogues of
      binary codes.  Vertical arrows relate Kingdoms to each other; e.g., the
      Qubit Kingdom descends from the Modular-qudit Kingdom because qubits are a
      special case of qudits.}\label{ref9}\end{flmFloat}

Codes are related by parent-child relations (directed) and cousin relations
(symmetric).  A child code is a special case of its parent.  Cousin relations
capture other types of connections.  These include quantum analogs of
classical codes (e.g., \flmRefsCref{code:reed_muller} and \flmRefsCref{code:quantum_reed_muller}
are cousins), and codes used in the construction of another code
(e.g., the \flmRefsHyperref{code:qubit_css}{Qubit CSS code}'s construction uses two
\flmRefsHyperref{code:binary_linear}{Binary Linear codes}).  All such relations can be
represented in a code graph (for example,
\flmHref{https://errorcorrectionzoo.org/code_graph}{this interactive version}).

\textbf{Primary hierarchy.}
The first parent listed in each code entry is the primary parent. It is chosen as the smallest code family containing the code that is defined over the same alphabet (or physical space, for quantum codes).
We choose this as the primary hierarchy because the structure and operations associated with a given alphabet determine how errors are detected and corrected, and therefore clearly distinguish different code types.
For example, the finite-field alphabet supports both addition and multiplication, while the ring of modular integers only fully supports addition.

Codes over the same alphabet are grouped into Kingdoms, which are in turn grouped into Domains (see \flmRefsCref{ref9}). The Classical Domain contains eight Kingdoms, with six of them corresponding to classical codes over bits, finite (a.k.a. Galois) fields, modular integers, real numbers, groups, and homogeneous spaces.

Classical codewords are \textit{elements of} an alphabet, whereas quantum codewords are continuously parameterized \textit{(wave-)functions over} that alphabet. The Quantum Domain includes six analogous Kingdoms for quantum codes defined on qubits, Galois qudits, modular qudits, bosonic modes, functions on groups, and functions on homogeneous spaces.
The Classical Domain includes two more Kingdoms for codes over spheres and matrices. The Quantum Domain is completed by Kingdoms of codes defined on spins and category-based spaces.
The Classical-Quantum (c-q) Domain, which organizes codes for transmission of classical information over quantum channels, contains one Kingdom each for binary and real-valued (a.k.a. analog) c-q codes.

Each Kingdom is defined by one or more root codes, which classify its codes by structure rather than alphabet. For example, the subsystem qubit code, one of the Qubit Kingdom's root codes, is the primary parent of all subsystem qubit codes.

Each Domain is likewise defined by root codes, and Kingdoms are related to Domains and to one another through the primary hierarchy of those root codes. For example, binary codes are children of \(q\)-ary codes over both Galois fields and modular integers for \(q=2\). Analogously, qubit codes are children of both Galois- and modular-qudit codes.

\textbf{Notable families.}
Complementing primary parents, secondary parents collect codes over different alphabets but with the same notable structure. 
For example, children of the cyclic code are all classical codes with cyclic symmetry, irrespective of their Kingdom. 
Notable families are defined for each Domain; analogously, there is a cyclic quantum-code family comprising cyclic quantum codes. 
Most codes in the Zoo are block codes --- subsets of \(n\) copies of a fixed alphabet (for classical codes), or subspaces of the corresponding wavefunction space (quantum codes).

Notable families also include codes with populated values for certain keys, such as codes that have been realized in real life (filled-in \EscVerb[formatcom=\flmFmtVRB{verba}]{realizations} key), or codes that admit fault-tolerant gadgets (filled-in \EscVerb[formatcom=\flmFmtVRB{verba}]{fault\_tolerance} key).

Notable code families are presented on the web version of the Error Correction
Zoo as \flmHref{https://errorcorrectionzoo.org/lists}{“code lists.”}




\section{Core Principles Behind the Zoo}

If a dentist's smile proves their skill, 
a site on error correction should prove its own resilience.
We follow strict reliability standards 
to provide a reliable source of information that can ensure long-term
accessibility.

\textbf{Reliability of the website.}
Inspired by the motto “Cool URLs don't
change”~\NoCaseChange{\protect\cite{cite10}}, we want to ensure that 
existing links and bookmarks to pages on the EC Zoo will work in 20 years.
Our compiled site is composed of static HTML pages, a
reliable standard for information preservation. We
have an explicit policy of never changing \EscVerb[formatcom=\flmFmtVRB{verba}]{code\_id} keys and URLs of pages
(barring exceptional circumstances or typos detected soon after the code's first online appearance).  Our domain name
‘\emph{errorcorrectionzoo.org}’ is not expected to change anytime soon.  

The Zoo's flexible design enables data to be processed and presented in other ways, which increases
the chances of absorption by human and machine friends alike. 
This note is uploaded to the arXiv to ensure long-term distribution and archival.
The Zoo's data resides in a git repository of standard YAML files which
is publicly hosted on \flmUrl{https://github.com/}; this choice ensures the Zoo's
data has high chances of preservation in an accessible form even if the website were to go offline.

\textbf{Evolving with the field.}
In order to “future-proof” the EC Zoo, we try to avoid making unnecessarily 
restrictive implicit assumptions in our design choices.  
Implicit assumptions
can seep in at the content level, at the structural level, as well as at the
technical level.  
Such choices can
prevent the site from reflecting an up-to-date picture of the field, leading to regrets about earlier design choices.  

For example, a poor choice of a code property field, say defining a numerical field for a code “distance,” could have required restructuring as the zoo grew to accommodate codes for which the
notion of “distance” is significantly subtler to define and for which any
numerical value would deserve an additional explanation.

Contents of code entries should use standard, modern notation
and jargon reflecting the relevant literature and that will be easy to
understand and follow in 10 years (an attempt to strictly standardize all the
notation used site-wide, for instance, would have been ill-advised).  
At a structural level, we avoided committing rigidly to how codes are organized, with the exception of the primary hierarchy by alphabet/physical space.
At a technical level, we made some design choices
that would facilitate integration of the EC Zoo with other
resources, databases, or software packages that might emerge as major standards
in the field.

\textbf{Structured data storage.} 
We reduce the overheads of updating the Zoo contents by ensuring that updates to
the YAML data files, along with the execution of appropriate github actions,
suffice to update the EC Zoo website.  This includes, among many
others, automatic creation of appropriate HTML files with cross-referenced
content, as well as compiling any necessary JavaScript configuration files for
the interactive code graph.  The pipeline is depicted in
\flmRefsCref{ref11}.  
At an intermediate stage, an in-memory object model
representing the data enables more elaborate processing, such as the execution
of standard graph algorithms on the code graph to find descendants, common
ancestors, etc.

\begin{flmFloat}{figure}{NumCap}\includegraphics[width=219.45600000000002bp,max width=\linewidth]{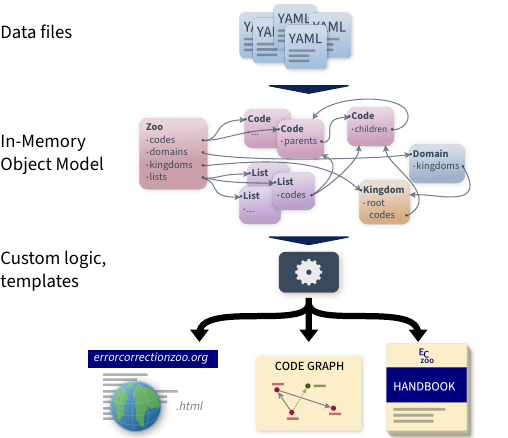}\caption{The EC Zoo processing pipeline.  The code data is
    maintained as a repository of structured YAML files, which are both
    human-editable and machine-readable.  The files are parsed into an in-memory
    object model, with JavaScript objects representing the zoo instance, code
    instances as well as lists, domains, and kingdoms.  The objects expose
    methods for computed properties, such as finding the parent kingdom of a
    code, exposing an API that enables powerful processing with custom code.
    Our code can generate our \flmHref{https://errorcorrectionzoo.org/}{HTML
      website}, static or interactive
    \flmHref{https://errorcorrectionzoo.org/code_graph}{code graphs},
    or even a print-ready handbook.}\label{ref11}\end{flmFloat}

At a technical level, the Zoo functionality is implemented in JavaScript and
Node.js, a language and framework close to web standards and with many available
libraries.
The Zoo relies on some Python and JavaScript software packages which were
developed and/or significantly improved as a byproduct of building the EC Zoo.  The package \emph{pylatexenc}, \emph{Flexible Latex-like Markup
  (FLM)} and \emph{ZooDb} are hosted on
\emph{github.com}~\NoCaseChange{\protect\cite{cite12}}.  The packages \emph{pylatexenc} and \emph{FLM} compile pseudo-LaTeX code
into an abstract syntax tree, enabling its further processing and translation
to HTML.  The package \emph{ZooDb} collects many of the defining core features
of how the EC Zoo is built, and is meant to enable the generation
of other similarly-scoped zoos.  \emph{ZooDb} is used, for instance, to build
the \emph{Quantum Channel Zoo}~\NoCaseChange{\protect\cite{cite13}}.

\textbf{Automated citation retrieval.} 
To facilitate the handling of the thousands of bibliographic references present
in the EC Zoo, our pipeline automatically generates full citations from any
arXiv ID or a DOI mentioned in the EC Zoo text.  This processing aligns with the
vision that automated semantic literature search and synthesis is the future.
The four citation types are an arXiv identifier, a DOI identifier, a predefined
list of references (used for certain widely cited references such as textbooks),
or by a manual in-place citation.  ArXiv and DOI identifiers result in an online
query to the \flmHref{https://info.arxiv.org/help/api/}{arXiv.org API} or to
\flmUrl{https://doi.org/} to fetch bibliographic details.  For instance, typing
“\EscVerb[formatcom=\flmFmtVRB{verbcode}]{\\cite\{arXiv\:0811\.4262\}}” automatically resolves to the citation of
the full published version of the paper \emph{B. Eastin and E. Knill,
  “Restrictions on Transversal Encoded Quantum Gate Sets”, Physical Review
  Letters 102, (2009) arXiv:0811.4262}.

\textbf{Continued relevance in the age of AI.} 
We use Codex~\NoCaseChange{\protect\cite{cite14}}, Copilot~\NoCaseChange{\protect\cite{cite15}}, Claude Code~\NoCaseChange{\protect\cite{cite16}}, and Gemini~\NoCaseChange{\protect\cite{cite17}} agentic coding systems to verify mathematical correctness, completeness of code relations, consistency of the code graph, accuracy of predefined references and manual citations, and spelling/grammar/formatting issues.
Among our numerous checks, we systematically asked agents to read each paper that was cited nine or more times in the EC Zoo, verify its citations for completeness and correctness, and add new citations in case important information was missing.  
The Zookeeper remains solely responsible for the accuracy and integrity of the content.

Given the right prompt, agents are making fewer mistakes, are not hallucinating like they were just a few years ago~\NoCaseChange{\protect\cite{cite7}}, and no longer require instructions on how to format entries into the EC Zoo repository.
Augmenting the EC Zoo pipeline with agent-assisted additions is a tantalizing way for the EC Zoo to scale up in proportion to the ever-growing fields it means to describe. 
All edits continue to be verified by the Zookeeper for now.  
Should AI models continue the same drastic pace of improvement, the EC Zoo could become fully automated without a loss in accuracy (cf. the \flmHref{https://www.qubitzoo.org/}{Qubit Zoo}).

At the same time, more and more humans are now primarily interfacing with agents directly instead of searching for results online.
Anyone can stand up a Wiki site using a single prompt~\NoCaseChange{\protect\cite{cite18}}.
It is therefore worth asking whether the EC Zoo, Wikipedia, and even the internet at large will very soon become second-order resources used by agents to summarize content for humans.
We leave open the question of whether the EC Zoo's big-picture features, such as its taxonomy, lists, and graphs, as well as its reliability and stability are sufficient to sustain its usefulness and impact.

\section{Outline of this Handbook}

We first list the Domains and Kingdoms and their root codes. We then describe
descendants and cousins of notable families. Following that, we print all code
entries of each of the three Domains in alphabetical order, organized by
Kingdom.  We accompany notable families with relevant subgraphs of the code
graph, and certain codes with descriptive figures or tables.  A legend for the
elements depicted in code graphs is presented in \flmRefsCref{ref19}.
The code index, list of contributors, and bibliography are printed last.

\begin{flmFloat}{figure}{NumCap}\includegraphics[width=219.45600000000002bp,max width=\linewidth]{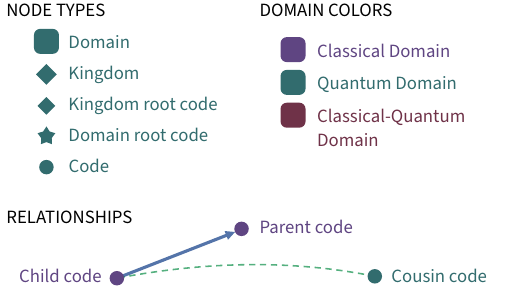}\caption{Elements that appear in a code graph}\label{ref19}\end{flmFloat}

\part{Overview of domains and kingdoms}

\onecolumngrid

All codes are organized in a taxonomy hierarchy (see \cref{ref8}).
Domains and kingdoms designate the underlying physical system on which the code is supported.
A code identifying a generic family of codes need not belong to any particular kingdom.

\section{Classical Domain}
\flmLDefLabelText{Classical Domain}{\label{domain:classical_domain}}

\begin{eczDomainSectionTop}

Codes for communication over classical channels


\eczhIncludeCodeGraph{Bare}{scale=0.6}{\columnwidth}{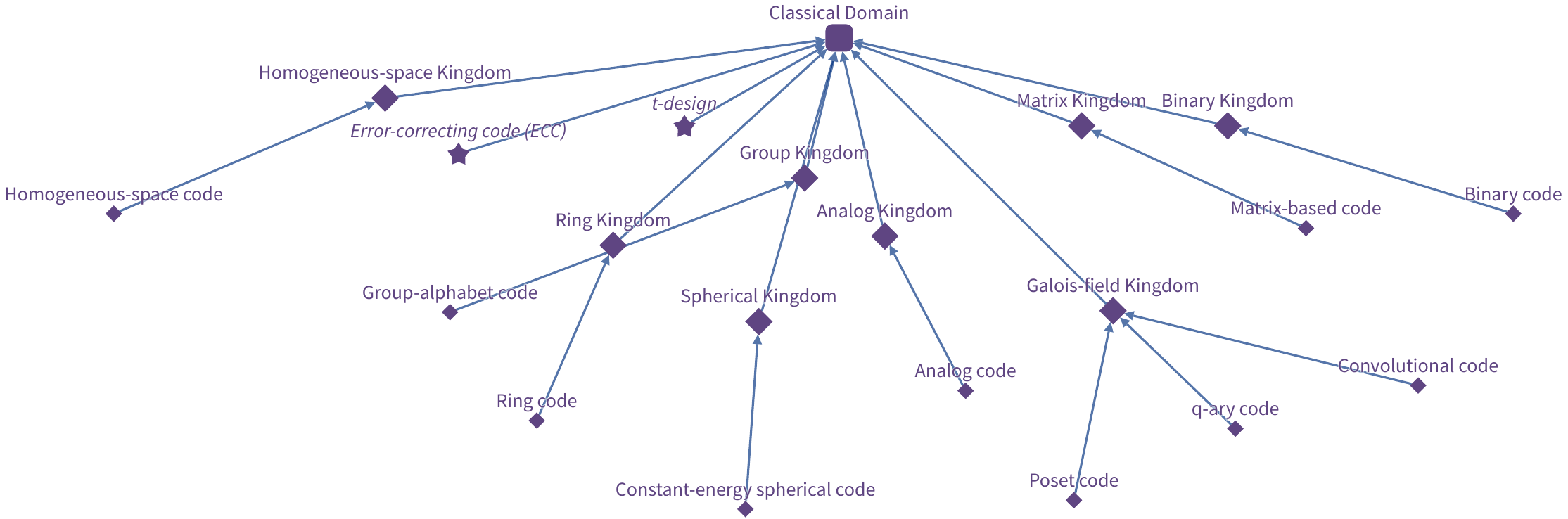}{Classical Domain}{}

\end{eczDomainSectionTop}

\codesection*{Root property codes}
\begin{eczListOfCodes}

\eczListOfCodesItem{ecc}{Error-correcting code (ECC)}
\eczhNoLabels{Code designed for transmission of classical information through classical channels.}

\eczListOfCodesItem{t-designs}{\(t\)-design}
\eczhNoLabels{A code whose codewords are uniformly distributed in a way that is useful for determining averages of polynomials over the code's underlying space \(X\).
In that way, the codewords form an approximation of the space.
A code is a design on \(X\) of \textit{strength} \(t\), i.e., a \(t\)-design on \(X\), if the average of any polynomial of degree up to \(t\) over its codewords is equal to the uniform average over all of \(X\).}

\end{eczListOfCodes}


\codesection*{Binary Kingdom}
\flmLDefLabelText{Binary Kingdom}{\label{kingdom:bits_into_bits}}

\begin{eczListOfCodes}
\eczListOfCodesItem{bits_into_bits}{Binary code}
\eczhNoLabels{Encodes \(K\) states (codewords) in \(n\) binary coordinates and has distance \(d\). Usually denoted as \((n,K,d)\). The distance is the minimum Hamming distance between a pair of distinct codewords.}
\end{eczListOfCodes}

\codesection*{Galois-field Kingdom}
\flmLDefLabelText{Galois-field Kingdom}{\label{kingdom:q-ary_digits_into_q-ary_digits}}

\begin{eczListOfCodes}
\eczListOfCodesItem{q-ary_digits_into_q-ary_digits}{\(q\)-ary code}
\eczhNoLabels{Encodes \(K\) states (codewords) in \(n\) \(q\)-ary coordinates over the field \(\mathbb{F}_q\), i.e., \(q\)-ary strings.
Error-correcting performance is quantified by some distance \(d\), which in turn is defined using a metric.
The default distance is the Hamming distance \(d\), i.e., the number of coordinates in which two distinct codewords differ; such codes are usually denoted as \((n,K,d)_q\). For linear codes, this is equivalently the Hamming weight of the lowest-weight nonzero codeword.
Unless stated otherwise, the distance for this class is the Hamming distance.},\eczListOfCodesItem{convolutional}{Convolutional code}
\eczhNoLabels{Infinite-block code that is formed using generator polynomials over a finite field. The encoder slides across contiguous subsets of the input string (like a convolutional neural network), evaluating the polynomials on that window to obtain parity bits. These parity bits are the encoded information.},\eczListOfCodesItem{poset}{Poset code}
\eczhNoLabels{Encodes \(K\) states (codewords) in \(n\) \(q\)-ary coordinates over the field \(\mathbb{F}_q\), with its distance evaluated in the poset metric.
}
\end{eczListOfCodes}

\codesection*{Matrix Kingdom}
\flmLDefLabelText{Matrix Kingdom}{\label{kingdom:matrices_into_matrices}}

\begin{eczListOfCodes}
\eczListOfCodesItem{matrices_into_matrices}{Matrix-based code}
\eczhNoLabels{Encodes \(K\) states (codewords) in an \(m\times n\) array of coordinates over a field (e.g., the Galois field \(\mathbb{F}_q\) or the complex numbers \(\mathbb{C}\)).}
\end{eczListOfCodes}

\codesection*{Analog Kingdom}
\flmLDefLabelText{Analog Kingdom}{\label{kingdom:analog}}

\begin{eczListOfCodes}
\eczListOfCodesItem{analog}{Analog code}
\eczhNoLabels{Encodes states (codewords) into continuous coordinates in the \(n\)-dimensional (real or complex) coordinate space (\(\mathbb{R}^n\) or \(\mathbb{C}^n\)).
Important subclasses include sphere packings, tilings, and modulation constellations.
The number of codewords may be infinite because the coordinate space is infinite, so various restricted versions have to be constructed in practice.
}
\end{eczListOfCodes}

\codesection*{Spherical Kingdom}
\flmLDefLabelText{Spherical Kingdom}{\label{kingdom:points_into_spheres}}

\begin{eczListOfCodes}
\eczListOfCodesItem{points_into_spheres}{Constant-energy spherical code}
\eczhNoLabels{Code whose codewords are points on a real or complex sphere whose radius squared is called the \textit{energy}.
Typically, only angular distances between points are relevant for code performance, so one can normalize codewords of a constant-energy code to obtain up a spherical code, i.e., a constant energy code with energy one.
}
\end{eczListOfCodes}

\codesection*{Ring Kingdom}
\flmLDefLabelText{Ring Kingdom}{\label{kingdom:rings_into_rings}}

\begin{eczListOfCodes}
\eczListOfCodesItem{rings_into_rings}{Ring code}
\eczhNoLabels{Encodes \(K\) states (codewords) in \(n\) coordinates over a finite ring \(R\).}
\end{eczListOfCodes}

\codesection*{Group Kingdom}
\flmLDefLabelText{Group Kingdom}{\label{kingdom:group_classical}}

\begin{eczListOfCodes}
\eczListOfCodesItem{group_classical}{Group-alphabet code}
\eczhNoLabels{Encodes \(K\) states (codewords) using symbols drawn from a group \(G\), typically with the group operation inducing a natural notion of \flmRefsHyperref{ref20}{translation} or symmetry on the alphabet. The number of codewords may be infinite for infinite groups, so various restricted versions have to be constructed in practice.
}
\end{eczListOfCodes}

\codesection*{Homogeneous-space Kingdom}
\flmLDefLabelText{Homogeneous-space Kingdom}{\label{kingdom:homogeneous_space_classical}}

\begin{eczListOfCodes}
\eczListOfCodesItem{homogeneous_space_classical}{Homogeneous-space code}
\eczhNoLabels{Encodes \(K\) states (codewords) into a homogeneous (a.k.a. coset) space \(G/H\), where \(G\) is a group and \(H\) is a subgroup of \(G\). The space is labeled by cosets of \(H\) in \(G\).
Notable groups include compact groups, locally compact Abelian groups, and finite groups. }
\end{eczListOfCodes}

\onecolumngrid

\section{Quantum Domain}
\flmLDefLabelText{Quantum Domain}{\label{domain:quantum_domain}}

\begin{eczDomainSectionTop}

Codes for quantum communication over quantum channels


\eczhIncludeCodeGraph{Bare}{scale=0.6}{\columnwidth}{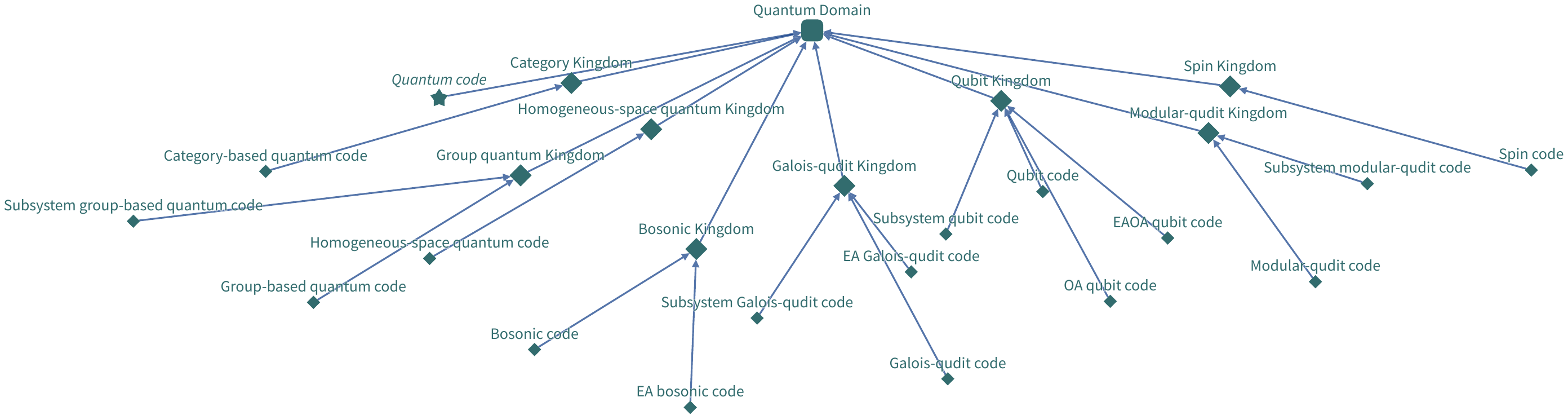}{Quantum Domain}{}

\end{eczDomainSectionTop}

\codesection*{Root property codes}
\begin{eczListOfCodes}

\eczListOfCodesItem{quantum_into_quantum}{Quantum code}
\eczhNoLabels{Code designed for transmission of quantum and/or classical information through a quantum channel for the purposes of robust storage, communication, or sensing. 
Transmission can be performed with side information or entanglement.}

\end{eczListOfCodes}


\codesection*{Qubit Kingdom}
\flmLDefLabelText{Qubit Kingdom}{\label{kingdom:qubits_into_qubits}}

\begin{eczListOfCodes}
\eczListOfCodesItem{qubits_into_qubits}{Qubit code}
\eczhNoLabels{Encodes \(K\)-dimensional Hilbert space into a \(2^n\)-dimensional (i.e., \(n\)-qubit) Hilbert space.
Usually denoted as \(\llparenthesis n,K\rrparenthesis \) or \(\llparenthesis n,K,d\rrparenthesis \), where \(d\) is the code's distance.},\eczListOfCodesItem{subsystem_qubits_into_qubits}{Subsystem qubit code}
\eczhNoLabels{Subsystem QECC encoding into a \(2^n\)-dimensional (i.e., \(n\)-qubit) Hilbert space.
},\eczListOfCodesItem{oa_qubits_into_qubits}{OA qubit code}
\eczhNoLabels{An OAQECC family that encompasses ordinary (i.e., subspace) qubit codes, subsystem qubit codes, and hybrid qubit codes using an operator-algebraic framework.},\eczListOfCodesItem{eaoa_qubits_into_qubits}{EAOA qubit code}
\eczhNoLabels{Entanglement-assisted qubit code in the operator-algebra framework.
This family encompasses ordinary entanglement-assisted subspace qubit codes, entanglement-assisted subsystem qubit codes, entanglement-assisted hybrid qubit codes, and their operator-algebra generalizations.}
\end{eczListOfCodes}

\codesection*{Modular-qudit Kingdom}
\flmLDefLabelText{Modular-qudit Kingdom}{\label{kingdom:qudits_into_qudits}}

\begin{eczListOfCodes}
\eczListOfCodesItem{qudits_into_qudits}{Modular-qudit code}
\eczhNoLabels{Encodes a \(K\)-dimensional Hilbert space into a \(q^n\)-dimensional (\(n\)-qudit) Hilbert space, with canonical qudit states \(|k\rangle\) labeled by elements \(k\) of the group \(\mathbb{Z}_q\) of integers \textit{modulo} \(q\).
Usually denoted as \(\llparenthesis n,K\rrparenthesis _{\mathbb{Z}_q}\) or \(\llparenthesis n,K,d\rrparenthesis _{\mathbb{Z}_q}\), whenever the code's distance \(d\) is defined, and with \(q=p\) when the dimension is prime.},\eczListOfCodesItem{subsystem_qudits_into_qudits}{Subsystem modular-qudit code}
\eczhNoLabels{Subsystem QECC encoding into a \(q^n\)-dimensional Hilbert space consisting of \(n\) modular qudits.
}
\end{eczListOfCodes}

\codesection*{Galois-qudit Kingdom}
\flmLDefLabelText{Galois-qudit Kingdom}{\label{kingdom:galois_into_galois}}

\begin{eczListOfCodes}
\eczListOfCodesItem{galois_into_galois}{Galois-qudit code}
\eczhNoLabels{Encodes \(K\)-dimensional Hilbert space into a \(q^n\)-dimensional (\(n\)-qudit) Hilbert space, with canonical qudit states \(|k\rangle\) labeled by elements \(k\) of the \textit{Galois field} \(\mathbb{F}_q\) and with \(q\) being a power of a prime \(p\).},\eczListOfCodesItem{subsystem_galois_into_galois}{Subsystem Galois-qudit code}
\eczhNoLabels{Subsystem QECC encoding into a \(q^n\)-dimensional Hilbert space consisting of \(n\) Galois qudits.
},\eczListOfCodesItem{ea_galois_into_galois}{EA Galois-qudit code}
\eczhNoLabels{Galois-qudit code designed to utilize pre-shared entanglement between sender and receiver.
}
\end{eczListOfCodes}

\codesection*{Bosonic Kingdom}
\flmLDefLabelText{Bosonic Kingdom}{\label{kingdom:oscillators}}

\begin{eczListOfCodes}
\eczListOfCodesItem{oscillators}{Bosonic code}
\eczhNoLabels{Encodes logical Hilbert space, finite- or infinite-dimensional, into a physical Hilbert space that contains at least one \textit{oscillator} (a.k.a. \textit{bosonic mode} or \textit{qumode}).},\eczListOfCodesItem{ea_oscillators}{EA bosonic code}
\eczhNoLabels{Bosonic code designed to utilize pre-shared entanglement between sender and receiver.
}
\end{eczListOfCodes}

\codesection*{Spin Kingdom}
\flmLDefLabelText{Spin Kingdom}{\label{kingdom:spins_into_spins}}

\begin{eczListOfCodes}
\eczListOfCodesItem{spins_into_spins}{Spin code}
\eczhNoLabels{Encodes a \(K\)-dimensional Hilbert space into a tensor-product or direct sum of factors, with each factor spanned by states of a quantum mechanical spin or, more generally, an irreducible representation of a compact Lie group.}
\end{eczListOfCodes}

\codesection*{Group quantum Kingdom}
\flmLDefLabelText{Group quantum Kingdom}{\label{kingdom:group_quantum}}

\begin{eczListOfCodes}
\eczListOfCodesItem{group_quantum}{Group-based quantum code}
\eczhNoLabels{Encodes a \textit{logical} Hilbert space, finite- or infinite-dimensional, into a \textit{physical} Hilbert space of \(L^2\)-normalizable functions on a second-countable unimodular group \(G\), i.e., a \(G\)\textit{-valued qudit} or \(G\)-qudit.
In other words, a group-valued qudit is a vector space whose canonical basis states \(|g\rangle\) are labeled by elements \(g\) of a group \(G\).
For \(K\)-dimensional logical subspace and for block codes defined on groups \(G^{n}\), can be denoted as \(\llparenthesis n,K\rrparenthesis _G\).
When the logical subspace is the Hilbert space of \(L^2\)-normalizable functions on \(G^{ k}\), can be denoted as \(\llbracket n,k\rrbracket _G\).
Ideal codewords may not be normalizable, depending on whether \(G\) is continuous and/or noncompact, so approximate versions have to be constructed in practice.},\eczListOfCodesItem{subsystem_group_quantum}{Subsystem group-based quantum code}
\eczhNoLabels{Group-based quantum code whose codespace admits a tensor-product decomposition into logical and gauge factors.
}
\end{eczListOfCodes}

\codesection*{Homogeneous-space quantum Kingdom}
\flmLDefLabelText{Homogeneous-space quantum Kingdom}{\label{kingdom:homogeneous_space_quantum}}

\begin{eczListOfCodes}
\eczListOfCodesItem{homogeneous_space_quantum}{Homogeneous-space quantum code}
\eczhNoLabels{Encodes a \textit{logical} Hilbert space, finite- or infinite-dimensional, into a \textit{physical} Hilbert space of \(L^2\)-normalizable functions on a homogeneous space \(G/H\) or, more generally, induced representations whose base space is \(G/H\) \NoCaseChange{\protect\cite{cite21,cite22,cite23,cite24,cite25}}. Here, \(G\) is a second-countable unimodular group, and \(H\) is a closed subgroup of \(G\).
}
\end{eczListOfCodes}

\codesection*{Category Kingdom}
\flmLDefLabelText{Category Kingdom}{\label{kingdom:category_quantum}}

\begin{eczListOfCodes}
\eczListOfCodesItem{category_quantum}{Category-based quantum code}
\eczhNoLabels{Encodes a finite-dimensional \textit{logical} Hilbert space into a \textit{physical} Hilbert space associated with a category.
The categories of interest are typically fusion categories, which subsume all finite groups and many state spaces associated with topological codes. 
Codes on modular fusion categories are often associated with a particular topological quantum field theory (TQFT), as the data of such theories is described by such categories.
}
\end{eczListOfCodes}

\onecolumngrid

\section{Classical-quantum Domain}
\flmLDefLabelText{Classical-quantum Domain}{\label{domain:classical_into_quantum_domain}}

\begin{eczDomainSectionTop}

Codes for classical communication over quantum channels


\eczhIncludeCodeGraph{Bare}{scale=0.6}{\columnwidth}{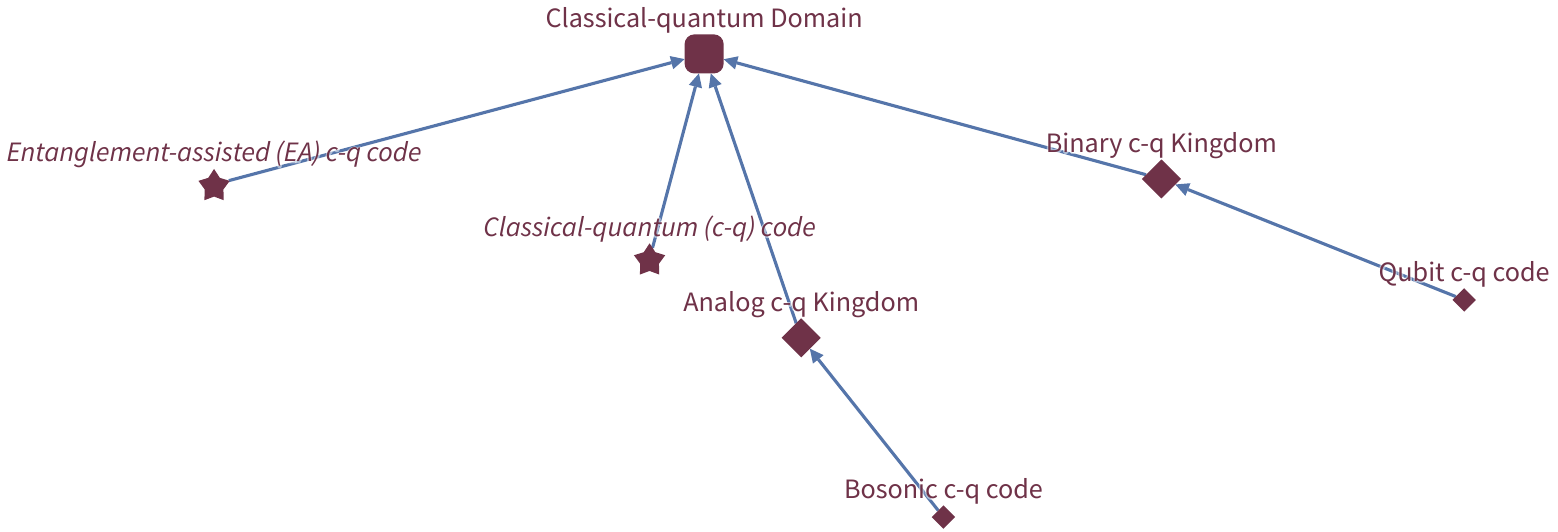}{Classical-quantum Domain}{}

\end{eczDomainSectionTop}

\codesection*{Root property codes}
\begin{eczListOfCodes}

\eczListOfCodesItem{classical_into_quantum}{Classical-quantum (c-q) code}
\eczhNoLabels{Code designed specifically for transmission of classical information through non-classical channels, e.g., quantum channels, hybrid classical-quantum channels, or channels with classical inputs and quantum outputs. 
Such codes include maps from a classical alphabet into a quantum Hilbert space.
}

\eczListOfCodesItem{ea_classical_into_quantum}{Entanglement-assisted (EA) c-q code}
\eczhNoLabels{Classical-quantum code whose encoding and decoding utilize pre-shared entanglement between sender and receiver.
The sender encodes classical information into quantum systems sent through a quantum channel, while the receiver decodes using the channel outputs together with retained halves of pre-shared entangled states.
}

\end{eczListOfCodes}


\codesection*{Binary c-q Kingdom}
\flmLDefLabelText{Binary c-q Kingdom}{\label{kingdom:qubit_classical_into_quantum}}

\begin{eczListOfCodes}
\eczListOfCodesItem{qubit_classical_into_quantum}{Qubit c-q code}
\eczhNoLabels{A qubit code designed for transmission of classical information in the form of bits through non-classical channels.
}
\end{eczListOfCodes}

\codesection*{Analog c-q Kingdom}
\flmLDefLabelText{Analog c-q Kingdom}{\label{kingdom:bosonic_classical_into_quantum}}

\begin{eczListOfCodes}
\eczListOfCodesItem{bosonic_classical_into_quantum}{Bosonic c-q code}
\eczhNoLabels{Bosonic code designed for transmission of classical information through non-classical channels.
Encodes classical symbols into bosonic quantum states for transmission over a quantum channel and decoding with a quantum-enhanced \textit{receiver}.
This entry includes bosonic c-q modulation formats and is distinct from a \flmRefsHyperref{code:modulation}{classical modulation scheme}, which maps classical symbols into classical electromagnetic signals for transmission over classical channels.
A bosonic c-q modulation format instead treats the transmitted signals as quantum states and allows the receiver to use quantum measurements.
}
\end{eczListOfCodes}

\onecolumngrid

\vspace{1.5em}

\part{Overview of notable code families}
\twocolumngrid
\onecolumngrid 
\eczlistdomainsection{Classical code families}

\eczcodelist{ag}{Algebraic-geometry codes}%

\eczhCodeListAutoDescription{All descendants of \flmRefsCref{code:ag}.}%

\eczhIncludeCodeGraph{Bare}{scale=0.5}{\columnwidth}{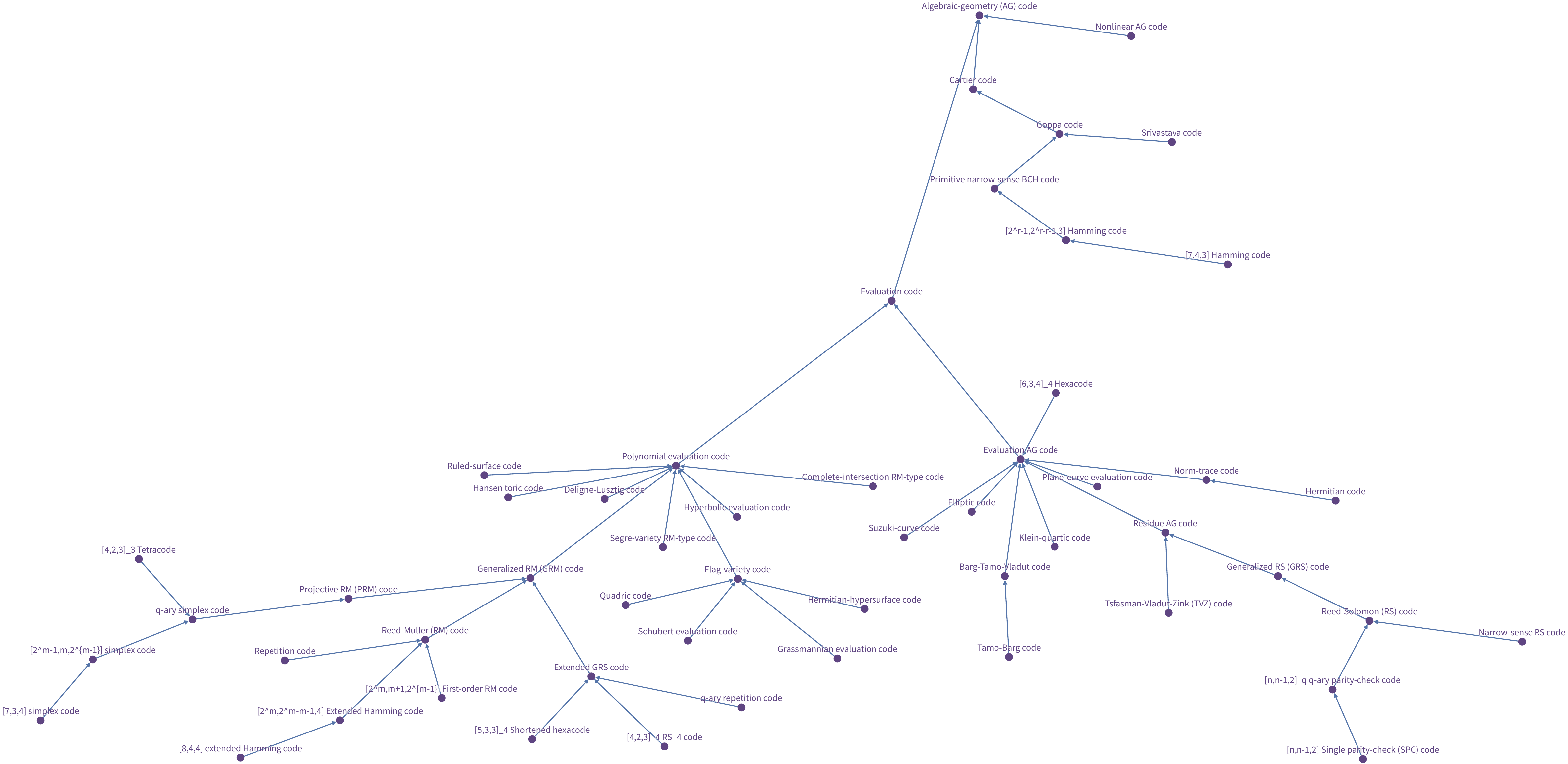}{Algebraic-geometry codes}{https://errorcorrectionzoo.org/code_graph#J\%7B\%22displayMode\%22\%3A\%22subset\%22\%2C\%22modeSubsetOptions\%22\%3A\%7B\%22codeIds\%22\%3A\%5B\%22ag\%22\%2C\%22tamo_barg_vladut\%22\%2C\%22cartier\%22\%2C\%22complete_intersections\%22\%2C\%22deligne_lusztig\%22\%2C\%22elliptic\%22\%2C\%22evaluation\%22\%2C\%22evaluation_varieties\%22\%2C\%22extended_reed_solomon\%22\%2C\%22flag_variety\%22\%2C\%22generalized_reed_muller\%22\%2C\%22generalized_reed_solomon\%22\%2C\%22goppa\%22\%2C\%22grassmannian_variety\%22\%2C\%22toric_classical\%22\%2C\%22hermitian\%22\%2C\%22hermitian_hypersurface\%22\%2C\%22cascaded_reed_solomon\%22\%2C\%22klein_quartic\%22\%2C\%22narrow_sense_reed_solomon\%22\%2C\%22nonlinear_ag\%22\%2C\%22norm_trace\%22\%2C\%22plane_curve\%22\%2C\%22evaluation_polynomial\%22\%2C\%22narrow_sense_q-ary_bch\%22\%2C\%22projective_reed_muller\%22\%2C\%22quadric\%22\%2C\%22reed_muller\%22\%2C\%22reed_solomon\%22\%2C\%22repetition\%22\%2C\%22residue\%22\%2C\%22ruled_surface\%22\%2C\%22schubert\%22\%2C\%22serge\%22\%2C\%22srivastava\%22\%2C\%22suzuki\%22\%2C\%22tamo_barg\%22\%2C\%22shimura\%22\%2C\%22extended_hamming\%22\%2C\%22biorthogonal\%22\%2C\%22simplex\%22\%2C\%22hamming\%22\%2C\%22tetracode\%22\%2C\%22reed_solomon_4\%22\%2C\%22shortened_hexacode\%22\%2C\%22hexacode\%22\%2C\%22simplex734\%22\%2C\%22hamming743\%22\%2C\%22hamming844\%22\%2C\%22parity_check\%22\%2C\%22q-ary_parity_check\%22\%2C\%22q-ary_repetition\%22\%2C\%22q-ary_simplex\%22\%5D\%2C\%22reusePreviousLayoutPositions\%22\%3Afalse\%2C\%22showIntermediateConnectingNodes\%22\%3Atrue\%2C\%22connectingNodesMaxDepth\%22\%3A15\%2C\%22connectingNodesPathMaxLength\%22\%3A20\%2C\%22connectingNodesMaxExtraDepth\%22\%3A3\%2C\%22connectingNodesOnlyKeepPathsWithAdditionalLength\%22\%3A1\%2C\%22connectingNodesToDomainsAndKingdoms\%22\%3Afalse\%2C\%22connectingNodesEdgeLengthsByType\%22\%3A\%7B\%22primaryParent\%22\%3A1\%2C\%22secondaryParent\%22\%3A4\%2C\%22cousin\%22\%3A6\%7D\%2C\%22nodeIds\%22\%3A\%5B\%5D\%7D\%2C\%22highlightImportantNodes\%22\%3A\%7B\%22highlightImportantNodes\%22\%3Afalse\%2C\%22highlightPrimaryParents\%22\%3Afalse\%2C\%22highlightRootConnectingEdges\%22\%3Afalse\%7D\%7D}

\begingroup
\small
\eczhBreakableDashes
\renewcommand\arraystretch{1.05}
\edef\myxtraspc{\eczListAddVSpaceXtraXtraStretch}
\begin{tabularx}{\linewidth}{>{\raggedright\arraybackslash}p{\eczListColWidth{name}} >{\hsize=1.0000\hsize }X}
\toprule
\eczListColTitle{Code} & \eczListColTitle{Description} \\
\midrule
\endfirsthead
\toprule
\eczListColTitleContinued{Code} & \eczListColTitleContinued{Description} \\
\midrule
\endhead
\bottomrule
\endfoot
\eczhRefIndex{code:ag}%
\eczhListValue{\flmRefsHyperref{code:ag}{Algebraic-geometry (AG) code}} & \eczhListValue{Binary or \(q\)-ary code or subcode constructed from an algebraic curve of some genus over a finite field via the evaluation construction, the residue construction, or more general constructions that yield nonlinear codes \NoCaseChange{\protect\cite[{Defs. 15.3.1 and 15.3.2}]{cite26}\protect\cite[{Sec. 15.4.4}]{cite26}}. Linear AG codes from the first two constructions are also called \textit{geometric Goppa codes}.}\\ 
\addlinespace[\myxtraspc]
\eczhRefIndex{code:tamo_barg_vladut}%
\eczhListValue{\flmRefsHyperref{code:tamo_barg_vladut}{Barg-Tamo-Vladut code}} & \eczhListValue{Evaluation AG code on algebraic curves built from a Galois cover \(\phi:Y\to X\), where the recovery sets are fibres over rational points of \(X\) that split completely in the cover \NoCaseChange{\protect\cite[{Def. 15.9.10}]{cite26}\protect\cite[{Thm. 15.9.14}]{cite26}}.
The Barg-Tamo-Vladut construction generalizes the Tamo-Barg construction from \(PG(1,q)\) to longer AG codes, and variants can be built with higher local distance or availability \(2\) via fibre products of curves \NoCaseChange{\protect\cite[{Thm. 15.9.19}]{cite26}\protect\cite[{Thm. 15.9.21}]{cite26}}.}\\ 
\addlinespace[\myxtraspc]
\eczhRefIndex{code:cartier}%
\eczhListValue{\flmRefsHyperref{code:cartier}{Cartier code}} & \eczhListValue{A generalization of the Goppa codes to codes defined from curves of nonzero genus. Each code is a subcode of a certain residue AG code and is constructed using the Cartier operator.}\\ 
\addlinespace[\myxtraspc]
\eczhRefIndex{code:complete_intersections}%
\eczhListValue{\flmRefsHyperref{code:complete_intersections}{Complete-intersection RM-type code}} & \eczhListValue{Evaluation code of polynomials evaluated on points lying on a complete intersection.}\\ 
\addlinespace[\myxtraspc]
\eczhRefIndex{code:deligne_lusztig}%
\eczhListValue{\flmRefsHyperref{code:deligne_lusztig}{Deligne-Lusztig code}} & \eczhListValue{Evaluation code of polynomials evaluated on points lying on a Deligne-Lusztig variety, often a Deligne-Lusztig curve in the classical one-dimensional cases.}\\ 
\addlinespace[\myxtraspc]
\eczhRefIndex{code:elliptic}%
\eczhListValue{\flmRefsHyperref{code:elliptic}{Elliptic code}} & \eczhListValue{Evaluation AG code of rational functions evaluated on points lying on an elliptic curve, i.e., a curve of genus one.}\\ 
\addlinespace[\myxtraspc]
\eczhRefIndex{code:evaluation}%
\eczhListValue{\flmRefsHyperref{code:evaluation}{Evaluation AG code}} & \eczhListValue{Evaluation code over \(\mathbb{F}_q\) on a set of points \({\cal P} = \left( P_1,P_2,\cdots,P_n \right)\) lying on an algebraic curve \(\cal X\) defined over \(\mathbb{F}_q\), where the corresponding vector space \(L\) of functions \(f\) consists of certain rational functions (or, in special cases, polynomials).}\\ 
\addlinespace[\myxtraspc]
\eczhRefIndex{code:evaluation_varieties}%
\eczhListValue{\flmRefsHyperref{code:evaluation_varieties}{Evaluation code}} & \eczhListValue{Code whose codewords are evaluations of functions at certain fixed points. Code properties can be inferred from the structure of the functions and the underlying geometric object containing the points, often using results from algebraic geometry.}\\ 
\addlinespace[\myxtraspc]
\eczhRefIndex{code:extended_reed_solomon}%
\eczhListValue{\flmRefsHyperref{code:extended_reed_solomon}{Extended GRS code}} & \eczhListValue{A GRS code extended by one extra coordinate to form an \([n+1,k,n-k+2]_q\) MDS code. In projective language, this corresponds to adding one more evaluation point, often interpreted as the point at infinity; in suitable equivalent descriptions, one may instead use an affine point such as \(0\). The case when \(n=q-1\), multipliers \(v_i=1\), and \(\alpha_i\) are \(i-1\)st powers of a primitive \(n\)th root of unity is an \textit{extended narrow-sense RS code}.}\\ 
\addlinespace[\myxtraspc]
\eczhRefIndex{code:flag_variety}%
\eczhListValue{\flmRefsHyperref{code:flag_variety}{Flag-variety code}} & \eczhListValue{Evaluation code of polynomials evaluated on points lying on a flag variety.}\\ 
\addlinespace[\myxtraspc]
\eczhRefIndex{code:generalized_reed_muller}%
\eczhListValue{\flmRefsHyperref{code:generalized_reed_muller}{Generalized RM (GRM) code}} & \eczhListValue{Extensions of RM codes to \(q\)-ary coordinates that can be described as multivariate polynomial evaluation codes over affine or projective space.}\\ 
\addlinespace[\myxtraspc]
\eczhRefIndex{code:generalized_reed_solomon}%
\eczhListValue{\flmRefsHyperref{code:generalized_reed_solomon}{Generalized RS (GRS) code}} & \eczhListValue{An \([n,k,n-k+1]_q\) MDS code that is a modification of the RS code where codeword polynomials are multiplied by additional prefactors \NoCaseChange{\protect\cite[{Def. 15.3.19}]{cite26}}.}\\ 
\addlinespace[\myxtraspc]
\eczhRefIndex{code:goppa}%
\eczhListValue{\flmRefsHyperref{code:goppa}{Goppa code}} & \eczhListValue{A linear \(q\)-ary code defined from a polynomial \(G(x)\) over an extension field and a set of evaluation points \(L\) avoiding the roots of \(G\).
Goppa codes form a central family of alternant codes, admit efficient algebraic decoding algorithms, and include the binary Goppa codes used in the McEliece cryptosystem.
When the base field equals the coefficient field, they coincide with residue AG codes on \(PG(1,q^m)\); in general, classical Goppa codes are subfield subcodes of such AG codes \NoCaseChange{\protect\cite[{Rem. 15.3.27}]{cite26}\protect\cite[{Thm. 15.3.28}]{cite26}}.}\\ 
\addlinespace[\myxtraspc]
\eczhRefIndex{code:grassmannian_variety}%
\eczhListValue{\flmRefsHyperref{code:grassmannian_variety}{Grassmannian evaluation code}} & \eczhListValue{Evaluation code of polynomials evaluated on points lying on a finite-field Grassmannian embedded into projective space using the Plucker embedding \NoCaseChange{\protect\cite{cite27,cite28}}.}\\ 
\addlinespace[\myxtraspc]
\eczhRefIndex{code:toric_classical}%
\eczhListValue{\flmRefsHyperref{code:toric_classical}{Hansen toric code}} & \eczhListValue{Evaluation code of a linear space of polynomials evaluated on points lying on an affine or projective toric variety. If the space is taken to be all polynomials up to some degree, the code is called a \textit{toric RM-type code} of that degree.}\\ 
\addlinespace[\myxtraspc]
\eczhRefIndex{code:hermitian}%
\eczhListValue{\flmRefsHyperref{code:hermitian}{Hermitian code}} & \eczhListValue{Evaluation AG code of rational functions on a Hermitian curve over \(\mathbb{F}_{q^2}\).}\\ 
\addlinespace[\myxtraspc]
\eczhRefIndex{code:hermitian_hypersurface}%
\eczhListValue{\flmRefsHyperref{code:hermitian_hypersurface}{Hermitian-hypersurface code}} & \eczhListValue{Evaluation code of polynomials evaluated on points lying on a Hermitian hypersurface.}\\ 
\addlinespace[\myxtraspc]
\eczhRefIndex{code:cascaded_reed_solomon}%
\eczhListValue{\flmRefsHyperref{code:cascaded_reed_solomon}{Hyperbolic evaluation code}} & \eczhListValue{An evaluation code over polynomials in two variables. 
Generator matrices are determined in Ref. \NoCaseChange{\protect\cite{cite29}} following initial formulations of the codes as generalized concatenations of RS codes \NoCaseChange{\protect\cite{cite30,cite31}}; see \NoCaseChange{\protect\cite[{Exam. 4.26}]{cite32}}.}\\ 
\addlinespace[\myxtraspc]
\eczhRefIndex{code:klein_quartic}%
\eczhListValue{\flmRefsHyperref{code:klein_quartic}{Klein-quartic code}} & \eczhListValue{Evaluation AG code over \(\mathbb{F}_8\) of rational functions evaluated on points lying on the Klein quartic, which is defined by the equation \(x^3 y + y^3 z + z^3 x = 0\) \NoCaseChange{\protect\cite[{Ex. 2.75}]{cite32}}.}\\ 
\addlinespace[\myxtraspc]
\eczhRefIndex{code:narrow_sense_reed_solomon}%
\eczhListValue{\flmRefsHyperref{code:narrow_sense_reed_solomon}{Narrow-sense RS code}} & \eczhListValue{An \([q-1,k,n-k+1]_q\) RS code whose points \(\alpha_i\) are all \((i-1)\)st powers of a \flmRefsHyperref{ref33}{primitive} element \(\alpha\) of \(\mathbb{F}_q\).}\\ 
\addlinespace[\myxtraspc]
\eczhRefIndex{code:nonlinear_ag}%
\eczhListValue{\flmRefsHyperref{code:nonlinear_ag}{Nonlinear AG code}} & \eczhListValue{Nonlinear \(q\)-ary code constructed by evaluating functions, and in some constructions derivatives of functions, on an algebraic curve \NoCaseChange{\protect\cite[{Sec. 15.4.4}]{cite26}}.}\\ 
\addlinespace[\myxtraspc]
\eczhRefIndex{code:norm_trace}%
\eczhListValue{\flmRefsHyperref{code:norm_trace}{Norm-trace code}} & \eczhListValue{Evaluation AG code of rational functions evaluated on points lying on a Miura-Kamiya curve in either affine or projective space.
The family is named as such because the equations defining the curves can be expressed in terms of the \flmRefsHyperref{ref33}{field norm and field trace}.}\\ 
\addlinespace[\myxtraspc]
\eczhRefIndex{code:plane_curve}%
\eczhListValue{\flmRefsHyperref{code:plane_curve}{Plane-curve evaluation code}} & \eczhListValue{Evaluation AG code of bivariate polynomials of some finite maximum degree, evaluated at points lying on an affine or projective plane curve.}\\ 
\addlinespace[\myxtraspc]
\eczhRefIndex{code:evaluation_polynomial}%
\eczhListValue{\flmRefsHyperref{code:evaluation_polynomial}{Polynomial evaluation code}} & \eczhListValue{Evaluation code of polynomials (or, more generally, rational functions) at points \({\cal P} = \left( P_1,P_2,\cdots,P_n \right)\) on an algebraic variety \(\cal X\) of dimension greater than one (i.e., not an algebraic curve).}\\ 
\addlinespace[\myxtraspc]
\eczhRefIndex{code:narrow_sense_q-ary_bch}%
\eczhListValue{\flmRefsHyperref{code:narrow_sense_q-ary_bch}{Primitive narrow-sense BCH code}} & \eczhListValue{A \(q\)-ary BCH code for \(b=1\) and for \(n=q^r-1\) for some \(r\geq 2\).}\\ 
\addlinespace[\myxtraspc]
\eczhRefIndex{code:projective_reed_muller}%
\eczhListValue{\flmRefsHyperref{code:projective_reed_muller}{Projective RM (PRM) code}} & \eczhListValue{Evaluation code obtained by evaluating homogeneous polynomials on the points of the projective space \(PG(m,q)\), equivalently on representatives of the nonzero vectors in \(\mathbb{F}_q^{m+1}\) whose leftmost nonzero coordinate is one.}\\ 
\addlinespace[\myxtraspc]
\eczhRefIndex{code:quadric}%
\eczhListValue{\flmRefsHyperref{code:quadric}{Quadric code}} & \eczhListValue{Evaluation code of polynomials evaluated on points lying on a quadric hypersurface.}\\ 
\addlinespace[\myxtraspc]
\eczhRefIndex{code:reed_muller}%
\eczhListValue{\flmRefsHyperref{code:reed_muller}{Reed-Muller (RM) code}} & \eczhListValue{Member of the RM\((r,m)\) family of linear binary codes derived from multivariate polynomials. The code parameters are \([2^m,\sum_{j=0}^{r} {m \choose j},2^{m-r}]\), where \(r\) is the \textit{order} of the code satisfying \(0\leq r\leq m\).
First-order RM codes are also called biorthogonal codes, while \(m\)th order RM codes are also called \textit{universe} codes.
\textit{Punctured RM codes} RM\(^*(r,m)\) are obtained from RM codes by deleting one coordinate from each codeword.}\\ 
\addlinespace[\myxtraspc]
\eczhRefIndex{code:reed_solomon}%
\eczhListValue{\flmRefsHyperref{code:reed_solomon}{Reed-Solomon (RS) code}} & \eczhListValue{An \([n,k,n-k+1]_q\) linear code based on polynomials over \(\mathbb{F}_q\).}\\ 
\addlinespace[\myxtraspc]
\eczhRefIndex{code:repetition}%
\eczhListValue{\flmRefsHyperref{code:repetition}{Repetition code}} & \eczhListValue{\([n,1,n]\) binary linear code encoding one bit of information into an \(n\)-bit string.
Majority decoding requires \(n\) to be odd in order to avoid ties.
The idea is to increase the code distance by repeating the logical information several times. It is a \((n,1)\)-Hamming code.
Its automorphism group is \(S_n\).}\\ 
\addlinespace[\myxtraspc]
\eczhRefIndex{code:residue}%
\eczhListValue{\flmRefsHyperref{code:residue}{Residue AG code}} & \eczhListValue{Linear \(q\)-ary code defined using a set of \(\mathbb{F}_q\)-rational points \({\cal P} = \left( P_1,P_2,\cdots,P_n \right)\) on an algebraic curve \(\cal X\) and a linear space \(\Omega\) of certain rational differential forms \(\omega\) \NoCaseChange{\protect\cite[{Def. 15.3.2}]{cite26}}.}\\ 
\addlinespace[\myxtraspc]
\eczhRefIndex{code:ruled_surface}%
\eczhListValue{\flmRefsHyperref{code:ruled_surface}{Ruled-surface code}} & \eczhListValue{Evaluation code obtained by evaluating global sections of a line bundle, or equivalently suitable polynomial functions, on rational points of a ruled surface over a finite field. Such codes extend algebraic-geometry constructions from curves to certain projective surfaces \NoCaseChange{\protect\cite{cite34,cite35}}.}\\ 
\addlinespace[\myxtraspc]
\eczhRefIndex{code:schubert}%
\eczhListValue{\flmRefsHyperref{code:schubert}{Schubert evaluation code}} & \eczhListValue{Evaluation code of polynomials evaluated on points lying on a Schubert variety.}\\ 
\addlinespace[\myxtraspc]
\eczhRefIndex{code:serge}%
\eczhListValue{\flmRefsHyperref{code:serge}{Segre-variety RM-type code}} & \eczhListValue{Evaluation code of multihomogeneous polynomials evaluated on points of a Segre variety, i.e., on the Segre embedding of a product of projective spaces. These codes are Reed-Muller-type analogues adapted to product projective geometries \NoCaseChange{\protect\cite{cite36}}.}\\ 
\addlinespace[\myxtraspc]
\eczhRefIndex{code:srivastava}%
\eczhListValue{\flmRefsHyperref{code:srivastava}{Srivastava code}} & \eczhListValue{A special case of a generalized Srivastava code for \(z_j = \alpha_j^{\mu}\) for some \(\mu\) and \(t=1\).}\\ 
\addlinespace[\myxtraspc]
\eczhRefIndex{code:suzuki}%
\eczhListValue{\flmRefsHyperref{code:suzuki}{Suzuki-curve code}} & \eczhListValue{Evaluation AG code of rational functions evaluated on points lying on a Suzuki curve.}\\ 
\addlinespace[\myxtraspc]
\eczhRefIndex{code:tamo_barg}%
\eczhListValue{\flmRefsHyperref{code:tamo_barg}{Tamo-Barg code}} & \eczhListValue{A family of \(q\)-ary polynomial evaluation codes that are optimal LRCs and for which \(q\) is comparable to \(n\).}\\ 
\addlinespace[\myxtraspc]
\eczhRefIndex{code:shimura}%
\eczhListValue{\flmRefsHyperref{code:shimura}{Tsfasman-Vladut-Zink (TVZ) code}} & \eczhListValue{Member of a family of AG codes obtained from algebraic curves via the residue or evaluation construction. Sequences of curves with many rational points, such as Drinfeld modular curves, classical modular curves, or Garcia-Stichtenoth curves, yield the asymptotic parameters of the TVZ bound \NoCaseChange{\protect\cite[{Sec. 15.4.2}]{cite26}}.}\\ 
\addlinespace[\myxtraspc]
\eczhRefIndex{code:extended_hamming}%
\eczhListValue{\flmRefsHyperref{code:extended_hamming}{\([2^m,2^m-m-1,4]\) Extended Hamming code}} & \eczhListValue{Member of an infinite family of RM\((m-2,m)\) codes with parameters \([2^m,2^m-m-1, 4]\) for \(m \geq 2\) that are extensions of the Hamming codes by a parity-check bit.}\\ 
\addlinespace[\myxtraspc]
\eczhRefIndex{code:biorthogonal}%
\eczhListValue{\flmRefsHyperref{code:biorthogonal}{\([2^m,m+1,2^{m-1}]\) First-order RM code}} & \eczhListValue{A member of the family of first-order RM codes.
Its codewords are the rows of the \(2^m\)-dimensional Hadamard matrix \(H\) and its negation \(-H\) with the mapping \(+1\to 0\) and \(-1\to 1\).
The family is self-orthogonal for \(m \geq 3\) \NoCaseChange{\protect\cite{cite37}}.
They form a \((2^m,2^{m+1})\) biorthogonal spherical code under the \flmRefsHyperref{ref38}{antipodal mapping}.}\\ 
\addlinespace[\myxtraspc]
\eczhRefIndex{code:simplex}%
\eczhListValue{\flmRefsHyperref{code:simplex}{\([2^m-1,m,2^{m-1}]\) simplex code}} & \eczhListValue{A member of the equidistant code family dual to the \([2^m-1,2^m-m-1,3]\) Hamming family.}\\ 
\addlinespace[\myxtraspc]
\eczhRefIndex{code:hamming}%
\eczhListValue{\flmRefsHyperref{code:hamming}{\([2^r-1,2^r-r-1,3]\) Hamming code}} & \eczhListValue{Member of an infinite family of perfect linear codes with parameters \([2^r-1,2^r-r-1, 3]\) for \(r \geq 2\).
Their \(r \times (2^r-1) \) parity-check matrix \(H\) has all possible nonzero \(r\)-bit strings as its columns.
Adding a parity check yields the \([2^r,2^r-r-1, 4]\) extended Hamming code.}\\ 
\addlinespace[\myxtraspc]
\eczhRefIndex{code:tetracode}%
\eczhListValue{\flmRefsHyperref{code:tetracode}{\([4,2,3]_3\) Tetracode}} & \eczhListValue{The \([4,2,3]_3\) ternary self-dual MDS code that has connections to lattices \NoCaseChange{\protect\cite{cite39}}. Its weight enumerator is the Gleason polynomial \(g_4\) \NoCaseChange{\protect\cite[{Rem. 4.2.6}]{cite40}}.}\\ 
\addlinespace[\myxtraspc]
\eczhRefIndex{code:reed_solomon_4}%
\eczhListValue{\flmRefsHyperref{code:reed_solomon_4}{\([4,2,3]_4\) RS\(_4\) code}} & \eczhListValue{A Type II Euclidean self-dual extended RS code that is the smallest quaternary extended QR code \NoCaseChange{\protect\cite[{pg. 296}]{cite41}\protect\cite[{Sec. 2.4.2}]{cite42}}.
Puncturing the \([4,2,3]_4\) RS\(_4\) code yields the \([3,2,2]_4\) shortened RS\(_4\) code, which is an RS code \NoCaseChange{\protect\cite[{pg. 295}]{cite41}}.}\\ 
\addlinespace[\myxtraspc]
\eczhRefIndex{code:shortened_hexacode}%
\eczhListValue{\flmRefsHyperref{code:shortened_hexacode}{\([5,3,3]_4\) Shortened hexacode}} & \eczhListValue{A perfect \([5,3,3]_4\) quaternary Hamming code that is the result of puncturing the hexacode \NoCaseChange{\protect\cite{cite43}}.}\\ 
\addlinespace[\myxtraspc]
\eczhRefIndex{code:hexacode}%
\eczhListValue{\flmRefsHyperref{code:hexacode}{\([6,3,4]_4\) Hexacode}} & \eczhListValue{The \([6,3,4]_4\) Hermitian self-dual MDS code that has connections to projective geometry, lattices \NoCaseChange{\protect\cite{cite39}}, and conformal field theory \NoCaseChange{\protect\cite{cite44}}. Its weight enumerator is the Gleason polynomial \(g_7\) \NoCaseChange{\protect\cite[{Rem. 4.2.6}]{cite40}}.}\\ 
\addlinespace[\myxtraspc]
\eczhRefIndex{code:simplex734}%
\eczhListValue{\flmRefsHyperref{code:simplex734}{\([7,3,4]\) simplex code}} & \eczhListValue{Second-smallest nontrivial member of the simplex-code family.
The columns of its generator matrix are in one-to-one correspondence with the elements of the projective space \(PG(2,2)\), with each column being a chosen representative of the corresponding element.
The codewords form a \((8,9)\) simplex spherical code under the \flmRefsHyperref{ref38}{antipodal mapping}.
As a simplex code, it is equidistant: every nonzero codeword has Hamming weight \(4\).}\\ 
\addlinespace[\myxtraspc]
\eczhRefIndex{code:hamming743}%
\eczhListValue{\flmRefsHyperref{code:hamming743}{\([7,4,3]\) Hamming code}} & \eczhListValue{Second-smallest member of the Hamming code family.}\\ 
\addlinespace[\myxtraspc]
\eczhRefIndex{code:hamming844}%
\eczhListValue{\flmRefsHyperref{code:hamming844}{\([8,4,4]\) extended Hamming code}} & \eczhListValue{Extension of the \([7,4,3]\) Hamming code by a parity-check bit.
The smallest doubly even self-dual code, and the unique Type II code of length \(8\) \NoCaseChange{\protect\cite[{Rem. 4.3.10}]{cite40}}.}\\ 
\addlinespace[\myxtraspc]
\eczhRefIndex{code:parity_check}%
\eczhListValue{\flmRefsHyperref{code:parity_check}{\([n,n-1,2]\) Single parity-check (SPC) code}} & \eczhListValue{An \([n,n-1,2]\) linear binary code whose codewords consist of the message string appended with a \textit{parity-check bit} or \textit{parity bit} such that the parity (i.e., sum over all coordinates of each codeword) is zero.
If the Hamming weight of a message is odd (even), then the parity bit is one (zero).
This code requires only one extra bit of overhead and is therefore inexpensive.
Its codewords are all even-weight binary strings, and its parity-check matrix is a row vector of all ones.
Its automorphism group is \(S_n\).}\\ 
\addlinespace[\myxtraspc]
\eczhRefIndex{code:q-ary_parity_check}%
\eczhListValue{\flmRefsHyperref{code:q-ary_parity_check}{\([n,n-1,2]_q\) \(q\)-ary parity-check code}} & \eczhListValue{An \([n,n-1,2]_q\) linear \(q\)-ary code whose codewords consist of the message string appended with a \textit{parity-check} or \textit{zero-sum check digit} such that the sum over all coordinates of each codeword is zero.}\\ 
\addlinespace[\myxtraspc]
\eczhRefIndex{code:q-ary_repetition}%
\eczhListValue{\flmRefsHyperref{code:q-ary_repetition}{\(q\)-ary repetition code}} & \eczhListValue{An \([n,1,n]_q\) code consisting of codewords \((j,j,\cdots,j)\) for \(j \in \mathbb{F}_q\).}\\ 
\addlinespace[\myxtraspc]
\eczhRefIndex{code:q-ary_simplex}%
\eczhListValue{\flmRefsHyperref{code:q-ary_simplex}{\(q\)-ary simplex code}} & \eczhListValue{An \([n,m,q^{m-1}]_q\) equidistant projective code with \(n=\frac{q^m-1}{q-1}\), denoted as \(S(q,m)\). The columns of the generator matrix are in one-to-one correspondence with the elements of the projective space \(PG(m-1,q)\), with each column being a chosen representative of the corresponding element.
All nonzero simplex codewords have a constant weight of \(q^{m-1}\) \NoCaseChange{\protect\cite{cite45,cite46}}.}\\ 
\end{tabularx}\endgroup
\eczcodelist{binary_linear}{Binary linear codes
}%

\eczhCodeListAutoDescription{All descendants of \flmRefsCref{code:binary_linear}.}%

\eczhIncludeCodeGraph{Bare}{scale=0.5}{\columnwidth}{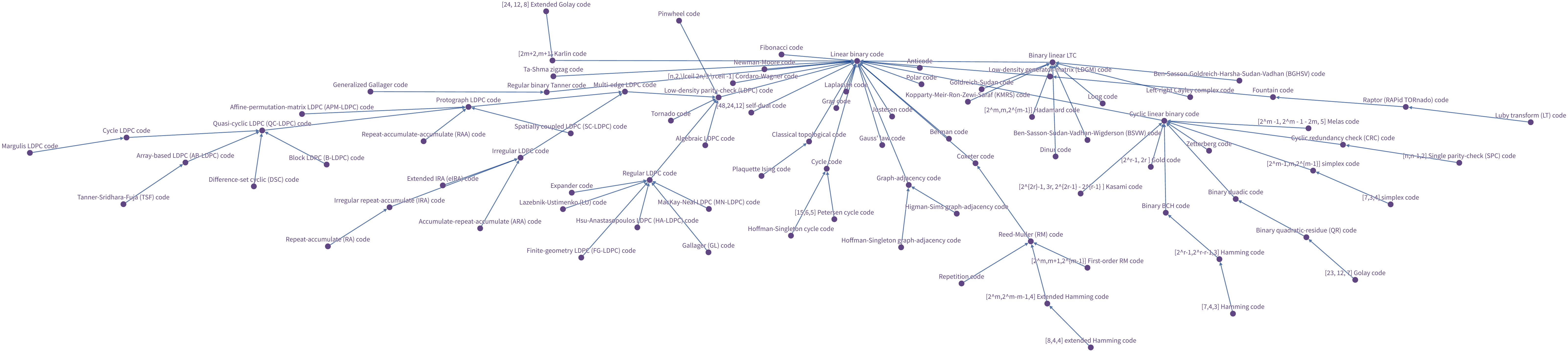}{Binary linear codes}{https://errorcorrectionzoo.org/code_graph#J\%7B\%22displayMode\%22\%3A\%22subset\%22\%2C\%22modeSubsetOptions\%22\%3A\%7B\%22codeIds\%22\%3A\%5B\%22ara\%22\%2C\%22apm_ldpc\%22\%2C\%22algebraic_ldpc\%22\%2C\%22anticode\%22\%2C\%22array_ldpc\%22\%2C\%22bsghsv-ltc\%22\%2C\%22bssvw-ltc\%22\%2C\%22berman\%22\%2C\%22bch\%22\%2C\%22binary_duadic\%22\%2C\%22binary_ltc\%22\%2C\%22binary_quad_residue\%22\%2C\%22b_ldpc\%22\%2C\%22topological_classical\%22\%2C\%22coxeter\%22\%2C\%22cycle_ldpc\%22\%2C\%22homological_classical\%22\%2C\%22binary_cyclic\%22\%2C\%22crc\%22\%2C\%22difference_set\%22\%2C\%22dinur\%22\%2C\%22expander\%22\%2C\%22extended_ira\%22\%2C\%22fibonacci_model\%22\%2C\%22pg_ldpc\%22\%2C\%22fountain\%22\%2C\%22gallager\%22\%2C\%22gauss_law\%22\%2C\%22generalized_gallager\%22\%2C\%22gs-ltc\%22\%2C\%22graph\%22\%2C\%22gray\%22\%2C\%22higman-sims_graph\%22\%2C\%22hoffman-singleton\%22\%2C\%22hoffman-singleton_graph\%22\%2C\%22ha_ldpc\%22\%2C\%22irregular_ldpc\%22\%2C\%22ira\%22\%2C\%22justesen\%22\%2C\%22kmrs-ltc\%22\%2C\%22laplacian\%22\%2C\%22lu_ldpc\%22\%2C\%22lr-cayley-complex\%22\%2C\%22binary_linear\%22\%2C\%22long\%22\%2C\%22ldgm\%22\%2C\%22ldpc\%22\%2C\%22luby_transform\%22\%2C\%22mn_ldpc\%22\%2C\%22margulis_ldpc\%22\%2C\%22multi_edge_ldpc\%22\%2C\%22newman_moore\%22\%2C\%22pinwheel\%22\%2C\%22plaquette_ising\%22\%2C\%22polar\%22\%2C\%22protograph_ldpc\%22\%2C\%22qc_ldpc\%22\%2C\%22raptor\%22\%2C\%22reed_muller\%22\%2C\%22regular_ldpc\%22\%2C\%22regular_binary_tanner\%22\%2C\%22ra\%22\%2C\%22raa\%22\%2C\%22repetition\%22\%2C\%22sc_ldpc\%22\%2C\%22ta-shma\%22\%2C\%22tsf\%22\%2C\%22tornado\%22\%2C\%22zetterberg\%22\%2C\%22petersen\%22\%2C\%22golay\%22\%2C\%22extended_golay\%22\%2C\%22melas\%22\%2C\%22extended_hamming\%22\%2C\%22biorthogonal\%22\%2C\%22hadamard\%22\%2C\%22simplex\%22\%2C\%22gold\%22\%2C\%22hamming\%22\%2C\%22kasami\%22\%2C\%22karlin\%22\%2C\%22self_dual_48_24_12\%22\%2C\%22simplex734\%22\%2C\%22hamming743\%22\%2C\%22hamming844\%22\%2C\%22cordaro_wagner\%22\%2C\%22parity_check\%22\%5D\%2C\%22reusePreviousLayoutPositions\%22\%3Afalse\%2C\%22showIntermediateConnectingNodes\%22\%3Atrue\%2C\%22connectingNodesMaxDepth\%22\%3A15\%2C\%22connectingNodesPathMaxLength\%22\%3A20\%2C\%22connectingNodesMaxExtraDepth\%22\%3A3\%2C\%22connectingNodesOnlyKeepPathsWithAdditionalLength\%22\%3A1\%2C\%22connectingNodesToDomainsAndKingdoms\%22\%3Afalse\%2C\%22connectingNodesEdgeLengthsByType\%22\%3A\%7B\%22primaryParent\%22\%3A1\%2C\%22secondaryParent\%22\%3A4\%2C\%22cousin\%22\%3A6\%7D\%2C\%22nodeIds\%22\%3A\%5B\%5D\%7D\%2C\%22highlightImportantNodes\%22\%3A\%7B\%22highlightImportantNodes\%22\%3Afalse\%2C\%22highlightPrimaryParents\%22\%3Afalse\%2C\%22highlightRootConnectingEdges\%22\%3Afalse\%7D\%7D}

\begingroup
\small
\eczhBreakableDashes
\renewcommand\arraystretch{1.05}
\edef\myxtraspc{\eczListAddVSpaceXtraXtraStretch}
\begin{tabularx}{\linewidth}{>{\raggedright\arraybackslash}p{\eczListColWidth{name}} >{\hsize=1.0000\hsize }X}
\toprule
\eczListColTitle{Code} & \eczListColTitle{Description} \\
\midrule
\endfirsthead
\toprule
\eczListColTitleContinued{Code} & \eczListColTitleContinued{Description} \\
\midrule
\endhead
\bottomrule
\endfoot
\eczhRefIndex{code:ara}%
\eczhListValue{\flmRefsHyperref{code:ara}{Accumulate-repeat-accumulate (ARA) code}} & \eczhListValue{A generalization of the RA code in which the outer repetition-code encoding step is augmented with an accumulator acting on a fraction of the incoming bits.
In addition, the code may be punctured after the final accumulating step.}\\ 
\addlinespace[\myxtraspc]
\eczhRefIndex{code:apm_ldpc}%
\eczhListValue{\flmRefsHyperref{code:apm_ldpc}{Affine-permutation-matrix LDPC (APM-LDPC) code}} & \eczhListValue{LDPC code whose parity-check matrix can be put into the form of a block matrix consisting of permutation submatrices representing the affine permutation group or the zero submatrix.
Given a cyclic group \(\mathbb{Z}_r\), the affine permutation group is \(\mathbb{Z}_r \rtimes \mathbb{Z}_r^{\times}\), where \(\mathbb{Z}_r^{\times}\) is the multiplicative group of integers modulo \(r\).
Such codes are often constructed by \flmRefsHyperref{ref47}{lifting} certain protographs into such block matrices \NoCaseChange{\protect\cite{cite48}}.}\\ 
\addlinespace[\myxtraspc]
\eczhRefIndex{code:algebraic_ldpc}%
\eczhListValue{\flmRefsHyperref{code:algebraic_ldpc}{Algebraic LDPC code}} & \eczhListValue{LDPC code whose parity check matrix is constructed explicitly (i.e., non-randomly) from a particular graph \NoCaseChange{\protect\cite{cite49,cite50}} or an algebraic structure such as a combinatorial design \NoCaseChange{\protect\cite{cite51,cite52,cite53}}, balanced incomplete block design \NoCaseChange{\protect\cite{cite54}}, a partial geometry \NoCaseChange{\protect\cite{cite55}}, a generalized polygon \NoCaseChange{\protect\cite{cite56,cite57}}, or a Latin square \NoCaseChange{\protect\cite{cite58,cite59,cite60}}.
The extra structure and/or symmetry \NoCaseChange{\protect\cite{cite61}} of these codes can often be used to gain a better understanding of their properties.}\\ 
\addlinespace[\myxtraspc]
\eczhRefIndex{code:anticode}%
\eczhListValue{\flmRefsHyperref{code:anticode}{Anticode}} & \eczhListValue{Linear binary code for which the distance between any two codewords is less than or equal to some value \(\delta\) called the maximum distance. Anticodes can be used to construct codes that saturate the Griesmer bound; see Refs. \NoCaseChange{\protect\cite{cite62,cite63,cite41}} for more details.}\\ 
\addlinespace[\myxtraspc]
\eczhRefIndex{code:array_ldpc}%
\eczhListValue{\flmRefsHyperref{code:array_ldpc}{Array-based LDPC (AB-LDPC) code}} & \eczhListValue{QC-LDPC code constructed deterministically from a disk array code known as a B-code.
Its parity-check matrix admits a compact representation \NoCaseChange{\protect\cite{cite64}} and is related to RS codes.}\\ 
\addlinespace[\myxtraspc]
\eczhRefIndex{code:bsghsv-ltc}%
\eczhListValue{\flmRefsHyperref{code:bsghsv-ltc}{Ben-Sasson-Goldreich-Harsha-Sudan-Vadhan (BGHSV) code}} & \eczhListValue{A member of a family of locally testable \([n,k,d]\) codes with \(n = k^{1+\epsilon}\) and query complexity of \flmRefsHyperref{ref65}{order} \(O(1/\epsilon)\), for any fixed \(\epsilon > 0\).}\\ 
\addlinespace[\myxtraspc]
\eczhRefIndex{code:bssvw-ltc}%
\eczhListValue{\flmRefsHyperref{code:bssvw-ltc}{Ben-Sasson-Sudan-Vadhan-Wigderson (BSVW) code}} & \eczhListValue{Locally testable \([n,k,d]\) code with \(n = k \cdot 2^{\tilde{O}(\sqrt{\log k})}\) and asymptotically constant query complexity, where \(\tilde{O}(f)=O(f\cdot (\log f)^c)\) for some fixed constant \(c\).}\\ 
\addlinespace[\myxtraspc]
\eczhRefIndex{code:berman}%
\eczhListValue{\flmRefsHyperref{code:berman}{Berman code}} & \eczhListValue{A member of a family of codes that is recursively constructed from the single parity-check code via a construction that is similar to the \((u|u+v)\) construction \NoCaseChange{\protect\cite{cite66}}.
Berman codes include RM codes as a special case.}\\ 
\addlinespace[\myxtraspc]
\eczhRefIndex{code:bch}%
\eczhListValue{\flmRefsHyperref{code:bch}{Binary BCH code}} & \eczhListValue{Cyclic binary code of odd length \(n\) whose zeroes are consecutive powers of a primitive \(n\)th root of unity \(\alpha\) (see \flmRefsCref{ref67}). More precisely, the generator polynomial of a BCH code of \textit{designed distance} \(\delta\geq 1\) is the lowest-degree monic polynomial with zeroes \(\{\alpha^b,\alpha^{b+1},\cdots,\alpha^{b+\delta-2}\}\) for some \(b\geq 0\). BCH codes are called \textit{narrow-sense} when \(b=1\), and are called \textit{primitive} when \(n=2^r-1\) for some \(r\geq 2\).}\\ 
\addlinespace[\myxtraspc]
\eczhRefIndex{code:binary_duadic}%
\eczhListValue{\flmRefsHyperref{code:binary_duadic}{Binary duadic code}} & \eczhListValue{Member of a pair of cyclic linear binary codes that satisfy certain relations, depending on whether the pair is \textit{even-like} or \textit{odd-like} duadic. Binary duadic codes generalize binary quadratic-residue codes \NoCaseChange{\protect\cite[{Sec. 2.7}]{cite68}}. Duadic codes exist for lengths \(n\) that are products of powers of primes, with each prime being \(\pm 1\) modulo \(8\) \NoCaseChange{\protect\cite{cite69}}.}\\ 
\addlinespace[\myxtraspc]
\eczhRefIndex{code:binary_ltc}%
\eczhListValue{\flmRefsHyperref{code:binary_ltc}{Binary linear LTC}} & \eczhListValue{A binary linear code \(C\) of length \(n\) that is a \((u,R)\)-LTC with query complexity \(u\) and soundness \(R>0\).}\\ 
\addlinespace[\myxtraspc]
\eczhRefIndex{code:binary_quad_residue}%
\eczhListValue{\flmRefsHyperref{code:binary_quad_residue}{Binary quadratic-residue (QR) code}} & \eczhListValue{Member of a quadruple of cyclic binary codes of prime length \(n=8m\pm 1\) for \(m\geq 1\) constructed using quadratic residues and nonresidues of \(n\) \NoCaseChange{\protect\cite[{Def. 3.2.8}]{cite70}}.}\\ 
\addlinespace[\myxtraspc]
\eczhRefIndex{code:b_ldpc}%
\eczhListValue{\flmRefsHyperref{code:b_ldpc}{Block LDPC (B-LDPC) code}} & \eczhListValue{Member of a particular class of irregular QC-LDPC codes with efficient encoders.}\\ 
\addlinespace[\myxtraspc]
\eczhRefIndex{code:topological_classical}%
\eczhListValue{\flmRefsHyperref{code:topological_classical}{Classical topological code}} & \eczhListValue{Classical code defined on a two-dimensional lattice and inspired by geometrically local topological quantum codes, such as the surface code or color code.}\\ 
\addlinespace[\myxtraspc]
\eczhRefIndex{code:coxeter}%
\eczhListValue{\flmRefsHyperref{code:coxeter}{Coxeter code}} & \eczhListValue{A member of a family of codes that generalize RM codes in a group-theoretic sense.}\\ 
\addlinespace[\myxtraspc]
\eczhRefIndex{code:cycle_ldpc}%
\eczhListValue{\flmRefsHyperref{code:cycle_ldpc}{Cycle LDPC code}} & \eczhListValue{An LDPC code whose parity-check matrix forms the incidence matrix of a graph, i.e., has weight-two columns.}\\ 
\addlinespace[\myxtraspc]
\eczhRefIndex{code:homological_classical}%
\eczhListValue{\flmRefsHyperref{code:homological_classical}{Cycle code}} & \eczhListValue{A code whose parity-check matrix is obtained from the incidence matrix of a graph over \(\mathbb{F}_2\).
This code's properties are derived from the size two chain complex associated with the graph.
Not every binary linear code is homological, but there exist homological families that asymptotically saturate the Hamming bound \NoCaseChange{\protect\cite{cite71}}.}\\ 
\addlinespace[\myxtraspc]
\eczhRefIndex{code:binary_cyclic}%
\eczhListValue{\flmRefsHyperref{code:binary_cyclic}{Cyclic linear binary code}} & \eczhListValue{A binary code of length \(n\) is cyclic if, for each codeword \(c_1 c_2 \cdots c_n\), the cyclically shifted string \(c_n c_1 \cdots c_{n-1}\) is also a codeword. A cyclic code is called \textit{primitive} when \(n=2^r-1\) for some \(r\geq 2\).}\\ 
\addlinespace[\myxtraspc]
\eczhRefIndex{code:crc}%
\eczhListValue{\flmRefsHyperref{code:crc}{Cyclic redundancy check (CRC) code}} & \eczhListValue{A generalization of the single parity-check code in which the generalization of the parity bit is the remainder of the data string (mapped into a polynomial via the \flmRefsCref{ref67}) divided by some generator polynomial.
A notable family of codes is referred to as \textit{CRC-}(\(m-1\)), where \(m\) is the length of the generator polynomial.}\\ 
\addlinespace[\myxtraspc]
\eczhRefIndex{code:difference_set}%
\eczhListValue{\flmRefsHyperref{code:difference_set}{Difference-set cyclic (DSC) code}} & \eczhListValue{Cyclic LDPC code constructed deterministically from a difference set.
Certain DSC codes satisfy more redundant constraints than Gallager codes and thus can outperform them \NoCaseChange{\protect\cite{cite72}}.}\\ 
\addlinespace[\myxtraspc]
\eczhRefIndex{code:dinur}%
\eczhListValue{\flmRefsHyperref{code:dinur}{Dinur code}} & \eczhListValue{Member of an infinite family of locally testable \([n,n/\text{polylog}(n),d]\) codes with vanishing rate. Code construction relies on tensor-product codes \NoCaseChange{\protect\cite{cite73}}.}\\ 
\addlinespace[\myxtraspc]
\eczhRefIndex{code:expander}%
\eczhListValue{\flmRefsHyperref{code:expander}{Expander code}} & \eczhListValue{LDPC code whose parity-check matrix is derived from the adjacency matrix of a bipartite expander graph \NoCaseChange{\protect\cite{cite74}} such as a Ramanujan graph or a Cayley graph of a projective special linear group over a finite field \NoCaseChange{\protect\cite{cite75,cite76}}.
Expander codes admit efficient encoding and decoding algorithms and yield an explicit (i.e., non-random) asymptotically good LDPC code family.}\\ 
\addlinespace[\myxtraspc]
\eczhRefIndex{code:extended_ira}%
\eczhListValue{\flmRefsHyperref{code:extended_ira}{Extended IRA (eIRA) code}} & \eczhListValue{A generalization of the IRA code in which the outer LDGM code is replaced by a random sparse matrix containing no weight-two columns.}\\ 
\addlinespace[\myxtraspc]
\eczhRefIndex{code:fibonacci_model}%
\eczhListValue{\flmRefsHyperref{code:fibonacci_model}{Fibonacci code}} & \eczhListValue{Quantum-inspired binary linear code defined on an \(L\times L/2\) lattice with one bit on each site, where \(L=2^N\) for an integer \(N\geq 2\). The codewords are defined to satisfy the condition that, for each lattice site \((x,y)\), the bits on \((x,y)\), \((x+1,y)\), \((x-1,y)\) and \((x,y+1)\) (where the lattice is taken to be periodic in both directions) contain an even number of \(1\)'s.}\\ 
\addlinespace[\myxtraspc]
\eczhRefIndex{code:pg_ldpc}%
\eczhListValue{\flmRefsHyperref{code:pg_ldpc}{Finite-geometry LDPC (FG-LDPC) code}} & \eczhListValue{LDPC code whose parity-check matrix is the incidence matrix of points and hyperplanes in either a Euclidean or a projective geometry.
Such codes are called \textit{Euclidean-geometry LDPC (EG-LDPC)} and \textit{projective-geometry LDPC (PG-LDPC)}, respectively.
Such constructions have been generalized to incidence matrices of hyperplanes of different dimensions \NoCaseChange{\protect\cite{cite77}}.}\\ 
\addlinespace[\myxtraspc]
\eczhRefIndex{code:fountain}%
\eczhListValue{\flmRefsHyperref{code:fountain}{Fountain code}} & \eczhListValue{Code based on the idea of generating an endless stream of custom encoded packets for the receiver. The code is designed so that the receiver can recover the original transmission of size \(Kl\) bits after receiving at least \(K\) packets each of \(l\) bits.}\\ 
\addlinespace[\myxtraspc]
\eczhRefIndex{code:gallager}%
\eczhListValue{\flmRefsHyperref{code:gallager}{Gallager (GL) code}} & \eczhListValue{The first LDPC code.
The rows of the parity-check matrix of this regular code are divided into equal subsets, and columns in the first subset are randomly permuted to yield the corresponding rows in subsequent subsets.}\\ 
\addlinespace[\myxtraspc]
\eczhRefIndex{code:gauss_law}%
\eczhListValue{\flmRefsHyperref{code:gauss_law}{Gauss' law code}} & \eczhListValue{An \([m+Dm,Dm,3]\) linear binary code for \(m\geq 3^D\), defined by the Gauss' law constraint of a \(D\)-dimensional fermionic \(\mathbb{Z}_2\) gauge theory \NoCaseChange{\protect\cite[{Thm. 1}]{cite78}}.
The code can be rephrased as a distance-one stabilizer code whose stabilizers consist of gauge-group elements.
It can be concatenated to form a stabilizer code for fault-tolerant quantum simulation of the underlying gauge theory \NoCaseChange{\protect\cite{cite79,cite78}}.}\\ 
\addlinespace[\myxtraspc]
\eczhRefIndex{code:generalized_gallager}%
\eczhListValue{\flmRefsHyperref{code:generalized_gallager}{Generalized Gallager code}} & \eczhListValue{A LDPC code that generalizes the Gallager codes using the Tanner construction.
While Gallager-code parity-check matrices consist of repetition-code submatrices that are randomly permuted, generalized Gallager-code matrices instead use more general binary linear component codes.}\\ 
\addlinespace[\myxtraspc]
\eczhRefIndex{code:gs-ltc}%
\eczhListValue{\flmRefsHyperref{code:gs-ltc}{Goldreich-Sudan code}} & \eczhListValue{Locally testable \([n,k,d]\) code with \(n = k^{1+O(1/u)}\) and distance of \flmRefsHyperref{ref65}{order} \(\Omega(n)\) for query complexity \(u\). The same work also presented a probabilistic construction of codes of size \(k^{1+o(1)}\).}\\ 
\addlinespace[\myxtraspc]
\eczhRefIndex{code:graph}%
\eczhListValue{\flmRefsHyperref{code:graph}{Graph-adjacency code}} & \eczhListValue{Binary linear code whose generator matrix is that of the row space of the adjacency matrix of a strongly regular graph.
Given an adjacency matrix \(A\), the generator matrix is either \(G=A\) or \(G=(I|A)\), where \(I\) is the identity matrix.}\\ 
\addlinespace[\myxtraspc]
\eczhRefIndex{code:gray}%
\eczhListValue{\flmRefsHyperref{code:gray}{Gray code}} & \eczhListValue{The first Gray code \NoCaseChange{\protect\cite{cite80}}, now called the \textit{binary reflected Gray code}, is a trivial code that orders length-\(n\) binary strings such that nearest-neighbor strings differ by only one digit via what is known as the \flmTerm{term}{ref81}{}{Gray map}.}\\ 
\addlinespace[\myxtraspc]
\eczhRefIndex{code:higman-sims_graph}%
\eczhListValue{\flmRefsHyperref{code:higman-sims_graph}{Higman-Sims graph-adjacency code}} & \eczhListValue{A graph-based code whose generator matrix is the row space of the adjacency matrix of the Higman-Sims graph, yielding a \([100,22,22]\) code \(C_{HS}\) whose dual is a \([100,78,6]\) code \NoCaseChange{\protect\cite[{Table IV}]{cite82}}.}\\ 
\addlinespace[\myxtraspc]
\eczhRefIndex{code:hoffman-singleton}%
\eczhListValue{\flmRefsHyperref{code:hoffman-singleton}{Hoffman-Singleton cycle code}} & \eczhListValue{A \([50,21,12]\) cycle code whose parity-check matrix is the incidence matrix of the Hoffman-Singleton graph \NoCaseChange{\protect\cite{cite83}}.
Its dual is a \([50,29,8]\) code \NoCaseChange{\protect\cite[{Table II}]{cite82}}.}\\ 
\addlinespace[\myxtraspc]
\eczhRefIndex{code:hoffman-singleton_graph}%
\eczhListValue{\flmRefsHyperref{code:hoffman-singleton_graph}{Hoffman-Singleton graph-adjacency code}} & \eczhListValue{A graph-based code whose generator matrix is the row space of the adjacency matrix of the Hoffman-Singleton graph \NoCaseChange{\protect\cite{cite83}}.
This \([50,22,7]\) code is dual to a \([50,28,8]\) code \NoCaseChange{\protect\cite[{Table III}]{cite82}}.}\\ 
\addlinespace[\myxtraspc]
\eczhRefIndex{code:ha_ldpc}%
\eczhListValue{\flmRefsHyperref{code:ha_ldpc}{Hsu-Anastasopoulos LDPC (HA-LDPC) code}} & \eczhListValue{A regular LDPC code obtained from a concatenation of a certain random regular LDPC code and a certain random LDGM code.
An \((l,r,g)\)-HA-LDPC code can be written using punctured LDPC and LDGM parts, and it is dual to the corresponding \((r,l,g)\)-MN-LDPC code \NoCaseChange{\protect\cite{cite84}}.
Using rate-one LDGM codes eliminates high-weight codewords while admitting an amount of low-weight codewords that asymptotically vanishes, allowing code families to achieve the \flmRefsHyperref{ref85}{GV bound} with high probability.}\\ 
\addlinespace[\myxtraspc]
\eczhRefIndex{code:irregular_ldpc}%
\eczhListValue{\flmRefsHyperref{code:irregular_ldpc}{Irregular LDPC code}} & \eczhListValue{An LDPC code whose parity-check matrix has a variable number of entries in each row or column.}\\ 
\addlinespace[\myxtraspc]
\eczhRefIndex{code:ira}%
\eczhListValue{\flmRefsHyperref{code:ira}{Irregular repeat-accumulate (IRA) code}} & \eczhListValue{A generalization of the RA code in which the outer 1-in-3 repetition encoding step is replaced by an LDGM code.
A simple version is when different bits in the RA block are repeated a different number of times.}\\ 
\addlinespace[\myxtraspc]
\eczhRefIndex{code:justesen}%
\eczhListValue{\flmRefsHyperref{code:justesen}{Justesen code}} & \eczhListValue{Binary linear code resulting from generalized concatenation of an outer RS code with multiple inner codes sampled from the Wozencraft ensemble, i.e., \(N\) distinct binary inner codes of dimension \(m\) and length \(2m\).
Justesen codes were among the first explicit asymptotically good codes \NoCaseChange{\protect\cite[{Sec. 3.3.3}]{cite70}}.}\\ 
\addlinespace[\myxtraspc]
\eczhRefIndex{code:kmrs-ltc}%
\eczhListValue{\flmRefsHyperref{code:kmrs-ltc}{Kopparty-Meir-Ron-Zewi-Saraf (KMRS) code}} & \eczhListValue{Member of a family of locally testable binary linear codes with constant rate, constant relative distance, and subpolynomial query complexity \(u = (\log n)^{O(\log \log n)}\).
Later work by Gopi, Kopparty, Oliveira, Ron-Zewi, and Saraf \NoCaseChange{\protect\cite{cite86}} showed that related concatenated codes achieve the \flmRefsHyperref{ref85}{GV bound}.}\\ 
\addlinespace[\myxtraspc]
\eczhRefIndex{code:laplacian}%
\eczhListValue{\flmRefsHyperref{code:laplacian}{Laplacian code}} & \eczhListValue{A binary linear code whose parity-check matrix is the graph Laplacian reduced mod 2.
For an undirected graph \(\Gamma\) with degree matrix \(D\) and adjacency matrix \(A\), the parity-check matrix is the symmetric matrix \(H=(D-A)\bmod 2\).}\\ 
\addlinespace[\myxtraspc]
\eczhRefIndex{code:lu_ldpc}%
\eczhListValue{\flmRefsHyperref{code:lu_ldpc}{Lazebnik-Ustimenko (LU) code}} & \eczhListValue{LDPC code whose Tanner graph comes from a particular family of \(q\)-regular graphs \NoCaseChange{\protect\cite{cite87}} of known girth and relatively large stopping sets.}\\ 
\addlinespace[\myxtraspc]
\eczhRefIndex{code:lr-cayley-complex}%
\eczhListValue{\flmRefsHyperref{code:lr-cayley-complex}{Left-right Cayley complex code}} & \eczhListValue{Binary code constructed on a left-right Cayley complex using a pair of base codes \(C_A,C_B\) and an expander graph \NoCaseChange{\protect\cite{cite74}}. Bits live on the squares of the complex, while local constraints are imposed on edges; for a fixed graph vertex, the incident symbols form a codeword of the tensor code \(C_A \otimes C_B\). A family of such codes is one of the first \(c^3\)-LTCs \NoCaseChange{\protect\cite{cite88}}.}\\ 
\addlinespace[\myxtraspc]
\eczhRefIndex{code:binary_linear}%
\eczhListValue{\flmRefsHyperref{code:binary_linear}{Linear binary code}} & \eczhListValue{An \((n,2^k,d)\) linear code is denoted as \([n,k]\) or \([n,k,d]\), where \(k\) is the code's dimension, and where \(d\) is the code's distance. Its codewords form a linear subspace, i.e., for any codewords \(x,y\), \(x+y\) is also a codeword. A code that is not linear is called \textit{nonlinear}.}\\ 
\addlinespace[\myxtraspc]
\eczhRefIndex{code:long}%
\eczhListValue{\flmRefsHyperref{code:long}{Long code}} & \eczhListValue{Nonlinear locally testable code of extremely large length that is not practical, but is useful for certain probabilistically checkable proof (PCP) constructions \NoCaseChange{\protect\cite{cite89}}.}\\ 
\addlinespace[\myxtraspc]
\eczhRefIndex{code:ldgm}%
\eczhListValue{\flmRefsHyperref{code:ldgm}{Low-density generator-matrix (LDGM) code}} & \eczhListValue{Binary linear code with a sparse generator matrix.
Alternatively, a member of an infinite family of \([n,k,d]\) codes for which the number of nonzero entries in each row and column of the generator matrix are both bounded by a constant as \(n\to\infty\). The dual of an LDGM code has a sparse parity-check matrix and is called an LDPC code.}\\ 
\addlinespace[\myxtraspc]
\eczhRefIndex{code:ldpc}%
\eczhListValue{\flmRefsHyperref{code:ldpc}{Low-density parity-check (LDPC) code}} & \eczhListValue{A binary linear code with a sparse parity-check matrix.
Often a member of an infinite family of \([n,k,d]\) codes for which the numbers of nonzero entries in each row and in each column of the parity-check matrix are both bounded above by a constant as \(n\to\infty\).}\\ 
\addlinespace[\myxtraspc]
\eczhRefIndex{code:luby_transform}%
\eczhListValue{\flmRefsHyperref{code:luby_transform}{Luby transform (LT) code}} & \eczhListValue{Erasure codes based on fountain codes. They improve on random linear fountain codes by having a much more efficient encoding and decoding algorithm.}\\ 
\addlinespace[\myxtraspc]
\eczhRefIndex{code:mn_ldpc}%
\eczhListValue{\flmRefsHyperref{code:mn_ldpc}{MacKay-Neal LDPC (MN-LDPC) code}} & \eczhListValue{A code whose parity-check matrix is constructed non-deterministically via the MacKay-Neal prescription.}\\ 
\addlinespace[\myxtraspc]
\eczhRefIndex{code:margulis_ldpc}%
\eczhListValue{\flmRefsHyperref{code:margulis_ldpc}{Margulis LDPC code}} & \eczhListValue{Member of a class of LDPC codes deterministically constructed from explicit sparse regular expander graphs.
The underlying Margulis-Gabber-Galil graph family provides explicit expanders \NoCaseChange{\protect\cite{cite49,cite90}}, yielding deterministic sparse parity-check matrices.
Related explicit LDPC constructions \NoCaseChange{\protect\cite{cite91}} utilize Ramanujan graphs \NoCaseChange{\protect\cite{cite75,cite76}}.}\\ 
\addlinespace[\myxtraspc]
\eczhRefIndex{code:multi_edge_ldpc}%
\eczhListValue{\flmRefsHyperref{code:multi_edge_ldpc}{Multi-edge LDPC code}} & \eczhListValue{Irregular LDPC code whose construction generalizes those of the original examples of irregular LDPC as well as RA codes.}\\ 
\addlinespace[\myxtraspc]
\eczhRefIndex{code:newman_moore}%
\eczhListValue{\flmRefsHyperref{code:newman_moore}{Newman-Moore code}} & \eczhListValue{Member of a family of \([L^2,O(L),O(L^{\frac{\log 3}{\log 2}})]\) binary linear codes on \(L\times L\) square lattices that form the ground-state subspace of a class of exactly solvable spin-glass models with three-body interactions.
The codewords resemble the Sierpinski triangle on a square lattice, which can be generated by a cellular automaton \NoCaseChange{\protect\cite{cite92}}.}\\ 
\addlinespace[\myxtraspc]
\eczhRefIndex{code:pinwheel}%
\eczhListValue{\flmRefsHyperref{code:pinwheel}{Pinwheel code}} & \eczhListValue{A geometrically local binary LDPC code defined on planar graphs obtained from the pinwheel tiling \NoCaseChange{\protect\cite{cite93}}.
Both bits and checks live on vertices of the graph.
If \(L_N\) is the graph Laplacian at generation \(N\), the undepleted check matrix is \(\tilde H_N=(L_N-\mathbb{I})\bmod 2\), and the actual parity-check matrix \(H_N\) is obtained by removing an evenly spaced fraction of boundary checks.}\\ 
\addlinespace[\myxtraspc]
\eczhRefIndex{code:plaquette_ising}%
\eczhListValue{\flmRefsHyperref{code:plaquette_ising}{Plaquette Ising code}} & \eczhListValue{Classical code defined on a cubic lattice in usually two or three dimensions whose parity checks are applied on the four vertices of each square.}\\ 
\addlinespace[\myxtraspc]
\eczhRefIndex{code:polar}%
\eczhListValue{\flmRefsHyperref{code:polar}{Polar code}} & \eczhListValue{A member of a family of binary linear codes whose encoding is based on picking the "most reliable bits" at the output of a noise channel.}\\ 
\addlinespace[\myxtraspc]
\eczhRefIndex{code:protograph_ldpc}%
\eczhListValue{\flmRefsHyperref{code:protograph_ldpc}{Protograph LDPC code}} & \eczhListValue{Binary version of a \(q\)-ary protograph LDPC code.
Its parity-check matrix can be written as a block matrix whose blocks are either sums of permutation matrices or the zero matrix.}\\ 
\addlinespace[\myxtraspc]
\eczhRefIndex{code:qc_ldpc}%
\eczhListValue{\flmRefsHyperref{code:qc_ldpc}{Quasi-cyclic LDPC (QC-LDPC) code}} & \eczhListValue{LDPC code that can be put into quasi-cyclic form.
Its parity check matrix can be put into the form of a block matrix consisting of either circulant permutation sub-matrices or the zero sub-matrix.
Such codes are often constructed by \flmRefsHyperref{ref47}{lifting} certain protographs into such block matrices \NoCaseChange{\protect\cite{cite48}}.
Their simple structure makes them useful for several wireless communication standards.}\\ 
\addlinespace[\myxtraspc]
\eczhRefIndex{code:raptor}%
\eczhListValue{\flmRefsHyperref{code:raptor}{Raptor (RAPid TORnado) code}} & \eczhListValue{Raptor codes are concatenated erasure codes with two layers: an outer \textit{pre-code} and a Luby-Transform (LT) inner code.
The pre-code is a linear binary erasure code, which is applied first to the input to create some redundant data.
The LT code is then applied to the intermediate symbols from the pre-code to generate final output symbols to be transmitted.}\\ 
\addlinespace[\myxtraspc]
\eczhRefIndex{code:reed_muller}%
\eczhListValue{\flmRefsHyperref{code:reed_muller}{Reed-Muller (RM) code}} & \eczhListValue{Member of the RM\((r,m)\) family of linear binary codes derived from multivariate polynomials. The code parameters are \([2^m,\sum_{j=0}^{r} {m \choose j},2^{m-r}]\), where \(r\) is the \textit{order} of the code satisfying \(0\leq r\leq m\).
First-order RM codes are also called biorthogonal codes, while \(m\)th order RM codes are also called \textit{universe} codes.
\textit{Punctured RM codes} RM\(^*(r,m)\) are obtained from RM codes by deleting one coordinate from each codeword.}\\ 
\addlinespace[\myxtraspc]
\eczhRefIndex{code:regular_ldpc}%
\eczhListValue{\flmRefsHyperref{code:regular_ldpc}{Regular LDPC code}} & \eczhListValue{An LDPC code whose parity-check matrix has a fixed number of ones in each row and each column.
Such a code is called \((j,k)\)-regular if each column has weight \(j\) (variable-node degree) and each row has weight \(k\) (check-node degree).
If the parity-check matrix has \(n\) columns and \(m\) rows, then regularity implies \(nj = mk\).}\\ 
\addlinespace[\myxtraspc]
\eczhRefIndex{code:regular_binary_tanner}%
\eczhListValue{\flmRefsHyperref{code:regular_binary_tanner}{Regular binary Tanner code}} & \eczhListValue{A binary Tanner code defined on a regular bipartite graph, with the inner code being the same for all vertices.}\\ 
\addlinespace[\myxtraspc]
\eczhRefIndex{code:ra}%
\eczhListValue{\flmRefsHyperref{code:ra}{Repeat-accumulate (RA) code}} & \eczhListValue{An LDPC code whose parity-check matrix has weight-two columns arranged in a step-like pattern for its last columns \NoCaseChange{\protect\cite{cite94}}.}\\ 
\addlinespace[\myxtraspc]
\eczhRefIndex{code:raa}%
\eczhListValue{\flmRefsHyperref{code:raa}{Repeat-accumulate-accumulate (RAA) code}} & \eczhListValue{Generalization of the RA code in which two accumulators and permutations are used.}\\ 
\addlinespace[\myxtraspc]
\eczhRefIndex{code:repetition}%
\eczhListValue{\flmRefsHyperref{code:repetition}{Repetition code}} & \eczhListValue{\([n,1,n]\) binary linear code encoding one bit of information into an \(n\)-bit string.
Majority decoding requires \(n\) to be odd in order to avoid ties.
The idea is to increase the code distance by repeating the logical information several times. It is a \((n,1)\)-Hamming code.
Its automorphism group is \(S_n\).}\\ 
\addlinespace[\myxtraspc]
\eczhRefIndex{code:sc_ldpc}%
\eczhListValue{\flmRefsHyperref{code:sc_ldpc}{Spatially coupled LDPC (SC-LDPC) code}} & \eczhListValue{An LDPC code whose parity-check matrix is constructed by "spatially" coupling several copies of a regular LDPC parity-check matrix in chain-like fashion (or, more generally, in grid-like fashion) to yield a band matrix.
A finite-length chain is then capped by imposing either open boundary conditions (yielding \textit{non-tail-biting} SC-LDPC codes) or periodic boundary conditions (yielding \textit{tail-biting} SC-LDPC codes); sometimes extra \textit{terminating vertices} are added to the ends of the chain.
Matrices corresponding to translationally invariant chains are called \textit{time-invariant}, and otherwise are called \textit{time-varying}.
These codes can be constructed, e.g., using the \flmRefsHyperref{ref47}{lifting} procedure or using edge-cutting vectors \NoCaseChange{\protect\cite{cite95}}.
Spatial coupling can also be applied to MN-LDPC and HA-LDPC protographs, yielding bounded-density SC-MN and SC-HA families \NoCaseChange{\protect\cite{cite84}}.}\\ 
\addlinespace[\myxtraspc]
\eczhRefIndex{code:ta-shma}%
\eczhListValue{\flmRefsHyperref{code:ta-shma}{Ta-Shma zigzag code}} & \eczhListValue{Member of a family of \(\epsilon\)-balanced codes that nearly achieves the \flmRefsHyperref{ref85}{asymptotic GV bound}. The codes have relative distance \(\frac{1}{2}-\frac{\epsilon}{2}\) and rate of \flmRefsHyperref{ref65}{order} \(\Omega (\epsilon^{2+\beta})\) for \(\beta\to 0\) as \(n\to\infty\) \NoCaseChange{\protect\cite{cite96}}.}\\ 
\addlinespace[\myxtraspc]
\eczhRefIndex{code:tsf}%
\eczhListValue{\flmRefsHyperref{code:tsf}{Tanner-Sridhara-Fuja (TSF) code}} & \eczhListValue{Array QC-LDPC code constructed from a cyclically shifted identity matrix; see \NoCaseChange{\protect\cite[{Exam. 21.6.5}]{cite97}}.}\\ 
\addlinespace[\myxtraspc]
\eczhRefIndex{code:tornado}%
\eczhListValue{\flmRefsHyperref{code:tornado}{Tornado code}} & \eczhListValue{Linear binary erasure code that is a precursor to fountain codes and is built from a multilayer cascade of sparse bipartite graphs. Its encoding and decoding operations involve only XOR gates \NoCaseChange{\protect\cite{cite98,cite99}\protect\cite[{Sec. 30.2}]{cite100}}.}\\ 
\addlinespace[\myxtraspc]
\eczhRefIndex{code:zetterberg}%
\eczhListValue{\flmRefsHyperref{code:zetterberg}{Zetterberg code}} & \eczhListValue{Family of binary cyclic \([2^{2s}+1,2^{2s}-4s+1]\) codes with distance \(d\geq 5\) generated by the minimal polynomial \(g_s(x)\) of \(\alpha\) over \(\mathbb{F}_2\), where \(\alpha\) is a primitive \(n\)th root of unity in the field \(\mathbb{F}_{2^{4s}}\). They are quasi-perfect codes and are one of the best known families of double-error correcting binary linear codes.}\\ 
\addlinespace[\myxtraspc]
\eczhRefIndex{code:petersen}%
\eczhListValue{\flmRefsHyperref{code:petersen}{\([15,6,5]\) Petersen cycle code}} & \eczhListValue{A \([15,6,5]\) cycle code whose parity-check matrix is the incidence matrix of the Petersen graph.
The Petersen graph can be thought of as a dodecahedron with antipodes identified \NoCaseChange{\protect\cite[{Appx. A.2.1}]{cite101}}.}\\ 
\addlinespace[\myxtraspc]
\eczhRefIndex{code:golay}%
\eczhListValue{\flmRefsHyperref{code:golay}{\([23, 12, 7]\) Golay code}} & \eczhListValue{A \([23, 12, 7]\) perfect binary linear code with connections to various areas of mathematics, e.g., lattices \NoCaseChange{\protect\cite{cite39}} and sporadic simple groups \NoCaseChange{\protect\cite{cite41}}.
Up to equivalence, it is unique for its parameters \NoCaseChange{\protect\cite{cite102}}.
The dual of the Golay code is its \([23,11,8]\) even-weight subcode \NoCaseChange{\protect\cite{cite103,cite104}}.}\\ 
\addlinespace[\myxtraspc]
\eczhRefIndex{code:extended_golay}%
\eczhListValue{\flmRefsHyperref{code:extended_golay}{\([24, 12, 8]\) Extended Golay code}} & \eczhListValue{A self-dual \([24, 12, 8]\) code that is obtained from the Golay code by adding a parity check.
Equivalently, puncturing any coordinate yields the \([23,12,7]\) Golay code.
Up to equivalence, it is unique for its parameters \NoCaseChange{\protect\cite{cite102}}, and it is the unique \([24,12,8]\) extremal Type II code \NoCaseChange{\protect\cite[{Rems. 4.3.10 and 4.3.11}]{cite40}}.}\\ 
\addlinespace[\myxtraspc]
\eczhRefIndex{code:melas}%
\eczhListValue{\flmRefsHyperref{code:melas}{\([2^m -1, 2^m - 1 - 2m, 5]\) Melas code}} & \eczhListValue{Cyclic linear code whose generator polynomial is \(g(x) = p(x)p(x)^{\star}\), where \(p(x)\) is a primitive polynomial of degree \(m\) that is the minimal polynomial over \(\mathbb{F}_2\) of an element \(\alpha\) of order \(2^m -1\) in \(\mathbb{F}_{2^m}\), \(m\) is odd and greater than five, and '\(\star\)' denotes reciprocation \NoCaseChange{\protect\cite{cite105}}.}\\ 
\addlinespace[\myxtraspc]
\eczhRefIndex{code:extended_hamming}%
\eczhListValue{\flmRefsHyperref{code:extended_hamming}{\([2^m,2^m-m-1,4]\) Extended Hamming code}} & \eczhListValue{Member of an infinite family of RM\((m-2,m)\) codes with parameters \([2^m,2^m-m-1, 4]\) for \(m \geq 2\) that are extensions of the Hamming codes by a parity-check bit.}\\ 
\addlinespace[\myxtraspc]
\eczhRefIndex{code:biorthogonal}%
\eczhListValue{\flmRefsHyperref{code:biorthogonal}{\([2^m,m+1,2^{m-1}]\) First-order RM code}} & \eczhListValue{A member of the family of first-order RM codes.
Its codewords are the rows of the \(2^m\)-dimensional Hadamard matrix \(H\) and its negation \(-H\) with the mapping \(+1\to 0\) and \(-1\to 1\).
The family is self-orthogonal for \(m \geq 3\) \NoCaseChange{\protect\cite{cite37}}.
They form a \((2^m,2^{m+1})\) biorthogonal spherical code under the \flmRefsHyperref{ref38}{antipodal mapping}.}\\ 
\addlinespace[\myxtraspc]
\eczhRefIndex{code:hadamard}%
\eczhListValue{\flmRefsHyperref{code:hadamard}{\([2^m,m,2^{m-1}]\) Hadamard code}} & \eczhListValue{An \([2^m,m,2^{m-1}]\) balanced binary code.
The \([2^m,m+1,2^{m-1}]\) augmented Hadamard code is the first-order RM code (a.k.a. RM\((1,m)\)), while the \([2^m-1,m,2^{m-1}]\) shortened Hadamard code is the simplex code (a.k.a. RM\(^*(1,m)\)).}\\ 
\addlinespace[\myxtraspc]
\eczhRefIndex{code:simplex}%
\eczhListValue{\flmRefsHyperref{code:simplex}{\([2^m-1,m,2^{m-1}]\) simplex code}} & \eczhListValue{A member of the equidistant code family dual to the \([2^m-1,2^m-m-1,3]\) Hamming family.}\\ 
\addlinespace[\myxtraspc]
\eczhRefIndex{code:gold}%
\eczhListValue{\flmRefsHyperref{code:gold}{\([2^r-1, 2r ]\) Gold code}} & \eczhListValue{A cyclic binary linear code characterized by the generator polynomial of degree \(r\) of two maximum-period sequences of period \(2^r-1\) with absolute cross-correlation \( \leq 2^{(r+2)/2}\). Gold codewords are generated using \(m\)-sequences \(x\) and \(y\), which are codewords of simplex codes with check polynomials of degree \(r\) \NoCaseChange{\protect\cite{cite106}}.}\\ 
\addlinespace[\myxtraspc]
\eczhRefIndex{code:hamming}%
\eczhListValue{\flmRefsHyperref{code:hamming}{\([2^r-1,2^r-r-1,3]\) Hamming code}} & \eczhListValue{Member of an infinite family of perfect linear codes with parameters \([2^r-1,2^r-r-1, 3]\) for \(r \geq 2\).
Their \(r \times (2^r-1) \) parity-check matrix \(H\) has all possible nonzero \(r\)-bit strings as its columns.
Adding a parity check yields the \([2^r,2^r-r-1, 4]\) extended Hamming code.}\\ 
\addlinespace[\myxtraspc]
\eczhRefIndex{code:kasami}%
\eczhListValue{\flmRefsHyperref{code:kasami}{\([2^{2r}-1, 3r, 2^{2r-1} - 2^{r-1} ]\) Kasami code}} & \eczhListValue{Member of the family of \([2^{2r}-1, 3r, 2^{2r-1} - 2^{r-1} ]\) cyclic binary linear codes. 
Gold and Kasami codes are both constructed by picking a set of cyclically unrelated sequences of binary linear codes with low cross-correlation \NoCaseChange{\protect\cite{cite107,cite108}}.}\\ 
\addlinespace[\myxtraspc]
\eczhRefIndex{code:karlin}%
\eczhListValue{\flmRefsHyperref{code:karlin}{\([2m+2,m+1]\) Karlin code}} & \eczhListValue{Member of the family of \([2m+2,m+1]\) double circulant codes such that \(m\) is prime of the form \(8k+3\) for some \(k\), and \(2m+2\) is a multiple of eight.
See \NoCaseChange{\protect\cite[{Ch. 16}]{cite41}} for their generator matrix.
Karlin codes can be mapped to extended cyclic and extended quadratic-residue codes over \(\mathbb{F}_4\) \NoCaseChange{\protect\cite{cite109,cite110}\protect\cite[{Ch. 16}]{cite41}\protect\cite[{Sec. 2.4.2}]{cite42}} by identifying \((0,\omega,\bar{\omega},1)\) with \((00),(10),(01),(11)\) \NoCaseChange{\protect\cite{cite109}}.}\\ 
\addlinespace[\myxtraspc]
\eczhRefIndex{code:self_dual_48_24_12}%
\eczhListValue{\flmRefsHyperref{code:self_dual_48_24_12}{\([48,24,12]\) self-dual code}} & \eczhListValue{An extended quadratic-residue code that is the unique self-dual doubly even \([48,24,12]\) code. It is extremal Type II, and its automorphism group is \(PSL(2,47)\) \NoCaseChange{\protect\cite{cite111}\protect\cite[{Rem. 4.3.11}]{cite40}}.}\\ 
\addlinespace[\myxtraspc]
\eczhRefIndex{code:simplex734}%
\eczhListValue{\flmRefsHyperref{code:simplex734}{\([7,3,4]\) simplex code}} & \eczhListValue{Second-smallest nontrivial member of the simplex-code family.
The columns of its generator matrix are in one-to-one correspondence with the elements of the projective space \(PG(2,2)\), with each column being a chosen representative of the corresponding element.
The codewords form a \((8,9)\) simplex spherical code under the \flmRefsHyperref{ref38}{antipodal mapping}.
As a simplex code, it is equidistant: every nonzero codeword has Hamming weight \(4\).}\\ 
\addlinespace[\myxtraspc]
\eczhRefIndex{code:hamming743}%
\eczhListValue{\flmRefsHyperref{code:hamming743}{\([7,4,3]\) Hamming code}} & \eczhListValue{Second-smallest member of the Hamming code family.}\\ 
\addlinespace[\myxtraspc]
\eczhRefIndex{code:hamming844}%
\eczhListValue{\flmRefsHyperref{code:hamming844}{\([8,4,4]\) extended Hamming code}} & \eczhListValue{Extension of the \([7,4,3]\) Hamming code by a parity-check bit.
The smallest doubly even self-dual code, and the unique Type II code of length \(8\) \NoCaseChange{\protect\cite[{Rem. 4.3.10}]{cite40}}.}\\ 
\addlinespace[\myxtraspc]
\eczhRefIndex{code:cordaro_wagner}%
\eczhListValue{\flmRefsHyperref{code:cordaro_wagner}{\([n,2,\lceil 2n/3 \rceil -1]\) Cordaro-Wagner code}} & \eczhListValue{A linear binary code with \(k=2\) that is optimal among binary linear \([n,2,d]\) codes.}\\ 
\addlinespace[\myxtraspc]
\eczhRefIndex{code:parity_check}%
\eczhListValue{\flmRefsHyperref{code:parity_check}{\([n,n-1,2]\) Single parity-check (SPC) code}} & \eczhListValue{An \([n,n-1,2]\) linear binary code whose codewords consist of the message string appended with a \textit{parity-check bit} or \textit{parity bit} such that the parity (i.e., sum over all coordinates of each codeword) is zero.
If the Hamming weight of a message is odd (even), then the parity bit is one (zero).
This code requires only one extra bit of overhead and is therefore inexpensive.
Its codewords are all even-weight binary strings, and its parity-check matrix is a row vector of all ones.
Its automorphism group is \(S_n\).}\\ 
\end{tabularx}\endgroup
\eczcodelist{self_dual}{Classical self-dual objects}%

\eczhCodeListAutoDescription{Union of:
\begin{itemize}\item codes that are descendants of \flmRefsCref{code:self_dual_over_rings}
\item codes that are descendants of \flmRefsCref{code:self_dual_additive}
\item codes that are descendants of \flmRefsCref{code:self_dual_polytope}
\item codes that are descendants of \flmRefsCref{code:self_dual_lattice}
\end{itemize}}%

\eczhIncludeCodeGraph{Bare}{scale=0.5}{\columnwidth}{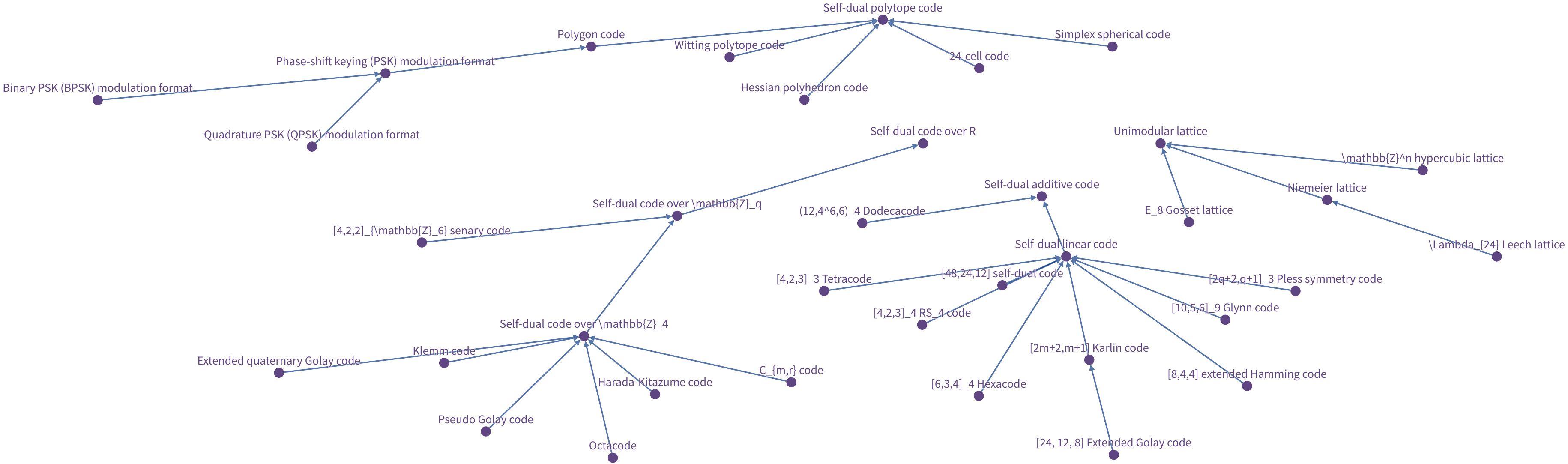}{Classical self-dual objects}{https://errorcorrectionzoo.org/code_graph#J\%7B\%22displayMode\%22\%3A\%22subset\%22\%2C\%22modeSubsetOptions\%22\%3A\%7B\%22codeIds\%22\%3A\%5B\%2224cell\%22\%2C\%22bpsk\%22\%2C\%22quaternary_golay\%22\%2C\%22harada_kitazume\%22\%2C\%22hessian_polyhedron\%22\%2C\%22klemm\%22\%2C\%22niemeier\%22\%2C\%22octacode\%22\%2C\%22psk\%22\%2C\%22polygon\%22\%2C\%22pseudo_golay\%22\%2C\%22qpsk\%22\%2C\%22self_dual_additive\%22\%2C\%22self_dual_over_rings\%22\%2C\%22self_dual_over_z4\%22\%2C\%22self_dual_over_zq\%22\%2C\%22self_dual\%22\%2C\%22self_dual_polytope\%22\%2C\%22simplex_spherical\%22\%2C\%22self_dual_lattice\%22\%2C\%22witting_polytope\%22\%2C\%22dodecacode\%22\%2C\%22cmr\%22\%2C\%22eeight\%22\%2C\%22glynn\%22\%2C\%22extended_golay\%22\%2C\%22karlin\%22\%2C\%22pless_symmetry\%22\%2C\%22self_dual_z6\%22\%2C\%22tetracode\%22\%2C\%22reed_solomon_4\%22\%2C\%22self_dual_48_24_12\%22\%2C\%22hexacode\%22\%2C\%22hamming844\%22\%2C\%22leech\%22\%2C\%22hypercubic\%22\%5D\%2C\%22reusePreviousLayoutPositions\%22\%3Afalse\%2C\%22showIntermediateConnectingNodes\%22\%3Atrue\%2C\%22connectingNodesMaxDepth\%22\%3A15\%2C\%22connectingNodesPathMaxLength\%22\%3A20\%2C\%22connectingNodesMaxExtraDepth\%22\%3A3\%2C\%22connectingNodesOnlyKeepPathsWithAdditionalLength\%22\%3A1\%2C\%22connectingNodesToDomainsAndKingdoms\%22\%3Afalse\%2C\%22connectingNodesEdgeLengthsByType\%22\%3A\%7B\%22primaryParent\%22\%3A1\%2C\%22secondaryParent\%22\%3A4\%2C\%22cousin\%22\%3A6\%7D\%2C\%22nodeIds\%22\%3A\%5B\%5D\%7D\%2C\%22highlightImportantNodes\%22\%3A\%7B\%22highlightImportantNodes\%22\%3Afalse\%2C\%22highlightPrimaryParents\%22\%3Afalse\%2C\%22highlightRootConnectingEdges\%22\%3Afalse\%7D\%7D}

\begingroup
\small
\eczhBreakableDashes
\renewcommand\arraystretch{1.05}
\edef\myxtraspc{\eczListAddVSpaceXtraXtraStretch}
\begin{tabularx}{\linewidth}{>{\raggedright\arraybackslash}p{\eczListColWidth{name}} >{\hsize=1.0000\hsize }X}
\toprule
\eczListColTitle{Code} & \eczListColTitle{Description} \\
\midrule
\endfirsthead
\toprule
\eczListColTitleContinued{Code} & \eczListColTitleContinued{Description} \\
\midrule
\endhead
\bottomrule
\endfoot
\eczhRefIndex{code:24cell}%
\eczhListValue{\flmRefsHyperref{code:24cell}{24-cell code}} & \eczhListValue{Spherical \((4,24,1)\) code whose codewords are the vertices of the 24-cell.
Codewords form the minimal lattice-shell code of the \(D_4\) lattice.}\\ 
\addlinespace[\myxtraspc]
\eczhRefIndex{code:bpsk}%
\eczhListValue{\flmRefsHyperref{code:bpsk}{Binary PSK (BPSK) modulation format}} & \eczhListValue{Encodes one bit of information into a constellation of antipodal points \(\pm\alpha\) for complex \(\alpha\).
These points are typically associated with two phases of an electromagnetic signal.}\\ 
\addlinespace[\myxtraspc]
\eczhRefIndex{code:quaternary_golay}%
\eczhListValue{\flmRefsHyperref{code:quaternary_golay}{Extended quaternary Golay code}} & \eczhListValue{An extended quadratic residue quaternary linear \((24,4^{12},12)_{\mathbb{Z}_4}\) code that is a quaternary version of the Golay code.
The code has Lee distance 12, Hamming distance 8, and Euclidean distance 16 \NoCaseChange{\protect\cite{cite112}}. 
Under the \flmTerm{term}{ref81}{}{Gray map}, the code yields a formally self-dual binary \((48,2^{24},12)\) code whose distance distribution is the \flmRefsHyperref{ref113}{MacWilliams transform} of the distance distribution of its dual code \NoCaseChange{\protect\cite{cite112}}.
The lattice \(A(C)/2\) obtained from the code is the Leech lattice \NoCaseChange{\protect\cite{cite112}}.}\\ 
\addlinespace[\myxtraspc]
\eczhRefIndex{code:harada_kitazume}%
\eczhListValue{\flmRefsHyperref{code:harada_kitazume}{Harada-Kitazume code}} & \eczhListValue{A member of a family of extremal Type II self-dual codes over \(\mathbb{Z}_4\) that yield all Niemeier lattices via \flmTerm{term}{ref114}{}{Construction \(A_4\)}.}\\ 
\addlinespace[\myxtraspc]
\eczhRefIndex{code:hessian_polyhedron}%
\eczhListValue{\flmRefsHyperref{code:hessian_polyhedron}{Hessian polyhedron code}} & \eczhListValue{Spherical \((6,27,3/2)\) code whose codewords are the vertices of the Hessian complex polyhedron and the \(2_{21}\) polytope.
Two copies of the code yield the \((6,54,1)\) \textit{double Hessian polyhedron} (a.k.a. diplo-Schläfli) code.
The code can be obtained from the Schläfli graph \NoCaseChange{\protect\cite[{Ch. 9}]{cite115}}.
The (antipodal pairs of) points of the (double) Hessian polyhedron correspond to the 27 lines on a smooth cubic surface in \(\mathbb{C}P^3\) \NoCaseChange{\protect\cite{cite116,cite117,cite118,cite119,cite120}}.}\\ 
\addlinespace[\myxtraspc]
\eczhRefIndex{code:klemm}%
\eczhListValue{\flmRefsHyperref{code:klemm}{Klemm code}} & \eczhListValue{A member of a family of self-dual linear \((4m,4^1 2^{4m-2})_{\mathbb{Z}_4}\) codes.
Its generator matrix consists of a sum of the generator matrix of the repetition code and twice the generator matrix of the SPC code \NoCaseChange{\protect\cite{cite121}}.}\\ 
\addlinespace[\myxtraspc]
\eczhRefIndex{code:niemeier}%
\eczhListValue{\flmRefsHyperref{code:niemeier}{Niemeier lattice}} & \eczhListValue{One of the 24 positive-definite even unimodular lattices of rank 24.
The 24 lattices are \(D_{24}\), \(D_{16}E_8\), \(E_8^3\), \(A_{24}\), \(D_{12}^2\), \(A_{17}E_7\), \(D_{10}E_7^2\), \(A_{15}D_9\), \(D_8^3\), \(A_{12}^2\), \(A_{11}D_7E_6\), \(E_6^4\), \(A_9^2D_6\), \(D_6^4\), \(A_8^3\), \(A_7^2D_5^2\), \(A_6^4\), \(A_5^4D_4\), \(D_4^6\), \(A_4^6\), \(A_3^8\), \(A_2^{12}\), \(A_1^{24}\), and \(\Lambda_{24}\) (the Leech lattice) \NoCaseChange{\protect\cite[{Table 16.1}]{cite39}}.}\\ 
\addlinespace[\myxtraspc]
\eczhRefIndex{code:octacode}%
\eczhListValue{\flmRefsHyperref{code:octacode}{Octacode}} & \eczhListValue{The unique self-dual linear \((8,4^4,6)_{\mathbb{Z}_4}\) code of Euclidean distance 8.
Its shortened version is called the \((7,4^3,6)_{\mathbb{Z}_4}\) \textit{heptacode}.}\\ 
\addlinespace[\myxtraspc]
\eczhRefIndex{code:psk}%
\eczhListValue{\flmRefsHyperref{code:psk}{Phase-shift keying (PSK) modulation format}} & \eczhListValue{A \(q\)-ary phase-shift keying (\(q\)-PSK) encodes one \(q\)-ary digit of information into a constellation of \(q\) points distributed equidistantly on a circle in \(\mathbb{C}\) or, equivalently, \(\mathbb{R}^2\).}\\ 
\addlinespace[\myxtraspc]
\eczhRefIndex{code:polygon}%
\eczhListValue{\flmRefsHyperref{code:polygon}{Polygon code}} & \eczhListValue{Spherical \((1,q,4\sin^2 \frac{\pi}{q})\) code for any \(q\geq2\) whose codewords are the vertices of a \(q\)-gon. Special cases include the line segment (\(q=2\)), triangle (\(q=3\)), square (\(q=4\)), pentagon (\(q=5\)), and hexagon (\(q=6\)).}\\ 
\addlinespace[\myxtraspc]
\eczhRefIndex{code:pseudo_golay}%
\eczhListValue{\flmRefsHyperref{code:pseudo_golay}{Pseudo Golay code}} & \eczhListValue{Any one of 13 quaternary extremal Type II self-dual linear codes over \(\mathbb{Z}_4\) of length 24 whose mod-two reduction (mapping \(0,1,2,3\) to \(0,1,0,1\)) is the Golay code \NoCaseChange{\protect\cite[{Thm. 11}]{cite122}}.
Each code has Lee distance 12, Hamming distance 8, and Euclidean distance 16 \NoCaseChange{\protect\cite[{Thm. 9}]{cite122}}.}\\ 
\addlinespace[\myxtraspc]
\eczhRefIndex{code:qpsk}%
\eczhListValue{\flmRefsHyperref{code:qpsk}{Quadrature PSK (QPSK) modulation format}} & \eczhListValue{A quaternary encoding into a constellation of four points distributed equidistantly on a circle.
For the case of \(\pi/4\)-QPSK, the constellation is \(\{e^{\pm i\frac{\pi}{4}},e^{\pm i\frac{3\pi}{4}}\}\).}\\ 
\addlinespace[\myxtraspc]
\eczhRefIndex{code:self_dual_additive}%
\eczhListValue{\flmRefsHyperref{code:self_dual_additive}{Self-dual additive code}} & \eczhListValue{An additive \(q\)-ary code \(C \subseteq \mathbb{F}_q^n\) that is equal to its dual, \(C^\perp = C\), where the dual is defined with respect to some inner product, usually the trace-Hermitian inner product.}\\ 
\addlinespace[\myxtraspc]
\eczhRefIndex{code:self_dual_over_rings}%
\eczhListValue{\flmRefsHyperref{code:self_dual_over_rings}{Self-dual code over \(R\)}} & \eczhListValue{An additive linear code \(C\) over a ring \(R\) that is equal to its dual, \(C^\perp = C\), where the dual is defined with respect to some inner product.}\\ 
\addlinespace[\myxtraspc]
\eczhRefIndex{code:self_dual_over_z4}%
\eczhListValue{\flmRefsHyperref{code:self_dual_over_z4}{Self-dual code over \(\mathbb{Z}_4\)}} & \eczhListValue{A linear code \(C\) over \(\mathbb{Z}_4\) that is equal to its dual, \(C^\perp = C\), where the dual is defined with respect to the standard inner product.
The code contains \(2^n\) codewords \NoCaseChange{\protect\cite[{Corr. 1.3}]{cite123}}.}\\ 
\addlinespace[\myxtraspc]
\eczhRefIndex{code:self_dual_over_zq}%
\eczhListValue{\flmRefsHyperref{code:self_dual_over_zq}{Self-dual code over \(\mathbb{Z}_q\)}} & \eczhListValue{A linear code \(C\) over \(\mathbb{Z}_q\) that is equal to its dual, \(C^\perp = C\), where the dual is defined with respect to the standard inner product.}\\ 
\addlinespace[\myxtraspc]
\eczhRefIndex{code:self_dual}%
\eczhListValue{\flmRefsHyperref{code:self_dual}{Self-dual linear code}} & \eczhListValue{An \([n,n/2]_q\) code that is equal to its dual, \(C^\perp = C\), where the dual is defined with respect to an inner product, most commonly either Euclidean or Hermitian.
Self-dual codes exist only for even lengths and have dimension \(k=n/2\).
A code that is equivalent to its dual is called \textit{isodual}.
Any self-dual code is isodual, and hence formally self-dual \NoCaseChange{\protect\cite[{Rem. 4.2.2}]{cite40}}.}\\ 
\addlinespace[\myxtraspc]
\eczhRefIndex{code:self_dual_polytope}%
\eczhListValue{\flmRefsHyperref{code:self_dual_polytope}{Self-dual polytope code}} & \eczhListValue{A spherical code whose codewords are the vertices of a self-dual polytope.}\\ 
\addlinespace[\myxtraspc]
\eczhRefIndex{code:simplex_spherical}%
\eczhListValue{\flmRefsHyperref{code:simplex_spherical}{Simplex spherical code}} & \eczhListValue{Spherical \((n,n+1,2+2/n)\) code whose codewords are all permutations of the \(n+1\)-dimensional vector \((1,1,\cdots,1,-n)\), up to normalization, forming an \(n\)-simplex.
Codewords are all equidistant and their components add up to zero.
Simplex spherical codewords in 2 (3, 4) dimensions form the vertices of a triangle (tetrahedron, 5-cell).
In general, the code makes up the vertices of an \(n\)-simplex.
The union of a simplex and its antipodal simplex forms the vertices of a bi-simplex, which has \(2(n+1)\) vertices.}\\ 
\addlinespace[\myxtraspc]
\eczhRefIndex{code:self_dual_lattice}%
\eczhListValue{\flmRefsHyperref{code:self_dual_lattice}{Unimodular lattice}} & \eczhListValue{A lattice, scaled to be integral, that is equal to its dual, \(L^\perp = L\).
Unimodular lattices have \(\det L = \pm 1\).}\\ 
\addlinespace[\myxtraspc]
\eczhRefIndex{code:witting_polytope}%
\eczhListValue{\flmRefsHyperref{code:witting_polytope}{Witting polytope code}} & \eczhListValue{Spherical \((8,240,1)\) code whose codewords are the vertices of the Witting complex polytope, the \(4_{21}\) polytope, and the minimal lattice-shell code of the \(E_8\) lattice.
The code is optimal and unique up to equivalence \NoCaseChange{\protect\cite{cite124,cite39,cite125}}.
Antipodal pairs of points of the \(4_{21}\) polytope code correspond to the 120 tritangent planes of a canonical sextic curve in \(\mathbb{C}P^3\) \NoCaseChange{\protect\cite{cite117,cite118,cite119,cite120}}.}\\ 
\addlinespace[\myxtraspc]
\eczhRefIndex{code:dodecacode}%
\eczhListValue{\flmRefsHyperref{code:dodecacode}{\((12,4^6,6)_4\) Dodecacode}} & \eczhListValue{The unique trace-Hermitian self-dual additive \((12,4^6,6)_4\) code.
Its codewords are cyclic permutations of \((\omega 10100100101)\), where \(\mathbb{F}_4=\{0,1,\omega,\bar{\omega}\}\) is the \flmRefsHyperref{ref33}{quaternary Galois field} \NoCaseChange{\protect\cite[{Sec. 2.4.8}]{cite42}}.
Another generator matrix can be found in \NoCaseChange{\protect\cite[{Exam. 9.10.8}]{cite126}}.}\\ 
\addlinespace[\myxtraspc]
\eczhRefIndex{code:cmr}%
\eczhListValue{\flmRefsHyperref{code:cmr}{\(C_{m,r}\) code}} & \eczhListValue{A member of a family of Type IV self-dual quaternary linear codes over \(\mathbb{Z}_4\) generated by \(\textnormal{RM}(r,m) + 2\textnormal{RM}(m-r-1,m)\) for \(3r \leq m-1\) \NoCaseChange{\protect\cite{cite121}}.}\\ 
\addlinespace[\myxtraspc]
\eczhRefIndex{code:eeight}%
\eczhListValue{\flmRefsHyperref{code:eeight}{\(E_8\) Gosset lattice}} & \eczhListValue{Even unimodular BW lattice in dimension \(8\), consisting of the Cayley integral octonions rescaled by \(\sqrt{2}\).
The lattice corresponds to the \([8,4,4]\) Hamming code via \flmTerm{term}{ref127}{}{Construction A}.}\\ 
\addlinespace[\myxtraspc]
\eczhRefIndex{code:glynn}%
\eczhListValue{\flmRefsHyperref{code:glynn}{\([10,5,6]_9\) Glynn code}} & \eczhListValue{The unique trace-Hermitian self-dual \([10,5,6]_9\) code, constructed using a 10-arc in \(PG(4,9)\) that is not a rational curve.}\\ 
\addlinespace[\myxtraspc]
\eczhRefIndex{code:extended_golay}%
\eczhListValue{\flmRefsHyperref{code:extended_golay}{\([24, 12, 8]\) Extended Golay code}} & \eczhListValue{A self-dual \([24, 12, 8]\) code that is obtained from the Golay code by adding a parity check.
Equivalently, puncturing any coordinate yields the \([23,12,7]\) Golay code.
Up to equivalence, it is unique for its parameters \NoCaseChange{\protect\cite{cite102}}, and it is the unique \([24,12,8]\) extremal Type II code \NoCaseChange{\protect\cite[{Rems. 4.3.10 and 4.3.11}]{cite40}}.}\\ 
\addlinespace[\myxtraspc]
\eczhRefIndex{code:karlin}%
\eczhListValue{\flmRefsHyperref{code:karlin}{\([2m+2,m+1]\) Karlin code}} & \eczhListValue{Member of the family of \([2m+2,m+1]\) double circulant codes such that \(m\) is prime of the form \(8k+3\) for some \(k\), and \(2m+2\) is a multiple of eight.
See \NoCaseChange{\protect\cite[{Ch. 16}]{cite41}} for their generator matrix.
Karlin codes can be mapped to extended cyclic and extended quadratic-residue codes over \(\mathbb{F}_4\) \NoCaseChange{\protect\cite{cite109,cite110}\protect\cite[{Ch. 16}]{cite41}\protect\cite[{Sec. 2.4.2}]{cite42}} by identifying \((0,\omega,\bar{\omega},1)\) with \((00),(10),(01),(11)\) \NoCaseChange{\protect\cite{cite109}}.}\\ 
\addlinespace[\myxtraspc]
\eczhRefIndex{code:pless_symmetry}%
\eczhListValue{\flmRefsHyperref{code:pless_symmetry}{\([2q+2,q+1]_3\) Pless symmetry code}} & \eczhListValue{A member of a family of self-dual ternary \([2q+2,q+1]_3\) codes for any power of an odd prime satisfying \(q \equiv 2\) modulo 3.}\\ 
\addlinespace[\myxtraspc]
\eczhRefIndex{code:self_dual_z6}%
\eczhListValue{\flmRefsHyperref{code:self_dual_z6}{\([4,2,2]_{\mathbb{Z}_6}\) senary code}} & \eczhListValue{A self-dual code over \(\mathbb{Z}_6\) that is one of two such codes, up to permutations \NoCaseChange{\protect\cite{cite128}}.}\\ 
\addlinespace[\myxtraspc]
\eczhRefIndex{code:tetracode}%
\eczhListValue{\flmRefsHyperref{code:tetracode}{\([4,2,3]_3\) Tetracode}} & \eczhListValue{The \([4,2,3]_3\) ternary self-dual MDS code that has connections to lattices \NoCaseChange{\protect\cite{cite39}}. Its weight enumerator is the Gleason polynomial \(g_4\) \NoCaseChange{\protect\cite[{Rem. 4.2.6}]{cite40}}.}\\ 
\addlinespace[\myxtraspc]
\eczhRefIndex{code:reed_solomon_4}%
\eczhListValue{\flmRefsHyperref{code:reed_solomon_4}{\([4,2,3]_4\) RS\(_4\) code}} & \eczhListValue{A Type II Euclidean self-dual extended RS code that is the smallest quaternary extended QR code \NoCaseChange{\protect\cite[{pg. 296}]{cite41}\protect\cite[{Sec. 2.4.2}]{cite42}}.
Puncturing the \([4,2,3]_4\) RS\(_4\) code yields the \([3,2,2]_4\) shortened RS\(_4\) code, which is an RS code \NoCaseChange{\protect\cite[{pg. 295}]{cite41}}.}\\ 
\addlinespace[\myxtraspc]
\eczhRefIndex{code:self_dual_48_24_12}%
\eczhListValue{\flmRefsHyperref{code:self_dual_48_24_12}{\([48,24,12]\) self-dual code}} & \eczhListValue{An extended quadratic-residue code that is the unique self-dual doubly even \([48,24,12]\) code. It is extremal Type II, and its automorphism group is \(PSL(2,47)\) \NoCaseChange{\protect\cite{cite111}\protect\cite[{Rem. 4.3.11}]{cite40}}.}\\ 
\addlinespace[\myxtraspc]
\eczhRefIndex{code:hexacode}%
\eczhListValue{\flmRefsHyperref{code:hexacode}{\([6,3,4]_4\) Hexacode}} & \eczhListValue{The \([6,3,4]_4\) Hermitian self-dual MDS code that has connections to projective geometry, lattices \NoCaseChange{\protect\cite{cite39}}, and conformal field theory \NoCaseChange{\protect\cite{cite44}}. Its weight enumerator is the Gleason polynomial \(g_7\) \NoCaseChange{\protect\cite[{Rem. 4.2.6}]{cite40}}.}\\ 
\addlinespace[\myxtraspc]
\eczhRefIndex{code:hamming844}%
\eczhListValue{\flmRefsHyperref{code:hamming844}{\([8,4,4]\) extended Hamming code}} & \eczhListValue{Extension of the \([7,4,3]\) Hamming code by a parity-check bit.
The smallest doubly even self-dual code, and the unique Type II code of length \(8\) \NoCaseChange{\protect\cite[{Rem. 4.3.10}]{cite40}}.}\\ 
\addlinespace[\myxtraspc]
\eczhRefIndex{code:leech}%
\eczhListValue{\flmRefsHyperref{code:leech}{\(\Lambda_{24}\) Leech lattice}} & \eczhListValue{Even unimodular lattice in 24 dimensions that exhibits optimal packing.
Its automorphism group is the Conway group Co\(_0\).}\\ 
\addlinespace[\myxtraspc]
\eczhRefIndex{code:hypercubic}%
\eczhListValue{\flmRefsHyperref{code:hypercubic}{\(\mathbb{Z}^n\) hypercubic lattice}} & \eczhListValue{Lattice-based code consisting of all integer vectors in \(n\) dimensions.
Its generator matrix is the \(n\)-dimensional identity matrix.
Its automorphism group consists of all coordinate permutations and sign changes.}\\ 
\end{tabularx}\endgroup
\eczcodelist{combinatorial_design}{Combinatorial designs and friends}%

\eczhCodeListAutoDescription{All descendants and cousins of \flmRefsCref{code:combinatorial_design}.}%

\eczhIncludeCodeGraph{Bare}{scale=0.5}{\columnwidth}{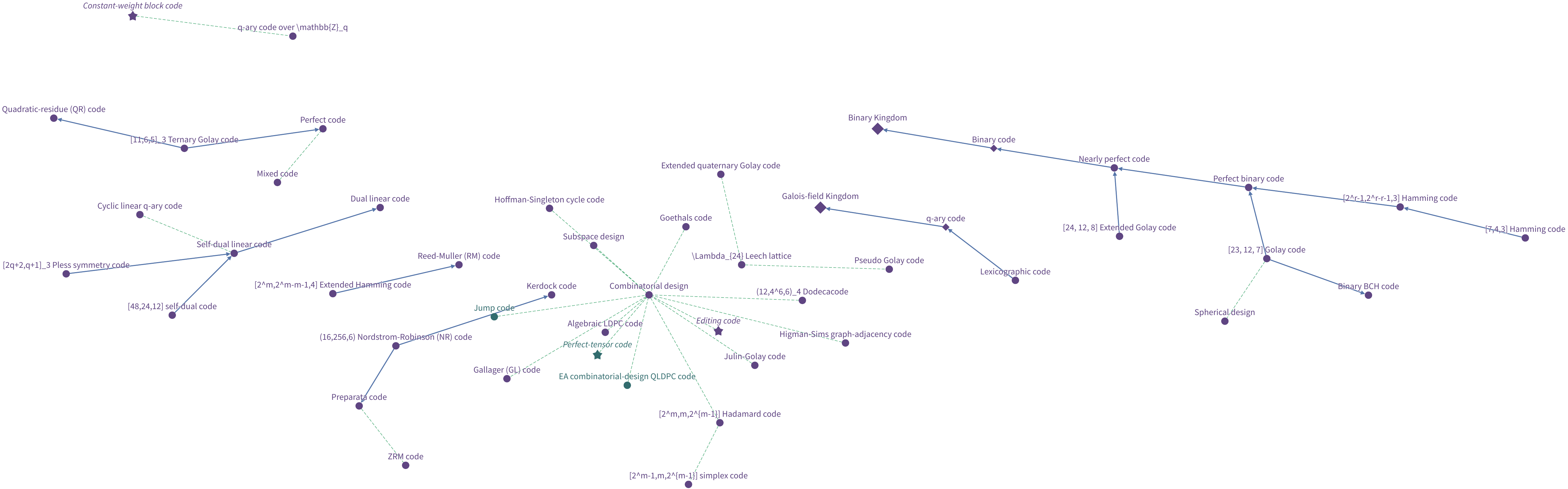}{Combinatorial designs and friends}{https://errorcorrectionzoo.org/code_graph#J\%7B\%22displayMode\%22\%3A\%22subset\%22\%2C\%22modeSubsetOptions\%22\%3A\%7B\%22codeIds\%22\%3A\%5B\%22algebraic_ldpc\%22\%2C\%22bch\%22\%2C\%22bits_into_bits\%22\%2C\%22combinatorial_design\%22\%2C\%22q-ary_constant_weight\%22\%2C\%22q-ary_cyclic\%22\%2C\%22dual\%22\%2C\%22ea_design_qldpc\%22\%2C\%22insertion_deletion\%22\%2C\%22quaternary_golay\%22\%2C\%22gallager\%22\%2C\%22goethals\%22\%2C\%22higman-sims_graph\%22\%2C\%22hoffman-singleton\%22\%2C\%22julin12\%22\%2C\%22jump\%22\%2C\%22kerdock\%22\%2C\%22lexicographic\%22\%2C\%22mixed\%22\%2C\%22nearly_perfect\%22\%2C\%22perfect_binary\%22\%2C\%22perfect\%22\%2C\%22ame\%22\%2C\%22preparata\%22\%2C\%22pseudo_golay\%22\%2C\%22q-ary_quad_residue\%22\%2C\%22reed_muller\%22\%2C\%22self_dual\%22\%2C\%22spherical_design\%22\%2C\%22subspace_design\%22\%2C\%22zrm\%22\%2C\%22dodecacode\%22\%2C\%22nordstrom_robinson\%22\%2C\%22ternary_golay\%22\%2C\%22golay\%22\%2C\%22extended_golay\%22\%2C\%22extended_hamming\%22\%2C\%22hadamard\%22\%2C\%22simplex\%22\%2C\%22hamming\%22\%2C\%22pless_symmetry\%22\%2C\%22self_dual_48_24_12\%22\%2C\%22hamming743\%22\%2C\%22leech\%22\%2C\%22q-ary_digits_into_q-ary_digits\%22\%2C\%22q-ary_over_zq\%22\%5D\%2C\%22reusePreviousLayoutPositions\%22\%3Afalse\%2C\%22showIntermediateConnectingNodes\%22\%3Atrue\%2C\%22connectingNodesMaxDepth\%22\%3A15\%2C\%22connectingNodesPathMaxLength\%22\%3A20\%2C\%22connectingNodesMaxExtraDepth\%22\%3A3\%2C\%22connectingNodesOnlyKeepPathsWithAdditionalLength\%22\%3A1\%2C\%22connectingNodesToDomainsAndKingdoms\%22\%3Afalse\%2C\%22connectingNodesEdgeLengthsByType\%22\%3A\%7B\%22primaryParent\%22\%3A1\%2C\%22secondaryParent\%22\%3A4\%2C\%22cousin\%22\%3A6\%7D\%2C\%22nodeIds\%22\%3A\%5B\%22k_bits_into_bits\%22\%2C\%22k_q-ary_digits_into_q-ary_digits\%22\%5D\%7D\%2C\%22highlightImportantNodes\%22\%3A\%7B\%22highlightImportantNodes\%22\%3Afalse\%2C\%22highlightPrimaryParents\%22\%3Afalse\%2C\%22highlightRootConnectingEdges\%22\%3Afalse\%7D\%7D}

\begingroup
\small
\eczhBreakableDashes
\renewcommand\arraystretch{1.05}
\edef\myxtraspc{\eczListAddVSpaceXtraXtraStretch}
\begin{tabularx}{\linewidth}{>{\raggedright\arraybackslash}p{\eczListColWidth{name}} >{\hsize=1.0000\hsize }X}
\toprule
\eczListColTitle{Code} & \eczListColTitle{Relation} \\
\midrule
\endfirsthead
\toprule
\eczListColTitleContinued{Code} & \eczListColTitleContinued{Relation} \\
\midrule
\endhead
\bottomrule
\endfoot
\eczhRefIndex{code:algebraic_ldpc}%
\eczhListValue{\flmRefsHyperref{code:algebraic_ldpc}{Algebraic LDPC code}} & \eczhListValue{Combinatorial designs can be used to construct explicit LDPC codes \NoCaseChange{\protect\cite{cite51,cite52,cite53}}.}\\ 
\addlinespace[\myxtraspc]
\eczhRefIndex{code:bch}%
\eczhListValue{\flmRefsHyperref{code:bch}{Binary BCH code}} & \eczhListValue{A family of BCH codes supports an infinite family of combinatorial 4-designs \NoCaseChange{\protect\cite{cite129,cite130}}.}\\ 
\addlinespace[\myxtraspc]
\eczhRefIndex{code:bits_into_bits}%
\eczhListValue{\flmRefsHyperref{code:bits_into_bits}{Binary code}} & \eczhListValue{If the \flmRefsHyperref{ref113}{number} of a code is less than or equal to its \flmRefsHyperref{ref113}{dual distance}, then some sets of fixed-weight codewords form a combinatorial design \NoCaseChange{\protect\cite[{Thm. 6.7}]{cite41}}.}\\ 
\addlinespace[\myxtraspc]
\eczhRefIndex{code:combinatorial_design}%
\eczhListValue{\flmRefsHyperref{code:combinatorial_design}{Combinatorial design}} & \eczhListValue{\eczListValueNA }\\ 
\addlinespace[\myxtraspc]
\eczhRefIndex{code:q-ary_constant_weight}%
\eczhListValue{\flmRefsHyperref{code:q-ary_constant_weight}{Constant-weight block code}} & \eczhListValue{Optimal constant-weight codes over \(\mathbb{Z}_q\) can be constructed \NoCaseChange{\protect\cite{cite131}} from a generalization of combinatorial designs to \(q\)-ary alphabets \NoCaseChange{\protect\cite{cite132,cite133}}.}\\ 
\addlinespace[\myxtraspc]
\eczhRefIndex{code:q-ary_cyclic}%
\eczhListValue{\flmRefsHyperref{code:q-ary_cyclic}{Cyclic linear \(q\)-ary code}} & \eczhListValue{Two families of cyclic \(q\)-ary codes support an infinite family of combinatorial 3-designs \NoCaseChange{\protect\cite{cite134}}.
The supports of all fixed-weight codewords of a \(q\)-ary cyclic code support a combinatorial \(1\)-design \NoCaseChange{\protect\cite[{Corr. 5.2.4}]{cite135}}.}\\ 
\addlinespace[\myxtraspc]
\eczhRefIndex{code:dual}%
\eczhListValue{\flmRefsHyperref{code:dual}{Dual linear code}} & \eczhListValue{Linear codes and their duals are related to combinatorial designs via the Assmus-Mattson theorem \NoCaseChange{\protect\cite{cite136,cite137}} (see \NoCaseChange{\protect\cite[{Sec. 5.4}]{cite135}}).}\\ 
\addlinespace[\myxtraspc]
\eczhRefIndex{code:ea_design_qldpc}%
\eczhListValue{\flmRefsHyperref{code:ea_design_qldpc}{EA combinatorial-design QLDPC code}} & \eczhListValue{Combinatorial designs can be used to construct EA QLDPC codes \NoCaseChange{\protect\cite{cite138}}.}\\ 
\addlinespace[\myxtraspc]
\eczhRefIndex{code:insertion_deletion}%
\eczhListValue{\flmRefsHyperref{code:insertion_deletion}{Editing code}} & \eczhListValue{Perfect deletion correcting codes can be constructed using combinatorial design theory \NoCaseChange{\protect\cite{cite139,cite140}}.}\\ 
\addlinespace[\myxtraspc]
\eczhRefIndex{code:quaternary_golay}%
\eczhListValue{\flmRefsHyperref{code:quaternary_golay}{Extended quaternary Golay code}} & \eczhListValue{Supports of codewords of any fixed symmetrized type of the extended quaternary Golay code form a 5-design \NoCaseChange{\protect\cite{cite141,cite142,cite143}}.}\\ 
\addlinespace[\myxtraspc]
\eczhRefIndex{code:gallager}%
\eczhListValue{\flmRefsHyperref{code:gallager}{Gallager (GL) code}} & \eczhListValue{Some Steiner systems can be used to construct Gallager codes \NoCaseChange{\protect\cite{cite72}}.}\\ 
\addlinespace[\myxtraspc]
\eczhRefIndex{code:goethals}%
\eczhListValue{\flmRefsHyperref{code:goethals}{Goethals code}} & \eczhListValue{Goethals codes form an infinite family of nonlinear binary codes supporting 3-designs \NoCaseChange{\protect\cite[{Table 5.1}]{cite135}}.}\\ 
\addlinespace[\myxtraspc]
\eczhRefIndex{code:higman-sims_graph}%
\eczhListValue{\flmRefsHyperref{code:higman-sims_graph}{Higman-Sims graph-adjacency code}} & \eczhListValue{The 4125 codewords of weight 36 of the \([100,22,32]\) code \(C_{100}\) form a \(2\)-\((100,36,525)\) design, which can be used for majority decoding of single errors in \(C_{100}^\perp\) \NoCaseChange{\protect\cite[{Rem. 1.7}]{cite82}}.}\\ 
\addlinespace[\myxtraspc]
\eczhRefIndex{code:hoffman-singleton}%
\eczhListValue{\flmRefsHyperref{code:hoffman-singleton}{Hoffman-Singleton cycle code}} & \eczhListValue{The incidence matrix of the Hoffman-Singleton graph can be converted into a \(2\)-\((50,14,13)\) design \NoCaseChange{\protect\cite[{Prop. 1.1}]{cite82}}.}\\ 
\addlinespace[\myxtraspc]
\eczhRefIndex{code:julin12}%
\eczhListValue{\flmRefsHyperref{code:julin12}{Julin-Golay code}} & \eczhListValue{Julin-Golay codes are constructed from the Steiner system \(S(5,6,12)\) arising from the extended \((12,132,4)\) code \NoCaseChange{\protect\cite[{pgs. 70-72}]{cite41}}.}\\ 
\addlinespace[\myxtraspc]
\eczhRefIndex{code:jump}%
\eczhListValue{\flmRefsHyperref{code:jump}{Jump code}} & \eczhListValue{Certain types of combinatorial designs can be used to obtain jump codes \NoCaseChange{\protect\cite{cite144,cite145,cite146}}.}\\ 
\addlinespace[\myxtraspc]
\eczhRefIndex{code:kerdock}%
\eczhListValue{\flmRefsHyperref{code:kerdock}{Kerdock code}} & \eczhListValue{Kerdock codes form an infinite family of nonlinear binary codes supporting 3-designs \NoCaseChange{\protect\cite[{Rem. 5.5.6}]{cite135}}.}\\ 
\addlinespace[\myxtraspc]
\eczhRefIndex{code:lexicographic}%
\eczhListValue{\flmRefsHyperref{code:lexicographic}{Lexicographic code}} & \eczhListValue{Some lexicodes yield Steiner systems \NoCaseChange{\protect\cite{cite147}}.}\\ 
\addlinespace[\myxtraspc]
\eczhRefIndex{code:mixed}%
\eczhListValue{\flmRefsHyperref{code:mixed}{Mixed code}} & \eczhListValue{Combinatorial designs have been generalized to mixed alphabets \NoCaseChange{\protect\cite{cite148}}.}\\ 
\addlinespace[\myxtraspc]
\eczhRefIndex{code:nearly_perfect}%
\eczhListValue{\flmRefsHyperref{code:nearly_perfect}{Nearly perfect code}} & \eczhListValue{The minimum-weight codewords in a nearly perfect code containing the zero vector support a \(t\)-\((n,2t+1,\lfloor (n-t)/(t+1) \rfloor)\) design, while the minimum-weight codewords in the extended code support a \((t+1)\)-\((n+1,2t+2,\lfloor (n-t)/(t+1) \rfloor)\) design \NoCaseChange{\protect\cite[{Thm. 5.5.4}]{cite135}}.}\\ 
\addlinespace[\myxtraspc]
\eczhRefIndex{code:perfect_binary}%
\eczhListValue{\flmRefsHyperref{code:perfect_binary}{Perfect binary code}} & \eczhListValue{If a perfect binary code contains the zero vector, then its minimum-weight codewords support a Steiner \((t+1)\)-\((n,2t+1,1)\) design, while the minimum-weight codewords in the extended code support a Steiner \((t+2)\)-\((n+1,2t+2,1)\) design \NoCaseChange{\protect\cite[{Thm. 5.3.1(b)}]{cite135}}.}\\ 
\addlinespace[\myxtraspc]
\eczhRefIndex{code:perfect}%
\eczhListValue{\flmRefsHyperref{code:perfect}{Perfect code}} & \eczhListValue{Perfect codes and combinatorial designs are related \NoCaseChange{\protect\cite{cite149,cite150}}.}\\ 
\addlinespace[\myxtraspc]
\eczhRefIndex{code:ame}%
\eczhListValue{\flmRefsHyperref{code:ame}{Perfect-tensor code}} & \eczhListValue{Combinatorial designs and \(d\)-uniform quantum states are related \NoCaseChange{\protect\cite{cite151,cite152,cite153}}.}\\ 
\addlinespace[\myxtraspc]
\eczhRefIndex{code:preparata}%
\eczhListValue{\flmRefsHyperref{code:preparata}{Preparata code}} & \eczhListValue{Preparata codewords of each weight form 3-designs, and the minimum-weight codewords yield infinite families of 4-designs, including Steiner 4-designs with block sizes 5 and 6 \NoCaseChange{\protect\cite[{Rem. 5.5.6 and Thms. 5.5.7, 5.5.11}]{cite135}\protect\cite[{pg. 471}]{cite41}}.}\\ 
\addlinespace[\myxtraspc]
\eczhRefIndex{code:pseudo_golay}%
\eczhListValue{\flmRefsHyperref{code:pseudo_golay}{Pseudo Golay code}} & \eczhListValue{Supports of codewords of any fixed symmetrized type of pseudo Golay codes form a 5-design \NoCaseChange{\protect\cite{cite141,cite142,cite143}}.}\\ 
\addlinespace[\myxtraspc]
\eczhRefIndex{code:q-ary_quad_residue}%
\eczhListValue{\flmRefsHyperref{code:q-ary_quad_residue}{Quadratic-residue (QR) code}} & \eczhListValue{The supports of fixed-weight codewords of certain \(q\)-ary QR codes support combinatorial designs \NoCaseChange{\protect\cite{cite149,cite136,cite154}}, including \(3\)-designs \NoCaseChange{\protect\cite{cite155}}.}\\ 
\addlinespace[\myxtraspc]
\eczhRefIndex{code:reed_muller}%
\eczhListValue{\flmRefsHyperref{code:reed_muller}{Reed-Muller (RM) code}} & \eczhListValue{Fixed-weight RM codewords of weight less than \(2^m\) support combinatorial 3-designs \NoCaseChange{\protect\cite[{Exam. 5.2.7}]{cite135}}.}\\ 
\addlinespace[\myxtraspc]
\eczhRefIndex{code:self_dual}%
\eczhListValue{\flmRefsHyperref{code:self_dual}{Self-dual linear code}} & \eczhListValue{Self-dual extremal codes yield combinatorial \(\leq 5\)-designs using the Assmus-Mattson theorem \NoCaseChange{\protect\cite{cite136}} (see \NoCaseChange{\protect\cite[{Sec. 5.4}]{cite135}}).
See \NoCaseChange{\protect\cite[{Table 1.61, pg. 683}]{cite156}} for a table of combinatorial designs obtained from self-dual codes.}\\ 
\addlinespace[\myxtraspc]
\eczhRefIndex{code:spherical_design}%
\eczhListValue{\flmRefsHyperref{code:spherical_design}{Spherical design}} & \eczhListValue{Spherical designs can be thought of as Euclidean analogues of combinatorial designs \NoCaseChange{\protect\cite{cite157}}.}\\ 
\addlinespace[\myxtraspc]
\eczhRefIndex{code:subspace_design}%
\eczhListValue{\flmRefsHyperref{code:subspace_design}{Subspace design}} & \eczhListValue{Combinatorial designs are designs in Johnson space, the space of all size-\(w\) subsets of a set with \(n\) elements. The \(q\)-Johnson spaces generalize this notion to subspaces and reduce to Johnson spaces at \(q=1\). In other words, combinatorial designs are designs over spaces of subsets, while subspace designs are designs over spaces of subspaces.}\\ 
\addlinespace[\myxtraspc]
\eczhRefIndex{code:zrm}%
\eczhListValue{\flmRefsHyperref{code:zrm}{ZRM code}} & \eczhListValue{The weight-four codewords of the binary image of the dual of ZRM\((1,m)\) form a Steiner system that is identical to that formed by the weight-four codewords of an extended Hamming code \NoCaseChange{\protect\cite{cite158}}.}\\ 
\addlinespace[\myxtraspc]
\eczhRefIndex{code:dodecacode}%
\eczhListValue{\flmRefsHyperref{code:dodecacode}{\((12,4^6,6)_4\) Dodecacode}} & \eczhListValue{There exists a \(5\)-\((12, 6, 3)\) design in the dodecacode, and a \(3\)-\((11, 5, 4)\) design in the shortened dodecacode \NoCaseChange{\protect\cite{cite159}}.}\\ 
\addlinespace[\myxtraspc]
\eczhRefIndex{code:nordstrom_robinson}%
\eczhListValue{\flmRefsHyperref{code:nordstrom_robinson}{\((16,256,6)\) Nordstrom-Robinson (NR) code}} & \eczhListValue{NR codewords give \(3\)-\((16, 6, 4)\), \(3\)-\((16, 8, 3)\), and \(3\)-\((16, 10, 24)\) designs, while the punctured code of length \(15\) and minimum distance \(5\) meets the Johnson bound and supports \(2\)-designs \NoCaseChange{\protect\cite[{Exam. 5.5.5}]{cite135}\protect\cite[{pg. 164}]{cite41}}.}\\ 
\addlinespace[\myxtraspc]
\eczhRefIndex{code:ternary_golay}%
\eczhListValue{\flmRefsHyperref{code:ternary_golay}{\([11,6,5]_3\) Ternary Golay code}} & \eczhListValue{The supports of the weight-five codewords of the ternary Golay code and the weight-six codewords of the extended ternary Golay code support the Steiner systems \(S(4,5,11)\) and \(S(5,6,12)\), respectively \NoCaseChange{\protect\cite{cite160,cite154}\protect\cite[{pg. 89}]{cite39}}. The latter blocks are called hexads.}\\ 
\addlinespace[\myxtraspc]
\eczhRefIndex{code:golay}%
\eczhListValue{\flmRefsHyperref{code:golay}{\([23, 12, 7]\) Golay code}} & \eczhListValue{The supports of the weight-seven codewords of the Golay code support the Steiner system \(S(4,7,23)\) \NoCaseChange{\protect\cite{cite160,cite154}\protect\cite[{pg. 89}]{cite39}}.}\\ 
\addlinespace[\myxtraspc]
\eczhRefIndex{code:extended_golay}%
\eczhListValue{\flmRefsHyperref{code:extended_golay}{\([24, 12, 8]\) Extended Golay code}} & \eczhListValue{The supports of the weight-eight codewords of the extended Golay code support the Steiner system \(S(5,8,24)\) \NoCaseChange{\protect\cite{cite160,cite154}\protect\cite[{pg. 89}]{cite39}\protect\cite[{Ch. 10, pg. 276}]{cite39}}. Its blocks are called octads.}\\ 
\addlinespace[\myxtraspc]
\eczhRefIndex{code:extended_hamming}%
\eczhListValue{\flmRefsHyperref{code:extended_hamming}{\([2^m,2^m-m-1,4]\) Extended Hamming code}} & \eczhListValue{Weight-four codewords of the \([2^r,2^r-r-1, 4]\) extended Hamming code support the Steiner system \(S(3,4,2^r)\) \NoCaseChange{\protect\cite[{pg. 89}]{cite39}}.}\\ 
\addlinespace[\myxtraspc]
\eczhRefIndex{code:hadamard}%
\eczhListValue{\flmRefsHyperref{code:hadamard}{\([2^m,m,2^{m-1}]\) Hadamard code}} & \eczhListValue{\textit{Hadamard designs} are combinatorial designs constructed from Hadamard matrices \NoCaseChange{\protect\cite{cite161,cite162}}; see Ref. \NoCaseChange{\protect\cite{cite163}}.}\\ 
\addlinespace[\myxtraspc]
\eczhRefIndex{code:simplex}%
\eczhListValue{\flmRefsHyperref{code:simplex}{\([2^m-1,m,2^{m-1}]\) simplex code}} & \eczhListValue{Simplex codewords form a 2-design \NoCaseChange{\protect\cite[{pg. 166}]{cite41}}.}\\ 
\addlinespace[\myxtraspc]
\eczhRefIndex{code:hamming}%
\eczhListValue{\flmRefsHyperref{code:hamming}{\([2^r-1,2^r-r-1,3]\) Hamming code}} & \eczhListValue{Weight-three codewords of the \([2^r-1,2^r-r-1, 3]\) Hamming code support the Steiner system \(S(2,3,2^r-1)\) \NoCaseChange{\protect\cite[{pg. 89}]{cite39}}.}\\ 
\addlinespace[\myxtraspc]
\eczhRefIndex{code:pless_symmetry}%
\eczhListValue{\flmRefsHyperref{code:pless_symmetry}{\([2q+2,q+1]_3\) Pless symmetry code}} & \eczhListValue{The supports of fixed-weight codewords of certain Pless symmetry codes support combinatorial designs \NoCaseChange{\protect\cite{cite164,cite165,cite154}}.}\\ 
\addlinespace[\myxtraspc]
\eczhRefIndex{code:self_dual_48_24_12}%
\eczhListValue{\flmRefsHyperref{code:self_dual_48_24_12}{\([48,24,12]\) self-dual code}} & \eczhListValue{Fixed-weight codewords of extremal Type II codes of length divisible by \(24\) form combinatorial 5-designs \NoCaseChange{\protect\cite[{Thm. 4.3.16(a)}]{cite40}}. There are several designs associated with this code \NoCaseChange{\protect\cite{cite166}}.}\\ 
\addlinespace[\myxtraspc]
\eczhRefIndex{code:hamming743}%
\eczhListValue{\flmRefsHyperref{code:hamming743}{\([7,4,3]\) Hamming code}} & \eczhListValue{Weight-three and weight-four codewords of the \([7,4,3]\) Hamming code support combinatorial \(2\)-\((7,3,1)\) and \(2\)-\((7,4,2)\) designs, respectively \NoCaseChange{\protect\cite[{Exam. 5.2.5}]{cite135}}.}\\ 
\addlinespace[\myxtraspc]
\eczhRefIndex{code:leech}%
\eczhListValue{\flmRefsHyperref{code:leech}{\(\Lambda_{24}\) Leech lattice}} & \eczhListValue{The Leech lattice is completely determined by the Steiner system \(S(5,8,24)\) formed by the octads of the extended Golay code \NoCaseChange{\protect\cite[{Ch. 12, pg. 335, Thm. 6}]{cite39}}.}\\ 
\addlinespace[\myxtraspc]
\eczhRefIndex{code:q-ary_digits_into_q-ary_digits}%
\eczhListValue{\flmRefsHyperref{code:q-ary_digits_into_q-ary_digits}{\(q\)-ary code}} & \eczhListValue{Designs can be constructed from \(q\)-ary codes by taking the supports of a subset of codewords of constant weight.}\\ 
\addlinespace[\myxtraspc]
\eczhRefIndex{code:q-ary_over_zq}%
\eczhListValue{\flmRefsHyperref{code:q-ary_over_zq}{\(q\)-ary code over \(\mathbb{Z}_q\)}} & \eczhListValue{Optimal constant-weight codes over \(\mathbb{Z}_q\) can be constructed \NoCaseChange{\protect\cite{cite131}} from a generalization of combinatorial designs to \(q\)-ary alphabets \NoCaseChange{\protect\cite{cite132,cite133}}.}\\ 
\end{tabularx}\endgroup
\eczcodelist{constant_weight}{Constant-weight codes and friends
}%

\eczhCodeListAutoDescription{Union of:
\begin{itemize}\item codes that are descendants of \flmRefsCref{code:q-ary_constant_weight}
\item codes that are cousins of \flmRefsCref{code:q-ary_constant_weight}
\item codes that are cousins of \flmRefsCref{code:constant_weight}
\end{itemize}}%

\eczhIncludeCodeGraph{Bare}{scale=0.5}{\columnwidth}{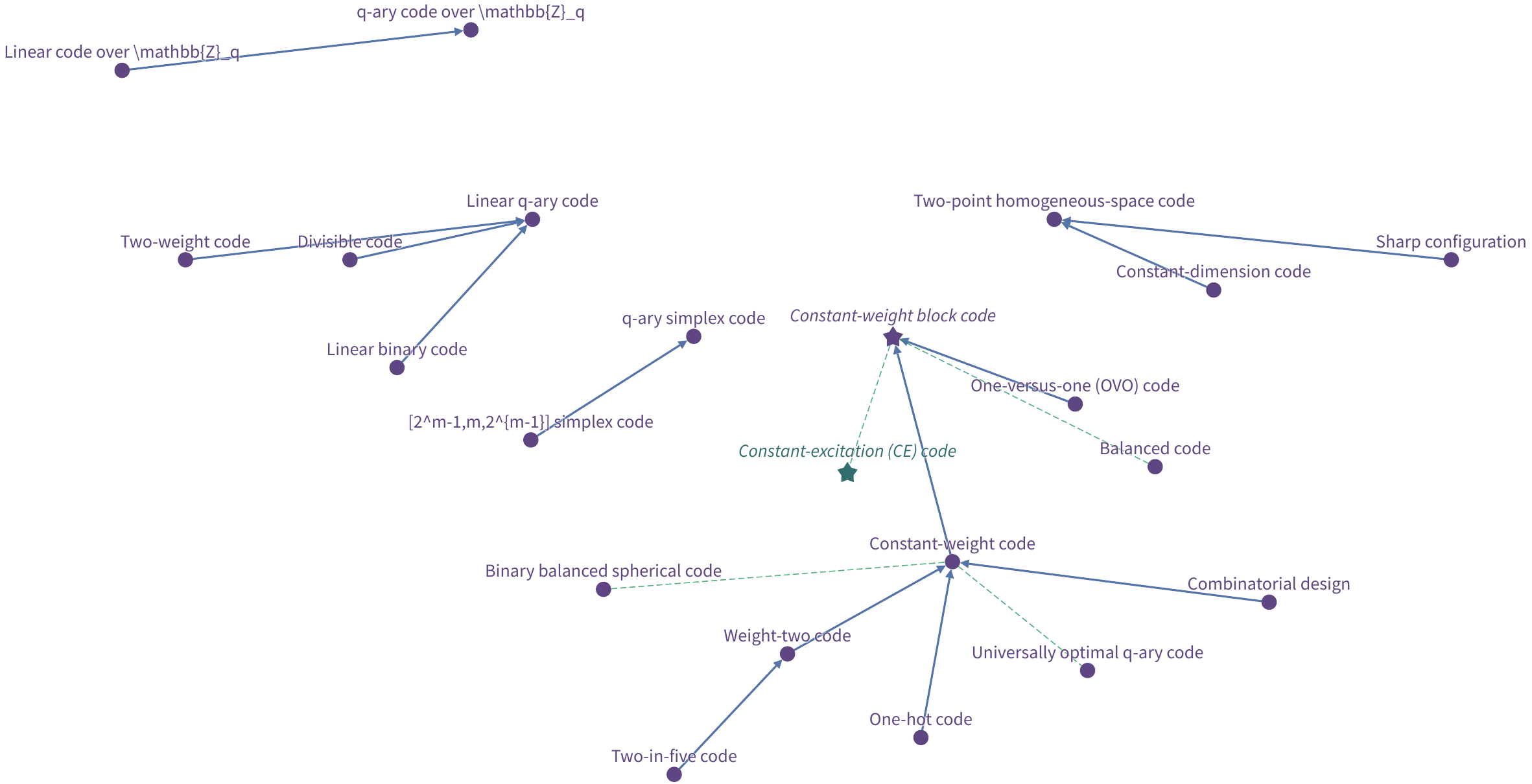}{Constant-weight codes and friends}{https://errorcorrectionzoo.org/code_graph#J\%7B\%22displayMode\%22\%3A\%22subset\%22\%2C\%22modeSubsetOptions\%22\%3A\%7B\%22codeIds\%22\%3A\%5B\%22balanced\%22\%2C\%22binary_balanced\%22\%2C\%22combinatorial_design\%22\%2C\%22finite_grassmann\%22\%2C\%22constant_excitation\%22\%2C\%22q-ary_constant_weight\%22\%2C\%22constant_weight\%22\%2C\%22divisible\%22\%2C\%22q-ary_linear\%22\%2C\%22binary_linear\%22\%2C\%22q-ary_linear_over_zq\%22\%2C\%22one_hot\%22\%2C\%22one_vs_one\%22\%2C\%22delsarte_optimal\%22\%2C\%22two_in_five\%22\%2C\%222pt_homogeneous\%22\%2C\%22two_weight\%22\%2C\%22univ_opt_q-ary\%22\%2C\%22weight_two\%22\%2C\%22simplex\%22\%2C\%22q-ary_over_zq\%22\%2C\%22q-ary_simplex\%22\%5D\%2C\%22reusePreviousLayoutPositions\%22\%3Afalse\%2C\%22showIntermediateConnectingNodes\%22\%3Atrue\%2C\%22connectingNodesMaxDepth\%22\%3A15\%2C\%22connectingNodesPathMaxLength\%22\%3A20\%2C\%22connectingNodesMaxExtraDepth\%22\%3A3\%2C\%22connectingNodesOnlyKeepPathsWithAdditionalLength\%22\%3A1\%2C\%22connectingNodesToDomainsAndKingdoms\%22\%3Afalse\%2C\%22connectingNodesEdgeLengthsByType\%22\%3A\%7B\%22primaryParent\%22\%3A1\%2C\%22secondaryParent\%22\%3A4\%2C\%22cousin\%22\%3A6\%7D\%2C\%22nodeIds\%22\%3A\%5B\%5D\%7D\%2C\%22highlightImportantNodes\%22\%3A\%7B\%22highlightImportantNodes\%22\%3Afalse\%2C\%22highlightPrimaryParents\%22\%3Afalse\%2C\%22highlightRootConnectingEdges\%22\%3Afalse\%7D\%7D}

\begingroup
\small
\eczhBreakableDashes
\renewcommand\arraystretch{1.05}
\edef\myxtraspc{\eczListAddVSpaceXtraXtraStretch}
\begin{tabularx}{\linewidth}{>{\raggedright\arraybackslash}p{\eczListColWidth{name}} >{\hsize=1.0000\hsize }X}
\toprule
\eczListColTitle{Code} & \eczListColTitle{Description} \\
\midrule
\endfirsthead
\toprule
\eczListColTitleContinued{Code} & \eczListColTitleContinued{Description} \\
\midrule
\endhead
\bottomrule
\endfoot
\eczhRefIndex{code:balanced}%
\eczhListValue{\flmRefsHyperref{code:balanced}{Balanced code}} & \eczhListValue{An even-length-\(n\) \(q\)-ary code whose nonzero codewords all have a Hamming weight of \(n/2\).
A code is \(\epsilon\)\textit{-balanced} if the relative weight (i.e., weight divided by \(n\)) of all nonzero codewords lies in the interval \([\frac{1-\epsilon}{2},\frac{1+\epsilon}{2}]\).
A code is \(\gamma\)\textit{-unbiased} if the relative weight lies in the interval \((\frac{1}{2}-\frac{1}{n^{\gamma}},\frac{1}{2}+\frac{1}{n^{\gamma}})\).}\\ 
\addlinespace[\myxtraspc]
\eczhRefIndex{code:binary_balanced}%
\eczhListValue{\flmRefsHyperref{code:binary_balanced}{Binary balanced spherical code}} & \eczhListValue{An \((n-1,K,\frac{nd}{nw-w^2})\) spherical code obtained from a constant-weight-\(w\) binary \((n,K,d)\) code via the component-wise binary balanced mapping.}\\ 
\addlinespace[\myxtraspc]
\eczhRefIndex{code:combinatorial_design}%
\eczhListValue{\flmRefsHyperref{code:combinatorial_design}{Combinatorial design}} & \eczhListValue{A constant-weight binary code that is mapped into a combinatorial \(t\)-design.}\\ 
\addlinespace[\myxtraspc]
\eczhRefIndex{code:finite_grassmann}%
\eczhListValue{\flmRefsHyperref{code:finite_grassmann}{Constant-dimension code}} & \eczhListValue{A subspace code whose codewords are \(k\)-dimensional subspaces of \(\mathbb{F}_q^n\) for fixed \(k\).
Constant-dimension codes are equivalent to linear authentication codes \NoCaseChange{\protect\cite[{Thm. 4.1}]{cite167}}.}\\ 
\addlinespace[\myxtraspc]
\eczhRefIndex{code:constant_excitation}%
\eczhListValue{\flmRefsHyperref{code:constant_excitation}{Constant-excitation (CE) code}} & \eczhListValue{Code whose codewords lie in an eigenspace of fixed total energy or fixed total excitation number for the underlying quantum system.
For qubit codes, such a Hamiltonian is often the \textit{total spin Hamiltonian}, \(H=\sum_i Z_i\).
For spin-\(S\) codes, this generalizes to \(H=\sum_i J_z^{(i)}\), where \(J_z\) is the spin-\(S\) \(Z\)-operator.
For bosonic (and, similarly, for fermion) codes, such as Fock-state codes, codewords are often in an eigenspace with eigenvalue \(N>0\) of the \textit{total excitation} or \textit{energy Hamiltonian}, \(H=\sum_i \hat{n}_i\).}\\ 
\addlinespace[\myxtraspc]
\eczhRefIndex{code:q-ary_constant_weight}%
\eczhListValue{\flmRefsHyperref{code:q-ary_constant_weight}{Constant-weight block code}} & \eczhListValue{A block code whose codewords all have the same number of nonzero coordinates.
Code constructions exist for codes over fields \NoCaseChange{\protect\cite{cite168}} or rings \NoCaseChange{\protect\cite{cite169}}.}\\ 
\addlinespace[\myxtraspc]
\eczhRefIndex{code:constant_weight}%
\eczhListValue{\flmRefsHyperref{code:constant_weight}{Constant-weight code}} & \eczhListValue{A binary code whose codewords are all constrained to have the same Hamming weight \(w\). In the linear setting, this usually refers to all nonzero codewords having the same weight, since every linear code contains the zero codeword.}\\ 
\addlinespace[\myxtraspc]
\eczhRefIndex{code:divisible}%
\eczhListValue{\flmRefsHyperref{code:divisible}{Divisible code}} & \eczhListValue{A linear \(q\)-ary block code is \(\Delta\)-divisible if the Hamming weight of each of its codewords is divisible by divisor \(\Delta\).
A \(2\)-divisible (\(4\)-divisible, \(8\)-divisible) code is called \textit{even} (\textit{doubly even}, \textit{triply even}) \NoCaseChange{\protect\cite{cite170,cite39}}.
A code is called \textit{singly even} if all codewords are even and at least one has weight equal to 2 modulo 4.
More generally, a code is \(m\)\textit{-even} if it is \(2^{m}\)-divisible.}\\ 
\addlinespace[\myxtraspc]
\eczhRefIndex{code:q-ary_linear}%
\eczhListValue{\flmRefsHyperref{code:q-ary_linear}{Linear \(q\)-ary code}} & \eczhListValue{An \((n,K,d)_q\) linear code is denoted as \([n,k,d]_q\), where \(k=\log_q K\) is an integer. Its codewords form a linear subspace, i.e., for any codewords \(x,y\), \(\alpha x+ \beta y\) is also a codeword for any field elements \(\alpha,\beta \in \mathbb{F}_q\).
This extra structure yields much information about their properties, making them a large and well-studied subset of codes.}\\ 
\addlinespace[\myxtraspc]
\eczhRefIndex{code:binary_linear}%
\eczhListValue{\flmRefsHyperref{code:binary_linear}{Linear binary code}} & \eczhListValue{An \((n,2^k,d)\) linear code is denoted as \([n,k]\) or \([n,k,d]\), where \(k\) is the code's dimension, and where \(d\) is the code's distance. Its codewords form a linear subspace, i.e., for any codewords \(x,y\), \(x+y\) is also a codeword. A code that is not linear is called \textit{nonlinear}.}\\ 
\addlinespace[\myxtraspc]
\eczhRefIndex{code:q-ary_linear_over_zq}%
\eczhListValue{\flmRefsHyperref{code:q-ary_linear_over_zq}{Linear code over \(\mathbb{Z}_q\)}} & \eczhListValue{A code encoding \(K\) states (codewords) in \(n\) coordinates over the ring \(\mathbb{Z}_q\) of integers modulo \(q\) that forms an Abelian subgroup of \(\mathbb{Z}_q^n\) under addition.
Since addition of \(m\) identical elements is equivalent to multiplying by \(m\), linear codes over \(\mathbb{Z}_q\) are automatically closed under scalar multiplication.
More technically, linear codes over \(\mathbb{Z}_q\) are submodules of \(\mathbb{Z}_q^n\).}\\ 
\addlinespace[\myxtraspc]
\eczhRefIndex{code:one_hot}%
\eczhListValue{\flmRefsHyperref{code:one_hot}{One-hot code}} & \eczhListValue{A nonlinear binary code whose codewords are all those with Hamming weight one. The reverse of this code, where all codewords have Hamming weight \(n-1\) is called a \textit{one-cold} code.}\\ 
\addlinespace[\myxtraspc]
\eczhRefIndex{code:one_vs_one}%
\eczhListValue{\flmRefsHyperref{code:one_vs_one}{One-versus-one (OVO) code}} & \eczhListValue{A length-\(n\) ternary code over \(\{\pm 1,0\}\) whose generator matrix has columns with one \(+1\), one \(-1\), and the rest of the entries zero.}\\ 
\addlinespace[\myxtraspc]
\eczhRefIndex{code:delsarte_optimal}%
\eczhListValue{\flmRefsHyperref{code:delsarte_optimal}{Sharp configuration}} & \eczhListValue{A code \(W\) in a compact connected two-point homogeneous space with \(m=l(W)\) distinct distances such that either \(r(W) \geq 2m-1\), or \(r(W)=2m-2>0\) and \(W\) is diametrical \NoCaseChange{\protect\cite{cite171}}.}\\ 
\addlinespace[\myxtraspc]
\eczhRefIndex{code:two_in_five}%
\eczhListValue{\flmRefsHyperref{code:two_in_five}{Two-in-five code}} & \eczhListValue{A nonlinear binary code consisting of the 10 weight-two five-bit strings, thereby providing an encoding for the decimal digits 0 through 9.}\\ 
\addlinespace[\myxtraspc]
\eczhRefIndex{code:2pt_homogeneous}%
\eczhListValue{\flmRefsHyperref{code:2pt_homogeneous}{Two-point homogeneous-space code}} & \eczhListValue{Encodes \(K\) states (codewords) into a two-point homogeneous space \(G/H\), i.e., a homogeneous space with a \(G\)-invariant metric whose symmetry group acts transitively on pairs of points at any fixed distance.
In the compact connected case, these are precisely the rank-one symmetric spaces. Finite examples tag along as the corresponding combinatorial analogues, where pairs of points are indistinguishable up to the symmetry group once their mutual distance is fixed.}\\ 
\addlinespace[\myxtraspc]
\eczhRefIndex{code:two_weight}%
\eczhListValue{\flmRefsHyperref{code:two_weight}{Two-weight code}} & \eczhListValue{A linear \(q\)-ary code whose codewords all have one of two possible nonzero Hamming weights \NoCaseChange{\protect\cite[{Def. 19.1}]{cite172}}.}\\ 
\addlinespace[\myxtraspc]
\eczhRefIndex{code:univ_opt_q-ary}%
\eczhListValue{\flmRefsHyperref{code:univ_opt_q-ary}{Universally optimal \(q\)-ary code}} & \eczhListValue{A \(q\)-ary code that (weakly) minimizes all completely monotonic potentials on Hamming space \NoCaseChange{\protect\cite{cite173}}.
Equivalently, its binomial moments are minimal among all codes with the same size and block length \NoCaseChange{\protect\cite[{Lemma 4}]{cite173}}.}\\ 
\addlinespace[\myxtraspc]
\eczhRefIndex{code:weight_two}%
\eczhListValue{\flmRefsHyperref{code:weight_two}{Weight-two code}} & \eczhListValue{A length-\(n\) binary code whose codewords all have Hamming weight two. Such codes provide slightly extra redundancy for storage of small-scale information such as ZIP codes or decimal digits.}\\ 
\addlinespace[\myxtraspc]
\eczhRefIndex{code:simplex}%
\eczhListValue{\flmRefsHyperref{code:simplex}{\([2^m-1,m,2^{m-1}]\) simplex code}} & \eczhListValue{A member of the equidistant code family dual to the \([2^m-1,2^m-m-1,3]\) Hamming family.}\\ 
\addlinespace[\myxtraspc]
\eczhRefIndex{code:q-ary_over_zq}%
\eczhListValue{\flmRefsHyperref{code:q-ary_over_zq}{\(q\)-ary code over \(\mathbb{Z}_q\)}} & \eczhListValue{A code encoding \(K\) states (codewords) in \(n\) coordinates over the ring \(\mathbb{Z}_q\) of integers modulo \(q\).}\\ 
\addlinespace[\myxtraspc]
\eczhRefIndex{code:q-ary_simplex}%
\eczhListValue{\flmRefsHyperref{code:q-ary_simplex}{\(q\)-ary simplex code}} & \eczhListValue{An \([n,m,q^{m-1}]_q\) equidistant projective code with \(n=\frac{q^m-1}{q-1}\), denoted as \(S(q,m)\). The columns of the generator matrix are in one-to-one correspondence with the elements of the projective space \(PG(m-1,q)\), with each column being a chosen representative of the corresponding element.
All nonzero simplex codewords have a constant weight of \(q^{m-1}\) \NoCaseChange{\protect\cite{cite45,cite46}}.}\\ 
\end{tabularx}\endgroup
\eczcodelist{cyclic}{Cyclic codes}%

\eczhCodeListAutoDescription{All descendants of \flmRefsCref{code:cyclic}.}%

\eczhIncludeCodeGraph{Bare}{scale=0.5}{\columnwidth}{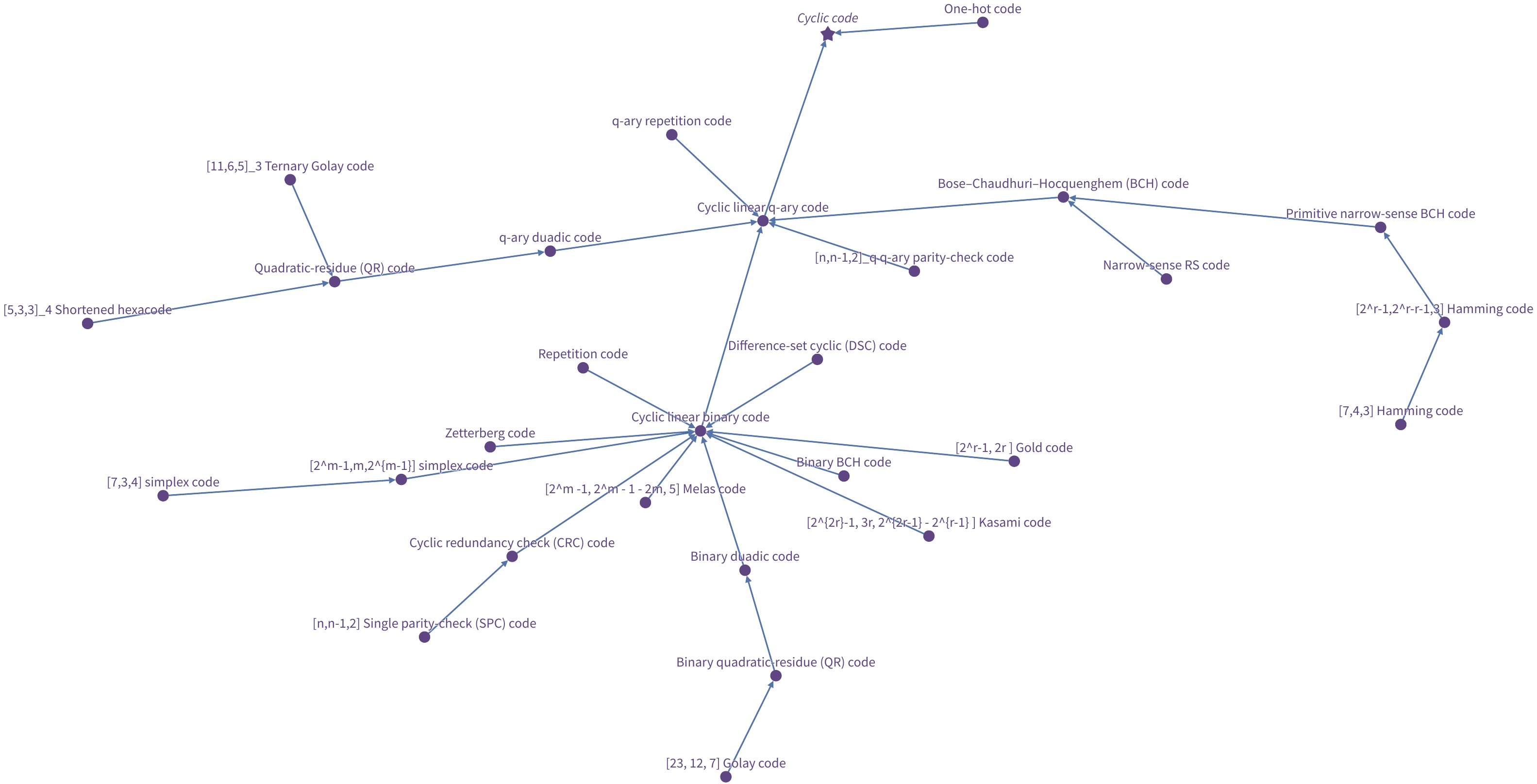}{Cyclic codes}{https://errorcorrectionzoo.org/code_graph#J\%7B\%22displayMode\%22\%3A\%22subset\%22\%2C\%22modeSubsetOptions\%22\%3A\%7B\%22codeIds\%22\%3A\%5B\%22bch\%22\%2C\%22binary_duadic\%22\%2C\%22binary_quad_residue\%22\%2C\%22q-ary_bch\%22\%2C\%22cyclic\%22\%2C\%22q-ary_cyclic\%22\%2C\%22binary_cyclic\%22\%2C\%22crc\%22\%2C\%22difference_set\%22\%2C\%22narrow_sense_reed_solomon\%22\%2C\%22one_hot\%22\%2C\%22narrow_sense_q-ary_bch\%22\%2C\%22q-ary_quad_residue\%22\%2C\%22repetition\%22\%2C\%22zetterberg\%22\%2C\%22ternary_golay\%22\%2C\%22golay\%22\%2C\%22melas\%22\%2C\%22simplex\%22\%2C\%22gold\%22\%2C\%22hamming\%22\%2C\%22kasami\%22\%2C\%22shortened_hexacode\%22\%2C\%22simplex734\%22\%2C\%22hamming743\%22\%2C\%22parity_check\%22\%2C\%22q-ary_parity_check\%22\%2C\%22q-ary_duadic\%22\%2C\%22q-ary_repetition\%22\%5D\%2C\%22reusePreviousLayoutPositions\%22\%3Afalse\%2C\%22showIntermediateConnectingNodes\%22\%3Atrue\%2C\%22connectingNodesMaxDepth\%22\%3A15\%2C\%22connectingNodesPathMaxLength\%22\%3A20\%2C\%22connectingNodesMaxExtraDepth\%22\%3A3\%2C\%22connectingNodesOnlyKeepPathsWithAdditionalLength\%22\%3A1\%2C\%22connectingNodesToDomainsAndKingdoms\%22\%3Afalse\%2C\%22connectingNodesEdgeLengthsByType\%22\%3A\%7B\%22primaryParent\%22\%3A1\%2C\%22secondaryParent\%22\%3A4\%2C\%22cousin\%22\%3A6\%7D\%2C\%22nodeIds\%22\%3A\%5B\%5D\%7D\%2C\%22highlightImportantNodes\%22\%3A\%7B\%22highlightImportantNodes\%22\%3Afalse\%2C\%22highlightPrimaryParents\%22\%3Afalse\%2C\%22highlightRootConnectingEdges\%22\%3Afalse\%7D\%7D}

\begingroup
\small
\eczhBreakableDashes
\renewcommand\arraystretch{1.05}
\edef\myxtraspc{\eczListAddVSpaceXtraXtraStretch}
\begin{tabularx}{\linewidth}{>{\raggedright\arraybackslash}p{\eczListColWidth{name}} >{\hsize=1.0000\hsize }X}
\toprule
\eczListColTitle{Code} & \eczListColTitle{Description} \\
\midrule
\endfirsthead
\toprule
\eczListColTitleContinued{Code} & \eczListColTitleContinued{Description} \\
\midrule
\endhead
\bottomrule
\endfoot
\eczhRefIndex{code:bch}%
\eczhListValue{\flmRefsHyperref{code:bch}{Binary BCH code}} & \eczhListValue{Cyclic binary code of odd length \(n\) whose zeroes are consecutive powers of a primitive \(n\)th root of unity \(\alpha\) (see \flmRefsCref{ref67}). More precisely, the generator polynomial of a BCH code of \textit{designed distance} \(\delta\geq 1\) is the lowest-degree monic polynomial with zeroes \(\{\alpha^b,\alpha^{b+1},\cdots,\alpha^{b+\delta-2}\}\) for some \(b\geq 0\). BCH codes are called \textit{narrow-sense} when \(b=1\), and are called \textit{primitive} when \(n=2^r-1\) for some \(r\geq 2\).}\\ 
\addlinespace[\myxtraspc]
\eczhRefIndex{code:binary_duadic}%
\eczhListValue{\flmRefsHyperref{code:binary_duadic}{Binary duadic code}} & \eczhListValue{Member of a pair of cyclic linear binary codes that satisfy certain relations, depending on whether the pair is \textit{even-like} or \textit{odd-like} duadic. Binary duadic codes generalize binary quadratic-residue codes \NoCaseChange{\protect\cite[{Sec. 2.7}]{cite68}}. Duadic codes exist for lengths \(n\) that are products of powers of primes, with each prime being \(\pm 1\) modulo \(8\) \NoCaseChange{\protect\cite{cite69}}.}\\ 
\addlinespace[\myxtraspc]
\eczhRefIndex{code:binary_quad_residue}%
\eczhListValue{\flmRefsHyperref{code:binary_quad_residue}{Binary quadratic-residue (QR) code}} & \eczhListValue{Member of a quadruple of cyclic binary codes of prime length \(n=8m\pm 1\) for \(m\geq 1\) constructed using quadratic residues and nonresidues of \(n\) \NoCaseChange{\protect\cite[{Def. 3.2.8}]{cite70}}.}\\ 
\addlinespace[\myxtraspc]
\eczhRefIndex{code:q-ary_bch}%
\eczhListValue{\flmRefsHyperref{code:q-ary_bch}{Bose–Chaudhuri–Hocquenghem (BCH) code}} & \eczhListValue{A cyclic \(q\)-ary code, with \(n\) and \(q\) relatively prime, whose zeroes are consecutive powers of a primitive \(n\)th root of unity \(\alpha\).}\\ 
\addlinespace[\myxtraspc]
\eczhRefIndex{code:cyclic}%
\eczhListValue{\flmRefsHyperref{code:cyclic}{Cyclic code}} & \eczhListValue{A block code of length \(n\) over an alphabet is cyclic if, for each codeword \(c_1 c_2 \cdots c_n\), the cyclically shifted string \(c_n c_1 \cdots c_{n-1}\) is also a codeword.}\\ 
\addlinespace[\myxtraspc]
\eczhRefIndex{code:q-ary_cyclic}%
\eczhListValue{\flmRefsHyperref{code:q-ary_cyclic}{Cyclic linear \(q\)-ary code}} & \eczhListValue{A \(q\)-ary code of length \(n\) is cyclic if, for each codeword \(c_1 c_2 \cdots c_n\), the cyclically shifted string \(c_n c_1 \cdots c_{n-1}\) is also a codeword. A cyclic code is called \textit{primitive} when \(n=q^r-1\) for some \(r\geq 2\). A \textit{shortened cyclic code} is obtained from a cyclic code by taking only codewords with the first \(j\) zero entries, and deleting those zeroes.}\\ 
\addlinespace[\myxtraspc]
\eczhRefIndex{code:binary_cyclic}%
\eczhListValue{\flmRefsHyperref{code:binary_cyclic}{Cyclic linear binary code}} & \eczhListValue{A binary code of length \(n\) is cyclic if, for each codeword \(c_1 c_2 \cdots c_n\), the cyclically shifted string \(c_n c_1 \cdots c_{n-1}\) is also a codeword. A cyclic code is called \textit{primitive} when \(n=2^r-1\) for some \(r\geq 2\).}\\ 
\addlinespace[\myxtraspc]
\eczhRefIndex{code:crc}%
\eczhListValue{\flmRefsHyperref{code:crc}{Cyclic redundancy check (CRC) code}} & \eczhListValue{A generalization of the single parity-check code in which the generalization of the parity bit is the remainder of the data string (mapped into a polynomial via the \flmRefsCref{ref67}) divided by some generator polynomial.
A notable family of codes is referred to as \textit{CRC-}(\(m-1\)), where \(m\) is the length of the generator polynomial.}\\ 
\addlinespace[\myxtraspc]
\eczhRefIndex{code:difference_set}%
\eczhListValue{\flmRefsHyperref{code:difference_set}{Difference-set cyclic (DSC) code}} & \eczhListValue{Cyclic LDPC code constructed deterministically from a difference set.
Certain DSC codes satisfy more redundant constraints than Gallager codes and thus can outperform them \NoCaseChange{\protect\cite{cite72}}.}\\ 
\addlinespace[\myxtraspc]
\eczhRefIndex{code:narrow_sense_reed_solomon}%
\eczhListValue{\flmRefsHyperref{code:narrow_sense_reed_solomon}{Narrow-sense RS code}} & \eczhListValue{An \([q-1,k,n-k+1]_q\) RS code whose points \(\alpha_i\) are all \((i-1)\)st powers of a \flmRefsHyperref{ref33}{primitive} element \(\alpha\) of \(\mathbb{F}_q\).}\\ 
\addlinespace[\myxtraspc]
\eczhRefIndex{code:one_hot}%
\eczhListValue{\flmRefsHyperref{code:one_hot}{One-hot code}} & \eczhListValue{A nonlinear binary code whose codewords are all those with Hamming weight one. The reverse of this code, where all codewords have Hamming weight \(n-1\) is called a \textit{one-cold} code.}\\ 
\addlinespace[\myxtraspc]
\eczhRefIndex{code:narrow_sense_q-ary_bch}%
\eczhListValue{\flmRefsHyperref{code:narrow_sense_q-ary_bch}{Primitive narrow-sense BCH code}} & \eczhListValue{A \(q\)-ary BCH code for \(b=1\) and for \(n=q^r-1\) for some \(r\geq 2\).}\\ 
\addlinespace[\myxtraspc]
\eczhRefIndex{code:q-ary_quad_residue}%
\eczhListValue{\flmRefsHyperref{code:q-ary_quad_residue}{Quadratic-residue (QR) code}} & \eczhListValue{Member of a quadruple of cyclic \(q\)-ary codes of prime length \(n\) where \(q\) is prime and a quadratic-residue modulo \(n\) \NoCaseChange{\protect\cite[{Def. 3.2.8}]{cite70}}.
The codes are constructed using quadratic residues and nonresidues of \(n\).
The definition extends to prime-power alphabet sizes and to prime-power lengths \NoCaseChange{\protect\cite{cite174,cite175}\protect\cite[{Rem. 3.2.9}]{cite70}}.
A quadratic-residue code of prime length \(p\) has dimension \((p+1)/2\) \NoCaseChange{\protect\cite[{Sec. 3.2.1}]{cite70}}.}\\ 
\addlinespace[\myxtraspc]
\eczhRefIndex{code:repetition}%
\eczhListValue{\flmRefsHyperref{code:repetition}{Repetition code}} & \eczhListValue{\([n,1,n]\) binary linear code encoding one bit of information into an \(n\)-bit string.
Majority decoding requires \(n\) to be odd in order to avoid ties.
The idea is to increase the code distance by repeating the logical information several times. It is a \((n,1)\)-Hamming code.
Its automorphism group is \(S_n\).}\\ 
\addlinespace[\myxtraspc]
\eczhRefIndex{code:zetterberg}%
\eczhListValue{\flmRefsHyperref{code:zetterberg}{Zetterberg code}} & \eczhListValue{Family of binary cyclic \([2^{2s}+1,2^{2s}-4s+1]\) codes with distance \(d\geq 5\) generated by the minimal polynomial \(g_s(x)\) of \(\alpha\) over \(\mathbb{F}_2\), where \(\alpha\) is a primitive \(n\)th root of unity in the field \(\mathbb{F}_{2^{4s}}\). They are quasi-perfect codes and are one of the best known families of double-error correcting binary linear codes.}\\ 
\addlinespace[\myxtraspc]
\eczhRefIndex{code:ternary_golay}%
\eczhListValue{\flmRefsHyperref{code:ternary_golay}{\([11,6,5]_3\) Ternary Golay code}} & \eczhListValue{A \([11,6,5]_3\) perfect ternary linear code with connections to various areas of mathematics, e.g., lattices \NoCaseChange{\protect\cite{cite39}} and sporadic simple groups \NoCaseChange{\protect\cite{cite41}}.
Adding a parity bit to the code results in the self-dual \([12,6,6]_3\) \textit{extended ternary Golay code}, whose weight enumerator is the Gleason polynomial \(g_5\) \NoCaseChange{\protect\cite[{Rem. 4.2.6}]{cite40}}.
Up to equivalence, both codes are unique for their respective parameters \NoCaseChange{\protect\cite{cite102}}.
The dual of the ternary Golay code is a \([11,5,6]_3\) projective two-weight subcode \NoCaseChange{\protect\cite[{Exam. 19.3.2}]{cite172}}.}\\ 
\addlinespace[\myxtraspc]
\eczhRefIndex{code:golay}%
\eczhListValue{\flmRefsHyperref{code:golay}{\([23, 12, 7]\) Golay code}} & \eczhListValue{A \([23, 12, 7]\) perfect binary linear code with connections to various areas of mathematics, e.g., lattices \NoCaseChange{\protect\cite{cite39}} and sporadic simple groups \NoCaseChange{\protect\cite{cite41}}.
Up to equivalence, it is unique for its parameters \NoCaseChange{\protect\cite{cite102}}.
The dual of the Golay code is its \([23,11,8]\) even-weight subcode \NoCaseChange{\protect\cite{cite103,cite104}}.}\\ 
\addlinespace[\myxtraspc]
\eczhRefIndex{code:melas}%
\eczhListValue{\flmRefsHyperref{code:melas}{\([2^m -1, 2^m - 1 - 2m, 5]\) Melas code}} & \eczhListValue{Cyclic linear code whose generator polynomial is \(g(x) = p(x)p(x)^{\star}\), where \(p(x)\) is a primitive polynomial of degree \(m\) that is the minimal polynomial over \(\mathbb{F}_2\) of an element \(\alpha\) of order \(2^m -1\) in \(\mathbb{F}_{2^m}\), \(m\) is odd and greater than five, and '\(\star\)' denotes reciprocation \NoCaseChange{\protect\cite{cite105}}.}\\ 
\addlinespace[\myxtraspc]
\eczhRefIndex{code:simplex}%
\eczhListValue{\flmRefsHyperref{code:simplex}{\([2^m-1,m,2^{m-1}]\) simplex code}} & \eczhListValue{A member of the equidistant code family dual to the \([2^m-1,2^m-m-1,3]\) Hamming family.}\\ 
\addlinespace[\myxtraspc]
\eczhRefIndex{code:gold}%
\eczhListValue{\flmRefsHyperref{code:gold}{\([2^r-1, 2r ]\) Gold code}} & \eczhListValue{A cyclic binary linear code characterized by the generator polynomial of degree \(r\) of two maximum-period sequences of period \(2^r-1\) with absolute cross-correlation \( \leq 2^{(r+2)/2}\). Gold codewords are generated using \(m\)-sequences \(x\) and \(y\), which are codewords of simplex codes with check polynomials of degree \(r\) \NoCaseChange{\protect\cite{cite106}}.}\\ 
\addlinespace[\myxtraspc]
\eczhRefIndex{code:hamming}%
\eczhListValue{\flmRefsHyperref{code:hamming}{\([2^r-1,2^r-r-1,3]\) Hamming code}} & \eczhListValue{Member of an infinite family of perfect linear codes with parameters \([2^r-1,2^r-r-1, 3]\) for \(r \geq 2\).
Their \(r \times (2^r-1) \) parity-check matrix \(H\) has all possible nonzero \(r\)-bit strings as its columns.
Adding a parity check yields the \([2^r,2^r-r-1, 4]\) extended Hamming code.}\\ 
\addlinespace[\myxtraspc]
\eczhRefIndex{code:kasami}%
\eczhListValue{\flmRefsHyperref{code:kasami}{\([2^{2r}-1, 3r, 2^{2r-1} - 2^{r-1} ]\) Kasami code}} & \eczhListValue{Member of the family of \([2^{2r}-1, 3r, 2^{2r-1} - 2^{r-1} ]\) cyclic binary linear codes. 
Gold and Kasami codes are both constructed by picking a set of cyclically unrelated sequences of binary linear codes with low cross-correlation \NoCaseChange{\protect\cite{cite107,cite108}}.}\\ 
\addlinespace[\myxtraspc]
\eczhRefIndex{code:shortened_hexacode}%
\eczhListValue{\flmRefsHyperref{code:shortened_hexacode}{\([5,3,3]_4\) Shortened hexacode}} & \eczhListValue{A perfect \([5,3,3]_4\) quaternary Hamming code that is the result of puncturing the hexacode \NoCaseChange{\protect\cite{cite43}}.}\\ 
\addlinespace[\myxtraspc]
\eczhRefIndex{code:simplex734}%
\eczhListValue{\flmRefsHyperref{code:simplex734}{\([7,3,4]\) simplex code}} & \eczhListValue{Second-smallest nontrivial member of the simplex-code family.
The columns of its generator matrix are in one-to-one correspondence with the elements of the projective space \(PG(2,2)\), with each column being a chosen representative of the corresponding element.
The codewords form a \((8,9)\) simplex spherical code under the \flmRefsHyperref{ref38}{antipodal mapping}.
As a simplex code, it is equidistant: every nonzero codeword has Hamming weight \(4\).}\\ 
\addlinespace[\myxtraspc]
\eczhRefIndex{code:hamming743}%
\eczhListValue{\flmRefsHyperref{code:hamming743}{\([7,4,3]\) Hamming code}} & \eczhListValue{Second-smallest member of the Hamming code family.}\\ 
\addlinespace[\myxtraspc]
\eczhRefIndex{code:parity_check}%
\eczhListValue{\flmRefsHyperref{code:parity_check}{\([n,n-1,2]\) Single parity-check (SPC) code}} & \eczhListValue{An \([n,n-1,2]\) linear binary code whose codewords consist of the message string appended with a \textit{parity-check bit} or \textit{parity bit} such that the parity (i.e., sum over all coordinates of each codeword) is zero.
If the Hamming weight of a message is odd (even), then the parity bit is one (zero).
This code requires only one extra bit of overhead and is therefore inexpensive.
Its codewords are all even-weight binary strings, and its parity-check matrix is a row vector of all ones.
Its automorphism group is \(S_n\).}\\ 
\addlinespace[\myxtraspc]
\eczhRefIndex{code:q-ary_parity_check}%
\eczhListValue{\flmRefsHyperref{code:q-ary_parity_check}{\([n,n-1,2]_q\) \(q\)-ary parity-check code}} & \eczhListValue{An \([n,n-1,2]_q\) linear \(q\)-ary code whose codewords consist of the message string appended with a \textit{parity-check} or \textit{zero-sum check digit} such that the sum over all coordinates of each codeword is zero.}\\ 
\addlinespace[\myxtraspc]
\eczhRefIndex{code:q-ary_duadic}%
\eczhListValue{\flmRefsHyperref{code:q-ary_duadic}{\(q\)-ary duadic code}} & \eczhListValue{Member of a pair of cyclic linear \(q\)-ary codes that satisfy certain relations, depending on whether the pair is \textit{even-like} or \textit{odd-like} duadic.
Duadic codes exist only when \(q\) is a square modulo \(n\) \NoCaseChange{\protect\cite{cite69}}.}\\ 
\addlinespace[\myxtraspc]
\eczhRefIndex{code:q-ary_repetition}%
\eczhListValue{\flmRefsHyperref{code:q-ary_repetition}{\(q\)-ary repetition code}} & \eczhListValue{An \([n,1,n]_q\) code consisting of codewords \((j,j,\cdots,j)\) for \(j \in \mathbb{F}_q\).}\\ 
\end{tabularx}\endgroup
\eczcodelist{evaluation}{Evaluation codes
}%

\eczhCodeListAutoDescription{All descendants of \flmRefsCref{code:evaluation_varieties}.}%

\eczhIncludeCodeGraph{Bare}{scale=0.5}{\columnwidth}{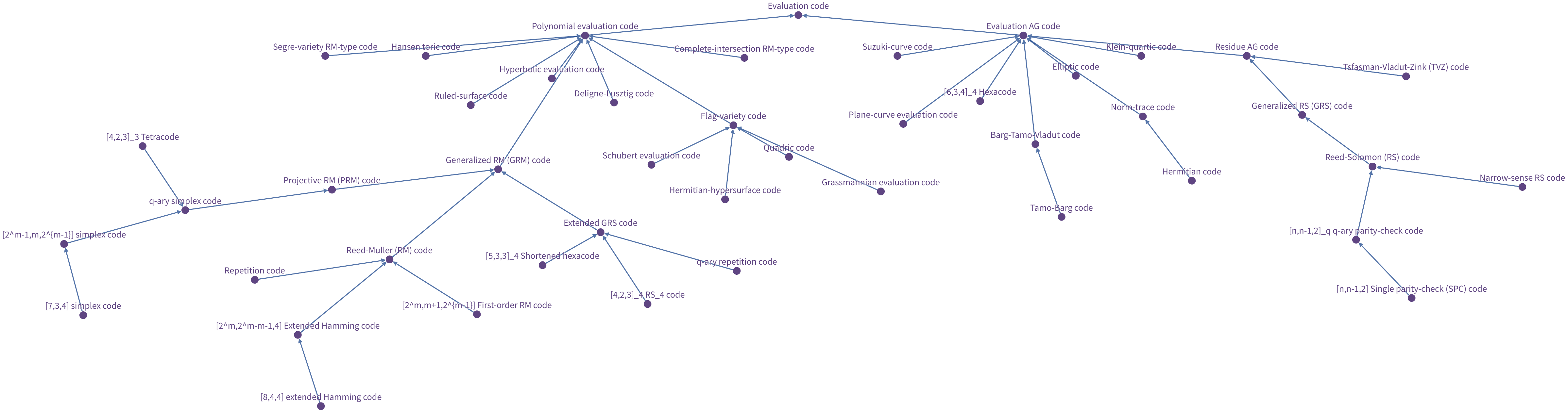}{Evaluation codes}{https://errorcorrectionzoo.org/code_graph#J\%7B\%22displayMode\%22\%3A\%22subset\%22\%2C\%22modeSubsetOptions\%22\%3A\%7B\%22codeIds\%22\%3A\%5B\%22tamo_barg_vladut\%22\%2C\%22complete_intersections\%22\%2C\%22deligne_lusztig\%22\%2C\%22elliptic\%22\%2C\%22evaluation\%22\%2C\%22evaluation_varieties\%22\%2C\%22extended_reed_solomon\%22\%2C\%22flag_variety\%22\%2C\%22generalized_reed_muller\%22\%2C\%22generalized_reed_solomon\%22\%2C\%22grassmannian_variety\%22\%2C\%22toric_classical\%22\%2C\%22hermitian\%22\%2C\%22hermitian_hypersurface\%22\%2C\%22cascaded_reed_solomon\%22\%2C\%22klein_quartic\%22\%2C\%22narrow_sense_reed_solomon\%22\%2C\%22norm_trace\%22\%2C\%22plane_curve\%22\%2C\%22evaluation_polynomial\%22\%2C\%22projective_reed_muller\%22\%2C\%22quadric\%22\%2C\%22reed_muller\%22\%2C\%22reed_solomon\%22\%2C\%22repetition\%22\%2C\%22residue\%22\%2C\%22ruled_surface\%22\%2C\%22schubert\%22\%2C\%22serge\%22\%2C\%22suzuki\%22\%2C\%22tamo_barg\%22\%2C\%22shimura\%22\%2C\%22extended_hamming\%22\%2C\%22biorthogonal\%22\%2C\%22simplex\%22\%2C\%22tetracode\%22\%2C\%22reed_solomon_4\%22\%2C\%22shortened_hexacode\%22\%2C\%22hexacode\%22\%2C\%22simplex734\%22\%2C\%22hamming844\%22\%2C\%22parity_check\%22\%2C\%22q-ary_parity_check\%22\%2C\%22q-ary_repetition\%22\%2C\%22q-ary_simplex\%22\%5D\%2C\%22reusePreviousLayoutPositions\%22\%3Afalse\%2C\%22showIntermediateConnectingNodes\%22\%3Atrue\%2C\%22connectingNodesMaxDepth\%22\%3A15\%2C\%22connectingNodesPathMaxLength\%22\%3A20\%2C\%22connectingNodesMaxExtraDepth\%22\%3A3\%2C\%22connectingNodesOnlyKeepPathsWithAdditionalLength\%22\%3A1\%2C\%22connectingNodesToDomainsAndKingdoms\%22\%3Afalse\%2C\%22connectingNodesEdgeLengthsByType\%22\%3A\%7B\%22primaryParent\%22\%3A1\%2C\%22secondaryParent\%22\%3A4\%2C\%22cousin\%22\%3A6\%7D\%2C\%22nodeIds\%22\%3A\%5B\%5D\%7D\%2C\%22highlightImportantNodes\%22\%3A\%7B\%22highlightImportantNodes\%22\%3Afalse\%2C\%22highlightPrimaryParents\%22\%3Afalse\%2C\%22highlightRootConnectingEdges\%22\%3Afalse\%7D\%7D}

\begingroup
\small
\eczhBreakableDashes
\renewcommand\arraystretch{1.05}
\edef\myxtraspc{\eczListAddVSpaceXtraXtraStretch}
\begin{tabularx}{\linewidth}{>{\raggedright\arraybackslash}p{\eczListColWidth{name}} >{\hsize=1.0000\hsize }X}
\toprule
\eczListColTitle{Code} & \eczListColTitle{Description} \\
\midrule
\endfirsthead
\toprule
\eczListColTitleContinued{Code} & \eczListColTitleContinued{Description} \\
\midrule
\endhead
\bottomrule
\endfoot
\eczhRefIndex{code:tamo_barg_vladut}%
\eczhListValue{\flmRefsHyperref{code:tamo_barg_vladut}{Barg-Tamo-Vladut code}} & \eczhListValue{Evaluation AG code on algebraic curves built from a Galois cover \(\phi:Y\to X\), where the recovery sets are fibres over rational points of \(X\) that split completely in the cover \NoCaseChange{\protect\cite[{Def. 15.9.10}]{cite26}\protect\cite[{Thm. 15.9.14}]{cite26}}.
The Barg-Tamo-Vladut construction generalizes the Tamo-Barg construction from \(PG(1,q)\) to longer AG codes, and variants can be built with higher local distance or availability \(2\) via fibre products of curves \NoCaseChange{\protect\cite[{Thm. 15.9.19}]{cite26}\protect\cite[{Thm. 15.9.21}]{cite26}}.}\\ 
\addlinespace[\myxtraspc]
\eczhRefIndex{code:complete_intersections}%
\eczhListValue{\flmRefsHyperref{code:complete_intersections}{Complete-intersection RM-type code}} & \eczhListValue{Evaluation code of polynomials evaluated on points lying on a complete intersection.}\\ 
\addlinespace[\myxtraspc]
\eczhRefIndex{code:deligne_lusztig}%
\eczhListValue{\flmRefsHyperref{code:deligne_lusztig}{Deligne-Lusztig code}} & \eczhListValue{Evaluation code of polynomials evaluated on points lying on a Deligne-Lusztig variety, often a Deligne-Lusztig curve in the classical one-dimensional cases.}\\ 
\addlinespace[\myxtraspc]
\eczhRefIndex{code:elliptic}%
\eczhListValue{\flmRefsHyperref{code:elliptic}{Elliptic code}} & \eczhListValue{Evaluation AG code of rational functions evaluated on points lying on an elliptic curve, i.e., a curve of genus one.}\\ 
\addlinespace[\myxtraspc]
\eczhRefIndex{code:evaluation}%
\eczhListValue{\flmRefsHyperref{code:evaluation}{Evaluation AG code}} & \eczhListValue{Evaluation code over \(\mathbb{F}_q\) on a set of points \({\cal P} = \left( P_1,P_2,\cdots,P_n \right)\) lying on an algebraic curve \(\cal X\) defined over \(\mathbb{F}_q\), where the corresponding vector space \(L\) of functions \(f\) consists of certain rational functions (or, in special cases, polynomials).}\\ 
\addlinespace[\myxtraspc]
\eczhRefIndex{code:evaluation_varieties}%
\eczhListValue{\flmRefsHyperref{code:evaluation_varieties}{Evaluation code}} & \eczhListValue{Code whose codewords are evaluations of functions at certain fixed points. Code properties can be inferred from the structure of the functions and the underlying geometric object containing the points, often using results from algebraic geometry.}\\ 
\addlinespace[\myxtraspc]
\eczhRefIndex{code:extended_reed_solomon}%
\eczhListValue{\flmRefsHyperref{code:extended_reed_solomon}{Extended GRS code}} & \eczhListValue{A GRS code extended by one extra coordinate to form an \([n+1,k,n-k+2]_q\) MDS code. In projective language, this corresponds to adding one more evaluation point, often interpreted as the point at infinity; in suitable equivalent descriptions, one may instead use an affine point such as \(0\). The case when \(n=q-1\), multipliers \(v_i=1\), and \(\alpha_i\) are \(i-1\)st powers of a primitive \(n\)th root of unity is an \textit{extended narrow-sense RS code}.}\\ 
\addlinespace[\myxtraspc]
\eczhRefIndex{code:flag_variety}%
\eczhListValue{\flmRefsHyperref{code:flag_variety}{Flag-variety code}} & \eczhListValue{Evaluation code of polynomials evaluated on points lying on a flag variety.}\\ 
\addlinespace[\myxtraspc]
\eczhRefIndex{code:generalized_reed_muller}%
\eczhListValue{\flmRefsHyperref{code:generalized_reed_muller}{Generalized RM (GRM) code}} & \eczhListValue{Extensions of RM codes to \(q\)-ary coordinates that can be described as multivariate polynomial evaluation codes over affine or projective space.}\\ 
\addlinespace[\myxtraspc]
\eczhRefIndex{code:generalized_reed_solomon}%
\eczhListValue{\flmRefsHyperref{code:generalized_reed_solomon}{Generalized RS (GRS) code}} & \eczhListValue{An \([n,k,n-k+1]_q\) MDS code that is a modification of the RS code where codeword polynomials are multiplied by additional prefactors \NoCaseChange{\protect\cite[{Def. 15.3.19}]{cite26}}.}\\ 
\addlinespace[\myxtraspc]
\eczhRefIndex{code:grassmannian_variety}%
\eczhListValue{\flmRefsHyperref{code:grassmannian_variety}{Grassmannian evaluation code}} & \eczhListValue{Evaluation code of polynomials evaluated on points lying on a finite-field Grassmannian embedded into projective space using the Plucker embedding \NoCaseChange{\protect\cite{cite27,cite28}}.}\\ 
\addlinespace[\myxtraspc]
\eczhRefIndex{code:toric_classical}%
\eczhListValue{\flmRefsHyperref{code:toric_classical}{Hansen toric code}} & \eczhListValue{Evaluation code of a linear space of polynomials evaluated on points lying on an affine or projective toric variety. If the space is taken to be all polynomials up to some degree, the code is called a \textit{toric RM-type code} of that degree.}\\ 
\addlinespace[\myxtraspc]
\eczhRefIndex{code:hermitian}%
\eczhListValue{\flmRefsHyperref{code:hermitian}{Hermitian code}} & \eczhListValue{Evaluation AG code of rational functions on a Hermitian curve over \(\mathbb{F}_{q^2}\).}\\ 
\addlinespace[\myxtraspc]
\eczhRefIndex{code:hermitian_hypersurface}%
\eczhListValue{\flmRefsHyperref{code:hermitian_hypersurface}{Hermitian-hypersurface code}} & \eczhListValue{Evaluation code of polynomials evaluated on points lying on a Hermitian hypersurface.}\\ 
\addlinespace[\myxtraspc]
\eczhRefIndex{code:cascaded_reed_solomon}%
\eczhListValue{\flmRefsHyperref{code:cascaded_reed_solomon}{Hyperbolic evaluation code}} & \eczhListValue{An evaluation code over polynomials in two variables. 
Generator matrices are determined in Ref. \NoCaseChange{\protect\cite{cite29}} following initial formulations of the codes as generalized concatenations of RS codes \NoCaseChange{\protect\cite{cite30,cite31}}; see \NoCaseChange{\protect\cite[{Exam. 4.26}]{cite32}}.}\\ 
\addlinespace[\myxtraspc]
\eczhRefIndex{code:klein_quartic}%
\eczhListValue{\flmRefsHyperref{code:klein_quartic}{Klein-quartic code}} & \eczhListValue{Evaluation AG code over \(\mathbb{F}_8\) of rational functions evaluated on points lying on the Klein quartic, which is defined by the equation \(x^3 y + y^3 z + z^3 x = 0\) \NoCaseChange{\protect\cite[{Ex. 2.75}]{cite32}}.}\\ 
\addlinespace[\myxtraspc]
\eczhRefIndex{code:narrow_sense_reed_solomon}%
\eczhListValue{\flmRefsHyperref{code:narrow_sense_reed_solomon}{Narrow-sense RS code}} & \eczhListValue{An \([q-1,k,n-k+1]_q\) RS code whose points \(\alpha_i\) are all \((i-1)\)st powers of a \flmRefsHyperref{ref33}{primitive} element \(\alpha\) of \(\mathbb{F}_q\).}\\ 
\addlinespace[\myxtraspc]
\eczhRefIndex{code:norm_trace}%
\eczhListValue{\flmRefsHyperref{code:norm_trace}{Norm-trace code}} & \eczhListValue{Evaluation AG code of rational functions evaluated on points lying on a Miura-Kamiya curve in either affine or projective space.
The family is named as such because the equations defining the curves can be expressed in terms of the \flmRefsHyperref{ref33}{field norm and field trace}.}\\ 
\addlinespace[\myxtraspc]
\eczhRefIndex{code:plane_curve}%
\eczhListValue{\flmRefsHyperref{code:plane_curve}{Plane-curve evaluation code}} & \eczhListValue{Evaluation AG code of bivariate polynomials of some finite maximum degree, evaluated at points lying on an affine or projective plane curve.}\\ 
\addlinespace[\myxtraspc]
\eczhRefIndex{code:evaluation_polynomial}%
\eczhListValue{\flmRefsHyperref{code:evaluation_polynomial}{Polynomial evaluation code}} & \eczhListValue{Evaluation code of polynomials (or, more generally, rational functions) at points \({\cal P} = \left( P_1,P_2,\cdots,P_n \right)\) on an algebraic variety \(\cal X\) of dimension greater than one (i.e., not an algebraic curve).}\\ 
\addlinespace[\myxtraspc]
\eczhRefIndex{code:projective_reed_muller}%
\eczhListValue{\flmRefsHyperref{code:projective_reed_muller}{Projective RM (PRM) code}} & \eczhListValue{Evaluation code obtained by evaluating homogeneous polynomials on the points of the projective space \(PG(m,q)\), equivalently on representatives of the nonzero vectors in \(\mathbb{F}_q^{m+1}\) whose leftmost nonzero coordinate is one.}\\ 
\addlinespace[\myxtraspc]
\eczhRefIndex{code:quadric}%
\eczhListValue{\flmRefsHyperref{code:quadric}{Quadric code}} & \eczhListValue{Evaluation code of polynomials evaluated on points lying on a quadric hypersurface.}\\ 
\addlinespace[\myxtraspc]
\eczhRefIndex{code:reed_muller}%
\eczhListValue{\flmRefsHyperref{code:reed_muller}{Reed-Muller (RM) code}} & \eczhListValue{Member of the RM\((r,m)\) family of linear binary codes derived from multivariate polynomials. The code parameters are \([2^m,\sum_{j=0}^{r} {m \choose j},2^{m-r}]\), where \(r\) is the \textit{order} of the code satisfying \(0\leq r\leq m\).
First-order RM codes are also called biorthogonal codes, while \(m\)th order RM codes are also called \textit{universe} codes.
\textit{Punctured RM codes} RM\(^*(r,m)\) are obtained from RM codes by deleting one coordinate from each codeword.}\\ 
\addlinespace[\myxtraspc]
\eczhRefIndex{code:reed_solomon}%
\eczhListValue{\flmRefsHyperref{code:reed_solomon}{Reed-Solomon (RS) code}} & \eczhListValue{An \([n,k,n-k+1]_q\) linear code based on polynomials over \(\mathbb{F}_q\).}\\ 
\addlinespace[\myxtraspc]
\eczhRefIndex{code:repetition}%
\eczhListValue{\flmRefsHyperref{code:repetition}{Repetition code}} & \eczhListValue{\([n,1,n]\) binary linear code encoding one bit of information into an \(n\)-bit string.
Majority decoding requires \(n\) to be odd in order to avoid ties.
The idea is to increase the code distance by repeating the logical information several times. It is a \((n,1)\)-Hamming code.
Its automorphism group is \(S_n\).}\\ 
\addlinespace[\myxtraspc]
\eczhRefIndex{code:residue}%
\eczhListValue{\flmRefsHyperref{code:residue}{Residue AG code}} & \eczhListValue{Linear \(q\)-ary code defined using a set of \(\mathbb{F}_q\)-rational points \({\cal P} = \left( P_1,P_2,\cdots,P_n \right)\) on an algebraic curve \(\cal X\) and a linear space \(\Omega\) of certain rational differential forms \(\omega\) \NoCaseChange{\protect\cite[{Def. 15.3.2}]{cite26}}.}\\ 
\addlinespace[\myxtraspc]
\eczhRefIndex{code:ruled_surface}%
\eczhListValue{\flmRefsHyperref{code:ruled_surface}{Ruled-surface code}} & \eczhListValue{Evaluation code obtained by evaluating global sections of a line bundle, or equivalently suitable polynomial functions, on rational points of a ruled surface over a finite field. Such codes extend algebraic-geometry constructions from curves to certain projective surfaces \NoCaseChange{\protect\cite{cite34,cite35}}.}\\ 
\addlinespace[\myxtraspc]
\eczhRefIndex{code:schubert}%
\eczhListValue{\flmRefsHyperref{code:schubert}{Schubert evaluation code}} & \eczhListValue{Evaluation code of polynomials evaluated on points lying on a Schubert variety.}\\ 
\addlinespace[\myxtraspc]
\eczhRefIndex{code:serge}%
\eczhListValue{\flmRefsHyperref{code:serge}{Segre-variety RM-type code}} & \eczhListValue{Evaluation code of multihomogeneous polynomials evaluated on points of a Segre variety, i.e., on the Segre embedding of a product of projective spaces. These codes are Reed-Muller-type analogues adapted to product projective geometries \NoCaseChange{\protect\cite{cite36}}.}\\ 
\addlinespace[\myxtraspc]
\eczhRefIndex{code:suzuki}%
\eczhListValue{\flmRefsHyperref{code:suzuki}{Suzuki-curve code}} & \eczhListValue{Evaluation AG code of rational functions evaluated on points lying on a Suzuki curve.}\\ 
\addlinespace[\myxtraspc]
\eczhRefIndex{code:tamo_barg}%
\eczhListValue{\flmRefsHyperref{code:tamo_barg}{Tamo-Barg code}} & \eczhListValue{A family of \(q\)-ary polynomial evaluation codes that are optimal LRCs and for which \(q\) is comparable to \(n\).}\\ 
\addlinespace[\myxtraspc]
\eczhRefIndex{code:shimura}%
\eczhListValue{\flmRefsHyperref{code:shimura}{Tsfasman-Vladut-Zink (TVZ) code}} & \eczhListValue{Member of a family of AG codes obtained from algebraic curves via the residue or evaluation construction. Sequences of curves with many rational points, such as Drinfeld modular curves, classical modular curves, or Garcia-Stichtenoth curves, yield the asymptotic parameters of the TVZ bound \NoCaseChange{\protect\cite[{Sec. 15.4.2}]{cite26}}.}\\ 
\addlinespace[\myxtraspc]
\eczhRefIndex{code:extended_hamming}%
\eczhListValue{\flmRefsHyperref{code:extended_hamming}{\([2^m,2^m-m-1,4]\) Extended Hamming code}} & \eczhListValue{Member of an infinite family of RM\((m-2,m)\) codes with parameters \([2^m,2^m-m-1, 4]\) for \(m \geq 2\) that are extensions of the Hamming codes by a parity-check bit.}\\ 
\addlinespace[\myxtraspc]
\eczhRefIndex{code:biorthogonal}%
\eczhListValue{\flmRefsHyperref{code:biorthogonal}{\([2^m,m+1,2^{m-1}]\) First-order RM code}} & \eczhListValue{A member of the family of first-order RM codes.
Its codewords are the rows of the \(2^m\)-dimensional Hadamard matrix \(H\) and its negation \(-H\) with the mapping \(+1\to 0\) and \(-1\to 1\).
The family is self-orthogonal for \(m \geq 3\) \NoCaseChange{\protect\cite{cite37}}.
They form a \((2^m,2^{m+1})\) biorthogonal spherical code under the \flmRefsHyperref{ref38}{antipodal mapping}.}\\ 
\addlinespace[\myxtraspc]
\eczhRefIndex{code:simplex}%
\eczhListValue{\flmRefsHyperref{code:simplex}{\([2^m-1,m,2^{m-1}]\) simplex code}} & \eczhListValue{A member of the equidistant code family dual to the \([2^m-1,2^m-m-1,3]\) Hamming family.}\\ 
\addlinespace[\myxtraspc]
\eczhRefIndex{code:tetracode}%
\eczhListValue{\flmRefsHyperref{code:tetracode}{\([4,2,3]_3\) Tetracode}} & \eczhListValue{The \([4,2,3]_3\) ternary self-dual MDS code that has connections to lattices \NoCaseChange{\protect\cite{cite39}}. Its weight enumerator is the Gleason polynomial \(g_4\) \NoCaseChange{\protect\cite[{Rem. 4.2.6}]{cite40}}.}\\ 
\addlinespace[\myxtraspc]
\eczhRefIndex{code:reed_solomon_4}%
\eczhListValue{\flmRefsHyperref{code:reed_solomon_4}{\([4,2,3]_4\) RS\(_4\) code}} & \eczhListValue{A Type II Euclidean self-dual extended RS code that is the smallest quaternary extended QR code \NoCaseChange{\protect\cite[{pg. 296}]{cite41}\protect\cite[{Sec. 2.4.2}]{cite42}}.
Puncturing the \([4,2,3]_4\) RS\(_4\) code yields the \([3,2,2]_4\) shortened RS\(_4\) code, which is an RS code \NoCaseChange{\protect\cite[{pg. 295}]{cite41}}.}\\ 
\addlinespace[\myxtraspc]
\eczhRefIndex{code:shortened_hexacode}%
\eczhListValue{\flmRefsHyperref{code:shortened_hexacode}{\([5,3,3]_4\) Shortened hexacode}} & \eczhListValue{A perfect \([5,3,3]_4\) quaternary Hamming code that is the result of puncturing the hexacode \NoCaseChange{\protect\cite{cite43}}.}\\ 
\addlinespace[\myxtraspc]
\eczhRefIndex{code:hexacode}%
\eczhListValue{\flmRefsHyperref{code:hexacode}{\([6,3,4]_4\) Hexacode}} & \eczhListValue{The \([6,3,4]_4\) Hermitian self-dual MDS code that has connections to projective geometry, lattices \NoCaseChange{\protect\cite{cite39}}, and conformal field theory \NoCaseChange{\protect\cite{cite44}}. Its weight enumerator is the Gleason polynomial \(g_7\) \NoCaseChange{\protect\cite[{Rem. 4.2.6}]{cite40}}.}\\ 
\addlinespace[\myxtraspc]
\eczhRefIndex{code:simplex734}%
\eczhListValue{\flmRefsHyperref{code:simplex734}{\([7,3,4]\) simplex code}} & \eczhListValue{Second-smallest nontrivial member of the simplex-code family.
The columns of its generator matrix are in one-to-one correspondence with the elements of the projective space \(PG(2,2)\), with each column being a chosen representative of the corresponding element.
The codewords form a \((8,9)\) simplex spherical code under the \flmRefsHyperref{ref38}{antipodal mapping}.
As a simplex code, it is equidistant: every nonzero codeword has Hamming weight \(4\).}\\ 
\addlinespace[\myxtraspc]
\eczhRefIndex{code:hamming844}%
\eczhListValue{\flmRefsHyperref{code:hamming844}{\([8,4,4]\) extended Hamming code}} & \eczhListValue{Extension of the \([7,4,3]\) Hamming code by a parity-check bit.
The smallest doubly even self-dual code, and the unique Type II code of length \(8\) \NoCaseChange{\protect\cite[{Rem. 4.3.10}]{cite40}}.}\\ 
\addlinespace[\myxtraspc]
\eczhRefIndex{code:parity_check}%
\eczhListValue{\flmRefsHyperref{code:parity_check}{\([n,n-1,2]\) Single parity-check (SPC) code}} & \eczhListValue{An \([n,n-1,2]\) linear binary code whose codewords consist of the message string appended with a \textit{parity-check bit} or \textit{parity bit} such that the parity (i.e., sum over all coordinates of each codeword) is zero.
If the Hamming weight of a message is odd (even), then the parity bit is one (zero).
This code requires only one extra bit of overhead and is therefore inexpensive.
Its codewords are all even-weight binary strings, and its parity-check matrix is a row vector of all ones.
Its automorphism group is \(S_n\).}\\ 
\addlinespace[\myxtraspc]
\eczhRefIndex{code:q-ary_parity_check}%
\eczhListValue{\flmRefsHyperref{code:q-ary_parity_check}{\([n,n-1,2]_q\) \(q\)-ary parity-check code}} & \eczhListValue{An \([n,n-1,2]_q\) linear \(q\)-ary code whose codewords consist of the message string appended with a \textit{parity-check} or \textit{zero-sum check digit} such that the sum over all coordinates of each codeword is zero.}\\ 
\addlinespace[\myxtraspc]
\eczhRefIndex{code:q-ary_repetition}%
\eczhListValue{\flmRefsHyperref{code:q-ary_repetition}{\(q\)-ary repetition code}} & \eczhListValue{An \([n,1,n]_q\) code consisting of codewords \((j,j,\cdots,j)\) for \(j \in \mathbb{F}_q\).}\\ 
\addlinespace[\myxtraspc]
\eczhRefIndex{code:q-ary_simplex}%
\eczhListValue{\flmRefsHyperref{code:q-ary_simplex}{\(q\)-ary simplex code}} & \eczhListValue{An \([n,m,q^{m-1}]_q\) equidistant projective code with \(n=\frac{q^m-1}{q-1}\), denoted as \(S(q,m)\). The columns of the generator matrix are in one-to-one correspondence with the elements of the projective space \(PG(m-1,q)\), with each column being a chosen representative of the corresponding element.
All nonzero simplex codewords have a constant weight of \(q^{m-1}\) \NoCaseChange{\protect\cite{cite45,cite46}}.}\\ 
\end{tabularx}\endgroup
\eczcodelist{lattice}{Lattices}%

\eczhCodeListAutoDescription{All descendants of \flmRefsCref{code:points_into_lattices}.}%

\eczhIncludeCodeGraph{Bare}{scale=0.5}{\columnwidth}{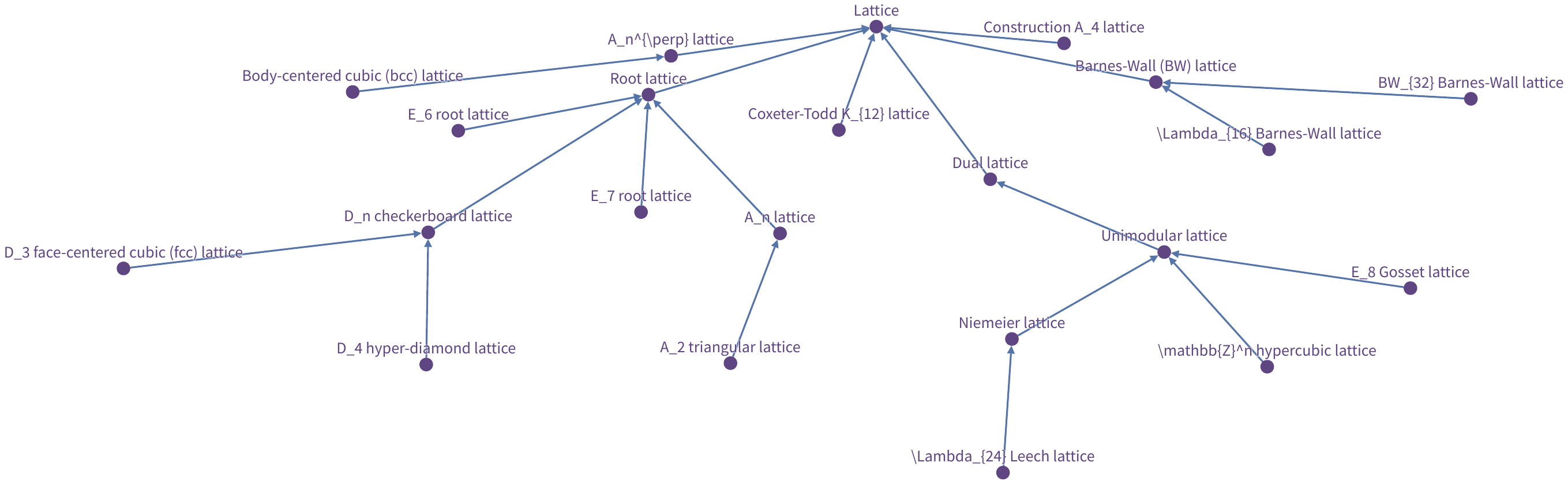}{Lattices}{https://errorcorrectionzoo.org/code_graph#J\%7B\%22displayMode\%22\%3A\%22subset\%22\%2C\%22modeSubsetOptions\%22\%3A\%7B\%22codeIds\%22\%3A\%5B\%22barnes_wall\%22\%2C\%22bcc\%22\%2C\%22construction_a4\%22\%2C\%22coxeter_todd\%22\%2C\%22dual_lattice\%22\%2C\%22points_into_lattices\%22\%2C\%22niemeier\%22\%2C\%22root\%22\%2C\%22self_dual_lattice\%22\%2C\%22hexagonal\%22\%2C\%22an\%22\%2C\%22an_dual\%22\%2C\%22bw32\%22\%2C\%22dthree\%22\%2C\%22dfour\%22\%2C\%22dn\%22\%2C\%22esix\%22\%2C\%22eseven\%22\%2C\%22eeight\%22\%2C\%22lambda16\%22\%2C\%22leech\%22\%2C\%22hypercubic\%22\%5D\%2C\%22reusePreviousLayoutPositions\%22\%3Afalse\%2C\%22showIntermediateConnectingNodes\%22\%3Atrue\%2C\%22connectingNodesMaxDepth\%22\%3A15\%2C\%22connectingNodesPathMaxLength\%22\%3A20\%2C\%22connectingNodesMaxExtraDepth\%22\%3A3\%2C\%22connectingNodesOnlyKeepPathsWithAdditionalLength\%22\%3A1\%2C\%22connectingNodesToDomainsAndKingdoms\%22\%3Afalse\%2C\%22connectingNodesEdgeLengthsByType\%22\%3A\%7B\%22primaryParent\%22\%3A1\%2C\%22secondaryParent\%22\%3A4\%2C\%22cousin\%22\%3A6\%7D\%2C\%22nodeIds\%22\%3A\%5B\%5D\%7D\%2C\%22highlightImportantNodes\%22\%3A\%7B\%22highlightImportantNodes\%22\%3Afalse\%2C\%22highlightPrimaryParents\%22\%3Afalse\%2C\%22highlightRootConnectingEdges\%22\%3Afalse\%7D\%7D}

\begingroup
\small
\eczhBreakableDashes
\renewcommand\arraystretch{1.05}
\edef\myxtraspc{\eczListAddVSpaceXtraXtraStretch}
\begin{tabularx}{\linewidth}{>{\raggedright\arraybackslash}p{\eczListColWidth{name}} >{\hsize=1.0000\hsize }X}
\toprule
\eczListColTitle{Code} & \eczListColTitle{Description} \\
\midrule
\endfirsthead
\toprule
\eczListColTitleContinued{Code} & \eczListColTitleContinued{Description} \\
\midrule
\endhead
\bottomrule
\endfoot
\eczhRefIndex{code:barnes_wall}%
\eczhListValue{\flmRefsHyperref{code:barnes_wall}{Barnes-Wall (BW) lattice}} & \eczhListValue{Member of a family of \(2^{m+1}\)-dimensional lattices, denoted as BW\(_{2^{m+1}}\), that are the densest lattices known.
Members include the integer square lattice \(\mathbb{Z}^2\), \(D_4\), the Gosset \(E_8\) lattice, and the \(\Lambda_{16}\) lattice, corresponding to \(m\in\{0,1,2,3\}\), respectively.}\\ 
\addlinespace[\myxtraspc]
\eczhRefIndex{code:bcc}%
\eczhListValue{\flmRefsHyperref{code:bcc}{Body-centered cubic (bcc) lattice}} & \eczhListValue{Three-dimensional lattice consisting of all points \((x,y,z)\) whose integer components are either all even or all odd.}\\ 
\addlinespace[\myxtraspc]
\eczhRefIndex{code:construction_a4}%
\eczhListValue{\flmRefsHyperref{code:construction_a4}{Construction \(A_4\) lattice}} & \eczhListValue{A lattice that is constructed from a linear code over \(\mathbb{Z}_4\) using \flmTerm{term}{ref114}{}{Construction \(A_4\)}.}\\ 
\addlinespace[\myxtraspc]
\eczhRefIndex{code:coxeter_todd}%
\eczhListValue{\flmRefsHyperref{code:coxeter_todd}{Coxeter-Todd \(K_{12}\) lattice}} & \eczhListValue{Even integral lattice in dimension \(12\) that gives the densest known lattice packing.
Its automorphism group was discovered by Mitchell \NoCaseChange{\protect\cite{cite176}}.
As a real lattice, \(K_{12}\) is equivalent, up to rescaling, to its dual \(K_{12}^{\perp}\) \NoCaseChange{\protect\cite[{Ch. 4, pg. 128}]{cite39}}.
For more details, see \NoCaseChange{\protect\cite{cite177}\protect\cite[{Sec. 4.9}]{cite39}}.}\\ 
\addlinespace[\myxtraspc]
\eczhRefIndex{code:dual_lattice}%
\eczhListValue{\flmRefsHyperref{code:dual_lattice}{Dual lattice}} & \eczhListValue{For any dimensional lattice \(L\), the dual lattice is the set of vectors whose inner products with the elements of \(L\) are integers.}\\ 
\addlinespace[\myxtraspc]
\eczhRefIndex{code:points_into_lattices}%
\eczhListValue{\flmRefsHyperref{code:points_into_lattices}{Lattice}} & \eczhListValue{Encodes states (codewords) in coordinates of an \(n\)-dimensional lattice, i.e., a discrete set of points in Euclidean space \(\mathbb{R}^n\) that forms a group under vector addition when translated so that one point is at the origin. The number of codewords may be infinite because the ambient Euclidean space is unbounded, so various restricted versions have to be constructed in practice. Since lattices are closed under addition, lattice-based codes can be thought of as linear codes over the reals.}\\ 
\addlinespace[\myxtraspc]
\eczhRefIndex{code:niemeier}%
\eczhListValue{\flmRefsHyperref{code:niemeier}{Niemeier lattice}} & \eczhListValue{One of the 24 positive-definite even unimodular lattices of rank 24.
The 24 lattices are \(D_{24}\), \(D_{16}E_8\), \(E_8^3\), \(A_{24}\), \(D_{12}^2\), \(A_{17}E_7\), \(D_{10}E_7^2\), \(A_{15}D_9\), \(D_8^3\), \(A_{12}^2\), \(A_{11}D_7E_6\), \(E_6^4\), \(A_9^2D_6\), \(D_6^4\), \(A_8^3\), \(A_7^2D_5^2\), \(A_6^4\), \(A_5^4D_4\), \(D_4^6\), \(A_4^6\), \(A_3^8\), \(A_2^{12}\), \(A_1^{24}\), and \(\Lambda_{24}\) (the Leech lattice) \NoCaseChange{\protect\cite[{Table 16.1}]{cite39}}.}\\ 
\addlinespace[\myxtraspc]
\eczhRefIndex{code:root}%
\eczhListValue{\flmRefsHyperref{code:root}{Root lattice}} & \eczhListValue{A lattice that is symmetric under a specific crystallographic reflection group; see \NoCaseChange{\protect\cite[{Table 4.1}]{cite39}} for the list of crystallographic reflection groups and their corresponding root lattices.
The root-lattice family consists of lattices \(A_n\), \(\mathbb{Z}^n\), or \(D_n\) for dimension \(n\), or \(E_{i}\) for \(i\in\{6,7,8\}\).
Their generator matrices can be taken to be the root matrices of the corresponding reflection groups.}\\ 
\addlinespace[\myxtraspc]
\eczhRefIndex{code:self_dual_lattice}%
\eczhListValue{\flmRefsHyperref{code:self_dual_lattice}{Unimodular lattice}} & \eczhListValue{A lattice, scaled to be integral, that is equal to its dual, \(L^\perp = L\).
Unimodular lattices have \(\det L = \pm 1\).}\\ 
\addlinespace[\myxtraspc]
\eczhRefIndex{code:hexagonal}%
\eczhListValue{\flmRefsHyperref{code:hexagonal}{\(A_2\) triangular lattice}} & \eczhListValue{Two-dimensional lattice that corresponds to the triangular tiling and that exhibits optimal packing, solving the packing, kissing, covering and quantization problems.
As a tiling, its dual (whose points lie at the centers of each triangle) is the honeycomb tiling.}\\ 
\addlinespace[\myxtraspc]
\eczhRefIndex{code:an}%
\eczhListValue{\flmRefsHyperref{code:an}{\(A_n\) lattice}} & \eczhListValue{Lattice-based \(n\)-dimensional code that can be simply defined in \(n+1\) dimensions as the set of integer vectors \(x\) lying in the hyperplane \(x_0+x_1+\cdots+x_{n} = 0\).}\\ 
\addlinespace[\myxtraspc]
\eczhRefIndex{code:an_dual}%
\eczhListValue{\flmRefsHyperref{code:an_dual}{\(A_n^{\perp}\) lattice}} & \eczhListValue{Lattice-based \(n\)-dimensional code whose codewords form the dual of the \(A_n\) lattice.}\\ 
\addlinespace[\myxtraspc]
\eczhRefIndex{code:bw32}%
\eczhListValue{\flmRefsHyperref{code:bw32}{\(BW_{32}\) Barnes-Wall lattice}} & \eczhListValue{BW lattice in dimension \(32\).}\\ 
\addlinespace[\myxtraspc]
\eczhRefIndex{code:dthree}%
\eczhListValue{\flmRefsHyperref{code:dthree}{\(D_3\) face-centered cubic (fcc) lattice}} & \eczhListValue{Laminated three-dimensional lattice consisting of layers of triangular lattices.}\\ 
\addlinespace[\myxtraspc]
\eczhRefIndex{code:dfour}%
\eczhListValue{\flmRefsHyperref{code:dfour}{\(D_4\) hyper-diamond lattice}} & \eczhListValue{BW lattice in dimension \(4\), which is the lattice corresponding to the \([4,3,2]\) SPC code via \flmTerm{term}{ref127}{}{Construction A}.
The lattice points form the \(\{3,3,4,3\}\) tessellation of 4-space \NoCaseChange{\protect\cite[{pg. 136}]{cite178}}.}\\ 
\addlinespace[\myxtraspc]
\eczhRefIndex{code:dn}%
\eczhListValue{\flmRefsHyperref{code:dn}{\(D_n\) checkerboard lattice}} & \eczhListValue{Lattice consisting of all points whose coordinates add up to an even integer.}\\ 
\addlinespace[\myxtraspc]
\eczhRefIndex{code:esix}%
\eczhListValue{\flmRefsHyperref{code:esix}{\(E_6\) root lattice}} & \eczhListValue{Exceptional root lattice in dimension \(6\).}\\ 
\addlinespace[\myxtraspc]
\eczhRefIndex{code:eseven}%
\eczhListValue{\flmRefsHyperref{code:eseven}{\(E_7\) root lattice}} & \eczhListValue{Exceptional root lattice in dimension \(7\).}\\ 
\addlinespace[\myxtraspc]
\eczhRefIndex{code:eeight}%
\eczhListValue{\flmRefsHyperref{code:eeight}{\(E_8\) Gosset lattice}} & \eczhListValue{Even unimodular BW lattice in dimension \(8\), consisting of the Cayley integral octonions rescaled by \(\sqrt{2}\).
The lattice corresponds to the \([8,4,4]\) Hamming code via \flmTerm{term}{ref127}{}{Construction A}.}\\ 
\addlinespace[\myxtraspc]
\eczhRefIndex{code:lambda16}%
\eczhListValue{\flmRefsHyperref{code:lambda16}{\(\Lambda_{16}\) Barnes-Wall lattice}} & \eczhListValue{BW lattice in dimension \(16\).}\\ 
\addlinespace[\myxtraspc]
\eczhRefIndex{code:leech}%
\eczhListValue{\flmRefsHyperref{code:leech}{\(\Lambda_{24}\) Leech lattice}} & \eczhListValue{Even unimodular lattice in 24 dimensions that exhibits optimal packing.
Its automorphism group is the Conway group Co\(_0\).}\\ 
\addlinespace[\myxtraspc]
\eczhRefIndex{code:hypercubic}%
\eczhListValue{\flmRefsHyperref{code:hypercubic}{\(\mathbb{Z}^n\) hypercubic lattice}} & \eczhListValue{Lattice-based code consisting of all integer vectors in \(n\) dimensions.
Its generator matrix is the \(n\)-dimensional identity matrix.
Its automorphism group consists of all coordinate permutations and sign changes.}\\ 
\end{tabularx}\endgroup
\eczcodelist{ldpc}{LDPC codes
}%

\eczhCodeListAutoDescription{All descendants of \flmRefsCref{code:q-ary_ldpc}.}%

\eczhIncludeCodeGraph{Bare}{scale=0.5}{\columnwidth}{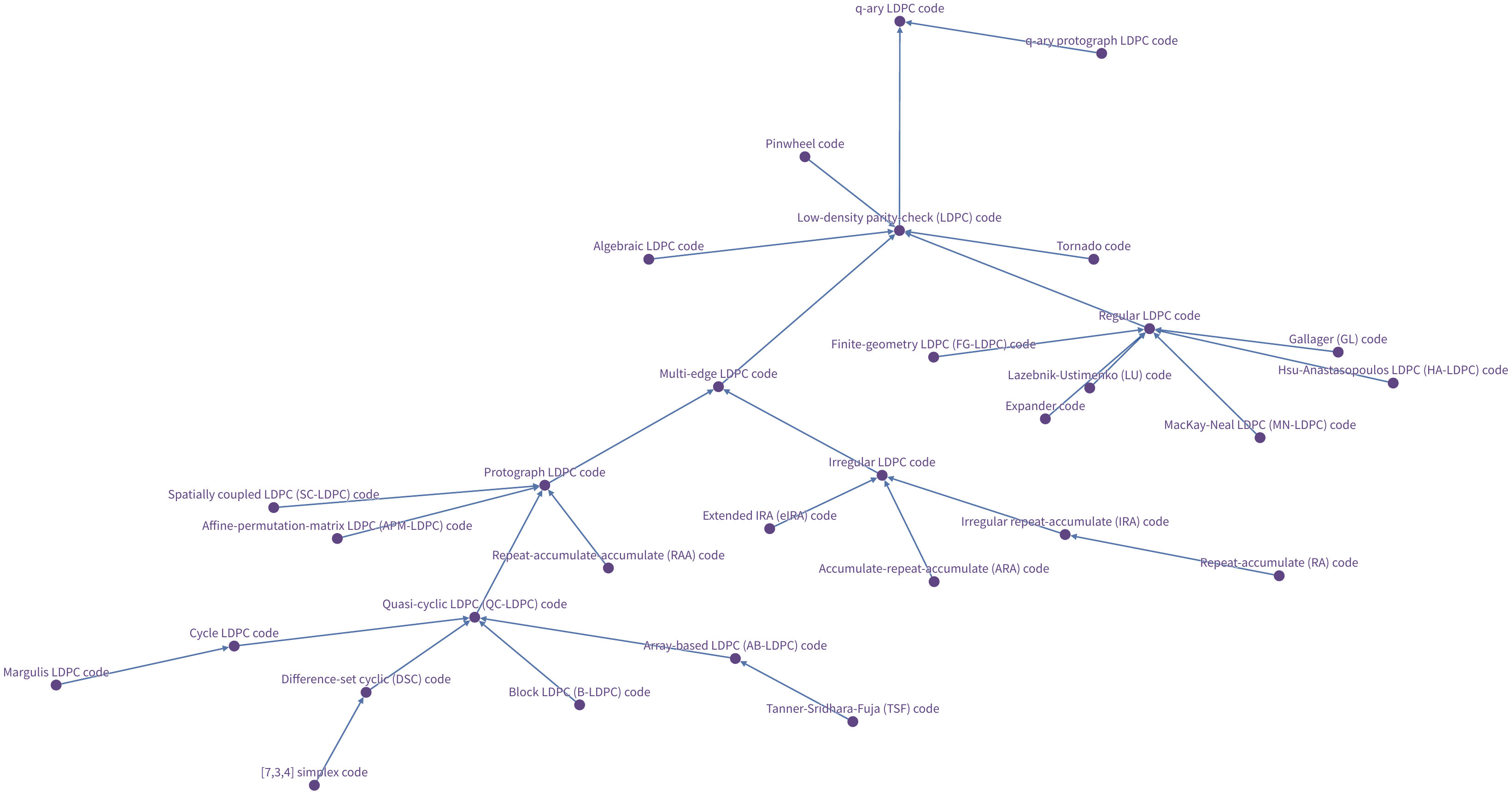}{LDPC codes}{https://errorcorrectionzoo.org/code_graph#J\%7B\%22displayMode\%22\%3A\%22subset\%22\%2C\%22modeSubsetOptions\%22\%3A\%7B\%22codeIds\%22\%3A\%5B\%22ara\%22\%2C\%22apm_ldpc\%22\%2C\%22algebraic_ldpc\%22\%2C\%22array_ldpc\%22\%2C\%22b_ldpc\%22\%2C\%22cycle_ldpc\%22\%2C\%22difference_set\%22\%2C\%22expander\%22\%2C\%22extended_ira\%22\%2C\%22pg_ldpc\%22\%2C\%22gallager\%22\%2C\%22ha_ldpc\%22\%2C\%22irregular_ldpc\%22\%2C\%22ira\%22\%2C\%22lu_ldpc\%22\%2C\%22ldpc\%22\%2C\%22mn_ldpc\%22\%2C\%22margulis_ldpc\%22\%2C\%22multi_edge_ldpc\%22\%2C\%22pinwheel\%22\%2C\%22protograph_ldpc\%22\%2C\%22qc_ldpc\%22\%2C\%22regular_ldpc\%22\%2C\%22ra\%22\%2C\%22raa\%22\%2C\%22sc_ldpc\%22\%2C\%22tsf\%22\%2C\%22tornado\%22\%2C\%22simplex734\%22\%2C\%22q-ary_ldpc\%22\%2C\%22q-ary_protograph_ldpc\%22\%5D\%2C\%22reusePreviousLayoutPositions\%22\%3Afalse\%2C\%22showIntermediateConnectingNodes\%22\%3Atrue\%2C\%22connectingNodesMaxDepth\%22\%3A15\%2C\%22connectingNodesPathMaxLength\%22\%3A20\%2C\%22connectingNodesMaxExtraDepth\%22\%3A3\%2C\%22connectingNodesOnlyKeepPathsWithAdditionalLength\%22\%3A1\%2C\%22connectingNodesToDomainsAndKingdoms\%22\%3Afalse\%2C\%22connectingNodesEdgeLengthsByType\%22\%3A\%7B\%22primaryParent\%22\%3A1\%2C\%22secondaryParent\%22\%3A4\%2C\%22cousin\%22\%3A6\%7D\%2C\%22nodeIds\%22\%3A\%5B\%5D\%7D\%2C\%22highlightImportantNodes\%22\%3A\%7B\%22highlightImportantNodes\%22\%3Afalse\%2C\%22highlightPrimaryParents\%22\%3Afalse\%2C\%22highlightRootConnectingEdges\%22\%3Afalse\%7D\%7D}

\begingroup
\small
\eczhBreakableDashes
\renewcommand\arraystretch{1.05}
\edef\myxtraspc{\eczListAddVSpaceXtraXtraStretch}
\begin{tabularx}{\linewidth}{>{\raggedright\arraybackslash}p{\eczListColWidth{name}} >{\hsize=1.0000\hsize }X}
\toprule
\eczListColTitle{Code} & \eczListColTitle{Description} \\
\midrule
\endfirsthead
\toprule
\eczListColTitleContinued{Code} & \eczListColTitleContinued{Description} \\
\midrule
\endhead
\bottomrule
\endfoot
\eczhRefIndex{code:ara}%
\eczhListValue{\flmRefsHyperref{code:ara}{Accumulate-repeat-accumulate (ARA) code}} & \eczhListValue{A generalization of the RA code in which the outer repetition-code encoding step is augmented with an accumulator acting on a fraction of the incoming bits.
In addition, the code may be punctured after the final accumulating step.}\\ 
\addlinespace[\myxtraspc]
\eczhRefIndex{code:apm_ldpc}%
\eczhListValue{\flmRefsHyperref{code:apm_ldpc}{Affine-permutation-matrix LDPC (APM-LDPC) code}} & \eczhListValue{LDPC code whose parity-check matrix can be put into the form of a block matrix consisting of permutation submatrices representing the affine permutation group or the zero submatrix.
Given a cyclic group \(\mathbb{Z}_r\), the affine permutation group is \(\mathbb{Z}_r \rtimes \mathbb{Z}_r^{\times}\), where \(\mathbb{Z}_r^{\times}\) is the multiplicative group of integers modulo \(r\).
Such codes are often constructed by \flmRefsHyperref{ref47}{lifting} certain protographs into such block matrices \NoCaseChange{\protect\cite{cite48}}.}\\ 
\addlinespace[\myxtraspc]
\eczhRefIndex{code:algebraic_ldpc}%
\eczhListValue{\flmRefsHyperref{code:algebraic_ldpc}{Algebraic LDPC code}} & \eczhListValue{LDPC code whose parity check matrix is constructed explicitly (i.e., non-randomly) from a particular graph \NoCaseChange{\protect\cite{cite49,cite50}} or an algebraic structure such as a combinatorial design \NoCaseChange{\protect\cite{cite51,cite52,cite53}}, balanced incomplete block design \NoCaseChange{\protect\cite{cite54}}, a partial geometry \NoCaseChange{\protect\cite{cite55}}, a generalized polygon \NoCaseChange{\protect\cite{cite56,cite57}}, or a Latin square \NoCaseChange{\protect\cite{cite58,cite59,cite60}}.
The extra structure and/or symmetry \NoCaseChange{\protect\cite{cite61}} of these codes can often be used to gain a better understanding of their properties.}\\ 
\addlinespace[\myxtraspc]
\eczhRefIndex{code:array_ldpc}%
\eczhListValue{\flmRefsHyperref{code:array_ldpc}{Array-based LDPC (AB-LDPC) code}} & \eczhListValue{QC-LDPC code constructed deterministically from a disk array code known as a B-code.
Its parity-check matrix admits a compact representation \NoCaseChange{\protect\cite{cite64}} and is related to RS codes.}\\ 
\addlinespace[\myxtraspc]
\eczhRefIndex{code:b_ldpc}%
\eczhListValue{\flmRefsHyperref{code:b_ldpc}{Block LDPC (B-LDPC) code}} & \eczhListValue{Member of a particular class of irregular QC-LDPC codes with efficient encoders.}\\ 
\addlinespace[\myxtraspc]
\eczhRefIndex{code:cycle_ldpc}%
\eczhListValue{\flmRefsHyperref{code:cycle_ldpc}{Cycle LDPC code}} & \eczhListValue{An LDPC code whose parity-check matrix forms the incidence matrix of a graph, i.e., has weight-two columns.}\\ 
\addlinespace[\myxtraspc]
\eczhRefIndex{code:difference_set}%
\eczhListValue{\flmRefsHyperref{code:difference_set}{Difference-set cyclic (DSC) code}} & \eczhListValue{Cyclic LDPC code constructed deterministically from a difference set.
Certain DSC codes satisfy more redundant constraints than Gallager codes and thus can outperform them \NoCaseChange{\protect\cite{cite72}}.}\\ 
\addlinespace[\myxtraspc]
\eczhRefIndex{code:expander}%
\eczhListValue{\flmRefsHyperref{code:expander}{Expander code}} & \eczhListValue{LDPC code whose parity-check matrix is derived from the adjacency matrix of a bipartite expander graph \NoCaseChange{\protect\cite{cite74}} such as a Ramanujan graph or a Cayley graph of a projective special linear group over a finite field \NoCaseChange{\protect\cite{cite75,cite76}}.
Expander codes admit efficient encoding and decoding algorithms and yield an explicit (i.e., non-random) asymptotically good LDPC code family.}\\ 
\addlinespace[\myxtraspc]
\eczhRefIndex{code:extended_ira}%
\eczhListValue{\flmRefsHyperref{code:extended_ira}{Extended IRA (eIRA) code}} & \eczhListValue{A generalization of the IRA code in which the outer LDGM code is replaced by a random sparse matrix containing no weight-two columns.}\\ 
\addlinespace[\myxtraspc]
\eczhRefIndex{code:pg_ldpc}%
\eczhListValue{\flmRefsHyperref{code:pg_ldpc}{Finite-geometry LDPC (FG-LDPC) code}} & \eczhListValue{LDPC code whose parity-check matrix is the incidence matrix of points and hyperplanes in either a Euclidean or a projective geometry.
Such codes are called \textit{Euclidean-geometry LDPC (EG-LDPC)} and \textit{projective-geometry LDPC (PG-LDPC)}, respectively.
Such constructions have been generalized to incidence matrices of hyperplanes of different dimensions \NoCaseChange{\protect\cite{cite77}}.}\\ 
\addlinespace[\myxtraspc]
\eczhRefIndex{code:gallager}%
\eczhListValue{\flmRefsHyperref{code:gallager}{Gallager (GL) code}} & \eczhListValue{The first LDPC code.
The rows of the parity-check matrix of this regular code are divided into equal subsets, and columns in the first subset are randomly permuted to yield the corresponding rows in subsequent subsets.}\\ 
\addlinespace[\myxtraspc]
\eczhRefIndex{code:ha_ldpc}%
\eczhListValue{\flmRefsHyperref{code:ha_ldpc}{Hsu-Anastasopoulos LDPC (HA-LDPC) code}} & \eczhListValue{A regular LDPC code obtained from a concatenation of a certain random regular LDPC code and a certain random LDGM code.
An \((l,r,g)\)-HA-LDPC code can be written using punctured LDPC and LDGM parts, and it is dual to the corresponding \((r,l,g)\)-MN-LDPC code \NoCaseChange{\protect\cite{cite84}}.
Using rate-one LDGM codes eliminates high-weight codewords while admitting an amount of low-weight codewords that asymptotically vanishes, allowing code families to achieve the \flmRefsHyperref{ref85}{GV bound} with high probability.}\\ 
\addlinespace[\myxtraspc]
\eczhRefIndex{code:irregular_ldpc}%
\eczhListValue{\flmRefsHyperref{code:irregular_ldpc}{Irregular LDPC code}} & \eczhListValue{An LDPC code whose parity-check matrix has a variable number of entries in each row or column.}\\ 
\addlinespace[\myxtraspc]
\eczhRefIndex{code:ira}%
\eczhListValue{\flmRefsHyperref{code:ira}{Irregular repeat-accumulate (IRA) code}} & \eczhListValue{A generalization of the RA code in which the outer 1-in-3 repetition encoding step is replaced by an LDGM code.
A simple version is when different bits in the RA block are repeated a different number of times.}\\ 
\addlinespace[\myxtraspc]
\eczhRefIndex{code:lu_ldpc}%
\eczhListValue{\flmRefsHyperref{code:lu_ldpc}{Lazebnik-Ustimenko (LU) code}} & \eczhListValue{LDPC code whose Tanner graph comes from a particular family of \(q\)-regular graphs \NoCaseChange{\protect\cite{cite87}} of known girth and relatively large stopping sets.}\\ 
\addlinespace[\myxtraspc]
\eczhRefIndex{code:ldpc}%
\eczhListValue{\flmRefsHyperref{code:ldpc}{Low-density parity-check (LDPC) code}} & \eczhListValue{A binary linear code with a sparse parity-check matrix.
Often a member of an infinite family of \([n,k,d]\) codes for which the numbers of nonzero entries in each row and in each column of the parity-check matrix are both bounded above by a constant as \(n\to\infty\).}\\ 
\addlinespace[\myxtraspc]
\eczhRefIndex{code:mn_ldpc}%
\eczhListValue{\flmRefsHyperref{code:mn_ldpc}{MacKay-Neal LDPC (MN-LDPC) code}} & \eczhListValue{A code whose parity-check matrix is constructed non-deterministically via the MacKay-Neal prescription.}\\ 
\addlinespace[\myxtraspc]
\eczhRefIndex{code:margulis_ldpc}%
\eczhListValue{\flmRefsHyperref{code:margulis_ldpc}{Margulis LDPC code}} & \eczhListValue{Member of a class of LDPC codes deterministically constructed from explicit sparse regular expander graphs.
The underlying Margulis-Gabber-Galil graph family provides explicit expanders \NoCaseChange{\protect\cite{cite49,cite90}}, yielding deterministic sparse parity-check matrices.
Related explicit LDPC constructions \NoCaseChange{\protect\cite{cite91}} utilize Ramanujan graphs \NoCaseChange{\protect\cite{cite75,cite76}}.}\\ 
\addlinespace[\myxtraspc]
\eczhRefIndex{code:multi_edge_ldpc}%
\eczhListValue{\flmRefsHyperref{code:multi_edge_ldpc}{Multi-edge LDPC code}} & \eczhListValue{Irregular LDPC code whose construction generalizes those of the original examples of irregular LDPC as well as RA codes.}\\ 
\addlinespace[\myxtraspc]
\eczhRefIndex{code:pinwheel}%
\eczhListValue{\flmRefsHyperref{code:pinwheel}{Pinwheel code}} & \eczhListValue{A geometrically local binary LDPC code defined on planar graphs obtained from the pinwheel tiling \NoCaseChange{\protect\cite{cite93}}.
Both bits and checks live on vertices of the graph.
If \(L_N\) is the graph Laplacian at generation \(N\), the undepleted check matrix is \(\tilde H_N=(L_N-\mathbb{I})\bmod 2\), and the actual parity-check matrix \(H_N\) is obtained by removing an evenly spaced fraction of boundary checks.}\\ 
\addlinespace[\myxtraspc]
\eczhRefIndex{code:protograph_ldpc}%
\eczhListValue{\flmRefsHyperref{code:protograph_ldpc}{Protograph LDPC code}} & \eczhListValue{Binary version of a \(q\)-ary protograph LDPC code.
Its parity-check matrix can be written as a block matrix whose blocks are either sums of permutation matrices or the zero matrix.}\\ 
\addlinespace[\myxtraspc]
\eczhRefIndex{code:qc_ldpc}%
\eczhListValue{\flmRefsHyperref{code:qc_ldpc}{Quasi-cyclic LDPC (QC-LDPC) code}} & \eczhListValue{LDPC code that can be put into quasi-cyclic form.
Its parity check matrix can be put into the form of a block matrix consisting of either circulant permutation sub-matrices or the zero sub-matrix.
Such codes are often constructed by \flmRefsHyperref{ref47}{lifting} certain protographs into such block matrices \NoCaseChange{\protect\cite{cite48}}.
Their simple structure makes them useful for several wireless communication standards.}\\ 
\addlinespace[\myxtraspc]
\eczhRefIndex{code:regular_ldpc}%
\eczhListValue{\flmRefsHyperref{code:regular_ldpc}{Regular LDPC code}} & \eczhListValue{An LDPC code whose parity-check matrix has a fixed number of ones in each row and each column.
Such a code is called \((j,k)\)-regular if each column has weight \(j\) (variable-node degree) and each row has weight \(k\) (check-node degree).
If the parity-check matrix has \(n\) columns and \(m\) rows, then regularity implies \(nj = mk\).}\\ 
\addlinespace[\myxtraspc]
\eczhRefIndex{code:ra}%
\eczhListValue{\flmRefsHyperref{code:ra}{Repeat-accumulate (RA) code}} & \eczhListValue{An LDPC code whose parity-check matrix has weight-two columns arranged in a step-like pattern for its last columns \NoCaseChange{\protect\cite{cite94}}.}\\ 
\addlinespace[\myxtraspc]
\eczhRefIndex{code:raa}%
\eczhListValue{\flmRefsHyperref{code:raa}{Repeat-accumulate-accumulate (RAA) code}} & \eczhListValue{Generalization of the RA code in which two accumulators and permutations are used.}\\ 
\addlinespace[\myxtraspc]
\eczhRefIndex{code:sc_ldpc}%
\eczhListValue{\flmRefsHyperref{code:sc_ldpc}{Spatially coupled LDPC (SC-LDPC) code}} & \eczhListValue{An LDPC code whose parity-check matrix is constructed by "spatially" coupling several copies of a regular LDPC parity-check matrix in chain-like fashion (or, more generally, in grid-like fashion) to yield a band matrix.
A finite-length chain is then capped by imposing either open boundary conditions (yielding \textit{non-tail-biting} SC-LDPC codes) or periodic boundary conditions (yielding \textit{tail-biting} SC-LDPC codes); sometimes extra \textit{terminating vertices} are added to the ends of the chain.
Matrices corresponding to translationally invariant chains are called \textit{time-invariant}, and otherwise are called \textit{time-varying}.
These codes can be constructed, e.g., using the \flmRefsHyperref{ref47}{lifting} procedure or using edge-cutting vectors \NoCaseChange{\protect\cite{cite95}}.
Spatial coupling can also be applied to MN-LDPC and HA-LDPC protographs, yielding bounded-density SC-MN and SC-HA families \NoCaseChange{\protect\cite{cite84}}.}\\ 
\addlinespace[\myxtraspc]
\eczhRefIndex{code:tsf}%
\eczhListValue{\flmRefsHyperref{code:tsf}{Tanner-Sridhara-Fuja (TSF) code}} & \eczhListValue{Array QC-LDPC code constructed from a cyclically shifted identity matrix; see \NoCaseChange{\protect\cite[{Exam. 21.6.5}]{cite97}}.}\\ 
\addlinespace[\myxtraspc]
\eczhRefIndex{code:tornado}%
\eczhListValue{\flmRefsHyperref{code:tornado}{Tornado code}} & \eczhListValue{Linear binary erasure code that is a precursor to fountain codes and is built from a multilayer cascade of sparse bipartite graphs. Its encoding and decoding operations involve only XOR gates \NoCaseChange{\protect\cite{cite98,cite99}\protect\cite[{Sec. 30.2}]{cite100}}.}\\ 
\addlinespace[\myxtraspc]
\eczhRefIndex{code:simplex734}%
\eczhListValue{\flmRefsHyperref{code:simplex734}{\([7,3,4]\) simplex code}} & \eczhListValue{Second-smallest nontrivial member of the simplex-code family.
The columns of its generator matrix are in one-to-one correspondence with the elements of the projective space \(PG(2,2)\), with each column being a chosen representative of the corresponding element.
The codewords form a \((8,9)\) simplex spherical code under the \flmRefsHyperref{ref38}{antipodal mapping}.
As a simplex code, it is equidistant: every nonzero codeword has Hamming weight \(4\).}\\ 
\addlinespace[\myxtraspc]
\eczhRefIndex{code:q-ary_ldpc}%
\eczhListValue{\flmRefsHyperref{code:q-ary_ldpc}{\(q\)-ary LDPC code}} & \eczhListValue{A \(q\)-ary linear code with a sparse parity-check matrix.
Alternatively, a member of an infinite family of \([n,k,d]_q\) codes for which the number of nonzero entries in each row and column of the parity-check matrix are both bounded above by a constant as \(n\to\infty\).}\\ 
\addlinespace[\myxtraspc]
\eczhRefIndex{code:q-ary_protograph_ldpc}%
\eczhListValue{\flmRefsHyperref{code:q-ary_protograph_ldpc}{\(q\)-ary protograph LDPC code}} & \eczhListValue{A \(q\)-ary LDPC code whose parity-check matrix is constructed using the \flmRefsHyperref{ref47}{lifting} procedure applied to the incidence matrix of a sparse graph called, in this context, a \textit{protograph}.
An ability to assign non-binary edge weight called \textit{edge scaling} can also be used in code construction.}\\ 
\end{tabularx}\endgroup
\eczcodelist{lcc}{Locally correctable codes and friends}%

\eczhCodeListAutoDescription{Union of:
\begin{itemize}\item codes that are descendants of \flmRefsCref{code:lcc}
\item codes that are cousins of \flmRefsCref{code:lcc}
\item codes that are cousins of \flmRefsCref{code:q-ary_lcc}
\end{itemize}}%

\eczhIncludeCodeGraph{Bare}{scale=0.5}{\columnwidth}{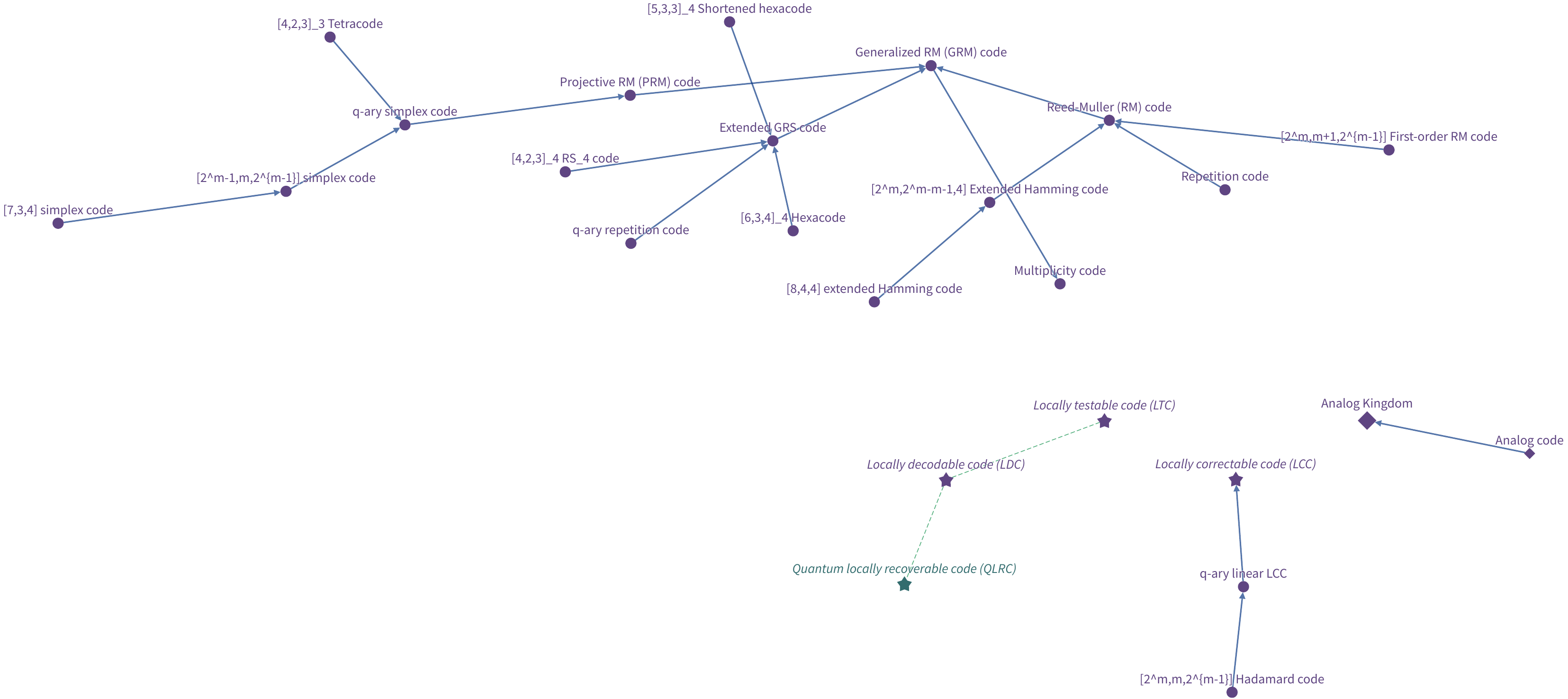}{Locally correctable codes and friends}{https://errorcorrectionzoo.org/code_graph#J\%7B\%22displayMode\%22\%3A\%22subset\%22\%2C\%22modeSubsetOptions\%22\%3A\%7B\%22codeIds\%22\%3A\%5B\%22analog\%22\%2C\%22extended_reed_solomon\%22\%2C\%22generalized_reed_muller\%22\%2C\%22lcc\%22\%2C\%22ldc\%22\%2C\%22ltc\%22\%2C\%22multiplicity\%22\%2C\%22projective_reed_muller\%22\%2C\%22quantum_locally_recoverable\%22\%2C\%22reed_muller\%22\%2C\%22repetition\%22\%2C\%22extended_hamming\%22\%2C\%22biorthogonal\%22\%2C\%22hadamard\%22\%2C\%22simplex\%22\%2C\%22tetracode\%22\%2C\%22reed_solomon_4\%22\%2C\%22shortened_hexacode\%22\%2C\%22hexacode\%22\%2C\%22simplex734\%22\%2C\%22hamming844\%22\%2C\%22q-ary_lcc\%22\%2C\%22q-ary_repetition\%22\%2C\%22q-ary_simplex\%22\%5D\%2C\%22reusePreviousLayoutPositions\%22\%3Afalse\%2C\%22showIntermediateConnectingNodes\%22\%3Atrue\%2C\%22connectingNodesMaxDepth\%22\%3A15\%2C\%22connectingNodesPathMaxLength\%22\%3A20\%2C\%22connectingNodesMaxExtraDepth\%22\%3A3\%2C\%22connectingNodesOnlyKeepPathsWithAdditionalLength\%22\%3A1\%2C\%22connectingNodesToDomainsAndKingdoms\%22\%3Afalse\%2C\%22connectingNodesEdgeLengthsByType\%22\%3A\%7B\%22primaryParent\%22\%3A1\%2C\%22secondaryParent\%22\%3A4\%2C\%22cousin\%22\%3A6\%7D\%2C\%22nodeIds\%22\%3A\%5B\%22k_analog\%22\%5D\%7D\%2C\%22highlightImportantNodes\%22\%3A\%7B\%22highlightImportantNodes\%22\%3Afalse\%2C\%22highlightPrimaryParents\%22\%3Afalse\%2C\%22highlightRootConnectingEdges\%22\%3Afalse\%7D\%7D}

\begingroup
\small
\eczhBreakableDashes
\renewcommand\arraystretch{1.05}
\edef\myxtraspc{\eczListAddVSpaceXtraXtraStretch}
\begin{tabularx}{\linewidth}{>{\raggedright\arraybackslash}p{\eczListColWidth{name}} >{\hsize=1.0000\hsize }X}
\toprule
\eczListColTitle{Code} & \eczListColTitle{Description} \\
\midrule
\endfirsthead
\toprule
\eczListColTitleContinued{Code} & \eczListColTitleContinued{Description} \\
\midrule
\endhead
\bottomrule
\endfoot
\eczhRefIndex{code:analog}%
\eczhListValue{\flmRefsHyperref{code:analog}{Analog code}} & \eczhListValue{Encodes states (codewords) into continuous coordinates in the \(n\)-dimensional (real or complex) coordinate space (\(\mathbb{R}^n\) or \(\mathbb{C}^n\)).
Important subclasses include sphere packings, tilings, and modulation constellations.
The number of codewords may be infinite because the coordinate space is infinite, so various restricted versions have to be constructed in practice.}\\ 
\addlinespace[\myxtraspc]
\eczhRefIndex{code:extended_reed_solomon}%
\eczhListValue{\flmRefsHyperref{code:extended_reed_solomon}{Extended GRS code}} & \eczhListValue{A GRS code extended by one extra coordinate to form an \([n+1,k,n-k+2]_q\) MDS code. In projective language, this corresponds to adding one more evaluation point, often interpreted as the point at infinity; in suitable equivalent descriptions, one may instead use an affine point such as \(0\). The case when \(n=q-1\), multipliers \(v_i=1\), and \(\alpha_i\) are \(i-1\)st powers of a primitive \(n\)th root of unity is an \textit{extended narrow-sense RS code}.}\\ 
\addlinespace[\myxtraspc]
\eczhRefIndex{code:generalized_reed_muller}%
\eczhListValue{\flmRefsHyperref{code:generalized_reed_muller}{Generalized RM (GRM) code}} & \eczhListValue{Extensions of RM codes to \(q\)-ary coordinates that can be described as multivariate polynomial evaluation codes over affine or projective space.}\\ 
\addlinespace[\myxtraspc]
\eczhRefIndex{code:lcc}%
\eczhListValue{\flmRefsHyperref{code:lcc}{Locally correctable code (LCC)}} & \eczhListValue{Recall that a block code encodes a length-\(k\) message into a length-\(n\) codeword, which is then sent through a noise channel to yield a received word.
Informally, an LCC is a block code for which one can recover any coordinate of a codeword from at most \(r\) coordinates of the received word (assuming the received word is within some tolerated corruption rate \(\delta\)).}\\ 
\addlinespace[\myxtraspc]
\eczhRefIndex{code:ldc}%
\eczhListValue{\flmRefsHyperref{code:ldc}{Locally decodable code (LDC)}} & \eczhListValue{Recall that a block code encodes a length-\(k\) message into a length-\(n\) codeword, which is then sent through a noise channel to yield a received word.
Informally, an LDC is a block code for which one can recover any coordinate of the message from at most \(r\) coordinates of the received word (assuming the received word is within some tolerated corruption rate \(\delta\)).
Efficiency of the decoding is quantified by the code's \textit{query complexity} \(r\), and decoding is performed by sampling subsets of \(r\) bits.}\\ 
\addlinespace[\myxtraspc]
\eczhRefIndex{code:ltc}%
\eczhListValue{\flmRefsHyperref{code:ltc}{Locally testable code (LTC)}} & \eczhListValue{Code for which one can efficiently check whether a given string is a codeword or is far from a codeword. Efficiency of the verification is quantified by the code's \textit{query complexity} \(u\), while effectiveness is quantified by the code's \textit{soundness} \(R\).}\\ 
\addlinespace[\myxtraspc]
\eczhRefIndex{code:multiplicity}%
\eczhListValue{\flmRefsHyperref{code:multiplicity}{Multiplicity code}} & \eczhListValue{A generalization of an \(m\)-variate polynomial evaluation code based on evaluating polynomials together with their Hasse derivatives up to order \(s-1\) at all points in \(\mathbb{F}_q^m\).
Originally proposed for coding using the Rosenbloom-Tsfasman metric \NoCaseChange{\protect\cite{cite179}}.
Univariate (\(m=1\)) \NoCaseChange{\protect\cite{cite179,cite180}} and multivariate (\(m>1\)) \NoCaseChange{\protect\cite{cite181}} codes have been proposed.}\\ 
\addlinespace[\myxtraspc]
\eczhRefIndex{code:projective_reed_muller}%
\eczhListValue{\flmRefsHyperref{code:projective_reed_muller}{Projective RM (PRM) code}} & \eczhListValue{Evaluation code obtained by evaluating homogeneous polynomials on the points of the projective space \(PG(m,q)\), equivalently on representatives of the nonzero vectors in \(\mathbb{F}_q^{m+1}\) whose leftmost nonzero coordinate is one.}\\ 
\addlinespace[\myxtraspc]
\eczhRefIndex{code:quantum_locally_recoverable}%
\eczhListValue{\flmRefsHyperref{code:quantum_locally_recoverable}{Quantum locally recoverable code (QLRC)}} & \eczhListValue{A QLRC of locality \(r\) is a block quantum code whose code states can be recovered after a single erasure by accessing at most \(r-1\) other subsystems and applying a recovery map.}\\ 
\addlinespace[\myxtraspc]
\eczhRefIndex{code:reed_muller}%
\eczhListValue{\flmRefsHyperref{code:reed_muller}{Reed-Muller (RM) code}} & \eczhListValue{Member of the RM\((r,m)\) family of linear binary codes derived from multivariate polynomials. The code parameters are \([2^m,\sum_{j=0}^{r} {m \choose j},2^{m-r}]\), where \(r\) is the \textit{order} of the code satisfying \(0\leq r\leq m\).
First-order RM codes are also called biorthogonal codes, while \(m\)th order RM codes are also called \textit{universe} codes.
\textit{Punctured RM codes} RM\(^*(r,m)\) are obtained from RM codes by deleting one coordinate from each codeword.}\\ 
\addlinespace[\myxtraspc]
\eczhRefIndex{code:repetition}%
\eczhListValue{\flmRefsHyperref{code:repetition}{Repetition code}} & \eczhListValue{\([n,1,n]\) binary linear code encoding one bit of information into an \(n\)-bit string.
Majority decoding requires \(n\) to be odd in order to avoid ties.
The idea is to increase the code distance by repeating the logical information several times. It is a \((n,1)\)-Hamming code.
Its automorphism group is \(S_n\).}\\ 
\addlinespace[\myxtraspc]
\eczhRefIndex{code:extended_hamming}%
\eczhListValue{\flmRefsHyperref{code:extended_hamming}{\([2^m,2^m-m-1,4]\) Extended Hamming code}} & \eczhListValue{Member of an infinite family of RM\((m-2,m)\) codes with parameters \([2^m,2^m-m-1, 4]\) for \(m \geq 2\) that are extensions of the Hamming codes by a parity-check bit.}\\ 
\addlinespace[\myxtraspc]
\eczhRefIndex{code:biorthogonal}%
\eczhListValue{\flmRefsHyperref{code:biorthogonal}{\([2^m,m+1,2^{m-1}]\) First-order RM code}} & \eczhListValue{A member of the family of first-order RM codes.
Its codewords are the rows of the \(2^m\)-dimensional Hadamard matrix \(H\) and its negation \(-H\) with the mapping \(+1\to 0\) and \(-1\to 1\).
The family is self-orthogonal for \(m \geq 3\) \NoCaseChange{\protect\cite{cite37}}.
They form a \((2^m,2^{m+1})\) biorthogonal spherical code under the \flmRefsHyperref{ref38}{antipodal mapping}.}\\ 
\addlinespace[\myxtraspc]
\eczhRefIndex{code:hadamard}%
\eczhListValue{\flmRefsHyperref{code:hadamard}{\([2^m,m,2^{m-1}]\) Hadamard code}} & \eczhListValue{An \([2^m,m,2^{m-1}]\) balanced binary code.
The \([2^m,m+1,2^{m-1}]\) augmented Hadamard code is the first-order RM code (a.k.a. RM\((1,m)\)), while the \([2^m-1,m,2^{m-1}]\) shortened Hadamard code is the simplex code (a.k.a. RM\(^*(1,m)\)).}\\ 
\addlinespace[\myxtraspc]
\eczhRefIndex{code:simplex}%
\eczhListValue{\flmRefsHyperref{code:simplex}{\([2^m-1,m,2^{m-1}]\) simplex code}} & \eczhListValue{A member of the equidistant code family dual to the \([2^m-1,2^m-m-1,3]\) Hamming family.}\\ 
\addlinespace[\myxtraspc]
\eczhRefIndex{code:tetracode}%
\eczhListValue{\flmRefsHyperref{code:tetracode}{\([4,2,3]_3\) Tetracode}} & \eczhListValue{The \([4,2,3]_3\) ternary self-dual MDS code that has connections to lattices \NoCaseChange{\protect\cite{cite39}}. Its weight enumerator is the Gleason polynomial \(g_4\) \NoCaseChange{\protect\cite[{Rem. 4.2.6}]{cite40}}.}\\ 
\addlinespace[\myxtraspc]
\eczhRefIndex{code:reed_solomon_4}%
\eczhListValue{\flmRefsHyperref{code:reed_solomon_4}{\([4,2,3]_4\) RS\(_4\) code}} & \eczhListValue{A Type II Euclidean self-dual extended RS code that is the smallest quaternary extended QR code \NoCaseChange{\protect\cite[{pg. 296}]{cite41}\protect\cite[{Sec. 2.4.2}]{cite42}}.
Puncturing the \([4,2,3]_4\) RS\(_4\) code yields the \([3,2,2]_4\) shortened RS\(_4\) code, which is an RS code \NoCaseChange{\protect\cite[{pg. 295}]{cite41}}.}\\ 
\addlinespace[\myxtraspc]
\eczhRefIndex{code:shortened_hexacode}%
\eczhListValue{\flmRefsHyperref{code:shortened_hexacode}{\([5,3,3]_4\) Shortened hexacode}} & \eczhListValue{A perfect \([5,3,3]_4\) quaternary Hamming code that is the result of puncturing the hexacode \NoCaseChange{\protect\cite{cite43}}.}\\ 
\addlinespace[\myxtraspc]
\eczhRefIndex{code:hexacode}%
\eczhListValue{\flmRefsHyperref{code:hexacode}{\([6,3,4]_4\) Hexacode}} & \eczhListValue{The \([6,3,4]_4\) Hermitian self-dual MDS code that has connections to projective geometry, lattices \NoCaseChange{\protect\cite{cite39}}, and conformal field theory \NoCaseChange{\protect\cite{cite44}}. Its weight enumerator is the Gleason polynomial \(g_7\) \NoCaseChange{\protect\cite[{Rem. 4.2.6}]{cite40}}.}\\ 
\addlinespace[\myxtraspc]
\eczhRefIndex{code:simplex734}%
\eczhListValue{\flmRefsHyperref{code:simplex734}{\([7,3,4]\) simplex code}} & \eczhListValue{Second-smallest nontrivial member of the simplex-code family.
The columns of its generator matrix are in one-to-one correspondence with the elements of the projective space \(PG(2,2)\), with each column being a chosen representative of the corresponding element.
The codewords form a \((8,9)\) simplex spherical code under the \flmRefsHyperref{ref38}{antipodal mapping}.
As a simplex code, it is equidistant: every nonzero codeword has Hamming weight \(4\).}\\ 
\addlinespace[\myxtraspc]
\eczhRefIndex{code:hamming844}%
\eczhListValue{\flmRefsHyperref{code:hamming844}{\([8,4,4]\) extended Hamming code}} & \eczhListValue{Extension of the \([7,4,3]\) Hamming code by a parity-check bit.
The smallest doubly even self-dual code, and the unique Type II code of length \(8\) \NoCaseChange{\protect\cite[{Rem. 4.3.10}]{cite40}}.}\\ 
\addlinespace[\myxtraspc]
\eczhRefIndex{code:q-ary_lcc}%
\eczhListValue{\flmRefsHyperref{code:q-ary_lcc}{\(q\)-ary linear LCC}} & \eczhListValue{A \(q\)-ary linear code for which one can recover any coordinate of a codeword from at most \(r\) coordinates of a received word (assuming the corruption rate is within some tolerated threshold \(\delta\)).}\\ 
\addlinespace[\myxtraspc]
\eczhRefIndex{code:q-ary_repetition}%
\eczhListValue{\flmRefsHyperref{code:q-ary_repetition}{\(q\)-ary repetition code}} & \eczhListValue{An \([n,1,n]_q\) code consisting of codewords \((j,j,\cdots,j)\) for \(j \in \mathbb{F}_q\).}\\ 
\addlinespace[\myxtraspc]
\eczhRefIndex{code:q-ary_simplex}%
\eczhListValue{\flmRefsHyperref{code:q-ary_simplex}{\(q\)-ary simplex code}} & \eczhListValue{An \([n,m,q^{m-1}]_q\) equidistant projective code with \(n=\frac{q^m-1}{q-1}\), denoted as \(S(q,m)\). The columns of the generator matrix are in one-to-one correspondence with the elements of the projective space \(PG(m-1,q)\), with each column being a chosen representative of the corresponding element.
All nonzero simplex codewords have a constant weight of \(q^{m-1}\) \NoCaseChange{\protect\cite{cite45,cite46}}.}\\ 
\end{tabularx}\endgroup
\eczcodelist{lrc}{Locally recoverable codes}%

\eczhCodeListAutoDescription{All descendants of \flmRefsCref{code:locally_recoverable}.}%

\eczhIncludeCodeGraph{Bare}{scale=0.5}{\columnwidth}{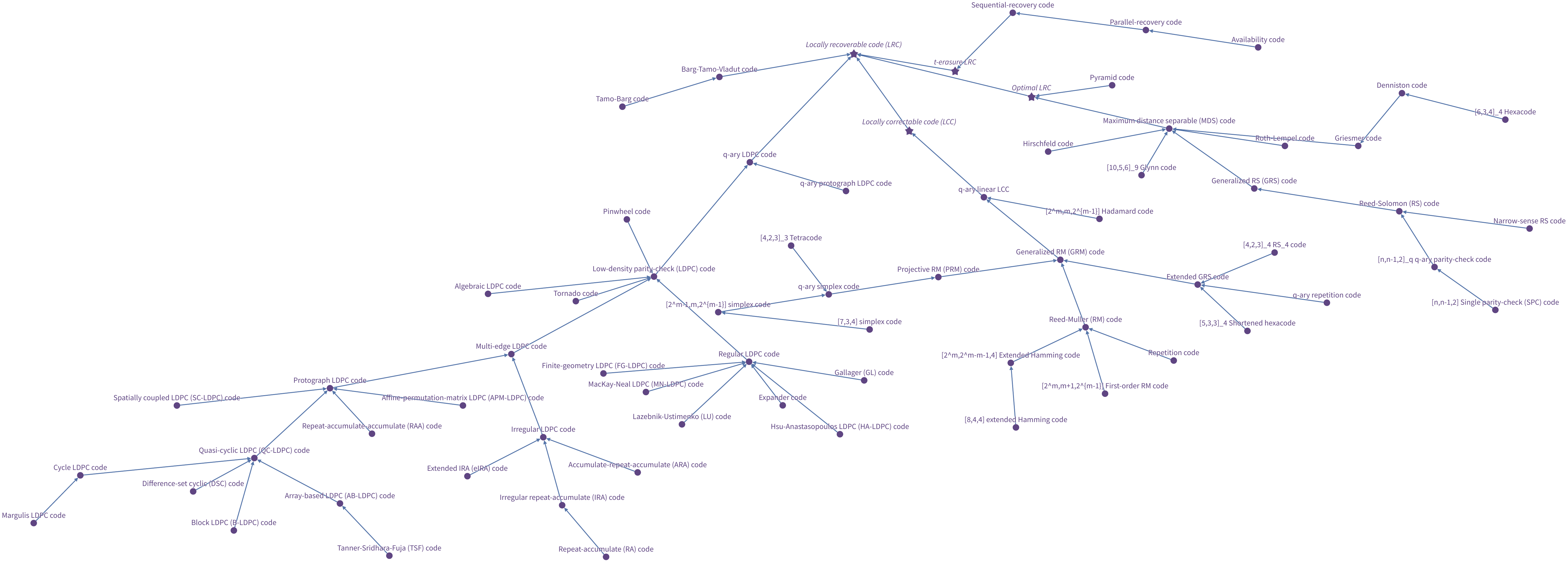}{Locally recoverable codes}{https://errorcorrectionzoo.org/code_graph#J\%7B\%22displayMode\%22\%3A\%22subset\%22\%2C\%22modeSubsetOptions\%22\%3A\%7B\%22codeIds\%22\%3A\%5B\%22ara\%22\%2C\%22apm_ldpc\%22\%2C\%22algebraic_ldpc\%22\%2C\%22array_ldpc\%22\%2C\%22codes_with_availability\%22\%2C\%22tamo_barg_vladut\%22\%2C\%22b_ldpc\%22\%2C\%22cycle_ldpc\%22\%2C\%22denniston\%22\%2C\%22difference_set\%22\%2C\%22expander\%22\%2C\%22extended_reed_solomon\%22\%2C\%22extended_ira\%22\%2C\%22pg_ldpc\%22\%2C\%22gallager\%22\%2C\%22generalized_reed_muller\%22\%2C\%22generalized_reed_solomon\%22\%2C\%22griesmer\%22\%2C\%22hirschfeld\%22\%2C\%22ha_ldpc\%22\%2C\%22irregular_ldpc\%22\%2C\%22ira\%22\%2C\%22lu_ldpc\%22\%2C\%22lcc\%22\%2C\%22locally_recoverable\%22\%2C\%22ldpc\%22\%2C\%22mn_ldpc\%22\%2C\%22margulis_ldpc\%22\%2C\%22mds\%22\%2C\%22multi_edge_ldpc\%22\%2C\%22narrow_sense_reed_solomon\%22\%2C\%22optimal_lrc\%22\%2C\%22parallel_recovery\%22\%2C\%22pinwheel\%22\%2C\%22projective_reed_muller\%22\%2C\%22protograph_ldpc\%22\%2C\%22pyramid\%22\%2C\%22qc_ldpc\%22\%2C\%22reed_muller\%22\%2C\%22reed_solomon\%22\%2C\%22regular_ldpc\%22\%2C\%22ra\%22\%2C\%22raa\%22\%2C\%22repetition\%22\%2C\%22roth_lempel\%22\%2C\%22sequential_recovery\%22\%2C\%22sc_ldpc\%22\%2C\%22tamo_barg\%22\%2C\%22tsf\%22\%2C\%22tornado\%22\%2C\%22glynn\%22\%2C\%22extended_hamming\%22\%2C\%22biorthogonal\%22\%2C\%22hadamard\%22\%2C\%22simplex\%22\%2C\%22tetracode\%22\%2C\%22reed_solomon_4\%22\%2C\%22shortened_hexacode\%22\%2C\%22hexacode\%22\%2C\%22simplex734\%22\%2C\%22hamming844\%22\%2C\%22parity_check\%22\%2C\%22q-ary_parity_check\%22\%2C\%22q-ary_ldpc\%22\%2C\%22q-ary_lcc\%22\%2C\%22q-ary_protograph_ldpc\%22\%2C\%22q-ary_repetition\%22\%2C\%22q-ary_simplex\%22\%2C\%22multiple_erasure_lrc\%22\%5D\%2C\%22reusePreviousLayoutPositions\%22\%3Afalse\%2C\%22showIntermediateConnectingNodes\%22\%3Atrue\%2C\%22connectingNodesMaxDepth\%22\%3A15\%2C\%22connectingNodesPathMaxLength\%22\%3A20\%2C\%22connectingNodesMaxExtraDepth\%22\%3A3\%2C\%22connectingNodesOnlyKeepPathsWithAdditionalLength\%22\%3A1\%2C\%22connectingNodesToDomainsAndKingdoms\%22\%3Afalse\%2C\%22connectingNodesEdgeLengthsByType\%22\%3A\%7B\%22primaryParent\%22\%3A1\%2C\%22secondaryParent\%22\%3A4\%2C\%22cousin\%22\%3A6\%7D\%2C\%22nodeIds\%22\%3A\%5B\%5D\%7D\%2C\%22highlightImportantNodes\%22\%3A\%7B\%22highlightImportantNodes\%22\%3Afalse\%2C\%22highlightPrimaryParents\%22\%3Afalse\%2C\%22highlightRootConnectingEdges\%22\%3Afalse\%7D\%7D}

\begingroup
\small
\eczhBreakableDashes
\renewcommand\arraystretch{1.05}
\edef\myxtraspc{\eczListAddVSpaceXtraXtraStretch}
\begin{tabularx}{\linewidth}{>{\raggedright\arraybackslash}p{\eczListColWidth{name}} >{\hsize=1.0000\hsize }X}
\toprule
\eczListColTitle{Code} & \eczListColTitle{Description} \\
\midrule
\endfirsthead
\toprule
\eczListColTitleContinued{Code} & \eczListColTitleContinued{Description} \\
\midrule
\endhead
\bottomrule
\endfoot
\eczhRefIndex{code:ara}%
\eczhListValue{\flmRefsHyperref{code:ara}{Accumulate-repeat-accumulate (ARA) code}} & \eczhListValue{A generalization of the RA code in which the outer repetition-code encoding step is augmented with an accumulator acting on a fraction of the incoming bits.
In addition, the code may be punctured after the final accumulating step.}\\ 
\addlinespace[\myxtraspc]
\eczhRefIndex{code:apm_ldpc}%
\eczhListValue{\flmRefsHyperref{code:apm_ldpc}{Affine-permutation-matrix LDPC (APM-LDPC) code}} & \eczhListValue{LDPC code whose parity-check matrix can be put into the form of a block matrix consisting of permutation submatrices representing the affine permutation group or the zero submatrix.
Given a cyclic group \(\mathbb{Z}_r\), the affine permutation group is \(\mathbb{Z}_r \rtimes \mathbb{Z}_r^{\times}\), where \(\mathbb{Z}_r^{\times}\) is the multiplicative group of integers modulo \(r\).
Such codes are often constructed by \flmRefsHyperref{ref47}{lifting} certain protographs into such block matrices \NoCaseChange{\protect\cite{cite48}}.}\\ 
\addlinespace[\myxtraspc]
\eczhRefIndex{code:algebraic_ldpc}%
\eczhListValue{\flmRefsHyperref{code:algebraic_ldpc}{Algebraic LDPC code}} & \eczhListValue{LDPC code whose parity check matrix is constructed explicitly (i.e., non-randomly) from a particular graph \NoCaseChange{\protect\cite{cite49,cite50}} or an algebraic structure such as a combinatorial design \NoCaseChange{\protect\cite{cite51,cite52,cite53}}, balanced incomplete block design \NoCaseChange{\protect\cite{cite54}}, a partial geometry \NoCaseChange{\protect\cite{cite55}}, a generalized polygon \NoCaseChange{\protect\cite{cite56,cite57}}, or a Latin square \NoCaseChange{\protect\cite{cite58,cite59,cite60}}.
The extra structure and/or symmetry \NoCaseChange{\protect\cite{cite61}} of these codes can often be used to gain a better understanding of their properties.}\\ 
\addlinespace[\myxtraspc]
\eczhRefIndex{code:array_ldpc}%
\eczhListValue{\flmRefsHyperref{code:array_ldpc}{Array-based LDPC (AB-LDPC) code}} & \eczhListValue{QC-LDPC code constructed deterministically from a disk array code known as a B-code.
Its parity-check matrix admits a compact representation \NoCaseChange{\protect\cite{cite64}} and is related to RS codes.}\\ 
\addlinespace[\myxtraspc]
\eczhRefIndex{code:codes_with_availability}%
\eczhListValue{\flmRefsHyperref{code:codes_with_availability}{Availability code}} & \eczhListValue{A \(t\)-availability parallel-recovery code is a code such that coordinates can be recovered in multiple ways.
The availability of a locally recoverable code is the minimum, over all coordinates, of the number of recovery sets for that coordinate \NoCaseChange{\protect\cite[{Def. 15.9.20}]{cite26}}.
That way, the code accommodates nodes that may be inaccessible during the recovery procedure.}\\ 
\addlinespace[\myxtraspc]
\eczhRefIndex{code:tamo_barg_vladut}%
\eczhListValue{\flmRefsHyperref{code:tamo_barg_vladut}{Barg-Tamo-Vladut code}} & \eczhListValue{Evaluation AG code on algebraic curves built from a Galois cover \(\phi:Y\to X\), where the recovery sets are fibres over rational points of \(X\) that split completely in the cover \NoCaseChange{\protect\cite[{Def. 15.9.10}]{cite26}\protect\cite[{Thm. 15.9.14}]{cite26}}.
The Barg-Tamo-Vladut construction generalizes the Tamo-Barg construction from \(PG(1,q)\) to longer AG codes, and variants can be built with higher local distance or availability \(2\) via fibre products of curves \NoCaseChange{\protect\cite[{Thm. 15.9.19}]{cite26}\protect\cite[{Thm. 15.9.21}]{cite26}}.}\\ 
\addlinespace[\myxtraspc]
\eczhRefIndex{code:b_ldpc}%
\eczhListValue{\flmRefsHyperref{code:b_ldpc}{Block LDPC (B-LDPC) code}} & \eczhListValue{Member of a particular class of irregular QC-LDPC codes with efficient encoders.}\\ 
\addlinespace[\myxtraspc]
\eczhRefIndex{code:cycle_ldpc}%
\eczhListValue{\flmRefsHyperref{code:cycle_ldpc}{Cycle LDPC code}} & \eczhListValue{An LDPC code whose parity-check matrix forms the incidence matrix of a graph, i.e., has weight-two columns.}\\ 
\addlinespace[\myxtraspc]
\eczhRefIndex{code:denniston}%
\eczhListValue{\flmRefsHyperref{code:denniston}{Denniston code}} & \eczhListValue{Projective code that is part of a family of \([2^{a+i}+2^i-2^a,3,2^{a+i}-2^a]_{2^a}\) codes for \(i < a\) constructed using Denniston arcs \NoCaseChange{\protect\cite[{Sec. 19.7.3}]{cite172}}.}\\ 
\addlinespace[\myxtraspc]
\eczhRefIndex{code:difference_set}%
\eczhListValue{\flmRefsHyperref{code:difference_set}{Difference-set cyclic (DSC) code}} & \eczhListValue{Cyclic LDPC code constructed deterministically from a difference set.
Certain DSC codes satisfy more redundant constraints than Gallager codes and thus can outperform them \NoCaseChange{\protect\cite{cite72}}.}\\ 
\addlinespace[\myxtraspc]
\eczhRefIndex{code:expander}%
\eczhListValue{\flmRefsHyperref{code:expander}{Expander code}} & \eczhListValue{LDPC code whose parity-check matrix is derived from the adjacency matrix of a bipartite expander graph \NoCaseChange{\protect\cite{cite74}} such as a Ramanujan graph or a Cayley graph of a projective special linear group over a finite field \NoCaseChange{\protect\cite{cite75,cite76}}.
Expander codes admit efficient encoding and decoding algorithms and yield an explicit (i.e., non-random) asymptotically good LDPC code family.}\\ 
\addlinespace[\myxtraspc]
\eczhRefIndex{code:extended_reed_solomon}%
\eczhListValue{\flmRefsHyperref{code:extended_reed_solomon}{Extended GRS code}} & \eczhListValue{A GRS code extended by one extra coordinate to form an \([n+1,k,n-k+2]_q\) MDS code. In projective language, this corresponds to adding one more evaluation point, often interpreted as the point at infinity; in suitable equivalent descriptions, one may instead use an affine point such as \(0\). The case when \(n=q-1\), multipliers \(v_i=1\), and \(\alpha_i\) are \(i-1\)st powers of a primitive \(n\)th root of unity is an \textit{extended narrow-sense RS code}.}\\ 
\addlinespace[\myxtraspc]
\eczhRefIndex{code:extended_ira}%
\eczhListValue{\flmRefsHyperref{code:extended_ira}{Extended IRA (eIRA) code}} & \eczhListValue{A generalization of the IRA code in which the outer LDGM code is replaced by a random sparse matrix containing no weight-two columns.}\\ 
\addlinespace[\myxtraspc]
\eczhRefIndex{code:pg_ldpc}%
\eczhListValue{\flmRefsHyperref{code:pg_ldpc}{Finite-geometry LDPC (FG-LDPC) code}} & \eczhListValue{LDPC code whose parity-check matrix is the incidence matrix of points and hyperplanes in either a Euclidean or a projective geometry.
Such codes are called \textit{Euclidean-geometry LDPC (EG-LDPC)} and \textit{projective-geometry LDPC (PG-LDPC)}, respectively.
Such constructions have been generalized to incidence matrices of hyperplanes of different dimensions \NoCaseChange{\protect\cite{cite77}}.}\\ 
\addlinespace[\myxtraspc]
\eczhRefIndex{code:gallager}%
\eczhListValue{\flmRefsHyperref{code:gallager}{Gallager (GL) code}} & \eczhListValue{The first LDPC code.
The rows of the parity-check matrix of this regular code are divided into equal subsets, and columns in the first subset are randomly permuted to yield the corresponding rows in subsequent subsets.}\\ 
\addlinespace[\myxtraspc]
\eczhRefIndex{code:generalized_reed_muller}%
\eczhListValue{\flmRefsHyperref{code:generalized_reed_muller}{Generalized RM (GRM) code}} & \eczhListValue{Extensions of RM codes to \(q\)-ary coordinates that can be described as multivariate polynomial evaluation codes over affine or projective space.}\\ 
\addlinespace[\myxtraspc]
\eczhRefIndex{code:generalized_reed_solomon}%
\eczhListValue{\flmRefsHyperref{code:generalized_reed_solomon}{Generalized RS (GRS) code}} & \eczhListValue{An \([n,k,n-k+1]_q\) MDS code that is a modification of the RS code where codeword polynomials are multiplied by additional prefactors \NoCaseChange{\protect\cite[{Def. 15.3.19}]{cite26}}.}\\ 
\addlinespace[\myxtraspc]
\eczhRefIndex{code:griesmer}%
\eczhListValue{\flmRefsHyperref{code:griesmer}{Griesmer code}} & \eczhListValue{A type of \(q\)-ary code whose parameters satisfy the Griesmer bound with equality.}\\ 
\addlinespace[\myxtraspc]
\eczhRefIndex{code:hirschfeld}%
\eczhListValue{\flmRefsHyperref{code:hirschfeld}{Hirschfeld code}} & \eczhListValue{A \([q+1,4,q-2]_q\) projective geometry code for non-prime \(q\) that is an example of an MDS code that is not an RS code; see \NoCaseChange{\protect\cite[{Exam. 7.6}]{cite182}} for the generator matrix.}\\ 
\addlinespace[\myxtraspc]
\eczhRefIndex{code:ha_ldpc}%
\eczhListValue{\flmRefsHyperref{code:ha_ldpc}{Hsu-Anastasopoulos LDPC (HA-LDPC) code}} & \eczhListValue{A regular LDPC code obtained from a concatenation of a certain random regular LDPC code and a certain random LDGM code.
An \((l,r,g)\)-HA-LDPC code can be written using punctured LDPC and LDGM parts, and it is dual to the corresponding \((r,l,g)\)-MN-LDPC code \NoCaseChange{\protect\cite{cite84}}.
Using rate-one LDGM codes eliminates high-weight codewords while admitting an amount of low-weight codewords that asymptotically vanishes, allowing code families to achieve the \flmRefsHyperref{ref85}{GV bound} with high probability.}\\ 
\addlinespace[\myxtraspc]
\eczhRefIndex{code:irregular_ldpc}%
\eczhListValue{\flmRefsHyperref{code:irregular_ldpc}{Irregular LDPC code}} & \eczhListValue{An LDPC code whose parity-check matrix has a variable number of entries in each row or column.}\\ 
\addlinespace[\myxtraspc]
\eczhRefIndex{code:ira}%
\eczhListValue{\flmRefsHyperref{code:ira}{Irregular repeat-accumulate (IRA) code}} & \eczhListValue{A generalization of the RA code in which the outer 1-in-3 repetition encoding step is replaced by an LDGM code.
A simple version is when different bits in the RA block are repeated a different number of times.}\\ 
\addlinespace[\myxtraspc]
\eczhRefIndex{code:lu_ldpc}%
\eczhListValue{\flmRefsHyperref{code:lu_ldpc}{Lazebnik-Ustimenko (LU) code}} & \eczhListValue{LDPC code whose Tanner graph comes from a particular family of \(q\)-regular graphs \NoCaseChange{\protect\cite{cite87}} of known girth and relatively large stopping sets.}\\ 
\addlinespace[\myxtraspc]
\eczhRefIndex{code:lcc}%
\eczhListValue{\flmRefsHyperref{code:lcc}{Locally correctable code (LCC)}} & \eczhListValue{Recall that a block code encodes a length-\(k\) message into a length-\(n\) codeword, which is then sent through a noise channel to yield a received word.
Informally, an LCC is a block code for which one can recover any coordinate of a codeword from at most \(r\) coordinates of the received word (assuming the received word is within some tolerated corruption rate \(\delta\)).}\\ 
\addlinespace[\myxtraspc]
\eczhRefIndex{code:locally_recoverable}%
\eczhListValue{\flmRefsHyperref{code:locally_recoverable}{Locally recoverable code (LRC)}} & \eczhListValue{A block code for which one can recover any coordinate of a codeword from at most \(r\) other coordinates of the codeword \NoCaseChange{\protect\cite[{Def. 15.9.3}]{cite26}}.}\\ 
\addlinespace[\myxtraspc]
\eczhRefIndex{code:ldpc}%
\eczhListValue{\flmRefsHyperref{code:ldpc}{Low-density parity-check (LDPC) code}} & \eczhListValue{A binary linear code with a sparse parity-check matrix.
Often a member of an infinite family of \([n,k,d]\) codes for which the numbers of nonzero entries in each row and in each column of the parity-check matrix are both bounded above by a constant as \(n\to\infty\).}\\ 
\addlinespace[\myxtraspc]
\eczhRefIndex{code:mn_ldpc}%
\eczhListValue{\flmRefsHyperref{code:mn_ldpc}{MacKay-Neal LDPC (MN-LDPC) code}} & \eczhListValue{A code whose parity-check matrix is constructed non-deterministically via the MacKay-Neal prescription.}\\ 
\addlinespace[\myxtraspc]
\eczhRefIndex{code:margulis_ldpc}%
\eczhListValue{\flmRefsHyperref{code:margulis_ldpc}{Margulis LDPC code}} & \eczhListValue{Member of a class of LDPC codes deterministically constructed from explicit sparse regular expander graphs.
The underlying Margulis-Gabber-Galil graph family provides explicit expanders \NoCaseChange{\protect\cite{cite49,cite90}}, yielding deterministic sparse parity-check matrices.
Related explicit LDPC constructions \NoCaseChange{\protect\cite{cite91}} utilize Ramanujan graphs \NoCaseChange{\protect\cite{cite75,cite76}}.}\\ 
\addlinespace[\myxtraspc]
\eczhRefIndex{code:mds}%
\eczhListValue{\flmRefsHyperref{code:mds}{Maximum distance separable (MDS) code}} & \eczhListValue{A \(q\)-ary linear code whose parameters satisfy the Singleton bound with equality.}\\ 
\addlinespace[\myxtraspc]
\eczhRefIndex{code:multi_edge_ldpc}%
\eczhListValue{\flmRefsHyperref{code:multi_edge_ldpc}{Multi-edge LDPC code}} & \eczhListValue{Irregular LDPC code whose construction generalizes those of the original examples of irregular LDPC as well as RA codes.}\\ 
\addlinespace[\myxtraspc]
\eczhRefIndex{code:narrow_sense_reed_solomon}%
\eczhListValue{\flmRefsHyperref{code:narrow_sense_reed_solomon}{Narrow-sense RS code}} & \eczhListValue{An \([q-1,k,n-k+1]_q\) RS code whose points \(\alpha_i\) are all \((i-1)\)st powers of a \flmRefsHyperref{ref33}{primitive} element \(\alpha\) of \(\mathbb{F}_q\).}\\ 
\addlinespace[\myxtraspc]
\eczhRefIndex{code:optimal_lrc}%
\eczhListValue{\flmRefsHyperref{code:optimal_lrc}{Optimal LRC}} & \eczhListValue{An LRC whose parameters saturate a generalized Singleton bound.}\\ 
\addlinespace[\myxtraspc]
\eczhRefIndex{code:parallel_recovery}%
\eczhListValue{\flmRefsHyperref{code:parallel_recovery}{Parallel-recovery code}} & \eczhListValue{A \(t\)-erasure LRC whose coordinate erasures can be recovered in parallel.}\\ 
\addlinespace[\myxtraspc]
\eczhRefIndex{code:pinwheel}%
\eczhListValue{\flmRefsHyperref{code:pinwheel}{Pinwheel code}} & \eczhListValue{A geometrically local binary LDPC code defined on planar graphs obtained from the pinwheel tiling \NoCaseChange{\protect\cite{cite93}}.
Both bits and checks live on vertices of the graph.
If \(L_N\) is the graph Laplacian at generation \(N\), the undepleted check matrix is \(\tilde H_N=(L_N-\mathbb{I})\bmod 2\), and the actual parity-check matrix \(H_N\) is obtained by removing an evenly spaced fraction of boundary checks.}\\ 
\addlinespace[\myxtraspc]
\eczhRefIndex{code:projective_reed_muller}%
\eczhListValue{\flmRefsHyperref{code:projective_reed_muller}{Projective RM (PRM) code}} & \eczhListValue{Evaluation code obtained by evaluating homogeneous polynomials on the points of the projective space \(PG(m,q)\), equivalently on representatives of the nonzero vectors in \(\mathbb{F}_q^{m+1}\) whose leftmost nonzero coordinate is one.}\\ 
\addlinespace[\myxtraspc]
\eczhRefIndex{code:protograph_ldpc}%
\eczhListValue{\flmRefsHyperref{code:protograph_ldpc}{Protograph LDPC code}} & \eczhListValue{Binary version of a \(q\)-ary protograph LDPC code.
Its parity-check matrix can be written as a block matrix whose blocks are either sums of permutation matrices or the zero matrix.}\\ 
\addlinespace[\myxtraspc]
\eczhRefIndex{code:pyramid}%
\eczhListValue{\flmRefsHyperref{code:pyramid}{Pyramid code}} & \eczhListValue{An LRC whose generator matrix is that of an RS code in standard form, but some of whose columns are split into multiple columns; see \NoCaseChange{\protect\cite[{Sec. 31.3.1.1}]{cite183}} for an example.}\\ 
\addlinespace[\myxtraspc]
\eczhRefIndex{code:qc_ldpc}%
\eczhListValue{\flmRefsHyperref{code:qc_ldpc}{Quasi-cyclic LDPC (QC-LDPC) code}} & \eczhListValue{LDPC code that can be put into quasi-cyclic form.
Its parity check matrix can be put into the form of a block matrix consisting of either circulant permutation sub-matrices or the zero sub-matrix.
Such codes are often constructed by \flmRefsHyperref{ref47}{lifting} certain protographs into such block matrices \NoCaseChange{\protect\cite{cite48}}.
Their simple structure makes them useful for several wireless communication standards.}\\ 
\addlinespace[\myxtraspc]
\eczhRefIndex{code:reed_muller}%
\eczhListValue{\flmRefsHyperref{code:reed_muller}{Reed-Muller (RM) code}} & \eczhListValue{Member of the RM\((r,m)\) family of linear binary codes derived from multivariate polynomials. The code parameters are \([2^m,\sum_{j=0}^{r} {m \choose j},2^{m-r}]\), where \(r\) is the \textit{order} of the code satisfying \(0\leq r\leq m\).
First-order RM codes are also called biorthogonal codes, while \(m\)th order RM codes are also called \textit{universe} codes.
\textit{Punctured RM codes} RM\(^*(r,m)\) are obtained from RM codes by deleting one coordinate from each codeword.}\\ 
\addlinespace[\myxtraspc]
\eczhRefIndex{code:reed_solomon}%
\eczhListValue{\flmRefsHyperref{code:reed_solomon}{Reed-Solomon (RS) code}} & \eczhListValue{An \([n,k,n-k+1]_q\) linear code based on polynomials over \(\mathbb{F}_q\).}\\ 
\addlinespace[\myxtraspc]
\eczhRefIndex{code:regular_ldpc}%
\eczhListValue{\flmRefsHyperref{code:regular_ldpc}{Regular LDPC code}} & \eczhListValue{An LDPC code whose parity-check matrix has a fixed number of ones in each row and each column.
Such a code is called \((j,k)\)-regular if each column has weight \(j\) (variable-node degree) and each row has weight \(k\) (check-node degree).
If the parity-check matrix has \(n\) columns and \(m\) rows, then regularity implies \(nj = mk\).}\\ 
\addlinespace[\myxtraspc]
\eczhRefIndex{code:ra}%
\eczhListValue{\flmRefsHyperref{code:ra}{Repeat-accumulate (RA) code}} & \eczhListValue{An LDPC code whose parity-check matrix has weight-two columns arranged in a step-like pattern for its last columns \NoCaseChange{\protect\cite{cite94}}.}\\ 
\addlinespace[\myxtraspc]
\eczhRefIndex{code:raa}%
\eczhListValue{\flmRefsHyperref{code:raa}{Repeat-accumulate-accumulate (RAA) code}} & \eczhListValue{Generalization of the RA code in which two accumulators and permutations are used.}\\ 
\addlinespace[\myxtraspc]
\eczhRefIndex{code:repetition}%
\eczhListValue{\flmRefsHyperref{code:repetition}{Repetition code}} & \eczhListValue{\([n,1,n]\) binary linear code encoding one bit of information into an \(n\)-bit string.
Majority decoding requires \(n\) to be odd in order to avoid ties.
The idea is to increase the code distance by repeating the logical information several times. It is a \((n,1)\)-Hamming code.
Its automorphism group is \(S_n\).}\\ 
\addlinespace[\myxtraspc]
\eczhRefIndex{code:roth_lempel}%
\eczhListValue{\flmRefsHyperref{code:roth_lempel}{Roth-Lempel code}} & \eczhListValue{Member of a \(q\)-ary linear code family that includes many examples of MDS codes that are not GRS codes.}\\ 
\addlinespace[\myxtraspc]
\eczhRefIndex{code:sequential_recovery}%
\eczhListValue{\flmRefsHyperref{code:sequential_recovery}{Sequential-recovery code}} & \eczhListValue{A \(t\)-erasure LRC whose erased coordinates are recovered sequentially.}\\ 
\addlinespace[\myxtraspc]
\eczhRefIndex{code:sc_ldpc}%
\eczhListValue{\flmRefsHyperref{code:sc_ldpc}{Spatially coupled LDPC (SC-LDPC) code}} & \eczhListValue{An LDPC code whose parity-check matrix is constructed by "spatially" coupling several copies of a regular LDPC parity-check matrix in chain-like fashion (or, more generally, in grid-like fashion) to yield a band matrix.
A finite-length chain is then capped by imposing either open boundary conditions (yielding \textit{non-tail-biting} SC-LDPC codes) or periodic boundary conditions (yielding \textit{tail-biting} SC-LDPC codes); sometimes extra \textit{terminating vertices} are added to the ends of the chain.
Matrices corresponding to translationally invariant chains are called \textit{time-invariant}, and otherwise are called \textit{time-varying}.
These codes can be constructed, e.g., using the \flmRefsHyperref{ref47}{lifting} procedure or using edge-cutting vectors \NoCaseChange{\protect\cite{cite95}}.
Spatial coupling can also be applied to MN-LDPC and HA-LDPC protographs, yielding bounded-density SC-MN and SC-HA families \NoCaseChange{\protect\cite{cite84}}.}\\ 
\addlinespace[\myxtraspc]
\eczhRefIndex{code:tamo_barg}%
\eczhListValue{\flmRefsHyperref{code:tamo_barg}{Tamo-Barg code}} & \eczhListValue{A family of \(q\)-ary polynomial evaluation codes that are optimal LRCs and for which \(q\) is comparable to \(n\).}\\ 
\addlinespace[\myxtraspc]
\eczhRefIndex{code:tsf}%
\eczhListValue{\flmRefsHyperref{code:tsf}{Tanner-Sridhara-Fuja (TSF) code}} & \eczhListValue{Array QC-LDPC code constructed from a cyclically shifted identity matrix; see \NoCaseChange{\protect\cite[{Exam. 21.6.5}]{cite97}}.}\\ 
\addlinespace[\myxtraspc]
\eczhRefIndex{code:tornado}%
\eczhListValue{\flmRefsHyperref{code:tornado}{Tornado code}} & \eczhListValue{Linear binary erasure code that is a precursor to fountain codes and is built from a multilayer cascade of sparse bipartite graphs. Its encoding and decoding operations involve only XOR gates \NoCaseChange{\protect\cite{cite98,cite99}\protect\cite[{Sec. 30.2}]{cite100}}.}\\ 
\addlinespace[\myxtraspc]
\eczhRefIndex{code:glynn}%
\eczhListValue{\flmRefsHyperref{code:glynn}{\([10,5,6]_9\) Glynn code}} & \eczhListValue{The unique trace-Hermitian self-dual \([10,5,6]_9\) code, constructed using a 10-arc in \(PG(4,9)\) that is not a rational curve.}\\ 
\addlinespace[\myxtraspc]
\eczhRefIndex{code:extended_hamming}%
\eczhListValue{\flmRefsHyperref{code:extended_hamming}{\([2^m,2^m-m-1,4]\) Extended Hamming code}} & \eczhListValue{Member of an infinite family of RM\((m-2,m)\) codes with parameters \([2^m,2^m-m-1, 4]\) for \(m \geq 2\) that are extensions of the Hamming codes by a parity-check bit.}\\ 
\addlinespace[\myxtraspc]
\eczhRefIndex{code:biorthogonal}%
\eczhListValue{\flmRefsHyperref{code:biorthogonal}{\([2^m,m+1,2^{m-1}]\) First-order RM code}} & \eczhListValue{A member of the family of first-order RM codes.
Its codewords are the rows of the \(2^m\)-dimensional Hadamard matrix \(H\) and its negation \(-H\) with the mapping \(+1\to 0\) and \(-1\to 1\).
The family is self-orthogonal for \(m \geq 3\) \NoCaseChange{\protect\cite{cite37}}.
They form a \((2^m,2^{m+1})\) biorthogonal spherical code under the \flmRefsHyperref{ref38}{antipodal mapping}.}\\ 
\addlinespace[\myxtraspc]
\eczhRefIndex{code:hadamard}%
\eczhListValue{\flmRefsHyperref{code:hadamard}{\([2^m,m,2^{m-1}]\) Hadamard code}} & \eczhListValue{An \([2^m,m,2^{m-1}]\) balanced binary code.
The \([2^m,m+1,2^{m-1}]\) augmented Hadamard code is the first-order RM code (a.k.a. RM\((1,m)\)), while the \([2^m-1,m,2^{m-1}]\) shortened Hadamard code is the simplex code (a.k.a. RM\(^*(1,m)\)).}\\ 
\addlinespace[\myxtraspc]
\eczhRefIndex{code:simplex}%
\eczhListValue{\flmRefsHyperref{code:simplex}{\([2^m-1,m,2^{m-1}]\) simplex code}} & \eczhListValue{A member of the equidistant code family dual to the \([2^m-1,2^m-m-1,3]\) Hamming family.}\\ 
\addlinespace[\myxtraspc]
\eczhRefIndex{code:tetracode}%
\eczhListValue{\flmRefsHyperref{code:tetracode}{\([4,2,3]_3\) Tetracode}} & \eczhListValue{The \([4,2,3]_3\) ternary self-dual MDS code that has connections to lattices \NoCaseChange{\protect\cite{cite39}}. Its weight enumerator is the Gleason polynomial \(g_4\) \NoCaseChange{\protect\cite[{Rem. 4.2.6}]{cite40}}.}\\ 
\addlinespace[\myxtraspc]
\eczhRefIndex{code:reed_solomon_4}%
\eczhListValue{\flmRefsHyperref{code:reed_solomon_4}{\([4,2,3]_4\) RS\(_4\) code}} & \eczhListValue{A Type II Euclidean self-dual extended RS code that is the smallest quaternary extended QR code \NoCaseChange{\protect\cite[{pg. 296}]{cite41}\protect\cite[{Sec. 2.4.2}]{cite42}}.
Puncturing the \([4,2,3]_4\) RS\(_4\) code yields the \([3,2,2]_4\) shortened RS\(_4\) code, which is an RS code \NoCaseChange{\protect\cite[{pg. 295}]{cite41}}.}\\ 
\addlinespace[\myxtraspc]
\eczhRefIndex{code:shortened_hexacode}%
\eczhListValue{\flmRefsHyperref{code:shortened_hexacode}{\([5,3,3]_4\) Shortened hexacode}} & \eczhListValue{A perfect \([5,3,3]_4\) quaternary Hamming code that is the result of puncturing the hexacode \NoCaseChange{\protect\cite{cite43}}.}\\ 
\addlinespace[\myxtraspc]
\eczhRefIndex{code:hexacode}%
\eczhListValue{\flmRefsHyperref{code:hexacode}{\([6,3,4]_4\) Hexacode}} & \eczhListValue{The \([6,3,4]_4\) Hermitian self-dual MDS code that has connections to projective geometry, lattices \NoCaseChange{\protect\cite{cite39}}, and conformal field theory \NoCaseChange{\protect\cite{cite44}}. Its weight enumerator is the Gleason polynomial \(g_7\) \NoCaseChange{\protect\cite[{Rem. 4.2.6}]{cite40}}.}\\ 
\addlinespace[\myxtraspc]
\eczhRefIndex{code:simplex734}%
\eczhListValue{\flmRefsHyperref{code:simplex734}{\([7,3,4]\) simplex code}} & \eczhListValue{Second-smallest nontrivial member of the simplex-code family.
The columns of its generator matrix are in one-to-one correspondence with the elements of the projective space \(PG(2,2)\), with each column being a chosen representative of the corresponding element.
The codewords form a \((8,9)\) simplex spherical code under the \flmRefsHyperref{ref38}{antipodal mapping}.
As a simplex code, it is equidistant: every nonzero codeword has Hamming weight \(4\).}\\ 
\addlinespace[\myxtraspc]
\eczhRefIndex{code:hamming844}%
\eczhListValue{\flmRefsHyperref{code:hamming844}{\([8,4,4]\) extended Hamming code}} & \eczhListValue{Extension of the \([7,4,3]\) Hamming code by a parity-check bit.
The smallest doubly even self-dual code, and the unique Type II code of length \(8\) \NoCaseChange{\protect\cite[{Rem. 4.3.10}]{cite40}}.}\\ 
\addlinespace[\myxtraspc]
\eczhRefIndex{code:parity_check}%
\eczhListValue{\flmRefsHyperref{code:parity_check}{\([n,n-1,2]\) Single parity-check (SPC) code}} & \eczhListValue{An \([n,n-1,2]\) linear binary code whose codewords consist of the message string appended with a \textit{parity-check bit} or \textit{parity bit} such that the parity (i.e., sum over all coordinates of each codeword) is zero.
If the Hamming weight of a message is odd (even), then the parity bit is one (zero).
This code requires only one extra bit of overhead and is therefore inexpensive.
Its codewords are all even-weight binary strings, and its parity-check matrix is a row vector of all ones.
Its automorphism group is \(S_n\).}\\ 
\addlinespace[\myxtraspc]
\eczhRefIndex{code:q-ary_parity_check}%
\eczhListValue{\flmRefsHyperref{code:q-ary_parity_check}{\([n,n-1,2]_q\) \(q\)-ary parity-check code}} & \eczhListValue{An \([n,n-1,2]_q\) linear \(q\)-ary code whose codewords consist of the message string appended with a \textit{parity-check} or \textit{zero-sum check digit} such that the sum over all coordinates of each codeword is zero.}\\ 
\addlinespace[\myxtraspc]
\eczhRefIndex{code:q-ary_ldpc}%
\eczhListValue{\flmRefsHyperref{code:q-ary_ldpc}{\(q\)-ary LDPC code}} & \eczhListValue{A \(q\)-ary linear code with a sparse parity-check matrix.
Alternatively, a member of an infinite family of \([n,k,d]_q\) codes for which the number of nonzero entries in each row and column of the parity-check matrix are both bounded above by a constant as \(n\to\infty\).}\\ 
\addlinespace[\myxtraspc]
\eczhRefIndex{code:q-ary_lcc}%
\eczhListValue{\flmRefsHyperref{code:q-ary_lcc}{\(q\)-ary linear LCC}} & \eczhListValue{A \(q\)-ary linear code for which one can recover any coordinate of a codeword from at most \(r\) coordinates of a received word (assuming the corruption rate is within some tolerated threshold \(\delta\)).}\\ 
\addlinespace[\myxtraspc]
\eczhRefIndex{code:q-ary_protograph_ldpc}%
\eczhListValue{\flmRefsHyperref{code:q-ary_protograph_ldpc}{\(q\)-ary protograph LDPC code}} & \eczhListValue{A \(q\)-ary LDPC code whose parity-check matrix is constructed using the \flmRefsHyperref{ref47}{lifting} procedure applied to the incidence matrix of a sparse graph called, in this context, a \textit{protograph}.
An ability to assign non-binary edge weight called \textit{edge scaling} can also be used in code construction.}\\ 
\addlinespace[\myxtraspc]
\eczhRefIndex{code:q-ary_repetition}%
\eczhListValue{\flmRefsHyperref{code:q-ary_repetition}{\(q\)-ary repetition code}} & \eczhListValue{An \([n,1,n]_q\) code consisting of codewords \((j,j,\cdots,j)\) for \(j \in \mathbb{F}_q\).}\\ 
\addlinespace[\myxtraspc]
\eczhRefIndex{code:q-ary_simplex}%
\eczhListValue{\flmRefsHyperref{code:q-ary_simplex}{\(q\)-ary simplex code}} & \eczhListValue{An \([n,m,q^{m-1}]_q\) equidistant projective code with \(n=\frac{q^m-1}{q-1}\), denoted as \(S(q,m)\). The columns of the generator matrix are in one-to-one correspondence with the elements of the projective space \(PG(m-1,q)\), with each column being a chosen representative of the corresponding element.
All nonzero simplex codewords have a constant weight of \(q^{m-1}\) \NoCaseChange{\protect\cite{cite45,cite46}}.}\\ 
\addlinespace[\myxtraspc]
\eczhRefIndex{code:multiple_erasure_lrc}%
\eczhListValue{\flmRefsHyperref{code:multiple_erasure_lrc}{\(t\)-erasure LRC}} & \eczhListValue{A code which admits local recoverability against more than one coordinate erasure, typically up to some specified number \(t\) of erasures per local group.}\\ 
\end{tabularx}\endgroup
\eczcodelist{ltc}{Locally testable codes and friends}%

\eczhCodeListAutoDescription{Union of:
\begin{itemize}\item codes that are descendants of \flmRefsCref{code:ltc}
\item codes that are cousins of \flmRefsCref{code:ltc}
\item codes that are cousins of \flmRefsCref{code:q-ary_ltc}
\item codes that are cousins of \flmRefsCref{code:binary_ltc}
\end{itemize}}%

\eczhIncludeCodeGraph{Bare}{scale=0.5}{\columnwidth}{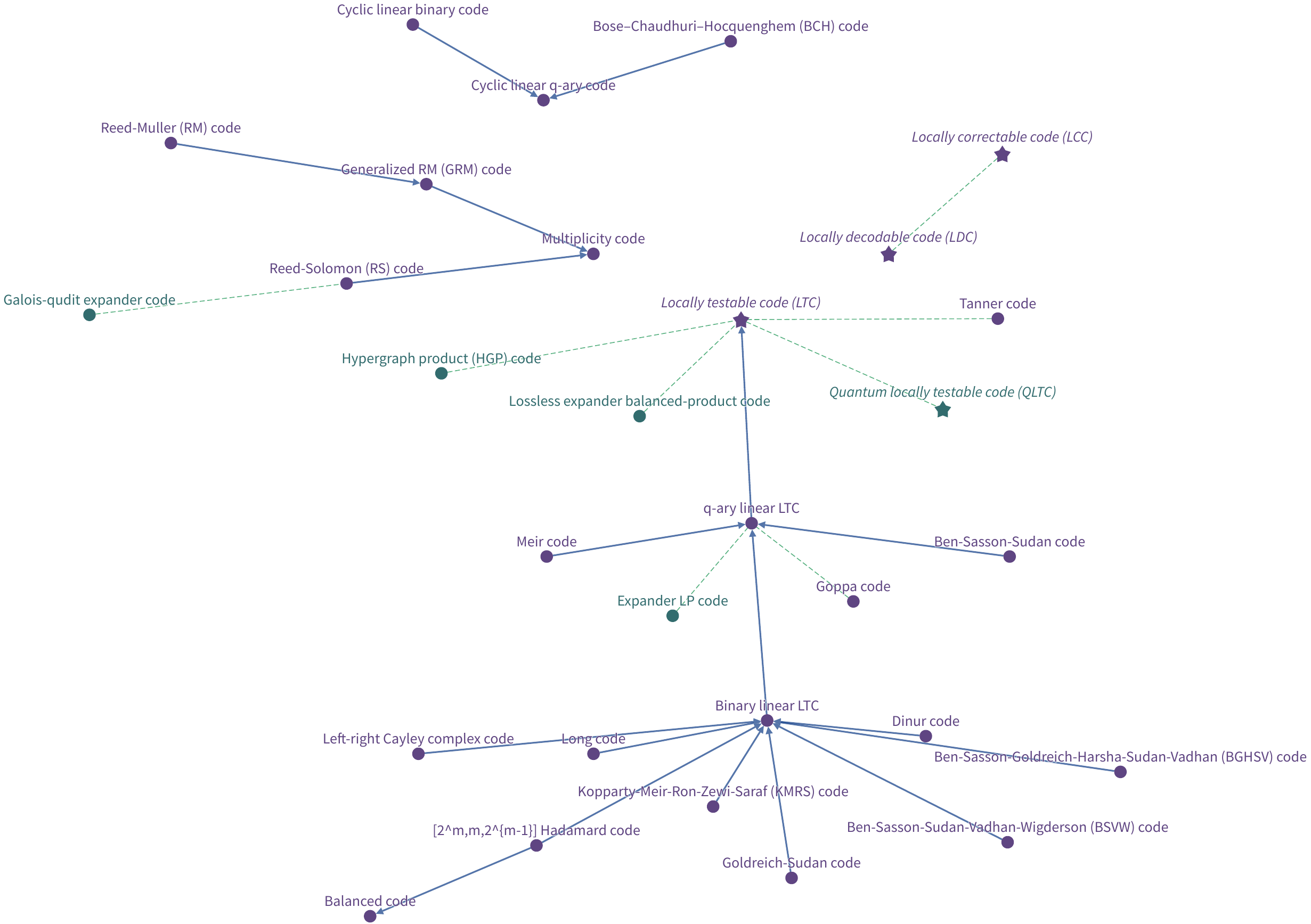}{Locally testable codes and friends}{https://errorcorrectionzoo.org/code_graph#J\%7B\%22displayMode\%22\%3A\%22subset\%22\%2C\%22modeSubsetOptions\%22\%3A\%7B\%22codeIds\%22\%3A\%5B\%22balanced\%22\%2C\%22bsghsv-ltc\%22\%2C\%22bs-ltc\%22\%2C\%22bssvw-ltc\%22\%2C\%22binary_ltc\%22\%2C\%22q-ary_bch\%22\%2C\%22q-ary_cyclic\%22\%2C\%22binary_cyclic\%22\%2C\%22dinur\%22\%2C\%22expander_lifted_product\%22\%2C\%22galois_expander\%22\%2C\%22generalized_reed_muller\%22\%2C\%22gs-ltc\%22\%2C\%22goppa\%22\%2C\%22hypergraph_product\%22\%2C\%22kmrs-ltc\%22\%2C\%22lr-cayley-complex\%22\%2C\%22lcc\%22\%2C\%22ldc\%22\%2C\%22ltc\%22\%2C\%22long\%22\%2C\%22lossless_expander\%22\%2C\%22meir\%22\%2C\%22multiplicity\%22\%2C\%22qltc\%22\%2C\%22reed_muller\%22\%2C\%22reed_solomon\%22\%2C\%22tanner\%22\%2C\%22hadamard\%22\%2C\%22q-ary_ltc\%22\%5D\%2C\%22reusePreviousLayoutPositions\%22\%3Afalse\%2C\%22showIntermediateConnectingNodes\%22\%3Atrue\%2C\%22connectingNodesMaxDepth\%22\%3A15\%2C\%22connectingNodesPathMaxLength\%22\%3A20\%2C\%22connectingNodesMaxExtraDepth\%22\%3A3\%2C\%22connectingNodesOnlyKeepPathsWithAdditionalLength\%22\%3A1\%2C\%22connectingNodesToDomainsAndKingdoms\%22\%3Afalse\%2C\%22connectingNodesEdgeLengthsByType\%22\%3A\%7B\%22primaryParent\%22\%3A1\%2C\%22secondaryParent\%22\%3A4\%2C\%22cousin\%22\%3A6\%7D\%2C\%22nodeIds\%22\%3A\%5B\%5D\%7D\%2C\%22highlightImportantNodes\%22\%3A\%7B\%22highlightImportantNodes\%22\%3Afalse\%2C\%22highlightPrimaryParents\%22\%3Afalse\%2C\%22highlightRootConnectingEdges\%22\%3Afalse\%7D\%7D}

\begingroup
\small
\eczhBreakableDashes
\renewcommand\arraystretch{1.05}
\edef\myxtraspc{\eczListAddVSpaceXtraXtraStretch}
\begin{tabularx}{\linewidth}{>{\raggedright\arraybackslash}p{\eczListColWidth{name}} >{\hsize=1.0000\hsize }X}
\toprule
\eczListColTitle{Code} & \eczListColTitle{Description} \\
\midrule
\endfirsthead
\toprule
\eczListColTitleContinued{Code} & \eczListColTitleContinued{Description} \\
\midrule
\endhead
\bottomrule
\endfoot
\eczhRefIndex{code:balanced}%
\eczhListValue{\flmRefsHyperref{code:balanced}{Balanced code}} & \eczhListValue{An even-length-\(n\) \(q\)-ary code whose nonzero codewords all have a Hamming weight of \(n/2\).
A code is \(\epsilon\)\textit{-balanced} if the relative weight (i.e., weight divided by \(n\)) of all nonzero codewords lies in the interval \([\frac{1-\epsilon}{2},\frac{1+\epsilon}{2}]\).
A code is \(\gamma\)\textit{-unbiased} if the relative weight lies in the interval \((\frac{1}{2}-\frac{1}{n^{\gamma}},\frac{1}{2}+\frac{1}{n^{\gamma}})\).}\\ 
\addlinespace[\myxtraspc]
\eczhRefIndex{code:bsghsv-ltc}%
\eczhListValue{\flmRefsHyperref{code:bsghsv-ltc}{Ben-Sasson-Goldreich-Harsha-Sudan-Vadhan (BGHSV) code}} & \eczhListValue{A member of a family of locally testable \([n,k,d]\) codes with \(n = k^{1+\epsilon}\) and query complexity of \flmRefsHyperref{ref65}{order} \(O(1/\epsilon)\), for any fixed \(\epsilon > 0\).}\\ 
\addlinespace[\myxtraspc]
\eczhRefIndex{code:bs-ltc}%
\eczhListValue{\flmRefsHyperref{code:bs-ltc}{Ben-Sasson-Sudan code}} & \eczhListValue{Locally testable \([n,k/2,d]_{2^m}\) code with \(k\) a power of two, \(n = k \log^{c} k\), and query complexity \(\log^{c}k\) for some universal constant \(c\).}\\ 
\addlinespace[\myxtraspc]
\eczhRefIndex{code:bssvw-ltc}%
\eczhListValue{\flmRefsHyperref{code:bssvw-ltc}{Ben-Sasson-Sudan-Vadhan-Wigderson (BSVW) code}} & \eczhListValue{Locally testable \([n,k,d]\) code with \(n = k \cdot 2^{\tilde{O}(\sqrt{\log k})}\) and asymptotically constant query complexity, where \(\tilde{O}(f)=O(f\cdot (\log f)^c)\) for some fixed constant \(c\).}\\ 
\addlinespace[\myxtraspc]
\eczhRefIndex{code:binary_ltc}%
\eczhListValue{\flmRefsHyperref{code:binary_ltc}{Binary linear LTC}} & \eczhListValue{A binary linear code \(C\) of length \(n\) that is a \((u,R)\)-LTC with query complexity \(u\) and soundness \(R>0\).}\\ 
\addlinespace[\myxtraspc]
\eczhRefIndex{code:q-ary_bch}%
\eczhListValue{\flmRefsHyperref{code:q-ary_bch}{Bose–Chaudhuri–Hocquenghem (BCH) code}} & \eczhListValue{A cyclic \(q\)-ary code, with \(n\) and \(q\) relatively prime, whose zeroes are consecutive powers of a primitive \(n\)th root of unity \(\alpha\).}\\ 
\addlinespace[\myxtraspc]
\eczhRefIndex{code:q-ary_cyclic}%
\eczhListValue{\flmRefsHyperref{code:q-ary_cyclic}{Cyclic linear \(q\)-ary code}} & \eczhListValue{A \(q\)-ary code of length \(n\) is cyclic if, for each codeword \(c_1 c_2 \cdots c_n\), the cyclically shifted string \(c_n c_1 \cdots c_{n-1}\) is also a codeword. A cyclic code is called \textit{primitive} when \(n=q^r-1\) for some \(r\geq 2\). A \textit{shortened cyclic code} is obtained from a cyclic code by taking only codewords with the first \(j\) zero entries, and deleting those zeroes.}\\ 
\addlinespace[\myxtraspc]
\eczhRefIndex{code:binary_cyclic}%
\eczhListValue{\flmRefsHyperref{code:binary_cyclic}{Cyclic linear binary code}} & \eczhListValue{A binary code of length \(n\) is cyclic if, for each codeword \(c_1 c_2 \cdots c_n\), the cyclically shifted string \(c_n c_1 \cdots c_{n-1}\) is also a codeword. A cyclic code is called \textit{primitive} when \(n=2^r-1\) for some \(r\geq 2\).}\\ 
\addlinespace[\myxtraspc]
\eczhRefIndex{code:dinur}%
\eczhListValue{\flmRefsHyperref{code:dinur}{Dinur code}} & \eczhListValue{Member of an infinite family of locally testable \([n,n/\text{polylog}(n),d]\) codes with vanishing rate. Code construction relies on tensor-product codes \NoCaseChange{\protect\cite{cite73}}.}\\ 
\addlinespace[\myxtraspc]
\eczhRefIndex{code:expander_lifted_product}%
\eczhListValue{\flmRefsHyperref{code:expander_lifted_product}{Expander LP code}} & \eczhListValue{Family of \(G\)-lifted product codes constructed using two classical \flmRefsHyperref{code:expander}{expander codes}, equivalently two regular \flmRefsHyperref{code:tanner}{Tanner codes} defined on the same expander graph \NoCaseChange{\protect\cite{cite74}}. For certain parameters, this construction yields the first asymptotically good QLDPC codes. Classical codes resulting from the same lifted-product complexes are one of the first two families of \(c^3\)-LTCs \NoCaseChange{\protect\cite{cite184}}.}\\ 
\addlinespace[\myxtraspc]
\eczhRefIndex{code:galois_expander}%
\eczhListValue{\flmRefsHyperref{code:galois_expander}{Galois-qudit expander code}} & \eczhListValue{Galois-qudit CSS code obtained from tensor products of chain complexes associated with an explicit family of expander codes with Reed-Solomon local checks.}\\ 
\addlinespace[\myxtraspc]
\eczhRefIndex{code:generalized_reed_muller}%
\eczhListValue{\flmRefsHyperref{code:generalized_reed_muller}{Generalized RM (GRM) code}} & \eczhListValue{Extensions of RM codes to \(q\)-ary coordinates that can be described as multivariate polynomial evaluation codes over affine or projective space.}\\ 
\addlinespace[\myxtraspc]
\eczhRefIndex{code:gs-ltc}%
\eczhListValue{\flmRefsHyperref{code:gs-ltc}{Goldreich-Sudan code}} & \eczhListValue{Locally testable \([n,k,d]\) code with \(n = k^{1+O(1/u)}\) and distance of \flmRefsHyperref{ref65}{order} \(\Omega(n)\) for query complexity \(u\). The same work also presented a probabilistic construction of codes of size \(k^{1+o(1)}\).}\\ 
\addlinespace[\myxtraspc]
\eczhRefIndex{code:goppa}%
\eczhListValue{\flmRefsHyperref{code:goppa}{Goppa code}} & \eczhListValue{A linear \(q\)-ary code defined from a polynomial \(G(x)\) over an extension field and a set of evaluation points \(L\) avoiding the roots of \(G\).
Goppa codes form a central family of alternant codes, admit efficient algebraic decoding algorithms, and include the binary Goppa codes used in the McEliece cryptosystem.
When the base field equals the coefficient field, they coincide with residue AG codes on \(PG(1,q^m)\); in general, classical Goppa codes are subfield subcodes of such AG codes \NoCaseChange{\protect\cite[{Rem. 15.3.27}]{cite26}\protect\cite[{Thm. 15.3.28}]{cite26}}.}\\ 
\addlinespace[\myxtraspc]
\eczhRefIndex{code:hypergraph_product}%
\eczhListValue{\flmRefsHyperref{code:hypergraph_product}{Hypergraph product (HGP) code}} & \eczhListValue{A member of a family of CSS codes whose stabilizer generator matrix is obtained from a hypergraph product of two classical linear binary codes.}\\ 
\addlinespace[\myxtraspc]
\eczhRefIndex{code:kmrs-ltc}%
\eczhListValue{\flmRefsHyperref{code:kmrs-ltc}{Kopparty-Meir-Ron-Zewi-Saraf (KMRS) code}} & \eczhListValue{Member of a family of locally testable binary linear codes with constant rate, constant relative distance, and subpolynomial query complexity \(u = (\log n)^{O(\log \log n)}\).
Later work by Gopi, Kopparty, Oliveira, Ron-Zewi, and Saraf \NoCaseChange{\protect\cite{cite86}} showed that related concatenated codes achieve the \flmRefsHyperref{ref85}{GV bound}.}\\ 
\addlinespace[\myxtraspc]
\eczhRefIndex{code:lr-cayley-complex}%
\eczhListValue{\flmRefsHyperref{code:lr-cayley-complex}{Left-right Cayley complex code}} & \eczhListValue{Binary code constructed on a left-right Cayley complex using a pair of base codes \(C_A,C_B\) and an expander graph \NoCaseChange{\protect\cite{cite74}}. Bits live on the squares of the complex, while local constraints are imposed on edges; for a fixed graph vertex, the incident symbols form a codeword of the tensor code \(C_A \otimes C_B\). A family of such codes is one of the first \(c^3\)-LTCs \NoCaseChange{\protect\cite{cite88}}.}\\ 
\addlinespace[\myxtraspc]
\eczhRefIndex{code:lcc}%
\eczhListValue{\flmRefsHyperref{code:lcc}{Locally correctable code (LCC)}} & \eczhListValue{Recall that a block code encodes a length-\(k\) message into a length-\(n\) codeword, which is then sent through a noise channel to yield a received word.
Informally, an LCC is a block code for which one can recover any coordinate of a codeword from at most \(r\) coordinates of the received word (assuming the received word is within some tolerated corruption rate \(\delta\)).}\\ 
\addlinespace[\myxtraspc]
\eczhRefIndex{code:ldc}%
\eczhListValue{\flmRefsHyperref{code:ldc}{Locally decodable code (LDC)}} & \eczhListValue{Recall that a block code encodes a length-\(k\) message into a length-\(n\) codeword, which is then sent through a noise channel to yield a received word.
Informally, an LDC is a block code for which one can recover any coordinate of the message from at most \(r\) coordinates of the received word (assuming the received word is within some tolerated corruption rate \(\delta\)).
Efficiency of the decoding is quantified by the code's \textit{query complexity} \(r\), and decoding is performed by sampling subsets of \(r\) bits.}\\ 
\addlinespace[\myxtraspc]
\eczhRefIndex{code:ltc}%
\eczhListValue{\flmRefsHyperref{code:ltc}{Locally testable code (LTC)}} & \eczhListValue{Code for which one can efficiently check whether a given string is a codeword or is far from a codeword. Efficiency of the verification is quantified by the code's \textit{query complexity} \(u\), while effectiveness is quantified by the code's \textit{soundness} \(R\).}\\ 
\addlinespace[\myxtraspc]
\eczhRefIndex{code:long}%
\eczhListValue{\flmRefsHyperref{code:long}{Long code}} & \eczhListValue{Nonlinear locally testable code of extremely large length that is not practical, but is useful for certain probabilistically checkable proof (PCP) constructions \NoCaseChange{\protect\cite{cite89}}.}\\ 
\addlinespace[\myxtraspc]
\eczhRefIndex{code:lossless_expander}%
\eczhListValue{\flmRefsHyperref{code:lossless_expander}{Lossless expander balanced-product code}} & \eczhListValue{QLDPC code constructed by taking the balanced product of lossless expander graphs.
Using one part of a quantum-code chain complex constructed with one-sided loss expanders \NoCaseChange{\protect\cite{cite185}} yields a \(c^3\)-LTC \NoCaseChange{\protect\cite{cite186}}.
Using two-sided expanders \NoCaseChange{\protect\cite{cite187}} yields an asymptotically good QLDPC code family \NoCaseChange{\protect\cite{cite188}}.}\\ 
\addlinespace[\myxtraspc]
\eczhRefIndex{code:meir}%
\eczhListValue{\flmRefsHyperref{code:meir}{Meir code}} & \eczhListValue{Locally testable \([n,k,d]_q\) code with query complexity \(\text{poly}(\log k)\) and rejection ratio \(R/n = 1/\text{poly}(\log k)\). The code construction is probabilistic and combinatorial.}\\ 
\addlinespace[\myxtraspc]
\eczhRefIndex{code:multiplicity}%
\eczhListValue{\flmRefsHyperref{code:multiplicity}{Multiplicity code}} & \eczhListValue{A generalization of an \(m\)-variate polynomial evaluation code based on evaluating polynomials together with their Hasse derivatives up to order \(s-1\) at all points in \(\mathbb{F}_q^m\).
Originally proposed for coding using the Rosenbloom-Tsfasman metric \NoCaseChange{\protect\cite{cite179}}.
Univariate (\(m=1\)) \NoCaseChange{\protect\cite{cite179,cite180}} and multivariate (\(m>1\)) \NoCaseChange{\protect\cite{cite181}} codes have been proposed.}\\ 
\addlinespace[\myxtraspc]
\eczhRefIndex{code:qltc}%
\eczhListValue{\flmRefsHyperref{code:qltc}{Quantum locally testable code (QLTC)}} & \eczhListValue{A local commuting-projector Hamiltonian-based block quantum code which has a nonzero average-energy penalty for creating large errors. Informally, states that are far away from the codespace of a QLTC have to be excited states of a number of the code's local projectors that scales linearly with \(n\).}\\ 
\addlinespace[\myxtraspc]
\eczhRefIndex{code:reed_muller}%
\eczhListValue{\flmRefsHyperref{code:reed_muller}{Reed-Muller (RM) code}} & \eczhListValue{Member of the RM\((r,m)\) family of linear binary codes derived from multivariate polynomials. The code parameters are \([2^m,\sum_{j=0}^{r} {m \choose j},2^{m-r}]\), where \(r\) is the \textit{order} of the code satisfying \(0\leq r\leq m\).
First-order RM codes are also called biorthogonal codes, while \(m\)th order RM codes are also called \textit{universe} codes.
\textit{Punctured RM codes} RM\(^*(r,m)\) are obtained from RM codes by deleting one coordinate from each codeword.}\\ 
\addlinespace[\myxtraspc]
\eczhRefIndex{code:reed_solomon}%
\eczhListValue{\flmRefsHyperref{code:reed_solomon}{Reed-Solomon (RS) code}} & \eczhListValue{An \([n,k,n-k+1]_q\) linear code based on polynomials over \(\mathbb{F}_q\).}\\ 
\addlinespace[\myxtraspc]
\eczhRefIndex{code:tanner}%
\eczhListValue{\flmRefsHyperref{code:tanner}{Tanner code}} & \eczhListValue{A linear \(q\)-ary code defined on a bipartite graph similar to a Tanner graph.
This \textit{generalized Tanner graph} consists of variable nodes and constraint nodes, with the generalization being that the constraint nodes of degree \(r\) now represent a linear code of length \(r\).}\\ 
\addlinespace[\myxtraspc]
\eczhRefIndex{code:hadamard}%
\eczhListValue{\flmRefsHyperref{code:hadamard}{\([2^m,m,2^{m-1}]\) Hadamard code}} & \eczhListValue{An \([2^m,m,2^{m-1}]\) balanced binary code.
The \([2^m,m+1,2^{m-1}]\) augmented Hadamard code is the first-order RM code (a.k.a. RM\((1,m)\)), while the \([2^m-1,m,2^{m-1}]\) shortened Hadamard code is the simplex code (a.k.a. RM\(^*(1,m)\)).}\\ 
\addlinespace[\myxtraspc]
\eczhRefIndex{code:q-ary_ltc}%
\eczhListValue{\flmRefsHyperref{code:q-ary_ltc}{\(q\)-ary linear LTC}} & \eczhListValue{A \(q\)-ary linear code \(C\) of length \(n\) that is a \((u,R)\)-LTC with query complexity \(u\) and soundness \(R>0\).}\\ 
\end{tabularx}\endgroup
\eczcodelist{mds}{MDS codes and generalizations
}%

\eczhCodeListAutoDescription{Union of:
\begin{itemize}\item codes that are descendants of \flmRefsCref{code:mds}
\item codes that are descendants of \flmRefsCref{code:mds_array}
\item codes that are descendants of \flmRefsCref{code:maximum_rank_distance}
\item codes that are descendants of \flmRefsCref{code:maximum_sum_rank_distance}
\end{itemize}}%

\eczhIncludeCodeGraph{Bare}{scale=0.5}{\columnwidth}{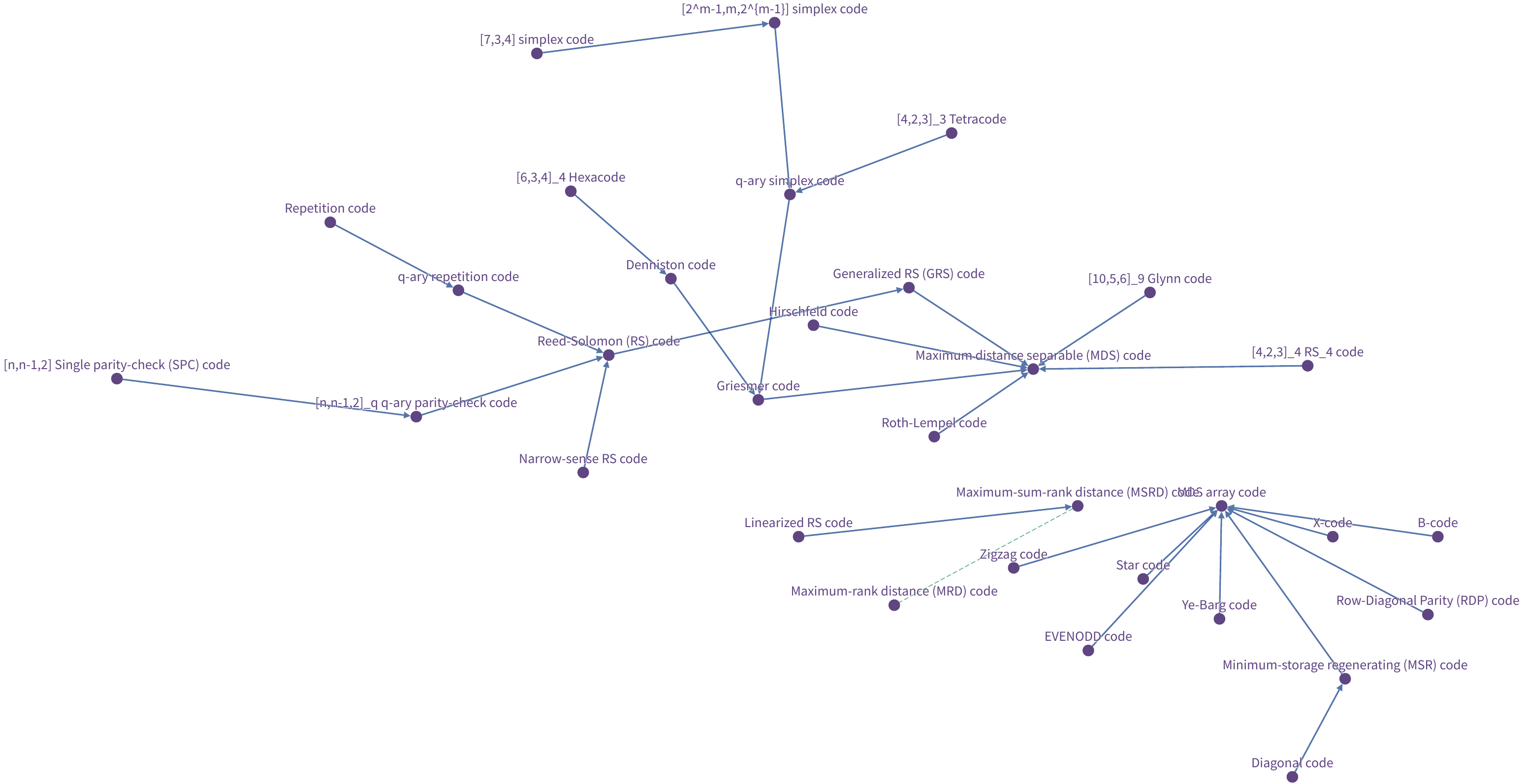}{MDS codes and generalizations}{https://errorcorrectionzoo.org/code_graph#J\%7B\%22displayMode\%22\%3A\%22subset\%22\%2C\%22modeSubsetOptions\%22\%3A\%7B\%22codeIds\%22\%3A\%5B\%22b_array\%22\%2C\%22denniston\%22\%2C\%22diagonal\%22\%2C\%22evenodd\%22\%2C\%22generalized_reed_solomon\%22\%2C\%22griesmer\%22\%2C\%22hirschfeld\%22\%2C\%22linearized_reed_solomon\%22\%2C\%22mds_array\%22\%2C\%22mds\%22\%2C\%22maximum_rank_distance\%22\%2C\%22maximum_sum_rank_distance\%22\%2C\%22msr\%22\%2C\%22narrow_sense_reed_solomon\%22\%2C\%22reed_solomon\%22\%2C\%22repetition\%22\%2C\%22roth_lempel\%22\%2C\%22rdp\%22\%2C\%22star\%22\%2C\%22x_array\%22\%2C\%22ye_barg\%22\%2C\%22zigzag\%22\%2C\%22glynn\%22\%2C\%22simplex\%22\%2C\%22tetracode\%22\%2C\%22reed_solomon_4\%22\%2C\%22hexacode\%22\%2C\%22simplex734\%22\%2C\%22parity_check\%22\%2C\%22q-ary_parity_check\%22\%2C\%22q-ary_repetition\%22\%2C\%22q-ary_simplex\%22\%5D\%2C\%22reusePreviousLayoutPositions\%22\%3Afalse\%2C\%22showIntermediateConnectingNodes\%22\%3Atrue\%2C\%22connectingNodesMaxDepth\%22\%3A15\%2C\%22connectingNodesPathMaxLength\%22\%3A20\%2C\%22connectingNodesMaxExtraDepth\%22\%3A3\%2C\%22connectingNodesOnlyKeepPathsWithAdditionalLength\%22\%3A1\%2C\%22connectingNodesToDomainsAndKingdoms\%22\%3Afalse\%2C\%22connectingNodesEdgeLengthsByType\%22\%3A\%7B\%22primaryParent\%22\%3A1\%2C\%22secondaryParent\%22\%3A4\%2C\%22cousin\%22\%3A6\%7D\%2C\%22nodeIds\%22\%3A\%5B\%5D\%7D\%2C\%22highlightImportantNodes\%22\%3A\%7B\%22highlightImportantNodes\%22\%3Afalse\%2C\%22highlightPrimaryParents\%22\%3Afalse\%2C\%22highlightRootConnectingEdges\%22\%3Afalse\%7D\%7D}

\begingroup
\small
\eczhBreakableDashes
\renewcommand\arraystretch{1.05}
\edef\myxtraspc{\eczListAddVSpaceXtraXtraStretch}
\begin{tabularx}{\linewidth}{>{\raggedright\arraybackslash}p{\eczListColWidth{name}} >{\hsize=1.0000\hsize }X}
\toprule
\eczListColTitle{Code} & \eczListColTitle{Description} \\
\midrule
\endfirsthead
\toprule
\eczListColTitleContinued{Code} & \eczListColTitleContinued{Description} \\
\midrule
\endhead
\bottomrule
\endfoot
\eczhRefIndex{code:b_array}%
\eczhListValue{\flmRefsHyperref{code:b_array}{B-code}} & \eczhListValue{Binary MDS block array code \(\mathcal{B}_2(m)\) on \((m-1)\times m\) arrays with horizontal and toroidal diagonal parity checks \NoCaseChange{\protect\cite{cite189}}.}\\ 
\addlinespace[\myxtraspc]
\eczhRefIndex{code:denniston}%
\eczhListValue{\flmRefsHyperref{code:denniston}{Denniston code}} & \eczhListValue{Projective code that is part of a family of \([2^{a+i}+2^i-2^a,3,2^{a+i}-2^a]_{2^a}\) codes for \(i < a\) constructed using Denniston arcs \NoCaseChange{\protect\cite[{Sec. 19.7.3}]{cite172}}.}\\ 
\addlinespace[\myxtraspc]
\eczhRefIndex{code:diagonal}%
\eczhListValue{\flmRefsHyperref{code:diagonal}{Diagonal code}} & \eczhListValue{Member of an explicit family of high-rate \([n,k,d,\alpha, \beta = \frac{\alpha}{d-k+1}, M=k\alpha]\) MSR codes for any \(r\) and \(n\).
Such codes can optimally repair any \(f\) failed nodes from any \(d\) helper nodes for all \(d\), \(1 \le f \le r\) and \(k \le d \le n-f\) simultaneously.
These codes can be constructed over any base field \(\mathbb{F}_q\) as long as \(|\mathbb{F}_q| \ge sn\), where \(s = \text{lcm}(1,2,\cdots,r)\).}\\ 
\addlinespace[\myxtraspc]
\eczhRefIndex{code:evenodd}%
\eczhListValue{\flmRefsHyperref{code:evenodd}{EVENODD code}} & \eczhListValue{Binary array code \(\mathcal{EO}_2(m)\) with independent horizontal and diagonal parity columns, designed to retain optimal double-erasure protection while simplifying small updates \NoCaseChange{\protect\cite{cite189}}.}\\ 
\addlinespace[\myxtraspc]
\eczhRefIndex{code:generalized_reed_solomon}%
\eczhListValue{\flmRefsHyperref{code:generalized_reed_solomon}{Generalized RS (GRS) code}} & \eczhListValue{An \([n,k,n-k+1]_q\) MDS code that is a modification of the RS code where codeword polynomials are multiplied by additional prefactors \NoCaseChange{\protect\cite[{Def. 15.3.19}]{cite26}}.}\\ 
\addlinespace[\myxtraspc]
\eczhRefIndex{code:griesmer}%
\eczhListValue{\flmRefsHyperref{code:griesmer}{Griesmer code}} & \eczhListValue{A type of \(q\)-ary code whose parameters satisfy the Griesmer bound with equality.}\\ 
\addlinespace[\myxtraspc]
\eczhRefIndex{code:hirschfeld}%
\eczhListValue{\flmRefsHyperref{code:hirschfeld}{Hirschfeld code}} & \eczhListValue{A \([q+1,4,q-2]_q\) projective geometry code for non-prime \(q\) that is an example of an MDS code that is not an RS code; see \NoCaseChange{\protect\cite[{Exam. 7.6}]{cite182}} for the generator matrix.}\\ 
\addlinespace[\myxtraspc]
\eczhRefIndex{code:linearized_reed_solomon}%
\eczhListValue{\flmRefsHyperref{code:linearized_reed_solomon}{Linearized RS code}} & \eczhListValue{A code obtained by linearizing a skew RS code, i.e., by translating evaluations of skew polynomials into operator evaluations over blocks.}\\ 
\addlinespace[\myxtraspc]
\eczhRefIndex{code:mds_array}%
\eczhListValue{\flmRefsHyperref{code:mds_array}{MDS array code}} & \eczhListValue{An \((n,k,m)\) array code whose codewords can be recovered by any \(k\) out of \(n\) nodes, where each node stores a length-\(m\) column of the codeword. MDS array codes are MDS codes when each matrix codeword is treated as a vector by converting each column into a single coordinate via subpacketization.}\\ 
\addlinespace[\myxtraspc]
\eczhRefIndex{code:mds}%
\eczhListValue{\flmRefsHyperref{code:mds}{Maximum distance separable (MDS) code}} & \eczhListValue{A \(q\)-ary linear code whose parameters satisfy the Singleton bound with equality.}\\ 
\addlinespace[\myxtraspc]
\eczhRefIndex{code:maximum_rank_distance}%
\eczhListValue{\flmRefsHyperref{code:maximum_rank_distance}{Maximum-rank distance (MRD) code}} & \eczhListValue{A rank-metric code whose parameters satisfy the rank-metric Singleton-like bound with equality.}\\ 
\addlinespace[\myxtraspc]
\eczhRefIndex{code:maximum_sum_rank_distance}%
\eczhListValue{\flmRefsHyperref{code:maximum_sum_rank_distance}{Maximum-sum-rank distance (MSRD) code}} & \eczhListValue{A sum-rank-metric code whose parameters satisfy the sum-rank-metric Singleton bound with equality.}\\ 
\addlinespace[\myxtraspc]
\eczhRefIndex{code:msr}%
\eczhListValue{\flmRefsHyperref{code:msr}{Minimum-storage regenerating (MSR) code}} & \eczhListValue{A regenerating code that corresponds to the minimum-storage extreme point of the storage-bandwidth trade-off curve, characterized by \(\alpha = (d-k+1)\beta\).}\\ 
\addlinespace[\myxtraspc]
\eczhRefIndex{code:narrow_sense_reed_solomon}%
\eczhListValue{\flmRefsHyperref{code:narrow_sense_reed_solomon}{Narrow-sense RS code}} & \eczhListValue{An \([q-1,k,n-k+1]_q\) RS code whose points \(\alpha_i\) are all \((i-1)\)st powers of a \flmRefsHyperref{ref33}{primitive} element \(\alpha\) of \(\mathbb{F}_q\).}\\ 
\addlinespace[\myxtraspc]
\eczhRefIndex{code:reed_solomon}%
\eczhListValue{\flmRefsHyperref{code:reed_solomon}{Reed-Solomon (RS) code}} & \eczhListValue{An \([n,k,n-k+1]_q\) linear code based on polynomials over \(\mathbb{F}_q\).}\\ 
\addlinespace[\myxtraspc]
\eczhRefIndex{code:repetition}%
\eczhListValue{\flmRefsHyperref{code:repetition}{Repetition code}} & \eczhListValue{\([n,1,n]\) binary linear code encoding one bit of information into an \(n\)-bit string.
Majority decoding requires \(n\) to be odd in order to avoid ties.
The idea is to increase the code distance by repeating the logical information several times. It is a \((n,1)\)-Hamming code.
Its automorphism group is \(S_n\).}\\ 
\addlinespace[\myxtraspc]
\eczhRefIndex{code:roth_lempel}%
\eczhListValue{\flmRefsHyperref{code:roth_lempel}{Roth-Lempel code}} & \eczhListValue{Member of a \(q\)-ary linear code family that includes many examples of MDS codes that are not GRS codes.}\\ 
\addlinespace[\myxtraspc]
\eczhRefIndex{code:rdp}%
\eczhListValue{\flmRefsHyperref{code:rdp}{Row-Diagonal Parity (RDP) code}} & \eczhListValue{An MDS array code protecting against double erasures.}\\ 
\addlinespace[\myxtraspc]
\eczhRefIndex{code:star}%
\eczhListValue{\flmRefsHyperref{code:star}{Star code}} & \eczhListValue{An MDS array code protecting against triple erasures.}\\ 
\addlinespace[\myxtraspc]
\eczhRefIndex{code:x_array}%
\eczhListValue{\flmRefsHyperref{code:x_array}{X-code}} & \eczhListValue{An MDS array code with a simple geometrical construction that achieves optimal encoding and update complexity.}\\ 
\addlinespace[\myxtraspc]
\eczhRefIndex{code:ye_barg}%
\eczhListValue{\flmRefsHyperref{code:ye_barg}{Ye-Barg code}} & \eczhListValue{A member of an explicit family of MDS array codes with the optimal access property.
The constructions of Ye and Barg achieve optimal repair bandwidth for single-node repair and include optimal-access variants with nearly optimal sub-packetization \NoCaseChange{\protect\cite{cite190,cite191}}.}\\ 
\addlinespace[\myxtraspc]
\eczhRefIndex{code:zigzag}%
\eczhListValue{\flmRefsHyperref{code:zigzag}{Zigzag code}} & \eczhListValue{An MDS array code correcting two erasures with optimal rebuilding ratio; see Ref. \NoCaseChange{\protect\cite{cite192}} for definitions.}\\ 
\addlinespace[\myxtraspc]
\eczhRefIndex{code:glynn}%
\eczhListValue{\flmRefsHyperref{code:glynn}{\([10,5,6]_9\) Glynn code}} & \eczhListValue{The unique trace-Hermitian self-dual \([10,5,6]_9\) code, constructed using a 10-arc in \(PG(4,9)\) that is not a rational curve.}\\ 
\addlinespace[\myxtraspc]
\eczhRefIndex{code:simplex}%
\eczhListValue{\flmRefsHyperref{code:simplex}{\([2^m-1,m,2^{m-1}]\) simplex code}} & \eczhListValue{A member of the equidistant code family dual to the \([2^m-1,2^m-m-1,3]\) Hamming family.}\\ 
\addlinespace[\myxtraspc]
\eczhRefIndex{code:tetracode}%
\eczhListValue{\flmRefsHyperref{code:tetracode}{\([4,2,3]_3\) Tetracode}} & \eczhListValue{The \([4,2,3]_3\) ternary self-dual MDS code that has connections to lattices \NoCaseChange{\protect\cite{cite39}}. Its weight enumerator is the Gleason polynomial \(g_4\) \NoCaseChange{\protect\cite[{Rem. 4.2.6}]{cite40}}.}\\ 
\addlinespace[\myxtraspc]
\eczhRefIndex{code:reed_solomon_4}%
\eczhListValue{\flmRefsHyperref{code:reed_solomon_4}{\([4,2,3]_4\) RS\(_4\) code}} & \eczhListValue{A Type II Euclidean self-dual extended RS code that is the smallest quaternary extended QR code \NoCaseChange{\protect\cite[{pg. 296}]{cite41}\protect\cite[{Sec. 2.4.2}]{cite42}}.
Puncturing the \([4,2,3]_4\) RS\(_4\) code yields the \([3,2,2]_4\) shortened RS\(_4\) code, which is an RS code \NoCaseChange{\protect\cite[{pg. 295}]{cite41}}.}\\ 
\addlinespace[\myxtraspc]
\eczhRefIndex{code:hexacode}%
\eczhListValue{\flmRefsHyperref{code:hexacode}{\([6,3,4]_4\) Hexacode}} & \eczhListValue{The \([6,3,4]_4\) Hermitian self-dual MDS code that has connections to projective geometry, lattices \NoCaseChange{\protect\cite{cite39}}, and conformal field theory \NoCaseChange{\protect\cite{cite44}}. Its weight enumerator is the Gleason polynomial \(g_7\) \NoCaseChange{\protect\cite[{Rem. 4.2.6}]{cite40}}.}\\ 
\addlinespace[\myxtraspc]
\eczhRefIndex{code:simplex734}%
\eczhListValue{\flmRefsHyperref{code:simplex734}{\([7,3,4]\) simplex code}} & \eczhListValue{Second-smallest nontrivial member of the simplex-code family.
The columns of its generator matrix are in one-to-one correspondence with the elements of the projective space \(PG(2,2)\), with each column being a chosen representative of the corresponding element.
The codewords form a \((8,9)\) simplex spherical code under the \flmRefsHyperref{ref38}{antipodal mapping}.
As a simplex code, it is equidistant: every nonzero codeword has Hamming weight \(4\).}\\ 
\addlinespace[\myxtraspc]
\eczhRefIndex{code:parity_check}%
\eczhListValue{\flmRefsHyperref{code:parity_check}{\([n,n-1,2]\) Single parity-check (SPC) code}} & \eczhListValue{An \([n,n-1,2]\) linear binary code whose codewords consist of the message string appended with a \textit{parity-check bit} or \textit{parity bit} such that the parity (i.e., sum over all coordinates of each codeword) is zero.
If the Hamming weight of a message is odd (even), then the parity bit is one (zero).
This code requires only one extra bit of overhead and is therefore inexpensive.
Its codewords are all even-weight binary strings, and its parity-check matrix is a row vector of all ones.
Its automorphism group is \(S_n\).}\\ 
\addlinespace[\myxtraspc]
\eczhRefIndex{code:q-ary_parity_check}%
\eczhListValue{\flmRefsHyperref{code:q-ary_parity_check}{\([n,n-1,2]_q\) \(q\)-ary parity-check code}} & \eczhListValue{An \([n,n-1,2]_q\) linear \(q\)-ary code whose codewords consist of the message string appended with a \textit{parity-check} or \textit{zero-sum check digit} such that the sum over all coordinates of each codeword is zero.}\\ 
\addlinespace[\myxtraspc]
\eczhRefIndex{code:q-ary_repetition}%
\eczhListValue{\flmRefsHyperref{code:q-ary_repetition}{\(q\)-ary repetition code}} & \eczhListValue{An \([n,1,n]_q\) code consisting of codewords \((j,j,\cdots,j)\) for \(j \in \mathbb{F}_q\).}\\ 
\addlinespace[\myxtraspc]
\eczhRefIndex{code:q-ary_simplex}%
\eczhListValue{\flmRefsHyperref{code:q-ary_simplex}{\(q\)-ary simplex code}} & \eczhListValue{An \([n,m,q^{m-1}]_q\) equidistant projective code with \(n=\frac{q^m-1}{q-1}\), denoted as \(S(q,m)\). The columns of the generator matrix are in one-to-one correspondence with the elements of the projective space \(PG(m-1,q)\), with each column being a chosen representative of the corresponding element.
All nonzero simplex codewords have a constant weight of \(q^{m-1}\) \NoCaseChange{\protect\cite{cite45,cite46}}.}\\ 
\end{tabularx}\endgroup
\eczcodelist{modulation}{Modulation schemes}%

\eczhCodeListAutoDescription{All descendants of \flmRefsCref{code:modulation}.}%

\eczhIncludeCodeGraph{Bare}{scale=0.5}{\columnwidth}{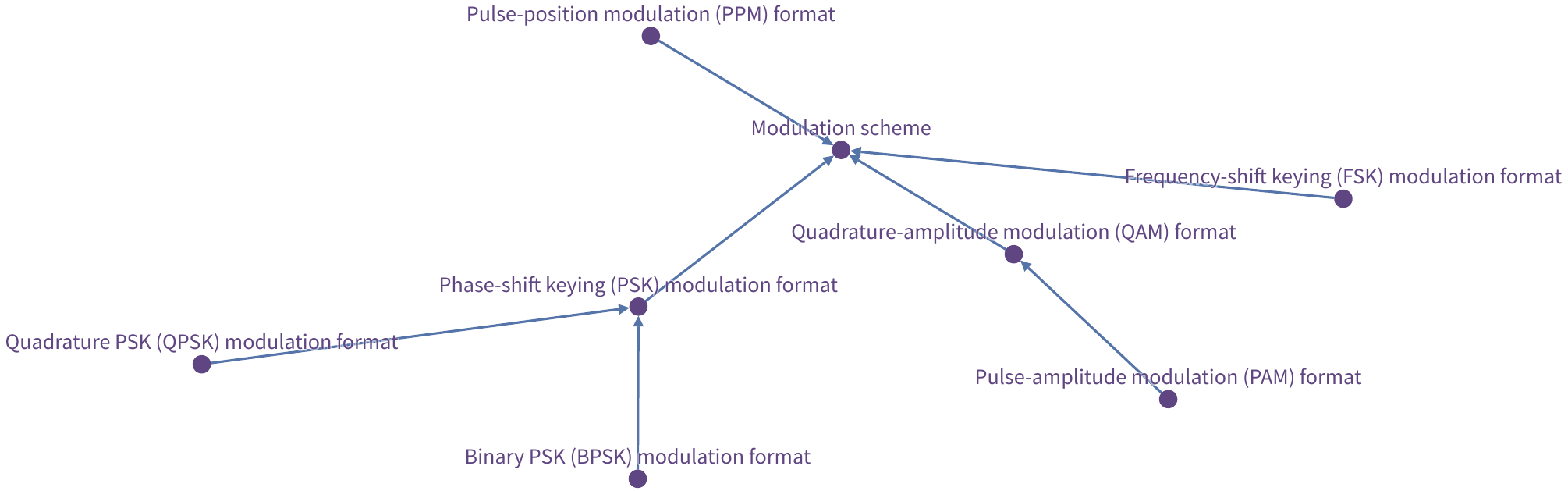}{Modulation schemes}{https://errorcorrectionzoo.org/code_graph#J\%7B\%22displayMode\%22\%3A\%22subset\%22\%2C\%22modeSubsetOptions\%22\%3A\%7B\%22codeIds\%22\%3A\%5B\%22bpsk\%22\%2C\%22fsk\%22\%2C\%22modulation\%22\%2C\%22psk\%22\%2C\%22pam\%22\%2C\%22ppm\%22\%2C\%22qpsk\%22\%2C\%22qam\%22\%5D\%2C\%22reusePreviousLayoutPositions\%22\%3Afalse\%2C\%22showIntermediateConnectingNodes\%22\%3Atrue\%2C\%22connectingNodesMaxDepth\%22\%3A15\%2C\%22connectingNodesPathMaxLength\%22\%3A20\%2C\%22connectingNodesMaxExtraDepth\%22\%3A3\%2C\%22connectingNodesOnlyKeepPathsWithAdditionalLength\%22\%3A1\%2C\%22connectingNodesToDomainsAndKingdoms\%22\%3Afalse\%2C\%22connectingNodesEdgeLengthsByType\%22\%3A\%7B\%22primaryParent\%22\%3A1\%2C\%22secondaryParent\%22\%3A4\%2C\%22cousin\%22\%3A6\%7D\%2C\%22nodeIds\%22\%3A\%5B\%5D\%7D\%2C\%22highlightImportantNodes\%22\%3A\%7B\%22highlightImportantNodes\%22\%3Afalse\%2C\%22highlightPrimaryParents\%22\%3Afalse\%2C\%22highlightRootConnectingEdges\%22\%3Afalse\%7D\%7D}

\begingroup
\small
\eczhBreakableDashes
\renewcommand\arraystretch{1.05}
\edef\myxtraspc{\eczListAddVSpaceXtraXtraStretch}
\begin{tabularx}{\linewidth}{>{\raggedright\arraybackslash}p{\eczListColWidth{name}} >{\hsize=1.0000\hsize }X}
\toprule
\eczListColTitle{Code} & \eczListColTitle{Description} \\
\midrule
\endfirsthead
\toprule
\eczListColTitleContinued{Code} & \eczListColTitleContinued{Description} \\
\midrule
\endhead
\bottomrule
\endfoot
\eczhRefIndex{code:bpsk}%
\eczhListValue{\flmRefsHyperref{code:bpsk}{Binary PSK (BPSK) modulation format}} & \eczhListValue{Encodes one bit of information into a constellation of antipodal points \(\pm\alpha\) for complex \(\alpha\).
These points are typically associated with two phases of an electromagnetic signal.}\\ 
\addlinespace[\myxtraspc]
\eczhRefIndex{code:fsk}%
\eczhListValue{\flmRefsHyperref{code:fsk}{Frequency-shift keying (FSK) modulation format}} & \eczhListValue{A \(q\)-ary frequency-shift keying (\(q\)-FSK) encodes one \(q\)-ary digit of information into signals with \(q\) different frequencies.
In its standard orthogonal form, each symbol is carried by one of \(q\) approximately orthogonal tones over a fixed symbol interval.}\\ 
\addlinespace[\myxtraspc]
\eczhRefIndex{code:modulation}%
\eczhListValue{\flmRefsHyperref{code:modulation}{Modulation scheme}} & \eczhListValue{A sphere packing mapped into a time-dependent electromagnetic signal \NoCaseChange{\protect\cite{cite193,cite194}}.
There is a close relation between abstract real-space encodings and modulation schemes, and certain simple sphere packings are often synonymous with their corresponding modulation schemes.}\\ 
\addlinespace[\myxtraspc]
\eczhRefIndex{code:psk}%
\eczhListValue{\flmRefsHyperref{code:psk}{Phase-shift keying (PSK) modulation format}} & \eczhListValue{A \(q\)-ary phase-shift keying (\(q\)-PSK) encodes one \(q\)-ary digit of information into a constellation of \(q\) points distributed equidistantly on a circle in \(\mathbb{C}\) or, equivalently, \(\mathbb{R}^2\).}\\ 
\addlinespace[\myxtraspc]
\eczhRefIndex{code:pam}%
\eczhListValue{\flmRefsHyperref{code:pam}{Pulse-amplitude modulation (PAM) format}} & \eczhListValue{Encodes a \(q\)-ary digit into a constellation of equally spaced points on the real line.
A standard \(q\)-PAM constellation can be written as \(\{(2i-q-1)\alpha\}_{i=1}^{q}\) for some real scaling factor \(\alpha\); for \(q=8\), this yields \(\{ \pm \alpha,\pm 3\alpha,\pm 5\alpha, \pm 7\alpha \}\).
The points in the constellation are typically associated with one quadrature of an electromagnetic signal.}\\ 
\addlinespace[\myxtraspc]
\eczhRefIndex{code:ppm}%
\eczhListValue{\flmRefsHyperref{code:ppm}{Pulse-position modulation (PPM) format}} & \eczhListValue{A modulation code with \(q\) equal-energy signals in which each codeword has one pulse in one of \(q\) time slots and zeros elsewhere.}\\ 
\addlinespace[\myxtraspc]
\eczhRefIndex{code:qpsk}%
\eczhListValue{\flmRefsHyperref{code:qpsk}{Quadrature PSK (QPSK) modulation format}} & \eczhListValue{A quaternary encoding into a constellation of four points distributed equidistantly on a circle.
For the case of \(\pi/4\)-QPSK, the constellation is \(\{e^{\pm i\frac{\pi}{4}},e^{\pm i\frac{3\pi}{4}}\}\).}\\ 
\addlinespace[\myxtraspc]
\eczhRefIndex{code:qam}%
\eczhListValue{\flmRefsHyperref{code:qam}{Quadrature-amplitude modulation (QAM) format}} & \eczhListValue{Encodes into a finite set of points in \(\mathbb{R}^{2}\), often treated as \(\mathbb{C}\).
Each point is associated with a complex amplitude of an electromagnetic signal, so information is encoded jointly in the in-phase and quadrature components \NoCaseChange{\protect\cite[{Ch. 16}]{cite194}}.}\\ 
\end{tabularx}\endgroup
\eczcodelist{q-ary_linear}{Non-binary linear codes
}%

\eczhCodeListAutoDescription{Codes that are descendants of \flmRefsCref{code:q-ary_linear} and not descendants of \flmRefsCref{code:binary_linear}.}%

\eczhIncludeCodeGraph{Bare}{scale=0.5}{\columnwidth}{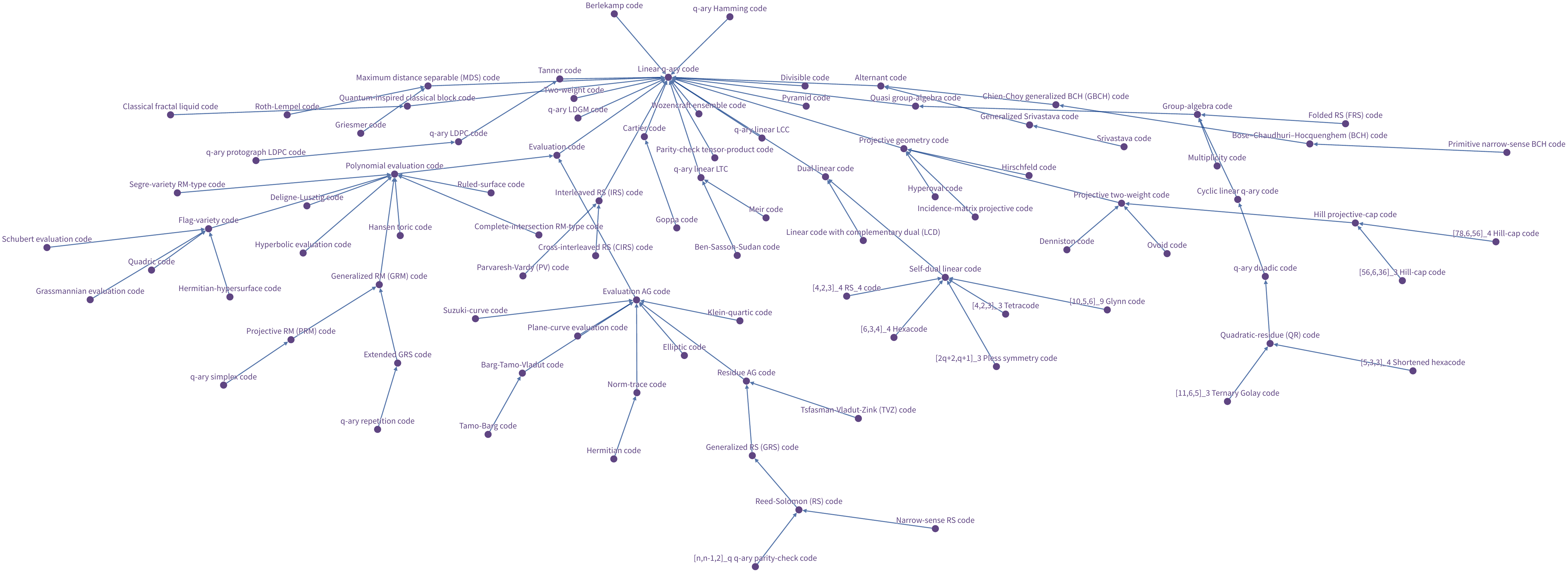}{Non-binary linear codes}{https://errorcorrectionzoo.org/code_graph#J\%7B\%22displayMode\%22\%3A\%22subset\%22\%2C\%22modeSubsetOptions\%22\%3A\%7B\%22codeIds\%22\%3A\%5B\%22alternant\%22\%2C\%22tamo_barg_vladut\%22\%2C\%22bs-ltc\%22\%2C\%22berlekamp\%22\%2C\%22q-ary_bch\%22\%2C\%22cartier\%22\%2C\%22gbch\%22\%2C\%22classical_fractal_liquid\%22\%2C\%22complete_intersections\%22\%2C\%22cross_interleaved_reed_solomon\%22\%2C\%22q-ary_cyclic\%22\%2C\%22deligne_lusztig\%22\%2C\%22denniston\%22\%2C\%22divisible\%22\%2C\%22dual\%22\%2C\%22elliptic\%22\%2C\%22evaluation\%22\%2C\%22evaluation_varieties\%22\%2C\%22extended_reed_solomon\%22\%2C\%22flag_variety\%22\%2C\%22folded_reed_solomon\%22\%2C\%22generalized_reed_muller\%22\%2C\%22generalized_reed_solomon\%22\%2C\%22generalized_srivastava\%22\%2C\%22goppa\%22\%2C\%22grassmannian_variety\%22\%2C\%22griesmer\%22\%2C\%22group\%22\%2C\%22toric_classical\%22\%2C\%22hermitian\%22\%2C\%22hermitian_hypersurface\%22\%2C\%22hill_cap\%22\%2C\%22hirschfeld\%22\%2C\%22cascaded_reed_solomon\%22\%2C\%22hyperoval\%22\%2C\%22incidence_matrix\%22\%2C\%22interleaved_reed_solomon\%22\%2C\%22klein_quartic\%22\%2C\%22q-ary_linear\%22\%2C\%22lcd\%22\%2C\%22mds\%22\%2C\%22meir\%22\%2C\%22multiplicity\%22\%2C\%22narrow_sense_reed_solomon\%22\%2C\%22norm_trace\%22\%2C\%22bose_qvist\%22\%2C\%22parity_check_tensor\%22\%2C\%22parvaresh_vardy\%22\%2C\%22plane_curve\%22\%2C\%22evaluation_polynomial\%22\%2C\%22narrow_sense_q-ary_bch\%22\%2C\%22projective_reed_muller\%22\%2C\%22projective\%22\%2C\%22projective_two_weight\%22\%2C\%22pyramid\%22\%2C\%22q-ary_quad_residue\%22\%2C\%22quadric\%22\%2C\%22quantum_inspired\%22\%2C\%22quasi_group\%22\%2C\%22reed_solomon\%22\%2C\%22residue\%22\%2C\%22roth_lempel\%22\%2C\%22ruled_surface\%22\%2C\%22schubert\%22\%2C\%22serge\%22\%2C\%22self_dual\%22\%2C\%22srivastava\%22\%2C\%22suzuki\%22\%2C\%22tamo_barg\%22\%2C\%22tanner\%22\%2C\%22shimura\%22\%2C\%22two_weight\%22\%2C\%22wozencraft\%22\%2C\%22glynn\%22\%2C\%22ternary_golay\%22\%2C\%22pless_symmetry\%22\%2C\%22tetracode\%22\%2C\%22reed_solomon_4\%22\%2C\%22shortened_hexacode\%22\%2C\%22hill_56_6_36\%22\%2C\%22hexacode\%22\%2C\%22hill_78_6_56\%22\%2C\%22q-ary_parity_check\%22\%2C\%22q-ary_hamming\%22\%2C\%22q-ary_ldgm\%22\%2C\%22q-ary_ldpc\%22\%2C\%22q-ary_duadic\%22\%2C\%22q-ary_lcc\%22\%2C\%22q-ary_ltc\%22\%2C\%22q-ary_protograph_ldpc\%22\%2C\%22q-ary_repetition\%22\%2C\%22q-ary_simplex\%22\%5D\%2C\%22reusePreviousLayoutPositions\%22\%3Afalse\%2C\%22showIntermediateConnectingNodes\%22\%3Atrue\%2C\%22connectingNodesMaxDepth\%22\%3A15\%2C\%22connectingNodesPathMaxLength\%22\%3A20\%2C\%22connectingNodesMaxExtraDepth\%22\%3A3\%2C\%22connectingNodesOnlyKeepPathsWithAdditionalLength\%22\%3A1\%2C\%22connectingNodesToDomainsAndKingdoms\%22\%3Afalse\%2C\%22connectingNodesEdgeLengthsByType\%22\%3A\%7B\%22primaryParent\%22\%3A1\%2C\%22secondaryParent\%22\%3A4\%2C\%22cousin\%22\%3A6\%7D\%2C\%22nodeIds\%22\%3A\%5B\%5D\%7D\%2C\%22highlightImportantNodes\%22\%3A\%7B\%22highlightImportantNodes\%22\%3Afalse\%2C\%22highlightPrimaryParents\%22\%3Afalse\%2C\%22highlightRootConnectingEdges\%22\%3Afalse\%7D\%7D}

\begingroup
\small
\eczhBreakableDashes
\renewcommand\arraystretch{1.05}
\edef\myxtraspc{\eczListAddVSpaceXtraXtraStretch}
\begin{tabularx}{\linewidth}{>{\raggedright\arraybackslash}p{\eczListColWidth{name}} >{\hsize=1.0000\hsize }X}
\toprule
\eczListColTitle{Code} & \eczListColTitle{Description} \\
\midrule
\endfirsthead
\toprule
\eczListColTitleContinued{Code} & \eczListColTitleContinued{Description} \\
\midrule
\endhead
\bottomrule
\endfoot
\eczhRefIndex{code:alternant}%
\eczhListValue{\flmRefsHyperref{code:alternant}{Alternant code}} & \eczhListValue{Given a length-\(n\) GRS code \(C\) over \(\mathbb{F}_{q^m}\), an alternant code is the \(\mathbb{F}_q\)-\flmRefsHyperref{ref33}{subfield} subcode of the dual of \(C\); see \NoCaseChange{\protect\cite[{Ch. 12}]{cite41}}. Its parity-check matrix is an alternant matrix.}\\ 
\addlinespace[\myxtraspc]
\eczhRefIndex{code:tamo_barg_vladut}%
\eczhListValue{\flmRefsHyperref{code:tamo_barg_vladut}{Barg-Tamo-Vladut code}} & \eczhListValue{Evaluation AG code on algebraic curves built from a Galois cover \(\phi:Y\to X\), where the recovery sets are fibres over rational points of \(X\) that split completely in the cover \NoCaseChange{\protect\cite[{Def. 15.9.10}]{cite26}\protect\cite[{Thm. 15.9.14}]{cite26}}.
The Barg-Tamo-Vladut construction generalizes the Tamo-Barg construction from \(PG(1,q)\) to longer AG codes, and variants can be built with higher local distance or availability \(2\) via fibre products of curves \NoCaseChange{\protect\cite[{Thm. 15.9.19}]{cite26}\protect\cite[{Thm. 15.9.21}]{cite26}}.}\\ 
\addlinespace[\myxtraspc]
\eczhRefIndex{code:bs-ltc}%
\eczhListValue{\flmRefsHyperref{code:bs-ltc}{Ben-Sasson-Sudan code}} & \eczhListValue{Locally testable \([n,k/2,d]_{2^m}\) code with \(k\) a power of two, \(n = k \log^{c} k\), and query complexity \(\log^{c}k\) for some universal constant \(c\).}\\ 
\addlinespace[\myxtraspc]
\eczhRefIndex{code:berlekamp}%
\eczhListValue{\flmRefsHyperref{code:berlekamp}{Berlekamp code}} & \eczhListValue{A linear \(p\)-ary code (for prime \(p\)) that has Lee distance 5 and whose construction resembles that of RS codes.
It is obtained by first constructing an RS-like parity-check matrix out of a certain \flmRefsHyperref{ref33}{field extension} of \(\mathbb{F}_p\) and then taking the \flmRefsHyperref{ref33}{subfield} subcode of the corresponding code; see \NoCaseChange{\protect\cite[{Ch. 10.6}]{cite195}}.}\\ 
\addlinespace[\myxtraspc]
\eczhRefIndex{code:q-ary_bch}%
\eczhListValue{\flmRefsHyperref{code:q-ary_bch}{Bose–Chaudhuri–Hocquenghem (BCH) code}} & \eczhListValue{A cyclic \(q\)-ary code, with \(n\) and \(q\) relatively prime, whose zeroes are consecutive powers of a primitive \(n\)th root of unity \(\alpha\).}\\ 
\addlinespace[\myxtraspc]
\eczhRefIndex{code:cartier}%
\eczhListValue{\flmRefsHyperref{code:cartier}{Cartier code}} & \eczhListValue{A generalization of the Goppa codes to codes defined from curves of nonzero genus. Each code is a subcode of a certain residue AG code and is constructed using the Cartier operator.}\\ 
\addlinespace[\myxtraspc]
\eczhRefIndex{code:gbch}%
\eczhListValue{\flmRefsHyperref{code:gbch}{Chien-Choy generalized BCH (GBCH) code}} & \eczhListValue{An \([n,k\geq n-rm, d\geq r+1]_q\) alternant code defined using two polynomials \(P(x),G(x)\) that are relatively prime to \(x^n-1\), with \(\deg P \leq n-1\) and \(r = \deg G \leq n-1\).}\\ 
\addlinespace[\myxtraspc]
\eczhRefIndex{code:classical_fractal_liquid}%
\eczhListValue{\flmRefsHyperref{code:classical_fractal_liquid}{Classical fractal liquid code}} & \eczhListValue{Member of a family of \([L^D,O(L^{D-1}),O(L^{D-\epsilon})]_p\) linear codes on \(D\)-dimensional square lattices of side length \(L\) and for prime \(p\) and \(\epsilon > 0\) that is based on \(p\)-ary generalizations of the Sierpinski triangle.}\\ 
\addlinespace[\myxtraspc]
\eczhRefIndex{code:complete_intersections}%
\eczhListValue{\flmRefsHyperref{code:complete_intersections}{Complete-intersection RM-type code}} & \eczhListValue{Evaluation code of polynomials evaluated on points lying on a complete intersection.}\\ 
\addlinespace[\myxtraspc]
\eczhRefIndex{code:cross_interleaved_reed_solomon}%
\eczhListValue{\flmRefsHyperref{code:cross_interleaved_reed_solomon}{Cross-interleaved RS (CIRS) code}} & \eczhListValue{An IRS code constructed from two shortened RS codes and two forms of interleaving. In a 2D array visualization, one component code runs vertically and the other diagonally \NoCaseChange{\protect\cite{cite189}}.}\\ 
\addlinespace[\myxtraspc]
\eczhRefIndex{code:q-ary_cyclic}%
\eczhListValue{\flmRefsHyperref{code:q-ary_cyclic}{Cyclic linear \(q\)-ary code}} & \eczhListValue{A \(q\)-ary code of length \(n\) is cyclic if, for each codeword \(c_1 c_2 \cdots c_n\), the cyclically shifted string \(c_n c_1 \cdots c_{n-1}\) is also a codeword. A cyclic code is called \textit{primitive} when \(n=q^r-1\) for some \(r\geq 2\). A \textit{shortened cyclic code} is obtained from a cyclic code by taking only codewords with the first \(j\) zero entries, and deleting those zeroes.}\\ 
\addlinespace[\myxtraspc]
\eczhRefIndex{code:deligne_lusztig}%
\eczhListValue{\flmRefsHyperref{code:deligne_lusztig}{Deligne-Lusztig code}} & \eczhListValue{Evaluation code of polynomials evaluated on points lying on a Deligne-Lusztig variety, often a Deligne-Lusztig curve in the classical one-dimensional cases.}\\ 
\addlinespace[\myxtraspc]
\eczhRefIndex{code:denniston}%
\eczhListValue{\flmRefsHyperref{code:denniston}{Denniston code}} & \eczhListValue{Projective code that is part of a family of \([2^{a+i}+2^i-2^a,3,2^{a+i}-2^a]_{2^a}\) codes for \(i < a\) constructed using Denniston arcs \NoCaseChange{\protect\cite[{Sec. 19.7.3}]{cite172}}.}\\ 
\addlinespace[\myxtraspc]
\eczhRefIndex{code:divisible}%
\eczhListValue{\flmRefsHyperref{code:divisible}{Divisible code}} & \eczhListValue{A linear \(q\)-ary block code is \(\Delta\)-divisible if the Hamming weight of each of its codewords is divisible by divisor \(\Delta\).
A \(2\)-divisible (\(4\)-divisible, \(8\)-divisible) code is called \textit{even} (\textit{doubly even}, \textit{triply even}) \NoCaseChange{\protect\cite{cite170,cite39}}.
A code is called \textit{singly even} if all codewords are even and at least one has weight equal to 2 modulo 4.
More generally, a code is \(m\)\textit{-even} if it is \(2^{m}\)-divisible.}\\ 
\addlinespace[\myxtraspc]
\eczhRefIndex{code:dual}%
\eczhListValue{\flmRefsHyperref{code:dual}{Dual linear code}} & \eczhListValue{For any \([n,k]_q\) linear code \(C\), the dual code is the set of \(q\)-ary strings that are orthogonal to the codewords of \(C\) under a particular inner product.}\\ 
\addlinespace[\myxtraspc]
\eczhRefIndex{code:elliptic}%
\eczhListValue{\flmRefsHyperref{code:elliptic}{Elliptic code}} & \eczhListValue{Evaluation AG code of rational functions evaluated on points lying on an elliptic curve, i.e., a curve of genus one.}\\ 
\addlinespace[\myxtraspc]
\eczhRefIndex{code:evaluation}%
\eczhListValue{\flmRefsHyperref{code:evaluation}{Evaluation AG code}} & \eczhListValue{Evaluation code over \(\mathbb{F}_q\) on a set of points \({\cal P} = \left( P_1,P_2,\cdots,P_n \right)\) lying on an algebraic curve \(\cal X\) defined over \(\mathbb{F}_q\), where the corresponding vector space \(L\) of functions \(f\) consists of certain rational functions (or, in special cases, polynomials).}\\ 
\addlinespace[\myxtraspc]
\eczhRefIndex{code:evaluation_varieties}%
\eczhListValue{\flmRefsHyperref{code:evaluation_varieties}{Evaluation code}} & \eczhListValue{Code whose codewords are evaluations of functions at certain fixed points. Code properties can be inferred from the structure of the functions and the underlying geometric object containing the points, often using results from algebraic geometry.}\\ 
\addlinespace[\myxtraspc]
\eczhRefIndex{code:extended_reed_solomon}%
\eczhListValue{\flmRefsHyperref{code:extended_reed_solomon}{Extended GRS code}} & \eczhListValue{A GRS code extended by one extra coordinate to form an \([n+1,k,n-k+2]_q\) MDS code. In projective language, this corresponds to adding one more evaluation point, often interpreted as the point at infinity; in suitable equivalent descriptions, one may instead use an affine point such as \(0\). The case when \(n=q-1\), multipliers \(v_i=1\), and \(\alpha_i\) are \(i-1\)st powers of a primitive \(n\)th root of unity is an \textit{extended narrow-sense RS code}.}\\ 
\addlinespace[\myxtraspc]
\eczhRefIndex{code:flag_variety}%
\eczhListValue{\flmRefsHyperref{code:flag_variety}{Flag-variety code}} & \eczhListValue{Evaluation code of polynomials evaluated on points lying on a flag variety.}\\ 
\addlinespace[\myxtraspc]
\eczhRefIndex{code:folded_reed_solomon}%
\eczhListValue{\flmRefsHyperref{code:folded_reed_solomon}{Folded RS (FRS) code}} & \eczhListValue{A code obtained from an RS code by bundling consecutive symbols into larger alphabet symbols. This preserves the algebraic structure of the parent RS code while lowering the block length over an extension alphabet, and it is a key ingredient in capacity-approaching list-decoding constructions.}\\ 
\addlinespace[\myxtraspc]
\eczhRefIndex{code:generalized_reed_muller}%
\eczhListValue{\flmRefsHyperref{code:generalized_reed_muller}{Generalized RM (GRM) code}} & \eczhListValue{Extensions of RM codes to \(q\)-ary coordinates that can be described as multivariate polynomial evaluation codes over affine or projective space.}\\ 
\addlinespace[\myxtraspc]
\eczhRefIndex{code:generalized_reed_solomon}%
\eczhListValue{\flmRefsHyperref{code:generalized_reed_solomon}{Generalized RS (GRS) code}} & \eczhListValue{An \([n,k,n-k+1]_q\) MDS code that is a modification of the RS code where codeword polynomials are multiplied by additional prefactors \NoCaseChange{\protect\cite[{Def. 15.3.19}]{cite26}}.}\\ 
\addlinespace[\myxtraspc]
\eczhRefIndex{code:generalized_srivastava}%
\eczhListValue{\flmRefsHyperref{code:generalized_srivastava}{Generalized Srivastava code}} & \eczhListValue{An \([n,k \geq n-mst,d \geq st+1 ]_q\) alternant code defined for \(n+s\) distinct elements \(\alpha_1,\alpha_2,\cdots,\alpha_n,w_1,w_2,\cdots,w_s\) and \(n\) nonzero elements \(z_1,z_2,\cdots,z_n\) of \(\mathbb{F}_{q^m}\).}\\ 
\addlinespace[\myxtraspc]
\eczhRefIndex{code:goppa}%
\eczhListValue{\flmRefsHyperref{code:goppa}{Goppa code}} & \eczhListValue{A linear \(q\)-ary code defined from a polynomial \(G(x)\) over an extension field and a set of evaluation points \(L\) avoiding the roots of \(G\).
Goppa codes form a central family of alternant codes, admit efficient algebraic decoding algorithms, and include the binary Goppa codes used in the McEliece cryptosystem.
When the base field equals the coefficient field, they coincide with residue AG codes on \(PG(1,q^m)\); in general, classical Goppa codes are subfield subcodes of such AG codes \NoCaseChange{\protect\cite[{Rem. 15.3.27}]{cite26}\protect\cite[{Thm. 15.3.28}]{cite26}}.}\\ 
\addlinespace[\myxtraspc]
\eczhRefIndex{code:grassmannian_variety}%
\eczhListValue{\flmRefsHyperref{code:grassmannian_variety}{Grassmannian evaluation code}} & \eczhListValue{Evaluation code of polynomials evaluated on points lying on a finite-field Grassmannian embedded into projective space using the Plucker embedding \NoCaseChange{\protect\cite{cite27,cite28}}.}\\ 
\addlinespace[\myxtraspc]
\eczhRefIndex{code:griesmer}%
\eczhListValue{\flmRefsHyperref{code:griesmer}{Griesmer code}} & \eczhListValue{A type of \(q\)-ary code whose parameters satisfy the Griesmer bound with equality.}\\ 
\addlinespace[\myxtraspc]
\eczhRefIndex{code:group}%
\eczhListValue{\flmRefsHyperref{code:group}{Group-algebra code}} & \eczhListValue{An \( [n,k]_q \) code associated with a finite group \(G\) of order \(n\), viewed as an ideal in the group algebra \(\mathbb{F}_q[G]\) \NoCaseChange{\protect\cite[{Def. 16.4.3}]{cite196}}.
Equivalently, after identifying the \(n\) coordinate positions of each codeword with elements of \(G\), the code is invariant under the regular action of \(G\) and thus becomes a \(G\)-submodule of \(\mathbb{F}_q^n\) \NoCaseChange{\protect\cite{cite198}\protect\cite[{Lemma 2.3}]{cite197}}.
A group-algebra code for an Abelian group is called an \textit{Abelian group-algebra code}.}\\ 
\addlinespace[\myxtraspc]
\eczhRefIndex{code:toric_classical}%
\eczhListValue{\flmRefsHyperref{code:toric_classical}{Hansen toric code}} & \eczhListValue{Evaluation code of a linear space of polynomials evaluated on points lying on an affine or projective toric variety. If the space is taken to be all polynomials up to some degree, the code is called a \textit{toric RM-type code} of that degree.}\\ 
\addlinespace[\myxtraspc]
\eczhRefIndex{code:hermitian}%
\eczhListValue{\flmRefsHyperref{code:hermitian}{Hermitian code}} & \eczhListValue{Evaluation AG code of rational functions on a Hermitian curve over \(\mathbb{F}_{q^2}\).}\\ 
\addlinespace[\myxtraspc]
\eczhRefIndex{code:hermitian_hypersurface}%
\eczhListValue{\flmRefsHyperref{code:hermitian_hypersurface}{Hermitian-hypersurface code}} & \eczhListValue{Evaluation code of polynomials evaluated on points lying on a Hermitian hypersurface.}\\ 
\addlinespace[\myxtraspc]
\eczhRefIndex{code:hill_cap}%
\eczhListValue{\flmRefsHyperref{code:hill_cap}{Hill projective-cap code}} & \eczhListValue{Member of a projective code family that contains two \(q\)-ary sharp configurations \NoCaseChange{\protect\cite[{Table 12.1}]{cite199}} and that is constructed using projective caps.}\\ 
\addlinespace[\myxtraspc]
\eczhRefIndex{code:hirschfeld}%
\eczhListValue{\flmRefsHyperref{code:hirschfeld}{Hirschfeld code}} & \eczhListValue{A \([q+1,4,q-2]_q\) projective geometry code for non-prime \(q\) that is an example of an MDS code that is not an RS code; see \NoCaseChange{\protect\cite[{Exam. 7.6}]{cite182}} for the generator matrix.}\\ 
\addlinespace[\myxtraspc]
\eczhRefIndex{code:cascaded_reed_solomon}%
\eczhListValue{\flmRefsHyperref{code:cascaded_reed_solomon}{Hyperbolic evaluation code}} & \eczhListValue{An evaluation code over polynomials in two variables. 
Generator matrices are determined in Ref. \NoCaseChange{\protect\cite{cite29}} following initial formulations of the codes as generalized concatenations of RS codes \NoCaseChange{\protect\cite{cite30,cite31}}; see \NoCaseChange{\protect\cite[{Exam. 4.26}]{cite32}}.}\\ 
\addlinespace[\myxtraspc]
\eczhRefIndex{code:hyperoval}%
\eczhListValue{\flmRefsHyperref{code:hyperoval}{Hyperoval code}} & \eczhListValue{Projective code constructed from a hyperoval in the projective plane \(PG(2,q)\), where \(q\) is even. Since a hyperoval is a set of \(q+2\) points with no three collinear, the corresponding projective code has parameters \([q+2,3,q]_q\) \NoCaseChange{\protect\cite[{Exam. 19.2.1}]{cite172}}; the \([6,3,4]_4\) hexacode is the smallest example.}\\ 
\addlinespace[\myxtraspc]
\eczhRefIndex{code:incidence_matrix}%
\eczhListValue{\flmRefsHyperref{code:incidence_matrix}{Incidence-matrix projective code}} & \eczhListValue{A projective code whose generator matrix is the incidence matrix of points and hyperplanes in a projective space.
This construction has been generalized to incidence matrices of other structures \NoCaseChange{\protect\cite{cite200,cite201}\protect\cite[{Sec. 14.4}]{cite202}}.
More generally, columns of a code's parity-check matrix can also be organized as an incidence matrix.}\\ 
\addlinespace[\myxtraspc]
\eczhRefIndex{code:interleaved_reed_solomon}%
\eczhListValue{\flmRefsHyperref{code:interleaved_reed_solomon}{Interleaved RS (IRS) code}} & \eczhListValue{A modification of RS codes where multiple polynomials are used to define each codeword.}\\ 
\addlinespace[\myxtraspc]
\eczhRefIndex{code:klein_quartic}%
\eczhListValue{\flmRefsHyperref{code:klein_quartic}{Klein-quartic code}} & \eczhListValue{Evaluation AG code over \(\mathbb{F}_8\) of rational functions evaluated on points lying on the Klein quartic, which is defined by the equation \(x^3 y + y^3 z + z^3 x = 0\) \NoCaseChange{\protect\cite[{Ex. 2.75}]{cite32}}.}\\ 
\addlinespace[\myxtraspc]
\eczhRefIndex{code:q-ary_linear}%
\eczhListValue{\flmRefsHyperref{code:q-ary_linear}{Linear \(q\)-ary code}} & \eczhListValue{An \((n,K,d)_q\) linear code is denoted as \([n,k,d]_q\), where \(k=\log_q K\) is an integer. Its codewords form a linear subspace, i.e., for any codewords \(x,y\), \(\alpha x+ \beta y\) is also a codeword for any field elements \(\alpha,\beta \in \mathbb{F}_q\).
This extra structure yields much information about their properties, making them a large and well-studied subset of codes.}\\ 
\addlinespace[\myxtraspc]
\eczhRefIndex{code:lcd}%
\eczhListValue{\flmRefsHyperref{code:lcd}{Linear code with complementary dual (LCD)}} & \eczhListValue{A linear code \(C\) admits a complementary dual if \(C\) and its dual code \(C^{\perp}\) do not share any nonzero codewords. Equivalently, \(\mathbb{F}_q^n = C \oplus C^{\perp}\).}\\ 
\addlinespace[\myxtraspc]
\eczhRefIndex{code:mds}%
\eczhListValue{\flmRefsHyperref{code:mds}{Maximum distance separable (MDS) code}} & \eczhListValue{A \(q\)-ary linear code whose parameters satisfy the Singleton bound with equality.}\\ 
\addlinespace[\myxtraspc]
\eczhRefIndex{code:meir}%
\eczhListValue{\flmRefsHyperref{code:meir}{Meir code}} & \eczhListValue{Locally testable \([n,k,d]_q\) code with query complexity \(\text{poly}(\log k)\) and rejection ratio \(R/n = 1/\text{poly}(\log k)\). The code construction is probabilistic and combinatorial.}\\ 
\addlinespace[\myxtraspc]
\eczhRefIndex{code:multiplicity}%
\eczhListValue{\flmRefsHyperref{code:multiplicity}{Multiplicity code}} & \eczhListValue{A generalization of an \(m\)-variate polynomial evaluation code based on evaluating polynomials together with their Hasse derivatives up to order \(s-1\) at all points in \(\mathbb{F}_q^m\).
Originally proposed for coding using the Rosenbloom-Tsfasman metric \NoCaseChange{\protect\cite{cite179}}.
Univariate (\(m=1\)) \NoCaseChange{\protect\cite{cite179,cite180}} and multivariate (\(m>1\)) \NoCaseChange{\protect\cite{cite181}} codes have been proposed.}\\ 
\addlinespace[\myxtraspc]
\eczhRefIndex{code:narrow_sense_reed_solomon}%
\eczhListValue{\flmRefsHyperref{code:narrow_sense_reed_solomon}{Narrow-sense RS code}} & \eczhListValue{An \([q-1,k,n-k+1]_q\) RS code whose points \(\alpha_i\) are all \((i-1)\)st powers of a \flmRefsHyperref{ref33}{primitive} element \(\alpha\) of \(\mathbb{F}_q\).}\\ 
\addlinespace[\myxtraspc]
\eczhRefIndex{code:norm_trace}%
\eczhListValue{\flmRefsHyperref{code:norm_trace}{Norm-trace code}} & \eczhListValue{Evaluation AG code of rational functions evaluated on points lying on a Miura-Kamiya curve in either affine or projective space.
The family is named as such because the equations defining the curves can be expressed in terms of the \flmRefsHyperref{ref33}{field norm and field trace}.}\\ 
\addlinespace[\myxtraspc]
\eczhRefIndex{code:bose_qvist}%
\eczhListValue{\flmRefsHyperref{code:bose_qvist}{Ovoid code}} & \eczhListValue{Member of a \([q^2+1,4,q^2-q]_q\) projective two-weight code family obtained from ovoids in \(\mathrm{PG}(3,q)\).
If the columns of a generator matrix are the \(q^2+1\) points of an ovoid, then every hyperplane meets the ovoid in either \(1\) or \(q+1\) points, yielding the two nonzero weights \(q^2\) and \(q^2-q\).
See \NoCaseChange{\protect\cite[{pg. 107}]{cite203}\protect\cite[{pg. 192}]{cite62}} for further details.}\\ 
\addlinespace[\myxtraspc]
\eczhRefIndex{code:parity_check_tensor}%
\eczhListValue{\flmRefsHyperref{code:parity_check_tensor}{Parity-check tensor-product code}} & \eczhListValue{A \(q\)-ary linear code constructed out of two \(q\)-ary linear codes with parity-check matrices \(H_A,H_B\) such that its parity-check matrix is \(H_A \otimes H_B\).  
Its dual has parity-check matrix \(G_A\otimes G_B\), where \(G_{A,B}\) are the generator matrices of the two underlying codes \NoCaseChange{\protect\cite{cite204}}.}\\ 
\addlinespace[\myxtraspc]
\eczhRefIndex{code:parvaresh_vardy}%
\eczhListValue{\flmRefsHyperref{code:parvaresh_vardy}{Parvaresh-Vardy (PV) code}} & \eczhListValue{An IRS code with additional algebraic relations (a.k.a. correlations) between the codeword polynomials \(\{f^{(j)}\}_{j=1}^{t}\). These relations yield a list decoder that achieves list-decoding capacity.}\\ 
\addlinespace[\myxtraspc]
\eczhRefIndex{code:plane_curve}%
\eczhListValue{\flmRefsHyperref{code:plane_curve}{Plane-curve evaluation code}} & \eczhListValue{Evaluation AG code of bivariate polynomials of some finite maximum degree, evaluated at points lying on an affine or projective plane curve.}\\ 
\addlinespace[\myxtraspc]
\eczhRefIndex{code:evaluation_polynomial}%
\eczhListValue{\flmRefsHyperref{code:evaluation_polynomial}{Polynomial evaluation code}} & \eczhListValue{Evaluation code of polynomials (or, more generally, rational functions) at points \({\cal P} = \left( P_1,P_2,\cdots,P_n \right)\) on an algebraic variety \(\cal X\) of dimension greater than one (i.e., not an algebraic curve).}\\ 
\addlinespace[\myxtraspc]
\eczhRefIndex{code:narrow_sense_q-ary_bch}%
\eczhListValue{\flmRefsHyperref{code:narrow_sense_q-ary_bch}{Primitive narrow-sense BCH code}} & \eczhListValue{A \(q\)-ary BCH code for \(b=1\) and for \(n=q^r-1\) for some \(r\geq 2\).}\\ 
\addlinespace[\myxtraspc]
\eczhRefIndex{code:projective_reed_muller}%
\eczhListValue{\flmRefsHyperref{code:projective_reed_muller}{Projective RM (PRM) code}} & \eczhListValue{Evaluation code obtained by evaluating homogeneous polynomials on the points of the projective space \(PG(m,q)\), equivalently on representatives of the nonzero vectors in \(\mathbb{F}_q^{m+1}\) whose leftmost nonzero coordinate is one.}\\ 
\addlinespace[\myxtraspc]
\eczhRefIndex{code:projective}%
\eczhListValue{\flmRefsHyperref{code:projective}{Projective geometry code}} & \eczhListValue{Linear \(q\)-ary \([n,k,d]\) code whose generator matrix \(G\) does not contain any repeated columns or the zero column.
That way, each column corresponds to a distinct point in the projective space \(PG(k-1,q)\) arising from a \(k\)-dimensional vector space over \(\mathbb{F}_q\). A choice of \(k\) linearly independent columns determines an \textit{information set}.
Columns of a code's parity-check matrix can similarly correspond to points in projective space. This formulation yields connections to projective geometry, which can be applied to determine code properties.}\\ 
\addlinespace[\myxtraspc]
\eczhRefIndex{code:projective_two_weight}%
\eczhListValue{\flmRefsHyperref{code:projective_two_weight}{Projective two-weight code}} & \eczhListValue{A projective code whose codewords all have one of two possible nonzero Hamming weights.}\\ 
\addlinespace[\myxtraspc]
\eczhRefIndex{code:pyramid}%
\eczhListValue{\flmRefsHyperref{code:pyramid}{Pyramid code}} & \eczhListValue{An LRC whose generator matrix is that of an RS code in standard form, but some of whose columns are split into multiple columns; see \NoCaseChange{\protect\cite[{Sec. 31.3.1.1}]{cite183}} for an example.}\\ 
\addlinespace[\myxtraspc]
\eczhRefIndex{code:q-ary_quad_residue}%
\eczhListValue{\flmRefsHyperref{code:q-ary_quad_residue}{Quadratic-residue (QR) code}} & \eczhListValue{Member of a quadruple of cyclic \(q\)-ary codes of prime length \(n\) where \(q\) is prime and a quadratic-residue modulo \(n\) \NoCaseChange{\protect\cite[{Def. 3.2.8}]{cite70}}.
The codes are constructed using quadratic residues and nonresidues of \(n\).
The definition extends to prime-power alphabet sizes and to prime-power lengths \NoCaseChange{\protect\cite{cite174,cite175}\protect\cite[{Rem. 3.2.9}]{cite70}}.
A quadratic-residue code of prime length \(p\) has dimension \((p+1)/2\) \NoCaseChange{\protect\cite[{Sec. 3.2.1}]{cite70}}.}\\ 
\addlinespace[\myxtraspc]
\eczhRefIndex{code:quadric}%
\eczhListValue{\flmRefsHyperref{code:quadric}{Quadric code}} & \eczhListValue{Evaluation code of polynomials evaluated on points lying on a quadric hypersurface.}\\ 
\addlinespace[\myxtraspc]
\eczhRefIndex{code:quantum_inspired}%
\eczhListValue{\flmRefsHyperref{code:quantum_inspired}{Quantum-inspired classical block code}} & \eczhListValue{A \(q\)-ary linear code whose construction was inspired by a quantum code.}\\ 
\addlinespace[\myxtraspc]
\eczhRefIndex{code:quasi_group}%
\eczhListValue{\flmRefsHyperref{code:quasi_group}{Quasi group-algebra code}} & \eczhListValue{A \(q\)-ary linear code based on a finite group \( G \) of order \(n/\ell\) for some \textit{index} \(\ell\).
The code is a right submodule of the direct sum of \(\ell\) copies of the \flmRefsHyperref{ref205}{group algebra} \(\mathbb{F}_q G\).
A quasi group-algebra code for an Abelian group is called an \textit{Abelian quasi group-algebra code}.}\\ 
\addlinespace[\myxtraspc]
\eczhRefIndex{code:reed_solomon}%
\eczhListValue{\flmRefsHyperref{code:reed_solomon}{Reed-Solomon (RS) code}} & \eczhListValue{An \([n,k,n-k+1]_q\) linear code based on polynomials over \(\mathbb{F}_q\).}\\ 
\addlinespace[\myxtraspc]
\eczhRefIndex{code:residue}%
\eczhListValue{\flmRefsHyperref{code:residue}{Residue AG code}} & \eczhListValue{Linear \(q\)-ary code defined using a set of \(\mathbb{F}_q\)-rational points \({\cal P} = \left( P_1,P_2,\cdots,P_n \right)\) on an algebraic curve \(\cal X\) and a linear space \(\Omega\) of certain rational differential forms \(\omega\) \NoCaseChange{\protect\cite[{Def. 15.3.2}]{cite26}}.}\\ 
\addlinespace[\myxtraspc]
\eczhRefIndex{code:roth_lempel}%
\eczhListValue{\flmRefsHyperref{code:roth_lempel}{Roth-Lempel code}} & \eczhListValue{Member of a \(q\)-ary linear code family that includes many examples of MDS codes that are not GRS codes.}\\ 
\addlinespace[\myxtraspc]
\eczhRefIndex{code:ruled_surface}%
\eczhListValue{\flmRefsHyperref{code:ruled_surface}{Ruled-surface code}} & \eczhListValue{Evaluation code obtained by evaluating global sections of a line bundle, or equivalently suitable polynomial functions, on rational points of a ruled surface over a finite field. Such codes extend algebraic-geometry constructions from curves to certain projective surfaces \NoCaseChange{\protect\cite{cite34,cite35}}.}\\ 
\addlinespace[\myxtraspc]
\eczhRefIndex{code:schubert}%
\eczhListValue{\flmRefsHyperref{code:schubert}{Schubert evaluation code}} & \eczhListValue{Evaluation code of polynomials evaluated on points lying on a Schubert variety.}\\ 
\addlinespace[\myxtraspc]
\eczhRefIndex{code:serge}%
\eczhListValue{\flmRefsHyperref{code:serge}{Segre-variety RM-type code}} & \eczhListValue{Evaluation code of multihomogeneous polynomials evaluated on points of a Segre variety, i.e., on the Segre embedding of a product of projective spaces. These codes are Reed-Muller-type analogues adapted to product projective geometries \NoCaseChange{\protect\cite{cite36}}.}\\ 
\addlinespace[\myxtraspc]
\eczhRefIndex{code:self_dual}%
\eczhListValue{\flmRefsHyperref{code:self_dual}{Self-dual linear code}} & \eczhListValue{An \([n,n/2]_q\) code that is equal to its dual, \(C^\perp = C\), where the dual is defined with respect to an inner product, most commonly either Euclidean or Hermitian.
Self-dual codes exist only for even lengths and have dimension \(k=n/2\).
A code that is equivalent to its dual is called \textit{isodual}.
Any self-dual code is isodual, and hence formally self-dual \NoCaseChange{\protect\cite[{Rem. 4.2.2}]{cite40}}.}\\ 
\addlinespace[\myxtraspc]
\eczhRefIndex{code:srivastava}%
\eczhListValue{\flmRefsHyperref{code:srivastava}{Srivastava code}} & \eczhListValue{A special case of a generalized Srivastava code for \(z_j = \alpha_j^{\mu}\) for some \(\mu\) and \(t=1\).}\\ 
\addlinespace[\myxtraspc]
\eczhRefIndex{code:suzuki}%
\eczhListValue{\flmRefsHyperref{code:suzuki}{Suzuki-curve code}} & \eczhListValue{Evaluation AG code of rational functions evaluated on points lying on a Suzuki curve.}\\ 
\addlinespace[\myxtraspc]
\eczhRefIndex{code:tamo_barg}%
\eczhListValue{\flmRefsHyperref{code:tamo_barg}{Tamo-Barg code}} & \eczhListValue{A family of \(q\)-ary polynomial evaluation codes that are optimal LRCs and for which \(q\) is comparable to \(n\).}\\ 
\addlinespace[\myxtraspc]
\eczhRefIndex{code:tanner}%
\eczhListValue{\flmRefsHyperref{code:tanner}{Tanner code}} & \eczhListValue{A linear \(q\)-ary code defined on a bipartite graph similar to a Tanner graph.
This \textit{generalized Tanner graph} consists of variable nodes and constraint nodes, with the generalization being that the constraint nodes of degree \(r\) now represent a linear code of length \(r\).}\\ 
\addlinespace[\myxtraspc]
\eczhRefIndex{code:shimura}%
\eczhListValue{\flmRefsHyperref{code:shimura}{Tsfasman-Vladut-Zink (TVZ) code}} & \eczhListValue{Member of a family of AG codes obtained from algebraic curves via the residue or evaluation construction. Sequences of curves with many rational points, such as Drinfeld modular curves, classical modular curves, or Garcia-Stichtenoth curves, yield the asymptotic parameters of the TVZ bound \NoCaseChange{\protect\cite[{Sec. 15.4.2}]{cite26}}.}\\ 
\addlinespace[\myxtraspc]
\eczhRefIndex{code:two_weight}%
\eczhListValue{\flmRefsHyperref{code:two_weight}{Two-weight code}} & \eczhListValue{A linear \(q\)-ary code whose codewords all have one of two possible nonzero Hamming weights \NoCaseChange{\protect\cite[{Def. 19.1}]{cite172}}.}\\ 
\addlinespace[\myxtraspc]
\eczhRefIndex{code:wozencraft}%
\eczhListValue{\flmRefsHyperref{code:wozencraft}{Wozencraft ensemble code}} & \eczhListValue{A code that is part of the Wozencraft ensemble, a set of codes most of whose members achieve the \flmRefsHyperref{ref85}{GV bound}.}\\ 
\addlinespace[\myxtraspc]
\eczhRefIndex{code:glynn}%
\eczhListValue{\flmRefsHyperref{code:glynn}{\([10,5,6]_9\) Glynn code}} & \eczhListValue{The unique trace-Hermitian self-dual \([10,5,6]_9\) code, constructed using a 10-arc in \(PG(4,9)\) that is not a rational curve.}\\ 
\addlinespace[\myxtraspc]
\eczhRefIndex{code:ternary_golay}%
\eczhListValue{\flmRefsHyperref{code:ternary_golay}{\([11,6,5]_3\) Ternary Golay code}} & \eczhListValue{A \([11,6,5]_3\) perfect ternary linear code with connections to various areas of mathematics, e.g., lattices \NoCaseChange{\protect\cite{cite39}} and sporadic simple groups \NoCaseChange{\protect\cite{cite41}}.
Adding a parity bit to the code results in the self-dual \([12,6,6]_3\) \textit{extended ternary Golay code}, whose weight enumerator is the Gleason polynomial \(g_5\) \NoCaseChange{\protect\cite[{Rem. 4.2.6}]{cite40}}.
Up to equivalence, both codes are unique for their respective parameters \NoCaseChange{\protect\cite{cite102}}.
The dual of the ternary Golay code is a \([11,5,6]_3\) projective two-weight subcode \NoCaseChange{\protect\cite[{Exam. 19.3.2}]{cite172}}.}\\ 
\addlinespace[\myxtraspc]
\eczhRefIndex{code:pless_symmetry}%
\eczhListValue{\flmRefsHyperref{code:pless_symmetry}{\([2q+2,q+1]_3\) Pless symmetry code}} & \eczhListValue{A member of a family of self-dual ternary \([2q+2,q+1]_3\) codes for any power of an odd prime satisfying \(q \equiv 2\) modulo 3.}\\ 
\addlinespace[\myxtraspc]
\eczhRefIndex{code:tetracode}%
\eczhListValue{\flmRefsHyperref{code:tetracode}{\([4,2,3]_3\) Tetracode}} & \eczhListValue{The \([4,2,3]_3\) ternary self-dual MDS code that has connections to lattices \NoCaseChange{\protect\cite{cite39}}. Its weight enumerator is the Gleason polynomial \(g_4\) \NoCaseChange{\protect\cite[{Rem. 4.2.6}]{cite40}}.}\\ 
\addlinespace[\myxtraspc]
\eczhRefIndex{code:reed_solomon_4}%
\eczhListValue{\flmRefsHyperref{code:reed_solomon_4}{\([4,2,3]_4\) RS\(_4\) code}} & \eczhListValue{A Type II Euclidean self-dual extended RS code that is the smallest quaternary extended QR code \NoCaseChange{\protect\cite[{pg. 296}]{cite41}\protect\cite[{Sec. 2.4.2}]{cite42}}.
Puncturing the \([4,2,3]_4\) RS\(_4\) code yields the \([3,2,2]_4\) shortened RS\(_4\) code, which is an RS code \NoCaseChange{\protect\cite[{pg. 295}]{cite41}}.}\\ 
\addlinespace[\myxtraspc]
\eczhRefIndex{code:shortened_hexacode}%
\eczhListValue{\flmRefsHyperref{code:shortened_hexacode}{\([5,3,3]_4\) Shortened hexacode}} & \eczhListValue{A perfect \([5,3,3]_4\) quaternary Hamming code that is the result of puncturing the hexacode \NoCaseChange{\protect\cite{cite43}}.}\\ 
\addlinespace[\myxtraspc]
\eczhRefIndex{code:hill_56_6_36}%
\eczhListValue{\flmRefsHyperref{code:hill_56_6_36}{\([56,6,36]_3\) Hill-cap code}} & \eczhListValue{Projective two-weight ternary code based on the Games graph \NoCaseChange{\protect\cite{cite206}\protect\cite[{Table 19.1}]{cite172}} and Hill's 56-cap \NoCaseChange{\protect\cite{cite207}}.
Its automorphism group contains \(PSL(3,4)\) \NoCaseChange{\protect\cite{cite208}}.}\\ 
\addlinespace[\myxtraspc]
\eczhRefIndex{code:hexacode}%
\eczhListValue{\flmRefsHyperref{code:hexacode}{\([6,3,4]_4\) Hexacode}} & \eczhListValue{The \([6,3,4]_4\) Hermitian self-dual MDS code that has connections to projective geometry, lattices \NoCaseChange{\protect\cite{cite39}}, and conformal field theory \NoCaseChange{\protect\cite{cite44}}. Its weight enumerator is the Gleason polynomial \(g_7\) \NoCaseChange{\protect\cite[{Rem. 4.2.6}]{cite40}}.}\\ 
\addlinespace[\myxtraspc]
\eczhRefIndex{code:hill_78_6_56}%
\eczhListValue{\flmRefsHyperref{code:hill_78_6_56}{\([78,6,56]_4\) Hill-cap code}} & \eczhListValue{Projective two-weight quaternary code based on a cap that corresponds to a strongly regular graph \NoCaseChange{\protect\cite[{Table 7.1}]{cite206}}.}\\ 
\addlinespace[\myxtraspc]
\eczhRefIndex{code:q-ary_parity_check}%
\eczhListValue{\flmRefsHyperref{code:q-ary_parity_check}{\([n,n-1,2]_q\) \(q\)-ary parity-check code}} & \eczhListValue{An \([n,n-1,2]_q\) linear \(q\)-ary code whose codewords consist of the message string appended with a \textit{parity-check} or \textit{zero-sum check digit} such that the sum over all coordinates of each codeword is zero.}\\ 
\addlinespace[\myxtraspc]
\eczhRefIndex{code:q-ary_hamming}%
\eczhListValue{\flmRefsHyperref{code:q-ary_hamming}{\(q\)-ary Hamming code}} & \eczhListValue{Member of an infinite family of perfect linear \(q\)-ary codes with parameters \([(q^r-1)/(q-1),(q^r-1)/(q-1)-r, 3]_q\) for \(r \geq 2\) \NoCaseChange{\protect\cite[{(3.1)}]{cite70}}.
These are precisely the nontrivial perfect linear codes over \(\mathbb{F}_q\) \NoCaseChange{\protect\cite[{Thm. 3.3.1}]{cite70}}.}\\ 
\addlinespace[\myxtraspc]
\eczhRefIndex{code:q-ary_ldgm}%
\eczhListValue{\flmRefsHyperref{code:q-ary_ldgm}{\(q\)-ary LDGM code}} & \eczhListValue{\(q\)-ary linear code with a sparse generator matrix.
Alternatively, a member of an infinite family of \([n,k,d]_q\) codes for which the number of nonzero entries in each row and column of the generator matrix are both bounded by a constant as \(n\to\infty\).}\\ 
\addlinespace[\myxtraspc]
\eczhRefIndex{code:q-ary_ldpc}%
\eczhListValue{\flmRefsHyperref{code:q-ary_ldpc}{\(q\)-ary LDPC code}} & \eczhListValue{A \(q\)-ary linear code with a sparse parity-check matrix.
Alternatively, a member of an infinite family of \([n,k,d]_q\) codes for which the number of nonzero entries in each row and column of the parity-check matrix are both bounded above by a constant as \(n\to\infty\).}\\ 
\addlinespace[\myxtraspc]
\eczhRefIndex{code:q-ary_duadic}%
\eczhListValue{\flmRefsHyperref{code:q-ary_duadic}{\(q\)-ary duadic code}} & \eczhListValue{Member of a pair of cyclic linear \(q\)-ary codes that satisfy certain relations, depending on whether the pair is \textit{even-like} or \textit{odd-like} duadic.
Duadic codes exist only when \(q\) is a square modulo \(n\) \NoCaseChange{\protect\cite{cite69}}.}\\ 
\addlinespace[\myxtraspc]
\eczhRefIndex{code:q-ary_lcc}%
\eczhListValue{\flmRefsHyperref{code:q-ary_lcc}{\(q\)-ary linear LCC}} & \eczhListValue{A \(q\)-ary linear code for which one can recover any coordinate of a codeword from at most \(r\) coordinates of a received word (assuming the corruption rate is within some tolerated threshold \(\delta\)).}\\ 
\addlinespace[\myxtraspc]
\eczhRefIndex{code:q-ary_ltc}%
\eczhListValue{\flmRefsHyperref{code:q-ary_ltc}{\(q\)-ary linear LTC}} & \eczhListValue{A \(q\)-ary linear code \(C\) of length \(n\) that is a \((u,R)\)-LTC with query complexity \(u\) and soundness \(R>0\).}\\ 
\addlinespace[\myxtraspc]
\eczhRefIndex{code:q-ary_protograph_ldpc}%
\eczhListValue{\flmRefsHyperref{code:q-ary_protograph_ldpc}{\(q\)-ary protograph LDPC code}} & \eczhListValue{A \(q\)-ary LDPC code whose parity-check matrix is constructed using the \flmRefsHyperref{ref47}{lifting} procedure applied to the incidence matrix of a sparse graph called, in this context, a \textit{protograph}.
An ability to assign non-binary edge weight called \textit{edge scaling} can also be used in code construction.}\\ 
\addlinespace[\myxtraspc]
\eczhRefIndex{code:q-ary_repetition}%
\eczhListValue{\flmRefsHyperref{code:q-ary_repetition}{\(q\)-ary repetition code}} & \eczhListValue{An \([n,1,n]_q\) code consisting of codewords \((j,j,\cdots,j)\) for \(j \in \mathbb{F}_q\).}\\ 
\addlinespace[\myxtraspc]
\eczhRefIndex{code:q-ary_simplex}%
\eczhListValue{\flmRefsHyperref{code:q-ary_simplex}{\(q\)-ary simplex code}} & \eczhListValue{An \([n,m,q^{m-1}]_q\) equidistant projective code with \(n=\frac{q^m-1}{q-1}\), denoted as \(S(q,m)\). The columns of the generator matrix are in one-to-one correspondence with the elements of the projective space \(PG(m-1,q)\), with each column being a chosen representative of the corresponding element.
All nonzero simplex codewords have a constant weight of \(q^{m-1}\) \NoCaseChange{\protect\cite{cite45,cite46}}.}\\ 
\end{tabularx}\endgroup
\eczcodelist{orthogonal_array}{Orthogonal arrays and friends}%

\eczhCodeListAutoDescription{All descendants and cousins of \flmRefsCref{code:orthogonal_array}.}%

\eczhIncludeCodeGraph{Bare}{scale=0.5}{\columnwidth}{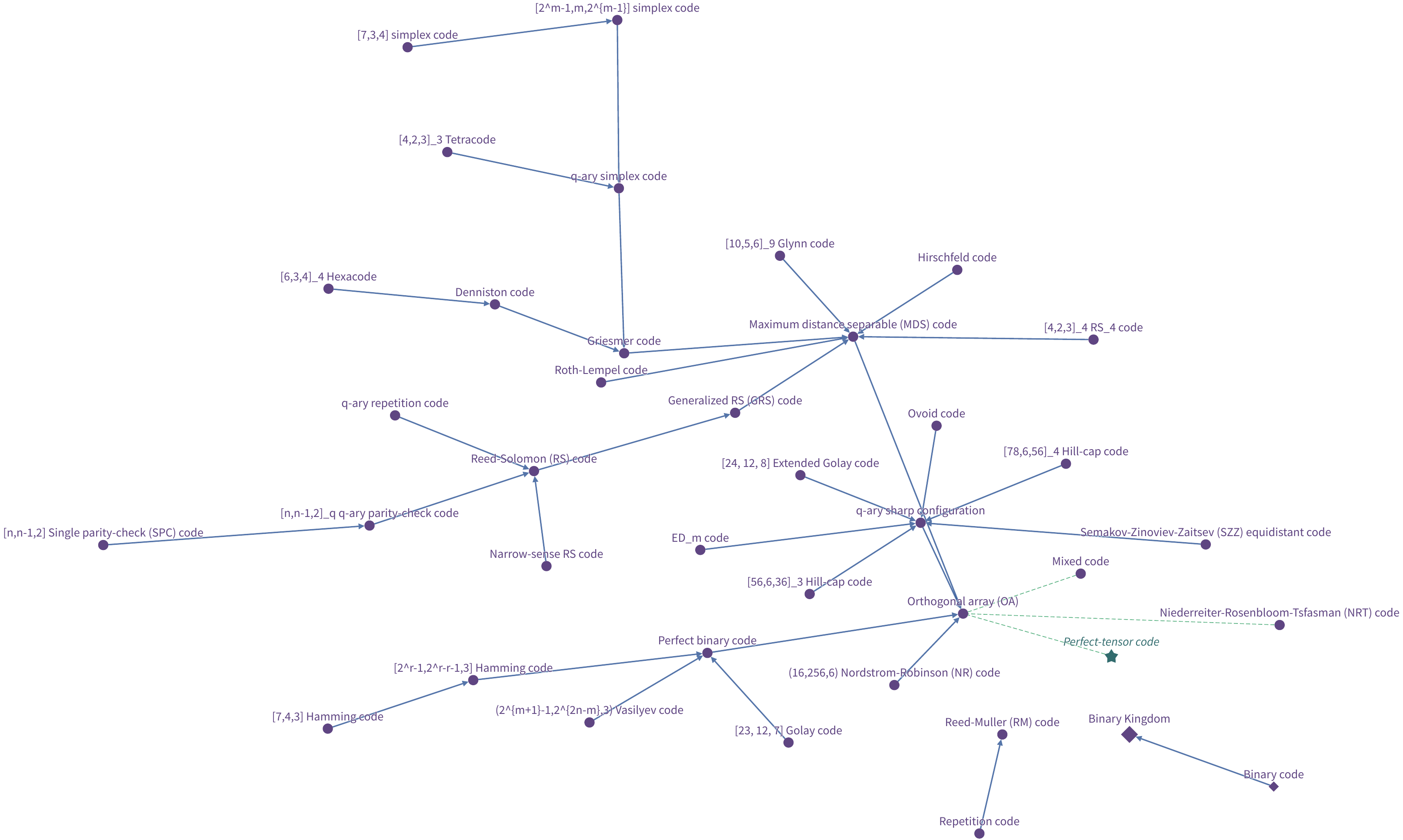}{Orthogonal arrays and friends}{https://errorcorrectionzoo.org/code_graph#J\%7B\%22displayMode\%22\%3A\%22subset\%22\%2C\%22modeSubsetOptions\%22\%3A\%7B\%22codeIds\%22\%3A\%5B\%22bits_into_bits\%22\%2C\%22denniston\%22\%2C\%22generalized_reed_solomon\%22\%2C\%22griesmer\%22\%2C\%22hirschfeld\%22\%2C\%22mds\%22\%2C\%22mixed\%22\%2C\%22narrow_sense_reed_solomon\%22\%2C\%22nrt\%22\%2C\%22orthogonal_array\%22\%2C\%22bose_qvist\%22\%2C\%22perfect_binary\%22\%2C\%22ame\%22\%2C\%22reed_muller\%22\%2C\%22reed_solomon\%22\%2C\%22repetition\%22\%2C\%22roth_lempel\%22\%2C\%22semakov_zinoviev_zaitsev\%22\%2C\%22nordstrom_robinson\%22\%2C\%22vasilyev\%22\%2C\%22semakov_zinoviev\%22\%2C\%22glynn\%22\%2C\%22golay\%22\%2C\%22extended_golay\%22\%2C\%22simplex\%22\%2C\%22hamming\%22\%2C\%22tetracode\%22\%2C\%22reed_solomon_4\%22\%2C\%22hill_56_6_36\%22\%2C\%22hexacode\%22\%2C\%22simplex734\%22\%2C\%22hamming743\%22\%2C\%22hill_78_6_56\%22\%2C\%22parity_check\%22\%2C\%22q-ary_parity_check\%22\%2C\%22q-ary_repetition\%22\%2C\%22delsarte_optimal_q-ary\%22\%2C\%22q-ary_simplex\%22\%5D\%2C\%22reusePreviousLayoutPositions\%22\%3Afalse\%2C\%22showIntermediateConnectingNodes\%22\%3Atrue\%2C\%22connectingNodesMaxDepth\%22\%3A15\%2C\%22connectingNodesPathMaxLength\%22\%3A20\%2C\%22connectingNodesMaxExtraDepth\%22\%3A3\%2C\%22connectingNodesOnlyKeepPathsWithAdditionalLength\%22\%3A1\%2C\%22connectingNodesToDomainsAndKingdoms\%22\%3Afalse\%2C\%22connectingNodesEdgeLengthsByType\%22\%3A\%7B\%22primaryParent\%22\%3A1\%2C\%22secondaryParent\%22\%3A4\%2C\%22cousin\%22\%3A6\%7D\%2C\%22nodeIds\%22\%3A\%5B\%22k_bits_into_bits\%22\%5D\%7D\%2C\%22highlightImportantNodes\%22\%3A\%7B\%22highlightImportantNodes\%22\%3Afalse\%2C\%22highlightPrimaryParents\%22\%3Afalse\%2C\%22highlightRootConnectingEdges\%22\%3Afalse\%7D\%7D}

\begingroup
\small
\eczhBreakableDashes
\renewcommand\arraystretch{1.05}
\edef\myxtraspc{\eczListAddVSpaceXtraXtraStretch}
\begin{tabularx}{\linewidth}{>{\raggedright\arraybackslash}p{\eczListColWidth{name}} >{\hsize=0.7463\hsize }X >{\hsize=0.2537\hsize }X}
\toprule
\eczListColTitle{Code} & \eczListColTitle{Description} & \eczListColTitle{Relation} \\
\midrule
\endfirsthead
\toprule
\eczListColTitleContinued{Code} & \eczListColTitleContinued{Description} & \eczListColTitleContinued{Relation} \\
\midrule
\endhead
\bottomrule
\endfoot
\eczhRefIndex{code:bits_into_bits}%
\eczhListValue{\flmRefsHyperref{code:bits_into_bits}{Binary code}} & \eczhListValue{Encodes \(K\) states (codewords) in \(n\) binary coordinates and has distance \(d\). Usually denoted as \((n,K,d)\). The distance is the minimum Hamming distance between a pair of distinct codewords.} & \eczhListValue{An \((n,K)\) binary code with \flmRefsHyperref{ref113}{dual distance} \(d^{\perp}\) is an OA\(_{K/2^{d^{\perp}-1}}(d^{\perp}-1,n,2)\) \NoCaseChange{\protect\cite{cite209}\protect\cite[{Ch. 5}]{cite41}}.}\\ 
\addlinespace[\myxtraspc]
\eczhRefIndex{code:denniston}%
\eczhListValue{\flmRefsHyperref{code:denniston}{Denniston code}} & \eczhListValue{Projective code that is part of a family of \([2^{a+i}+2^i-2^a,3,2^{a+i}-2^a]_{2^a}\) codes for \(i < a\) constructed using Denniston arcs \NoCaseChange{\protect\cite[{Sec. 19.7.3}]{cite172}}.} & \eczhListValue{\eczListValueNA }\\ 
\addlinespace[\myxtraspc]
\eczhRefIndex{code:generalized_reed_solomon}%
\eczhListValue{\flmRefsHyperref{code:generalized_reed_solomon}{Generalized RS (GRS) code}} & \eczhListValue{An \([n,k,n-k+1]_q\) MDS code that is a modification of the RS code where codeword polynomials are multiplied by additional prefactors \NoCaseChange{\protect\cite[{Def. 15.3.19}]{cite26}}.} & \eczhListValue{\eczListValueNA }\\ 
\addlinespace[\myxtraspc]
\eczhRefIndex{code:griesmer}%
\eczhListValue{\flmRefsHyperref{code:griesmer}{Griesmer code}} & \eczhListValue{A type of \(q\)-ary code whose parameters satisfy the Griesmer bound with equality.} & \eczhListValue{\eczListValueNA }\\ 
\addlinespace[\myxtraspc]
\eczhRefIndex{code:hirschfeld}%
\eczhListValue{\flmRefsHyperref{code:hirschfeld}{Hirschfeld code}} & \eczhListValue{A \([q+1,4,q-2]_q\) projective geometry code for non-prime \(q\) that is an example of an MDS code that is not an RS code; see \NoCaseChange{\protect\cite[{Exam. 7.6}]{cite182}} for the generator matrix.} & \eczhListValue{\eczListValueNA }\\ 
\addlinespace[\myxtraspc]
\eczhRefIndex{code:mds}%
\eczhListValue{\flmRefsHyperref{code:mds}{Maximum distance separable (MDS) code}} & \eczhListValue{A \(q\)-ary linear code whose parameters satisfy the Singleton bound with equality.} & \eczhListValue{An MDS code is an OA\(_{1}(k,n,q)\) \NoCaseChange{\protect\cite[{Thm. 3.3.19}]{cite70}}.}\\ 
\addlinespace[\myxtraspc]
\eczhRefIndex{code:mixed}%
\eczhListValue{\flmRefsHyperref{code:mixed}{Mixed code}} & \eczhListValue{Encodes \(K\) states (codewords) in a string of two or more coordinates, each of which takes values in one of two or more possible groups.} & \eczhListValue{Orthogonal arrays generalized to mixed alphabets are called mixed-level orthogonal arrays \NoCaseChange{\protect\cite{cite210,cite211}} (see \NoCaseChange{\protect\cite[{Ch. 9}]{cite212}}). See Ref. \NoCaseChange{\protect\cite{cite213}} for bounds on mixed orthogonal arrays.}\\ 
\addlinespace[\myxtraspc]
\eczhRefIndex{code:narrow_sense_reed_solomon}%
\eczhListValue{\flmRefsHyperref{code:narrow_sense_reed_solomon}{Narrow-sense RS code}} & \eczhListValue{An \([q-1,k,n-k+1]_q\) RS code whose points \(\alpha_i\) are all \((i-1)\)st powers of a \flmRefsHyperref{ref33}{primitive} element \(\alpha\) of \(\mathbb{F}_q\).} & \eczhListValue{\eczListValueNA }\\ 
\addlinespace[\myxtraspc]
\eczhRefIndex{code:nrt}%
\eczhListValue{\flmRefsHyperref{code:nrt}{Niederreiter-Rosenbloom-Tsfasman (NRT) code}} & \eczhListValue{A poset code based on a partial ordering of \([n]\), i.e., \(1\leq 2\leq \cdots \leq n\).} & \eczhListValue{There exist orthogonal arrays in ordered Hamming space \NoCaseChange{\protect\cite{cite214,cite215}}.}\\ 
\addlinespace[\myxtraspc]
\eczhRefIndex{code:orthogonal_array}%
\eczhListValue{\flmRefsHyperref{code:orthogonal_array}{Orthogonal array (OA)}} & \eczhListValue{An orthogonal array, or OA\(_{\lambda}(t,n,q)\), of \textit{strength} \(t\) with \(q\) \textit{levels} and \(n\) \textit{constraints} is a set of \(q\)-ary strings such that any subset of \(t\) coordinates contains every length-\(t\) string an equal number of times \(\lambda\), which is the \textit{index} of the array \NoCaseChange{\protect\cite[{Def. 3.3.18}]{cite70}}.} & \eczhListValue{\eczListValueNA }\\ 
\addlinespace[\myxtraspc]
\eczhRefIndex{code:bose_qvist}%
\eczhListValue{\flmRefsHyperref{code:bose_qvist}{Ovoid code}} & \eczhListValue{Member of a \([q^2+1,4,q^2-q]_q\) projective two-weight code family obtained from ovoids in \(\mathrm{PG}(3,q)\).
If the columns of a generator matrix are the \(q^2+1\) points of an ovoid, then every hyperplane meets the ovoid in either \(1\) or \(q+1\) points, yielding the two nonzero weights \(q^2\) and \(q^2-q\).
See \NoCaseChange{\protect\cite[{pg. 107}]{cite203}\protect\cite[{pg. 192}]{cite62}} for further details.} & \eczhListValue{\eczListValueNA }\\ 
\addlinespace[\myxtraspc]
\eczhRefIndex{code:perfect_binary}%
\eczhListValue{\flmRefsHyperref{code:perfect_binary}{Perfect binary code}} & \eczhListValue{A type of binary code whose parameters satisfy the Hamming bound with equality.} & \eczhListValue{Perfect distance-three binary codes of length \(n =2^m-1\) are equivalent to binary orthogonal arrays of strength \(t = 2^{m-1}-1\) \NoCaseChange{\protect\cite{cite216,cite217,cite218}}.}\\ 
\addlinespace[\myxtraspc]
\eczhRefIndex{code:ame}%
\eczhListValue{\flmRefsHyperref{code:ame}{Perfect-tensor code}} & \eczhListValue{Block quantum code encoding one subsystem into an odd number \(n\) subsystems whose encoding isometry is a perfect tensor.
This code stems from an AME\((n,q)\) \flmRefsHyperref{ref219}{AME state}, or equivalently, a \(\llparenthesis n+1,1,\lfloor (n+1)/2 \rfloor + 1\rrparenthesis \) code.} & \eczhListValue{Orthogonal arrays and \(d\)-uniform quantum states are related \NoCaseChange{\protect\cite{cite220,cite152,cite221,cite222,cite223,cite224}}.}\\ 
\addlinespace[\myxtraspc]
\eczhRefIndex{code:reed_muller}%
\eczhListValue{\flmRefsHyperref{code:reed_muller}{Reed-Muller (RM) code}} & \eczhListValue{Member of the RM\((r,m)\) family of linear binary codes derived from multivariate polynomials. The code parameters are \([2^m,\sum_{j=0}^{r} {m \choose j},2^{m-r}]\), where \(r\) is the \textit{order} of the code satisfying \(0\leq r\leq m\).
First-order RM codes are also called biorthogonal codes, while \(m\)th order RM codes are also called \textit{universe} codes.
\textit{Punctured RM codes} RM\(^*(r,m)\) are obtained from RM codes by deleting one coordinate from each codeword.} & \eczhListValue{RM codes are related to orthogonal arrays \NoCaseChange{\protect\cite[{Exam. 10.57}]{cite225}}.}\\ 
\addlinespace[\myxtraspc]
\eczhRefIndex{code:reed_solomon}%
\eczhListValue{\flmRefsHyperref{code:reed_solomon}{Reed-Solomon (RS) code}} & \eczhListValue{An \([n,k,n-k+1]_q\) linear code based on polynomials over \(\mathbb{F}_q\).} & \eczhListValue{\eczListValueNA }\\ 
\addlinespace[\myxtraspc]
\eczhRefIndex{code:repetition}%
\eczhListValue{\flmRefsHyperref{code:repetition}{Repetition code}} & \eczhListValue{\([n,1,n]\) binary linear code encoding one bit of information into an \(n\)-bit string.
Majority decoding requires \(n\) to be odd in order to avoid ties.
The idea is to increase the code distance by repeating the logical information several times. It is a \((n,1)\)-Hamming code.
Its automorphism group is \(S_n\).} & \eczhListValue{\eczListValueNA }\\ 
\addlinespace[\myxtraspc]
\eczhRefIndex{code:roth_lempel}%
\eczhListValue{\flmRefsHyperref{code:roth_lempel}{Roth-Lempel code}} & \eczhListValue{Member of a \(q\)-ary linear code family that includes many examples of MDS codes that are not GRS codes.} & \eczhListValue{\eczListValueNA }\\ 
\addlinespace[\myxtraspc]
\eczhRefIndex{code:semakov_zinoviev_zaitsev}%
\eczhListValue{\flmRefsHyperref{code:semakov_zinoviev_zaitsev}{Semakov-Zinoviev-Zaitsev (SZZ) equidistant code}} & \eczhListValue{Member of a nonlinear \(q\)-ary code family that is related to affine resolvable block designs and that is universally optimal.} & \eczhListValue{\eczListValueNA }\\ 
\addlinespace[\myxtraspc]
\eczhRefIndex{code:nordstrom_robinson}%
\eczhListValue{\flmRefsHyperref{code:nordstrom_robinson}{\((16,256,6)\) Nordstrom-Robinson (NR) code}} & \eczhListValue{A nonlinear \((16,256,6)\) binary code that is the smallest Kerdock code and the smallest Preparata code.
The size of this code is larger than the largest possible linear code with the same length and distance.} & \eczhListValue{The NR code is an orthogonal array of strength \(5\) \NoCaseChange{\protect\cite[{pg. 141}]{cite41}}.}\\ 
\addlinespace[\myxtraspc]
\eczhRefIndex{code:vasilyev}%
\eczhListValue{\flmRefsHyperref{code:vasilyev}{\((2^{m+1}-1,2^{2n-m},3)\) Vasilyev code}} & \eczhListValue{Member of an infinite \((2^{m+1}-1,2^{2n-m},3)\) family of perfect nonlinear codes for any \(m \geq 3\).
Constructed by applying a modification of the \((u|u+v)\) construction to a perfect \((2^m-1,2^{n-m},3)\) code, not necessarily linear \NoCaseChange{\protect\cite[{pg. 77}]{cite41}}.} & \eczhListValue{\eczListValueNA }\\ 
\addlinespace[\myxtraspc]
\eczhRefIndex{code:semakov_zinoviev}%
\eczhListValue{\flmRefsHyperref{code:semakov_zinoviev}{\(ED_m\) code}} & \eczhListValue{Member of a family of nonlinear \( (c\frac{qt-1}{(t-1,q-1)},qt,ct\frac{q-1}{(t-1,q-1)}) \) \(q\)-ary codes, where \(c,t\geq 1\) and \((a,b)\) is the greatest common divisor of \(a\) and \(b\).
Such codes are universally optimal and are related to resolvable block designs.} & \eczhListValue{\eczListValueNA }\\ 
\addlinespace[\myxtraspc]
\eczhRefIndex{code:glynn}%
\eczhListValue{\flmRefsHyperref{code:glynn}{\([10,5,6]_9\) Glynn code}} & \eczhListValue{The unique trace-Hermitian self-dual \([10,5,6]_9\) code, constructed using a 10-arc in \(PG(4,9)\) that is not a rational curve.} & \eczhListValue{\eczListValueNA }\\ 
\addlinespace[\myxtraspc]
\eczhRefIndex{code:golay}%
\eczhListValue{\flmRefsHyperref{code:golay}{\([23, 12, 7]\) Golay code}} & \eczhListValue{A \([23, 12, 7]\) perfect binary linear code with connections to various areas of mathematics, e.g., lattices \NoCaseChange{\protect\cite{cite39}} and sporadic simple groups \NoCaseChange{\protect\cite{cite41}}.
Up to equivalence, it is unique for its parameters \NoCaseChange{\protect\cite{cite102}}.
The dual of the Golay code is its \([23,11,8]\) even-weight subcode \NoCaseChange{\protect\cite{cite103,cite104}}.} & \eczhListValue{\eczListValueNA }\\ 
\addlinespace[\myxtraspc]
\eczhRefIndex{code:extended_golay}%
\eczhListValue{\flmRefsHyperref{code:extended_golay}{\([24, 12, 8]\) Extended Golay code}} & \eczhListValue{A self-dual \([24, 12, 8]\) code that is obtained from the Golay code by adding a parity check.
Equivalently, puncturing any coordinate yields the \([23,12,7]\) Golay code.
Up to equivalence, it is unique for its parameters \NoCaseChange{\protect\cite{cite102}}, and it is the unique \([24,12,8]\) extremal Type II code \NoCaseChange{\protect\cite[{Rems. 4.3.10 and 4.3.11}]{cite40}}.} & \eczhListValue{The extended Golay code is an orthogonal array of strength 7 \NoCaseChange{\protect\cite[{Exam. 1}]{cite226}}.}\\ 
\addlinespace[\myxtraspc]
\eczhRefIndex{code:simplex}%
\eczhListValue{\flmRefsHyperref{code:simplex}{\([2^m-1,m,2^{m-1}]\) simplex code}} & \eczhListValue{A member of the equidistant code family dual to the \([2^m-1,2^m-m-1,3]\) Hamming family.} & \eczhListValue{\eczListValueNA }\\ 
\addlinespace[\myxtraspc]
\eczhRefIndex{code:hamming}%
\eczhListValue{\flmRefsHyperref{code:hamming}{\([2^r-1,2^r-r-1,3]\) Hamming code}} & \eczhListValue{Member of an infinite family of perfect linear codes with parameters \([2^r-1,2^r-r-1, 3]\) for \(r \geq 2\).
Their \(r \times (2^r-1) \) parity-check matrix \(H\) has all possible nonzero \(r\)-bit strings as its columns.
Adding a parity check yields the \([2^r,2^r-r-1, 4]\) extended Hamming code.} & \eczhListValue{\eczListValueNA }\\ 
\addlinespace[\myxtraspc]
\eczhRefIndex{code:tetracode}%
\eczhListValue{\flmRefsHyperref{code:tetracode}{\([4,2,3]_3\) Tetracode}} & \eczhListValue{The \([4,2,3]_3\) ternary self-dual MDS code that has connections to lattices \NoCaseChange{\protect\cite{cite39}}. Its weight enumerator is the Gleason polynomial \(g_4\) \NoCaseChange{\protect\cite[{Rem. 4.2.6}]{cite40}}.} & \eczhListValue{\eczListValueNA }\\ 
\addlinespace[\myxtraspc]
\eczhRefIndex{code:reed_solomon_4}%
\eczhListValue{\flmRefsHyperref{code:reed_solomon_4}{\([4,2,3]_4\) RS\(_4\) code}} & \eczhListValue{A Type II Euclidean self-dual extended RS code that is the smallest quaternary extended QR code \NoCaseChange{\protect\cite[{pg. 296}]{cite41}\protect\cite[{Sec. 2.4.2}]{cite42}}.
Puncturing the \([4,2,3]_4\) RS\(_4\) code yields the \([3,2,2]_4\) shortened RS\(_4\) code, which is an RS code \NoCaseChange{\protect\cite[{pg. 295}]{cite41}}.} & \eczhListValue{\eczListValueNA }\\ 
\addlinespace[\myxtraspc]
\eczhRefIndex{code:hill_56_6_36}%
\eczhListValue{\flmRefsHyperref{code:hill_56_6_36}{\([56,6,36]_3\) Hill-cap code}} & \eczhListValue{Projective two-weight ternary code based on the Games graph \NoCaseChange{\protect\cite{cite206}\protect\cite[{Table 19.1}]{cite172}} and Hill's 56-cap \NoCaseChange{\protect\cite{cite207}}.
Its automorphism group contains \(PSL(3,4)\) \NoCaseChange{\protect\cite{cite208}}.} & \eczhListValue{\eczListValueNA }\\ 
\addlinespace[\myxtraspc]
\eczhRefIndex{code:hexacode}%
\eczhListValue{\flmRefsHyperref{code:hexacode}{\([6,3,4]_4\) Hexacode}} & \eczhListValue{The \([6,3,4]_4\) Hermitian self-dual MDS code that has connections to projective geometry, lattices \NoCaseChange{\protect\cite{cite39}}, and conformal field theory \NoCaseChange{\protect\cite{cite44}}. Its weight enumerator is the Gleason polynomial \(g_7\) \NoCaseChange{\protect\cite[{Rem. 4.2.6}]{cite40}}.} & \eczhListValue{\eczListValueNA }\\ 
\addlinespace[\myxtraspc]
\eczhRefIndex{code:simplex734}%
\eczhListValue{\flmRefsHyperref{code:simplex734}{\([7,3,4]\) simplex code}} & \eczhListValue{Second-smallest nontrivial member of the simplex-code family.
The columns of its generator matrix are in one-to-one correspondence with the elements of the projective space \(PG(2,2)\), with each column being a chosen representative of the corresponding element.
The codewords form a \((8,9)\) simplex spherical code under the \flmRefsHyperref{ref38}{antipodal mapping}.
As a simplex code, it is equidistant: every nonzero codeword has Hamming weight \(4\).} & \eczhListValue{\eczListValueNA }\\ 
\addlinespace[\myxtraspc]
\eczhRefIndex{code:hamming743}%
\eczhListValue{\flmRefsHyperref{code:hamming743}{\([7,4,3]\) Hamming code}} & \eczhListValue{Second-smallest member of the Hamming code family.} & \eczhListValue{\eczListValueNA }\\ 
\addlinespace[\myxtraspc]
\eczhRefIndex{code:hill_78_6_56}%
\eczhListValue{\flmRefsHyperref{code:hill_78_6_56}{\([78,6,56]_4\) Hill-cap code}} & \eczhListValue{Projective two-weight quaternary code based on a cap that corresponds to a strongly regular graph \NoCaseChange{\protect\cite[{Table 7.1}]{cite206}}.} & \eczhListValue{\eczListValueNA }\\ 
\addlinespace[\myxtraspc]
\eczhRefIndex{code:parity_check}%
\eczhListValue{\flmRefsHyperref{code:parity_check}{\([n,n-1,2]\) Single parity-check (SPC) code}} & \eczhListValue{An \([n,n-1,2]\) linear binary code whose codewords consist of the message string appended with a \textit{parity-check bit} or \textit{parity bit} such that the parity (i.e., sum over all coordinates of each codeword) is zero.
If the Hamming weight of a message is odd (even), then the parity bit is one (zero).
This code requires only one extra bit of overhead and is therefore inexpensive.
Its codewords are all even-weight binary strings, and its parity-check matrix is a row vector of all ones.
Its automorphism group is \(S_n\).} & \eczhListValue{\eczListValueNA }\\ 
\addlinespace[\myxtraspc]
\eczhRefIndex{code:q-ary_parity_check}%
\eczhListValue{\flmRefsHyperref{code:q-ary_parity_check}{\([n,n-1,2]_q\) \(q\)-ary parity-check code}} & \eczhListValue{An \([n,n-1,2]_q\) linear \(q\)-ary code whose codewords consist of the message string appended with a \textit{parity-check} or \textit{zero-sum check digit} such that the sum over all coordinates of each codeword is zero.} & \eczhListValue{\eczListValueNA }\\ 
\addlinespace[\myxtraspc]
\eczhRefIndex{code:q-ary_repetition}%
\eczhListValue{\flmRefsHyperref{code:q-ary_repetition}{\(q\)-ary repetition code}} & \eczhListValue{An \([n,1,n]_q\) code consisting of codewords \((j,j,\cdots,j)\) for \(j \in \mathbb{F}_q\).} & \eczhListValue{\eczListValueNA }\\ 
\addlinespace[\myxtraspc]
\eczhRefIndex{code:delsarte_optimal_q-ary}%
\eczhListValue{\flmRefsHyperref{code:delsarte_optimal_q-ary}{\(q\)-ary sharp configuration}} & \eczhListValue{A \(q\)-ary code that admits \(m\) different distances between distinct codewords and forms a design of strength \(2m-1\) or greater.} & \eczhListValue{\eczListValueNA }\\ 
\addlinespace[\myxtraspc]
\eczhRefIndex{code:q-ary_simplex}%
\eczhListValue{\flmRefsHyperref{code:q-ary_simplex}{\(q\)-ary simplex code}} & \eczhListValue{An \([n,m,q^{m-1}]_q\) equidistant projective code with \(n=\frac{q^m-1}{q-1}\), denoted as \(S(q,m)\). The columns of the generator matrix are in one-to-one correspondence with the elements of the projective space \(PG(m-1,q)\), with each column being a chosen representative of the corresponding element.
All nonzero simplex codewords have a constant weight of \(q^{m-1}\) \NoCaseChange{\protect\cite{cite45,cite46}}.} & \eczhListValue{\eczListValueNA }\\ 
\end{tabularx}\endgroup
\eczcodelist{perfect}{Perfect codes and friends
}%

\eczhCodeListAutoDescription{All descendants and cousins of \flmRefsCref{code:perfect}.}%

\eczhIncludeCodeGraph{Bare}{scale=0.5}{\columnwidth}{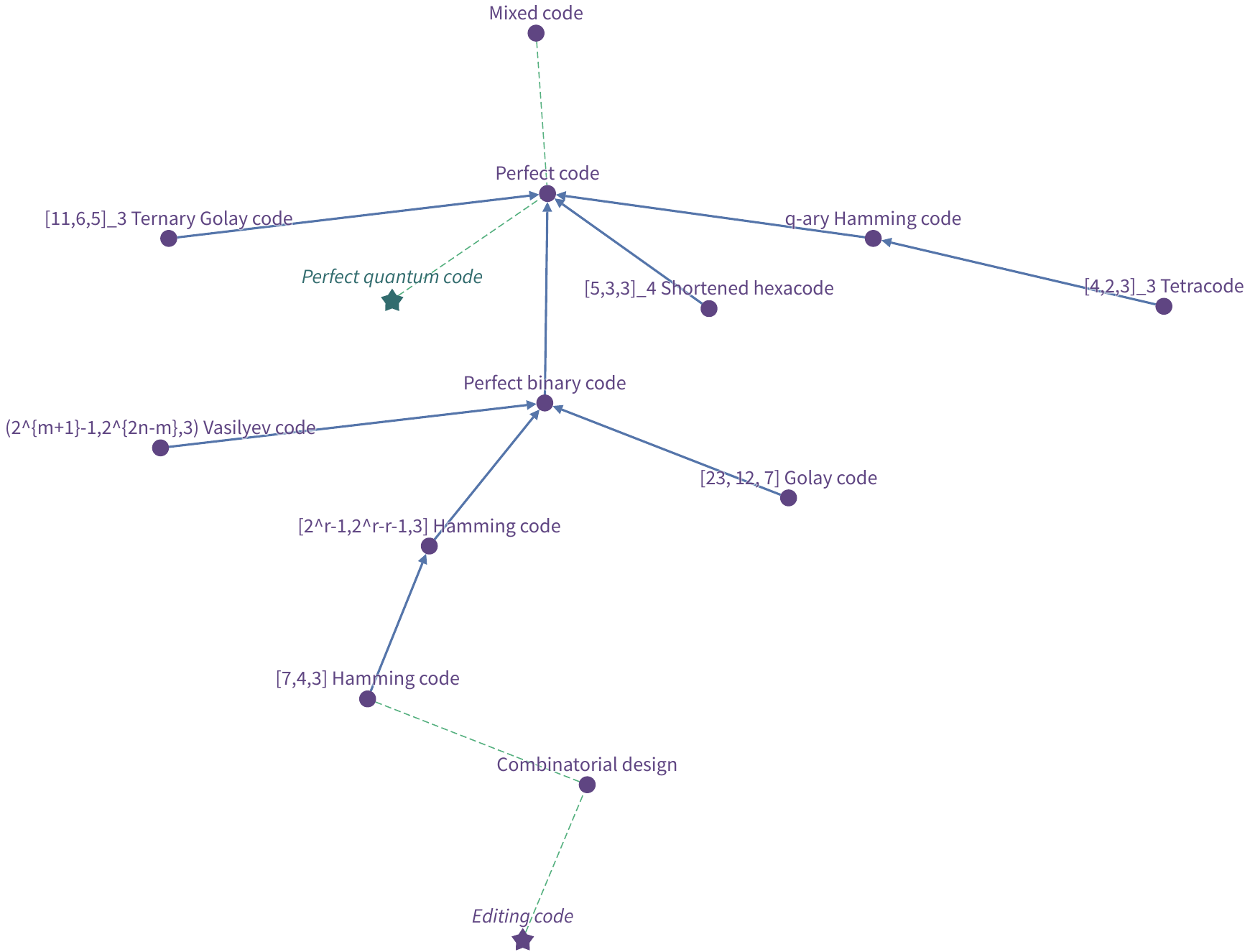}{Perfect codes and friends}{https://errorcorrectionzoo.org/code_graph#J\%7B\%22displayMode\%22\%3A\%22subset\%22\%2C\%22modeSubsetOptions\%22\%3A\%7B\%22codeIds\%22\%3A\%5B\%22combinatorial_design\%22\%2C\%22insertion_deletion\%22\%2C\%22mixed\%22\%2C\%22perfect_binary\%22\%2C\%22perfect\%22\%2C\%22quantum_perfect\%22\%2C\%22vasilyev\%22\%2C\%22ternary_golay\%22\%2C\%22golay\%22\%2C\%22hamming\%22\%2C\%22tetracode\%22\%2C\%22shortened_hexacode\%22\%2C\%22hamming743\%22\%2C\%22q-ary_hamming\%22\%5D\%2C\%22reusePreviousLayoutPositions\%22\%3Afalse\%2C\%22showIntermediateConnectingNodes\%22\%3Atrue\%2C\%22connectingNodesMaxDepth\%22\%3A15\%2C\%22connectingNodesPathMaxLength\%22\%3A20\%2C\%22connectingNodesMaxExtraDepth\%22\%3A3\%2C\%22connectingNodesOnlyKeepPathsWithAdditionalLength\%22\%3A1\%2C\%22connectingNodesToDomainsAndKingdoms\%22\%3Afalse\%2C\%22connectingNodesEdgeLengthsByType\%22\%3A\%7B\%22primaryParent\%22\%3A1\%2C\%22secondaryParent\%22\%3A4\%2C\%22cousin\%22\%3A6\%7D\%2C\%22nodeIds\%22\%3A\%5B\%5D\%7D\%2C\%22highlightImportantNodes\%22\%3A\%7B\%22highlightImportantNodes\%22\%3Afalse\%2C\%22highlightPrimaryParents\%22\%3Afalse\%2C\%22highlightRootConnectingEdges\%22\%3Afalse\%7D\%7D}

\begingroup
\small
\eczhBreakableDashes
\renewcommand\arraystretch{1.05}
\edef\myxtraspc{\eczListAddVSpaceXtraXtraStretch}
\begin{tabularx}{\linewidth}{>{\raggedright\arraybackslash}p{\eczListColWidth{name}} >{\hsize=1.0000\hsize }X}
\toprule
\eczListColTitle{Code} & \eczListColTitle{Description} \\
\midrule
\endfirsthead
\toprule
\eczListColTitleContinued{Code} & \eczListColTitleContinued{Description} \\
\midrule
\endhead
\bottomrule
\endfoot
\eczhRefIndex{code:combinatorial_design}%
\eczhListValue{\flmRefsHyperref{code:combinatorial_design}{Combinatorial design}} & \eczhListValue{A constant-weight binary code that is mapped into a combinatorial \(t\)-design.}\\ 
\addlinespace[\myxtraspc]
\eczhRefIndex{code:insertion_deletion}%
\eczhListValue{\flmRefsHyperref{code:insertion_deletion}{Editing code}} & \eczhListValue{A block code designed to protect against insertions, where a new symbol is added somewhere within the string, and deletions, where a symbol at an unknown location is erased.}\\ 
\addlinespace[\myxtraspc]
\eczhRefIndex{code:mixed}%
\eczhListValue{\flmRefsHyperref{code:mixed}{Mixed code}} & \eczhListValue{Encodes \(K\) states (codewords) in a string of two or more coordinates, each of which takes values in one of two or more possible groups.}\\ 
\addlinespace[\myxtraspc]
\eczhRefIndex{code:perfect_binary}%
\eczhListValue{\flmRefsHyperref{code:perfect_binary}{Perfect binary code}} & \eczhListValue{A type of binary code whose parameters satisfy the Hamming bound with equality.}\\ 
\addlinespace[\myxtraspc]
\eczhRefIndex{code:perfect}%
\eczhListValue{\flmRefsHyperref{code:perfect}{Perfect code}} & \eczhListValue{A type of \(q\)-ary code whose parameters satisfy the Hamming bound with equality.}\\ 
\addlinespace[\myxtraspc]
\eczhRefIndex{code:quantum_perfect}%
\eczhListValue{\flmRefsHyperref{code:quantum_perfect}{Perfect quantum code}} & \eczhListValue{A type of block quantum code whose parameters satisfy the quantum Hamming bound with equality.}\\ 
\addlinespace[\myxtraspc]
\eczhRefIndex{code:vasilyev}%
\eczhListValue{\flmRefsHyperref{code:vasilyev}{\((2^{m+1}-1,2^{2n-m},3)\) Vasilyev code}} & \eczhListValue{Member of an infinite \((2^{m+1}-1,2^{2n-m},3)\) family of perfect nonlinear codes for any \(m \geq 3\).
Constructed by applying a modification of the \((u|u+v)\) construction to a perfect \((2^m-1,2^{n-m},3)\) code, not necessarily linear \NoCaseChange{\protect\cite[{pg. 77}]{cite41}}.}\\ 
\addlinespace[\myxtraspc]
\eczhRefIndex{code:ternary_golay}%
\eczhListValue{\flmRefsHyperref{code:ternary_golay}{\([11,6,5]_3\) Ternary Golay code}} & \eczhListValue{A \([11,6,5]_3\) perfect ternary linear code with connections to various areas of mathematics, e.g., lattices \NoCaseChange{\protect\cite{cite39}} and sporadic simple groups \NoCaseChange{\protect\cite{cite41}}.
Adding a parity bit to the code results in the self-dual \([12,6,6]_3\) \textit{extended ternary Golay code}, whose weight enumerator is the Gleason polynomial \(g_5\) \NoCaseChange{\protect\cite[{Rem. 4.2.6}]{cite40}}.
Up to equivalence, both codes are unique for their respective parameters \NoCaseChange{\protect\cite{cite102}}.
The dual of the ternary Golay code is a \([11,5,6]_3\) projective two-weight subcode \NoCaseChange{\protect\cite[{Exam. 19.3.2}]{cite172}}.}\\ 
\addlinespace[\myxtraspc]
\eczhRefIndex{code:golay}%
\eczhListValue{\flmRefsHyperref{code:golay}{\([23, 12, 7]\) Golay code}} & \eczhListValue{A \([23, 12, 7]\) perfect binary linear code with connections to various areas of mathematics, e.g., lattices \NoCaseChange{\protect\cite{cite39}} and sporadic simple groups \NoCaseChange{\protect\cite{cite41}}.
Up to equivalence, it is unique for its parameters \NoCaseChange{\protect\cite{cite102}}.
The dual of the Golay code is its \([23,11,8]\) even-weight subcode \NoCaseChange{\protect\cite{cite103,cite104}}.}\\ 
\addlinespace[\myxtraspc]
\eczhRefIndex{code:hamming}%
\eczhListValue{\flmRefsHyperref{code:hamming}{\([2^r-1,2^r-r-1,3]\) Hamming code}} & \eczhListValue{Member of an infinite family of perfect linear codes with parameters \([2^r-1,2^r-r-1, 3]\) for \(r \geq 2\).
Their \(r \times (2^r-1) \) parity-check matrix \(H\) has all possible nonzero \(r\)-bit strings as its columns.
Adding a parity check yields the \([2^r,2^r-r-1, 4]\) extended Hamming code.}\\ 
\addlinespace[\myxtraspc]
\eczhRefIndex{code:tetracode}%
\eczhListValue{\flmRefsHyperref{code:tetracode}{\([4,2,3]_3\) Tetracode}} & \eczhListValue{The \([4,2,3]_3\) ternary self-dual MDS code that has connections to lattices \NoCaseChange{\protect\cite{cite39}}. Its weight enumerator is the Gleason polynomial \(g_4\) \NoCaseChange{\protect\cite[{Rem. 4.2.6}]{cite40}}.}\\ 
\addlinespace[\myxtraspc]
\eczhRefIndex{code:shortened_hexacode}%
\eczhListValue{\flmRefsHyperref{code:shortened_hexacode}{\([5,3,3]_4\) Shortened hexacode}} & \eczhListValue{A perfect \([5,3,3]_4\) quaternary Hamming code that is the result of puncturing the hexacode \NoCaseChange{\protect\cite{cite43}}.}\\ 
\addlinespace[\myxtraspc]
\eczhRefIndex{code:hamming743}%
\eczhListValue{\flmRefsHyperref{code:hamming743}{\([7,4,3]\) Hamming code}} & \eczhListValue{Second-smallest member of the Hamming code family.}\\ 
\addlinespace[\myxtraspc]
\eczhRefIndex{code:q-ary_hamming}%
\eczhListValue{\flmRefsHyperref{code:q-ary_hamming}{\(q\)-ary Hamming code}} & \eczhListValue{Member of an infinite family of perfect linear \(q\)-ary codes with parameters \([(q^r-1)/(q-1),(q^r-1)/(q-1)-r, 3]_q\) for \(r \geq 2\) \NoCaseChange{\protect\cite[{(3.1)}]{cite70}}.
These are precisely the nontrivial perfect linear codes over \(\mathbb{F}_q\) \NoCaseChange{\protect\cite[{Thm. 3.3.1}]{cite70}}.}\\ 
\end{tabularx}\endgroup
\eczcodelist{polytope}{Polytopes}%

\eczhCodeListAutoDescription{All descendants of \flmRefsCref{code:polytope}.}%

\eczhIncludeCodeGraph{Bare}{scale=0.5}{\columnwidth}{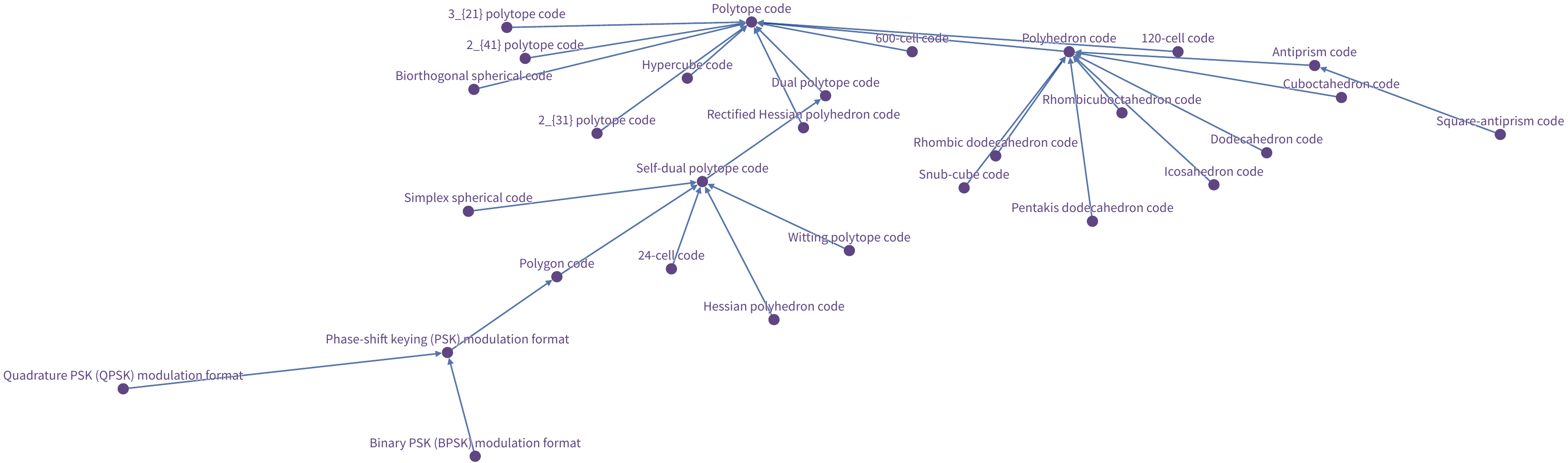}{Polytopes}{https://errorcorrectionzoo.org/code_graph#J\%7B\%22displayMode\%22\%3A\%22subset\%22\%2C\%22modeSubsetOptions\%22\%3A\%7B\%22codeIds\%22\%3A\%5B\%22120cell\%22\%2C\%2224cell\%22\%2C\%22600cell\%22\%2C\%22antiprism\%22\%2C\%22bpsk\%22\%2C\%22biorthogonal_spherical\%22\%2C\%22cubeoctahedron\%22\%2C\%22dodecahedron\%22\%2C\%22dual_polytope\%22\%2C\%22hessian_polyhedron\%22\%2C\%22hypercube\%22\%2C\%22icosahedron\%22\%2C\%22pentakis_dodecahedron\%22\%2C\%22psk\%22\%2C\%22polygon\%22\%2C\%22polyhedron\%22\%2C\%22polytope\%22\%2C\%22qpsk\%22\%2C\%22rect_hessian_polyhedron\%22\%2C\%22rhombic_dodecahedron\%22\%2C\%22rhombicuboctahedron\%22\%2C\%22self_dual_polytope\%22\%2C\%22simplex_spherical\%22\%2C\%22snub_cube\%22\%2C\%22square_antiprism\%22\%2C\%22witting_polytope\%22\%2C\%22231_polytope\%22\%2C\%22241_polytope\%22\%2C\%22hess_polytope\%22\%5D\%2C\%22reusePreviousLayoutPositions\%22\%3Afalse\%2C\%22showIntermediateConnectingNodes\%22\%3Atrue\%2C\%22connectingNodesMaxDepth\%22\%3A15\%2C\%22connectingNodesPathMaxLength\%22\%3A20\%2C\%22connectingNodesMaxExtraDepth\%22\%3A3\%2C\%22connectingNodesOnlyKeepPathsWithAdditionalLength\%22\%3A1\%2C\%22connectingNodesToDomainsAndKingdoms\%22\%3Afalse\%2C\%22connectingNodesEdgeLengthsByType\%22\%3A\%7B\%22primaryParent\%22\%3A1\%2C\%22secondaryParent\%22\%3A4\%2C\%22cousin\%22\%3A6\%7D\%2C\%22nodeIds\%22\%3A\%5B\%5D\%7D\%2C\%22highlightImportantNodes\%22\%3A\%7B\%22highlightImportantNodes\%22\%3Afalse\%2C\%22highlightPrimaryParents\%22\%3Afalse\%2C\%22highlightRootConnectingEdges\%22\%3Afalse\%7D\%7D}

\begingroup
\small
\eczhBreakableDashes
\renewcommand\arraystretch{1.05}
\edef\myxtraspc{\eczListAddVSpaceXtraXtraStretch}
\begin{tabularx}{\linewidth}{>{\raggedright\arraybackslash}p{\eczListColWidth{name}} >{\hsize=1.0000\hsize }X}
\toprule
\eczListColTitle{Code} & \eczListColTitle{Description} \\
\midrule
\endfirsthead
\toprule
\eczListColTitleContinued{Code} & \eczListColTitleContinued{Description} \\
\midrule
\endhead
\bottomrule
\endfoot
\eczhRefIndex{code:120cell}%
\eczhListValue{\flmRefsHyperref{code:120cell}{120-cell code}} & \eczhListValue{Spherical \((4,600,(7-3\sqrt{5})/4)\) code whose codewords are the vertices of the 120-cell.
See \NoCaseChange{\protect\cite{cite178}\protect\cite[{Table 1}]{cite227}\protect\cite[{Table 3}]{cite228}} for explicit realizations of its 600 codewords.}\\ 
\addlinespace[\myxtraspc]
\eczhRefIndex{code:24cell}%
\eczhListValue{\flmRefsHyperref{code:24cell}{24-cell code}} & \eczhListValue{Spherical \((4,24,1)\) code whose codewords are the vertices of the 24-cell.
Codewords form the minimal lattice-shell code of the \(D_4\) lattice.}\\ 
\addlinespace[\myxtraspc]
\eczhRefIndex{code:600cell}%
\eczhListValue{\flmRefsHyperref{code:600cell}{600-cell code}} & \eczhListValue{Spherical \((4,120,(3-\sqrt{5})/2)\) code whose codewords are the vertices of the 600-cell.
See \NoCaseChange{\protect\cite[{Table 1}]{cite229}\protect\cite[{Table 3}]{cite228}} for realizations of the 120 codewords.
A realization of the 600-cell can be given in terms of icosians, which are quaternion coordinates of the 120 elements of the binary icosahedral group \(2I \cong 2.A_5\) (a.k.a. the icosian group) \NoCaseChange{\protect\cite{cite230}\protect\cite[{Ch. 8, pg. 207}]{cite39}}.}\\ 
\addlinespace[\myxtraspc]
\eczhRefIndex{code:antiprism}%
\eczhListValue{\flmRefsHyperref{code:antiprism}{Antiprism code}} & \eczhListValue{Spherical \((3,2q)\) code for \(q \geq 2\) whose codewords are the vertices of a \(q\)-antiprism.}\\ 
\addlinespace[\myxtraspc]
\eczhRefIndex{code:bpsk}%
\eczhListValue{\flmRefsHyperref{code:bpsk}{Binary PSK (BPSK) modulation format}} & \eczhListValue{Encodes one bit of information into a constellation of antipodal points \(\pm\alpha\) for complex \(\alpha\).
These points are typically associated with two phases of an electromagnetic signal.}\\ 
\addlinespace[\myxtraspc]
\eczhRefIndex{code:biorthogonal_spherical}%
\eczhListValue{\flmRefsHyperref{code:biorthogonal_spherical}{Biorthogonal spherical code}} & \eczhListValue{Spherical \((n,2n,2)\) code whose codewords are all permutations of the \(n\)-dimensional vectors \((0,0,\cdots,0,\pm1)\), up to normalization.
The code makes up the vertices of an \(n\)-orthoplex (a.k.a. hyperoctahedron or cross polytope).}\\ 
\addlinespace[\myxtraspc]
\eczhRefIndex{code:cubeoctahedron}%
\eczhListValue{\flmRefsHyperref{code:cubeoctahedron}{Cuboctahedron code}} & \eczhListValue{Spherical \((3,12,1)\) code whose codewords are the vertices of the cuboctahedron.
Codewords form the minimal lattice-shell code of the \(D_3\) face-centered cubic (fcc) lattice.}\\ 
\addlinespace[\myxtraspc]
\eczhRefIndex{code:dodecahedron}%
\eczhListValue{\flmRefsHyperref{code:dodecahedron}{Dodecahedron code}} & \eczhListValue{Spherical \((3,20,2-2\sqrt{5}/3)\) code whose codewords are the vertices of the dodecahedron (alternatively, the centers of the faces of a icosahedron, the dodecahedron's dual polytope).}\\ 
\addlinespace[\myxtraspc]
\eczhRefIndex{code:dual_polytope}%
\eczhListValue{\flmRefsHyperref{code:dual_polytope}{Dual polytope code}} & \eczhListValue{For a spherical code whose codewords are vertices of a convex polytope, the dual code consists of codewords corresponding to the facets of the original polytope, i.e., to the vertices of the polar dual polytope. For regular polytopes, these dual codewords can be represented by the normalized centers of the facets of the original polytope. The dual codewords make up the vertices of the polytope dual to the original polytope.}\\ 
\addlinespace[\myxtraspc]
\eczhRefIndex{code:hessian_polyhedron}%
\eczhListValue{\flmRefsHyperref{code:hessian_polyhedron}{Hessian polyhedron code}} & \eczhListValue{Spherical \((6,27,3/2)\) code whose codewords are the vertices of the Hessian complex polyhedron and the \(2_{21}\) polytope.
Two copies of the code yield the \((6,54,1)\) \textit{double Hessian polyhedron} (a.k.a. diplo-Schläfli) code.
The code can be obtained from the Schläfli graph \NoCaseChange{\protect\cite[{Ch. 9}]{cite115}}.
The (antipodal pairs of) points of the (double) Hessian polyhedron correspond to the 27 lines on a smooth cubic surface in \(\mathbb{C}P^3\) \NoCaseChange{\protect\cite{cite116,cite117,cite118,cite119,cite120}}.}\\ 
\addlinespace[\myxtraspc]
\eczhRefIndex{code:hypercube}%
\eczhListValue{\flmRefsHyperref{code:hypercube}{Hypercube code}} & \eczhListValue{Spherical \((n,2^n,4/n)\) code whose codewords are vertices of an \(n\)-cube, i.e., all permutations and negations of the vector \((1,1,\cdots,1)\), up to normalization.}\\ 
\addlinespace[\myxtraspc]
\eczhRefIndex{code:icosahedron}%
\eczhListValue{\flmRefsHyperref{code:icosahedron}{Icosahedron code}} & \eczhListValue{Spherical \((3,12,2-2/\sqrt{5})\) code whose codewords are the vertices of the icosahedron (alternatively, the centers of the faces of a dodecahedron, the icosahedron's dual polytope).}\\ 
\addlinespace[\myxtraspc]
\eczhRefIndex{code:pentakis_dodecahedron}%
\eczhListValue{\flmRefsHyperref{code:pentakis_dodecahedron}{Pentakis dodecahedron code}} & \eczhListValue{Spherical \((3,32,(9-\sqrt{5})/6)\) code whose codewords are the vertices of the pentakis dodecahedron, the convex hull of the icosahedron and dodecahedron.}\\ 
\addlinespace[\myxtraspc]
\eczhRefIndex{code:psk}%
\eczhListValue{\flmRefsHyperref{code:psk}{Phase-shift keying (PSK) modulation format}} & \eczhListValue{A \(q\)-ary phase-shift keying (\(q\)-PSK) encodes one \(q\)-ary digit of information into a constellation of \(q\) points distributed equidistantly on a circle in \(\mathbb{C}\) or, equivalently, \(\mathbb{R}^2\).}\\ 
\addlinespace[\myxtraspc]
\eczhRefIndex{code:polygon}%
\eczhListValue{\flmRefsHyperref{code:polygon}{Polygon code}} & \eczhListValue{Spherical \((1,q,4\sin^2 \frac{\pi}{q})\) code for any \(q\geq2\) whose codewords are the vertices of a \(q\)-gon. Special cases include the line segment (\(q=2\)), triangle (\(q=3\)), square (\(q=4\)), pentagon (\(q=5\)), and hexagon (\(q=6\)).}\\ 
\addlinespace[\myxtraspc]
\eczhRefIndex{code:polyhedron}%
\eczhListValue{\flmRefsHyperref{code:polyhedron}{Polyhedron code}} & \eczhListValue{A polytope code in three dimensions, i.e., a spherical code whose codewords form vertices of a polyhedron.}\\ 
\addlinespace[\myxtraspc]
\eczhRefIndex{code:polytope}%
\eczhListValue{\flmRefsHyperref{code:polytope}{Polytope code}} & \eczhListValue{Spherical code whose codewords are the vertices of a polytope, i.e., a geometrical figure bounded by lines, planes, and hyperplanes in either real \NoCaseChange{\protect\cite{cite178}} or complex \NoCaseChange{\protect\cite{cite231}} space.
A polytope in two (three, four) dimensions is called a polygon (polyhedron, polychoron).}\\ 
\addlinespace[\myxtraspc]
\eczhRefIndex{code:qpsk}%
\eczhListValue{\flmRefsHyperref{code:qpsk}{Quadrature PSK (QPSK) modulation format}} & \eczhListValue{A quaternary encoding into a constellation of four points distributed equidistantly on a circle.
For the case of \(\pi/4\)-QPSK, the constellation is \(\{e^{\pm i\frac{\pi}{4}},e^{\pm i\frac{3\pi}{4}}\}\).}\\ 
\addlinespace[\myxtraspc]
\eczhRefIndex{code:rect_hessian_polyhedron}%
\eczhListValue{\flmRefsHyperref{code:rect_hessian_polyhedron}{Rectified Hessian polyhedron code}} & \eczhListValue{Spherical \((6,72,1)\) code whose codewords are the vertices of the rectified Hessian complex polyhedron and the \(1_{22}\) polytope.
Codewords form the minimal lattice-shell code of the \(E_6\) lattice, i.e., the 72 roots of \(E_6\) after normalization to the unit sphere.
See \NoCaseChange{\protect\cite[{pg. 127}]{cite231}\protect\cite[{pg. 126}]{cite39}} for realizations of the 72 codewords.}\\ 
\addlinespace[\myxtraspc]
\eczhRefIndex{code:rhombic_dodecahedron}%
\eczhListValue{\flmRefsHyperref{code:rhombic_dodecahedron}{Rhombic dodecahedron code}} & \eczhListValue{Spherical \((3,14,2-2/\sqrt{3})\) code whose codewords are the normalized vertices of the rhombic dodecahedron.
Equivalently, the codewords are the union of the vertices of a cube and an octahedron on the unit sphere.}\\ 
\addlinespace[\myxtraspc]
\eczhRefIndex{code:rhombicuboctahedron}%
\eczhListValue{\flmRefsHyperref{code:rhombicuboctahedron}{Rhombicuboctahedron code}} & \eczhListValue{Spherical \((3,24,4/(5+2 \sqrt{2}\rrparenthesis \) code whose codewords are the vertices of the rhombicuboctahedron.}\\ 
\addlinespace[\myxtraspc]
\eczhRefIndex{code:self_dual_polytope}%
\eczhListValue{\flmRefsHyperref{code:self_dual_polytope}{Self-dual polytope code}} & \eczhListValue{A spherical code whose codewords are the vertices of a self-dual polytope.}\\ 
\addlinespace[\myxtraspc]
\eczhRefIndex{code:simplex_spherical}%
\eczhListValue{\flmRefsHyperref{code:simplex_spherical}{Simplex spherical code}} & \eczhListValue{Spherical \((n,n+1,2+2/n)\) code whose codewords are all permutations of the \(n+1\)-dimensional vector \((1,1,\cdots,1,-n)\), up to normalization, forming an \(n\)-simplex.
Codewords are all equidistant and their components add up to zero.
Simplex spherical codewords in 2 (3, 4) dimensions form the vertices of a triangle (tetrahedron, 5-cell).
In general, the code makes up the vertices of an \(n\)-simplex.
The union of a simplex and its antipodal simplex forms the vertices of a bi-simplex, which has \(2(n+1)\) vertices.}\\ 
\addlinespace[\myxtraspc]
\eczhRefIndex{code:snub_cube}%
\eczhListValue{\flmRefsHyperref{code:snub_cube}{Snub-cube code}} & \eczhListValue{Spherical \((3,24,\frac{2(t^2+1)}{3t^2+2t+2})\) code whose codewords are the vertices of a snub cube, normalized to lie on the unit sphere.
Here, \(t \approx 1.839\) is the tribonacci constant, the real root of \(t^3-t^2-t-1=0\), and the minimum distance squared is approximately \(0.55384\).}\\ 
\addlinespace[\myxtraspc]
\eczhRefIndex{code:square_antiprism}%
\eczhListValue{\flmRefsHyperref{code:square_antiprism}{Square-antiprism code}} & \eczhListValue{Spherical \((3,8,4(4-\sqrt{2})/7)\) code whose codewords are the vertices of the square antiprism \NoCaseChange{\protect\cite[{pg. 72}]{cite115}}.}\\ 
\addlinespace[\myxtraspc]
\eczhRefIndex{code:witting_polytope}%
\eczhListValue{\flmRefsHyperref{code:witting_polytope}{Witting polytope code}} & \eczhListValue{Spherical \((8,240,1)\) code whose codewords are the vertices of the Witting complex polytope, the \(4_{21}\) polytope, and the minimal lattice-shell code of the \(E_8\) lattice.
The code is optimal and unique up to equivalence \NoCaseChange{\protect\cite{cite124,cite39,cite125}}.
Antipodal pairs of points of the \(4_{21}\) polytope code correspond to the 120 tritangent planes of a canonical sextic curve in \(\mathbb{C}P^3\) \NoCaseChange{\protect\cite{cite117,cite118,cite119,cite120}}.}\\ 
\addlinespace[\myxtraspc]
\eczhRefIndex{code:231_polytope}%
\eczhListValue{\flmRefsHyperref{code:231_polytope}{\(2_{31}\) polytope code}} & \eczhListValue{An antipodal spherical \((7,126,1)\) code whose codewords are the vertices of the smallest shell of the \(E_7\) lattice \NoCaseChange{\protect\cite{cite232}}.}\\ 
\addlinespace[\myxtraspc]
\eczhRefIndex{code:241_polytope}%
\eczhListValue{\flmRefsHyperref{code:241_polytope}{\(2_{41}\) polytope code}} & \eczhListValue{An antipodal spherical \((8,2160,1/2)\) code whose codewords are the vertices of the second-smallest shell of the \(E_8\) lattice \NoCaseChange{\protect\cite{cite232}\protect\cite[{Table 10.3}]{cite115}}.}\\ 
\addlinespace[\myxtraspc]
\eczhRefIndex{code:hess_polytope}%
\eczhListValue{\flmRefsHyperref{code:hess_polytope}{\(3_{21}\) polytope code}} & \eczhListValue{Spherical \((7,56,1/3)\) code whose codewords are the vertices of the \(3_{21}\) polytope (a.k.a. the Hess polytope).
The vertices form the kissing configuration of the Witting polytope code.
The 1-skeleton of this polytope is the Gosset graph \NoCaseChange{\protect\cite{cite178}}.
The code is optimal and unique up to equivalence \NoCaseChange{\protect\cite{cite124,cite39,cite125}}.}\\ 
\end{tabularx}\endgroup
\eczcodelist{projective}{Projective codes}%

\eczhCodeListAutoDescription{All descendants of \flmRefsCref{code:projective}.}%

\eczhIncludeCodeGraph{Bare}{scale=0.5}{\columnwidth}{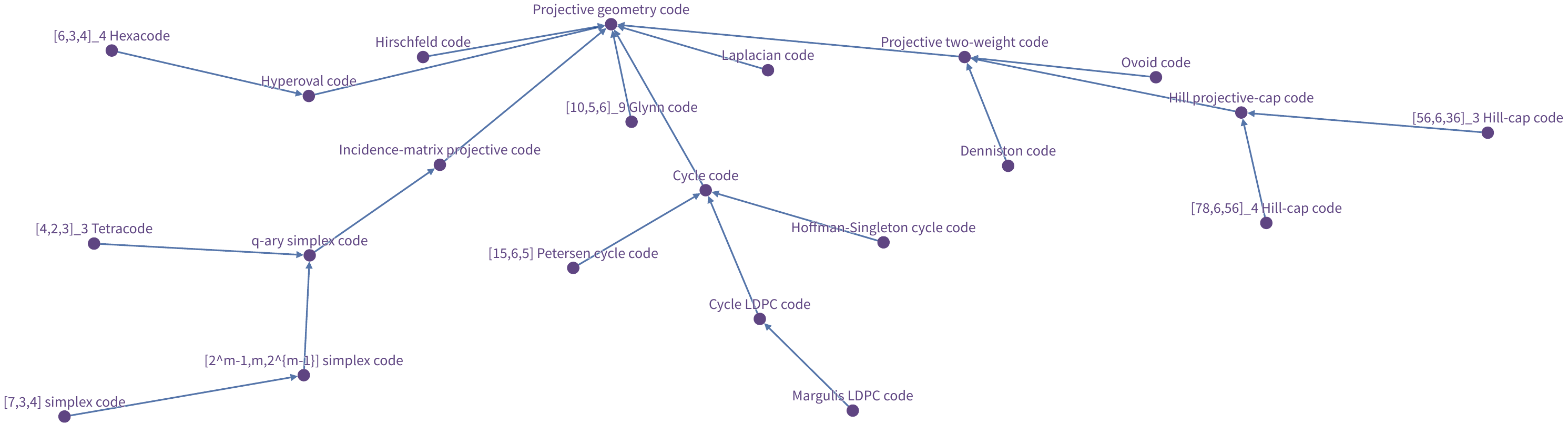}{Projective codes}{https://errorcorrectionzoo.org/code_graph#J\%7B\%22displayMode\%22\%3A\%22subset\%22\%2C\%22modeSubsetOptions\%22\%3A\%7B\%22codeIds\%22\%3A\%5B\%22cycle_ldpc\%22\%2C\%22homological_classical\%22\%2C\%22denniston\%22\%2C\%22hill_cap\%22\%2C\%22hirschfeld\%22\%2C\%22hoffman-singleton\%22\%2C\%22hyperoval\%22\%2C\%22incidence_matrix\%22\%2C\%22laplacian\%22\%2C\%22margulis_ldpc\%22\%2C\%22bose_qvist\%22\%2C\%22projective\%22\%2C\%22projective_two_weight\%22\%2C\%22glynn\%22\%2C\%22petersen\%22\%2C\%22simplex\%22\%2C\%22tetracode\%22\%2C\%22hill_56_6_36\%22\%2C\%22hexacode\%22\%2C\%22simplex734\%22\%2C\%22hill_78_6_56\%22\%2C\%22q-ary_simplex\%22\%5D\%2C\%22reusePreviousLayoutPositions\%22\%3Afalse\%2C\%22showIntermediateConnectingNodes\%22\%3Atrue\%2C\%22connectingNodesMaxDepth\%22\%3A15\%2C\%22connectingNodesPathMaxLength\%22\%3A20\%2C\%22connectingNodesMaxExtraDepth\%22\%3A3\%2C\%22connectingNodesOnlyKeepPathsWithAdditionalLength\%22\%3A1\%2C\%22connectingNodesToDomainsAndKingdoms\%22\%3Afalse\%2C\%22connectingNodesEdgeLengthsByType\%22\%3A\%7B\%22primaryParent\%22\%3A1\%2C\%22secondaryParent\%22\%3A4\%2C\%22cousin\%22\%3A6\%7D\%2C\%22nodeIds\%22\%3A\%5B\%5D\%7D\%2C\%22highlightImportantNodes\%22\%3A\%7B\%22highlightImportantNodes\%22\%3Afalse\%2C\%22highlightPrimaryParents\%22\%3Afalse\%2C\%22highlightRootConnectingEdges\%22\%3Afalse\%7D\%7D}

\begingroup
\small
\eczhBreakableDashes
\renewcommand\arraystretch{1.05}
\edef\myxtraspc{\eczListAddVSpaceXtraXtraStretch}
\begin{tabularx}{\linewidth}{>{\raggedright\arraybackslash}p{\eczListColWidth{name}} >{\hsize=1.0000\hsize }X}
\toprule
\eczListColTitle{Code} & \eczListColTitle{Description} \\
\midrule
\endfirsthead
\toprule
\eczListColTitleContinued{Code} & \eczListColTitleContinued{Description} \\
\midrule
\endhead
\bottomrule
\endfoot
\eczhRefIndex{code:cycle_ldpc}%
\eczhListValue{\flmRefsHyperref{code:cycle_ldpc}{Cycle LDPC code}} & \eczhListValue{An LDPC code whose parity-check matrix forms the incidence matrix of a graph, i.e., has weight-two columns.}\\ 
\addlinespace[\myxtraspc]
\eczhRefIndex{code:homological_classical}%
\eczhListValue{\flmRefsHyperref{code:homological_classical}{Cycle code}} & \eczhListValue{A code whose parity-check matrix is obtained from the incidence matrix of a graph over \(\mathbb{F}_2\).
This code's properties are derived from the size two chain complex associated with the graph.
Not every binary linear code is homological, but there exist homological families that asymptotically saturate the Hamming bound \NoCaseChange{\protect\cite{cite71}}.}\\ 
\addlinespace[\myxtraspc]
\eczhRefIndex{code:denniston}%
\eczhListValue{\flmRefsHyperref{code:denniston}{Denniston code}} & \eczhListValue{Projective code that is part of a family of \([2^{a+i}+2^i-2^a,3,2^{a+i}-2^a]_{2^a}\) codes for \(i < a\) constructed using Denniston arcs \NoCaseChange{\protect\cite[{Sec. 19.7.3}]{cite172}}.}\\ 
\addlinespace[\myxtraspc]
\eczhRefIndex{code:hill_cap}%
\eczhListValue{\flmRefsHyperref{code:hill_cap}{Hill projective-cap code}} & \eczhListValue{Member of a projective code family that contains two \(q\)-ary sharp configurations \NoCaseChange{\protect\cite[{Table 12.1}]{cite199}} and that is constructed using projective caps.}\\ 
\addlinespace[\myxtraspc]
\eczhRefIndex{code:hirschfeld}%
\eczhListValue{\flmRefsHyperref{code:hirschfeld}{Hirschfeld code}} & \eczhListValue{A \([q+1,4,q-2]_q\) projective geometry code for non-prime \(q\) that is an example of an MDS code that is not an RS code; see \NoCaseChange{\protect\cite[{Exam. 7.6}]{cite182}} for the generator matrix.}\\ 
\addlinespace[\myxtraspc]
\eczhRefIndex{code:hoffman-singleton}%
\eczhListValue{\flmRefsHyperref{code:hoffman-singleton}{Hoffman-Singleton cycle code}} & \eczhListValue{A \([50,21,12]\) cycle code whose parity-check matrix is the incidence matrix of the Hoffman-Singleton graph \NoCaseChange{\protect\cite{cite83}}.
Its dual is a \([50,29,8]\) code \NoCaseChange{\protect\cite[{Table II}]{cite82}}.}\\ 
\addlinespace[\myxtraspc]
\eczhRefIndex{code:hyperoval}%
\eczhListValue{\flmRefsHyperref{code:hyperoval}{Hyperoval code}} & \eczhListValue{Projective code constructed from a hyperoval in the projective plane \(PG(2,q)\), where \(q\) is even. Since a hyperoval is a set of \(q+2\) points with no three collinear, the corresponding projective code has parameters \([q+2,3,q]_q\) \NoCaseChange{\protect\cite[{Exam. 19.2.1}]{cite172}}; the \([6,3,4]_4\) hexacode is the smallest example.}\\ 
\addlinespace[\myxtraspc]
\eczhRefIndex{code:incidence_matrix}%
\eczhListValue{\flmRefsHyperref{code:incidence_matrix}{Incidence-matrix projective code}} & \eczhListValue{A projective code whose generator matrix is the incidence matrix of points and hyperplanes in a projective space.
This construction has been generalized to incidence matrices of other structures \NoCaseChange{\protect\cite{cite200,cite201}\protect\cite[{Sec. 14.4}]{cite202}}.
More generally, columns of a code's parity-check matrix can also be organized as an incidence matrix.}\\ 
\addlinespace[\myxtraspc]
\eczhRefIndex{code:laplacian}%
\eczhListValue{\flmRefsHyperref{code:laplacian}{Laplacian code}} & \eczhListValue{A binary linear code whose parity-check matrix is the graph Laplacian reduced mod 2.
For an undirected graph \(\Gamma\) with degree matrix \(D\) and adjacency matrix \(A\), the parity-check matrix is the symmetric matrix \(H=(D-A)\bmod 2\).}\\ 
\addlinespace[\myxtraspc]
\eczhRefIndex{code:margulis_ldpc}%
\eczhListValue{\flmRefsHyperref{code:margulis_ldpc}{Margulis LDPC code}} & \eczhListValue{Member of a class of LDPC codes deterministically constructed from explicit sparse regular expander graphs.
The underlying Margulis-Gabber-Galil graph family provides explicit expanders \NoCaseChange{\protect\cite{cite49,cite90}}, yielding deterministic sparse parity-check matrices.
Related explicit LDPC constructions \NoCaseChange{\protect\cite{cite91}} utilize Ramanujan graphs \NoCaseChange{\protect\cite{cite75,cite76}}.}\\ 
\addlinespace[\myxtraspc]
\eczhRefIndex{code:bose_qvist}%
\eczhListValue{\flmRefsHyperref{code:bose_qvist}{Ovoid code}} & \eczhListValue{Member of a \([q^2+1,4,q^2-q]_q\) projective two-weight code family obtained from ovoids in \(\mathrm{PG}(3,q)\).
If the columns of a generator matrix are the \(q^2+1\) points of an ovoid, then every hyperplane meets the ovoid in either \(1\) or \(q+1\) points, yielding the two nonzero weights \(q^2\) and \(q^2-q\).
See \NoCaseChange{\protect\cite[{pg. 107}]{cite203}\protect\cite[{pg. 192}]{cite62}} for further details.}\\ 
\addlinespace[\myxtraspc]
\eczhRefIndex{code:projective}%
\eczhListValue{\flmRefsHyperref{code:projective}{Projective geometry code}} & \eczhListValue{Linear \(q\)-ary \([n,k,d]\) code whose generator matrix \(G\) does not contain any repeated columns or the zero column.
That way, each column corresponds to a distinct point in the projective space \(PG(k-1,q)\) arising from a \(k\)-dimensional vector space over \(\mathbb{F}_q\). A choice of \(k\) linearly independent columns determines an \textit{information set}.
Columns of a code's parity-check matrix can similarly correspond to points in projective space. This formulation yields connections to projective geometry, which can be applied to determine code properties.}\\ 
\addlinespace[\myxtraspc]
\eczhRefIndex{code:projective_two_weight}%
\eczhListValue{\flmRefsHyperref{code:projective_two_weight}{Projective two-weight code}} & \eczhListValue{A projective code whose codewords all have one of two possible nonzero Hamming weights.}\\ 
\addlinespace[\myxtraspc]
\eczhRefIndex{code:glynn}%
\eczhListValue{\flmRefsHyperref{code:glynn}{\([10,5,6]_9\) Glynn code}} & \eczhListValue{The unique trace-Hermitian self-dual \([10,5,6]_9\) code, constructed using a 10-arc in \(PG(4,9)\) that is not a rational curve.}\\ 
\addlinespace[\myxtraspc]
\eczhRefIndex{code:petersen}%
\eczhListValue{\flmRefsHyperref{code:petersen}{\([15,6,5]\) Petersen cycle code}} & \eczhListValue{A \([15,6,5]\) cycle code whose parity-check matrix is the incidence matrix of the Petersen graph.
The Petersen graph can be thought of as a dodecahedron with antipodes identified \NoCaseChange{\protect\cite[{Appx. A.2.1}]{cite101}}.}\\ 
\addlinespace[\myxtraspc]
\eczhRefIndex{code:simplex}%
\eczhListValue{\flmRefsHyperref{code:simplex}{\([2^m-1,m,2^{m-1}]\) simplex code}} & \eczhListValue{A member of the equidistant code family dual to the \([2^m-1,2^m-m-1,3]\) Hamming family.}\\ 
\addlinespace[\myxtraspc]
\eczhRefIndex{code:tetracode}%
\eczhListValue{\flmRefsHyperref{code:tetracode}{\([4,2,3]_3\) Tetracode}} & \eczhListValue{The \([4,2,3]_3\) ternary self-dual MDS code that has connections to lattices \NoCaseChange{\protect\cite{cite39}}. Its weight enumerator is the Gleason polynomial \(g_4\) \NoCaseChange{\protect\cite[{Rem. 4.2.6}]{cite40}}.}\\ 
\addlinespace[\myxtraspc]
\eczhRefIndex{code:hill_56_6_36}%
\eczhListValue{\flmRefsHyperref{code:hill_56_6_36}{\([56,6,36]_3\) Hill-cap code}} & \eczhListValue{Projective two-weight ternary code based on the Games graph \NoCaseChange{\protect\cite{cite206}\protect\cite[{Table 19.1}]{cite172}} and Hill's 56-cap \NoCaseChange{\protect\cite{cite207}}.
Its automorphism group contains \(PSL(3,4)\) \NoCaseChange{\protect\cite{cite208}}.}\\ 
\addlinespace[\myxtraspc]
\eczhRefIndex{code:hexacode}%
\eczhListValue{\flmRefsHyperref{code:hexacode}{\([6,3,4]_4\) Hexacode}} & \eczhListValue{The \([6,3,4]_4\) Hermitian self-dual MDS code that has connections to projective geometry, lattices \NoCaseChange{\protect\cite{cite39}}, and conformal field theory \NoCaseChange{\protect\cite{cite44}}. Its weight enumerator is the Gleason polynomial \(g_7\) \NoCaseChange{\protect\cite[{Rem. 4.2.6}]{cite40}}.}\\ 
\addlinespace[\myxtraspc]
\eczhRefIndex{code:simplex734}%
\eczhListValue{\flmRefsHyperref{code:simplex734}{\([7,3,4]\) simplex code}} & \eczhListValue{Second-smallest nontrivial member of the simplex-code family.
The columns of its generator matrix are in one-to-one correspondence with the elements of the projective space \(PG(2,2)\), with each column being a chosen representative of the corresponding element.
The codewords form a \((8,9)\) simplex spherical code under the \flmRefsHyperref{ref38}{antipodal mapping}.
As a simplex code, it is equidistant: every nonzero codeword has Hamming weight \(4\).}\\ 
\addlinespace[\myxtraspc]
\eczhRefIndex{code:hill_78_6_56}%
\eczhListValue{\flmRefsHyperref{code:hill_78_6_56}{\([78,6,56]_4\) Hill-cap code}} & \eczhListValue{Projective two-weight quaternary code based on a cap that corresponds to a strongly regular graph \NoCaseChange{\protect\cite[{Table 7.1}]{cite206}}.}\\ 
\addlinespace[\myxtraspc]
\eczhRefIndex{code:q-ary_simplex}%
\eczhListValue{\flmRefsHyperref{code:q-ary_simplex}{\(q\)-ary simplex code}} & \eczhListValue{An \([n,m,q^{m-1}]_q\) equidistant projective code with \(n=\frac{q^m-1}{q-1}\), denoted as \(S(q,m)\). The columns of the generator matrix are in one-to-one correspondence with the elements of the projective space \(PG(m-1,q)\), with each column being a chosen representative of the corresponding element.
All nonzero simplex codewords have a constant weight of \(q^{m-1}\) \NoCaseChange{\protect\cite{cite45,cite46}}.}\\ 
\end{tabularx}\endgroup
\eczcodelist{quantum_inspired}{Quantum-inspired classical codes and friends
}%

\eczhCodeListAutoDescription{All descendants and cousins of \flmRefsCref{code:quantum_inspired}.}%

\eczhIncludeCodeGraph{Bare}{scale=0.5}{\columnwidth}{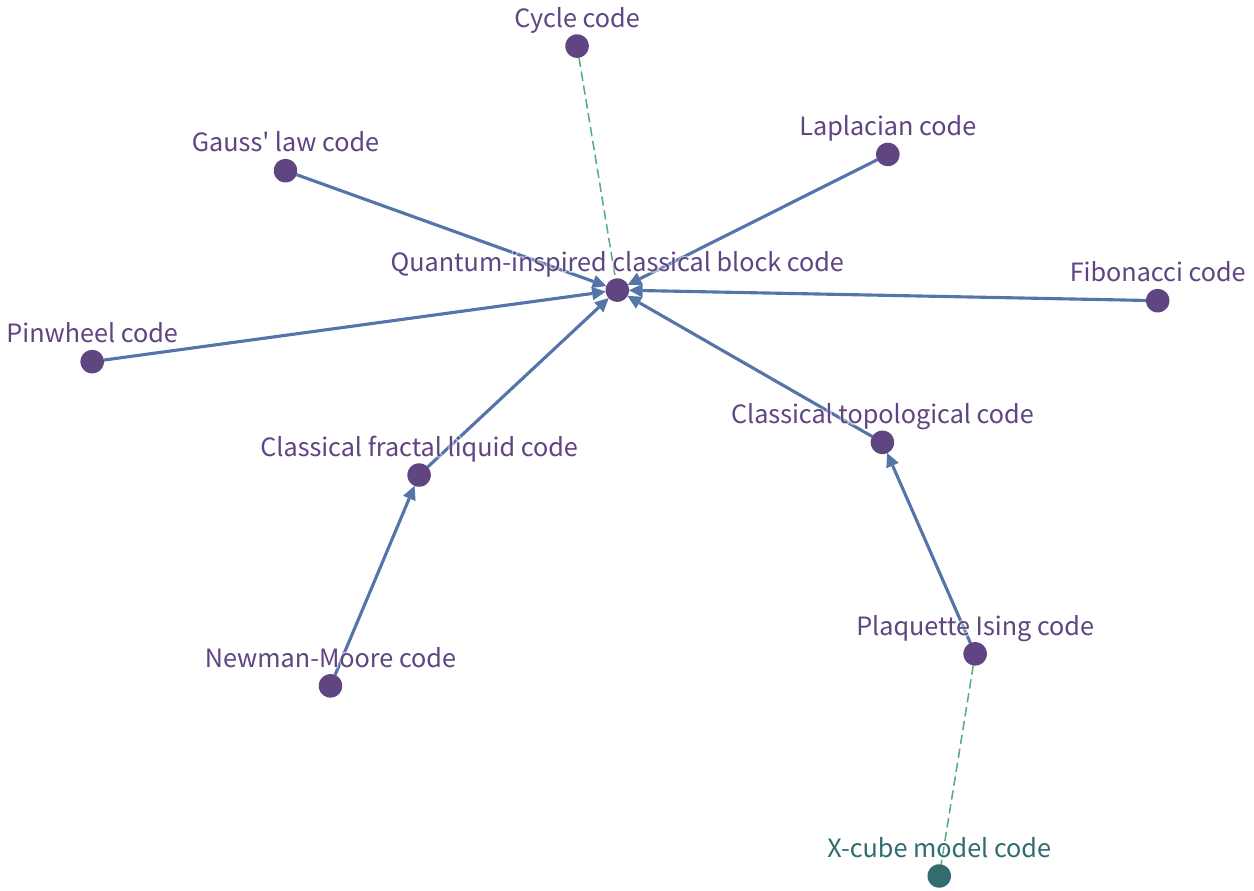}{Quantum-inspired classical codes and friends}{https://errorcorrectionzoo.org/code_graph#J\%7B\%22displayMode\%22\%3A\%22subset\%22\%2C\%22modeSubsetOptions\%22\%3A\%7B\%22codeIds\%22\%3A\%5B\%22classical_fractal_liquid\%22\%2C\%22topological_classical\%22\%2C\%22homological_classical\%22\%2C\%22fibonacci_model\%22\%2C\%22gauss_law\%22\%2C\%22laplacian\%22\%2C\%22newman_moore\%22\%2C\%22pinwheel\%22\%2C\%22plaquette_ising\%22\%2C\%22quantum_inspired\%22\%2C\%22xcube\%22\%5D\%2C\%22reusePreviousLayoutPositions\%22\%3Afalse\%2C\%22showIntermediateConnectingNodes\%22\%3Atrue\%2C\%22connectingNodesMaxDepth\%22\%3A15\%2C\%22connectingNodesPathMaxLength\%22\%3A20\%2C\%22connectingNodesMaxExtraDepth\%22\%3A3\%2C\%22connectingNodesOnlyKeepPathsWithAdditionalLength\%22\%3A1\%2C\%22connectingNodesToDomainsAndKingdoms\%22\%3Afalse\%2C\%22connectingNodesEdgeLengthsByType\%22\%3A\%7B\%22primaryParent\%22\%3A1\%2C\%22secondaryParent\%22\%3A4\%2C\%22cousin\%22\%3A6\%7D\%2C\%22nodeIds\%22\%3A\%5B\%5D\%7D\%2C\%22highlightImportantNodes\%22\%3A\%7B\%22highlightImportantNodes\%22\%3Afalse\%2C\%22highlightPrimaryParents\%22\%3Afalse\%2C\%22highlightRootConnectingEdges\%22\%3Afalse\%7D\%7D}

\begingroup
\small
\eczhBreakableDashes
\renewcommand\arraystretch{1.05}
\edef\myxtraspc{\eczListAddVSpaceXtraXtraStretch}
\begin{tabularx}{\linewidth}{>{\raggedright\arraybackslash}p{\eczListColWidth{name}} >{\hsize=1.0000\hsize }X}
\toprule
\eczListColTitle{Code} & \eczListColTitle{Description} \\
\midrule
\endfirsthead
\toprule
\eczListColTitleContinued{Code} & \eczListColTitleContinued{Description} \\
\midrule
\endhead
\bottomrule
\endfoot
\eczhRefIndex{code:classical_fractal_liquid}%
\eczhListValue{\flmRefsHyperref{code:classical_fractal_liquid}{Classical fractal liquid code}} & \eczhListValue{Member of a family of \([L^D,O(L^{D-1}),O(L^{D-\epsilon})]_p\) linear codes on \(D\)-dimensional square lattices of side length \(L\) and for prime \(p\) and \(\epsilon > 0\) that is based on \(p\)-ary generalizations of the Sierpinski triangle.}\\ 
\addlinespace[\myxtraspc]
\eczhRefIndex{code:topological_classical}%
\eczhListValue{\flmRefsHyperref{code:topological_classical}{Classical topological code}} & \eczhListValue{Classical code defined on a two-dimensional lattice and inspired by geometrically local topological quantum codes, such as the surface code or color code.}\\ 
\addlinespace[\myxtraspc]
\eczhRefIndex{code:homological_classical}%
\eczhListValue{\flmRefsHyperref{code:homological_classical}{Cycle code}} & \eczhListValue{A code whose parity-check matrix is obtained from the incidence matrix of a graph over \(\mathbb{F}_2\).
This code's properties are derived from the size two chain complex associated with the graph.
Not every binary linear code is homological, but there exist homological families that asymptotically saturate the Hamming bound \NoCaseChange{\protect\cite{cite71}}.}\\ 
\addlinespace[\myxtraspc]
\eczhRefIndex{code:fibonacci_model}%
\eczhListValue{\flmRefsHyperref{code:fibonacci_model}{Fibonacci code}} & \eczhListValue{Quantum-inspired binary linear code defined on an \(L\times L/2\) lattice with one bit on each site, where \(L=2^N\) for an integer \(N\geq 2\). The codewords are defined to satisfy the condition that, for each lattice site \((x,y)\), the bits on \((x,y)\), \((x+1,y)\), \((x-1,y)\) and \((x,y+1)\) (where the lattice is taken to be periodic in both directions) contain an even number of \(1\)'s.}\\ 
\addlinespace[\myxtraspc]
\eczhRefIndex{code:gauss_law}%
\eczhListValue{\flmRefsHyperref{code:gauss_law}{Gauss' law code}} & \eczhListValue{An \([m+Dm,Dm,3]\) linear binary code for \(m\geq 3^D\), defined by the Gauss' law constraint of a \(D\)-dimensional fermionic \(\mathbb{Z}_2\) gauge theory \NoCaseChange{\protect\cite[{Thm. 1}]{cite78}}.
The code can be rephrased as a distance-one stabilizer code whose stabilizers consist of gauge-group elements.
It can be concatenated to form a stabilizer code for fault-tolerant quantum simulation of the underlying gauge theory \NoCaseChange{\protect\cite{cite79,cite78}}.}\\ 
\addlinespace[\myxtraspc]
\eczhRefIndex{code:laplacian}%
\eczhListValue{\flmRefsHyperref{code:laplacian}{Laplacian code}} & \eczhListValue{A binary linear code whose parity-check matrix is the graph Laplacian reduced mod 2.
For an undirected graph \(\Gamma\) with degree matrix \(D\) and adjacency matrix \(A\), the parity-check matrix is the symmetric matrix \(H=(D-A)\bmod 2\).}\\ 
\addlinespace[\myxtraspc]
\eczhRefIndex{code:newman_moore}%
\eczhListValue{\flmRefsHyperref{code:newman_moore}{Newman-Moore code}} & \eczhListValue{Member of a family of \([L^2,O(L),O(L^{\frac{\log 3}{\log 2}})]\) binary linear codes on \(L\times L\) square lattices that form the ground-state subspace of a class of exactly solvable spin-glass models with three-body interactions.
The codewords resemble the Sierpinski triangle on a square lattice, which can be generated by a cellular automaton \NoCaseChange{\protect\cite{cite92}}.}\\ 
\addlinespace[\myxtraspc]
\eczhRefIndex{code:pinwheel}%
\eczhListValue{\flmRefsHyperref{code:pinwheel}{Pinwheel code}} & \eczhListValue{A geometrically local binary LDPC code defined on planar graphs obtained from the pinwheel tiling \NoCaseChange{\protect\cite{cite93}}.
Both bits and checks live on vertices of the graph.
If \(L_N\) is the graph Laplacian at generation \(N\), the undepleted check matrix is \(\tilde H_N=(L_N-\mathbb{I})\bmod 2\), and the actual parity-check matrix \(H_N\) is obtained by removing an evenly spaced fraction of boundary checks.}\\ 
\addlinespace[\myxtraspc]
\eczhRefIndex{code:plaquette_ising}%
\eczhListValue{\flmRefsHyperref{code:plaquette_ising}{Plaquette Ising code}} & \eczhListValue{Classical code defined on a cubic lattice in usually two or three dimensions whose parity checks are applied on the four vertices of each square.}\\ 
\addlinespace[\myxtraspc]
\eczhRefIndex{code:quantum_inspired}%
\eczhListValue{\flmRefsHyperref{code:quantum_inspired}{Quantum-inspired classical block code}} & \eczhListValue{A \(q\)-ary linear code whose construction was inspired by a quantum code.}\\ 
\addlinespace[\myxtraspc]
\eczhRefIndex{code:xcube}%
\eczhListValue{\flmRefsHyperref{code:xcube}{X-cube model code}} & \eczhListValue{A foliated type-I fracton CSS code on a cubic lattice with qubits on edges, cube stabilizers, and three cross-shaped vertex stabilizers for each vertex \NoCaseChange{\protect\cite{cite233}}.
It supports a subextensive number of logical qubits.}\\ 
\end{tabularx}\endgroup
\eczcodelist{realizations}{Realized classical codes
}%

\eczhCodeListAutoDescription{All codes in \flmRefsCref{domain:classical_domain} with \emph{Realizations}.}%

\eczhIncludeCodeGraph{Bare}{scale=0.5}{\columnwidth}{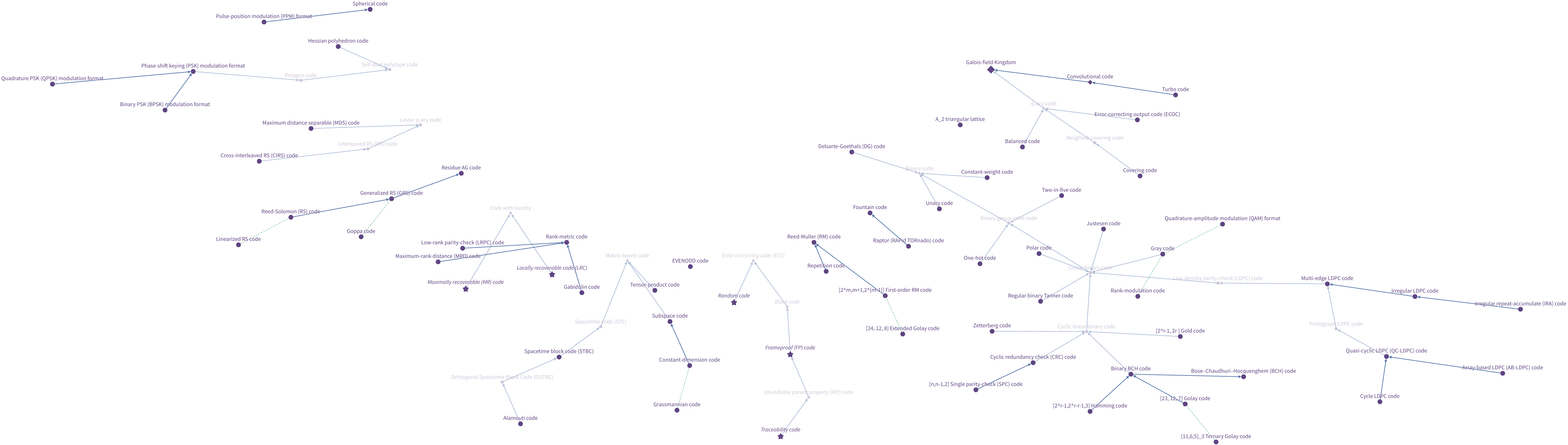}{Realized classical codes}{https://errorcorrectionzoo.org/code_graph#J\%7B\%22displayMode\%22\%3A\%22subset\%22\%2C\%22modeSubsetOptions\%22\%3A\%7B\%22codeIds\%22\%3A\%5B\%22alamouti\%22\%2C\%22array_ldpc\%22\%2C\%22balanced\%22\%2C\%22bch\%22\%2C\%22bpsk\%22\%2C\%22q-ary_bch\%22\%2C\%22finite_grassmann\%22\%2C\%22constant_weight\%22\%2C\%22convolutional\%22\%2C\%22covering\%22\%2C\%22cross_interleaved_reed_solomon\%22\%2C\%22cycle_ldpc\%22\%2C\%22crc\%22\%2C\%22delsarte_goethals\%22\%2C\%22evenodd\%22\%2C\%22ecoc\%22\%2C\%22fountain\%22\%2C\%22frameproof\%22\%2C\%22gabidulin\%22\%2C\%22generalized_reed_solomon\%22\%2C\%22goppa\%22\%2C\%22grassmannian\%22\%2C\%22gray\%22\%2C\%22hessian_polyhedron\%22\%2C\%22irregular_ldpc\%22\%2C\%22ira\%22\%2C\%22justesen\%22\%2C\%22linearized_reed_solomon\%22\%2C\%22locally_recoverable\%22\%2C\%22lrpc\%22\%2C\%22maximally_recoverable\%22\%2C\%22mds\%22\%2C\%22maximum_rank_distance\%22\%2C\%22multi_edge_ldpc\%22\%2C\%22one_hot\%22\%2C\%22psk\%22\%2C\%22polar\%22\%2C\%22ppm\%22\%2C\%22qpsk\%22\%2C\%22qam\%22\%2C\%22qc_ldpc\%22\%2C\%22random\%22\%2C\%22rank_metric\%22\%2C\%22rank_modulation\%22\%2C\%22raptor\%22\%2C\%22reed_muller\%22\%2C\%22reed_solomon\%22\%2C\%22regular_binary_tanner\%22\%2C\%22repetition\%22\%2C\%22residue\%22\%2C\%22spacetime_block\%22\%2C\%22spherical\%22\%2C\%22subspace\%22\%2C\%22tensor\%22\%2C\%22traceability\%22\%2C\%22turbo\%22\%2C\%22two_in_five\%22\%2C\%22unary\%22\%2C\%22zetterberg\%22\%2C\%22hexagonal\%22\%2C\%22ternary_golay\%22\%2C\%22golay\%22\%2C\%22extended_golay\%22\%2C\%22biorthogonal\%22\%2C\%22gold\%22\%2C\%22hamming\%22\%2C\%22parity_check\%22\%5D\%2C\%22reusePreviousLayoutPositions\%22\%3Afalse\%2C\%22showIntermediateConnectingNodes\%22\%3Atrue\%2C\%22connectingNodesMaxDepth\%22\%3A15\%2C\%22connectingNodesPathMaxLength\%22\%3A20\%2C\%22connectingNodesMaxExtraDepth\%22\%3A3\%2C\%22connectingNodesOnlyKeepPathsWithAdditionalLength\%22\%3A1\%2C\%22connectingNodesToDomainsAndKingdoms\%22\%3Afalse\%2C\%22connectingNodesEdgeLengthsByType\%22\%3A\%7B\%22primaryParent\%22\%3A1\%2C\%22secondaryParent\%22\%3A4\%2C\%22cousin\%22\%3A6\%7D\%2C\%22nodeIds\%22\%3A\%5B\%22k_q-ary_digits_into_q-ary_digits\%22\%5D\%7D\%2C\%22highlightImportantNodes\%22\%3A\%7B\%22highlightImportantNodes\%22\%3Afalse\%2C\%22highlightPrimaryParents\%22\%3Afalse\%2C\%22highlightRootConnectingEdges\%22\%3Afalse\%7D\%7D}

\begingroup
\small
\eczhBreakableDashes
\renewcommand\arraystretch{1.05}
\edef\myxtraspc{\eczListAddVSpaceXtraXtraStretch}
\begin{tabularx}{\linewidth}{>{\eczlistsmallsize\raggedright\arraybackslash}p{\eczListColWidth{name}} >{\hsize=1.0000\hsize \eczlistsmallsize}X}
\toprule
\eczListColTitle{Name} & \eczListColTitle{Realization(s)} \\
\midrule
\endfirsthead
\toprule
\eczListColTitleContinued{Name} & \eczListColTitleContinued{Realization(s)} \\
\midrule
\endhead
\bottomrule
\endfoot
\eczhRefIndex{code:alamouti}%
\eczhListValue{\flmRefsHyperref{code:alamouti}{Alamouti code}} & \eczhListValue{\begin{eczvaluelist}
\item\relax Wireless standards since: 3G, LTE, LTE-Advanced, and 5G.
\item\relax Wireless communication: IEEE 802.11n, IEEE 802.11ad, IEEE 802.11ay, etc.
\end{eczvaluelist}
}\\ 
\addlinespace[\myxtraspc]
\eczhRefIndex{code:array_ldpc}%
\eczhListValue{\flmRefsHyperref{code:array_ldpc}{Array-based LDPC (AB-LDPC) code}} & \eczhListValue{\begin{eczvaluelist}
\item\relax Certain AB-LDPC codes have been proposed to be used for DSL transmission \NoCaseChange{\protect\cite{cite234}}.
\end{eczvaluelist}
}\\ 
\addlinespace[\myxtraspc]
\eczhRefIndex{code:balanced}%
\eczhListValue{\flmRefsHyperref{code:balanced}{Balanced code}} & \eczhListValue{\begin{eczvaluelist}
\item\relax Balanced length-eight code, known as a 6b/8b encoding, used for balancing direct current in a communications system \NoCaseChange{\protect\cite{cite235}}
\end{eczvaluelist}
}\\ 
\addlinespace[\myxtraspc]
\eczhRefIndex{code:bch}%
\eczhListValue{\flmRefsHyperref{code:bch}{Binary BCH code}} & \eczhListValue{\begin{eczvaluelist}
\item\relax Satellite communication \NoCaseChange{\protect\cite{cite236}}
\end{eczvaluelist}
}\\ 
\addlinespace[\myxtraspc]
\eczhRefIndex{code:bpsk}%
\eczhListValue{\flmRefsHyperref{code:bpsk}{Binary PSK (BPSK) modulation format}} & \eczhListValue{\begin{eczvaluelist}
\item\relax Telephone-line modems throughout 1950s and 1960s: Bell 103 and 202, as well as international standards V.21 \NoCaseChange{\protect\cite{cite237}} and V.23 \NoCaseChange{\protect\cite{cite238}}.
\end{eczvaluelist}
}\\ 
\addlinespace[\myxtraspc]
\eczhRefIndex{code:q-ary_bch}%
\eczhListValue{\flmRefsHyperref{code:q-ary_bch}{Bose–Chaudhuri–Hocquenghem (BCH) code}} & \eczhListValue{\begin{eczvaluelist}
\item\relax DVDs, disk drives, and two-dimensional bar codes \NoCaseChange{\protect\cite{cite239}}.
\end{eczvaluelist}
}\\ 
\addlinespace[\myxtraspc]
\eczhRefIndex{code:finite_grassmann}%
\eczhListValue{\flmRefsHyperref{code:finite_grassmann}{Constant-dimension code}} & \eczhListValue{\begin{eczvaluelist}
\item\relax Linear authentication \NoCaseChange{\protect\cite{cite167}}.
\end{eczvaluelist}
}\\ 
\addlinespace[\myxtraspc]
\eczhRefIndex{code:constant_weight}%
\eczhListValue{\flmRefsHyperref{code:constant_weight}{Constant-weight code}} & \eczhListValue{\begin{eczvaluelist}
\item\relax Radio-network frequency hopping \NoCaseChange{\protect\cite{cite240}}.
\end{eczvaluelist}
}\\ 
\addlinespace[\myxtraspc]
\eczhRefIndex{code:convolutional}%
\eczhListValue{\flmRefsHyperref{code:convolutional}{Convolutional code}} & \eczhListValue{\begin{eczvaluelist}
\item\relax A type of convolutional code used in Real-time Application networks \NoCaseChange{\protect\cite{cite241}}.
\item\relax Mobile and radio communications (3G networks) use convolutional codes concatenated with RS codes to obtain suitable performance \NoCaseChange{\protect\cite{cite242}}.
\item\relax A convolutional code with rate 1/2 was used for deep-space and satellite communication \NoCaseChange{\protect\cite{cite243}}
\end{eczvaluelist}
}\\ 
\addlinespace[\myxtraspc]
\eczhRefIndex{code:covering}%
\eczhListValue{\flmRefsHyperref{code:covering}{Covering code}} & \eczhListValue{\begin{eczvaluelist}
\item\relax Data compression, both with and without distortion constraints, can be phrased in terms of covering codes \NoCaseChange{\protect\cite{cite244}}.
\item\relax Football-pool problem: finding the smallest number of bets on a set of matches needed to guarantee at least one bet has at most \(\rho\) errors \NoCaseChange{\protect\cite{cite245,cite246}}.
\end{eczvaluelist}
}\\ 
\addlinespace[\myxtraspc]
\eczhRefIndex{code:cross_interleaved_reed_solomon}%
\eczhListValue{\flmRefsHyperref{code:cross_interleaved_reed_solomon}{Cross-interleaved RS (CIRS) code}} & \eczhListValue{\begin{eczvaluelist}
\item\relax Compact discs (CDs); see \NoCaseChange{\protect\cite{cite189}\protect\cite[{Sec. 5.6}]{cite126}\protect\cite[{Ch. 4}]{cite247}}.
\end{eczvaluelist}
}\\ 
\addlinespace[\myxtraspc]
\eczhRefIndex{code:cycle_ldpc}%
\eczhListValue{\flmRefsHyperref{code:cycle_ldpc}{Cycle LDPC code}} & \eczhListValue{\begin{eczvaluelist}
\item\relax Cycle LDPC codes have been proposed to be used for MIMO channels \NoCaseChange{\protect\cite{cite248}}.
\end{eczvaluelist}
}\\ 
\addlinespace[\myxtraspc]
\eczhRefIndex{code:crc}%
\eczhListValue{\flmRefsHyperref{code:crc}{Cyclic redundancy check (CRC) code}} & \eczhListValue{\begin{eczvaluelist}
\item\relax CRC-16 and CRC-32 are used in data transmission, e.g., IEEE 802.16e, IEEE 802.3 \NoCaseChange{\protect\cite{cite249}} and TCP/IP communication \NoCaseChange{\protect\cite[{Sec. 2.3.3}]{cite250}}.
\end{eczvaluelist}
}\\ 
\addlinespace[\myxtraspc]
\eczhRefIndex{code:delsarte_goethals}%
\eczhListValue{\flmRefsHyperref{code:delsarte_goethals}{Delsarte-Goethals (DG) code}} & \eczhListValue{\begin{eczvaluelist}
\item\relax Space-time signaling \NoCaseChange{\protect\cite{cite251}}.
\item\relax Compressed neighbor discovery in a network \NoCaseChange{\protect\cite{cite252}}.
\end{eczvaluelist}
}\\ 
\addlinespace[\myxtraspc]
\eczhRefIndex{code:evenodd}%
\eczhListValue{\flmRefsHyperref{code:evenodd}{EVENODD code}} & \eczhListValue{\begin{eczvaluelist}
\item\relax Can be implemented on standard RAID-5 controllers without extra finite-field hardware \NoCaseChange{\protect\cite{cite189}}.
\end{eczvaluelist}
}\\ 
\addlinespace[\myxtraspc]
\eczhRefIndex{code:ecoc}%
\eczhListValue{\flmRefsHyperref{code:ecoc}{Error-correcting output code (ECOC)}} & \eczhListValue{\begin{eczvaluelist}
\item\relax Multiclass problems in machine learning, relevant to facial recognition \NoCaseChange{\protect\cite{cite253,cite254}}, text recognition \NoCaseChange{\protect\cite{cite255}}, or digit classification \NoCaseChange{\protect\cite{cite256}}.
\end{eczvaluelist}
}\\ 
\addlinespace[\myxtraspc]
\eczhRefIndex{code:fountain}%
\eczhListValue{\flmRefsHyperref{code:fountain}{Fountain code}} & \eczhListValue{\begin{eczvaluelist}
\item\relax Designed for servers sending data to many recipients, such as during broadcasting or file distribution \NoCaseChange{\protect\cite{cite257,cite258}}.
\item\relax DNA storage \NoCaseChange{\protect\cite{cite259}}.
\end{eczvaluelist}
}\\ 
\addlinespace[\myxtraspc]
\eczhRefIndex{code:frameproof}%
\eczhListValue{\flmRefsHyperref{code:frameproof}{Frameproof (FP) code}} & \eczhListValue{\begin{eczvaluelist}
\item\relax FP codes are utilized in digital fingerprinting and watermarking \NoCaseChange{\protect\cite{cite260}}.
\end{eczvaluelist}
}\\ 
\addlinespace[\myxtraspc]
\eczhRefIndex{code:gabidulin}%
\eczhListValue{\flmRefsHyperref{code:gabidulin}{Gabidulin code}} & \eczhListValue{\begin{eczvaluelist}
\item\relax Public-key cryptosystems \NoCaseChange{\protect\cite{cite261,cite262}}.
\item\relax Digital watermarking. The Gabidulin code provides efficient correction against luminance tampering and image-slicing distortion due to the consistency of the rank against alterations such as column swapping \NoCaseChange{\protect\cite{cite263}}.
\end{eczvaluelist}
}\\ 
\addlinespace[\myxtraspc]
\eczhRefIndex{code:generalized_reed_solomon}%
\eczhListValue{\flmRefsHyperref{code:generalized_reed_solomon}{Generalized RS (GRS) code}} & \eczhListValue{\begin{eczvaluelist}
\item\relax Commonly used in mass storage systems such as CDs, DVDs, QR codes etc.
\item\relax Various cloud storage systems \NoCaseChange{\protect\cite{cite264}}.
\item\relax A variation of the McEliece public-key cryptosystem \NoCaseChange{\protect\cite{cite265,cite266}} by Niederreiter \NoCaseChange{\protect\cite{cite267}} replaced the generator matrix by the parity check matrix of a GRS code. This was proven to be insecure since the public key exposes the algebraic structure of code \NoCaseChange{\protect\cite{cite268}}. More recent works focus on methods to mask the algebraic structure using subcodes of GRS codes \NoCaseChange{\protect\cite{cite261}}. For example, a key-recovery attack was developed in Ref. \NoCaseChange{\protect\cite{cite269}} for a variant of masking method proposed in Ref. \NoCaseChange{\protect\cite{cite270}}.
\end{eczvaluelist}
}\\ 
\addlinespace[\myxtraspc]
\eczhRefIndex{code:goppa}%
\eczhListValue{\flmRefsHyperref{code:goppa}{Goppa code}} & \eczhListValue{\begin{eczvaluelist}
\item\relax The McEliece public-key cryptosystem \NoCaseChange{\protect\cite{cite265,cite266}}. The protocol relies on the assumptions that Goppa-code generator matrices are hard to distinguish from random linear codes. However, there is an algorithm distinguishing between the two code classes in a time subexponential in \(n\) \NoCaseChange{\protect\cite{cite271}}.
\end{eczvaluelist}
}\\ 
\addlinespace[\myxtraspc]
\eczhRefIndex{code:grassmannian}%
\eczhListValue{\flmRefsHyperref{code:grassmannian}{Grassmannian code}} & \eczhListValue{\begin{eczvaluelist}
\item\relax Multiple description coding \NoCaseChange{\protect\cite{cite272}}.
\item\relax Digital fingerprinting \NoCaseChange{\protect\cite{cite273}}.
\end{eczvaluelist}
}\\ 
\addlinespace[\myxtraspc]
\eczhRefIndex{code:gray}%
\eczhListValue{\flmRefsHyperref{code:gray}{Gray code}} & \eczhListValue{\begin{eczvaluelist}
\item\relax Three-dimensional imaging \NoCaseChange{\protect\cite{cite274}}.
\item\relax Broadcasting and communication \NoCaseChange{\protect\cite{cite275}}.
\end{eczvaluelist}
}\\ 
\addlinespace[\myxtraspc]
\eczhRefIndex{code:hessian_polyhedron}%
\eczhListValue{\flmRefsHyperref{code:hessian_polyhedron}{Hessian polyhedron code}} & \eczhListValue{\begin{eczvaluelist}
\item\relax Quantum mechanical SIC-POVMs \NoCaseChange{\protect\cite{cite276}}.
\end{eczvaluelist}
}\\ 
\addlinespace[\myxtraspc]
\eczhRefIndex{code:irregular_ldpc}%
\eczhListValue{\flmRefsHyperref{code:irregular_ldpc}{Irregular LDPC code}} & \eczhListValue{\begin{eczvaluelist}
\item\relax Satellite communication after concatenating with a modulation scheme \NoCaseChange{\protect\cite{cite277}}.
\end{eczvaluelist}
}\\ 
\addlinespace[\myxtraspc]
\eczhRefIndex{code:ira}%
\eczhListValue{\flmRefsHyperref{code:ira}{Irregular repeat-accumulate (IRA) code}} & \eczhListValue{\begin{eczvaluelist}
\item\relax LDPC codes used for digital satellite video broadcasting per the DVB-S2 standard \NoCaseChange{\protect\cite{cite278,cite279}} utilize IRA code features and were subject to litigation; see Ref. \NoCaseChange{\protect\cite{cite280}}.
\item\relax Apple and Broadcom Wi-Fi devices utilize IRA encoding and decoding features and are subject to ongoing litigation; see Ref. \NoCaseChange{\protect\cite{cite280}}.
\end{eczvaluelist}
}\\ 
\addlinespace[\myxtraspc]
\eczhRefIndex{code:justesen}%
\eczhListValue{\flmRefsHyperref{code:justesen}{Justesen code}} & \eczhListValue{\begin{eczvaluelist}
\item\relax Generating small-bias sample spaces, i.e., probability distributions that parity functions cannot typically distinguish from the uniform distribution \NoCaseChange{\protect\cite{cite281}}.
\end{eczvaluelist}
}\\ 
\addlinespace[\myxtraspc]
\eczhRefIndex{code:linearized_reed_solomon}%
\eczhListValue{\flmRefsHyperref{code:linearized_reed_solomon}{Linearized RS code}} & \eczhListValue{\begin{eczvaluelist}
\item\relax Network coding \NoCaseChange{\protect\cite{cite282}}.
\item\relax Code-based cryptography \NoCaseChange{\protect\cite{cite283,cite284}}.
\end{eczvaluelist}
}\\ 
\addlinespace[\myxtraspc]
\eczhRefIndex{code:locally_recoverable}%
\eczhListValue{\flmRefsHyperref{code:locally_recoverable}{Locally recoverable code (LRC)}} & \eczhListValue{\begin{eczvaluelist}
\item\relax An \((18,14,7)\) LRC has been used in the Windows Azure cloud storage system \NoCaseChange{\protect\cite{cite285}}; see also \NoCaseChange{\protect\cite[{31.3.1.2}]{cite183}}.
\item\relax Facebook f4 BLOB cloud storage system \NoCaseChange{\protect\cite{cite286}}
\end{eczvaluelist}
}\\ 
\addlinespace[\myxtraspc]
\eczhRefIndex{code:lrpc}%
\eczhListValue{\flmRefsHyperref{code:lrpc}{Low-rank parity-check (LRPC) code}} & \eczhListValue{\begin{eczvaluelist}
\item\relax Cryptosystem \NoCaseChange{\protect\cite{cite287}} that is a rank-metric analogue of NTRU \NoCaseChange{\protect\cite{cite288}} and MDPC \NoCaseChange{\protect\cite{cite289}} cryptosystems.
\item\relax Post-quantum cryptography \NoCaseChange{\protect\cite{cite290}}.
\end{eczvaluelist}
}\\ 
\addlinespace[\myxtraspc]
\eczhRefIndex{code:maximally_recoverable}%
\eczhListValue{\flmRefsHyperref{code:maximally_recoverable}{Maximally recoverable (MR) code}} & \eczhListValue{\begin{eczvaluelist}
\item\relax RAID storage \NoCaseChange{\protect\cite{cite291}}.
\end{eczvaluelist}
}\\ 
\addlinespace[\myxtraspc]
\eczhRefIndex{code:mds}%
\eczhListValue{\flmRefsHyperref{code:mds}{Maximum distance separable (MDS) code}} & \eczhListValue{\begin{eczvaluelist}
\item\relax Automatic repeat request (ARQ) data transmission protocols \NoCaseChange{\protect\cite[{Ch. 7}]{cite247}}.
\end{eczvaluelist}
}\\ 
\addlinespace[\myxtraspc]
\eczhRefIndex{code:maximum_rank_distance}%
\eczhListValue{\flmRefsHyperref{code:maximum_rank_distance}{Maximum-rank distance (MRD) code}} & \eczhListValue{\begin{eczvaluelist}
\item\relax Useful for error and erasure correction in network coding \NoCaseChange{\protect\cite{cite292,cite293}}.
\end{eczvaluelist}
}\\ 
\addlinespace[\myxtraspc]
\eczhRefIndex{code:multi_edge_ldpc}%
\eczhListValue{\flmRefsHyperref{code:multi_edge_ldpc}{Multi-edge LDPC code}} & \eczhListValue{\begin{eczvaluelist}
\item\relax Quantum key distribution \NoCaseChange{\protect\cite{cite294,cite295,cite296}}.
\end{eczvaluelist}
}\\ 
\addlinespace[\myxtraspc]
\eczhRefIndex{code:one_hot}%
\eczhListValue{\flmRefsHyperref{code:one_hot}{One-hot code}} & \eczhListValue{\begin{eczvaluelist}
\item\relax The bi-quinary code, a combination of one-hot 1-in-2 and 1-in-5 one-hot codes to encode decimal digits, was used in several early computers \NoCaseChange{\protect\cite[{Ch. 27}]{cite297}}.
\item\relax Marking the state of a finite automaton \NoCaseChange{\protect\cite{cite298}}.
\item\relax Encoding the output of the different classes of a classifier neural network \NoCaseChange{\protect\cite{cite299}}. One-hot codes are the primary codes used in multiclass classification \NoCaseChange{\protect\cite{cite300,cite301,cite302,cite303}}.
\end{eczvaluelist}
}\\ 
\addlinespace[\myxtraspc]
\eczhRefIndex{code:psk}%
\eczhListValue{\flmRefsHyperref{code:psk}{Phase-shift keying (PSK) modulation format}} & \eczhListValue{\begin{eczvaluelist}
\item\relax Telephone-line modems: 1967 Milgo 4400/48 and international standard V.27 used 8-PSK \NoCaseChange{\protect\cite{cite304}}.
\end{eczvaluelist}
}\\ 
\addlinespace[\myxtraspc]
\eczhRefIndex{code:polar}%
\eczhListValue{\flmRefsHyperref{code:polar}{Polar code}} & \eczhListValue{\begin{eczvaluelist}
\item\relax Code control channels for the 5G NR (New Radio) interfaces \NoCaseChange{\protect\cite{cite305}}.
\end{eczvaluelist}
}\\ 
\addlinespace[\myxtraspc]
\eczhRefIndex{code:ppm}%
\eczhListValue{\flmRefsHyperref{code:ppm}{Pulse-position modulation (PPM) format}} & \eczhListValue{\begin{eczvaluelist}
\item\relax Greek hydraulic semaphore system \NoCaseChange{\protect\cite{cite306,cite307}}.
\item\relax Telegraph time-division multiplexing.
\item\relax Radio-control, fiber-optic communications, and deep-space communications.
\end{eczvaluelist}
}\\ 
\addlinespace[\myxtraspc]
\eczhRefIndex{code:qpsk}%
\eczhListValue{\flmRefsHyperref{code:qpsk}{Quadrature PSK (QPSK) modulation format}} & \eczhListValue{\begin{eczvaluelist}
\item\relax Japanese and North American digital cellular and personal systems \NoCaseChange{\protect\cite{cite308}}.
\item\relax Telephone-line modems: 1962 Bell 201 and international standard V.24 \NoCaseChange{\protect\cite{cite309}}.
\end{eczvaluelist}
}\\ 
\addlinespace[\myxtraspc]
\eczhRefIndex{code:qam}%
\eczhListValue{\flmRefsHyperref{code:qam}{Quadrature-amplitude modulation (QAM) format}} & \eczhListValue{\begin{eczvaluelist}
\item\relax Optical communication (e.g., Ref. \NoCaseChange{\protect\cite{cite310}}).
\item\relax Telephone-line modems: 1971 Codex 9600C and international standard V.29 used 16-QAM \NoCaseChange{\protect\cite{cite311}}.
\end{eczvaluelist}
}\\ 
\addlinespace[\myxtraspc]
\eczhRefIndex{code:qc_ldpc}%
\eczhListValue{\flmRefsHyperref{code:qc_ldpc}{Quasi-cyclic LDPC (QC-LDPC) code}} & \eczhListValue{\begin{eczvaluelist}
\item\relax 5G NR cellular communication for the traffic channel \NoCaseChange{\protect\cite{cite312,cite313}}.
\item\relax Wireless communication: WiMAX (IEEE 802.16e) \NoCaseChange{\protect\cite{cite314,cite315,cite316}}, WiFi 4 (IEEE 802.11n) \NoCaseChange{\protect\cite{cite317}}, and WPAN (IEEE 802.15.3c) \NoCaseChange{\protect\cite{cite318}}.
\end{eczvaluelist}
}\\ 
\addlinespace[\myxtraspc]
\eczhRefIndex{code:random}%
\eczhListValue{\flmRefsHyperref{code:random}{Random code}} & \eczhListValue{\begin{eczvaluelist}
\item\relax Distributed storage systems \NoCaseChange{\protect\cite{cite319}}.
\item\relax Classical and quantum cryptography based on the learning-with-errors problem, which is related to decoding a random linear code \NoCaseChange{\protect\cite{cite320}}.
\item\relax Random codes can be used to realize secure computation \NoCaseChange{\protect\cite{cite321}}.
\end{eczvaluelist}
}\\ 
\addlinespace[\myxtraspc]
\eczhRefIndex{code:rank_metric}%
\eczhListValue{\flmRefsHyperref{code:rank_metric}{Rank-metric code}} & \eczhListValue{\begin{eczvaluelist}
\item\relax Identity-Based Encryption \NoCaseChange{\protect\cite{cite322}}.
\item\relax Digital watermarking \NoCaseChange{\protect\cite{cite323}}.
\item\relax Network coding and streaming media broadcasting \NoCaseChange{\protect\cite{cite324}}.
\end{eczvaluelist}
}\\ 
\addlinespace[\myxtraspc]
\eczhRefIndex{code:rank_modulation}%
\eczhListValue{\flmRefsHyperref{code:rank_modulation}{Rank-modulation code}} & \eczhListValue{\begin{eczvaluelist}
\item\relax Electronic devices where charges can either increase in an individual cell or decrease in a block of adjacent cells, e.g., flash memories \NoCaseChange{\protect\cite{cite325}}.
\end{eczvaluelist}
}\\ 
\addlinespace[\myxtraspc]
\eczhRefIndex{code:raptor}%
\eczhListValue{\flmRefsHyperref{code:raptor}{Raptor (RAPid TORnado) code}} & \eczhListValue{\begin{eczvaluelist}
\item\relax Two versions of Raptor codes have been standardized by IETF: \flmHref{https://datatracker.ietf.org/doc/html/rfc5053}{R10} and the more recent \flmHref{https://tools.ietf.org/html/rfc6330}{RaptorQ}. RaptorQ is used in mobile multimedia broadcasts as specified in ETSI technical specifications. It is also used in the mobile \flmHref{https://www.atsc.org/wp-content/uploads/2016/01/A331S33-174r6-Signaling-Delivery-Sync-FEC.pdf}{Next Gen TV} standard.
\item\relax Raptor codes are useful in scenarios where erasure (i.e. weak signal or noisy channel) is common, such as in military or disaster scenarios.
\end{eczvaluelist}
}\\ 
\addlinespace[\myxtraspc]
\eczhRefIndex{code:reed_muller}%
\eczhListValue{\flmRefsHyperref{code:reed_muller}{Reed-Muller (RM) code}} & \eczhListValue{\begin{eczvaluelist}
\item\relax Deep-space communication \NoCaseChange{\protect\cite{cite326,cite327}}.
\end{eczvaluelist}
}\\ 
\addlinespace[\myxtraspc]
\eczhRefIndex{code:reed_solomon}%
\eczhListValue{\flmRefsHyperref{code:reed_solomon}{Reed-Solomon (RS) code}} & \eczhListValue{\begin{eczvaluelist}
\item\relax RS Product Code (RSPC) was used in DVDs \NoCaseChange{\protect\cite[{Ch. 4}]{cite247}}.
\item\relax DSL technologies and their variants against impluse noise \NoCaseChange{\protect\cite{cite328}}.
\item\relax Cryptographic primitives based on the hardness of decoding RS codes for more than \(1-\sqrt{k/n}+\epsilon\) errors. This is equivalent to the polynomial reconstruction problem \NoCaseChange{\protect\cite{cite329}}.
\item\relax RS codes as outer codes concatenated with convolutional codes are used indirectly in space exploration programs such as Voyager and Galileo. RS codes were part of a telemetry channel coding standard issued by the Consultative Committee for Space Data Systems \NoCaseChange{\protect\cite[{Ch. 3}]{cite247}}.
\item\relax Automatic repeat request (ARQ) data transmission protocols \NoCaseChange{\protect\cite[{Ch. 7}]{cite247}}.
\item\relax Slow-frequency-hop spread-spectrum transmission \NoCaseChange{\protect\cite[{Chs. 8-9}]{cite247}}.
\item\relax RS codes over \(q=2^m\) are used in RAID 6 \NoCaseChange{\protect\cite{cite330,cite331}}; see \NoCaseChange{\protect\cite{cite189}}.
\item\relax Coded sharding designs in blockchains to increase efficiency \NoCaseChange{\protect\cite{cite332}}.
\item\relax Used in QR-Codes to retrieve damaged barcodes \NoCaseChange{\protect\cite{cite333}}.
\item\relax Wireless communication systems such as 3G, DVB, and WiMAX \NoCaseChange{\protect\cite{cite334}}.
\item\relax Correcting pooled testing results for SARS-CoV-2 \NoCaseChange{\protect\cite{cite335}}.
\item\relax DNA storage \NoCaseChange{\protect\cite{cite336}}.
\end{eczvaluelist}
}\\ 
\addlinespace[\myxtraspc]
\eczhRefIndex{code:regular_binary_tanner}%
\eczhListValue{\flmRefsHyperref{code:regular_binary_tanner}{Regular binary Tanner code}} & \eczhListValue{\begin{eczvaluelist}
\item\relax First hardware implementation was done using a semi-systolic decoding architecture \NoCaseChange{\protect\cite{cite337}}.
\end{eczvaluelist}
}\\ 
\addlinespace[\myxtraspc]
\eczhRefIndex{code:repetition}%
\eczhListValue{\flmRefsHyperref{code:repetition}{Repetition code}} & \eczhListValue{\begin{eczvaluelist}
\item\relax Repetition codes, in conjunction with other codes, were used in magnetic disks \NoCaseChange{\protect\cite{cite338}}.
\item\relax Communication protocols such as FlexRay \NoCaseChange{\protect\cite{cite339}}.'
\end{eczvaluelist}
}\\ 
\addlinespace[\myxtraspc]
\eczhRefIndex{code:residue}%
\eczhListValue{\flmRefsHyperref{code:residue}{Residue AG code}} & \eczhListValue{\begin{eczvaluelist}
\item\relax Improvements over the McEliece public-key cryptosystem to linear AG codes on curves of arbitrary genus \NoCaseChange{\protect\cite{cite340}}. Only the \flmRefsHyperref{ref33}{subfield} subcode proposal remains resilient to attacks \NoCaseChange{\protect\cite[{Sec. 15.7.5.3}]{cite26}}.
\item\relax Algebraic-geometric secret-sharing schemes \NoCaseChange{\protect\cite{cite341}}.
\end{eczvaluelist}
}\\ 
\addlinespace[\myxtraspc]
\eczhRefIndex{code:spacetime_block}%
\eczhListValue{\flmRefsHyperref{code:spacetime_block}{Spacetime block code (STBC)}} & \eczhListValue{\begin{eczvaluelist}
\item\relax High data-rate wireless communication, e.g., WiMAX (IEEE 802.16m) \NoCaseChange{\protect\cite{cite342,cite343,cite344}}.
\end{eczvaluelist}
}\\ 
\addlinespace[\myxtraspc]
\eczhRefIndex{code:spherical}%
\eczhListValue{\flmRefsHyperref{code:spherical}{Spherical code}} & \eczhListValue{\begin{eczvaluelist}
\item\relax Spherical codes are relevant to modern Hopfield networks \NoCaseChange{\protect\cite{cite345,cite346}}
\end{eczvaluelist}
}\\ 
\addlinespace[\myxtraspc]
\eczhRefIndex{code:subspace}%
\eczhListValue{\flmRefsHyperref{code:subspace}{Subspace code}} & \eczhListValue{\begin{eczvaluelist}
\item\relax Packet-based transmission over linear networks \NoCaseChange{\protect\cite{cite292,cite293,cite347}}.
\end{eczvaluelist}
}\\ 
\addlinespace[\myxtraspc]
\eczhRefIndex{code:tensor}%
\eczhListValue{\flmRefsHyperref{code:tensor}{Tensor-product code}} & \eczhListValue{\begin{eczvaluelist}
\item\relax Construction can be used in magnetic recording by taking the tensor product of an RS code and a parity-check code \NoCaseChange{\protect\cite{cite348}}.
\end{eczvaluelist}
}\\ 
\addlinespace[\myxtraspc]
\eczhRefIndex{code:traceability}%
\eczhListValue{\flmRefsHyperref{code:traceability}{Traceability code}} & \eczhListValue{\begin{eczvaluelist}
\item\relax Broadcast messages, pay-per-view movies, and protecting copyrighted online material \NoCaseChange{\protect\cite{cite349}}.
\end{eczvaluelist}
}\\ 
\addlinespace[\myxtraspc]
\eczhRefIndex{code:turbo}%
\eczhListValue{\flmRefsHyperref{code:turbo}{Turbo code}} & \eczhListValue{\begin{eczvaluelist}
\item\relax Recommended by Consultative Committee on Space Data Systems (CCSDS) for telemetry and telecommand \NoCaseChange{\protect\cite{cite350,cite351}}.
\item\relax Several standards related to wireless communication, including W-CDMA, DVB-RCS, TD-SCDMA, 802.16, and CDMA2000 \NoCaseChange{\protect\cite{cite352}}.
\end{eczvaluelist}
}\\ 
\addlinespace[\myxtraspc]
\eczhRefIndex{code:two_in_five}%
\eczhListValue{\flmRefsHyperref{code:two_in_five}{Two-in-five code}} & \eczhListValue{\begin{eczvaluelist}
\item\relax Used in the United States Postal Service's POSTNET barcode system as well as the Postal Alphanumeric Encoding Technique (PLANET).
\item\relax Forms the numerical part of the \textit{Code 39} barcode encoding.
\item\relax Early IBM computers \NoCaseChange{\protect\cite{cite353,cite354}}.
\end{eczvaluelist}
}\\ 
\addlinespace[\myxtraspc]
\eczhRefIndex{code:unary}%
\eczhListValue{\flmRefsHyperref{code:unary}{Unary code}} & \eczhListValue{\begin{eczvaluelist}
\item\relax Neural networks \NoCaseChange{\protect\cite{cite355}}.
\item\relax Birdsong production \NoCaseChange{\protect\cite{cite356}}.
\end{eczvaluelist}
}\\ 
\addlinespace[\myxtraspc]
\eczhRefIndex{code:zetterberg}%
\eczhListValue{\flmRefsHyperref{code:zetterberg}{Zetterberg code}} & \eczhListValue{\begin{eczvaluelist}
\item\relax Code used to provide better protection of data transmission with its double error correcting capacity \NoCaseChange{\protect\cite{cite357}}.
\end{eczvaluelist}
}\\ 
\addlinespace[\myxtraspc]
\eczhRefIndex{code:hexagonal}%
\eczhListValue{\flmRefsHyperref{code:hexagonal}{\(A_2\) triangular lattice}} & \eczhListValue{\begin{eczvaluelist}
\item\relax Wireless communication \NoCaseChange{\protect\cite{cite358,cite359}}.
\end{eczvaluelist}
}\\ 
\addlinespace[\myxtraspc]
\eczhRefIndex{code:ternary_golay}%
\eczhListValue{\flmRefsHyperref{code:ternary_golay}{\([11,6,5]_3\) Ternary Golay code}} & \eczhListValue{\begin{eczvaluelist}
\item\relax Code used in football pools with at least one good bet \NoCaseChange{\protect\cite{cite360,cite246}}. In fact, the code was originally constructed by Juhani Virtakallio and published in the Finnish football pool magazine Veikkaaja \NoCaseChange{\protect\cite{cite361,cite246,cite354}}.
\item\relax Proofs of the quantum mechanical Kochen-Specker theorem \NoCaseChange{\protect\cite{cite362}}.
\end{eczvaluelist}
}\\ 
\addlinespace[\myxtraspc]
\eczhRefIndex{code:golay}%
\eczhListValue{\flmRefsHyperref{code:golay}{\([23, 12, 7]\) Golay code}} & \eczhListValue{\begin{eczvaluelist}
\item\relax Proofs of the quantum mechanical Kochen-Specker theorem \NoCaseChange{\protect\cite{cite362}}.
\end{eczvaluelist}
}\\ 
\addlinespace[\myxtraspc]
\eczhRefIndex{code:extended_golay}%
\eczhListValue{\flmRefsHyperref{code:extended_golay}{\([24, 12, 8]\) Extended Golay code}} & \eczhListValue{\begin{eczvaluelist}
\item\relax Voyager 1 and 2 spacecraft, transmitting hundreds of color pictures of Jupiter and Saturn in their 1979, 1980, and 1981 fly-bys \NoCaseChange{\protect\cite{cite363}}.
\item\relax American military standards for automatic link establishment in high frequency radio systems \NoCaseChange{\protect\cite{cite364}}.
\end{eczvaluelist}
}\\ 
\addlinespace[\myxtraspc]
\eczhRefIndex{code:biorthogonal}%
\eczhListValue{\flmRefsHyperref{code:biorthogonal}{\([2^m,m+1,2^{m-1}]\) First-order RM code}} & \eczhListValue{\begin{eczvaluelist}
\item\relax The \([32, 6, 16]\) RM\((1,5)\) code was used for the 1971 Mariner 9 spacecraft \NoCaseChange{\protect\cite{cite365}}.
\end{eczvaluelist}
}\\ 
\addlinespace[\myxtraspc]
\eczhRefIndex{code:gold}%
\eczhListValue{\flmRefsHyperref{code:gold}{\([2^r-1, 2r ]\) Gold code}} & \eczhListValue{\begin{eczvaluelist}
\item\relax Used for synchronization purposes in telecommunication \NoCaseChange{\protect\cite{cite366}}
\item\relax GPS C/A for satellite navigation \NoCaseChange{\protect\cite{cite367}}.
\end{eczvaluelist}
}\\ 
\addlinespace[\myxtraspc]
\eczhRefIndex{code:hamming}%
\eczhListValue{\flmRefsHyperref{code:hamming}{\([2^r-1,2^r-r-1,3]\) Hamming code}} & \eczhListValue{\begin{eczvaluelist}
\item\relax Commonly used when error rates are very low, for example, computer RAM or integrated circuits \NoCaseChange{\protect\cite{cite368}}.
\item\relax Hamming-code based matrix embedding used in steganography \NoCaseChange{\protect\cite{cite369,cite370}}.
\end{eczvaluelist}
}\\ 
\addlinespace[\myxtraspc]
\eczhRefIndex{code:parity_check}%
\eczhListValue{\flmRefsHyperref{code:parity_check}{\([n,n-1,2]\) Single parity-check (SPC) code}} & \eczhListValue{\begin{eczvaluelist}
\item\relax Can be realized on almost every communication device. SPCs are some of the earliest error-correcting codes \NoCaseChange{\protect\cite[{Ch. 27}]{cite297}}.
\end{eczvaluelist}
}\\ 
\end{tabularx}\endgroup
\eczcodelist{small}{Small-distance classical codes and friends
}%

\eczhCodeListAutoDescription{All descendants and cousins of \flmRefsCref{code:small_distance}.}%

\eczhIncludeCodeGraph{Bare}{scale=0.5}{\columnwidth}{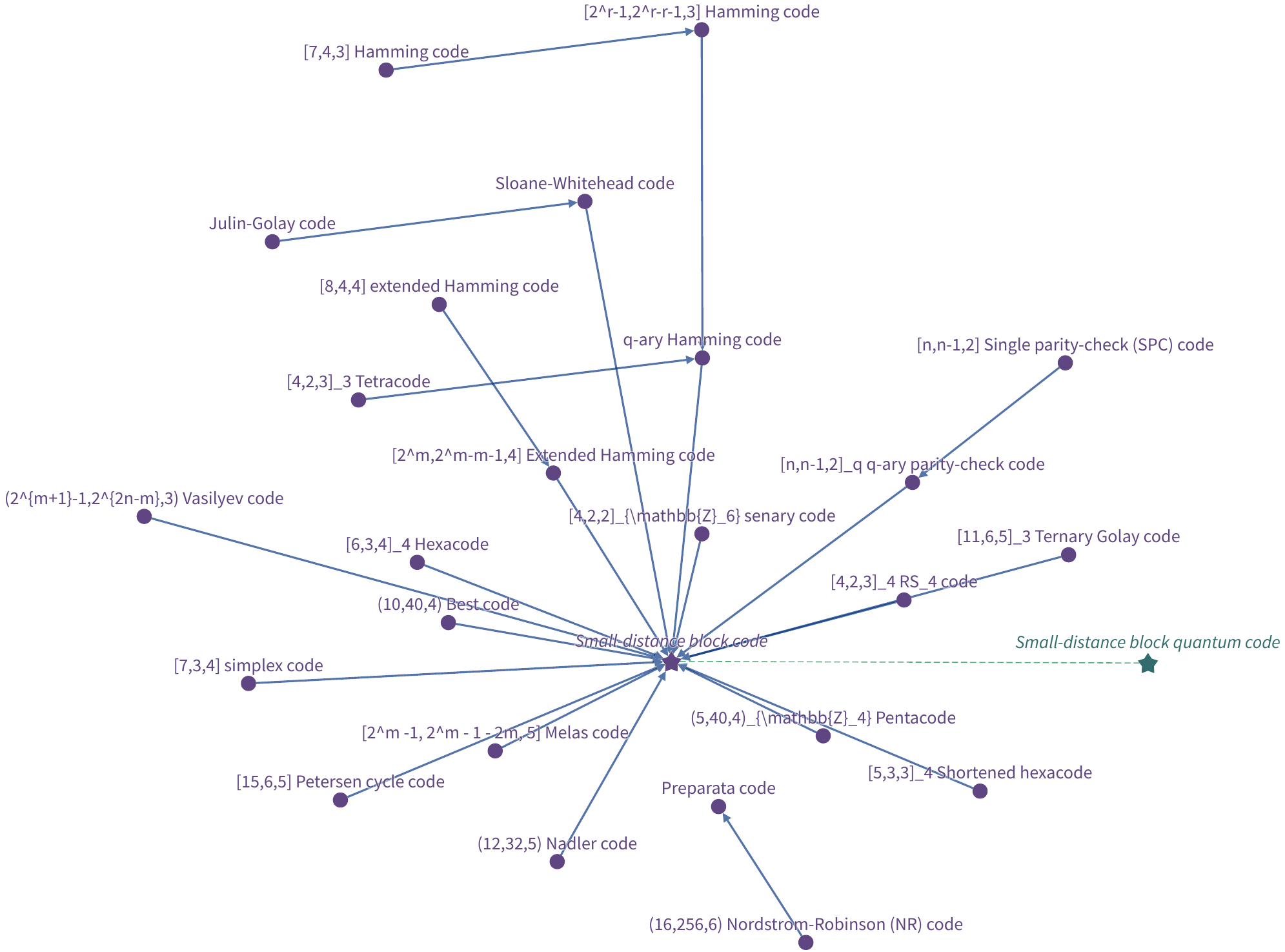}{Small-distance classical codes and friends}{https://errorcorrectionzoo.org/code_graph#J\%7B\%22displayMode\%22\%3A\%22subset\%22\%2C\%22modeSubsetOptions\%22\%3A\%7B\%22codeIds\%22\%3A\%5B\%22julin12\%22\%2C\%22preparata\%22\%2C\%22sloane_whitehead\%22\%2C\%22small_distance\%22\%2C\%22small_distance_quantum\%22\%2C\%22best\%22\%2C\%22nadler\%22\%2C\%22nordstrom_robinson\%22\%2C\%22vasilyev\%22\%2C\%22pentacode\%22\%2C\%22ternary_golay\%22\%2C\%22petersen\%22\%2C\%22melas\%22\%2C\%22extended_hamming\%22\%2C\%22hamming\%22\%2C\%22self_dual_z6\%22\%2C\%22tetracode\%22\%2C\%22reed_solomon_4\%22\%2C\%22shortened_hexacode\%22\%2C\%22hexacode\%22\%2C\%22simplex734\%22\%2C\%22hamming743\%22\%2C\%22hamming844\%22\%2C\%22parity_check\%22\%2C\%22q-ary_parity_check\%22\%2C\%22q-ary_hamming\%22\%5D\%2C\%22reusePreviousLayoutPositions\%22\%3Afalse\%2C\%22showIntermediateConnectingNodes\%22\%3Atrue\%2C\%22connectingNodesMaxDepth\%22\%3A15\%2C\%22connectingNodesPathMaxLength\%22\%3A20\%2C\%22connectingNodesMaxExtraDepth\%22\%3A3\%2C\%22connectingNodesOnlyKeepPathsWithAdditionalLength\%22\%3A1\%2C\%22connectingNodesToDomainsAndKingdoms\%22\%3Afalse\%2C\%22connectingNodesEdgeLengthsByType\%22\%3A\%7B\%22primaryParent\%22\%3A1\%2C\%22secondaryParent\%22\%3A4\%2C\%22cousin\%22\%3A6\%7D\%2C\%22nodeIds\%22\%3A\%5B\%5D\%7D\%2C\%22highlightImportantNodes\%22\%3A\%7B\%22highlightImportantNodes\%22\%3Afalse\%2C\%22highlightPrimaryParents\%22\%3Afalse\%2C\%22highlightRootConnectingEdges\%22\%3Afalse\%7D\%7D}

\begingroup
\small
\eczhBreakableDashes
\renewcommand\arraystretch{1.05}
\edef\myxtraspc{\eczListAddVSpaceXtraXtraStretch}
\begin{tabularx}{\linewidth}{>{\raggedright\arraybackslash}p{\eczListColWidth{name}} >{\hsize=1.0000\hsize }X}
\toprule
\eczListColTitle{Code} & \eczListColTitle{Description} \\
\midrule
\endfirsthead
\toprule
\eczListColTitleContinued{Code} & \eczListColTitleContinued{Description} \\
\midrule
\endhead
\bottomrule
\endfoot
\eczhRefIndex{code:julin12}%
\eczhListValue{\flmRefsHyperref{code:julin12}{Julin-Golay code}} & \eczhListValue{One of several nonlinear binary \((12,144,4)\) codes based on the Steiner system \(S(5,6,12)\) \NoCaseChange{\protect\cite{cite371,cite372}\protect\cite[{Sec. 2.7}]{cite41}\protect\cite[{Sec. 4}]{cite373}}
or their shortened versions, the nonlinear \((11,72,4)\), \((10,38,4)\), and \((9,20,4)\) Julin-Golay codes.
Several of these codes contain more codewords than linear codes of the same length and distance and yield non-lattice sphere-packings that hold records in 9 and 11 dimensions.}\\ 
\addlinespace[\myxtraspc]
\eczhRefIndex{code:preparata}%
\eczhListValue{\flmRefsHyperref{code:preparata}{Preparata code}} & \eczhListValue{A nonlinear binary \((2^{m+1}, 2^{2^{m+1}-2m-2}, 6)\) code where \(m\) is odd.
Puncturing a Preparata code yields the \textit{shortened Preparata code} with parameters \((2^{m+1}-1, 2^{2^{m+1}-2m-2}, 5)\).}\\ 
\addlinespace[\myxtraspc]
\eczhRefIndex{code:sloane_whitehead}%
\eczhListValue{\flmRefsHyperref{code:sloane_whitehead}{Sloane-Whitehead code}} & \eczhListValue{Member of an infinite \((n,\lambda\cdot 2^{n-m-1},3)\) nonlinear code family for any \(n\) satisfying \(2^m \leq n < 3\cdot 2^{m-1}\) for some \(m\) and for \(\lambda\in\{20/16,19/16,18/16\}\).
Such a code has more codewords than any linear code with the same length and distance.
The code is constructed by applying the \((u|u+v)\) construction recursively to the Julin-Golay codes \NoCaseChange{\protect\cite{cite374}\protect\cite[{Secs. 2.7 and 2.9}]{cite41}}.}\\ 
\addlinespace[\myxtraspc]
\eczhRefIndex{code:small_distance}%
\eczhListValue{\flmRefsHyperref{code:small_distance}{Small-distance block code}} & \eczhListValue{A block code of length \(n\) that either detects or corrects errors on at most two coordinates, i.e., has distance \(d \leq 5\).}\\ 
\addlinespace[\myxtraspc]
\eczhRefIndex{code:small_distance_quantum}%
\eczhListValue{\flmRefsHyperref{code:small_distance_quantum}{Small-distance block quantum code}} & \eczhListValue{A block quantum code on \(n\) subsystems that either detects or corrects errors on at most two subsystems, i.e., have distance \(\leq 5\).}\\ 
\addlinespace[\myxtraspc]
\eczhRefIndex{code:best}%
\eczhListValue{\flmRefsHyperref{code:best}{\((10,40,4)\) Best code}} & \eczhListValue{Binary nonlinear \((10,40,4)\) code that is unique \NoCaseChange{\protect\cite{cite375}}.
Under \flmTerm{term}{ref127}{}{Construction A}, this code yields \(P_{10c}\), a non-lattice sphere packing that is the densest known in 10 dimensions \NoCaseChange{\protect\cite{cite376}\protect\cite[{pg. 140}]{cite39}}.}\\ 
\addlinespace[\myxtraspc]
\eczhRefIndex{code:nadler}%
\eczhListValue{\flmRefsHyperref{code:nadler}{\((12,32,5)\) Nadler code}} & \eczhListValue{A nonlinear \((12,32,5)\) binary code that is the largest double-error-correcting code.}\\ 
\addlinespace[\myxtraspc]
\eczhRefIndex{code:nordstrom_robinson}%
\eczhListValue{\flmRefsHyperref{code:nordstrom_robinson}{\((16,256,6)\) Nordstrom-Robinson (NR) code}} & \eczhListValue{A nonlinear \((16,256,6)\) binary code that is the smallest Kerdock code and the smallest Preparata code.
The size of this code is larger than the largest possible linear code with the same length and distance.}\\ 
\addlinespace[\myxtraspc]
\eczhRefIndex{code:vasilyev}%
\eczhListValue{\flmRefsHyperref{code:vasilyev}{\((2^{m+1}-1,2^{2n-m},3)\) Vasilyev code}} & \eczhListValue{Member of an infinite \((2^{m+1}-1,2^{2n-m},3)\) family of perfect nonlinear codes for any \(m \geq 3\).
Constructed by applying a modification of the \((u|u+v)\) construction to a perfect \((2^m-1,2^{n-m},3)\) code, not necessarily linear \NoCaseChange{\protect\cite[{pg. 77}]{cite41}}.}\\ 
\addlinespace[\myxtraspc]
\eczhRefIndex{code:pentacode}%
\eczhListValue{\flmRefsHyperref{code:pentacode}{\((5,40,4)_{\mathbb{Z}_4}\) Pentacode}} & \eczhListValue{Nonlinear \((5,40,4)_{\mathbb{Z}_4}\) code over \(\mathbb{Z}_4\) whose codewords are all cyclic permutations and negations of the strings \(01112\), \(03110\), \(21310\), and \(21132\).}\\ 
\addlinespace[\myxtraspc]
\eczhRefIndex{code:ternary_golay}%
\eczhListValue{\flmRefsHyperref{code:ternary_golay}{\([11,6,5]_3\) Ternary Golay code}} & \eczhListValue{A \([11,6,5]_3\) perfect ternary linear code with connections to various areas of mathematics, e.g., lattices \NoCaseChange{\protect\cite{cite39}} and sporadic simple groups \NoCaseChange{\protect\cite{cite41}}.
Adding a parity bit to the code results in the self-dual \([12,6,6]_3\) \textit{extended ternary Golay code}, whose weight enumerator is the Gleason polynomial \(g_5\) \NoCaseChange{\protect\cite[{Rem. 4.2.6}]{cite40}}.
Up to equivalence, both codes are unique for their respective parameters \NoCaseChange{\protect\cite{cite102}}.
The dual of the ternary Golay code is a \([11,5,6]_3\) projective two-weight subcode \NoCaseChange{\protect\cite[{Exam. 19.3.2}]{cite172}}.}\\ 
\addlinespace[\myxtraspc]
\eczhRefIndex{code:petersen}%
\eczhListValue{\flmRefsHyperref{code:petersen}{\([15,6,5]\) Petersen cycle code}} & \eczhListValue{A \([15,6,5]\) cycle code whose parity-check matrix is the incidence matrix of the Petersen graph.
The Petersen graph can be thought of as a dodecahedron with antipodes identified \NoCaseChange{\protect\cite[{Appx. A.2.1}]{cite101}}.}\\ 
\addlinespace[\myxtraspc]
\eczhRefIndex{code:melas}%
\eczhListValue{\flmRefsHyperref{code:melas}{\([2^m -1, 2^m - 1 - 2m, 5]\) Melas code}} & \eczhListValue{Cyclic linear code whose generator polynomial is \(g(x) = p(x)p(x)^{\star}\), where \(p(x)\) is a primitive polynomial of degree \(m\) that is the minimal polynomial over \(\mathbb{F}_2\) of an element \(\alpha\) of order \(2^m -1\) in \(\mathbb{F}_{2^m}\), \(m\) is odd and greater than five, and '\(\star\)' denotes reciprocation \NoCaseChange{\protect\cite{cite105}}.}\\ 
\addlinespace[\myxtraspc]
\eczhRefIndex{code:extended_hamming}%
\eczhListValue{\flmRefsHyperref{code:extended_hamming}{\([2^m,2^m-m-1,4]\) Extended Hamming code}} & \eczhListValue{Member of an infinite family of RM\((m-2,m)\) codes with parameters \([2^m,2^m-m-1, 4]\) for \(m \geq 2\) that are extensions of the Hamming codes by a parity-check bit.}\\ 
\addlinespace[\myxtraspc]
\eczhRefIndex{code:hamming}%
\eczhListValue{\flmRefsHyperref{code:hamming}{\([2^r-1,2^r-r-1,3]\) Hamming code}} & \eczhListValue{Member of an infinite family of perfect linear codes with parameters \([2^r-1,2^r-r-1, 3]\) for \(r \geq 2\).
Their \(r \times (2^r-1) \) parity-check matrix \(H\) has all possible nonzero \(r\)-bit strings as its columns.
Adding a parity check yields the \([2^r,2^r-r-1, 4]\) extended Hamming code.}\\ 
\addlinespace[\myxtraspc]
\eczhRefIndex{code:self_dual_z6}%
\eczhListValue{\flmRefsHyperref{code:self_dual_z6}{\([4,2,2]_{\mathbb{Z}_6}\) senary code}} & \eczhListValue{A self-dual code over \(\mathbb{Z}_6\) that is one of two such codes, up to permutations \NoCaseChange{\protect\cite{cite128}}.}\\ 
\addlinespace[\myxtraspc]
\eczhRefIndex{code:tetracode}%
\eczhListValue{\flmRefsHyperref{code:tetracode}{\([4,2,3]_3\) Tetracode}} & \eczhListValue{The \([4,2,3]_3\) ternary self-dual MDS code that has connections to lattices \NoCaseChange{\protect\cite{cite39}}. Its weight enumerator is the Gleason polynomial \(g_4\) \NoCaseChange{\protect\cite[{Rem. 4.2.6}]{cite40}}.}\\ 
\addlinespace[\myxtraspc]
\eczhRefIndex{code:reed_solomon_4}%
\eczhListValue{\flmRefsHyperref{code:reed_solomon_4}{\([4,2,3]_4\) RS\(_4\) code}} & \eczhListValue{A Type II Euclidean self-dual extended RS code that is the smallest quaternary extended QR code \NoCaseChange{\protect\cite[{pg. 296}]{cite41}\protect\cite[{Sec. 2.4.2}]{cite42}}.
Puncturing the \([4,2,3]_4\) RS\(_4\) code yields the \([3,2,2]_4\) shortened RS\(_4\) code, which is an RS code \NoCaseChange{\protect\cite[{pg. 295}]{cite41}}.}\\ 
\addlinespace[\myxtraspc]
\eczhRefIndex{code:shortened_hexacode}%
\eczhListValue{\flmRefsHyperref{code:shortened_hexacode}{\([5,3,3]_4\) Shortened hexacode}} & \eczhListValue{A perfect \([5,3,3]_4\) quaternary Hamming code that is the result of puncturing the hexacode \NoCaseChange{\protect\cite{cite43}}.}\\ 
\addlinespace[\myxtraspc]
\eczhRefIndex{code:hexacode}%
\eczhListValue{\flmRefsHyperref{code:hexacode}{\([6,3,4]_4\) Hexacode}} & \eczhListValue{The \([6,3,4]_4\) Hermitian self-dual MDS code that has connections to projective geometry, lattices \NoCaseChange{\protect\cite{cite39}}, and conformal field theory \NoCaseChange{\protect\cite{cite44}}. Its weight enumerator is the Gleason polynomial \(g_7\) \NoCaseChange{\protect\cite[{Rem. 4.2.6}]{cite40}}.}\\ 
\addlinespace[\myxtraspc]
\eczhRefIndex{code:simplex734}%
\eczhListValue{\flmRefsHyperref{code:simplex734}{\([7,3,4]\) simplex code}} & \eczhListValue{Second-smallest nontrivial member of the simplex-code family.
The columns of its generator matrix are in one-to-one correspondence with the elements of the projective space \(PG(2,2)\), with each column being a chosen representative of the corresponding element.
The codewords form a \((8,9)\) simplex spherical code under the \flmRefsHyperref{ref38}{antipodal mapping}.
As a simplex code, it is equidistant: every nonzero codeword has Hamming weight \(4\).}\\ 
\addlinespace[\myxtraspc]
\eczhRefIndex{code:hamming743}%
\eczhListValue{\flmRefsHyperref{code:hamming743}{\([7,4,3]\) Hamming code}} & \eczhListValue{Second-smallest member of the Hamming code family.}\\ 
\addlinespace[\myxtraspc]
\eczhRefIndex{code:hamming844}%
\eczhListValue{\flmRefsHyperref{code:hamming844}{\([8,4,4]\) extended Hamming code}} & \eczhListValue{Extension of the \([7,4,3]\) Hamming code by a parity-check bit.
The smallest doubly even self-dual code, and the unique Type II code of length \(8\) \NoCaseChange{\protect\cite[{Rem. 4.3.10}]{cite40}}.}\\ 
\addlinespace[\myxtraspc]
\eczhRefIndex{code:parity_check}%
\eczhListValue{\flmRefsHyperref{code:parity_check}{\([n,n-1,2]\) Single parity-check (SPC) code}} & \eczhListValue{An \([n,n-1,2]\) linear binary code whose codewords consist of the message string appended with a \textit{parity-check bit} or \textit{parity bit} such that the parity (i.e., sum over all coordinates of each codeword) is zero.
If the Hamming weight of a message is odd (even), then the parity bit is one (zero).
This code requires only one extra bit of overhead and is therefore inexpensive.
Its codewords are all even-weight binary strings, and its parity-check matrix is a row vector of all ones.
Its automorphism group is \(S_n\).}\\ 
\addlinespace[\myxtraspc]
\eczhRefIndex{code:q-ary_parity_check}%
\eczhListValue{\flmRefsHyperref{code:q-ary_parity_check}{\([n,n-1,2]_q\) \(q\)-ary parity-check code}} & \eczhListValue{An \([n,n-1,2]_q\) linear \(q\)-ary code whose codewords consist of the message string appended with a \textit{parity-check} or \textit{zero-sum check digit} such that the sum over all coordinates of each codeword is zero.}\\ 
\addlinespace[\myxtraspc]
\eczhRefIndex{code:q-ary_hamming}%
\eczhListValue{\flmRefsHyperref{code:q-ary_hamming}{\(q\)-ary Hamming code}} & \eczhListValue{Member of an infinite family of perfect linear \(q\)-ary codes with parameters \([(q^r-1)/(q-1),(q^r-1)/(q-1)-r, 3]_q\) for \(r \geq 2\) \NoCaseChange{\protect\cite[{(3.1)}]{cite70}}.
These are precisely the nontrivial perfect linear codes over \(\mathbb{F}_q\) \NoCaseChange{\protect\cite[{Thm. 3.3.1}]{cite70}}.}\\ 
\end{tabularx}\endgroup
\eczcodelist{spherical_design}{Spherical designs}%

\eczhCodeListAutoDescription{All descendants of \flmRefsCref{code:spherical_design}.}%

\eczhIncludeCodeGraph{Bare}{scale=0.5}{\columnwidth}{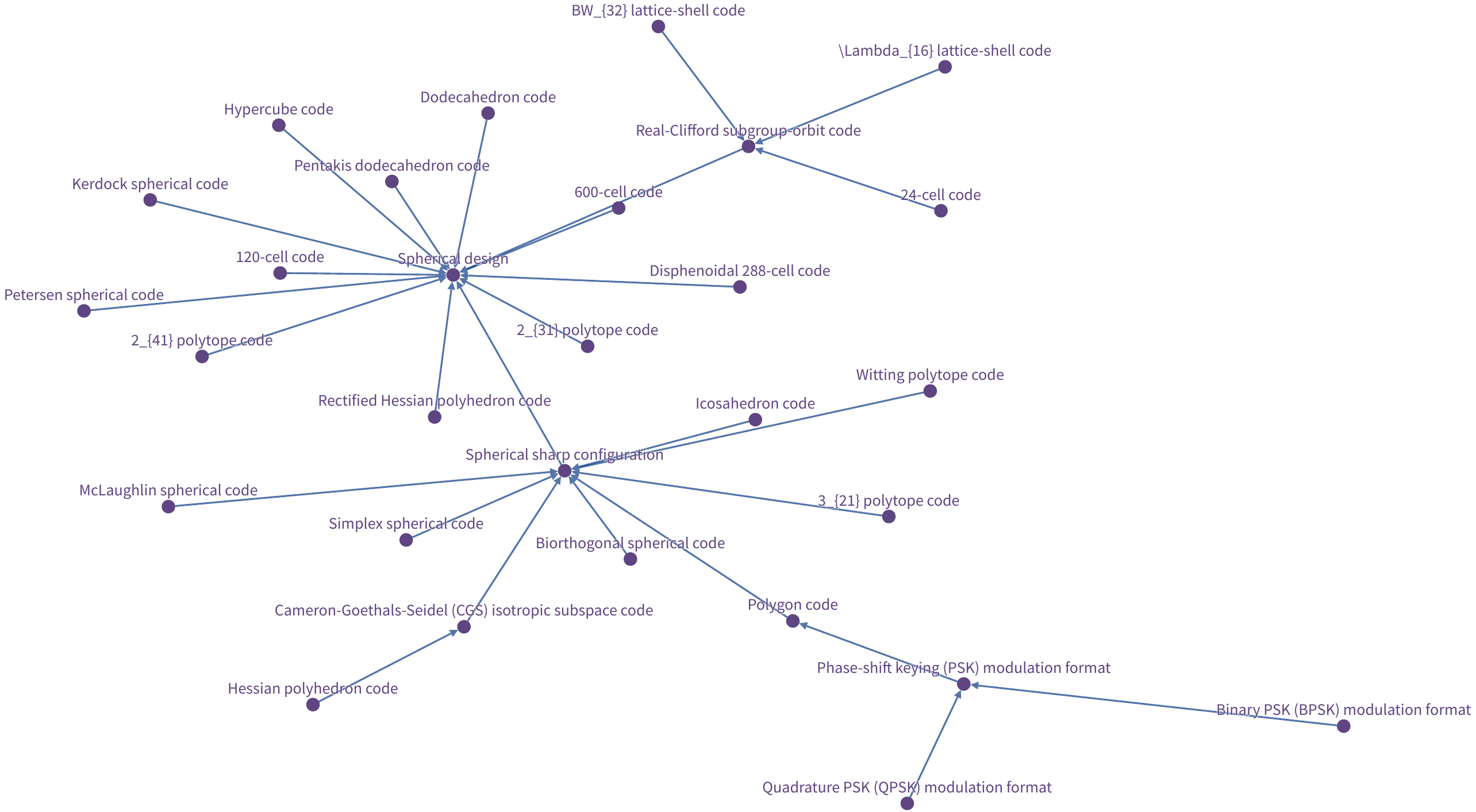}{Spherical designs}{https://errorcorrectionzoo.org/code_graph#J\%7B\%22displayMode\%22\%3A\%22subset\%22\%2C\%22modeSubsetOptions\%22\%3A\%7B\%22codeIds\%22\%3A\%5B\%22120cell\%22\%2C\%2224cell\%22\%2C\%22600cell\%22\%2C\%22bpsk\%22\%2C\%22biorthogonal_spherical\%22\%2C\%22cgs_spherical\%22\%2C\%22disphenoidal288cell\%22\%2C\%22dodecahedron\%22\%2C\%22hessian_polyhedron\%22\%2C\%22hypercube\%22\%2C\%22icosahedron\%22\%2C\%22kerdock_spherical\%22\%2C\%22mclaughlin\%22\%2C\%22pentakis_dodecahedron\%22\%2C\%22petersen_spherical\%22\%2C\%22psk\%22\%2C\%22polygon\%22\%2C\%22qpsk\%22\%2C\%22sidelnikov\%22\%2C\%22rect_hessian_polyhedron\%22\%2C\%22simplex_spherical\%22\%2C\%22spherical_design\%22\%2C\%22sharp_config\%22\%2C\%22witting_polytope\%22\%2C\%22231_polytope\%22\%2C\%22241_polytope\%22\%2C\%22hess_polytope\%22\%2C\%22bw32_shell\%22\%2C\%22lambda16_shell\%22\%5D\%2C\%22reusePreviousLayoutPositions\%22\%3Afalse\%2C\%22showIntermediateConnectingNodes\%22\%3Atrue\%2C\%22connectingNodesMaxDepth\%22\%3A15\%2C\%22connectingNodesPathMaxLength\%22\%3A20\%2C\%22connectingNodesMaxExtraDepth\%22\%3A3\%2C\%22connectingNodesOnlyKeepPathsWithAdditionalLength\%22\%3A1\%2C\%22connectingNodesToDomainsAndKingdoms\%22\%3Afalse\%2C\%22connectingNodesEdgeLengthsByType\%22\%3A\%7B\%22primaryParent\%22\%3A1\%2C\%22secondaryParent\%22\%3A4\%2C\%22cousin\%22\%3A6\%7D\%2C\%22nodeIds\%22\%3A\%5B\%5D\%7D\%2C\%22highlightImportantNodes\%22\%3A\%7B\%22highlightImportantNodes\%22\%3Afalse\%2C\%22highlightPrimaryParents\%22\%3Afalse\%2C\%22highlightRootConnectingEdges\%22\%3Afalse\%7D\%7D}

\begingroup
\small
\eczhBreakableDashes
\renewcommand\arraystretch{1.05}
\edef\myxtraspc{\eczListAddVSpaceXtraXtraStretch}
\begin{tabularx}{\linewidth}{>{\raggedright\arraybackslash}p{\eczListColWidth{name}} >{\hsize=1.0000\hsize }X}
\toprule
\eczListColTitle{Code} & \eczListColTitle{Relation} \\
\midrule
\endfirsthead
\toprule
\eczListColTitleContinued{Code} & \eczListColTitleContinued{Relation} \\
\midrule
\endhead
\bottomrule
\endfoot
\eczhRefIndex{code:120cell}%
\eczhListValue{\flmRefsHyperref{code:120cell}{120-cell code}} & \eczhListValue{The code forms a spherical 11-design because its vertices can be divided into five 600-cells, each of which forms said design.}\\ 
\addlinespace[\myxtraspc]
\eczhRefIndex{code:24cell}%
\eczhListValue{\flmRefsHyperref{code:24cell}{24-cell code}} & \eczhListValue{The 24-cell code is a spherical 5-design \NoCaseChange{\protect\cite{cite377}}.}\\ 
\addlinespace[\myxtraspc]
\eczhRefIndex{code:600cell}%
\eczhListValue{\flmRefsHyperref{code:600cell}{600-cell code}} & \eczhListValue{The 600-cell code forms a spherical 11-design that is unique up to equivalence \NoCaseChange{\protect\cite{cite378}}.}\\ 
\addlinespace[\myxtraspc]
\eczhRefIndex{code:bpsk}%
\eczhListValue{\flmRefsHyperref{code:bpsk}{Binary PSK (BPSK) modulation format}} & \eczhListValue{\eczListValueNA }\\ 
\addlinespace[\myxtraspc]
\eczhRefIndex{code:biorthogonal_spherical}%
\eczhListValue{\flmRefsHyperref{code:biorthogonal_spherical}{Biorthogonal spherical code}} & \eczhListValue{Biorthogonal spherical codes are the only tight spherical 3-designs \NoCaseChange{\protect\cite[{Tab. 9.3}]{cite115}}. A suitable weighted union of the vertices of a hypercube and an orthoplex forms a weighted spherical 5-design in dimensions \(\geq 3\) \NoCaseChange{\protect\cite[{Sec. 8.6, Ex. 5-2}]{cite379}\protect\cite[{Exam. 2.6}]{cite380}}.}\\ 
\addlinespace[\myxtraspc]
\eczhRefIndex{code:cgs_spherical}%
\eczhListValue{\flmRefsHyperref{code:cgs_spherical}{Cameron-Goethals-Seidel (CGS) isotropic subspace code}} & \eczhListValue{\eczListValueNA }\\ 
\addlinespace[\myxtraspc]
\eczhRefIndex{code:disphenoidal288cell}%
\eczhListValue{\flmRefsHyperref{code:disphenoidal288cell}{Disphenoidal 288-cell code}} & \eczhListValue{The disphenoidal 288-cell code forms a spherical 7-design \NoCaseChange{\protect\cite{cite381}}.}\\ 
\addlinespace[\myxtraspc]
\eczhRefIndex{code:dodecahedron}%
\eczhListValue{\flmRefsHyperref{code:dodecahedron}{Dodecahedron code}} & \eczhListValue{The dodecahedron code forms a spherical 5-design \NoCaseChange{\protect\cite{cite382}}.}\\ 
\addlinespace[\myxtraspc]
\eczhRefIndex{code:hessian_polyhedron}%
\eczhListValue{\flmRefsHyperref{code:hessian_polyhedron}{Hessian polyhedron code}} & \eczhListValue{The Hessian polytope code forms a tight spherical 4-design \NoCaseChange{\protect\cite[{Exam. 7.3}]{cite383}}. The double Hessian polytope code forms a spherical 5-design \NoCaseChange{\protect\cite{cite384}}.}\\ 
\addlinespace[\myxtraspc]
\eczhRefIndex{code:hypercube}%
\eczhListValue{\flmRefsHyperref{code:hypercube}{Hypercube code}} & \eczhListValue{Hypercube codes form spherical 3-designs. The weighted union of the vertices of a hypercube and an orthoplex form a weighted spherical 5-design in dimensions \(\geq 3\) \NoCaseChange{\protect\cite[{Sec. 8.6, Ex. 5-2}]{cite379}\protect\cite[{Exam. 2.6}]{cite380}}.}\\ 
\addlinespace[\myxtraspc]
\eczhRefIndex{code:icosahedron}%
\eczhListValue{\flmRefsHyperref{code:icosahedron}{Icosahedron code}} & \eczhListValue{The icosahedron code forms a unique tight spherical 5-design \NoCaseChange{\protect\cite{cite385}\protect\cite[{Exam. 9.6.1}]{cite115}}.}\\ 
\addlinespace[\myxtraspc]
\eczhRefIndex{code:kerdock_spherical}%
\eczhListValue{\flmRefsHyperref{code:kerdock_spherical}{Kerdock spherical code}} & \eczhListValue{Kerdock spherical codes form spherical 3-designs because their codewords are unions of \(2^{2r-1}+1\) orthoplexes \NoCaseChange{\protect\cite{cite386}}.}\\ 
\addlinespace[\myxtraspc]
\eczhRefIndex{code:mclaughlin}%
\eczhListValue{\flmRefsHyperref{code:mclaughlin}{McLaughlin spherical code}} & \eczhListValue{Both McLaughlin spherical codes are sharp configurations \NoCaseChange{\protect\cite{cite119,cite387}}. The \((22,275,1/6)\) code is a unique and tight spherical 4-design, while the \((23,552,1/5)\) code is a unique and tight spherical 5-design; see Ref. \NoCaseChange{\protect\cite[{Appx. A}]{cite119}}.}\\ 
\addlinespace[\myxtraspc]
\eczhRefIndex{code:pentakis_dodecahedron}%
\eczhListValue{\flmRefsHyperref{code:pentakis_dodecahedron}{Pentakis dodecahedron code}} & \eczhListValue{Vertices of the pentakis dodecahedron form a weighted spherical 9-design \NoCaseChange{\protect\cite{cite388,cite389}\protect\cite[{Exam. 2.5}]{cite380}}.}\\ 
\addlinespace[\myxtraspc]
\eczhRefIndex{code:petersen_spherical}%
\eczhListValue{\flmRefsHyperref{code:petersen_spherical}{Petersen spherical code}} & \eczhListValue{The Petersen spherical code forms a spherical 2-design \NoCaseChange{\protect\cite{cite390}}.}\\ 
\addlinespace[\myxtraspc]
\eczhRefIndex{code:psk}%
\eczhListValue{\flmRefsHyperref{code:psk}{Phase-shift keying (PSK) modulation format}} & \eczhListValue{\eczListValueNA }\\ 
\addlinespace[\myxtraspc]
\eczhRefIndex{code:polygon}%
\eczhListValue{\flmRefsHyperref{code:polygon}{Polygon code}} & \eczhListValue{A \(q\)-gon is a tight spherical \(q-1\) design.}\\ 
\addlinespace[\myxtraspc]
\eczhRefIndex{code:qpsk}%
\eczhListValue{\flmRefsHyperref{code:qpsk}{Quadrature PSK (QPSK) modulation format}} & \eczhListValue{\eczListValueNA }\\ 
\addlinespace[\myxtraspc]
\eczhRefIndex{code:sidelnikov}%
\eczhListValue{\flmRefsHyperref{code:sidelnikov}{Real-Clifford subgroup-orbit code}} & \eczhListValue{The orbit of any point under the real Clifford subgroup is a spherical 7-design \NoCaseChange{\protect\cite{cite391}}, and some are 11-designs \NoCaseChange{\protect\cite{cite392}}.}\\ 
\addlinespace[\myxtraspc]
\eczhRefIndex{code:rect_hessian_polyhedron}%
\eczhListValue{\flmRefsHyperref{code:rect_hessian_polyhedron}{Rectified Hessian polyhedron code}} & \eczhListValue{The rectified Hessian polyhedron code forms a spherical 5-design \NoCaseChange{\protect\cite{cite393}}.}\\ 
\addlinespace[\myxtraspc]
\eczhRefIndex{code:simplex_spherical}%
\eczhListValue{\flmRefsHyperref{code:simplex_spherical}{Simplex spherical code}} & \eczhListValue{Simplex spherical codes are the only tight spherical 2-designs \NoCaseChange{\protect\cite[{Tab. 9.3}]{cite115}}. The bi-simplex is a spherical 3-design since antipodal codes have zero averages over odd-degree polynomials.}\\ 
\addlinespace[\myxtraspc]
\eczhRefIndex{code:spherical_design}%
\eczhListValue{\flmRefsHyperref{code:spherical_design}{Spherical design}} & \eczhListValue{\eczListValueNA }\\ 
\addlinespace[\myxtraspc]
\eczhRefIndex{code:sharp_config}%
\eczhListValue{\flmRefsHyperref{code:sharp_config}{Spherical sharp configuration}} & \eczhListValue{Spherical sharp configurations are spherical designs of strength \(2m-1\) for some \(m\).}\\ 
\addlinespace[\myxtraspc]
\eczhRefIndex{code:witting_polytope}%
\eczhListValue{\flmRefsHyperref{code:witting_polytope}{Witting polytope code}} & \eczhListValue{The Witting polytope code forms a tight spherical 7-design \NoCaseChange{\protect\cite{cite124}\protect\cite[{Ch. 14}]{cite39}}.}\\ 
\addlinespace[\myxtraspc]
\eczhRefIndex{code:231_polytope}%
\eczhListValue{\flmRefsHyperref{code:231_polytope}{\(2_{31}\) polytope code}} & \eczhListValue{The 126 vertices of the \(2_{31}\) polytope form a spherical 5-design \NoCaseChange{\protect\cite{cite384}}.}\\ 
\addlinespace[\myxtraspc]
\eczhRefIndex{code:241_polytope}%
\eczhListValue{\flmRefsHyperref{code:241_polytope}{\(2_{41}\) polytope code}} & \eczhListValue{The \(2_{41}\) polytope code forms a spherical 7-design \NoCaseChange{\protect\cite{cite232}}.}\\ 
\addlinespace[\myxtraspc]
\eczhRefIndex{code:hess_polytope}%
\eczhListValue{\flmRefsHyperref{code:hess_polytope}{\(3_{21}\) polytope code}} & \eczhListValue{The \(3_{21}\) polytope code forms a tight spherical 5-design \NoCaseChange{\protect\cite{cite385,cite124}\protect\cite[{Ch. 14}]{cite39}\protect\cite[{Table 1}]{cite119}}.}\\ 
\addlinespace[\myxtraspc]
\eczhRefIndex{code:bw32_shell}%
\eczhListValue{\flmRefsHyperref{code:bw32_shell}{\(BW_{32}\) lattice-shell code}} & \eczhListValue{\eczListValueNA }\\ 
\addlinespace[\myxtraspc]
\eczhRefIndex{code:lambda16_shell}%
\eczhListValue{\flmRefsHyperref{code:lambda16_shell}{\(\Lambda_{16}\) lattice-shell code}} & \eczhListValue{\eczListValueNA }\\ 
\end{tabularx}\endgroup
\eczcodelist{univ_opt}{Universally optimal codes}%

\eczhCodeListAutoDescription{All descendants of \flmRefsCref{code:univ_opt}.}%

\eczhIncludeCodeGraph{Bare}{scale=0.5}{\columnwidth}{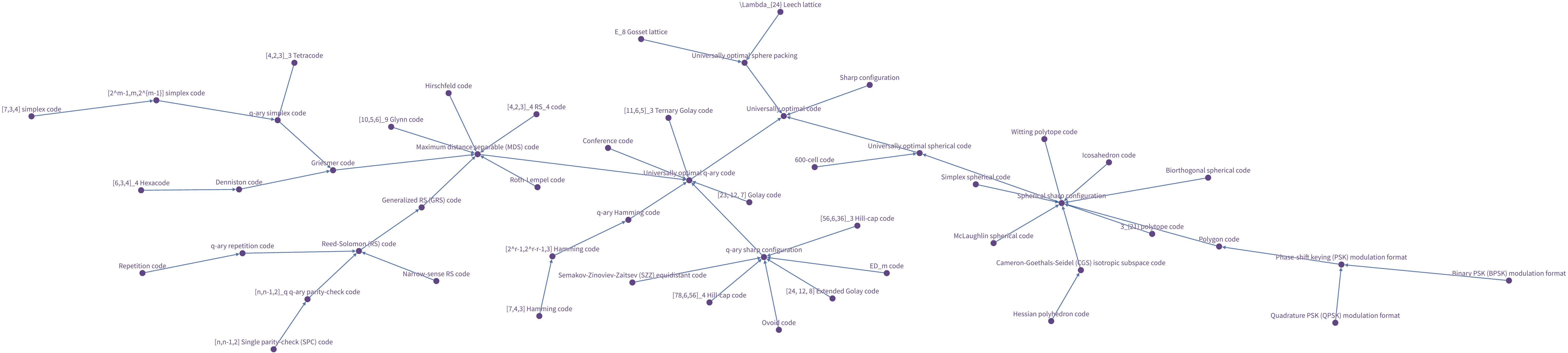}{Universally optimal codes}{https://errorcorrectionzoo.org/code_graph#J\%7B\%22displayMode\%22\%3A\%22subset\%22\%2C\%22modeSubsetOptions\%22\%3A\%7B\%22codeIds\%22\%3A\%5B\%22600cell\%22\%2C\%22bpsk\%22\%2C\%22biorthogonal_spherical\%22\%2C\%22cgs_spherical\%22\%2C\%22conference\%22\%2C\%22denniston\%22\%2C\%22generalized_reed_solomon\%22\%2C\%22griesmer\%22\%2C\%22hessian_polyhedron\%22\%2C\%22hirschfeld\%22\%2C\%22icosahedron\%22\%2C\%22mds\%22\%2C\%22mclaughlin\%22\%2C\%22narrow_sense_reed_solomon\%22\%2C\%22bose_qvist\%22\%2C\%22psk\%22\%2C\%22polygon\%22\%2C\%22qpsk\%22\%2C\%22reed_solomon\%22\%2C\%22repetition\%22\%2C\%22roth_lempel\%22\%2C\%22semakov_zinoviev_zaitsev\%22\%2C\%22delsarte_optimal\%22\%2C\%22simplex_spherical\%22\%2C\%22sharp_config\%22\%2C\%22univ_opt_q-ary\%22\%2C\%22univ_opt\%22\%2C\%22univ_opt_analog\%22\%2C\%22univ_opt_spherical\%22\%2C\%22witting_polytope\%22\%2C\%22hess_polytope\%22\%2C\%22semakov_zinoviev\%22\%2C\%22eeight\%22\%2C\%22glynn\%22\%2C\%22ternary_golay\%22\%2C\%22golay\%22\%2C\%22extended_golay\%22\%2C\%22simplex\%22\%2C\%22hamming\%22\%2C\%22tetracode\%22\%2C\%22reed_solomon_4\%22\%2C\%22hill_56_6_36\%22\%2C\%22hexacode\%22\%2C\%22simplex734\%22\%2C\%22hamming743\%22\%2C\%22hill_78_6_56\%22\%2C\%22parity_check\%22\%2C\%22q-ary_parity_check\%22\%2C\%22leech\%22\%2C\%22q-ary_hamming\%22\%2C\%22q-ary_repetition\%22\%2C\%22delsarte_optimal_q-ary\%22\%2C\%22q-ary_simplex\%22\%5D\%2C\%22reusePreviousLayoutPositions\%22\%3Afalse\%2C\%22showIntermediateConnectingNodes\%22\%3Atrue\%2C\%22connectingNodesMaxDepth\%22\%3A15\%2C\%22connectingNodesPathMaxLength\%22\%3A20\%2C\%22connectingNodesMaxExtraDepth\%22\%3A3\%2C\%22connectingNodesOnlyKeepPathsWithAdditionalLength\%22\%3A1\%2C\%22connectingNodesToDomainsAndKingdoms\%22\%3Afalse\%2C\%22connectingNodesEdgeLengthsByType\%22\%3A\%7B\%22primaryParent\%22\%3A1\%2C\%22secondaryParent\%22\%3A4\%2C\%22cousin\%22\%3A6\%7D\%2C\%22nodeIds\%22\%3A\%5B\%5D\%7D\%2C\%22highlightImportantNodes\%22\%3A\%7B\%22highlightImportantNodes\%22\%3Afalse\%2C\%22highlightPrimaryParents\%22\%3Afalse\%2C\%22highlightRootConnectingEdges\%22\%3Afalse\%7D\%7D}

\begingroup
\small
\eczhBreakableDashes
\renewcommand\arraystretch{1.05}
\edef\myxtraspc{\eczListAddVSpaceXtraXtraStretch}
\begin{tabularx}{\linewidth}{>{\raggedright\arraybackslash}p{\eczListColWidth{name}} >{\hsize=1.0000\hsize }X}
\toprule
\eczListColTitle{Code} & \eczListColTitle{Description} \\
\midrule
\endfirsthead
\toprule
\eczListColTitleContinued{Code} & \eczListColTitleContinued{Description} \\
\midrule
\endhead
\bottomrule
\endfoot
\eczhRefIndex{code:600cell}%
\eczhListValue{\flmRefsHyperref{code:600cell}{600-cell code}} & \eczhListValue{Spherical \((4,120,(3-\sqrt{5})/2)\) code whose codewords are the vertices of the 600-cell.
See \NoCaseChange{\protect\cite[{Table 1}]{cite229}\protect\cite[{Table 3}]{cite228}} for realizations of the 120 codewords.
A realization of the 600-cell can be given in terms of icosians, which are quaternion coordinates of the 120 elements of the binary icosahedral group \(2I \cong 2.A_5\) (a.k.a. the icosian group) \NoCaseChange{\protect\cite{cite230}\protect\cite[{Ch. 8, pg. 207}]{cite39}}.}\\ 
\addlinespace[\myxtraspc]
\eczhRefIndex{code:bpsk}%
\eczhListValue{\flmRefsHyperref{code:bpsk}{Binary PSK (BPSK) modulation format}} & \eczhListValue{Encodes one bit of information into a constellation of antipodal points \(\pm\alpha\) for complex \(\alpha\).
These points are typically associated with two phases of an electromagnetic signal.}\\ 
\addlinespace[\myxtraspc]
\eczhRefIndex{code:biorthogonal_spherical}%
\eczhListValue{\flmRefsHyperref{code:biorthogonal_spherical}{Biorthogonal spherical code}} & \eczhListValue{Spherical \((n,2n,2)\) code whose codewords are all permutations of the \(n\)-dimensional vectors \((0,0,\cdots,0,\pm1)\), up to normalization.
The code makes up the vertices of an \(n\)-orthoplex (a.k.a. hyperoctahedron or cross polytope).}\\ 
\addlinespace[\myxtraspc]
\eczhRefIndex{code:cgs_spherical}%
\eczhListValue{\flmRefsHyperref{code:cgs_spherical}{Cameron-Goethals-Seidel (CGS) isotropic subspace code}} & \eczhListValue{Member of a \((q(q^2-q+1),(q+1)(q^3+1),2-2/q^2)\) family of spherical codes for any prime-power \(q\).
Constructed from generalized quadrangles, which in this case correspond to sets of totally isotropic points and lines in the projective space \(PG(5,q)\) \NoCaseChange{\protect\cite[{Exam. 9.4.5}]{cite115}}.
There exist multiple distinct spherical codes using this construction for \(q>3\) \NoCaseChange{\protect\cite{cite119,cite386}}.}\\ 
\addlinespace[\myxtraspc]
\eczhRefIndex{code:conference}%
\eczhListValue{\flmRefsHyperref{code:conference}{Conference code}} & \eczhListValue{A member of the family of \((n,2n+2,(n-1)/2)\) nonlinear binary codes for \(n=1\) modulo 4 that are constructed from conference matrices.}\\ 
\addlinespace[\myxtraspc]
\eczhRefIndex{code:denniston}%
\eczhListValue{\flmRefsHyperref{code:denniston}{Denniston code}} & \eczhListValue{Projective code that is part of a family of \([2^{a+i}+2^i-2^a,3,2^{a+i}-2^a]_{2^a}\) codes for \(i < a\) constructed using Denniston arcs \NoCaseChange{\protect\cite[{Sec. 19.7.3}]{cite172}}.}\\ 
\addlinespace[\myxtraspc]
\eczhRefIndex{code:generalized_reed_solomon}%
\eczhListValue{\flmRefsHyperref{code:generalized_reed_solomon}{Generalized RS (GRS) code}} & \eczhListValue{An \([n,k,n-k+1]_q\) MDS code that is a modification of the RS code where codeword polynomials are multiplied by additional prefactors \NoCaseChange{\protect\cite[{Def. 15.3.19}]{cite26}}.}\\ 
\addlinespace[\myxtraspc]
\eczhRefIndex{code:griesmer}%
\eczhListValue{\flmRefsHyperref{code:griesmer}{Griesmer code}} & \eczhListValue{A type of \(q\)-ary code whose parameters satisfy the Griesmer bound with equality.}\\ 
\addlinespace[\myxtraspc]
\eczhRefIndex{code:hessian_polyhedron}%
\eczhListValue{\flmRefsHyperref{code:hessian_polyhedron}{Hessian polyhedron code}} & \eczhListValue{Spherical \((6,27,3/2)\) code whose codewords are the vertices of the Hessian complex polyhedron and the \(2_{21}\) polytope.
Two copies of the code yield the \((6,54,1)\) \textit{double Hessian polyhedron} (a.k.a. diplo-Schläfli) code.
The code can be obtained from the Schläfli graph \NoCaseChange{\protect\cite[{Ch. 9}]{cite115}}.
The (antipodal pairs of) points of the (double) Hessian polyhedron correspond to the 27 lines on a smooth cubic surface in \(\mathbb{C}P^3\) \NoCaseChange{\protect\cite{cite116,cite117,cite118,cite119,cite120}}.}\\ 
\addlinespace[\myxtraspc]
\eczhRefIndex{code:hirschfeld}%
\eczhListValue{\flmRefsHyperref{code:hirschfeld}{Hirschfeld code}} & \eczhListValue{A \([q+1,4,q-2]_q\) projective geometry code for non-prime \(q\) that is an example of an MDS code that is not an RS code; see \NoCaseChange{\protect\cite[{Exam. 7.6}]{cite182}} for the generator matrix.}\\ 
\addlinespace[\myxtraspc]
\eczhRefIndex{code:icosahedron}%
\eczhListValue{\flmRefsHyperref{code:icosahedron}{Icosahedron code}} & \eczhListValue{Spherical \((3,12,2-2/\sqrt{5})\) code whose codewords are the vertices of the icosahedron (alternatively, the centers of the faces of a dodecahedron, the icosahedron's dual polytope).}\\ 
\addlinespace[\myxtraspc]
\eczhRefIndex{code:mds}%
\eczhListValue{\flmRefsHyperref{code:mds}{Maximum distance separable (MDS) code}} & \eczhListValue{A \(q\)-ary linear code whose parameters satisfy the Singleton bound with equality.}\\ 
\addlinespace[\myxtraspc]
\eczhRefIndex{code:mclaughlin}%
\eczhListValue{\flmRefsHyperref{code:mclaughlin}{McLaughlin spherical code}} & \eczhListValue{The \((22,275,1/6)\) or \((23,552,1/5)\) code associated with the McLaughlin graph and the Leech lattice.
See Ref. \NoCaseChange{\protect\cite{cite387}} for explicit constructions of and relations between both codes.}\\ 
\addlinespace[\myxtraspc]
\eczhRefIndex{code:narrow_sense_reed_solomon}%
\eczhListValue{\flmRefsHyperref{code:narrow_sense_reed_solomon}{Narrow-sense RS code}} & \eczhListValue{An \([q-1,k,n-k+1]_q\) RS code whose points \(\alpha_i\) are all \((i-1)\)st powers of a \flmRefsHyperref{ref33}{primitive} element \(\alpha\) of \(\mathbb{F}_q\).}\\ 
\addlinespace[\myxtraspc]
\eczhRefIndex{code:bose_qvist}%
\eczhListValue{\flmRefsHyperref{code:bose_qvist}{Ovoid code}} & \eczhListValue{Member of a \([q^2+1,4,q^2-q]_q\) projective two-weight code family obtained from ovoids in \(\mathrm{PG}(3,q)\).
If the columns of a generator matrix are the \(q^2+1\) points of an ovoid, then every hyperplane meets the ovoid in either \(1\) or \(q+1\) points, yielding the two nonzero weights \(q^2\) and \(q^2-q\).
See \NoCaseChange{\protect\cite[{pg. 107}]{cite203}\protect\cite[{pg. 192}]{cite62}} for further details.}\\ 
\addlinespace[\myxtraspc]
\eczhRefIndex{code:psk}%
\eczhListValue{\flmRefsHyperref{code:psk}{Phase-shift keying (PSK) modulation format}} & \eczhListValue{A \(q\)-ary phase-shift keying (\(q\)-PSK) encodes one \(q\)-ary digit of information into a constellation of \(q\) points distributed equidistantly on a circle in \(\mathbb{C}\) or, equivalently, \(\mathbb{R}^2\).}\\ 
\addlinespace[\myxtraspc]
\eczhRefIndex{code:polygon}%
\eczhListValue{\flmRefsHyperref{code:polygon}{Polygon code}} & \eczhListValue{Spherical \((1,q,4\sin^2 \frac{\pi}{q})\) code for any \(q\geq2\) whose codewords are the vertices of a \(q\)-gon. Special cases include the line segment (\(q=2\)), triangle (\(q=3\)), square (\(q=4\)), pentagon (\(q=5\)), and hexagon (\(q=6\)).}\\ 
\addlinespace[\myxtraspc]
\eczhRefIndex{code:qpsk}%
\eczhListValue{\flmRefsHyperref{code:qpsk}{Quadrature PSK (QPSK) modulation format}} & \eczhListValue{A quaternary encoding into a constellation of four points distributed equidistantly on a circle.
For the case of \(\pi/4\)-QPSK, the constellation is \(\{e^{\pm i\frac{\pi}{4}},e^{\pm i\frac{3\pi}{4}}\}\).}\\ 
\addlinespace[\myxtraspc]
\eczhRefIndex{code:reed_solomon}%
\eczhListValue{\flmRefsHyperref{code:reed_solomon}{Reed-Solomon (RS) code}} & \eczhListValue{An \([n,k,n-k+1]_q\) linear code based on polynomials over \(\mathbb{F}_q\).}\\ 
\addlinespace[\myxtraspc]
\eczhRefIndex{code:repetition}%
\eczhListValue{\flmRefsHyperref{code:repetition}{Repetition code}} & \eczhListValue{\([n,1,n]\) binary linear code encoding one bit of information into an \(n\)-bit string.
Majority decoding requires \(n\) to be odd in order to avoid ties.
The idea is to increase the code distance by repeating the logical information several times. It is a \((n,1)\)-Hamming code.
Its automorphism group is \(S_n\).}\\ 
\addlinespace[\myxtraspc]
\eczhRefIndex{code:roth_lempel}%
\eczhListValue{\flmRefsHyperref{code:roth_lempel}{Roth-Lempel code}} & \eczhListValue{Member of a \(q\)-ary linear code family that includes many examples of MDS codes that are not GRS codes.}\\ 
\addlinespace[\myxtraspc]
\eczhRefIndex{code:semakov_zinoviev_zaitsev}%
\eczhListValue{\flmRefsHyperref{code:semakov_zinoviev_zaitsev}{Semakov-Zinoviev-Zaitsev (SZZ) equidistant code}} & \eczhListValue{Member of a nonlinear \(q\)-ary code family that is related to affine resolvable block designs and that is universally optimal.}\\ 
\addlinespace[\myxtraspc]
\eczhRefIndex{code:delsarte_optimal}%
\eczhListValue{\flmRefsHyperref{code:delsarte_optimal}{Sharp configuration}} & \eczhListValue{A code \(W\) in a compact connected two-point homogeneous space with \(m=l(W)\) distinct distances such that either \(r(W) \geq 2m-1\), or \(r(W)=2m-2>0\) and \(W\) is diametrical \NoCaseChange{\protect\cite{cite171}}.}\\ 
\addlinespace[\myxtraspc]
\eczhRefIndex{code:simplex_spherical}%
\eczhListValue{\flmRefsHyperref{code:simplex_spherical}{Simplex spherical code}} & \eczhListValue{Spherical \((n,n+1,2+2/n)\) code whose codewords are all permutations of the \(n+1\)-dimensional vector \((1,1,\cdots,1,-n)\), up to normalization, forming an \(n\)-simplex.
Codewords are all equidistant and their components add up to zero.
Simplex spherical codewords in 2 (3, 4) dimensions form the vertices of a triangle (tetrahedron, 5-cell).
In general, the code makes up the vertices of an \(n\)-simplex.
The union of a simplex and its antipodal simplex forms the vertices of a bi-simplex, which has \(2(n+1)\) vertices.}\\ 
\addlinespace[\myxtraspc]
\eczhRefIndex{code:sharp_config}%
\eczhListValue{\flmRefsHyperref{code:sharp_config}{Spherical sharp configuration}} & \eczhListValue{A spherical code that is a spherical design of strength \(2m-1\) for some \(m\) and that has \(m\) distances between distinct points.
All known spherical sharp configurations are either obtained from the Leech or \(E_8\) lattice, certain regular polytopes, or are CGS isotropic subspace spherical codes \NoCaseChange{\protect\cite[{Table 1}]{cite394}} (see also \NoCaseChange{\protect\cite[{Table 9.1}]{cite171}}).}\\ 
\addlinespace[\myxtraspc]
\eczhRefIndex{code:univ_opt_q-ary}%
\eczhListValue{\flmRefsHyperref{code:univ_opt_q-ary}{Universally optimal \(q\)-ary code}} & \eczhListValue{A \(q\)-ary code that (weakly) minimizes all completely monotonic potentials on Hamming space \NoCaseChange{\protect\cite{cite173}}.
Equivalently, its binomial moments are minimal among all codes with the same size and block length \NoCaseChange{\protect\cite[{Lemma 4}]{cite173}}.}\\ 
\addlinespace[\myxtraspc]
\eczhRefIndex{code:univ_opt}%
\eczhListValue{\flmRefsHyperref{code:univ_opt}{Universally optimal code}} & \eczhListValue{A code that minimizes, among all codes of the same cardinality, a large family of potential functions.}\\ 
\addlinespace[\myxtraspc]
\eczhRefIndex{code:univ_opt_analog}%
\eczhListValue{\flmRefsHyperref{code:univ_opt_analog}{Universally optimal sphere packing}} & \eczhListValue{A periodic sphere packing that (weakly) minimizes all completely monotonic potentials of square Euclidean distance among all periodic packings of the same density.}\\ 
\addlinespace[\myxtraspc]
\eczhRefIndex{code:univ_opt_spherical}%
\eczhListValue{\flmRefsHyperref{code:univ_opt_spherical}{Universally optimal spherical code}} & \eczhListValue{A spherical code that (weakly) minimizes all completely monotonic potentials on the sphere for its cardinality. See \NoCaseChange{\protect\cite{cite395}\protect\cite[{Sec. 12.4}]{cite199}} for further discussion.
In particular, all spherical sharp configurations and the 600-cell are universally optimal \NoCaseChange{\protect\cite[{Thm. 1.2}]{cite119}}.}\\ 
\addlinespace[\myxtraspc]
\eczhRefIndex{code:witting_polytope}%
\eczhListValue{\flmRefsHyperref{code:witting_polytope}{Witting polytope code}} & \eczhListValue{Spherical \((8,240,1)\) code whose codewords are the vertices of the Witting complex polytope, the \(4_{21}\) polytope, and the minimal lattice-shell code of the \(E_8\) lattice.
The code is optimal and unique up to equivalence \NoCaseChange{\protect\cite{cite124,cite39,cite125}}.
Antipodal pairs of points of the \(4_{21}\) polytope code correspond to the 120 tritangent planes of a canonical sextic curve in \(\mathbb{C}P^3\) \NoCaseChange{\protect\cite{cite117,cite118,cite119,cite120}}.}\\ 
\addlinespace[\myxtraspc]
\eczhRefIndex{code:hess_polytope}%
\eczhListValue{\flmRefsHyperref{code:hess_polytope}{\(3_{21}\) polytope code}} & \eczhListValue{Spherical \((7,56,1/3)\) code whose codewords are the vertices of the \(3_{21}\) polytope (a.k.a. the Hess polytope).
The vertices form the kissing configuration of the Witting polytope code.
The 1-skeleton of this polytope is the Gosset graph \NoCaseChange{\protect\cite{cite178}}.
The code is optimal and unique up to equivalence \NoCaseChange{\protect\cite{cite124,cite39,cite125}}.}\\ 
\addlinespace[\myxtraspc]
\eczhRefIndex{code:semakov_zinoviev}%
\eczhListValue{\flmRefsHyperref{code:semakov_zinoviev}{\(ED_m\) code}} & \eczhListValue{Member of a family of nonlinear \( (c\frac{qt-1}{(t-1,q-1)},qt,ct\frac{q-1}{(t-1,q-1)}) \) \(q\)-ary codes, where \(c,t\geq 1\) and \((a,b)\) is the greatest common divisor of \(a\) and \(b\).
Such codes are universally optimal and are related to resolvable block designs.}\\ 
\addlinespace[\myxtraspc]
\eczhRefIndex{code:eeight}%
\eczhListValue{\flmRefsHyperref{code:eeight}{\(E_8\) Gosset lattice}} & \eczhListValue{Even unimodular BW lattice in dimension \(8\), consisting of the Cayley integral octonions rescaled by \(\sqrt{2}\).
The lattice corresponds to the \([8,4,4]\) Hamming code via \flmTerm{term}{ref127}{}{Construction A}.}\\ 
\addlinespace[\myxtraspc]
\eczhRefIndex{code:glynn}%
\eczhListValue{\flmRefsHyperref{code:glynn}{\([10,5,6]_9\) Glynn code}} & \eczhListValue{The unique trace-Hermitian self-dual \([10,5,6]_9\) code, constructed using a 10-arc in \(PG(4,9)\) that is not a rational curve.}\\ 
\addlinespace[\myxtraspc]
\eczhRefIndex{code:ternary_golay}%
\eczhListValue{\flmRefsHyperref{code:ternary_golay}{\([11,6,5]_3\) Ternary Golay code}} & \eczhListValue{A \([11,6,5]_3\) perfect ternary linear code with connections to various areas of mathematics, e.g., lattices \NoCaseChange{\protect\cite{cite39}} and sporadic simple groups \NoCaseChange{\protect\cite{cite41}}.
Adding a parity bit to the code results in the self-dual \([12,6,6]_3\) \textit{extended ternary Golay code}, whose weight enumerator is the Gleason polynomial \(g_5\) \NoCaseChange{\protect\cite[{Rem. 4.2.6}]{cite40}}.
Up to equivalence, both codes are unique for their respective parameters \NoCaseChange{\protect\cite{cite102}}.
The dual of the ternary Golay code is a \([11,5,6]_3\) projective two-weight subcode \NoCaseChange{\protect\cite[{Exam. 19.3.2}]{cite172}}.}\\ 
\addlinespace[\myxtraspc]
\eczhRefIndex{code:golay}%
\eczhListValue{\flmRefsHyperref{code:golay}{\([23, 12, 7]\) Golay code}} & \eczhListValue{A \([23, 12, 7]\) perfect binary linear code with connections to various areas of mathematics, e.g., lattices \NoCaseChange{\protect\cite{cite39}} and sporadic simple groups \NoCaseChange{\protect\cite{cite41}}.
Up to equivalence, it is unique for its parameters \NoCaseChange{\protect\cite{cite102}}.
The dual of the Golay code is its \([23,11,8]\) even-weight subcode \NoCaseChange{\protect\cite{cite103,cite104}}.}\\ 
\addlinespace[\myxtraspc]
\eczhRefIndex{code:extended_golay}%
\eczhListValue{\flmRefsHyperref{code:extended_golay}{\([24, 12, 8]\) Extended Golay code}} & \eczhListValue{A self-dual \([24, 12, 8]\) code that is obtained from the Golay code by adding a parity check.
Equivalently, puncturing any coordinate yields the \([23,12,7]\) Golay code.
Up to equivalence, it is unique for its parameters \NoCaseChange{\protect\cite{cite102}}, and it is the unique \([24,12,8]\) extremal Type II code \NoCaseChange{\protect\cite[{Rems. 4.3.10 and 4.3.11}]{cite40}}.}\\ 
\addlinespace[\myxtraspc]
\eczhRefIndex{code:simplex}%
\eczhListValue{\flmRefsHyperref{code:simplex}{\([2^m-1,m,2^{m-1}]\) simplex code}} & \eczhListValue{A member of the equidistant code family dual to the \([2^m-1,2^m-m-1,3]\) Hamming family.}\\ 
\addlinespace[\myxtraspc]
\eczhRefIndex{code:hamming}%
\eczhListValue{\flmRefsHyperref{code:hamming}{\([2^r-1,2^r-r-1,3]\) Hamming code}} & \eczhListValue{Member of an infinite family of perfect linear codes with parameters \([2^r-1,2^r-r-1, 3]\) for \(r \geq 2\).
Their \(r \times (2^r-1) \) parity-check matrix \(H\) has all possible nonzero \(r\)-bit strings as its columns.
Adding a parity check yields the \([2^r,2^r-r-1, 4]\) extended Hamming code.}\\ 
\addlinespace[\myxtraspc]
\eczhRefIndex{code:tetracode}%
\eczhListValue{\flmRefsHyperref{code:tetracode}{\([4,2,3]_3\) Tetracode}} & \eczhListValue{The \([4,2,3]_3\) ternary self-dual MDS code that has connections to lattices \NoCaseChange{\protect\cite{cite39}}. Its weight enumerator is the Gleason polynomial \(g_4\) \NoCaseChange{\protect\cite[{Rem. 4.2.6}]{cite40}}.}\\ 
\addlinespace[\myxtraspc]
\eczhRefIndex{code:reed_solomon_4}%
\eczhListValue{\flmRefsHyperref{code:reed_solomon_4}{\([4,2,3]_4\) RS\(_4\) code}} & \eczhListValue{A Type II Euclidean self-dual extended RS code that is the smallest quaternary extended QR code \NoCaseChange{\protect\cite[{pg. 296}]{cite41}\protect\cite[{Sec. 2.4.2}]{cite42}}.
Puncturing the \([4,2,3]_4\) RS\(_4\) code yields the \([3,2,2]_4\) shortened RS\(_4\) code, which is an RS code \NoCaseChange{\protect\cite[{pg. 295}]{cite41}}.}\\ 
\addlinespace[\myxtraspc]
\eczhRefIndex{code:hill_56_6_36}%
\eczhListValue{\flmRefsHyperref{code:hill_56_6_36}{\([56,6,36]_3\) Hill-cap code}} & \eczhListValue{Projective two-weight ternary code based on the Games graph \NoCaseChange{\protect\cite{cite206}\protect\cite[{Table 19.1}]{cite172}} and Hill's 56-cap \NoCaseChange{\protect\cite{cite207}}.
Its automorphism group contains \(PSL(3,4)\) \NoCaseChange{\protect\cite{cite208}}.}\\ 
\addlinespace[\myxtraspc]
\eczhRefIndex{code:hexacode}%
\eczhListValue{\flmRefsHyperref{code:hexacode}{\([6,3,4]_4\) Hexacode}} & \eczhListValue{The \([6,3,4]_4\) Hermitian self-dual MDS code that has connections to projective geometry, lattices \NoCaseChange{\protect\cite{cite39}}, and conformal field theory \NoCaseChange{\protect\cite{cite44}}. Its weight enumerator is the Gleason polynomial \(g_7\) \NoCaseChange{\protect\cite[{Rem. 4.2.6}]{cite40}}.}\\ 
\addlinespace[\myxtraspc]
\eczhRefIndex{code:simplex734}%
\eczhListValue{\flmRefsHyperref{code:simplex734}{\([7,3,4]\) simplex code}} & \eczhListValue{Second-smallest nontrivial member of the simplex-code family.
The columns of its generator matrix are in one-to-one correspondence with the elements of the projective space \(PG(2,2)\), with each column being a chosen representative of the corresponding element.
The codewords form a \((8,9)\) simplex spherical code under the \flmRefsHyperref{ref38}{antipodal mapping}.
As a simplex code, it is equidistant: every nonzero codeword has Hamming weight \(4\).}\\ 
\addlinespace[\myxtraspc]
\eczhRefIndex{code:hamming743}%
\eczhListValue{\flmRefsHyperref{code:hamming743}{\([7,4,3]\) Hamming code}} & \eczhListValue{Second-smallest member of the Hamming code family.}\\ 
\addlinespace[\myxtraspc]
\eczhRefIndex{code:hill_78_6_56}%
\eczhListValue{\flmRefsHyperref{code:hill_78_6_56}{\([78,6,56]_4\) Hill-cap code}} & \eczhListValue{Projective two-weight quaternary code based on a cap that corresponds to a strongly regular graph \NoCaseChange{\protect\cite[{Table 7.1}]{cite206}}.}\\ 
\addlinespace[\myxtraspc]
\eczhRefIndex{code:parity_check}%
\eczhListValue{\flmRefsHyperref{code:parity_check}{\([n,n-1,2]\) Single parity-check (SPC) code}} & \eczhListValue{An \([n,n-1,2]\) linear binary code whose codewords consist of the message string appended with a \textit{parity-check bit} or \textit{parity bit} such that the parity (i.e., sum over all coordinates of each codeword) is zero.
If the Hamming weight of a message is odd (even), then the parity bit is one (zero).
This code requires only one extra bit of overhead and is therefore inexpensive.
Its codewords are all even-weight binary strings, and its parity-check matrix is a row vector of all ones.
Its automorphism group is \(S_n\).}\\ 
\addlinespace[\myxtraspc]
\eczhRefIndex{code:q-ary_parity_check}%
\eczhListValue{\flmRefsHyperref{code:q-ary_parity_check}{\([n,n-1,2]_q\) \(q\)-ary parity-check code}} & \eczhListValue{An \([n,n-1,2]_q\) linear \(q\)-ary code whose codewords consist of the message string appended with a \textit{parity-check} or \textit{zero-sum check digit} such that the sum over all coordinates of each codeword is zero.}\\ 
\addlinespace[\myxtraspc]
\eczhRefIndex{code:leech}%
\eczhListValue{\flmRefsHyperref{code:leech}{\(\Lambda_{24}\) Leech lattice}} & \eczhListValue{Even unimodular lattice in 24 dimensions that exhibits optimal packing.
Its automorphism group is the Conway group Co\(_0\).}\\ 
\addlinespace[\myxtraspc]
\eczhRefIndex{code:q-ary_hamming}%
\eczhListValue{\flmRefsHyperref{code:q-ary_hamming}{\(q\)-ary Hamming code}} & \eczhListValue{Member of an infinite family of perfect linear \(q\)-ary codes with parameters \([(q^r-1)/(q-1),(q^r-1)/(q-1)-r, 3]_q\) for \(r \geq 2\) \NoCaseChange{\protect\cite[{(3.1)}]{cite70}}.
These are precisely the nontrivial perfect linear codes over \(\mathbb{F}_q\) \NoCaseChange{\protect\cite[{Thm. 3.3.1}]{cite70}}.}\\ 
\addlinespace[\myxtraspc]
\eczhRefIndex{code:q-ary_repetition}%
\eczhListValue{\flmRefsHyperref{code:q-ary_repetition}{\(q\)-ary repetition code}} & \eczhListValue{An \([n,1,n]_q\) code consisting of codewords \((j,j,\cdots,j)\) for \(j \in \mathbb{F}_q\).}\\ 
\addlinespace[\myxtraspc]
\eczhRefIndex{code:delsarte_optimal_q-ary}%
\eczhListValue{\flmRefsHyperref{code:delsarte_optimal_q-ary}{\(q\)-ary sharp configuration}} & \eczhListValue{A \(q\)-ary code that admits \(m\) different distances between distinct codewords and forms a design of strength \(2m-1\) or greater.}\\ 
\addlinespace[\myxtraspc]
\eczhRefIndex{code:q-ary_simplex}%
\eczhListValue{\flmRefsHyperref{code:q-ary_simplex}{\(q\)-ary simplex code}} & \eczhListValue{An \([n,m,q^{m-1}]_q\) equidistant projective code with \(n=\frac{q^m-1}{q-1}\), denoted as \(S(q,m)\). The columns of the generator matrix are in one-to-one correspondence with the elements of the projective space \(PG(m-1,q)\), with each column being a chosen representative of the corresponding element.
All nonzero simplex codewords have a constant weight of \(q^{m-1}\) \NoCaseChange{\protect\cite{cite45,cite46}}.}\\ 
\end{tabularx}\endgroup
\eczlistdomainsection{Quantum code families}

\eczcodelist{stabilizer_1d}{1D stabilizer codes
}%

\eczhCodeListAutoDescription{All descendants of \flmRefsCref{code:1d_stabilizer}.}%

\eczhIncludeCodeGraph{Bare}{scale=0.5}{\columnwidth}{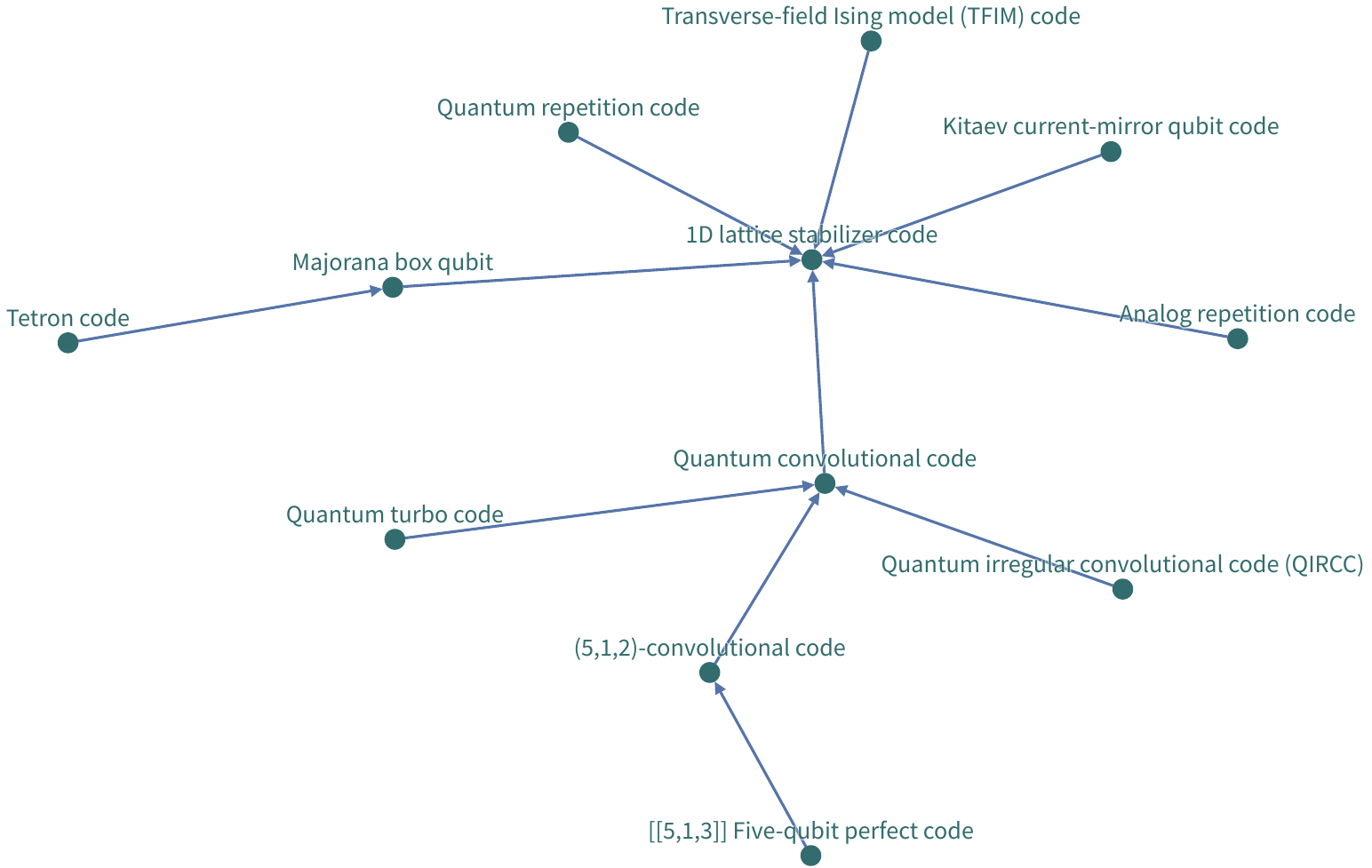}{1D stabilizer codes}{https://errorcorrectionzoo.org/code_graph#J\%7B\%22displayMode\%22\%3A\%22subset\%22\%2C\%22modeSubsetOptions\%22\%3A\%7B\%22codeIds\%22\%3A\%5B\%221d_stabilizer\%22\%2C\%22analog_repetition\%22\%2C\%22current_mirror\%22\%2C\%22mbq\%22\%2C\%22quantum_convolutional\%22\%2C\%22quantum_irregular_convolutional\%22\%2C\%22quantum_repetition\%22\%2C\%22quantum_turbo\%22\%2C\%22tetron\%22\%2C\%22tfim\%22\%2C\%22stab_5_1_2_convolutional\%22\%2C\%22stab_5_1_3\%22\%5D\%2C\%22reusePreviousLayoutPositions\%22\%3Afalse\%2C\%22showIntermediateConnectingNodes\%22\%3Atrue\%2C\%22connectingNodesMaxDepth\%22\%3A15\%2C\%22connectingNodesPathMaxLength\%22\%3A20\%2C\%22connectingNodesMaxExtraDepth\%22\%3A3\%2C\%22connectingNodesOnlyKeepPathsWithAdditionalLength\%22\%3A1\%2C\%22connectingNodesToDomainsAndKingdoms\%22\%3Afalse\%2C\%22connectingNodesEdgeLengthsByType\%22\%3A\%7B\%22primaryParent\%22\%3A1\%2C\%22secondaryParent\%22\%3A4\%2C\%22cousin\%22\%3A6\%7D\%2C\%22nodeIds\%22\%3A\%5B\%5D\%7D\%2C\%22highlightImportantNodes\%22\%3A\%7B\%22highlightImportantNodes\%22\%3Afalse\%2C\%22highlightPrimaryParents\%22\%3Afalse\%2C\%22highlightRootConnectingEdges\%22\%3Afalse\%7D\%7D}

\begingroup
\small
\eczhBreakableDashes
\renewcommand\arraystretch{1.05}
\edef\myxtraspc{\eczListAddVSpaceXtraXtraStretch}
\begin{tabularx}{\linewidth}{>{\raggedright\arraybackslash}p{\eczListColWidth{name}} >{\hsize=1.0000\hsize }X}
\toprule
\eczListColTitle{Code} & \eczListColTitle{Description} \\
\midrule
\endfirsthead
\toprule
\eczListColTitleContinued{Code} & \eczListColTitleContinued{Description} \\
\midrule
\endhead
\bottomrule
\endfoot
\eczhRefIndex{code:1d_stabilizer}%
\eczhListValue{\flmRefsHyperref{code:1d_stabilizer}{1D lattice stabilizer code}} & \eczhListValue{Lattice stabilizer code in one Euclidean dimension, using either the ordinary block notion of locality or the fermionic/Majorana notion of locality.}\\ 
\addlinespace[\myxtraspc]
\eczhRefIndex{code:analog_repetition}%
\eczhListValue{\flmRefsHyperref{code:analog_repetition}{Analog repetition code}} & \eczhListValue{An \(\llbracket n,1\rrbracket _{\mathbb{R}}\) analog stabilizer version of the quantum repetition code, encoding the position states of one mode into an odd number \(n\) of modes.}\\ 
\addlinespace[\myxtraspc]
\eczhRefIndex{code:current_mirror}%
\eczhListValue{\flmRefsHyperref{code:current_mirror}{Kitaev current-mirror qubit code}} & \eczhListValue{Member of the family of \(\llbracket 2n,(0,2),(2,n)\rrbracket _{\mathbb{Z}}\) homological rotor codes storing a logical qubit on a thin Möbius strip.
The ideal code can be obtained from a Josephson-junction \NoCaseChange{\protect\cite{cite396}} system \NoCaseChange{\protect\cite{cite397}}.}\\ 
\addlinespace[\myxtraspc]
\eczhRefIndex{code:mbq}%
\eczhListValue{\flmRefsHyperref{code:mbq}{Majorana box qubit}} & \eczhListValue{A family of Majorana stabilizer codes obtained by fixing the total fermion parity of \(n\) fermionic modes, equivalently \(2n\) Majorana zero modes, within the ground-state subspace of \(n\) Kitaev Majorana chain Hamiltonians.
The resulting positive-parity subspace encodes \(n-1\) logical qubits and has Majorana distance \(2\).}\\ 
\addlinespace[\myxtraspc]
\eczhRefIndex{code:quantum_convolutional}%
\eczhListValue{\flmRefsHyperref{code:quantum_convolutional}{Quantum convolutional code}} & \eczhListValue{1D translationally invariant qubit stabilizer code whose stabilizer group can be partitioned into constant-size subsets of constant support and of constant overlap between neighboring sets.
Initially formulated as a quantum analogue of convolutional codes, which were designed to protect a continuous and never-ending stream of information.
Precise formulations sometimes begin with a finite-dimensional lattice, with the intent to take the thermodynamic limit; logical dimension can be infinite as well.}\\ 
\addlinespace[\myxtraspc]
\eczhRefIndex{code:quantum_irregular_convolutional}%
\eczhListValue{\flmRefsHyperref{code:quantum_irregular_convolutional}{Quantum irregular convolutional code (QIRCC)}} & \eczhListValue{Quantum convolutional code whose stabilizer group consists of different constant-size subsets.}\\ 
\addlinespace[\myxtraspc]
\eczhRefIndex{code:quantum_repetition}%
\eczhListValue{\flmRefsHyperref{code:quantum_repetition}{Quantum repetition code}} & \eczhListValue{Encodes \(1\) qubit into \(n\) qubits according to \(|0\rangle\to|\phi_0\rangle^{\otimes n}\) and \(|1\rangle\to|\phi_1\rangle^{\otimes n}\). The code is called a \textit{bit-flip} code when \(|\phi_i\rangle = |i\rangle\), and a \textit{phase-flip} code when \(|\phi_0\rangle = |+\rangle\) and \(|\phi_1\rangle = |-\rangle\).
This repetition-style encoding does not clone an arbitrary quantum state; instead, it extends the copying of computational-basis states linearly to entangled codewords  \NoCaseChange{\protect\cite[{Ch. 2}]{cite398}}.}\\ 
\addlinespace[\myxtraspc]
\eczhRefIndex{code:quantum_turbo}%
\eczhListValue{\flmRefsHyperref{code:quantum_turbo}{Quantum turbo code}} & \eczhListValue{A quantum version of the turbo code, obtained from an interleaved serial quantum concatenation \NoCaseChange{\protect\cite[{Def. 30}]{cite399}} of quantum convolutional codes.
The interleaver induces long-range entanglement and can increase the minimum distance relative to the constituent convolutional codes \NoCaseChange{\protect\cite{cite400}}.}\\ 
\addlinespace[\myxtraspc]
\eczhRefIndex{code:tetron}%
\eczhListValue{\flmRefsHyperref{code:tetron}{Tetron code}} & \eczhListValue{A \(\llbracket 2,1,2\rrbracket _{f}\) Majorana box qubit encoding a logical qubit into four Majorana modes, equivalently into the fixed-total-parity sector of two physical fermionic modes.
Four Majorana zero modes are the smallest aggregate that supports a qubit in a fixed fermion-parity sector \NoCaseChange{\protect\cite{cite401}}.
This code can be concatenated with various qubit codes such as surface codes and color codes.
Four-boundary Majorana surface-code patches are logical tetrons, i.e., higher-distance analogues of this physical tetron block \NoCaseChange{\protect\cite{cite402}}.}\\ 
\addlinespace[\myxtraspc]
\eczhRefIndex{code:tfim}%
\eczhListValue{\flmRefsHyperref{code:tfim}{Transverse-field Ising model (TFIM) code}} & \eczhListValue{A 1D translationally invariant stabilizer code whose encoding is a constant-depth circuit of nearest-neighbor gates on alternating even and odd bonds that consist of transverse-field Ising Hamiltonian interactions. The code allows for perfect state transfer of arbitrary distance using local operations and classical communications (LOCC).}\\ 
\addlinespace[\myxtraspc]
\eczhRefIndex{code:stab_5_1_2_convolutional}%
\eczhListValue{\flmRefsHyperref{code:stab_5_1_2_convolutional}{\((5,1,2)\)-convolutional code}} & \eczhListValue{Family of quantum convolutional codes that are 1D lattice generalizations of the five-qubit perfect code, with the former's lattice-translation symmetry being the extension of the latter's cyclic permutation symmetry.}\\ 
\addlinespace[\myxtraspc]
\eczhRefIndex{code:stab_5_1_3}%
\eczhListValue{\flmRefsHyperref{code:stab_5_1_3}{\(\llbracket 5,1,3\rrbracket \) Five-qubit perfect code}} & \eczhListValue{Five-qubit cyclic stabilizer code that is the smallest qubit stabilizer code to correct a single-qubit error.}\\ 
\end{tabularx}\endgroup
\eczcodelist{stabilizer_2d}{2D stabilizer codes
}%

\eczhCodeListAutoDescription{All descendants of \flmRefsCref{code:2d_stabilizer}.}%

\eczhIncludeCodeGraph{Bare}{scale=0.5}{\columnwidth}{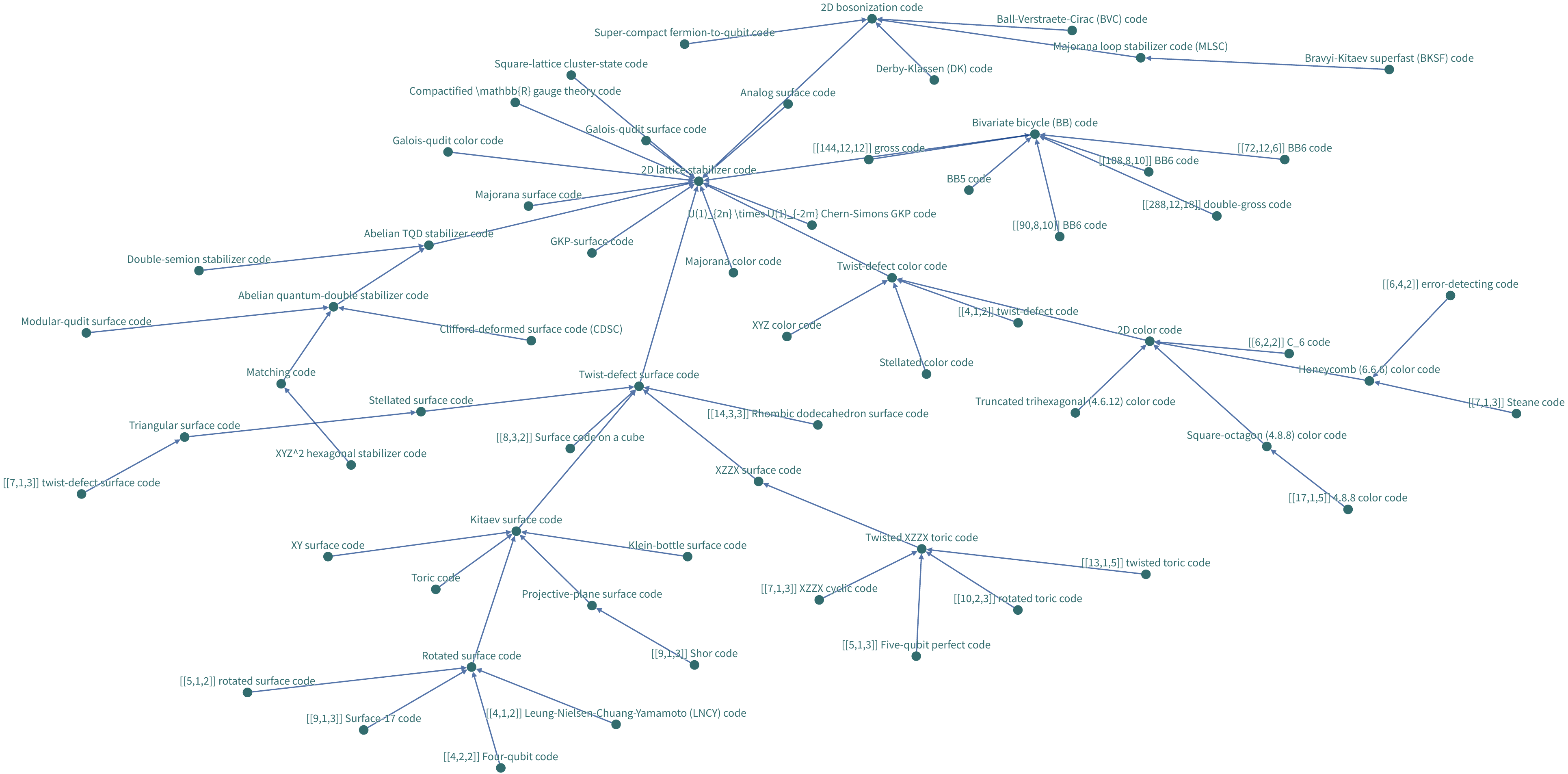}{2D stabilizer codes}{https://errorcorrectionzoo.org/code_graph#J\%7B\%22displayMode\%22\%3A\%22subset\%22\%2C\%22modeSubsetOptions\%22\%3A\%7B\%22codeIds\%22\%3A\%5B\%222d_bosonization\%22\%2C\%222d_color\%22\%2C\%222d_stabilizer\%22\%2C\%22tqd_abelian_stabilizer\%22\%2C\%22quantum_double_abelian\%22\%2C\%22analog_surface\%22\%2C\%22bb5\%22\%2C\%22bvc\%22\%2C\%22qcga\%22\%2C\%22bksf\%22\%2C\%22clifford-deformed_surface\%22\%2C\%22compactified_r\%22\%2C\%22derby_klassen\%22\%2C\%22double_semion\%22\%2C\%22gkp_surface_concatenated\%22\%2C\%22galois_color\%22\%2C\%22galois_topological\%22\%2C\%22triangular_color\%22\%2C\%22surface\%22\%2C\%22klein_bottle\%22\%2C\%22majorana_color\%22\%2C\%22mlsc\%22\%2C\%22majorana_surface\%22\%2C\%22matching\%22\%2C\%22qudit_surface\%22\%2C\%22real_projective_plane\%22\%2C\%22rotated_surface\%22\%2C\%22square_lattice_cluster\%22\%2C\%22488_color\%22\%2C\%22stellated_color\%22\%2C\%22stellated_surface\%22\%2C\%22super_compact\%22\%2C\%22toric\%22\%2C\%22triangle_surface\%22\%2C\%224612_color\%22\%2C\%22twist_defect_color\%22\%2C\%22twist_defect_surface\%22\%2C\%22twisted_xzzx\%22\%2C\%22xysurface\%22\%2C\%22xyz_color\%22\%2C\%22xyz_hexagonal\%22\%2C\%22xzzx\%22\%2C\%22chern_simons_gkp\%22\%2C\%22xzzx_10_2_3\%22\%2C\%22bb108\%22\%2C\%22stab_13_1_5\%22\%2C\%22rhombic_dodecahedron_surface\%22\%2C\%22gross\%22\%2C\%22stab_17_1_5\%22\%2C\%22bb288\%22\%2C\%22css_4_1_2\%22\%2C\%22stab_4_1_2\%22\%2C\%22stab_4_2_2\%22\%2C\%22stab_5_1_2\%22\%2C\%22stab_5_1_3\%22\%2C\%22stab_6_2_2\%22\%2C\%22stab_6_4_2\%22\%2C\%22steane\%22\%2C\%22xzzx_7_1_3\%22\%2C\%22twist_defect_7_1_3\%22\%2C\%22bb72\%22\%2C\%22cubic_surface\%22\%2C\%22shor_nine\%22\%2C\%22surface-17\%22\%2C\%22bb90\%22\%5D\%2C\%22reusePreviousLayoutPositions\%22\%3Afalse\%2C\%22showIntermediateConnectingNodes\%22\%3Atrue\%2C\%22connectingNodesMaxDepth\%22\%3A15\%2C\%22connectingNodesPathMaxLength\%22\%3A20\%2C\%22connectingNodesMaxExtraDepth\%22\%3A3\%2C\%22connectingNodesOnlyKeepPathsWithAdditionalLength\%22\%3A1\%2C\%22connectingNodesToDomainsAndKingdoms\%22\%3Afalse\%2C\%22connectingNodesEdgeLengthsByType\%22\%3A\%7B\%22primaryParent\%22\%3A1\%2C\%22secondaryParent\%22\%3A4\%2C\%22cousin\%22\%3A6\%7D\%2C\%22nodeIds\%22\%3A\%5B\%5D\%7D\%2C\%22highlightImportantNodes\%22\%3A\%7B\%22highlightImportantNodes\%22\%3Afalse\%2C\%22highlightPrimaryParents\%22\%3Afalse\%2C\%22highlightRootConnectingEdges\%22\%3Afalse\%7D\%7D}

\begingroup
\small
\eczhBreakableDashes
\renewcommand\arraystretch{1.05}
\edef\myxtraspc{\eczListAddVSpaceXtraXtraStretch}
\begin{tabularx}{\linewidth}{>{\raggedright\arraybackslash}p{\eczListColWidth{name}} >{\hsize=1.0000\hsize }X}
\toprule
\eczListColTitle{Code} & \eczListColTitle{Description} \\
\midrule
\endfirsthead
\toprule
\eczListColTitleContinued{Code} & \eczListColTitleContinued{Description} \\
\midrule
\endhead
\bottomrule
\endfoot
\eczhRefIndex{code:2d_bosonization}%
\eczhListValue{\flmRefsHyperref{code:2d_bosonization}{2D bosonization code}} & \eczhListValue{A mapping between a 2D lattice quadratic Hamiltonian of Majorana modes and a 2D lattice of qubits.
The original exact 2D bosonization code \NoCaseChange{\protect\cite{cite403}} is a stabilizer code whose generators are products of plaquettes and stars of the surface code, with gauge constraints that project onto a toric-code-like subspace with emergent fermions \NoCaseChange{\protect\cite{cite403,cite404}}.
Finite-depth generalized local unitary Clifford circuits generate a family of equivalent local encodings with qubit-to-fermion ratio \(r = 1 + \frac{1}{2k}\) for any positive integer \(k\); the square-lattice compact encoding with \(r=1.5\) and the super-compact encoding with \(r=1.25\) are explicit examples \NoCaseChange{\protect\cite{cite404}}.}\\ 
\addlinespace[\myxtraspc]
\eczhRefIndex{code:2d_color}%
\eczhListValue{\flmRefsHyperref{code:2d_color}{2D color code}} & \eczhListValue{Color code defined on a graph embedded in a two-dimensional surface.
Each face hosts two stabilizer generators, a Pauli-\(X\) and a Pauli-\(Z\) string acting on all the qubits of the face.}\\ 
\addlinespace[\myxtraspc]
\eczhRefIndex{code:2d_stabilizer}%
\eczhListValue{\flmRefsHyperref{code:2d_stabilizer}{2D lattice stabilizer code}} & \eczhListValue{Lattice stabilizer code in two Euclidean dimensions, using either the ordinary block notion of locality or the fermionic/Majorana notion of locality.}\\ 
\addlinespace[\myxtraspc]
\eczhRefIndex{code:tqd_abelian_stabilizer}%
\eczhListValue{\flmRefsHyperref{code:tqd_abelian_stabilizer}{Abelian TQD stabilizer code}} & \eczhListValue{Modular-qudit stabilizer code whose codewords realize a 2D Abelian twisted-quantum-double topological order on composite-dimensional qudits.
For every finite Abelian group \(G=\prod_i \mathbb{Z}_{N_i}\) and every product of Type-I and Type-II cocycles, there is a Pauli stabilizer Hamiltonian realizing the corresponding Abelian TQD \NoCaseChange{\protect\cite{cite405}}.
Equivalently, these codes exhaust the 2D Abelian topological orders that admit gapped boundaries \NoCaseChange{\protect\cite{cite405,cite406}}.}\\ 
\addlinespace[\myxtraspc]
\eczhRefIndex{code:quantum_double_abelian}%
\eczhListValue{\flmRefsHyperref{code:quantum_double_abelian}{Abelian quantum-double stabilizer code}} & \eczhListValue{Modular-qudit stabilizer code whose codewords realize 2D modular gapped Abelian topological order with trivial cocycle.
The corresponding anyon theory is defined by an Abelian group.
The \(G=\mathbb{Z}_2\) instance on a torus is the toric code, and cyclic-group instances reduce to modular-qudit surface codes.
All such codes can be realized by a stack of modular-qudit surface codes because all finite Abelian groups are direct products of cyclic groups.}\\ 
\addlinespace[\myxtraspc]
\eczhRefIndex{code:analog_surface}%
\eczhListValue{\flmRefsHyperref{code:analog_surface}{Analog surface code}} & \eczhListValue{An analog CSS version of the Kitaev surface code realizing a phase of 2D \(\mathbb{R}\) gauge theory.}\\ 
\addlinespace[\myxtraspc]
\eczhRefIndex{code:bb5}%
\eczhListValue{\flmRefsHyperref{code:bb5}{BB5 code}} & \eczhListValue{A BB code with weight-five stabilizer generators
(contrasting with the weight-six checks of standard BB codes), designed and
benchmarked for long chains of trapped ions \NoCaseChange{\protect\cite{cite407}}.}\\ 
\addlinespace[\myxtraspc]
\eczhRefIndex{code:bvc}%
\eczhListValue{\flmRefsHyperref{code:bvc}{Ball-Verstraete-Cirac (BVC) code}} & \eczhListValue{A 2D fermion-into-qubit encoding that builds upon the JW transformation by eliminating the weight-\(O(n)\) non-local \(Z\)-type string at the expense of introducing an auxiliary qubit per site and local gauge constraints.
See \NoCaseChange{\protect\cite[{Sec. IV.B}]{cite404}} for details.}\\ 
\addlinespace[\myxtraspc]
\eczhRefIndex{code:qcga}%
\eczhListValue{\flmRefsHyperref{code:qcga}{Bivariate bicycle (BB) code}} & \eczhListValue{One of several Abelian 2BGA codes which admit time-optimal syndrome measurement circuits that can be implemented in a two-layer
architecture, a generalization of the square-lattice architecture
optimal for the surface codes.
Codes can be classified by the weight of their checks, e.g., by BB\(w\) where \(w\) is the check weight.}\\ 
\addlinespace[\myxtraspc]
\eczhRefIndex{code:bksf}%
\eczhListValue{\flmRefsHyperref{code:bksf}{Bravyi-Kitaev superfast (BKSF) code}} & \eczhListValue{A single-error-detecting fermion-into-qubit encoding defined on a 2D qubit lattice whose stabilizers are associated with loops in the lattice.
For the square-lattice edge ordering used in Ref. \NoCaseChange{\protect\cite{cite404}}, the BKSF logical operators coincide with exact 2D bosonization on the dual lattice after relabeling \(X\) and \(Y\).
The code can be generalized to a single error-correcting code (i.e., with distance three) on graphs of degree \(\geq 6\) \NoCaseChange{\protect\cite{cite408}}.}\\ 
\addlinespace[\myxtraspc]
\eczhRefIndex{code:clifford-deformed_surface}%
\eczhListValue{\flmRefsHyperref{code:clifford-deformed_surface}{Clifford-deformed surface code (CDSC)}} & \eczhListValue{A generally non-CSS derivative of the surface code defined by applying a translationally invariant constant-depth \flmRefsHyperref{ref409}{Clifford circuit} to the original (CSS) surface code.
Unlike the surface code, CDSCs include codes whose thresholds and subthreshold performance are enhanced under noise biased towards dephasing.
Examples of CDSCs include the XY code, XZZX code, and random CDSCs.}\\ 
\addlinespace[\myxtraspc]
\eczhRefIndex{code:compactified_r}%
\eczhListValue{\flmRefsHyperref{code:compactified_r}{Compactified \(\mathbb{R}\) gauge theory code}} & \eczhListValue{An integer-homology bosonic CSS code realizing 2D \(U(1)\) gauge theory on bosonic modes.
The code can be obtained from the analog surface code by \flmRefsHyperref{ref410}{condensing} certain anyons \NoCaseChange{\protect\cite{cite411}}. 
This results in a pinning of each mode to the space of periodic functions, which is the Hilbert space of a physical rotor, and can be thought of as compactification of the 2D \(\mathbb{R}\) gauge theory phase realized by the analog surface code.}\\ 
\addlinespace[\myxtraspc]
\eczhRefIndex{code:derby_klassen}%
\eczhListValue{\flmRefsHyperref{code:derby_klassen}{Derby-Klassen (DK) code}} & \eczhListValue{A fermion-into-qubit code defined on regular tilings with maximum degree 4 whose stabilizers are associated with loops in the tiling.
The code outperforms several other encodings in terms of encoding rate \NoCaseChange{\protect\cite[{Table I}]{cite412}}.
It has been extended for models with several modes per site \NoCaseChange{\protect\cite{cite413}}.}\\ 
\addlinespace[\myxtraspc]
\eczhRefIndex{code:double_semion}%
\eczhListValue{\flmRefsHyperref{code:double_semion}{Double-semion stabilizer code}} & \eczhListValue{A 2D lattice modular-qudit stabilizer code with qudit dimension \(q=4\) that realizes the 2D double semion topological phase.
The code can be obtained from a \(\mathbb{Z}_4\) toric-code ground state by \flmRefsHyperref{ref410}{condensing} the emergent boson \(e^2 m^2\); in the stabilizer construction this condensation is implemented by two-body measurements \NoCaseChange{\protect\cite{cite405,cite414}}.
Its ground-state subspace can be mapped to that of the double-semion string-net model by a finite-depth quantum circuit with ancillas \NoCaseChange{\protect\cite{cite405}}.}\\ 
\addlinespace[\myxtraspc]
\eczhRefIndex{code:gkp_surface_concatenated}%
\eczhListValue{\flmRefsHyperref{code:gkp_surface_concatenated}{GKP-surface code}} & \eczhListValue{A concatenated code whose outer code is a GKP code and whose inner code is a surface code, including toric surface-code variants \NoCaseChange{\protect\cite{cite415,cite416}}, rotated surface codes \NoCaseChange{\protect\cite{cite417,cite418,cite419,cite420}}, and XZZX surface codes \NoCaseChange{\protect\cite{cite421}}.}\\ 
\addlinespace[\myxtraspc]
\eczhRefIndex{code:galois_color}%
\eczhListValue{\flmRefsHyperref{code:galois_color}{Galois-qudit color code}} & \eczhListValue{Extension of the color code to 2D lattices of Galois qudits.}\\ 
\addlinespace[\myxtraspc]
\eczhRefIndex{code:galois_topological}%
\eczhListValue{\flmRefsHyperref{code:galois_topological}{Galois-qudit surface code}} & \eczhListValue{Extension of the surface code to 2D lattices of Galois qudits.}\\ 
\addlinespace[\myxtraspc]
\eczhRefIndex{code:triangular_color}%
\eczhListValue{\flmRefsHyperref{code:triangular_color}{Honeycomb (6.6.6) color code}} & \eczhListValue{2D color code defined on a (typically triangular) patch of the 6.6.6 (honeycomb) tiling.
The usual triangular patch has three differently colored boundaries, encodes one logical qubit, and is local-Clifford equivalent to a folded surface/toric code with two smooth and two rough boundaries \NoCaseChange{\protect\cite{cite422}}.}\\ 
\addlinespace[\myxtraspc]
\eczhRefIndex{code:surface}%
\eczhListValue{\flmRefsHyperref{code:surface}{Kitaev surface code}} & \eczhListValue{A family of Abelian topological \flmRefsHyperref{code:css}{CSS stabilizer} codes
whose generators are few-body \(X\)-type and \(Z\)-type Pauli strings
associated to the stars and plaquettes, respectively, of a cellulation of a
two-dimensional surface (with a qubit located at each edge of the
cellulation).
Codewords correspond to ground states of the surface code Hamiltonian, and error operators create or annihilate pairs of anyonic charges or vortices.}\\ 
\addlinespace[\myxtraspc]
\eczhRefIndex{code:klein_bottle}%
\eczhListValue{\flmRefsHyperref{code:klein_bottle}{Klein-bottle surface code}} & \eczhListValue{A family of Kitaev surface codes on the non-orientable Klein bottle.}\\ 
\addlinespace[\myxtraspc]
\eczhRefIndex{code:majorana_color}%
\eczhListValue{\flmRefsHyperref{code:majorana_color}{Majorana color code}} & \eczhListValue{A fermionic analogue of a 2D color code.}\\ 
\addlinespace[\myxtraspc]
\eczhRefIndex{code:mlsc}%
\eczhListValue{\flmRefsHyperref{code:mlsc}{Majorana loop stabilizer code (MLSC)}} & \eczhListValue{A single-error-correcting fermion-into-qubit encoding defined on a 2D qubit lattice whose stabilizers are associated with loops in the lattice.}\\ 
\addlinespace[\myxtraspc]
\eczhRefIndex{code:majorana_surface}%
\eczhListValue{\flmRefsHyperref{code:majorana_surface}{Majorana surface code}} & \eczhListValue{Fermionic analogue of the surface code defined on a three-colorable 2D tiling whose face operators are non-overlapping even-Majorana stabilizers.
Open patches with four or six alternating colored boundaries encode logical tetrons or hexons.
The uniform 4.8.8, 6.6.6, and 4.6.12 tilings yield families with tetron, hexon, or dodecon building blocks and with twist-based lattice surgery supporting minimal-overhead logical Clifford gates \NoCaseChange{\protect\cite{cite402}}.}\\ 
\addlinespace[\myxtraspc]
\eczhRefIndex{code:matching}%
\eczhListValue{\flmRefsHyperref{code:matching}{Matching code}} & \eczhListValue{Member of a class of qubit stabilizer codes based on the Abelian phase of the Kitaev honeycomb model.}\\ 
\addlinespace[\myxtraspc]
\eczhRefIndex{code:qudit_surface}%
\eczhListValue{\flmRefsHyperref{code:qudit_surface}{Modular-qudit surface code}} & \eczhListValue{Extension of the surface code to prime-dimensional \NoCaseChange{\protect\cite{cite423,cite424}} and more general modular qudits.
Stabilizer generators are few-body \(X\)-type and \(Z\)-type Pauli strings associated to the stars and plaquettes, respectively, of a tessellation of a two-dimensional surface.
Since qudits have more than one \(X\) and \(Z\)-type operator, various sets of stabilizer generators can be defined.}\\ 
\addlinespace[\myxtraspc]
\eczhRefIndex{code:real_projective_plane}%
\eczhListValue{\flmRefsHyperref{code:real_projective_plane}{Projective-plane surface code}} & \eczhListValue{A family of Kitaev surface codes on the non-orientable 2-dimensional compact manifold \(\mathbb{R}P^2\) (in contrast to a genus-\(g\) surface).
Whereas genus-\(g\) surface codes require \(2g\) logical qubits, qubit codes on \(\mathbb{R}P^2\) are made from a single logical qubit.}\\ 
\addlinespace[\myxtraspc]
\eczhRefIndex{code:rotated_surface}%
\eczhListValue{\flmRefsHyperref{code:rotated_surface}{Rotated surface code}} & \eczhListValue{Variant of the surface code obtained by taking the medial graph of the surface code lattice (treated as a graph) and applying a procedure to construct the check operators \NoCaseChange{\protect\cite{cite425,cite426}\protect\cite[{Fig. 8}]{cite427}}.
On a square lattice, this amounts to a rotation by 45 degrees such that qubits are on vertices, and both \(X\)- and \(Z\)-type check operators occupy plaquettes in an alternating checkerboard pattern.}\\ 
\addlinespace[\myxtraspc]
\eczhRefIndex{code:square_lattice_cluster}%
\eczhListValue{\flmRefsHyperref{code:square_lattice_cluster}{Square-lattice cluster-state code}} & \eczhListValue{A code based on the cluster state on a square lattice that was used in the first proposal for MBQC \NoCaseChange{\protect\cite{cite428,cite429}}.
In the one-way model, the pre-entangled square-lattice cluster is a universal resource, and the computation is carried out entirely by adaptive single-qubit measurements.}\\ 
\addlinespace[\myxtraspc]
\eczhRefIndex{code:488_color}%
\eczhListValue{\flmRefsHyperref{code:488_color}{Square-octagon (4.8.8) color code}} & \eczhListValue{2D color code defined on a patch of the 4.8.8 (square-octagon) tiling, which itself is obtained by applying a fattening procedure to the square lattice \NoCaseChange{\protect\cite{cite430}}.
An equivalent description uses the Tetrakis square tiling (a.k.a. the Union Jack lattice), which is dual to the 4.8.8 lattice \NoCaseChange{\protect\cite{cite431}}.
Among the three semiregular triangular 2D color-code families, the 4.8.8 family uses the fewest physical qubits for a given distance and is the only one of the three with transversal implementations of the full Clifford group \NoCaseChange{\protect\cite{cite432}}.}\\ 
\addlinespace[\myxtraspc]
\eczhRefIndex{code:stellated_color}%
\eczhListValue{\flmRefsHyperref{code:stellated_color}{Stellated color code}} & \eczhListValue{A non-CSS color-code family on a lattice patch with a single central puncture that hosts a twist defect connected to the boundary by a domain wall.}\\ 
\addlinespace[\myxtraspc]
\eczhRefIndex{code:stellated_surface}%
\eczhListValue{\flmRefsHyperref{code:stellated_surface}{Stellated surface code}} & \eczhListValue{A twist-defect surface-code family parameterized by a rotational symmetry order \(s\), with a central toric-code twist connected to the boundary by a domain wall.
The \(s=3\) member is the triangular surface code \NoCaseChange{\protect\cite{cite433}}.}\\ 
\addlinespace[\myxtraspc]
\eczhRefIndex{code:super_compact}%
\eczhListValue{\flmRefsHyperref{code:super_compact}{Super-compact fermion-to-qubit code}} & \eczhListValue{A 2D fermion-into-qubit encoding on the square lattice obtained from exact 2D bosonization by a finite-depth generalized local unitary Clifford circuit, followed by re-pairing of Majorana modes and a slight lattice deformation.
The code uses \(1.25\) qubits per fermion, improving on the square-lattice compact encoding with ratio \(r=1.5\).
Its fermion-parity, hopping, and stabilizer operators have weights \(1\)-\(2\), \(2\)-\(6\), and \(12\), respectively \NoCaseChange{\protect\cite[{Table I}]{cite404}}.}\\ 
\addlinespace[\myxtraspc]
\eczhRefIndex{code:toric}%
\eczhListValue{\flmRefsHyperref{code:toric}{Toric code}} & \eczhListValue{Version of the Kitaev surface code on a square lattice with periodic boundary conditions, encoding two logical qubits.
Being the first manifestation of the surface code, "toric code" is often an alternative name for the general construction.
\textit{Twisted toric code} \NoCaseChange{\protect\cite[{Fig. 8}]{cite434}} refers to the construction on a torus with twisted (a.k.a. shifted) boundary conditions.
In the original Hamiltonian construction, open Pauli-\(Z\) and Pauli-\(X\) strings create pairs of electric charges and magnetic vortices, and braiding one type around the other yields the nontrivial Abelian anyonic phase \NoCaseChange{\protect\cite{cite423}}.}\\ 
\addlinespace[\myxtraspc]
\eczhRefIndex{code:triangle_surface}%
\eczhListValue{\flmRefsHyperref{code:triangle_surface}{Triangular surface code}} & \eczhListValue{A member of a twist-defect surface code family with a single central twist whose planar layout fits within a triangle.
Triangle codes can be viewed as three conjoined surface-code patches projected into two dimensions, with weight-four plaquette stabilizers and weight-two edge stabilizers \NoCaseChange{\protect\cite{cite433}}.
Symmetric distance-\(d\) triangle codes use \(3d^2/4+1/4\) data qubits, i.e., about \(25\%\) fewer than the rotated surface code for a given odd distance.
Logical \(\overline{X}\), \(\overline{Y}\), and \(\overline{Z}\) operators can be supported on the three sides of the triangle, enabling initialization and measurement in any Pauli basis.}\\ 
\addlinespace[\myxtraspc]
\eczhRefIndex{code:4612_color}%
\eczhListValue{\flmRefsHyperref{code:4612_color}{Truncated trihexagonal (4.6.12) color code}} & \eczhListValue{2D color code defined on a (typically triangular) patch of the 4.6.12 (truncated trihexagonal or square-hexagon-dodecagon) tiling.}\\ 
\addlinespace[\myxtraspc]
\eczhRefIndex{code:twist_defect_color}%
\eczhListValue{\flmRefsHyperref{code:twist_defect_color}{Twist-defect color code}} & \eczhListValue{A non-CSS extension of the 2D color code whose non-CSS stabilizer generators are associated with twist defects of the associated lattice.
These twists terminate domain walls that permute color labels, Pauli labels, or interchange the two.}\\ 
\addlinespace[\myxtraspc]
\eczhRefIndex{code:twist_defect_surface}%
\eczhListValue{\flmRefsHyperref{code:twist_defect_surface}{Twist-defect surface code}} & \eczhListValue{A non-CSS extension of the 2D surface-code construction whose non-CSS stabilizer generators are associated with twist defects of the associated lattice.
A related construction \NoCaseChange{\protect\cite{cite435}} doubles the number of qubits in the lattice via \flmRefsHyperref{ref436}{symplectic doubling}.}\\ 
\addlinespace[\myxtraspc]
\eczhRefIndex{code:twisted_xzzx}%
\eczhListValue{\flmRefsHyperref{code:twisted_xzzx}{Twisted XZZX toric code}} & \eczhListValue{A cyclic code that can be thought of as the XZZX toric code with shifted (a.k.a twisted) boundary conditions.
Admits a set of stabilizer generators that are equivalent to cyclic shifts of a particular weight-four \(XZZX\) Pauli string.}\\ 
\addlinespace[\myxtraspc]
\eczhRefIndex{code:xysurface}%
\eczhListValue{\flmRefsHyperref{code:xysurface}{XY surface code}} & \eczhListValue{A variant of the surface code whose generators are \(XXXX\)  and \(YYYY\), obtained by mapping \(Z \to Y\) in the surface code.
While this code is equivalent to a CSS surface code with the same distance, other properties like noise-bias performance can differ significantly.}\\ 
\addlinespace[\myxtraspc]
\eczhRefIndex{code:xyz_color}%
\eczhListValue{\flmRefsHyperref{code:xyz_color}{XYZ color code}} & \eczhListValue{A variant of the 6.6.6 color code whose generators are \(XZXZXZ\) and \(ZYZYZY\) Pauli strings associated to each hexagonal in the hexagonal (6.6.6) tiling. 
A further variation called the \textit{domain wall color code} admits generators of the form \(XXXZZZ\) and \(ZZZXXX\) \NoCaseChange{\protect\cite{cite437}}.
While such codes are equivalent to CSS color codes with the same distance, other properties like noise-bias performance can differ significantly.}\\ 
\addlinespace[\myxtraspc]
\eczhRefIndex{code:xyz_hexagonal}%
\eczhListValue{\flmRefsHyperref{code:xyz_hexagonal}{XYZ\(^2\) hexagonal stabilizer code}} & \eczhListValue{An instance of the matching code based on the Kitaev honeycomb model. It is described on a honeycomb tiling with \(XYZXYZ\) stabilizers on each hexagonal plaquette. Each vertical pair of qubits has an \(XX\), \(YY\), or \(ZZ\) link stabilizer depending on the orientation of the plaquette stabilizers.}\\ 
\addlinespace[\myxtraspc]
\eczhRefIndex{code:xzzx}%
\eczhListValue{\flmRefsHyperref{code:xzzx}{XZZX surface code}} & \eczhListValue{A variant of the rotated surface code whose generators are \(XZZX\) Pauli strings associated, clockwise, to the vertices of each face of a two-dimensional lattice (with a qubit located at each vertex of the tessellation).}\\ 
\addlinespace[\myxtraspc]
\eczhRefIndex{code:chern_simons_gkp}%
\eczhListValue{\flmRefsHyperref{code:chern_simons_gkp}{\(U(1)_{2n} \times U(1)_{-2m}\) Chern-Simons GKP code}} & \eczhListValue{A non-CSS multimode GKP code defined on a 2D mode lattice that encodes a qudit logical space and whose excitations are characterized by the \(U(1)_{2n} \times U(1)_{-2m}\) Chern-Simons theory.
The code can be obtained from the analog surface code by \flmRefsHyperref{ref410}{condensing} certain anyons \NoCaseChange{\protect\cite{cite411}}.}\\ 
\addlinespace[\myxtraspc]
\eczhRefIndex{code:xzzx_10_2_3}%
\eczhListValue{\flmRefsHyperref{code:xzzx_10_2_3}{\(\llbracket 10,2,3\rrbracket \) rotated toric code}} & \eczhListValue{Rotated toric code that is the CSS form of the twisted XZZX toric code with parameters \(a=1\), \(b=3\) \NoCaseChange{\protect\cite{cite438,cite427}},
related to the XZZX form by Hadamard on the \(B\)-sublattice.
It is also the \flmRefsHyperref{ref436}{symplectic double} (a.k.a. genus-one double cover) of the \(\llbracket 5,1,3\rrbracket \) five-qubit perfect code \NoCaseChange{\protect\cite{cite439,cite435}}, the \flmRefsHyperref{ref436}{symplectic double} of the \(\llbracket 5,1,2\rrbracket \) rotated surface code \NoCaseChange{\protect\cite{cite435}}, and a BCC code \NoCaseChange{\protect\cite{cite440}}.}\\ 
\addlinespace[\myxtraspc]
\eczhRefIndex{code:bb108}%
\eczhListValue{\flmRefsHyperref{code:bb108}{\(\llbracket 108,8,10\rrbracket \) BB6 code}} & \eczhListValue{A bivariate bicycle (BB) code with parameters \(\llbracket 108,8,10\rrbracket \) and weight-six stabilizer generators \NoCaseChange{\protect\cite{cite441}}.}\\ 
\addlinespace[\myxtraspc]
\eczhRefIndex{code:stab_13_1_5}%
\eczhListValue{\flmRefsHyperref{code:stab_13_1_5}{\(\llbracket 13,1,5\rrbracket \) twisted toric code}} & \eczhListValue{Thirteen-qubit twisted toric code whose stabilizer tableau consists of cyclic permutations of the \(XZZX\)-type Pauli string \(XIZZIXIIIIIII\).
The code can be thought of as a small twisted XZZX code \NoCaseChange{\protect\cite[{Exam. 11 and Fig. 3}]{cite438}}.}\\ 
\addlinespace[\myxtraspc]
\eczhRefIndex{code:rhombic_dodecahedron_surface}%
\eczhListValue{\flmRefsHyperref{code:rhombic_dodecahedron_surface}{\(\llbracket 14,3,3\rrbracket \) Rhombic dodecahedron surface code}} & \eczhListValue{A \(\llbracket 14,3,3\rrbracket \) twist-defect surface code whose qubits lie on the vertices of a rhombic dodecahedron.
Its non-CSS nature is due to twist defects \NoCaseChange{\protect\cite{cite442}} stemming from the geometry of the polytope.
A local-Clifford-equivalent clean realization has only \(X\)- and \(Z\)-type operators on its four-valent vertices, and its symplectic double is a \(\llbracket 28,6,3\rrbracket \) genus-three code \NoCaseChange{\protect\cite{cite435}}.}\\ 
\addlinespace[\myxtraspc]
\eczhRefIndex{code:gross}%
\eczhListValue{\flmRefsHyperref{code:gross}{\(\llbracket 144,12,12\rrbracket \) gross code}} & \eczhListValue{A BB code which requires less physical and ancilla qubits (for syndrome extraction) than the surface code with the same number of logical qubits and distance.
The gross code is equivalent to 8 copies of the surface code via a constant-depth Clifford circuit, and is an element of a larger family of 2D stabilizer codes \NoCaseChange{\protect\cite{cite443}}.
The name stems from the fact that a gross is a dozen dozen.}\\ 
\addlinespace[\myxtraspc]
\eczhRefIndex{code:stab_17_1_5}%
\eczhListValue{\flmRefsHyperref{code:stab_17_1_5}{\(\llbracket 17,1,5\rrbracket \) 4.8.8 color code}} & \eczhListValue{Seventeen-qubit doubly even 2D color code that admits a transversal implementation of the logical Clifford group.
The smallest distance-five CSS code has \(n=17\) \NoCaseChange{\protect\cite{cite444}}.
It is also a normal self-dual CSS code whose transversal Hadamard acts logically, making it suitable as an inner code for fifth-order magic-state distillation \NoCaseChange{\protect\cite{cite101}}.}\\ 
\addlinespace[\myxtraspc]
\eczhRefIndex{code:bb288}%
\eczhListValue{\flmRefsHyperref{code:bb288}{\(\llbracket 288,12,18\rrbracket \) double-gross code}} & \eczhListValue{A bivariate bicycle (BB) code with parameters \(\llbracket 288,12,18\rrbracket \) and weight-six stabilizer generators \NoCaseChange{\protect\cite{cite441}}.}\\ 
\addlinespace[\myxtraspc]
\eczhRefIndex{code:css_4_1_2}%
\eczhListValue{\flmRefsHyperref{code:css_4_1_2}{\(\llbracket 4,1,2\rrbracket \) Leung-Nielsen-Chuang-Yamamoto (LNCY) code}} & \eczhListValue{A four-qubit CSS stabilizer code that is the only qubit CSS code with such parameters.}\\ 
\addlinespace[\myxtraspc]
\eczhRefIndex{code:stab_4_1_2}%
\eczhListValue{\flmRefsHyperref{code:stab_4_1_2}{\(\llbracket 4,1,2\rrbracket \) twist-defect code}} & \eczhListValue{A four-qubit non-CSS stabilizer code that can be interpreted as the smallest triangular color code with \(x\)-, \(y\)-, and \(z\)-type Pauli boundaries \NoCaseChange{\protect\cite[{Fig. 7}]{cite445}}, and equivalently as a small twist-defect surface code on a tetrahedron inscribed in a sphere \NoCaseChange{\protect\cite{cite435}}.
It is the only non-CSS qubit stabilizer code with parameters \(\llbracket 4,1,2\rrbracket \).
The code admits weight-three stabilizer generators and weight-two logical Pauli \(X,Y,Z\) operators.}\\ 
\addlinespace[\myxtraspc]
\eczhRefIndex{code:stab_4_2_2}%
\eczhListValue{\flmRefsHyperref{code:stab_4_2_2}{\(\llbracket 4,2,2\rrbracket \) Four-qubit code}} & \eczhListValue{A four-qubit hyperbolic self-dual CSS stabilizer code that is the smallest two-logical-qubit stabilizer code to detect a single-qubit error.
It is unique for its parameters \NoCaseChange{\protect\cite[{Thm. 8}]{cite446}}.}\\ 
\addlinespace[\myxtraspc]
\eczhRefIndex{code:stab_5_1_2}%
\eczhListValue{\flmRefsHyperref{code:stab_5_1_2}{\(\llbracket 5,1,2\rrbracket \) rotated surface code}} & \eczhListValue{A rotated surface code on one rung of a ladder, with one qubit on the rung, and four qubits surrounding it. This is the smallest code that implements a fault-tolerant logical \(S\) gate using a diagonal depth-one Clifford circuit \NoCaseChange{\protect\cite{cite447}}.}\\ 
\addlinespace[\myxtraspc]
\eczhRefIndex{code:stab_5_1_3}%
\eczhListValue{\flmRefsHyperref{code:stab_5_1_3}{\(\llbracket 5,1,3\rrbracket \) Five-qubit perfect code}} & \eczhListValue{Five-qubit cyclic stabilizer code that is the smallest qubit stabilizer code to correct a single-qubit error.}\\ 
\addlinespace[\myxtraspc]
\eczhRefIndex{code:stab_6_2_2}%
\eczhListValue{\flmRefsHyperref{code:stab_6_2_2}{\(\llbracket 6,2,2\rrbracket \) \(C_6\) code}} & \eczhListValue{Error-detecting normal self-dual CSS code on three qubit pairs that encodes a logical qubit pair and detects any error acting on one pair \NoCaseChange{\protect\cite{cite448}}.
In Knill's \(C_4/C_6\) architecture, this code is used at the second and higher concatenation levels.}\\ 
\addlinespace[\myxtraspc]
\eczhRefIndex{code:stab_6_4_2}%
\eczhListValue{\flmRefsHyperref{code:stab_6_4_2}{\(\llbracket 6,4,2\rrbracket \) error-detecting code}} & \eczhListValue{Self-complementary six-qubit code with rate \(2/3\) that is unique for its parameters, up to equivalence \NoCaseChange{\protect\cite[{Tab. III}]{cite449}}.
Concatenations of this code with itself yield the \(\llbracket 6^r,4^r,2^r\rrbracket \) level-\(r\) \textit{many-hypercube} code \NoCaseChange{\protect\cite{cite450}}.}\\ 
\addlinespace[\myxtraspc]
\eczhRefIndex{code:steane}%
\eczhListValue{\flmRefsHyperref{code:steane}{\(\llbracket 7,1,3\rrbracket \) Steane code}} & \eczhListValue{A \(\llbracket 7,1,3\rrbracket \) self-dual CSS code that is the smallest qubit CSS code to correct a single-qubit error \NoCaseChange{\protect\cite{cite451}}.
The code is constructed using the classical binary \([7,4,3]\) Hamming code for protecting against both \(X\) and \(Z\) errors.}\\ 
\addlinespace[\myxtraspc]
\eczhRefIndex{code:xzzx_7_1_3}%
\eczhListValue{\flmRefsHyperref{code:xzzx_7_1_3}{\(\llbracket 7,1,3\rrbracket \) XZZX cyclic code}} & \eczhListValue{A \(\llbracket 7,1,3\rrbracket \) cyclic non-CSS code whose stabilizer generators are the seven cyclic shifts of the weight-four Pauli string \(XZIZXII\),
any six of which are independent.
The non-identity support of this generator is \(XZZX\) (at positions 0, 1, 3, 4), making this a member of the twisted XZZX code family.
It is one of sixteen distinct indecomposable \(\llbracket 7,1,3\rrbracket \) codes \NoCaseChange{\protect\cite{cite452}}.}\\ 
\addlinespace[\myxtraspc]
\eczhRefIndex{code:twist_defect_7_1_3}%
\eczhListValue{\flmRefsHyperref{code:twist_defect_7_1_3}{\(\llbracket 7,1,3\rrbracket \) twist-defect surface code}} & \eczhListValue{A \(\llbracket 7,1,3\rrbracket \) code (different from the Steane code) that is a small example of a twist-defect surface code.}\\ 
\addlinespace[\myxtraspc]
\eczhRefIndex{code:bb72}%
\eczhListValue{\flmRefsHyperref{code:bb72}{\(\llbracket 72,12,6\rrbracket \) BB6 code}} & \eczhListValue{A bivariate bicycle (BB) code with parameters \(\llbracket 72,12,6\rrbracket \) and weight-six stabilizer generators \NoCaseChange{\protect\cite{cite441}}.}\\ 
\addlinespace[\myxtraspc]
\eczhRefIndex{code:cubic_surface}%
\eczhListValue{\flmRefsHyperref{code:cubic_surface}{\(\llbracket 8,3,2\rrbracket \) Surface code on a cube}} & \eczhListValue{An \(\llbracket 8,3,2\rrbracket \) twist-defect surface code whose qubits lie on the vertices of a cube.
It is obtained by three-coloring the faces of a cube and placing \(X\), \(Y\), and \(Z\) stabilizer generators on each pair of faces of the same color.
Its non-CSS nature is due to twist defects \NoCaseChange{\protect\cite{cite442}} stemming from the geometry of the polytope.}\\ 
\addlinespace[\myxtraspc]
\eczhRefIndex{code:shor_nine}%
\eczhListValue{\flmRefsHyperref{code:shor_nine}{\(\llbracket 9,1,3\rrbracket \) Shor code}} & \eczhListValue{Nine-qubit \flmRefsHyperref{code:css}{CSS code} that is the first quantum error-correcting code \NoCaseChange{\protect\cite[{ID 8802}]{cite453}}.
Among indecomposable \(\llbracket 9,1,3\rrbracket \) CSS codes, the Shor code has the largest automorphism group \NoCaseChange{\protect\cite{cite454}}.}\\ 
\addlinespace[\myxtraspc]
\eczhRefIndex{code:surface-17}%
\eczhListValue{\flmRefsHyperref{code:surface-17}{\(\llbracket 9,1,3\rrbracket \) Surface-17 code}} & \eczhListValue{A \(\llbracket 9,1,3\rrbracket \) rotated surface code named for the sum of its 9 data qubits and 8 syndrome qubits.
It is one of the four inequivalent CSS gauge fixings of the nine-qubit Bacon-Shor code \NoCaseChange{\protect\cite{cite454}}.
It uses the smallest number of qubits to perform fault-tolerant error correction on a surface code with parallel syndrome extraction.}\\ 
\addlinespace[\myxtraspc]
\eczhRefIndex{code:bb90}%
\eczhListValue{\flmRefsHyperref{code:bb90}{\(\llbracket 90,8,10\rrbracket \) BB6 code}} & \eczhListValue{A bivariate bicycle (BB) code with parameters \(\llbracket 90,8,10\rrbracket \) and weight-six stabilizer generators \NoCaseChange{\protect\cite{cite441}}.}\\ 
\end{tabularx}\endgroup
\eczcodelist{stabilizer_3d}{3D stabilizer codes
}%

\eczhCodeListAutoDescription{All descendants of \flmRefsCref{code:3d_stabilizer}.}%

\eczhIncludeCodeGraph{Bare}{scale=0.5}{\columnwidth}{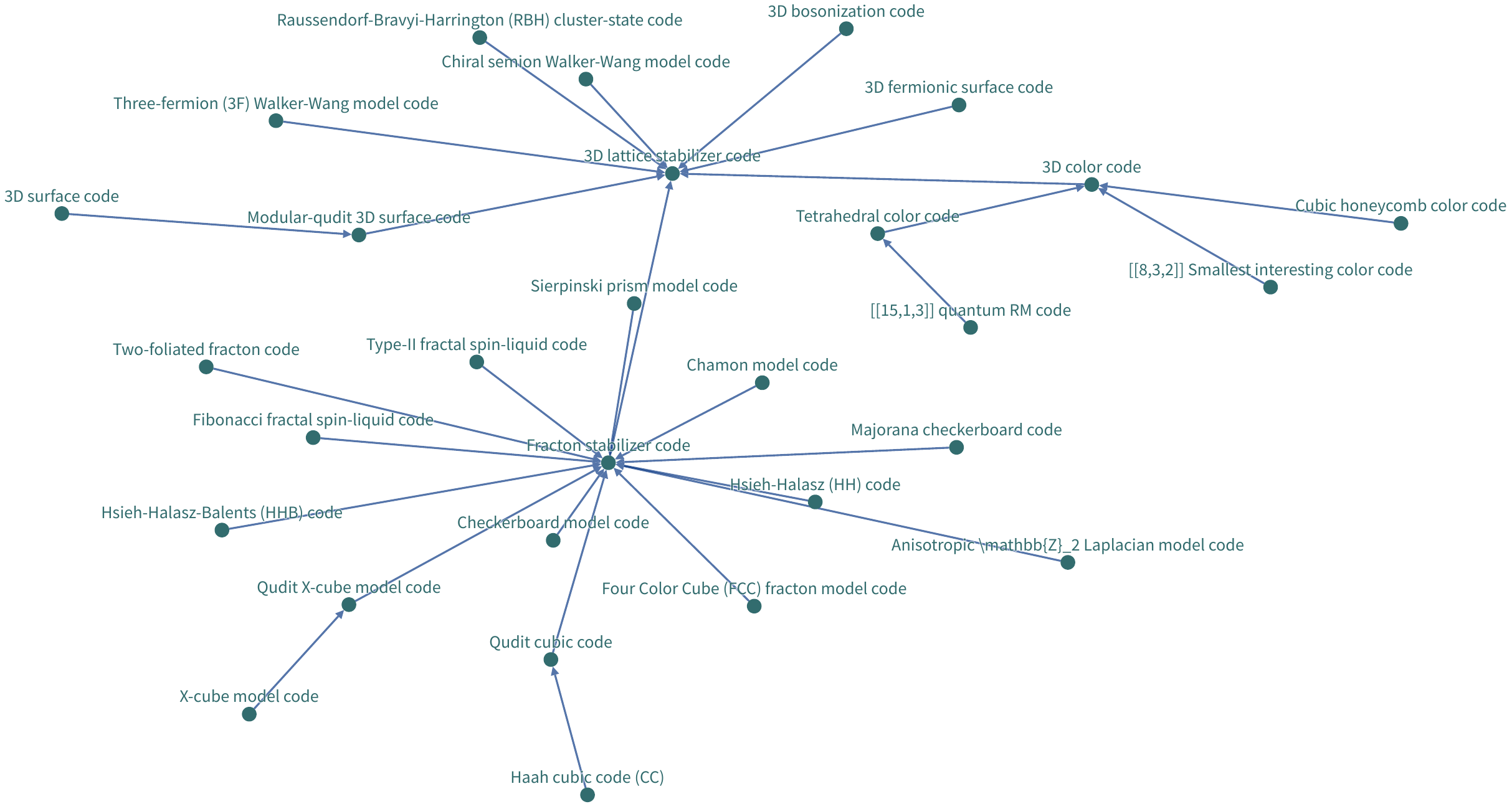}{3D stabilizer codes}{https://errorcorrectionzoo.org/code_graph#J\%7B\%22displayMode\%22\%3A\%22subset\%22\%2C\%22modeSubsetOptions\%22\%3A\%7B\%22codeIds\%22\%3A\%5B\%223d_bosonization\%22\%2C\%223d_color\%22\%2C\%223d_fermionic_surface\%22\%2C\%223d_stabilizer\%22\%2C\%223d_surface\%22\%2C\%22anisotropic_z2_laplacian\%22\%2C\%22chamon\%22\%2C\%22checkerboard\%22\%2C\%223d_semion\%22\%2C\%22cubic_honeycomb_color\%22\%2C\%22fibonacci_fractal_liquid\%22\%2C\%22fcc_fracton\%22\%2C\%22fracton\%22\%2C\%22haah_cubic\%22\%2C\%22hh_fracton\%22\%2C\%22hhb_fracton\%22\%2C\%22majorana_checkerboard\%22\%2C\%22qudit_3d_surface\%22\%2C\%22qudit_xcube\%22\%2C\%22qudit_cubic\%22\%2C\%22rbh\%22\%2C\%22sierpinsky_fractal_liquid\%22\%2C\%22tetrahedral_color\%22\%2C\%22three_fermion\%22\%2C\%22two_foliated\%22\%2C\%22fractal_liquid\%22\%2C\%22xcube\%22\%2C\%22stab_15_1_3\%22\%2C\%22stab_8_3_2\%22\%5D\%2C\%22reusePreviousLayoutPositions\%22\%3Afalse\%2C\%22showIntermediateConnectingNodes\%22\%3Atrue\%2C\%22connectingNodesMaxDepth\%22\%3A15\%2C\%22connectingNodesPathMaxLength\%22\%3A20\%2C\%22connectingNodesMaxExtraDepth\%22\%3A3\%2C\%22connectingNodesOnlyKeepPathsWithAdditionalLength\%22\%3A1\%2C\%22connectingNodesToDomainsAndKingdoms\%22\%3Afalse\%2C\%22connectingNodesEdgeLengthsByType\%22\%3A\%7B\%22primaryParent\%22\%3A1\%2C\%22secondaryParent\%22\%3A4\%2C\%22cousin\%22\%3A6\%7D\%2C\%22nodeIds\%22\%3A\%5B\%5D\%7D\%2C\%22highlightImportantNodes\%22\%3A\%7B\%22highlightImportantNodes\%22\%3Afalse\%2C\%22highlightPrimaryParents\%22\%3Afalse\%2C\%22highlightRootConnectingEdges\%22\%3Afalse\%7D\%7D}

\begingroup
\small
\eczhBreakableDashes
\renewcommand\arraystretch{1.05}
\edef\myxtraspc{\eczListAddVSpaceXtraXtraStretch}
\begin{tabularx}{\linewidth}{>{\raggedright\arraybackslash}p{\eczListColWidth{name}} >{\hsize=1.0000\hsize }X}
\toprule
\eczListColTitle{Code} & \eczListColTitle{Description} \\
\midrule
\endfirsthead
\toprule
\eczListColTitleContinued{Code} & \eczListColTitleContinued{Description} \\
\midrule
\endhead
\bottomrule
\endfoot
\eczhRefIndex{code:3d_bosonization}%
\eczhListValue{\flmRefsHyperref{code:3d_bosonization}{3D bosonization code}} & \eczhListValue{A mapping from a 3D lattice quadratic Hamiltonian of Majorana modes to a lattice of qubits which realizes a \(\mathbb{Z}_2\) gauge theory with a particular Gauss law.}\\ 
\addlinespace[\myxtraspc]
\eczhRefIndex{code:3d_color}%
\eczhListValue{\flmRefsHyperref{code:3d_color}{3D color code}} & \eczhListValue{Color code defined on a four-valent, four-colorable 3-colex in a 3-manifold.
In the original colex realization, qubits sit on vertices, \(X\)-type stabilizers are attached to 3-cells, and \(Z\)-type stabilizers are attached to faces \NoCaseChange{\protect\cite{cite430}}.}\\ 
\addlinespace[\myxtraspc]
\eczhRefIndex{code:3d_fermionic_surface}%
\eczhListValue{\flmRefsHyperref{code:3d_fermionic_surface}{3D fermionic surface code}} & \eczhListValue{A non-CSS variant of the 3D Kitaev surface code that realizes \(\mathbb{Z}_2\) gauge theory with an emergent fermion, i.e., the fermionic-charge bosonic-loop (FcBl) phase \NoCaseChange{\protect\cite{cite455}}.
The model can be defined on a cubic lattice in several ways \NoCaseChange{\protect\cite[{Eq. (D45-46)}]{cite456}}.
Realizations on other lattices also exist \NoCaseChange{\protect\cite{cite457}}, and the phase of this code also exists in the 3D Kitaev honeycomb model \NoCaseChange{\protect\cite{cite458}}.}\\ 
\addlinespace[\myxtraspc]
\eczhRefIndex{code:3d_stabilizer}%
\eczhListValue{\flmRefsHyperref{code:3d_stabilizer}{3D lattice stabilizer code}} & \eczhListValue{Lattice stabilizer code in three Euclidean dimensions, using either the ordinary block notion of locality or the fermionic/Majorana notion of locality.}\\ 
\addlinespace[\myxtraspc]
\eczhRefIndex{code:3d_surface}%
\eczhListValue{\flmRefsHyperref{code:3d_surface}{3D surface code}} & \eczhListValue{A generalization of the Kitaev surface code defined on a 3D cubic lattice.
Qubits are placed on edges, \(Z\)-type stabilizer generators are placed on square plaquettes oriented in all three directions, and \(X\)-type stabilizers are placed on the six edges neighboring every vertex \NoCaseChange{\protect\cite{cite459}}.}\\ 
\addlinespace[\myxtraspc]
\eczhRefIndex{code:anisotropic_z2_laplacian}%
\eczhListValue{\flmRefsHyperref{code:anisotropic_z2_laplacian}{Anisotropic \(\mathbb{Z}_2\) Laplacian model code}} & \eczhListValue{A graph-based analogue of a Type-I fracton phase with lineon-like restricted mobility \NoCaseChange{\protect\cite{cite460,cite461}}.}\\ 
\addlinespace[\myxtraspc]
\eczhRefIndex{code:chamon}%
\eczhListValue{\flmRefsHyperref{code:chamon}{Chamon model code}} & \eczhListValue{A foliated type-I fracton non-CSS code defined on a cubic lattice using one weight-eight stabilizer generator acting on the eight vertices of each cube in the lattice \NoCaseChange{\protect\cite[{Eq. (D38)}]{cite456}}.}\\ 
\addlinespace[\myxtraspc]
\eczhRefIndex{code:checkerboard}%
\eczhListValue{\flmRefsHyperref{code:checkerboard}{Checkerboard model code}} & \eczhListValue{A foliated type-I fracton code defined on a cubic lattice that admits weight-eight  \(X\)- and \(Z\)-type stabilizer generators on the eight vertices of each cube in the lattice.
A tetrahedral Ising model can be used to obtain the checkerboard model by gauging \NoCaseChange{\protect\cite{cite462,cite463,cite233,cite464,cite465,cite466,cite467,cite468,cite469,cite470}} its subsystem symmetry \NoCaseChange{\protect\cite{cite233}}.
In that construction, the checkerboard model is self-dual under exchange of \(X\)- and \(Z\)-type stabilizers, and its composites include dimension-1 and dimension-2 excitations, i.e., lineons and planons in later terminology, with anyonic mutual and self-statistics \NoCaseChange{\protect\cite{cite233}}.}\\ 
\addlinespace[\myxtraspc]
\eczhRefIndex{code:3d_semion}%
\eczhListValue{\flmRefsHyperref{code:3d_semion}{Chiral semion Walker-Wang model code}} & \eczhListValue{A 3D lattice modular-qudit stabilizer code with qudit dimension \(q=4\) whose low-energy excitations on boundaries realize the chiral semion topological order.
The model admits 2D chiral semion topological order at one of its surfaces \NoCaseChange{\protect\cite{cite471,cite472}}.
The corresponding phase can also be realized via a non-stabilizer Hamiltonian \NoCaseChange{\protect\cite{cite473}}.}\\ 
\addlinespace[\myxtraspc]
\eczhRefIndex{code:cubic_honeycomb_color}%
\eczhListValue{\flmRefsHyperref{code:cubic_honeycomb_color}{Cubic honeycomb color code}} & \eczhListValue{3D color code defined on a four-colorable bitruncated cubic honeycomb uniform tiling.}\\ 
\addlinespace[\myxtraspc]
\eczhRefIndex{code:fibonacci_fractal_liquid}%
\eczhListValue{\flmRefsHyperref{code:fibonacci_fractal_liquid}{Fibonacci fractal spin-liquid code}} & \eczhListValue{A fractal type-I fracton CSS code defined on a cubic lattice \NoCaseChange{\protect\cite[{Eq. (D23)}]{cite456}}.}\\ 
\addlinespace[\myxtraspc]
\eczhRefIndex{code:fcc_fracton}%
\eczhListValue{\flmRefsHyperref{code:fcc_fracton}{Four Color Cube (FCC) fracton model code}} & \eczhListValue{A fracton code obtained from four coupled X-cube models using p-membrane condensation.
A modular-qudit generalization has been proposed \NoCaseChange{\protect\cite{cite474}}.}\\ 
\addlinespace[\myxtraspc]
\eczhRefIndex{code:fracton}%
\eczhListValue{\flmRefsHyperref{code:fracton}{Fracton stabilizer code}} & \eczhListValue{A 3D modular-qudit stabilizer code whose codewords make up the ground-state space of a Hamiltonian in a fracton phase.
Unlike topological phases, whose excitations can move in any direction, fracton phases are characterized by excitations whose movement is restricted.}\\ 
\addlinespace[\myxtraspc]
\eczhRefIndex{code:haah_cubic}%
\eczhListValue{\flmRefsHyperref{code:haah_cubic}{Haah cubic code (CC)}} & \eczhListValue{A 3D lattice stabilizer code on a length-\(L\) cubic lattice with one or two qubits per site.
Admits two types of stabilizer generators with support on each cube of the lattice.
In the non-CSS case, these two are related by spatial inversion.
For CSS codes, we require that the product of all corner operators is the identity.
We lastly require that there are no non-trivial string operators, meaning that single-site operators are a phase, and any period one logical operator \(l \in \mathsf{S}^{\perp}\) is just a phase.}\\ 
\addlinespace[\myxtraspc]
\eczhRefIndex{code:hh_fracton}%
\eczhListValue{\flmRefsHyperref{code:hh_fracton}{Hsieh-Halasz (HH) code}} & \eczhListValue{Member of one of two families of fracton codes, named HH-I and HH-II, defined on a cubic lattice with two qubits per site.
HH-I (HH-II) is a CSS (non-CSS) stabilizer code family, with the former identified as a foliated type-I fracton code that is decomposable into two separate lattice models \NoCaseChange{\protect\cite{cite456}}.
The sorting analysis of Ref. \NoCaseChange{\protect\cite{cite456}} leaves HH-II inconclusive, consistent with either a fractal type-I or a type-II fracton phase.}\\ 
\addlinespace[\myxtraspc]
\eczhRefIndex{code:hhb_fracton}%
\eczhListValue{\flmRefsHyperref{code:hhb_fracton}{Hsieh-Halasz-Balents (HHB) code}} & \eczhListValue{Member of one of two families of fracton codes, named HHB model A and B, defined on a cubic lattice with two qubits per site.
Both are expected to be foliated type-I fracton codes \NoCaseChange{\protect\cite[{Eqs. (D42-D43)}]{cite456}}.}\\ 
\addlinespace[\myxtraspc]
\eczhRefIndex{code:majorana_checkerboard}%
\eczhListValue{\flmRefsHyperref{code:majorana_checkerboard}{Majorana checkerboard code}} & \eczhListValue{A Majorana analogue of the X-cube model defined on a cubic lattice.
The code admits weight-eight Majorana stabilizer generators on the eight vertices of each cube of a checkerboard sublattice.}\\ 
\addlinespace[\myxtraspc]
\eczhRefIndex{code:qudit_3d_surface}%
\eczhListValue{\flmRefsHyperref{code:qudit_3d_surface}{Modular-qudit 3D surface code}} & \eczhListValue{A generalization of the 3D surface code to modular qudits.
Qudits are placed on edges, \(Z\)-type stabilizer generators are placed on square plaquettes oriented in all three directions, and \(X\)-type stabilizers are placed on the six edges neighboring every vertex \NoCaseChange{\protect\cite{cite459}}.}\\ 
\addlinespace[\myxtraspc]
\eczhRefIndex{code:qudit_xcube}%
\eczhListValue{\flmRefsHyperref{code:qudit_xcube}{Qudit X-cube model code}} & \eczhListValue{Generalization of the X-cube model code to modular qudits.}\\ 
\addlinespace[\myxtraspc]
\eczhRefIndex{code:qudit_cubic}%
\eczhListValue{\flmRefsHyperref{code:qudit_cubic}{Qudit cubic code}} & \eczhListValue{Generalization of the Haah cubic code to modular qudits.}\\ 
\addlinespace[\myxtraspc]
\eczhRefIndex{code:rbh}%
\eczhListValue{\flmRefsHyperref{code:rbh}{Raussendorf-Bravyi-Harrington (RBH) cluster-state code}} & \eczhListValue{A three-dimensional cluster-state code defined on the bcc lattice (i.e., a cubic lattice with qubits on edges and faces).}\\ 
\addlinespace[\myxtraspc]
\eczhRefIndex{code:sierpinsky_fractal_liquid}%
\eczhListValue{\flmRefsHyperref{code:sierpinsky_fractal_liquid}{Sierpinski prism model code}} & \eczhListValue{A fractal type-I fracton CSS code defined on a cubic lattice \NoCaseChange{\protect\cite[{Eq. (D22)}]{cite456}}.
The code admits an excitation-moving operator shaped like a Sierpinski triangle \NoCaseChange{\protect\cite[{Fig. 2}]{cite456}}.}\\ 
\addlinespace[\myxtraspc]
\eczhRefIndex{code:tetrahedral_color}%
\eczhListValue{\flmRefsHyperref{code:tetrahedral_color}{Tetrahedral color code}} & \eczhListValue{A 3D color code defined on a colored tetrahedron cut from a suitably colored BCC lattice \NoCaseChange{\protect\cite{cite475}}.
Qubits are placed on tetrahedra, on the triangles covering the tetrahedron faces, on the edges along the tetrahedron edges, and on the tetrahedron vertices.
The code has both string-like and sheet-like logical operators \NoCaseChange{\protect\cite{cite476}}.}\\ 
\addlinespace[\myxtraspc]
\eczhRefIndex{code:three_fermion}%
\eczhListValue{\flmRefsHyperref{code:three_fermion}{Three-fermion (3F) Walker-Wang model code}} & \eczhListValue{A 3D lattice stabilizer code whose bulk realizes a 3D time-reversal SPT order \NoCaseChange{\protect\cite{cite477}} and whose gapped boundary supports the 2D three-fermion (3F) topological order.
The code can be used as a resource state for fault-tolerant MBQC \NoCaseChange{\protect\cite{cite478}}.}\\ 
\addlinespace[\myxtraspc]
\eczhRefIndex{code:two_foliated}%
\eczhListValue{\flmRefsHyperref{code:two_foliated}{Two-foliated fracton code}} & \eczhListValue{A type-I fracton code obtained by gauging \NoCaseChange{\protect\cite{cite462,cite463,cite233,cite464,cite465,cite466,cite467,cite468,cite469,cite470}} a 3D paramagnet with planar subsystem symmetries in two directions.
In that construction, the gauge charges are lineons and the flux excitations are also lineons moving in the same direction, yielding the anisotropic lineon model \NoCaseChange{\protect\cite[{Sec. 4.1.2}]{cite467}}.}\\ 
\addlinespace[\myxtraspc]
\eczhRefIndex{code:fractal_liquid}%
\eczhListValue{\flmRefsHyperref{code:fractal_liquid}{Type-II fractal spin-liquid code}} & \eczhListValue{A type-II fracton prime-qudit CSS code defined on a cubic lattice \NoCaseChange{\protect\cite[{Eqs. (D9-D10)}]{cite456}}.}\\ 
\addlinespace[\myxtraspc]
\eczhRefIndex{code:xcube}%
\eczhListValue{\flmRefsHyperref{code:xcube}{X-cube model code}} & \eczhListValue{A foliated type-I fracton CSS code on a cubic lattice with qubits on edges, cube stabilizers, and three cross-shaped vertex stabilizers for each vertex \NoCaseChange{\protect\cite{cite233}}.
It supports a subextensive number of logical qubits.}\\ 
\addlinespace[\myxtraspc]
\eczhRefIndex{code:stab_15_1_3}%
\eczhListValue{\flmRefsHyperref{code:stab_15_1_3}{\(\llbracket 15,1,3\rrbracket \) quantum RM code}} & \eczhListValue{A \(\llbracket 15,1,3\rrbracket \) quantum Reed-Muller code that is most easily thought of as a tetrahedral 3D color code.
It can be constructed as a CSS code from the \([15,5,8]\) punctured Reed-Muller code and its even subcode, which explains its transversal \(T^\dagger\) gate \NoCaseChange{\protect\cite{cite398}}.}\\ 
\addlinespace[\myxtraspc]
\eczhRefIndex{code:stab_8_3_2}%
\eczhListValue{\flmRefsHyperref{code:stab_8_3_2}{\(\llbracket 8,3,2\rrbracket \) Smallest interesting color code}} & \eczhListValue{Smallest 3D color code whose physical qubits lie on vertices of a cube and which admits a (weakly) transversal \(CCZ\) gate.}\\ 
\end{tabularx}\endgroup
\eczcodelist{stabilizer_4d}{4D stabilizer codes
}%

\eczhCodeListAutoDescription{All descendants of \flmRefsCref{code:4d_stabilizer}.}%

\eczhIncludeCodeGraph{Bare}{scale=0.5}{\columnwidth}{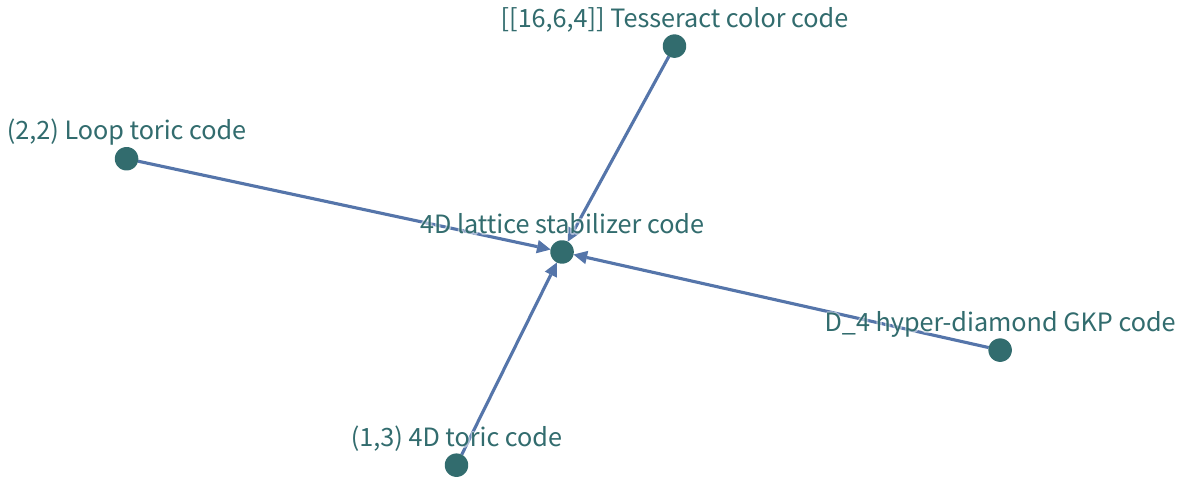}{4D stabilizer codes}{https://errorcorrectionzoo.org/code_graph#J\%7B\%22displayMode\%22\%3A\%22subset\%22\%2C\%22modeSubsetOptions\%22\%3A\%7B\%22codeIds\%22\%3A\%5B\%224d_stabilizer\%22\%2C\%224d_13_surface\%22\%2C\%224d_surface\%22\%2C\%22dfour_gkp\%22\%2C\%22stab_16_6_4\%22\%5D\%2C\%22reusePreviousLayoutPositions\%22\%3Afalse\%2C\%22showIntermediateConnectingNodes\%22\%3Atrue\%2C\%22connectingNodesMaxDepth\%22\%3A15\%2C\%22connectingNodesPathMaxLength\%22\%3A20\%2C\%22connectingNodesMaxExtraDepth\%22\%3A3\%2C\%22connectingNodesOnlyKeepPathsWithAdditionalLength\%22\%3A1\%2C\%22connectingNodesToDomainsAndKingdoms\%22\%3Afalse\%2C\%22connectingNodesEdgeLengthsByType\%22\%3A\%7B\%22primaryParent\%22\%3A1\%2C\%22secondaryParent\%22\%3A4\%2C\%22cousin\%22\%3A6\%7D\%2C\%22nodeIds\%22\%3A\%5B\%5D\%7D\%2C\%22highlightImportantNodes\%22\%3A\%7B\%22highlightImportantNodes\%22\%3Afalse\%2C\%22highlightPrimaryParents\%22\%3Afalse\%2C\%22highlightRootConnectingEdges\%22\%3Afalse\%7D\%7D}

\begingroup
\small
\eczhBreakableDashes
\renewcommand\arraystretch{1.05}
\edef\myxtraspc{\eczListAddVSpaceXtraXtraStretch}
\begin{tabularx}{\linewidth}{>{\raggedright\arraybackslash}p{\eczListColWidth{name}} >{\hsize=1.0000\hsize }X}
\toprule
\eczListColTitle{Code} & \eczListColTitle{Description} \\
\midrule
\endfirsthead
\toprule
\eczListColTitleContinued{Code} & \eczListColTitleContinued{Description} \\
\midrule
\endhead
\bottomrule
\endfoot
\eczhRefIndex{code:4d_stabilizer}%
\eczhListValue{\flmRefsHyperref{code:4d_stabilizer}{4D lattice stabilizer code}} & \eczhListValue{Lattice stabilizer code in four Euclidean dimensions, using either the ordinary block notion of locality or the fermionic/Majorana notion of locality.}\\ 
\addlinespace[\myxtraspc]
\eczhRefIndex{code:4d_13_surface}%
\eczhListValue{\flmRefsHyperref{code:4d_13_surface}{\((1,3)\) 4D toric code}} & \eczhListValue{A generalization of the Kitaev surface code defined on a 4D lattice.
The code is called a \((1,3)\) toric code because it admits 1D \(Z\)-type and 3D \(X\)-type logical operators.
In the hypercubic lattice version, qubits are placed on edges, each \(Z\)-type stabilizer generator is supported on cubes on the boundary of a hypercube, and \(X\)-type stabilizers are placed on the edges neighboring every vertex \NoCaseChange{\protect\cite{cite479}}.}\\ 
\addlinespace[\myxtraspc]
\eczhRefIndex{code:4d_surface}%
\eczhListValue{\flmRefsHyperref{code:4d_surface}{\((2,2)\) Loop toric code}} & \eczhListValue{A generalization of the Kitaev surface code defined on a 4D lattice.
The code is called a \((2,2)\) toric code because it admits 2D membrane \(Z\)-type and \(X\)-type logical operators.
Both types of operators create 1D (i.e., loop) excitations at their edges.
The code serves as a self-correcting quantum memory \NoCaseChange{\protect\cite{cite480,cite481}}.}\\ 
\addlinespace[\myxtraspc]
\eczhRefIndex{code:dfour_gkp}%
\eczhListValue{\flmRefsHyperref{code:dfour_gkp}{\(D_4\) hyper-diamond GKP code}} & \eczhListValue{Two-mode GKP qubit-into-oscillator code based on the \(D_4\) hyper-diamond lattice \NoCaseChange{\protect\cite{cite482}}.}\\ 
\addlinespace[\myxtraspc]
\eczhRefIndex{code:stab_16_6_4}%
\eczhListValue{\flmRefsHyperref{code:stab_16_6_4}{\(\llbracket 16,6,4\rrbracket \) Tesseract color code}} & \eczhListValue{A (hyperbolic self-dual CSS) 4D color code defined on a tesseract, with stabilizer generators of both types supported on each cube. 
A \(\llbracket 16,4,2,4\rrbracket \) tesseract subsystem code can be obtained from this code by using two logical qubits as gauge qubits \NoCaseChange{\protect\cite{cite483}}.}\\ 
\end{tabularx}\endgroup
\eczcodelist{good_qldpc}{Asymptotically good QLDPC codes and friends
}%

\eczhCodeListAutoDescription{All cousins of \flmRefsCref{code:good_qldpc}.}%

\eczhIncludeCodeGraph{Bare}{scale=0.5}{\columnwidth}{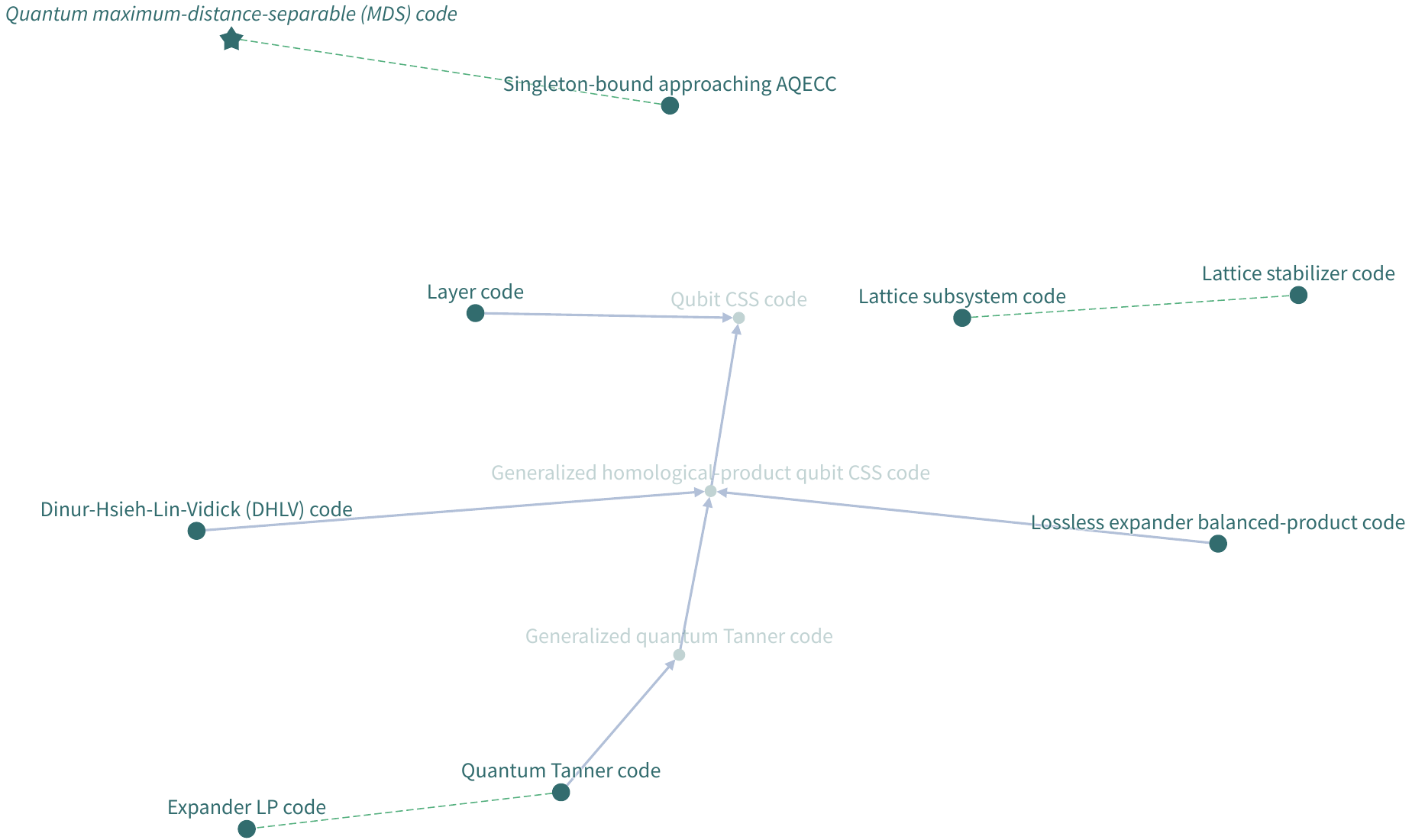}{Asymptotically good QLDPC codes and friends}{https://errorcorrectionzoo.org/code_graph#J\%7B\%22displayMode\%22\%3A\%22subset\%22\%2C\%22modeSubsetOptions\%22\%3A\%7B\%22codeIds\%22\%3A\%5B\%22dhlv\%22\%2C\%22expander_lifted_product\%22\%2C\%22translationally_invariant_stabilizer\%22\%2C\%22translationally_invariant_subsystem\%22\%2C\%22layer\%22\%2C\%22lossless_expander\%22\%2C\%22quantum_tanner\%22\%2C\%22quantum_mds\%22\%2C\%22quantum_singleton\%22\%5D\%2C\%22reusePreviousLayoutPositions\%22\%3Afalse\%2C\%22showIntermediateConnectingNodes\%22\%3Atrue\%2C\%22connectingNodesMaxDepth\%22\%3A15\%2C\%22connectingNodesPathMaxLength\%22\%3A20\%2C\%22connectingNodesMaxExtraDepth\%22\%3A3\%2C\%22connectingNodesOnlyKeepPathsWithAdditionalLength\%22\%3A1\%2C\%22connectingNodesToDomainsAndKingdoms\%22\%3Afalse\%2C\%22connectingNodesEdgeLengthsByType\%22\%3A\%7B\%22primaryParent\%22\%3A1\%2C\%22secondaryParent\%22\%3A4\%2C\%22cousin\%22\%3A6\%7D\%2C\%22nodeIds\%22\%3A\%5B\%5D\%7D\%2C\%22highlightImportantNodes\%22\%3A\%7B\%22highlightImportantNodes\%22\%3Afalse\%2C\%22highlightPrimaryParents\%22\%3Afalse\%2C\%22highlightRootConnectingEdges\%22\%3Afalse\%7D\%7D}

\begingroup
\small
\eczhBreakableDashes
\renewcommand\arraystretch{1.05}
\edef\myxtraspc{\eczListAddVSpaceXtraXtraStretch}
\begin{tabularx}{\linewidth}{>{\raggedright\arraybackslash}p{\eczListColWidth{name}} >{\hsize=1.0000\hsize }X}
\toprule
\eczListColTitle{Code} & \eczListColTitle{Relation} \\
\midrule
\endfirsthead
\toprule
\eczListColTitleContinued{Code} & \eczListColTitleContinued{Relation} \\
\midrule
\endhead
\bottomrule
\endfoot
\eczhRefIndex{code:dhlv}%
\eczhListValue{\flmRefsHyperref{code:dhlv}{Dinur-Hsieh-Lin-Vidick (DHLV) code}} & \eczhListValue{DHLV code construction yields asymptotically good QLDPC codes.}\\ 
\addlinespace[\myxtraspc]
\eczhRefIndex{code:expander_lifted_product}%
\eczhListValue{\flmRefsHyperref{code:expander_lifted_product}{Expander LP code}} & \eczhListValue{Lifted products of certain classical Tanner codes are the first asymptotically good QLDPC codes.}\\ 
\addlinespace[\myxtraspc]
\eczhRefIndex{code:translationally_invariant_stabilizer}%
\eczhListValue{\flmRefsHyperref{code:translationally_invariant_stabilizer}{Lattice stabilizer code}} & \eczhListValue{Chain complexes describing some QLDPC codes \NoCaseChange{\protect\cite{cite484,cite485}}, and, more generally, CSS codes \NoCaseChange{\protect\cite{cite486}} can be 'lifted' into higher-dimensional manifolds admitting some notion of geometric locality. Applying this procedure to good QLDPC codes yields \(\llbracket n,n^{1-2/D},n^{1-1/D}\rrbracket \) lattice stabilizer codes in \(D\) spatial dimensions that saturate the \flmRefsHyperref{ref487}{BPT bound} \NoCaseChange{\protect\cite{cite488,cite485,cite489}}.}\\ 
\addlinespace[\myxtraspc]
\eczhRefIndex{code:translationally_invariant_subsystem}%
\eczhListValue{\flmRefsHyperref{code:translationally_invariant_subsystem}{Lattice subsystem code}} & \eczhListValue{An \(\llbracket n,k,d\rrbracket \) qubit stabilizer code can be converted into an \flmRefsHyperref{ref65}{order} \(\llbracket O(\ell \delta n),k,\Omega(d/w)\rrbracket \) subsystem qubit stabilizer code with weight-three gauge operators via the wire-code mapping \NoCaseChange{\protect\cite{cite490}}, which uses \flmRefsHyperref{ref491}{weight reduction}. 
Here, \(w\) and \(\delta\) are the weight and degree of the input code's Tanner graph, while \(\ell\) is the length of the longest edge of a particular embedding of that graph.
Applying this procedure to good QLDPC codes and using an embedding into \(D\)-dimensional Euclidean space yields lattice subsystem codes whose logical-qubit number and distance both scale as \(\Theta(n^{1-1/D})\) as functions of block length \(n\), saturating the \flmRefsHyperref{ref492}{subsystem BT bound} \NoCaseChange{\protect\cite{cite490}}.}\\ 
\addlinespace[\myxtraspc]
\eczhRefIndex{code:layer}%
\eczhListValue{\flmRefsHyperref{code:layer}{Layer code}} & \eczhListValue{Layer codes achieve the 3D \flmRefsHyperref{ref487}{BPT bound}, with parameters  \(\llbracket n,\Theta(n^{1/3}),\Theta(n^{1/3})\rrbracket \), when asymptotically good QLDPC codes are used in the construction.}\\ 
\addlinespace[\myxtraspc]
\eczhRefIndex{code:lossless_expander}%
\eczhListValue{\flmRefsHyperref{code:lossless_expander}{Lossless expander balanced-product code}} & \eczhListValue{Taking a balanced product of two-sided expanders \NoCaseChange{\protect\cite{cite187}} yields an asymptotically good QLDPC code family \NoCaseChange{\protect\cite{cite188}}.}\\ 
\addlinespace[\myxtraspc]
\eczhRefIndex{code:quantum_tanner}%
\eczhListValue{\flmRefsHyperref{code:quantum_tanner}{Quantum Tanner code}} & \eczhListValue{Quantum Tanner code construction yields asymptotically good QLDPC codes.}\\ 
\addlinespace[\myxtraspc]
\eczhRefIndex{code:quantum_mds}%
\eczhListValue{\flmRefsHyperref{code:quantum_mds}{Quantum maximum-distance-separable (MDS) code}} & \eczhListValue{AEL distance amplification \NoCaseChange{\protect\cite{cite493,cite494}} can be used to construct asymptotically good QLDPC codes that approach the quantum Singleton bound \NoCaseChange{\protect\cite[{Corr. 5.3}]{cite495}}.}\\ 
\addlinespace[\myxtraspc]
\eczhRefIndex{code:quantum_singleton}%
\eczhListValue{\flmRefsHyperref{code:quantum_singleton}{Singleton-bound approaching AQECC}} & \eczhListValue{AEL distance amplification \NoCaseChange{\protect\cite{cite493,cite494}} can be used to construct constant-alphabet QLDPC CSS codes of any target rate \(R\) and relative distance \((1-R-\gamma)/2\) that are decodable in linear time up to half that distance \NoCaseChange{\protect\cite[{Corr. 5.3}]{cite495}}. The AEL distance-amplification framework also yields constant-alphabet approximate quantum codes that decode nearly up to the quantum Singleton bound \NoCaseChange{\protect\cite{cite495}}.}\\ 
\end{tabularx}\endgroup
\eczcodelist{fock_state}{Bosonic Fock-state codes
}%

\eczhCodeListAutoDescription{All descendants of \flmRefsCref{code:fock_state}.}%

\eczhIncludeCodeGraph{Bare}{scale=0.5}{\columnwidth}{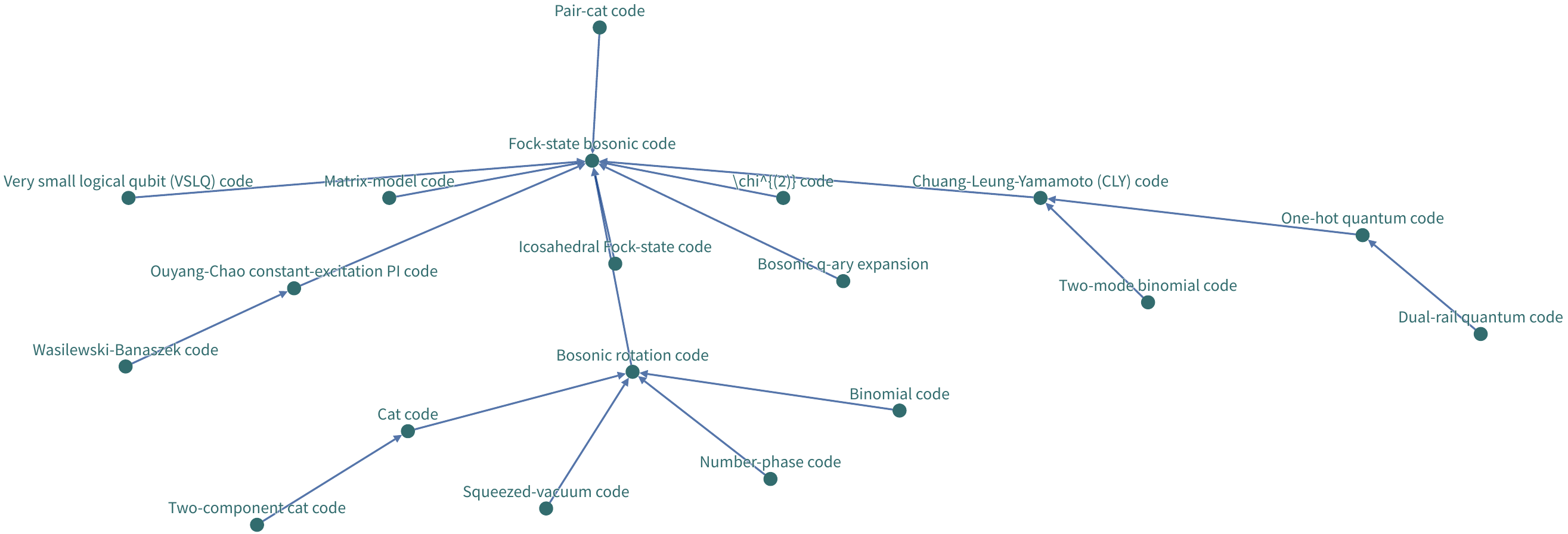}{Bosonic Fock-state codes}{https://errorcorrectionzoo.org/code_graph#J\%7B\%22displayMode\%22\%3A\%22subset\%22\%2C\%22modeSubsetOptions\%22\%3A\%7B\%22codeIds\%22\%3A\%5B\%22binomial\%22\%2C\%22bosonic_q-ary_expansion\%22\%2C\%22bosonic_rotation\%22\%2C\%22cat\%22\%2C\%22chuang-leung-yamamoto\%22\%2C\%22dual_rail\%22\%2C\%22fock_state\%22\%2C\%22icosahedral_fock\%22\%2C\%22matrix_qm\%22\%2C\%22number_phase\%22\%2C\%22one_hot_quantum\%22\%2C\%22constant_excitation_permutation_invariant\%22\%2C\%22paircat\%22\%2C\%22squeezed_vacuum\%22\%2C\%22two-legged-cat\%22\%2C\%22two-mode_binomial\%22\%2C\%22very-small-logical-qubit\%22\%2C\%22wasilewski-banaszek\%22\%2C\%22chi2\%22\%5D\%2C\%22reusePreviousLayoutPositions\%22\%3Afalse\%2C\%22showIntermediateConnectingNodes\%22\%3Atrue\%2C\%22connectingNodesMaxDepth\%22\%3A15\%2C\%22connectingNodesPathMaxLength\%22\%3A20\%2C\%22connectingNodesMaxExtraDepth\%22\%3A3\%2C\%22connectingNodesOnlyKeepPathsWithAdditionalLength\%22\%3A1\%2C\%22connectingNodesToDomainsAndKingdoms\%22\%3Afalse\%2C\%22connectingNodesEdgeLengthsByType\%22\%3A\%7B\%22primaryParent\%22\%3A1\%2C\%22secondaryParent\%22\%3A4\%2C\%22cousin\%22\%3A6\%7D\%2C\%22nodeIds\%22\%3A\%5B\%5D\%7D\%2C\%22highlightImportantNodes\%22\%3A\%7B\%22highlightImportantNodes\%22\%3Afalse\%2C\%22highlightPrimaryParents\%22\%3Afalse\%2C\%22highlightRootConnectingEdges\%22\%3Afalse\%7D\%7D}

\begingroup
\small
\eczhBreakableDashes
\renewcommand\arraystretch{1.05}
\edef\myxtraspc{\eczListAddVSpaceXtraXtraStretch}
\begin{tabularx}{\linewidth}{>{\raggedright\arraybackslash}p{\eczListColWidth{name}} >{\hsize=1.0000\hsize }X}
\toprule
\eczListColTitle{Code} & \eczListColTitle{Description} \\
\midrule
\endfirsthead
\toprule
\eczListColTitleContinued{Code} & \eczListColTitleContinued{Description} \\
\midrule
\endhead
\bottomrule
\endfoot
\eczhRefIndex{code:binomial}%
\eczhListValue{\flmRefsHyperref{code:binomial}{Binomial code}} & \eczhListValue{Bosonic rotation codes designed to approximately protect against errors consisting of powers of raising and lowering operators up to some maximum power. Binomial codes can be thought of as spin-coherent states embedded into an oscillator \NoCaseChange{\protect\cite{cite496}}.}\\ 
\addlinespace[\myxtraspc]
\eczhRefIndex{code:bosonic_q-ary_expansion}%
\eczhListValue{\flmRefsHyperref{code:bosonic_q-ary_expansion}{Bosonic \(q\)-ary expansion}} & \eczhListValue{A one-to-one mapping between basis states on \(n\) prime-dimensional qudits (of dimension \(q=p\)) and the subspace of the first \(p^n\) single-mode Fock states.
While this mapping offers a way to map qudits into a single mode, noise models for the two code families induce different notions of locality and thus qualitatively different physical interpretations \NoCaseChange{\protect\cite{cite497}}.}\\ 
\addlinespace[\myxtraspc]
\eczhRefIndex{code:bosonic_rotation}%
\eczhListValue{\flmRefsHyperref{code:bosonic_rotation}{Bosonic rotation code}} & \eczhListValue{A single-mode Fock-state bosonic code whose codespace is preserved by a phase-space rotation by a multiple of \(2\pi/N\) for some \(N\). The rotation symmetry ensures that encoded states have support only on every \(N^{\textrm{th}}\) Fock state. For example, single-mode Fock-state codes for \(N=2\) encoding a qubit admit basis states that are, respectively, supported on Fock state sets \(\{|0\rangle,|4\rangle,|8\rangle,\cdots\}\) and \(\{|2\rangle,|6\rangle,|10\rangle,\cdots\}\).}\\ 
\addlinespace[\myxtraspc]
\eczhRefIndex{code:cat}%
\eczhListValue{\flmRefsHyperref{code:cat}{Cat code}} & \eczhListValue{Rotation-symmetric bosonic Fock-state code encoding a \(q\)-dimensional qudit into one oscillator which utilizes a constellation of \(q(S+1)\) coherent states distributed equidistantly around a circle in phase space of radius \(\alpha\).}\\ 
\addlinespace[\myxtraspc]
\eczhRefIndex{code:chuang-leung-yamamoto}%
\eczhListValue{\flmRefsHyperref{code:chuang-leung-yamamoto}{Chuang-Leung-Yamamoto (CLY) code}} & \eczhListValue{Bosonic Fock-state code that encodes \(k\) qubits into \(n\) oscillators, with each oscillator restricted to having at most \(N\) excitations. Codewords are superpositions of oscillator Fock states which have exactly \(N\) total excitations, and are either uniform (i.e., balanced) superpositions or unbalanced superpositions.}\\ 
\addlinespace[\myxtraspc]
\eczhRefIndex{code:dual_rail}%
\eczhListValue{\flmRefsHyperref{code:dual_rail}{Dual-rail quantum code}} & \eczhListValue{Two-mode bosonic code encoding a logical qubit in Fock states with one excitation.
The logical-zero state is represented by \(|10\rangle\), while the logical-one state is represented by \(|01\rangle\).
This encoding is often realized in temporal or spatial modes, corresponding to a \textit{time-bin} or \textit{frequency-bin} encoding.
Two different types of photon polarization can also be used.}\\ 
\addlinespace[\myxtraspc]
\eczhRefIndex{code:fock_state}%
\eczhListValue{\flmRefsHyperref{code:fock_state}{Fock-state bosonic code}} & \eczhListValue{Qudit-into-oscillator code whose protection against \flmRefsHyperref{ref498}{AD} noise (i.e., photon loss) stems from the use of disjoint sets of Fock states for the construction of each code basis state. The simplest example is the dual-rail code, which has codewords consisting of single Fock states \(|10\rangle\) and \(|01\rangle\). This code can detect a single loss error since a loss operator in either mode maps one of the codewords to a different Fock state \(|00\rangle\). More involved codewords consist of several well-separated Fock states such that multiple loss events can be detected and corrected.}\\ 
\addlinespace[\myxtraspc]
\eczhRefIndex{code:icosahedral_fock}%
\eczhListValue{\flmRefsHyperref{code:icosahedral_fock}{Icosahedral Fock-state code}} & \eczhListValue{A constant-excitation Fock-state code designed to realize the \(2I\) group of gates using Gaussian rotations.
It is obtained from the corresponding icosahedral spin code via the \flmRefsHyperref{ref499}{simplex mapping} between spin and constant-excitation Fock spaces \NoCaseChange{\protect\cite{cite500}}.}\\ 
\addlinespace[\myxtraspc]
\eczhRefIndex{code:matrix_qm}%
\eczhListValue{\flmRefsHyperref{code:matrix_qm}{Matrix-model code}} & \eczhListValue{Multimode Fock-state bosonic approximate code derived from a matrix model, i.e., a bosonic theory with a large non-Abelian gauge group.
The model's degrees of freedom are matrix-valued bosons \(a\), each consisting of \(N^2\) harmonic oscillator modes and subject to an \(SU(N)\) gauge symmetry.}\\ 
\addlinespace[\myxtraspc]
\eczhRefIndex{code:number_phase}%
\eczhListValue{\flmRefsHyperref{code:number_phase}{Number-phase code}} & \eczhListValue{Bosonic rotation code consisting of superpositions of Pegg-Barnett phase states \NoCaseChange{\protect\cite{cite501}}.}\\ 
\addlinespace[\myxtraspc]
\eczhRefIndex{code:one_hot_quantum}%
\eczhListValue{\flmRefsHyperref{code:one_hot_quantum}{One-hot quantum code}} & \eczhListValue{Encoding of a \(q\)-dimensional qudit into the single-excitation subspace of \(q\) modes. The \(j\)th logical state is the multi-mode Fock state with one photon in mode \(j\) and zero photons in the other modes.
This code is useful for encoding and performing operations on qudits in multiple modes \NoCaseChange{\protect\cite{cite502,cite503,cite504,cite505,cite506}}.}\\ 
\addlinespace[\myxtraspc]
\eczhRefIndex{code:constant_excitation_permutation_invariant}%
\eczhListValue{\flmRefsHyperref{code:constant_excitation_permutation_invariant}{Ouyang-Chao constant-excitation PI code}} & \eczhListValue{A constant-excitation PI Fock-state code whose construction is based on integer partitions.}\\ 
\addlinespace[\myxtraspc]
\eczhRefIndex{code:paircat}%
\eczhListValue{\flmRefsHyperref{code:paircat}{Pair-cat code}} & \eczhListValue{Two- or higher-mode extension of cat codes whose codewords are right eigenstates of powers of products of the modes' lowering operators. Many gadgets for cat codes have two-mode pair-cat analogues, with the advantage being that such gates can be done in parallel with a dissipative error-correction process.}\\ 
\addlinespace[\myxtraspc]
\eczhRefIndex{code:squeezed_vacuum}%
\eczhListValue{\flmRefsHyperref{code:squeezed_vacuum}{Squeezed-vacuum code}} & \eczhListValue{A squeezed Fock-state code constructed from a coherent superposition of \(m\) squeezed vacuum states, each squeezed along equiangular axes in phase space.}\\ 
\addlinespace[\myxtraspc]
\eczhRefIndex{code:two-legged-cat}%
\eczhListValue{\flmRefsHyperref{code:two-legged-cat}{Two-component cat code}} & \eczhListValue{Code whose codespace is spanned by two coherent states \(\left|\pm\alpha\right\rangle\) for nonzero complex \(\alpha\).}\\ 
\addlinespace[\myxtraspc]
\eczhRefIndex{code:two-mode_binomial}%
\eczhListValue{\flmRefsHyperref{code:two-mode_binomial}{Two-mode binomial code}} & \eczhListValue{Two-mode constant-energy CLY code whose coefficients are square-roots of binomial coefficients.}\\ 
\addlinespace[\myxtraspc]
\eczhRefIndex{code:very-small-logical-qubit}%
\eczhListValue{\flmRefsHyperref{code:very-small-logical-qubit}{Very small logical qubit (VSLQ) code}} & \eczhListValue{A code consisting of two logical codewords \(|\pm\rangle \propto (|0\rangle\pm|2\rangle)(|0\rangle\pm|2\rangle)\), where the total Hilbert space is the tensor product of two transmon qudits (whose ground states \(|0\rangle\) and second excited states \(|2\rangle\) are used in the codewords).
Since the code is intended to protect against losses, the qutrits can equivalently be thought of as oscillator Fock-state subspaces.}\\ 
\addlinespace[\myxtraspc]
\eczhRefIndex{code:wasilewski-banaszek}%
\eczhListValue{\flmRefsHyperref{code:wasilewski-banaszek}{Wasilewski-Banaszek code}} & \eczhListValue{Three-oscillator constant-excitation Fock-state code encoding a single logical qubit.}\\ 
\addlinespace[\myxtraspc]
\eczhRefIndex{code:chi2}%
\eczhListValue{\flmRefsHyperref{code:chi2}{\(\chi^{(2)}\) code}} & \eczhListValue{A \(3n\)-mode bosonic Fock-state code that requires only linear optics and the \(\chi^{(2)}\) optical nonlinear interaction for encoding, decoding, and logical gates.
Codewords lie in Fock-state subspaces that are invariant under Hermitian combinations of the \(\chi^{(2)}\) nonlinearities \(abc^\dagger\) and \(i abc^\dagger\), where \(a\), \(b\), and \(c\) are lowering operators acting on one of the \(n\) triples of modes on which the codes are defined.
Codewords are also \(+1\) eigenstates of stabilizer-like \textit{symmetry operators}, and photon parities are error syndromes.}\\ 
\end{tabularx}\endgroup
\eczcodelist{oscillator_stabilizer}{Bosonic stabilizer codes
}%

\eczhCodeListAutoDescription{All descendants of \flmRefsCref{code:oscillator_stabilizer}.}%

\eczhIncludeCodeGraph{Bare}{scale=0.5}{\columnwidth}{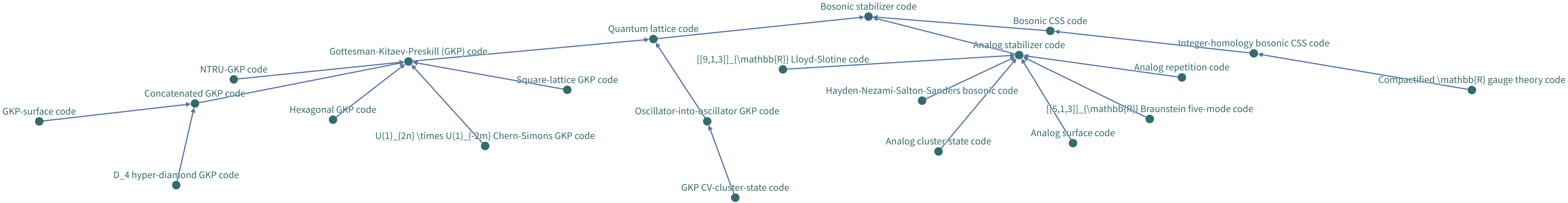}{Bosonic stabilizer codes}{https://errorcorrectionzoo.org/code_graph#J\%7B\%22displayMode\%22\%3A\%22subset\%22\%2C\%22modeSubsetOptions\%22\%3A\%7B\%22codeIds\%22\%3A\%5B\%22cv_cluster_state\%22\%2C\%22analog_repetition\%22\%2C\%22analog_stabilizer\%22\%2C\%22analog_surface\%22\%2C\%22oscillator_css\%22\%2C\%22oscillator_stabilizer\%22\%2C\%22compactified_r\%22\%2C\%22gkp_concatenated\%22\%2C\%22gkp-cluster-state\%22\%2C\%22gkp_surface_concatenated\%22\%2C\%22multimodegkp\%22\%2C\%22hnss\%22\%2C\%22hexagonal_gkp\%22\%2C\%22homological_cv\%22\%2C\%22ntru_gkp\%22\%2C\%22gkp-stabilizer\%22\%2C\%22quantum_lattice\%22\%2C\%22gkp\%22\%2C\%22dfour_gkp\%22\%2C\%22chern_simons_gkp\%22\%2C\%22braunstein\%22\%2C\%22lloyd_slotine\%22\%5D\%2C\%22reusePreviousLayoutPositions\%22\%3Afalse\%2C\%22showIntermediateConnectingNodes\%22\%3Atrue\%2C\%22connectingNodesMaxDepth\%22\%3A15\%2C\%22connectingNodesPathMaxLength\%22\%3A20\%2C\%22connectingNodesMaxExtraDepth\%22\%3A3\%2C\%22connectingNodesOnlyKeepPathsWithAdditionalLength\%22\%3A1\%2C\%22connectingNodesToDomainsAndKingdoms\%22\%3Afalse\%2C\%22connectingNodesEdgeLengthsByType\%22\%3A\%7B\%22primaryParent\%22\%3A1\%2C\%22secondaryParent\%22\%3A4\%2C\%22cousin\%22\%3A6\%7D\%2C\%22nodeIds\%22\%3A\%5B\%5D\%7D\%2C\%22highlightImportantNodes\%22\%3A\%7B\%22highlightImportantNodes\%22\%3Afalse\%2C\%22highlightPrimaryParents\%22\%3Afalse\%2C\%22highlightRootConnectingEdges\%22\%3Afalse\%7D\%7D}

\begingroup
\small
\eczhBreakableDashes
\renewcommand\arraystretch{1.05}
\edef\myxtraspc{\eczListAddVSpaceXtraXtraStretch}
\begin{tabularx}{\linewidth}{>{\raggedright\arraybackslash}p{\eczListColWidth{name}} >{\hsize=1.0000\hsize }X}
\toprule
\eczListColTitle{Code} & \eczListColTitle{Description} \\
\midrule
\endfirsthead
\toprule
\eczListColTitleContinued{Code} & \eczListColTitleContinued{Description} \\
\midrule
\endhead
\bottomrule
\endfoot
\eczhRefIndex{code:cv_cluster_state}%
\eczhListValue{\flmRefsHyperref{code:cv_cluster_state}{Analog cluster-state code}} & \eczhListValue{A code based on a continuous-variable (CV), or analog, cluster state.
Such a state can be used to perform MBQC of logical modes, which substitutes the temporal dimension necessary for decoding a conventional code with a spatial dimension.
The exact analog cluster state is non-normalizable, so approximate constructions have to be considered.}\\ 
\addlinespace[\myxtraspc]
\eczhRefIndex{code:analog_repetition}%
\eczhListValue{\flmRefsHyperref{code:analog_repetition}{Analog repetition code}} & \eczhListValue{An \(\llbracket n,1\rrbracket _{\mathbb{R}}\) analog stabilizer version of the quantum repetition code, encoding the position states of one mode into an odd number \(n\) of modes.}\\ 
\addlinespace[\myxtraspc]
\eczhRefIndex{code:analog_stabilizer}%
\eczhListValue{\flmRefsHyperref{code:analog_stabilizer}{Analog stabilizer code}} & \eczhListValue{An oscillator-into-oscillator stabilizer code encoding logical oscillator modes into \(n\) physical modes. If the code is defined by \(r\) independent nullifiers, then it is denoted by \(\llbracket n,n-r\rrbracket _{\mathbb{R}}\) and encodes \(k=n-r\) logical modes \NoCaseChange{\protect\cite{cite507}}.
Any analog stabilizer state can be thought of as a pure Gaussian state that has been infinitely squeezed on all modes \NoCaseChange{\protect\cite{cite507}}.}\\ 
\addlinespace[\myxtraspc]
\eczhRefIndex{code:analog_surface}%
\eczhListValue{\flmRefsHyperref{code:analog_surface}{Analog surface code}} & \eczhListValue{An analog CSS version of the Kitaev surface code realizing a phase of 2D \(\mathbb{R}\) gauge theory.}\\ 
\addlinespace[\myxtraspc]
\eczhRefIndex{code:oscillator_css}%
\eczhListValue{\flmRefsHyperref{code:oscillator_css}{Bosonic CSS code}} & \eczhListValue{Bosonic stabilizer code admitting a set of stabilizer generators that are either position or momentum displacements.}\\ 
\addlinespace[\myxtraspc]
\eczhRefIndex{code:oscillator_stabilizer}%
\eczhListValue{\flmRefsHyperref{code:oscillator_stabilizer}{Bosonic stabilizer code}} & \eczhListValue{Bosonic code whose codespace is defined as the common \(+1\) eigenspace of a group of mutually commuting displacement operators.
Displacements form the stabilizers of the code, and have continuous eigenvalues, in contrast with the discrete set of eigenvalues of qubit stabilizers.
As a result, exact codewords are non-normalizable, so approximate constructions have to be considered.
Stabilizer groups are any locally compact Abelian subgroups of \(\mathbb{R}^n\), can themselves contain discrete or continuous subgroups, and can admit logical qudit and/or oscillator logical subspaces.}\\ 
\addlinespace[\myxtraspc]
\eczhRefIndex{code:compactified_r}%
\eczhListValue{\flmRefsHyperref{code:compactified_r}{Compactified \(\mathbb{R}\) gauge theory code}} & \eczhListValue{An integer-homology bosonic CSS code realizing 2D \(U(1)\) gauge theory on bosonic modes.
The code can be obtained from the analog surface code by \flmRefsHyperref{ref410}{condensing} certain anyons \NoCaseChange{\protect\cite{cite411}}. 
This results in a pinning of each mode to the space of periodic functions, which is the Hilbert space of a physical rotor, and can be thought of as compactification of the 2D \(\mathbb{R}\) gauge theory phase realized by the analog surface code.}\\ 
\addlinespace[\myxtraspc]
\eczhRefIndex{code:gkp_concatenated}%
\eczhListValue{\flmRefsHyperref{code:gkp_concatenated}{Concatenated GKP code}} & \eczhListValue{A concatenated code whose outer code is a GKP code. In other words, a bosonic code that can be thought of as a concatenation of an arbitrary inner code and another bosonic outer code. Most examples encode physical qubits of an inner stabilizer code into the square-lattice GKP code.}\\ 
\addlinespace[\myxtraspc]
\eczhRefIndex{code:gkp-cluster-state}%
\eczhListValue{\flmRefsHyperref{code:gkp-cluster-state}{GKP CV-cluster-state code}} & \eczhListValue{A cluster-state code that utilizes a generalized analog cluster state with some of its physical modes initialized in GKP (resource) states.
Alternatively, it can be thought of as a multimode GKP code whose encoding consists of initializing \(k\) modes in momentum states (or, in the normalizable case, squeezed vacua), \(n-k\) modes in (normalizable) GKP states, and applying a Gaussian circuit consisting of two-body gates \(e^{i V_{jk} \hat{x}_j \hat{x}_k }\) for some angles \(V_{jk}\).
The code provides a way to perform fault-tolerant MBQC, with the required number \(n-k\) of GKP-encoded physical modes determined by the particular protocol \NoCaseChange{\protect\cite{cite508,cite509,cite415,cite510}}.}\\ 
\addlinespace[\myxtraspc]
\eczhRefIndex{code:gkp_surface_concatenated}%
\eczhListValue{\flmRefsHyperref{code:gkp_surface_concatenated}{GKP-surface code}} & \eczhListValue{A concatenated code whose outer code is a GKP code and whose inner code is a surface code, including toric surface-code variants \NoCaseChange{\protect\cite{cite415,cite416}}, rotated surface codes \NoCaseChange{\protect\cite{cite417,cite418,cite419,cite420}}, and XZZX surface codes \NoCaseChange{\protect\cite{cite421}}.}\\ 
\addlinespace[\myxtraspc]
\eczhRefIndex{code:multimodegkp}%
\eczhListValue{\flmRefsHyperref{code:multimodegkp}{Gottesman-Kitaev-Preskill (GKP) code}} & \eczhListValue{Quantum lattice code for a non-degenerate lattice, thereby admitting a finite-dimensional logical subspace.
Codes on \(n\) modes can be constructed from lattices with \(2n\)-dimensional full-rank Gram matrices \(A\).
Any GKP code can be generated from a Gram matrix in standard form via a Gaussian unitary transformation \NoCaseChange{\protect\cite[{Corr. 1}]{cite511}}.}\\ 
\addlinespace[\myxtraspc]
\eczhRefIndex{code:hnss}%
\eczhListValue{\flmRefsHyperref{code:hnss}{Hayden-Nezami-Salton-Sanders bosonic code}} & \eczhListValue{An \(\llbracket n,1\rrbracket _{\mathbb{R}}\) analog CSS code defined using homological structures associated with an \(n-1\) simplex. Relevant to the study of spacetime replication of quantum information \NoCaseChange{\protect\cite{cite512}}.}\\ 
\addlinespace[\myxtraspc]
\eczhRefIndex{code:hexagonal_gkp}%
\eczhListValue{\flmRefsHyperref{code:hexagonal_gkp}{Hexagonal GKP code}} & \eczhListValue{Single-mode GKP qudit-into-oscillator code based on the triangular lattice. Offers the best error correction against displacement noise in a single mode due to the optimal packing of the underlying lattice \NoCaseChange{\protect\cite[{Sec. VI}]{cite513}}.}\\ 
\addlinespace[\myxtraspc]
\eczhRefIndex{code:homological_cv}%
\eczhListValue{\flmRefsHyperref{code:homological_cv}{Integer-homology bosonic CSS code}} & \eczhListValue{A bosonic stabilizer code whose physical modes have been restricted, via a single GKP stabilizer, from the space of functions on the real line to the space of periodic functions.
This restriction effectively realizes a rotor on each physical mode, allowing one to construct homological rotor codes out of displacement stabilizer groups.
The stabilizer group is continuous, but contains discrete components in the form of the single-mode GKP stabilizers.
The homology group of the logical operators has a torsion component because the chain complexes are defined over the ring of integers, which yields codes with finite logical dimension.}\\ 
\addlinespace[\myxtraspc]
\eczhRefIndex{code:ntru_gkp}%
\eczhListValue{\flmRefsHyperref{code:ntru_gkp}{NTRU-GKP code}} & \eczhListValue{Multi-mode GKP code whose underlying lattice is utilized in variations of the NTRU cryptosystem \NoCaseChange{\protect\cite{cite288}}.
Randomized constructions yield constant-rate GKP code families whose largest decodable displacement length scales as \(O(\sqrt{n})\) with high probability.}\\ 
\addlinespace[\myxtraspc]
\eczhRefIndex{code:gkp-stabilizer}%
\eczhListValue{\flmRefsHyperref{code:gkp-stabilizer}{Oscillator-into-oscillator GKP code}} & \eczhListValue{Multimode GKP code with an infinite-dimensional logical space. Can be obtained by considering an \(n\)-mode GKP code with a finite-dimensional logical space, removing stabilizers such that the logical space becomes infinite dimensional, and applying a Gaussian circuit.}\\ 
\addlinespace[\myxtraspc]
\eczhRefIndex{code:quantum_lattice}%
\eczhListValue{\flmRefsHyperref{code:quantum_lattice}{Quantum lattice code}} & \eczhListValue{Bosonic stabilizer code on \(n\) bosonic modes whose stabilizer group is an infinite countable group of oscillator displacement operators which implement lattice translations in phase space.}\\ 
\addlinespace[\myxtraspc]
\eczhRefIndex{code:gkp}%
\eczhListValue{\flmRefsHyperref{code:gkp}{Square-lattice GKP code}} & \eczhListValue{Single-mode GKP qudit-into-oscillator CSS code based on the rectangular lattice.
Its stabilizer generators are oscillator displacement operators \(\hat{S}_q(2\alpha)=e^{-2i\alpha \hat{p}}\) and \(\hat{S}_p(2\beta)=e^{2i\beta \hat{x}}\).
To ensure \(\hat{S}_q(2\alpha)\) and \(\hat{S}_p(2\beta)\) generate a stabilizer group that is Abelian, there is a constraint that \(\alpha\beta=2q\pi\) where \(q\) is an integer denoting the logical dimension.}\\ 
\addlinespace[\myxtraspc]
\eczhRefIndex{code:dfour_gkp}%
\eczhListValue{\flmRefsHyperref{code:dfour_gkp}{\(D_4\) hyper-diamond GKP code}} & \eczhListValue{Two-mode GKP qubit-into-oscillator code based on the \(D_4\) hyper-diamond lattice \NoCaseChange{\protect\cite{cite482}}.}\\ 
\addlinespace[\myxtraspc]
\eczhRefIndex{code:chern_simons_gkp}%
\eczhListValue{\flmRefsHyperref{code:chern_simons_gkp}{\(U(1)_{2n} \times U(1)_{-2m}\) Chern-Simons GKP code}} & \eczhListValue{A non-CSS multimode GKP code defined on a 2D mode lattice that encodes a qudit logical space and whose excitations are characterized by the \(U(1)_{2n} \times U(1)_{-2m}\) Chern-Simons theory.
The code can be obtained from the analog surface code by \flmRefsHyperref{ref410}{condensing} certain anyons \NoCaseChange{\protect\cite{cite411}}.}\\ 
\addlinespace[\myxtraspc]
\eczhRefIndex{code:braunstein}%
\eczhListValue{\flmRefsHyperref{code:braunstein}{\(\llbracket 5,1,3\rrbracket _{\mathbb{R}}\) Braunstein five-mode code}} & \eczhListValue{An analog stabilizer version of the five-qubit perfect code, encoding one mode into five and correcting arbitrary errors on any one mode.}\\ 
\addlinespace[\myxtraspc]
\eczhRefIndex{code:lloyd_slotine}%
\eczhListValue{\flmRefsHyperref{code:lloyd_slotine}{\(\llbracket 9,1,3\rrbracket _{\mathbb{R}}\) Lloyd-Slotine code}} & \eczhListValue{An analog stabilizer version of Shor's nine-qubit code, encoding one mode into nine and correcting arbitrary errors on any one mode.}\\ 
\end{tabularx}\endgroup
\eczcodelist{quantum_concatenated}{Concatenated quantum codes
}%

\eczhCodeListAutoDescription{All descendants of \flmRefsCref{code:quantum_concatenated}.}%

\eczhIncludeCodeGraph{Bare}{scale=0.5}{\columnwidth}{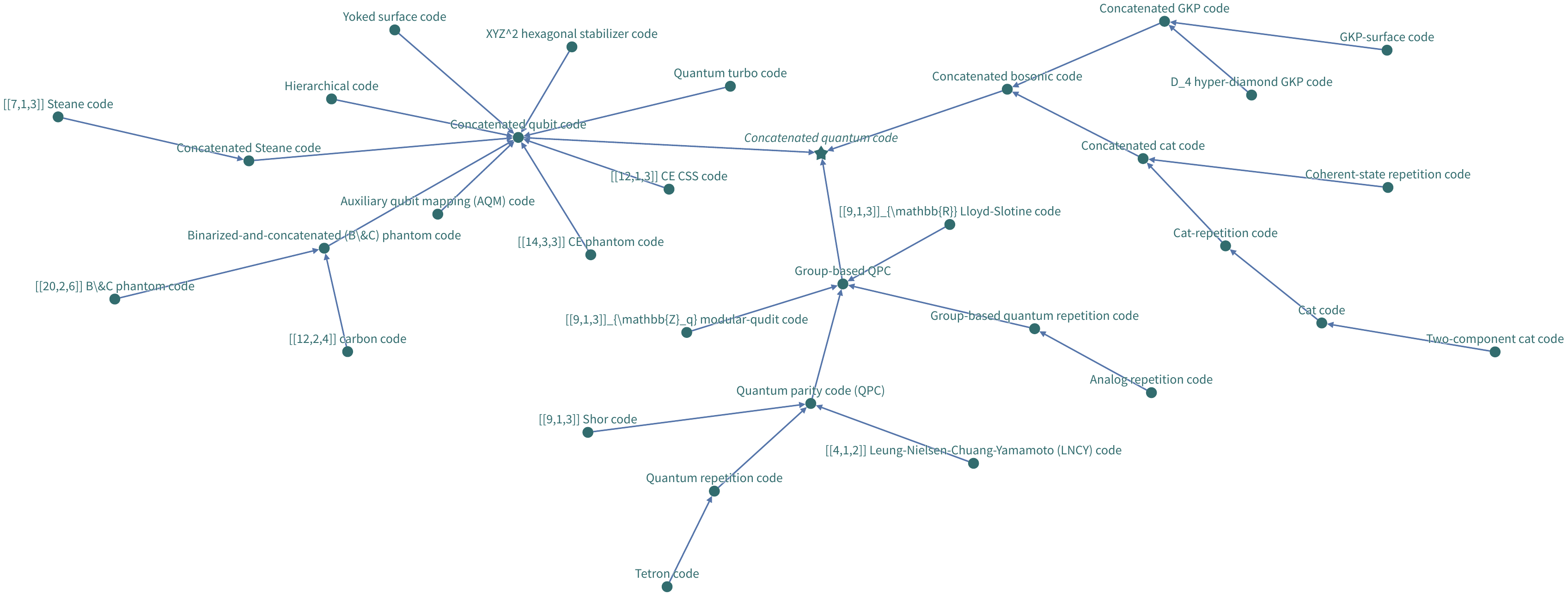}{Concatenated quantum codes}{https://errorcorrectionzoo.org/code_graph#J\%7B\%22displayMode\%22\%3A\%22subset\%22\%2C\%22modeSubsetOptions\%22\%3A\%7B\%22codeIds\%22\%3A\%5B\%22analog_repetition\%22\%2C\%22aqm\%22\%2C\%22bc_phantom\%22\%2C\%22cat\%22\%2C\%22cat_repetition\%22\%2C\%22coherent_state_repetition\%22\%2C\%22gkp_concatenated\%22\%2C\%22concatenated_steane\%22\%2C\%22oscillators_concatenated\%22\%2C\%22cat_concatenated\%22\%2C\%22quantum_concatenated\%22\%2C\%22qubit_concatenated\%22\%2C\%22gkp_surface_concatenated\%22\%2C\%22group_quantum_parity\%22\%2C\%22group_quantum_repetition\%22\%2C\%22hierarchical\%22\%2C\%22quantum_parity\%22\%2C\%22quantum_repetition\%22\%2C\%22quantum_turbo\%22\%2C\%22tetron\%22\%2C\%22two-legged-cat\%22\%2C\%22xyz_hexagonal\%22\%2C\%22yoked_surface\%22\%2C\%22dfour_gkp\%22\%2C\%22css_12_1_3\%22\%2C\%22carbon\%22\%2C\%22phantom_14_3_3\%22\%2C\%22stab_20_2_6\%22\%2C\%22css_4_1_2\%22\%2C\%22steane\%22\%2C\%22shor_nine\%22\%2C\%22lloyd_slotine\%22\%2C\%22stab_9_1_3\%22\%5D\%2C\%22reusePreviousLayoutPositions\%22\%3Afalse\%2C\%22showIntermediateConnectingNodes\%22\%3Atrue\%2C\%22connectingNodesMaxDepth\%22\%3A15\%2C\%22connectingNodesPathMaxLength\%22\%3A20\%2C\%22connectingNodesMaxExtraDepth\%22\%3A3\%2C\%22connectingNodesOnlyKeepPathsWithAdditionalLength\%22\%3A1\%2C\%22connectingNodesToDomainsAndKingdoms\%22\%3Afalse\%2C\%22connectingNodesEdgeLengthsByType\%22\%3A\%7B\%22primaryParent\%22\%3A1\%2C\%22secondaryParent\%22\%3A4\%2C\%22cousin\%22\%3A6\%7D\%2C\%22nodeIds\%22\%3A\%5B\%5D\%7D\%2C\%22highlightImportantNodes\%22\%3A\%7B\%22highlightImportantNodes\%22\%3Afalse\%2C\%22highlightPrimaryParents\%22\%3Afalse\%2C\%22highlightRootConnectingEdges\%22\%3Afalse\%7D\%7D}

\begingroup
\small
\eczhBreakableDashes
\renewcommand\arraystretch{1.05}
\edef\myxtraspc{\eczListAddVSpaceXtraXtraStretch}
\begin{tabularx}{\linewidth}{>{\raggedright\arraybackslash}p{\eczListColWidth{name}} >{\hsize=1.0000\hsize }X}
\toprule
\eczListColTitle{Code} & \eczListColTitle{Description} \\
\midrule
\endfirsthead
\toprule
\eczListColTitleContinued{Code} & \eczListColTitleContinued{Description} \\
\midrule
\endhead
\bottomrule
\endfoot
\eczhRefIndex{code:analog_repetition}%
\eczhListValue{\flmRefsHyperref{code:analog_repetition}{Analog repetition code}} & \eczhListValue{An \(\llbracket n,1\rrbracket _{\mathbb{R}}\) analog stabilizer version of the quantum repetition code, encoding the position states of one mode into an odd number \(n\) of modes.}\\ 
\addlinespace[\myxtraspc]
\eczhRefIndex{code:aqm}%
\eczhListValue{\flmRefsHyperref{code:aqm}{Auxiliary qubit mapping (AQM) code}} & \eczhListValue{A concatenation of the JW transformation code with a qubit stabilizer code.}\\ 
\addlinespace[\myxtraspc]
\eczhRefIndex{code:bc_phantom}%
\eczhListValue{\flmRefsHyperref{code:bc_phantom}{Binarized-and-concatenated (B\&C) phantom code}} & \eczhListValue{Member of a family of \(k=2\) CSS phantom codes obtained from a \(q=4\) Galois-qudit CSS code by binarizing each \(\mathbb{F}_4\) qudit into two qubits and then concatenating each qubit pair with the \(\llbracket 4,2,2\rrbracket \) code \NoCaseChange{\protect\cite{cite514}}.}\\ 
\addlinespace[\myxtraspc]
\eczhRefIndex{code:cat}%
\eczhListValue{\flmRefsHyperref{code:cat}{Cat code}} & \eczhListValue{Rotation-symmetric bosonic Fock-state code encoding a \(q\)-dimensional qudit into one oscillator which utilizes a constellation of \(q(S+1)\) coherent states distributed equidistantly around a circle in phase space of radius \(\alpha\).}\\ 
\addlinespace[\myxtraspc]
\eczhRefIndex{code:cat_repetition}%
\eczhListValue{\flmRefsHyperref{code:cat_repetition}{Cat-repetition code}} & \eczhListValue{A concatenated qubit-into-\(n\)-mode code obtained by encoding each qubit of a quantum repetition code into a two-component cat code in its cat-state basis.}\\ 
\addlinespace[\myxtraspc]
\eczhRefIndex{code:coherent_state_repetition}%
\eczhListValue{\flmRefsHyperref{code:coherent_state_repetition}{Coherent-state repetition code}} & \eczhListValue{A concatenated qubit-into-\(n\)-mode code (for odd \(n\)) whose inner code is a quantum repetition code and whose outer code is the two-component cat code in its coherent-state basis.}\\ 
\addlinespace[\myxtraspc]
\eczhRefIndex{code:gkp_concatenated}%
\eczhListValue{\flmRefsHyperref{code:gkp_concatenated}{Concatenated GKP code}} & \eczhListValue{A concatenated code whose outer code is a GKP code. In other words, a bosonic code that can be thought of as a concatenation of an arbitrary inner code and another bosonic outer code. Most examples encode physical qubits of an inner stabilizer code into the square-lattice GKP code.}\\ 
\addlinespace[\myxtraspc]
\eczhRefIndex{code:concatenated_steane}%
\eczhListValue{\flmRefsHyperref{code:concatenated_steane}{Concatenated Steane code}} & \eczhListValue{A member of the family of \(\llbracket 7^m,1,3^m\rrbracket \) CSS codes, each of which is a recursive level-\(m\) concatenation of the Steane code.
This family is one of the first to admit a \flmRefsHyperref{ref515}{concatenated threshold} \NoCaseChange{\protect\cite{cite516,cite517,cite518,cite519,cite520}}.}\\ 
\addlinespace[\myxtraspc]
\eczhRefIndex{code:oscillators_concatenated}%
\eczhListValue{\flmRefsHyperref{code:oscillators_concatenated}{Concatenated bosonic code}} & \eczhListValue{A concatenated code whose outer code is a bosonic code. In other words, a bosonic code that can be thought of as a concatenation of a possibly non-bosonic inner code and a bosonic outer code.}\\ 
\addlinespace[\myxtraspc]
\eczhRefIndex{code:cat_concatenated}%
\eczhListValue{\flmRefsHyperref{code:cat_concatenated}{Concatenated cat code}} & \eczhListValue{A concatenated code obtained by encoding the physical qubits of an inner qubit code into cat-code states.
Most examples concatenate a qubit stabilizer code with the two-component cat code in its cat-state basis.}\\ 
\addlinespace[\myxtraspc]
\eczhRefIndex{code:quantum_concatenated}%
\eczhListValue{\flmRefsHyperref{code:quantum_concatenated}{Concatenated quantum code}} & \eczhListValue{A combination of two quantum codes, an inner code \(C_{\text{in}}\) and an outer code \(C_{\text{out}}\), where the physical subspace used for the inner code consists of the logical subspace of the outer code.
In other words, one first encodes in the inner code, and then encodes each of its physical registers in the outer code.
An inner \(C_{\text{in}} = \llparenthesis n_1,K,d_1\rrparenthesis _{q_1}\) and outer \(C_{\text{out}} = \llparenthesis n_2,q_1,d_2\rrparenthesis _{q_2}\) block quantum code yield an \(\llparenthesis n_1 n_2, K, d \geq d_1d_2\rrparenthesis _{q_2}\) concatenated block quantum code \NoCaseChange{\protect\cite{cite398}}.}\\ 
\addlinespace[\myxtraspc]
\eczhRefIndex{code:qubit_concatenated}%
\eczhListValue{\flmRefsHyperref{code:qubit_concatenated}{Concatenated qubit code}} & \eczhListValue{A concatenated code whose outer code is a qubit code. In other words, a qubit code that can be thought of as a concatenation of an inner qubit code and an outer qubit code.
An inner \(C_{\text{in}} = \llparenthesis n_1,K,d_1\rrparenthesis \) and outer \(C_{\text{out}} = \llparenthesis n_2,2,d_2\rrparenthesis \) qubit code yield an \(\llparenthesis n_1 n_2, K, d \geq d_1d_2\rrparenthesis \) concatenated qubit code.}\\ 
\addlinespace[\myxtraspc]
\eczhRefIndex{code:gkp_surface_concatenated}%
\eczhListValue{\flmRefsHyperref{code:gkp_surface_concatenated}{GKP-surface code}} & \eczhListValue{A concatenated code whose outer code is a GKP code and whose inner code is a surface code, including toric surface-code variants \NoCaseChange{\protect\cite{cite415,cite416}}, rotated surface codes \NoCaseChange{\protect\cite{cite417,cite418,cite419,cite420}}, and XZZX surface codes \NoCaseChange{\protect\cite{cite421}}.}\\ 
\addlinespace[\myxtraspc]
\eczhRefIndex{code:group_quantum_parity}%
\eczhListValue{\flmRefsHyperref{code:group_quantum_parity}{Group-based QPC}} & \eczhListValue{An \(\llbracket m r,1,\min(m,r)\rrbracket _G\) generalization of the QPC.}\\ 
\addlinespace[\myxtraspc]
\eczhRefIndex{code:group_quantum_repetition}%
\eczhListValue{\flmRefsHyperref{code:group_quantum_repetition}{Group-based quantum repetition code}} & \eczhListValue{An \(\llbracket n,1\rrbracket _G\) generalization of the quantum repetition code.}\\ 
\addlinespace[\myxtraspc]
\eczhRefIndex{code:hierarchical}%
\eczhListValue{\flmRefsHyperref{code:hierarchical}{Hierarchical code}} & \eczhListValue{Member of a family of \(\llbracket n,k,d\rrbracket \) qubit stabilizer codes resulting from a concatenation of a constant-rate \flmRefsHyperref{code:qldpc}{QLDPC code} with a \flmRefsHyperref{code:rotated_surface}{rotated surface code}.
Concatenation allows for syndrome extraction to be performed on a 2D geometry while maintaining a threshold at the expense of a logarithmically vanishing rate.
The growing syndrome extraction circuit depth allows known bounds in the literature to be weakened \NoCaseChange{\protect\cite{cite521,cite522}}.}\\ 
\addlinespace[\myxtraspc]
\eczhRefIndex{code:quantum_parity}%
\eczhListValue{\flmRefsHyperref{code:quantum_parity}{Quantum parity code (QPC)}} & \eczhListValue{A \(\llbracket m_1 m_2,1,\min(m_1,m_2)\rrbracket \) CSS code family obtained from concatenating an \(m_1\)-qubit bit-flip repetition code with an \(m_2\)-qubit phase-flip repetition code.}\\ 
\addlinespace[\myxtraspc]
\eczhRefIndex{code:quantum_repetition}%
\eczhListValue{\flmRefsHyperref{code:quantum_repetition}{Quantum repetition code}} & \eczhListValue{Encodes \(1\) qubit into \(n\) qubits according to \(|0\rangle\to|\phi_0\rangle^{\otimes n}\) and \(|1\rangle\to|\phi_1\rangle^{\otimes n}\). The code is called a \textit{bit-flip} code when \(|\phi_i\rangle = |i\rangle\), and a \textit{phase-flip} code when \(|\phi_0\rangle = |+\rangle\) and \(|\phi_1\rangle = |-\rangle\).
This repetition-style encoding does not clone an arbitrary quantum state; instead, it extends the copying of computational-basis states linearly to entangled codewords  \NoCaseChange{\protect\cite[{Ch. 2}]{cite398}}.}\\ 
\addlinespace[\myxtraspc]
\eczhRefIndex{code:quantum_turbo}%
\eczhListValue{\flmRefsHyperref{code:quantum_turbo}{Quantum turbo code}} & \eczhListValue{A quantum version of the turbo code, obtained from an interleaved serial quantum concatenation \NoCaseChange{\protect\cite[{Def. 30}]{cite399}} of quantum convolutional codes.
The interleaver induces long-range entanglement and can increase the minimum distance relative to the constituent convolutional codes \NoCaseChange{\protect\cite{cite400}}.}\\ 
\addlinespace[\myxtraspc]
\eczhRefIndex{code:tetron}%
\eczhListValue{\flmRefsHyperref{code:tetron}{Tetron code}} & \eczhListValue{A \(\llbracket 2,1,2\rrbracket _{f}\) Majorana box qubit encoding a logical qubit into four Majorana modes, equivalently into the fixed-total-parity sector of two physical fermionic modes.
Four Majorana zero modes are the smallest aggregate that supports a qubit in a fixed fermion-parity sector \NoCaseChange{\protect\cite{cite401}}.
This code can be concatenated with various qubit codes such as surface codes and color codes.
Four-boundary Majorana surface-code patches are logical tetrons, i.e., higher-distance analogues of this physical tetron block \NoCaseChange{\protect\cite{cite402}}.}\\ 
\addlinespace[\myxtraspc]
\eczhRefIndex{code:two-legged-cat}%
\eczhListValue{\flmRefsHyperref{code:two-legged-cat}{Two-component cat code}} & \eczhListValue{Code whose codespace is spanned by two coherent states \(\left|\pm\alpha\right\rangle\) for nonzero complex \(\alpha\).}\\ 
\addlinespace[\myxtraspc]
\eczhRefIndex{code:xyz_hexagonal}%
\eczhListValue{\flmRefsHyperref{code:xyz_hexagonal}{XYZ\(^2\) hexagonal stabilizer code}} & \eczhListValue{An instance of the matching code based on the Kitaev honeycomb model. It is described on a honeycomb tiling with \(XYZXYZ\) stabilizers on each hexagonal plaquette. Each vertical pair of qubits has an \(XX\), \(YY\), or \(ZZ\) link stabilizer depending on the orientation of the plaquette stabilizers.}\\ 
\addlinespace[\myxtraspc]
\eczhRefIndex{code:yoked_surface}%
\eczhListValue{\flmRefsHyperref{code:yoked_surface}{Yoked surface code}} & \eczhListValue{Member of a family of \(\llbracket n,k,d\rrbracket \) qubit CSS codes resulting from a concatenation of a \flmRefsHyperref{code:qmdpc}{QMDPC code} with a \flmRefsHyperref{code:rotated_surface}{rotated surface code}.
Concatenation does not impose additional connectivity constraints and can triple the number of logical qubits per physical qubit when compared to the original surface code.
Concatenation with 1D (2D) QMDPC yields codes with twice (four times) the distance.
Using the concatenation convention of the Zoo, the stabilizer generators of the inner QMDPC code are referred to as \textit{yokes} in this context; the cited paper \NoCaseChange{\protect\cite{cite523}} uses the opposite inner/outer terminology.}\\ 
\addlinespace[\myxtraspc]
\eczhRefIndex{code:dfour_gkp}%
\eczhListValue{\flmRefsHyperref{code:dfour_gkp}{\(D_4\) hyper-diamond GKP code}} & \eczhListValue{Two-mode GKP qubit-into-oscillator code based on the \(D_4\) hyper-diamond lattice \NoCaseChange{\protect\cite{cite482}}.}\\ 
\addlinespace[\myxtraspc]
\eczhRefIndex{code:css_12_1_3}%
\eczhListValue{\flmRefsHyperref{code:css_12_1_3}{\(\llbracket 12,1,3\rrbracket \) CE CSS code}} & \eczhListValue{Twelve-qubit constant-excitation (CE) CSS code that encodes one logical qubit with distance three.
It is the smallest CE CSS code that corrects a single-qubit error \NoCaseChange{\protect\cite{cite524}}.
Codewords lie in a fixed Hamming-weight subspace, making the code immune to coherent noise in the form of transversal \(Z\)-rotations.}\\ 
\addlinespace[\myxtraspc]
\eczhRefIndex{code:carbon}%
\eczhListValue{\flmRefsHyperref{code:carbon}{\(\llbracket 12,2,4\rrbracket \) carbon code}} & \eczhListValue{Twelve-qubit CSS code based on Knill's \(C_4/C_6\) scheme \NoCaseChange{\protect\cite{cite525}}.
Using the concatenation convention of the Zoo, the carbon code can be viewed as a block concatenation with inner code \(\llbracket 4,2,2\rrbracket \) and outer code \(C_6\): three inner \(\llbracket 4,2,2\rrbracket \) blocks encode six intermediate qubits, which are then encoded into two logical qubits by the outer \(\llbracket 6,2,2\rrbracket \) code.}\\ 
\addlinespace[\myxtraspc]
\eczhRefIndex{code:phantom_14_3_3}%
\eczhListValue{\flmRefsHyperref{code:phantom_14_3_3}{\(\llbracket 14,3,3\rrbracket \) CE phantom code}} & \eczhListValue{CSS phantom code obtained by concatenating the \(\llbracket 7,3,(d_X=3,d_Z=2)\rrbracket \) punctured hypercube code with the two-qubit phase-flip repetition code.
The code is equivalent to the \(\llbracket 14,3,3\rrbracket \) constant-excitation (CE) CSS code obtained by applying dual-rail concatenation to the \(\llbracket 7,3,2\rrbracket \) punctured hypercube code, up to single-qubit Clifford gates, a physical-qubit permutation, and a Pauli frame \NoCaseChange{\protect\cite{cite524}}.}\\ 
\addlinespace[\myxtraspc]
\eczhRefIndex{code:stab_20_2_6}%
\eczhListValue{\flmRefsHyperref{code:stab_20_2_6}{\(\llbracket 20,2,6\rrbracket \) B\&C phantom code}} & \eczhListValue{Self-dual CSS code on 20 physical qubits encoding two logical qubits with distance 6.
The code is obtained by binarizing the \(\llbracket 5,1,3\rrbracket _4\) code in the self-dual normal basis \(\{\omega,\omega^2\}\) to the \(\llbracket 10,2,3\rrbracket \) binarized Galois-qudit code and then concatenating each qubit pair with the \(\llbracket 4,2,2\rrbracket \) code \NoCaseChange{\protect\cite{cite514}}.}\\ 
\addlinespace[\myxtraspc]
\eczhRefIndex{code:css_4_1_2}%
\eczhListValue{\flmRefsHyperref{code:css_4_1_2}{\(\llbracket 4,1,2\rrbracket \) Leung-Nielsen-Chuang-Yamamoto (LNCY) code}} & \eczhListValue{A four-qubit CSS stabilizer code that is the only qubit CSS code with such parameters.}\\ 
\addlinespace[\myxtraspc]
\eczhRefIndex{code:steane}%
\eczhListValue{\flmRefsHyperref{code:steane}{\(\llbracket 7,1,3\rrbracket \) Steane code}} & \eczhListValue{A \(\llbracket 7,1,3\rrbracket \) self-dual CSS code that is the smallest qubit CSS code to correct a single-qubit error \NoCaseChange{\protect\cite{cite451}}.
The code is constructed using the classical binary \([7,4,3]\) Hamming code for protecting against both \(X\) and \(Z\) errors.}\\ 
\addlinespace[\myxtraspc]
\eczhRefIndex{code:shor_nine}%
\eczhListValue{\flmRefsHyperref{code:shor_nine}{\(\llbracket 9,1,3\rrbracket \) Shor code}} & \eczhListValue{Nine-qubit \flmRefsHyperref{code:css}{CSS code} that is the first quantum error-correcting code \NoCaseChange{\protect\cite[{ID 8802}]{cite453}}.
Among indecomposable \(\llbracket 9,1,3\rrbracket \) CSS codes, the Shor code has the largest automorphism group \NoCaseChange{\protect\cite{cite454}}.}\\ 
\addlinespace[\myxtraspc]
\eczhRefIndex{code:lloyd_slotine}%
\eczhListValue{\flmRefsHyperref{code:lloyd_slotine}{\(\llbracket 9,1,3\rrbracket _{\mathbb{R}}\) Lloyd-Slotine code}} & \eczhListValue{An analog stabilizer version of Shor's nine-qubit code, encoding one mode into nine and correcting arbitrary errors on any one mode.}\\ 
\addlinespace[\myxtraspc]
\eczhRefIndex{code:stab_9_1_3}%
\eczhListValue{\flmRefsHyperref{code:stab_9_1_3}{\(\llbracket 9,1,3\rrbracket _{\mathbb{Z}_q}\) modular-qudit code}} & \eczhListValue{Modular-qudit CSS code that generalizes the \(\llbracket 9,1,3\rrbracket \) Shor code to \(q\)-level systems.}\\ 
\end{tabularx}\endgroup
\eczcodelist{constant_excitation}{Constant-excitation quantum codes
}%

\eczhCodeListAutoDescription{All descendants of \flmRefsCref{code:constant_excitation}.}%

\eczhIncludeCodeGraph{Bare}{scale=0.5}{\columnwidth}{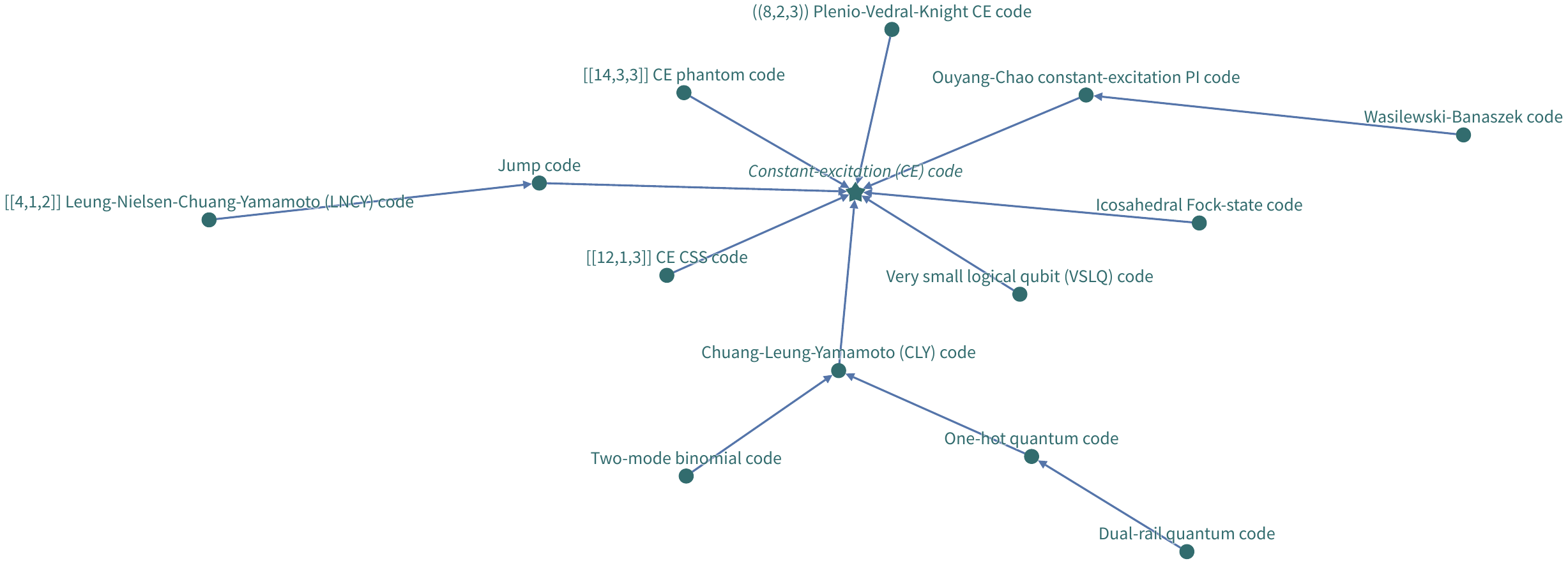}{Constant-excitation quantum codes}{https://errorcorrectionzoo.org/code_graph#J\%7B\%22displayMode\%22\%3A\%22subset\%22\%2C\%22modeSubsetOptions\%22\%3A\%7B\%22codeIds\%22\%3A\%5B\%22chuang-leung-yamamoto\%22\%2C\%22constant_excitation\%22\%2C\%22dual_rail\%22\%2C\%22icosahedral_fock\%22\%2C\%22jump\%22\%2C\%22one_hot_quantum\%22\%2C\%22constant_excitation_permutation_invariant\%22\%2C\%22two-mode_binomial\%22\%2C\%22very-small-logical-qubit\%22\%2C\%22wasilewski-banaszek\%22\%2C\%22qubit_8_1_3\%22\%2C\%22css_12_1_3\%22\%2C\%22phantom_14_3_3\%22\%2C\%22css_4_1_2\%22\%5D\%2C\%22reusePreviousLayoutPositions\%22\%3Afalse\%2C\%22showIntermediateConnectingNodes\%22\%3Atrue\%2C\%22connectingNodesMaxDepth\%22\%3A15\%2C\%22connectingNodesPathMaxLength\%22\%3A20\%2C\%22connectingNodesMaxExtraDepth\%22\%3A3\%2C\%22connectingNodesOnlyKeepPathsWithAdditionalLength\%22\%3A1\%2C\%22connectingNodesToDomainsAndKingdoms\%22\%3Afalse\%2C\%22connectingNodesEdgeLengthsByType\%22\%3A\%7B\%22primaryParent\%22\%3A1\%2C\%22secondaryParent\%22\%3A4\%2C\%22cousin\%22\%3A6\%7D\%2C\%22nodeIds\%22\%3A\%5B\%5D\%7D\%2C\%22highlightImportantNodes\%22\%3A\%7B\%22highlightImportantNodes\%22\%3Afalse\%2C\%22highlightPrimaryParents\%22\%3Afalse\%2C\%22highlightRootConnectingEdges\%22\%3Afalse\%7D\%7D}

\begingroup
\small
\eczhBreakableDashes
\renewcommand\arraystretch{1.05}
\edef\myxtraspc{\eczListAddVSpaceXtraXtraStretch}
\begin{tabularx}{\linewidth}{>{\raggedright\arraybackslash}p{\eczListColWidth{name}} >{\hsize=1.0000\hsize }X}
\toprule
\eczListColTitle{Code} & \eczListColTitle{Description} \\
\midrule
\endfirsthead
\toprule
\eczListColTitleContinued{Code} & \eczListColTitleContinued{Description} \\
\midrule
\endhead
\bottomrule
\endfoot
\eczhRefIndex{code:chuang-leung-yamamoto}%
\eczhListValue{\flmRefsHyperref{code:chuang-leung-yamamoto}{Chuang-Leung-Yamamoto (CLY) code}} & \eczhListValue{Bosonic Fock-state code that encodes \(k\) qubits into \(n\) oscillators, with each oscillator restricted to having at most \(N\) excitations. Codewords are superpositions of oscillator Fock states which have exactly \(N\) total excitations, and are either uniform (i.e., balanced) superpositions or unbalanced superpositions.}\\ 
\addlinespace[\myxtraspc]
\eczhRefIndex{code:constant_excitation}%
\eczhListValue{\flmRefsHyperref{code:constant_excitation}{Constant-excitation (CE) code}} & \eczhListValue{Code whose codewords lie in an eigenspace of fixed total energy or fixed total excitation number for the underlying quantum system.
For qubit codes, such a Hamiltonian is often the \textit{total spin Hamiltonian}, \(H=\sum_i Z_i\).
For spin-\(S\) codes, this generalizes to \(H=\sum_i J_z^{(i)}\), where \(J_z\) is the spin-\(S\) \(Z\)-operator.
For bosonic (and, similarly, for fermion) codes, such as Fock-state codes, codewords are often in an eigenspace with eigenvalue \(N>0\) of the \textit{total excitation} or \textit{energy Hamiltonian}, \(H=\sum_i \hat{n}_i\).}\\ 
\addlinespace[\myxtraspc]
\eczhRefIndex{code:dual_rail}%
\eczhListValue{\flmRefsHyperref{code:dual_rail}{Dual-rail quantum code}} & \eczhListValue{Two-mode bosonic code encoding a logical qubit in Fock states with one excitation.
The logical-zero state is represented by \(|10\rangle\), while the logical-one state is represented by \(|01\rangle\).
This encoding is often realized in temporal or spatial modes, corresponding to a \textit{time-bin} or \textit{frequency-bin} encoding.
Two different types of photon polarization can also be used.}\\ 
\addlinespace[\myxtraspc]
\eczhRefIndex{code:icosahedral_fock}%
\eczhListValue{\flmRefsHyperref{code:icosahedral_fock}{Icosahedral Fock-state code}} & \eczhListValue{A constant-excitation Fock-state code designed to realize the \(2I\) group of gates using Gaussian rotations.
It is obtained from the corresponding icosahedral spin code via the \flmRefsHyperref{ref499}{simplex mapping} between spin and constant-excitation Fock spaces \NoCaseChange{\protect\cite{cite500}}.}\\ 
\addlinespace[\myxtraspc]
\eczhRefIndex{code:jump}%
\eczhListValue{\flmRefsHyperref{code:jump}{Jump code}} & \eczhListValue{A CE code designed to detect and correct \flmRefsHyperref{ref498}{AD} errors.
An \(\llparenthesis n,K\rrparenthesis \) jump code is denoted as \(\llparenthesis n,K,t\rrparenthesis _w\) (which conflicts with modular-qudit notation), where \(t\) is the maximum number of qubits that can be corrected after each one has undergone a jump error \(|0\rangle\langle 1|\), and where each codeword is a uniform superposition of qubit basis states with Hamming weight \(w\).}\\ 
\addlinespace[\myxtraspc]
\eczhRefIndex{code:one_hot_quantum}%
\eczhListValue{\flmRefsHyperref{code:one_hot_quantum}{One-hot quantum code}} & \eczhListValue{Encoding of a \(q\)-dimensional qudit into the single-excitation subspace of \(q\) modes. The \(j\)th logical state is the multi-mode Fock state with one photon in mode \(j\) and zero photons in the other modes.
This code is useful for encoding and performing operations on qudits in multiple modes \NoCaseChange{\protect\cite{cite502,cite503,cite504,cite505,cite506}}.}\\ 
\addlinespace[\myxtraspc]
\eczhRefIndex{code:constant_excitation_permutation_invariant}%
\eczhListValue{\flmRefsHyperref{code:constant_excitation_permutation_invariant}{Ouyang-Chao constant-excitation PI code}} & \eczhListValue{A constant-excitation PI Fock-state code whose construction is based on integer partitions.}\\ 
\addlinespace[\myxtraspc]
\eczhRefIndex{code:two-mode_binomial}%
\eczhListValue{\flmRefsHyperref{code:two-mode_binomial}{Two-mode binomial code}} & \eczhListValue{Two-mode constant-energy CLY code whose coefficients are square-roots of binomial coefficients.}\\ 
\addlinespace[\myxtraspc]
\eczhRefIndex{code:very-small-logical-qubit}%
\eczhListValue{\flmRefsHyperref{code:very-small-logical-qubit}{Very small logical qubit (VSLQ) code}} & \eczhListValue{A code consisting of two logical codewords \(|\pm\rangle \propto (|0\rangle\pm|2\rangle)(|0\rangle\pm|2\rangle)\), where the total Hilbert space is the tensor product of two transmon qudits (whose ground states \(|0\rangle\) and second excited states \(|2\rangle\) are used in the codewords).
Since the code is intended to protect against losses, the qutrits can equivalently be thought of as oscillator Fock-state subspaces.}\\ 
\addlinespace[\myxtraspc]
\eczhRefIndex{code:wasilewski-banaszek}%
\eczhListValue{\flmRefsHyperref{code:wasilewski-banaszek}{Wasilewski-Banaszek code}} & \eczhListValue{Three-oscillator constant-excitation Fock-state code encoding a single logical qubit.}\\ 
\addlinespace[\myxtraspc]
\eczhRefIndex{code:qubit_8_1_3}%
\eczhListValue{\flmRefsHyperref{code:qubit_8_1_3}{\(\llparenthesis 8,2,3\rrparenthesis \) Plenio-Vedral-Knight CE code}} & \eczhListValue{An eight-qubit single-error-correcting code that is the first CE code.
Each logical state is a superposition of computational basis states with four excitations.}\\ 
\addlinespace[\myxtraspc]
\eczhRefIndex{code:css_12_1_3}%
\eczhListValue{\flmRefsHyperref{code:css_12_1_3}{\(\llbracket 12,1,3\rrbracket \) CE CSS code}} & \eczhListValue{Twelve-qubit constant-excitation (CE) CSS code that encodes one logical qubit with distance three.
It is the smallest CE CSS code that corrects a single-qubit error \NoCaseChange{\protect\cite{cite524}}.
Codewords lie in a fixed Hamming-weight subspace, making the code immune to coherent noise in the form of transversal \(Z\)-rotations.}\\ 
\addlinespace[\myxtraspc]
\eczhRefIndex{code:phantom_14_3_3}%
\eczhListValue{\flmRefsHyperref{code:phantom_14_3_3}{\(\llbracket 14,3,3\rrbracket \) CE phantom code}} & \eczhListValue{CSS phantom code obtained by concatenating the \(\llbracket 7,3,(d_X=3,d_Z=2)\rrbracket \) punctured hypercube code with the two-qubit phase-flip repetition code.
The code is equivalent to the \(\llbracket 14,3,3\rrbracket \) constant-excitation (CE) CSS code obtained by applying dual-rail concatenation to the \(\llbracket 7,3,2\rrbracket \) punctured hypercube code, up to single-qubit Clifford gates, a physical-qubit permutation, and a Pauli frame \NoCaseChange{\protect\cite{cite524}}.}\\ 
\addlinespace[\myxtraspc]
\eczhRefIndex{code:css_4_1_2}%
\eczhListValue{\flmRefsHyperref{code:css_4_1_2}{\(\llbracket 4,1,2\rrbracket \) Leung-Nielsen-Chuang-Yamamoto (LNCY) code}} & \eczhListValue{A four-qubit CSS stabilizer code that is the only qubit CSS code with such parameters.}\\ 
\end{tabularx}\endgroup
\eczcodelist{quantum_cyclic}{Cyclic quantum codes
}%

\eczhCodeListAutoDescription{All descendants of \flmRefsCref{code:quantum_cyclic}.}%

\eczhIncludeCodeGraph{Bare}{scale=0.5}{\columnwidth}{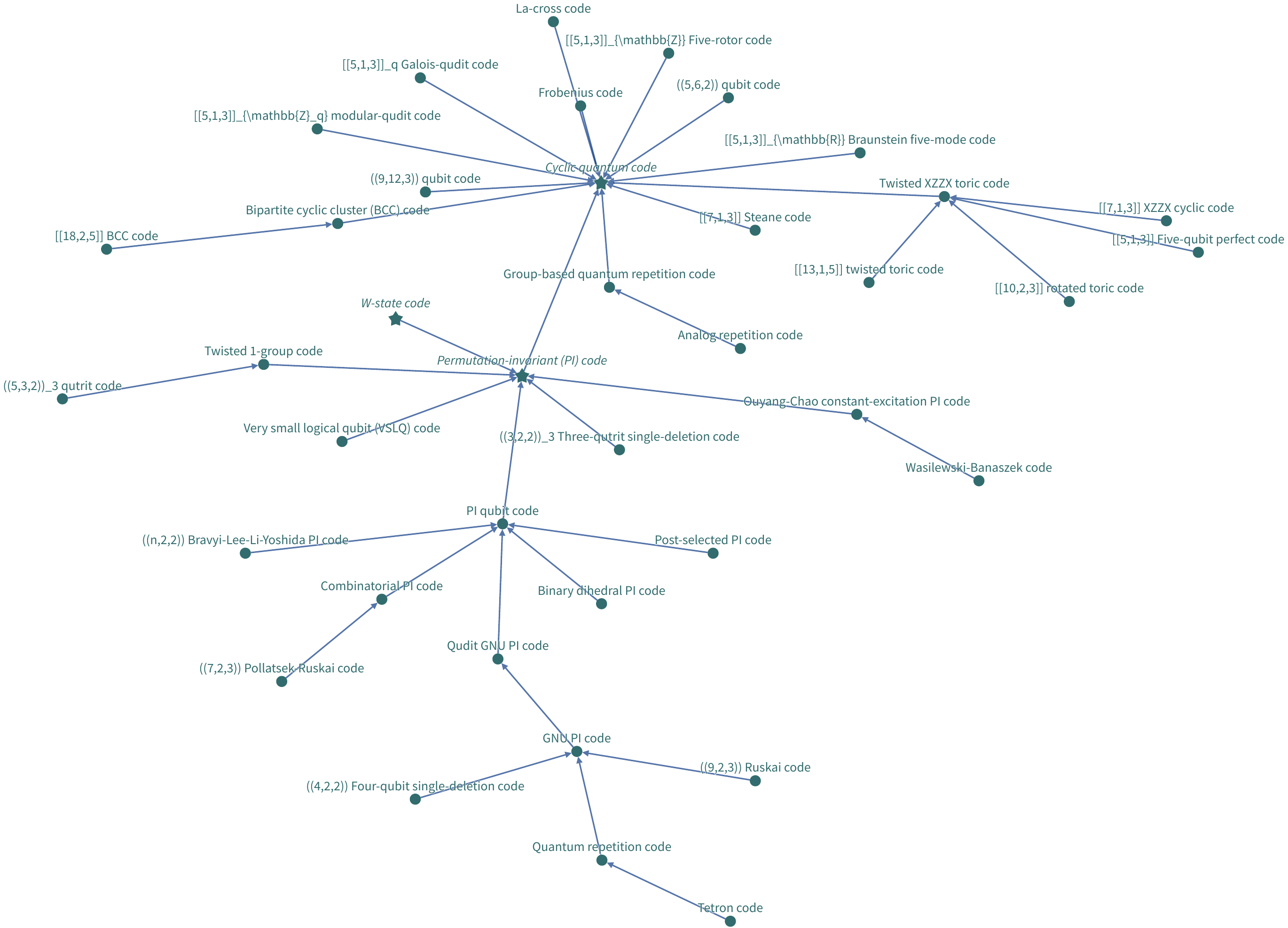}{Cyclic quantum codes}{https://errorcorrectionzoo.org/code_graph#J\%7B\%22displayMode\%22\%3A\%22subset\%22\%2C\%22modeSubsetOptions\%22\%3A\%7B\%22codeIds\%22\%3A\%5B\%22analog_repetition\%22\%2C\%22binary_dihedral_permutation_invariant\%22\%2C\%22bipartite_cyclic_cluster\%22\%2C\%22combinatorial_permutation_invariant\%22\%2C\%22quantum_cyclic\%22\%2C\%22frobenius\%22\%2C\%22gnu_permutation_invariant\%22\%2C\%22group_quantum_repetition\%22\%2C\%22lacross\%22\%2C\%22constant_excitation_permutation_invariant\%22\%2C\%22qubit_permutation_invariant\%22\%2C\%22permutation_invariant\%22\%2C\%22ampdamp_post_selected\%22\%2C\%22quantum_repetition\%22\%2C\%22qudit_gnu_permutation_invariant\%22\%2C\%22tetron\%22\%2C\%22twisted_xzzx\%22\%2C\%22t_group\%22\%2C\%22very-small-logical-qubit\%22\%2C\%22w_state\%22\%2C\%22wasilewski-banaszek\%22\%2C\%22three_qutrit_permutation_invariant\%22\%2C\%22four_qubit_permutation_invariant\%22\%2C\%22su3_sigma360\%22\%2C\%22qubit_5_6_2\%22\%2C\%22icosahedral_permutation_invariant\%22\%2C\%22qubit_9_12_3\%22\%2C\%22ruskai\%22\%2C\%22unentangled_permutation_invariant\%22\%2C\%22xzzx_10_2_3\%22\%2C\%22stab_13_1_5\%22\%2C\%22stab_18_2_5\%22\%2C\%22stab_5_1_3\%22\%2C\%22galois_5_1_3\%22\%2C\%22braunstein\%22\%2C\%22qudit_5_1_3\%22\%2C\%22rotor_5_1_3\%22\%2C\%22steane\%22\%2C\%22xzzx_7_1_3\%22\%5D\%2C\%22reusePreviousLayoutPositions\%22\%3Afalse\%2C\%22showIntermediateConnectingNodes\%22\%3Atrue\%2C\%22connectingNodesMaxDepth\%22\%3A15\%2C\%22connectingNodesPathMaxLength\%22\%3A20\%2C\%22connectingNodesMaxExtraDepth\%22\%3A3\%2C\%22connectingNodesOnlyKeepPathsWithAdditionalLength\%22\%3A1\%2C\%22connectingNodesToDomainsAndKingdoms\%22\%3Afalse\%2C\%22connectingNodesEdgeLengthsByType\%22\%3A\%7B\%22primaryParent\%22\%3A1\%2C\%22secondaryParent\%22\%3A4\%2C\%22cousin\%22\%3A6\%7D\%2C\%22nodeIds\%22\%3A\%5B\%5D\%7D\%2C\%22highlightImportantNodes\%22\%3A\%7B\%22highlightImportantNodes\%22\%3Afalse\%2C\%22highlightPrimaryParents\%22\%3Afalse\%2C\%22highlightRootConnectingEdges\%22\%3Afalse\%7D\%7D}

\begingroup
\small
\eczhBreakableDashes
\renewcommand\arraystretch{1.05}
\edef\myxtraspc{\eczListAddVSpaceXtraXtraStretch}
\begin{tabularx}{\linewidth}{>{\raggedright\arraybackslash}p{\eczListColWidth{name}} >{\hsize=1.0000\hsize }X}
\toprule
\eczListColTitle{Code} & \eczListColTitle{Description} \\
\midrule
\endfirsthead
\toprule
\eczListColTitleContinued{Code} & \eczListColTitleContinued{Description} \\
\midrule
\endhead
\bottomrule
\endfoot
\eczhRefIndex{code:analog_repetition}%
\eczhListValue{\flmRefsHyperref{code:analog_repetition}{Analog repetition code}} & \eczhListValue{An \(\llbracket n,1\rrbracket _{\mathbb{R}}\) analog stabilizer version of the quantum repetition code, encoding the position states of one mode into an odd number \(n\) of modes.}\\ 
\addlinespace[\myxtraspc]
\eczhRefIndex{code:binary_dihedral_permutation_invariant}%
\eczhListValue{\flmRefsHyperref{code:binary_dihedral_permutation_invariant}{Binary dihedral PI code}} & \eczhListValue{Multi-qubit PI code designed to realize gates from the binary dihedral group transversally.
Can also be interpreted as a single-spin code.
The codespace projection is a projection onto an irrep of the \textit{binary dihedral group} \( \mathsf{BD}_{2N} = \langle\omega I, X, P\rangle \) of order \(8N\), where \( \omega \) is a \( 2N \)th root of unity, and \( P = \text{diag} ( 1, \omega^2) \).}\\ 
\addlinespace[\myxtraspc]
\eczhRefIndex{code:bipartite_cyclic_cluster}%
\eczhListValue{\flmRefsHyperref{code:bipartite_cyclic_cluster}{Bipartite cyclic cluster (BCC) code}} & \eczhListValue{Cyclic CSS code constructed from a bipartite cluster state with cyclic invariance,
emphasizing simplicity of state preparation over simplicity of stabilizers.}\\ 
\addlinespace[\myxtraspc]
\eczhRefIndex{code:combinatorial_permutation_invariant}%
\eczhListValue{\flmRefsHyperref{code:combinatorial_permutation_invariant}{Combinatorial PI code}} & \eczhListValue{A member of a family of PI quantum codes whose correction properties are derived from solving a family of combinatorial identities.
The code encodes one logical qubit in superpositions of \flmRefsHyperref{ref526}{Dicke states} whose coefficients are square roots of ratios of binomial coefficients.}\\ 
\addlinespace[\myxtraspc]
\eczhRefIndex{code:quantum_cyclic}%
\eczhListValue{\flmRefsHyperref{code:quantum_cyclic}{Cyclic quantum code}} & \eczhListValue{A block quantum code such that cyclic permutations of the subsystems leave the codespace invariant. In other words, the automorphism group of the code contains the cyclic group \(\mathbb{Z}_n\).}\\ 
\addlinespace[\myxtraspc]
\eczhRefIndex{code:frobenius}%
\eczhListValue{\flmRefsHyperref{code:frobenius}{Frobenius code}} & \eczhListValue{A cyclic prime-qudit stabilizer code whose length \(n\) divides \(p^t + 1\) for some positive integer \(t\).}\\ 
\addlinespace[\myxtraspc]
\eczhRefIndex{code:gnu_permutation_invariant}%
\eczhListValue{\flmRefsHyperref{code:gnu_permutation_invariant}{GNU PI code}} & \eczhListValue{PI code whose codewords can be expressed as superpositions of \flmRefsHyperref{ref526}{Dicke states} with coefficients are square-roots of the binomial distribution.}\\ 
\addlinespace[\myxtraspc]
\eczhRefIndex{code:group_quantum_repetition}%
\eczhListValue{\flmRefsHyperref{code:group_quantum_repetition}{Group-based quantum repetition code}} & \eczhListValue{An \(\llbracket n,1\rrbracket _G\) generalization of the quantum repetition code.}\\ 
\addlinespace[\myxtraspc]
\eczhRefIndex{code:lacross}%
\eczhListValue{\flmRefsHyperref{code:lacross}{La-cross code}} & \eczhListValue{Code constructed using a hypergraph product of two copies of a cyclic LDPC code.
The construction uses cyclic LDPC codes with \flmRefsHyperref{ref67}{generating polynomials} \(1+x+x^k\) for some \(k\).
Using a length-\(n\) seed code yields an \(\llbracket 2n^2,2k^2\rrbracket \) family for periodic boundary conditions and an \(\llbracket (n-k)^2+n^2,k^2\rrbracket \) family for open boundary conditions.}\\ 
\addlinespace[\myxtraspc]
\eczhRefIndex{code:constant_excitation_permutation_invariant}%
\eczhListValue{\flmRefsHyperref{code:constant_excitation_permutation_invariant}{Ouyang-Chao constant-excitation PI code}} & \eczhListValue{A constant-excitation PI Fock-state code whose construction is based on integer partitions.}\\ 
\addlinespace[\myxtraspc]
\eczhRefIndex{code:qubit_permutation_invariant}%
\eczhListValue{\flmRefsHyperref{code:qubit_permutation_invariant}{PI qubit code}} & \eczhListValue{Block quantum code defined on two-dimensional subsystems such that any permutation of the subsystems leaves any codeword invariant.}\\ 
\addlinespace[\myxtraspc]
\eczhRefIndex{code:permutation_invariant}%
\eczhListValue{\flmRefsHyperref{code:permutation_invariant}{Permutation-invariant (PI) code}} & \eczhListValue{Block quantum code such that any permutation of the subsystems leaves any codeword invariant.
In other words, the automorphism group of the code contains the symmetric group \(S_n\).}\\ 
\addlinespace[\myxtraspc]
\eczhRefIndex{code:ampdamp_post_selected}%
\eczhListValue{\flmRefsHyperref{code:ampdamp_post_selected}{Post-selected PI code}} & \eczhListValue{PI qubit code whose recovery succeeds at protecting against \flmRefsHyperref{ref498}{AD} errors with a success probability less than one.}\\ 
\addlinespace[\myxtraspc]
\eczhRefIndex{code:quantum_repetition}%
\eczhListValue{\flmRefsHyperref{code:quantum_repetition}{Quantum repetition code}} & \eczhListValue{Encodes \(1\) qubit into \(n\) qubits according to \(|0\rangle\to|\phi_0\rangle^{\otimes n}\) and \(|1\rangle\to|\phi_1\rangle^{\otimes n}\). The code is called a \textit{bit-flip} code when \(|\phi_i\rangle = |i\rangle\), and a \textit{phase-flip} code when \(|\phi_0\rangle = |+\rangle\) and \(|\phi_1\rangle = |-\rangle\).
This repetition-style encoding does not clone an arbitrary quantum state; instead, it extends the copying of computational-basis states linearly to entangled codewords  \NoCaseChange{\protect\cite[{Ch. 2}]{cite398}}.}\\ 
\addlinespace[\myxtraspc]
\eczhRefIndex{code:qudit_gnu_permutation_invariant}%
\eczhListValue{\flmRefsHyperref{code:qudit_gnu_permutation_invariant}{Qudit GNU PI code}} & \eczhListValue{Extension of the GNU PI codes to those encoding logical qudits into physical qubits.
Codewords can be expressed as superpositions of \flmRefsHyperref{ref526}{Dicke states} with coefficients given by square roots of polynomial coefficients, with the case of binomial coefficients reducing to the GNU PI codes.}\\ 
\addlinespace[\myxtraspc]
\eczhRefIndex{code:tetron}%
\eczhListValue{\flmRefsHyperref{code:tetron}{Tetron code}} & \eczhListValue{A \(\llbracket 2,1,2\rrbracket _{f}\) Majorana box qubit encoding a logical qubit into four Majorana modes, equivalently into the fixed-total-parity sector of two physical fermionic modes.
Four Majorana zero modes are the smallest aggregate that supports a qubit in a fixed fermion-parity sector \NoCaseChange{\protect\cite{cite401}}.
This code can be concatenated with various qubit codes such as surface codes and color codes.
Four-boundary Majorana surface-code patches are logical tetrons, i.e., higher-distance analogues of this physical tetron block \NoCaseChange{\protect\cite{cite402}}.}\\ 
\addlinespace[\myxtraspc]
\eczhRefIndex{code:twisted_xzzx}%
\eczhListValue{\flmRefsHyperref{code:twisted_xzzx}{Twisted XZZX toric code}} & \eczhListValue{A cyclic code that can be thought of as the XZZX toric code with shifted (a.k.a twisted) boundary conditions.
Admits a set of stabilizer generators that are equivalent to cyclic shifts of a particular weight-four \(XZZX\) Pauli string.}\\ 
\addlinespace[\myxtraspc]
\eczhRefIndex{code:t_group}%
\eczhListValue{\flmRefsHyperref{code:t_group}{Twisted \(1\)-group code}} & \eczhListValue{Block group-representation code realizing particular irreps of particular groups such that a distance of two is automatically guaranteed.
Groups which admit irreps with this property are called \textit{twisted (unitary) \(1\)-groups} and include the binary icosahedral group \(2I\), the \(\Sigma(360\phi)\) subgroup of \(SU(3)\), the family \(\{PSp(2b, 3), b \geq 1\}\), and the alternating groups \(A_{5,6}\).
Groups whose irreps are images of the appropriate irreps of twisted \(1\)-groups also yield such properties, e.g., the binary tetrahedral group \(2T\) or qutrit Pauli group \(\Sigma(72\phi)\).}\\ 
\addlinespace[\myxtraspc]
\eczhRefIndex{code:very-small-logical-qubit}%
\eczhListValue{\flmRefsHyperref{code:very-small-logical-qubit}{Very small logical qubit (VSLQ) code}} & \eczhListValue{A code consisting of two logical codewords \(|\pm\rangle \propto (|0\rangle\pm|2\rangle)(|0\rangle\pm|2\rangle)\), where the total Hilbert space is the tensor product of two transmon qudits (whose ground states \(|0\rangle\) and second excited states \(|2\rangle\) are used in the codewords).
Since the code is intended to protect against losses, the qutrits can equivalently be thought of as oscillator Fock-state subspaces.}\\ 
\addlinespace[\myxtraspc]
\eczhRefIndex{code:w_state}%
\eczhListValue{\flmRefsHyperref{code:w_state}{W-state code}} & \eczhListValue{Approximate block quantum code whose encoding resembles the structure of the
W state~\NoCaseChange{\protect\cite{cite527}}.
This code enables universal quantum computation with transversal gates.}\\ 
\addlinespace[\myxtraspc]
\eczhRefIndex{code:wasilewski-banaszek}%
\eczhListValue{\flmRefsHyperref{code:wasilewski-banaszek}{Wasilewski-Banaszek code}} & \eczhListValue{Three-oscillator constant-excitation Fock-state code encoding a single logical qubit.}\\ 
\addlinespace[\myxtraspc]
\eczhRefIndex{code:three_qutrit_permutation_invariant}%
\eczhListValue{\flmRefsHyperref{code:three_qutrit_permutation_invariant}{\(\llparenthesis 3,2,2\rrparenthesis _3\) Three-qutrit single-deletion code}} & \eczhListValue{Three-qutrit PI code that is the smallest qutrit PI code to correct one deletion error.}\\ 
\addlinespace[\myxtraspc]
\eczhRefIndex{code:four_qubit_permutation_invariant}%
\eczhListValue{\flmRefsHyperref{code:four_qubit_permutation_invariant}{\(\llparenthesis 4,2,2\rrparenthesis \) Four-qubit single-deletion code}} & \eczhListValue{Four-qubit PI code that is the smallest qubit code to correct one deletion error.}\\ 
\addlinespace[\myxtraspc]
\eczhRefIndex{code:su3_sigma360}%
\eczhListValue{\flmRefsHyperref{code:su3_sigma360}{\(\llparenthesis 5,3,2\rrparenthesis _3\) qutrit code}} & \eczhListValue{Smallest qutrit block code realizing the \(\Sigma(360\phi)=3.A_6\) subgroup of \(SU(3)\) transversally.
The next smallest code is \(\llparenthesis 7,3,2\rrparenthesis _3\).}\\ 
\addlinespace[\myxtraspc]
\eczhRefIndex{code:qubit_5_6_2}%
\eczhListValue{\flmRefsHyperref{code:qubit_5_6_2}{\(\llparenthesis 5,6,2\rrparenthesis \) qubit code}} & \eczhListValue{Five-qubit cyclic CWS code detecting a single-qubit error.
This code has a logical subspace whose dimension is larger than that of the \(\llbracket 5,2,2\rrbracket \) code, the best five-qubit stabilizer code with the same distance \NoCaseChange{\protect\cite{cite452}}.}\\ 
\addlinespace[\myxtraspc]
\eczhRefIndex{code:icosahedral_permutation_invariant}%
\eczhListValue{\flmRefsHyperref{code:icosahedral_permutation_invariant}{\(\llparenthesis 7,2,3\rrparenthesis \) Pollatsek-Ruskai code}} & \eczhListValue{Seven-qubit PI code that realizes gates from the binary icosahedral group transversally.
Can also be interpreted as a spin-\(7/2\) single-spin code.
The codespace projection is a projection onto an irrep of the binary icosahedral group \(2I\).
See Ref. \NoCaseChange{\protect\cite{cite528}} for other non-PI codes realizing \(2I\) gates transversally.}\\ 
\addlinespace[\myxtraspc]
\eczhRefIndex{code:qubit_9_12_3}%
\eczhListValue{\flmRefsHyperref{code:qubit_9_12_3}{\(\llparenthesis 9,12,3\rrparenthesis \) qubit code}} & \eczhListValue{Nine-qubit cyclic CWS code correcting a single-qubit error.
This code has a logical subspace whose dimension is larger than that of the \(\llbracket 9,3,3\rrbracket \) code, the best nine-qubit stabilizer code with the same distance \NoCaseChange{\protect\cite{cite449}}.}\\ 
\addlinespace[\myxtraspc]
\eczhRefIndex{code:ruskai}%
\eczhListValue{\flmRefsHyperref{code:ruskai}{\(\llparenthesis 9,2,3\rrparenthesis \) Ruskai code}} & \eczhListValue{Nine-qubit PI code that protects against single-qubit errors as well as two-qubit errors arising from exchange processes.}\\ 
\addlinespace[\myxtraspc]
\eczhRefIndex{code:unentangled_permutation_invariant}%
\eczhListValue{\flmRefsHyperref{code:unentangled_permutation_invariant}{\(\llparenthesis n,2,2\rrparenthesis \) Bravyi-Lee-Li-Yoshida PI code}} & \eczhListValue{PI distance-two code on \(n\geq4\) qubits whose degree of entanglement vanishes asymptotically with \(n\) \NoCaseChange{\protect\cite[{Appx. D}]{cite529}} (cf. \NoCaseChange{\protect\cite{cite530}}).}\\ 
\addlinespace[\myxtraspc]
\eczhRefIndex{code:xzzx_10_2_3}%
\eczhListValue{\flmRefsHyperref{code:xzzx_10_2_3}{\(\llbracket 10,2,3\rrbracket \) rotated toric code}} & \eczhListValue{Rotated toric code that is the CSS form of the twisted XZZX toric code with parameters \(a=1\), \(b=3\) \NoCaseChange{\protect\cite{cite438,cite427}},
related to the XZZX form by Hadamard on the \(B\)-sublattice.
It is also the \flmRefsHyperref{ref436}{symplectic double} (a.k.a. genus-one double cover) of the \(\llbracket 5,1,3\rrbracket \) five-qubit perfect code \NoCaseChange{\protect\cite{cite439,cite435}}, the \flmRefsHyperref{ref436}{symplectic double} of the \(\llbracket 5,1,2\rrbracket \) rotated surface code \NoCaseChange{\protect\cite{cite435}}, and a BCC code \NoCaseChange{\protect\cite{cite440}}.}\\ 
\addlinespace[\myxtraspc]
\eczhRefIndex{code:stab_13_1_5}%
\eczhListValue{\flmRefsHyperref{code:stab_13_1_5}{\(\llbracket 13,1,5\rrbracket \) twisted toric code}} & \eczhListValue{Thirteen-qubit twisted toric code whose stabilizer tableau consists of cyclic permutations of the \(XZZX\)-type Pauli string \(XIZZIXIIIIIII\).
The code can be thought of as a small twisted XZZX code \NoCaseChange{\protect\cite[{Exam. 11 and Fig. 3}]{cite438}}.}\\ 
\addlinespace[\myxtraspc]
\eczhRefIndex{code:stab_18_2_5}%
\eczhListValue{\flmRefsHyperref{code:stab_18_2_5}{\(\llbracket 18,2,5\rrbracket \) BCC code}} & \eczhListValue{BCC code on 18 qubits encoding 2 logical qubits with distance 5, found by computer search \NoCaseChange{\protect\cite{cite440}}.}\\ 
\addlinespace[\myxtraspc]
\eczhRefIndex{code:stab_5_1_3}%
\eczhListValue{\flmRefsHyperref{code:stab_5_1_3}{\(\llbracket 5,1,3\rrbracket \) Five-qubit perfect code}} & \eczhListValue{Five-qubit cyclic stabilizer code that is the smallest qubit stabilizer code to correct a single-qubit error.}\\ 
\addlinespace[\myxtraspc]
\eczhRefIndex{code:galois_5_1_3}%
\eczhListValue{\flmRefsHyperref{code:galois_5_1_3}{\(\llbracket 5,1,3\rrbracket _q\) Galois-qudit code}} & \eczhListValue{True stabilizer code that generalizes the five-qubit perfect code to Galois qudits of prime-power dimension \(q=p^m\). It has \(4m\) stabilizer generators expressed as \(X_{\gamma} Z_{\gamma} Z_{-\gamma} X_{-\gamma} I\) and its cyclic permutations, with \(\gamma\) iterating over basis elements of \(\mathbb{F}_q\) over \(\mathbb{F}_p\).}\\ 
\addlinespace[\myxtraspc]
\eczhRefIndex{code:braunstein}%
\eczhListValue{\flmRefsHyperref{code:braunstein}{\(\llbracket 5,1,3\rrbracket _{\mathbb{R}}\) Braunstein five-mode code}} & \eczhListValue{An analog stabilizer version of the five-qubit perfect code, encoding one mode into five and correcting arbitrary errors on any one mode.}\\ 
\addlinespace[\myxtraspc]
\eczhRefIndex{code:qudit_5_1_3}%
\eczhListValue{\flmRefsHyperref{code:qudit_5_1_3}{\(\llbracket 5,1,3\rrbracket _{\mathbb{Z}_q}\) modular-qudit code}} & \eczhListValue{Modular-qudit stabilizer code that generalizes the five-qubit perfect code using properties of the multiplicative group \(\mathbb{Z}_q\) \NoCaseChange{\protect\cite{cite531}}; see also \NoCaseChange{\protect\cite[{Thm. 13}]{cite532}}. It has four stabilizer generators consisting of \(X Z Z^\dagger X^\dagger I\) and its cyclic permutations.}\\ 
\addlinespace[\myxtraspc]
\eczhRefIndex{code:rotor_5_1_3}%
\eczhListValue{\flmRefsHyperref{code:rotor_5_1_3}{\(\llbracket 5,1,3\rrbracket _{\mathbb{Z}}\) Five-rotor code}} & \eczhListValue{Extension of the five-qubit stabilizer code to the integer alphabet, i.e., the angular momentum states of a rotor. The code is \(U(1)\)-covariant and ideal codewords are not normalizable.}\\ 
\addlinespace[\myxtraspc]
\eczhRefIndex{code:steane}%
\eczhListValue{\flmRefsHyperref{code:steane}{\(\llbracket 7,1,3\rrbracket \) Steane code}} & \eczhListValue{A \(\llbracket 7,1,3\rrbracket \) self-dual CSS code that is the smallest qubit CSS code to correct a single-qubit error \NoCaseChange{\protect\cite{cite451}}.
The code is constructed using the classical binary \([7,4,3]\) Hamming code for protecting against both \(X\) and \(Z\) errors.}\\ 
\addlinespace[\myxtraspc]
\eczhRefIndex{code:xzzx_7_1_3}%
\eczhListValue{\flmRefsHyperref{code:xzzx_7_1_3}{\(\llbracket 7,1,3\rrbracket \) XZZX cyclic code}} & \eczhListValue{A \(\llbracket 7,1,3\rrbracket \) cyclic non-CSS code whose stabilizer generators are the seven cyclic shifts of the weight-four Pauli string \(XZIZXII\),
any six of which are independent.
The non-identity support of this generator is \(XZZX\) (at positions 0, 1, 3, 4), making this a member of the twisted XZZX code family.
It is one of sixteen distinct indecomposable \(\llbracket 7,1,3\rrbracket \) codes \NoCaseChange{\protect\cite{cite452}}.}\\ 
\end{tabularx}\endgroup
\eczcodelist{dynamic_gen}{Dynamically generated quantum codes and friends
}%

\eczhCodeListAutoDescription{All descendants and cousins of \flmRefsCref{code:dynamic_gen}.}%

\eczhIncludeCodeGraph{Bare}{scale=0.5}{\columnwidth}{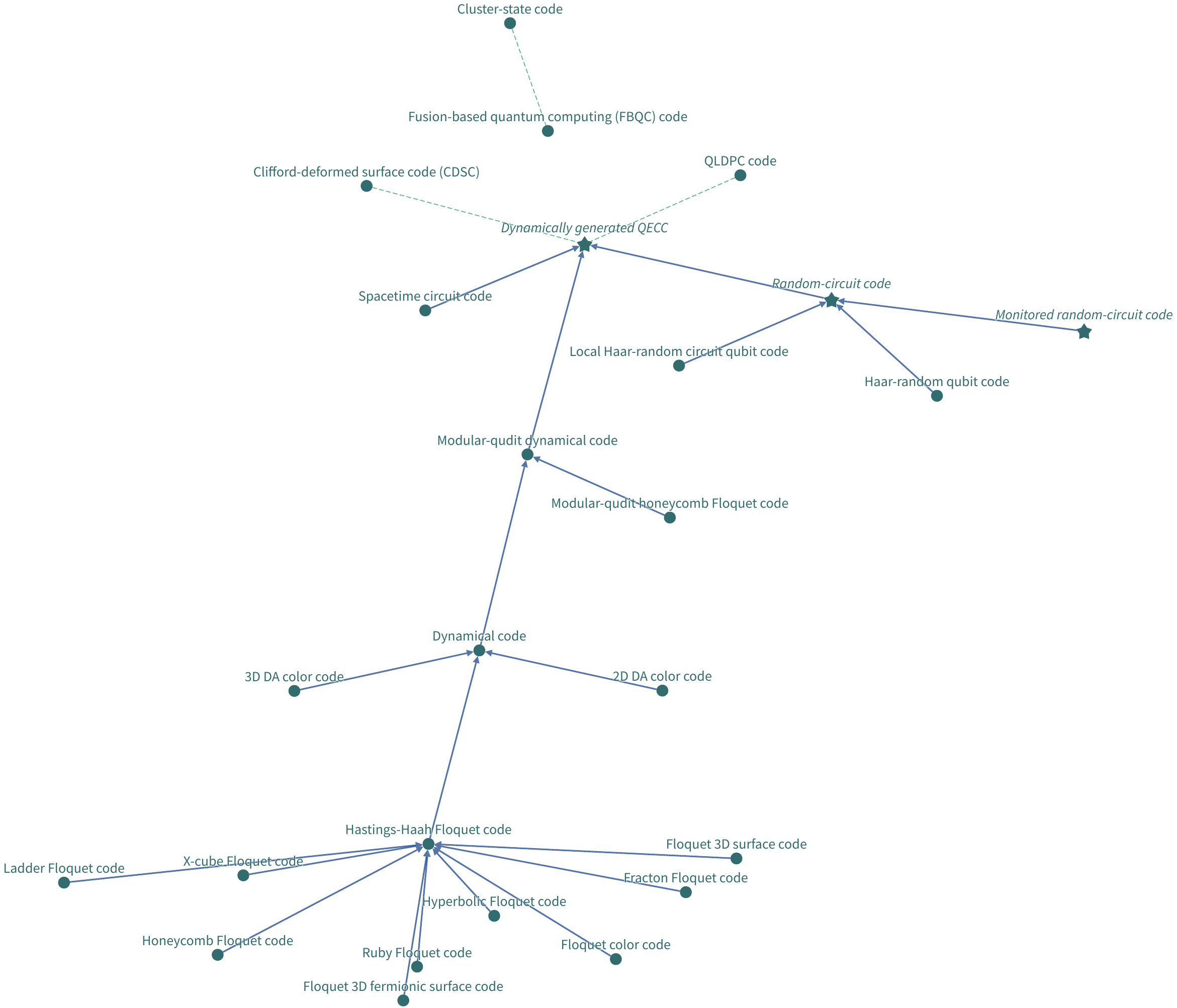}{Dynamically generated quantum codes and friends}{https://errorcorrectionzoo.org/code_graph#J\%7B\%22displayMode\%22\%3A\%22subset\%22\%2C\%22modeSubsetOptions\%22\%3A\%7B\%22codeIds\%22\%3A\%5B\%22da_color_2d\%22\%2C\%22da_color_3d\%22\%2C\%22clifford-deformed_surface\%22\%2C\%22cluster_state\%22\%2C\%22da\%22\%2C\%22dynamic_gen\%22\%2C\%22floquet_3d_fermionic_surface\%22\%2C\%22floquet_3d_surface\%22\%2C\%22floquet_color\%22\%2C\%22floquet_fracton\%22\%2C\%22fusion\%22\%2C\%22haar_random\%22\%2C\%22floquet\%22\%2C\%22honeycomb_floquet\%22\%2C\%22hyperbolic_floquet\%22\%2C\%22ladder\%22\%2C\%22local_haar_random\%22\%2C\%22qudit_da\%22\%2C\%22qudit_honeycomb\%22\%2C\%22monitored_random_circuits\%22\%2C\%22general_qldpc\%22\%2C\%22random_circuit\%22\%2C\%22floquet_xyz_ruby\%22\%2C\%22spacetime_circuit\%22\%2C\%22floquet_xcube\%22\%5D\%2C\%22reusePreviousLayoutPositions\%22\%3Afalse\%2C\%22showIntermediateConnectingNodes\%22\%3Atrue\%2C\%22connectingNodesMaxDepth\%22\%3A15\%2C\%22connectingNodesPathMaxLength\%22\%3A20\%2C\%22connectingNodesMaxExtraDepth\%22\%3A3\%2C\%22connectingNodesOnlyKeepPathsWithAdditionalLength\%22\%3A1\%2C\%22connectingNodesToDomainsAndKingdoms\%22\%3Afalse\%2C\%22connectingNodesEdgeLengthsByType\%22\%3A\%7B\%22primaryParent\%22\%3A1\%2C\%22secondaryParent\%22\%3A4\%2C\%22cousin\%22\%3A6\%7D\%2C\%22nodeIds\%22\%3A\%5B\%5D\%7D\%2C\%22highlightImportantNodes\%22\%3A\%7B\%22highlightImportantNodes\%22\%3Afalse\%2C\%22highlightPrimaryParents\%22\%3Afalse\%2C\%22highlightRootConnectingEdges\%22\%3Afalse\%7D\%7D}

\begingroup
\small
\eczhBreakableDashes
\renewcommand\arraystretch{1.05}
\edef\myxtraspc{\eczListAddVSpaceXtraXtraStretch}
\begin{tabularx}{\linewidth}{>{\raggedright\arraybackslash}p{\eczListColWidth{name}} >{\hsize=1.0000\hsize }X}
\toprule
\eczListColTitle{Code} & \eczListColTitle{Description} \\
\midrule
\endfirsthead
\toprule
\eczListColTitleContinued{Code} & \eczListColTitleContinued{Description} \\
\midrule
\endhead
\bottomrule
\endfoot
\eczhRefIndex{code:da_color_2d}%
\eczhListValue{\flmRefsHyperref{code:da_color_2d}{2D DA color code}} & \eczhListValue{A 2D dynamical code constructed aperiodically that utilizes measurement sequences to encode logical information with automorphisms of the 2D color code.
The code is assembled from short measurement sequences that can realize all 72 automorphisms of the 2D color code.
On a stack of \(N\) triangular patches with a Pauli boundary, the code encodes \(N\) logical qubits.}\\ 
\addlinespace[\myxtraspc]
\eczhRefIndex{code:da_color_3d}%
\eczhListValue{\flmRefsHyperref{code:da_color_3d}{3D DA color code}} & \eczhListValue{A 3D dynamical code constructed aperiodically that utilizes measurement sequences to encode logical information with automorphisms of the 3D color code.
The code represents the first step towards universal quantum computation with dynamical automorphism codes.}\\ 
\addlinespace[\myxtraspc]
\eczhRefIndex{code:clifford-deformed_surface}%
\eczhListValue{\flmRefsHyperref{code:clifford-deformed_surface}{Clifford-deformed surface code (CDSC)}} & \eczhListValue{A generally non-CSS derivative of the surface code defined by applying a translationally invariant constant-depth \flmRefsHyperref{ref409}{Clifford circuit} to the original (CSS) surface code.
Unlike the surface code, CDSCs include codes whose thresholds and subthreshold performance are enhanced under noise biased towards dephasing.
Examples of CDSCs include the XY code, XZZX code, and random CDSCs.}\\ 
\addlinespace[\myxtraspc]
\eczhRefIndex{code:cluster_state}%
\eczhListValue{\flmRefsHyperref{code:cluster_state}{Cluster-state code}} & \eczhListValue{A code based on a cluster state (a.k.a. graph state) and often used in measurement-based quantum computation (MBQC) \NoCaseChange{\protect\cite{cite428,cite429}} (a.k.a. one-way quantum processing), which substitutes the temporal dimension necessary for decoding a conventional code with a spatial dimension.
This is done by encoding the computation into the features of the cluster state's graph.}\\ 
\addlinespace[\myxtraspc]
\eczhRefIndex{code:da}%
\eczhListValue{\flmRefsHyperref{code:da}{Dynamical code}} & \eczhListValue{Dynamically generated stabilizer-based code whose (not necessarily periodic) sequence of few-body measurements implements state initialization, logical gates and error detection.}\\ 
\addlinespace[\myxtraspc]
\eczhRefIndex{code:dynamic_gen}%
\eczhListValue{\flmRefsHyperref{code:dynamic_gen}{Dynamically generated QECC}} & \eczhListValue{Block quantum code whose natural definition is in terms of a many-body scaling limit of a local dynamical process.
Such processes, which are often non-deterministic, update the code structure and can include random unitary evolution or non-commuting projective measurements.}\\ 
\addlinespace[\myxtraspc]
\eczhRefIndex{code:floquet_3d_fermionic_surface}%
\eczhListValue{\flmRefsHyperref{code:floquet_3d_fermionic_surface}{Floquet 3D fermionic surface code}} & \eczhListValue{A 3D Floquet code on a trivalent lattice whose weight-two checks are the \(XX\), \(YY\), and \(ZZ\) edge terms of the 3D Kitaev honeycomb model \NoCaseChange{\protect\cite{cite458,cite533}}.}\\ 
\addlinespace[\myxtraspc]
\eczhRefIndex{code:floquet_3d_surface}%
\eczhListValue{\flmRefsHyperref{code:floquet_3d_surface}{Floquet 3D surface code}} & \eczhListValue{A 3D Floquet code on a truncated cubic honeycomb with pairs of physical qubits on vertices.
It is constructed from three stacks of square-octagon Floquet toric codes, coupled by interlayer \(YY\) measurements in a coupled-layer construction \NoCaseChange{\protect\cite{cite534,cite533}}.}\\ 
\addlinespace[\myxtraspc]
\eczhRefIndex{code:floquet_color}%
\eczhListValue{\flmRefsHyperref{code:floquet_color}{Floquet color code}} & \eczhListValue{2D Floquet code on a trivalent 2D lattice whose parent topological phase is the \(\mathbb{Z}_2\times\mathbb{Z}_2\) 2D color-code phase and whose measurements cycle logical quantum information between the nine \(\mathbb{Z}_2\) surface-code \flmRefsHyperref{ref410}{condensed phases} of the parent phase.
The code's ISG is the stabilizer group of one of the nine surface codes.}\\ 
\addlinespace[\myxtraspc]
\eczhRefIndex{code:floquet_fracton}%
\eczhListValue{\flmRefsHyperref{code:floquet_fracton}{Fracton Floquet code}} & \eczhListValue{3D Floquet code whose qubits are placed on vertices of a truncated cubic honeycomb.
Its weight-two check operators are placed on edges of each truncated cube, while weight-three check operators are placed on each triangle.
Its ISG can be that of the X-cube model code or the checkerboard model code.
On a three-torus of size \(L_x \times L_y \times L_z\), the code consists of \(n= 48L_xL_yL_z\) physical qubits and encodes \(k= 2(L_x+L_y+L_z)-6\) logical qubits.}\\ 
\addlinespace[\myxtraspc]
\eczhRefIndex{code:fusion}%
\eczhListValue{\flmRefsHyperref{code:fusion}{Fusion-based quantum computing (FBQC) code}} & \eczhListValue{Code whose codewords are resource states used in an FBQC scheme.}\\ 
\addlinespace[\myxtraspc]
\eczhRefIndex{code:haar_random}%
\eczhListValue{\flmRefsHyperref{code:haar_random}{Haar-random qubit code}} & \eczhListValue{Haar-random codewords are generated in a process involving averaging over unitary operations distributed according to the Haar measure. Haar-random codes are used to prove statements about the capacity of a quantum channel to transmit quantum information \NoCaseChange{\protect\cite{cite535}}, but encoding and decoding in such \(n\)-qubit codes quickly becomes impractical as \(n\to\infty\).}\\ 
\addlinespace[\myxtraspc]
\eczhRefIndex{code:floquet}%
\eczhListValue{\flmRefsHyperref{code:floquet}{Hastings-Haah Floquet code}} & \eczhListValue{Dynamical code whose sequence of check-operator measurements is periodic.
The original Hastings-Haah construction introduced periodic measurement schedules that dynamically generate logical qubits even when the underlying subsystem code has fewer or no logical qubits \NoCaseChange{\protect\cite{cite536}}.
Its basic examples are the 2D honeycomb Floquet code and the 1D ladder Floquet code.}\\ 
\addlinespace[\myxtraspc]
\eczhRefIndex{code:honeycomb_floquet}%
\eczhListValue{\flmRefsHyperref{code:honeycomb_floquet}{Honeycomb Floquet code}} & \eczhListValue{2D Floquet code based on the Kitaev honeycomb model \NoCaseChange{\protect\cite{cite537}} whose logical qubits are generated through a particular sequence of measurements.
A CSS version of the code has been proposed which loosens the restriction of which sequences to use \NoCaseChange{\protect\cite{cite538}}.
The code has also been generalized to arbitrary non-chiral, Abelian topological order \NoCaseChange{\protect\cite{cite539}}.}\\ 
\addlinespace[\myxtraspc]
\eczhRefIndex{code:hyperbolic_floquet}%
\eczhListValue{\flmRefsHyperref{code:hyperbolic_floquet}{Hyperbolic Floquet code}} & \eczhListValue{Floquet code whose check-operators correspond to edges of a hyperbolic lattice of degree 3.}\\ 
\addlinespace[\myxtraspc]
\eczhRefIndex{code:ladder}%
\eczhListValue{\flmRefsHyperref{code:ladder}{Ladder Floquet code}} & \eczhListValue{1D Floquet code defined on a ladder qubit geometry, with one qubit per vertex.
The check operators consist of \(ZZ\) checks on each rung and alternating \(XX\) and \(YY\) checks on the legs.
The period-four measurement schedule measures \(ZZ\), \(XX\), \(ZZ\), and \(YY\) in rounds \(0,1,2,3\) mod \(4\), respectively, dynamically generating one logical qubit \NoCaseChange{\protect\cite{cite536}}.}\\ 
\addlinespace[\myxtraspc]
\eczhRefIndex{code:local_haar_random}%
\eczhListValue{\flmRefsHyperref{code:local_haar_random}{Local Haar-random circuit qubit code}} & \eczhListValue{An \(n\)-qubit code whose codewords are a pair of approximately locally indistinguishable states produced by starting with any two orthogonal \(n\)-qubit states and acting with a random unitary circuit of depth polynomial in \(n\).
Two states are \textit{locally indistinguishable} if they cannot be distinguished by local measurements. A single layer of the encoding circuit is composed of about \(n/2\) two-qubit nearest-neighbor gates run in parallel, with each gate drawn randomly from the Haar distribution on two-qubit unitaries.}\\ 
\addlinespace[\myxtraspc]
\eczhRefIndex{code:qudit_da}%
\eczhListValue{\flmRefsHyperref{code:qudit_da}{Modular-qudit dynamical code}} & \eczhListValue{Dynamically generated stabilizer-based modular-qudit code whose (not necessarily periodic) sequence of few-body measurements implements state initialization, logical gates and error detection.}\\ 
\addlinespace[\myxtraspc]
\eczhRefIndex{code:qudit_honeycomb}%
\eczhListValue{\flmRefsHyperref{code:qudit_honeycomb}{Modular-qudit honeycomb Floquet code}} & \eczhListValue{A modular-qudit extension of the honeycomb Floquet code.}\\ 
\addlinespace[\myxtraspc]
\eczhRefIndex{code:monitored_random_circuits}%
\eczhListValue{\flmRefsHyperref{code:monitored_random_circuits}{Monitored random-circuit code}} & \eczhListValue{Error-correcting code arising from a monitored random circuit. Such a circuit is described by a series of intermittent random local projective Pauli measurements with random unitary time-evolution operators.}\\ 
\addlinespace[\myxtraspc]
\eczhRefIndex{code:general_qldpc}%
\eczhListValue{\flmRefsHyperref{code:general_qldpc}{QLDPC code}} & \eczhListValue{Member of a family of stabilizer codes for which the number of sites participating in each stabilizer generator and the number of stabilizer generators that each site participates in are both bounded by a constant as \(n\to\infty\).
Sometimes, the two parameters are explicitly stated: each site of an \((l,w)\)\textit{-regular QLDPC code} is acted on by \(\leq l\) generators of weight \(\leq w\).}\\ 
\addlinespace[\myxtraspc]
\eczhRefIndex{code:random_circuit}%
\eczhListValue{\flmRefsHyperref{code:random_circuit}{Random-circuit code}} & \eczhListValue{Code whose encoding is naturally constructed by randomly sampling from a large set of quantum circuits. Examples include short random Clifford circuits that define good quantum error-correcting codes \NoCaseChange{\protect\cite{cite540}} and monitored random circuits whose mixed phase dynamically generates error-protected subspaces with nonzero channel-capacity density on polynomial timescales \NoCaseChange{\protect\cite{cite541}}.}\\ 
\addlinespace[\myxtraspc]
\eczhRefIndex{code:floquet_xyz_ruby}%
\eczhListValue{\flmRefsHyperref{code:floquet_xyz_ruby}{Ruby Floquet code}} & \eczhListValue{2D Floquet code whose qubits are placed on vertices of a ruby tiling, with weight-two Pauli check operators on \(x\)-, \(y\)-, and \(z\)-labeled edges \NoCaseChange{\protect\cite{cite542}}.
The code admits two different measurement schedules, the XYZ ruby schedule and the color-code schedule.}\\ 
\addlinespace[\myxtraspc]
\eczhRefIndex{code:spacetime_circuit}%
\eczhListValue{\flmRefsHyperref{code:spacetime_circuit}{Spacetime circuit code}} & \eczhListValue{Qubit stabilizer code constructed from a \flmRefsHyperref{ref409}{Clifford circuit}, i.e., a circuit made up of \flmRefsHyperref{ref409}{Clifford gates} and Pauli measurements, in order to detect and correct circuit faults.
The code utilizes redundancy in the measurement outcomes of a circuit to correct circuit faults, which correspond to Pauli errors of the code.}\\ 
\addlinespace[\myxtraspc]
\eczhRefIndex{code:floquet_xcube}%
\eczhListValue{\flmRefsHyperref{code:floquet_xcube}{X-cube Floquet code}} & \eczhListValue{A 3D Floquet code on the truncated cubic honeycomb, built from coupled layers of square-octagon Floquet toric codes.}\\ 
\end{tabularx}\endgroup
\eczcodelist{eaqecc}{Entanglement-assisted quantum codes and friends
}%

\eczhCodeListAutoDescription{All descendants and cousins of \flmRefsCref{code:eaqecc}.}%

\eczhIncludeCodeGraph{Bare}{scale=0.5}{\columnwidth}{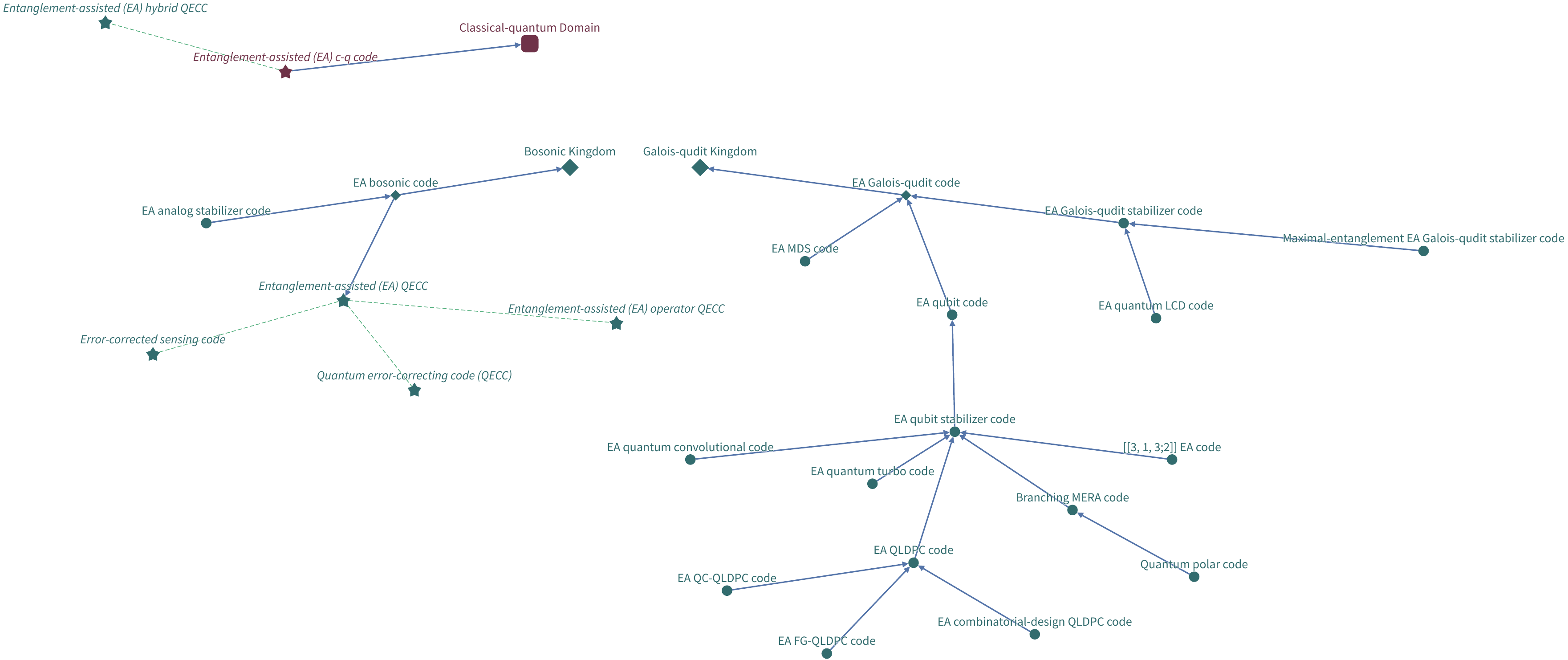}{Entanglement-assisted quantum codes and friends}{https://errorcorrectionzoo.org/code_graph#J\%7B\%22displayMode\%22\%3A\%22subset\%22\%2C\%22modeSubsetOptions\%22\%3A\%7B\%22codeIds\%22\%3A\%5B\%22branching_mera\%22\%2C\%22ea_pg_qldpc\%22\%2C\%22ea_galois_into_galois\%22\%2C\%22ea_galois_stabilizer\%22\%2C\%22ea_mds\%22\%2C\%22ea_qc_qldpc\%22\%2C\%22ea_qldpc\%22\%2C\%22ea_analog_stabilizer\%22\%2C\%22ea_oscillators\%22\%2C\%22ea_design_qldpc\%22\%2C\%22ea_quantum_lcd\%22\%2C\%22ea_quantum_convolutional\%22\%2C\%22ea_turbo\%22\%2C\%22ea_qubits_into_qubits\%22\%2C\%22eastab\%22\%2C\%22eaqecc\%22\%2C\%22ea_classical_into_quantum\%22\%2C\%22eacq\%22\%2C\%22eaoecc\%22\%2C\%22metopt\%22\%2C\%22maximal_entanglement_galois_stabilizer\%22\%2C\%22qecc\%22\%2C\%22quantum_polar\%22\%2C\%22ea_3_1_3-2\%22\%5D\%2C\%22reusePreviousLayoutPositions\%22\%3Afalse\%2C\%22showIntermediateConnectingNodes\%22\%3Atrue\%2C\%22connectingNodesMaxDepth\%22\%3A15\%2C\%22connectingNodesPathMaxLength\%22\%3A20\%2C\%22connectingNodesMaxExtraDepth\%22\%3A3\%2C\%22connectingNodesOnlyKeepPathsWithAdditionalLength\%22\%3A1\%2C\%22connectingNodesToDomainsAndKingdoms\%22\%3Afalse\%2C\%22connectingNodesEdgeLengthsByType\%22\%3A\%7B\%22primaryParent\%22\%3A1\%2C\%22secondaryParent\%22\%3A4\%2C\%22cousin\%22\%3A6\%7D\%2C\%22nodeIds\%22\%3A\%5B\%22k_galois_into_galois\%22\%2C\%22k_oscillators\%22\%2C\%22d_classical_into_quantum_domain\%22\%5D\%7D\%2C\%22highlightImportantNodes\%22\%3A\%7B\%22highlightImportantNodes\%22\%3Afalse\%2C\%22highlightPrimaryParents\%22\%3Afalse\%2C\%22highlightRootConnectingEdges\%22\%3Afalse\%7D\%7D}

\begingroup
\small
\eczhBreakableDashes
\renewcommand\arraystretch{1.05}
\edef\myxtraspc{\eczListAddVSpaceXtraXtraStretch}
\begin{tabularx}{\linewidth}{>{\raggedright\arraybackslash}p{\eczListColWidth{name}} >{\hsize=1.0000\hsize }X}
\toprule
\eczListColTitle{Code} & \eczListColTitle{Description} \\
\midrule
\endfirsthead
\toprule
\eczListColTitleContinued{Code} & \eczListColTitleContinued{Description} \\
\midrule
\endhead
\bottomrule
\endfoot
\eczhRefIndex{code:branching_mera}%
\eczhListValue{\flmRefsHyperref{code:branching_mera}{Branching MERA code}} & \eczhListValue{Qubit stabilizer code whose encoding circuit corresponds to a branching MERA \NoCaseChange{\protect\cite{cite543}} tensor network.
These codes generalize quantum polar codes by reinstating the disentanglers omitted in the branching-tree tensor-network construction \NoCaseChange{\protect\cite{cite400}}.}\\ 
\addlinespace[\myxtraspc]
\eczhRefIndex{code:ea_pg_qldpc}%
\eczhListValue{\flmRefsHyperref{code:ea_pg_qldpc}{EA FG-QLDPC code}} & \eczhListValue{One of several EA QLDPC code families constructed from finite-geometry LDPC (FG-LDPC) codes.
The construction includes families whose entanglement-consumption rate \(c/n\) decreases with block length \(n\) \NoCaseChange{\protect\cite{cite544}}.
Two such FG-based families require only one ebit (\(c=1\)) independent of code length \NoCaseChange{\protect\cite{cite544}}.}\\ 
\addlinespace[\myxtraspc]
\eczhRefIndex{code:ea_galois_into_galois}%
\eczhListValue{\flmRefsHyperref{code:ea_galois_into_galois}{EA Galois-qudit code}} & \eczhListValue{Galois-qudit code designed to utilize pre-shared entanglement between sender and receiver.}\\ 
\addlinespace[\myxtraspc]
\eczhRefIndex{code:ea_galois_stabilizer}%
\eczhListValue{\flmRefsHyperref{code:ea_galois_stabilizer}{EA Galois-qudit stabilizer code}} & \eczhListValue{A Galois-qudit stabilizer code constructed using a variation of the stabilizer formalism designed to utilize pre-shared entanglement between sender and receiver.
A code is typically denoted as \(\llbracket n,k;e\rrbracket _q\) or \(\llbracket n,k,d;e\rrbracket _q\), where \(d\) is the distance of the EA code and \(e\) is the number of required pre-shared maximally entangled Galois-qudit states.}\\ 
\addlinespace[\myxtraspc]
\eczhRefIndex{code:ea_mds}%
\eczhListValue{\flmRefsHyperref{code:ea_mds}{EA MDS code}} & \eczhListValue{EA Galois-qudit code whose parameters make the EAQECC Singleton bound \NoCaseChange{\protect\cite[{Thm. 6}]{cite545}} become an equality.}\\ 
\addlinespace[\myxtraspc]
\eczhRefIndex{code:ea_qc_qldpc}%
\eczhListValue{\flmRefsHyperref{code:ea_qc_qldpc}{EA QC-QLDPC code}} & \eczhListValue{One of several EA QLDPC code families constructed from classical QC-LDPC codes with girth at least six.
The entanglement assistance removes the dual-containing constraint in the CSS construction, avoiding many 4-cycles while retaining SPA decoding \NoCaseChange{\protect\cite{cite546}}.}\\ 
\addlinespace[\myxtraspc]
\eczhRefIndex{code:ea_qldpc}%
\eczhListValue{\flmRefsHyperref{code:ea_qldpc}{EA QLDPC code}} & \eczhListValue{EA qubit stabilizer code for which the number of sites participating in each stabilizer generator and the number of stabilizer generators that each site participates in are both bounded by a constant \(w\) as \(n\to\infty\).}\\ 
\addlinespace[\myxtraspc]
\eczhRefIndex{code:ea_analog_stabilizer}%
\eczhListValue{\flmRefsHyperref{code:ea_analog_stabilizer}{EA analog stabilizer code}} & \eczhListValue{Constructed using a variation of the analog stabilizer formalism designed to utilize pre-shared entanglement between sender and receiver.}\\ 
\addlinespace[\myxtraspc]
\eczhRefIndex{code:ea_oscillators}%
\eczhListValue{\flmRefsHyperref{code:ea_oscillators}{EA bosonic code}} & \eczhListValue{Bosonic code designed to utilize pre-shared entanglement between sender and receiver.}\\ 
\addlinespace[\myxtraspc]
\eczhRefIndex{code:ea_design_qldpc}%
\eczhListValue{\flmRefsHyperref{code:ea_design_qldpc}{EA combinatorial-design QLDPC code}} & \eczhListValue{One of several EA QLDPC code families constructed from combinatorial designs.}\\ 
\addlinespace[\myxtraspc]
\eczhRefIndex{code:ea_quantum_lcd}%
\eczhListValue{\flmRefsHyperref{code:ea_quantum_lcd}{EA quantum LCD code}} & \eczhListValue{An EA Galois-qudit stabilizer code constructed from an LCD code.
This family includes the first asymptotically good EA Galois-qudit codes.}\\ 
\addlinespace[\myxtraspc]
\eczhRefIndex{code:ea_quantum_convolutional}%
\eczhListValue{\flmRefsHyperref{code:ea_quantum_convolutional}{EA quantum convolutional code}} & \eczhListValue{A quantum convolutional code designed to utilize pre-shared entanglement between sender and receiver \NoCaseChange{\protect\cite{cite547,cite548,cite549}}.
Entanglement assistance removes the self-orthogonality constraint that ordinary quantum convolutional codes inherit from the stabilizer formalism, allowing arbitrary classical convolutional codes to be imported into quantum ones \NoCaseChange{\protect\cite{cite549}}.
In some constructions, the additional ebits also reduce the memory requirements of the encoding circuit \NoCaseChange{\protect\cite{cite550}}.}\\ 
\addlinespace[\myxtraspc]
\eczhRefIndex{code:ea_turbo}%
\eczhListValue{\flmRefsHyperref{code:ea_turbo}{EA quantum turbo code}} & \eczhListValue{A quantum turbo code which uses pre-shared entanglement.
This allows its encoder to be both recursive and non-catastrophic.}\\ 
\addlinespace[\myxtraspc]
\eczhRefIndex{code:ea_qubits_into_qubits}%
\eczhListValue{\flmRefsHyperref{code:ea_qubits_into_qubits}{EA qubit code}} & \eczhListValue{Qubit code designed to utilize pre-shared entanglement between sender and receiver.}\\ 
\addlinespace[\myxtraspc]
\eczhRefIndex{code:eastab}%
\eczhListValue{\flmRefsHyperref{code:eastab}{EA qubit stabilizer code}} & \eczhListValue{A code constructed using a variation of the stabilizer formalism designed to utilize pre-shared entanglement between sender and receiver.
A code is typically denoted as \(\llbracket n,k;e\rrbracket \) or \(\llbracket n,k,d;e\rrbracket \), where \(d\) is the distance of the EA code and \(e\) is the number of required pre-shared maximally entangled Bell states (ebits).
While other entangled states can be used, there is always a choice of generators such that Bell states suffice while still using the fewest ebits.}\\ 
\addlinespace[\myxtraspc]
\eczhRefIndex{code:eaqecc}%
\eczhListValue{\flmRefsHyperref{code:eaqecc}{Entanglement-assisted (EA) QECC}} & \eczhListValue{QECC whose encoding and decoding utilize pre-shared entanglement between sender and receiver.}\\ 
\addlinespace[\myxtraspc]
\eczhRefIndex{code:ea_classical_into_quantum}%
\eczhListValue{\flmRefsHyperref{code:ea_classical_into_quantum}{Entanglement-assisted (EA) c-q code}} & \eczhListValue{Classical-quantum code whose encoding and decoding utilize pre-shared entanglement between sender and receiver.
The sender encodes classical information into quantum systems sent through a quantum channel, while the receiver decodes using the channel outputs together with retained halves of pre-shared entangled states.}\\ 
\addlinespace[\myxtraspc]
\eczhRefIndex{code:eacq}%
\eczhListValue{\flmRefsHyperref{code:eacq}{Entanglement-assisted (EA) hybrid QECC}} & \eczhListValue{Code that encodes quantum and classical information and requires pre-shared
entanglement for transmission.}\\ 
\addlinespace[\myxtraspc]
\eczhRefIndex{code:eaoecc}%
\eczhListValue{\flmRefsHyperref{code:eaoecc}{Entanglement-assisted (EA) operator QECC}} & \eczhListValue{Subsystem QECC whose encoding and decoding utilize pre-shared entanglement between sender and receiver.}\\ 
\addlinespace[\myxtraspc]
\eczhRefIndex{code:metopt}%
\eczhListValue{\flmRefsHyperref{code:metopt}{Error-corrected sensing code}} & \eczhListValue{Code that can be obtained via an optimization procedure that ensures correction against a set \(\cal{E}\) of errors as well as guaranteeing optimal precision in locally estimating a parameter using a noiseless ancilla. For tensor-product spaces consisting of \(n\) subsystems (e.g., qubits, modular qudits, or Galois qudits), the procedure can yield a code whose parameter estimation precision satisfies \textit{Heisenberg scaling}, i.e., scales quadratically with the number \(n\) of subsystems.}\\ 
\addlinespace[\myxtraspc]
\eczhRefIndex{code:maximal_entanglement_galois_stabilizer}%
\eczhListValue{\flmRefsHyperref{code:maximal_entanglement_galois_stabilizer}{Maximal-entanglement EA Galois-qudit stabilizer code}} & \eczhListValue{An \(\llbracket n,k,d;e\rrbracket _q\) EA Galois-qudit stabilizer code for which \(e = n-k\), i.e., the number of required pre-shared maximally entangled Galois-qudit pairs saturates the defining maximal-entanglement condition.}\\ 
\addlinespace[\myxtraspc]
\eczhRefIndex{code:qecc}%
\eczhListValue{\flmRefsHyperref{code:qecc}{Quantum error-correcting code (QECC)}} & \eczhListValue{Encodes quantum information in a (\textit{logical}) subspace of a
(\textit{physical}) Hilbert space such that it is possible to recover said
information from errors that act as linear maps on the physical space.
The state space of a QECC is contained in the space of complex \(L^2\)-normalizable functions of some configuration space, which usually corresponds to the alphabet of a classical code.}\\ 
\addlinespace[\myxtraspc]
\eczhRefIndex{code:quantum_polar}%
\eczhListValue{\flmRefsHyperref{code:quantum_polar}{Quantum polar code}} & \eczhListValue{Entanglement-assisted CSS code utilized in a quantum polar coding scheme producing entangled pairs of qubits between sender and receiver. In such a scheme, the amplitude and phase information of a quantum state is handled in complementary fashion \NoCaseChange{\protect\cite{cite551}} using an encoding based on classical polar codes. Variants of the initial scheme have been developed for degradable channels \NoCaseChange{\protect\cite{cite552}} and extended to arbitrary channels \NoCaseChange{\protect\cite{cite553}}.}\\ 
\addlinespace[\myxtraspc]
\eczhRefIndex{code:ea_3_1_3-2}%
\eczhListValue{\flmRefsHyperref{code:ea_3_1_3-2}{\(\llbracket 3, 1, 3;2\rrbracket \) EA code}} & \eczhListValue{Distance-three EA stabilizer code encoding one logical qubit and using two ebits.
It is the smallest example of an EA code correcting an arbitrary single-qubit error.}\\ 
\end{tabularx}\endgroup
\eczcodelist{fermions_into_qubits}{Fermion-into-qubit encodings
}%

\eczhCodeListAutoDescription{All descendants of \flmRefsCref{code:fermions_into_qubits}.}%

\eczhIncludeCodeGraph{Bare}{scale=0.5}{\columnwidth}{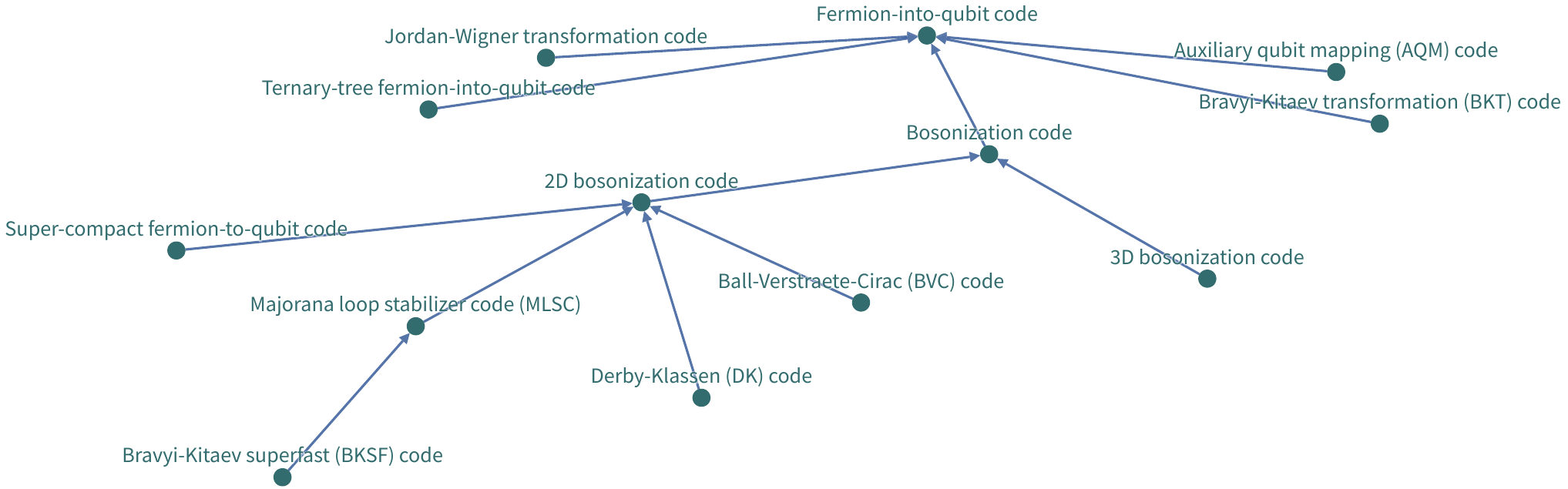}{Fermion-into-qubit encodings}{https://errorcorrectionzoo.org/code_graph#J\%7B\%22displayMode\%22\%3A\%22subset\%22\%2C\%22modeSubsetOptions\%22\%3A\%7B\%22codeIds\%22\%3A\%5B\%222d_bosonization\%22\%2C\%223d_bosonization\%22\%2C\%22aqm\%22\%2C\%22bvc\%22\%2C\%22bosonization\%22\%2C\%22bksf\%22\%2C\%22bkt\%22\%2C\%22derby_klassen\%22\%2C\%22fermions_into_qubits\%22\%2C\%22jw\%22\%2C\%22mlsc\%22\%2C\%22super_compact\%22\%2C\%22ternary_tree_fermion\%22\%5D\%2C\%22reusePreviousLayoutPositions\%22\%3Afalse\%2C\%22showIntermediateConnectingNodes\%22\%3Atrue\%2C\%22connectingNodesMaxDepth\%22\%3A15\%2C\%22connectingNodesPathMaxLength\%22\%3A20\%2C\%22connectingNodesMaxExtraDepth\%22\%3A3\%2C\%22connectingNodesOnlyKeepPathsWithAdditionalLength\%22\%3A1\%2C\%22connectingNodesToDomainsAndKingdoms\%22\%3Afalse\%2C\%22connectingNodesEdgeLengthsByType\%22\%3A\%7B\%22primaryParent\%22\%3A1\%2C\%22secondaryParent\%22\%3A4\%2C\%22cousin\%22\%3A6\%7D\%2C\%22nodeIds\%22\%3A\%5B\%5D\%7D\%2C\%22highlightImportantNodes\%22\%3A\%7B\%22highlightImportantNodes\%22\%3Afalse\%2C\%22highlightPrimaryParents\%22\%3Afalse\%2C\%22highlightRootConnectingEdges\%22\%3Afalse\%7D\%7D}

\begingroup
\small
\eczhBreakableDashes
\renewcommand\arraystretch{1.05}
\edef\myxtraspc{\eczListAddVSpaceXtraXtraStretch}
\begin{tabularx}{\linewidth}{>{\raggedright\arraybackslash}p{\eczListColWidth{name}} >{\hsize=1.0000\hsize }X}
\toprule
\eczListColTitle{Code} & \eczListColTitle{Description} \\
\midrule
\endfirsthead
\toprule
\eczListColTitleContinued{Code} & \eczListColTitleContinued{Description} \\
\midrule
\endhead
\bottomrule
\endfoot
\eczhRefIndex{code:2d_bosonization}%
\eczhListValue{\flmRefsHyperref{code:2d_bosonization}{2D bosonization code}} & \eczhListValue{A mapping between a 2D lattice quadratic Hamiltonian of Majorana modes and a 2D lattice of qubits.
The original exact 2D bosonization code \NoCaseChange{\protect\cite{cite403}} is a stabilizer code whose generators are products of plaquettes and stars of the surface code, with gauge constraints that project onto a toric-code-like subspace with emergent fermions \NoCaseChange{\protect\cite{cite403,cite404}}.
Finite-depth generalized local unitary Clifford circuits generate a family of equivalent local encodings with qubit-to-fermion ratio \(r = 1 + \frac{1}{2k}\) for any positive integer \(k\); the square-lattice compact encoding with \(r=1.5\) and the super-compact encoding with \(r=1.25\) are explicit examples \NoCaseChange{\protect\cite{cite404}}.}\\ 
\addlinespace[\myxtraspc]
\eczhRefIndex{code:3d_bosonization}%
\eczhListValue{\flmRefsHyperref{code:3d_bosonization}{3D bosonization code}} & \eczhListValue{A mapping from a 3D lattice quadratic Hamiltonian of Majorana modes to a lattice of qubits which realizes a \(\mathbb{Z}_2\) gauge theory with a particular Gauss law.}\\ 
\addlinespace[\myxtraspc]
\eczhRefIndex{code:aqm}%
\eczhListValue{\flmRefsHyperref{code:aqm}{Auxiliary qubit mapping (AQM) code}} & \eczhListValue{A concatenation of the JW transformation code with a qubit stabilizer code.}\\ 
\addlinespace[\myxtraspc]
\eczhRefIndex{code:bvc}%
\eczhListValue{\flmRefsHyperref{code:bvc}{Ball-Verstraete-Cirac (BVC) code}} & \eczhListValue{A 2D fermion-into-qubit encoding that builds upon the JW transformation by eliminating the weight-\(O(n)\) non-local \(Z\)-type string at the expense of introducing an auxiliary qubit per site and local gauge constraints.
See \NoCaseChange{\protect\cite[{Sec. IV.B}]{cite404}} for details.}\\ 
\addlinespace[\myxtraspc]
\eczhRefIndex{code:bosonization}%
\eczhListValue{\flmRefsHyperref{code:bosonization}{Bosonization code}} & \eczhListValue{A mapping that maps a \(D\)-dimensional lattice quadratic Hamiltonian of Majorana modes into a lattice of qubits.
The resulting qubit code can realize various topological phases, depending on the initial Majorana-mode Hamiltonian and its symmetries.}\\ 
\addlinespace[\myxtraspc]
\eczhRefIndex{code:bksf}%
\eczhListValue{\flmRefsHyperref{code:bksf}{Bravyi-Kitaev superfast (BKSF) code}} & \eczhListValue{A single-error-detecting fermion-into-qubit encoding defined on a 2D qubit lattice whose stabilizers are associated with loops in the lattice.
For the square-lattice edge ordering used in Ref. \NoCaseChange{\protect\cite{cite404}}, the BKSF logical operators coincide with exact 2D bosonization on the dual lattice after relabeling \(X\) and \(Y\).
The code can be generalized to a single error-correcting code (i.e., with distance three) on graphs of degree \(\geq 6\) \NoCaseChange{\protect\cite{cite408}}.}\\ 
\addlinespace[\myxtraspc]
\eczhRefIndex{code:bkt}%
\eczhListValue{\flmRefsHyperref{code:bkt}{Bravyi-Kitaev transformation (BKT) code}} & \eczhListValue{A fermion-into-qubit encoding that maps Majorana operators into Pauli strings of weight \(\lceil \log_2(n+1) \rceil\).
The code can be reformulated in terms of Fenwick trees \NoCaseChange{\protect\cite{cite554}}, and the Pauli-string weight can be further optimized to yield the \textit{segmented Bravyi-Kitaev (SBK) transformation code} \NoCaseChange{\protect\cite{cite555}} (see also Ref. \NoCaseChange{\protect\cite{cite556}}).}\\ 
\addlinespace[\myxtraspc]
\eczhRefIndex{code:derby_klassen}%
\eczhListValue{\flmRefsHyperref{code:derby_klassen}{Derby-Klassen (DK) code}} & \eczhListValue{A fermion-into-qubit code defined on regular tilings with maximum degree 4 whose stabilizers are associated with loops in the tiling.
The code outperforms several other encodings in terms of encoding rate \NoCaseChange{\protect\cite[{Table I}]{cite412}}.
It has been extended for models with several modes per site \NoCaseChange{\protect\cite{cite413}}.}\\ 
\addlinespace[\myxtraspc]
\eczhRefIndex{code:fermions_into_qubits}%
\eczhListValue{\flmRefsHyperref{code:fermions_into_qubits}{Fermion-into-qubit code}} & \eczhListValue{Qubit stabilizer code encoding a logical fermionic Hilbert space into a physical space of \(n\) qubits.
Such codes are primarily intended for simulating fermionic systems on quantum computers, and some of them have error-detecting, correcting, and transmuting properties.}\\ 
\addlinespace[\myxtraspc]
\eczhRefIndex{code:jw}%
\eczhListValue{\flmRefsHyperref{code:jw}{Jordan-Wigner transformation code}} & \eczhListValue{A mapping between qubit Pauli strings and Majorana operators that can be thought of as a trivial \(\llbracket n,n\rrbracket \) code.
The mapping is best described as converting a chain of \(n\) qubits into a chain of \(2n\) Majorana modes (i.e., \(n\) fermionic modes).
It maps Majorana operators into Pauli strings of weight \(O(n)\).}\\ 
\addlinespace[\myxtraspc]
\eczhRefIndex{code:mlsc}%
\eczhListValue{\flmRefsHyperref{code:mlsc}{Majorana loop stabilizer code (MLSC)}} & \eczhListValue{A single-error-correcting fermion-into-qubit encoding defined on a 2D qubit lattice whose stabilizers are associated with loops in the lattice.}\\ 
\addlinespace[\myxtraspc]
\eczhRefIndex{code:super_compact}%
\eczhListValue{\flmRefsHyperref{code:super_compact}{Super-compact fermion-to-qubit code}} & \eczhListValue{A 2D fermion-into-qubit encoding on the square lattice obtained from exact 2D bosonization by a finite-depth generalized local unitary Clifford circuit, followed by re-pairing of Majorana modes and a slight lattice deformation.
The code uses \(1.25\) qubits per fermion, improving on the square-lattice compact encoding with ratio \(r=1.5\).
Its fermion-parity, hopping, and stabilizer operators have weights \(1\)-\(2\), \(2\)-\(6\), and \(12\), respectively \NoCaseChange{\protect\cite[{Table I}]{cite404}}.}\\ 
\addlinespace[\myxtraspc]
\eczhRefIndex{code:ternary_tree_fermion}%
\eczhListValue{\flmRefsHyperref{code:ternary_tree_fermion}{Ternary-tree fermion-into-qubit code}} & \eczhListValue{A fermion-into-qubit encoding defined on ternary trees that maps Majorana operators into Pauli strings of weight \(\lceil \log_3 (2n+1) \rceil\).}\\ 
\end{tabularx}\endgroup
\eczcodelist{fermion}{Fermionic codes
}%

\eczhCodeListAutoDescription{Union of:
\begin{itemize}\item codes that are descendants of \flmRefsCref{code:fermions}
\item codes that are descendants of \flmRefsCref{code:majorana_subsystem}
\end{itemize}}%

\eczhIncludeCodeGraph{Bare}{scale=0.5}{\columnwidth}{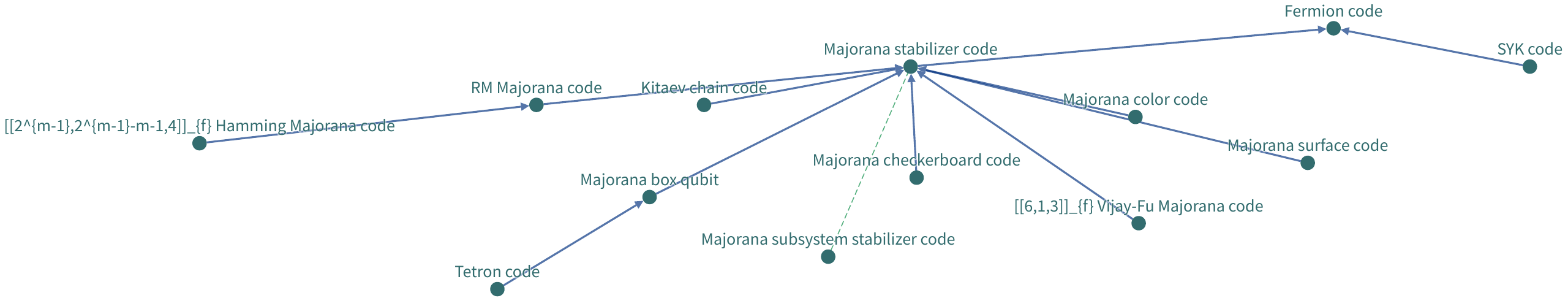}{Fermionic codes}{https://errorcorrectionzoo.org/code_graph#J\%7B\%22displayMode\%22\%3A\%22subset\%22\%2C\%22modeSubsetOptions\%22\%3A\%7B\%22codeIds\%22\%3A\%5B\%22fermions\%22\%2C\%22kitaev_chain\%22\%2C\%22mbq\%22\%2C\%22majorana_checkerboard\%22\%2C\%22majorana_color\%22\%2C\%22majorana_stab\%22\%2C\%22majorana_subsystem\%22\%2C\%22majorana_surface\%22\%2C\%22majorana_reed_muller\%22\%2C\%22syk\%22\%2C\%22tetron\%22\%2C\%22majorana_hamming\%22\%2C\%22majorana_6_1_3\%22\%5D\%2C\%22reusePreviousLayoutPositions\%22\%3Afalse\%2C\%22showIntermediateConnectingNodes\%22\%3Atrue\%2C\%22connectingNodesMaxDepth\%22\%3A15\%2C\%22connectingNodesPathMaxLength\%22\%3A20\%2C\%22connectingNodesMaxExtraDepth\%22\%3A3\%2C\%22connectingNodesOnlyKeepPathsWithAdditionalLength\%22\%3A1\%2C\%22connectingNodesToDomainsAndKingdoms\%22\%3Afalse\%2C\%22connectingNodesEdgeLengthsByType\%22\%3A\%7B\%22primaryParent\%22\%3A1\%2C\%22secondaryParent\%22\%3A4\%2C\%22cousin\%22\%3A6\%7D\%2C\%22nodeIds\%22\%3A\%5B\%5D\%7D\%2C\%22highlightImportantNodes\%22\%3A\%7B\%22highlightImportantNodes\%22\%3Afalse\%2C\%22highlightPrimaryParents\%22\%3Afalse\%2C\%22highlightRootConnectingEdges\%22\%3Afalse\%7D\%7D}

\begingroup
\small
\eczhBreakableDashes
\renewcommand\arraystretch{1.05}
\edef\myxtraspc{\eczListAddVSpaceXtraXtraStretch}
\begin{tabularx}{\linewidth}{>{\raggedright\arraybackslash}p{\eczListColWidth{name}} >{\hsize=1.0000\hsize }X}
\toprule
\eczListColTitle{Code} & \eczListColTitle{Description} \\
\midrule
\endfirsthead
\toprule
\eczListColTitleContinued{Code} & \eczListColTitleContinued{Description} \\
\midrule
\endhead
\bottomrule
\endfoot
\eczhRefIndex{code:fermions}%
\eczhListValue{\flmRefsHyperref{code:fermions}{Fermion code}} & \eczhListValue{Finite-dimensional quantum error-correcting code encoding a logical qudit or fermionic Hilbert space into a physical Fock space of fermionic modes.
Codes are typically described using Majorana operators, which are linear combinations of fermionic creation and annihilation operators \NoCaseChange{\protect\cite{cite557}}.
Majorana operators may either be considered individually or paired in various ways into creation and annihilation operators to yield fermionic modes.
They form a Clifford algebra and can be interpreted as Ising anyons in certain contexts.}\\ 
\addlinespace[\myxtraspc]
\eczhRefIndex{code:kitaev_chain}%
\eczhListValue{\flmRefsHyperref{code:kitaev_chain}{Kitaev chain code}} & \eczhListValue{A Majorana stabilizer code obtained from the ground-state subspace of the Kitaev Majorana chain in its fermionic topological phase \NoCaseChange{\protect\cite{cite558}}. Its codespace is stabilized by nearest-neighbor Majorana bilinears, while two unpaired edge Majoranas furnish one logical fermionic mode. Under parity-preserving noise, it behaves as the Majorana analogue of the repetition code \NoCaseChange{\protect\cite{cite559}}.}\\ 
\addlinespace[\myxtraspc]
\eczhRefIndex{code:mbq}%
\eczhListValue{\flmRefsHyperref{code:mbq}{Majorana box qubit}} & \eczhListValue{A family of Majorana stabilizer codes obtained by fixing the total fermion parity of \(n\) fermionic modes, equivalently \(2n\) Majorana zero modes, within the ground-state subspace of \(n\) Kitaev Majorana chain Hamiltonians.
The resulting positive-parity subspace encodes \(n-1\) logical qubits and has Majorana distance \(2\).}\\ 
\addlinespace[\myxtraspc]
\eczhRefIndex{code:majorana_checkerboard}%
\eczhListValue{\flmRefsHyperref{code:majorana_checkerboard}{Majorana checkerboard code}} & \eczhListValue{A Majorana analogue of the X-cube model defined on a cubic lattice.
The code admits weight-eight Majorana stabilizer generators on the eight vertices of each cube of a checkerboard sublattice.}\\ 
\addlinespace[\myxtraspc]
\eczhRefIndex{code:majorana_color}%
\eczhListValue{\flmRefsHyperref{code:majorana_color}{Majorana color code}} & \eczhListValue{A fermionic analogue of a 2D color code.}\\ 
\addlinespace[\myxtraspc]
\eczhRefIndex{code:majorana_stab}%
\eczhListValue{\flmRefsHyperref{code:majorana_stab}{Majorana stabilizer code}} & \eczhListValue{A stabilizer code whose stabilizers are products of an even number of Majorana fermion operators, analogous to Pauli strings for a traditional stabilizer code and referred to as \textit{Majorana stabilizers}.
The codespace is the mutual \(+1\) eigenspace of all Majorana stabilizers.}\\ 
\addlinespace[\myxtraspc]
\eczhRefIndex{code:majorana_subsystem}%
\eczhListValue{\flmRefsHyperref{code:majorana_subsystem}{Majorana subsystem stabilizer code}} & \eczhListValue{A Majorana subsystem code with some of its logical qubits denoted as \textit{gauge} qubits and not used for storage of logical information.}\\ 
\addlinespace[\myxtraspc]
\eczhRefIndex{code:majorana_surface}%
\eczhListValue{\flmRefsHyperref{code:majorana_surface}{Majorana surface code}} & \eczhListValue{Fermionic analogue of the surface code defined on a three-colorable 2D tiling whose face operators are non-overlapping even-Majorana stabilizers.
Open patches with four or six alternating colored boundaries encode logical tetrons or hexons.
The uniform 4.8.8, 6.6.6, and 4.6.12 tilings yield families with tetron, hexon, or dodecon building blocks and with twist-based lattice surgery supporting minimal-overhead logical Clifford gates \NoCaseChange{\protect\cite{cite402}}.}\\ 
\addlinespace[\myxtraspc]
\eczhRefIndex{code:majorana_reed_muller}%
\eczhListValue{\flmRefsHyperref{code:majorana_reed_muller}{RM Majorana code}} & \eczhListValue{A Majorana stabilizer code constructed from a self-orthogonal RM code.
These codes have the additional property that the global fermion parity is fixed in the codespace. 
Logical measurements are reduced to parity measurements of some subset of Majorana fermions in the code.}\\ 
\addlinespace[\myxtraspc]
\eczhRefIndex{code:syk}%
\eczhListValue{\flmRefsHyperref{code:syk}{SYK code}} & \eczhListValue{Approximate \(n\)-fermionic code whose codewords are low-energy states of the Sachdev-Ye-Kitaev (SYK) Hamiltonian \NoCaseChange{\protect\cite{cite560,cite561}} or other low-rank SYK models \NoCaseChange{\protect\cite{cite562,cite563}}.}\\ 
\addlinespace[\myxtraspc]
\eczhRefIndex{code:tetron}%
\eczhListValue{\flmRefsHyperref{code:tetron}{Tetron code}} & \eczhListValue{A \(\llbracket 2,1,2\rrbracket _{f}\) Majorana box qubit encoding a logical qubit into four Majorana modes, equivalently into the fixed-total-parity sector of two physical fermionic modes.
Four Majorana zero modes are the smallest aggregate that supports a qubit in a fixed fermion-parity sector \NoCaseChange{\protect\cite{cite401}}.
This code can be concatenated with various qubit codes such as surface codes and color codes.
Four-boundary Majorana surface-code patches are logical tetrons, i.e., higher-distance analogues of this physical tetron block \NoCaseChange{\protect\cite{cite402}}.}\\ 
\addlinespace[\myxtraspc]
\eczhRefIndex{code:majorana_hamming}%
\eczhListValue{\flmRefsHyperref{code:majorana_hamming}{\(\llbracket 2^{m-1},2^{m-1}-m-1,4\rrbracket _{f}\) Hamming Majorana code}} & \eczhListValue{A member of the \(\llbracket 2^{m-1},2^{m-1}-m-1,4\rrbracket _{f}\) family of Majorana stabilizer codes for \(m \geq 3\) constructed from a self-orthogonal first-order RM code (whose dual is the extended Hamming code).
A shortened \(\llbracket 2^{m-1}-1,2^{m-1}-m-2,3\rrbracket _{f}\) version can also be defined \NoCaseChange{\protect\cite[{Prop. 2.5.1}]{cite564}}.
The logical subspace of the \(\llbracket 8,3,4\rrbracket _{f}\) Hamming Majorana code is a Cartan subspace of the \(E_8\) Lie algebra \NoCaseChange{\protect\cite{cite565}}.}\\ 
\addlinespace[\myxtraspc]
\eczhRefIndex{code:majorana_6_1_3}%
\eczhListValue{\flmRefsHyperref{code:majorana_6_1_3}{\(\llbracket 6,1,3\rrbracket _{f}\) Vijay-Fu Majorana code}} & \eczhListValue{A Majorana stabilizer code encoding a logical fermion into six physical fermions.
This code is the shortest code correcting single fermion-parity flips \NoCaseChange{\protect\cite{cite566}}.}\\ 
\end{tabularx}\endgroup
\eczcodelist{fracton}{Fracton codes and friends
}%

\eczhCodeListAutoDescription{All descendants and cousins of \flmRefsCref{code:fracton}.}%

\eczhIncludeCodeGraph{Bare}{scale=0.5}{\columnwidth}{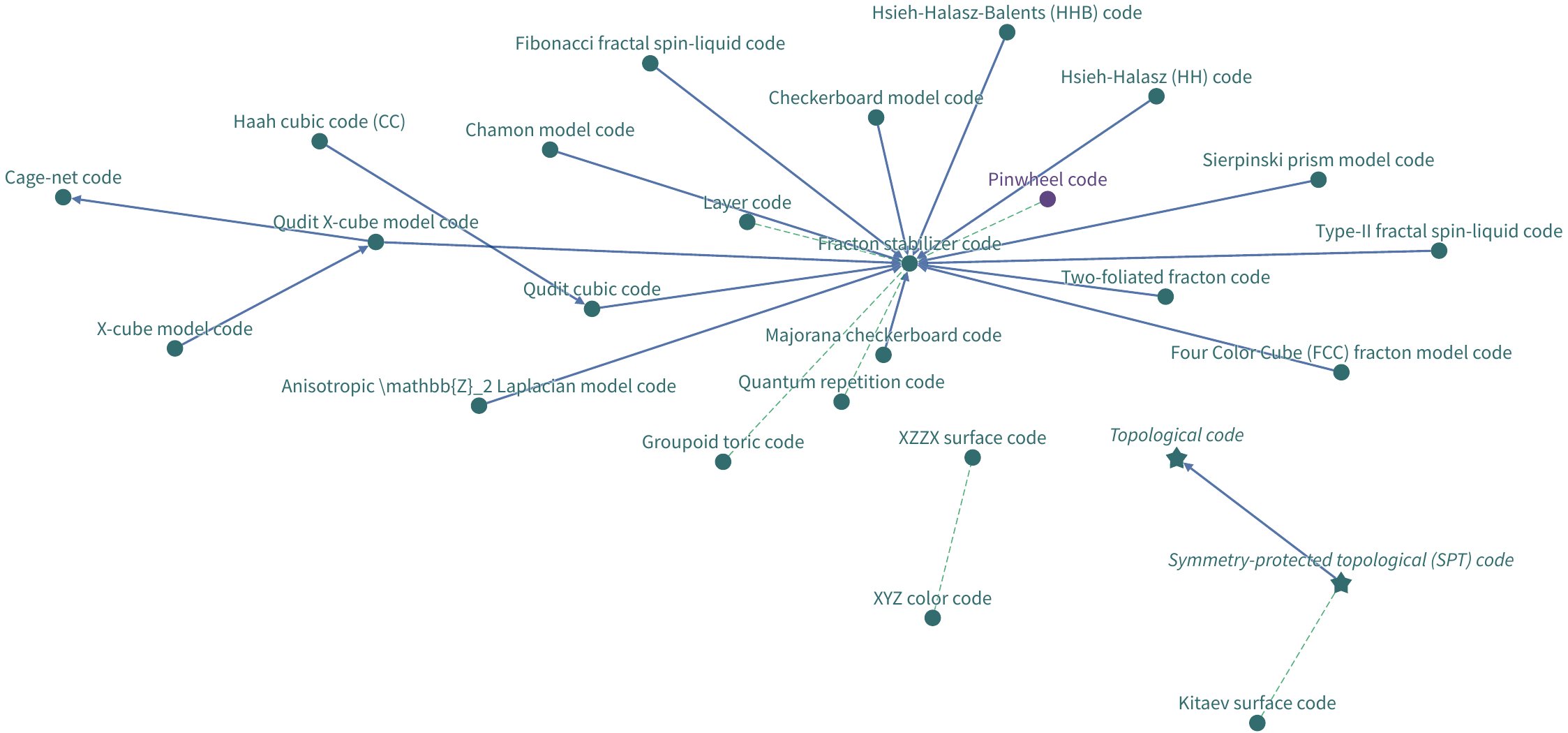}{Fracton codes and friends}{https://errorcorrectionzoo.org/code_graph#J\%7B\%22displayMode\%22\%3A\%22subset\%22\%2C\%22modeSubsetOptions\%22\%3A\%7B\%22codeIds\%22\%3A\%5B\%22anisotropic_z2_laplacian\%22\%2C\%22cage_net\%22\%2C\%22chamon\%22\%2C\%22checkerboard\%22\%2C\%22fibonacci_fractal_liquid\%22\%2C\%22fcc_fracton\%22\%2C\%22fracton\%22\%2C\%22groupoid_surface\%22\%2C\%22haah_cubic\%22\%2C\%22hh_fracton\%22\%2C\%22hhb_fracton\%22\%2C\%22surface\%22\%2C\%22layer\%22\%2C\%22majorana_checkerboard\%22\%2C\%22pinwheel\%22\%2C\%22quantum_repetition\%22\%2C\%22qudit_xcube\%22\%2C\%22qudit_cubic\%22\%2C\%22sierpinsky_fractal_liquid\%22\%2C\%22spt\%22\%2C\%22topological\%22\%2C\%22two_foliated\%22\%2C\%22fractal_liquid\%22\%2C\%22xcube\%22\%2C\%22xyz_color\%22\%2C\%22xzzx\%22\%5D\%2C\%22reusePreviousLayoutPositions\%22\%3Afalse\%2C\%22showIntermediateConnectingNodes\%22\%3Atrue\%2C\%22connectingNodesMaxDepth\%22\%3A15\%2C\%22connectingNodesPathMaxLength\%22\%3A20\%2C\%22connectingNodesMaxExtraDepth\%22\%3A3\%2C\%22connectingNodesOnlyKeepPathsWithAdditionalLength\%22\%3A1\%2C\%22connectingNodesToDomainsAndKingdoms\%22\%3Afalse\%2C\%22connectingNodesEdgeLengthsByType\%22\%3A\%7B\%22primaryParent\%22\%3A1\%2C\%22secondaryParent\%22\%3A4\%2C\%22cousin\%22\%3A6\%7D\%2C\%22nodeIds\%22\%3A\%5B\%5D\%7D\%2C\%22highlightImportantNodes\%22\%3A\%7B\%22highlightImportantNodes\%22\%3Afalse\%2C\%22highlightPrimaryParents\%22\%3Afalse\%2C\%22highlightRootConnectingEdges\%22\%3Afalse\%7D\%7D}

\begingroup
\small
\eczhBreakableDashes
\renewcommand\arraystretch{1.05}
\edef\myxtraspc{\eczListAddVSpaceXtraXtraStretch}
\begin{tabularx}{\linewidth}{>{\raggedright\arraybackslash}p{\eczListColWidth{name}} >{\hsize=1.0000\hsize }X}
\toprule
\eczListColTitle{Code} & \eczListColTitle{Description} \\
\midrule
\endfirsthead
\toprule
\eczListColTitleContinued{Code} & \eczListColTitleContinued{Description} \\
\midrule
\endhead
\bottomrule
\endfoot
\eczhRefIndex{code:anisotropic_z2_laplacian}%
\eczhListValue{\flmRefsHyperref{code:anisotropic_z2_laplacian}{Anisotropic \(\mathbb{Z}_2\) Laplacian model code}} & \eczhListValue{A graph-based analogue of a Type-I fracton phase with lineon-like restricted mobility \NoCaseChange{\protect\cite{cite460,cite461}}.}\\ 
\addlinespace[\myxtraspc]
\eczhRefIndex{code:cage_net}%
\eczhListValue{\flmRefsHyperref{code:cage_net}{Cage-net code}} & \eczhListValue{A commuting-projector code family obtained by coupling layers of two-dimensional topological orders and condensing extended one-dimensional flux strings \NoCaseChange{\protect\cite{cite567}}.
A modern lattice realization starts from isotropic stacks of \(G\)-graded string-net models \NoCaseChange{\protect\cite{cite568}}.
The family includes stabilizer examples such as the \(\mathbb{Z}_2\) string-membrane-net realization of the X-cube model \NoCaseChange{\protect\cite{cite569}} as well as non-stabilizer examples with non-Abelian restricted-mobility excitations \NoCaseChange{\protect\cite{cite567}}.
String-membrane-net and cage-net constructions can realize the same fracton phases; for isotropic stacks, this equivalence can be understood via generalized local unitaries \NoCaseChange{\protect\cite{cite568}}.
The cage-net construction can be used to realize various fracton phases, stabilizer and otherwise.}\\ 
\addlinespace[\myxtraspc]
\eczhRefIndex{code:chamon}%
\eczhListValue{\flmRefsHyperref{code:chamon}{Chamon model code}} & \eczhListValue{A foliated type-I fracton non-CSS code defined on a cubic lattice using one weight-eight stabilizer generator acting on the eight vertices of each cube in the lattice \NoCaseChange{\protect\cite[{Eq. (D38)}]{cite456}}.}\\ 
\addlinespace[\myxtraspc]
\eczhRefIndex{code:checkerboard}%
\eczhListValue{\flmRefsHyperref{code:checkerboard}{Checkerboard model code}} & \eczhListValue{A foliated type-I fracton code defined on a cubic lattice that admits weight-eight  \(X\)- and \(Z\)-type stabilizer generators on the eight vertices of each cube in the lattice.
A tetrahedral Ising model can be used to obtain the checkerboard model by gauging \NoCaseChange{\protect\cite{cite462,cite463,cite233,cite464,cite465,cite466,cite467,cite468,cite469,cite470}} its subsystem symmetry \NoCaseChange{\protect\cite{cite233}}.
In that construction, the checkerboard model is self-dual under exchange of \(X\)- and \(Z\)-type stabilizers, and its composites include dimension-1 and dimension-2 excitations, i.e., lineons and planons in later terminology, with anyonic mutual and self-statistics \NoCaseChange{\protect\cite{cite233}}.}\\ 
\addlinespace[\myxtraspc]
\eczhRefIndex{code:fibonacci_fractal_liquid}%
\eczhListValue{\flmRefsHyperref{code:fibonacci_fractal_liquid}{Fibonacci fractal spin-liquid code}} & \eczhListValue{A fractal type-I fracton CSS code defined on a cubic lattice \NoCaseChange{\protect\cite[{Eq. (D23)}]{cite456}}.}\\ 
\addlinespace[\myxtraspc]
\eczhRefIndex{code:fcc_fracton}%
\eczhListValue{\flmRefsHyperref{code:fcc_fracton}{Four Color Cube (FCC) fracton model code}} & \eczhListValue{A fracton code obtained from four coupled X-cube models using p-membrane condensation.
A modular-qudit generalization has been proposed \NoCaseChange{\protect\cite{cite474}}.}\\ 
\addlinespace[\myxtraspc]
\eczhRefIndex{code:fracton}%
\eczhListValue{\flmRefsHyperref{code:fracton}{Fracton stabilizer code}} & \eczhListValue{A 3D modular-qudit stabilizer code whose codewords make up the ground-state space of a Hamiltonian in a fracton phase.
Unlike topological phases, whose excitations can move in any direction, fracton phases are characterized by excitations whose movement is restricted.}\\ 
\addlinespace[\myxtraspc]
\eczhRefIndex{code:groupoid_surface}%
\eczhListValue{\flmRefsHyperref{code:groupoid_surface}{Groupoid toric code}} & \eczhListValue{Extension of the Kitaev surface code from Abelian groups to groupoids, i.e., multi-fusion categories in which every morphism is an isomorphism \NoCaseChange{\protect\cite{cite570}}.
Some models admit fracton-like features such as extensive ground-state degeneracy and excitations with restricted mobility.
The robustness of these features has not yet been established.}\\ 
\addlinespace[\myxtraspc]
\eczhRefIndex{code:haah_cubic}%
\eczhListValue{\flmRefsHyperref{code:haah_cubic}{Haah cubic code (CC)}} & \eczhListValue{A 3D lattice stabilizer code on a length-\(L\) cubic lattice with one or two qubits per site.
Admits two types of stabilizer generators with support on each cube of the lattice.
In the non-CSS case, these two are related by spatial inversion.
For CSS codes, we require that the product of all corner operators is the identity.
We lastly require that there are no non-trivial string operators, meaning that single-site operators are a phase, and any period one logical operator \(l \in \mathsf{S}^{\perp}\) is just a phase.}\\ 
\addlinespace[\myxtraspc]
\eczhRefIndex{code:hh_fracton}%
\eczhListValue{\flmRefsHyperref{code:hh_fracton}{Hsieh-Halasz (HH) code}} & \eczhListValue{Member of one of two families of fracton codes, named HH-I and HH-II, defined on a cubic lattice with two qubits per site.
HH-I (HH-II) is a CSS (non-CSS) stabilizer code family, with the former identified as a foliated type-I fracton code that is decomposable into two separate lattice models \NoCaseChange{\protect\cite{cite456}}.
The sorting analysis of Ref. \NoCaseChange{\protect\cite{cite456}} leaves HH-II inconclusive, consistent with either a fractal type-I or a type-II fracton phase.}\\ 
\addlinespace[\myxtraspc]
\eczhRefIndex{code:hhb_fracton}%
\eczhListValue{\flmRefsHyperref{code:hhb_fracton}{Hsieh-Halasz-Balents (HHB) code}} & \eczhListValue{Member of one of two families of fracton codes, named HHB model A and B, defined on a cubic lattice with two qubits per site.
Both are expected to be foliated type-I fracton codes \NoCaseChange{\protect\cite[{Eqs. (D42-D43)}]{cite456}}.}\\ 
\addlinespace[\myxtraspc]
\eczhRefIndex{code:surface}%
\eczhListValue{\flmRefsHyperref{code:surface}{Kitaev surface code}} & \eczhListValue{A family of Abelian topological \flmRefsHyperref{code:css}{CSS stabilizer} codes
whose generators are few-body \(X\)-type and \(Z\)-type Pauli strings
associated to the stars and plaquettes, respectively, of a cellulation of a
two-dimensional surface (with a qubit located at each edge of the
cellulation).
Codewords correspond to ground states of the surface code Hamiltonian, and error operators create or annihilate pairs of anyonic charges or vortices.}\\ 
\addlinespace[\myxtraspc]
\eczhRefIndex{code:layer}%
\eczhListValue{\flmRefsHyperref{code:layer}{Layer code}} & \eczhListValue{Member of a family of qubit QLDPC CSS codes with stabilizer generator weights \(\leq 6\) that are obtained by coupling layers of 2D surface codes according to the Tanner graph of a QLDPC code (or a more general qubit stabilizer code).
Geometric locality is maintained because, instead of being concatenated, each pair of parallel surface-code squares is fused (or quasi-concatenated) with perpendicular surface-code squares via lattice surgery.}\\ 
\addlinespace[\myxtraspc]
\eczhRefIndex{code:majorana_checkerboard}%
\eczhListValue{\flmRefsHyperref{code:majorana_checkerboard}{Majorana checkerboard code}} & \eczhListValue{A Majorana analogue of the X-cube model defined on a cubic lattice.
The code admits weight-eight Majorana stabilizer generators on the eight vertices of each cube of a checkerboard sublattice.}\\ 
\addlinespace[\myxtraspc]
\eczhRefIndex{code:pinwheel}%
\eczhListValue{\flmRefsHyperref{code:pinwheel}{Pinwheel code}} & \eczhListValue{A geometrically local binary LDPC code defined on planar graphs obtained from the pinwheel tiling \NoCaseChange{\protect\cite{cite93}}.
Both bits and checks live on vertices of the graph.
If \(L_N\) is the graph Laplacian at generation \(N\), the undepleted check matrix is \(\tilde H_N=(L_N-\mathbb{I})\bmod 2\), and the actual parity-check matrix \(H_N\) is obtained by removing an evenly spaced fraction of boundary checks.}\\ 
\addlinespace[\myxtraspc]
\eczhRefIndex{code:quantum_repetition}%
\eczhListValue{\flmRefsHyperref{code:quantum_repetition}{Quantum repetition code}} & \eczhListValue{Encodes \(1\) qubit into \(n\) qubits according to \(|0\rangle\to|\phi_0\rangle^{\otimes n}\) and \(|1\rangle\to|\phi_1\rangle^{\otimes n}\). The code is called a \textit{bit-flip} code when \(|\phi_i\rangle = |i\rangle\), and a \textit{phase-flip} code when \(|\phi_0\rangle = |+\rangle\) and \(|\phi_1\rangle = |-\rangle\).
This repetition-style encoding does not clone an arbitrary quantum state; instead, it extends the copying of computational-basis states linearly to entangled codewords  \NoCaseChange{\protect\cite[{Ch. 2}]{cite398}}.}\\ 
\addlinespace[\myxtraspc]
\eczhRefIndex{code:qudit_xcube}%
\eczhListValue{\flmRefsHyperref{code:qudit_xcube}{Qudit X-cube model code}} & \eczhListValue{Generalization of the X-cube model code to modular qudits.}\\ 
\addlinespace[\myxtraspc]
\eczhRefIndex{code:qudit_cubic}%
\eczhListValue{\flmRefsHyperref{code:qudit_cubic}{Qudit cubic code}} & \eczhListValue{Generalization of the Haah cubic code to modular qudits.}\\ 
\addlinespace[\myxtraspc]
\eczhRefIndex{code:sierpinsky_fractal_liquid}%
\eczhListValue{\flmRefsHyperref{code:sierpinsky_fractal_liquid}{Sierpinski prism model code}} & \eczhListValue{A fractal type-I fracton CSS code defined on a cubic lattice \NoCaseChange{\protect\cite[{Eq. (D22)}]{cite456}}.
The code admits an excitation-moving operator shaped like a Sierpinski triangle \NoCaseChange{\protect\cite[{Fig. 2}]{cite456}}.}\\ 
\addlinespace[\myxtraspc]
\eczhRefIndex{code:spt}%
\eczhListValue{\flmRefsHyperref{code:spt}{Symmetry-protected topological (SPT) code}} & \eczhListValue{A code whose codewords form the ground-state or low-energy subspace of a code Hamiltonian realizing symmetry-protected topological (SPT) order.}\\ 
\addlinespace[\myxtraspc]
\eczhRefIndex{code:topological}%
\eczhListValue{\flmRefsHyperref{code:topological}{Topological code}} & \eczhListValue{A code whose codewords form the ground-state or low-energy subspace of a (typically geometrically local) code Hamiltonian realizing a topological phase.
A topological phase may be \textit{bosonic} or \textit{fermionic}, i.e., constructed out of underlying subsystems whose operators commute or anti-commute with each other, respectively.
Unless otherwise noted, the phases discussed are bosonic.}\\ 
\addlinespace[\myxtraspc]
\eczhRefIndex{code:two_foliated}%
\eczhListValue{\flmRefsHyperref{code:two_foliated}{Two-foliated fracton code}} & \eczhListValue{A type-I fracton code obtained by gauging \NoCaseChange{\protect\cite{cite462,cite463,cite233,cite464,cite465,cite466,cite467,cite468,cite469,cite470}} a 3D paramagnet with planar subsystem symmetries in two directions.
In that construction, the gauge charges are lineons and the flux excitations are also lineons moving in the same direction, yielding the anisotropic lineon model \NoCaseChange{\protect\cite[{Sec. 4.1.2}]{cite467}}.}\\ 
\addlinespace[\myxtraspc]
\eczhRefIndex{code:fractal_liquid}%
\eczhListValue{\flmRefsHyperref{code:fractal_liquid}{Type-II fractal spin-liquid code}} & \eczhListValue{A type-II fracton prime-qudit CSS code defined on a cubic lattice \NoCaseChange{\protect\cite[{Eqs. (D9-D10)}]{cite456}}.}\\ 
\addlinespace[\myxtraspc]
\eczhRefIndex{code:xcube}%
\eczhListValue{\flmRefsHyperref{code:xcube}{X-cube model code}} & \eczhListValue{A foliated type-I fracton CSS code on a cubic lattice with qubits on edges, cube stabilizers, and three cross-shaped vertex stabilizers for each vertex \NoCaseChange{\protect\cite{cite233}}.
It supports a subextensive number of logical qubits.}\\ 
\addlinespace[\myxtraspc]
\eczhRefIndex{code:xyz_color}%
\eczhListValue{\flmRefsHyperref{code:xyz_color}{XYZ color code}} & \eczhListValue{A variant of the 6.6.6 color code whose generators are \(XZXZXZ\) and \(ZYZYZY\) Pauli strings associated to each hexagonal in the hexagonal (6.6.6) tiling. 
A further variation called the \textit{domain wall color code} admits generators of the form \(XXXZZZ\) and \(ZZZXXX\) \NoCaseChange{\protect\cite{cite437}}.
While such codes are equivalent to CSS color codes with the same distance, other properties like noise-bias performance can differ significantly.}\\ 
\addlinespace[\myxtraspc]
\eczhRefIndex{code:xzzx}%
\eczhListValue{\flmRefsHyperref{code:xzzx}{XZZX surface code}} & \eczhListValue{A variant of the rotated surface code whose generators are \(XZZX\) Pauli strings associated, clockwise, to the vertices of each face of a two-dimensional lattice (with a qubit located at each vertex of the tessellation).}\\ 
\end{tabularx}\endgroup
\eczcodelist{hamiltonian}{Hamiltonian-based codes
}%

\eczhCodeListAutoDescription{Codes that are descendants of \flmRefsCref{code:hamiltonian} and not descendants of any of \flmRefsCref{code:stabilizer}, \flmRefsCref{code:constant_excitation}.}%

\eczhIncludeCodeGraph{Bare}{scale=0.5}{\columnwidth}{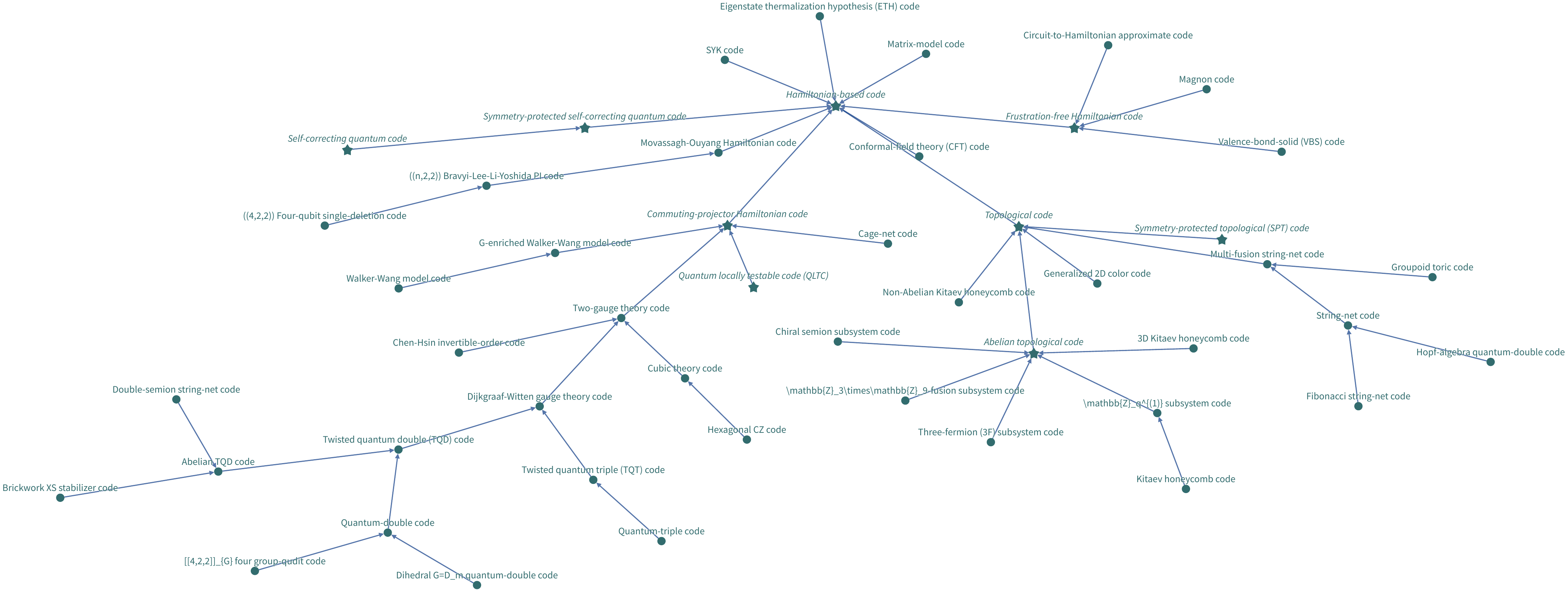}{Hamiltonian-based codes}{https://errorcorrectionzoo.org/code_graph#J\%7B\%22displayMode\%22\%3A\%22subset\%22\%2C\%22modeSubsetOptions\%22\%3A\%7B\%22codeIds\%22\%3A\%5B\%223d_kitaev_honeycomb\%22\%2C\%22tqd_abelian\%22\%2C\%22topological_abelian\%22\%2C\%22brickwork\%22\%2C\%22cage_net\%22\%2C\%22invertible\%22\%2C\%22semion\%22\%2C\%22circuit_to_hamiltonian\%22\%2C\%22commuting_projector\%22\%2C\%22cft\%22\%2C\%22cubic_theory\%22\%2C\%22quantum_double_dihedral\%22\%2C\%22dijkgraaf_witten\%22\%2C\%22double_semion_string_net\%22\%2C\%22eth\%22\%2C\%22fibonacci\%22\%2C\%22frustration_free\%22\%2C\%22generalized_color\%22\%2C\%22groupoid_surface\%22\%2C\%22hamiltonian\%22\%2C\%22hexagonal_cz\%22\%2C\%22hopf_quantum_double\%22\%2C\%22kitaev_honeycomb\%22\%2C\%22mps\%22\%2C\%22matrix_qm\%22\%2C\%22movassagh_ouyang\%22\%2C\%22enriched_string_net\%22\%2C\%22nonabelian_kitaev_honeycomb\%22\%2C\%22qltc\%22\%2C\%22quantum_double\%22\%2C\%22quantum_triple\%22\%2C\%22syk\%22\%2C\%22self_correct\%22\%2C\%22string_net\%22\%2C\%22symmetry_protected_self_correct\%22\%2C\%22spt\%22\%2C\%22subsystem_three_fermion\%22\%2C\%22topological\%22\%2C\%22tqd\%22\%2C\%22tqt\%22\%2C\%22yetter_gauge_theory\%22\%2C\%22vbs\%22\%2C\%22walker_wang\%22\%2C\%22four_qubit_permutation_invariant\%22\%2C\%22unentangled_permutation_invariant\%22\%2C\%22enriched_walker_wang\%22\%2C\%22group_4_2_2\%22\%2C\%22zthree_znine\%22\%2C\%22qudit_znone\%22\%5D\%2C\%22reusePreviousLayoutPositions\%22\%3Afalse\%2C\%22showIntermediateConnectingNodes\%22\%3Atrue\%2C\%22connectingNodesMaxDepth\%22\%3A15\%2C\%22connectingNodesPathMaxLength\%22\%3A20\%2C\%22connectingNodesMaxExtraDepth\%22\%3A3\%2C\%22connectingNodesOnlyKeepPathsWithAdditionalLength\%22\%3A1\%2C\%22connectingNodesToDomainsAndKingdoms\%22\%3Afalse\%2C\%22connectingNodesEdgeLengthsByType\%22\%3A\%7B\%22primaryParent\%22\%3A1\%2C\%22secondaryParent\%22\%3A4\%2C\%22cousin\%22\%3A6\%7D\%2C\%22nodeIds\%22\%3A\%5B\%5D\%7D\%2C\%22highlightImportantNodes\%22\%3A\%7B\%22highlightImportantNodes\%22\%3Afalse\%2C\%22highlightPrimaryParents\%22\%3Afalse\%2C\%22highlightRootConnectingEdges\%22\%3Afalse\%7D\%7D}

\begingroup
\small
\eczhBreakableDashes
\renewcommand\arraystretch{1.05}
\edef\myxtraspc{\eczListAddVSpaceXtraXtraStretch}
\begin{tabularx}{\linewidth}{>{\raggedright\arraybackslash}p{\eczListColWidth{name}} >{\hsize=1.0000\hsize }X}
\toprule
\eczListColTitle{Code} & \eczListColTitle{Description} \\
\midrule
\endfirsthead
\toprule
\eczListColTitleContinued{Code} & \eczListColTitleContinued{Description} \\
\midrule
\endhead
\bottomrule
\endfoot
\eczhRefIndex{code:3d_kitaev_honeycomb}%
\eczhListValue{\flmRefsHyperref{code:3d_kitaev_honeycomb}{3D Kitaev honeycomb code}} & \eczhListValue{3D subsystem stabilizer code whose Hamiltonian is a 3D generalization of the Kitaev honeycomb model.
One of the phases realized by the 3D Kitaev honeycomb Hamiltonian is that of the 3D fermionic surface code \NoCaseChange{\protect\cite{cite458}}.}\\ 
\addlinespace[\myxtraspc]
\eczhRefIndex{code:tqd_abelian}%
\eczhListValue{\flmRefsHyperref{code:tqd_abelian}{Abelian TQD code}} & \eczhListValue{TQD code whose codewords realize a 2D Abelian twisted-quantum-double topological order.
For Abelian TQDs, the corresponding anyon theory is defined by an Abelian group and a group cocycle built from Type-I, Type-II, or Type-III 3-cocycles \NoCaseChange{\protect\cite{cite571,cite405,cite572}}.
Abelian TQDs with Type-I and -II cocycles account for all 2D Abelian topological orders that admit gapped boundaries \NoCaseChange{\protect\cite{cite573}}.
Abelian TQDs with Type-III cocycles may admit non-Abelian topological orders.}\\ 
\addlinespace[\myxtraspc]
\eczhRefIndex{code:topological_abelian}%
\eczhListValue{\flmRefsHyperref{code:topological_abelian}{Abelian topological code}} & \eczhListValue{Code whose codewords realize topological order associated with an Abelian anyon theory.
In 2D, this is equivalent to a unitary braided fusion category which is also an Abelian group under fusion \NoCaseChange{\protect\cite{cite574}}.
Unless otherwise noted, the phases discussed are bosonic.}\\ 
\addlinespace[\myxtraspc]
\eczhRefIndex{code:brickwork}%
\eczhListValue{\flmRefsHyperref{code:brickwork}{Brickwork \(XS\) stabilizer code}} & \eczhListValue{An \(XS\) stabilizer code that realizes the topological order of the Type-III \(G=\mathbb{Z}^3_2\) TQD model \NoCaseChange{\protect\cite{cite575,cite576}}, which is the same topological order as the \(G=D_4\) quantum double \NoCaseChange{\protect\cite{cite577}}.
Its qubits are placed on a 2D square lattice, and the stabilizers are defined using two overlapping rectangular tilings.}\\ 
\addlinespace[\myxtraspc]
\eczhRefIndex{code:cage_net}%
\eczhListValue{\flmRefsHyperref{code:cage_net}{Cage-net code}} & \eczhListValue{A commuting-projector code family obtained by coupling layers of two-dimensional topological orders and condensing extended one-dimensional flux strings \NoCaseChange{\protect\cite{cite567}}.
A modern lattice realization starts from isotropic stacks of \(G\)-graded string-net models \NoCaseChange{\protect\cite{cite568}}.
The family includes stabilizer examples such as the \(\mathbb{Z}_2\) string-membrane-net realization of the X-cube model \NoCaseChange{\protect\cite{cite569}} as well as non-stabilizer examples with non-Abelian restricted-mobility excitations \NoCaseChange{\protect\cite{cite567}}.
String-membrane-net and cage-net constructions can realize the same fracton phases; for isotropic stacks, this equivalence can be understood via generalized local unitaries \NoCaseChange{\protect\cite{cite568}}.
The cage-net construction can be used to realize various fracton phases, stabilizer and otherwise.}\\ 
\addlinespace[\myxtraspc]
\eczhRefIndex{code:invertible}%
\eczhListValue{\flmRefsHyperref{code:invertible}{Chen-Hsin invertible-order code}} & \eczhListValue{A geometrically local commuting-projector code that realizes beyond-group-cohomology invertible topological phases in arbitrary dimensions.
Its code Hamiltonian terms include Pauli-\(Z\) operators and products of Pauli-\(X\) operators and \(CZ\) gates \NoCaseChange{\protect\cite[{Eq. (3.25)}]{cite578}}.
Instances of the code in 4D realize the 3D \(\mathbb{Z}_2\) gauge theory with fermionic charge and either bosonic (FcBl) or fermionic (FcFl) loop excitations at their boundaries \NoCaseChange{\protect\cite{cite579,cite455}}; see Ref. \NoCaseChange{\protect\cite{cite580}} for a different lattice-model formulation of the FcBl boundary code.}\\ 
\addlinespace[\myxtraspc]
\eczhRefIndex{code:semion}%
\eczhListValue{\flmRefsHyperref{code:semion}{Chiral semion subsystem code}} & \eczhListValue{Modular-qudit subsystem stabilizer code with qudit dimension \(q=4\) that is characterized by the chiral semion topological phase.
The code admits a set of geometrically local stabilizer generators on a torus.}\\ 
\addlinespace[\myxtraspc]
\eczhRefIndex{code:circuit_to_hamiltonian}%
\eczhListValue{\flmRefsHyperref{code:circuit_to_hamiltonian}{Circuit-to-Hamiltonian approximate code}} & \eczhListValue{Approximate qubit block code that forms the ground-state space of a frustration-free Hamiltonian with non-commuting terms.
Its distance and logical-qubit number are both of \flmRefsHyperref{ref65}{order} \(\Omega(n/\log^5 n)\) \NoCaseChange{\protect\cite[{Thm. 3.1}]{cite581}}.
The code is an approximate non-stabilizer QLWC code since the Hamiltonian consists of non-commuting 9-local non-Pauli projectors, with each qubit acted on by \flmRefsHyperref{ref65}{order} \(O( \text{polylog}(n) )\) projectors.}\\ 
\addlinespace[\myxtraspc]
\eczhRefIndex{code:commuting_projector}%
\eczhListValue{\flmRefsHyperref{code:commuting_projector}{Commuting-projector Hamiltonian code}} & \eczhListValue{Hamiltonian-based code whose Hamiltonian terms can be expressed as orthogonal projectors (i.e., Hermitian operators with eigenvalues 0 or 1) that commute with each other.}\\ 
\addlinespace[\myxtraspc]
\eczhRefIndex{code:cft}%
\eczhListValue{\flmRefsHyperref{code:cft}{Conformal-field theory (CFT) code}} & \eczhListValue{Approximate code whose codewords lie in the low-energy subspace of a conformal field theory, e.g., the quantum Ising model at its critical point \NoCaseChange{\protect\cite{cite582,cite583}}.
Its encoding is argued to perform source coding (i.e., compression) as well as channel coding (i.e., error correction) \NoCaseChange{\protect\cite{cite582}}.}\\ 
\addlinespace[\myxtraspc]
\eczhRefIndex{code:cubic_theory}%
\eczhListValue{\flmRefsHyperref{code:cubic_theory}{Cubic theory code}} & \eczhListValue{A geometrically local commuting-projector code family defined on triangulations in arbitrary spatial dimensions.
Its Hamiltonian contains Pauli-\(Z\) flux terms and non-Pauli Gauss-law terms built from products of Pauli-\(X\) operators and \(CZ\) gates.
These commuting non-Pauli stabilizers realize higher-form \(\mathbb{Z}_2^3\) gauge theories with Abelian electric excitations and non-Abelian magnetic excitations.}\\ 
\addlinespace[\myxtraspc]
\eczhRefIndex{code:quantum_double_dihedral}%
\eczhListValue{\flmRefsHyperref{code:quantum_double_dihedral}{Dihedral \(G=D_m\) quantum-double code}} & \eczhListValue{Quantum-double code whose codewords realize topological order associated with the dihedral group \(D_m\) of order \(2m\).
For \(m \geq 3\), these codes are non-Abelian, with the simplest case given by \(D_3=S_3\), the permutation group on three objects.
On an oriented lattice, each edge hosts a \(2m\)-dimensional group qudit, and the codespace is the ground-state subspace of the corresponding quantum double Hamiltonian \NoCaseChange{\protect\cite{cite423}}.}\\ 
\addlinespace[\myxtraspc]
\eczhRefIndex{code:dijkgraaf_witten}%
\eczhListValue{\flmRefsHyperref{code:dijkgraaf_witten}{Dijkgraaf-Witten gauge theory code}} & \eczhListValue{A code whose codewords realize \(D\)-dimensional lattice Dijkgraaf-Witten gauge theory \NoCaseChange{\protect\cite{cite584,cite585}} for a finite group \(G\) and a \(D+1\)-cocycle \(\omega\) in the cohomology class \(H^{D+1}( G, U(1) )\).
When the cocycle is non-trivial, the gauge theory is called a \textit{twisted gauge theory}.
There exist lattice-model formulations in arbitrary spatial dimension \NoCaseChange{\protect\cite{cite586}}.
Boundaries and excitations have been studied for arbitrary dimension \NoCaseChange{\protect\cite{cite587}}.}\\ 
\addlinespace[\myxtraspc]
\eczhRefIndex{code:double_semion_string_net}%
\eczhListValue{\flmRefsHyperref{code:double_semion_string_net}{Double-semion string-net code}} & \eczhListValue{An \(XS\) stabilizer code that realizes the 2D double semion topological phase.
The model can be extended to other spatial dimensions \NoCaseChange{\protect\cite{cite588}}.}\\ 
\addlinespace[\myxtraspc]
\eczhRefIndex{code:eth}%
\eczhListValue{\flmRefsHyperref{code:eth}{Eigenstate thermalization hypothesis (ETH) code}} & \eczhListValue{An \(n\)-qubit approximate code whose codespace is formed by eigenstates of a translationally-invariant quantum many-body system which satisfies the Eigenstate Thermalization Hypothesis (ETH).
ETH ensures that codewords cannot be locally distinguished in the thermodynamic limit.
Relevant many-body systems include 1D non-interacting spin chains or frustration-free systems such as Motzkin chains and Heisenberg models.}\\ 
\addlinespace[\myxtraspc]
\eczhRefIndex{code:fibonacci}%
\eczhListValue{\flmRefsHyperref{code:fibonacci}{Fibonacci string-net code}} & \eczhListValue{Quantum error correcting code associated with the Levin-Wen string-net model with the Fibonacci input category, admitting two types of encodings.}\\ 
\addlinespace[\myxtraspc]
\eczhRefIndex{code:frustration_free}%
\eczhListValue{\flmRefsHyperref{code:frustration_free}{Frustration-free Hamiltonian code}} & \eczhListValue{Hamiltonian-based code whose Hamiltonian is frustration free, i.e., whose ground states minimize the energy of each term.}\\ 
\addlinespace[\myxtraspc]
\eczhRefIndex{code:generalized_color}%
\eczhListValue{\flmRefsHyperref{code:generalized_color}{Generalized 2D color code}} & \eczhListValue{Member of a family of non-Abelian 2D topological codes, defined by a finite group \( G \), that serves as a generalization of the 2D color code (for which \(G=\mathbb{Z}_2 \times \mathbb{Z}_2\)).
Hamiltonian terms are built from \flmRefsHyperref{ref20}{group-based right- and left-multiplication \(X\)-type operators together with \(Z\)-type operators}.}\\ 
\addlinespace[\myxtraspc]
\eczhRefIndex{code:groupoid_surface}%
\eczhListValue{\flmRefsHyperref{code:groupoid_surface}{Groupoid toric code}} & \eczhListValue{Extension of the Kitaev surface code from Abelian groups to groupoids, i.e., multi-fusion categories in which every morphism is an isomorphism \NoCaseChange{\protect\cite{cite570}}.
Some models admit fracton-like features such as extensive ground-state degeneracy and excitations with restricted mobility.
The robustness of these features has not yet been established.}\\ 
\addlinespace[\myxtraspc]
\eczhRefIndex{code:hamiltonian}%
\eczhListValue{\flmRefsHyperref{code:hamiltonian}{Hamiltonian-based code}} & \eczhListValue{Code whose codespace corresponds to a set of energy eigenstates of a quantum-mechanical Hamiltonian, i.e., a Hermitian operator whose expectation value measures the energy of its underlying physical system.
The codespace is typically a set of low-energy eigenstates or ground states, but can include subspaces of arbitrarily high energy.
Hamiltonians whose eigenstates are the canonical basis elements are called \textit{classical}; otherwise, a Hamiltonian is called \textit{quantum}.}\\ 
\addlinespace[\myxtraspc]
\eczhRefIndex{code:hexagonal_cz}%
\eczhListValue{\flmRefsHyperref{code:hexagonal_cz}{Hexagonal \(CZ\) code}} & \eczhListValue{A hexagonal-lattice realization of the \(2+1\)D \(l=m=n=1\) cubic theory / Type-III \(\mathbb{Z}_2^3\) twisted quantum double phase.
Its stabilizers are products of Pauli-\(Z\) operators and \(CZ\) gates \NoCaseChange{\protect\cite[{Fig. 6}]{cite575}\protect\cite[{Fig. 3}]{cite589}}.
The ground-state subspace of the hexagonal \(CZ\) code realizes the topological order of the Type-III \(G=\mathbb{Z}^3_2\) Abelian TQD model \NoCaseChange{\protect\cite{cite575,cite576}}, which is the same topological order as the \(G=D_4\) non-Abelian quantum double \NoCaseChange{\protect\cite{cite577}}.
The stabilizers include \(CZ\) operators acting on hexagonal loops, but a reduced version exists where only two \(CZ\) gates act on each loop \NoCaseChange{\protect\cite{cite589}}.}\\ 
\addlinespace[\myxtraspc]
\eczhRefIndex{code:hopf_quantum_double}%
\eczhListValue{\flmRefsHyperref{code:hopf_quantum_double}{Hopf-algebra quantum-double code}} & \eczhListValue{Code whose codewords realize 2D gapped topological order defined on qudits valued in a Hopf algebra \(H\).
The code Hamiltonian is a generalization \NoCaseChange{\protect\cite{cite590,cite591}} of the quantum double model from group algebras to Hopf algebras, as anticipated by Kitaev \NoCaseChange{\protect\cite{cite423}}.
Boundaries of these models have been examined \NoCaseChange{\protect\cite{cite592,cite593}}.}\\ 
\addlinespace[\myxtraspc]
\eczhRefIndex{code:kitaev_honeycomb}%
\eczhListValue{\flmRefsHyperref{code:kitaev_honeycomb}{Kitaev honeycomb code}} & \eczhListValue{Subsystem qubit stabilizer code underlying the Kitaev honeycomb model \NoCaseChange{\protect\cite{cite537,cite594}}.
Its gauge generators are the two-qubit \(XX\), \(YY\), and \(ZZ\) link operators on the three edge types of the honeycomb lattice \NoCaseChange{\protect\cite[{Sec. 3.2}]{cite594}}.
Its stabilizer group is generated by loop operators, and syndrome extraction can be reduced to ordered measurements of the two-qubit link operators \NoCaseChange{\protect\cite[{Sec. 3.2}]{cite594}}.
This is the \(q=2\) instance of the \(\mathbb{Z}_q^{(1)}\) subsystem code and does not encode any logical qubits \NoCaseChange{\protect\cite{cite594}\protect\cite[{Sec. 7.3}]{cite414}}.}\\ 
\addlinespace[\myxtraspc]
\eczhRefIndex{code:mps}%
\eczhListValue{\flmRefsHyperref{code:mps}{Magnon code}} & \eczhListValue{An \(n\)-spin approximate code whose codespace of \(k=\Omega(\log n)\) qubits is efficiently described in terms of particular matrix product states or Bethe ansatz tensor networks.
Magnon codewords are low-energy excited states of the frustration-free Heisenberg-XXX model Hamiltonian \NoCaseChange{\protect\cite{cite595}}.}\\ 
\addlinespace[\myxtraspc]
\eczhRefIndex{code:matrix_qm}%
\eczhListValue{\flmRefsHyperref{code:matrix_qm}{Matrix-model code}} & \eczhListValue{Multimode Fock-state bosonic approximate code derived from a matrix model, i.e., a bosonic theory with a large non-Abelian gauge group.
The model's degrees of freedom are matrix-valued bosons \(a\), each consisting of \(N^2\) harmonic oscillator modes and subject to an \(SU(N)\) gauge symmetry.}\\ 
\addlinespace[\myxtraspc]
\eczhRefIndex{code:movassagh_ouyang}%
\eczhListValue{\flmRefsHyperref{code:movassagh_ouyang}{Movassagh-Ouyang Hamiltonian code}} & \eczhListValue{This is a family of codes derived via an algorithm that takes as input \textit{any} binary classical code and outputs a quantum code (note that this framework can be extended to \(q\)-ary codes).  
The algorithm is probabilistic but succeeds almost surely if the classical code is random. 
An explicit code construction does exist for linear distance codes encoding one logical qubit using Radon's theorem \NoCaseChange{\protect\cite{cite596,cite597}}. 
For finite rate codes, there is no rigorous proof that the construction algorithm succeeds, and approximate constructions are described instead.}\\ 
\addlinespace[\myxtraspc]
\eczhRefIndex{code:enriched_string_net}%
\eczhListValue{\flmRefsHyperref{code:enriched_string_net}{Multi-fusion string-net code}} & \eczhListValue{Family of codes resulting from the string-net construction but whose input is a unitary multi-fusion category (as opposed to a unitary fusion category).}\\ 
\addlinespace[\myxtraspc]
\eczhRefIndex{code:nonabelian_kitaev_honeycomb}%
\eczhListValue{\flmRefsHyperref{code:nonabelian_kitaev_honeycomb}{Non-Abelian Kitaev honeycomb code}} & \eczhListValue{Code whose logical subspace in the gapped non-Abelian phase of the Kitaev honeycomb model with a magnetic field is labeled by different fusion outcomes of Ising anyons \NoCaseChange{\protect\cite{cite537}}.}\\ 
\addlinespace[\myxtraspc]
\eczhRefIndex{code:qltc}%
\eczhListValue{\flmRefsHyperref{code:qltc}{Quantum locally testable code (QLTC)}} & \eczhListValue{A local commuting-projector Hamiltonian-based block quantum code which has a nonzero average-energy penalty for creating large errors. Informally, states that are far away from the codespace of a QLTC have to be excited states of a number of the code's local projectors that scales linearly with \(n\).}\\ 
\addlinespace[\myxtraspc]
\eczhRefIndex{code:quantum_double}%
\eczhListValue{\flmRefsHyperref{code:quantum_double}{Quantum-double code}} & \eczhListValue{Group-based code whose codewords realize 2D modular gapped topological order defined by a finite group \(G\).
The code's generators are few-body operators associated to the stars and plaquettes, respectively, of a tessellation of a two-dimensional surface (with a qudit of dimension \( |G| \) located at each edge of the tessellation).
The original Hamiltonian can be re-expressed via \flmRefsHyperref{ref20}{group-based right- and left-multiplication \(X\)-type as well as \(Z\)-type error} operators \NoCaseChange{\protect\cite[{Sec. 3.3}]{cite598}}.}\\ 
\addlinespace[\myxtraspc]
\eczhRefIndex{code:quantum_triple}%
\eczhListValue{\flmRefsHyperref{code:quantum_triple}{Quantum-triple code}} & \eczhListValue{Group-based code whose codewords realize 3D topological order defined by a finite group \(G\).}\\ 
\addlinespace[\myxtraspc]
\eczhRefIndex{code:syk}%
\eczhListValue{\flmRefsHyperref{code:syk}{SYK code}} & \eczhListValue{Approximate \(n\)-fermionic code whose codewords are low-energy states of the Sachdev-Ye-Kitaev (SYK) Hamiltonian \NoCaseChange{\protect\cite{cite560,cite561}} or other low-rank SYK models \NoCaseChange{\protect\cite{cite562,cite563}}.}\\ 
\addlinespace[\myxtraspc]
\eczhRefIndex{code:self_correct}%
\eczhListValue{\flmRefsHyperref{code:self_correct}{Self-correcting quantum code}} & \eczhListValue{A block quantum code that forms the ground-state subspace of an \(n\)-body geometrically local Hamiltonian whose logical information is recoverable for arbitrarily long times in the \(n\to\infty\) limit after interaction with a sufficiently cold thermal environment.
Typically, one also requires a decoder whose decoding time scales polynomially with \(n\) and a finite energy density.}\\ 
\addlinespace[\myxtraspc]
\eczhRefIndex{code:string_net}%
\eczhListValue{\flmRefsHyperref{code:string_net}{String-net code}} & \eczhListValue{A non-stabilizer commuting-projector code whose codewords realize a 2D topological order rendered by a Turaev-Viro topological field theory.
The corresponding anyon theory is defined by a (multiplicity-free) unitary fusion category \( \mathcal{C} \).
The code is defined on a cell decomposition dual to a triangulation of a two-dimensional surface, with a qudit of dimension \( |\mathcal{C}| \) located at each edge of the decomposition.
For modular input categories, the emergent anyon theory is the doubled category \( \mathcal{C}\otimes\mathcal{C}^{*} \) \NoCaseChange{\protect\cite{cite599}}.
These models realize local topological order (LTO) \NoCaseChange{\protect\cite{cite600}}.}\\ 
\addlinespace[\myxtraspc]
\eczhRefIndex{code:symmetry_protected_self_correct}%
\eczhListValue{\flmRefsHyperref{code:symmetry_protected_self_correct}{Symmetry-protected self-correcting quantum code}} & \eczhListValue{A code which admits a restricted notion of thermal stability against symmetric perturbations, i.e., perturbations that commute with a set of operators forming a group \(G\) called the \textit{symmetry group}.}\\ 
\addlinespace[\myxtraspc]
\eczhRefIndex{code:spt}%
\eczhListValue{\flmRefsHyperref{code:spt}{Symmetry-protected topological (SPT) code}} & \eczhListValue{A code whose codewords form the ground-state or low-energy subspace of a code Hamiltonian realizing symmetry-protected topological (SPT) order.}\\ 
\addlinespace[\myxtraspc]
\eczhRefIndex{code:subsystem_three_fermion}%
\eczhListValue{\flmRefsHyperref{code:subsystem_three_fermion}{Three-fermion (3F) subsystem code}} & \eczhListValue{2D subsystem stabilizer code whose low-energy excitations realize the three-fermion anyon theory \NoCaseChange{\protect\cite{cite601,cite602,cite603}}.
One version uses two qubits at each site \NoCaseChange{\protect\cite{cite414}}, while other manifestations utilize a single qubit per site and only weight-two (two-body) interactions \NoCaseChange{\protect\cite{cite602,cite604}}.
All are expected to be equivalent to each other via a local constant-depth \flmRefsHyperref{ref409}{Clifford circuit}.}\\ 
\addlinespace[\myxtraspc]
\eczhRefIndex{code:topological}%
\eczhListValue{\flmRefsHyperref{code:topological}{Topological code}} & \eczhListValue{A code whose codewords form the ground-state or low-energy subspace of a (typically geometrically local) code Hamiltonian realizing a topological phase.
A topological phase may be \textit{bosonic} or \textit{fermionic}, i.e., constructed out of underlying subsystems whose operators commute or anti-commute with each other, respectively.
Unless otherwise noted, the phases discussed are bosonic.}\\ 
\addlinespace[\myxtraspc]
\eczhRefIndex{code:tqd}%
\eczhListValue{\flmRefsHyperref{code:tqd}{Twisted quantum double (TQD) code}} & \eczhListValue{Code whose codewords realize a 2D topological order rendered by a Chern-Simons topological field theory.
The corresponding anyon theory is defined by a finite group \(G\) and a 3-cocycle \(\omega\in H^3( G, U(1) )\) \NoCaseChange{\protect\cite{cite605,cite571,cite405}}.
Canonical TQD models \NoCaseChange{\protect\cite{cite571}} are defined on group-valued qudits.}\\ 
\addlinespace[\myxtraspc]
\eczhRefIndex{code:tqt}%
\eczhListValue{\flmRefsHyperref{code:tqt}{Twisted quantum triple (TQT) code}} & \eczhListValue{Group-based code realizing a 3D topological order rendered by a Dijkgraaf-Witten gauge theory.
The corresponding anyon theory is defined by a finite group \(G\) and a Type-IV four-cocycle \(\omega\).
Canonical TQT models \NoCaseChange{\protect\cite{cite586,cite606}} and other formulations whose ground states are in the same phase are all defined on group-valued qudits.}\\ 
\addlinespace[\myxtraspc]
\eczhRefIndex{code:yetter_gauge_theory}%
\eczhListValue{\flmRefsHyperref{code:yetter_gauge_theory}{Two-gauge theory code}} & \eczhListValue{A code whose codewords realize lattice two-gauge theory \NoCaseChange{\protect\cite{cite607,cite608,cite609,cite610,cite611,cite612,cite613,cite614,cite615}} for a finite \textit{two-group} (a.k.a. a \textit{crossed module}) in arbitrary spatial dimension.
There exist several lattice-model formulations in arbitrary spatial dimension \NoCaseChange{\protect\cite{cite616,cite617}} as well as explicitly in 3D \NoCaseChange{\protect\cite{cite618,cite619,cite620,cite621}} and 4D \NoCaseChange{\protect\cite{cite621}}, with the 3D case realizing the Yetter model \NoCaseChange{\protect\cite{cite622,cite623,cite624,cite625}}.}\\ 
\addlinespace[\myxtraspc]
\eczhRefIndex{code:vbs}%
\eczhListValue{\flmRefsHyperref{code:vbs}{Valence-bond-solid (VBS) code}} & \eczhListValue{A member of an approximate \(q\)-dimensional spin-code family whose codespace is described in terms of \(SU(q)\) valence-bond-solid (VBS) \NoCaseChange{\protect\cite{cite626}} matrix product states with various boundary conditions.
The codes become exact when either \(n\) or \(q\) go to infinity.
The original work on these codes studied the \(q=2\) case \NoCaseChange{\protect\cite{cite627}}.}\\ 
\addlinespace[\myxtraspc]
\eczhRefIndex{code:walker_wang}%
\eczhListValue{\flmRefsHyperref{code:walker_wang}{Walker-Wang model code}} & \eczhListValue{A non-stabilizer commuting-projector 3D topological code defined by a unitary braided fusion category \( \mathcal{C} \) (also known as a unitary premodular category).
The code is defined on a cubic lattice that is resolved to be trivalent, with a qudit of dimension \( |\mathcal{C}| \) located at each edge.
The codespace is the ground-state subspace of the Walker-Wang model Hamiltonian \NoCaseChange{\protect\cite{cite628}} and realizes the Crane-Yetter model \NoCaseChange{\protect\cite{cite629,cite630,cite631}}.
A single-state version of the code provides a resource state for MBQC \NoCaseChange{\protect\cite{cite478}}.}\\ 
\addlinespace[\myxtraspc]
\eczhRefIndex{code:four_qubit_permutation_invariant}%
\eczhListValue{\flmRefsHyperref{code:four_qubit_permutation_invariant}{\(\llparenthesis 4,2,2\rrparenthesis \) Four-qubit single-deletion code}} & \eczhListValue{Four-qubit PI code that is the smallest qubit code to correct one deletion error.}\\ 
\addlinespace[\myxtraspc]
\eczhRefIndex{code:unentangled_permutation_invariant}%
\eczhListValue{\flmRefsHyperref{code:unentangled_permutation_invariant}{\(\llparenthesis n,2,2\rrparenthesis \) Bravyi-Lee-Li-Yoshida PI code}} & \eczhListValue{PI distance-two code on \(n\geq4\) qubits whose degree of entanglement vanishes asymptotically with \(n\) \NoCaseChange{\protect\cite[{Appx. D}]{cite529}} (cf. \NoCaseChange{\protect\cite{cite530}}).}\\ 
\addlinespace[\myxtraspc]
\eczhRefIndex{code:enriched_walker_wang}%
\eczhListValue{\flmRefsHyperref{code:enriched_walker_wang}{\(G\)-enriched Walker-Wang model code}} & \eczhListValue{A 3D topological code defined by a unitary \(G\)-crossed braided fusion category \( \mathcal{C} \) \NoCaseChange{\protect\cite{cite632,cite633}}, where \(G\) is a finite group.
The model realizes TQFTs that include two-gauge theories, those behind Walker-Wang models, as well as the Kashaev TQFT \NoCaseChange{\protect\cite{cite634,cite635}}.
It has been generalized to include domain walls \NoCaseChange{\protect\cite{cite636}}.}\\ 
\addlinespace[\myxtraspc]
\eczhRefIndex{code:group_4_2_2}%
\eczhListValue{\flmRefsHyperref{code:group_4_2_2}{\(\llbracket 4,2,2\rrbracket _{G}\) four group-qudit code}} & \eczhListValue{\(\llbracket 4,2,2\rrbracket _{G}\) group quantum code that is an extension of the four-qubit code to group-valued qudits.}\\ 
\addlinespace[\myxtraspc]
\eczhRefIndex{code:zthree_znine}%
\eczhListValue{\flmRefsHyperref{code:zthree_znine}{\(\mathbb{Z}_3\times\mathbb{Z}_9\)-fusion subsystem code}} & \eczhListValue{Modular-qudit 2D subsystem stabilizer code whose low-energy excitations realize a non-modular anyon theory with \(\mathbb{Z}_3\times\mathbb{Z}_9\) fusion rules.
Encodes two qutrits when put on a torus.}\\ 
\addlinespace[\myxtraspc]
\eczhRefIndex{code:qudit_znone}%
\eczhListValue{\flmRefsHyperref{code:qudit_znone}{\(\mathbb{Z}_q^{(1)}\) subsystem code}} & \eczhListValue{Modular-qudit subsystem code, based on the Kitaev honeycomb model \NoCaseChange{\protect\cite{cite537}} and its generalization \NoCaseChange{\protect\cite{cite637}}, that is characterized by the \(\mathbb{Z}_q^{(1)}\) anyon theory \NoCaseChange{\protect\cite{cite638}}, which is modular for odd prime \(q\) and non-modular otherwise. Encodes a single \(q\)-dimensional qudit when put on a torus for odd \(q\), and a \(q/2\)-dimensional qudit for even \(q\). This code can be constructed using geometrically local gauge generators, but does not admit geometrically local stabilizer generators. For \(q=2\), the code reduces to the subsystem code underlying the Kitaev honeycomb model code as well as the honeycomb Floquet code.}\\ 
\end{tabularx}\endgroup
\eczcodelist{holographic}{Holographic codes
}%

\eczhCodeListAutoDescription{All descendants of \flmRefsCref{code:holographic}.}%

\eczhIncludeCodeGraph{Bare}{scale=0.5}{\columnwidth}{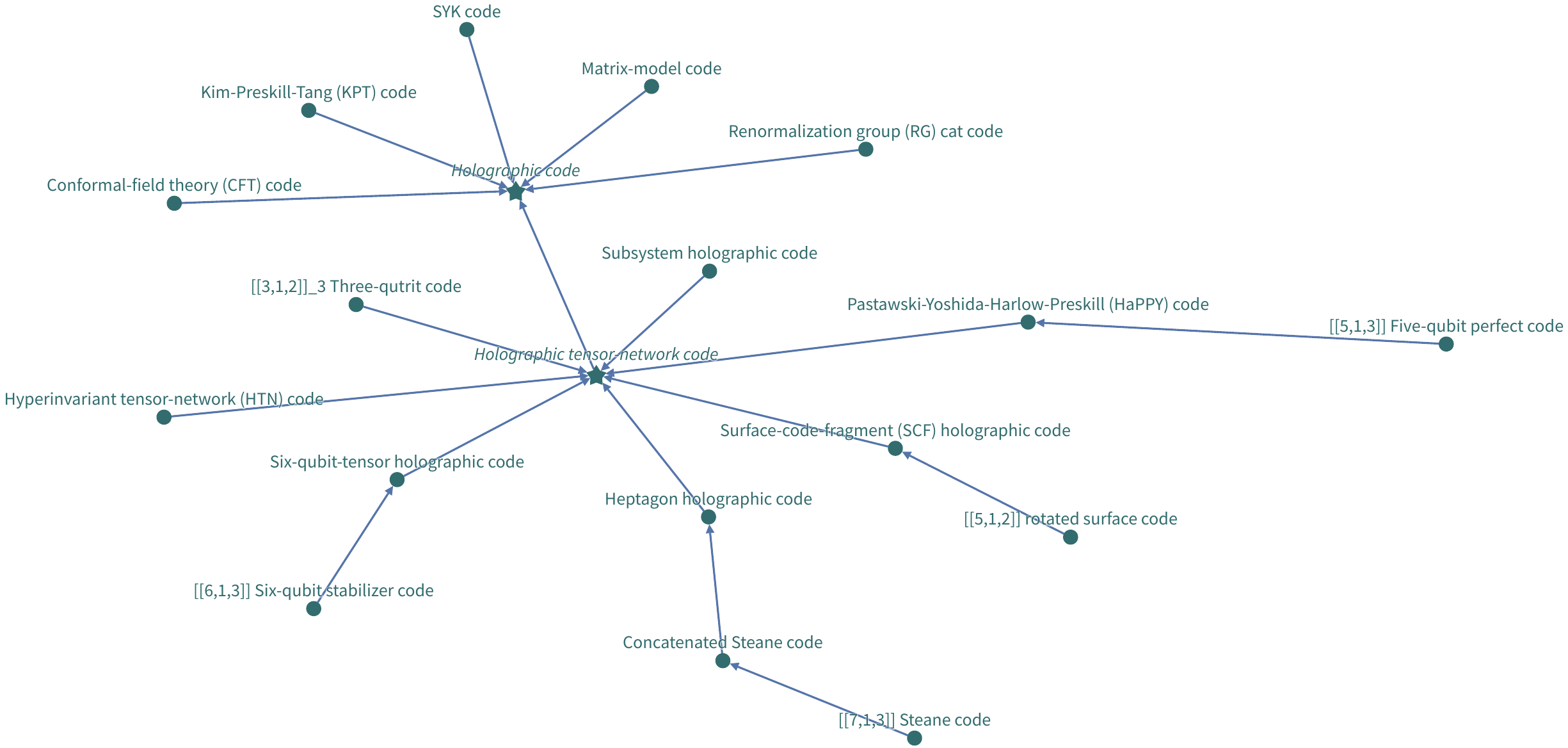}{Holographic codes}{https://errorcorrectionzoo.org/code_graph#J\%7B\%22displayMode\%22\%3A\%22subset\%22\%2C\%22modeSubsetOptions\%22\%3A\%7B\%22codeIds\%22\%3A\%5B\%22concatenated_steane\%22\%2C\%22cft\%22\%2C\%22holographic_steane\%22\%2C\%22holographic\%22\%2C\%22holographic_tensor\%22\%2C\%22holographic_hyperinvariant\%22\%2C\%22kpt\%22\%2C\%22matrix_qm\%22\%2C\%22happy\%22\%2C\%22rg_cat\%22\%2C\%22syk\%22\%2C\%22holographic_6_1_3\%22\%2C\%22holographic_subsystem\%22\%2C\%22holographic_5_1_2\%22\%2C\%22stab_3_1_2\%22\%2C\%22stab_5_1_2\%22\%2C\%22stab_5_1_3\%22\%2C\%22stab_6_1_3\%22\%2C\%22steane\%22\%5D\%2C\%22reusePreviousLayoutPositions\%22\%3Afalse\%2C\%22showIntermediateConnectingNodes\%22\%3Atrue\%2C\%22connectingNodesMaxDepth\%22\%3A15\%2C\%22connectingNodesPathMaxLength\%22\%3A20\%2C\%22connectingNodesMaxExtraDepth\%22\%3A3\%2C\%22connectingNodesOnlyKeepPathsWithAdditionalLength\%22\%3A1\%2C\%22connectingNodesToDomainsAndKingdoms\%22\%3Afalse\%2C\%22connectingNodesEdgeLengthsByType\%22\%3A\%7B\%22primaryParent\%22\%3A1\%2C\%22secondaryParent\%22\%3A4\%2C\%22cousin\%22\%3A6\%7D\%2C\%22nodeIds\%22\%3A\%5B\%5D\%7D\%2C\%22highlightImportantNodes\%22\%3A\%7B\%22highlightImportantNodes\%22\%3Afalse\%2C\%22highlightPrimaryParents\%22\%3Afalse\%2C\%22highlightRootConnectingEdges\%22\%3Afalse\%7D\%7D}

\begingroup
\small
\eczhBreakableDashes
\renewcommand\arraystretch{1.05}
\edef\myxtraspc{\eczListAddVSpaceXtraXtraStretch}
\begin{tabularx}{\linewidth}{>{\raggedright\arraybackslash}p{\eczListColWidth{name}} >{\hsize=1.0000\hsize }X}
\toprule
\eczListColTitle{Code} & \eczListColTitle{Description} \\
\midrule
\endfirsthead
\toprule
\eczListColTitleContinued{Code} & \eczListColTitleContinued{Description} \\
\midrule
\endhead
\bottomrule
\endfoot
\eczhRefIndex{code:concatenated_steane}%
\eczhListValue{\flmRefsHyperref{code:concatenated_steane}{Concatenated Steane code}} & \eczhListValue{A member of the family of \(\llbracket 7^m,1,3^m\rrbracket \) CSS codes, each of which is a recursive level-\(m\) concatenation of the Steane code.
This family is one of the first to admit a \flmRefsHyperref{ref515}{concatenated threshold} \NoCaseChange{\protect\cite{cite516,cite517,cite518,cite519,cite520}}.}\\ 
\addlinespace[\myxtraspc]
\eczhRefIndex{code:cft}%
\eczhListValue{\flmRefsHyperref{code:cft}{Conformal-field theory (CFT) code}} & \eczhListValue{Approximate code whose codewords lie in the low-energy subspace of a conformal field theory, e.g., the quantum Ising model at its critical point \NoCaseChange{\protect\cite{cite582,cite583}}.
Its encoding is argued to perform source coding (i.e., compression) as well as channel coding (i.e., error correction) \NoCaseChange{\protect\cite{cite582}}.}\\ 
\addlinespace[\myxtraspc]
\eczhRefIndex{code:holographic_steane}%
\eczhListValue{\flmRefsHyperref{code:holographic_steane}{Heptagon holographic code}} & \eczhListValue{Holographic tensor-network code constructed out of a network of encoding isometries of the Steane code.
Depending on how the isometry tensors are contracted, there is a zero-rate and a finite-rate code family.}\\ 
\addlinespace[\myxtraspc]
\eczhRefIndex{code:holographic}%
\eczhListValue{\flmRefsHyperref{code:holographic}{Holographic code}} & \eczhListValue{Block quantum code whose features serve to model aspects of the AdS/CFT holographic duality and, more generally, quantum gravity.}\\ 
\addlinespace[\myxtraspc]
\eczhRefIndex{code:holographic_tensor}%
\eczhListValue{\flmRefsHyperref{code:holographic_tensor}{Holographic tensor-network code}} & \eczhListValue{Quantum Lego code whose encoding isometry forms a holographic tensor network, i.e., a tensor network associated with a tiling of hyperbolic space.
Physical qubits are associated with uncontracted tensor legs at the boundary of the tessellation, while logical qubits are associated with uncontracted legs in the bulk.
The number of layers emanating from the central point of the tiling is the \textit{radius} of the code.}\\ 
\addlinespace[\myxtraspc]
\eczhRefIndex{code:holographic_hyperinvariant}%
\eczhListValue{\flmRefsHyperref{code:holographic_hyperinvariant}{Hyperinvariant tensor-network (HTN) code}} & \eczhListValue{Holographic tensor-network code constructed out of a hyperinvariant tensor network \NoCaseChange{\protect\cite{cite639}}, i.e., a MERA-like network admitting a hyperbolic geometry.
The network is defined using two layers A and B, with constituent tensors satisfying isometry conditions (a.k.a. multitensor constraints).}\\ 
\addlinespace[\myxtraspc]
\eczhRefIndex{code:kpt}%
\eczhListValue{\flmRefsHyperref{code:kpt}{Kim-Preskill-Tang (KPT) code}} & \eczhListValue{An approximate quantum error-correcting code that protects the encoded interior of a black hole from computationally bounded exterior observers.
Under the assumption that the Hawking radiation emitted by an old black hole is pseudorandom, there exists a subspace of the radiation system that encodes the black hole interior, entangled with the late outgoing Hawking quanta.
The logical operators of this code, called ghost operators in \NoCaseChange{\protect\cite{cite640}}, commute with efficient operations acting on the radiation, protecting the interior up to corrections exponentially small in the black hole's entropy.
The construction is state dependent: the encoding depends on the state that collapsed to form the black hole \NoCaseChange{\protect\cite{cite640}}.}\\ 
\addlinespace[\myxtraspc]
\eczhRefIndex{code:matrix_qm}%
\eczhListValue{\flmRefsHyperref{code:matrix_qm}{Matrix-model code}} & \eczhListValue{Multimode Fock-state bosonic approximate code derived from a matrix model, i.e., a bosonic theory with a large non-Abelian gauge group.
The model's degrees of freedom are matrix-valued bosons \(a\), each consisting of \(N^2\) harmonic oscillator modes and subject to an \(SU(N)\) gauge symmetry.}\\ 
\addlinespace[\myxtraspc]
\eczhRefIndex{code:happy}%
\eczhListValue{\flmRefsHyperref{code:happy}{Pastawski-Yoshida-Harlow-Preskill (HaPPY) code}} & \eczhListValue{Holographic code constructed from six-leg five-qubit \flmRefsHyperref{ref219}{perfect tensors} placed on hyperbolic pentagon and hexagon tilings.
The code serves as a minimal model for several aspects of the AdS/CFT holographic duality \NoCaseChange{\protect\cite{cite641}} and potentially a dS/CFT duality \NoCaseChange{\protect\cite{cite642}}.}\\ 
\addlinespace[\myxtraspc]
\eczhRefIndex{code:rg_cat}%
\eczhListValue{\flmRefsHyperref{code:rg_cat}{Renormalization group (RG) cat code}} & \eczhListValue{Code whose codespace is spanned by \(q\) field-theoretic coherent states which are flowing under the renormalization group (RG) flow of massive free fields. The code approximately protects against displacements that represent local (i.e., short-distance, ultraviolet, or UV) operators. Intuitively, this is because RG cat codewords represent non-local (i.e., long-distance) degrees of freedom, which should only be excitable by acting on a macroscopically large number of short-distance degrees of freedom.}\\ 
\addlinespace[\myxtraspc]
\eczhRefIndex{code:syk}%
\eczhListValue{\flmRefsHyperref{code:syk}{SYK code}} & \eczhListValue{Approximate \(n\)-fermionic code whose codewords are low-energy states of the Sachdev-Ye-Kitaev (SYK) Hamiltonian \NoCaseChange{\protect\cite{cite560,cite561}} or other low-rank SYK models \NoCaseChange{\protect\cite{cite562,cite563}}.}\\ 
\addlinespace[\myxtraspc]
\eczhRefIndex{code:holographic_6_1_3}%
\eczhListValue{\flmRefsHyperref{code:holographic_6_1_3}{Six-qubit-tensor holographic code}} & \eczhListValue{Holographic tensor-network code constructed out of a network of encoding isometries of the \(\llbracket 6,1,3\rrbracket \) six-qubit stabilizer code.
The structure of the isometry is similar to that of the heptagon holographic code since both isometries are rank-six tensors, but the isometry in this case is neither a \flmRefsHyperref{ref219}{perfect tensor} nor a \flmRefsHyperref{code:block_perfect}{planar-perfect tensor}.}\\ 
\addlinespace[\myxtraspc]
\eczhRefIndex{code:holographic_subsystem}%
\eczhListValue{\flmRefsHyperref{code:holographic_subsystem}{Subsystem holographic code}} & \eczhListValue{A holographic tensor-network code constructed out of alternating isometries of the five-qubit and \(\llbracket 4,1,1,2\rrbracket \) Bacon-Shor codes.}\\ 
\addlinespace[\myxtraspc]
\eczhRefIndex{code:holographic_5_1_2}%
\eczhListValue{\flmRefsHyperref{code:holographic_5_1_2}{Surface-code-fragment (SCF) holographic code}} & \eczhListValue{Holographic tensor-network code constructed out of a network of encoding isometries of the \(\llbracket 5,1,2\rrbracket \) rotated surface code.
The structure of the isometry is similar to that of the HaPPY code since both isometries are rank-six tensors.
In the case of the SCF holographic code, the isometry is only a \flmRefsHyperref{code:block_perfect}{planar-perfect tensor} (as opposed to a \flmRefsHyperref{ref219}{perfect tensor}).}\\ 
\addlinespace[\myxtraspc]
\eczhRefIndex{code:stab_3_1_2}%
\eczhListValue{\flmRefsHyperref{code:stab_3_1_2}{\(\llbracket 3,1,2\rrbracket _3\) Three-qutrit code}} & \eczhListValue{A \(\llbracket 3,1,2\rrbracket _3\) prime-qudit CSS code that is the smallest qutrit stabilizer code to detect a single-qutrit error.
It has stabilizer generators \(ZZZ\) and \(XXX\). The code defines a quantum secret-sharing scheme and serves as a minimal model for the AdS/CFT holographic duality. It is also the smallest non-trivial instance of a quantum maximum distance separable code (QMDS), saturating the quantum Singleton bound.}\\ 
\addlinespace[\myxtraspc]
\eczhRefIndex{code:stab_5_1_2}%
\eczhListValue{\flmRefsHyperref{code:stab_5_1_2}{\(\llbracket 5,1,2\rrbracket \) rotated surface code}} & \eczhListValue{A rotated surface code on one rung of a ladder, with one qubit on the rung, and four qubits surrounding it. This is the smallest code that implements a fault-tolerant logical \(S\) gate using a diagonal depth-one Clifford circuit \NoCaseChange{\protect\cite{cite447}}.}\\ 
\addlinespace[\myxtraspc]
\eczhRefIndex{code:stab_5_1_3}%
\eczhListValue{\flmRefsHyperref{code:stab_5_1_3}{\(\llbracket 5,1,3\rrbracket \) Five-qubit perfect code}} & \eczhListValue{Five-qubit cyclic stabilizer code that is the smallest qubit stabilizer code to correct a single-qubit error.}\\ 
\addlinespace[\myxtraspc]
\eczhRefIndex{code:stab_6_1_3}%
\eczhListValue{\flmRefsHyperref{code:stab_6_1_3}{\(\llbracket 6,1,3\rrbracket \) Six-qubit stabilizer code}} & \eczhListValue{A degenerate, non-trivial \(\llbracket 6,1,3\rrbracket \) stabilizer code.
It is one of two six-qubit distance-three codes that are unique up to equivalence \NoCaseChange{\protect\cite{cite449}}, with the other code being decomposable and an extension of the five-qubit code \NoCaseChange{\protect\cite{cite451}\protect\cite[{ID 87}]{cite453}}.
The code admits fault-tolerant syndrome extraction using only one ancilla per stabilizer generator measurement.}\\ 
\addlinespace[\myxtraspc]
\eczhRefIndex{code:steane}%
\eczhListValue{\flmRefsHyperref{code:steane}{\(\llbracket 7,1,3\rrbracket \) Steane code}} & \eczhListValue{A \(\llbracket 7,1,3\rrbracket \) self-dual CSS code that is the smallest qubit CSS code to correct a single-qubit error \NoCaseChange{\protect\cite{cite451}}.
The code is constructed using the classical binary \([7,4,3]\) Hamming code for protecting against both \(X\) and \(Z\) errors.}\\ 
\end{tabularx}\endgroup
\eczcodelist{qubit_translationally_invariant}{Lattice qubit stabilizer codes
}%

\eczhCodeListAutoDescription{Codes that are descendants of all of \flmRefsCref{code:translationally_invariant_stabilizer}, \flmRefsCref{code:qubit_stabilizer}.}%

\eczhIncludeCodeGraph{Bare}{scale=0.5}{\columnwidth}{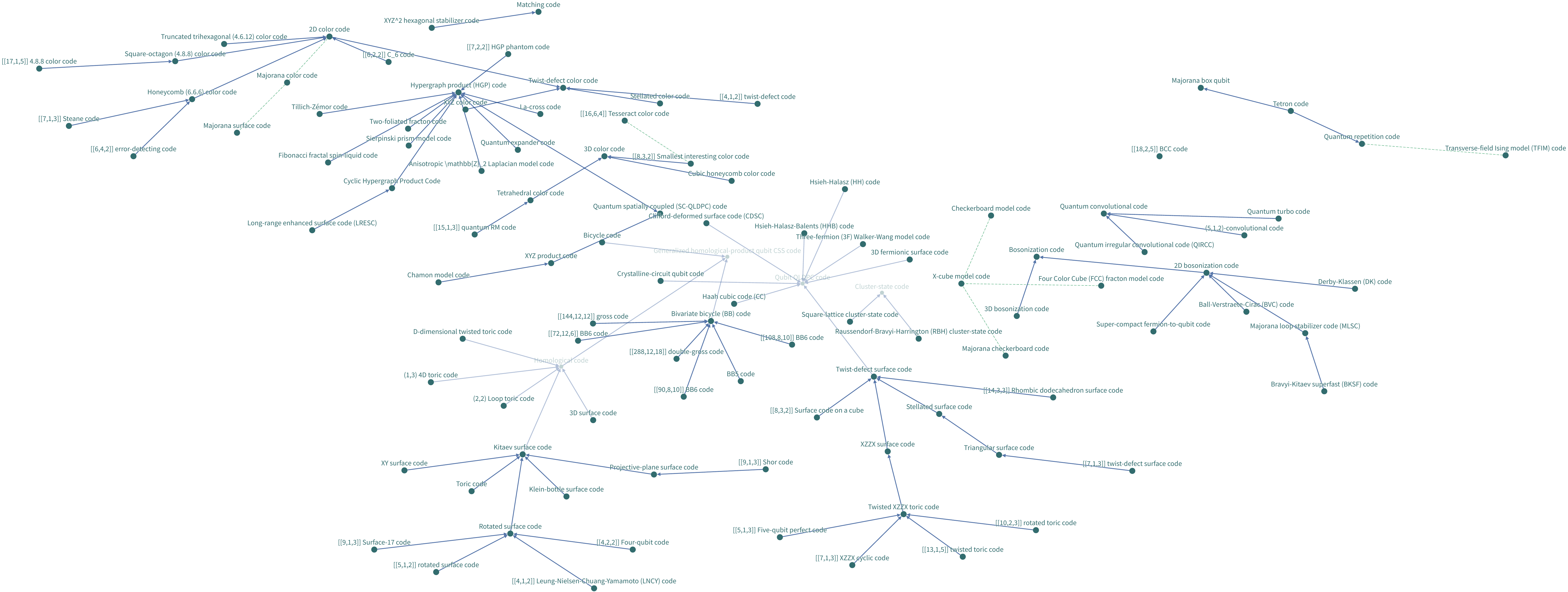}{Lattice qubit stabilizer codes}{https://errorcorrectionzoo.org/code_graph#J\%7B\%22displayMode\%22\%3A\%22subset\%22\%2C\%22modeSubsetOptions\%22\%3A\%7B\%22codeIds\%22\%3A\%5B\%222d_bosonization\%22\%2C\%222d_color\%22\%2C\%223d_bosonization\%22\%2C\%223d_color\%22\%2C\%223d_fermionic_surface\%22\%2C\%223d_surface\%22\%2C\%22anisotropic_z2_laplacian\%22\%2C\%22bb5\%22\%2C\%22bvc\%22\%2C\%22bicycle\%22\%2C\%22qcga\%22\%2C\%22bosonization\%22\%2C\%22bksf\%22\%2C\%22chamon\%22\%2C\%22checkerboard\%22\%2C\%22clifford-deformed_surface\%22\%2C\%22crystalline_dynamic_gen\%22\%2C\%22cubic_honeycomb_color\%22\%2C\%22cyclic_hgp\%22\%2C\%22derby_klassen\%22\%2C\%22fibonacci_fractal_liquid\%22\%2C\%22fcc_fracton\%22\%2C\%22haah_cubic\%22\%2C\%22triangular_color\%22\%2C\%22hh_fracton\%22\%2C\%22hhb_fracton\%22\%2C\%22hypergraph_product\%22\%2C\%22surface\%22\%2C\%22klein_bottle\%22\%2C\%22lacross\%22\%2C\%22lresc\%22\%2C\%22mbq\%22\%2C\%22majorana_checkerboard\%22\%2C\%22majorana_color\%22\%2C\%22mlsc\%22\%2C\%22majorana_surface\%22\%2C\%22matching\%22\%2C\%22real_projective_plane\%22\%2C\%22quantum_convolutional\%22\%2C\%22quantum_expander\%22\%2C\%22quantum_irregular_convolutional\%22\%2C\%22quantum_repetition\%22\%2C\%22sc_qldpc\%22\%2C\%22quantum_turbo\%22\%2C\%22rbh\%22\%2C\%22rotated_surface\%22\%2C\%22sierpinsky_fractal_liquid\%22\%2C\%22square_lattice_cluster\%22\%2C\%22488_color\%22\%2C\%22stellated_color\%22\%2C\%22stellated_surface\%22\%2C\%22super_compact\%22\%2C\%22tetrahedral_color\%22\%2C\%22tetron\%22\%2C\%22three_fermion\%22\%2C\%22tillichzemor\%22\%2C\%22toric\%22\%2C\%22tfim\%22\%2C\%22triangle_surface\%22\%2C\%224612_color\%22\%2C\%22twist_defect_color\%22\%2C\%22twist_defect_surface\%22\%2C\%22twisted_xzzx\%22\%2C\%22two_foliated\%22\%2C\%22xcube\%22\%2C\%22xysurface\%22\%2C\%22xyz_color\%22\%2C\%22xyz_product\%22\%2C\%22xyz_hexagonal\%22\%2C\%22xzzx\%22\%2C\%224d_13_surface\%22\%2C\%224d_surface\%22\%2C\%22stab_5_1_2_convolutional\%22\%2C\%22higher_dimensional_toric\%22\%2C\%22xzzx_10_2_3\%22\%2C\%22bb108\%22\%2C\%22stab_13_1_5\%22\%2C\%22rhombic_dodecahedron_surface\%22\%2C\%22gross\%22\%2C\%22stab_15_1_3\%22\%2C\%22stab_16_6_4\%22\%2C\%22stab_17_1_5\%22\%2C\%22stab_18_2_5\%22\%2C\%22bb288\%22\%2C\%22css_4_1_2\%22\%2C\%22stab_4_1_2\%22\%2C\%22stab_4_2_2\%22\%2C\%22stab_5_1_2\%22\%2C\%22stab_5_1_3\%22\%2C\%22stab_6_2_2\%22\%2C\%22stab_6_4_2\%22\%2C\%22steane\%22\%2C\%22xzzx_7_1_3\%22\%2C\%22twist_defect_7_1_3\%22\%2C\%22hgp_7_2_2\%22\%2C\%22bb72\%22\%2C\%22stab_8_3_2\%22\%2C\%22cubic_surface\%22\%2C\%22shor_nine\%22\%2C\%22surface-17\%22\%2C\%22bb90\%22\%5D\%2C\%22reusePreviousLayoutPositions\%22\%3Afalse\%2C\%22showIntermediateConnectingNodes\%22\%3Atrue\%2C\%22connectingNodesMaxDepth\%22\%3A15\%2C\%22connectingNodesPathMaxLength\%22\%3A20\%2C\%22connectingNodesMaxExtraDepth\%22\%3A3\%2C\%22connectingNodesOnlyKeepPathsWithAdditionalLength\%22\%3A1\%2C\%22connectingNodesToDomainsAndKingdoms\%22\%3Afalse\%2C\%22connectingNodesEdgeLengthsByType\%22\%3A\%7B\%22primaryParent\%22\%3A1\%2C\%22secondaryParent\%22\%3A4\%2C\%22cousin\%22\%3A6\%7D\%2C\%22nodeIds\%22\%3A\%5B\%5D\%7D\%2C\%22highlightImportantNodes\%22\%3A\%7B\%22highlightImportantNodes\%22\%3Afalse\%2C\%22highlightPrimaryParents\%22\%3Afalse\%2C\%22highlightRootConnectingEdges\%22\%3Afalse\%7D\%7D}

\begingroup
\small
\eczhBreakableDashes
\renewcommand\arraystretch{1.05}
\edef\myxtraspc{\eczListAddVSpaceXtraXtraStretch}
\begin{tabularx}{\linewidth}{>{\raggedright\arraybackslash}p{\eczListColWidth{name}} >{\hsize=1.0000\hsize }X}
\toprule
\eczListColTitle{Code} & \eczListColTitle{Description} \\
\midrule
\endfirsthead
\toprule
\eczListColTitleContinued{Code} & \eczListColTitleContinued{Description} \\
\midrule
\endhead
\bottomrule
\endfoot
\eczhRefIndex{code:2d_bosonization}%
\eczhListValue{\flmRefsHyperref{code:2d_bosonization}{2D bosonization code}} & \eczhListValue{A mapping between a 2D lattice quadratic Hamiltonian of Majorana modes and a 2D lattice of qubits.
The original exact 2D bosonization code \NoCaseChange{\protect\cite{cite403}} is a stabilizer code whose generators are products of plaquettes and stars of the surface code, with gauge constraints that project onto a toric-code-like subspace with emergent fermions \NoCaseChange{\protect\cite{cite403,cite404}}.
Finite-depth generalized local unitary Clifford circuits generate a family of equivalent local encodings with qubit-to-fermion ratio \(r = 1 + \frac{1}{2k}\) for any positive integer \(k\); the square-lattice compact encoding with \(r=1.5\) and the super-compact encoding with \(r=1.25\) are explicit examples \NoCaseChange{\protect\cite{cite404}}.}\\ 
\addlinespace[\myxtraspc]
\eczhRefIndex{code:2d_color}%
\eczhListValue{\flmRefsHyperref{code:2d_color}{2D color code}} & \eczhListValue{Color code defined on a graph embedded in a two-dimensional surface.
Each face hosts two stabilizer generators, a Pauli-\(X\) and a Pauli-\(Z\) string acting on all the qubits of the face.}\\ 
\addlinespace[\myxtraspc]
\eczhRefIndex{code:3d_bosonization}%
\eczhListValue{\flmRefsHyperref{code:3d_bosonization}{3D bosonization code}} & \eczhListValue{A mapping from a 3D lattice quadratic Hamiltonian of Majorana modes to a lattice of qubits which realizes a \(\mathbb{Z}_2\) gauge theory with a particular Gauss law.}\\ 
\addlinespace[\myxtraspc]
\eczhRefIndex{code:3d_color}%
\eczhListValue{\flmRefsHyperref{code:3d_color}{3D color code}} & \eczhListValue{Color code defined on a four-valent, four-colorable 3-colex in a 3-manifold.
In the original colex realization, qubits sit on vertices, \(X\)-type stabilizers are attached to 3-cells, and \(Z\)-type stabilizers are attached to faces \NoCaseChange{\protect\cite{cite430}}.}\\ 
\addlinespace[\myxtraspc]
\eczhRefIndex{code:3d_fermionic_surface}%
\eczhListValue{\flmRefsHyperref{code:3d_fermionic_surface}{3D fermionic surface code}} & \eczhListValue{A non-CSS variant of the 3D Kitaev surface code that realizes \(\mathbb{Z}_2\) gauge theory with an emergent fermion, i.e., the fermionic-charge bosonic-loop (FcBl) phase \NoCaseChange{\protect\cite{cite455}}.
The model can be defined on a cubic lattice in several ways \NoCaseChange{\protect\cite[{Eq. (D45-46)}]{cite456}}.
Realizations on other lattices also exist \NoCaseChange{\protect\cite{cite457}}, and the phase of this code also exists in the 3D Kitaev honeycomb model \NoCaseChange{\protect\cite{cite458}}.}\\ 
\addlinespace[\myxtraspc]
\eczhRefIndex{code:3d_surface}%
\eczhListValue{\flmRefsHyperref{code:3d_surface}{3D surface code}} & \eczhListValue{A generalization of the Kitaev surface code defined on a 3D cubic lattice.
Qubits are placed on edges, \(Z\)-type stabilizer generators are placed on square plaquettes oriented in all three directions, and \(X\)-type stabilizers are placed on the six edges neighboring every vertex \NoCaseChange{\protect\cite{cite459}}.}\\ 
\addlinespace[\myxtraspc]
\eczhRefIndex{code:anisotropic_z2_laplacian}%
\eczhListValue{\flmRefsHyperref{code:anisotropic_z2_laplacian}{Anisotropic \(\mathbb{Z}_2\) Laplacian model code}} & \eczhListValue{A graph-based analogue of a Type-I fracton phase with lineon-like restricted mobility \NoCaseChange{\protect\cite{cite460,cite461}}.}\\ 
\addlinespace[\myxtraspc]
\eczhRefIndex{code:bb5}%
\eczhListValue{\flmRefsHyperref{code:bb5}{BB5 code}} & \eczhListValue{A BB code with weight-five stabilizer generators
(contrasting with the weight-six checks of standard BB codes), designed and
benchmarked for long chains of trapped ions \NoCaseChange{\protect\cite{cite407}}.}\\ 
\addlinespace[\myxtraspc]
\eczhRefIndex{code:bvc}%
\eczhListValue{\flmRefsHyperref{code:bvc}{Ball-Verstraete-Cirac (BVC) code}} & \eczhListValue{A 2D fermion-into-qubit encoding that builds upon the JW transformation by eliminating the weight-\(O(n)\) non-local \(Z\)-type string at the expense of introducing an auxiliary qubit per site and local gauge constraints.
See \NoCaseChange{\protect\cite[{Sec. IV.B}]{cite404}} for details.}\\ 
\addlinespace[\myxtraspc]
\eczhRefIndex{code:bicycle}%
\eczhListValue{\flmRefsHyperref{code:bicycle}{Bicycle code}} & \eczhListValue{A CSS code whose stabilizer generator matrix blocks are \(H_{X}=H_{Z}=(A|A^T)\), where \(A\) is a circulant matrix.
The fact that \(A\) commutes with its transpose ensures that the CSS condition is satisfied.
Bicycle codes are the first QLDPC codes.}\\ 
\addlinespace[\myxtraspc]
\eczhRefIndex{code:qcga}%
\eczhListValue{\flmRefsHyperref{code:qcga}{Bivariate bicycle (BB) code}} & \eczhListValue{One of several Abelian 2BGA codes which admit time-optimal syndrome measurement circuits that can be implemented in a two-layer
architecture, a generalization of the square-lattice architecture
optimal for the surface codes.
Codes can be classified by the weight of their checks, e.g., by BB\(w\) where \(w\) is the check weight.}\\ 
\addlinespace[\myxtraspc]
\eczhRefIndex{code:bosonization}%
\eczhListValue{\flmRefsHyperref{code:bosonization}{Bosonization code}} & \eczhListValue{A mapping that maps a \(D\)-dimensional lattice quadratic Hamiltonian of Majorana modes into a lattice of qubits.
The resulting qubit code can realize various topological phases, depending on the initial Majorana-mode Hamiltonian and its symmetries.}\\ 
\addlinespace[\myxtraspc]
\eczhRefIndex{code:bksf}%
\eczhListValue{\flmRefsHyperref{code:bksf}{Bravyi-Kitaev superfast (BKSF) code}} & \eczhListValue{A single-error-detecting fermion-into-qubit encoding defined on a 2D qubit lattice whose stabilizers are associated with loops in the lattice.
For the square-lattice edge ordering used in Ref. \NoCaseChange{\protect\cite{cite404}}, the BKSF logical operators coincide with exact 2D bosonization on the dual lattice after relabeling \(X\) and \(Y\).
The code can be generalized to a single error-correcting code (i.e., with distance three) on graphs of degree \(\geq 6\) \NoCaseChange{\protect\cite{cite408}}.}\\ 
\addlinespace[\myxtraspc]
\eczhRefIndex{code:chamon}%
\eczhListValue{\flmRefsHyperref{code:chamon}{Chamon model code}} & \eczhListValue{A foliated type-I fracton non-CSS code defined on a cubic lattice using one weight-eight stabilizer generator acting on the eight vertices of each cube in the lattice \NoCaseChange{\protect\cite[{Eq. (D38)}]{cite456}}.}\\ 
\addlinespace[\myxtraspc]
\eczhRefIndex{code:checkerboard}%
\eczhListValue{\flmRefsHyperref{code:checkerboard}{Checkerboard model code}} & \eczhListValue{A foliated type-I fracton code defined on a cubic lattice that admits weight-eight  \(X\)- and \(Z\)-type stabilizer generators on the eight vertices of each cube in the lattice.
A tetrahedral Ising model can be used to obtain the checkerboard model by gauging \NoCaseChange{\protect\cite{cite462,cite463,cite233,cite464,cite465,cite466,cite467,cite468,cite469,cite470}} its subsystem symmetry \NoCaseChange{\protect\cite{cite233}}.
In that construction, the checkerboard model is self-dual under exchange of \(X\)- and \(Z\)-type stabilizers, and its composites include dimension-1 and dimension-2 excitations, i.e., lineons and planons in later terminology, with anyonic mutual and self-statistics \NoCaseChange{\protect\cite{cite233}}.}\\ 
\addlinespace[\myxtraspc]
\eczhRefIndex{code:clifford-deformed_surface}%
\eczhListValue{\flmRefsHyperref{code:clifford-deformed_surface}{Clifford-deformed surface code (CDSC)}} & \eczhListValue{A generally non-CSS derivative of the surface code defined by applying a translationally invariant constant-depth \flmRefsHyperref{ref409}{Clifford circuit} to the original (CSS) surface code.
Unlike the surface code, CDSCs include codes whose thresholds and subthreshold performance are enhanced under noise biased towards dephasing.
Examples of CDSCs include the XY code, XZZX code, and random CDSCs.}\\ 
\addlinespace[\myxtraspc]
\eczhRefIndex{code:crystalline_dynamic_gen}%
\eczhListValue{\flmRefsHyperref{code:crystalline_dynamic_gen}{Crystalline-circuit qubit code}} & \eczhListValue{Code dynamically generated by constant-depth unitary \flmRefsHyperref{ref409}{Clifford circuits} defined on a lattice with some crystalline symmetry. A notable example is the circuit defined on a rotated square lattice with vertices corresponding to iSWAP gates and edges decorated by \(R_X[\pi/2]\), a single-qubit rotation by \(\pi/2\) around the \(X\)-axis. This circuit is invariant under space-time translations by a unit cell \((T, a)\) and all transformations of the square lattice point group \(D_4\).}\\ 
\addlinespace[\myxtraspc]
\eczhRefIndex{code:cubic_honeycomb_color}%
\eczhListValue{\flmRefsHyperref{code:cubic_honeycomb_color}{Cubic honeycomb color code}} & \eczhListValue{3D color code defined on a four-colorable bitruncated cubic honeycomb uniform tiling.}\\ 
\addlinespace[\myxtraspc]
\eczhRefIndex{code:cyclic_hgp}%
\eczhListValue{\flmRefsHyperref{code:cyclic_hgp}{Cyclic Hypergraph Product Code}} & \eczhListValue{The hypergraph product code constructed using two low-weight circulant matrices. The code family \(\mathrm{C2}\) is the product of a cyclic LDPC code with itself, and the family \(\mathrm{CxR}\) is the product of the cyclic code with a repetition code.}\\ 
\addlinespace[\myxtraspc]
\eczhRefIndex{code:derby_klassen}%
\eczhListValue{\flmRefsHyperref{code:derby_klassen}{Derby-Klassen (DK) code}} & \eczhListValue{A fermion-into-qubit code defined on regular tilings with maximum degree 4 whose stabilizers are associated with loops in the tiling.
The code outperforms several other encodings in terms of encoding rate \NoCaseChange{\protect\cite[{Table I}]{cite412}}.
It has been extended for models with several modes per site \NoCaseChange{\protect\cite{cite413}}.}\\ 
\addlinespace[\myxtraspc]
\eczhRefIndex{code:fibonacci_fractal_liquid}%
\eczhListValue{\flmRefsHyperref{code:fibonacci_fractal_liquid}{Fibonacci fractal spin-liquid code}} & \eczhListValue{A fractal type-I fracton CSS code defined on a cubic lattice \NoCaseChange{\protect\cite[{Eq. (D23)}]{cite456}}.}\\ 
\addlinespace[\myxtraspc]
\eczhRefIndex{code:fcc_fracton}%
\eczhListValue{\flmRefsHyperref{code:fcc_fracton}{Four Color Cube (FCC) fracton model code}} & \eczhListValue{A fracton code obtained from four coupled X-cube models using p-membrane condensation.
A modular-qudit generalization has been proposed \NoCaseChange{\protect\cite{cite474}}.}\\ 
\addlinespace[\myxtraspc]
\eczhRefIndex{code:haah_cubic}%
\eczhListValue{\flmRefsHyperref{code:haah_cubic}{Haah cubic code (CC)}} & \eczhListValue{A 3D lattice stabilizer code on a length-\(L\) cubic lattice with one or two qubits per site.
Admits two types of stabilizer generators with support on each cube of the lattice.
In the non-CSS case, these two are related by spatial inversion.
For CSS codes, we require that the product of all corner operators is the identity.
We lastly require that there are no non-trivial string operators, meaning that single-site operators are a phase, and any period one logical operator \(l \in \mathsf{S}^{\perp}\) is just a phase.}\\ 
\addlinespace[\myxtraspc]
\eczhRefIndex{code:triangular_color}%
\eczhListValue{\flmRefsHyperref{code:triangular_color}{Honeycomb (6.6.6) color code}} & \eczhListValue{2D color code defined on a (typically triangular) patch of the 6.6.6 (honeycomb) tiling.
The usual triangular patch has three differently colored boundaries, encodes one logical qubit, and is local-Clifford equivalent to a folded surface/toric code with two smooth and two rough boundaries \NoCaseChange{\protect\cite{cite422}}.}\\ 
\addlinespace[\myxtraspc]
\eczhRefIndex{code:hh_fracton}%
\eczhListValue{\flmRefsHyperref{code:hh_fracton}{Hsieh-Halasz (HH) code}} & \eczhListValue{Member of one of two families of fracton codes, named HH-I and HH-II, defined on a cubic lattice with two qubits per site.
HH-I (HH-II) is a CSS (non-CSS) stabilizer code family, with the former identified as a foliated type-I fracton code that is decomposable into two separate lattice models \NoCaseChange{\protect\cite{cite456}}.
The sorting analysis of Ref. \NoCaseChange{\protect\cite{cite456}} leaves HH-II inconclusive, consistent with either a fractal type-I or a type-II fracton phase.}\\ 
\addlinespace[\myxtraspc]
\eczhRefIndex{code:hhb_fracton}%
\eczhListValue{\flmRefsHyperref{code:hhb_fracton}{Hsieh-Halasz-Balents (HHB) code}} & \eczhListValue{Member of one of two families of fracton codes, named HHB model A and B, defined on a cubic lattice with two qubits per site.
Both are expected to be foliated type-I fracton codes \NoCaseChange{\protect\cite[{Eqs. (D42-D43)}]{cite456}}.}\\ 
\addlinespace[\myxtraspc]
\eczhRefIndex{code:hypergraph_product}%
\eczhListValue{\flmRefsHyperref{code:hypergraph_product}{Hypergraph product (HGP) code}} & \eczhListValue{A member of a family of CSS codes whose stabilizer generator matrix is obtained from a hypergraph product of two classical linear binary codes.}\\ 
\addlinespace[\myxtraspc]
\eczhRefIndex{code:surface}%
\eczhListValue{\flmRefsHyperref{code:surface}{Kitaev surface code}} & \eczhListValue{A family of Abelian topological \flmRefsHyperref{code:css}{CSS stabilizer} codes
whose generators are few-body \(X\)-type and \(Z\)-type Pauli strings
associated to the stars and plaquettes, respectively, of a cellulation of a
two-dimensional surface (with a qubit located at each edge of the
cellulation).
Codewords correspond to ground states of the surface code Hamiltonian, and error operators create or annihilate pairs of anyonic charges or vortices.}\\ 
\addlinespace[\myxtraspc]
\eczhRefIndex{code:klein_bottle}%
\eczhListValue{\flmRefsHyperref{code:klein_bottle}{Klein-bottle surface code}} & \eczhListValue{A family of Kitaev surface codes on the non-orientable Klein bottle.}\\ 
\addlinespace[\myxtraspc]
\eczhRefIndex{code:lacross}%
\eczhListValue{\flmRefsHyperref{code:lacross}{La-cross code}} & \eczhListValue{Code constructed using a hypergraph product of two copies of a cyclic LDPC code.
The construction uses cyclic LDPC codes with \flmRefsHyperref{ref67}{generating polynomials} \(1+x+x^k\) for some \(k\).
Using a length-\(n\) seed code yields an \(\llbracket 2n^2,2k^2\rrbracket \) family for periodic boundary conditions and an \(\llbracket (n-k)^2+n^2,k^2\rrbracket \) family for open boundary conditions.}\\ 
\addlinespace[\myxtraspc]
\eczhRefIndex{code:lresc}%
\eczhListValue{\flmRefsHyperref{code:lresc}{Long-range enhanced surface code (LRESC)}} & \eczhListValue{Code constructed using a hypergraph product of two copies of a concatenated LDPC-repetition seed code.
This family interpolates between surface codes and hypergraph codes since the hypergraph product of two repetition codes yields the planar surface code.
The construction uses small \([3,2,2]\) and \([6,2,4]\) LDPC codes concatenated with \([4,1,4]\) and \([2,1,2]\) repetition codes, respectively.
An example using a \([5,2,3]\) code is also presented.}\\ 
\addlinespace[\myxtraspc]
\eczhRefIndex{code:mbq}%
\eczhListValue{\flmRefsHyperref{code:mbq}{Majorana box qubit}} & \eczhListValue{A family of Majorana stabilizer codes obtained by fixing the total fermion parity of \(n\) fermionic modes, equivalently \(2n\) Majorana zero modes, within the ground-state subspace of \(n\) Kitaev Majorana chain Hamiltonians.
The resulting positive-parity subspace encodes \(n-1\) logical qubits and has Majorana distance \(2\).}\\ 
\addlinespace[\myxtraspc]
\eczhRefIndex{code:majorana_checkerboard}%
\eczhListValue{\flmRefsHyperref{code:majorana_checkerboard}{Majorana checkerboard code}} & \eczhListValue{A Majorana analogue of the X-cube model defined on a cubic lattice.
The code admits weight-eight Majorana stabilizer generators on the eight vertices of each cube of a checkerboard sublattice.}\\ 
\addlinespace[\myxtraspc]
\eczhRefIndex{code:majorana_color}%
\eczhListValue{\flmRefsHyperref{code:majorana_color}{Majorana color code}} & \eczhListValue{A fermionic analogue of a 2D color code.}\\ 
\addlinespace[\myxtraspc]
\eczhRefIndex{code:mlsc}%
\eczhListValue{\flmRefsHyperref{code:mlsc}{Majorana loop stabilizer code (MLSC)}} & \eczhListValue{A single-error-correcting fermion-into-qubit encoding defined on a 2D qubit lattice whose stabilizers are associated with loops in the lattice.}\\ 
\addlinespace[\myxtraspc]
\eczhRefIndex{code:majorana_surface}%
\eczhListValue{\flmRefsHyperref{code:majorana_surface}{Majorana surface code}} & \eczhListValue{Fermionic analogue of the surface code defined on a three-colorable 2D tiling whose face operators are non-overlapping even-Majorana stabilizers.
Open patches with four or six alternating colored boundaries encode logical tetrons or hexons.
The uniform 4.8.8, 6.6.6, and 4.6.12 tilings yield families with tetron, hexon, or dodecon building blocks and with twist-based lattice surgery supporting minimal-overhead logical Clifford gates \NoCaseChange{\protect\cite{cite402}}.}\\ 
\addlinespace[\myxtraspc]
\eczhRefIndex{code:matching}%
\eczhListValue{\flmRefsHyperref{code:matching}{Matching code}} & \eczhListValue{Member of a class of qubit stabilizer codes based on the Abelian phase of the Kitaev honeycomb model.}\\ 
\addlinespace[\myxtraspc]
\eczhRefIndex{code:real_projective_plane}%
\eczhListValue{\flmRefsHyperref{code:real_projective_plane}{Projective-plane surface code}} & \eczhListValue{A family of Kitaev surface codes on the non-orientable 2-dimensional compact manifold \(\mathbb{R}P^2\) (in contrast to a genus-\(g\) surface).
Whereas genus-\(g\) surface codes require \(2g\) logical qubits, qubit codes on \(\mathbb{R}P^2\) are made from a single logical qubit.}\\ 
\addlinespace[\myxtraspc]
\eczhRefIndex{code:quantum_convolutional}%
\eczhListValue{\flmRefsHyperref{code:quantum_convolutional}{Quantum convolutional code}} & \eczhListValue{1D translationally invariant qubit stabilizer code whose stabilizer group can be partitioned into constant-size subsets of constant support and of constant overlap between neighboring sets.
Initially formulated as a quantum analogue of convolutional codes, which were designed to protect a continuous and never-ending stream of information.
Precise formulations sometimes begin with a finite-dimensional lattice, with the intent to take the thermodynamic limit; logical dimension can be infinite as well.}\\ 
\addlinespace[\myxtraspc]
\eczhRefIndex{code:quantum_expander}%
\eczhListValue{\flmRefsHyperref{code:quantum_expander}{Quantum expander code}} & \eczhListValue{CSS code constructed from a hypergraph product of bipartite expander graphs \NoCaseChange{\protect\cite{cite74}} with bounded left and right vertex degrees. For every bipartite graph there is an associated matrix (the parity check matrix) with columns indexed by the left vertices, rows indexed by the right vertices, and 1 entries whenever a left and right vertex are connected. This matrix can serve as the parity check matrix of a classical code. Two bipartite expander graphs can be used to construct a quantum CSS code (the quantum expander code) via the hypergraph product of their parity check matrices.}\\ 
\addlinespace[\myxtraspc]
\eczhRefIndex{code:quantum_irregular_convolutional}%
\eczhListValue{\flmRefsHyperref{code:quantum_irregular_convolutional}{Quantum irregular convolutional code (QIRCC)}} & \eczhListValue{Quantum convolutional code whose stabilizer group consists of different constant-size subsets.}\\ 
\addlinespace[\myxtraspc]
\eczhRefIndex{code:quantum_repetition}%
\eczhListValue{\flmRefsHyperref{code:quantum_repetition}{Quantum repetition code}} & \eczhListValue{Encodes \(1\) qubit into \(n\) qubits according to \(|0\rangle\to|\phi_0\rangle^{\otimes n}\) and \(|1\rangle\to|\phi_1\rangle^{\otimes n}\). The code is called a \textit{bit-flip} code when \(|\phi_i\rangle = |i\rangle\), and a \textit{phase-flip} code when \(|\phi_0\rangle = |+\rangle\) and \(|\phi_1\rangle = |-\rangle\).
This repetition-style encoding does not clone an arbitrary quantum state; instead, it extends the copying of computational-basis states linearly to entangled codewords  \NoCaseChange{\protect\cite[{Ch. 2}]{cite398}}.}\\ 
\addlinespace[\myxtraspc]
\eczhRefIndex{code:sc_qldpc}%
\eczhListValue{\flmRefsHyperref{code:sc_qldpc}{Quantum spatially coupled (SC-QLDPC) code}} & \eczhListValue{QLDPC code whose stabilizer generator matrix resembles the parity-check matrix of SC-LDPC codes.
There exist CSS \NoCaseChange{\protect\cite{cite643}} and stabilizer constructions \NoCaseChange{\protect\cite{cite644}}.
In either case, the stabilizer generator matrix is constructed by "spatially" coupling sub-matrix blocks in chain-like fashion (or, more generally, in grid-like fashion) to yield a band matrix.
The sub-matrix blocks have to satisfy certain conditions amongst themselves so that the resulting band matrix is a stabilizer generator matrix.
Matrices corresponding to translationally invariant chains are called \textit{time-invariant}, and otherwise are called \textit{time-varying}.}\\ 
\addlinespace[\myxtraspc]
\eczhRefIndex{code:quantum_turbo}%
\eczhListValue{\flmRefsHyperref{code:quantum_turbo}{Quantum turbo code}} & \eczhListValue{A quantum version of the turbo code, obtained from an interleaved serial quantum concatenation \NoCaseChange{\protect\cite[{Def. 30}]{cite399}} of quantum convolutional codes.
The interleaver induces long-range entanglement and can increase the minimum distance relative to the constituent convolutional codes \NoCaseChange{\protect\cite{cite400}}.}\\ 
\addlinespace[\myxtraspc]
\eczhRefIndex{code:rbh}%
\eczhListValue{\flmRefsHyperref{code:rbh}{Raussendorf-Bravyi-Harrington (RBH) cluster-state code}} & \eczhListValue{A three-dimensional cluster-state code defined on the bcc lattice (i.e., a cubic lattice with qubits on edges and faces).}\\ 
\addlinespace[\myxtraspc]
\eczhRefIndex{code:rotated_surface}%
\eczhListValue{\flmRefsHyperref{code:rotated_surface}{Rotated surface code}} & \eczhListValue{Variant of the surface code obtained by taking the medial graph of the surface code lattice (treated as a graph) and applying a procedure to construct the check operators \NoCaseChange{\protect\cite{cite425,cite426}\protect\cite[{Fig. 8}]{cite427}}.
On a square lattice, this amounts to a rotation by 45 degrees such that qubits are on vertices, and both \(X\)- and \(Z\)-type check operators occupy plaquettes in an alternating checkerboard pattern.}\\ 
\addlinespace[\myxtraspc]
\eczhRefIndex{code:sierpinsky_fractal_liquid}%
\eczhListValue{\flmRefsHyperref{code:sierpinsky_fractal_liquid}{Sierpinski prism model code}} & \eczhListValue{A fractal type-I fracton CSS code defined on a cubic lattice \NoCaseChange{\protect\cite[{Eq. (D22)}]{cite456}}.
The code admits an excitation-moving operator shaped like a Sierpinski triangle \NoCaseChange{\protect\cite[{Fig. 2}]{cite456}}.}\\ 
\addlinespace[\myxtraspc]
\eczhRefIndex{code:square_lattice_cluster}%
\eczhListValue{\flmRefsHyperref{code:square_lattice_cluster}{Square-lattice cluster-state code}} & \eczhListValue{A code based on the cluster state on a square lattice that was used in the first proposal for MBQC \NoCaseChange{\protect\cite{cite428,cite429}}.
In the one-way model, the pre-entangled square-lattice cluster is a universal resource, and the computation is carried out entirely by adaptive single-qubit measurements.}\\ 
\addlinespace[\myxtraspc]
\eczhRefIndex{code:488_color}%
\eczhListValue{\flmRefsHyperref{code:488_color}{Square-octagon (4.8.8) color code}} & \eczhListValue{2D color code defined on a patch of the 4.8.8 (square-octagon) tiling, which itself is obtained by applying a fattening procedure to the square lattice \NoCaseChange{\protect\cite{cite430}}.
An equivalent description uses the Tetrakis square tiling (a.k.a. the Union Jack lattice), which is dual to the 4.8.8 lattice \NoCaseChange{\protect\cite{cite431}}.
Among the three semiregular triangular 2D color-code families, the 4.8.8 family uses the fewest physical qubits for a given distance and is the only one of the three with transversal implementations of the full Clifford group \NoCaseChange{\protect\cite{cite432}}.}\\ 
\addlinespace[\myxtraspc]
\eczhRefIndex{code:stellated_color}%
\eczhListValue{\flmRefsHyperref{code:stellated_color}{Stellated color code}} & \eczhListValue{A non-CSS color-code family on a lattice patch with a single central puncture that hosts a twist defect connected to the boundary by a domain wall.}\\ 
\addlinespace[\myxtraspc]
\eczhRefIndex{code:stellated_surface}%
\eczhListValue{\flmRefsHyperref{code:stellated_surface}{Stellated surface code}} & \eczhListValue{A twist-defect surface-code family parameterized by a rotational symmetry order \(s\), with a central toric-code twist connected to the boundary by a domain wall.
The \(s=3\) member is the triangular surface code \NoCaseChange{\protect\cite{cite433}}.}\\ 
\addlinespace[\myxtraspc]
\eczhRefIndex{code:super_compact}%
\eczhListValue{\flmRefsHyperref{code:super_compact}{Super-compact fermion-to-qubit code}} & \eczhListValue{A 2D fermion-into-qubit encoding on the square lattice obtained from exact 2D bosonization by a finite-depth generalized local unitary Clifford circuit, followed by re-pairing of Majorana modes and a slight lattice deformation.
The code uses \(1.25\) qubits per fermion, improving on the square-lattice compact encoding with ratio \(r=1.5\).
Its fermion-parity, hopping, and stabilizer operators have weights \(1\)-\(2\), \(2\)-\(6\), and \(12\), respectively \NoCaseChange{\protect\cite[{Table I}]{cite404}}.}\\ 
\addlinespace[\myxtraspc]
\eczhRefIndex{code:tetrahedral_color}%
\eczhListValue{\flmRefsHyperref{code:tetrahedral_color}{Tetrahedral color code}} & \eczhListValue{A 3D color code defined on a colored tetrahedron cut from a suitably colored BCC lattice \NoCaseChange{\protect\cite{cite475}}.
Qubits are placed on tetrahedra, on the triangles covering the tetrahedron faces, on the edges along the tetrahedron edges, and on the tetrahedron vertices.
The code has both string-like and sheet-like logical operators \NoCaseChange{\protect\cite{cite476}}.}\\ 
\addlinespace[\myxtraspc]
\eczhRefIndex{code:tetron}%
\eczhListValue{\flmRefsHyperref{code:tetron}{Tetron code}} & \eczhListValue{A \(\llbracket 2,1,2\rrbracket _{f}\) Majorana box qubit encoding a logical qubit into four Majorana modes, equivalently into the fixed-total-parity sector of two physical fermionic modes.
Four Majorana zero modes are the smallest aggregate that supports a qubit in a fixed fermion-parity sector \NoCaseChange{\protect\cite{cite401}}.
This code can be concatenated with various qubit codes such as surface codes and color codes.
Four-boundary Majorana surface-code patches are logical tetrons, i.e., higher-distance analogues of this physical tetron block \NoCaseChange{\protect\cite{cite402}}.}\\ 
\addlinespace[\myxtraspc]
\eczhRefIndex{code:three_fermion}%
\eczhListValue{\flmRefsHyperref{code:three_fermion}{Three-fermion (3F) Walker-Wang model code}} & \eczhListValue{A 3D lattice stabilizer code whose bulk realizes a 3D time-reversal SPT order \NoCaseChange{\protect\cite{cite477}} and whose gapped boundary supports the 2D three-fermion (3F) topological order.
The code can be used as a resource state for fault-tolerant MBQC \NoCaseChange{\protect\cite{cite478}}.}\\ 
\addlinespace[\myxtraspc]
\eczhRefIndex{code:tillichzemor}%
\eczhListValue{\flmRefsHyperref{code:tillichzemor}{Tillich-Zémor code}} & \eczhListValue{A family of \(\llbracket n^2 + m^2, (n - \text{rank}([C \mid M]) )^2 + (m - \text{rank}([C \mid M]^\top) )^2, d\rrbracket \) quantum LDPC codes constructed via the hypergraph product of two classical \((n, m, r)\)-structured LDPC seed codes}\\ 
\addlinespace[\myxtraspc]
\eczhRefIndex{code:toric}%
\eczhListValue{\flmRefsHyperref{code:toric}{Toric code}} & \eczhListValue{Version of the Kitaev surface code on a square lattice with periodic boundary conditions, encoding two logical qubits.
Being the first manifestation of the surface code, "toric code" is often an alternative name for the general construction.
\textit{Twisted toric code} \NoCaseChange{\protect\cite[{Fig. 8}]{cite434}} refers to the construction on a torus with twisted (a.k.a. shifted) boundary conditions.
In the original Hamiltonian construction, open Pauli-\(Z\) and Pauli-\(X\) strings create pairs of electric charges and magnetic vortices, and braiding one type around the other yields the nontrivial Abelian anyonic phase \NoCaseChange{\protect\cite{cite423}}.}\\ 
\addlinespace[\myxtraspc]
\eczhRefIndex{code:tfim}%
\eczhListValue{\flmRefsHyperref{code:tfim}{Transverse-field Ising model (TFIM) code}} & \eczhListValue{A 1D translationally invariant stabilizer code whose encoding is a constant-depth circuit of nearest-neighbor gates on alternating even and odd bonds that consist of transverse-field Ising Hamiltonian interactions. The code allows for perfect state transfer of arbitrary distance using local operations and classical communications (LOCC).}\\ 
\addlinespace[\myxtraspc]
\eczhRefIndex{code:triangle_surface}%
\eczhListValue{\flmRefsHyperref{code:triangle_surface}{Triangular surface code}} & \eczhListValue{A member of a twist-defect surface code family with a single central twist whose planar layout fits within a triangle.
Triangle codes can be viewed as three conjoined surface-code patches projected into two dimensions, with weight-four plaquette stabilizers and weight-two edge stabilizers \NoCaseChange{\protect\cite{cite433}}.
Symmetric distance-\(d\) triangle codes use \(3d^2/4+1/4\) data qubits, i.e., about \(25\%\) fewer than the rotated surface code for a given odd distance.
Logical \(\overline{X}\), \(\overline{Y}\), and \(\overline{Z}\) operators can be supported on the three sides of the triangle, enabling initialization and measurement in any Pauli basis.}\\ 
\addlinespace[\myxtraspc]
\eczhRefIndex{code:4612_color}%
\eczhListValue{\flmRefsHyperref{code:4612_color}{Truncated trihexagonal (4.6.12) color code}} & \eczhListValue{2D color code defined on a (typically triangular) patch of the 4.6.12 (truncated trihexagonal or square-hexagon-dodecagon) tiling.}\\ 
\addlinespace[\myxtraspc]
\eczhRefIndex{code:twist_defect_color}%
\eczhListValue{\flmRefsHyperref{code:twist_defect_color}{Twist-defect color code}} & \eczhListValue{A non-CSS extension of the 2D color code whose non-CSS stabilizer generators are associated with twist defects of the associated lattice.
These twists terminate domain walls that permute color labels, Pauli labels, or interchange the two.}\\ 
\addlinespace[\myxtraspc]
\eczhRefIndex{code:twist_defect_surface}%
\eczhListValue{\flmRefsHyperref{code:twist_defect_surface}{Twist-defect surface code}} & \eczhListValue{A non-CSS extension of the 2D surface-code construction whose non-CSS stabilizer generators are associated with twist defects of the associated lattice.
A related construction \NoCaseChange{\protect\cite{cite435}} doubles the number of qubits in the lattice via \flmRefsHyperref{ref436}{symplectic doubling}.}\\ 
\addlinespace[\myxtraspc]
\eczhRefIndex{code:twisted_xzzx}%
\eczhListValue{\flmRefsHyperref{code:twisted_xzzx}{Twisted XZZX toric code}} & \eczhListValue{A cyclic code that can be thought of as the XZZX toric code with shifted (a.k.a twisted) boundary conditions.
Admits a set of stabilizer generators that are equivalent to cyclic shifts of a particular weight-four \(XZZX\) Pauli string.}\\ 
\addlinespace[\myxtraspc]
\eczhRefIndex{code:two_foliated}%
\eczhListValue{\flmRefsHyperref{code:two_foliated}{Two-foliated fracton code}} & \eczhListValue{A type-I fracton code obtained by gauging \NoCaseChange{\protect\cite{cite462,cite463,cite233,cite464,cite465,cite466,cite467,cite468,cite469,cite470}} a 3D paramagnet with planar subsystem symmetries in two directions.
In that construction, the gauge charges are lineons and the flux excitations are also lineons moving in the same direction, yielding the anisotropic lineon model \NoCaseChange{\protect\cite[{Sec. 4.1.2}]{cite467}}.}\\ 
\addlinespace[\myxtraspc]
\eczhRefIndex{code:xcube}%
\eczhListValue{\flmRefsHyperref{code:xcube}{X-cube model code}} & \eczhListValue{A foliated type-I fracton CSS code on a cubic lattice with qubits on edges, cube stabilizers, and three cross-shaped vertex stabilizers for each vertex \NoCaseChange{\protect\cite{cite233}}.
It supports a subextensive number of logical qubits.}\\ 
\addlinespace[\myxtraspc]
\eczhRefIndex{code:xysurface}%
\eczhListValue{\flmRefsHyperref{code:xysurface}{XY surface code}} & \eczhListValue{A variant of the surface code whose generators are \(XXXX\)  and \(YYYY\), obtained by mapping \(Z \to Y\) in the surface code.
While this code is equivalent to a CSS surface code with the same distance, other properties like noise-bias performance can differ significantly.}\\ 
\addlinespace[\myxtraspc]
\eczhRefIndex{code:xyz_color}%
\eczhListValue{\flmRefsHyperref{code:xyz_color}{XYZ color code}} & \eczhListValue{A variant of the 6.6.6 color code whose generators are \(XZXZXZ\) and \(ZYZYZY\) Pauli strings associated to each hexagonal in the hexagonal (6.6.6) tiling. 
A further variation called the \textit{domain wall color code} admits generators of the form \(XXXZZZ\) and \(ZZZXXX\) \NoCaseChange{\protect\cite{cite437}}.
While such codes are equivalent to CSS color codes with the same distance, other properties like noise-bias performance can differ significantly.}\\ 
\addlinespace[\myxtraspc]
\eczhRefIndex{code:xyz_product}%
\eczhListValue{\flmRefsHyperref{code:xyz_product}{XYZ product code}} & \eczhListValue{A non-CSS QLDPC code obtained from a three-fold variant of the hypergraph product applied to three classical binary codes with parity-check matrices \(H_1,H_2,H_3\).  Unlike CSS three-fold hypergraph product codes, the third input code acts through Pauli-\(Y\) checks \NoCaseChange{\protect\cite{cite645}}.
Under mild assumptions, the code dimension is determined by a tensor Sylvester equation over \(\mathbb{F}_2\), and the minimum-distance problem reduces up to constant factors to how closely a related inhomogeneous tensor Sylvester equation can be satisfied \NoCaseChange{\protect\cite{cite645}}.  When the underlying classical codes are repetition codes, the construction yields the \flmRefsHyperref{code:chamon}{Chamon model code}.}\\ 
\addlinespace[\myxtraspc]
\eczhRefIndex{code:xyz_hexagonal}%
\eczhListValue{\flmRefsHyperref{code:xyz_hexagonal}{XYZ\(^2\) hexagonal stabilizer code}} & \eczhListValue{An instance of the matching code based on the Kitaev honeycomb model. It is described on a honeycomb tiling with \(XYZXYZ\) stabilizers on each hexagonal plaquette. Each vertical pair of qubits has an \(XX\), \(YY\), or \(ZZ\) link stabilizer depending on the orientation of the plaquette stabilizers.}\\ 
\addlinespace[\myxtraspc]
\eczhRefIndex{code:xzzx}%
\eczhListValue{\flmRefsHyperref{code:xzzx}{XZZX surface code}} & \eczhListValue{A variant of the rotated surface code whose generators are \(XZZX\) Pauli strings associated, clockwise, to the vertices of each face of a two-dimensional lattice (with a qubit located at each vertex of the tessellation).}\\ 
\addlinespace[\myxtraspc]
\eczhRefIndex{code:4d_13_surface}%
\eczhListValue{\flmRefsHyperref{code:4d_13_surface}{\((1,3)\) 4D toric code}} & \eczhListValue{A generalization of the Kitaev surface code defined on a 4D lattice.
The code is called a \((1,3)\) toric code because it admits 1D \(Z\)-type and 3D \(X\)-type logical operators.
In the hypercubic lattice version, qubits are placed on edges, each \(Z\)-type stabilizer generator is supported on cubes on the boundary of a hypercube, and \(X\)-type stabilizers are placed on the edges neighboring every vertex \NoCaseChange{\protect\cite{cite479}}.}\\ 
\addlinespace[\myxtraspc]
\eczhRefIndex{code:4d_surface}%
\eczhListValue{\flmRefsHyperref{code:4d_surface}{\((2,2)\) Loop toric code}} & \eczhListValue{A generalization of the Kitaev surface code defined on a 4D lattice.
The code is called a \((2,2)\) toric code because it admits 2D membrane \(Z\)-type and \(X\)-type logical operators.
Both types of operators create 1D (i.e., loop) excitations at their edges.
The code serves as a self-correcting quantum memory \NoCaseChange{\protect\cite{cite480,cite481}}.}\\ 
\addlinespace[\myxtraspc]
\eczhRefIndex{code:stab_5_1_2_convolutional}%
\eczhListValue{\flmRefsHyperref{code:stab_5_1_2_convolutional}{\((5,1,2)\)-convolutional code}} & \eczhListValue{Family of quantum convolutional codes that are 1D lattice generalizations of the five-qubit perfect code, with the former's lattice-translation symmetry being the extension of the latter's cyclic permutation symmetry.}\\ 
\addlinespace[\myxtraspc]
\eczhRefIndex{code:higher_dimensional_toric}%
\eczhListValue{\flmRefsHyperref{code:higher_dimensional_toric}{\(D\)-dimensional twisted toric code}} & \eczhListValue{Extension of the Kitaev toric code to higher-dimensional lattices with regular or shifted (a.k.a. twisted) boundary conditions.
Such boundary conditions yield qubit geometries that are tori \(\mathbb{R}^D/\Lambda\), where \(\Lambda\) is an arbitrary \(D\)-dimensional lattice.
Picking a hypercubic lattice yields the ordinary \(D\)-dimensional toric code.}\\ 
\addlinespace[\myxtraspc]
\eczhRefIndex{code:xzzx_10_2_3}%
\eczhListValue{\flmRefsHyperref{code:xzzx_10_2_3}{\(\llbracket 10,2,3\rrbracket \) rotated toric code}} & \eczhListValue{Rotated toric code that is the CSS form of the twisted XZZX toric code with parameters \(a=1\), \(b=3\) \NoCaseChange{\protect\cite{cite438,cite427}},
related to the XZZX form by Hadamard on the \(B\)-sublattice.
It is also the \flmRefsHyperref{ref436}{symplectic double} (a.k.a. genus-one double cover) of the \(\llbracket 5,1,3\rrbracket \) five-qubit perfect code \NoCaseChange{\protect\cite{cite439,cite435}}, the \flmRefsHyperref{ref436}{symplectic double} of the \(\llbracket 5,1,2\rrbracket \) rotated surface code \NoCaseChange{\protect\cite{cite435}}, and a BCC code \NoCaseChange{\protect\cite{cite440}}.}\\ 
\addlinespace[\myxtraspc]
\eczhRefIndex{code:bb108}%
\eczhListValue{\flmRefsHyperref{code:bb108}{\(\llbracket 108,8,10\rrbracket \) BB6 code}} & \eczhListValue{A bivariate bicycle (BB) code with parameters \(\llbracket 108,8,10\rrbracket \) and weight-six stabilizer generators \NoCaseChange{\protect\cite{cite441}}.}\\ 
\addlinespace[\myxtraspc]
\eczhRefIndex{code:stab_13_1_5}%
\eczhListValue{\flmRefsHyperref{code:stab_13_1_5}{\(\llbracket 13,1,5\rrbracket \) twisted toric code}} & \eczhListValue{Thirteen-qubit twisted toric code whose stabilizer tableau consists of cyclic permutations of the \(XZZX\)-type Pauli string \(XIZZIXIIIIIII\).
The code can be thought of as a small twisted XZZX code \NoCaseChange{\protect\cite[{Exam. 11 and Fig. 3}]{cite438}}.}\\ 
\addlinespace[\myxtraspc]
\eczhRefIndex{code:rhombic_dodecahedron_surface}%
\eczhListValue{\flmRefsHyperref{code:rhombic_dodecahedron_surface}{\(\llbracket 14,3,3\rrbracket \) Rhombic dodecahedron surface code}} & \eczhListValue{A \(\llbracket 14,3,3\rrbracket \) twist-defect surface code whose qubits lie on the vertices of a rhombic dodecahedron.
Its non-CSS nature is due to twist defects \NoCaseChange{\protect\cite{cite442}} stemming from the geometry of the polytope.
A local-Clifford-equivalent clean realization has only \(X\)- and \(Z\)-type operators on its four-valent vertices, and its symplectic double is a \(\llbracket 28,6,3\rrbracket \) genus-three code \NoCaseChange{\protect\cite{cite435}}.}\\ 
\addlinespace[\myxtraspc]
\eczhRefIndex{code:gross}%
\eczhListValue{\flmRefsHyperref{code:gross}{\(\llbracket 144,12,12\rrbracket \) gross code}} & \eczhListValue{A BB code which requires less physical and ancilla qubits (for syndrome extraction) than the surface code with the same number of logical qubits and distance.
The gross code is equivalent to 8 copies of the surface code via a constant-depth Clifford circuit, and is an element of a larger family of 2D stabilizer codes \NoCaseChange{\protect\cite{cite443}}.
The name stems from the fact that a gross is a dozen dozen.}\\ 
\addlinespace[\myxtraspc]
\eczhRefIndex{code:stab_15_1_3}%
\eczhListValue{\flmRefsHyperref{code:stab_15_1_3}{\(\llbracket 15,1,3\rrbracket \) quantum RM code}} & \eczhListValue{A \(\llbracket 15,1,3\rrbracket \) quantum Reed-Muller code that is most easily thought of as a tetrahedral 3D color code.
It can be constructed as a CSS code from the \([15,5,8]\) punctured Reed-Muller code and its even subcode, which explains its transversal \(T^\dagger\) gate \NoCaseChange{\protect\cite{cite398}}.}\\ 
\addlinespace[\myxtraspc]
\eczhRefIndex{code:stab_16_6_4}%
\eczhListValue{\flmRefsHyperref{code:stab_16_6_4}{\(\llbracket 16,6,4\rrbracket \) Tesseract color code}} & \eczhListValue{A (hyperbolic self-dual CSS) 4D color code defined on a tesseract, with stabilizer generators of both types supported on each cube. 
A \(\llbracket 16,4,2,4\rrbracket \) tesseract subsystem code can be obtained from this code by using two logical qubits as gauge qubits \NoCaseChange{\protect\cite{cite483}}.}\\ 
\addlinespace[\myxtraspc]
\eczhRefIndex{code:stab_17_1_5}%
\eczhListValue{\flmRefsHyperref{code:stab_17_1_5}{\(\llbracket 17,1,5\rrbracket \) 4.8.8 color code}} & \eczhListValue{Seventeen-qubit doubly even 2D color code that admits a transversal implementation of the logical Clifford group.
The smallest distance-five CSS code has \(n=17\) \NoCaseChange{\protect\cite{cite444}}.
It is also a normal self-dual CSS code whose transversal Hadamard acts logically, making it suitable as an inner code for fifth-order magic-state distillation \NoCaseChange{\protect\cite{cite101}}.}\\ 
\addlinespace[\myxtraspc]
\eczhRefIndex{code:stab_18_2_5}%
\eczhListValue{\flmRefsHyperref{code:stab_18_2_5}{\(\llbracket 18,2,5\rrbracket \) BCC code}} & \eczhListValue{BCC code on 18 qubits encoding 2 logical qubits with distance 5, found by computer search \NoCaseChange{\protect\cite{cite440}}.}\\ 
\addlinespace[\myxtraspc]
\eczhRefIndex{code:bb288}%
\eczhListValue{\flmRefsHyperref{code:bb288}{\(\llbracket 288,12,18\rrbracket \) double-gross code}} & \eczhListValue{A bivariate bicycle (BB) code with parameters \(\llbracket 288,12,18\rrbracket \) and weight-six stabilizer generators \NoCaseChange{\protect\cite{cite441}}.}\\ 
\addlinespace[\myxtraspc]
\eczhRefIndex{code:css_4_1_2}%
\eczhListValue{\flmRefsHyperref{code:css_4_1_2}{\(\llbracket 4,1,2\rrbracket \) Leung-Nielsen-Chuang-Yamamoto (LNCY) code}} & \eczhListValue{A four-qubit CSS stabilizer code that is the only qubit CSS code with such parameters.}\\ 
\addlinespace[\myxtraspc]
\eczhRefIndex{code:stab_4_1_2}%
\eczhListValue{\flmRefsHyperref{code:stab_4_1_2}{\(\llbracket 4,1,2\rrbracket \) twist-defect code}} & \eczhListValue{A four-qubit non-CSS stabilizer code that can be interpreted as the smallest triangular color code with \(x\)-, \(y\)-, and \(z\)-type Pauli boundaries \NoCaseChange{\protect\cite[{Fig. 7}]{cite445}}, and equivalently as a small twist-defect surface code on a tetrahedron inscribed in a sphere \NoCaseChange{\protect\cite{cite435}}.
It is the only non-CSS qubit stabilizer code with parameters \(\llbracket 4,1,2\rrbracket \).
The code admits weight-three stabilizer generators and weight-two logical Pauli \(X,Y,Z\) operators.}\\ 
\addlinespace[\myxtraspc]
\eczhRefIndex{code:stab_4_2_2}%
\eczhListValue{\flmRefsHyperref{code:stab_4_2_2}{\(\llbracket 4,2,2\rrbracket \) Four-qubit code}} & \eczhListValue{A four-qubit hyperbolic self-dual CSS stabilizer code that is the smallest two-logical-qubit stabilizer code to detect a single-qubit error.
It is unique for its parameters \NoCaseChange{\protect\cite[{Thm. 8}]{cite446}}.}\\ 
\addlinespace[\myxtraspc]
\eczhRefIndex{code:stab_5_1_2}%
\eczhListValue{\flmRefsHyperref{code:stab_5_1_2}{\(\llbracket 5,1,2\rrbracket \) rotated surface code}} & \eczhListValue{A rotated surface code on one rung of a ladder, with one qubit on the rung, and four qubits surrounding it. This is the smallest code that implements a fault-tolerant logical \(S\) gate using a diagonal depth-one Clifford circuit \NoCaseChange{\protect\cite{cite447}}.}\\ 
\addlinespace[\myxtraspc]
\eczhRefIndex{code:stab_5_1_3}%
\eczhListValue{\flmRefsHyperref{code:stab_5_1_3}{\(\llbracket 5,1,3\rrbracket \) Five-qubit perfect code}} & \eczhListValue{Five-qubit cyclic stabilizer code that is the smallest qubit stabilizer code to correct a single-qubit error.}\\ 
\addlinespace[\myxtraspc]
\eczhRefIndex{code:stab_6_2_2}%
\eczhListValue{\flmRefsHyperref{code:stab_6_2_2}{\(\llbracket 6,2,2\rrbracket \) \(C_6\) code}} & \eczhListValue{Error-detecting normal self-dual CSS code on three qubit pairs that encodes a logical qubit pair and detects any error acting on one pair \NoCaseChange{\protect\cite{cite448}}.
In Knill's \(C_4/C_6\) architecture, this code is used at the second and higher concatenation levels.}\\ 
\addlinespace[\myxtraspc]
\eczhRefIndex{code:stab_6_4_2}%
\eczhListValue{\flmRefsHyperref{code:stab_6_4_2}{\(\llbracket 6,4,2\rrbracket \) error-detecting code}} & \eczhListValue{Self-complementary six-qubit code with rate \(2/3\) that is unique for its parameters, up to equivalence \NoCaseChange{\protect\cite[{Tab. III}]{cite449}}.
Concatenations of this code with itself yield the \(\llbracket 6^r,4^r,2^r\rrbracket \) level-\(r\) \textit{many-hypercube} code \NoCaseChange{\protect\cite{cite450}}.}\\ 
\addlinespace[\myxtraspc]
\eczhRefIndex{code:steane}%
\eczhListValue{\flmRefsHyperref{code:steane}{\(\llbracket 7,1,3\rrbracket \) Steane code}} & \eczhListValue{A \(\llbracket 7,1,3\rrbracket \) self-dual CSS code that is the smallest qubit CSS code to correct a single-qubit error \NoCaseChange{\protect\cite{cite451}}.
The code is constructed using the classical binary \([7,4,3]\) Hamming code for protecting against both \(X\) and \(Z\) errors.}\\ 
\addlinespace[\myxtraspc]
\eczhRefIndex{code:xzzx_7_1_3}%
\eczhListValue{\flmRefsHyperref{code:xzzx_7_1_3}{\(\llbracket 7,1,3\rrbracket \) XZZX cyclic code}} & \eczhListValue{A \(\llbracket 7,1,3\rrbracket \) cyclic non-CSS code whose stabilizer generators are the seven cyclic shifts of the weight-four Pauli string \(XZIZXII\),
any six of which are independent.
The non-identity support of this generator is \(XZZX\) (at positions 0, 1, 3, 4), making this a member of the twisted XZZX code family.
It is one of sixteen distinct indecomposable \(\llbracket 7,1,3\rrbracket \) codes \NoCaseChange{\protect\cite{cite452}}.}\\ 
\addlinespace[\myxtraspc]
\eczhRefIndex{code:twist_defect_7_1_3}%
\eczhListValue{\flmRefsHyperref{code:twist_defect_7_1_3}{\(\llbracket 7,1,3\rrbracket \) twist-defect surface code}} & \eczhListValue{A \(\llbracket 7,1,3\rrbracket \) code (different from the Steane code) that is a small example of a twist-defect surface code.}\\ 
\addlinespace[\myxtraspc]
\eczhRefIndex{code:hgp_7_2_2}%
\eczhListValue{\flmRefsHyperref{code:hgp_7_2_2}{\(\llbracket 7,2,2\rrbracket \) HGP phantom code}} & \eczhListValue{Smallest member of a family of CSS phantom HGP codes obtained from the hypergraph product of a classical simplex code and a repetition code.}\\ 
\addlinespace[\myxtraspc]
\eczhRefIndex{code:bb72}%
\eczhListValue{\flmRefsHyperref{code:bb72}{\(\llbracket 72,12,6\rrbracket \) BB6 code}} & \eczhListValue{A bivariate bicycle (BB) code with parameters \(\llbracket 72,12,6\rrbracket \) and weight-six stabilizer generators \NoCaseChange{\protect\cite{cite441}}.}\\ 
\addlinespace[\myxtraspc]
\eczhRefIndex{code:stab_8_3_2}%
\eczhListValue{\flmRefsHyperref{code:stab_8_3_2}{\(\llbracket 8,3,2\rrbracket \) Smallest interesting color code}} & \eczhListValue{Smallest 3D color code whose physical qubits lie on vertices of a cube and which admits a (weakly) transversal \(CCZ\) gate.}\\ 
\addlinespace[\myxtraspc]
\eczhRefIndex{code:cubic_surface}%
\eczhListValue{\flmRefsHyperref{code:cubic_surface}{\(\llbracket 8,3,2\rrbracket \) Surface code on a cube}} & \eczhListValue{An \(\llbracket 8,3,2\rrbracket \) twist-defect surface code whose qubits lie on the vertices of a cube.
It is obtained by three-coloring the faces of a cube and placing \(X\), \(Y\), and \(Z\) stabilizer generators on each pair of faces of the same color.
Its non-CSS nature is due to twist defects \NoCaseChange{\protect\cite{cite442}} stemming from the geometry of the polytope.}\\ 
\addlinespace[\myxtraspc]
\eczhRefIndex{code:shor_nine}%
\eczhListValue{\flmRefsHyperref{code:shor_nine}{\(\llbracket 9,1,3\rrbracket \) Shor code}} & \eczhListValue{Nine-qubit \flmRefsHyperref{code:css}{CSS code} that is the first quantum error-correcting code \NoCaseChange{\protect\cite[{ID 8802}]{cite453}}.
Among indecomposable \(\llbracket 9,1,3\rrbracket \) CSS codes, the Shor code has the largest automorphism group \NoCaseChange{\protect\cite{cite454}}.}\\ 
\addlinespace[\myxtraspc]
\eczhRefIndex{code:surface-17}%
\eczhListValue{\flmRefsHyperref{code:surface-17}{\(\llbracket 9,1,3\rrbracket \) Surface-17 code}} & \eczhListValue{A \(\llbracket 9,1,3\rrbracket \) rotated surface code named for the sum of its 9 data qubits and 8 syndrome qubits.
It is one of the four inequivalent CSS gauge fixings of the nine-qubit Bacon-Shor code \NoCaseChange{\protect\cite{cite454}}.
It uses the smallest number of qubits to perform fault-tolerant error correction on a surface code with parallel syndrome extraction.}\\ 
\addlinespace[\myxtraspc]
\eczhRefIndex{code:bb90}%
\eczhListValue{\flmRefsHyperref{code:bb90}{\(\llbracket 90,8,10\rrbracket \) BB6 code}} & \eczhListValue{A bivariate bicycle (BB) code with parameters \(\llbracket 90,8,10\rrbracket \) and weight-six stabilizer generators \NoCaseChange{\protect\cite{cite441}}.}\\ 
\end{tabularx}\endgroup
\eczcodelist{single_subsystem}{Monolithic quantum codes
}%

\eczhCodeListAutoDescription{All descendants of \flmRefsCref{code:single_subsystem}.}%

\eczhIncludeCodeGraph{Bare}{scale=0.5}{\columnwidth}{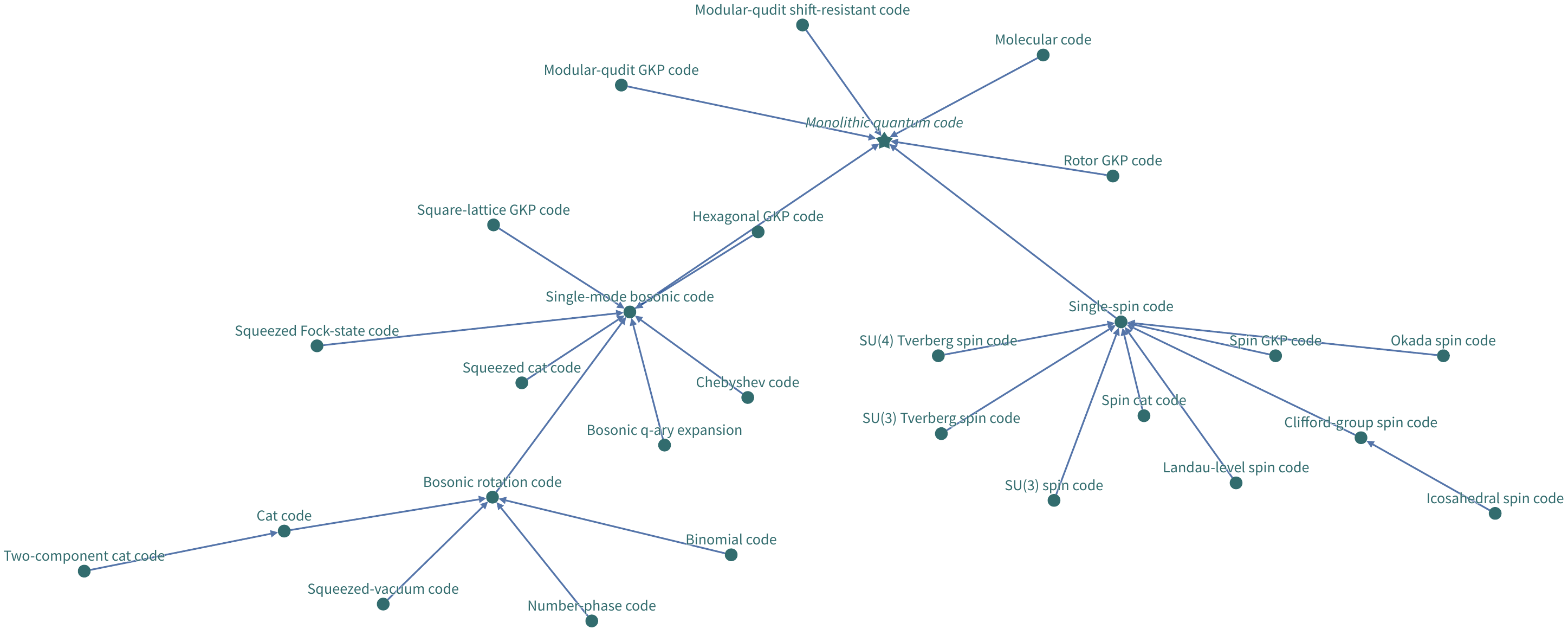}{Monolithic quantum codes}{https://errorcorrectionzoo.org/code_graph#J\%7B\%22displayMode\%22\%3A\%22subset\%22\%2C\%22modeSubsetOptions\%22\%3A\%7B\%22codeIds\%22\%3A\%5B\%22binomial\%22\%2C\%22bosonic_q-ary_expansion\%22\%2C\%22bosonic_rotation\%22\%2C\%22cat\%22\%2C\%22chebyshev\%22\%2C\%22j_gross\%22\%2C\%22hexagonal_gkp\%22\%2C\%22icosahedral_spin\%22\%2C\%22landau_level\%22\%2C\%22qudit_gkp\%22\%2C\%22qudit_sign\%22\%2C\%22molecular\%22\%2C\%22single_subsystem\%22\%2C\%22number_phase\%22\%2C\%22okada\%22\%2C\%22rotor_gkp\%22\%2C\%22single-mode\%22\%2C\%22single_spin\%22\%2C\%22spin_gkp\%22\%2C\%22spin_cat\%22\%2C\%22gkp\%22\%2C\%22squeezed_fock_state\%22\%2C\%22squeezed_cat\%22\%2C\%22squeezed_vacuum\%22\%2C\%22two-legged-cat\%22\%2C\%22su3_tverberg_spin\%22\%2C\%22su3_spin\%22\%2C\%22su4_tverberg_spin\%22\%5D\%2C\%22reusePreviousLayoutPositions\%22\%3Afalse\%2C\%22showIntermediateConnectingNodes\%22\%3Atrue\%2C\%22connectingNodesMaxDepth\%22\%3A15\%2C\%22connectingNodesPathMaxLength\%22\%3A20\%2C\%22connectingNodesMaxExtraDepth\%22\%3A3\%2C\%22connectingNodesOnlyKeepPathsWithAdditionalLength\%22\%3A1\%2C\%22connectingNodesToDomainsAndKingdoms\%22\%3Afalse\%2C\%22connectingNodesEdgeLengthsByType\%22\%3A\%7B\%22primaryParent\%22\%3A1\%2C\%22secondaryParent\%22\%3A4\%2C\%22cousin\%22\%3A6\%7D\%2C\%22nodeIds\%22\%3A\%5B\%5D\%7D\%2C\%22highlightImportantNodes\%22\%3A\%7B\%22highlightImportantNodes\%22\%3Afalse\%2C\%22highlightPrimaryParents\%22\%3Afalse\%2C\%22highlightRootConnectingEdges\%22\%3Afalse\%7D\%7D}

\begingroup
\small
\eczhBreakableDashes
\renewcommand\arraystretch{1.05}
\edef\myxtraspc{\eczListAddVSpaceXtraXtraStretch}
\begin{tabularx}{\linewidth}{>{\raggedright\arraybackslash}p{\eczListColWidth{name}} >{\hsize=1.0000\hsize }X}
\toprule
\eczListColTitle{Code} & \eczListColTitle{Description} \\
\midrule
\endfirsthead
\toprule
\eczListColTitleContinued{Code} & \eczListColTitleContinued{Description} \\
\midrule
\endhead
\bottomrule
\endfoot
\eczhRefIndex{code:binomial}%
\eczhListValue{\flmRefsHyperref{code:binomial}{Binomial code}} & \eczhListValue{Bosonic rotation codes designed to approximately protect against errors consisting of powers of raising and lowering operators up to some maximum power. Binomial codes can be thought of as spin-coherent states embedded into an oscillator \NoCaseChange{\protect\cite{cite496}}.}\\ 
\addlinespace[\myxtraspc]
\eczhRefIndex{code:bosonic_q-ary_expansion}%
\eczhListValue{\flmRefsHyperref{code:bosonic_q-ary_expansion}{Bosonic \(q\)-ary expansion}} & \eczhListValue{A one-to-one mapping between basis states on \(n\) prime-dimensional qudits (of dimension \(q=p\)) and the subspace of the first \(p^n\) single-mode Fock states.
While this mapping offers a way to map qudits into a single mode, noise models for the two code families induce different notions of locality and thus qualitatively different physical interpretations \NoCaseChange{\protect\cite{cite497}}.}\\ 
\addlinespace[\myxtraspc]
\eczhRefIndex{code:bosonic_rotation}%
\eczhListValue{\flmRefsHyperref{code:bosonic_rotation}{Bosonic rotation code}} & \eczhListValue{A single-mode Fock-state bosonic code whose codespace is preserved by a phase-space rotation by a multiple of \(2\pi/N\) for some \(N\). The rotation symmetry ensures that encoded states have support only on every \(N^{\textrm{th}}\) Fock state. For example, single-mode Fock-state codes for \(N=2\) encoding a qubit admit basis states that are, respectively, supported on Fock state sets \(\{|0\rangle,|4\rangle,|8\rangle,\cdots\}\) and \(\{|2\rangle,|6\rangle,|10\rangle,\cdots\}\).}\\ 
\addlinespace[\myxtraspc]
\eczhRefIndex{code:cat}%
\eczhListValue{\flmRefsHyperref{code:cat}{Cat code}} & \eczhListValue{Rotation-symmetric bosonic Fock-state code encoding a \(q\)-dimensional qudit into one oscillator which utilizes a constellation of \(q(S+1)\) coherent states distributed equidistantly around a circle in phase space of radius \(\alpha\).}\\ 
\addlinespace[\myxtraspc]
\eczhRefIndex{code:chebyshev}%
\eczhListValue{\flmRefsHyperref{code:chebyshev}{Chebyshev code}} & \eczhListValue{Single-mode bosonic Fock-state code that can be used for error-corrected sensing of a signal Hamiltonian \({\hat n}^s\), where \({\hat n}\) is the occupation number operator.}\\ 
\addlinespace[\myxtraspc]
\eczhRefIndex{code:j_gross}%
\eczhListValue{\flmRefsHyperref{code:j_gross}{Clifford-group spin code}} & \eczhListValue{A single-spin code designed to realize a discrete group of gates using \(SU(2)\) rotations.
Codewords are subspaces of a spin's Hilbert space that house irreducible representations (irreps) of a discrete subgroup of \(SU(2)\).}\\ 
\addlinespace[\myxtraspc]
\eczhRefIndex{code:hexagonal_gkp}%
\eczhListValue{\flmRefsHyperref{code:hexagonal_gkp}{Hexagonal GKP code}} & \eczhListValue{Single-mode GKP qudit-into-oscillator code based on the triangular lattice. Offers the best error correction against displacement noise in a single mode due to the optimal packing of the underlying lattice \NoCaseChange{\protect\cite[{Sec. VI}]{cite513}}.}\\ 
\addlinespace[\myxtraspc]
\eczhRefIndex{code:icosahedral_spin}%
\eczhListValue{\flmRefsHyperref{code:icosahedral_spin}{Icosahedral spin code}} & \eczhListValue{A spin-\(7/2\) single-spin code designed to realize the binary icosahedral group \(2I\) using \(SU(2)\) rotations \NoCaseChange{\protect\cite{cite646}}.
The codespace is a two-dimensional irrep subspace obtained by restricting the spin-\(7/2\) representation of \(SU(2)\) to the subgroup \(2I\).
Under the \flmRefsHyperref{ref526}{Dicke-state mapping}, this code is equivalent to the \(\llparenthesis 7,2,3\rrparenthesis \) Pollatsek-Ruskai permutation-invariant code \NoCaseChange{\protect\cite{cite646,cite647}}.}\\ 
\addlinespace[\myxtraspc]
\eczhRefIndex{code:landau_level}%
\eczhListValue{\flmRefsHyperref{code:landau_level}{Landau-level spin code}} & \eczhListValue{Approximate quantum code that encodes a qudit in the finite-dimensional Hilbert space of a single spin, i.e., a spherical Landau level.
Codewords are approximately orthogonal spin coherent states whose orientations are spaced maximally far apart along a great circle (equator) of the sphere.
The larger the spin, the better the performance.}\\ 
\addlinespace[\myxtraspc]
\eczhRefIndex{code:qudit_gkp}%
\eczhListValue{\flmRefsHyperref{code:qudit_gkp}{Modular-qudit GKP code}} & \eczhListValue{Modular-qudit analogue of the GKP code.
Encodes a qudit into a larger qudit and protects against Pauli shifts up to some maximum value.}\\ 
\addlinespace[\myxtraspc]
\eczhRefIndex{code:qudit_sign}%
\eczhListValue{\flmRefsHyperref{code:qudit_sign}{Modular-qudit shift-resistant code}} & \eczhListValue{Monolithic code encoding a qubit into a single modular qudit and protecting against either \(Z\)-type or \(X\)-type modular-qudit Pauli shifts.}\\ 
\addlinespace[\myxtraspc]
\eczhRefIndex{code:molecular}%
\eczhListValue{\flmRefsHyperref{code:molecular}{Molecular code}} & \eczhListValue{Approximate quantum code that encodes a finite-dimensional logical space into the Hilbert space of \(L^2\)-normalizable functions on \(SO(3)\), i.e., rotational states of an asymmetric rigid body such as a polyatomic molecule.}\\ 
\addlinespace[\myxtraspc]
\eczhRefIndex{code:single_subsystem}%
\eczhListValue{\flmRefsHyperref{code:single_subsystem}{Monolithic quantum code}} & \eczhListValue{A code constructed in a single quantum system, i.e., a physical space that is \textit{not} treated as a tensor product of \(n\) identical subsystems.
Examples include codes in a single qudit, spin, oscillator, or molecule.}\\ 
\addlinespace[\myxtraspc]
\eczhRefIndex{code:number_phase}%
\eczhListValue{\flmRefsHyperref{code:number_phase}{Number-phase code}} & \eczhListValue{Bosonic rotation code consisting of superpositions of Pegg-Barnett phase states \NoCaseChange{\protect\cite{cite501}}.}\\ 
\addlinespace[\myxtraspc]
\eczhRefIndex{code:okada}%
\eczhListValue{\flmRefsHyperref{code:okada}{Okada spin code}} & \eczhListValue{Non-diagonal \(SU(2)\) single-spin code in the spin-\(J = 3m\) irrep for integer \(m \geq 1\), encoding a logical \((2m+1)\)-dimensional space.
The construction uses a \textit{non-diagonal} subspace (one for which the projected error space \(P_{\mathcal{B}}\mathcal{E}P_{\mathcal{B}}\) is block-diagonal rather than diagonal) to exceed the dimension bound achievable by the Tverberg theorem construction \NoCaseChange{\protect\cite{cite648}}.}\\ 
\addlinespace[\myxtraspc]
\eczhRefIndex{code:rotor_gkp}%
\eczhListValue{\flmRefsHyperref{code:rotor_gkp}{Rotor GKP code}} & \eczhListValue{GKP code protecting against small angular position and momentum shifts of a planar rotor.}\\ 
\addlinespace[\myxtraspc]
\eczhRefIndex{code:single-mode}%
\eczhListValue{\flmRefsHyperref{code:single-mode}{Single-mode bosonic code}} & \eczhListValue{Encodes a \(K\)-dimensional Hilbert space into a single bosonic mode. A trivial single-mode code encoding a qubit into the first two Fock states \(\{|0\rangle,|1\rangle\}\) is called the \textit{single-rail} encoding \NoCaseChange{\protect\cite{cite649,cite650}}.}\\ 
\addlinespace[\myxtraspc]
\eczhRefIndex{code:single_spin}%
\eczhListValue{\flmRefsHyperref{code:single_spin}{Single-spin code}} & \eczhListValue{An encoding into a monolithic (i.e. non-tensor-product) Hilbert space that houses an irreducible representation of \(SU(2)\) or, more generally, another Lie group.
In some cases, this space can be thought of as the permutation invariant subspace of a particular tensor-product space.}\\ 
\addlinespace[\myxtraspc]
\eczhRefIndex{code:spin_gkp}%
\eczhListValue{\flmRefsHyperref{code:spin_gkp}{Spin GKP code}} & \eczhListValue{An analogue of the single-mode GKP code designed for atomic ensembles. It was designed using the Holstein-Primakoff mapping \NoCaseChange{\protect\cite{cite651,cite652,cite653}} to pull back the phase-space structure of a bosonic system to the compact phase space of a quantum spin. A different construction emerges depending on which particular expression for GKP codewords is pulled back.}\\ 
\addlinespace[\myxtraspc]
\eczhRefIndex{code:spin_cat}%
\eczhListValue{\flmRefsHyperref{code:spin_cat}{Spin cat code}} & \eczhListValue{An analogue of the two-component cat code for a large spin, which is often realized in the PI subspace of atomic ensembles.}\\ 
\addlinespace[\myxtraspc]
\eczhRefIndex{code:gkp}%
\eczhListValue{\flmRefsHyperref{code:gkp}{Square-lattice GKP code}} & \eczhListValue{Single-mode GKP qudit-into-oscillator CSS code based on the rectangular lattice.
Its stabilizer generators are oscillator displacement operators \(\hat{S}_q(2\alpha)=e^{-2i\alpha \hat{p}}\) and \(\hat{S}_p(2\beta)=e^{2i\beta \hat{x}}\).
To ensure \(\hat{S}_q(2\alpha)\) and \(\hat{S}_p(2\beta)\) generate a stabilizer group that is Abelian, there is a constraint that \(\alpha\beta=2q\pi\) where \(q\) is an integer denoting the logical dimension.}\\ 
\addlinespace[\myxtraspc]
\eczhRefIndex{code:squeezed_fock_state}%
\eczhListValue{\flmRefsHyperref{code:squeezed_fock_state}{Squeezed Fock-state code}} & \eczhListValue{Approximate bosonic code that encodes a qubit into a superposition of one or a few squeezed Fock states, some of which are the result of a photon-number resolving measurement \NoCaseChange{\protect\cite{cite654}}.}\\ 
\addlinespace[\myxtraspc]
\eczhRefIndex{code:squeezed_cat}%
\eczhListValue{\flmRefsHyperref{code:squeezed_cat}{Squeezed cat code}} & \eczhListValue{Two-component cat code whose two coherent states have been squeezed in a direction perpendicular to the segment formed by the two coherent state values \(\pm\alpha\).}\\ 
\addlinespace[\myxtraspc]
\eczhRefIndex{code:squeezed_vacuum}%
\eczhListValue{\flmRefsHyperref{code:squeezed_vacuum}{Squeezed-vacuum code}} & \eczhListValue{A squeezed Fock-state code constructed from a coherent superposition of \(m\) squeezed vacuum states, each squeezed along equiangular axes in phase space.}\\ 
\addlinespace[\myxtraspc]
\eczhRefIndex{code:two-legged-cat}%
\eczhListValue{\flmRefsHyperref{code:two-legged-cat}{Two-component cat code}} & \eczhListValue{Code whose codespace is spanned by two coherent states \(\left|\pm\alpha\right\rangle\) for nonzero complex \(\alpha\).}\\ 
\addlinespace[\myxtraspc]
\eczhRefIndex{code:su3_tverberg_spin}%
\eczhListValue{\flmRefsHyperref{code:su3_tverberg_spin}{\(SU(3)\) Tverberg spin code}} & \eczhListValue{\(SU(3)\) single-spin code family obtained from the two-step Tverberg construction \NoCaseChange{\protect\cite{cite648}} in the totally symmetric \(N\)-particle irrep \(\mathcal{H}=\mathrm{Sym}^N(\mathbb{C}^3)\) of \(\mathfrak{su}(3)\), which has dimension \(\binom{N+2}{2}\).
Its weight basis is indexed by the \flmRefsHyperref{ref655}{discrete simplex} \(\Delta_{3,N}\), whose centered form is the triangular \(A_2\) lattice.}\\ 
\addlinespace[\myxtraspc]
\eczhRefIndex{code:su3_spin}%
\eczhListValue{\flmRefsHyperref{code:su3_spin}{\(SU(3)\) spin code}} & \eczhListValue{An extension of Clifford single-spin codes to the group \(SU(3)\), whose codespace is a projection onto a particular irrep of a subgroup of \(SU(3)\) of an underlying spin that houses some particular irrep of \(SU(3)\).}\\ 
\addlinespace[\myxtraspc]
\eczhRefIndex{code:su4_tverberg_spin}%
\eczhListValue{\flmRefsHyperref{code:su4_tverberg_spin}{\(SU(4)\) Tverberg spin code}} & \eczhListValue{Single-spin code family in the totally symmetric \(N\)-particle irrep \(\mathcal{H}=\mathrm{Sym}^N(\mathbb{C}^4)\) of \(\mathfrak{su}(4)\), whose weight diagram is the \flmRefsHyperref{ref655}{discrete simplex} \(\Delta_{4,N}\), equivalently its centered tetrahedral realization.
The construction uses the two-step Tverberg-theorem method \NoCaseChange{\protect\cite{cite648}}: first choose an intermediate subspace from a distance-two subset of \(\Delta_{4,N}\), then combine basis states whose convex hulls contain the origin.}\\ 
\end{tabularx}\endgroup
\eczcodelist{oa_qubit}{Operator-algebra qubit codes
}%

\eczhCodeListAutoDescription{Codes that are descendants of \flmRefsCref{code:oa_qubits_into_qubits} and not descendants of \flmRefsCref{code:qubits_into_qubits}.}%

\eczhIncludeCodeGraph{Bare}{scale=0.5}{\columnwidth}{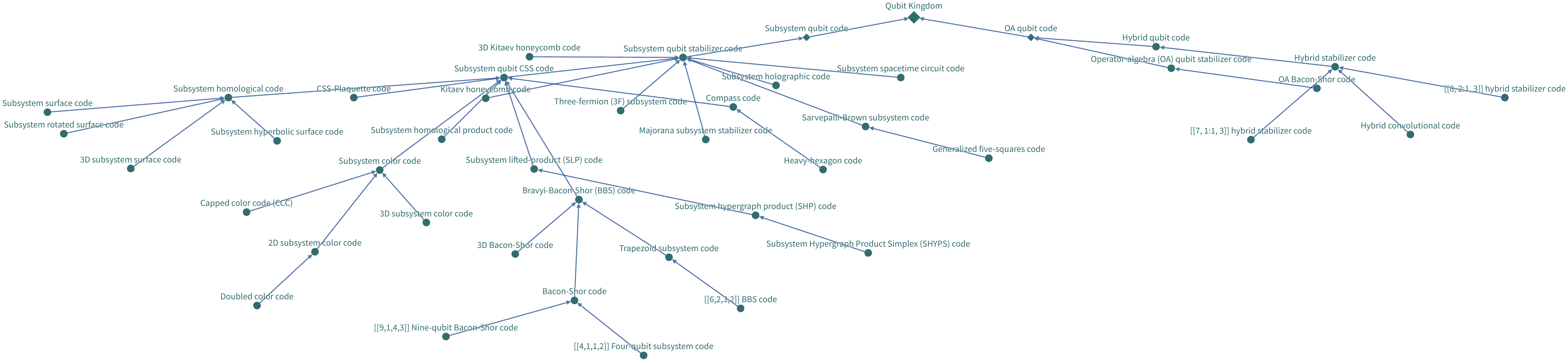}{Operator-algebra qubit codes}{https://errorcorrectionzoo.org/code_graph#J\%7B\%22displayMode\%22\%3A\%22subset\%22\%2C\%22modeSubsetOptions\%22\%3A\%7B\%22codeIds\%22\%3A\%5B\%222d_subsystem_color\%22\%2C\%223d_bacon_shor\%22\%2C\%223d_kitaev_honeycomb\%22\%2C\%223d_subsystem_color\%22\%2C\%223d_subsystem_surface\%22\%2C\%22bacon_shor\%22\%2C\%22bravyi_bacon_shor\%22\%2C\%22css_plaquette\%22\%2C\%22capped_color\%22\%2C\%22compass_model\%22\%2C\%22doubled_color\%22\%2C\%22five_squares\%22\%2C\%22heavy_hex\%22\%2C\%22hybrid_convolutional\%22\%2C\%22hybrid_qubits_into_qubits\%22\%2C\%22hybrid_stabilizer\%22\%2C\%22kitaev_honeycomb\%22\%2C\%22majorana_subsystem\%22\%2C\%22hybrid_bacon_shor\%22\%2C\%22oa_qubits_into_qubits\%22\%2C\%22qubit_stabilizer_oaqecc\%22\%2C\%22subsystem_hypergraph\%22\%2C\%22shyps\%22\%2C\%22subsystem_color\%22\%2C\%22holographic_subsystem\%22\%2C\%22subsystem_higher_dimensional_surface\%22\%2C\%22subsystem_product\%22\%2C\%22subsystem_hyperbolic_surface\%22\%2C\%22subsystem_quantum_parity\%22\%2C\%22subsystem_lifted_product\%22\%2C\%22qubit_subsystem_css\%22\%2C\%22subsystem_qubits_into_qubits\%22\%2C\%22qubit_subsystem_stabilizer\%22\%2C\%22subsystem_rotated_surface\%22\%2C\%22subsystem_spacetime_circuit\%22\%2C\%22subsystem_surface\%22\%2C\%22subsystem_three_fermion\%22\%2C\%22trapezoid\%22\%2C\%22bacon_shor_4\%22\%2C\%22bravyi_bacon_shor_6\%22\%2C\%22hybrid_7_1-1_3\%22\%2C\%22hybrid_8_2-1_3\%22\%2C\%22bacon_shor_9\%22\%5D\%2C\%22reusePreviousLayoutPositions\%22\%3Afalse\%2C\%22showIntermediateConnectingNodes\%22\%3Atrue\%2C\%22connectingNodesMaxDepth\%22\%3A15\%2C\%22connectingNodesPathMaxLength\%22\%3A20\%2C\%22connectingNodesMaxExtraDepth\%22\%3A3\%2C\%22connectingNodesOnlyKeepPathsWithAdditionalLength\%22\%3A1\%2C\%22connectingNodesToDomainsAndKingdoms\%22\%3Afalse\%2C\%22connectingNodesEdgeLengthsByType\%22\%3A\%7B\%22primaryParent\%22\%3A1\%2C\%22secondaryParent\%22\%3A4\%2C\%22cousin\%22\%3A6\%7D\%2C\%22nodeIds\%22\%3A\%5B\%22k_qubits_into_qubits\%22\%2C\%22k_qubits_into_qubits\%22\%5D\%7D\%2C\%22highlightImportantNodes\%22\%3A\%7B\%22highlightImportantNodes\%22\%3Afalse\%2C\%22highlightPrimaryParents\%22\%3Afalse\%2C\%22highlightRootConnectingEdges\%22\%3Afalse\%7D\%7D}

\begingroup
\small
\eczhBreakableDashes
\renewcommand\arraystretch{1.05}
\edef\myxtraspc{\eczListAddVSpaceXtraXtraStretch}
\begin{tabularx}{\linewidth}{>{\raggedright\arraybackslash}p{\eczListColWidth{name}} >{\hsize=1.0000\hsize }X}
\toprule
\eczListColTitle{Code} & \eczListColTitle{Description} \\
\midrule
\endfirsthead
\toprule
\eczListColTitleContinued{Code} & \eczListColTitleContinued{Description} \\
\midrule
\endhead
\bottomrule
\endfoot
\eczhRefIndex{code:2d_subsystem_color}%
\eczhListValue{\flmRefsHyperref{code:2d_subsystem_color}{2D subsystem color code}} & \eczhListValue{A subsystem version of the 2D color code.
The original topological subsystem-code example is defined on the Union Jack lattice \NoCaseChange{\protect\cite{cite604}}; the square-octagon-lattice hypergraph construction of \NoCaseChange{\protect\cite{cite594}} reproduces the same code from a complementary viewpoint.}\\ 
\addlinespace[\myxtraspc]
\eczhRefIndex{code:3d_bacon_shor}%
\eczhListValue{\flmRefsHyperref{code:3d_bacon_shor}{3D Bacon-Shor code}} & \eczhListValue{Generalization of the Bacon-Shor code to three dimensions that was conjectured to be a self-correcting memory.
It is defined on a cubic lattice and admits sheet-like stabilizer generators.}\\ 
\addlinespace[\myxtraspc]
\eczhRefIndex{code:3d_kitaev_honeycomb}%
\eczhListValue{\flmRefsHyperref{code:3d_kitaev_honeycomb}{3D Kitaev honeycomb code}} & \eczhListValue{3D subsystem stabilizer code whose Hamiltonian is a 3D generalization of the Kitaev honeycomb model.
One of the phases realized by the 3D Kitaev honeycomb Hamiltonian is that of the 3D fermionic surface code \NoCaseChange{\protect\cite{cite458}}.}\\ 
\addlinespace[\myxtraspc]
\eczhRefIndex{code:3d_subsystem_color}%
\eczhListValue{\flmRefsHyperref{code:3d_subsystem_color}{3D subsystem color code}} & \eczhListValue{A subsystem version of the 3D color code defined on a 3-colex.}\\ 
\addlinespace[\myxtraspc]
\eczhRefIndex{code:3d_subsystem_surface}%
\eczhListValue{\flmRefsHyperref{code:3d_subsystem_surface}{3D subsystem surface code}} & \eczhListValue{Subsystem generalization of the surface code on a 3D cubic lattice with gauge-group generators of weight at most three.}\\ 
\addlinespace[\myxtraspc]
\eczhRefIndex{code:bacon_shor}%
\eczhListValue{\flmRefsHyperref{code:bacon_shor}{Bacon-Shor code}} & \eczhListValue{Subsystem CSS code defined on an \(m_1 \times m_2\) lattice of qubits that generalizes the \(\llbracket 9,1,3\rrbracket \) (subspace) Shor code.
It is said to be \textit{symmetric} when \(m_1=m_2\), and \textit{asymmetric} otherwise.}\\ 
\addlinespace[\myxtraspc]
\eczhRefIndex{code:bravyi_bacon_shor}%
\eczhListValue{\flmRefsHyperref{code:bravyi_bacon_shor}{Bravyi-Bacon-Shor (BBS) code}} & \eczhListValue{A CSS subsystem stabilizer code generalizing Bacon-Shor codes to a larger set of qubit geometries.
Defined through a binary matrix \(A\) such that physical qubits live on sites \((i,j)\) whenever \(A_{i,j}=1\).
The gauge group is generated by 2-qubit operators, including \(XX\) interactions between any two qubits sharing a column in \(A\), and \(ZZ\) interactions between any two qubits sharing a row.}\\ 
\addlinespace[\myxtraspc]
\eczhRefIndex{code:css_plaquette}%
\eczhListValue{\flmRefsHyperref{code:css_plaquette}{CSS-Plaquette code}} & \eczhListValue{Generalization of the Bacon-Shor code to three dimensions, defined on a cubic lattice and admitting string-like stabilizer generators.}\\ 
\addlinespace[\myxtraspc]
\eczhRefIndex{code:capped_color}%
\eczhListValue{\flmRefsHyperref{code:capped_color}{Capped color code (CCC)}} & \eczhListValue{A non-geometrically local subsystem color code consisting of two layers of 2D color code stacked together and topped (or capped) by a single qubit.
Gauge fixing yields two types of codes, capped color codes in H or T form.
Layers of 2D color codes can also be stacked together in a recursive construction, yielding \textit{recursive capped color codes} (RCCCs).}\\ 
\addlinespace[\myxtraspc]
\eczhRefIndex{code:compass_model}%
\eczhListValue{\flmRefsHyperref{code:compass_model}{Compass code}} & \eczhListValue{Subspace or subsystem CSS code defined by gauge-fixing the Bacon-Shor code, i.e., the code whose gauge group consists of terms in the compass model Hamiltonian \NoCaseChange{\protect\cite{cite656,cite657,cite658}} on a square lattice.
Families of random codes perform well against biased noise and spatially dependent (i.e., asymmetric) noise.}\\ 
\addlinespace[\myxtraspc]
\eczhRefIndex{code:doubled_color}%
\eczhListValue{\flmRefsHyperref{code:doubled_color}{Doubled color code}} & \eczhListValue{Family of \(\llbracket 2t^3+8t^2+6t-1,1,2t+1\rrbracket \) subsystem color codes (with \(t\geq 1\)), constructed using a generalization of the doubling transformation \NoCaseChange{\protect\cite{cite659}}, that admit a Clifford + \(T\) transversal gate set using gauge fixing.}\\ 
\addlinespace[\myxtraspc]
\eczhRefIndex{code:five_squares}%
\eczhListValue{\flmRefsHyperref{code:five_squares}{Generalized five-squares code}} & \eczhListValue{Member of a family of subsystem codes that are generalizations \NoCaseChange{\protect\cite{cite660,cite661}} of a code defined on a three-valent hypergraph associated with the five-squares lattice \NoCaseChange{\protect\cite{cite594}}.
The original five-squares code is a 2D topological subsystem code with local two-qubit gauge generators; on a torus, it encodes two logical qubits \NoCaseChange{\protect\cite{cite594}}.}\\ 
\addlinespace[\myxtraspc]
\eczhRefIndex{code:heavy_hex}%
\eczhListValue{\flmRefsHyperref{code:heavy_hex}{Heavy-hexagon code}} & \eczhListValue{Subsystem stabilizer code on the heavy-hexagonal point set that combines Bacon-Shor and surface-code stabilizers.
Encodes one logical qubit into \(n=(5d^2-2d-1)/2\) physical qubits with distance \(d\).
The heavy-hexagonal point set allows for low degree (at most 3) connectivity between all the data and ancilla qubits, which is suitable for fixed-frequency transmon qubits subject to frequency collision errors.
The code can be split into a surface and a Bacon-Shor code, with the idling qubits of one code serving as the physical qubits of the other \NoCaseChange{\protect\cite{cite662}}.}\\ 
\addlinespace[\myxtraspc]
\eczhRefIndex{code:hybrid_convolutional}%
\eczhListValue{\flmRefsHyperref{code:hybrid_convolutional}{Hybrid convolutional code}} & \eczhListValue{A quantum convolutional code which protects both quantum and classical information.}\\ 
\addlinespace[\myxtraspc]
\eczhRefIndex{code:hybrid_qubits_into_qubits}%
\eczhListValue{\flmRefsHyperref{code:hybrid_qubits_into_qubits}{Hybrid qubit code}} & \eczhListValue{A qubit code which stores both quantum and classical information.
Usually denoted as \(\llparenthesis n,K:M\rrparenthesis \) or \(\llparenthesis n,K:M,d\rrparenthesis \), where \(K\) is the dimension of the underlying quantum code, \(M\) is the size of the classical code, and \(d\) is the distance.}\\ 
\addlinespace[\myxtraspc]
\eczhRefIndex{code:hybrid_stabilizer}%
\eczhListValue{\flmRefsHyperref{code:hybrid_stabilizer}{Hybrid stabilizer code}} & \eczhListValue{A qubit stabilizer code which stores both quantum and classical information.
Usually denoted as \(\llbracket n,k:c\rrbracket \) or \(\llbracket n,k:c,d\rrbracket \), where \(k\) (\(c\)) is the number of encoded qubits (classical bits), and where \(d\) is the distance.}\\ 
\addlinespace[\myxtraspc]
\eczhRefIndex{code:kitaev_honeycomb}%
\eczhListValue{\flmRefsHyperref{code:kitaev_honeycomb}{Kitaev honeycomb code}} & \eczhListValue{Subsystem qubit stabilizer code underlying the Kitaev honeycomb model \NoCaseChange{\protect\cite{cite537,cite594}}.
Its gauge generators are the two-qubit \(XX\), \(YY\), and \(ZZ\) link operators on the three edge types of the honeycomb lattice \NoCaseChange{\protect\cite[{Sec. 3.2}]{cite594}}.
Its stabilizer group is generated by loop operators, and syndrome extraction can be reduced to ordered measurements of the two-qubit link operators \NoCaseChange{\protect\cite[{Sec. 3.2}]{cite594}}.
This is the \(q=2\) instance of the \(\mathbb{Z}_q^{(1)}\) subsystem code and does not encode any logical qubits \NoCaseChange{\protect\cite{cite594}\protect\cite[{Sec. 7.3}]{cite414}}.}\\ 
\addlinespace[\myxtraspc]
\eczhRefIndex{code:majorana_subsystem}%
\eczhListValue{\flmRefsHyperref{code:majorana_subsystem}{Majorana subsystem stabilizer code}} & \eczhListValue{A Majorana subsystem code with some of its logical qubits denoted as \textit{gauge} qubits and not used for storage of logical information.}\\ 
\addlinespace[\myxtraspc]
\eczhRefIndex{code:hybrid_bacon_shor}%
\eczhListValue{\flmRefsHyperref{code:hybrid_bacon_shor}{OA Bacon-Shor code}} & \eczhListValue{Family of OA qubit stabilizer codes derived from Bacon-Shor subsystem codes by using their extra gauge structure to store classical information.}\\ 
\addlinespace[\myxtraspc]
\eczhRefIndex{code:oa_qubits_into_qubits}%
\eczhListValue{\flmRefsHyperref{code:oa_qubits_into_qubits}{OA qubit code}} & \eczhListValue{An OAQECC family that encompasses ordinary (i.e., subspace) qubit codes, subsystem qubit codes, and hybrid qubit codes using an operator-algebraic framework.}\\ 
\addlinespace[\myxtraspc]
\eczhRefIndex{code:qubit_stabilizer_oaqecc}%
\eczhListValue{\flmRefsHyperref{code:qubit_stabilizer_oaqecc}{Operator-algebra (OA) qubit stabilizer code}} & \eczhListValue{An OAQECC in which the commutant \(\mathcal{A}'\) of the logical algebra \(\mathcal{A}\) arises as the \flmRefsHyperref{ref205}{group algebra} of a subgroup \(\mathsf{G}\) of the \(n\)-qubit \flmRefsHyperref{ref663}{Pauli group} \(\mathsf{P}_n\).}\\ 
\addlinespace[\myxtraspc]
\eczhRefIndex{code:subsystem_hypergraph}%
\eczhListValue{\flmRefsHyperref{code:subsystem_hypergraph}{Sarvepalli-Brown subsystem code}} & \eczhListValue{Member of a family of non-CSS subsystem codes constructed from hypergraphs that satisfy certain assumptions \NoCaseChange{\protect\cite[{Construction C}]{cite660}}.}\\ 
\addlinespace[\myxtraspc]
\eczhRefIndex{code:shyps}%
\eczhListValue{\flmRefsHyperref{code:shyps}{Subsystem Hypergraph Product Simplex (SHYPS) code}} & \eczhListValue{Family of quantum LDPC codes obtained by combining the \flmRefsHyperref{code:subsystem_quantum_parity}{subsystem hypergraph product code} construction with classical \flmRefsHyperref{code:simplex}{simplex codes}. 
The results are CSS subsystem codes with weight-three gauge generators and code parameters \(\llbracket n=(2^r − 1)^2, k=r^2, d=2^{r-1}\rrbracket \) for \(r \geq 3\).}\\ 
\addlinespace[\myxtraspc]
\eczhRefIndex{code:subsystem_color}%
\eczhListValue{\flmRefsHyperref{code:subsystem_color}{Subsystem color code}} & \eczhListValue{A subsystem version of the color code.}\\ 
\addlinespace[\myxtraspc]
\eczhRefIndex{code:holographic_subsystem}%
\eczhListValue{\flmRefsHyperref{code:holographic_subsystem}{Subsystem holographic code}} & \eczhListValue{A holographic tensor-network code constructed out of alternating isometries of the five-qubit and \(\llbracket 4,1,1,2\rrbracket \) Bacon-Shor codes.}\\ 
\addlinespace[\myxtraspc]
\eczhRefIndex{code:subsystem_higher_dimensional_surface}%
\eczhListValue{\flmRefsHyperref{code:subsystem_higher_dimensional_surface}{Subsystem homological code}} & \eczhListValue{A subsystem CSS code that is a subsystem version of the homological code, defined on cellulations of manifolds in arbitrary dimensions.
Gauge-group generators are of lower weight than the stabilizers of the corresponding surface code, enabling fault-tolerant syndrome extraction with simpler circuits.
The stabilizer group may contain generators of unbounded weight, distinguishing these codes from stabilizer codes with bounded-weight generators for which some logical qubits were re-assigned to be gauge qubits.}\\ 
\addlinespace[\myxtraspc]
\eczhRefIndex{code:subsystem_product}%
\eczhListValue{\flmRefsHyperref{code:subsystem_product}{Subsystem homological product code}} & \eczhListValue{A CSS subsystem code constructed from a product of two (subspace) CSS codes.
The case for qubits is formulated below, but these codes have also been extended to Galois qudits \NoCaseChange{\protect\cite{cite664}}.}\\ 
\addlinespace[\myxtraspc]
\eczhRefIndex{code:subsystem_hyperbolic_surface}%
\eczhListValue{\flmRefsHyperref{code:subsystem_hyperbolic_surface}{Subsystem hyperbolic surface code}} & \eczhListValue{Subsystem generalization of the surface code on a 2D hyperbolic tessellation with gauge-group generators of weight at most three.
An \(\{r,4\}\) hyperbolic tessellation with \(E\) edges yields a \(\llbracket 3E/2,(1/2-2/r)E+2,(1-2/r)E,d\rrbracket \) subsystem code.}\\ 
\addlinespace[\myxtraspc]
\eczhRefIndex{code:subsystem_quantum_parity}%
\eczhListValue{\flmRefsHyperref{code:subsystem_quantum_parity}{Subsystem hypergraph product (SHP) code}} & \eczhListValue{A CSS subsystem version of the generalized Shor code that has the same parameters as the subspace version, but requires fewer stabilizer measurements, resulting in a simpler error recovery routine.
The code can also be thought of as a subsystem version of an HGP code because two such codes reduce to an HGP code upon gauge fixing \NoCaseChange{\protect\cite[{Sec. III}]{cite665}}.
The code can be obtained from a generalized Shor code by removing certain stabilizers that do not affect the code distance.}\\ 
\addlinespace[\myxtraspc]
\eczhRefIndex{code:subsystem_lifted_product}%
\eczhListValue{\flmRefsHyperref{code:subsystem_lifted_product}{Subsystem lifted-product (SLP) code}} & \eczhListValue{Member of a family of subsystem CSS codes constructed from a subsystem hypergraph product of a \flmRefsHyperref{ref47}{lifted} binary linear code.}\\ 
\addlinespace[\myxtraspc]
\eczhRefIndex{code:qubit_subsystem_css}%
\eczhListValue{\flmRefsHyperref{code:qubit_subsystem_css}{Subsystem qubit CSS code}} & \eczhListValue{Subsystem qubit stabilizer code which admits a set of gauge-group generators which consist of either all-\(Z\) or all-\(X\) Pauli strings.
This ensures that the code's stabilizer group is also CSS.}\\ 
\addlinespace[\myxtraspc]
\eczhRefIndex{code:subsystem_qubits_into_qubits}%
\eczhListValue{\flmRefsHyperref{code:subsystem_qubits_into_qubits}{Subsystem qubit code}} & \eczhListValue{Subsystem QECC encoding into a \(2^n\)-dimensional (i.e., \(n\)-qubit) Hilbert space.}\\ 
\addlinespace[\myxtraspc]
\eczhRefIndex{code:qubit_subsystem_stabilizer}%
\eczhListValue{\flmRefsHyperref{code:qubit_subsystem_stabilizer}{Subsystem qubit stabilizer code}} & \eczhListValue{A stabilizer code with some of its logical qubits denoted as \textit{gauge} qubits and not used for storage of logical information.
Note that this doesn't lead to new codes but does lead to new error correction and fault tolerance procedures.
Subsystem codes are denoted by \(\llbracket n,k,g,d\rrbracket \), similar to stabilizer codes, but with an extra parameter \(g\) denoting the number of gauge qubits.}\\ 
\addlinespace[\myxtraspc]
\eczhRefIndex{code:subsystem_rotated_surface}%
\eczhListValue{\flmRefsHyperref{code:subsystem_rotated_surface}{Subsystem rotated surface code}} & \eczhListValue{Subsystem version of the rotated surface code.}\\ 
\addlinespace[\myxtraspc]
\eczhRefIndex{code:subsystem_spacetime_circuit}%
\eczhListValue{\flmRefsHyperref{code:subsystem_spacetime_circuit}{Subsystem spacetime circuit code}} & \eczhListValue{Subsystem stabilizer code obtained from a spacetime circuit code by \flmRefsHyperref{ref666}{gauging out} logical operators that correspond to circuit faults with trivial effect \NoCaseChange{\protect\cite[{Sec. 5.4}]{cite667}}.
In the original circuit-to-code construction, each circuit element is replaced by low-weight gauge generators enforcing its input-output relations, yielding subsystem codes from restricted Clifford postselection circuits \NoCaseChange{\protect\cite{cite668}}.}\\ 
\addlinespace[\myxtraspc]
\eczhRefIndex{code:subsystem_surface}%
\eczhListValue{\flmRefsHyperref{code:subsystem_surface}{Subsystem surface code}} & \eczhListValue{Subsystem version of the surface code defined on a square lattice with qubits placed at every vertex and center of every edge.
Its gauge checks are weight-three triangle operators of type \(XXX\) and \(ZZZ\) \NoCaseChange{\protect\cite{cite669}}.}\\ 
\addlinespace[\myxtraspc]
\eczhRefIndex{code:subsystem_three_fermion}%
\eczhListValue{\flmRefsHyperref{code:subsystem_three_fermion}{Three-fermion (3F) subsystem code}} & \eczhListValue{2D subsystem stabilizer code whose low-energy excitations realize the three-fermion anyon theory \NoCaseChange{\protect\cite{cite601,cite602,cite603}}.
One version uses two qubits at each site \NoCaseChange{\protect\cite{cite414}}, while other manifestations utilize a single qubit per site and only weight-two (two-body) interactions \NoCaseChange{\protect\cite{cite602,cite604}}.
All are expected to be equivalent to each other via a local constant-depth \flmRefsHyperref{ref409}{Clifford circuit}.}\\ 
\addlinespace[\myxtraspc]
\eczhRefIndex{code:trapezoid}%
\eczhListValue{\flmRefsHyperref{code:trapezoid}{Trapezoid subsystem code}} & \eczhListValue{A member of a family of BBS codes with weight-two (two-body) gauge generators designed to suppress errors in adiabatic quantum computation.}\\ 
\addlinespace[\myxtraspc]
\eczhRefIndex{code:bacon_shor_4}%
\eczhListValue{\flmRefsHyperref{code:bacon_shor_4}{\(\llbracket 4,1,1,2\rrbracket \) Four-qubit subsystem code}} & \eczhListValue{Error-detecting four-qubit subsystem stabilizer code encoding one logical qubit and one gauge qubit.}\\ 
\addlinespace[\myxtraspc]
\eczhRefIndex{code:bravyi_bacon_shor_6}%
\eczhListValue{\flmRefsHyperref{code:bravyi_bacon_shor_6}{\(\llbracket 6,2,3,2\rrbracket \) BBS code}} & \eczhListValue{Error-detecting six-qubit BBS subsystem code with parameters \(\llbracket 6,2,3,2\rrbracket \) that can suppress errors in adiabatic quantum computation \NoCaseChange{\protect\cite{cite670}}.}\\ 
\addlinespace[\myxtraspc]
\eczhRefIndex{code:hybrid_7_1-1_3}%
\eczhListValue{\flmRefsHyperref{code:hybrid_7_1-1_3}{\(\llbracket 7, 1:1, 3\rrbracket \) hybrid stabilizer code}} & \eczhListValue{A distance-three seven-qubit hybrid stabilizer code storing one qubit and one classical bit.
Admits a stabilizer generator set with a weight-two generator, which delineates the underlying classical code \NoCaseChange{\protect\cite[{Eq. (3)}]{cite671}}.}\\ 
\addlinespace[\myxtraspc]
\eczhRefIndex{code:hybrid_8_2-1_3}%
\eczhListValue{\flmRefsHyperref{code:hybrid_8_2-1_3}{\(\llbracket 8, 2:1, 3\rrbracket \) hybrid stabilizer code}} & \eczhListValue{A code obtained from the \(\llbracket 8,3,3\rrbracket \) Gottesman code by using one of its logical qubits as a classical bit.
One can also use two logical qubits as classical bits, obtaining an \(\llbracket 8,1:2,3\rrbracket \) hybrid stabilizer code.}\\ 
\addlinespace[\myxtraspc]
\eczhRefIndex{code:bacon_shor_9}%
\eczhListValue{\flmRefsHyperref{code:bacon_shor_9}{\(\llbracket 9,1,4,3\rrbracket \) Nine-qubit Bacon-Shor code}} & \eczhListValue{Error-correcting nine-qubit subsystem stabilizer code encoding one logical qubit and four gauge qubits.
There are exactly four inequivalent CSS gauge fixings of the code, including the \flmRefsHyperref{code:shor_nine}{Shor code} and the \flmRefsHyperref{code:surface-17}{surface-17 code} \NoCaseChange{\protect\cite{cite454}}.}\\ 
\end{tabularx}\endgroup
\eczcodelist{quantum_perfect}{Perfect quantum codes and friends
}%

\eczhCodeListAutoDescription{All descendants and cousins of \flmRefsCref{code:quantum_perfect}.}%

\eczhIncludeCodeGraph{Bare}{scale=0.5}{\columnwidth}{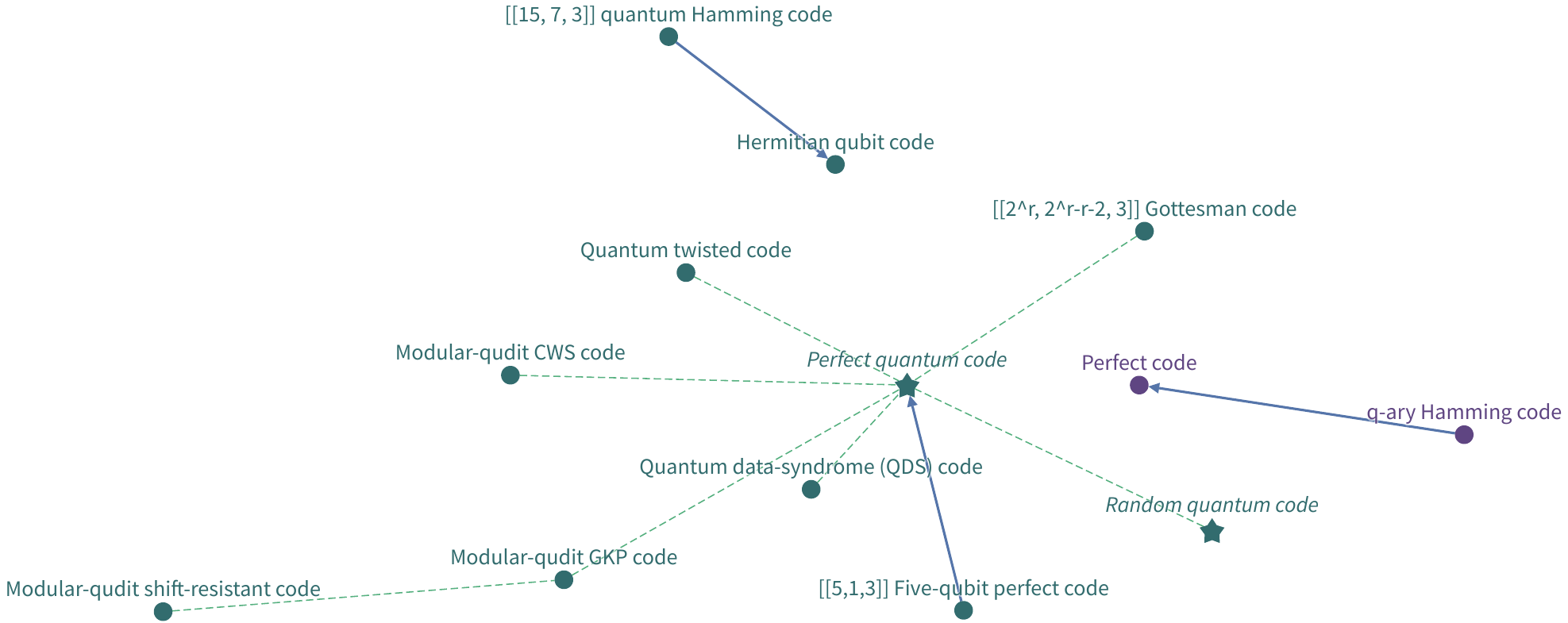}{Perfect quantum codes and friends}{https://errorcorrectionzoo.org/code_graph#J\%7B\%22displayMode\%22\%3A\%22subset\%22\%2C\%22modeSubsetOptions\%22\%3A\%7B\%22codeIds\%22\%3A\%5B\%22stabilizer_over_gf4\%22\%2C\%22qudit_cws\%22\%2C\%22qudit_gkp\%22\%2C\%22qudit_sign\%22\%2C\%22perfect\%22\%2C\%22quantum_perfect\%22\%2C\%22data_syndrome\%22\%2C\%22quantum_twisted\%22\%2C\%22quantum_random\%22\%2C\%22stab_15_7_3\%22\%2C\%22quantum_hamming\%22\%2C\%22stab_5_1_3\%22\%2C\%22q-ary_hamming\%22\%5D\%2C\%22reusePreviousLayoutPositions\%22\%3Afalse\%2C\%22showIntermediateConnectingNodes\%22\%3Atrue\%2C\%22connectingNodesMaxDepth\%22\%3A15\%2C\%22connectingNodesPathMaxLength\%22\%3A20\%2C\%22connectingNodesMaxExtraDepth\%22\%3A3\%2C\%22connectingNodesOnlyKeepPathsWithAdditionalLength\%22\%3A1\%2C\%22connectingNodesToDomainsAndKingdoms\%22\%3Afalse\%2C\%22connectingNodesEdgeLengthsByType\%22\%3A\%7B\%22primaryParent\%22\%3A1\%2C\%22secondaryParent\%22\%3A4\%2C\%22cousin\%22\%3A6\%7D\%2C\%22nodeIds\%22\%3A\%5B\%5D\%7D\%2C\%22highlightImportantNodes\%22\%3A\%7B\%22highlightImportantNodes\%22\%3Afalse\%2C\%22highlightPrimaryParents\%22\%3Afalse\%2C\%22highlightRootConnectingEdges\%22\%3Afalse\%7D\%7D}

\begingroup
\small
\eczhBreakableDashes
\renewcommand\arraystretch{1.05}
\edef\myxtraspc{\eczListAddVSpaceXtraXtraStretch}
\begin{tabularx}{\linewidth}{>{\raggedright\arraybackslash}p{\eczListColWidth{name}} >{\hsize=1.0000\hsize }X}
\toprule
\eczListColTitle{Code} & \eczListColTitle{Description} \\
\midrule
\endfirsthead
\toprule
\eczListColTitleContinued{Code} & \eczListColTitleContinued{Description} \\
\midrule
\endhead
\bottomrule
\endfoot
\eczhRefIndex{code:stabilizer_over_gf4}%
\eczhListValue{\flmRefsHyperref{code:stabilizer_over_gf4}{Hermitian qubit code}} & \eczhListValue{A qubit stabilizer code constructed from a Hermitian self-orthogonal linear quaternary code using the Hermitian construction.}\\ 
\addlinespace[\myxtraspc]
\eczhRefIndex{code:qudit_cws}%
\eczhListValue{\flmRefsHyperref{code:qudit_cws}{Modular-qudit CWS code}} & \eczhListValue{A CWS code for modular qudits, defined using a modular-qudit cluster state and a set of modular-qudit \(Z\)-type Pauli strings defined by a \(q\)-ary classical code over \(\mathbb{Z}_q\).}\\ 
\addlinespace[\myxtraspc]
\eczhRefIndex{code:qudit_gkp}%
\eczhListValue{\flmRefsHyperref{code:qudit_gkp}{Modular-qudit GKP code}} & \eczhListValue{Modular-qudit analogue of the GKP code.
Encodes a qudit into a larger qudit and protects against Pauli shifts up to some maximum value.}\\ 
\addlinespace[\myxtraspc]
\eczhRefIndex{code:qudit_sign}%
\eczhListValue{\flmRefsHyperref{code:qudit_sign}{Modular-qudit shift-resistant code}} & \eczhListValue{Monolithic code encoding a qubit into a single modular qudit and protecting against either \(Z\)-type or \(X\)-type modular-qudit Pauli shifts.}\\ 
\addlinespace[\myxtraspc]
\eczhRefIndex{code:perfect}%
\eczhListValue{\flmRefsHyperref{code:perfect}{Perfect code}} & \eczhListValue{A type of \(q\)-ary code whose parameters satisfy the Hamming bound with equality.}\\ 
\addlinespace[\myxtraspc]
\eczhRefIndex{code:quantum_perfect}%
\eczhListValue{\flmRefsHyperref{code:quantum_perfect}{Perfect quantum code}} & \eczhListValue{A type of block quantum code whose parameters satisfy the quantum Hamming bound with equality.}\\ 
\addlinespace[\myxtraspc]
\eczhRefIndex{code:data_syndrome}%
\eczhListValue{\flmRefsHyperref{code:data_syndrome}{Quantum data-syndrome (QDS) code}} & \eczhListValue{Stabilizer code designed to correct both data qubit errors and syndrome measurement errors simultaneously due to extra redundancy in its stabilizer generators.}\\ 
\addlinespace[\myxtraspc]
\eczhRefIndex{code:quantum_twisted}%
\eczhListValue{\flmRefsHyperref{code:quantum_twisted}{Quantum twisted code}} & \eczhListValue{Hermitian stabilizer code constructed from twisted BCH codes.}\\ 
\addlinespace[\myxtraspc]
\eczhRefIndex{code:quantum_random}%
\eczhListValue{\flmRefsHyperref{code:quantum_random}{Random quantum code}} & \eczhListValue{Quantum code whose construction is non-deterministic in some way, i.e., codes that utilize an element of randomness somewhere in their construction. Members of this class range from fully non-deterministic codes (e.g., random-circuit codes), to codes whose multi-step construction is deterministic with the exception of a single step (e.g., expander lifted-product codes).}\\ 
\addlinespace[\myxtraspc]
\eczhRefIndex{code:stab_15_7_3}%
\eczhListValue{\flmRefsHyperref{code:stab_15_7_3}{\(\llbracket 15, 7, 3\rrbracket \) quantum Hamming code}} & \eczhListValue{Self-dual quantum Hamming code that admits permutation-based CZ logical gates.
The code is constructed using the CSS construction from the \([15,11,3]\) Hamming code and its \([15,4,8]\) dual simplex code.}\\ 
\addlinespace[\myxtraspc]
\eczhRefIndex{code:quantum_hamming}%
\eczhListValue{\flmRefsHyperref{code:quantum_hamming}{\(\llbracket 2^r, 2^r-r-2, 3\rrbracket \) Gottesman code}} & \eczhListValue{A family of \flmRefsHyperref{ref672}{pure} \NoCaseChange{\protect\cite{cite449}} non-CSS stabilizer codes of distance \(3\) that saturate the asymptotic quantum Hamming bound.}\\ 
\addlinespace[\myxtraspc]
\eczhRefIndex{code:stab_5_1_3}%
\eczhListValue{\flmRefsHyperref{code:stab_5_1_3}{\(\llbracket 5,1,3\rrbracket \) Five-qubit perfect code}} & \eczhListValue{Five-qubit cyclic stabilizer code that is the smallest qubit stabilizer code to correct a single-qubit error.}\\ 
\addlinespace[\myxtraspc]
\eczhRefIndex{code:q-ary_hamming}%
\eczhListValue{\flmRefsHyperref{code:q-ary_hamming}{\(q\)-ary Hamming code}} & \eczhListValue{Member of an infinite family of perfect linear \(q\)-ary codes with parameters \([(q^r-1)/(q-1),(q^r-1)/(q-1)-r, 3]_q\) for \(r \geq 2\) \NoCaseChange{\protect\cite[{(3.1)}]{cite70}}.
These are precisely the nontrivial perfect linear codes over \(\mathbb{F}_q\) \NoCaseChange{\protect\cite[{Thm. 3.3.1}]{cite70}}.}\\ 
\end{tabularx}\endgroup
\eczcodelist{subsystem_qldpc}{QLDPC subsystem codes
}%

\eczhCodeListAutoDescription{All descendants of \flmRefsCref{code:sparse_subsystem}.}%

\eczhIncludeCodeGraph{Bare}{scale=0.5}{\columnwidth}{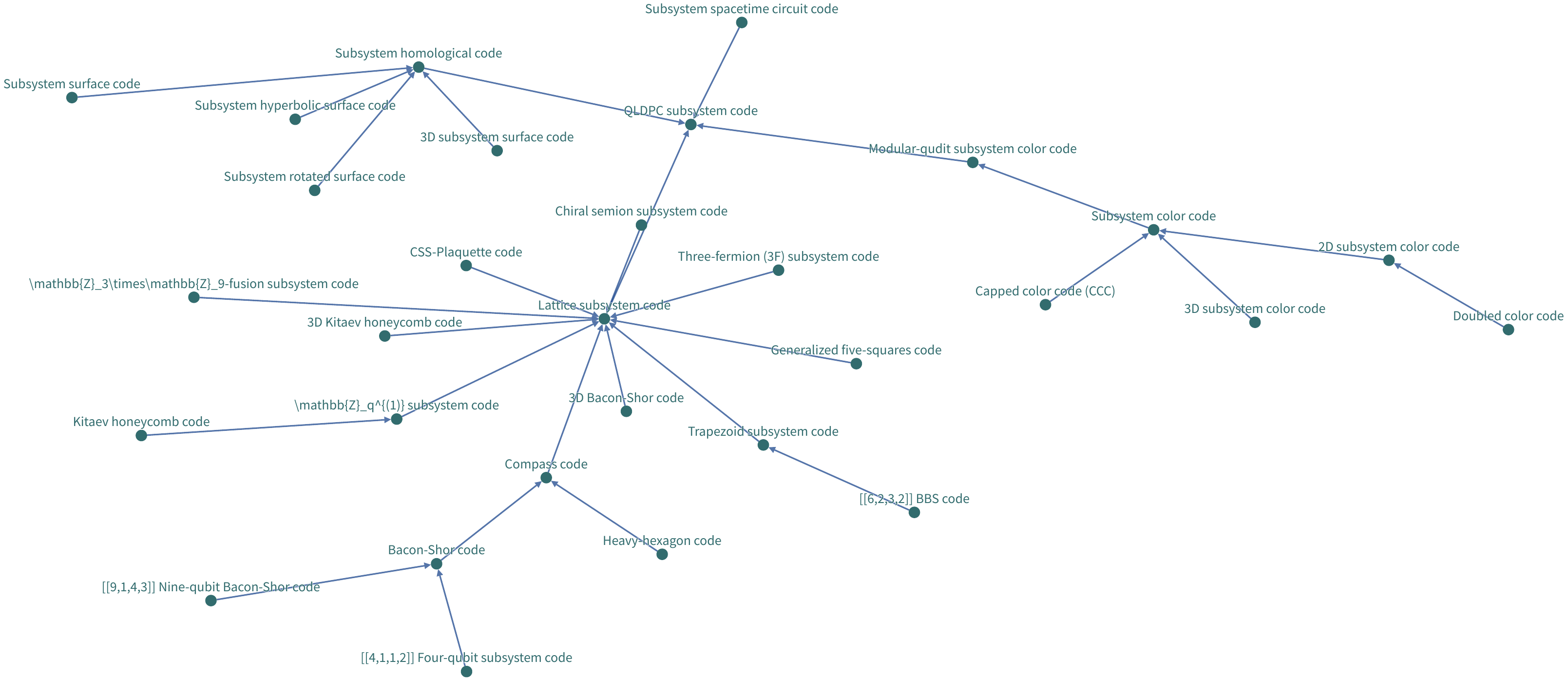}{QLDPC subsystem codes}{https://errorcorrectionzoo.org/code_graph#J\%7B\%22displayMode\%22\%3A\%22subset\%22\%2C\%22modeSubsetOptions\%22\%3A\%7B\%22codeIds\%22\%3A\%5B\%222d_subsystem_color\%22\%2C\%223d_bacon_shor\%22\%2C\%223d_kitaev_honeycomb\%22\%2C\%223d_subsystem_color\%22\%2C\%223d_subsystem_surface\%22\%2C\%22bacon_shor\%22\%2C\%22css_plaquette\%22\%2C\%22capped_color\%22\%2C\%22semion\%22\%2C\%22compass_model\%22\%2C\%22doubled_color\%22\%2C\%22five_squares\%22\%2C\%22heavy_hex\%22\%2C\%22kitaev_honeycomb\%22\%2C\%22translationally_invariant_subsystem\%22\%2C\%22qudit_subsystem_color\%22\%2C\%22sparse_subsystem\%22\%2C\%22subsystem_color\%22\%2C\%22subsystem_higher_dimensional_surface\%22\%2C\%22subsystem_hyperbolic_surface\%22\%2C\%22subsystem_rotated_surface\%22\%2C\%22subsystem_spacetime_circuit\%22\%2C\%22subsystem_surface\%22\%2C\%22subsystem_three_fermion\%22\%2C\%22trapezoid\%22\%2C\%22bacon_shor_4\%22\%2C\%22bravyi_bacon_shor_6\%22\%2C\%22bacon_shor_9\%22\%2C\%22zthree_znine\%22\%2C\%22qudit_znone\%22\%5D\%2C\%22reusePreviousLayoutPositions\%22\%3Afalse\%2C\%22showIntermediateConnectingNodes\%22\%3Atrue\%2C\%22connectingNodesMaxDepth\%22\%3A15\%2C\%22connectingNodesPathMaxLength\%22\%3A20\%2C\%22connectingNodesMaxExtraDepth\%22\%3A3\%2C\%22connectingNodesOnlyKeepPathsWithAdditionalLength\%22\%3A1\%2C\%22connectingNodesToDomainsAndKingdoms\%22\%3Afalse\%2C\%22connectingNodesEdgeLengthsByType\%22\%3A\%7B\%22primaryParent\%22\%3A1\%2C\%22secondaryParent\%22\%3A4\%2C\%22cousin\%22\%3A6\%7D\%2C\%22nodeIds\%22\%3A\%5B\%5D\%7D\%2C\%22highlightImportantNodes\%22\%3A\%7B\%22highlightImportantNodes\%22\%3Afalse\%2C\%22highlightPrimaryParents\%22\%3Afalse\%2C\%22highlightRootConnectingEdges\%22\%3Afalse\%7D\%7D}

\begingroup
\small
\eczhBreakableDashes
\renewcommand\arraystretch{1.05}
\edef\myxtraspc{\eczListAddVSpaceXtraXtraStretch}
\begin{tabularx}{\linewidth}{>{\raggedright\arraybackslash}p{\eczListColWidth{name}} >{\hsize=1.0000\hsize }X}
\toprule
\eczListColTitle{Code} & \eczListColTitle{Description} \\
\midrule
\endfirsthead
\toprule
\eczListColTitleContinued{Code} & \eczListColTitleContinued{Description} \\
\midrule
\endhead
\bottomrule
\endfoot
\eczhRefIndex{code:2d_subsystem_color}%
\eczhListValue{\flmRefsHyperref{code:2d_subsystem_color}{2D subsystem color code}} & \eczhListValue{A subsystem version of the 2D color code.
The original topological subsystem-code example is defined on the Union Jack lattice \NoCaseChange{\protect\cite{cite604}}; the square-octagon-lattice hypergraph construction of \NoCaseChange{\protect\cite{cite594}} reproduces the same code from a complementary viewpoint.}\\ 
\addlinespace[\myxtraspc]
\eczhRefIndex{code:3d_bacon_shor}%
\eczhListValue{\flmRefsHyperref{code:3d_bacon_shor}{3D Bacon-Shor code}} & \eczhListValue{Generalization of the Bacon-Shor code to three dimensions that was conjectured to be a self-correcting memory.
It is defined on a cubic lattice and admits sheet-like stabilizer generators.}\\ 
\addlinespace[\myxtraspc]
\eczhRefIndex{code:3d_kitaev_honeycomb}%
\eczhListValue{\flmRefsHyperref{code:3d_kitaev_honeycomb}{3D Kitaev honeycomb code}} & \eczhListValue{3D subsystem stabilizer code whose Hamiltonian is a 3D generalization of the Kitaev honeycomb model.
One of the phases realized by the 3D Kitaev honeycomb Hamiltonian is that of the 3D fermionic surface code \NoCaseChange{\protect\cite{cite458}}.}\\ 
\addlinespace[\myxtraspc]
\eczhRefIndex{code:3d_subsystem_color}%
\eczhListValue{\flmRefsHyperref{code:3d_subsystem_color}{3D subsystem color code}} & \eczhListValue{A subsystem version of the 3D color code defined on a 3-colex.}\\ 
\addlinespace[\myxtraspc]
\eczhRefIndex{code:3d_subsystem_surface}%
\eczhListValue{\flmRefsHyperref{code:3d_subsystem_surface}{3D subsystem surface code}} & \eczhListValue{Subsystem generalization of the surface code on a 3D cubic lattice with gauge-group generators of weight at most three.}\\ 
\addlinespace[\myxtraspc]
\eczhRefIndex{code:bacon_shor}%
\eczhListValue{\flmRefsHyperref{code:bacon_shor}{Bacon-Shor code}} & \eczhListValue{Subsystem CSS code defined on an \(m_1 \times m_2\) lattice of qubits that generalizes the \(\llbracket 9,1,3\rrbracket \) (subspace) Shor code.
It is said to be \textit{symmetric} when \(m_1=m_2\), and \textit{asymmetric} otherwise.}\\ 
\addlinespace[\myxtraspc]
\eczhRefIndex{code:css_plaquette}%
\eczhListValue{\flmRefsHyperref{code:css_plaquette}{CSS-Plaquette code}} & \eczhListValue{Generalization of the Bacon-Shor code to three dimensions, defined on a cubic lattice and admitting string-like stabilizer generators.}\\ 
\addlinespace[\myxtraspc]
\eczhRefIndex{code:capped_color}%
\eczhListValue{\flmRefsHyperref{code:capped_color}{Capped color code (CCC)}} & \eczhListValue{A non-geometrically local subsystem color code consisting of two layers of 2D color code stacked together and topped (or capped) by a single qubit.
Gauge fixing yields two types of codes, capped color codes in H or T form.
Layers of 2D color codes can also be stacked together in a recursive construction, yielding \textit{recursive capped color codes} (RCCCs).}\\ 
\addlinespace[\myxtraspc]
\eczhRefIndex{code:semion}%
\eczhListValue{\flmRefsHyperref{code:semion}{Chiral semion subsystem code}} & \eczhListValue{Modular-qudit subsystem stabilizer code with qudit dimension \(q=4\) that is characterized by the chiral semion topological phase.
The code admits a set of geometrically local stabilizer generators on a torus.}\\ 
\addlinespace[\myxtraspc]
\eczhRefIndex{code:compass_model}%
\eczhListValue{\flmRefsHyperref{code:compass_model}{Compass code}} & \eczhListValue{Subspace or subsystem CSS code defined by gauge-fixing the Bacon-Shor code, i.e., the code whose gauge group consists of terms in the compass model Hamiltonian \NoCaseChange{\protect\cite{cite656,cite657,cite658}} on a square lattice.
Families of random codes perform well against biased noise and spatially dependent (i.e., asymmetric) noise.}\\ 
\addlinespace[\myxtraspc]
\eczhRefIndex{code:doubled_color}%
\eczhListValue{\flmRefsHyperref{code:doubled_color}{Doubled color code}} & \eczhListValue{Family of \(\llbracket 2t^3+8t^2+6t-1,1,2t+1\rrbracket \) subsystem color codes (with \(t\geq 1\)), constructed using a generalization of the doubling transformation \NoCaseChange{\protect\cite{cite659}}, that admit a Clifford + \(T\) transversal gate set using gauge fixing.}\\ 
\addlinespace[\myxtraspc]
\eczhRefIndex{code:five_squares}%
\eczhListValue{\flmRefsHyperref{code:five_squares}{Generalized five-squares code}} & \eczhListValue{Member of a family of subsystem codes that are generalizations \NoCaseChange{\protect\cite{cite660,cite661}} of a code defined on a three-valent hypergraph associated with the five-squares lattice \NoCaseChange{\protect\cite{cite594}}.
The original five-squares code is a 2D topological subsystem code with local two-qubit gauge generators; on a torus, it encodes two logical qubits \NoCaseChange{\protect\cite{cite594}}.}\\ 
\addlinespace[\myxtraspc]
\eczhRefIndex{code:heavy_hex}%
\eczhListValue{\flmRefsHyperref{code:heavy_hex}{Heavy-hexagon code}} & \eczhListValue{Subsystem stabilizer code on the heavy-hexagonal point set that combines Bacon-Shor and surface-code stabilizers.
Encodes one logical qubit into \(n=(5d^2-2d-1)/2\) physical qubits with distance \(d\).
The heavy-hexagonal point set allows for low degree (at most 3) connectivity between all the data and ancilla qubits, which is suitable for fixed-frequency transmon qubits subject to frequency collision errors.
The code can be split into a surface and a Bacon-Shor code, with the idling qubits of one code serving as the physical qubits of the other \NoCaseChange{\protect\cite{cite662}}.}\\ 
\addlinespace[\myxtraspc]
\eczhRefIndex{code:kitaev_honeycomb}%
\eczhListValue{\flmRefsHyperref{code:kitaev_honeycomb}{Kitaev honeycomb code}} & \eczhListValue{Subsystem qubit stabilizer code underlying the Kitaev honeycomb model \NoCaseChange{\protect\cite{cite537,cite594}}.
Its gauge generators are the two-qubit \(XX\), \(YY\), and \(ZZ\) link operators on the three edge types of the honeycomb lattice \NoCaseChange{\protect\cite[{Sec. 3.2}]{cite594}}.
Its stabilizer group is generated by loop operators, and syndrome extraction can be reduced to ordered measurements of the two-qubit link operators \NoCaseChange{\protect\cite[{Sec. 3.2}]{cite594}}.
This is the \(q=2\) instance of the \(\mathbb{Z}_q^{(1)}\) subsystem code and does not encode any logical qubits \NoCaseChange{\protect\cite{cite594}\protect\cite[{Sec. 7.3}]{cite414}}.}\\ 
\addlinespace[\myxtraspc]
\eczhRefIndex{code:translationally_invariant_subsystem}%
\eczhListValue{\flmRefsHyperref{code:translationally_invariant_subsystem}{Lattice subsystem code}} & \eczhListValue{A geometrically local qubit, modular-qudit, or Galois-qudit subsystem stabilizer code with qudits organized on a lattice modeled by the additive group \(\mathbb{Z}^D\) for spatial dimension \(D\), using either the ordinary block notion of locality or the fermionic/Majorana notion of locality.
On an infinite lattice, its gauge group is generated by few-site Pauli operators and their translations, in which case the code is called \textit{translationally invariant subsystem code}.
The stabilizer group may contain generators of unbounded weight, distinguishing these codes from stabilizer codes with bounded-weight generators for which some logical qubits were re-assigned to be gauge qubits.}\\ 
\addlinespace[\myxtraspc]
\eczhRefIndex{code:qudit_subsystem_color}%
\eczhListValue{\flmRefsHyperref{code:qudit_subsystem_color}{Modular-qudit subsystem color code}} & \eczhListValue{An extension of subsystem color codes to modular qudits.
Codes are defined analogously to qubit subsystem color codes, but a directionality is required in order to make the modular-qudit stabilizers commute \NoCaseChange{\protect\cite[{Sec. VII}]{cite673}}.}\\ 
\addlinespace[\myxtraspc]
\eczhRefIndex{code:sparse_subsystem}%
\eczhListValue{\flmRefsHyperref{code:sparse_subsystem}{QLDPC subsystem code}} & \eczhListValue{Member of a family of subsystem stabilizer codes for which the number of sites participating in each gauge generator and the number of gauge generators that each site participates in are both bounded by a constant as \(n\to\infty\).
The stabilizer group may contain generators of unbounded weight, distinguishing these codes from stabilizer codes with bounded-weight generators for which some logical qubits were re-assigned to be gauge qubits.}\\ 
\addlinespace[\myxtraspc]
\eczhRefIndex{code:subsystem_color}%
\eczhListValue{\flmRefsHyperref{code:subsystem_color}{Subsystem color code}} & \eczhListValue{A subsystem version of the color code.}\\ 
\addlinespace[\myxtraspc]
\eczhRefIndex{code:subsystem_higher_dimensional_surface}%
\eczhListValue{\flmRefsHyperref{code:subsystem_higher_dimensional_surface}{Subsystem homological code}} & \eczhListValue{A subsystem CSS code that is a subsystem version of the homological code, defined on cellulations of manifolds in arbitrary dimensions.
Gauge-group generators are of lower weight than the stabilizers of the corresponding surface code, enabling fault-tolerant syndrome extraction with simpler circuits.
The stabilizer group may contain generators of unbounded weight, distinguishing these codes from stabilizer codes with bounded-weight generators for which some logical qubits were re-assigned to be gauge qubits.}\\ 
\addlinespace[\myxtraspc]
\eczhRefIndex{code:subsystem_hyperbolic_surface}%
\eczhListValue{\flmRefsHyperref{code:subsystem_hyperbolic_surface}{Subsystem hyperbolic surface code}} & \eczhListValue{Subsystem generalization of the surface code on a 2D hyperbolic tessellation with gauge-group generators of weight at most three.
An \(\{r,4\}\) hyperbolic tessellation with \(E\) edges yields a \(\llbracket 3E/2,(1/2-2/r)E+2,(1-2/r)E,d\rrbracket \) subsystem code.}\\ 
\addlinespace[\myxtraspc]
\eczhRefIndex{code:subsystem_rotated_surface}%
\eczhListValue{\flmRefsHyperref{code:subsystem_rotated_surface}{Subsystem rotated surface code}} & \eczhListValue{Subsystem version of the rotated surface code.}\\ 
\addlinespace[\myxtraspc]
\eczhRefIndex{code:subsystem_spacetime_circuit}%
\eczhListValue{\flmRefsHyperref{code:subsystem_spacetime_circuit}{Subsystem spacetime circuit code}} & \eczhListValue{Subsystem stabilizer code obtained from a spacetime circuit code by \flmRefsHyperref{ref666}{gauging out} logical operators that correspond to circuit faults with trivial effect \NoCaseChange{\protect\cite[{Sec. 5.4}]{cite667}}.
In the original circuit-to-code construction, each circuit element is replaced by low-weight gauge generators enforcing its input-output relations, yielding subsystem codes from restricted Clifford postselection circuits \NoCaseChange{\protect\cite{cite668}}.}\\ 
\addlinespace[\myxtraspc]
\eczhRefIndex{code:subsystem_surface}%
\eczhListValue{\flmRefsHyperref{code:subsystem_surface}{Subsystem surface code}} & \eczhListValue{Subsystem version of the surface code defined on a square lattice with qubits placed at every vertex and center of every edge.
Its gauge checks are weight-three triangle operators of type \(XXX\) and \(ZZZ\) \NoCaseChange{\protect\cite{cite669}}.}\\ 
\addlinespace[\myxtraspc]
\eczhRefIndex{code:subsystem_three_fermion}%
\eczhListValue{\flmRefsHyperref{code:subsystem_three_fermion}{Three-fermion (3F) subsystem code}} & \eczhListValue{2D subsystem stabilizer code whose low-energy excitations realize the three-fermion anyon theory \NoCaseChange{\protect\cite{cite601,cite602,cite603}}.
One version uses two qubits at each site \NoCaseChange{\protect\cite{cite414}}, while other manifestations utilize a single qubit per site and only weight-two (two-body) interactions \NoCaseChange{\protect\cite{cite602,cite604}}.
All are expected to be equivalent to each other via a local constant-depth \flmRefsHyperref{ref409}{Clifford circuit}.}\\ 
\addlinespace[\myxtraspc]
\eczhRefIndex{code:trapezoid}%
\eczhListValue{\flmRefsHyperref{code:trapezoid}{Trapezoid subsystem code}} & \eczhListValue{A member of a family of BBS codes with weight-two (two-body) gauge generators designed to suppress errors in adiabatic quantum computation.}\\ 
\addlinespace[\myxtraspc]
\eczhRefIndex{code:bacon_shor_4}%
\eczhListValue{\flmRefsHyperref{code:bacon_shor_4}{\(\llbracket 4,1,1,2\rrbracket \) Four-qubit subsystem code}} & \eczhListValue{Error-detecting four-qubit subsystem stabilizer code encoding one logical qubit and one gauge qubit.}\\ 
\addlinespace[\myxtraspc]
\eczhRefIndex{code:bravyi_bacon_shor_6}%
\eczhListValue{\flmRefsHyperref{code:bravyi_bacon_shor_6}{\(\llbracket 6,2,3,2\rrbracket \) BBS code}} & \eczhListValue{Error-detecting six-qubit BBS subsystem code with parameters \(\llbracket 6,2,3,2\rrbracket \) that can suppress errors in adiabatic quantum computation \NoCaseChange{\protect\cite{cite670}}.}\\ 
\addlinespace[\myxtraspc]
\eczhRefIndex{code:bacon_shor_9}%
\eczhListValue{\flmRefsHyperref{code:bacon_shor_9}{\(\llbracket 9,1,4,3\rrbracket \) Nine-qubit Bacon-Shor code}} & \eczhListValue{Error-correcting nine-qubit subsystem stabilizer code encoding one logical qubit and four gauge qubits.
There are exactly four inequivalent CSS gauge fixings of the code, including the \flmRefsHyperref{code:shor_nine}{Shor code} and the \flmRefsHyperref{code:surface-17}{surface-17 code} \NoCaseChange{\protect\cite{cite454}}.}\\ 
\addlinespace[\myxtraspc]
\eczhRefIndex{code:zthree_znine}%
\eczhListValue{\flmRefsHyperref{code:zthree_znine}{\(\mathbb{Z}_3\times\mathbb{Z}_9\)-fusion subsystem code}} & \eczhListValue{Modular-qudit 2D subsystem stabilizer code whose low-energy excitations realize a non-modular anyon theory with \(\mathbb{Z}_3\times\mathbb{Z}_9\) fusion rules.
Encodes two qutrits when put on a torus.}\\ 
\addlinespace[\myxtraspc]
\eczhRefIndex{code:qudit_znone}%
\eczhListValue{\flmRefsHyperref{code:qudit_znone}{\(\mathbb{Z}_q^{(1)}\) subsystem code}} & \eczhListValue{Modular-qudit subsystem code, based on the Kitaev honeycomb model \NoCaseChange{\protect\cite{cite537}} and its generalization \NoCaseChange{\protect\cite{cite637}}, that is characterized by the \(\mathbb{Z}_q^{(1)}\) anyon theory \NoCaseChange{\protect\cite{cite638}}, which is modular for odd prime \(q\) and non-modular otherwise. Encodes a single \(q\)-dimensional qudit when put on a torus for odd \(q\), and a \(q/2\)-dimensional qudit for even \(q\). This code can be constructed using geometrically local gauge generators, but does not admit geometrically local stabilizer generators. For \(q=2\), the code reduces to the subsystem code underlying the Kitaev honeycomb model code as well as the honeycomb Floquet code.}\\ 
\end{tabularx}\endgroup
\eczcodelist{homological}{Quantum codes based on homological products
}%

\eczhCodeListAutoDescription{All descendants of \flmRefsCref{code:generalized_homological_product}.}%

\eczhIncludeCodeGraph{Bare}{scale=0.5}{\columnwidth}{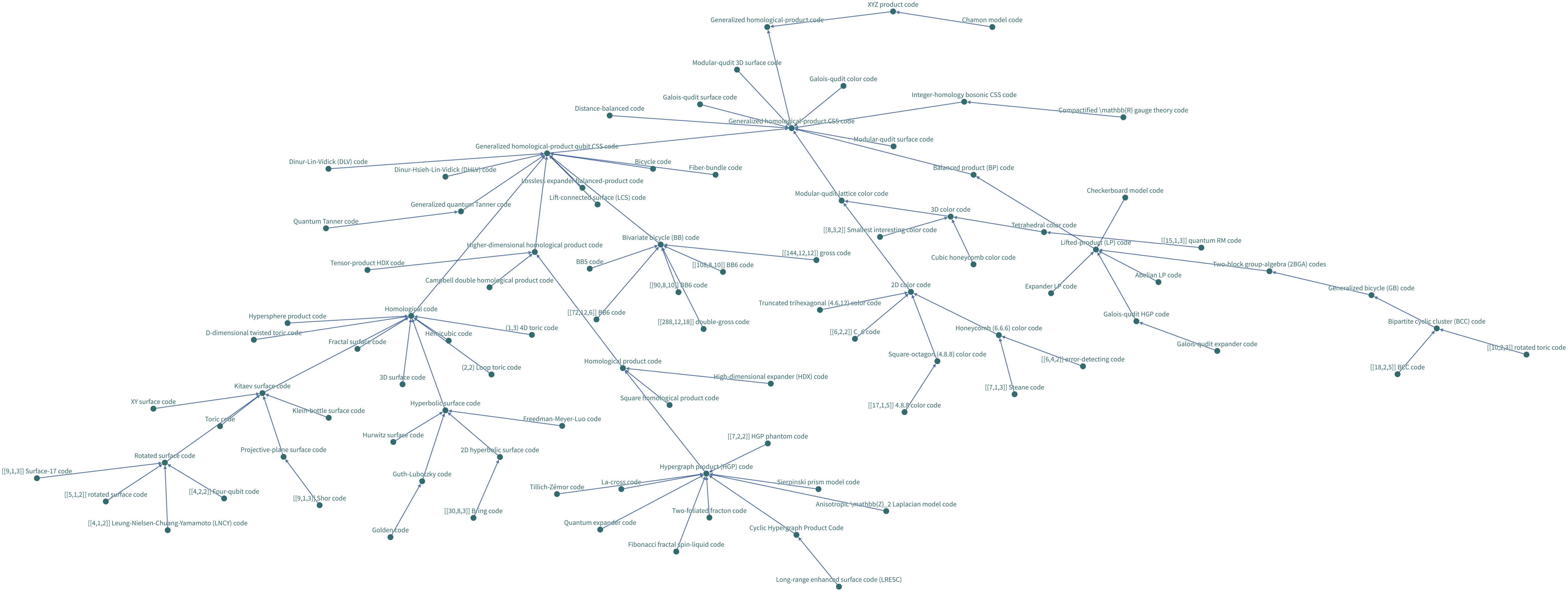}{Quantum codes based on homological products}{https://errorcorrectionzoo.org/code_graph#J\%7B\%22displayMode\%22\%3A\%22subset\%22\%2C\%22modeSubsetOptions\%22\%3A\%7B\%22codeIds\%22\%3A\%5B\%222d_color\%22\%2C\%22two_dimensional_hyperbolic_surface\%22\%2C\%223d_color\%22\%2C\%223d_surface\%22\%2C\%22abelian_lifted_product\%22\%2C\%22anisotropic_z2_laplacian\%22\%2C\%22bb5\%22\%2C\%22balanced_product\%22\%2C\%22bicycle\%22\%2C\%22bipartite_cyclic_cluster\%22\%2C\%22qcga\%22\%2C\%22double_homological_product\%22\%2C\%22chamon\%22\%2C\%22checkerboard\%22\%2C\%22compactified_r\%22\%2C\%22cubic_honeycomb_color\%22\%2C\%22cyclic_hgp\%22\%2C\%22dhlv\%22\%2C\%22dlv\%22\%2C\%22distance_balanced\%22\%2C\%22expander_lifted_product\%22\%2C\%22fiber_bundle\%22\%2C\%22fibonacci_fractal_liquid\%22\%2C\%22fractal_surface\%22\%2C\%22freedman_meyer_luo\%22\%2C\%22galois_hypergraph_product\%22\%2C\%22galois_color\%22\%2C\%22galois_expander\%22\%2C\%22galois_topological\%22\%2C\%22generalized_bicycle\%22\%2C\%22generalized_homological_product_css\%22\%2C\%22generalized_homological_product\%22\%2C\%22qubit_generalized_homological_product_css\%22\%2C\%22generalized_quantum_tanner\%22\%2C\%22golden\%22\%2C\%22four_dimensional_hyperbolic\%22\%2C\%22hemicubic\%22\%2C\%22ramanujan_tensor_product\%22\%2C\%22multisector_hypergraph\%22\%2C\%22higher_dimensional_surface\%22\%2C\%22homological_product\%22\%2C\%22triangular_color\%22\%2C\%22hurwitz_surface\%22\%2C\%22hyperbolic_surface\%22\%2C\%22hypergraph_product\%22\%2C\%22hypersphere_product\%22\%2C\%22homological_cv\%22\%2C\%22surface\%22\%2C\%22klein_bottle\%22\%2C\%22lacross\%22\%2C\%22lcs\%22\%2C\%22lifted_product\%22\%2C\%22lresc\%22\%2C\%22lossless_expander\%22\%2C\%22qudit_3d_surface\%22\%2C\%22qudit_color\%22\%2C\%22qudit_surface\%22\%2C\%22real_projective_plane\%22\%2C\%22quantum_tanner\%22\%2C\%22quantum_expander\%22\%2C\%22rotated_surface\%22\%2C\%22sierpinsky_fractal_liquid\%22\%2C\%22square_homological_product\%22\%2C\%22488_color\%22\%2C\%22iterated_ramanujan\%22\%2C\%22tetrahedral_color\%22\%2C\%22tillichzemor\%22\%2C\%22toric\%22\%2C\%224612_color\%22\%2C\%222bga\%22\%2C\%22two_foliated\%22\%2C\%22xysurface\%22\%2C\%22xyz_product\%22\%2C\%224d_13_surface\%22\%2C\%224d_surface\%22\%2C\%22higher_dimensional_toric\%22\%2C\%22xzzx_10_2_3\%22\%2C\%22bb108\%22\%2C\%22gross\%22\%2C\%22stab_15_1_3\%22\%2C\%22stab_17_1_5\%22\%2C\%22stab_18_2_5\%22\%2C\%22bb288\%22\%2C\%22stellated_dodecahedron_css\%22\%2C\%22css_4_1_2\%22\%2C\%22stab_4_2_2\%22\%2C\%22stab_5_1_2\%22\%2C\%22stab_6_2_2\%22\%2C\%22stab_6_4_2\%22\%2C\%22steane\%22\%2C\%22hgp_7_2_2\%22\%2C\%22bb72\%22\%2C\%22stab_8_3_2\%22\%2C\%22shor_nine\%22\%2C\%22surface-17\%22\%2C\%22bb90\%22\%5D\%2C\%22reusePreviousLayoutPositions\%22\%3Afalse\%2C\%22showIntermediateConnectingNodes\%22\%3Atrue\%2C\%22connectingNodesMaxDepth\%22\%3A15\%2C\%22connectingNodesPathMaxLength\%22\%3A20\%2C\%22connectingNodesMaxExtraDepth\%22\%3A3\%2C\%22connectingNodesOnlyKeepPathsWithAdditionalLength\%22\%3A1\%2C\%22connectingNodesToDomainsAndKingdoms\%22\%3Afalse\%2C\%22connectingNodesEdgeLengthsByType\%22\%3A\%7B\%22primaryParent\%22\%3A1\%2C\%22secondaryParent\%22\%3A4\%2C\%22cousin\%22\%3A6\%7D\%2C\%22nodeIds\%22\%3A\%5B\%5D\%7D\%2C\%22highlightImportantNodes\%22\%3A\%7B\%22highlightImportantNodes\%22\%3Afalse\%2C\%22highlightPrimaryParents\%22\%3Afalse\%2C\%22highlightRootConnectingEdges\%22\%3Afalse\%7D\%7D}

\begingroup
\small
\eczhBreakableDashes
\renewcommand\arraystretch{1.05}
\edef\myxtraspc{\eczListAddVSpaceXtraXtraStretch}
\begin{tabularx}{\linewidth}{>{\raggedright\arraybackslash}p{\eczListColWidth{name}} >{\hsize=1.0000\hsize }X}
\toprule
\eczListColTitle{Code} & \eczListColTitle{Description} \\
\midrule
\endfirsthead
\toprule
\eczListColTitleContinued{Code} & \eczListColTitleContinued{Description} \\
\midrule
\endhead
\bottomrule
\endfoot
\eczhRefIndex{code:2d_color}%
\eczhListValue{\flmRefsHyperref{code:2d_color}{2D color code}} & \eczhListValue{Color code defined on a graph embedded in a two-dimensional surface.
Each face hosts two stabilizer generators, a Pauli-\(X\) and a Pauli-\(Z\) string acting on all the qubits of the face.}\\ 
\addlinespace[\myxtraspc]
\eczhRefIndex{code:two_dimensional_hyperbolic_surface}%
\eczhListValue{\flmRefsHyperref{code:two_dimensional_hyperbolic_surface}{2D hyperbolic surface code}} & \eczhListValue{Hyperbolic surface codes based on a tessellation of a closed 2D manifold with a hyperbolic geometry (i.e., non-Euclidean geometry, e.g., saddle surfaces when defined on a 2D plane).}\\ 
\addlinespace[\myxtraspc]
\eczhRefIndex{code:3d_color}%
\eczhListValue{\flmRefsHyperref{code:3d_color}{3D color code}} & \eczhListValue{Color code defined on a four-valent, four-colorable 3-colex in a 3-manifold.
In the original colex realization, qubits sit on vertices, \(X\)-type stabilizers are attached to 3-cells, and \(Z\)-type stabilizers are attached to faces \NoCaseChange{\protect\cite{cite430}}.}\\ 
\addlinespace[\myxtraspc]
\eczhRefIndex{code:3d_surface}%
\eczhListValue{\flmRefsHyperref{code:3d_surface}{3D surface code}} & \eczhListValue{A generalization of the Kitaev surface code defined on a 3D cubic lattice.
Qubits are placed on edges, \(Z\)-type stabilizer generators are placed on square plaquettes oriented in all three directions, and \(X\)-type stabilizers are placed on the six edges neighboring every vertex \NoCaseChange{\protect\cite{cite459}}.}\\ 
\addlinespace[\myxtraspc]
\eczhRefIndex{code:abelian_lifted_product}%
\eczhListValue{\flmRefsHyperref{code:abelian_lifted_product}{Abelian LP code}} & \eczhListValue{A lifted-product code whose lift group \(G\) is Abelian.
The case of \(G\) being a cyclic group is a GB code (a.k.a. a quasi-cyclic LP code) \NoCaseChange{\protect\cite[{Sec. III.E}]{cite674}}.
A particular family with \(G=\mathbb{Z}_{\ell}\) yields codes with parameters \(\llbracket n,k=\Theta(\log n),d=\Theta(n/\log n)\rrbracket \) \NoCaseChange{\protect\cite{cite674}}.}\\ 
\addlinespace[\myxtraspc]
\eczhRefIndex{code:anisotropic_z2_laplacian}%
\eczhListValue{\flmRefsHyperref{code:anisotropic_z2_laplacian}{Anisotropic \(\mathbb{Z}_2\) Laplacian model code}} & \eczhListValue{A graph-based analogue of a Type-I fracton phase with lineon-like restricted mobility \NoCaseChange{\protect\cite{cite460,cite461}}.}\\ 
\addlinespace[\myxtraspc]
\eczhRefIndex{code:bb5}%
\eczhListValue{\flmRefsHyperref{code:bb5}{BB5 code}} & \eczhListValue{A BB code with weight-five stabilizer generators
(contrasting with the weight-six checks of standard BB codes), designed and
benchmarked for long chains of trapped ions \NoCaseChange{\protect\cite{cite407}}.}\\ 
\addlinespace[\myxtraspc]
\eczhRefIndex{code:balanced_product}%
\eczhListValue{\flmRefsHyperref{code:balanced_product}{Balanced product (BP) code}} & \eczhListValue{Family of CSS quantum codes obtained from two classical-code chain complexes that share a common group symmetry.
The balanced product can be understood as taking the usual tensor or hypergraph product and then quotienting by the shared symmetry action.
This can reduce the overall number of physical qubits \(n\) while, in favorable cases, preserving the number of encoded qubits and the code distance, thereby improving the encoding rate \(k/n\) and normalized distance \(d/n\) compared to the underlying tensor or hypergraph product.}\\ 
\addlinespace[\myxtraspc]
\eczhRefIndex{code:bicycle}%
\eczhListValue{\flmRefsHyperref{code:bicycle}{Bicycle code}} & \eczhListValue{A CSS code whose stabilizer generator matrix blocks are \(H_{X}=H_{Z}=(A|A^T)\), where \(A\) is a circulant matrix.
The fact that \(A\) commutes with its transpose ensures that the CSS condition is satisfied.
Bicycle codes are the first QLDPC codes.}\\ 
\addlinespace[\myxtraspc]
\eczhRefIndex{code:bipartite_cyclic_cluster}%
\eczhListValue{\flmRefsHyperref{code:bipartite_cyclic_cluster}{Bipartite cyclic cluster (BCC) code}} & \eczhListValue{Cyclic CSS code constructed from a bipartite cluster state with cyclic invariance,
emphasizing simplicity of state preparation over simplicity of stabilizers.}\\ 
\addlinespace[\myxtraspc]
\eczhRefIndex{code:qcga}%
\eczhListValue{\flmRefsHyperref{code:qcga}{Bivariate bicycle (BB) code}} & \eczhListValue{One of several Abelian 2BGA codes which admit time-optimal syndrome measurement circuits that can be implemented in a two-layer
architecture, a generalization of the square-lattice architecture
optimal for the surface codes.
Codes can be classified by the weight of their checks, e.g., by BB\(w\) where \(w\) is the check weight.}\\ 
\addlinespace[\myxtraspc]
\eczhRefIndex{code:double_homological_product}%
\eczhListValue{\flmRefsHyperref{code:double_homological_product}{Campbell double homological product code}} & \eczhListValue{A multi-dimensional HGP code derived from two applications of the hypergraph product to a classical code, resulting in a length-\(4\) chain complex.
The construction method allows for the use of two different classical codes as inputs, with Ref. \NoCaseChange{\protect\cite{cite675}} assuming identical input codes for simplicity.}\\ 
\addlinespace[\myxtraspc]
\eczhRefIndex{code:chamon}%
\eczhListValue{\flmRefsHyperref{code:chamon}{Chamon model code}} & \eczhListValue{A foliated type-I fracton non-CSS code defined on a cubic lattice using one weight-eight stabilizer generator acting on the eight vertices of each cube in the lattice \NoCaseChange{\protect\cite[{Eq. (D38)}]{cite456}}.}\\ 
\addlinespace[\myxtraspc]
\eczhRefIndex{code:checkerboard}%
\eczhListValue{\flmRefsHyperref{code:checkerboard}{Checkerboard model code}} & \eczhListValue{A foliated type-I fracton code defined on a cubic lattice that admits weight-eight  \(X\)- and \(Z\)-type stabilizer generators on the eight vertices of each cube in the lattice.
A tetrahedral Ising model can be used to obtain the checkerboard model by gauging \NoCaseChange{\protect\cite{cite462,cite463,cite233,cite464,cite465,cite466,cite467,cite468,cite469,cite470}} its subsystem symmetry \NoCaseChange{\protect\cite{cite233}}.
In that construction, the checkerboard model is self-dual under exchange of \(X\)- and \(Z\)-type stabilizers, and its composites include dimension-1 and dimension-2 excitations, i.e., lineons and planons in later terminology, with anyonic mutual and self-statistics \NoCaseChange{\protect\cite{cite233}}.}\\ 
\addlinespace[\myxtraspc]
\eczhRefIndex{code:compactified_r}%
\eczhListValue{\flmRefsHyperref{code:compactified_r}{Compactified \(\mathbb{R}\) gauge theory code}} & \eczhListValue{An integer-homology bosonic CSS code realizing 2D \(U(1)\) gauge theory on bosonic modes.
The code can be obtained from the analog surface code by \flmRefsHyperref{ref410}{condensing} certain anyons \NoCaseChange{\protect\cite{cite411}}. 
This results in a pinning of each mode to the space of periodic functions, which is the Hilbert space of a physical rotor, and can be thought of as compactification of the 2D \(\mathbb{R}\) gauge theory phase realized by the analog surface code.}\\ 
\addlinespace[\myxtraspc]
\eczhRefIndex{code:cubic_honeycomb_color}%
\eczhListValue{\flmRefsHyperref{code:cubic_honeycomb_color}{Cubic honeycomb color code}} & \eczhListValue{3D color code defined on a four-colorable bitruncated cubic honeycomb uniform tiling.}\\ 
\addlinespace[\myxtraspc]
\eczhRefIndex{code:cyclic_hgp}%
\eczhListValue{\flmRefsHyperref{code:cyclic_hgp}{Cyclic Hypergraph Product Code}} & \eczhListValue{The hypergraph product code constructed using two low-weight circulant matrices. The code family \(\mathrm{C2}\) is the product of a cyclic LDPC code with itself, and the family \(\mathrm{CxR}\) is the product of the cyclic code with a repetition code.}\\ 
\addlinespace[\myxtraspc]
\eczhRefIndex{code:dhlv}%
\eczhListValue{\flmRefsHyperref{code:dhlv}{Dinur-Hsieh-Lin-Vidick (DHLV) code}} & \eczhListValue{A family of asymptotically good QLDPC codes which are related to expander LP codes in that the roles of the check operators and physical qubits are exchanged.}\\ 
\addlinespace[\myxtraspc]
\eczhRefIndex{code:dlv}%
\eczhListValue{\flmRefsHyperref{code:dlv}{Dinur-Lin-Vidick (DLV) code}} & \eczhListValue{Member of a family of codes constructed using cubical chain complexes, which are \(t\)-order extensions of the complexes underlying expander codes (\(t=1\)) and expander lifted-product codes (\(t=2\)).}\\ 
\addlinespace[\myxtraspc]
\eczhRefIndex{code:distance_balanced}%
\eczhListValue{\flmRefsHyperref{code:distance_balanced}{Distance-balanced code}} & \eczhListValue{Galois-qudit CSS code obtained from a CSS code by increasing the smaller of the \(X\)- and \(Z\)-distances using a homological-product-based balancing step or one of its generalizations.
The initial code is said to be \textit{unbalanced}, i.e., tailored to noise biased toward either bit- or phase-flip errors, and the procedure can result in a code that treats both types of errors on a more equal footing.}\\ 
\addlinespace[\myxtraspc]
\eczhRefIndex{code:expander_lifted_product}%
\eczhListValue{\flmRefsHyperref{code:expander_lifted_product}{Expander LP code}} & \eczhListValue{Family of \(G\)-lifted product codes constructed using two classical \flmRefsHyperref{code:expander}{expander codes}, equivalently two regular \flmRefsHyperref{code:tanner}{Tanner codes} defined on the same expander graph \NoCaseChange{\protect\cite{cite74}}. For certain parameters, this construction yields the first asymptotically good QLDPC codes. Classical codes resulting from the same lifted-product complexes are one of the first two families of \(c^3\)-LTCs \NoCaseChange{\protect\cite{cite184}}.}\\ 
\addlinespace[\myxtraspc]
\eczhRefIndex{code:fiber_bundle}%
\eczhListValue{\flmRefsHyperref{code:fiber_bundle}{Fiber-bundle code}} & \eczhListValue{A CSS code constructed by combining one code as the base and another as the fiber of a fiber bundle.
In particular, taking a random LDPC code as the base and a cyclic repetition code as the fiber yields, after distance balancing, a QLDPC code with distance of \flmRefsHyperref{ref65}{order} \(\Omega( n^{3/5}/\text{polylog}(n) )\) and rate of \flmRefsHyperref{ref65}{order} \(\Omega( n^{-2/5}/\text{polylog}(n) )\).}\\ 
\addlinespace[\myxtraspc]
\eczhRefIndex{code:fibonacci_fractal_liquid}%
\eczhListValue{\flmRefsHyperref{code:fibonacci_fractal_liquid}{Fibonacci fractal spin-liquid code}} & \eczhListValue{A fractal type-I fracton CSS code defined on a cubic lattice \NoCaseChange{\protect\cite[{Eq. (D23)}]{cite456}}.}\\ 
\addlinespace[\myxtraspc]
\eczhRefIndex{code:fractal_surface}%
\eczhListValue{\flmRefsHyperref{code:fractal_surface}{Fractal surface code}} & \eczhListValue{Kitaev surface code on a fractal geometry, which is obtained by removing qubits from the surface code on a cubic lattice.
A related construction, the \textit{fractal product code}, is a hypergraph product of two classical codes defined on a Sierpinski carpet graph \NoCaseChange{\protect\cite{cite676}}. 
The underlying classical codes form classical self-correcting memories \NoCaseChange{\protect\cite{cite677,cite678,cite679}}.}\\ 
\addlinespace[\myxtraspc]
\eczhRefIndex{code:freedman_meyer_luo}%
\eczhListValue{\flmRefsHyperref{code:freedman_meyer_luo}{Freedman-Meyer-Luo code}} & \eczhListValue{Hyperbolic surface code constructed using cellulation of a Riemannian Manifold \(M\) exhibiting systolic freedom \NoCaseChange{\protect\cite{cite680}}. Codes derived from such manifolds can achieve distances scaling better than \(\sqrt{n}\), something that is impossible using closed 2D surfaces or 2D surfaces with boundaries \NoCaseChange{\protect\cite{cite681}}. Improved codes are obtained by studying a weak family of Riemann metrics on closed 4-dimensional manifolds \(S^2\otimes S^2\) with the \(\mathbb{Z}_2\)-homology.}\\ 
\addlinespace[\myxtraspc]
\eczhRefIndex{code:galois_hypergraph_product}%
\eczhListValue{\flmRefsHyperref{code:galois_hypergraph_product}{Galois-qudit HGP code}} & \eczhListValue{A member of a family of Galois-qudit CSS codes whose stabilizer generator matrix is obtained from a hypergraph product of two classical linear \(q\)-ary codes.}\\ 
\addlinespace[\myxtraspc]
\eczhRefIndex{code:galois_color}%
\eczhListValue{\flmRefsHyperref{code:galois_color}{Galois-qudit color code}} & \eczhListValue{Extension of the color code to 2D lattices of Galois qudits.}\\ 
\addlinespace[\myxtraspc]
\eczhRefIndex{code:galois_expander}%
\eczhListValue{\flmRefsHyperref{code:galois_expander}{Galois-qudit expander code}} & \eczhListValue{Galois-qudit CSS code obtained from tensor products of chain complexes associated with an explicit family of expander codes with Reed-Solomon local checks.}\\ 
\addlinespace[\myxtraspc]
\eczhRefIndex{code:galois_topological}%
\eczhListValue{\flmRefsHyperref{code:galois_topological}{Galois-qudit surface code}} & \eczhListValue{Extension of the surface code to 2D lattices of Galois qudits.}\\ 
\addlinespace[\myxtraspc]
\eczhRefIndex{code:generalized_bicycle}%
\eczhListValue{\flmRefsHyperref{code:generalized_bicycle}{Generalized bicycle (GB) code}} & \eczhListValue{A quasi-cyclic Galois-qudit CSS code constructed using a generalized version of the bicycle ansatz \NoCaseChange{\protect\cite{cite682}} from a pair of equivalent index-two quasi-cyclic linear codes.
Equivalently, the codes can be constructed via the lifted-product construction for \(G\) being a cyclic group \NoCaseChange{\protect\cite[{Sec. III.E}]{cite674}}.}\\ 
\addlinespace[\myxtraspc]
\eczhRefIndex{code:generalized_homological_product_css}%
\eczhListValue{\flmRefsHyperref{code:generalized_homological_product_css}{Generalized homological-product CSS code}} & \eczhListValue{CSS code whose properties are determined from an underlying chain complex, which often consists of some type of product of other chain complexes.}\\ 
\addlinespace[\myxtraspc]
\eczhRefIndex{code:generalized_homological_product}%
\eczhListValue{\flmRefsHyperref{code:generalized_homological_product}{Generalized homological-product code}} & \eczhListValue{Stabilizer code whose properties are determined from an underlying chain complex, which often consists of some type of product of other chain complexes.
The \flmRefsCref{ref683} yields an interpretation of codes in terms of chain complexes, thus allowing for the use of various products from homology in constructing codes.}\\ 
\addlinespace[\myxtraspc]
\eczhRefIndex{code:qubit_generalized_homological_product_css}%
\eczhListValue{\flmRefsHyperref{code:qubit_generalized_homological_product_css}{Generalized homological-product qubit CSS code}} & \eczhListValue{A qubit CSS code whose properties are determined from an underlying chain complex via the \flmRefsHyperref{ref683}{qubit CSS-to-homology correspondence}. This complex often consists of some type of product of other chain complexes.}\\ 
\addlinespace[\myxtraspc]
\eczhRefIndex{code:generalized_quantum_tanner}%
\eczhListValue{\flmRefsHyperref{code:generalized_quantum_tanner}{Generalized quantum Tanner code}} & \eczhListValue{An extension of quantum Tanner codes to codes constructed from two commuting regular graphs with the same vertex set.
This allows for code construction using finite sets and Schreier graphs, yielding a broader family of square complexes.}\\ 
\addlinespace[\myxtraspc]
\eczhRefIndex{code:golden}%
\eczhListValue{\flmRefsHyperref{code:golden}{Golden code}} & \eczhListValue{Variant of the Guth-Lubotzky hyperbolic surface code that uses regular tessellations of 4-dimensional hyperbolic space.}\\ 
\addlinespace[\myxtraspc]
\eczhRefIndex{code:four_dimensional_hyperbolic}%
\eczhListValue{\flmRefsHyperref{code:four_dimensional_hyperbolic}{Guth-Lubotzky code}} & \eczhListValue{Homological linear-rate code based on cellulations of certain 4D hyperbolic manifolds with particular homology and systolic properties.}\\ 
\addlinespace[\myxtraspc]
\eczhRefIndex{code:hemicubic}%
\eczhListValue{\flmRefsHyperref{code:hemicubic}{Hemicubic code}} & \eczhListValue{Homological code constructed out of cubes in high dimensions.
The hemicubic code family has asymptotically diminishing soundness that scales as \flmRefsHyperref{ref65}{order} \(\Omega(1/\log n)\), locality of stabilizer generators scaling as \flmRefsHyperref{ref65}{order} \(O(\log n)\), and distance of \flmRefsHyperref{ref65}{order} \(\Theta(\sqrt{n})\).}\\ 
\addlinespace[\myxtraspc]
\eczhRefIndex{code:ramanujan_tensor_product}%
\eczhListValue{\flmRefsHyperref{code:ramanujan_tensor_product}{High-dimensional expander (HDX) code}} & \eczhListValue{CSS code obtained by applying the generalized distance-balancing/product construction of Ref. \NoCaseChange{\protect\cite{cite684}} to a Ramanujan-complex quantum code and an asymptotically good classical LDPC code.}\\ 
\addlinespace[\myxtraspc]
\eczhRefIndex{code:multisector_hypergraph}%
\eczhListValue{\flmRefsHyperref{code:multisector_hypergraph}{Higher-dimensional homological product code}} & \eczhListValue{A qubit CSS code formulated using a tensor product of two or more chain complexes, each of length one or greater.
The number of chain complexes participating in the product is the \textit{dimension} of the code.
When all chain complexes are length-one, meaning that they correspond to classical codes, the code is called a \textit{higher-dimensional HGP code} (a.k.a. multi-sector HGP code or iterative HGP code).}\\ 
\addlinespace[\myxtraspc]
\eczhRefIndex{code:higher_dimensional_surface}%
\eczhListValue{\flmRefsHyperref{code:higher_dimensional_surface}{Homological code}} & \eczhListValue{A CSS extension of the Kitaev surface code to arbitrary manifolds.
The version on a Euclidean manifold of some fixed dimension is called the \(D\)\textit{-dimensional "surface"} or \(D\)\textit{-dimensional toric} code.}\\ 
\addlinespace[\myxtraspc]
\eczhRefIndex{code:homological_product}%
\eczhListValue{\flmRefsHyperref{code:homological_product}{Homological product code}} & \eczhListValue{CSS code formulated using the tensor product of two chain complexes of length one or greater (see \flmRefsCref{ref683}).}\\ 
\addlinespace[\myxtraspc]
\eczhRefIndex{code:triangular_color}%
\eczhListValue{\flmRefsHyperref{code:triangular_color}{Honeycomb (6.6.6) color code}} & \eczhListValue{2D color code defined on a (typically triangular) patch of the 6.6.6 (honeycomb) tiling.
The usual triangular patch has three differently colored boundaries, encodes one logical qubit, and is local-Clifford equivalent to a folded surface/toric code with two smooth and two rough boundaries \NoCaseChange{\protect\cite{cite422}}.}\\ 
\addlinespace[\myxtraspc]
\eczhRefIndex{code:hurwitz_surface}%
\eczhListValue{\flmRefsHyperref{code:hurwitz_surface}{Hurwitz surface code}} & \eczhListValue{Homological code constructed on triangulations of Hurwitz surfaces.}\\ 
\addlinespace[\myxtraspc]
\eczhRefIndex{code:hyperbolic_surface}%
\eczhListValue{\flmRefsHyperref{code:hyperbolic_surface}{Hyperbolic surface code}} & \eczhListValue{An extension of the Kitaev surface code construction to hyperbolic manifolds.
Given a cellulation of a hyperbolic manifold of arbitrary dimension, qubits are put on \(i\)-dimensional faces, \(X\)-type stabilizers are associated with \((i-1)\)-faces, while \(Z\)-type stabilizers are associated with \(i+1\)-faces.}\\ 
\addlinespace[\myxtraspc]
\eczhRefIndex{code:hypergraph_product}%
\eczhListValue{\flmRefsHyperref{code:hypergraph_product}{Hypergraph product (HGP) code}} & \eczhListValue{A member of a family of CSS codes whose stabilizer generator matrix is obtained from a hypergraph product of two classical linear binary codes.}\\ 
\addlinespace[\myxtraspc]
\eczhRefIndex{code:hypersphere_product}%
\eczhListValue{\flmRefsHyperref{code:hypersphere_product}{Hypersphere product code}} & \eczhListValue{Homological code based on products of hyperspheres.
The hypersphere product code family has asymptotically diminishing soundness that scales as \flmRefsHyperref{ref65}{order} \(O(1/\log (n)^2)\), locality of stabilizer generators scaling as \flmRefsHyperref{ref65}{order} \(O(\log n/ \log\log n)\), and distance of \flmRefsHyperref{ref65}{order} \(\Theta(\sqrt{n})\).}\\ 
\addlinespace[\myxtraspc]
\eczhRefIndex{code:homological_cv}%
\eczhListValue{\flmRefsHyperref{code:homological_cv}{Integer-homology bosonic CSS code}} & \eczhListValue{A bosonic stabilizer code whose physical modes have been restricted, via a single GKP stabilizer, from the space of functions on the real line to the space of periodic functions.
This restriction effectively realizes a rotor on each physical mode, allowing one to construct homological rotor codes out of displacement stabilizer groups.
The stabilizer group is continuous, but contains discrete components in the form of the single-mode GKP stabilizers.
The homology group of the logical operators has a torsion component because the chain complexes are defined over the ring of integers, which yields codes with finite logical dimension.}\\ 
\addlinespace[\myxtraspc]
\eczhRefIndex{code:surface}%
\eczhListValue{\flmRefsHyperref{code:surface}{Kitaev surface code}} & \eczhListValue{A family of Abelian topological \flmRefsHyperref{code:css}{CSS stabilizer} codes
whose generators are few-body \(X\)-type and \(Z\)-type Pauli strings
associated to the stars and plaquettes, respectively, of a cellulation of a
two-dimensional surface (with a qubit located at each edge of the
cellulation).
Codewords correspond to ground states of the surface code Hamiltonian, and error operators create or annihilate pairs of anyonic charges or vortices.}\\ 
\addlinespace[\myxtraspc]
\eczhRefIndex{code:klein_bottle}%
\eczhListValue{\flmRefsHyperref{code:klein_bottle}{Klein-bottle surface code}} & \eczhListValue{A family of Kitaev surface codes on the non-orientable Klein bottle.}\\ 
\addlinespace[\myxtraspc]
\eczhRefIndex{code:lacross}%
\eczhListValue{\flmRefsHyperref{code:lacross}{La-cross code}} & \eczhListValue{Code constructed using a hypergraph product of two copies of a cyclic LDPC code.
The construction uses cyclic LDPC codes with \flmRefsHyperref{ref67}{generating polynomials} \(1+x+x^k\) for some \(k\).
Using a length-\(n\) seed code yields an \(\llbracket 2n^2,2k^2\rrbracket \) family for periodic boundary conditions and an \(\llbracket (n-k)^2+n^2,k^2\rrbracket \) family for open boundary conditions.}\\ 
\addlinespace[\myxtraspc]
\eczhRefIndex{code:lcs}%
\eczhListValue{\flmRefsHyperref{code:lcs}{Lift-connected surface (LCS) code}} & \eczhListValue{Member of one of several families of lifted-product codes that consist of sparsely interconnected stacks of surface codes.}\\ 
\addlinespace[\myxtraspc]
\eczhRefIndex{code:lifted_product}%
\eczhListValue{\flmRefsHyperref{code:lifted_product}{Lifted-product (LP) code}} & \eczhListValue{Galois-qudit code that utilizes the notion of a lifted product in its construction. Lifted products of certain classical Tanner codes are the first (asymptotically) \textit{good QLDPC codes}.}\\ 
\addlinespace[\myxtraspc]
\eczhRefIndex{code:lresc}%
\eczhListValue{\flmRefsHyperref{code:lresc}{Long-range enhanced surface code (LRESC)}} & \eczhListValue{Code constructed using a hypergraph product of two copies of a concatenated LDPC-repetition seed code.
This family interpolates between surface codes and hypergraph codes since the hypergraph product of two repetition codes yields the planar surface code.
The construction uses small \([3,2,2]\) and \([6,2,4]\) LDPC codes concatenated with \([4,1,4]\) and \([2,1,2]\) repetition codes, respectively.
An example using a \([5,2,3]\) code is also presented.}\\ 
\addlinespace[\myxtraspc]
\eczhRefIndex{code:lossless_expander}%
\eczhListValue{\flmRefsHyperref{code:lossless_expander}{Lossless expander balanced-product code}} & \eczhListValue{QLDPC code constructed by taking the balanced product of lossless expander graphs.
Using one part of a quantum-code chain complex constructed with one-sided loss expanders \NoCaseChange{\protect\cite{cite185}} yields a \(c^3\)-LTC \NoCaseChange{\protect\cite{cite186}}.
Using two-sided expanders \NoCaseChange{\protect\cite{cite187}} yields an asymptotically good QLDPC code family \NoCaseChange{\protect\cite{cite188}}.}\\ 
\addlinespace[\myxtraspc]
\eczhRefIndex{code:qudit_3d_surface}%
\eczhListValue{\flmRefsHyperref{code:qudit_3d_surface}{Modular-qudit 3D surface code}} & \eczhListValue{A generalization of the 3D surface code to modular qudits.
Qudits are placed on edges, \(Z\)-type stabilizer generators are placed on square plaquettes oriented in all three directions, and \(X\)-type stabilizers are placed on the six edges neighboring every vertex \NoCaseChange{\protect\cite{cite459}}.}\\ 
\addlinespace[\myxtraspc]
\eczhRefIndex{code:qudit_color}%
\eczhListValue{\flmRefsHyperref{code:qudit_color}{Modular-qudit lattice color code}} & \eczhListValue{Extension of the color code to lattices of modular qudits.
Codes are defined analogously to qubit color codes on suitable lattices of any spatial dimension, but a directionality is required in order to make the modular-qudit stabilizers commute.
This can be done by puncturing a hyperspherical lattice \NoCaseChange{\protect\cite{cite475}} or constructing a star-bipartition; see \NoCaseChange{\protect\cite[{Sec. III}]{cite673}}.
Logical dimension is determined by the genus of the underlying surface (for closed surfaces), types of boundaries (for open surfaces), and/or any twist defects present.}\\ 
\addlinespace[\myxtraspc]
\eczhRefIndex{code:qudit_surface}%
\eczhListValue{\flmRefsHyperref{code:qudit_surface}{Modular-qudit surface code}} & \eczhListValue{Extension of the surface code to prime-dimensional \NoCaseChange{\protect\cite{cite423,cite424}} and more general modular qudits.
Stabilizer generators are few-body \(X\)-type and \(Z\)-type Pauli strings associated to the stars and plaquettes, respectively, of a tessellation of a two-dimensional surface.
Since qudits have more than one \(X\) and \(Z\)-type operator, various sets of stabilizer generators can be defined.}\\ 
\addlinespace[\myxtraspc]
\eczhRefIndex{code:real_projective_plane}%
\eczhListValue{\flmRefsHyperref{code:real_projective_plane}{Projective-plane surface code}} & \eczhListValue{A family of Kitaev surface codes on the non-orientable 2-dimensional compact manifold \(\mathbb{R}P^2\) (in contrast to a genus-\(g\) surface).
Whereas genus-\(g\) surface codes require \(2g\) logical qubits, qubit codes on \(\mathbb{R}P^2\) are made from a single logical qubit.}\\ 
\addlinespace[\myxtraspc]
\eczhRefIndex{code:quantum_tanner}%
\eczhListValue{\flmRefsHyperref{code:quantum_tanner}{Quantum Tanner code}} & \eczhListValue{Member of a family of QLDPC codes based on two compatible classical Tanner codes defined on a two-dimensional Cayley complex, a complex constructed from Cayley graphs of groups.
For certain choices of codes and complex, the resulting codes have asymptotically good parameters.
See Ref. \NoCaseChange{\protect\cite{cite685}} for explicit instances based on dihedral groups.
This construction has been generalized to Schreier graphs \NoCaseChange{\protect\cite{cite686}}.}\\ 
\addlinespace[\myxtraspc]
\eczhRefIndex{code:quantum_expander}%
\eczhListValue{\flmRefsHyperref{code:quantum_expander}{Quantum expander code}} & \eczhListValue{CSS code constructed from a hypergraph product of bipartite expander graphs \NoCaseChange{\protect\cite{cite74}} with bounded left and right vertex degrees. For every bipartite graph there is an associated matrix (the parity check matrix) with columns indexed by the left vertices, rows indexed by the right vertices, and 1 entries whenever a left and right vertex are connected. This matrix can serve as the parity check matrix of a classical code. Two bipartite expander graphs can be used to construct a quantum CSS code (the quantum expander code) via the hypergraph product of their parity check matrices.}\\ 
\addlinespace[\myxtraspc]
\eczhRefIndex{code:rotated_surface}%
\eczhListValue{\flmRefsHyperref{code:rotated_surface}{Rotated surface code}} & \eczhListValue{Variant of the surface code obtained by taking the medial graph of the surface code lattice (treated as a graph) and applying a procedure to construct the check operators \NoCaseChange{\protect\cite{cite425,cite426}\protect\cite[{Fig. 8}]{cite427}}.
On a square lattice, this amounts to a rotation by 45 degrees such that qubits are on vertices, and both \(X\)- and \(Z\)-type check operators occupy plaquettes in an alternating checkerboard pattern.}\\ 
\addlinespace[\myxtraspc]
\eczhRefIndex{code:sierpinsky_fractal_liquid}%
\eczhListValue{\flmRefsHyperref{code:sierpinsky_fractal_liquid}{Sierpinski prism model code}} & \eczhListValue{A fractal type-I fracton CSS code defined on a cubic lattice \NoCaseChange{\protect\cite[{Eq. (D22)}]{cite456}}.
The code admits an excitation-moving operator shaped like a Sierpinski triangle \NoCaseChange{\protect\cite[{Fig. 2}]{cite456}}.}\\ 
\addlinespace[\myxtraspc]
\eczhRefIndex{code:square_homological_product}%
\eczhListValue{\flmRefsHyperref{code:square_homological_product}{Square homological product code}} & \eczhListValue{Homological product code whose underlying quantum-code boundary operators are square matrices (see \flmRefsCref{ref683}).}\\ 
\addlinespace[\myxtraspc]
\eczhRefIndex{code:488_color}%
\eczhListValue{\flmRefsHyperref{code:488_color}{Square-octagon (4.8.8) color code}} & \eczhListValue{2D color code defined on a patch of the 4.8.8 (square-octagon) tiling, which itself is obtained by applying a fattening procedure to the square lattice \NoCaseChange{\protect\cite{cite430}}.
An equivalent description uses the Tetrakis square tiling (a.k.a. the Union Jack lattice), which is dual to the 4.8.8 lattice \NoCaseChange{\protect\cite{cite431}}.
Among the three semiregular triangular 2D color-code families, the 4.8.8 family uses the fewest physical qubits for a given distance and is the only one of the three with transversal implementations of the full Clifford group \NoCaseChange{\protect\cite{cite432}}.}\\ 
\addlinespace[\myxtraspc]
\eczhRefIndex{code:iterated_ramanujan}%
\eczhListValue{\flmRefsHyperref{code:iterated_ramanujan}{Tensor-product HDX code}} & \eczhListValue{A code constructed in a similar way as the HDX code, but utilizing iterated homological products of \textit{multiple} Ramanujan complexes and then applying distance balancing. For any fixed tensor-power parameter \(c\), these yield explicit QLDPC codes with distance scaling as \(\sqrt{n}\log^{c} n\), improving on the original HDX construction by replacing a single logarithmic enhancement with arbitrarily high fixed polylogarithmic enhancement. The utility of such tensor products comes from the fact that one of the Ramanujan complexes is a \textit{collective cosystolic expander} as opposed to just a cosystolic expander.}\\ 
\addlinespace[\myxtraspc]
\eczhRefIndex{code:tetrahedral_color}%
\eczhListValue{\flmRefsHyperref{code:tetrahedral_color}{Tetrahedral color code}} & \eczhListValue{A 3D color code defined on a colored tetrahedron cut from a suitably colored BCC lattice \NoCaseChange{\protect\cite{cite475}}.
Qubits are placed on tetrahedra, on the triangles covering the tetrahedron faces, on the edges along the tetrahedron edges, and on the tetrahedron vertices.
The code has both string-like and sheet-like logical operators \NoCaseChange{\protect\cite{cite476}}.}\\ 
\addlinespace[\myxtraspc]
\eczhRefIndex{code:tillichzemor}%
\eczhListValue{\flmRefsHyperref{code:tillichzemor}{Tillich-Zémor code}} & \eczhListValue{A family of \(\llbracket n^2 + m^2, (n - \text{rank}([C \mid M]) )^2 + (m - \text{rank}([C \mid M]^\top) )^2, d\rrbracket \) quantum LDPC codes constructed via the hypergraph product of two classical \((n, m, r)\)-structured LDPC seed codes}\\ 
\addlinespace[\myxtraspc]
\eczhRefIndex{code:toric}%
\eczhListValue{\flmRefsHyperref{code:toric}{Toric code}} & \eczhListValue{Version of the Kitaev surface code on a square lattice with periodic boundary conditions, encoding two logical qubits.
Being the first manifestation of the surface code, "toric code" is often an alternative name for the general construction.
\textit{Twisted toric code} \NoCaseChange{\protect\cite[{Fig. 8}]{cite434}} refers to the construction on a torus with twisted (a.k.a. shifted) boundary conditions.
In the original Hamiltonian construction, open Pauli-\(Z\) and Pauli-\(X\) strings create pairs of electric charges and magnetic vortices, and braiding one type around the other yields the nontrivial Abelian anyonic phase \NoCaseChange{\protect\cite{cite423}}.}\\ 
\addlinespace[\myxtraspc]
\eczhRefIndex{code:4612_color}%
\eczhListValue{\flmRefsHyperref{code:4612_color}{Truncated trihexagonal (4.6.12) color code}} & \eczhListValue{2D color code defined on a (typically triangular) patch of the 4.6.12 (truncated trihexagonal or square-hexagon-dodecagon) tiling.}\\ 
\addlinespace[\myxtraspc]
\eczhRefIndex{code:2bga}%
\eczhListValue{\flmRefsHyperref{code:2bga}{Two-block group-algebra (2BGA) codes}} & \eczhListValue{2BGA codes are the one-by-one, or smallest, \flmRefsHyperref{code:lifted_product}{LP codes}:
\(LP(a,b)\) is defined by a pair of \flmRefsHyperref{ref205}{group algebra} elements
\(a,b\in \mathbb{F}_q[G]\), where \(G\) is a finite group.
If \(|G|=\ell\), then the code has length \(n=2\ell\).}\\ 
\addlinespace[\myxtraspc]
\eczhRefIndex{code:two_foliated}%
\eczhListValue{\flmRefsHyperref{code:two_foliated}{Two-foliated fracton code}} & \eczhListValue{A type-I fracton code obtained by gauging \NoCaseChange{\protect\cite{cite462,cite463,cite233,cite464,cite465,cite466,cite467,cite468,cite469,cite470}} a 3D paramagnet with planar subsystem symmetries in two directions.
In that construction, the gauge charges are lineons and the flux excitations are also lineons moving in the same direction, yielding the anisotropic lineon model \NoCaseChange{\protect\cite[{Sec. 4.1.2}]{cite467}}.}\\ 
\addlinespace[\myxtraspc]
\eczhRefIndex{code:xysurface}%
\eczhListValue{\flmRefsHyperref{code:xysurface}{XY surface code}} & \eczhListValue{A variant of the surface code whose generators are \(XXXX\)  and \(YYYY\), obtained by mapping \(Z \to Y\) in the surface code.
While this code is equivalent to a CSS surface code with the same distance, other properties like noise-bias performance can differ significantly.}\\ 
\addlinespace[\myxtraspc]
\eczhRefIndex{code:xyz_product}%
\eczhListValue{\flmRefsHyperref{code:xyz_product}{XYZ product code}} & \eczhListValue{A non-CSS QLDPC code obtained from a three-fold variant of the hypergraph product applied to three classical binary codes with parity-check matrices \(H_1,H_2,H_3\).  Unlike CSS three-fold hypergraph product codes, the third input code acts through Pauli-\(Y\) checks \NoCaseChange{\protect\cite{cite645}}.
Under mild assumptions, the code dimension is determined by a tensor Sylvester equation over \(\mathbb{F}_2\), and the minimum-distance problem reduces up to constant factors to how closely a related inhomogeneous tensor Sylvester equation can be satisfied \NoCaseChange{\protect\cite{cite645}}.  When the underlying classical codes are repetition codes, the construction yields the \flmRefsHyperref{code:chamon}{Chamon model code}.}\\ 
\addlinespace[\myxtraspc]
\eczhRefIndex{code:4d_13_surface}%
\eczhListValue{\flmRefsHyperref{code:4d_13_surface}{\((1,3)\) 4D toric code}} & \eczhListValue{A generalization of the Kitaev surface code defined on a 4D lattice.
The code is called a \((1,3)\) toric code because it admits 1D \(Z\)-type and 3D \(X\)-type logical operators.
In the hypercubic lattice version, qubits are placed on edges, each \(Z\)-type stabilizer generator is supported on cubes on the boundary of a hypercube, and \(X\)-type stabilizers are placed on the edges neighboring every vertex \NoCaseChange{\protect\cite{cite479}}.}\\ 
\addlinespace[\myxtraspc]
\eczhRefIndex{code:4d_surface}%
\eczhListValue{\flmRefsHyperref{code:4d_surface}{\((2,2)\) Loop toric code}} & \eczhListValue{A generalization of the Kitaev surface code defined on a 4D lattice.
The code is called a \((2,2)\) toric code because it admits 2D membrane \(Z\)-type and \(X\)-type logical operators.
Both types of operators create 1D (i.e., loop) excitations at their edges.
The code serves as a self-correcting quantum memory \NoCaseChange{\protect\cite{cite480,cite481}}.}\\ 
\addlinespace[\myxtraspc]
\eczhRefIndex{code:higher_dimensional_toric}%
\eczhListValue{\flmRefsHyperref{code:higher_dimensional_toric}{\(D\)-dimensional twisted toric code}} & \eczhListValue{Extension of the Kitaev toric code to higher-dimensional lattices with regular or shifted (a.k.a. twisted) boundary conditions.
Such boundary conditions yield qubit geometries that are tori \(\mathbb{R}^D/\Lambda\), where \(\Lambda\) is an arbitrary \(D\)-dimensional lattice.
Picking a hypercubic lattice yields the ordinary \(D\)-dimensional toric code.}\\ 
\addlinespace[\myxtraspc]
\eczhRefIndex{code:xzzx_10_2_3}%
\eczhListValue{\flmRefsHyperref{code:xzzx_10_2_3}{\(\llbracket 10,2,3\rrbracket \) rotated toric code}} & \eczhListValue{Rotated toric code that is the CSS form of the twisted XZZX toric code with parameters \(a=1\), \(b=3\) \NoCaseChange{\protect\cite{cite438,cite427}},
related to the XZZX form by Hadamard on the \(B\)-sublattice.
It is also the \flmRefsHyperref{ref436}{symplectic double} (a.k.a. genus-one double cover) of the \(\llbracket 5,1,3\rrbracket \) five-qubit perfect code \NoCaseChange{\protect\cite{cite439,cite435}}, the \flmRefsHyperref{ref436}{symplectic double} of the \(\llbracket 5,1,2\rrbracket \) rotated surface code \NoCaseChange{\protect\cite{cite435}}, and a BCC code \NoCaseChange{\protect\cite{cite440}}.}\\ 
\addlinespace[\myxtraspc]
\eczhRefIndex{code:bb108}%
\eczhListValue{\flmRefsHyperref{code:bb108}{\(\llbracket 108,8,10\rrbracket \) BB6 code}} & \eczhListValue{A bivariate bicycle (BB) code with parameters \(\llbracket 108,8,10\rrbracket \) and weight-six stabilizer generators \NoCaseChange{\protect\cite{cite441}}.}\\ 
\addlinespace[\myxtraspc]
\eczhRefIndex{code:gross}%
\eczhListValue{\flmRefsHyperref{code:gross}{\(\llbracket 144,12,12\rrbracket \) gross code}} & \eczhListValue{A BB code which requires less physical and ancilla qubits (for syndrome extraction) than the surface code with the same number of logical qubits and distance.
The gross code is equivalent to 8 copies of the surface code via a constant-depth Clifford circuit, and is an element of a larger family of 2D stabilizer codes \NoCaseChange{\protect\cite{cite443}}.
The name stems from the fact that a gross is a dozen dozen.}\\ 
\addlinespace[\myxtraspc]
\eczhRefIndex{code:stab_15_1_3}%
\eczhListValue{\flmRefsHyperref{code:stab_15_1_3}{\(\llbracket 15,1,3\rrbracket \) quantum RM code}} & \eczhListValue{A \(\llbracket 15,1,3\rrbracket \) quantum Reed-Muller code that is most easily thought of as a tetrahedral 3D color code.
It can be constructed as a CSS code from the \([15,5,8]\) punctured Reed-Muller code and its even subcode, which explains its transversal \(T^\dagger\) gate \NoCaseChange{\protect\cite{cite398}}.}\\ 
\addlinespace[\myxtraspc]
\eczhRefIndex{code:stab_17_1_5}%
\eczhListValue{\flmRefsHyperref{code:stab_17_1_5}{\(\llbracket 17,1,5\rrbracket \) 4.8.8 color code}} & \eczhListValue{Seventeen-qubit doubly even 2D color code that admits a transversal implementation of the logical Clifford group.
The smallest distance-five CSS code has \(n=17\) \NoCaseChange{\protect\cite{cite444}}.
It is also a normal self-dual CSS code whose transversal Hadamard acts logically, making it suitable as an inner code for fifth-order magic-state distillation \NoCaseChange{\protect\cite{cite101}}.}\\ 
\addlinespace[\myxtraspc]
\eczhRefIndex{code:stab_18_2_5}%
\eczhListValue{\flmRefsHyperref{code:stab_18_2_5}{\(\llbracket 18,2,5\rrbracket \) BCC code}} & \eczhListValue{BCC code on 18 qubits encoding 2 logical qubits with distance 5, found by computer search \NoCaseChange{\protect\cite{cite440}}.}\\ 
\addlinespace[\myxtraspc]
\eczhRefIndex{code:bb288}%
\eczhListValue{\flmRefsHyperref{code:bb288}{\(\llbracket 288,12,18\rrbracket \) double-gross code}} & \eczhListValue{A bivariate bicycle (BB) code with parameters \(\llbracket 288,12,18\rrbracket \) and weight-six stabilizer generators \NoCaseChange{\protect\cite{cite441}}.}\\ 
\addlinespace[\myxtraspc]
\eczhRefIndex{code:stellated_dodecahedron_css}%
\eczhListValue{\flmRefsHyperref{code:stellated_dodecahedron_css}{\(\llbracket 30,8,3\rrbracket \) Bring code}} & \eczhListValue{A \(\llbracket 30,8,3\rrbracket \) hyperbolic surface code on a quotient of the \(\{5,5\}\) hyperbolic tiling called Bring's curve.
Its qubits and stabilizer generators lie on the vertices of the small stellated dodecahedron. It admits a set of weight-five stabilizer generators.}\\ 
\addlinespace[\myxtraspc]
\eczhRefIndex{code:css_4_1_2}%
\eczhListValue{\flmRefsHyperref{code:css_4_1_2}{\(\llbracket 4,1,2\rrbracket \) Leung-Nielsen-Chuang-Yamamoto (LNCY) code}} & \eczhListValue{A four-qubit CSS stabilizer code that is the only qubit CSS code with such parameters.}\\ 
\addlinespace[\myxtraspc]
\eczhRefIndex{code:stab_4_2_2}%
\eczhListValue{\flmRefsHyperref{code:stab_4_2_2}{\(\llbracket 4,2,2\rrbracket \) Four-qubit code}} & \eczhListValue{A four-qubit hyperbolic self-dual CSS stabilizer code that is the smallest two-logical-qubit stabilizer code to detect a single-qubit error.
It is unique for its parameters \NoCaseChange{\protect\cite[{Thm. 8}]{cite446}}.}\\ 
\addlinespace[\myxtraspc]
\eczhRefIndex{code:stab_5_1_2}%
\eczhListValue{\flmRefsHyperref{code:stab_5_1_2}{\(\llbracket 5,1,2\rrbracket \) rotated surface code}} & \eczhListValue{A rotated surface code on one rung of a ladder, with one qubit on the rung, and four qubits surrounding it. This is the smallest code that implements a fault-tolerant logical \(S\) gate using a diagonal depth-one Clifford circuit \NoCaseChange{\protect\cite{cite447}}.}\\ 
\addlinespace[\myxtraspc]
\eczhRefIndex{code:stab_6_2_2}%
\eczhListValue{\flmRefsHyperref{code:stab_6_2_2}{\(\llbracket 6,2,2\rrbracket \) \(C_6\) code}} & \eczhListValue{Error-detecting normal self-dual CSS code on three qubit pairs that encodes a logical qubit pair and detects any error acting on one pair \NoCaseChange{\protect\cite{cite448}}.
In Knill's \(C_4/C_6\) architecture, this code is used at the second and higher concatenation levels.}\\ 
\addlinespace[\myxtraspc]
\eczhRefIndex{code:stab_6_4_2}%
\eczhListValue{\flmRefsHyperref{code:stab_6_4_2}{\(\llbracket 6,4,2\rrbracket \) error-detecting code}} & \eczhListValue{Self-complementary six-qubit code with rate \(2/3\) that is unique for its parameters, up to equivalence \NoCaseChange{\protect\cite[{Tab. III}]{cite449}}.
Concatenations of this code with itself yield the \(\llbracket 6^r,4^r,2^r\rrbracket \) level-\(r\) \textit{many-hypercube} code \NoCaseChange{\protect\cite{cite450}}.}\\ 
\addlinespace[\myxtraspc]
\eczhRefIndex{code:steane}%
\eczhListValue{\flmRefsHyperref{code:steane}{\(\llbracket 7,1,3\rrbracket \) Steane code}} & \eczhListValue{A \(\llbracket 7,1,3\rrbracket \) self-dual CSS code that is the smallest qubit CSS code to correct a single-qubit error \NoCaseChange{\protect\cite{cite451}}.
The code is constructed using the classical binary \([7,4,3]\) Hamming code for protecting against both \(X\) and \(Z\) errors.}\\ 
\addlinespace[\myxtraspc]
\eczhRefIndex{code:hgp_7_2_2}%
\eczhListValue{\flmRefsHyperref{code:hgp_7_2_2}{\(\llbracket 7,2,2\rrbracket \) HGP phantom code}} & \eczhListValue{Smallest member of a family of CSS phantom HGP codes obtained from the hypergraph product of a classical simplex code and a repetition code.}\\ 
\addlinespace[\myxtraspc]
\eczhRefIndex{code:bb72}%
\eczhListValue{\flmRefsHyperref{code:bb72}{\(\llbracket 72,12,6\rrbracket \) BB6 code}} & \eczhListValue{A bivariate bicycle (BB) code with parameters \(\llbracket 72,12,6\rrbracket \) and weight-six stabilizer generators \NoCaseChange{\protect\cite{cite441}}.}\\ 
\addlinespace[\myxtraspc]
\eczhRefIndex{code:stab_8_3_2}%
\eczhListValue{\flmRefsHyperref{code:stab_8_3_2}{\(\llbracket 8,3,2\rrbracket \) Smallest interesting color code}} & \eczhListValue{Smallest 3D color code whose physical qubits lie on vertices of a cube and which admits a (weakly) transversal \(CCZ\) gate.}\\ 
\addlinespace[\myxtraspc]
\eczhRefIndex{code:shor_nine}%
\eczhListValue{\flmRefsHyperref{code:shor_nine}{\(\llbracket 9,1,3\rrbracket \) Shor code}} & \eczhListValue{Nine-qubit \flmRefsHyperref{code:css}{CSS code} that is the first quantum error-correcting code \NoCaseChange{\protect\cite[{ID 8802}]{cite453}}.
Among indecomposable \(\llbracket 9,1,3\rrbracket \) CSS codes, the Shor code has the largest automorphism group \NoCaseChange{\protect\cite{cite454}}.}\\ 
\addlinespace[\myxtraspc]
\eczhRefIndex{code:surface-17}%
\eczhListValue{\flmRefsHyperref{code:surface-17}{\(\llbracket 9,1,3\rrbracket \) Surface-17 code}} & \eczhListValue{A \(\llbracket 9,1,3\rrbracket \) rotated surface code named for the sum of its 9 data qubits and 8 syndrome qubits.
It is one of the four inequivalent CSS gauge fixings of the nine-qubit Bacon-Shor code \NoCaseChange{\protect\cite{cite454}}.
It uses the smallest number of qubits to perform fault-tolerant error correction on a surface code with parallel syndrome extraction.}\\ 
\addlinespace[\myxtraspc]
\eczhRefIndex{code:bb90}%
\eczhListValue{\flmRefsHyperref{code:bb90}{\(\llbracket 90,8,10\rrbracket \) BB6 code}} & \eczhListValue{A bivariate bicycle (BB) code with parameters \(\llbracket 90,8,10\rrbracket \) and weight-six stabilizer generators \NoCaseChange{\protect\cite{cite441}}.}\\ 
\end{tabularx}\endgroup
\eczcodelist{quantum_magic}{Quantum codes with magic-state yield parameters
}%

\eczhCodeListAutoDescription{All codes with \emph{Magic} that are not descendants of \flmRefsCref{code:ecc}.}%

\eczhIncludeCodeGraph{Bare}{scale=0.5}{\columnwidth}{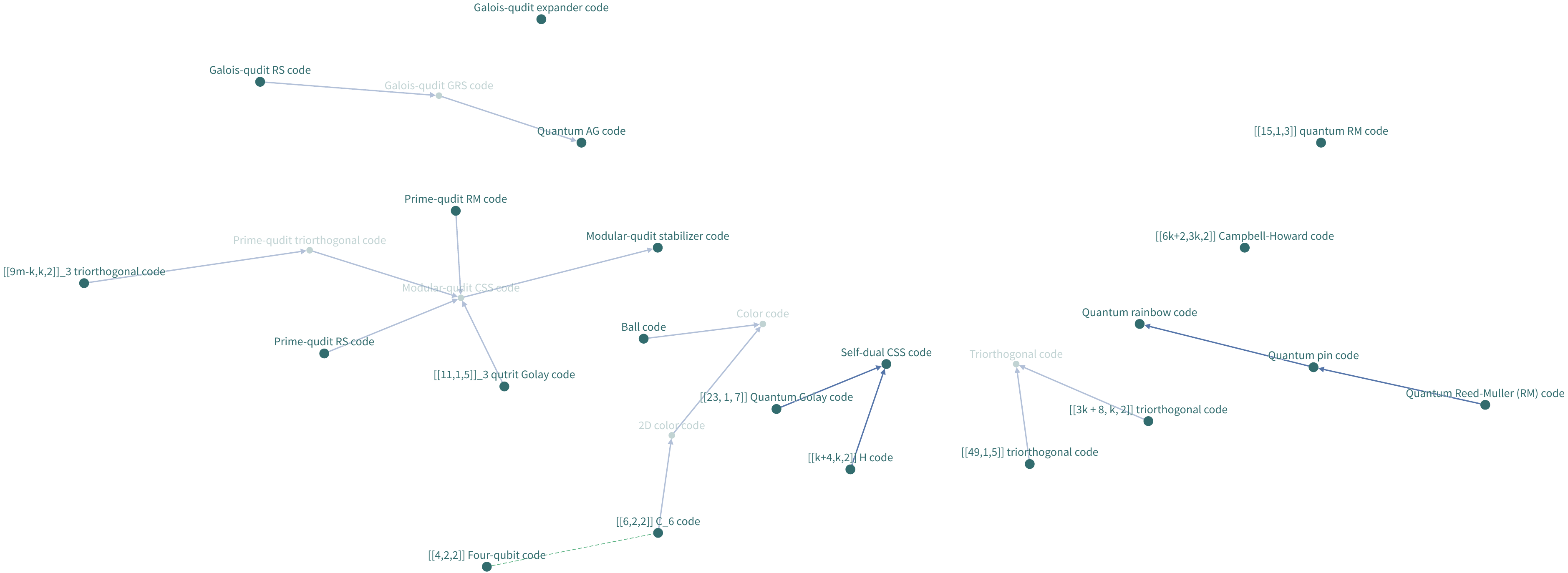}{Quantum codes with magic-state yield parameters}{https://errorcorrectionzoo.org/code_graph#J\%7B\%22displayMode\%22\%3A\%22subset\%22\%2C\%22modeSubsetOptions\%22\%3A\%7B\%22codeIds\%22\%3A\%5B\%22ball_color\%22\%2C\%22galois_polynomial\%22\%2C\%22galois_expander\%22\%2C\%22qudit_stabilizer\%22\%2C\%22qudit_reed_muller\%22\%2C\%22polynomial\%22\%2C\%22quantum_ag\%22\%2C\%22quantum_reed_muller\%22\%2C\%22quantum_pin\%22\%2C\%22quantum_rainbow\%22\%2C\%22self_dual_css\%22\%2C\%22qutrit_golay\%22\%2C\%22stab_15_1_3\%22\%2C\%22qubit_golay\%22\%2C\%22small_triorthogonal\%22\%2C\%22stab_4_2_2\%22\%2C\%22stab_49_1_5\%22\%2C\%22stab_6_2_2\%22\%2C\%22campbell_howard\%22\%2C\%22qutrit_small_triorthogonal\%22\%2C\%22quantum_h\%22\%5D\%2C\%22reusePreviousLayoutPositions\%22\%3Afalse\%2C\%22showIntermediateConnectingNodes\%22\%3Atrue\%2C\%22connectingNodesMaxDepth\%22\%3A15\%2C\%22connectingNodesPathMaxLength\%22\%3A20\%2C\%22connectingNodesMaxExtraDepth\%22\%3A3\%2C\%22connectingNodesOnlyKeepPathsWithAdditionalLength\%22\%3A1\%2C\%22connectingNodesToDomainsAndKingdoms\%22\%3Afalse\%2C\%22connectingNodesEdgeLengthsByType\%22\%3A\%7B\%22primaryParent\%22\%3A1\%2C\%22secondaryParent\%22\%3A4\%2C\%22cousin\%22\%3A6\%7D\%2C\%22nodeIds\%22\%3A\%5B\%5D\%7D\%2C\%22highlightImportantNodes\%22\%3A\%7B\%22highlightImportantNodes\%22\%3Afalse\%2C\%22highlightPrimaryParents\%22\%3Afalse\%2C\%22highlightRootConnectingEdges\%22\%3Afalse\%7D\%7D}

\begingroup
\small
\eczhBreakableDashes
\renewcommand\arraystretch{1.05}
\edef\myxtraspc{\eczListAddVSpaceXtraXtraStretch}
\begin{tabularx}{\linewidth}{>{\eczlistsmallsize\raggedright\arraybackslash}p{\eczListColWidth{name}} >{\hsize=1.0000\hsize \eczlistsmallsize}X}
\toprule
\eczListColTitle{Name} & \eczListColTitle{Magic-state yield parameter} \\
\midrule
\endfirsthead
\toprule
\eczListColTitleContinued{Name} & \eczListColTitleContinued{Magic-state yield parameter} \\
\midrule
\endhead
\bottomrule
\endfoot
\eczhRefIndex{code:ball_color}%
\eczhListValue{\flmRefsHyperref{code:ball_color}{Ball code}} & \eczhListValue{The 3D ball codes on duals of the truncated octahedron, truncated cuboctahedron, and truncated icosidodecahedron have \(\gamma\) close to one \NoCaseChange{\protect\cite{cite687}}.}\\ 
\addlinespace[\myxtraspc]
\eczhRefIndex{code:galois_polynomial}%
\eczhListValue{\flmRefsHyperref{code:galois_polynomial}{Galois-qudit RS code}} & \eczhListValue{Punctured RS codes can be used for magic-state distillation with a spacetime overhead of \((\log \frac{1}{\epsilon})^{\gamma}\), with the magic-state scaling exponent \(\gamma \to 0\) with decreasing \(\epsilon\) \NoCaseChange{\protect\cite{cite688}}.}\\ 
\addlinespace[\myxtraspc]
\eczhRefIndex{code:galois_expander}%
\eczhListValue{\flmRefsHyperref{code:galois_expander}{Galois-qudit expander code}} & \eczhListValue{For every integer \(r\geq 2\) and every \(\epsilon>0\), the construction yields \(\llbracket N,K\geq N^{1-\epsilon},D\geq N^{1/r}/\operatorname{poly}(\log N)\rrbracket _q\) QLDPC Galois-qudit quantum expander codes with transversal \(C^{r-1} Z\) gates and stabilizer weight \(w\leq\operatorname{poly}(\log N)\) \NoCaseChange{\protect\cite{cite689}}. This construction allows for arbitrarily small magic-state yield parameter \(\gamma\).}\\ 
\addlinespace[\myxtraspc]
\eczhRefIndex{code:qudit_stabilizer}%
\eczhListValue{\flmRefsHyperref{code:qudit_stabilizer}{Modular-qudit stabilizer code}} & \eczhListValue{The \textit{magic-state yield parameter} \(\gamma = \log_d(n/k)\) quantifies the overhead cost of magic-state distillation per the original protocol \NoCaseChange{\protect\cite{cite690,cite691}}.}\\ 
\addlinespace[\myxtraspc]
\eczhRefIndex{code:qudit_reed_muller}%
\eczhListValue{\flmRefsHyperref{code:qudit_reed_muller}{Prime-qudit RM code}} & \eczhListValue{An odd-prime-qudit CSS code family constructed from first-order punctured GRM codes can be used for qudit magic-state distillation; see \NoCaseChange{\protect\cite[{Table I}]{cite692}} for yields.}\\ 
\addlinespace[\myxtraspc]
\eczhRefIndex{code:polynomial}%
\eczhListValue{\flmRefsHyperref{code:polynomial}{Prime-qudit RS code}} & \eczhListValue{Triorthogonal \(p\)-dimensional prime-qudit RS codes achieve a magic-state yield parameter \(\gamma = O(1/\log p)\) \NoCaseChange{\protect\cite{cite693}}.}\\ 
\addlinespace[\myxtraspc]
\eczhRefIndex{code:quantum_ag}%
\eczhListValue{\flmRefsHyperref{code:quantum_ag}{Quantum AG code}} & \eczhListValue{By defining a generalization of triorthogonal matrices to Galois qudits of dimension \(q=2^m\), one can construct an asymptotically good family of quantum AG codes that admits a diagonal transversal gate at the third level of the \flmTerm{term}{ref694}{}{Clifford hierarchy} and attains a zero magic-state yield parameter, \(\gamma = 0\) \NoCaseChange{\protect\cite{cite695}}. This code can be treated as a qubit code by decomposing each Galois qudit into a Kronecker product of \(m\) qubits; see \NoCaseChange{\protect\cite{cite696,cite398,cite698,cite699,cite700}\protect\cite[{Sec. 5.3}]{cite697}}. Two other asymptotically good families \NoCaseChange{\protect\cite{cite699,cite698}} admit a transversal \(CCZ\) gate (a different diagonal gate at the third level of the \flmTerm{term}{ref694}{}{Clifford hierarchy}) and achieve \(\gamma \to 0\) with constant alphabet size.}\\ 
\addlinespace[\myxtraspc]
\eczhRefIndex{code:quantum_reed_muller}%
\eczhListValue{\flmRefsHyperref{code:quantum_reed_muller}{Quantum Reed-Muller (RM) code}} & \eczhListValue{The family constructed out of shortened RM codes with parameters \(\llbracket \sum_{i=w+1}^m \binom{m}{i}, \sum_{i=0}^{w} \binom{m}{i}, \sum_{i=w+1}^{r+1} \binom{r+1}{i}\rrbracket \) for integers \(m > 2r\) and \(r > w \geq 0\) yields protocols with an exponent of \(\gamma < 0.678\), with the fewest-resource protocol with \(\gamma < 1\) requiring a code with parameters \(\{r,w,m\} = \{19,14,3r+1\}\) such that \(n \approx 2^{58}\) qubits \NoCaseChange{\protect\cite[{Corr. 1}]{cite701}}. This refutes a conjecture that no protocol could achieve \(\gamma < 1\) \NoCaseChange{\protect\cite{cite691}}.}\\ 
\addlinespace[\myxtraspc]
\eczhRefIndex{code:quantum_pin}%
\eczhListValue{\flmRefsHyperref{code:quantum_pin}{Quantum pin code}} & \eczhListValue{A family of punctured pin codes admits \(\gamma \approx 1.6\) \NoCaseChange{\protect\cite[{Table VII}]{cite702}}.}\\ 
\addlinespace[\myxtraspc]
\eczhRefIndex{code:quantum_rainbow}%
\eczhListValue{\flmRefsHyperref{code:quantum_rainbow}{Quantum rainbow code}} & \eczhListValue{Hypergraph products of color codes yield quantum rainbow codes with growing distance and transversal gates in the \flmTerm{term}{ref694}{}{Clifford hierarchy}. In particular, utilizing this construction for quasi-hyperbolic color codes \NoCaseChange{\protect\cite{cite703}} yields an \(\llbracket n,O(n),O(\log n)\rrbracket \) triorthogonal code family satisfying the necessary conditions for the magic-state yield parameter \(\gamma\) to become arbitrarily small \NoCaseChange{\protect\cite{cite704}}.}\\ 
\addlinespace[\myxtraspc]
\eczhRefIndex{code:self_dual_css}%
\eczhListValue{\flmRefsHyperref{code:self_dual_css}{Self-dual CSS code}} & \eczhListValue{Normal and hyperbolic self-dual CSS codes yield magic-state distillation protocols with asymptotically constant space overhead and yield parameter \(\gamma \to 1^{+}\) \NoCaseChange{\protect\cite{cite705}\protect\cite[{Thms. 4.1 and 4.2}]{cite101}}.}\\ 
\addlinespace[\myxtraspc]
\eczhRefIndex{code:qutrit_golay}%
\eczhListValue{\flmRefsHyperref{code:qutrit_golay}{\(\llbracket 11,1,5\rrbracket _3\) qutrit Golay code}} & \eczhListValue{Magic-state distillation scaling exponent \(\gamma=\log_3(1728\times 11) \approx 8.97\), where the \(1728\) factor comes from the fact that one round of distillation succeeds with probability \(\approx 1/1728\) \NoCaseChange{\protect\cite{cite706}}.}\\ 
\addlinespace[\myxtraspc]
\eczhRefIndex{code:stab_15_1_3}%
\eczhListValue{\flmRefsHyperref{code:stab_15_1_3}{\(\llbracket 15,1,3\rrbracket \) quantum RM code}} & \eczhListValue{Magic-state yield parameter \( \gamma= \log_d (n/k)\approx 2.47\) \NoCaseChange{\protect\cite{cite101}\protect\cite[{Box 2}]{cite707}}.}\\ 
\addlinespace[\myxtraspc]
\eczhRefIndex{code:qubit_golay}%
\eczhListValue{\flmRefsHyperref{code:qubit_golay}{\(\llbracket 23, 1, 7\rrbracket \) Quantum Golay code}} & \eczhListValue{Magic-state distillation scaling exponent \(\gamma=\log_2 23 \approx 4.52\)\NoCaseChange{\protect\cite{cite706}}.}\\ 
\addlinespace[\myxtraspc]
\eczhRefIndex{code:small_triorthogonal}%
\eczhListValue{\flmRefsHyperref{code:small_triorthogonal}{\(\llbracket 3k + 8, k, 2\rrbracket \) triorthogonal code}} & \eczhListValue{The family yields the asymptotic exponent \(\gamma = \log_2 \frac{3k+8}{k} \to \log_2 3 \approx 1.6\) for sufficiently large \(k\) \NoCaseChange{\protect\cite[{Box 2}]{cite707}}; see \NoCaseChange{\protect\cite[{Table V}]{cite705}}.}\\ 
\addlinespace[\myxtraspc]
\eczhRefIndex{code:stab_4_2_2}%
\eczhListValue{\flmRefsHyperref{code:stab_4_2_2}{\(\llbracket 4,2,2\rrbracket \) Four-qubit code}} & \eczhListValue{Various magic-state distillation protocols exist for the \(\llbracket 4,2,2\rrbracket \) qubit code and the \(C_6\) code in what are known as Meier-Eastin-Knill (MEK) protocols \NoCaseChange{\protect\cite{cite708,cite101}}. In the inner/outer-code formulation of Ref. \NoCaseChange{\protect\cite{cite101}}, the \(\llbracket 4,2,2\rrbracket \) code is a hyperbolic self-dual inner code for quadratic distillation. For example, the magic-state yield parameter is \(\gamma = \log_2 5 \approx 2.322\) for a protocol using the \(\llbracket 10,2,2\rrbracket \) code \NoCaseChange{\protect\cite[{Box 2}]{cite707}}; see also \NoCaseChange{\protect\cite[{Table IV}]{cite705}}.}\\ 
\addlinespace[\myxtraspc]
\eczhRefIndex{code:stab_49_1_5}%
\eczhListValue{\flmRefsHyperref{code:stab_49_1_5}{\(\llbracket 49,1,5\rrbracket \) triorthogonal code}} & \eczhListValue{The code yields an exponent \(\gamma = \log 49 / \log 5 \approx 2.42\).}\\ 
\addlinespace[\myxtraspc]
\eczhRefIndex{code:stab_6_2_2}%
\eczhListValue{\flmRefsHyperref{code:stab_6_2_2}{\(\llbracket 6,2,2\rrbracket \) \(C_6\) code}} & \eczhListValue{Various magic-state distillation protocols exist for the \(\llbracket 4,2,2\rrbracket \) qubit code and the \(C_6\) code in what are known as Meier-Eastin-Knill (MEK) protocols \NoCaseChange{\protect\cite{cite708}}. For example, the magic-state yield parameter is \(\gamma = \log_2 5 \approx 2.322\) for a protocol using the \(\llbracket 10,2,2\rrbracket \) code \NoCaseChange{\protect\cite[{Box 2}]{cite707}}; see also \NoCaseChange{\protect\cite[{Table IV}]{cite705}}.}\\ 
\addlinespace[\myxtraspc]
\eczhRefIndex{code:campbell_howard}%
\eczhListValue{\flmRefsHyperref{code:campbell_howard}{\(\llbracket 6k+2,3k,2\rrbracket \) Campbell-Howard code}} & \eczhListValue{A total of \(r\) rounds of magic-state distillation yields a magic-state yield parameter \(\gamma\to 1^{+}\) as \(k,r\rightarrow \infty\). This matches the Bravyi-Haah conjectured lower bound \(\gamma \geq 1\) for concatenated triorthogonal-matrix protocols \NoCaseChange{\protect\cite[{Sec. VI}]{cite691}}.}\\ 
\addlinespace[\myxtraspc]
\eczhRefIndex{code:qutrit_small_triorthogonal}%
\eczhListValue{\flmRefsHyperref{code:qutrit_small_triorthogonal}{\(\llbracket 9m-k,k,2\rrbracket _3\) triorthogonal code}} & \eczhListValue{For \(k = 3m-2\), the family yields the magic-state yield parameter \(\gamma = \log_2 (2+\frac{6}{3m-2}) \to 1\) as \(m\to\infty\) \NoCaseChange{\protect\cite{cite709}}.}\\ 
\addlinespace[\myxtraspc]
\eczhRefIndex{code:quantum_h}%
\eczhListValue{\flmRefsHyperref{code:quantum_h}{\(\llbracket k+4,k,2\rrbracket \) H code}} & \eczhListValue{A total of \(r\) rounds of magic-state distillation yields a magic-state yield parameter \(\gamma\to 1^{+}\) as \(k,r\rightarrow \infty\); see \NoCaseChange{\protect\cite[{Box 2}]{cite707}}. This matches the Bravyi-Haah conjectured lower bound \(\gamma \geq 1\) for concatenated triorthogonal-matrix protocols \NoCaseChange{\protect\cite[{Sec. VI}]{cite691}}.}\\ 
\end{tabularx}\endgroup
\eczcodelist{quantum_transversal}{Quantum codes with transversal or permutation-based gates
}%

\eczhCodeListAutoDescription{All codes with \emph{Transversal and Permutation-Based Gates} that are not descendants of \flmRefsCref{code:ecc}.}%

\eczhIncludeCodeGraph{Bare}{scale=0.5}{\columnwidth}{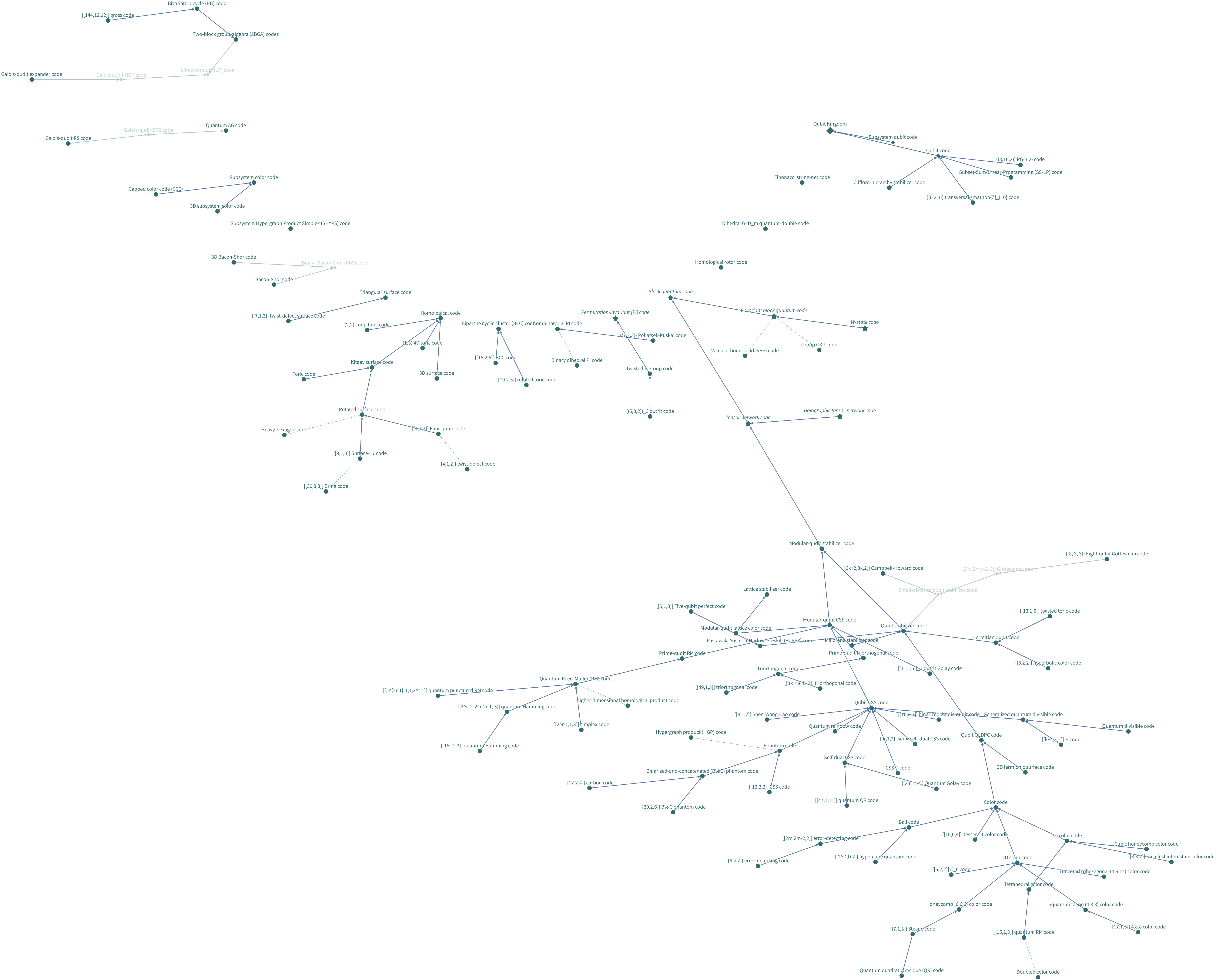}{Quantum codes with transversal or permutation-based gates}{https://errorcorrectionzoo.org/code_graph#J\%7B\%22displayMode\%22\%3A\%22subset\%22\%2C\%22modeSubsetOptions\%22\%3A\%7B\%22codeIds\%22\%3A\%5B\%222d_color\%22\%2C\%223d_bacon_shor\%22\%2C\%223d_color\%22\%2C\%223d_fermionic_surface\%22\%2C\%223d_subsystem_color\%22\%2C\%223d_surface\%22\%2C\%22bacon_shor\%22\%2C\%22ball_color\%22\%2C\%22bc_phantom\%22\%2C\%22binary_dihedral_permutation_invariant\%22\%2C\%22bipartite_cyclic_cluster\%22\%2C\%22qcga\%22\%2C\%22block_quantum\%22\%2C\%22css-t\%22\%2C\%22capped_color\%22\%2C\%22clifford_hierarchy\%22\%2C\%22color\%22\%2C\%22combinatorial_permutation_invariant\%22\%2C\%22covariant\%22\%2C\%22cubic_honeycomb_color\%22\%2C\%22quantum_double_dihedral\%22\%2C\%22doubled_color\%22\%2C\%22fibonacci\%22\%2C\%22galois_polynomial\%22\%2C\%22galois_expander\%22\%2C\%22generalized_quantum_divisible\%22\%2C\%22group_gkp\%22\%2C\%22heavy_hex\%22\%2C\%22stabilizer_over_gf4\%22\%2C\%22multisector_hypergraph\%22\%2C\%22holographic_tensor\%22\%2C\%22higher_dimensional_surface\%22\%2C\%22homological_rotor\%22\%2C\%22triangular_color\%22\%2C\%22hypergraph_product\%22\%2C\%22surface\%22\%2C\%22translationally_invariant_stabilizer\%22\%2C\%22majorana_stab\%22\%2C\%22qudit_css\%22\%2C\%22qudit_color\%22\%2C\%22qudit_stabilizer\%22\%2C\%22happy\%22\%2C\%22permutation_invariant\%22\%2C\%22phantom\%22\%2C\%22qudit_reed_muller\%22\%2C\%22qudit_triorthogonal\%22\%2C\%22quantum_ag\%22\%2C\%22quantum_reed_muller\%22\%2C\%22quantum_divisible\%22\%2C\%22galois_quad_residue\%22\%2C\%22quantum_rainbow\%22\%2C\%22qubit_css\%22\%2C\%22qldpc\%22\%2C\%22qubits_into_qubits\%22\%2C\%22qubit_stabilizer\%22\%2C\%22rotated_surface\%22\%2C\%22self_dual_css\%22\%2C\%22488_color\%22\%2C\%22sslp\%22\%2C\%22shyps\%22\%2C\%22subsystem_color\%22\%2C\%22subsystem_qubits_into_qubits\%22\%2C\%22quantum_lego\%22\%2C\%22tetrahedral_color\%22\%2C\%22toric\%22\%2C\%22triangle_surface\%22\%2C\%22quantum_triorthogonal\%22\%2C\%224612_color\%22\%2C\%22t_group\%22\%2C\%222bga\%22\%2C\%22vbs\%22\%2C\%22w_state\%22\%2C\%22su3_sigma360\%22\%2C\%22qubit_6_2_3\%22\%2C\%22icosahedral_permutation_invariant\%22\%2C\%22qubit_8_4_2\%22\%2C\%224d_13_surface\%22\%2C\%224d_surface\%22\%2C\%22stab_10_2_3\%22\%2C\%22xzzx_10_2_3\%22\%2C\%22qutrit_golay\%22\%2C\%22stab_12_2_2\%22\%2C\%22carbon\%22\%2C\%22stab_13_1_5\%22\%2C\%22gross\%22\%2C\%22stab_15_7_3\%22\%2C\%22stab_15_1_3\%22\%2C\%22stab_16_6_4\%22\%2C\%22stab_17_1_5\%22\%2C\%22stab_18_2_5\%22\%2C\%22stab_20_2_6\%22\%2C\%22qubit_golay\%22\%2C\%22hypercube_quantum\%22\%2C\%22quantum_hamming_css\%22\%2C\%22diagonal_clifford\%22\%2C\%22single_qubit_clifford\%22\%2C\%22iceberg\%22\%2C\%22stellated_dodecahedron_css\%22\%2C\%22small_triorthogonal\%22\%2C\%22stab_4_1_2\%22\%2C\%22stab_4_2_2\%22\%2C\%22stab_47_1_11\%22\%2C\%22stab_49_1_5\%22\%2C\%22stab_5_1_3\%22\%2C\%22css_6_1_2\%22\%2C\%22stab_6_2_2\%22\%2C\%22stab_6_4_2\%22\%2C\%22campbell_howard\%22\%2C\%22steane\%22\%2C\%22twist_defect_7_1_3\%22\%2C\%22stab_8_3_3\%22\%2C\%22stab_8_1_2\%22\%2C\%22stab_8_2_2\%22\%2C\%22stab_8_3_2\%22\%2C\%22surface-17\%22\%2C\%22quantum_h\%22\%5D\%2C\%22reusePreviousLayoutPositions\%22\%3Afalse\%2C\%22showIntermediateConnectingNodes\%22\%3Atrue\%2C\%22connectingNodesMaxDepth\%22\%3A15\%2C\%22connectingNodesPathMaxLength\%22\%3A20\%2C\%22connectingNodesMaxExtraDepth\%22\%3A3\%2C\%22connectingNodesOnlyKeepPathsWithAdditionalLength\%22\%3A1\%2C\%22connectingNodesToDomainsAndKingdoms\%22\%3Afalse\%2C\%22connectingNodesEdgeLengthsByType\%22\%3A\%7B\%22primaryParent\%22\%3A1\%2C\%22secondaryParent\%22\%3A4\%2C\%22cousin\%22\%3A6\%7D\%2C\%22nodeIds\%22\%3A\%5B\%22k_qubits_into_qubits\%22\%2C\%22k_qubits_into_qubits\%22\%5D\%7D\%2C\%22highlightImportantNodes\%22\%3A\%7B\%22highlightImportantNodes\%22\%3Afalse\%2C\%22highlightPrimaryParents\%22\%3Afalse\%2C\%22highlightRootConnectingEdges\%22\%3Afalse\%7D\%7D}

\begingroup
\small
\eczhBreakableDashes
\renewcommand\arraystretch{1.05}
\edef\myxtraspc{\eczListAddVSpaceXtraXtraStretch}
\begin{tabularx}{\linewidth}{>{\eczlistsmallsize\raggedright\arraybackslash}p{\eczListColWidth{name}} >{\hsize=1.0000\hsize \eczlistsmallsize}X}
\toprule
\eczListColTitle{Name} & \eczListColTitle{Transversal gates} \\
\midrule
\endfirsthead
\toprule
\eczListColTitleContinued{Name} & \eczListColTitleContinued{Transversal gates} \\
\midrule
\endhead
\bottomrule
\endfoot
\eczhRefIndex{code:2d_color}%
\eczhListValue{\flmRefsHyperref{code:2d_color}{2D color code}} & \eczhListValue{\begin{eczvaluelist}
\item\relax CNOT gate because the code is CSS.
\item\relax Hadamard gates for any qubit geometry which yields a self-dual CSS code.
\item\relax Certain triangular 2D color codes on suitably chosen lattices admit transversal implementations of the full Clifford group, including \(H\), \(S^\dagger\), and \(CNOT\), without selective addressing \NoCaseChange{\protect\cite{cite710}}.
\end{eczvaluelist}
}\\ 
\addlinespace[\myxtraspc]
\eczhRefIndex{code:3d_bacon_shor}%
\eczhListValue{\flmRefsHyperref{code:3d_bacon_shor}{3D Bacon-Shor code}} & \eczhListValue{\begin{eczvaluelist}
\item\relax Logical \(CCZ\) gates on three code blocks of different orientations \NoCaseChange{\protect\cite{cite711}}.
\end{eczvaluelist}
}\\ 
\addlinespace[\myxtraspc]
\eczhRefIndex{code:3d_color}%
\eczhListValue{\flmRefsHyperref{code:3d_color}{3D color code}} & \eczhListValue{\begin{eczvaluelist}
\item\relax Transversal action of \(T\) gates on color codes on general 3-manifolds realizes a \(CCZ\) gate on three logical qubits and is related to a topological invariant that is called the triple intersection number; this gate is related to the fact that this code admits a cup product structure \NoCaseChange{\protect\cite{cite703}}.
\item\relax Transversal \(S\) gate on color codes on general 3-manifolds corresponds to a higher-form symmetry \NoCaseChange{\protect\cite{cite703}}.
\item\relax Universal transversal gates can be achieved using lattice surgery or code deformation \NoCaseChange{\protect\cite{cite712,cite713}}.
\item\relax Families of 3D color codes on quasi-hyperbolic, fibre-bundle, and Torelli mapping-torus 3-manifolds support collective logical \(CCZ\) gates via transversal \(T\) and individually addressable, parallelizable logical \(CZ\) gates via transversal \(S\) on codimension-1 submanifolds. Their rate-distance scalings are \(O(1/\log n)\) with \(d=O(\log n)\), \(O(1/\log^2 n)\) with \(d=\Omega(\log^2 n)\), and \(O(1)\) with distance scaling unknown, respectively \NoCaseChange{\protect\cite{cite703}}.
\end{eczvaluelist}
}\\ 
\addlinespace[\myxtraspc]
\eczhRefIndex{code:3d_fermionic_surface}%
\eczhListValue{\flmRefsHyperref{code:3d_fermionic_surface}{3D fermionic surface code}} & \eczhListValue{\begin{eczvaluelist}
\item\relax \(CCZ\) and \(CS\) gates can be obtained for the fermionic 3D surface code on certain manifolds by circuits that can be interpreted as moving and spreading lattice realizations of Kitaev chain and \(p+ip\) defects \NoCaseChange{\protect\cite{cite714}}.
\end{eczvaluelist}
}\\ 
\addlinespace[\myxtraspc]
\eczhRefIndex{code:3d_subsystem_color}%
\eczhListValue{\flmRefsHyperref{code:3d_subsystem_color}{3D subsystem color code}} & \eczhListValue{\begin{eczvaluelist}
\item\relax For the \((1,1)\) member, \(CNOT\) and Hadamard are transversal; gauge fixing to the \((1,2)\) code enables a transversal \(R_3\) gate, yielding a universal gate set without encoded ancillas \NoCaseChange{\protect\cite{cite475}}.
\end{eczvaluelist}
}\\ 
\addlinespace[\myxtraspc]
\eczhRefIndex{code:3d_surface}%
\eczhListValue{\flmRefsHyperref{code:3d_surface}{3D surface code}} & \eczhListValue{\begin{eczvaluelist}
\item\relax For a stack of three 3D surface codes on the same rectified cubic lattice, pairwise logical \(CZ\) and triple logical \(CCZ\) gates are transversal \NoCaseChange{\protect\cite{cite715}}.
\end{eczvaluelist}
}\\ 
\addlinespace[\myxtraspc]
\eczhRefIndex{code:bacon_shor}%
\eczhListValue{\flmRefsHyperref{code:bacon_shor}{Bacon-Shor code}} & \eczhListValue{\begin{eczvaluelist}
\item\relax Logical Hadamard is transversal in symmetric Bacon-Shor codes up to a qubit permutation \NoCaseChange{\protect\cite{cite716}} and can be implemented with teleportation \NoCaseChange{\protect\cite{cite717}}.
\item\relax Bacon-Shor codes on an \(m \times m^k\) lattice admit transversal \(k\)-qubit-controlled \(Z\) gates \NoCaseChange{\protect\cite{cite711}}.
\end{eczvaluelist}
}\\ 
\addlinespace[\myxtraspc]
\eczhRefIndex{code:ball_color}%
\eczhListValue{\flmRefsHyperref{code:ball_color}{Ball code}} & \eczhListValue{\begin{eczvaluelist}
\item\relax The 3D ball codes on duals of the truncated octahedron, truncated cuboctahedron, and truncated icosidodecahedron admit logical \(CCZ\)-type gates implemented by physical \(T\) and \(T^\dagger\) gates \NoCaseChange{\protect\cite[{Exam. 4, Thm. 5}]{cite687}}.
\end{eczvaluelist}
}\\ 
\addlinespace[\myxtraspc]
\eczhRefIndex{code:bc_phantom}%
\eczhListValue{\flmRefsHyperref{code:bc_phantom}{Binarized-and-concatenated (B\&C) phantom code}} & \eczhListValue{\begin{eczvaluelist}
\item\relax Ordinary CSS self-duality of the starting \(q=4\) Galois-qudit CSS code yields a permutation-assisted logical Hadamard on the resulting qubit code. Hermitian self-duality, meaning that the \(Z\)-type check space is the Frobenius conjugate of the \(X\)-type check space, additionally yields a logical \(CZ\) gate \NoCaseChange{\protect\cite{cite514}}.
\end{eczvaluelist}
}\\ 
\addlinespace[\myxtraspc]
\eczhRefIndex{code:binary_dihedral_permutation_invariant}%
\eczhListValue{\flmRefsHyperref{code:binary_dihedral_permutation_invariant}{Binary dihedral PI code}} & \eczhListValue{\begin{eczvaluelist}
\item\relax Binary dihedral group gates can be realized transversally, which include subgroups of any level of the \flmTerm{term}{ref694}{}{Clifford hierarchy} and subgroups which cannot be realized by any qubit stabilizer code.
\end{eczvaluelist}
}\\ 
\addlinespace[\myxtraspc]
\eczhRefIndex{code:bipartite_cyclic_cluster}%
\eczhListValue{\flmRefsHyperref{code:bipartite_cyclic_cluster}{Bipartite cyclic cluster (BCC) code}} & \eczhListValue{\begin{eczvaluelist}
\item\relax For \(k=2\), all BCC codes admit a transversal Hadamard-SWAP logical gate via qubit permutation \((m,1) \mapsto (-m,2)\) followed by Hadamard on all qubits \NoCaseChange{\protect\cite{cite440}}.
\item\relax The \(\llbracket d^2+1,2,d\rrbracket \) BCC codes also admit a logical Hadamard without SWAP via \(m \mapsto md \pmod{n}\) followed by Hadamard \NoCaseChange{\protect\cite{cite440}}.
\end{eczvaluelist}
}\\ 
\addlinespace[\myxtraspc]
\eczhRefIndex{code:qcga}%
\eczhListValue{\flmRefsHyperref{code:qcga}{Bivariate bicycle (BB) code}} & \eczhListValue{\begin{eczvaluelist}
\item\relax Logical Pauli operators and fold-transversal gates studied in Ref. \NoCaseChange{\protect\cite{cite718,cite719}}.
\end{eczvaluelist}
}\\ 
\addlinespace[\myxtraspc]
\eczhRefIndex{code:block_quantum}%
\eczhListValue{\flmRefsHyperref{code:block_quantum}{Block quantum code}} & \eczhListValue{\begin{eczvaluelist}
\item\relax \begin{defterm}{Eastin-Knill theorem}\label{ref720}\label{ref721} \textit{Transversal gates} are logical gates on block codes that can be realized as tensor products of unitary operations acting on subsets of subsystems whose size is independent of \(n\). For subsets of size one, gates are sometimes called \textit{strongly transversal} if the single-subsystem unitaries are identical, and \textit{weakly transversal} otherwise. A universal gate set for a finite-dimensional block quantum code cannot be transversal for any code that detects single-block errors due to the Eastin-Knill theorem \NoCaseChange{\protect\cite{cite722}}. \end{defterm}
\item\relax A qudit code of length \(n\) with permutation automorphism subgroups \(N\triangleleft G\leq \mathrm{PAut}(Q)\) and simple non-Abelian quotient \(G/N\) must satisfy \(n\geq \mu(G/N)\) \NoCaseChange{\protect\cite[{Thm. S5}]{cite723}}.
\end{eczvaluelist}
}\\ 
\addlinespace[\myxtraspc]
\eczhRefIndex{code:css-t}%
\eczhListValue{\flmRefsHyperref{code:css-t}{CSS-T code}} & \eczhListValue{\begin{eczvaluelist}
\item\relax A physical transversal \(T\) gate is either the identity (up to a global phase) or a logical gate \NoCaseChange{\protect\cite{cite724}}.
\end{eczvaluelist}
}\\ 
\addlinespace[\myxtraspc]
\eczhRefIndex{code:capped_color}%
\eczhListValue{\flmRefsHyperref{code:capped_color}{Capped color code (CCC)}} & \eczhListValue{\begin{eczvaluelist}
\item\relax Capped color codes in H (T) form admit a transversal Hadamard (T) gate.
\end{eczvaluelist}
}\\ 
\addlinespace[\myxtraspc]
\eczhRefIndex{code:clifford_hierarchy}%
\eczhListValue{\flmRefsHyperref{code:clifford_hierarchy}{Clifford-hierarchy stabilizer code}} & \eczhListValue{\begin{eczvaluelist}
\item\relax A transversal logical \(\text{diag}(1, e^{i2\pi/2^D})\) gate at the \(D\)th level of the Clifford hierarchy in \((D-1)\) spatial dimensions \NoCaseChange{\protect\cite{cite725}}.
\end{eczvaluelist}
}\\ 
\addlinespace[\myxtraspc]
\eczhRefIndex{code:color}%
\eczhListValue{\flmRefsHyperref{code:color}{Color code}} & \eczhListValue{\begin{eczvaluelist}
\item\relax Some color codes on \(D\)-dimensional lattices can transversally implement a gate at the \(D\)th level of the \flmTerm{term}{ref694}{}{Clifford hierarchy} in the form of a \(Z\)-rotation by angle \(\pi/2^{D-1}\) \NoCaseChange{\protect\cite[{Fig. 3}]{cite726}}.
\end{eczvaluelist}
}\\ 
\addlinespace[\myxtraspc]
\eczhRefIndex{code:combinatorial_permutation_invariant}%
\eczhListValue{\flmRefsHyperref{code:combinatorial_permutation_invariant}{Combinatorial PI code}} & \eczhListValue{\begin{eczvaluelist}
\item\relax A class of combinatorial PI codes called \((b,g,1)\)-codes has been identified that admits logical gates in the diagonal \flmTerm{term}{ref694}{}{Clifford hierarchy} from transversal \(Z\)-axis rotations \NoCaseChange{\protect\cite{cite727}}.
\end{eczvaluelist}
}\\ 
\addlinespace[\myxtraspc]
\eczhRefIndex{code:covariant}%
\eczhListValue{\flmRefsHyperref{code:covariant}{Covariant block quantum code}} & \eczhListValue{\begin{eczvaluelist}
\item\relax \(G\)-covariant codes defined on a tensor product space consisting of \(n\) subsystems are equivalent to codes with a transversal gate set realizing \(G\).
\end{eczvaluelist}
}\\ 
\addlinespace[\myxtraspc]
\eczhRefIndex{code:cubic_honeycomb_color}%
\eczhListValue{\flmRefsHyperref{code:cubic_honeycomb_color}{Cubic honeycomb color code}} & \eczhListValue{\begin{eczvaluelist}
\item\relax A code family on a truncated cube with particular boundary conditions admits a transversal control-\(S\) gate via physical \(T\) and \(T^{\dagger}\) gates \NoCaseChange{\protect\cite{cite728}}.
\end{eczvaluelist}
}\\ 
\addlinespace[\myxtraspc]
\eczhRefIndex{code:quantum_double_dihedral}%
\eczhListValue{\flmRefsHyperref{code:quantum_double_dihedral}{Dihedral \(G=D_m\) quantum-double code}} & \eczhListValue{\begin{eczvaluelist}
\item\relax On a triangular patch encoding one logical qubit, a transversal \(T^{1/N}=\mathrm{diag}(1,e^{i\pi/(4N)})\) gate can be implemented when \(G=D_{4N}\) \NoCaseChange{\protect\cite{cite729,cite730}}.
\end{eczvaluelist}
}\\ 
\addlinespace[\myxtraspc]
\eczhRefIndex{code:doubled_color}%
\eczhListValue{\flmRefsHyperref{code:doubled_color}{Doubled color code}} & \eczhListValue{\begin{eczvaluelist}
\item\relax Doubled color codes are triply even, so they yield a transversal \(T\) gate \NoCaseChange{\protect\cite{cite731}}. Using gauge fixing, the codes admit a Clifford + \(T\) transversal gate set.
\end{eczvaluelist}
}\\ 
\addlinespace[\myxtraspc]
\eczhRefIndex{code:fibonacci}%
\eczhListValue{\flmRefsHyperref{code:fibonacci}{Fibonacci string-net code}} & \eczhListValue{\begin{eczvaluelist}
\item\relax A universal transversal gate set could be implemented in a folded version of this code using the techniques introduced in Ref \NoCaseChange{\protect\cite{cite732}}.
\end{eczvaluelist}
}\\ 
\addlinespace[\myxtraspc]
\eczhRefIndex{code:galois_polynomial}%
\eczhListValue{\flmRefsHyperref{code:galois_polynomial}{Galois-qudit RS code}} & \eczhListValue{\begin{eczvaluelist}
\item\relax There exists an \flmRefsHyperref{ref65}{order} \(\llbracket n,\Theta(n),\Theta(n)\rrbracket _{n^2}\) punctured RS code family that admits transversal \(CCZ\) gates for any three logical qubits \NoCaseChange{\protect\cite{cite733}}. This code can be treated as a qubit code by decomposing each Galois qudit into a Kronecker product of several qubits; see \NoCaseChange{\protect\cite{cite696,cite398,cite698,cite699,cite700}\protect\cite[{Sec. 5.3}]{cite697}}. This yields a qubit code family that is asymptotically good up to poly-logarithmic factors \NoCaseChange{\protect\cite{cite733}}.
\item\relax Quantum RS codes can be concatenated with quantum multiplication-friendly codes to yield CSS codes over a constant prime-power alphabet \(q\) admitting a transversal \(CCZ_q\) gate (and, for \(q \geq 5\), also a transversal \(U_q\) gate) with \flmRefsHyperref{ref65}{nearly linear} code dimension and distance, \(k,d = \Omega(n/2^{O(\log^* n)})\), where \(\log^* n\) is the iterated logarithm (the number of times \(\log\) must be applied to \(n\) before the result is \(\leq 1\)) \NoCaseChange{\protect\cite{cite699}}.
\end{eczvaluelist}
}\\ 
\addlinespace[\myxtraspc]
\eczhRefIndex{code:galois_expander}%
\eczhListValue{\flmRefsHyperref{code:galois_expander}{Galois-qudit expander code}} & \eczhListValue{\begin{eczvaluelist}
\item\relax For every integer \(r\geq 2\), there are QLDPC Galois-qudit quantum expander codes with transversal \(C^{r-1} Z\) gates and parameters \(\llbracket N,K\geq N^{1-\epsilon},D\geq N^{1/r}/\operatorname{poly}(\log N)\rrbracket _q\) \NoCaseChange{\protect\cite{cite689}}. By decomposing each Galois qudit into a Kronecker product of qubits, this yields an explicit qubit CSS QLDPC family with parameters \(\llbracket N,K\geq N^{1-\epsilon},D\geq N^{1/r}/\operatorname{poly}(\log N)\rrbracket \), stabilizer weight \(\operatorname{poly}(\log N)\), and transversal \(C^{r-1}Z\) gates acting on \(N^{1-\epsilon}\) disjoint logical \(r\)-tuples.
\end{eczvaluelist}
}\\ 
\addlinespace[\myxtraspc]
\eczhRefIndex{code:generalized_quantum_divisible}%
\eczhListValue{\flmRefsHyperref{code:generalized_quantum_divisible}{Generalized quantum divisible code}} & \eczhListValue{\begin{eczvaluelist}
\item\relax A level-\(\nu\) generalized quantum divisible code admits a diagonal transversal gate at the \(\nu\)th level of the \flmTerm{term}{ref694}{}{Clifford hierarchy} \NoCaseChange{\protect\cite[{Lemma V.3}]{cite734}}.
\end{eczvaluelist}
}\\ 
\addlinespace[\myxtraspc]
\eczhRefIndex{code:group_gkp}%
\eczhListValue{\flmRefsHyperref{code:group_gkp}{Group GKP code}} & \eczhListValue{\begin{eczvaluelist}
\item\relax Group-GKP codes corresponding to the \(G^{k_1} \subseteq G^{k_2} \subset G^{n}\) group construction admit \(X\)-type logical group-multiplication gates, and are thus covariant with respect to the induced \(G^{k_2}\)-action \NoCaseChange{\protect\cite{cite735}}.
\end{eczvaluelist}
}\\ 
\addlinespace[\myxtraspc]
\eczhRefIndex{code:heavy_hex}%
\eczhListValue{\flmRefsHyperref{code:heavy_hex}{Heavy-hexagon code}} & \eczhListValue{\begin{eczvaluelist}
\item\relax CNOT gates are transversal for this code. However, for most architectures, all logical gates would be implemented using lattice surgery methods.
\end{eczvaluelist}
}\\ 
\addlinespace[\myxtraspc]
\eczhRefIndex{code:stabilizer_over_gf4}%
\eczhListValue{\flmRefsHyperref{code:stabilizer_over_gf4}{Hermitian qubit code}} & \eczhListValue{\begin{eczvaluelist}
\item\relax All code automorphisms lie in the \flmRefsHyperref{ref409}{Clifford group} \NoCaseChange{\protect\cite[{Corr. 16}]{cite446}}, so transversal physical gates implement only Clifford logical gates.
\item\relax Transversal \(SH\) and \(HS\) "facet" gates (a.k.a. \(M_3\) gates) which cyclically permute Paulis as \(X \to Y\), \(Y \to Z\), and \(Z \to X\) \NoCaseChange{\protect\cite[{Sec. 8.2}]{cite736}}.
\item\relax The three-block transversal gate mapping each physical \(X \to XYZ\) and each \(Z \to ZXY\) implements a logical gate \NoCaseChange{\protect\cite{cite737}\protect\cite[{Exam. 2}]{cite532}}.
\item\relax A qubit stabilizer code is Hermitian if and only if a transversal \(R\) gate leaves the stabilizer group invariant \NoCaseChange{\protect\cite{cite454}}.
\item\relax Hermitian qubit codes admit the group \(U(\ell,\mathbb{F}_4)\) of diagonal transversal gates on \(\ell\) codeblocks \NoCaseChange{\protect\cite{cite738}}.
\end{eczvaluelist}
}\\ 
\addlinespace[\myxtraspc]
\eczhRefIndex{code:multisector_hypergraph}%
\eczhListValue{\flmRefsHyperref{code:multisector_hypergraph}{Higher-dimensional homological product code}} & \eczhListValue{\begin{eczvaluelist}
\item\relax Transversal gates for multi-dimensional HGP codes of all dimensions lie in the Clifford group \NoCaseChange{\protect\cite{cite739}}.
\end{eczvaluelist}
}\\ 
\addlinespace[\myxtraspc]
\eczhRefIndex{code:holographic_tensor}%
\eczhListValue{\flmRefsHyperref{code:holographic_tensor}{Holographic tensor-network code}} & \eczhListValue{\begin{eczvaluelist}
\item\relax There exist holographic approximate codes with arbitrary transversal gate sets for any compact Lie group \NoCaseChange{\protect\cite{cite468}}. However, for sufficiently localized logical subsystems of holographic stabilizer codes, the set of transversally implementable logical operations is contained in the \flmRefsHyperref{ref409}{Clifford group} \NoCaseChange{\protect\cite{cite740}}.
\end{eczvaluelist}
}\\ 
\addlinespace[\myxtraspc]
\eczhRefIndex{code:higher_dimensional_surface}%
\eczhListValue{\flmRefsHyperref{code:higher_dimensional_surface}{Homological code}} & \eczhListValue{\begin{eczvaluelist}
\item\relax Locality preserving operations can be determined for stacks of homological codes in any dimension \NoCaseChange{\protect\cite{cite741}}.
\end{eczvaluelist}
}\\ 
\addlinespace[\myxtraspc]
\eczhRefIndex{code:homological_rotor}%
\eczhListValue{\flmRefsHyperref{code:homological_rotor}{Homological rotor code}} & \eczhListValue{\begin{eczvaluelist}
\item\relax All generalized Pauli gates are realized transversally.
\end{eczvaluelist}
}\\ 
\addlinespace[\myxtraspc]
\eczhRefIndex{code:triangular_color}%
\eczhListValue{\flmRefsHyperref{code:triangular_color}{Honeycomb (6.6.6) color code}} & \eczhListValue{\begin{eczvaluelist}
\item\relax CNOT gate because the code is CSS.
\item\relax Hadamard gates for any qubit geometry which yields a self-dual CSS code.
\end{eczvaluelist}
}\\ 
\addlinespace[\myxtraspc]
\eczhRefIndex{code:hypergraph_product}%
\eczhListValue{\flmRefsHyperref{code:hypergraph_product}{Hypergraph product (HGP) code}} & \eczhListValue{\begin{eczvaluelist}
\item\relax Hadamard (up to logical SWAP gates) and control-\(Z\) on all logical qubits \NoCaseChange{\protect\cite{cite742}}.
\item\relax Patch-transversal gates inherited from the automorphism group of the underlying classical codes \NoCaseChange{\protect\cite[{Appx. D}]{cite743}}.
\item\relax Orientation-preserving constant-depth circuits can only implement gates in the Clifford group \NoCaseChange{\protect\cite{cite739}}.
\item\relax Permutation-based gates (automorphism gadgets) can be inherited from the underlying classical codes in the hypergraph construction \NoCaseChange{\protect\cite{cite744}}.
\end{eczvaluelist}
}\\ 
\addlinespace[\myxtraspc]
\eczhRefIndex{code:surface}%
\eczhListValue{\flmRefsHyperref{code:surface}{Kitaev surface code}} & \eczhListValue{\begin{eczvaluelist}
\item\relax Folded surface codes, which are local-Clifford equivalent to triangular color codes with three differently colored boundaries, admit transversal \flmRefsHyperref{ref409}{Clifford gates} \NoCaseChange{\protect\cite{cite422,cite745}}.
\item\relax Fold-transversal initialization of the \(|Y\rangle\) logical state \NoCaseChange{\protect\cite{cite745,cite746,cite747,cite748}}.

\end{eczvaluelist}
}\\ 
\addlinespace[\myxtraspc]
\eczhRefIndex{code:translationally_invariant_stabilizer}%
\eczhListValue{\flmRefsHyperref{code:translationally_invariant_stabilizer}{Lattice stabilizer code}} & \eczhListValue{\begin{eczvaluelist}
\item\relax Fold-transversal logical gates can be obtained from crystalline symmetries \NoCaseChange{\protect\cite{cite749}}.
\end{eczvaluelist}
}\\ 
\addlinespace[\myxtraspc]
\eczhRefIndex{code:majorana_stab}%
\eczhListValue{\flmRefsHyperref{code:majorana_stab}{Majorana stabilizer code}} & \eczhListValue{\begin{eczvaluelist}
\item\relax Transversal Clifford operations are discussed in Ref. \NoCaseChange{\protect\cite{cite750}}.
\end{eczvaluelist}
}\\ 
\addlinespace[\myxtraspc]
\eczhRefIndex{code:qudit_css}%
\eczhListValue{\flmRefsHyperref{code:qudit_css}{Modular-qudit CSS code}} & \eczhListValue{\begin{eczvaluelist}
\item\relax Modular-qudit generalizations of CNOT gates are transversal interblock gates for all modular-qudit CSS codes.
\item\relax Scalar multiplication physical gates, which map \(|x\rangle \mapsto |ax\rangle\) for \(a \in \mathbb{Z}_q^\times\), the mutiplicative group modulo \(q\).
\end{eczvaluelist}
}\\ 
\addlinespace[\myxtraspc]
\eczhRefIndex{code:qudit_color}%
\eczhListValue{\flmRefsHyperref{code:qudit_color}{Modular-qudit lattice color code}} & \eczhListValue{\begin{eczvaluelist}
\item\relax Some modular-qudit lattice color codes on \(D\)-dimensional lattices can transversally implement a gate at the \((D-1)\)st level of the \flmRefsHyperref{ref751}{qudit Clifford hierarchy} \NoCaseChange{\protect\cite{cite673}}.
\end{eczvaluelist}
}\\ 
\addlinespace[\myxtraspc]
\eczhRefIndex{code:qudit_stabilizer}%
\eczhListValue{\flmRefsHyperref{code:qudit_stabilizer}{Modular-qudit stabilizer code}} & \eczhListValue{\begin{eczvaluelist}
\item\relax All qudit stabilizer codes realize modular qudit Pauli transformations transversally.
\end{eczvaluelist}
}\\ 
\addlinespace[\myxtraspc]
\eczhRefIndex{code:happy}%
\eczhListValue{\flmRefsHyperref{code:happy}{Pastawski-Yoshida-Harlow-Preskill (HaPPY) code}} & \eczhListValue{\begin{eczvaluelist}
\item\relax Any transversal gate of the five-qubit code is a transversal gate of the HaPPY code since the HaPPY code is constructed from five-qubit encoding tensors, which are covariant under such gates.
\item\relax For locality-preserving physical gates on the boundary, the set of transversally implementable logical operations in the bulk is strictly contained in the \flmRefsHyperref{ref409}{Clifford group} \NoCaseChange{\protect\cite{cite740}}.
\end{eczvaluelist}
}\\ 
\addlinespace[\myxtraspc]
\eczhRefIndex{code:permutation_invariant}%
\eczhListValue{\flmRefsHyperref{code:permutation_invariant}{Permutation-invariant (PI) code}} & \eczhListValue{\begin{eczvaluelist}
\item\relax Qudit Dicke states are in one-to-one correspondence with points on the \flmRefsHyperref{ref655}{discrete simplex} \(\Delta_{q,n}\), which houses the totally symmetric irrep of \(SU(q)\) \NoCaseChange{\protect\cite{cite500}}. Any transversal gates of the form \(U^{\otimes N}\), with \(U \in SU(q)\), will implement logical operations in a subgroup of \(SU(q)\).
\end{eczvaluelist}
}\\ 
\addlinespace[\myxtraspc]
\eczhRefIndex{code:phantom}%
\eczhListValue{\flmRefsHyperref{code:phantom}{Phantom code}} & \eczhListValue{\begin{eczvaluelist}
\item\relax For CSS phantom codes, interblock \(\overline{\mathrm{CNOT}}\) gates are transversal. Combining transversal interblock CNOTs with in-block permutation CNOTs implements any logical CNOT circuit on \(2^a\) phantom-code blocks in physical depth at most \(4(2^a-1)\), up to a residual logical-qubit permutation; for unidirectional CNOT circuits, the bound is \(2(2^a-1)\) while preserving logical-qubit order \NoCaseChange{\protect\cite{cite514}}.
\item\relax A stabilizer code supporting a logical gate by qubit permutations cannot admit any strictly transversal logical gate that does not commute with that permutation-implemented logical gate, ruling out strictly transversal implementations of several gates on phantom codes \NoCaseChange{\protect\cite{cite514}}.
\item\relax Additional logical Clifford and non-Clifford gates can arise from code automorphisms combining local Cliffords and qubit permutations, from fold-diagonal gates using patterned one- and two-qubit diagonal interactions, and from non-uniform diagonal single-qubit rotations \NoCaseChange{\protect\cite{cite514}}.
\end{eczvaluelist}
}\\ 
\addlinespace[\myxtraspc]
\eczhRefIndex{code:qudit_reed_muller}%
\eczhListValue{\flmRefsHyperref{code:qudit_reed_muller}{Prime-qudit RM code}} & \eczhListValue{\begin{eczvaluelist}
\item\relax An odd-prime-qudit CSS code family constructed from first-order punctured GRM codes transversally implements a diagonal gate at any level of the \flmRefsHyperref{ref751}{qudit Clifford hierarchy} \NoCaseChange{\protect\cite{cite692,cite752}}.
\end{eczvaluelist}
}\\ 
\addlinespace[\myxtraspc]
\eczhRefIndex{code:qudit_triorthogonal}%
\eczhListValue{\flmRefsHyperref{code:qudit_triorthogonal}{Prime-qudit triorthogonal code}} & \eczhListValue{\begin{eczvaluelist}
\item\relax Admits a transversal gate from the third level of the \flmRefsHyperref{ref751}{qudit Clifford hierarchy} \NoCaseChange{\protect\cite{cite693}}.
\end{eczvaluelist}
}\\ 
\addlinespace[\myxtraspc]
\eczhRefIndex{code:quantum_ag}%
\eczhListValue{\flmRefsHyperref{code:quantum_ag}{Quantum AG code}} & \eczhListValue{\begin{eczvaluelist}
\item\relax There exist three asymptotically good code families \NoCaseChange{\protect\cite{cite695,cite699,cite698}} that admit a diagonal transversal gate at the third level of the \flmTerm{term}{ref694}{}{Clifford hierarchy}.
\item\relax By decomposing each Galois qudit into a Kronecker product of qubits, the family of Ref. \NoCaseChange{\protect\cite{cite698}} yields an explicit asymptotically good qubit CSS code family with parameters \(\llbracket N,K=\Theta(N),D=\Theta(N)\rrbracket \) on which \(\overline{CCZ}^{\otimes K}\) is realized by a transversal application of physical \(CCZ\) gates on a constant fraction of qubits.
\item\relax There exists an asymptotically good code family that admits three-Galois-qudit non-Clifford gates for any three logical Galois qudits \NoCaseChange{\protect\cite{cite733}}.
\end{eczvaluelist}
}\\ 
\addlinespace[\myxtraspc]
\eczhRefIndex{code:quantum_reed_muller}%
\eczhListValue{\flmRefsHyperref{code:quantum_reed_muller}{Quantum Reed-Muller (RM) code}} & \eczhListValue{\begin{eczvaluelist}
\item\relax Stabilizer generators can be defined as Pauli strings acting on subsets of qubits corresponding to subcubes of the Hamming \(n\)-cube (a.k.a. Boolean hypercube) \NoCaseChange{\protect\cite{cite753}}. Transversal \(Z\)-rotations by angles \(\pi/2^k\) acting on subcubes can implement logical multi-controlled-\(Z\) gates \NoCaseChange{\protect\cite{cite753}}.
\item\relax The \(\llbracket 2^m,{m \choose r}, 2^{\min(r,m-r)}\rrbracket \) family, where \(r\) divides \(m\), admits diagonal gates in the form of \(Z\)-rotations by angle \(\pi/2^{m/r}\) \NoCaseChange{\protect\cite{cite754,cite755,cite756}\protect\cite[{Exam. 8 and Thm. 19}]{cite724}}. Of these, the self-dual sub-family for \(m=2r\) admits logical Clifford group gates via permutations, transversal gates, and fold-transversal gates with the help of ancillas \NoCaseChange{\protect\cite{cite757}}; the ancilla requirement was later removed \NoCaseChange{\protect\cite{cite758}}.
\item\relax The \(\llbracket 64,15,4\rrbracket \) member is highlighted in Ref. \NoCaseChange{\protect\cite{cite759}} as a candidate for transversal IQP sampling because a transversal \(T\) gate implements a logical circuit consisting of 15 \(CCZ\) gates.
\item\relax The family constructed out of shortened RM codes with parameters \(\llbracket \sum_{i=w+1}^m \binom{m}{i}, \sum_{i=0}^{w} \binom{m}{i}, \sum_{i=w+1}^{r+1} \binom{r+1}{i}\rrbracket \) for integers \(m > 2r\) and \(r > w \geq 0\) admits a transversal gate at the \(\nu\)th level in the hierarchy whenever \(m > \nu r\) \NoCaseChange{\protect\cite[{Thm. 1}]{cite701}}.
\end{eczvaluelist}
}\\ 
\addlinespace[\myxtraspc]
\eczhRefIndex{code:quantum_divisible}%
\eczhListValue{\flmRefsHyperref{code:quantum_divisible}{Quantum divisible code}} & \eczhListValue{\begin{eczvaluelist}
\item\relax A self-dual weakly doubly even \(\llbracket n,1,d\rrbracket \) CSS code admits a partitioned transversal physical \(S\) gate that realizes \(\overline{S}^m\), where \(m=|M^+|-|M^-| \pmod 4\); for odd \(m\), together with transversal Hadamard and CNOT, this yields the full logical Clifford group transversally \NoCaseChange{\protect\cite{cite731}\protect\cite[{Lemma 4}]{cite760}}.
\item\relax A weakly triply even \(\llbracket n,1,d\rrbracket \) CSS code with a strongly transversal logical \(X\) gate admits a partitioned transversal physical \(T\) gate that realizes \(\overline{T}^m\), where \(m=|M^+|-|M^-| \pmod 8\) \NoCaseChange{\protect\cite{cite731}\protect\cite[{Lemma 3}]{cite760}}.
\item\relax If the \(X\)-type stabilizers of a CSS code form an \(\nu\)-even classical code, and if all \(X\)-type logicals are \((\nu-1)\)-even, then the code admits a diagonal transversal gate in the \(\nu\)th level of the \flmTerm{term}{ref694}{}{Clifford hierarchy} \NoCaseChange{\protect\cite[{Prop. 8}]{cite702}}.
\end{eczvaluelist}
}\\ 
\addlinespace[\myxtraspc]
\eczhRefIndex{code:galois_quad_residue}%
\eczhListValue{\flmRefsHyperref{code:galois_quad_residue}{Quantum quadratic-residue (QR) code}} & \eczhListValue{\begin{eczvaluelist}
\item\relax Qubit quantum QR codes admit transversal implementations of the \flmRefsHyperref{ref409}{single-qubit Clifford group} \NoCaseChange{\protect\cite{cite760}}. They yield a family of high-distance triorthogonal codes \NoCaseChange{\protect\cite{cite760}} via the doubling transformation \NoCaseChange{\protect\cite{cite731}}; such codes admit transversal implementations of the \(T\) gate.
\end{eczvaluelist}
}\\ 
\addlinespace[\myxtraspc]
\eczhRefIndex{code:quantum_rainbow}%
\eczhListValue{\flmRefsHyperref{code:quantum_rainbow}{Quantum rainbow code}} & \eczhListValue{\begin{eczvaluelist}
\item\relax Hypergraph products of color codes yield quantum rainbow codes with growing distance and transversal gates in the \flmTerm{term}{ref694}{}{Clifford hierarchy} \NoCaseChange{\protect\cite{cite704}}.
\end{eczvaluelist}
}\\ 
\addlinespace[\myxtraspc]
\eczhRefIndex{code:qubit_css}%
\eczhListValue{\flmRefsHyperref{code:qubit_css}{Qubit CSS code}} & \eczhListValue{\begin{eczvaluelist}
\item\relax Transversal CNOT gates preserve the logical subspace, up to \(X\)-type Paulis, iff a qubit stabilizer code is CSS \NoCaseChange{\protect\cite{cite761,cite398}}. The Paulis are necessary for when the code is stabilized by stabilizers with a minus in front of them, e.g., \(-XXXX\) and \(ZZZZ\).
\item\relax \textit{Fold-transversal} \NoCaseChange{\protect\cite{cite422,cite745,cite762}} \flmRefsHyperref{ref409}{Clifford gates} are transversal gates combined with qubit permutations. Some of these can be obtained from automorphism groups of the underlying classical codes \NoCaseChange{\protect\cite[{Thms. 2-3}]{cite763}}.
\item\relax Necessary and sufficient conditions for diagonal physical gates on a CSS code to induce logical gates in the \flmTerm{term}{ref694}{}{Clifford hierarchy} have been formulated \NoCaseChange{\protect\cite{cite766}\protect\cite[{Thm. 9}]{cite764}\protect\cite[{Thm. 5}]{cite765}}. There are routines that can determine what diagonal gates in the \flmTerm{term}{ref694}{}{Clifford hierarchy} are realized by a code \NoCaseChange{\protect\cite{cite767,cite768}}.
\item\relax CSS code families with asymptotic rate \(> 1/3\) and distance at \(\geq 3\) do not admit logical qubit permutations from physical permutations \NoCaseChange{\protect\cite{cite769}}.
\item\relax Diagonal transversal \flmRefsHyperref{ref409}{Clifford gates} on \(\ell\) codeblocks of a CSS code form \(GL(\ell,\mathbb{F}_2)\) for non-self-dual CSS codes, \(U(\ell,R_8)\) for \textit{semi-self-dual CSS codes} (i.e., CSS codes whose \(X\)-type stabilizers are contained in the \(Z\)-type stabilizers), and \(Sp(2\ell,\mathbb{F}_2)\) for self-dual CSS codes \NoCaseChange{\protect\cite{cite738}}.
\item\relax Diagonal transversal gate groups can be defined using a set of equations \NoCaseChange{\protect\cite{cite770}}.
\end{eczvaluelist}
}\\ 
\addlinespace[\myxtraspc]
\eczhRefIndex{code:qldpc}%
\eczhListValue{\flmRefsHyperref{code:qldpc}{Qubit QLDPC code}} & \eczhListValue{\begin{eczvaluelist}
\item\relax There are recipes to determine transversal gates for asymmetric qubit QLDPC codes \NoCaseChange{\protect\cite{cite771}}.
\end{eczvaluelist}
}\\ 
\addlinespace[\myxtraspc]
\eczhRefIndex{code:qubits_into_qubits}%
\eczhListValue{\flmRefsHyperref{code:qubits_into_qubits}{Qubit code}} & \eczhListValue{\begin{eczvaluelist}
\item\relax A qubit code is \(U\)-\textit{quasi-transversal} if it can realize the logical gate \(U\) in the third level of the \flmTerm{term}{ref694}{}{Clifford hierarchy} using the physical gate \(C T^{\otimes n}\), where \(C\) is some Clifford gate \NoCaseChange{\protect\cite[{Def. 4}]{cite754}}.
\item\relax If a qubit code \(Q\) of length \(n\) has compact subgroups \(N\triangleleft G\leq \mathrm{Aut}(Q)\) such that \(G/N\) is finite, non-Abelian, simple, and not \(A_5\), then \(n\) is at least the minimal permutation degree \(\mu(G/N)\) \NoCaseChange{\protect\cite[{Thm. 1}]{cite723}}.
\end{eczvaluelist}
}\\ 
\addlinespace[\myxtraspc]
\eczhRefIndex{code:qubit_stabilizer}%
\eczhListValue{\flmRefsHyperref{code:qubit_stabilizer}{Qubit stabilizer code}} & \eczhListValue{\begin{eczvaluelist}
\item\relax Logical Pauli transformations \NoCaseChange{\protect\cite{cite736,cite772,cite773}}.
\item\relax The four-block transversal gate mapping each \(X \to IXXX\) and each \(Z \to IZZZ\) implements the same logical gate on all qubits \NoCaseChange{\protect\cite{cite736}\protect\cite[{Exam. 1}]{cite532}}. More generally, in the \(t\)-block setting, the commutant of the \(t\)-fold action of the \(n\)-qubit Clifford group contains invertible operators realizing the stochastic orthogonal group \(O_t(2)\), yielding inter-block symmetries in tensor powers of stabilizer constructions \NoCaseChange{\protect\cite{cite774}}.
\item\relax Transversal logical gates are in a finite level of the \flmTerm{term}{ref694}{}{Clifford hierarchy}, which is shown using stabilizer \textit{disjointness} \NoCaseChange{\protect\cite{cite775}} (see also \NoCaseChange{\protect\cite{cite776,cite777}}).
Transversal gates for \(n\in\{1,2\}\) are semi-Clifford \NoCaseChange{\protect\cite{cite778}}.

\item\relax No stabilizer code can implement a classical universal gate set transversally \NoCaseChange{\protect\cite{cite779}}.
\item\relax Fold-transversal gates have been extended from qubit CSS codes to qubit stabilizer codes, and there is an algorithm to determine them from the stabilizer group \NoCaseChange{\protect\cite{cite719}}.
\item\relax Computation can be sped up substantially for codes that admit transversal measurements of logical \(X\) and \(Z\) \NoCaseChange{\protect\cite{cite780}}.
\item\relax Diagonal transversal \flmRefsHyperref{ref409}{Clifford gates} on multiple code blocks must form one of six families of matrix groups: \(O(\ell,\mathbb{F}_2)\) generically, \(U(\ell,\mathbb{F}_4)\) for Hermitian qubit codes, \(GL(\ell,\mathbb{F}_2)\) for non-self-dual CSS codes, \(O( \ell, \mathbb{F}_2[x]/(x^2) )\) for self-dual non-CSS codes, \(U(\ell,R_8)\) for intermediate semi-self-dual CSS or self-dual semi-CSS cases, and \(Sp(2\ell,\mathbb{F}_2)\) for self-dual CSS codes \NoCaseChange{\protect\cite{cite738}}. There are two \(\llbracket 8,1,3\rrbracket \) self-dual non-CSS codes; see QECDB \NoCaseChange{\protect\cite{cite781}}.
\item\relax An entangling diagonal transversal two-qubit \flmRefsHyperref{ref409}{Clifford gate} exists only for codes equivalent to CSS or self-dual codes; canonical representatives are the inter-block CNOT in the CSS case and a \(Y\)-controlled-\(Y\) gate, locally Clifford equivalent to \(CZ\), in the self-dual case \NoCaseChange{\protect\cite{cite738}}.
\item\relax Implementing the full Clifford group on \(k\) logical qubits requires \(k\)-fold transversal gates; in particular, stabilizer codes cannot admit a transversal (1-fold) implementation of the full \flmRefsHyperref{ref409}{Clifford group} on more than one logical qubit, a fold-transversal (2-fold) implementation on more than two logical qubits, or a code-automorphism implementation on more than one logical qubit \NoCaseChange{\protect\cite{cite782}}. These bounds are tight: \(k\) copies of the \(\llbracket 7,1,3\rrbracket \) Steane code form a \(\llbracket 7k,k,3\rrbracket \) code admitting a \(k\)-fold transversal \flmRefsHyperref{ref409}{Clifford group} \NoCaseChange{\protect\cite{cite782}}.
\end{eczvaluelist}
}\\ 
\addlinespace[\myxtraspc]
\eczhRefIndex{code:rotated_surface}%
\eczhListValue{\flmRefsHyperref{code:rotated_surface}{Rotated surface code}} & \eczhListValue{\begin{eczvaluelist}
\item\relax Fold-transversal \(S\) gate \NoCaseChange{\protect\cite{cite783,cite748}}.
\end{eczvaluelist}
}\\ 
\addlinespace[\myxtraspc]
\eczhRefIndex{code:self_dual_css}%
\eczhListValue{\flmRefsHyperref{code:self_dual_css}{Self-dual CSS code}} & \eczhListValue{\begin{eczvaluelist}
\item\relax Self-dual CSS codes admit a transversal Hadamard gate. There are criteria for when such codes realize logical gates from tensor products of \(S\) and \(S^{\dagger}\) gates \NoCaseChange{\protect\cite{cite784}}.
\item\relax Diagonal transversal \flmRefsHyperref{ref409}{Clifford gates} on \(\ell\) codeblocks of a CSS code form \(Sp(2\ell,\mathbb{F}_2)\) for self-dual CSS codes \NoCaseChange{\protect\cite{cite738}}.
\item\relax A self-dual weakly doubly even \(\llbracket n,1,d\rrbracket \) CSS code admits a partitioned transversal physical \(S\) gate that realizes \(\overline{S}^m\), where \(m=|M^+|-|M^-| \pmod 4\); for odd \(m\), together with transversal Hadamard and CNOT, this yields the full logical Clifford group transversally \NoCaseChange{\protect\cite{cite731}\protect\cite[{Lemma 4}]{cite760}}.
\end{eczvaluelist}
}\\ 
\addlinespace[\myxtraspc]
\eczhRefIndex{code:488_color}%
\eczhListValue{\flmRefsHyperref{code:488_color}{Square-octagon (4.8.8) color code}} & \eczhListValue{\begin{eczvaluelist}
\item\relax CNOT gate because the code is CSS.
\item\relax Hadamard gates for any qubit geometry which yields a self-dual CSS code.
\item\relax Transversal \(S\) gate \NoCaseChange{\protect\cite{cite710,cite432}}.
\item\relax Transversal logical \flmRefsHyperref{ref409}{Clifford gates} in the Union Jack formulation \NoCaseChange{\protect\cite{cite431}}.
\item\relax Single-qubit Clifford and CNOT gates between qubits encoded in holes in the lattice can be implemented via braiding \NoCaseChange{\protect\cite{cite713}}.
\end{eczvaluelist}
}\\ 
\addlinespace[\myxtraspc]
\eczhRefIndex{code:sslp}%
\eczhListValue{\flmRefsHyperref{code:sslp}{Subset-Sum-Linear-Programming (SS-LP) code}} & \eczhListValue{\begin{eczvaluelist}
\item\relax SS-LP codes are optimized to admit diagonal gates transversally and include \(\llparenthesis 7,2,3\rrparenthesis \) codes that realize the \(\mathsf{BD}_{16}\) and \(\mathsf{BD}_{32}\) groups transversally, yielding \(T\) and \(\sqrt{T}\) gates, respectively. Larger codes include an \(\llparenthesis 8,2,3\rrparenthesis \) code that transversally realizes \(\mathsf{BD}_{64}\).
\end{eczvaluelist}
}\\ 
\addlinespace[\myxtraspc]
\eczhRefIndex{code:shyps}%
\eczhListValue{\flmRefsHyperref{code:shyps}{Subsystem Hypergraph Product Simplex (SHYPS) code}} & \eczhListValue{\begin{eczvaluelist}
\item\relax Cross-block transversal CNOT gates \(\prod_{i=1}^n CNOT_{i, n+(\sigma_1\otimes\sigma_2)(i)}\) for \(\sigma_1, \sigma_2\) automorphisms of the classical simplex code. These operators implement a generating set of cross-block logical CNOT gates.
\item\relax In-block phase-type fold-transversal gates with \(Sp(2k,\mathbb{F}_2)\) representation \(\begin{pmatrix} I & (\sigma \otimes \sigma^T)\tau \\ 0 & 1\end{pmatrix}\) for \(\sigma\) an automorphism of the classical simplex code, and \(\tau\) a self-inverse permutation which acts like \(\tau (e_i \otimes e_j) = e_j \otimes e_i\) for the canonical basis \(\{e_i\}\) of \(\mathbb{F}_2^{\sqrt{n}}\). These operators implement a generating set of logical in-block diagonal gates \NoCaseChange{\protect\cite{cite785}}.
\item\relax Cross-block phase-type fold-transversal gates \(\prod_{i=1}^n CZ_{i, n+(\sigma_1\otimes\sigma_2)\tau(i)}\) for \(\sigma_1, \sigma_2\) automorphisms of the classical simplex code and \(\tau\) as above. These operators implement a generating set of logical cross-block diagonal gates \NoCaseChange{\protect\cite{cite785}}.
\item\relax Fold-transversal Hadamard gate \(H^{\otimes n} \tau \), with \(\tau\) as above. Implements logical Hadamard-SWAP operator \(H^{\otimes k} \tau_k \), with \(\tau_k\) defined analogously to \(\tau\) \NoCaseChange{\protect\cite{cite785}}.
\end{eczvaluelist}
}\\ 
\addlinespace[\myxtraspc]
\eczhRefIndex{code:subsystem_color}%
\eczhListValue{\flmRefsHyperref{code:subsystem_color}{Subsystem color code}} & \eczhListValue{\begin{eczvaluelist}
\item\relax For a \(D\)-dimensional \((d,e)\) gauge color code, \(CNOT\) is transversal, Hadamard is transversal when \(d=e\), and \(R_n=\operatorname{diag}(1,e^{2\pi i/2^n})\) is transversal whenever \(D \geq n(D-e)\) \NoCaseChange{\protect\cite{cite475}}.
\end{eczvaluelist}
}\\ 
\addlinespace[\myxtraspc]
\eczhRefIndex{code:subsystem_qubits_into_qubits}%
\eczhListValue{\flmRefsHyperref{code:subsystem_qubits_into_qubits}{Subsystem qubit code}} & \eczhListValue{\begin{eczvaluelist}
\item\relax If a subsystem qubit code \(Q\) of length \(n\) has compact subgroups \(N\triangleleft G\leq \mathrm{Aut}(Q)\) such that \(G/N\) is finite, non-Abelian, simple, and not \(A_5\), then \(n\) is at least the minimal permutation degree \(\mu(G/N)\) \NoCaseChange{\protect\cite[{Thm. 1}]{cite723}}.
\end{eczvaluelist}
}\\ 
\addlinespace[\myxtraspc]
\eczhRefIndex{code:quantum_lego}%
\eczhListValue{\flmRefsHyperref{code:quantum_lego}{Tensor-network code}} & \eczhListValue{\begin{eczvaluelist}
\item\relax The quantum Lego framework yields an \(\llbracket 8,1,2\rrbracket \) stabilizer code that admits a transversal logical \(T\) gate that originates from that of a trivial (distance-one) \(\llbracket 7,1\rrbracket \) code. This code, in turn, is obtained from the \(\llbracket 15,1,3\rrbracket \) code \NoCaseChange{\protect\cite{cite786}}.
\end{eczvaluelist}
}\\ 
\addlinespace[\myxtraspc]
\eczhRefIndex{code:tetrahedral_color}%
\eczhListValue{\flmRefsHyperref{code:tetrahedral_color}{Tetrahedral color code}} & \eczhListValue{\begin{eczvaluelist}
\item\relax A \(\llbracket 5d^3-12d^2+16,3,d\rrbracket \) close relative of this code admits a logical \(CCZ\) gate via single-qubit rotations; for this family, stabilizers remain asymptotically constant-weight and can be gauge-reduced to weight at most six \NoCaseChange{\protect\cite{cite759}}.
\end{eczvaluelist}
}\\ 
\addlinespace[\myxtraspc]
\eczhRefIndex{code:toric}%
\eczhListValue{\flmRefsHyperref{code:toric}{Toric code}} & \eczhListValue{\begin{eczvaluelist}
\item\relax Transversal logical Pauli gates correspond to Pauli strings on non-trivial loops of the torus.
\end{eczvaluelist}
}\\ 
\addlinespace[\myxtraspc]
\eczhRefIndex{code:triangle_surface}%
\eczhListValue{\flmRefsHyperref{code:triangle_surface}{Triangular surface code}} & \eczhListValue{\begin{eczvaluelist}
\item\relax Triangle codes admit transversal order-three single-qubit gates in the \flmRefsHyperref{ref409}{Clifford group}, e.g., \(\bar{SH}\) \NoCaseChange{\protect\cite{cite433}}.
\end{eczvaluelist}
}\\ 
\addlinespace[\myxtraspc]
\eczhRefIndex{code:quantum_triorthogonal}%
\eczhListValue{\flmRefsHyperref{code:quantum_triorthogonal}{Triorthogonal code}} & \eczhListValue{\begin{eczvaluelist}
\item\relax Transversal action of \(T\) gates on all qubits, followed by a particular pattern of \(CZ\) and \(S\) gates, will realize a logical \(T\) gate \NoCaseChange{\protect\cite[{Lemma 2}]{cite691}\protect\cite[{Rem. 3}]{cite760}}. When an additional condition on logical-\(X\) operators is satisfied, the \(CZ\) and \(S\) gates are no longer necessary \NoCaseChange{\protect\cite[{Thm. 14}]{cite724}}.
\item\relax Triorthogonality is necessary but not sufficient for physical transversal \(T\) gates on each qubit to realize the identity logical gate \NoCaseChange{\protect\cite[{Thm. 12}]{cite724}}.
\item\relax Certain codes realize controlled-controlled-\(Z\) gates \NoCaseChange{\protect\cite{cite787}}, realized via physical \(CCZ\) gates on three code blocks.
\item\relax Triorthogonal codes realizing logical \(T\) gates using only physical \(T\) gates can be paired up with self-dual CSS codes to yield a transversal CNOT gate \NoCaseChange{\protect\cite{cite788}}.
\end{eczvaluelist}
}\\ 
\addlinespace[\myxtraspc]
\eczhRefIndex{code:4612_color}%
\eczhListValue{\flmRefsHyperref{code:4612_color}{Truncated trihexagonal (4.6.12) color code}} & \eczhListValue{\begin{eczvaluelist}
\item\relax CNOT gate because the code is CSS.
\item\relax Hadamard gates for any qubit geometry which yields a self-dual CSS code.
\end{eczvaluelist}
}\\ 
\addlinespace[\myxtraspc]
\eczhRefIndex{code:t_group}%
\eczhListValue{\flmRefsHyperref{code:t_group}{Twisted \(1\)-group code}} & \eczhListValue{\begin{eczvaluelist}
\item\relax All gates in the underlying twisted \(1\)-group. See \NoCaseChange{\protect\cite[{Table II}]{cite789}} for other notable groups including the sporadic groups.
\end{eczvaluelist}
}\\ 
\addlinespace[\myxtraspc]
\eczhRefIndex{code:2bga}%
\eczhListValue{\flmRefsHyperref{code:2bga}{Two-block group-algebra (2BGA) codes}} & \eczhListValue{\begin{eczvaluelist}
\item\relax Logical Pauli operators and fold-transversal gates have been studied \NoCaseChange{\protect\cite{cite718,cite719}}.
\end{eczvaluelist}
}\\ 
\addlinespace[\myxtraspc]
\eczhRefIndex{code:vbs}%
\eczhListValue{\flmRefsHyperref{code:vbs}{Valence-bond-solid (VBS) code}} & \eczhListValue{\begin{eczvaluelist}
\item\relax Two classes of (approximate) VBS codes have \(SU(q)\) transversal gates \NoCaseChange{\protect\cite[{Tab. III}]{cite790}}.
\end{eczvaluelist}
}\\ 
\addlinespace[\myxtraspc]
\eczhRefIndex{code:w_state}%
\eczhListValue{\flmRefsHyperref{code:w_state}{W-state code}} & \eczhListValue{\begin{eczvaluelist}
\item\relax All logical gates can be implemented transversally. The logical unitary \(U_L\) can be performed with the physical unitary \(U_L\otimes U_L\otimes\cdots\otimes U_L\), where on the physical space \(U_L\) is taken to act trivially on \(\ket\perp\), i.e., \( U_L\ket\perp = \ket\perp\).
\end{eczvaluelist}
}\\ 
\addlinespace[\myxtraspc]
\eczhRefIndex{code:su3_sigma360}%
\eczhListValue{\flmRefsHyperref{code:su3_sigma360}{\(\llparenthesis 5,3,2\rrparenthesis _3\) qutrit code}} & \eczhListValue{\begin{eczvaluelist}
\item\relax \(\Sigma(360\phi)=3.A_6\) group gates can be realized transversally.
\end{eczvaluelist}
}\\ 
\addlinespace[\myxtraspc]
\eczhRefIndex{code:qubit_6_2_3}%
\eczhListValue{\flmRefsHyperref{code:qubit_6_2_3}{\(\llparenthesis 6,2,3\rrparenthesis \) transversal-\(\mathbb{Z}_{10}\) code}} & \eczhListValue{\begin{eczvaluelist}
\item\relax The \(\mathbb{Z}_{10}\) group can be realized transversally \NoCaseChange{\protect\cite{cite528}}. This is the smallest known distance-three code supporting a transversal gate outside of the \flmRefsHyperref{ref409}{Clifford group}.
\end{eczvaluelist}
}\\ 
\addlinespace[\myxtraspc]
\eczhRefIndex{code:icosahedral_permutation_invariant}%
\eczhListValue{\flmRefsHyperref{code:icosahedral_permutation_invariant}{\(\llparenthesis 7,2,3\rrparenthesis \) Pollatsek-Ruskai code}} & \eczhListValue{\begin{eczvaluelist}
\item\relax Binary icosahedral group \(2I\) gates can be realized transversally \NoCaseChange{\protect\cite{cite647}}. See Ref. \NoCaseChange{\protect\cite{cite528}} for other non-PI codes realizing \(2I\) gates transversally.
\end{eczvaluelist}
}\\ 
\addlinespace[\myxtraspc]
\eczhRefIndex{code:qubit_8_4_2}%
\eczhListValue{\flmRefsHyperref{code:qubit_8_4_2}{\(\llparenthesis 8,16,2\rrparenthesis \) \(PG(3,2)\) code}} & \eczhListValue{\begin{eczvaluelist}
\item\relax Even physical-qubit permutations act as \(GL(4,\mathbb{F}_2)\) on the logical basis, and odd permutations extend the permutation automorphism group to the full symmetric group \(S_8\) \NoCaseChange{\protect\cite{cite723}}.
\item\relax \(T^{\otimes 8}\) is a transversal non-Clifford gate implementing \(2\ket{\overline{0}}\bra{\overline{0}}-I\) on the logical subspace \NoCaseChange{\protect\cite{cite723}}.
\item\relax A specific odd permutation implements a non-Clifford logical involution, and the full permutation automorphism group is \(S_8\) \NoCaseChange{\protect\cite{cite723}}.
\end{eczvaluelist}
}\\ 
\addlinespace[\myxtraspc]
\eczhRefIndex{code:4d_13_surface}%
\eczhListValue{\flmRefsHyperref{code:4d_13_surface}{\((1,3)\) 4D toric code}} & \eczhListValue{\begin{eczvaluelist}
\item\relax Logical \(CCCZ\) gate on a hyper-diamond lattice \NoCaseChange{\protect\cite{cite479}}.
\end{eczvaluelist}
}\\ 
\addlinespace[\myxtraspc]
\eczhRefIndex{code:4d_surface}%
\eczhListValue{\flmRefsHyperref{code:4d_surface}{\((2,2)\) Loop toric code}} & \eczhListValue{\begin{eczvaluelist}
\item\relax Only logical \flmRefsHyperref{ref409}{Clifford gates} can be implemented transversally when defined on a hypercubic lattice \NoCaseChange{\protect\cite{cite479}}.
\end{eczvaluelist}
}\\ 
\addlinespace[\myxtraspc]
\eczhRefIndex{code:stab_10_2_3}%
\eczhListValue{\flmRefsHyperref{code:stab_10_2_3}{\(\llbracket 10,2,3\rrbracket \) binarized Galois-qudit code}} & \eczhListValue{\begin{eczvaluelist}
\item\relax The code is permutation-equivalent to its Hadamard dual: transversal physical Hadamards followed by swapping the two qubits in each binarized \(\mathbb{F}_4\) coordinate preserve the codespace and implement a logical \(H^{\otimes 2}\) up to logical \(\mathrm{SWAP}\). This is the binarized version of the permutation-assisted Hadamard available from plain self-duality of the starting \(q=4\) Galois-qudit CSS code \NoCaseChange{\protect\cite{cite514}}.
\end{eczvaluelist}
}\\ 
\addlinespace[\myxtraspc]
\eczhRefIndex{code:xzzx_10_2_3}%
\eczhListValue{\flmRefsHyperref{code:xzzx_10_2_3}{\(\llbracket 10,2,3\rrbracket \) rotated toric code}} & \eczhListValue{\begin{eczvaluelist}
\item\relax Transversal Hadamard-SWAP: qubit permutation \(m \mapsto -m\) (mod 10) paired with sublattice swap \(A \leftrightarrow B\), followed by Hadamard on all qubits \NoCaseChange{\protect\cite{cite440}}.
\item\relax Logical Hadamard without SWAP via \(m \mapsto 3m\) (mod 10) followed by Hadamard \NoCaseChange{\protect\cite{cite440}}.
\end{eczvaluelist}
}\\ 
\addlinespace[\myxtraspc]
\eczhRefIndex{code:qutrit_golay}%
\eczhListValue{\flmRefsHyperref{code:qutrit_golay}{\(\llbracket 11,1,5\rrbracket _3\) qutrit Golay code}} & \eczhListValue{\begin{eczvaluelist}
\item\relax All single-qutrit encoded Clifford gates \NoCaseChange{\protect\cite{cite706}}.
\end{eczvaluelist}
}\\ 
\addlinespace[\myxtraspc]
\eczhRefIndex{code:stab_12_2_2}%
\eczhListValue{\flmRefsHyperref{code:stab_12_2_2}{\(\llbracket 12,2,2\rrbracket \) CSS code}} & \eczhListValue{\begin{eczvaluelist}
\item\relax Logical \(CS_{01}\) gate via transversal application of \(T\) on qubits \(0,3,4,7,8,11\), \(T^3\) on qubits \(1,2,5,6\), and \(T^7 = T^{\dagger}\) on qubits \(9,10\) \NoCaseChange{\protect\cite[{Fig. 6}]{cite767}}.
\end{eczvaluelist}
}\\ 
\addlinespace[\myxtraspc]
\eczhRefIndex{code:carbon}%
\eczhListValue{\flmRefsHyperref{code:carbon}{\(\llbracket 12,2,4\rrbracket \) carbon code}} & \eczhListValue{\begin{eczvaluelist}
\item\relax Two-block CNOT gates are transversal because the code is CSS.
\item\relax Automorphism groups of the underlying classical codes can yield transversal \flmRefsHyperref{ref409}{Clifford gates} when combined with qubit permutations \NoCaseChange{\protect\cite{cite763}}. In particular, logical Hadamard is realized by a transversal physical Hadamard followed by a qubit permutation, and a logical one-block CNOT is implemented by a qubit permutation \NoCaseChange{\protect\cite{cite448,cite525}}.
\end{eczvaluelist}
}\\ 
\addlinespace[\myxtraspc]
\eczhRefIndex{code:stab_13_1_5}%
\eczhListValue{\flmRefsHyperref{code:stab_13_1_5}{\(\llbracket 13,1,5\rrbracket \) twisted toric code}} & \eczhListValue{\begin{eczvaluelist}
\item\relax No non-Pauli transversal gates.
\end{eczvaluelist}
}\\ 
\addlinespace[\myxtraspc]
\eczhRefIndex{code:gross}%
\eczhListValue{\flmRefsHyperref{code:gross}{\(\llbracket 144,12,12\rrbracket \) gross code}} & \eczhListValue{\begin{eczvaluelist}
\item\relax Logical Pauli operators and fold-transversal gates studied in Ref. \NoCaseChange{\protect\cite{cite719}}.
\end{eczvaluelist}
}\\ 
\addlinespace[\myxtraspc]
\eczhRefIndex{code:stab_15_7_3}%
\eczhListValue{\flmRefsHyperref{code:stab_15_7_3}{\(\llbracket 15, 7, 3\rrbracket \) quantum Hamming code}} & \eczhListValue{\begin{eczvaluelist}
\item\relax CNOT gate because it is a CSS code.
\item\relax In a suitable logical basis, single-qubit Clifford operations applied transversally yield the corresponding \flmRefsHyperref{ref409}{Clifford gates} on each of the seven logical qubits \NoCaseChange{\protect\cite{cite791}}.
\item\relax Automorphism groups of the underlying classical codes can yield transversal \flmRefsHyperref{ref409}{Clifford gates} when combined with qubit permutations \NoCaseChange{\protect\cite[{Sec. IV.A}]{cite763}}.
\item\relax Transversal interblock \(CCZ\) gate \NoCaseChange{\protect\cite{cite787}}.
\end{eczvaluelist}
}\\ 
\addlinespace[\myxtraspc]
\eczhRefIndex{code:stab_15_1_3}%
\eczhListValue{\flmRefsHyperref{code:stab_15_1_3}{\(\llbracket 15,1,3\rrbracket \) quantum RM code}} & \eczhListValue{\begin{eczvaluelist}
\item\relax A transversal logical \(T\) is implemented by applying a \(T^\dagger\) gate on every qubit \NoCaseChange{\protect\cite{cite792,cite793,cite707}}. This is the smallest qubit stabilizer code with a (strongly) transversal gate outside of the \flmRefsHyperref{ref409}{Clifford group} \NoCaseChange{\protect\cite{cite794}}.
\item\relax Transversal logical \(CS\) gates between code blocks, together with transversal \(T\) and CNOT, can be used to realize IQP-like sampling architectures on a hypercube connectivity graph \NoCaseChange{\protect\cite{cite759}}.
\end{eczvaluelist}
}\\ 
\addlinespace[\myxtraspc]
\eczhRefIndex{code:stab_16_6_4}%
\eczhListValue{\flmRefsHyperref{code:stab_16_6_4}{\(\llbracket 16,6,4\rrbracket \) Tesseract color code}} & \eczhListValue{\begin{eczvaluelist}
\item\relax Global transversal \(S\) implements a logical circuit composed of \(CZ\) and \(Z\) gates \NoCaseChange{\protect\cite{cite753,cite724}}
\item\relax Transversal Hadamard can be chosen to swap three pairs of logical qubits, allowing even-weight Hadamard-product measurements in magic-state distillation protocols \NoCaseChange{\protect\cite[{Sec. III}]{cite101}}.
\end{eczvaluelist}
}\\ 
\addlinespace[\myxtraspc]
\eczhRefIndex{code:stab_17_1_5}%
\eczhListValue{\flmRefsHyperref{code:stab_17_1_5}{\(\llbracket 17,1,5\rrbracket \) 4.8.8 color code}} & \eczhListValue{\begin{eczvaluelist}
\item\relax Transversal implementation of the logical Clifford group.
\end{eczvaluelist}
}\\ 
\addlinespace[\myxtraspc]
\eczhRefIndex{code:stab_18_2_5}%
\eczhListValue{\flmRefsHyperref{code:stab_18_2_5}{\(\llbracket 18,2,5\rrbracket \) BCC code}} & \eczhListValue{\begin{eczvaluelist}
\item\relax Transversal Hadamard-SWAP logical gate (inherited from BCC code structure) \NoCaseChange{\protect\cite{cite440}}.
\item\relax Logical Hadamard without SWAP: the code is self-dual, so transversal physical Hadamard implements logical \(H^{\otimes 2}\) \NoCaseChange{\protect\cite{cite440}}.
\end{eczvaluelist}
}\\ 
\addlinespace[\myxtraspc]
\eczhRefIndex{code:stab_20_2_6}%
\eczhListValue{\flmRefsHyperref{code:stab_20_2_6}{\(\llbracket 20,2,6\rrbracket \) B\&C phantom code}} & \eczhListValue{\begin{eczvaluelist}
\item\relax Logical CNOT gates in both directions between the two logical qubits are realized by qubit permutations within a code block \NoCaseChange{\protect\cite{cite514,cite795}}.
\item\relax Fold-diagonal logical \(SS\) gates are available from the self-duality of the outer \(\llbracket 5,1,3\rrbracket _4\) code and the inner \(\llbracket 4,2,2\rrbracket \) layer \NoCaseChange{\protect\cite{cite514}}.
\end{eczvaluelist}
}\\ 
\addlinespace[\myxtraspc]
\eczhRefIndex{code:qubit_golay}%
\eczhListValue{\flmRefsHyperref{code:qubit_golay}{\(\llbracket 23, 1, 7\rrbracket \) Quantum Golay code}} & \eczhListValue{\begin{eczvaluelist}
\item\relax \flmRefsHyperref{ref409}{Single-qubit Clifford group} by choosing \(\overline{U}=U^{\otimes 23}\) for every Clifford unitary \(U\) \NoCaseChange{\protect\cite{cite796}}.
\end{eczvaluelist}
}\\ 
\addlinespace[\myxtraspc]
\eczhRefIndex{code:hypercube_quantum}%
\eczhListValue{\flmRefsHyperref{code:hypercube_quantum}{\(\llbracket 2^D,D,2\rrbracket \) hypercube quantum code}} & \eczhListValue{\begin{eczvaluelist}
\item\relax  \(CZ\), \(CCZ\), and generalized \(CZ\) gates at the \((D-1)\)-st level of the Clifford hierarchy \NoCaseChange{\protect\cite{cite797}\protect\cite[{Exam. 6.10}]{cite798}}. CNOT and SWAP gates can be realized by qubit permutations \NoCaseChange{\protect\cite{cite759}}.
\end{eczvaluelist}
}\\ 
\addlinespace[\myxtraspc]
\eczhRefIndex{code:quantum_hamming_css}%
\eczhListValue{\flmRefsHyperref{code:quantum_hamming_css}{\(\llbracket 2^r-1, 2^r-2r-1, 3\rrbracket \) quantum Hamming code}} & \eczhListValue{\begin{eczvaluelist}
\item\relax Pauli, Hadamard, and CNOT gates.
\end{eczvaluelist}
}\\ 
\addlinespace[\myxtraspc]
\eczhRefIndex{code:diagonal_clifford}%
\eczhListValue{\flmRefsHyperref{code:diagonal_clifford}{\(\llbracket 2^r-1,1,3\rrbracket \) simplex code}} & \eczhListValue{\begin{eczvaluelist}
\item\relax Each code transversally implements a diagonal gate at the \((r-1)\)st level of the \flmTerm{term}{ref694}{}{Clifford hierarchy} in the form of a \(Z\)-rotation by angle \(-\pi/2^{r-1}\) \NoCaseChange{\protect\cite{cite799,cite800}}. These are the smallest distance-three qubit stabilizer codes with such a (strongly) transversal gate \NoCaseChange{\protect\cite{cite794}}.
\end{eczvaluelist}
}\\ 
\addlinespace[\myxtraspc]
\eczhRefIndex{code:single_qubit_clifford}%
\eczhListValue{\flmRefsHyperref{code:single_qubit_clifford}{\(\llbracket 2^{2r-1}-1,1,2^r-1\rrbracket \) quantum punctured RM code}} & \eczhListValue{\begin{eczvaluelist}
\item\relax All single-qubit gates in the \flmRefsHyperref{ref409}{Clifford group}.
\end{eczvaluelist}
}\\ 
\addlinespace[\myxtraspc]
\eczhRefIndex{code:iceberg}%
\eczhListValue{\flmRefsHyperref{code:iceberg}{\(\llbracket 2m,2m-2,2\rrbracket \) error-detecting code}} & \eczhListValue{\begin{eczvaluelist}
\item\relax Transveral CNOT gates can be performed by first teleporting qubits into different code blocks \NoCaseChange{\protect\cite[{Sec. VII}]{cite737}}.
\item\relax For \(2m\) being a multiple of four, this code houses a transversal representation of the single-qubit Clifford group \NoCaseChange{\protect\cite[{Sec. 3.1}]{cite801}}. Its \(n\)-block version houses a transversal representation of the \(n\)-qubit Clifford group \NoCaseChange{\protect\cite[{Secs. 3.1 and 3.4}]{cite801}}. The commutant of transversal representations of the Clifford group contains qubit permutations and projections onto the \(\llbracket 2m,2m-2,2\rrbracket \) error-detecting code \NoCaseChange{\protect\cite{cite802}}.
\end{eczvaluelist}
}\\ 
\addlinespace[\myxtraspc]
\eczhRefIndex{code:stellated_dodecahedron_css}%
\eczhListValue{\flmRefsHyperref{code:stellated_dodecahedron_css}{\(\llbracket 30,8,3\rrbracket \) Bring code}} & \eczhListValue{\begin{eczvaluelist}
\item\relax \flmRefsHyperref{ref409}{Clifford group} of four of the eight logical qubits can be implemented by transversal gates combined with qubit permutations \NoCaseChange{\protect\cite{cite762}}.
\end{eczvaluelist}
}\\ 
\addlinespace[\myxtraspc]
\eczhRefIndex{code:small_triorthogonal}%
\eczhListValue{\flmRefsHyperref{code:small_triorthogonal}{\(\llbracket 3k + 8, k, 2\rrbracket \) triorthogonal code}} & \eczhListValue{\begin{eczvaluelist}
\item\relax The code admits a transversal \(T\) gate \NoCaseChange{\protect\cite[{Lemma 2}]{cite691}}.
\end{eczvaluelist}
}\\ 
\addlinespace[\myxtraspc]
\eczhRefIndex{code:stab_4_1_2}%
\eczhListValue{\flmRefsHyperref{code:stab_4_1_2}{\(\llbracket 4,1,2\rrbracket \) twist-defect code}} & \eczhListValue{\begin{eczvaluelist}
\item\relax Weight-two transversal logical Pauli \(X,Y,Z\) operations \NoCaseChange{\protect\cite{cite803}}.
\end{eczvaluelist}
}\\ 
\addlinespace[\myxtraspc]
\eczhRefIndex{code:stab_4_2_2}%
\eczhListValue{\flmRefsHyperref{code:stab_4_2_2}{\(\llbracket 4,2,2\rrbracket \) Four-qubit code}} & \eczhListValue{\begin{eczvaluelist}
\item\relax A tensor product of Hadamard gates applies a Hadamard gate to both logical qubits, and a tensor product of \(S=\sqrt{Z}\) gates applies a \(CZ\) gate followed by a logical \(Z\) on both qubits \NoCaseChange{\protect\cite{cite804}} (see also \NoCaseChange{\protect\cite{cite805}}). A logical \(CZ\) gate is then realized by \(\sqrt{Z}\otimes\sqrt{Z}^{\dagger}\otimes\sqrt{Z}^{\dagger}\otimes\sqrt{Z}\). With a different logical basis, transversal Hadamard swaps the two logical qubits, enabling control-SWAP and \(H^{\otimes 2}\)-measurement routines for quadratic magic-state distillation \NoCaseChange{\protect\cite[{Sec. I.B.1}]{cite101}}.
\item\relax This code is the only four-qubit subspace to house a transversal representation of the single-qubit Clifford group but not of the single-qubit unitary group \NoCaseChange{\protect\cite[{Eq. (104)}]{cite801}}. Equivalently, its projector is the only extra generator of the fourth tensor-power Clifford commutant beyond qubit permutations \NoCaseChange{\protect\cite[{Thm. 1}]{cite801}}. Its \(n\)-block version is a \(\llbracket 4n,2n,2\rrbracket \) code, which houses a signed permutation representation of the \(n\)-qubit Clifford group \NoCaseChange{\protect\cite[{Sec. 3.1 and Appx. B}]{cite801}}.
\end{eczvaluelist}
}\\ 
\addlinespace[\myxtraspc]
\eczhRefIndex{code:stab_47_1_11}%
\eczhListValue{\flmRefsHyperref{code:stab_47_1_11}{\(\llbracket 47,1,11\rrbracket \) quantum QR code}} & \eczhListValue{\begin{eczvaluelist}
\item\relax All logical \flmRefsHyperref{ref409}{Clifford gates} are realized transversally \NoCaseChange{\protect\cite{cite760}}.
\end{eczvaluelist}
}\\ 
\addlinespace[\myxtraspc]
\eczhRefIndex{code:stab_49_1_5}%
\eczhListValue{\flmRefsHyperref{code:stab_49_1_5}{\(\llbracket 49,1,5\rrbracket \) triorthogonal code}} & \eczhListValue{\begin{eczvaluelist}
\item\relax The code admits a transversal \(T\) gate \NoCaseChange{\protect\cite[{Appx. B}]{cite691}}.
\end{eczvaluelist}
}\\ 
\addlinespace[\myxtraspc]
\eczhRefIndex{code:stab_5_1_3}%
\eczhListValue{\flmRefsHyperref{code:stab_5_1_3}{\(\llbracket 5,1,3\rrbracket \) Five-qubit perfect code}} & \eczhListValue{\begin{eczvaluelist}
\item\relax A non-Pauli Hadamard-phase "facet" gate \(SH\) and three-qubit Clifford operation \(M_3\) \NoCaseChange{\protect\cite{cite737,cite736}}. These realize the \(2T\) binary tetrahedral subgroup of \(SU(2)\).
\item\relax The entire logical Clifford group can be realized using fold-transversal gates \NoCaseChange{\protect\cite{cite806,cite807,cite719}}.
\item\relax The code does not admit any \flmRefsHyperref{ref409}{non-Clifford} transversal gates \NoCaseChange{\protect\cite{cite446}}; in particular, see \NoCaseChange{\protect\cite{cite808}} for the case of collective \(Z\) rotations.
\item\relax Transversal gates can be interpreted as monodromies under a particular notion of parallel transport \NoCaseChange{\protect\cite[{Exam. 6.4.2}]{cite809}}.
\end{eczvaluelist}
}\\ 
\addlinespace[\myxtraspc]
\eczhRefIndex{code:css_6_1_2}%
\eczhListValue{\flmRefsHyperref{code:css_6_1_2}{\(\llbracket 6,1,2\rrbracket \) semi-self-dual CSS code}} & \eczhListValue{\begin{eczvaluelist}
\item\relax The physical gate \(S^{\dagger}_{0}S^{\dagger}_{1}S^{\dagger}_{2}S_{3}S_{4}S_{5}\) implements the logical action \(\bar{S}\) \NoCaseChange{\protect\cite[{ID 59}]{cite453}}.
\item\relax In the equivalent self-dual semi-CSS presentation obtained by fixing one logical qubit of the \(\llbracket 6,2,2\rrbracket \) \(C_6\) code to \(|Y^{-}\rangle_L\), transversal gates support a fault-tolerant magic-state-preparation circuit \NoCaseChange{\protect\cite{cite738}\protect\cite[{ID 59}]{cite453}}.
\end{eczvaluelist}
}\\ 
\addlinespace[\myxtraspc]
\eczhRefIndex{code:stab_6_2_2}%
\eczhListValue{\flmRefsHyperref{code:stab_6_2_2}{\(\llbracket 6,2,2\rrbracket \) \(C_6\) code}} & \eczhListValue{\begin{eczvaluelist}
\item\relax Transversal physical Hadamards preserve the codespace because the displayed \(X\)- and \(Z\)-check spaces coincide, making this a normal self-dual qubit CSS code \NoCaseChange{\protect\cite{cite810}}.
\end{eczvaluelist}
}\\ 
\addlinespace[\myxtraspc]
\eczhRefIndex{code:stab_6_4_2}%
\eczhListValue{\flmRefsHyperref{code:stab_6_4_2}{\(\llbracket 6,4,2\rrbracket \) error-detecting code}} & \eczhListValue{\begin{eczvaluelist}
\item\relax CNOT and Hadamard gates \NoCaseChange{\protect\cite[{Appx. B}]{cite450}}.
\item\relax A \(CZ\) gate implemented by transversal \(S\) and \(S^{\dagger}\) \NoCaseChange{\protect\cite{cite805}}; see also \NoCaseChange{\protect\cite{cite687}}.
\end{eczvaluelist}
}\\ 
\addlinespace[\myxtraspc]
\eczhRefIndex{code:campbell_howard}%
\eczhListValue{\flmRefsHyperref{code:campbell_howard}{\(\llbracket 6k+2,3k,2\rrbracket \) Campbell-Howard code}} & \eczhListValue{\begin{eczvaluelist}
\item\relax Quasi-transversal \(CCZ^{\otimes k}\) gates \NoCaseChange{\protect\cite{cite754}}.
\end{eczvaluelist}
}\\ 
\addlinespace[\myxtraspc]
\eczhRefIndex{code:steane}%
\eczhListValue{\flmRefsHyperref{code:steane}{\(\llbracket 7,1,3\rrbracket \) Steane code}} & \eczhListValue{\begin{eczvaluelist}
\item\relax The \flmRefsHyperref{ref409}{single-qubit Clifford group} \NoCaseChange{\protect\cite{cite761,cite776}}. More generally, \(k\) copies of the Steane code form a \(\llbracket 7k,k,3\rrbracket \) code that admits a \(k\)-fold transversal implementation of the full \flmRefsHyperref{ref409}{Clifford group} on all \(k\) logical qubits, showing the tightness of a no-go theorem that requires at least \(k\)-fold transversal gadgets for the full Clifford group \NoCaseChange{\protect\cite{cite782}}.
\end{eczvaluelist}
}\\ 
\addlinespace[\myxtraspc]
\eczhRefIndex{code:twist_defect_7_1_3}%
\eczhListValue{\flmRefsHyperref{code:twist_defect_7_1_3}{\(\llbracket 7,1,3\rrbracket \) twist-defect surface code}} & \eczhListValue{\begin{eczvaluelist}
\item\relax Admits certain transversal order-three single-qubit \flmRefsHyperref{ref409}{Clifford} gates (e.g., \(SH\)) \NoCaseChange{\protect\cite{cite433}}.
\end{eczvaluelist}
}\\ 
\addlinespace[\myxtraspc]
\eczhRefIndex{code:stab_8_3_3}%
\eczhListValue{\flmRefsHyperref{code:stab_8_3_3}{\(\llbracket 8, 3, 3\rrbracket \) Eight-qubit Gottesman code}} & \eczhListValue{\begin{eczvaluelist}
\item\relax Permutation-based gates \NoCaseChange{\protect\cite[{Sec. IV.D}]{cite763}}.
\item\relax No gates outside of the \flmRefsHyperref{ref663}{Pauli group} were found in Ref. \NoCaseChange{\protect\cite{cite805}}.
\end{eczvaluelist}
}\\ 
\addlinespace[\myxtraspc]
\eczhRefIndex{code:stab_8_1_2}%
\eczhListValue{\flmRefsHyperref{code:stab_8_1_2}{\(\llbracket 8,1,2\rrbracket \) Shen-Wang-Cao code}} & \eczhListValue{\begin{eczvaluelist}
\item\relax Logical \(T\) gate \(\bar{T}=T^{\otimes 6}\otimes K\) via physical \(T\) gates on the first six qubits and the two-qubit gate \(K\propto\cos(\pi/8)(I\otimes I)-i\sin(\pi/8)(Z\otimes Z)\propto\mathrm{diag}(1,e^{i\pi/4},e^{i\pi/4},1)\) on qubits 7 and 8 \NoCaseChange{\protect\cite{cite786}}.
\end{eczvaluelist}
}\\ 
\addlinespace[\myxtraspc]
\eczhRefIndex{code:stab_8_2_2}%
\eczhListValue{\flmRefsHyperref{code:stab_8_2_2}{\(\llbracket 8,2,2\rrbracket \) hyperbolic color code}} & \eczhListValue{\begin{eczvaluelist}
\item\relax Applying transversal \(S\) and \(S^{\dagger}\), \(\sqrt{X}\), and Hadamard gates yields various logical gates \NoCaseChange{\protect\cite{cite805}}. For instance, the physical gate \(S^{\dagger}_{2}S^{\dagger}_{3}S_{4}S_{5}S_{6}S_{7}CZ_{01}\) implements the logical action \(\bar{S}_{0} \bar{Z}_{1} \overline{CZ}_{01}\).
\end{eczvaluelist}
}\\ 
\addlinespace[\myxtraspc]
\eczhRefIndex{code:stab_8_3_2}%
\eczhListValue{\flmRefsHyperref{code:stab_8_3_2}{\(\llbracket 8,3,2\rrbracket \) Smallest interesting color code}} & \eczhListValue{\begin{eczvaluelist}
\item\relax \(CZ\) gates between any two logical qubits \NoCaseChange{\protect\cite{cite805}} and (weakly) transversal \(CCZ\) gate \NoCaseChange{\protect\cite{cite422,cite797,cite805}}.
\end{eczvaluelist}
}\\ 
\addlinespace[\myxtraspc]
\eczhRefIndex{code:surface-17}%
\eczhListValue{\flmRefsHyperref{code:surface-17}{\(\llbracket 9,1,3\rrbracket \) Surface-17 code}} & \eczhListValue{\begin{eczvaluelist}
\item\relax Pauli gates, CNOT gate, and \(H\) gate (with relabeling).
\end{eczvaluelist}
}\\ 
\addlinespace[\myxtraspc]
\eczhRefIndex{code:quantum_h}%
\eczhListValue{\flmRefsHyperref{code:quantum_h}{\(\llbracket k+4,k,2\rrbracket \) H code}} & \eczhListValue{\begin{eczvaluelist}
\item\relax Hadamard and \(TXT^{\dagger}\) gates, with the latter Clifford-equivalent to Hadamard, and where \(T=\exp(i\pi(I-Z)/8)\) is the \(\pi/8\)-rotation gate.
\end{eczvaluelist}
}\\ 
\end{tabularx}\endgroup
\eczcodelist{css}{Quantum CSS codes (non-qubit)
}%

\eczhCodeListAutoDescription{Codes that are descendants of \flmRefsCref{code:css} and not descendants of \flmRefsCref{code:qubit_css}.}%

\eczhIncludeCodeGraph{Bare}{scale=0.5}{\columnwidth}{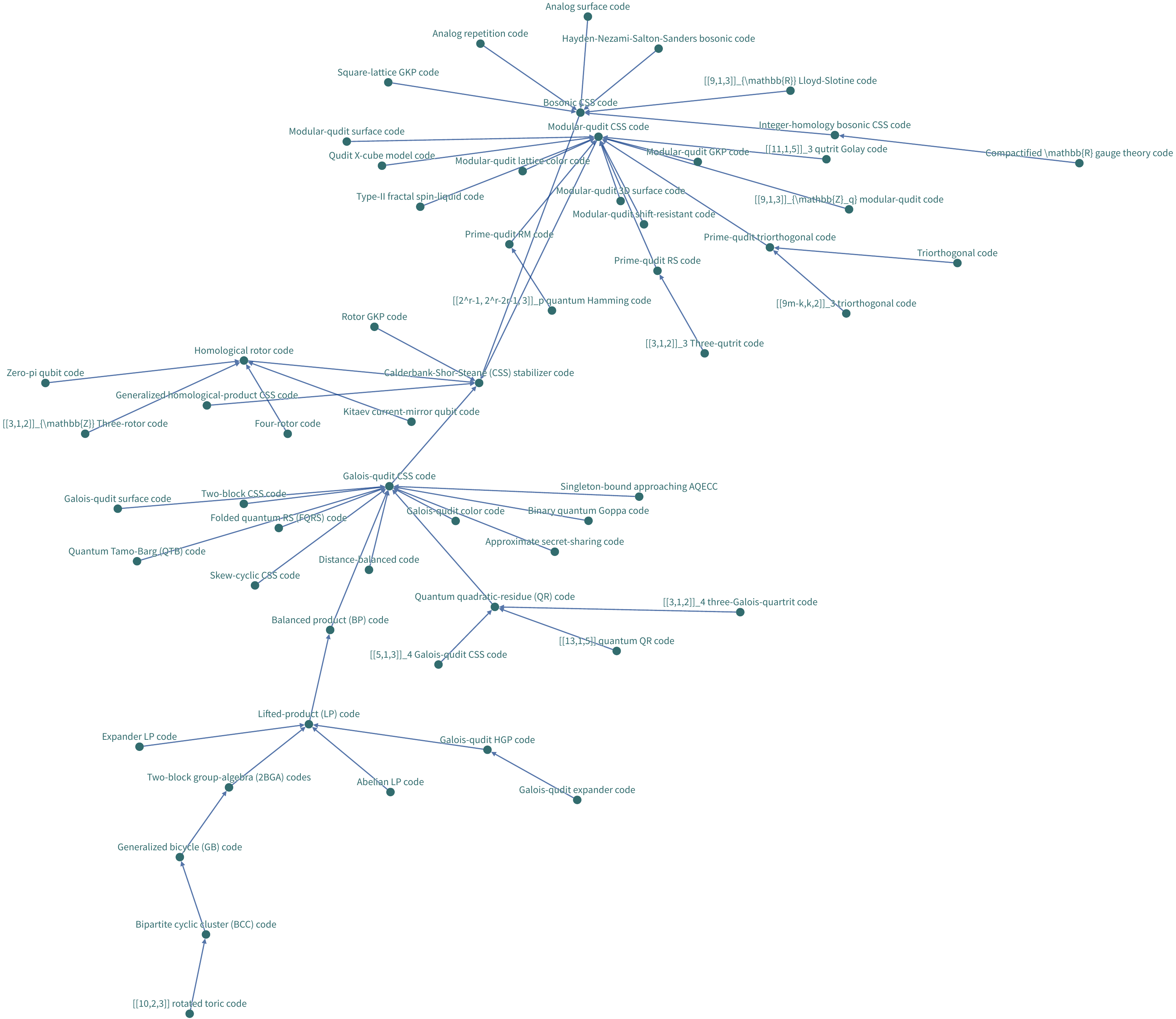}{Quantum CSS codes (non-qubit)}{https://errorcorrectionzoo.org/code_graph#J\%7B\%22displayMode\%22\%3A\%22subset\%22\%2C\%22modeSubsetOptions\%22\%3A\%7B\%22codeIds\%22\%3A\%5B\%22abelian_lifted_product\%22\%2C\%22analog_repetition\%22\%2C\%22analog_surface\%22\%2C\%22quantum_secret_sharing\%22\%2C\%22balanced_product\%22\%2C\%22binary_quantum_goppa\%22\%2C\%22bipartite_cyclic_cluster\%22\%2C\%22oscillator_css\%22\%2C\%22css\%22\%2C\%22compactified_r\%22\%2C\%22distance_balanced\%22\%2C\%22expander_lifted_product\%22\%2C\%22galois_fqrs\%22\%2C\%22rotor_4_2_2\%22\%2C\%22galois_css\%22\%2C\%22galois_hypergraph_product\%22\%2C\%22galois_color\%22\%2C\%22galois_expander\%22\%2C\%22galois_topological\%22\%2C\%22generalized_bicycle\%22\%2C\%22generalized_homological_product_css\%22\%2C\%22hnss\%22\%2C\%22homological_rotor\%22\%2C\%22homological_cv\%22\%2C\%22current_mirror\%22\%2C\%22lifted_product\%22\%2C\%22qudit_3d_surface\%22\%2C\%22qudit_css\%22\%2C\%22qudit_gkp\%22\%2C\%22qudit_color\%22\%2C\%22qudit_sign\%22\%2C\%22qudit_surface\%22\%2C\%22qudit_reed_muller\%22\%2C\%22polynomial\%22\%2C\%22qudit_triorthogonal\%22\%2C\%22quantum_tamo_barg\%22\%2C\%22galois_quad_residue\%22\%2C\%22qudit_xcube\%22\%2C\%22rotor_gkp\%22\%2C\%22quantum_singleton\%22\%2C\%22skew-cyclic_galois_css\%22\%2C\%22gkp\%22\%2C\%22quantum_triorthogonal\%22\%2C\%22two_block_quantum\%22\%2C\%222bga\%22\%2C\%22fractal_liquid\%22\%2C\%22zero_pi\%22\%2C\%22xzzx_10_2_3\%22\%2C\%22qutrit_golay\%22\%2C\%22quad_residue_13_1_5\%22\%2C\%22qudit_hamming_css\%22\%2C\%22stab_3_1_2\%22\%2C\%22galois_3_1_2\%22\%2C\%22rotor_3_1_2\%22\%2C\%22css_5_1_3\%22\%2C\%22lloyd_slotine\%22\%2C\%22stab_9_1_3\%22\%2C\%22qutrit_small_triorthogonal\%22\%5D\%2C\%22reusePreviousLayoutPositions\%22\%3Afalse\%2C\%22showIntermediateConnectingNodes\%22\%3Atrue\%2C\%22connectingNodesMaxDepth\%22\%3A15\%2C\%22connectingNodesPathMaxLength\%22\%3A20\%2C\%22connectingNodesMaxExtraDepth\%22\%3A3\%2C\%22connectingNodesOnlyKeepPathsWithAdditionalLength\%22\%3A1\%2C\%22connectingNodesToDomainsAndKingdoms\%22\%3Afalse\%2C\%22connectingNodesEdgeLengthsByType\%22\%3A\%7B\%22primaryParent\%22\%3A1\%2C\%22secondaryParent\%22\%3A4\%2C\%22cousin\%22\%3A6\%7D\%2C\%22nodeIds\%22\%3A\%5B\%5D\%7D\%2C\%22highlightImportantNodes\%22\%3A\%7B\%22highlightImportantNodes\%22\%3Afalse\%2C\%22highlightPrimaryParents\%22\%3Afalse\%2C\%22highlightRootConnectingEdges\%22\%3Afalse\%7D\%7D}

\begingroup
\small
\eczhBreakableDashes
\renewcommand\arraystretch{1.05}
\edef\myxtraspc{\eczListAddVSpaceXtraXtraStretch}
\begin{tabularx}{\linewidth}{>{\raggedright\arraybackslash}p{\eczListColWidth{name}} >{\hsize=1.0000\hsize }X}
\toprule
\eczListColTitle{Code} & \eczListColTitle{Description} \\
\midrule
\endfirsthead
\toprule
\eczListColTitleContinued{Code} & \eczListColTitleContinued{Description} \\
\midrule
\endhead
\bottomrule
\endfoot
\eczhRefIndex{code:abelian_lifted_product}%
\eczhListValue{\flmRefsHyperref{code:abelian_lifted_product}{Abelian LP code}} & \eczhListValue{A lifted-product code whose lift group \(G\) is Abelian.
The case of \(G\) being a cyclic group is a GB code (a.k.a. a quasi-cyclic LP code) \NoCaseChange{\protect\cite[{Sec. III.E}]{cite674}}.
A particular family with \(G=\mathbb{Z}_{\ell}\) yields codes with parameters \(\llbracket n,k=\Theta(\log n),d=\Theta(n/\log n)\rrbracket \) \NoCaseChange{\protect\cite{cite674}}.}\\ 
\addlinespace[\myxtraspc]
\eczhRefIndex{code:analog_repetition}%
\eczhListValue{\flmRefsHyperref{code:analog_repetition}{Analog repetition code}} & \eczhListValue{An \(\llbracket n,1\rrbracket _{\mathbb{R}}\) analog stabilizer version of the quantum repetition code, encoding the position states of one mode into an odd number \(n\) of modes.}\\ 
\addlinespace[\myxtraspc]
\eczhRefIndex{code:analog_surface}%
\eczhListValue{\flmRefsHyperref{code:analog_surface}{Analog surface code}} & \eczhListValue{An analog CSS version of the Kitaev surface code realizing a phase of 2D \(\mathbb{R}\) gauge theory.}\\ 
\addlinespace[\myxtraspc]
\eczhRefIndex{code:quantum_secret_sharing}%
\eczhListValue{\flmRefsHyperref{code:quantum_secret_sharing}{Approximate secret-sharing code}} & \eczhListValue{A family of \( \llbracket n,k,d\rrbracket _q \) CSS codes approximately correcting errors on up to \(\lfloor (n-1)/2 \rfloor\) Galois qudits, i.e., with approximate distance approaching the no-cloning bound \(n/2\). Constructed using a \flmRefsHyperref{ref811}{non-degenerate} CSS code, such as a polynomial quantum code, and a classical authentication scheme. The code can be viewed as a \(t\)-error-tolerant secret sharing scheme. Since the code yields a small logical subspace using large registers that contain both classical and quantum information, it is not useful for practical error correction problems, but instead demonstrates the power of approximate quantum error correction.}\\ 
\addlinespace[\myxtraspc]
\eczhRefIndex{code:balanced_product}%
\eczhListValue{\flmRefsHyperref{code:balanced_product}{Balanced product (BP) code}} & \eczhListValue{Family of CSS quantum codes obtained from two classical-code chain complexes that share a common group symmetry.
The balanced product can be understood as taking the usual tensor or hypergraph product and then quotienting by the shared symmetry action.
This can reduce the overall number of physical qubits \(n\) while, in favorable cases, preserving the number of encoded qubits and the code distance, thereby improving the encoding rate \(k/n\) and normalized distance \(d/n\) compared to the underlying tensor or hypergraph product.}\\ 
\addlinespace[\myxtraspc]
\eczhRefIndex{code:binary_quantum_goppa}%
\eczhListValue{\flmRefsHyperref{code:binary_quantum_goppa}{Binary quantum Goppa code}} & \eczhListValue{A quantum AG code obtained from algebraic-geometric Goppa codes via the Galois-qudit CSS construction.}\\ 
\addlinespace[\myxtraspc]
\eczhRefIndex{code:bipartite_cyclic_cluster}%
\eczhListValue{\flmRefsHyperref{code:bipartite_cyclic_cluster}{Bipartite cyclic cluster (BCC) code}} & \eczhListValue{Cyclic CSS code constructed from a bipartite cluster state with cyclic invariance,
emphasizing simplicity of state preparation over simplicity of stabilizers.}\\ 
\addlinespace[\myxtraspc]
\eczhRefIndex{code:oscillator_css}%
\eczhListValue{\flmRefsHyperref{code:oscillator_css}{Bosonic CSS code}} & \eczhListValue{Bosonic stabilizer code admitting a set of stabilizer generators that are either position or momentum displacements.}\\ 
\addlinespace[\myxtraspc]
\eczhRefIndex{code:css}%
\eczhListValue{\flmRefsHyperref{code:css}{Calderbank-Shor-Steane (CSS) stabilizer code}} & \eczhListValue{A stabilizer code admitting a set of stabilizer generators that are either \(Z\)-type or \(X\)-type operators.
The two sets of stabilizer generators can often be related to parts of a chain complex over the appropriate ring or field.}\\ 
\addlinespace[\myxtraspc]
\eczhRefIndex{code:compactified_r}%
\eczhListValue{\flmRefsHyperref{code:compactified_r}{Compactified \(\mathbb{R}\) gauge theory code}} & \eczhListValue{An integer-homology bosonic CSS code realizing 2D \(U(1)\) gauge theory on bosonic modes.
The code can be obtained from the analog surface code by \flmRefsHyperref{ref410}{condensing} certain anyons \NoCaseChange{\protect\cite{cite411}}. 
This results in a pinning of each mode to the space of periodic functions, which is the Hilbert space of a physical rotor, and can be thought of as compactification of the 2D \(\mathbb{R}\) gauge theory phase realized by the analog surface code.}\\ 
\addlinespace[\myxtraspc]
\eczhRefIndex{code:distance_balanced}%
\eczhListValue{\flmRefsHyperref{code:distance_balanced}{Distance-balanced code}} & \eczhListValue{Galois-qudit CSS code obtained from a CSS code by increasing the smaller of the \(X\)- and \(Z\)-distances using a homological-product-based balancing step or one of its generalizations.
The initial code is said to be \textit{unbalanced}, i.e., tailored to noise biased toward either bit- or phase-flip errors, and the procedure can result in a code that treats both types of errors on a more equal footing.}\\ 
\addlinespace[\myxtraspc]
\eczhRefIndex{code:expander_lifted_product}%
\eczhListValue{\flmRefsHyperref{code:expander_lifted_product}{Expander LP code}} & \eczhListValue{Family of \(G\)-lifted product codes constructed using two classical \flmRefsHyperref{code:expander}{expander codes}, equivalently two regular \flmRefsHyperref{code:tanner}{Tanner codes} defined on the same expander graph \NoCaseChange{\protect\cite{cite74}}. For certain parameters, this construction yields the first asymptotically good QLDPC codes. Classical codes resulting from the same lifted-product complexes are one of the first two families of \(c^3\)-LTCs \NoCaseChange{\protect\cite{cite184}}.}\\ 
\addlinespace[\myxtraspc]
\eczhRefIndex{code:galois_fqrs}%
\eczhListValue{\flmRefsHyperref{code:galois_fqrs}{Folded quantum RS (FQRS) code}} & \eczhListValue{CSS code on \(q^m\)-dimensional Galois-qudits that is constructed from folded RS (FRS) codes (i.e., an RS code whose coordinates have been grouped together) via the Galois-qudit CSS construction.
This code is used to construct Singleton-bound approaching approximate quantum codes.}\\ 
\addlinespace[\myxtraspc]
\eczhRefIndex{code:rotor_4_2_2}%
\eczhListValue{\flmRefsHyperref{code:rotor_4_2_2}{Four-rotor code}} & \eczhListValue{\(\llbracket 4,2,2\rrbracket _{\mathbb Z}\) CSS rotor code that is an extension of the four-qubit code to the integer alphabet, i.e., the angular momentum states of a rotor.}\\ 
\addlinespace[\myxtraspc]
\eczhRefIndex{code:galois_css}%
\eczhListValue{\flmRefsHyperref{code:galois_css}{Galois-qudit CSS code}} & \eczhListValue{An \(\llbracket n,k,d\rrbracket _q \) Galois-qudit true stabilizer code admitting a set of stabilizer generators that
are either \(Z\)-type or \(X\)-type Galois-qudit Pauli strings.
Codes can be defined from chain complexes over \(\mathbb{F}_q\) via an extension of \flmRefsHyperref{ref683}{qubit CSS-to-homology correspondence} to Galois qudits.}\\ 
\addlinespace[\myxtraspc]
\eczhRefIndex{code:galois_hypergraph_product}%
\eczhListValue{\flmRefsHyperref{code:galois_hypergraph_product}{Galois-qudit HGP code}} & \eczhListValue{A member of a family of Galois-qudit CSS codes whose stabilizer generator matrix is obtained from a hypergraph product of two classical linear \(q\)-ary codes.}\\ 
\addlinespace[\myxtraspc]
\eczhRefIndex{code:galois_color}%
\eczhListValue{\flmRefsHyperref{code:galois_color}{Galois-qudit color code}} & \eczhListValue{Extension of the color code to 2D lattices of Galois qudits.}\\ 
\addlinespace[\myxtraspc]
\eczhRefIndex{code:galois_expander}%
\eczhListValue{\flmRefsHyperref{code:galois_expander}{Galois-qudit expander code}} & \eczhListValue{Galois-qudit CSS code obtained from tensor products of chain complexes associated with an explicit family of expander codes with Reed-Solomon local checks.}\\ 
\addlinespace[\myxtraspc]
\eczhRefIndex{code:galois_topological}%
\eczhListValue{\flmRefsHyperref{code:galois_topological}{Galois-qudit surface code}} & \eczhListValue{Extension of the surface code to 2D lattices of Galois qudits.}\\ 
\addlinespace[\myxtraspc]
\eczhRefIndex{code:generalized_bicycle}%
\eczhListValue{\flmRefsHyperref{code:generalized_bicycle}{Generalized bicycle (GB) code}} & \eczhListValue{A quasi-cyclic Galois-qudit CSS code constructed using a generalized version of the bicycle ansatz \NoCaseChange{\protect\cite{cite682}} from a pair of equivalent index-two quasi-cyclic linear codes.
Equivalently, the codes can be constructed via the lifted-product construction for \(G\) being a cyclic group \NoCaseChange{\protect\cite[{Sec. III.E}]{cite674}}.}\\ 
\addlinespace[\myxtraspc]
\eczhRefIndex{code:generalized_homological_product_css}%
\eczhListValue{\flmRefsHyperref{code:generalized_homological_product_css}{Generalized homological-product CSS code}} & \eczhListValue{CSS code whose properties are determined from an underlying chain complex, which often consists of some type of product of other chain complexes.}\\ 
\addlinespace[\myxtraspc]
\eczhRefIndex{code:hnss}%
\eczhListValue{\flmRefsHyperref{code:hnss}{Hayden-Nezami-Salton-Sanders bosonic code}} & \eczhListValue{An \(\llbracket n,1\rrbracket _{\mathbb{R}}\) analog CSS code defined using homological structures associated with an \(n-1\) simplex. Relevant to the study of spacetime replication of quantum information \NoCaseChange{\protect\cite{cite512}}.}\\ 
\addlinespace[\myxtraspc]
\eczhRefIndex{code:homological_rotor}%
\eczhListValue{\flmRefsHyperref{code:homological_rotor}{Homological rotor code}} & \eczhListValue{A CSS rotor code stabilized by a group of rotor \(X\)-type and \(Z\)-type generalized Pauli operators.
Codes are formulated using an extension of the \flmRefsHyperref{ref683}{qubit CSS-to-homology correspondence} to rotors.
The homology group of the logical operators has a torsion component because the chain complexes are defined over the ring of integers, which yields codes with finite logical dimension, i.e., encoding logical qudits instead of only logical rotors.}\\ 
\addlinespace[\myxtraspc]
\eczhRefIndex{code:homological_cv}%
\eczhListValue{\flmRefsHyperref{code:homological_cv}{Integer-homology bosonic CSS code}} & \eczhListValue{A bosonic stabilizer code whose physical modes have been restricted, via a single GKP stabilizer, from the space of functions on the real line to the space of periodic functions.
This restriction effectively realizes a rotor on each physical mode, allowing one to construct homological rotor codes out of displacement stabilizer groups.
The stabilizer group is continuous, but contains discrete components in the form of the single-mode GKP stabilizers.
The homology group of the logical operators has a torsion component because the chain complexes are defined over the ring of integers, which yields codes with finite logical dimension.}\\ 
\addlinespace[\myxtraspc]
\eczhRefIndex{code:current_mirror}%
\eczhListValue{\flmRefsHyperref{code:current_mirror}{Kitaev current-mirror qubit code}} & \eczhListValue{Member of the family of \(\llbracket 2n,(0,2),(2,n)\rrbracket _{\mathbb{Z}}\) homological rotor codes storing a logical qubit on a thin Möbius strip.
The ideal code can be obtained from a Josephson-junction \NoCaseChange{\protect\cite{cite396}} system \NoCaseChange{\protect\cite{cite397}}.}\\ 
\addlinespace[\myxtraspc]
\eczhRefIndex{code:lifted_product}%
\eczhListValue{\flmRefsHyperref{code:lifted_product}{Lifted-product (LP) code}} & \eczhListValue{Galois-qudit code that utilizes the notion of a lifted product in its construction. Lifted products of certain classical Tanner codes are the first (asymptotically) \textit{good QLDPC codes}.}\\ 
\addlinespace[\myxtraspc]
\eczhRefIndex{code:qudit_3d_surface}%
\eczhListValue{\flmRefsHyperref{code:qudit_3d_surface}{Modular-qudit 3D surface code}} & \eczhListValue{A generalization of the 3D surface code to modular qudits.
Qudits are placed on edges, \(Z\)-type stabilizer generators are placed on square plaquettes oriented in all three directions, and \(X\)-type stabilizers are placed on the six edges neighboring every vertex \NoCaseChange{\protect\cite{cite459}}.}\\ 
\addlinespace[\myxtraspc]
\eczhRefIndex{code:qudit_css}%
\eczhListValue{\flmRefsHyperref{code:qudit_css}{Modular-qudit CSS code}} & \eczhListValue{An \(\llparenthesis n,K,d\rrparenthesis _q\) modular-qudit stabilizer code admitting a set of stabilizer generators that
are either \(Z\)-type or \(X\)-type Pauli strings.
Codes can be defined from two classical codes and/or chain complexes over the ring \(\mathbb{Z}_q\) via an extension of \flmRefsHyperref{ref683}{qubit CSS-to-homology correspondence} to modular qudits \NoCaseChange{\protect\cite{cite71}}.
The homology group of the logical operators has a torsion component because the chain complexes are defined over a ring, which yields codes whose logical dimension is not a power of \(q\).}\\ 
\addlinespace[\myxtraspc]
\eczhRefIndex{code:qudit_gkp}%
\eczhListValue{\flmRefsHyperref{code:qudit_gkp}{Modular-qudit GKP code}} & \eczhListValue{Modular-qudit analogue of the GKP code.
Encodes a qudit into a larger qudit and protects against Pauli shifts up to some maximum value.}\\ 
\addlinespace[\myxtraspc]
\eczhRefIndex{code:qudit_color}%
\eczhListValue{\flmRefsHyperref{code:qudit_color}{Modular-qudit lattice color code}} & \eczhListValue{Extension of the color code to lattices of modular qudits.
Codes are defined analogously to qubit color codes on suitable lattices of any spatial dimension, but a directionality is required in order to make the modular-qudit stabilizers commute.
This can be done by puncturing a hyperspherical lattice \NoCaseChange{\protect\cite{cite475}} or constructing a star-bipartition; see \NoCaseChange{\protect\cite[{Sec. III}]{cite673}}.
Logical dimension is determined by the genus of the underlying surface (for closed surfaces), types of boundaries (for open surfaces), and/or any twist defects present.}\\ 
\addlinespace[\myxtraspc]
\eczhRefIndex{code:qudit_sign}%
\eczhListValue{\flmRefsHyperref{code:qudit_sign}{Modular-qudit shift-resistant code}} & \eczhListValue{Monolithic code encoding a qubit into a single modular qudit and protecting against either \(Z\)-type or \(X\)-type modular-qudit Pauli shifts.}\\ 
\addlinespace[\myxtraspc]
\eczhRefIndex{code:qudit_surface}%
\eczhListValue{\flmRefsHyperref{code:qudit_surface}{Modular-qudit surface code}} & \eczhListValue{Extension of the surface code to prime-dimensional \NoCaseChange{\protect\cite{cite423,cite424}} and more general modular qudits.
Stabilizer generators are few-body \(X\)-type and \(Z\)-type Pauli strings associated to the stars and plaquettes, respectively, of a tessellation of a two-dimensional surface.
Since qudits have more than one \(X\) and \(Z\)-type operator, various sets of stabilizer generators can be defined.}\\ 
\addlinespace[\myxtraspc]
\eczhRefIndex{code:qudit_reed_muller}%
\eczhListValue{\flmRefsHyperref{code:qudit_reed_muller}{Prime-qudit RM code}} & \eczhListValue{Modular-qudit stabilizer code constructed from GRM codes or their duals via the modular-qudit CSS construction.}\\ 
\addlinespace[\myxtraspc]
\eczhRefIndex{code:polynomial}%
\eczhListValue{\flmRefsHyperref{code:polynomial}{Prime-qudit RS code}} & \eczhListValue{Prime-qudit CSS code constructed using two RS codes.}\\ 
\addlinespace[\myxtraspc]
\eczhRefIndex{code:qudit_triorthogonal}%
\eczhListValue{\flmRefsHyperref{code:qudit_triorthogonal}{Prime-qudit triorthogonal code}} & \eczhListValue{An \(m \times n\) matrix over \(\mathbb{F}_p=\mathbb{Z}_p\) is triorthogonal if its rows \(r_1, \ldots, r_m\) satisfy \(|r_i \cdot r_j| = 0\) and \(|r_i \cdot r_j \cdot r_k| = 0\) modulo \(p\), where addition and multiplication are done on \(\mathbb{F}_p\).
The triorthogonal prime-qudit CSS code associated with the matrix is constructed by mapping nonzero entries in self-orthogonal rows to \(X\) operators, and \(Z\) operators for each row in the orthogonal complement \NoCaseChange{\protect\cite{cite693,cite709}}.}\\ 
\addlinespace[\myxtraspc]
\eczhRefIndex{code:quantum_tamo_barg}%
\eczhListValue{\flmRefsHyperref{code:quantum_tamo_barg}{Quantum Tamo-Barg (QTB) code}} & \eczhListValue{A member of a family of Galois-qudit CSS codes whose underlying classical codes consist of Tamo-Barg codes together with specific low-weight codewords.
Folded versions of QTB codes, or \textit{FQTB codes}, defined on qudits whose dimension depends on \(n\), yield explicit examples of QLRCs of arbitrary locality \(r\) \NoCaseChange{\protect\cite[{Cor. 64}]{cite812}}.}\\ 
\addlinespace[\myxtraspc]
\eczhRefIndex{code:galois_quad_residue}%
\eczhListValue{\flmRefsHyperref{code:galois_quad_residue}{Quantum quadratic-residue (QR) code}} & \eczhListValue{Galois-qudit \(\llbracket n,1\rrbracket _q\) \flmRefsHyperref{ref672}{pure} self-dual Galois-qudit CSS code constructed from a dual-containing QR code via the Galois-qudit CSS construction.
For \(q\) not divisible by \(n\), its distance satisfies \(d^2-d+1 \geq n\) when \(n \equiv 3\) modulo 4 \NoCaseChange{\protect\cite[{Thm. 40}]{cite813}} and \(d \geq \sqrt{n}\) when \(n\equiv 1\) modulo 4 \NoCaseChange{\protect\cite[{Thm. 41}]{cite813}}.}\\ 
\addlinespace[\myxtraspc]
\eczhRefIndex{code:qudit_xcube}%
\eczhListValue{\flmRefsHyperref{code:qudit_xcube}{Qudit X-cube model code}} & \eczhListValue{Generalization of the X-cube model code to modular qudits.}\\ 
\addlinespace[\myxtraspc]
\eczhRefIndex{code:rotor_gkp}%
\eczhListValue{\flmRefsHyperref{code:rotor_gkp}{Rotor GKP code}} & \eczhListValue{GKP code protecting against small angular position and momentum shifts of a planar rotor.}\\ 
\addlinespace[\myxtraspc]
\eczhRefIndex{code:quantum_singleton}%
\eczhListValue{\flmRefsHyperref{code:quantum_singleton}{Singleton-bound approaching AQECC}} & \eczhListValue{A member of an approximate quantum code family of rate \(R\) that can tolerate adversarial errors nearly saturating the quantum Singleton bound of \((1-R)/2\).
The formulation of such codes relies on a notion of \textit{quantum list decoding} \NoCaseChange{\protect\cite{cite814,cite495}}.}\\ 
\addlinespace[\myxtraspc]
\eczhRefIndex{code:skew-cyclic_galois_css}%
\eczhListValue{\flmRefsHyperref{code:skew-cyclic_galois_css}{Skew-cyclic CSS code}} & \eczhListValue{A member of a family of Galois-qudit CSS codes constructed from skew-cyclic classical codes over rings \NoCaseChange{\protect\cite[{Thm. 5.8}]{cite815}}.
See related study \NoCaseChange{\protect\cite{cite816}} that uses cyclic codes over rings.}\\ 
\addlinespace[\myxtraspc]
\eczhRefIndex{code:gkp}%
\eczhListValue{\flmRefsHyperref{code:gkp}{Square-lattice GKP code}} & \eczhListValue{Single-mode GKP qudit-into-oscillator CSS code based on the rectangular lattice.
Its stabilizer generators are oscillator displacement operators \(\hat{S}_q(2\alpha)=e^{-2i\alpha \hat{p}}\) and \(\hat{S}_p(2\beta)=e^{2i\beta \hat{x}}\).
To ensure \(\hat{S}_q(2\alpha)\) and \(\hat{S}_p(2\beta)\) generate a stabilizer group that is Abelian, there is a constraint that \(\alpha\beta=2q\pi\) where \(q\) is an integer denoting the logical dimension.}\\ 
\addlinespace[\myxtraspc]
\eczhRefIndex{code:quantum_triorthogonal}%
\eczhListValue{\flmRefsHyperref{code:quantum_triorthogonal}{Triorthogonal code}} & \eczhListValue{Qubit CSS code whose \(X\)-type logicals and stabilizer generators form a triorthogonal matrix (defined below) in the \flmRefsHyperref{ref817}{symplectic representation}.}\\ 
\addlinespace[\myxtraspc]
\eczhRefIndex{code:two_block_quantum}%
\eczhListValue{\flmRefsHyperref{code:two_block_quantum}{Two-block CSS code}} & \eczhListValue{Galois-qudit CSS code whose stabilizer generator matrices \(H_X=(A_1,B_1)\) and \(H_Z=(B^T_2,-A^T_2)\), are constructed from four matrices satisfying \(A_1 B_2 - B_1 A_2 = 0\).
In the case the two pairs are equal, we have \(H_X=(A,B)\) and \(H_Z=(B^T,-A^T)\), constructed from a pair of square commuting matrices \(A\) and \(B\).}\\ 
\addlinespace[\myxtraspc]
\eczhRefIndex{code:2bga}%
\eczhListValue{\flmRefsHyperref{code:2bga}{Two-block group-algebra (2BGA) codes}} & \eczhListValue{2BGA codes are the one-by-one, or smallest, \flmRefsHyperref{code:lifted_product}{LP codes}:
\(LP(a,b)\) is defined by a pair of \flmRefsHyperref{ref205}{group algebra} elements
\(a,b\in \mathbb{F}_q[G]\), where \(G\) is a finite group.
If \(|G|=\ell\), then the code has length \(n=2\ell\).}\\ 
\addlinespace[\myxtraspc]
\eczhRefIndex{code:fractal_liquid}%
\eczhListValue{\flmRefsHyperref{code:fractal_liquid}{Type-II fractal spin-liquid code}} & \eczhListValue{A type-II fracton prime-qudit CSS code defined on a cubic lattice \NoCaseChange{\protect\cite[{Eqs. (D9-D10)}]{cite456}}.}\\ 
\addlinespace[\myxtraspc]
\eczhRefIndex{code:zero_pi}%
\eczhListValue{\flmRefsHyperref{code:zero_pi}{Zero-pi qubit code}} & \eczhListValue{A \(\llbracket 2,(0,2),(2,1)\rrbracket _{\mathbb{Z}}\) homological rotor code on the smallest tiling of the projective plane \(\mathbb{R}P^2\).
The ideal code can be obtained from a four-rotor Josephson-junction \NoCaseChange{\protect\cite{cite396}} system after a choice of grounding \NoCaseChange{\protect\cite{cite397}}.}\\ 
\addlinespace[\myxtraspc]
\eczhRefIndex{code:xzzx_10_2_3}%
\eczhListValue{\flmRefsHyperref{code:xzzx_10_2_3}{\(\llbracket 10,2,3\rrbracket \) rotated toric code}} & \eczhListValue{Rotated toric code that is the CSS form of the twisted XZZX toric code with parameters \(a=1\), \(b=3\) \NoCaseChange{\protect\cite{cite438,cite427}},
related to the XZZX form by Hadamard on the \(B\)-sublattice.
It is also the \flmRefsHyperref{ref436}{symplectic double} (a.k.a. genus-one double cover) of the \(\llbracket 5,1,3\rrbracket \) five-qubit perfect code \NoCaseChange{\protect\cite{cite439,cite435}}, the \flmRefsHyperref{ref436}{symplectic double} of the \(\llbracket 5,1,2\rrbracket \) rotated surface code \NoCaseChange{\protect\cite{cite435}}, and a BCC code \NoCaseChange{\protect\cite{cite440}}.}\\ 
\addlinespace[\myxtraspc]
\eczhRefIndex{code:qutrit_golay}%
\eczhListValue{\flmRefsHyperref{code:qutrit_golay}{\(\llbracket 11,1,5\rrbracket _3\) qutrit Golay code}} & \eczhListValue{An \(\llbracket 11,1,5\rrbracket _3\) code constructed from the ternary Golay code via the CSS construction.
The code's stabilizer generator matrix blocks \(H_{X}\) and \(H_{Z}\) are both the generator matrix of the ternary Golay code.}\\ 
\addlinespace[\myxtraspc]
\eczhRefIndex{code:quad_residue_13_1_5}%
\eczhListValue{\flmRefsHyperref{code:quad_residue_13_1_5}{\(\llbracket 13,1,5\rrbracket \) quantum QR code}} & \eczhListValue{Thirteen-qubit cyclic Hermitian qubit code derived from a quaternary quadratic-residue code using the Hermitian construction \NoCaseChange{\protect\cite{cite819}\protect\cite[{pg. 11}]{cite818}}.
The code admits a stabilizer tableau whose rows are cyclic permutations of the Pauli string \(XXZZIZIIIZIZZ\) \NoCaseChange{\protect\cite[{ID 67067199e766ad364e845262}]{cite781}}.}\\ 
\addlinespace[\myxtraspc]
\eczhRefIndex{code:qudit_hamming_css}%
\eczhListValue{\flmRefsHyperref{code:qudit_hamming_css}{\(\llbracket 2^r-1, 2^r-2r-1, 3\rrbracket _p\) quantum Hamming code}} & \eczhListValue{A family of CSS codes extending \flmRefsHyperref{code:quantum_hamming_css}{quantum Hamming codes} to prime qudits of dimension \(p\) by expressing the qubit code stabilizers in local-dimension-invariant (LDI) form \NoCaseChange{\protect\cite{cite820}}.}\\ 
\addlinespace[\myxtraspc]
\eczhRefIndex{code:stab_3_1_2}%
\eczhListValue{\flmRefsHyperref{code:stab_3_1_2}{\(\llbracket 3,1,2\rrbracket _3\) Three-qutrit code}} & \eczhListValue{A \(\llbracket 3,1,2\rrbracket _3\) prime-qudit CSS code that is the smallest qutrit stabilizer code to detect a single-qutrit error.
It has stabilizer generators \(ZZZ\) and \(XXX\). The code defines a quantum secret-sharing scheme and serves as a minimal model for the AdS/CFT holographic duality. It is also the smallest non-trivial instance of a quantum maximum distance separable code (QMDS), saturating the quantum Singleton bound.}\\ 
\addlinespace[\myxtraspc]
\eczhRefIndex{code:galois_3_1_2}%
\eczhListValue{\flmRefsHyperref{code:galois_3_1_2}{\(\llbracket 3,1,2\rrbracket _4\) three-Galois-quartrit code}} & \eczhListValue{Three-Galois-qudit CSS code over \(\mathbb{F}_4=\{0,1,\omega,\omega^2\}\) that encodes one logical Galois qudit and detects a single-qudit error.}\\ 
\addlinespace[\myxtraspc]
\eczhRefIndex{code:rotor_3_1_2}%
\eczhListValue{\flmRefsHyperref{code:rotor_3_1_2}{\(\llbracket 3,1,2\rrbracket _{\mathbb{Z}}\) Three-rotor code}} & \eczhListValue{\(\llbracket 3,1,2\rrbracket _{\mathbb Z}\) rotor code that is an extension of the \(\llbracket 3,1,2\rrbracket _3\) qutrit CSS code to the integer alphabet, i.e., the angular momentum states of a rotor.}\\ 
\addlinespace[\myxtraspc]
\eczhRefIndex{code:css_5_1_3}%
\eczhListValue{\flmRefsHyperref{code:css_5_1_3}{\(\llbracket 5,1,3\rrbracket _4\) Galois-qudit CSS code}} & \eczhListValue{Five-Galois-qudit CSS code over \(\mathbb{F}_4=\{0,1,\omega,\omega^2\}\) that encodes one logical Galois qudit and corrects a single-qudit error.}\\ 
\addlinespace[\myxtraspc]
\eczhRefIndex{code:lloyd_slotine}%
\eczhListValue{\flmRefsHyperref{code:lloyd_slotine}{\(\llbracket 9,1,3\rrbracket _{\mathbb{R}}\) Lloyd-Slotine code}} & \eczhListValue{An analog stabilizer version of Shor's nine-qubit code, encoding one mode into nine and correcting arbitrary errors on any one mode.}\\ 
\addlinespace[\myxtraspc]
\eczhRefIndex{code:stab_9_1_3}%
\eczhListValue{\flmRefsHyperref{code:stab_9_1_3}{\(\llbracket 9,1,3\rrbracket _{\mathbb{Z}_q}\) modular-qudit code}} & \eczhListValue{Modular-qudit CSS code that generalizes the \(\llbracket 9,1,3\rrbracket \) Shor code to \(q\)-level systems.}\\ 
\addlinespace[\myxtraspc]
\eczhRefIndex{code:qutrit_small_triorthogonal}%
\eczhListValue{\flmRefsHyperref{code:qutrit_small_triorthogonal}{\(\llbracket 9m-k,k,2\rrbracket _3\) triorthogonal code}} & \eczhListValue{Member of the \(\llbracket 9m-k,k,2\rrbracket _3\) family of triorthogonal qutrit codes (for \(k\leq 3m-2\)) that admit a transversal diagonal gate in the third level of the \flmRefsHyperref{ref751}{qudit Clifford hierarchy} and that are relevant for magic-state distillation.}\\ 
\end{tabularx}\endgroup
\eczcodelist{general_qldpc}{Quantum LDPC codes
}%

\eczhCodeListAutoDescription{Codes that are descendants of \flmRefsCref{code:general_qldpc} and not descendants of \flmRefsCref{code:qldpc}.}%

\eczhIncludeCodeGraph{Bare}{scale=0.5}{\columnwidth}{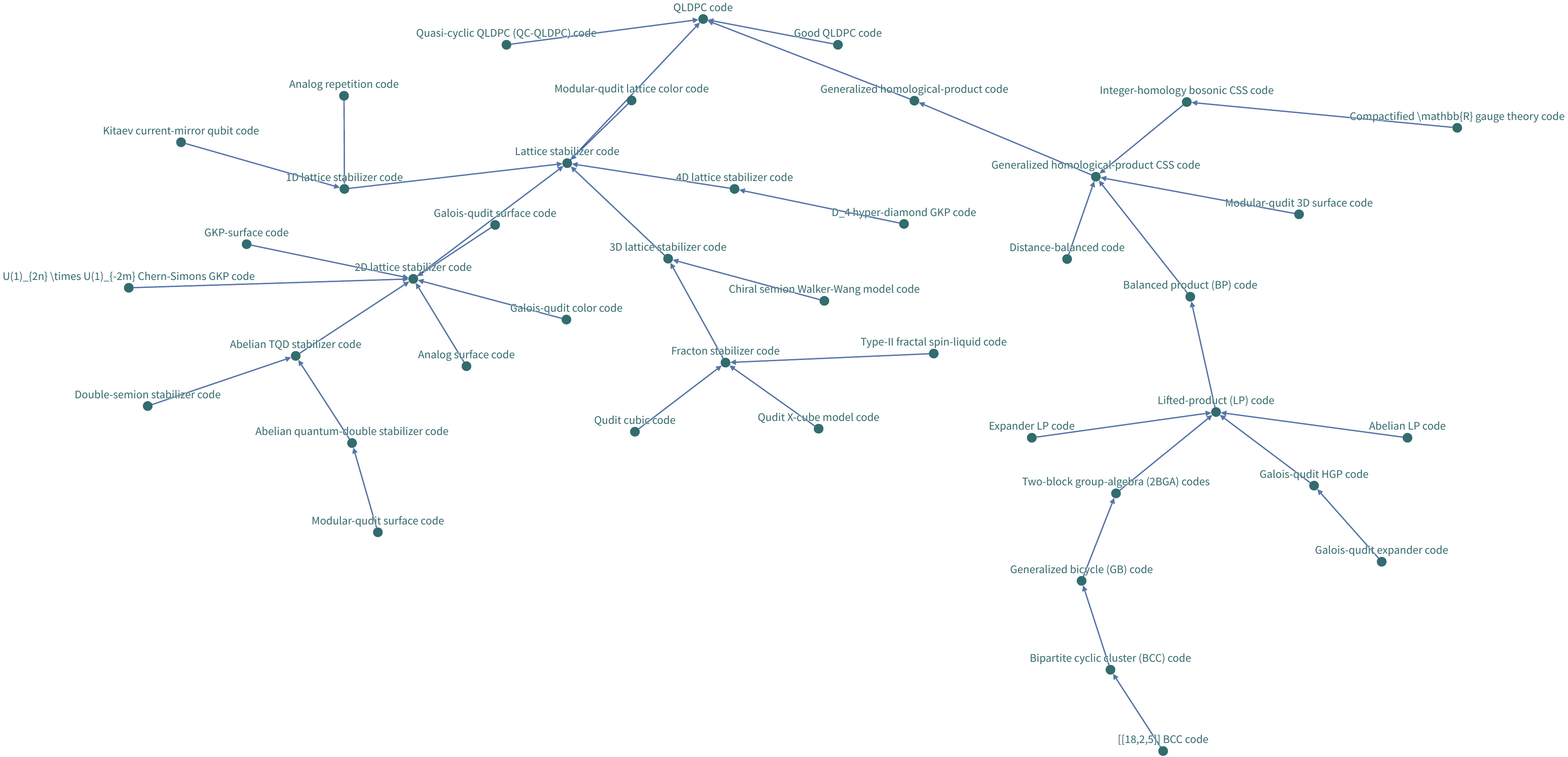}{Quantum LDPC codes}{https://errorcorrectionzoo.org/code_graph#J\%7B\%22displayMode\%22\%3A\%22subset\%22\%2C\%22modeSubsetOptions\%22\%3A\%7B\%22codeIds\%22\%3A\%5B\%221d_stabilizer\%22\%2C\%222d_stabilizer\%22\%2C\%223d_stabilizer\%22\%2C\%224d_stabilizer\%22\%2C\%22abelian_lifted_product\%22\%2C\%22tqd_abelian_stabilizer\%22\%2C\%22quantum_double_abelian\%22\%2C\%22analog_repetition\%22\%2C\%22analog_surface\%22\%2C\%22balanced_product\%22\%2C\%22bipartite_cyclic_cluster\%22\%2C\%223d_semion\%22\%2C\%22compactified_r\%22\%2C\%22distance_balanced\%22\%2C\%22double_semion\%22\%2C\%22expander_lifted_product\%22\%2C\%22fracton\%22\%2C\%22gkp_surface_concatenated\%22\%2C\%22galois_hypergraph_product\%22\%2C\%22galois_color\%22\%2C\%22galois_expander\%22\%2C\%22galois_topological\%22\%2C\%22generalized_bicycle\%22\%2C\%22generalized_homological_product_css\%22\%2C\%22generalized_homological_product\%22\%2C\%22good_qldpc\%22\%2C\%22homological_cv\%22\%2C\%22current_mirror\%22\%2C\%22translationally_invariant_stabilizer\%22\%2C\%22lifted_product\%22\%2C\%22qudit_3d_surface\%22\%2C\%22qudit_color\%22\%2C\%22qudit_surface\%22\%2C\%22general_qldpc\%22\%2C\%22quasi_cyclic_qldpc\%22\%2C\%22qudit_xcube\%22\%2C\%22qudit_cubic\%22\%2C\%222bga\%22\%2C\%22fractal_liquid\%22\%2C\%22dfour_gkp\%22\%2C\%22chern_simons_gkp\%22\%2C\%22stab_18_2_5\%22\%5D\%2C\%22reusePreviousLayoutPositions\%22\%3Afalse\%2C\%22showIntermediateConnectingNodes\%22\%3Atrue\%2C\%22connectingNodesMaxDepth\%22\%3A15\%2C\%22connectingNodesPathMaxLength\%22\%3A20\%2C\%22connectingNodesMaxExtraDepth\%22\%3A3\%2C\%22connectingNodesOnlyKeepPathsWithAdditionalLength\%22\%3A1\%2C\%22connectingNodesToDomainsAndKingdoms\%22\%3Afalse\%2C\%22connectingNodesEdgeLengthsByType\%22\%3A\%7B\%22primaryParent\%22\%3A1\%2C\%22secondaryParent\%22\%3A4\%2C\%22cousin\%22\%3A6\%7D\%2C\%22nodeIds\%22\%3A\%5B\%5D\%7D\%2C\%22highlightImportantNodes\%22\%3A\%7B\%22highlightImportantNodes\%22\%3Afalse\%2C\%22highlightPrimaryParents\%22\%3Afalse\%2C\%22highlightRootConnectingEdges\%22\%3Afalse\%7D\%7D}

\begingroup
\small
\eczhBreakableDashes
\renewcommand\arraystretch{1.05}
\edef\myxtraspc{\eczListAddVSpaceXtraXtraStretch}
\begin{tabularx}{\linewidth}{>{\raggedright\arraybackslash}p{\eczListColWidth{name}} >{\hsize=1.0000\hsize }X}
\toprule
\eczListColTitle{Code} & \eczListColTitle{Description} \\
\midrule
\endfirsthead
\toprule
\eczListColTitleContinued{Code} & \eczListColTitleContinued{Description} \\
\midrule
\endhead
\bottomrule
\endfoot
\eczhRefIndex{code:1d_stabilizer}%
\eczhListValue{\flmRefsHyperref{code:1d_stabilizer}{1D lattice stabilizer code}} & \eczhListValue{Lattice stabilizer code in one Euclidean dimension, using either the ordinary block notion of locality or the fermionic/Majorana notion of locality.}\\ 
\addlinespace[\myxtraspc]
\eczhRefIndex{code:2d_stabilizer}%
\eczhListValue{\flmRefsHyperref{code:2d_stabilizer}{2D lattice stabilizer code}} & \eczhListValue{Lattice stabilizer code in two Euclidean dimensions, using either the ordinary block notion of locality or the fermionic/Majorana notion of locality.}\\ 
\addlinespace[\myxtraspc]
\eczhRefIndex{code:3d_stabilizer}%
\eczhListValue{\flmRefsHyperref{code:3d_stabilizer}{3D lattice stabilizer code}} & \eczhListValue{Lattice stabilizer code in three Euclidean dimensions, using either the ordinary block notion of locality or the fermionic/Majorana notion of locality.}\\ 
\addlinespace[\myxtraspc]
\eczhRefIndex{code:4d_stabilizer}%
\eczhListValue{\flmRefsHyperref{code:4d_stabilizer}{4D lattice stabilizer code}} & \eczhListValue{Lattice stabilizer code in four Euclidean dimensions, using either the ordinary block notion of locality or the fermionic/Majorana notion of locality.}\\ 
\addlinespace[\myxtraspc]
\eczhRefIndex{code:abelian_lifted_product}%
\eczhListValue{\flmRefsHyperref{code:abelian_lifted_product}{Abelian LP code}} & \eczhListValue{A lifted-product code whose lift group \(G\) is Abelian.
The case of \(G\) being a cyclic group is a GB code (a.k.a. a quasi-cyclic LP code) \NoCaseChange{\protect\cite[{Sec. III.E}]{cite674}}.
A particular family with \(G=\mathbb{Z}_{\ell}\) yields codes with parameters \(\llbracket n,k=\Theta(\log n),d=\Theta(n/\log n)\rrbracket \) \NoCaseChange{\protect\cite{cite674}}.}\\ 
\addlinespace[\myxtraspc]
\eczhRefIndex{code:tqd_abelian_stabilizer}%
\eczhListValue{\flmRefsHyperref{code:tqd_abelian_stabilizer}{Abelian TQD stabilizer code}} & \eczhListValue{Modular-qudit stabilizer code whose codewords realize a 2D Abelian twisted-quantum-double topological order on composite-dimensional qudits.
For every finite Abelian group \(G=\prod_i \mathbb{Z}_{N_i}\) and every product of Type-I and Type-II cocycles, there is a Pauli stabilizer Hamiltonian realizing the corresponding Abelian TQD \NoCaseChange{\protect\cite{cite405}}.
Equivalently, these codes exhaust the 2D Abelian topological orders that admit gapped boundaries \NoCaseChange{\protect\cite{cite405,cite406}}.}\\ 
\addlinespace[\myxtraspc]
\eczhRefIndex{code:quantum_double_abelian}%
\eczhListValue{\flmRefsHyperref{code:quantum_double_abelian}{Abelian quantum-double stabilizer code}} & \eczhListValue{Modular-qudit stabilizer code whose codewords realize 2D modular gapped Abelian topological order with trivial cocycle.
The corresponding anyon theory is defined by an Abelian group.
The \(G=\mathbb{Z}_2\) instance on a torus is the toric code, and cyclic-group instances reduce to modular-qudit surface codes.
All such codes can be realized by a stack of modular-qudit surface codes because all finite Abelian groups are direct products of cyclic groups.}\\ 
\addlinespace[\myxtraspc]
\eczhRefIndex{code:analog_repetition}%
\eczhListValue{\flmRefsHyperref{code:analog_repetition}{Analog repetition code}} & \eczhListValue{An \(\llbracket n,1\rrbracket _{\mathbb{R}}\) analog stabilizer version of the quantum repetition code, encoding the position states of one mode into an odd number \(n\) of modes.}\\ 
\addlinespace[\myxtraspc]
\eczhRefIndex{code:analog_surface}%
\eczhListValue{\flmRefsHyperref{code:analog_surface}{Analog surface code}} & \eczhListValue{An analog CSS version of the Kitaev surface code realizing a phase of 2D \(\mathbb{R}\) gauge theory.}\\ 
\addlinespace[\myxtraspc]
\eczhRefIndex{code:balanced_product}%
\eczhListValue{\flmRefsHyperref{code:balanced_product}{Balanced product (BP) code}} & \eczhListValue{Family of CSS quantum codes obtained from two classical-code chain complexes that share a common group symmetry.
The balanced product can be understood as taking the usual tensor or hypergraph product and then quotienting by the shared symmetry action.
This can reduce the overall number of physical qubits \(n\) while, in favorable cases, preserving the number of encoded qubits and the code distance, thereby improving the encoding rate \(k/n\) and normalized distance \(d/n\) compared to the underlying tensor or hypergraph product.}\\ 
\addlinespace[\myxtraspc]
\eczhRefIndex{code:bipartite_cyclic_cluster}%
\eczhListValue{\flmRefsHyperref{code:bipartite_cyclic_cluster}{Bipartite cyclic cluster (BCC) code}} & \eczhListValue{Cyclic CSS code constructed from a bipartite cluster state with cyclic invariance,
emphasizing simplicity of state preparation over simplicity of stabilizers.}\\ 
\addlinespace[\myxtraspc]
\eczhRefIndex{code:3d_semion}%
\eczhListValue{\flmRefsHyperref{code:3d_semion}{Chiral semion Walker-Wang model code}} & \eczhListValue{A 3D lattice modular-qudit stabilizer code with qudit dimension \(q=4\) whose low-energy excitations on boundaries realize the chiral semion topological order.
The model admits 2D chiral semion topological order at one of its surfaces \NoCaseChange{\protect\cite{cite471,cite472}}.
The corresponding phase can also be realized via a non-stabilizer Hamiltonian \NoCaseChange{\protect\cite{cite473}}.}\\ 
\addlinespace[\myxtraspc]
\eczhRefIndex{code:compactified_r}%
\eczhListValue{\flmRefsHyperref{code:compactified_r}{Compactified \(\mathbb{R}\) gauge theory code}} & \eczhListValue{An integer-homology bosonic CSS code realizing 2D \(U(1)\) gauge theory on bosonic modes.
The code can be obtained from the analog surface code by \flmRefsHyperref{ref410}{condensing} certain anyons \NoCaseChange{\protect\cite{cite411}}. 
This results in a pinning of each mode to the space of periodic functions, which is the Hilbert space of a physical rotor, and can be thought of as compactification of the 2D \(\mathbb{R}\) gauge theory phase realized by the analog surface code.}\\ 
\addlinespace[\myxtraspc]
\eczhRefIndex{code:distance_balanced}%
\eczhListValue{\flmRefsHyperref{code:distance_balanced}{Distance-balanced code}} & \eczhListValue{Galois-qudit CSS code obtained from a CSS code by increasing the smaller of the \(X\)- and \(Z\)-distances using a homological-product-based balancing step or one of its generalizations.
The initial code is said to be \textit{unbalanced}, i.e., tailored to noise biased toward either bit- or phase-flip errors, and the procedure can result in a code that treats both types of errors on a more equal footing.}\\ 
\addlinespace[\myxtraspc]
\eczhRefIndex{code:double_semion}%
\eczhListValue{\flmRefsHyperref{code:double_semion}{Double-semion stabilizer code}} & \eczhListValue{A 2D lattice modular-qudit stabilizer code with qudit dimension \(q=4\) that realizes the 2D double semion topological phase.
The code can be obtained from a \(\mathbb{Z}_4\) toric-code ground state by \flmRefsHyperref{ref410}{condensing} the emergent boson \(e^2 m^2\); in the stabilizer construction this condensation is implemented by two-body measurements \NoCaseChange{\protect\cite{cite405,cite414}}.
Its ground-state subspace can be mapped to that of the double-semion string-net model by a finite-depth quantum circuit with ancillas \NoCaseChange{\protect\cite{cite405}}.}\\ 
\addlinespace[\myxtraspc]
\eczhRefIndex{code:expander_lifted_product}%
\eczhListValue{\flmRefsHyperref{code:expander_lifted_product}{Expander LP code}} & \eczhListValue{Family of \(G\)-lifted product codes constructed using two classical \flmRefsHyperref{code:expander}{expander codes}, equivalently two regular \flmRefsHyperref{code:tanner}{Tanner codes} defined on the same expander graph \NoCaseChange{\protect\cite{cite74}}. For certain parameters, this construction yields the first asymptotically good QLDPC codes. Classical codes resulting from the same lifted-product complexes are one of the first two families of \(c^3\)-LTCs \NoCaseChange{\protect\cite{cite184}}.}\\ 
\addlinespace[\myxtraspc]
\eczhRefIndex{code:fracton}%
\eczhListValue{\flmRefsHyperref{code:fracton}{Fracton stabilizer code}} & \eczhListValue{A 3D modular-qudit stabilizer code whose codewords make up the ground-state space of a Hamiltonian in a fracton phase.
Unlike topological phases, whose excitations can move in any direction, fracton phases are characterized by excitations whose movement is restricted.}\\ 
\addlinespace[\myxtraspc]
\eczhRefIndex{code:gkp_surface_concatenated}%
\eczhListValue{\flmRefsHyperref{code:gkp_surface_concatenated}{GKP-surface code}} & \eczhListValue{A concatenated code whose outer code is a GKP code and whose inner code is a surface code, including toric surface-code variants \NoCaseChange{\protect\cite{cite415,cite416}}, rotated surface codes \NoCaseChange{\protect\cite{cite417,cite418,cite419,cite420}}, and XZZX surface codes \NoCaseChange{\protect\cite{cite421}}.}\\ 
\addlinespace[\myxtraspc]
\eczhRefIndex{code:galois_hypergraph_product}%
\eczhListValue{\flmRefsHyperref{code:galois_hypergraph_product}{Galois-qudit HGP code}} & \eczhListValue{A member of a family of Galois-qudit CSS codes whose stabilizer generator matrix is obtained from a hypergraph product of two classical linear \(q\)-ary codes.}\\ 
\addlinespace[\myxtraspc]
\eczhRefIndex{code:galois_color}%
\eczhListValue{\flmRefsHyperref{code:galois_color}{Galois-qudit color code}} & \eczhListValue{Extension of the color code to 2D lattices of Galois qudits.}\\ 
\addlinespace[\myxtraspc]
\eczhRefIndex{code:galois_expander}%
\eczhListValue{\flmRefsHyperref{code:galois_expander}{Galois-qudit expander code}} & \eczhListValue{Galois-qudit CSS code obtained from tensor products of chain complexes associated with an explicit family of expander codes with Reed-Solomon local checks.}\\ 
\addlinespace[\myxtraspc]
\eczhRefIndex{code:galois_topological}%
\eczhListValue{\flmRefsHyperref{code:galois_topological}{Galois-qudit surface code}} & \eczhListValue{Extension of the surface code to 2D lattices of Galois qudits.}\\ 
\addlinespace[\myxtraspc]
\eczhRefIndex{code:generalized_bicycle}%
\eczhListValue{\flmRefsHyperref{code:generalized_bicycle}{Generalized bicycle (GB) code}} & \eczhListValue{A quasi-cyclic Galois-qudit CSS code constructed using a generalized version of the bicycle ansatz \NoCaseChange{\protect\cite{cite682}} from a pair of equivalent index-two quasi-cyclic linear codes.
Equivalently, the codes can be constructed via the lifted-product construction for \(G\) being a cyclic group \NoCaseChange{\protect\cite[{Sec. III.E}]{cite674}}.}\\ 
\addlinespace[\myxtraspc]
\eczhRefIndex{code:generalized_homological_product_css}%
\eczhListValue{\flmRefsHyperref{code:generalized_homological_product_css}{Generalized homological-product CSS code}} & \eczhListValue{CSS code whose properties are determined from an underlying chain complex, which often consists of some type of product of other chain complexes.}\\ 
\addlinespace[\myxtraspc]
\eczhRefIndex{code:generalized_homological_product}%
\eczhListValue{\flmRefsHyperref{code:generalized_homological_product}{Generalized homological-product code}} & \eczhListValue{Stabilizer code whose properties are determined from an underlying chain complex, which often consists of some type of product of other chain complexes.
The \flmRefsCref{ref683} yields an interpretation of codes in terms of chain complexes, thus allowing for the use of various products from homology in constructing codes.}\\ 
\addlinespace[\myxtraspc]
\eczhRefIndex{code:good_qldpc}%
\eczhListValue{\flmRefsHyperref{code:good_qldpc}{Good QLDPC code}} & \eczhListValue{Also called \textit{asymptotically good QLDPC codes}. A family of QLDPC codes \(\llbracket n_i,k_i,d_i\rrbracket \) whose asymptotic rate \(\lim_{i\to\infty} k_i/n_i\) and asymptotic distance \(\lim_{i\to\infty} d_i/n_i\) are both positive.}\\ 
\addlinespace[\myxtraspc]
\eczhRefIndex{code:homological_cv}%
\eczhListValue{\flmRefsHyperref{code:homological_cv}{Integer-homology bosonic CSS code}} & \eczhListValue{A bosonic stabilizer code whose physical modes have been restricted, via a single GKP stabilizer, from the space of functions on the real line to the space of periodic functions.
This restriction effectively realizes a rotor on each physical mode, allowing one to construct homological rotor codes out of displacement stabilizer groups.
The stabilizer group is continuous, but contains discrete components in the form of the single-mode GKP stabilizers.
The homology group of the logical operators has a torsion component because the chain complexes are defined over the ring of integers, which yields codes with finite logical dimension.}\\ 
\addlinespace[\myxtraspc]
\eczhRefIndex{code:current_mirror}%
\eczhListValue{\flmRefsHyperref{code:current_mirror}{Kitaev current-mirror qubit code}} & \eczhListValue{Member of the family of \(\llbracket 2n,(0,2),(2,n)\rrbracket _{\mathbb{Z}}\) homological rotor codes storing a logical qubit on a thin Möbius strip.
The ideal code can be obtained from a Josephson-junction \NoCaseChange{\protect\cite{cite396}} system \NoCaseChange{\protect\cite{cite397}}.}\\ 
\addlinespace[\myxtraspc]
\eczhRefIndex{code:translationally_invariant_stabilizer}%
\eczhListValue{\flmRefsHyperref{code:translationally_invariant_stabilizer}{Lattice stabilizer code}} & \eczhListValue{A geometrically local stabilizer code with sites organized on a lattice modeled by the additive group \(\mathbb{Z}^D\) for spatial dimension \(D\), using either the ordinary block notion of locality or the fermionic/Majorana notion of locality.
On an infinite lattice, its stabilizer group is generated by few-site Pauli-type operators and their translations, in which case the code is called \textit{translationally invariant stabilizer code}.
Boundary conditions have to be imposed on the lattice in order to obtain finite-dimensional versions.
Lattice defects and boundaries between different codes can also be introduced.}\\ 
\addlinespace[\myxtraspc]
\eczhRefIndex{code:lifted_product}%
\eczhListValue{\flmRefsHyperref{code:lifted_product}{Lifted-product (LP) code}} & \eczhListValue{Galois-qudit code that utilizes the notion of a lifted product in its construction. Lifted products of certain classical Tanner codes are the first (asymptotically) \textit{good QLDPC codes}.}\\ 
\addlinespace[\myxtraspc]
\eczhRefIndex{code:qudit_3d_surface}%
\eczhListValue{\flmRefsHyperref{code:qudit_3d_surface}{Modular-qudit 3D surface code}} & \eczhListValue{A generalization of the 3D surface code to modular qudits.
Qudits are placed on edges, \(Z\)-type stabilizer generators are placed on square plaquettes oriented in all three directions, and \(X\)-type stabilizers are placed on the six edges neighboring every vertex \NoCaseChange{\protect\cite{cite459}}.}\\ 
\addlinespace[\myxtraspc]
\eczhRefIndex{code:qudit_color}%
\eczhListValue{\flmRefsHyperref{code:qudit_color}{Modular-qudit lattice color code}} & \eczhListValue{Extension of the color code to lattices of modular qudits.
Codes are defined analogously to qubit color codes on suitable lattices of any spatial dimension, but a directionality is required in order to make the modular-qudit stabilizers commute.
This can be done by puncturing a hyperspherical lattice \NoCaseChange{\protect\cite{cite475}} or constructing a star-bipartition; see \NoCaseChange{\protect\cite[{Sec. III}]{cite673}}.
Logical dimension is determined by the genus of the underlying surface (for closed surfaces), types of boundaries (for open surfaces), and/or any twist defects present.}\\ 
\addlinespace[\myxtraspc]
\eczhRefIndex{code:qudit_surface}%
\eczhListValue{\flmRefsHyperref{code:qudit_surface}{Modular-qudit surface code}} & \eczhListValue{Extension of the surface code to prime-dimensional \NoCaseChange{\protect\cite{cite423,cite424}} and more general modular qudits.
Stabilizer generators are few-body \(X\)-type and \(Z\)-type Pauli strings associated to the stars and plaquettes, respectively, of a tessellation of a two-dimensional surface.
Since qudits have more than one \(X\) and \(Z\)-type operator, various sets of stabilizer generators can be defined.}\\ 
\addlinespace[\myxtraspc]
\eczhRefIndex{code:general_qldpc}%
\eczhListValue{\flmRefsHyperref{code:general_qldpc}{QLDPC code}} & \eczhListValue{Member of a family of stabilizer codes for which the number of sites participating in each stabilizer generator and the number of stabilizer generators that each site participates in are both bounded by a constant as \(n\to\infty\).
Sometimes, the two parameters are explicitly stated: each site of an \((l,w)\)\textit{-regular QLDPC code} is acted on by \(\leq l\) generators of weight \(\leq w\).}\\ 
\addlinespace[\myxtraspc]
\eczhRefIndex{code:quasi_cyclic_qldpc}%
\eczhListValue{\flmRefsHyperref{code:quasi_cyclic_qldpc}{Quasi-cyclic QLDPC (QC-QLDPC) code}} & \eczhListValue{A QLDPC code such that cyclic shifts of the subsystems by a fixed \(\ell\geq 1\) leave the codespace invariant.
Stabilizer generator matrices of such codes can be put into block form, where each nonzero block is a circulant matrix \NoCaseChange{\protect\cite{cite821,cite822}}.}\\ 
\addlinespace[\myxtraspc]
\eczhRefIndex{code:qudit_xcube}%
\eczhListValue{\flmRefsHyperref{code:qudit_xcube}{Qudit X-cube model code}} & \eczhListValue{Generalization of the X-cube model code to modular qudits.}\\ 
\addlinespace[\myxtraspc]
\eczhRefIndex{code:qudit_cubic}%
\eczhListValue{\flmRefsHyperref{code:qudit_cubic}{Qudit cubic code}} & \eczhListValue{Generalization of the Haah cubic code to modular qudits.}\\ 
\addlinespace[\myxtraspc]
\eczhRefIndex{code:2bga}%
\eczhListValue{\flmRefsHyperref{code:2bga}{Two-block group-algebra (2BGA) codes}} & \eczhListValue{2BGA codes are the one-by-one, or smallest, \flmRefsHyperref{code:lifted_product}{LP codes}:
\(LP(a,b)\) is defined by a pair of \flmRefsHyperref{ref205}{group algebra} elements
\(a,b\in \mathbb{F}_q[G]\), where \(G\) is a finite group.
If \(|G|=\ell\), then the code has length \(n=2\ell\).}\\ 
\addlinespace[\myxtraspc]
\eczhRefIndex{code:fractal_liquid}%
\eczhListValue{\flmRefsHyperref{code:fractal_liquid}{Type-II fractal spin-liquid code}} & \eczhListValue{A type-II fracton prime-qudit CSS code defined on a cubic lattice \NoCaseChange{\protect\cite[{Eqs. (D9-D10)}]{cite456}}.}\\ 
\addlinespace[\myxtraspc]
\eczhRefIndex{code:dfour_gkp}%
\eczhListValue{\flmRefsHyperref{code:dfour_gkp}{\(D_4\) hyper-diamond GKP code}} & \eczhListValue{Two-mode GKP qubit-into-oscillator code based on the \(D_4\) hyper-diamond lattice \NoCaseChange{\protect\cite{cite482}}.}\\ 
\addlinespace[\myxtraspc]
\eczhRefIndex{code:chern_simons_gkp}%
\eczhListValue{\flmRefsHyperref{code:chern_simons_gkp}{\(U(1)_{2n} \times U(1)_{-2m}\) Chern-Simons GKP code}} & \eczhListValue{A non-CSS multimode GKP code defined on a 2D mode lattice that encodes a qudit logical space and whose excitations are characterized by the \(U(1)_{2n} \times U(1)_{-2m}\) Chern-Simons theory.
The code can be obtained from the analog surface code by \flmRefsHyperref{ref410}{condensing} certain anyons \NoCaseChange{\protect\cite{cite411}}.}\\ 
\addlinespace[\myxtraspc]
\eczhRefIndex{code:stab_18_2_5}%
\eczhListValue{\flmRefsHyperref{code:stab_18_2_5}{\(\llbracket 18,2,5\rrbracket \) BCC code}} & \eczhListValue{BCC code on 18 qubits encoding 2 logical qubits with distance 5, found by computer search \NoCaseChange{\protect\cite{cite440}}.}\\ 
\end{tabularx}\endgroup
\eczcodelist{qltc}{Quantum locally testable codes and friends
}%

\eczhCodeListAutoDescription{All descendants and cousins of \flmRefsCref{code:qltc}.}%

\eczhIncludeCodeGraph{Bare}{scale=0.5}{\columnwidth}{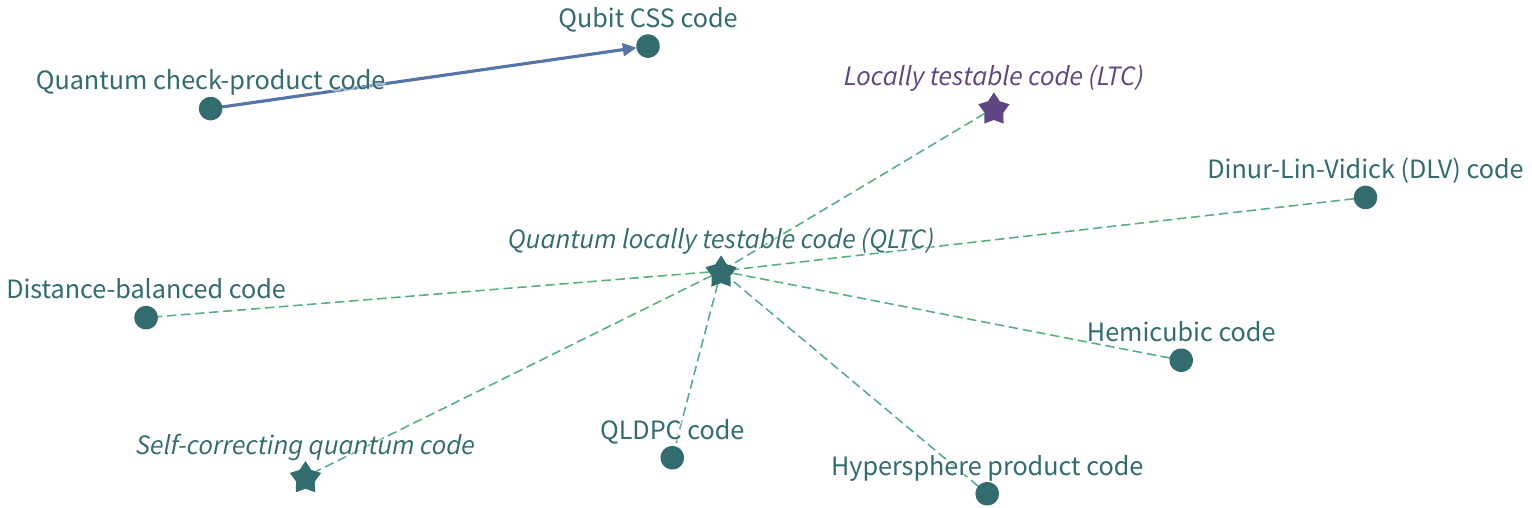}{Quantum locally testable codes and friends}{https://errorcorrectionzoo.org/code_graph#J\%7B\%22displayMode\%22\%3A\%22subset\%22\%2C\%22modeSubsetOptions\%22\%3A\%7B\%22codeIds\%22\%3A\%5B\%22dlv\%22\%2C\%22distance_balanced\%22\%2C\%22hemicubic\%22\%2C\%22hypersphere_product\%22\%2C\%22ltc\%22\%2C\%22general_qldpc\%22\%2C\%22check_product\%22\%2C\%22qltc\%22\%2C\%22qubit_css\%22\%2C\%22self_correct\%22\%5D\%2C\%22reusePreviousLayoutPositions\%22\%3Afalse\%2C\%22showIntermediateConnectingNodes\%22\%3Atrue\%2C\%22connectingNodesMaxDepth\%22\%3A15\%2C\%22connectingNodesPathMaxLength\%22\%3A20\%2C\%22connectingNodesMaxExtraDepth\%22\%3A3\%2C\%22connectingNodesOnlyKeepPathsWithAdditionalLength\%22\%3A1\%2C\%22connectingNodesToDomainsAndKingdoms\%22\%3Afalse\%2C\%22connectingNodesEdgeLengthsByType\%22\%3A\%7B\%22primaryParent\%22\%3A1\%2C\%22secondaryParent\%22\%3A4\%2C\%22cousin\%22\%3A6\%7D\%2C\%22nodeIds\%22\%3A\%5B\%5D\%7D\%2C\%22highlightImportantNodes\%22\%3A\%7B\%22highlightImportantNodes\%22\%3Afalse\%2C\%22highlightPrimaryParents\%22\%3Afalse\%2C\%22highlightRootConnectingEdges\%22\%3Afalse\%7D\%7D}

\begingroup
\small
\eczhBreakableDashes
\renewcommand\arraystretch{1.05}
\edef\myxtraspc{\eczListAddVSpaceXtraXtraStretch}
\begin{tabularx}{\linewidth}{>{\raggedright\arraybackslash}p{\eczListColWidth{name}} >{\hsize=1.0000\hsize }X}
\toprule
\eczListColTitle{Code} & \eczListColTitle{Description} \\
\midrule
\endfirsthead
\toprule
\eczListColTitleContinued{Code} & \eczListColTitleContinued{Description} \\
\midrule
\endhead
\bottomrule
\endfoot
\eczhRefIndex{code:dlv}%
\eczhListValue{\flmRefsHyperref{code:dlv}{Dinur-Lin-Vidick (DLV) code}} & \eczhListValue{Member of a family of codes constructed using cubical chain complexes, which are \(t\)-order extensions of the complexes underlying expander codes (\(t=1\)) and expander lifted-product codes (\(t=2\)).}\\ 
\addlinespace[\myxtraspc]
\eczhRefIndex{code:distance_balanced}%
\eczhListValue{\flmRefsHyperref{code:distance_balanced}{Distance-balanced code}} & \eczhListValue{Galois-qudit CSS code obtained from a CSS code by increasing the smaller of the \(X\)- and \(Z\)-distances using a homological-product-based balancing step or one of its generalizations.
The initial code is said to be \textit{unbalanced}, i.e., tailored to noise biased toward either bit- or phase-flip errors, and the procedure can result in a code that treats both types of errors on a more equal footing.}\\ 
\addlinespace[\myxtraspc]
\eczhRefIndex{code:hemicubic}%
\eczhListValue{\flmRefsHyperref{code:hemicubic}{Hemicubic code}} & \eczhListValue{Homological code constructed out of cubes in high dimensions.
The hemicubic code family has asymptotically diminishing soundness that scales as \flmRefsHyperref{ref65}{order} \(\Omega(1/\log n)\), locality of stabilizer generators scaling as \flmRefsHyperref{ref65}{order} \(O(\log n)\), and distance of \flmRefsHyperref{ref65}{order} \(\Theta(\sqrt{n})\).}\\ 
\addlinespace[\myxtraspc]
\eczhRefIndex{code:hypersphere_product}%
\eczhListValue{\flmRefsHyperref{code:hypersphere_product}{Hypersphere product code}} & \eczhListValue{Homological code based on products of hyperspheres.
The hypersphere product code family has asymptotically diminishing soundness that scales as \flmRefsHyperref{ref65}{order} \(O(1/\log (n)^2)\), locality of stabilizer generators scaling as \flmRefsHyperref{ref65}{order} \(O(\log n/ \log\log n)\), and distance of \flmRefsHyperref{ref65}{order} \(\Theta(\sqrt{n})\).}\\ 
\addlinespace[\myxtraspc]
\eczhRefIndex{code:ltc}%
\eczhListValue{\flmRefsHyperref{code:ltc}{Locally testable code (LTC)}} & \eczhListValue{Code for which one can efficiently check whether a given string is a codeword or is far from a codeword. Efficiency of the verification is quantified by the code's \textit{query complexity} \(u\), while effectiveness is quantified by the code's \textit{soundness} \(R\).}\\ 
\addlinespace[\myxtraspc]
\eczhRefIndex{code:general_qldpc}%
\eczhListValue{\flmRefsHyperref{code:general_qldpc}{QLDPC code}} & \eczhListValue{Member of a family of stabilizer codes for which the number of sites participating in each stabilizer generator and the number of stabilizer generators that each site participates in are both bounded by a constant as \(n\to\infty\).
Sometimes, the two parameters are explicitly stated: each site of an \((l,w)\)\textit{-regular QLDPC code} is acted on by \(\leq l\) generators of weight \(\leq w\).}\\ 
\addlinespace[\myxtraspc]
\eczhRefIndex{code:check_product}%
\eczhListValue{\flmRefsHyperref{code:check_product}{Quantum check-product code}} & \eczhListValue{CSS code constructed from an extension of the check product (between two classical codes) to a product between a classical and a quantum code.}\\ 
\addlinespace[\myxtraspc]
\eczhRefIndex{code:qltc}%
\eczhListValue{\flmRefsHyperref{code:qltc}{Quantum locally testable code (QLTC)}} & \eczhListValue{A local commuting-projector Hamiltonian-based block quantum code which has a nonzero average-energy penalty for creating large errors. Informally, states that are far away from the codespace of a QLTC have to be excited states of a number of the code's local projectors that scales linearly with \(n\).}\\ 
\addlinespace[\myxtraspc]
\eczhRefIndex{code:qubit_css}%
\eczhListValue{\flmRefsHyperref{code:qubit_css}{Qubit CSS code}} & \eczhListValue{An \(\llbracket n,k,d\rrbracket \) stabilizer code admitting a set of stabilizer generators that are either \(Z\)-type or \(X\)-type Pauli strings.
Codes can be defined from two classical codes and/or chain complexes over \(\mathbb{Z}_2\) per the \flmRefsHyperref{ref683}{qubit CSS-to-homology correspondence} below.}\\ 
\addlinespace[\myxtraspc]
\eczhRefIndex{code:self_correct}%
\eczhListValue{\flmRefsHyperref{code:self_correct}{Self-correcting quantum code}} & \eczhListValue{A block quantum code that forms the ground-state subspace of an \(n\)-body geometrically local Hamiltonian whose logical information is recoverable for arbitrarily long times in the \(n\to\infty\) limit after interaction with a sufficiently cold thermal environment.
Typically, one also requires a decoder whose decoding time scales polynomially with \(n\) and a finite energy density.}\\ 
\end{tabularx}\endgroup
\eczcodelist{quantum_mds}{Quantum MDS codes and friends
}%

\eczhCodeListAutoDescription{Union of:
\begin{itemize}\item codes that are descendants of \flmRefsCref{code:quantum_mds}
\item codes that are cousins of \flmRefsCref{code:quantum_mds}
\item codes that are descendants of \flmRefsCref{code:ea_mds}
\item codes that are cousins of \flmRefsCref{code:ea_mds}
\end{itemize}}%

\eczhIncludeCodeGraph{Bare}{scale=0.5}{\columnwidth}{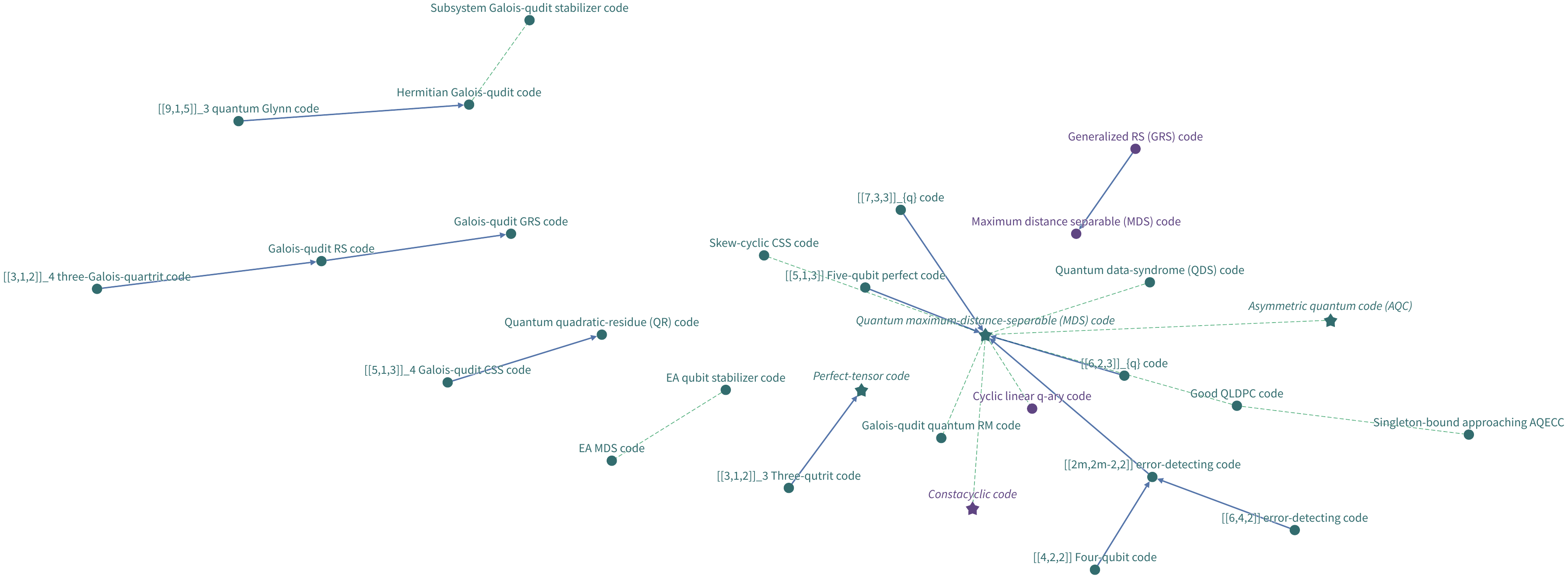}{Quantum MDS codes and friends}{https://errorcorrectionzoo.org/code_graph#J\%7B\%22displayMode\%22\%3A\%22subset\%22\%2C\%22modeSubsetOptions\%22\%3A\%7B\%22codeIds\%22\%3A\%5B\%22asymmetric_qecc\%22\%2C\%22constacyclic\%22\%2C\%22q-ary_cyclic\%22\%2C\%22ea_mds\%22\%2C\%22eastab\%22\%2C\%22galois_grs\%22\%2C\%22galois_polynomial\%22\%2C\%22galois_reed_muller\%22\%2C\%22generalized_reed_solomon\%22\%2C\%22good_qldpc\%22\%2C\%22stabilizer_over_gfqsq\%22\%2C\%22mds\%22\%2C\%22ame\%22\%2C\%22data_syndrome\%22\%2C\%22quantum_mds\%22\%2C\%22galois_quad_residue\%22\%2C\%22quantum_singleton\%22\%2C\%22skew-cyclic_galois_css\%22\%2C\%22galois_subsystem_stabilizer\%22\%2C\%22iceberg\%22\%2C\%22stab_3_1_2\%22\%2C\%22galois_3_1_2\%22\%2C\%22stab_4_2_2\%22\%2C\%22stab_5_1_3\%22\%2C\%22css_5_1_3\%22\%2C\%22galois_6_2_3\%22\%2C\%22stab_6_4_2\%22\%2C\%22galois_7_3_3\%22\%2C\%22stab_9_1_5\%22\%5D\%2C\%22reusePreviousLayoutPositions\%22\%3Afalse\%2C\%22showIntermediateConnectingNodes\%22\%3Atrue\%2C\%22connectingNodesMaxDepth\%22\%3A15\%2C\%22connectingNodesPathMaxLength\%22\%3A20\%2C\%22connectingNodesMaxExtraDepth\%22\%3A3\%2C\%22connectingNodesOnlyKeepPathsWithAdditionalLength\%22\%3A1\%2C\%22connectingNodesToDomainsAndKingdoms\%22\%3Afalse\%2C\%22connectingNodesEdgeLengthsByType\%22\%3A\%7B\%22primaryParent\%22\%3A1\%2C\%22secondaryParent\%22\%3A4\%2C\%22cousin\%22\%3A6\%7D\%2C\%22nodeIds\%22\%3A\%5B\%5D\%7D\%2C\%22highlightImportantNodes\%22\%3A\%7B\%22highlightImportantNodes\%22\%3Afalse\%2C\%22highlightPrimaryParents\%22\%3Afalse\%2C\%22highlightRootConnectingEdges\%22\%3Afalse\%7D\%7D}

\begingroup
\small
\eczhBreakableDashes
\renewcommand\arraystretch{1.05}
\edef\myxtraspc{\eczListAddVSpaceXtraXtraStretch}
\begin{tabularx}{\linewidth}{>{\raggedright\arraybackslash}p{\eczListColWidth{name}} >{\hsize=1.0000\hsize }X}
\toprule
\eczListColTitle{Code} & \eczListColTitle{Description} \\
\midrule
\endfirsthead
\toprule
\eczListColTitleContinued{Code} & \eczListColTitleContinued{Description} \\
\midrule
\endhead
\bottomrule
\endfoot
\eczhRefIndex{code:asymmetric_qecc}%
\eczhListValue{\flmRefsHyperref{code:asymmetric_qecc}{Asymmetric quantum code (AQC)}} & \eczhListValue{Quantum systems can be roughly characterized by two types of noise, a bit-flip noise that maps canonical basis states into each other, and a phase-flip noise that induces relative phases between superpositions of such basis states.
A code cannot protect against both types of noise arbitrarily well, and there is a tradeoff between the two types of protection.
An AQC is one that performs much better against one type of noise than the other type.
Such codes typically have tunable distances against each noise type and include CSS codes, GKP codes, and QSCs.}\\ 
\addlinespace[\myxtraspc]
\eczhRefIndex{code:constacyclic}%
\eczhListValue{\flmRefsHyperref{code:constacyclic}{Constacyclic code}} & \eczhListValue{A block code \(C\) of length \(n\) over an alphabet \(R\) is \(\alpha\)-constacyclic (or \(\alpha\)-twisted) if, for each string \(c_1 c_2 \cdots c_n\in C\), the string \(\alpha c_n, c_1, \cdots, c_{n-1} \in C\) \NoCaseChange{\protect\cite[{Def. 3.2.7}]{cite70}}.
A \(-1\)-constacyclic code is called \textit{negacyclic}.}\\ 
\addlinespace[\myxtraspc]
\eczhRefIndex{code:q-ary_cyclic}%
\eczhListValue{\flmRefsHyperref{code:q-ary_cyclic}{Cyclic linear \(q\)-ary code}} & \eczhListValue{A \(q\)-ary code of length \(n\) is cyclic if, for each codeword \(c_1 c_2 \cdots c_n\), the cyclically shifted string \(c_n c_1 \cdots c_{n-1}\) is also a codeword. A cyclic code is called \textit{primitive} when \(n=q^r-1\) for some \(r\geq 2\). A \textit{shortened cyclic code} is obtained from a cyclic code by taking only codewords with the first \(j\) zero entries, and deleting those zeroes.}\\ 
\addlinespace[\myxtraspc]
\eczhRefIndex{code:ea_mds}%
\eczhListValue{\flmRefsHyperref{code:ea_mds}{EA MDS code}} & \eczhListValue{EA Galois-qudit code whose parameters make the EAQECC Singleton bound \NoCaseChange{\protect\cite[{Thm. 6}]{cite545}} become an equality.}\\ 
\addlinespace[\myxtraspc]
\eczhRefIndex{code:eastab}%
\eczhListValue{\flmRefsHyperref{code:eastab}{EA qubit stabilizer code}} & \eczhListValue{A code constructed using a variation of the stabilizer formalism designed to utilize pre-shared entanglement between sender and receiver.
A code is typically denoted as \(\llbracket n,k;e\rrbracket \) or \(\llbracket n,k,d;e\rrbracket \), where \(d\) is the distance of the EA code and \(e\) is the number of required pre-shared maximally entangled Bell states (ebits).
While other entangled states can be used, there is always a choice of generators such that Bell states suffice while still using the fewest ebits.}\\ 
\addlinespace[\myxtraspc]
\eczhRefIndex{code:galois_grs}%
\eczhListValue{\flmRefsHyperref{code:galois_grs}{Galois-qudit GRS code}} & \eczhListValue{A true \(q\)-Galois-qudit stabilizer code constructed from GRS codes via either the Hermitian construction \NoCaseChange{\protect\cite{cite823,cite824,cite825}} or the Galois-qudit CSS construction \NoCaseChange{\protect\cite{cite826,cite827}}.}\\ 
\addlinespace[\myxtraspc]
\eczhRefIndex{code:galois_polynomial}%
\eczhListValue{\flmRefsHyperref{code:galois_polynomial}{Galois-qudit RS code}} & \eczhListValue{A Galois-qudit CSS code family (with \(q>n\)) constructed using two RS codes over \(\mathbb{F}_q\).}\\ 
\addlinespace[\myxtraspc]
\eczhRefIndex{code:galois_reed_muller}%
\eczhListValue{\flmRefsHyperref{code:galois_reed_muller}{Galois-qudit quantum RM code}} & \eczhListValue{True Galois-qudit stabilizer code constructed from generalized Reed-Muller (GRM) codes via the Galois-qudit Hermitian construction, the Galois-qudit CSS construction, or directly from their parity-check matrices \NoCaseChange{\protect\cite{cite828}\protect\cite[{Sec. 4.2}]{cite829}}.}\\ 
\addlinespace[\myxtraspc]
\eczhRefIndex{code:generalized_reed_solomon}%
\eczhListValue{\flmRefsHyperref{code:generalized_reed_solomon}{Generalized RS (GRS) code}} & \eczhListValue{An \([n,k,n-k+1]_q\) MDS code that is a modification of the RS code where codeword polynomials are multiplied by additional prefactors \NoCaseChange{\protect\cite[{Def. 15.3.19}]{cite26}}.}\\ 
\addlinespace[\myxtraspc]
\eczhRefIndex{code:good_qldpc}%
\eczhListValue{\flmRefsHyperref{code:good_qldpc}{Good QLDPC code}} & \eczhListValue{Also called \textit{asymptotically good QLDPC codes}. A family of QLDPC codes \(\llbracket n_i,k_i,d_i\rrbracket \) whose asymptotic rate \(\lim_{i\to\infty} k_i/n_i\) and asymptotic distance \(\lim_{i\to\infty} d_i/n_i\) are both positive.}\\ 
\addlinespace[\myxtraspc]
\eczhRefIndex{code:stabilizer_over_gfqsq}%
\eczhListValue{\flmRefsHyperref{code:stabilizer_over_gfqsq}{Hermitian Galois-qudit code}} & \eczhListValue{An \(\llbracket n,k,d\rrbracket _q\) true Galois-qudit stabilizer code constructed from a Hermitian self-orthogonal linear code over \(\mathbb{F}_{q^2}\) using the one-to-one correspondence between the Galois-qudit Pauli matrices and elements of the Galois field \(\mathbb{F}_{q^2}\).}\\ 
\addlinespace[\myxtraspc]
\eczhRefIndex{code:mds}%
\eczhListValue{\flmRefsHyperref{code:mds}{Maximum distance separable (MDS) code}} & \eczhListValue{A \(q\)-ary linear code whose parameters satisfy the Singleton bound with equality.}\\ 
\addlinespace[\myxtraspc]
\eczhRefIndex{code:ame}%
\eczhListValue{\flmRefsHyperref{code:ame}{Perfect-tensor code}} & \eczhListValue{Block quantum code encoding one subsystem into an odd number \(n\) subsystems whose encoding isometry is a perfect tensor.
This code stems from an AME\((n,q)\) \flmRefsHyperref{ref219}{AME state}, or equivalently, a \(\llparenthesis n+1,1,\lfloor (n+1)/2 \rfloor + 1\rrparenthesis \) code.}\\ 
\addlinespace[\myxtraspc]
\eczhRefIndex{code:data_syndrome}%
\eczhListValue{\flmRefsHyperref{code:data_syndrome}{Quantum data-syndrome (QDS) code}} & \eczhListValue{Stabilizer code designed to correct both data qubit errors and syndrome measurement errors simultaneously due to extra redundancy in its stabilizer generators.}\\ 
\addlinespace[\myxtraspc]
\eczhRefIndex{code:quantum_mds}%
\eczhListValue{\flmRefsHyperref{code:quantum_mds}{Quantum maximum-distance-separable (MDS) code}} & \eczhListValue{A type of block quantum code whose parameters satisfy the quantum Singleton bound with equality.}\\ 
\addlinespace[\myxtraspc]
\eczhRefIndex{code:galois_quad_residue}%
\eczhListValue{\flmRefsHyperref{code:galois_quad_residue}{Quantum quadratic-residue (QR) code}} & \eczhListValue{Galois-qudit \(\llbracket n,1\rrbracket _q\) \flmRefsHyperref{ref672}{pure} self-dual Galois-qudit CSS code constructed from a dual-containing QR code via the Galois-qudit CSS construction.
For \(q\) not divisible by \(n\), its distance satisfies \(d^2-d+1 \geq n\) when \(n \equiv 3\) modulo 4 \NoCaseChange{\protect\cite[{Thm. 40}]{cite813}} and \(d \geq \sqrt{n}\) when \(n\equiv 1\) modulo 4 \NoCaseChange{\protect\cite[{Thm. 41}]{cite813}}.}\\ 
\addlinespace[\myxtraspc]
\eczhRefIndex{code:quantum_singleton}%
\eczhListValue{\flmRefsHyperref{code:quantum_singleton}{Singleton-bound approaching AQECC}} & \eczhListValue{A member of an approximate quantum code family of rate \(R\) that can tolerate adversarial errors nearly saturating the quantum Singleton bound of \((1-R)/2\).
The formulation of such codes relies on a notion of \textit{quantum list decoding} \NoCaseChange{\protect\cite{cite814,cite495}}.}\\ 
\addlinespace[\myxtraspc]
\eczhRefIndex{code:skew-cyclic_galois_css}%
\eczhListValue{\flmRefsHyperref{code:skew-cyclic_galois_css}{Skew-cyclic CSS code}} & \eczhListValue{A member of a family of Galois-qudit CSS codes constructed from skew-cyclic classical codes over rings \NoCaseChange{\protect\cite[{Thm. 5.8}]{cite815}}.
See related study \NoCaseChange{\protect\cite{cite816}} that uses cyclic codes over rings.}\\ 
\addlinespace[\myxtraspc]
\eczhRefIndex{code:galois_subsystem_stabilizer}%
\eczhListValue{\flmRefsHyperref{code:galois_subsystem_stabilizer}{Subsystem Galois-qudit stabilizer code}} & \eczhListValue{Galois-qudit generalization of a subsystem qubit stabilizer code.
Can be obtained by taking a Galois-qudit stabilizer code and assigning some of its logical qudits to be gauge qudits.}\\ 
\addlinespace[\myxtraspc]
\eczhRefIndex{code:iceberg}%
\eczhListValue{\flmRefsHyperref{code:iceberg}{\(\llbracket 2m,2m-2,2\rrbracket \) error-detecting code}} & \eczhListValue{Self-complementary and self-dual CSS code for \(m\geq 2\) with generators \(\{XX\cdots X, ZZ\cdots Z\} \) acting on all \(2m\) physical qubits.
The code is constructed via the CSS construction from an SPC code and a repetition code \NoCaseChange{\protect\cite[{Sec. III}]{cite773}}.
This is the highest-rate distance-two code when an even number of qubits is used \NoCaseChange{\protect\cite{cite449}}.}\\ 
\addlinespace[\myxtraspc]
\eczhRefIndex{code:stab_3_1_2}%
\eczhListValue{\flmRefsHyperref{code:stab_3_1_2}{\(\llbracket 3,1,2\rrbracket _3\) Three-qutrit code}} & \eczhListValue{A \(\llbracket 3,1,2\rrbracket _3\) prime-qudit CSS code that is the smallest qutrit stabilizer code to detect a single-qutrit error.
It has stabilizer generators \(ZZZ\) and \(XXX\). The code defines a quantum secret-sharing scheme and serves as a minimal model for the AdS/CFT holographic duality. It is also the smallest non-trivial instance of a quantum maximum distance separable code (QMDS), saturating the quantum Singleton bound.}\\ 
\addlinespace[\myxtraspc]
\eczhRefIndex{code:galois_3_1_2}%
\eczhListValue{\flmRefsHyperref{code:galois_3_1_2}{\(\llbracket 3,1,2\rrbracket _4\) three-Galois-quartrit code}} & \eczhListValue{Three-Galois-qudit CSS code over \(\mathbb{F}_4=\{0,1,\omega,\omega^2\}\) that encodes one logical Galois qudit and detects a single-qudit error.}\\ 
\addlinespace[\myxtraspc]
\eczhRefIndex{code:stab_4_2_2}%
\eczhListValue{\flmRefsHyperref{code:stab_4_2_2}{\(\llbracket 4,2,2\rrbracket \) Four-qubit code}} & \eczhListValue{A four-qubit hyperbolic self-dual CSS stabilizer code that is the smallest two-logical-qubit stabilizer code to detect a single-qubit error.
It is unique for its parameters \NoCaseChange{\protect\cite[{Thm. 8}]{cite446}}.}\\ 
\addlinespace[\myxtraspc]
\eczhRefIndex{code:stab_5_1_3}%
\eczhListValue{\flmRefsHyperref{code:stab_5_1_3}{\(\llbracket 5,1,3\rrbracket \) Five-qubit perfect code}} & \eczhListValue{Five-qubit cyclic stabilizer code that is the smallest qubit stabilizer code to correct a single-qubit error.}\\ 
\addlinespace[\myxtraspc]
\eczhRefIndex{code:css_5_1_3}%
\eczhListValue{\flmRefsHyperref{code:css_5_1_3}{\(\llbracket 5,1,3\rrbracket _4\) Galois-qudit CSS code}} & \eczhListValue{Five-Galois-qudit CSS code over \(\mathbb{F}_4=\{0,1,\omega,\omega^2\}\) that encodes one logical Galois qudit and corrects a single-qudit error.}\\ 
\addlinespace[\myxtraspc]
\eczhRefIndex{code:galois_6_2_3}%
\eczhListValue{\flmRefsHyperref{code:galois_6_2_3}{\(\llbracket 6,2,3\rrbracket _{q}\) code}} & \eczhListValue{Six-qudit MDS error-correcting code defined for Galois-qudit dimension \(q=3\) \NoCaseChange{\protect\cite{cite830}}, \(q=2^2\) \NoCaseChange{\protect\cite{cite831}}, and \(q \geq 5\) \NoCaseChange{\protect\cite{cite830}\protect\cite[{Exam. 33}]{cite813}}.
This code cannot exist for qubits (\(q=2\)).}\\ 
\addlinespace[\myxtraspc]
\eczhRefIndex{code:stab_6_4_2}%
\eczhListValue{\flmRefsHyperref{code:stab_6_4_2}{\(\llbracket 6,4,2\rrbracket \) error-detecting code}} & \eczhListValue{Self-complementary six-qubit code with rate \(2/3\) that is unique for its parameters, up to equivalence \NoCaseChange{\protect\cite[{Tab. III}]{cite449}}.
Concatenations of this code with itself yield the \(\llbracket 6^r,4^r,2^r\rrbracket \) level-\(r\) \textit{many-hypercube} code \NoCaseChange{\protect\cite{cite450}}.}\\ 
\addlinespace[\myxtraspc]
\eczhRefIndex{code:galois_7_3_3}%
\eczhListValue{\flmRefsHyperref{code:galois_7_3_3}{\(\llbracket 7,3,3\rrbracket _{q}\) code}} & \eczhListValue{Seven-qudit MDS error-detecting code defined for Galois-qudit dimension \(q=3\) \NoCaseChange{\protect\cite{cite830}} and \(q \geq 7\) \NoCaseChange{\protect\cite{cite830}\protect\cite[{Exam. 33}]{cite813}}.
This code cannot exist for qubits (\(q=2\)).}\\ 
\addlinespace[\myxtraspc]
\eczhRefIndex{code:stab_9_1_5}%
\eczhListValue{\flmRefsHyperref{code:stab_9_1_5}{\(\llbracket 9,1,5\rrbracket _3\) quantum Glynn code}} & \eczhListValue{Nine-qutrit \flmRefsHyperref{ref672}{pure} Hermitian code that is the smallest qutrit stabilizer code to correct two-qutrit errors.}\\ 
\end{tabularx}\endgroup
\eczcodelist{quantum_reed_muller}{Quantum Reed-Muller codes
}%

\eczhCodeListAutoDescription{All descendants of \flmRefsCref{code:galois_reed_muller}.}%

\eczhIncludeCodeGraph{Bare}{scale=0.5}{\columnwidth}{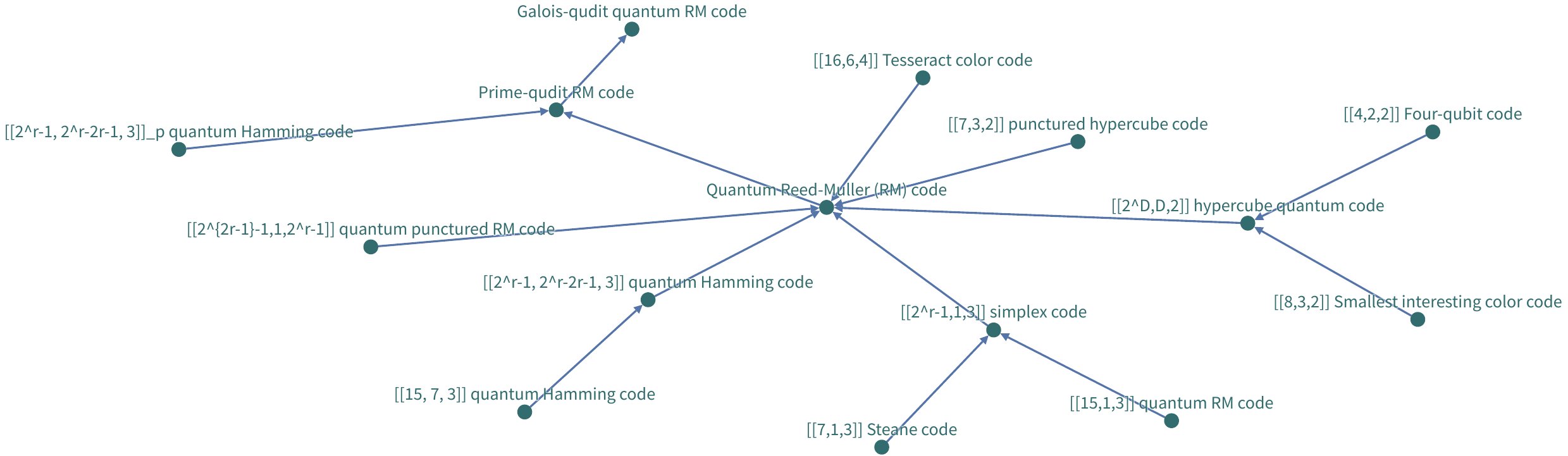}{Quantum Reed-Muller codes}{https://errorcorrectionzoo.org/code_graph#J\%7B\%22displayMode\%22\%3A\%22subset\%22\%2C\%22modeSubsetOptions\%22\%3A\%7B\%22codeIds\%22\%3A\%5B\%22galois_reed_muller\%22\%2C\%22qudit_reed_muller\%22\%2C\%22quantum_reed_muller\%22\%2C\%22stab_15_7_3\%22\%2C\%22stab_15_1_3\%22\%2C\%22stab_16_6_4\%22\%2C\%22hypercube_quantum\%22\%2C\%22quantum_hamming_css\%22\%2C\%22qudit_hamming_css\%22\%2C\%22diagonal_clifford\%22\%2C\%22single_qubit_clifford\%22\%2C\%22stab_4_2_2\%22\%2C\%22steane\%22\%2C\%22xz_7_3_2\%22\%2C\%22stab_8_3_2\%22\%5D\%2C\%22reusePreviousLayoutPositions\%22\%3Afalse\%2C\%22showIntermediateConnectingNodes\%22\%3Atrue\%2C\%22connectingNodesMaxDepth\%22\%3A15\%2C\%22connectingNodesPathMaxLength\%22\%3A20\%2C\%22connectingNodesMaxExtraDepth\%22\%3A3\%2C\%22connectingNodesOnlyKeepPathsWithAdditionalLength\%22\%3A1\%2C\%22connectingNodesToDomainsAndKingdoms\%22\%3Afalse\%2C\%22connectingNodesEdgeLengthsByType\%22\%3A\%7B\%22primaryParent\%22\%3A1\%2C\%22secondaryParent\%22\%3A4\%2C\%22cousin\%22\%3A6\%7D\%2C\%22nodeIds\%22\%3A\%5B\%5D\%7D\%2C\%22highlightImportantNodes\%22\%3A\%7B\%22highlightImportantNodes\%22\%3Afalse\%2C\%22highlightPrimaryParents\%22\%3Afalse\%2C\%22highlightRootConnectingEdges\%22\%3Afalse\%7D\%7D}

\begingroup
\small
\eczhBreakableDashes
\renewcommand\arraystretch{1.05}
\edef\myxtraspc{\eczListAddVSpaceXtraXtraStretch}
\begin{tabularx}{\linewidth}{>{\raggedright\arraybackslash}p{\eczListColWidth{name}} >{\hsize=1.0000\hsize }X}
\toprule
\eczListColTitle{Code} & \eczListColTitle{Description} \\
\midrule
\endfirsthead
\toprule
\eczListColTitleContinued{Code} & \eczListColTitleContinued{Description} \\
\midrule
\endhead
\bottomrule
\endfoot
\eczhRefIndex{code:galois_reed_muller}%
\eczhListValue{\flmRefsHyperref{code:galois_reed_muller}{Galois-qudit quantum RM code}} & \eczhListValue{True Galois-qudit stabilizer code constructed from generalized Reed-Muller (GRM) codes via the Galois-qudit Hermitian construction, the Galois-qudit CSS construction, or directly from their parity-check matrices \NoCaseChange{\protect\cite{cite828}\protect\cite[{Sec. 4.2}]{cite829}}.}\\ 
\addlinespace[\myxtraspc]
\eczhRefIndex{code:qudit_reed_muller}%
\eczhListValue{\flmRefsHyperref{code:qudit_reed_muller}{Prime-qudit RM code}} & \eczhListValue{Modular-qudit stabilizer code constructed from GRM codes or their duals via the modular-qudit CSS construction.}\\ 
\addlinespace[\myxtraspc]
\eczhRefIndex{code:quantum_reed_muller}%
\eczhListValue{\flmRefsHyperref{code:quantum_reed_muller}{Quantum Reed-Muller (RM) code}} & \eczhListValue{A CSS code formed from a classical RM code or its punctured/shortened versions.
Such codes often admit transversal logical gates in the \flmTerm{term}{ref694}{}{Clifford hierarchy}.}\\ 
\addlinespace[\myxtraspc]
\eczhRefIndex{code:stab_15_7_3}%
\eczhListValue{\flmRefsHyperref{code:stab_15_7_3}{\(\llbracket 15, 7, 3\rrbracket \) quantum Hamming code}} & \eczhListValue{Self-dual quantum Hamming code that admits permutation-based CZ logical gates.
The code is constructed using the CSS construction from the \([15,11,3]\) Hamming code and its \([15,4,8]\) dual simplex code.}\\ 
\addlinespace[\myxtraspc]
\eczhRefIndex{code:stab_15_1_3}%
\eczhListValue{\flmRefsHyperref{code:stab_15_1_3}{\(\llbracket 15,1,3\rrbracket \) quantum RM code}} & \eczhListValue{A \(\llbracket 15,1,3\rrbracket \) quantum Reed-Muller code that is most easily thought of as a tetrahedral 3D color code.
It can be constructed as a CSS code from the \([15,5,8]\) punctured Reed-Muller code and its even subcode, which explains its transversal \(T^\dagger\) gate \NoCaseChange{\protect\cite{cite398}}.}\\ 
\addlinespace[\myxtraspc]
\eczhRefIndex{code:stab_16_6_4}%
\eczhListValue{\flmRefsHyperref{code:stab_16_6_4}{\(\llbracket 16,6,4\rrbracket \) Tesseract color code}} & \eczhListValue{A (hyperbolic self-dual CSS) 4D color code defined on a tesseract, with stabilizer generators of both types supported on each cube. 
A \(\llbracket 16,4,2,4\rrbracket \) tesseract subsystem code can be obtained from this code by using two logical qubits as gauge qubits \NoCaseChange{\protect\cite{cite483}}.}\\ 
\addlinespace[\myxtraspc]
\eczhRefIndex{code:hypercube_quantum}%
\eczhListValue{\flmRefsHyperref{code:hypercube_quantum}{\(\llbracket 2^D,D,2\rrbracket \) hypercube quantum code}} & \eczhListValue{Member of a family of codes defined by placing qubits on a \(D\)-dimensional hypercube, \(Z\)-stabilizers on all two-dimensional faces, and an \(X\)-stabilizer on all vertices.
These codes realize gates at the \((D-1)\)-st level of the Clifford hierarchy.
The measured physical bit string can be post-processed into both a logical output string and stabilizer checks, enabling end-of-circuit error detection directly from classical samples \NoCaseChange{\protect\cite{cite759}}.
Puncturing the \(\llbracket 2^D,D,2\rrbracket \) hypercube quantum code yields the \(\llbracket 2^D-1,D,2\rrbracket \) punctured-hypercube family.}\\ 
\addlinespace[\myxtraspc]
\eczhRefIndex{code:quantum_hamming_css}%
\eczhListValue{\flmRefsHyperref{code:quantum_hamming_css}{\(\llbracket 2^r-1, 2^r-2r-1, 3\rrbracket \) quantum Hamming code}} & \eczhListValue{Member of a family of self-dual CSS codes constructed from \([2^r-1,2^r-r-1,3]=C_X=C_Z\) Hamming codes and their duals, the simplex codes.
The code's stabilizer generator matrix blocks \(H_{X}\) and \(H_{Z}\) are both the generator matrix for a simplex code.
The weight of each stabilizer generator is \(2^{r-1}\).}\\ 
\addlinespace[\myxtraspc]
\eczhRefIndex{code:qudit_hamming_css}%
\eczhListValue{\flmRefsHyperref{code:qudit_hamming_css}{\(\llbracket 2^r-1, 2^r-2r-1, 3\rrbracket _p\) quantum Hamming code}} & \eczhListValue{A family of CSS codes extending \flmRefsHyperref{code:quantum_hamming_css}{quantum Hamming codes} to prime qudits of dimension \(p\) by expressing the qubit code stabilizers in local-dimension-invariant (LDI) form \NoCaseChange{\protect\cite{cite820}}.}\\ 
\addlinespace[\myxtraspc]
\eczhRefIndex{code:diagonal_clifford}%
\eczhListValue{\flmRefsHyperref{code:diagonal_clifford}{\(\llbracket 2^r-1,1,3\rrbracket \) simplex code}} & \eczhListValue{Member of a color-code family constructed from a punctured first-order RM\((1,m=r)\) code and its even subcode for \(r \geq 3\).
Each code transversally implements a diagonal gate at the \((r-1)\)st level of the \flmTerm{term}{ref694}{}{Clifford hierarchy} \NoCaseChange{\protect\cite{cite799,cite800}}.
Each code is a color code defined on a simplex in \(r-1\) dimensions \NoCaseChange{\protect\cite{cite475,cite832}}, where qubits are placed on the vertices, edges, and faces as well as on the simplex itself.}\\ 
\addlinespace[\myxtraspc]
\eczhRefIndex{code:single_qubit_clifford}%
\eczhListValue{\flmRefsHyperref{code:single_qubit_clifford}{\(\llbracket 2^{2r-1}-1,1,2^r-1\rrbracket \) quantum punctured RM code}} & \eczhListValue{A quantum Reed-Muller code constructed from a punctured self-dual RM code and its even subcode for \(r \geq 2\).}\\ 
\addlinespace[\myxtraspc]
\eczhRefIndex{code:stab_4_2_2}%
\eczhListValue{\flmRefsHyperref{code:stab_4_2_2}{\(\llbracket 4,2,2\rrbracket \) Four-qubit code}} & \eczhListValue{A four-qubit hyperbolic self-dual CSS stabilizer code that is the smallest two-logical-qubit stabilizer code to detect a single-qubit error.
It is unique for its parameters \NoCaseChange{\protect\cite[{Thm. 8}]{cite446}}.}\\ 
\addlinespace[\myxtraspc]
\eczhRefIndex{code:steane}%
\eczhListValue{\flmRefsHyperref{code:steane}{\(\llbracket 7,1,3\rrbracket \) Steane code}} & \eczhListValue{A \(\llbracket 7,1,3\rrbracket \) self-dual CSS code that is the smallest qubit CSS code to correct a single-qubit error \NoCaseChange{\protect\cite{cite451}}.
The code is constructed using the classical binary \([7,4,3]\) Hamming code for protecting against both \(X\) and \(Z\) errors.}\\ 
\addlinespace[\myxtraspc]
\eczhRefIndex{code:xz_7_3_2}%
\eczhListValue{\flmRefsHyperref{code:xz_7_3_2}{\(\llbracket 7,3,2\rrbracket \) punctured hypercube code}} & \eczhListValue{A seven-qubit \flmRefsHyperref{ref672}{pure} CSS code that corrects a single \(X\) or a single \(Z\) error, but not a single \(Y\) error.
It is the punctured version of the \(\llbracket 8,3,2\rrbracket \) hypercube quantum code and is a phantom code \NoCaseChange{\protect\cite{cite514}}.}\\ 
\addlinespace[\myxtraspc]
\eczhRefIndex{code:stab_8_3_2}%
\eczhListValue{\flmRefsHyperref{code:stab_8_3_2}{\(\llbracket 8,3,2\rrbracket \) Smallest interesting color code}} & \eczhListValue{Smallest 3D color code whose physical qubits lie on vertices of a cube and which admits a (weakly) transversal \(CCZ\) gate.}\\ 
\end{tabularx}\endgroup
\eczcodelist{qldpc}{Qubit QLDPC codes (non-lattice)
}%

\eczhCodeListAutoDescription{Codes that are descendants of \flmRefsCref{code:qldpc} and not descendants of \flmRefsCref{code:translationally_invariant_stabilizer}.}%

\eczhIncludeCodeGraph{Bare}{scale=0.5}{\columnwidth}{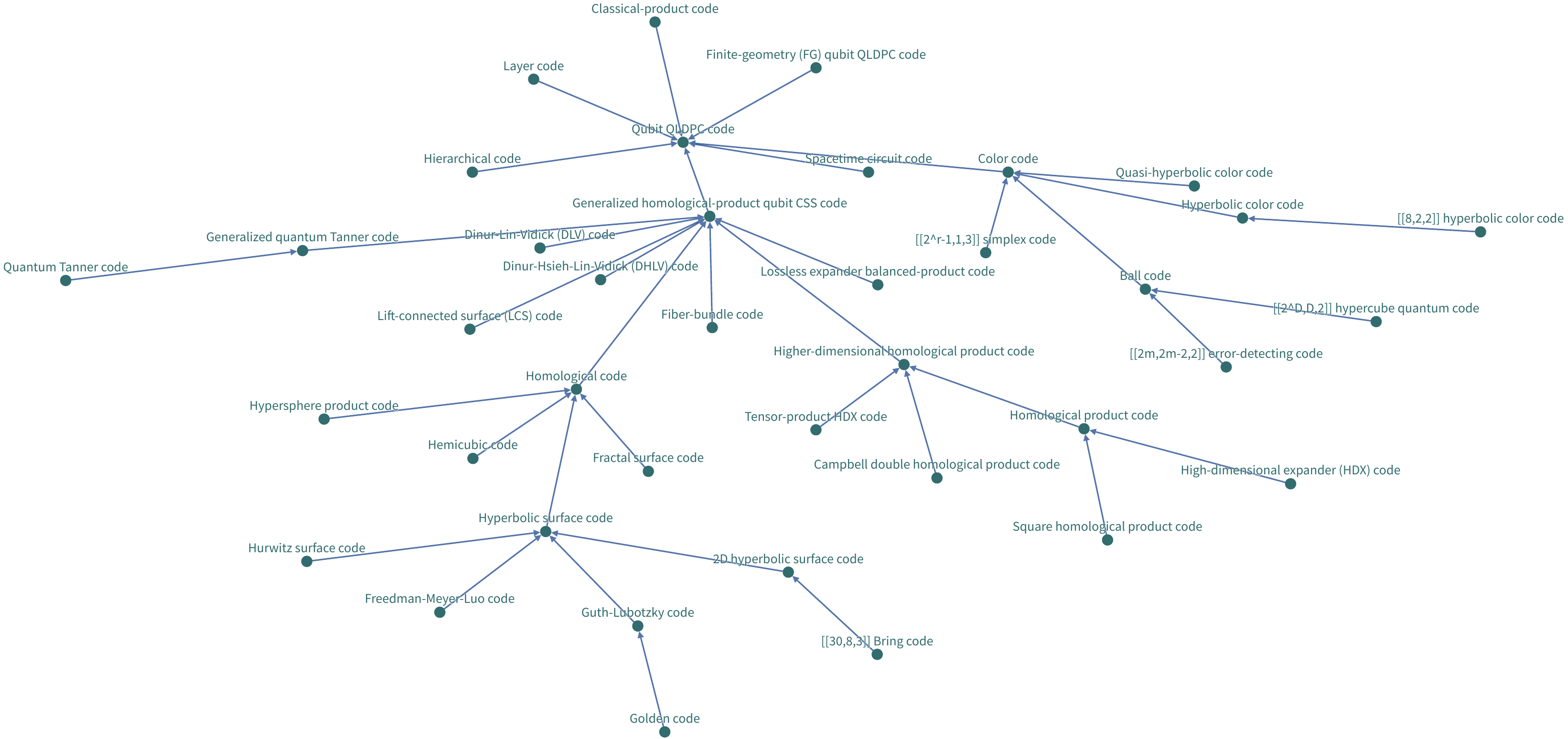}{Qubit QLDPC codes (non-lattice)}{https://errorcorrectionzoo.org/code_graph#J\%7B\%22displayMode\%22\%3A\%22subset\%22\%2C\%22modeSubsetOptions\%22\%3A\%7B\%22codeIds\%22\%3A\%5B\%22two_dimensional_hyperbolic_surface\%22\%2C\%22ball_color\%22\%2C\%22double_homological_product\%22\%2C\%22classical_product\%22\%2C\%22color\%22\%2C\%22dhlv\%22\%2C\%22dlv\%22\%2C\%22fiber_bundle\%22\%2C\%22pg_qldpc\%22\%2C\%22fractal_surface\%22\%2C\%22freedman_meyer_luo\%22\%2C\%22qubit_generalized_homological_product_css\%22\%2C\%22generalized_quantum_tanner\%22\%2C\%22golden\%22\%2C\%22four_dimensional_hyperbolic\%22\%2C\%22hemicubic\%22\%2C\%22hierarchical\%22\%2C\%22ramanujan_tensor_product\%22\%2C\%22multisector_hypergraph\%22\%2C\%22higher_dimensional_surface\%22\%2C\%22homological_product\%22\%2C\%22hurwitz_surface\%22\%2C\%22hyperbolic_color\%22\%2C\%22hyperbolic_surface\%22\%2C\%22hypersphere_product\%22\%2C\%22layer\%22\%2C\%22lcs\%22\%2C\%22lossless_expander\%22\%2C\%22quantum_tanner\%22\%2C\%22quasi_hyperbolic_color\%22\%2C\%22qldpc\%22\%2C\%22spacetime_circuit\%22\%2C\%22square_homological_product\%22\%2C\%22iterated_ramanujan\%22\%2C\%22hypercube_quantum\%22\%2C\%22diagonal_clifford\%22\%2C\%22iceberg\%22\%2C\%22stellated_dodecahedron_css\%22\%2C\%22stab_8_2_2\%22\%5D\%2C\%22reusePreviousLayoutPositions\%22\%3Afalse\%2C\%22showIntermediateConnectingNodes\%22\%3Atrue\%2C\%22connectingNodesMaxDepth\%22\%3A15\%2C\%22connectingNodesPathMaxLength\%22\%3A20\%2C\%22connectingNodesMaxExtraDepth\%22\%3A3\%2C\%22connectingNodesOnlyKeepPathsWithAdditionalLength\%22\%3A1\%2C\%22connectingNodesToDomainsAndKingdoms\%22\%3Afalse\%2C\%22connectingNodesEdgeLengthsByType\%22\%3A\%7B\%22primaryParent\%22\%3A1\%2C\%22secondaryParent\%22\%3A4\%2C\%22cousin\%22\%3A6\%7D\%2C\%22nodeIds\%22\%3A\%5B\%5D\%7D\%2C\%22highlightImportantNodes\%22\%3A\%7B\%22highlightImportantNodes\%22\%3Afalse\%2C\%22highlightPrimaryParents\%22\%3Afalse\%2C\%22highlightRootConnectingEdges\%22\%3Afalse\%7D\%7D}

\begingroup
\small
\eczhBreakableDashes
\renewcommand\arraystretch{1.05}
\edef\myxtraspc{\eczListAddVSpaceXtraXtraStretch}
\begin{tabularx}{\linewidth}{>{\raggedright\arraybackslash}p{\eczListColWidth{name}} >{\hsize=1.0000\hsize }X}
\toprule
\eczListColTitle{Code} & \eczListColTitle{Description} \\
\midrule
\endfirsthead
\toprule
\eczListColTitleContinued{Code} & \eczListColTitleContinued{Description} \\
\midrule
\endhead
\bottomrule
\endfoot
\eczhRefIndex{code:two_dimensional_hyperbolic_surface}%
\eczhListValue{\flmRefsHyperref{code:two_dimensional_hyperbolic_surface}{2D hyperbolic surface code}} & \eczhListValue{Hyperbolic surface codes based on a tessellation of a closed 2D manifold with a hyperbolic geometry (i.e., non-Euclidean geometry, e.g., saddle surfaces when defined on a 2D plane).}\\ 
\addlinespace[\myxtraspc]
\eczhRefIndex{code:ball_color}%
\eczhListValue{\flmRefsHyperref{code:ball_color}{Ball code}} & \eczhListValue{A distance-two color code defined on a colorable \(D\)-ball, equivalently on a \(D\)-colex with boundary \NoCaseChange{\protect\cite[{Appx. A}]{cite687}}.
In the morphing construction of Ref. \NoCaseChange{\protect\cite{cite687}}, ball codes arise as the child codes associated with the morphed ball-like regions.
This family includes hypercube codes (defined on balls constructed from hyperoctahedra) and 3D ball codes (defined on duals of certain Archimedean solids).}\\ 
\addlinespace[\myxtraspc]
\eczhRefIndex{code:double_homological_product}%
\eczhListValue{\flmRefsHyperref{code:double_homological_product}{Campbell double homological product code}} & \eczhListValue{A multi-dimensional HGP code derived from two applications of the hypergraph product to a classical code, resulting in a length-\(4\) chain complex.
The construction method allows for the use of two different classical codes as inputs, with Ref. \NoCaseChange{\protect\cite{cite675}} assuming identical input codes for simplicity.}\\ 
\addlinespace[\myxtraspc]
\eczhRefIndex{code:classical_product}%
\eczhListValue{\flmRefsHyperref{code:classical_product}{Classical-product code}} & \eczhListValue{A QLDPC qubit CSS code constructed by separately constructing the \(X\) and \(Z\) check matrices using product constructions from classical codes. A particular \(\llbracket 512,174,8\rrbracket \) code performed well \NoCaseChange{\protect\cite{cite204}} against erasure and depolarizing noise when compared to other notable CSS codes, such as the asymptotically good quantum Tanner codes. These codes have been generalized to the \textit{intersecting subset code} family \NoCaseChange{\protect\cite{cite833}}.}\\ 
\addlinespace[\myxtraspc]
\eczhRefIndex{code:color}%
\eczhListValue{\flmRefsHyperref{code:color}{Color code}} & \eczhListValue{Member of a family of qubit CSS codes defined on particular \(D\)-dimensional graphs.}\\ 
\addlinespace[\myxtraspc]
\eczhRefIndex{code:dhlv}%
\eczhListValue{\flmRefsHyperref{code:dhlv}{Dinur-Hsieh-Lin-Vidick (DHLV) code}} & \eczhListValue{A family of asymptotically good QLDPC codes which are related to expander LP codes in that the roles of the check operators and physical qubits are exchanged.}\\ 
\addlinespace[\myxtraspc]
\eczhRefIndex{code:dlv}%
\eczhListValue{\flmRefsHyperref{code:dlv}{Dinur-Lin-Vidick (DLV) code}} & \eczhListValue{Member of a family of codes constructed using cubical chain complexes, which are \(t\)-order extensions of the complexes underlying expander codes (\(t=1\)) and expander lifted-product codes (\(t=2\)).}\\ 
\addlinespace[\myxtraspc]
\eczhRefIndex{code:fiber_bundle}%
\eczhListValue{\flmRefsHyperref{code:fiber_bundle}{Fiber-bundle code}} & \eczhListValue{A CSS code constructed by combining one code as the base and another as the fiber of a fiber bundle.
In particular, taking a random LDPC code as the base and a cyclic repetition code as the fiber yields, after distance balancing, a QLDPC code with distance of \flmRefsHyperref{ref65}{order} \(\Omega( n^{3/5}/\text{polylog}(n) )\) and rate of \flmRefsHyperref{ref65}{order} \(\Omega( n^{-2/5}/\text{polylog}(n) )\).}\\ 
\addlinespace[\myxtraspc]
\eczhRefIndex{code:pg_qldpc}%
\eczhListValue{\flmRefsHyperref{code:pg_qldpc}{Finite-geometry (FG) qubit QLDPC code}} & \eczhListValue{CSS code constructed from linear binary codes whose parity-check or generator matrices are incidence matrices of points, hyperplanes, or other structures in finite geometries.
These codes can be interpreted as quantum versions of FG-LDPC codes, but some of them \NoCaseChange{\protect\cite{cite834,cite835}} are not strictly QLDPC.}\\ 
\addlinespace[\myxtraspc]
\eczhRefIndex{code:fractal_surface}%
\eczhListValue{\flmRefsHyperref{code:fractal_surface}{Fractal surface code}} & \eczhListValue{Kitaev surface code on a fractal geometry, which is obtained by removing qubits from the surface code on a cubic lattice.
A related construction, the \textit{fractal product code}, is a hypergraph product of two classical codes defined on a Sierpinski carpet graph \NoCaseChange{\protect\cite{cite676}}. 
The underlying classical codes form classical self-correcting memories \NoCaseChange{\protect\cite{cite677,cite678,cite679}}.}\\ 
\addlinespace[\myxtraspc]
\eczhRefIndex{code:freedman_meyer_luo}%
\eczhListValue{\flmRefsHyperref{code:freedman_meyer_luo}{Freedman-Meyer-Luo code}} & \eczhListValue{Hyperbolic surface code constructed using cellulation of a Riemannian Manifold \(M\) exhibiting systolic freedom \NoCaseChange{\protect\cite{cite680}}. Codes derived from such manifolds can achieve distances scaling better than \(\sqrt{n}\), something that is impossible using closed 2D surfaces or 2D surfaces with boundaries \NoCaseChange{\protect\cite{cite681}}. Improved codes are obtained by studying a weak family of Riemann metrics on closed 4-dimensional manifolds \(S^2\otimes S^2\) with the \(\mathbb{Z}_2\)-homology.}\\ 
\addlinespace[\myxtraspc]
\eczhRefIndex{code:qubit_generalized_homological_product_css}%
\eczhListValue{\flmRefsHyperref{code:qubit_generalized_homological_product_css}{Generalized homological-product qubit CSS code}} & \eczhListValue{A qubit CSS code whose properties are determined from an underlying chain complex via the \flmRefsHyperref{ref683}{qubit CSS-to-homology correspondence}. This complex often consists of some type of product of other chain complexes.}\\ 
\addlinespace[\myxtraspc]
\eczhRefIndex{code:generalized_quantum_tanner}%
\eczhListValue{\flmRefsHyperref{code:generalized_quantum_tanner}{Generalized quantum Tanner code}} & \eczhListValue{An extension of quantum Tanner codes to codes constructed from two commuting regular graphs with the same vertex set.
This allows for code construction using finite sets and Schreier graphs, yielding a broader family of square complexes.}\\ 
\addlinespace[\myxtraspc]
\eczhRefIndex{code:golden}%
\eczhListValue{\flmRefsHyperref{code:golden}{Golden code}} & \eczhListValue{Variant of the Guth-Lubotzky hyperbolic surface code that uses regular tessellations of 4-dimensional hyperbolic space.}\\ 
\addlinespace[\myxtraspc]
\eczhRefIndex{code:four_dimensional_hyperbolic}%
\eczhListValue{\flmRefsHyperref{code:four_dimensional_hyperbolic}{Guth-Lubotzky code}} & \eczhListValue{Homological linear-rate code based on cellulations of certain 4D hyperbolic manifolds with particular homology and systolic properties.}\\ 
\addlinespace[\myxtraspc]
\eczhRefIndex{code:hemicubic}%
\eczhListValue{\flmRefsHyperref{code:hemicubic}{Hemicubic code}} & \eczhListValue{Homological code constructed out of cubes in high dimensions.
The hemicubic code family has asymptotically diminishing soundness that scales as \flmRefsHyperref{ref65}{order} \(\Omega(1/\log n)\), locality of stabilizer generators scaling as \flmRefsHyperref{ref65}{order} \(O(\log n)\), and distance of \flmRefsHyperref{ref65}{order} \(\Theta(\sqrt{n})\).}\\ 
\addlinespace[\myxtraspc]
\eczhRefIndex{code:hierarchical}%
\eczhListValue{\flmRefsHyperref{code:hierarchical}{Hierarchical code}} & \eczhListValue{Member of a family of \(\llbracket n,k,d\rrbracket \) qubit stabilizer codes resulting from a concatenation of a constant-rate \flmRefsHyperref{code:qldpc}{QLDPC code} with a \flmRefsHyperref{code:rotated_surface}{rotated surface code}.
Concatenation allows for syndrome extraction to be performed on a 2D geometry while maintaining a threshold at the expense of a logarithmically vanishing rate.
The growing syndrome extraction circuit depth allows known bounds in the literature to be weakened \NoCaseChange{\protect\cite{cite521,cite522}}.}\\ 
\addlinespace[\myxtraspc]
\eczhRefIndex{code:ramanujan_tensor_product}%
\eczhListValue{\flmRefsHyperref{code:ramanujan_tensor_product}{High-dimensional expander (HDX) code}} & \eczhListValue{CSS code obtained by applying the generalized distance-balancing/product construction of Ref. \NoCaseChange{\protect\cite{cite684}} to a Ramanujan-complex quantum code and an asymptotically good classical LDPC code.}\\ 
\addlinespace[\myxtraspc]
\eczhRefIndex{code:multisector_hypergraph}%
\eczhListValue{\flmRefsHyperref{code:multisector_hypergraph}{Higher-dimensional homological product code}} & \eczhListValue{A qubit CSS code formulated using a tensor product of two or more chain complexes, each of length one or greater.
The number of chain complexes participating in the product is the \textit{dimension} of the code.
When all chain complexes are length-one, meaning that they correspond to classical codes, the code is called a \textit{higher-dimensional HGP code} (a.k.a. multi-sector HGP code or iterative HGP code).}\\ 
\addlinespace[\myxtraspc]
\eczhRefIndex{code:higher_dimensional_surface}%
\eczhListValue{\flmRefsHyperref{code:higher_dimensional_surface}{Homological code}} & \eczhListValue{A CSS extension of the Kitaev surface code to arbitrary manifolds.
The version on a Euclidean manifold of some fixed dimension is called the \(D\)\textit{-dimensional "surface"} or \(D\)\textit{-dimensional toric} code.}\\ 
\addlinespace[\myxtraspc]
\eczhRefIndex{code:homological_product}%
\eczhListValue{\flmRefsHyperref{code:homological_product}{Homological product code}} & \eczhListValue{CSS code formulated using the tensor product of two chain complexes of length one or greater (see \flmRefsCref{ref683}).}\\ 
\addlinespace[\myxtraspc]
\eczhRefIndex{code:hurwitz_surface}%
\eczhListValue{\flmRefsHyperref{code:hurwitz_surface}{Hurwitz surface code}} & \eczhListValue{Homological code constructed on triangulations of Hurwitz surfaces.}\\ 
\addlinespace[\myxtraspc]
\eczhRefIndex{code:hyperbolic_color}%
\eczhListValue{\flmRefsHyperref{code:hyperbolic_color}{Hyperbolic color code}} & \eczhListValue{An extension of the color code construction to hyperbolic manifolds.
As opposed to there being only three types of uniform three-valent and three-colorable lattice tilings in the 2D Euclidean plane, there is an infinite number of admissible hyperbolic tilings in the 2D hyperbolic plane \NoCaseChange{\protect\cite{cite836}}.
Certain double covers of hyperbolic tilings also yield admissible tilings \NoCaseChange{\protect\cite{cite837}}.
Other admissible hyperbolic tilings can be obtained via a fattening procedure \NoCaseChange{\protect\cite{cite430}}; see also a construction based on the more general quantum pin codes \NoCaseChange{\protect\cite{cite702}}.}\\ 
\addlinespace[\myxtraspc]
\eczhRefIndex{code:hyperbolic_surface}%
\eczhListValue{\flmRefsHyperref{code:hyperbolic_surface}{Hyperbolic surface code}} & \eczhListValue{An extension of the Kitaev surface code construction to hyperbolic manifolds.
Given a cellulation of a hyperbolic manifold of arbitrary dimension, qubits are put on \(i\)-dimensional faces, \(X\)-type stabilizers are associated with \((i-1)\)-faces, while \(Z\)-type stabilizers are associated with \(i+1\)-faces.}\\ 
\addlinespace[\myxtraspc]
\eczhRefIndex{code:hypersphere_product}%
\eczhListValue{\flmRefsHyperref{code:hypersphere_product}{Hypersphere product code}} & \eczhListValue{Homological code based on products of hyperspheres.
The hypersphere product code family has asymptotically diminishing soundness that scales as \flmRefsHyperref{ref65}{order} \(O(1/\log (n)^2)\), locality of stabilizer generators scaling as \flmRefsHyperref{ref65}{order} \(O(\log n/ \log\log n)\), and distance of \flmRefsHyperref{ref65}{order} \(\Theta(\sqrt{n})\).}\\ 
\addlinespace[\myxtraspc]
\eczhRefIndex{code:layer}%
\eczhListValue{\flmRefsHyperref{code:layer}{Layer code}} & \eczhListValue{Member of a family of qubit QLDPC CSS codes with stabilizer generator weights \(\leq 6\) that are obtained by coupling layers of 2D surface codes according to the Tanner graph of a QLDPC code (or a more general qubit stabilizer code).
Geometric locality is maintained because, instead of being concatenated, each pair of parallel surface-code squares is fused (or quasi-concatenated) with perpendicular surface-code squares via lattice surgery.}\\ 
\addlinespace[\myxtraspc]
\eczhRefIndex{code:lcs}%
\eczhListValue{\flmRefsHyperref{code:lcs}{Lift-connected surface (LCS) code}} & \eczhListValue{Member of one of several families of lifted-product codes that consist of sparsely interconnected stacks of surface codes.}\\ 
\addlinespace[\myxtraspc]
\eczhRefIndex{code:lossless_expander}%
\eczhListValue{\flmRefsHyperref{code:lossless_expander}{Lossless expander balanced-product code}} & \eczhListValue{QLDPC code constructed by taking the balanced product of lossless expander graphs.
Using one part of a quantum-code chain complex constructed with one-sided loss expanders \NoCaseChange{\protect\cite{cite185}} yields a \(c^3\)-LTC \NoCaseChange{\protect\cite{cite186}}.
Using two-sided expanders \NoCaseChange{\protect\cite{cite187}} yields an asymptotically good QLDPC code family \NoCaseChange{\protect\cite{cite188}}.}\\ 
\addlinespace[\myxtraspc]
\eczhRefIndex{code:quantum_tanner}%
\eczhListValue{\flmRefsHyperref{code:quantum_tanner}{Quantum Tanner code}} & \eczhListValue{Member of a family of QLDPC codes based on two compatible classical Tanner codes defined on a two-dimensional Cayley complex, a complex constructed from Cayley graphs of groups.
For certain choices of codes and complex, the resulting codes have asymptotically good parameters.
See Ref. \NoCaseChange{\protect\cite{cite685}} for explicit instances based on dihedral groups.
This construction has been generalized to Schreier graphs \NoCaseChange{\protect\cite{cite686}}.}\\ 
\addlinespace[\myxtraspc]
\eczhRefIndex{code:quasi_hyperbolic_color}%
\eczhListValue{\flmRefsHyperref{code:quasi_hyperbolic_color}{Quasi-hyperbolic color code}} & \eczhListValue{An extension of the color code construction to quasi-hyperbolic 3-manifolds, e.g., a product of a 2D hyperbolic surface and a circle.}\\ 
\addlinespace[\myxtraspc]
\eczhRefIndex{code:qldpc}%
\eczhListValue{\flmRefsHyperref{code:qldpc}{Qubit QLDPC code}} & \eczhListValue{Member of a family of \(\llbracket n,k,d\rrbracket \) qubit stabilizer codes for which the number of sites participating in each stabilizer generator and the number of stabilizer generators that each site participates in are both bounded by a constant \(w\) as \(n\to\infty\).
The code can be denoted by \(\llbracket n,k,d,w\rrbracket \).
Sometimes, the two parameters are explicitly stated: each site of an \((l,w)\)\textit{-regular qubit QLDPC code} is acted on by \(\leq l\) generators of weight \(\leq w\).}\\ 
\addlinespace[\myxtraspc]
\eczhRefIndex{code:spacetime_circuit}%
\eczhListValue{\flmRefsHyperref{code:spacetime_circuit}{Spacetime circuit code}} & \eczhListValue{Qubit stabilizer code constructed from a \flmRefsHyperref{ref409}{Clifford circuit}, i.e., a circuit made up of \flmRefsHyperref{ref409}{Clifford gates} and Pauli measurements, in order to detect and correct circuit faults.
The code utilizes redundancy in the measurement outcomes of a circuit to correct circuit faults, which correspond to Pauli errors of the code.}\\ 
\addlinespace[\myxtraspc]
\eczhRefIndex{code:square_homological_product}%
\eczhListValue{\flmRefsHyperref{code:square_homological_product}{Square homological product code}} & \eczhListValue{Homological product code whose underlying quantum-code boundary operators are square matrices (see \flmRefsCref{ref683}).}\\ 
\addlinespace[\myxtraspc]
\eczhRefIndex{code:iterated_ramanujan}%
\eczhListValue{\flmRefsHyperref{code:iterated_ramanujan}{Tensor-product HDX code}} & \eczhListValue{A code constructed in a similar way as the HDX code, but utilizing iterated homological products of \textit{multiple} Ramanujan complexes and then applying distance balancing. For any fixed tensor-power parameter \(c\), these yield explicit QLDPC codes with distance scaling as \(\sqrt{n}\log^{c} n\), improving on the original HDX construction by replacing a single logarithmic enhancement with arbitrarily high fixed polylogarithmic enhancement. The utility of such tensor products comes from the fact that one of the Ramanujan complexes is a \textit{collective cosystolic expander} as opposed to just a cosystolic expander.}\\ 
\addlinespace[\myxtraspc]
\eczhRefIndex{code:hypercube_quantum}%
\eczhListValue{\flmRefsHyperref{code:hypercube_quantum}{\(\llbracket 2^D,D,2\rrbracket \) hypercube quantum code}} & \eczhListValue{Member of a family of codes defined by placing qubits on a \(D\)-dimensional hypercube, \(Z\)-stabilizers on all two-dimensional faces, and an \(X\)-stabilizer on all vertices.
These codes realize gates at the \((D-1)\)-st level of the Clifford hierarchy.
The measured physical bit string can be post-processed into both a logical output string and stabilizer checks, enabling end-of-circuit error detection directly from classical samples \NoCaseChange{\protect\cite{cite759}}.
Puncturing the \(\llbracket 2^D,D,2\rrbracket \) hypercube quantum code yields the \(\llbracket 2^D-1,D,2\rrbracket \) punctured-hypercube family.}\\ 
\addlinespace[\myxtraspc]
\eczhRefIndex{code:diagonal_clifford}%
\eczhListValue{\flmRefsHyperref{code:diagonal_clifford}{\(\llbracket 2^r-1,1,3\rrbracket \) simplex code}} & \eczhListValue{Member of a color-code family constructed from a punctured first-order RM\((1,m=r)\) code and its even subcode for \(r \geq 3\).
Each code transversally implements a diagonal gate at the \((r-1)\)st level of the \flmTerm{term}{ref694}{}{Clifford hierarchy} \NoCaseChange{\protect\cite{cite799,cite800}}.
Each code is a color code defined on a simplex in \(r-1\) dimensions \NoCaseChange{\protect\cite{cite475,cite832}}, where qubits are placed on the vertices, edges, and faces as well as on the simplex itself.}\\ 
\addlinespace[\myxtraspc]
\eczhRefIndex{code:iceberg}%
\eczhListValue{\flmRefsHyperref{code:iceberg}{\(\llbracket 2m,2m-2,2\rrbracket \) error-detecting code}} & \eczhListValue{Self-complementary and self-dual CSS code for \(m\geq 2\) with generators \(\{XX\cdots X, ZZ\cdots Z\} \) acting on all \(2m\) physical qubits.
The code is constructed via the CSS construction from an SPC code and a repetition code \NoCaseChange{\protect\cite[{Sec. III}]{cite773}}.
This is the highest-rate distance-two code when an even number of qubits is used \NoCaseChange{\protect\cite{cite449}}.}\\ 
\addlinespace[\myxtraspc]
\eczhRefIndex{code:stellated_dodecahedron_css}%
\eczhListValue{\flmRefsHyperref{code:stellated_dodecahedron_css}{\(\llbracket 30,8,3\rrbracket \) Bring code}} & \eczhListValue{A \(\llbracket 30,8,3\rrbracket \) hyperbolic surface code on a quotient of the \(\{5,5\}\) hyperbolic tiling called Bring's curve.
Its qubits and stabilizer generators lie on the vertices of the small stellated dodecahedron. It admits a set of weight-five stabilizer generators.}\\ 
\addlinespace[\myxtraspc]
\eczhRefIndex{code:stab_8_2_2}%
\eczhListValue{\flmRefsHyperref{code:stab_8_2_2}{\(\llbracket 8,2,2\rrbracket \) hyperbolic color code}} & \eczhListValue{An \(\llbracket 8,2,2\rrbracket \) hyperbolic color code defined on the projective plane. It is a self-dual CSS code.}\\ 
\end{tabularx}\endgroup
\eczcodelist{self_correct}{Self-correcting quantum codes and friends
}%

\eczhCodeListAutoDescription{All descendants and cousins of \flmRefsCref{code:self_correct}.}%

\eczhIncludeCodeGraph{Bare}{scale=0.5}{\columnwidth}{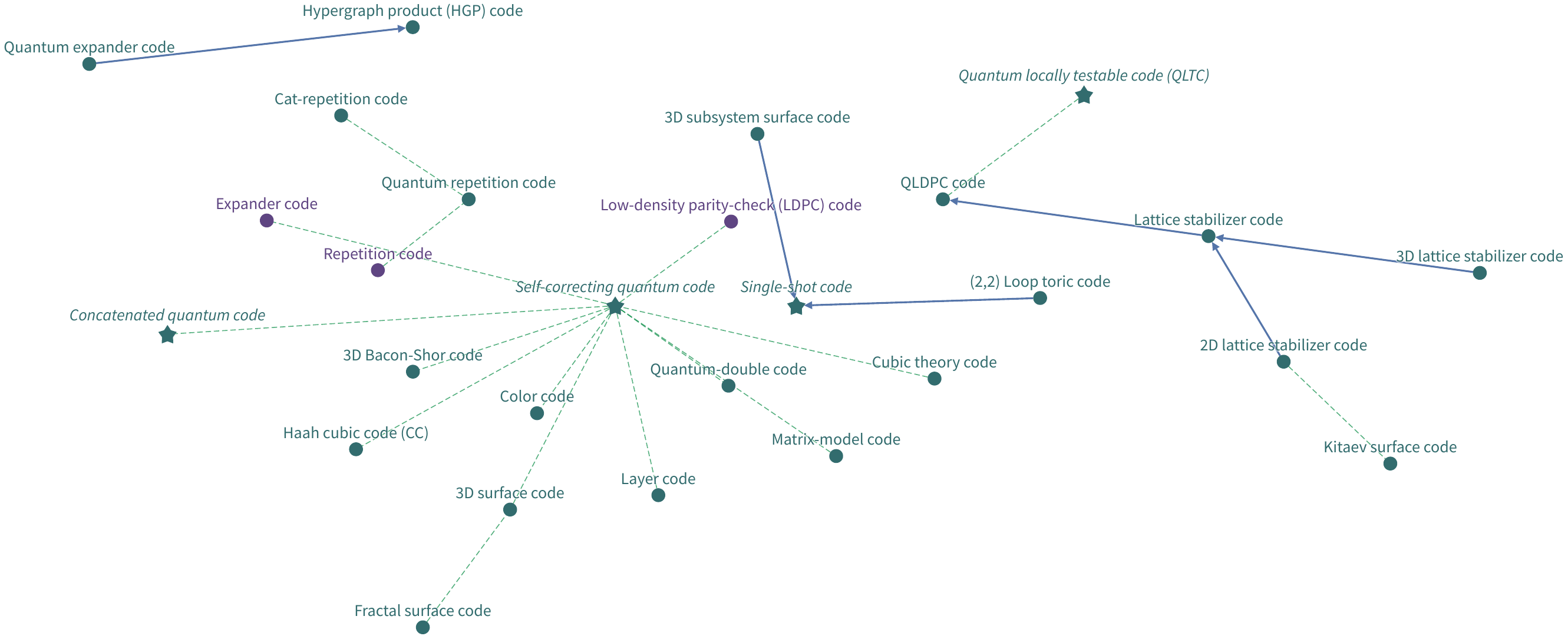}{Self-correcting quantum codes and friends}{https://errorcorrectionzoo.org/code_graph#J\%7B\%22displayMode\%22\%3A\%22subset\%22\%2C\%22modeSubsetOptions\%22\%3A\%7B\%22codeIds\%22\%3A\%5B\%222d_stabilizer\%22\%2C\%223d_bacon_shor\%22\%2C\%223d_stabilizer\%22\%2C\%223d_subsystem_surface\%22\%2C\%223d_surface\%22\%2C\%22cat_repetition\%22\%2C\%22color\%22\%2C\%22quantum_concatenated\%22\%2C\%22cubic_theory\%22\%2C\%22expander\%22\%2C\%22fractal_surface\%22\%2C\%22haah_cubic\%22\%2C\%22hypergraph_product\%22\%2C\%22surface\%22\%2C\%22translationally_invariant_stabilizer\%22\%2C\%22layer\%22\%2C\%22ldpc\%22\%2C\%22matrix_qm\%22\%2C\%22general_qldpc\%22\%2C\%22quantum_expander\%22\%2C\%22qltc\%22\%2C\%22quantum_repetition\%22\%2C\%22quantum_double\%22\%2C\%22repetition\%22\%2C\%22self_correct\%22\%2C\%22single_shot\%22\%2C\%224d_surface\%22\%5D\%2C\%22reusePreviousLayoutPositions\%22\%3Afalse\%2C\%22showIntermediateConnectingNodes\%22\%3Atrue\%2C\%22connectingNodesMaxDepth\%22\%3A15\%2C\%22connectingNodesPathMaxLength\%22\%3A20\%2C\%22connectingNodesMaxExtraDepth\%22\%3A3\%2C\%22connectingNodesOnlyKeepPathsWithAdditionalLength\%22\%3A1\%2C\%22connectingNodesToDomainsAndKingdoms\%22\%3Afalse\%2C\%22connectingNodesEdgeLengthsByType\%22\%3A\%7B\%22primaryParent\%22\%3A1\%2C\%22secondaryParent\%22\%3A4\%2C\%22cousin\%22\%3A6\%7D\%2C\%22nodeIds\%22\%3A\%5B\%5D\%7D\%2C\%22highlightImportantNodes\%22\%3A\%7B\%22highlightImportantNodes\%22\%3Afalse\%2C\%22highlightPrimaryParents\%22\%3Afalse\%2C\%22highlightRootConnectingEdges\%22\%3Afalse\%7D\%7D}

\begingroup
\small
\eczhBreakableDashes
\renewcommand\arraystretch{1.05}
\edef\myxtraspc{\eczListAddVSpaceXtraXtraStretch}
\begin{tabularx}{\linewidth}{>{\raggedright\arraybackslash}p{\eczListColWidth{name}} >{\hsize=1.0000\hsize }X}
\toprule
\eczListColTitle{Code} & \eczListColTitle{Description} \\
\midrule
\endfirsthead
\toprule
\eczListColTitleContinued{Code} & \eczListColTitleContinued{Description} \\
\midrule
\endhead
\bottomrule
\endfoot
\eczhRefIndex{code:2d_stabilizer}%
\eczhListValue{\flmRefsHyperref{code:2d_stabilizer}{2D lattice stabilizer code}} & \eczhListValue{Lattice stabilizer code in two Euclidean dimensions, using either the ordinary block notion of locality or the fermionic/Majorana notion of locality.}\\ 
\addlinespace[\myxtraspc]
\eczhRefIndex{code:3d_bacon_shor}%
\eczhListValue{\flmRefsHyperref{code:3d_bacon_shor}{3D Bacon-Shor code}} & \eczhListValue{Generalization of the Bacon-Shor code to three dimensions that was conjectured to be a self-correcting memory.
It is defined on a cubic lattice and admits sheet-like stabilizer generators.}\\ 
\addlinespace[\myxtraspc]
\eczhRefIndex{code:3d_stabilizer}%
\eczhListValue{\flmRefsHyperref{code:3d_stabilizer}{3D lattice stabilizer code}} & \eczhListValue{Lattice stabilizer code in three Euclidean dimensions, using either the ordinary block notion of locality or the fermionic/Majorana notion of locality.}\\ 
\addlinespace[\myxtraspc]
\eczhRefIndex{code:3d_subsystem_surface}%
\eczhListValue{\flmRefsHyperref{code:3d_subsystem_surface}{3D subsystem surface code}} & \eczhListValue{Subsystem generalization of the surface code on a 3D cubic lattice with gauge-group generators of weight at most three.}\\ 
\addlinespace[\myxtraspc]
\eczhRefIndex{code:3d_surface}%
\eczhListValue{\flmRefsHyperref{code:3d_surface}{3D surface code}} & \eczhListValue{A generalization of the Kitaev surface code defined on a 3D cubic lattice.
Qubits are placed on edges, \(Z\)-type stabilizer generators are placed on square plaquettes oriented in all three directions, and \(X\)-type stabilizers are placed on the six edges neighboring every vertex \NoCaseChange{\protect\cite{cite459}}.}\\ 
\addlinespace[\myxtraspc]
\eczhRefIndex{code:cat_repetition}%
\eczhListValue{\flmRefsHyperref{code:cat_repetition}{Cat-repetition code}} & \eczhListValue{A concatenated qubit-into-\(n\)-mode code obtained by encoding each qubit of a quantum repetition code into a two-component cat code in its cat-state basis.}\\ 
\addlinespace[\myxtraspc]
\eczhRefIndex{code:color}%
\eczhListValue{\flmRefsHyperref{code:color}{Color code}} & \eczhListValue{Member of a family of qubit CSS codes defined on particular \(D\)-dimensional graphs.}\\ 
\addlinespace[\myxtraspc]
\eczhRefIndex{code:quantum_concatenated}%
\eczhListValue{\flmRefsHyperref{code:quantum_concatenated}{Concatenated quantum code}} & \eczhListValue{A combination of two quantum codes, an inner code \(C_{\text{in}}\) and an outer code \(C_{\text{out}}\), where the physical subspace used for the inner code consists of the logical subspace of the outer code.
In other words, one first encodes in the inner code, and then encodes each of its physical registers in the outer code.
An inner \(C_{\text{in}} = \llparenthesis n_1,K,d_1\rrparenthesis _{q_1}\) and outer \(C_{\text{out}} = \llparenthesis n_2,q_1,d_2\rrparenthesis _{q_2}\) block quantum code yield an \(\llparenthesis n_1 n_2, K, d \geq d_1d_2\rrparenthesis _{q_2}\) concatenated block quantum code \NoCaseChange{\protect\cite{cite398}}.}\\ 
\addlinespace[\myxtraspc]
\eczhRefIndex{code:cubic_theory}%
\eczhListValue{\flmRefsHyperref{code:cubic_theory}{Cubic theory code}} & \eczhListValue{A geometrically local commuting-projector code family defined on triangulations in arbitrary spatial dimensions.
Its Hamiltonian contains Pauli-\(Z\) flux terms and non-Pauli Gauss-law terms built from products of Pauli-\(X\) operators and \(CZ\) gates.
These commuting non-Pauli stabilizers realize higher-form \(\mathbb{Z}_2^3\) gauge theories with Abelian electric excitations and non-Abelian magnetic excitations.}\\ 
\addlinespace[\myxtraspc]
\eczhRefIndex{code:expander}%
\eczhListValue{\flmRefsHyperref{code:expander}{Expander code}} & \eczhListValue{LDPC code whose parity-check matrix is derived from the adjacency matrix of a bipartite expander graph \NoCaseChange{\protect\cite{cite74}} such as a Ramanujan graph or a Cayley graph of a projective special linear group over a finite field \NoCaseChange{\protect\cite{cite75,cite76}}.
Expander codes admit efficient encoding and decoding algorithms and yield an explicit (i.e., non-random) asymptotically good LDPC code family.}\\ 
\addlinespace[\myxtraspc]
\eczhRefIndex{code:fractal_surface}%
\eczhListValue{\flmRefsHyperref{code:fractal_surface}{Fractal surface code}} & \eczhListValue{Kitaev surface code on a fractal geometry, which is obtained by removing qubits from the surface code on a cubic lattice.
A related construction, the \textit{fractal product code}, is a hypergraph product of two classical codes defined on a Sierpinski carpet graph \NoCaseChange{\protect\cite{cite676}}. 
The underlying classical codes form classical self-correcting memories \NoCaseChange{\protect\cite{cite677,cite678,cite679}}.}\\ 
\addlinespace[\myxtraspc]
\eczhRefIndex{code:haah_cubic}%
\eczhListValue{\flmRefsHyperref{code:haah_cubic}{Haah cubic code (CC)}} & \eczhListValue{A 3D lattice stabilizer code on a length-\(L\) cubic lattice with one or two qubits per site.
Admits two types of stabilizer generators with support on each cube of the lattice.
In the non-CSS case, these two are related by spatial inversion.
For CSS codes, we require that the product of all corner operators is the identity.
We lastly require that there are no non-trivial string operators, meaning that single-site operators are a phase, and any period one logical operator \(l \in \mathsf{S}^{\perp}\) is just a phase.}\\ 
\addlinespace[\myxtraspc]
\eczhRefIndex{code:hypergraph_product}%
\eczhListValue{\flmRefsHyperref{code:hypergraph_product}{Hypergraph product (HGP) code}} & \eczhListValue{A member of a family of CSS codes whose stabilizer generator matrix is obtained from a hypergraph product of two classical linear binary codes.}\\ 
\addlinespace[\myxtraspc]
\eczhRefIndex{code:surface}%
\eczhListValue{\flmRefsHyperref{code:surface}{Kitaev surface code}} & \eczhListValue{A family of Abelian topological \flmRefsHyperref{code:css}{CSS stabilizer} codes
whose generators are few-body \(X\)-type and \(Z\)-type Pauli strings
associated to the stars and plaquettes, respectively, of a cellulation of a
two-dimensional surface (with a qubit located at each edge of the
cellulation).
Codewords correspond to ground states of the surface code Hamiltonian, and error operators create or annihilate pairs of anyonic charges or vortices.}\\ 
\addlinespace[\myxtraspc]
\eczhRefIndex{code:translationally_invariant_stabilizer}%
\eczhListValue{\flmRefsHyperref{code:translationally_invariant_stabilizer}{Lattice stabilizer code}} & \eczhListValue{A geometrically local stabilizer code with sites organized on a lattice modeled by the additive group \(\mathbb{Z}^D\) for spatial dimension \(D\), using either the ordinary block notion of locality or the fermionic/Majorana notion of locality.
On an infinite lattice, its stabilizer group is generated by few-site Pauli-type operators and their translations, in which case the code is called \textit{translationally invariant stabilizer code}.
Boundary conditions have to be imposed on the lattice in order to obtain finite-dimensional versions.
Lattice defects and boundaries between different codes can also be introduced.}\\ 
\addlinespace[\myxtraspc]
\eczhRefIndex{code:layer}%
\eczhListValue{\flmRefsHyperref{code:layer}{Layer code}} & \eczhListValue{Member of a family of qubit QLDPC CSS codes with stabilizer generator weights \(\leq 6\) that are obtained by coupling layers of 2D surface codes according to the Tanner graph of a QLDPC code (or a more general qubit stabilizer code).
Geometric locality is maintained because, instead of being concatenated, each pair of parallel surface-code squares is fused (or quasi-concatenated) with perpendicular surface-code squares via lattice surgery.}\\ 
\addlinespace[\myxtraspc]
\eczhRefIndex{code:ldpc}%
\eczhListValue{\flmRefsHyperref{code:ldpc}{Low-density parity-check (LDPC) code}} & \eczhListValue{A binary linear code with a sparse parity-check matrix.
Often a member of an infinite family of \([n,k,d]\) codes for which the numbers of nonzero entries in each row and in each column of the parity-check matrix are both bounded above by a constant as \(n\to\infty\).}\\ 
\addlinespace[\myxtraspc]
\eczhRefIndex{code:matrix_qm}%
\eczhListValue{\flmRefsHyperref{code:matrix_qm}{Matrix-model code}} & \eczhListValue{Multimode Fock-state bosonic approximate code derived from a matrix model, i.e., a bosonic theory with a large non-Abelian gauge group.
The model's degrees of freedom are matrix-valued bosons \(a\), each consisting of \(N^2\) harmonic oscillator modes and subject to an \(SU(N)\) gauge symmetry.}\\ 
\addlinespace[\myxtraspc]
\eczhRefIndex{code:general_qldpc}%
\eczhListValue{\flmRefsHyperref{code:general_qldpc}{QLDPC code}} & \eczhListValue{Member of a family of stabilizer codes for which the number of sites participating in each stabilizer generator and the number of stabilizer generators that each site participates in are both bounded by a constant as \(n\to\infty\).
Sometimes, the two parameters are explicitly stated: each site of an \((l,w)\)\textit{-regular QLDPC code} is acted on by \(\leq l\) generators of weight \(\leq w\).}\\ 
\addlinespace[\myxtraspc]
\eczhRefIndex{code:quantum_expander}%
\eczhListValue{\flmRefsHyperref{code:quantum_expander}{Quantum expander code}} & \eczhListValue{CSS code constructed from a hypergraph product of bipartite expander graphs \NoCaseChange{\protect\cite{cite74}} with bounded left and right vertex degrees. For every bipartite graph there is an associated matrix (the parity check matrix) with columns indexed by the left vertices, rows indexed by the right vertices, and 1 entries whenever a left and right vertex are connected. This matrix can serve as the parity check matrix of a classical code. Two bipartite expander graphs can be used to construct a quantum CSS code (the quantum expander code) via the hypergraph product of their parity check matrices.}\\ 
\addlinespace[\myxtraspc]
\eczhRefIndex{code:qltc}%
\eczhListValue{\flmRefsHyperref{code:qltc}{Quantum locally testable code (QLTC)}} & \eczhListValue{A local commuting-projector Hamiltonian-based block quantum code which has a nonzero average-energy penalty for creating large errors. Informally, states that are far away from the codespace of a QLTC have to be excited states of a number of the code's local projectors that scales linearly with \(n\).}\\ 
\addlinespace[\myxtraspc]
\eczhRefIndex{code:quantum_repetition}%
\eczhListValue{\flmRefsHyperref{code:quantum_repetition}{Quantum repetition code}} & \eczhListValue{Encodes \(1\) qubit into \(n\) qubits according to \(|0\rangle\to|\phi_0\rangle^{\otimes n}\) and \(|1\rangle\to|\phi_1\rangle^{\otimes n}\). The code is called a \textit{bit-flip} code when \(|\phi_i\rangle = |i\rangle\), and a \textit{phase-flip} code when \(|\phi_0\rangle = |+\rangle\) and \(|\phi_1\rangle = |-\rangle\).
This repetition-style encoding does not clone an arbitrary quantum state; instead, it extends the copying of computational-basis states linearly to entangled codewords  \NoCaseChange{\protect\cite[{Ch. 2}]{cite398}}.}\\ 
\addlinespace[\myxtraspc]
\eczhRefIndex{code:quantum_double}%
\eczhListValue{\flmRefsHyperref{code:quantum_double}{Quantum-double code}} & \eczhListValue{Group-based code whose codewords realize 2D modular gapped topological order defined by a finite group \(G\).
The code's generators are few-body operators associated to the stars and plaquettes, respectively, of a tessellation of a two-dimensional surface (with a qudit of dimension \( |G| \) located at each edge of the tessellation).
The original Hamiltonian can be re-expressed via \flmRefsHyperref{ref20}{group-based right- and left-multiplication \(X\)-type as well as \(Z\)-type error} operators \NoCaseChange{\protect\cite[{Sec. 3.3}]{cite598}}.}\\ 
\addlinespace[\myxtraspc]
\eczhRefIndex{code:repetition}%
\eczhListValue{\flmRefsHyperref{code:repetition}{Repetition code}} & \eczhListValue{\([n,1,n]\) binary linear code encoding one bit of information into an \(n\)-bit string.
Majority decoding requires \(n\) to be odd in order to avoid ties.
The idea is to increase the code distance by repeating the logical information several times. It is a \((n,1)\)-Hamming code.
Its automorphism group is \(S_n\).}\\ 
\addlinespace[\myxtraspc]
\eczhRefIndex{code:self_correct}%
\eczhListValue{\flmRefsHyperref{code:self_correct}{Self-correcting quantum code}} & \eczhListValue{A block quantum code that forms the ground-state subspace of an \(n\)-body geometrically local Hamiltonian whose logical information is recoverable for arbitrarily long times in the \(n\to\infty\) limit after interaction with a sufficiently cold thermal environment.
Typically, one also requires a decoder whose decoding time scales polynomially with \(n\) and a finite energy density.}\\ 
\addlinespace[\myxtraspc]
\eczhRefIndex{code:single_shot}%
\eczhListValue{\flmRefsHyperref{code:single_shot}{Single-shot code}} & \eczhListValue{Block quantum qudit code whose error-syndrome weights increase linearly with the distance of the error state to the code space.}\\ 
\addlinespace[\myxtraspc]
\eczhRefIndex{code:4d_surface}%
\eczhListValue{\flmRefsHyperref{code:4d_surface}{\((2,2)\) Loop toric code}} & \eczhListValue{A generalization of the Kitaev surface code defined on a 4D lattice.
The code is called a \((2,2)\) toric code because it admits 2D membrane \(Z\)-type and \(X\)-type logical operators.
Both types of operators create 1D (i.e., loop) excitations at their edges.
The code serves as a self-correcting quantum memory \NoCaseChange{\protect\cite{cite480,cite481}}.}\\ 
\end{tabularx}\endgroup
\eczcodelist{single_shot}{Single-shot codes and friends
}%

\eczhCodeListAutoDescription{All descendants and cousins of \flmRefsCref{code:single_shot}.}%

\eczhIncludeCodeGraph{Bare}{scale=0.5}{\columnwidth}{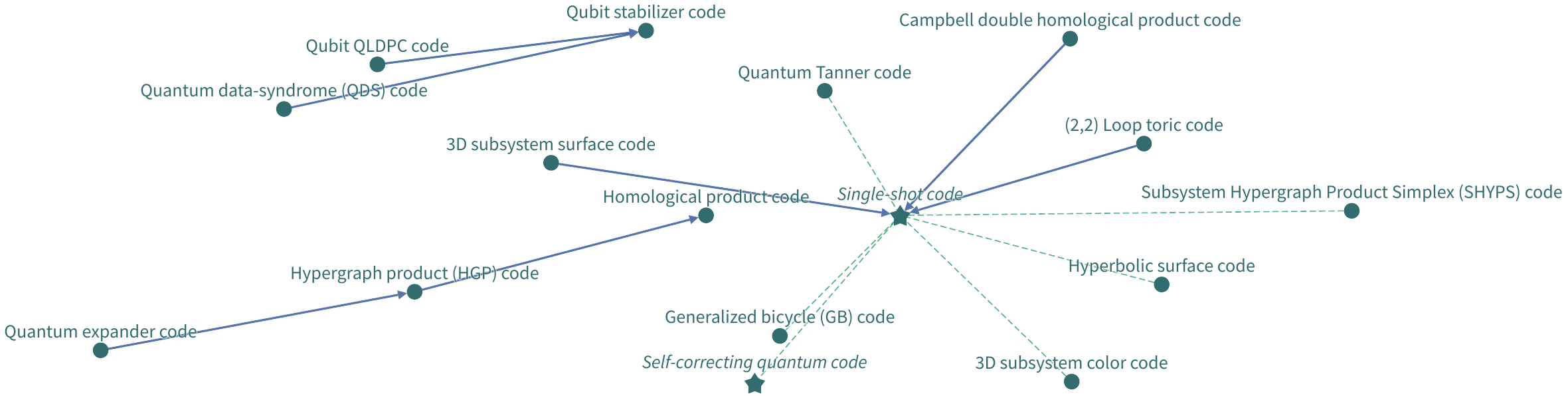}{Single-shot codes and friends}{https://errorcorrectionzoo.org/code_graph#J\%7B\%22displayMode\%22\%3A\%22subset\%22\%2C\%22modeSubsetOptions\%22\%3A\%7B\%22codeIds\%22\%3A\%5B\%223d_subsystem_color\%22\%2C\%223d_subsystem_surface\%22\%2C\%22double_homological_product\%22\%2C\%22generalized_bicycle\%22\%2C\%22homological_product\%22\%2C\%22hyperbolic_surface\%22\%2C\%22hypergraph_product\%22\%2C\%22quantum_tanner\%22\%2C\%22data_syndrome\%22\%2C\%22quantum_expander\%22\%2C\%22qldpc\%22\%2C\%22qubit_stabilizer\%22\%2C\%22self_correct\%22\%2C\%22single_shot\%22\%2C\%22shyps\%22\%2C\%224d_surface\%22\%5D\%2C\%22reusePreviousLayoutPositions\%22\%3Afalse\%2C\%22showIntermediateConnectingNodes\%22\%3Atrue\%2C\%22connectingNodesMaxDepth\%22\%3A15\%2C\%22connectingNodesPathMaxLength\%22\%3A20\%2C\%22connectingNodesMaxExtraDepth\%22\%3A3\%2C\%22connectingNodesOnlyKeepPathsWithAdditionalLength\%22\%3A1\%2C\%22connectingNodesToDomainsAndKingdoms\%22\%3Afalse\%2C\%22connectingNodesEdgeLengthsByType\%22\%3A\%7B\%22primaryParent\%22\%3A1\%2C\%22secondaryParent\%22\%3A4\%2C\%22cousin\%22\%3A6\%7D\%2C\%22nodeIds\%22\%3A\%5B\%5D\%7D\%2C\%22highlightImportantNodes\%22\%3A\%7B\%22highlightImportantNodes\%22\%3Afalse\%2C\%22highlightPrimaryParents\%22\%3Afalse\%2C\%22highlightRootConnectingEdges\%22\%3Afalse\%7D\%7D}

\begingroup
\small
\eczhBreakableDashes
\renewcommand\arraystretch{1.05}
\edef\myxtraspc{\eczListAddVSpaceXtraXtraStretch}
\begin{tabularx}{\linewidth}{>{\raggedright\arraybackslash}p{\eczListColWidth{name}} >{\hsize=1.0000\hsize }X}
\toprule
\eczListColTitle{Code} & \eczListColTitle{Relation} \\
\midrule
\endfirsthead
\toprule
\eczListColTitleContinued{Code} & \eczListColTitleContinued{Relation} \\
\midrule
\endhead
\bottomrule
\endfoot
\eczhRefIndex{code:3d_subsystem_color}%
\eczhListValue{\flmRefsHyperref{code:3d_subsystem_color}{3D subsystem color code}} & \eczhListValue{The 3D subsystem color code defined on the cube-truncated rhombic dodecahedral honeycomb, i.e., a tessellation of cubes and chamfered cubes (a.k.a. tetratruncated rhombic dodecahedra) \NoCaseChange{\protect\cite[{Fig. 1}]{cite832}}, is a single-shot code \NoCaseChange{\protect\cite{cite838,cite832}}.}\\ 
\addlinespace[\myxtraspc]
\eczhRefIndex{code:3d_subsystem_surface}%
\eczhListValue{\flmRefsHyperref{code:3d_subsystem_surface}{3D subsystem surface code}} & \eczhListValue{The 3D subsystem surface code is a single-shot code \NoCaseChange{\protect\cite{cite839,cite840}}; see Ref. \NoCaseChange{\protect\cite{cite841}} for an alternative formulation.}\\ 
\addlinespace[\myxtraspc]
\eczhRefIndex{code:double_homological_product}%
\eczhListValue{\flmRefsHyperref{code:double_homological_product}{Campbell double homological product code}} & \eczhListValue{For a minimal input chain complex associated with a classical \([n,k,d]\) code, the Campbell double homological product code is a single-shot code with \(d_{\text{ss}}=\infty\), \((d,f)\)-soundness for \(f(x)=x^3/4\), and check redundancy bounded by \(<2\) \NoCaseChange{\protect\cite{cite675}}.}\\ 
\addlinespace[\myxtraspc]
\eczhRefIndex{code:generalized_bicycle}%
\eczhListValue{\flmRefsHyperref{code:generalized_bicycle}{Generalized bicycle (GB) code}} & \eczhListValue{A qubit GB code \(\llbracket n,k,d\rrbracket _2\) has \(k\) non-trivial relations between the syndrome bits, which is expected to help with operation in a fault-tolerant regime (in the presence of syndrome measurement errors). See Ref. \NoCaseChange{\protect\cite{cite842}} for many examples of such codes. There is numerical evidence that a particular family is single shot \NoCaseChange{\protect\cite{cite843}}.}\\ 
\addlinespace[\myxtraspc]
\eczhRefIndex{code:homological_product}%
\eczhListValue{\flmRefsHyperref{code:homological_product}{Homological product code}} & \eczhListValue{It is conjectured that a particular class of codes called three-dimensional product codes is single shot \NoCaseChange{\protect\cite{cite844}}.}\\ 
\addlinespace[\myxtraspc]
\eczhRefIndex{code:hyperbolic_surface}%
\eczhListValue{\flmRefsHyperref{code:hyperbolic_surface}{Hyperbolic surface code}} & \eczhListValue{A 4D hyperbolic surface code can be decoded with the Hastings decoder \NoCaseChange{\protect\cite{cite845}} in time \(O(n\log n)\) and with a logical error scaling inverse polynomially with \(n\).}\\ 
\addlinespace[\myxtraspc]
\eczhRefIndex{code:hypergraph_product}%
\eczhListValue{\flmRefsHyperref{code:hypergraph_product}{Hypergraph product (HGP) code}} & \eczhListValue{Two-fold application of the hypergraph product to a pair of binary linear codes yields single-shot QLDPC codes that exploit redundancy in their stabilizer generators \NoCaseChange{\protect\cite{cite675}}.}\\ 
\addlinespace[\myxtraspc]
\eczhRefIndex{code:quantum_tanner}%
\eczhListValue{\flmRefsHyperref{code:quantum_tanner}{Quantum Tanner code}} & \eczhListValue{Certain quantum Tanner codes facilitate single-shot decoding \NoCaseChange{\protect\cite{cite846}}.}\\ 
\addlinespace[\myxtraspc]
\eczhRefIndex{code:data_syndrome}%
\eczhListValue{\flmRefsHyperref{code:data_syndrome}{Quantum data-syndrome (QDS) code}} & \eczhListValue{QDS codes are closely related to single-shot codes because both use redundant syndrome information to suppress measurement errors in a single round of syndrome extraction \NoCaseChange{\protect\cite{cite675}}.}\\ 
\addlinespace[\myxtraspc]
\eczhRefIndex{code:quantum_expander}%
\eczhListValue{\flmRefsHyperref{code:quantum_expander}{Quantum expander code}} & \eczhListValue{Quantum expander codes are single-shot \NoCaseChange{\protect\cite{cite847}}.}\\ 
\addlinespace[\myxtraspc]
\eczhRefIndex{code:qldpc}%
\eczhListValue{\flmRefsHyperref{code:qldpc}{Qubit QLDPC code}} & \eczhListValue{Qubit QLDPC codes satisfying linear confinement are single shot \NoCaseChange{\protect\cite{cite844}}. Any code that admits a local greedy decoder also satisfies linear confinement, and so is single shot \NoCaseChange{\protect\cite{cite848}}.}\\ 
\addlinespace[\myxtraspc]
\eczhRefIndex{code:qubit_stabilizer}%
\eczhListValue{\flmRefsHyperref{code:qubit_stabilizer}{Qubit stabilizer code}} & \eczhListValue{Any stabilizer code can be single shot if sufficiently non-local high-weight stabilizer generators are used for syndrome measurements.  These can be obtained with a Gaussian elimination procedure \NoCaseChange{\protect\cite{cite675}}.}\\ 
\addlinespace[\myxtraspc]
\eczhRefIndex{code:self_correct}%
\eczhListValue{\flmRefsHyperref{code:self_correct}{Self-correcting quantum code}} & \eczhListValue{The presence of an energy barrier (i.e., confinement) is sufficient for a code to be single shot, and is also conjectured to be necessary for a code to be a self-correcting memory. Linear confinement of QLDPC (LDPC) codes implies (classical) self-correction \NoCaseChange{\protect\cite{cite849}}.}\\ 
\addlinespace[\myxtraspc]
\eczhRefIndex{code:single_shot}%
\eczhListValue{\flmRefsHyperref{code:single_shot}{Single-shot code}} & \eczhListValue{\eczListValueNA }\\ 
\addlinespace[\myxtraspc]
\eczhRefIndex{code:shyps}%
\eczhListValue{\flmRefsHyperref{code:shyps}{Subsystem Hypergraph Product Simplex (SHYPS) code}} & \eczhListValue{SHYPS codes exhibit practical single-shot signatures, including logical error-rate stability under small-window sliding-window decoding and constant single-shot distance \(d_{\mathrm{ss}}=3\), which supports using one syndrome-extraction round between logical generators \NoCaseChange{\protect\cite{cite785}}.}\\ 
\addlinespace[\myxtraspc]
\eczhRefIndex{code:4d_surface}%
\eczhListValue{\flmRefsHyperref{code:4d_surface}{\((2,2)\) Loop toric code}} & \eczhListValue{Single-shot QEC has been realized using the \(\llbracket 33,1,4\rrbracket \) loop toric code on the Quantinuum H2 device \NoCaseChange{\protect\cite{cite850}}.}\\ 
\end{tabularx}\endgroup
\eczcodelist{small_quantum}{Small-distance quantum codes and friends (non-qubit-stabilizer)
}%

\eczhCodeListAutoDescription{Union of:
\begin{itemize}\item codes that are descendants of \flmRefsCref{code:small_distance_quantum} and not descendants of \flmRefsCref{code:small_distance_qubit_stabilizer}
\item codes that are cousins of \flmRefsCref{code:small_distance_quantum}
\end{itemize}}%

\eczhIncludeCodeGraph{Bare}{scale=0.5}{\columnwidth}{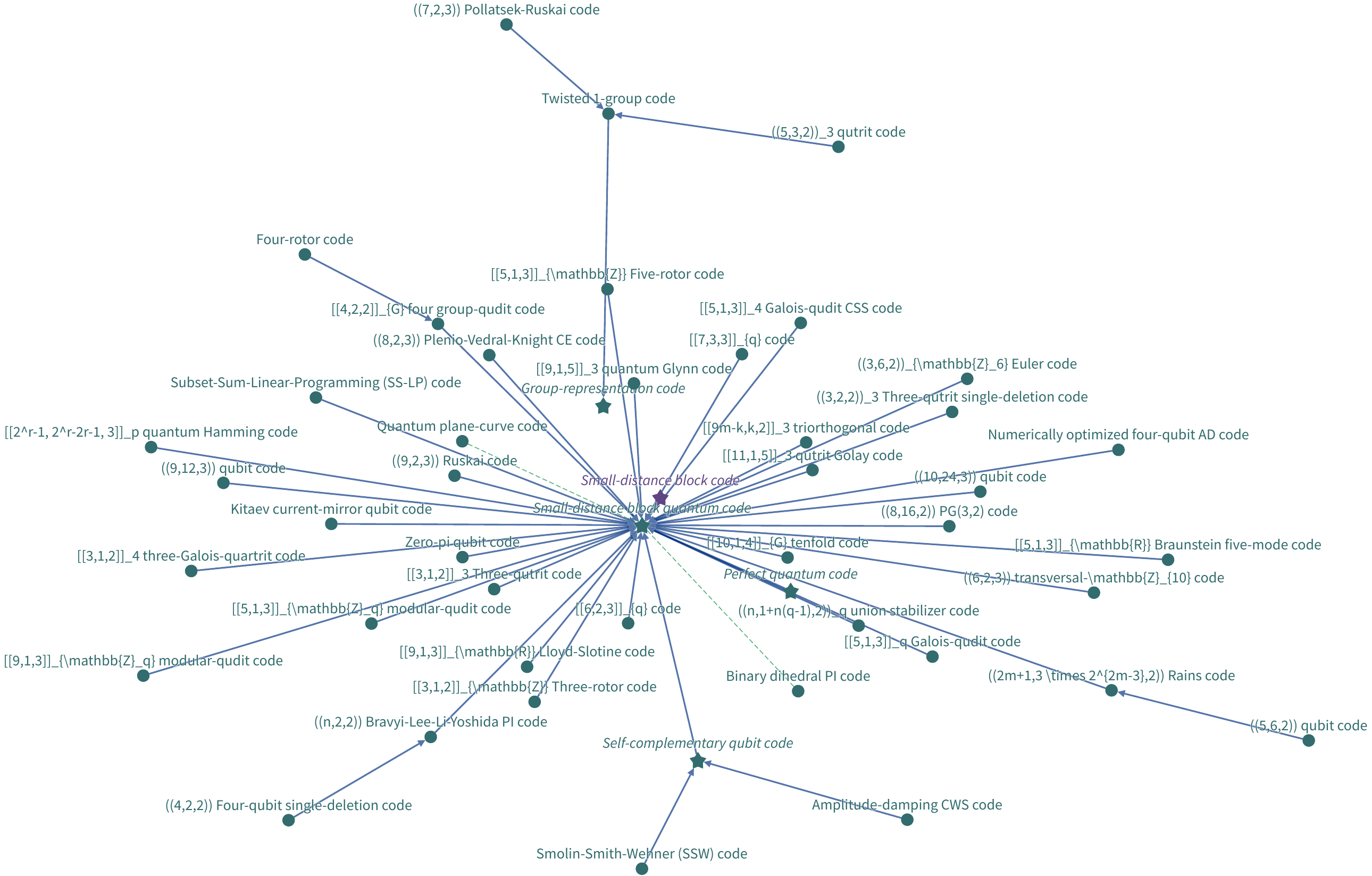}{Small-distance quantum codes and friends (non-qubit-stabilizer)}{https://errorcorrectionzoo.org/code_graph#J\%7B\%22displayMode\%22\%3A\%22subset\%22\%2C\%22modeSubsetOptions\%22\%3A\%7B\%22codeIds\%22\%3A\%5B\%22ampdamp_cws\%22\%2C\%22binary_dihedral_permutation_invariant\%22\%2C\%22rotor_4_2_2\%22\%2C\%22group_representation\%22\%2C\%22current_mirror\%22\%2C\%22ampdamp_numopt\%22\%2C\%22quantum_perfect\%22\%2C\%22quantum_plane_curve\%22\%2C\%22self_complementary\%22\%2C\%22small_distance\%22\%2C\%22small_distance_quantum\%22\%2C\%22ssw\%22\%2C\%22sslp\%22\%2C\%22t_group\%22\%2C\%22zero_pi\%22\%2C\%22qubit_10_24_3\%22\%2C\%22rains\%22\%2C\%22three_qutrit_permutation_invariant\%22\%2C\%22qudit_3_6_2\%22\%2C\%22four_qubit_permutation_invariant\%22\%2C\%22su3_sigma360\%22\%2C\%22qubit_5_6_2\%22\%2C\%22qubit_6_2_3\%22\%2C\%22icosahedral_permutation_invariant\%22\%2C\%22qubit_8_4_2\%22\%2C\%22qubit_8_1_3\%22\%2C\%22qubit_9_12_3\%22\%2C\%22ruskai\%22\%2C\%22arvind\%22\%2C\%22unentangled_permutation_invariant\%22\%2C\%22group_10_1_4\%22\%2C\%22qutrit_golay\%22\%2C\%22qudit_hamming_css\%22\%2C\%22stab_3_1_2\%22\%2C\%22galois_3_1_2\%22\%2C\%22rotor_3_1_2\%22\%2C\%22group_4_2_2\%22\%2C\%22css_5_1_3\%22\%2C\%22galois_5_1_3\%22\%2C\%22braunstein\%22\%2C\%22qudit_5_1_3\%22\%2C\%22rotor_5_1_3\%22\%2C\%22galois_6_2_3\%22\%2C\%22galois_7_3_3\%22\%2C\%22lloyd_slotine\%22\%2C\%22stab_9_1_3\%22\%2C\%22stab_9_1_5\%22\%2C\%22qutrit_small_triorthogonal\%22\%5D\%2C\%22reusePreviousLayoutPositions\%22\%3Afalse\%2C\%22showIntermediateConnectingNodes\%22\%3Atrue\%2C\%22connectingNodesMaxDepth\%22\%3A15\%2C\%22connectingNodesPathMaxLength\%22\%3A20\%2C\%22connectingNodesMaxExtraDepth\%22\%3A3\%2C\%22connectingNodesOnlyKeepPathsWithAdditionalLength\%22\%3A1\%2C\%22connectingNodesToDomainsAndKingdoms\%22\%3Afalse\%2C\%22connectingNodesEdgeLengthsByType\%22\%3A\%7B\%22primaryParent\%22\%3A1\%2C\%22secondaryParent\%22\%3A4\%2C\%22cousin\%22\%3A6\%7D\%2C\%22nodeIds\%22\%3A\%5B\%5D\%7D\%2C\%22highlightImportantNodes\%22\%3A\%7B\%22highlightImportantNodes\%22\%3Afalse\%2C\%22highlightPrimaryParents\%22\%3Afalse\%2C\%22highlightRootConnectingEdges\%22\%3Afalse\%7D\%7D}

\begingroup
\small
\eczhBreakableDashes
\renewcommand\arraystretch{1.05}
\edef\myxtraspc{\eczListAddVSpaceXtraXtraStretch}
\begin{tabularx}{\linewidth}{>{\raggedright\arraybackslash}p{\eczListColWidth{name}} >{\hsize=1.0000\hsize }X}
\toprule
\eczListColTitle{Code} & \eczListColTitle{Description} \\
\midrule
\endfirsthead
\toprule
\eczListColTitleContinued{Code} & \eczListColTitleContinued{Description} \\
\midrule
\endhead
\bottomrule
\endfoot
\eczhRefIndex{code:ampdamp_cws}%
\eczhListValue{\flmRefsHyperref{code:ampdamp_cws}{Amplitude-damping CWS code}} & \eczhListValue{Self-complementary CWS code that is designed to detect and correct \flmRefsHyperref{ref498}{AD} errors.}\\ 
\addlinespace[\myxtraspc]
\eczhRefIndex{code:binary_dihedral_permutation_invariant}%
\eczhListValue{\flmRefsHyperref{code:binary_dihedral_permutation_invariant}{Binary dihedral PI code}} & \eczhListValue{Multi-qubit PI code designed to realize gates from the binary dihedral group transversally.
Can also be interpreted as a single-spin code.
The codespace projection is a projection onto an irrep of the \textit{binary dihedral group} \( \mathsf{BD}_{2N} = \langle\omega I, X, P\rangle \) of order \(8N\), where \( \omega \) is a \( 2N \)th root of unity, and \( P = \text{diag} ( 1, \omega^2) \).}\\ 
\addlinespace[\myxtraspc]
\eczhRefIndex{code:rotor_4_2_2}%
\eczhListValue{\flmRefsHyperref{code:rotor_4_2_2}{Four-rotor code}} & \eczhListValue{\(\llbracket 4,2,2\rrbracket _{\mathbb Z}\) CSS rotor code that is an extension of the four-qubit code to the integer alphabet, i.e., the angular momentum states of a rotor.}\\ 
\addlinespace[\myxtraspc]
\eczhRefIndex{code:group_representation}%
\eczhListValue{\flmRefsHyperref{code:group_representation}{Group-representation code}} & \eczhListValue{Code whose projector is onto an irreducible representation of a subgroup \(G\) of a group of canonical or distinguished unitary operations, e.g., transversal gates in the case of block quantum codes, Gaussian operations in the case of bosonic codes, or \(SU(2)\) operations in the case of single-spin codes.}\\ 
\addlinespace[\myxtraspc]
\eczhRefIndex{code:current_mirror}%
\eczhListValue{\flmRefsHyperref{code:current_mirror}{Kitaev current-mirror qubit code}} & \eczhListValue{Member of the family of \(\llbracket 2n,(0,2),(2,n)\rrbracket _{\mathbb{Z}}\) homological rotor codes storing a logical qubit on a thin Möbius strip.
The ideal code can be obtained from a Josephson-junction \NoCaseChange{\protect\cite{cite396}} system \NoCaseChange{\protect\cite{cite397}}.}\\ 
\addlinespace[\myxtraspc]
\eczhRefIndex{code:ampdamp_numopt}%
\eczhListValue{\flmRefsHyperref{code:ampdamp_numopt}{Numerically optimized four-qubit AD code}} & \eczhListValue{One of several four-qubit codes that can (approximately) correct a single \flmRefsHyperref{ref498}{AD} error with higher fidelity than the \(\llbracket 4,1,2\rrbracket \) subcodes of the \(\llbracket 4,2,2\rrbracket \) code.}\\ 
\addlinespace[\myxtraspc]
\eczhRefIndex{code:quantum_perfect}%
\eczhListValue{\flmRefsHyperref{code:quantum_perfect}{Perfect quantum code}} & \eczhListValue{A type of block quantum code whose parameters satisfy the quantum Hamming bound with equality.}\\ 
\addlinespace[\myxtraspc]
\eczhRefIndex{code:quantum_plane_curve}%
\eczhListValue{\flmRefsHyperref{code:quantum_plane_curve}{Quantum plane-curve code}} & \eczhListValue{Quantum AG code constructed from plane-curve codes via the Galois-qudit Hermitian construction.
Code parameters are \(\llbracket q^3,q^3+q^2-3q-2r,r+2q-q^2\rrbracket _q\), where \(r\) is an integer satisfying \(q^2 - 2 \leq r \leq q^2 + q - 3\), and where the underlying plane curve is \(y^q + y = x^{q-1}\).}\\ 
\addlinespace[\myxtraspc]
\eczhRefIndex{code:self_complementary}%
\eczhListValue{\flmRefsHyperref{code:self_complementary}{Self-complementary qubit code}} & \eczhListValue{A qubit code which admits a basis of codewords of the form \(|c\rangle+|\overline{c}\rangle\), where \(c\) is a bitstring and \(\overline{c}\) is its negation a.k.a. complement. 
Their codewords generalize the two-qubit Bell states and three-qubit GHZ states and are often called \textit{(qubit) cat states} or \textit{poor-man's GHZ states}.
Such codes were originally pointed out to perform well against \flmRefsHyperref{ref498}{AD} noise \NoCaseChange{\protect\cite{cite851}}.}\\ 
\addlinespace[\myxtraspc]
\eczhRefIndex{code:small_distance}%
\eczhListValue{\flmRefsHyperref{code:small_distance}{Small-distance block code}} & \eczhListValue{A block code of length \(n\) that either detects or corrects errors on at most two coordinates, i.e., has distance \(d \leq 5\).}\\ 
\addlinespace[\myxtraspc]
\eczhRefIndex{code:small_distance_quantum}%
\eczhListValue{\flmRefsHyperref{code:small_distance_quantum}{Small-distance block quantum code}} & \eczhListValue{A block quantum code on \(n\) subsystems that either detects or corrects errors on at most two subsystems, i.e., have distance \(\leq 5\).}\\ 
\addlinespace[\myxtraspc]
\eczhRefIndex{code:ssw}%
\eczhListValue{\flmRefsHyperref{code:ssw}{Smolin-Smith-Wehner (SSW) code}} & \eczhListValue{A family of \(\llparenthesis n=4k+2l+3,M_{k,l},2\rrparenthesis \) self-complementary CWS codes, where \(M_{k,l} \approx 2^{n-2}(1-\sqrt{2/(\pi(n-1\rrparenthesis })\).
For \(n \geq 11\), these codes have a logical subspace whose dimension is larger than that of the largest stabilizer code for the same \(n\) and \(d\).
Ref. \NoCaseChange{\protect\cite{cite452}} augments a star-graph-based subfamily \(\llparenthesis 4n+1,M_n,2\rrparenthesis \) by one additional graph-state basis word, yielding \(\llparenthesis 4n+1,M_n+1,2\rrparenthesis \) codes with \(M_n=2^{4n-1}-\frac{1}{2}\binom{4n}{2n}\).
In the CWS description of Ref. \NoCaseChange{\protect\cite{cite852}}, the underlying stabilizer state is locally Clifford-equivalent to a GHZ state and its standard-form graph is a star graph.}\\ 
\addlinespace[\myxtraspc]
\eczhRefIndex{code:sslp}%
\eczhListValue{\flmRefsHyperref{code:sslp}{Subset-Sum-Linear-Programming (SS-LP) code}} & \eczhListValue{Qubit block quantum code that encodes a logical qubit and that is constructed using the Subset-Sum-Linear-Programming (SS-LP) numerical construction.
SS-LP codes are optimized to admit diagonal gates transversally and include \(\llparenthesis 7,2,3\rrparenthesis \) codes that realize the \(\mathsf{BD}_{16}\) and \(\mathsf{BD}_{32}\) groups transversally, yielding \(T\) and \(\sqrt{T}\) gates, respectively. Larger codes include an \(\llparenthesis 8,2,3\rrparenthesis \) code that transversally realizes \(\mathsf{BD}_{64}\).}\\ 
\addlinespace[\myxtraspc]
\eczhRefIndex{code:t_group}%
\eczhListValue{\flmRefsHyperref{code:t_group}{Twisted \(1\)-group code}} & \eczhListValue{Block group-representation code realizing particular irreps of particular groups such that a distance of two is automatically guaranteed.
Groups which admit irreps with this property are called \textit{twisted (unitary) \(1\)-groups} and include the binary icosahedral group \(2I\), the \(\Sigma(360\phi)\) subgroup of \(SU(3)\), the family \(\{PSp(2b, 3), b \geq 1\}\), and the alternating groups \(A_{5,6}\).
Groups whose irreps are images of the appropriate irreps of twisted \(1\)-groups also yield such properties, e.g., the binary tetrahedral group \(2T\) or qutrit Pauli group \(\Sigma(72\phi)\).}\\ 
\addlinespace[\myxtraspc]
\eczhRefIndex{code:zero_pi}%
\eczhListValue{\flmRefsHyperref{code:zero_pi}{Zero-pi qubit code}} & \eczhListValue{A \(\llbracket 2,(0,2),(2,1)\rrbracket _{\mathbb{Z}}\) homological rotor code on the smallest tiling of the projective plane \(\mathbb{R}P^2\).
The ideal code can be obtained from a four-rotor Josephson-junction \NoCaseChange{\protect\cite{cite396}} system after a choice of grounding \NoCaseChange{\protect\cite{cite397}}.}\\ 
\addlinespace[\myxtraspc]
\eczhRefIndex{code:qubit_10_24_3}%
\eczhListValue{\flmRefsHyperref{code:qubit_10_24_3}{\(\llparenthesis 10,24,3\rrparenthesis \) qubit code}} & \eczhListValue{Ten-qubit nonadditive CWS code saturating the linear-programming bound for one-error-correcting codes on ten qubits \NoCaseChange{\protect\cite{cite452}}.
It was constructed using the coding-clique graph-theoretic framework introduced in \NoCaseChange{\protect\cite{cite452}}, which unifies additive and nonadditive code constructions.}\\ 
\addlinespace[\myxtraspc]
\eczhRefIndex{code:rains}%
\eczhListValue{\flmRefsHyperref{code:rains}{\(\llparenthesis 2m+1,3 \times 2^{2m-3},2\rrparenthesis \) Rains code}} & \eczhListValue{Member of a family of \flmRefsHyperref{ref672}{pure} odd-length distance-two CWS codes with parameters \(\llparenthesis 2m+1,3 \times 2^{2m-3},2\rrparenthesis \) for all \(m \geq 2\), constructed recursively from the \(\llparenthesis 5,6,2\rrparenthesis \) code \NoCaseChange{\protect\cite{cite854,cite855}\protect\cite[{Lem. 5 and Thm. 4}]{cite446}\protect\cite[{Exam. 8}]{cite853}}.}\\ 
\addlinespace[\myxtraspc]
\eczhRefIndex{code:three_qutrit_permutation_invariant}%
\eczhListValue{\flmRefsHyperref{code:three_qutrit_permutation_invariant}{\(\llparenthesis 3,2,2\rrparenthesis _3\) Three-qutrit single-deletion code}} & \eczhListValue{Three-qutrit PI code that is the smallest qutrit PI code to correct one deletion error.}\\ 
\addlinespace[\myxtraspc]
\eczhRefIndex{code:qudit_3_6_2}%
\eczhListValue{\flmRefsHyperref{code:qudit_3_6_2}{\(\llparenthesis 3,6,2\rrparenthesis _{\mathbb{Z}_6}\) Euler code}} & \eczhListValue{Three-qudit error-detecting code with logical dimension \(K=6\) that is obtained from a particular \flmRefsHyperref{ref219}{AME state} that serves as a solution of a quantum analogue of the classical problem of 36 officers of Euler.
The code is obtained from a \(\llparenthesis 4,1,3\rrparenthesis _{\mathbb{Z}_6}\) code.}\\ 
\addlinespace[\myxtraspc]
\eczhRefIndex{code:four_qubit_permutation_invariant}%
\eczhListValue{\flmRefsHyperref{code:four_qubit_permutation_invariant}{\(\llparenthesis 4,2,2\rrparenthesis \) Four-qubit single-deletion code}} & \eczhListValue{Four-qubit PI code that is the smallest qubit code to correct one deletion error.}\\ 
\addlinespace[\myxtraspc]
\eczhRefIndex{code:su3_sigma360}%
\eczhListValue{\flmRefsHyperref{code:su3_sigma360}{\(\llparenthesis 5,3,2\rrparenthesis _3\) qutrit code}} & \eczhListValue{Smallest qutrit block code realizing the \(\Sigma(360\phi)=3.A_6\) subgroup of \(SU(3)\) transversally.
The next smallest code is \(\llparenthesis 7,3,2\rrparenthesis _3\).}\\ 
\addlinespace[\myxtraspc]
\eczhRefIndex{code:qubit_5_6_2}%
\eczhListValue{\flmRefsHyperref{code:qubit_5_6_2}{\(\llparenthesis 5,6,2\rrparenthesis \) qubit code}} & \eczhListValue{Five-qubit cyclic CWS code detecting a single-qubit error.
This code has a logical subspace whose dimension is larger than that of the \(\llbracket 5,2,2\rrbracket \) code, the best five-qubit stabilizer code with the same distance \NoCaseChange{\protect\cite{cite452}}.}\\ 
\addlinespace[\myxtraspc]
\eczhRefIndex{code:qubit_6_2_3}%
\eczhListValue{\flmRefsHyperref{code:qubit_6_2_3}{\(\llparenthesis 6,2,3\rrparenthesis \) transversal-\(\mathbb{Z}_{10}\) code}} & \eczhListValue{Six-qubit code that realizes gates from the group \(\mathbb{Z}_{10}\) transversally.
This is the smallest known distance-three code supporting a transversal gate outside of the \flmRefsHyperref{ref409}{Clifford group}.
See Ref. \NoCaseChange{\protect\cite{cite528}} for the codewords.}\\ 
\addlinespace[\myxtraspc]
\eczhRefIndex{code:icosahedral_permutation_invariant}%
\eczhListValue{\flmRefsHyperref{code:icosahedral_permutation_invariant}{\(\llparenthesis 7,2,3\rrparenthesis \) Pollatsek-Ruskai code}} & \eczhListValue{Seven-qubit PI code that realizes gates from the binary icosahedral group transversally.
Can also be interpreted as a spin-\(7/2\) single-spin code.
The codespace projection is a projection onto an irrep of the binary icosahedral group \(2I\).
See Ref. \NoCaseChange{\protect\cite{cite528}} for other non-PI codes realizing \(2I\) gates transversally.}\\ 
\addlinespace[\myxtraspc]
\eczhRefIndex{code:qubit_8_4_2}%
\eczhListValue{\flmRefsHyperref{code:qubit_8_4_2}{\(\llparenthesis 8,16,2\rrparenthesis \) \(PG(3,2)\) code}} & \eczhListValue{Eight-qubit code encoding four logical qubits whose logical basis consists of a GHZ state together with fifteen states built from the incidence geometry of the projective space \(PG(3,2)\).}\\ 
\addlinespace[\myxtraspc]
\eczhRefIndex{code:qubit_8_1_3}%
\eczhListValue{\flmRefsHyperref{code:qubit_8_1_3}{\(\llparenthesis 8,2,3\rrparenthesis \) Plenio-Vedral-Knight CE code}} & \eczhListValue{An eight-qubit single-error-correcting code that is the first CE code.
Each logical state is a superposition of computational basis states with four excitations.}\\ 
\addlinespace[\myxtraspc]
\eczhRefIndex{code:qubit_9_12_3}%
\eczhListValue{\flmRefsHyperref{code:qubit_9_12_3}{\(\llparenthesis 9,12,3\rrparenthesis \) qubit code}} & \eczhListValue{Nine-qubit cyclic CWS code correcting a single-qubit error.
This code has a logical subspace whose dimension is larger than that of the \(\llbracket 9,3,3\rrbracket \) code, the best nine-qubit stabilizer code with the same distance \NoCaseChange{\protect\cite{cite449}}.}\\ 
\addlinespace[\myxtraspc]
\eczhRefIndex{code:ruskai}%
\eczhListValue{\flmRefsHyperref{code:ruskai}{\(\llparenthesis 9,2,3\rrparenthesis \) Ruskai code}} & \eczhListValue{Nine-qubit PI code that protects against single-qubit errors as well as two-qubit errors arising from exchange processes.}\\ 
\addlinespace[\myxtraspc]
\eczhRefIndex{code:arvind}%
\eczhListValue{\flmRefsHyperref{code:arvind}{\(\llparenthesis n,1+n(q-1),2\rrparenthesis _q\) union stabilizer code}} & \eczhListValue{Member of a family of \(\llparenthesis n,1+n(q-1),2\rrparenthesis _q\) Galois-qudit union stabilizer codes for odd \(n\).}\\ 
\addlinespace[\myxtraspc]
\eczhRefIndex{code:unentangled_permutation_invariant}%
\eczhListValue{\flmRefsHyperref{code:unentangled_permutation_invariant}{\(\llparenthesis n,2,2\rrparenthesis \) Bravyi-Lee-Li-Yoshida PI code}} & \eczhListValue{PI distance-two code on \(n\geq4\) qubits whose degree of entanglement vanishes asymptotically with \(n\) \NoCaseChange{\protect\cite[{Appx. D}]{cite529}} (cf. \NoCaseChange{\protect\cite{cite530}}).}\\ 
\addlinespace[\myxtraspc]
\eczhRefIndex{code:group_10_1_4}%
\eczhListValue{\flmRefsHyperref{code:group_10_1_4}{\(\llbracket 10,1,4\rrbracket _{G}\) tenfold code}} & \eczhListValue{A \(\llbracket 10,1,4\rrbracket _{G}\) group code for finite Abelian \(G\).
The code is defined using a graph that is closely related to the \(\llbracket 5,1,3\rrbracket \) code.}\\ 
\addlinespace[\myxtraspc]
\eczhRefIndex{code:qutrit_golay}%
\eczhListValue{\flmRefsHyperref{code:qutrit_golay}{\(\llbracket 11,1,5\rrbracket _3\) qutrit Golay code}} & \eczhListValue{An \(\llbracket 11,1,5\rrbracket _3\) code constructed from the ternary Golay code via the CSS construction.
The code's stabilizer generator matrix blocks \(H_{X}\) and \(H_{Z}\) are both the generator matrix of the ternary Golay code.}\\ 
\addlinespace[\myxtraspc]
\eczhRefIndex{code:qudit_hamming_css}%
\eczhListValue{\flmRefsHyperref{code:qudit_hamming_css}{\(\llbracket 2^r-1, 2^r-2r-1, 3\rrbracket _p\) quantum Hamming code}} & \eczhListValue{A family of CSS codes extending \flmRefsHyperref{code:quantum_hamming_css}{quantum Hamming codes} to prime qudits of dimension \(p\) by expressing the qubit code stabilizers in local-dimension-invariant (LDI) form \NoCaseChange{\protect\cite{cite820}}.}\\ 
\addlinespace[\myxtraspc]
\eczhRefIndex{code:stab_3_1_2}%
\eczhListValue{\flmRefsHyperref{code:stab_3_1_2}{\(\llbracket 3,1,2\rrbracket _3\) Three-qutrit code}} & \eczhListValue{A \(\llbracket 3,1,2\rrbracket _3\) prime-qudit CSS code that is the smallest qutrit stabilizer code to detect a single-qutrit error.
It has stabilizer generators \(ZZZ\) and \(XXX\). The code defines a quantum secret-sharing scheme and serves as a minimal model for the AdS/CFT holographic duality. It is also the smallest non-trivial instance of a quantum maximum distance separable code (QMDS), saturating the quantum Singleton bound.}\\ 
\addlinespace[\myxtraspc]
\eczhRefIndex{code:galois_3_1_2}%
\eczhListValue{\flmRefsHyperref{code:galois_3_1_2}{\(\llbracket 3,1,2\rrbracket _4\) three-Galois-quartrit code}} & \eczhListValue{Three-Galois-qudit CSS code over \(\mathbb{F}_4=\{0,1,\omega,\omega^2\}\) that encodes one logical Galois qudit and detects a single-qudit error.}\\ 
\addlinespace[\myxtraspc]
\eczhRefIndex{code:rotor_3_1_2}%
\eczhListValue{\flmRefsHyperref{code:rotor_3_1_2}{\(\llbracket 3,1,2\rrbracket _{\mathbb{Z}}\) Three-rotor code}} & \eczhListValue{\(\llbracket 3,1,2\rrbracket _{\mathbb Z}\) rotor code that is an extension of the \(\llbracket 3,1,2\rrbracket _3\) qutrit CSS code to the integer alphabet, i.e., the angular momentum states of a rotor.}\\ 
\addlinespace[\myxtraspc]
\eczhRefIndex{code:group_4_2_2}%
\eczhListValue{\flmRefsHyperref{code:group_4_2_2}{\(\llbracket 4,2,2\rrbracket _{G}\) four group-qudit code}} & \eczhListValue{\(\llbracket 4,2,2\rrbracket _{G}\) group quantum code that is an extension of the four-qubit code to group-valued qudits.}\\ 
\addlinespace[\myxtraspc]
\eczhRefIndex{code:css_5_1_3}%
\eczhListValue{\flmRefsHyperref{code:css_5_1_3}{\(\llbracket 5,1,3\rrbracket _4\) Galois-qudit CSS code}} & \eczhListValue{Five-Galois-qudit CSS code over \(\mathbb{F}_4=\{0,1,\omega,\omega^2\}\) that encodes one logical Galois qudit and corrects a single-qudit error.}\\ 
\addlinespace[\myxtraspc]
\eczhRefIndex{code:galois_5_1_3}%
\eczhListValue{\flmRefsHyperref{code:galois_5_1_3}{\(\llbracket 5,1,3\rrbracket _q\) Galois-qudit code}} & \eczhListValue{True stabilizer code that generalizes the five-qubit perfect code to Galois qudits of prime-power dimension \(q=p^m\). It has \(4m\) stabilizer generators expressed as \(X_{\gamma} Z_{\gamma} Z_{-\gamma} X_{-\gamma} I\) and its cyclic permutations, with \(\gamma\) iterating over basis elements of \(\mathbb{F}_q\) over \(\mathbb{F}_p\).}\\ 
\addlinespace[\myxtraspc]
\eczhRefIndex{code:braunstein}%
\eczhListValue{\flmRefsHyperref{code:braunstein}{\(\llbracket 5,1,3\rrbracket _{\mathbb{R}}\) Braunstein five-mode code}} & \eczhListValue{An analog stabilizer version of the five-qubit perfect code, encoding one mode into five and correcting arbitrary errors on any one mode.}\\ 
\addlinespace[\myxtraspc]
\eczhRefIndex{code:qudit_5_1_3}%
\eczhListValue{\flmRefsHyperref{code:qudit_5_1_3}{\(\llbracket 5,1,3\rrbracket _{\mathbb{Z}_q}\) modular-qudit code}} & \eczhListValue{Modular-qudit stabilizer code that generalizes the five-qubit perfect code using properties of the multiplicative group \(\mathbb{Z}_q\) \NoCaseChange{\protect\cite{cite531}}; see also \NoCaseChange{\protect\cite[{Thm. 13}]{cite532}}. It has four stabilizer generators consisting of \(X Z Z^\dagger X^\dagger I\) and its cyclic permutations.}\\ 
\addlinespace[\myxtraspc]
\eczhRefIndex{code:rotor_5_1_3}%
\eczhListValue{\flmRefsHyperref{code:rotor_5_1_3}{\(\llbracket 5,1,3\rrbracket _{\mathbb{Z}}\) Five-rotor code}} & \eczhListValue{Extension of the five-qubit stabilizer code to the integer alphabet, i.e., the angular momentum states of a rotor. The code is \(U(1)\)-covariant and ideal codewords are not normalizable.}\\ 
\addlinespace[\myxtraspc]
\eczhRefIndex{code:galois_6_2_3}%
\eczhListValue{\flmRefsHyperref{code:galois_6_2_3}{\(\llbracket 6,2,3\rrbracket _{q}\) code}} & \eczhListValue{Six-qudit MDS error-correcting code defined for Galois-qudit dimension \(q=3\) \NoCaseChange{\protect\cite{cite830}}, \(q=2^2\) \NoCaseChange{\protect\cite{cite831}}, and \(q \geq 5\) \NoCaseChange{\protect\cite{cite830}\protect\cite[{Exam. 33}]{cite813}}.
This code cannot exist for qubits (\(q=2\)).}\\ 
\addlinespace[\myxtraspc]
\eczhRefIndex{code:galois_7_3_3}%
\eczhListValue{\flmRefsHyperref{code:galois_7_3_3}{\(\llbracket 7,3,3\rrbracket _{q}\) code}} & \eczhListValue{Seven-qudit MDS error-detecting code defined for Galois-qudit dimension \(q=3\) \NoCaseChange{\protect\cite{cite830}} and \(q \geq 7\) \NoCaseChange{\protect\cite{cite830}\protect\cite[{Exam. 33}]{cite813}}.
This code cannot exist for qubits (\(q=2\)).}\\ 
\addlinespace[\myxtraspc]
\eczhRefIndex{code:lloyd_slotine}%
\eczhListValue{\flmRefsHyperref{code:lloyd_slotine}{\(\llbracket 9,1,3\rrbracket _{\mathbb{R}}\) Lloyd-Slotine code}} & \eczhListValue{An analog stabilizer version of Shor's nine-qubit code, encoding one mode into nine and correcting arbitrary errors on any one mode.}\\ 
\addlinespace[\myxtraspc]
\eczhRefIndex{code:stab_9_1_3}%
\eczhListValue{\flmRefsHyperref{code:stab_9_1_3}{\(\llbracket 9,1,3\rrbracket _{\mathbb{Z}_q}\) modular-qudit code}} & \eczhListValue{Modular-qudit CSS code that generalizes the \(\llbracket 9,1,3\rrbracket \) Shor code to \(q\)-level systems.}\\ 
\addlinespace[\myxtraspc]
\eczhRefIndex{code:stab_9_1_5}%
\eczhListValue{\flmRefsHyperref{code:stab_9_1_5}{\(\llbracket 9,1,5\rrbracket _3\) quantum Glynn code}} & \eczhListValue{Nine-qutrit \flmRefsHyperref{ref672}{pure} Hermitian code that is the smallest qutrit stabilizer code to correct two-qutrit errors.}\\ 
\addlinespace[\myxtraspc]
\eczhRefIndex{code:qutrit_small_triorthogonal}%
\eczhListValue{\flmRefsHyperref{code:qutrit_small_triorthogonal}{\(\llbracket 9m-k,k,2\rrbracket _3\) triorthogonal code}} & \eczhListValue{Member of the \(\llbracket 9m-k,k,2\rrbracket _3\) family of triorthogonal qutrit codes (for \(k\leq 3m-2\)) that admit a transversal diagonal gate in the third level of the \flmRefsHyperref{ref751}{qudit Clifford hierarchy} and that are relevant for magic-state distillation.}\\ 
\end{tabularx}\endgroup
\eczcodelist{small_qubit_stabilizer}{Small-distance qubit stabilizer codes and friends
}%

\eczhCodeListAutoDescription{All descendants and cousins of \flmRefsCref{code:small_distance_qubit_stabilizer}.}%

\eczhIncludeCodeGraph{Bare}{scale=0.5}{\columnwidth}{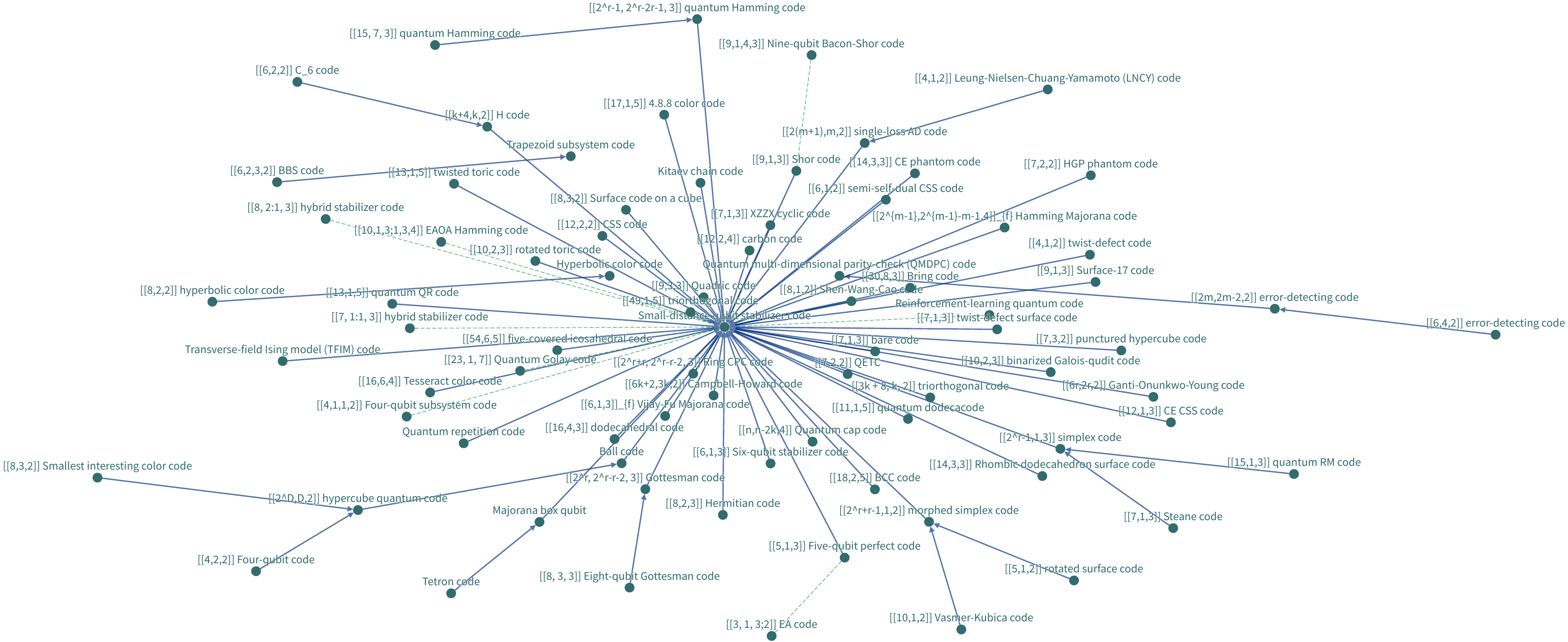}{Small-distance qubit stabilizer codes and friends}{https://errorcorrectionzoo.org/code_graph#J\%7B\%22displayMode\%22\%3A\%22subset\%22\%2C\%22modeSubsetOptions\%22\%3A\%7B\%22codeIds\%22\%3A\%5B\%22ball_color\%22\%2C\%22hyperbolic_color\%22\%2C\%22kitaev_chain\%22\%2C\%22mbq\%22\%2C\%22qmdpc\%22\%2C\%22quantum_repetition\%22\%2C\%22reinforcement_learning\%22\%2C\%22small_distance_qubit_stabilizer\%22\%2C\%22tetron\%22\%2C\%22tfim\%22\%2C\%22trapezoid\%22\%2C\%22stab_10_1_2\%22\%2C\%22eaoa_hamming\%22\%2C\%22stab_10_2_3\%22\%2C\%22xzzx_10_2_3\%22\%2C\%22stab_11_1_5\%22\%2C\%22css_12_1_3\%22\%2C\%22stab_12_2_2\%22\%2C\%22carbon\%22\%2C\%22quad_residue_13_1_5\%22\%2C\%22stab_13_1_5\%22\%2C\%22phantom_14_3_3\%22\%2C\%22rhombic_dodecahedron_surface\%22\%2C\%22stab_15_7_3\%22\%2C\%22stab_15_1_3\%22\%2C\%22quantum_dodecahedron\%22\%2C\%22stab_16_6_4\%22\%2C\%22stab_17_1_5\%22\%2C\%22stab_18_2_5\%22\%2C\%22ampdamp_stabilizer\%22\%2C\%22qubit_golay\%22\%2C\%22hypercube_quantum\%22\%2C\%22ring_cpc\%22\%2C\%22morphed_diagonal_clifford\%22\%2C\%22quantum_hamming\%22\%2C\%22quantum_hamming_css\%22\%2C\%22diagonal_clifford\%22\%2C\%22majorana_hamming\%22\%2C\%22iceberg\%22\%2C\%22ea_3_1_3-2\%22\%2C\%22stellated_dodecahedron_css\%22\%2C\%22small_triorthogonal\%22\%2C\%22bacon_shor_4\%22\%2C\%22css_4_1_2\%22\%2C\%22stab_4_1_2\%22\%2C\%22stab_4_2_2\%22\%2C\%22stab_49_1_5\%22\%2C\%22stab_5_1_2\%22\%2C\%22stab_5_1_3\%22\%2C\%22quantum_icosahedron\%22\%2C\%22css_6_1_2\%22\%2C\%22stab_6_1_3\%22\%2C\%22majorana_6_1_3\%22\%2C\%22stab_6_2_2\%22\%2C\%22bravyi_bacon_shor_6\%22\%2C\%22stab_6_4_2\%22\%2C\%22campbell_howard\%22\%2C\%22goy\%22\%2C\%22hybrid_7_1-1_3\%22\%2C\%22steane\%22\%2C\%22xzzx_7_1_3\%22\%2C\%22bare_7_1_3\%22\%2C\%22twist_defect_7_1_3\%22\%2C\%22hgp_7_2_2\%22\%2C\%22qetc_7_2\%22\%2C\%22xz_7_3_2\%22\%2C\%22hybrid_8_2-1_3\%22\%2C\%22stab_8_3_3\%22\%2C\%22stab_8_1_2\%22\%2C\%22stab_8_2_2\%22\%2C\%22stab_8_2_3\%22\%2C\%22stab_8_3_2\%22\%2C\%22cubic_surface\%22\%2C\%22shor_nine\%22\%2C\%22surface-17\%22\%2C\%22bacon_shor_9\%22\%2C\%22stab_9_3_3\%22\%2C\%22quantum_h\%22\%2C\%22quantum_cap\%22\%5D\%2C\%22reusePreviousLayoutPositions\%22\%3Afalse\%2C\%22showIntermediateConnectingNodes\%22\%3Atrue\%2C\%22connectingNodesMaxDepth\%22\%3A15\%2C\%22connectingNodesPathMaxLength\%22\%3A20\%2C\%22connectingNodesMaxExtraDepth\%22\%3A3\%2C\%22connectingNodesOnlyKeepPathsWithAdditionalLength\%22\%3A1\%2C\%22connectingNodesToDomainsAndKingdoms\%22\%3Afalse\%2C\%22connectingNodesEdgeLengthsByType\%22\%3A\%7B\%22primaryParent\%22\%3A1\%2C\%22secondaryParent\%22\%3A4\%2C\%22cousin\%22\%3A6\%7D\%2C\%22nodeIds\%22\%3A\%5B\%5D\%7D\%2C\%22highlightImportantNodes\%22\%3A\%7B\%22highlightImportantNodes\%22\%3Afalse\%2C\%22highlightPrimaryParents\%22\%3Afalse\%2C\%22highlightRootConnectingEdges\%22\%3Afalse\%7D\%7D}

\begingroup
\small
\eczhBreakableDashes
\renewcommand\arraystretch{1.05}
\edef\myxtraspc{\eczListAddVSpaceXtraXtraStretch}
\begin{tabularx}{\linewidth}{>{\raggedright\arraybackslash}p{\eczListColWidth{name}} >{\hsize=1.0000\hsize }X}
\toprule
\eczListColTitle{Code} & \eczListColTitle{Description} \\
\midrule
\endfirsthead
\toprule
\eczListColTitleContinued{Code} & \eczListColTitleContinued{Description} \\
\midrule
\endhead
\bottomrule
\endfoot
\eczhRefIndex{code:ball_color}%
\eczhListValue{\flmRefsHyperref{code:ball_color}{Ball code}} & \eczhListValue{A distance-two color code defined on a colorable \(D\)-ball, equivalently on a \(D\)-colex with boundary \NoCaseChange{\protect\cite[{Appx. A}]{cite687}}.
In the morphing construction of Ref. \NoCaseChange{\protect\cite{cite687}}, ball codes arise as the child codes associated with the morphed ball-like regions.
This family includes hypercube codes (defined on balls constructed from hyperoctahedra) and 3D ball codes (defined on duals of certain Archimedean solids).}\\ 
\addlinespace[\myxtraspc]
\eczhRefIndex{code:hyperbolic_color}%
\eczhListValue{\flmRefsHyperref{code:hyperbolic_color}{Hyperbolic color code}} & \eczhListValue{An extension of the color code construction to hyperbolic manifolds.
As opposed to there being only three types of uniform three-valent and three-colorable lattice tilings in the 2D Euclidean plane, there is an infinite number of admissible hyperbolic tilings in the 2D hyperbolic plane \NoCaseChange{\protect\cite{cite836}}.
Certain double covers of hyperbolic tilings also yield admissible tilings \NoCaseChange{\protect\cite{cite837}}.
Other admissible hyperbolic tilings can be obtained via a fattening procedure \NoCaseChange{\protect\cite{cite430}}; see also a construction based on the more general quantum pin codes \NoCaseChange{\protect\cite{cite702}}.}\\ 
\addlinespace[\myxtraspc]
\eczhRefIndex{code:kitaev_chain}%
\eczhListValue{\flmRefsHyperref{code:kitaev_chain}{Kitaev chain code}} & \eczhListValue{A Majorana stabilizer code obtained from the ground-state subspace of the Kitaev Majorana chain in its fermionic topological phase \NoCaseChange{\protect\cite{cite558}}. Its codespace is stabilized by nearest-neighbor Majorana bilinears, while two unpaired edge Majoranas furnish one logical fermionic mode. Under parity-preserving noise, it behaves as the Majorana analogue of the repetition code \NoCaseChange{\protect\cite{cite559}}.}\\ 
\addlinespace[\myxtraspc]
\eczhRefIndex{code:mbq}%
\eczhListValue{\flmRefsHyperref{code:mbq}{Majorana box qubit}} & \eczhListValue{A family of Majorana stabilizer codes obtained by fixing the total fermion parity of \(n\) fermionic modes, equivalently \(2n\) Majorana zero modes, within the ground-state subspace of \(n\) Kitaev Majorana chain Hamiltonians.
The resulting positive-parity subspace encodes \(n-1\) logical qubits and has Majorana distance \(2\).}\\ 
\addlinespace[\myxtraspc]
\eczhRefIndex{code:qmdpc}%
\eczhListValue{\flmRefsHyperref{code:qmdpc}{Quantum multi-dimensional parity-check (QMDPC) code}} & \eczhListValue{High-rate low-distance CSS code whose qubits lie on a \(D\)-dimensional rectangle, with \(X\)-type stabilizer generators defined on each \(D-1\)-dimensional rectangle.
The \(Z\)-type stabilizer generators are defined via permutations in order to commute with the \(X\)-type generators.}\\ 
\addlinespace[\myxtraspc]
\eczhRefIndex{code:quantum_repetition}%
\eczhListValue{\flmRefsHyperref{code:quantum_repetition}{Quantum repetition code}} & \eczhListValue{Encodes \(1\) qubit into \(n\) qubits according to \(|0\rangle\to|\phi_0\rangle^{\otimes n}\) and \(|1\rangle\to|\phi_1\rangle^{\otimes n}\). The code is called a \textit{bit-flip} code when \(|\phi_i\rangle = |i\rangle\), and a \textit{phase-flip} code when \(|\phi_0\rangle = |+\rangle\) and \(|\phi_1\rangle = |-\rangle\).
This repetition-style encoding does not clone an arbitrary quantum state; instead, it extends the copying of computational-basis states linearly to entangled codewords  \NoCaseChange{\protect\cite[{Ch. 2}]{cite398}}.}\\ 
\addlinespace[\myxtraspc]
\eczhRefIndex{code:reinforcement_learning}%
\eczhListValue{\flmRefsHyperref{code:reinforcement_learning}{Reinforcement-learning quantum code}} & \eczhListValue{An approximate qubit code obtained from a numerical optimization involving a reinforcement learning agent.}\\ 
\addlinespace[\myxtraspc]
\eczhRefIndex{code:small_distance_qubit_stabilizer}%
\eczhListValue{\flmRefsHyperref{code:small_distance_qubit_stabilizer}{Small-distance qubit stabilizer code}} & \eczhListValue{A qubit stabilizer code that either detects or corrects errors on at most two subsystems, i.e., has distance \(\leq 5\).}\\ 
\addlinespace[\myxtraspc]
\eczhRefIndex{code:tetron}%
\eczhListValue{\flmRefsHyperref{code:tetron}{Tetron code}} & \eczhListValue{A \(\llbracket 2,1,2\rrbracket _{f}\) Majorana box qubit encoding a logical qubit into four Majorana modes, equivalently into the fixed-total-parity sector of two physical fermionic modes.
Four Majorana zero modes are the smallest aggregate that supports a qubit in a fixed fermion-parity sector \NoCaseChange{\protect\cite{cite401}}.
This code can be concatenated with various qubit codes such as surface codes and color codes.
Four-boundary Majorana surface-code patches are logical tetrons, i.e., higher-distance analogues of this physical tetron block \NoCaseChange{\protect\cite{cite402}}.}\\ 
\addlinespace[\myxtraspc]
\eczhRefIndex{code:tfim}%
\eczhListValue{\flmRefsHyperref{code:tfim}{Transverse-field Ising model (TFIM) code}} & \eczhListValue{A 1D translationally invariant stabilizer code whose encoding is a constant-depth circuit of nearest-neighbor gates on alternating even and odd bonds that consist of transverse-field Ising Hamiltonian interactions. The code allows for perfect state transfer of arbitrary distance using local operations and classical communications (LOCC).}\\ 
\addlinespace[\myxtraspc]
\eczhRefIndex{code:trapezoid}%
\eczhListValue{\flmRefsHyperref{code:trapezoid}{Trapezoid subsystem code}} & \eczhListValue{A member of a family of BBS codes with weight-two (two-body) gauge generators designed to suppress errors in adiabatic quantum computation.}\\ 
\addlinespace[\myxtraspc]
\eczhRefIndex{code:stab_10_1_2}%
\eczhListValue{\flmRefsHyperref{code:stab_10_1_2}{\(\llbracket 10,1,2\rrbracket \) Vasmer-Kubica code}} & \eczhListValue{A stabilizer code obtained by morphing the \(\llbracket 15,1,3\rrbracket \) quantum Reed-Muller code on a subset whose child code is the \(\llbracket 8,3,2\rrbracket \) smallest interesting color code \NoCaseChange{\protect\cite{cite687}}.
It is the smallest known stabilizer code with a fault-tolerant logical \(T\) gate, implemented via physical \(T\), \(T^{\dagger}\), and \(CCZ\) gates \NoCaseChange{\protect\cite{cite687}}.}\\ 
\addlinespace[\myxtraspc]
\eczhRefIndex{code:eaoa_hamming}%
\eczhListValue{\flmRefsHyperref{code:eaoa_hamming}{\(\llbracket 10,1,3;1,3,4\rrbracket \) EAOA Hamming code}} & \eczhListValue{An EAOA qubit stabilizer code constructed from the dual of a \([10,6,3]\) code obtained by shortening the classical \([15,11,3]\) Hamming code at five positions.
In the notation of the parent entry, the example of Ref. \NoCaseChange{\protect\cite{cite856}} is a \(\llbracket 10,1,3;1,3,4\rrbracket \) code: it encodes one logical qubit and four classical strings (equivalently, two classical bits), while retaining one gauge qubit and using three ebits.}\\ 
\addlinespace[\myxtraspc]
\eczhRefIndex{code:stab_10_2_3}%
\eczhListValue{\flmRefsHyperref{code:stab_10_2_3}{\(\llbracket 10,2,3\rrbracket \) binarized Galois-qudit code}} & \eczhListValue{CSS code obtained by binarizing a \(\llbracket 5,1,3\rrbracket _4\) Galois-qudit CSS code in the self-dual normal basis \(\{\omega,\omega^2\}\).}\\ 
\addlinespace[\myxtraspc]
\eczhRefIndex{code:xzzx_10_2_3}%
\eczhListValue{\flmRefsHyperref{code:xzzx_10_2_3}{\(\llbracket 10,2,3\rrbracket \) rotated toric code}} & \eczhListValue{Rotated toric code that is the CSS form of the twisted XZZX toric code with parameters \(a=1\), \(b=3\) \NoCaseChange{\protect\cite{cite438,cite427}},
related to the XZZX form by Hadamard on the \(B\)-sublattice.
It is also the \flmRefsHyperref{ref436}{symplectic double} (a.k.a. genus-one double cover) of the \(\llbracket 5,1,3\rrbracket \) five-qubit perfect code \NoCaseChange{\protect\cite{cite439,cite435}}, the \flmRefsHyperref{ref436}{symplectic double} of the \(\llbracket 5,1,2\rrbracket \) rotated surface code \NoCaseChange{\protect\cite{cite435}}, and a BCC code \NoCaseChange{\protect\cite{cite440}}.}\\ 
\addlinespace[\myxtraspc]
\eczhRefIndex{code:stab_11_1_5}%
\eczhListValue{\flmRefsHyperref{code:stab_11_1_5}{\(\llbracket 11,1,5\rrbracket \) quantum dodecacode}} & \eczhListValue{Eleven-qubit \flmRefsHyperref{ref672}{pure} stabilizer code that is the smallest qubit stabilizer code to correct two-qubit errors.
It can be obtained from the non-additive dodecacode by puncturing \NoCaseChange{\protect\cite[{Table IV}]{cite449}}.}\\ 
\addlinespace[\myxtraspc]
\eczhRefIndex{code:css_12_1_3}%
\eczhListValue{\flmRefsHyperref{code:css_12_1_3}{\(\llbracket 12,1,3\rrbracket \) CE CSS code}} & \eczhListValue{Twelve-qubit constant-excitation (CE) CSS code that encodes one logical qubit with distance three.
It is the smallest CE CSS code that corrects a single-qubit error \NoCaseChange{\protect\cite{cite524}}.
Codewords lie in a fixed Hamming-weight subspace, making the code immune to coherent noise in the form of transversal \(Z\)-rotations.}\\ 
\addlinespace[\myxtraspc]
\eczhRefIndex{code:stab_12_2_2}%
\eczhListValue{\flmRefsHyperref{code:stab_12_2_2}{\(\llbracket 12,2,2\rrbracket \) CSS code}} & \eczhListValue{CSS code that admits a logical \(CS\) gate via application of physical \(T\) and \(T^{\dagger}\) gates.}\\ 
\addlinespace[\myxtraspc]
\eczhRefIndex{code:carbon}%
\eczhListValue{\flmRefsHyperref{code:carbon}{\(\llbracket 12,2,4\rrbracket \) carbon code}} & \eczhListValue{Twelve-qubit CSS code based on Knill's \(C_4/C_6\) scheme \NoCaseChange{\protect\cite{cite525}}.
Using the concatenation convention of the Zoo, the carbon code can be viewed as a block concatenation with inner code \(\llbracket 4,2,2\rrbracket \) and outer code \(C_6\): three inner \(\llbracket 4,2,2\rrbracket \) blocks encode six intermediate qubits, which are then encoded into two logical qubits by the outer \(\llbracket 6,2,2\rrbracket \) code.}\\ 
\addlinespace[\myxtraspc]
\eczhRefIndex{code:quad_residue_13_1_5}%
\eczhListValue{\flmRefsHyperref{code:quad_residue_13_1_5}{\(\llbracket 13,1,5\rrbracket \) quantum QR code}} & \eczhListValue{Thirteen-qubit cyclic Hermitian qubit code derived from a quaternary quadratic-residue code using the Hermitian construction \NoCaseChange{\protect\cite{cite819}\protect\cite[{pg. 11}]{cite818}}.
The code admits a stabilizer tableau whose rows are cyclic permutations of the Pauli string \(XXZZIZIIIZIZZ\) \NoCaseChange{\protect\cite[{ID 67067199e766ad364e845262}]{cite781}}.}\\ 
\addlinespace[\myxtraspc]
\eczhRefIndex{code:stab_13_1_5}%
\eczhListValue{\flmRefsHyperref{code:stab_13_1_5}{\(\llbracket 13,1,5\rrbracket \) twisted toric code}} & \eczhListValue{Thirteen-qubit twisted toric code whose stabilizer tableau consists of cyclic permutations of the \(XZZX\)-type Pauli string \(XIZZIXIIIIIII\).
The code can be thought of as a small twisted XZZX code \NoCaseChange{\protect\cite[{Exam. 11 and Fig. 3}]{cite438}}.}\\ 
\addlinespace[\myxtraspc]
\eczhRefIndex{code:phantom_14_3_3}%
\eczhListValue{\flmRefsHyperref{code:phantom_14_3_3}{\(\llbracket 14,3,3\rrbracket \) CE phantom code}} & \eczhListValue{CSS phantom code obtained by concatenating the \(\llbracket 7,3,(d_X=3,d_Z=2)\rrbracket \) punctured hypercube code with the two-qubit phase-flip repetition code.
The code is equivalent to the \(\llbracket 14,3,3\rrbracket \) constant-excitation (CE) CSS code obtained by applying dual-rail concatenation to the \(\llbracket 7,3,2\rrbracket \) punctured hypercube code, up to single-qubit Clifford gates, a physical-qubit permutation, and a Pauli frame \NoCaseChange{\protect\cite{cite524}}.}\\ 
\addlinespace[\myxtraspc]
\eczhRefIndex{code:rhombic_dodecahedron_surface}%
\eczhListValue{\flmRefsHyperref{code:rhombic_dodecahedron_surface}{\(\llbracket 14,3,3\rrbracket \) Rhombic dodecahedron surface code}} & \eczhListValue{A \(\llbracket 14,3,3\rrbracket \) twist-defect surface code whose qubits lie on the vertices of a rhombic dodecahedron.
Its non-CSS nature is due to twist defects \NoCaseChange{\protect\cite{cite442}} stemming from the geometry of the polytope.
A local-Clifford-equivalent clean realization has only \(X\)- and \(Z\)-type operators on its four-valent vertices, and its symplectic double is a \(\llbracket 28,6,3\rrbracket \) genus-three code \NoCaseChange{\protect\cite{cite435}}.}\\ 
\addlinespace[\myxtraspc]
\eczhRefIndex{code:stab_15_7_3}%
\eczhListValue{\flmRefsHyperref{code:stab_15_7_3}{\(\llbracket 15, 7, 3\rrbracket \) quantum Hamming code}} & \eczhListValue{Self-dual quantum Hamming code that admits permutation-based CZ logical gates.
The code is constructed using the CSS construction from the \([15,11,3]\) Hamming code and its \([15,4,8]\) dual simplex code.}\\ 
\addlinespace[\myxtraspc]
\eczhRefIndex{code:stab_15_1_3}%
\eczhListValue{\flmRefsHyperref{code:stab_15_1_3}{\(\llbracket 15,1,3\rrbracket \) quantum RM code}} & \eczhListValue{A \(\llbracket 15,1,3\rrbracket \) quantum Reed-Muller code that is most easily thought of as a tetrahedral 3D color code.
It can be constructed as a CSS code from the \([15,5,8]\) punctured Reed-Muller code and its even subcode, which explains its transversal \(T^\dagger\) gate \NoCaseChange{\protect\cite{cite398}}.}\\ 
\addlinespace[\myxtraspc]
\eczhRefIndex{code:quantum_dodecahedron}%
\eczhListValue{\flmRefsHyperref{code:quantum_dodecahedron}{\(\llbracket 16,4,3\rrbracket \) dodecahedral code}} & \eczhListValue{A \(\llbracket 16,4,3\rrbracket \) non-CSS qubit stabilizer code whose \flmRefsHyperref{ref857}{encoder-respecting form} is the graph of vertices of a dodecahedron \NoCaseChange{\protect\cite{cite858}}.}\\ 
\addlinespace[\myxtraspc]
\eczhRefIndex{code:stab_16_6_4}%
\eczhListValue{\flmRefsHyperref{code:stab_16_6_4}{\(\llbracket 16,6,4\rrbracket \) Tesseract color code}} & \eczhListValue{A (hyperbolic self-dual CSS) 4D color code defined on a tesseract, with stabilizer generators of both types supported on each cube. 
A \(\llbracket 16,4,2,4\rrbracket \) tesseract subsystem code can be obtained from this code by using two logical qubits as gauge qubits \NoCaseChange{\protect\cite{cite483}}.}\\ 
\addlinespace[\myxtraspc]
\eczhRefIndex{code:stab_17_1_5}%
\eczhListValue{\flmRefsHyperref{code:stab_17_1_5}{\(\llbracket 17,1,5\rrbracket \) 4.8.8 color code}} & \eczhListValue{Seventeen-qubit doubly even 2D color code that admits a transversal implementation of the logical Clifford group.
The smallest distance-five CSS code has \(n=17\) \NoCaseChange{\protect\cite{cite444}}.
It is also a normal self-dual CSS code whose transversal Hadamard acts logically, making it suitable as an inner code for fifth-order magic-state distillation \NoCaseChange{\protect\cite{cite101}}.}\\ 
\addlinespace[\myxtraspc]
\eczhRefIndex{code:stab_18_2_5}%
\eczhListValue{\flmRefsHyperref{code:stab_18_2_5}{\(\llbracket 18,2,5\rrbracket \) BCC code}} & \eczhListValue{BCC code on 18 qubits encoding 2 logical qubits with distance 5, found by computer search \NoCaseChange{\protect\cite{cite440}}.}\\ 
\addlinespace[\myxtraspc]
\eczhRefIndex{code:ampdamp_stabilizer}%
\eczhListValue{\flmRefsHyperref{code:ampdamp_stabilizer}{\(\llbracket 2(m+1),m,2\rrbracket \) single-loss AD code}} & \eczhListValue{A member of a class of \(\llbracket 2(m+1),m,2\rrbracket \) CSS codes for \(m\geq 1\) that generalizes the \(\llbracket 4,1,2\rrbracket \) approximate amplitude-damping code of Ref. \NoCaseChange{\protect\cite{cite859}}.
Its \(Z\)-type generators are \(m+1\) pairwise products, with each qubit participating in only one check; the single \(X\)-type generator is the all-\(X\) string.}\\ 
\addlinespace[\myxtraspc]
\eczhRefIndex{code:qubit_golay}%
\eczhListValue{\flmRefsHyperref{code:qubit_golay}{\(\llbracket 23, 1, 7\rrbracket \) Quantum Golay code}} & \eczhListValue{A \(\llbracket 23, 1, 7\rrbracket \) self-dual CSS code with eleven stabilizer generators of each type, and with each generator being weight eight.}\\ 
\addlinespace[\myxtraspc]
\eczhRefIndex{code:hypercube_quantum}%
\eczhListValue{\flmRefsHyperref{code:hypercube_quantum}{\(\llbracket 2^D,D,2\rrbracket \) hypercube quantum code}} & \eczhListValue{Member of a family of codes defined by placing qubits on a \(D\)-dimensional hypercube, \(Z\)-stabilizers on all two-dimensional faces, and an \(X\)-stabilizer on all vertices.
These codes realize gates at the \((D-1)\)-st level of the Clifford hierarchy.
The measured physical bit string can be post-processed into both a logical output string and stabilizer checks, enabling end-of-circuit error detection directly from classical samples \NoCaseChange{\protect\cite{cite759}}.
Puncturing the \(\llbracket 2^D,D,2\rrbracket \) hypercube quantum code yields the \(\llbracket 2^D-1,D,2\rrbracket \) punctured-hypercube family.}\\ 
\addlinespace[\myxtraspc]
\eczhRefIndex{code:ring_cpc}%
\eczhListValue{\flmRefsHyperref{code:ring_cpc}{\(\llbracket 2^r+r, 2^r-r-2, 3\rrbracket \) Ring CPC code}} & \eczhListValue{A family of \(\llbracket 2^r+r, 2^r-r-2, 3\rrbracket \) CPC codes for \(r \geq 3\) whose matrices are based on the shortened version of the \([2^r-1,2^r-r-1,3]\) Hamming code.
See \NoCaseChange{\protect\cite[{Thm. 4}]{cite860}} for their stabilizer generator matrix.}\\ 
\addlinespace[\myxtraspc]
\eczhRefIndex{code:morphed_diagonal_clifford}%
\eczhListValue{\flmRefsHyperref{code:morphed_diagonal_clifford}{\(\llbracket 2^r+r-1,1,2\rrbracket \) morphed simplex code}} & \eczhListValue{A member of a family of codes obtained by morphing the \(\llbracket 2^{r+1}-1,1,3\rrbracket \) simplex codes on a region whose child code is a \(\llbracket 2^r,r,2\rrbracket \) hypercube code \NoCaseChange{\protect\cite[{Appx. C}]{cite687}}.
The morphing process replaces a subset of qubits with their logical qubits, yielding a code with parameters \(\llbracket 2^r+r-1,1,2\rrbracket \) that inherits a diagonal gate at the \((r-1)\)st level of the Clifford hierarchy from the parent code.}\\ 
\addlinespace[\myxtraspc]
\eczhRefIndex{code:quantum_hamming}%
\eczhListValue{\flmRefsHyperref{code:quantum_hamming}{\(\llbracket 2^r, 2^r-r-2, 3\rrbracket \) Gottesman code}} & \eczhListValue{A family of \flmRefsHyperref{ref672}{pure} \NoCaseChange{\protect\cite{cite449}} non-CSS stabilizer codes of distance \(3\) that saturate the asymptotic quantum Hamming bound.}\\ 
\addlinespace[\myxtraspc]
\eczhRefIndex{code:quantum_hamming_css}%
\eczhListValue{\flmRefsHyperref{code:quantum_hamming_css}{\(\llbracket 2^r-1, 2^r-2r-1, 3\rrbracket \) quantum Hamming code}} & \eczhListValue{Member of a family of self-dual CSS codes constructed from \([2^r-1,2^r-r-1,3]=C_X=C_Z\) Hamming codes and their duals, the simplex codes.
The code's stabilizer generator matrix blocks \(H_{X}\) and \(H_{Z}\) are both the generator matrix for a simplex code.
The weight of each stabilizer generator is \(2^{r-1}\).}\\ 
\addlinespace[\myxtraspc]
\eczhRefIndex{code:diagonal_clifford}%
\eczhListValue{\flmRefsHyperref{code:diagonal_clifford}{\(\llbracket 2^r-1,1,3\rrbracket \) simplex code}} & \eczhListValue{Member of a color-code family constructed from a punctured first-order RM\((1,m=r)\) code and its even subcode for \(r \geq 3\).
Each code transversally implements a diagonal gate at the \((r-1)\)st level of the \flmTerm{term}{ref694}{}{Clifford hierarchy} \NoCaseChange{\protect\cite{cite799,cite800}}.
Each code is a color code defined on a simplex in \(r-1\) dimensions \NoCaseChange{\protect\cite{cite475,cite832}}, where qubits are placed on the vertices, edges, and faces as well as on the simplex itself.}\\ 
\addlinespace[\myxtraspc]
\eczhRefIndex{code:majorana_hamming}%
\eczhListValue{\flmRefsHyperref{code:majorana_hamming}{\(\llbracket 2^{m-1},2^{m-1}-m-1,4\rrbracket _{f}\) Hamming Majorana code}} & \eczhListValue{A member of the \(\llbracket 2^{m-1},2^{m-1}-m-1,4\rrbracket _{f}\) family of Majorana stabilizer codes for \(m \geq 3\) constructed from a self-orthogonal first-order RM code (whose dual is the extended Hamming code).
A shortened \(\llbracket 2^{m-1}-1,2^{m-1}-m-2,3\rrbracket _{f}\) version can also be defined \NoCaseChange{\protect\cite[{Prop. 2.5.1}]{cite564}}.
The logical subspace of the \(\llbracket 8,3,4\rrbracket _{f}\) Hamming Majorana code is a Cartan subspace of the \(E_8\) Lie algebra \NoCaseChange{\protect\cite{cite565}}.}\\ 
\addlinespace[\myxtraspc]
\eczhRefIndex{code:iceberg}%
\eczhListValue{\flmRefsHyperref{code:iceberg}{\(\llbracket 2m,2m-2,2\rrbracket \) error-detecting code}} & \eczhListValue{Self-complementary and self-dual CSS code for \(m\geq 2\) with generators \(\{XX\cdots X, ZZ\cdots Z\} \) acting on all \(2m\) physical qubits.
The code is constructed via the CSS construction from an SPC code and a repetition code \NoCaseChange{\protect\cite[{Sec. III}]{cite773}}.
This is the highest-rate distance-two code when an even number of qubits is used \NoCaseChange{\protect\cite{cite449}}.}\\ 
\addlinespace[\myxtraspc]
\eczhRefIndex{code:ea_3_1_3-2}%
\eczhListValue{\flmRefsHyperref{code:ea_3_1_3-2}{\(\llbracket 3, 1, 3;2\rrbracket \) EA code}} & \eczhListValue{Distance-three EA stabilizer code encoding one logical qubit and using two ebits.
It is the smallest example of an EA code correcting an arbitrary single-qubit error.}\\ 
\addlinespace[\myxtraspc]
\eczhRefIndex{code:stellated_dodecahedron_css}%
\eczhListValue{\flmRefsHyperref{code:stellated_dodecahedron_css}{\(\llbracket 30,8,3\rrbracket \) Bring code}} & \eczhListValue{A \(\llbracket 30,8,3\rrbracket \) hyperbolic surface code on a quotient of the \(\{5,5\}\) hyperbolic tiling called Bring's curve.
Its qubits and stabilizer generators lie on the vertices of the small stellated dodecahedron. It admits a set of weight-five stabilizer generators.}\\ 
\addlinespace[\myxtraspc]
\eczhRefIndex{code:small_triorthogonal}%
\eczhListValue{\flmRefsHyperref{code:small_triorthogonal}{\(\llbracket 3k + 8, k, 2\rrbracket \) triorthogonal code}} & \eczhListValue{Member of the \(\llbracket 3k + 8, k, 2\rrbracket \) family (for even \(k\)) of triorthogonal and quantum divisible codes that admit a transversal \(T\) gate and are relevant for magic-state distillation \NoCaseChange{\protect\cite{cite691}\protect\cite[{Sec. VI.C}]{cite734}}.}\\ 
\addlinespace[\myxtraspc]
\eczhRefIndex{code:bacon_shor_4}%
\eczhListValue{\flmRefsHyperref{code:bacon_shor_4}{\(\llbracket 4,1,1,2\rrbracket \) Four-qubit subsystem code}} & \eczhListValue{Error-detecting four-qubit subsystem stabilizer code encoding one logical qubit and one gauge qubit.}\\ 
\addlinespace[\myxtraspc]
\eczhRefIndex{code:css_4_1_2}%
\eczhListValue{\flmRefsHyperref{code:css_4_1_2}{\(\llbracket 4,1,2\rrbracket \) Leung-Nielsen-Chuang-Yamamoto (LNCY) code}} & \eczhListValue{A four-qubit CSS stabilizer code that is the only qubit CSS code with such parameters.}\\ 
\addlinespace[\myxtraspc]
\eczhRefIndex{code:stab_4_1_2}%
\eczhListValue{\flmRefsHyperref{code:stab_4_1_2}{\(\llbracket 4,1,2\rrbracket \) twist-defect code}} & \eczhListValue{A four-qubit non-CSS stabilizer code that can be interpreted as the smallest triangular color code with \(x\)-, \(y\)-, and \(z\)-type Pauli boundaries \NoCaseChange{\protect\cite[{Fig. 7}]{cite445}}, and equivalently as a small twist-defect surface code on a tetrahedron inscribed in a sphere \NoCaseChange{\protect\cite{cite435}}.
It is the only non-CSS qubit stabilizer code with parameters \(\llbracket 4,1,2\rrbracket \).
The code admits weight-three stabilizer generators and weight-two logical Pauli \(X,Y,Z\) operators.}\\ 
\addlinespace[\myxtraspc]
\eczhRefIndex{code:stab_4_2_2}%
\eczhListValue{\flmRefsHyperref{code:stab_4_2_2}{\(\llbracket 4,2,2\rrbracket \) Four-qubit code}} & \eczhListValue{A four-qubit hyperbolic self-dual CSS stabilizer code that is the smallest two-logical-qubit stabilizer code to detect a single-qubit error.
It is unique for its parameters \NoCaseChange{\protect\cite[{Thm. 8}]{cite446}}.}\\ 
\addlinespace[\myxtraspc]
\eczhRefIndex{code:stab_49_1_5}%
\eczhListValue{\flmRefsHyperref{code:stab_49_1_5}{\(\llbracket 49,1,5\rrbracket \) triorthogonal code}} & \eczhListValue{Triorthogonal and quantum divisible code which is the smallest distance-five stabilizer code to admit a transversal \(T\) gate \NoCaseChange{\protect\cite{cite659}\protect\cite[{Appx. B}]{cite691}}.
It is one example of a level-three generalized divisible code obtainable from the doubling transformation \NoCaseChange{\protect\cite[{Sec. VI.D}]{cite734}}.
Its \(X\)-type stabilizers form a triply even linear binary code in the \flmRefsHyperref{ref817}{symplectic representation}.}\\ 
\addlinespace[\myxtraspc]
\eczhRefIndex{code:stab_5_1_2}%
\eczhListValue{\flmRefsHyperref{code:stab_5_1_2}{\(\llbracket 5,1,2\rrbracket \) rotated surface code}} & \eczhListValue{A rotated surface code on one rung of a ladder, with one qubit on the rung, and four qubits surrounding it. This is the smallest code that implements a fault-tolerant logical \(S\) gate using a diagonal depth-one Clifford circuit \NoCaseChange{\protect\cite{cite447}}.}\\ 
\addlinespace[\myxtraspc]
\eczhRefIndex{code:stab_5_1_3}%
\eczhListValue{\flmRefsHyperref{code:stab_5_1_3}{\(\llbracket 5,1,3\rrbracket \) Five-qubit perfect code}} & \eczhListValue{Five-qubit cyclic stabilizer code that is the smallest qubit stabilizer code to correct a single-qubit error.}\\ 
\addlinespace[\myxtraspc]
\eczhRefIndex{code:quantum_icosahedron}%
\eczhListValue{\flmRefsHyperref{code:quantum_icosahedron}{\(\llbracket 54,6,5\rrbracket \) five-covered icosahedral code}} & \eczhListValue{A \(\llbracket 54,6,5\rrbracket \) qubit stabilizer code whose \flmRefsHyperref{ref857}{encoder-respecting form} is the graph of a five-cover of the icosahedron \NoCaseChange{\protect\cite{cite858}}.
The covering-space construction avoids the weight-three logical operators that occur for the bare icosahedral graph \NoCaseChange{\protect\cite{cite858}}.}\\ 
\addlinespace[\myxtraspc]
\eczhRefIndex{code:css_6_1_2}%
\eczhListValue{\flmRefsHyperref{code:css_6_1_2}{\(\llbracket 6,1,2\rrbracket \) semi-self-dual CSS code}} & \eczhListValue{A six-qubit CSS stabilizer code that is an example of a semi-self-dual CSS code, i.e., a CSS code whose \(X\)-type stabilizers are contained in the \(Z\)-type stabilizers \NoCaseChange{\protect\cite{cite738}}.}\\ 
\addlinespace[\myxtraspc]
\eczhRefIndex{code:stab_6_1_3}%
\eczhListValue{\flmRefsHyperref{code:stab_6_1_3}{\(\llbracket 6,1,3\rrbracket \) Six-qubit stabilizer code}} & \eczhListValue{A degenerate, non-trivial \(\llbracket 6,1,3\rrbracket \) stabilizer code.
It is one of two six-qubit distance-three codes that are unique up to equivalence \NoCaseChange{\protect\cite{cite449}}, with the other code being decomposable and an extension of the five-qubit code \NoCaseChange{\protect\cite{cite451}\protect\cite[{ID 87}]{cite453}}.
The code admits fault-tolerant syndrome extraction using only one ancilla per stabilizer generator measurement.}\\ 
\addlinespace[\myxtraspc]
\eczhRefIndex{code:majorana_6_1_3}%
\eczhListValue{\flmRefsHyperref{code:majorana_6_1_3}{\(\llbracket 6,1,3\rrbracket _{f}\) Vijay-Fu Majorana code}} & \eczhListValue{A Majorana stabilizer code encoding a logical fermion into six physical fermions.
This code is the shortest code correcting single fermion-parity flips \NoCaseChange{\protect\cite{cite566}}.}\\ 
\addlinespace[\myxtraspc]
\eczhRefIndex{code:stab_6_2_2}%
\eczhListValue{\flmRefsHyperref{code:stab_6_2_2}{\(\llbracket 6,2,2\rrbracket \) \(C_6\) code}} & \eczhListValue{Error-detecting normal self-dual CSS code on three qubit pairs that encodes a logical qubit pair and detects any error acting on one pair \NoCaseChange{\protect\cite{cite448}}.
In Knill's \(C_4/C_6\) architecture, this code is used at the second and higher concatenation levels.}\\ 
\addlinespace[\myxtraspc]
\eczhRefIndex{code:bravyi_bacon_shor_6}%
\eczhListValue{\flmRefsHyperref{code:bravyi_bacon_shor_6}{\(\llbracket 6,2,3,2\rrbracket \) BBS code}} & \eczhListValue{Error-detecting six-qubit BBS subsystem code with parameters \(\llbracket 6,2,3,2\rrbracket \) that can suppress errors in adiabatic quantum computation \NoCaseChange{\protect\cite{cite670}}.}\\ 
\addlinespace[\myxtraspc]
\eczhRefIndex{code:stab_6_4_2}%
\eczhListValue{\flmRefsHyperref{code:stab_6_4_2}{\(\llbracket 6,4,2\rrbracket \) error-detecting code}} & \eczhListValue{Self-complementary six-qubit code with rate \(2/3\) that is unique for its parameters, up to equivalence \NoCaseChange{\protect\cite[{Tab. III}]{cite449}}.
Concatenations of this code with itself yield the \(\llbracket 6^r,4^r,2^r\rrbracket \) level-\(r\) \textit{many-hypercube} code \NoCaseChange{\protect\cite{cite450}}.}\\ 
\addlinespace[\myxtraspc]
\eczhRefIndex{code:campbell_howard}%
\eczhListValue{\flmRefsHyperref{code:campbell_howard}{\(\llbracket 6k+2,3k,2\rrbracket \) Campbell-Howard code}} & \eczhListValue{Family of \(\llbracket 6k+2,3k,2\rrbracket \) qubit stabilizer codes with quasi-transversal \(CCZ^{\otimes k}\) gates that are relevant to magic-state distillation.
In the synthillation framework, these distance-two codes realize batches of logical \(CCZ\) gates using physical \(T\) gates followed by a Clifford correction.}\\ 
\addlinespace[\myxtraspc]
\eczhRefIndex{code:goy}%
\eczhListValue{\flmRefsHyperref{code:goy}{\(\llbracket 6r,2r,2\rrbracket \) Ganti-Onunkwo-Young code}} & \eczhListValue{A member of the family of \(\llbracket 6r,2r,2\rrbracket \) CSS codes designed to suppress errors in adiabatic quantum computation.
All but two of its stabilizer generators are weight-two (two-body), and the remaining two are weight-\(4r\).}\\ 
\addlinespace[\myxtraspc]
\eczhRefIndex{code:hybrid_7_1-1_3}%
\eczhListValue{\flmRefsHyperref{code:hybrid_7_1-1_3}{\(\llbracket 7, 1:1, 3\rrbracket \) hybrid stabilizer code}} & \eczhListValue{A distance-three seven-qubit hybrid stabilizer code storing one qubit and one classical bit.
Admits a stabilizer generator set with a weight-two generator, which delineates the underlying classical code \NoCaseChange{\protect\cite[{Eq. (3)}]{cite671}}.}\\ 
\addlinespace[\myxtraspc]
\eczhRefIndex{code:steane}%
\eczhListValue{\flmRefsHyperref{code:steane}{\(\llbracket 7,1,3\rrbracket \) Steane code}} & \eczhListValue{A \(\llbracket 7,1,3\rrbracket \) self-dual CSS code that is the smallest qubit CSS code to correct a single-qubit error \NoCaseChange{\protect\cite{cite451}}.
The code is constructed using the classical binary \([7,4,3]\) Hamming code for protecting against both \(X\) and \(Z\) errors.}\\ 
\addlinespace[\myxtraspc]
\eczhRefIndex{code:xzzx_7_1_3}%
\eczhListValue{\flmRefsHyperref{code:xzzx_7_1_3}{\(\llbracket 7,1,3\rrbracket \) XZZX cyclic code}} & \eczhListValue{A \(\llbracket 7,1,3\rrbracket \) cyclic non-CSS code whose stabilizer generators are the seven cyclic shifts of the weight-four Pauli string \(XZIZXII\),
any six of which are independent.
The non-identity support of this generator is \(XZZX\) (at positions 0, 1, 3, 4), making this a member of the twisted XZZX code family.
It is one of sixteen distinct indecomposable \(\llbracket 7,1,3\rrbracket \) codes \NoCaseChange{\protect\cite{cite452}}.}\\ 
\addlinespace[\myxtraspc]
\eczhRefIndex{code:bare_7_1_3}%
\eczhListValue{\flmRefsHyperref{code:bare_7_1_3}{\(\llbracket 7,1,3\rrbracket \) bare code}} & \eczhListValue{A \(\llbracket 7,1,3\rrbracket \) code that admits fault-tolerant syndrome extraction using only one ancilla per stabilizer generator measurement.}\\ 
\addlinespace[\myxtraspc]
\eczhRefIndex{code:twist_defect_7_1_3}%
\eczhListValue{\flmRefsHyperref{code:twist_defect_7_1_3}{\(\llbracket 7,1,3\rrbracket \) twist-defect surface code}} & \eczhListValue{A \(\llbracket 7,1,3\rrbracket \) code (different from the Steane code) that is a small example of a twist-defect surface code.}\\ 
\addlinespace[\myxtraspc]
\eczhRefIndex{code:hgp_7_2_2}%
\eczhListValue{\flmRefsHyperref{code:hgp_7_2_2}{\(\llbracket 7,2,2\rrbracket \) HGP phantom code}} & \eczhListValue{Smallest member of a family of CSS phantom HGP codes obtained from the hypergraph product of a classical simplex code and a repetition code.}\\ 
\addlinespace[\myxtraspc]
\eczhRefIndex{code:qetc_7_2}%
\eczhListValue{\flmRefsHyperref{code:qetc_7_2}{\(\llbracket 7,2,2\rrbracket \) QETC}} & \eczhListValue{Seven-qubit \flmRefsHyperref{ref672}{pure} QETC that transmutes all single-qubit Pauli errors to logical phase errors.}\\ 
\addlinespace[\myxtraspc]
\eczhRefIndex{code:xz_7_3_2}%
\eczhListValue{\flmRefsHyperref{code:xz_7_3_2}{\(\llbracket 7,3,2\rrbracket \) punctured hypercube code}} & \eczhListValue{A seven-qubit \flmRefsHyperref{ref672}{pure} CSS code that corrects a single \(X\) or a single \(Z\) error, but not a single \(Y\) error.
It is the punctured version of the \(\llbracket 8,3,2\rrbracket \) hypercube quantum code and is a phantom code \NoCaseChange{\protect\cite{cite514}}.}\\ 
\addlinespace[\myxtraspc]
\eczhRefIndex{code:hybrid_8_2-1_3}%
\eczhListValue{\flmRefsHyperref{code:hybrid_8_2-1_3}{\(\llbracket 8, 2:1, 3\rrbracket \) hybrid stabilizer code}} & \eczhListValue{A code obtained from the \(\llbracket 8,3,3\rrbracket \) Gottesman code by using one of its logical qubits as a classical bit.
One can also use two logical qubits as classical bits, obtaining an \(\llbracket 8,1:2,3\rrbracket \) hybrid stabilizer code.}\\ 
\addlinespace[\myxtraspc]
\eczhRefIndex{code:stab_8_3_3}%
\eczhListValue{\flmRefsHyperref{code:stab_8_3_3}{\(\llbracket 8, 3, 3\rrbracket \) Eight-qubit Gottesman code}} & \eczhListValue{Eight-qubit \flmRefsHyperref{ref811}{non-degenerate} code that can be obtained from a modified CSS construction using the \([8,4,4]\) extended Hamming code and a \([8,7,2]\) even-weight code \NoCaseChange{\protect\cite{cite861}}.
The modification introduces signs between the codewords.}\\ 
\addlinespace[\myxtraspc]
\eczhRefIndex{code:stab_8_1_2}%
\eczhListValue{\flmRefsHyperref{code:stab_8_1_2}{\(\llbracket 8,1,2\rrbracket \) Shen-Wang-Cao code}} & \eczhListValue{A stabilizer code that admits a logical \(T\) gate via application of physical \(T\) gates and a \(CZ\)-like gate.}\\ 
\addlinespace[\myxtraspc]
\eczhRefIndex{code:stab_8_2_2}%
\eczhListValue{\flmRefsHyperref{code:stab_8_2_2}{\(\llbracket 8,2,2\rrbracket \) hyperbolic color code}} & \eczhListValue{An \(\llbracket 8,2,2\rrbracket \) hyperbolic color code defined on the projective plane. It is a self-dual CSS code.}\\ 
\addlinespace[\myxtraspc]
\eczhRefIndex{code:stab_8_2_3}%
\eczhListValue{\flmRefsHyperref{code:stab_8_2_3}{\(\llbracket 8,2,3\rrbracket \) Hermitian code}} & \eczhListValue{A non-CSS Hermitian eight-qubit stabilizer code that is the only Hermitian code and that has the largest automorphism group among the 20 inequivalent \(\llbracket 8,2,3\rrbracket \) codes \NoCaseChange{\protect\cite{cite454}}.
The code has an exceptionally short fault-tolerant syndrome measurement sequence \NoCaseChange{\protect\cite{cite862,cite454}}.}\\ 
\addlinespace[\myxtraspc]
\eczhRefIndex{code:stab_8_3_2}%
\eczhListValue{\flmRefsHyperref{code:stab_8_3_2}{\(\llbracket 8,3,2\rrbracket \) Smallest interesting color code}} & \eczhListValue{Smallest 3D color code whose physical qubits lie on vertices of a cube and which admits a (weakly) transversal \(CCZ\) gate.}\\ 
\addlinespace[\myxtraspc]
\eczhRefIndex{code:cubic_surface}%
\eczhListValue{\flmRefsHyperref{code:cubic_surface}{\(\llbracket 8,3,2\rrbracket \) Surface code on a cube}} & \eczhListValue{An \(\llbracket 8,3,2\rrbracket \) twist-defect surface code whose qubits lie on the vertices of a cube.
It is obtained by three-coloring the faces of a cube and placing \(X\), \(Y\), and \(Z\) stabilizer generators on each pair of faces of the same color.
Its non-CSS nature is due to twist defects \NoCaseChange{\protect\cite{cite442}} stemming from the geometry of the polytope.}\\ 
\addlinespace[\myxtraspc]
\eczhRefIndex{code:shor_nine}%
\eczhListValue{\flmRefsHyperref{code:shor_nine}{\(\llbracket 9,1,3\rrbracket \) Shor code}} & \eczhListValue{Nine-qubit \flmRefsHyperref{code:css}{CSS code} that is the first quantum error-correcting code \NoCaseChange{\protect\cite[{ID 8802}]{cite453}}.
Among indecomposable \(\llbracket 9,1,3\rrbracket \) CSS codes, the Shor code has the largest automorphism group \NoCaseChange{\protect\cite{cite454}}.}\\ 
\addlinespace[\myxtraspc]
\eczhRefIndex{code:surface-17}%
\eczhListValue{\flmRefsHyperref{code:surface-17}{\(\llbracket 9,1,3\rrbracket \) Surface-17 code}} & \eczhListValue{A \(\llbracket 9,1,3\rrbracket \) rotated surface code named for the sum of its 9 data qubits and 8 syndrome qubits.
It is one of the four inequivalent CSS gauge fixings of the nine-qubit Bacon-Shor code \NoCaseChange{\protect\cite{cite454}}.
It uses the smallest number of qubits to perform fault-tolerant error correction on a surface code with parallel syndrome extraction.}\\ 
\addlinespace[\myxtraspc]
\eczhRefIndex{code:bacon_shor_9}%
\eczhListValue{\flmRefsHyperref{code:bacon_shor_9}{\(\llbracket 9,1,4,3\rrbracket \) Nine-qubit Bacon-Shor code}} & \eczhListValue{Error-correcting nine-qubit subsystem stabilizer code encoding one logical qubit and four gauge qubits.
There are exactly four inequivalent CSS gauge fixings of the code, including the \flmRefsHyperref{code:shor_nine}{Shor code} and the \flmRefsHyperref{code:surface-17}{surface-17 code} \NoCaseChange{\protect\cite{cite454}}.}\\ 
\addlinespace[\myxtraspc]
\eczhRefIndex{code:stab_9_3_3}%
\eczhListValue{\flmRefsHyperref{code:stab_9_3_3}{\(\llbracket 9,3,3\rrbracket \) Quadric code}} & \eczhListValue{Nine-qubit \flmRefsHyperref{ref672}{pure} Hermitian qubit code constructed from the almost MDS \([9,3,6]_4\) Hermitian self-orthogonal code.
It is the only \flmRefsHyperref{ref672}{pure} Hermitian code with its parameters and is the highest-distance qubit stabilizer code for its \(n\) and \(k\).}\\ 
\addlinespace[\myxtraspc]
\eczhRefIndex{code:quantum_h}%
\eczhListValue{\flmRefsHyperref{code:quantum_h}{\(\llbracket k+4,k,2\rrbracket \) H code}} & \eczhListValue{Family of \(\llbracket k+4,k,2\rrbracket \) self-dual CSS codes (for even \(k\)) with transversal Hadamard gates that are relevant to magic state distillation.
The four stabilizer generators are \(X_1X_2X_3X_4\), \(Z_1Z_2Z_3Z_4\), \(X_1X_2X_5X_6...X_{k+4}\), and \(Z_1Z_2Z_5Z_6...Z_{k+4}\).}\\ 
\addlinespace[\myxtraspc]
\eczhRefIndex{code:quantum_cap}%
\eczhListValue{\flmRefsHyperref{code:quantum_cap}{\(\llbracket n,n-2k,4\rrbracket \) Quantum cap code}} & \eczhListValue{A distance-four \flmRefsHyperref{ref672}{pure} Hermitian qubit code constructed from a Hermitian self-orthogonal \([n,k]_4\) code associated with an \(n\)-cap in \(PG(k-1,4)\).}\\ 
\end{tabularx}\endgroup
\eczcodelist{stabilizer}{Stabilizer codes (non-CSS, non-qubit)
}%

\eczhCodeListAutoDescription{Codes that are descendants of \flmRefsCref{code:stabilizer} and not descendants of any of \flmRefsCref{code:css}, \flmRefsCref{code:qubit_stabilizer}.}%

\eczhIncludeCodeGraph{Bare}{scale=0.5}{\columnwidth}{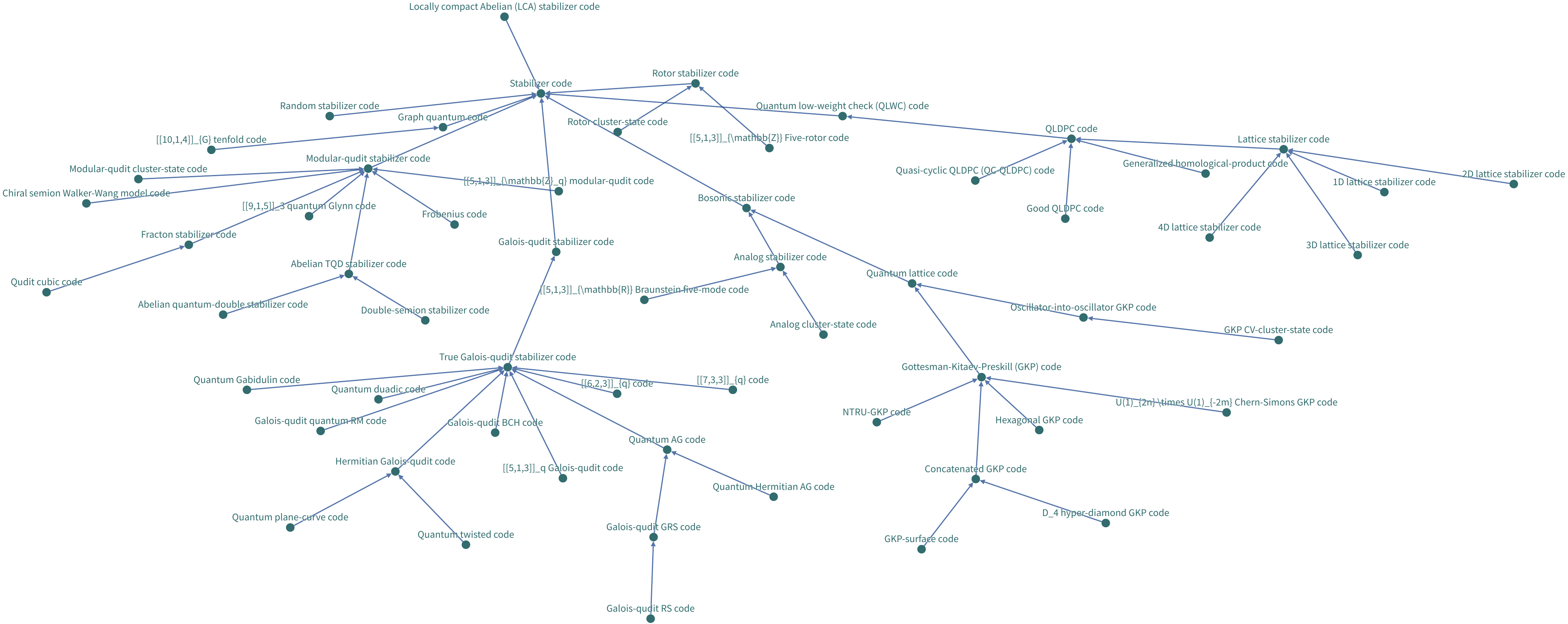}{Stabilizer codes (non-CSS, non-qubit)}{https://errorcorrectionzoo.org/code_graph#J\%7B\%22displayMode\%22\%3A\%22subset\%22\%2C\%22modeSubsetOptions\%22\%3A\%7B\%22codeIds\%22\%3A\%5B\%221d_stabilizer\%22\%2C\%222d_stabilizer\%22\%2C\%223d_stabilizer\%22\%2C\%224d_stabilizer\%22\%2C\%22tqd_abelian_stabilizer\%22\%2C\%22quantum_double_abelian\%22\%2C\%22cv_cluster_state\%22\%2C\%22analog_stabilizer\%22\%2C\%22oscillator_stabilizer\%22\%2C\%223d_semion\%22\%2C\%22gkp_concatenated\%22\%2C\%22double_semion\%22\%2C\%22fracton\%22\%2C\%22frobenius\%22\%2C\%22gkp-cluster-state\%22\%2C\%22gkp_surface_concatenated\%22\%2C\%22galois_bch\%22\%2C\%22galois_grs\%22\%2C\%22galois_polynomial\%22\%2C\%22galois_reed_muller\%22\%2C\%22galois_stabilizer\%22\%2C\%22generalized_homological_product\%22\%2C\%22good_qldpc\%22\%2C\%22multimodegkp\%22\%2C\%22graph_quantum\%22\%2C\%22stabilizer_over_gfqsq\%22\%2C\%22hexagonal_gkp\%22\%2C\%22translationally_invariant_stabilizer\%22\%2C\%22lca_stabilizer\%22\%2C\%22qudit_cluster_state\%22\%2C\%22qudit_stabilizer\%22\%2C\%22ntru_gkp\%22\%2C\%22gkp-stabilizer\%22\%2C\%22general_qldpc\%22\%2C\%22quantum_ag\%22\%2C\%22quantum_gabidulin\%22\%2C\%22quantum_hermitian_ag\%22\%2C\%22galois_duadic\%22\%2C\%22quantum_lattice\%22\%2C\%22qlwc\%22\%2C\%22quantum_plane_curve\%22\%2C\%22quantum_twisted\%22\%2C\%22quasi_cyclic_qldpc\%22\%2C\%22qudit_cubic\%22\%2C\%22random_stabilizer\%22\%2C\%22rotor_cluster\%22\%2C\%22rotor_stabilizer\%22\%2C\%22stabilizer\%22\%2C\%22galois_true_stabilizer\%22\%2C\%22dfour_gkp\%22\%2C\%22chern_simons_gkp\%22\%2C\%22group_10_1_4\%22\%2C\%22galois_5_1_3\%22\%2C\%22braunstein\%22\%2C\%22qudit_5_1_3\%22\%2C\%22rotor_5_1_3\%22\%2C\%22galois_6_2_3\%22\%2C\%22galois_7_3_3\%22\%2C\%22stab_9_1_5\%22\%5D\%2C\%22reusePreviousLayoutPositions\%22\%3Afalse\%2C\%22showIntermediateConnectingNodes\%22\%3Atrue\%2C\%22connectingNodesMaxDepth\%22\%3A15\%2C\%22connectingNodesPathMaxLength\%22\%3A20\%2C\%22connectingNodesMaxExtraDepth\%22\%3A3\%2C\%22connectingNodesOnlyKeepPathsWithAdditionalLength\%22\%3A1\%2C\%22connectingNodesToDomainsAndKingdoms\%22\%3Afalse\%2C\%22connectingNodesEdgeLengthsByType\%22\%3A\%7B\%22primaryParent\%22\%3A1\%2C\%22secondaryParent\%22\%3A4\%2C\%22cousin\%22\%3A6\%7D\%2C\%22nodeIds\%22\%3A\%5B\%5D\%7D\%2C\%22highlightImportantNodes\%22\%3A\%7B\%22highlightImportantNodes\%22\%3Afalse\%2C\%22highlightPrimaryParents\%22\%3Afalse\%2C\%22highlightRootConnectingEdges\%22\%3Afalse\%7D\%7D}

\begingroup
\small
\eczhBreakableDashes
\renewcommand\arraystretch{1.05}
\edef\myxtraspc{\eczListAddVSpaceXtraXtraStretch}
\begin{tabularx}{\linewidth}{>{\raggedright\arraybackslash}p{\eczListColWidth{name}} >{\hsize=1.0000\hsize }X}
\toprule
\eczListColTitle{Code} & \eczListColTitle{Description} \\
\midrule
\endfirsthead
\toprule
\eczListColTitleContinued{Code} & \eczListColTitleContinued{Description} \\
\midrule
\endhead
\bottomrule
\endfoot
\eczhRefIndex{code:1d_stabilizer}%
\eczhListValue{\flmRefsHyperref{code:1d_stabilizer}{1D lattice stabilizer code}} & \eczhListValue{Lattice stabilizer code in one Euclidean dimension, using either the ordinary block notion of locality or the fermionic/Majorana notion of locality.}\\ 
\addlinespace[\myxtraspc]
\eczhRefIndex{code:2d_stabilizer}%
\eczhListValue{\flmRefsHyperref{code:2d_stabilizer}{2D lattice stabilizer code}} & \eczhListValue{Lattice stabilizer code in two Euclidean dimensions, using either the ordinary block notion of locality or the fermionic/Majorana notion of locality.}\\ 
\addlinespace[\myxtraspc]
\eczhRefIndex{code:3d_stabilizer}%
\eczhListValue{\flmRefsHyperref{code:3d_stabilizer}{3D lattice stabilizer code}} & \eczhListValue{Lattice stabilizer code in three Euclidean dimensions, using either the ordinary block notion of locality or the fermionic/Majorana notion of locality.}\\ 
\addlinespace[\myxtraspc]
\eczhRefIndex{code:4d_stabilizer}%
\eczhListValue{\flmRefsHyperref{code:4d_stabilizer}{4D lattice stabilizer code}} & \eczhListValue{Lattice stabilizer code in four Euclidean dimensions, using either the ordinary block notion of locality or the fermionic/Majorana notion of locality.}\\ 
\addlinespace[\myxtraspc]
\eczhRefIndex{code:tqd_abelian_stabilizer}%
\eczhListValue{\flmRefsHyperref{code:tqd_abelian_stabilizer}{Abelian TQD stabilizer code}} & \eczhListValue{Modular-qudit stabilizer code whose codewords realize a 2D Abelian twisted-quantum-double topological order on composite-dimensional qudits.
For every finite Abelian group \(G=\prod_i \mathbb{Z}_{N_i}\) and every product of Type-I and Type-II cocycles, there is a Pauli stabilizer Hamiltonian realizing the corresponding Abelian TQD \NoCaseChange{\protect\cite{cite405}}.
Equivalently, these codes exhaust the 2D Abelian topological orders that admit gapped boundaries \NoCaseChange{\protect\cite{cite405,cite406}}.}\\ 
\addlinespace[\myxtraspc]
\eczhRefIndex{code:quantum_double_abelian}%
\eczhListValue{\flmRefsHyperref{code:quantum_double_abelian}{Abelian quantum-double stabilizer code}} & \eczhListValue{Modular-qudit stabilizer code whose codewords realize 2D modular gapped Abelian topological order with trivial cocycle.
The corresponding anyon theory is defined by an Abelian group.
The \(G=\mathbb{Z}_2\) instance on a torus is the toric code, and cyclic-group instances reduce to modular-qudit surface codes.
All such codes can be realized by a stack of modular-qudit surface codes because all finite Abelian groups are direct products of cyclic groups.}\\ 
\addlinespace[\myxtraspc]
\eczhRefIndex{code:cv_cluster_state}%
\eczhListValue{\flmRefsHyperref{code:cv_cluster_state}{Analog cluster-state code}} & \eczhListValue{A code based on a continuous-variable (CV), or analog, cluster state.
Such a state can be used to perform MBQC of logical modes, which substitutes the temporal dimension necessary for decoding a conventional code with a spatial dimension.
The exact analog cluster state is non-normalizable, so approximate constructions have to be considered.}\\ 
\addlinespace[\myxtraspc]
\eczhRefIndex{code:analog_stabilizer}%
\eczhListValue{\flmRefsHyperref{code:analog_stabilizer}{Analog stabilizer code}} & \eczhListValue{An oscillator-into-oscillator stabilizer code encoding logical oscillator modes into \(n\) physical modes. If the code is defined by \(r\) independent nullifiers, then it is denoted by \(\llbracket n,n-r\rrbracket _{\mathbb{R}}\) and encodes \(k=n-r\) logical modes \NoCaseChange{\protect\cite{cite507}}.
Any analog stabilizer state can be thought of as a pure Gaussian state that has been infinitely squeezed on all modes \NoCaseChange{\protect\cite{cite507}}.}\\ 
\addlinespace[\myxtraspc]
\eczhRefIndex{code:oscillator_stabilizer}%
\eczhListValue{\flmRefsHyperref{code:oscillator_stabilizer}{Bosonic stabilizer code}} & \eczhListValue{Bosonic code whose codespace is defined as the common \(+1\) eigenspace of a group of mutually commuting displacement operators.
Displacements form the stabilizers of the code, and have continuous eigenvalues, in contrast with the discrete set of eigenvalues of qubit stabilizers.
As a result, exact codewords are non-normalizable, so approximate constructions have to be considered.
Stabilizer groups are any locally compact Abelian subgroups of \(\mathbb{R}^n\), can themselves contain discrete or continuous subgroups, and can admit logical qudit and/or oscillator logical subspaces.}\\ 
\addlinespace[\myxtraspc]
\eczhRefIndex{code:3d_semion}%
\eczhListValue{\flmRefsHyperref{code:3d_semion}{Chiral semion Walker-Wang model code}} & \eczhListValue{A 3D lattice modular-qudit stabilizer code with qudit dimension \(q=4\) whose low-energy excitations on boundaries realize the chiral semion topological order.
The model admits 2D chiral semion topological order at one of its surfaces \NoCaseChange{\protect\cite{cite471,cite472}}.
The corresponding phase can also be realized via a non-stabilizer Hamiltonian \NoCaseChange{\protect\cite{cite473}}.}\\ 
\addlinespace[\myxtraspc]
\eczhRefIndex{code:gkp_concatenated}%
\eczhListValue{\flmRefsHyperref{code:gkp_concatenated}{Concatenated GKP code}} & \eczhListValue{A concatenated code whose outer code is a GKP code. In other words, a bosonic code that can be thought of as a concatenation of an arbitrary inner code and another bosonic outer code. Most examples encode physical qubits of an inner stabilizer code into the square-lattice GKP code.}\\ 
\addlinespace[\myxtraspc]
\eczhRefIndex{code:double_semion}%
\eczhListValue{\flmRefsHyperref{code:double_semion}{Double-semion stabilizer code}} & \eczhListValue{A 2D lattice modular-qudit stabilizer code with qudit dimension \(q=4\) that realizes the 2D double semion topological phase.
The code can be obtained from a \(\mathbb{Z}_4\) toric-code ground state by \flmRefsHyperref{ref410}{condensing} the emergent boson \(e^2 m^2\); in the stabilizer construction this condensation is implemented by two-body measurements \NoCaseChange{\protect\cite{cite405,cite414}}.
Its ground-state subspace can be mapped to that of the double-semion string-net model by a finite-depth quantum circuit with ancillas \NoCaseChange{\protect\cite{cite405}}.}\\ 
\addlinespace[\myxtraspc]
\eczhRefIndex{code:fracton}%
\eczhListValue{\flmRefsHyperref{code:fracton}{Fracton stabilizer code}} & \eczhListValue{A 3D modular-qudit stabilizer code whose codewords make up the ground-state space of a Hamiltonian in a fracton phase.
Unlike topological phases, whose excitations can move in any direction, fracton phases are characterized by excitations whose movement is restricted.}\\ 
\addlinespace[\myxtraspc]
\eczhRefIndex{code:frobenius}%
\eczhListValue{\flmRefsHyperref{code:frobenius}{Frobenius code}} & \eczhListValue{A cyclic prime-qudit stabilizer code whose length \(n\) divides \(p^t + 1\) for some positive integer \(t\).}\\ 
\addlinespace[\myxtraspc]
\eczhRefIndex{code:gkp-cluster-state}%
\eczhListValue{\flmRefsHyperref{code:gkp-cluster-state}{GKP CV-cluster-state code}} & \eczhListValue{A cluster-state code that utilizes a generalized analog cluster state with some of its physical modes initialized in GKP (resource) states.
Alternatively, it can be thought of as a multimode GKP code whose encoding consists of initializing \(k\) modes in momentum states (or, in the normalizable case, squeezed vacua), \(n-k\) modes in (normalizable) GKP states, and applying a Gaussian circuit consisting of two-body gates \(e^{i V_{jk} \hat{x}_j \hat{x}_k }\) for some angles \(V_{jk}\).
The code provides a way to perform fault-tolerant MBQC, with the required number \(n-k\) of GKP-encoded physical modes determined by the particular protocol \NoCaseChange{\protect\cite{cite508,cite509,cite415,cite510}}.}\\ 
\addlinespace[\myxtraspc]
\eczhRefIndex{code:gkp_surface_concatenated}%
\eczhListValue{\flmRefsHyperref{code:gkp_surface_concatenated}{GKP-surface code}} & \eczhListValue{A concatenated code whose outer code is a GKP code and whose inner code is a surface code, including toric surface-code variants \NoCaseChange{\protect\cite{cite415,cite416}}, rotated surface codes \NoCaseChange{\protect\cite{cite417,cite418,cite419,cite420}}, and XZZX surface codes \NoCaseChange{\protect\cite{cite421}}.}\\ 
\addlinespace[\myxtraspc]
\eczhRefIndex{code:galois_bch}%
\eczhListValue{\flmRefsHyperref{code:galois_bch}{Galois-qudit BCH code}} & \eczhListValue{True Galois-qudit stabilizer code constructed from BCH codes via either the Hermitian construction or the Galois-qudit CSS construction.
Parameters can be improved by applying \flmRefsCref{ref863} \NoCaseChange{\protect\cite{cite864}}, e.g., as in Ref. \NoCaseChange{\protect\cite{cite865}}.}\\ 
\addlinespace[\myxtraspc]
\eczhRefIndex{code:galois_grs}%
\eczhListValue{\flmRefsHyperref{code:galois_grs}{Galois-qudit GRS code}} & \eczhListValue{A true \(q\)-Galois-qudit stabilizer code constructed from GRS codes via either the Hermitian construction \NoCaseChange{\protect\cite{cite823,cite824,cite825}} or the Galois-qudit CSS construction \NoCaseChange{\protect\cite{cite826,cite827}}.}\\ 
\addlinespace[\myxtraspc]
\eczhRefIndex{code:galois_polynomial}%
\eczhListValue{\flmRefsHyperref{code:galois_polynomial}{Galois-qudit RS code}} & \eczhListValue{A Galois-qudit CSS code family (with \(q>n\)) constructed using two RS codes over \(\mathbb{F}_q\).}\\ 
\addlinespace[\myxtraspc]
\eczhRefIndex{code:galois_reed_muller}%
\eczhListValue{\flmRefsHyperref{code:galois_reed_muller}{Galois-qudit quantum RM code}} & \eczhListValue{True Galois-qudit stabilizer code constructed from generalized Reed-Muller (GRM) codes via the Galois-qudit Hermitian construction, the Galois-qudit CSS construction, or directly from their parity-check matrices \NoCaseChange{\protect\cite{cite828}\protect\cite[{Sec. 4.2}]{cite829}}.}\\ 
\addlinespace[\myxtraspc]
\eczhRefIndex{code:galois_stabilizer}%
\eczhListValue{\flmRefsHyperref{code:galois_stabilizer}{Galois-qudit stabilizer code}} & \eczhListValue{An \(\llparenthesis n,K,d\rrparenthesis _q\) Galois-qudit code whose logical subspace is the joint eigenspace of commuting Galois-qudit Pauli operators forming the code's stabilizer group \(\mathsf{S}\). Traditionally, the logical subspace is the joint \(+1\) eigenspace, and the stabilizer group does not contain \(e^{i \phi} I\) for any \(\phi \neq 0\). The distance \(d\) is the minimum weight of a Galois-qudit Pauli string that implements a nontrivial logical operation in the code.}\\ 
\addlinespace[\myxtraspc]
\eczhRefIndex{code:generalized_homological_product}%
\eczhListValue{\flmRefsHyperref{code:generalized_homological_product}{Generalized homological-product code}} & \eczhListValue{Stabilizer code whose properties are determined from an underlying chain complex, which often consists of some type of product of other chain complexes.
The \flmRefsCref{ref683} yields an interpretation of codes in terms of chain complexes, thus allowing for the use of various products from homology in constructing codes.}\\ 
\addlinespace[\myxtraspc]
\eczhRefIndex{code:good_qldpc}%
\eczhListValue{\flmRefsHyperref{code:good_qldpc}{Good QLDPC code}} & \eczhListValue{Also called \textit{asymptotically good QLDPC codes}. A family of QLDPC codes \(\llbracket n_i,k_i,d_i\rrbracket \) whose asymptotic rate \(\lim_{i\to\infty} k_i/n_i\) and asymptotic distance \(\lim_{i\to\infty} d_i/n_i\) are both positive.}\\ 
\addlinespace[\myxtraspc]
\eczhRefIndex{code:multimodegkp}%
\eczhListValue{\flmRefsHyperref{code:multimodegkp}{Gottesman-Kitaev-Preskill (GKP) code}} & \eczhListValue{Quantum lattice code for a non-degenerate lattice, thereby admitting a finite-dimensional logical subspace.
Codes on \(n\) modes can be constructed from lattices with \(2n\)-dimensional full-rank Gram matrices \(A\).
Any GKP code can be generated from a Gram matrix in standard form via a Gaussian unitary transformation \NoCaseChange{\protect\cite[{Corr. 1}]{cite511}}.}\\ 
\addlinespace[\myxtraspc]
\eczhRefIndex{code:graph_quantum}%
\eczhListValue{\flmRefsHyperref{code:graph_quantum}{Graph quantum code}} & \eczhListValue{A stabilizer code on tensor products of \(G\)-valued qudits for Abelian \(G\) whose encoding isometry is defined using a graph \NoCaseChange{\protect\cite[{Eqs. (4-5)}]{cite866}}.
An analytical form of the codewords exists in terms of the adjacency matrix of the graph and bicharacters of the Abelian group \NoCaseChange{\protect\cite{cite866}}; see \NoCaseChange{\protect\cite[{Eq. (1)}]{cite867}}.
A graph quantum code for \(G=\mathbb{Z}_2\) contains a cluster state as one of its codewords and reduces to a cluster state when its logical dimension is one \NoCaseChange{\protect\cite{cite868}}.}\\ 
\addlinespace[\myxtraspc]
\eczhRefIndex{code:stabilizer_over_gfqsq}%
\eczhListValue{\flmRefsHyperref{code:stabilizer_over_gfqsq}{Hermitian Galois-qudit code}} & \eczhListValue{An \(\llbracket n,k,d\rrbracket _q\) true Galois-qudit stabilizer code constructed from a Hermitian self-orthogonal linear code over \(\mathbb{F}_{q^2}\) using the one-to-one correspondence between the Galois-qudit Pauli matrices and elements of the Galois field \(\mathbb{F}_{q^2}\).}\\ 
\addlinespace[\myxtraspc]
\eczhRefIndex{code:hexagonal_gkp}%
\eczhListValue{\flmRefsHyperref{code:hexagonal_gkp}{Hexagonal GKP code}} & \eczhListValue{Single-mode GKP qudit-into-oscillator code based on the triangular lattice. Offers the best error correction against displacement noise in a single mode due to the optimal packing of the underlying lattice \NoCaseChange{\protect\cite[{Sec. VI}]{cite513}}.}\\ 
\addlinespace[\myxtraspc]
\eczhRefIndex{code:translationally_invariant_stabilizer}%
\eczhListValue{\flmRefsHyperref{code:translationally_invariant_stabilizer}{Lattice stabilizer code}} & \eczhListValue{A geometrically local stabilizer code with sites organized on a lattice modeled by the additive group \(\mathbb{Z}^D\) for spatial dimension \(D\), using either the ordinary block notion of locality or the fermionic/Majorana notion of locality.
On an infinite lattice, its stabilizer group is generated by few-site Pauli-type operators and their translations, in which case the code is called \textit{translationally invariant stabilizer code}.
Boundary conditions have to be imposed on the lattice in order to obtain finite-dimensional versions.
Lattice defects and boundaries between different codes can also be introduced.}\\ 
\addlinespace[\myxtraspc]
\eczhRefIndex{code:lca_stabilizer}%
\eczhListValue{\flmRefsHyperref{code:lca_stabilizer}{Locally compact Abelian (LCA) stabilizer code}} & \eczhListValue{A mixed oscillator stabilizer code whose codewords are quantum lattice states defined on any number of qudits and a nonzero number of oscillators.
Its stabilizers are countably infinite subgroups of the qudit Pauli and oscillator displacement groups.
Codewords are entangled across the qudit-oscillator bipartition.}\\ 
\addlinespace[\myxtraspc]
\eczhRefIndex{code:qudit_cluster_state}%
\eczhListValue{\flmRefsHyperref{code:qudit_cluster_state}{Modular-qudit cluster-state code}} & \eczhListValue{A code based on a modular-qudit cluster state.}\\ 
\addlinespace[\myxtraspc]
\eczhRefIndex{code:qudit_stabilizer}%
\eczhListValue{\flmRefsHyperref{code:qudit_stabilizer}{Modular-qudit stabilizer code}} & \eczhListValue{An \(\llparenthesis n,K,d\rrparenthesis _q\) modular-qudit code whose logical subspace is the joint eigenspace of commuting qudit Pauli operators forming the code's stabilizer group \(\mathsf{S}\) \NoCaseChange{\protect\cite[{Sec. 3.6}]{cite736}}.
Traditionally, the logical subspace is the joint \(+1\) eigenspace, and the stabilizer group does not contain \(e^{i \phi} I\) for any \(\phi \neq 0\).
The distance \(d\) is the minimum weight of a qudit Pauli string that implements a nontrivial logical operation in the code.}\\ 
\addlinespace[\myxtraspc]
\eczhRefIndex{code:ntru_gkp}%
\eczhListValue{\flmRefsHyperref{code:ntru_gkp}{NTRU-GKP code}} & \eczhListValue{Multi-mode GKP code whose underlying lattice is utilized in variations of the NTRU cryptosystem \NoCaseChange{\protect\cite{cite288}}.
Randomized constructions yield constant-rate GKP code families whose largest decodable displacement length scales as \(O(\sqrt{n})\) with high probability.}\\ 
\addlinespace[\myxtraspc]
\eczhRefIndex{code:gkp-stabilizer}%
\eczhListValue{\flmRefsHyperref{code:gkp-stabilizer}{Oscillator-into-oscillator GKP code}} & \eczhListValue{Multimode GKP code with an infinite-dimensional logical space. Can be obtained by considering an \(n\)-mode GKP code with a finite-dimensional logical space, removing stabilizers such that the logical space becomes infinite dimensional, and applying a Gaussian circuit.}\\ 
\addlinespace[\myxtraspc]
\eczhRefIndex{code:general_qldpc}%
\eczhListValue{\flmRefsHyperref{code:general_qldpc}{QLDPC code}} & \eczhListValue{Member of a family of stabilizer codes for which the number of sites participating in each stabilizer generator and the number of stabilizer generators that each site participates in are both bounded by a constant as \(n\to\infty\).
Sometimes, the two parameters are explicitly stated: each site of an \((l,w)\)\textit{-regular QLDPC code} is acted on by \(\leq l\) generators of weight \(\leq w\).}\\ 
\addlinespace[\myxtraspc]
\eczhRefIndex{code:quantum_ag}%
\eczhListValue{\flmRefsHyperref{code:quantum_ag}{Quantum AG code}} & \eczhListValue{True Galois-qudit stabilizer code constructed from evaluation AG codes via the Galois-qudit Hermitian construction or the Galois-qudit CSS construction.}\\ 
\addlinespace[\myxtraspc]
\eczhRefIndex{code:quantum_gabidulin}%
\eczhListValue{\flmRefsHyperref{code:quantum_gabidulin}{Quantum Gabidulin code}} & \eczhListValue{A Galois-qudit stabilizer code over \(n\) Galois qudits of dimension \(q = 2^n \) that is useful in protecting against faults in qubit Clifford circuits acting on stacked quantum memories.
This code can be treated as a code on an \(n\times n\) qubit stacked memory by decomposing each Galois qudit into a Kronecker product of \(n\) qubits; see \NoCaseChange{\protect\cite{cite696,cite398,cite698,cite699,cite700}\protect\cite[{Sec. 5.3}]{cite697}}.}\\ 
\addlinespace[\myxtraspc]
\eczhRefIndex{code:quantum_hermitian_ag}%
\eczhListValue{\flmRefsHyperref{code:quantum_hermitian_ag}{Quantum Hermitian AG code}} & \eczhListValue{Quantum AG code constructed from Hermitian AG codes via the Galois-qudit Hermitian construction or the Galois-qudit CSS construction.
The underlying classical codes can be constructed from one-point \NoCaseChange{\protect\cite{cite869}} or two-point \NoCaseChange{\protect\cite{cite870}} Hermitian codes on Hermitian curves (see also Ref. \NoCaseChange{\protect\cite{cite871}}).
In parameter ranges where two-point Hermitian codes improve on one-point codes, the resulting quantum codes can also have improved parameters \NoCaseChange{\protect\cite{cite870}}.}\\ 
\addlinespace[\myxtraspc]
\eczhRefIndex{code:galois_duadic}%
\eczhListValue{\flmRefsHyperref{code:galois_duadic}{Quantum duadic code}} & \eczhListValue{True Galois-qudit stabilizer code constructed from \(q\)-ary duadic codes via the Hermitian construction or the Galois-qudit CSS construction.
Large subclasses of quantum duadic codes are degenerate \NoCaseChange{\protect\cite[{Sec. 5.5}]{cite872}}.}\\ 
\addlinespace[\myxtraspc]
\eczhRefIndex{code:quantum_lattice}%
\eczhListValue{\flmRefsHyperref{code:quantum_lattice}{Quantum lattice code}} & \eczhListValue{Bosonic stabilizer code on \(n\) bosonic modes whose stabilizer group is an infinite countable group of oscillator displacement operators which implement lattice translations in phase space.}\\ 
\addlinespace[\myxtraspc]
\eczhRefIndex{code:qlwc}%
\eczhListValue{\flmRefsHyperref{code:qlwc}{Quantum low-weight check (QLWC) code}} & \eczhListValue{Member of a family of \(\llbracket n,k,d\rrbracket \) stabilizer codes for which the number of sites participating in each stabilizer generator is bounded by a constant as \(n\to\infty\).}\\ 
\addlinespace[\myxtraspc]
\eczhRefIndex{code:quantum_plane_curve}%
\eczhListValue{\flmRefsHyperref{code:quantum_plane_curve}{Quantum plane-curve code}} & \eczhListValue{Quantum AG code constructed from plane-curve codes via the Galois-qudit Hermitian construction.
Code parameters are \(\llbracket q^3,q^3+q^2-3q-2r,r+2q-q^2\rrbracket _q\), where \(r\) is an integer satisfying \(q^2 - 2 \leq r \leq q^2 + q - 3\), and where the underlying plane curve is \(y^q + y = x^{q-1}\).}\\ 
\addlinespace[\myxtraspc]
\eczhRefIndex{code:quantum_twisted}%
\eczhListValue{\flmRefsHyperref{code:quantum_twisted}{Quantum twisted code}} & \eczhListValue{Hermitian stabilizer code constructed from twisted BCH codes.}\\ 
\addlinespace[\myxtraspc]
\eczhRefIndex{code:quasi_cyclic_qldpc}%
\eczhListValue{\flmRefsHyperref{code:quasi_cyclic_qldpc}{Quasi-cyclic QLDPC (QC-QLDPC) code}} & \eczhListValue{A QLDPC code such that cyclic shifts of the subsystems by a fixed \(\ell\geq 1\) leave the codespace invariant.
Stabilizer generator matrices of such codes can be put into block form, where each nonzero block is a circulant matrix \NoCaseChange{\protect\cite{cite821,cite822}}.}\\ 
\addlinespace[\myxtraspc]
\eczhRefIndex{code:qudit_cubic}%
\eczhListValue{\flmRefsHyperref{code:qudit_cubic}{Qudit cubic code}} & \eczhListValue{Generalization of the Haah cubic code to modular qudits.}\\ 
\addlinespace[\myxtraspc]
\eczhRefIndex{code:random_stabilizer}%
\eczhListValue{\flmRefsHyperref{code:random_stabilizer}{Random stabilizer code}} & \eczhListValue{An \(n\)-qubit, modular-qudit, or Galois-qudit stabilizer code whose construction is non-deterministic.
Since stabilizer encoders are Clifford circuits, such codes can be thought of as arising from random Clifford circuits.}\\ 
\addlinespace[\myxtraspc]
\eczhRefIndex{code:rotor_cluster}%
\eczhListValue{\flmRefsHyperref{code:rotor_cluster}{Rotor cluster-state code}} & \eczhListValue{Rotor analogue of the qubit and analog cluster-state codes.
The exact rotor cluster state is non-normalizable, so approximate constructions have to be considered.
Defined from a real-valued weighted adjacency matrix of a graph \NoCaseChange{\protect\cite{cite507}}.}\\ 
\addlinespace[\myxtraspc]
\eczhRefIndex{code:rotor_stabilizer}%
\eczhListValue{\flmRefsHyperref{code:rotor_stabilizer}{Rotor stabilizer code}} & \eczhListValue{Rotor code whose codespace is defined as the common \(+1\) eigenspace of a group of mutually commuting rotor generalized Pauli operators.
The stabilizer group can be either discrete or continuous, corresponding to modular or linear constraints on angular positions and momenta.
Both cases can yield finite or infinite logical dimension.
Exact codewords are non-normalizable, so approximate constructions have to be considered.}\\ 
\addlinespace[\myxtraspc]
\eczhRefIndex{code:stabilizer}%
\eczhListValue{\flmRefsHyperref{code:stabilizer}{Stabilizer code}} & \eczhListValue{A code whose logical subspace is the joint eigenspace (usually with eigenvalue \(+1\)) of a set of commuting unitary Pauli-type operators forming the code's stabilizer group.
They can be block codes defined on tensor-product spaces of qubits or qudits, or non-block codes defined on single sufficiently large Hilbert spaces such as bosonic modes or other Abelian group spaces.}\\ 
\addlinespace[\myxtraspc]
\eczhRefIndex{code:galois_true_stabilizer}%
\eczhListValue{\flmRefsHyperref{code:galois_true_stabilizer}{True Galois-qudit stabilizer code}} & \eczhListValue{A \(\llbracket n,k,d\rrbracket _q\) stabilizer code whose stabilizer's \flmRefsHyperref{ref873}{Galois symplectic representation} forms a linear subspace. In other words, the set of \(q\)-ary vectors representing the stabilizer group is closed under both addition and multiplication by elements of \(\mathbb{F}_q\). In contrast, Galois-qudit stabilizer codes admit sets of vectors that are closed under addition only.}\\ 
\addlinespace[\myxtraspc]
\eczhRefIndex{code:dfour_gkp}%
\eczhListValue{\flmRefsHyperref{code:dfour_gkp}{\(D_4\) hyper-diamond GKP code}} & \eczhListValue{Two-mode GKP qubit-into-oscillator code based on the \(D_4\) hyper-diamond lattice \NoCaseChange{\protect\cite{cite482}}.}\\ 
\addlinespace[\myxtraspc]
\eczhRefIndex{code:chern_simons_gkp}%
\eczhListValue{\flmRefsHyperref{code:chern_simons_gkp}{\(U(1)_{2n} \times U(1)_{-2m}\) Chern-Simons GKP code}} & \eczhListValue{A non-CSS multimode GKP code defined on a 2D mode lattice that encodes a qudit logical space and whose excitations are characterized by the \(U(1)_{2n} \times U(1)_{-2m}\) Chern-Simons theory.
The code can be obtained from the analog surface code by \flmRefsHyperref{ref410}{condensing} certain anyons \NoCaseChange{\protect\cite{cite411}}.}\\ 
\addlinespace[\myxtraspc]
\eczhRefIndex{code:group_10_1_4}%
\eczhListValue{\flmRefsHyperref{code:group_10_1_4}{\(\llbracket 10,1,4\rrbracket _{G}\) tenfold code}} & \eczhListValue{A \(\llbracket 10,1,4\rrbracket _{G}\) group code for finite Abelian \(G\).
The code is defined using a graph that is closely related to the \(\llbracket 5,1,3\rrbracket \) code.}\\ 
\addlinespace[\myxtraspc]
\eczhRefIndex{code:galois_5_1_3}%
\eczhListValue{\flmRefsHyperref{code:galois_5_1_3}{\(\llbracket 5,1,3\rrbracket _q\) Galois-qudit code}} & \eczhListValue{True stabilizer code that generalizes the five-qubit perfect code to Galois qudits of prime-power dimension \(q=p^m\). It has \(4m\) stabilizer generators expressed as \(X_{\gamma} Z_{\gamma} Z_{-\gamma} X_{-\gamma} I\) and its cyclic permutations, with \(\gamma\) iterating over basis elements of \(\mathbb{F}_q\) over \(\mathbb{F}_p\).}\\ 
\addlinespace[\myxtraspc]
\eczhRefIndex{code:braunstein}%
\eczhListValue{\flmRefsHyperref{code:braunstein}{\(\llbracket 5,1,3\rrbracket _{\mathbb{R}}\) Braunstein five-mode code}} & \eczhListValue{An analog stabilizer version of the five-qubit perfect code, encoding one mode into five and correcting arbitrary errors on any one mode.}\\ 
\addlinespace[\myxtraspc]
\eczhRefIndex{code:qudit_5_1_3}%
\eczhListValue{\flmRefsHyperref{code:qudit_5_1_3}{\(\llbracket 5,1,3\rrbracket _{\mathbb{Z}_q}\) modular-qudit code}} & \eczhListValue{Modular-qudit stabilizer code that generalizes the five-qubit perfect code using properties of the multiplicative group \(\mathbb{Z}_q\) \NoCaseChange{\protect\cite{cite531}}; see also \NoCaseChange{\protect\cite[{Thm. 13}]{cite532}}. It has four stabilizer generators consisting of \(X Z Z^\dagger X^\dagger I\) and its cyclic permutations.}\\ 
\addlinespace[\myxtraspc]
\eczhRefIndex{code:rotor_5_1_3}%
\eczhListValue{\flmRefsHyperref{code:rotor_5_1_3}{\(\llbracket 5,1,3\rrbracket _{\mathbb{Z}}\) Five-rotor code}} & \eczhListValue{Extension of the five-qubit stabilizer code to the integer alphabet, i.e., the angular momentum states of a rotor. The code is \(U(1)\)-covariant and ideal codewords are not normalizable.}\\ 
\addlinespace[\myxtraspc]
\eczhRefIndex{code:galois_6_2_3}%
\eczhListValue{\flmRefsHyperref{code:galois_6_2_3}{\(\llbracket 6,2,3\rrbracket _{q}\) code}} & \eczhListValue{Six-qudit MDS error-correcting code defined for Galois-qudit dimension \(q=3\) \NoCaseChange{\protect\cite{cite830}}, \(q=2^2\) \NoCaseChange{\protect\cite{cite831}}, and \(q \geq 5\) \NoCaseChange{\protect\cite{cite830}\protect\cite[{Exam. 33}]{cite813}}.
This code cannot exist for qubits (\(q=2\)).}\\ 
\addlinespace[\myxtraspc]
\eczhRefIndex{code:galois_7_3_3}%
\eczhListValue{\flmRefsHyperref{code:galois_7_3_3}{\(\llbracket 7,3,3\rrbracket _{q}\) code}} & \eczhListValue{Seven-qudit MDS error-detecting code defined for Galois-qudit dimension \(q=3\) \NoCaseChange{\protect\cite{cite830}} and \(q \geq 7\) \NoCaseChange{\protect\cite{cite830}\protect\cite[{Exam. 33}]{cite813}}.
This code cannot exist for qubits (\(q=2\)).}\\ 
\addlinespace[\myxtraspc]
\eczhRefIndex{code:stab_9_1_5}%
\eczhListValue{\flmRefsHyperref{code:stab_9_1_5}{\(\llbracket 9,1,5\rrbracket _3\) quantum Glynn code}} & \eczhListValue{Nine-qutrit \flmRefsHyperref{ref672}{pure} Hermitian code that is the smallest qutrit stabilizer code to correct two-qutrit errors.}\\ 
\end{tabularx}\endgroup
\eczcodelist{subsystem}{Subsystem codes (non-qubit)
}%

\eczhCodeListAutoDescription{Codes that are descendants of \flmRefsCref{code:oecc} and not descendants of any of \flmRefsCref{code:subsystem_qubits_into_qubits}.}%

\eczhIncludeCodeGraph{Bare}{scale=0.5}{\columnwidth}{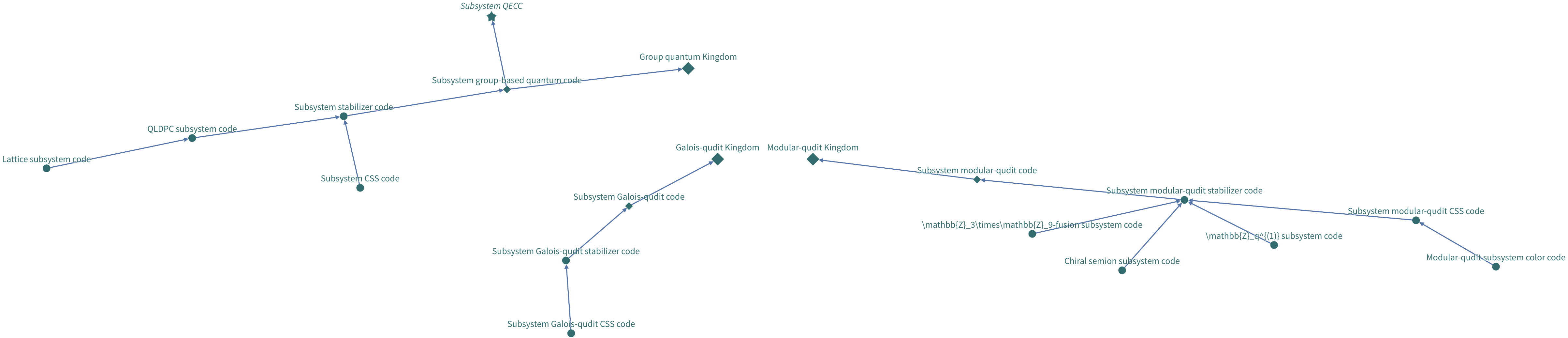}{Subsystem codes (non-qubit)}{https://errorcorrectionzoo.org/code_graph#J\%7B\%22displayMode\%22\%3A\%22subset\%22\%2C\%22modeSubsetOptions\%22\%3A\%7B\%22codeIds\%22\%3A\%5B\%22semion\%22\%2C\%22translationally_invariant_subsystem\%22\%2C\%22qudit_subsystem_color\%22\%2C\%22sparse_subsystem\%22\%2C\%22subsystem_css\%22\%2C\%22galois_subsystem_css\%22\%2C\%22subsystem_galois_into_galois\%22\%2C\%22galois_subsystem_stabilizer\%22\%2C\%22oecc\%22\%2C\%22subsystem_group_quantum\%22\%2C\%22qudit_subsystem_css\%22\%2C\%22subsystem_qudits_into_qudits\%22\%2C\%22qudit_subsystem_stabilizer\%22\%2C\%22subsystem_stabilizer\%22\%2C\%22zthree_znine\%22\%2C\%22qudit_znone\%22\%5D\%2C\%22reusePreviousLayoutPositions\%22\%3Afalse\%2C\%22showIntermediateConnectingNodes\%22\%3Atrue\%2C\%22connectingNodesMaxDepth\%22\%3A15\%2C\%22connectingNodesPathMaxLength\%22\%3A20\%2C\%22connectingNodesMaxExtraDepth\%22\%3A3\%2C\%22connectingNodesOnlyKeepPathsWithAdditionalLength\%22\%3A1\%2C\%22connectingNodesToDomainsAndKingdoms\%22\%3Afalse\%2C\%22connectingNodesEdgeLengthsByType\%22\%3A\%7B\%22primaryParent\%22\%3A1\%2C\%22secondaryParent\%22\%3A4\%2C\%22cousin\%22\%3A6\%7D\%2C\%22nodeIds\%22\%3A\%5B\%22k_galois_into_galois\%22\%2C\%22k_group_quantum\%22\%2C\%22k_qudits_into_qudits\%22\%5D\%7D\%2C\%22highlightImportantNodes\%22\%3A\%7B\%22highlightImportantNodes\%22\%3Afalse\%2C\%22highlightPrimaryParents\%22\%3Afalse\%2C\%22highlightRootConnectingEdges\%22\%3Afalse\%7D\%7D}

\begingroup
\small
\eczhBreakableDashes
\renewcommand\arraystretch{1.05}
\edef\myxtraspc{\eczListAddVSpaceXtraXtraStretch}
\begin{tabularx}{\linewidth}{>{\raggedright\arraybackslash}p{\eczListColWidth{name}} >{\hsize=1.0000\hsize }X}
\toprule
\eczListColTitle{Code} & \eczListColTitle{Description} \\
\midrule
\endfirsthead
\toprule
\eczListColTitleContinued{Code} & \eczListColTitleContinued{Description} \\
\midrule
\endhead
\bottomrule
\endfoot
\eczhRefIndex{code:semion}%
\eczhListValue{\flmRefsHyperref{code:semion}{Chiral semion subsystem code}} & \eczhListValue{Modular-qudit subsystem stabilizer code with qudit dimension \(q=4\) that is characterized by the chiral semion topological phase.
The code admits a set of geometrically local stabilizer generators on a torus.}\\ 
\addlinespace[\myxtraspc]
\eczhRefIndex{code:translationally_invariant_subsystem}%
\eczhListValue{\flmRefsHyperref{code:translationally_invariant_subsystem}{Lattice subsystem code}} & \eczhListValue{A geometrically local qubit, modular-qudit, or Galois-qudit subsystem stabilizer code with qudits organized on a lattice modeled by the additive group \(\mathbb{Z}^D\) for spatial dimension \(D\), using either the ordinary block notion of locality or the fermionic/Majorana notion of locality.
On an infinite lattice, its gauge group is generated by few-site Pauli operators and their translations, in which case the code is called \textit{translationally invariant subsystem code}.
The stabilizer group may contain generators of unbounded weight, distinguishing these codes from stabilizer codes with bounded-weight generators for which some logical qubits were re-assigned to be gauge qubits.}\\ 
\addlinespace[\myxtraspc]
\eczhRefIndex{code:qudit_subsystem_color}%
\eczhListValue{\flmRefsHyperref{code:qudit_subsystem_color}{Modular-qudit subsystem color code}} & \eczhListValue{An extension of subsystem color codes to modular qudits.
Codes are defined analogously to qubit subsystem color codes, but a directionality is required in order to make the modular-qudit stabilizers commute \NoCaseChange{\protect\cite[{Sec. VII}]{cite673}}.}\\ 
\addlinespace[\myxtraspc]
\eczhRefIndex{code:sparse_subsystem}%
\eczhListValue{\flmRefsHyperref{code:sparse_subsystem}{QLDPC subsystem code}} & \eczhListValue{Member of a family of subsystem stabilizer codes for which the number of sites participating in each gauge generator and the number of gauge generators that each site participates in are both bounded by a constant as \(n\to\infty\).
The stabilizer group may contain generators of unbounded weight, distinguishing these codes from stabilizer codes with bounded-weight generators for which some logical qubits were re-assigned to be gauge qubits.}\\ 
\addlinespace[\myxtraspc]
\eczhRefIndex{code:subsystem_css}%
\eczhListValue{\flmRefsHyperref{code:subsystem_css}{Subsystem CSS code}} & \eczhListValue{A subsystem stabilizer code admitting a set of gauge-group generators that are either \(Z\)-type or \(X\)-type operators.
This ensures that the associated stabilizer group is also CSS.}\\ 
\addlinespace[\myxtraspc]
\eczhRefIndex{code:galois_subsystem_css}%
\eczhListValue{\flmRefsHyperref{code:galois_subsystem_css}{Subsystem Galois-qudit CSS code}} & \eczhListValue{Galois-qudit subsystem stabilizer code which admits a set of gauge-group generators which consist of either all-\(Z\) or all-\(X\) Galois-qudit Pauli strings.}\\ 
\addlinespace[\myxtraspc]
\eczhRefIndex{code:subsystem_galois_into_galois}%
\eczhListValue{\flmRefsHyperref{code:subsystem_galois_into_galois}{Subsystem Galois-qudit code}} & \eczhListValue{Subsystem QECC encoding into a \(q^n\)-dimensional Hilbert space consisting of \(n\) Galois qudits.}\\ 
\addlinespace[\myxtraspc]
\eczhRefIndex{code:galois_subsystem_stabilizer}%
\eczhListValue{\flmRefsHyperref{code:galois_subsystem_stabilizer}{Subsystem Galois-qudit stabilizer code}} & \eczhListValue{Galois-qudit generalization of a subsystem qubit stabilizer code.
Can be obtained by taking a Galois-qudit stabilizer code and assigning some of its logical qudits to be gauge qudits.}\\ 
\addlinespace[\myxtraspc]
\eczhRefIndex{code:oecc}%
\eczhListValue{\flmRefsHyperref{code:oecc}{Subsystem QECC}} & \eczhListValue{A quantum code which encodes quantum information in a tensor factor of a subspace that is decomposed into a tensor product of subsystems.}\\ 
\addlinespace[\myxtraspc]
\eczhRefIndex{code:subsystem_group_quantum}%
\eczhListValue{\flmRefsHyperref{code:subsystem_group_quantum}{Subsystem group-based quantum code}} & \eczhListValue{Group-based quantum code whose codespace admits a tensor-product decomposition into logical and gauge factors.}\\ 
\addlinespace[\myxtraspc]
\eczhRefIndex{code:qudit_subsystem_css}%
\eczhListValue{\flmRefsHyperref{code:qudit_subsystem_css}{Subsystem modular-qudit CSS code}} & \eczhListValue{Modular-qudit subsystem stabilizer code which admits a set of gauge-group generators which consist of either all-\(Z\) or all-\(X\) modular-qudit Pauli strings.
This ensures that the code's stabilizer group is also CSS.}\\ 
\addlinespace[\myxtraspc]
\eczhRefIndex{code:subsystem_qudits_into_qudits}%
\eczhListValue{\flmRefsHyperref{code:subsystem_qudits_into_qudits}{Subsystem modular-qudit code}} & \eczhListValue{Subsystem QECC encoding into a \(q^n\)-dimensional Hilbert space consisting of \(n\) modular qudits.}\\ 
\addlinespace[\myxtraspc]
\eczhRefIndex{code:qudit_subsystem_stabilizer}%
\eczhListValue{\flmRefsHyperref{code:qudit_subsystem_stabilizer}{Subsystem modular-qudit stabilizer code}} & \eczhListValue{Modular-qudit generalization of a subsystem qubit stabilizer code.
Can be obtained by taking a modular-qudit stabilizer code and assigning some of its logical qudits to be gauge qudits.
For composite qudit dimensions, such codes need not encode an integer number of qudits.}\\ 
\addlinespace[\myxtraspc]
\eczhRefIndex{code:subsystem_stabilizer}%
\eczhListValue{\flmRefsHyperref{code:subsystem_stabilizer}{Subsystem stabilizer code}} & \eczhListValue{A subsystem code that is derived from a stabilizer code by assigning some factors of the stabilizer code's logical tensor-product structure to gauge degrees of freedom.}\\ 
\addlinespace[\myxtraspc]
\eczhRefIndex{code:zthree_znine}%
\eczhListValue{\flmRefsHyperref{code:zthree_znine}{\(\mathbb{Z}_3\times\mathbb{Z}_9\)-fusion subsystem code}} & \eczhListValue{Modular-qudit 2D subsystem stabilizer code whose low-energy excitations realize a non-modular anyon theory with \(\mathbb{Z}_3\times\mathbb{Z}_9\) fusion rules.
Encodes two qutrits when put on a torus.}\\ 
\addlinespace[\myxtraspc]
\eczhRefIndex{code:qudit_znone}%
\eczhListValue{\flmRefsHyperref{code:qudit_znone}{\(\mathbb{Z}_q^{(1)}\) subsystem code}} & \eczhListValue{Modular-qudit subsystem code, based on the Kitaev honeycomb model \NoCaseChange{\protect\cite{cite537}} and its generalization \NoCaseChange{\protect\cite{cite637}}, that is characterized by the \(\mathbb{Z}_q^{(1)}\) anyon theory \NoCaseChange{\protect\cite{cite638}}, which is modular for odd prime \(q\) and non-modular otherwise. Encodes a single \(q\)-dimensional qudit when put on a torus for odd \(q\), and a \(q/2\)-dimensional qudit for even \(q\). This code can be constructed using geometrically local gauge generators, but does not admit geometrically local stabilizer generators. For \(q=2\), the code reduces to the subsystem code underlying the Kitaev honeycomb model code as well as the honeycomb Floquet code.}\\ 
\end{tabularx}\endgroup
\eczcodelist{topological}{Topological codes
}%

\eczhCodeListAutoDescription{All descendants of \flmRefsCref{code:topological}.}%

\eczhIncludeCodeGraph{Bare}{scale=0.5}{\columnwidth}{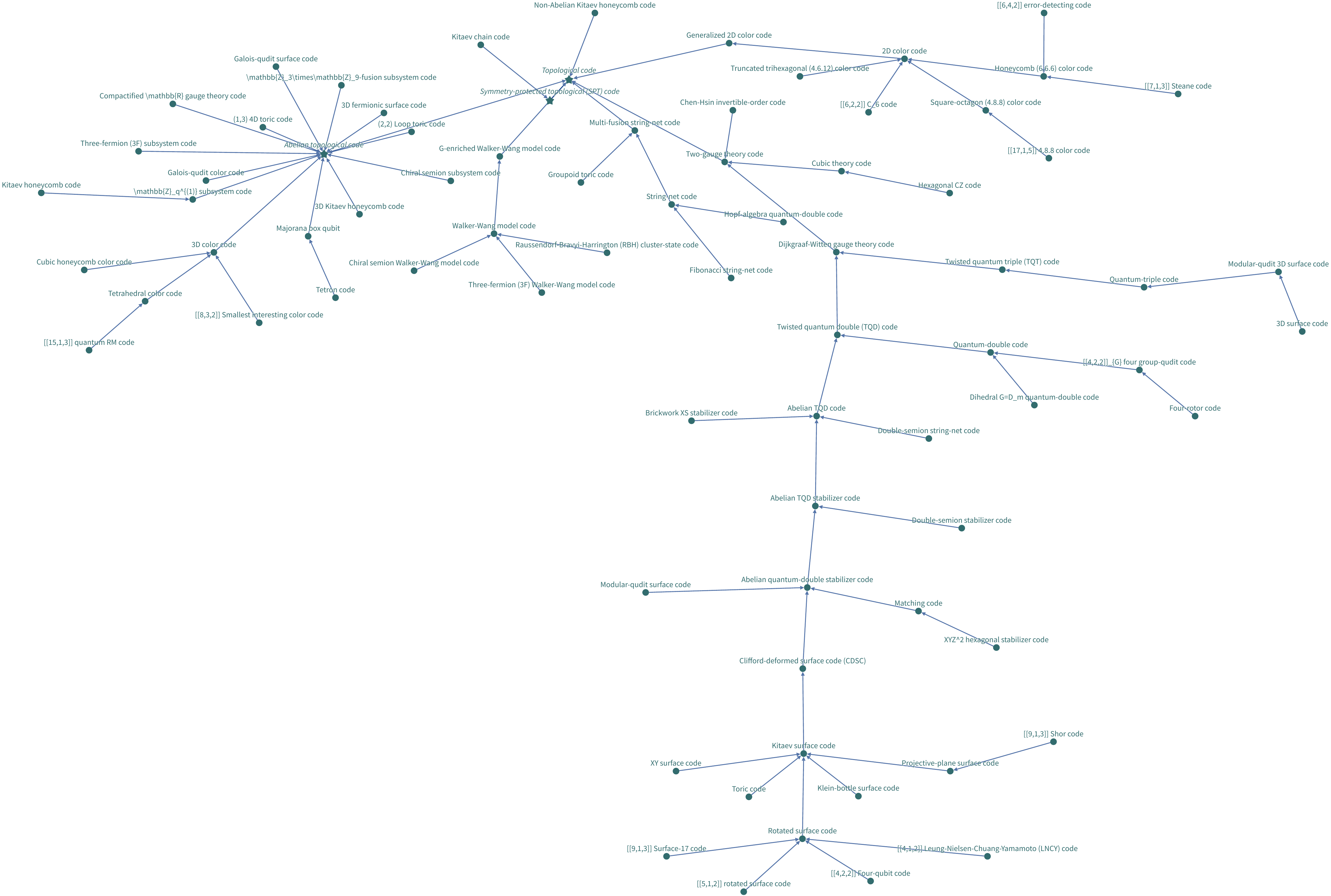}{Topological codes}{https://errorcorrectionzoo.org/code_graph#J\%7B\%22displayMode\%22\%3A\%22subset\%22\%2C\%22modeSubsetOptions\%22\%3A\%7B\%22codeIds\%22\%3A\%5B\%222d_color\%22\%2C\%223d_kitaev_honeycomb\%22\%2C\%223d_color\%22\%2C\%223d_fermionic_surface\%22\%2C\%223d_surface\%22\%2C\%22tqd_abelian\%22\%2C\%22tqd_abelian_stabilizer\%22\%2C\%22quantum_double_abelian\%22\%2C\%22topological_abelian\%22\%2C\%22brickwork\%22\%2C\%22invertible\%22\%2C\%223d_semion\%22\%2C\%22semion\%22\%2C\%22clifford-deformed_surface\%22\%2C\%22compactified_r\%22\%2C\%22cubic_honeycomb_color\%22\%2C\%22cubic_theory\%22\%2C\%22quantum_double_dihedral\%22\%2C\%22dijkgraaf_witten\%22\%2C\%22double_semion\%22\%2C\%22double_semion_string_net\%22\%2C\%22fibonacci\%22\%2C\%22rotor_4_2_2\%22\%2C\%22galois_color\%22\%2C\%22galois_topological\%22\%2C\%22generalized_color\%22\%2C\%22groupoid_surface\%22\%2C\%22hexagonal_cz\%22\%2C\%22triangular_color\%22\%2C\%22hopf_quantum_double\%22\%2C\%22kitaev_chain\%22\%2C\%22kitaev_honeycomb\%22\%2C\%22surface\%22\%2C\%22klein_bottle\%22\%2C\%22mbq\%22\%2C\%22matching\%22\%2C\%22qudit_3d_surface\%22\%2C\%22qudit_surface\%22\%2C\%22enriched_string_net\%22\%2C\%22nonabelian_kitaev_honeycomb\%22\%2C\%22real_projective_plane\%22\%2C\%22quantum_double\%22\%2C\%22quantum_triple\%22\%2C\%22rbh\%22\%2C\%22rotated_surface\%22\%2C\%22488_color\%22\%2C\%22string_net\%22\%2C\%22spt\%22\%2C\%22tetrahedral_color\%22\%2C\%22tetron\%22\%2C\%22three_fermion\%22\%2C\%22subsystem_three_fermion\%22\%2C\%22topological\%22\%2C\%22toric\%22\%2C\%224612_color\%22\%2C\%22tqd\%22\%2C\%22tqt\%22\%2C\%22yetter_gauge_theory\%22\%2C\%22walker_wang\%22\%2C\%22xysurface\%22\%2C\%22xyz_hexagonal\%22\%2C\%224d_13_surface\%22\%2C\%224d_surface\%22\%2C\%22enriched_walker_wang\%22\%2C\%22stab_15_1_3\%22\%2C\%22stab_17_1_5\%22\%2C\%22css_4_1_2\%22\%2C\%22stab_4_2_2\%22\%2C\%22group_4_2_2\%22\%2C\%22stab_5_1_2\%22\%2C\%22stab_6_2_2\%22\%2C\%22stab_6_4_2\%22\%2C\%22steane\%22\%2C\%22stab_8_3_2\%22\%2C\%22shor_nine\%22\%2C\%22surface-17\%22\%2C\%22zthree_znine\%22\%2C\%22qudit_znone\%22\%5D\%2C\%22reusePreviousLayoutPositions\%22\%3Afalse\%2C\%22showIntermediateConnectingNodes\%22\%3Atrue\%2C\%22connectingNodesMaxDepth\%22\%3A15\%2C\%22connectingNodesPathMaxLength\%22\%3A20\%2C\%22connectingNodesMaxExtraDepth\%22\%3A3\%2C\%22connectingNodesOnlyKeepPathsWithAdditionalLength\%22\%3A1\%2C\%22connectingNodesToDomainsAndKingdoms\%22\%3Afalse\%2C\%22connectingNodesEdgeLengthsByType\%22\%3A\%7B\%22primaryParent\%22\%3A1\%2C\%22secondaryParent\%22\%3A4\%2C\%22cousin\%22\%3A6\%7D\%2C\%22nodeIds\%22\%3A\%5B\%5D\%7D\%2C\%22highlightImportantNodes\%22\%3A\%7B\%22highlightImportantNodes\%22\%3Afalse\%2C\%22highlightPrimaryParents\%22\%3Afalse\%2C\%22highlightRootConnectingEdges\%22\%3Afalse\%7D\%7D}

\begingroup
\small
\eczhBreakableDashes
\renewcommand\arraystretch{1.05}
\edef\myxtraspc{\eczListAddVSpaceXtraXtraStretch}
\begin{tabularx}{\linewidth}{>{\raggedright\arraybackslash}p{\eczListColWidth{name}} >{\hsize=1.0000\hsize }X}
\toprule
\eczListColTitle{Code} & \eczListColTitle{Description} \\
\midrule
\endfirsthead
\toprule
\eczListColTitleContinued{Code} & \eczListColTitleContinued{Description} \\
\midrule
\endhead
\bottomrule
\endfoot
\eczhRefIndex{code:2d_color}%
\eczhListValue{\flmRefsHyperref{code:2d_color}{2D color code}} & \eczhListValue{Color code defined on a graph embedded in a two-dimensional surface.
Each face hosts two stabilizer generators, a Pauli-\(X\) and a Pauli-\(Z\) string acting on all the qubits of the face.}\\ 
\addlinespace[\myxtraspc]
\eczhRefIndex{code:3d_kitaev_honeycomb}%
\eczhListValue{\flmRefsHyperref{code:3d_kitaev_honeycomb}{3D Kitaev honeycomb code}} & \eczhListValue{3D subsystem stabilizer code whose Hamiltonian is a 3D generalization of the Kitaev honeycomb model.
One of the phases realized by the 3D Kitaev honeycomb Hamiltonian is that of the 3D fermionic surface code \NoCaseChange{\protect\cite{cite458}}.}\\ 
\addlinespace[\myxtraspc]
\eczhRefIndex{code:3d_color}%
\eczhListValue{\flmRefsHyperref{code:3d_color}{3D color code}} & \eczhListValue{Color code defined on a four-valent, four-colorable 3-colex in a 3-manifold.
In the original colex realization, qubits sit on vertices, \(X\)-type stabilizers are attached to 3-cells, and \(Z\)-type stabilizers are attached to faces \NoCaseChange{\protect\cite{cite430}}.}\\ 
\addlinespace[\myxtraspc]
\eczhRefIndex{code:3d_fermionic_surface}%
\eczhListValue{\flmRefsHyperref{code:3d_fermionic_surface}{3D fermionic surface code}} & \eczhListValue{A non-CSS variant of the 3D Kitaev surface code that realizes \(\mathbb{Z}_2\) gauge theory with an emergent fermion, i.e., the fermionic-charge bosonic-loop (FcBl) phase \NoCaseChange{\protect\cite{cite455}}.
The model can be defined on a cubic lattice in several ways \NoCaseChange{\protect\cite[{Eq. (D45-46)}]{cite456}}.
Realizations on other lattices also exist \NoCaseChange{\protect\cite{cite457}}, and the phase of this code also exists in the 3D Kitaev honeycomb model \NoCaseChange{\protect\cite{cite458}}.}\\ 
\addlinespace[\myxtraspc]
\eczhRefIndex{code:3d_surface}%
\eczhListValue{\flmRefsHyperref{code:3d_surface}{3D surface code}} & \eczhListValue{A generalization of the Kitaev surface code defined on a 3D cubic lattice.
Qubits are placed on edges, \(Z\)-type stabilizer generators are placed on square plaquettes oriented in all three directions, and \(X\)-type stabilizers are placed on the six edges neighboring every vertex \NoCaseChange{\protect\cite{cite459}}.}\\ 
\addlinespace[\myxtraspc]
\eczhRefIndex{code:tqd_abelian}%
\eczhListValue{\flmRefsHyperref{code:tqd_abelian}{Abelian TQD code}} & \eczhListValue{TQD code whose codewords realize a 2D Abelian twisted-quantum-double topological order.
For Abelian TQDs, the corresponding anyon theory is defined by an Abelian group and a group cocycle built from Type-I, Type-II, or Type-III 3-cocycles \NoCaseChange{\protect\cite{cite571,cite405,cite572}}.
Abelian TQDs with Type-I and -II cocycles account for all 2D Abelian topological orders that admit gapped boundaries \NoCaseChange{\protect\cite{cite573}}.
Abelian TQDs with Type-III cocycles may admit non-Abelian topological orders.}\\ 
\addlinespace[\myxtraspc]
\eczhRefIndex{code:tqd_abelian_stabilizer}%
\eczhListValue{\flmRefsHyperref{code:tqd_abelian_stabilizer}{Abelian TQD stabilizer code}} & \eczhListValue{Modular-qudit stabilizer code whose codewords realize a 2D Abelian twisted-quantum-double topological order on composite-dimensional qudits.
For every finite Abelian group \(G=\prod_i \mathbb{Z}_{N_i}\) and every product of Type-I and Type-II cocycles, there is a Pauli stabilizer Hamiltonian realizing the corresponding Abelian TQD \NoCaseChange{\protect\cite{cite405}}.
Equivalently, these codes exhaust the 2D Abelian topological orders that admit gapped boundaries \NoCaseChange{\protect\cite{cite405,cite406}}.}\\ 
\addlinespace[\myxtraspc]
\eczhRefIndex{code:quantum_double_abelian}%
\eczhListValue{\flmRefsHyperref{code:quantum_double_abelian}{Abelian quantum-double stabilizer code}} & \eczhListValue{Modular-qudit stabilizer code whose codewords realize 2D modular gapped Abelian topological order with trivial cocycle.
The corresponding anyon theory is defined by an Abelian group.
The \(G=\mathbb{Z}_2\) instance on a torus is the toric code, and cyclic-group instances reduce to modular-qudit surface codes.
All such codes can be realized by a stack of modular-qudit surface codes because all finite Abelian groups are direct products of cyclic groups.}\\ 
\addlinespace[\myxtraspc]
\eczhRefIndex{code:topological_abelian}%
\eczhListValue{\flmRefsHyperref{code:topological_abelian}{Abelian topological code}} & \eczhListValue{Code whose codewords realize topological order associated with an Abelian anyon theory.
In 2D, this is equivalent to a unitary braided fusion category which is also an Abelian group under fusion \NoCaseChange{\protect\cite{cite574}}.
Unless otherwise noted, the phases discussed are bosonic.}\\ 
\addlinespace[\myxtraspc]
\eczhRefIndex{code:brickwork}%
\eczhListValue{\flmRefsHyperref{code:brickwork}{Brickwork \(XS\) stabilizer code}} & \eczhListValue{An \(XS\) stabilizer code that realizes the topological order of the Type-III \(G=\mathbb{Z}^3_2\) TQD model \NoCaseChange{\protect\cite{cite575,cite576}}, which is the same topological order as the \(G=D_4\) quantum double \NoCaseChange{\protect\cite{cite577}}.
Its qubits are placed on a 2D square lattice, and the stabilizers are defined using two overlapping rectangular tilings.}\\ 
\addlinespace[\myxtraspc]
\eczhRefIndex{code:invertible}%
\eczhListValue{\flmRefsHyperref{code:invertible}{Chen-Hsin invertible-order code}} & \eczhListValue{A geometrically local commuting-projector code that realizes beyond-group-cohomology invertible topological phases in arbitrary dimensions.
Its code Hamiltonian terms include Pauli-\(Z\) operators and products of Pauli-\(X\) operators and \(CZ\) gates \NoCaseChange{\protect\cite[{Eq. (3.25)}]{cite578}}.
Instances of the code in 4D realize the 3D \(\mathbb{Z}_2\) gauge theory with fermionic charge and either bosonic (FcBl) or fermionic (FcFl) loop excitations at their boundaries \NoCaseChange{\protect\cite{cite579,cite455}}; see Ref. \NoCaseChange{\protect\cite{cite580}} for a different lattice-model formulation of the FcBl boundary code.}\\ 
\addlinespace[\myxtraspc]
\eczhRefIndex{code:3d_semion}%
\eczhListValue{\flmRefsHyperref{code:3d_semion}{Chiral semion Walker-Wang model code}} & \eczhListValue{A 3D lattice modular-qudit stabilizer code with qudit dimension \(q=4\) whose low-energy excitations on boundaries realize the chiral semion topological order.
The model admits 2D chiral semion topological order at one of its surfaces \NoCaseChange{\protect\cite{cite471,cite472}}.
The corresponding phase can also be realized via a non-stabilizer Hamiltonian \NoCaseChange{\protect\cite{cite473}}.}\\ 
\addlinespace[\myxtraspc]
\eczhRefIndex{code:semion}%
\eczhListValue{\flmRefsHyperref{code:semion}{Chiral semion subsystem code}} & \eczhListValue{Modular-qudit subsystem stabilizer code with qudit dimension \(q=4\) that is characterized by the chiral semion topological phase.
The code admits a set of geometrically local stabilizer generators on a torus.}\\ 
\addlinespace[\myxtraspc]
\eczhRefIndex{code:clifford-deformed_surface}%
\eczhListValue{\flmRefsHyperref{code:clifford-deformed_surface}{Clifford-deformed surface code (CDSC)}} & \eczhListValue{A generally non-CSS derivative of the surface code defined by applying a translationally invariant constant-depth \flmRefsHyperref{ref409}{Clifford circuit} to the original (CSS) surface code.
Unlike the surface code, CDSCs include codes whose thresholds and subthreshold performance are enhanced under noise biased towards dephasing.
Examples of CDSCs include the XY code, XZZX code, and random CDSCs.}\\ 
\addlinespace[\myxtraspc]
\eczhRefIndex{code:compactified_r}%
\eczhListValue{\flmRefsHyperref{code:compactified_r}{Compactified \(\mathbb{R}\) gauge theory code}} & \eczhListValue{An integer-homology bosonic CSS code realizing 2D \(U(1)\) gauge theory on bosonic modes.
The code can be obtained from the analog surface code by \flmRefsHyperref{ref410}{condensing} certain anyons \NoCaseChange{\protect\cite{cite411}}. 
This results in a pinning of each mode to the space of periodic functions, which is the Hilbert space of a physical rotor, and can be thought of as compactification of the 2D \(\mathbb{R}\) gauge theory phase realized by the analog surface code.}\\ 
\addlinespace[\myxtraspc]
\eczhRefIndex{code:cubic_honeycomb_color}%
\eczhListValue{\flmRefsHyperref{code:cubic_honeycomb_color}{Cubic honeycomb color code}} & \eczhListValue{3D color code defined on a four-colorable bitruncated cubic honeycomb uniform tiling.}\\ 
\addlinespace[\myxtraspc]
\eczhRefIndex{code:cubic_theory}%
\eczhListValue{\flmRefsHyperref{code:cubic_theory}{Cubic theory code}} & \eczhListValue{A geometrically local commuting-projector code family defined on triangulations in arbitrary spatial dimensions.
Its Hamiltonian contains Pauli-\(Z\) flux terms and non-Pauli Gauss-law terms built from products of Pauli-\(X\) operators and \(CZ\) gates.
These commuting non-Pauli stabilizers realize higher-form \(\mathbb{Z}_2^3\) gauge theories with Abelian electric excitations and non-Abelian magnetic excitations.}\\ 
\addlinespace[\myxtraspc]
\eczhRefIndex{code:quantum_double_dihedral}%
\eczhListValue{\flmRefsHyperref{code:quantum_double_dihedral}{Dihedral \(G=D_m\) quantum-double code}} & \eczhListValue{Quantum-double code whose codewords realize topological order associated with the dihedral group \(D_m\) of order \(2m\).
For \(m \geq 3\), these codes are non-Abelian, with the simplest case given by \(D_3=S_3\), the permutation group on three objects.
On an oriented lattice, each edge hosts a \(2m\)-dimensional group qudit, and the codespace is the ground-state subspace of the corresponding quantum double Hamiltonian \NoCaseChange{\protect\cite{cite423}}.}\\ 
\addlinespace[\myxtraspc]
\eczhRefIndex{code:dijkgraaf_witten}%
\eczhListValue{\flmRefsHyperref{code:dijkgraaf_witten}{Dijkgraaf-Witten gauge theory code}} & \eczhListValue{A code whose codewords realize \(D\)-dimensional lattice Dijkgraaf-Witten gauge theory \NoCaseChange{\protect\cite{cite584,cite585}} for a finite group \(G\) and a \(D+1\)-cocycle \(\omega\) in the cohomology class \(H^{D+1}( G, U(1) )\).
When the cocycle is non-trivial, the gauge theory is called a \textit{twisted gauge theory}.
There exist lattice-model formulations in arbitrary spatial dimension \NoCaseChange{\protect\cite{cite586}}.
Boundaries and excitations have been studied for arbitrary dimension \NoCaseChange{\protect\cite{cite587}}.}\\ 
\addlinespace[\myxtraspc]
\eczhRefIndex{code:double_semion}%
\eczhListValue{\flmRefsHyperref{code:double_semion}{Double-semion stabilizer code}} & \eczhListValue{A 2D lattice modular-qudit stabilizer code with qudit dimension \(q=4\) that realizes the 2D double semion topological phase.
The code can be obtained from a \(\mathbb{Z}_4\) toric-code ground state by \flmRefsHyperref{ref410}{condensing} the emergent boson \(e^2 m^2\); in the stabilizer construction this condensation is implemented by two-body measurements \NoCaseChange{\protect\cite{cite405,cite414}}.
Its ground-state subspace can be mapped to that of the double-semion string-net model by a finite-depth quantum circuit with ancillas \NoCaseChange{\protect\cite{cite405}}.}\\ 
\addlinespace[\myxtraspc]
\eczhRefIndex{code:double_semion_string_net}%
\eczhListValue{\flmRefsHyperref{code:double_semion_string_net}{Double-semion string-net code}} & \eczhListValue{An \(XS\) stabilizer code that realizes the 2D double semion topological phase.
The model can be extended to other spatial dimensions \NoCaseChange{\protect\cite{cite588}}.}\\ 
\addlinespace[\myxtraspc]
\eczhRefIndex{code:fibonacci}%
\eczhListValue{\flmRefsHyperref{code:fibonacci}{Fibonacci string-net code}} & \eczhListValue{Quantum error correcting code associated with the Levin-Wen string-net model with the Fibonacci input category, admitting two types of encodings.}\\ 
\addlinespace[\myxtraspc]
\eczhRefIndex{code:rotor_4_2_2}%
\eczhListValue{\flmRefsHyperref{code:rotor_4_2_2}{Four-rotor code}} & \eczhListValue{\(\llbracket 4,2,2\rrbracket _{\mathbb Z}\) CSS rotor code that is an extension of the four-qubit code to the integer alphabet, i.e., the angular momentum states of a rotor.}\\ 
\addlinespace[\myxtraspc]
\eczhRefIndex{code:galois_color}%
\eczhListValue{\flmRefsHyperref{code:galois_color}{Galois-qudit color code}} & \eczhListValue{Extension of the color code to 2D lattices of Galois qudits.}\\ 
\addlinespace[\myxtraspc]
\eczhRefIndex{code:galois_topological}%
\eczhListValue{\flmRefsHyperref{code:galois_topological}{Galois-qudit surface code}} & \eczhListValue{Extension of the surface code to 2D lattices of Galois qudits.}\\ 
\addlinespace[\myxtraspc]
\eczhRefIndex{code:generalized_color}%
\eczhListValue{\flmRefsHyperref{code:generalized_color}{Generalized 2D color code}} & \eczhListValue{Member of a family of non-Abelian 2D topological codes, defined by a finite group \( G \), that serves as a generalization of the 2D color code (for which \(G=\mathbb{Z}_2 \times \mathbb{Z}_2\)).
Hamiltonian terms are built from \flmRefsHyperref{ref20}{group-based right- and left-multiplication \(X\)-type operators together with \(Z\)-type operators}.}\\ 
\addlinespace[\myxtraspc]
\eczhRefIndex{code:groupoid_surface}%
\eczhListValue{\flmRefsHyperref{code:groupoid_surface}{Groupoid toric code}} & \eczhListValue{Extension of the Kitaev surface code from Abelian groups to groupoids, i.e., multi-fusion categories in which every morphism is an isomorphism \NoCaseChange{\protect\cite{cite570}}.
Some models admit fracton-like features such as extensive ground-state degeneracy and excitations with restricted mobility.
The robustness of these features has not yet been established.}\\ 
\addlinespace[\myxtraspc]
\eczhRefIndex{code:hexagonal_cz}%
\eczhListValue{\flmRefsHyperref{code:hexagonal_cz}{Hexagonal \(CZ\) code}} & \eczhListValue{A hexagonal-lattice realization of the \(2+1\)D \(l=m=n=1\) cubic theory / Type-III \(\mathbb{Z}_2^3\) twisted quantum double phase.
Its stabilizers are products of Pauli-\(Z\) operators and \(CZ\) gates \NoCaseChange{\protect\cite[{Fig. 6}]{cite575}\protect\cite[{Fig. 3}]{cite589}}.
The ground-state subspace of the hexagonal \(CZ\) code realizes the topological order of the Type-III \(G=\mathbb{Z}^3_2\) Abelian TQD model \NoCaseChange{\protect\cite{cite575,cite576}}, which is the same topological order as the \(G=D_4\) non-Abelian quantum double \NoCaseChange{\protect\cite{cite577}}.
The stabilizers include \(CZ\) operators acting on hexagonal loops, but a reduced version exists where only two \(CZ\) gates act on each loop \NoCaseChange{\protect\cite{cite589}}.}\\ 
\addlinespace[\myxtraspc]
\eczhRefIndex{code:triangular_color}%
\eczhListValue{\flmRefsHyperref{code:triangular_color}{Honeycomb (6.6.6) color code}} & \eczhListValue{2D color code defined on a (typically triangular) patch of the 6.6.6 (honeycomb) tiling.
The usual triangular patch has three differently colored boundaries, encodes one logical qubit, and is local-Clifford equivalent to a folded surface/toric code with two smooth and two rough boundaries \NoCaseChange{\protect\cite{cite422}}.}\\ 
\addlinespace[\myxtraspc]
\eczhRefIndex{code:hopf_quantum_double}%
\eczhListValue{\flmRefsHyperref{code:hopf_quantum_double}{Hopf-algebra quantum-double code}} & \eczhListValue{Code whose codewords realize 2D gapped topological order defined on qudits valued in a Hopf algebra \(H\).
The code Hamiltonian is a generalization \NoCaseChange{\protect\cite{cite590,cite591}} of the quantum double model from group algebras to Hopf algebras, as anticipated by Kitaev \NoCaseChange{\protect\cite{cite423}}.
Boundaries of these models have been examined \NoCaseChange{\protect\cite{cite592,cite593}}.}\\ 
\addlinespace[\myxtraspc]
\eczhRefIndex{code:kitaev_chain}%
\eczhListValue{\flmRefsHyperref{code:kitaev_chain}{Kitaev chain code}} & \eczhListValue{A Majorana stabilizer code obtained from the ground-state subspace of the Kitaev Majorana chain in its fermionic topological phase \NoCaseChange{\protect\cite{cite558}}. Its codespace is stabilized by nearest-neighbor Majorana bilinears, while two unpaired edge Majoranas furnish one logical fermionic mode. Under parity-preserving noise, it behaves as the Majorana analogue of the repetition code \NoCaseChange{\protect\cite{cite559}}.}\\ 
\addlinespace[\myxtraspc]
\eczhRefIndex{code:kitaev_honeycomb}%
\eczhListValue{\flmRefsHyperref{code:kitaev_honeycomb}{Kitaev honeycomb code}} & \eczhListValue{Subsystem qubit stabilizer code underlying the Kitaev honeycomb model \NoCaseChange{\protect\cite{cite537,cite594}}.
Its gauge generators are the two-qubit \(XX\), \(YY\), and \(ZZ\) link operators on the three edge types of the honeycomb lattice \NoCaseChange{\protect\cite[{Sec. 3.2}]{cite594}}.
Its stabilizer group is generated by loop operators, and syndrome extraction can be reduced to ordered measurements of the two-qubit link operators \NoCaseChange{\protect\cite[{Sec. 3.2}]{cite594}}.
This is the \(q=2\) instance of the \(\mathbb{Z}_q^{(1)}\) subsystem code and does not encode any logical qubits \NoCaseChange{\protect\cite{cite594}\protect\cite[{Sec. 7.3}]{cite414}}.}\\ 
\addlinespace[\myxtraspc]
\eczhRefIndex{code:surface}%
\eczhListValue{\flmRefsHyperref{code:surface}{Kitaev surface code}} & \eczhListValue{A family of Abelian topological \flmRefsHyperref{code:css}{CSS stabilizer} codes
whose generators are few-body \(X\)-type and \(Z\)-type Pauli strings
associated to the stars and plaquettes, respectively, of a cellulation of a
two-dimensional surface (with a qubit located at each edge of the
cellulation).
Codewords correspond to ground states of the surface code Hamiltonian, and error operators create or annihilate pairs of anyonic charges or vortices.}\\ 
\addlinespace[\myxtraspc]
\eczhRefIndex{code:klein_bottle}%
\eczhListValue{\flmRefsHyperref{code:klein_bottle}{Klein-bottle surface code}} & \eczhListValue{A family of Kitaev surface codes on the non-orientable Klein bottle.}\\ 
\addlinespace[\myxtraspc]
\eczhRefIndex{code:mbq}%
\eczhListValue{\flmRefsHyperref{code:mbq}{Majorana box qubit}} & \eczhListValue{A family of Majorana stabilizer codes obtained by fixing the total fermion parity of \(n\) fermionic modes, equivalently \(2n\) Majorana zero modes, within the ground-state subspace of \(n\) Kitaev Majorana chain Hamiltonians.
The resulting positive-parity subspace encodes \(n-1\) logical qubits and has Majorana distance \(2\).}\\ 
\addlinespace[\myxtraspc]
\eczhRefIndex{code:matching}%
\eczhListValue{\flmRefsHyperref{code:matching}{Matching code}} & \eczhListValue{Member of a class of qubit stabilizer codes based on the Abelian phase of the Kitaev honeycomb model.}\\ 
\addlinespace[\myxtraspc]
\eczhRefIndex{code:qudit_3d_surface}%
\eczhListValue{\flmRefsHyperref{code:qudit_3d_surface}{Modular-qudit 3D surface code}} & \eczhListValue{A generalization of the 3D surface code to modular qudits.
Qudits are placed on edges, \(Z\)-type stabilizer generators are placed on square plaquettes oriented in all three directions, and \(X\)-type stabilizers are placed on the six edges neighboring every vertex \NoCaseChange{\protect\cite{cite459}}.}\\ 
\addlinespace[\myxtraspc]
\eczhRefIndex{code:qudit_surface}%
\eczhListValue{\flmRefsHyperref{code:qudit_surface}{Modular-qudit surface code}} & \eczhListValue{Extension of the surface code to prime-dimensional \NoCaseChange{\protect\cite{cite423,cite424}} and more general modular qudits.
Stabilizer generators are few-body \(X\)-type and \(Z\)-type Pauli strings associated to the stars and plaquettes, respectively, of a tessellation of a two-dimensional surface.
Since qudits have more than one \(X\) and \(Z\)-type operator, various sets of stabilizer generators can be defined.}\\ 
\addlinespace[\myxtraspc]
\eczhRefIndex{code:enriched_string_net}%
\eczhListValue{\flmRefsHyperref{code:enriched_string_net}{Multi-fusion string-net code}} & \eczhListValue{Family of codes resulting from the string-net construction but whose input is a unitary multi-fusion category (as opposed to a unitary fusion category).}\\ 
\addlinespace[\myxtraspc]
\eczhRefIndex{code:nonabelian_kitaev_honeycomb}%
\eczhListValue{\flmRefsHyperref{code:nonabelian_kitaev_honeycomb}{Non-Abelian Kitaev honeycomb code}} & \eczhListValue{Code whose logical subspace in the gapped non-Abelian phase of the Kitaev honeycomb model with a magnetic field is labeled by different fusion outcomes of Ising anyons \NoCaseChange{\protect\cite{cite537}}.}\\ 
\addlinespace[\myxtraspc]
\eczhRefIndex{code:real_projective_plane}%
\eczhListValue{\flmRefsHyperref{code:real_projective_plane}{Projective-plane surface code}} & \eczhListValue{A family of Kitaev surface codes on the non-orientable 2-dimensional compact manifold \(\mathbb{R}P^2\) (in contrast to a genus-\(g\) surface).
Whereas genus-\(g\) surface codes require \(2g\) logical qubits, qubit codes on \(\mathbb{R}P^2\) are made from a single logical qubit.}\\ 
\addlinespace[\myxtraspc]
\eczhRefIndex{code:quantum_double}%
\eczhListValue{\flmRefsHyperref{code:quantum_double}{Quantum-double code}} & \eczhListValue{Group-based code whose codewords realize 2D modular gapped topological order defined by a finite group \(G\).
The code's generators are few-body operators associated to the stars and plaquettes, respectively, of a tessellation of a two-dimensional surface (with a qudit of dimension \( |G| \) located at each edge of the tessellation).
The original Hamiltonian can be re-expressed via \flmRefsHyperref{ref20}{group-based right- and left-multiplication \(X\)-type as well as \(Z\)-type error} operators \NoCaseChange{\protect\cite[{Sec. 3.3}]{cite598}}.}\\ 
\addlinespace[\myxtraspc]
\eczhRefIndex{code:quantum_triple}%
\eczhListValue{\flmRefsHyperref{code:quantum_triple}{Quantum-triple code}} & \eczhListValue{Group-based code whose codewords realize 3D topological order defined by a finite group \(G\).}\\ 
\addlinespace[\myxtraspc]
\eczhRefIndex{code:rbh}%
\eczhListValue{\flmRefsHyperref{code:rbh}{Raussendorf-Bravyi-Harrington (RBH) cluster-state code}} & \eczhListValue{A three-dimensional cluster-state code defined on the bcc lattice (i.e., a cubic lattice with qubits on edges and faces).}\\ 
\addlinespace[\myxtraspc]
\eczhRefIndex{code:rotated_surface}%
\eczhListValue{\flmRefsHyperref{code:rotated_surface}{Rotated surface code}} & \eczhListValue{Variant of the surface code obtained by taking the medial graph of the surface code lattice (treated as a graph) and applying a procedure to construct the check operators \NoCaseChange{\protect\cite{cite425,cite426}\protect\cite[{Fig. 8}]{cite427}}.
On a square lattice, this amounts to a rotation by 45 degrees such that qubits are on vertices, and both \(X\)- and \(Z\)-type check operators occupy plaquettes in an alternating checkerboard pattern.}\\ 
\addlinespace[\myxtraspc]
\eczhRefIndex{code:488_color}%
\eczhListValue{\flmRefsHyperref{code:488_color}{Square-octagon (4.8.8) color code}} & \eczhListValue{2D color code defined on a patch of the 4.8.8 (square-octagon) tiling, which itself is obtained by applying a fattening procedure to the square lattice \NoCaseChange{\protect\cite{cite430}}.
An equivalent description uses the Tetrakis square tiling (a.k.a. the Union Jack lattice), which is dual to the 4.8.8 lattice \NoCaseChange{\protect\cite{cite431}}.
Among the three semiregular triangular 2D color-code families, the 4.8.8 family uses the fewest physical qubits for a given distance and is the only one of the three with transversal implementations of the full Clifford group \NoCaseChange{\protect\cite{cite432}}.}\\ 
\addlinespace[\myxtraspc]
\eczhRefIndex{code:string_net}%
\eczhListValue{\flmRefsHyperref{code:string_net}{String-net code}} & \eczhListValue{A non-stabilizer commuting-projector code whose codewords realize a 2D topological order rendered by a Turaev-Viro topological field theory.
The corresponding anyon theory is defined by a (multiplicity-free) unitary fusion category \( \mathcal{C} \).
The code is defined on a cell decomposition dual to a triangulation of a two-dimensional surface, with a qudit of dimension \( |\mathcal{C}| \) located at each edge of the decomposition.
For modular input categories, the emergent anyon theory is the doubled category \( \mathcal{C}\otimes\mathcal{C}^{*} \) \NoCaseChange{\protect\cite{cite599}}.
These models realize local topological order (LTO) \NoCaseChange{\protect\cite{cite600}}.}\\ 
\addlinespace[\myxtraspc]
\eczhRefIndex{code:spt}%
\eczhListValue{\flmRefsHyperref{code:spt}{Symmetry-protected topological (SPT) code}} & \eczhListValue{A code whose codewords form the ground-state or low-energy subspace of a code Hamiltonian realizing symmetry-protected topological (SPT) order.}\\ 
\addlinespace[\myxtraspc]
\eczhRefIndex{code:tetrahedral_color}%
\eczhListValue{\flmRefsHyperref{code:tetrahedral_color}{Tetrahedral color code}} & \eczhListValue{A 3D color code defined on a colored tetrahedron cut from a suitably colored BCC lattice \NoCaseChange{\protect\cite{cite475}}.
Qubits are placed on tetrahedra, on the triangles covering the tetrahedron faces, on the edges along the tetrahedron edges, and on the tetrahedron vertices.
The code has both string-like and sheet-like logical operators \NoCaseChange{\protect\cite{cite476}}.}\\ 
\addlinespace[\myxtraspc]
\eczhRefIndex{code:tetron}%
\eczhListValue{\flmRefsHyperref{code:tetron}{Tetron code}} & \eczhListValue{A \(\llbracket 2,1,2\rrbracket _{f}\) Majorana box qubit encoding a logical qubit into four Majorana modes, equivalently into the fixed-total-parity sector of two physical fermionic modes.
Four Majorana zero modes are the smallest aggregate that supports a qubit in a fixed fermion-parity sector \NoCaseChange{\protect\cite{cite401}}.
This code can be concatenated with various qubit codes such as surface codes and color codes.
Four-boundary Majorana surface-code patches are logical tetrons, i.e., higher-distance analogues of this physical tetron block \NoCaseChange{\protect\cite{cite402}}.}\\ 
\addlinespace[\myxtraspc]
\eczhRefIndex{code:three_fermion}%
\eczhListValue{\flmRefsHyperref{code:three_fermion}{Three-fermion (3F) Walker-Wang model code}} & \eczhListValue{A 3D lattice stabilizer code whose bulk realizes a 3D time-reversal SPT order \NoCaseChange{\protect\cite{cite477}} and whose gapped boundary supports the 2D three-fermion (3F) topological order.
The code can be used as a resource state for fault-tolerant MBQC \NoCaseChange{\protect\cite{cite478}}.}\\ 
\addlinespace[\myxtraspc]
\eczhRefIndex{code:subsystem_three_fermion}%
\eczhListValue{\flmRefsHyperref{code:subsystem_three_fermion}{Three-fermion (3F) subsystem code}} & \eczhListValue{2D subsystem stabilizer code whose low-energy excitations realize the three-fermion anyon theory \NoCaseChange{\protect\cite{cite601,cite602,cite603}}.
One version uses two qubits at each site \NoCaseChange{\protect\cite{cite414}}, while other manifestations utilize a single qubit per site and only weight-two (two-body) interactions \NoCaseChange{\protect\cite{cite602,cite604}}.
All are expected to be equivalent to each other via a local constant-depth \flmRefsHyperref{ref409}{Clifford circuit}.}\\ 
\addlinespace[\myxtraspc]
\eczhRefIndex{code:topological}%
\eczhListValue{\flmRefsHyperref{code:topological}{Topological code}} & \eczhListValue{A code whose codewords form the ground-state or low-energy subspace of a (typically geometrically local) code Hamiltonian realizing a topological phase.
A topological phase may be \textit{bosonic} or \textit{fermionic}, i.e., constructed out of underlying subsystems whose operators commute or anti-commute with each other, respectively.
Unless otherwise noted, the phases discussed are bosonic.}\\ 
\addlinespace[\myxtraspc]
\eczhRefIndex{code:toric}%
\eczhListValue{\flmRefsHyperref{code:toric}{Toric code}} & \eczhListValue{Version of the Kitaev surface code on a square lattice with periodic boundary conditions, encoding two logical qubits.
Being the first manifestation of the surface code, "toric code" is often an alternative name for the general construction.
\textit{Twisted toric code} \NoCaseChange{\protect\cite[{Fig. 8}]{cite434}} refers to the construction on a torus with twisted (a.k.a. shifted) boundary conditions.
In the original Hamiltonian construction, open Pauli-\(Z\) and Pauli-\(X\) strings create pairs of electric charges and magnetic vortices, and braiding one type around the other yields the nontrivial Abelian anyonic phase \NoCaseChange{\protect\cite{cite423}}.}\\ 
\addlinespace[\myxtraspc]
\eczhRefIndex{code:4612_color}%
\eczhListValue{\flmRefsHyperref{code:4612_color}{Truncated trihexagonal (4.6.12) color code}} & \eczhListValue{2D color code defined on a (typically triangular) patch of the 4.6.12 (truncated trihexagonal or square-hexagon-dodecagon) tiling.}\\ 
\addlinespace[\myxtraspc]
\eczhRefIndex{code:tqd}%
\eczhListValue{\flmRefsHyperref{code:tqd}{Twisted quantum double (TQD) code}} & \eczhListValue{Code whose codewords realize a 2D topological order rendered by a Chern-Simons topological field theory.
The corresponding anyon theory is defined by a finite group \(G\) and a 3-cocycle \(\omega\in H^3( G, U(1) )\) \NoCaseChange{\protect\cite{cite605,cite571,cite405}}.
Canonical TQD models \NoCaseChange{\protect\cite{cite571}} are defined on group-valued qudits.}\\ 
\addlinespace[\myxtraspc]
\eczhRefIndex{code:tqt}%
\eczhListValue{\flmRefsHyperref{code:tqt}{Twisted quantum triple (TQT) code}} & \eczhListValue{Group-based code realizing a 3D topological order rendered by a Dijkgraaf-Witten gauge theory.
The corresponding anyon theory is defined by a finite group \(G\) and a Type-IV four-cocycle \(\omega\).
Canonical TQT models \NoCaseChange{\protect\cite{cite586,cite606}} and other formulations whose ground states are in the same phase are all defined on group-valued qudits.}\\ 
\addlinespace[\myxtraspc]
\eczhRefIndex{code:yetter_gauge_theory}%
\eczhListValue{\flmRefsHyperref{code:yetter_gauge_theory}{Two-gauge theory code}} & \eczhListValue{A code whose codewords realize lattice two-gauge theory \NoCaseChange{\protect\cite{cite607,cite608,cite609,cite610,cite611,cite612,cite613,cite614,cite615}} for a finite \textit{two-group} (a.k.a. a \textit{crossed module}) in arbitrary spatial dimension.
There exist several lattice-model formulations in arbitrary spatial dimension \NoCaseChange{\protect\cite{cite616,cite617}} as well as explicitly in 3D \NoCaseChange{\protect\cite{cite618,cite619,cite620,cite621}} and 4D \NoCaseChange{\protect\cite{cite621}}, with the 3D case realizing the Yetter model \NoCaseChange{\protect\cite{cite622,cite623,cite624,cite625}}.}\\ 
\addlinespace[\myxtraspc]
\eczhRefIndex{code:walker_wang}%
\eczhListValue{\flmRefsHyperref{code:walker_wang}{Walker-Wang model code}} & \eczhListValue{A non-stabilizer commuting-projector 3D topological code defined by a unitary braided fusion category \( \mathcal{C} \) (also known as a unitary premodular category).
The code is defined on a cubic lattice that is resolved to be trivalent, with a qudit of dimension \( |\mathcal{C}| \) located at each edge.
The codespace is the ground-state subspace of the Walker-Wang model Hamiltonian \NoCaseChange{\protect\cite{cite628}} and realizes the Crane-Yetter model \NoCaseChange{\protect\cite{cite629,cite630,cite631}}.
A single-state version of the code provides a resource state for MBQC \NoCaseChange{\protect\cite{cite478}}.}\\ 
\addlinespace[\myxtraspc]
\eczhRefIndex{code:xysurface}%
\eczhListValue{\flmRefsHyperref{code:xysurface}{XY surface code}} & \eczhListValue{A variant of the surface code whose generators are \(XXXX\)  and \(YYYY\), obtained by mapping \(Z \to Y\) in the surface code.
While this code is equivalent to a CSS surface code with the same distance, other properties like noise-bias performance can differ significantly.}\\ 
\addlinespace[\myxtraspc]
\eczhRefIndex{code:xyz_hexagonal}%
\eczhListValue{\flmRefsHyperref{code:xyz_hexagonal}{XYZ\(^2\) hexagonal stabilizer code}} & \eczhListValue{An instance of the matching code based on the Kitaev honeycomb model. It is described on a honeycomb tiling with \(XYZXYZ\) stabilizers on each hexagonal plaquette. Each vertical pair of qubits has an \(XX\), \(YY\), or \(ZZ\) link stabilizer depending on the orientation of the plaquette stabilizers.}\\ 
\addlinespace[\myxtraspc]
\eczhRefIndex{code:4d_13_surface}%
\eczhListValue{\flmRefsHyperref{code:4d_13_surface}{\((1,3)\) 4D toric code}} & \eczhListValue{A generalization of the Kitaev surface code defined on a 4D lattice.
The code is called a \((1,3)\) toric code because it admits 1D \(Z\)-type and 3D \(X\)-type logical operators.
In the hypercubic lattice version, qubits are placed on edges, each \(Z\)-type stabilizer generator is supported on cubes on the boundary of a hypercube, and \(X\)-type stabilizers are placed on the edges neighboring every vertex \NoCaseChange{\protect\cite{cite479}}.}\\ 
\addlinespace[\myxtraspc]
\eczhRefIndex{code:4d_surface}%
\eczhListValue{\flmRefsHyperref{code:4d_surface}{\((2,2)\) Loop toric code}} & \eczhListValue{A generalization of the Kitaev surface code defined on a 4D lattice.
The code is called a \((2,2)\) toric code because it admits 2D membrane \(Z\)-type and \(X\)-type logical operators.
Both types of operators create 1D (i.e., loop) excitations at their edges.
The code serves as a self-correcting quantum memory \NoCaseChange{\protect\cite{cite480,cite481}}.}\\ 
\addlinespace[\myxtraspc]
\eczhRefIndex{code:enriched_walker_wang}%
\eczhListValue{\flmRefsHyperref{code:enriched_walker_wang}{\(G\)-enriched Walker-Wang model code}} & \eczhListValue{A 3D topological code defined by a unitary \(G\)-crossed braided fusion category \( \mathcal{C} \) \NoCaseChange{\protect\cite{cite632,cite633}}, where \(G\) is a finite group.
The model realizes TQFTs that include two-gauge theories, those behind Walker-Wang models, as well as the Kashaev TQFT \NoCaseChange{\protect\cite{cite634,cite635}}.
It has been generalized to include domain walls \NoCaseChange{\protect\cite{cite636}}.}\\ 
\addlinespace[\myxtraspc]
\eczhRefIndex{code:stab_15_1_3}%
\eczhListValue{\flmRefsHyperref{code:stab_15_1_3}{\(\llbracket 15,1,3\rrbracket \) quantum RM code}} & \eczhListValue{A \(\llbracket 15,1,3\rrbracket \) quantum Reed-Muller code that is most easily thought of as a tetrahedral 3D color code.
It can be constructed as a CSS code from the \([15,5,8]\) punctured Reed-Muller code and its even subcode, which explains its transversal \(T^\dagger\) gate \NoCaseChange{\protect\cite{cite398}}.}\\ 
\addlinespace[\myxtraspc]
\eczhRefIndex{code:stab_17_1_5}%
\eczhListValue{\flmRefsHyperref{code:stab_17_1_5}{\(\llbracket 17,1,5\rrbracket \) 4.8.8 color code}} & \eczhListValue{Seventeen-qubit doubly even 2D color code that admits a transversal implementation of the logical Clifford group.
The smallest distance-five CSS code has \(n=17\) \NoCaseChange{\protect\cite{cite444}}.
It is also a normal self-dual CSS code whose transversal Hadamard acts logically, making it suitable as an inner code for fifth-order magic-state distillation \NoCaseChange{\protect\cite{cite101}}.}\\ 
\addlinespace[\myxtraspc]
\eczhRefIndex{code:css_4_1_2}%
\eczhListValue{\flmRefsHyperref{code:css_4_1_2}{\(\llbracket 4,1,2\rrbracket \) Leung-Nielsen-Chuang-Yamamoto (LNCY) code}} & \eczhListValue{A four-qubit CSS stabilizer code that is the only qubit CSS code with such parameters.}\\ 
\addlinespace[\myxtraspc]
\eczhRefIndex{code:stab_4_2_2}%
\eczhListValue{\flmRefsHyperref{code:stab_4_2_2}{\(\llbracket 4,2,2\rrbracket \) Four-qubit code}} & \eczhListValue{A four-qubit hyperbolic self-dual CSS stabilizer code that is the smallest two-logical-qubit stabilizer code to detect a single-qubit error.
It is unique for its parameters \NoCaseChange{\protect\cite[{Thm. 8}]{cite446}}.}\\ 
\addlinespace[\myxtraspc]
\eczhRefIndex{code:group_4_2_2}%
\eczhListValue{\flmRefsHyperref{code:group_4_2_2}{\(\llbracket 4,2,2\rrbracket _{G}\) four group-qudit code}} & \eczhListValue{\(\llbracket 4,2,2\rrbracket _{G}\) group quantum code that is an extension of the four-qubit code to group-valued qudits.}\\ 
\addlinespace[\myxtraspc]
\eczhRefIndex{code:stab_5_1_2}%
\eczhListValue{\flmRefsHyperref{code:stab_5_1_2}{\(\llbracket 5,1,2\rrbracket \) rotated surface code}} & \eczhListValue{A rotated surface code on one rung of a ladder, with one qubit on the rung, and four qubits surrounding it. This is the smallest code that implements a fault-tolerant logical \(S\) gate using a diagonal depth-one Clifford circuit \NoCaseChange{\protect\cite{cite447}}.}\\ 
\addlinespace[\myxtraspc]
\eczhRefIndex{code:stab_6_2_2}%
\eczhListValue{\flmRefsHyperref{code:stab_6_2_2}{\(\llbracket 6,2,2\rrbracket \) \(C_6\) code}} & \eczhListValue{Error-detecting normal self-dual CSS code on three qubit pairs that encodes a logical qubit pair and detects any error acting on one pair \NoCaseChange{\protect\cite{cite448}}.
In Knill's \(C_4/C_6\) architecture, this code is used at the second and higher concatenation levels.}\\ 
\addlinespace[\myxtraspc]
\eczhRefIndex{code:stab_6_4_2}%
\eczhListValue{\flmRefsHyperref{code:stab_6_4_2}{\(\llbracket 6,4,2\rrbracket \) error-detecting code}} & \eczhListValue{Self-complementary six-qubit code with rate \(2/3\) that is unique for its parameters, up to equivalence \NoCaseChange{\protect\cite[{Tab. III}]{cite449}}.
Concatenations of this code with itself yield the \(\llbracket 6^r,4^r,2^r\rrbracket \) level-\(r\) \textit{many-hypercube} code \NoCaseChange{\protect\cite{cite450}}.}\\ 
\addlinespace[\myxtraspc]
\eczhRefIndex{code:steane}%
\eczhListValue{\flmRefsHyperref{code:steane}{\(\llbracket 7,1,3\rrbracket \) Steane code}} & \eczhListValue{A \(\llbracket 7,1,3\rrbracket \) self-dual CSS code that is the smallest qubit CSS code to correct a single-qubit error \NoCaseChange{\protect\cite{cite451}}.
The code is constructed using the classical binary \([7,4,3]\) Hamming code for protecting against both \(X\) and \(Z\) errors.}\\ 
\addlinespace[\myxtraspc]
\eczhRefIndex{code:stab_8_3_2}%
\eczhListValue{\flmRefsHyperref{code:stab_8_3_2}{\(\llbracket 8,3,2\rrbracket \) Smallest interesting color code}} & \eczhListValue{Smallest 3D color code whose physical qubits lie on vertices of a cube and which admits a (weakly) transversal \(CCZ\) gate.}\\ 
\addlinespace[\myxtraspc]
\eczhRefIndex{code:shor_nine}%
\eczhListValue{\flmRefsHyperref{code:shor_nine}{\(\llbracket 9,1,3\rrbracket \) Shor code}} & \eczhListValue{Nine-qubit \flmRefsHyperref{code:css}{CSS code} that is the first quantum error-correcting code \NoCaseChange{\protect\cite[{ID 8802}]{cite453}}.
Among indecomposable \(\llbracket 9,1,3\rrbracket \) CSS codes, the Shor code has the largest automorphism group \NoCaseChange{\protect\cite{cite454}}.}\\ 
\addlinespace[\myxtraspc]
\eczhRefIndex{code:surface-17}%
\eczhListValue{\flmRefsHyperref{code:surface-17}{\(\llbracket 9,1,3\rrbracket \) Surface-17 code}} & \eczhListValue{A \(\llbracket 9,1,3\rrbracket \) rotated surface code named for the sum of its 9 data qubits and 8 syndrome qubits.
It is one of the four inequivalent CSS gauge fixings of the nine-qubit Bacon-Shor code \NoCaseChange{\protect\cite{cite454}}.
It uses the smallest number of qubits to perform fault-tolerant error correction on a surface code with parallel syndrome extraction.}\\ 
\addlinespace[\myxtraspc]
\eczhRefIndex{code:zthree_znine}%
\eczhListValue{\flmRefsHyperref{code:zthree_znine}{\(\mathbb{Z}_3\times\mathbb{Z}_9\)-fusion subsystem code}} & \eczhListValue{Modular-qudit 2D subsystem stabilizer code whose low-energy excitations realize a non-modular anyon theory with \(\mathbb{Z}_3\times\mathbb{Z}_9\) fusion rules.
Encodes two qutrits when put on a torus.}\\ 
\addlinespace[\myxtraspc]
\eczhRefIndex{code:qudit_znone}%
\eczhListValue{\flmRefsHyperref{code:qudit_znone}{\(\mathbb{Z}_q^{(1)}\) subsystem code}} & \eczhListValue{Modular-qudit subsystem code, based on the Kitaev honeycomb model \NoCaseChange{\protect\cite{cite537}} and its generalization \NoCaseChange{\protect\cite{cite637}}, that is characterized by the \(\mathbb{Z}_q^{(1)}\) anyon theory \NoCaseChange{\protect\cite{cite638}}, which is modular for odd prime \(q\) and non-modular otherwise. Encodes a single \(q\)-dimensional qudit when put on a torus for odd \(q\), and a \(q/2\)-dimensional qudit for even \(q\). This code can be constructed using geometrically local gauge generators, but does not admit geometrically local stabilizer generators. For \(q=2\), the code reduces to the subsystem code underlying the Kitaev honeycomb model code as well as the honeycomb Floquet code.}\\ 
\end{tabularx}\endgroup
\eczlistdomainsection{Classical-into-quantum code families}

\eczcodelist{coherent_state_c-q}{Coherent-state c-q codes}%

\eczhCodeListAutoDescription{All descendants of \flmRefsCref{code:coherent_state_c-q}.}%

\eczhIncludeCodeGraph{Bare}{scale=0.5}{\columnwidth}{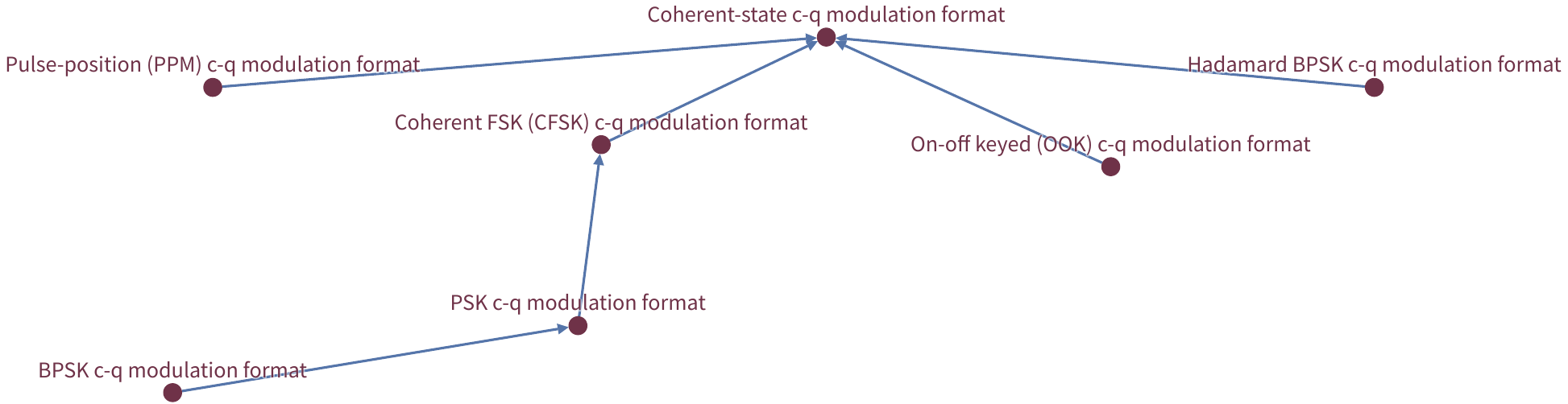}{Coherent-state c-q codes}{https://errorcorrectionzoo.org/code_graph#J\%7B\%22displayMode\%22\%3A\%22subset\%22\%2C\%22modeSubsetOptions\%22\%3A\%7B\%22codeIds\%22\%3A\%5B\%22quantum_bpsk\%22\%2C\%22quantum_fsk\%22\%2C\%22coherent_state_c-q\%22\%2C\%22quantum_hadamard_bpsk\%22\%2C\%22quantum_ook\%22\%2C\%22quantum_psk\%22\%2C\%22quantum_ppm\%22\%5D\%2C\%22reusePreviousLayoutPositions\%22\%3Afalse\%2C\%22showIntermediateConnectingNodes\%22\%3Atrue\%2C\%22connectingNodesMaxDepth\%22\%3A15\%2C\%22connectingNodesPathMaxLength\%22\%3A20\%2C\%22connectingNodesMaxExtraDepth\%22\%3A3\%2C\%22connectingNodesOnlyKeepPathsWithAdditionalLength\%22\%3A1\%2C\%22connectingNodesToDomainsAndKingdoms\%22\%3Afalse\%2C\%22connectingNodesEdgeLengthsByType\%22\%3A\%7B\%22primaryParent\%22\%3A1\%2C\%22secondaryParent\%22\%3A4\%2C\%22cousin\%22\%3A6\%7D\%2C\%22nodeIds\%22\%3A\%5B\%5D\%7D\%2C\%22highlightImportantNodes\%22\%3A\%7B\%22highlightImportantNodes\%22\%3Afalse\%2C\%22highlightPrimaryParents\%22\%3Afalse\%2C\%22highlightRootConnectingEdges\%22\%3Afalse\%7D\%7D}

\begingroup
\small
\eczhBreakableDashes
\renewcommand\arraystretch{1.05}
\edef\myxtraspc{\eczListAddVSpaceXtraXtraStretch}
\begin{tabularx}{\linewidth}{>{\raggedright\arraybackslash}p{\eczListColWidth{name}} >{\hsize=1.0000\hsize }X}
\toprule
\eczListColTitle{Code} & \eczListColTitle{Description} \\
\midrule
\endfirsthead
\toprule
\eczListColTitleContinued{Code} & \eczListColTitleContinued{Description} \\
\midrule
\endhead
\bottomrule
\endfoot
\eczhRefIndex{code:quantum_bpsk}%
\eczhListValue{\flmRefsHyperref{code:quantum_bpsk}{BPSK c-q modulation format}} & \eczhListValue{Coherent-state c-q binary code encoding into two coherent states \(|\pm\alpha\rangle\) for complex \(\alpha\). A shifted version, with codewords \(\{|0\rangle,|\alpha\rangle\}\), is called binary amplitude modulation (BAM), The three-state subcode \(\{|\alpha,\alpha\rangle,|-\alpha,\alpha\rangle,|\alpha,-\alpha\rangle\}\) of two-mode BPSK is called the \textit{single-degeneracy code} \NoCaseChange{\protect\cite{cite874}}.}\\ 
\addlinespace[\myxtraspc]
\eczhRefIndex{code:quantum_fsk}%
\eczhListValue{\flmRefsHyperref{code:quantum_fsk}{Coherent FSK (CFSK) c-q modulation format}} & \eczhListValue{\flmRefsHyperref{code:coherent_state_c-q}{Coherent-state c-q code} encoding into coherent states that are frequency-shifted with certain initial relative phase.}\\ 
\addlinespace[\myxtraspc]
\eczhRefIndex{code:coherent_state_c-q}%
\eczhListValue{\flmRefsHyperref{code:coherent_state_c-q}{Coherent-state c-q modulation format}} & \eczhListValue{Bosonic c-q code whose codewords form a constellation of coherent states.
Encodes classical symbols into coherent states for transmission over a quantum channel and decoding with a quantum-enhanced \textit{receiver}.}\\ 
\addlinespace[\myxtraspc]
\eczhRefIndex{code:quantum_hadamard_bpsk}%
\eczhListValue{\flmRefsHyperref{code:quantum_hadamard_bpsk}{Hadamard BPSK c-q modulation format}} & \eczhListValue{Multimode coherent-state c-q code that is a concatenation of a Hadamard code with a BPSK c-q code.
Its codewords are \(n\)-mode coherent states whose components \(\pm\alpha\) are arranged according to rows of a Hadamard matrix.}\\ 
\addlinespace[\myxtraspc]
\eczhRefIndex{code:quantum_ook}%
\eczhListValue{\flmRefsHyperref{code:quantum_ook}{On-off keyed (OOK) c-q modulation format}} & \eczhListValue{Coherent-state c-q binary code whose encoding is either in the vacuum \(|0\rangle\) or in a nonzero coherent state \(|\alpha\rangle\).}\\ 
\addlinespace[\myxtraspc]
\eczhRefIndex{code:quantum_psk}%
\eczhListValue{\flmRefsHyperref{code:quantum_psk}{PSK c-q modulation format}} & \eczhListValue{Coherent-state c-q \(q\)-ary code whose \(j\)th codeword corresponds to a coherent state whose phase is the \(j\)th multiple of \(2\pi/q\). These states are also called geometrically uniform states (GUS) \NoCaseChange{\protect\cite{cite875}}.}\\ 
\addlinespace[\myxtraspc]
\eczhRefIndex{code:quantum_ppm}%
\eczhListValue{\flmRefsHyperref{code:quantum_ppm}{Pulse-position (PPM) c-q modulation format}} & \eczhListValue{A \(q\)-PPM c-q code is a coherent-state c-q code whose \(j\)th codeword corresponds to a tensor-product state of zero-amplitude coherent states at all modes except mode \(j\).
For example, a 3-PPM encoding corresponds to the three-mode states \(|\alpha\rangle|0\rangle|0\rangle\), \(|0\rangle|\alpha\rangle|0\rangle\), and \(|0\rangle|0\rangle|\alpha\rangle\) for some complex \(\alpha\).
The dual of a PPM code is obtained by the exchange \(0\leftrightarrow\alpha\).}\\ 
\end{tabularx}\endgroup
\part{Codes in the Classical Domain}
\onecolumngrid

\begin{eczEpigraph}
\begin{quote}
\flmQuoteSetup{quote}%
It has been only in the past two or three years that the theory of error-correcting codes has advanced sufficiently to provide the basis for practical systems. Computer technology also has continued to advance, decreasing the cost and increasing the speed of equipment used to implement error correction. At the same time, there is a growing need for communication channels of extreme reliability for computers and automatic control systems. As the need grows and as coding theory and computer technology continue to develop, error detection and error correction will play an essential role in the success of complex systems.
\flmQuoteAttributed{W. Wesley Peterson, 1962}
\end{quote}
\end{eczEpigraph}

\section{Property codes}

\twocolumngrid

\eczcode{t-designs}{\(t\)-design}{}
\codefieldsection{Alternative Names}
\begin{eczvaluelist}
\item\relax Quadrature
\item\relax Cubature
\item\relax Averaging set
\end{eczvaluelist}
\eczhIndexCodeAliasName{t-designs}{Quadrature}
\eczhIndexCodeAliasName{t-designs}{Cubature}
\eczhIndexCodeAliasName{t-designs}{Averaging set}
\codefieldsection{Description}
A code whose codewords are uniformly distributed in a way that is useful for determining averages of polynomials over the code's underlying space \(X\).
In that way, the codewords form an approximation of the space.
A code is a design on \(X\) of \textit{strength} \(t\), i.e., a \(t\)-design on \(X\), if the average of any polynomial of degree up to \(t\) over its codewords is equal to the uniform average over all of \(X\).

As such, a design can be used to determine the average of degree-\(\leq t\) polynomials \(p\) over \(X\),
\flmMathEnvironment{align}{}{
  \int_{X}\textnormal{d}xp(x)={\textstyle \frac{1}{|D|}}\sum_{x\in D}p(x)~,
}
where the integral is over \(X\) (given some measure \(d x\)), while the sum is over the design \(D\subset X\).
A \textit{weighted design} is a design for which each term \(p(x)\) in the above sum must be multiplied by a weight \(w(x)\) in order to be equal to the left-hand side.
The most well-known examples of weighted designs are exact \textit{Gaussian quadrature} or \textit{cubature} formulas for integration over the reals \NoCaseChange{\protect\cite{cite379,cite876,cite877,cite878,cite879}}, \(X = \mathbb{R}^n\) (with appropriate measure); these tend to be weighted designs.

Fixed-weight codewords of a binary code can form a design on \(X\) being a Johnson space \(J(n,w)\), i.e., the space of length-\(n\) binary strings of weight \(w\).
Such a design is called a \flmRefsHyperref{code:combinatorial_design}{combinatorial design} (a.k.a. block design or covering design) \NoCaseChange{\protect\cite{cite880}}, which includes Steiner systems as a special case.
Other designs exist when \(X\) is \(q\)-ary Hamming space (where they are called \flmRefsHyperref{code:orthogonal_array}{orthogonal arrays}), ordered Hamming space \NoCaseChange{\protect\cite{cite214,cite215}}, \(q\)-Johnson space \NoCaseChange{\protect\cite{cite881,cite882}} (where they are called \flmRefsHyperref{code:subspace_design}{subspace designs}), or a sphere \NoCaseChange{\protect\cite{cite385}} (where they are called \flmRefsHyperref{code:spherical_design}{spherical designs}).

Complex projective designs are designs on complex projective space, i.e., the space of all pure quantum states \NoCaseChange{\protect\cite{cite883,cite884,cite885}}.
Generalizations to infinite-dimensional spaces yield rigged designs, or more colloquially, continuous-variable (CV) designs \NoCaseChange{\protect\cite{cite886}}, which can be used as operator-valued measures for the space of bosonic quantum states (i.e., Schwartz space over the reals).

Designs also exist on groups.
Designs on the unitary (projective unitary) group are called \flmRefsHyperref{code:unitary_design}{strong unitary (unitary) designs} \NoCaseChange{\protect\cite{cite887,cite888,cite889,cite890}}, while \(t\)-designs on the permutation group are called permutation \(t\)-designs \NoCaseChange{\protect\cite{cite891}} (a.k.a. \(t\)-wise independent permutations).

Other notable designs not included below include torus designs \NoCaseChange{\protect\cite{cite892,cite893}}, simplex designs \NoCaseChange{\protect\cite{cite894,cite895,cite896,cite897}},  quantum-channel designs \NoCaseChange{\protect\cite{cite898}}, designs on vertex operator algebras (a.k.a. conformal designs) \NoCaseChange{\protect\cite{cite899}}, and fractional designs \NoCaseChange{\protect\cite{cite900}}.
Existence has been proven for combinatorial designs \NoCaseChange{\protect\cite{cite901,cite902,cite903,cite904,cite905}}, subspace designs \NoCaseChange{\protect\cite{cite906,cite907}}, as well as designs on continuous topological spaces \NoCaseChange{\protect\cite{cite908,cite909,cite910,cite911}}.

\codefieldsection{Notes}
\begin{eczvaluelist}
\item\relax See books \NoCaseChange{\protect\cite{cite379,cite163}} for tables of various designs.
\end{eczvaluelist}
\codefieldsection{Children}
\begin{eczvaluelist}
\item\relax
\flmRefsHyperref[eczindexfamilyrel]{code:subspace_design}{Subspace design} --- Subspace designs are designs on the finite-field Grassmannian (a.k.a. \(q\)-Johnson space or \(q\)-Johnson association scheme) \NoCaseChange{\protect\cite{cite912}\protect\cite[{Sec. 8.6}]{cite913}}.
\item\relax
\flmRefsHyperref[eczindexfamilyrel]{code:unitary_design}{Unitary \(t\)-design} --- Unitary \(t\)-designs are designs on the unitary group \(U(N)\).
\item\relax
\flmRefsHyperref[eczindexfamilyrel]{code:delsarte_optimal}{Sharp configuration} --- Sharp configurations attain a universal bound expressed in terms of the minimal distance, the number of distances between codewords, and the strength of the design formed by the codewords.
\item\relax
\flmRefsHyperref[eczindexfamilyrel]{code:orthogonal_array}{Orthogonal array (OA)} --- Orthogonal arrays are designs on Hamming space \(\mathbb{F}_q^n\) (a.k.a. the Hamming association scheme) \NoCaseChange{\protect\cite{cite880,cite912,cite914}\protect\cite[{Exam. 1}]{cite226}}; see also Ref. \NoCaseChange{\protect\cite{cite915}}.
\item\relax
\flmRefsHyperref[eczindexfamilyrel]{code:spherical_design}{Spherical design} --- Spherical designs are designs on real or complex spheres.
\end{eczvaluelist}
\codefieldsection{Cousins}
\begin{eczvaluelist}
\item\relax
\flmRefsHyperref[eczindexfamilyrel]{code:ecc}{Error-correcting code (ECC)} --- ECCs and \(t\)-designs on two-point homogeneous spaces are intimately related via association schemes \NoCaseChange{\protect\cite{cite226,cite916}}.
\item\relax
\flmRefsHyperref[eczindexfamilyrel]{code:2pt_homogeneous}{Two-point homogeneous-space code} --- Designs exist on compact connected two-point homogeneous spaces \NoCaseChange{\protect\cite{cite226,cite914,cite917}}. ECCs and \(t\)-designs on two-point homogeneous spaces are intimately related via association schemes \NoCaseChange{\protect\cite{cite226,cite916}}.
\item\relax
\flmRefsHyperref[eczindexfamilyrel]{code:barnes_wall}{Barnes-Wall (BW) lattice} --- BW lattices support Grassmannian 6-designs \NoCaseChange{\protect\cite{cite918}}.
\item\relax
\flmRefsHyperref[eczindexfamilyrel]{code:binary_permutation}{Code in permutations} --- The \(GA(n,\mathbb{F}_q)\) group is a permutation 2-design for general \(q\), and a 3-design for \(q=2\). This follows from the fact that the group acts transitively on ordered pairs of distinct points, and on ordered triples of distinct points for \(q=2\) \NoCaseChange{\protect\cite{cite919}}.
\item\relax
\flmRefsHyperref[eczindexfamilyrel]{code:complex_projective}{Complex projective space code} --- Pure quantum states in an \((N+1)\)-dimensional Hilbert space are parameterized by points in the complex projective space \(\mathbb{C}P^N\). As such, complex projective designs are designs on the space of pure quantum states \NoCaseChange{\protect\cite{cite883,cite884,cite885}}. Symmetric informationally complete quantum measurements (SIC-POVMs) \NoCaseChange{\protect\cite{cite920,cite883}} and mutually unbiased bases (MUBs) \NoCaseChange{\protect\cite{cite921,cite922,cite923,cite924,cite925,cite926}} are important examples of such designs.
\item\relax
\flmRefsHyperref[eczindexfamilyrel]{code:grassmannian}{Grassmannian code} --- Designs have been formulated on Grassmannians \NoCaseChange{\protect\cite{cite927,cite928,cite918,cite911,cite929}}.
\item\relax
\flmRefsHyperref[eczindexfamilyrel]{code:poset}{Poset code} --- Designs exist on ordered Hamming space \NoCaseChange{\protect\cite{cite214,cite215}}.
\item\relax
\flmRefsHyperref[eczindexfamilyrel]{code:esix_shell}{\(E_6\) lattice-shell code} --- The 36 antipodal pairs of the smallest \(E_6\) lattice shell form a 2-design in \(\mathbb{R}P^5\) \NoCaseChange{\protect\cite{cite119}}.
\item\relax
\flmRefsHyperref[eczindexfamilyrel]{code:polygon}{Polygon code} --- For even \(q\), the \(q/2\) sets of antipodal pairs of a \(q\)-gon form a tight design on the projective line \(\mathbb{R}P^1\) \NoCaseChange{\protect\cite{cite917}}.
\item\relax
\flmRefsHyperref[eczindexfamilyrel]{code:24cell}{24-cell code} --- The 12 antipodal pairs of the 24-cell code form a sharp configuration and a 2-design in \(\mathbb{R}P^3\) \NoCaseChange{\protect\cite{cite119}}.
\item\relax
\flmRefsHyperref[eczindexfamilyrel]{code:231_polytope}{\(2_{31}\) polytope code} --- The 63 antipodal pairs of vertices of the \(2_{31}\) polytope form a 2-design in \(\mathbb{R}P^6\) \NoCaseChange{\protect\cite{cite119}}.
\item\relax
\flmRefsHyperref[eczindexfamilyrel]{code:witting_polytope}{Witting polytope code} --- Antipodal pairs of points of the Witting polytope code form a 3-design in \(\mathbb{R}P^7\) \NoCaseChange{\protect\cite{cite119}}.
\item\relax
\flmRefsHyperref[eczindexfamilyrel]{code:coherent_constellation}{Coherent-state constellation code} --- Coherent-state constellation codes consisting of points from a Gaussian quadrature rule can be concatenated with quantum polar codes to achieve the Gaussian coherent information of the thermal noise channel \NoCaseChange{\protect\cite{cite930,cite931}}.
\item\relax
\flmRefsHyperref[eczindexfamilyrel]{code:number_phase}{Number-phase code} --- Pegg-Barnett phase states undergoing Kerr evolution, together with Fock states, form a rigged 2-design for a single mode \NoCaseChange{\protect\cite{cite886}}.
\item\relax
\flmRefsHyperref[eczindexfamilyrel]{code:oscillators}{Bosonic code} --- Gaussian states, under a particular measure, do not form rigged two-designs \NoCaseChange{\protect\cite{cite932}}.
\item\relax
\flmRefsHyperref[eczindexfamilyrel]{code:multimodegkp}{Gottesman-Kitaev-Preskill (GKP) code} --- GKP states on \(n\) modes and their displaced versions for all possible lattices form a rigged 2-design for all \(n\) \NoCaseChange{\protect\cite{cite933}}.
\item\relax
\flmRefsHyperref[eczindexfamilyrel]{code:cluster_state}{Cluster-state code} --- Kerdock codes correspond to cluster states, and the corresponding Clifford-group automorphisms of this set form a particular group \NoCaseChange{\protect\cite{cite934}} that is a unitary 2-design on \(U(2^n)\) \NoCaseChange{\protect\cite{cite935}}. As such, cluster states form complex projective 2-designs on \(\mathbb{C}P^{2^n-1}\). These are useful in matrix-vector multiplication \NoCaseChange{\protect\cite{cite936}}.
\item\relax
\flmRefsHyperref[eczindexfamilyrel]{code:qubit_stabilizer}{Qubit stabilizer code} --- Stabilizer states on \(n\) qubits form complex projective 3-designs, but not 4-designs, on \(\mathbb{C}P^{2^n-1}\) \NoCaseChange{\protect\cite{cite937}}. The \flmRefsHyperref{ref409}{Clifford group} is a unitary 2-design \NoCaseChange{\protect\cite{cite938}} and a 3-design \NoCaseChange{\protect\cite{cite940,cite941}\protect\cite[{Thm. 1.6(B)}]{cite939}\protect\cite[{pg. 191}]{cite42}} on \(U(2^n)\). The \(\llbracket 2m,2m-2,2\rrbracket \) code for \(2m\) being a multiple of four obstructs the Clifford group from being a 4-design \NoCaseChange{\protect\cite{cite801}}.
\item\relax
\flmRefsHyperref[eczindexfamilyrel]{code:qudit_stabilizer}{Modular-qudit stabilizer code} --- Stabilizer states on \(n\) prime-dimensional qudits form complex projective 2-designs on \(\mathbb{C}P^{p^n-1}\), and they form 3-designs if and only if \(p=2\) \NoCaseChange{\protect\cite{cite937}}. The prime-qudit Clifford group is a unitary 2-design on \(U(p^n)\) \NoCaseChange{\protect\cite{cite942}}.
\item\relax
\flmRefsHyperref[eczindexfamilyrel]{code:galois_stabilizer}{Galois-qudit stabilizer code} --- Stabilizer states on \(n\) Galois qubits form 2-designs on complex projective spaces \(\mathbb{C}P^{p^{mn}}\) \NoCaseChange{\protect\cite{cite943}}.
\end{eczvaluelist}
\eczhbkcontributors{ Greg Kuperberg, Alexander Barg, \eczhuVVA }
\endeczcode

\eczcode{multiple_erasure_lrc}{\(t\)-erasure LRC}{}
\codefieldsection{Alternative Names}
\begin{eczvaluelist}
\item\relax Multiple-erasure LRC
\end{eczvaluelist}
\eczhIndexCodeAliasName{multiple_erasure_lrc}{Multiple-erasure LRC}
\codefieldsection{Description}
A code which admits local recoverability against more than one coordinate erasure, typically up to some specified number \(t\) of erasures per local group.

\codefieldsection{Parent}
\begin{eczvaluelist}
\item\relax
\flmRefsHyperref[eczindexfamilyrel]{code:locally_recoverable}{Locally recoverable code (LRC)} --- A \(t\)-erasure LRC is locally recoverable by definition since \(t > 1\).
\end{eczvaluelist}
\codefieldsection{Child}
\begin{eczvaluelist}
\item\relax
\flmRefsHyperref[eczindexfamilyrel]{code:sequential_recovery}{Sequential-recovery code}\end{eczvaluelist}
\eczhbkcontributors{ \eczhuVVA }
\endeczcode

\eczcode{batch}{Batch code}{~\NoCaseChange{\protect\cite{cite944}}}
\codefieldsection{Description}
Binary code designed for minimizing the total amount of storage and the worst-case maximal load on any device in a distributed system.

An \((n,N,k,m,t)\) batch code encodes a length-\(n\) string \(x_1,\cdots,x_n\) into an \(m\)-tuple of strings of total length \(N\) (also called buckets), such that for each \(k\)-tuple of distinct indices \(i_1,i_2,...,i_k\), the entries \(x_{i_1},...,x_{i_k}\) can be decoded by reading at most \(t\) symbols from each bucket.
If each bucket of a batch code contains a single symbol, then the \((n,N,k,m)\) batch code is \textit{primitive}.

If, for any multiset \(i_1, i_2, ..., i_k \in [n]\), there is a partition of buckets into subsets \(S_1, ..., S_k \subset [m]\) such that each \(x_{i_j}\) can be recovered by reading at most one symbol from each bucket in \(S_j\), then the \((n, N, k, m)\) code is a \textit{multiset batch code}.

\codefieldsection{Protection}
The Gadget Lemma states that any \((n,N,k,m)\) batch code at \(t=1\) can be transformed into a multiset \((rn,rN,k,m)\) batch code for any positive integer \(r\) \NoCaseChange{\protect\cite{cite945}}.

Combining two batch codes \(C_1\) and \(C_2\), which are \((n_1,N_1,k_1,m_1)\) and \((n_2,N_2,k_2,m_2)\) batch codes respectively, yields a composite batch code \(C_1\otimes C_2\), which is an \((n_1, m_1N_2, k_1 k_2, m_1 m_2)\) batch code.

\codefieldsection{Notes}
\begin{eczvaluelist}
\item\relax See Ref. \NoCaseChange{\protect\cite{cite946}}.
\end{eczvaluelist}
\codefieldsection{Parents}
\begin{eczvaluelist}
\item\relax
\flmRefsHyperref[eczindexfamilyrel]{code:block}{Block code}\item\relax
\flmRefsHyperref[eczindexfamilyrel]{code:ecc_finite}{Finite-dimensional error-correcting code (ECC)}\end{eczvaluelist}
\codefieldsection{Cousins}
\begin{eczvaluelist}
\item\relax
\flmRefsHyperref[eczindexfamilyrel]{code:locally_recoverable}{Locally recoverable code (LRC)} --- A systematic batch code with restricted size of reconstruction sets can recover any query of \(t\) information symbols with recovery sets of size \(r\) \NoCaseChange{\protect\cite{cite947,cite948}}.
\item\relax
\flmRefsHyperref[eczindexfamilyrel]{code:ldc}{Locally decodable code (LDC)} --- Batch codes and LDCs are related \NoCaseChange{\protect\cite{cite944,cite949}\protect\cite[{Ch. 10.3}]{cite946}}.
\item\relax
\flmRefsHyperref[eczindexfamilyrel]{code:hamming}{\([2^r-1,2^r-r-1,3]\) Hamming code} --- Hamming codes can be used to construct batch codes \NoCaseChange{\protect\cite{cite950}\protect\cite[{Exam. 10.9}]{cite946}}.
\item\relax
\flmRefsHyperref[eczindexfamilyrel]{code:generalized_reed_muller}{Generalized RM (GRM) code} --- GRM codes can be used to construct batch codes \NoCaseChange{\protect\cite{cite950}}.
\item\relax
\flmRefsHyperref[eczindexfamilyrel]{code:multiplicity}{Multiplicity code} --- Multiplicity codes can be used to construct batch codes \NoCaseChange{\protect\cite{cite951}}.
\item\relax
\flmRefsHyperref[eczindexfamilyrel]{code:pir}{Private information retrieval (PIR) code} --- Batch and PIR codes are related \NoCaseChange{\protect\cite{cite949,cite952}}.
\end{eczvaluelist}
\eczhbkcontributors{ Yijia Xu, \eczhuVVA }
\endeczcode

\eczcode{block}{Block code}{}
\codefieldsection{Description}
A code intended to encode a piece, or block, of a data stream on a \textit{block} of \(n\) symbols, with each symbol taking values from a fixed alphabet \(\Sigma\).

The overall alphabet of the code is \(\Sigma^n\), and \(n\) is called the \textit{length} of the code \NoCaseChange{\protect\cite[{Ch. 3}]{cite953}}.
In some cases, only a subset of \(\Sigma^n\) is available to use for constructing the code.
For example, in the case of spherical codes, one is constrained to \(n\)-dimensional real vectors on the unit sphere.

An alternative more stringent definition of the code (not used here) is in terms of a map encoding logical information from \(\Sigma^k\) into \(\Sigma^n\), yielding an \((n,k,d)_{\Sigma}\) block code, where \(d\) is the code distance.

\codefieldsection{Protection}
Block codes protect from errors acting on a few of the \(n\) symbols. A block code with \textit{distance} \(d\) detects errors acting on up to \(d-1\) symbols, and corrects erasure errors on up to \(d-1\) symbols.

\codefieldsection{Rate}
Ideal decoding error is suppressed exponentially with the number of subsystems \(n\), and the exponent has been studied in Ref. \NoCaseChange{\protect\cite{cite954,cite955,cite956}}.
\codefieldsection{Decoding}
\begin{eczvaluelist}
\item\relax Decoding an error-correcting code is equivalent to finding the ground state of a statistical mechanical model \NoCaseChange{\protect\cite{cite957}}.
\end{eczvaluelist}
\codefieldsection{Notes}
\begin{eczvaluelist}
\item\relax \begin{defterm}{Asymptotic notation}\label{ref958}\label{ref65}
We are often interested in how parameters of particular infinite block-code families scale with increasing block length \(n\), necessitating the use of asymptotic or Bachmann–Landau notation; see the book \NoCaseChange{\protect\cite{cite959}}. The table below summarizes the notation used throughout the EC Zoo for relating functions \(f,g\) that both grow with \(n\).
\begin{flmFloat}{table}{NumCap}\flmCellsBeginCenter
\long\def\flmTempTypesetThisTable#1{%
\begin{tblr}{#1,
  hspan=minimal,
  cell{1}{1}={}{c, font={\flmCellsHeaderFont}},
  cell{1}{2}={}{c, font={\flmCellsHeaderFont}},
  cell{2}{1}={}{c},
  cell{2}{2}={}{c},
  cell{3}{1}={}{c},
  cell{3}{2}={}{c},
  cell{4}{1}={}{c},
  cell{4}{2}={}{c},
  hline{2}={1}{.4pt,solid},
  hline{2}={2}{.4pt,solid}}%
\toprule
relation & behavior\\

  \(f(n)=O( g(n) )\) & \(\phantom{c_{2}\geq}{\displaystyle \underset{n\to\infty}{\lim}\frac{|f(n)|}{|g(n)|}}\leq c\)
      \\

  \(f(n)=\Omega( g(n) )\) & \(0<{\displaystyle \underset{n\to\infty}{\lim}\frac{|f(n)|}{|g(n)|}}\phantom{\leq c_{2}}\)
      \\

  \(f(n)=\Theta( g(n) )\) & \(c_{1}\leq{\displaystyle \underset{n\to\infty}{\lim}\frac{|f(n)|}{|g(n)|}}\leq c_{2}\)
  \\
\bottomrule
\end{tblr}%
}%
\def\flmTmpMaxW{\dimexpr 0.96\linewidth\relax}%
\setbox0=\hbox{\flmTempTypesetThisTable{colspec={cc}}}%
\ifdim\wd0<\flmTmpMaxW\relax
  \leavevmode\box0 
\else
  \flmTempTypesetThisTable{width=\flmTmpMaxW,colspec={X[-1]X[-1]}}
\fi
\flmCellsEndCenter \caption{Asymptotic notation relating functions \(f,g\) that both grow with system size \(n\); \(c,c_1,c_2\) are positive \(n\)-independent constants. In some cases, \(\lim\) should be replaced with \(\limsup\).}\label{ref960}\end{flmFloat}
\end{defterm}

\end{eczvaluelist}
\codefieldsection{Parent}
\begin{eczvaluelist}
\item\relax
\flmRefsHyperref[eczindexfamilyrel]{code:ecc}{Error-correcting code (ECC)}\end{eczvaluelist}
\codefieldsection{Children}
\begin{eczvaluelist}
\item\relax
\flmRefsHyperref[eczindexfamilyrel]{code:analog}{Analog code}\item\relax
\flmRefsHyperref[eczindexfamilyrel]{code:group_linear}{Linear code over \(G\)} --- Linear codes over \(G\) are linear block codes with \(\Sigma=G\).
\item\relax
\flmRefsHyperref[eczindexfamilyrel]{code:matrices_into_matrices}{Matrix-based code}\item\relax
\flmRefsHyperref[eczindexfamilyrel]{code:checksum}{Checksum code}\item\relax
\flmRefsHyperref[eczindexfamilyrel]{code:frameproof}{Frameproof (FP) code}\item\relax
\flmRefsHyperref[eczindexfamilyrel]{code:distributed_storage}{Distributed-storage code}\item\relax
\flmRefsHyperref[eczindexfamilyrel]{code:ldc}{Locally decodable code (LDC)}\item\relax
\flmRefsHyperref[eczindexfamilyrel]{code:ltc}{Locally testable code (LTC)}\item\relax
\flmRefsHyperref[eczindexfamilyrel]{code:batch}{Batch code}\item\relax
\flmRefsHyperref[eczindexfamilyrel]{code:pir}{Private information retrieval (PIR) code}\item\relax
\flmRefsHyperref[eczindexfamilyrel]{code:insertion_deletion}{Editing code}\item\relax
\flmRefsHyperref[eczindexfamilyrel]{code:q-ary_constant_weight}{Constant-weight block code}\item\relax
\flmRefsHyperref[eczindexfamilyrel]{code:small_distance}{Small-distance block code}\item\relax
\flmRefsHyperref[eczindexfamilyrel]{code:reversible}{Reversible code}\item\relax
\flmRefsHyperref[eczindexfamilyrel]{code:skew_cyclic}{Skew-cyclic code}\item\relax
\flmRefsHyperref[eczindexfamilyrel]{code:quasi_twisted}{Quasi-twisted code}\item\relax
\flmRefsHyperref[eczindexfamilyrel]{code:rings_into_rings}{Ring code} --- Ring codes are block codes with \(\Sigma=R\).
\end{eczvaluelist}
\codefieldsection{Cousin}
\begin{eczvaluelist}
\item\relax
\flmRefsHyperref[eczindexfamilyrel]{code:block_quantum}{Block quantum code} --- Block quantum codes are quantum analogues of block codes.
\end{eczvaluelist}
\eczhbkcontributors{ \eczhuVVA }
\endeczcode

\eczcode{checksum}{Checksum code}{}
\codefieldsection{Description}
An error-detecting block code that appends to a message a shorter string, called a checksum, that is computed from the message.

\codefieldsection{Protection}
The checksum can be recalculated from the received message and compared with the appended checksum in order to detect certain transmission or storage errors.
\codefieldsection{Parents}
\begin{eczvaluelist}
\item\relax
\flmRefsHyperref[eczindexfamilyrel]{code:block}{Block code}\item\relax
\flmRefsHyperref[eczindexfamilyrel]{code:ecc_finite}{Finite-dimensional error-correcting code (ECC)}\end{eczvaluelist}
\codefieldsection{Children}
\begin{eczvaluelist}
\item\relax
\flmRefsHyperref[eczindexfamilyrel]{code:crc}{Cyclic redundancy check (CRC) code}\item\relax
\flmRefsHyperref[eczindexfamilyrel]{code:isbn}{International Standard Book Number (ISBN) code} --- The last digit of an ISBN-10 string is a check digit computed modulo 11 \NoCaseChange{\protect\cite{cite961}}.
\item\relax
\flmRefsHyperref[eczindexfamilyrel]{code:q-ary_parity_check}{\([n,n-1,2]_q\) \(q\)-ary parity-check code} --- The \(q\)-ary parity check code is a simple example of a checksum code, with the parity of the message being the checksum.
\item\relax
\flmRefsHyperref[eczindexfamilyrel]{code:upc}{Universal Product Code (UPC)} --- The last digit of a UPC barcode is a base-10 check digit computed modulo 10 from an alternating weighted sum of the preceding digits \NoCaseChange{\protect\cite{cite962}}.
\end{eczvaluelist}
\eczhbkcontributors{ \eczhuVVA }
\endeczcode

\eczcode{circular_dna}{Circular DNA code}{~\NoCaseChange{\protect\cite{cite963}}}
\codefieldsection{Description}
A DNA code that consists of subsets of three-letter DNA sequences that can be used to express any DNA string (up to some length) when the string is written in a circle.

\codefieldsection{Notes}
\begin{eczvaluelist}
\item\relax Review of circular codes \NoCaseChange{\protect\cite{cite964}}.
\end{eczvaluelist}
\codefieldsection{Parent}
\begin{eczvaluelist}
\item\relax
\flmRefsHyperref[eczindexfamilyrel]{code:dna}{DNA storage code}\end{eczvaluelist}
\eczhbkcontributors{ \eczhuVVA }
\endeczcode

\eczcode{code_with_locality}{Code with locality}{}
\codefieldsection{Description}
A code has \((r,\delta)\) locality if each codeword coordinate belongs to a repair group of size at most \(r+\delta-1\) whose restricted code has minimum distance at least \(\delta\) \NoCaseChange{\protect\cite[{Sec. 31.3.4.5}]{cite183}}.
Equivalently, given a codeword \(c\) and coordinate \(c_i\), there exists a coordinate set \(S_i\) of size \(\leq r+\delta-1\) containing \(i\) such that the restriction \(C_{|S_i}\) has minimum distance at least \(\delta\).

\codefieldsection{Protection}
There is a generalized Singleton minimum distance bound \NoCaseChange{\protect\cite{cite965}},
\flmMathEnvironment{align}{}{
     d\leq n-k+1-(\left\lceil k/r\right\rceil -1)(\delta-1)~,
}
with codes saturating this bound being \textit{optimal codes with locality}.
The \(\delta=2\) case recovers optimal LRCs.

\codefieldsection{Parent}
\begin{eczvaluelist}
\item\relax
\flmRefsHyperref[eczindexfamilyrel]{code:distributed_storage}{Distributed-storage code}\end{eczvaluelist}
\codefieldsection{Children}
\begin{eczvaluelist}
\item\relax
\flmRefsHyperref[eczindexfamilyrel]{code:locally_recoverable}{Locally recoverable code (LRC)} --- An LRC of locality \(r\) is a code with \((r,2)\) locality \NoCaseChange{\protect\cite[{Sec. 31.3.4.5}]{cite183}}.
\item\relax
\flmRefsHyperref[eczindexfamilyrel]{code:maximally_recoverable}{Maximally recoverable (MR) code}\end{eczvaluelist}
\eczhbkcontributors{ \eczhuVVA }
\endeczcode

\eczcode{concatenated}{Concatenated code}{~\NoCaseChange{\protect\cite{cite966}}}
\codefieldsection{Alternative Names}
\begin{eczvaluelist}
\item\relax Serially concatenated code
\end{eczvaluelist}
\eczhIndexCodeAliasName{concatenated}{Serially concatenated code}
\codefieldsection{Description}
A code whose encoding mapping is a composition of two mappings: first the message set is mapped onto the code space of the outer code, then each coordinate of the outer code is mapped onto the code space of the inner code.
In the basic construction, the outer code's alphabet is the finite field \(\mathbb{F}_{p^m}\) and the \(m\)-dimensional inner code is over the field \(\mathbb{F}_p\). The construction is not limited to linear codes.

\codefieldsection{Rate}
There exist bounds on distance and rate of concatenated codes with a fixed outer and random inner code \NoCaseChange{\protect\cite{cite967,cite968}}.
\codefieldsection{Decoding}
\begin{eczvaluelist}
\item\relax Generalized minimum-distance decoder \NoCaseChange{\protect\cite{cite969}}.
\end{eczvaluelist}
\codefieldsection{Parent}
\begin{eczvaluelist}
\item\relax
\flmRefsHyperref[eczindexfamilyrel]{code:generalized_concatenated}{Generalized concatenated code (GCC)}\end{eczvaluelist}
\codefieldsection{Children}
\begin{eczvaluelist}
\item\relax
\flmRefsHyperref[eczindexfamilyrel]{code:ira}{Irregular repeat-accumulate (IRA) code} --- IRA codes can be interpreted as serial concatenated codes \NoCaseChange{\protect\cite{cite970}}.
\item\relax
\flmRefsHyperref[eczindexfamilyrel]{code:binary_balanced}{Binary balanced spherical code} --- A binary balanced spherical code can be thought of as a concatenation of a constant-weight binary outer code with a shifted and scaled BPSK-like inner code.
\item\relax
\flmRefsHyperref[eczindexfamilyrel]{code:polyphase}{Polyphase code} --- A polyphase code can be thought of as a concatenation of a \(q\)-ary outer code with a PSK inner code.
\end{eczvaluelist}
\codefieldsection{Cousins}
\begin{eczvaluelist}
\item\relax
\flmRefsHyperref[eczindexfamilyrel]{code:quantum_concatenated}{Concatenated quantum code} --- Quantum codes can be concatenated with classical codes to yield good quantum codes \NoCaseChange{\protect\cite{cite971}}.
\item\relax
\flmRefsHyperref[eczindexfamilyrel]{code:ha_ldpc}{Hsu-Anastasopoulos LDPC (HA-LDPC) code} --- HA-LDPC codes are a concatenation of an LDPC and an LDGM code.
\item\relax
\flmRefsHyperref[eczindexfamilyrel]{code:tensor}{Tensor-product code} --- Tensor-product codes can be viewed both as serial or parallel concatenated codes \NoCaseChange{\protect\cite{cite972}}.
\item\relax
\flmRefsHyperref[eczindexfamilyrel]{code:generalized_reed_solomon}{Generalized RS (GRS) code} --- Concatenations of GRS codes with random linear codes almost surely attain the \flmRefsHyperref{ref85}{GV bound} \NoCaseChange{\protect\cite{cite973}}.
\item\relax
\flmRefsHyperref[eczindexfamilyrel]{code:lresc}{Long-range enhanced surface code (LRESC)} --- LRESCs are constructed using a hypergraph product of two copies of a concatenated LDPC-repetition seed code.
\end{eczvaluelist}
\eczhbkcontributors{ Alexander Barg, \eczhuVVA }
\endeczcode

\eczcode{constacyclic}{Constacyclic code}{}
\codefieldsection{Alternative Names}
\begin{eczvaluelist}
\item\relax Twisted code
\end{eczvaluelist}
\eczhIndexCodeAliasName{constacyclic}{Twisted code}
\codefieldsection{Description}
A block code \(C\) of length \(n\) over an alphabet \(R\) is \(\alpha\)-constacyclic (or \(\alpha\)-twisted) if, for each string \(c_1 c_2 \cdots c_n\in C\), the string \(\alpha c_n, c_1, \cdots, c_{n-1} \in C\) \NoCaseChange{\protect\cite[{Def. 3.2.7}]{cite70}}.
A \(-1\)-constacyclic code is called \textit{negacyclic}.

\codefieldsection{Parent}
\begin{eczvaluelist}
\item\relax
\flmRefsHyperref[eczindexfamilyrel]{code:quasi_twisted}{Quasi-twisted code} --- Quasi-twisted codes with \(\ell=1\) are constacyclic.
\end{eczvaluelist}
\codefieldsection{Children}
\begin{eczvaluelist}
\item\relax
\flmRefsHyperref[eczindexfamilyrel]{code:cyclic}{Cyclic code} --- Constacyclic codes with \(\alpha=1\) are cyclic.
\item\relax
\flmRefsHyperref[eczindexfamilyrel]{code:berlekamp}{Berlekamp code} --- Berlekamp codes are negacyclic \NoCaseChange{\protect\cite[{Ch. 9}]{cite974}}.
\end{eczvaluelist}
\codefieldsection{Cousins}
\begin{eczvaluelist}
\item\relax
\flmRefsHyperref[eczindexfamilyrel]{code:quantum_mds}{Quantum maximum-distance-separable (MDS) code} --- Many quantum MDS codes are constructed from Hermitian self-orthogonal codes over \(\mathbb{F}_{q^2}\) using the Hermitian construction \NoCaseChange{\protect\cite{cite975,cite976,cite977,cite978}}, in particular from cyclic \NoCaseChange{\protect\cite{cite979}}, constacyclic \NoCaseChange{\protect\cite{cite980,cite981,cite978}}, and negacyclic \NoCaseChange{\protect\cite{cite982}} codes.
\item\relax
\flmRefsHyperref[eczindexfamilyrel]{code:stabilizer_over_gf4}{Hermitian qubit code} --- Duadic constacyclic codes yield many examples of Hermitian qubit codes \NoCaseChange{\protect\cite{cite983}}.
\end{eczvaluelist}
\eczhbkcontributors{ \eczhuVVA }
\endeczcode

\eczcode{q-ary_constant_weight}{Constant-weight block code}{~\NoCaseChange{\protect\cite{cite984}}}
\codefieldsection{Alternative Names}
\begin{eczvaluelist}
\item\relax One-weight block code
\end{eczvaluelist}
\eczhIndexCodeAliasName{q-ary_constant_weight}{One-weight block code}
\codefieldsection{Description}
A block code whose codewords all have the same number of nonzero coordinates.
Code constructions exist for codes over fields \NoCaseChange{\protect\cite{cite168}} or rings \NoCaseChange{\protect\cite{cite169}}.

The set of all weight-\(w\) \(q\)-ary strings of length \(n\) forms the \textit{nonbinary Johnson space}, a finite symmetric space \(G/H\) with \(G = S_{q-1} \wr S_n\) \NoCaseChange{\protect\cite{cite984}\protect\cite[{Sec. 8.8}]{cite913}\protect\cite[{Table 3}]{cite985}}. The number of such strings is \({n \choose w} (q-1)^w\).

\codefieldsection{Protection}
Upper bounds on the largest possible number of codewords, given the length \(n\), distance \(d\), and weight \(w\), can be derived using linear- and semidefinite-programming methods on the nonbinary Johnson association scheme \NoCaseChange{\protect\cite{cite984,cite985}}.

\codefieldsection{Parents}
\begin{eczvaluelist}
\item\relax
\flmRefsHyperref[eczindexfamilyrel]{code:block}{Block code}\item\relax
\flmRefsHyperref[eczindexfamilyrel]{code:ecc_finite}{Finite-dimensional error-correcting code (ECC)}\item\relax
\flmRefsHyperref[eczindexfamilyrel]{code:symmetric_space}{Symmetric-space code} --- The set of all weight-\(w\) \(q\)-ary strings of length \(n\) forms the \textit{nonbinary Johnson space} (a.k.a. \(q\)-ary Johnson space), a finite symmetric space \(G/H\) with \(G = S_{q-1} \wr S_n\) \NoCaseChange{\protect\cite{cite984}\protect\cite[{Sec. 8.8}]{cite913}\protect\cite[{Table 3}]{cite985}}. The number of such strings is \({n \choose w} (q-1)^w\). This reduces to the Johnson space for \(q=2\).
\end{eczvaluelist}
\codefieldsection{Children}
\begin{eczvaluelist}
\item\relax
\flmRefsHyperref[eczindexfamilyrel]{code:constant_weight}{Constant-weight code} --- The set of all weight-\(w\) binary strings of length \(n\) forms the \textit{Johnson space} \(J(n,w)\), a finite two-point homogeneous space \(G/H\) with \(G = S_n\) and \(H = S_w \times S_{n-w}\) \NoCaseChange{\protect\cite{cite880,cite986,cite912,cite171}\protect\cite[{Sec. 4.2.1}]{cite987}\protect\cite[{Table 2}]{cite985}}. This is a special case of the nonbinary Johnson space for \(q=2\).
\item\relax
\flmRefsHyperref[eczindexfamilyrel]{code:one_vs_one}{One-versus-one (OVO) code}\end{eczvaluelist}
\codefieldsection{Cousins}
\begin{eczvaluelist}
\item\relax
\flmRefsHyperref[eczindexfamilyrel]{code:constant_excitation}{Constant-excitation (CE) code} --- Constant-weight codes are classical analogues of qubit constant-excitation codes.
\item\relax
\flmRefsHyperref[eczindexfamilyrel]{code:q-ary_linear_over_zq}{Linear code over \(\mathbb{Z}_q\)} --- Constant-weight linear codes over \(\mathbb{Z}_q\) have been classified \NoCaseChange{\protect\cite{cite169}}.
\item\relax
\flmRefsHyperref[eczindexfamilyrel]{code:q-ary_over_zq}{\(q\)-ary code over \(\mathbb{Z}_q\)} --- Optimal constant-weight codes over \(\mathbb{Z}_q\) can be constructed \NoCaseChange{\protect\cite{cite131}} from a generalization of combinatorial designs to \(q\)-ary alphabets \NoCaseChange{\protect\cite{cite132,cite133}}.
\item\relax
\flmRefsHyperref[eczindexfamilyrel]{code:combinatorial_design}{Combinatorial design} --- Optimal constant-weight codes over \(\mathbb{Z}_q\) can be constructed \NoCaseChange{\protect\cite{cite131}} from a generalization of combinatorial designs to \(q\)-ary alphabets \NoCaseChange{\protect\cite{cite132,cite133}}.
\item\relax
\flmRefsHyperref[eczindexfamilyrel]{code:q-ary_linear}{Linear \(q\)-ary code} --- Linear \(q\)-ary codes cannot be constant weight, but can have nonzero codewords with constant weight. All such codes are equidistant, and Bonisoli's theorem states that any equidistant linear code is a direct sum of \(q\)-ary simplex codes \NoCaseChange{\protect\cite{cite988}} (see also Refs. \NoCaseChange{\protect\cite{cite45,cite46}}).
\item\relax
\flmRefsHyperref[eczindexfamilyrel]{code:q-ary_simplex}{\(q\)-ary simplex code} --- Linear \(q\)-ary codes cannot be constant weight, but can have nonzero codewords with constant weight. All such codes are equidistant, and Bonisoli's theorem states that any equidistant linear code is a direct sum of \(q\)-ary simplex codes \NoCaseChange{\protect\cite{cite988}} (see also Refs. \NoCaseChange{\protect\cite{cite45,cite46}}).
\item\relax
\flmRefsHyperref[eczindexfamilyrel]{code:balanced}{Balanced code} --- Balanced codes are not automatically constant-weight because they may contain the zero codeword.
\item\relax
\flmRefsHyperref[eczindexfamilyrel]{code:two_weight}{Two-weight code} --- Each codeword of a constant-weight (two-weight) code has one (two) possible Hamming weight(s).
\end{eczvaluelist}
\eczhbkcontributors{ \eczhuVVA }
\endeczcode

\eczcode{cyclic}{Cyclic code}{~\NoCaseChange{\protect\cite{cite989,cite990,cite991,cite992,cite993}}}
\codefieldsection{Description}
A block code of length \(n\) over an alphabet is cyclic if, for each codeword \(c_1 c_2 \cdots c_n\), the cyclically shifted string \(c_n c_1 \cdots c_{n-1}\) is also a codeword.

\codefieldsection{Notes}
\begin{eczvaluelist}
\item\relax Cyclic codes over finite fields, together with defining sets and basic distance bounds, are reviewed in \NoCaseChange{\protect\cite[{Secs. 2.1-2.4}]{cite68}}.
\end{eczvaluelist}
\codefieldsection{Parents}
\begin{eczvaluelist}
\item\relax
\flmRefsHyperref[eczindexfamilyrel]{code:quasi_cyclic}{Quasi-cyclic code} --- Quasi-cyclic codes with \(\ell=1\) are cyclic.
\item\relax
\flmRefsHyperref[eczindexfamilyrel]{code:constacyclic}{Constacyclic code} --- Constacyclic codes with \(\alpha=1\) are cyclic.
\item\relax
\flmRefsHyperref[eczindexfamilyrel]{code:skew_cyclic}{Skew-cyclic code} --- Skew-cyclic codes with \(\theta\) trivial are cyclic.
\end{eczvaluelist}
\codefieldsection{Children}
\begin{eczvaluelist}
\item\relax
\flmRefsHyperref[eczindexfamilyrel]{code:one_hot}{One-hot code} --- The one-hot code is a cyclic non-linear binary code.
\item\relax
\flmRefsHyperref[eczindexfamilyrel]{code:q-ary_cyclic}{Cyclic linear \(q\)-ary code}\end{eczvaluelist}
\codefieldsection{Cousins}
\begin{eczvaluelist}
\item\relax
\flmRefsHyperref[eczindexfamilyrel]{code:octacode}{Octacode} --- The heptacode is a cyclic code over \(\mathbb{Z}_4\) with generator polynomial \(x^3+3x^2+2x+3\) \NoCaseChange{\protect\cite{cite42}}.
\item\relax
\flmRefsHyperref[eczindexfamilyrel]{code:lattice_shell}{Lattice-shell code} --- Lattice-shell codewords are often permutations of a particular set of reference vectors, meaning that a cyclic permutation of a codeword yields another codeword.
\item\relax
\flmRefsHyperref[eczindexfamilyrel]{code:quantum_cyclic}{Cyclic quantum code} --- Cyclic quantum codes are quantum analogues of cyclic codes.
\end{eczvaluelist}
\eczhbkcontributors{ Nolan Coble, \eczhuVVA }
\endeczcode

\eczcode{distributed_storage}{Distributed-storage code}{}
\codefieldsection{Description}
Block code designed to encode information into spatial nodes such that it is possible to recover said information after failure of some nodes by accessing the remaining \textit{helper nodes} with minimal bandwidth.

\codefieldsection{Protection}
Typically designed to protect a distributed storage system against the failure of a single node or multiple nodes.
\codefieldsection{Notes}
\begin{eczvaluelist}
\item\relax See Refs. \NoCaseChange{\protect\cite{cite183,cite946}}.
\end{eczvaluelist}
\codefieldsection{Parents}
\begin{eczvaluelist}
\item\relax
\flmRefsHyperref[eczindexfamilyrel]{code:block}{Block code}\item\relax
\flmRefsHyperref[eczindexfamilyrel]{code:ecc_finite}{Finite-dimensional error-correcting code (ECC)}\end{eczvaluelist}
\codefieldsection{Children}
\begin{eczvaluelist}
\item\relax
\flmRefsHyperref[eczindexfamilyrel]{code:array}{Array code}\item\relax
\flmRefsHyperref[eczindexfamilyrel]{code:sum_rank_metric}{Sum-rank-metric code} --- Sum-rank-metric codes are useful for distributed storage \NoCaseChange{\protect\cite{cite994}}.
\item\relax
\flmRefsHyperref[eczindexfamilyrel]{code:code_with_locality}{Code with locality}\end{eczvaluelist}
\codefieldsection{Cousins}
\begin{eczvaluelist}
\item\relax
\flmRefsHyperref[eczindexfamilyrel]{code:fountain}{Fountain code} --- There are proposals \NoCaseChange{\protect\cite{cite995,cite996}} adapting fountain codes to distributed storage systems.
\item\relax
\flmRefsHyperref[eczindexfamilyrel]{code:generalized_reed_solomon}{Generalized RS (GRS) code} --- GRS codes are used in various cloud storage systems \NoCaseChange{\protect\cite{cite264}}.
\end{eczvaluelist}
\eczhbkcontributors{ Fengxing Zhu, \eczhuVVA }
\endeczcode

\eczcode{dna}{DNA storage code}{~\NoCaseChange{\protect\cite{cite997}}}
\codefieldsection{Description}
Code that was designed (or that can be applied) to encode information into the four-base-pair alphabet of a DNA molecule.

There exist several proposals tailored specifically for DNA storage \NoCaseChange{\protect\cite{cite997,cite998,cite999,cite1000,cite1001}}.

\codefieldsection{Protection}
Noise affecting DNA molecules can include insertions and deletions \NoCaseChange{\protect\cite{cite1002}}. The DNA data storage channel has been characterized \NoCaseChange{\protect\cite{cite997,cite1003,cite1002}}.

\codefieldsection{Notes}
\begin{eczvaluelist}
\item\relax Review of DNA-based coding \NoCaseChange{\protect\cite{cite1004}}.
\item\relax Communicating with DNA may be simplified to a graph-based communication model \NoCaseChange{\protect\cite{cite1005}}.
\end{eczvaluelist}
\codefieldsection{Parent}
\begin{eczvaluelist}
\item\relax
\flmRefsHyperref[eczindexfamilyrel]{code:insertion_deletion}{Editing code} --- DNA codes can typically handle base-pair insertions and deletions.
\end{eczvaluelist}
\codefieldsection{Child}
\begin{eczvaluelist}
\item\relax
\flmRefsHyperref[eczindexfamilyrel]{code:circular_dna}{Circular DNA code}\end{eczvaluelist}
\codefieldsection{Cousins}
\begin{eczvaluelist}
\item\relax
\flmRefsHyperref[eczindexfamilyrel]{code:ldpc}{Low-density parity-check (LDPC) code} --- LDPC codes are potentially relevant for DNA storage \NoCaseChange{\protect\cite{cite1006}}.
\item\relax
\flmRefsHyperref[eczindexfamilyrel]{code:simplex_discrete}{Simplex integer-based code} --- Simplex integer-based codewords are intended to model multisets of DNA molecules \NoCaseChange{\protect\cite{cite1007}}. Points in a \flmRefsHyperref{ref655}{discrete simplex} are in one-to-one correspondence to multisets because their coordinates denote the multiplicity of each element in a given multiset \NoCaseChange{\protect\cite{cite1007}}.
\item\relax
\flmRefsHyperref[eczindexfamilyrel]{code:fountain}{Fountain code} --- Fountain codes have been used for DNA storage \NoCaseChange{\protect\cite{cite259}}.
\item\relax
\flmRefsHyperref[eczindexfamilyrel]{code:reed_solomon}{Reed-Solomon (RS) code} --- RS codes have been used for DNA storage \NoCaseChange{\protect\cite{cite336}}.
\end{eczvaluelist}
\eczhbkcontributors{ \eczhuVVA }
\endeczcode

\eczcode{insertion_deletion}{Editing code}{~\NoCaseChange{\protect\cite{cite1008}}}
\codefieldsection{Alternative Names}
\begin{eczvaluelist}
\item\relax Insertion and deletion code
\end{eczvaluelist}
\eczhIndexCodeAliasName{insertion_deletion}{Insertion and deletion code}
\codefieldsection{Description}
A block code designed to protect against insertions, where a new symbol is added somewhere within the string, and deletions, where a symbol at an unknown location is erased.

\codefieldsection{Protection}
The metric measuring distance of a received word to the nearest codeword is the \textit{Levenshtein deletion distance}: given vectors \(u,v\), this distance is one-half the smallest number of deletions and insertions needed to change \(u\) to \(v\).
A code \(C\) corrects \(e\) deletions if all codewords are separated by at least \(e+1\) in the deletion distance \NoCaseChange{\protect\cite{cite1009}}.
Similar distances, collectively called \textit{editing distances}, can be defined for insertions and related operations \NoCaseChange{\protect\cite[{Sec. 22.7}]{cite1010}}.

\codefieldsection{Rate}
An asymptotically good linear code against bit-flip errors can be converted into an asymptotically good code against insertion-deletion errors \NoCaseChange{\protect\cite{cite1011}}.
\codefieldsection{Notes}
\begin{eczvaluelist}
\item\relax See Refs. \NoCaseChange{\protect\cite{cite1012,cite1013}\protect\cite[{Sec. 22.7}]{cite1010}} for more details.
\end{eczvaluelist}
\codefieldsection{Parents}
\begin{eczvaluelist}
\item\relax
\flmRefsHyperref[eczindexfamilyrel]{code:block}{Block code}\item\relax
\flmRefsHyperref[eczindexfamilyrel]{code:ecc_finite}{Finite-dimensional error-correcting code (ECC)}\end{eczvaluelist}
\codefieldsection{Children}
\begin{eczvaluelist}
\item\relax
\flmRefsHyperref[eczindexfamilyrel]{code:vt_single_deletion}{Varshamov-Tenengolts (VT) code}\item\relax
\flmRefsHyperref[eczindexfamilyrel]{code:dna}{DNA storage code} --- DNA codes can typically handle base-pair insertions and deletions.
\end{eczvaluelist}
\codefieldsection{Cousins}
\begin{eczvaluelist}
\item\relax
\flmRefsHyperref[eczindexfamilyrel]{code:permutation_invariant}{Permutation-invariant (PI) code} --- PI codes of distance \(d\) can protect against \(d-1\) (quantum) deletion errors.
\item\relax
\flmRefsHyperref[eczindexfamilyrel]{code:combinatorial_design}{Combinatorial design} --- Perfect deletion correcting codes can be constructed using combinatorial design theory \NoCaseChange{\protect\cite{cite139,cite140}}.
\item\relax
\flmRefsHyperref[eczindexfamilyrel]{code:perfect}{Perfect code} --- Perfect deletion correcting codes can be constructed using combinatorial design theory \NoCaseChange{\protect\cite{cite139,cite140}}.
\end{eczvaluelist}
\eczhbkcontributors{ Fengxing Zhu, \eczhuVVA }
\endeczcode

\eczcode{ecc}{Error-correcting code (ECC)}{}
\codefieldsection{Description}
Code designed for transmission of classical information through classical channels.

A code is a subset of a set or \textit{alphabet} \(\Sigma\), with each element called a \textit{codeword}. 
An error-correcting code consists of \(K\) codewords over an alphabet with \(N\) elements such that it is possible to recover the codewords from errors \(E\) from some error set \(\mathcal{E}\).
The table below lists the most common alphabets, along with names of the corresponding codewords of a block code, i.e., a code on \(n\) copies of the alphabet.
  \begin{flmFloat}{table}{NumCap}\flmCellsBeginCenter
\long\def\flmTempTypesetThisTable#1{%
\begin{tblr}{#1,
  hspan=minimal,
  cell{1}{1}={}{c, font={\flmCellsHeaderFont}},
  cell{1}{2}={}{c, font={\flmCellsHeaderFont}},
  cell{2}{1}={}{c},
  cell{2}{2}={}{c},
  cell{3}{1}={}{c},
  cell{3}{2}={}{c},
  cell{4}{1}={}{c},
  cell{4}{2}={}{c},
  cell{5}{1}={}{c},
  cell{5}{2}={}{c},
  cell{6}{1}={}{c},
  cell{6}{2}={}{c},
  cell{7}{1}={}{c},
  cell{7}{2}={}{c},
  hline{2}={1}{.4pt,solid},
  hline{2}={2}{.4pt,solid}}%
\toprule
alphabet \(\Sigma\) & codewords\\

    \(\mathbb{Z}_{2}=\mathbb{F}_2\) & bitstrings
        \\

    \(\mathbb{F}_q\) & \(q\)-ary strings
        \\

    \(\mathbb{Z}_{q}\) & \(q\)-ary strings over \(\mathbb{Z}_{q}\)
        \\

    \(\mathbb{R}\) & sphere packings
        \\

    \(G\) & group elements
        \\

    \(G/H\) & cosets
    \\
\bottomrule
\end{tblr}%
}%
\def\flmTmpMaxW{\dimexpr 0.96\linewidth\relax}%
\setbox0=\hbox{\flmTempTypesetThisTable{colspec={cc}}}%
\ifdim\wd0<\flmTmpMaxW\relax
  \leavevmode\box0 
\else
  \flmTempTypesetThisTable{width=\flmTmpMaxW,colspec={X[-1]X[-1]}}
\fi
\flmCellsEndCenter \caption{Table listing the most common alphabets used in ECCs. Here, \(\mathbb{F}_q\) is a \flmRefsHyperref{ref33}{finite field}, \(G\) is a group, and \(H\) is a subgroup of \(G\).}\label{ref1014}\end{flmFloat}

\subsection{Finite-field alphabet}
\begin{defterm}{Finite fields}\label{ref1015}\label{ref33}
The most common and useful \NoCaseChange{\protect\cite{cite1016}} alphabets used in block codes are Galois or finite fields \(\mathbb{F}_q\), which are sets of \(q\) elements closed under addition and multiplication.
They are finite analogues of the real or complex numbers, and a unique field exists for every power \(q=p^m\) of a prime \(p\).
The prime-field case reduces to \(\mathbb{Z}_p\), a group under addition that is promoted to a field by defining multiplication modulo \(p\); the case \(p=2\) yields the binary field \(\mathbb{Z}_2\).
Every finite field comes with a 0 element (additive identity), a 1 element (multiplicative identity), and additive (multiplicative) inverses for all (nonzero) elements.  
An element whose powers exhaust all nonzero field elements is called \textit{primitive}.
Fields come with a trace operation, the \textit{field trace}, which maps elements \(\gamma \in \mathbb{F}_q\) to elements of \(\mathbb{F}_p\) as
\flmMathEnvironment{align}{}{
  \text{tr}(\gamma)=\sum_{k=0}^{m-1}\gamma^{p^{k}}~.
}
The field trace can be thought of as an averaging over the field's Galois group, which is the cyclic group generated by \(\gamma\to\gamma^p\) \NoCaseChange{\protect\cite[{pg. 113}]{cite41}}.
Fields also come with a \textit{field norm},
\flmMathEnvironment{align}{}{
  N(\gamma)=\prod_{k=0}^{m-1}\gamma^{p^{k}}=\gamma^{(p^{m}-1)/(p-1)}~.
}
In the case of the complex numbers, analogues of the field trace and field norm are the real part and  norm squared of a complex number, respectively.

Any field \(\mathbb{F}_{q=p^m}\) can be thought of as an \(m\)-dimensional vector space over \(\mathbb{F}_p\) a.k.a. the \(m\)th \textit{extension} of \(\mathbb{F}_p\) (similar to the complex numbers being an extension of the reals).
Conversely, \(\mathbb{F}_p\) is an example of a \textit{subfield} of \(\mathbb{F}_q\).
Certain field elements are chosen to be the \textit{basis} of \(\mathbb{F}_q\) over \(\mathbb{F}_p\), and all other elements are expressed as linear combinations of these basis elements.
More generally, elements of fields such as \(\mathbb{F}_{p^{ml}}\) can be written as \(m\)-dimensional vectors over \(\mathbb{F}_{p^l}\) or \((m\times l)\)-dimensional matrices over \(\mathbb{F}_p\). 
This idea is used to convert between ordinary block codes and matrix-based codes such as disk array codes and rank-metric codes.
The field norm and field trace can likewise be defined for fields \(\mathbb{F}_{q^m}\) that are extensions of \(\mathbb{F}_q\) for non-prime \(q\).
\end{defterm}

An example of a field is the quaternary Galois field \(\mathbb{F}_4 = \{0,1,\omega, \omega^2=\bar{\omega}\}\) with \(p=m=2\).
In this case, \(\omega\) can be interpreted as a third root of unity, but more formally it is defined as a solution to the polynomial equation \(1+x+x^2=0\).
Field elements can be represented as two-dimensional vectors with binary elements, \(\mathbb{F}_4=\mathbb{F}_2^2\), using the basis \(1\cong(1,0)\) and \(\omega\cong(0,1)\):
\flmMathEnvironment{align}{}{
  0&\leftrightarrow(0,0)\cong0\cdot1+0\cdot\omega\\1&\leftrightarrow(0,1)\cong0\cdot1+1\cdot\omega\\\omega&\leftrightarrow(1,1)\cong1\cdot1+1\cdot\omega\\\bar{\omega}&\leftrightarrow(1,0)\cong1\cdot1+0\cdot\omega~.
}
In this way, the field elements form the Klein four group \(\mathbb{Z}_2\times\mathbb{Z}_2\) under addition.
One can check that the trace operation, \(\text{tr}(\gamma) = \gamma + \gamma^2\), outputs either 0 or 1 for any element \(\gamma\in \mathbb{F}_4\).

\codefieldsection{Rate}
The Shannon channel capacity (the maximum of the mutual information over input distributions) is the highest rate of information transmission through a classical (i.e., non-quantum) channel with arbitrarily small error rate \NoCaseChange{\protect\cite{cite1}}.
The fault-tolerant capacity is the capacity for the more general case where the encoding and decoding maps are also assumed to undergo noise \NoCaseChange{\protect\cite{cite1017}}.

Corrections to the capacity and tradeoff between decoding error, code rate and code length are determined using small \NoCaseChange{\protect\cite{cite1018,cite1019,cite1020}}, moderate \NoCaseChange{\protect\cite{cite1021,cite1022,cite1023}} and large \NoCaseChange{\protect\cite{cite1024,cite1025,cite1026,cite1027}} deviation analysis.
Sometimes the difference from the asymptotic rate at finite block length can be characterized by the \textit{channel dispersion} \NoCaseChange{\protect\cite{cite1020,cite1028}}.
Doeblin coefficients \NoCaseChange{\protect\cite{cite1029}} for classical channels have been studied \NoCaseChange{\protect\cite{cite1030}}.

\codefieldsection{Notes}
\begin{eczvaluelist}
\item\relax See Ref. \NoCaseChange{\protect\cite{cite1031}} for a list of open problems in coding theory.
\item\relax See Refs. \NoCaseChange{\protect\cite{cite1032,cite1033}} for reviews of coding theory.
\item\relax Classical systems such as RAM \NoCaseChange{\protect\cite{cite1034}}, HPC systems \NoCaseChange{\protect\cite{cite1035,cite1036,cite1037}}, and data centers \NoCaseChange{\protect\cite{cite1038}} suffer from noise.
\end{eczvaluelist}
\codefieldsection{Parent}
\begin{eczvaluelist}
\item\relax
\flmRefsHyperref[eczindexfamilyrel]{code:classical_into_quantum}{Classical-quantum (c-q) code} --- Any ECC can be embedded into a quantum Hilbert space, and thus passed through a quantum channel, by associating elements of the alphabet with basis vectors in a Hilbert space over the complex numbers. In other words, classical codewords are elements of an alphabet, while quantum codewords are functions on the alphabet. Classical codes can be unified with quantum codes using various algebraic frameworks \NoCaseChange{\protect\cite{cite1039,cite1040}}.
\end{eczvaluelist}
\codefieldsection{Children}
\begin{eczvaluelist}
\item\relax
\flmRefsHyperref[eczindexfamilyrel]{code:homogeneous_space_classical}{Homogeneous-space code}\item\relax
\flmRefsHyperref[eczindexfamilyrel]{code:block}{Block code}\item\relax
\flmRefsHyperref[eczindexfamilyrel]{code:generalized_concatenated}{Generalized concatenated code (GCC)}\item\relax
\flmRefsHyperref[eczindexfamilyrel]{code:parallel_concatenated}{Parallel concatenated code}\item\relax
\flmRefsHyperref[eczindexfamilyrel]{code:ecc_finite}{Finite-dimensional error-correcting code (ECC)}\item\relax
\flmRefsHyperref[eczindexfamilyrel]{code:group_orbit}{Group-orbit code} --- Not all codes are group-orbit codes, and more generally one can classify codewords into orbits of the automorphism group \NoCaseChange{\protect\cite{cite1041}}.
\item\relax
\flmRefsHyperref[eczindexfamilyrel]{code:random}{Random code}\item\relax
\flmRefsHyperref[eczindexfamilyrel]{code:convolutional}{Convolutional code}\end{eczvaluelist}
\codefieldsection{Cousins}
\begin{eczvaluelist}
\item\relax
\flmRefsHyperref[eczindexfamilyrel]{code:2pt_homogeneous}{Two-point homogeneous-space code} --- ECCs and \(t\)-designs on two-point homogeneous spaces are intimately related via association schemes \NoCaseChange{\protect\cite{cite226,cite916}}.
\item\relax
\flmRefsHyperref[eczindexfamilyrel]{code:t-designs}{\(t\)-design} --- ECCs and \(t\)-designs on two-point homogeneous spaces are intimately related via association schemes \NoCaseChange{\protect\cite{cite226,cite916}}.
\item\relax
\flmRefsHyperref[eczindexfamilyrel]{code:qecc}{Quantum error-correcting code (QECC)} --- Quantum information cannot be copied using a linear process \NoCaseChange{\protect\cite{cite1042}}, so one cannot send several copies of a quantum state through a channel as can be done for classical information. The \flmTerm{term}{ref1043}{}{Knill-Laflamme conditions} can similarly be formulated for classical codes \NoCaseChange{\protect\cite[{Sec. 3}]{cite1044}}, although they are not as widely used as those for quantum codes.
\end{eczvaluelist}
\eczhbkcontributors{ \eczhuVVA }
\endeczcode

\eczcode{ecc_finite}{Finite-dimensional error-correcting code (ECC)}{~\NoCaseChange{\protect\cite{cite1}}}
\codefieldsection{Description}
An error-correcting code defined over a finite alphabet.

\codefieldsection{Protection}
A code corrects errors associated with a noise channel if it is possible to recover any codeword after its coordinates have been altered during transmission through the channel. 

More technically, an error-correcting code can be specified by an \textit{encoder} \(e:[1\cdots K]\to\Sigma^n\) together with a set \(\mathcal{E}\) of correctable errors \(E:\Sigma^n\to Y\) such that there exists a \textit{decoder} \(d:Y\to[1\cdots K]\) satisfying \(d( E( e(x) ) )=x\) for all \(E\in\mathcal{E}\) and messages \(x\in[1\cdots K]\) \NoCaseChange{\protect\cite{cite398}}.

Finite ECCs can also be defined by axiomatically defining their encoding functions \NoCaseChange{\protect\cite{cite45}}.

\codefieldsection{Decoding}
\begin{eczvaluelist}
\item\relax Capacity-achieving Guessing Random Additive Noise Decoding (GRAND) \NoCaseChange{\protect\cite{cite1045}} (see also \NoCaseChange{\protect\cite{cite1046}}).
\end{eczvaluelist}
\codefieldsection{Notes}
\begin{eczvaluelist}
\item\relax The modern theory of error-correcting codes is rooted in the foundational work of C. Shannon \NoCaseChange{\protect\cite{cite1}}, but error-correcting codes have been used prior to that work \NoCaseChange{\protect\cite{cite246}}.
\item\relax Boolean networks, designed to model gene regulatory networks, generically develop error-correcting codes when they are evolved to perform computations \NoCaseChange{\protect\cite{cite1047}}.
\end{eczvaluelist}
\codefieldsection{Parent}
\begin{eczvaluelist}
\item\relax
\flmRefsHyperref[eczindexfamilyrel]{code:ecc}{Error-correcting code (ECC)}\end{eczvaluelist}
\codefieldsection{Children}
\begin{eczvaluelist}
\item\relax
\flmRefsHyperref[eczindexfamilyrel]{code:matrices_into_matrices}{Matrix-based code}\item\relax
\flmRefsHyperref[eczindexfamilyrel]{code:checksum}{Checksum code}\item\relax
\flmRefsHyperref[eczindexfamilyrel]{code:frameproof}{Frameproof (FP) code}\item\relax
\flmRefsHyperref[eczindexfamilyrel]{code:distributed_storage}{Distributed-storage code}\item\relax
\flmRefsHyperref[eczindexfamilyrel]{code:ldc}{Locally decodable code (LDC)}\item\relax
\flmRefsHyperref[eczindexfamilyrel]{code:ltc}{Locally testable code (LTC)}\item\relax
\flmRefsHyperref[eczindexfamilyrel]{code:batch}{Batch code}\item\relax
\flmRefsHyperref[eczindexfamilyrel]{code:pir}{Private information retrieval (PIR) code}\item\relax
\flmRefsHyperref[eczindexfamilyrel]{code:insertion_deletion}{Editing code}\item\relax
\flmRefsHyperref[eczindexfamilyrel]{code:q-ary_constant_weight}{Constant-weight block code}\item\relax
\flmRefsHyperref[eczindexfamilyrel]{code:small_distance}{Small-distance block code}\item\relax
\flmRefsHyperref[eczindexfamilyrel]{code:rings_into_rings}{Ring code}\end{eczvaluelist}
\codefieldsection{Cousin}
\begin{eczvaluelist}
\item\relax
\flmRefsHyperref[eczindexfamilyrel]{code:qecc_finite}{Finite-dimensional quantum error-correcting code} --- Finite-dimensional QECCs are quantum analogues of finite-dimensional classical ECCs.
\end{eczvaluelist}
\eczhbkcontributors{ \eczhuVVA }
\endeczcode

\eczcode{frameproof}{Frameproof (FP) code}{~\NoCaseChange{\protect\cite{cite1048,cite1049}}}
\codefieldsection{Description}
A block code designed to prevent a group of users from framing another user outside of the group for creating an unauthorized copy of data.
FP codes help to provide software protection from the illegal distribution and copying of computer software and copyrighted materials.
These codes help protect products of distributors as well as other naive users from being framed for illegal activity \NoCaseChange{\protect\cite{cite1050}}.

A \(c\)\textit{-separating} code has the property that, for any two disjoint sets that each contain at most \(c\) codewords, there is at least one position where the set of symbols of each set are disjoint \NoCaseChange{\protect\cite{cite1051}}.
Separating codes are, equivalently, codes with the \textit{secure FP (SFP)} property \NoCaseChange{\protect\cite{cite1049}}.

Let us define \(\Gamma = \{w^{(1)}, \dots, w^{(n)}\} \subseteq \mathbb{F}_2^{l}\) as an ( \(l,n\) )-code.
Each codeword \(w^{(i)}\) corresponds to a user \(u_i\).
Let \(C\) be a group of users.
A bit in position \(i\) is undetectable for the group \(C\) when the words assigned to the users in the group match at the same position \(i\).
The feasible set of the group \(C\), denoted \(F(C;\Gamma)\) or \(F(C)\), for some user \(u \in C\) contains all of the codewords that match the group's set of undetectable bits.
Finally, if every subset \(S \subset \Gamma\) of size at most \(c\) satisfies \(F(S)\cap\Gamma = S\), then \(\Gamma\) is a \(c\)\textit{-FP} code \NoCaseChange{\protect\cite{cite1050}}.

Any \(c\)-FP code must be of length at least \(c\) \NoCaseChange{\protect\cite{cite1050}}.
A length-\(l\) \(q\)-ary \(c\)-FP code has at most \(tq^{\lceil l/c \rceil} + O(q^{\lceil l/c \rceil - 1})\) codewords, where \(t\) is an integer between \(1\) and \(c\), and \(t \equiv l\) modulo \(c\) \NoCaseChange{\protect\cite{cite260}}.

\codefieldsection{Rate}
FP codes tend to have large minimum distance and low rate \NoCaseChange{\protect\cite{cite1050}}. Specifically, for any positive integers \(n\) and \(c\), if \(l = 16c^{2}\log n\), then there exists a \(c\)-FP \( (l,n) \)-code which has rate \((\log n)/l\) \NoCaseChange{\protect\cite{cite1050}}. See Ref. \NoCaseChange{\protect\cite{cite1052}} for other bounds on FP codes.
\codefieldsection{Realizations}
\begin{eczvaluelist}
\item\relax FP codes are utilized in digital fingerprinting and watermarking \NoCaseChange{\protect\cite{cite260}}.
\end{eczvaluelist}
\codefieldsection{Parents}
\begin{eczvaluelist}
\item\relax
\flmRefsHyperref[eczindexfamilyrel]{code:block}{Block code}\item\relax
\flmRefsHyperref[eczindexfamilyrel]{code:ecc_finite}{Finite-dimensional error-correcting code (ECC)}\end{eczvaluelist}
\codefieldsection{Child}
\begin{eczvaluelist}
\item\relax
\flmRefsHyperref[eczindexfamilyrel]{code:ipp}{Identifiable parent property (IPP) code} --- A \(t\)-IPP code is a \(t\)-SFP code, which further implies it is a \(t\)-FP code; see \NoCaseChange{\protect\cite{cite1053}\protect\cite[{Sec. I.C}]{cite349}}.
\end{eczvaluelist}
\codefieldsection{Cousins}
\begin{eczvaluelist}
\item\relax
\flmRefsHyperref[eczindexfamilyrel]{code:evaluation}{Evaluation AG code} --- Asymptotic bounds on FP codes can be formulated using evaluation AG codes \NoCaseChange{\protect\cite{cite1054,cite1055}}.
A sufficient condition for an evaluation AG code to be FP can be recast as an instance of the Riemann-Roch equation \NoCaseChange{\protect\cite[{Sec. 15.8.2}]{cite26}}.
AG-based constructions of binary \((2,1)\)-separating systems can beat a random-coding lower bound \NoCaseChange{\protect\cite[{Thm. 15.8.13}]{cite26}}.

\item\relax
\flmRefsHyperref[eczindexfamilyrel]{code:kerdock}{Kerdock code} --- Kerdock codes of sufficient order are separating \NoCaseChange{\protect\cite{cite1056,cite1057}}.
\end{eczvaluelist}
\eczhbkcontributors{ Raley Roberts, \eczhuVVA }
\endeczcode

\eczcode{generalized_concatenated}{Generalized concatenated code (GCC)}{~\NoCaseChange{\protect\cite{cite1058,cite1059}}}
\codefieldsection{Alternative Names}
\begin{eczvaluelist}
\item\relax Cascade code
\end{eczvaluelist}
\eczhIndexCodeAliasName{generalized_concatenated}{Cascade code}
\codefieldsection{Description}
A code that combines multiple outer codes of the same length and (possibly) different dimensions with a single inner code; see Refs. \NoCaseChange{\protect\cite{cite1059}\protect\cite[{Ch. 18}]{cite41}}.

\codefieldsection{Notes}
\begin{eczvaluelist}
\item\relax Introductions to GCCs \NoCaseChange{\protect\cite{cite1060,cite1061}}.
\end{eczvaluelist}
\codefieldsection{Parent}
\begin{eczvaluelist}
\item\relax
\flmRefsHyperref[eczindexfamilyrel]{code:ecc}{Error-correcting code (ECC)}\end{eczvaluelist}
\codefieldsection{Children}
\begin{eczvaluelist}
\item\relax
\flmRefsHyperref[eczindexfamilyrel]{code:justesen}{Justesen code} --- Justesen codes can be considered as a generalized concatenation of an outer RS code with \(N\) distinct binary inner codes.
\item\relax
\flmRefsHyperref[eczindexfamilyrel]{code:polar}{Polar code} --- Polar codes can be represented as generalized concatenations of their kernels.
\item\relax
\flmRefsHyperref[eczindexfamilyrel]{code:concatenated}{Concatenated code}\end{eczvaluelist}
\codefieldsection{Cousin}
\begin{eczvaluelist}
\item\relax
\flmRefsHyperref[eczindexfamilyrel]{code:concatenated_c-q}{Concatenated c-q code} --- Concatenated c-q codes are c-q analogues of generalized concatenated codes.
\end{eczvaluelist}
\eczhbkcontributors{ Alexander Barg, \eczhuVVA }
\endeczcode

\eczcode{group_orbit}{Group-orbit code}{}
\codefieldsection{Description}
Code whose set of codewords forms an orbit of some reference codeword under a subgroup of the \textit{automorphism group}, i.e., the group of distance-preserving transformations on the metric space defined with the code's alphabet.

\codefieldsection{Parent}
\begin{eczvaluelist}
\item\relax
\flmRefsHyperref[eczindexfamilyrel]{code:ecc}{Error-correcting code (ECC)} --- Not all codes are group-orbit codes, and more generally one can classify codewords into orbits of the automorphism group \NoCaseChange{\protect\cite{cite1041}}.
\end{eczvaluelist}
\codefieldsection{Children}
\begin{eczvaluelist}
\item\relax
\flmRefsHyperref[eczindexfamilyrel]{code:binary_group_orbit}{Binary group-orbit code} --- Binary group-orbit codes are group-orbit codes in Hamming space.
\item\relax
\flmRefsHyperref[eczindexfamilyrel]{code:group_linear}{Linear code over \(G\)} --- The set of codewords of a linear code over \(G\) can be thought of as an orbit of a particular codeword under the group formed by the code. However, group orbit codes do not have to be linear \NoCaseChange{\protect\cite[{Remark 8.4.3}]{cite115}}.
\item\relax
\flmRefsHyperref[eczindexfamilyrel]{code:spacetime_group}{Multi-channel group-orbit code}\item\relax
\flmRefsHyperref[eczindexfamilyrel]{code:slepian_group}{Slepian group-orbit code} --- Slepian group-orbit codes are group-orbit codes on spheres.
\end{eczvaluelist}
\codefieldsection{Cousin}
\begin{eczvaluelist}
\item\relax
\flmRefsHyperref[eczindexfamilyrel]{code:group}{Group-algebra code} --- A group-algebra code admits a regular, i.e., free and transitive, action on coordinates by a subgroup of its permutation automorphism group \NoCaseChange{\protect\cite[{Thm. 16.4.7}]{cite196}}. This differs from a group-orbit code, whose defining group action is transitive on codewords.
\end{eczvaluelist}
\eczhbkcontributors{ \eczhuVVA }
\endeczcode

\eczcode{ipp}{Identifiable parent property (IPP) code}{~\NoCaseChange{\protect\cite{cite1062}}}
\codefieldsection{Description}
A code that is embedded in copyrighted content in order to detect unauthorized redistribution of said content by pirates.
IPP codes are designed to detect pirates even when segments of content are mixed together so as to conceal the pirates' identities.

A unique codeword of an IPP code is hidden into each copy of the content (say, a movie or videogame) such that reselling a copy would expose the seller as a pirate.

To overcome this, a coalition of \(t\) pirates can mix their copies together so as to obfuscate their identities.
The \textit{descendant} bit-string corresponding to the new mixed copy therefore contains subsets of codeword coordinates of the copies that were used in the mixing.

More technically, let \(B=\{c^{(1)},\cdots,c^{(t)}\}\) be a subset of \(t\) IPP codewords, corresponding to the copies of the pirate coalition.
Then, \(a\) is called a \textit{descendant} of \(B\) if every coordinate of \(a\) also exists as a coordinate (in the same position) of at least one of the codewords in \(B\).
In other words, for all coordinates \(i\), there exists some \(c \in B\) such that \(a_i = c_i\).

A \(t\)-IPP code has the property that, given any descendant, a parent codeword can always be identified by examining all possible \(t\)-subsets of the code \NoCaseChange{\protect\cite{cite1053}}.

\codefieldsection{Rate}
A hypergraph approach can be used to show that, for any \(t \leq q\), there exists a sequence of IPP codes with asymptotically non-vanishing rate \NoCaseChange{\protect\cite{cite1063}}.
\codefieldsection{Parent}
\begin{eczvaluelist}
\item\relax
\flmRefsHyperref[eczindexfamilyrel]{code:frameproof}{Frameproof (FP) code} --- A \(t\)-IPP code is a \(t\)-SFP code, which further implies it is a \(t\)-FP code; see \NoCaseChange{\protect\cite{cite1053}\protect\cite[{Sec. I.C}]{cite349}}.
\end{eczvaluelist}
\codefieldsection{Child}
\begin{eczvaluelist}
\item\relax
\flmRefsHyperref[eczindexfamilyrel]{code:traceability}{Traceability code} --- Traceability codes allow for detection of parents of pirated descendant copies by only determining the closest codeword to the descendant; see \NoCaseChange{\protect\cite[{Lemma 1.3}]{cite349}}.
\end{eczvaluelist}
\eczhbkcontributors{ Fengxing Zhu, \eczhuVVA }
\endeczcode

\eczcode{lcc}{Locally correctable code (LCC)}{}
\codefieldsection{Description}
Recall that a block code encodes a length-\(k\) message into a length-\(n\) codeword, which is then sent through a noise channel to yield a received word.
Informally, an LCC is a block code for which one can recover any coordinate of a codeword from at most \(r\) coordinates of the received word (assuming the received word is within some tolerated corruption rate \(\delta\)).

Modified versions of local correctability include \textit{relaxed local correctability} \NoCaseChange{\protect\cite{cite1064}}.

\codefieldsection{Protection}
Three-query LCCs have to have length that is superpolynomial in the message length \NoCaseChange{\protect\cite{cite1065}}.

\codefieldsection{Notes}
\begin{eczvaluelist}
\item\relax See \NoCaseChange{\protect\cite{cite1066}} and Ref. \NoCaseChange{\protect\cite{cite1067}} for an introduction to LDCs and LCCs.
\item\relax See a popular summary of the result about three-query LCCs in \flmHref{https://www.quantamagazine.org/magical-error-correction-scheme-proved-inherently-inefficient-20240109}{Quanta Magazine}.
\end{eczvaluelist}
\codefieldsection{Parent}
\begin{eczvaluelist}
\item\relax
\flmRefsHyperref[eczindexfamilyrel]{code:locally_recoverable}{Locally recoverable code (LRC)} --- LRCs recover a coordinate from a small number of other coordinates of an uncorrupted codeword, while LCCs allow recovery from a corrupted received word within the decoding radius. Since an uncorrupted codeword is a special case of a received word, any LCC is also an LRC.
\end{eczvaluelist}
\codefieldsection{Child}
\begin{eczvaluelist}
\item\relax
\flmRefsHyperref[eczindexfamilyrel]{code:q-ary_lcc}{\(q\)-ary linear LCC}\end{eczvaluelist}
\codefieldsection{Cousins}
\begin{eczvaluelist}
\item\relax
\flmRefsHyperref[eczindexfamilyrel]{code:ldc}{Locally decodable code (LDC)} --- Any family of LCCs can be converted to a family of LDCs whose rate differs by a constant \NoCaseChange{\protect\cite{cite1068}}; see \NoCaseChange{\protect\cite[{Sec. 2.4.1}]{cite1067}}.
\item\relax
\flmRefsHyperref[eczindexfamilyrel]{code:ltc}{Locally testable code (LTC)} --- There are relations between LDCs and LTCs \NoCaseChange{\protect\cite{cite1069}}.
\item\relax
\flmRefsHyperref[eczindexfamilyrel]{code:quantum_locally_recoverable}{Quantum locally recoverable code (QLRC)} --- A quantum code cannot admit two disjoint local recovery sets for the same qudit unless that qudit is fixed, ruling out a natural quantum analogue of LCCs \NoCaseChange{\protect\cite[{Thm. 74}]{cite812}}.
\item\relax
\flmRefsHyperref[eczindexfamilyrel]{code:analog}{Analog code} --- LCCs can also be defined over the real or complex numbers, and there are no complex 2-query LCCs \NoCaseChange{\protect\cite{cite1070}}.
\end{eczvaluelist}
\eczhbkcontributors{ \eczhuVVA }
\endeczcode

\eczcode{ldc}{Locally decodable code (LDC)}{~\NoCaseChange{\protect\cite{cite1071}}}
\codefieldsection{Description}
Recall that a block code encodes a length-\(k\) message into a length-\(n\) codeword, which is then sent through a noise channel to yield a received word.
Informally, an LDC is a block code for which one can recover any coordinate of the message from at most \(r\) coordinates of the received word (assuming the received word is within some tolerated corruption rate \(\delta\)).
Efficiency of the decoding is quantified by the code's \textit{query complexity} \(r\), and decoding is performed by sampling subsets of \(r\) bits.

LDCs have applications in computational complexity theory and cryptography \NoCaseChange{\protect\cite{cite1073,cite1074,cite1075}\protect\cite[{Sec. 17.4}]{cite1072}}.

Modified versions of local decodability include \textit{relaxed local decodability} \NoCaseChange{\protect\cite{cite1076}}.

\codefieldsection{Rate}
Families of LDCs with query complexity \(r=2\) need \(n\) to scale exponentially with \(k\) \NoCaseChange{\protect\cite{cite1077,cite1078}}.
\codefieldsection{Decoding}
\begin{eczvaluelist}
\item\relax LDCs admit decoders whose runtime scales polylogarithmically with \(n\).
\end{eczvaluelist}
\codefieldsection{Notes}
\begin{eczvaluelist}
\item\relax See \NoCaseChange{\protect\cite{cite1066}} and Ref. \NoCaseChange{\protect\cite{cite1067,cite946}} for introductions to LDCs and LCCs.
\end{eczvaluelist}
\codefieldsection{Parents}
\begin{eczvaluelist}
\item\relax
\flmRefsHyperref[eczindexfamilyrel]{code:block}{Block code}\item\relax
\flmRefsHyperref[eczindexfamilyrel]{code:ecc_finite}{Finite-dimensional error-correcting code (ECC)}\end{eczvaluelist}
\codefieldsection{Cousins}
\begin{eczvaluelist}
\item\relax
\flmRefsHyperref[eczindexfamilyrel]{code:ltc}{Locally testable code (LTC)} --- There are relations between LDCs and LTCs \NoCaseChange{\protect\cite{cite1069}}.
\item\relax
\flmRefsHyperref[eczindexfamilyrel]{code:quantum_locally_recoverable}{Quantum locally recoverable code (QLRC)} --- There are quantum counterparts of LDCs, but they can be transformed into (classical) LDCs which can be decoded well on average \NoCaseChange{\protect\cite{cite1079}}.
\item\relax
\flmRefsHyperref[eczindexfamilyrel]{code:expander}{Expander code} --- Expander codes are locally decodable provided that the inner code satisfies certain properties; there exist code families with rate approaching one \NoCaseChange{\protect\cite{cite1080}}.
\item\relax
\flmRefsHyperref[eczindexfamilyrel]{code:lcc}{Locally correctable code (LCC)} --- Any family of LCCs can be converted to a family of LDCs whose rate differs by a constant \NoCaseChange{\protect\cite{cite1068}}; see \NoCaseChange{\protect\cite[{Sec. 2.4.1}]{cite1067}}.
\item\relax
\flmRefsHyperref[eczindexfamilyrel]{code:batch}{Batch code} --- Batch codes and LDCs are related \NoCaseChange{\protect\cite{cite944,cite949}\protect\cite[{Ch. 10.3}]{cite946}}.
\item\relax
\flmRefsHyperref[eczindexfamilyrel]{code:pir}{Private information retrieval (PIR) code} --- Any \textit{smooth} LDC yields a PIR scheme \NoCaseChange{\protect\cite{cite1081}}; see also Ref. \NoCaseChange{\protect\cite{cite1082}}.
\item\relax
\flmRefsHyperref[eczindexfamilyrel]{code:q-ary_lcc}{\(q\)-ary linear LCC} --- Linear LCCs can be converted into LDCs with the same locality \(r\) \NoCaseChange{\protect\cite[{Sec. 2.4.1}]{cite1067}}.
\end{eczvaluelist}
\eczhbkcontributors{ \eczhuVVA }
\endeczcode

\eczcode{locally_recoverable}{Locally recoverable code (LRC)}{}
\codefieldsection{Alternative Names}
\begin{eczvaluelist}
\item\relax Locally repairable code
\end{eczvaluelist}
\eczhIndexCodeAliasName{locally_recoverable}{Locally repairable code}
\codefieldsection{Description}
A block code for which one can recover any coordinate of a codeword from at most \(r\) other coordinates of the codeword \NoCaseChange{\protect\cite[{Def. 15.9.3}]{cite26}}.

An LRC of locality \(r\) is a block code for which, given a codeword \(c\) and coordinate \(c_i\), \(c_i\) can be recovered from at most \(r\) other coordinates of \(c\).
An \(r\)-locally recoverable code of length \(n\) and dimension \(k\) is denoted as an \((n,k,r)\) LRC.
The definition can be generalized to \(t\)-LRC, if every coordinate is recoverable from \(t\) disjoint subsets of coordinates; the corresponding \textit{availability} is the minimum number of recovery sets per coordinate \NoCaseChange{\protect\cite[{Def. 15.9.20}]{cite26}}.

\codefieldsection{Protection}
A Singleton-like bound is
\flmMathEnvironment{align}{}{
  d \leq n-k-\left\lceil\frac{k}{r}\right\rceil+2
}
for an \((n,k,r)\) LRC \NoCaseChange{\protect\cite[{Thm. 15.9.8}]{cite26}}.
A study on the parameters of \(t\)-LRCs, together with known bounds, can be found in Ref. \NoCaseChange{\protect\cite{cite1083}}.
See Ref. \NoCaseChange{\protect\cite{cite1084}} for more bounds on locally recoverable distributed storage codes.

\codefieldsection{Rate}
The rate of an \((n,k,r)\) LRC satisfies 
\flmMathEnvironment{align}{}{
\frac{k}{n}\leq\frac{r}{r+1}~.
}
Various asymptotic lower \NoCaseChange{\protect\cite{cite1085,cite1086}} and upper bounds \NoCaseChange{\protect\cite{cite1087}} exist.

\codefieldsection{Realizations}
\begin{eczvaluelist}
\item\relax An \((18,14,7)\) LRC has been used in the Windows Azure cloud storage system \NoCaseChange{\protect\cite{cite285}}; see also \NoCaseChange{\protect\cite[{31.3.1.2}]{cite183}}.
\item\relax Facebook f4 BLOB cloud storage system \NoCaseChange{\protect\cite{cite286}}
\end{eczvaluelist}
\codefieldsection{Notes}
\begin{eczvaluelist}
\item\relax See Ref. \NoCaseChange{\protect\cite{cite946}}.
\end{eczvaluelist}
\codefieldsection{Parent}
\begin{eczvaluelist}
\item\relax
\flmRefsHyperref[eczindexfamilyrel]{code:code_with_locality}{Code with locality} --- An LRC of locality \(r\) is a code with \((r,2)\) locality \NoCaseChange{\protect\cite[{Sec. 31.3.4.5}]{cite183}}.
\end{eczvaluelist}
\codefieldsection{Children}
\begin{eczvaluelist}
\item\relax
\flmRefsHyperref[eczindexfamilyrel]{code:lcc}{Locally correctable code (LCC)} --- LRCs recover a coordinate from a small number of other coordinates of an uncorrupted codeword, while LCCs allow recovery from a corrupted received word within the decoding radius. Since an uncorrupted codeword is a special case of a received word, any LCC is also an LRC.
\item\relax
\flmRefsHyperref[eczindexfamilyrel]{code:multiple_erasure_lrc}{\(t\)-erasure LRC} --- A \(t\)-erasure LRC is locally recoverable by definition since \(t > 1\).
\item\relax
\flmRefsHyperref[eczindexfamilyrel]{code:optimal_lrc}{Optimal LRC}\item\relax
\flmRefsHyperref[eczindexfamilyrel]{code:tamo_barg_vladut}{Barg-Tamo-Vladut code} --- Barg-Tamo-Vladut codes form a family of locally recoverable codes obtained from algebraic curves \NoCaseChange{\protect\cite[{Thm. 15.9.14}]{cite26}}.
\item\relax
\flmRefsHyperref[eczindexfamilyrel]{code:q-ary_ldpc}{\(q\)-ary LDPC code} --- LDPC codes are linear LRCs whose locality is the maximum number of nonzero entries in a row of the parity-check matrix \NoCaseChange{\protect\cite{cite812}}.
\end{eczvaluelist}
\codefieldsection{Cousins}
\begin{eczvaluelist}
\item\relax
\flmRefsHyperref[eczindexfamilyrel]{code:q-ary_linear}{Linear \(q\)-ary code} --- A \(q\)-ary linear code is an LRC of locality \(r\) if each coordinate participates in at least one parity check of weight \(\leq r\) \NoCaseChange{\protect\cite{cite812}\protect\cite[{Sec. 31.3.4.5}]{cite183}}.
\item\relax
\flmRefsHyperref[eczindexfamilyrel]{code:hadamard}{\([2^m,m,2^{m-1}]\) Hadamard code} --- The Hadamard code is an LRC with \(r=3\) \NoCaseChange{\protect\cite{cite812}}.
\item\relax
\flmRefsHyperref[eczindexfamilyrel]{code:regenerating}{Regenerating code (RGC)} --- RGCs and LRCs are related via the group repair with optimal access problem \NoCaseChange{\protect\cite{cite191}}.
\item\relax
\flmRefsHyperref[eczindexfamilyrel]{code:linearized_reed_solomon}{Linearized RS code} --- Linearized RS codes can be used to construct locally recoverable codes \NoCaseChange{\protect\cite{cite994}}.
\item\relax
\flmRefsHyperref[eczindexfamilyrel]{code:batch}{Batch code} --- A systematic batch code with restricted size of reconstruction sets can recover any query of \(t\) information symbols with recovery sets of size \(r\) \NoCaseChange{\protect\cite{cite947,cite948}}.
\item\relax
\flmRefsHyperref[eczindexfamilyrel]{code:pir}{Private information retrieval (PIR) code} --- LRCs and PIR codes are related \NoCaseChange{\protect\cite{cite1088,cite952}}: LRCs are designed to recover a codeword coordinate from a small set of other codeword coordinates, while PIR codes are designed to recover from many disjoint sets of arbitrary size \NoCaseChange{\protect\cite{cite1088}}.
\item\relax
\flmRefsHyperref[eczindexfamilyrel]{code:codes_with_availability}{Availability code} --- Availability is the number of recovery sets available for each coordinate of an LRC \NoCaseChange{\protect\cite[{Def. 15.9.20}]{cite26}}.
\item\relax
\flmRefsHyperref[eczindexfamilyrel]{code:q-ary_repetition}{\(q\)-ary repetition code} --- The \(q\)-ary repetition code is an LRC with \(r=2\) \NoCaseChange{\protect\cite{cite812}}.
\item\relax
\flmRefsHyperref[eczindexfamilyrel]{code:quantum_locally_recoverable}{Quantum locally recoverable code (QLRC)} --- QLRCs are quantum analogues of LRCs.
\end{eczvaluelist}
\eczhbkcontributors{ Mustafa Doger, \eczhuVVA }
\endeczcode

\eczcode{ltc}{Locally testable code (LTC)}{~\NoCaseChange{\protect\cite{cite1089,cite1090,cite1091,cite1092}}}
\codefieldsection{Description}
Code for which one can efficiently check whether a given string is a codeword or is far from a codeword. Efficiency of the verification is quantified by the code's \textit{query complexity} \(u\), while effectiveness is quantified by the code's \textit{soundness} \(R\).

Typically, one looks at how \(R\) scales with increasing code size for infinite families of codes, defining LTC families as those for which the soundness is asymptotically constant. Such LTC families with asymptotically constant distance, rate, and query complexity are called \(c^3\)\textit{-LTCs}. The first two such families are classical codes arising from the \flmRefsHyperref{code:expander_lifted_product}{expander lifted-product} quantum code construction and \flmRefsHyperref{code:lr-cayley-complex}{left-right Cayley complex} codes.

A technical definition for codes over binary alphabets is provided as follows; for general alphabets, see Ref. \NoCaseChange{\protect\cite{cite1093}}.
The idea behind LTCs is to be able to reliably test whether a given bit-string \(x\) is in the code by only sampling subsets of \(u\) bits.
To have something to check against, we first have to define a collection of length-\(u\) subsets \(S\) of bit locations that are called \textit{allowed local views}.
A code is an LTC if the following two conditions are satisfied \NoCaseChange{\protect\cite[{Thm. 1.1}]{cite88}}.

First, if \(x\) is a codeword, then all of its restrictions \(x|_S\) to the subsets \(S\) are allowed local views,
\flmMathEnvironment{align}{}{
  x\in C \Rightarrow x|_S \in \{\text{allowed local views}\}~.
}
This guarantees that codewords can indeed be determined from this limited sampling procedure.

Second, the probability that a given restriction is not an allowed local view is lower-bounded by the relative distance to the code,
\flmMathEnvironment{align}{}{
  \text{Pr}_S (x|_S\text{ not allowed local view}) \geq \frac{R}{n} D(x,C)~,
}
where \(D(x,C)\) is the Hamming distance between \(x\) and the closest codeword to \(x\). This condition ensures that strings \(x\) can be deemed to be not in the codespace with high probability, i.e., with probability increasing as \(x\) gets farther from the code.

\codefieldsection{Notes}
\begin{eczvaluelist}
\item\relax LTCs first appeared implicitly in works on probabilistically checkable proofs (PCPs) \NoCaseChange{\protect\cite{cite1094,cite1095,cite1096,cite1089,cite1097,cite1098}}; see Ref. \NoCaseChange{\protect\cite{cite1093}} for a review.
\end{eczvaluelist}
\codefieldsection{Parents}
\begin{eczvaluelist}
\item\relax
\flmRefsHyperref[eczindexfamilyrel]{code:block}{Block code}\item\relax
\flmRefsHyperref[eczindexfamilyrel]{code:ecc_finite}{Finite-dimensional error-correcting code (ECC)}\end{eczvaluelist}
\codefieldsection{Child}
\begin{eczvaluelist}
\item\relax
\flmRefsHyperref[eczindexfamilyrel]{code:q-ary_ltc}{\(q\)-ary linear LTC}\end{eczvaluelist}
\codefieldsection{Cousins}
\begin{eczvaluelist}
\item\relax
\flmRefsHyperref[eczindexfamilyrel]{code:qltc}{Quantum locally testable code (QLTC)} --- QLTCs are quantum analogues of LTCs.
\item\relax
\flmRefsHyperref[eczindexfamilyrel]{code:ldc}{Locally decodable code (LDC)} --- There are relations between LDCs and LTCs \NoCaseChange{\protect\cite{cite1069}}.
\item\relax
\flmRefsHyperref[eczindexfamilyrel]{code:lcc}{Locally correctable code (LCC)} --- There are relations between LDCs and LTCs \NoCaseChange{\protect\cite{cite1069}}.
\item\relax
\flmRefsHyperref[eczindexfamilyrel]{code:multiplicity}{Multiplicity code} --- Some multiplicity codes are locally testable by an appropriate test \NoCaseChange{\protect\cite{cite1099,cite1100}}.
\item\relax
\flmRefsHyperref[eczindexfamilyrel]{code:tanner}{Tanner code} --- Tanner codes can be generalized to \textit{sheaf codes}, whose local codes satisfy a certain hierarchy. This allows for a way to formulate and understand many generalized homological-product CSS codes \NoCaseChange{\protect\cite{cite1101}} and LTCs \NoCaseChange{\protect\cite{cite1102}}.
\item\relax
\flmRefsHyperref[eczindexfamilyrel]{code:balanced}{Balanced code} --- Random low-rate unbiased linear codes are LTCs \NoCaseChange{\protect\cite{cite1103}}.
\item\relax
\flmRefsHyperref[eczindexfamilyrel]{code:lossless_expander}{Lossless expander balanced-product code} --- Using one part of a quantum-code chain complex constructed with one-sided loss expanders yields a \(c^3\)-LTC \NoCaseChange{\protect\cite{cite186}}.
\item\relax
\flmRefsHyperref[eczindexfamilyrel]{code:hypergraph_product}{Hypergraph product (HGP) code} --- Applying the hypergraph product to an LTC yields a code which provides an explicit example of \textit{No Low-Error Trivial States (NLETS)} \NoCaseChange{\protect\cite{cite1104}}.
\end{eczvaluelist}
\eczhbkcontributors{ \eczhuVVA }
\endeczcode

\eczcode{maximally_recoverable}{Maximally recoverable (MR) code}{~\NoCaseChange{\protect\cite{cite1105,cite1106}}}
\codefieldsection{Alternative Names}
\begin{eczvaluelist}
\item\relax Partial MDS code
\end{eczvaluelist}
\eczhIndexCodeAliasName{maximally_recoverable}{Partial MDS code}
\codefieldsection{Description}
A code with \((r,\delta)\) locality such that puncturing any \(\delta-1\) coordinates from each local \([r+\delta-1,r,\delta]\) code yields an MDS code.

\codefieldsection{Realizations}
\begin{eczvaluelist}
\item\relax RAID storage \NoCaseChange{\protect\cite{cite291}}.
\end{eczvaluelist}
\codefieldsection{Parent}
\begin{eczvaluelist}
\item\relax
\flmRefsHyperref[eczindexfamilyrel]{code:code_with_locality}{Code with locality}\end{eczvaluelist}
\eczhbkcontributors{ \eczhuVVA }
\endeczcode

\eczcode{optimal_lrc}{Optimal LRC}{~\NoCaseChange{\protect\cite{cite1107,cite1108}}}
\codefieldsection{Description}
An LRC whose parameters saturate a generalized Singleton bound.

A \((n,k,r)\) LRC with distance \(d\) is optimal if its parameters \(n\), \(k\), \(d\), and \(q\) are such that the generalized Singleton bound
\flmMathEnvironment{align}{}{
d\leq n-k-\left \lceil\frac{k}{r}\right \rceil+2\label{ref1109}
}
becomes an equality.
When \(k=r\), the generalized Singleton bound becomes the Singleton bound.

The generalized Singleton bound \eqref{ref1109} does not account for \(q\)-ary alphabet size.
A more general bound (including the nonlinear case) is given in Ref. \NoCaseChange{\protect\cite{cite1086}}.

\codefieldsection{Parent}
\begin{eczvaluelist}
\item\relax
\flmRefsHyperref[eczindexfamilyrel]{code:locally_recoverable}{Locally recoverable code (LRC)}\end{eczvaluelist}
\codefieldsection{Children}
\begin{eczvaluelist}
\item\relax
\flmRefsHyperref[eczindexfamilyrel]{code:pyramid}{Pyramid code}\item\relax
\flmRefsHyperref[eczindexfamilyrel]{code:tamo_barg}{Tamo-Barg code} --- For \(r \mid k\), Tamo-Barg codes meet the LRC Singleton bound \NoCaseChange{\protect\cite[{Sec. 15.9.3}]{cite26}}.
\item\relax
\flmRefsHyperref[eczindexfamilyrel]{code:mds}{Maximum distance separable (MDS) code} --- The generalized Singleton bound becomes the Singleton bound for \(k=r\), so optimal LRCs with that constraint are MDS.
\end{eczvaluelist}
\eczhbkcontributors{ \eczhuVVA }
\endeczcode

\eczcode{parallel_concatenated}{Parallel concatenated code}{}
\codefieldsection{Description}
A code obtained by encoding the same information sequence using two or more component encoders in parallel, often with one branch preceded by an interleaver. Turbo codes are a standard example \NoCaseChange{\protect\cite{cite972}}.

\codefieldsection{Parent}
\begin{eczvaluelist}
\item\relax
\flmRefsHyperref[eczindexfamilyrel]{code:ecc}{Error-correcting code (ECC)}\end{eczvaluelist}
\codefieldsection{Children}
\begin{eczvaluelist}
\item\relax
\flmRefsHyperref[eczindexfamilyrel]{code:turbo}{Turbo code}\item\relax
\flmRefsHyperref[eczindexfamilyrel]{code:tanner}{Tanner code} --- Tanner codes are examples of parallel concatenation \NoCaseChange{\protect\cite{cite972}}.
\end{eczvaluelist}
\codefieldsection{Cousin}
\begin{eczvaluelist}
\item\relax
\flmRefsHyperref[eczindexfamilyrel]{code:tensor}{Tensor-product code} --- Tensor-product codes can be viewed both as serial or parallel concatenated codes \NoCaseChange{\protect\cite{cite972}}.
\end{eczvaluelist}
\eczhbkcontributors{ \eczhuVVA }
\endeczcode

\eczcode{pir}{Private information retrieval (PIR) code}{~\NoCaseChange{\protect\cite{cite1110,cite1111}}}
\codefieldsection{Description}
A code used to obtain information from several servers privately, i.e., without the servers knowing what information was obtained.

A \(k\)\textit{-server PIR code} is a block code for which one can recover any coordinate of a codeword from \(k\) disjoint sets of coordinates of the codeword \NoCaseChange{\protect\cite{cite1088}}.

\codefieldsection{Notes}
\begin{eczvaluelist}
\item\relax See Ref. \NoCaseChange{\protect\cite{cite946,cite1111}}.
\end{eczvaluelist}
\codefieldsection{Parents}
\begin{eczvaluelist}
\item\relax
\flmRefsHyperref[eczindexfamilyrel]{code:block}{Block code}\item\relax
\flmRefsHyperref[eczindexfamilyrel]{code:ecc_finite}{Finite-dimensional error-correcting code (ECC)}\end{eczvaluelist}
\codefieldsection{Cousins}
\begin{eczvaluelist}
\item\relax
\flmRefsHyperref[eczindexfamilyrel]{code:multiplicity}{Multiplicity code} --- Multiplicity codes can be used to construct PIR codes \NoCaseChange{\protect\cite{cite951}}.
\item\relax
\flmRefsHyperref[eczindexfamilyrel]{code:batch}{Batch code} --- Batch and PIR codes are related \NoCaseChange{\protect\cite{cite949,cite952}}.
\item\relax
\flmRefsHyperref[eczindexfamilyrel]{code:locally_recoverable}{Locally recoverable code (LRC)} --- LRCs and PIR codes are related \NoCaseChange{\protect\cite{cite1088,cite952}}: LRCs are designed to recover a codeword coordinate from a small set of other codeword coordinates, while PIR codes are designed to recover from many disjoint sets of arbitrary size \NoCaseChange{\protect\cite{cite1088}}.
\item\relax
\flmRefsHyperref[eczindexfamilyrel]{code:ldc}{Locally decodable code (LDC)} --- Any \textit{smooth} LDC yields a PIR scheme \NoCaseChange{\protect\cite{cite1081}}; see also Ref. \NoCaseChange{\protect\cite{cite1082}}.
\item\relax
\flmRefsHyperref[eczindexfamilyrel]{code:reed_solomon}{Reed-Solomon (RS) code} --- RS codes can be used to construct PIR codes \NoCaseChange{\protect\cite{cite1112}}.
\end{eczvaluelist}
\eczhbkcontributors{ \eczhuVVA }
\endeczcode

\eczcode{quasi_cyclic}{Quasi-cyclic code}{~\NoCaseChange{\protect\cite{cite1113}}}
\codefieldsection{Description}
A block code of length \(n\) is quasi-cyclic if, for each codeword \(c_1 \cdots c_{\ell} c_{\ell+1} \cdots c_n\), the cyclic shift by \(\ell\) positions, \(c_{n-\ell+1} \cdots c_n c_1 \cdots c_{n-\ell}\), is also a codeword.
Codes for which \(\ell = 1\) are cyclic, while codes for which \(\ell = 2\) are called \textit{double circulant}.

The generator of an \([mn_0,mk_0]\) quasi-cyclic linear code is representable as a block matrix of \(m \times m\) circulant matrices \NoCaseChange{\protect\cite{cite41,cite1114}}.

Quasi-cyclic codes can also be understood in terms of cyclic-shift automorphisms.
Cyclic codes are invariant under a shift by one position, while quasi-cyclic codes are invariant under a shift by \(\ell\) positions.

\codefieldsection{Notes}
\begin{eczvaluelist}
\item\relax A database of quasi-cyclic codes with searchable parameters such as block length and dimension is constructed and displayed \flmHref{http://www.tec.hkr.se/~chen/research/codes/qc.htm}{here}.
\item\relax See \NoCaseChange{\protect\cite[{Ch. 16}]{cite41}} for a review of double circulant codes.
\end{eczvaluelist}
\codefieldsection{Parent}
\begin{eczvaluelist}
\item\relax
\flmRefsHyperref[eczindexfamilyrel]{code:quasi_twisted}{Quasi-twisted code} --- Quasi-twisted codes with \(\alpha=1\) are quasi-cyclic.
\end{eczvaluelist}
\codefieldsection{Children}
\begin{eczvaluelist}
\item\relax
\flmRefsHyperref[eczindexfamilyrel]{code:karlin}{\([2m+2,m+1]\) Karlin code} --- Karlin codes can be mapped to extended cyclic and extended quadratic-residue codes over \(\mathbb{F}_4\) \NoCaseChange{\protect\cite{cite109,cite110}\protect\cite[{Ch. 16}]{cite41}\protect\cite[{Sec. 2.4.2}]{cite42}} by identifying \((0,\omega,\bar{\omega},1)\) with \((00),(10),(01),(11)\) \NoCaseChange{\protect\cite{cite109}}.
\item\relax
\flmRefsHyperref[eczindexfamilyrel]{code:qc_ldpc}{Quasi-cyclic LDPC (QC-LDPC) code}\item\relax
\flmRefsHyperref[eczindexfamilyrel]{code:cyclic}{Cyclic code} --- Quasi-cyclic codes with \(\ell=1\) are cyclic.
\item\relax
\flmRefsHyperref[eczindexfamilyrel]{code:pless_symmetry}{\([2q+2,q+1]_3\) Pless symmetry code} --- Pless symmetry codes are double circulant \NoCaseChange{\protect\cite[{pg. 510}]{cite41}}.
\end{eczvaluelist}
\codefieldsection{Cousins}
\begin{eczvaluelist}
\item\relax
\flmRefsHyperref[eczindexfamilyrel]{code:quasi_group}{Quasi group-algebra code} --- A quasi group-algebra code for \(G\) being the cyclic group is a quasi-cyclic \(q\)-ary linear code \NoCaseChange{\protect\cite[{pg. 4}]{cite1115}}.
\item\relax
\flmRefsHyperref[eczindexfamilyrel]{code:self_dual}{Self-dual linear code} --- Quasi-cyclic self-dual constructions include double circulant codes and, in odd characteristic, their negacirculant analogs such as double negacirculant and four-negacirculant codes \NoCaseChange{\protect\cite[{Sec. 4.4}]{cite40}}.
\item\relax
\flmRefsHyperref[eczindexfamilyrel]{code:convolutional}{Convolutional code} --- Quasi-cyclic codes can be \textit{unwrapped} to obtain convolutional codes \NoCaseChange{\protect\cite{cite1116,cite1117,cite1118,cite1119,cite1120,cite1121,cite1122}}.
\item\relax
\flmRefsHyperref[eczindexfamilyrel]{code:sc_qldpc}{Quantum spatially coupled (SC-QLDPC) code} --- Quasi-cyclic binary code parity-check matrices can be used as sub-matrices to define a 1D SC-QLDPC code \NoCaseChange{\protect\cite{cite643}}.
\item\relax
\flmRefsHyperref[eczindexfamilyrel]{code:quantum_divisible}{Quantum divisible code} --- Certain double circulant codes can be used to construct doubly even \(\llbracket 55,1,11\rrbracket \) and \(\llbracket 87,1,15\rrbracket \) codes \NoCaseChange{\protect\cite{cite1123}}.
\item\relax
\flmRefsHyperref[eczindexfamilyrel]{code:binary_quad_residue}{Binary quadratic-residue (QR) code} --- Binary QR codes are equivalent to double circulant codes for all \(n<200\) except 89 and 167 \NoCaseChange{\protect\cite{cite1124}}.
\item\relax
\flmRefsHyperref[eczindexfamilyrel]{code:skew_cyclic}{Skew-cyclic code} --- Under certain conditions, there is an equivalent quasi-cyclic or cyclic code for every skew-cyclic code \NoCaseChange{\protect\cite{cite1125}}.
\item\relax
\flmRefsHyperref[eczindexfamilyrel]{code:quantum_quasi_cyclic}{Quasi-cyclic quantum code} --- Quasi-cyclic quantum codes are quantum analogues of quasi-cyclic codes.
\end{eczvaluelist}
\eczhbkcontributors{ Micah Shaw, Nolan Coble, \eczhuVVA }
\endeczcode

\eczcode{quasi_twisted}{Quasi-twisted code}{}
\codefieldsection{Description}
A block code of length \(n\) is \(\alpha\)-quasi-twisted if, for each codeword \(c_1 \cdots c_{\ell} c_{\ell+1} \cdots c_n\) with \(\ell\mid n\), the string \(\alpha c_{n-\ell+1}, \alpha c_{n-\ell+2}, \cdots, \alpha c_n, c_1, c_2, \cdots, c_{n-\ell}\) is also a codeword \NoCaseChange{\protect\cite[{Def. 3.2.7}]{cite70}\protect\cite[{Sec. 17.7}]{cite1126}}.

\codefieldsection{Parent}
\begin{eczvaluelist}
\item\relax
\flmRefsHyperref[eczindexfamilyrel]{code:block}{Block code}\end{eczvaluelist}
\codefieldsection{Children}
\begin{eczvaluelist}
\item\relax
\flmRefsHyperref[eczindexfamilyrel]{code:quasi_cyclic}{Quasi-cyclic code} --- Quasi-twisted codes with \(\alpha=1\) are quasi-cyclic.
\item\relax
\flmRefsHyperref[eczindexfamilyrel]{code:constacyclic}{Constacyclic code} --- Quasi-twisted codes with \(\ell=1\) are constacyclic.
\item\relax
\flmRefsHyperref[eczindexfamilyrel]{code:hill_56_6_36}{\([56,6,36]_3\) Hill-cap code} --- The \([56,6,36]_3\) Hill-cap code is quasi-twisted \NoCaseChange{\protect\cite{cite208}}.
\end{eczvaluelist}
\eczhbkcontributors{ Connor Clayton, \eczhuVVA }
\endeczcode

\eczcode{random}{Random code}{~\NoCaseChange{\protect\cite{cite1}}}
\codefieldsection{Description}
Code whose construction is non-deterministic in some way, i.e., codes that utilize an element of randomness somewhere in their construction. Members of this class range from fully non-deterministic codes to codes whose multi-step construction is deterministic except for a single step.

Typically, random codes are selected with a uniform distribution from some ensemble of codes. 
For example, a random binary code is a set of \(K\) codewords of length \(n\) chosen uniformly from the ensemble of all \(2^n\) bit-strings. Each bit in each codeword is chosen independently and uniformly. 
For another example, a random binary linear code is generated from \(k\) randomly chosen generators of length \(n\). Equivalently, it is defined by a random \(k \times n\) binary generator matrix whose entries are chosen independently and uniformly.

In both of the above random code constructions, the ensemble size scales exponentially with \(n\). A common convention is to think of the resulting code constructions as effectively explicit (as opposed to random) in cases where the ensemble size is independent of \(n\) or even when the size scales polynomially with \(n\).

\codefieldsection{Rate}
Typical random codes (TRC) or typical random linear codes (TLC) refer to codes in the respective ensemble that satisfy a certain minimum-distance property. The relative fraction of typical codes in the ensemble approaches one as \(n\) goes to infinity \NoCaseChange{\protect\cite{cite1}} (see also Ref. \NoCaseChange{\protect\cite{cite1127}}). Asymptotically, given relative distance \(\delta=d/n\), the maximum rate for a TRC is \(R=\frac{1}{2}R_{GV}(\delta)\), where \(R_{GV}(\delta)=1-h(\delta)\) is the Gilbert-Varshamov (GV) bound and \(h(\delta)\) is the binary entropy function. The maximum rate for a TLC is \(R=R_{GV}(\delta)\), meaning that TLCs achieve the \flmRefsHyperref{ref85}{asymptotic GV bound}.
\codefieldsection{Decoding}
\begin{eczvaluelist}
\item\relax Ball-collision decoding \NoCaseChange{\protect\cite{cite1128}}.
\item\relax Information set decoding (ISD) \NoCaseChange{\protect\cite{cite1129}} and Finiasz and Sendrier (FS-ISD) decoding \NoCaseChange{\protect\cite{cite1130}}.
\end{eczvaluelist}
\codefieldsection{Realizations}
\begin{eczvaluelist}
\item\relax Distributed storage systems \NoCaseChange{\protect\cite{cite319}}.
\item\relax Classical and quantum cryptography based on the learning-with-errors problem, which is related to decoding a random linear code \NoCaseChange{\protect\cite{cite320}}.
\item\relax Random codes can be used to realize secure computation \NoCaseChange{\protect\cite{cite321}}.
\end{eczvaluelist}
\codefieldsection{Notes}
\begin{eczvaluelist}
\item\relax Shannon's pioneering work \NoCaseChange{\protect\cite{cite1}} analyzes the distance distribution of the code given a rate. Given \(n\) and rate \(R\), the minimum distance of a TRC is governed asymptotically by the \flmRefsHyperref{ref85}{GV bound} \(d=n\delta_{GV}(2R)\), where \(\delta_{GV}(x)=h^{-1}(1-x)\) for \(0\leq x \leq 1\). For a TLC, the minimum distance is similarly governed by \(d=n\delta_{GV}(R)\).
\end{eczvaluelist}
\codefieldsection{Parent}
\begin{eczvaluelist}
\item\relax
\flmRefsHyperref[eczindexfamilyrel]{code:ecc}{Error-correcting code (ECC)}\end{eczvaluelist}
\codefieldsection{Children}
\begin{eczvaluelist}
\item\relax
\flmRefsHyperref[eczindexfamilyrel]{code:justesen}{Justesen code} --- The required inner codes are obtained by random sampling from the Wozencraft ensemble, whose length scales logarithmically with \(n\).
\item\relax
\flmRefsHyperref[eczindexfamilyrel]{code:mn_ldpc}{MacKay-Neal LDPC (MN-LDPC) code}\item\relax
\flmRefsHyperref[eczindexfamilyrel]{code:meir}{Meir code} --- Meir codes are defined using a probabilistic combinatorial construction.
\end{eczvaluelist}
\codefieldsection{Cousins}
\begin{eczvaluelist}
\item\relax
\flmRefsHyperref[eczindexfamilyrel]{code:fountain}{Fountain code} --- Fountain codes are typically generated from random sparse encoding ensembles.
\item\relax
\flmRefsHyperref[eczindexfamilyrel]{code:gs-ltc}{Goldreich-Sudan code} --- Randomness enters the probabilistic constructions associated with Goldreich-Sudan LTCs.
\item\relax
\flmRefsHyperref[eczindexfamilyrel]{code:ldpc}{Low-density parity-check (LDPC) code} --- LDPC codes are often constructed nondeterministically.
\item\relax
\flmRefsHyperref[eczindexfamilyrel]{code:generalized_reed_solomon}{Generalized RS (GRS) code} --- Concatenations of GRS codes with random linear codes almost surely attain the \flmRefsHyperref{ref85}{GV bound} \NoCaseChange{\protect\cite{cite973}}.
\item\relax
\flmRefsHyperref[eczindexfamilyrel]{code:quantum_random}{Random quantum code} --- Random quantum codes are quantum analogues of random classical codes.
\item\relax
\flmRefsHyperref[eczindexfamilyrel]{code:expander_lifted_product}{Expander LP code} --- Expander lifted-product codes are quantum CSS codes that utilize short classical codes in their construction which need to satisfy some properties \NoCaseChange{\protect\cite[{Lemma 10}]{cite184}}. It is shown that such codes exist, but they are not explicitly constructed. Such codes can be obtained by repeated random sampling or by performing a search of all codes of desired length. Nevertheless, since the length of the desired short codes does not scale with \(n\), this construction is effectively explicit.
\end{eczvaluelist}
\eczhbkcontributors{ Xiao Xiao, \eczhuVVA }
\endeczcode

\eczcode{reversible}{Reversible code}{}
\codefieldsection{Description}
A block code of length \(n\) over an alphabet is reversible if, for each codeword \(c_1 c_2 \cdots c_n\), the reversed string \(c_n \cdots c_2 c_1\) is also a codeword.

\codefieldsection{Notes}
\begin{eczvaluelist}
\item\relax Reversible cyclic codes are studied in \NoCaseChange{\protect\cite[{Sec. 2.10}]{cite68}}.
\end{eczvaluelist}
\codefieldsection{Parent}
\begin{eczvaluelist}
\item\relax
\flmRefsHyperref[eczindexfamilyrel]{code:block}{Block code}\end{eczvaluelist}
\codefieldsection{Child}
\begin{eczvaluelist}
\item\relax
\flmRefsHyperref[eczindexfamilyrel]{code:melas}{\([2^m -1, 2^m - 1 - 2m, 5]\) Melas code}\end{eczvaluelist}
\codefieldsection{Cousins}
\begin{eczvaluelist}
\item\relax
\flmRefsHyperref[eczindexfamilyrel]{code:q-ary_cyclic}{Cyclic linear \(q\)-ary code} --- A reversible cyclic code is a cyclic code with self-reciprocal generator polynomial and is an LCD code \NoCaseChange{\protect\cite[{Thm. 2.10.3}]{cite68}}.
\item\relax
\flmRefsHyperref[eczindexfamilyrel]{code:lcd}{Linear code with complementary dual (LCD)} --- A reversible cyclic code is a cyclic code with self-reciprocal generator polynomial and is an LCD code \NoCaseChange{\protect\cite[{Thm. 2.10.3}]{cite68}}.
\item\relax
\flmRefsHyperref[eczindexfamilyrel]{code:quantum_tensor_product}{Quantum tensor-product code} --- Reversible cyclic codes can be used to construct quantum tensor-product codes \NoCaseChange{\protect\cite{cite1131}}.
\end{eczvaluelist}
\eczhbkcontributors{ \eczhuVVA }
\endeczcode

\eczcode{skew_cyclic}{Skew-cyclic code}{~\NoCaseChange{\protect\cite{cite1132}}}
\codefieldsection{Description}
A block code \(C\) of length \(n\) over an alphabet \(R\) is skew-cyclic if there exists an automorphism, \(\theta\), of \(R\), such that for each string \(c_1 c_2 \cdots c_n\in C\), the skew-cyclically shifted string \(\theta(c_n) \theta(c_1) \cdots \theta(c_{n-1})\in C\). We say that \(C\) is a \(\theta\)-cyclic code over \(R\).
\codefieldsection{Notes}
\begin{eczvaluelist}
\item\relax Computer algebra software is used to find most codes of this type. Ref. \NoCaseChange{\protect\cite{cite1132}} gives several examples of codes, which have slightly improved minimum distance for some \((n,k)\) codes.
\end{eczvaluelist}
\codefieldsection{Parent}
\begin{eczvaluelist}
\item\relax
\flmRefsHyperref[eczindexfamilyrel]{code:block}{Block code}\end{eczvaluelist}
\codefieldsection{Child}
\begin{eczvaluelist}
\item\relax
\flmRefsHyperref[eczindexfamilyrel]{code:cyclic}{Cyclic code} --- Skew-cyclic codes with \(\theta\) trivial are cyclic.
\end{eczvaluelist}
\codefieldsection{Cousins}
\begin{eczvaluelist}
\item\relax
\flmRefsHyperref[eczindexfamilyrel]{code:quasi_cyclic}{Quasi-cyclic code} --- Under certain conditions, there is an equivalent quasi-cyclic or cyclic code for every skew-cyclic code \NoCaseChange{\protect\cite{cite1125}}.
\item\relax
\flmRefsHyperref[eczindexfamilyrel]{code:q-ary_bch}{Bose–Chaudhuri–Hocquenghem (BCH) code} --- There exist skew-BCH codes, which are skew-cyclic versions of BCH codes \NoCaseChange{\protect\cite{cite1132}}.
\item\relax
\flmRefsHyperref[eczindexfamilyrel]{code:skew-cyclic_galois_css}{Skew-cyclic CSS code} --- Skew-cyclic CSS codes are constructed from classical skew-cyclic codes over rings.
\end{eczvaluelist}
\eczhbkcontributors{ Nolan Coble, \eczhuVVA }
\endeczcode

\eczcode{small_distance}{Small-distance block code}{}
\codefieldsection{Description}
A block code of length \(n\) that either detects or corrects errors on at most two coordinates, i.e., has distance \(d \leq 5\).

See Ref. \NoCaseChange{\protect\cite{cite1133}} for a comparison of short block codes.

\codefieldsection{Parents}
\begin{eczvaluelist}
\item\relax
\flmRefsHyperref[eczindexfamilyrel]{code:block}{Block code}\item\relax
\flmRefsHyperref[eczindexfamilyrel]{code:ecc_finite}{Finite-dimensional error-correcting code (ECC)}\end{eczvaluelist}
\codefieldsection{Children}
\begin{eczvaluelist}
\item\relax
\flmRefsHyperref[eczindexfamilyrel]{code:melas}{\([2^m -1, 2^m - 1 - 2m, 5]\) Melas code}\item\relax
\flmRefsHyperref[eczindexfamilyrel]{code:petersen}{\([15,6,5]\) Petersen cycle code}\item\relax
\flmRefsHyperref[eczindexfamilyrel]{code:nadler}{\((12,32,5)\) Nadler code}\item\relax
\flmRefsHyperref[eczindexfamilyrel]{code:best}{\((10,40,4)\) Best code}\item\relax
\flmRefsHyperref[eczindexfamilyrel]{code:sloane_whitehead}{Sloane-Whitehead code}\item\relax
\flmRefsHyperref[eczindexfamilyrel]{code:vasilyev}{\((2^{m+1}-1,2^{2n-m},3)\) Vasilyev code}\item\relax
\flmRefsHyperref[eczindexfamilyrel]{code:simplex734}{\([7,3,4]\) simplex code}\item\relax
\flmRefsHyperref[eczindexfamilyrel]{code:extended_hamming}{\([2^m,2^m-m-1,4]\) Extended Hamming code}\item\relax
\flmRefsHyperref[eczindexfamilyrel]{code:hexacode}{\([6,3,4]_4\) Hexacode}\item\relax
\flmRefsHyperref[eczindexfamilyrel]{code:q-ary_hamming}{\(q\)-ary Hamming code}\item\relax
\flmRefsHyperref[eczindexfamilyrel]{code:q-ary_parity_check}{\([n,n-1,2]_q\) \(q\)-ary parity-check code}\item\relax
\flmRefsHyperref[eczindexfamilyrel]{code:reed_solomon_4}{\([4,2,3]_4\) RS\(_4\) code}\item\relax
\flmRefsHyperref[eczindexfamilyrel]{code:shortened_hexacode}{\([5,3,3]_4\) Shortened hexacode}\item\relax
\flmRefsHyperref[eczindexfamilyrel]{code:ternary_golay}{\([11,6,5]_3\) Ternary Golay code}\item\relax
\flmRefsHyperref[eczindexfamilyrel]{code:self_dual_z6}{\([4,2,2]_{\mathbb{Z}_6}\) senary code}\item\relax
\flmRefsHyperref[eczindexfamilyrel]{code:pentacode}{\((5,40,4)_{\mathbb{Z}_4}\) Pentacode}\end{eczvaluelist}
\codefieldsection{Cousins}
\begin{eczvaluelist}
\item\relax
\flmRefsHyperref[eczindexfamilyrel]{code:small_distance_quantum}{Small-distance block quantum code} --- Small-distance block quantum codes are quantum analogues of small-distance block codes.
\item\relax
\flmRefsHyperref[eczindexfamilyrel]{code:preparata}{Preparata code} --- Shortened Preparata codes form an infinite family of small-distance block codes with minimum distance \(5\).
\item\relax
\flmRefsHyperref[eczindexfamilyrel]{code:nordstrom_robinson}{\((16,256,6)\) Nordstrom-Robinson (NR) code} --- The NR code can be shortened to produce unique \((15, 256, 5)\), \((14, 128, 5)\), and \((13, 64, 5)\) codes \NoCaseChange{\protect\cite[{pg. 74}]{cite41}}.
\end{eczvaluelist}
\eczhbkcontributors{ \eczhuVVA }
\endeczcode

\eczcode{traceability}{Traceability code}{~\NoCaseChange{\protect\cite{cite1134}}}
\codefieldsection{Description}
An IPP code with which it is possible to detect a parent of a given pirated descendant by finding the closest codeword to that descendant.

Codes with strong traceability trace at least one member of a group that has constructed a pirate decoder (i.e., a generic pirate decryption process \NoCaseChange{\protect\cite{cite1134}}).
A code with weak traceability has the ability to ensure that no group is able to frame another user \NoCaseChange{\protect\cite{cite349}}.

\codefieldsection{Rate}
Suppose \(n\) is the number of users, \(k\) is the number of users known by the pirates, and \(p\) is the probability that the pirates cannot be traced.
An open (public) resilient scheme using a hash function has users' personal keys consisting of \(O(k^{2}\log n)\) decryption keys, which is the number of decryptions needed to reveal the information.
The amount of data redundancy overhead is about \(O(k^{4}\log n)\) \NoCaseChange{\protect\cite{cite1134}}.

A secret resilient scheme using a hash function has users' personal keys consisting of \(O( k \log(n/p) )\) decryption keys, which is the number of decryptions needed to reveal the information.
The amount of data redundancy overhead is about \(O( k^{2} \log(n/p) )\) \NoCaseChange{\protect\cite{cite1134}}.

A threshold (secret) scheme using a hash function that is successful against pirates which decrypt with probability \(> q\) has users' personal keys consisting of \((4k/3q)\log(n/p)\) decryption keys (note that this is the same as in the secret resilient scheme above).
These types of schemes only need \flmRefsHyperref{ref65}{order} \(O(1)\) decryption operations performed by users to decrypt the information successfully.
Finally, the amount of data redundancy overhead is 4k encrypted keys, a large improvement compared to the above \NoCaseChange{\protect\cite{cite1134}}.

\codefieldsection{Realizations}
\begin{eczvaluelist}
\item\relax Broadcast messages, pay-per-view movies, and protecting copyrighted online material \NoCaseChange{\protect\cite{cite349}}.
\end{eczvaluelist}
\codefieldsection{Notes}
\begin{eczvaluelist}
\item\relax Note that in the feature section above, the hash function maps the users into a set of \(2k^{2}\) decryption keys \NoCaseChange{\protect\cite{cite1134}}.
\item\relax For code tables, see Refs. \NoCaseChange{\protect\cite{cite1134,cite349}}.
\end{eczvaluelist}
\codefieldsection{Parent}
\begin{eczvaluelist}
\item\relax
\flmRefsHyperref[eczindexfamilyrel]{code:ipp}{Identifiable parent property (IPP) code} --- Traceability codes allow for detection of parents of pirated descendant copies by only determining the closest codeword to the descendant; see \NoCaseChange{\protect\cite[{Lemma 1.3}]{cite349}}.
\end{eczvaluelist}
\codefieldsection{Cousin}
\begin{eczvaluelist}
\item\relax
\flmRefsHyperref[eczindexfamilyrel]{code:q-ary_digits_into_q-ary_digits}{\(q\)-ary code} --- A \(q\)-ary code with distance \(d \geq n(1-1/t^2)\) has the \(t\)-traceability property \NoCaseChange{\protect\cite[{Thm. 4.3}]{cite349}}.
\end{eczvaluelist}
\eczhbkcontributors{ Raley Roberts, \eczhuVVA }
\endeczcode

\onecolumngrid
\clearpage

\section{Binary Kingdom}

\begin{eczEpigraph}
\begin{quote}
\flmQuoteSetup{quote}%
A message with content and clarity\\
Has gotten to be quite a rarity.\\
\indent To combat the terror\\
\indent Of serious error,\\
Use bits of appropriate parity.
\flmQuoteAttributed{Solomon W. Golomb}
\end{quote}
\end{eczEpigraph}

\twocolumngrid

\eczcode{best}{\((10,40,4)\) Best code}{~\NoCaseChange{\protect\cite{cite1135,cite1136}}}
\eczhIndexCodeAliasName{best}{Best code}
\codefieldsection{Description}
Binary nonlinear \((10,40,4)\) code that is unique \NoCaseChange{\protect\cite{cite375}}.
Under \flmTerm{term}{ref127}{}{Construction A}, this code yields \(P_{10c}\), a non-lattice sphere packing that is the densest known in 10 dimensions \NoCaseChange{\protect\cite{cite376}\protect\cite[{pg. 140}]{cite39}}.

\codefieldsection{Parents}
\begin{eczvaluelist}
\item\relax
\flmRefsHyperref[eczindexfamilyrel]{code:bits_into_bits}{Binary code}\item\relax
\flmRefsHyperref[eczindexfamilyrel]{code:small_distance}{Small-distance block code}\end{eczvaluelist}
\codefieldsection{Cousins}
\begin{eczvaluelist}
\item\relax
\flmRefsHyperref[eczindexfamilyrel]{code:construction_a}{Construction A code} --- Using \flmTerm{term}{ref127}{}{Construction A}, the Best code yields \(P_{10c}\), a non-lattice sphere packing in 10 dimensions that is the densest known \NoCaseChange{\protect\cite{cite376}\protect\cite[{pg. 140}]{cite39}}.
\item\relax
\flmRefsHyperref[eczindexfamilyrel]{code:sphere_packing}{Sphere packing} --- Using \flmTerm{term}{ref127}{}{Construction A}, the Best code yields \(P_{10c}\), a non-lattice sphere packing in 10 dimensions that is the densest known \NoCaseChange{\protect\cite{cite376}\protect\cite[{pg. 140}]{cite39}}.
\item\relax
\flmRefsHyperref[eczindexfamilyrel]{code:gray}{Gray code} --- Codewords of the Best code can be obtained by applying the Gray map to the pentacode \NoCaseChange{\protect\cite[{Sec. 2}]{cite373}}.
\item\relax
\flmRefsHyperref[eczindexfamilyrel]{code:pentacode}{\((5,40,4)_{\mathbb{Z}_4}\) Pentacode} --- Codewords of the Best code can be obtained by applying the Gray map to the pentacode \NoCaseChange{\protect\cite[{Sec. 2}]{cite373}}.
\end{eczvaluelist}
\eczhbkcontributors{ \eczhuVVA }
\endeczcode

\eczcode{nadler}{\((12,32,5)\) Nadler code}{~\NoCaseChange{\protect\cite{cite1137}}}
\eczhIndexCodeAliasName{nadler}{Nadler code}
\codefieldsection{Description}
A nonlinear \((12,32,5)\) binary code that is the largest double-error-correcting code.

An inequivalent code with the same parameters was constructed by Van Lint \NoCaseChange{\protect\cite{cite1138}} and presented in \NoCaseChange{\protect\cite[{Ch. 2, Sec. 8}]{cite41}}.
Its automorphism group is a subgroup of \(S_3\times S_4\) \NoCaseChange{\protect\cite{cite1138}}.

While the two codes are inequivalent, the \((13,64,5)\) extension of both codes is unique and is a shortened NR code \NoCaseChange{\protect\cite{cite1139}}.

\codefieldsection{Parents}
\begin{eczvaluelist}
\item\relax
\flmRefsHyperref[eczindexfamilyrel]{code:bits_into_bits}{Binary code}\item\relax
\flmRefsHyperref[eczindexfamilyrel]{code:small_distance}{Small-distance block code}\end{eczvaluelist}
\codefieldsection{Cousin}
\begin{eczvaluelist}
\item\relax
\flmRefsHyperref[eczindexfamilyrel]{code:nordstrom_robinson}{\((16,256,6)\) Nordstrom-Robinson (NR) code} --- The NR code is an extension of the Nadler code \NoCaseChange{\protect\cite{cite1138,cite1139}}.
\end{eczvaluelist}
\eczhbkcontributors{ \eczhuVVA }
\endeczcode

\eczcode{nordstrom_robinson}{\((16,256,6)\) Nordstrom-Robinson (NR) code}{~\NoCaseChange{\protect\cite{cite1140,cite1141}}}
\codefieldsection{Alternative Names}
\begin{eczvaluelist}
\item\relax Semakov-Zinoviev code
\end{eczvaluelist}
\eczhIndexCodeAliasName{nordstrom_robinson}{Nordstrom-Robinson (NR) code}
\eczhIndexCodeAliasName{nordstrom_robinson}{Semakov-Zinoviev code}
\codefieldsection{Description}
A nonlinear \((16,256,6)\) binary code that is the smallest Kerdock code and the smallest Preparata code.
The size of this code is larger than the largest possible linear code with the same length and distance.

The code can be shortened to produce optimal \((15, 128, 6)\), \((14, 64, 6)\) and \((13, 32, 6)\) codes, as well as unique \((15, 256, 5)\), \((15, 128, 6)\), \((14, 128, 5)\), \((14, 64, 6)\), \((13, 64, 5)\) and \((13, 32, 6)\) codes \NoCaseChange{\protect\cite[{pg. 74}]{cite41}}.
Further shortening yields the \((12, 32, 5)\) Nadler code \NoCaseChange{\protect\cite{cite1138}}.

The automorphism group of the code is \(\mathbb{Z}_2^4 \times A_7\) \NoCaseChange{\protect\cite{cite1142,cite1143,cite1144}\protect\cite[{pg. 478}]{cite41}}.

\codefieldsection{Parents}
\begin{eczvaluelist}
\item\relax
\flmRefsHyperref[eczindexfamilyrel]{code:kerdock}{Kerdock code} --- The NR code is the smallest Kerdock code.
\item\relax
\flmRefsHyperref[eczindexfamilyrel]{code:preparata}{Preparata code} --- The NR code is the smallest Preparata code.
\item\relax
\flmRefsHyperref[eczindexfamilyrel]{code:orthogonal_array}{Orthogonal array (OA)} --- The NR code is an orthogonal array of strength \(5\) \NoCaseChange{\protect\cite[{pg. 141}]{cite41}}.
\end{eczvaluelist}
\codefieldsection{Cousins}
\begin{eczvaluelist}
\item\relax
\flmRefsHyperref[eczindexfamilyrel]{code:octacode}{Octacode} --- The NR code is the image of the octacode under the \flmTerm{term}{ref81}{}{Gray map} \NoCaseChange{\protect\cite{cite1146,cite123}\protect\cite[{Sec. 6.3}]{cite1145}\protect\cite[{Thm. 12}]{cite158}}. The \((14, 64, 6)\) shortened NR code is the image of the heptacode under the \flmTerm{term}{ref81}{}{Gray map} \NoCaseChange{\protect\cite[{Exam. 5}]{cite1147}}.
\item\relax
\flmRefsHyperref[eczindexfamilyrel]{code:biorthogonal}{\([2^m,m+1,2^{m-1}]\) First-order RM code} --- The NR code is the union of eight cosets of a linear \([16,5,8]\) code, i.e., the first-order Reed-Muller (biorthogonal) code \NoCaseChange{\protect\cite[{pgs. 76 and 476}]{cite41}}.
\item\relax
\flmRefsHyperref[eczindexfamilyrel]{code:extended_golay}{\([24, 12, 8]\) Extended Golay code} --- The NR code can be constructed using the extended Golay code by first selecting a set of codewords satisfying certain conditions and then deleting specific coordinates \NoCaseChange{\protect\cite[{pg. 73}]{cite41}}.
\item\relax
\flmRefsHyperref[eczindexfamilyrel]{code:self_dual}{Self-dual linear code} --- The NR code is self-dual in that its distance distribution is invariant under the \flmRefsHyperref{ref113}{MacWilliams transform} \NoCaseChange{\protect\cite{cite1148}}.
It maps to the octacode, a self-dual code over \(\mathbb{Z}_4\) under the \flmTerm{term}{ref81}{}{Gray map} \NoCaseChange{\protect\cite{cite1149,cite1146,cite123}\protect\cite[{Sec. 6.3}]{cite1145}}.

\item\relax
\flmRefsHyperref[eczindexfamilyrel]{code:nearly_perfect}{Nearly perfect code} --- The punctured \((15,256,5)\) NR code saturates the Johnson bound and is therefore nearly perfect \NoCaseChange{\protect\cite[{Exam. 5.5.5}]{cite135}}.
\item\relax
\flmRefsHyperref[eczindexfamilyrel]{code:combinatorial_design}{Combinatorial design} --- NR codewords give \(3\)-\((16, 6, 4)\), \(3\)-\((16, 8, 3)\), and \(3\)-\((16, 10, 24)\) designs, while the punctured code of length \(15\) and minimum distance \(5\) meets the Johnson bound and supports \(2\)-designs \NoCaseChange{\protect\cite[{Exam. 5.5.5}]{cite135}\protect\cite[{pg. 164}]{cite41}}.
\item\relax
\flmRefsHyperref[eczindexfamilyrel]{code:small_distance}{Small-distance block code} --- The NR code can be shortened to produce unique \((15, 256, 5)\), \((14, 128, 5)\), and \((13, 64, 5)\) codes \NoCaseChange{\protect\cite[{pg. 74}]{cite41}}.
\item\relax
\flmRefsHyperref[eczindexfamilyrel]{code:nadler}{\((12,32,5)\) Nadler code} --- The NR code is an extension of the Nadler code \NoCaseChange{\protect\cite{cite1138,cite1139}}.
\end{eczvaluelist}
\eczhbkcontributors{ Madhura Pankaja, \eczhuVVA }
\endeczcode

\eczcode{vasilyev}{\((2^{m+1}-1,2^{2n-m},3)\) Vasilyev code}{~\NoCaseChange{\protect\cite{cite1150}}}
\eczhIndexCodeAliasName{vasilyev}{Vasilyev code}
\codefieldsection{Description}
Member of an infinite \((2^{m+1}-1,2^{2n-m},3)\) family of perfect nonlinear codes for any \(m \geq 3\).
Constructed by applying a modification of the \((u|u+v)\) construction to a perfect \((2^m-1,2^{n-m},3)\) code, not necessarily linear \NoCaseChange{\protect\cite[{pg. 77}]{cite41}}.

The automorphism group of these codes is always nontrivial \NoCaseChange{\protect\cite{cite1151}}.

\codefieldsection{Parents}
\begin{eczvaluelist}
\item\relax
\flmRefsHyperref[eczindexfamilyrel]{code:perfect_binary}{Perfect binary code} --- Vasilyev codes are perfect nonlinear binary codes and are inequivalent to any linear code.
\item\relax
\flmRefsHyperref[eczindexfamilyrel]{code:uplusv}{\((u|u+v)\)-construction code}\item\relax
\flmRefsHyperref[eczindexfamilyrel]{code:small_distance}{Small-distance block code}\end{eczvaluelist}
\eczhbkcontributors{ \eczhuVVA }
\endeczcode

\eczcode{petersen}{\([15,6,5]\) Petersen cycle code}{~\NoCaseChange{\protect\cite{cite1152}}}
\eczhIndexCodeAliasName{petersen}{Petersen cycle code}
\codefieldsection{Description}
A \([15,6,5]\) cycle code whose parity-check matrix is the incidence matrix of the Petersen graph.
The Petersen graph can be thought of as a dodecahedron with antipodes identified \NoCaseChange{\protect\cite[{Appx. A.2.1}]{cite101}}.

\codefieldsection{Gates}
\begin{eczvaluelist}
\item\relax In the inner/outer-code framework for magic-state distillation, it is the smallest outer code with weight-three checks that is \((4,2)\)-sensitive, giving a fifth-order protocol on 15 magic states \NoCaseChange{\protect\cite[{Appx. A.2.1}]{cite101}}.
\end{eczvaluelist}
\codefieldsection{Notes}
\begin{eczvaluelist}
\item\relax See \NoCaseChange{\protect\cite[{Example 10.11.7}]{cite1153}} for more details.
\end{eczvaluelist}
\codefieldsection{Parents}
\begin{eczvaluelist}
\item\relax
\flmRefsHyperref[eczindexfamilyrel]{code:homological_classical}{Cycle code}\item\relax
\flmRefsHyperref[eczindexfamilyrel]{code:small_distance}{Small-distance block code}\end{eczvaluelist}
\codefieldsection{Cousins}
\begin{eczvaluelist}
\item\relax
\flmRefsHyperref[eczindexfamilyrel]{code:dodecahedron}{Dodecahedron code} --- The Petersen graph can be thought of as a dodecahedron with antipodes identified \NoCaseChange{\protect\cite[{Appx. A.2.1}]{cite101}}.
\item\relax
\flmRefsHyperref[eczindexfamilyrel]{code:petersen_spherical}{Petersen spherical code} --- The Petersen spherical code is a spherical realization associated with the Petersen cycle code.
\end{eczvaluelist}
\eczhbkcontributors{ \eczhuVVA }
\endeczcode

\eczcode{kasami}{\([2^{2r}-1, 3r, 2^{2r-1} - 2^{r-1} ]\) Kasami code}{~\NoCaseChange{\protect\cite{cite1154}}}
\eczhIndexCodeAliasName{kasami}{Kasami code}
\codefieldsection{Description}
Member of the family of \([2^{2r}-1, 3r, 2^{2r-1} - 2^{r-1} ]\) cyclic binary linear codes. 
Gold and Kasami codes are both constructed by picking a set of cyclically unrelated sequences of binary linear codes with low cross-correlation \NoCaseChange{\protect\cite{cite107,cite108}}.

\codefieldsection{Parent}
\begin{eczvaluelist}
\item\relax
\flmRefsHyperref[eczindexfamilyrel]{code:binary_cyclic}{Cyclic linear binary code}\end{eczvaluelist}
\codefieldsection{Cousins}
\begin{eczvaluelist}
\item\relax
\flmRefsHyperref[eczindexfamilyrel]{code:gold}{\([2^r-1, 2r ]\) Gold code} --- Gold and Kasami codes are both constructed by picking a set of cyclically unrelated sequences of binary linear codes with low cross-correlation \NoCaseChange{\protect\cite{cite107,cite108}}.
\item\relax
\flmRefsHyperref[eczindexfamilyrel]{code:griesmer}{Griesmer code} --- Kasami codes satisfy the Griesmer bound for certain parameters \NoCaseChange{\protect\cite{cite1155}}.
\end{eczvaluelist}
\eczhbkcontributors{ \eczhuVVA }
\endeczcode

\eczcode{melas}{\([2^m -1, 2^m - 1 - 2m, 5]\) Melas code}{~\NoCaseChange{\protect\cite{cite1156,cite1157}}}
\eczhIndexCodeAliasName{melas}{Melas code}
\codefieldsection{Description}
Cyclic linear code whose generator polynomial is \(g(x) = p(x)p(x)^{\star}\), where \(p(x)\) is a primitive polynomial of degree \(m\) that is the minimal polynomial over \(\mathbb{F}_2\) of an element \(\alpha\) of order \(2^m -1\) in \(\mathbb{F}_{2^m}\), \(m\) is odd and greater than five, and '\(\star\)' denotes reciprocation \NoCaseChange{\protect\cite{cite105}}.
\codefieldsection{Decoding}
\begin{eczvaluelist}
\item\relax Algebraic decoder \NoCaseChange{\protect\cite{cite105}}.
\end{eczvaluelist}
\codefieldsection{Parents}
\begin{eczvaluelist}
\item\relax
\flmRefsHyperref[eczindexfamilyrel]{code:binary_cyclic}{Cyclic linear binary code}\item\relax
\flmRefsHyperref[eczindexfamilyrel]{code:reversible}{Reversible code}\item\relax
\flmRefsHyperref[eczindexfamilyrel]{code:small_distance}{Small-distance block code}\end{eczvaluelist}
\codefieldsection{Cousin}
\begin{eczvaluelist}
\item\relax
\flmRefsHyperref[eczindexfamilyrel]{code:quaternary_over_z4}{Linear code over \(\mathbb{Z}_4\)} --- The even-weight subcode of the Melas code can be lifted to a linear code over \(\mathbb{Z}_4\) \NoCaseChange{\protect\cite{cite105}}.
\end{eczvaluelist}
\eczhbkcontributors{ Khalil Guy, \eczhuVVA }
\endeczcode

\eczcode{simplex}{\([2^m-1,m,2^{m-1}]\) simplex code}{~\NoCaseChange{\protect\cite{cite1158,cite1}}}
\codefieldsection{Alternative Names}
\begin{eczvaluelist}
\item\relax Shortened Hadamard code
\item\relax RM\(^*(1,m)\) code
\item\relax Maximum-length feedback-shift-register code
\end{eczvaluelist}
\eczhIndexCodeAliasName{simplex}{simplex code}
\eczhIndexCodeAliasName{simplex}{Shortened Hadamard code}
\eczhIndexCodeAliasName{simplex}{RM\(^*(1,m)\) code}
\eczhIndexCodeAliasName{simplex}{Maximum-length feedback-shift-register code}
\codefieldsection{Description}
A member of the equidistant code family dual to the \([2^m-1,2^m-m-1,3]\) Hamming family.

The columns of its generator matrix are in one-to-one correspondence with the elements of the projective space \(PG(m-1,2)\), with each column being a chosen representative of the corresponding element.
Simplex codes saturate the Plotkin bound and hence have nonzero codewords all of the same weight, \(2^{m-1}\) \NoCaseChange{\protect\cite[{Th. 11(a)}]{cite41}\protect\cite[{Thm. 1.10.5}]{cite1159}}.
The codewords form a \((2^m,2^m+1)\) simplex spherical code under the \flmRefsHyperref{ref38}{antipodal mapping}.

A punctured simplex code is known as a \textit{MacDonald code} \NoCaseChange{\protect\cite{cite1160}}, with parameters \([2^m-2^u,m,2^{m-1}-2^{u-1}]\) for \(u \leq m-1\) \NoCaseChange{\protect\cite{cite1161}}.

The automorphism group of the code is \(GL(m,\mathbb{F}_2)\) \NoCaseChange{\protect\cite{cite41}}.

\codefieldsection{Protection}
Simplex codes saturate the Plotkin bound \NoCaseChange{\protect\cite[{pg. 43}]{cite41}}.

\codefieldsection{Decoding}
\begin{eczvaluelist}
\item\relax Serial orthogonal decoder \NoCaseChange{\protect\cite{cite1162,cite1163}}
\item\relax Quantum decoder \NoCaseChange{\protect\cite{cite1164}}.
\end{eczvaluelist}
\codefieldsection{Notes}
\begin{eczvaluelist}
\item\relax Simplex codes are equivalent to irreducible cyclic codes \(C(q,n,1)\) when \(\gcd(q-1,m)=1\) \NoCaseChange{\protect\cite[{Exam. 2.5.1}]{cite68}}.
\end{eczvaluelist}
\codefieldsection{Parents}
\begin{eczvaluelist}
\item\relax
\flmRefsHyperref[eczindexfamilyrel]{code:binary_cyclic}{Cyclic linear binary code} --- Simplex codes can be realized as maximal-length feedback-shift-register codes, and are therefore cyclic \NoCaseChange{\protect\cite[{pgs. 89 and 216}]{cite41}}.
\item\relax
\flmRefsHyperref[eczindexfamilyrel]{code:q-ary_simplex}{\(q\)-ary simplex code} --- \(q\)-ary simplex codes reduce to simplex codes for \(q=2\).
\end{eczvaluelist}
\codefieldsection{Child}
\begin{eczvaluelist}
\item\relax
\flmRefsHyperref[eczindexfamilyrel]{code:simplex734}{\([7,3,4]\) simplex code}\end{eczvaluelist}
\codefieldsection{Cousins}
\begin{eczvaluelist}
\item\relax
\flmRefsHyperref[eczindexfamilyrel]{code:combinatorial_design}{Combinatorial design} --- Simplex codewords form a 2-design \NoCaseChange{\protect\cite[{pg. 166}]{cite41}}.
\item\relax
\flmRefsHyperref[eczindexfamilyrel]{code:binary_linear}{Linear binary code} --- Linear binary codes cannot be constant weight, but can have nonzero codewords with constant weight. All such codes are equidistant, and Bonisoli's theorem states that any equidistant linear binary code is a direct sum of simplex codes \NoCaseChange{\protect\cite{cite988}} (see also Refs. \NoCaseChange{\protect\cite{cite45,cite46}}).
\item\relax
\flmRefsHyperref[eczindexfamilyrel]{code:hamming}{\([2^r-1,2^r-r-1,3]\) Hamming code} --- Hamming and simplex codes are dual to each other.
\item\relax
\flmRefsHyperref[eczindexfamilyrel]{code:dual}{Dual linear code} --- Hamming and simplex codes are dual to each other.
\item\relax
\flmRefsHyperref[eczindexfamilyrel]{code:simplex_spherical}{Simplex spherical code} --- Binary simplex codes map to \((2^m,2^m+1)\) simplex spherical codes under the \flmRefsHyperref{ref38}{antipodal mapping} \NoCaseChange{\protect\cite[{Sec. 6.5.2}]{cite1165}\protect\cite[{pg. 18}]{cite115}}. In other words, simplex (simplex spherical) codes form simplices in Hamming (Euclidean) space.
\item\relax
\flmRefsHyperref[eczindexfamilyrel]{code:biorthogonal}{\([2^m,m+1,2^{m-1}]\) First-order RM code} --- First-order RM codes and simplex codes are interconvertible via shortening and lengthening \NoCaseChange{\protect\cite[{pg. 31}]{cite41}}. Punctured first-order RM codes and simplex codes are interconvertible via expurgation and augmentation \NoCaseChange{\protect\cite[{pg. 31}]{cite41}}.
\item\relax
\flmRefsHyperref[eczindexfamilyrel]{code:constant_weight}{Constant-weight code} --- Linear binary codes cannot be constant weight, but can have nonzero codewords with constant weight. All such codes are equidistant, and Bonisoli's theorem states that any equidistant linear binary code is a direct sum of simplex codes \NoCaseChange{\protect\cite{cite988}} (see also Refs. \NoCaseChange{\protect\cite{cite45,cite46}}).
\item\relax
\flmRefsHyperref[eczindexfamilyrel]{code:reed_muller}{Reed-Muller (RM) code} --- Simplex are equivalent to RM\(^*(1,m)\).
\item\relax
\flmRefsHyperref[eczindexfamilyrel]{code:simplex_discrete}{Simplex integer-based code} --- Codewords of simplex integer-based codes are restricted to lie in a \flmRefsHyperref{ref655}{discrete simplex}.
\item\relax
\flmRefsHyperref[eczindexfamilyrel]{code:gold}{\([2^r-1, 2r ]\) Gold code} --- Simplex codes are used to make gold codes. The dual of a Gold code is the intersection of the duals of the simplex codes used to construct it \NoCaseChange{\protect\cite{cite1166}}.
\item\relax
\flmRefsHyperref[eczindexfamilyrel]{code:hadamard}{\([2^m,m,2^{m-1}]\) Hadamard code} --- The \([2^m-1,m,2^{m-1}]\) shortened Hadamard code is the simplex code (a.k.a. RM\(^*(1,m)\)).
\item\relax
\flmRefsHyperref[eczindexfamilyrel]{code:hgp_7_2_2}{\(\llbracket 7,2,2\rrbracket \) HGP phantom code} --- This code is constructed from the hypergraph product of the \([3,2,2]\) simplex code and the \([2,1,2]\) repetition code \NoCaseChange{\protect\cite{cite514}}.
\item\relax
\flmRefsHyperref[eczindexfamilyrel]{code:quantum_hamming_css}{\(\llbracket 2^r-1, 2^r-2r-1, 3\rrbracket \) quantum Hamming code} --- Quantum Hamming codes result from applying the CSS construction to Hamming codes and their duals the simplex codes.
\item\relax
\flmRefsHyperref[eczindexfamilyrel]{code:shyps}{Subsystem Hypergraph Product Simplex (SHYPS) code} --- SHYPS code gauge generator matrices are constructed from hypergraph products of simplex codes \NoCaseChange{\protect\cite{cite785}}.
\end{eczvaluelist}
\eczhbkcontributors{ Yi-Ting (Rick) Tu, \eczhuVVA }
\endeczcode

\eczcode{extended_hamming}{\([2^m,2^m-m-1,4]\) Extended Hamming code}{~\NoCaseChange{\protect\cite{cite1,cite1167,cite1168}}}
\codefieldsection{Alternative Names}
\begin{eczvaluelist}
\item\relax RM\((m-2,m)\) code
\end{eczvaluelist}
\eczhIndexCodeAliasName{extended_hamming}{Extended Hamming code}
\eczhIndexCodeAliasName{extended_hamming}{RM\((m-2,m)\) code}
\codefieldsection{Description}
Member of an infinite family of RM\((m-2,m)\) codes with parameters \([2^m,2^m-m-1, 4]\) for \(m \geq 2\) that are extensions of the Hamming codes by a parity-check bit.

\codefieldsection{Parents}
\begin{eczvaluelist}
\item\relax
\flmRefsHyperref[eczindexfamilyrel]{code:reed_muller}{Reed-Muller (RM) code} --- Extended Hamming codes are RM\((m-2,m)\) codes.
\item\relax
\flmRefsHyperref[eczindexfamilyrel]{code:small_distance}{Small-distance block code}\end{eczvaluelist}
\codefieldsection{Child}
\begin{eczvaluelist}
\item\relax
\flmRefsHyperref[eczindexfamilyrel]{code:hamming844}{\([8,4,4]\) extended Hamming code}\end{eczvaluelist}
\codefieldsection{Cousins}
\begin{eczvaluelist}
\item\relax
\flmRefsHyperref[eczindexfamilyrel]{code:karlin}{\([2m+2,m+1]\) Karlin code} --- The extended Hamming code is equivalent to the Karlin double circulant code for \(m=3\) \NoCaseChange{\protect\cite[{Ch. 16}]{cite41}}.
\item\relax
\flmRefsHyperref[eczindexfamilyrel]{code:univ_opt_q-ary}{Universally optimal \(q\)-ary code} --- Several extended Hamming codes are LP universally optimal codes \NoCaseChange{\protect\cite{cite173}}.
\item\relax
\flmRefsHyperref[eczindexfamilyrel]{code:biorthogonal}{\([2^m,m+1,2^{m-1}]\) First-order RM code} --- Extended Hamming and first-order RM codes are dual to each other.
\item\relax
\flmRefsHyperref[eczindexfamilyrel]{code:dual}{Dual linear code} --- Extended Hamming and first-order RM codes are dual to each other.
\item\relax
\flmRefsHyperref[eczindexfamilyrel]{code:incidence_matrix}{Incidence-matrix projective code} --- Columns of an extended Hamming code's parity-check matrix correspond to points in \(PG(m-1,2)\) that lie in the complement of a hyperplane \NoCaseChange{\protect\cite[{pg. 182}]{cite961}}.
\item\relax
\flmRefsHyperref[eczindexfamilyrel]{code:combinatorial_design}{Combinatorial design} --- Weight-four codewords of the \([2^r,2^r-r-1, 4]\) extended Hamming code support the Steiner system \(S(3,4,2^r)\) \NoCaseChange{\protect\cite[{pg. 89}]{cite39}}.
\item\relax
\flmRefsHyperref[eczindexfamilyrel]{code:preparata}{Preparata code} --- Any code with the same parameters as the Preparata code must be a distance invariant subcode of a (possibly nonlinear) code with the same parameters as the extended Hamming code \NoCaseChange{\protect\cite{cite1169,cite1170}}.
\item\relax
\flmRefsHyperref[eczindexfamilyrel]{code:hamming}{\([2^r-1,2^r-r-1,3]\) Hamming code} --- Extended Hamming codes are extensions of Hamming codes by a parity-check bit. Puncturing extended Hamming codes yields the Hamming codes.
\item\relax
\flmRefsHyperref[eczindexfamilyrel]{code:zrm}{ZRM code} --- The weight-four codewords of the binary image of the dual of ZRM\((1,m)\) form a Steiner system that is identical to that formed by the weight-four codewords of an extended Hamming code \NoCaseChange{\protect\cite{cite158}}.
\item\relax
\flmRefsHyperref[eczindexfamilyrel]{code:majorana_hamming}{\(\llbracket 2^{m-1},2^{m-1}-m-1,4\rrbracket _{f}\) Hamming Majorana code} --- A Hamming Majorana code is constructed from a first-order RM code (whose dual is the extended Hamming code).
\end{eczvaluelist}
\eczhbkcontributors{ Dhruv Devulapalli, \eczhuVVA }
\endeczcode

\eczcode{hadamard}{\([2^m,m,2^{m-1}]\) Hadamard code}{}
\codefieldsection{Alternative Names}
\begin{eczvaluelist}
\item\relax Walsh code
\item\relax Walsh-Hadamard code
\end{eczvaluelist}
\eczhIndexCodeAliasName{hadamard}{Hadamard code}
\eczhIndexCodeAliasName{hadamard}{Walsh code}
\eczhIndexCodeAliasName{hadamard}{Walsh-Hadamard code}
\codefieldsection{Description}
An \([2^m,m,2^{m-1}]\) balanced binary code.
The \([2^m,m+1,2^{m-1}]\) augmented Hadamard code is the first-order RM code (a.k.a. RM\((1,m)\)), while the \([2^m-1,m,2^{m-1}]\) shortened Hadamard code is the simplex code (a.k.a. RM\(^*(1,m)\)).

\codefieldsection{Notes}
\begin{eczvaluelist}
\item\relax Review of Hadamard matrices \NoCaseChange{\protect\cite{cite1171}}.
\end{eczvaluelist}
\codefieldsection{Parents}
\begin{eczvaluelist}
\item\relax
\flmRefsHyperref[eczindexfamilyrel]{code:binary_ltc}{Binary linear LTC} --- The Hadamard code is the first code to be identified as a (three-query) LTC \NoCaseChange{\protect\cite{cite1172,cite1094}}.
\item\relax
\flmRefsHyperref[eczindexfamilyrel]{code:balanced}{Balanced code} --- Each nonzero Hadamard codeword has length \(2^m\) and Hamming weight of \(2^{m-1}\), making this code balanced.
\item\relax
\flmRefsHyperref[eczindexfamilyrel]{code:q-ary_lcc}{\(q\)-ary linear LCC} --- Hadamard codes are two-query LDCs and LCCs \NoCaseChange{\protect\cite{cite1073,cite1067}}.
\end{eczvaluelist}
\codefieldsection{Cousins}
\begin{eczvaluelist}
\item\relax
\flmRefsHyperref[eczindexfamilyrel]{code:locally_recoverable}{Locally recoverable code (LRC)} --- The Hadamard code is an LRC with \(r=3\) \NoCaseChange{\protect\cite{cite812}}.
\item\relax
\flmRefsHyperref[eczindexfamilyrel]{code:long}{Long code} --- The Hadamard code is a subcode of the long code and can be obtained by restricting the long-code construction to only linear functions.
\item\relax
\flmRefsHyperref[eczindexfamilyrel]{code:binary_quad_residue}{Binary quadratic-residue (QR) code} --- For Hadamard matrices obtained from the Paley construction, the linear span of the resulting Hadamard codes yields quadratic-residue codes \NoCaseChange{\protect\cite[{pg. 49}]{cite41}}.
\item\relax
\flmRefsHyperref[eczindexfamilyrel]{code:reed_muller}{Reed-Muller (RM) code} --- The \([2^m,m+1,2^{m-1}]\) augmented Hadamard code is the first-order RM code (a.k.a. RM\((1,m)\)). The \([2^m-1,m,2^{m-1}]\) shortened Hadamard code is the simplex code (a.k.a. RM\(^*(1,m)\)). Rows of a Hadamard matrix forming a Prometheus orthonormal set (PONS) are codewords of a coset of RM\((1,m)\) in RM\((2,m)\) \NoCaseChange{\protect\cite{cite1173}}.
\item\relax
\flmRefsHyperref[eczindexfamilyrel]{code:simplex}{\([2^m-1,m,2^{m-1}]\) simplex code} --- The \([2^m-1,m,2^{m-1}]\) shortened Hadamard code is the simplex code (a.k.a. RM\(^*(1,m)\)).
\item\relax
\flmRefsHyperref[eczindexfamilyrel]{code:biorthogonal}{\([2^m,m+1,2^{m-1}]\) First-order RM code} --- The \([2^m,m+1,2^{m-1}]\) augmented Hadamard code is the first-order RM code (a.k.a. RM\((1,m)\)). Rows of a Hadamard matrix forming a Prometheus orthonormal set (PONS) are codewords of a coset of RM\((1,m)\) in RM\((2,m)\) \NoCaseChange{\protect\cite{cite1173}}.
\item\relax
\flmRefsHyperref[eczindexfamilyrel]{code:delsarte_optimal_q-ary}{\(q\)-ary sharp configuration} --- The augmented binary Hadamard code family is listed among the sharp configurations in \NoCaseChange{\protect\cite[{Table 12.1}]{cite199}}.
\item\relax
\flmRefsHyperref[eczindexfamilyrel]{code:univ_opt_q-ary}{Universally optimal \(q\)-ary code} --- Several punctured Hadamard codes are LP universally optimal codes \NoCaseChange{\protect\cite{cite173}}.
\item\relax
\flmRefsHyperref[eczindexfamilyrel]{code:combinatorial_design}{Combinatorial design} --- \textit{Hadamard designs} are combinatorial designs constructed from Hadamard matrices \NoCaseChange{\protect\cite{cite161,cite162}}; see Ref. \NoCaseChange{\protect\cite{cite163}}.
\item\relax
\flmRefsHyperref[eczindexfamilyrel]{code:polar}{Polar code} --- The generator matrices of RM and polar codes are different submatrices of Kronecker products of Hadamard matrices \NoCaseChange{\protect\cite{cite1174,cite365}}. There are families interpolating between the two codes \NoCaseChange{\protect\cite{cite1175}}.
\item\relax
\flmRefsHyperref[eczindexfamilyrel]{code:difference_set}{Difference-set cyclic (DSC) code} --- \textit{Hadamard difference sets} are difference sets constructed from Hadamard matrices \NoCaseChange{\protect\cite[{Ch. 6}]{cite1176}}.
\item\relax
\flmRefsHyperref[eczindexfamilyrel]{code:ecoc}{Error-correcting output code (ECOC)} --- Hadamard codes and subcodes can be used as ECOCs \NoCaseChange{\protect\cite{cite1177,cite1178,cite1179}}.
\item\relax
\flmRefsHyperref[eczindexfamilyrel]{code:quantum_hadamard_bpsk}{Hadamard BPSK c-q modulation format} --- The Hadamard BPSK c-q code can be thought of as a concatenation of the Hadamard binary linear code with BPSK for the purposes of transmission of classical information over quantum channels.
\end{eczvaluelist}
\eczhbkcontributors{ Alexander Barg, Dhruv Devulapalli, \eczhuVVA }
\endeczcode

\eczcode{biorthogonal}{\([2^m,m+1,2^{m-1}]\) First-order RM code}{}
\codefieldsection{Alternative Names}
\begin{eczvaluelist}
\item\relax Biorthogonal code
\item\relax RM\((1,m)\) code
\item\relax Augmented Hadamard code
\end{eczvaluelist}
\eczhIndexCodeAliasName{biorthogonal}{First-order RM code}
\eczhIndexCodeAliasName{biorthogonal}{Biorthogonal code}
\eczhIndexCodeAliasName{biorthogonal}{RM\((1,m)\) code}
\eczhIndexCodeAliasName{biorthogonal}{Augmented Hadamard code}
\codefieldsection{Description}
A member of the family of first-order RM codes.
Its codewords are the rows of the \(2^m\)-dimensional Hadamard matrix \(H\) and its negation \(-H\) with the mapping \(+1\to 0\) and \(-1\to 1\).
The family is self-orthogonal for \(m \geq 3\) \NoCaseChange{\protect\cite{cite37}}.
They form a \((2^m,2^{m+1})\) biorthogonal spherical code under the \flmRefsHyperref{ref38}{antipodal mapping}.

The automorphism group of the code is \(GA(m,\mathbb{F}_2)\) \NoCaseChange{\protect\cite{cite41}}.

\codefieldsection{Decoding}
\begin{eczvaluelist}
\item\relax First-order RM codes admit specialized decoders, such as the Fast Hadamard Transform decoder \NoCaseChange{\protect\cite{cite327}}.
\item\relax Maximum-likelihood decoding is practical via the Green Machine decoder \NoCaseChange{\protect\cite[{Ch. 14}]{cite41}}.
\end{eczvaluelist}
\codefieldsection{Realizations}
\begin{eczvaluelist}
\item\relax The \([32, 6, 16]\) RM\((1,5)\) code was used for the 1971 Mariner 9 spacecraft \NoCaseChange{\protect\cite{cite365}}.
\end{eczvaluelist}
\codefieldsection{Notes}
\begin{eczvaluelist}
\item\relax See Ref. \NoCaseChange{\protect\cite{cite1180}} for the weight distribution of the \(2^{26}\) cosets of the \([32,6]\) first-order RM code.
\end{eczvaluelist}
\codefieldsection{Parent}
\begin{eczvaluelist}
\item\relax
\flmRefsHyperref[eczindexfamilyrel]{code:reed_muller}{Reed-Muller (RM) code} --- First-order RM codes are RM\((1,m)\) codes.
\end{eczvaluelist}
\codefieldsection{Child}
\begin{eczvaluelist}
\item\relax
\flmRefsHyperref[eczindexfamilyrel]{code:hamming844}{\([8,4,4]\) extended Hamming code} --- The \([8,4,4]\) extended Hamming code is the first-order RM\((1,3)\) code.
\end{eczvaluelist}
\codefieldsection{Cousins}
\begin{eczvaluelist}
\item\relax
\flmRefsHyperref[eczindexfamilyrel]{code:biorthogonal_spherical}{Biorthogonal spherical code} --- Each first-order RM code maps to a \((2^m,2^{m+1})\) biorthogonal spherical code under the \flmRefsHyperref{ref38}{antipodal mapping} \NoCaseChange{\protect\cite{cite1181}\protect\cite[{Sec. 6.4}]{cite1165}\protect\cite[{pg. 19}]{cite115}}. In other words, first-order RM (biorthogonal spherical) codes form orthoplexes in Hamming (Euclidean) space.
\item\relax
\flmRefsHyperref[eczindexfamilyrel]{code:lambda16}{\(\Lambda_{16}\) Barnes-Wall lattice} --- Applying Construction B to the first-order RM\((1,4)\) code yields the \(\Lambda_{16}\) Barnes-Wall lattice \NoCaseChange{\protect\cite[{Ch. 4, pg. 130}]{cite39}\protect\cite[{Exam. 10.7.2}]{cite115}}.
\item\relax
\flmRefsHyperref[eczindexfamilyrel]{code:extended_golay}{\([24, 12, 8]\) Extended Golay code} --- The first-order RM\((1,4)\) Reed-Muller code is a subcode of the extended Golay code \NoCaseChange{\protect\cite[{Ch. 5, pg. 146}]{cite39}}.
\item\relax
\flmRefsHyperref[eczindexfamilyrel]{code:kerdock}{Kerdock code} --- Kerdock code is a subcode of a second-order RM Code \NoCaseChange{\protect\cite[{pg. 457}]{cite41}}.
It consists of a number of cosets of RM\((2,m)\) created by quotienting with first-order RM\((1,m)\) codes.

\item\relax
\flmRefsHyperref[eczindexfamilyrel]{code:nordstrom_robinson}{\((16,256,6)\) Nordstrom-Robinson (NR) code} --- The NR code is the union of eight cosets of a linear \([16,5,8]\) code, i.e., the first-order Reed-Muller (biorthogonal) code \NoCaseChange{\protect\cite[{pgs. 76 and 476}]{cite41}}.
\item\relax
\flmRefsHyperref[eczindexfamilyrel]{code:levenshtein}{Levenshtein code} --- First-order RM codes and Levenshtein codes are both constructed using Hadamard matrices.
\item\relax
\flmRefsHyperref[eczindexfamilyrel]{code:hadamard}{\([2^m,m,2^{m-1}]\) Hadamard code} --- The \([2^m,m+1,2^{m-1}]\) augmented Hadamard code is the first-order RM code (a.k.a. RM\((1,m)\)). Rows of a Hadamard matrix forming a Prometheus orthonormal set (PONS) are codewords of a coset of RM\((1,m)\) in RM\((2,m)\) \NoCaseChange{\protect\cite{cite1173}}.
\item\relax
\flmRefsHyperref[eczindexfamilyrel]{code:simplex}{\([2^m-1,m,2^{m-1}]\) simplex code} --- First-order RM codes and simplex codes are interconvertible via shortening and lengthening \NoCaseChange{\protect\cite[{pg. 31}]{cite41}}. Punctured first-order RM codes and simplex codes are interconvertible via expurgation and augmentation \NoCaseChange{\protect\cite[{pg. 31}]{cite41}}.
\item\relax
\flmRefsHyperref[eczindexfamilyrel]{code:extended_hamming}{\([2^m,2^m-m-1,4]\) Extended Hamming code} --- Extended Hamming and first-order RM codes are dual to each other.
\item\relax
\flmRefsHyperref[eczindexfamilyrel]{code:majorana_hamming}{\(\llbracket 2^{m-1},2^{m-1}-m-1,4\rrbracket _{f}\) Hamming Majorana code} --- A Hamming Majorana code is constructed from a first-order RM code (whose dual is the extended Hamming code).
\item\relax
\flmRefsHyperref[eczindexfamilyrel]{code:quantum_hamming}{\(\llbracket 2^r, 2^r-r-2, 3\rrbracket \) Gottesman code} --- Gottesman codes can be obtained from a modified CSS construction \NoCaseChange{\protect\cite{cite1182,cite861}} with a \([2^r,r+1,2^{r-1}] = C_2^{\perp}\) first-order RM code and a \([2^r,2^r-1,2] = C_1\) even-weight code \NoCaseChange{\protect\cite{cite1182,cite861}}.
\item\relax
\flmRefsHyperref[eczindexfamilyrel]{code:stab_16_6_4}{\(\llbracket 16,6,4\rrbracket \) Tesseract color code} --- The tesseract color code is constructed from the \([16,5,8]\) first-order RM code via the CSS construction \NoCaseChange{\protect\cite{cite101,cite757}}.
\item\relax
\flmRefsHyperref[eczindexfamilyrel]{code:quantum_divisible}{Quantum divisible code} --- Quantum divisible codes can be constructed out of first-order RM\((1,m)\) codes \NoCaseChange{\protect\cite{cite765}}.
\item\relax
\flmRefsHyperref[eczindexfamilyrel]{code:diagonal_clifford}{\(\llbracket 2^r-1,1,3\rrbracket \) simplex code} --- The \(\llbracket 2^r-1,1,3\rrbracket \) simplex code is constructed with a punctured first-order RM code and its even subcode.
\end{eczvaluelist}
\eczhbkcontributors{ \eczhuVVA }
\endeczcode

\eczcode{gold}{\([2^r-1, 2r ]\) Gold code}{~\NoCaseChange{\protect\cite{cite106}}}
\eczhIndexCodeAliasName{gold}{Gold code}
\codefieldsection{Description}
A cyclic binary linear code characterized by the generator polynomial of degree \(r\) of two maximum-period sequences of period \(2^r-1\) with absolute cross-correlation \( \leq 2^{(r+2)/2}\). Gold codewords are generated using \(m\)-sequences \(x\) and \(y\), which are codewords of simplex codes with check polynomials of degree \(r\) \NoCaseChange{\protect\cite{cite106}}.
\codefieldsection{Encoding}
\begin{eczvaluelist}
\item\relax Information bits are initialized in the shift registers of the two \(m\)-sequences \(x\) and \(y\).
\end{eczvaluelist}
\codefieldsection{Decoding}
\begin{eczvaluelist}
\item\relax General decoding is done by building a sparse parity check matrix, followed by applying an iterative message passing algorithm. \NoCaseChange{\protect\cite{cite1183}}.
\end{eczvaluelist}
\codefieldsection{Realizations}
\begin{eczvaluelist}
\item\relax Used for synchronization purposes in telecommunication \NoCaseChange{\protect\cite{cite366}}
\item\relax GPS C/A for satellite navigation \NoCaseChange{\protect\cite{cite367}}.
\end{eczvaluelist}
\codefieldsection{Parent}
\begin{eczvaluelist}
\item\relax
\flmRefsHyperref[eczindexfamilyrel]{code:binary_cyclic}{Cyclic linear binary code}\end{eczvaluelist}
\codefieldsection{Cousins}
\begin{eczvaluelist}
\item\relax
\flmRefsHyperref[eczindexfamilyrel]{code:simplex}{\([2^m-1,m,2^{m-1}]\) simplex code} --- Simplex codes are used to make gold codes. The dual of a Gold code is the intersection of the duals of the simplex codes used to construct it \NoCaseChange{\protect\cite{cite1166}}.
\item\relax
\flmRefsHyperref[eczindexfamilyrel]{code:kasami}{\([2^{2r}-1, 3r, 2^{2r-1} - 2^{r-1} ]\) Kasami code} --- Gold and Kasami codes are both constructed by picking a set of cyclically unrelated sequences of binary linear codes with low cross-correlation \NoCaseChange{\protect\cite{cite107,cite108}}.
\end{eczvaluelist}
\eczhbkcontributors{ Khalil Guy, \eczhuVVA }
\endeczcode

\eczcode{hamming}{\([2^r-1,2^r-r-1,3]\) Hamming code}{~\NoCaseChange{\protect\cite{cite1167,cite1168}}}
\codefieldsection{Alternative Names}
\begin{eczvaluelist}
\item\relax RM\(^*(r-2,r)\) code
\end{eczvaluelist}
\eczhIndexCodeAliasName{hamming}{Hamming code}
\eczhIndexCodeAliasName{hamming}{RM\(^*(r-2,r)\) code}
\codefieldsection{Description}
Member of an infinite family of perfect linear codes with parameters \([2^r-1,2^r-r-1, 3]\) for \(r \geq 2\).
Their \(r \times (2^r-1) \) parity-check matrix \(H\) has all possible nonzero \(r\)-bit strings as its columns.
Adding a parity check yields the \([2^r,2^r-r-1, 4]\) extended Hamming code.

\codefieldsection{Protection}
Can detect 1-bit and 2-bit errors, and can correct 1-bit errors.
\codefieldsection{Rate}
Asymptotic rate \(k/n = 1-\frac{\log n}{n} \to 1\) and normalized distance \(d/n \to 0\).
\codefieldsection{Realizations}
\begin{eczvaluelist}
\item\relax Commonly used when error rates are very low, for example, computer RAM or integrated circuits \NoCaseChange{\protect\cite{cite368}}.
\item\relax Hamming-code based matrix embedding used in steganography \NoCaseChange{\protect\cite{cite369,cite370}}.
\end{eczvaluelist}
\codefieldsection{Notes}
\begin{eczvaluelist}
\item\relax See Kaiserslautern database \NoCaseChange{\protect\cite{cite1184}} for explicit codes.
\item\relax See Ref. \NoCaseChange{\protect\cite[{Sec. 1.10}]{cite1159}} for an introduction to Hamming codes, including their perfectness (Thm. 1.10.4).
\item\relax Binary Hamming codes are duals of irreducible cyclic codes \(C(2,n,1)\) \NoCaseChange{\protect\cite[{Exam. 2.5.1}]{cite68}}.
\end{eczvaluelist}
\codefieldsection{Parents}
\begin{eczvaluelist}
\item\relax
\flmRefsHyperref[eczindexfamilyrel]{code:perfect_binary}{Perfect binary code}\item\relax
\flmRefsHyperref[eczindexfamilyrel]{code:q-ary_hamming}{\(q\)-ary Hamming code} --- The \(q\)-ary Hamming codes reduce to the Hamming codes at \(q=2\).
\item\relax
\flmRefsHyperref[eczindexfamilyrel]{code:narrow_sense_q-ary_bch}{Primitive narrow-sense BCH code} --- Binary Hamming codes are binary primitive narrow-sense BCH codes \NoCaseChange{\protect\cite[{Corr. 5.1.5}]{cite126}}. Binary Hamming codes can be written in cyclic form \NoCaseChange{\protect\cite[{Thm. 12.22}]{cite961}}.
\item\relax
\flmRefsHyperref[eczindexfamilyrel]{code:bch}{Binary BCH code} --- Binary Hamming codes are binary primitive narrow-sense BCH codes \NoCaseChange{\protect\cite[{Corr. 5.1.5}]{cite126}}. Binary Hamming codes can be written in cyclic form \NoCaseChange{\protect\cite[{Thm. 12.22}]{cite961}}.
\item\relax
\flmRefsHyperref[eczindexfamilyrel]{code:lexicographic}{Lexicographic code} --- Hamming codes are lexicodes \NoCaseChange{\protect\cite{cite147}}.
\end{eczvaluelist}
\codefieldsection{Child}
\begin{eczvaluelist}
\item\relax
\flmRefsHyperref[eczindexfamilyrel]{code:hamming743}{\([7,4,3]\) Hamming code}\end{eczvaluelist}
\codefieldsection{Cousins}
\begin{eczvaluelist}
\item\relax
\flmRefsHyperref[eczindexfamilyrel]{code:univ_opt_q-ary}{Universally optimal \(q\)-ary code} --- Binary Hamming codes and several of their extended, punctured, and shortened versions are LP universally optimal codes \NoCaseChange{\protect\cite{cite173}}.
\item\relax
\flmRefsHyperref[eczindexfamilyrel]{code:binary_quad_residue}{Binary quadratic-residue (QR) code} --- The \([7,4,3]\) Hamming code is a binary quadratic-residue code \NoCaseChange{\protect\cite[{Ex. 3.2.10}]{cite70}}.
\item\relax
\flmRefsHyperref[eczindexfamilyrel]{code:constantin_rao}{Constantin-Rao (CR) code} --- The nonlinear CR codes for \(G = \mathbb{Z}_2^r\) reduce to Hamming codes at lengths \(n = 2^r - 1\) \NoCaseChange{\protect\cite{cite1185}}; see Ref. \NoCaseChange{\protect\cite{cite1186}}.
\item\relax
\flmRefsHyperref[eczindexfamilyrel]{code:extended_hamming}{\([2^m,2^m-m-1,4]\) Extended Hamming code} --- Extended Hamming codes are extensions of Hamming codes by a parity-check bit. Puncturing extended Hamming codes yields the Hamming codes.
\item\relax
\flmRefsHyperref[eczindexfamilyrel]{code:reed_muller}{Reed-Muller (RM) code} --- Binary Hamming codes are equivalent to RM\(^*(r-2,r)\).
\item\relax
\flmRefsHyperref[eczindexfamilyrel]{code:nearly_perfect}{Nearly perfect code} --- Shortened Hamming codes \([2^r-2,2^r-r-2,3]\) are nearly perfect \NoCaseChange{\protect\cite[{pg. 533}]{cite41}}.
\item\relax
\flmRefsHyperref[eczindexfamilyrel]{code:combinatorial_design}{Combinatorial design} --- Weight-three codewords of the \([2^r-1,2^r-r-1, 3]\) Hamming code support the Steiner system \(S(2,3,2^r-1)\) \NoCaseChange{\protect\cite[{pg. 89}]{cite39}}.
\item\relax
\flmRefsHyperref[eczindexfamilyrel]{code:preparata}{Preparata code} --- The union of a shortened Preparata code and some of its translates forms a Hamming code \NoCaseChange{\protect\cite[{pg. 475}]{cite41}}.

\item\relax
\flmRefsHyperref[eczindexfamilyrel]{code:repetition}{Repetition code} --- The triple repetition code \([3,1,3]\) is the smallest Hamming code.
\item\relax
\flmRefsHyperref[eczindexfamilyrel]{code:simplex}{\([2^m-1,m,2^{m-1}]\) simplex code} --- Hamming and simplex codes are dual to each other.
\item\relax
\flmRefsHyperref[eczindexfamilyrel]{code:batch}{Batch code} --- Hamming codes can be used to construct batch codes \NoCaseChange{\protect\cite{cite950}\protect\cite[{Exam. 10.9}]{cite946}}.
\item\relax
\flmRefsHyperref[eczindexfamilyrel]{code:ampdamp}{Amplitude-damping (AD) code} --- Ref. \NoCaseChange{\protect\cite{cite1187}} presents a \(\llbracket 7,3\rrbracket \) qubit stabilizer code for a single \flmRefsHyperref{ref498}{AD} error based on the classical \([7,4,3]\) Hamming code.
\item\relax
\flmRefsHyperref[eczindexfamilyrel]{code:eaoa_hamming}{\(\llbracket 10,1,3;1,3,4\rrbracket \) EAOA Hamming code} --- The dual of a \([10,6,3]\) code obtained by shortening the \([15,11,3]\) Hamming code at five positions can be used to construct \(\llbracket 10,1,3;1,3,4\rrbracket \) EAOA Hamming code \NoCaseChange{\protect\cite{cite856}}.
\item\relax
\flmRefsHyperref[eczindexfamilyrel]{code:kls}{Khesin-Lu-Shor code} --- The \flmRefsHyperref{ref857}{encoder-respecting form} of the \(\llbracket m 2^m / (m+1), 2^m / (m+1), d(m)\rrbracket \) Khesin-Lu-Shor code is the graph of a hypercube in \(m = 2^r - 1\) dimensions, and input nodes in the graph are codewords of the \([2^r-1,2^r-r-1,3]\) Hamming code \NoCaseChange{\protect\cite{cite858}}.
\item\relax
\flmRefsHyperref[eczindexfamilyrel]{code:quantum_hamming}{\(\llbracket 2^r, 2^r-r-2, 3\rrbracket \) Gottesman code} --- \(\llbracket 2^r, 2^r-r-2, 3\rrbracket \) Gottesman codes are analogues of Hamming codes in that they saturate the asymptotic Hamming bound.
\item\relax
\flmRefsHyperref[eczindexfamilyrel]{code:ring_cpc}{\(\llbracket 2^r+r, 2^r-r-2, 3\rrbracket \) Ring CPC code} --- The ring CPC code is obtained from the shortened Hamming code via the CPC construction \NoCaseChange{\protect\cite{cite860}}.
\item\relax
\flmRefsHyperref[eczindexfamilyrel]{code:cpc}{Coherent-parity-check (CPC) code} --- \textit{Tripartite CPC codes} are constructed from Hamming codes via the CPC construction \NoCaseChange{\protect\cite[{Thm. 4}]{cite860}}.
\item\relax
\flmRefsHyperref[eczindexfamilyrel]{code:quantum_hamming_css}{\(\llbracket 2^r-1, 2^r-2r-1, 3\rrbracket \) quantum Hamming code} --- Quantum Hamming codes result from applying the CSS construction to Hamming codes and their duals the simplex codes.
\end{eczvaluelist}
\eczhbkcontributors{ Dhruv Devulapalli, \eczhuVVA }
\endeczcode

\eczcode{golay}{\([23, 12, 7]\) Golay code}{~\NoCaseChange{\protect\cite{cite1168}}}
\eczhIndexCodeAliasName{golay}{Golay code}
\codefieldsection{Description}
A \([23, 12, 7]\) perfect binary linear code with connections to various areas of mathematics, e.g., lattices \NoCaseChange{\protect\cite{cite39}} and sporadic simple groups \NoCaseChange{\protect\cite{cite41}}.
Up to equivalence, it is unique for its parameters \NoCaseChange{\protect\cite{cite102}}.
The dual of the Golay code is its \([23,11,8]\) even-weight subcode \NoCaseChange{\protect\cite{cite103,cite104}}.

Shortening the Golay code yields \textit{shortened Golay codes} \NoCaseChange{\protect\cite{cite1041}} like the \([22,10,8]\), \([22,11,7]\), and \([22,12,6]\) codes.
The quadruply shortened Golay code is optimal \NoCaseChange{\protect\cite{cite1188}}.

To construct the Golay code, one can use the great dodecahedron to generate codewords by placing message bits on the faces and calculating the parity bits that live on the 12 vertices of the inner icosahedron.
Its generator matrix is \NoCaseChange{\protect\cite[{Table II}]{cite1032}}
\flmMathEnvironment{align}{}{
  \left(\begin{smallmatrix}
  0 & 1 & 0 & 0 & 1 & 0 & 0 & 1 & 1 & 1 & 1 & 1 & 0 & 0 & 0 & 0 & 0 & 0 & 0 & 0 & 0 & 0 & 1\\
  1 & 0 & 0 & 1 & 0 & 0 & 1 & 1 & 1 & 1 & 1 & 0 & 0 & 0 & 0 & 0 & 0 & 0 & 0 & 0 & 0 & 1 & 0\\
  0 & 1 & 1 & 0 & 1 & 1 & 1 & 0 & 0 & 0 & 1 & 1 & 0 & 0 & 0 & 0 & 0 & 0 & 0 & 0 & 1 & 0 & 0\\
  1 & 1 & 0 & 1 & 1 & 1 & 0 & 0 & 0 & 1 & 1 & 0 & 0 & 0 & 0 & 0 & 0 & 0 & 0 & 1 & 0 & 0 & 0\\
  1 & 1 & 1 & 1 & 0 & 0 & 0 & 1 & 0 & 0 & 1 & 1 & 0 & 0 & 0 & 0 & 0 & 0 & 1 & 0 & 0 & 0 & 0\\
  1 & 0 & 1 & 0 & 1 & 0 & 1 & 1 & 1 & 0 & 0 & 1 & 0 & 0 & 0 & 0 & 0 & 1 & 0 & 0 & 0 & 0 & 0\\
  0 & 0 & 0 & 1 & 1 & 1 & 1 & 0 & 1 & 1 & 0 & 1 & 0 & 0 & 0 & 0 & 1 & 0 & 0 & 0 & 0 & 0 & 0\\
  0 & 0 & 1 & 1 & 1 & 1 & 0 & 1 & 1 & 0 & 1 & 0 & 0 & 0 & 0 & 1 & 0 & 0 & 0 & 0 & 0 & 0 & 0\\
  0 & 1 & 1 & 1 & 1 & 0 & 1 & 1 & 0 & 1 & 0 & 0 & 0 & 0 & 1 & 0 & 0 & 0 & 0 & 0 & 0 & 0 & 0\\
  1 & 1 & 1 & 1 & 0 & 1 & 1 & 0 & 1 & 0 & 0 & 0 & 0 & 1 & 0 & 0 & 0 & 0 & 0 & 0 & 0 & 0 & 0\\
  1 & 0 & 1 & 0 & 0 & 1 & 0 & 0 & 1 & 1 & 1 & 1 & 1 & 0 & 0 & 0 & 0 & 0 & 0 & 0 & 0 & 0 & 0
  \end{smallmatrix}\right)~.
}

The automorphism group of the Golay code is the Mathieu group \(\mathcal{M}_{23}\), a sporadic simple group.
The automorphism group of several shortened Golay codes is \(\mathcal{M}_{22}\) \NoCaseChange{\protect\cite{cite1041}}.

\codefieldsection{Decoding}
\begin{eczvaluelist}
\item\relax The Golay code has a trellis representation and can thus be decoded using trellis decoding \NoCaseChange{\protect\cite{cite1189,cite1190}}.
\item\relax Bounded-distance decoder requiring at most 121 real operations \NoCaseChange{\protect\cite{cite1191}}.
\end{eczvaluelist}
\codefieldsection{Realizations}
\begin{eczvaluelist}
\item\relax Proofs of the quantum mechanical Kochen-Specker theorem \NoCaseChange{\protect\cite{cite362}}.
\end{eczvaluelist}
\codefieldsection{Notes}
\begin{eczvaluelist}
\item\relax See Ref. \NoCaseChange{\protect\cite[{Sec. 1.13}]{cite1159}\protect\cite[{Secs. 2.5 and 2.7}]{cite68}} for an introduction to Golay codes.
\end{eczvaluelist}
\codefieldsection{Parents}
\begin{eczvaluelist}
\item\relax
\flmRefsHyperref[eczindexfamilyrel]{code:perfect_binary}{Perfect binary code} --- The Golay code is perfect \NoCaseChange{\protect\cite[{Thm. 12.3.3 and Def. 12.3.4}]{cite199}}.
\item\relax
\flmRefsHyperref[eczindexfamilyrel]{code:binary_quad_residue}{Binary quadratic-residue (QR) code} --- The Golay code is a binary quadratic residue code with generator polynomial \(r(x)\) over \(\mathbb{F}_2\) with length \(n=23\) \NoCaseChange{\protect\cite[{Ex. 3.2.10}]{cite70}\protect\cite[{Ch. 16}]{cite41}}.
\item\relax
\flmRefsHyperref[eczindexfamilyrel]{code:bch}{Binary BCH code} --- The Golay code is equivalent to a BCH code with Bose distance 5 \NoCaseChange{\protect\cite[{Ch. 20}]{cite41}}.
\item\relax
\flmRefsHyperref[eczindexfamilyrel]{code:univ_opt_q-ary}{Universally optimal \(q\)-ary code} --- The Golay code and several of its extended, shortened, and punctured versions are LP universally optimal codes \NoCaseChange{\protect\cite{cite173}}.
\end{eczvaluelist}
\codefieldsection{Cousins}
\begin{eczvaluelist}
\item\relax
\flmRefsHyperref[eczindexfamilyrel]{code:dual}{Dual linear code} --- The dual of the Golay code is its \([23,11,8]\) even-weight subcode \NoCaseChange{\protect\cite{cite103,cite104}}.
\item\relax
\flmRefsHyperref[eczindexfamilyrel]{code:delsarte_optimal_q-ary}{\(q\)-ary sharp configuration} --- The dual \([23,11,8]\) even-weight subcode of the Golay code is a sharp configuration \NoCaseChange{\protect\cite[{Table 12.1}]{cite199}}.
\item\relax
\flmRefsHyperref[eczindexfamilyrel]{code:spherical_design}{Spherical design} --- The dual of the Golay code forms a spherical 3-design under the \flmRefsHyperref{ref38}{antipodal mapping} \NoCaseChange{\protect\cite[{Exam. 9.3}]{cite385}}.
\item\relax
\flmRefsHyperref[eczindexfamilyrel]{code:combinatorial_design}{Combinatorial design} --- The supports of the weight-seven codewords of the Golay code support the Steiner system \(S(4,7,23)\) \NoCaseChange{\protect\cite{cite160,cite154}\protect\cite[{pg. 89}]{cite39}}.
\item\relax
\flmRefsHyperref[eczindexfamilyrel]{code:extended_golay}{\([24, 12, 8]\) Extended Golay code} --- The extended Golay code is an extension of the Golay code by a parity-check bit.
\item\relax
\flmRefsHyperref[eczindexfamilyrel]{code:hexacode}{\([6,3,4]_4\) Hexacode} --- There is a connection between automorphisms of the even Golay code and the holomorph of the hexacode \NoCaseChange{\protect\cite{cite44}}. The hexacode is often referred to as the extended Golay code over \(\mathbb{F}_4\) \NoCaseChange{\protect\cite{cite126}}.
\item\relax
\flmRefsHyperref[eczindexfamilyrel]{code:shortened_hexacode}{\([5,3,3]_4\) Shortened hexacode} --- The shortened hexacode is often referred to as the Golay code over \(\mathbb{F}_4\) \NoCaseChange{\protect\cite{cite126}}.
\item\relax
\flmRefsHyperref[eczindexfamilyrel]{code:ternary_golay}{\([11,6,5]_3\) Ternary Golay code} --- The ternary Golay code is the ternary counterpart of the binary Golay code.
\item\relax
\flmRefsHyperref[eczindexfamilyrel]{code:mclaughlin}{McLaughlin spherical code} --- The McLaughlin spherical code can be constructed from length-23 Golay codewords \NoCaseChange{\protect\cite{cite387}}.
\item\relax
\flmRefsHyperref[eczindexfamilyrel]{code:qubit_golay}{\(\llbracket 23, 1, 7\rrbracket \) Quantum Golay code} --- The qubit Golay code is a CSS code constructed with the Golay code.
\item\relax
\flmRefsHyperref[eczindexfamilyrel]{code:stellated_dodecahedron_css}{\(\llbracket 30,8,3\rrbracket \) Bring code} --- The automorphism group of the parity-check matrix of the Golay code is the same as a certain automorphism group of the Bring code \NoCaseChange{\protect\cite{cite762}}.
\end{eczvaluelist}
\eczhbkcontributors{ Vikram Elijah Amin, Noah Berthusen, \eczhuVVA }
\endeczcode

\eczcode{extended_golay}{\([24, 12, 8]\) Extended Golay code}{~\NoCaseChange{\protect\cite{cite1168}}}
\eczhIndexCodeAliasName{extended_golay}{Extended Golay code}
\codefieldsection{Description}
A self-dual \([24, 12, 8]\) code that is obtained from the Golay code by adding a parity check.
Equivalently, puncturing any coordinate yields the \([23,12,7]\) Golay code.
Up to equivalence, it is unique for its parameters \NoCaseChange{\protect\cite{cite102}}, and it is the unique \([24,12,8]\) extremal Type II code \NoCaseChange{\protect\cite[{Rems. 4.3.10 and 4.3.11}]{cite40}}.

The automorphism group of the extended Golay code is the Mathieu group \(\mathcal{M}_{24}\), a sporadic simple group \NoCaseChange{\protect\cite[{Rem. 4.3.11}]{cite40}}.

\codefieldsection{Decoding}
\begin{eczvaluelist}
\item\relax Majority decoding \NoCaseChange{\protect\cite{cite1143}}.
\item\relax Decoder using the hexacode \NoCaseChange{\protect\cite{cite1192}}.
\item\relax The extended Golay code has a trellis representation and can thus be decoded using trellis decoding \NoCaseChange{\protect\cite{cite1189,cite1190}}.
\end{eczvaluelist}
\codefieldsection{Realizations}
\begin{eczvaluelist}
\item\relax Voyager 1 and 2 spacecraft, transmitting hundreds of color pictures of Jupiter and Saturn in their 1979, 1980, and 1981 fly-bys \NoCaseChange{\protect\cite{cite363}}.
\item\relax American military standards for automatic link establishment in high frequency radio systems \NoCaseChange{\protect\cite{cite364}}.
\end{eczvaluelist}
\codefieldsection{Notes}
\begin{eczvaluelist}
\item\relax See Ref. \NoCaseChange{\protect\cite[{Sec. 1.13}]{cite1159}\protect\cite[{Sec. 2.5}]{cite68}} for an introduction to Golay codes.
\end{eczvaluelist}
\codefieldsection{Parents}
\begin{eczvaluelist}
\item\relax
\flmRefsHyperref[eczindexfamilyrel]{code:nearly_perfect}{Nearly perfect code} --- The extended Golay code is nearly perfect.
\item\relax
\flmRefsHyperref[eczindexfamilyrel]{code:delsarte_optimal_q-ary}{\(q\)-ary sharp configuration} --- The extended Golay code is a sharp configuration \NoCaseChange{\protect\cite[{Table 12.1}]{cite199}}.
\item\relax
\flmRefsHyperref[eczindexfamilyrel]{code:karlin}{\([2m+2,m+1]\) Karlin code} --- The extended Golay code is equivalent to the Karlin double circulant code for \(m=11\) \NoCaseChange{\protect\cite[{Ch. 16}]{cite41}}.
\item\relax
\flmRefsHyperref[eczindexfamilyrel]{code:quasi_group}{Quasi group-algebra code} --- The extended Golay code is a quasi group-algebra code for various groups \NoCaseChange{\protect\cite{cite1193,cite1194,cite1115}}.
\item\relax
\flmRefsHyperref[eczindexfamilyrel]{code:lexicographic}{Lexicographic code} --- The extended Golay code is a lexicode \NoCaseChange{\protect\cite{cite1195,cite147}\protect\cite[{pg. 327}]{cite41}}.
\end{eczvaluelist}
\codefieldsection{Cousins}
\begin{eczvaluelist}
\item\relax
\flmRefsHyperref[eczindexfamilyrel]{code:orthogonal_array}{Orthogonal array (OA)} --- The extended Golay code is an orthogonal array of strength 7 \NoCaseChange{\protect\cite[{Exam. 1}]{cite226}}.
\item\relax
\flmRefsHyperref[eczindexfamilyrel]{code:self_dual}{Self-dual linear code} --- The extended Golay code is the unique \([24,12,8]\) code, and in particular the unique self-dual doubly even code with those parameters \NoCaseChange{\protect\cite{cite102}\protect\cite[{Rem. 4.3.11}]{cite40}}.
\item\relax
\flmRefsHyperref[eczindexfamilyrel]{code:biorthogonal}{\([2^m,m+1,2^{m-1}]\) First-order RM code} --- The first-order RM\((1,4)\) Reed-Muller code is a subcode of the extended Golay code \NoCaseChange{\protect\cite[{Ch. 5, pg. 146}]{cite39}}.
\item\relax
\flmRefsHyperref[eczindexfamilyrel]{code:hamming844}{\([8,4,4]\) extended Hamming code} --- The extended Golay code can be constructed from two suitably chosen extended Hamming \([8,4,4]\) codes using the \(|a+x|b+x|a+b+x|\) construction \NoCaseChange{\protect\cite[{pg. 588}]{cite41}}.
\item\relax
\flmRefsHyperref[eczindexfamilyrel]{code:binary_quad_residue}{Binary quadratic-residue (QR) code} --- The extended Golay code is an extended binary quadratic-residue code \NoCaseChange{\protect\cite[{Ch. 16}]{cite41}}.
\item\relax
\flmRefsHyperref[eczindexfamilyrel]{code:icosahedron}{Icosahedron code} --- The parity bits of the extended Golay code can be visualized to lie on the vertices of the icosahedron; see \NoCaseChange{\protect\cite{cite1196}} for more details. To construct the code, one can use the great dodecahedron to generate codewords by placing message bits on the faces and calculating the parity bits that live on the 12 vertices of the inner icosahedron.
\item\relax
\flmRefsHyperref[eczindexfamilyrel]{code:dodecahedron}{Dodecahedron code} --- The parity bits of the extended Golay code can be visualized to lie on the vertices of the icosahedron; see \NoCaseChange{\protect\cite{cite1196}} for more details. To construct the code, one can use the great dodecahedron to generate codewords by placing message bits on the faces and calculating the parity bits that live on the 12 vertices of the inner icosahedron.
\item\relax
\flmRefsHyperref[eczindexfamilyrel]{code:golay}{\([23, 12, 7]\) Golay code} --- The extended Golay code is an extension of the Golay code by a parity-check bit.
\item\relax
\flmRefsHyperref[eczindexfamilyrel]{code:group}{Group-algebra code} --- The extended Golay code is a group-algebra code for various groups \NoCaseChange{\protect\cite{cite1197,cite1194,cite1193}}; see \NoCaseChange{\protect\cite{cite1115}\protect\cite[{Exam. 16.5.1}]{cite196}}.
\item\relax
\flmRefsHyperref[eczindexfamilyrel]{code:leech}{\(\Lambda_{24}\) Leech lattice} --- Two copies of the 24-dimensional packing obtained from the extended Golay code can be fitted together without overlap to form the Leech lattice \NoCaseChange{\protect\cite[{Ch. 5, pg. 145}]{cite39}}. Half of the lattice can be obtained using Construction \(B^{\star}\) \NoCaseChange{\protect\cite[{Exam. 10.7.3}]{cite115}}.
\item\relax
\flmRefsHyperref[eczindexfamilyrel]{code:niemeier}{Niemeier lattice} --- The extended Golay code is the glue code for the Niemeier lattice \(A_1^{24}\) \NoCaseChange{\protect\cite[{Ch. 16, pg. 408}]{cite39}}.
\item\relax
\flmRefsHyperref[eczindexfamilyrel]{code:combinatorial_design}{Combinatorial design} --- The supports of the weight-eight codewords of the extended Golay code support the Steiner system \(S(5,8,24)\) \NoCaseChange{\protect\cite{cite160,cite154}\protect\cite[{pg. 89}]{cite39}\protect\cite[{Ch. 10, pg. 276}]{cite39}}. Its blocks are called octads.
\item\relax
\flmRefsHyperref[eczindexfamilyrel]{code:nordstrom_robinson}{\((16,256,6)\) Nordstrom-Robinson (NR) code} --- The NR code can be constructed using the extended Golay code by first selecting a set of codewords satisfying certain conditions and then deleting specific coordinates \NoCaseChange{\protect\cite[{pg. 73}]{cite41}}.
\item\relax
\flmRefsHyperref[eczindexfamilyrel]{code:hexacode}{\([6,3,4]_4\) Hexacode} --- Extended Golay codewords can be obtained from hexacodewords \NoCaseChange{\protect\cite{cite39}}. The hexacode can be used to decode the extended Golay code \NoCaseChange{\protect\cite{cite1192}}.
\item\relax
\flmRefsHyperref[eczindexfamilyrel]{code:ternary_golay}{\([11,6,5]_3\) Ternary Golay code} --- The extended ternary Golay code is the ternary counterpart of the extended binary Golay code.
\item\relax
\flmRefsHyperref[eczindexfamilyrel]{code:pseudo_golay}{Pseudo Golay code} --- The mod-two reduction (mapping \(0,1,2,3\) to \(0,1,0,1\)) of all pseudo Golay codes yields the extended Golay code; see Ref. \NoCaseChange{\protect\cite{cite1198}}.
\item\relax
\flmRefsHyperref[eczindexfamilyrel]{code:quaternary_golay}{Extended quaternary Golay code} --- Codewords of the extended quaternary Golay code with entries 0 and 2 are of the form \(2c\), where \(c\) is a codeword of the extended Golay code. Its mod-two reduction (mapping \(0,1,2,3\) to \(0,1,0,1\)) also yields the extended Golay code \NoCaseChange{\protect\cite{cite112,cite1198}}. The quaternary Golay code can be constructed from the extended Golay code by Hensel lifting to \(\mathbb{Z}_4\) \NoCaseChange{\protect\cite{cite112,cite1199}\protect\cite[{3rd Ed., pg. xxxiii}]{cite39}}.
\end{eczvaluelist}
\eczhbkcontributors{ Vikram Elijah Amin, \eczhuVVA }
\endeczcode

\eczcode{karlin}{\([2m+2,m+1]\) Karlin code}{~\NoCaseChange{\protect\cite{cite1200}}}
\eczhIndexCodeAliasName{karlin}{Karlin code}
\codefieldsection{Description}
Member of the family of \([2m+2,m+1]\) double circulant codes such that \(m\) is prime of the form \(8k+3\) for some \(k\), and \(2m+2\) is a multiple of eight.
See \NoCaseChange{\protect\cite[{Ch. 16}]{cite41}} for their generator matrix.
Karlin codes can be mapped to extended cyclic and extended quadratic-residue codes over \(\mathbb{F}_4\) \NoCaseChange{\protect\cite{cite109,cite110}\protect\cite[{Ch. 16}]{cite41}\protect\cite[{Sec. 2.4.2}]{cite42}} by identifying \((0,\omega,\bar{\omega},1)\) with \((00),(10),(01),(11)\) \NoCaseChange{\protect\cite{cite109}}.

\codefieldsection{Parents}
\begin{eczvaluelist}
\item\relax
\flmRefsHyperref[eczindexfamilyrel]{code:binary_linear}{Linear binary code}\item\relax
\flmRefsHyperref[eczindexfamilyrel]{code:quasi_cyclic}{Quasi-cyclic code} --- Karlin codes can be mapped to extended cyclic and extended quadratic-residue codes over \(\mathbb{F}_4\) \NoCaseChange{\protect\cite{cite109,cite110}\protect\cite[{Ch. 16}]{cite41}\protect\cite[{Sec. 2.4.2}]{cite42}} by identifying \((0,\omega,\bar{\omega},1)\) with \((00),(10),(01),(11)\) \NoCaseChange{\protect\cite{cite109}}.
\item\relax
\flmRefsHyperref[eczindexfamilyrel]{code:self_dual}{Self-dual linear code} --- Karlin codes are Euclidean self-dual doubly even codes \NoCaseChange{\protect\cite[{Ch. 16}]{cite41}}, and some of them are extremal \NoCaseChange{\protect\cite{cite1201,cite1202}}.
\end{eczvaluelist}
\codefieldsection{Child}
\begin{eczvaluelist}
\item\relax
\flmRefsHyperref[eczindexfamilyrel]{code:extended_golay}{\([24, 12, 8]\) Extended Golay code} --- The extended Golay code is equivalent to the Karlin double circulant code for \(m=11\) \NoCaseChange{\protect\cite[{Ch. 16}]{cite41}}.
\end{eczvaluelist}
\codefieldsection{Cousins}
\begin{eczvaluelist}
\item\relax
\flmRefsHyperref[eczindexfamilyrel]{code:q-ary_quad_residue}{Quadratic-residue (QR) code} --- Karlin codes can be mapped to extended cyclic and extended quadratic-residue codes over \(\mathbb{F}_4\) \NoCaseChange{\protect\cite{cite109,cite110}\protect\cite[{Ch. 16}]{cite41}\protect\cite[{Sec. 2.4.2}]{cite42}} by identifying \((0,\omega,\bar{\omega},1)\) with \((00),(10),(01),(11)\) \NoCaseChange{\protect\cite{cite109}}.
\item\relax
\flmRefsHyperref[eczindexfamilyrel]{code:extended_hamming}{\([2^m,2^m-m-1,4]\) Extended Hamming code} --- The extended Hamming code is equivalent to the Karlin double circulant code for \(m=3\) \NoCaseChange{\protect\cite[{Ch. 16}]{cite41}}.
\end{eczvaluelist}
\eczhbkcontributors{ \eczhuVVA }
\endeczcode

\eczcode{self_dual_48_24_12}{\([48,24,12]\) self-dual code}{}
\eczhIndexCodeAliasName{self_dual_48_24_12}{self-dual code}
\codefieldsection{Description}
An extended quadratic-residue code that is the unique self-dual doubly even \([48,24,12]\) code. It is extremal Type II, and its automorphism group is \(PSL(2,47)\) \NoCaseChange{\protect\cite{cite111}\protect\cite[{Rem. 4.3.11}]{cite40}}.

\codefieldsection{Parents}
\begin{eczvaluelist}
\item\relax
\flmRefsHyperref[eczindexfamilyrel]{code:binary_linear}{Linear binary code}\item\relax
\flmRefsHyperref[eczindexfamilyrel]{code:self_dual}{Self-dual linear code} --- The \([48,24,12]\) code is the unique self-dual doubly even code with those parameters \NoCaseChange{\protect\cite{cite111}\protect\cite[{Rem. 4.3.11}]{cite40}}.
\item\relax
\flmRefsHyperref[eczindexfamilyrel]{code:divisible}{Divisible code} --- The \([48,24,12]\) code is doubly even and hence Type II \NoCaseChange{\protect\cite[{Rem. 4.1.10}]{cite40}}; it is the unique self-dual doubly even code with those parameters \NoCaseChange{\protect\cite{cite111}}.
\end{eczvaluelist}
\codefieldsection{Cousins}
\begin{eczvaluelist}
\item\relax
\flmRefsHyperref[eczindexfamilyrel]{code:binary_quad_residue}{Binary quadratic-residue (QR) code} --- The \([48,24,12]\) self-dual code is an extended quadratic-residue code \NoCaseChange{\protect\cite[{Ch. 16}]{cite41}}.
\item\relax
\flmRefsHyperref[eczindexfamilyrel]{code:combinatorial_design}{Combinatorial design} --- Fixed-weight codewords of extremal Type II codes of length divisible by \(24\) form combinatorial 5-designs \NoCaseChange{\protect\cite[{Thm. 4.3.16(a)}]{cite40}}. There are several designs associated with this code \NoCaseChange{\protect\cite{cite166}}.
\item\relax
\flmRefsHyperref[eczindexfamilyrel]{code:group}{Group-algebra code} --- The \([48,24,12]\) self-dual code is a group code for \(G\) being a dihedral group \NoCaseChange{\protect\cite{cite1203}\protect\cite[{Exam. 16.5.1}]{cite196}}.
\item\relax
\flmRefsHyperref[eczindexfamilyrel]{code:stab_47_1_11}{\(\llbracket 47,1,11\rrbracket \) quantum QR code} --- Applying the puncture-and-CSS construction to the \([48,24,12]\) self-dual doubly even quadratic-residue code yields the \(\llbracket 47,1,11\rrbracket \) quantum QR code \NoCaseChange{\protect\cite{cite760}}.
\end{eczvaluelist}
\eczhbkcontributors{ \eczhuVVA }
\endeczcode

\eczcode{simplex734}{\([7,3,4]\) simplex code}{}
\codefieldsection{Alternative Names}
\begin{eczvaluelist}
\item\relax RM\(^*(1,3)\) code
\item\relax Little Hamming code
\end{eczvaluelist}
\eczhIndexCodeAliasName{simplex734}{simplex code}
\eczhIndexCodeAliasName{simplex734}{RM\(^*(1,3)\) code}
\eczhIndexCodeAliasName{simplex734}{Little Hamming code}
\codefieldsection{Description}
Second-smallest nontrivial member of the simplex-code family.
The columns of its generator matrix are in one-to-one correspondence with the elements of the projective space \(PG(2,2)\), with each column being a chosen representative of the corresponding element.
The codewords form a \((8,9)\) simplex spherical code under the \flmRefsHyperref{ref38}{antipodal mapping}.
As a simplex code, it is equidistant: every nonzero codeword has Hamming weight \(4\).

Its generator matrix is
\flmMathEnvironment{align}{}{
\left(\begin{array}{ccccccc}
 1 & 0 & 1 & 1 & 1 & 0 & 0 \\
 1 & 1 & 1 & 0 & 0 & 1 & 0 \\
 0 & 1 & 1 & 1 & 0 & 0 & 1
\end{array}\right)~.
}
The automorphism group of the code is \(GL(3,\mathbb{F}_2)\cong PSL(2,\mathbb{F}_7)\), the second-smallest non-abelian finite simple group.

\codefieldsection{Protection}
Being a simplex code, it saturates the Plotkin bound \NoCaseChange{\protect\cite[{pg. 43}]{cite41}}.

\codefieldsection{Parents}
\begin{eczvaluelist}
\item\relax
\flmRefsHyperref[eczindexfamilyrel]{code:simplex}{\([2^m-1,m,2^{m-1}]\) simplex code}\item\relax
\flmRefsHyperref[eczindexfamilyrel]{code:difference_set}{Difference-set cyclic (DSC) code} --- The \([7,3,4]\) simplex code is the smallest difference-set cyclic code, arising from the lines of the projective plane \(PG(2,2)\) \NoCaseChange{\protect\cite[{pg. 397}]{cite41}}.
\item\relax
\flmRefsHyperref[eczindexfamilyrel]{code:small_distance}{Small-distance block code}\end{eczvaluelist}
\codefieldsection{Cousins}
\begin{eczvaluelist}
\item\relax
\flmRefsHyperref[eczindexfamilyrel]{code:incidence_matrix}{Incidence-matrix projective code} --- The \([7,3,4]\) simplex code is the smallest difference-set cyclic code, arising from the lines of the projective plane \(PG(2,2)\) \NoCaseChange{\protect\cite[{pg. 397}]{cite41}}.
\item\relax
\flmRefsHyperref[eczindexfamilyrel]{code:hamming743}{\([7,4,3]\) Hamming code} --- The \([7,3,4]\) simplex code is the dual of the Hamming code and also its even-weight subcode \NoCaseChange{\protect\cite{cite103,cite104}}.
\item\relax
\flmRefsHyperref[eczindexfamilyrel]{code:eseven}{\(E_7\) root lattice} --- The \([7,3,4]\) simplex code yields the \(E_7\) root lattice via \flmTerm{term}{ref127}{}{Construction A} \NoCaseChange{\protect\cite{cite1204}\protect\cite[{Exam. 10.5.3}]{cite115}\protect\cite[{pg. 138}]{cite39}}.
\item\relax
\flmRefsHyperref[eczindexfamilyrel]{code:octacode}{Octacode} --- Codewords of the heptacode with entries 0 and 2 are of the form \(2c\), where \(c\) is a codeword of the \([7,3,4]\) simplex code \NoCaseChange{\protect\cite[{Exam. 5}]{cite1147}}.
\end{eczvaluelist}
\eczhbkcontributors{ \eczhuVVA }
\endeczcode

\eczcode{hamming743}{\([7,4,3]\) Hamming code}{~\NoCaseChange{\protect\cite{cite1,cite1167,cite1168}}}
\eczhIndexCodeAliasName{hamming743}{Hamming code}
\codefieldsection{Description}
Second-smallest member of the Hamming code family.

Its generator matrix is
\flmMathEnvironment{align}{}{
\left(\begin{array}{ccccccc}
  1 & 0 & 0 & 0 & 1 & 1 & 0\\
  0 & 1 & 0 & 0 & 1 & 0 & 1\\
  0 & 0 & 1 & 0 & 0 & 1 & 1\\
  0 & 0 & 0 & 1 & 1 & 1 & 1
\end{array}\right)~.
}
Up to equivalence, this is the only nontrivial length-seven perfect binary code containing the zero vector.
The automorphism group of the code is \(GL(3,\mathbb{F}_2)\), the second-smallest simple group.

The Hamming code can be extended by a parity-check bit to yield the \([8,4,4]\) extended Hamming code, the smallest doubly even self-dual code.
It can be shortened to yield the \([6,3,3]\) \textit{shortened Hamming code}.
The \([7,3,4]\) simplex code is the dual of the Hamming code and also its even-weight subcode \NoCaseChange{\protect\cite{cite103,cite104}}.

\codefieldsection{Protection}
Can detect 1-bit and 2-bit errors, and can correct 1-bit errors.
\codefieldsection{Notes}
\begin{eczvaluelist}
\item\relax Shannon used this code as an example of a code achieving the channel capacity of a symmetric channel in which blocks of 7 bits are either transmitted without error or with exactly one bit flipped, with each of the eight outcomes being equally likely (so that the equivocation is \(\log_2 8 = 3\) bits and the capacity is \(4\) bits per block) \NoCaseChange{\protect\cite{cite1}}.
\end{eczvaluelist}
\codefieldsection{Parents}
\begin{eczvaluelist}
\item\relax
\flmRefsHyperref[eczindexfamilyrel]{code:hamming}{\([2^r-1,2^r-r-1,3]\) Hamming code}\item\relax
\flmRefsHyperref[eczindexfamilyrel]{code:binary_quad_residue}{Binary quadratic-residue (QR) code} --- The \([7,4,3]\) Hamming code is a quadratic-residue code with generator polynomial \(1+x+x^3\) \NoCaseChange{\protect\cite{cite41}}.
\end{eczvaluelist}
\codefieldsection{Cousins}
\begin{eczvaluelist}
\item\relax
\flmRefsHyperref[eczindexfamilyrel]{code:incidence_matrix}{Incidence-matrix projective code} --- The \([7,4,3]\) Hamming code parity-check matrix corresponds to points in the Fano plane \(PG(2,2)\) \NoCaseChange{\protect\cite[{Exam. 21.4.2}]{cite97}}.
\item\relax
\flmRefsHyperref[eczindexfamilyrel]{code:hamming844}{\([8,4,4]\) extended Hamming code} --- The Hamming code can be extended by a parity-check bit to yield the \([8,4,4]\) extended Hamming code, the smallest doubly even self-dual code.
\item\relax
\flmRefsHyperref[eczindexfamilyrel]{code:dual}{Dual linear code} --- The \([7,3,4]\) simplex code is the dual of the Hamming code and also its even-weight subcode \NoCaseChange{\protect\cite{cite103,cite104}}.
\item\relax
\flmRefsHyperref[eczindexfamilyrel]{code:steane}{\(\llbracket 7,1,3\rrbracket \) Steane code} --- The Steane code is constructed from the \([7,4,3]\) classical Hamming code via the CSS construction.
\item\relax
\flmRefsHyperref[eczindexfamilyrel]{code:griesmer}{Griesmer code} --- Starting with the \([6,3,3]\) shortened Hamming code and applying the \((u|u+v)\) construction recursively using the repetition code yields a family of \([2^m,m+1,2^{m-1}]\) codes for \(m\geq1\) that saturate the Griesmer bound \NoCaseChange{\protect\cite[{pg. 90}]{cite62}}.
\item\relax
\flmRefsHyperref[eczindexfamilyrel]{code:eseven}{\(E_7\) root lattice} --- The \([7,4,3]\) Hamming code yields the \(E_7^{\perp}\) lattice via \flmTerm{term}{ref127}{}{Construction A} \NoCaseChange{\protect\cite{cite1204}}.
\item\relax
\flmRefsHyperref[eczindexfamilyrel]{code:combinatorial_design}{Combinatorial design} --- Weight-three and weight-four codewords of the \([7,4,3]\) Hamming code support combinatorial \(2\)-\((7,3,1)\) and \(2\)-\((7,4,2)\) designs, respectively \NoCaseChange{\protect\cite[{Exam. 5.2.5}]{cite135}}.
\item\relax
\flmRefsHyperref[eczindexfamilyrel]{code:simplex734}{\([7,3,4]\) simplex code} --- The \([7,3,4]\) simplex code is the dual of the Hamming code and also its even-weight subcode \NoCaseChange{\protect\cite{cite103,cite104}}.
\item\relax
\flmRefsHyperref[eczindexfamilyrel]{code:uplusv}{\((u|u+v)\)-construction code} --- Starting with the \([6,3,3]\) shortened Hamming code and applying the \((u|u+v)\) construction recursively using the repetition code yields a family of \([2^m,m+1,2^{m-1}]\) codes for \(m\geq1\) that saturate the Griesmer bound \NoCaseChange{\protect\cite[{pg. 90}]{cite62}}.
\item\relax
\flmRefsHyperref[eczindexfamilyrel]{code:octacode}{Octacode} --- The heptacode can be obtained by Hensel-lifting the \([7,4,3]\) Hamming code to \(\mathbb{Z}_4\) \NoCaseChange{\protect\cite{cite1199,cite158}}.
\item\relax
\flmRefsHyperref[eczindexfamilyrel]{code:xz_7_3_2}{\(\llbracket 7,3,2\rrbracket \) punctured hypercube code} --- The \(\llbracket 7,3,2\rrbracket \) punctured hypercube code \(H_X\) check matrix is the parity-check matrix of the \([7,4,3]\) Hamming code, while its \(H_Z\) matrix is that of the SPC code.
\end{eczvaluelist}
\eczhbkcontributors{ \eczhuVVA }
\endeczcode

\eczcode{cordaro_wagner}{\([n,2,\lceil 2n/3 \rceil -1]\) Cordaro-Wagner code}{~\NoCaseChange{\protect\cite{cite1205}}}
\eczhIndexCodeAliasName{cordaro_wagner}{Cordaro-Wagner code}
\codefieldsection{Description}
A linear binary code with \(k=2\) that is optimal among binary linear \([n,2,d]\) codes.

\codefieldsection{Notes}
\begin{eczvaluelist}
\item\relax Implementation in Komm Python software library for communication systems \NoCaseChange{\protect\cite{cite1206}}.
\end{eczvaluelist}
\codefieldsection{Parent}
\begin{eczvaluelist}
\item\relax
\flmRefsHyperref[eczindexfamilyrel]{code:binary_linear}{Linear binary code}\end{eczvaluelist}
\eczhbkcontributors{ \eczhuVVA }
\endeczcode

\eczcode{parity_check}{\([n,n-1,2]\) Single parity-check (SPC) code}{}
\codefieldsection{Alternative Names}
\begin{eczvaluelist}
\item\relax Sum-zero code
\item\relax Zero-sum code
\item\relax Even-weight code
\end{eczvaluelist}
\eczhIndexCodeAliasName{parity_check}{Single parity-check (SPC) code}
\eczhIndexCodeAliasName{parity_check}{Sum-zero code}
\eczhIndexCodeAliasName{parity_check}{Zero-sum code}
\eczhIndexCodeAliasName{parity_check}{Even-weight code}
\codefieldsection{Description}
An \([n,n-1,2]\) linear binary code whose codewords consist of the message string appended with a \textit{parity-check bit} or \textit{parity bit} such that the parity (i.e., sum over all coordinates of each codeword) is zero.
If the Hamming weight of a message is odd (even), then the parity bit is one (zero).
This code requires only one extra bit of overhead and is therefore inexpensive.
Its codewords are all even-weight binary strings, and its parity-check matrix is a row vector of all ones.
Its automorphism group is \(S_n\).

\codefieldsection{Protection}
This code cannot correct errors; it can only detect a single-bit error.
\codefieldsection{Rate}
The code rate is \(\frac{n-1}{n}\to 1\) as \(n\to\infty\).
\codefieldsection{Decoding}
\begin{eczvaluelist}
\item\relax If the receiver finds that the parity information of a codeword disagrees with the parity bit, then the receiver will discard the information and request a resend.
\item\relax Wagner's rule yields a procedure that is linear in \(n\) \NoCaseChange{\protect\cite{cite1207}} (see \NoCaseChange{\protect\cite[{Sec. 29.7.2}]{cite194}} for a description).
\end{eczvaluelist}
\codefieldsection{Realizations}
\begin{eczvaluelist}
\item\relax Can be realized on almost every communication device. SPCs are some of the earliest error-correcting codes \NoCaseChange{\protect\cite[{Ch. 27}]{cite297}}.
\end{eczvaluelist}
\codefieldsection{Notes}
\begin{eczvaluelist}
\item\relax A parity-check code is the mod-two version of the casting out nines procedure \NoCaseChange{\protect\cite{cite1208,cite1209}}.
\end{eczvaluelist}
\codefieldsection{Parents}
\begin{eczvaluelist}
\item\relax
\flmRefsHyperref[eczindexfamilyrel]{code:crc}{Cyclic redundancy check (CRC) code} --- A CRC using the divisor 11 is a single parity-check code \NoCaseChange{\protect\cite[{Sec. 2.3.3}]{cite250}}.
\item\relax
\flmRefsHyperref[eczindexfamilyrel]{code:q-ary_parity_check}{\([n,n-1,2]_q\) \(q\)-ary parity-check code}\item\relax
\flmRefsHyperref[eczindexfamilyrel]{code:nearly_perfect}{Nearly perfect code}\item\relax
\flmRefsHyperref[eczindexfamilyrel]{code:divisible}{Divisible code} --- Binary SPCs are two-divisible.
\item\relax
\flmRefsHyperref[eczindexfamilyrel]{code:lexicographic}{Lexicographic code} --- SPCs are lexicodes \NoCaseChange{\protect\cite{cite147}}.
\item\relax
\flmRefsHyperref[eczindexfamilyrel]{code:delsarte_optimal_q-ary}{\(q\)-ary sharp configuration} --- The SPC code is a binary sharp configuration \NoCaseChange{\protect\cite[{Table 12.1}]{cite199}}.
\end{eczvaluelist}
\codefieldsection{Cousins}
\begin{eczvaluelist}
\item\relax
\flmRefsHyperref[eczindexfamilyrel]{code:reed_muller}{Reed-Muller (RM) code} --- Binary SPC codes of length \(2^m\) are RM\((m-1,m)\) codes.
\item\relax
\flmRefsHyperref[eczindexfamilyrel]{code:repetition}{Repetition code} --- Binary SPCs and repetition codes are dual to each other.
\item\relax
\flmRefsHyperref[eczindexfamilyrel]{code:dual}{Dual linear code} --- Binary SPCs and repetition codes are dual to each other.
\item\relax
\flmRefsHyperref[eczindexfamilyrel]{code:binary_linear}{Linear binary code} --- Any \([n,k,d]\) code with odd distance can be \textit{extended} to an \([n+1,k,d+1]\) code by adding a bit storing the sum of codeword coordinates.
\item\relax
\flmRefsHyperref[eczindexfamilyrel]{code:ldgm}{Low-density generator-matrix (LDGM) code} --- Concatenated SPCs are LDGM \NoCaseChange{\protect\cite{cite1210}}.
\item\relax
\flmRefsHyperref[eczindexfamilyrel]{code:stab_4_2_2}{\(\llbracket 4,2,2\rrbracket \) Four-qubit code} --- The \(\llbracket 4,2,2\rrbracket \) code is constructed from the \([4,3,2]\) SPC code via the CSS construction.
\item\relax
\flmRefsHyperref[eczindexfamilyrel]{code:dfour}{\(D_4\) hyper-diamond lattice} --- The \(D_4\) lattice is constructed out of the \([4,3,2]\) SPC code via \flmTerm{term}{ref127}{}{Construction A} \NoCaseChange{\protect\cite[{pg. 138}]{cite39}}.
\item\relax
\flmRefsHyperref[eczindexfamilyrel]{code:dn}{\(D_n\) checkerboard lattice} --- \([n,n-1,2]\) SPC codes yield \(D_n\) checkerboard lattices via \flmTerm{term}{ref127}{}{Construction A} \NoCaseChange{\protect\cite[{Exam. 10.5.1}]{cite115}\protect\cite[{pg. 138}]{cite39}}.
\item\relax
\flmRefsHyperref[eczindexfamilyrel]{code:klemm}{Klemm code} --- The generator matrix of the Klemm code consists of a sum of the generator matrix of the repetition code and twice the generator matrix of the SPC code \NoCaseChange{\protect\cite{cite121}}.
\item\relax
\flmRefsHyperref[eczindexfamilyrel]{code:iceberg}{\(\llbracket 2m,2m-2,2\rrbracket \) error-detecting code} --- The \(\llbracket 2m,2m-2,2\rrbracket \) error-detecting code is constructed via the CSS construction from an SPC code and its dual repetition code \NoCaseChange{\protect\cite[{Sec. III}]{cite773}}.
\item\relax
\flmRefsHyperref[eczindexfamilyrel]{code:xz_7_3_2}{\(\llbracket 7,3,2\rrbracket \) punctured hypercube code} --- The \(\llbracket 7,3,2\rrbracket \) punctured hypercube code \(H_X\) check matrix is the parity-check matrix of the \([7,4,3]\) Hamming code, while its \(H_Z\) matrix is that of the SPC code.
\item\relax
\flmRefsHyperref[eczindexfamilyrel]{code:stab_8_3_2}{\(\llbracket 8,3,2\rrbracket \) Smallest interesting color code} --- The \(\llbracket 8,3,2\rrbracket \) hypercube code \(H_X\) check matrix is the parity-check matrix of the \([8,4,4]\) extended Hamming code, while its \(H_Z\) matrix is that of the SPC code.
\item\relax
\flmRefsHyperref[eczindexfamilyrel]{code:classical_product}{Classical-product code} --- SPC codes are used as component codes in classical-product code constructions.
\end{eczvaluelist}
\eczhbkcontributors{ Yijia Xu, \eczhuVVA }
\endeczcode

\eczcode{ara}{Accumulate-repeat-accumulate (ARA) code}{~\NoCaseChange{\protect\cite{cite1211}}}
\codefieldsection{Description}
A generalization of the RA code in which the outer repetition-code encoding step is augmented with an accumulator acting on a fraction of the incoming bits.
In addition, the code may be punctured after the final accumulating step.

\codefieldsection{Parents}
\begin{eczvaluelist}
\item\relax
\flmRefsHyperref[eczindexfamilyrel]{code:irregular_ldpc}{Irregular LDPC code}\item\relax
\flmRefsHyperref[eczindexfamilyrel]{code:protograph_ldpc}{Protograph LDPC code} --- ARA codes can be formulated as protograph LDPC codes \NoCaseChange{\protect\cite{cite1212}}.
\end{eczvaluelist}
\codefieldsection{Child}
\begin{eczvaluelist}
\item\relax
\flmRefsHyperref[eczindexfamilyrel]{code:ra}{Repeat-accumulate (RA) code} --- ARA codes with no pre-encoding accumulator and no post-accumulator puncturing reduce to RA codes.
\end{eczvaluelist}
\eczhbkcontributors{ \eczhuVVA }
\endeczcode

\eczcode{apm_ldpc}{Affine-permutation-matrix LDPC (APM-LDPC) code}{~\NoCaseChange{\protect\cite{cite1213}}}
\codefieldsection{Description}
LDPC code whose parity-check matrix can be put into the form of a block matrix consisting of permutation submatrices representing the affine permutation group or the zero submatrix.
Given a cyclic group \(\mathbb{Z}_r\), the affine permutation group is \(\mathbb{Z}_r \rtimes \mathbb{Z}_r^{\times}\), where \(\mathbb{Z}_r^{\times}\) is the multiplicative group of integers modulo \(r\).
Such codes are often constructed by \flmRefsHyperref{ref47}{lifting} certain protographs into such block matrices \NoCaseChange{\protect\cite{cite48}}.

\codefieldsection{Parent}
\begin{eczvaluelist}
\item\relax
\flmRefsHyperref[eczindexfamilyrel]{code:protograph_ldpc}{Protograph LDPC code} --- Parity-check matrices of APM-LDPC codes can be put into block form where the nonzero blocks are permutation matrices representing the affine permutation group \(\mathbb{Z}_r \rtimes \mathbb{Z}_r^{\times}\).
\end{eczvaluelist}
\codefieldsection{Cousin}
\begin{eczvaluelist}
\item\relax
\flmRefsHyperref[eczindexfamilyrel]{code:2bga}{Two-block group-algebra (2BGA) codes} --- APM-LDPC codes can be used to construct non-Abelian 2BGA codes based on the affine permutation group \NoCaseChange{\protect\cite{cite1214}}.
\end{eczvaluelist}
\eczhbkcontributors{ \eczhuVVA }
\endeczcode

\eczcode{algebraic_ldpc}{Algebraic LDPC code}{}
\codefieldsection{Description}
LDPC code whose parity check matrix is constructed explicitly (i.e., non-randomly) from a particular graph \NoCaseChange{\protect\cite{cite49,cite50}} or an algebraic structure such as a combinatorial design \NoCaseChange{\protect\cite{cite51,cite52,cite53}}, balanced incomplete block design \NoCaseChange{\protect\cite{cite54}}, a partial geometry \NoCaseChange{\protect\cite{cite55}}, a generalized polygon \NoCaseChange{\protect\cite{cite56,cite57}}, or a Latin square \NoCaseChange{\protect\cite{cite58,cite59,cite60}}.
The extra structure and/or symmetry \NoCaseChange{\protect\cite{cite61}} of these codes can often be used to gain a better understanding of their properties.

\codefieldsection{Parent}
\begin{eczvaluelist}
\item\relax
\flmRefsHyperref[eczindexfamilyrel]{code:ldpc}{Low-density parity-check (LDPC) code}\end{eczvaluelist}
\codefieldsection{Children}
\begin{eczvaluelist}
\item\relax
\flmRefsHyperref[eczindexfamilyrel]{code:array_ldpc}{Array-based LDPC (AB-LDPC) code}\item\relax
\flmRefsHyperref[eczindexfamilyrel]{code:b_ldpc}{Block LDPC (B-LDPC) code}\item\relax
\flmRefsHyperref[eczindexfamilyrel]{code:difference_set}{Difference-set cyclic (DSC) code}\item\relax
\flmRefsHyperref[eczindexfamilyrel]{code:lu_ldpc}{Lazebnik-Ustimenko (LU) code}\item\relax
\flmRefsHyperref[eczindexfamilyrel]{code:margulis_ldpc}{Margulis LDPC code}\item\relax
\flmRefsHyperref[eczindexfamilyrel]{code:pg_ldpc}{Finite-geometry LDPC (FG-LDPC) code}\end{eczvaluelist}
\codefieldsection{Cousins}
\begin{eczvaluelist}
\item\relax
\flmRefsHyperref[eczindexfamilyrel]{code:combinatorial_design}{Combinatorial design} --- Combinatorial designs can be used to construct explicit LDPC codes \NoCaseChange{\protect\cite{cite51,cite52,cite53}}.
\item\relax
\flmRefsHyperref[eczindexfamilyrel]{code:protograph_ldpc}{Protograph LDPC code} --- Some deterministic protograph LDPC codes \NoCaseChange{\protect\cite{cite1215}} can be obtained from the theory of voltage graphs \NoCaseChange{\protect\cite{cite1216,cite1217}}.
\item\relax
\flmRefsHyperref[eczindexfamilyrel]{code:qldpc}{Qubit QLDPC code} --- Algebraic LDPC codes made from Latin squares can be used to make qubit QLDPC codes \NoCaseChange{\protect\cite[{Ch. 15}]{cite872}}.
\end{eczvaluelist}
\eczhbkcontributors{ \eczhuVVA }
\endeczcode

\eczcode{anticode}{Anticode}{~\NoCaseChange{\protect\cite{cite1218,cite1219}}}
\codefieldsection{Description}
Linear binary code for which the distance between any two codewords is less than or equal to some value \(\delta\) called the maximum distance. Anticodes can be used to construct codes that saturate the Griesmer bound; see Refs. \NoCaseChange{\protect\cite{cite62,cite63,cite41}} for more details.
\codefieldsection{Parent}
\begin{eczvaluelist}
\item\relax
\flmRefsHyperref[eczindexfamilyrel]{code:binary_linear}{Linear binary code}\end{eczvaluelist}
\codefieldsection{Cousins}
\begin{eczvaluelist}
\item\relax
\flmRefsHyperref[eczindexfamilyrel]{code:griesmer}{Griesmer code} --- Several anticode (e.g., \NoCaseChange{\protect\cite{cite1220,cite1221}}) and related \NoCaseChange{\protect\cite{cite1222}} constructions saturate the Griesmer bound; see Refs. \NoCaseChange{\protect\cite{cite62,cite63,cite41}} for more details.
\item\relax
\flmRefsHyperref[eczindexfamilyrel]{code:antipode}{Antipode sphere packing} --- The antipode and anticode constructions are morally similar \NoCaseChange{\protect\cite{cite39}}.
\item\relax
\flmRefsHyperref[eczindexfamilyrel]{code:projective}{Projective geometry code} --- There is a relation between anticodes and minihypers \NoCaseChange{\protect\cite[{pg. 295}]{cite62}}.
\end{eczvaluelist}
\eczhbkcontributors{ \eczhuVVA }
\endeczcode

\eczcode{array_ldpc}{Array-based LDPC (AB-LDPC) code}{~\NoCaseChange{\protect\cite{cite1223,cite1224}}}
\codefieldsection{Description}
QC-LDPC code constructed deterministically from a disk array code known as a B-code.
Its parity-check matrix admits a compact representation \NoCaseChange{\protect\cite{cite64}} and is related to RS codes.

\codefieldsection{Realizations}
\begin{eczvaluelist}
\item\relax Certain AB-LDPC codes have been proposed to be used for DSL transmission \NoCaseChange{\protect\cite{cite234}}.
\end{eczvaluelist}
\codefieldsection{Parents}
\begin{eczvaluelist}
\item\relax
\flmRefsHyperref[eczindexfamilyrel]{code:qc_ldpc}{Quasi-cyclic LDPC (QC-LDPC) code}\item\relax
\flmRefsHyperref[eczindexfamilyrel]{code:algebraic_ldpc}{Algebraic LDPC code}\end{eczvaluelist}
\codefieldsection{Child}
\begin{eczvaluelist}
\item\relax
\flmRefsHyperref[eczindexfamilyrel]{code:tsf}{Tanner-Sridhara-Fuja (TSF) code}\end{eczvaluelist}
\codefieldsection{Cousin}
\begin{eczvaluelist}
\item\relax
\flmRefsHyperref[eczindexfamilyrel]{code:b_array}{B-code} --- AB-LDPC codes are constructed from certain classes of B-codes. B-codes can be viewed as binary codes by mapping their ring elements to permutation matrices (cf. \flmRefsHyperref{ref47}{lifting}). The resulting codes turn out to be LDPC \NoCaseChange{\protect\cite{cite1224}}.
\end{eczvaluelist}
\eczhbkcontributors{ \eczhuVVA }
\endeczcode

\eczcode{bsghsv-ltc}{Ben-Sasson-Goldreich-Harsha-Sudan-Vadhan (BGHSV) code}{~\NoCaseChange{\protect\cite{cite1076}}}
\codefieldsection{Description}
A member of a family of locally testable \([n,k,d]\) codes with \(n = k^{1+\epsilon}\) and query complexity of \flmRefsHyperref{ref65}{order} \(O(1/\epsilon)\), for any fixed \(\epsilon > 0\).

\codefieldsection{Parent}
\begin{eczvaluelist}
\item\relax
\flmRefsHyperref[eczindexfamilyrel]{code:binary_ltc}{Binary linear LTC}\end{eczvaluelist}
\eczhbkcontributors{ \eczhuVVA }
\endeczcode

\eczcode{bssvw-ltc}{Ben-Sasson-Sudan-Vadhan-Wigderson (BSVW) code}{~\NoCaseChange{\protect\cite{cite1225}}}
\codefieldsection{Description}
Locally testable \([n,k,d]\) code with \(n = k \cdot 2^{\tilde{O}(\sqrt{\log k})}\) and asymptotically constant query complexity, where \(\tilde{O}(f)=O(f\cdot (\log f)^c)\) for some fixed constant \(c\).

\codefieldsection{Parent}
\begin{eczvaluelist}
\item\relax
\flmRefsHyperref[eczindexfamilyrel]{code:binary_ltc}{Binary linear LTC}\end{eczvaluelist}
\eczhbkcontributors{ \eczhuVVA }
\endeczcode

\eczcode{berman}{Berman code}{~\NoCaseChange{\protect\cite{cite1226,cite1227,cite66}}}
\codefieldsection{Description}
A member of a family of codes that is recursively constructed from the single parity-check code via a construction that is similar to the \((u|u+v)\) construction \NoCaseChange{\protect\cite{cite66}}.
Berman codes include RM codes as a special case.

\codefieldsection{Rate}
Achieve capacity of the BEC and BMS channels \NoCaseChange{\protect\cite{cite66}}.
\codefieldsection{Parent}
\begin{eczvaluelist}
\item\relax
\flmRefsHyperref[eczindexfamilyrel]{code:binary_linear}{Linear binary code}\end{eczvaluelist}
\codefieldsection{Child}
\begin{eczvaluelist}
\item\relax
\flmRefsHyperref[eczindexfamilyrel]{code:reed_muller}{Reed-Muller (RM) code} --- Berman codes include RM codes as a special case \NoCaseChange{\protect\cite{cite66}}.
\end{eczvaluelist}
\codefieldsection{Cousin}
\begin{eczvaluelist}
\item\relax
\flmRefsHyperref[eczindexfamilyrel]{code:uplusv}{\((u|u+v)\)-construction code} --- Berman codes are recursively constructed via a construction that is similar to the \((u|u+v)\) construction \NoCaseChange{\protect\cite{cite66}}.
\end{eczvaluelist}
\eczhbkcontributors{ \eczhuVVA }
\endeczcode

\eczcode{bch}{Binary BCH code}{~\NoCaseChange{\protect\cite{cite1228,cite1229,cite1230}}}
\codefieldsection{Description}
Cyclic binary code of odd length \(n\) whose zeroes are consecutive powers of a primitive \(n\)th root of unity \(\alpha\) (see \flmRefsCref{ref67}). More precisely, the generator polynomial of a BCH code of \textit{designed distance} \(\delta\geq 1\) is the lowest-degree monic polynomial with zeroes \(\{\alpha^b,\alpha^{b+1},\cdots,\alpha^{b+\delta-2}\}\) for some \(b\geq 0\). BCH codes are called \textit{narrow-sense} when \(b=1\), and are called \textit{primitive} when \(n=2^r-1\) for some \(r\geq 2\).

The code dimension is related to the \textit{multiplicative order} of \(2\) modulo \(n\), i.e., the smallest integer \(m\) such that \(n\) divides \(2^m-1\). The dimension of a BCH code with \(\delta=2t+1\) is at least \(n-mt\). The field \(\mathbb{F}_{2^m}\) is the smallest field containing the above root of unity \(\alpha\), and is the splitting field of the polynomial \(x^n-1\) (see \flmRefsCref{ref67}).

\codefieldsection{Protection}
By the BCH bound, a BCH code with designed distance \(\delta\) has true distance \(d\geq\delta\). BCH codes with different designed distances may coincide, and the largest possible designed distance for a given code is the \textit{Bose distance}; the true distance may still be larger than that.

\codefieldsection{Rate}
Primitive BCH codes are asymptotically bad \NoCaseChange{\protect\cite[{Thm. 2.6.3}]{cite68}}.
\codefieldsection{Decoding}
\begin{eczvaluelist}
\item\relax Peterson decoder with runtime of \flmRefsHyperref{ref65}{order} \(O(n^3)\) \NoCaseChange{\protect\cite{cite1231,cite1232}} (see exposition in Ref. \NoCaseChange{\protect\cite{cite1233}}).
\item\relax Berlekamp-Massey decoder with runtime of \flmRefsHyperref{ref65}{order} \(O(n^2)\) \NoCaseChange{\protect\cite{cite1234,cite1235}} and modification by Burton \NoCaseChange{\protect\cite{cite1236}}; see also \NoCaseChange{\protect\cite{cite993,cite1024}}.
\item\relax Sugiyama et al. modification of the extended Euclidean algorithm \NoCaseChange{\protect\cite{cite1237,cite1238}}.
\item\relax Guruswami-Sudan list decoder \NoCaseChange{\protect\cite{cite1239,cite1240}}.
\end{eczvaluelist}
\codefieldsection{Realizations}
\begin{eczvaluelist}
\item\relax Satellite communication \NoCaseChange{\protect\cite{cite236}}
\end{eczvaluelist}
\codefieldsection{Notes}
\begin{eczvaluelist}
\item\relax See books \NoCaseChange{\protect\cite{cite41,cite1241,cite126}\protect\cite[{Sec. 1.14}]{cite1159}\protect\cite[{Secs. 2.6-2.6.3}]{cite68}} for expositions on BCH codes and code tables.
\item\relax See Kaiserslautern database \NoCaseChange{\protect\cite{cite1184}} for explicit codes.
\item\relax See corresponding MinT database entry \NoCaseChange{\protect\cite{cite1242}}.
\end{eczvaluelist}
\codefieldsection{Parents}
\begin{eczvaluelist}
\item\relax
\flmRefsHyperref[eczindexfamilyrel]{code:binary_cyclic}{Cyclic linear binary code}\item\relax
\flmRefsHyperref[eczindexfamilyrel]{code:q-ary_bch}{Bose–Chaudhuri–Hocquenghem (BCH) code}\end{eczvaluelist}
\codefieldsection{Children}
\begin{eczvaluelist}
\item\relax
\flmRefsHyperref[eczindexfamilyrel]{code:golay}{\([23, 12, 7]\) Golay code} --- The Golay code is equivalent to a BCH code with Bose distance 5 \NoCaseChange{\protect\cite[{Ch. 20}]{cite41}}.
\item\relax
\flmRefsHyperref[eczindexfamilyrel]{code:hamming}{\([2^r-1,2^r-r-1,3]\) Hamming code} --- Binary Hamming codes are binary primitive narrow-sense BCH codes \NoCaseChange{\protect\cite[{Corr. 5.1.5}]{cite126}}. Binary Hamming codes can be written in cyclic form \NoCaseChange{\protect\cite[{Thm. 12.22}]{cite961}}.
\end{eczvaluelist}
\codefieldsection{Cousins}
\begin{eczvaluelist}
\item\relax
\flmRefsHyperref[eczindexfamilyrel]{code:quasi_perfect}{Quasi-perfect code} --- Only double error-correcting BCH codes \([2^m-1,n-2m,5]\) are quasi-perfect \NoCaseChange{\protect\cite{cite1243,cite1244}} (see also \NoCaseChange{\protect\cite[{Ch. 9}]{cite41}}).
\item\relax
\flmRefsHyperref[eczindexfamilyrel]{code:griesmer}{Griesmer code} --- The \([15,5,7]\) BCH code extended with a parity check saturates the Griesmer bound \NoCaseChange{\protect\cite[{pg. 157}]{cite62}}.
\item\relax
\flmRefsHyperref[eczindexfamilyrel]{code:combinatorial_design}{Combinatorial design} --- A family of BCH codes supports an infinite family of combinatorial 4-designs \NoCaseChange{\protect\cite{cite129,cite130}}.
\item\relax
\flmRefsHyperref[eczindexfamilyrel]{code:preparata}{Preparata code} --- Preparata codes contain twice as many codewords as the extended double-error-correcting BCH codes of the same length and minimum distance, and have the greatest possible number of codewords for this minimum distance \NoCaseChange{\protect\cite{cite1245}\protect\cite[{pg. 475}]{cite41}}.
\item\relax
\flmRefsHyperref[eczindexfamilyrel]{code:reed_muller}{Reed-Muller (RM) code} --- RM\(^*(r,m)\) codes are equivalent to subcodes of BCH codes of designed distance \(2^{m-r}-1\), while RM\((r,m)\) are subcodes of extended BCH codes of the same designed distance \NoCaseChange{\protect\cite[{Ch. 13}]{cite41}}.
\item\relax
\flmRefsHyperref[eczindexfamilyrel]{code:ecoc}{Error-correcting output code (ECOC)} --- BCH codes can be used as ECOCs \NoCaseChange{\protect\cite{cite1177}}.
\item\relax
\flmRefsHyperref[eczindexfamilyrel]{code:quantum_bch}{Qubit BCH code} --- Binary BCH codes are used to construct a subset of qubit BCH codes via the CSS construction.
\item\relax
\flmRefsHyperref[eczindexfamilyrel]{code:quantum_synchronizable}{Quantum synchronizable code} --- BCH codes can be used to construct quantum synchronizable codes via the CSS construction \NoCaseChange{\protect\cite{cite1246}}.
\item\relax
\flmRefsHyperref[eczindexfamilyrel]{code:generalized_bicycle}{Generalized bicycle (GB) code} --- There exist examples of GB codes whose syndromes are protected by small BCH codes \NoCaseChange{\protect\cite{cite1247}}.
\end{eczvaluelist}
\eczhbkcontributors{ Muhammad Junaid Aftab, Nolan Coble, Manasi Shingane, \eczhuVVA }
\endeczcode

\eczcode{bits_into_bits}{Binary code}{}

\codefieldsection{Kingdom root code}
for the \flmRefsHyperref{kingdom:bits_into_bits}{Binary Kingdom}.
\codefieldsection{Description}
Encodes \(K\) states (codewords) in \(n\) binary coordinates and has distance \(d\). Usually denoted as \((n,K,d)\). The distance is the minimum Hamming distance between a pair of distinct codewords.

The coordinate permutation group \(S_n\) of order \(n!\) is formed by \(n\)-dimensional matrices with a 1 in each row and column \NoCaseChange{\protect\cite[{Ch. 8}]{cite41}\protect\cite[{Ch. 3}]{cite39}}.
The group of isometries of Hamming space is the hyperoctahedral group \(\mathbb{Z}_2\wr S_n=\mathbb{Z}_2^n\rtimes S_n\), i.e., the permutation group together with the group formed by the action of binary space on itself (under addition).
Two binary codes are \textit{equivalent} if the codewords of one code can be mapped into those of the other under a hyperoctahedral group element \NoCaseChange{\protect\cite[{Def. 1.8.8}]{cite1159}\protect\cite[{Sec. 3.2}]{cite70}}.
Determining equivalence of two codes can be done by putting each in a canonical form and mapping to a graph isomorphism problem \NoCaseChange{\protect\cite{cite1248}\protect\cite[{Def. 3.2.18 and Sec. 3.2.2}]{cite70}}.

\codefieldsection{Protection}
A binary code \(C\) \textit{corrects} \(t\) errors in the Hamming distance if
\flmMathEnvironment{align}{}{
  \forall x \in C~,~D(x,x+y) < D(x' , x+y)
}
for all codewords \(x' \neq x\) and all \(y\) such that \(|y|=t\), where \(D\) is the \textit{Hamming distance} and \(|y| = D(y,0) \). A code corrects \(t\) errors if and only if \(d \geq 2t+1\), i.e., a code corrects errors on \(t \leq \left\lfloor (d-1)/2 \right\rfloor\) coordinates.
The number of correctable errors is called the \textit{decoding radius}, and it is upper bounded by half of the distance.
In addition, a code detects errors on up to \(d-1\) coordinates, and corrects erasure errors on up to \(d-1\) coordinates.

A binary code \(C\) is \textit{distance invariant} if it has the same Hamming weight distribution as that of its translates \(c + C\) for all codewords \(c\).

\codefieldsection{Rate}
The rate of a binary code is usually defined as \(R=\frac{1}{n}\log_{2}K\) bits per symbol. The maximum rate of a binary code with linear distance is upper bounded by the McEliece, Rodemich, Rumsey and Welch (MRRW) bound \NoCaseChange{\protect\cite{cite1249}} (see Refs. \NoCaseChange{\protect\cite{cite1250,cite1251,cite1252,cite1253,cite1254,cite1255}} for other proofs).
\codefieldsection{Gates}
\begin{eczvaluelist}
\item\relax The group of reversible binary operations is the permutation group \(S_{2^n}\) of all possible \(n\)-bit strings. Reversible gate sets have been classified \NoCaseChange{\protect\cite{cite1256}}.
\end{eczvaluelist}
\codefieldsection{Decoding}
\begin{eczvaluelist}
\item\relax For few-bit codes (\(n\) is small), decoding can be based on a lookup table. For infinite code families, the size of such a table scales exponentially with \(n\), so approximate decoding algorithms scaling polynomially with \(n\) have to be used. The decoder determining the most likely error given a noise channel is called the \textit{maximum-likelihood decoder}.
\item\relax Given a received string \(x\) and an error bound \(e\), a \textit{list decoder} returns a list of all codewords that are at most \(e\) from \(x\) in Hamming distance. The number of codewords in a neighborhood of \(x\) has to be polynomial in \(n\) in order for this decoder to run in time polynomial in \(n\).
\end{eczvaluelist}
\codefieldsection{Parents}
\begin{eczvaluelist}
\item\relax
\flmRefsHyperref[eczindexfamilyrel]{code:q-ary_digits_into_q-ary_digits}{\(q\)-ary code} --- A \(q\)-ary code reduces to a binary code at \(q=2\). Ternary computing may be more applicable than binary computing to cryptographic schemes \NoCaseChange{\protect\cite{cite1257,cite1258}}.
\item\relax
\flmRefsHyperref[eczindexfamilyrel]{code:q-ary_over_zq}{\(q\)-ary code over \(\mathbb{Z}_q\)} --- A \(q\)-ary code over \(\mathbb{Z}_q\) reduces to a binary code at \(q=2\). Ternary computing may be more applicable than binary computing to cryptographic schemes \NoCaseChange{\protect\cite{cite1257,cite1258}}.
\end{eczvaluelist}
\codefieldsection{Children}
\begin{eczvaluelist}
\item\relax
\flmRefsHyperref[eczindexfamilyrel]{code:binary_group_orbit}{Binary group-orbit code}\item\relax
\flmRefsHyperref[eczindexfamilyrel]{code:constant_weight}{Constant-weight code}\item\relax
\flmRefsHyperref[eczindexfamilyrel]{code:nearly_perfect}{Nearly perfect code}\item\relax
\flmRefsHyperref[eczindexfamilyrel]{code:unary}{Unary code}\item\relax
\flmRefsHyperref[eczindexfamilyrel]{code:conference}{Conference code}\item\relax
\flmRefsHyperref[eczindexfamilyrel]{code:constantin_rao}{Constantin-Rao (CR) code}\item\relax
\flmRefsHyperref[eczindexfamilyrel]{code:hergert}{Hergert code}\item\relax
\flmRefsHyperref[eczindexfamilyrel]{code:delsarte_goethals}{Delsarte-Goethals (DG) code}\item\relax
\flmRefsHyperref[eczindexfamilyrel]{code:nadler}{\((12,32,5)\) Nadler code}\item\relax
\flmRefsHyperref[eczindexfamilyrel]{code:levenshtein}{Levenshtein code}\item\relax
\flmRefsHyperref[eczindexfamilyrel]{code:best}{\((10,40,4)\) Best code}\item\relax
\flmRefsHyperref[eczindexfamilyrel]{code:sloane_whitehead}{Sloane-Whitehead code}\item\relax
\flmRefsHyperref[eczindexfamilyrel]{code:superimposed}{Superimposed code}\end{eczvaluelist}
\codefieldsection{Cousins}
\begin{eczvaluelist}
\item\relax
\flmRefsHyperref[eczindexfamilyrel]{code:orthogonal_array}{Orthogonal array (OA)} --- An \((n,K)\) binary code with \flmRefsHyperref{ref113}{dual distance} \(d^{\perp}\) is an OA\(_{K/2^{d^{\perp}-1}}(d^{\perp}-1,n,2)\) \NoCaseChange{\protect\cite{cite209}\protect\cite[{Ch. 5}]{cite41}}.
\item\relax
\flmRefsHyperref[eczindexfamilyrel]{code:qubit_classical_into_quantum}{Qubit c-q code} --- Any binary code can be embedded into a qubit Hilbert space, and thus passed through a qubit channel, by associating length-\(n\) bitstrings with basis vectors in a Hilbert space over \(\mathbb{Z}_2^n\). For example, a bit of information can be embedded into a two-dimensional vector space by associating the two bit values with two basis vectors for the space.
\item\relax
\flmRefsHyperref[eczindexfamilyrel]{code:qubits_into_qubits}{Qubit code} --- Qubit codes are quantum counterparts of binary codes.
\item\relax
\flmRefsHyperref[eczindexfamilyrel]{code:construction_a}{Construction A code} --- Each binary code yields a sphere packing under \flmTerm{term}{ref127}{}{Construction A}.
\item\relax
\flmRefsHyperref[eczindexfamilyrel]{code:combinatorial_design}{Combinatorial design} --- If the \flmRefsHyperref{ref113}{number} of a code is less than or equal to its \flmRefsHyperref{ref113}{dual distance}, then some sets of fixed-weight codewords form a combinatorial design \NoCaseChange{\protect\cite[{Thm. 6.7}]{cite41}}.
\item\relax
\flmRefsHyperref[eczindexfamilyrel]{code:sum_rank_metric}{Sum-rank-metric code} --- The sum-rank metric generalizes both the Hamming metric and the rank metric \NoCaseChange{\protect\cite{cite1259}}.
\item\relax
\flmRefsHyperref[eczindexfamilyrel]{code:hypercube}{Hypercube code} --- Binary strings are elements of the Hamming \(n\)-cube (a.k.a. Boolean hypercube).
\item\relax
\flmRefsHyperref[eczindexfamilyrel]{code:binary_antipodal}{Binary antipodal code} --- Binary antipodal codes are spherical codes obtained from binary codes via the \flmRefsHyperref{ref38}{antipodal mapping}.
\item\relax
\flmRefsHyperref[eczindexfamilyrel]{code:fock_state}{Fock-state bosonic code} --- Fock-state code distance is a natural extension of Hamming distance between binary strings.
\item\relax
\flmRefsHyperref[eczindexfamilyrel]{code:spt}{Symmetry-protected topological (SPT) code} --- SPT orders may be used for encoding classical information \NoCaseChange{\protect\cite{cite1260}}.
\item\relax
\flmRefsHyperref[eczindexfamilyrel]{code:movassagh_ouyang}{Movassagh-Ouyang Hamiltonian code} --- Movassagh-Ouyang codes are constructed from classical binary codes.
\item\relax
\flmRefsHyperref[eczindexfamilyrel]{code:self_complementary}{Self-complementary qubit code} --- A binary code is called \textit{self-complementary} if, for each codeword \(c\), its negation \(\overline{c}\) is also a codeword \NoCaseChange{\protect\cite{cite1261}}. Any self-complementary \((n,K,d > 1)\) classical code yields an \(\llparenthesis n,K/2,2\rrparenthesis \) self-complementary quantum code whose quantum codewords are superpositions of the classical codewords and their complements \NoCaseChange{\protect\cite[{Lemma 1}]{cite1262}}. Self-complementary classical code parameters are governed by the Gray-Rankin bound \NoCaseChange{\protect\cite{cite1263}}.
\end{eczvaluelist}
\eczhbkcontributors{ \eczhuVVA }
\endeczcode

\eczcode{binary_duadic}{Binary duadic code}{~\NoCaseChange{\protect\cite{cite1264}}}
\codefieldsection{Description}
Member of a pair of cyclic linear binary codes that satisfy certain relations, depending on whether the pair is \textit{even-like} or \textit{odd-like} duadic. Binary duadic codes generalize binary quadratic-residue codes \NoCaseChange{\protect\cite[{Sec. 2.7}]{cite68}}. Duadic codes exist for lengths \(n\) that are products of powers of primes, with each prime being \(\pm 1\) modulo \(8\) \NoCaseChange{\protect\cite{cite69}}.

Duadic codes come in two pairs, an even-like duadic pair and an odd-like duadic pair. All codewords in an even-like pair are \textit{even-like}, i.e., \(\sum_i c_i = 0\). By contrast, an odd-like pair is not even-like, i.e., it contains at least one codeword with \(\sum_i c_i = 1\).

Duadic code pairs can be defined in terms of their idempotent generators (see \flmRefsCref{ref67}).
A pair of even-like codes \(C_1\) and \(C_2\) with respective idempotents \(e_1\) and \(e_2\) is an \textit{even-like duadic pair} if (1) \(e_1(x)+e_2(x)=1-\frac{1}{n}(1+x+x^2+\cdots+x^{n-1})\) and (2) there exists a multiplier \(\mu\) such that \(C_1 \mu=C_2\) and \(C_2 \mu=C_1\).

There is an odd-like duadic pair \(\{D_1,D_2\}\) associated with the even-like pair \(\{C_1, C_2\}\), where \(1-e_2(x)\) generates \(D_1\) and \(1-e_1(x)\) generates \(D_2\). The even-pair codes are \([n,\frac{n-1}{2}]\) codes while the odd-pair codes are \([n,\frac{n+1}{2}]\) codes.

Families of binary duadic codes were constructed in Ref. \NoCaseChange{\protect\cite{cite1265}}.

\codefieldsection{Protection}
For odd-like duadic codes, the common minimum odd-like weight \(d_o\) satisfies \(d_o^2 \geq n\); if the splitting is given by \(\mu=-1\), then \(d_o^2-d_o+1 \geq n\) \NoCaseChange{\protect\cite[{Thm. 2.7.1}]{cite68}}.
\codefieldsection{Notes}
\begin{eczvaluelist}
\item\relax Reviews of duadic codes \NoCaseChange{\protect\cite{cite69,cite126}\protect\cite[{Sec. 2.7}]{cite68}}.
\end{eczvaluelist}
\codefieldsection{Parents}
\begin{eczvaluelist}
\item\relax
\flmRefsHyperref[eczindexfamilyrel]{code:binary_cyclic}{Cyclic linear binary code}\item\relax
\flmRefsHyperref[eczindexfamilyrel]{code:q-ary_duadic}{\(q\)-ary duadic code}\end{eczvaluelist}
\codefieldsection{Child}
\begin{eczvaluelist}
\item\relax
\flmRefsHyperref[eczindexfamilyrel]{code:binary_quad_residue}{Binary quadratic-residue (QR) code} --- QR codes are duadic codes of prime length satisfying certain relations \NoCaseChange{\protect\cite{cite69}}.
\end{eczvaluelist}
\codefieldsection{Cousin}
\begin{eczvaluelist}
\item\relax
\flmRefsHyperref[eczindexfamilyrel]{code:reed_muller}{Reed-Muller (RM) code} --- Certain punctured RM codes, such as RM\(^*(2,5)\) \NoCaseChange{\protect\cite[{Table 6.2}]{cite126}} and codes of order \((m-1)/2\) for odd \(m\) \NoCaseChange{\protect\cite{cite1266}}, are duadic.
\end{eczvaluelist}
\eczhbkcontributors{ Yijia Xu, \eczhuVVA }
\endeczcode

\eczcode{binary_group_orbit}{Binary group-orbit code}{~\NoCaseChange{\protect\cite{cite1267,cite1268}}}
\codefieldsection{Description}
Binary length-\(n\) code whose codewords correspond to points in an orbit of some \textit{initial vector} under a \textit{generating group} \(G\), where \(G\) is a subgroup of the Hamming-space isometry group generated by coordinate permutations and bitwise translations.

\codefieldsection{Parents}
\begin{eczvaluelist}
\item\relax
\flmRefsHyperref[eczindexfamilyrel]{code:bits_into_bits}{Binary code}\item\relax
\flmRefsHyperref[eczindexfamilyrel]{code:group_orbit}{Group-orbit code} --- Binary group-orbit codes are group-orbit codes in Hamming space.
\end{eczvaluelist}
\codefieldsection{Children}
\begin{eczvaluelist}
\item\relax
\flmRefsHyperref[eczindexfamilyrel]{code:binary_linear}{Linear binary code} --- The set of codewords of a binary linear code can be thought of as an orbit of a particular codeword under the translation group formed by the code \NoCaseChange{\protect\cite[{Thm. 8.4.2}]{cite115}}. However, binary group-orbit codes do not have to be linear; see \NoCaseChange{\protect\cite[{Remark 8.4.3}]{cite115}}.
\item\relax
\flmRefsHyperref[eczindexfamilyrel]{code:two_in_five}{Two-in-five code} --- The two-in-five code is a binary group-orbit code with group \(S_5\).
\item\relax
\flmRefsHyperref[eczindexfamilyrel]{code:one_hot}{One-hot code} --- The one-hot code is a binary group-orbit code with the cyclic permutation group \(\mathbb{Z}_n\).
\end{eczvaluelist}
\codefieldsection{Cousin}
\begin{eczvaluelist}
\item\relax
\flmRefsHyperref[eczindexfamilyrel]{code:slepian_group}{Slepian group-orbit code} --- Binary group-orbit codes can be mapped into Slepian group-orbit codes via various mappings \NoCaseChange{\protect\cite[{Ch. 8}]{cite115}}.
\end{eczvaluelist}
\eczhbkcontributors{ \eczhuVVA }
\endeczcode

\eczcode{binary_ltc}{Binary linear LTC}{}
\codefieldsection{Description}
A binary linear code \(C\) of length \(n\) that is a \((u,R)\)-LTC with query complexity \(u\) and soundness \(R>0\).

More technically, the code is a \((u,R)\)-LTC if the rows of its parity-check matrix \(H\in \mathbb{F}_2^{r\times n}\) have weight at most \(u\) and if
\flmMathEnvironment{align}{}{
  \frac{1}{r}\operatorname{wt}(H x) \geq \frac{R}{n} D(x,C)
}
holds for any bitstring \(x\), where \(D(x,C)\) is the Hamming distance between \(x\) and the closest codeword to \(x\) \NoCaseChange{\protect\cite[{Def. 11}]{cite1269}}.

\codefieldsection{Parents}
\begin{eczvaluelist}
\item\relax
\flmRefsHyperref[eczindexfamilyrel]{code:binary_linear}{Linear binary code} --- Linear binary codes with distances \(\frac{1}{2}n-\sqrt{t n}\) for some \(t\) are called almost-orthogonal and are locally testable with query complexity of \flmRefsHyperref{ref65}{order} \(O(t)\) \NoCaseChange{\protect\cite{cite1270}}. This was later improved to codes with distance \(\frac{1}{2}n-O(n^{1-\gamma})\) for any positive \(\gamma\) \NoCaseChange{\protect\cite{cite1271}}, provided that the number of codewords is polynomial in \(n\).
\item\relax
\flmRefsHyperref[eczindexfamilyrel]{code:q-ary_ltc}{\(q\)-ary linear LTC}\end{eczvaluelist}
\codefieldsection{Children}
\begin{eczvaluelist}
\item\relax
\flmRefsHyperref[eczindexfamilyrel]{code:bsghsv-ltc}{Ben-Sasson-Goldreich-Harsha-Sudan-Vadhan (BGHSV) code}\item\relax
\flmRefsHyperref[eczindexfamilyrel]{code:bssvw-ltc}{Ben-Sasson-Sudan-Vadhan-Wigderson (BSVW) code}\item\relax
\flmRefsHyperref[eczindexfamilyrel]{code:dinur}{Dinur code}\item\relax
\flmRefsHyperref[eczindexfamilyrel]{code:gs-ltc}{Goldreich-Sudan code} --- Goldreich-Sudan codes resulted from what is often referred to as the first systematic study of LTCs.
\item\relax
\flmRefsHyperref[eczindexfamilyrel]{code:kmrs-ltc}{Kopparty-Meir-Ron-Zewi-Saraf (KMRS) code}\item\relax
\flmRefsHyperref[eczindexfamilyrel]{code:long}{Long code}\item\relax
\flmRefsHyperref[eczindexfamilyrel]{code:lr-cayley-complex}{Left-right Cayley complex code} --- Left-right Cayley complex codes yield one of the first two families of \(c^3\)-LTCs.
\item\relax
\flmRefsHyperref[eczindexfamilyrel]{code:hadamard}{\([2^m,m,2^{m-1}]\) Hadamard code} --- The Hadamard code is the first code to be identified as a (three-query) LTC \NoCaseChange{\protect\cite{cite1172,cite1094}}.
\end{eczvaluelist}
\codefieldsection{Cousins}
\begin{eczvaluelist}
\item\relax
\flmRefsHyperref[eczindexfamilyrel]{code:binary_cyclic}{Cyclic linear binary code} --- Cyclic linear codes cannot be \(c^3\)-LTCs \NoCaseChange{\protect\cite{cite1272}}. Codeword symmetries are in general an obstruction to achieving such LTCs \NoCaseChange{\protect\cite{cite1273}}.
\item\relax
\flmRefsHyperref[eczindexfamilyrel]{code:reed_muller}{Reed-Muller (RM) code} --- RM codes can be LTCs in the low- \NoCaseChange{\protect\cite{cite1274,cite1275}} and high-error \NoCaseChange{\protect\cite{cite1276}} regimes; see also \NoCaseChange{\protect\cite{cite1277}}.
\end{eczvaluelist}
\eczhbkcontributors{ \eczhuVVA }
\endeczcode

\eczcode{binary_quad_residue}{Binary quadratic-residue (QR) code}{}
\codefieldsection{Description}
Member of a quadruple of cyclic binary codes of prime length \(n=8m\pm 1\) for \(m\geq 1\) constructed using quadratic residues and nonresidues of \(n\) \NoCaseChange{\protect\cite[{Def. 3.2.8}]{cite70}}.

The roots of the generator polynomial \(r(x)\) of the first code (see \flmRefsCref{ref67}) are all of the inequivalent quadratic residues of \(n\), and the second code's generator polynomial is \((x-1)r(x)\). The roots of the generator polynomial \(a(x)\) of the third code are all inequivalent nonresidues of \(n\), and the fourth code's generator polynomial is \((x-1)a(x)\). The codes corresponding to polynomials \(r,a\) are often called \textit{augmented} quadratic-residue codes, while the remaining codes are called \textit{expurgated}.

The extended versions of odd-like binary quadratic-residue codes have automorphism groups containing \(PSL(2,p)\) by the Gleason-Prange theorem \NoCaseChange{\protect\cite[{Thm. 3.2.11}]{cite70}}; the extensions of the \([7,4,3]\) Hamming and \([23,12,7]\) Golay codes are exceptional examples with larger automorphism groups \NoCaseChange{\protect\cite[{Rem. 3.2.12}]{cite70}}.

\codefieldsection{Protection}
For odd-like quadratic-residue codes of prime length, the common minimum distance \(d\) satisfies \(d^2 \geq n\); if \(-1\) is a quadratic non-residue, then \(d^2-d+1 \geq n\) \NoCaseChange{\protect\cite[{Thm. 2.7.4}]{cite68}}.
\codefieldsection{Decoding}
\begin{eczvaluelist}
\item\relax Algebraic decoder \NoCaseChange{\protect\cite{cite1278}}.
\end{eczvaluelist}
\codefieldsection{Notes}
\begin{eczvaluelist}
\item\relax Introduction of quadratic-residue codes in Refs. \NoCaseChange{\protect\cite{cite41,cite126}\protect\cite[{Sec. 2.7}]{cite68}}.
\end{eczvaluelist}
\codefieldsection{Parents}
\begin{eczvaluelist}
\item\relax
\flmRefsHyperref[eczindexfamilyrel]{code:binary_duadic}{Binary duadic code} --- QR codes are duadic codes of prime length satisfying certain relations \NoCaseChange{\protect\cite{cite69}}.
\item\relax
\flmRefsHyperref[eczindexfamilyrel]{code:q-ary_quad_residue}{Quadratic-residue (QR) code}\end{eczvaluelist}
\codefieldsection{Children}
\begin{eczvaluelist}
\item\relax
\flmRefsHyperref[eczindexfamilyrel]{code:golay}{\([23, 12, 7]\) Golay code} --- The Golay code is a binary quadratic residue code with generator polynomial \(r(x)\) over \(\mathbb{F}_2\) with length \(n=23\) \NoCaseChange{\protect\cite[{Ex. 3.2.10}]{cite70}\protect\cite[{Ch. 16}]{cite41}}.
\item\relax
\flmRefsHyperref[eczindexfamilyrel]{code:hamming743}{\([7,4,3]\) Hamming code} --- The \([7,4,3]\) Hamming code is a quadratic-residue code with generator polynomial \(1+x+x^3\) \NoCaseChange{\protect\cite{cite41}}.
\end{eczvaluelist}
\codefieldsection{Cousins}
\begin{eczvaluelist}
\item\relax
\flmRefsHyperref[eczindexfamilyrel]{code:divisible}{Divisible code} --- Extended binary quadratic residue codes of length \(8m\) are self-dual doubly even codes \NoCaseChange{\protect\cite[{pg. 82}]{cite39}}.
\item\relax
\flmRefsHyperref[eczindexfamilyrel]{code:self_dual}{Self-dual linear code} --- The length-\(72\) extended binary quadratic-residue code is self-dual but not extremal \NoCaseChange{\protect\cite[{Rem. 4.3.10}]{cite40}}.
\item\relax
\flmRefsHyperref[eczindexfamilyrel]{code:lexicographic}{Lexicographic code} --- The \([18,9,6]\) binary QR code is a lexicode \NoCaseChange{\protect\cite{cite147}}.
\item\relax
\flmRefsHyperref[eczindexfamilyrel]{code:quasi_cyclic}{Quasi-cyclic code} --- Binary QR codes are equivalent to double circulant codes for all \(n<200\) except 89 and 167 \NoCaseChange{\protect\cite{cite1124}}.
\item\relax
\flmRefsHyperref[eczindexfamilyrel]{code:extended_golay}{\([24, 12, 8]\) Extended Golay code} --- The extended Golay code is an extended binary quadratic-residue code \NoCaseChange{\protect\cite[{Ch. 16}]{cite41}}.
\item\relax
\flmRefsHyperref[eczindexfamilyrel]{code:self_dual_48_24_12}{\([48,24,12]\) self-dual code} --- The \([48,24,12]\) self-dual code is an extended quadratic-residue code \NoCaseChange{\protect\cite[{Ch. 16}]{cite41}}.
\item\relax
\flmRefsHyperref[eczindexfamilyrel]{code:hadamard}{\([2^m,m,2^{m-1}]\) Hadamard code} --- For Hadamard matrices obtained from the Paley construction, the linear span of the resulting Hadamard codes yields quadratic-residue codes \NoCaseChange{\protect\cite[{pg. 49}]{cite41}}.
\item\relax
\flmRefsHyperref[eczindexfamilyrel]{code:hamming}{\([2^r-1,2^r-r-1,3]\) Hamming code} --- The \([7,4,3]\) Hamming code is a binary quadratic-residue code \NoCaseChange{\protect\cite[{Ex. 3.2.10}]{cite70}}.
\item\relax
\flmRefsHyperref[eczindexfamilyrel]{code:hamming844}{\([8,4,4]\) extended Hamming code} --- The \([8,4,4]\) extended Hamming code is an extended quadratic-residue code \NoCaseChange{\protect\cite{cite41}}.
\item\relax
\flmRefsHyperref[eczindexfamilyrel]{code:uplusv}{\((u|u+v)\)-construction code} --- The \((u|u+v)\) construction can be used to obtain nonlinear binary quadratic-residue codes \NoCaseChange{\protect\cite{cite374}}.
\item\relax
\flmRefsHyperref[eczindexfamilyrel]{code:quantum_synchronizable}{Quantum synchronizable code} --- Binary QR codes can be used to construct quantum synchronizable codes via the CSS construction \NoCaseChange{\protect\cite{cite1279}}.
\end{eczvaluelist}
\eczhbkcontributors{ Yijia Xu, \eczhuVVA }
\endeczcode

\eczcode{b_ldpc}{Block LDPC (B-LDPC) code}{~\NoCaseChange{\protect\cite{cite1280}}}
\codefieldsection{Description}
Member of a particular class of irregular QC-LDPC codes with efficient encoders.

\codefieldsection{Encoding}
\begin{eczvaluelist}
\item\relax Efficient encoder \NoCaseChange{\protect\cite{cite1280}}.
\end{eczvaluelist}
\codefieldsection{Parents}
\begin{eczvaluelist}
\item\relax
\flmRefsHyperref[eczindexfamilyrel]{code:qc_ldpc}{Quasi-cyclic LDPC (QC-LDPC) code}\item\relax
\flmRefsHyperref[eczindexfamilyrel]{code:irregular_ldpc}{Irregular LDPC code}\item\relax
\flmRefsHyperref[eczindexfamilyrel]{code:algebraic_ldpc}{Algebraic LDPC code}\end{eczvaluelist}
\eczhbkcontributors{ \eczhuVVA }
\endeczcode

\eczcode{topological_classical}{Classical topological code}{~\NoCaseChange{\protect\cite{cite71,cite1281,cite1282}}}
\codefieldsection{Description}
Classical code defined on a two-dimensional lattice and inspired by geometrically local topological quantum codes, such as the surface code or color code.

\codefieldsection{Parents}
\begin{eczvaluelist}
\item\relax
\flmRefsHyperref[eczindexfamilyrel]{code:binary_linear}{Linear binary code}\item\relax
\flmRefsHyperref[eczindexfamilyrel]{code:quantum_inspired}{Quantum-inspired classical block code}\end{eczvaluelist}
\codefieldsection{Child}
\begin{eczvaluelist}
\item\relax
\flmRefsHyperref[eczindexfamilyrel]{code:plaquette_ising}{Plaquette Ising code} --- The 2D plaquette Ising model can be constructed by coupling layers of 1D \(\mathbb{Z}_2\) lattice gauge theory \NoCaseChange{\protect\cite{cite1283}}. A field-theoretic description of the 2D plaquette Ising model can be obtained by coupling layers of 1D gauge theory \NoCaseChange{\protect\cite{cite568}}.
\end{eczvaluelist}
\codefieldsection{Cousin}
\begin{eczvaluelist}
\item\relax
\flmRefsHyperref[eczindexfamilyrel]{code:topological_abelian}{Abelian topological code} --- Some topological orders have classical analogues that can be used for error correction.
\end{eczvaluelist}
\eczhbkcontributors{ \eczhuVVA }
\endeczcode

\eczcode{combinatorial_design}{Combinatorial design}{}
\codefieldsection{Alternative Names}
\begin{eczvaluelist}
\item\relax Block design
\item\relax Covering design
\item\relax Support design
\end{eczvaluelist}
\eczhIndexCodeAliasName{combinatorial_design}{Block design}
\eczhIndexCodeAliasName{combinatorial_design}{Covering design}
\eczhIndexCodeAliasName{combinatorial_design}{Support design}
\codefieldsection{Description}
A constant-weight binary code that is mapped into a combinatorial \(t\)-design.

The mapping proceeds as follows for a length-\(n\) code with codewords of constant weight \(w\).
A codeword \(c\) corresponds to a \textit{block} of the design, with the codeword's \(j\)th coordinate labeling whether or not element \(j\) is contained in the block.
There are a total of \(n\) \textit{elements}, and the constant weight of the code implies that each block contains a fixed number \(w\) of elements.
The code supports an \(S(t,w,n)\) \textit{Steiner system} if each subset of \(t\leq w\) elements is contained in exactly one block.
More generally, the code supports a \textit{combinatorial \(t\)-design}, or, more precisely, a \(t\)-\((n,w,\lambda)\)-design, if each such \(t\)-subset is contained in exactly \(\lambda \geq 1\) blocks.
A combinatorial 2-design with two block intersection sizes is called a \textit{quasi-symmetric design}.
A \(3\)-\((q^d+1,q+1,1)\) combinatorial design is sometimes called a \textit{Witt design} (a.k.a. a \textit{spherical geometry design}) \NoCaseChange{\protect\cite{cite1284,cite1286}\protect\cite[{Remark 6.10}]{cite1285}}.

For example, the seven codewords of weight \(w=3\) of the \([7,4,3]\) Hamming code support a \(2\)-\((7,3,1)\)-design a.k.a. an \(S(2,3,7)\) Steiner system.
The codeword \(1000110\) corresponds to a block containing elements 1, 5, and 6.
Similarly, the other six codewords correspond to blocks 257, 367, 147, 246, 345, and 123.
Each pair of elements is contained in exactly one block.

Combinatorial \(t\)-designs exist for all \(t\) \NoCaseChange{\protect\cite{cite901,cite902,cite903,cite904,cite905}}.
Existence of certain quasi-symmetric designs has also been established \NoCaseChange{\protect\cite{cite1287}}.

\codefieldsection{Notes}
\begin{eczvaluelist}
\item\relax See \NoCaseChange{\protect\cite{cite1288,cite135,cite154}} for reviews on combinatorial designs.
\item\relax See \NoCaseChange{\protect\cite{cite1285,cite163,cite1289,cite225}} for books on combinatorial designs.
\end{eczvaluelist}
\codefieldsection{Parent}
\begin{eczvaluelist}
\item\relax
\flmRefsHyperref[eczindexfamilyrel]{code:constant_weight}{Constant-weight code}\end{eczvaluelist}
\codefieldsection{Cousins}
\begin{eczvaluelist}
\item\relax
\flmRefsHyperref[eczindexfamilyrel]{code:bits_into_bits}{Binary code} --- If the \flmRefsHyperref{ref113}{number} of a code is less than or equal to its \flmRefsHyperref{ref113}{dual distance}, then some sets of fixed-weight codewords form a combinatorial design \NoCaseChange{\protect\cite[{Thm. 6.7}]{cite41}}.
\item\relax
\flmRefsHyperref[eczindexfamilyrel]{code:q-ary_digits_into_q-ary_digits}{\(q\)-ary code} --- Designs can be constructed from \(q\)-ary codes by taking the supports of a subset of codewords of constant weight.
\item\relax
\flmRefsHyperref[eczindexfamilyrel]{code:reed_muller}{Reed-Muller (RM) code} --- Fixed-weight RM codewords of weight less than \(2^m\) support combinatorial 3-designs \NoCaseChange{\protect\cite[{Exam. 5.2.7}]{cite135}}.
\item\relax
\flmRefsHyperref[eczindexfamilyrel]{code:hamming743}{\([7,4,3]\) Hamming code} --- Weight-three and weight-four codewords of the \([7,4,3]\) Hamming code support combinatorial \(2\)-\((7,3,1)\) and \(2\)-\((7,4,2)\) designs, respectively \NoCaseChange{\protect\cite[{Exam. 5.2.5}]{cite135}}.
\item\relax
\flmRefsHyperref[eczindexfamilyrel]{code:hamming}{\([2^r-1,2^r-r-1,3]\) Hamming code} --- Weight-three codewords of the \([2^r-1,2^r-r-1, 3]\) Hamming code support the Steiner system \(S(2,3,2^r-1)\) \NoCaseChange{\protect\cite[{pg. 89}]{cite39}}.
\item\relax
\flmRefsHyperref[eczindexfamilyrel]{code:extended_hamming}{\([2^m,2^m-m-1,4]\) Extended Hamming code} --- Weight-four codewords of the \([2^r,2^r-r-1, 4]\) extended Hamming code support the Steiner system \(S(3,4,2^r)\) \NoCaseChange{\protect\cite[{pg. 89}]{cite39}}.
\item\relax
\flmRefsHyperref[eczindexfamilyrel]{code:q-ary_quad_residue}{Quadratic-residue (QR) code} --- The supports of fixed-weight codewords of certain \(q\)-ary QR codes support combinatorial designs \NoCaseChange{\protect\cite{cite149,cite136,cite154}}, including \(3\)-designs \NoCaseChange{\protect\cite{cite155}}.
\item\relax
\flmRefsHyperref[eczindexfamilyrel]{code:pless_symmetry}{\([2q+2,q+1]_3\) Pless symmetry code} --- The supports of fixed-weight codewords of certain Pless symmetry codes support combinatorial designs \NoCaseChange{\protect\cite{cite164,cite165,cite154}}.
\item\relax
\flmRefsHyperref[eczindexfamilyrel]{code:golay}{\([23, 12, 7]\) Golay code} --- The supports of the weight-seven codewords of the Golay code support the Steiner system \(S(4,7,23)\) \NoCaseChange{\protect\cite{cite160,cite154}\protect\cite[{pg. 89}]{cite39}}.
\item\relax
\flmRefsHyperref[eczindexfamilyrel]{code:extended_golay}{\([24, 12, 8]\) Extended Golay code} --- The supports of the weight-eight codewords of the extended Golay code support the Steiner system \(S(5,8,24)\) \NoCaseChange{\protect\cite{cite160,cite154}\protect\cite[{pg. 89}]{cite39}\protect\cite[{Ch. 10, pg. 276}]{cite39}}. Its blocks are called octads.
\item\relax
\flmRefsHyperref[eczindexfamilyrel]{code:ternary_golay}{\([11,6,5]_3\) Ternary Golay code} --- The supports of the weight-five codewords of the ternary Golay code and the weight-six codewords of the extended ternary Golay code support the Steiner systems \(S(4,5,11)\) and \(S(5,6,12)\), respectively \NoCaseChange{\protect\cite{cite160,cite154}\protect\cite[{pg. 89}]{cite39}}. The latter blocks are called hexads.
\item\relax
\flmRefsHyperref[eczindexfamilyrel]{code:perfect}{Perfect code} --- Perfect codes and combinatorial designs are related \NoCaseChange{\protect\cite{cite149,cite150}}.
\item\relax
\flmRefsHyperref[eczindexfamilyrel]{code:dual}{Dual linear code} --- Linear codes and their duals are related to combinatorial designs via the Assmus-Mattson theorem \NoCaseChange{\protect\cite{cite136,cite137}} (see \NoCaseChange{\protect\cite[{Sec. 5.4}]{cite135}}).
\item\relax
\flmRefsHyperref[eczindexfamilyrel]{code:self_dual}{Self-dual linear code} --- Self-dual extremal codes yield combinatorial \(\leq 5\)-designs using the Assmus-Mattson theorem \NoCaseChange{\protect\cite{cite136}} (see \NoCaseChange{\protect\cite[{Sec. 5.4}]{cite135}}).
See \NoCaseChange{\protect\cite[{Table 1.61, pg. 683}]{cite156}} for a table of combinatorial designs obtained from self-dual codes.

\item\relax
\flmRefsHyperref[eczindexfamilyrel]{code:gallager}{Gallager (GL) code} --- Some Steiner systems can be used to construct Gallager codes \NoCaseChange{\protect\cite{cite72}}.
\item\relax
\flmRefsHyperref[eczindexfamilyrel]{code:algebraic_ldpc}{Algebraic LDPC code} --- Combinatorial designs can be used to construct explicit LDPC codes \NoCaseChange{\protect\cite{cite51,cite52,cite53}}.
\item\relax
\flmRefsHyperref[eczindexfamilyrel]{code:hadamard}{\([2^m,m,2^{m-1}]\) Hadamard code} --- \textit{Hadamard designs} are combinatorial designs constructed from Hadamard matrices \NoCaseChange{\protect\cite{cite161,cite162}}; see Ref. \NoCaseChange{\protect\cite{cite163}}.
\item\relax
\flmRefsHyperref[eczindexfamilyrel]{code:leech}{\(\Lambda_{24}\) Leech lattice} --- The Leech lattice is completely determined by the Steiner system \(S(5,8,24)\) formed by the octads of the extended Golay code \NoCaseChange{\protect\cite[{Ch. 12, pg. 335, Thm. 6}]{cite39}}.
\item\relax
\flmRefsHyperref[eczindexfamilyrel]{code:nearly_perfect}{Nearly perfect code} --- The minimum-weight codewords in a nearly perfect code containing the zero vector support a \(t\)-\((n,2t+1,\lfloor (n-t)/(t+1) \rfloor)\) design, while the minimum-weight codewords in the extended code support a \((t+1)\)-\((n+1,2t+2,\lfloor (n-t)/(t+1) \rfloor)\) design \NoCaseChange{\protect\cite[{Thm. 5.5.4}]{cite135}}.
\item\relax
\flmRefsHyperref[eczindexfamilyrel]{code:perfect_binary}{Perfect binary code} --- If a perfect binary code contains the zero vector, then its minimum-weight codewords support a Steiner \((t+1)\)-\((n,2t+1,1)\) design, while the minimum-weight codewords in the extended code support a Steiner \((t+2)\)-\((n+1,2t+2,1)\) design \NoCaseChange{\protect\cite[{Thm. 5.3.1(b)}]{cite135}}.
\item\relax
\flmRefsHyperref[eczindexfamilyrel]{code:bch}{Binary BCH code} --- A family of BCH codes supports an infinite family of combinatorial 4-designs \NoCaseChange{\protect\cite{cite129,cite130}}.
\item\relax
\flmRefsHyperref[eczindexfamilyrel]{code:self_dual_48_24_12}{\([48,24,12]\) self-dual code} --- Fixed-weight codewords of extremal Type II codes of length divisible by \(24\) form combinatorial 5-designs \NoCaseChange{\protect\cite[{Thm. 4.3.16(a)}]{cite40}}. There are several designs associated with this code \NoCaseChange{\protect\cite{cite166}}.
\item\relax
\flmRefsHyperref[eczindexfamilyrel]{code:higman-sims_graph}{Higman-Sims graph-adjacency code} --- The 4125 codewords of weight 36 of the \([100,22,32]\) code \(C_{100}\) form a \(2\)-\((100,36,525)\) design, which can be used for majority decoding of single errors in \(C_{100}^\perp\) \NoCaseChange{\protect\cite[{Rem. 1.7}]{cite82}}.
\item\relax
\flmRefsHyperref[eczindexfamilyrel]{code:hoffman-singleton}{Hoffman-Singleton cycle code} --- The incidence matrix of the Hoffman-Singleton graph can be converted into a \(2\)-\((50,14,13)\) design \NoCaseChange{\protect\cite[{Prop. 1.1}]{cite82}}.
\item\relax
\flmRefsHyperref[eczindexfamilyrel]{code:goethals}{Goethals code} --- Goethals codes form an infinite family of nonlinear binary codes supporting 3-designs \NoCaseChange{\protect\cite[{Table 5.1}]{cite135}}.
\item\relax
\flmRefsHyperref[eczindexfamilyrel]{code:preparata}{Preparata code} --- Preparata codewords of each weight form 3-designs, and the minimum-weight codewords yield infinite families of 4-designs, including Steiner 4-designs with block sizes 5 and 6 \NoCaseChange{\protect\cite[{Rem. 5.5.6 and Thms. 5.5.7, 5.5.11}]{cite135}\protect\cite[{pg. 471}]{cite41}}.
\item\relax
\flmRefsHyperref[eczindexfamilyrel]{code:kerdock}{Kerdock code} --- Kerdock codes form an infinite family of nonlinear binary codes supporting 3-designs \NoCaseChange{\protect\cite[{Rem. 5.5.6}]{cite135}}.
\item\relax
\flmRefsHyperref[eczindexfamilyrel]{code:nordstrom_robinson}{\((16,256,6)\) Nordstrom-Robinson (NR) code} --- NR codewords give \(3\)-\((16, 6, 4)\), \(3\)-\((16, 8, 3)\), and \(3\)-\((16, 10, 24)\) designs, while the punctured code of length \(15\) and minimum distance \(5\) meets the Johnson bound and supports \(2\)-designs \NoCaseChange{\protect\cite[{Exam. 5.5.5}]{cite135}\protect\cite[{pg. 164}]{cite41}}.
\item\relax
\flmRefsHyperref[eczindexfamilyrel]{code:julin12}{Julin-Golay code} --- Julin-Golay codes are constructed from the Steiner system \(S(5,6,12)\) arising from the extended \((12,132,4)\) code \NoCaseChange{\protect\cite[{pgs. 70-72}]{cite41}}.
\item\relax
\flmRefsHyperref[eczindexfamilyrel]{code:simplex}{\([2^m-1,m,2^{m-1}]\) simplex code} --- Simplex codewords form a 2-design \NoCaseChange{\protect\cite[{pg. 166}]{cite41}}.
\item\relax
\flmRefsHyperref[eczindexfamilyrel]{code:mixed}{Mixed code} --- Combinatorial designs have been generalized to mixed alphabets \NoCaseChange{\protect\cite{cite148}}.
\item\relax
\flmRefsHyperref[eczindexfamilyrel]{code:subspace_design}{Subspace design} --- Combinatorial designs are designs in Johnson space, the space of all size-\(w\) subsets of a set with \(n\) elements. The \(q\)-Johnson spaces generalize this notion to subspaces and reduce to Johnson spaces at \(q=1\). In other words, combinatorial designs are designs over spaces of subsets, while subspace designs are designs over spaces of subspaces.
\item\relax
\flmRefsHyperref[eczindexfamilyrel]{code:insertion_deletion}{Editing code} --- Perfect deletion correcting codes can be constructed using combinatorial design theory \NoCaseChange{\protect\cite{cite139,cite140}}.
\item\relax
\flmRefsHyperref[eczindexfamilyrel]{code:q-ary_constant_weight}{Constant-weight block code} --- Optimal constant-weight codes over \(\mathbb{Z}_q\) can be constructed \NoCaseChange{\protect\cite{cite131}} from a generalization of combinatorial designs to \(q\)-ary alphabets \NoCaseChange{\protect\cite{cite132,cite133}}.
\item\relax
\flmRefsHyperref[eczindexfamilyrel]{code:dodecacode}{\((12,4^6,6)_4\) Dodecacode} --- There exists a \(5\)-\((12, 6, 3)\) design in the dodecacode, and a \(3\)-\((11, 5, 4)\) design in the shortened dodecacode \NoCaseChange{\protect\cite{cite159}}.
\item\relax
\flmRefsHyperref[eczindexfamilyrel]{code:q-ary_cyclic}{Cyclic linear \(q\)-ary code} --- Two families of cyclic \(q\)-ary codes support an infinite family of combinatorial 3-designs \NoCaseChange{\protect\cite{cite134}}.
The supports of all fixed-weight codewords of a \(q\)-ary cyclic code support a combinatorial \(1\)-design \NoCaseChange{\protect\cite[{Corr. 5.2.4}]{cite135}}.

\item\relax
\flmRefsHyperref[eczindexfamilyrel]{code:lexicographic}{Lexicographic code} --- Some lexicodes yield Steiner systems \NoCaseChange{\protect\cite{cite147}}.
\item\relax
\flmRefsHyperref[eczindexfamilyrel]{code:zrm}{ZRM code} --- The weight-four codewords of the binary image of the dual of ZRM\((1,m)\) form a Steiner system that is identical to that formed by the weight-four codewords of an extended Hamming code \NoCaseChange{\protect\cite{cite158}}.
\item\relax
\flmRefsHyperref[eczindexfamilyrel]{code:pseudo_golay}{Pseudo Golay code} --- Supports of codewords of any fixed symmetrized type of pseudo Golay codes form a 5-design \NoCaseChange{\protect\cite{cite141,cite142,cite143}}.
\item\relax
\flmRefsHyperref[eczindexfamilyrel]{code:quaternary_golay}{Extended quaternary Golay code} --- Supports of codewords of any fixed symmetrized type of the extended quaternary Golay code form a 5-design \NoCaseChange{\protect\cite{cite141,cite142,cite143}}.
\item\relax
\flmRefsHyperref[eczindexfamilyrel]{code:q-ary_over_zq}{\(q\)-ary code over \(\mathbb{Z}_q\)} --- Optimal constant-weight codes over \(\mathbb{Z}_q\) can be constructed \NoCaseChange{\protect\cite{cite131}} from a generalization of combinatorial designs to \(q\)-ary alphabets \NoCaseChange{\protect\cite{cite132,cite133}}.
\item\relax
\flmRefsHyperref[eczindexfamilyrel]{code:spherical_design}{Spherical design} --- Spherical designs can be thought of as Euclidean analogues of combinatorial designs \NoCaseChange{\protect\cite{cite157}}.
\item\relax
\flmRefsHyperref[eczindexfamilyrel]{code:ame}{Perfect-tensor code} --- Combinatorial designs and \(d\)-uniform quantum states are related \NoCaseChange{\protect\cite{cite151,cite152,cite153}}.
\item\relax
\flmRefsHyperref[eczindexfamilyrel]{code:ea_design_qldpc}{EA combinatorial-design QLDPC code} --- Combinatorial designs can be used to construct EA QLDPC codes \NoCaseChange{\protect\cite{cite138}}.
\item\relax
\flmRefsHyperref[eczindexfamilyrel]{code:jump}{Jump code} --- Certain types of combinatorial designs can be used to obtain jump codes \NoCaseChange{\protect\cite{cite144,cite145,cite146}}.
\end{eczvaluelist}
\eczhbkcontributors{ \eczhuVVA }
\endeczcode

\eczcode{conference}{Conference code}{~\NoCaseChange{\protect\cite{cite1290}\protect\cite[{pg. 55}]{cite41}}}
\codefieldsection{Description}
A member of the family of \((n,2n+2,(n-1)/2)\) nonlinear binary codes for \(n=1\) modulo 4 that are constructed from conference matrices.

A \textit{conference matrix} \(H\) is a symmetric \(n+1\)-dimensional matrix with zero on its diagonal and \(\pm 1\) elsewhere that satisfies \(H H^T = n I_{n+1} \), where \(I_n\) is the \(n\)-dimensional identity matrix.
By multiplying rows and columns by \(-1\), \(H\) can be \textit{normalized} to the form \(\left(\begin{smallmatrix}0 & f\\
f^{T} & J
\end{smallmatrix}\right)\), where \(J\) is an \(n\)-dimensional matrix satisfying \(J^{T}J=nI_{n}-F\) for some matrix \(F\) satisfying \(JF=FJ=0\).
The code is made up of the \(2n\) rows of the two matrices \(\frac{1}{2}\left(I+F\pm J\right)\) along with the all-zeroes and all-ones vectors.

\codefieldsection{Parents}
\begin{eczvaluelist}
\item\relax
\flmRefsHyperref[eczindexfamilyrel]{code:bits_into_bits}{Binary code}\item\relax
\flmRefsHyperref[eczindexfamilyrel]{code:univ_opt_q-ary}{Universally optimal \(q\)-ary code} --- Conference codes are LP universally optimal codes \NoCaseChange{\protect\cite{cite173}}.
\end{eczvaluelist}
\eczhbkcontributors{ \eczhuVVA }
\endeczcode

\eczcode{constant_weight}{Constant-weight code}{}
\codefieldsection{Alternative Names}
\begin{eczvaluelist}
\item\relax One-weight code
\end{eczvaluelist}
\eczhIndexCodeAliasName{constant_weight}{One-weight code}
\codefieldsection{Description}
A binary code whose codewords are all constrained to have the same Hamming weight \(w\). In the linear setting, this usually refers to all nonzero codewords having the same weight, since every linear code contains the zero codeword.

The \textit{complement} of a constant-weight code is a constant-weight code obtained by interchanging zeroes and ones in the codewords.
Constant-weight codes that contain all strings of some fixed Hamming weight are known as \textit{\(m\)-in-\(n\)} or \({n \choose m}\) codes.

The set of all weight-\(w\) binary strings of length \(n\) forms the \textit{Johnson space} \(J(n,w)\), a finite two-point homogeneous space \(G/H\) with \(G = S_n\) and \(H = S_w \times S_{n-w}\) \NoCaseChange{\protect\cite{cite880,cite986,cite912,cite171}\protect\cite[{Sec. 4.2.1}]{cite987}\protect\cite[{Table 2}]{cite985}}. In these coordinates, the Johnson metric is half the Hamming distance between characteristic vectors \NoCaseChange{\protect\cite{cite171}}. The number of such binary strings is the binomial coefficient \(n \choose w\).

Nontrivial linear binary codes cannot have all codewords of the same weight, but they can have all nonzero codewords of the same weight. All such codes are equidistant, and Bonisoli''s theorem states that any equidistant linear binary code is a direct sum of simplex codes \NoCaseChange{\protect\cite{cite988}} (see also Refs. \NoCaseChange{\protect\cite{cite45,cite46}}).

\codefieldsection{Protection}
See Ref. \NoCaseChange{\protect\cite{cite1291}} for upper bounds on dimension. See book \NoCaseChange{\protect\cite{cite41}} for (Johnson) bounds on size.

\codefieldsection{Realizations}
\begin{eczvaluelist}
\item\relax Radio-network frequency hopping \NoCaseChange{\protect\cite{cite240}}.
\end{eczvaluelist}
\codefieldsection{Notes}
\begin{eczvaluelist}
\item\relax Tables of constant-weight codes for \(n \leq 28\) \NoCaseChange{\protect\cite{cite1292}} and \(n > 28\) \NoCaseChange{\protect\cite{cite240}}.
\end{eczvaluelist}
\codefieldsection{Parents}
\begin{eczvaluelist}
\item\relax
\flmRefsHyperref[eczindexfamilyrel]{code:bits_into_bits}{Binary code}\item\relax
\flmRefsHyperref[eczindexfamilyrel]{code:q-ary_constant_weight}{Constant-weight block code} --- The set of all weight-\(w\) binary strings of length \(n\) forms the \textit{Johnson space} \(J(n,w)\), a finite two-point homogeneous space \(G/H\) with \(G = S_n\) and \(H = S_w \times S_{n-w}\) \NoCaseChange{\protect\cite{cite880,cite986,cite912,cite171}\protect\cite[{Sec. 4.2.1}]{cite987}\protect\cite[{Table 2}]{cite985}}. This is a special case of the nonbinary Johnson space for \(q=2\).
\end{eczvaluelist}
\codefieldsection{Children}
\begin{eczvaluelist}
\item\relax
\flmRefsHyperref[eczindexfamilyrel]{code:combinatorial_design}{Combinatorial design}\item\relax
\flmRefsHyperref[eczindexfamilyrel]{code:weight_two}{Weight-two code}\item\relax
\flmRefsHyperref[eczindexfamilyrel]{code:one_hot}{One-hot code}\end{eczvaluelist}
\codefieldsection{Cousins}
\begin{eczvaluelist}
\item\relax
\flmRefsHyperref[eczindexfamilyrel]{code:2pt_homogeneous}{Two-point homogeneous-space code} --- The set of all weight-\(w\) binary strings of length \(n\) forms the \textit{Johnson space} \(J(n,w)\), a finite two-point homogeneous space \(G/H\) with \(G = S_n\) and \(H = S_w \times S_{n-w}\) \NoCaseChange{\protect\cite{cite880,cite986,cite912,cite171}\protect\cite[{Sec. 4.2.1}]{cite987}\protect\cite[{Table 2}]{cite985}}.
\item\relax
\flmRefsHyperref[eczindexfamilyrel]{code:finite_grassmann}{Constant-dimension code} --- Codewords of length \(n\) and weight \(w\) are in one-to-one correspondence with subsets of \(n\) objects with \(w\) elements. The \(q\)-Johnson spaces generalize this notion to subspaces and reduce to Johnson spaces at \(q=1\). In other words, \((q=2)\)-Johnson space is not the same as (binary) Johnson space since the former indexes subspaces, while the latter indexes subsets.
\item\relax
\flmRefsHyperref[eczindexfamilyrel]{code:delsarte_optimal}{Sharp configuration} --- See \NoCaseChange{\protect\cite[{Table 8.4}]{cite171}} for lists of constant-weight binary codes in Johnson spaces that are Delsarte codes, and hence maximum codes in that metric space.
\item\relax
\flmRefsHyperref[eczindexfamilyrel]{code:univ_opt_q-ary}{Universally optimal \(q\)-ary code} --- Some constant-weight codes, such as simplex codes, are also universally optimal binary codes.
\item\relax
\flmRefsHyperref[eczindexfamilyrel]{code:divisible}{Divisible code} --- Codes whose codewords have a constant weight of \(m\) are automatically \(m\)-divisible. However, divisible codes are linear by definition while constant-weight codes do not have to be.
\item\relax
\flmRefsHyperref[eczindexfamilyrel]{code:binary_linear}{Linear binary code} --- Nontrivial linear binary codes cannot have all codewords of the same weight, but they can have all nonzero codewords of the same weight. All such codes are equidistant, and Bonisoli's theorem states that any equidistant linear binary code is a direct sum of simplex codes \NoCaseChange{\protect\cite{cite988}} (see also Refs. \NoCaseChange{\protect\cite{cite45,cite46}}).
\item\relax
\flmRefsHyperref[eczindexfamilyrel]{code:simplex}{\([2^m-1,m,2^{m-1}]\) simplex code} --- Linear binary codes cannot be constant weight, but can have nonzero codewords with constant weight. All such codes are equidistant, and Bonisoli's theorem states that any equidistant linear binary code is a direct sum of simplex codes \NoCaseChange{\protect\cite{cite988}} (see also Refs. \NoCaseChange{\protect\cite{cite45,cite46}}).
\item\relax
\flmRefsHyperref[eczindexfamilyrel]{code:binary_balanced}{Binary balanced spherical code} --- Binary balanced spherical codes are obtained from constant-weight binary codes.
\end{eczvaluelist}
\eczhbkcontributors{ Micah Shaw, \eczhuVVA }
\endeczcode

\eczcode{constantin_rao}{Constantin-Rao (CR) code}{~\NoCaseChange{\protect\cite{cite1293}}}
\codefieldsection{Description}
A nonlinear single-asymmetric-error code that generalize VT codes and that is constructed from an Abelian group.

A CR code for an Abelian group \(G\) of order \(n+1\) and fixed group element \(g\) consists of all binary strings \(c=c_1c_2\cdots c_n\) that satisfy \(\sum_{i=1}^n c_i g_i = g\) \NoCaseChange{\protect\cite[{Def. 1.3}]{cite1186}}.
Here, addition is the group operation, the multiplication \(1 g_i = g_i\), and \(0 g_i = g_0\) is the identity element. 

CR codes can be generalized to the \(q\)-ary case and also to codes correcting more than one asymmetric error \NoCaseChange{\protect\cite{cite1185}}.
\codefieldsection{Protection}
Protect against single errors induced by the asymmetric noise channel.
Codes for some groups, and in particular, the VT codes, also protect against single deletions and insertions \NoCaseChange{\protect\cite{cite1294}}.

\codefieldsection{Rate}
CR codes for particular groups have higher rates than distance-one codes under the binary asymmetric channel for all lengths except \(n = 2^r - 1\), in which case CR codes reduce to Hamming codes \NoCaseChange{\protect\cite{cite1185}}; see Ref. \NoCaseChange{\protect\cite{cite1186}}.
Size analysis is presented in Refs. \NoCaseChange{\protect\cite{cite1295,cite1296}}.

\codefieldsection{Parent}
\begin{eczvaluelist}
\item\relax
\flmRefsHyperref[eczindexfamilyrel]{code:bits_into_bits}{Binary code}\end{eczvaluelist}
\codefieldsection{Child}
\begin{eczvaluelist}
\item\relax
\flmRefsHyperref[eczindexfamilyrel]{code:vt_single_deletion}{Varshamov-Tenengolts (VT) code} --- CR codes for \(G=\mathbb{Z}_{n+1}\) reduce to VT codes.
\end{eczvaluelist}
\codefieldsection{Cousins}
\begin{eczvaluelist}
\item\relax
\flmRefsHyperref[eczindexfamilyrel]{code:q-ary_digits_into_q-ary_digits}{\(q\)-ary code} --- CR codes, and their special cases the VT codes, can be converted to ternary codes with nice structure via a \textit{binary-to-ternary} map \(00\to 0\), \(11\to 0\), \(01\to 1\), and \(10\to 2\) \NoCaseChange{\protect\cite{cite1186}}.
\item\relax
\flmRefsHyperref[eczindexfamilyrel]{code:q-ary_linear_over_zq}{Linear code over \(\mathbb{Z}_q\)} --- CR codes, and their special cases the VT codes, can be converted to ternary codes with nice structure via a \textit{binary-to-ternary} map \(00\to 0\), \(11\to 0\), \(01\to 1\), and \(10\to 2\) \NoCaseChange{\protect\cite{cite1186}}.
\item\relax
\flmRefsHyperref[eczindexfamilyrel]{code:hamming}{\([2^r-1,2^r-r-1,3]\) Hamming code} --- The nonlinear CR codes for \(G = \mathbb{Z}_2^r\) reduce to Hamming codes at lengths \(n = 2^r - 1\) \NoCaseChange{\protect\cite{cite1185}}; see Ref. \NoCaseChange{\protect\cite{cite1186}}.
\item\relax
\flmRefsHyperref[eczindexfamilyrel]{code:ampdamp_cws}{Amplitude-damping CWS code} --- Amplitude-damping CWS codes can be obtained from CR codes \NoCaseChange{\protect\cite{cite1297}}.
\end{eczvaluelist}
\eczhbkcontributors{ \eczhuVVA }
\endeczcode

\eczcode{coxeter}{Coxeter code}{~\NoCaseChange{\protect\cite{cite1298}}}
\codefieldsection{Description}
A member of a family of codes that generalize RM codes in a group-theoretic sense.

An RM\((r,m)\) code is spanned by indicators of all subcubes of dimension \(m-r\) in the \(m\)-dimensional cube (this is a redundant generating set \NoCaseChange{\protect\cite{cite753}}), i.e., by the cosets of rank-\((m-r)\) subgroups of \(\mathbb{Z}_2^m\).
For a finite Coxeter group \(W\) with \(m\) generators, the Coxeter code \(C_W(r)\) of order \(r\) with \(-1 \leq r \leq m\) is similarly spanned by indicators of all the cosets of rank-\((m-r)\) parabolic subgroups of \(W\).

The dimension \(\dim( C_W(r) )=\sum_{i=0}^r 
\left\langle \begin{smallmatrix}W\\ i \end{smallmatrix}\right\rangle \), 
where \(\left\langle \begin{smallmatrix}W\\ i \end{smallmatrix}\right\rangle \) is the \(W\)-Eulerian number, i.e., the count of elements in \(W\) with descent number equal to \(i\).

\codefieldsection{Protection}
The family of Coxeter codes is closed under duality, \(( C_W(r) )^{\perp}=C_W(m-r-1)\).
For \(q < r\), \(C_W(q) \subsetneq C_W(r)\).
The distance \(d( C_W(r) ) \geq 2^{m-r}\).

\codefieldsection{Parent}
\begin{eczvaluelist}
\item\relax
\flmRefsHyperref[eczindexfamilyrel]{code:binary_linear}{Linear binary code}\end{eczvaluelist}
\codefieldsection{Child}
\begin{eczvaluelist}
\item\relax
\flmRefsHyperref[eczindexfamilyrel]{code:reed_muller}{Reed-Muller (RM) code} --- An RM\((r,m)\) code is spanned by indicators of all subcubes of dimension \(m-r\) in the \(m\)-dimensional cube (this is a redundant generating set \NoCaseChange{\protect\cite{cite753}}), i.e., by the cosets of rank-\((m-r)\) subgroups of \(\mathbb{Z}_2^m\). For a finite Coxeter group \(W\) with \(m\) generators, the Coxeter code \(C_W(r)\) of order \(r\) with \(-1 \leq r \leq m\) is similarly spanned by indicators of all the cosets of rank-\((m-r)\) parabolic subgroups of \(W\).
\end{eczvaluelist}
\codefieldsection{Cousin}
\begin{eczvaluelist}
\item\relax
\flmRefsHyperref[eczindexfamilyrel]{code:qubit_css}{Qubit CSS code} --- Coxeter codes can be used to make qubit CSS codes \NoCaseChange{\protect\cite{cite1298}}.
\end{eczvaluelist}
\eczhbkcontributors{ Alexander Barg, \eczhuVVA }
\endeczcode

\eczcode{homological_classical}{Cycle code}{~\NoCaseChange{\protect\cite{cite1299,cite1300,cite1301,cite1302,cite1152,cite71,cite1303}}}
\codefieldsection{Alternative Names}
\begin{eczvaluelist}
\item\relax Graph theoretic code
\item\relax Graph homology code
\item\relax Graph code
\end{eczvaluelist}
\eczhIndexCodeAliasName{homological_classical}{Graph theoretic code}
\eczhIndexCodeAliasName{homological_classical}{Graph homology code}
\eczhIndexCodeAliasName{homological_classical}{Graph code}
\codefieldsection{Description}
A code whose parity-check matrix is obtained from the incidence matrix of a graph over \(\mathbb{F}_2\).
This code's properties are derived from the size two chain complex associated with the graph.
Not every binary linear code is homological, but there exist homological families that asymptotically saturate the Hamming bound \NoCaseChange{\protect\cite{cite71}}.

Let \(G\) be a finite, undirected, connected graph without loops and multiple edges.
Let \(|V|\) be its number of vertices and \(|E|\) be its number of edges.
Its incidence matrix is a \(|V| \times |E|\) matrix whose rows are indexed by the vertices of \(G\), and whose columns are indexed by the edges of \(G\).
The matrix element \(H_{ij}=1\) if vertex \(i\) belongs to the edge \(j\) and \(H_{ij}=0\) otherwise.
Over \(\mathbb{F}_2\), one row is redundant for a connected graph, so a parity-check matrix of the cycle code can be taken to be any full-rank set of \(|V|-1\) rows.

\codefieldsection{Protection}
More generally, an \([n,k,d]\) cycle code constructed out of a connected graph has \(n = |E|\) and \(k = 1 - \chi(G) = 1-|V|+|E|\), where \(\chi(G)\) is the Euler characteristic of the graph \NoCaseChange{\protect\cite{cite1304}}.
The code distance is equal to the shortest size of a cycle, guaranteed to exist since \(G\) is not a tree.

\codefieldsection{Parents}
\begin{eczvaluelist}
\item\relax
\flmRefsHyperref[eczindexfamilyrel]{code:binary_linear}{Linear binary code}\item\relax
\flmRefsHyperref[eczindexfamilyrel]{code:projective}{Projective geometry code} --- Incidence matrices of graphs have no repeated columns since that would correspond to multi-edges. Therefore, cycle codes can be interpreted as projective codes.
\end{eczvaluelist}
\codefieldsection{Children}
\begin{eczvaluelist}
\item\relax
\flmRefsHyperref[eczindexfamilyrel]{code:hoffman-singleton}{Hoffman-Singleton cycle code}\item\relax
\flmRefsHyperref[eczindexfamilyrel]{code:petersen}{\([15,6,5]\) Petersen cycle code}\item\relax
\flmRefsHyperref[eczindexfamilyrel]{code:cycle_ldpc}{Cycle LDPC code}\end{eczvaluelist}
\codefieldsection{Cousins}
\begin{eczvaluelist}
\item\relax
\flmRefsHyperref[eczindexfamilyrel]{code:graph}{Graph-adjacency code} --- Graph-adjacency (cycle) codes' generator (parity-check) matrices are defined using adjacency (incidence) matrices of graphs.
\item\relax
\flmRefsHyperref[eczindexfamilyrel]{code:quantum_inspired}{Quantum-inspired classical block code} --- Cycle codes have been known in classical coding theory, and have been rediscovered in the quantum context; see Ref. \NoCaseChange{\protect\cite{cite1304}} for a brief exposition.
\item\relax
\flmRefsHyperref[eczindexfamilyrel]{code:generalized_homological_product}{Generalized homological-product code} --- Cycle codes have been known in classical coding theory, and have been rediscovered in the quantum context; see Ref. \NoCaseChange{\protect\cite{cite1304}} for a brief exposition.
\item\relax
\flmRefsHyperref[eczindexfamilyrel]{code:higher_dimensional_surface}{Homological code} --- Cycle codes feature in generalizations of the surface code \NoCaseChange{\protect\cite{cite1304}}.
\item\relax
\flmRefsHyperref[eczindexfamilyrel]{code:perfect_binary}{Perfect binary code} --- A family of cycle codes saturate the asymptotic Hamming bound \NoCaseChange{\protect\cite{cite71}}.
\item\relax
\flmRefsHyperref[eczindexfamilyrel]{code:qubit_css}{Qubit CSS code} --- Cycle codes, including the Petersen cycle and Hoffman-Singleton cycle codes, feature in magic-state distillation protocols \NoCaseChange{\protect\cite[{Appx. A.2.1}]{cite101}\protect\cite[{Sec. VII.A}]{cite705}}.
\end{eczvaluelist}
\eczhbkcontributors{ Fengxing Zhu, Seyed Sajjad Nezhadi, \eczhuVVA }
\endeczcode

\eczcode{cycle_ldpc}{Cycle LDPC code}{~\NoCaseChange{\protect\cite{cite1152}}}
\codefieldsection{Description}
An LDPC code whose parity-check matrix forms the incidence matrix of a graph, i.e., has weight-two columns.

\codefieldsection{Protection}
The minimum distance of a cycle LDPC code is \(d\geq g/2\), where \(g\) is the girth of the code's Tanner graph \NoCaseChange{\protect\cite[{Remark 21.2.13}]{cite97}}.

\codefieldsection{Rate}
Cycle codes are not asymptotically good \NoCaseChange{\protect\cite{cite1305}}.
\codefieldsection{Encoding}
\begin{eczvaluelist}
\item\relax Linear-time encoder \NoCaseChange{\protect\cite{cite1306}}.
\end{eczvaluelist}
\codefieldsection{Realizations}
\begin{eczvaluelist}
\item\relax Cycle LDPC codes have been proposed to be used for MIMO channels \NoCaseChange{\protect\cite{cite248}}.
\end{eczvaluelist}
\codefieldsection{Parents}
\begin{eczvaluelist}
\item\relax
\flmRefsHyperref[eczindexfamilyrel]{code:qc_ldpc}{Quasi-cyclic LDPC (QC-LDPC) code} --- Cycle LDPC codes form a class of regular QC LDPC codes \NoCaseChange{\protect\cite{cite1307}}.
\item\relax
\flmRefsHyperref[eczindexfamilyrel]{code:regular_ldpc}{Regular LDPC code} --- Cycle LDPC codes form a class of regular QC LDPC codes \NoCaseChange{\protect\cite{cite1307}}.
\item\relax
\flmRefsHyperref[eczindexfamilyrel]{code:homological_classical}{Cycle code}\end{eczvaluelist}
\codefieldsection{Child}
\begin{eczvaluelist}
\item\relax
\flmRefsHyperref[eczindexfamilyrel]{code:margulis_ldpc}{Margulis LDPC code} --- Margulis LDPC codes are examples of cycle codes for particular large-girth graphs \NoCaseChange{\protect\cite{cite1304}}.
\end{eczvaluelist}
\eczhbkcontributors{ Fengxing Zhu, \eczhuVVA }
\endeczcode

\eczcode{binary_cyclic}{Cyclic linear binary code}{}
\codefieldsection{Description}
A binary code of length \(n\) is cyclic if, for each codeword \(c_1 c_2 \cdots c_n\), the cyclically shifted string \(c_n c_1 \cdots c_{n-1}\) is also a codeword. A cyclic code is called \textit{primitive} when \(n=2^r-1\) for some \(r\geq 2\). 

A \textit{shortened cyclic code} is obtained from a cyclic code by taking only codewords with the first \(j\) zero entries, and deleting those zeroes.
For odd \(n\), cyclic binary codes can also be described by defining sets of roots of generator polynomials, and by equivalent generating-idempotent and trace constructions \NoCaseChange{\protect\cite[{Secs. 2.3 and 2.4}]{cite68}}.

\codefieldsection{Protection}
The BCH and Hartmann-Tzeng bounds give lower bounds on the distance of cyclic binary codes \NoCaseChange{\protect\cite[{Sec. 2.4}]{cite68}}; the shift bound \NoCaseChange{\protect\cite{cite1308}} gives another lower bound.
\codefieldsection{Decoding}
\begin{eczvaluelist}
\item\relax Meggitt decoder \NoCaseChange{\protect\cite{cite357}}.
\item\relax Information set decoding (ISD) \NoCaseChange{\protect\cite{cite1309}}, a probabilistic decoding strategy that essentially tries to guess an information set of \(k\) correct positions in the received word, where \(k\) is the code dimension. Then, an error vector is constructed to map the received word onto the nearest codeword, assuming those \(k\) positions are error-free. When the Hamming weight of the error vector is low enough, that codeword is assumed to be the intended transmission.
\item\relax Permutation decoding \NoCaseChange{\protect\cite{cite1310}}.
\end{eczvaluelist}
\codefieldsection{Notes}
\begin{eczvaluelist}
\item\relax See \NoCaseChange{\protect\cite[{Ch. 7}]{cite41}\protect\cite[{Secs. 2.1-2.4}]{cite68}} for expositions on cyclic codes.
\item\relax See Ref. \NoCaseChange{\protect\cite{cite1311}} for tables of the best binary cyclic codes of odd lengths 101 to 127.
\end{eczvaluelist}
\codefieldsection{Parents}
\begin{eczvaluelist}
\item\relax
\flmRefsHyperref[eczindexfamilyrel]{code:binary_linear}{Linear binary code}\item\relax
\flmRefsHyperref[eczindexfamilyrel]{code:q-ary_cyclic}{Cyclic linear \(q\)-ary code}\end{eczvaluelist}
\codefieldsection{Children}
\begin{eczvaluelist}
\item\relax
\flmRefsHyperref[eczindexfamilyrel]{code:bch}{Binary BCH code}\item\relax
\flmRefsHyperref[eczindexfamilyrel]{code:binary_duadic}{Binary duadic code}\item\relax
\flmRefsHyperref[eczindexfamilyrel]{code:gold}{\([2^r-1, 2r ]\) Gold code}\item\relax
\flmRefsHyperref[eczindexfamilyrel]{code:kasami}{\([2^{2r}-1, 3r, 2^{2r-1} - 2^{r-1} ]\) Kasami code}\item\relax
\flmRefsHyperref[eczindexfamilyrel]{code:melas}{\([2^m -1, 2^m - 1 - 2m, 5]\) Melas code}\item\relax
\flmRefsHyperref[eczindexfamilyrel]{code:zetterberg}{Zetterberg code}\item\relax
\flmRefsHyperref[eczindexfamilyrel]{code:crc}{Cyclic redundancy check (CRC) code}\item\relax
\flmRefsHyperref[eczindexfamilyrel]{code:repetition}{Repetition code} --- The repetition code is cyclic with generator polynomial \(1+x+\cdots+x^{n-1}\).
\item\relax
\flmRefsHyperref[eczindexfamilyrel]{code:simplex}{\([2^m-1,m,2^{m-1}]\) simplex code} --- Simplex codes can be realized as maximal-length feedback-shift-register codes, and are therefore cyclic \NoCaseChange{\protect\cite[{pgs. 89 and 216}]{cite41}}.
\item\relax
\flmRefsHyperref[eczindexfamilyrel]{code:difference_set}{Difference-set cyclic (DSC) code}\end{eczvaluelist}
\codefieldsection{Cousins}
\begin{eczvaluelist}
\item\relax
\flmRefsHyperref[eczindexfamilyrel]{code:binary_ltc}{Binary linear LTC} --- Cyclic linear codes cannot be \(c^3\)-LTCs \NoCaseChange{\protect\cite{cite1272}}. Codeword symmetries are in general an obstruction to achieving such LTCs \NoCaseChange{\protect\cite{cite1273}}.
\item\relax
\flmRefsHyperref[eczindexfamilyrel]{code:one_hot}{One-hot code} --- The one-hot code is a cyclic non-linear binary code.
\item\relax
\flmRefsHyperref[eczindexfamilyrel]{code:reed_muller}{Reed-Muller (RM) code} --- Punctured RM codes, i.e., RM codes with nonzero evaluation points, are cyclic \NoCaseChange{\protect\cite{cite1312,cite1313}\protect\cite[{Ch. 13, Thm. 11}]{cite41}\protect\cite[{Sec. 2.8}]{cite68}\protect\cite[{pg. 52}]{cite1314}}, making RM codes extended cyclic codes.
\item\relax
\flmRefsHyperref[eczindexfamilyrel]{code:majorana_stab}{Majorana stabilizer code} --- Cyclic binary linear codes can be used to construct translation-invariant Majorana stabilizer codes, provided that they are also self-orthogonal \NoCaseChange{\protect\cite{cite566}}.
\item\relax
\flmRefsHyperref[eczindexfamilyrel]{code:css-t}{CSS-T code} --- Binary cyclic and extended cyclic codes can be used to construct CSS-T codes via the CSS construction \NoCaseChange{\protect\cite{cite1315}}.
\item\relax
\flmRefsHyperref[eczindexfamilyrel]{code:multisector_hypergraph}{Higher-dimensional homological product code} --- Higher-dimensional homological-product codes can be constructed out of CSS codes that in turn stem from cyclic codes \NoCaseChange{\protect\cite[{Sec. 4.2}]{cite835}}.
\item\relax
\flmRefsHyperref[eczindexfamilyrel]{code:quantum_synchronizable}{Quantum synchronizable code} --- The original construction of quantum synchronizable codes is based on pairs of binary cyclic codes satisfying \(C^\perp \subseteq C \subset D\) \NoCaseChange{\protect\cite{cite1246}}.
\item\relax
\flmRefsHyperref[eczindexfamilyrel]{code:rotated_surface}{Rotated surface code} --- Periodic checkerboard or rotated-toric codes on the same lattice can be obtained from hypergraph products of two cyclic linear binary codes with palindromic check polynomials \NoCaseChange{\protect\cite[{Sec. IV.D}]{cite1316}}.
\end{eczvaluelist}
\eczhbkcontributors{ Mazin Karjikar, \eczhuVVA }
\endeczcode

\eczcode{crc}{Cyclic redundancy check (CRC) code}{~\NoCaseChange{\protect\cite{cite1317,cite993,cite1241}}}
\codefieldsection{Alternative Names}
\begin{eczvaluelist}
\item\relax Frame check sequence (FCS)
\end{eczvaluelist}
\eczhIndexCodeAliasName{crc}{Frame check sequence (FCS)}
\codefieldsection{Description}
A generalization of the single parity-check code in which the generalization of the parity bit is the remainder of the data string (mapped into a polynomial via the \flmRefsCref{ref67}) divided by some generator polynomial.
A notable family of codes is referred to as \textit{CRC-}(\(m-1\)), where \(m\) is the length of the generator polynomial.

\codefieldsection{Protection}
Detects any burst error up to the length of the generator polynomial.
\codefieldsection{Decoding}
\begin{eczvaluelist}
\item\relax GRAND \NoCaseChange{\protect\cite{cite1318}}.
\end{eczvaluelist}
\codefieldsection{Realizations}
\begin{eczvaluelist}
\item\relax CRC-16 and CRC-32 are used in data transmission, e.g., IEEE 802.16e, IEEE 802.3 \NoCaseChange{\protect\cite{cite249}} and TCP/IP communication \NoCaseChange{\protect\cite[{Sec. 2.3.3}]{cite250}}.
\end{eczvaluelist}
\codefieldsection{Notes}
\begin{eczvaluelist}
\item\relax See Ref. \NoCaseChange{\protect\cite{cite1319}} and book \NoCaseChange{\protect\cite{cite1320}} for introductions to CRC codes.
\item\relax See Refs. \NoCaseChange{\protect\cite{cite1321,cite1322}} for exhaustive lists of CRC polynomials, as well as the \flmHref{https://users.ece.cmu.edu/~koopman/crc/crc32.html}{CRC Polynomial Zoo website} by Philip Koopman.
\end{eczvaluelist}
\codefieldsection{Parents}
\begin{eczvaluelist}
\item\relax
\flmRefsHyperref[eczindexfamilyrel]{code:binary_cyclic}{Cyclic linear binary code}\item\relax
\flmRefsHyperref[eczindexfamilyrel]{code:checksum}{Checksum code}\end{eczvaluelist}
\codefieldsection{Child}
\begin{eczvaluelist}
\item\relax
\flmRefsHyperref[eczindexfamilyrel]{code:parity_check}{\([n,n-1,2]\) Single parity-check (SPC) code} --- A CRC using the divisor 11 is a single parity-check code \NoCaseChange{\protect\cite[{Sec. 2.3.3}]{cite250}}.
\end{eczvaluelist}
\codefieldsection{Cousin}
\begin{eczvaluelist}
\item\relax
\flmRefsHyperref[eczindexfamilyrel]{code:polar}{Polar code} --- CRC codes concatenated with polar codes yield improved performance of the SCL polar-code decoder \NoCaseChange{\protect\cite{cite1323,cite1324,cite1325}}.
\end{eczvaluelist}
\eczhbkcontributors{ Gage Erwin, \eczhuVVA }
\endeczcode

\eczcode{delsarte_goethals}{Delsarte-Goethals (DG) code}{~\NoCaseChange{\protect\cite{cite1326}}}
\codefieldsection{Description}
Member of a family of \((2^{2t+2},2^{(2t+1)(t-r+2)+2t+3},2^{2t+1}-2^{2t+1-r})\) binary nonlinear codes for parameters \(r \geq 1\) and \(m = 2t+2 \geq 4\), denoted by DG\((m,r)\), that generalizes the Kerdock code.

The code DG\((m,r)\) is a nonlinear subcode of the second-order Reed-Muller code RM\((2,m)\), and equals RM\((2,m)\) at \(r=1\) \NoCaseChange{\protect\cite[{pg. 461}]{cite41}}.
The code is the union of certain cosets of RM\((1,m)\) in RM\((2,m)\) that are specified by bilinear forms \NoCaseChange{\protect\cite{cite1326}}.
The code DG\((m,r+1)\) is a union of disjoint translates of DG\((m,r)\).

While DG\((m,r)\) is not generally linear, it is the Gray map image of a certain extended cyclic linear code over \(\mathbb{Z_4}\) \NoCaseChange{\protect\cite{cite158}}.
These codes are distance invariant \NoCaseChange{\protect\cite{cite158}}, so the distance and weight distributions are the same.

Their automorphism groups are determined in Ref. \NoCaseChange{\protect\cite{cite1327}}.

\codefieldsection{Decoding}
\begin{eczvaluelist}
\item\relax Since the equivalent \(\mathbb{Z_4}\) codes are extended cyclic codes, efficient encoding and decoding is possible \NoCaseChange{\protect\cite{cite158}}.
\end{eczvaluelist}
\codefieldsection{Realizations}
\begin{eczvaluelist}
\item\relax Space-time signaling \NoCaseChange{\protect\cite{cite251}}.
\item\relax Compressed neighbor discovery in a network \NoCaseChange{\protect\cite{cite252}}.
\end{eczvaluelist}
\codefieldsection{Parent}
\begin{eczvaluelist}
\item\relax
\flmRefsHyperref[eczindexfamilyrel]{code:bits_into_bits}{Binary code}\end{eczvaluelist}
\codefieldsection{Child}
\begin{eczvaluelist}
\item\relax
\flmRefsHyperref[eczindexfamilyrel]{code:kerdock}{Kerdock code} --- A Kerdock code of length \(2^m\) is equivalent to DG\((m,m/2)\) and is a subcode of DG\((m,r)\) \NoCaseChange{\protect\cite[{pg. 461}]{cite41}}.
\end{eczvaluelist}
\codefieldsection{Cousins}
\begin{eczvaluelist}
\item\relax
\flmRefsHyperref[eczindexfamilyrel]{code:quaternary_over_z4}{Linear code over \(\mathbb{Z}_4\)} --- DG codes can be seen, via the \flmTerm{term}{ref81}{}{Gray map}, as extended linear cyclic codes over \(\mathbb{Z}_4\) \NoCaseChange{\protect\cite{cite158}}.
\item\relax
\flmRefsHyperref[eczindexfamilyrel]{code:gray}{Gray code} --- DG codes can be seen, via the \flmTerm{term}{ref81}{}{Gray map}, as extended linear cyclic codes over \(\mathbb{Z}_4\) \NoCaseChange{\protect\cite{cite158}}.
\item\relax
\flmRefsHyperref[eczindexfamilyrel]{code:reed_muller}{Reed-Muller (RM) code} --- The code DG\((m,r)\) is a subcode of the second-order Reed-Muller code RM\((2,m)\), and equals RM\((2,m)\) at \(r=1\) \NoCaseChange{\protect\cite[{pg. 461}]{cite41}}. The code is the union of certain cosets of the first-order RM\((1,m)\) code in RM\((2,m)\) that are specified by bilinear forms \NoCaseChange{\protect\cite{cite1326}}.
\item\relax
\flmRefsHyperref[eczindexfamilyrel]{code:goethals}{Goethals code} --- Goethals codes for a given \(m\) are duals of DG\((m,1/2(m-2\rrparenthesis \) codes in that their distance distribution is equal to the \flmRefsHyperref{ref113}{MacWilliams transform} of the distance distribution of the DG codes \NoCaseChange{\protect\cite{cite1328}\protect\cite[{pg. 476}]{cite41}}.
However, the two codes are images of a pair of mutually dual linear codes over \(\mathbb{Z}_4\) under the \flmTerm{term}{ref81}{}{Gray map}  \NoCaseChange{\protect\cite{cite158}}.

\item\relax
\flmRefsHyperref[eczindexfamilyrel]{code:hergert}{Hergert code} --- Hergert codes are duals of DG codes in that their distance distribution is equal to the \flmRefsHyperref{ref113}{MacWilliams transform} of the distance distribution of DG codes \NoCaseChange{\protect\cite{cite1328}}. However, the two codes are images of a pair of mutually dual linear codes over \(\mathbb{Z}_4\) under the \flmTerm{term}{ref81}{}{Gray map} \NoCaseChange{\protect\cite{cite158,cite123}}.
\end{eczvaluelist}
\eczhbkcontributors{ Madhura Pankaja, \eczhuVVA }
\endeczcode

\eczcode{difference_set}{Difference-set cyclic (DSC) code}{~\NoCaseChange{\protect\cite{cite1329}}}
\codefieldsection{Description}
Cyclic LDPC code constructed deterministically from a difference set.
Certain DSC codes satisfy more redundant constraints than Gallager codes and thus can outperform them \NoCaseChange{\protect\cite{cite72}}.

\codefieldsection{Notes}
\begin{eczvaluelist}
\item\relax See book \NoCaseChange{\protect\cite[{Ch. 6}]{cite1176}} for a general theory of linear codes made from difference sets.
\end{eczvaluelist}
\codefieldsection{Parents}
\begin{eczvaluelist}
\item\relax
\flmRefsHyperref[eczindexfamilyrel]{code:qc_ldpc}{Quasi-cyclic LDPC (QC-LDPC) code}\item\relax
\flmRefsHyperref[eczindexfamilyrel]{code:algebraic_ldpc}{Algebraic LDPC code}\item\relax
\flmRefsHyperref[eczindexfamilyrel]{code:binary_cyclic}{Cyclic linear binary code}\end{eczvaluelist}
\codefieldsection{Child}
\begin{eczvaluelist}
\item\relax
\flmRefsHyperref[eczindexfamilyrel]{code:simplex734}{\([7,3,4]\) simplex code} --- The \([7,3,4]\) simplex code is the smallest difference-set cyclic code, arising from the lines of the projective plane \(PG(2,2)\) \NoCaseChange{\protect\cite[{pg. 397}]{cite41}}.
\end{eczvaluelist}
\codefieldsection{Cousins}
\begin{eczvaluelist}
\item\relax
\flmRefsHyperref[eczindexfamilyrel]{code:hadamard}{\([2^m,m,2^{m-1}]\) Hadamard code} --- \textit{Hadamard difference sets} are difference sets constructed from Hadamard matrices \NoCaseChange{\protect\cite[{Ch. 6}]{cite1176}}.
\item\relax
\flmRefsHyperref[eczindexfamilyrel]{code:hyperoval}{Hyperoval code} --- Hyperoval difference sets yield DSC codes \NoCaseChange{\protect\cite{cite1330}\protect\cite[{Ch. 6}]{cite1176}}.
\item\relax
\flmRefsHyperref[eczindexfamilyrel]{code:generalized_reed_muller}{Generalized RM (GRM) code} --- DSC codes can be \flmRefsHyperref{ref33}{subfield} subcodes of GRM codes, and vice versa \NoCaseChange{\protect\cite[{Thm. 6.14}]{cite1176}}.
\item\relax
\flmRefsHyperref[eczindexfamilyrel]{code:incidence_matrix}{Incidence-matrix projective code} --- Planar difference-set cyclic codes arise from the incidence vectors of the lines of a projective plane \(PG(2,p^s)\) \NoCaseChange{\protect\cite[{pgs. 397-398}]{cite41}}.
\end{eczvaluelist}
\eczhbkcontributors{ \eczhuVVA }
\endeczcode

\eczcode{dinur}{Dinur code}{~\NoCaseChange{\protect\cite{cite1331}}}
\codefieldsection{Description}
Member of an infinite family of locally testable \([n,n/\text{polylog}(n),d]\) codes with vanishing rate. Code construction relies on tensor-product codes \NoCaseChange{\protect\cite{cite73}}.

\codefieldsection{Parent}
\begin{eczvaluelist}
\item\relax
\flmRefsHyperref[eczindexfamilyrel]{code:binary_ltc}{Binary linear LTC}\end{eczvaluelist}
\eczhbkcontributors{ \eczhuVVA }
\endeczcode

\eczcode{expander}{Expander code}{~\NoCaseChange{\protect\cite{cite1332}}}
\codefieldsection{Alternative Names}
\begin{eczvaluelist}
\item\relax Sipser-Spielman code
\end{eczvaluelist}
\eczhIndexCodeAliasName{expander}{Sipser-Spielman code}
\codefieldsection{Description}
LDPC code whose parity-check matrix is derived from the adjacency matrix of a bipartite expander graph \NoCaseChange{\protect\cite{cite74}} such as a Ramanujan graph or a Cayley graph of a projective special linear group over a finite field \NoCaseChange{\protect\cite{cite75,cite76}}.
Expander codes admit efficient encoding and decoding algorithms and yield an explicit (i.e., non-random) asymptotically good LDPC code family.

The rate and distance of the expander code depend on specific parameters of the corresponding graph.
A (\(n, m, D, \gamma, \alpha\)) bipartite expander graph is defined as a \(D\)-left-regular graph with \(n\) left nodes and \(m\) right nodes such that for any subset of left nodes \(S\) of size at most \(\gamma n\), the neighborhood \(N(S)\) has size at least \(\alpha|S|\). Given a (\(n, m, D, \gamma, (1-\epsilon)D\)) expander graph, the corresponding expander code has rate at least \(1 - m/n\) and distance at least \(2(1-\epsilon)\gamma n\) for any \(\epsilon < 1/2\).
Explicit constructions for expander graphs \NoCaseChange{\protect\cite{cite74}} with any ratio \(n/m\) are known where \(D = \text{polylog}(n/m)\), \(\gamma = \Omega(1/D)\) and arbitrary \(\epsilon\) \NoCaseChange{\protect\cite{cite185}}.

\codefieldsection{Protection}
There exist minimum distance bounds \NoCaseChange{\protect\cite{cite1332,cite1333}} as well as bounds on decoding performance \NoCaseChange{\protect\cite{cite1334,cite1335,cite1336}} in terms of features of the expander graph.

\codefieldsection{Rate}
The rate is at least \(1 - m/n\), where \(n\) is the number of left nodes and \(m\) is the number of right nodes in the bipartite expander graph. Expander codes yield an explicit (i.e., non-random) asymptotically good LDPC code family \NoCaseChange{\protect\cite{cite1332}}.
\codefieldsection{Encoding}
\begin{eczvaluelist}
\item\relax Multiplication by generator matrix with runtime \(O(n^2)\)
\end{eczvaluelist}
\codefieldsection{Decoding}
\begin{eczvaluelist}
\item\relax Decoding can be done in \(O(n)\) runtime using a greedy \textit{flip decoder} \NoCaseChange{\protect\cite{cite1332}} (see also \NoCaseChange{\protect\cite{cite1337}}). The algorithm consists of flipping a bit of the received word if it will result in a greater number of satisfied parity checks. This is repeated until a codeword is reached.
\item\relax 'Find erasures and Decode' a.k.a. Viderman's algorithm correcting \flmRefsHyperref{ref65}{order} \(\Omega(n)\) errors in \flmRefsHyperref{ref65}{order} \(O(n)\) time \NoCaseChange{\protect\cite{cite1338}}.
\end{eczvaluelist}
\codefieldsection{Fault Tolerance}
\begin{eczvaluelist}
\item\relax The flip decoding algorithm is fault tolerant against parity-check errors \NoCaseChange{\protect\cite{cite1339}}; see also \NoCaseChange{\protect\cite{cite1340}}.
\end{eczvaluelist}
\codefieldsection{Parent}
\begin{eczvaluelist}
\item\relax
\flmRefsHyperref[eczindexfamilyrel]{code:regular_ldpc}{Regular LDPC code} --- Expander codes yield an explicit (i.e., non-random) asymptotically good LDPC code family \NoCaseChange{\protect\cite{cite1332}}.
\end{eczvaluelist}
\codefieldsection{Cousins}
\begin{eczvaluelist}
\item\relax
\flmRefsHyperref[eczindexfamilyrel]{code:ldc}{Locally decodable code (LDC)} --- Expander codes are locally decodable provided that the inner code satisfies certain properties; there exist code families with rate approaching one \NoCaseChange{\protect\cite{cite1080}}.
\item\relax
\flmRefsHyperref[eczindexfamilyrel]{code:lr-cayley-complex}{Left-right Cayley complex code} --- Left-right Cayley complex codes can be viewed as Tanner-like codes on expander graphs \NoCaseChange{\protect\cite{cite74}}, but with bits defined on squares and constraints on edges (as opposed to edges and vertices, respectively, for expander codes). Expander codes are also typically not locally testable \NoCaseChange{\protect\cite{cite1341}}.
\item\relax
\flmRefsHyperref[eczindexfamilyrel]{code:q-ary_ldpc}{\(q\)-ary LDPC code} --- Asymptotically good non-binary expander codes can be constructed \NoCaseChange{\protect\cite{cite1343}\protect\cite[{Thm. 7.16}]{cite1342}} by generalizing the originally binary expander constructions \NoCaseChange{\protect\cite{cite1332,cite1344}}.
\item\relax
\flmRefsHyperref[eczindexfamilyrel]{code:self_correct}{Self-correcting quantum code} --- Constant-rate random (quantum) expander codes are self-correcting (quantum) memories, but have no thermodynamic phase transitions \NoCaseChange{\protect\cite{cite849}}.
\item\relax
\flmRefsHyperref[eczindexfamilyrel]{code:quantum_expander}{Quantum expander code} --- Quantum expander codes are quantum analogues of expander codes.
\item\relax
\flmRefsHyperref[eczindexfamilyrel]{code:galois_expander}{Galois-qudit expander code} --- The explicit expander-code construction of \NoCaseChange{\protect\cite{cite689}} yields \(\llbracket N,K\geq N^{1-\epsilon},D\geq N^{1/r}/\operatorname{poly}(\log N)\rrbracket _q\) QLDPC Galois-qudit quantum expander codes with transversal \(C^{r-1} Z\) gates. Balanced products of the same expander-code complexes also yield \([n,k\geq n^{1-\epsilon},d\geq n/\operatorname{poly}(\log n)]_q\) LTCs exhibiting the multiplication property.
\item\relax
\flmRefsHyperref[eczindexfamilyrel]{code:expander_lifted_product}{Expander LP code} --- Expander LP codes are lifted products of expander codes with different local codes \NoCaseChange{\protect\cite{cite184}}.
\end{eczvaluelist}
\eczhbkcontributors{ Jon Nelson, \eczhuVVA }
\endeczcode

\eczcode{extended_ira}{Extended IRA (eIRA) code}{~\NoCaseChange{\protect\cite{cite1345,cite1346,cite1347}}}
\codefieldsection{Description}
A generalization of the IRA code in which the outer LDGM code is replaced by a random sparse matrix containing no weight-two columns.

\codefieldsection{Parent}
\begin{eczvaluelist}
\item\relax
\flmRefsHyperref[eczindexfamilyrel]{code:irregular_ldpc}{Irregular LDPC code}\end{eczvaluelist}
\eczhbkcontributors{ \eczhuVVA }
\endeczcode

\eczcode{fibonacci_model}{Fibonacci code}{~\NoCaseChange{\protect\cite{cite1348}}}
\codefieldsection{Description}
Quantum-inspired binary linear code defined on an \(L\times L/2\) lattice with one bit on each site, where \(L=2^N\) for an integer \(N\geq 2\). The codewords are defined to satisfy the condition that, for each lattice site \((x,y)\), the bits on \((x,y)\), \((x+1,y)\), \((x-1,y)\) and \((x,y+1)\) (where the lattice is taken to be periodic in both directions) contain an even number of \(1\)'s.

\codefieldsection{Protection}
Protects against small weight errors and string-like errors. The code distance is more than \(L\), but the exact value is unknown.

\codefieldsection{Encoding}
\begin{eczvaluelist}
\item\relax The codewords can be generated using a cellular automaton of length \(L\) (periodic). The \(2^L\) possible initial states correspond to the \(2^L\) codewords. For each generation, the state of each cell is the XOR sum of that cell and its two neighbors in the previous generation. After \(L/2-1\) generations, the entire history generated by the automaton corresponds to a codeword, where the initial state is the first row of the lattice, the first generation is the second row, etc. In polynomial language, the update rule is \(f(x)=1+x+x^2\) over \(\mathbb{F}_2\), so the resulting fractal has dimension \(\log_2(1+\sqrt{5})\) \NoCaseChange{\protect\cite{cite1348}}.
\end{eczvaluelist}
\codefieldsection{Decoding}
\begin{eczvaluelist}
\item\relax An efficient algorithm based on minimum-weight perfect matching \NoCaseChange{\protect\cite{cite1349}}, which can correct high-weight errors that span rows and columns of the 2D lattice, with failure rate decaying super-exponentially with \(L\).
\end{eczvaluelist}
\codefieldsection{Parents}
\begin{eczvaluelist}
\item\relax
\flmRefsHyperref[eczindexfamilyrel]{code:binary_linear}{Linear binary code}\item\relax
\flmRefsHyperref[eczindexfamilyrel]{code:quantum_inspired}{Quantum-inspired classical block code}\end{eczvaluelist}
\codefieldsection{Cousins}
\begin{eczvaluelist}
\item\relax
\flmRefsHyperref[eczindexfamilyrel]{code:haah_cubic}{Haah cubic code (CC)} --- The Fibonacci code is designed to mimic the fractal properties of (quantum) Haah cubic code so that studying the former can help us toward the development of an efficient algorithm for the latter \NoCaseChange{\protect\cite{cite1349}}.
\item\relax
\flmRefsHyperref[eczindexfamilyrel]{code:fibonacci_fractal_liquid}{Fibonacci fractal spin-liquid code} --- The Fibonacci fractal spin-liquid code is a hypergraph product of the repetition code and the Fibonacci code \NoCaseChange{\protect\cite{cite1348}}, and can be formulated directly as a BP code \NoCaseChange{\protect\cite{cite1350}}.
\end{eczvaluelist}
\eczhbkcontributors{ Yi-Ting (Rick) Tu, \eczhuVVA }
\endeczcode

\eczcode{pg_ldpc}{Finite-geometry LDPC (FG-LDPC) code}{~\NoCaseChange{\protect\cite{cite1351}}}
\codefieldsection{Description}
LDPC code whose parity-check matrix is the incidence matrix of points and hyperplanes in either a Euclidean or a projective geometry.
Such codes are called \textit{Euclidean-geometry LDPC (EG-LDPC)} and \textit{projective-geometry LDPC (PG-LDPC)}, respectively.
Such constructions have been generalized to incidence matrices of hyperplanes of different dimensions \NoCaseChange{\protect\cite{cite77}}.

\codefieldsection{Parents}
\begin{eczvaluelist}
\item\relax
\flmRefsHyperref[eczindexfamilyrel]{code:regular_ldpc}{Regular LDPC code}\item\relax
\flmRefsHyperref[eczindexfamilyrel]{code:algebraic_ldpc}{Algebraic LDPC code}\end{eczvaluelist}
\codefieldsection{Cousins}
\begin{eczvaluelist}
\item\relax
\flmRefsHyperref[eczindexfamilyrel]{code:qc_ldpc}{Quasi-cyclic LDPC (QC-LDPC) code} --- Many FG-LDPC codes can be put into quasi-cyclic form \NoCaseChange{\protect\cite{cite1351,cite77}\protect\cite[{pg. 286}]{cite1352}}.
\item\relax
\flmRefsHyperref[eczindexfamilyrel]{code:incidence_matrix}{Incidence-matrix projective code} --- The parity-check matrix of a PG-LDPC code is the incidence matrix of points and hyperplanes in a projective space.
\item\relax
\flmRefsHyperref[eczindexfamilyrel]{code:generalized_reed_muller}{Generalized RM (GRM) code} --- Some EG-LDPC codes are duals of \flmRefsHyperref{ref33}{subfield} subcodes of GRM codes \NoCaseChange{\protect\cite[{pg. 448}]{cite1353}}.
\item\relax
\flmRefsHyperref[eczindexfamilyrel]{code:asymmetric_qecc}{Asymmetric quantum code (AQC)} --- FG-LDPC codes can be used to construct asymmetric CSS codes \NoCaseChange{\protect\cite{cite1355}\protect\cite[{Lemma 4.1}]{cite1354}}.
\item\relax
\flmRefsHyperref[eczindexfamilyrel]{code:ea_pg_qldpc}{EA FG-QLDPC code} --- EA FG-QLDPC codes are entanglement-assisted quantum analogues of finite-geometry LDPC codes.
\item\relax
\flmRefsHyperref[eczindexfamilyrel]{code:pg_qldpc}{Finite-geometry (FG) qubit QLDPC code} --- Quantum versions of PG-LDPC and EG-LDPC codes can be constructed via the CSS construction \NoCaseChange{\protect\cite{cite1356,cite834}}.
\item\relax
\flmRefsHyperref[eczindexfamilyrel]{code:abelian_lifted_product}{Abelian LP code} --- FG-LDPC codes can be used to construct Abelian LP codes \NoCaseChange{\protect\cite{cite1357}}.
\end{eczvaluelist}
\eczhbkcontributors{ \eczhuVVA }
\endeczcode

\eczcode{fountain}{Fountain code}{~\NoCaseChange{\protect\cite{cite98}}}
\codefieldsection{Description}
Code based on the idea of generating an endless stream of custom encoded packets for the receiver. The code is designed so that the receiver can recover the original transmission of size \(Kl\) bits after receiving at least \(K\) packets each of \(l\) bits.

The simplest example of a fountain code is the random linear fountain code. Take some message of size \(Kl\) and split into \(K\) packets, \(p_0, p_1, ..., p_K\). For each packet \(\hat{p}_n\) to be transmitted do the following: Generate \(K\) random bits \(G_{nk}\) and let \(\hat{p}_n\) be the bitwise sum of the source packets when \(G_{nk}\) is 1,
\flmMathEnvironment{align}{}{
\hat{p}_n = \sum_{k=1}^K p_k G_{kn}~.
}
Error correction can then be applied to each packet.

\codefieldsection{Protection}
Designed to protect against erasures during broadcasting of information by a sender to multiple receivers.
\codefieldsection{Rate}
Random linear fountain codes approach the Shannon limit as the file size \(K\) increases. However, they do not have a fixed encoding rate.
\codefieldsection{Decoding}
\begin{eczvaluelist}
\item\relax Invert the fragment generator matrix resulting from the continuous encoding process. If exactly \(K\) packets are received, then the probability of decoding correctly is \(0.289\). Extra packets increase this probability exponentially. The decoding runtime is dominated by the matrix inversion step, which takes \flmRefsHyperref{ref65}{order} \(O(n^3)\) time.
\end{eczvaluelist}
\codefieldsection{Realizations}
\begin{eczvaluelist}
\item\relax Designed for servers sending data to many recipients, such as during broadcasting or file distribution \NoCaseChange{\protect\cite{cite257,cite258}}.
\item\relax DNA storage \NoCaseChange{\protect\cite{cite259}}.
\end{eczvaluelist}
\codefieldsection{Notes}
\begin{eczvaluelist}
\item\relax Review on fountain codes can be found in Refs. \NoCaseChange{\protect\cite{cite1358,cite1359,cite946}}.
\end{eczvaluelist}
\codefieldsection{Parent}
\begin{eczvaluelist}
\item\relax
\flmRefsHyperref[eczindexfamilyrel]{code:ldgm}{Low-density generator-matrix (LDGM) code}\end{eczvaluelist}
\codefieldsection{Child}
\begin{eczvaluelist}
\item\relax
\flmRefsHyperref[eczindexfamilyrel]{code:raptor}{Raptor (RAPid TORnado) code}\end{eczvaluelist}
\codefieldsection{Cousins}
\begin{eczvaluelist}
\item\relax
\flmRefsHyperref[eczindexfamilyrel]{code:random}{Random code} --- Fountain codes are typically generated from random sparse encoding ensembles.
\item\relax
\flmRefsHyperref[eczindexfamilyrel]{code:distributed_storage}{Distributed-storage code} --- There are proposals \NoCaseChange{\protect\cite{cite995,cite996}} adapting fountain codes to distributed storage systems.
\item\relax
\flmRefsHyperref[eczindexfamilyrel]{code:dna}{DNA storage code} --- Fountain codes have been used for DNA storage \NoCaseChange{\protect\cite{cite259}}.
\item\relax
\flmRefsHyperref[eczindexfamilyrel]{code:tornado}{Tornado code} --- Tornado codes, the precursor to fountain codes, are much slower to encode and decode in the low-rate regime applicable to scalable data transmission \NoCaseChange{\protect\cite{cite1359,cite257}}.
\end{eczvaluelist}
\eczhbkcontributors{ Noah Berthusen, \eczhuVVA }
\endeczcode

\eczcode{gallager}{Gallager (GL) code}{~\NoCaseChange{\protect\cite{cite1360,cite1361}}}
\codefieldsection{Description}
The first LDPC code.
The rows of the parity-check matrix of this regular code are divided into equal subsets, and columns in the first subset are randomly permuted to yield the corresponding rows in subsequent subsets.

For example, the parity-check matrix
\flmMathEnvironment{align}{}{
\begin{pmatrix}
1 & 1 & 0 & 0\\
0 & 0 & 1 & 1\\
1 & 0 & 1 & 0\\
0 & 1 & 0 & 1
\end{pmatrix}
}
contains two subsets, each consisting of two rows, and the last two rows are obtained from the first two by exchanging the second and third columns.

\codefieldsection{Protection}
With high probability, random GL codes have minimum distance that grows linearly with block length \NoCaseChange{\protect\cite{cite1361}}. There exist GL codes that are able to correct errors of weight less than \(c n\) for some constant \(c\) in \flmRefsHyperref{ref65}{order} \(O(n\log n)\) decoding operations \NoCaseChange{\protect\cite{cite1362}}.
\codefieldsection{Rate}
GL codes nearly achieve Shannon capacity against binary-input additive Gaussian white noise using iterative decoding \NoCaseChange{\protect\cite{cite1363,cite1364}}. GL codes can outperform RS codes at short block length \NoCaseChange{\protect\cite{cite72}}.
\codefieldsection{Parents}
\begin{eczvaluelist}
\item\relax
\flmRefsHyperref[eczindexfamilyrel]{code:regular_ldpc}{Regular LDPC code} --- GL codes are the first LDPC codes.
\item\relax
\flmRefsHyperref[eczindexfamilyrel]{code:generalized_gallager}{Generalized Gallager code}\end{eczvaluelist}
\codefieldsection{Cousin}
\begin{eczvaluelist}
\item\relax
\flmRefsHyperref[eczindexfamilyrel]{code:combinatorial_design}{Combinatorial design} --- Some Steiner systems can be used to construct Gallager codes \NoCaseChange{\protect\cite{cite72}}.
\end{eczvaluelist}
\eczhbkcontributors{ \eczhuVVA }
\endeczcode

\eczcode{gauss_law}{Gauss' law code}{~\NoCaseChange{\protect\cite{cite79,cite78}}}
\codefieldsection{Description}
An \([m+Dm,Dm,3]\) linear binary code for \(m\geq 3^D\), defined by the Gauss' law constraint of a \(D\)-dimensional fermionic \(\mathbb{Z}_2\) gauge theory \NoCaseChange{\protect\cite[{Thm. 1}]{cite78}}.
The code can be rephrased as a distance-one stabilizer code whose stabilizers consist of gauge-group elements.
It can be concatenated to form a stabilizer code for fault-tolerant quantum simulation of the underlying gauge theory \NoCaseChange{\protect\cite{cite79,cite78}}.

\codefieldsection{Parents}
\begin{eczvaluelist}
\item\relax
\flmRefsHyperref[eczindexfamilyrel]{code:binary_linear}{Linear binary code}\item\relax
\flmRefsHyperref[eczindexfamilyrel]{code:quantum_inspired}{Quantum-inspired classical block code}\end{eczvaluelist}
\codefieldsection{Cousins}
\begin{eczvaluelist}
\item\relax
\flmRefsHyperref[eczindexfamilyrel]{code:topological_abelian}{Abelian topological code} --- Gauge-group elements of a \(D\)-dimensional fermionic \(\mathbb{Z}_2\) gauge theory can arise from a single-error-correcting linear binary code \NoCaseChange{\protect\cite[{Thm. 1}]{cite78}}. There is a general correspondence between stabilizer codes and gauge theory, with the stabilizer group playing the role of the gauge group \NoCaseChange{\protect\cite{cite1365}}, and with the Gauss' law code being a specific example.
\item\relax
\flmRefsHyperref[eczindexfamilyrel]{code:fermions_into_qubits}{Fermion-into-qubit code} --- Gauge-group elements of a \(D\)-dimensional fermionic \(\mathbb{Z}_2\) gauge theory can arise from a single-error-correcting linear binary code \NoCaseChange{\protect\cite[{Thm. 1}]{cite78}}.
\item\relax
\flmRefsHyperref[eczindexfamilyrel]{code:qubit_concatenated}{Concatenated qubit code} --- The Gauss' law code can be concatenated to form a stabilizer code for fault-tolerant quantum simulation of the underlying gauge theory \NoCaseChange{\protect\cite{cite79,cite78}}.
\item\relax
\flmRefsHyperref[eczindexfamilyrel]{code:iceberg}{\(\llbracket 2m,2m-2,2\rrbracket \) error-detecting code} --- The iceberg code can be used for robustly simulating \(SU(2)\) gauge theories \NoCaseChange{\protect\cite{cite1366}}.
\end{eczvaluelist}
\eczhbkcontributors{ \eczhuVVA }
\endeczcode

\eczcode{generalized_gallager}{Generalized Gallager code}{~\NoCaseChange{\protect\cite{cite1367}}}
\codefieldsection{Description}
A LDPC code that generalizes the Gallager codes using the Tanner construction.
While Gallager-code parity-check matrices consist of repetition-code submatrices that are randomly permuted, generalized Gallager-code matrices instead use more general binary linear component codes.

\codefieldsection{Parent}
\begin{eczvaluelist}
\item\relax
\flmRefsHyperref[eczindexfamilyrel]{code:regular_binary_tanner}{Regular binary Tanner code}\end{eczvaluelist}
\codefieldsection{Child}
\begin{eczvaluelist}
\item\relax
\flmRefsHyperref[eczindexfamilyrel]{code:gallager}{Gallager (GL) code}\end{eczvaluelist}
\eczhbkcontributors{ \eczhuVVA }
\endeczcode

\eczcode{goethals}{Goethals code}{~\NoCaseChange{\protect\cite{cite1368}}}
\codefieldsection{Description}
Member of a family of \((2^m,2^{2^m-3m+1},8)\) binary nonlinear codes for \(m \geq 6\) that generalizes the Preparata codes.
The code can be constructed as a disjoint union of cosets of a certain linear code \NoCaseChange{\protect\cite[{Ch. 15}]{cite41}}.

\codefieldsection{Rate}
The rate is \({2^m -3m +1}/2^m\), going to 1 as block length goes to infinity.
\codefieldsection{Parent}
\begin{eczvaluelist}
\item\relax
\flmRefsHyperref[eczindexfamilyrel]{code:hergert}{Hergert code} --- Goethals codes are equivalent to Hergert codes for \(r=3\) \NoCaseChange{\protect\cite[{Thm. 2}]{cite1326}}.
\end{eczvaluelist}
\codefieldsection{Cousins}
\begin{eczvaluelist}
\item\relax
\flmRefsHyperref[eczindexfamilyrel]{code:combinatorial_design}{Combinatorial design} --- Goethals codes form an infinite family of nonlinear binary codes supporting 3-designs \NoCaseChange{\protect\cite[{Table 5.1}]{cite135}}.
\item\relax
\flmRefsHyperref[eczindexfamilyrel]{code:delsarte_goethals}{Delsarte-Goethals (DG) code} --- Goethals codes for a given \(m\) are duals of DG\((m,1/2(m-2\rrparenthesis \) codes in that their distance distribution is equal to the \flmRefsHyperref{ref113}{MacWilliams transform} of the distance distribution of the DG codes \NoCaseChange{\protect\cite{cite1328}\protect\cite[{pg. 476}]{cite41}}.
However, the two codes are images of a pair of mutually dual linear codes over \(\mathbb{Z}_4\) under the \flmTerm{term}{ref81}{}{Gray map}  \NoCaseChange{\protect\cite{cite158}}.

\item\relax
\flmRefsHyperref[eczindexfamilyrel]{code:quantum_goethals_preparata}{\(\llparenthesis 2^m,2^{2^m−5m+1},8\rrparenthesis \) Goethals-Preparata code} --- The \(\llparenthesis 2^m,2^{2^m−5m+1},8\rrparenthesis \) Goethals-Preparata code is constructed using the classical Goethals and Preparata codes \NoCaseChange{\protect\cite{cite1369,cite1370}}. A construction using the \(\mathbb{Z}_4\) versions of the Goethals and Preparata codes and the \flmTerm{term}{ref81}{}{Gray map} yields qubit code families with similar parameters \NoCaseChange{\protect\cite{cite1371}}.
\end{eczvaluelist}
\eczhbkcontributors{ Madhura Pankaja, \eczhuVVA }
\endeczcode

\eczcode{gs-ltc}{Goldreich-Sudan code}{~\NoCaseChange{\protect\cite{cite1372}}}
\codefieldsection{Description}
Locally testable \([n,k,d]\) code with \(n = k^{1+O(1/u)}\) and distance of \flmRefsHyperref{ref65}{order} \(\Omega(n)\) for query complexity \(u\). The same work also presented a probabilistic construction of codes of size \(k^{1+o(1)}\).

\codefieldsection{Parent}
\begin{eczvaluelist}
\item\relax
\flmRefsHyperref[eczindexfamilyrel]{code:binary_ltc}{Binary linear LTC} --- Goldreich-Sudan codes resulted from what is often referred to as the first systematic study of LTCs.
\end{eczvaluelist}
\codefieldsection{Cousin}
\begin{eczvaluelist}
\item\relax
\flmRefsHyperref[eczindexfamilyrel]{code:random}{Random code} --- Randomness enters the probabilistic constructions associated with Goldreich-Sudan LTCs.
\end{eczvaluelist}
\eczhbkcontributors{ \eczhuVVA }
\endeczcode

\eczcode{graph}{Graph-adjacency code}{~\NoCaseChange{\protect\cite{cite82,cite1373}}}
\codefieldsection{Description}
Binary linear code whose generator matrix is that of the row space of the adjacency matrix of a strongly regular graph.
Given an adjacency matrix \(A\), the generator matrix is either \(G=A\) or \(G=(I|A)\), where \(I\) is the identity matrix.

Codes based on strongly regular graphs are sometimes optimal or nearly optimal for their length and size \NoCaseChange{\protect\cite{cite82}}.

\codefieldsection{Parent}
\begin{eczvaluelist}
\item\relax
\flmRefsHyperref[eczindexfamilyrel]{code:binary_linear}{Linear binary code}\end{eczvaluelist}
\codefieldsection{Children}
\begin{eczvaluelist}
\item\relax
\flmRefsHyperref[eczindexfamilyrel]{code:higman-sims_graph}{Higman-Sims graph-adjacency code}\item\relax
\flmRefsHyperref[eczindexfamilyrel]{code:hoffman-singleton_graph}{Hoffman-Singleton graph-adjacency code}\end{eczvaluelist}
\codefieldsection{Cousins}
\begin{eczvaluelist}
\item\relax
\flmRefsHyperref[eczindexfamilyrel]{code:homological_classical}{Cycle code} --- Graph-adjacency (cycle) codes' generator (parity-check) matrices are defined using adjacency (incidence) matrices of graphs.
\item\relax
\flmRefsHyperref[eczindexfamilyrel]{code:stabilizer_over_gf4}{Hermitian qubit code} --- Bounds on self-dual \(\llbracket n,0,d\rrbracket \) Hermitian codes based on graphs have been derived \NoCaseChange{\protect\cite{cite1373}}.
\end{eczvaluelist}
\eczhbkcontributors{ \eczhuVVA }
\endeczcode

\eczcode{gray}{Gray code}{~\NoCaseChange{\protect\cite{cite80,cite1374,cite1375}}}
\codefieldsection{Description}
The first Gray code \NoCaseChange{\protect\cite{cite80}}, now called the \textit{binary reflected Gray code}, is a trivial code that orders length-\(n\) binary strings such that nearest-neighbor strings differ by only one digit via what is known as the \flmTerm{term}{ref81}{}{Gray map}.

\begin{defterm}{Gray map}\label{ref81}
The Gray map converts a quaternary string over \(\mathbb{Z}_4\) into a binary string such that the Hamming distance of the binary representation is one between any two consecutive quaternary digits. 
For a single digit, the mapping is \(0\to 00\), \(1\to 01\), \(2\to 11\), and \(3\to 10\).
This differs from the usual binary expansion of the natural numbers, which maps \(0\to 00\), \(1\to 01\), \(2\to 10\), and \(3\to 11\).
A linear code \(C\) over \(\mathbb{Z}_4\) can be mapped, via the Gray map, to a distance invariant binary code \NoCaseChange{\protect\cite[{Thm. 2}]{cite158}}. 
The binary code is linear if and only if doubling the component-wise product of any two codewords in \(C\) yields another codeword in \(C\) \NoCaseChange{\protect\cite[{Sec. 6.3}]{cite1145}\protect\cite[{Thm. 12.2.3}]{cite126}}.
The map converts Lee to Hamming weight, and Lee distance to Hamming distance \NoCaseChange{\protect\cite[{Sec. 6.3}]{cite1145}\protect\cite[{Thm. 3.1 and Prop. 3.3}]{cite123}}.
\end{defterm}

Gray codes have been generalized such that nearest-neighbor strings differ by only one digit when the strings are arranged in higher-dimensional hypercubes \NoCaseChange{\protect\cite{cite1374}}.
Further generalizations called \textit{combinatorial Gray codes} \NoCaseChange{\protect\cite{cite1375}} refer to methods to organize combinatorial objects such that successive objects differ in some particular way.
Particular \(q\)-ary extensions \NoCaseChange{\protect\cite{cite1376}} of Gray codes may be useful in digital imaging and signal processing.

\codefieldsection{Encoding}
\begin{eczvaluelist}
\item\relax Efficient encoder for binary reflected Gray code \NoCaseChange{\protect\cite{cite1377}}.
\end{eczvaluelist}
\codefieldsection{Realizations}
\begin{eczvaluelist}
\item\relax Three-dimensional imaging \NoCaseChange{\protect\cite{cite274}}.
\item\relax Broadcasting and communication \NoCaseChange{\protect\cite{cite275}}.
\end{eczvaluelist}
\codefieldsection{Notes}
\begin{eczvaluelist}
\item\relax See Refs. \NoCaseChange{\protect\cite{cite1378,cite1379,cite1380}} reviews of various Gray codes.
\end{eczvaluelist}
\codefieldsection{Parent}
\begin{eczvaluelist}
\item\relax
\flmRefsHyperref[eczindexfamilyrel]{code:binary_linear}{Linear binary code} --- A linear code \(C\) over \(\mathbb{Z}_4\) can be mapped, via the \flmTerm{term}{ref81}{}{Gray map}, to a binary code. The binary code is linear if and only if doubling the component-wise product of any two codewords in \(C\) yields another codeword in \(C\) \NoCaseChange{\protect\cite[{Thm. 12.2.3}]{cite126}}.
\end{eczvaluelist}
\codefieldsection{Cousins}
\begin{eczvaluelist}
\item\relax
\flmRefsHyperref[eczindexfamilyrel]{code:qubits_into_qubits}{Qubit code} --- Gray codes are useful for optimizing qubit unitary circuits \NoCaseChange{\protect\cite{cite1381}} and for encoding qudits in multiple qubits \NoCaseChange{\protect\cite{cite506}}.
\item\relax
\flmRefsHyperref[eczindexfamilyrel]{code:quaternary_over_z4}{Linear code over \(\mathbb{Z}_4\)} --- A linear code \(C\) over \(\mathbb{Z}_4\) can be mapped, via the \flmTerm{term}{ref81}{}{Gray map}, to a binary code. The binary code is linear if and only if doubling the component-wise product of any two codewords in \(C\) yields another codeword in \(C\) \NoCaseChange{\protect\cite[{Thm. 12.2.3}]{cite126}}. More specifically, a linear quaternary code over \(\mathbb{Z}_4\) of length \(n\), type \(4^{k_1}2^{k_2}\), and minimum Lee weight \(d\) maps under the \flmTerm{term}{ref81}{}{Gray map} to a binary code of length \(2n\), cardinality \(2^{2k_1+k_2}\), and minimum Hamming weight \(d\) \NoCaseChange{\protect\cite[{Sec. 6.3}]{cite1145}}.
\item\relax
\flmRefsHyperref[eczindexfamilyrel]{code:qam}{Quadrature-amplitude modulation (QAM) format} --- 2D Gray codes are often concatenated with \(n=1\) lattice-based QAM codes so that the Hamming distance between the bitstrings encoded into the points is a discretized version of the Euclidean distance between the points.
\item\relax
\flmRefsHyperref[eczindexfamilyrel]{code:hergert}{Hergert code} --- Hergert codes can be seen, via the \flmTerm{term}{ref81}{}{Gray map}, as linear codes over \(\mathbb{Z}_4\) \NoCaseChange{\protect\cite{cite158,cite123}}.
\item\relax
\flmRefsHyperref[eczindexfamilyrel]{code:preparata}{Preparata code} --- The binary Preparata code is the Gray-map image of the quaternary code QRM\((m-2,m)\) \NoCaseChange{\protect\cite[{Thm. 19}]{cite158}}.
\item\relax
\flmRefsHyperref[eczindexfamilyrel]{code:delsarte_goethals}{Delsarte-Goethals (DG) code} --- DG codes can be seen, via the \flmTerm{term}{ref81}{}{Gray map}, as extended linear cyclic codes over \(\mathbb{Z}_4\) \NoCaseChange{\protect\cite{cite158}}.
\item\relax
\flmRefsHyperref[eczindexfamilyrel]{code:kerdock}{Kerdock code} --- The binary Kerdock code is the Gray-map image of the quaternary code QRM\((1,m)\), an extended cyclic code over \(\mathbb{Z}_4\) \NoCaseChange{\protect\cite[{Thm. 19}]{cite158}} (see also Ref. \NoCaseChange{\protect\cite{cite1382}}).
\item\relax
\flmRefsHyperref[eczindexfamilyrel]{code:best}{\((10,40,4)\) Best code} --- Codewords of the Best code can be obtained by applying the Gray map to the pentacode \NoCaseChange{\protect\cite[{Sec. 2}]{cite373}}.
\item\relax
\flmRefsHyperref[eczindexfamilyrel]{code:julin12}{Julin-Golay code} --- Julin codes can be obtained from simple nonlinear codes over \(\mathbb{Z}_4\) using the Gray map \NoCaseChange{\protect\cite{cite373}}.
\item\relax
\flmRefsHyperref[eczindexfamilyrel]{code:rank_modulation}{Rank-modulation code} --- The rank-modulation Gray code is an extension of the original binary Gray code to a code on the permutation group \NoCaseChange{\protect\cite{cite325}}.
\item\relax
\flmRefsHyperref[eczindexfamilyrel]{code:zrm}{ZRM code} --- The image of the ZRM\((r,m-1)\) code under the \flmTerm{term}{ref81}{}{Gray map} is the RM\((r,m)\) code \NoCaseChange{\protect\cite[{Thm. 7}]{cite158}}.
\item\relax
\flmRefsHyperref[eczindexfamilyrel]{code:self_dual_over_z4}{Self-dual code over \(\mathbb{Z}_4\)} --- Under the \flmTerm{term}{ref81}{}{Gray map}, any self-dual code over \(\mathbb{Z}_4\) maps to a formally self-dual binary code \NoCaseChange{\protect\cite{cite112}}.
\item\relax
\flmRefsHyperref[eczindexfamilyrel]{code:psk}{Phase-shift keying (PSK) modulation format} --- 1D Gray codes are often concatenated with PSKs so that the Hamming distance between the bitstrings encoded into the points is a discretized version of the Euclidean distance between the points.
\item\relax
\flmRefsHyperref[eczindexfamilyrel]{code:quantum_goethals_preparata}{\(\llparenthesis 2^m,2^{2^m−5m+1},8\rrparenthesis \) Goethals-Preparata code} --- A construction using the \(\mathbb{Z}_4\) versions of the Goethals and Preparata codes and the \flmTerm{term}{ref81}{}{Gray map} yields qubit code families with similar parameters as the \(\llparenthesis 2^m,2^{2^m−5m+1},8\rrparenthesis \) Goethals-Preparata code \NoCaseChange{\protect\cite{cite1371}}.
\end{eczvaluelist}
\eczhbkcontributors{ \eczhuVVA }
\endeczcode

\eczcode{hergert}{Hergert code}{~\NoCaseChange{\protect\cite{cite1326}}}
\codefieldsection{Alternative Names}
\begin{eczvaluelist}
\item\relax Goethals-Delsarte (GD) code
\end{eczvaluelist}
\eczhIndexCodeAliasName{hergert}{Goethals-Delsarte (GD) code}
\codefieldsection{Description}
A nonlinear subcode of an RM code that is a formal dual of a nonlinear DG code in the sense that its distance distribution is equal to the \flmRefsHyperref{ref113}{MacWilliams transform} of the distance distribution of the corresponding DG code.

The Hergert code for \( m/2 \geq r \geq 2 \) is a nonlinear subcode of RM\((m-2,m)\), and equals RM\((m-2,m)\) at \(r=1\) \NoCaseChange{\protect\cite[{Thm. 2}]{cite1326}}.
For each DG\((m,r)\) code, the Hergert code is defined as the union of cosets of RM\((m-3,m)\) in RM\((m-2,m)\), with coset representatives obtained by applying a particular linear bijection to the coset representatives of the DG code \NoCaseChange{\protect\cite{cite1326}}.

\codefieldsection{Decoding}
\begin{eczvaluelist}
\item\relax Since the equivalent \(\mathbb{Z}_4\) codes are extended cyclic codes, efficient encoding and decoding is possible \NoCaseChange{\protect\cite{cite158,cite1383}}.
\end{eczvaluelist}
\codefieldsection{Parent}
\begin{eczvaluelist}
\item\relax
\flmRefsHyperref[eczindexfamilyrel]{code:bits_into_bits}{Binary code}\end{eczvaluelist}
\codefieldsection{Children}
\begin{eczvaluelist}
\item\relax
\flmRefsHyperref[eczindexfamilyrel]{code:goethals}{Goethals code} --- Goethals codes are equivalent to Hergert codes for \(r=3\) \NoCaseChange{\protect\cite[{Thm. 2}]{cite1326}}.
\item\relax
\flmRefsHyperref[eczindexfamilyrel]{code:preparata}{Preparata code} --- Preparata codes are equivalent to Hergert codes for \(r=2\) \NoCaseChange{\protect\cite[{Thm. 2}]{cite1326}}.
\end{eczvaluelist}
\codefieldsection{Cousins}
\begin{eczvaluelist}
\item\relax
\flmRefsHyperref[eczindexfamilyrel]{code:delsarte_goethals}{Delsarte-Goethals (DG) code} --- Hergert codes are duals of DG codes in that their distance distribution is equal to the \flmRefsHyperref{ref113}{MacWilliams transform} of the distance distribution of DG codes \NoCaseChange{\protect\cite{cite1328}}. However, the two codes are images of a pair of mutually dual linear codes over \(\mathbb{Z}_4\) under the \flmTerm{term}{ref81}{}{Gray map} \NoCaseChange{\protect\cite{cite158,cite123}}.
\item\relax
\flmRefsHyperref[eczindexfamilyrel]{code:dual}{Dual linear code} --- Hergert codes are duals of DG codes in that their distance distribution is equal to the \flmRefsHyperref{ref113}{MacWilliams transform} of the distance distribution of DG codes \NoCaseChange{\protect\cite{cite1328}}. However, the two codes are images of a pair of mutually dual linear codes over \(\mathbb{Z}_4\) under the \flmTerm{term}{ref81}{}{Gray map} \NoCaseChange{\protect\cite{cite158,cite123}}.
\item\relax
\flmRefsHyperref[eczindexfamilyrel]{code:quaternary_over_z4}{Linear code over \(\mathbb{Z}_4\)} --- Hergert codes can be seen, via the \flmTerm{term}{ref81}{}{Gray map}, as linear codes over \(\mathbb{Z}_4\) \NoCaseChange{\protect\cite{cite158,cite123}}.
\item\relax
\flmRefsHyperref[eczindexfamilyrel]{code:gray}{Gray code} --- Hergert codes can be seen, via the \flmTerm{term}{ref81}{}{Gray map}, as linear codes over \(\mathbb{Z}_4\) \NoCaseChange{\protect\cite{cite158,cite123}}.
\end{eczvaluelist}
\eczhbkcontributors{ \eczhuVVA }
\endeczcode

\eczcode{higman-sims_graph}{Higman-Sims graph-adjacency code}{~\NoCaseChange{\protect\cite{cite82,cite1373}}}
\codefieldsection{Description}
A graph-based code whose generator matrix is the row space of the adjacency matrix of the Higman-Sims graph, yielding a \([100,22,22]\) code \(C_{HS}\) whose dual is a \([100,78,6]\) code \NoCaseChange{\protect\cite[{Table IV}]{cite82}}.

A related \([100,22,32]\) code \(C_{100}\), invariant under the Higman-Sims simple group \(HS\), is obtained by restricting a \([176,22,50]\) code invariant under the simple group \(Co_3\) \NoCaseChange{\protect\cite{cite1384}} to the 100 nonzero coordinates of a fixed minimum-weight codeword.
Its dual is an optimal \([100,78,8]\) code \NoCaseChange{\protect\cite[{Table VI}]{cite82}}.
The full automorphism groups are \(\mathrm{Aut}(C_{HS}) = 2\cdot HS\) and \(\mathrm{Aut}(C_{100}) = HS\) \NoCaseChange{\protect\cite[{Rem. 1.5}]{cite82}}.
The codes \(C_{HS}\) and \(C_{100}\) intersect in their doubly-even-weight subcodes, which have dimension 21 \NoCaseChange{\protect\cite[{Rem. 1.5}]{cite82}}.

\codefieldsection{Decoding}
\begin{eczvaluelist}
\item\relax The rows of the adjacency matrix can be used as orthogonal parity checks enabling majority decoding of \(C_{HS}^\perp\) up to two errors \NoCaseChange{\protect\cite[{Prop. 1.4}]{cite82}}.
\end{eczvaluelist}
\codefieldsection{Parent}
\begin{eczvaluelist}
\item\relax
\flmRefsHyperref[eczindexfamilyrel]{code:graph}{Graph-adjacency code}\end{eczvaluelist}
\codefieldsection{Cousins}
\begin{eczvaluelist}
\item\relax
\flmRefsHyperref[eczindexfamilyrel]{code:combinatorial_design}{Combinatorial design} --- The 4125 codewords of weight 36 of the \([100,22,32]\) code \(C_{100}\) form a \(2\)-\((100,36,525)\) design, which can be used for majority decoding of single errors in \(C_{100}^\perp\) \NoCaseChange{\protect\cite[{Rem. 1.7}]{cite82}}.
\item\relax
\flmRefsHyperref[eczindexfamilyrel]{code:leech}{\(\Lambda_{24}\) Leech lattice} --- The Higman-Sims graph occurs in the Leech lattice \NoCaseChange{\protect\cite{cite39}}.
\item\relax
\flmRefsHyperref[eczindexfamilyrel]{code:2pt_homogeneous}{Two-point homogeneous-space code} --- The Higman-Sims graph is distance-transitive, hence it is a finite two-point homogeneous space \NoCaseChange{\protect\cite{cite1385}}.
\end{eczvaluelist}
\eczhbkcontributors{ Andrey Boris Khesin, \eczhuVVA }
\endeczcode

\eczcode{hoffman-singleton}{Hoffman-Singleton cycle code}{~\NoCaseChange{\protect\cite{cite82,cite1373}}}
\codefieldsection{Description}
A \([50,21,12]\) cycle code whose parity-check matrix is the incidence matrix of the Hoffman-Singleton graph \NoCaseChange{\protect\cite{cite83}}.
Its dual is a \([50,29,8]\) code \NoCaseChange{\protect\cite[{Table II}]{cite82}}.

The Hoffman-Singleton graph-adjacency code \([50,22,7]\) contains this code as a subcode of codimension 1 \NoCaseChange{\protect\cite{cite82}}.
The dual \([50,29,8]\) code admits majority decoding correcting up to two errors via Rudolph's algorithm applied to the \(2\text{-}(50,14,13)\) design derived from the Hoffman-Singleton graph \NoCaseChange{\protect\cite[{Cor. 1.3}]{cite82}}.

\codefieldsection{Parent}
\begin{eczvaluelist}
\item\relax
\flmRefsHyperref[eczindexfamilyrel]{code:homological_classical}{Cycle code}\end{eczvaluelist}
\codefieldsection{Cousins}
\begin{eczvaluelist}
\item\relax
\flmRefsHyperref[eczindexfamilyrel]{code:hoffman-singleton_graph}{Hoffman-Singleton graph-adjacency code} --- The Hoffman-Singleton cycle code and graph-adjacency code are both derived from the Hoffman-Singleton graph.
\item\relax
\flmRefsHyperref[eczindexfamilyrel]{code:combinatorial_design}{Combinatorial design} --- The incidence matrix of the Hoffman-Singleton graph can be converted into a \(2\)-\((50,14,13)\) design \NoCaseChange{\protect\cite[{Prop. 1.1}]{cite82}}.
\end{eczvaluelist}
\eczhbkcontributors{ \eczhuVVA }
\endeczcode

\eczcode{hoffman-singleton_graph}{Hoffman-Singleton graph-adjacency code}{~\NoCaseChange{\protect\cite{cite82,cite1373}}}
\codefieldsection{Description}
A graph-based code whose generator matrix is the row space of the adjacency matrix of the Hoffman-Singleton graph \NoCaseChange{\protect\cite{cite83}}.
This \([50,22,7]\) code is dual to a \([50,28,8]\) code \NoCaseChange{\protect\cite[{Table III}]{cite82}}.

\codefieldsection{Decoding}
\begin{eczvaluelist}
\item\relax The rows of the adjacency matrix serve as orthogonal parity checks enabling majority decoding of the \([50,28,8]\) dual code up to three errors (its full error-correcting capacity) \NoCaseChange{\protect\cite{cite82}}.
\end{eczvaluelist}
\codefieldsection{Parent}
\begin{eczvaluelist}
\item\relax
\flmRefsHyperref[eczindexfamilyrel]{code:graph}{Graph-adjacency code}\end{eczvaluelist}
\codefieldsection{Cousins}
\begin{eczvaluelist}
\item\relax
\flmRefsHyperref[eczindexfamilyrel]{code:2pt_homogeneous}{Two-point homogeneous-space code} --- The Hoffman-Singleton graph is distance-transitive, hence it is a finite two-point homogeneous space \NoCaseChange{\protect\cite{cite1385}}.
\item\relax
\flmRefsHyperref[eczindexfamilyrel]{code:hoffman-singleton}{Hoffman-Singleton cycle code} --- The Hoffman-Singleton cycle code and graph-adjacency code are both derived from the Hoffman-Singleton graph.
\end{eczvaluelist}
\eczhbkcontributors{ \eczhuVVA }
\endeczcode

\eczcode{ha_ldpc}{Hsu-Anastasopoulos LDPC (HA-LDPC) code}{~\NoCaseChange{\protect\cite{cite1386}}}
\codefieldsection{Description}
A regular LDPC code obtained from a concatenation of a certain random regular LDPC code and a certain random LDGM code.
An \((l,r,g)\)-HA-LDPC code can be written using punctured LDPC and LDGM parts, and it is dual to the corresponding \((r,l,g)\)-MN-LDPC code \NoCaseChange{\protect\cite{cite84}}.
Using rate-one LDGM codes eliminates high-weight codewords while admitting an amount of low-weight codewords that asymptotically vanishes, allowing code families to achieve the \flmRefsHyperref{ref85}{GV bound} with high probability.

An \((l,r,g)\)-HA-LDPC code can be written using punctured LDPC and LDGM parts, and it is dual to the corresponding \((r,l,g)\)-MN-LDPC code \NoCaseChange{\protect\cite{cite84}}.

\codefieldsection{Rate}
HA-LDPC codes achieve capacity on the BEC channel under BP decoding \NoCaseChange{\protect\cite{cite1386}} and the memoryless binary-input output-symmetric (MBIOS) channels under ML decoding \NoCaseChange{\protect\cite{cite1386}} and under MAP decoding \NoCaseChange{\protect\cite{cite84}}. Bounded-density spatially coupled HA-LDPC codes have BEC BP thresholds close to the Shannon limit \NoCaseChange{\protect\cite{cite84}}. They also achieve the \flmRefsHyperref{ref85}{GV bound} with asymptotically high probability when the concatenation is with a rate-one LDGM code \NoCaseChange{\protect\cite{cite1386}}.
\codefieldsection{Parents}
\begin{eczvaluelist}
\item\relax
\flmRefsHyperref[eczindexfamilyrel]{code:regular_ldpc}{Regular LDPC code}\item\relax
\flmRefsHyperref[eczindexfamilyrel]{code:multi_edge_ldpc}{Multi-edge LDPC code} --- HA-LDPC codes can be formulated as multi-edge LDPC codes \NoCaseChange{\protect\cite{cite84}}.
\end{eczvaluelist}
\codefieldsection{Cousins}
\begin{eczvaluelist}
\item\relax
\flmRefsHyperref[eczindexfamilyrel]{code:ldgm}{Low-density generator-matrix (LDGM) code} --- HA-LDPC codes are a concatenation of an LDPC and an LDGM code.
\item\relax
\flmRefsHyperref[eczindexfamilyrel]{code:concatenated}{Concatenated code} --- HA-LDPC codes are a concatenation of an LDPC and an LDGM code.
\item\relax
\flmRefsHyperref[eczindexfamilyrel]{code:mn_ldpc}{MacKay-Neal LDPC (MN-LDPC) code} --- HA-LDPC and MN-LDPC codes are dual to each other \NoCaseChange{\protect\cite{cite84}}.
\item\relax
\flmRefsHyperref[eczindexfamilyrel]{code:dual}{Dual linear code} --- HA-LDPC and MN-LDPC codes are dual to each other \NoCaseChange{\protect\cite{cite84}}.
\item\relax
\flmRefsHyperref[eczindexfamilyrel]{code:sc_ldpc}{Spatially coupled LDPC (SC-LDPC) code} --- Spatial coupling of HA-LDPC protographs yields bounded-density SC-HA codes with BEC BP thresholds close to the Shannon limit \NoCaseChange{\protect\cite{cite84}}.
\end{eczvaluelist}
\eczhbkcontributors{ \eczhuVVA }
\endeczcode

\eczcode{irregular_ldpc}{Irregular LDPC code}{~\NoCaseChange{\protect\cite{cite1387,cite1388}}}
\codefieldsection{Description}
An LDPC code whose parity-check matrix has a variable number of entries in each row or column.

\codefieldsection{Rate}
Nearly achieve capacity against binary-input additive Gaussian white noise using iterative decoding \NoCaseChange{\protect\cite{cite1389,cite1390}}. Such sequences have sublinearly growing distance per block length \NoCaseChange{\protect\cite{cite1391}}.
\codefieldsection{Realizations}
\begin{eczvaluelist}
\item\relax Satellite communication after concatenating with a modulation scheme \NoCaseChange{\protect\cite{cite277}}.
\end{eczvaluelist}
\codefieldsection{Notes}
\begin{eczvaluelist}
\item\relax Useful tools for designing irregular LDPC codes can be found in Refs. \NoCaseChange{\protect\cite{cite1392,cite1393}}.
\end{eczvaluelist}
\codefieldsection{Parent}
\begin{eczvaluelist}
\item\relax
\flmRefsHyperref[eczindexfamilyrel]{code:multi_edge_ldpc}{Multi-edge LDPC code} --- The multi-edge code construction generalizes several of the original examples of irregular LDPC codes. Irregular LDPC codes can be formulated as multi-edge LDPC codes \NoCaseChange{\protect\cite[{Sec. XI}]{cite1394}}.
\end{eczvaluelist}
\codefieldsection{Children}
\begin{eczvaluelist}
\item\relax
\flmRefsHyperref[eczindexfamilyrel]{code:ara}{Accumulate-repeat-accumulate (ARA) code}\item\relax
\flmRefsHyperref[eczindexfamilyrel]{code:extended_ira}{Extended IRA (eIRA) code}\item\relax
\flmRefsHyperref[eczindexfamilyrel]{code:ira}{Irregular repeat-accumulate (IRA) code}\item\relax
\flmRefsHyperref[eczindexfamilyrel]{code:b_ldpc}{Block LDPC (B-LDPC) code}\end{eczvaluelist}
\codefieldsection{Cousin}
\begin{eczvaluelist}
\item\relax
\flmRefsHyperref[eczindexfamilyrel]{code:regular_ldpc}{Regular LDPC code} --- Irregular LDPC codes have variable node and check node degrees, while regular LDPC codes have fixed node degrees. Irregular LDPC codes with optimized degree distributions can outperform regular ones under iterative decoding \NoCaseChange{\protect\cite{cite1364,cite1395}}.
\end{eczvaluelist}
\eczhbkcontributors{ \eczhuVVA }
\endeczcode

\eczcode{ira}{Irregular repeat-accumulate (IRA) code}{~\NoCaseChange{\protect\cite{cite1396,cite1397,cite280}}}
\codefieldsection{Description}
A generalization of the RA code in which the outer 1-in-3 repetition encoding step is replaced by an LDGM code.
A simple version is when different bits in the RA block are repeated a different number of times.

IRA codes can be optimized against various noise channels \NoCaseChange{\protect\cite{cite1398}}.

\codefieldsection{Rate}
IRA codes nearly achieve the Shannon capacity of the binary erasure channel using iterative decoding \NoCaseChange{\protect\cite{cite1396}}. Puncturing lessens the decoding complexity while still allowing sequences of codes to achieve capacity \NoCaseChange{\protect\cite{cite1399}}.
\codefieldsection{Encoding}
\begin{eczvaluelist}
\item\relax One linear-time encoder for a \textit{systematic} IRA code consists of first encoding into an \([n,k]\) LDGM binary linear code, applying a random permutation, and then applying an accumulator to obtain \flmMathEnvironment{align}{}{ (u_{1},u_{1}+u_{2},\cdots,u_{1}+\cdots+u_{N})~, } where \(N\) is the length of the permuted LDGM output.
\end{eczvaluelist}
\codefieldsection{Decoding}
\begin{eczvaluelist}
\item\relax Linear-time decoder \NoCaseChange{\protect\cite{cite1396}}.
\end{eczvaluelist}
\codefieldsection{Realizations}
\begin{eczvaluelist}
\item\relax LDPC codes used for digital satellite video broadcasting per the DVB-S2 standard \NoCaseChange{\protect\cite{cite278,cite279}} utilize IRA code features and were subject to litigation; see Ref. \NoCaseChange{\protect\cite{cite280}}.
\item\relax Apple and Broadcom Wi-Fi devices utilize IRA encoding and decoding features and are subject to ongoing litigation; see Ref. \NoCaseChange{\protect\cite{cite280}}.
\end{eczvaluelist}
\codefieldsection{Parents}
\begin{eczvaluelist}
\item\relax
\flmRefsHyperref[eczindexfamilyrel]{code:irregular_ldpc}{Irregular LDPC code}\item\relax
\flmRefsHyperref[eczindexfamilyrel]{code:protograph_ldpc}{Protograph LDPC code} --- IRA codes can be formulated as protograph LDPC codes \NoCaseChange{\protect\cite{cite1212}}.
\item\relax
\flmRefsHyperref[eczindexfamilyrel]{code:concatenated}{Concatenated code} --- IRA codes can be interpreted as serial concatenated codes \NoCaseChange{\protect\cite{cite970}}.
\end{eczvaluelist}
\codefieldsection{Child}
\begin{eczvaluelist}
\item\relax
\flmRefsHyperref[eczindexfamilyrel]{code:ra}{Repeat-accumulate (RA) code} --- IRA codes for which the outer code is a 1-in-3 repetition code reduce to RA codes.
\end{eczvaluelist}
\codefieldsection{Cousins}
\begin{eczvaluelist}
\item\relax
\flmRefsHyperref[eczindexfamilyrel]{code:ldgm}{Low-density generator-matrix (LDGM) code} --- IRA codes replace the outer 1-in-3 repetition encoding step in RA codes with an LDGM code.
\item\relax
\flmRefsHyperref[eczindexfamilyrel]{code:mn_ldpc}{MacKay-Neal LDPC (MN-LDPC) code} --- MN-LDPC and IRA codes intersect for certain parameters \NoCaseChange{\protect\cite{cite1400}}.
\end{eczvaluelist}
\eczhbkcontributors{ \eczhuVVA }
\endeczcode

\eczcode{julin12}{Julin-Golay code}{~\NoCaseChange{\protect\cite{cite1401,cite1402,cite376}}}
\codefieldsection{Description}
One of several nonlinear binary \((12,144,4)\) codes based on the Steiner system \(S(5,6,12)\) \NoCaseChange{\protect\cite{cite371,cite372}\protect\cite[{Sec. 2.7}]{cite41}\protect\cite[{Sec. 4}]{cite373}}
or their shortened versions, the nonlinear \((11,72,4)\), \((10,38,4)\), and \((9,20,4)\) Julin-Golay codes.
Several of these codes contain more codewords than linear codes of the same length and distance and yield non-lattice sphere-packings that hold records in 9 and 11 dimensions.

The codewords of the length-12 codes are 132 distinct mod-two pairwise row sums of an \(11\)-dimensional matrix derived from the \(12\)-dimensional Hadamard matrix \(H\) along with their negations, 6 mutually disjoint codewords of weight two, and 6 codewords of weight 10 whose complements are mutually disjoint.

Using \flmTerm{term}{ref127}{}{Construction A}, some Julin-Golay length-12 codes yield 12-dimensional non-lattice sphere packings, collectively called \(P_{12a}\), with kissing number 840 \NoCaseChange{\protect\cite{cite376}\protect\cite[{pg. 139}]{cite39}}.
This is the highest known kissing number in that dimension.
The length-11 code yields \(P_{11a}\), a non-lattice sphere packing that is the densest known in 11 dimensions.
The length-9 code yields a non-lattice sphere packing called \(P_{9a}\) with kissing number 306, the highest known in 9 dimensions.

The Julin-Golay length-12 codes are not to be confused with the Best \((12,144,4)\) code \NoCaseChange{\protect\cite{cite1135}}, which is not based on a Steiner system \NoCaseChange{\protect\cite[{Sec. 3}]{cite373}}.

\codefieldsection{Parent}
\begin{eczvaluelist}
\item\relax
\flmRefsHyperref[eczindexfamilyrel]{code:sloane_whitehead}{Sloane-Whitehead code} --- Julin-Golay codes are the starting codes for the Sloane-Whitehead construction \NoCaseChange{\protect\cite{cite374}\protect\cite[{Secs. 2.7 and 2.9}]{cite41}}.
\end{eczvaluelist}
\codefieldsection{Cousins}
\begin{eczvaluelist}
\item\relax
\flmRefsHyperref[eczindexfamilyrel]{code:combinatorial_design}{Combinatorial design} --- Julin-Golay codes are constructed from the Steiner system \(S(5,6,12)\) arising from the extended \((12,132,4)\) code \NoCaseChange{\protect\cite[{pgs. 70-72}]{cite41}}.
\item\relax
\flmRefsHyperref[eczindexfamilyrel]{code:sphere_packing}{Sphere packing} --- Using \flmTerm{term}{ref127}{}{Construction A}, the Julin-Golay codes yield non-lattice sphere-packings that hold records in 9 and 11 dimensions.
\item\relax
\flmRefsHyperref[eczindexfamilyrel]{code:construction_a}{Construction A code} --- Using \flmTerm{term}{ref127}{}{Construction A}, the Julin-Golay codes yield non-lattice sphere-packings that hold records in 9 and 11 dimensions.
\item\relax
\flmRefsHyperref[eczindexfamilyrel]{code:q-ary_over_zq}{\(q\)-ary code over \(\mathbb{Z}_q\)} --- Julin codes can be obtained from simple nonlinear codes over \(\mathbb{Z}_4\) using the Gray map \NoCaseChange{\protect\cite{cite373}}.
\item\relax
\flmRefsHyperref[eczindexfamilyrel]{code:gray}{Gray code} --- Julin codes can be obtained from simple nonlinear codes over \(\mathbb{Z}_4\) using the Gray map \NoCaseChange{\protect\cite{cite373}}.
\end{eczvaluelist}
\eczhbkcontributors{ \eczhuVVA }
\endeczcode

\eczcode{justesen}{Justesen code}{~\NoCaseChange{\protect\cite{cite1403}}}
\codefieldsection{Description}
Binary linear code resulting from generalized concatenation of an outer RS code with multiple inner codes sampled from the Wozencraft ensemble, i.e., \(N\) distinct binary inner codes of dimension \(m\) and length \(2m\).
Justesen codes were among the first explicit asymptotically good codes \NoCaseChange{\protect\cite[{Sec. 3.3.3}]{cite70}}.

Justesen codes are parameterized by \(m\), with length \(n=2mN\) and dimension \(k=mK\), where \((N=2^m-1,K)\) is the RS outer code over \(\mathbb{F}_{2^m}\).
Further modifications have improved code parameters \NoCaseChange{\protect\cite{cite1404,cite1405,cite1406}}.

\codefieldsection{Rate}
The first asymptotically good codes.
Rate is \(R_m=k/n=K/2N\geq R\) and the relative minimum distance satisfies \(\delta_m=d_m/n\geq 0.11(1-2R)\), where \(K=\left\lceil 2NR\right\rceil\) for asymptotic rate \(0<R<1/2\); see \NoCaseChange{\protect\cite[{pg. 311}]{cite41}}.

The code can be improved and the range of \(R\) extended from 0 to 1 by \textit{puncturing}, i.e., by erasing \(s\) digits from each inner codeword. This yields a code of length \(n=(2m-s)N\) and rate \(R=mk/(2m-s)N\). The lower bound on the relative distance of the punctured code approaches \(d_m/n=(1-R/r)H^{-1}(1-r)\) as \(m\) goes to infinity, where \(r\) is the maximum of 1/2 and the solution to \(R=r^2/( 1+\log( 1-h^{-1}(1-r) ) )\), and \(h\) is the binary entropy function.

\codefieldsection{Decoding}
\begin{eczvaluelist}
\item\relax Generalized minimum distance decoding \NoCaseChange{\protect\cite{cite1403}}.
\end{eczvaluelist}
\codefieldsection{Realizations}
\begin{eczvaluelist}
\item\relax Generating small-bias sample spaces, i.e., probability distributions that parity functions cannot typically distinguish from the uniform distribution \NoCaseChange{\protect\cite{cite281}}.
\end{eczvaluelist}
\codefieldsection{Parents}
\begin{eczvaluelist}
\item\relax
\flmRefsHyperref[eczindexfamilyrel]{code:binary_linear}{Linear binary code}\item\relax
\flmRefsHyperref[eczindexfamilyrel]{code:generalized_concatenated}{Generalized concatenated code (GCC)} --- Justesen codes can be considered as a generalized concatenation of an outer RS code with \(N\) distinct binary inner codes.
\item\relax
\flmRefsHyperref[eczindexfamilyrel]{code:random}{Random code} --- The required inner codes are obtained by random sampling from the Wozencraft ensemble, whose length scales logarithmically with \(n\).
\end{eczvaluelist}
\codefieldsection{Cousins}
\begin{eczvaluelist}
\item\relax
\flmRefsHyperref[eczindexfamilyrel]{code:reed_solomon}{Reed-Solomon (RS) code} --- An RS code is the outer code of Justesen codes.
\item\relax
\flmRefsHyperref[eczindexfamilyrel]{code:wozencraft}{Wozencraft ensemble code} --- Wozencraft ensemble forms the inner codes of Justesen codes.
\item\relax
\flmRefsHyperref[eczindexfamilyrel]{code:q-ary_bch}{Bose–Chaudhuri–Hocquenghem (BCH) code} --- Using more general BCH codes instead of RS codes can improve the parameters of the Justesen codes \NoCaseChange{\protect\cite{cite1404}}.
\item\relax
\flmRefsHyperref[eczindexfamilyrel]{code:movassagh_ouyang}{Movassagh-Ouyang Hamiltonian code} --- Justesen codes can be used to build a family of \(n\)-qubit Movassagh-Ouyang Hamiltonian spin codes encoding one logical qubit with linear distance. These codes form the ground-state subspace of a frustration-free geometrically local Hamiltonian \NoCaseChange{\protect\cite{cite1407}}.
\end{eczvaluelist}
\eczhbkcontributors{ Xiao Xiao, \eczhuVVA }
\endeczcode

\eczcode{kerdock}{Kerdock code}{~\NoCaseChange{\protect\cite{cite1408}}}
\codefieldsection{Description}
Binary nonlinear \((2^m, 2^{2m}, 2^{m-1} - 2^{(m-2)/2})\) for even \(m\) consisting of the first-order Reed-Muller code RM\((1,m)\) with maximum-rank cosets of RM\((1,m)\) in RM\((2,m)\).

The size of code book, \(|2^{m-1}||\text{RM}(1,m)|\), is twice the size of the largest possible linear code with the same length and distance.
The relative minimum distance tends to \(\frac{1}{2}\) as \(m\) becomes large.

Let the matrices \(A = [a_{ij}]\) constitute the Kerdock set \(K\), i.e., a set of symmetric binary matrices with zero diagonal entries with the property that differences of distinct matrices in the set have full rank.
Define Boolean functions of the form \(Q(X) + l(X) + b\), where \(Q(X) = \sum_{1 \leq i < j \leq m} a_{ij}X_{i}X_{j}\), the function \(l(X)\) is linear in \(m\) variables, and \(b\) is a bit.
Codewords are formed as evaluations of these functions over \(\mathbb{F}_2^{m}\) in the variable \(X \in \mathbb{F}_2^{m}\).

The automorphism group of these codes is \(\Gamma A(1,2^{m-1})\times\mathbb{F}_2\) \NoCaseChange{\protect\cite{cite1409}}.

\codefieldsection{Rate}
The transmission rate is \(2m/2^m\) which tends to 0 as \(m\) becomes large, hence these codes are asymptotically poor.
\codefieldsection{Decoding}
\begin{eczvaluelist}
\item\relax Soft decision decoding involves extending the Fast Hadamard Transform decoding algorithm for the first-order RM code to Kerdock code \NoCaseChange{\protect\cite{cite158}}.
\item\relax Complexity of soft decision decoding algorithm: \(4^m\) multiplications and \(m4^m\) additions \NoCaseChange{\protect\cite{cite158}}.
\end{eczvaluelist}
\codefieldsection{Notes}
\begin{eczvaluelist}
\item\relax See corresponding MinT database entry \NoCaseChange{\protect\cite{cite1410}}.
\end{eczvaluelist}
\codefieldsection{Parent}
\begin{eczvaluelist}
\item\relax
\flmRefsHyperref[eczindexfamilyrel]{code:delsarte_goethals}{Delsarte-Goethals (DG) code} --- A Kerdock code of length \(2^m\) is equivalent to DG\((m,m/2)\) and is a subcode of DG\((m,r)\) \NoCaseChange{\protect\cite[{pg. 461}]{cite41}}.
\end{eczvaluelist}
\codefieldsection{Child}
\begin{eczvaluelist}
\item\relax
\flmRefsHyperref[eczindexfamilyrel]{code:nordstrom_robinson}{\((16,256,6)\) Nordstrom-Robinson (NR) code} --- The NR code is the smallest Kerdock code.
\end{eczvaluelist}
\codefieldsection{Cousins}
\begin{eczvaluelist}
\item\relax
\flmRefsHyperref[eczindexfamilyrel]{code:reed_muller}{Reed-Muller (RM) code} --- Kerdock code is a subcode of a second-order RM Code \NoCaseChange{\protect\cite[{pg. 457}]{cite41}}.
It consists of a number of cosets of RM\((2,m)\) created by quotienting with first-order RM\((1,m)\) codes.

\item\relax
\flmRefsHyperref[eczindexfamilyrel]{code:combinatorial_design}{Combinatorial design} --- Kerdock codes form an infinite family of nonlinear binary codes supporting 3-designs \NoCaseChange{\protect\cite[{Rem. 5.5.6}]{cite135}}.
\item\relax
\flmRefsHyperref[eczindexfamilyrel]{code:biorthogonal}{\([2^m,m+1,2^{m-1}]\) First-order RM code} --- Kerdock code is a subcode of a second-order RM Code \NoCaseChange{\protect\cite[{pg. 457}]{cite41}}.
It consists of a number of cosets of RM\((2,m)\) created by quotienting with first-order RM\((1,m)\) codes.

\item\relax
\flmRefsHyperref[eczindexfamilyrel]{code:preparata}{Preparata code} --- Preparata codes are duals of Kerdock codes in that their distance distribution is equal to the \flmRefsHyperref{ref113}{MacWilliams transform} of the distance distribution of Kerdock codes \NoCaseChange{\protect\cite{cite1148}}.
However, the two codes are images of a pair of mutually dual linear codes over \(\mathbb{Z}_4\) under the \flmTerm{term}{ref81}{}{Gray map}  \NoCaseChange{\protect\cite{cite1149}\protect\cite[{Sec. 6.3}]{cite1145}}.

\item\relax
\flmRefsHyperref[eczindexfamilyrel]{code:cluster_state}{Cluster-state code} --- Kerdock codes correspond to cluster states, and the corresponding Clifford-group automorphisms of this set form a particular group \NoCaseChange{\protect\cite{cite934}} that is a unitary 2-design on \(U(2^n)\) \NoCaseChange{\protect\cite{cite935}}. As such, cluster states form complex projective on 2-designs \(\mathbb{C}P^{2^n}\). These are useful in matrix-vector multiplication \NoCaseChange{\protect\cite{cite936}}.
\item\relax
\flmRefsHyperref[eczindexfamilyrel]{code:unitary_design}{Unitary \(t\)-design} --- Kerdock codes correspond to cluster states, and the corresponding Clifford-group automorphisms of this set form a particular group \NoCaseChange{\protect\cite{cite934}} that is a unitary 2-design on \(U(2^n)\) \NoCaseChange{\protect\cite{cite935}}. As such, cluster states form complex projective on 2-designs \(\mathbb{C}P^{2^n}\). These are useful in matrix-vector multiplication \NoCaseChange{\protect\cite{cite936}}.
\item\relax
\flmRefsHyperref[eczindexfamilyrel]{code:complex_projective}{Complex projective space code} --- Kerdock codes correspond to cluster states, and the corresponding Clifford-group automorphisms of this set form a particular group \NoCaseChange{\protect\cite{cite934}} that is a unitary 2-design on \(U(2^n)\) \NoCaseChange{\protect\cite{cite935}}. As such, cluster states form complex projective on 2-designs \(\mathbb{C}P^{2^n}\). These are useful in matrix-vector multiplication \NoCaseChange{\protect\cite{cite936}}.
\item\relax
\flmRefsHyperref[eczindexfamilyrel]{code:univ_opt_q-ary}{Universally optimal \(q\)-ary code} --- Kerdock codes are asymptotically close to the LP and universal-energy bounds \NoCaseChange{\protect\cite[{Exam. 12.3.25}]{cite199}}.
\item\relax
\flmRefsHyperref[eczindexfamilyrel]{code:quaternary_reed_muller}{Quaternary RM (QRM) code} --- The binary Kerdock code is the Gray-map image of the quaternary code QRM\((1,m)\), an extended cyclic code over \(\mathbb{Z}_4\) \NoCaseChange{\protect\cite[{Thm. 19}]{cite158}} (see also Ref. \NoCaseChange{\protect\cite{cite1382}}).
\item\relax
\flmRefsHyperref[eczindexfamilyrel]{code:gray}{Gray code} --- The binary Kerdock code is the Gray-map image of the quaternary code QRM\((1,m)\), an extended cyclic code over \(\mathbb{Z}_4\) \NoCaseChange{\protect\cite[{Thm. 19}]{cite158}} (see also Ref. \NoCaseChange{\protect\cite{cite1382}}).
\item\relax
\flmRefsHyperref[eczindexfamilyrel]{code:zrm}{ZRM code} --- Each Kerdock code is contained in a corresponding ZRM\((2,m)\) code \NoCaseChange{\protect\cite{cite158}}.
\item\relax
\flmRefsHyperref[eczindexfamilyrel]{code:real_projective}{Real projective space code} --- The 12 sets of antipodal pairs of the 24-cell code form a sharp configuration in the projective space \(\mathbb{R}P^3\) \NoCaseChange{\protect\cite{cite119}}. This is a special case of a family of real projective plane codes, constructed using Kerdock codes \NoCaseChange{\protect\cite{cite1411}} (cf. \NoCaseChange{\protect\cite{cite917}}).
\item\relax
\flmRefsHyperref[eczindexfamilyrel]{code:clifford_group}{Clifford group} --- Kerdock codes correspond to cluster states, and the corresponding Clifford-group automorphisms of this set form a particular group \NoCaseChange{\protect\cite{cite934}} that is a unitary 2-design on \(U(2^n)\) \NoCaseChange{\protect\cite{cite935}}. As such, cluster states form complex projective 2-designs on \(\mathbb{C}P^{2^n}\). These are useful in matrix-vector multiplication \NoCaseChange{\protect\cite{cite936}}.
\item\relax
\flmRefsHyperref[eczindexfamilyrel]{code:frameproof}{Frameproof (FP) code} --- Kerdock codes of sufficient order are separating \NoCaseChange{\protect\cite{cite1056,cite1057}}.
\item\relax
\flmRefsHyperref[eczindexfamilyrel]{code:24cell}{24-cell code} --- The 12 antipodal pairs of the 24-cell code form a sharp configuration and a 2-design in \(\mathbb{R}P^3\) \NoCaseChange{\protect\cite{cite119}}. This is a special case of a family of real projective plane codes, constructed using Kerdock codes \NoCaseChange{\protect\cite{cite1411}} (cf. \NoCaseChange{\protect\cite{cite917}}).
\item\relax
\flmRefsHyperref[eczindexfamilyrel]{code:kerdock_spherical}{Kerdock spherical code} --- Kerdock spherical codes can be obtained from Kerdock codes using the \flmRefsHyperref{ref38}{antipodal mapping} \NoCaseChange{\protect\cite[{pg. 157}]{cite115}}.
\end{eczvaluelist}
\eczhbkcontributors{ Shuubham Ojha, \eczhuVVA }
\endeczcode

\eczcode{kmrs-ltc}{Kopparty-Meir-Ron-Zewi-Saraf (KMRS) code}{~\NoCaseChange{\protect\cite{cite1412,cite86}}}
\codefieldsection{Description}
Member of a family of locally testable binary linear codes with constant rate, constant relative distance, and subpolynomial query complexity \(u = (\log n)^{O(\log \log n)}\).
Later work by Gopi, Kopparty, Oliveira, Ron-Zewi, and Saraf \NoCaseChange{\protect\cite{cite86}} showed that related concatenated codes achieve the \flmRefsHyperref{ref85}{GV bound}.

\codefieldsection{Parent}
\begin{eczvaluelist}
\item\relax
\flmRefsHyperref[eczindexfamilyrel]{code:binary_ltc}{Binary linear LTC}\end{eczvaluelist}
\eczhbkcontributors{ \eczhuVVA }
\endeczcode

\eczcode{laplacian}{Laplacian code}{~\NoCaseChange{\protect\cite{cite1350}}}
\codefieldsection{Description}
A binary linear code whose parity-check matrix is the graph Laplacian reduced mod 2.
For an undirected graph \(\Gamma\) with degree matrix \(D\) and adjacency matrix \(A\), the parity-check matrix is the symmetric matrix \(H=(D-A)\bmod 2\).

\codefieldsection{Protection}
A connected graph always contributes at least one logical bit.
For generic sparse bounded-degree graph ensembles, the logical dimension stays \(O(1)\), while highly symmetric graphs such as square lattices or complete graphs can have much larger rank deficiency \NoCaseChange{\protect\cite{cite1350}}.

\codefieldsection{Rate}
Generic sparse-graph Laplacian-code families have vanishing rate with only \(O(1)\) logical bits, whereas more symmetric families can exhibit enhanced rank deficiency \NoCaseChange{\protect\cite{cite1350}}.
\codefieldsection{Parents}
\begin{eczvaluelist}
\item\relax
\flmRefsHyperref[eczindexfamilyrel]{code:binary_linear}{Linear binary code}\item\relax
\flmRefsHyperref[eczindexfamilyrel]{code:projective}{Projective geometry code} --- Incidence matrices of graphs have no repeated columns since that would correspond to multi-edges. Therefore, Laplacian codes can be interpreted as projective codes.
\item\relax
\flmRefsHyperref[eczindexfamilyrel]{code:quantum_inspired}{Quantum-inspired classical block code}\end{eczvaluelist}
\codefieldsection{Cousins}
\begin{eczvaluelist}
\item\relax
\flmRefsHyperref[eczindexfamilyrel]{code:ldpc}{Low-density parity-check (LDPC) code} --- Laplacian codes on bounded-degree graph families are classical LDPC codes.
\item\relax
\flmRefsHyperref[eczindexfamilyrel]{code:pinwheel}{Pinwheel code} --- The pinwheel code is derived from the graph Laplacian of the pinwheel tiling, with a fraction of boundary checks removed.
\item\relax
\flmRefsHyperref[eczindexfamilyrel]{code:anisotropic_z2_laplacian}{Anisotropic \(\mathbb{Z}_2\) Laplacian model code} --- The anisotropic \(\mathbb{Z}_2\) Laplacian model is the hypergraph product of a cyclic repetition code and a Laplacian code \NoCaseChange{\protect\cite{cite1350}}.
\end{eczvaluelist}
\eczhbkcontributors{ \eczhuVVA }
\endeczcode

\eczcode{lu_ldpc}{Lazebnik-Ustimenko (LU) code}{~\NoCaseChange{\protect\cite{cite87,cite1413}}}
\codefieldsection{Description}
LDPC code whose Tanner graph comes from a particular family of \(q\)-regular graphs \NoCaseChange{\protect\cite{cite87}} of known girth and relatively large stopping sets.

\codefieldsection{Parents}
\begin{eczvaluelist}
\item\relax
\flmRefsHyperref[eczindexfamilyrel]{code:regular_ldpc}{Regular LDPC code}\item\relax
\flmRefsHyperref[eczindexfamilyrel]{code:algebraic_ldpc}{Algebraic LDPC code}\end{eczvaluelist}
\eczhbkcontributors{ \eczhuVVA }
\endeczcode

\eczcode{lr-cayley-complex}{Left-right Cayley complex code}{~\NoCaseChange{\protect\cite{cite88}}}
\codefieldsection{Description}
Binary code constructed on a left-right Cayley complex using a pair of base codes \(C_A,C_B\) and an expander graph \NoCaseChange{\protect\cite{cite74}}. Bits live on the squares of the complex, while local constraints are imposed on edges; for a fixed graph vertex, the incident symbols form a codeword of the tensor code \(C_A \otimes C_B\). A family of such codes is one of the first \(c^3\)-LTCs \NoCaseChange{\protect\cite{cite88}}.

\codefieldsection{Parent}
\begin{eczvaluelist}
\item\relax
\flmRefsHyperref[eczindexfamilyrel]{code:binary_ltc}{Binary linear LTC} --- Left-right Cayley complex codes yield one of the first two families of \(c^3\)-LTCs.
\end{eczvaluelist}
\codefieldsection{Cousins}
\begin{eczvaluelist}
\item\relax
\flmRefsHyperref[eczindexfamilyrel]{code:tensor}{Tensor-product code} --- Left-right Cayley complex codewords for a fixed graph vertex are codewords of a tensor code.
\item\relax
\flmRefsHyperref[eczindexfamilyrel]{code:expander}{Expander code} --- Left-right Cayley complex codes can be viewed as Tanner-like codes on expander graphs \NoCaseChange{\protect\cite{cite74}}, but with bits defined on squares and constraints on edges (as opposed to edges and vertices, respectively, for expander codes). Expander codes are also typically not locally testable \NoCaseChange{\protect\cite{cite1341}}.
\item\relax
\flmRefsHyperref[eczindexfamilyrel]{code:balanced_product}{Balanced product (BP) code} --- Left-right Cayley complexes can be obtained via a balanced product of \(G\)-graphs \NoCaseChange{\protect\cite{cite88}}.
\item\relax
\flmRefsHyperref[eczindexfamilyrel]{code:quantum_tanner}{Quantum Tanner code} --- Applying the CSS construction to two left-right Cayley complex codes yields quantum Tanner codes, and one can simultaneously prove a linear distance for the quantum code and local testability for one of its constituent classical codes \NoCaseChange{\protect\cite{cite1414}}.
\end{eczvaluelist}
\eczhbkcontributors{ \eczhuVVA }
\endeczcode

\eczcode{levenshtein}{Levenshtein code}{~\NoCaseChange{\protect\cite{cite1415}}}
\codefieldsection{Description}
Binary codes constructed from combining two codes \(A'\) constructed out of Hadamard matrices.

Let \(H_n\) be a normalized Hadamard matrix. The generator matrix for an \((n-1,n,n/2)\) code \(A_n\) is obtained by taking \(H_n\), replacing the +1's by 0's and the -1's by 1's, and deleting the first column. Taking only the codewords of \(A_n\) which begin with 0 and deleting the leading 0 yields the generator matrix of an \((n-2,n/2,n/2)\) code \(A_n'\).

Next, apply the following way of combining codes. Suppose we have an \((n_1,M_1,d_1)\) code \(C_1\) and an \((n_2,M_2,d_2)\) code \(C_2\). The combined \((an_1+bn_2,\min(M_1,M_2),ad_1+bd_2)\) code \(a C_1\oplus b C_2\) may be constructed by pasting \(a\) copies of \(C_1\) and \(b\) copies of \(C_2\) together and omitting the last \(|M_1-M_2|\) rows. Applying this to construct a Levenshtein code with length \(n\) and distance \(d\), define \(k=\lfloor d/(2d-n)\rfloor\), \(a=d(2k+1)-n(k+1)\), and \(b=kn-d(2k-1)\). If \(n\) is even, construct \(\frac{a}{2}A_{4k}'\oplus\frac{b}{2}A_{4k+4}'\). If \(n\) is odd and \(k\) is even, construct \(aA_{2k}\oplus\frac{b}{2}A_{4k+4}'\). If \(n\) is odd and \(k\) is odd, construct \(\frac{a}{2}A_{4k}'\oplus b A_{2k+2}\).

\codefieldsection{Protection}
Levenshtein codes meet the Plotkin bound \(K\leq 2\left\lfloor\frac{d}{2d-n}\right\rfloor\), where \(K\) is the number of codewords, \(d\) is the distance, and \(n\) is the length, and with the assumption that the Hadamard matrices for such parameters exist. The general proof depends on the correctness of Hadamard's conjecture \NoCaseChange{\protect\cite{cite41}}.

\codefieldsection{Parent}
\begin{eczvaluelist}
\item\relax
\flmRefsHyperref[eczindexfamilyrel]{code:bits_into_bits}{Binary code}\end{eczvaluelist}
\codefieldsection{Cousin}
\begin{eczvaluelist}
\item\relax
\flmRefsHyperref[eczindexfamilyrel]{code:biorthogonal}{\([2^m,m+1,2^{m-1}]\) First-order RM code} --- First-order RM codes and Levenshtein codes are both constructed using Hadamard matrices.
\end{eczvaluelist}
\eczhbkcontributors{ Richard Barney, \eczhuVVA }
\endeczcode

\eczcode{binary_linear}{Linear binary code}{}
\codefieldsection{Description}
An \((n,2^k,d)\) linear code is denoted as \([n,k]\) or \([n,k,d]\), where \(k\) is the code's dimension, and where \(d\) is the code's distance. Its codewords form a linear subspace, i.e., for any codewords \(x,y\), \(x+y\) is also a codeword. A code that is not linear is called \textit{nonlinear}.

Linear codes can be defined in terms of a \textit{generator matrix} \(G\), whose rows form a basis for the \(k\)-dimensional codespace. Given a message \(x\), the corresponding encoded codeword is \(G^T x\). The generator matrix can be reduced via coordinate permutations to its \textit{standard form} or \textit{systematic form} \(G = [I_k~A]\), where \(I_k\) is a \(k\times k\) identity matrix and \(A\) is a \(k \times (n-k)\) binary matrix.

The \textit{automorphism group} of a linear binary code is the largest subgroup of coordinate permutations that maps the code onto itself.
Two linear binary codes are \textit{equivalent} if the codewords of one code can be mapped into those of the other under a permutation \NoCaseChange{\protect\cite[{Remark 3.2.1}]{cite70}\protect\cite[{Ch. 8}]{cite41}\protect\cite[{Ch. 3}]{cite39}}.

\codefieldsection{Protection}
Distance \(d\) of a linear code is the number of nonzero entries in the (nonzero) codeword with the smallest such number. Corrects any error set for which no two elements of the set add up to a codeword.

Linear codes admit a \textit{parity check matrix} \(H\), whose columns make up a set of \textit{parity checks}, i.e., a maximal linearly independent set of vectors that are in the kernel of \(G\). It follows that
\flmMathEnvironment{align}{}{
  G H^{\text{T}} = 0 \mod 2~.
}

The decision problem corresponding to finding the minimum distance is also \(NP\)-complete \NoCaseChange{\protect\cite{cite1416}}, and approximating the weight enumerator is \(\#P\)-complete \NoCaseChange{\protect\cite{cite1417}}.

There are several standard procedures for increasing or decreasing the length of an \([n,k,d]\) code \NoCaseChange{\protect\cite[{Ch. 1}]{cite41}}:
\begin{enumerate}[(1)]\item \textit{Puncturing}: removing a coordinate to yield a code whose length is shorter by one and whose distance is \(\geq d-1\).
\item \textit{Expurgating}: removing odd-weight codewords of a non-even-weight code to yield a code whose dimension is \(k-1\).
\item \textit{Augmenting}: adding the all-ones codeword to a code without it to yield a code whose dimension is \(k+1\).
\item \textit{Lengthening}: adding the all-ones codeword and then adding a parity check to yield a code whose size and dimension increase by one.
\item \textit{Shortening}: keeping only codewords which have a zero in a fixed coordinate and removing that coordinate to yield a code whose length is shorter by one.
\end{enumerate}

\codefieldsection{Rate}
A family of linear codes \(C_i = [n_i,k_i,d_i]\) is \textit{asymptotically good} if the asymptotic rate \(\lim_{i\to\infty} k_i/n_i\) and asymptotic distance \(\lim_{i\to\infty} d_i/n_i\) are both positive. Nearly all good linear binary codes for the asymmetric channel are also good for the symmetric channel \NoCaseChange{\protect\cite{cite1418}}; this is not the case for non-binary codes \NoCaseChange{\protect\cite{cite1186}}.

Binary linear codes on \(D\)-dimensional Euclidean lattices are limited by the \textit{classical Bravyi-Poulin-Terhal (BPT) bound} \NoCaseChange{\protect\cite{cite1419,cite1420}}, which states that \(d = O(n^{1-1/D})\) and that \(k d^{1/D} = O(n)\) (using \flmRefsHyperref{ref65}{asymptotic notation}). 
This bound is the classical analogue of the \flmRefsHyperref{ref487}{BPT bound} for qubit stabilizer codes and the \flmRefsHyperref{ref492}{subsystem BT bound} for subsystem qubit stabilizer codes.

\codefieldsection{Decoding}
\begin{eczvaluelist}
\item\relax Decoding an arbitrary linear binary code is \(NP\)-complete \NoCaseChange{\protect\cite{cite1421,cite1422}}.
\item\relax Slepian's standard-array decoding \NoCaseChange{\protect\cite{cite1268}}.
\item\relax Recursive maximum likelihood decoding \NoCaseChange{\protect\cite{cite1423}}.
\item\relax Deep learning \NoCaseChange{\protect\cite{cite1424}} and a transformer graph neural net (GNN) for soft decoding \NoCaseChange{\protect\cite{cite1425}}.
\item\relax Chase decoding, which uses channel measurement information \NoCaseChange{\protect\cite{cite1426}}.
\end{eczvaluelist}
\codefieldsection{Notes}
\begin{eczvaluelist}
\item\relax Tables of bounds and examples of linear codes for various \(n\) and \(k\), extending code tables by A. E. Brouwer \NoCaseChange{\protect\cite{cite1427}}, are maintained by M. Grassl at this \flmHref{https://www.codetables.de/}{website}.
\end{eczvaluelist}
\codefieldsection{Parents}
\begin{eczvaluelist}
\item\relax
\flmRefsHyperref[eczindexfamilyrel]{code:binary_group_orbit}{Binary group-orbit code} --- The set of codewords of a binary linear code can be thought of as an orbit of a particular codeword under the translation group formed by the code \NoCaseChange{\protect\cite[{Thm. 8.4.2}]{cite115}}. However, binary group-orbit codes do not have to be linear; see \NoCaseChange{\protect\cite[{Remark 8.4.3}]{cite115}}.
\item\relax
\flmRefsHyperref[eczindexfamilyrel]{code:q-ary_linear}{Linear \(q\)-ary code} --- Linear binary codes are linear \(q\)-ary codes for \(q=2\).
\item\relax
\flmRefsHyperref[eczindexfamilyrel]{code:q-ary_linear_over_zq}{Linear code over \(\mathbb{Z}_q\)} --- Linear binary codes are linear \(q\)-ary codes over \(\mathbb{Z}_q\) for \(q=2\).
\end{eczvaluelist}
\codefieldsection{Children}
\begin{eczvaluelist}
\item\relax
\flmRefsHyperref[eczindexfamilyrel]{code:anticode}{Anticode}\item\relax
\flmRefsHyperref[eczindexfamilyrel]{code:binary_cyclic}{Cyclic linear binary code}\item\relax
\flmRefsHyperref[eczindexfamilyrel]{code:karlin}{\([2m+2,m+1]\) Karlin code}\item\relax
\flmRefsHyperref[eczindexfamilyrel]{code:self_dual_48_24_12}{\([48,24,12]\) self-dual code}\item\relax
\flmRefsHyperref[eczindexfamilyrel]{code:cordaro_wagner}{\([n,2,\lceil 2n/3 \rceil -1]\) Cordaro-Wagner code}\item\relax
\flmRefsHyperref[eczindexfamilyrel]{code:graph}{Graph-adjacency code}\item\relax
\flmRefsHyperref[eczindexfamilyrel]{code:homological_classical}{Cycle code}\item\relax
\flmRefsHyperref[eczindexfamilyrel]{code:laplacian}{Laplacian code}\item\relax
\flmRefsHyperref[eczindexfamilyrel]{code:justesen}{Justesen code}\item\relax
\flmRefsHyperref[eczindexfamilyrel]{code:binary_ltc}{Binary linear LTC} --- Linear binary codes with distances \(\frac{1}{2}n-\sqrt{t n}\) for some \(t\) are called almost-orthogonal and are locally testable with query complexity of \flmRefsHyperref{ref65}{order} \(O(t)\) \NoCaseChange{\protect\cite{cite1270}}. This was later improved to codes with distance \(\frac{1}{2}n-O(n^{1-\gamma})\) for any positive \(\gamma\) \NoCaseChange{\protect\cite{cite1271}}, provided that the number of codewords is polynomial in \(n\).
\item\relax
\flmRefsHyperref[eczindexfamilyrel]{code:gray}{Gray code} --- A linear code \(C\) over \(\mathbb{Z}_4\) can be mapped, via the \flmTerm{term}{ref81}{}{Gray map}, to a binary code. The binary code is linear if and only if doubling the component-wise product of any two codewords in \(C\) yields another codeword in \(C\) \NoCaseChange{\protect\cite[{Thm. 12.2.3}]{cite126}}.
\item\relax
\flmRefsHyperref[eczindexfamilyrel]{code:fibonacci_model}{Fibonacci code}\item\relax
\flmRefsHyperref[eczindexfamilyrel]{code:gauss_law}{Gauss' law code}\item\relax
\flmRefsHyperref[eczindexfamilyrel]{code:newman_moore}{Newman-Moore code}\item\relax
\flmRefsHyperref[eczindexfamilyrel]{code:topological_classical}{Classical topological code}\item\relax
\flmRefsHyperref[eczindexfamilyrel]{code:berman}{Berman code}\item\relax
\flmRefsHyperref[eczindexfamilyrel]{code:coxeter}{Coxeter code}\item\relax
\flmRefsHyperref[eczindexfamilyrel]{code:polar}{Polar code}\item\relax
\flmRefsHyperref[eczindexfamilyrel]{code:ta-shma}{Ta-Shma zigzag code}\item\relax
\flmRefsHyperref[eczindexfamilyrel]{code:ldgm}{Low-density generator-matrix (LDGM) code}\item\relax
\flmRefsHyperref[eczindexfamilyrel]{code:ldpc}{Low-density parity-check (LDPC) code}\item\relax
\flmRefsHyperref[eczindexfamilyrel]{code:regular_binary_tanner}{Regular binary Tanner code}\end{eczvaluelist}
\codefieldsection{Cousins}
\begin{eczvaluelist}
\item\relax
\flmRefsHyperref[eczindexfamilyrel]{code:frustration_free}{Frustration-free Hamiltonian code} --- Parity-check constraints defining a binary linear code can be encoded into a classical Ising model Hamiltonian, a commuting-projector model whose terms contain products of Pauli \(Z\) matrices participating in each parity check. Such Ising models are also frustration-free since the codewords satisfy all parity checks.
\item\relax
\flmRefsHyperref[eczindexfamilyrel]{code:commuting_projector}{Commuting-projector Hamiltonian code} --- Parity-check constraints defining a binary linear code can be encoded into a classical Ising model Hamiltonian, a commuting-projector model whose terms contain products of Pauli \(Z\) matrices participating in each parity check. Such Ising models are also frustration-free since the codewords satisfy all parity checks.
\item\relax
\flmRefsHyperref[eczindexfamilyrel]{code:construction_a}{Construction A code} --- Every binary linear code yields a lattice under \flmTerm{term}{ref127}{}{Construction A}.
\item\relax
\flmRefsHyperref[eczindexfamilyrel]{code:constant_weight}{Constant-weight code} --- Nontrivial linear binary codes cannot have all codewords of the same weight, but they can have all nonzero codewords of the same weight. All such codes are equidistant, and Bonisoli's theorem states that any equidistant linear binary code is a direct sum of simplex codes \NoCaseChange{\protect\cite{cite988}} (see also Refs. \NoCaseChange{\protect\cite{cite45,cite46}}).
\item\relax
\flmRefsHyperref[eczindexfamilyrel]{code:parity_check}{\([n,n-1,2]\) Single parity-check (SPC) code} --- Any \([n,k,d]\) code with odd distance can be \textit{extended} to an \([n+1,k,d+1]\) code by adding a bit storing the sum of codeword coordinates.
\item\relax
\flmRefsHyperref[eczindexfamilyrel]{code:simplex}{\([2^m-1,m,2^{m-1}]\) simplex code} --- Linear binary codes cannot be constant weight, but can have nonzero codewords with constant weight. All such codes are equidistant, and Bonisoli's theorem states that any equidistant linear binary code is a direct sum of simplex codes \NoCaseChange{\protect\cite{cite988}} (see also Refs. \NoCaseChange{\protect\cite{cite45,cite46}}).
\item\relax
\flmRefsHyperref[eczindexfamilyrel]{code:lexicographic}{Lexicographic code} --- Binary lexicodes are linear \NoCaseChange{\protect\cite{cite147}}.
\item\relax
\flmRefsHyperref[eczindexfamilyrel]{code:quaternary_over_z4}{Linear code over \(\mathbb{Z}_4\)} --- A linear quaternary code over \(\mathbb{Z}_4\) of length \(n\), type \(4^{k_1}2^{k_2}\), and minimum Lee weight \(d\) maps under the \flmTerm{term}{ref81}{}{Gray map} to a binary code of length \(2n\), cardinality \(2^{2k_1+k_2}\), and minimum Hamming weight \(d\) \NoCaseChange{\protect\cite[{Sec. 6.3}]{cite1145}}.
\item\relax
\flmRefsHyperref[eczindexfamilyrel]{code:slepian_group}{Slepian group-orbit code} --- Any length-\(n\) binary linear code can be used to define a diagonal subgroup of \(n\)-dimensional rotation matrices with \(\pm 1\) on the diagonals via the \flmRefsHyperref{ref38}{antipodal mapping} \(0\to+1\) and \(1\to-1\). The orbit of this subgroup yields the corresponding Slepian group-orbit code; see \NoCaseChange{\protect\cite[{Thm. 8.5.2}]{cite115}}.
\item\relax
\flmRefsHyperref[eczindexfamilyrel]{code:bpsk}{Binary PSK (BPSK) modulation format} --- Concatenating binary linear codes with BPSK yields a standard way of digitizing the analog AWGN channel \NoCaseChange{\protect\cite[{Ch. 29}]{cite194}}.
\item\relax
\flmRefsHyperref[eczindexfamilyrel]{code:lca_stabilizer}{Locally compact Abelian (LCA) stabilizer code} --- Linear binary codes can be used to construct LCA stabilizer codes \NoCaseChange{\protect\cite{cite1428}}.
\item\relax
\flmRefsHyperref[eczindexfamilyrel]{code:spacetime_circuit}{Spacetime circuit code} --- The set of measurement outcomes of a \flmRefsHyperref{ref409}{Clifford circuit} can be made into a classical binary linear code.
Error syndromes of the spacetime circuit code can be used to obtain the parity checks of the outcome code.

\item\relax
\flmRefsHyperref[eczindexfamilyrel]{code:eastab}{EA qubit stabilizer code} --- Any linear binary code can be used to construct an EA qubit stabilizer code \NoCaseChange{\protect\cite{cite1429,cite1430,cite1431}}.
\item\relax
\flmRefsHyperref[eczindexfamilyrel]{code:majorana_stab}{Majorana stabilizer code} --- When constructing a Majorana stabilizer code from a self-orthogonal classical code with an odd number of bits and generator matrix \(G\), a more complex procedure must be applied to ensure that the fermion code has an even number of Majorana zero modes, and thus a physical Hilbert space \NoCaseChange{\protect\cite{cite1432,cite566}}. Rather than taking \(G\) to be the stabilizer matrix as in the even case, we take \(G\oplus G\). This is a concatenation of classical codes as in the CSS construction and it yields a mapping \([2n-1,k,d]\rightarrow \llbracket 2n-1,2n-1-k,d^\perp\rrbracket _f\). This procedure may be further generalized by concatenating two different self-orthogonal classical codes with an odd number of bits, as is often done in the CSS construction.
\item\relax
\flmRefsHyperref[eczindexfamilyrel]{code:cpc}{Coherent-parity-check (CPC) code} --- The CPC Construction uses two binary linear codes.
\item\relax
\flmRefsHyperref[eczindexfamilyrel]{code:data_syndrome}{Quantum data-syndrome (QDS) code} --- The QDS code construction employs a particular binary linear code to provide protection against syndrome measurement errors.
\item\relax
\flmRefsHyperref[eczindexfamilyrel]{code:css-t}{CSS-T code} --- CSS-T codes are constructed from a pair of linear binary codes via the CSS construction, with the pair satisfying certain conditions \NoCaseChange{\protect\cite{cite1315}}.
\item\relax
\flmRefsHyperref[eczindexfamilyrel]{code:multisector_hypergraph}{Higher-dimensional homological product code} --- \(D\)-dimensional HGP codes are constructed using a hypergraph product of \(D\) linear binary codes.
\item\relax
\flmRefsHyperref[eczindexfamilyrel]{code:xyz_product}{XYZ product code} --- The XYZ product code is a non-CSS three-fold variant of the hypergraph product built from three classical linear binary codes \NoCaseChange{\protect\cite{cite645}}.
\item\relax
\flmRefsHyperref[eczindexfamilyrel]{code:qubit_css}{Qubit CSS code} --- The CSS construction uses two related binary linear codes, \(C_X\) and \(C_Z\).
\item\relax
\flmRefsHyperref[eczindexfamilyrel]{code:qubit_stabilizer}{Qubit stabilizer code} --- Qubit stabilizer codes are the closest quantum analogues of binary linear codes because addition modulo two corresponds to multiplication of stabilizers in the quantum case. 
Any binary linear code can be thought of as a qubit stabilizer code with \(Z\)-type stabilizer generators \NoCaseChange{\protect\cite{cite1434}\protect\cite[{Table I}]{cite1433}}. 
The stabilizer generators are extracted from rows of the parity-check matrix, while logical \(X\) Paulis correspond to rows of the generator matrix.
States close to the equal superposition of all bitstrings within Hamming distance \(b\) of a binary linear code can be prepared efficiently \NoCaseChange{\protect\cite{cite1435}}.
Binary linear codes can be used for error-corrected entanglement distillation protocols \NoCaseChange{\protect\cite{cite1436}}.

\item\relax
\flmRefsHyperref[eczindexfamilyrel]{code:2d_color}{2D color code} --- As CSS codes, variants of the 2D color code are constructed out of self-dual classical codes on cubic planar graphs \NoCaseChange{\protect\cite{cite1437}}.
\item\relax
\flmRefsHyperref[eczindexfamilyrel]{code:fractal_surface}{Fractal surface code} --- The fractal product code is a hypergraph product of two classical codes defined on a Sierpinski carpet graph \NoCaseChange{\protect\cite{cite676}}.
\item\relax
\flmRefsHyperref[eczindexfamilyrel]{code:subsystem_lifted_product}{Subsystem lifted-product (SLP) code} --- SLP codes are constructed from a subsystem hypergraph product of a \flmRefsHyperref{ref47}{lifted} binary linear code.
\end{eczvaluelist}
\eczhbkcontributors{ \eczhuVVA }
\endeczcode

\eczcode{long}{Long code}{~\NoCaseChange{\protect\cite{cite1438,cite1439,cite1440}}}
\codefieldsection{Description}
Nonlinear locally testable code of extremely large length that is not practical, but is useful for certain probabilistically checkable proof (PCP) constructions \NoCaseChange{\protect\cite{cite89}}. 

For \(x\in\mathbb{F}_2^k\), the long-code encoding of \(x\) is the binary string
\flmMathEnvironment{align}{}{
  \mathrm{Long}(x)=\left(f(x)\right)_{f:\mathbb{F}_2^k\to\mathbb{F}_2}~,
}
whose coordinates are indexed by all Boolean functions on \(\mathbb{F}_2^k\).
Thus, the code has length \(2^{2^k}\) and \(2^k\) codewords.

\codefieldsection{Protection}
Any two distinct codewords differ on exactly half of the coordinates, so the minimum distance is \(2^{2^k-1}\).

\codefieldsection{Parent}
\begin{eczvaluelist}
\item\relax
\flmRefsHyperref[eczindexfamilyrel]{code:binary_ltc}{Binary linear LTC}\end{eczvaluelist}
\codefieldsection{Cousin}
\begin{eczvaluelist}
\item\relax
\flmRefsHyperref[eczindexfamilyrel]{code:hadamard}{\([2^m,m,2^{m-1}]\) Hadamard code} --- The Hadamard code is a subcode of the long code and can be obtained by restricting the long-code construction to only linear functions.
\end{eczvaluelist}
\eczhbkcontributors{ \eczhuVVA }
\endeczcode

\eczcode{ldgm}{Low-density generator-matrix (LDGM) code}{}
\codefieldsection{Description}
Binary linear code with a sparse generator matrix.
Alternatively, a member of an infinite family of \([n,k,d]\) codes for which the number of nonzero entries in each row and column of the generator matrix are both bounded by a constant as \(n\to\infty\). The dual of an LDGM code has a sparse parity-check matrix and is called an LDPC code.

\codefieldsection{Rate}
Certain LDGM codes come close to achieving Shannon capacity \NoCaseChange{\protect\cite{cite1441}}.
\codefieldsection{Parents}
\begin{eczvaluelist}
\item\relax
\flmRefsHyperref[eczindexfamilyrel]{code:binary_linear}{Linear binary code}\item\relax
\flmRefsHyperref[eczindexfamilyrel]{code:q-ary_ldgm}{\(q\)-ary LDGM code}\end{eczvaluelist}
\codefieldsection{Child}
\begin{eczvaluelist}
\item\relax
\flmRefsHyperref[eczindexfamilyrel]{code:fountain}{Fountain code}\end{eczvaluelist}
\codefieldsection{Cousins}
\begin{eczvaluelist}
\item\relax
\flmRefsHyperref[eczindexfamilyrel]{code:dual}{Dual linear code} --- The dual of an LDPC code has a sparse generator matrix and is called an LDGM code.
\item\relax
\flmRefsHyperref[eczindexfamilyrel]{code:parity_check}{\([n,n-1,2]\) Single parity-check (SPC) code} --- Concatenated SPCs are LDGM \NoCaseChange{\protect\cite{cite1210}}.
\item\relax
\flmRefsHyperref[eczindexfamilyrel]{code:ira}{Irregular repeat-accumulate (IRA) code} --- IRA codes replace the outer 1-in-3 repetition encoding step in RA codes with an LDGM code.
\item\relax
\flmRefsHyperref[eczindexfamilyrel]{code:ldpc}{Low-density parity-check (LDPC) code} --- The dual of an LDPC code has a sparse generator matrix and is called an LDGM code.
\item\relax
\flmRefsHyperref[eczindexfamilyrel]{code:ha_ldpc}{Hsu-Anastasopoulos LDPC (HA-LDPC) code} --- HA-LDPC codes are a concatenation of an LDPC and an LDGM code.
\item\relax
\flmRefsHyperref[eczindexfamilyrel]{code:mn_ldpc}{MacKay-Neal LDPC (MN-LDPC) code} --- \((l,r,1\))-MN-LDPC codes are LDGM \NoCaseChange{\protect\cite{cite84}}.
\item\relax
\flmRefsHyperref[eczindexfamilyrel]{code:qldpc}{Qubit QLDPC code} --- LDGM codes can yield CSS \NoCaseChange{\protect\cite{cite1442,cite1443,cite1444,cite1445}} and non-CSS \NoCaseChange{\protect\cite{cite1446,cite1447}} qubit QLDPC codes. Some of the LDGM-based CSS codes have \(n\)-independent minimum distance and no code capacity threshold \NoCaseChange{\protect\cite[{Sec. 4.2}]{cite1448}}.
\end{eczvaluelist}
\eczhbkcontributors{ \eczhuVVA }
\endeczcode

\eczcode{ldpc}{Low-density parity-check (LDPC) code}{~\NoCaseChange{\protect\cite{cite1360,cite1361,cite1449}}}
\codefieldsection{Alternative Names}
\begin{eczvaluelist}
\item\relax Sparse graph code
\end{eczvaluelist}
\eczhIndexCodeAliasName{ldpc}{Sparse graph code}
\codefieldsection{Description}
A binary linear code with a sparse parity-check matrix.
Often a member of an infinite family of \([n,k,d]\) codes for which the numbers of nonzero entries in each row and in each column of the parity-check matrix are both bounded above by a constant as \(n\to\infty\).

An LDPC code is \((j,k)\)-\textit{regular} if the parity-check matrix has a fixed number of \(j\) nonzero entries in each column and \(k\) entries in each row; otherwise, the LDPC code is \textit{irregular}.
Irregular LDPC codes are characterized by the fractions \(v_j\) of columns of weight \(j\) as well as the fractions \(h_j\) of rows of weight \(j\); these are collectively called the \textit{degree distribution}.

A \textit{parity check} is performed by taking the inner product of a row of the parity-check matrix with a codeword that has been affected by a noise channel.
A parity check yields either zero (the parity constraint is satisfied) or one (the parity constraint is violated).

In alternative conventions (not used here), LDPC codes are referred to as \textit{simple LDPC} codes, as opposed to generalized LDPC codes, also called Tanner codes.

\codefieldsection{Rate}
Some LDPC codes achieve the Shannon capacity of the binary symmetric channel under maximum-likelihood decoding \NoCaseChange{\protect\cite{cite1364,cite1360,cite1450}}. Other LDPC codes achieve capacity for smaller block lengths under belief-propagation decoding \NoCaseChange{\protect\cite{cite1451}}. Random LDPC codes achieve list-decoding capacity \NoCaseChange{\protect\cite{cite1452}}.
\codefieldsection{Encoding}
\begin{eczvaluelist}
\item\relax Almost linear-time encoder based on transforming the parity-check matrix into upper triangular form \NoCaseChange{\protect\cite{cite1453}}.
\end{eczvaluelist}
\codefieldsection{Decoding}
\begin{eczvaluelist}
\item\relax Gallager introduced a hard-decision iterative decoder that flips any digit contained in more than a fixed number of unsatisfied parity checks \NoCaseChange{\protect\cite[{Ch. 4}]{cite1361}}.
\item\relax Gallager also introduced an iterative probabilistic decoding procedure on the computation tree; this is a precursor to modern message-passing, belief-propagation (BP), and sum-product decoding \NoCaseChange{\protect\cite{cite1454,cite1455,cite1456,cite1395}\protect\cite[{Ch. 4}]{cite1361}} (see also \NoCaseChange{\protect\cite{cite1241,cite1457}}).
\item\relax Soft-decision Sum-Product Algorithm (SPA) \NoCaseChange{\protect\cite{cite1456,cite1395}} and its simplification the Min-Sum Algorithm (MSA) \NoCaseChange{\protect\cite{cite1458}}.
\item\relax Linear programming \NoCaseChange{\protect\cite{cite1459,cite1460,cite1461}}.
\item\relax Iterative decoder using stochastic computation \NoCaseChange{\protect\cite{cite1462}}.
\item\relax Iterative LDPC decoders can get stuck at \textit{stopping sets} of their Tanner graphs \NoCaseChange{\protect\cite{cite1463}}, with decoder performance improving with the size of the smallest stopping set; see \NoCaseChange{\protect\cite[{Sec. 21.3.1}]{cite97}} for more details. The smallest stopping set size can reach the minimum distance of the code \NoCaseChange{\protect\cite{cite1464}}.
\item\relax Ensembles of random LDPC codes under iterative decoders are subject to the \textit{concentration theorem} \NoCaseChange{\protect\cite{cite1395,cite1389}}; see \NoCaseChange{\protect\cite[{Thm. 21.7.1}]{cite97}} for the case of the BEC.
\item\relax Reinforcement learning \NoCaseChange{\protect\cite{cite1465}}.
\item\relax Quantum-enhanced BP decoding \NoCaseChange{\protect\cite{cite1466}}.
\end{eczvaluelist}
\codefieldsection{Notes}
\begin{eczvaluelist}
\item\relax The potential of LDPC codes was noted by Margulis \NoCaseChange{\protect\cite{cite49}}, but realized by the broader community \NoCaseChange{\protect\cite{cite1363,cite1364}} much later after their discovery by Gallager \NoCaseChange{\protect\cite{cite1360,cite1361}}.
\item\relax See books \NoCaseChange{\protect\cite{cite1467,cite1468,cite946}} and reviews \NoCaseChange{\protect\cite{cite1469,cite1470,cite1471,cite1472}} for introductions to LDPC codes, belief-propagation decoding, and connections to statistical mechanics. Other introductory references include Refs. \NoCaseChange{\protect\cite{cite1473,cite94,cite1474,cite1475}} as well as a review of LDPC codes circa 2005 \NoCaseChange{\protect\cite{cite1476}}.
\item\relax See Kaiserslautern database \NoCaseChange{\protect\cite{cite1184}} for explicit representatives of several classes of LDPC codes, including \(q\)-ary, WiMAX, multi-edge, and spatially-coupled.
\item\relax See pretty-good-codes database \NoCaseChange{\protect\cite{cite1477}} for explicit representatives and benchmarking.
\item\relax See the Encyclopedia of sparse graph codes for explicit representatives \NoCaseChange{\protect\cite{cite1478}}
\item\relax LDPC codes have been considered for quantum key distribution \NoCaseChange{\protect\cite{cite1479}}.
\item\relax Codes have been benchmarked using AFF3CT toolbox \NoCaseChange{\protect\cite{cite1480}}.
\item\relax LDPC Python software library for decoding LDPC and QLDPC codes \NoCaseChange{\protect\cite{cite1481,cite1482}}.
\end{eczvaluelist}
\codefieldsection{Parents}
\begin{eczvaluelist}
\item\relax
\flmRefsHyperref[eczindexfamilyrel]{code:binary_linear}{Linear binary code}\item\relax
\flmRefsHyperref[eczindexfamilyrel]{code:q-ary_ldpc}{\(q\)-ary LDPC code}\end{eczvaluelist}
\codefieldsection{Children}
\begin{eczvaluelist}
\item\relax
\flmRefsHyperref[eczindexfamilyrel]{code:tornado}{Tornado code} --- Tornado codes are LDPC codes that use a highly irregular weight distribution for their underlying graphs \NoCaseChange{\protect\cite{cite257}}.
\item\relax
\flmRefsHyperref[eczindexfamilyrel]{code:pinwheel}{Pinwheel code}\item\relax
\flmRefsHyperref[eczindexfamilyrel]{code:algebraic_ldpc}{Algebraic LDPC code}\item\relax
\flmRefsHyperref[eczindexfamilyrel]{code:multi_edge_ldpc}{Multi-edge LDPC code}\item\relax
\flmRefsHyperref[eczindexfamilyrel]{code:regular_ldpc}{Regular LDPC code}\end{eczvaluelist}
\codefieldsection{Cousins}
\begin{eczvaluelist}
\item\relax
\flmRefsHyperref[eczindexfamilyrel]{code:tensor}{Tensor-product code} --- Tensor products of random LDPC codes are robustly testable \NoCaseChange{\protect\cite{cite1483,cite1484}}.
\item\relax
\flmRefsHyperref[eczindexfamilyrel]{code:ldgm}{Low-density generator-matrix (LDGM) code} --- The dual of an LDPC code has a sparse generator matrix and is called an LDGM code.
\item\relax
\flmRefsHyperref[eczindexfamilyrel]{code:dual}{Dual linear code} --- The dual of an LDPC code has a sparse generator matrix and is called an LDGM code.
\item\relax
\flmRefsHyperref[eczindexfamilyrel]{code:random}{Random code} --- LDPC codes are often constructed nondeterministically.
\item\relax
\flmRefsHyperref[eczindexfamilyrel]{code:hamiltonian}{Hamiltonian-based code} --- There are relations between LDPC codes and statistical mechanical models of spin glasses \NoCaseChange{\protect\cite{cite1485,cite1467,cite1468,cite946}}.
\item\relax
\flmRefsHyperref[eczindexfamilyrel]{code:laplacian}{Laplacian code} --- Laplacian codes on bounded-degree graph families are classical LDPC codes.
\item\relax
\flmRefsHyperref[eczindexfamilyrel]{code:lrpc}{Low-rank parity-check (LRPC) code} --- LRPC codes are rank-metric analogues of LDPC codes \NoCaseChange{\protect\cite{cite287}}.
\item\relax
\flmRefsHyperref[eczindexfamilyrel]{code:dna}{DNA storage code} --- LDPC codes are potentially relevant for DNA storage \NoCaseChange{\protect\cite{cite1006}}.
\item\relax
\flmRefsHyperref[eczindexfamilyrel]{code:ld_convolutional}{LDPC convolutional code (LDPC-CC)} --- LDPC-CCs are convolutional analogues of LDPC codes.
\item\relax
\flmRefsHyperref[eczindexfamilyrel]{code:lhz}{Lechner-Hauke-Zoller (LHZ) code} --- The LHZ code is an LDPC c-q code designed to convert the long-range interactions of a quantum annealer into local constraints.
\item\relax
\flmRefsHyperref[eczindexfamilyrel]{code:cat_concatenated}{Concatenated cat code} --- Cat codes have been concatenated with LDPC codes (treated as qubit stabilizer codes) \NoCaseChange{\protect\cite{cite1486}}.
\item\relax
\flmRefsHyperref[eczindexfamilyrel]{code:self_correct}{Self-correcting quantum code} --- Linear confinement of QLDPC (LDPC) codes implies (classical) self-correction \NoCaseChange{\protect\cite{cite849}}.
\item\relax
\flmRefsHyperref[eczindexfamilyrel]{code:spacetime_circuit}{Spacetime circuit code} --- There is an equivalence between \flmRefsHyperref{ref409}{Clifford circuits} and LDPC codes with bit-check symmetry \NoCaseChange{\protect\cite{cite1487}}.
\item\relax
\flmRefsHyperref[eczindexfamilyrel]{code:ea_qldpc}{EA QLDPC code} --- There exist necessary and sufficient conditions for an EA QLDPC code consuming \(e=1\) ebit that is obtainable from a pair of LDPC codes \NoCaseChange{\protect\cite{cite1488}}.
\item\relax
\flmRefsHyperref[eczindexfamilyrel]{code:lresc}{Long-range enhanced surface code (LRESC)} --- LRESCs are constructed using a hypergraph product of two copies of a concatenated LDPC-repetition seed code.
\item\relax
\flmRefsHyperref[eczindexfamilyrel]{code:qldpc}{Qubit QLDPC code} --- Qubit QLDPC codes are quantum analogues of binary LDPC codes.
\end{eczvaluelist}
\eczhbkcontributors{ Armin Gerami, \eczhuVVA }
\endeczcode

\eczcode{luby_transform}{Luby transform (LT) code}{~\NoCaseChange{\protect\cite{cite1489}}}
\codefieldsection{Description}
Erasure codes based on fountain codes. They improve on random linear fountain codes by having a much more efficient encoding and decoding algorithm.

LT codes can be constructed as follows. First, randomly choose a degree \(d_n\) from a degree distribution depending on total size \(K\). Then, randomly choose \(d_n\) distinct source packets and let the packet to be transmitted \(\hat{p}_n\) be the bitwise sum of the chosen input packets. This forms a graph connecting encoded packets to source packets.

\codefieldsection{Decoding}
\begin{eczvaluelist}
\item\relax Sum-Product Algorithm (SPA), often called a peeling decoder \NoCaseChange{\protect\cite{cite1490,cite1491}}, similar to belief propagation \NoCaseChange{\protect\cite{cite1492}}.
\end{eczvaluelist}
\codefieldsection{Parent}
\begin{eczvaluelist}
\item\relax
\flmRefsHyperref[eczindexfamilyrel]{code:raptor}{Raptor (RAPid TORnado) code} --- Raptor codes using a trivial pre-code are LT codes. Typically, Raptor codes have constant-sized overhead but are faster to decode.
\end{eczvaluelist}
\eczhbkcontributors{ Thomas Wrona, Noah Berthusen, \eczhuVVA }
\endeczcode

\eczcode{mn_ldpc}{MacKay-Neal LDPC (MN-LDPC) code}{~\NoCaseChange{\protect\cite{cite1493,cite1364}}}
\codefieldsection{Description}
A code whose parity-check matrix is constructed non-deterministically via the MacKay-Neal prescription.

An \((l,r,g)\)-MN-LDPC code is a non-systematic two-edge-type LDPC code whose parity-check matrix is of the form \((H_1~H_2)\), where \(H_1\) is a random binary matrix of column weight \(l\) and row weight \(r\), and \(H_2\) is a random binary matrix of column and row weight \(g\); the bits corresponding to \(H_1\) are punctured \NoCaseChange{\protect\cite{cite84}}.

\codefieldsection{Rate}
Certain sequences of optimally decoded codes can nearly achieve the Shannon capacity \NoCaseChange{\protect\cite{cite1493,cite1364}}. A sequence of codes achieves the capacity of memoryless binary-input symmetric-output channels under MAP decoding \NoCaseChange{\protect\cite{cite84}}. Standard MN-LDPC codes have no BP threshold, while bounded-density spatially coupled MN-LDPC codes have BEC BP thresholds close to the Shannon limit \NoCaseChange{\protect\cite{cite84}}.
\codefieldsection{Decoding}
\begin{eczvaluelist}
\item\relax Free-energy minimization and a BP decoder \NoCaseChange{\protect\cite{cite1493}}.
\end{eczvaluelist}
\codefieldsection{Parents}
\begin{eczvaluelist}
\item\relax
\flmRefsHyperref[eczindexfamilyrel]{code:regular_ldpc}{Regular LDPC code} --- MN-LDPC codes re-invigorated the study of LDPC codes about 30 years after their discovery.
\item\relax
\flmRefsHyperref[eczindexfamilyrel]{code:multi_edge_ldpc}{Multi-edge LDPC code} --- MN-LDPC codes can be formulated as multi-edge LDPC codes \NoCaseChange{\protect\cite{cite84}}.
\item\relax
\flmRefsHyperref[eczindexfamilyrel]{code:random}{Random code}\end{eczvaluelist}
\codefieldsection{Cousins}
\begin{eczvaluelist}
\item\relax
\flmRefsHyperref[eczindexfamilyrel]{code:ldgm}{Low-density generator-matrix (LDGM) code} --- \((l,r,1\))-MN-LDPC codes are LDGM \NoCaseChange{\protect\cite{cite84}}.
\item\relax
\flmRefsHyperref[eczindexfamilyrel]{code:ira}{Irregular repeat-accumulate (IRA) code} --- MN-LDPC and IRA codes intersect for certain parameters \NoCaseChange{\protect\cite{cite1400}}.
\item\relax
\flmRefsHyperref[eczindexfamilyrel]{code:ha_ldpc}{Hsu-Anastasopoulos LDPC (HA-LDPC) code} --- HA-LDPC and MN-LDPC codes are dual to each other \NoCaseChange{\protect\cite{cite84}}.
\item\relax
\flmRefsHyperref[eczindexfamilyrel]{code:sc_ldpc}{Spatially coupled LDPC (SC-LDPC) code} --- Spatial coupling of MN-LDPC protographs yields bounded-density SC-MN codes with BEC BP thresholds close to the Shannon limit \NoCaseChange{\protect\cite{cite84}}.
\end{eczvaluelist}
\eczhbkcontributors{ \eczhuVVA }
\endeczcode

\eczcode{margulis_ldpc}{Margulis LDPC code}{~\NoCaseChange{\protect\cite{cite49,cite90}}}
\codefieldsection{Alternative Names}
\begin{eczvaluelist}
\item\relax Margulis-Gabber-Galil LDPC code
\end{eczvaluelist}
\eczhIndexCodeAliasName{margulis_ldpc}{Margulis-Gabber-Galil LDPC code}
\codefieldsection{Description}
Member of a class of LDPC codes deterministically constructed from explicit sparse regular expander graphs.
The underlying Margulis-Gabber-Galil graph family provides explicit expanders \NoCaseChange{\protect\cite{cite49,cite90}}, yielding deterministic sparse parity-check matrices.
Related explicit LDPC constructions \NoCaseChange{\protect\cite{cite91}} utilize Ramanujan graphs \NoCaseChange{\protect\cite{cite75,cite76}}.

\codefieldsection{Parents}
\begin{eczvaluelist}
\item\relax
\flmRefsHyperref[eczindexfamilyrel]{code:cycle_ldpc}{Cycle LDPC code} --- Margulis LDPC codes are examples of cycle codes for particular large-girth graphs \NoCaseChange{\protect\cite{cite1304}}.
\item\relax
\flmRefsHyperref[eczindexfamilyrel]{code:algebraic_ldpc}{Algebraic LDPC code}\end{eczvaluelist}
\eczhbkcontributors{ \eczhuVVA }
\endeczcode

\eczcode{multi_edge_ldpc}{Multi-edge LDPC code}{~\NoCaseChange{\protect\cite{cite1394}}}
\codefieldsection{Description}
Irregular LDPC code whose construction generalizes those of the original examples of irregular LDPC as well as RA codes.

\codefieldsection{Realizations}
\begin{eczvaluelist}
\item\relax Quantum key distribution \NoCaseChange{\protect\cite{cite294,cite295,cite296}}.
\end{eczvaluelist}
\codefieldsection{Parent}
\begin{eczvaluelist}
\item\relax
\flmRefsHyperref[eczindexfamilyrel]{code:ldpc}{Low-density parity-check (LDPC) code}\end{eczvaluelist}
\codefieldsection{Children}
\begin{eczvaluelist}
\item\relax
\flmRefsHyperref[eczindexfamilyrel]{code:irregular_ldpc}{Irregular LDPC code} --- The multi-edge code construction generalizes several of the original examples of irregular LDPC codes. Irregular LDPC codes can be formulated as multi-edge LDPC codes \NoCaseChange{\protect\cite[{Sec. XI}]{cite1394}}.
\item\relax
\flmRefsHyperref[eczindexfamilyrel]{code:protograph_ldpc}{Protograph LDPC code} --- LDPC codes based on protographs can be formulated as multi-edge LDPC codes \NoCaseChange{\protect\cite{cite1494}}.
\item\relax
\flmRefsHyperref[eczindexfamilyrel]{code:ha_ldpc}{Hsu-Anastasopoulos LDPC (HA-LDPC) code} --- HA-LDPC codes can be formulated as multi-edge LDPC codes \NoCaseChange{\protect\cite{cite84}}.
\item\relax
\flmRefsHyperref[eczindexfamilyrel]{code:mn_ldpc}{MacKay-Neal LDPC (MN-LDPC) code} --- MN-LDPC codes can be formulated as multi-edge LDPC codes \NoCaseChange{\protect\cite{cite84}}.
\end{eczvaluelist}
\eczhbkcontributors{ \eczhuVVA }
\endeczcode

\eczcode{nearly_perfect}{Nearly perfect code}{~\NoCaseChange{\protect\cite{cite1495,cite1496,cite1169}}}
\codefieldsection{Description}
A binary code whose parameters satisfy the Johnson bound with equality.

An \((n,K,2t+1)\) binary code is nearly perfect if parameters \(n\), \(K\), and \(t\) are such that the Johnson bound \NoCaseChange{\protect\cite{cite1495}},
\flmMathEnvironment{align}{}{
  \frac{{n \choose t}\left(\frac{n-t}{t+1}-\left\lfloor \frac{n-t}{t+1}\right\rfloor \right)}{\left\lfloor \frac{n}{t+1}\right\rfloor }+\sum_{j=0}^{t}{n \choose j}\leq2^{n}/K
}
becomes an equality \NoCaseChange{\protect\cite[{Sec. 2.3.5}]{cite126}\protect\cite[{Ch. 17}]{cite41}}.
All nearly perfect binary codes are either perfect or correspond to shortened Preparata codes or a particular family of distance-three nonlinear codes \NoCaseChange{\protect\cite{cite1497}}.

Similar definitions can be made for \(q\)-ary codes, but all nearly perfect \(q\)-ary codes must be perfect \NoCaseChange{\protect\cite{cite1498,cite1499}}.

\codefieldsection{Parents}
\begin{eczvaluelist}
\item\relax
\flmRefsHyperref[eczindexfamilyrel]{code:bits_into_bits}{Binary code}\item\relax
\flmRefsHyperref[eczindexfamilyrel]{code:quasi_perfect}{Quasi-perfect code} --- Nearly perfect codes are quasi-perfect \NoCaseChange{\protect\cite[{pg. 533}]{cite41}}.
\end{eczvaluelist}
\codefieldsection{Children}
\begin{eczvaluelist}
\item\relax
\flmRefsHyperref[eczindexfamilyrel]{code:perfect_binary}{Perfect binary code} --- Perfect binary codes are nearly perfect, and \(t+1\) divides \(n-t\) for such codes. In addition, any perfect code can be extended to a nearly perfect code.
\item\relax
\flmRefsHyperref[eczindexfamilyrel]{code:extended_golay}{\([24, 12, 8]\) Extended Golay code} --- The extended Golay code is nearly perfect.
\item\relax
\flmRefsHyperref[eczindexfamilyrel]{code:parity_check}{\([n,n-1,2]\) Single parity-check (SPC) code}\item\relax
\flmRefsHyperref[eczindexfamilyrel]{code:repetition}{Repetition code}\end{eczvaluelist}
\codefieldsection{Cousins}
\begin{eczvaluelist}
\item\relax
\flmRefsHyperref[eczindexfamilyrel]{code:combinatorial_design}{Combinatorial design} --- The minimum-weight codewords in a nearly perfect code containing the zero vector support a \(t\)-\((n,2t+1,\lfloor (n-t)/(t+1) \rfloor)\) design, while the minimum-weight codewords in the extended code support a \((t+1)\)-\((n+1,2t+2,\lfloor (n-t)/(t+1) \rfloor)\) design \NoCaseChange{\protect\cite[{Thm. 5.5.4}]{cite135}}.
\item\relax
\flmRefsHyperref[eczindexfamilyrel]{code:preparata}{Preparata code} --- Shortened Preparata codes are uniformly packed and nearly perfect \NoCaseChange{\protect\cite{cite1496}}. For any word \(u\) and shortened Preparata codebook \(C\) with \(d(u, C) > 2\), we have that \(u\) has distance 2 or 3 to exactly \(\left\lfloor (2^{m+1}-1)/3\right\rfloor\) codewords.
\item\relax
\flmRefsHyperref[eczindexfamilyrel]{code:nordstrom_robinson}{\((16,256,6)\) Nordstrom-Robinson (NR) code} --- The punctured \((15,256,5)\) NR code saturates the Johnson bound and is therefore nearly perfect \NoCaseChange{\protect\cite[{Exam. 5.5.5}]{cite135}}.
\item\relax
\flmRefsHyperref[eczindexfamilyrel]{code:hamming}{\([2^r-1,2^r-r-1,3]\) Hamming code} --- Shortened Hamming codes \([2^r-2,2^r-r-2,3]\) are nearly perfect \NoCaseChange{\protect\cite[{pg. 533}]{cite41}}.
\end{eczvaluelist}
\eczhbkcontributors{ \eczhuVVA }
\endeczcode

\eczcode{newman_moore}{Newman-Moore code}{~\NoCaseChange{\protect\cite{cite1500}}}
\codefieldsection{Description}
Member of a family of \([L^2,O(L),O(L^{\frac{\log 3}{\log 2}})]\) binary linear codes on \(L\times L\) square lattices that form the ground-state subspace of a class of exactly solvable spin-glass models with three-body interactions.
The codewords resemble the Sierpinski triangle on a square lattice, which can be generated by a cellular automaton \NoCaseChange{\protect\cite{cite92}}.

The code is not cyclic, but admits a bivariate \flmRefsHyperref{ref67}{polynomial representation} \(1+x+y\) for its generator matrix.

\codefieldsection{Protection}
Code parameters nearly saturate the classical version of the \flmRefsHyperref{ref487}{BPT bound}, based on numerical simulations and analytical arguments \NoCaseChange{\protect\cite[{Appx. A}]{cite1419}}.
\codefieldsection{Decoding}
\begin{eczvaluelist}
\item\relax Efficient decoder \NoCaseChange{\protect\cite{cite92}}.
\end{eczvaluelist}
\codefieldsection{Parents}
\begin{eczvaluelist}
\item\relax
\flmRefsHyperref[eczindexfamilyrel]{code:binary_linear}{Linear binary code}\item\relax
\flmRefsHyperref[eczindexfamilyrel]{code:classical_fractal_liquid}{Classical fractal liquid code}\end{eczvaluelist}
\codefieldsection{Cousin}
\begin{eczvaluelist}
\item\relax
\flmRefsHyperref[eczindexfamilyrel]{code:sierpinsky_fractal_liquid}{Sierpinski prism model code} --- The Sierpinski prism model code is a hypergraph product of the repetition code and the Newman-Moore code \NoCaseChange{\protect\cite{cite1501,cite1350}}.
\end{eczvaluelist}
\eczhbkcontributors{ \eczhuVVA }
\endeczcode

\eczcode{one_hot}{One-hot code}{~\NoCaseChange{\protect\cite{cite1502}}}
\codefieldsection{Alternative Names}
\begin{eczvaluelist}
\item\relax One-vs-all (OVA) code
\item\relax One-against-all (1AA) code
\item\relax One-vs-rest (OvR) code
\item\relax \(1\)-in-\(n\) code
\end{eczvaluelist}
\eczhIndexCodeAliasName{one_hot}{One-vs-all (OVA) code}
\eczhIndexCodeAliasName{one_hot}{One-against-all (1AA) code}
\eczhIndexCodeAliasName{one_hot}{One-vs-rest (OvR) code}
\eczhIndexCodeAliasName{one_hot}{\(1\)-in-\(n\) code}
\codefieldsection{Description}
A nonlinear binary code whose codewords are all those with Hamming weight one. The reverse of this code, where all codewords have Hamming weight \(n-1\) is called a \textit{one-cold} code.

\codefieldsection{Realizations}
\begin{eczvaluelist}
\item\relax The bi-quinary code, a combination of one-hot 1-in-2 and 1-in-5 one-hot codes to encode decimal digits, was used in several early computers \NoCaseChange{\protect\cite[{Ch. 27}]{cite297}}.
\item\relax Marking the state of a finite automaton \NoCaseChange{\protect\cite{cite298}}.
\item\relax Encoding the output of the different classes of a classifier neural network \NoCaseChange{\protect\cite{cite299}}. One-hot codes are the primary codes used in multiclass classification \NoCaseChange{\protect\cite{cite300,cite301,cite302,cite303}}.
\end{eczvaluelist}
\codefieldsection{Parents}
\begin{eczvaluelist}
\item\relax
\flmRefsHyperref[eczindexfamilyrel]{code:binary_group_orbit}{Binary group-orbit code} --- The one-hot code is a binary group-orbit code with the cyclic permutation group \(\mathbb{Z}_n\).
\item\relax
\flmRefsHyperref[eczindexfamilyrel]{code:constant_weight}{Constant-weight code}\item\relax
\flmRefsHyperref[eczindexfamilyrel]{code:cyclic}{Cyclic code} --- The one-hot code is a cyclic non-linear binary code.
\item\relax
\flmRefsHyperref[eczindexfamilyrel]{code:ecoc}{Error-correcting output code (ECOC)} --- One-hot codes are the primary codes used in multiclass classification \NoCaseChange{\protect\cite{cite300,cite301,cite302,cite303}}.
\end{eczvaluelist}
\codefieldsection{Cousins}
\begin{eczvaluelist}
\item\relax
\flmRefsHyperref[eczindexfamilyrel]{code:ppm}{Pulse-position modulation (PPM) format} --- The PPM code is a continuous analogue of the one-hot code.
\item\relax
\flmRefsHyperref[eczindexfamilyrel]{code:quantum_ppm}{Pulse-position (PPM) c-q modulation format} --- The PPM c-q code is a continuous analogue of the one-hot code designed for transmission through quantum channels.
\item\relax
\flmRefsHyperref[eczindexfamilyrel]{code:binary_cyclic}{Cyclic linear binary code} --- The one-hot code is a cyclic non-linear binary code.
\item\relax
\flmRefsHyperref[eczindexfamilyrel]{code:one_hot_quantum}{One-hot quantum code} --- The one-hot quantum code is the quantum version of the one-hot code.
\end{eczvaluelist}
\eczhbkcontributors{ \eczhuVVA }
\endeczcode

\eczcode{perfect_binary}{Perfect binary code}{}
\codefieldsection{Description}
A type of binary code whose parameters satisfy the Hamming bound with equality.

An \((n,K,2t+1)\) binary code is perfect if parameters \(n\), \(K\), and \(t\) are such that the binary Hamming (a.k.a. sphere-packing) bound
\flmMathEnvironment{align}{}{
\sum_{j=0}^{t} {n \choose j} \leq 2^{n}/K
}
becomes an equality. For example, for a code with one logical bit (\(K=2\)) and \(t=1\), the bound becomes \(n+1 \leq 2^{n-1}\).
Perfect codes are those for which balls of Hamming radius \(t\) exactly fill the space of all \(n\) binary strings.

Any perfect linear binary code is either a binary repetition code, a binary Hamming code, or the binary Golay code \NoCaseChange{\protect\cite{cite1499}\protect\cite[{Thm. 3.3.1}]{cite70}}.

For codes with \(K=2^k\), one can work out an asymptotic Hamming bound in the large-\(n,k,t\) limit,
\flmMathEnvironment{align}{}{
\frac{k}{n}\leq 1-h(t/n),
}
where \(h\) is the binary entropy function.

There are many inequivalent nonlinear perfect binary codes \NoCaseChange{\protect\cite{cite1150,cite1503,cite1504,cite217,cite1505}}; for example, there are 5983 inequivalent perfect distance-three codes of length 15 \NoCaseChange{\protect\cite{cite1506}}.
Nonlinear perfect codes can have arbitrary finite groups as their automorphism groups \NoCaseChange{\protect\cite{cite1507}}, including the trivial group \NoCaseChange{\protect\cite{cite1508,cite1509}}.
The automorphism group of a distance-three perfect binary code is no greater than the automorphism of the Hamming code of the same length \NoCaseChange{\protect\cite{cite1510,cite1511,cite1512,cite1513}}.

\codefieldsection{Parents}
\begin{eczvaluelist}
\item\relax
\flmRefsHyperref[eczindexfamilyrel]{code:nearly_perfect}{Nearly perfect code} --- Perfect binary codes are nearly perfect, and \(t+1\) divides \(n-t\) for such codes. In addition, any perfect code can be extended to a nearly perfect code.
\item\relax
\flmRefsHyperref[eczindexfamilyrel]{code:perfect}{Perfect code}\item\relax
\flmRefsHyperref[eczindexfamilyrel]{code:orthogonal_array}{Orthogonal array (OA)} --- Perfect distance-three binary codes of length \(n =2^m-1\) are equivalent to binary orthogonal arrays of strength \(t = 2^{m-1}-1\) \NoCaseChange{\protect\cite{cite216,cite217,cite218}}.
\end{eczvaluelist}
\codefieldsection{Children}
\begin{eczvaluelist}
\item\relax
\flmRefsHyperref[eczindexfamilyrel]{code:golay}{\([23, 12, 7]\) Golay code} --- The Golay code is perfect \NoCaseChange{\protect\cite[{Thm. 12.3.3 and Def. 12.3.4}]{cite199}}.
\item\relax
\flmRefsHyperref[eczindexfamilyrel]{code:vasilyev}{\((2^{m+1}-1,2^{2n-m},3)\) Vasilyev code} --- Vasilyev codes are perfect nonlinear binary codes and are inequivalent to any linear code.
\item\relax
\flmRefsHyperref[eczindexfamilyrel]{code:hamming}{\([2^r-1,2^r-r-1,3]\) Hamming code}\end{eczvaluelist}
\codefieldsection{Cousins}
\begin{eczvaluelist}
\item\relax
\flmRefsHyperref[eczindexfamilyrel]{code:combinatorial_design}{Combinatorial design} --- If a perfect binary code contains the zero vector, then its minimum-weight codewords support a Steiner \((t+1)\)-\((n,2t+1,1)\) design, while the minimum-weight codewords in the extended code support a Steiner \((t+2)\)-\((n+1,2t+2,1)\) design \NoCaseChange{\protect\cite[{Thm. 5.3.1(b)}]{cite135}}.
\item\relax
\flmRefsHyperref[eczindexfamilyrel]{code:homological_classical}{Cycle code} --- A family of cycle codes saturate the asymptotic Hamming bound \NoCaseChange{\protect\cite{cite71}}.
\item\relax
\flmRefsHyperref[eczindexfamilyrel]{code:repetition}{Repetition code} --- Repetition codes are trivially perfect for odd \(n\) \NoCaseChange{\protect\cite[{Def. 12.3.4}]{cite199}\protect\cite[{pg. 180}]{cite41}}.
\end{eczvaluelist}
\eczhbkcontributors{ Mustafa Doger, \eczhuVVA }
\endeczcode

\eczcode{plaquette_ising}{Plaquette Ising code}{~\NoCaseChange{\protect\cite{cite1514,cite1515,cite1516}}}
\codefieldsection{Alternative Names}
\begin{eczvaluelist}
\item\relax Right-angle water ice code
\item\relax Xu-Moore code
\end{eczvaluelist}
\eczhIndexCodeAliasName{plaquette_ising}{Right-angle water ice code}
\eczhIndexCodeAliasName{plaquette_ising}{Xu-Moore code}
\codefieldsection{Description}
Classical code defined on a cubic lattice in usually two or three dimensions whose parity checks are applied on the four vertices of each square.

\codefieldsection{Protection}
The 2D code has parameters \([L^2,2L-1,L]\) on a square lattice of size \(L\) \NoCaseChange{\protect\cite{cite1517}}.
The 3D version has been studied in Ref. \NoCaseChange{\protect\cite{cite1518}}.

\codefieldsection{Parent}
\begin{eczvaluelist}
\item\relax
\flmRefsHyperref[eczindexfamilyrel]{code:topological_classical}{Classical topological code} --- The 2D plaquette Ising model can be constructed by coupling layers of 1D \(\mathbb{Z}_2\) lattice gauge theory \NoCaseChange{\protect\cite{cite1283}}. A field-theoretic description of the 2D plaquette Ising model can be obtained by coupling layers of 1D gauge theory \NoCaseChange{\protect\cite{cite568}}.
\end{eczvaluelist}
\codefieldsection{Cousins}
\begin{eczvaluelist}
\item\relax
\flmRefsHyperref[eczindexfamilyrel]{code:rotated_surface}{Rotated surface code} --- The 2D plaquette Ising model can be thought of as the rotated surface code whose \(X\)-type stabilizer generators have been converted to \(Z\)-type stabilizer generators.
\item\relax
\flmRefsHyperref[eczindexfamilyrel]{code:bacon_shor}{Bacon-Shor code} --- Ungauging \NoCaseChange{\protect\cite{cite462,cite463,cite233,cite464,cite465,cite466,cite467,cite468,cite469,cite470}} the \(Z\)-type symmetries of the 2D Bacon-Shor gauge Hamiltonian yields the Xu-Moore model, with emergent horizontal and vertical rigid \(X\)-type symmetries \NoCaseChange{\protect\cite{cite466}}.
\item\relax
\flmRefsHyperref[eczindexfamilyrel]{code:xcube}{X-cube model code} --- The 3D plaquette Ising model can be used to obtain the X-cube model by gauging \NoCaseChange{\protect\cite{cite462,cite463,cite233,cite464,cite465,cite466,cite467,cite468,cite469,cite470}} its subsystem symmetry \NoCaseChange{\protect\cite{cite233}}.
\item\relax
\flmRefsHyperref[eczindexfamilyrel]{code:topological_abelian}{Abelian topological code} --- The 2D plaquette Ising model can be constructed by coupling layers of 1D \(\mathbb{Z}_2\) lattice gauge theory \NoCaseChange{\protect\cite{cite1283}}. A field-theoretic description of the 2D plaquette Ising model can be obtained by coupling layers of 1D gauge theory \NoCaseChange{\protect\cite{cite568}}.
\item\relax
\flmRefsHyperref[eczindexfamilyrel]{code:translationally_invariant_stabilizer}{Lattice stabilizer code}\item\relax
\flmRefsHyperref[eczindexfamilyrel]{code:two_foliated}{Two-foliated fracton code} --- The two-foliated fracton code is a hypergraph product of the repetition code and the plaquette Ising code on a square lattice with periodic boundary conditions \NoCaseChange{\protect\cite{cite1517}}.
\end{eczvaluelist}
\eczhbkcontributors{ \eczhuVVA }
\endeczcode

\eczcode{polar}{Polar code}{~\NoCaseChange{\protect\cite{cite1519,cite1174}}}
\codefieldsection{Description}
A member of a family of binary linear codes whose encoding is based on picking the "most reliable bits" at the output of a noise channel.

In its basic version, a binary linear polar code encodes \(K\) message bits into \(N=2^n\) bits. The linear transformation that defines the code is given by the matrix \(G^{(n)}=B_N G^{\otimes n}\), where \(B_N\) is a certain \(N\times N\) permutation matrix, and \(G^{\otimes n}\) is the \(n\)th Kronecker power of the \(2\times 2\) \textit{kernel matrix} \(G=\left(\begin{smallmatrix}1 & 0\\ 1 & 1 \end{smallmatrix}\right)\). To encode \(K\) message bits, one forms an \(N\)-vector \(u\) in which \(K\) coordinates represent the message bits. The remaining \(N-K\) coordinates are set to some fixed values and are said to be \textit{frozen}. The codeword \(x \in \{0,1\}^N\) is obtained as \(x=u G^{(n)}\).

The choice of the frozen coordinates depends on the communication channel, and they correspond to the least reliable bits on the output of the channel under a particular decoding procedure called successive cancellation decoding. If the communication channel is input-symmetric, the values of the frozen bits can be set to zero.

There are multiple variants of the above basic construction, in particular relying on other kernel matrices \NoCaseChange{\protect\cite{cite1520}}.
The codes can be defined for nonbinary alphabets, and they can be adjusted to support tasks such as lossless and lossy compression, successive refinement, communication over the multiple access channel, communication over the wiretap channel, and many others.

The affine automorphism group of polar codes is the lower-triangular affine group \NoCaseChange{\protect\cite{cite1521,cite1522}}.

\codefieldsection{Protection}
Protects against various types of noise in the communication channel, for instance, errors, erasures, or other types of noise. Distance plays no role in the analysis of its properties, and is much lower than the largest possible value given \(K,N\). Polar codes often need to be tailored for a given channel, but universal constructions also exist \NoCaseChange{\protect\cite{cite1523}}.
\codefieldsection{Rate}
Supports reliable transmission at rates \(K/N\) approaching the Shannon capacity of the channel under the successive cancellation decoder \NoCaseChange{\protect\cite{cite1174,cite1524}}; see also Refs. \NoCaseChange{\protect\cite{cite1525,cite1526}}.
\codefieldsection{Decoding}
\begin{eczvaluelist}
\item\relax Successive cancellation (SC) decoder \NoCaseChange{\protect\cite{cite1174}}.
\item\relax Successive cancellation list (SCL) decoder \NoCaseChange{\protect\cite{cite1324}} and a modification utilizing sequence repetition (SR-List) \NoCaseChange{\protect\cite{cite1527}}.
\item\relax Soft cancellation (SCAN) decoder \NoCaseChange{\protect\cite{cite1528,cite1529}}.
\item\relax Belief propagation (BP) decoder \NoCaseChange{\protect\cite{cite1530}}.
\item\relax Noisy quantum gate-based decoder \NoCaseChange{\protect\cite{cite1531}}.
\end{eczvaluelist}
\codefieldsection{Realizations}
\begin{eczvaluelist}
\item\relax Code control channels for the 5G NR (New Radio) interfaces \NoCaseChange{\protect\cite{cite305}}.
\end{eczvaluelist}
\codefieldsection{Notes}
\begin{eczvaluelist}
\item\relax For more details, see Refs. \NoCaseChange{\protect\cite{cite1532,cite946}}.
\item\relax See Kaiserslautern database \NoCaseChange{\protect\cite{cite1184}} and the pretty-good-codes database \NoCaseChange{\protect\cite{cite1477}} for explicit representatives and benchmarking.
\item\relax Codes have been benchmarked using AFF3CT toolbox \NoCaseChange{\protect\cite{cite1480}}.
\item\relax Polar codes are also useful for source coding \NoCaseChange{\protect\cite{cite1533}}.
\end{eczvaluelist}
\codefieldsection{Parents}
\begin{eczvaluelist}
\item\relax
\flmRefsHyperref[eczindexfamilyrel]{code:binary_linear}{Linear binary code}\item\relax
\flmRefsHyperref[eczindexfamilyrel]{code:generalized_concatenated}{Generalized concatenated code (GCC)} --- Polar codes can be represented as generalized concatenations of their kernels.
\end{eczvaluelist}
\codefieldsection{Cousins}
\begin{eczvaluelist}
\item\relax
\flmRefsHyperref[eczindexfamilyrel]{code:reed_muller}{Reed-Muller (RM) code} --- The generator matrices of RM and polar codes are different submatrices of Kronecker products of Hadamard matrices \NoCaseChange{\protect\cite{cite1174,cite365}}. There are families interpolating between the two codes \NoCaseChange{\protect\cite{cite1175}}.
\item\relax
\flmRefsHyperref[eczindexfamilyrel]{code:hadamard}{\([2^m,m,2^{m-1}]\) Hadamard code} --- The generator matrices of RM and polar codes are different submatrices of Kronecker products of Hadamard matrices \NoCaseChange{\protect\cite{cite1174,cite365}}. There are families interpolating between the two codes \NoCaseChange{\protect\cite{cite1175}}.
\item\relax
\flmRefsHyperref[eczindexfamilyrel]{code:polar_for_quantum}{Polar c-q code} --- Quantum-classical polar codes generalize polar codes for transmission through channels with quantum output.
\item\relax
\flmRefsHyperref[eczindexfamilyrel]{code:crc}{Cyclic redundancy check (CRC) code} --- CRC codes concatenated with polar codes yield improved performance of the SCL polar-code decoder \NoCaseChange{\protect\cite{cite1323,cite1324,cite1325}}.
\item\relax
\flmRefsHyperref[eczindexfamilyrel]{code:branching_mera}{Branching MERA code} --- Classical versions of branching MERA codes can be thought of as extensions of polar codes \NoCaseChange{\protect\cite{cite1534,cite1535}}.
\item\relax
\flmRefsHyperref[eczindexfamilyrel]{code:quantum_polar}{Quantum polar code} --- Without entanglement assistance, quantum polar codes are CSS codes constructed out of polar codes.
\end{eczvaluelist}
\eczhbkcontributors{ Alexander Barg, \eczhuVVA }
\endeczcode

\eczcode{preparata}{Preparata code}{~\NoCaseChange{\protect\cite{cite1245}}}
\codefieldsection{Description}
A nonlinear binary \((2^{m+1}, 2^{2^{m+1}-2m-2}, 6)\) code where \(m\) is odd.
Puncturing a Preparata code yields the \textit{shortened Preparata code} with parameters \((2^{m+1}-1, 2^{2^{m+1}-2m-2}, 5)\).

The automorphism groups of Preparata and shortened Preparata codes are determined in Refs. \NoCaseChange{\protect\cite{cite1536,cite1537,cite1148}}.

\codefieldsection{Rate}
The rate is \(\frac{2^{m+1}-2m-2}{2^{m+1}}\), tending to 1 as \(m\) becomes large.
\codefieldsection{Notes}
\begin{eczvaluelist}
\item\relax See corresponding MinT database entry \NoCaseChange{\protect\cite{cite1538}}.
\end{eczvaluelist}
\codefieldsection{Parent}
\begin{eczvaluelist}
\item\relax
\flmRefsHyperref[eczindexfamilyrel]{code:hergert}{Hergert code} --- Preparata codes are equivalent to Hergert codes for \(r=2\) \NoCaseChange{\protect\cite[{Thm. 2}]{cite1326}}.
\end{eczvaluelist}
\codefieldsection{Child}
\begin{eczvaluelist}
\item\relax
\flmRefsHyperref[eczindexfamilyrel]{code:nordstrom_robinson}{\((16,256,6)\) Nordstrom-Robinson (NR) code} --- The NR code is the smallest Preparata code.
\end{eczvaluelist}
\codefieldsection{Cousins}
\begin{eczvaluelist}
\item\relax
\flmRefsHyperref[eczindexfamilyrel]{code:small_distance}{Small-distance block code} --- Shortened Preparata codes form an infinite family of small-distance block codes with minimum distance \(5\).
\item\relax
\flmRefsHyperref[eczindexfamilyrel]{code:nearly_perfect}{Nearly perfect code} --- Shortened Preparata codes are uniformly packed and nearly perfect \NoCaseChange{\protect\cite{cite1496}}. For any word \(u\) and shortened Preparata codebook \(C\) with \(d(u, C) > 2\), we have that \(u\) has distance 2 or 3 to exactly \(\left\lfloor (2^{m+1}-1)/3\right\rfloor\) codewords.
\item\relax
\flmRefsHyperref[eczindexfamilyrel]{code:quasi_perfect}{Quasi-perfect code} --- Shortened Preparata codes are quasi-perfect \NoCaseChange{\protect\cite[{pg. 475}]{cite41}}.
\item\relax
\flmRefsHyperref[eczindexfamilyrel]{code:reed_muller}{Reed-Muller (RM) code} --- Preparata codes are nonlinear subcodes of second-order Reed-Muller codes, and shortened Preparata codes are obtained from them by puncturing.
\item\relax
\flmRefsHyperref[eczindexfamilyrel]{code:bch}{Binary BCH code} --- Preparata codes contain twice as many codewords as the extended double-error-correcting BCH codes of the same length and minimum distance, and have the greatest possible number of codewords for this minimum distance \NoCaseChange{\protect\cite{cite1245}\protect\cite[{pg. 475}]{cite41}}.
\item\relax
\flmRefsHyperref[eczindexfamilyrel]{code:hamming}{\([2^r-1,2^r-r-1,3]\) Hamming code} --- The union of a shortened Preparata code and some of its translates forms a Hamming code \NoCaseChange{\protect\cite[{pg. 475}]{cite41}}.

\item\relax
\flmRefsHyperref[eczindexfamilyrel]{code:extended_hamming}{\([2^m,2^m-m-1,4]\) Extended Hamming code} --- Any code with the same parameters as the Preparata code must be a distance invariant subcode of a (possibly nonlinear) code with the same parameters as the extended Hamming code \NoCaseChange{\protect\cite{cite1169,cite1170}}.
\item\relax
\flmRefsHyperref[eczindexfamilyrel]{code:combinatorial_design}{Combinatorial design} --- Preparata codewords of each weight form 3-designs, and the minimum-weight codewords yield infinite families of 4-designs, including Steiner 4-designs with block sizes 5 and 6 \NoCaseChange{\protect\cite[{Rem. 5.5.6 and Thms. 5.5.7, 5.5.11}]{cite135}\protect\cite[{pg. 471}]{cite41}}.
\item\relax
\flmRefsHyperref[eczindexfamilyrel]{code:quaternary_reed_muller}{Quaternary RM (QRM) code} --- The binary Preparata code is the Gray-map image of the quaternary code QRM\((m-2,m)\) \NoCaseChange{\protect\cite[{Thm. 19}]{cite158}}.
\item\relax
\flmRefsHyperref[eczindexfamilyrel]{code:gray}{Gray code} --- The binary Preparata code is the Gray-map image of the quaternary code QRM\((m-2,m)\) \NoCaseChange{\protect\cite[{Thm. 19}]{cite158}}.
\item\relax
\flmRefsHyperref[eczindexfamilyrel]{code:zrm}{ZRM code} --- Each Preparata code is contained in a corresponding dual of ZRM\((1,m)\) \NoCaseChange{\protect\cite{cite158}}.
\item\relax
\flmRefsHyperref[eczindexfamilyrel]{code:kerdock}{Kerdock code} --- Preparata codes are duals of Kerdock codes in that their distance distribution is equal to the \flmRefsHyperref{ref113}{MacWilliams transform} of the distance distribution of Kerdock codes \NoCaseChange{\protect\cite{cite1148}}.
However, the two codes are images of a pair of mutually dual linear codes over \(\mathbb{Z}_4\) under the \flmTerm{term}{ref81}{}{Gray map}  \NoCaseChange{\protect\cite{cite1149}\protect\cite[{Sec. 6.3}]{cite1145}}.

\item\relax
\flmRefsHyperref[eczindexfamilyrel]{code:quantum_goethals_preparata}{\(\llparenthesis 2^m,2^{2^m−5m+1},8\rrparenthesis \) Goethals-Preparata code} --- The \(\llparenthesis 2^m,2^{2^m−5m+1},8\rrparenthesis \) Goethals-Preparata code is constructed using the classical Goethals and Preparata codes \NoCaseChange{\protect\cite{cite1369,cite1370}}. A construction using the \(\mathbb{Z}_4\) versions of the Goethals and Preparata codes and the \flmTerm{term}{ref81}{}{Gray map} yields qubit code families with similar parameters \NoCaseChange{\protect\cite{cite1371}}.
\end{eczvaluelist}
\eczhbkcontributors{ Shuubham Ojha, \eczhuVVA }
\endeczcode

\eczcode{protograph_ldpc}{Protograph LDPC code}{~\NoCaseChange{\protect\cite{cite1539,cite1212,cite1540,cite1541}}}
\codefieldsection{Description}
Binary version of a \(q\)-ary protograph LDPC code.
Its parity-check matrix can be written as a block matrix whose blocks are either sums of permutation matrices or the zero matrix.

\codefieldsection{Protection}
The minimum distance of protograph codes is bounded by a function of the number of commuting permutation-matrix blocks \NoCaseChange{\protect\cite{cite72}}.

\codefieldsection{Notes}
\begin{eczvaluelist}
\item\relax For reviews on protograph LDPC codes, see Ref. \NoCaseChange{\protect\cite{cite1542}}.
\end{eczvaluelist}
\codefieldsection{Parents}
\begin{eczvaluelist}
\item\relax
\flmRefsHyperref[eczindexfamilyrel]{code:multi_edge_ldpc}{Multi-edge LDPC code} --- LDPC codes based on protographs can be formulated as multi-edge LDPC codes \NoCaseChange{\protect\cite{cite1494}}.
\item\relax
\flmRefsHyperref[eczindexfamilyrel]{code:q-ary_protograph_ldpc}{\(q\)-ary protograph LDPC code}\end{eczvaluelist}
\codefieldsection{Children}
\begin{eczvaluelist}
\item\relax
\flmRefsHyperref[eczindexfamilyrel]{code:ara}{Accumulate-repeat-accumulate (ARA) code} --- ARA codes can be formulated as protograph LDPC codes \NoCaseChange{\protect\cite{cite1212}}.
\item\relax
\flmRefsHyperref[eczindexfamilyrel]{code:ira}{Irregular repeat-accumulate (IRA) code} --- IRA codes can be formulated as protograph LDPC codes \NoCaseChange{\protect\cite{cite1212}}.
\item\relax
\flmRefsHyperref[eczindexfamilyrel]{code:raa}{Repeat-accumulate-accumulate (RAA) code} --- RAA codes can be formulated as protograph LDPC codes \NoCaseChange{\protect\cite{cite1543}}.
\item\relax
\flmRefsHyperref[eczindexfamilyrel]{code:apm_ldpc}{Affine-permutation-matrix LDPC (APM-LDPC) code} --- Parity-check matrices of APM-LDPC codes can be put into block form where the nonzero blocks are permutation matrices representing the affine permutation group \(\mathbb{Z}_r \rtimes \mathbb{Z}_r^{\times}\).
\item\relax
\flmRefsHyperref[eczindexfamilyrel]{code:qc_ldpc}{Quasi-cyclic LDPC (QC-LDPC) code} --- Parity-check matrices of QC-LDPC codes can be put into block form where the nonzero blocks are circulant permutation matrices representing a cyclic group.
\item\relax
\flmRefsHyperref[eczindexfamilyrel]{code:sc_ldpc}{Spatially coupled LDPC (SC-LDPC) code} --- SC-LDPC codes can be interpreted as protograph LDPC codes \NoCaseChange{\protect\cite{cite1544}}.
\end{eczvaluelist}
\codefieldsection{Cousin}
\begin{eczvaluelist}
\item\relax
\flmRefsHyperref[eczindexfamilyrel]{code:algebraic_ldpc}{Algebraic LDPC code} --- Some deterministic protograph LDPC codes \NoCaseChange{\protect\cite{cite1215}} can be obtained from the theory of voltage graphs \NoCaseChange{\protect\cite{cite1216,cite1217}}.
\end{eczvaluelist}
\eczhbkcontributors{ \eczhuVVA }
\endeczcode

\eczcode{qc_ldpc}{Quasi-cyclic LDPC (QC-LDPC) code}{~\NoCaseChange{\protect\cite{cite1545,cite1546,cite61,cite1547,cite1548,cite1549,cite1550}\protect\cite[{Appx. C}]{cite1361}}}
\codefieldsection{Description}
LDPC code that can be put into quasi-cyclic form.
Its parity check matrix can be put into the form of a block matrix consisting of either circulant permutation sub-matrices or the zero sub-matrix.
Such codes are often constructed by \flmRefsHyperref{ref47}{lifting} certain protographs into such block matrices \NoCaseChange{\protect\cite{cite48}}.
Their simple structure makes them useful for several wireless communication standards.

\codefieldsection{Protection}
Minimum-distance upper bounds exist \NoCaseChange{\protect\cite{cite72,cite1551}}.

\codefieldsection{Realizations}
\begin{eczvaluelist}
\item\relax 5G NR cellular communication for the traffic channel \NoCaseChange{\protect\cite{cite312,cite313}}.
\item\relax Wireless communication: WiMAX (IEEE 802.16e) \NoCaseChange{\protect\cite{cite314,cite315,cite316}}, WiFi 4 (IEEE 802.11n) \NoCaseChange{\protect\cite{cite317}}, and WPAN (IEEE 802.15.3c) \NoCaseChange{\protect\cite{cite318}}.
\end{eczvaluelist}
\codefieldsection{Parents}
\begin{eczvaluelist}
\item\relax
\flmRefsHyperref[eczindexfamilyrel]{code:protograph_ldpc}{Protograph LDPC code} --- Parity-check matrices of QC-LDPC codes can be put into block form where the nonzero blocks are circulant permutation matrices representing a cyclic group.
\item\relax
\flmRefsHyperref[eczindexfamilyrel]{code:quasi_cyclic}{Quasi-cyclic code}\end{eczvaluelist}
\codefieldsection{Children}
\begin{eczvaluelist}
\item\relax
\flmRefsHyperref[eczindexfamilyrel]{code:array_ldpc}{Array-based LDPC (AB-LDPC) code}\item\relax
\flmRefsHyperref[eczindexfamilyrel]{code:b_ldpc}{Block LDPC (B-LDPC) code}\item\relax
\flmRefsHyperref[eczindexfamilyrel]{code:difference_set}{Difference-set cyclic (DSC) code}\item\relax
\flmRefsHyperref[eczindexfamilyrel]{code:cycle_ldpc}{Cycle LDPC code} --- Cycle LDPC codes form a class of regular QC LDPC codes \NoCaseChange{\protect\cite{cite1307}}.
\end{eczvaluelist}
\codefieldsection{Cousins}
\begin{eczvaluelist}
\item\relax
\flmRefsHyperref[eczindexfamilyrel]{code:ld_convolutional}{LDPC convolutional code (LDPC-CC)} --- QC-LDPC codes can be \textit{unwrapped} to obtain LDPC-CCs by expressing each circulant matrix block as a power of some generating circulant matrix and replacing that generating matrix by the shift operator of the convolutional code \NoCaseChange{\protect\cite{cite1307}}.
\item\relax
\flmRefsHyperref[eczindexfamilyrel]{code:ra}{Repeat-accumulate (RA) code} --- There exist quasi-cyclic versions of RA codes \NoCaseChange{\protect\cite{cite1552}}.
\item\relax
\flmRefsHyperref[eczindexfamilyrel]{code:pg_ldpc}{Finite-geometry LDPC (FG-LDPC) code} --- Many FG-LDPC codes can be put into quasi-cyclic form \NoCaseChange{\protect\cite{cite1351,cite77}\protect\cite[{pg. 286}]{cite1352}}.
\item\relax
\flmRefsHyperref[eczindexfamilyrel]{code:quasi_cyclic_qldpc}{Quasi-cyclic QLDPC (QC-QLDPC) code} --- QC-QLDPC codes are quantum counterparts of QC-LDPC codes. QC-LDPC codes can be used to make qubit QLDPC codes using various non-CSS constructions \NoCaseChange{\protect\cite{cite1553}}. There exist explicit constructions of both whose parity-check (stabilizer generator) matrices have column weight 2 and girth 12 \NoCaseChange{\protect\cite{cite1554}}.
\item\relax
\flmRefsHyperref[eczindexfamilyrel]{code:ea_qc_qldpc}{EA QC-QLDPC code} --- EA QC-QLDPC codes are entanglement-assisted quantum analogues of QC-LDPC codes.
\item\relax
\flmRefsHyperref[eczindexfamilyrel]{code:multisector_hypergraph}{Higher-dimensional homological product code} --- Higher-dimensional hypergraph-product codes can be constructed out of QC-LDPC codes \NoCaseChange{\protect\cite[{Table III}]{cite1555}}.
\item\relax
\flmRefsHyperref[eczindexfamilyrel]{code:cyclic_hgp}{Cyclic Hypergraph Product Code} --- A classical cyclic LDPC code with parameters \([n,k,d]\) yields a \(\mathrm{C2}\) code with parameters \(\llbracket 2n^2,2k^2,d\rrbracket \) and a \(\mathrm{CxR}\) code with parameters \(\llbracket 2nd,2k,d\rrbracket \).
\item\relax
\flmRefsHyperref[eczindexfamilyrel]{code:lacross}{La-cross code} --- La-cross codes are constructed using a hypergraph product of a cyclic LDPC code with itself.
\item\relax
\flmRefsHyperref[eczindexfamilyrel]{code:abelian_lifted_product}{Abelian LP code} --- QC-LDPC codes can be \flmRefsHyperref{ref47}{lifted} to yield various Abelian LP codes \NoCaseChange{\protect\cite{cite1556,cite1357,cite843}}. Conversely, the Abelian LP construction yields notable families of QC-LDPC codes \NoCaseChange{\protect\cite{cite1557}}.
\end{eczvaluelist}
\eczhbkcontributors{ \eczhuVVA }
\endeczcode

\eczcode{raptor}{Raptor (RAPid TORnado) code}{~\NoCaseChange{\protect\cite{cite257,cite1558}}}
\codefieldsection{Description}
Raptor codes are concatenated erasure codes with two layers: an outer \textit{pre-code} and a Luby-Transform (LT) inner code.
The pre-code is a linear binary erasure code, which is applied first to the input to create some redundant data.
The LT code is then applied to the intermediate symbols from the pre-code to generate final output symbols to be transmitted.

The parameters for a Raptor code are \( ( k, C, \Omega(x) ) \), with \(C\) being the pre-code with dimension \( k \), and \( \Omega(x) \) being the degree distribution for the LT code.

The times to encode and decode source blocks are both linear. The space requirement is \(1/R \), where \(R\) is the rate of the pre-code. Raptor codes with the simplest output distribution (LT code) are called \textit{pre-code-only} (PCO).

\codefieldsection{Protection}
As a type of fountain code, a Raptor code is designed to correct erasures. The error probability of Raptor codes is measured in terms of its overhead, which is how many additional symbols are received above the dimension of the input \(k\). This relationship can vary widely depending on the input pre-code and degree distribution. For a well-designed degree distribution, the error probability of a Raptor code is directly related to the error probability of the pre-code's decoder. In other words, if there is a linear time decoder for the pre-code that has subexponentially small error probability, then the Raptor code's error probability will decrease exponentially with increasing overhead (past the \(n-k\) overhead symbols necessary for the pre-code).
\codefieldsection{Decoding}
\begin{eczvaluelist}
\item\relax Raptor codes can be decoded using inactivation decoding \NoCaseChange{\protect\cite{cite1559}}, a combination of belief-propagation and Gaussian elimination decoding.
\end{eczvaluelist}
\codefieldsection{Realizations}
\begin{eczvaluelist}
\item\relax Two versions of Raptor codes have been standardized by IETF: \flmHref{https://datatracker.ietf.org/doc/html/rfc5053}{R10} and the more recent \flmHref{https://tools.ietf.org/html/rfc6330}{RaptorQ}. RaptorQ is used in mobile multimedia broadcasts as specified in ETSI technical specifications. It is also used in the mobile \flmHref{https://www.atsc.org/wp-content/uploads/2016/01/A331S33-174r6-Signaling-Delivery-Sync-FEC.pdf}{Next Gen TV} standard.
\item\relax Raptor codes are useful in scenarios where erasure (i.e. weak signal or noisy channel) is common, such as in military or disaster scenarios.
\end{eczvaluelist}
\codefieldsection{Notes}
\begin{eczvaluelist}
\item\relax There is an open source RaptorQ implementation in \flmHref{https://openrq-team.github.io/openrq/}{Java} and \flmHref{https://github.com/cberner/raptorq}{Rust}.
\end{eczvaluelist}
\codefieldsection{Parent}
\begin{eczvaluelist}
\item\relax
\flmRefsHyperref[eczindexfamilyrel]{code:fountain}{Fountain code}\end{eczvaluelist}
\codefieldsection{Child}
\begin{eczvaluelist}
\item\relax
\flmRefsHyperref[eczindexfamilyrel]{code:luby_transform}{Luby transform (LT) code} --- Raptor codes using a trivial pre-code are LT codes. Typically, Raptor codes have constant-sized overhead but are faster to decode.
\end{eczvaluelist}
\codefieldsection{Cousin}
\begin{eczvaluelist}
\item\relax
\flmRefsHyperref[eczindexfamilyrel]{code:tornado}{Tornado code} --- Tornado codes, which can be used as a pre-code for raptor codes, also use a multi-layer approach where redundant symbols are created by one code for another code to use as input.
\end{eczvaluelist}
\eczhbkcontributors{ Thomas Wrona, \eczhuVVA }
\endeczcode

\eczcode{reed_muller}{Reed-Muller (RM) code}{~\NoCaseChange{\protect\cite{cite1560,cite1561,cite1562}}}
\codefieldsection{Description}
Member of the RM\((r,m)\) family of linear binary codes derived from multivariate polynomials. The code parameters are \([2^m,\sum_{j=0}^{r} {m \choose j},2^{m-r}]\), where \(r\) is the \textit{order} of the code satisfying \(0\leq r\leq m\).
First-order RM codes are also called biorthogonal codes, while \(m\)th order RM codes are also called \textit{universe} codes.
\textit{Punctured RM codes} RM\(^*(r,m)\) are obtained from RM codes by deleting one coordinate from each codeword.

Generator matrices of RM codes are constructed using the \((u|u+v)\) construction by starting from the \(2^m\)-dimensional matrix \(F^{(m)}=\left(\begin{smallmatrix}1 & 0\\
1 & 1
\end{smallmatrix}\right)^{\otimes m}\), labeling its rows top-to-bottom from \(0\) to \(2^m-1\), converting the labels to binary strings of length \(m\), and deleting all rows whose labels have a Hamming weight less than \(m-r\). The recursive nature of the tensor product in the matrix \(F^{(m)}\) implies that RM\((r,m)\) is a subcode of RM\((r+1,m)\).

Another way to generate RM codewords is to list all outcomes of all polynomials of \(m\) binary variables of degree at most \(r\) \NoCaseChange{\protect\cite{cite1563}} (see also \NoCaseChange{\protect\cite[{Ch. 13}]{cite41}}).

The automorphism group of the RM\((r,m)\) (RM\(^*(r,m)\)) code is \(GA(m,\mathbb{F}_2)\) (\(GL(m,\mathbb{F}_2)\)) for \(1 \leq r \leq m-2\) \NoCaseChange{\protect\cite{cite41}}.
For \(m\geq 5\) and \(m=3\), the only binary linear codes of length \(2^m-1\) whose automorphism group contains \(GL(m,\mathbb{F}_2)\) are the punctured and shortened Reed-Muller codes \NoCaseChange{\protect\cite{cite1564,cite723}}.

\codefieldsection{Protection}
The \textit{Schwartz-Zippel lemma} provides a distance lower bound on RM codes, extending the degree-based distance argument familiar from RS codes.
There is a relation between RM code performance against correlated generalizations of multiple-access channels (MACs) and quantum RM code performance against Pauli channels \NoCaseChange{\protect\cite{cite1565}}.

\codefieldsection{Rate}
Achieves capacity of the binary erasure channel \NoCaseChange{\protect\cite{cite1566,cite1567}}, the binary memoryless symmetric (BMS) channel under bitwise maximum-a-posteriori decoding \NoCaseChange{\protect\cite{cite1568}} (see also Ref. \NoCaseChange{\protect\cite{cite1569}}), and the binary symmetric channel (BSC), solving a long-standing conjecture \NoCaseChange{\protect\cite{cite1570}}.
\codefieldsection{Decoding}
\begin{eczvaluelist}
\item\relax Reed decoder with \(r+1\)-step majority decoding corrects \(\frac{1}{2}(2^{m-r}-1)\) errors \NoCaseChange{\protect\cite{cite1560}} (see also \NoCaseChange{\protect\cite[{Ch. 13}]{cite41}}).
\item\relax Sequential code-reduction decoding \NoCaseChange{\protect\cite{cite1571}}.
\item\relax Matrix factorization can be used to decode an RM\((n,n-3)\) code \NoCaseChange{\protect\cite{cite1572}}; see \NoCaseChange{\protect\cite{cite754}}.
\end{eczvaluelist}
\codefieldsection{Realizations}
\begin{eczvaluelist}
\item\relax Deep-space communication \NoCaseChange{\protect\cite{cite326,cite327}}.
\end{eczvaluelist}
\codefieldsection{Notes}
\begin{eczvaluelist}
\item\relax See \NoCaseChange{\protect\cite{cite1573}\protect\cite[{Chs. 13-15}]{cite41}\protect\cite[{Sec. 1.11}]{cite1159}\protect\cite[{Secs. 2.8 and 2.9}]{cite68}} for details of RM codes, punctured RM codes, and related cyclic generalizations.
\item\relax Review on theory and algorithms of RM codes \NoCaseChange{\protect\cite{cite1563}}.
\end{eczvaluelist}
\codefieldsection{Parents}
\begin{eczvaluelist}
\item\relax
\flmRefsHyperref[eczindexfamilyrel]{code:coxeter}{Coxeter code} --- An RM\((r,m)\) code is spanned by indicators of all subcubes of dimension \(m-r\) in the \(m\)-dimensional cube (this is a redundant generating set \NoCaseChange{\protect\cite{cite753}}), i.e., by the cosets of rank-\((m-r)\) subgroups of \(\mathbb{Z}_2^m\). For a finite Coxeter group \(W\) with \(m\) generators, the Coxeter code \(C_W(r)\) of order \(r\) with \(-1 \leq r \leq m\) is similarly spanned by indicators of all the cosets of rank-\((m-r)\) parabolic subgroups of \(W\).
\item\relax
\flmRefsHyperref[eczindexfamilyrel]{code:berman}{Berman code} --- Berman codes include RM codes as a special case \NoCaseChange{\protect\cite{cite66}}.
\item\relax
\flmRefsHyperref[eczindexfamilyrel]{code:uplusv}{\((u|u+v)\)-construction code} --- All RM codes can be constructed via the \((u|u+v)\) construction \NoCaseChange{\protect\cite[{Ch. 13}]{cite41}}.
\item\relax
\flmRefsHyperref[eczindexfamilyrel]{code:generalized_reed_muller}{Generalized RM (GRM) code} --- Binary GRM codes are RM codes.
\item\relax
\flmRefsHyperref[eczindexfamilyrel]{code:cascaded_reed_solomon}{Hyperbolic evaluation code} --- RM codes are special cases of hyperbolic evaluation codes \NoCaseChange{\protect\cite[{Thm. 3 proof}]{cite29}}.
\item\relax
\flmRefsHyperref[eczindexfamilyrel]{code:divisible}{Divisible code} --- An RM\((r,m)\) code is \(2^{\left\lceil m/r\right\rceil-1}\)-divisible, according to McEliece's theorem \NoCaseChange{\protect\cite{cite1574,cite1575}}.
\end{eczvaluelist}
\codefieldsection{Children}
\begin{eczvaluelist}
\item\relax
\flmRefsHyperref[eczindexfamilyrel]{code:biorthogonal}{\([2^m,m+1,2^{m-1}]\) First-order RM code} --- First-order RM codes are RM\((1,m)\) codes.
\item\relax
\flmRefsHyperref[eczindexfamilyrel]{code:repetition}{Repetition code} --- Repetition codes are RM\((0,m)\) codes.
\item\relax
\flmRefsHyperref[eczindexfamilyrel]{code:extended_hamming}{\([2^m,2^m-m-1,4]\) Extended Hamming code} --- Extended Hamming codes are RM\((m-2,m)\) codes.
\end{eczvaluelist}
\codefieldsection{Cousins}
\begin{eczvaluelist}
\item\relax
\flmRefsHyperref[eczindexfamilyrel]{code:group}{Group-algebra code} --- RM codes are group-algebra codes \NoCaseChange{\protect\cite{cite1576,cite1577}\protect\cite[{Exam. 16.4.11}]{cite196}}. Consider a binary vector space of dimension \( m \). Under addition, this forms a finite group with \( 2^m \) elements known as an elementary Abelian 2-group -- the direct product of \( m \) two-element cyclic groups \( \mathbb{Z}_2 \times \dots \times \mathbb{Z}_2 \). Denote this group by \( G_m \). Let \( J \) be the Jacobson radical of the \flmRefsHyperref{ref205}{group algebra} \( \mathbb{F}_2 G_m \). RM\((r,m)\) codes correspond to the ideal \( J^{m-r} \). The length of the code is \( |G_m| = 2^m \), the distance is \( 2^{m-r} \), and the dimension is \( \sum_{i=0}^r {m \choose i} \). A similar construction exists for choices of a prime \( p\neq 2 \).
\item\relax
\flmRefsHyperref[eczindexfamilyrel]{code:bch}{Binary BCH code} --- RM\(^*(r,m)\) codes are equivalent to subcodes of BCH codes of designed distance \(2^{m-r}-1\), while RM\((r,m)\) are subcodes of extended BCH codes of the same designed distance \NoCaseChange{\protect\cite[{Ch. 13}]{cite41}}.
\item\relax
\flmRefsHyperref[eczindexfamilyrel]{code:quaternary_over_z4}{Linear code over \(\mathbb{Z}_4\)} --- Binary Reed-Muller codes are images of linear quaternary codes over \(\mathbb{Z}_4\) under the Gray map \NoCaseChange{\protect\cite[{Sec. 6.3}]{cite1145}}.
\item\relax
\flmRefsHyperref[eczindexfamilyrel]{code:dual}{Dual linear code} --- The codes RM\((r,m)\) and RM\((m-r-1,m)\) are dual to each other, with the case \(m = 2r+1\) being self dual.
\item\relax
\flmRefsHyperref[eczindexfamilyrel]{code:self_dual}{Self-dual linear code} --- The codes RM\((r,m)\) and RM\((m-r-1,m)\) are dual to each other, with the case \(m = 2r+1\) being self dual.
\item\relax
\flmRefsHyperref[eczindexfamilyrel]{code:binary_duadic}{Binary duadic code} --- Certain punctured RM codes, such as RM\(^*(2,5)\) \NoCaseChange{\protect\cite[{Table 6.2}]{cite126}} and codes of order \((m-1)/2\) for odd \(m\) \NoCaseChange{\protect\cite{cite1266}}, are duadic.
\item\relax
\flmRefsHyperref[eczindexfamilyrel]{code:binary_cyclic}{Cyclic linear binary code} --- Punctured RM codes, i.e., RM codes with nonzero evaluation points, are cyclic \NoCaseChange{\protect\cite{cite1312,cite1313}\protect\cite[{Ch. 13, Thm. 11}]{cite41}\protect\cite[{Sec. 2.8}]{cite68}\protect\cite[{pg. 52}]{cite1314}}, making RM codes extended cyclic codes.
\item\relax
\flmRefsHyperref[eczindexfamilyrel]{code:binary_ltc}{Binary linear LTC} --- RM codes can be LTCs in the low- \NoCaseChange{\protect\cite{cite1274,cite1275}} and high-error \NoCaseChange{\protect\cite{cite1276}} regimes; see also \NoCaseChange{\protect\cite{cite1277}}.
\item\relax
\flmRefsHyperref[eczindexfamilyrel]{code:qubits_into_qubits}{Qubit code} --- Optimizing T gates in a qubit circuit that uses CNOT and T gates is equivalent to decoding a particular RM code \NoCaseChange{\protect\cite{cite1578}}.
\item\relax
\flmRefsHyperref[eczindexfamilyrel]{code:orthogonal_array}{Orthogonal array (OA)} --- RM codes are related to orthogonal arrays \NoCaseChange{\protect\cite[{Exam. 10.57}]{cite225}}.
\item\relax
\flmRefsHyperref[eczindexfamilyrel]{code:barnes_wall}{Barnes-Wall (BW) lattice} --- BW lattices are lattice analogues of RM codes in that both can be constructed recursively via a \(|u|u+v|\) construction \NoCaseChange{\protect\cite{cite1579,cite1580}}.

\item\relax
\flmRefsHyperref[eczindexfamilyrel]{code:combinatorial_design}{Combinatorial design} --- Fixed-weight RM codewords of weight less than \(2^m\) support combinatorial 3-designs \NoCaseChange{\protect\cite[{Exam. 5.2.7}]{cite135}}.
\item\relax
\flmRefsHyperref[eczindexfamilyrel]{code:parity_check}{\([n,n-1,2]\) Single parity-check (SPC) code} --- Binary SPC codes of length \(2^m\) are RM\((m-1,m)\) codes.
\item\relax
\flmRefsHyperref[eczindexfamilyrel]{code:preparata}{Preparata code} --- Preparata codes are nonlinear subcodes of second-order Reed-Muller codes, and shortened Preparata codes are obtained from them by puncturing.
\item\relax
\flmRefsHyperref[eczindexfamilyrel]{code:delsarte_goethals}{Delsarte-Goethals (DG) code} --- The code DG\((m,r)\) is a subcode of the second-order Reed-Muller code RM\((2,m)\), and equals RM\((2,m)\) at \(r=1\) \NoCaseChange{\protect\cite[{pg. 461}]{cite41}}. The code is the union of certain cosets of the first-order RM\((1,m)\) code in RM\((2,m)\) that are specified by bilinear forms \NoCaseChange{\protect\cite{cite1326}}.
\item\relax
\flmRefsHyperref[eczindexfamilyrel]{code:kerdock}{Kerdock code} --- Kerdock code is a subcode of a second-order RM Code \NoCaseChange{\protect\cite[{pg. 457}]{cite41}}.
It consists of a number of cosets of RM\((2,m)\) created by quotienting with first-order RM\((1,m)\) codes.

\item\relax
\flmRefsHyperref[eczindexfamilyrel]{code:hadamard}{\([2^m,m,2^{m-1}]\) Hadamard code} --- The \([2^m,m+1,2^{m-1}]\) augmented Hadamard code is the first-order RM code (a.k.a. RM\((1,m)\)). The \([2^m-1,m,2^{m-1}]\) shortened Hadamard code is the simplex code (a.k.a. RM\(^*(1,m)\)). Rows of a Hadamard matrix forming a Prometheus orthonormal set (PONS) are codewords of a coset of RM\((1,m)\) in RM\((2,m)\) \NoCaseChange{\protect\cite{cite1173}}.
\item\relax
\flmRefsHyperref[eczindexfamilyrel]{code:simplex}{\([2^m-1,m,2^{m-1}]\) simplex code} --- Simplex are equivalent to RM\(^*(1,m)\).
\item\relax
\flmRefsHyperref[eczindexfamilyrel]{code:hamming}{\([2^r-1,2^r-r-1,3]\) Hamming code} --- Binary Hamming codes are equivalent to RM\(^*(r-2,r)\).
\item\relax
\flmRefsHyperref[eczindexfamilyrel]{code:polar}{Polar code} --- The generator matrices of RM and polar codes are different submatrices of Kronecker products of Hadamard matrices \NoCaseChange{\protect\cite{cite1174,cite365}}. There are families interpolating between the two codes \NoCaseChange{\protect\cite{cite1175}}.
\item\relax
\flmRefsHyperref[eczindexfamilyrel]{code:cmr}{\(C_{m,r}\) code} --- The \(C_{m,r}\) code is generated by \(\textnormal{RM}(r,m) + 2\textnormal{RM}(m-r-1,m)\) for \(3r \leq m-1\) \NoCaseChange{\protect\cite{cite121}}.
\item\relax
\flmRefsHyperref[eczindexfamilyrel]{code:quaternary_reed_muller}{Quaternary RM (QRM) code} --- The mod-two reduction of the QRM\((r,m)\) code is the RM\((r,m)\) code \NoCaseChange{\protect\cite[{Thm. 19}]{cite158}}.
\item\relax
\flmRefsHyperref[eczindexfamilyrel]{code:zrm}{ZRM code} --- The ZRM code is generated by \(\textnormal{RM}(r-1,m-1) + 2\textnormal{RM}(r,m-1)\) \NoCaseChange{\protect\cite[{Thm. 7}]{cite158}}. 
The image of the ZRM\((r,m-1)\) code under the \flmTerm{term}{ref81}{}{Gray map} is the RM\((r,m)\) code \NoCaseChange{\protect\cite[{Thm. 7}]{cite158}}.

\item\relax
\flmRefsHyperref[eczindexfamilyrel]{code:klemm}{Klemm code} --- The Klemm code at \(m=4\) is generated by \(\textnormal{RM}(0,4) + 2\textnormal{RM}(3,4)\) \NoCaseChange{\protect\cite{cite1581}}.
\item\relax
\flmRefsHyperref[eczindexfamilyrel]{code:majorana_reed_muller}{RM Majorana code} --- RM Majorana codes are constructed from self-orthogonal RM codes.
\item\relax
\flmRefsHyperref[eczindexfamilyrel]{code:quantum_reed_muller}{Quantum Reed-Muller (RM) code} --- Quantum RM codes are constructed from RM codes via the CSS construction. There is a relation between RM code performance against correlated generalizations of multiple-access channels (MACs) and quantum RM code performance against Pauli channels \NoCaseChange{\protect\cite{cite1565}}.
\end{eczvaluelist}
\eczhbkcontributors{ Anqi Gong, \eczhuVVA }
\endeczcode

\eczcode{regular_binary_tanner}{Regular binary Tanner code}{~\NoCaseChange{\protect\cite{cite1582}}}
\codefieldsection{Alternative Names}
\begin{eczvaluelist}
\item\relax Regular binary GLDPC code
\end{eczvaluelist}
\eczhIndexCodeAliasName{regular_binary_tanner}{Regular binary GLDPC code}
\codefieldsection{Description}
A binary Tanner code defined on a regular bipartite graph, with the inner code being the same for all vertices.

An alternative definition maps the variable nodes of the generalized Tanner graph to edges of a regular graph \(G\) and the constraint nodes to vertices of \(G\).
Bits are placed on edges of \(G\) such that each subsequence of bits corresponding to edges incident on any vertex belongs to some \textit{short} binary linear code \(C_0\).

More technically, let \(G(V,E)\) be a \(\Delta\)-regular (not necessarily bipartite) graph with number of vertices \(|V| = n \) and number of edges \(|E| = N = n\Delta/2\). Let \(C_0\) be a linear binary code of length \(\Delta\) and rate \(R_0\). The Tanner code \(T(G,C_0)\), whose bits are placed on edges of the graph, consists of the following codewords:
\flmMathEnvironment{align}{}{
\left\{ c \in \mathbb{F}_2^{N}\,\text{s.t. }\forall v\in V,\left.c\right|_{E(v)}\in C_{0}\right\} ~,
}
where \(\left.c\right|_{E(v)}\) is the subsequence formed by the \(\Delta\) bits located on the edges incident on the vertex \(v\). The dimension of \(T\) is at least \(N -n(\Delta -\Delta R_0) = N(2R_0-1)\geq 0\) whenever \(R_0 \geq \frac{1}{2}\).

\codefieldsection{Protection}
Minimum-distance bounds in terms of graph and short-code parameters include the bit-oriented and parity-oriented eigenvalue bounds \NoCaseChange{\protect\cite{cite1583}} and related extensions \NoCaseChange{\protect\cite{cite1584}}. Related eigenvalue techniques also yield lower bounds on stopping-set size and pseudocodeword weight for expander/LDPC realizations \NoCaseChange{\protect\cite{cite1585}}.

\codefieldsection{Rate}
For a short code \(C_0\) with rate \(R_0\), the Tanner code has rate \(R \geq 2R_0-1\). If \(C_0\) satisfies the \flmRefsHyperref{ref85}{GV bound}, i.e., \(R_0 \geq 1-h(\delta_0)\), then \(R \geq 1-2h(\delta_0)\), where \(\delta_0\) is the relative distance of \(C_0\) and \(h\) is the binary entropy function.
\codefieldsection{Encoding}
\begin{eczvaluelist}
\item\relax Quadratic algorithm: This technique works for all linear block codes and encodes using matrix multiplication \NoCaseChange{\protect\cite{cite1586}}.
\item\relax Using the non-Abelian Fast Fourier Transform and exploiting the symmetry of the underlying graph, an encoding algorithm that requires \(O(n^{4/3})\) has been devised in \NoCaseChange{\protect\cite{cite1586}}.
\item\relax A modified construction yields codes that may be encoded in linear time yet maintain similar performance \NoCaseChange{\protect\cite{cite1339}}.
\end{eczvaluelist}
\codefieldsection{Decoding}
\begin{eczvaluelist}
\item\relax Parallel decoding algorithm corrects a fraction \(\delta_0^2/48\) of errors for Tanner codes \NoCaseChange{\protect\cite{cite1332}}. A modification of said algorithm improves the fraction to \(\delta_0^2/4\) with no extra cost to complexity \NoCaseChange{\protect\cite{cite1344}}.
\item\relax Soft-decision linear-time decoder correcting errors almost up to half of the Blokh-Zyablov bound \NoCaseChange{\protect\cite{cite972}}.
\end{eczvaluelist}
\codefieldsection{Realizations}
\begin{eczvaluelist}
\item\relax First hardware implementation was done using a semi-systolic decoding architecture \NoCaseChange{\protect\cite{cite337}}.
\end{eczvaluelist}
\codefieldsection{Parents}
\begin{eczvaluelist}
\item\relax
\flmRefsHyperref[eczindexfamilyrel]{code:binary_linear}{Linear binary code}\item\relax
\flmRefsHyperref[eczindexfamilyrel]{code:tanner}{Tanner code} --- Regular binary Tanner codes are binary Tanner codes defined on regular sparse bipartite graphs, with the inner code being the same for all vertices.
\end{eczvaluelist}
\codefieldsection{Children}
\begin{eczvaluelist}
\item\relax
\flmRefsHyperref[eczindexfamilyrel]{code:generalized_gallager}{Generalized Gallager code}\item\relax
\flmRefsHyperref[eczindexfamilyrel]{code:regular_ldpc}{Regular LDPC code} --- Regular LDPC codes are regular binary Tanner codes defined on sparse graphs whose constraint nodes represent parity-check codes.
\end{eczvaluelist}
\codefieldsection{Cousins}
\begin{eczvaluelist}
\item\relax
\flmRefsHyperref[eczindexfamilyrel]{code:dhlv}{Dinur-Hsieh-Lin-Vidick (DHLV) code} --- Regular binary Tanner codes are used in constructing quantum DHLV codes.
\item\relax
\flmRefsHyperref[eczindexfamilyrel]{code:quantum_tanner}{Quantum Tanner code} --- Regular binary Tanner codes are used in constructing quantum Tanner codes.
\end{eczvaluelist}
\eczhbkcontributors{ Xiaozhen Fu, \eczhuVVA }
\endeczcode

\eczcode{regular_ldpc}{Regular LDPC code}{~\NoCaseChange{\protect\cite{cite1360,cite1361}}}
\codefieldsection{Description}
An LDPC code whose parity-check matrix has a fixed number of ones in each row and each column.
Such a code is called \((j,k)\)-regular if each column has weight \(j\) (variable-node degree) and each row has weight \(k\) (check-node degree).
If the parity-check matrix has \(n\) columns and \(m\) rows, then regularity implies \(nj = mk\).

\codefieldsection{Protection}
Random \((j,k)\)-regular LDPC codes have minimum distance that grows linearly with block length for any fixed \(j \geq 3\) (with \(k > j\)) \NoCaseChange{\protect\cite{cite1361,cite1587}}.
In contrast, \((2,k)\)-regular LDPC codes have minimum distance growing at most logarithmically with block length \NoCaseChange{\protect\cite{cite1361}}.

\codefieldsection{Rate}
An ensemble of \((j,k)\)-regular LDPC codes has a \textit{design rate} \(1 - j/k\). This is a lower bound on the true rate, is the true rate if all constraints are linearly independent, and equals the rate asymptotically \NoCaseChange{\protect\cite{cite1361}\protect\cite[{Sec. 3.4}]{cite1473}}.

\codefieldsection{Decoding}
\begin{eczvaluelist}
\item\relax The BP decoding threshold for regular codes is strictly below the Shannon capacity for any fixed \(j,k\) \NoCaseChange{\protect\cite[{Sec. 3.11}]{cite1473}}.
\end{eczvaluelist}
\codefieldsection{Parents}
\begin{eczvaluelist}
\item\relax
\flmRefsHyperref[eczindexfamilyrel]{code:ldpc}{Low-density parity-check (LDPC) code}\item\relax
\flmRefsHyperref[eczindexfamilyrel]{code:regular_binary_tanner}{Regular binary Tanner code} --- Regular LDPC codes are regular binary Tanner codes defined on sparse graphs whose constraint nodes represent parity-check codes.
\end{eczvaluelist}
\codefieldsection{Children}
\begin{eczvaluelist}
\item\relax
\flmRefsHyperref[eczindexfamilyrel]{code:cycle_ldpc}{Cycle LDPC code} --- Cycle LDPC codes form a class of regular QC LDPC codes \NoCaseChange{\protect\cite{cite1307}}.
\item\relax
\flmRefsHyperref[eczindexfamilyrel]{code:expander}{Expander code} --- Expander codes yield an explicit (i.e., non-random) asymptotically good LDPC code family \NoCaseChange{\protect\cite{cite1332}}.
\item\relax
\flmRefsHyperref[eczindexfamilyrel]{code:gallager}{Gallager (GL) code} --- GL codes are the first LDPC codes.
\item\relax
\flmRefsHyperref[eczindexfamilyrel]{code:ha_ldpc}{Hsu-Anastasopoulos LDPC (HA-LDPC) code}\item\relax
\flmRefsHyperref[eczindexfamilyrel]{code:lu_ldpc}{Lazebnik-Ustimenko (LU) code}\item\relax
\flmRefsHyperref[eczindexfamilyrel]{code:mn_ldpc}{MacKay-Neal LDPC (MN-LDPC) code} --- MN-LDPC codes re-invigorated the study of LDPC codes about 30 years after their discovery.
\item\relax
\flmRefsHyperref[eczindexfamilyrel]{code:pg_ldpc}{Finite-geometry LDPC (FG-LDPC) code}\end{eczvaluelist}
\codefieldsection{Cousin}
\begin{eczvaluelist}
\item\relax
\flmRefsHyperref[eczindexfamilyrel]{code:irregular_ldpc}{Irregular LDPC code} --- Irregular LDPC codes have variable node and check node degrees, while regular LDPC codes have fixed node degrees. Irregular LDPC codes with optimized degree distributions can outperform regular ones under iterative decoding \NoCaseChange{\protect\cite{cite1364,cite1395}}.
\end{eczvaluelist}
\eczhbkcontributors{ \eczhuVVA }
\endeczcode

\eczcode{ra}{Repeat-accumulate (RA) code}{~\NoCaseChange{\protect\cite{cite1588}}}
\codefieldsection{Description}
An LDPC code whose parity-check matrix has weight-two columns arranged in a step-like pattern for its last columns \NoCaseChange{\protect\cite{cite94}}.

\codefieldsection{Protection}
Minimum-distance upper bounds \NoCaseChange{\protect\cite{cite1589,cite1552}}.

\codefieldsection{Rate}
RA codes are not asymptotically good \NoCaseChange{\protect\cite{cite1590}}.
\codefieldsection{Encoding}
\begin{eczvaluelist}
\item\relax An encoder for an RA code acting on a string \((c_1c_2\cdots c_K)\) of logical bits begins by repeating each bit three times to obtain the length-\(3K\) bitstring \((c_1 c_1 c_1 c_2 c_2 c_2 \cdots c_K c_K c_K)\), permuting using a random permutation to obtain a bitstring \(u\), and applying the mod-two accumulated sum (or \textit{accumulator}) to obtain \NoCaseChange{\protect\cite[{Ch. 49}]{cite1474}} \flmMathEnvironment{align}{}{ (u_{1},u_{1}+u_{2},\cdots,u_{1}+\cdots+u_{3K})~. } The first repeating step is effectively using a 1-in-3 repetition code, which can be thought of as the outer code in this concatenated construction.
\end{eczvaluelist}
\codefieldsection{Parents}
\begin{eczvaluelist}
\item\relax
\flmRefsHyperref[eczindexfamilyrel]{code:ira}{Irregular repeat-accumulate (IRA) code} --- IRA codes for which the outer code is a 1-in-3 repetition code reduce to RA codes.
\item\relax
\flmRefsHyperref[eczindexfamilyrel]{code:ara}{Accumulate-repeat-accumulate (ARA) code} --- ARA codes with no pre-encoding accumulator and no post-accumulator puncturing reduce to RA codes.
\item\relax
\flmRefsHyperref[eczindexfamilyrel]{code:raa}{Repeat-accumulate-accumulate (RAA) code} --- RAA codes with no second permutation and accumulator reduce to RA codes.
\end{eczvaluelist}
\codefieldsection{Cousin}
\begin{eczvaluelist}
\item\relax
\flmRefsHyperref[eczindexfamilyrel]{code:qc_ldpc}{Quasi-cyclic LDPC (QC-LDPC) code} --- There exist quasi-cyclic versions of RA codes \NoCaseChange{\protect\cite{cite1552}}.
\end{eczvaluelist}
\eczhbkcontributors{ \eczhuVVA }
\endeczcode

\eczcode{raa}{Repeat-accumulate-accumulate (RAA) code}{~\NoCaseChange{\protect\cite{cite1590}}}
\codefieldsection{Description}
Generalization of the RA code in which two accumulators and permutations are used.

\codefieldsection{Rate}
Some sequences of non-deterministic RAA codes are asymptotically good \NoCaseChange{\protect\cite{cite1590,cite1591}}.
\codefieldsection{Encoding}
\begin{eczvaluelist}
\item\relax An encoder for an RAA code is the same as that for the RA code, followed by a second round of permutation and accumulation.
\end{eczvaluelist}
\codefieldsection{Parent}
\begin{eczvaluelist}
\item\relax
\flmRefsHyperref[eczindexfamilyrel]{code:protograph_ldpc}{Protograph LDPC code} --- RAA codes can be formulated as protograph LDPC codes \NoCaseChange{\protect\cite{cite1543}}.
\end{eczvaluelist}
\codefieldsection{Child}
\begin{eczvaluelist}
\item\relax
\flmRefsHyperref[eczindexfamilyrel]{code:ra}{Repeat-accumulate (RA) code} --- RAA codes with no second permutation and accumulator reduce to RA codes.
\end{eczvaluelist}
\eczhbkcontributors{ \eczhuVVA }
\endeczcode

\eczcode{repetition}{Repetition code}{}
\codefieldsection{Description}
\([n,1,n]\) binary linear code encoding one bit of information into an \(n\)-bit string.
Majority decoding requires \(n\) to be odd in order to avoid ties.
The idea is to increase the code distance by repeating the logical information several times. It is a \((n,1)\)-Hamming code.
Its automorphism group is \(S_n\).

\codefieldsection{Protection}
Detects errors on up to \(n-1\) coordinates, corrects errors on up to \(\frac{n-1}{2}\) coordinates, and corrects erasure errors on up to \(n-1\) coordinates. The generator matrix is \(G=\left(\begin{smallmatrix}1 & 1&\cdots& 1 & 1 \end{smallmatrix}\right)\).
\codefieldsection{Rate}
Code rate is \(\frac{1}{n}\), code distance is \(n\).
\codefieldsection{Decoding}
\begin{eczvaluelist}
\item\relax Calculate the Hamming weight \(d_H\) of an error word. If \(d_H\leq \frac{n-1}{2}\), decode the code as 0. If \(d_H\geq \frac{n+1}{2}\), decode the code as 1.
\item\relax Local automaton decoder for the repetition code on a 2D lattice based on Toom's rule \NoCaseChange{\protect\cite{cite1592,cite1593,cite1594,cite1595}}.
\item\relax Local automaton decoder for the repetition code on a 1D lattice by Gacs that is translation-invariant, that does not require synchronization of local clocks, and that has a constant encoding rate \NoCaseChange{\protect\cite{cite1596,cite1597,cite1598,cite1599}}.
\item\relax Local automaton decoder for the repetition code on a 1D lattice by Tsirelson \NoCaseChange{\protect\cite{cite1600}}.
\item\relax Local automaton decoder obtained from reinforcement learning \NoCaseChange{\protect\cite{cite1601}}.
\end{eczvaluelist}
\codefieldsection{Fault Tolerance}
\begin{eczvaluelist}
\item\relax Triple modular redundancy (TMR) error-correction protocol \NoCaseChange{\protect\cite{cite1602}} for fault-tolerant memory operations and classical gate operations; see \NoCaseChange{\protect\cite[{Secs. 2.6 and 2.7}]{cite497}} for a pedagogical explanation.
\end{eczvaluelist}
\codefieldsection{Threshold}
\begin{eczvaluelist}
\item\relax Suppose each bit has probability \(p\) of being received correctly, independent for each bit. The probability that a repetition code is received correctly is \(\sum_{k=0}^{(n-1)/2}\frac{n!}{k!(n-k)!}p^{n-k}(1-p)^{k}\). If \(\frac{1}{2}\leq p\), then one can always increase the probability of success by increasing the number of physical bits \(n\); see \NoCaseChange{\protect\cite[{Sec. 2.2.1}]{cite497}} for a pedagogical explanation.
\item\relax The first threshold theorem for classical circuits was proven by von Neumann \NoCaseChange{\protect\cite{cite1603}} using cellular automata \NoCaseChange{\protect\cite{cite1604,cite1603}} and which spurred the study of noisy classical circuits \NoCaseChange{\protect\cite{cite1605,cite1606,cite1607,cite1608}}.
\end{eczvaluelist}
\codefieldsection{Realizations}
\begin{eczvaluelist}
\item\relax Repetition codes, in conjunction with other codes, were used in magnetic disks \NoCaseChange{\protect\cite{cite338}}.
\item\relax Communication protocols such as FlexRay \NoCaseChange{\protect\cite{cite339}}.'
\end{eczvaluelist}
\codefieldsection{Parents}
\begin{eczvaluelist}
\item\relax
\flmRefsHyperref[eczindexfamilyrel]{code:reed_muller}{Reed-Muller (RM) code} --- Repetition codes are RM\((0,m)\) codes.
\item\relax
\flmRefsHyperref[eczindexfamilyrel]{code:nearly_perfect}{Nearly perfect code}\item\relax
\flmRefsHyperref[eczindexfamilyrel]{code:binary_cyclic}{Cyclic linear binary code} --- The repetition code is cyclic with generator polynomial \(1+x+\cdots+x^{n-1}\).
\item\relax
\flmRefsHyperref[eczindexfamilyrel]{code:q-ary_repetition}{\(q\)-ary repetition code} --- \(q\)-ary repetition code reduce to repetition codes for \(q=2\).
\end{eczvaluelist}
\codefieldsection{Cousins}
\begin{eczvaluelist}
\item\relax
\flmRefsHyperref[eczindexfamilyrel]{code:mds}{Maximum distance separable (MDS) code} --- Binary repetition codes are trivial MDS codes \NoCaseChange{\protect\cite[{Thm. 12.3.1}]{cite199}\protect\cite[{Sec. 3.3.2}]{cite70}}.
\item\relax
\flmRefsHyperref[eczindexfamilyrel]{code:perfect_binary}{Perfect binary code} --- Repetition codes are trivially perfect for odd \(n\) \NoCaseChange{\protect\cite[{Def. 12.3.4}]{cite199}\protect\cite[{pg. 180}]{cite41}}.
\item\relax
\flmRefsHyperref[eczindexfamilyrel]{code:quantum_repetition}{Quantum repetition code} --- A quantum repetition code can be thought of as a classical \([n,1,n]\) repetition code embedded in a quantum system.
\item\relax
\flmRefsHyperref[eczindexfamilyrel]{code:hamming}{\([2^r-1,2^r-r-1,3]\) Hamming code} --- The triple repetition code \([3,1,3]\) is the smallest Hamming code.
\item\relax
\flmRefsHyperref[eczindexfamilyrel]{code:eeight}{\(E_8\) Gosset lattice} --- The \([8,1,8]\) repetition code can be used to obtain the \(E_8\) Gosset lattice \NoCaseChange{\protect\cite[{Exam. 10.7.1}]{cite115}}.
\item\relax
\flmRefsHyperref[eczindexfamilyrel]{code:dfour}{\(D_4\) hyper-diamond lattice} --- Construction \(A_c\) applied to the binary repetition code \(\{00,11\}\) over the Gaussian integers yields a Gaussian lattice whose corresponding real lattice is \(D_4\) \NoCaseChange{\protect\cite[{Ch. 7, pg. 202}]{cite39}}.
\item\relax
\flmRefsHyperref[eczindexfamilyrel]{code:parity_check}{\([n,n-1,2]\) Single parity-check (SPC) code} --- Binary SPCs and repetition codes are dual to each other.
\item\relax
\flmRefsHyperref[eczindexfamilyrel]{code:pinwheel}{Pinwheel code} --- The hypergraph product of a pinwheel code with a cyclic repetition code yields a local Type-I fracton model in three dimensions \NoCaseChange{\protect\cite{cite1350}}.
\item\relax
\flmRefsHyperref[eczindexfamilyrel]{code:klemm}{Klemm code} --- The generator matrix of the Klemm code consists of a sum of the generator matrix of the repetition code and twice the generator matrix of the SPC code \NoCaseChange{\protect\cite{cite121}}.
\item\relax
\flmRefsHyperref[eczindexfamilyrel]{code:compactified_r}{Compactified \(\mathbb{R}\) gauge theory code} --- The compactified \(\mathbb{R}\) gauge theory code is constructed from a hypergraph product of two repetition codes over the integers.
\item\relax
\flmRefsHyperref[eczindexfamilyrel]{code:tiger_surface}{Tiger surface code} --- The tiger surface code is constructed from a hypergraph product of two repetition codes over the integers.
\item\relax
\flmRefsHyperref[eczindexfamilyrel]{code:self_correct}{Self-correcting quantum code} --- The repetition code associated with the 2D classical Ising model is a self-correcting classical memory \NoCaseChange{\protect\cite{cite1609}\protect\cite[{Sec. V.A}]{cite1610}}.
\item\relax
\flmRefsHyperref[eczindexfamilyrel]{code:iceberg}{\(\llbracket 2m,2m-2,2\rrbracket \) error-detecting code} --- The \(\llbracket 2m,2m-2,2\rrbracket \) error-detecting code is constructed via the CSS construction from an SPC code and its dual repetition code \NoCaseChange{\protect\cite[{Sec. III}]{cite773}}.
\item\relax
\flmRefsHyperref[eczindexfamilyrel]{code:hgp_7_2_2}{\(\llbracket 7,2,2\rrbracket \) HGP phantom code} --- This code is constructed from the hypergraph product of the \([3,2,2]\) simplex code and the \([2,1,2]\) repetition code \NoCaseChange{\protect\cite{cite514}}.
\item\relax
\flmRefsHyperref[eczindexfamilyrel]{code:anisotropic_z2_laplacian}{Anisotropic \(\mathbb{Z}_2\) Laplacian model code} --- The anisotropic \(\mathbb{Z}_2\) Laplacian model is the hypergraph product of a cyclic repetition code and a Laplacian code \NoCaseChange{\protect\cite{cite1350}}.
\item\relax
\flmRefsHyperref[eczindexfamilyrel]{code:chamon}{Chamon model code} --- The Chamon model code can be obtained from an XYZ product of three repetition codes \NoCaseChange{\protect\cite{cite1611}}, in a construction different from the 3D surface code; see \NoCaseChange{\protect\cite[{Sec. 3.4}]{cite645}}.
\item\relax
\flmRefsHyperref[eczindexfamilyrel]{code:fibonacci_fractal_liquid}{Fibonacci fractal spin-liquid code} --- The Fibonacci fractal spin-liquid code is a hypergraph product of the repetition code and the Fibonacci code \NoCaseChange{\protect\cite{cite1348}}, and can be formulated directly as a BP code \NoCaseChange{\protect\cite{cite1350}}.
\item\relax
\flmRefsHyperref[eczindexfamilyrel]{code:sierpinsky_fractal_liquid}{Sierpinski prism model code} --- The Sierpinski prism model code is a hypergraph product of the repetition code and the Newman-Moore code \NoCaseChange{\protect\cite{cite1501,cite1350}}.
\item\relax
\flmRefsHyperref[eczindexfamilyrel]{code:two_foliated}{Two-foliated fracton code} --- The two-foliated fracton code is a hypergraph product of the repetition code and the plaquette Ising code on a square lattice with periodic boundary conditions \NoCaseChange{\protect\cite{cite1517}}.
\item\relax
\flmRefsHyperref[eczindexfamilyrel]{code:lresc}{Long-range enhanced surface code (LRESC)} --- LRESCs are constructed using a hypergraph product of two copies of a concatenated LDPC-repetition seed code.
\item\relax
\flmRefsHyperref[eczindexfamilyrel]{code:surface}{Kitaev surface code} --- The planar surface code on a square lattice can be obtained from a hypergraph product of two repetition codes with appropriate boundary checks.
\item\relax
\flmRefsHyperref[eczindexfamilyrel]{code:toric}{Toric code} --- The toric code can be obtained from a hypergraph product of two repetition codes \NoCaseChange{\protect\cite[{Exam. 6}]{cite1316}}. Other hypergraph products of two repetition codes yield the related \(\llbracket 2d^2-2d+1,1,d\rrbracket \) CSS code family \NoCaseChange{\protect\cite[{Exam. 5}]{cite1316}}.
\item\relax
\flmRefsHyperref[eczindexfamilyrel]{code:3d_surface}{3D surface code} --- The 3D planar and toric code on a cubic lattice can be obtained from a hypergraph product of three repetition codes \NoCaseChange{\protect\cite{cite1613}\protect\cite[{Exam. A.1}]{cite1612}}.
\item\relax
\flmRefsHyperref[eczindexfamilyrel]{code:4d_13_surface}{\((1,3)\) 4D toric code} --- The 4D \((1,3)\) planar (toric) code on a hypercubic lattice can be obtained from a particular choice of chain complex from a hypergraph product of four repetition codes \NoCaseChange{\protect\cite{cite1613}}.
\item\relax
\flmRefsHyperref[eczindexfamilyrel]{code:4d_surface}{\((2,2)\) Loop toric code} --- The 4D loop planar (toric) code on a hypercubic lattice can be obtained from a particular choice of chain complex from a hypergraph product of four repetition codes \NoCaseChange{\protect\cite{cite1613}}.
\item\relax
\flmRefsHyperref[eczindexfamilyrel]{code:xzzx}{XZZX surface code} --- The Chamon model code can be obtained from an XYZ product of three repetition codes \NoCaseChange{\protect\cite{cite1611}}; see \NoCaseChange{\protect\cite[{Sec. 3.4}]{cite645}}. Using only two repetition codes in the analogous 2D construction yields the XZZX code, making it a 2D analogue of the Chamon code \NoCaseChange{\protect\cite[{Sec. 2}]{cite645}}.
\end{eczvaluelist}
\eczhbkcontributors{ Bao Bach, Yijia Xu, \eczhuVVA }
\endeczcode

\eczcode{sloane_whitehead}{Sloane-Whitehead code}{~\NoCaseChange{\protect\cite{cite374}}}
\codefieldsection{Description}
Member of an infinite \((n,\lambda\cdot 2^{n-m-1},3)\) nonlinear code family for any \(n\) satisfying \(2^m \leq n < 3\cdot 2^{m-1}\) for some \(m\) and for \(\lambda\in\{20/16,19/16,18/16\}\).
Such a code has more codewords than any linear code with the same length and distance.
The code is constructed by applying the \((u|u+v)\) construction recursively to the Julin-Golay codes \NoCaseChange{\protect\cite{cite374}\protect\cite[{Secs. 2.7 and 2.9}]{cite41}}.

\codefieldsection{Parents}
\begin{eczvaluelist}
\item\relax
\flmRefsHyperref[eczindexfamilyrel]{code:bits_into_bits}{Binary code}\item\relax
\flmRefsHyperref[eczindexfamilyrel]{code:uplusv}{\((u|u+v)\)-construction code}\item\relax
\flmRefsHyperref[eczindexfamilyrel]{code:small_distance}{Small-distance block code}\end{eczvaluelist}
\codefieldsection{Child}
\begin{eczvaluelist}
\item\relax
\flmRefsHyperref[eczindexfamilyrel]{code:julin12}{Julin-Golay code} --- Julin-Golay codes are the starting codes for the Sloane-Whitehead construction \NoCaseChange{\protect\cite{cite374}\protect\cite[{Secs. 2.7 and 2.9}]{cite41}}.
\end{eczvaluelist}
\eczhbkcontributors{ \eczhuVVA }
\endeczcode

\eczcode{sc_ldpc}{Spatially coupled LDPC (SC-LDPC) code}{~\NoCaseChange{\protect\cite{cite1614,cite1615,cite1616,cite1617,cite1618}}}
\codefieldsection{Description}
An LDPC code whose parity-check matrix is constructed by "spatially" coupling several copies of a regular LDPC parity-check matrix in chain-like fashion (or, more generally, in grid-like fashion) to yield a band matrix.
A finite-length chain is then capped by imposing either open boundary conditions (yielding \textit{non-tail-biting} SC-LDPC codes) or periodic boundary conditions (yielding \textit{tail-biting} SC-LDPC codes); sometimes extra \textit{terminating vertices} are added to the ends of the chain.
Matrices corresponding to translationally invariant chains are called \textit{time-invariant}, and otherwise are called \textit{time-varying}.
These codes can be constructed, e.g., using the \flmRefsHyperref{ref47}{lifting} procedure or using edge-cutting vectors \NoCaseChange{\protect\cite{cite95}}.
Spatial coupling can also be applied to MN-LDPC and HA-LDPC protographs, yielding bounded-density SC-MN and SC-HA families \NoCaseChange{\protect\cite{cite84}}.

\codefieldsection{Protection}
SC-LDPCs sometimes outperform other LDPC constructions \NoCaseChange{\protect\cite{cite1619,cite1620}}.

\codefieldsection{Rate}
Spatial coupling of LDPC codes can increase the achievable rate against BEC, coming close to the capacity \NoCaseChange{\protect\cite{cite1614,cite1617,cite1621}}. SC-LDPC codes achieve capacity of the binary memoryless symmetric (BMS) channel \NoCaseChange{\protect\cite{cite1451,cite1622}}. Spatial coupling of MN-LDPC and HA-LDPC codes yields bounded-density SC-MN and SC-HA families whose BEC BP thresholds are empirically close to the Shannon limit \NoCaseChange{\protect\cite{cite84}}.
\codefieldsection{Parent}
\begin{eczvaluelist}
\item\relax
\flmRefsHyperref[eczindexfamilyrel]{code:protograph_ldpc}{Protograph LDPC code} --- SC-LDPC codes can be interpreted as protograph LDPC codes \NoCaseChange{\protect\cite{cite1544}}.
\end{eczvaluelist}
\codefieldsection{Cousins}
\begin{eczvaluelist}
\item\relax
\flmRefsHyperref[eczindexfamilyrel]{code:ld_convolutional}{LDPC convolutional code (LDPC-CC)} --- Infinite-block versions of SC-LDPC are LDPC-CCs.
\item\relax
\flmRefsHyperref[eczindexfamilyrel]{code:mn_ldpc}{MacKay-Neal LDPC (MN-LDPC) code} --- Spatial coupling of MN-LDPC protographs yields bounded-density SC-MN codes with BEC BP thresholds close to the Shannon limit \NoCaseChange{\protect\cite{cite84}}.
\item\relax
\flmRefsHyperref[eczindexfamilyrel]{code:ha_ldpc}{Hsu-Anastasopoulos LDPC (HA-LDPC) code} --- Spatial coupling of HA-LDPC protographs yields bounded-density SC-HA codes with BEC BP thresholds close to the Shannon limit \NoCaseChange{\protect\cite{cite84}}.
\item\relax
\flmRefsHyperref[eczindexfamilyrel]{code:sc_qldpc}{Quantum spatially coupled (SC-QLDPC) code} --- SC-QLDPC code stabilizer-generator matrices have similar block form as the parity-check matrices of SC-LDPC codes.
\end{eczvaluelist}
\eczhbkcontributors{ \eczhuVVA }
\endeczcode

\eczcode{superimposed}{Superimposed code}{~\NoCaseChange{\protect\cite{cite1623,cite1624,cite1625,cite1626}}}
\codefieldsection{Description}
A set of binary strings with the property that the bitwise OR of any subset of at most \(s\) codewords uniquely identifies that subset, for some prescribed strength \(s\).
Equivalently, in the associated family of supports, no codeword support is contained in the union of \(s\) others \NoCaseChange{\protect\cite{cite1626}\protect\cite[{Thm. 1.2}]{cite1627}}.

\codefieldsection{Notes}
\begin{eczvaluelist}
\item\relax See Refs. \NoCaseChange{\protect\cite{cite1628,cite1629,cite1630}}.
\end{eczvaluelist}
\codefieldsection{Parent}
\begin{eczvaluelist}
\item\relax
\flmRefsHyperref[eczindexfamilyrel]{code:bits_into_bits}{Binary code}\end{eczvaluelist}
\eczhbkcontributors{ \eczhuVVA }
\endeczcode

\eczcode{ta-shma}{Ta-Shma zigzag code}{~\NoCaseChange{\protect\cite{cite1631}}}
\codefieldsection{Description}
Member of a family of \(\epsilon\)-balanced codes that nearly achieves the \flmRefsHyperref{ref85}{asymptotic GV bound}. The codes have relative distance \(\frac{1}{2}-\frac{\epsilon}{2}\) and rate of \flmRefsHyperref{ref65}{order} \(\Omega (\epsilon^{2+\beta})\) for \(\beta\to 0\) as \(n\to\infty\) \NoCaseChange{\protect\cite{cite96}}.

\codefieldsection{Decoding}
\begin{eczvaluelist}
\item\relax Unique and list decoders \NoCaseChange{\protect\cite{cite96}}.
\end{eczvaluelist}
\codefieldsection{Parent}
\begin{eczvaluelist}
\item\relax
\flmRefsHyperref[eczindexfamilyrel]{code:binary_linear}{Linear binary code}\end{eczvaluelist}
\codefieldsection{Cousin}
\begin{eczvaluelist}
\item\relax
\flmRefsHyperref[eczindexfamilyrel]{code:balanced}{Balanced code} --- Ta-Shma codes are \(\epsilon\)-balanced.
\end{eczvaluelist}
\eczhbkcontributors{ \eczhuVVA }
\endeczcode

\eczcode{tsf}{Tanner-Sridhara-Fuja (TSF) code}{~\NoCaseChange{\protect\cite{cite61}}}
\codefieldsection{Description}
Array QC-LDPC code constructed from a cyclically shifted identity matrix; see \NoCaseChange{\protect\cite[{Exam. 21.6.5}]{cite97}}.

\codefieldsection{Parent}
\begin{eczvaluelist}
\item\relax
\flmRefsHyperref[eczindexfamilyrel]{code:array_ldpc}{Array-based LDPC (AB-LDPC) code}\end{eczvaluelist}
\eczhbkcontributors{ \eczhuVVA }
\endeczcode

\eczcode{tornado}{Tornado code}{~\NoCaseChange{\protect\cite{cite98,cite1388,cite99}}}
\codefieldsection{Description}
Linear binary erasure code that is a precursor to fountain codes and is built from a multilayer cascade of sparse bipartite graphs. Its encoding and decoding operations involve only XOR gates \NoCaseChange{\protect\cite{cite98,cite99}\protect\cite[{Sec. 30.2}]{cite100}}.

\codefieldsection{Rate}
Tornado-code ensembles can come arbitrarily close to the capacity of the binary erasure channel \NoCaseChange{\protect\cite{cite99}}.
\codefieldsection{Encoding}
\begin{eczvaluelist}
\item\relax Linear-time encoder \NoCaseChange{\protect\cite{cite99}}.
\end{eczvaluelist}
\codefieldsection{Decoding}
\begin{eczvaluelist}
\item\relax Linear-time peeling decoder \NoCaseChange{\protect\cite{cite99}}. This decoder either terminates when it has removed a given erasure pattern or when it is stuck in a \textit{stopping set}.
\end{eczvaluelist}
\codefieldsection{Parent}
\begin{eczvaluelist}
\item\relax
\flmRefsHyperref[eczindexfamilyrel]{code:ldpc}{Low-density parity-check (LDPC) code} --- Tornado codes are LDPC codes that use a highly irregular weight distribution for their underlying graphs \NoCaseChange{\protect\cite{cite257}}.
\end{eczvaluelist}
\codefieldsection{Cousins}
\begin{eczvaluelist}
\item\relax
\flmRefsHyperref[eczindexfamilyrel]{code:fountain}{Fountain code} --- Tornado codes, the precursor to fountain codes, are much slower to encode and decode in the low-rate regime applicable to scalable data transmission \NoCaseChange{\protect\cite{cite1359,cite257}}.
\item\relax
\flmRefsHyperref[eczindexfamilyrel]{code:raptor}{Raptor (RAPid TORnado) code} --- Tornado codes, which can be used as a pre-code for raptor codes, also use a multi-layer approach where redundant symbols are created by one code for another code to use as input.
\end{eczvaluelist}
\eczhbkcontributors{ Thomas Wrona, \eczhuVVA }
\endeczcode

\eczcode{two_in_five}{Two-in-five code}{~\NoCaseChange{\protect\cite{cite1632}}}
\codefieldsection{Alternative Names}
\begin{eczvaluelist}
\item\relax Two-out-of-five code
\end{eczvaluelist}
\eczhIndexCodeAliasName{two_in_five}{Two-out-of-five code}
\codefieldsection{Description}
A nonlinear binary code consisting of the 10 weight-two five-bit strings, thereby providing an encoding for the decimal digits 0 through 9. 

\codefieldsection{Protection}
Detects some single bit-flips as well as unidirectional errors using the fact that each codeword has weight two. The code fails for any sequence of flips that maintains the constant weight.

\codefieldsection{Realizations}
\begin{eczvaluelist}
\item\relax Used in the United States Postal Service's POSTNET barcode system as well as the Postal Alphanumeric Encoding Technique (PLANET).
\item\relax Forms the numerical part of the \textit{Code 39} barcode encoding.
\item\relax Early IBM computers \NoCaseChange{\protect\cite{cite353,cite354}}.
\end{eczvaluelist}
\codefieldsection{Parents}
\begin{eczvaluelist}
\item\relax
\flmRefsHyperref[eczindexfamilyrel]{code:binary_group_orbit}{Binary group-orbit code} --- The two-in-five code is a binary group-orbit code with group \(S_5\).
\item\relax
\flmRefsHyperref[eczindexfamilyrel]{code:weight_two}{Weight-two code}\end{eczvaluelist}
\eczhbkcontributors{ \eczhuVVA }
\endeczcode

\eczcode{unary}{Unary code}{}
\codefieldsection{Alternative Names}
\begin{eczvaluelist}
\item\relax Thermometer code
\end{eczvaluelist}
\eczhIndexCodeAliasName{unary}{Thermometer code}
\codefieldsection{Description}
Trivial code that encodes integers \(0\) through \(n\) into binary strings of length \(n\).
The \(i\)th codeword is a string consisting of \(i\) ones followed by \(n-i\) zeroes.

\codefieldsection{Realizations}
\begin{eczvaluelist}
\item\relax Neural networks \NoCaseChange{\protect\cite{cite355}}.
\item\relax Birdsong production \NoCaseChange{\protect\cite{cite356}}.
\end{eczvaluelist}
\codefieldsection{Parent}
\begin{eczvaluelist}
\item\relax
\flmRefsHyperref[eczindexfamilyrel]{code:bits_into_bits}{Binary code}\end{eczvaluelist}
\codefieldsection{Cousin}
\begin{eczvaluelist}
\item\relax
\flmRefsHyperref[eczindexfamilyrel]{code:qlwc}{Quantum low-weight check (QLWC) code} --- A family of approximate non-stabilizer qubit QLWC codes with linear distance and rate has been constructed \NoCaseChange{\protect\cite{cite1633}} using unary codes that arise from the Feynman-Kitaev clock construction \NoCaseChange{\protect\cite{cite1634}}.
\end{eczvaluelist}
\eczhbkcontributors{ \eczhuVVA }
\endeczcode

\eczcode{vt_single_deletion}{Varshamov-Tenengolts (VT) code}{~\NoCaseChange{\protect\cite{cite1635,cite1636}}}
\codefieldsection{Description}
Nearly optimal binary deletion-correcting code and code for the asymmetric channel. 

Given integers \(n\geq 1\) and \(a\in\{0,1,\dots,n\}\), the associated Varshamov-Tenengolts code \(C_{n,a}\) corresponds to the set
\flmMathEnvironment{align}{}{
C_{n,a}=\left\{x\in\{0,1\}^n: \sum_{i=1}^n i~x_i = a\mod (n+1) \right\}.
}

By adapting a construction of Tenengolts \NoCaseChange{\protect\cite{cite1637}}, VT codes can be modified to yield slightly longer linear codes \NoCaseChange{\protect\cite{cite1009}}.
VT codes can be generalized to the \(q\)-ary case \NoCaseChange{\protect\cite{cite1638,cite1639,cite1012}}.
\codefieldsection{Protection}
Corrects a single asymmetric error (a \(0\) mapped to a \(1\)), a single deletion, or a single insertion of an arbitrary bit in an arbitrary position for any choice of \(a\).
\codefieldsection{Rate}
Has asymptotic rate \(1\). The redundancy is about \(\log_2(n+1)\) bits, making the construction nearly optimal for single-deletion correction.
\codefieldsection{Decoding}
\begin{eczvaluelist}
\item\relax Decoder based on checksums \(\sum_{i=1}^n i~x_i^{\prime}\) of corrupted codewords \(x_i^{\prime}\) \NoCaseChange{\protect\cite{cite1636,cite1640}}.
\end{eczvaluelist}
\codefieldsection{Parents}
\begin{eczvaluelist}
\item\relax
\flmRefsHyperref[eczindexfamilyrel]{code:constantin_rao}{Constantin-Rao (CR) code} --- CR codes for \(G=\mathbb{Z}_{n+1}\) reduce to VT codes.
\item\relax
\flmRefsHyperref[eczindexfamilyrel]{code:insertion_deletion}{Editing code}\end{eczvaluelist}
\eczhbkcontributors{ João Ribeiro, \eczhuVVA }
\endeczcode

\eczcode{weight_two}{Weight-two code}{~\NoCaseChange{\protect\cite{cite1632}}}
\codefieldsection{Description}
A length-\(n\) binary code whose codewords all have Hamming weight two. Such codes provide slightly extra redundancy for storage of small-scale information such as ZIP codes or decimal digits.

\codefieldsection{Parent}
\begin{eczvaluelist}
\item\relax
\flmRefsHyperref[eczindexfamilyrel]{code:constant_weight}{Constant-weight code}\end{eczvaluelist}
\codefieldsection{Child}
\begin{eczvaluelist}
\item\relax
\flmRefsHyperref[eczindexfamilyrel]{code:two_in_five}{Two-in-five code}\end{eczvaluelist}
\codefieldsection{Cousin}
\begin{eczvaluelist}
\item\relax
\flmRefsHyperref[eczindexfamilyrel]{code:two_weight}{Two-weight code} --- Codewords of two-weight codes have one of two possible Hamming weights, while those of weight-two codes have Hamming weight two.
\end{eczvaluelist}
\eczhbkcontributors{ \eczhuVVA }
\endeczcode

\eczcode{zetterberg}{Zetterberg code}{~\NoCaseChange{\protect\cite{cite1641}}}
\codefieldsection{Description}
Family of binary cyclic \([2^{2s}+1,2^{2s}-4s+1]\) codes with distance \(d\geq 5\) generated by the minimal polynomial \(g_s(x)\) of \(\alpha\) over \(\mathbb{F}_2\), where \(\alpha\) is a primitive \(n\)th root of unity in the field \(\mathbb{F}_{2^{4s}}\). They are quasi-perfect codes and are one of the best known families of double-error correcting binary linear codes.
\codefieldsection{Protection}
Correct at least all weight-2 errors.
\codefieldsection{Rate}
The rate is given by \(1-\frac{4s}{n}\), which is asymptotically good, with a minimum distance of 5.
\codefieldsection{Decoding}
\begin{eczvaluelist}
\item\relax Kallquist first described an algebraic decoding theorem \NoCaseChange{\protect\cite{cite1642}}. A faster version was later provided in Ref. \NoCaseChange{\protect\cite{cite1643}} and further modified in Ref. \NoCaseChange{\protect\cite{cite1644}}.
\end{eczvaluelist}
\codefieldsection{Realizations}
\begin{eczvaluelist}
\item\relax Code used to provide better protection of data transmission with its double error correcting capacity \NoCaseChange{\protect\cite{cite357}}.
\end{eczvaluelist}
\codefieldsection{Parents}
\begin{eczvaluelist}
\item\relax
\flmRefsHyperref[eczindexfamilyrel]{code:binary_cyclic}{Cyclic linear binary code}\item\relax
\flmRefsHyperref[eczindexfamilyrel]{code:quasi_perfect}{Quasi-perfect code} --- Zetterberg codes are quasi-perfect, with each \(n\)-bit string at most three bit-flips away from a codeword \NoCaseChange{\protect\cite{cite1643}}.
\end{eczvaluelist}
\eczhbkcontributors{ Xiaozhen Fu, \eczhuVVA }
\endeczcode

\onecolumngrid
\clearpage

\section{Galois-field Kingdom}

\begin{eczEpigraph}
\begin{quote}
\flmQuoteSetup{quote}%
In Galois fields, full of flowers\\
primitive elements dance for hours\\
climbing sequentially through the trees\\
and shouting occasional parities\\

The syndromes like ghosts in the misty damp\\
feed the smoldering fires of the Berlekamp\\
and high-flying exponents sometimes are downed\\
on the jagged peaks of the Gilbert bound.
\flmQuoteAttributed{S. B. Weinstein}
\end{quote}
\end{eczEpigraph}

\twocolumngrid

\eczcode{dodecacode}{\((12,4^6,6)_4\) Dodecacode}{~\NoCaseChange{\protect\cite{cite449}}}
\eczhIndexCodeAliasName{dodecacode}{Dodecacode}
\codefieldsection{Description}
The unique trace-Hermitian self-dual additive \((12,4^6,6)_4\) code.
Its codewords are cyclic permutations of \((\omega 10100100101)\), where \(\mathbb{F}_4=\{0,1,\omega,\bar{\omega}\}\) is the \flmRefsHyperref{ref33}{quaternary Galois field} \NoCaseChange{\protect\cite[{Sec. 2.4.8}]{cite42}}.
Another generator matrix can be found in \NoCaseChange{\protect\cite[{Exam. 9.10.8}]{cite126}}.

The dodecacode is a self-dual additive code, and there is no self-dual linear code with the same parameters \NoCaseChange{\protect\cite{cite1645}}.

Puncturing the code yields the \((11,4^6,5)_4\) additive code known as the \textit{punctured} or \textit{shortened dodecacode} \NoCaseChange{\protect\cite{cite1646}}.

\codefieldsection{Parent}
\begin{eczvaluelist}
\item\relax
\flmRefsHyperref[eczindexfamilyrel]{code:self_dual_additive}{Self-dual additive code} --- The dodecacode is trace-Hermitian self-dual additive.
\end{eczvaluelist}
\codefieldsection{Cousins}
\begin{eczvaluelist}
\item\relax
\flmRefsHyperref[eczindexfamilyrel]{code:combinatorial_design}{Combinatorial design} --- There exists a \(5\)-\((12, 6, 3)\) design in the dodecacode, and a \(3\)-\((11, 5, 4)\) design in the shortened dodecacode \NoCaseChange{\protect\cite{cite159}}.
\item\relax
\flmRefsHyperref[eczindexfamilyrel]{code:stab_11_1_5}{\(\llbracket 11,1,5\rrbracket \) quantum dodecacode} --- The dodecacode corresponds to a \(\llbracket 12,0,6\rrbracket \) quantum code in the \flmRefsHyperref{code:qubit_stabilizer}{\(\mathbb{F}_4\) representation}  \NoCaseChange{\protect\cite{cite449}}. The \(\llbracket 11,1,5\rrbracket \) quantum dodecacode code corresponds to the shortened dodecacode \NoCaseChange{\protect\cite{cite449,cite1647}}. A \flmRefsHyperref{ref672}{pure} \(\llbracket 10,1,4\rrbracket \) quantum code can be obtained from the doubly punctured dodecacode \NoCaseChange{\protect\cite{cite1647}}. These codes are not obtained from the Hermitian construction since none of the classical codes are linear.
\item\relax
\flmRefsHyperref[eczindexfamilyrel]{code:uniformly_packed}{Uniformly packed code} --- The punctured dodecacode code is uniformly packed \NoCaseChange{\protect\cite{cite1646}}.
\end{eczvaluelist}
\eczhbkcontributors{ \eczhuVVA }
\endeczcode

\eczcode{uplusv}{\((u|u+v)\)-construction code}{~\NoCaseChange{\protect\cite{cite1648,cite374}}}
\codefieldsection{Description}
Code constructed using a concatenation procedure that takes in two \(q\)-ary codes \(C_1,C_2\) and produces a new code whose codewords are \((u|u+v)\) for all \(u\in C_1\) and \(v\in C_2\).
If the two codes have parameters \((n,K_1,d_1)\) and \((n,K_2,d_2)\), then the output code is a \((2n,K_1 K_2, \min\{2d_1,d_2\})\) code \NoCaseChange{\protect\cite[{Thm. 5.10}]{cite62}\protect\cite[{pg. 76}]{cite41}}. When \(C_1\) and \(C_2\) are linear \([n,k_1,d_1]_q\) and \([n,k_2,d_2]_q\) codes, the resulting code is linear with parameters \([2n,k_1+k_2,\min\{2d_1,d_2\}]_q\).

\codefieldsection{Parent}
\begin{eczvaluelist}
\item\relax
\flmRefsHyperref[eczindexfamilyrel]{code:matrix_product}{Matrix-product code}\end{eczvaluelist}
\codefieldsection{Children}
\begin{eczvaluelist}
\item\relax
\flmRefsHyperref[eczindexfamilyrel]{code:sloane_whitehead}{Sloane-Whitehead code}\item\relax
\flmRefsHyperref[eczindexfamilyrel]{code:vasilyev}{\((2^{m+1}-1,2^{2n-m},3)\) Vasilyev code}\item\relax
\flmRefsHyperref[eczindexfamilyrel]{code:reed_muller}{Reed-Muller (RM) code} --- All RM codes can be constructed via the \((u|u+v)\) construction \NoCaseChange{\protect\cite[{Ch. 13}]{cite41}}.
\end{eczvaluelist}
\codefieldsection{Cousins}
\begin{eczvaluelist}
\item\relax
\flmRefsHyperref[eczindexfamilyrel]{code:hamming743}{\([7,4,3]\) Hamming code} --- Starting with the \([6,3,3]\) shortened Hamming code and applying the \((u|u+v)\) construction recursively using the repetition code yields a family of \([2^m,m+1,2^{m-1}]\) codes for \(m\geq1\) that saturate the Griesmer bound \NoCaseChange{\protect\cite[{pg. 90}]{cite62}}.
\item\relax
\flmRefsHyperref[eczindexfamilyrel]{code:binary_quad_residue}{Binary quadratic-residue (QR) code} --- The \((u|u+v)\) construction can be used to obtain nonlinear binary quadratic-residue codes \NoCaseChange{\protect\cite{cite374}}.
\item\relax
\flmRefsHyperref[eczindexfamilyrel]{code:berman}{Berman code} --- Berman codes are recursively constructed via a construction that is similar to the \((u|u+v)\) construction \NoCaseChange{\protect\cite{cite66}}.
\end{eczvaluelist}
\eczhbkcontributors{ \eczhuVVA }
\endeczcode

\eczcode{glynn}{\([10,5,6]_9\) Glynn code}{~\NoCaseChange{\protect\cite{cite1649}}}
\eczhIndexCodeAliasName{glynn}{Glynn code}
\codefieldsection{Description}
The unique trace-Hermitian self-dual \([10,5,6]_9\) code, constructed using a 10-arc in \(PG(4,9)\) that is not a rational curve.

The Glynn code is the unique trace-Hermitian self-dual code for its parameters, and is not Euclidean self-dual \NoCaseChange{\protect\cite{cite1650,cite1651,cite1652}}.

\codefieldsection{Parents}
\begin{eczvaluelist}
\item\relax
\flmRefsHyperref[eczindexfamilyrel]{code:self_dual}{Self-dual linear code} --- The Glynn code is trace-Hermitian self-dual, and is not Euclidean self-dual \NoCaseChange{\protect\cite{cite1650,cite1651,cite1652}}.
\item\relax
\flmRefsHyperref[eczindexfamilyrel]{code:projective}{Projective geometry code} --- The Glynn code is constructed using a 10-arc in \(PG(4,9)\) that is not a rational curve.
\item\relax
\flmRefsHyperref[eczindexfamilyrel]{code:mds}{Maximum distance separable (MDS) code} --- The Glynn code is a rare example of an MDS code that is not related to an RS code.
\end{eczvaluelist}
\codefieldsection{Cousins}
\begin{eczvaluelist}
\item\relax
\flmRefsHyperref[eczindexfamilyrel]{code:reed_solomon}{Reed-Solomon (RS) code} --- The only other inequivalent \([10,5,6]_9\) code is an RS code, which is the unique Euclidean self-dual code for its parameters, and which is not Hermitian self-dual \NoCaseChange{\protect\cite{cite1650,cite1651,cite1652}}.
\item\relax
\flmRefsHyperref[eczindexfamilyrel]{code:stab_9_1_5}{\(\llbracket 9,1,5\rrbracket _3\) quantum Glynn code} --- Applying the \flmRefsHyperref{code:stabilizer_over_gfqsq}{Hermitian construction} to the Glynn code yields a \(\llbracket 10,0,6\rrbracket _3\) state \NoCaseChange{\protect\cite{cite1653,cite1654}}. The \(\llbracket 9,1,5\rrbracket _3\) quantum Glynn code can be obtained by applying the \flmRefsHyperref{code:stabilizer_over_gfqsq}{Hermitian construction} to the shortened Glynn code \NoCaseChange{\protect\cite[{Corr. 4}]{cite1653}} (cf. \NoCaseChange{\protect\cite[{Exam. 7}]{cite1655}}).
\end{eczvaluelist}
\eczhbkcontributors{ \eczhuVVA }
\endeczcode

\eczcode{ternary_golay}{\([11,6,5]_3\) Ternary Golay code}{~\NoCaseChange{\protect\cite{cite361,cite1168}}}
\eczhIndexCodeAliasName{ternary_golay}{Ternary Golay code}
\codefieldsection{Description}
A \([11,6,5]_3\) perfect ternary linear code with connections to various areas of mathematics, e.g., lattices \NoCaseChange{\protect\cite{cite39}} and sporadic simple groups \NoCaseChange{\protect\cite{cite41}}.
Adding a parity bit to the code results in the self-dual \([12,6,6]_3\) \textit{extended ternary Golay code}, whose weight enumerator is the Gleason polynomial \(g_5\) \NoCaseChange{\protect\cite[{Rem. 4.2.6}]{cite40}}.
Up to equivalence, both codes are unique for their respective parameters \NoCaseChange{\protect\cite{cite102}}.
The dual of the ternary Golay code is a \([11,5,6]_3\) projective two-weight subcode \NoCaseChange{\protect\cite[{Exam. 19.3.2}]{cite172}}.

A generator matrix for the ternary Golay code is
\flmMathEnvironment{align}{}{
\left(\begin{array}{ccccccccccc}
  1 & 0 & 0 & 0 & 0 & 0 & 1 & 1 & 1 & 1 & 1 \\
  0 & 1 & 0 & 0 & 0 & 0 & 1 & 1 & 2 & 2 & 0 \\
  0 & 0 & 1 & 0 & 0 & 0 & 1 & 2 & 1 & 0 & 2 \\
  0 & 0 & 0 & 1 & 0 & 0 & 2 & 1 & 0 & 1 & 2 \\
  0 & 0 & 0 & 0 & 1 & 0 & 2 & 0 & 1 & 2 & 1 \\
  0 & 0 & 0 & 0 & 0 & 1 & 0 & 2 & 2 & 1 & 1
\end{array}\right)~.
}

The permutation automorphism group of the ternary Golay code is the Mathieu group \(\mathcal{M}_{11}\), while the monomial automorphism group of the extended ternary Golay code is the double cover \(2.\mathcal{M}_{12}\); these are closely related to two sporadic simple groups.

\codefieldsection{Decoding}
\begin{eczvaluelist}
\item\relax Decoder for the extended ternary Golay code using the tetracode \NoCaseChange{\protect\cite{cite1192}}.
\end{eczvaluelist}
\codefieldsection{Realizations}
\begin{eczvaluelist}
\item\relax Code used in football pools with at least one good bet \NoCaseChange{\protect\cite{cite360,cite246}}. In fact, the code was originally constructed by Juhani Virtakallio and published in the Finnish football pool magazine Veikkaaja \NoCaseChange{\protect\cite{cite361,cite246,cite354}}.
\item\relax Proofs of the quantum mechanical Kochen-Specker theorem \NoCaseChange{\protect\cite{cite362}}.
\end{eczvaluelist}
\codefieldsection{Notes}
\begin{eczvaluelist}
\item\relax The ternary Golay code is a quadratic-residue code \NoCaseChange{\protect\cite[{Sec. 2.7}]{cite68}}.
\end{eczvaluelist}
\codefieldsection{Parents}
\begin{eczvaluelist}
\item\relax
\flmRefsHyperref[eczindexfamilyrel]{code:perfect}{Perfect code} --- The ternary Golay code is perfect \NoCaseChange{\protect\cite[{Thm. 12.3.3 and Def. 12.3.4}]{cite199}}.
\item\relax
\flmRefsHyperref[eczindexfamilyrel]{code:q-ary_quad_residue}{Quadratic-residue (QR) code} --- The ternary Golay code is a quadratic-residue code over \(\mathbb{F}_3\) with residue set \(Q = \{1, 3, 4, 5, 9\} \) and generator polynomial \(x^5 + x^4 - x^3 + x^2 - 1\) \NoCaseChange{\protect\cite[{Ex. 3.2.10}]{cite70}\protect\cite[{Ch. 16}]{cite41}}.
\item\relax
\flmRefsHyperref[eczindexfamilyrel]{code:q-ary_linear_over_zq}{Linear code over \(\mathbb{Z}_q\)}\item\relax
\flmRefsHyperref[eczindexfamilyrel]{code:univ_opt_q-ary}{Universally optimal \(q\)-ary code} --- The ternary Golay code and several of its extended, shortened, and punctured versions are LP universally optimal codes \NoCaseChange{\protect\cite{cite173}}. The \([11,6,5]_3\) code is universally optimal but not sharp.
\item\relax
\flmRefsHyperref[eczindexfamilyrel]{code:small_distance}{Small-distance block code}\end{eczvaluelist}
\codefieldsection{Cousins}
\begin{eczvaluelist}
\item\relax
\flmRefsHyperref[eczindexfamilyrel]{code:golay}{\([23, 12, 7]\) Golay code} --- The ternary Golay code is the ternary counterpart of the binary Golay code.
\item\relax
\flmRefsHyperref[eczindexfamilyrel]{code:extended_golay}{\([24, 12, 8]\) Extended Golay code} --- The extended ternary Golay code is the ternary counterpart of the extended binary Golay code.
\item\relax
\flmRefsHyperref[eczindexfamilyrel]{code:self_dual}{Self-dual linear code} --- The extended ternary Golay code is self-dual, i.e., a Type III code in the terminology of \NoCaseChange{\protect\cite[{Rems. 4.2.6 and 4.3.2}]{cite40}}.
\item\relax
\flmRefsHyperref[eczindexfamilyrel]{code:self_dual_over_zq}{Self-dual code over \(\mathbb{Z}_q\)} --- The extended ternary Golay code is self-dual \NoCaseChange{\protect\cite[{Rem. 4.2.6}]{cite40}}.
\item\relax
\flmRefsHyperref[eczindexfamilyrel]{code:projective}{Projective geometry code} --- The extended ternary Golay code admits a projective geometric construction \NoCaseChange{\protect\cite[{pg. 296}]{cite62}}.
\item\relax
\flmRefsHyperref[eczindexfamilyrel]{code:divisible}{Divisible code} --- The extended ternary Golay code is 3-divisible because ternary self-dual codes are Type III \NoCaseChange{\protect\cite[{Thm. 4.1.9}]{cite40}}.
\item\relax
\flmRefsHyperref[eczindexfamilyrel]{code:delsarte_optimal_q-ary}{\(q\)-ary sharp configuration} --- The dual \([11,5,6]_3\) code of the ternary Golay code is the length-\(11\) ternary Golay-family sharp configuration in \NoCaseChange{\protect\cite[{Table 12.1}]{cite199}}. The extended ternary Golay code is the corresponding sharp configuration at length \(12\).
\item\relax
\flmRefsHyperref[eczindexfamilyrel]{code:projective_two_weight}{Projective two-weight code} --- The dual of the ternary Golay code is a projective two-weight subcode \NoCaseChange{\protect\cite{cite1656,cite1657}\protect\cite[{Exam. 19.3.2}]{cite172}\protect\cite[{Table 7.1}]{cite206}}.
\item\relax
\flmRefsHyperref[eczindexfamilyrel]{code:leech}{\(\Lambda_{24}\) Leech lattice} --- A 12-dimensional complex version of the Leech lattice can be obtained from the ternary Golay code \NoCaseChange{\protect\cite{cite1658,cite1659}\protect\cite[{pg. 200}]{cite39}}.
\item\relax
\flmRefsHyperref[eczindexfamilyrel]{code:niemeier}{Niemeier lattice} --- The extended ternary Golay code is the glue code for the Niemeier lattice \(A^{12}_2\) \NoCaseChange{\protect\cite[{Ch. 16, pg. 408}]{cite39}}.
\item\relax
\flmRefsHyperref[eczindexfamilyrel]{code:combinatorial_design}{Combinatorial design} --- The supports of the weight-five codewords of the ternary Golay code and the weight-six codewords of the extended ternary Golay code support the Steiner systems \(S(4,5,11)\) and \(S(5,6,12)\), respectively \NoCaseChange{\protect\cite{cite160,cite154}\protect\cite[{pg. 89}]{cite39}}. The latter blocks are called hexads.
\item\relax
\flmRefsHyperref[eczindexfamilyrel]{code:pless_symmetry}{\([2q+2,q+1]_3\) Pless symmetry code} --- The Pless symmetry code for \(p=5\) is the extended ternary Golay code.
\item\relax
\flmRefsHyperref[eczindexfamilyrel]{code:tetracode}{\([4,2,3]_3\) Tetracode} --- Extended ternary Golay codewords can be obtained from tetracodewords \NoCaseChange{\protect\cite{cite39}}. The tetracode can be used to decode the extended ternary Golay code \NoCaseChange{\protect\cite{cite1192}}.
\item\relax
\flmRefsHyperref[eczindexfamilyrel]{code:qutrit_golay}{\(\llbracket 11,1,5\rrbracket _3\) qutrit Golay code} --- The qutrit Golay code is a CSS code constructed from the ternary Golay code.
\end{eczvaluelist}
\eczhbkcontributors{ Vikram Elijah Amin, Shashank Sule, \eczhuVVA }
\endeczcode

\eczcode{pless_symmetry}{\([2q+2,q+1]_3\) Pless symmetry code}{~\NoCaseChange{\protect\cite{cite1660,cite165}}}
\codefieldsection{Alternative Names}
\begin{eczvaluelist}
\item\relax Pless double circulant code
\end{eczvaluelist}
\eczhIndexCodeAliasName{pless_symmetry}{Pless symmetry code}
\eczhIndexCodeAliasName{pless_symmetry}{Pless double circulant code}
\codefieldsection{Description}
A member of a family of self-dual ternary \([2q+2,q+1]_3\) codes for any power of an odd prime satisfying \(q \equiv 2\) modulo 3.

The code's generator matrix is \(G = [I | S_q]\), where \(I\) is the \((q+1)\)-dimensional identity matrix, and where the matrix \(S_q\) is shown in the following image (with \(q=p\)).
There, \(\chi(0)=0\), \(\chi(x)=1\) if \(x\) is a square in \(\mathbb{F}_q\), and \(\chi(x)=-1\) if \(x\) is not a square in \(\mathbb{F}_q\).
\begin{flmFloat}{table}{Bare}\includegraphics[width=252.74970000000002bp,max width=\linewidth]{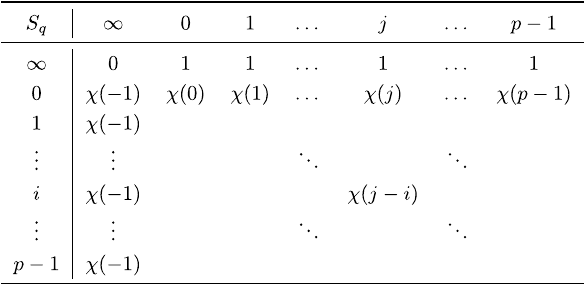}\end{flmFloat}

See \NoCaseChange{\protect\cite[{Sec. 10.5}]{cite126}\protect\cite[{pg. 87}]{cite39}} for more details.

\codefieldsection{Rate}
Achieve capacity of the binary erasure channel; see Ref. \NoCaseChange{\protect\cite{cite1661}}.
\codefieldsection{Parents}
\begin{eczvaluelist}
\item\relax
\flmRefsHyperref[eczindexfamilyrel]{code:self_dual}{Self-dual linear code}\item\relax
\flmRefsHyperref[eczindexfamilyrel]{code:self_dual_over_zq}{Self-dual code over \(\mathbb{Z}_q\)}\item\relax
\flmRefsHyperref[eczindexfamilyrel]{code:quasi_cyclic}{Quasi-cyclic code} --- Pless symmetry codes are double circulant \NoCaseChange{\protect\cite[{pg. 510}]{cite41}}.
\end{eczvaluelist}
\codefieldsection{Cousins}
\begin{eczvaluelist}
\item\relax
\flmRefsHyperref[eczindexfamilyrel]{code:q-ary_quad_residue}{Quadratic-residue (QR) code} --- Pless symmetry codes for lengths 24, 48, and 60 they have the same Hamming weight enumerators as the corresponding extended QR codes but are not equivalent \NoCaseChange{\protect\cite[{pg. 511}]{cite41}}.
\item\relax
\flmRefsHyperref[eczindexfamilyrel]{code:ternary_golay}{\([11,6,5]_3\) Ternary Golay code} --- The Pless symmetry code for \(p=5\) is the extended ternary Golay code.
\item\relax
\flmRefsHyperref[eczindexfamilyrel]{code:combinatorial_design}{Combinatorial design} --- The supports of fixed-weight codewords of certain Pless symmetry codes support combinatorial designs \NoCaseChange{\protect\cite{cite164,cite165,cite154}}.
\end{eczvaluelist}
\eczhbkcontributors{ Connor Clayton, \eczhuVVA }
\endeczcode

\eczcode{tetracode}{\([4,2,3]_3\) Tetracode}{~\NoCaseChange{\protect\cite{cite39}}}
\eczhIndexCodeAliasName{tetracode}{Tetracode}
\codefieldsection{Description}
The \([4,2,3]_3\) ternary self-dual MDS code that has connections to lattices \NoCaseChange{\protect\cite{cite39}}. Its weight enumerator is the Gleason polynomial \(g_4\) \NoCaseChange{\protect\cite[{Rem. 4.2.6}]{cite40}}.

A generator matrix is
\flmMathEnvironment{align}{}{
  \begin{pmatrix}1 & 0 & 1 & 1\\
  0 & 1 & 1 & 2
  \end{pmatrix}~,
}
where \(\mathbb{F}_3 = \{0,1,2\}\).

\codefieldsection{Notes}
\begin{eczvaluelist}
\item\relax See corresponding MinT database entry \NoCaseChange{\protect\cite{cite1662}}.
\end{eczvaluelist}
\codefieldsection{Parents}
\begin{eczvaluelist}
\item\relax
\flmRefsHyperref[eczindexfamilyrel]{code:self_dual}{Self-dual linear code} --- The tetracode is Euclidean self-dual, i.e., Type III in the terminology of \NoCaseChange{\protect\cite[{Rems. 4.2.6 and 4.3.2}]{cite40}}.
\item\relax
\flmRefsHyperref[eczindexfamilyrel]{code:self_dual_over_zq}{Self-dual code over \(\mathbb{Z}_q\)}\item\relax
\flmRefsHyperref[eczindexfamilyrel]{code:q-ary_simplex}{\(q\)-ary simplex code} --- The tetracode is equivalent to \(S(3,2)\).
\item\relax
\flmRefsHyperref[eczindexfamilyrel]{code:q-ary_hamming}{\(q\)-ary Hamming code} --- The tetracode is equivalent to the \(r=2\) \(3\)-ary Hamming code.
\item\relax
\flmRefsHyperref[eczindexfamilyrel]{code:extended_reed_solomon}{Extended GRS code} --- The tetracode is an extended RS code \NoCaseChange{\protect\cite[{pg. 81}]{cite39}}.
\item\relax
\flmRefsHyperref[eczindexfamilyrel]{code:lexicographic}{Lexicographic code} --- The tetracode is a lexicode \NoCaseChange{\protect\cite{cite147}}.
\end{eczvaluelist}
\codefieldsection{Cousins}
\begin{eczvaluelist}
\item\relax
\flmRefsHyperref[eczindexfamilyrel]{code:mds}{Maximum distance separable (MDS) code} --- The tetracode is a unique MDS code \NoCaseChange{\protect\cite{cite1663,cite1664}}.
\item\relax
\flmRefsHyperref[eczindexfamilyrel]{code:ternary_golay}{\([11,6,5]_3\) Ternary Golay code} --- Extended ternary Golay codewords can be obtained from tetracodewords \NoCaseChange{\protect\cite{cite39}}. The tetracode can be used to decode the extended ternary Golay code \NoCaseChange{\protect\cite{cite1192}}.
\item\relax
\flmRefsHyperref[eczindexfamilyrel]{code:eeight}{\(E_8\) Gosset lattice} --- The \([4,2,3]_3\) tetracode can be used to obtain the \(E_8\) Gosset lattice \NoCaseChange{\protect\cite[{Exam. 10.5.5}]{cite115}\protect\cite[{Ch. 7, pg. 200}]{cite39}}.
\item\relax
\flmRefsHyperref[eczindexfamilyrel]{code:niemeier}{Niemeier lattice} --- The tetracode is the glue code for the Niemeier lattice \(E_6^4\) \NoCaseChange{\protect\cite[{Ch. 16, pg. 408}]{cite39}}.
\end{eczvaluelist}
\eczhbkcontributors{ \eczhuVVA }
\endeczcode

\eczcode{reed_solomon_4}{\([4,2,3]_4\) RS\(_4\) code}{}
\codefieldsection{Alternative Names}
\begin{eczvaluelist}
\item\relax \(XQ(\mathbb{F}_4,3)\)
\end{eczvaluelist}
\eczhIndexCodeAliasName{reed_solomon_4}{RS\(_4\) code}
\eczhIndexCodeAliasName{reed_solomon_4}{\(XQ(\mathbb{F}_4,3)\)}
\codefieldsection{Description}
A Type II Euclidean self-dual extended RS code that is the smallest quaternary extended QR code \NoCaseChange{\protect\cite[{pg. 296}]{cite41}\protect\cite[{Sec. 2.4.2}]{cite42}}.
Puncturing the \([4,2,3]_4\) RS\(_4\) code yields the \([3,2,2]_4\) shortened RS\(_4\) code, which is an RS code \NoCaseChange{\protect\cite[{pg. 295}]{cite41}}.

A generator matrix for the code is
\flmMathEnvironment{align}{}{
  \begin{pmatrix}
  1 & 1 & 1 & 1 \\
  0 & 1 & \omega & \omega^2
  \end{pmatrix}~,
}
where \(\mathbb{F}_4 = \{0,1,\omega, \bar{\omega}\}\) is the \flmRefsHyperref{ref33}{quaternary Galois field}.

The automorphism group of the code is \(3.S_4\) \NoCaseChange{\protect\cite[{Sec. 2.4.2}]{cite42}}.

\codefieldsection{Parents}
\begin{eczvaluelist}
\item\relax
\flmRefsHyperref[eczindexfamilyrel]{code:self_dual}{Self-dual linear code} --- The RS\(_4\) is the smallest Type II Euclidean self-dual code \NoCaseChange{\protect\cite[{Sec. 2.4.2}]{cite42}}.
\item\relax
\flmRefsHyperref[eczindexfamilyrel]{code:extended_reed_solomon}{Extended GRS code} --- The RS\(_4\) is an extended RS code \NoCaseChange{\protect\cite[{pg. 296}]{cite41}}.
\item\relax
\flmRefsHyperref[eczindexfamilyrel]{code:mds}{Maximum distance separable (MDS) code}\item\relax
\flmRefsHyperref[eczindexfamilyrel]{code:small_distance}{Small-distance block code}\end{eczvaluelist}
\codefieldsection{Cousins}
\begin{eczvaluelist}
\item\relax
\flmRefsHyperref[eczindexfamilyrel]{code:q-ary_quad_residue}{Quadratic-residue (QR) code} --- The RS\(_4\) code is the smallest quaternary extended QR code \NoCaseChange{\protect\cite[{Sec. 2.4.2}]{cite42}}. The shortened RS\(_4\) code is the smallest quaternary QR code.
\item\relax
\flmRefsHyperref[eczindexfamilyrel]{code:hamming844}{\([8,4,4]\) extended Hamming code} --- The RS\(_4\) code can be mapped to the \([8,4,4]\) extended Hamming code \NoCaseChange{\protect\cite[{Sec. 2.4.2}]{cite42}} by identifying \((0,\omega,\bar{\omega},1)\) with \((00),(10),(01),(11)\) \NoCaseChange{\protect\cite{cite109}}.
\item\relax
\flmRefsHyperref[eczindexfamilyrel]{code:galois_3_1_2}{\(\llbracket 3,1,2\rrbracket _4\) three-Galois-quartrit code} --- The \(\llbracket 3,1,2\rrbracket _4\) code is constructed from the shortened RS\(_4\) code \NoCaseChange{\protect\cite{cite514}}.
\end{eczvaluelist}
\eczhbkcontributors{ \eczhuVVA }
\endeczcode

\eczcode{shortened_hexacode}{\([5,3,3]_4\) Shortened hexacode}{~\NoCaseChange{\protect\cite{cite1665,cite39}}}
\codefieldsection{Alternative Names}
\begin{eczvaluelist}
\item\relax Shorter hexacode
\item\relax Golay code over \(\mathbb{F}_4\)
\end{eczvaluelist}
\eczhIndexCodeAliasName{shortened_hexacode}{Shortened hexacode}
\eczhIndexCodeAliasName{shortened_hexacode}{Shorter hexacode}
\eczhIndexCodeAliasName{shortened_hexacode}{Golay code over \(\mathbb{F}_4\)}
\codefieldsection{Description}
A perfect \([5,3,3]_4\) quaternary Hamming code that is the result of puncturing the hexacode \NoCaseChange{\protect\cite{cite43}}.

\codefieldsection{Parents}
\begin{eczvaluelist}
\item\relax
\flmRefsHyperref[eczindexfamilyrel]{code:perfect}{Perfect code} --- The shortened hexacode is perfect \NoCaseChange{\protect\cite[{Exer. 578}]{cite126}}.
\item\relax
\flmRefsHyperref[eczindexfamilyrel]{code:q-ary_quad_residue}{Quadratic-residue (QR) code} --- The shortened hexacode is an odd-like quadratic-residue code \NoCaseChange{\protect\cite[{Exam. 6.6.8}]{cite126}}.
\item\relax
\flmRefsHyperref[eczindexfamilyrel]{code:extended_reed_solomon}{Extended GRS code} --- The shortened hexacode is a doubly extended narrow-sense RS code \NoCaseChange{\protect\cite[{pg. 82}]{cite39}}.
\item\relax
\flmRefsHyperref[eczindexfamilyrel]{code:small_distance}{Small-distance block code}\end{eczvaluelist}
\codefieldsection{Cousins}
\begin{eczvaluelist}
\item\relax
\flmRefsHyperref[eczindexfamilyrel]{code:hexacode}{\([6,3,4]_4\) Hexacode} --- The shortened hexacode is obtained by puncturing the hexacode.
\item\relax
\flmRefsHyperref[eczindexfamilyrel]{code:self_dual}{Self-dual linear code} --- The hexacode and the shortened hexacode are extremal \NoCaseChange{\protect\cite[{Tab. 9.14}]{cite126}\protect\cite[{Tm. 12}]{cite43}}.
\item\relax
\flmRefsHyperref[eczindexfamilyrel]{code:stab_5_1_3}{\(\llbracket 5,1,3\rrbracket \) Five-qubit perfect code} --- The five-qubit code can be obtained either by applying the \flmRefsHyperref{code:stabilizer_over_gf4}{qubit Hermitian construction} to the shortened hexacode \NoCaseChange{\protect\cite[{Exam. A}]{cite1666}} or by tracing out a qubit of the \(\llbracket 6,0,4\rrbracket \) code \NoCaseChange{\protect\cite[{Appx. A}]{cite1667}}.
\item\relax
\flmRefsHyperref[eczindexfamilyrel]{code:golay}{\([23, 12, 7]\) Golay code} --- The shortened hexacode is often referred to as the Golay code over \(\mathbb{F}_4\) \NoCaseChange{\protect\cite{cite126}}.
\item\relax
\flmRefsHyperref[eczindexfamilyrel]{code:reed_solomon}{Reed-Solomon (RS) code} --- The dual of the shortened hexacode code is a \([5,2,4]_4\) doubly extended RS code \NoCaseChange{\protect\cite[{Exam. A}]{cite1666}}.
\item\relax
\flmRefsHyperref[eczindexfamilyrel]{code:css_5_1_3}{\(\llbracket 5,1,3\rrbracket _4\) Galois-qudit CSS code} --- The \(\llbracket 5,1,3\rrbracket _4\) code is obtained from the shortened hexacode \NoCaseChange{\protect\cite{cite514}}.
\end{eczvaluelist}
\eczhbkcontributors{ \eczhuVVA }
\endeczcode

\eczcode{hill_56_6_36}{\([56,6,36]_3\) Hill-cap code}{~\NoCaseChange{\protect\cite{cite207}}}
\eczhIndexCodeAliasName{hill_56_6_36}{Hill-cap code}
\codefieldsection{Description}
Projective two-weight ternary code based on the Games graph \NoCaseChange{\protect\cite{cite206}\protect\cite[{Table 19.1}]{cite172}} and Hill's 56-cap \NoCaseChange{\protect\cite{cite207}}.
Its automorphism group contains \(PSL(3,4)\) \NoCaseChange{\protect\cite{cite208}}.

\codefieldsection{Parents}
\begin{eczvaluelist}
\item\relax
\flmRefsHyperref[eczindexfamilyrel]{code:hill_cap}{Hill projective-cap code}\item\relax
\flmRefsHyperref[eczindexfamilyrel]{code:quasi_twisted}{Quasi-twisted code} --- The \([56,6,36]_3\) Hill-cap code is quasi-twisted \NoCaseChange{\protect\cite{cite208}}.
\item\relax
\flmRefsHyperref[eczindexfamilyrel]{code:q-ary_linear_over_zq}{Linear code over \(\mathbb{Z}_q\)}\item\relax
\flmRefsHyperref[eczindexfamilyrel]{code:delsarte_optimal_q-ary}{\(q\)-ary sharp configuration} --- The \([56,6,36]_3\) Hill-cap code is a \(q\)-ary sharp configurations \NoCaseChange{\protect\cite[{Table 12.1}]{cite199}}.
\end{eczvaluelist}
\eczhbkcontributors{ \eczhuVVA }
\endeczcode

\eczcode{hexacode}{\([6,3,4]_4\) Hexacode}{~\NoCaseChange{\protect\cite{cite1665,cite39}}}
\codefieldsection{Alternative Names}
\begin{eczvaluelist}
\item\relax Extended Golay code over \(\mathbb{F}_4\)
\end{eczvaluelist}
\eczhIndexCodeAliasName{hexacode}{Hexacode}
\eczhIndexCodeAliasName{hexacode}{Extended Golay code over \(\mathbb{F}_4\)}
\codefieldsection{Description}
The \([6,3,4]_4\) Hermitian self-dual MDS code that has connections to projective geometry, lattices \NoCaseChange{\protect\cite{cite39}}, and conformal field theory \NoCaseChange{\protect\cite{cite44}}. Its weight enumerator is the Gleason polynomial \(g_7\) \NoCaseChange{\protect\cite[{Rem. 4.2.6}]{cite40}}.

A generator matrix for the hexacode is \NoCaseChange{\protect\cite[{Exam. 1.7.8}]{cite126}}
\flmMathEnvironment{align}{}{
  \begin{pmatrix}
  1 & 0 & 0 & 1 & 1 & \omega\\
  0 & 1 & 0 & 1 & \omega & 1\\
  0 & 0 & 1 & \omega & 1 & 1
  \end{pmatrix}~,
}
where \(\mathbb{F}_4 = \{0,1,\omega, \bar{\omega}\}\) is the \flmRefsHyperref{ref33}{quaternary Galois field}.

The permutation automorphism group of the hexacode is \(A_5 \cong I\), and the automorphism group is \(3.A_6 \cong \Sigma(360\phi)\). This is expanded to \(3.S_6\) if one includes the Galois automorphism \(\omega \leftrightarrow \omega^2\) \NoCaseChange{\protect\cite[{Exam. 1.7.8}]{cite126}\protect\cite[{pg. 520}]{cite41}\protect\cite[{Ch. 3, pg. 82}]{cite39}}.

\codefieldsection{Decoding}
\begin{eczvaluelist}
\item\relax Bounded-distance decoder requiring at most 34 real operations \NoCaseChange{\protect\cite{cite1191}}.
\end{eczvaluelist}
\codefieldsection{Notes}
\begin{eczvaluelist}
\item\relax See \NoCaseChange{\protect\cite[{Sec. 10.3}]{cite126}} for an exposition.
\item\relax See corresponding MinT database entry \NoCaseChange{\protect\cite{cite1668}}.
\end{eczvaluelist}
\codefieldsection{Parents}
\begin{eczvaluelist}
\item\relax
\flmRefsHyperref[eczindexfamilyrel]{code:self_dual}{Self-dual linear code} --- The hexacode is Hermitian self-dual \NoCaseChange{\protect\cite[{Rem. 4.2.6}]{cite40}} and, as a result, is also trace-Hermitian self-dual additive \NoCaseChange{\protect\cite[{Sec. 9.10}]{cite126}}. The hexacode and the shortened hexacode are extremal \NoCaseChange{\protect\cite[{Tab. 9.14}]{cite126}\protect\cite[{Tm. 12}]{cite43}}.
\item\relax
\flmRefsHyperref[eczindexfamilyrel]{code:hyperoval}{Hyperoval code} --- Columns of hexacode's generator matrix represent the six points of a hyperoval in the projective plane \(PG(2,4)\), an example of a two-weight code \NoCaseChange{\protect\cite[{pg. 289}]{cite62}\protect\cite[{Exam. 19.2.1}]{cite172}}.
\item\relax
\flmRefsHyperref[eczindexfamilyrel]{code:evaluation}{Evaluation AG code} --- The hexacode is an evaluation AG code over the \flmRefsHyperref{ref33}{quaternary Galois field} \(\mathbb{F}_4 = \{0,1,\omega, \bar{\omega}\}\) with \(\cal X\) defined by \(x^2 y + \omega y^2 z + \bar{\omega} z^2 x = 0\) \NoCaseChange{\protect\cite[{Exam. 2.77}]{cite32}}.
\item\relax
\flmRefsHyperref[eczindexfamilyrel]{code:denniston}{Denniston code} --- A version of the hexacode is recovered for Denniston code parameters \(i=1\) and \(a=2\) \NoCaseChange{\protect\cite{cite62}}.
\item\relax
\flmRefsHyperref[eczindexfamilyrel]{code:extended_reed_solomon}{Extended GRS code} --- The hexacode is a triply extended RS code \NoCaseChange{\protect\cite[{pg. 82}]{cite39}}.
\item\relax
\flmRefsHyperref[eczindexfamilyrel]{code:lexicographic}{Lexicographic code} --- Hexacodewords can be arranged in an order from smallest to largest, with each codeword differing at four places from the next \NoCaseChange{\protect\cite{cite1669}\protect\cite[{pg. 327}]{cite41}}.
\item\relax
\flmRefsHyperref[eczindexfamilyrel]{code:small_distance}{Small-distance block code}\end{eczvaluelist}
\codefieldsection{Cousins}
\begin{eczvaluelist}
\item\relax
\flmRefsHyperref[eczindexfamilyrel]{code:mds}{Maximum distance separable (MDS) code} --- The hexacode is an MDS code \NoCaseChange{\protect\cite[{Exer. 578}]{cite126}}.
\item\relax
\flmRefsHyperref[eczindexfamilyrel]{code:q-ary_quad_residue}{Quadratic-residue (QR) code} --- The hexacode is the smallest example of an extended quadratic-residue code of Type \(4^H\) \NoCaseChange{\protect\cite[{Sec. 2.4.6}]{cite42}\protect\cite[{Exer. 363}]{cite126}}.
\item\relax
\flmRefsHyperref[eczindexfamilyrel]{code:q-ary_hamming}{\(q\)-ary Hamming code} --- The hexacode is an extended quaternary Hamming code \NoCaseChange{\protect\cite[{Exer. 578}]{cite126}}.
\item\relax
\flmRefsHyperref[eczindexfamilyrel]{code:stab_5_1_3}{\(\llbracket 5,1,3\rrbracket \) Five-qubit perfect code} --- Applying the \flmRefsHyperref{code:stabilizer_over_gf4}{qubit Hermitian construction} to the hexacode yields the \(\llbracket 6,0,4\rrbracket \) quantum hexacode \NoCaseChange{\protect\cite{cite1670}}, and tracing out any one qubit of that code yields the \(\llbracket 5,1,3\rrbracket \) five-qubit code \NoCaseChange{\protect\cite[{Cor. 9 and Cor. 10}]{cite446}}.
\item\relax
\flmRefsHyperref[eczindexfamilyrel]{code:extended_golay}{\([24, 12, 8]\) Extended Golay code} --- Extended Golay codewords can be obtained from hexacodewords \NoCaseChange{\protect\cite{cite39}}. The hexacode can be used to decode the extended Golay code \NoCaseChange{\protect\cite{cite1192}}.
\item\relax
\flmRefsHyperref[eczindexfamilyrel]{code:golay}{\([23, 12, 7]\) Golay code} --- There is a connection between automorphisms of the even Golay code and the holomorph of the hexacode \NoCaseChange{\protect\cite{cite44}}. The hexacode is often referred to as the extended Golay code over \(\mathbb{F}_4\) \NoCaseChange{\protect\cite{cite126}}.
\item\relax
\flmRefsHyperref[eczindexfamilyrel]{code:spherical_design}{Spherical design} --- The hexacode is a complex spherical 3-design when embedded into the complex sphere via the polyphase mapping \NoCaseChange{\protect\cite{cite1671}}.
\item\relax
\flmRefsHyperref[eczindexfamilyrel]{code:polyphase}{Polyphase code} --- The hexacode is a complex spherical 3-design when embedded into the complex sphere via the polyphase mapping \NoCaseChange{\protect\cite{cite1671}}.
\item\relax
\flmRefsHyperref[eczindexfamilyrel]{code:coxeter_todd}{Coxeter-Todd \(K_{12}\) lattice} --- The hexacode can be used to obtain the Coxeter-Todd \(K_{12}\) lattice \NoCaseChange{\protect\cite[{Exam. 10.5.6}]{cite115}\protect\cite[{Ch. 7, pg. 198}]{cite39}}.
\item\relax
\flmRefsHyperref[eczindexfamilyrel]{code:niemeier}{Niemeier lattice} --- The hexacode is the glue code for the Niemeier lattice \(D_4^6\) \NoCaseChange{\protect\cite[{Ch. 16, pg. 408}]{cite39}}.
\item\relax
\flmRefsHyperref[eczindexfamilyrel]{code:shortened_hexacode}{\([5,3,3]_4\) Shortened hexacode} --- The shortened hexacode is obtained by puncturing the hexacode.
\end{eczvaluelist}
\eczhbkcontributors{ Simon Burton, \eczhuVVA }
\endeczcode

\eczcode{hill_78_6_56}{\([78,6,56]_4\) Hill-cap code}{~\NoCaseChange{\protect\cite{cite1672}}}
\eczhIndexCodeAliasName{hill_78_6_56}{Hill-cap code}
\codefieldsection{Description}
Projective two-weight quaternary code based on a cap that corresponds to a strongly regular graph \NoCaseChange{\protect\cite[{Table 7.1}]{cite206}}.

\codefieldsection{Parents}
\begin{eczvaluelist}
\item\relax
\flmRefsHyperref[eczindexfamilyrel]{code:hill_cap}{Hill projective-cap code}\item\relax
\flmRefsHyperref[eczindexfamilyrel]{code:delsarte_optimal_q-ary}{\(q\)-ary sharp configuration} --- The \([78,6,56]_4\) Hill-cap code is a \(q\)-ary sharp configuration \NoCaseChange{\protect\cite[{Table 12.1}]{cite199}}.
\end{eczvaluelist}
\eczhbkcontributors{ \eczhuVVA }
\endeczcode

\eczcode{hamming844}{\([8,4,4]\) extended Hamming code}{~\NoCaseChange{\protect\cite{cite1,cite1167,cite1168}}}
\codefieldsection{Alternative Names}
\begin{eczvaluelist}
\item\relax \([8,4,4]\) \(e_8\) code
\end{eczvaluelist}
\eczhIndexCodeAliasName{hamming844}{extended Hamming code}
\eczhIndexCodeAliasName{hamming844}{\([8,4,4]\) \(e_8\) code}
\codefieldsection{Description}
Extension of the \([7,4,3]\) Hamming code by a parity-check bit.
The smallest doubly even self-dual code, and the unique Type II code of length \(8\) \NoCaseChange{\protect\cite[{Rem. 4.3.10}]{cite40}}.

A generator matrix is
\flmMathEnvironment{align}{}{
\begin{pmatrix}
1 & 1 & 1 & 1 & 1 & 1 & 1 & 1 \\
0 & 0 & 0 & 0 & 1 & 1 & 1 & 1 \\
0 & 0 & 1 & 1 & 0 & 0 & 1 & 1 \\
0 & 1 & 0 & 1 & 0 & 1 & 0 & 1
\end{pmatrix}~,
}
equivalent to the standard form
\flmMathEnvironment{align}{}{
\begin{pmatrix}
1 & 0 & 0 & 0 & 0 & 1 & 1 & 1 \\
0 & 1 & 0 & 0 & 1 & 0 & 1 & 1 \\
0 & 0 & 1 & 0 & 1 & 1 & 0 & 1 \\
0 & 0 & 0 & 1 & 1 & 1 & 1 & 0
\end{pmatrix}~.
}
Its automorphism group is \(GA(3,\mathbb{F}_2)\) \NoCaseChange{\protect\cite{cite1115}}.

\codefieldsection{Parents}
\begin{eczvaluelist}
\item\relax
\flmRefsHyperref[eczindexfamilyrel]{code:self_dual}{Self-dual linear code} --- The \([8,4,4]\) extended Hamming code is the smallest doubly even self-dual code, and the unique Type II code of length \(8\) \NoCaseChange{\protect\cite[{Rem. 4.3.10}]{cite40}}.
\item\relax
\flmRefsHyperref[eczindexfamilyrel]{code:extended_hamming}{\([2^m,2^m-m-1,4]\) Extended Hamming code}\item\relax
\flmRefsHyperref[eczindexfamilyrel]{code:biorthogonal}{\([2^m,m+1,2^{m-1}]\) First-order RM code} --- The \([8,4,4]\) extended Hamming code is the first-order RM\((1,3)\) code.
\end{eczvaluelist}
\codefieldsection{Cousins}
\begin{eczvaluelist}
\item\relax
\flmRefsHyperref[eczindexfamilyrel]{code:group}{Group-algebra code} --- The \([8,4,4]\) extended Hamming code is a group-algebra code for the group \(\mathbb{Z}_2 \times \mathbb{Z}_4\) \NoCaseChange{\protect\cite{cite1115}}.
\item\relax
\flmRefsHyperref[eczindexfamilyrel]{code:binary_quad_residue}{Binary quadratic-residue (QR) code} --- The \([8,4,4]\) extended Hamming code is an extended quadratic-residue code \NoCaseChange{\protect\cite{cite41}}.
\item\relax
\flmRefsHyperref[eczindexfamilyrel]{code:divisible}{Divisible code} --- The \([8,4,4]\) extended Hamming code is the smallest doubly even self-dual code, and the unique Type II code of length \(8\) \NoCaseChange{\protect\cite[{Rem. 4.3.10}]{cite40}}.
\item\relax
\flmRefsHyperref[eczindexfamilyrel]{code:eeight}{\(E_8\) Gosset lattice} --- The \([8,4,4]\) extended Hamming code yields the \(E_8\) Gosset lattice via \flmTerm{term}{ref127}{}{Construction A} \NoCaseChange{\protect\cite[{Exam. 10.5.2}]{cite115}\protect\cite[{pg. 138}]{cite39}}.
\item\relax
\flmRefsHyperref[eczindexfamilyrel]{code:extended_golay}{\([24, 12, 8]\) Extended Golay code} --- The extended Golay code can be constructed from two suitably chosen extended Hamming \([8,4,4]\) codes using the \(|a+x|b+x|a+b+x|\) construction \NoCaseChange{\protect\cite[{pg. 588}]{cite41}}.
\item\relax
\flmRefsHyperref[eczindexfamilyrel]{code:hamming743}{\([7,4,3]\) Hamming code} --- The Hamming code can be extended by a parity-check bit to yield the \([8,4,4]\) extended Hamming code, the smallest doubly even self-dual code.
\item\relax
\flmRefsHyperref[eczindexfamilyrel]{code:reed_solomon_4}{\([4,2,3]_4\) RS\(_4\) code} --- The RS\(_4\) code can be mapped to the \([8,4,4]\) extended Hamming code \NoCaseChange{\protect\cite[{Sec. 2.4.2}]{cite42}} by identifying \((0,\omega,\bar{\omega},1)\) with \((00),(10),(01),(11)\) \NoCaseChange{\protect\cite{cite109}}.
\item\relax
\flmRefsHyperref[eczindexfamilyrel]{code:klemm}{Klemm code} --- The binary image of the \(m=1\) Klemm code under the \flmTerm{term}{ref81}{}{Gray map} is the \([8,4,4]\) extended Hamming code \NoCaseChange{\protect\cite[{Exam. 3.2}]{cite123}}.
\item\relax
\flmRefsHyperref[eczindexfamilyrel]{code:octacode}{Octacode} --- The mod-two reduction of the octacode is the \([8,4,4]\) extended Hamming code \NoCaseChange{\protect\cite{cite42}}. The octacode can be obtained by Hensel-lifting the \([8,4,4]\) extended Hamming code to \(\mathbb{Z}_4\) \NoCaseChange{\protect\cite{cite1199}}.
\item\relax
\flmRefsHyperref[eczindexfamilyrel]{code:stab_8_3_2}{\(\llbracket 8,3,2\rrbracket \) Smallest interesting color code} --- The \(\llbracket 8,3,2\rrbracket \) hypercube code \(H_X\) check matrix is the parity-check matrix of the \([8,4,4]\) extended Hamming code, while its \(H_Z\) matrix is that of the SPC code.
\item\relax
\flmRefsHyperref[eczindexfamilyrel]{code:stab_8_3_3}{\(\llbracket 8, 3, 3\rrbracket \) Eight-qubit Gottesman code} --- The \(\llbracket 8, 3, 3\rrbracket \) code is obtained via a modified CSS construction from the \([8,4,4]\) extended Hamming code.
\end{eczvaluelist}
\eczhbkcontributors{ \eczhuVVA }
\endeczcode

\eczcode{q-ary_parity_check}{\([n,n-1,2]_q\) \(q\)-ary parity-check code}{}
\codefieldsection{Alternative Names}
\begin{eczvaluelist}
\item\relax \(q\)-ary sum-zero code
\item\relax \(q\)-ary zero-sum code
\end{eczvaluelist}
\eczhIndexCodeAliasName{q-ary_parity_check}{\(q\)-ary parity-check code}
\eczhIndexCodeAliasName{q-ary_parity_check}{\(q\)-ary sum-zero code}
\eczhIndexCodeAliasName{q-ary_parity_check}{\(q\)-ary zero-sum code}
\codefieldsection{Description}
An \([n,n-1,2]_q\) linear \(q\)-ary code whose codewords consist of the message string appended with a \textit{parity-check} or \textit{zero-sum check digit} such that the sum over all coordinates of each codeword is zero.
\codefieldsection{Parents}
\begin{eczvaluelist}
\item\relax
\flmRefsHyperref[eczindexfamilyrel]{code:reed_solomon}{Reed-Solomon (RS) code} --- RS codes for \(k=n-1\) are parity-check codes \NoCaseChange{\protect\cite{cite1673}}.
\item\relax
\flmRefsHyperref[eczindexfamilyrel]{code:q-ary_cyclic}{Cyclic linear \(q\)-ary code} --- Since permutations preserve coordinate sums, the cyclic permutation of an SPC codeword is another codeword. The generator polynomial of the code is \(x-1\).
\item\relax
\flmRefsHyperref[eczindexfamilyrel]{code:checksum}{Checksum code} --- The \(q\)-ary parity check code is a simple example of a checksum code, with the parity of the message being the checksum.
\item\relax
\flmRefsHyperref[eczindexfamilyrel]{code:small_distance}{Small-distance block code}\end{eczvaluelist}
\codefieldsection{Child}
\begin{eczvaluelist}
\item\relax
\flmRefsHyperref[eczindexfamilyrel]{code:parity_check}{\([n,n-1,2]\) Single parity-check (SPC) code}\end{eczvaluelist}
\codefieldsection{Cousins}
\begin{eczvaluelist}
\item\relax
\flmRefsHyperref[eczindexfamilyrel]{code:mds}{Maximum distance separable (MDS) code}\item\relax
\flmRefsHyperref[eczindexfamilyrel]{code:q-ary_ldgm}{\(q\)-ary LDGM code} --- Concatenated \(q\)-ary parity-check codes are LDGM \NoCaseChange{\protect\cite{cite1210}}.
\end{eczvaluelist}
\eczhbkcontributors{ \eczhuVVA }
\endeczcode

\eczcode{semakov_zinoviev}{\(ED_m\) code}{~\NoCaseChange{\protect\cite{cite1674}}}
\codefieldsection{Alternative Names}
\begin{eczvaluelist}
\item\relax Equidistant code with maximal distance
\end{eczvaluelist}
\eczhIndexCodeAliasName{semakov_zinoviev}{code}
\eczhIndexCodeAliasName{semakov_zinoviev}{Equidistant code with maximal distance}
\codefieldsection{Description}
Member of a family of nonlinear \( (c\frac{qt-1}{(t-1,q-1)},qt,ct\frac{q-1}{(t-1,q-1)}) \) \(q\)-ary codes, where \(c,t\geq 1\) and \((a,b)\) is the greatest common divisor of \(a\) and \(b\).
Such codes are universally optimal and are related to resolvable block designs.

\codefieldsection{Parent}
\begin{eczvaluelist}
\item\relax
\flmRefsHyperref[eczindexfamilyrel]{code:delsarte_optimal_q-ary}{\(q\)-ary sharp configuration} --- The \(ED_m\) code is a \(q\)-ary sharp configuration \NoCaseChange{\protect\cite[{Table 12.1}]{cite199}}.
\end{eczvaluelist}
\eczhbkcontributors{ \eczhuVVA }
\endeczcode

\eczcode{q-ary_digits_into_q-ary_digits}{\(q\)-ary code}{}

\codefieldsection{Kingdom root code}
for the \flmRefsHyperref{kingdom:q-ary_digits_into_q-ary_digits}{Galois-field Kingdom}.
\codefieldsection{Description}
Encodes \(K\) states (codewords) in \(n\) \(q\)-ary coordinates over the field \(\mathbb{F}_q\), i.e., \(q\)-ary strings.
Error-correcting performance is quantified by some distance \(d\), which in turn is defined using a metric.
The default distance is the Hamming distance \(d\), i.e., the number of coordinates in which two distinct codewords differ; such codes are usually denoted as \((n,K,d)_q\). For linear codes, this is equivalently the Hamming weight of the lowest-weight nonzero codeword.
Unless stated otherwise, the distance for this class is the Hamming distance.

Two \(q\)-ary codes are \textit{equivalent} if the codewords of one code can be mapped into those of the other under a combination of a coordinate permutation and a permutation of the elements of each coordinate. 
The full group of such composite permutations is \(S_q \wr S_n\) \NoCaseChange{\protect\cite[{Def. 1.8.8}]{cite1159}\protect\cite[{Sec. 3.2}]{cite70}}.

\codefieldsection{Protection}
A code detects errors on up to \(d-1\) coordinates, corrects erasure errors on up to \(d-1\) coordinates, and corrects general errors on up to \(\left\lfloor (d-1)/2 \right\rfloor\) coordinates.
Often, the relative Hamming distance \(\delta=d/n\) is used to compare codes of different lengths.
A code is \textit{equidistant} if the Hamming distance between any two codewords is \(d\). 

An analogue of the \flmTerm{term}{ref81}{}{Gray map} and Lee weight can be defined for codes over \(\mathbb{F}_4\) by identifying \((0,\omega,\bar{\omega},1)\) with \((00),(10),(01),(11)\) \NoCaseChange{\protect\cite{cite109}}.

\subsection{Noise channels}

Noise channels include the symmetric noise channel, asymmetric noise channels \NoCaseChange{\protect\cite{cite1418,cite1675,cite1293,cite1185,cite1676}}, as well as insertion/deletion noise and its generalization to substring edits \NoCaseChange{\protect\cite{cite1677}}.

\subsection{Weight enumerator and four fundamental parameters}
\begin{defterm}{Weight enumerator}\label{ref1678}\label{ref113}
Determining protection and bounds on code parameters can also be done using the code's \textit{weight enumerator} (cf. \flmRefsHyperref{ref672}{quantum weight enumerators}),
  \flmMathEnvironment{align}{}{
  \begin{split}
    A(x,y)&=\sum_{j=0}^{n}A_{j}x^{n-j}y^{j}\\
    A_{j}&=\text{number of wt-}j\text{ codewords}~.
  \end{split}
  }
The weight enumerator and its Fourier transform, the \textit{dual weight enumerator}, satisfy the \textit{MacWilliams identity} \NoCaseChange{\protect\cite{cite1679,cite1680}}; see \NoCaseChange{\protect\cite[{Ch. 5}]{cite41}}.
The transform between the two enumerators is called the \textit{MacWilliams transform}.
Computing the weight enumerator is \(\#P\)-hard \NoCaseChange{\protect\cite{cite1417}}.

The distance of the code is the index \(j\) of the first nonzero coefficient \(A_j\), while the \textit{dual distance} is the index of the first nonzero coefficient of the dual weight enumerator.
The \textit{number} of the code is the number of nonzero \(A_j\)'s, corresponding to the number of distinct nonzero distances between codewords.
The \textit{external distance} is the number of nonzero coefficients of the dual weight enumerator.
The distance, dual distance, number and external distance make up the \textit{four fundamental parameters} of a code \NoCaseChange{\protect\cite{cite216}\protect\cite[{Ch. 5}]{cite41}}.

Other types of weight enumerators include the Hamming weight enumerator, Lee weight enumerator, joint weight enumerator, split weight enumerator, and biweight enumerator \NoCaseChange{\protect\cite{cite41}}.
\end{defterm}

\subsection{Bounds on code parameters}
Bounds on the parameters of an \((n,K,d)_q\) code include the Hamming a.k.a. sphere-packing bound, Singleton bound, Gilbert-Varshamov (GV) bound, Griesmer bound, Plotkin bound, Johnson bound, and various linear programming (LP) bounds; see \NoCaseChange{\protect\cite[{Sec. 1.9}]{cite1159}}.
A code whose parameters attain the Hamming bound (Singleton bound, Griesmer bound, Johnson bound, Delsarte LP bound) is called a perfect code (an MDS code, a Griesmer code, a nearly perfect code, an \flmRefsHyperref{code:univ_opt_q-ary}{LP universally optimal code}).

\begin{defterm}{Gilbert-Varshamov (GV) bound}\label{ref1681}\label{ref85}
The Gilbert-Varshamov \NoCaseChange{\protect\cite{cite1682,cite1683}}, or Gilbert-Shannon-Varshamov, bound states that the maximum size \(K\) of a \(q\)-ary code with distance \(d\) satisfies
\flmMathEnvironment{align}{}{
K\sum_{j=0}^{d-1}{n \choose j}(q-1)^{j}\geq q^{n}~.
}
In other words, if the left-hand side of the above is less than or equal to the right-hand side, then a code with such parameters exists.
The GV bound gives rise to the \textit{asymptotic GV bound} (i.e., GV bound in the \(n\to\infty\) limit), expressed in terms of the maximum achievable rate \(R\) and relative distance \(\delta\),
\flmMathEnvironment{align}{}{
  R\geq 1-h_{q}(\delta)~,
}
where \(h_q\) is the \(q\)-ary entropy function,
\flmMathEnvironment{align}{}{
  h_{q}(\delta)=-\delta\log_{q}\frac{\delta}{q-1}-(1-\delta)\log_{q}(1-\delta)~.
}
\end{defterm}

\codefieldsection{Rate}
The rate of a \(q\)-ary code is usually defined as \(R=\frac{1}{n}\log_q K\) dits per symbol. The maximum rate of a \(q\)-ary code with linear distance is lower bounded by the \flmRefsHyperref{ref85}{asymptotic GV bound} and upper bounded by the \(q\)-ary version of the MRRW LP bound \NoCaseChange{\protect\cite{cite1684}}.
\codefieldsection{Decoding}
\begin{eczvaluelist}
\item\relax For small \(n\), decoding can be based on a lookup table. For infinite code families, the size of such a table scales exponentially with \(n\), so approximate decoding algorithms scaling polynomially with \(n\) have to be used. The decoder determining the most likely error given a noise channel is called the \textit{maximum-likelihood decoder}.
\item\relax Given a received string \(x\) and an error bound \(e\), a \textit{list decoder} returns a list of all codewords that are at most \(e\) from \(x\). The number of codewords in a neighborhood of \(x\) has to be polynomial in \(n\) in order for this decoder to run in time polynomial in \(n\).
\end{eczvaluelist}
\codefieldsection{Threshold}
\begin{eczvaluelist}
\item\relax Threshold for large-alphabet circuits is higher than for Boolean circuits \NoCaseChange{\protect\cite{cite1685}}.
\end{eczvaluelist}
\codefieldsection{Notes}
\begin{eczvaluelist}
\item\relax Tables of bounds and examples of linear codes for various \(n\) and \(k\), extending code tables by Brouwer \NoCaseChange{\protect\cite{cite1427}}, are maintained by M. Grassl at this \flmHref{https://www.codetables.de/}{website}.
\end{eczvaluelist}
\codefieldsection{Parents}
\begin{eczvaluelist}
\item\relax
\flmRefsHyperref[eczindexfamilyrel]{code:rings_into_rings}{Ring code} --- Galois fields are rings. Codes over algebraic number fields have also been studied \NoCaseChange{\protect\cite{cite1686,cite1687,cite1688}}.
\item\relax
\flmRefsHyperref[eczindexfamilyrel]{code:matrices_into_matrices}{Matrix-based code} --- Matrix-based codes over \(\mathbb{F}_q\) whose codewords are vectors reduce to \(q\)-ary codes. Elements of fields such as \(\mathbb{F}_{p^{ml}}\) can be written as \(m\)-dimensional vectors over \(\mathbb{F}_{p^{l}}\) or \((m\times l)\)-dimensional matrices over \(\mathbb{F}_p\). This idea is used to convert between ordinary block codes and matrix-based codes such as disk array codes and rank-metric codes.
\item\relax
\flmRefsHyperref[eczindexfamilyrel]{code:2pt_homogeneous}{Two-point homogeneous-space code} --- Hamming space can be regarded as a finite two-point homogeneous space \(G/H\) where \(G = S_q \wr S_n\) is its isometry group \NoCaseChange{\protect\cite[{Sec. 5.3}]{cite987}\protect\cite[{Table 2}]{cite985}}.
\end{eczvaluelist}
\codefieldsection{Children}
\begin{eczvaluelist}
\item\relax
\flmRefsHyperref[eczindexfamilyrel]{code:bits_into_bits}{Binary code} --- A \(q\)-ary code reduces to a binary code at \(q=2\). Ternary computing may be more applicable than binary computing to cryptographic schemes \NoCaseChange{\protect\cite{cite1257,cite1258}}.
\item\relax
\flmRefsHyperref[eczindexfamilyrel]{code:q-ary_additive}{Additive \(q\)-ary code}\item\relax
\flmRefsHyperref[eczindexfamilyrel]{code:ag}{Algebraic-geometry (AG) code}\item\relax
\flmRefsHyperref[eczindexfamilyrel]{code:sequential_recovery}{Sequential-recovery code}\item\relax
\flmRefsHyperref[eczindexfamilyrel]{code:isbn}{International Standard Book Number (ISBN) code} --- The last digit of an ISBN-10 string is a check digit computed modulo 11 \NoCaseChange{\protect\cite{cite961}}.
\item\relax
\flmRefsHyperref[eczindexfamilyrel]{code:ecoc}{Error-correcting output code (ECOC)}\item\relax
\flmRefsHyperref[eczindexfamilyrel]{code:lexicographic}{Lexicographic code}\item\relax
\flmRefsHyperref[eczindexfamilyrel]{code:orthogonal_array}{Orthogonal array (OA)} --- There is a relation between \(q\)-ary codes and orthogonal arrays which is phrased in terms of the codes' \flmRefsHyperref{ref113}{dual distance} \NoCaseChange{\protect\cite[{Thm. 4.5}]{cite216}\protect\cite[{Thm. 4.9}]{cite212}}.
\item\relax
\flmRefsHyperref[eczindexfamilyrel]{code:weighed_covering}{Weighted-covering code}\item\relax
\flmRefsHyperref[eczindexfamilyrel]{code:completely_regular}{Completely regular code}\item\relax
\flmRefsHyperref[eczindexfamilyrel]{code:matrix_product}{Matrix-product code}\item\relax
\flmRefsHyperref[eczindexfamilyrel]{code:univ_opt_q-ary}{Universally optimal \(q\)-ary code}\item\relax
\flmRefsHyperref[eczindexfamilyrel]{code:balanced}{Balanced code}\end{eczvaluelist}
\codefieldsection{Cousins}
\begin{eczvaluelist}
\item\relax
\flmRefsHyperref[eczindexfamilyrel]{code:galois_into_galois}{Galois-qudit code} --- Galois-qudit codes are quantum counterparts of \(q\)-ary codes.
\item\relax
\flmRefsHyperref[eczindexfamilyrel]{code:q-ary_over_zq}{\(q\)-ary code over \(\mathbb{Z}_q\)} --- \(q\)-ary codes for \(q=p\) prime are \(p\)-ary codes over \(\mathbb{Z}_p \cong \mathbb{F}_p\).
\item\relax
\flmRefsHyperref[eczindexfamilyrel]{code:combinatorial_design}{Combinatorial design} --- Designs can be constructed from \(q\)-ary codes by taking the supports of a subset of codewords of constant weight.
\item\relax
\flmRefsHyperref[eczindexfamilyrel]{code:constantin_rao}{Constantin-Rao (CR) code} --- CR codes, and their special cases the VT codes, can be converted to ternary codes with nice structure via a \textit{binary-to-ternary} map \(00\to 0\), \(11\to 0\), \(01\to 1\), and \(10\to 2\) \NoCaseChange{\protect\cite{cite1186}}.
\item\relax
\flmRefsHyperref[eczindexfamilyrel]{code:traceability}{Traceability code} --- A \(q\)-ary code with distance \(d \geq n(1-1/t^2)\) has the \(t\)-traceability property \NoCaseChange{\protect\cite[{Thm. 4.3}]{cite349}}.
\item\relax
\flmRefsHyperref[eczindexfamilyrel]{code:convolutional}{Convolutional code} --- Convolutional codes for finite block size reduce to \(q\)-ary codes.
\item\relax
\flmRefsHyperref[eczindexfamilyrel]{code:polyphase}{Polyphase code} --- Polyphase codes are spherical codes that can be obtained from \(q\)-ary codes.
\end{eczvaluelist}
\eczhbkcontributors{ \eczhuVVA }
\endeczcode

\eczcode{q-ary_duadic}{\(q\)-ary duadic code}{~\NoCaseChange{\protect\cite{cite1689,cite69,cite1690,cite1691}}}
\codefieldsection{Description}
Member of a pair of cyclic linear \(q\)-ary codes that satisfy certain relations, depending on whether the pair is \textit{even-like} or \textit{odd-like} duadic.
Duadic codes exist only when \(q\) is a square modulo \(n\) \NoCaseChange{\protect\cite{cite69}}.

Duadic codes come in two pairs, an even-like duadic pair and an odd-like duadic pair. All codewords in an even-like pair are \textit{even-like}, i.e., \(\sum_i c_i = 0\). By contrast, an odd-like pair is not even-like, i.e., it contains at least one codeword with \(\sum_i c_i \neq 0\).

Duadic code pairs can be defined in terms of their idempotent generators (see \flmRefsCref{ref67}).
A pair of even-like codes \(C_1\) and \(C_2\) with respective idempotents \(e_1\) and \(e_2\) is an \textit{even-like duadic pair} if (1) \(e_1(x)+e_2(x)=1-\frac{1}{n}(1+x+x^2+\cdots+x^{n-1})\) and (2) there exists a multiplier \(\mu\) such that \(C_1 \mu=C_2\) and \(C_2 \mu=C_1\).

There is an odd-like duadic pair \(\{D_1,D_2\}\) associated with the even-like pair \(\{C_1, C_2\}\), where \(1-e_2(x)\) generates \(D_1\) and \(1-e_1(x)\) generates \(D_2\). The even-pair codes are \([n,\frac{n-1}{2}]_q\) codes while the odd-pair codes are \([n,\frac{n+1}{2}]_q\) codes.

\codefieldsection{Protection}
For odd-like duadic codes, the common minimum odd-like weight \(d_o\) satisfies \(d_o^2 \geq n\); if the splitting is given by \(\mu=-1\), then \(d_o^2-d_o+1 \geq n\) \NoCaseChange{\protect\cite[{Thm. 2.7.1}]{cite68}}.
\codefieldsection{Notes}
\begin{eczvaluelist}
\item\relax Reviews of duadic codes \NoCaseChange{\protect\cite{cite69,cite126}\protect\cite[{Sec. 2.7}]{cite68}}.
\end{eczvaluelist}
\codefieldsection{Parent}
\begin{eczvaluelist}
\item\relax
\flmRefsHyperref[eczindexfamilyrel]{code:q-ary_cyclic}{Cyclic linear \(q\)-ary code}\end{eczvaluelist}
\codefieldsection{Children}
\begin{eczvaluelist}
\item\relax
\flmRefsHyperref[eczindexfamilyrel]{code:binary_duadic}{Binary duadic code}\item\relax
\flmRefsHyperref[eczindexfamilyrel]{code:q-ary_quad_residue}{Quadratic-residue (QR) code} --- QR codes are duadic codes of prime length satisfying certain relations \NoCaseChange{\protect\cite{cite69}}.
\end{eczvaluelist}
\codefieldsection{Cousins}
\begin{eczvaluelist}
\item\relax
\flmRefsHyperref[eczindexfamilyrel]{code:self_dual}{Self-dual linear code} --- Under certain conditions, extended odd-like duadic codes are self-dual \NoCaseChange{\protect\cite[{Sec. 2.7}]{cite68}}.
\item\relax
\flmRefsHyperref[eczindexfamilyrel]{code:projective}{Projective geometry code} --- The weight-five codewords of the \([21,5,6]_4\) quaternary duadic code support a projective plane of order 4 \NoCaseChange{\protect\cite[{Table 6.3}]{cite126}}.
\item\relax
\flmRefsHyperref[eczindexfamilyrel]{code:galois_duadic}{Quantum duadic code} --- Quantum duadic codes are quantum analogues of \(q\)-ary duadic codes.
\end{eczvaluelist}
\eczhbkcontributors{ \eczhuVVA }
\endeczcode

\eczcode{q-ary_hamming}{\(q\)-ary Hamming code}{~\NoCaseChange{\protect\cite{cite1692,cite1168}}}
\codefieldsection{Description}
Member of an infinite family of perfect linear \(q\)-ary codes with parameters \([(q^r-1)/(q-1),(q^r-1)/(q-1)-r, 3]_q\) for \(r \geq 2\) \NoCaseChange{\protect\cite[{(3.1)}]{cite70}}.
These are precisely the nontrivial perfect linear codes over \(\mathbb{F}_q\) \NoCaseChange{\protect\cite[{Thm. 3.3.1}]{cite70}}.

The automorphism group is known \NoCaseChange{\protect\cite{cite1693}}.

\codefieldsection{Protection}
Can detect up to two symbol errors and correct one symbol error.
\codefieldsection{Notes}
\begin{eczvaluelist}
\item\relax Example 2.2.2 studies the subfield subcode \(H_{3,2^2}|_{\mathbb{F}_2}\), a \([21,16,3]_2\) code, of a \(q\)-ary Hamming code over an extension field \NoCaseChange{\protect\cite[{Exam. 2.2.2}]{cite68}}.
\end{eczvaluelist}
\codefieldsection{Parents}
\begin{eczvaluelist}
\item\relax
\flmRefsHyperref[eczindexfamilyrel]{code:perfect}{Perfect code}\item\relax
\flmRefsHyperref[eczindexfamilyrel]{code:q-ary_linear}{Linear \(q\)-ary code}\item\relax
\flmRefsHyperref[eczindexfamilyrel]{code:univ_opt_q-ary}{Universally optimal \(q\)-ary code} --- Hamming codes and their once-punctured and once-shortened versions are LP universally optimal codes \NoCaseChange{\protect\cite{cite173}}.
\item\relax
\flmRefsHyperref[eczindexfamilyrel]{code:small_distance}{Small-distance block code}\end{eczvaluelist}
\codefieldsection{Children}
\begin{eczvaluelist}
\item\relax
\flmRefsHyperref[eczindexfamilyrel]{code:hamming}{\([2^r-1,2^r-r-1,3]\) Hamming code} --- The \(q\)-ary Hamming codes reduce to the Hamming codes at \(q=2\).
\item\relax
\flmRefsHyperref[eczindexfamilyrel]{code:tetracode}{\([4,2,3]_3\) Tetracode} --- The tetracode is equivalent to the \(r=2\) \(3\)-ary Hamming code.
\end{eczvaluelist}
\codefieldsection{Cousins}
\begin{eczvaluelist}
\item\relax
\flmRefsHyperref[eczindexfamilyrel]{code:incidence_matrix}{Incidence-matrix projective code} --- Columns of a Hamming parity-check matrix correspond to 1D subspaces of \(\mathbb{F}_q^r\).
\item\relax
\flmRefsHyperref[eczindexfamilyrel]{code:q-ary_bch}{Bose–Chaudhuri–Hocquenghem (BCH) code} --- When \(\gcd(r,q-1)=1\), \(q\)-ary Hamming codes are narrow-sense BCH codes \NoCaseChange{\protect\cite[{Exam. 16.4.10}]{cite196}\protect\cite[{Thm. 5.1.4}]{cite126}}, which are cyclic \NoCaseChange{\protect\cite[{pg. 194}]{cite41}\protect\cite[{Exam. 2.5.1}]{cite68}}.
\item\relax
\flmRefsHyperref[eczindexfamilyrel]{code:hexacode}{\([6,3,4]_4\) Hexacode} --- The hexacode is an extended quaternary Hamming code \NoCaseChange{\protect\cite[{Exer. 578}]{cite126}}.
\item\relax
\flmRefsHyperref[eczindexfamilyrel]{code:q-ary_simplex}{\(q\)-ary simplex code} --- \(q\)-ary Hamming and \(q\)-ary simplex codes are dual to each other \NoCaseChange{\protect\cite[{pg. 45}]{cite1314}}.
\item\relax
\flmRefsHyperref[eczindexfamilyrel]{code:quantum_perfect}{Perfect quantum code} --- For qubits (\(q=2\)), the only nontrivial perfect codes are the stabilizer code family \(\llbracket (4^r-1)/3, (4^r-1)/3 - 2r, 3\rrbracket \) for \(r \geq 2\), obtained from Hamming codes over \(\mathbb{F}_4\) via the Hermitian construction \NoCaseChange{\protect\cite{cite1694,cite449}}. These codes are related to partial spreads in projective geometry \NoCaseChange{\protect\cite{cite1695}}.
\end{eczvaluelist}
\eczhbkcontributors{ \eczhuVVA }
\endeczcode

\eczcode{q-ary_ldgm}{\(q\)-ary LDGM code}{}
\codefieldsection{Description}
\(q\)-ary linear code with a sparse generator matrix.
Alternatively, a member of an infinite family of \([n,k,d]_q\) codes for which the number of nonzero entries in each row and column of the generator matrix are both bounded by a constant as \(n\to\infty\).

\codefieldsection{Parent}
\begin{eczvaluelist}
\item\relax
\flmRefsHyperref[eczindexfamilyrel]{code:q-ary_linear}{Linear \(q\)-ary code}\end{eczvaluelist}
\codefieldsection{Child}
\begin{eczvaluelist}
\item\relax
\flmRefsHyperref[eczindexfamilyrel]{code:ldgm}{Low-density generator-matrix (LDGM) code}\end{eczvaluelist}
\codefieldsection{Cousins}
\begin{eczvaluelist}
\item\relax
\flmRefsHyperref[eczindexfamilyrel]{code:dual}{Dual linear code} --- The dual of a \(q\)-ary LDPC code has a sparse generator matrix and is called a \(q\)-ary LDGM code.
\item\relax
\flmRefsHyperref[eczindexfamilyrel]{code:q-ary_parity_check}{\([n,n-1,2]_q\) \(q\)-ary parity-check code} --- Concatenated \(q\)-ary parity-check codes are LDGM \NoCaseChange{\protect\cite{cite1210}}.
\item\relax
\flmRefsHyperref[eczindexfamilyrel]{code:q-ary_ldpc}{\(q\)-ary LDPC code} --- The dual of a \(q\)-ary LDPC code has a sparse generator matrix and is called a \(q\)-ary LDGM code.
\end{eczvaluelist}
\eczhbkcontributors{ \eczhuVVA }
\endeczcode

\eczcode{q-ary_ldpc}{\(q\)-ary LDPC code}{~\NoCaseChange{\protect\cite{cite1696}\protect\cite[{Ch. 5}]{cite1361}}}
\codefieldsection{Alternative Names}
\begin{eczvaluelist}
\item\relax Non-binary LDPC (NBDPC) code
\end{eczvaluelist}
\eczhIndexCodeAliasName{q-ary_ldpc}{Non-binary LDPC (NBDPC) code}
\codefieldsection{Description}
A \(q\)-ary linear code with a sparse parity-check matrix.
Alternatively, a member of an infinite family of \([n,k,d]_q\) codes for which the number of nonzero entries in each row and column of the parity-check matrix are both bounded above by a constant as \(n\to\infty\).

A \textit{parity check} is performed by taking the inner product of a row of the parity-check matrix with a codeword that has been affected by a noise channel.
A parity check yields either zero (a satisfied check) or a nonzero field element (an unsatisfied check).
Despite the fact that there is more than one nonzero outcome, \(q\)-ary linear codes with sparse parity-check matrices are also called LDPC codes.

\codefieldsection{Protection}
For fixed \(j \geq 3\), typical \((n,j,k)\) \(q\)-ary LDPC codes have minimum distance growing linearly with block length, extending Gallager's binary analysis to arbitrary alphabet sizes \NoCaseChange{\protect\cite[{Ch. 5}]{cite1361}}.
Non-binary cycle LDPC codes for \(q\geq 32\) can exhibit good performance \NoCaseChange{\protect\cite{cite1697,cite1698,cite1699}}.

\codefieldsection{Rate}
An \((n,j,k)\) \(q\)-ary LDPC code has rate at least \(1-j/k\) by parity-check counting, and Gallager extended the corresponding regular-ensemble analysis to arbitrary alphabet sizes \NoCaseChange{\protect\cite[{Ch. 5}]{cite1361}}. Asymptotically good non-binary expander codes can be constructed \NoCaseChange{\protect\cite{cite1343}\protect\cite[{Thm. 7.16}]{cite1342}} by generalizing the originally binary expander constructions \NoCaseChange{\protect\cite{cite1332,cite1344}}.
\codefieldsection{Decoding}
\begin{eczvaluelist}
\item\relax Gallager generalized his iterative probabilistic decoding procedure to arbitrary alphabet sizes, with symbol-update rules expressed in terms of modulo-\(q\) check sums \NoCaseChange{\protect\cite[{Secs. 5.3--5.4}]{cite1361}}.
\end{eczvaluelist}
\codefieldsection{Parents}
\begin{eczvaluelist}
\item\relax
\flmRefsHyperref[eczindexfamilyrel]{code:tanner}{Tanner code} --- \(q\)-ary LDPC codes are \(q\)-ary Tanner codes on sparse bipartite graphs whose constraint nodes represent \(q\)-ary parity-check codes.
\item\relax
\flmRefsHyperref[eczindexfamilyrel]{code:locally_recoverable}{Locally recoverable code (LRC)} --- LDPC codes are linear LRCs whose locality is the maximum number of nonzero entries in a row of the parity-check matrix \NoCaseChange{\protect\cite{cite812}}.
\end{eczvaluelist}
\codefieldsection{Children}
\begin{eczvaluelist}
\item\relax
\flmRefsHyperref[eczindexfamilyrel]{code:ldpc}{Low-density parity-check (LDPC) code}\item\relax
\flmRefsHyperref[eczindexfamilyrel]{code:q-ary_protograph_ldpc}{\(q\)-ary protograph LDPC code}\end{eczvaluelist}
\codefieldsection{Cousins}
\begin{eczvaluelist}
\item\relax
\flmRefsHyperref[eczindexfamilyrel]{code:q-ary_ldgm}{\(q\)-ary LDGM code} --- The dual of a \(q\)-ary LDPC code has a sparse generator matrix and is called a \(q\)-ary LDGM code.
\item\relax
\flmRefsHyperref[eczindexfamilyrel]{code:dual}{Dual linear code} --- The dual of a \(q\)-ary LDPC code has a sparse generator matrix and is called a \(q\)-ary LDGM code.
\item\relax
\flmRefsHyperref[eczindexfamilyrel]{code:expander}{Expander code} --- Asymptotically good non-binary expander codes can be constructed \NoCaseChange{\protect\cite{cite1343}\protect\cite[{Thm. 7.16}]{cite1342}} by generalizing the originally binary expander constructions \NoCaseChange{\protect\cite{cite1332,cite1344}}.
\item\relax
\flmRefsHyperref[eczindexfamilyrel]{code:general_qldpc}{QLDPC code} --- Galois-qudit QLDPC codes are quantum analogues of \(q\)-ary LDPC codes.
\end{eczvaluelist}
\eczhbkcontributors{ \eczhuVVA }
\endeczcode

\eczcode{q-ary_lcc}{\(q\)-ary linear LCC}{}
\codefieldsection{Description}
A \(q\)-ary linear code for which one can recover any coordinate of a codeword from at most \(r\) coordinates of a received word (assuming the corruption rate is within some tolerated threshold \(\delta\)).

\codefieldsection{Parents}
\begin{eczvaluelist}
\item\relax
\flmRefsHyperref[eczindexfamilyrel]{code:q-ary_linear}{Linear \(q\)-ary code}\item\relax
\flmRefsHyperref[eczindexfamilyrel]{code:lcc}{Locally correctable code (LCC)}\end{eczvaluelist}
\codefieldsection{Children}
\begin{eczvaluelist}
\item\relax
\flmRefsHyperref[eczindexfamilyrel]{code:hadamard}{\([2^m,m,2^{m-1}]\) Hadamard code} --- Hadamard codes are two-query LDCs and LCCs \NoCaseChange{\protect\cite{cite1073,cite1067}}.
\item\relax
\flmRefsHyperref[eczindexfamilyrel]{code:generalized_reed_muller}{Generalized RM (GRM) code} --- GRM codes are LDCs and LCCs \NoCaseChange{\protect\cite{cite1073,cite1067}}.
\end{eczvaluelist}
\codefieldsection{Cousins}
\begin{eczvaluelist}
\item\relax
\flmRefsHyperref[eczindexfamilyrel]{code:ldc}{Locally decodable code (LDC)} --- Linear LCCs can be converted into LDCs with the same locality \(r\) \NoCaseChange{\protect\cite[{Sec. 2.4.1}]{cite1067}}.
\item\relax
\flmRefsHyperref[eczindexfamilyrel]{code:multiplicity}{Multiplicity code} --- There exist multiplicity codes with rate arbitrarily close to one that are locally decodable and locally correctable from a constant error fraction \NoCaseChange{\protect\cite{cite181}}.
\end{eczvaluelist}
\eczhbkcontributors{ \eczhuVVA }
\endeczcode

\eczcode{q-ary_ltc}{\(q\)-ary linear LTC}{}
\codefieldsection{Description}
A \(q\)-ary linear code \(C\) of length \(n\) that is a \((u,R)\)-LTC with query complexity \(u\) and soundness \(R>0\).

More technically, the code is a \((u,R)\)-LTC if the rows of its parity-check matrix \(H\in \mathbb{F}_q^{r\times n}\) have weight at most \(u\) and if
\flmMathEnvironment{align}{}{
  \frac{1}{r}\operatorname{wt}(H x) \geq \frac{R}{n} D(x,C)
}
holds for any \(q\)-ary string \(x\), where \(D(x,C)\) is the \(q\)-ary Hamming distance between \(x\) and the closest codeword to \(x\) \NoCaseChange{\protect\cite[{Def. 11}]{cite1269}}.
A code satisfying the above constraint without the weight-\(u\) restriction is called an \(R\)\textit{-testable code} \NoCaseChange{\protect\cite{cite1700}}.

\codefieldsection{Notes}
\begin{eczvaluelist}
\item\relax Testable binary codes admit weakly stable presentations of their corresponding groups; this property is relevant to the disproof of the Connes embedding conjecture \NoCaseChange{\protect\cite{cite1700}}.
\end{eczvaluelist}
\codefieldsection{Parents}
\begin{eczvaluelist}
\item\relax
\flmRefsHyperref[eczindexfamilyrel]{code:q-ary_linear}{Linear \(q\)-ary code} --- Linear \(q\)-ary codes with distances \(\frac{1}{2}n-\sqrt{t n}\) for some \(t\) are called almost-orthogonal and are locally testable with query complexity of \flmRefsHyperref{ref65}{order} \(O(t)\) \NoCaseChange{\protect\cite{cite1270}}. This was later improved to codes with distance \(\frac{1}{2}n-O(n^{1-\gamma})\) for any positive \(\gamma\) \NoCaseChange{\protect\cite{cite1271}}, provided that the number of codewords is polynomial in \(n\).
\item\relax
\flmRefsHyperref[eczindexfamilyrel]{code:ltc}{Locally testable code (LTC)}\end{eczvaluelist}
\codefieldsection{Children}
\begin{eczvaluelist}
\item\relax
\flmRefsHyperref[eczindexfamilyrel]{code:binary_ltc}{Binary linear LTC}\item\relax
\flmRefsHyperref[eczindexfamilyrel]{code:bs-ltc}{Ben-Sasson-Sudan code}\item\relax
\flmRefsHyperref[eczindexfamilyrel]{code:meir}{Meir code} --- Meir codes stand out in that they are based on a combinatorial construction, while other LTCs often use algebraic tools.
\end{eczvaluelist}
\codefieldsection{Cousins}
\begin{eczvaluelist}
\item\relax
\flmRefsHyperref[eczindexfamilyrel]{code:reed_solomon}{Reed-Solomon (RS) code} --- RS codes can be used to construct LTCs encoding \(k\) bits with length \(k \text{polylog}(k)\) and query complexity \(\text{polylog}(k)\) \NoCaseChange{\protect\cite{cite1701}}.
\item\relax
\flmRefsHyperref[eczindexfamilyrel]{code:goppa}{Goppa code} --- Goppa codes are locally testable \NoCaseChange{\protect\cite{cite1270}}.
\item\relax
\flmRefsHyperref[eczindexfamilyrel]{code:generalized_reed_muller}{Generalized RM (GRM) code} --- GRM codes for \(r<q\) can be LTCs in the low-error \NoCaseChange{\protect\cite{cite1089,cite1098}} and high-error \NoCaseChange{\protect\cite{cite1702,cite1703}} regimes. They admit weakly stable presentations of their corresponding groups \NoCaseChange{\protect\cite{cite1700}}.
\item\relax
\flmRefsHyperref[eczindexfamilyrel]{code:q-ary_bch}{Bose–Chaudhuri–Hocquenghem (BCH) code} --- Duals of BCH codes are locally testable \NoCaseChange{\protect\cite{cite1270}}.
\item\relax
\flmRefsHyperref[eczindexfamilyrel]{code:q-ary_cyclic}{Cyclic linear \(q\)-ary code} --- Cyclic linear codes cannot be \(c^3\)-LTCs \NoCaseChange{\protect\cite{cite1272}}. Codeword symmetries are in general an obstruction to achieving such LTCs \NoCaseChange{\protect\cite{cite1273}}.
\item\relax
\flmRefsHyperref[eczindexfamilyrel]{code:galois_expander}{Galois-qudit expander code} --- Balanced products of the RS-based expander-code complexes in \NoCaseChange{\protect\cite{cite689}} yield \([n,k\geq n^{1-\epsilon},d\geq n/\operatorname{poly}(\log n)]_q\) LTCs exhibiting the multiplication property.
\item\relax
\flmRefsHyperref[eczindexfamilyrel]{code:expander_lifted_product}{Expander LP code} --- Classical codes resulting from the expander lifted-product construction are one of the first two families of \(c^3\)-LTCs \NoCaseChange{\protect\cite{cite184}}.
\end{eczvaluelist}
\eczhbkcontributors{ \eczhuVVA }
\endeczcode

\eczcode{q-ary_protograph_ldpc}{\(q\)-ary protograph LDPC code}{~\NoCaseChange{\protect\cite{cite1704,cite1705,cite1706,cite1707}}}
\codefieldsection{Description}
A \(q\)-ary LDPC code whose parity-check matrix is constructed using the \flmRefsHyperref{ref47}{lifting} procedure applied to the incidence matrix of a sparse graph called, in this context, a \textit{protograph}.
An ability to assign non-binary edge weight called \textit{edge scaling} can also be used in code construction.

\begin{defterm}{Lifting}\label{ref1708}\label{ref47}
Given the incidence matrix \(A\) of a protograph, each nonzero entry is replaced by a sum of \(\ell\)-dimensional permutation matrices while each zero entry is replaced by the \(\ell\)-dimensional zero matrix.
The resulting matrix is called a \textit{lift} of \(A\).
The permutation matrices can be chosen randomly or deterministically, with a deterministic lift also called a \textit{permutation voltage assignment} in the theory of voltage graphs \NoCaseChange{\protect\cite{cite1216,cite1217}}.

The matrices can come from a group \(G\) or its \flmRefsHyperref{ref205}{group algebra} \(\mathbb{F}_q G\), in which case the lift is often called a \(G\)\textit{-lift}.
In this case, matrix entries of a \(\mathbb{F}_q\)-valued matrix \(A\) are substituted with matrices forming the regular representation of \(\mathbb{F}_q G\) according to some rule.

For example, the lift of a binary two-dimensional incidence matrix using two-dimensional permutation matrices associated with the group \(\mathbb{Z}_2\) is as follows:
\flmMathEnvironment{align}{}{
  \begin{pmatrix}1 & 1\\
  0 & 1
  \end{pmatrix}\to\left(\begin{smallmatrix}0 & 1 & 0 & 1\\
  1 & 0 & 1 & 0\\
  0 & 0 & 1 & 0\\
  0 & 0 & 0 & 1
  \end{smallmatrix}\right)~.
}
Here, the two nonzero entries in the top row are replaced by the exchange permutation while the bottom nonzero entry is replaced by the trivial permutation.
\end{defterm}

\codefieldsection{Protection}
Minimum distance bounds \NoCaseChange{\protect\cite{cite1709}}.

\codefieldsection{Parent}
\begin{eczvaluelist}
\item\relax
\flmRefsHyperref[eczindexfamilyrel]{code:q-ary_ldpc}{\(q\)-ary LDPC code}\end{eczvaluelist}
\codefieldsection{Child}
\begin{eczvaluelist}
\item\relax
\flmRefsHyperref[eczindexfamilyrel]{code:protograph_ldpc}{Protograph LDPC code}\end{eczvaluelist}
\eczhbkcontributors{ \eczhuVVA }
\endeczcode

\eczcode{q-ary_repetition}{\(q\)-ary repetition code}{}
\codefieldsection{Description}
An \([n,1,n]_q\) code consisting of codewords \((j,j,\cdots,j)\) for \(j \in \mathbb{F}_q\).
\codefieldsection{Protection}
Detects up to \(n-1\) symbol errors, corrects up to \(\left\lfloor (n-1)/2\right\rfloor\) symbol errors by majority vote, and corrects up to \(n-1\) erasures.
\codefieldsection{Rate}
Code rate is \(1/n\), and code distance is \(n\).
\codefieldsection{Decoding}
\begin{eczvaluelist}
\item\relax The receiver can use majority vote to recover the information.
\end{eczvaluelist}
\codefieldsection{Parents}
\begin{eczvaluelist}
\item\relax
\flmRefsHyperref[eczindexfamilyrel]{code:extended_reed_solomon}{Extended GRS code} --- \(q\)-ary repetition codes can be thought of as extended RS codes \NoCaseChange{\protect\cite{cite1673}}. GRM\(_q(0,m)\) codes are evaluations of all zero-degree polynomials on \(\mathbb{F}_q^m\), which are just the \(q\) constant polynomials, so \(q\)-ary repetition codes are GRM\(_q(0,m)\) codes.
\item\relax
\flmRefsHyperref[eczindexfamilyrel]{code:reed_solomon}{Reed-Solomon (RS) code} --- \(q\)-ary repetition codes can be thought of as RS codes \NoCaseChange{\protect\cite{cite1673}}.
\item\relax
\flmRefsHyperref[eczindexfamilyrel]{code:q-ary_cyclic}{Cyclic linear \(q\)-ary code} --- The \(q\)-ary repetition code is cyclic with generator polynomial \(1+x+\cdots+x^{n-1}\).
\item\relax
\flmRefsHyperref[eczindexfamilyrel]{code:delsarte_optimal_q-ary}{\(q\)-ary sharp configuration} --- The \(q\)-ary repetition code is a \(q\)-ary sharp configuration \NoCaseChange{\protect\cite[{Table 12.1}]{cite199}}.
\end{eczvaluelist}
\codefieldsection{Child}
\begin{eczvaluelist}
\item\relax
\flmRefsHyperref[eczindexfamilyrel]{code:repetition}{Repetition code} --- \(q\)-ary repetition code reduce to repetition codes for \(q=2\).
\end{eczvaluelist}
\codefieldsection{Cousins}
\begin{eczvaluelist}
\item\relax
\flmRefsHyperref[eczindexfamilyrel]{code:locally_recoverable}{Locally recoverable code (LRC)} --- The \(q\)-ary repetition code is an LRC with \(r=2\) \NoCaseChange{\protect\cite{cite812}}.
\item\relax
\flmRefsHyperref[eczindexfamilyrel]{code:lambda16}{\(\Lambda_{16}\) Barnes-Wall lattice} --- The \(\Lambda_{16}\) Barnes-Wall lattice can be obtained from the \([4,1,4]\) repetition code over \(\mathbb{F}_9\) via Quebbemann's construction \NoCaseChange{\protect\cite[{Ch. 8, pg. 219}]{cite39}}.
\item\relax
\flmRefsHyperref[eczindexfamilyrel]{code:coxeter_todd}{Coxeter-Todd \(K_{12}\) lattice} --- Applying Construction \(B_c\) to the ternary repetition code of length \(6\) over the Eisenstein integers yields the Coxeter-Todd \(K_{12}\) lattice \NoCaseChange{\protect\cite[{Ch. 7, pg. 200}]{cite39}}.
\item\relax
\flmRefsHyperref[eczindexfamilyrel]{code:esix}{\(E_6\) root lattice} --- The \([3,1,3]_3\) ternary repetition code can be used to obtain the \(E_6\) root lattice \NoCaseChange{\protect\cite[{Exam. 10.5.4}]{cite115}\protect\cite[{Ch. 7, pg. 200}]{cite39}}.
\end{eczvaluelist}
\eczhbkcontributors{ \eczhuVVA }
\endeczcode

\eczcode{delsarte_optimal_q-ary}{\(q\)-ary sharp configuration}{~\NoCaseChange{\protect\cite{cite171,cite914,cite173}}}
\codefieldsection{Alternative Names}
\begin{eczvaluelist}
\item\relax Delsarte code in Hamming space
\item\relax \(q\)-ary Delsarte code
\end{eczvaluelist}
\eczhIndexCodeAliasName{delsarte_optimal_q-ary}{Delsarte code in Hamming space}
\eczhIndexCodeAliasName{delsarte_optimal_q-ary}{\(q\)-ary Delsarte code}
\codefieldsection{Description}
A \(q\)-ary code that admits \(m\) different distances between distinct codewords and forms a design of strength \(2m-1\) or greater.

\codefieldsection{Parents}
\begin{eczvaluelist}
\item\relax
\flmRefsHyperref[eczindexfamilyrel]{code:univ_opt_q-ary}{Universally optimal \(q\)-ary code} --- All \(q\)-ary sharp configurations are universally optimal \(q\)-ary codes \NoCaseChange{\protect\cite{cite173}}, but the converse is not true.
\item\relax
\flmRefsHyperref[eczindexfamilyrel]{code:delsarte_optimal}{Sharp configuration}\item\relax
\flmRefsHyperref[eczindexfamilyrel]{code:orthogonal_array}{Orthogonal array (OA)}\end{eczvaluelist}
\codefieldsection{Children}
\begin{eczvaluelist}
\item\relax
\flmRefsHyperref[eczindexfamilyrel]{code:extended_golay}{\([24, 12, 8]\) Extended Golay code} --- The extended Golay code is a sharp configuration \NoCaseChange{\protect\cite[{Table 12.1}]{cite199}}.
\item\relax
\flmRefsHyperref[eczindexfamilyrel]{code:parity_check}{\([n,n-1,2]\) Single parity-check (SPC) code} --- The SPC code is a binary sharp configuration \NoCaseChange{\protect\cite[{Table 12.1}]{cite199}}.
\item\relax
\flmRefsHyperref[eczindexfamilyrel]{code:q-ary_repetition}{\(q\)-ary repetition code} --- The \(q\)-ary repetition code is a \(q\)-ary sharp configuration \NoCaseChange{\protect\cite[{Table 12.1}]{cite199}}.
\item\relax
\flmRefsHyperref[eczindexfamilyrel]{code:bose_qvist}{Ovoid code} --- The ovoid code is a \(q\)-ary sharp configuration \NoCaseChange{\protect\cite[{Table 12.1}]{cite199}}.
\item\relax
\flmRefsHyperref[eczindexfamilyrel]{code:denniston}{Denniston code} --- The Denniston code is a \(q\)-ary sharp configuration \NoCaseChange{\protect\cite[{Table 12.1}]{cite199}}.
\item\relax
\flmRefsHyperref[eczindexfamilyrel]{code:hill_56_6_36}{\([56,6,36]_3\) Hill-cap code} --- The \([56,6,36]_3\) Hill-cap code is a \(q\)-ary sharp configurations \NoCaseChange{\protect\cite[{Table 12.1}]{cite199}}.
\item\relax
\flmRefsHyperref[eczindexfamilyrel]{code:hill_78_6_56}{\([78,6,56]_4\) Hill-cap code} --- The \([78,6,56]_4\) Hill-cap code is a \(q\)-ary sharp configuration \NoCaseChange{\protect\cite[{Table 12.1}]{cite199}}.
\item\relax
\flmRefsHyperref[eczindexfamilyrel]{code:semakov_zinoviev}{\(ED_m\) code} --- The \(ED_m\) code is a \(q\)-ary sharp configuration \NoCaseChange{\protect\cite[{Table 12.1}]{cite199}}.
\item\relax
\flmRefsHyperref[eczindexfamilyrel]{code:semakov_zinoviev_zaitsev}{Semakov-Zinoviev-Zaitsev (SZZ) equidistant code} --- The SZZ equidistant code is a \(q\)-ary sharp configuration \NoCaseChange{\protect\cite[{Table 12.1}]{cite199}}.
\end{eczvaluelist}
\codefieldsection{Cousins}
\begin{eczvaluelist}
\item\relax
\flmRefsHyperref[eczindexfamilyrel]{code:golay}{\([23, 12, 7]\) Golay code} --- The dual \([23,11,8]\) even-weight subcode of the Golay code is a sharp configuration \NoCaseChange{\protect\cite[{Table 12.1}]{cite199}}.
\item\relax
\flmRefsHyperref[eczindexfamilyrel]{code:hadamard}{\([2^m,m,2^{m-1}]\) Hadamard code} --- The augmented binary Hadamard code family is listed among the sharp configurations in \NoCaseChange{\protect\cite[{Table 12.1}]{cite199}}.
\item\relax
\flmRefsHyperref[eczindexfamilyrel]{code:ternary_golay}{\([11,6,5]_3\) Ternary Golay code} --- The dual \([11,5,6]_3\) code of the ternary Golay code is the length-\(11\) ternary Golay-family sharp configuration in \NoCaseChange{\protect\cite[{Table 12.1}]{cite199}}. The extended ternary Golay code is the corresponding sharp configuration at length \(12\).
\item\relax
\flmRefsHyperref[eczindexfamilyrel]{code:hyperoval}{Hyperoval code} --- Codes based on hyperovals in \(PG(2,q)\) are \(q\)-ary sharp configurations \NoCaseChange{\protect\cite[{Table 12.1}]{cite199}}.
\end{eczvaluelist}
\eczhbkcontributors{ Alexander Barg, \eczhuVVA }
\endeczcode

\eczcode{q-ary_simplex}{\(q\)-ary simplex code}{~\NoCaseChange{\protect\cite{cite1158,cite1}}}
\codefieldsection{Alternative Names}
\begin{eczvaluelist}
\item\relax First-order PRM code
\item\relax \(q\)-ary maximum-length feedback-shift-register code
\end{eczvaluelist}
\eczhIndexCodeAliasName{q-ary_simplex}{First-order PRM code}
\eczhIndexCodeAliasName{q-ary_simplex}{\(q\)-ary maximum-length feedback-shift-register code}
\codefieldsection{Description}
An \([n,m,q^{m-1}]_q\) equidistant projective code with \(n=\frac{q^m-1}{q-1}\), denoted as \(S(q,m)\). The columns of the generator matrix are in one-to-one correspondence with the elements of the projective space \(PG(m-1,q)\), with each column being a chosen representative of the corresponding element.
All nonzero simplex codewords have a constant weight of \(q^{m-1}\) \NoCaseChange{\protect\cite{cite45,cite46}}.

The dual of a \(q\)-ary simplex code is the \([n,n-m,3]_q\) \(q\)-ary Hamming code.
A punctured simplex code is known as a \textit{\(q\)-ary MacDonald code} \NoCaseChange{\protect\cite{cite1160}}, with parameters \([\frac{q^m-q^u}{q-1},m,q^{m-1}-q^{u-1}]_q\) for \(u \leq m-1\) \NoCaseChange{\protect\cite{cite1161}}.

\codefieldsection{Protection}
The simplex codes saturate the Plotkin bound \NoCaseChange{\protect\cite[{Sec. 1.2.2}]{cite1314}}.

\codefieldsection{Decoding}
\begin{eczvaluelist}
\item\relax Permutation decoder for simplex \NoCaseChange{\protect\cite{cite1710}} and MacDonald \NoCaseChange{\protect\cite{cite1711}} codes.
\end{eczvaluelist}
\codefieldsection{Notes}
\begin{eczvaluelist}
\item\relax See corresponding MinT database entry \NoCaseChange{\protect\cite{cite1712}}.
\item\relax Simplex codes are equivalent to irreducible cyclic codes \(C(q,n,1)\) when \(\gcd(q-1,m)=1\) \NoCaseChange{\protect\cite[{Exam. 2.5.1}]{cite68}}.
\end{eczvaluelist}
\codefieldsection{Parents}
\begin{eczvaluelist}
\item\relax
\flmRefsHyperref[eczindexfamilyrel]{code:projective_reed_muller}{Projective RM (PRM) code} --- The \(q\)-ary simplex codes are first-order PRM codes \NoCaseChange{\protect\cite[{Sec. 1.2.2}]{cite1314}}.
\item\relax
\flmRefsHyperref[eczindexfamilyrel]{code:incidence_matrix}{Incidence-matrix projective code} --- Columns of a simplex code's generator matrix correspond to 1D subspaces of \(\mathbb{F}_q^m\), i.e., to points of \(PG(m-1,q)\).
\item\relax
\flmRefsHyperref[eczindexfamilyrel]{code:griesmer}{Griesmer code} --- Simplex codes saturate the Griesmer bound (\NoCaseChange{\protect\cite{cite62}}, Exer. 5.1.11).
\end{eczvaluelist}
\codefieldsection{Children}
\begin{eczvaluelist}
\item\relax
\flmRefsHyperref[eczindexfamilyrel]{code:simplex}{\([2^m-1,m,2^{m-1}]\) simplex code} --- \(q\)-ary simplex codes reduce to simplex codes for \(q=2\).
\item\relax
\flmRefsHyperref[eczindexfamilyrel]{code:tetracode}{\([4,2,3]_3\) Tetracode} --- The tetracode is equivalent to \(S(3,2)\).
\end{eczvaluelist}
\codefieldsection{Cousins}
\begin{eczvaluelist}
\item\relax
\flmRefsHyperref[eczindexfamilyrel]{code:univ_opt_q-ary}{Universally optimal \(q\)-ary code} --- Simplex codes and their once-punctured versions are LP universally optimal codes \NoCaseChange{\protect\cite{cite173}}.
\item\relax
\flmRefsHyperref[eczindexfamilyrel]{code:q-ary_linear}{Linear \(q\)-ary code} --- Linear \(q\)-ary codes cannot be constant weight, but can have nonzero codewords with constant weight. All such codes are equidistant, and Bonisoli's theorem states that any equidistant linear code is a direct sum of \(q\)-ary simplex codes \NoCaseChange{\protect\cite{cite988}} (see also Refs. \NoCaseChange{\protect\cite{cite45,cite46}}).
\item\relax
\flmRefsHyperref[eczindexfamilyrel]{code:q-ary_constant_weight}{Constant-weight block code} --- Linear \(q\)-ary codes cannot be constant weight, but can have nonzero codewords with constant weight. All such codes are equidistant, and Bonisoli's theorem states that any equidistant linear code is a direct sum of \(q\)-ary simplex codes \NoCaseChange{\protect\cite{cite988}} (see also Refs. \NoCaseChange{\protect\cite{cite45,cite46}}).
\item\relax
\flmRefsHyperref[eczindexfamilyrel]{code:group}{Group-algebra code} --- Over a prime field \(\mathbb{F}_p\), simplex codes with parameters \([(p^m-1)/(p-1),m,p^{m-1}]_p\) and \(\gcd(m,p-1)=1\) are group-algebra codes \NoCaseChange{\protect\cite[{Exam. 16.8.2}]{cite196}}.
\item\relax
\flmRefsHyperref[eczindexfamilyrel]{code:q-ary_hamming}{\(q\)-ary Hamming code} --- \(q\)-ary Hamming and \(q\)-ary simplex codes are dual to each other \NoCaseChange{\protect\cite[{pg. 45}]{cite1314}}.
\item\relax
\flmRefsHyperref[eczindexfamilyrel]{code:dual}{Dual linear code} --- \(q\)-ary Hamming and \(q\)-ary simplex codes are dual to each other \NoCaseChange{\protect\cite[{pg. 45}]{cite1314}}
\item\relax
\flmRefsHyperref[eczindexfamilyrel]{code:two_weight}{Two-weight code} --- \(q\)-ary MacDonald codes are two-weight codes with weights \(q^{m-1}-q^{u-1}\) and \(q^{m-1}\) \NoCaseChange{\protect\cite{cite1161}}.
\item\relax
\flmRefsHyperref[eczindexfamilyrel]{code:extended_reed_solomon}{Extended GRS code} --- \(q\)-ary simplex codes for \(m=2\) can be thought of as extended RS codes \NoCaseChange{\protect\cite{cite1673}}.
\end{eczvaluelist}
\eczhbkcontributors{ \eczhuVVA }
\endeczcode

\eczcode{q-ary_additive}{Additive \(q\)-ary code}{}
\codefieldsection{Description}
A \(q\)-ary code whose codewords are closed under addition, i.e., for any codewords \(x,y\), \(x+y\) is also a codeword.
If \(q=p^m\), then additive closure already implies \(\mathbb{F}_p\)-linearity because multiplying a codeword by any \(\lambda\in\mathbb{F}_p\) is equivalent to adding that codeword to itself \(\lambda\) times.

\codefieldsection{Parents}
\begin{eczvaluelist}
\item\relax
\flmRefsHyperref[eczindexfamilyrel]{code:q-ary_digits_into_q-ary_digits}{\(q\)-ary code}\item\relax
\flmRefsHyperref[eczindexfamilyrel]{code:group_linear}{Linear code over \(G\)} --- Additive \(q\)-ary codes are linear over the additive group \(G=\mathbb{F}_q\). If \(q=p^m\), they are always \(\mathbb{F}_p\)-linear, but for \(m>1\) they need not be \(\mathbb{F}_q\)-linear.
\end{eczvaluelist}
\codefieldsection{Children}
\begin{eczvaluelist}
\item\relax
\flmRefsHyperref[eczindexfamilyrel]{code:twisted_bch}{Twisted BCH code}\item\relax
\flmRefsHyperref[eczindexfamilyrel]{code:dual_additive}{Dual additive code}\item\relax
\flmRefsHyperref[eczindexfamilyrel]{code:q-ary_linear}{Linear \(q\)-ary code} --- For \(q>2\), additive codes need not be linear since linearity also requires closure under multiplication.
\end{eczvaluelist}
\codefieldsection{Cousins}
\begin{eczvaluelist}
\item\relax
\flmRefsHyperref[eczindexfamilyrel]{code:pentacode}{\((5,40,4)_{\mathbb{Z}_4}\) Pentacode} --- A close relative of the pentacode is an additive quaternary code \NoCaseChange{\protect\cite{cite1713}}.
\item\relax
\flmRefsHyperref[eczindexfamilyrel]{code:eastab}{EA qubit stabilizer code} --- There is a relation between quaternary additive codes and EA qubit stabilizer codes \NoCaseChange{\protect\cite{cite1714}}.
\item\relax
\flmRefsHyperref[eczindexfamilyrel]{code:galois_stabilizer}{Galois-qudit stabilizer code} --- Galois-qudit stabilizer codes are the closest quantum analogues of additive codes over \(\mathbb{F}_q\) because addition in the field corresponds to multiplication of stabilizers in the quantum case.
\end{eczvaluelist}
\eczhbkcontributors{ Shuubham Ojha, \eczhuVVA }
\endeczcode

\eczcode{ag}{Algebraic-geometry (AG) code}{~\NoCaseChange{\protect\cite{cite1715,cite1716,cite1717}}}
\codefieldsection{Description}
Binary or \(q\)-ary code or subcode constructed from an algebraic curve of some genus over a finite field via the evaluation construction, the residue construction, or more general constructions that yield nonlinear codes \NoCaseChange{\protect\cite[{Defs. 15.3.1 and 15.3.2}]{cite26}\protect\cite[{Sec. 15.4.4}]{cite26}}. Linear AG codes from the first two constructions are also called \textit{geometric Goppa codes}.

In alternative conventions (not used here), AG codes are restricted to be linear and/or include \flmRefsHyperref{code:evaluation_varieties}{evaluation} codes defined using algebraic varieties more general than curves.

\codefieldsection{Rate}
Several sequences of linear AG codes beat the \flmRefsHyperref{ref85}{GV bound} and/or are asymptotically good \NoCaseChange{\protect\cite{cite1718,cite1719,cite1720}\protect\cite[{Thm. 15.4.3}]{cite26}} (see Ref. \NoCaseChange{\protect\cite{cite32}} for details). For square \(q\), there exist sequences of linear AG codes whose asymptotic rate and relative distance satisfy \flmMathEnvironment{align}{}{ \frac{k}{n} \geq 1 - \frac{d}{n} - \frac{1}{\sqrt{q}-1}~, } which follows from the Drinfeld-Vladut bound \NoCaseChange{\protect\cite{cite1721}}. Nonlinear AG codes can outperform this bound \NoCaseChange{\protect\cite[{Thm. 15.4.6}]{cite26}}.
\codefieldsection{Notes}
\begin{eczvaluelist}
\item\relax See book by Goppa \NoCaseChange{\protect\cite{cite1722}}.
\end{eczvaluelist}
\codefieldsection{Parent}
\begin{eczvaluelist}
\item\relax
\flmRefsHyperref[eczindexfamilyrel]{code:q-ary_digits_into_q-ary_digits}{\(q\)-ary code}\end{eczvaluelist}
\codefieldsection{Children}
\begin{eczvaluelist}
\item\relax
\flmRefsHyperref[eczindexfamilyrel]{code:evaluation_varieties}{Evaluation code} --- Evaluation codes on algebraic varieties are AG codes. The AG-code literature has mostly focused on codes on algebraic curves, i.e., one-dimensional varieties.
\item\relax
\flmRefsHyperref[eczindexfamilyrel]{code:nonlinear_ag}{Nonlinear AG code}\item\relax
\flmRefsHyperref[eczindexfamilyrel]{code:cartier}{Cartier code} --- Every Cartier code is contained in a \flmRefsHyperref{ref33}{subfield} subcode of a residue AG code. Cartier codes share similar asymptotic properties to \flmRefsHyperref{ref33}{subfield} subcodes of residue AG codes, with both families admitting sequences of codes that achieve the \flmRefsHyperref{ref85}{GV bound}.
\end{eczvaluelist}
\codefieldsection{Cousins}
\begin{eczvaluelist}
\item\relax
\flmRefsHyperref[eczindexfamilyrel]{code:mds}{Maximum distance separable (MDS) code} --- Near MDS \([n,k,d]_{p^m}\) AG codes exist when \(n,p,m\) satisfy certain relations according to the Tsfasman-Vladut bound \NoCaseChange{\protect\cite{cite1312,cite1313,cite1314,cite63}}.
\item\relax
\flmRefsHyperref[eczindexfamilyrel]{code:ea_galois_stabilizer}{EA Galois-qudit stabilizer code} --- Certain AG codes can be used to construct EA Galois-qudit stabilizer codes \NoCaseChange{\protect\cite{cite1723}}.
\end{eczvaluelist}
\eczhbkcontributors{ Alexander Barg, \eczhuVVA }
\endeczcode

\eczcode{alternant}{Alternant code}{~\NoCaseChange{\protect\cite{cite1724,cite1725,cite1726,cite1727}}}
\codefieldsection{Description}
Given a length-\(n\) GRS code \(C\) over \(\mathbb{F}_{q^m}\), an alternant code is the \(\mathbb{F}_q\)-\flmRefsHyperref{ref33}{subfield} subcode of the dual of \(C\); see \NoCaseChange{\protect\cite[{Ch. 12}]{cite41}}. Its parity-check matrix is an alternant matrix.
\codefieldsection{Decoding}
\begin{eczvaluelist}
\item\relax Variation of the Berlekamp-Welch algorithm \NoCaseChange{\protect\cite{cite1728}}.
\item\relax Euclidean algorithm; see \NoCaseChange{\protect\cite[{Ch. 12}]{cite41}} for more details.
\item\relax Guruswami-Sudan list decoder \NoCaseChange{\protect\cite{cite1239,cite1240}}.
\end{eczvaluelist}
\codefieldsection{Notes}
\begin{eczvaluelist}
\item\relax See \NoCaseChange{\protect\cite[{Ch. 12}]{cite41}} for more details.
\end{eczvaluelist}
\codefieldsection{Parent}
\begin{eczvaluelist}
\item\relax
\flmRefsHyperref[eczindexfamilyrel]{code:q-ary_linear}{Linear \(q\)-ary code}\end{eczvaluelist}
\codefieldsection{Children}
\begin{eczvaluelist}
\item\relax
\flmRefsHyperref[eczindexfamilyrel]{code:goppa}{Goppa code} --- Goppa codes are a special case of alternant codes \NoCaseChange{\protect\cite[{Ch. 12}]{cite41}}.
\item\relax
\flmRefsHyperref[eczindexfamilyrel]{code:gbch}{Chien-Choy generalized BCH (GBCH) code} --- GBCH codes are a special case of alternant codes \NoCaseChange{\protect\cite[{Ch. 12}]{cite41}}.
\item\relax
\flmRefsHyperref[eczindexfamilyrel]{code:generalized_srivastava}{Generalized Srivastava code} --- Generalized Srivastava codes are a special case of alternant codes \NoCaseChange{\protect\cite[{Ch. 12}]{cite41}}.
\end{eczvaluelist}
\codefieldsection{Cousins}
\begin{eczvaluelist}
\item\relax
\flmRefsHyperref[eczindexfamilyrel]{code:generalized_reed_solomon}{Generalized RS (GRS) code} --- Alternant codes are \flmRefsHyperref{ref33}{subfield} subcodes of GRS codes \NoCaseChange{\protect\cite{cite1727}}.
\item\relax
\flmRefsHyperref[eczindexfamilyrel]{code:berlekamp}{Berlekamp code} --- Berlekamp codes reduce to narrow-sense alternant codes for \(p=2\) \NoCaseChange{\protect\cite[{Ch. 10.6}]{cite195}}.
\item\relax
\flmRefsHyperref[eczindexfamilyrel]{code:qubit_css}{Qubit CSS code} --- Alternant codes used in the CSS construction yield quantum codes that asymptotically achieve the \flmRefsHyperref{ref1729}{quantum GV bound} \NoCaseChange{\protect\cite{cite1730}}.
\end{eczvaluelist}
\eczhbkcontributors{ Khalil Guy, Manasi Shingane, \eczhuVVA }
\endeczcode

\eczcode{codes_with_availability}{Availability code}{~\NoCaseChange{\protect\cite{cite1731,cite1732}}}
\codefieldsection{Description}
A \(t\)-availability parallel-recovery code is a code such that coordinates can be recovered in multiple ways.
The availability of a locally recoverable code is the minimum, over all coordinates, of the number of recovery sets for that coordinate \NoCaseChange{\protect\cite[{Def. 15.9.20}]{cite26}}.
That way, the code accommodates nodes that may be inaccessible during the recovery procedure.

\codefieldsection{Parent}
\begin{eczvaluelist}
\item\relax
\flmRefsHyperref[eczindexfamilyrel]{code:parallel_recovery}{Parallel-recovery code}\end{eczvaluelist}
\codefieldsection{Cousins}
\begin{eczvaluelist}
\item\relax
\flmRefsHyperref[eczindexfamilyrel]{code:locally_recoverable}{Locally recoverable code (LRC)} --- Availability is the number of recovery sets available for each coordinate of an LRC \NoCaseChange{\protect\cite[{Def. 15.9.20}]{cite26}}.
\item\relax
\flmRefsHyperref[eczindexfamilyrel]{code:tamo_barg_vladut}{Barg-Tamo-Vladut code} --- Fibre-product constructions of Barg-Tamo-Vladut codes yield LRCs with availability \(2\) \NoCaseChange{\protect\cite[{Thm. 15.9.21}]{cite26}}.
\end{eczvaluelist}
\eczhbkcontributors{ \eczhuVVA }
\endeczcode

\eczcode{balanced}{Balanced code}{~\NoCaseChange{\protect\cite{cite1733}}}
\codefieldsection{Description}
An even-length-\(n\) \(q\)-ary code whose nonzero codewords all have a Hamming weight of \(n/2\).
A code is \(\epsilon\)\textit{-balanced} if the relative weight (i.e., weight divided by \(n\)) of all nonzero codewords lies in the interval \([\frac{1-\epsilon}{2},\frac{1+\epsilon}{2}]\).
A code is \(\gamma\)\textit{-unbiased} if the relative weight lies in the interval \((\frac{1}{2}-\frac{1}{n^{\gamma}},\frac{1}{2}+\frac{1}{n^{\gamma}})\).

\codefieldsection{Protection}
Can detect unidirectional errors, such as a zero going to a one.
\codefieldsection{Encoding}
\begin{eczvaluelist}
\item\relax Efficient encoder \NoCaseChange{\protect\cite{cite1733}}.
\end{eczvaluelist}
\codefieldsection{Decoding}
\begin{eczvaluelist}
\item\relax Efficient decoder \NoCaseChange{\protect\cite{cite1733,cite1734,cite1735}}.
\end{eczvaluelist}
\codefieldsection{Realizations}
\begin{eczvaluelist}
\item\relax Balanced length-eight code, known as a 6b/8b encoding, used for balancing direct current in a communications system \NoCaseChange{\protect\cite{cite235}}
\end{eczvaluelist}
\codefieldsection{Parent}
\begin{eczvaluelist}
\item\relax
\flmRefsHyperref[eczindexfamilyrel]{code:q-ary_digits_into_q-ary_digits}{\(q\)-ary code}\end{eczvaluelist}
\codefieldsection{Child}
\begin{eczvaluelist}
\item\relax
\flmRefsHyperref[eczindexfamilyrel]{code:hadamard}{\([2^m,m,2^{m-1}]\) Hadamard code} --- Each nonzero Hadamard codeword has length \(2^m\) and Hamming weight of \(2^{m-1}\), making this code balanced.
\end{eczvaluelist}
\codefieldsection{Cousins}
\begin{eczvaluelist}
\item\relax
\flmRefsHyperref[eczindexfamilyrel]{code:q-ary_constant_weight}{Constant-weight block code} --- Balanced codes are not automatically constant-weight because they may contain the zero codeword.
\item\relax
\flmRefsHyperref[eczindexfamilyrel]{code:ltc}{Locally testable code (LTC)} --- Random low-rate unbiased linear codes are LTCs \NoCaseChange{\protect\cite{cite1103}}.
\item\relax
\flmRefsHyperref[eczindexfamilyrel]{code:ta-shma}{Ta-Shma zigzag code} --- Ta-Shma codes are \(\epsilon\)-balanced.
\end{eczvaluelist}
\eczhbkcontributors{ \eczhuVVA }
\endeczcode

\eczcode{tamo_barg_vladut}{Barg-Tamo-Vladut code}{~\NoCaseChange{\protect\cite{cite1736,cite1085}}}
\codefieldsection{Description}
Evaluation AG code on algebraic curves built from a Galois cover \(\phi:Y\to X\), where the recovery sets are fibres over rational points of \(X\) that split completely in the cover \NoCaseChange{\protect\cite[{Def. 15.9.10}]{cite26}\protect\cite[{Thm. 15.9.14}]{cite26}}.
The Barg-Tamo-Vladut construction generalizes the Tamo-Barg construction from \(PG(1,q)\) to longer AG codes, and variants can be built with higher local distance or availability \(2\) via fibre products of curves \NoCaseChange{\protect\cite[{Thm. 15.9.19}]{cite26}\protect\cite[{Thm. 15.9.21}]{cite26}}.

\codefieldsection{Rate}
Barg-Tamo-Vladut codes on asymptotically maximal curves improve upon the asymptotic LRC GV bound \NoCaseChange{\protect\cite{cite1085}}.
\codefieldsection{Parents}
\begin{eczvaluelist}
\item\relax
\flmRefsHyperref[eczindexfamilyrel]{code:evaluation}{Evaluation AG code} --- Barg-Tamo-Vladut codes are evaluation AG codes on algebraic curves built from curve covers \NoCaseChange{\protect\cite[{Thm. 15.9.14}]{cite26}}.
\item\relax
\flmRefsHyperref[eczindexfamilyrel]{code:locally_recoverable}{Locally recoverable code (LRC)} --- Barg-Tamo-Vladut codes form a family of locally recoverable codes obtained from algebraic curves \NoCaseChange{\protect\cite[{Thm. 15.9.14}]{cite26}}.
\end{eczvaluelist}
\codefieldsection{Child}
\begin{eczvaluelist}
\item\relax
\flmRefsHyperref[eczindexfamilyrel]{code:tamo_barg}{Tamo-Barg code} --- Tamo-Barg codes are the \(PG(1,q)\)/genus-zero instance of the covering-based construction that later yields Barg-Tamo-Vladut codes \NoCaseChange{\protect\cite[{Sec. 15.9.4}]{cite26}}.
\end{eczvaluelist}
\codefieldsection{Cousins}
\begin{eczvaluelist}
\item\relax
\flmRefsHyperref[eczindexfamilyrel]{code:codes_with_availability}{Availability code} --- Fibre-product constructions of Barg-Tamo-Vladut codes yield LRCs with availability \(2\) \NoCaseChange{\protect\cite[{Thm. 15.9.21}]{cite26}}.
\item\relax
\flmRefsHyperref[eczindexfamilyrel]{code:hermitian}{Hermitian code} --- Hermitian-curve examples of these codes are given in \NoCaseChange{\protect\cite[{Rem. 15.9.16}]{cite26}}.
\end{eczvaluelist}
\eczhbkcontributors{ \eczhuVVA }
\endeczcode

\eczcode{bs-ltc}{Ben-Sasson-Sudan code}{~\NoCaseChange{\protect\cite{cite1737}}}
\codefieldsection{Description}
Locally testable \([n,k/2,d]_{2^m}\) code with \(k\) a power of two, \(n = k \log^{c} k\), and query complexity \(\log^{c}k\) for some universal constant \(c\).

\codefieldsection{Parent}
\begin{eczvaluelist}
\item\relax
\flmRefsHyperref[eczindexfamilyrel]{code:q-ary_ltc}{\(q\)-ary linear LTC}\end{eczvaluelist}
\eczhbkcontributors{ \eczhuVVA }
\endeczcode

\eczcode{q-ary_bch}{Bose–Chaudhuri–Hocquenghem (BCH) code}{~\NoCaseChange{\protect\cite{cite1738}}}
\codefieldsection{Description}
A cyclic \(q\)-ary code, with \(n\) and \(q\) relatively prime, whose zeroes are consecutive powers of a primitive \(n\)th root of unity \(\alpha\). 

More precisely, the generator polynomial of a BCH code of \textit{designed distance} \(\delta\geq 1\) is the lowest-degree monic polynomial with zeroes \(\{\alpha^b,\alpha^{b+1},\cdots,\alpha^{b+\delta-2}\}\) for some \(b\geq 0\). BCH codes are called \textit{narrow-sense} when \(b=1\), and are called \textit{primitive} when \(n=q^r-1\) for some \(r\geq 2\).
More general BCH codes can be defined for zeroes of the form \(\{\alpha^b,\alpha^{b+s},\alpha^{b+2s},\cdots,\alpha^{b+(\delta-2)s}\}\), where gcd\((s,n)=1\).

The code dimension is related to the \textit{multiplicative order} of \(q\) modulo \(n\), i.e., the smallest integer \(m\) such that \(n\) divides \(q^m-1\). The dimension of a BCH code is at least \(n-m(\delta-1)\). The field \(\mathbb{F}_{q^m}\) is the smallest field containing the above root of unity \(\alpha\), and is the splitting field of the polynomial \(x^n-1\) (see \flmRefsCref{ref67}).

\codefieldsection{Protection}
By the BCH bound, a BCH code with designed distance \(\delta\) has true minimum distance \(d\geq\delta\). BCH codes with different designed distances may coincide, and the largest possible designed distance for a given code is the \textit{Bose distance}; the true distance may still be larger than that.

\codefieldsection{Rate}
Primitive BCH codes are asymptotically bad \NoCaseChange{\protect\cite[{Thm. 2.6.3}]{cite68}\protect\cite[{pg.~269}]{cite41}}.
\codefieldsection{Decoding}
\begin{eczvaluelist}
\item\relax Berlekamp-Massey decoder with runtime of \flmRefsHyperref{ref65}{order} \(O(n^2)\) \NoCaseChange{\protect\cite{cite1739,cite1234,cite1235}} and modification by Burton \NoCaseChange{\protect\cite{cite1236}}; see also \NoCaseChange{\protect\cite{cite993,cite1024}}.
\item\relax Gorenstein-Peterson-Zierler decoder with runtime of \flmRefsHyperref{ref65}{order} \(O(n^3)\) \NoCaseChange{\protect\cite{cite1231,cite1738}} (see exposition in Ref. \NoCaseChange{\protect\cite{cite1233}}).
\item\relax Sugiyama et al. modification of the extended Euclidean algorithm \NoCaseChange{\protect\cite{cite1237,cite1238}}.
\end{eczvaluelist}
\codefieldsection{Realizations}
\begin{eczvaluelist}
\item\relax DVDs, disk drives, and two-dimensional bar codes \NoCaseChange{\protect\cite{cite239}}.
\end{eczvaluelist}
\codefieldsection{Notes}
\begin{eczvaluelist}
\item\relax See books \NoCaseChange{\protect\cite{cite41,cite1241,cite126}\protect\cite[{Secs. 2.6-2.6.3}]{cite68}} for expositions on BCH codes and code tables.
\item\relax See Kaiserslautern database \NoCaseChange{\protect\cite{cite1184}} for explicit codes.
\item\relax See corresponding MinT database entry \NoCaseChange{\protect\cite{cite1242}}.
\end{eczvaluelist}
\codefieldsection{Parents}
\begin{eczvaluelist}
\item\relax
\flmRefsHyperref[eczindexfamilyrel]{code:gbch}{Chien-Choy generalized BCH (GBCH) code} --- \(q\)-ary BCH codes are a special case of GBCH codes \NoCaseChange{\protect\cite[{Ch. 12}]{cite41}}.
\item\relax
\flmRefsHyperref[eczindexfamilyrel]{code:q-ary_cyclic}{Cyclic linear \(q\)-ary code}\end{eczvaluelist}
\codefieldsection{Children}
\begin{eczvaluelist}
\item\relax
\flmRefsHyperref[eczindexfamilyrel]{code:bch}{Binary BCH code}\item\relax
\flmRefsHyperref[eczindexfamilyrel]{code:narrow_sense_reed_solomon}{Narrow-sense RS code} --- Narrow-sense RS codes are \(q\)-ary BCH codes \NoCaseChange{\protect\cite[{Remark 15.3.21}]{cite26}\protect\cite[{Thms. 5.2.1 and 5.2.3}]{cite126}}. Their minimal distance is equal to their designed distance \NoCaseChange{\protect\cite[{pg. 81}]{cite39}}.
\item\relax
\flmRefsHyperref[eczindexfamilyrel]{code:narrow_sense_q-ary_bch}{Primitive narrow-sense BCH code} --- BCH codes are called narrow-sense when \(b=1\), and are called primitive when \(n=q^r-1\) for some \(r\geq 2\).
\end{eczvaluelist}
\codefieldsection{Cousins}
\begin{eczvaluelist}
\item\relax
\flmRefsHyperref[eczindexfamilyrel]{code:twisted_bch}{Twisted BCH code} --- Twisted BCH codes are obtained from \(q\)-ary BCH codes via various lengthening and extension procedures.
\item\relax
\flmRefsHyperref[eczindexfamilyrel]{code:q-ary_ltc}{\(q\)-ary linear LTC} --- Duals of BCH codes are locally testable \NoCaseChange{\protect\cite{cite1270}}.
\item\relax
\flmRefsHyperref[eczindexfamilyrel]{code:q-ary_linear_over_zq}{Linear code over \(\mathbb{Z}_q\)} --- BCH codes for \(q=p\) prime field can also work as codes over the Lee metric \NoCaseChange{\protect\cite{cite1740}}.
\item\relax
\flmRefsHyperref[eczindexfamilyrel]{code:justesen}{Justesen code} --- Using more general BCH codes instead of RS codes can improve the parameters of the Justesen codes \NoCaseChange{\protect\cite{cite1404}}.
\item\relax
\flmRefsHyperref[eczindexfamilyrel]{code:skew_cyclic}{Skew-cyclic code} --- There exist skew-BCH codes, which are skew-cyclic versions of BCH codes \NoCaseChange{\protect\cite{cite1132}}.
\item\relax
\flmRefsHyperref[eczindexfamilyrel]{code:reed_solomon}{Reed-Solomon (RS) code} --- An RS code can be represented as a union of cosets, with each coset being an interleaver of several binary BCH codes \NoCaseChange{\protect\cite{cite1741}}.
BCH codes are \flmRefsHyperref{ref33}{subfield} subcodes of RS codes, while primitive RS codes are primitive BCH codes \NoCaseChange{\protect\cite[{Sec. 2.6}]{cite68}}.

\item\relax
\flmRefsHyperref[eczindexfamilyrel]{code:q-ary_hamming}{\(q\)-ary Hamming code} --- When \(\gcd(r,q-1)=1\), \(q\)-ary Hamming codes are narrow-sense BCH codes \NoCaseChange{\protect\cite[{Exam. 16.4.10}]{cite196}\protect\cite[{Thm. 5.1.4}]{cite126}}, which are cyclic \NoCaseChange{\protect\cite[{pg. 194}]{cite41}\protect\cite[{Exam. 2.5.1}]{cite68}}.
\item\relax
\flmRefsHyperref[eczindexfamilyrel]{code:quantum_bch}{Qubit BCH code} --- BCH codes are used to construct qubit BCH codes via the CSS construction or the Hermitian construction.
\item\relax
\flmRefsHyperref[eczindexfamilyrel]{code:qubit_subsystem_stabilizer}{Subsystem qubit stabilizer code} --- BCH codes yield subsystem stabilizer codes via the subsystem extension of the Hermitian construction to subsystem codes \NoCaseChange{\protect\cite[{Exam. 1}]{cite1742}}.
\item\relax
\flmRefsHyperref[eczindexfamilyrel]{code:galois_bch}{Galois-qudit BCH code} --- Galois-qudit BCH codes are quantum analogues of q-ary BCH codes.
\end{eczvaluelist}
\eczhbkcontributors{ Muhammad Junaid Aftab, \eczhuVVA }
\endeczcode

\eczcode{cartier}{Cartier code}{~\NoCaseChange{\protect\cite{cite1743}}}
\codefieldsection{Description}
A generalization of the Goppa codes to codes defined from curves of nonzero genus. Each code is a subcode of a certain residue AG code and is constructed using the Cartier operator.
\codefieldsection{Rate}
Cartier codes share similar asymptotic properties to \flmRefsHyperref{ref33}{subfield} subcodes of residue AG codes, with both families admitting sequences of codes that achieve the \flmRefsHyperref{ref85}{GV bound}.
\codefieldsection{Parents}
\begin{eczvaluelist}
\item\relax
\flmRefsHyperref[eczindexfamilyrel]{code:q-ary_linear}{Linear \(q\)-ary code}\item\relax
\flmRefsHyperref[eczindexfamilyrel]{code:ag}{Algebraic-geometry (AG) code} --- Every Cartier code is contained in a \flmRefsHyperref{ref33}{subfield} subcode of a residue AG code. Cartier codes share similar asymptotic properties to \flmRefsHyperref{ref33}{subfield} subcodes of residue AG codes, with both families admitting sequences of codes that achieve the \flmRefsHyperref{ref85}{GV bound}.
\end{eczvaluelist}
\codefieldsection{Child}
\begin{eczvaluelist}
\item\relax
\flmRefsHyperref[eczindexfamilyrel]{code:goppa}{Goppa code} --- Goppa codes are Cartier codes from a curve of genus zero \NoCaseChange{\protect\cite{cite1743}}.
\end{eczvaluelist}
\codefieldsection{Cousin}
\begin{eczvaluelist}
\item\relax
\flmRefsHyperref[eczindexfamilyrel]{code:residue}{Residue AG code} --- Every Cartier code is contained in a \flmRefsHyperref{ref33}{subfield} subcode of a residue AG code. Cartier codes share similar asymptotic properties to \flmRefsHyperref{ref33}{subfield} subcodes of residue AG codes, with both families admitting sequences of codes that achieve the \flmRefsHyperref{ref85}{GV bound}.
\end{eczvaluelist}
\eczhbkcontributors{ \eczhuVVA }
\endeczcode

\eczcode{gbch}{Chien-Choy generalized BCH (GBCH) code}{~\NoCaseChange{\protect\cite{cite1744}}}
\codefieldsection{Description}
An \([n,k\geq n-rm, d\geq r+1]_q\) alternant code defined using two polynomials \(P(x),G(x)\) that are relatively prime to \(x^n-1\), with \(\deg P \leq n-1\) and \(r = \deg G \leq n-1\).

See \NoCaseChange{\protect\cite[{Ch. 12}]{cite41}} for the parity-check matrix.

\codefieldsection{Parent}
\begin{eczvaluelist}
\item\relax
\flmRefsHyperref[eczindexfamilyrel]{code:alternant}{Alternant code} --- GBCH codes are a special case of alternant codes \NoCaseChange{\protect\cite[{Ch. 12}]{cite41}}.
\end{eczvaluelist}
\codefieldsection{Children}
\begin{eczvaluelist}
\item\relax
\flmRefsHyperref[eczindexfamilyrel]{code:q-ary_bch}{Bose–Chaudhuri–Hocquenghem (BCH) code} --- \(q\)-ary BCH codes are a special case of GBCH codes \NoCaseChange{\protect\cite[{Ch. 12}]{cite41}}.
\item\relax
\flmRefsHyperref[eczindexfamilyrel]{code:srivastava}{Srivastava code} --- Generalized Srivastava codes are a special case of GBCH codes \NoCaseChange{\protect\cite[{Ch. 12}]{cite41}}.
\end{eczvaluelist}
\codefieldsection{Cousin}
\begin{eczvaluelist}
\item\relax
\flmRefsHyperref[eczindexfamilyrel]{code:goppa}{Goppa code} --- In the binary case, GBCH\((z^{n-1},G)\) is the Goppa code \(\Gamma(L,G)\) where \(L\) consists of the \(n\)th roots of unity \NoCaseChange{\protect\cite[{pg. 360}]{cite41}}.
\end{eczvaluelist}
\eczhbkcontributors{ \eczhuVVA }
\endeczcode

\eczcode{classical_fractal_liquid}{Classical fractal liquid code}{~\NoCaseChange{\protect\cite{cite1745,cite1348}}}
\codefieldsection{Description}
Member of a family of \([L^D,O(L^{D-1}),O(L^{D-\epsilon})]_p\) linear codes on \(D\)-dimensional square lattices of side length \(L\) and for prime \(p\) and \(\epsilon > 0\) that is based on \(p\)-ary generalizations of the Sierpinski triangle.

\codefieldsection{Protection}
Parameters of some code families saturate the classical version of the \flmRefsHyperref{ref487}{BPT bound} \NoCaseChange{\protect\cite{cite1745}}.
\codefieldsection{Encoding}
\begin{eczvaluelist}
\item\relax Ground-state configurations are generated by linear cellular automata, exhibit discrete scale symmetries, and can flow in limit cycles under real-space renormalization-group transformations \NoCaseChange{\protect\cite{cite1348}}.
\end{eczvaluelist}
\codefieldsection{Parents}
\begin{eczvaluelist}
\item\relax
\flmRefsHyperref[eczindexfamilyrel]{code:quantum_inspired}{Quantum-inspired classical block code}\item\relax
\flmRefsHyperref[eczindexfamilyrel]{code:q-ary_linear_over_zq}{Linear code over \(\mathbb{Z}_q\)}\end{eczvaluelist}
\codefieldsection{Child}
\begin{eczvaluelist}
\item\relax
\flmRefsHyperref[eczindexfamilyrel]{code:newman_moore}{Newman-Moore code}\end{eczvaluelist}
\codefieldsection{Cousin}
\begin{eczvaluelist}
\item\relax
\flmRefsHyperref[eczindexfamilyrel]{code:commuting_projector}{Commuting-projector Hamiltonian code} --- Classical fractal liquid codewords form the ground-state space of a class of exactly solvable spin-glass Ising models with three-body interactions.
\end{eczvaluelist}
\eczhbkcontributors{ \eczhuVVA }
\endeczcode

\eczcode{complete_intersections}{Complete-intersection RM-type code}{~\NoCaseChange{\protect\cite{cite1746}}}
\codefieldsection{Description}
Evaluation code of polynomials evaluated on points lying on a complete intersection.

\codefieldsection{Protection}
Distance bounds formulated in Ref. \NoCaseChange{\protect\cite{cite1747}}.
\codefieldsection{Parent}
\begin{eczvaluelist}
\item\relax
\flmRefsHyperref[eczindexfamilyrel]{code:evaluation_polynomial}{Polynomial evaluation code} --- Complete-intersection RM-type codes are polynomial evaluation codes with \(\cal X\) being a complete intersection.
\end{eczvaluelist}
\eczhbkcontributors{ \eczhuVVA }
\endeczcode

\eczcode{completely_regular}{Completely regular code}{~\NoCaseChange{\protect\cite{cite880}}}
\codefieldsection{Description}
A code \(C\) is completely regular if the \flmRefsHyperref{ref113}{weight distribution} of any coset \(e+C\) depends only on the distance \(d(e,C)\) of \(e\) to \(C\) \NoCaseChange{\protect\cite{cite1748}}. 

\codefieldsection{Notes}
\begin{eczvaluelist}
\item\relax See review \NoCaseChange{\protect\cite{cite1748}} on completely regular codes.
\end{eczvaluelist}
\codefieldsection{Parent}
\begin{eczvaluelist}
\item\relax
\flmRefsHyperref[eczindexfamilyrel]{code:q-ary_digits_into_q-ary_digits}{\(q\)-ary code}\end{eczvaluelist}
\codefieldsection{Children}
\begin{eczvaluelist}
\item\relax
\flmRefsHyperref[eczindexfamilyrel]{code:perfect}{Perfect code} --- Perfect codes and extended perfect codes are completely regular \NoCaseChange{\protect\cite{cite1748}}.
\item\relax
\flmRefsHyperref[eczindexfamilyrel]{code:uniformly_packed}{Uniformly packed code} --- Uniformly packed codes are completely regular \NoCaseChange{\protect\cite{cite1749}\protect\cite[{Prop. 2.6}]{cite1748}}.
\end{eczvaluelist}
\eczhbkcontributors{ \eczhuVVA }
\endeczcode

\eczcode{convolutional}{Convolutional code}{~\NoCaseChange{\protect\cite{cite1750}}}
\codefieldsection{Alternative Names}
\begin{eczvaluelist}
\item\relax Tree code
\end{eczvaluelist}
\eczhIndexCodeAliasName{convolutional}{Tree code}

\codefieldsection{Kingdom root code}
for the \flmRefsHyperref{kingdom:q-ary_digits_into_q-ary_digits}{Galois-field Kingdom}.
\codefieldsection{Description}
Infinite-block code that is formed using generator polynomials over a finite field. The encoder slides across contiguous subsets of the input string (like a convolutional neural network), evaluating the polynomials on that window to obtain parity bits. These parity bits are the encoded information.
\codefieldsection{Protection}
There are linear programming bounds for convolutional codes \NoCaseChange{\protect\cite{cite1751}}.

\codefieldsection{Rate}
Depends on the polynomials used. Using puncturing \NoCaseChange{\protect\cite{cite1752}}, the rate for the code can be increased from \(\frac{1}{t}\) to \(\frac{s}{t}\), where \(t\) is the number of output bits, and \(s\) depends on the puncturing done. This is done by deleting some pieces of the encoder output such that maximum-likelihood decoders remain effective.
\codefieldsection{Encoding}
\begin{eczvaluelist}
\item\relax Evaluation on the generator polynomials. Can be implemented with a small number of XOR gates
\end{eczvaluelist}
\codefieldsection{Decoding}
\begin{eczvaluelist}
\item\relax Decoders based on the Viterbi algorithm (trellis decoding) were developed first, which result in the most-likely codeword for the encoded bits \NoCaseChange{\protect\cite{cite1753}}.
\item\relax BCJR decoder, also a trellis-based decoder \NoCaseChange{\protect\cite{cite1754}}.
\end{eczvaluelist}
\codefieldsection{Realizations}
\begin{eczvaluelist}
\item\relax A type of convolutional code used in Real-time Application networks \NoCaseChange{\protect\cite{cite241}}.
\item\relax Mobile and radio communications (3G networks) use convolutional codes concatenated with RS codes to obtain suitable performance \NoCaseChange{\protect\cite{cite242}}.
\item\relax A convolutional code with rate 1/2 was used for deep-space and satellite communication \NoCaseChange{\protect\cite{cite243}}
\end{eczvaluelist}
\codefieldsection{Notes}
\begin{eczvaluelist}
\item\relax There are connections between convolutional codes and statistical mechanical models \NoCaseChange{\protect\cite{cite1755,cite1467,cite1468}}.
\item\relax See books \NoCaseChange{\protect\cite{cite1756,cite993}} for introductions to convolutional codes.
\end{eczvaluelist}
\codefieldsection{Parent}
\begin{eczvaluelist}
\item\relax
\flmRefsHyperref[eczindexfamilyrel]{code:ecc}{Error-correcting code (ECC)}\end{eczvaluelist}
\codefieldsection{Children}
\begin{eczvaluelist}
\item\relax
\flmRefsHyperref[eczindexfamilyrel]{code:irregular_convolutional}{Irregular convolutional code (IRCC)}\item\relax
\flmRefsHyperref[eczindexfamilyrel]{code:ld_convolutional}{LDPC convolutional code (LDPC-CC)}\item\relax
\flmRefsHyperref[eczindexfamilyrel]{code:turbo}{Turbo code}\end{eczvaluelist}
\codefieldsection{Cousins}
\begin{eczvaluelist}
\item\relax
\flmRefsHyperref[eczindexfamilyrel]{code:q-ary_digits_into_q-ary_digits}{\(q\)-ary code} --- Convolutional codes for finite block size reduce to \(q\)-ary codes.
\item\relax
\flmRefsHyperref[eczindexfamilyrel]{code:quantum_convolutional}{Quantum convolutional code} --- Quantum convolutional codes are quantum analogues of convolutional codes.
\item\relax
\flmRefsHyperref[eczindexfamilyrel]{code:reed_solomon}{Reed-Solomon (RS) code} --- Convolutional codes can be constructed from \NoCaseChange{\protect\cite{cite1757}} and concatenated with \NoCaseChange{\protect\cite{cite242}} RS codes.
\item\relax
\flmRefsHyperref[eczindexfamilyrel]{code:binary_permutation}{Code in permutations} --- Convolutional codes in permutations have been constructed \NoCaseChange{\protect\cite{cite1758}}.
\item\relax
\flmRefsHyperref[eczindexfamilyrel]{code:quasi_cyclic}{Quasi-cyclic code} --- Quasi-cyclic codes can be \textit{unwrapped} to obtain convolutional codes \NoCaseChange{\protect\cite{cite1116,cite1117,cite1118,cite1119,cite1120,cite1121,cite1122}}.
\item\relax
\flmRefsHyperref[eczindexfamilyrel]{code:ea_quantum_convolutional}{EA quantum convolutional code} --- EA quantum convolutional codes are entanglement-assisted quantum analogues of convolutional codes.
\item\relax
\flmRefsHyperref[eczindexfamilyrel]{code:hybrid_convolutional}{Hybrid convolutional code} --- Hybrid convolutional codes are hybrid c-q analogues of convolutional codes.
\end{eczvaluelist}
\eczhbkcontributors{ Benjamin Quiring, \eczhuVVA }
\endeczcode

\eczcode{covering}{Covering code}{}
\codefieldsection{Description}
A \(q\)-ary code \(C\) is \(\rho\)-covering if \(\forall v \in \mathbb{F}_q^{n}\), there is a codeword \(c \in C\) such that the Hamming distance \(D(c,v)\leq \rho\). More generally, a covering code in a metric space is covering if the union of balls of some radius centered at the codewords covers the entire space.

The \textit{covering radius} \(\rho(C)\) is the smallest non-negative integer \(\rho\) such that \(C\) is \(\rho\)-covering, i.e.
\flmMathEnvironment{align}{}{
  \rho(C)=\max_{{v\in \mathbb{F}_q^{n}}}\min_{{c\in C}}d(v,c)~.
}
For a linear code \([n,k]_q\), the covering radius is the minimum number of columns of the code's parity check matrix which cover \(\mathbb{F}_q^{n-k}\).

The covering radius satisfies various inequalities. A code \(C\) with distance \(d\) satisfies the relation
\flmMathEnvironment{align}{}{
  \rho(C)\geq \frac{|d-1|}{2}~.\label{ref1759}
}
Linear \([n,k]_q\) codes also satisfy the \textit{redundancy bound}
\flmMathEnvironment{align}{}{
  \rho(C)\leq n-k
}
and the \textit{sphere covering bound}
\flmMathEnvironment{align}{}{
  \rho(C)\leq \min{\left(p~\bigg\rvert \sum_{i=0}^{p} {n \choose i}(q-1)^{i}|C| \geq q^{n}\right)}~.\label{ref1760}
}
A code is perfect iff it satisfies Eqs. \eqref{ref1759} and \eqref{ref1760} with equality.

In general, finding the covering radius of a code is \(NP\)-hard \NoCaseChange{\protect\cite{cite1761}}. Complexity analysis as well as an extensive study of bounds can be found in Ref. \NoCaseChange{\protect\cite{cite244}}.

\codefieldsection{Realizations}
\begin{eczvaluelist}
\item\relax Data compression, both with and without distortion constraints, can be phrased in terms of covering codes \NoCaseChange{\protect\cite{cite244}}.
\item\relax Football-pool problem: finding the smallest number of bets on a set of matches needed to guarantee at least one bet has at most \(\rho\) errors \NoCaseChange{\protect\cite{cite245,cite246}}.
\end{eczvaluelist}
\codefieldsection{Notes}
\begin{eczvaluelist}
\item\relax See book \NoCaseChange{\protect\cite{cite244}} for an exposition on covering codes.
\end{eczvaluelist}
\codefieldsection{Parent}
\begin{eczvaluelist}
\item\relax
\flmRefsHyperref[eczindexfamilyrel]{code:weighed_covering}{Weighted-covering code} --- An \(m\)-weighted covering code for \(m_j=1\) is a covering code of covering radius at most \(r\) \NoCaseChange{\protect\cite[{Ch. 13}]{cite244}}.
\end{eczvaluelist}
\codefieldsection{Child}
\begin{eczvaluelist}
\item\relax
\flmRefsHyperref[eczindexfamilyrel]{code:perfect}{Perfect code} --- Perfect codes are covering codes with the minimum number of codewords.
\end{eczvaluelist}
\codefieldsection{Cousin}
\begin{eczvaluelist}
\item\relax
\flmRefsHyperref[eczindexfamilyrel]{code:binary-ternary}{Binary-ternary mixed code} --- See Ref. \NoCaseChange{\protect\cite{cite1762}} for bounds on binary-ternary mixed covering codes.
\end{eczvaluelist}
\eczhbkcontributors{ Mustafa Doger, \eczhuVVA }
\endeczcode

\eczcode{cross_interleaved_reed_solomon}{Cross-interleaved RS (CIRS) code}{~\NoCaseChange{\protect\cite{cite1763,cite1764}}}
\codefieldsection{Description}
An IRS code constructed from two shortened RS codes and two forms of interleaving. In a 2D array visualization, one component code runs vertically and the other diagonally \NoCaseChange{\protect\cite{cite189}}.

\codefieldsection{Protection}
Can correct burst errors by first decoding one RS component and then treating residual failures as erasures for the other component \NoCaseChange{\protect\cite{cite189}}.

\codefieldsection{Realizations}
\begin{eczvaluelist}
\item\relax Compact discs (CDs); see \NoCaseChange{\protect\cite{cite189}\protect\cite[{Sec. 5.6}]{cite126}\protect\cite[{Ch. 4}]{cite247}}.
\end{eczvaluelist}
\codefieldsection{Parent}
\begin{eczvaluelist}
\item\relax
\flmRefsHyperref[eczindexfamilyrel]{code:interleaved_reed_solomon}{Interleaved RS (IRS) code}\end{eczvaluelist}
\codefieldsection{Cousin}
\begin{eczvaluelist}
\item\relax
\flmRefsHyperref[eczindexfamilyrel]{code:array}{Array code} --- The CIRS code can also be visualized as a 2D array code \NoCaseChange{\protect\cite{cite189}}.
\end{eczvaluelist}
\eczhbkcontributors{ \eczhuVVA }
\endeczcode

\eczcode{q-ary_cyclic}{Cyclic linear \(q\)-ary code}{}
\codefieldsection{Description}
A \(q\)-ary code of length \(n\) is cyclic if, for each codeword \(c_1 c_2 \cdots c_n\), the cyclically shifted string \(c_n c_1 \cdots c_{n-1}\) is also a codeword. A cyclic code is called \textit{primitive} when \(n=q^r-1\) for some \(r\geq 2\). A \textit{shortened cyclic code} is obtained from a cyclic code by taking only codewords with the first \(j\) zero entries, and deleting those zeroes.

\subsection{Cyclic-to-polynomial correspondence}

\begin{defterm}{Cyclic-to-polynomial correspondence}\label{ref1765}\label{ref67}
Cyclic linear \(q\)-ary codes and their properties can be naturally formulated using the theory of polynomials.
Codewords \(c_1 c_2 \cdots c_n\) of a cyclic \(q\)-ary code can be thought of as coefficients in a polynomial \(c_1+c_2 x+\cdots+c_n x^{n-1}\) in the set of polynomials with \(q\)-ary coefficients \(\mathbb{F}_q[x]\).
Polynomials corresponding to codewords of a linear cyclic code form an ideal (i.e., are closed under multiplication and addition) in the ring \(\mathbb{F}_q[x]/(x^n-1)\) (i.e., the set of equivalence classes of polynomials congruent modulo \(x^n-1\)).
Multiplication of a codeword polynomial \(c(x)\) by \(x\) in such a ring corresponds to a cyclic shift of the corresponding codeword string.
\end{defterm}

Codeword polynomials of a cyclic code can be generated, via multiplication, by a \textit{generator polynomial} \(g(x)\).
A particular generator polynomial \(e(x)\) has the additional property of being \textit{idempotent}, i.e., \(e(x)^2=e(x)\). Given a generator polynomial, the corresponding \textit{check polynomial} \(h(x)=(x^n-1)/g(x)\) yields zero when multiplying a codeword polynomial. Its coefficients correspond to the code's parity check matrix.

Since the generator polynomial \(g(x)\) is a polynomial over \(\mathbb{F}_q\), it can be factorized over some potentially larger \textit{splitting field} (just like \(x^2+1\) can be factorized over the complex numbers but not the reals).
Whenever \(q\) and \(n\) are relatively prime, cyclic codes can also be defined in terms of roots of \(g(x)\).
Such roots are called \textit{zeroes} of the code, and they are all powers of a primitive \(n\)th root of unity because \(g(x)\) is a divisor of \(x^n-1\).
Since the generator polynomial generates all codeword polynomials \(c(x)\) by multiplication by \(x\), its zeroes are also zeroes of those polynomials.
In the coprime case, cyclic codes also admit unique generating idempotents, parity-check polynomials, and trace representations \NoCaseChange{\protect\cite[{Secs. 2.3 and 2.4}]{cite68}}.
In the same setting, an \textit{irreducible cyclic code} \(C(q,n,i)\) is the length-\(n\) cyclic code over \(\mathbb{F}_q\) whose parity-check polynomial is the minimal polynomial \(M_{\alpha^i}(x)\) of \(\alpha^i\), where \(\alpha\) is a primitive \(n\)th root of unity. Such codes are also called \textit{minimal cyclic codes} because they correspond to minimal ideals of \(\mathbb{F}_q[x]/(x^n-1)\) \NoCaseChange{\protect\cite[{Sec. 2.5}]{cite68}}.

\codefieldsection{Protection}
The BCH and Hartmann-Tzeng bounds give lower bounds on the distance of cyclic \(q\)-ary codes \NoCaseChange{\protect\cite[{Sec. 2.4}]{cite68}}; the shift bound \NoCaseChange{\protect\cite{cite1308}} gives another lower bound.
\codefieldsection{Decoding}
\begin{eczvaluelist}
\item\relax Meggitt decoder \NoCaseChange{\protect\cite{cite357}}.
\item\relax Information set decoding (ISD) \NoCaseChange{\protect\cite{cite1309}}, a probabilistic decoding strategy that essentially tries to guess an information set of \(k\) correct positions in the received word, where \(k\) is the code dimension. Then, an error vector is constructed to map the received word onto the nearest codeword, assuming those \(k\) positions are error-free. When the Hamming weight of the error vector is low enough, that codeword is assumed to be the intended transmission.
\item\relax Permutation decoding \NoCaseChange{\protect\cite{cite1310}}.
\end{eczvaluelist}
\codefieldsection{Notes}
\begin{eczvaluelist}
\item\relax See \NoCaseChange{\protect\cite[{Ch. 7}]{cite41}\protect\cite[{Sec. 1.12}]{cite1159}\protect\cite[{Secs. 2.1-2.4}]{cite68}} for expositions.
\item\relax Any nontrivial \(q\)-ary linear code invariant under the general affine group \(GA(m,\mathbb{F}_{q^r})\) for some \(r > 1\) is equivalent to an extended cyclic code \NoCaseChange{\protect\cite{cite1766}}.
\end{eczvaluelist}
\codefieldsection{Parents}
\begin{eczvaluelist}
\item\relax
\flmRefsHyperref[eczindexfamilyrel]{code:group}{Group-algebra code} --- A length-\(n\) cyclic \(q\)-ary linear code is an Abelian group-algebra code for the cyclic group with \(n\) elements \( \mathbb{Z}_n \) \NoCaseChange{\protect\cite[{Exam. 16.4.9}]{cite196}}.
\item\relax
\flmRefsHyperref[eczindexfamilyrel]{code:cyclic}{Cyclic code}\end{eczvaluelist}
\codefieldsection{Children}
\begin{eczvaluelist}
\item\relax
\flmRefsHyperref[eczindexfamilyrel]{code:binary_cyclic}{Cyclic linear binary code}\item\relax
\flmRefsHyperref[eczindexfamilyrel]{code:q-ary_bch}{Bose–Chaudhuri–Hocquenghem (BCH) code}\item\relax
\flmRefsHyperref[eczindexfamilyrel]{code:q-ary_parity_check}{\([n,n-1,2]_q\) \(q\)-ary parity-check code} --- Since permutations preserve coordinate sums, the cyclic permutation of an SPC codeword is another codeword. The generator polynomial of the code is \(x-1\).
\item\relax
\flmRefsHyperref[eczindexfamilyrel]{code:q-ary_repetition}{\(q\)-ary repetition code} --- The \(q\)-ary repetition code is cyclic with generator polynomial \(1+x+\cdots+x^{n-1}\).
\item\relax
\flmRefsHyperref[eczindexfamilyrel]{code:q-ary_duadic}{\(q\)-ary duadic code}\end{eczvaluelist}
\codefieldsection{Cousins}
\begin{eczvaluelist}
\item\relax
\flmRefsHyperref[eczindexfamilyrel]{code:q-ary_ltc}{\(q\)-ary linear LTC} --- Cyclic linear codes cannot be \(c^3\)-LTCs \NoCaseChange{\protect\cite{cite1272}}. Codeword symmetries are in general an obstruction to achieving such LTCs \NoCaseChange{\protect\cite{cite1273}}.
\item\relax
\flmRefsHyperref[eczindexfamilyrel]{code:self_dual}{Self-dual linear code} --- See Refs. \NoCaseChange{\protect\cite{cite1767,cite1768}} for tables of cyclic self-dual codes.
\item\relax
\flmRefsHyperref[eczindexfamilyrel]{code:combinatorial_design}{Combinatorial design} --- Two families of cyclic \(q\)-ary codes support an infinite family of combinatorial 3-designs \NoCaseChange{\protect\cite{cite134}}.
The supports of all fixed-weight codewords of a \(q\)-ary cyclic code support a combinatorial \(1\)-design \NoCaseChange{\protect\cite[{Corr. 5.2.4}]{cite135}}.

\item\relax
\flmRefsHyperref[eczindexfamilyrel]{code:reversible}{Reversible code} --- A reversible cyclic code is a cyclic code with self-reciprocal generator polynomial and is an LCD code \NoCaseChange{\protect\cite[{Thm. 2.10.3}]{cite68}}.
\item\relax
\flmRefsHyperref[eczindexfamilyrel]{code:reed_solomon}{Reed-Solomon (RS) code} --- If the length divides \(q-1\), then it is possible to construct a cyclic RS code. Such codes are Abelian group-algebra codes \NoCaseChange{\protect\cite[{Exam. 16.4.9}]{cite196}}.
\item\relax
\flmRefsHyperref[eczindexfamilyrel]{code:generalized_reed_muller}{Generalized RM (GRM) code} --- Punctured GRM codes, i.e., GRM codes with nonzero evaluation points, are cyclic, and their extensions recover GRM codes \NoCaseChange{\protect\cite[{Sec. 2.8}]{cite68}\protect\cite[{pg. 52}]{cite1314}}.
\item\relax
\flmRefsHyperref[eczindexfamilyrel]{code:lcd}{Linear code with complementary dual (LCD)} --- A cyclic \([n,k]\) code with generator polynomial \(g(x)\) is LCD if and only if \(g(x)\) is self-reciprocal and \(\gcd( g(x), (x^{n}-1)/g(x) )=1\) \NoCaseChange{\protect\cite[{Cor. 16.7.11}]{cite196}}.
\item\relax
\flmRefsHyperref[eczindexfamilyrel]{code:projective}{Projective geometry code} --- Every projective linear code is a punctured code of a special cyclic code \NoCaseChange{\protect\cite[{Thm. 2.3.6}]{cite68}}.
\item\relax
\flmRefsHyperref[eczindexfamilyrel]{code:quantum_mds}{Quantum maximum-distance-separable (MDS) code} --- Quantum MDS codes can be constructed from \(q\)-ary cyclic codes using the Hermitian construction \NoCaseChange{\protect\cite{cite979}}.
\item\relax
\flmRefsHyperref[eczindexfamilyrel]{code:quantum_tensor_product}{Quantum tensor-product code} --- Reversible cyclic codes can be used to construct quantum tensor-product codes \NoCaseChange{\protect\cite{cite1131}}.
\item\relax
\flmRefsHyperref[eczindexfamilyrel]{code:galois_css}{Galois-qudit CSS code} --- Galois CSS codes can be constructed using self-orthogonal \(q\)-ary cyclic codes \NoCaseChange{\protect\cite{cite1769}}.
\end{eczvaluelist}
\eczhbkcontributors{ Mazin Karjikar, \eczhuVVA }
\endeczcode

\eczcode{deligne_lusztig}{Deligne-Lusztig code}{~\NoCaseChange{\protect\cite{cite34,cite1770,cite1771,cite1772}}}
\codefieldsection{Description}
Evaluation code of polynomials evaluated on points lying on a Deligne-Lusztig variety, often a Deligne-Lusztig curve in the classical one-dimensional cases.

\codefieldsection{Parent}
\begin{eczvaluelist}
\item\relax
\flmRefsHyperref[eczindexfamilyrel]{code:evaluation_polynomial}{Polynomial evaluation code} --- Deligne-Lusztig codes are evaluation AG codes with \(\cal X\) a Deligne-Lusztig variety.
\end{eczvaluelist}
\eczhbkcontributors{ \eczhuVVA }
\endeczcode

\eczcode{denniston}{Denniston code}{~\NoCaseChange{\protect\cite{cite1773}}}
\codefieldsection{Description}
Projective code that is part of a family of \([2^{a+i}+2^i-2^a,3,2^{a+i}-2^a]_{2^a}\) codes for \(i < a\) constructed using Denniston arcs \NoCaseChange{\protect\cite[{Sec. 19.7.3}]{cite172}}.

\codefieldsection{Notes}
\begin{eczvaluelist}
\item\relax See corresponding MinT database entry \NoCaseChange{\protect\cite{cite1774}}.
\end{eczvaluelist}
\codefieldsection{Parents}
\begin{eczvaluelist}
\item\relax
\flmRefsHyperref[eczindexfamilyrel]{code:projective_two_weight}{Projective two-weight code} --- Denniston codes are projective two-weight codes on maximal arcs \NoCaseChange{\protect\cite{cite1774}\protect\cite[{Sec. 19.7.3}]{cite172}}.
\item\relax
\flmRefsHyperref[eczindexfamilyrel]{code:griesmer}{Griesmer code}\item\relax
\flmRefsHyperref[eczindexfamilyrel]{code:delsarte_optimal_q-ary}{\(q\)-ary sharp configuration} --- The Denniston code is a \(q\)-ary sharp configuration \NoCaseChange{\protect\cite[{Table 12.1}]{cite199}}.
\end{eczvaluelist}
\codefieldsection{Child}
\begin{eczvaluelist}
\item\relax
\flmRefsHyperref[eczindexfamilyrel]{code:hexacode}{\([6,3,4]_4\) Hexacode} --- A version of the hexacode is recovered for Denniston code parameters \(i=1\) and \(a=2\) \NoCaseChange{\protect\cite{cite62}}.
\end{eczvaluelist}
\codefieldsection{Cousin}
\begin{eczvaluelist}
\item\relax
\flmRefsHyperref[eczindexfamilyrel]{code:hyperoval}{Hyperoval code} --- Denniston codes for \(i=1\) realize the hyperoval case of maximal arcs in \(PG(2,2^a)\) \NoCaseChange{\protect\cite{cite1774}\protect\cite[{Sec. 19.7.3}]{cite172}}.
\end{eczvaluelist}
\eczhbkcontributors{ \eczhuVVA }
\endeczcode

\eczcode{divisible}{Divisible code}{~\NoCaseChange{\protect\cite{cite1775}}}
\codefieldsection{Description}
A linear \(q\)-ary block code is \(\Delta\)-divisible if the Hamming weight of each of its codewords is divisible by divisor \(\Delta\).
A \(2\)-divisible (\(4\)-divisible, \(8\)-divisible) code is called \textit{even} (\textit{doubly even}, \textit{triply even}) \NoCaseChange{\protect\cite{cite170,cite39}}.
A code is called \textit{singly even} if all codewords are even and at least one has weight equal to 2 modulo 4.
More generally, a code is \(m\)\textit{-even} if it is \(2^{m}\)-divisible.

\codefieldsection{Notes}
\begin{eczvaluelist}
\item\relax See Ref. \NoCaseChange{\protect\cite{cite659}} for an introduction to triply even binary linear codes and their construction from doubly even codes.
\item\relax Doubly even self-dual codes have been classified up to \(n\leq 40\) \NoCaseChange{\protect\cite{cite1776}}.
\item\relax There are ten maximal triply even codes of length 48 up to equivalence \NoCaseChange{\protect\cite{cite659}}.
\end{eczvaluelist}
\codefieldsection{Parent}
\begin{eczvaluelist}
\item\relax
\flmRefsHyperref[eczindexfamilyrel]{code:q-ary_linear}{Linear \(q\)-ary code}\end{eczvaluelist}
\codefieldsection{Children}
\begin{eczvaluelist}
\item\relax
\flmRefsHyperref[eczindexfamilyrel]{code:self_dual_48_24_12}{\([48,24,12]\) self-dual code} --- The \([48,24,12]\) code is doubly even and hence Type II \NoCaseChange{\protect\cite[{Rem. 4.1.10}]{cite40}}; it is the unique self-dual doubly even code with those parameters \NoCaseChange{\protect\cite{cite111}}.
\item\relax
\flmRefsHyperref[eczindexfamilyrel]{code:parity_check}{\([n,n-1,2]\) Single parity-check (SPC) code} --- Binary SPCs are two-divisible.
\item\relax
\flmRefsHyperref[eczindexfamilyrel]{code:reed_muller}{Reed-Muller (RM) code} --- An RM\((r,m)\) code is \(2^{\left\lceil m/r\right\rceil-1}\)-divisible, according to McEliece's theorem \NoCaseChange{\protect\cite{cite1574,cite1575}}.
\end{eczvaluelist}
\codefieldsection{Cousins}
\begin{eczvaluelist}
\item\relax
\flmRefsHyperref[eczindexfamilyrel]{code:group}{Group-algebra code} --- If \(C\) is a group code over a field of characteristic \(p\), then the monomial kernel \(K_M(C)\) has order dividing the weight of every codeword, and the \(p^{\prime}\)-part of the divisor of \(C\) equals the \(p^{\prime}\)-part of \(|K_M(C)|\) \NoCaseChange{\protect\cite[{Thm. 16.8.3}]{cite196}}.
\item\relax
\flmRefsHyperref[eczindexfamilyrel]{code:constant_weight}{Constant-weight code} --- Codes whose codewords have a constant weight of \(m\) are automatically \(m\)-divisible. However, divisible codes are linear by definition while constant-weight codes do not have to be.
\item\relax
\flmRefsHyperref[eczindexfamilyrel]{code:binary_quad_residue}{Binary quadratic-residue (QR) code} --- Extended binary quadratic residue codes of length \(8m\) are self-dual doubly even codes \NoCaseChange{\protect\cite[{pg. 82}]{cite39}}.
\item\relax
\flmRefsHyperref[eczindexfamilyrel]{code:hamming844}{\([8,4,4]\) extended Hamming code} --- The \([8,4,4]\) extended Hamming code is the smallest doubly even self-dual code, and the unique Type II code of length \(8\) \NoCaseChange{\protect\cite[{Rem. 4.3.10}]{cite40}}.
\item\relax
\flmRefsHyperref[eczindexfamilyrel]{code:self_dual}{Self-dual linear code} --- Binary self-orthogonal codes are even, doubly even binary codes are self-orthogonal, and binary self-dual codes split into singly-even Type I and doubly-even Type II families \NoCaseChange{\protect\cite[{Def. 4.1.6}]{cite40}\protect\cite[{Rems. 4.1.7 and 4.1.10}]{cite40}}. Ternary self-dual codes are 3-divisible and Hermitian self-dual quaternary codes are 2-divisible \NoCaseChange{\protect\cite[{Thm. 4.1.9}]{cite40}}.
\item\relax
\flmRefsHyperref[eczindexfamilyrel]{code:ternary_golay}{\([11,6,5]_3\) Ternary Golay code} --- The extended ternary Golay code is 3-divisible because ternary self-dual codes are Type III \NoCaseChange{\protect\cite[{Thm. 4.1.9}]{cite40}}.
\item\relax
\flmRefsHyperref[eczindexfamilyrel]{code:griesmer}{Griesmer code} --- If a \(p\)-ary Griesmer code with \(p\) prime is such that a power of \(p\) divides the distance, then the code is divisible by that power \NoCaseChange{\protect\cite{cite1777}}.
\item\relax
\flmRefsHyperref[eczindexfamilyrel]{code:two_weight}{Two-weight code} --- Two-weight codes are \(m\)-divisible, where \(m\) is the greatest common divisor of their two possible weights.
\item\relax
\flmRefsHyperref[eczindexfamilyrel]{code:stabilizer_over_gf4}{Hermitian qubit code} --- The commutation requirement for a Hermitian stabilizer code implies that its underlying Hermitian self-orthogonal linear code over \(\mathbb{F}_4\) is even; the converse is also true \NoCaseChange{\protect\cite[{Thm. 1.4.10}]{cite126}}.
\item\relax
\flmRefsHyperref[eczindexfamilyrel]{code:quantum_divisible}{Quantum divisible code} --- The \(X\)-type stabilizers of a level-\(\nu\) quantum divisible code form a \(\nu\)-even linear binary code.
\item\relax
\flmRefsHyperref[eczindexfamilyrel]{code:stab_49_1_5}{\(\llbracket 49,1,5\rrbracket \) triorthogonal code} --- The \(\llbracket 49,1,5\rrbracket \) triorthogonal code stabilizer generator matrix can be obtained from a triply even linear binary code \NoCaseChange{\protect\cite[{Appx. B}]{cite691}}.
\end{eczvaluelist}
\eczhbkcontributors{ \eczhuVVA }
\endeczcode

\eczcode{dual_additive}{Dual additive code}{}
\codefieldsection{Description}
For any \(q\)-ary additive code \(C\), the dual code is the set of \(q\)-ary strings that are orthogonal to the codewords of \(C\) under a particular inner product.

The dual additive (or orthogonal additive) code is
\flmMathEnvironment{align}{}{
C^\perp = \{ y\in \mathbb{F}_q^{n} ~|~ x \star y=0 \forall x\in C\},
}
where the \textit{trace inner product} is \(x\star y = \sum_{i=1}^n \text{tr}(x_i y_i)\) for coordinates \(x_i,y_i\) and for \(\textit{tr}\) being the \flmRefsHyperref{ref33}{field trace}.

A code that is contained in its dual, \(C \subseteq C^\perp\), is called \textit{self-orthogonal additive} or \textit{weakly self-dual additive}. A code that contains its dual, \(C^\perp \subseteq C\), is called \textit{dual-containing additive}. A code that is equal to its dual, \(C^\perp = C\), is called \textit{self-dual additive}. A code is dual-containing additive iff its dual is self-orthogonal additive.

An alternative definition of dual substitutes the trace inner product for the \textit{trace-Hermitian inner product}, \(x\star y \to \sum_{i=1}^n \text{tr}(x_i y^{p}_i)\).
Another extension for when \(q=p^2\), relevant to \flmRefsHyperref{code:stabilizer_over_gfqsq}{certain stabilizer codes} and reducing to the trace-Hermitian case for \(q=4\), is the \textit{trace-alternating inner product},
\flmMathEnvironment{align}{}{
  x\star y \to \sum_{i=1}^{n}\text{tr}\left(\frac{x_{i}y_{i}^{\sqrt{q}}-x_{i}^{\sqrt{q}}y_{i}}{\alpha-\alpha^{q}}\right)~,
}
where \(\{1,\alpha\}\) is a basis of \(\mathbb{F}_q\) over \(\mathbb{F}_{\sqrt{q}}\).
Self-dual additive codes with respect to the trace-Hermitian (trace-alternating) inner product are called \textit{trace Hermitian (trace-alternating) self-dual additive}; similar definitions hold for self-orthogonal additive and dual-containing additive.

\codefieldsection{Parent}
\begin{eczvaluelist}
\item\relax
\flmRefsHyperref[eczindexfamilyrel]{code:q-ary_additive}{Additive \(q\)-ary code}\end{eczvaluelist}
\codefieldsection{Child}
\begin{eczvaluelist}
\item\relax
\flmRefsHyperref[eczindexfamilyrel]{code:self_dual_additive}{Self-dual additive code}\end{eczvaluelist}
\codefieldsection{Cousins}
\begin{eczvaluelist}
\item\relax
\flmRefsHyperref[eczindexfamilyrel]{code:dual}{Dual linear code} --- Different inner products are typically used to define duals of linear and additive codes.
\item\relax
\flmRefsHyperref[eczindexfamilyrel]{code:dual_over_rings}{Dual linear code over \(R\)} --- Dual additive codes are additive analogues of dual linear codes over rings.
\item\relax
\flmRefsHyperref[eczindexfamilyrel]{code:qubit_stabilizer}{Qubit stabilizer code} --- Qubit stabilizer codes are in one-to-one correspondence with trace-Hermitian self-orthogonal additive quaternary codes of length \(n\) via the \flmRefsHyperref{ref1778}{\(\mathbb{F}_4\) representation}.
\item\relax
\flmRefsHyperref[eczindexfamilyrel]{code:galois_stabilizer}{Galois-qudit stabilizer code} --- Galois-qudit stabilizer codes are in one-to-one correspondence with trace-symplectic self-orthogonal additive codes of length \(2n\) over \(\mathbb{F}_q\) via the \flmRefsHyperref{ref873}{Galois symplectic representation} \NoCaseChange{\protect\cite{cite696}}. They are also in one-to-one correspondence with trace-alternating self-orthogonal additive codes of length \(n\) over \(\mathbb{F}_{q^2}\) via the \flmRefsHyperref{ref1779}{\(\mathbb{F}_{q^2}\) representation}.
\end{eczvaluelist}
\eczhbkcontributors{ \eczhuVVA }
\endeczcode

\eczcode{dual}{Dual linear code}{}
\codefieldsection{Alternative Names}
\begin{eczvaluelist}
\item\relax Orthogonal linear code
\end{eczvaluelist}
\eczhIndexCodeAliasName{dual}{Orthogonal linear code}
\codefieldsection{Description}
For any \([n,k]_q\) linear code \(C\), the dual code is the set of \(q\)-ary strings that are orthogonal to the codewords of \(C\) under a particular inner product.

The dual code is a linear code defined by
\flmMathEnvironment{align}{}{
C^\perp = \{ y\in \mathbb{F}_q^{n} ~|~ x\cdot y=0 \forall x\in C\},
}
where the \textit{ordinary}, \textit{standard}, \textit{Euclidean}, or \textit{symmetric} inner product is \(x\cdot y = \sum_{i=1}^n x_i y_i\) for coordinates \(x_i,y_i\).

A code that is contained in its dual, \(C \subseteq C^\perp\), is called \textit{self-orthogonal}, \textit{weakly self-dual}, or \textit{null}.
A self-orthogonal code is called \textit{maximal} if it is not contained in the dual of any other code.
A code that contains its dual, \(C^\perp \subseteq C\), is called \textit{dual-containing}.
A code that is equal to its dual, \(C^\perp = C\), is called self-dual.
A code that is equivalent to its dual is called iso-dual.
The intersection of a code and its dual, \(C \cap C^{\perp}\), is called the \textit{hull} of the code.
A code admits a complementary dual if \(C\) and \(C^{\perp}\) do not share any codewords; such codes are called LCD codes.
The dual of a dual code is the original code.
A code is dual-containing iff its dual is self-orthogonal.

The dual code \(C^\perp\) is the row space of the parity check matrix of \(C\). The dual code is the kernel of a generator matrix for \(C\), and \(\dim C^\perp = n-k\).
The automorphism group of a linear binary code and its dual are the same \NoCaseChange{\protect\cite[{pg. 230}]{cite41}}.

An alternative definition of dual substitutes the Euclidean inner product for the \textit{Hermitian inner product},
\flmMathEnvironment{align}{}{
  x\cdot y \to x\cdot \bar{y} = \sum_{i=1}^n x_i y^{p}_i~.
}
Self-dual codes with respect to the above product are called \textit{Hermitian self-dual}; similar definitions hold for self-orthogonal and dual-containing.

More general inner products can also be considered \NoCaseChange{\protect\cite{cite1780}}.

\codefieldsection{Protection}
The dual of an \([n,k,d] \) code is an \([n,n-k,d^{\perp}]\) code, where the \flmRefsHyperref{ref113}{dual distance} \(d^{\perp}\) is not always related to \(d\).
The generator matrix of \(C^\perp\) is the parity check matrix of \(C\), and vice versa.

The generator matrix of the Hermitian dual of a code with generator matrix \(G = [I_k~~A]\) is \([-\bar{A}^T~~I_{n-k}]\), where \(\bar{A}\) contains matrix elements of \(A\) raised to the \(p\)th power.
A code is Hermitian self-dual if and only if \(A \bar{A}^{T} = -I_{n/2}\).

\codefieldsection{Parents}
\begin{eczvaluelist}
\item\relax
\flmRefsHyperref[eczindexfamilyrel]{code:q-ary_linear}{Linear \(q\)-ary code}\item\relax
\flmRefsHyperref[eczindexfamilyrel]{code:dual_over_rings}{Dual linear code over \(R\)}\end{eczvaluelist}
\codefieldsection{Children}
\begin{eczvaluelist}
\item\relax
\flmRefsHyperref[eczindexfamilyrel]{code:lcd}{Linear code with complementary dual (LCD)}\item\relax
\flmRefsHyperref[eczindexfamilyrel]{code:self_dual}{Self-dual linear code}\end{eczvaluelist}
\codefieldsection{Cousins}
\begin{eczvaluelist}
\item\relax
\flmRefsHyperref[eczindexfamilyrel]{code:dual_lattice}{Dual lattice} --- Dual lattices are lattice analogues of dual codes.
\item\relax
\flmRefsHyperref[eczindexfamilyrel]{code:combinatorial_design}{Combinatorial design} --- Linear codes and their duals are related to combinatorial designs via the Assmus-Mattson theorem \NoCaseChange{\protect\cite{cite136,cite137}} (see \NoCaseChange{\protect\cite[{Sec. 5.4}]{cite135}}).
\item\relax
\flmRefsHyperref[eczindexfamilyrel]{code:golay}{\([23, 12, 7]\) Golay code} --- The dual of the Golay code is its \([23,11,8]\) even-weight subcode \NoCaseChange{\protect\cite{cite103,cite104}}.
\item\relax
\flmRefsHyperref[eczindexfamilyrel]{code:parity_check}{\([n,n-1,2]\) Single parity-check (SPC) code} --- Binary SPCs and repetition codes are dual to each other.
\item\relax
\flmRefsHyperref[eczindexfamilyrel]{code:hergert}{Hergert code} --- Hergert codes are duals of DG codes in that their distance distribution is equal to the \flmRefsHyperref{ref113}{MacWilliams transform} of the distance distribution of DG codes \NoCaseChange{\protect\cite{cite1328}}. However, the two codes are images of a pair of mutually dual linear codes over \(\mathbb{Z}_4\) under the \flmTerm{term}{ref81}{}{Gray map} \NoCaseChange{\protect\cite{cite158,cite123}}.
\item\relax
\flmRefsHyperref[eczindexfamilyrel]{code:simplex}{\([2^m-1,m,2^{m-1}]\) simplex code} --- Hamming and simplex codes are dual to each other.
\item\relax
\flmRefsHyperref[eczindexfamilyrel]{code:extended_hamming}{\([2^m,2^m-m-1,4]\) Extended Hamming code} --- Extended Hamming and first-order RM codes are dual to each other.
\item\relax
\flmRefsHyperref[eczindexfamilyrel]{code:hamming743}{\([7,4,3]\) Hamming code} --- The \([7,3,4]\) simplex code is the dual of the Hamming code and also its even-weight subcode \NoCaseChange{\protect\cite{cite103,cite104}}.
\item\relax
\flmRefsHyperref[eczindexfamilyrel]{code:reed_muller}{Reed-Muller (RM) code} --- The codes RM\((r,m)\) and RM\((m-r-1,m)\) are dual to each other, with the case \(m = 2r+1\) being self dual.
\item\relax
\flmRefsHyperref[eczindexfamilyrel]{code:ldgm}{Low-density generator-matrix (LDGM) code} --- The dual of an LDPC code has a sparse generator matrix and is called an LDGM code.
\item\relax
\flmRefsHyperref[eczindexfamilyrel]{code:ldpc}{Low-density parity-check (LDPC) code} --- The dual of an LDPC code has a sparse generator matrix and is called an LDGM code.
\item\relax
\flmRefsHyperref[eczindexfamilyrel]{code:ha_ldpc}{Hsu-Anastasopoulos LDPC (HA-LDPC) code} --- HA-LDPC and MN-LDPC codes are dual to each other \NoCaseChange{\protect\cite{cite84}}.
\item\relax
\flmRefsHyperref[eczindexfamilyrel]{code:generalized_reed_solomon}{Generalized RS (GRS) code} --- The dual of a GRS code is another GRS code \NoCaseChange{\protect\cite[{pg. 304}]{cite41}}.
\item\relax
\flmRefsHyperref[eczindexfamilyrel]{code:reed_solomon}{Reed-Solomon (RS) code} --- The dual of an RS code is an RS code \NoCaseChange{\protect\cite[{pg. 296}]{cite41}}.
\item\relax
\flmRefsHyperref[eczindexfamilyrel]{code:generalized_reed_muller}{Generalized RM (GRM) code} --- The dual of a GRM code is also a GRM code \NoCaseChange{\protect\cite{cite1781,cite1782}}.
\item\relax
\flmRefsHyperref[eczindexfamilyrel]{code:mds}{Maximum distance separable (MDS) code} --- A linear binary or \(q\)-ary \([n,k,n-k+1]\) code is MDS if and only if its dual \([n,n-k,k+1]\) is MDS \NoCaseChange{\protect\cite[{Thm. 1.9.13}]{cite1159}}.
\item\relax
\flmRefsHyperref[eczindexfamilyrel]{code:dual_additive}{Dual additive code} --- Different inner products are typically used to define duals of linear and additive codes.
\item\relax
\flmRefsHyperref[eczindexfamilyrel]{code:q-ary_simplex}{\(q\)-ary simplex code} --- \(q\)-ary Hamming and \(q\)-ary simplex codes are dual to each other \NoCaseChange{\protect\cite[{pg. 45}]{cite1314}}
\item\relax
\flmRefsHyperref[eczindexfamilyrel]{code:q-ary_ldgm}{\(q\)-ary LDGM code} --- The dual of a \(q\)-ary LDPC code has a sparse generator matrix and is called a \(q\)-ary LDGM code.
\item\relax
\flmRefsHyperref[eczindexfamilyrel]{code:q-ary_ldpc}{\(q\)-ary LDPC code} --- The dual of a \(q\)-ary LDPC code has a sparse generator matrix and is called a \(q\)-ary LDGM code.
\item\relax
\flmRefsHyperref[eczindexfamilyrel]{code:majorana_stab}{Majorana stabilizer code} --- Classical self-orthogonal codes can be used to construct Majorana stabilizer codes \NoCaseChange{\protect\cite{cite566,cite1783,cite1784}}. The direct relationship between the two codes follows from expressing the Majorana strings as binary vectors – akin to the \flmRefsHyperref{ref817}{symplectic representation} – and observing that the binary stabilizer matrix \(S\) for such a Majorana stabilizer code satisfies \(S\cdot S^T=0\) because it has commuting stabilizers, which is precisely the condition \(G\cdot G^T=0\) on the generator matrix \(G\) of a self-orthogonal classical code. A self-orthogonal classical code \(C\) with parameters \([2N,k,d]\) yields a Majorana stabilizer code with parameters \(\llbracket N,N-k,d^\perp\rrbracket _f\), where \(d^\perp\) is the code distance of the dual code \(C^\perp\).
\item\relax
\flmRefsHyperref[eczindexfamilyrel]{code:stabilizer_over_gf4}{Hermitian qubit code} --- Hermitian qubit codes are constructed from Hermitian self-orthogonal linear codes over \(\mathbb{F}_4\) via the \flmRefsHyperref{ref1778}{\(\mathbb{F}_4\) representation}.
\item\relax
\flmRefsHyperref[eczindexfamilyrel]{code:qubit_stabilizer}{Qubit stabilizer code} --- Qubit stabilizer codes are in one-to-one correspondence with symplectic self-orthogonal binary linear codes of length \(2n\) via the \flmRefsHyperref{ref817}{symplectic representation}.
\item\relax
\flmRefsHyperref[eczindexfamilyrel]{code:self_dual_css}{Self-dual CSS code} --- Self-dual CSS codes arise from dual-containing (equivalently, self-orthogonal a.k.a. weakly self-dual) binary linear codes.
\item\relax
\flmRefsHyperref[eczindexfamilyrel]{code:stabilizer_over_gfqsq}{Hermitian Galois-qudit code} --- Hermitian codes are constructed from Hermitian self-orthogonal linear codes over \(\mathbb{F}_{q^2}\) via the \flmRefsHyperref{ref1779}{\(\mathbb{F}_{q^2}\) representation}.
\end{eczvaluelist}
\eczhbkcontributors{ Micah Shaw, Dhruv Devulapalli, \eczhuVVA }
\endeczcode

\eczcode{elliptic}{Elliptic code}{}
\codefieldsection{Description}
Evaluation AG code of rational functions evaluated on points lying on an elliptic curve, i.e., a curve of genus one.

\codefieldsection{Parent}
\begin{eczvaluelist}
\item\relax
\flmRefsHyperref[eczindexfamilyrel]{code:evaluation}{Evaluation AG code} --- Elliptic codes are evaluation AG codes with \(\cal X\) being an elliptic curve, i.e., curve of genus one \NoCaseChange{\protect\cite{cite1312,cite1313,cite32}\protect\cite[{Ch. 3.2}]{cite1314}}.
\end{eczvaluelist}
\codefieldsection{Cousins}
\begin{eczvaluelist}
\item\relax
\flmRefsHyperref[eczindexfamilyrel]{code:mds}{Maximum distance separable (MDS) code} --- Elliptic codes can be MDS \NoCaseChange{\protect\cite[{Exam. 15.5.3}]{cite26}\protect\cite[{pg. 310}]{cite1314}\protect\cite[{Sec. 4.4.2}]{cite1312}}.
\item\relax
\flmRefsHyperref[eczindexfamilyrel]{code:quantum_ag}{Quantum AG code} --- Elliptic codes can be used to construct quantum AG codes \NoCaseChange{\protect\cite{cite871}}.
\end{eczvaluelist}
\eczhbkcontributors{ En-Jui Kuo, \eczhuVVA }
\endeczcode

\eczcode{ecoc}{Error-correcting output code (ECOC)}{~\NoCaseChange{\protect\cite{cite1177,cite1785}}}
\codefieldsection{Description}
A length-\(n\) binary or ternary (over \(\{\pm 1,0\}\)) block code used to convey information about classes to classifiers in multiclass machine learning.
Rows of the code's generator matrix denote different classes, while columns correspond to classifiers.
The \(\pm 1\) elements can be used to distinguish between a pair of chosen classes, while a zero entry corresponds to a classifier ignoring that particular class.

A \textit{data-driven ECOC (DECOC)} \NoCaseChange{\protect\cite{cite1786}} is an ECOC whose design takes into account features of the data that is to be classified.
Not always decoded using the Hamming metric.

\codefieldsection{Decoding}
\begin{eczvaluelist}
\item\relax Standard Hamming-distance decoding \NoCaseChange{\protect\cite{cite1502}}.
\item\relax Inverse Hamming decoding \NoCaseChange{\protect\cite{cite1787}}.
\item\relax Euclidean-distance decoding or attenuated Euclidean decoding \NoCaseChange{\protect\cite{cite1788}}.
\item\relax Loss-based decoding \NoCaseChange{\protect\cite{cite1785}}.
\item\relax Probabilistic-based decoding \NoCaseChange{\protect\cite{cite1789}}.
\end{eczvaluelist}
\codefieldsection{Realizations}
\begin{eczvaluelist}
\item\relax Multiclass problems in machine learning, relevant to facial recognition \NoCaseChange{\protect\cite{cite253,cite254}}, text recognition \NoCaseChange{\protect\cite{cite255}}, or digit classification \NoCaseChange{\protect\cite{cite256}}.
\end{eczvaluelist}
\codefieldsection{Notes}
\begin{eczvaluelist}
\item\relax See \NoCaseChange{\protect\cite{cite1790}\protect\cite[{Ch. 6}]{cite300}} for overviews of ECOCs.
\item\relax See \NoCaseChange{\protect\cite{cite1791}} for a library of ECOCs.
\end{eczvaluelist}
\codefieldsection{Parents}
\begin{eczvaluelist}
\item\relax
\flmRefsHyperref[eczindexfamilyrel]{code:q-ary_digits_into_q-ary_digits}{\(q\)-ary code}\item\relax
\flmRefsHyperref[eczindexfamilyrel]{code:q-ary_over_zq}{\(q\)-ary code over \(\mathbb{Z}_q\)}\end{eczvaluelist}
\codefieldsection{Children}
\begin{eczvaluelist}
\item\relax
\flmRefsHyperref[eczindexfamilyrel]{code:one_hot}{One-hot code} --- One-hot codes are the primary codes used in multiclass classification \NoCaseChange{\protect\cite{cite300,cite301,cite302,cite303}}.
\item\relax
\flmRefsHyperref[eczindexfamilyrel]{code:one_vs_one}{One-versus-one (OVO) code} --- One-vs-one codes are often used in multiclass classification because they separate the multiclass task into several two-class tasks \NoCaseChange{\protect\cite{cite300}}.
\end{eczvaluelist}
\codefieldsection{Cousins}
\begin{eczvaluelist}
\item\relax
\flmRefsHyperref[eczindexfamilyrel]{code:hadamard}{\([2^m,m,2^{m-1}]\) Hadamard code} --- Hadamard codes and subcodes can be used as ECOCs \NoCaseChange{\protect\cite{cite1177,cite1178,cite1179}}.
\item\relax
\flmRefsHyperref[eczindexfamilyrel]{code:bch}{Binary BCH code} --- BCH codes can be used as ECOCs \NoCaseChange{\protect\cite{cite1177}}.
\end{eczvaluelist}
\eczhbkcontributors{ \eczhuVVA }
\endeczcode

\eczcode{evaluation}{Evaluation AG code}{}
\codefieldsection{Description}
Evaluation code over \(\mathbb{F}_q\) on a set of points \({\cal P} = \left( P_1,P_2,\cdots,P_n \right)\) lying on an algebraic curve \(\cal X\) defined over \(\mathbb{F}_q\), where the corresponding vector space \(L\) of functions \(f\) consists of certain rational functions (or, in special cases, polynomials).

Codewords are evaluations of all functions at the specified points,
\flmMathEnvironment{align}{}{
  \left( f(P_1), f(P_2), \cdots, f(P_n) \right) \quad\quad\forall f\in L~.
}
The code is denoted as \(C_L({\cal X},{\cal P},D)\), where the divisor \(D\) (of degree less than \(n\)) determines which rational functions to use by prescribing features associated with their zeroes and poles. The original motivation for evaluation codes, which are generalizations of RS codes that expand both the types of functions used as well as the available evaluation points, was to increase code length while maintaining good distance and size.

The algebraic curve \(\cal X\) used for this construction is typically taken to be absolutely irreducible (i.e., irreducible over the algebraic closure of \(\mathbb{F}_q\)), and often smooth and projective after replacing an affine model by its nonsingular projective model.
The curve can be defined over affine space or projective space, which contains the affine coordinates as a subset and which can yield an increase in length.
If evaluations are made over projective coordinates, then the codewords are evaluations of homogeneous polynomials, and there are relations between such polynomials with polynomials over affine space. See Refs. \NoCaseChange{\protect\cite{cite1312,cite1313,cite1314,cite32,cite1792}} for more details.

Any AG code on \(PG(1,q)\) is diagonally equivalent to a one-point code \NoCaseChange{\protect\cite[{Rem. 15.3.25}]{cite26}}.

\codefieldsection{Protection}
A lower bound on distance is \(d \geq n- \deg (D)\) \NoCaseChange{\protect\cite[{Thm. 15.3.12}]{cite26}}. The order or Feng-Rao bound, a generalization of the shift bound for cyclic codes, gives a lower bound on the distance of evaluation AG codes \NoCaseChange{\protect\cite{cite1793,cite1794,cite1795}}. Connection to semigroups yields another bound \NoCaseChange{\protect\cite{cite1796,cite32}}.
\codefieldsection{Decoding}
\begin{eczvaluelist}
\item\relax Generalization of plane-curve decoder \NoCaseChange{\protect\cite{cite1797}}. Another decoder \NoCaseChange{\protect\cite{cite1798}} was later shown to be equivalent in Ref. \NoCaseChange{\protect\cite{cite1799}}. Application of several algorithms in parallel can be used to decode up to half the minimum distance \NoCaseChange{\protect\cite{cite1800,cite1801}}. Computational procedure implementing these decoders is based on an extension of the Berlekamp-Massey algorithm by Sakata \NoCaseChange{\protect\cite{cite1802,cite1803,cite1804}}.
\item\relax Decoder based on majority voting of unknown syndromes \NoCaseChange{\protect\cite{cite1793}} decodes up to half of the minimum distance \NoCaseChange{\protect\cite{cite1805}}.
\item\relax List decoders generalizing Sudan's RS decoder by Shokrollahi-Wasserman \NoCaseChange{\protect\cite{cite1806}} and Guruswami-Sudan \NoCaseChange{\protect\cite{cite1240}}.
\end{eczvaluelist}
\codefieldsection{Notes}
\begin{eczvaluelist}
\item\relax See Refs. \NoCaseChange{\protect\cite{cite32,cite1807,cite1808,cite26,cite1809}} for surveys and overviews of decoders.
\end{eczvaluelist}
\codefieldsection{Parent}
\begin{eczvaluelist}
\item\relax
\flmRefsHyperref[eczindexfamilyrel]{code:evaluation_varieties}{Evaluation code} --- Evaluation AG codes are evaluation codes for which \(\cal X\) is an algebraic curve, i.e., an algebraic variety of dimension one \NoCaseChange{\protect\cite{cite32}}.
\end{eczvaluelist}
\codefieldsection{Children}
\begin{eczvaluelist}
\item\relax
\flmRefsHyperref[eczindexfamilyrel]{code:elliptic}{Elliptic code} --- Elliptic codes are evaluation AG codes with \(\cal X\) being an elliptic curve, i.e., curve of genus one \NoCaseChange{\protect\cite{cite1312,cite1313,cite32}\protect\cite[{Ch. 3.2}]{cite1314}}.
\item\relax
\flmRefsHyperref[eczindexfamilyrel]{code:klein_quartic}{Klein-quartic code} --- Klein-quartic codes are evaluation AG codes with \(\cal X\) being the Klein quartic (\NoCaseChange{\protect\cite[{Ex. 2}]{cite32}}.75)\NoCaseChange{\protect\cite{cite1314}}.
\item\relax
\flmRefsHyperref[eczindexfamilyrel]{code:norm_trace}{Norm-trace code} --- Norm-trace codes are evaluation AG codes with \(\cal X\) being a Miura-Kamiya curve \NoCaseChange{\protect\cite{cite1810}}.
\item\relax
\flmRefsHyperref[eczindexfamilyrel]{code:plane_curve}{Plane-curve evaluation code} --- Plane-curve evaluation codes are evaluation AG codes of bivariate polynomials with \(\cal X\) being an affine plane curve \NoCaseChange{\protect\cite{cite1314}\protect\cite[{Thm. 2.27}]{cite32}}.
\item\relax
\flmRefsHyperref[eczindexfamilyrel]{code:suzuki}{Suzuki-curve code} --- Suzuki-curve codes are evaluation AG codes with \(\cal X\) being a Suzuki curve.
\item\relax
\flmRefsHyperref[eczindexfamilyrel]{code:residue}{Residue AG code} --- Any residue AG code of differential forms can be restated, up to diagonal equivalence, as an evaluation AG code of functions \NoCaseChange{\protect\cite{cite1312,cite1313,cite1314}\protect\cite[{Lemma 15.3.10}]{cite26}\protect\cite[{Thm. 2.72}]{cite32}}. Evaluation and residue AG codes are dual to each other \NoCaseChange{\protect\cite{cite32}\protect\cite[{Thm. 15.3.3}]{cite26}}.
\item\relax
\flmRefsHyperref[eczindexfamilyrel]{code:tamo_barg_vladut}{Barg-Tamo-Vladut code} --- Barg-Tamo-Vladut codes are evaluation AG codes on algebraic curves built from curve covers \NoCaseChange{\protect\cite[{Thm. 15.9.14}]{cite26}}.
\item\relax
\flmRefsHyperref[eczindexfamilyrel]{code:hexacode}{\([6,3,4]_4\) Hexacode} --- The hexacode is an evaluation AG code over the \flmRefsHyperref{ref33}{quaternary Galois field} \(\mathbb{F}_4 = \{0,1,\omega, \bar{\omega}\}\) with \(\cal X\) defined by \(x^2 y + \omega y^2 z + \bar{\omega} z^2 x = 0\) \NoCaseChange{\protect\cite[{Exam. 2.77}]{cite32}}.
\end{eczvaluelist}
\codefieldsection{Cousins}
\begin{eczvaluelist}
\item\relax
\flmRefsHyperref[eczindexfamilyrel]{code:q-ary_linear}{Linear \(q\)-ary code} --- The degree of the divisor for evaluation AG codes is restricted to be less than \(n\). When there is no restriction, any \(q\)-ary linear code can be formulated as an evaluation AG code \NoCaseChange{\protect\cite{cite1811}}.
\item\relax
\flmRefsHyperref[eczindexfamilyrel]{code:frameproof}{Frameproof (FP) code} --- Asymptotic bounds on FP codes can be formulated using evaluation AG codes \NoCaseChange{\protect\cite{cite1054,cite1055}}.
A sufficient condition for an evaluation AG code to be FP can be recast as an instance of the Riemann-Roch equation \NoCaseChange{\protect\cite[{Sec. 15.8.2}]{cite26}}.
AG-based constructions of binary \((2,1)\)-separating systems can beat a random-coding lower bound \NoCaseChange{\protect\cite[{Thm. 15.8.13}]{cite26}}.

\item\relax
\flmRefsHyperref[eczindexfamilyrel]{code:shimura}{Tsfasman-Vladut-Zink (TVZ) code} --- TVZ codes can also be formulated as evaluation AG codes on algebraic curves; these are dual to the corresponding residue AG codes.
\item\relax
\flmRefsHyperref[eczindexfamilyrel]{code:quantum_ag}{Quantum AG code} --- Quantum AG codes are quantum analogues of evaluation AG codes.
\end{eczvaluelist}
\eczhbkcontributors{ En-Jui Kuo, \eczhuVVA }
\endeczcode

\eczcode{evaluation_varieties}{Evaluation code}{~\NoCaseChange{\protect\cite{cite32,cite1312,cite1313,cite1314}}}
\codefieldsection{Description}
Code whose codewords are evaluations of functions at certain fixed points. Code properties can be inferred from the structure of the functions and the underlying geometric object containing the points, often using results from algebraic geometry.

Let \(\cal{X}\) be a geometric object that contains a subset \({\cal P} = \left( P_1,P_2,\cdots,P_n \right) \) consisting of \(n\) points \(P_j\). Let \(L\) be a vector space over \(\mathbb{F}_q\) of functions \(f\) that take values in \(\mathbb{F}_q\). Each \(f\in L\) yields a codeword of an evaluation code \(C_L({\cal X},{\cal P})\) of the form
\flmMathEnvironment{align}{}{
  \left( f(P_1), f(P_2), \cdots, f(P_n) \right) \quad\quad\forall f\in L~.
}
This is a linear \(q\)-ary code since the functions \(f\) take values in \(\mathbb{F}_q\) and form a vector space.

Examples of geometric objects \(\cal X\) include affine or projective spaces over \(\mathbb{F}_q\) as well as subsets of those spaces determined by some constraints. Prominent subsets are \textit{algebraic varieties}, which, for algebraically closed fields, are sets of solutions of systems of polynomial equations in either affine or projective space. 
Algebraic curves are algebraic varieties of dimension one \NoCaseChange{\protect\cite{cite32}}, and those used for this construction are sets of zeroes of one or more nontrivial polynomials forming a prime ideal.

The functions \(f\) are typically polynomials for the case of algebraic varieties, but can be promoted to rational functions to either define codes on projective coordinates and/or to determine code properties using results from algebraic geometry.
For example, any degree-\(k\) univariate polynomial \(\sum_j^{k} p_j x^j\) is \textit{homogenized} into a bivariate polynomial \(\sum_j^{k} p_j x^j y^{k-j}\) and divided by another bivariate polynomial of the same degree, giving rise to a \textit{homogeneous rational function}, or \textit{form}.
Similar homogenization can be done for multivariate polynomials by adding an extra variable as above.
Forms are the most general cases considered for evaluation codes since they encompass all polynomials via the reverse of the above procedure.

One can specify the space \(L\) by the number of variables input into the rational functions as well as their maximum degree. One can additionally select only functions that have zeroes at certain points with certain multiplicities. A bookkeeping device for this data is the \textit{divisor} \(D\), and the corresponding vector space of functions defined using the variety \(\cal X\) and the divisor is the \textit{Riemann-Roch space} \(L=L(D)\) \NoCaseChange{\protect\cite[{pg. 313}]{cite26}}.
Codes based on divisors with only one pole (of arbitrary order) are called \textit{one-point codes} \NoCaseChange{\protect\cite[{Remark 4.4}]{cite32}}.

\codefieldsection{Protection}
Properties of the underlying geometric object \(\cal X\) can be used to bound the code dimension \(k\) and distance \(d\). The order or Feng-Rao bound gives a lower bound on the distance of evaluation codes \NoCaseChange{\protect\cite{cite1812,cite1793,cite32}}; see \NoCaseChange{\protect\cite{cite1813}\protect\cite[{Ch. 4}]{cite32}} for more discussion.
\codefieldsection{Notes}
\begin{eczvaluelist}
\item\relax See books \NoCaseChange{\protect\cite{cite1312,cite1313,cite32,cite1814,cite1722,cite1815}} for more information.
\item\relax See LMFDB \NoCaseChange{\protect\cite{cite1816}} for a database of varieties.
\end{eczvaluelist}
\codefieldsection{Parents}
\begin{eczvaluelist}
\item\relax
\flmRefsHyperref[eczindexfamilyrel]{code:q-ary_linear}{Linear \(q\)-ary code} --- Evaluation codes are defined using polynomial or rational functions evaluated on a subset of affine or projective space. Given access to more general structures (i.e., morphisms of algebras), any \(q\)-ary linear code can be formulated as an evaluation code \NoCaseChange{\protect\cite[{Sec. 4.1}]{cite32}\protect\cite[{Prop. 1.1.4}]{cite1314}\protect\cite[{Prop. 1.1.4}]{cite1312}}.
\item\relax
\flmRefsHyperref[eczindexfamilyrel]{code:ag}{Algebraic-geometry (AG) code} --- Evaluation codes on algebraic varieties are AG codes. The AG-code literature has mostly focused on codes on algebraic curves, i.e., one-dimensional varieties.
\end{eczvaluelist}
\codefieldsection{Children}
\begin{eczvaluelist}
\item\relax
\flmRefsHyperref[eczindexfamilyrel]{code:evaluation}{Evaluation AG code} --- Evaluation AG codes are evaluation codes for which \(\cal X\) is an algebraic curve, i.e., an algebraic variety of dimension one \NoCaseChange{\protect\cite{cite32}}.
\item\relax
\flmRefsHyperref[eczindexfamilyrel]{code:evaluation_polynomial}{Polynomial evaluation code} --- Polynomial evaluation codes are evaluation codes for which \(\cal X\) is an algebraic variety of dimension greater than one.
\end{eczvaluelist}
\codefieldsection{Cousin}
\begin{eczvaluelist}
\item\relax
\flmRefsHyperref[eczindexfamilyrel]{code:projective}{Projective geometry code} --- Codewords of an evaluation code of multivariate polynomials up to degree one evaluated at points in projective space yield a projective code.
\end{eczvaluelist}
\eczhbkcontributors{ Alexander Barg, \eczhuVVA }
\endeczcode

\eczcode{extended_reed_solomon}{Extended GRS code}{}
\codefieldsection{Description}
A GRS code extended by one extra coordinate to form an \([n+1,k,n-k+2]_q\) MDS code. In projective language, this corresponds to adding one more evaluation point, often interpreted as the point at infinity; in suitable equivalent descriptions, one may instead use an affine point such as \(0\). The case when \(n=q-1\), multipliers \(v_i=1\), and \(\alpha_i\) are \(i-1\)st powers of a primitive \(n\)th root of unity is an \textit{extended narrow-sense RS code}.

An \([q-1,k,q-k]_q\) narrow-sense RS code can be extended twice by adding two evaluation points (of which one can be zero) to yield a \([q+1,k,q-k+2]_q\) \textit{doubly extended narrow-sense RS code} \NoCaseChange{\protect\cite[{Rem. 15.3.21}]{cite26}}.
The two extra columns sometimes correspond to evaluating at zero and infinity if one switches to projective coordinates, in which case the doubly extended GRS code is a projective-line evaluation code.
There also exist \textit{triply extended RS codes} with parameters \([q+2,3,q-1]_q\) or \([q+2,q-1,4]_q\) \NoCaseChange{\protect\cite{cite62}}.

Their automorphism groups have been identified \NoCaseChange{\protect\cite{cite1817}}.

\codefieldsection{Notes}
\begin{eczvaluelist}
\item\relax See corresponding MinT database entry \NoCaseChange{\protect\cite{cite1673}}.
\end{eczvaluelist}
\codefieldsection{Parent}
\begin{eczvaluelist}
\item\relax
\flmRefsHyperref[eczindexfamilyrel]{code:generalized_reed_muller}{Generalized RM (GRM) code} --- GRM codes for univariate polynomials (\(m=1\)) reduce to extended RS codes \NoCaseChange{\protect\cite{cite1813}}.
\end{eczvaluelist}
\codefieldsection{Children}
\begin{eczvaluelist}
\item\relax
\flmRefsHyperref[eczindexfamilyrel]{code:hexacode}{\([6,3,4]_4\) Hexacode} --- The hexacode is a triply extended RS code \NoCaseChange{\protect\cite[{pg. 82}]{cite39}}.
\item\relax
\flmRefsHyperref[eczindexfamilyrel]{code:q-ary_repetition}{\(q\)-ary repetition code} --- \(q\)-ary repetition codes can be thought of as extended RS codes \NoCaseChange{\protect\cite{cite1673}}. GRM\(_q(0,m)\) codes are evaluations of all zero-degree polynomials on \(\mathbb{F}_q^m\), which are just the \(q\) constant polynomials, so \(q\)-ary repetition codes are GRM\(_q(0,m)\) codes.
\item\relax
\flmRefsHyperref[eczindexfamilyrel]{code:reed_solomon_4}{\([4,2,3]_4\) RS\(_4\) code} --- The RS\(_4\) is an extended RS code \NoCaseChange{\protect\cite[{pg. 296}]{cite41}}.
\item\relax
\flmRefsHyperref[eczindexfamilyrel]{code:shortened_hexacode}{\([5,3,3]_4\) Shortened hexacode} --- The shortened hexacode is a doubly extended narrow-sense RS code \NoCaseChange{\protect\cite[{pg. 82}]{cite39}}.
\item\relax
\flmRefsHyperref[eczindexfamilyrel]{code:tetracode}{\([4,2,3]_3\) Tetracode} --- The tetracode is an extended RS code \NoCaseChange{\protect\cite[{pg. 81}]{cite39}}.
\end{eczvaluelist}
\codefieldsection{Cousins}
\begin{eczvaluelist}
\item\relax
\flmRefsHyperref[eczindexfamilyrel]{code:generalized_reed_solomon}{Generalized RS (GRS) code} --- Extended GRS codes can be thought of as GRS codes that include an evaluation point of zero.
\item\relax
\flmRefsHyperref[eczindexfamilyrel]{code:hyperoval}{Hyperoval code} --- Columns of parity-check matrices of triply extended RS codes consist of points of a hyperoval \NoCaseChange{\protect\cite[{Prop. 17.5}]{cite62}}.
\item\relax
\flmRefsHyperref[eczindexfamilyrel]{code:mds}{Maximum distance separable (MDS) code} --- A GRS code can be extended to an MDS code \NoCaseChange{\protect\cite[{Thm. 5.3.4}]{cite126}}. Extended and doubly extended narrow-sense RS codes are MDS \NoCaseChange{\protect\cite[{Thms. 5.3.2 and 5.3.4}]{cite126}}, and there is an equivalence between the two for odd prime \(q\) \NoCaseChange{\protect\cite{cite1818}}.
\item\relax
\flmRefsHyperref[eczindexfamilyrel]{code:narrow_sense_reed_solomon}{Narrow-sense RS code} --- A narrow-sense RS code can be extended once, twice, or three times.
\item\relax
\flmRefsHyperref[eczindexfamilyrel]{code:reed_solomon}{Reed-Solomon (RS) code} --- Extending an RS code by one evaluation point, often interpreted as the point at infinity, yields an extended RS code.
\item\relax
\flmRefsHyperref[eczindexfamilyrel]{code:roth_lempel}{Roth-Lempel code} --- Roth-Lempel codes are doubly extended RS codes.
\item\relax
\flmRefsHyperref[eczindexfamilyrel]{code:projective}{Projective geometry code} --- Columns of parity-check matrices of doubly extended narrow-sense RS codes consist of points of a normal rational curve \NoCaseChange{\protect\cite[{Def. 14.2.6}]{cite202}}.
\item\relax
\flmRefsHyperref[eczindexfamilyrel]{code:q-ary_simplex}{\(q\)-ary simplex code} --- \(q\)-ary simplex codes for \(m=2\) can be thought of as extended RS codes \NoCaseChange{\protect\cite{cite1673}}.
\end{eczvaluelist}
\eczhbkcontributors{ \eczhuVVA }
\endeczcode

\eczcode{flag_variety}{Flag-variety code}{~\NoCaseChange{\protect\cite{cite1819}}}
\codefieldsection{Description}
Evaluation code of polynomials evaluated on points lying on a flag variety.

\codefieldsection{Parent}
\begin{eczvaluelist}
\item\relax
\flmRefsHyperref[eczindexfamilyrel]{code:evaluation_polynomial}{Polynomial evaluation code} --- Flag-variety codes are polynomial evaluation codes with \(\cal X\) being a flag variety.
\end{eczvaluelist}
\codefieldsection{Children}
\begin{eczvaluelist}
\item\relax
\flmRefsHyperref[eczindexfamilyrel]{code:grassmannian_variety}{Grassmannian evaluation code} --- Grassmannian evaluation codes are flag-variety evaluation codes with the flag variety being a Grassmannian.
\item\relax
\flmRefsHyperref[eczindexfamilyrel]{code:hermitian_hypersurface}{Hermitian-hypersurface code} --- Hermitian-hypersurface codes are flag-variety evaluation codes with the flag variety being a Hermitian hypersurface.
\item\relax
\flmRefsHyperref[eczindexfamilyrel]{code:quadric}{Quadric code} --- Quadric codes are flag-variety evaluation codes with the flag variety being a quadric hypersurface.
\item\relax
\flmRefsHyperref[eczindexfamilyrel]{code:schubert}{Schubert evaluation code} --- Schubert evaluation codes are flag-variety evaluation codes with the flag variety being a Schubert variety.
\end{eczvaluelist}
\codefieldsection{Cousin}
\begin{eczvaluelist}
\item\relax
\flmRefsHyperref[eczindexfamilyrel]{code:homogeneous_space_classical}{Homogeneous-space code} --- The flag variety is a finite homogeneous space \NoCaseChange{\protect\cite{cite28}}.
\end{eczvaluelist}
\eczhbkcontributors{ \eczhuVVA }
\endeczcode

\eczcode{folded_reed_solomon}{Folded RS (FRS) code}{~\NoCaseChange{\protect\cite{cite1820}}}
\codefieldsection{Description}
A code obtained from an RS code by bundling consecutive symbols into larger alphabet symbols. This preserves the algebraic structure of the parent RS code while lowering the block length over an extension alphabet, and it is a key ingredient in capacity-approaching list-decoding constructions.

An \([n/m,k]_{q^m}\) FRS code is obtained from an \([n,k]_q\) RS code by grouping consecutive evaluations, thereby reducing the block length over a larger alphabet. In this case, the evaluation points are powers of a field element \(\gamma\), namely \(\gamma^0,\gamma^1,\ldots,\gamma^{n-1}\). Each codeword \(\mu\) of an \(m\)-folded RS code is a string of \(n/m\) symbols, with each symbol being a block of values of a polynomial \(f_\mu\) at consecutive powers of \(\gamma\),
\flmMathEnvironment{align}{}{
\begin{split}
  \mu\to&\Big(\left(f_{\mu}(\gamma^{0}),\cdots,f_{\mu}(\gamma^{m-1})\right),\left(f_{\mu}(\gamma^{m}),\cdots,f_{\mu}(\gamma^{2m-1})\right)\cdots\\&\cdots,\left(f_{\mu}(\gamma^{n-m}),\cdots,f_{\mu}(\gamma^{n-1})\right)\Big)~.
\end{split}
}

\codefieldsection{Rate}
FRS codes achieve relaxed generalized Singleton bounds \NoCaseChange{\protect\cite{cite1821}}.
\codefieldsection{Decoding}
\begin{eczvaluelist}
\item\relax Guruswami and Rudra \NoCaseChange{\protect\cite{cite1822,cite1823}} achieved list-decoding up to \(1-\frac{k}{n}-\epsilon\) fraction of errors using the Parvaresh-Vardy algorithm \NoCaseChange{\protect\cite{cite1824}}; see Ref. \NoCaseChange{\protect\cite{cite1825}} for a randomized construction.
\item\relax List-decoding works up to the Johnson bound using the Guruswami-Sudan algorithm \NoCaseChange{\protect\cite{cite1826}}.
\item\relax Folded RS codes, concatenated with suitable inner codes, can be efficiently list-decoded up to the Blokh-Zyablov bound \NoCaseChange{\protect\cite{cite1822,cite1827}}.
\end{eczvaluelist}
\codefieldsection{Notes}
\begin{eczvaluelist}
\item\relax See the book \NoCaseChange{\protect\cite{cite1828}} for an introduction to FRS codes.
\item\relax A class of FRS codes can be used in the Yamakawa-Zhandry quantum algorithm \NoCaseChange{\protect\cite{cite1829}}.
\end{eczvaluelist}
\codefieldsection{Parent}
\begin{eczvaluelist}
\item\relax
\flmRefsHyperref[eczindexfamilyrel]{code:group}{Group-algebra code} --- FRS codes are polynomial ideal codes whose corresponding polynomial is a product of the polynomials of the RS codes that are being folded \NoCaseChange{\protect\cite{cite1826}}.
\end{eczvaluelist}
\codefieldsection{Child}
\begin{eczvaluelist}
\item\relax
\flmRefsHyperref[eczindexfamilyrel]{code:reed_solomon}{Reed-Solomon (RS) code} --- An FRS code with no extra grouping (\(m=1\)) reduces to an RS code.
\end{eczvaluelist}
\codefieldsection{Cousins}
\begin{eczvaluelist}
\item\relax
\flmRefsHyperref[eczindexfamilyrel]{code:parvaresh_vardy}{Parvaresh-Vardy (PV) code} --- The specific relations imposed on the polynomials of PV codes allow for them to be expressed in a similar way as FRS codes, but with more redundancy. Folded RS codes can be list-decoded up to a higher fraction of errors.
\item\relax
\flmRefsHyperref[eczindexfamilyrel]{code:galois_fqrs}{Folded quantum RS (FQRS) code} --- Folded quantum RS codes are quantum analogues of folded RS codes.
\end{eczvaluelist}
\eczhbkcontributors{ \eczhuVVA }
\endeczcode

\eczcode{generalized_reed_muller}{Generalized RM (GRM) code}{~\NoCaseChange{\protect\cite{cite1830,cite1831,cite1781}}}
\codefieldsection{Alternative Names}
\begin{eczvaluelist}
\item\relax \(q\)-ary RM code
\end{eczvaluelist}
\eczhIndexCodeAliasName{generalized_reed_muller}{\(q\)-ary RM code}
\codefieldsection{Description}
Extensions of RM codes to \(q\)-ary coordinates that can be described as multivariate polynomial evaluation codes over affine or projective space.

Affine Reed-Muller codes are denoted by GRM\(_q(r,m)\), being of length \(n=q^m\) over \(\mathbb{F}_q\) with \(0\leq r\leq m(q-1)\). Their codewords are evaluations of the set of all degree-\(\leq r\) polynomials in \(m\) variables at the points of \(\mathbb{F}_q\).

Since \(\beta^q=\beta\) for any \(\beta\in \mathbb{F}_q\), the above definition is not injective. Replacing each factor in each polynomial as \(x^q\to x\), the above set reduces to the set of all degree-\(\leq r\) polynomials in \(m\) variables such that no term has an exponent \(q\) or higher on any variable.

Its automorphism group is the general affine group \(GA(m,\mathbb{F}_q)\) \NoCaseChange{\protect\cite{cite1832}}.
Any nontrivial \(q\)-ary linear code invariant under this group is equivalent to a GRM code \NoCaseChange{\protect\cite{cite1766}}.

\codefieldsection{Protection}
Code parameters for specific \(m,r\) are given in Refs. \NoCaseChange{\protect\cite{cite1312,cite1313}\protect\cite[{pg. 46}]{cite1314}}.
\codefieldsection{Rate}
GRM codes achieve capacity on sufficiently symmetric non-binary channels \NoCaseChange{\protect\cite{cite1833}}.
\codefieldsection{Notes}
\begin{eczvaluelist}
\item\relax See books \NoCaseChange{\protect\cite{cite1573,cite126,cite1834}} for details of GRM codes.
\end{eczvaluelist}
\codefieldsection{Parents}
\begin{eczvaluelist}
\item\relax
\flmRefsHyperref[eczindexfamilyrel]{code:evaluation_polynomial}{Polynomial evaluation code} --- GRM (PRM) codes are multivariate polynomial evaluation codes with \(\cal X\) being the entire \(m\)-dimensional affine (projective) space over \(\mathbb{F}_q\) \NoCaseChange{\protect\cite{cite1835,cite32}\protect\cite[{pgs. 44-46}]{cite1314}}.
\item\relax
\flmRefsHyperref[eczindexfamilyrel]{code:multiplicity}{Multiplicity code} --- Multivariate multiplicity codes of degree \(s=1\) reduce to GRM codes.
\item\relax
\flmRefsHyperref[eczindexfamilyrel]{code:matrix_product}{Matrix-product code} --- Applying a special case of the matrix-product procedure yields GRM codes \NoCaseChange{\protect\cite{cite1836}}.
\item\relax
\flmRefsHyperref[eczindexfamilyrel]{code:q-ary_lcc}{\(q\)-ary linear LCC} --- GRM codes are LDCs and LCCs \NoCaseChange{\protect\cite{cite1073,cite1067}}.
\end{eczvaluelist}
\codefieldsection{Children}
\begin{eczvaluelist}
\item\relax
\flmRefsHyperref[eczindexfamilyrel]{code:reed_muller}{Reed-Muller (RM) code} --- Binary GRM codes are RM codes.
\item\relax
\flmRefsHyperref[eczindexfamilyrel]{code:extended_reed_solomon}{Extended GRS code} --- GRM codes for univariate polynomials (\(m=1\)) reduce to extended RS codes \NoCaseChange{\protect\cite{cite1813}}.
\item\relax
\flmRefsHyperref[eczindexfamilyrel]{code:projective_reed_muller}{Projective RM (PRM) code} --- Nonzero codewords of minimum weight of a \(r\)th-order \(q\)-ary projective RM code correspond to algebraic hypersurfaces of degree \(r\) having the largest number of points in the projective space \(PG(m,q)\) \NoCaseChange{\protect\cite[{Thm. 14.3.3}]{cite202}}.
\end{eczvaluelist}
\codefieldsection{Cousins}
\begin{eczvaluelist}
\item\relax
\flmRefsHyperref[eczindexfamilyrel]{code:group}{Group-algebra code} --- GRM codes over prime-power fields are group-algebra codes \NoCaseChange{\protect\cite{cite1576,cite1577,cite1837}\protect\cite[{Exam. 16.4.11}]{cite196}}.
\item\relax
\flmRefsHyperref[eczindexfamilyrel]{code:q-ary_cyclic}{Cyclic linear \(q\)-ary code} --- Punctured GRM codes, i.e., GRM codes with nonzero evaluation points, are cyclic, and their extensions recover GRM codes \NoCaseChange{\protect\cite[{Sec. 2.8}]{cite68}\protect\cite[{pg. 52}]{cite1314}}.
\item\relax
\flmRefsHyperref[eczindexfamilyrel]{code:q-ary_ltc}{\(q\)-ary linear LTC} --- GRM codes for \(r<q\) can be LTCs in the low-error \NoCaseChange{\protect\cite{cite1089,cite1098}} and high-error \NoCaseChange{\protect\cite{cite1702,cite1703}} regimes. They admit weakly stable presentations of their corresponding groups \NoCaseChange{\protect\cite{cite1700}}.
\item\relax
\flmRefsHyperref[eczindexfamilyrel]{code:dual}{Dual linear code} --- The dual of a GRM code is also a GRM code \NoCaseChange{\protect\cite{cite1781,cite1782}}.
\item\relax
\flmRefsHyperref[eczindexfamilyrel]{code:self_dual}{Self-dual linear code} --- Certain GRM codes are self-dual; in general, the dual of GRM\(_q(r,m)\) is another GRM code, namely GRM\(_q(m(q-1)-r-1,m)\) \NoCaseChange{\protect\cite{cite1781,cite1782}}.
\item\relax
\flmRefsHyperref[eczindexfamilyrel]{code:difference_set}{Difference-set cyclic (DSC) code} --- DSC codes can be \flmRefsHyperref{ref33}{subfield} subcodes of GRM codes, and vice versa \NoCaseChange{\protect\cite[{Thm. 6.14}]{cite1176}}.
\item\relax
\flmRefsHyperref[eczindexfamilyrel]{code:pg_ldpc}{Finite-geometry LDPC (FG-LDPC) code} --- Some EG-LDPC codes are duals of \flmRefsHyperref{ref33}{subfield} subcodes of GRM codes \NoCaseChange{\protect\cite[{pg. 448}]{cite1353}}.
\item\relax
\flmRefsHyperref[eczindexfamilyrel]{code:batch}{Batch code} --- GRM codes can be used to construct batch codes \NoCaseChange{\protect\cite{cite950}}.
\item\relax
\flmRefsHyperref[eczindexfamilyrel]{code:quantum_convolutional}{Quantum convolutional code} --- GRM codes can be used to construct quantum convolutional codes \NoCaseChange{\protect\cite{cite1838,cite1839}\protect\cite[{Sec. 12.4}]{cite872}}.
\item\relax
\flmRefsHyperref[eczindexfamilyrel]{code:galois_reed_muller}{Galois-qudit quantum RM code} --- Generalized RM codes can be used to construct Galois-qudit RM codes via the Galois-qudit Hermitian construction, the Galois-qudit CSS construction, or directly from their parity-check matrices \NoCaseChange{\protect\cite{cite828}\protect\cite[{Sec. 4.2}]{cite829}}.
\item\relax
\flmRefsHyperref[eczindexfamilyrel]{code:galois_expander}{Galois-qudit expander code} --- The explicit expander-code construction of \NoCaseChange{\protect\cite{cite689}} contains planted GRM codewords.
\end{eczvaluelist}
\eczhbkcontributors{ \eczhuVVA }
\endeczcode

\eczcode{generalized_reed_solomon}{Generalized RS (GRS) code}{}
\codefieldsection{Description}
An \([n,k,n-k+1]_q\) MDS code that is a modification of the RS code where codeword polynomials are multiplied by additional prefactors \NoCaseChange{\protect\cite[{Def. 15.3.19}]{cite26}}.

Each message \(\mu\) is encoded into a string of values of the corresponding polynomial \(f_\mu\) at the points \(\alpha_i\), multiplied by a corresponding nonzero factor \(v_i \in \mathbb{F}_q\),
\flmMathEnvironment{align}{}{
  \mu\to\left( v_{1}f_{\mu}\left(\alpha_{1}\right),v_{2}f_{\mu}\left(\alpha_{2}\right),\cdots,v_{n}f_{\mu}\left(\alpha_{n}\right)\right)~.
}

The dual of a GRS code is another GRS code \NoCaseChange{\protect\cite[{pg. 304}]{cite41}}.

\codefieldsection{Protection}
The code can detect up to \(n-k\) erasures and can correct up to \(\left\lfloor (n-k)/2\right\rfloor\) errors.
\codefieldsection{Rate}
Concatenations of GRS codes with random linear codes almost surely attains the \flmRefsHyperref{ref85}{GV bound} \NoCaseChange{\protect\cite{cite973}}.
\codefieldsection{Decoding}
\begin{eczvaluelist}
\item\relax The decoding process of GRS codes reduces to the solution of a polynomial congruence equation, usually referred to as the key equation. Decoding schemes are based on applications of the Euclid algorithm to solve the key equation.
\item\relax Berlekamp-Massey decoder with runtime of \flmRefsHyperref{ref65}{order} \(O(n^2)\) \NoCaseChange{\protect\cite{cite1739,cite1234,cite1235}}.
\item\relax Guruswami-Sudan list decoder \NoCaseChange{\protect\cite{cite1239,cite1240}} and modification by Koetter-Vardy for soft-decision decoding \NoCaseChange{\protect\cite{cite1840}}.
\item\relax Hard-decision decoder for errors within the Singleton bound \NoCaseChange{\protect\cite{cite1841}}.
\end{eczvaluelist}
\codefieldsection{Realizations}
\begin{eczvaluelist}
\item\relax Commonly used in mass storage systems such as CDs, DVDs, QR codes etc.
\item\relax Various cloud storage systems \NoCaseChange{\protect\cite{cite264}}.
\item\relax A variation of the McEliece public-key cryptosystem \NoCaseChange{\protect\cite{cite265,cite266}} by Niederreiter \NoCaseChange{\protect\cite{cite267}} replaced the generator matrix by the parity check matrix of a GRS code. This was proven to be insecure since the public key exposes the algebraic structure of code \NoCaseChange{\protect\cite{cite268}}. More recent works focus on methods to mask the algebraic structure using subcodes of GRS codes \NoCaseChange{\protect\cite{cite261}}. For example, a key-recovery attack was developed in Ref. \NoCaseChange{\protect\cite{cite269}} for a variant of masking method proposed in Ref. \NoCaseChange{\protect\cite{cite270}}.
\end{eczvaluelist}
\codefieldsection{Parents}
\begin{eczvaluelist}
\item\relax
\flmRefsHyperref[eczindexfamilyrel]{code:residue}{Residue AG code} --- GRS (RS) codes are in one-to-one correspondence with both evaluation AG codes of univariate polynomials \(f\) \NoCaseChange{\protect\cite[{Thm. 15.3.24}]{cite26}} and residue AG codes of univariate differential forms \NoCaseChange{\protect\cite[{Prop. 15.3.26}]{cite26}}, with \(\cal X\) being the projective (affine) line \NoCaseChange{\protect\cite{cite1312,cite1313,cite32}\protect\cite[{Ch. 3.2}]{cite1314}}. The \(C_L\) and \(C_{\Omega}\) constructions yield the same family of codes on the projective line (up to diagonal equivalence).
\item\relax
\flmRefsHyperref[eczindexfamilyrel]{code:mds}{Maximum distance separable (MDS) code} --- GRS codes have distance \(n-k+1\), saturating the Singleton bound.
\end{eczvaluelist}
\codefieldsection{Child}
\begin{eczvaluelist}
\item\relax
\flmRefsHyperref[eczindexfamilyrel]{code:reed_solomon}{Reed-Solomon (RS) code} --- A GRS code for which all multipliers \(v_i\) are unity reduces to an RS code \NoCaseChange{\protect\cite[{Def. 15.3.19}]{cite26}}.
\end{eczvaluelist}
\codefieldsection{Cousins}
\begin{eczvaluelist}
\item\relax
\flmRefsHyperref[eczindexfamilyrel]{code:dual}{Dual linear code} --- The dual of a GRS code is another GRS code \NoCaseChange{\protect\cite[{pg. 304}]{cite41}}.
\item\relax
\flmRefsHyperref[eczindexfamilyrel]{code:distributed_storage}{Distributed-storage code} --- GRS codes are used in various cloud storage systems \NoCaseChange{\protect\cite{cite264}}.
\item\relax
\flmRefsHyperref[eczindexfamilyrel]{code:concatenated}{Concatenated code} --- Concatenations of GRS codes with random linear codes almost surely attain the \flmRefsHyperref{ref85}{GV bound} \NoCaseChange{\protect\cite{cite973}}.
\item\relax
\flmRefsHyperref[eczindexfamilyrel]{code:random}{Random code} --- Concatenations of GRS codes with random linear codes almost surely attain the \flmRefsHyperref{ref85}{GV bound} \NoCaseChange{\protect\cite{cite973}}.
\item\relax
\flmRefsHyperref[eczindexfamilyrel]{code:q-ary_linear}{Linear \(q\)-ary code} --- Concatenations of GRS codes with random linear codes almost surely attain the \flmRefsHyperref{ref85}{GV bound} \NoCaseChange{\protect\cite{cite973}}.
\item\relax
\flmRefsHyperref[eczindexfamilyrel]{code:hermitian}{Hermitian code} --- Hermitian codes are concatenated GRS codes \NoCaseChange{\protect\cite{cite1842}}.
\item\relax
\flmRefsHyperref[eczindexfamilyrel]{code:extended_reed_solomon}{Extended GRS code} --- Extended GRS codes can be thought of as GRS codes that include an evaluation point of zero.
\item\relax
\flmRefsHyperref[eczindexfamilyrel]{code:goppa}{Goppa code} --- Goppa codes are \(\mathbb{F}_q\)-\flmRefsHyperref{ref33}{subfield} subcodes of the dual of the GRS code over \(\mathbb{F}_{q^m}\) with evaluation points \(\alpha_i\) and factors \(v_i=G(\alpha_i)^{-1}\) \NoCaseChange{\protect\cite{cite32}\protect\cite[{pg. 523}]{cite126}}. Since GRS codes are also residue AG codes on \(PG(1,q^m)\) \NoCaseChange{\protect\cite[{Prop. 15.3.26}]{cite26}}, Goppa codes are subfield subcodes of residue AG codes \NoCaseChange{\protect\cite[{Rem. 15.3.27}]{cite26}\protect\cite[{Thm. 15.3.28}]{cite26}}; when the base field equals the ambient field, they coincide with such GRS codes.
\item\relax
\flmRefsHyperref[eczindexfamilyrel]{code:alternant}{Alternant code} --- Alternant codes are \flmRefsHyperref{ref33}{subfield} subcodes of GRS codes \NoCaseChange{\protect\cite{cite1727}}.
\item\relax
\flmRefsHyperref[eczindexfamilyrel]{code:generalized_srivastava}{Generalized Srivastava code} --- Generalized Srivastava codes for \(m=1\) are equivalent to GRS codes \NoCaseChange{\protect\cite{cite1843}\protect\cite[{pg. 359}]{cite41}}.
\item\relax
\flmRefsHyperref[eczindexfamilyrel]{code:quantum_mds}{Quantum maximum-distance-separable (MDS) code} --- Some quantum MDS codes are constructed from cyclic and constacyclic codes \NoCaseChange{\protect\cite{cite1653}} which are GRS codes \NoCaseChange{\protect\cite{cite1844,cite1845}}.
\item\relax
\flmRefsHyperref[eczindexfamilyrel]{code:quantum_convolutional}{Quantum convolutional code} --- GRS codes can be used to construct quantum convolutional codes \NoCaseChange{\protect\cite[{Ch. 12}]{cite872}}.
\item\relax
\flmRefsHyperref[eczindexfamilyrel]{code:galois_fqrs}{Folded quantum RS (FQRS) code} --- A folded quantum generalized RS (GRS) code can be constructed in similar fashion from GRS codes as FQRS codes are constructed from FRS codes \NoCaseChange{\protect\cite[{Sec. 3}]{cite495}}.
\item\relax
\flmRefsHyperref[eczindexfamilyrel]{code:galois_grs}{Galois-qudit GRS code} --- Galois-qudit GRS codes are quantum analogues of generalized RS codes.
\end{eczvaluelist}
\eczhbkcontributors{ Muhammad Junaid Aftab, Qingfeng (Kee) Wang, \eczhuVVA }
\endeczcode

\eczcode{generalized_srivastava}{Generalized Srivastava code}{~\NoCaseChange{\protect\cite{cite1846}}}
\codefieldsection{Description}
An \([n,k \geq n-mst,d \geq st+1 ]_q\) alternant code defined for \(n+s\) distinct elements \(\alpha_1,\alpha_2,\cdots,\alpha_n,w_1,w_2,\cdots,w_s\) and \(n\) nonzero elements \(z_1,z_2,\cdots,z_n\) of \(\mathbb{F}_{q^m}\).

The code's parity-check matrix is \NoCaseChange{\protect\cite[{pg. 358}]{cite41}}
\flmMathEnvironment{align}{}{
  H=\begin{pmatrix}
  H_{1}\\
  H_{2}\\
  \vdots\\
  H_{s}
  \end{pmatrix}~,
}
where, for \(l=1,\ldots,s\),
\flmMathEnvironment{align}{}{
  H_{l}=\begin{pmatrix}
  \frac{z_{1}}{\alpha_{1}-w_{l}} & \frac{z_{2}}{\alpha_{2}-w_{l}} & \cdots & \frac{z_{n}}{\alpha_{n}-w_{l}}\\
  \frac{z_{1}}{\left(\alpha_{1}-w_{l}\right)^{2}} & \frac{z_{2}}{\left(\alpha_{2}-w_{l}\right)^{2}} & \cdots & \frac{z_{n}}{\left(\alpha_{n}-w_{l}\right)^{2}}\\
  \vdots & \vdots & \ddots & \vdots\\
  \frac{z_{1}}{\left(\alpha_{1}-w_{l}\right)^{t}} & \frac{z_{2}}{\left(\alpha_{2}-w_{l}\right)^{t}} & \cdots & \frac{z_{n}}{\left(\alpha_{n}-w_{l}\right)^{t}}
  \end{pmatrix}~.
}

\codefieldsection{Parent}
\begin{eczvaluelist}
\item\relax
\flmRefsHyperref[eczindexfamilyrel]{code:alternant}{Alternant code} --- Generalized Srivastava codes are a special case of alternant codes \NoCaseChange{\protect\cite[{Ch. 12}]{cite41}}.
\end{eczvaluelist}
\codefieldsection{Child}
\begin{eczvaluelist}
\item\relax
\flmRefsHyperref[eczindexfamilyrel]{code:srivastava}{Srivastava code} --- A Srivastava code is a special case of a generalized Srivastava code for \(z_j = \alpha_j^{\mu}\) for some \(\mu\) and \(t=1\).
\end{eczvaluelist}
\codefieldsection{Cousins}
\begin{eczvaluelist}
\item\relax
\flmRefsHyperref[eczindexfamilyrel]{code:generalized_reed_solomon}{Generalized RS (GRS) code} --- Generalized Srivastava codes for \(m=1\) are equivalent to GRS codes \NoCaseChange{\protect\cite{cite1843}\protect\cite[{pg. 359}]{cite41}}.
\item\relax
\flmRefsHyperref[eczindexfamilyrel]{code:mds}{Maximum distance separable (MDS) code} --- Generalized Srivastava codes for \(m=1\) are MDS codes \NoCaseChange{\protect\cite[{pg. 359}]{cite41}}.
\item\relax
\flmRefsHyperref[eczindexfamilyrel]{code:narrow_sense_q-ary_bch}{Primitive narrow-sense BCH code} --- Binary primitive generalized Srivastava codes with \(z_i=1\) and \(s=1\) are primitive narrow-sense BCH codes \NoCaseChange{\protect\cite[{pg. 359}]{cite41}}.
\end{eczvaluelist}
\eczhbkcontributors{ \eczhuVVA }
\endeczcode

\eczcode{goppa}{Goppa code}{~\NoCaseChange{\protect\cite{cite1847,cite1848,cite1849}}}
\codefieldsection{Alternative Names}
\begin{eczvaluelist}
\item\relax LG code
\end{eczvaluelist}
\eczhIndexCodeAliasName{goppa}{LG code}
\codefieldsection{Description}
A linear \(q\)-ary code defined from a polynomial \(G(x)\) over an extension field and a set of evaluation points \(L\) avoiding the roots of \(G\).
Goppa codes form a central family of alternant codes, admit efficient algebraic decoding algorithms, and include the binary Goppa codes used in the McEliece cryptosystem.
When the base field equals the coefficient field, they coincide with residue AG codes on \(PG(1,q^m)\); in general, classical Goppa codes are subfield subcodes of such AG codes \NoCaseChange{\protect\cite[{Rem. 15.3.27}]{cite26}\protect\cite[{Thm. 15.3.28}]{cite26}}.

Let \(G(x)\in \mathbb{F}_{q^m}[x]\) be a polynomial of degree \(r\), and let \(L=\{\alpha_1,\ldots,\alpha_n\}\subseteq \mathbb{F}_{q^m}\) be a set of distinct elements such that \(G(\alpha_i)\neq 0\) for all \(i\).
The Goppa code \(\Gamma(L,G)\) is the \([n,k,d]_q\) linear code consisting of all vectors \(a=(a_1,\ldots,a_n)\in\mathbb{F}_q^n\) such that
\flmMathEnvironment{align}{}{
  \sum_{i=1}^n \frac{a_i}{x-\alpha_i} \equiv 0 \pmod{G(x)}~.
}
Their duals are evaluation codes that are sometimes called \textit{geometric RS codes} \NoCaseChange{\protect\cite[{Thm. 2.71}]{cite32}}.

\codefieldsection{Protection}
The length is \(n=|L|\), the dimension satisfies \(k \geq n-mr\) where \(r=\deg G(x)\), and the minimum distance satisfies \(d \geq r+1\).
\codefieldsection{Decoding}
\begin{eczvaluelist}
\item\relax Algebraic decoding algorithms \NoCaseChange{\protect\cite{cite1850}}. If \( \text{deg} G(x) = 2t \) , then there exists a \(t\)-correcting algebraic decoding algorithm for \( \Gamma(L,G) \).
\item\relax Sugiyama et al. modification of the extended Euclidean algorithm \NoCaseChange{\protect\cite{cite1237,cite1238}}.
\item\relax Binary Goppa codes can be decoded using an RS-based decoder \NoCaseChange{\protect\cite{cite1851}}.
\item\relax List decoder for binary Goppa codes \NoCaseChange{\protect\cite{cite1852}}.
\end{eczvaluelist}
\codefieldsection{Realizations}
\begin{eczvaluelist}
\item\relax The McEliece public-key cryptosystem \NoCaseChange{\protect\cite{cite265,cite266}}. The protocol relies on the assumptions that Goppa-code generator matrices are hard to distinguish from random linear codes. However, there is an algorithm distinguishing between the two code classes in a time subexponential in \(n\) \NoCaseChange{\protect\cite{cite271}}.
\end{eczvaluelist}
\codefieldsection{Notes}
\begin{eczvaluelist}
\item\relax GAP function \flmHref{https://www.gap-system.org/Manuals/pkg/guava/doc/chap5.html\#X7EE808BB7D1E487A}{GoppaCode(G,L)} takes in a polynomial \(G\) that satisfies the necessary conditions for a Goppa code and a list \(L\) that contains elements in \(\mathbb{F}_q\) that are not roots of \(G\). It returns a Goppa code.
\end{eczvaluelist}
\codefieldsection{Parents}
\begin{eczvaluelist}
\item\relax
\flmRefsHyperref[eczindexfamilyrel]{code:cartier}{Cartier code} --- Goppa codes are Cartier codes from a curve of genus zero \NoCaseChange{\protect\cite{cite1743}}.
\item\relax
\flmRefsHyperref[eczindexfamilyrel]{code:alternant}{Alternant code} --- Goppa codes are a special case of alternant codes \NoCaseChange{\protect\cite[{Ch. 12}]{cite41}}.
\end{eczvaluelist}
\codefieldsection{Children}
\begin{eczvaluelist}
\item\relax
\flmRefsHyperref[eczindexfamilyrel]{code:narrow_sense_q-ary_bch}{Primitive narrow-sense BCH code} --- Primitive narrow-sense BCH codes are Goppa codes with \(L=\{1,\alpha^{-1},\cdots,\alpha^{1-n}\}\) and \(G(x)=x^{\delta-1}\) \NoCaseChange{\protect\cite[{pg. 522}]{cite126}}.
\item\relax
\flmRefsHyperref[eczindexfamilyrel]{code:srivastava}{Srivastava code} --- Generalized Srivastava codes are a special case of Goppa codes \NoCaseChange{\protect\cite[{Ch. 12}]{cite41}}.
\end{eczvaluelist}
\codefieldsection{Cousins}
\begin{eczvaluelist}
\item\relax
\flmRefsHyperref[eczindexfamilyrel]{code:generalized_reed_solomon}{Generalized RS (GRS) code} --- Goppa codes are \(\mathbb{F}_q\)-\flmRefsHyperref{ref33}{subfield} subcodes of the dual of the GRS code over \(\mathbb{F}_{q^m}\) with evaluation points \(\alpha_i\) and factors \(v_i=G(\alpha_i)^{-1}\) \NoCaseChange{\protect\cite{cite32}\protect\cite[{pg. 523}]{cite126}}. Since GRS codes are also residue AG codes on \(PG(1,q^m)\) \NoCaseChange{\protect\cite[{Prop. 15.3.26}]{cite26}}, Goppa codes are subfield subcodes of residue AG codes \NoCaseChange{\protect\cite[{Rem. 15.3.27}]{cite26}\protect\cite[{Thm. 15.3.28}]{cite26}}; when the base field equals the ambient field, they coincide with such GRS codes.
\item\relax
\flmRefsHyperref[eczindexfamilyrel]{code:q-ary_ltc}{\(q\)-ary linear LTC} --- Goppa codes are locally testable \NoCaseChange{\protect\cite{cite1270}}.
\item\relax
\flmRefsHyperref[eczindexfamilyrel]{code:gbch}{Chien-Choy generalized BCH (GBCH) code} --- In the binary case, GBCH\((z^{n-1},G)\) is the Goppa code \(\Gamma(L,G)\) where \(L\) consists of the \(n\)th roots of unity \NoCaseChange{\protect\cite[{pg. 360}]{cite41}}.
\item\relax
\flmRefsHyperref[eczindexfamilyrel]{code:binary_quantum_goppa}{Binary quantum Goppa code} --- Classical Goppa codes over various algebraic curves are used to construct quantum Goppa codes.
\end{eczvaluelist}
\eczhbkcontributors{ Manasi Shingane, \eczhuVVA }
\endeczcode

\eczcode{grassmannian_variety}{Grassmannian evaluation code}{~\NoCaseChange{\protect\cite{cite1853,cite1854,cite1855,cite27}}}
\codefieldsection{Description}
Evaluation code of polynomials evaluated on points lying on a finite-field Grassmannian embedded into projective space using the Plucker embedding \NoCaseChange{\protect\cite{cite27,cite28}}.

\codefieldsection{Parent}
\begin{eczvaluelist}
\item\relax
\flmRefsHyperref[eczindexfamilyrel]{code:flag_variety}{Flag-variety code} --- Grassmannian evaluation codes are flag-variety evaluation codes with the flag variety being a Grassmannian.
\end{eczvaluelist}
\codefieldsection{Cousins}
\begin{eczvaluelist}
\item\relax
\flmRefsHyperref[eczindexfamilyrel]{code:griesmer}{Griesmer code} --- The \([35,6,16]\) Grassmannian evaluation code, whose points lie on the Grassmannian \({\mathbb{G}(2,4)}\), attains the Griesmer bound \NoCaseChange{\protect\cite{cite1813}}.
\item\relax
\flmRefsHyperref[eczindexfamilyrel]{code:finite_grassmann}{Constant-dimension code} --- Grassmannian evaluation codes and constant-dimension codes are both built from the finite-field Grassmannian: the former evaluate functions on its points, while the latter use its \(k\)-dimensional subspaces themselves as codewords \NoCaseChange{\protect\cite{cite27,cite28}}.
\item\relax
\flmRefsHyperref[eczindexfamilyrel]{code:schubert}{Schubert evaluation code} --- Schubert varieties are subvarieties of Grassmannians, and Schubert evaluation codes were initially constructed as a generalization of Grassmannian evaluation codes.
\end{eczvaluelist}
\eczhbkcontributors{ \eczhuVVA }
\endeczcode

\eczcode{griesmer}{Griesmer code}{~\NoCaseChange{\protect\cite{cite1856,cite1857,cite1858}}}
\codefieldsection{Description}
A type of \(q\)-ary code whose parameters satisfy the Griesmer bound with equality.

A \([n,k,d]_q\) code is a Griesmer code if parameters \(n\), \(k\), \(d\), and \(q\) are such that the Griesmer bound
\flmMathEnvironment{align}{}{
  n\geq\sum_{j=0}^{k-1}\left\lceil \frac{d}{q^{j}}\right\rceil ~,
}
where \(\left\lceil x\right\rceil \) is the ceiling function, becomes an equality.

An \([n,2,d]_q\) code exists if and only if it is not excluded by the Griesmer bound. Every Griesmer code is generated by its minimum-weight codewords \NoCaseChange{\protect\cite{cite1859}}.

\codefieldsection{Parent}
\begin{eczvaluelist}
\item\relax
\flmRefsHyperref[eczindexfamilyrel]{code:mds}{Maximum distance separable (MDS) code} --- Singleton bound implies the Griesmer bound.
\end{eczvaluelist}
\codefieldsection{Children}
\begin{eczvaluelist}
\item\relax
\flmRefsHyperref[eczindexfamilyrel]{code:denniston}{Denniston code}\item\relax
\flmRefsHyperref[eczindexfamilyrel]{code:q-ary_simplex}{\(q\)-ary simplex code} --- Simplex codes saturate the Griesmer bound (\NoCaseChange{\protect\cite{cite62}}, Exer. 5.1.11).
\end{eczvaluelist}
\codefieldsection{Cousins}
\begin{eczvaluelist}
\item\relax
\flmRefsHyperref[eczindexfamilyrel]{code:divisible}{Divisible code} --- If a \(p\)-ary Griesmer code with \(p\) prime is such that a power of \(p\) divides the distance, then the code is divisible by that power \NoCaseChange{\protect\cite{cite1777}}.
\item\relax
\flmRefsHyperref[eczindexfamilyrel]{code:anticode}{Anticode} --- Several anticode (e.g., \NoCaseChange{\protect\cite{cite1220,cite1221}}) and related \NoCaseChange{\protect\cite{cite1222}} constructions saturate the Griesmer bound; see Refs. \NoCaseChange{\protect\cite{cite62,cite63,cite41}} for more details.
\item\relax
\flmRefsHyperref[eczindexfamilyrel]{code:bch}{Binary BCH code} --- The \([15,5,7]\) BCH code extended with a parity check saturates the Griesmer bound \NoCaseChange{\protect\cite[{pg. 157}]{cite62}}.
\item\relax
\flmRefsHyperref[eczindexfamilyrel]{code:kasami}{\([2^{2r}-1, 3r, 2^{2r-1} - 2^{r-1} ]\) Kasami code} --- Kasami codes satisfy the Griesmer bound for certain parameters \NoCaseChange{\protect\cite{cite1155}}.
\item\relax
\flmRefsHyperref[eczindexfamilyrel]{code:hamming743}{\([7,4,3]\) Hamming code} --- Starting with the \([6,3,3]\) shortened Hamming code and applying the \((u|u+v)\) construction recursively using the repetition code yields a family of \([2^m,m+1,2^{m-1}]\) codes for \(m\geq1\) that saturate the Griesmer bound \NoCaseChange{\protect\cite[{pg. 90}]{cite62}}.
\item\relax
\flmRefsHyperref[eczindexfamilyrel]{code:projective_reed_muller}{Projective RM (PRM) code} --- PRM codes for \(r=1\) attain the Griesmer bound for all \(m\) \NoCaseChange{\protect\cite{cite1813}}.
\item\relax
\flmRefsHyperref[eczindexfamilyrel]{code:grassmannian_variety}{Grassmannian evaluation code} --- The \([35,6,16]\) Grassmannian evaluation code, whose points lie on the Grassmannian \({\mathbb{G}(2,4)}\), attains the Griesmer bound \NoCaseChange{\protect\cite{cite1813}}.
\item\relax
\flmRefsHyperref[eczindexfamilyrel]{code:projective}{Projective geometry code} --- Arcs corresponding to Griesmer codes are called Griesmer arcs \NoCaseChange{\protect\cite[{pg. 248}]{cite1860}}. There is a one-to-one correspondence between Griesmer codes and minihypers \NoCaseChange{\protect\cite{cite1861,cite1862}}; see \NoCaseChange{\protect\cite{cite1863}\protect\cite[{Sec. 14.2.4}]{cite202}} for more details.
\item\relax
\flmRefsHyperref[eczindexfamilyrel]{code:galois_css}{Galois-qudit CSS code} --- A quantum version of the Griesmer bound has been derived for Galois-qudit CSS codes \NoCaseChange{\protect\cite{cite1864}} and Galois-qudit stabilizer codes \NoCaseChange{\protect\cite{cite1865}}.
\end{eczvaluelist}
\eczhbkcontributors{ \eczhuVVA }
\endeczcode

\eczcode{group}{Group-algebra code}{~\NoCaseChange{\protect\cite{cite1576}}}
\codefieldsection{Alternative Names}
\begin{eczvaluelist}
\item\relax \(G\) code
\end{eczvaluelist}
\eczhIndexCodeAliasName{group}{\(G\) code}
\codefieldsection{Description}
An \( [n,k]_q \) code associated with a finite group \(G\) of order \(n\), viewed as an ideal in the group algebra \(\mathbb{F}_q[G]\) \NoCaseChange{\protect\cite[{Def. 16.4.3}]{cite196}}.
Equivalently, after identifying the \(n\) coordinate positions of each codeword with elements of \(G\), the code is invariant under the regular action of \(G\) and thus becomes a \(G\)-submodule of \(\mathbb{F}_q^n\) \NoCaseChange{\protect\cite{cite198}\protect\cite[{Lemma 2.3}]{cite197}}.
A group-algebra code for an Abelian group is called an \textit{Abelian group-algebra code}.

\subsection{Group algebra}\label{ref205}
For a given field \(\mathbb{F}_q\) and a finite group \(G\) of order
\(|G|=\ell\), the \textit{group algebra} (a ring) \(\mathbb{F}_q[G]\) is defined as an
\(\mathbb{F}_q\)-linear space of all formal sums
\flmMathEnvironment{align}{}{
  x\equiv \sum_{g\in G}x_g g,\quad x_g\in \mathbb{F}_q,\label{ref1866}
}
where group elements \(g\in G\) serve as basis vectors,
equipped with the product naturally associated with the group
operation,
\flmMathEnvironment{align}{}{
  ab=\sum_{g\in G}\biggl(\sum_{h\in G} a_h b_{h^{-1}g}\biggr) g, \quad a,b\in \mathbb{F}_q[G].\label{ref1867}
}
Semisimple group algebras can be decomposed into simple components via a Wedderburn-Artin decomposition \NoCaseChange{\protect\cite{cite1869}\protect\cite[{Thm. 4.4, p. 112}]{cite1868}}.

\subsection{Group-algebra code}
A group-algebra code is a \( k \)-dimensional linear subspace of the \flmRefsHyperref{ref205}{group algebra} of \( G\) with coefficients in \(\mathbb{F}_q\). The formal definition is that a group-algebra code is a left, right, or two-sided ideal in the \flmRefsHyperref{ref205}{group algebra} \( \mathbb{F}_q G \).

A linear code is a group-algebra code for a group \(G\) if and only if \(G\) is isomorphic to a regular subgroup of the code's permutation automorphism group \NoCaseChange{\protect\cite{cite1870}\protect\cite[{Thm. 16.4.7}]{cite196}}.

\codefieldsection{Notes}
\begin{eczvaluelist}
\item\relax See \NoCaseChange{\protect\cite[{Def. 16.3.1}]{cite196}\protect\cite[{Def. 16.4.3}]{cite196}\protect\cite[{pg. 58}]{cite1314}} for introductions to group algebras and group-algebra codes.
\item\relax Not all Abelian group-algebra codes are for cyclic groups (cyclic codes) or for elementary Abelian \( p \) groups (e.g. Reed Muller codes \NoCaseChange{\protect\cite{cite1226}}). For example, there is a binary code with parameters \( [45,13,16] \) which is an Abelian group-algebra code for the group \( G = \mathbb{Z}_3 \times \mathbb{Z}_{15} \).
\end{eczvaluelist}
\codefieldsection{Parent}
\begin{eczvaluelist}
\item\relax
\flmRefsHyperref[eczindexfamilyrel]{code:quasi_group}{Quasi group-algebra code} --- A quasi group-algebra code of index \(\ell=1\) is a group-algebra code.
\end{eczvaluelist}
\codefieldsection{Children}
\begin{eczvaluelist}
\item\relax
\flmRefsHyperref[eczindexfamilyrel]{code:multiplicity}{Multiplicity code} --- Multiplicity codes of order \(s\) are Abelian group-algebra codes whose corresponding polynomial that is modded out is \((x-\alpha_j)^s\) for each evaluation point \(\alpha_j\) \NoCaseChange{\protect\cite{cite1826}}.
\item\relax
\flmRefsHyperref[eczindexfamilyrel]{code:folded_reed_solomon}{Folded RS (FRS) code} --- FRS codes are polynomial ideal codes whose corresponding polynomial is a product of the polynomials of the RS codes that are being folded \NoCaseChange{\protect\cite{cite1826}}.
\item\relax
\flmRefsHyperref[eczindexfamilyrel]{code:q-ary_cyclic}{Cyclic linear \(q\)-ary code} --- A length-\(n\) cyclic \(q\)-ary linear code is an Abelian group-algebra code for the cyclic group with \(n\) elements \( \mathbb{Z}_n \) \NoCaseChange{\protect\cite[{Exam. 16.4.9}]{cite196}}.
\end{eczvaluelist}
\codefieldsection{Cousins}
\begin{eczvaluelist}
\item\relax
\flmRefsHyperref[eczindexfamilyrel]{code:group_orbit}{Group-orbit code} --- A group-algebra code admits a regular, i.e., free and transitive, action on coordinates by a subgroup of its permutation automorphism group \NoCaseChange{\protect\cite[{Thm. 16.4.7}]{cite196}}. This differs from a group-orbit code, whose defining group action is transitive on codewords.
\item\relax
\flmRefsHyperref[eczindexfamilyrel]{code:extended_golay}{\([24, 12, 8]\) Extended Golay code} --- The extended Golay code is a group-algebra code for various groups \NoCaseChange{\protect\cite{cite1197,cite1194,cite1193}}; see \NoCaseChange{\protect\cite{cite1115}\protect\cite[{Exam. 16.5.1}]{cite196}}.
\item\relax
\flmRefsHyperref[eczindexfamilyrel]{code:self_dual_48_24_12}{\([48,24,12]\) self-dual code} --- The \([48,24,12]\) self-dual code is a group code for \(G\) being a dihedral group \NoCaseChange{\protect\cite{cite1203}\protect\cite[{Exam. 16.5.1}]{cite196}}.
\item\relax
\flmRefsHyperref[eczindexfamilyrel]{code:hamming844}{\([8,4,4]\) extended Hamming code} --- The \([8,4,4]\) extended Hamming code is a group-algebra code for the group \(\mathbb{Z}_2 \times \mathbb{Z}_4\) \NoCaseChange{\protect\cite{cite1115}}.
\item\relax
\flmRefsHyperref[eczindexfamilyrel]{code:reed_muller}{Reed-Muller (RM) code} --- RM codes are group-algebra codes \NoCaseChange{\protect\cite{cite1576,cite1577}\protect\cite[{Exam. 16.4.11}]{cite196}}. Consider a binary vector space of dimension \( m \). Under addition, this forms a finite group with \( 2^m \) elements known as an elementary Abelian 2-group -- the direct product of \( m \) two-element cyclic groups \( \mathbb{Z}_2 \times \dots \times \mathbb{Z}_2 \). Denote this group by \( G_m \). Let \( J \) be the Jacobson radical of the \flmRefsHyperref{ref205}{group algebra} \( \mathbb{F}_2 G_m \). RM\((r,m)\) codes correspond to the ideal \( J^{m-r} \). The length of the code is \( |G_m| = 2^m \), the distance is \( 2^{m-r} \), and the dimension is \( \sum_{i=0}^r {m \choose i} \). A similar construction exists for choices of a prime \( p\neq 2 \).
\item\relax
\flmRefsHyperref[eczindexfamilyrel]{code:hermitian}{Hermitian code} --- Some Hermitian codes are group-algebra codes \NoCaseChange{\protect\cite{cite1871}\protect\cite[{Remark 16.4.14}]{cite196}}.
\item\relax
\flmRefsHyperref[eczindexfamilyrel]{code:klein_quartic}{Klein-quartic code} --- Some Klein-quartic codes are group-algebra codes \NoCaseChange{\protect\cite[{Remark 16.4.14}]{cite196}}.
\item\relax
\flmRefsHyperref[eczindexfamilyrel]{code:suzuki}{Suzuki-curve code} --- Some Suzuki-curve codes are group-algebra codes \NoCaseChange{\protect\cite[{Remark 16.4.14}]{cite196}}.
\item\relax
\flmRefsHyperref[eczindexfamilyrel]{code:generalized_reed_muller}{Generalized RM (GRM) code} --- GRM codes over prime-power fields are group-algebra codes \NoCaseChange{\protect\cite{cite1576,cite1577,cite1837}\protect\cite[{Exam. 16.4.11}]{cite196}}.
\item\relax
\flmRefsHyperref[eczindexfamilyrel]{code:lcd}{Linear code with complementary dual (LCD)} --- A group code \(C \leq \mathbb{F}_q G\) is LCD if and only if \(C=e \mathbb{F}_q G\) for an idempotent \(e\) satisfying \(e=\hat{e}\), and then \(C^{\perp}=(1-e)\mathbb{F}_q G\) \NoCaseChange{\protect\cite[{Thm. 16.7.6}]{cite196}}.
\item\relax
\flmRefsHyperref[eczindexfamilyrel]{code:self_dual}{Self-dual linear code} --- Self-dual group codes exist exactly when the base field has characteristic \(2\) and the underlying group has even order \NoCaseChange{\protect\cite[{Thm. 16.5.4}]{cite196}}.
\item\relax
\flmRefsHyperref[eczindexfamilyrel]{code:q-ary_simplex}{\(q\)-ary simplex code} --- Over a prime field \(\mathbb{F}_p\), simplex codes with parameters \([(p^m-1)/(p-1),m,p^{m-1}]_p\) and \(\gcd(m,p-1)=1\) are group-algebra codes \NoCaseChange{\protect\cite[{Exam. 16.8.2}]{cite196}}.
\item\relax
\flmRefsHyperref[eczindexfamilyrel]{code:divisible}{Divisible code} --- If \(C\) is a group code over a field of characteristic \(p\), then the monomial kernel \(K_M(C)\) has order dividing the weight of every codeword, and the \(p^{\prime}\)-part of the divisor of \(C\) equals the \(p^{\prime}\)-part of \(|K_M(C)|\) \NoCaseChange{\protect\cite[{Thm. 16.8.3}]{cite196}}.
\item\relax
\flmRefsHyperref[eczindexfamilyrel]{code:2bga}{Two-block group-algebra (2BGA) codes} --- A 2BGA code \(LP(a,b)\) is constructible as a hypergraph-product code when the support subgroups generated by \(a\) and \(b\) are disjoint. In that case, the commuting matrices simultaneously acquire hypergraph-product Kronecker-product form, and the code can be obtained from a pair of classical group-algebra codes \NoCaseChange{\protect\cite[{Statements 8 and 12}]{cite842}}.

\end{eczvaluelist}
\eczhbkcontributors{ Ian Teixeira, \eczhuVVA }
\endeczcode

\eczcode{toric_classical}{Hansen toric code}{~\NoCaseChange{\protect\cite{cite1872,cite1873}}}
\codefieldsection{Description}
Evaluation code of a linear space of polynomials evaluated on points lying on an affine or projective toric variety. If the space is taken to be all polynomials up to some degree, the code is called a \textit{toric RM-type code} of that degree.

\codefieldsection{Protection}
Parameters of toric RM-type codes and various generalizations have been determined in Refs. \NoCaseChange{\protect\cite{cite1874,cite1875,cite1876,cite1877,cite1878,cite1879}}.
\codefieldsection{Notes}
\begin{eczvaluelist}
\item\relax See Ref. \NoCaseChange{\protect\cite{cite1879}} for various examples and implementations in Magma.
\end{eczvaluelist}
\codefieldsection{Parent}
\begin{eczvaluelist}
\item\relax
\flmRefsHyperref[eczindexfamilyrel]{code:evaluation_polynomial}{Polynomial evaluation code} --- Hansen toric codes are polynomial evaluation codes with \(\cal X\) being a toric variety.
\end{eczvaluelist}
\codefieldsection{Cousin}
\begin{eczvaluelist}
\item\relax
\flmRefsHyperref[eczindexfamilyrel]{code:toric}{Toric code} --- The toric code is not to be confused with the CSS code constructed from a polynomial evaluation code on a toric variety \NoCaseChange{\protect\cite{cite1880}}.
\end{eczvaluelist}
\eczhbkcontributors{ \eczhuVVA }
\endeczcode

\eczcode{hermitian}{Hermitian code}{~\NoCaseChange{\protect\cite{cite1881,cite1809}\protect\cite[{Sec. 4.4.3}]{cite1312}}}
\codefieldsection{Alternative Names}
\begin{eczvaluelist}
\item\relax Hermitian AG code
\end{eczvaluelist}
\eczhIndexCodeAliasName{hermitian}{Hermitian AG code}
\codefieldsection{Description}
Evaluation AG code of rational functions on a Hermitian curve over \(\mathbb{F}_{q^2}\).

A standard affine model of the curve is
\flmMathEnvironment{align}{}{
  y^q+y=x^{q+1}~,
}
and this curve has \(q^3+1\) rational points and genus \(g=q(q-1)/2\).

Hermitian codes are usually defined as one-point AG codes \(C_L(D,mP_\infty)\), where \(D\) is the sum of the \(q^3\) affine rational points and \(P_\infty\) is the unique point at infinity.
They provide much longer code lengths than Reed-Solomon codes over the same alphabet, since the standard one-point construction has length \(q^3\) over \(\mathbb{F}_{q^2}\).

There exists a family of Hermitian codes with automorphism group \(PGL(2,\mathbb{F}_q)\) \NoCaseChange{\protect\cite{cite1882}}.

\codefieldsection{Protection}
Distance is determined by the divisor and the geometry of the Hermitian curve \NoCaseChange{\protect\cite{cite1883}}; see \NoCaseChange{\protect\cite[{Sec. 5.3}]{cite32}} for the general result. For the standard one-point code \(C_L(D,mP_\infty)\) of length \(n=q^3\), the designed distance satisfies \(d \geq n-m\).
\codefieldsection{Rate}
For the standard one-point code \(C_L(D,mP_\infty)\) with genus \(g=q(q-1)/2\), Riemann-Roch gives
\flmMathEnvironment{align}{}{
  k=m-g+1
}
whenever \(2g-2 < m < n=q^3\).

\codefieldsection{Decoding}
\begin{eczvaluelist}
\item\relax Unique decoding using syndromes and error locator ideals for polynomial evaluations. Note that Hermitian codes are linear codes so we can compute the syndrome of a received vector. Moreover, akin to the error-locator ideals found in decoding RS codes, for the multivariate case we must define an error locator ideal \(\Lambda \) such that the variety of this ideal over \(\mathbb{F}^{2}_q\) is exactly the set of errors. The Sakata algorithm uses these two ingredients to get a unique decoding procedure \NoCaseChange{\protect\cite{cite1802}}.
\end{eczvaluelist}
\codefieldsection{Notes}
\begin{eczvaluelist}
\item\relax Certain structured classical optimization problems can be mapped into decoding and list decoding Hermitian codes via the Decoded Quantum Interferomentry (DQI) algorithm \NoCaseChange{\protect\cite{cite1884}}.
\end{eczvaluelist}
\codefieldsection{Parent}
\begin{eczvaluelist}
\item\relax
\flmRefsHyperref[eczindexfamilyrel]{code:norm_trace}{Norm-trace code} --- Hermitian codes are evaluation AG codes with \(\cal X\) being a Hermitian curve \NoCaseChange{\protect\cite{cite1809}\protect\cite[{Exam. 2.74}]{cite32}}. This curve is maximal, meaning that Hermitian codes are evaluation AG codes with maximum possible length given a fixed genus. They are a special case of norm-trace codes \NoCaseChange{\protect\cite{cite1810}}.
\end{eczvaluelist}
\codefieldsection{Cousins}
\begin{eczvaluelist}
\item\relax
\flmRefsHyperref[eczindexfamilyrel]{code:generalized_reed_solomon}{Generalized RS (GRS) code} --- Hermitian codes are concatenated GRS codes \NoCaseChange{\protect\cite{cite1842}}.
\item\relax
\flmRefsHyperref[eczindexfamilyrel]{code:group}{Group-algebra code} --- Some Hermitian codes are group-algebra codes \NoCaseChange{\protect\cite{cite1871}\protect\cite[{Remark 16.4.14}]{cite196}}.
\item\relax
\flmRefsHyperref[eczindexfamilyrel]{code:hermitian_hypersurface}{Hermitian-hypersurface code} --- Hermitian-hypersurface codes reduce to Hermitian codes of polynomials when the hypersurface is a curve.
\item\relax
\flmRefsHyperref[eczindexfamilyrel]{code:tamo_barg_vladut}{Barg-Tamo-Vladut code} --- Hermitian-curve examples of these codes are given in \NoCaseChange{\protect\cite[{Rem. 15.9.16}]{cite26}}.
\item\relax
\flmRefsHyperref[eczindexfamilyrel]{code:quantum_hermitian_ag}{Quantum Hermitian AG code} --- Quantum Hermitian AG codes are quantum analogues of Hermitian codes.
\end{eczvaluelist}
\eczhbkcontributors{ Shashank Sule, \eczhuVVA }
\endeczcode

\eczcode{hermitian_hypersurface}{Hermitian-hypersurface code}{~\NoCaseChange{\protect\cite{cite1885}}}
\codefieldsection{Description}
Evaluation code of polynomials evaluated on points lying on a Hermitian hypersurface.

\codefieldsection{Parent}
\begin{eczvaluelist}
\item\relax
\flmRefsHyperref[eczindexfamilyrel]{code:flag_variety}{Flag-variety code} --- Hermitian-hypersurface codes are flag-variety evaluation codes with the flag variety being a Hermitian hypersurface.
\end{eczvaluelist}
\codefieldsection{Cousin}
\begin{eczvaluelist}
\item\relax
\flmRefsHyperref[eczindexfamilyrel]{code:hermitian}{Hermitian code} --- Hermitian-hypersurface codes reduce to Hermitian codes of polynomials when the hypersurface is a curve.
\end{eczvaluelist}
\eczhbkcontributors{ \eczhuVVA }
\endeczcode

\eczcode{hill_cap}{Hill projective-cap code}{~\NoCaseChange{\protect\cite{cite1886}}}
\codefieldsection{Description}
Member of a projective code family that contains two \(q\)-ary sharp configurations \NoCaseChange{\protect\cite[{Table 12.1}]{cite199}} and that is constructed using projective caps. 

\codefieldsection{Parent}
\begin{eczvaluelist}
\item\relax
\flmRefsHyperref[eczindexfamilyrel]{code:projective_two_weight}{Projective two-weight code} --- Hill projective-cap codes are projective two-weight codes on projective caps \NoCaseChange{\protect\cite[{Table 19.1}]{cite172}}.
\end{eczvaluelist}
\codefieldsection{Children}
\begin{eczvaluelist}
\item\relax
\flmRefsHyperref[eczindexfamilyrel]{code:hill_56_6_36}{\([56,6,36]_3\) Hill-cap code}\item\relax
\flmRefsHyperref[eczindexfamilyrel]{code:hill_78_6_56}{\([78,6,56]_4\) Hill-cap code}\end{eczvaluelist}
\eczhbkcontributors{ \eczhuVVA }
\endeczcode

\eczcode{hirschfeld}{Hirschfeld code}{~\NoCaseChange{\protect\cite{cite1887}}}
\codefieldsection{Description}
A \([q+1,4,q-2]_q\) projective geometry code for non-prime \(q\) that is an example of an MDS code that is not an RS code; see \NoCaseChange{\protect\cite[{Exam. 7.6}]{cite182}} for the generator matrix.

\codefieldsection{Parents}
\begin{eczvaluelist}
\item\relax
\flmRefsHyperref[eczindexfamilyrel]{code:projective}{Projective geometry code}\item\relax
\flmRefsHyperref[eczindexfamilyrel]{code:mds}{Maximum distance separable (MDS) code} --- The Hirschfeld code is a rare example of an MDS code that is not related to an RS code.
\end{eczvaluelist}
\eczhbkcontributors{ \eczhuVVA }
\endeczcode

\eczcode{cascaded_reed_solomon}{Hyperbolic evaluation code}{~\NoCaseChange{\protect\cite{cite30,cite31,cite29}}}
\codefieldsection{Alternative Names}
\begin{eczvaluelist}
\item\relax Hyperbolic cascaded RS code
\end{eczvaluelist}
\eczhIndexCodeAliasName{cascaded_reed_solomon}{Hyperbolic cascaded RS code}
\codefieldsection{Description}
An evaluation code over polynomials in two variables. 
Generator matrices are determined in Ref. \NoCaseChange{\protect\cite{cite29}} following initial formulations of the codes as generalized concatenations of RS codes \NoCaseChange{\protect\cite{cite30,cite31}}; see \NoCaseChange{\protect\cite[{Exam. 4.26}]{cite32}}. 

\codefieldsection{Protection}
Minimum distance is determined by an order bound \NoCaseChange{\protect\cite{cite29,cite1888}}.

\codefieldsection{Parent}
\begin{eczvaluelist}
\item\relax
\flmRefsHyperref[eczindexfamilyrel]{code:evaluation_polynomial}{Polynomial evaluation code} --- A hyperbolic evaluation code is an evaluation code over polynomials in two variables.
\end{eczvaluelist}
\codefieldsection{Child}
\begin{eczvaluelist}
\item\relax
\flmRefsHyperref[eczindexfamilyrel]{code:reed_muller}{Reed-Muller (RM) code} --- RM codes are special cases of hyperbolic evaluation codes \NoCaseChange{\protect\cite[{Thm. 3 proof}]{cite29}}.
\end{eczvaluelist}
\codefieldsection{Cousin}
\begin{eczvaluelist}
\item\relax
\flmRefsHyperref[eczindexfamilyrel]{code:reed_solomon}{Reed-Solomon (RS) code} --- Hyperbolic evaluation codes were initially formulated as generalized concatenations (a.k.a. cascades) of RS codes \NoCaseChange{\protect\cite{cite30,cite31}}.
\end{eczvaluelist}
\eczhbkcontributors{ \eczhuVVA }
\endeczcode

\eczcode{hyperoval}{Hyperoval code}{~\NoCaseChange{\protect\cite{cite1889}}}
\codefieldsection{Description}
Projective code constructed from a hyperoval in the projective plane \(PG(2,q)\), where \(q\) is even. Since a hyperoval is a set of \(q+2\) points with no three collinear, the corresponding projective code has parameters \([q+2,3,q]_q\) \NoCaseChange{\protect\cite[{Exam. 19.2.1}]{cite172}}; the \([6,3,4]_4\) hexacode is the smallest example.

\codefieldsection{Parent}
\begin{eczvaluelist}
\item\relax
\flmRefsHyperref[eczindexfamilyrel]{code:projective}{Projective geometry code}\end{eczvaluelist}
\codefieldsection{Child}
\begin{eczvaluelist}
\item\relax
\flmRefsHyperref[eczindexfamilyrel]{code:hexacode}{\([6,3,4]_4\) Hexacode} --- Columns of hexacode's generator matrix represent the six points of a hyperoval in the projective plane \(PG(2,4)\), an example of a two-weight code \NoCaseChange{\protect\cite[{pg. 289}]{cite62}\protect\cite[{Exam. 19.2.1}]{cite172}}.
\end{eczvaluelist}
\codefieldsection{Cousins}
\begin{eczvaluelist}
\item\relax
\flmRefsHyperref[eczindexfamilyrel]{code:delsarte_optimal_q-ary}{\(q\)-ary sharp configuration} --- Codes based on hyperovals in \(PG(2,q)\) are \(q\)-ary sharp configurations \NoCaseChange{\protect\cite[{Table 12.1}]{cite199}}.
\item\relax
\flmRefsHyperref[eczindexfamilyrel]{code:projective_two_weight}{Projective two-weight code} --- Codes based on hyperovals in \(PG(2,q)\) with even \(q\) are projective two-weight codes \NoCaseChange{\protect\cite{cite1656,cite1657}\protect\cite[{Exam. 19.2.1}]{cite172}\protect\cite[{Table 7.1}]{cite206}}.
\item\relax
\flmRefsHyperref[eczindexfamilyrel]{code:difference_set}{Difference-set cyclic (DSC) code} --- Hyperoval difference sets yield DSC codes \NoCaseChange{\protect\cite{cite1330}\protect\cite[{Ch. 6}]{cite1176}}.
\item\relax
\flmRefsHyperref[eczindexfamilyrel]{code:extended_reed_solomon}{Extended GRS code} --- Columns of parity-check matrices of triply extended RS codes consist of points of a hyperoval \NoCaseChange{\protect\cite[{Prop. 17.5}]{cite62}}.
\item\relax
\flmRefsHyperref[eczindexfamilyrel]{code:denniston}{Denniston code} --- Denniston codes for \(i=1\) realize the hyperoval case of maximal arcs in \(PG(2,2^a)\) \NoCaseChange{\protect\cite{cite1774}\protect\cite[{Sec. 19.7.3}]{cite172}}.
\end{eczvaluelist}
\eczhbkcontributors{ Alexander Barg, \eczhuVVA }
\endeczcode

\eczcode{incidence_matrix}{Incidence-matrix projective code}{~\NoCaseChange{\protect\cite{cite1890,cite1891,cite990}}}
\codefieldsection{Description}
A projective code whose generator matrix is the incidence matrix of points and hyperplanes in a projective space.
This construction has been generalized to incidence matrices of other structures \NoCaseChange{\protect\cite{cite200,cite201}\protect\cite[{Sec. 14.4}]{cite202}}.
More generally, columns of a code's parity-check matrix can also be organized as an incidence matrix.

\codefieldsection{Parent}
\begin{eczvaluelist}
\item\relax
\flmRefsHyperref[eczindexfamilyrel]{code:projective}{Projective geometry code}\end{eczvaluelist}
\codefieldsection{Child}
\begin{eczvaluelist}
\item\relax
\flmRefsHyperref[eczindexfamilyrel]{code:q-ary_simplex}{\(q\)-ary simplex code} --- Columns of a simplex code's generator matrix correspond to 1D subspaces of \(\mathbb{F}_q^m\), i.e., to points of \(PG(m-1,q)\).
\end{eczvaluelist}
\codefieldsection{Cousins}
\begin{eczvaluelist}
\item\relax
\flmRefsHyperref[eczindexfamilyrel]{code:simplex734}{\([7,3,4]\) simplex code} --- The \([7,3,4]\) simplex code is the smallest difference-set cyclic code, arising from the lines of the projective plane \(PG(2,2)\) \NoCaseChange{\protect\cite[{pg. 397}]{cite41}}.
\item\relax
\flmRefsHyperref[eczindexfamilyrel]{code:extended_hamming}{\([2^m,2^m-m-1,4]\) Extended Hamming code} --- Columns of an extended Hamming code's parity-check matrix correspond to points in \(PG(m-1,2)\) that lie in the complement of a hyperplane \NoCaseChange{\protect\cite[{pg. 182}]{cite961}}.
\item\relax
\flmRefsHyperref[eczindexfamilyrel]{code:hamming743}{\([7,4,3]\) Hamming code} --- The \([7,4,3]\) Hamming code parity-check matrix corresponds to points in the Fano plane \(PG(2,2)\) \NoCaseChange{\protect\cite[{Exam. 21.4.2}]{cite97}}.
\item\relax
\flmRefsHyperref[eczindexfamilyrel]{code:difference_set}{Difference-set cyclic (DSC) code} --- Planar difference-set cyclic codes arise from the incidence vectors of the lines of a projective plane \(PG(2,p^s)\) \NoCaseChange{\protect\cite[{pgs. 397-398}]{cite41}}.
\item\relax
\flmRefsHyperref[eczindexfamilyrel]{code:pg_ldpc}{Finite-geometry LDPC (FG-LDPC) code} --- The parity-check matrix of a PG-LDPC code is the incidence matrix of points and hyperplanes in a projective space.
\item\relax
\flmRefsHyperref[eczindexfamilyrel]{code:q-ary_hamming}{\(q\)-ary Hamming code} --- Columns of a Hamming parity-check matrix correspond to 1D subspaces of \(\mathbb{F}_q^r\).
\end{eczvaluelist}
\eczhbkcontributors{ \eczhuVVA }
\endeczcode

\eczcode{interleaved_reed_solomon}{Interleaved RS (IRS) code}{}
\codefieldsection{Description}
A modification of RS codes where multiple polynomials are used to define each codeword. 

Each codeword \(\mu\) of a \(t\)-interleaved RS code is a string of values of the corresponding set \(\{f_\mu^{(1)},f_\mu^{(2)},\cdots,f_\mu^{(t)}\}\) of \(t\) polynomials at the points \(\alpha_i\). The vector codewords can be arranged in an array whose rows are ordinary RS codes for each polynomial \(f_\mu^{(j)}\), yielding the encoding
\flmMathEnvironment{align}{}{
\mu\to\left(
\begin{array}{cccc}
  f_{\mu}^{(1)}\left(\alpha_{1}\right) & f_{\mu}^{(1)}\left(\alpha_{2}\right) & \cdots & f_{\mu}^{(1)}\left(\alpha_{n}\right)\\
  f_{\mu}^{(2)}\left(\alpha_{1}\right) & f_{\mu}^{(2)}\left(\alpha_{2}\right) &  & f_{\mu}^{(2)}\left(\alpha_{n}\right)\\
  \vdots &  & \ddots & \vdots\\
  f_{\mu}^{(t)}\left(\alpha_{1}\right) & f_{\mu}^{(t)}\left(\alpha_{2}\right) & \cdots & f_{\mu}^{(t)}\left(\alpha_{n}\right)
\end{array}\right)~.
}

\codefieldsection{Decoding}
\begin{eczvaluelist}
\item\relax Decoder that corrects up to \(1-\frac{2k+n}{3n}\) fraction of random errors \NoCaseChange{\protect\cite{cite1892}}.
\item\relax Decoder that corrects up to \(1-(\frac{k}{n})^{2/3}\) fraction of random errors \NoCaseChange{\protect\cite{cite1893}}.
\end{eczvaluelist}
\codefieldsection{Parent}
\begin{eczvaluelist}
\item\relax
\flmRefsHyperref[eczindexfamilyrel]{code:q-ary_linear}{Linear \(q\)-ary code} --- IRS codes are linear over \(\mathbb{F}_q\) but not necessarily over \(\mathbb{F}_{q^t}\).
\end{eczvaluelist}
\codefieldsection{Children}
\begin{eczvaluelist}
\item\relax
\flmRefsHyperref[eczindexfamilyrel]{code:cross_interleaved_reed_solomon}{Cross-interleaved RS (CIRS) code}\item\relax
\flmRefsHyperref[eczindexfamilyrel]{code:parvaresh_vardy}{Parvaresh-Vardy (PV) code} --- PV codes are IRS codes with specific algebraic relations between the codeword polynomials that allow for efficient list decoding.
\item\relax
\flmRefsHyperref[eczindexfamilyrel]{code:reed_solomon}{Reed-Solomon (RS) code} --- An IRS code utilizing one polynomial \(f\) reduces to an RS code.
\end{eczvaluelist}
\eczhbkcontributors{ \eczhuVVA }
\endeczcode

\eczcode{isbn}{International Standard Book Number (ISBN) code}{}
\codefieldsection{Description}
A 10-digit checksum code used worldwide to identify books in stores.
In the ISBN-10 convention, the modulo-\(11\) check digit can take values between \(0\) and \(10\), with \(10\) denoted by the symbol \(X\).

\codefieldsection{Protection}
The last digit of an ISBN-10 string is a check digit computed modulo 11 \NoCaseChange{\protect\cite{cite961}}.
\codefieldsection{Parents}
\begin{eczvaluelist}
\item\relax
\flmRefsHyperref[eczindexfamilyrel]{code:q-ary_digits_into_q-ary_digits}{\(q\)-ary code} --- The last digit of an ISBN-10 string is a check digit computed modulo 11 \NoCaseChange{\protect\cite{cite961}}.
\item\relax
\flmRefsHyperref[eczindexfamilyrel]{code:checksum}{Checksum code} --- The last digit of an ISBN-10 string is a check digit computed modulo 11 \NoCaseChange{\protect\cite{cite961}}.
\end{eczvaluelist}
\eczhbkcontributors{ \eczhuVVA }
\endeczcode

\eczcode{irregular_convolutional}{Irregular convolutional code (IRCC)}{~\NoCaseChange{\protect\cite{cite1894,cite1895}}}
\codefieldsection{Description}
A convolutional code whose parity-check matrix has a variable number of entries in each row or column.

\codefieldsection{Decoding}
\begin{eczvaluelist}
\item\relax Extrinsic information transfer (EXIT) charts \NoCaseChange{\protect\cite{cite1895}}.
\end{eczvaluelist}
\codefieldsection{Parent}
\begin{eczvaluelist}
\item\relax
\flmRefsHyperref[eczindexfamilyrel]{code:convolutional}{Convolutional code}\end{eczvaluelist}
\codefieldsection{Cousin}
\begin{eczvaluelist}
\item\relax
\flmRefsHyperref[eczindexfamilyrel]{code:quantum_irregular_convolutional}{Quantum irregular convolutional code (QIRCC)} --- Quantum irregular convolutional codes are quantum analogues of irregular convolutional codes.
\end{eczvaluelist}
\eczhbkcontributors{ \eczhuVVA }
\endeczcode

\eczcode{klein_quartic}{Klein-quartic code}{~\NoCaseChange{\protect\cite{cite1896}}}
\codefieldsection{Description}
Evaluation AG code over \(\mathbb{F}_8\) of rational functions evaluated on points lying on the Klein quartic, which is defined by the equation \(x^3 y + y^3 z + z^3 x = 0\) \NoCaseChange{\protect\cite[{Ex. 2.75}]{cite32}}.

\codefieldsection{Protection}
Dimension \(k=8\) and distance \(d \geq 13\). Concatenation with the \([4,3,2]\) single parity check code, conversion to a binary code by expressing \(\mathbb{F}_8\) elements as vectors over \(\mathbb{F}_2\), and puncturing yields a \([91,24,25]\) binary code that set the world record for codes of length 91 \NoCaseChange{\protect\cite{cite1897}}.
\codefieldsection{Parent}
\begin{eczvaluelist}
\item\relax
\flmRefsHyperref[eczindexfamilyrel]{code:evaluation}{Evaluation AG code} --- Klein-quartic codes are evaluation AG codes with \(\cal X\) being the Klein quartic (\NoCaseChange{\protect\cite[{Ex. 2}]{cite32}}.75)\NoCaseChange{\protect\cite{cite1314}}.
\end{eczvaluelist}
\codefieldsection{Cousin}
\begin{eczvaluelist}
\item\relax
\flmRefsHyperref[eczindexfamilyrel]{code:group}{Group-algebra code} --- Some Klein-quartic codes are group-algebra codes \NoCaseChange{\protect\cite[{Remark 16.4.14}]{cite196}}.
\end{eczvaluelist}
\eczhbkcontributors{ \eczhuVVA }
\endeczcode

\eczcode{ld_convolutional}{LDPC convolutional code (LDPC-CC)}{~\NoCaseChange{\protect\cite{cite1614,cite1615,cite1616}}}
\codefieldsection{Alternative Names}
\begin{eczvaluelist}
\item\relax Low-density convolutional (LDC) code
\end{eczvaluelist}
\eczhIndexCodeAliasName{ld_convolutional}{Low-density convolutional (LDC) code}
\codefieldsection{Description}
Convolutional code defined by an infinite low-density parity-check matrix.

\codefieldsection{Parent}
\begin{eczvaluelist}
\item\relax
\flmRefsHyperref[eczindexfamilyrel]{code:convolutional}{Convolutional code}\end{eczvaluelist}
\codefieldsection{Cousins}
\begin{eczvaluelist}
\item\relax
\flmRefsHyperref[eczindexfamilyrel]{code:ldpc}{Low-density parity-check (LDPC) code} --- LDPC-CCs are convolutional analogues of LDPC codes.
\item\relax
\flmRefsHyperref[eczindexfamilyrel]{code:qc_ldpc}{Quasi-cyclic LDPC (QC-LDPC) code} --- QC-LDPC codes can be \textit{unwrapped} to obtain LDPC-CCs by expressing each circulant matrix block as a power of some generating circulant matrix and replacing that generating matrix by the shift operator of the convolutional code \NoCaseChange{\protect\cite{cite1307}}.
\item\relax
\flmRefsHyperref[eczindexfamilyrel]{code:sc_ldpc}{Spatially coupled LDPC (SC-LDPC) code} --- Infinite-block versions of SC-LDPC are LDPC-CCs.
\end{eczvaluelist}
\eczhbkcontributors{ \eczhuVVA }
\endeczcode

\eczcode{lexicographic}{Lexicographic code}{~\NoCaseChange{\protect\cite{cite1898,cite147}}}
\codefieldsection{Description}
A \(q\)-ary code whose codewords are constructed greedily and iteratively by starting with the all-zero word and adding codewords whose distance from all previously chosen codewords is at least the desired minimum distance of the code.

\codefieldsection{Parent}
\begin{eczvaluelist}
\item\relax
\flmRefsHyperref[eczindexfamilyrel]{code:q-ary_digits_into_q-ary_digits}{\(q\)-ary code}\end{eczvaluelist}
\codefieldsection{Children}
\begin{eczvaluelist}
\item\relax
\flmRefsHyperref[eczindexfamilyrel]{code:extended_golay}{\([24, 12, 8]\) Extended Golay code} --- The extended Golay code is a lexicode \NoCaseChange{\protect\cite{cite1195,cite147}\protect\cite[{pg. 327}]{cite41}}.
\item\relax
\flmRefsHyperref[eczindexfamilyrel]{code:parity_check}{\([n,n-1,2]\) Single parity-check (SPC) code} --- SPCs are lexicodes \NoCaseChange{\protect\cite{cite147}}.
\item\relax
\flmRefsHyperref[eczindexfamilyrel]{code:hamming}{\([2^r-1,2^r-r-1,3]\) Hamming code} --- Hamming codes are lexicodes \NoCaseChange{\protect\cite{cite147}}.
\item\relax
\flmRefsHyperref[eczindexfamilyrel]{code:hexacode}{\([6,3,4]_4\) Hexacode} --- Hexacodewords can be arranged in an order from smallest to largest, with each codeword differing at four places from the next \NoCaseChange{\protect\cite{cite1669}\protect\cite[{pg. 327}]{cite41}}.
\item\relax
\flmRefsHyperref[eczindexfamilyrel]{code:tetracode}{\([4,2,3]_3\) Tetracode} --- The tetracode is a lexicode \NoCaseChange{\protect\cite{cite147}}.
\end{eczvaluelist}
\codefieldsection{Cousins}
\begin{eczvaluelist}
\item\relax
\flmRefsHyperref[eczindexfamilyrel]{code:binary_linear}{Linear binary code} --- Binary lexicodes are linear \NoCaseChange{\protect\cite{cite147}}.
\item\relax
\flmRefsHyperref[eczindexfamilyrel]{code:points_into_lattices}{Lattice} --- Lexicographic codes are \(q\)-ary analogues of laminated lattices \NoCaseChange{\protect\cite{cite43}\protect\cite[{pg. 162}]{cite41}}.
\item\relax
\flmRefsHyperref[eczindexfamilyrel]{code:combinatorial_design}{Combinatorial design} --- Some lexicodes yield Steiner systems \NoCaseChange{\protect\cite{cite147}}.
\item\relax
\flmRefsHyperref[eczindexfamilyrel]{code:binary_quad_residue}{Binary quadratic-residue (QR) code} --- The \([18,9,6]\) binary QR code is a lexicode \NoCaseChange{\protect\cite{cite147}}.
\end{eczvaluelist}
\eczhbkcontributors{ \eczhuVVA }
\endeczcode

\eczcode{q-ary_linear}{Linear \(q\)-ary code}{}
\codefieldsection{Description}
An \((n,K,d)_q\) linear code is denoted as \([n,k,d]_q\), where \(k=\log_q K\) is an integer. Its codewords form a linear subspace, i.e., for any codewords \(x,y\), \(\alpha x+ \beta y\) is also a codeword for any field elements \(\alpha,\beta \in \mathbb{F}_q\).
This extra structure yields much information about their properties, making them a large and well-studied subset of codes.

Linear codes can be defined in terms of a \textit{generator matrix} \(G\), whose rows form a basis for the \(k\)-dimensional codespace. Given a message \(x\), the corresponding encoded codeword is \(G^T x\). The generator matrix can be reduced via coordinate permutations to its \textit{standard form} or \textit{systematic form} \(G = [I_k~~A]\), where \(I_k\) is a \(k\times k\) identity matrix and \(A\) is a \(k \times (n-k)\) \(q\)-ary matrix.
The code also comes with a parity check matrix \(H\), whose columns make up a maximal linearly independent set of vectors that are in the kernel of \(G\).
For an error vector \(e\), the associated error syndrome is \(He\).

The \textit{monomial group} of order \((q-1)^n n!\) is formed by \(n\)-dimensional matrices with one nonzero field element in each row and column.
Two linear \(q\)-ary codes are (monomial) \textit{equivalent} if the codewords of one code can be mapped into those of the other under a monomial group element \NoCaseChange{\protect\cite[{Ch. 8}]{cite41}\protect\cite[{Ch. 3}]{cite39}}.
The \textit{automorphism group} of a linear \(q\)-ary code is the largest subgroup of the monomial group that maps the code onto itself.

\codefieldsection{Protection}
Distance \(d\) of a linear code is the number of nonzero entries in the (nonzero) codeword with the smallest such number.
A linear code corrects an error set precisely when the difference of any pair of distinct errors is not a codeword, equivalently when distinct errors have distinct syndromes under the parity check matrix.

\codefieldsection{Rate}
Any code admitting a two-transitive automorphism group achieves capacity under the binary erasure channel \NoCaseChange{\protect\cite{cite1566,cite1567,cite1661}}.
\codefieldsection{Decoding}
\begin{eczvaluelist}
\item\relax Maximum likelihood (ML) decoding. This algorithm decodes a received word to the most likely sent codeword based on the received word. ML decoding of reduced complexity is possible for virtually all \(q\)-ary linear codes \NoCaseChange{\protect\cite{cite1899}}.
\item\relax Optimal symbol-by-symbol decoding rule \NoCaseChange{\protect\cite{cite1900}}.
\item\relax Information set decoding (ISD) \NoCaseChange{\protect\cite{cite1901}}, a probabilistic decoding strategy that essentially tries to guess \(k\) correct positions in the received word, where \(k\) is the size of the code. Then, an error vector is constructed to map the received word onto the nearest codeword, assuming the \(k\) positions are error free. When the Hamming weight of the error vector is low enough, that codeword is assumed to be the intended transmission.
\item\relax Generalized minimum-distance decoder \NoCaseChange{\protect\cite{cite969}}.
\item\relax Soft-decision maximum-likelihood trellis-based decoder \NoCaseChange{\protect\cite{cite1902}}.
\item\relax Random linear codes over large fields are list-recoverable and list-decodable up to near-optimal rates \NoCaseChange{\protect\cite{cite1903}}.
\item\relax Extensions of algebraic-geometry decoders to linear codes \NoCaseChange{\protect\cite{cite1904,cite1905}}.
\end{eczvaluelist}
\codefieldsection{Notes}
\begin{eczvaluelist}
\item\relax The two extreme examples of codes are the \([n,0,n]_q\) \textit{zero code} and its dual the \([n,n]_q\) \textit{universe code}.
\item\relax University of Salzburg's \flmHref{https://web.archive.org/web/20221213223211/http://mint.sbg.ac.at/table.php?i=c}{MinT application} generates an optimal parameter table for a linear code \([n,k,d]_q\), contingent on an optional fluctuation of maximal Hamming code distance, rank, and length, along with other specifications.
\item\relax CodingTheory Julia software library \NoCaseChange{\protect\cite{cite1906}}.
\end{eczvaluelist}
\codefieldsection{Parents}
\begin{eczvaluelist}
\item\relax
\flmRefsHyperref[eczindexfamilyrel]{code:q-ary_additive}{Additive \(q\)-ary code} --- For \(q>2\), additive codes need not be linear since linearity also requires closure under multiplication.
\item\relax
\flmRefsHyperref[eczindexfamilyrel]{code:rings_linear}{\(R\)-linear code} --- Linear \(q\)-ary codes are \(\mathbb{F}_q\)-linear.
\end{eczvaluelist}
\codefieldsection{Children}
\begin{eczvaluelist}
\item\relax
\flmRefsHyperref[eczindexfamilyrel]{code:binary_linear}{Linear binary code} --- Linear binary codes are linear \(q\)-ary codes for \(q=2\).
\item\relax
\flmRefsHyperref[eczindexfamilyrel]{code:evaluation_varieties}{Evaluation code} --- Evaluation codes are defined using polynomial or rational functions evaluated on a subset of affine or projective space. Given access to more general structures (i.e., morphisms of algebras), any \(q\)-ary linear code can be formulated as an evaluation code \NoCaseChange{\protect\cite[{Sec. 4.1}]{cite32}\protect\cite[{Prop. 1.1.4}]{cite1314}\protect\cite[{Prop. 1.1.4}]{cite1312}}.
\item\relax
\flmRefsHyperref[eczindexfamilyrel]{code:interleaved_reed_solomon}{Interleaved RS (IRS) code} --- IRS codes are linear over \(\mathbb{F}_q\) but not necessarily over \(\mathbb{F}_{q^t}\).
\item\relax
\flmRefsHyperref[eczindexfamilyrel]{code:cartier}{Cartier code}\item\relax
\flmRefsHyperref[eczindexfamilyrel]{code:alternant}{Alternant code}\item\relax
\flmRefsHyperref[eczindexfamilyrel]{code:pyramid}{Pyramid code}\item\relax
\flmRefsHyperref[eczindexfamilyrel]{code:q-ary_ltc}{\(q\)-ary linear LTC} --- Linear \(q\)-ary codes with distances \(\frac{1}{2}n-\sqrt{t n}\) for some \(t\) are called almost-orthogonal and are locally testable with query complexity of \flmRefsHyperref{ref65}{order} \(O(t)\) \NoCaseChange{\protect\cite{cite1270}}. This was later improved to codes with distance \(\frac{1}{2}n-O(n^{1-\gamma})\) for any positive \(\gamma\) \NoCaseChange{\protect\cite{cite1271}}, provided that the number of codewords is polynomial in \(n\).
\item\relax
\flmRefsHyperref[eczindexfamilyrel]{code:mds}{Maximum distance separable (MDS) code}\item\relax
\flmRefsHyperref[eczindexfamilyrel]{code:q-ary_lcc}{\(q\)-ary linear LCC}\item\relax
\flmRefsHyperref[eczindexfamilyrel]{code:dual}{Dual linear code}\item\relax
\flmRefsHyperref[eczindexfamilyrel]{code:q-ary_hamming}{\(q\)-ary Hamming code}\item\relax
\flmRefsHyperref[eczindexfamilyrel]{code:quasi_group}{Quasi group-algebra code} --- A linear code is a quasi group-algebra code for a group \(G\) and index \(\ell\) if and only if \(G\) is isomorphic to a regular subgroup of the code's permutation automorphism group which acts freely of index \(\ell\) on the coordinates \NoCaseChange{\protect\cite[{Thm. 3.5}]{cite1115}}.
\item\relax
\flmRefsHyperref[eczindexfamilyrel]{code:parity_check_tensor}{Parity-check tensor-product code}\item\relax
\flmRefsHyperref[eczindexfamilyrel]{code:projective}{Projective geometry code} --- Columns of the generator matrix of a projective linear \([n,k]_q\) code correspond to distinct nonzero points in projective space. In general, linear codes admit repeating columns or columns proportional to each other. In that case, the columns correspond to a multiset of non-distinct nonzero points, and multisets are in one-to-one correspondence to arcs in projective space \NoCaseChange{\protect\cite[{Thm. 1.1}]{cite1860}}.
\item\relax
\flmRefsHyperref[eczindexfamilyrel]{code:quantum_inspired}{Quantum-inspired classical block code}\item\relax
\flmRefsHyperref[eczindexfamilyrel]{code:q-ary_ldgm}{\(q\)-ary LDGM code}\item\relax
\flmRefsHyperref[eczindexfamilyrel]{code:tanner}{Tanner code}\item\relax
\flmRefsHyperref[eczindexfamilyrel]{code:divisible}{Divisible code}\item\relax
\flmRefsHyperref[eczindexfamilyrel]{code:two_weight}{Two-weight code}\item\relax
\flmRefsHyperref[eczindexfamilyrel]{code:wozencraft}{Wozencraft ensemble code}\item\relax
\flmRefsHyperref[eczindexfamilyrel]{code:berlekamp}{Berlekamp code}\end{eczvaluelist}
\codefieldsection{Cousins}
\begin{eczvaluelist}
\item\relax
\flmRefsHyperref[eczindexfamilyrel]{code:q-ary_linear_over_zq}{Linear code over \(\mathbb{Z}_q\)} --- \(q\)-ary linear codes for \(q=p\) prime are linear \(p\)-ary codes over \(\mathbb{Z}_p \cong \mathbb{F}_p\).
\item\relax
\flmRefsHyperref[eczindexfamilyrel]{code:rank_modulation}{Rank-modulation code} --- Almost all linear \(q\)-ary codes can be converted to rank-modulation codes \NoCaseChange{\protect\cite{cite1907}}.
\item\relax
\flmRefsHyperref[eczindexfamilyrel]{code:gabidulin}{Gabidulin code} --- Gabidulin codes over \(\mathbb{F}_{q^N}\), when expressed as vectors over \(\mathbb{F}_{q^N}\), are linear \(q\)-ary codes.
\item\relax
\flmRefsHyperref[eczindexfamilyrel]{code:locally_recoverable}{Locally recoverable code (LRC)} --- A \(q\)-ary linear code is an LRC of locality \(r\) if each coordinate participates in at least one parity check of weight \(\leq r\) \NoCaseChange{\protect\cite{cite812}\protect\cite[{Sec. 31.3.4.5}]{cite183}}.
\item\relax
\flmRefsHyperref[eczindexfamilyrel]{code:q-ary_constant_weight}{Constant-weight block code} --- Linear \(q\)-ary codes cannot be constant weight, but can have nonzero codewords with constant weight. All such codes are equidistant, and Bonisoli's theorem states that any equidistant linear code is a direct sum of \(q\)-ary simplex codes \NoCaseChange{\protect\cite{cite988}} (see also Refs. \NoCaseChange{\protect\cite{cite45,cite46}}).
\item\relax
\flmRefsHyperref[eczindexfamilyrel]{code:evaluation}{Evaluation AG code} --- The degree of the divisor for evaluation AG codes is restricted to be less than \(n\). When there is no restriction, any \(q\)-ary linear code can be formulated as an evaluation AG code \NoCaseChange{\protect\cite{cite1811}}.
\item\relax
\flmRefsHyperref[eczindexfamilyrel]{code:generalized_reed_solomon}{Generalized RS (GRS) code} --- Concatenations of GRS codes with random linear codes almost surely attain the \flmRefsHyperref{ref85}{GV bound} \NoCaseChange{\protect\cite{cite973}}.
\item\relax
\flmRefsHyperref[eczindexfamilyrel]{code:q-ary_simplex}{\(q\)-ary simplex code} --- Linear \(q\)-ary codes cannot be constant weight, but can have nonzero codewords with constant weight. All such codes are equidistant, and Bonisoli's theorem states that any equidistant linear code is a direct sum of \(q\)-ary simplex codes \NoCaseChange{\protect\cite{cite988}} (see also Refs. \NoCaseChange{\protect\cite{cite45,cite46}}).
\item\relax
\flmRefsHyperref[eczindexfamilyrel]{code:eastab}{EA qubit stabilizer code} --- Any quaternary linear code can be used to construct an EA qubit stabilizer code \NoCaseChange{\protect\cite{cite1430}}.
\item\relax
\flmRefsHyperref[eczindexfamilyrel]{code:galois_css}{Galois-qudit CSS code} --- The Galois-qudit CSS construction uses two related \(q\)-ary linear codes, \(C_X\) and \(C_Z\).
\item\relax
\flmRefsHyperref[eczindexfamilyrel]{code:galois_true_stabilizer}{True Galois-qudit stabilizer code} --- A true Galois-qudit stabilizer code is the closest quantum analogue of a linear code over \(\mathbb{F}_q\) because the \(q\)-ary vectors corresponding to the \flmRefsHyperref{ref873}{Galois symplectic representation} of the stabilizers form a linear subspace.
\item\relax
\flmRefsHyperref[eczindexfamilyrel]{code:galois_hypergraph_product}{Galois-qudit HGP code} --- Galois-qudit HGP codes are constructed out of two classical linear \(q\)-ary codes.
\end{eczvaluelist}
\eczhbkcontributors{ Micah Shaw, \eczhuVVA }
\endeczcode

\eczcode{lcd}{Linear code with complementary dual (LCD)}{~\NoCaseChange{\protect\cite{cite1908}}}
\codefieldsection{Description}
A linear code \(C\) admits a complementary dual if \(C\) and its dual code \(C^{\perp}\) do not share any nonzero codewords. Equivalently, \(\mathbb{F}_q^n = C \oplus C^{\perp}\).

\codefieldsection{Protection}
Optimal binary and ternary LCD codes have been characterized \NoCaseChange{\protect\cite{cite1909}}. LP bounds have been derived \NoCaseChange{\protect\cite{cite1910}}.

\codefieldsection{Rate}
Asymptotically good LCD codes exist \NoCaseChange{\protect\cite{cite1908}}.
\codefieldsection{Decoding}
\begin{eczvaluelist}
\item\relax The decoding problem reduces to finding the nearest codeword in \(C\) given a word in \(C^{\perp}\) \NoCaseChange{\protect\cite{cite1908}}.
\end{eczvaluelist}
\codefieldsection{Parent}
\begin{eczvaluelist}
\item\relax
\flmRefsHyperref[eczindexfamilyrel]{code:dual}{Dual linear code}\end{eczvaluelist}
\codefieldsection{Cousins}
\begin{eczvaluelist}
\item\relax
\flmRefsHyperref[eczindexfamilyrel]{code:group}{Group-algebra code} --- A group code \(C \leq \mathbb{F}_q G\) is LCD if and only if \(C=e \mathbb{F}_q G\) for an idempotent \(e\) satisfying \(e=\hat{e}\), and then \(C^{\perp}=(1-e)\mathbb{F}_q G\) \NoCaseChange{\protect\cite[{Thm. 16.7.6}]{cite196}}.
\item\relax
\flmRefsHyperref[eczindexfamilyrel]{code:q-ary_cyclic}{Cyclic linear \(q\)-ary code} --- A cyclic \([n,k]\) code with generator polynomial \(g(x)\) is LCD if and only if \(g(x)\) is self-reciprocal and \(\gcd( g(x), (x^{n}-1)/g(x) )=1\) \NoCaseChange{\protect\cite[{Cor. 16.7.11}]{cite196}}.
\item\relax
\flmRefsHyperref[eczindexfamilyrel]{code:reversible}{Reversible code} --- A reversible cyclic code is a cyclic code with self-reciprocal generator polynomial and is an LCD code \NoCaseChange{\protect\cite[{Thm. 2.10.3}]{cite68}}.
\item\relax
\flmRefsHyperref[eczindexfamilyrel]{code:ea_quantum_lcd}{EA quantum LCD code} --- Asymptotically good maximal-entanglement EA Galois-qudit stabilizer codes can be constructed from LCD codes \NoCaseChange{\protect\cite{cite1911}}.
\item\relax
\flmRefsHyperref[eczindexfamilyrel]{code:maximal_entanglement_galois_stabilizer}{Maximal-entanglement EA Galois-qudit stabilizer code} --- Asymptotically good maximal-entanglement EA Galois-qudit stabilizer codes can be constructed from LCD codes \NoCaseChange{\protect\cite{cite1911}}.
\end{eczvaluelist}
\eczhbkcontributors{ \eczhuVVA }
\endeczcode

\eczcode{matrix_product}{Matrix-product code}{~\NoCaseChange{\protect\cite{cite1836}}}
\codefieldsection{Description}
Code constructed from \(M\) length-\(n\) \(q\)-ary codes \(C_1,\ldots,C_M\) and an \(M\times N\) \(q\)-ary matrix \(A\), yielding all matrix products \([c_1\,\cdots\,c_M]A\) with \(c_i\in C_i\) and \(N\geq M\). A prominent subclass is the case in which \(A\) is non-singular by columns (NSC).

\codefieldsection{Decoding}
\begin{eczvaluelist}
\item\relax Decoder up to half of the minimum distance for NSC codes \NoCaseChange{\protect\cite{cite1912}}.
\end{eczvaluelist}
\codefieldsection{Parent}
\begin{eczvaluelist}
\item\relax
\flmRefsHyperref[eczindexfamilyrel]{code:q-ary_digits_into_q-ary_digits}{\(q\)-ary code}\end{eczvaluelist}
\codefieldsection{Children}
\begin{eczvaluelist}
\item\relax
\flmRefsHyperref[eczindexfamilyrel]{code:generalized_reed_muller}{Generalized RM (GRM) code} --- Applying a special case of the matrix-product procedure yields GRM codes \NoCaseChange{\protect\cite{cite1836}}.
\item\relax
\flmRefsHyperref[eczindexfamilyrel]{code:uplusv}{\((u|u+v)\)-construction code}\end{eczvaluelist}
\codefieldsection{Cousin}
\begin{eczvaluelist}
\item\relax
\flmRefsHyperref[eczindexfamilyrel]{code:stabilizer_over_gfqsq}{Hermitian Galois-qudit code} --- Hermitian self-orthogonal matrix-product codes over \(\mathbb{F}_{q^2}\) can be used to construct quantum codes via the Hermitian construction \NoCaseChange{\protect\cite{cite1913,cite1914}}.
\end{eczvaluelist}
\eczhbkcontributors{ \eczhuVVA }
\endeczcode

\eczcode{mds}{Maximum distance separable (MDS) code}{~\NoCaseChange{\protect\cite{cite1915}}}
\codefieldsection{Description}
A \(q\)-ary linear code whose parameters satisfy the Singleton bound with equality.

An \([n,k,d]_q\) code is MDS if parameters \(n\), \(k\), \(d\), and \(q\) are such that the Singleton bound
\flmMathEnvironment{align}{}{
d \leq n-k+1
}
becomes an equality.
Equivalently, a code with minimum distance \(d\) and dual distance \(d'\) is MDS precisely when \(d+d'=n+2\) and all distances occur \NoCaseChange{\protect\cite[{Thm. 12.3.1}]{cite199}}.
A code is called \textit{almost MDS} (AMDS) when \(d=n-k\).
A bound for general (i.e., nonlinear or unrestricted) \(q\)-ary codes can also be formulated; see \NoCaseChange{\protect\cite[{Thm. 1.9.10}]{cite1159}}.
A code is \textit{near MDS} (NMDS) if both the code and its dual are AMDS.

The codes \( [n,1,n]_q, [n,n-1,2]_q, [n,n]_q \) for any \(q\) are MDS codes. These are called the \textit{trivial} MDS codes \NoCaseChange{\protect\cite[{Sec. 3.3.2}]{cite70}}. The only binary MDS codes are the trivial ones.
Many, but not all, \(q\)-ary MDS codes are related to RS codes and their extensions; see, e.g., \NoCaseChange{\protect\cite[{Prob. 11.11}]{cite195}}.

\codefieldsection{Protection}
Given \(n\) and \(k\), MDS codes have the highest distance possible of all codes and so have the best possible error-correction properties.
\codefieldsection{Realizations}
\begin{eczvaluelist}
\item\relax Automatic repeat request (ARQ) data transmission protocols \NoCaseChange{\protect\cite[{Ch. 7}]{cite247}}.
\end{eczvaluelist}
\codefieldsection{Notes}
\begin{eczvaluelist}
\item\relax See Refs. \NoCaseChange{\protect\cite{cite1916}\protect\cite[{Sec. 11.4 notes}]{cite195}\protect\cite[{Ch. 11 notes}]{cite41}} for more on MDS codes and the MDS conjecture.
\item\relax An MDS code with \(k=2\) is equivalent to a set of \(n-k\) mutually orthogonal Latin squares of order \(q\) \NoCaseChange{\protect\cite[{Ch. 11 notes}]{cite41}}.
\end{eczvaluelist}
\codefieldsection{Parents}
\begin{eczvaluelist}
\item\relax
\flmRefsHyperref[eczindexfamilyrel]{code:q-ary_linear}{Linear \(q\)-ary code}\item\relax
\flmRefsHyperref[eczindexfamilyrel]{code:optimal_lrc}{Optimal LRC} --- The generalized Singleton bound becomes the Singleton bound for \(k=r\), so optimal LRCs with that constraint are MDS.
\item\relax
\flmRefsHyperref[eczindexfamilyrel]{code:univ_opt_q-ary}{Universally optimal \(q\)-ary code} --- MDS codes are LP universally optimal codes \NoCaseChange{\protect\cite{cite1917,cite173}}.
\item\relax
\flmRefsHyperref[eczindexfamilyrel]{code:orthogonal_array}{Orthogonal array (OA)} --- An MDS code is an OA\(_{1}(k,n,q)\) \NoCaseChange{\protect\cite[{Thm. 3.3.19}]{cite70}}.
\end{eczvaluelist}
\codefieldsection{Children}
\begin{eczvaluelist}
\item\relax
\flmRefsHyperref[eczindexfamilyrel]{code:generalized_reed_solomon}{Generalized RS (GRS) code} --- GRS codes have distance \(n-k+1\), saturating the Singleton bound.
\item\relax
\flmRefsHyperref[eczindexfamilyrel]{code:roth_lempel}{Roth-Lempel code} --- Roth-Lempel codes are examples of MDS codes that are not GRS codes.
\item\relax
\flmRefsHyperref[eczindexfamilyrel]{code:glynn}{\([10,5,6]_9\) Glynn code} --- The Glynn code is a rare example of an MDS code that is not related to an RS code.
\item\relax
\flmRefsHyperref[eczindexfamilyrel]{code:hirschfeld}{Hirschfeld code} --- The Hirschfeld code is a rare example of an MDS code that is not related to an RS code.
\item\relax
\flmRefsHyperref[eczindexfamilyrel]{code:reed_solomon_4}{\([4,2,3]_4\) RS\(_4\) code}\item\relax
\flmRefsHyperref[eczindexfamilyrel]{code:griesmer}{Griesmer code} --- Singleton bound implies the Griesmer bound.
\end{eczvaluelist}
\codefieldsection{Cousins}
\begin{eczvaluelist}
\item\relax
\flmRefsHyperref[eczindexfamilyrel]{code:dual}{Dual linear code} --- A linear binary or \(q\)-ary \([n,k,n-k+1]\) code is MDS if and only if its dual \([n,n-k,k+1]\) is MDS \NoCaseChange{\protect\cite[{Thm. 1.9.13}]{cite1159}}.
\item\relax
\flmRefsHyperref[eczindexfamilyrel]{code:projective}{Projective geometry code} --- A linear code is MDS (almost MDS) if and only if columns of its parity-check matrix form an \(n\)-arc (\(n\)-track) in projective space \NoCaseChange{\protect\cite{cite1889,cite1918,cite1863,cite1919}}. The dual of a MDS code is an MDS code, so MDS codes are projective. All \([9,3]\) MDS codes have been tabulated \NoCaseChange{\protect\cite{cite1920}} in terms of 9-arcs in the projective plane.
\item\relax
\flmRefsHyperref[eczindexfamilyrel]{code:repetition}{Repetition code} --- Binary repetition codes are trivial MDS codes \NoCaseChange{\protect\cite[{Thm. 12.3.1}]{cite199}\protect\cite[{Sec. 3.3.2}]{cite70}}.
\item\relax
\flmRefsHyperref[eczindexfamilyrel]{code:mds_array}{MDS array code} --- MDS array codes are MDS codes when each matrix codeword is treated as a vector by converting each column into a single coordinate via subpacketization.
\item\relax
\flmRefsHyperref[eczindexfamilyrel]{code:matrix_computation}{Distributed computation code} --- The first matrix multiplication code encoded each entry of the matrices to be multiplied into an MDS code \NoCaseChange{\protect\cite{cite1921}}.
\item\relax
\flmRefsHyperref[eczindexfamilyrel]{code:maximum_sum_rank_distance}{Maximum-sum-rank distance (MSRD) code} --- MSRD codes generalize MDS codes from the Hamming metric to the sum-rank metric.
\item\relax
\flmRefsHyperref[eczindexfamilyrel]{code:maximum_rank_distance}{Maximum-rank distance (MRD) code} --- MRD codes are matrix-code analogues of MDS codes.
\item\relax
\flmRefsHyperref[eczindexfamilyrel]{code:ag}{Algebraic-geometry (AG) code} --- Near MDS \([n,k,d]_{p^m}\) AG codes exist when \(n,p,m\) satisfy certain relations according to the Tsfasman-Vladut bound \NoCaseChange{\protect\cite{cite1312,cite1313,cite1314,cite63}}.
\item\relax
\flmRefsHyperref[eczindexfamilyrel]{code:elliptic}{Elliptic code} --- Elliptic codes can be MDS \NoCaseChange{\protect\cite[{Exam. 15.5.3}]{cite26}\protect\cite[{pg. 310}]{cite1314}\protect\cite[{Sec. 4.4.2}]{cite1312}}.
\item\relax
\flmRefsHyperref[eczindexfamilyrel]{code:extended_reed_solomon}{Extended GRS code} --- A GRS code can be extended to an MDS code \NoCaseChange{\protect\cite[{Thm. 5.3.4}]{cite126}}. Extended and doubly extended narrow-sense RS codes are MDS \NoCaseChange{\protect\cite[{Thms. 5.3.2 and 5.3.4}]{cite126}}, and there is an equivalence between the two for odd prime \(q\) \NoCaseChange{\protect\cite{cite1818}}.
\item\relax
\flmRefsHyperref[eczindexfamilyrel]{code:narrow_sense_reed_solomon}{Narrow-sense RS code} --- Extended and doubly extended narrow-sense RS codes are MDS \NoCaseChange{\protect\cite[{Thms. 5.3.2 and 5.3.4}]{cite126}\protect\cite[{Ch. 11}]{cite195}}, and there is an equivalence between the two for odd prime \(q\) \NoCaseChange{\protect\cite{cite1818}}.
\item\relax
\flmRefsHyperref[eczindexfamilyrel]{code:reed_solomon}{Reed-Solomon (RS) code} --- Reed-Solomon codes are important examples of MDS codes, and for prime \(q\), every \([q+1,k,q-k+2]_q\) MDS code is Reed-Solomon; for non-prime \(q\), non-equivalent maximal-length MDS codes also exist \NoCaseChange{\protect\cite{cite1818}\protect\cite[{Sec. 3.3.2}]{cite70}\protect\cite[{Ch. 11}]{cite195}}.
\item\relax
\flmRefsHyperref[eczindexfamilyrel]{code:generalized_srivastava}{Generalized Srivastava code} --- Generalized Srivastava codes for \(m=1\) are MDS codes \NoCaseChange{\protect\cite[{pg. 359}]{cite41}}.
\item\relax
\flmRefsHyperref[eczindexfamilyrel]{code:hexacode}{\([6,3,4]_4\) Hexacode} --- The hexacode is an MDS code \NoCaseChange{\protect\cite[{Exer. 578}]{cite126}}.
\item\relax
\flmRefsHyperref[eczindexfamilyrel]{code:q-ary_parity_check}{\([n,n-1,2]_q\) \(q\)-ary parity-check code}\item\relax
\flmRefsHyperref[eczindexfamilyrel]{code:tetracode}{\([4,2,3]_3\) Tetracode} --- The tetracode is a unique MDS code \NoCaseChange{\protect\cite{cite1663,cite1664}}.
\item\relax
\flmRefsHyperref[eczindexfamilyrel]{code:ea_mixed_alphabet_reed_solomon}{EA mixed-alphabet Reed-Solomon c-q code} --- EA mixed-alphabet RS c-q codes can saturate a block-erasure bound and, in some parameter ranges, exceed the classical Singleton bound for ordinary \(q\)-ary codes \NoCaseChange{\protect\cite{cite1922}}.
\item\relax
\flmRefsHyperref[eczindexfamilyrel]{code:quantum_mds}{Quantum maximum-distance-separable (MDS) code} --- Quantum MDS codes are quantum analogues of MDS codes.
\item\relax
\flmRefsHyperref[eczindexfamilyrel]{code:ame}{Perfect-tensor code} --- MDS codes can be used to obtain cluster states that are AME with minimal support \NoCaseChange{\protect\cite{cite1923,cite1924,cite1925,cite151,cite1926,cite1927}}.
\item\relax
\flmRefsHyperref[eczindexfamilyrel]{code:quantum_tensor_product}{Quantum tensor-product code} --- MDS codes can be used to construct quantum tensor-product codes \NoCaseChange{\protect\cite{cite1131}}.
\item\relax
\flmRefsHyperref[eczindexfamilyrel]{code:ea_mds}{EA MDS code} --- MDS codes give rise to families of EA Galois-qudit codes that saturate the original (erroneous) EAQECC Singleton bound \NoCaseChange{\protect\cite{cite1928}}.
\end{eczvaluelist}
\eczhbkcontributors{ Markus Grassl, Eric Kubischta, \eczhuVVA }
\endeczcode

\eczcode{meir}{Meir code}{~\NoCaseChange{\protect\cite{cite1929}}}
\codefieldsection{Description}
Locally testable \([n,k,d]_q\) code with query complexity \(\text{poly}(\log k)\) and rejection ratio \(R/n = 1/\text{poly}(\log k)\). The code construction is probabilistic and combinatorial.

\codefieldsection{Parents}
\begin{eczvaluelist}
\item\relax
\flmRefsHyperref[eczindexfamilyrel]{code:q-ary_ltc}{\(q\)-ary linear LTC} --- Meir codes stand out in that they are based on a combinatorial construction, while other LTCs often use algebraic tools.
\item\relax
\flmRefsHyperref[eczindexfamilyrel]{code:random}{Random code} --- Meir codes are defined using a probabilistic combinatorial construction.
\end{eczvaluelist}
\eczhbkcontributors{ \eczhuVVA }
\endeczcode

\eczcode{multiplicity}{Multiplicity code}{~\NoCaseChange{\protect\cite{cite179,cite180,cite181}}}
\codefieldsection{Description}
A generalization of an \(m\)-variate polynomial evaluation code based on evaluating polynomials together with their Hasse derivatives up to order \(s-1\) at all points in \(\mathbb{F}_q^m\).
Originally proposed for coding using the Rosenbloom-Tsfasman metric \NoCaseChange{\protect\cite{cite179}}.
Univariate (\(m=1\)) \NoCaseChange{\protect\cite{cite179,cite180}} and multivariate (\(m>1\)) \NoCaseChange{\protect\cite{cite181}} codes have been proposed.

\codefieldsection{Protection}
The multiplicity Schwartz-Zippel Lemma provides a lower bound on code distance \NoCaseChange{\protect\cite{cite1930}}.

\codefieldsection{Rate}
Multiplicity codes achieve relaxed generalized Singleton bounds \NoCaseChange{\protect\cite{cite1821}}.
\codefieldsection{Decoding}
\begin{eczvaluelist}
\item\relax Multivariate multiplicity codes can be decoded up to half of the minimum distance in polynomial time \NoCaseChange{\protect\cite{cite1931,cite1932}}.
\item\relax Univariate \NoCaseChange{\protect\cite{cite180}} and multivariate \NoCaseChange{\protect\cite{cite1931}} multiplicity codes can be list-decoded up to the Johnson bound. Certain univariate code families achieve the list-decoding capacity for sufficiently large field characteristic  \NoCaseChange{\protect\cite{cite1933,cite1931}}.
\end{eczvaluelist}
\codefieldsection{Notes}
\begin{eczvaluelist}
\item\relax See Ref. \NoCaseChange{\protect\cite{cite1934}} for a review of multiplicity codes.
\end{eczvaluelist}
\codefieldsection{Parent}
\begin{eczvaluelist}
\item\relax
\flmRefsHyperref[eczindexfamilyrel]{code:group}{Group-algebra code} --- Multiplicity codes of order \(s\) are Abelian group-algebra codes whose corresponding polynomial that is modded out is \((x-\alpha_j)^s\) for each evaluation point \(\alpha_j\) \NoCaseChange{\protect\cite{cite1826}}.
\end{eczvaluelist}
\codefieldsection{Children}
\begin{eczvaluelist}
\item\relax
\flmRefsHyperref[eczindexfamilyrel]{code:reed_solomon}{Reed-Solomon (RS) code} --- Univariate multiplicity codes of degree \(s=1\) reduce to RS codes.
\item\relax
\flmRefsHyperref[eczindexfamilyrel]{code:generalized_reed_muller}{Generalized RM (GRM) code} --- Multivariate multiplicity codes of degree \(s=1\) reduce to GRM codes.
\end{eczvaluelist}
\codefieldsection{Cousins}
\begin{eczvaluelist}
\item\relax
\flmRefsHyperref[eczindexfamilyrel]{code:q-ary_lcc}{\(q\)-ary linear LCC} --- There exist multiplicity codes with rate arbitrarily close to one that are locally decodable and locally correctable from a constant error fraction \NoCaseChange{\protect\cite{cite181}}.
\item\relax
\flmRefsHyperref[eczindexfamilyrel]{code:ltc}{Locally testable code (LTC)} --- Some multiplicity codes are locally testable by an appropriate test \NoCaseChange{\protect\cite{cite1099,cite1100}}.
\item\relax
\flmRefsHyperref[eczindexfamilyrel]{code:evaluation_polynomial}{Polynomial evaluation code} --- A multiplicity code is a generalization of an \(m\)-variate polynomial evaluation code based on evaluating polynomials together with their Hasse derivatives up to order \(s-1\) at all points in \(\mathbb{F}_q^m\).
\item\relax
\flmRefsHyperref[eczindexfamilyrel]{code:batch}{Batch code} --- Multiplicity codes can be used to construct batch codes \NoCaseChange{\protect\cite{cite951}}.
\item\relax
\flmRefsHyperref[eczindexfamilyrel]{code:pir}{Private information retrieval (PIR) code} --- Multiplicity codes can be used to construct PIR codes \NoCaseChange{\protect\cite{cite951}}.
\end{eczvaluelist}
\eczhbkcontributors{ \eczhuVVA }
\endeczcode

\eczcode{narrow_sense_reed_solomon}{Narrow-sense RS code}{~\NoCaseChange{\protect\cite{cite1935,cite1665,cite1936}}}
\codefieldsection{Description}
An \([q-1,k,n-k+1]_q\) RS code whose points \(\alpha_i\) are all \((i-1)\)st powers of a \flmRefsHyperref{ref33}{primitive} element \(\alpha\) of \(\mathbb{F}_q\).

A narrow-sense RS code encodes a message \(\mu=\{\mu_0,\cdots,\mu_{k-1}\}\) into \(\{f_\mu(1),f_\mu(\alpha),\cdots,f_\mu(\alpha^{n-1})\}\) using a message-dependent polynomial
\flmMathEnvironment{align}{}{
f_\mu(x)=\mu_0+\mu_1 x + \cdots + \mu_{k-1}x^{k-1}.
}
In other words, each message \(\mu\) is encoded into a string of values of the corresponding polynomial \(f_\mu\) at the points \(\alpha^{i-1}\),
\flmMathEnvironment{align}{}{
  \mu\to\left( f_{\mu}\left(1\right),f_{\mu}\left(\alpha\right),\cdots,f_{\mu}\left(\alpha^{n-1}\right)\right) \,.
}

In an alternative convention (not used here), this code is called an RS code, and the general-root case is a generalized RS code.

\codefieldsection{Parents}
\begin{eczvaluelist}
\item\relax
\flmRefsHyperref[eczindexfamilyrel]{code:reed_solomon}{Reed-Solomon (RS) code} --- A narrow-sense RS is an RS code with length \(n=q-1\) whose points \(\alpha_i\) are all \((i-1)\)st powers of a primitive element of \(\mathbb{F}_q\).
\item\relax
\flmRefsHyperref[eczindexfamilyrel]{code:q-ary_bch}{Bose–Chaudhuri–Hocquenghem (BCH) code} --- Narrow-sense RS codes are \(q\)-ary BCH codes \NoCaseChange{\protect\cite[{Remark 15.3.21}]{cite26}\protect\cite[{Thms. 5.2.1 and 5.2.3}]{cite126}}. Their minimal distance is equal to their designed distance \NoCaseChange{\protect\cite[{pg. 81}]{cite39}}.
\end{eczvaluelist}
\codefieldsection{Cousins}
\begin{eczvaluelist}
\item\relax
\flmRefsHyperref[eczindexfamilyrel]{code:extended_reed_solomon}{Extended GRS code} --- A narrow-sense RS code can be extended once, twice, or three times.
\item\relax
\flmRefsHyperref[eczindexfamilyrel]{code:mds}{Maximum distance separable (MDS) code} --- Extended and doubly extended narrow-sense RS codes are MDS \NoCaseChange{\protect\cite[{Thms. 5.3.2 and 5.3.4}]{cite126}\protect\cite[{Ch. 11}]{cite195}}, and there is an equivalence between the two for odd prime \(q\) \NoCaseChange{\protect\cite{cite1818}}.
\item\relax
\flmRefsHyperref[eczindexfamilyrel]{code:projective}{Projective geometry code} --- Columns of parity-check matrices of doubly extended narrow-sense RS codes consist of points of a normal rational curve \NoCaseChange{\protect\cite[{Def. 14.2.6}]{cite202}}.
\end{eczvaluelist}
\eczhbkcontributors{ \eczhuVVA }
\endeczcode

\eczcode{nrt}{Niederreiter-Rosenbloom-Tsfasman (NRT) code}{~\NoCaseChange{\protect\cite{cite1937,cite1938,cite1939,cite179}}}
\codefieldsection{Description}
A poset code based on a partial ordering of \([n]\), i.e., \(1\leq 2\leq \cdots \leq n\).

\codefieldsection{Protection}
LP bounds are provided in Refs. \NoCaseChange{\protect\cite{cite215,cite1940}}.

\codefieldsection{Parent}
\begin{eczvaluelist}
\item\relax
\flmRefsHyperref[eczindexfamilyrel]{code:poset}{Poset code}\end{eczvaluelist}
\codefieldsection{Child}
\begin{eczvaluelist}
\item\relax
\flmRefsHyperref[eczindexfamilyrel]{code:reed_solomon_nrt}{RS NRT code}\end{eczvaluelist}
\codefieldsection{Cousin}
\begin{eczvaluelist}
\item\relax
\flmRefsHyperref[eczindexfamilyrel]{code:orthogonal_array}{Orthogonal array (OA)} --- There exist orthogonal arrays in ordered Hamming space \NoCaseChange{\protect\cite{cite214,cite215}}.
\end{eczvaluelist}
\eczhbkcontributors{ \eczhuVVA }
\endeczcode

\eczcode{nonlinear_ag}{Nonlinear AG code}{~\NoCaseChange{\protect\cite{cite1941,cite1942,cite1943,cite1944,cite1945}}}
\codefieldsection{Description}
Nonlinear \(q\)-ary code constructed by evaluating functions, and in some constructions derivatives of functions, on an algebraic curve \NoCaseChange{\protect\cite[{Sec. 15.4.4}]{cite26}}.

\codefieldsection{Rate}
Certain nonlinear code sequences beat the Tsfasman-Vladut-Zink bound, outperforming linear AG codes \NoCaseChange{\protect\cite[{Thm. 15.4.6}]{cite26}}.
\codefieldsection{Parent}
\begin{eczvaluelist}
\item\relax
\flmRefsHyperref[eczindexfamilyrel]{code:ag}{Algebraic-geometry (AG) code}\end{eczvaluelist}
\eczhbkcontributors{ \eczhuVVA }
\endeczcode

\eczcode{norm_trace}{Norm-trace code}{~\NoCaseChange{\protect\cite{cite1810}}}
\codefieldsection{Description}
Evaluation AG code of rational functions evaluated on points lying on a Miura-Kamiya curve in either affine or projective space.
The family is named as such because the equations defining the curves can be expressed in terms of the \flmRefsHyperref{ref33}{field norm and field trace}.

\codefieldsection{Protection}
Minimum distance is determined by an order bound \NoCaseChange{\protect\cite{cite1810}}.

\codefieldsection{Parent}
\begin{eczvaluelist}
\item\relax
\flmRefsHyperref[eczindexfamilyrel]{code:evaluation}{Evaluation AG code} --- Norm-trace codes are evaluation AG codes with \(\cal X\) being a Miura-Kamiya curve \NoCaseChange{\protect\cite{cite1810}}.
\end{eczvaluelist}
\codefieldsection{Child}
\begin{eczvaluelist}
\item\relax
\flmRefsHyperref[eczindexfamilyrel]{code:hermitian}{Hermitian code} --- Hermitian codes are evaluation AG codes with \(\cal X\) being a Hermitian curve \NoCaseChange{\protect\cite{cite1809}\protect\cite[{Exam. 2.74}]{cite32}}. This curve is maximal, meaning that Hermitian codes are evaluation AG codes with maximum possible length given a fixed genus. They are a special case of norm-trace codes \NoCaseChange{\protect\cite{cite1810}}.
\end{eczvaluelist}
\eczhbkcontributors{ \eczhuVVA }
\endeczcode

\eczcode{one_vs_one}{One-versus-one (OVO) code}{~\NoCaseChange{\protect\cite{cite1946,cite1947}}}
\codefieldsection{Alternative Names}
\begin{eczvaluelist}
\item\relax One-against-one (1A1) code
\end{eczvaluelist}
\eczhIndexCodeAliasName{one_vs_one}{One-against-one (1A1) code}
\codefieldsection{Description}
A length-\(n\) ternary code over \(\{\pm 1,0\}\) whose generator matrix has columns with one \(+1\), one \(-1\), and the rest of the entries zero.

See \NoCaseChange{\protect\cite[{Tab. 6.3}]{cite300}} for an example, whose generator matrix is
\flmMathEnvironment{align}{}{
\left(
\begin{array}{cccccc}
1 & 1 & 1 & 0 & 0 & 0 \\
-1 & 0 & 0 & 1 & 1 & 0 \\
0 & -1 & 0 & -1 & 0 & 1 \\
0 & 0 & -1 & 0 & -1 & -1
\end{array}
\right)~.
}

\codefieldsection{Parents}
\begin{eczvaluelist}
\item\relax
\flmRefsHyperref[eczindexfamilyrel]{code:ecoc}{Error-correcting output code (ECOC)} --- One-vs-one codes are often used in multiclass classification because they separate the multiclass task into several two-class tasks \NoCaseChange{\protect\cite{cite300}}.
\item\relax
\flmRefsHyperref[eczindexfamilyrel]{code:q-ary_constant_weight}{Constant-weight block code}\end{eczvaluelist}
\eczhbkcontributors{ \eczhuVVA }
\endeczcode

\eczcode{orthogonal_array}{Orthogonal array (OA)}{~\NoCaseChange{\protect\cite{cite1948,cite1949,cite1950}}}
\codefieldsection{Description}
An orthogonal array, or OA\(_{\lambda}(t,n,q)\), of \textit{strength} \(t\) with \(q\) \textit{levels} and \(n\) \textit{constraints} is a set of \(q\)-ary strings such that any subset of \(t\) coordinates contains every length-\(t\) string an equal number of times \(\lambda\), which is the \textit{index} of the array \NoCaseChange{\protect\cite[{Def. 3.3.18}]{cite70}}.

\codefieldsection{Notes}
\begin{eczvaluelist}
\item\relax See \NoCaseChange{\protect\cite{cite212}} for a book on orthogonal arrays.
\end{eczvaluelist}
\codefieldsection{Parents}
\begin{eczvaluelist}
\item\relax
\flmRefsHyperref[eczindexfamilyrel]{code:q-ary_digits_into_q-ary_digits}{\(q\)-ary code} --- There is a relation between \(q\)-ary codes and orthogonal arrays which is phrased in terms of the codes' \flmRefsHyperref{ref113}{dual distance} \NoCaseChange{\protect\cite[{Thm. 4.5}]{cite216}\protect\cite[{Thm. 4.9}]{cite212}}.
\item\relax
\flmRefsHyperref[eczindexfamilyrel]{code:t-designs}{\(t\)-design} --- Orthogonal arrays are designs on Hamming space \(\mathbb{F}_q^n\) (a.k.a. the Hamming association scheme) \NoCaseChange{\protect\cite{cite880,cite912,cite914}\protect\cite[{Exam. 1}]{cite226}}; see also Ref. \NoCaseChange{\protect\cite{cite915}}.
\end{eczvaluelist}
\codefieldsection{Children}
\begin{eczvaluelist}
\item\relax
\flmRefsHyperref[eczindexfamilyrel]{code:perfect_binary}{Perfect binary code} --- Perfect distance-three binary codes of length \(n =2^m-1\) are equivalent to binary orthogonal arrays of strength \(t = 2^{m-1}-1\) \NoCaseChange{\protect\cite{cite216,cite217,cite218}}.
\item\relax
\flmRefsHyperref[eczindexfamilyrel]{code:nordstrom_robinson}{\((16,256,6)\) Nordstrom-Robinson (NR) code} --- The NR code is an orthogonal array of strength \(5\) \NoCaseChange{\protect\cite[{pg. 141}]{cite41}}.
\item\relax
\flmRefsHyperref[eczindexfamilyrel]{code:mds}{Maximum distance separable (MDS) code} --- An MDS code is an OA\(_{1}(k,n,q)\) \NoCaseChange{\protect\cite[{Thm. 3.3.19}]{cite70}}.
\item\relax
\flmRefsHyperref[eczindexfamilyrel]{code:delsarte_optimal_q-ary}{\(q\)-ary sharp configuration}\end{eczvaluelist}
\codefieldsection{Cousins}
\begin{eczvaluelist}
\item\relax
\flmRefsHyperref[eczindexfamilyrel]{code:bits_into_bits}{Binary code} --- An \((n,K)\) binary code with \flmRefsHyperref{ref113}{dual distance} \(d^{\perp}\) is an OA\(_{K/2^{d^{\perp}-1}}(d^{\perp}-1,n,2)\) \NoCaseChange{\protect\cite{cite209}\protect\cite[{Ch. 5}]{cite41}}.
\item\relax
\flmRefsHyperref[eczindexfamilyrel]{code:extended_golay}{\([24, 12, 8]\) Extended Golay code} --- The extended Golay code is an orthogonal array of strength 7 \NoCaseChange{\protect\cite[{Exam. 1}]{cite226}}.
\item\relax
\flmRefsHyperref[eczindexfamilyrel]{code:reed_muller}{Reed-Muller (RM) code} --- RM codes are related to orthogonal arrays \NoCaseChange{\protect\cite[{Exam. 10.57}]{cite225}}.
\item\relax
\flmRefsHyperref[eczindexfamilyrel]{code:mixed}{Mixed code} --- Orthogonal arrays generalized to mixed alphabets are called mixed-level orthogonal arrays \NoCaseChange{\protect\cite{cite210,cite211}} (see \NoCaseChange{\protect\cite[{Ch. 9}]{cite212}}). See Ref. \NoCaseChange{\protect\cite{cite213}} for bounds on mixed orthogonal arrays.
\item\relax
\flmRefsHyperref[eczindexfamilyrel]{code:nrt}{Niederreiter-Rosenbloom-Tsfasman (NRT) code} --- There exist orthogonal arrays in ordered Hamming space \NoCaseChange{\protect\cite{cite214,cite215}}.
\item\relax
\flmRefsHyperref[eczindexfamilyrel]{code:ame}{Perfect-tensor code} --- Orthogonal arrays and \(d\)-uniform quantum states are related \NoCaseChange{\protect\cite{cite220,cite152,cite221,cite222,cite223,cite224}}.
\end{eczvaluelist}
\eczhbkcontributors{ \eczhuVVA }
\endeczcode

\eczcode{bose_qvist}{Ovoid code}{~\NoCaseChange{\protect\cite{cite1889,cite1951}}}
\codefieldsection{Description}
Member of a \([q^2+1,4,q^2-q]_q\) projective two-weight code family obtained from ovoids in \(\mathrm{PG}(3,q)\).
If the columns of a generator matrix are the \(q^2+1\) points of an ovoid, then every hyperplane meets the ovoid in either \(1\) or \(q+1\) points, yielding the two nonzero weights \(q^2\) and \(q^2-q\).
See \NoCaseChange{\protect\cite[{pg. 107}]{cite203}\protect\cite[{pg. 192}]{cite62}} for further details.

\codefieldsection{Parents}
\begin{eczvaluelist}
\item\relax
\flmRefsHyperref[eczindexfamilyrel]{code:projective_two_weight}{Projective two-weight code} --- The ovoid code is a two-weight projective code \NoCaseChange{\protect\cite{cite1656,cite1657}\protect\cite[{Table 7.1}]{cite206}}.
\item\relax
\flmRefsHyperref[eczindexfamilyrel]{code:delsarte_optimal_q-ary}{\(q\)-ary sharp configuration} --- The ovoid code is a \(q\)-ary sharp configuration \NoCaseChange{\protect\cite[{Table 12.1}]{cite199}}.
\end{eczvaluelist}
\codefieldsection{Cousin}
\begin{eczvaluelist}
\item\relax
\flmRefsHyperref[eczindexfamilyrel]{code:univ_opt_q-ary}{Universally optimal \(q\)-ary code} --- Several shortened and punctured versions of the ovoid code are LP universally optimal codes \NoCaseChange{\protect\cite{cite173}}.
\end{eczvaluelist}
\eczhbkcontributors{ Alexander Barg, \eczhuVVA }
\endeczcode

\eczcode{parallel_recovery}{Parallel-recovery code}{~\NoCaseChange{\protect\cite{cite1952}}}
\codefieldsection{Description}
A \(t\)-erasure LRC whose coordinate erasures can be recovered in parallel.

\codefieldsection{Parent}
\begin{eczvaluelist}
\item\relax
\flmRefsHyperref[eczindexfamilyrel]{code:sequential_recovery}{Sequential-recovery code}\end{eczvaluelist}
\codefieldsection{Child}
\begin{eczvaluelist}
\item\relax
\flmRefsHyperref[eczindexfamilyrel]{code:codes_with_availability}{Availability code}\end{eczvaluelist}
\eczhbkcontributors{ \eczhuVVA }
\endeczcode

\eczcode{parity_check_tensor}{Parity-check tensor-product code}{~\NoCaseChange{\protect\cite{cite1953}}}
\codefieldsection{Description}
A \(q\)-ary linear code constructed out of two \(q\)-ary linear codes with parity-check matrices \(H_A,H_B\) such that its parity-check matrix is \(H_A \otimes H_B\).  
Its dual has parity-check matrix \(G_A\otimes G_B\), where \(G_{A,B}\) are the generator matrices of the two underlying codes \NoCaseChange{\protect\cite{cite204}}.

\codefieldsection{Notes}
\begin{eczvaluelist}
\item\relax See Ref. \NoCaseChange{\protect\cite{cite1954}} for an exposition.
\end{eczvaluelist}
\codefieldsection{Parent}
\begin{eczvaluelist}
\item\relax
\flmRefsHyperref[eczindexfamilyrel]{code:q-ary_linear}{Linear \(q\)-ary code}\end{eczvaluelist}
\codefieldsection{Cousin}
\begin{eczvaluelist}
\item\relax
\flmRefsHyperref[eczindexfamilyrel]{code:tensor}{Tensor-product code} --- Tensor-product codewords (parity-check tensor-product parity-check matrices) are constructed via an outer product of the underlying codes (parity-check matrices).
\end{eczvaluelist}
\eczhbkcontributors{ \eczhuVVA }
\endeczcode

\eczcode{parvaresh_vardy}{Parvaresh-Vardy (PV) code}{~\NoCaseChange{\protect\cite{cite1824}}}
\codefieldsection{Alternative Names}
\begin{eczvaluelist}
\item\relax Correlated RS code
\end{eczvaluelist}
\eczhIndexCodeAliasName{parvaresh_vardy}{Correlated RS code}
\codefieldsection{Description}
An IRS code with additional algebraic relations (a.k.a. correlations) between the codeword polynomials \(\{f^{(j)}\}_{j=1}^{t}\). These relations yield a list decoder that achieves list-decoding capacity.

\codefieldsection{Decoding}
\begin{eczvaluelist}
\item\relax PV codes can be list-decoded up to \(1-(t k/n)^{1/(t+1)}\) fraction of errors. This result improves over the Guruswami-Sudan algorithm for ordinary \flmRefsHyperref{code:reed_solomon}{RS} codes, which list-decodes up to \(1-\sqrt{k/n}\) fraction of errors.
\end{eczvaluelist}
\codefieldsection{Parent}
\begin{eczvaluelist}
\item\relax
\flmRefsHyperref[eczindexfamilyrel]{code:interleaved_reed_solomon}{Interleaved RS (IRS) code} --- PV codes are IRS codes with specific algebraic relations between the codeword polynomials that allow for efficient list decoding.
\end{eczvaluelist}
\codefieldsection{Cousin}
\begin{eczvaluelist}
\item\relax
\flmRefsHyperref[eczindexfamilyrel]{code:folded_reed_solomon}{Folded RS (FRS) code} --- The specific relations imposed on the polynomials of PV codes allow for them to be expressed in a similar way as FRS codes, but with more redundancy. Folded RS codes can be list-decoded up to a higher fraction of errors.
\end{eczvaluelist}
\eczhbkcontributors{ \eczhuVVA }
\endeczcode

\eczcode{perfect}{Perfect code}{}
\codefieldsection{Description}
A type of \(q\)-ary code whose parameters satisfy the Hamming bound with equality.

An \((n,K,2t+1)_q\) code is perfect if parameters \(n\), \(K\), \(t\), and \(q\) are such that the Hamming (a.k.a. sphere-packing) bound
\flmMathEnvironment{align}{}{
\sum_{j=0}^{t}(q-1)^{j}{n \choose j}\leq q^{n}/K
}
becomes an equality.
In other words, the code's packing radius matches its covering radius.

For example, for a binary \(q=2\) code with one logical bit (\(K=2\)) and \(t=1\), the bound becomes \(n+1 \leq 2^{n-1}\).
Perfect codes are those for which balls of Hamming radius \(t\) exactly fill the space of all \(n\) \(q\)-ary strings.

Any perfect linear code over \(\mathbb{F}_q\) is either a repetition code, a Hamming code, or a binary or ternary Golay code \NoCaseChange{\protect\cite{cite1499}\protect\cite[{Thm. 3.3.1}]{cite70}}.
If \(q\) is a prime power, any nontrivial perfect unrestricted code that is not equivalent to a linear code has the same length, size, and minimum distance as a Hamming code \NoCaseChange{\protect\cite[{Thm. 3.3.1}]{cite70}}; see \NoCaseChange{\protect\cite[{pg. 100}]{cite135}} for more details.
There are many nonlinear perfect codes \NoCaseChange{\protect\cite{cite1955,cite1956,cite1957,cite1958,cite1959,cite1960,cite1961,cite1150,cite1503,cite1504,cite217,cite1505,cite1506}}.

\codefieldsection{Notes}
\begin{eczvaluelist}
\item\relax See Ref. \NoCaseChange{\protect\cite[{Def. 1.9.8}]{cite1159}} for an introduction to perfect codes.
\item\relax See book \NoCaseChange{\protect\cite{cite1962}} on perfect codes for more details.
\end{eczvaluelist}
\codefieldsection{Parents}
\begin{eczvaluelist}
\item\relax
\flmRefsHyperref[eczindexfamilyrel]{code:completely_regular}{Completely regular code} --- Perfect codes and extended perfect codes are completely regular \NoCaseChange{\protect\cite{cite1748}}.
\item\relax
\flmRefsHyperref[eczindexfamilyrel]{code:covering}{Covering code} --- Perfect codes are covering codes with the minimum number of codewords.
\end{eczvaluelist}
\codefieldsection{Children}
\begin{eczvaluelist}
\item\relax
\flmRefsHyperref[eczindexfamilyrel]{code:perfect_binary}{Perfect binary code}\item\relax
\flmRefsHyperref[eczindexfamilyrel]{code:q-ary_hamming}{\(q\)-ary Hamming code}\item\relax
\flmRefsHyperref[eczindexfamilyrel]{code:shortened_hexacode}{\([5,3,3]_4\) Shortened hexacode} --- The shortened hexacode is perfect \NoCaseChange{\protect\cite[{Exer. 578}]{cite126}}.
\item\relax
\flmRefsHyperref[eczindexfamilyrel]{code:ternary_golay}{\([11,6,5]_3\) Ternary Golay code} --- The ternary Golay code is perfect \NoCaseChange{\protect\cite[{Thm. 12.3.3 and Def. 12.3.4}]{cite199}}.
\end{eczvaluelist}
\codefieldsection{Cousins}
\begin{eczvaluelist}
\item\relax
\flmRefsHyperref[eczindexfamilyrel]{code:combinatorial_design}{Combinatorial design} --- Perfect codes and combinatorial designs are related \NoCaseChange{\protect\cite{cite149,cite150}}.
\item\relax
\flmRefsHyperref[eczindexfamilyrel]{code:mixed}{Mixed code} --- Perfect mixed codes with minimum distance \(3\) can be constructed from partitions of vector spaces \NoCaseChange{\protect\cite[{Thm. 3.3.13}]{cite70}}.
\item\relax
\flmRefsHyperref[eczindexfamilyrel]{code:insertion_deletion}{Editing code} --- Perfect deletion correcting codes can be constructed using combinatorial design theory \NoCaseChange{\protect\cite{cite139,cite140}}.
\item\relax
\flmRefsHyperref[eczindexfamilyrel]{code:quantum_perfect}{Perfect quantum code} --- A classical (quantum) perfect code saturates the classical (quantum) Hamming bound.
\end{eczvaluelist}
\eczhbkcontributors{ Mustafa Doger, \eczhuVVA }
\endeczcode

\eczcode{pinwheel}{Pinwheel code}{~\NoCaseChange{\protect\cite{cite1350}}}
\codefieldsection{Description}
A geometrically local binary LDPC code defined on planar graphs obtained from the pinwheel tiling \NoCaseChange{\protect\cite{cite93}}.
Both bits and checks live on vertices of the graph.
If \(L_N\) is the graph Laplacian at generation \(N\), the undepleted check matrix is \(\tilde H_N=(L_N-\mathbb{I})\bmod 2\), and the actual parity-check matrix \(H_N\) is obtained by removing an evenly spaced fraction of boundary checks.

The construction was introduced as a concrete aperiodic seed code whose hypergraph products realize local Type-I and Type-II fracton models.

\codefieldsection{Protection}
For boundary-depletion period \(p\), the family has \(k\approx \sqrt{n}/p\) and distance \(d\sim n\) \NoCaseChange{\protect\cite{cite1350}}, saturating the classical \flmRefsHyperref{ref487}{BPT bound} \(k\sqrt{d}=O(n)\).
The same work gives numerical evidence for confinement.

\codefieldsection{Rate}
The encoding rate vanishes asymptotically because \(k\sim\sqrt{n}\) while the block length grows with the substitution generation \NoCaseChange{\protect\cite{cite1350}}.
\codefieldsection{Parents}
\begin{eczvaluelist}
\item\relax
\flmRefsHyperref[eczindexfamilyrel]{code:quantum_inspired}{Quantum-inspired classical block code}\item\relax
\flmRefsHyperref[eczindexfamilyrel]{code:ldpc}{Low-density parity-check (LDPC) code}\end{eczvaluelist}
\codefieldsection{Cousins}
\begin{eczvaluelist}
\item\relax
\flmRefsHyperref[eczindexfamilyrel]{code:laplacian}{Laplacian code} --- The pinwheel code is derived from the graph Laplacian of the pinwheel tiling, with a fraction of boundary checks removed.
\item\relax
\flmRefsHyperref[eczindexfamilyrel]{code:hypergraph_product}{Hypergraph product (HGP) code} --- The hypergraph product of a pinwheel code with a cyclic repetition code yields a local Type-I fracton model in three dimensions, while the hypergraph product of two pinwheel codes yields a local Type-II fracton model in four dimensions \NoCaseChange{\protect\cite{cite1350}}.
\item\relax
\flmRefsHyperref[eczindexfamilyrel]{code:repetition}{Repetition code} --- The hypergraph product of a pinwheel code with a cyclic repetition code yields a local Type-I fracton model in three dimensions \NoCaseChange{\protect\cite{cite1350}}.
\item\relax
\flmRefsHyperref[eczindexfamilyrel]{code:fracton}{Fracton stabilizer code} --- The hypergraph product of a pinwheel code with a cyclic repetition code yields a local Type-I fracton model in three dimensions, while the hypergraph product of two pinwheel codes yields a local Type-II fracton model in four dimensions \NoCaseChange{\protect\cite{cite1350}}.
\end{eczvaluelist}
\eczhbkcontributors{ \eczhuVVA }
\endeczcode

\eczcode{plane_curve}{Plane-curve evaluation code}{~\NoCaseChange{\protect\cite{cite1797}}}
\codefieldsection{Description}
Evaluation AG code of bivariate polynomials of some finite maximum degree, evaluated at points lying on an affine or projective plane curve.

\codefieldsection{Protection}
Bezout's theorem yields parameters \([n,k,d]\), which depend on the polynomial used to define the plane curve as well as the maximum degree of the polynomials used for evaluation \NoCaseChange{\protect\cite[{pg. 883}]{cite32}}. Distance bounds can be derived from how the plane curve is embedded in the ambient projective space \NoCaseChange{\protect\cite[{Thm. 4.1}]{cite1963}}.
\codefieldsection{Decoding}
\begin{eczvaluelist}
\item\relax Generalization of the Peterson algorithm for BCH codes \NoCaseChange{\protect\cite{cite1797,cite1964,cite1965}}.
\end{eczvaluelist}
\codefieldsection{Parent}
\begin{eczvaluelist}
\item\relax
\flmRefsHyperref[eczindexfamilyrel]{code:evaluation}{Evaluation AG code} --- Plane-curve evaluation codes are evaluation AG codes of bivariate polynomials with \(\cal X\) being an affine plane curve \NoCaseChange{\protect\cite{cite1314}\protect\cite[{Thm. 2.27}]{cite32}}.
\end{eczvaluelist}
\codefieldsection{Cousin}
\begin{eczvaluelist}
\item\relax
\flmRefsHyperref[eczindexfamilyrel]{code:quantum_plane_curve}{Quantum plane-curve code} --- Quantum plane-curve codes are quantum analogues of plane-curve evaluation codes.
\end{eczvaluelist}
\eczhbkcontributors{ \eczhuVVA }
\endeczcode

\eczcode{evaluation_polynomial}{Polynomial evaluation code}{}
\codefieldsection{Description}
Evaluation code of polynomials (or, more generally, rational functions) at points \({\cal P} = \left( P_1,P_2,\cdots,P_n \right)\) on an algebraic variety \(\cal X\) of dimension greater than one (i.e., not an algebraic curve). 

Codewords are evaluations of a linear space \(L\) of rational functions \(f\),
\flmMathEnvironment{align}{}{
  \left( f(P_1), f(P_2), \cdots, f(P_n) \right)~.
}
If the space is taken to be all multivariate polynomials up to some degree, the code is called a \textit{Reed-Muller-type code} or \textit{RM-type code} of that order.

\codefieldsection{Notes}
\begin{eczvaluelist}
\item\relax See Refs. \NoCaseChange{\protect\cite{cite1813,cite1879}} for reviews.
\end{eczvaluelist}
\codefieldsection{Parent}
\begin{eczvaluelist}
\item\relax
\flmRefsHyperref[eczindexfamilyrel]{code:evaluation_varieties}{Evaluation code} --- Polynomial evaluation codes are evaluation codes for which \(\cal X\) is an algebraic variety of dimension greater than one.
\end{eczvaluelist}
\codefieldsection{Children}
\begin{eczvaluelist}
\item\relax
\flmRefsHyperref[eczindexfamilyrel]{code:cascaded_reed_solomon}{Hyperbolic evaluation code} --- A hyperbolic evaluation code is an evaluation code over polynomials in two variables.
\item\relax
\flmRefsHyperref[eczindexfamilyrel]{code:generalized_reed_muller}{Generalized RM (GRM) code} --- GRM (PRM) codes are multivariate polynomial evaluation codes with \(\cal X\) being the entire \(m\)-dimensional affine (projective) space over \(\mathbb{F}_q\) \NoCaseChange{\protect\cite{cite1835,cite32}\protect\cite[{pgs. 44-46}]{cite1314}}.
\item\relax
\flmRefsHyperref[eczindexfamilyrel]{code:complete_intersections}{Complete-intersection RM-type code} --- Complete-intersection RM-type codes are polynomial evaluation codes with \(\cal X\) being a complete intersection.
\item\relax
\flmRefsHyperref[eczindexfamilyrel]{code:deligne_lusztig}{Deligne-Lusztig code} --- Deligne-Lusztig codes are evaluation AG codes with \(\cal X\) a Deligne-Lusztig variety.
\item\relax
\flmRefsHyperref[eczindexfamilyrel]{code:flag_variety}{Flag-variety code} --- Flag-variety codes are polynomial evaluation codes with \(\cal X\) being a flag variety.
\item\relax
\flmRefsHyperref[eczindexfamilyrel]{code:ruled_surface}{Ruled-surface code} --- Ruled-surface codes are polynomial evaluation codes with \(\cal X\) being a ruled surface.
\item\relax
\flmRefsHyperref[eczindexfamilyrel]{code:serge}{Segre-variety RM-type code} --- Segre-variety RM-type codes are polynomial evaluation codes with \(\cal X\) being a Segre variety.
\item\relax
\flmRefsHyperref[eczindexfamilyrel]{code:toric_classical}{Hansen toric code} --- Hansen toric codes are polynomial evaluation codes with \(\cal X\) being a toric variety.
\end{eczvaluelist}
\codefieldsection{Cousin}
\begin{eczvaluelist}
\item\relax
\flmRefsHyperref[eczindexfamilyrel]{code:multiplicity}{Multiplicity code} --- A multiplicity code is a generalization of an \(m\)-variate polynomial evaluation code based on evaluating polynomials together with their Hasse derivatives up to order \(s-1\) at all points in \(\mathbb{F}_q^m\).
\end{eczvaluelist}
\eczhbkcontributors{ \eczhuVVA }
\endeczcode

\eczcode{poset}{Poset code}{~\NoCaseChange{\protect\cite{cite1966}}}

\codefieldsection{Kingdom root code}
for the \flmRefsHyperref{kingdom:q-ary_digits_into_q-ary_digits}{Galois-field Kingdom}.
\codefieldsection{Description}
Encodes \(K\) states (codewords) in \(n\) \(q\)-ary coordinates over the field \(\mathbb{F}_q\), with its distance evaluated in the poset metric.

\codefieldsection{Protection}
Poset codes are quantified with respect to the poset metric \NoCaseChange{\protect\cite{cite1010}}.
This metric is based on a partial ordering \(\leq\) on \([n]=\{1,2,\cdots,n\}\) and the notion of ideals generated by the support of a \(q\)-ary string.
An ideal \(\langle \text{supp}(x) \rangle\) generated by \(x\in \mathbb{F}_q^n\) contains all elements of \([n]\) that are less than or equal to some element in the support of \(x\) in the partial ordering.
The \textit{poset metric} between two strings \(x,y\) is then the cardinality of the ideal generated by the support of \(x-y\), \(d_P(x,y) = |\langle \text{supp}(x-y) \rangle|\).

Generalizations of various bounds for ordinary \(q\)-ary codes have been developed for poset codes, including generalizations of \flmRefsHyperref{ref113}{MacWilliams identities} \NoCaseChange{\protect\cite{cite215,cite1967}}; see \NoCaseChange{\protect\cite{cite1010}}.

\codefieldsection{Notes}
\begin{eczvaluelist}
\item\relax See book \NoCaseChange{\protect\cite{cite1968}} for more details.
\end{eczvaluelist}
\codefieldsection{Parent}
\begin{eczvaluelist}
\item\relax
\flmRefsHyperref[eczindexfamilyrel]{code:symmetric_space}{Symmetric-space code} --- Ordered Hamming space can be viewed as a finite symmetric space \NoCaseChange{\protect\cite{cite214,cite215}\protect\cite[{Sec. 4.2.3}]{cite987}\protect\cite[{Table 3}]{cite985}}.
\end{eczvaluelist}
\codefieldsection{Child}
\begin{eczvaluelist}
\item\relax
\flmRefsHyperref[eczindexfamilyrel]{code:nrt}{Niederreiter-Rosenbloom-Tsfasman (NRT) code}\end{eczvaluelist}
\codefieldsection{Cousins}
\begin{eczvaluelist}
\item\relax
\flmRefsHyperref[eczindexfamilyrel]{code:subspace}{Subspace code} --- Poset-code and subspace-code distance metric families intersect only at the Hamming metric \NoCaseChange{\protect\cite{cite1010}}.
\item\relax
\flmRefsHyperref[eczindexfamilyrel]{code:t-designs}{\(t\)-design} --- Designs exist on ordered Hamming space \NoCaseChange{\protect\cite{cite214,cite215}}.
\end{eczvaluelist}
\eczhbkcontributors{ \eczhuVVA }
\endeczcode

\eczcode{narrow_sense_q-ary_bch}{Primitive narrow-sense BCH code}{~\NoCaseChange{\protect\cite{cite1738}}}
\codefieldsection{Description}
A \(q\)-ary BCH code for \(b=1\) and for \(n=q^r-1\) for some \(r\geq 2\).

\codefieldsection{Protection}
For primitive narrow-sense BCH codes, the minimum weight equals the minimum odd-like weight \NoCaseChange{\protect\cite[{Thm. 2.6.4}]{cite68}}.
For designed distance \(q^h-1\), the true minimum distance is \(q^h-1\) \NoCaseChange{\protect\cite[{Thm. 2.6.5}]{cite68}}.

\codefieldsection{Parents}
\begin{eczvaluelist}
\item\relax
\flmRefsHyperref[eczindexfamilyrel]{code:q-ary_bch}{Bose–Chaudhuri–Hocquenghem (BCH) code} --- BCH codes are called narrow-sense when \(b=1\), and are called primitive when \(n=q^r-1\) for some \(r\geq 2\).
\item\relax
\flmRefsHyperref[eczindexfamilyrel]{code:goppa}{Goppa code} --- Primitive narrow-sense BCH codes are Goppa codes with \(L=\{1,\alpha^{-1},\cdots,\alpha^{1-n}\}\) and \(G(x)=x^{\delta-1}\) \NoCaseChange{\protect\cite[{pg. 522}]{cite126}}.
\end{eczvaluelist}
\codefieldsection{Child}
\begin{eczvaluelist}
\item\relax
\flmRefsHyperref[eczindexfamilyrel]{code:hamming}{\([2^r-1,2^r-r-1,3]\) Hamming code} --- Binary Hamming codes are binary primitive narrow-sense BCH codes \NoCaseChange{\protect\cite[{Corr. 5.1.5}]{cite126}}. Binary Hamming codes can be written in cyclic form \NoCaseChange{\protect\cite[{Thm. 12.22}]{cite961}}.
\end{eczvaluelist}
\codefieldsection{Cousins}
\begin{eczvaluelist}
\item\relax
\flmRefsHyperref[eczindexfamilyrel]{code:generalized_srivastava}{Generalized Srivastava code} --- Binary primitive generalized Srivastava codes with \(z_i=1\) and \(s=1\) are primitive narrow-sense BCH codes \NoCaseChange{\protect\cite[{pg. 359}]{cite41}}.
\item\relax
\flmRefsHyperref[eczindexfamilyrel]{code:data_syndrome}{Quantum data-syndrome (QDS) code} --- Primitive narrow-sense BCH codes can be used as the syndrome measurement codes of a QDS code \NoCaseChange{\protect\cite{cite1969}}. This construction requires fewer measurements than a previous general construction \NoCaseChange{\protect\cite{cite1970}}.
\end{eczvaluelist}
\eczhbkcontributors{ \eczhuVVA }
\endeczcode

\eczcode{projective}{Projective geometry code}{}
\codefieldsection{Description}
Linear \(q\)-ary \([n,k,d]\) code whose generator matrix \(G\) does not contain any repeated columns or the zero column.
That way, each column corresponds to a distinct point in the projective space \(PG(k-1,q)\) arising from a \(k\)-dimensional vector space over \(\mathbb{F}_q\). A choice of \(k\) linearly independent columns determines an \textit{information set}.
Columns of a code's parity-check matrix can similarly correspond to points in projective space. This formulation yields connections to projective geometry, which can be applied to determine code properties.

Recall that a linear code encodes a message \(h\) into a codeword \(c = hG\). The \(i\)th coordinate of a codeword is given by the dot product \(h \cdot G_{i}\), with \(G_{i}\) being the \(i\)th column of the generator matrix. The zero-coordinate condition \(h \cdot x = 0\) defines a hyperplane of points \(x\) with normal vector \(h\). Therefore, the Hamming weight of the corresponding codeword is the number of points \(G_i\) \textit{not} contained in that hyperplane.

In general, linear codes can admit repeated columns or columns proportional to each other. In that case, the columns correspond to a multiset of not-necessarily-distinct nonzero points of projective space \NoCaseChange{\protect\cite{cite1860,cite63}}. Multisets can also be used to construct parity-check matrices of linear codes.

\codefieldsection{Protection}
Distance \(d\) is \(n\) minus the maximum number of points that are contained in a hyperplane. For \(n \geq 3\), a code is projective if and only if the distance of its dual code is at least three.

The weight enumerator of the code comes from the Tutte polynomial associated with the projective code \NoCaseChange{\protect\cite{cite1971}}.

\codefieldsection{Notes}
\begin{eczvaluelist}
\item\relax See corresponding definition in \flmHref{https://web.archive.org/web/20240228180821/http://mint.sbg.ac.at/glossary.php\#GProjectiveCode}{MinT}.
\end{eczvaluelist}
\codefieldsection{Parent}
\begin{eczvaluelist}
\item\relax
\flmRefsHyperref[eczindexfamilyrel]{code:q-ary_linear}{Linear \(q\)-ary code} --- Columns of the generator matrix of a projective linear \([n,k]_q\) code correspond to distinct nonzero points in projective space. In general, linear codes admit repeating columns or columns proportional to each other. In that case, the columns correspond to a multiset of non-distinct nonzero points, and multisets are in one-to-one correspondence to arcs in projective space \NoCaseChange{\protect\cite[{Thm. 1.1}]{cite1860}}.
\end{eczvaluelist}
\codefieldsection{Children}
\begin{eczvaluelist}
\item\relax
\flmRefsHyperref[eczindexfamilyrel]{code:homological_classical}{Cycle code} --- Incidence matrices of graphs have no repeated columns since that would correspond to multi-edges. Therefore, cycle codes can be interpreted as projective codes.
\item\relax
\flmRefsHyperref[eczindexfamilyrel]{code:laplacian}{Laplacian code} --- Incidence matrices of graphs have no repeated columns since that would correspond to multi-edges. Therefore, Laplacian codes can be interpreted as projective codes.
\item\relax
\flmRefsHyperref[eczindexfamilyrel]{code:glynn}{\([10,5,6]_9\) Glynn code} --- The Glynn code is constructed using a 10-arc in \(PG(4,9)\) that is not a rational curve.
\item\relax
\flmRefsHyperref[eczindexfamilyrel]{code:hirschfeld}{Hirschfeld code}\item\relax
\flmRefsHyperref[eczindexfamilyrel]{code:hyperoval}{Hyperoval code}\item\relax
\flmRefsHyperref[eczindexfamilyrel]{code:incidence_matrix}{Incidence-matrix projective code}\item\relax
\flmRefsHyperref[eczindexfamilyrel]{code:projective_two_weight}{Projective two-weight code} --- Projective two-weight codes are projective codes by definition \NoCaseChange{\protect\cite[{Sec. 19.2}]{cite172}} (see also \NoCaseChange{\protect\cite{cite1972,cite203,cite1973}}).
\end{eczvaluelist}
\codefieldsection{Cousins}
\begin{eczvaluelist}
\item\relax
\flmRefsHyperref[eczindexfamilyrel]{code:q-ary_cyclic}{Cyclic linear \(q\)-ary code} --- Every projective linear code is a punctured code of a special cyclic code \NoCaseChange{\protect\cite[{Thm. 2.3.6}]{cite68}}.
\item\relax
\flmRefsHyperref[eczindexfamilyrel]{code:evaluation_varieties}{Evaluation code} --- Codewords of an evaluation code of multivariate polynomials up to degree one evaluated at points in projective space yield a projective code.
\item\relax
\flmRefsHyperref[eczindexfamilyrel]{code:extended_reed_solomon}{Extended GRS code} --- Columns of parity-check matrices of doubly extended narrow-sense RS codes consist of points of a normal rational curve \NoCaseChange{\protect\cite[{Def. 14.2.6}]{cite202}}.
\item\relax
\flmRefsHyperref[eczindexfamilyrel]{code:narrow_sense_reed_solomon}{Narrow-sense RS code} --- Columns of parity-check matrices of doubly extended narrow-sense RS codes consist of points of a normal rational curve \NoCaseChange{\protect\cite[{Def. 14.2.6}]{cite202}}.
\item\relax
\flmRefsHyperref[eczindexfamilyrel]{code:griesmer}{Griesmer code} --- Arcs corresponding to Griesmer codes are called Griesmer arcs \NoCaseChange{\protect\cite[{pg. 248}]{cite1860}}. There is a one-to-one correspondence between Griesmer codes and minihypers \NoCaseChange{\protect\cite{cite1861,cite1862}}; see \NoCaseChange{\protect\cite{cite1863}\protect\cite[{Sec. 14.2.4}]{cite202}} for more details.
\item\relax
\flmRefsHyperref[eczindexfamilyrel]{code:anticode}{Anticode} --- There is a relation between anticodes and minihypers \NoCaseChange{\protect\cite[{pg. 295}]{cite62}}.
\item\relax
\flmRefsHyperref[eczindexfamilyrel]{code:finite_grassmann}{Constant-dimension code} --- The projective plane \(PG(k-1,q)\) is a special case of the finite-field Grassmannian \NoCaseChange{\protect\cite{cite28}}.
\item\relax
\flmRefsHyperref[eczindexfamilyrel]{code:subspace}{Subspace code} --- Subspace codes are sets of subspaces of a projective space \(PG(n-1,q)\).
\item\relax
\flmRefsHyperref[eczindexfamilyrel]{code:projective_reed_muller}{Projective RM (PRM) code} --- Nonzero codewords of minimum weight of a \(r\)th-order \(q\)-ary projective RM code correspond to algebraic hypersurfaces of degree \(r\) having the largest number of points in the projective space \(PG(m,q)\) \NoCaseChange{\protect\cite[{Thm. 14.3.3}]{cite202}}.
\item\relax
\flmRefsHyperref[eczindexfamilyrel]{code:mds}{Maximum distance separable (MDS) code} --- A linear code is MDS (almost MDS) if and only if columns of its parity-check matrix form an \(n\)-arc (\(n\)-track) in projective space \NoCaseChange{\protect\cite{cite1889,cite1918,cite1863,cite1919}}. The dual of a MDS code is an MDS code, so MDS codes are projective. All \([9,3]\) MDS codes have been tabulated \NoCaseChange{\protect\cite{cite1920}} in terms of 9-arcs in the projective plane.
\item\relax
\flmRefsHyperref[eczindexfamilyrel]{code:ternary_golay}{\([11,6,5]_3\) Ternary Golay code} --- The extended ternary Golay code admits a projective geometric construction \NoCaseChange{\protect\cite[{pg. 296}]{cite62}}.
\item\relax
\flmRefsHyperref[eczindexfamilyrel]{code:q-ary_duadic}{\(q\)-ary duadic code} --- The weight-five codewords of the \([21,5,6]_4\) quaternary duadic code support a projective plane of order 4 \NoCaseChange{\protect\cite[{Table 6.3}]{cite126}}.
\item\relax
\flmRefsHyperref[eczindexfamilyrel]{code:two_weight}{Two-weight code} --- Projective two-weight codes correspond to projective 2-character sets and certain strongly regular graphs \NoCaseChange{\protect\cite{cite1972,cite203,cite1973}\protect\cite[{Sec. 19.3.3}]{cite172}\protect\cite[{Sec. 19.9.1}]{cite172}}.
\item\relax
\flmRefsHyperref[eczindexfamilyrel]{code:octacode}{Octacode} --- Columns of the heptacode's (octacode's) generator matrix represent the seven (eight) points of a hyperoval (8-arc) in the projective Hjelmslev plane \(PHG(2,\mathbb{Z}_4)\) (\(PHG(3,\mathbb{Z}_4)\)) \NoCaseChange{\protect\cite{cite1974}\protect\cite[{Exam. 5}]{cite1147}}.
\item\relax
\flmRefsHyperref[eczindexfamilyrel]{code:cgs_spherical}{Cameron-Goethals-Seidel (CGS) isotropic subspace code} --- CGS isotropic subspace codes are constructed from incidence matrices of \(PG(5,q)\) \NoCaseChange{\protect\cite[{Exam. 9.4.5}]{cite115}}.
\item\relax
\flmRefsHyperref[eczindexfamilyrel]{code:quantum_cap}{\(\llbracket n,n-2k,4\rrbracket \) Quantum cap code} --- A quantum cap code is a distance-four \flmRefsHyperref{ref672}{pure} Hermitian qubit code constructed by identifying its underlying Hermitian self-orthogonal \([n,k]_4\) code with a particular projective cap in \(PG(k-1,4)\).
\item\relax
\flmRefsHyperref[eczindexfamilyrel]{code:quantum_hamming}{\(\llbracket 2^r, 2^r-r-2, 3\rrbracket \) Gottesman code} --- Gottesman codes are related to partial spreads in projective geometry \NoCaseChange{\protect\cite{cite1695}}.
\item\relax
\flmRefsHyperref[eczindexfamilyrel]{code:stab_9_3_3}{\(\llbracket 9,3,3\rrbracket \) Quadric code} --- The \(\llbracket 9,3,3\rrbracket \) quadric code can be constructed from the elliptic quadric in \(PG(5, 2)\) \NoCaseChange{\protect\cite{cite1695}}.
\item\relax
\flmRefsHyperref[eczindexfamilyrel]{code:stabilizer_over_gf4}{Hermitian qubit code} --- There is an equivalence between \flmRefsHyperref{ref672}{pure} \(\llbracket n,n-2k\rrbracket \) Hermitian qubit codes and certain sets of points in projective space \(PG(k-1,4)\) \NoCaseChange{\protect\cite[{Thm. 2.8}]{cite1695}}.
\item\relax
\flmRefsHyperref[eczindexfamilyrel]{code:purity_testing}{Purity-testing stabilizer code} --- Purity-testing stabilizer codes are constructed from normal rational curves.
\item\relax
\flmRefsHyperref[eczindexfamilyrel]{code:pg_qldpc}{Finite-geometry (FG) qubit QLDPC code} --- PG-QLDPC codes are constructed from linear binary codes whose parity-check or generator matrices are incidence matrices of structures in finite geometries.
\item\relax
\flmRefsHyperref[eczindexfamilyrel]{code:qubit_stabilizer}{Qubit stabilizer code} --- \(\llbracket n,k,d\rrbracket \) qubit stabilizer codes with no weight-one stabilizers are equivalent to particular "quantum" sets of lines in projective space \(PG(n-k-1,2)\) \NoCaseChange{\protect\cite{cite1976,cite1977}\protect\cite[{Thm. 3.7}]{cite1975}}. This equivalence is stated in the case of \flmRefsHyperref{ref672}{pure} qubit stabilizer codes with distance two or greater in \NoCaseChange{\protect\cite[{Thm. 2.6}]{cite1695}}.
\item\relax
\flmRefsHyperref[eczindexfamilyrel]{code:galois_non_stabilizer}{Galois-qudit USt code} --- Galois-qudit USt codes can be obtained from lines in projective space \NoCaseChange{\protect\cite{cite1977,cite1978}}.
\end{eczvaluelist}
\eczhbkcontributors{ \eczhuVVA }
\endeczcode

\eczcode{projective_reed_muller}{Projective RM (PRM) code}{~\NoCaseChange{\protect\cite{cite1979,cite1980}}}
\codefieldsection{Description}
Evaluation code obtained by evaluating homogeneous polynomials on the points of the projective space \(PG(m,q)\), equivalently on representatives of the nonzero vectors in \(\mathbb{F}_q^{m+1}\) whose leftmost nonzero coordinate is one.

PRM codes include the codes PRM\(_q(r,m)\) for \(r<q\), which are injective evaluation codes with parameters \NoCaseChange{\protect\cite{cite1981}}
\flmMathEnvironment{align}{}{
  \left[ q^m+q^{m-1}+\cdots +1, {m+r \choose r},(q+1-r)q^{m-1} \right]~.
}

The dual of a PRM code is not always a PRM code \NoCaseChange{\protect\cite{cite1980,cite1982}}.

\codefieldsection{Parent}
\begin{eczvaluelist}
\item\relax
\flmRefsHyperref[eczindexfamilyrel]{code:generalized_reed_muller}{Generalized RM (GRM) code} --- Nonzero codewords of minimum weight of a \(r\)th-order \(q\)-ary projective RM code correspond to algebraic hypersurfaces of degree \(r\) having the largest number of points in the projective space \(PG(m,q)\) \NoCaseChange{\protect\cite[{Thm. 14.3.3}]{cite202}}.
\end{eczvaluelist}
\codefieldsection{Child}
\begin{eczvaluelist}
\item\relax
\flmRefsHyperref[eczindexfamilyrel]{code:q-ary_simplex}{\(q\)-ary simplex code} --- The \(q\)-ary simplex codes are first-order PRM codes \NoCaseChange{\protect\cite[{Sec. 1.2.2}]{cite1314}}.
\end{eczvaluelist}
\codefieldsection{Cousins}
\begin{eczvaluelist}
\item\relax
\flmRefsHyperref[eczindexfamilyrel]{code:projective}{Projective geometry code} --- Nonzero codewords of minimum weight of a \(r\)th-order \(q\)-ary projective RM code correspond to algebraic hypersurfaces of degree \(r\) having the largest number of points in the projective space \(PG(m,q)\) \NoCaseChange{\protect\cite[{Thm. 14.3.3}]{cite202}}.
\item\relax
\flmRefsHyperref[eczindexfamilyrel]{code:griesmer}{Griesmer code} --- PRM codes for \(r=1\) attain the Griesmer bound for all \(m\) \NoCaseChange{\protect\cite{cite1813}}.
\item\relax
\flmRefsHyperref[eczindexfamilyrel]{code:galois_reed_muller}{Galois-qudit quantum RM code} --- Projective RM codes can be used to construct Galois-qudit RM codes \NoCaseChange{\protect\cite{cite1839}\protect\cite[{Sec. 6.2}]{cite872}}.
\end{eczvaluelist}
\eczhbkcontributors{ \eczhuVVA }
\endeczcode

\eczcode{projective_two_weight}{Projective two-weight code}{}
\codefieldsection{Description}
A projective code whose codewords all have one of two possible nonzero Hamming weights.

There is a correspondence between projective two-weight codes, projective two-character sets, and certain strongly regular graphs \NoCaseChange{\protect\cite{cite1983}\protect\cite[{Sec. 19.3.3}]{cite172}}.
As such, projective two-weight codes have been classified and can be constructed out of quadrics \NoCaseChange{\protect\cite{cite1917,cite1984}}, maximal arcs and hyperovals, Baer spaces, or Hermitian quadrics \NoCaseChange{\protect\cite[{Secs. 19.7.1-19.7.5}]{cite172}}.
There are also several sporadic examples \NoCaseChange{\protect\cite[{Table 19.1}]{cite172}}.

\codefieldsection{Protection}
In a projective two-weight code, the difference between the two nonzero weights is a power of the characteristic \NoCaseChange{\protect\cite[{Sec. 19.3.6}]{cite172}}.

\codefieldsection{Parents}
\begin{eczvaluelist}
\item\relax
\flmRefsHyperref[eczindexfamilyrel]{code:projective}{Projective geometry code} --- Projective two-weight codes are projective codes by definition \NoCaseChange{\protect\cite[{Sec. 19.2}]{cite172}} (see also \NoCaseChange{\protect\cite{cite1972,cite203,cite1973}}).
\item\relax
\flmRefsHyperref[eczindexfamilyrel]{code:two_weight}{Two-weight code} --- Projective two-weight codes are two-weight codes by definition \NoCaseChange{\protect\cite[{Def. 19.1}]{cite172}} (see also \NoCaseChange{\protect\cite{cite1972,cite203,cite1973}}).
\end{eczvaluelist}
\codefieldsection{Children}
\begin{eczvaluelist}
\item\relax
\flmRefsHyperref[eczindexfamilyrel]{code:bose_qvist}{Ovoid code} --- The ovoid code is a two-weight projective code \NoCaseChange{\protect\cite{cite1656,cite1657}\protect\cite[{Table 7.1}]{cite206}}.
\item\relax
\flmRefsHyperref[eczindexfamilyrel]{code:denniston}{Denniston code} --- Denniston codes are projective two-weight codes on maximal arcs \NoCaseChange{\protect\cite{cite1774}\protect\cite[{Sec. 19.7.3}]{cite172}}.
\item\relax
\flmRefsHyperref[eczindexfamilyrel]{code:hill_cap}{Hill projective-cap code} --- Hill projective-cap codes are projective two-weight codes on projective caps \NoCaseChange{\protect\cite[{Table 19.1}]{cite172}}.
\end{eczvaluelist}
\codefieldsection{Cousins}
\begin{eczvaluelist}
\item\relax
\flmRefsHyperref[eczindexfamilyrel]{code:ternary_golay}{\([11,6,5]_3\) Ternary Golay code} --- The dual of the ternary Golay code is a projective two-weight subcode \NoCaseChange{\protect\cite{cite1656,cite1657}\protect\cite[{Exam. 19.3.2}]{cite172}\protect\cite[{Table 7.1}]{cite206}}.
\item\relax
\flmRefsHyperref[eczindexfamilyrel]{code:hyperoval}{Hyperoval code} --- Codes based on hyperovals in \(PG(2,q)\) with even \(q\) are projective two-weight codes \NoCaseChange{\protect\cite{cite1656,cite1657}\protect\cite[{Exam. 19.2.1}]{cite172}\protect\cite[{Table 7.1}]{cite206}}.
\end{eczvaluelist}
\eczhbkcontributors{ \eczhuVVA }
\endeczcode

\eczcode{pyramid}{Pyramid code}{~\NoCaseChange{\protect\cite{cite1985}}}
\codefieldsection{Description}
An LRC whose generator matrix is that of an RS code in standard form, but some of whose columns are split into multiple columns; see \NoCaseChange{\protect\cite[{Sec. 31.3.1.1}]{cite183}} for an example. 

\codefieldsection{Parents}
\begin{eczvaluelist}
\item\relax
\flmRefsHyperref[eczindexfamilyrel]{code:q-ary_linear}{Linear \(q\)-ary code}\item\relax
\flmRefsHyperref[eczindexfamilyrel]{code:optimal_lrc}{Optimal LRC}\end{eczvaluelist}
\codefieldsection{Cousin}
\begin{eczvaluelist}
\item\relax
\flmRefsHyperref[eczindexfamilyrel]{code:reed_solomon}{Reed-Solomon (RS) code} --- A pyramid code is an LRC whose generator matrix is that of an RS code in standard form, but some of whose columns are split into multiple columns.
\end{eczvaluelist}
\eczhbkcontributors{ \eczhuVVA }
\endeczcode

\eczcode{q-ary_quad_residue}{Quadratic-residue (QR) code}{}
\codefieldsection{Description}
Member of a quadruple of cyclic \(q\)-ary codes of prime length \(n\) where \(q\) is prime and a quadratic-residue modulo \(n\) \NoCaseChange{\protect\cite[{Def. 3.2.8}]{cite70}}.
The codes are constructed using quadratic residues and nonresidues of \(n\).
The definition extends to prime-power alphabet sizes and to prime-power lengths \NoCaseChange{\protect\cite{cite174,cite175}\protect\cite[{Rem. 3.2.9}]{cite70}}.
A quadratic-residue code of prime length \(p\) has dimension \((p+1)/2\) \NoCaseChange{\protect\cite[{Sec. 3.2.1}]{cite70}}.

The roots of the generator polynomial \(r(x)\) of the first code (see \flmRefsCref{ref67}) are all of the inequivalent quadratic residues of \(n\), and the second code's generator polynomial is \((x-1)r(x)\). The roots of the generator polynomial \(a(x)\) of the third code are all inequivalent nonresidues of \(n\), and the fourth code's generator polynomial is \((x-1)a(x)\). The codes corresponding to polynomials \(r,a\) are often called \textit{augmented} quadratic-residue codes, while the remaining codes are called \textit{expurgated}.

The extended versions of odd-like quadratic-residue codes have automorphism groups containing \(PSL(2,p)\) \NoCaseChange{\protect\cite[{Thm. 3.2.11}]{cite70}}; such codes are the only codes with these symmetries \NoCaseChange{\protect\cite{cite1986}}.

\codefieldsection{Protection}
For odd-like quadratic-residue codes of prime length, the common minimum distance \(d\) satisfies \(d^2 \geq n\); if \(-1\) is a quadratic non-residue, then \(d^2-d+1 \geq n\) \NoCaseChange{\protect\cite[{Thm. 2.7.4}]{cite68}}.
\codefieldsection{Rate}
Achieve capacity of the binary erasure channel; see Ref. \NoCaseChange{\protect\cite{cite1661}}.
\codefieldsection{Notes}
\begin{eczvaluelist}
\item\relax Introduction of quadratic-residue codes in Refs. \NoCaseChange{\protect\cite{cite41,cite126}\protect\cite[{Sec. 2.7}]{cite68}}.
\end{eczvaluelist}
\codefieldsection{Parent}
\begin{eczvaluelist}
\item\relax
\flmRefsHyperref[eczindexfamilyrel]{code:q-ary_duadic}{\(q\)-ary duadic code} --- QR codes are duadic codes of prime length satisfying certain relations \NoCaseChange{\protect\cite{cite69}}.
\end{eczvaluelist}
\codefieldsection{Children}
\begin{eczvaluelist}
\item\relax
\flmRefsHyperref[eczindexfamilyrel]{code:binary_quad_residue}{Binary quadratic-residue (QR) code}\item\relax
\flmRefsHyperref[eczindexfamilyrel]{code:shortened_hexacode}{\([5,3,3]_4\) Shortened hexacode} --- The shortened hexacode is an odd-like quadratic-residue code \NoCaseChange{\protect\cite[{Exam. 6.6.8}]{cite126}}.
\item\relax
\flmRefsHyperref[eczindexfamilyrel]{code:ternary_golay}{\([11,6,5]_3\) Ternary Golay code} --- The ternary Golay code is a quadratic-residue code over \(\mathbb{F}_3\) with residue set \(Q = \{1, 3, 4, 5, 9\} \) and generator polynomial \(x^5 + x^4 - x^3 + x^2 - 1\) \NoCaseChange{\protect\cite[{Ex. 3.2.10}]{cite70}\protect\cite[{Ch. 16}]{cite41}}.
\end{eczvaluelist}
\codefieldsection{Cousins}
\begin{eczvaluelist}
\item\relax
\flmRefsHyperref[eczindexfamilyrel]{code:combinatorial_design}{Combinatorial design} --- The supports of fixed-weight codewords of certain \(q\)-ary QR codes support combinatorial designs \NoCaseChange{\protect\cite{cite149,cite136,cite154}}, including \(3\)-designs \NoCaseChange{\protect\cite{cite155}}.
\item\relax
\flmRefsHyperref[eczindexfamilyrel]{code:karlin}{\([2m+2,m+1]\) Karlin code} --- Karlin codes can be mapped to extended cyclic and extended quadratic-residue codes over \(\mathbb{F}_4\) \NoCaseChange{\protect\cite{cite109,cite110}\protect\cite[{Ch. 16}]{cite41}\protect\cite[{Sec. 2.4.2}]{cite42}} by identifying \((0,\omega,\bar{\omega},1)\) with \((00),(10),(01),(11)\) \NoCaseChange{\protect\cite{cite109}}.
\item\relax
\flmRefsHyperref[eczindexfamilyrel]{code:pless_symmetry}{\([2q+2,q+1]_3\) Pless symmetry code} --- Pless symmetry codes for lengths 24, 48, and 60 they have the same Hamming weight enumerators as the corresponding extended QR codes but are not equivalent \NoCaseChange{\protect\cite[{pg. 511}]{cite41}}.
\item\relax
\flmRefsHyperref[eczindexfamilyrel]{code:hexacode}{\([6,3,4]_4\) Hexacode} --- The hexacode is the smallest example of an extended quadratic-residue code of Type \(4^H\) \NoCaseChange{\protect\cite[{Sec. 2.4.6}]{cite42}\protect\cite[{Exer. 363}]{cite126}}.
\item\relax
\flmRefsHyperref[eczindexfamilyrel]{code:reed_solomon_4}{\([4,2,3]_4\) RS\(_4\) code} --- The RS\(_4\) code is the smallest quaternary extended QR code \NoCaseChange{\protect\cite[{Sec. 2.4.2}]{cite42}}. The shortened RS\(_4\) code is the smallest quaternary QR code.
\item\relax
\flmRefsHyperref[eczindexfamilyrel]{code:octacode}{Octacode} --- The octacode is equivalent to the length-eight quaternary extended quadratic-residue code over \(\mathbb{Z}_4\) \NoCaseChange{\protect\cite{cite112}}.
\item\relax
\flmRefsHyperref[eczindexfamilyrel]{code:galois_quad_residue}{Quantum quadratic-residue (QR) code} --- Quantum quadratic-residue codes are quantum analogues of \(q\)-ary quadratic-residue codes.
\end{eczvaluelist}
\eczhbkcontributors{ \eczhuVVA }
\endeczcode

\eczcode{quadric}{Quadric code}{~\NoCaseChange{\protect\cite{cite1987,cite1988}}}
\codefieldsection{Description}
Evaluation code of polynomials evaluated on points lying on a quadric hypersurface.

\codefieldsection{Parent}
\begin{eczvaluelist}
\item\relax
\flmRefsHyperref[eczindexfamilyrel]{code:flag_variety}{Flag-variety code} --- Quadric codes are flag-variety evaluation codes with the flag variety being a quadric hypersurface.
\end{eczvaluelist}
\eczhbkcontributors{ \eczhuVVA }
\endeczcode

\eczcode{quantum_inspired}{Quantum-inspired classical block code}{}
\codefieldsection{Description}
A \(q\)-ary linear code whose construction was inspired by a quantum code.

\codefieldsection{Parent}
\begin{eczvaluelist}
\item\relax
\flmRefsHyperref[eczindexfamilyrel]{code:q-ary_linear}{Linear \(q\)-ary code}\end{eczvaluelist}
\codefieldsection{Children}
\begin{eczvaluelist}
\item\relax
\flmRefsHyperref[eczindexfamilyrel]{code:laplacian}{Laplacian code}\item\relax
\flmRefsHyperref[eczindexfamilyrel]{code:fibonacci_model}{Fibonacci code}\item\relax
\flmRefsHyperref[eczindexfamilyrel]{code:gauss_law}{Gauss' law code}\item\relax
\flmRefsHyperref[eczindexfamilyrel]{code:pinwheel}{Pinwheel code}\item\relax
\flmRefsHyperref[eczindexfamilyrel]{code:topological_classical}{Classical topological code}\item\relax
\flmRefsHyperref[eczindexfamilyrel]{code:classical_fractal_liquid}{Classical fractal liquid code}\end{eczvaluelist}
\codefieldsection{Cousins}
\begin{eczvaluelist}
\item\relax
\flmRefsHyperref[eczindexfamilyrel]{code:homological_classical}{Cycle code} --- Cycle codes have been known in classical coding theory, and have been rediscovered in the quantum context; see Ref. \NoCaseChange{\protect\cite{cite1304}} for a brief exposition.
\item\relax
\flmRefsHyperref[eczindexfamilyrel]{code:xcube}{X-cube model code} --- According to Ref. \NoCaseChange{\protect\cite{cite1349}}, a classical analogue of the X-cube model is the eight-vertex model \NoCaseChange{\protect\cite{cite1989,cite1990,cite1991}}.
\end{eczvaluelist}
\eczhbkcontributors{ \eczhuVVA }
\endeczcode

\eczcode{quasi_group}{Quasi group-algebra code}{}
\codefieldsection{Alternative Names}
\begin{eczvaluelist}
\item\relax Quasi-\(G\) code
\end{eczvaluelist}
\eczhIndexCodeAliasName{quasi_group}{Quasi-\(G\) code}
\codefieldsection{Description}
A \(q\)-ary linear code based on a finite group \( G \) of order \(n/\ell\) for some \textit{index} \(\ell\).
The code is a right submodule of the direct sum of \(\ell\) copies of the \flmRefsHyperref{ref205}{group algebra} \(\mathbb{F}_q G\).
A quasi group-algebra code for an Abelian group is called an \textit{Abelian quasi group-algebra code}.

\codefieldsection{Parent}
\begin{eczvaluelist}
\item\relax
\flmRefsHyperref[eczindexfamilyrel]{code:q-ary_linear}{Linear \(q\)-ary code} --- A linear code is a quasi group-algebra code for a group \(G\) and index \(\ell\) if and only if \(G\) is isomorphic to a regular subgroup of the code's permutation automorphism group which acts freely of index \(\ell\) on the coordinates \NoCaseChange{\protect\cite[{Thm. 3.5}]{cite1115}}.
\end{eczvaluelist}
\codefieldsection{Children}
\begin{eczvaluelist}
\item\relax
\flmRefsHyperref[eczindexfamilyrel]{code:extended_golay}{\([24, 12, 8]\) Extended Golay code} --- The extended Golay code is a quasi group-algebra code for various groups \NoCaseChange{\protect\cite{cite1193,cite1194,cite1115}}.
\item\relax
\flmRefsHyperref[eczindexfamilyrel]{code:group}{Group-algebra code} --- A quasi group-algebra code of index \(\ell=1\) is a group-algebra code.
\end{eczvaluelist}
\codefieldsection{Cousin}
\begin{eczvaluelist}
\item\relax
\flmRefsHyperref[eczindexfamilyrel]{code:quasi_cyclic}{Quasi-cyclic code} --- A quasi group-algebra code for \(G\) being the cyclic group is a quasi-cyclic \(q\)-ary linear code \NoCaseChange{\protect\cite[{pg. 4}]{cite1115}}.
\end{eczvaluelist}
\eczhbkcontributors{ Pavel Panteleev, \eczhuVVA }
\endeczcode

\eczcode{quasi_perfect}{Quasi-perfect code}{}
\codefieldsection{Description}
Perfect codes \((n,K,d)_q\) are those for which balls of Hamming radius \(t=\left\lfloor (d-1)/2\right\rfloor\) exactly fill the space of all \(n\) \(q\)-ary strings. \textit{Quasi-perfect codes} are those for which balls of Hamming radius \(t\) are disjoint, while balls of radius \(t+1\) cover the space with possible overlaps. In other words, any \(q\)-ary string is at most \(t+1\) bit flips away from a codeword of a quasi-perfect code.

\codefieldsection{Protection}
Correct errors of weight \(t\) as well as some errors of weight \(t+1\).
\codefieldsection{Parents}
\begin{eczvaluelist}
\item\relax
\flmRefsHyperref[eczindexfamilyrel]{code:uniformly_packed}{Uniformly packed code} --- Quasi-perfect codes are uniformly packed \NoCaseChange{\protect\cite[{Def. 2.5}]{cite1748}}.
\item\relax
\flmRefsHyperref[eczindexfamilyrel]{code:weighed_covering}{Weighted-covering code} --- A quasi-perfect code is a special case of an \(m\)-weighted covering code with \(m\)-covering radius \(r=t+1\) \NoCaseChange{\protect\cite[{Ch. 13}]{cite244}}.
\end{eczvaluelist}
\codefieldsection{Children}
\begin{eczvaluelist}
\item\relax
\flmRefsHyperref[eczindexfamilyrel]{code:nearly_perfect}{Nearly perfect code} --- Nearly perfect codes are quasi-perfect \NoCaseChange{\protect\cite[{pg. 533}]{cite41}}.
\item\relax
\flmRefsHyperref[eczindexfamilyrel]{code:zetterberg}{Zetterberg code} --- Zetterberg codes are quasi-perfect, with each \(n\)-bit string at most three bit-flips away from a codeword \NoCaseChange{\protect\cite{cite1643}}.
\end{eczvaluelist}
\codefieldsection{Cousins}
\begin{eczvaluelist}
\item\relax
\flmRefsHyperref[eczindexfamilyrel]{code:bch}{Binary BCH code} --- Only double error-correcting BCH codes \([2^m-1,n-2m,5]\) are quasi-perfect \NoCaseChange{\protect\cite{cite1243,cite1244}} (see also \NoCaseChange{\protect\cite[{Ch. 9}]{cite41}}).
\item\relax
\flmRefsHyperref[eczindexfamilyrel]{code:preparata}{Preparata code} --- Shortened Preparata codes are quasi-perfect \NoCaseChange{\protect\cite[{pg. 475}]{cite41}}.
\end{eczvaluelist}
\eczhbkcontributors{ \eczhuVVA }
\endeczcode

\eczcode{reed_solomon}{Reed-Solomon (RS) code}{~\NoCaseChange{\protect\cite{cite1935,cite1665,cite1936}}}
\codefieldsection{Description}
An \([n,k,n-k+1]_q\) linear code based on polynomials over \(\mathbb{F}_q\).

Let \(\{\alpha_1,\cdots,\alpha_n\}\) be \(n\) distinct points in \(\mathbb{F}_q\). An RS code encodes a message \(\mu=\{\mu_0,\cdots,\mu_{k-1}\}\) into \(\{f_\mu(\alpha_1),\cdots,f_\mu(\alpha_n)\}\) using a message-dependent polynomial
\flmMathEnvironment{align}{}{
f_\mu(x)=\mu_0+\mu_1 x + \cdots + \mu_{k-1}x^{k-1}.
}
In other words, each message \(\mu\) is encoded into a string of values of the corresponding polynomial \(f_\mu\) at the points \(\alpha_i\),
\flmMathEnvironment{align}{}{
  \mu\to\left( f_{\mu}\left(\alpha_{1}\right),f_{\mu}\left(\alpha_{2}\right),\cdots,f_{\mu}\left(\alpha_{n}\right)\right) \,.
}

The dual of an RS code is an RS code \NoCaseChange{\protect\cite[{pg. 296}]{cite41}}.

\codefieldsection{Protection}
Since each polynomial \(f_{\mu}\) is of degree less than \(k\), it has less than \(k\) roots; this is called the \textit{degree mantra}. Therefore, the polynomial can be determined from its values at \(k\) points. This means that RS codes can correct erasures on up to \(n-k\) registers. The resulting distance, \(d=n-k+1\), saturates the Singleton bound.
\codefieldsection{Encoding}
\begin{eczvaluelist}
\item\relax Bit-serial encoder \NoCaseChange{\protect\cite{cite1992}}.
\item\relax \([n,k,n-k+1]\) RS code requires an \flmRefsHyperref{ref65}{order} \(O(n^2)\) operations while encoding if a straightforward matrix multiplication is employed and \(k=O(n)\). Using the FFT algorithm, complexity of evaluating a polynomial at \(n\) roots of unity becomes \(O(n\log n)\). The FFT can be generalized to finite fields and rings, which is referred as Number-theoretic Transform (NTT). However, for some values of \(n\), which can not be factorized into small primes or do not have \(n\) roots of unity, the FFT algorithm fails. Independently developed by \NoCaseChange{\protect\cite{cite1993,cite1994}} and generalized in Ref. \NoCaseChange{\protect\cite{cite1995}}, the additive FFT solves this problem by evaluating the polynomial at \(n-1\) roots of unity when \(n\) is power of 2.
\end{eczvaluelist}
\codefieldsection{Decoding}
\begin{eczvaluelist}
\item\relax Decoding general RS codes is \(NP\)-hard \NoCaseChange{\protect\cite{cite1996}}.
\item\relax Although using iFFT has its counterpart iNNT for finite fields, the decoding is usually standard polynomial interpolation in \(k=O(n\log^2 n)\). However, in erasure decoding, encoded values are only erased in \(r\) points, which is a specific case of polynomial interpolation and can be done in \(O(n\log n)\) by computing product of the received polynomial and an erasure locator polynomial and using long division to find an original polynomial. The long division step can be omitted to increase speed further by only dividing the derivative of the product polynomial, and derivative of erasure locator polynomial evaluated at erasure locations.
\item\relax Berlekamp-Massey decoder with runtime of \flmRefsHyperref{ref65}{order} \(O(n^2)\) \NoCaseChange{\protect\cite{cite1234,cite1235}}.
\item\relax Gorenstein-Peterson-Zierler decoder with runtime of \flmRefsHyperref{ref65}{order} \(O(n^3)\) \NoCaseChange{\protect\cite{cite1231,cite1738}} (see exposition in Ref. \NoCaseChange{\protect\cite{cite1233}}).
\item\relax Berlekamp-Welch decoder with runtime of \flmRefsHyperref{ref65}{order} \(O(n^3)\) \NoCaseChange{\protect\cite{cite1997}} (see exposition in Ref. \NoCaseChange{\protect\cite{cite1998}}), assuming that \(t \geq (n+k)/2\).
\item\relax Sugiyama et al. modification of the extended Euclidean algorithm \NoCaseChange{\protect\cite{cite1237,cite1238,cite1999}}.
\item\relax Gao decoder using extended Euclidean algorithm \NoCaseChange{\protect\cite{cite2000}}.
\item\relax Fast-Fourier-transform decoder with runtime of \flmRefsHyperref{ref65}{order} \(O(n \text{polylog}n)\) \NoCaseChange{\protect\cite{cite2001}}.
\item\relax List decoders try to find a low-degree bivariate polynomial \(Q(x,y)\) such that evaluation of \(Q\) at \((\alpha_i,y_i)\) is zero. By choosing proper degrees, it can be shown such polynomial exists by drawing an analogy between evaluation of \(Q(\alpha_i,y_i)\) and solving a homogeneous linear equation (interpolation). Once this is done, one lists roots of \(y\) that agree at \(\geq t\) points. The breakthrough Sudan list-decoding algorithm corrects up to \(1-\sqrt{2R}\) fraction of errors asymptotically in \(n\) \NoCaseChange{\protect\cite{cite2002}}. Roth and Ruckenstein proposed a modified key equation that allows for correction of more than \(\left\lfloor (n-k)/2 \right\rfloor\) errors \NoCaseChange{\protect\cite{cite2003}}. The Guruswami-Sudan algorithm improved the Sudan algorithm to \(1-\sqrt{R}\) \NoCaseChange{\protect\cite{cite1240}}, meaning that RS codes achieve list-decoding capacity; see Ref. \NoCaseChange{\protect\cite{cite2004}} for bounds. It was later shown that generic RS codes achieve list-decoding capacity \NoCaseChange{\protect\cite{cite2005}}. A modification of the Guruswami-Sudan algorithm by Koetter and Vardy is used for soft-decision decoding \NoCaseChange{\protect\cite{cite1840}} (see also Ref. \NoCaseChange{\protect\cite{cite1741}}). Subcodes of RS codes whose evaluation points lie in a \flmRefsHyperref{ref33}{subfield} can be decoded up to the \(1-R\) \NoCaseChange{\protect\cite{cite882}}. List decoding of RS codes is known as noisy polynomial interpolation in cryptography \NoCaseChange{\protect\cite{cite2006}}.
\item\relax The ubiquity of RS codes has yielded off-the-shelf VLSI integrated-circuit decoding hardware \NoCaseChange{\protect\cite{cite2007}} (see also \NoCaseChange{\protect\cite[{Chs. 5 and 10}]{cite247}}).
\end{eczvaluelist}
\codefieldsection{Realizations}
\begin{eczvaluelist}
\item\relax RS Product Code (RSPC) was used in DVDs \NoCaseChange{\protect\cite[{Ch. 4}]{cite247}}.
\item\relax DSL technologies and their variants against impluse noise \NoCaseChange{\protect\cite{cite328}}.
\item\relax Cryptographic primitives based on the hardness of decoding RS codes for more than \(1-\sqrt{k/n}+\epsilon\) errors. This is equivalent to the polynomial reconstruction problem \NoCaseChange{\protect\cite{cite329}}.
\item\relax RS codes as outer codes concatenated with convolutional codes are used indirectly in space exploration programs such as Voyager and Galileo. RS codes were part of a telemetry channel coding standard issued by the Consultative Committee for Space Data Systems \NoCaseChange{\protect\cite[{Ch. 3}]{cite247}}.
\item\relax Automatic repeat request (ARQ) data transmission protocols \NoCaseChange{\protect\cite[{Ch. 7}]{cite247}}.
\item\relax Slow-frequency-hop spread-spectrum transmission \NoCaseChange{\protect\cite[{Chs. 8-9}]{cite247}}.
\item\relax RS codes over \(q=2^m\) are used in RAID 6 \NoCaseChange{\protect\cite{cite330,cite331}}; see \NoCaseChange{\protect\cite{cite189}}.
\item\relax Coded sharding designs in blockchains to increase efficiency \NoCaseChange{\protect\cite{cite332}}.
\item\relax Used in QR-Codes to retrieve damaged barcodes \NoCaseChange{\protect\cite{cite333}}.
\item\relax Wireless communication systems such as 3G, DVB, and WiMAX \NoCaseChange{\protect\cite{cite334}}.
\item\relax Correcting pooled testing results for SARS-CoV-2 \NoCaseChange{\protect\cite{cite335}}.
\item\relax DNA storage \NoCaseChange{\protect\cite{cite336}}.
\end{eczvaluelist}
\codefieldsection{Notes}
\begin{eczvaluelist}
\item\relax See Ref. \NoCaseChange{\protect\cite[{Sec. 1.14}]{cite1159}} for an introduction to Reed-Solomon codes.
\item\relax See Kaiserslautern database \NoCaseChange{\protect\cite{cite1184}} for explicit codes.
\item\relax See corresponding MinT database entry \NoCaseChange{\protect\cite{cite2008}}.
\item\relax Popular summary in \flmHref{https://www.quantamagazine.org/how-mathematical-curves-power-cryptography-20220919/}{Quanta Magazine}.
\item\relax Certain structured classical optimization problems can be mapped into decoding and list decoding RS codes via the Decoded Quantum Interferomentry (DQI) algorithm \NoCaseChange{\protect\cite{cite2009,cite2010,cite2011}}.
\end{eczvaluelist}
\codefieldsection{Parents}
\begin{eczvaluelist}
\item\relax
\flmRefsHyperref[eczindexfamilyrel]{code:generalized_reed_solomon}{Generalized RS (GRS) code} --- A GRS code for which all multipliers \(v_i\) are unity reduces to an RS code \NoCaseChange{\protect\cite[{Def. 15.3.19}]{cite26}}.
\item\relax
\flmRefsHyperref[eczindexfamilyrel]{code:interleaved_reed_solomon}{Interleaved RS (IRS) code} --- An IRS code utilizing one polynomial \(f\) reduces to an RS code.
\item\relax
\flmRefsHyperref[eczindexfamilyrel]{code:folded_reed_solomon}{Folded RS (FRS) code} --- An FRS code with no extra grouping (\(m=1\)) reduces to an RS code.
\item\relax
\flmRefsHyperref[eczindexfamilyrel]{code:multiplicity}{Multiplicity code} --- Univariate multiplicity codes of degree \(s=1\) reduce to RS codes.
\item\relax
\flmRefsHyperref[eczindexfamilyrel]{code:tamo_barg}{Tamo-Barg code} --- Tamo-Barg codes are derived from Reed-Solomon codes; for \(r \mid k\), they have parameters \([n,k,n-k-k/r+2]\) and are optimal with respect to the LRC Singleton bound, while the special case \(r=k\) reduces to an RS code \NoCaseChange{\protect\cite[{Sec. 15.9.3}]{cite26}}.
\item\relax
\flmRefsHyperref[eczindexfamilyrel]{code:reed_solomon_nrt}{RS NRT code} --- RS NRT codes reduce to RS codes when the NRT metric is equivalent to the Hamming metric \NoCaseChange{\protect\cite{cite2012}}.
\end{eczvaluelist}
\codefieldsection{Children}
\begin{eczvaluelist}
\item\relax
\flmRefsHyperref[eczindexfamilyrel]{code:narrow_sense_reed_solomon}{Narrow-sense RS code} --- A narrow-sense RS is an RS code with length \(n=q-1\) whose points \(\alpha_i\) are all \((i-1)\)st powers of a primitive element of \(\mathbb{F}_q\).
\item\relax
\flmRefsHyperref[eczindexfamilyrel]{code:q-ary_parity_check}{\([n,n-1,2]_q\) \(q\)-ary parity-check code} --- RS codes for \(k=n-1\) are parity-check codes \NoCaseChange{\protect\cite{cite1673}}.
\item\relax
\flmRefsHyperref[eczindexfamilyrel]{code:q-ary_repetition}{\(q\)-ary repetition code} --- \(q\)-ary repetition codes can be thought of as RS codes \NoCaseChange{\protect\cite{cite1673}}.
\end{eczvaluelist}
\codefieldsection{Cousins}
\begin{eczvaluelist}
\item\relax
\flmRefsHyperref[eczindexfamilyrel]{code:dual}{Dual linear code} --- The dual of an RS code is an RS code \NoCaseChange{\protect\cite[{pg. 296}]{cite41}}.
\item\relax
\flmRefsHyperref[eczindexfamilyrel]{code:extended_reed_solomon}{Extended GRS code} --- Extending an RS code by one evaluation point, often interpreted as the point at infinity, yields an extended RS code.
\item\relax
\flmRefsHyperref[eczindexfamilyrel]{code:dna}{DNA storage code} --- RS codes have been used for DNA storage \NoCaseChange{\protect\cite{cite336}}.
\item\relax
\flmRefsHyperref[eczindexfamilyrel]{code:mds}{Maximum distance separable (MDS) code} --- Reed-Solomon codes are important examples of MDS codes, and for prime \(q\), every \([q+1,k,q-k+2]_q\) MDS code is Reed-Solomon; for non-prime \(q\), non-equivalent maximal-length MDS codes also exist \NoCaseChange{\protect\cite{cite1818}\protect\cite[{Sec. 3.3.2}]{cite70}\protect\cite[{Ch. 11}]{cite195}}.
\item\relax
\flmRefsHyperref[eczindexfamilyrel]{code:q-ary_bch}{Bose–Chaudhuri–Hocquenghem (BCH) code} --- An RS code can be represented as a union of cosets, with each coset being an interleaver of several binary BCH codes \NoCaseChange{\protect\cite{cite1741}}.
BCH codes are \flmRefsHyperref{ref33}{subfield} subcodes of RS codes, while primitive RS codes are primitive BCH codes \NoCaseChange{\protect\cite[{Sec. 2.6}]{cite68}}.

\item\relax
\flmRefsHyperref[eczindexfamilyrel]{code:q-ary_cyclic}{Cyclic linear \(q\)-ary code} --- If the length divides \(q-1\), then it is possible to construct a cyclic RS code. Such codes are Abelian group-algebra codes \NoCaseChange{\protect\cite[{Exam. 16.4.9}]{cite196}}.
\item\relax
\flmRefsHyperref[eczindexfamilyrel]{code:tensor}{Tensor-product code} --- Tensor codes constructed from RS codes are robustly testable \NoCaseChange{\protect\cite{cite2013}}.
\item\relax
\flmRefsHyperref[eczindexfamilyrel]{code:q-ary_ltc}{\(q\)-ary linear LTC} --- RS codes can be used to construct LTCs encoding \(k\) bits with length \(k \text{polylog}(k)\) and query complexity \(\text{polylog}(k)\) \NoCaseChange{\protect\cite{cite1701}}.
\item\relax
\flmRefsHyperref[eczindexfamilyrel]{code:pir}{Private information retrieval (PIR) code} --- RS codes can be used to construct PIR codes \NoCaseChange{\protect\cite{cite1112}}.
\item\relax
\flmRefsHyperref[eczindexfamilyrel]{code:analog_reed_solomon}{Analog RS code} --- Analog RS codes are versions of RS codes over the real and complex numbers.
\item\relax
\flmRefsHyperref[eczindexfamilyrel]{code:justesen}{Justesen code} --- An RS code is the outer code of Justesen codes.
\item\relax
\flmRefsHyperref[eczindexfamilyrel]{code:array}{Array code} --- RS codes over \(q=2^m\) are used in RAID 6 \NoCaseChange{\protect\cite{cite330,cite331}}; see \NoCaseChange{\protect\cite{cite189}}.
\item\relax
\flmRefsHyperref[eczindexfamilyrel]{code:b_array}{B-code} --- B-codes can be interpreted as RS codes over polynomials whose symbols lie in Galois rings \NoCaseChange{\protect\cite{cite2014,cite1224}}.
\item\relax
\flmRefsHyperref[eczindexfamilyrel]{code:linearized_reed_solomon}{Linearized RS code} --- Choosing \(\sigma=\operatorname{Id}\) and \(\delta=0\) makes linearized RS codes coincide with conventional RS codes, and the sum-rank metric reduces to the Hamming metric \NoCaseChange{\protect\cite[{Ex. 36}]{cite1259}}.
\item\relax
\flmRefsHyperref[eczindexfamilyrel]{code:maximum_rank_distance}{Maximum-rank distance (MRD) code} --- MRD rank-metric codes can be thought of as matrix analogues of MDS RS codes as both constructions utilize a Vandermonde matrix \NoCaseChange{\protect\cite{cite292}}.
\item\relax
\flmRefsHyperref[eczindexfamilyrel]{code:twisted_bch}{Twisted BCH code} --- Some twisted BCH codes are \flmRefsHyperref{ref33}{subfield} subcodes of RS codes \NoCaseChange{\protect\cite{cite2015}}.
\item\relax
\flmRefsHyperref[eczindexfamilyrel]{code:cascaded_reed_solomon}{Hyperbolic evaluation code} --- Hyperbolic evaluation codes were initially formulated as generalized concatenations (a.k.a. cascades) of RS codes \NoCaseChange{\protect\cite{cite30,cite31}}.
\item\relax
\flmRefsHyperref[eczindexfamilyrel]{code:convolutional}{Convolutional code} --- Convolutional codes can be constructed from \NoCaseChange{\protect\cite{cite1757}} and concatenated with \NoCaseChange{\protect\cite{cite242}} RS codes.
\item\relax
\flmRefsHyperref[eczindexfamilyrel]{code:pyramid}{Pyramid code} --- A pyramid code is an LRC whose generator matrix is that of an RS code in standard form, but some of whose columns are split into multiple columns.
\item\relax
\flmRefsHyperref[eczindexfamilyrel]{code:glynn}{\([10,5,6]_9\) Glynn code} --- The only other inequivalent \([10,5,6]_9\) code is an RS code, which is the unique Euclidean self-dual code for its parameters, and which is not Hermitian self-dual \NoCaseChange{\protect\cite{cite1650,cite1651,cite1652}}.
\item\relax
\flmRefsHyperref[eczindexfamilyrel]{code:shortened_hexacode}{\([5,3,3]_4\) Shortened hexacode} --- The dual of the shortened hexacode code is a \([5,2,4]_4\) doubly extended RS code \NoCaseChange{\protect\cite[{Exam. A}]{cite1666}}.
\item\relax
\flmRefsHyperref[eczindexfamilyrel]{code:berlekamp}{Berlekamp code} --- Berlekamp codes are obtained by first constructing an RS-like parity-check matrix out of a certain \flmRefsHyperref{ref33}{field extension} of \(\mathbb{F}_p\) and then taking the \flmRefsHyperref{ref33}{subfield} subcode of the corresponding code; see \NoCaseChange{\protect\cite[{Ch. 10.6}]{cite195}}.
\item\relax
\flmRefsHyperref[eczindexfamilyrel]{code:ea_mixed_alphabet_reed_solomon}{EA mixed-alphabet Reed-Solomon c-q code} --- EA mixed-alphabet RS c-q codes use Reed-Solomon polynomial evaluation, but evaluate over both \(\mathbb{F}_q\) and selected representatives from \(\mathbb{F}_{q^2}\setminus\mathbb{F}_q\) to support direct and dense-coded channel uses \NoCaseChange{\protect\cite{cite1922}}.
\item\relax
\flmRefsHyperref[eczindexfamilyrel]{code:qubit_subsystem_stabilizer}{Subsystem qubit stabilizer code} --- Primitive RS codes yield subsystem stabilizer codes via the subsystem extension of the Hermitian construction to subsystem codes \NoCaseChange{\protect\cite[{Exam. 3}]{cite1742}}.
\item\relax
\flmRefsHyperref[eczindexfamilyrel]{code:quantum_secret_sharing}{Approximate secret-sharing code} --- The classical information in this code is encoded using an RS code.
\item\relax
\flmRefsHyperref[eczindexfamilyrel]{code:galois_polynomial}{Galois-qudit RS code} --- Galois-qudit RS codes are CSS codes constructed from RS codes.
\item\relax
\flmRefsHyperref[eczindexfamilyrel]{code:galois_expander}{Galois-qudit expander code} --- The explicit expander-code construction with Reed-Solomon local checks in \NoCaseChange{\protect\cite{cite689}} yields \(\llbracket N,K\geq N^{1-\epsilon},D\geq N^{1/r}/\operatorname{poly}(\log N)\rrbracket _q\) QLDPC Galois-qudit quantum expander codes with transversal \(C^{r-1} Z\) gates. Balanced products of the same RS-based complexes also yield \([n,k\geq n^{1-\epsilon},d\geq n/\operatorname{poly}(\log n)]_q\) LTCs exhibiting the multiplication property.
\end{eczvaluelist}
\eczhbkcontributors{ Mustafa Doger, \eczhuVVA }
\endeczcode

\eczcode{residue}{Residue AG code}{}
\codefieldsection{Alternative Names}
\begin{eczvaluelist}
\item\relax Differential code
\end{eczvaluelist}
\eczhIndexCodeAliasName{residue}{Differential code}
\codefieldsection{Description}
Linear \(q\)-ary code defined using a set of \(\mathbb{F}_q\)-rational points \({\cal P} = \left( P_1,P_2,\cdots,P_n \right)\) on an algebraic curve \(\cal X\) and a linear space \(\Omega\) of certain rational differential forms \(\omega\) \NoCaseChange{\protect\cite[{Def. 15.3.2}]{cite26}}.

Codewords are evaluations of residues of the differential forms at the specified points,
\flmMathEnvironment{align}{}{
  \left(\text{Res}_{P_{1}}(\omega),\text{Res}_{P_{2}}(\omega),\cdots,\text{Res}_{P_{n}}(\omega)\right)\quad\quad\forall\omega\in\Omega~.
}
The code is denoted as \(C_{\Omega}({\cal X},{\cal P},D)\), where the \textit{divisor} \(D\) determines which rational differential forms to use.

\codefieldsection{Protection}
Riemann-Roch theorem yields code length \(n\), dimension \(k\), and a lower bound on distance in terms of the divisor \(D\) and the genus of the curve \(\cal X\) \NoCaseChange{\protect\cite[{Cor. 15.3.13}]{cite26}}. Distance bounds can also be derived from how an algebraic curve \(\cal X\) is embedded in the ambient projective space \NoCaseChange{\protect\cite{cite1963}}.
\codefieldsection{Realizations}
\begin{eczvaluelist}
\item\relax Improvements over the McEliece public-key cryptosystem to linear AG codes on curves of arbitrary genus \NoCaseChange{\protect\cite{cite340}}. Only the \flmRefsHyperref{ref33}{subfield} subcode proposal remains resilient to attacks \NoCaseChange{\protect\cite[{Sec. 15.7.5.3}]{cite26}}.
\item\relax Algebraic-geometric secret-sharing schemes \NoCaseChange{\protect\cite{cite341}}.
\end{eczvaluelist}
\codefieldsection{Parent}
\begin{eczvaluelist}
\item\relax
\flmRefsHyperref[eczindexfamilyrel]{code:evaluation}{Evaluation AG code} --- Any residue AG code of differential forms can be restated, up to diagonal equivalence, as an evaluation AG code of functions \NoCaseChange{\protect\cite{cite1312,cite1313,cite1314}\protect\cite[{Lemma 15.3.10}]{cite26}\protect\cite[{Thm. 2.72}]{cite32}}. Evaluation and residue AG codes are dual to each other \NoCaseChange{\protect\cite{cite32}\protect\cite[{Thm. 15.3.3}]{cite26}}.
\end{eczvaluelist}
\codefieldsection{Children}
\begin{eczvaluelist}
\item\relax
\flmRefsHyperref[eczindexfamilyrel]{code:generalized_reed_solomon}{Generalized RS (GRS) code} --- GRS (RS) codes are in one-to-one correspondence with both evaluation AG codes of univariate polynomials \(f\) \NoCaseChange{\protect\cite[{Thm. 15.3.24}]{cite26}} and residue AG codes of univariate differential forms \NoCaseChange{\protect\cite[{Prop. 15.3.26}]{cite26}}, with \(\cal X\) being the projective (affine) line \NoCaseChange{\protect\cite{cite1312,cite1313,cite32}\protect\cite[{Ch. 3.2}]{cite1314}}. The \(C_L\) and \(C_{\Omega}\) constructions yield the same family of codes on the projective line (up to diagonal equivalence).
\item\relax
\flmRefsHyperref[eczindexfamilyrel]{code:shimura}{Tsfasman-Vladut-Zink (TVZ) code} --- TVZ codes can be formulated as residue AG codes on algebraic curves.
\end{eczvaluelist}
\codefieldsection{Cousin}
\begin{eczvaluelist}
\item\relax
\flmRefsHyperref[eczindexfamilyrel]{code:cartier}{Cartier code} --- Every Cartier code is contained in a \flmRefsHyperref{ref33}{subfield} subcode of a residue AG code. Cartier codes share similar asymptotic properties to \flmRefsHyperref{ref33}{subfield} subcodes of residue AG codes, with both families admitting sequences of codes that achieve the \flmRefsHyperref{ref85}{GV bound}.
\end{eczvaluelist}
\eczhbkcontributors{ \eczhuVVA }
\endeczcode

\eczcode{roth_lempel}{Roth-Lempel code}{~\NoCaseChange{\protect\cite{cite2016}}}
\codefieldsection{Description}
Member of a \(q\)-ary linear code family that includes many examples of MDS codes that are not GRS codes.

The generator matrix of a Roth-Lempel code is
  \flmMathEnvironment{align}{}{
        \left(\begin{array}{cccccc}
  \alpha_{1}^{k-1} & \alpha_{2}^{k-1} & \cdots & \alpha_{n}^{k-1} & 1 & 0\\
  \alpha_{1}^{k-2} & \alpha_{2}^{k-2} & \cdots & \alpha_{n}^{k-2} & 0 & 1\\
  \vdots & \vdots & \ddots & \vdots & \vdots & \vdots\\
  \alpha_{1}^{2} & \alpha_{2}^{2} & \cdots & \alpha_{n}^{2} & 0 & 0\\
  \alpha_{1} & \alpha_{2} & \cdots & \alpha_{n} & 0 & 0\\
  1 & 1 & \cdots & 1 & 0 & 0
  \end{array}\right)~,
}
where \(\{\alpha_j\}\) is a set of elements of \(\mathbb{F}_q\).
The code is MDS if no subset of \(k-1\) elements sums to zero.

\codefieldsection{Parent}
\begin{eczvaluelist}
\item\relax
\flmRefsHyperref[eczindexfamilyrel]{code:mds}{Maximum distance separable (MDS) code} --- Roth-Lempel codes are examples of MDS codes that are not GRS codes.
\end{eczvaluelist}
\codefieldsection{Cousin}
\begin{eczvaluelist}
\item\relax
\flmRefsHyperref[eczindexfamilyrel]{code:extended_reed_solomon}{Extended GRS code} --- Roth-Lempel codes are doubly extended RS codes.
\end{eczvaluelist}
\eczhbkcontributors{ \eczhuVVA }
\endeczcode

\eczcode{reed_solomon_nrt}{RS NRT code}{~\NoCaseChange{\protect\cite{cite179}}}
\codefieldsection{Description}
An NRT analogue of an RS code.

\codefieldsection{Notes}
\begin{eczvaluelist}
\item\relax See corresponding MinT database entry \NoCaseChange{\protect\cite{cite2012}}.
\end{eczvaluelist}
\codefieldsection{Parent}
\begin{eczvaluelist}
\item\relax
\flmRefsHyperref[eczindexfamilyrel]{code:nrt}{Niederreiter-Rosenbloom-Tsfasman (NRT) code}\end{eczvaluelist}
\codefieldsection{Child}
\begin{eczvaluelist}
\item\relax
\flmRefsHyperref[eczindexfamilyrel]{code:reed_solomon}{Reed-Solomon (RS) code} --- RS NRT codes reduce to RS codes when the NRT metric is equivalent to the Hamming metric \NoCaseChange{\protect\cite{cite2012}}.
\end{eczvaluelist}
\eczhbkcontributors{ \eczhuVVA }
\endeczcode

\eczcode{ruled_surface}{Ruled-surface code}{~\NoCaseChange{\protect\cite{cite34,cite35}}}
\codefieldsection{Description}
Evaluation code obtained by evaluating global sections of a line bundle, or equivalently suitable polynomial functions, on rational points of a ruled surface over a finite field. Such codes extend algebraic-geometry constructions from curves to certain projective surfaces \NoCaseChange{\protect\cite{cite34,cite35}}.

\codefieldsection{Parent}
\begin{eczvaluelist}
\item\relax
\flmRefsHyperref[eczindexfamilyrel]{code:evaluation_polynomial}{Polynomial evaluation code} --- Ruled-surface codes are polynomial evaluation codes with \(\cal X\) being a ruled surface.
\end{eczvaluelist}
\eczhbkcontributors{ \eczhuVVA }
\endeczcode

\eczcode{schubert}{Schubert evaluation code}{~\NoCaseChange{\protect\cite{cite2017,cite2018}}}
\codefieldsection{Description}
Evaluation code of polynomials evaluated on points lying on a Schubert variety.

\codefieldsection{Protection}
Minimum distance bounds computed in Refs. \NoCaseChange{\protect\cite{cite2018,cite2019,cite2020}}.
\codefieldsection{Parent}
\begin{eczvaluelist}
\item\relax
\flmRefsHyperref[eczindexfamilyrel]{code:flag_variety}{Flag-variety code} --- Schubert evaluation codes are flag-variety evaluation codes with the flag variety being a Schubert variety.
\end{eczvaluelist}
\codefieldsection{Cousin}
\begin{eczvaluelist}
\item\relax
\flmRefsHyperref[eczindexfamilyrel]{code:grassmannian_variety}{Grassmannian evaluation code} --- Schubert varieties are subvarieties of Grassmannians, and Schubert evaluation codes were initially constructed as a generalization of Grassmannian evaluation codes.
\end{eczvaluelist}
\eczhbkcontributors{ \eczhuVVA }
\endeczcode

\eczcode{serge}{Segre-variety RM-type code}{~\NoCaseChange{\protect\cite{cite36}}}
\codefieldsection{Description}
Evaluation code of multihomogeneous polynomials evaluated on points of a Segre variety, i.e., on the Segre embedding of a product of projective spaces. These codes are Reed-Muller-type analogues adapted to product projective geometries \NoCaseChange{\protect\cite{cite36}}.

\codefieldsection{Parent}
\begin{eczvaluelist}
\item\relax
\flmRefsHyperref[eczindexfamilyrel]{code:evaluation_polynomial}{Polynomial evaluation code} --- Segre-variety RM-type codes are polynomial evaluation codes with \(\cal X\) being a Segre variety.
\end{eczvaluelist}
\eczhbkcontributors{ \eczhuVVA }
\endeczcode

\eczcode{self_dual_additive}{Self-dual additive code}{}
\codefieldsection{Description}
An additive \(q\)-ary code \(C \subseteq \mathbb{F}_q^n\) that is equal to its dual, \(C^\perp = C\), where the dual is defined with respect to some inner product, usually the trace-Hermitian inner product.

Self-dual additive codes that contain at least one codeword of odd weight are called \textit{Type I additive}.
Even self-dual additive codes are called \textit{Type II additive}, existing only for even \(n\) \NoCaseChange{\protect\cite[{Sec. 9.10}]{cite126}}.
Type I (type II) additive codes with length up to seven (eight) have been classified \NoCaseChange{\protect\cite{cite43}}.
Much is known about codes up to length 16 \NoCaseChange{\protect\cite{cite2021}}.

Two self-dual additive codes are \textit{equivalent} if they can be mapped into each other by a map that preserves self-duality \NoCaseChange{\protect\cite{cite1645}}.

\codefieldsection{Protection}
The minimum distance of a trace-Hermitian self-dual additive code satisfies \NoCaseChange{\protect\cite{cite2023}\protect\cite[{Thm. 33}]{cite2022}}
\flmMathEnvironment{align}{}{
  d\leq\begin{cases}
  2\left\lfloor \frac{n}{6}\right\rfloor +1 & n\equiv0\text{ mod 6 and code is Type I additive}\\
  4\left\lfloor \frac{n}{6}\right\rfloor +3 & n\equiv5\text{ mod 6 and code is Type I additive}\\
  2\left\lfloor \frac{n}{6}\right\rfloor +2 & \text{otherwise for Type I additive codes}\\
  2\left\lfloor \frac{n}{6}\right\rfloor +2 & \text{code is Type II additive}
  \end{cases}~.
}
A self-dual additive code saturating the above inequality is called \textit{extremal additive}.

\codefieldsection{Parent}
\begin{eczvaluelist}
\item\relax
\flmRefsHyperref[eczindexfamilyrel]{code:dual_additive}{Dual additive code}\end{eczvaluelist}
\codefieldsection{Children}
\begin{eczvaluelist}
\item\relax
\flmRefsHyperref[eczindexfamilyrel]{code:self_dual}{Self-dual linear code} --- Self-dual linear codes with respect to some inner product are automatically self-dual additive under the same inner product since linear codes are additive. In addition, quaternary linear codes are Hermitian self-orthogonal (self-dual) iff they are trace-Hermitian self-orthogonal (self-dual) additive \NoCaseChange{\protect\cite[{Thm. 27.4.1}]{cite2024}\protect\cite[{Thm. 9.10.3}]{cite126}}.
\item\relax
\flmRefsHyperref[eczindexfamilyrel]{code:dodecacode}{\((12,4^6,6)_4\) Dodecacode} --- The dodecacode is trace-Hermitian self-dual additive.
\end{eczvaluelist}
\eczhbkcontributors{ \eczhuVVA }
\endeczcode

\eczcode{self_dual}{Self-dual linear code}{}
\codefieldsection{Description}
An \([n,n/2]_q\) code that is equal to its dual, \(C^\perp = C\), where the dual is defined with respect to an inner product, most commonly either Euclidean or Hermitian.
Self-dual codes exist only for even lengths and have dimension \(k=n/2\).
A code that is equivalent to its dual is called \textit{isodual}.
Any self-dual code is isodual, and hence formally self-dual \NoCaseChange{\protect\cite[{Rem. 4.2.2}]{cite40}}.

A binary singly even (doubly even) self-dual code is called Type I (Type II), a ternary self-dual code is called Type III, and a Hermitian self-dual code over \(\mathbb{F}_4\) is called Type IV \NoCaseChange{\protect\cite[{Rems. 4.1.10 and 4.3.2}]{cite40}}.
Type III codes are 3-divisible, while Type IV codes are 2-divisible \NoCaseChange{\protect\cite[{Thm. 4.1.9}]{cite40}}.
Self-dual doubly even binary codes exist iff \(8|n\), self-dual ternary codes exist iff \(4|n\), and Hermitian self-dual codes over \(\mathbb{F}_4\) exist iff \(n\) is even \NoCaseChange{\protect\cite[{Thm. 4.1.13}]{cite40}}.

\codefieldsection{Protection}
The generator matrix of the Hermitian dual of a code with generator matrix \(G = [I_k~~A]\) is \([-\bar{A}^T~~I_{n-k}]\), where \(\bar{A}\) contains matrix elements of \(A\) raised to the \(p\)th power.
A code is Hermitian self-dual if and only if \(A \bar{A}^{T} = -I_{n/2}\).

The minimum distance of a formally self-dual even binary code, a Type II binary self-dual code, a Type III ternary self-dual code, and a Type IV Hermitian self-dual code over \(\mathbb{F}_4\) satisfies
\flmMathEnvironment{align}{}{
  d\leq\begin{cases}
    2\left\lfloor \frac{n}{8}\right\rfloor +2\\
    4\left\lfloor \frac{n}{24}\right\rfloor +4\\
    3\left\lfloor \frac{n}{12}\right\rfloor +3\\
    2\left\lfloor \frac{n}{6}\right\rfloor +2
  \end{cases}
}
respectively \NoCaseChange{\protect\cite[{Thm. 4.3.1}]{cite40}}.
More generally, binary self-dual codes satisfy Rains's bound \(d\leq 4\lfloor n/24\rfloor+4\), except when \(n\equiv 22\) modulo \(24\), in which case \(d\leq 4\lfloor n/24\rfloor+6\) \NoCaseChange{\protect\cite[{Thm. 4.3.6}]{cite40}}.
A self-dual code meeting the relevant upper bound is called \textit{extremal} \NoCaseChange{\protect\cite[{Def. 4.3.7}]{cite40}}.

Fixed-weight codewords of extremal Type II codes of length divisible by \(24\) form combinatorial 5-designs \NoCaseChange{\protect\cite[{Thm. 4.3.16(a)}]{cite40}}.
The extended Golay code and the \([48,24,12]\) self-dual code are two such codes.
It is not yet known whether a \([72,36,16]\) self-dual code exists \NoCaseChange{\protect\cite[{Rem. 4.3.11}]{cite40}}; see also \NoCaseChange{\protect\cite{cite2025,cite1031,cite2023}}.

Doubly even self-dual codes have been classified up to \(n\leq 40\) \NoCaseChange{\protect\cite{cite1776}}. 
The \([22,11]\) and \([24,12]\) doubly even self-dual codes have been classified, and there are nine inequivalent codes with the latter parameters \NoCaseChange{\protect\cite{cite2026}}.

For ternary self-dual codes, see \NoCaseChange{\protect\cite{cite2027,cite2023}\protect\cite[{Remark 4.3.14}]{cite2022}}.
Ternary self-dual codes have been classified for \(n=24\) \NoCaseChange{\protect\cite{cite2028}} and, in the extremal case, for \(n=28\) \NoCaseChange{\protect\cite{cite2029}}.

\codefieldsection{Notes}
\begin{eczvaluelist}
\item\relax See books \NoCaseChange{\protect\cite{cite42,cite126}} for more on self-dual codes.
\item\relax See Refs. \NoCaseChange{\protect\cite{cite2030,cite2031}} for constructions of binary self-dual codes.
\item\relax See \flmHref{https://www.unilim.fr/pages_perso/philippe.gaborit/SD/index.html}{Tables of Self-Dual Codes} by P. Gaborit and A. Otmani for a database of self-dual codes over \(\mathbb{F}_2\), \(\mathbb{F}_3\), \(\mathbb{F}_4\) (Euclidean or Hermitian), \(\mathbb{F}_5\), and \(\mathbb{F}_7\). See also Ref. \NoCaseChange{\protect\cite{cite2032}}.
\item\relax See \flmHref{https://www.math.is.tohoku.ac.jp/~munemasa/selfdualcodes.htm}{Database of self-dual codes} by M. Harada and A. Munemasa for a database of self-dual codes over \(\mathbb{F}_2\), \(\mathbb{F}_3\), \(\mathbb{F}_5\), and \(\mathbb{F}_7\).
\end{eczvaluelist}
\codefieldsection{Parents}
\begin{eczvaluelist}
\item\relax
\flmRefsHyperref[eczindexfamilyrel]{code:dual}{Dual linear code}\item\relax
\flmRefsHyperref[eczindexfamilyrel]{code:self_dual_additive}{Self-dual additive code} --- Self-dual linear codes with respect to some inner product are automatically self-dual additive under the same inner product since linear codes are additive. In addition, quaternary linear codes are Hermitian self-orthogonal (self-dual) iff they are trace-Hermitian self-orthogonal (self-dual) additive \NoCaseChange{\protect\cite[{Thm. 27.4.1}]{cite2024}\protect\cite[{Thm. 9.10.3}]{cite126}}.
\item\relax
\flmRefsHyperref[eczindexfamilyrel]{code:self_dual_over_rings}{Self-dual code over \(R\)} --- Self-dual linear codes are over fields, which are also rings.
\end{eczvaluelist}
\codefieldsection{Children}
\begin{eczvaluelist}
\item\relax
\flmRefsHyperref[eczindexfamilyrel]{code:karlin}{\([2m+2,m+1]\) Karlin code} --- Karlin codes are Euclidean self-dual doubly even codes \NoCaseChange{\protect\cite[{Ch. 16}]{cite41}}, and some of them are extremal \NoCaseChange{\protect\cite{cite1201,cite1202}}.
\item\relax
\flmRefsHyperref[eczindexfamilyrel]{code:self_dual_48_24_12}{\([48,24,12]\) self-dual code} --- The \([48,24,12]\) code is the unique self-dual doubly even code with those parameters \NoCaseChange{\protect\cite{cite111}\protect\cite[{Rem. 4.3.11}]{cite40}}.
\item\relax
\flmRefsHyperref[eczindexfamilyrel]{code:hamming844}{\([8,4,4]\) extended Hamming code} --- The \([8,4,4]\) extended Hamming code is the smallest doubly even self-dual code, and the unique Type II code of length \(8\) \NoCaseChange{\protect\cite[{Rem. 4.3.10}]{cite40}}.
\item\relax
\flmRefsHyperref[eczindexfamilyrel]{code:pless_symmetry}{\([2q+2,q+1]_3\) Pless symmetry code}\item\relax
\flmRefsHyperref[eczindexfamilyrel]{code:glynn}{\([10,5,6]_9\) Glynn code} --- The Glynn code is trace-Hermitian self-dual, and is not Euclidean self-dual \NoCaseChange{\protect\cite{cite1650,cite1651,cite1652}}.
\item\relax
\flmRefsHyperref[eczindexfamilyrel]{code:hexacode}{\([6,3,4]_4\) Hexacode} --- The hexacode is Hermitian self-dual \NoCaseChange{\protect\cite[{Rem. 4.2.6}]{cite40}} and, as a result, is also trace-Hermitian self-dual additive \NoCaseChange{\protect\cite[{Sec. 9.10}]{cite126}}. The hexacode and the shortened hexacode are extremal \NoCaseChange{\protect\cite[{Tab. 9.14}]{cite126}\protect\cite[{Tm. 12}]{cite43}}.
\item\relax
\flmRefsHyperref[eczindexfamilyrel]{code:reed_solomon_4}{\([4,2,3]_4\) RS\(_4\) code} --- The RS\(_4\) is the smallest Type II Euclidean self-dual code \NoCaseChange{\protect\cite[{Sec. 2.4.2}]{cite42}}.
\item\relax
\flmRefsHyperref[eczindexfamilyrel]{code:tetracode}{\([4,2,3]_3\) Tetracode} --- The tetracode is Euclidean self-dual, i.e., Type III in the terminology of \NoCaseChange{\protect\cite[{Rems. 4.2.6 and 4.3.2}]{cite40}}.
\end{eczvaluelist}
\codefieldsection{Cousins}
\begin{eczvaluelist}
\item\relax
\flmRefsHyperref[eczindexfamilyrel]{code:divisible}{Divisible code} --- Binary self-orthogonal codes are even, doubly even binary codes are self-orthogonal, and binary self-dual codes split into singly-even Type I and doubly-even Type II families \NoCaseChange{\protect\cite[{Def. 4.1.6}]{cite40}\protect\cite[{Rems. 4.1.7 and 4.1.10}]{cite40}}. Ternary self-dual codes are 3-divisible and Hermitian self-dual quaternary codes are 2-divisible \NoCaseChange{\protect\cite[{Thm. 4.1.9}]{cite40}}.
\item\relax
\flmRefsHyperref[eczindexfamilyrel]{code:group}{Group-algebra code} --- Self-dual group codes exist exactly when the base field has characteristic \(2\) and the underlying group has even order \NoCaseChange{\protect\cite[{Thm. 16.5.4}]{cite196}}.
\item\relax
\flmRefsHyperref[eczindexfamilyrel]{code:cft}{Conformal-field theory (CFT) code} --- Even self-dual binary codes and even unimodular lattices define CFTs \NoCaseChange{\protect\cite{cite2033,cite2034,cite2035}}. Self-dual ternary codes define superconformal field theories (SCFTs) \NoCaseChange{\protect\cite{cite2036}}.
\item\relax
\flmRefsHyperref[eczindexfamilyrel]{code:niemeier}{Niemeier lattice} --- The nine inequivalent \([24,12]\) doubly even self-dual codes \NoCaseChange{\protect\cite{cite2026}} yield certain Niemeier lattices via \flmTerm{term}{ref127}{}{Construction A} \NoCaseChange{\protect\cite{cite2037}}. Niemeier lattices can be constructed from ternary self-dual codes of length 24 \NoCaseChange{\protect\cite{cite2038}}.
\item\relax
\flmRefsHyperref[eczindexfamilyrel]{code:self_dual_lattice}{Unimodular lattice} --- Unimodular lattices are lattice analogues of self-dual codes. There are several parallels between (doubly even) self-dual binary codes and (even) unimodular lattices \NoCaseChange{\protect\cite{cite39,cite42,cite2039}}. Even self-dual binary codes and even unimodular lattices define CFTs \NoCaseChange{\protect\cite{cite2033,cite2034,cite2035}}.
\item\relax
\flmRefsHyperref[eczindexfamilyrel]{code:combinatorial_design}{Combinatorial design} --- Self-dual extremal codes yield combinatorial \(\leq 5\)-designs using the Assmus-Mattson theorem \NoCaseChange{\protect\cite{cite136}} (see \NoCaseChange{\protect\cite[{Sec. 5.4}]{cite135}}).
See \NoCaseChange{\protect\cite[{Table 1.61, pg. 683}]{cite156}} for a table of combinatorial designs obtained from self-dual codes.

\item\relax
\flmRefsHyperref[eczindexfamilyrel]{code:binary_quad_residue}{Binary quadratic-residue (QR) code} --- The length-\(72\) extended binary quadratic-residue code is self-dual but not extremal \NoCaseChange{\protect\cite[{Rem. 4.3.10}]{cite40}}.
\item\relax
\flmRefsHyperref[eczindexfamilyrel]{code:extended_golay}{\([24, 12, 8]\) Extended Golay code} --- The extended Golay code is the unique \([24,12,8]\) code, and in particular the unique self-dual doubly even code with those parameters \NoCaseChange{\protect\cite{cite102}\protect\cite[{Rem. 4.3.11}]{cite40}}.
\item\relax
\flmRefsHyperref[eczindexfamilyrel]{code:nordstrom_robinson}{\((16,256,6)\) Nordstrom-Robinson (NR) code} --- The NR code is self-dual in that its distance distribution is invariant under the \flmRefsHyperref{ref113}{MacWilliams transform} \NoCaseChange{\protect\cite{cite1148}}.
It maps to the octacode, a self-dual code over \(\mathbb{Z}_4\) under the \flmTerm{term}{ref81}{}{Gray map} \NoCaseChange{\protect\cite{cite1149,cite1146,cite123}\protect\cite[{Sec. 6.3}]{cite1145}}.

\item\relax
\flmRefsHyperref[eczindexfamilyrel]{code:reed_muller}{Reed-Muller (RM) code} --- The codes RM\((r,m)\) and RM\((m-r-1,m)\) are dual to each other, with the case \(m = 2r+1\) being self dual.
\item\relax
\flmRefsHyperref[eczindexfamilyrel]{code:quasi_cyclic}{Quasi-cyclic code} --- Quasi-cyclic self-dual constructions include double circulant codes and, in odd characteristic, their negacirculant analogs such as double negacirculant and four-negacirculant codes \NoCaseChange{\protect\cite[{Sec. 4.4}]{cite40}}.
\item\relax
\flmRefsHyperref[eczindexfamilyrel]{code:generalized_reed_muller}{Generalized RM (GRM) code} --- Certain GRM codes are self-dual; in general, the dual of GRM\(_q(r,m)\) is another GRM code, namely GRM\(_q(m(q-1)-r-1,m)\) \NoCaseChange{\protect\cite{cite1781,cite1782}}.
\item\relax
\flmRefsHyperref[eczindexfamilyrel]{code:shortened_hexacode}{\([5,3,3]_4\) Shortened hexacode} --- The hexacode and the shortened hexacode are extremal \NoCaseChange{\protect\cite[{Tab. 9.14}]{cite126}\protect\cite[{Tm. 12}]{cite43}}.
\item\relax
\flmRefsHyperref[eczindexfamilyrel]{code:ternary_golay}{\([11,6,5]_3\) Ternary Golay code} --- The extended ternary Golay code is self-dual, i.e., a Type III code in the terminology of \NoCaseChange{\protect\cite[{Rems. 4.2.6 and 4.3.2}]{cite40}}.
\item\relax
\flmRefsHyperref[eczindexfamilyrel]{code:q-ary_cyclic}{Cyclic linear \(q\)-ary code} --- See Refs. \NoCaseChange{\protect\cite{cite1767,cite1768}} for tables of cyclic self-dual codes.
\item\relax
\flmRefsHyperref[eczindexfamilyrel]{code:q-ary_duadic}{\(q\)-ary duadic code} --- Under certain conditions, extended odd-like duadic codes are self-dual \NoCaseChange{\protect\cite[{Sec. 2.7}]{cite68}}.
\item\relax
\flmRefsHyperref[eczindexfamilyrel]{code:harada_kitazume}{Harada-Kitazume code} --- Codewords consisting of 0 and 2 of nine Harada-Kitazume codes are of the form \(2c\), where \(c\) is a codeword of one of the nine corresponding \([24,12]\) doubly even self-dual codes \NoCaseChange{\protect\cite{cite2037}}.
\item\relax
\flmRefsHyperref[eczindexfamilyrel]{code:self_dual_over_z4}{Self-dual code over \(\mathbb{Z}_4\)} --- Under the \flmTerm{term}{ref81}{}{Gray map}, any self-dual code over \(\mathbb{Z}_4\) maps to a formally self-dual binary code \NoCaseChange{\protect\cite{cite112}}.
\item\relax
\flmRefsHyperref[eczindexfamilyrel]{code:self_dual_polytope}{Self-dual polytope code} --- Self-dual polytope codes are spherical analogues of self-dual linear codes.
\item\relax
\flmRefsHyperref[eczindexfamilyrel]{code:jump}{Jump code} --- Iso-dual codes can be used to construct jump codes \NoCaseChange{\protect\cite{cite145}}.
\item\relax
\flmRefsHyperref[eczindexfamilyrel]{code:stabilizer_over_gf4}{Hermitian qubit code} --- Hermitian qubit codes are constructed from Hermitian self-orthogonal linear codes over \(\mathbb{F}_4\) via the \flmRefsHyperref{ref1778}{\(\mathbb{F}_4\) representation}. This relation yields bounds on self-dual codes over \(\mathbb{F}_4\) \NoCaseChange{\protect\cite{cite2040}}.
\item\relax
\flmRefsHyperref[eczindexfamilyrel]{code:quantum_triorthogonal}{Triorthogonal code} --- Self-dual binary codes can be used to construct triorthogonal codes \NoCaseChange{\protect\cite{cite2041}}.
\end{eczvaluelist}
\eczhbkcontributors{ \eczhuVVA }
\endeczcode

\eczcode{semakov_zinoviev_zaitsev}{Semakov-Zinoviev-Zaitsev (SZZ) equidistant code}{~\NoCaseChange{\protect\cite{cite2042}}}
\codefieldsection{Description}
Member of a nonlinear \(q\)-ary code family that is related to affine resolvable block designs and that is universally optimal.

\codefieldsection{Parent}
\begin{eczvaluelist}
\item\relax
\flmRefsHyperref[eczindexfamilyrel]{code:delsarte_optimal_q-ary}{\(q\)-ary sharp configuration} --- The SZZ equidistant code is a \(q\)-ary sharp configuration \NoCaseChange{\protect\cite[{Table 12.1}]{cite199}}.
\end{eczvaluelist}
\eczhbkcontributors{ \eczhuVVA }
\endeczcode

\eczcode{sequential_recovery}{Sequential-recovery code}{~\NoCaseChange{\protect\cite{cite2043,cite2044}}}
\codefieldsection{Description}
A \(t\)-erasure LRC whose erased coordinates are recovered sequentially.

\codefieldsection{Parents}
\begin{eczvaluelist}
\item\relax
\flmRefsHyperref[eczindexfamilyrel]{code:q-ary_digits_into_q-ary_digits}{\(q\)-ary code}\item\relax
\flmRefsHyperref[eczindexfamilyrel]{code:multiple_erasure_lrc}{\(t\)-erasure LRC}\end{eczvaluelist}
\codefieldsection{Child}
\begin{eczvaluelist}
\item\relax
\flmRefsHyperref[eczindexfamilyrel]{code:parallel_recovery}{Parallel-recovery code}\end{eczvaluelist}
\eczhbkcontributors{ \eczhuVVA }
\endeczcode

\eczcode{srivastava}{Srivastava code}{~\NoCaseChange{\protect\cite{cite974,cite1846}}}
\codefieldsection{Description}
A special case of a generalized Srivastava code for \(z_j = \alpha_j^{\mu}\) for some \(\mu\) and \(t=1\).

The code's parity-check matrix is
\flmMathEnvironment{align}{}{
  H=\begin{pmatrix}\frac{\alpha_{1}^{\mu}}{\alpha_{1}-w_{1}} & \frac{\alpha_{2}^{\mu}}{\alpha_{2}-w_{1}} & \cdots & \frac{\alpha_{n}^{\mu}}{\alpha_{n}-w_{1}}\\
  \frac{\alpha_{1}^{\mu}}{\alpha_{1}-w_{2}} & \frac{\alpha_{2}^{\mu}}{\alpha_{2}-w_{2}} & \cdots & \frac{\alpha_{n}^{\mu}}{\alpha_{n}-w_{2}}\\
  \vdots & \vdots & \ddots & \vdots\\
  \frac{\alpha_{1}^{\mu}}{\alpha_{1}-w_{s}} & \frac{\alpha_{2}^{\mu}}{\alpha_{2}-w_{s}} & \cdots & \frac{\alpha_{n}^{\mu}}{\alpha_{n}-w_{s}}
  \end{pmatrix}~.
}

\codefieldsection{Protection}
Dimension and minimum distance are found in Refs. \NoCaseChange{\protect\cite{cite1846,cite1724}}.

\codefieldsection{Parents}
\begin{eczvaluelist}
\item\relax
\flmRefsHyperref[eczindexfamilyrel]{code:generalized_srivastava}{Generalized Srivastava code} --- A Srivastava code is a special case of a generalized Srivastava code for \(z_j = \alpha_j^{\mu}\) for some \(\mu\) and \(t=1\).
\item\relax
\flmRefsHyperref[eczindexfamilyrel]{code:goppa}{Goppa code} --- Generalized Srivastava codes are a special case of Goppa codes \NoCaseChange{\protect\cite[{Ch. 12}]{cite41}}.
\item\relax
\flmRefsHyperref[eczindexfamilyrel]{code:gbch}{Chien-Choy generalized BCH (GBCH) code} --- Generalized Srivastava codes are a special case of GBCH codes \NoCaseChange{\protect\cite[{Ch. 12}]{cite41}}.
\end{eczvaluelist}
\eczhbkcontributors{ \eczhuVVA }
\endeczcode

\eczcode{suzuki}{Suzuki-curve code}{~\NoCaseChange{\protect\cite{cite2045}}}
\codefieldsection{Description}
Evaluation AG code of rational functions evaluated on points lying on a Suzuki curve.

\codefieldsection{Parent}
\begin{eczvaluelist}
\item\relax
\flmRefsHyperref[eczindexfamilyrel]{code:evaluation}{Evaluation AG code} --- Suzuki-curve codes are evaluation AG codes with \(\cal X\) being a Suzuki curve.
\end{eczvaluelist}
\codefieldsection{Cousin}
\begin{eczvaluelist}
\item\relax
\flmRefsHyperref[eczindexfamilyrel]{code:group}{Group-algebra code} --- Some Suzuki-curve codes are group-algebra codes \NoCaseChange{\protect\cite[{Remark 16.4.14}]{cite196}}.
\end{eczvaluelist}
\eczhbkcontributors{ Shashank Sule, \eczhuVVA }
\endeczcode

\eczcode{tamo_barg}{Tamo-Barg code}{~\NoCaseChange{\protect\cite{cite2046}}}
\codefieldsection{Description}
A family of \(q\)-ary polynomial evaluation codes that are optimal LRCs and for which \(q\) is comparable to \(n\).

Each length-\(k\) message \(\mu\) is encoded into a string of values of its corresponding polynomial \(f_\mu\) at points \(\alpha\) in some size-\(n\) set \(A\).
This polynomial can be written as
\flmMathEnvironment{align}{}{
  f_{\mu}\left(x\right)=\sum_{i=0}^{r-1}\sum_{j=0}^{k/r-1}\mu_{ij}x^{i}g(x)^{j}~,
}
where \(\mu\) has been reorganized into an \(r \times k/r\) matrix, and where \(g(x)\) is a degree-\((r+1)\) polynomial that is constant on elements of certain partitions of \(A\) \NoCaseChange{\protect\cite[{Sec. III.A}]{cite2046}}.
This construction assumes that \(r\) divides \(k\), but a simple extension to other parameters can be formulated.

\codefieldsection{Decoding}
\begin{eczvaluelist}
\item\relax Polynomial evaluation over \(r\) points \NoCaseChange{\protect\cite{cite2046}}.
\end{eczvaluelist}
\codefieldsection{Parents}
\begin{eczvaluelist}
\item\relax
\flmRefsHyperref[eczindexfamilyrel]{code:tamo_barg_vladut}{Barg-Tamo-Vladut code} --- Tamo-Barg codes are the \(PG(1,q)\)/genus-zero instance of the covering-based construction that later yields Barg-Tamo-Vladut codes \NoCaseChange{\protect\cite[{Sec. 15.9.4}]{cite26}}.
\item\relax
\flmRefsHyperref[eczindexfamilyrel]{code:optimal_lrc}{Optimal LRC} --- For \(r \mid k\), Tamo-Barg codes meet the LRC Singleton bound \NoCaseChange{\protect\cite[{Sec. 15.9.3}]{cite26}}.
\end{eczvaluelist}
\codefieldsection{Child}
\begin{eczvaluelist}
\item\relax
\flmRefsHyperref[eczindexfamilyrel]{code:reed_solomon}{Reed-Solomon (RS) code} --- Tamo-Barg codes are derived from Reed-Solomon codes; for \(r \mid k\), they have parameters \([n,k,n-k-k/r+2]\) and are optimal with respect to the LRC Singleton bound, while the special case \(r=k\) reduces to an RS code \NoCaseChange{\protect\cite[{Sec. 15.9.3}]{cite26}}.
\end{eczvaluelist}
\codefieldsection{Cousin}
\begin{eczvaluelist}
\item\relax
\flmRefsHyperref[eczindexfamilyrel]{code:quantum_tamo_barg}{Quantum Tamo-Barg (QTB) code} --- QTB codes are CSS codes constructed from Tamo-Barg codes.
\end{eczvaluelist}
\eczhbkcontributors{ \eczhuVVA }
\endeczcode

\eczcode{tanner}{Tanner code}{~\NoCaseChange{\protect\cite{cite1582}}}
\codefieldsection{Alternative Names}
\begin{eczvaluelist}
\item\relax Generalized LDPC (GLDPC) code
\end{eczvaluelist}
\eczhIndexCodeAliasName{tanner}{Generalized LDPC (GLDPC) code}
\codefieldsection{Description}
A linear \(q\)-ary code defined on a bipartite graph similar to a Tanner graph.
This \textit{generalized Tanner graph} consists of variable nodes and constraint nodes, with the generalization being that the constraint nodes of degree \(r\) now represent a linear code of length \(r\).

A string is in the Tanner code if all substrings satisfy their corresponding constraints, i.e., all substrings are members of the linear codes represented on the constraint nodes.

A code is called \textit{regular} if the corresponding bipartite graph is regular.
Expansion properties of the underlying graph can yield efficient decoding algorithms.

\codefieldsection{Decoding}
\begin{eczvaluelist}
\item\relax Min-sum and sum-product iterative decoders for binary Tanner codes \NoCaseChange{\protect\cite{cite2047,cite2048}}; see also \NoCaseChange{\protect\cite{cite2049,cite1238}}. These decoders can be improved using a probabilistic message-passing schedule \NoCaseChange{\protect\cite{cite2050}}.
\item\relax Any code can be put into normal form without significantly altering the underlying graph or the decoding complexity \NoCaseChange{\protect\cite{cite2051}}. For an algebraic viewpoint on decoding, see \NoCaseChange{\protect\cite{cite2052}}.
\item\relax Iterative decoding is optimal for Tanner graphs that are free of cycles \NoCaseChange{\protect\cite{cite2048}}. However, codes that admit cycle-free representations have bounds on their distances \NoCaseChange{\protect\cite{cite2053,cite1582}}; see \NoCaseChange{\protect\cite{cite97,cite2054}}.
\end{eczvaluelist}
\codefieldsection{Parents}
\begin{eczvaluelist}
\item\relax
\flmRefsHyperref[eczindexfamilyrel]{code:q-ary_linear}{Linear \(q\)-ary code}\item\relax
\flmRefsHyperref[eczindexfamilyrel]{code:parallel_concatenated}{Parallel concatenated code} --- Tanner codes are examples of parallel concatenation \NoCaseChange{\protect\cite{cite972}}.
\end{eczvaluelist}
\codefieldsection{Children}
\begin{eczvaluelist}
\item\relax
\flmRefsHyperref[eczindexfamilyrel]{code:regular_binary_tanner}{Regular binary Tanner code} --- Regular binary Tanner codes are binary Tanner codes defined on regular sparse bipartite graphs, with the inner code being the same for all vertices.
\item\relax
\flmRefsHyperref[eczindexfamilyrel]{code:q-ary_ldpc}{\(q\)-ary LDPC code} --- \(q\)-ary LDPC codes are \(q\)-ary Tanner codes on sparse bipartite graphs whose constraint nodes represent \(q\)-ary parity-check codes.
\end{eczvaluelist}
\codefieldsection{Cousins}
\begin{eczvaluelist}
\item\relax
\flmRefsHyperref[eczindexfamilyrel]{code:generalized_homological_product_css}{Generalized homological-product CSS code} --- Tanner codes can be generalized to \textit{sheaf codes}, whose local codes satisfy a certain hierarchy. This allows for a way to formulate and understand many generalized homological-product CSS codes \NoCaseChange{\protect\cite{cite1101}} and LTCs \NoCaseChange{\protect\cite{cite1102}}.
\item\relax
\flmRefsHyperref[eczindexfamilyrel]{code:ltc}{Locally testable code (LTC)} --- Tanner codes can be generalized to \textit{sheaf codes}, whose local codes satisfy a certain hierarchy. This allows for a way to formulate and understand many generalized homological-product CSS codes \NoCaseChange{\protect\cite{cite1101}} and LTCs \NoCaseChange{\protect\cite{cite1102}}.
\end{eczvaluelist}
\eczhbkcontributors{ \eczhuVVA }
\endeczcode

\eczcode{shimura}{Tsfasman-Vladut-Zink (TVZ) code}{~\NoCaseChange{\protect\cite{cite1718}}}
\codefieldsection{Description}
Member of a family of AG codes obtained from algebraic curves via the residue or evaluation construction. Sequences of curves with many rational points, such as Drinfeld modular curves, classical modular curves, or Garcia-Stichtenoth curves, yield the asymptotic parameters of the TVZ bound \NoCaseChange{\protect\cite[{Sec. 15.4.2}]{cite26}}.
\codefieldsection{Rate}
TVZ codes, either in the residue AG or evaluation AG constructions, exceed the \flmRefsHyperref{ref85}{asymptotic GV bound} \NoCaseChange{\protect\cite{cite1718}} (see also Ref. \NoCaseChange{\protect\cite{cite2055}}). For square \(q \geq 49\), these codes improve on the GV bound; for smaller \(q\), they do not \NoCaseChange{\protect\cite[{Sec. 15.4.2}]{cite26}}. Roughly speaking, this result implies that AG codes can outperform random codes.
\codefieldsection{Parent}
\begin{eczvaluelist}
\item\relax
\flmRefsHyperref[eczindexfamilyrel]{code:residue}{Residue AG code} --- TVZ codes can be formulated as residue AG codes on algebraic curves.
\end{eczvaluelist}
\codefieldsection{Cousins}
\begin{eczvaluelist}
\item\relax
\flmRefsHyperref[eczindexfamilyrel]{code:evaluation}{Evaluation AG code} --- TVZ codes can also be formulated as evaluation AG codes on algebraic curves; these are dual to the corresponding residue AG codes.
\item\relax
\flmRefsHyperref[eczindexfamilyrel]{code:quantum_ag}{Quantum AG code} --- The AG codes used in an asymptotically good construction of quantum AG codes with non-Clifford transversal gates \NoCaseChange{\protect\cite{cite699}} are those of the TVZ codes.
\end{eczvaluelist}
\eczhbkcontributors{ \eczhuVVA }
\endeczcode

\eczcode{turbo}{Turbo code}{~\NoCaseChange{\protect\cite{cite2056,cite2057}}}
\codefieldsection{Description}
Code obtained from a parallel concatenation of two or more convolutional codes with permutations \textit{interleaving} the individual encodings.

The choice of interleaver is important to the code design \NoCaseChange{\protect\cite{cite2058,cite2059}}.

\codefieldsection{Protection}
Parallel concatenated turbo codes have typical minimum distance with upper bound \(O( \log(n) )\). Truhachev, Lentmacher, and Zignagirov produced a sequence of turbo codes with minimum distance of \flmRefsHyperref{ref65}{order} \(\Theta(\log(n) )\). \NoCaseChange{\protect\cite{cite2060}}.
Various bounds on code parameters exist \NoCaseChange{\protect\cite{cite2061,cite2062}}.

\codefieldsection{Rate}
Turbo codes nearly achieve the Shannon capacity \NoCaseChange{\protect\cite{cite2057}}.
\codefieldsection{Decoding}
\begin{eczvaluelist}
\item\relax Turbo decoder \NoCaseChange{\protect\cite{cite2057}}, an instance of BP decoding \NoCaseChange{\protect\cite{cite2063}}.
\item\relax Maximum A Posteriori (MAP) decoder \NoCaseChange{\protect\cite{cite2064}} and a soft output derivative \NoCaseChange{\protect\cite{cite2065}}. The use of soft outputs can improve code performance \NoCaseChange{\protect\cite{cite2066}}.
\item\relax List decoding \NoCaseChange{\protect\cite{cite2067}}.
\item\relax VLSI integrated-circuit decoding hardware \NoCaseChange{\protect\cite{cite2068}}.
\item\relax Autoencoder \NoCaseChange{\protect\cite{cite2069}}.
\end{eczvaluelist}
\codefieldsection{Realizations}
\begin{eczvaluelist}
\item\relax Recommended by Consultative Committee on Space Data Systems (CCSDS) for telemetry and telecommand \NoCaseChange{\protect\cite{cite350,cite351}}.
\item\relax Several standards related to wireless communication, including W-CDMA, DVB-RCS, TD-SCDMA, 802.16, and CDMA2000 \NoCaseChange{\protect\cite{cite352}}.
\end{eczvaluelist}
\codefieldsection{Notes}
\begin{eczvaluelist}
\item\relax See Refs. \NoCaseChange{\protect\cite{cite2070,cite352,cite2071,cite2072,cite2073}} for reviews of turbo codes.
\item\relax See database \NoCaseChange{\protect\cite{cite2074}} for explicit codes.
\item\relax Codes have been benchmarked using AFF3CT toolbox \NoCaseChange{\protect\cite{cite1480}}.
\end{eczvaluelist}
\codefieldsection{Parents}
\begin{eczvaluelist}
\item\relax
\flmRefsHyperref[eczindexfamilyrel]{code:convolutional}{Convolutional code}\item\relax
\flmRefsHyperref[eczindexfamilyrel]{code:parallel_concatenated}{Parallel concatenated code}\end{eczvaluelist}
\codefieldsection{Cousins}
\begin{eczvaluelist}
\item\relax
\flmRefsHyperref[eczindexfamilyrel]{code:qam}{Quadrature-amplitude modulation (QAM) format} --- Turbo codes concatenated with QAM codes offer a substantial coding gain \NoCaseChange{\protect\cite{cite2075}}.
\item\relax
\flmRefsHyperref[eczindexfamilyrel]{code:bpsk}{Binary PSK (BPSK) modulation format} --- Turbo codes can be concatenated with BPSK codes \NoCaseChange{\protect\cite{cite2076}}.
\item\relax
\flmRefsHyperref[eczindexfamilyrel]{code:ea_turbo}{EA quantum turbo code} --- EA quantum turbo codes are entanglement-assisted quantum analogues of turbo codes.
\item\relax
\flmRefsHyperref[eczindexfamilyrel]{code:quantum_turbo}{Quantum turbo code} --- Quantum turbo codes are quantum analogues of turbo codes.
\end{eczvaluelist}
\eczhbkcontributors{ \eczhuVVA }
\endeczcode

\eczcode{twisted_bch}{Twisted BCH code}{~\NoCaseChange{\protect\cite{cite2015,cite2077,cite2078}}}
\codefieldsection{Description}
Additive or linear \(q\)-ary code obtained from BCH codes via various lengthening and extension procedures such as Construction X and Construction XX.

\codefieldsection{Parent}
\begin{eczvaluelist}
\item\relax
\flmRefsHyperref[eczindexfamilyrel]{code:q-ary_additive}{Additive \(q\)-ary code}\end{eczvaluelist}
\codefieldsection{Cousins}
\begin{eczvaluelist}
\item\relax
\flmRefsHyperref[eczindexfamilyrel]{code:reed_solomon}{Reed-Solomon (RS) code} --- Some twisted BCH codes are \flmRefsHyperref{ref33}{subfield} subcodes of RS codes \NoCaseChange{\protect\cite{cite2015}}.
\item\relax
\flmRefsHyperref[eczindexfamilyrel]{code:q-ary_bch}{Bose–Chaudhuri–Hocquenghem (BCH) code} --- Twisted BCH codes are obtained from \(q\)-ary BCH codes via various lengthening and extension procedures.
\item\relax
\flmRefsHyperref[eczindexfamilyrel]{code:quantum_twisted}{Quantum twisted code} --- Quantum twisted codes are quantum analogues of twisted BCH codes.
\end{eczvaluelist}
\eczhbkcontributors{ \eczhuVVA }
\endeczcode

\eczcode{two_weight}{Two-weight code}{}
\codefieldsection{Description}
A linear \(q\)-ary code whose codewords all have one of two possible nonzero Hamming weights \NoCaseChange{\protect\cite[{Def. 19.1}]{cite172}}.

\codefieldsection{Protection}
Bounds on two-weight codes are known \NoCaseChange{\protect\cite{cite2079}} along with complete \flmRefsHyperref{ref113}{weight enumerators} \NoCaseChange{\protect\cite{cite2080}}.
Sizes of maximal two-weight, three-weight, and four-weight binary codes have been determined \NoCaseChange{\protect\cite{cite2081}}.

There is a classification of two-weight codes of length 40 \NoCaseChange{\protect\cite{cite2082}} as well as of \([n,m,d]_q\) codes with the two weights \(n,d\) \NoCaseChange{\protect\cite{cite1973}}.

\codefieldsection{Notes}
\begin{eczvaluelist}
\item\relax See \NoCaseChange{\protect\cite{cite172,cite203}} for overviews of two-weight codes and Refs. \NoCaseChange{\protect\cite{cite2083,cite2084}} for examples.
\item\relax If the difference between the two nonzero weights is not a power of the characteristic, then the code is necessarily nonprojective \NoCaseChange{\protect\cite[{Sec. 19.8}]{cite172}}.
\item\relax There are relations between two- and three-weight codes and Bent functions \NoCaseChange{\protect\cite{cite2085}}; see \NoCaseChange{\protect\cite[{Ch. 18}]{cite2086}}.
\end{eczvaluelist}
\codefieldsection{Parent}
\begin{eczvaluelist}
\item\relax
\flmRefsHyperref[eczindexfamilyrel]{code:q-ary_linear}{Linear \(q\)-ary code}\end{eczvaluelist}
\codefieldsection{Child}
\begin{eczvaluelist}
\item\relax
\flmRefsHyperref[eczindexfamilyrel]{code:projective_two_weight}{Projective two-weight code} --- Projective two-weight codes are two-weight codes by definition \NoCaseChange{\protect\cite[{Def. 19.1}]{cite172}} (see also \NoCaseChange{\protect\cite{cite1972,cite203,cite1973}}).
\end{eczvaluelist}
\codefieldsection{Cousins}
\begin{eczvaluelist}
\item\relax
\flmRefsHyperref[eczindexfamilyrel]{code:divisible}{Divisible code} --- Two-weight codes are \(m\)-divisible, where \(m\) is the greatest common divisor of their two possible weights.
\item\relax
\flmRefsHyperref[eczindexfamilyrel]{code:projective}{Projective geometry code} --- Projective two-weight codes correspond to projective 2-character sets and certain strongly regular graphs \NoCaseChange{\protect\cite{cite1972,cite203,cite1973}\protect\cite[{Sec. 19.3.3}]{cite172}\protect\cite[{Sec. 19.9.1}]{cite172}}.
\item\relax
\flmRefsHyperref[eczindexfamilyrel]{code:weight_two}{Weight-two code} --- Codewords of two-weight codes have one of two possible Hamming weights, while those of weight-two codes have Hamming weight two.
\item\relax
\flmRefsHyperref[eczindexfamilyrel]{code:q-ary_constant_weight}{Constant-weight block code} --- Each codeword of a constant-weight (two-weight) code has one (two) possible Hamming weight(s).
\item\relax
\flmRefsHyperref[eczindexfamilyrel]{code:q-ary_simplex}{\(q\)-ary simplex code} --- \(q\)-ary MacDonald codes are two-weight codes with weights \(q^{m-1}-q^{u-1}\) and \(q^{m-1}\) \NoCaseChange{\protect\cite{cite1161}}.
\end{eczvaluelist}
\eczhbkcontributors{ \eczhuVVA }
\endeczcode

\eczcode{uniformly_packed}{Uniformly packed code}{~\NoCaseChange{\protect\cite{cite1169,cite2087,cite1749}}}
\codefieldsection{Description}
An \((n,K,2t+1)_q\) code is uniformly packed if its \flmRefsHyperref{ref113}{external distance} is equal to \(t+1\) \NoCaseChange{\protect\cite{cite41}}; see \NoCaseChange{\protect\cite[{Def. 2.5}]{cite1748}} for the case of even distance and other generalizations.

\codefieldsection{Parent}
\begin{eczvaluelist}
\item\relax
\flmRefsHyperref[eczindexfamilyrel]{code:completely_regular}{Completely regular code} --- Uniformly packed codes are completely regular \NoCaseChange{\protect\cite{cite1749}\protect\cite[{Prop. 2.6}]{cite1748}}.
\end{eczvaluelist}
\codefieldsection{Child}
\begin{eczvaluelist}
\item\relax
\flmRefsHyperref[eczindexfamilyrel]{code:quasi_perfect}{Quasi-perfect code} --- Quasi-perfect codes are uniformly packed \NoCaseChange{\protect\cite[{Def. 2.5}]{cite1748}}.
\end{eczvaluelist}
\codefieldsection{Cousin}
\begin{eczvaluelist}
\item\relax
\flmRefsHyperref[eczindexfamilyrel]{code:dodecacode}{\((12,4^6,6)_4\) Dodecacode} --- The punctured dodecacode code is uniformly packed \NoCaseChange{\protect\cite{cite1646}}.
\end{eczvaluelist}
\eczhbkcontributors{ \eczhuVVA }
\endeczcode

\eczcode{univ_opt_q-ary}{Universally optimal \(q\)-ary code}{~\NoCaseChange{\protect\cite{cite2088,cite2089,cite171,cite914,cite2090,cite2091,cite173}}}
\codefieldsection{Description}
A \(q\)-ary code that (weakly) minimizes all completely monotonic potentials on Hamming space \NoCaseChange{\protect\cite{cite173}}.
Equivalently, its binomial moments are minimal among all codes with the same size and block length \NoCaseChange{\protect\cite[{Lemma 4}]{cite173}}.

All codes that attain the linear programming (LP) bound by Delsarte \NoCaseChange{\protect\cite{cite209}} are universally optimal \NoCaseChange{\protect\cite{cite173}}.
Such codes are called \textit{LP universally optimal} or \textit{extremal}.
If universal optimality is proved by these LP bounds, then deleting any one codeword yields another universally optimal code \NoCaseChange{\protect\cite[{Thm. 2}]{cite173}}.
However, not all universally optimal codes attain the Delsarte LP bound (for example, the \([23,12,7]\) Golay code \NoCaseChange{\protect\cite{cite173}\protect\cite[{Table 12.1}]{cite199}}).
See \NoCaseChange{\protect\cite[{Table 12.1}]{cite199}\protect\cite[{Table 1}]{cite173}} for a list of (LP) universally optimal codes.
See \NoCaseChange{\protect\cite[{Sec. 12.3}]{cite199}} for further discussion.

All codes that attain the Levenshtein bound \NoCaseChange{\protect\cite{cite2088,cite2089,cite171,cite914}}, which estimates the solution to Delsarte's linear program, are universally optimal \NoCaseChange{\protect\cite{cite2090}}; see \NoCaseChange{\protect\cite[{Thm. 12.3.23}]{cite199}}.
However, not all universally optimal codes attain the Levenshtein bound.

\codefieldsection{Parents}
\begin{eczvaluelist}
\item\relax
\flmRefsHyperref[eczindexfamilyrel]{code:q-ary_digits_into_q-ary_digits}{\(q\)-ary code}\item\relax
\flmRefsHyperref[eczindexfamilyrel]{code:univ_opt}{Universally optimal code}\end{eczvaluelist}
\codefieldsection{Children}
\begin{eczvaluelist}
\item\relax
\flmRefsHyperref[eczindexfamilyrel]{code:golay}{\([23, 12, 7]\) Golay code} --- The Golay code and several of its extended, shortened, and punctured versions are LP universally optimal codes \NoCaseChange{\protect\cite{cite173}}.
\item\relax
\flmRefsHyperref[eczindexfamilyrel]{code:conference}{Conference code} --- Conference codes are LP universally optimal codes \NoCaseChange{\protect\cite{cite173}}.
\item\relax
\flmRefsHyperref[eczindexfamilyrel]{code:mds}{Maximum distance separable (MDS) code} --- MDS codes are LP universally optimal codes \NoCaseChange{\protect\cite{cite1917,cite173}}.
\item\relax
\flmRefsHyperref[eczindexfamilyrel]{code:q-ary_hamming}{\(q\)-ary Hamming code} --- Hamming codes and their once-punctured and once-shortened versions are LP universally optimal codes \NoCaseChange{\protect\cite{cite173}}.
\item\relax
\flmRefsHyperref[eczindexfamilyrel]{code:ternary_golay}{\([11,6,5]_3\) Ternary Golay code} --- The ternary Golay code and several of its extended, shortened, and punctured versions are LP universally optimal codes \NoCaseChange{\protect\cite{cite173}}. The \([11,6,5]_3\) code is universally optimal but not sharp.
\item\relax
\flmRefsHyperref[eczindexfamilyrel]{code:delsarte_optimal_q-ary}{\(q\)-ary sharp configuration} --- All \(q\)-ary sharp configurations are universally optimal \(q\)-ary codes \NoCaseChange{\protect\cite{cite173}}, but the converse is not true.
\end{eczvaluelist}
\codefieldsection{Cousins}
\begin{eczvaluelist}
\item\relax
\flmRefsHyperref[eczindexfamilyrel]{code:constant_weight}{Constant-weight code} --- Some constant-weight codes, such as simplex codes, are also universally optimal binary codes.
\item\relax
\flmRefsHyperref[eczindexfamilyrel]{code:kerdock}{Kerdock code} --- Kerdock codes are asymptotically close to the LP and universal-energy bounds \NoCaseChange{\protect\cite[{Exam. 12.3.25}]{cite199}}.
\item\relax
\flmRefsHyperref[eczindexfamilyrel]{code:hadamard}{\([2^m,m,2^{m-1}]\) Hadamard code} --- Several punctured Hadamard codes are LP universally optimal codes \NoCaseChange{\protect\cite{cite173}}.
\item\relax
\flmRefsHyperref[eczindexfamilyrel]{code:extended_hamming}{\([2^m,2^m-m-1,4]\) Extended Hamming code} --- Several extended Hamming codes are LP universally optimal codes \NoCaseChange{\protect\cite{cite173}}.
\item\relax
\flmRefsHyperref[eczindexfamilyrel]{code:hamming}{\([2^r-1,2^r-r-1,3]\) Hamming code} --- Binary Hamming codes and several of their extended, punctured, and shortened versions are LP universally optimal codes \NoCaseChange{\protect\cite{cite173}}.
\item\relax
\flmRefsHyperref[eczindexfamilyrel]{code:bose_qvist}{Ovoid code} --- Several shortened and punctured versions of the ovoid code are LP universally optimal codes \NoCaseChange{\protect\cite{cite173}}.
\item\relax
\flmRefsHyperref[eczindexfamilyrel]{code:q-ary_simplex}{\(q\)-ary simplex code} --- Simplex codes and their once-punctured versions are LP universally optimal codes \NoCaseChange{\protect\cite{cite173}}.
\end{eczvaluelist}
\eczhbkcontributors{ Alexander Barg, \eczhuVVA }
\endeczcode

\eczcode{weighed_covering}{Weighted-covering code}{}
\codefieldsection{Description}
A \(q\)-ary code for which balls of some radius centered at its codewords provide a weighted covering of the Hamming space.

Let the \textit{outer} or \textit{weight distribution} of a \(q\)-ary string \(x\) with respect to a \(q\)-ary code \(C\) be \(A(x) = \left( A_0(x),A_1(x),\cdots,A_n(x) \right)\), where
\flmMathEnvironment{align}{}{
  A_j(x) = \left|\{ c \in C~\text{such that}~ D(c,x)=j \}\right|~,
}
and \(D\) is the Hamming distance. Given a tuple \(m=(m_1,m_2,\cdots,m_n)\) of rational numbers, the \textit{\(m\)-density} of the code at \(x\) is
\flmMathEnvironment{align}{}{
  \theta(x) = \sum_{j=0}^n m_j A_j(x)~.
}

A code is an \textit{\(m\)-weighted covering} if \(\theta(x)\geq1\) for all strings \(x\in \mathbb{F}_q^n\).
The \textit{\(m\)-covering radius} \(r\) is the largest \(j\) for which \(m_j\) is nonzero.

\codefieldsection{Notes}
\begin{eczvaluelist}
\item\relax See book \NoCaseChange{\protect\cite{cite244}} for an exposition on weighted covering codes and generalized sphere-packing bounds.
\item\relax See \NoCaseChange{\protect\cite[{Table 7.5.18}]{cite953}} for tables of codes with particular weighted coverings.
\end{eczvaluelist}
\codefieldsection{Parent}
\begin{eczvaluelist}
\item\relax
\flmRefsHyperref[eczindexfamilyrel]{code:q-ary_digits_into_q-ary_digits}{\(q\)-ary code}\end{eczvaluelist}
\codefieldsection{Children}
\begin{eczvaluelist}
\item\relax
\flmRefsHyperref[eczindexfamilyrel]{code:covering}{Covering code} --- An \(m\)-weighted covering code for \(m_j=1\) is a covering code of covering radius at most \(r\) \NoCaseChange{\protect\cite[{Ch. 13}]{cite244}}.
\item\relax
\flmRefsHyperref[eczindexfamilyrel]{code:quasi_perfect}{Quasi-perfect code} --- A quasi-perfect code is a special case of an \(m\)-weighted covering code with \(m\)-covering radius \(r=t+1\) \NoCaseChange{\protect\cite[{Ch. 13}]{cite244}}.
\end{eczvaluelist}
\eczhbkcontributors{ \eczhuVVA }
\endeczcode

\eczcode{wozencraft}{Wozencraft ensemble code}{~\NoCaseChange{\protect\cite{cite2092}}}
\codefieldsection{Description}
A code that is part of the Wozencraft ensemble, a set of codes most of whose members achieve the \flmRefsHyperref{ref85}{GV bound}.

Each \([2k,k]_q\) code is defined by a parameter \(\alpha \in \mathbb{F}_{q^k}\) and consists of codewords \((x,\alpha x)\) for each message \(x \in \mathbb{F}_{q^k}\), where each element of \(\mathbb{F}_{q^k}\) is expressed over \(\mathbb{F}_q^k\) using a fixed basis.

\codefieldsection{Protection}
Bounds and constructions with \flmRefsHyperref{ref65}{order} \(\Omega(\sqrt{k})\) distance are provided in Ref. \NoCaseChange{\protect\cite{cite2093}}.

\codefieldsection{Rate}
Meets the \flmRefsHyperref{ref85}{GV bound} for most choices of \(\alpha\). Puncturing the code yields a higher rate with also meeting the \flmRefsHyperref{ref85}{GV bound}; see Ref. \NoCaseChange{\protect\cite{cite2093}}. These codes can be used to asymptotically improve over the \flmRefsHyperref{ref85}{GV bound} \NoCaseChange{\protect\cite{cite2094}}.
\codefieldsection{Parent}
\begin{eczvaluelist}
\item\relax
\flmRefsHyperref[eczindexfamilyrel]{code:q-ary_linear}{Linear \(q\)-ary code}\end{eczvaluelist}
\codefieldsection{Cousin}
\begin{eczvaluelist}
\item\relax
\flmRefsHyperref[eczindexfamilyrel]{code:justesen}{Justesen code} --- Wozencraft ensemble forms the inner codes of Justesen codes.
\end{eczvaluelist}
\eczhbkcontributors{ \eczhuVVA }
\endeczcode

\onecolumngrid
\clearpage

\section{Matrix Kingdom}

\begin{eczEpigraph}
\begin{quote}
\flmQuoteSetup{quote}%
I was told to think outside the box, so I switched to a higher-dimensional matrix.
\flmQuoteAttributed{Andrej and Elena Cherkaev}
\end{quote}
\end{eczEpigraph}

\twocolumngrid

\eczcode{alamouti}{Alamouti code}{~\NoCaseChange{\protect\cite{cite2095}}}
\codefieldsection{Description}
The simplest OSTBC, with \(n=2\) transmitting antennas, \(m=1\) receiving antenna, and \(t=2\) channel uses.

The transmit matrix is
\flmMathEnvironment{align}{}{
  \begin{pmatrix}c_{1} & c_{2}\\
  -c_{2}^{\star} & c_{1}^{\star}
  \end{pmatrix}~,
}
where \(c_i\) are complex numbers, and where \(\star\) denotes complex conjugation.

\codefieldsection{Rate}
The only complex OSTBC with full rate, i.e., one complex information symbol per channel use \NoCaseChange{\protect\cite{cite2096,cite2097}}.
\codefieldsection{Realizations}
\begin{eczvaluelist}
\item\relax Wireless standards since: 3G, LTE, LTE-Advanced, and 5G.
\item\relax Wireless communication: IEEE 802.11n, IEEE 802.11ad, IEEE 802.11ay, etc.
\end{eczvaluelist}
\codefieldsection{Parents}
\begin{eczvaluelist}
\item\relax
\flmRefsHyperref[eczindexfamilyrel]{code:orth_spacetime_block}{Orthogonal Spacetime Block Code (OSTBC)}\item\relax
\flmRefsHyperref[eczindexfamilyrel]{code:linear_spacetime}{Linear STC}\item\relax
\flmRefsHyperref[eczindexfamilyrel]{code:unitary}{Unitary code} --- Codewords of the Alamouti code are two-dimensional unitary matrices.
\end{eczvaluelist}
\eczhbkcontributors{ \eczhuVVA }
\endeczcode

\eczcode{array}{Array code}{}
\codefieldsection{Alternative Names}
\begin{eczvaluelist}
\item\relax RAID code
\item\relax Disk array code
\end{eczvaluelist}
\eczhIndexCodeAliasName{array}{RAID code}
\eczhIndexCodeAliasName{array}{Disk array code}
\codefieldsection{Description}
Matrix code over \(\mathbb{F}_q\) designed for use in distributed storage, with the first such application being a RAID-type array of hard drives such that information is protected against erasure of one or more hard drives.
Each column of a matrix codeword is typically treated as a single codeword coordinate via subpacketization (defined below) and represents a large data block.
Array codes are denoted by \((n,k,m)\), where \(n\) is the number of nodes, \(k\) is the number of nodes needed to recover the data, and \(m\) is the column dimension of each codeword a.k.a. the subpacketization level.

Classical array codes also include block and convolutional families whose parity checks run along rows, columns, or diagonals, with applications to burst correction, multitrack magnetic recording, and disk arrays \NoCaseChange{\protect\cite{cite189}}.

\codefieldsection{Protection}
Since they are designed to protect against corruption of nodes in a distributed storage network, many array codes are analyzed in terms of their erasure capabilities.

An analogue of the Hamming metric for array codes comes from counting the number of column positions in which two codewords differ.
Equivalently, one may treat each length-\(m\) column as a single symbol from an alphabet of size \(q^m\); this interpretation underlies the subpacketization viewpoint.
Array codes are typically \(\mathbb{F}_q\)-linear as matrix codes. In some constructions, after choosing an identification of columns with symbols from an extension field \(\mathbb{F}_{q^m}\), one also obtains an \(\mathbb{F}_{q^m}\)-linear code.

There are other notions of distance for array codes, including the rank metric and its generalization the sum-rank metric.

\codefieldsection{Notes}
\begin{eczvaluelist}
\item\relax See \NoCaseChange{\protect\cite{cite189}\protect\cite[{Ch. 14}]{cite2098}} for introductions.
\end{eczvaluelist}
\codefieldsection{Parents}
\begin{eczvaluelist}
\item\relax
\flmRefsHyperref[eczindexfamilyrel]{code:matrices_into_matrices}{Matrix-based code}\item\relax
\flmRefsHyperref[eczindexfamilyrel]{code:distributed_storage}{Distributed-storage code}\end{eczvaluelist}
\codefieldsection{Children}
\begin{eczvaluelist}
\item\relax
\flmRefsHyperref[eczindexfamilyrel]{code:bbv}{Generalized EVENODD code}\item\relax
\flmRefsHyperref[eczindexfamilyrel]{code:mds_array}{MDS array code}\item\relax
\flmRefsHyperref[eczindexfamilyrel]{code:regenerating}{Regenerating code (RGC)}\end{eczvaluelist}
\codefieldsection{Cousins}
\begin{eczvaluelist}
\item\relax
\flmRefsHyperref[eczindexfamilyrel]{code:tensor}{Tensor-product code} --- Classical block array codes form a subclass of product codes, i.e., tensor-product codes \NoCaseChange{\protect\cite{cite189}}.
\item\relax
\flmRefsHyperref[eczindexfamilyrel]{code:reed_solomon}{Reed-Solomon (RS) code} --- RS codes over \(q=2^m\) are used in RAID 6 \NoCaseChange{\protect\cite{cite330,cite331}}; see \NoCaseChange{\protect\cite{cite189}}.
\item\relax
\flmRefsHyperref[eczindexfamilyrel]{code:cross_interleaved_reed_solomon}{Cross-interleaved RS (CIRS) code} --- The CIRS code can also be visualized as a 2D array code \NoCaseChange{\protect\cite{cite189}}.
\end{eczvaluelist}
\eczhbkcontributors{ \eczhuVVA }
\endeczcode

\eczcode{b_array}{B-code}{~\NoCaseChange{\protect\cite{cite2014}}}
\codefieldsection{Description}
Binary MDS block array code \(\mathcal{B}_2(m)\) on \((m-1)\times m\) arrays with horizontal and toroidal diagonal parity checks \NoCaseChange{\protect\cite{cite189}}.

\codefieldsection{Protection}
Corrects any two erased columns if and only if \(m\) is prime, and can then also correct any single column error \NoCaseChange{\protect\cite{cite189}}.

\codefieldsection{Decoding}
\begin{eczvaluelist}
\item\relax Efficient decoding algorithm against erasures \NoCaseChange{\protect\cite{cite2014}}.
\end{eczvaluelist}
\codefieldsection{Parent}
\begin{eczvaluelist}
\item\relax
\flmRefsHyperref[eczindexfamilyrel]{code:mds_array}{MDS array code} --- B-codes are examples of MDS array codes \NoCaseChange{\protect\cite{cite192}}.
\end{eczvaluelist}
\codefieldsection{Cousins}
\begin{eczvaluelist}
\item\relax
\flmRefsHyperref[eczindexfamilyrel]{code:reed_solomon}{Reed-Solomon (RS) code} --- B-codes can be interpreted as RS codes over polynomials whose symbols lie in Galois rings \NoCaseChange{\protect\cite{cite2014,cite1224}}.
\item\relax
\flmRefsHyperref[eczindexfamilyrel]{code:array_ldpc}{Array-based LDPC (AB-LDPC) code} --- AB-LDPC codes are constructed from certain classes of B-codes. B-codes can be viewed as binary codes by mapping their ring elements to permutation matrices (cf. \flmRefsHyperref{ref47}{lifting}). The resulting codes turn out to be LDPC \NoCaseChange{\protect\cite{cite1224}}.
\item\relax
\flmRefsHyperref[eczindexfamilyrel]{code:evenodd}{EVENODD code} --- EVENODD avoids the recursive parity updates of \(\mathcal{B}_2(m)\) while retaining optimal two-erasure correction \NoCaseChange{\protect\cite{cite189}}.
\end{eczvaluelist}
\eczhbkcontributors{ \eczhuVVA }
\endeczcode

\eczcode{clifford_group}{Clifford group}{~\NoCaseChange{\protect\cite{cite2099,cite2100,cite2101,cite2102,cite449,cite2103,cite2104}}}
\codefieldsection{Alternative Names}
\begin{eczvaluelist}
\item\relax Clifford-Weil group
\item\relax Normalizer of a group of symplectic type
\end{eczvaluelist}
\eczhIndexCodeAliasName{clifford_group}{Clifford-Weil group}
\eczhIndexCodeAliasName{clifford_group}{Normalizer of a group of symplectic type}
\codefieldsection{Description}
The Clifford group on \(n\) qubits is a subgroup of the unitary group \(U(2^n)\) that is the normalizer of the \flmRefsHyperref{ref663}{Pauli group}, that forms a unitary 3-design, and that is closely related to the automorphism group of BW lattices.
The group features prominently in quantum information, with the rest of the entry given in that context.

\begin{defterm}{Clifford group}\label{ref2105}\label{ref409}
The \(n\)-qubit Clifford group consists of the Pauli group as well as elements that permute Pauli operators amongst themselves.
Up to any phases and Pauli strings, the group is equivalent to the symplectic group \(Sp(2n,\mathbb{Z}_2)\).
A Clifford circuit is a quantum circuit, defined on some qubit geometry and consisting of only Clifford gates. 
See Refs. \NoCaseChange{\protect\cite{cite2104,cite2106,cite2107,cite2108,cite398}} for generators, relations, and normal form. 
\end{defterm}

The Clifford group cannot be expressed as a semidirect product of the Pauli and symplectic groups \NoCaseChange{\protect\cite{cite435}}.
Restricting the group to real-valued elements yields the \textit{real Clifford group} \NoCaseChange{\protect\cite{cite2109,cite2110}}, and including complex conjugation yields the \textit{extended Clifford group} \NoCaseChange{\protect\cite{cite2111}}.

Single-qubit Clifford gates, together with Paulis, realize a group with \(192\) elements.
Modding out phases yields the \(48\)-element \(2O\) binary octahedral subgroup of \(SU(2)\).
Further modding out the Pauli group, which corresponds to the quaternion group \(Q\), yields the permutation group \(S_3\), which consists of permutations of the three non-identity single-qubit Pauli matrices.

The two-qubit Clifford group, modded out by the Pauli group and phases, is isomorphic to \(S_6\), and its subgroups have been classified \NoCaseChange{\protect\cite{cite2112}}.

The commutant of four-fold \NoCaseChange{\protect\cite{cite801,cite2113}} and higher transversal representations of the Clifford group consists of qubit permutations and projections onto the \(\llbracket 2m,2m-2,2\rrbracket \) error-detecting code \NoCaseChange{\protect\cite{cite802}}.
In particular, the fourth tensor-power commutant is generated by tensor-factor permutations and the projector onto this special stabilizer code \NoCaseChange{\protect\cite[{Thm. 1}]{cite801}}.
Although generic Clifford orbits are not projective 4-designs, exact projective 4-designs can be constructed from Clifford orbits, and random Clifford orbits are typically close to 4-designs \NoCaseChange{\protect\cite{cite801}}.

Clifford-group elements can be sampled efficiently \NoCaseChange{\protect\cite{cite2114}}.

\codefieldsection{Notes}
\begin{eczvaluelist}
\item\relax See \NoCaseChange{\protect\cite[{preface}]{cite42}} for a history of the Clifford group.
\end{eczvaluelist}
\codefieldsection{Parent}
\begin{eczvaluelist}
\item\relax
\flmRefsHyperref[eczindexfamilyrel]{code:unitary_design}{Unitary \(t\)-design} --- Stabilizer states on \(n\) qubits form 3-designs on complex projective spaces \(\mathbb{C}P^{2^n}\) \NoCaseChange{\protect\cite{cite937}}. The \flmRefsHyperref{ref409}{Clifford group} is a unitary 2-design \NoCaseChange{\protect\cite{cite938}} and a 3-design \NoCaseChange{\protect\cite{cite940,cite941}\protect\cite[{Thm. 1.6(B)}]{cite939}\protect\cite[{pg. 191}]{cite42}} on \(U(2^n)\). The \(\llbracket 2m,2m-2,2\rrbracket \) code when \(2m\) is a multiple of four obstructs the Clifford group from being a 4-design \NoCaseChange{\protect\cite{cite801}}.
\end{eczvaluelist}
\codefieldsection{Cousins}
\begin{eczvaluelist}
\item\relax
\flmRefsHyperref[eczindexfamilyrel]{code:qubit_stabilizer}{Qubit stabilizer code} --- Computing with \flmRefsHyperref{ref409}{Clifford gates}, Pauli measurements, and classical feedforward acting on stabilizer states only can be efficiently simulated on a classical computer by tracking stabilizer and logical generators, according to the \textit{Gottesman-Knill theorem} \NoCaseChange{\protect\cite{cite2115,cite2116}}. Stabilizer states can be mapped into the first lattice shell of a BW lattice over a cyclotomic field, while the \flmRefsHyperref{ref409}{Clifford group} is related to the symmetry group of the lattice \NoCaseChange{\protect\cite{cite2117}}.
\item\relax
\flmRefsHyperref[eczindexfamilyrel]{code:kerdock}{Kerdock code} --- Kerdock codes correspond to cluster states, and the corresponding Clifford-group automorphisms of this set form a particular group \NoCaseChange{\protect\cite{cite934}} that is a unitary 2-design on \(U(2^n)\) \NoCaseChange{\protect\cite{cite935}}. As such, cluster states form complex projective 2-designs on \(\mathbb{C}P^{2^n}\). These are useful in matrix-vector multiplication \NoCaseChange{\protect\cite{cite936}}.
\item\relax
\flmRefsHyperref[eczindexfamilyrel]{code:cluster_state}{Cluster-state code} --- Kerdock codes correspond to cluster states, and the corresponding Clifford-group automorphisms of this set form a particular group \NoCaseChange{\protect\cite{cite934}} that is a unitary 2-design on \(U(2^n)\) \NoCaseChange{\protect\cite{cite935}}. As such, cluster states form complex projective 2-designs on \(\mathbb{C}P^{2^n}\). These are useful in matrix-vector multiplication \NoCaseChange{\protect\cite{cite936}}.
\item\relax
\flmRefsHyperref[eczindexfamilyrel]{code:complex_projective}{Complex projective space code} --- Stabilizer states on \(n\) qubits form 3-designs on complex projective spaces \(\mathbb{C}P^{2^n}\) \NoCaseChange{\protect\cite{cite937}}. The \flmRefsHyperref{ref409}{Clifford group} is a unitary 2-design \NoCaseChange{\protect\cite{cite938}} and a 3-design \NoCaseChange{\protect\cite{cite940,cite941}\protect\cite[{Thm. 1.6(B)}]{cite939}\protect\cite[{pg. 191}]{cite42}} on \(U(2^n)\). The \(\llbracket 2m,2m-2,2\rrbracket \) code when \(2m\) is a multiple of four obstructs the Clifford group from being a 4-design \NoCaseChange{\protect\cite{cite801}}.
\item\relax
\flmRefsHyperref[eczindexfamilyrel]{code:iceberg}{\(\llbracket 2m,2m-2,2\rrbracket \) error-detecting code} --- Stabilizer states on \(n\) qubits form 3-designs on complex projective spaces \(\mathbb{C}P^{2^n}\) \NoCaseChange{\protect\cite{cite937}}. The \flmRefsHyperref{ref409}{Clifford group} is a unitary 2-design \NoCaseChange{\protect\cite{cite938}} and a 3-design \NoCaseChange{\protect\cite{cite940,cite941}\protect\cite[{Thm. 1.6(B)}]{cite939}\protect\cite[{pg. 191}]{cite42}} on \(U(2^n)\). The \(\llbracket 2m,2m-2,2\rrbracket \) code when \(2m\) is a multiple of four obstructs the Clifford group from being a 4-design \NoCaseChange{\protect\cite{cite801}}.
\item\relax
\flmRefsHyperref[eczindexfamilyrel]{code:barnes_wall}{Barnes-Wall (BW) lattice} --- Stabilizer states can be mapped into the first lattice shell of a BW lattice over a cyclotomic field, while the \flmRefsHyperref{ref409}{Clifford group} is related to the symmetry group of the lattice \NoCaseChange{\protect\cite{cite2117}}.
\item\relax
\flmRefsHyperref[eczindexfamilyrel]{code:sidelnikov}{Real-Clifford subgroup-orbit code} --- The automorphism group of BW lattices is a subgroup of index 2 of a \flmRefsHyperref{ref409}{real Clifford group} \NoCaseChange{\protect\cite{cite2109,cite2110}} (see \NoCaseChange{\protect\cite{cite2103,cite2117}} for an explanation).
\item\relax
\flmRefsHyperref[eczindexfamilyrel]{code:haar_random}{Haar-random qubit code} --- Approximating the random projections through \(t\)-designs is necessary in order to make the Haar-random qubit protocol practical. Replacing with random \flmRefsHyperref{ref409}{Clifford gates} is especially convenient since the \flmRefsHyperref{ref409}{Clifford group} forms a unitary 2-design and produces stabilizer codes.
\item\relax
\flmRefsHyperref[eczindexfamilyrel]{code:clifford_hierarchy}{Clifford-hierarchy stabilizer code} --- Clifford-hierarchy codes are joint eigenspaces of subsets of the \flmRefsHyperref{ref2118}{Clifford hierarchy}, whose second level is the \flmRefsHyperref{ref409}{Clifford group}.
\item\relax
\flmRefsHyperref[eczindexfamilyrel]{code:qubits_into_qubits}{Qubit code} --- Computing with \flmRefsHyperref{ref409}{Clifford gates}, Pauli measurements, and classical feedforward acting on stabilizer states only can be efficiently simulated on a classical computer by tracking stabilizer and logical generators, according to the \textit{Gottesman-Knill theorem} \NoCaseChange{\protect\cite{cite2115,cite2116}}.
There is a canonical form for \flmRefsHyperref{ref409}{Clifford circuits} \NoCaseChange{\protect\cite{cite2119,cite2120}} and many algorithms for simulating them \NoCaseChange{\protect\cite{cite2121,cite2122,cite2123}}.
Universal quantum computing can be achieved using \flmRefsHyperref{ref409}{Clifford gates} and a single type of \flmRefsHyperref{ref409}{non-Clifford} gate, such as the \(T\) gate \NoCaseChange{\protect\cite{cite2124}}.
More generally, the \textit{Solovay-Kitaev} theorem \NoCaseChange{\protect\cite{cite2125,cite1634}} states that any subset of gates that generates a dense subgroup of the full \(n\)-qubit gate group can be used to construct any gate to arbitrary accuracy (see \NoCaseChange{\protect\cite{cite2126}\protect\cite[{Appx. 3}]{cite2127}}). The task of approximating a desired gate by \flmRefsHyperref{ref409}{Clifford gates} and a fixed set of \flmRefsHyperref{ref409}{non-Clifford} gates is called \textit{gate compilation} or \textit{circuit synthesis}.

\end{eczvaluelist}
\eczhbkcontributors{ David Gross, \eczhuVVA }
\endeczcode

\eczcode{finite_grassmann}{Constant-dimension code}{~\NoCaseChange{\protect\cite{cite292}}}
\codefieldsection{Alternative Names}
\begin{eczvaluelist}
\item\relax Linear authentication code
\item\relax Code in the \(q\)-Johnson scheme
\item\relax Finite-field Grassmannian code
\item\relax Grassmannian variety code
\item\relax Finite Grassmannian code
\item\relax Discrete Grassmannian code
\end{eczvaluelist}
\eczhIndexCodeAliasName{finite_grassmann}{Linear authentication code}
\eczhIndexCodeAliasName{finite_grassmann}{Code in the \(q\)-Johnson scheme}
\eczhIndexCodeAliasName{finite_grassmann}{Finite-field Grassmannian code}
\eczhIndexCodeAliasName{finite_grassmann}{Grassmannian variety code}
\eczhIndexCodeAliasName{finite_grassmann}{Finite Grassmannian code}
\eczhIndexCodeAliasName{finite_grassmann}{Discrete Grassmannian code}
\codefieldsection{Description}
A subspace code whose codewords are \(k\)-dimensional subspaces of \(\mathbb{F}_q^n\) for fixed \(k\).
Constant-dimension codes are equivalent to linear authentication codes \NoCaseChange{\protect\cite[{Thm. 4.1}]{cite167}}.

Each \(k\)-dimensional subspace of \(\mathbb{F}_q^n\) is a point on the finite-field Grassmannian (a.k.a. Grassmannian variety or \(q\)-Johnson space). The number of such subspaces is a \(q\)-binomial coefficient \NoCaseChange{\protect\cite{cite2128}\protect\cite[{Sec. 8.6}]{cite913}}.

\codefieldsection{Protection}
Johnson-type bounds have been developed \NoCaseChange{\protect\cite{cite2129}}.

\codefieldsection{Realizations}
\begin{eczvaluelist}
\item\relax Linear authentication \NoCaseChange{\protect\cite{cite167}}.
\end{eczvaluelist}
\codefieldsection{Parents}
\begin{eczvaluelist}
\item\relax
\flmRefsHyperref[eczindexfamilyrel]{code:subspace}{Subspace code} --- Subspace codes of constant dimension reduce to constant-dimension codes.
\item\relax
\flmRefsHyperref[eczindexfamilyrel]{code:2pt_homogeneous}{Two-point homogeneous-space code} --- The finite-field Grassmannian (a.k.a. \(q\)-Johnson space) can be regarded as a finite two-point homogeneous space \(G/H\) where \(G = GL(n,\mathbb{F}_q)\) \NoCaseChange{\protect\cite{cite28}\protect\cite[{Sec. 4.2.1}]{cite987}\protect\cite[{Table 2}]{cite985}\protect\cite[{Ch. 9}]{cite39}\protect\cite[{Sec. 8.6}]{cite913}}.
\end{eczvaluelist}
\codefieldsection{Child}
\begin{eczvaluelist}
\item\relax
\flmRefsHyperref[eczindexfamilyrel]{code:subspace_design}{Subspace design}\end{eczvaluelist}
\codefieldsection{Cousins}
\begin{eczvaluelist}
\item\relax
\flmRefsHyperref[eczindexfamilyrel]{code:grassmannian}{Grassmannian code} --- The finite-field Grassmannian is a finite analogue of the compact Grassmannians.
\item\relax
\flmRefsHyperref[eczindexfamilyrel]{code:rank_metric}{Rank-metric code} --- Rank-metric codes can be lifted to make constant-dimension codes \NoCaseChange{\protect\cite{cite293,cite347}}; see review \NoCaseChange{\protect\cite{cite2128}}.
\item\relax
\flmRefsHyperref[eczindexfamilyrel]{code:constant_weight}{Constant-weight code} --- Codewords of length \(n\) and weight \(w\) are in one-to-one correspondence with subsets of \(n\) objects with \(w\) elements. The \(q\)-Johnson spaces generalize this notion to subspaces and reduce to Johnson spaces at \(q=1\). In other words, \((q=2)\)-Johnson space is not the same as (binary) Johnson space since the former indexes subspaces, while the latter indexes subsets.
\item\relax
\flmRefsHyperref[eczindexfamilyrel]{code:grassmannian_variety}{Grassmannian evaluation code} --- Grassmannian evaluation codes and constant-dimension codes are both built from the finite-field Grassmannian: the former evaluate functions on its points, while the latter use its \(k\)-dimensional subspaces themselves as codewords \NoCaseChange{\protect\cite{cite27,cite28}}.
\item\relax
\flmRefsHyperref[eczindexfamilyrel]{code:projective}{Projective geometry code} --- The projective plane \(PG(k-1,q)\) is a special case of the finite-field Grassmannian \NoCaseChange{\protect\cite{cite28}}.
\end{eczvaluelist}
\eczhbkcontributors{ \eczhuVVA }
\endeczcode

\eczcode{determinant}{Determinant code}{~\NoCaseChange{\protect\cite{cite2130}}}
\codefieldsection{Description}
Determinant codes give optimal exact repair regenerating codes for any \([n,k,d=k]\) at all the points of the storage bandwidth trade-off curve.
The codes are linear, and the exact regenerating property is provided based on fundamental properties of matrix determinants.
The field size \(q\) required for this code construction is linear in \(n\).

\codefieldsection{Decoding}
\begin{eczvaluelist}
\item\relax For exact repair, the interior points of the storage-bandwidth trade-off curve can be shown to be the convex hull of \(k\) corner points described by \((\alpha_m,\beta_m)= (\binom{k}{m},\binom{k-1}{m-1})\) for \(m\in\{1,2,\cdots,k\}\).
\end{eczvaluelist}
\codefieldsection{Parent}
\begin{eczvaluelist}
\item\relax
\flmRefsHyperref[eczindexfamilyrel]{code:regenerating}{Regenerating code (RGC)}\end{eczvaluelist}
\eczhbkcontributors{ Adway Patra, \eczhuVVA }
\endeczcode

\eczcode{diagonal}{Diagonal code}{~\NoCaseChange{\protect\cite{cite2131}}}
\codefieldsection{Description}
Member of an explicit family of high-rate \([n,k,d,\alpha, \beta = \frac{\alpha}{d-k+1}, M=k\alpha]\) MSR codes for any \(r\) and \(n\).
Such codes can optimally repair any \(f\) failed nodes from any \(d\) helper nodes for all \(d\), \(1 \le f \le r\) and \(k \le d \le n-f\) simultaneously.
These codes can be constructed over any base field \(\mathbb{F}_q\) as long as \(|\mathbb{F}_q| \ge sn\), where \(s = \text{lcm}(1,2,\cdots,r)\).

Let \(C \in \mathbb{F}_q^{\alpha \times n}\) be a codeword, with \(C_i\) being the \(i\)-th coordinate.
Then, the code is defined as
\flmMathEnvironment{equation}{}{
  \mathsf{C} = \left\{(C_1,C_2,\cdots,C_n) \,\middle|\, \sum_{i=1}^n A_i^{t-1}C_i = 0,\; t=1,2,\cdots,r \right\}~,
}
where the matrices \(A_i\) are diagonal \(\alpha \times \alpha\) matrices.

\codefieldsection{Parent}
\begin{eczvaluelist}
\item\relax
\flmRefsHyperref[eczindexfamilyrel]{code:msr}{Minimum-storage regenerating (MSR) code}\end{eczvaluelist}
\eczhbkcontributors{ Adway Patra, \eczhuVVA }
\endeczcode

\eczcode{matrix_computation}{Distributed computation code}{}
\codefieldsection{Description}
Encoding that provides extra redundancy for distributed matrix computation algorithms such as matrix multiplication. Parallelized algorithms distribute a desired computation over many nodes, and a key performance bottleneck is due to some nodes completing their individual tasks much later than other nodes. Matrix computation codes provide a layer of redundancy such that the computation can be performed without having all nodes finish their piece of the computation.

\codefieldsection{Protection}
Allows computation to complete without waiting for \textit{stragglers}, or nodes that either do not finish or finish their portion of the computation much later than all other nodes.
\codefieldsection{Parent}
\begin{eczvaluelist}
\item\relax
\flmRefsHyperref[eczindexfamilyrel]{code:matrices_into_matrices}{Matrix-based code}\end{eczvaluelist}
\codefieldsection{Cousin}
\begin{eczvaluelist}
\item\relax
\flmRefsHyperref[eczindexfamilyrel]{code:mds}{Maximum distance separable (MDS) code} --- The first matrix multiplication code encoded each entry of the matrices to be multiplied into an MDS code \NoCaseChange{\protect\cite{cite1921}}.
\end{eczvaluelist}
\eczhbkcontributors{ \eczhuVVA }
\endeczcode

\eczcode{evenodd}{EVENODD code}{~\NoCaseChange{\protect\cite{cite2132}}}
\codefieldsection{Description}
Binary array code \(\mathcal{EO}_2(m)\) with independent horizontal and diagonal parity columns, designed to retain optimal double-erasure protection while simplifying small updates \NoCaseChange{\protect\cite{cite189}}.

\codefieldsection{Protection}
Corrects any two erased columns whenever \(m\) is prime, and has minimum column distance three in that case \NoCaseChange{\protect\cite{cite189}}.

\codefieldsection{Decoding}
\begin{eczvaluelist}
\item\relax Efficient decoding algorithm against two erasures \NoCaseChange{\protect\cite{cite2132}}.
\end{eczvaluelist}
\codefieldsection{Realizations}
\begin{eczvaluelist}
\item\relax Can be implemented on standard RAID-5 controllers without extra finite-field hardware \NoCaseChange{\protect\cite{cite189}}.
\end{eczvaluelist}
\codefieldsection{Parents}
\begin{eczvaluelist}
\item\relax
\flmRefsHyperref[eczindexfamilyrel]{code:bbv}{Generalized EVENODD code} --- Generalized EVENODD codes reduce to EVENODD codes for \(r=2\) \NoCaseChange{\protect\cite{cite189}}.
\item\relax
\flmRefsHyperref[eczindexfamilyrel]{code:mds_array}{MDS array code} --- EVENODD codes are examples of MDS array codes \NoCaseChange{\protect\cite{cite192}}.
\end{eczvaluelist}
\codefieldsection{Cousin}
\begin{eczvaluelist}
\item\relax
\flmRefsHyperref[eczindexfamilyrel]{code:b_array}{B-code} --- EVENODD avoids the recursive parity updates of \(\mathcal{B}_2(m)\) while retaining optimal two-erasure correction \NoCaseChange{\protect\cite{cite189}}.
\end{eczvaluelist}
\eczhbkcontributors{ \eczhuVVA }
\endeczcode

\eczcode{gabidulin}{Gabidulin code}{~\NoCaseChange{\protect\cite{cite2133,cite2134,cite2135}}}
\codefieldsection{Alternative Names}
\begin{eczvaluelist}
\item\relax Vector rank-metric code
\item\relax Delsarte-Gabidulin code
\end{eczvaluelist}
\eczhIndexCodeAliasName{gabidulin}{Vector rank-metric code}
\eczhIndexCodeAliasName{gabidulin}{Delsarte-Gabidulin code}
\codefieldsection{Description}
A linear code over \(\mathbb{F}_{q^N}\) that corrects errors over rank metric instead of the traditional Hamming distance. Every element \(\mathbb{F}_{q^N}\) can be written as an \(N\)-dimensional vector with coefficients in \(\mathbb{F}_q\), and the rank of a set of elements is rank of the matrix formed by their coefficients.

Given \(X^n=\text{span}\{x_i\}\), an \(n\)-dimensional vector space over \(\mathbb{F}_{q^N}\) (where \(q\) is a power of a prime number), the \textit{rank metric} \(d(x, y)\) is defined via the rank norm, \(\mathrm{rank}( A(x) )\), where
\flmMathEnvironment{align}{}{
A(x) = \begin{pmatrix} a_{11} & a_{12} & \ldots & a_{1n} \\ a_{21} & a_{22} & \ldots & a_{2n} \\  a_{N1} & a_{N2} & \ldots & a_{Nn} \end{pmatrix}
}
and \(x_i = a_{1i} u_1 + a_{2i} u_2 + \ldots + a_{Ni}u_N \) for some fixed basis \(\{u_i\}_{i=1}^N\).

\codefieldsection{Protection}
Set of vectors \(\{x_1, x_2, \ldots, x_M\}\) determines a rank code with distance \(d=\min d(x_i, x_j)\). The code with distance \(d\) corrects all errors with rank of the error not greater than \(\lfloor (d-1)/2\rfloor\).
\codefieldsection{Decoding}
\begin{eczvaluelist}
\item\relax Fast decoder based on a transform-domain approach \NoCaseChange{\protect\cite{cite2136}}.
\item\relax Algebraic list decoder that decodes up to the Singleton bound \NoCaseChange{\protect\cite{cite882}}.
\end{eczvaluelist}
\codefieldsection{Realizations}
\begin{eczvaluelist}
\item\relax Public-key cryptosystems \NoCaseChange{\protect\cite{cite261,cite262}}.
\item\relax Digital watermarking. The Gabidulin code provides efficient correction against luminance tampering and image-slicing distortion due to the consistency of the rank against alterations such as column swapping \NoCaseChange{\protect\cite{cite263}}.
\end{eczvaluelist}
\codefieldsection{Parent}
\begin{eczvaluelist}
\item\relax
\flmRefsHyperref[eczindexfamilyrel]{code:rank_metric}{Rank-metric code} --- Gabidulin codes over \(\mathbb{F}_{q^N}\), when expressed as matrices over \(\mathbb{F}_q\), are rank-metric codes (see Def. 14 in Ref. \NoCaseChange{\protect\cite{cite2137}}). The reverse is not always true, i.e., not every rank-metric code is Gabidulin \NoCaseChange{\protect\cite[{Rm. 16}]{cite2137}}.
\end{eczvaluelist}
\codefieldsection{Cousins}
\begin{eczvaluelist}
\item\relax
\flmRefsHyperref[eczindexfamilyrel]{code:maximum_rank_distance}{Maximum-rank distance (MRD) code} --- Gabidulin codes over \(\mathbb{F}_{q^N}\) with maximum rank-distance, when expressed as matrices over \(\mathbb{F}_q\), are MRD codes.
\item\relax
\flmRefsHyperref[eczindexfamilyrel]{code:q-ary_linear}{Linear \(q\)-ary code} --- Gabidulin codes over \(\mathbb{F}_{q^N}\), when expressed as vectors over \(\mathbb{F}_{q^N}\), are linear \(q\)-ary codes.
\item\relax
\flmRefsHyperref[eczindexfamilyrel]{code:subspace}{Subspace code} --- Gabidulin codes can be used to construct asymptotically good subspace codes \NoCaseChange{\protect\cite{cite167,cite292}}.
\item\relax
\flmRefsHyperref[eczindexfamilyrel]{code:linearized_reed_solomon}{Linearized RS code} --- Choosing one conjugacy class with \(\sigma\neq\operatorname{Id}\) and \(\delta=0\) recovers Gabidulin codes, and the sum-rank metric reduces to the rank metric \NoCaseChange{\protect\cite[{Ex. 37}]{cite1259}}. Over \(\mathbb{F}_{q^m}\) viewed over \(\mathbb{F}_q\), linearized RS codes can have length up to \((q-1)m\), whereas Gabidulin codes over the same extension have maximum length \(m\) \NoCaseChange{\protect\cite[{Sec. 4.2}]{cite1259}}.
\item\relax
\flmRefsHyperref[eczindexfamilyrel]{code:quantum_gabidulin}{Quantum Gabidulin code} --- A quantum Gabidulin code is defined using two Gabidulin codes with associated parameters \(r,s\), respectively, such that \(r+s = n\) \NoCaseChange{\protect\cite{cite2138}}.
\end{eczvaluelist}
\eczhbkcontributors{ Micah Shaw, Marianna Podzorova, \eczhuVVA }
\endeczcode

\eczcode{bbv}{Generalized EVENODD code}{~\NoCaseChange{\protect\cite{cite2139}}}
\codefieldsection{Alternative Names}
\begin{eczvaluelist}
\item\relax Blaum-Bruck-Vardy array code
\end{eczvaluelist}
\eczhIndexCodeAliasName{bbv}{Blaum-Bruck-Vardy array code}
\codefieldsection{Description}
Generalized EVENODD code \(\mathcal{EO}_r(m)\) with one horizontal parity column and \(r-1\) independently encoded diagonal-parity columns of different slopes \NoCaseChange{\protect\cite{cite189}}.

\codefieldsection{Protection}
For prime \(m\), the \(r=3\) member has minimum column distance four and is therefore an MDS array code; for \(r \geq 4\), the generalized EVENODD family is not MDS in general \NoCaseChange{\protect\cite{cite189}}.

\codefieldsection{Parent}
\begin{eczvaluelist}
\item\relax
\flmRefsHyperref[eczindexfamilyrel]{code:array}{Array code}\end{eczvaluelist}
\codefieldsection{Child}
\begin{eczvaluelist}
\item\relax
\flmRefsHyperref[eczindexfamilyrel]{code:evenodd}{EVENODD code} --- Generalized EVENODD codes reduce to EVENODD codes for \(r=2\) \NoCaseChange{\protect\cite{cite189}}.
\end{eczvaluelist}
\codefieldsection{Cousin}
\begin{eczvaluelist}
\item\relax
\flmRefsHyperref[eczindexfamilyrel]{code:mds_array}{MDS array code} --- Generalized EVENODD codes for prime \(m\) and \(r=3\) are MDS array codes \NoCaseChange{\protect\cite{cite189}}.
\end{eczvaluelist}
\eczhbkcontributors{ \eczhuVVA }
\endeczcode

\eczcode{linear_spacetime}{Linear STC}{}
\codefieldsection{Description}
Spacetime code whose set of matrix codewords is closed under addition and subtraction.

\codefieldsection{Decoding}
\begin{eczvaluelist}
\item\relax Sphere decoder \NoCaseChange{\protect\cite{cite2140,cite2141,cite2142}}.
\end{eczvaluelist}
\codefieldsection{Parent}
\begin{eczvaluelist}
\item\relax
\flmRefsHyperref[eczindexfamilyrel]{code:spacetime_block}{Spacetime block code (STBC)}\end{eczvaluelist}
\codefieldsection{Child}
\begin{eczvaluelist}
\item\relax
\flmRefsHyperref[eczindexfamilyrel]{code:alamouti}{Alamouti code}\end{eczvaluelist}
\eczhbkcontributors{ \eczhuVVA }
\endeczcode

\eczcode{linearized_reed_solomon}{Linearized RS code}{~\NoCaseChange{\protect\cite{cite2143,cite2144,cite1259}}}
\codefieldsection{Description}
A code obtained by linearizing a skew RS code, i.e., by translating evaluations of skew polynomials into operator evaluations over blocks.

The general definition over arbitrary division rings was introduced in \NoCaseChange{\protect\cite{cite1259}}. Earlier finite-field subfamilies include skew-polynomial constructions and the pasting MDS construction \NoCaseChange{\protect\cite{cite2143,cite2144}}.

\codefieldsection{Protection}
Linearized RS codes satisfy \(d_{\text{SR}}=n-k+1\), so they are MSRD and attain the sum-rank Singleton bound \NoCaseChange{\protect\cite[{Prop. 34, Thm. 4}]{cite1259}}.
\codefieldsection{Decoding}
\begin{eczvaluelist}
\item\relax Berlekamp-Welch-type decoder \NoCaseChange{\protect\cite{cite2144}} and its sum-rank version \NoCaseChange{\protect\cite{cite282}}.
\end{eczvaluelist}
\codefieldsection{Realizations}
\begin{eczvaluelist}
\item\relax Network coding \NoCaseChange{\protect\cite{cite282}}.
\item\relax Code-based cryptography \NoCaseChange{\protect\cite{cite283,cite284}}.
\end{eczvaluelist}
\codefieldsection{Parent}
\begin{eczvaluelist}
\item\relax
\flmRefsHyperref[eczindexfamilyrel]{code:maximum_sum_rank_distance}{Maximum-sum-rank distance (MSRD) code}\end{eczvaluelist}
\codefieldsection{Cousins}
\begin{eczvaluelist}
\item\relax
\flmRefsHyperref[eczindexfamilyrel]{code:reed_solomon}{Reed-Solomon (RS) code} --- Choosing \(\sigma=\operatorname{Id}\) and \(\delta=0\) makes linearized RS codes coincide with conventional RS codes, and the sum-rank metric reduces to the Hamming metric \NoCaseChange{\protect\cite[{Ex. 36}]{cite1259}}.
\item\relax
\flmRefsHyperref[eczindexfamilyrel]{code:gabidulin}{Gabidulin code} --- Choosing one conjugacy class with \(\sigma\neq\operatorname{Id}\) and \(\delta=0\) recovers Gabidulin codes, and the sum-rank metric reduces to the rank metric \NoCaseChange{\protect\cite[{Ex. 37}]{cite1259}}. Over \(\mathbb{F}_{q^m}\) viewed over \(\mathbb{F}_q\), linearized RS codes can have length up to \((q-1)m\), whereas Gabidulin codes over the same extension have maximum length \(m\) \NoCaseChange{\protect\cite[{Sec. 4.2}]{cite1259}}.
\item\relax
\flmRefsHyperref[eczindexfamilyrel]{code:locally_recoverable}{Locally recoverable code (LRC)} --- Linearized RS codes can be used to construct locally recoverable codes \NoCaseChange{\protect\cite{cite994}}.
\end{eczvaluelist}
\eczhbkcontributors{ \eczhuVVA }
\endeczcode

\eczcode{lrpc}{Low-rank parity-check (LRPC) code}{~\NoCaseChange{\protect\cite{cite287}}}
\codefieldsection{Description}
An LRPC code of rank \(d\) is a rank-metric code that, when interpreted as a linear code over \(\mathbb{F}_{q^m}\), admits an \((n-k)\times n\) parity-check matrix whose entries span a subspace of \(\mathbb{F}_{q^m}\) that is at most \(d\)-dimensional.

\codefieldsection{Decoding}
\begin{eczvaluelist}
\item\relax Efficient probabilistic decoder \NoCaseChange{\protect\cite{cite287}}.
\item\relax Mixed decoder \NoCaseChange{\protect\cite{cite290}}.
\end{eczvaluelist}
\codefieldsection{Realizations}
\begin{eczvaluelist}
\item\relax Cryptosystem \NoCaseChange{\protect\cite{cite287}} that is a rank-metric analogue of NTRU \NoCaseChange{\protect\cite{cite288}} and MDPC \NoCaseChange{\protect\cite{cite289}} cryptosystems.
\item\relax Post-quantum cryptography \NoCaseChange{\protect\cite{cite290}}.
\end{eczvaluelist}
\codefieldsection{Parent}
\begin{eczvaluelist}
\item\relax
\flmRefsHyperref[eczindexfamilyrel]{code:rank_metric}{Rank-metric code}\end{eczvaluelist}
\codefieldsection{Cousin}
\begin{eczvaluelist}
\item\relax
\flmRefsHyperref[eczindexfamilyrel]{code:ldpc}{Low-density parity-check (LDPC) code} --- LRPC codes are rank-metric analogues of LDPC codes \NoCaseChange{\protect\cite{cite287}}.
\end{eczvaluelist}
\eczhbkcontributors{ Mazin Karjikar, \eczhuVVA }
\endeczcode

\eczcode{matrices_into_matrices}{Matrix-based code}{}
\codefieldsection{Alternative Names}
\begin{eczvaluelist}
\item\relax Two-dimensional code
\end{eczvaluelist}
\eczhIndexCodeAliasName{matrices_into_matrices}{Two-dimensional code}

\codefieldsection{Kingdom root code}
for the \flmRefsHyperref{kingdom:matrices_into_matrices}{Matrix Kingdom}.
\codefieldsection{Description}
Encodes \(K\) states (codewords) in an \(m\times n\) array of coordinates over a field (e.g., the Galois field \(\mathbb{F}_q\) or the complex numbers \(\mathbb{C}\)).
\codefieldsection{Parents}
\begin{eczvaluelist}
\item\relax
\flmRefsHyperref[eczindexfamilyrel]{code:block}{Block code}\item\relax
\flmRefsHyperref[eczindexfamilyrel]{code:ecc_finite}{Finite-dimensional error-correcting code (ECC)}\item\relax
\flmRefsHyperref[eczindexfamilyrel]{code:group_classical}{Group-alphabet code} --- Matrix-based code alphabets are additive groups.
\end{eczvaluelist}
\codefieldsection{Children}
\begin{eczvaluelist}
\item\relax
\flmRefsHyperref[eczindexfamilyrel]{code:array}{Array code}\item\relax
\flmRefsHyperref[eczindexfamilyrel]{code:matrix_computation}{Distributed computation code}\item\relax
\flmRefsHyperref[eczindexfamilyrel]{code:spacetime}{Spacetime code (STC)}\item\relax
\flmRefsHyperref[eczindexfamilyrel]{code:subspace}{Subspace code} --- Subspace codes are represented by generator matrices of subspaces of \(\mathbb{F}_q^n\).
\item\relax
\flmRefsHyperref[eczindexfamilyrel]{code:sum_rank_metric}{Sum-rank-metric code}\item\relax
\flmRefsHyperref[eczindexfamilyrel]{code:tensor}{Tensor-product code}\item\relax
\flmRefsHyperref[eczindexfamilyrel]{code:unitary}{Unitary code}\item\relax
\flmRefsHyperref[eczindexfamilyrel]{code:q-ary_digits_into_q-ary_digits}{\(q\)-ary code} --- Matrix-based codes over \(\mathbb{F}_q\) whose codewords are vectors reduce to \(q\)-ary codes. Elements of fields such as \(\mathbb{F}_{p^{ml}}\) can be written as \(m\)-dimensional vectors over \(\mathbb{F}_{p^{l}}\) or \((m\times l)\)-dimensional matrices over \(\mathbb{F}_p\). This idea is used to convert between ordinary block codes and matrix-based codes such as disk array codes and rank-metric codes.
\end{eczvaluelist}
\eczhbkcontributors{ \eczhuVVA }
\endeczcode

\eczcode{maximum_rank_distance}{Maximum-rank distance (MRD) code}{~\NoCaseChange{\protect\cite{cite2134,cite2133,cite2135}}}
\codefieldsection{Alternative Names}
\begin{eczvaluelist}
\item\relax Optimal rank-distance code
\end{eczvaluelist}
\eczhIndexCodeAliasName{maximum_rank_distance}{Optimal rank-distance code}
\codefieldsection{Description}
A rank-metric code whose parameters satisfy the rank-metric Singleton-like bound with equality.

An \([n\times m,k,d]_q\) rank-metric code is MRD if its parameters are such that the Singleton-like bound
\flmMathEnvironment{align}{}{
k \leq \max(n, m) (\min(n, m) - d + 1)
}
becomes an equality.

\codefieldsection{Realizations}
\begin{eczvaluelist}
\item\relax Useful for error and erasure correction in network coding \NoCaseChange{\protect\cite{cite292,cite293}}.
\end{eczvaluelist}
\codefieldsection{Parent}
\begin{eczvaluelist}
\item\relax
\flmRefsHyperref[eczindexfamilyrel]{code:rank_metric}{Rank-metric code}\end{eczvaluelist}
\codefieldsection{Cousins}
\begin{eczvaluelist}
\item\relax
\flmRefsHyperref[eczindexfamilyrel]{code:mds}{Maximum distance separable (MDS) code} --- MRD codes are matrix-code analogues of MDS codes.
\item\relax
\flmRefsHyperref[eczindexfamilyrel]{code:reed_solomon}{Reed-Solomon (RS) code} --- MRD rank-metric codes can be thought of as matrix analogues of MDS RS codes as both constructions utilize a Vandermonde matrix \NoCaseChange{\protect\cite{cite292}}.
\item\relax
\flmRefsHyperref[eczindexfamilyrel]{code:maximum_sum_rank_distance}{Maximum-sum-rank distance (MSRD) code} --- MSRD codes generalize MRD codes from the rank metric to the sum-rank metric.
\item\relax
\flmRefsHyperref[eczindexfamilyrel]{code:gabidulin}{Gabidulin code} --- Gabidulin codes over \(\mathbb{F}_{q^N}\) with maximum rank-distance, when expressed as matrices over \(\mathbb{F}_q\), are MRD codes.
\end{eczvaluelist}
\eczhbkcontributors{ Marianna Podzorova, \eczhuVVA }
\endeczcode

\eczcode{maximum_sum_rank_distance}{Maximum-sum-rank distance (MSRD) code}{~\NoCaseChange{\protect\cite{cite1259}}}
\codefieldsection{Alternative Names}
\begin{eczvaluelist}
\item\relax Optimal sum-rank-distance code
\end{eczvaluelist}
\eczhIndexCodeAliasName{maximum_sum_rank_distance}{Optimal sum-rank-distance code}
\codefieldsection{Description}
A sum-rank-metric code whose parameters satisfy the sum-rank-metric Singleton bound with equality.

An \([n\times m,k,d]_q\) code is MSRD if its parameters are such that the sum-rank-metric Singleton bound \NoCaseChange{\protect\cite[{Prop. 34}]{cite1259}},
\flmMathEnvironment{align}{}{
d_{\text{SR}}(C) \leq n - k + 1~,
}
becomes an equality, where \(d_{\text{SR}}\) is the sum-rank metric.

\codefieldsection{Parent}
\begin{eczvaluelist}
\item\relax
\flmRefsHyperref[eczindexfamilyrel]{code:sum_rank_metric}{Sum-rank-metric code}\end{eczvaluelist}
\codefieldsection{Child}
\begin{eczvaluelist}
\item\relax
\flmRefsHyperref[eczindexfamilyrel]{code:linearized_reed_solomon}{Linearized RS code}\end{eczvaluelist}
\codefieldsection{Cousins}
\begin{eczvaluelist}
\item\relax
\flmRefsHyperref[eczindexfamilyrel]{code:mds}{Maximum distance separable (MDS) code} --- MSRD codes generalize MDS codes from the Hamming metric to the sum-rank metric.
\item\relax
\flmRefsHyperref[eczindexfamilyrel]{code:maximum_rank_distance}{Maximum-rank distance (MRD) code} --- MSRD codes generalize MRD codes from the rank metric to the sum-rank metric.
\end{eczvaluelist}
\eczhbkcontributors{ \eczhuVVA }
\endeczcode

\eczcode{mds_array}{MDS array code}{}
\codefieldsection{Description}
An \((n,k,m)\) array code whose codewords can be recovered by any \(k\) out of \(n\) nodes, where each node stores a length-\(m\) column of the codeword. MDS array codes are MDS codes when each matrix codeword is treated as a vector by converting each column into a single coordinate via subpacketization.

\codefieldsection{Parent}
\begin{eczvaluelist}
\item\relax
\flmRefsHyperref[eczindexfamilyrel]{code:array}{Array code}\end{eczvaluelist}
\codefieldsection{Children}
\begin{eczvaluelist}
\item\relax
\flmRefsHyperref[eczindexfamilyrel]{code:evenodd}{EVENODD code} --- EVENODD codes are examples of MDS array codes \NoCaseChange{\protect\cite{cite192}}.
\item\relax
\flmRefsHyperref[eczindexfamilyrel]{code:b_array}{B-code} --- B-codes are examples of MDS array codes \NoCaseChange{\protect\cite{cite192}}.
\item\relax
\flmRefsHyperref[eczindexfamilyrel]{code:rdp}{Row-Diagonal Parity (RDP) code}\item\relax
\flmRefsHyperref[eczindexfamilyrel]{code:star}{Star code}\item\relax
\flmRefsHyperref[eczindexfamilyrel]{code:x_array}{X-code} --- X-codes are examples of MDS array codes \NoCaseChange{\protect\cite{cite192}}.
\item\relax
\flmRefsHyperref[eczindexfamilyrel]{code:ye_barg}{Ye-Barg code}\item\relax
\flmRefsHyperref[eczindexfamilyrel]{code:zigzag}{Zigzag code}\item\relax
\flmRefsHyperref[eczindexfamilyrel]{code:msr}{Minimum-storage regenerating (MSR) code} --- MSR codes are MDS array codes; e.g., see \NoCaseChange{\protect\cite{cite2145}}.
\end{eczvaluelist}
\codefieldsection{Cousins}
\begin{eczvaluelist}
\item\relax
\flmRefsHyperref[eczindexfamilyrel]{code:mds}{Maximum distance separable (MDS) code} --- MDS array codes are MDS codes when each matrix codeword is treated as a vector by converting each column into a single coordinate via subpacketization.
\item\relax
\flmRefsHyperref[eczindexfamilyrel]{code:bbv}{Generalized EVENODD code} --- Generalized EVENODD codes for prime \(m\) and \(r=3\) are MDS array codes \NoCaseChange{\protect\cite{cite189}}.
\end{eczvaluelist}
\eczhbkcontributors{ \eczhuVVA }
\endeczcode

\eczcode{mbr}{Minimum-bandwidth regenerating (MBR) code}{}
\codefieldsection{Description}
A regenerating code that corresponds to the minimum-bandwidth extreme point of the storage-bandwidth trade-off curve, characterized by \(\alpha = d\beta\).

\codefieldsection{Parent}
\begin{eczvaluelist}
\item\relax
\flmRefsHyperref[eczindexfamilyrel]{code:regenerating}{Regenerating code (RGC)} --- MBR codes are minimum-bandwidth extreme points in the storage-bandwidth trade-off curve and are characterized by \(\alpha = d\beta\).
\end{eczvaluelist}
\codefieldsection{Cousin}
\begin{eczvaluelist}
\item\relax
\flmRefsHyperref[eczindexfamilyrel]{code:product_matrix}{Product-matrix (PM) code} --- One of the two PM code constructions yields MBR codes for all \([n,k,d]\).
\end{eczvaluelist}
\eczhbkcontributors{ \eczhuVVA }
\endeczcode

\eczcode{msr}{Minimum-storage regenerating (MSR) code}{}
\codefieldsection{Description}
A regenerating code that corresponds to the minimum-storage extreme point of the storage-bandwidth trade-off curve, characterized by \(\alpha = (d-k+1)\beta\).

\codefieldsection{Parents}
\begin{eczvaluelist}
\item\relax
\flmRefsHyperref[eczindexfamilyrel]{code:regenerating}{Regenerating code (RGC)} --- MSR codes are extreme points in the storage-bandwidth trade-off curve and are characterised by \(\alpha = (d-k+1)\beta\).
\item\relax
\flmRefsHyperref[eczindexfamilyrel]{code:mds_array}{MDS array code} --- MSR codes are MDS array codes; e.g., see \NoCaseChange{\protect\cite{cite2145}}.
\end{eczvaluelist}
\codefieldsection{Child}
\begin{eczvaluelist}
\item\relax
\flmRefsHyperref[eczindexfamilyrel]{code:diagonal}{Diagonal code}\end{eczvaluelist}
\codefieldsection{Cousin}
\begin{eczvaluelist}
\item\relax
\flmRefsHyperref[eczindexfamilyrel]{code:product_matrix}{Product-matrix (PM) code} --- One of the two PM code constructions yields MSR codes for all \([n,k,d \ge 2k-2]\).
\end{eczvaluelist}
\eczhbkcontributors{ \eczhuVVA }
\endeczcode

\eczcode{spacetime_group}{Multi-channel group-orbit code}{~\NoCaseChange{\protect\cite{cite2146}}}
\codefieldsection{Description}
Extension of group-orbit codes to multi-antenna transmission, in which a finite matrix group acts on a reference \(T\times n\) spacetime matrix to generate the codebook.

\codefieldsection{Parents}
\begin{eczvaluelist}
\item\relax
\flmRefsHyperref[eczindexfamilyrel]{code:spacetime_block}{Spacetime block code (STBC)}\item\relax
\flmRefsHyperref[eczindexfamilyrel]{code:group_orbit}{Group-orbit code}\end{eczvaluelist}
\eczhbkcontributors{ \eczhuVVA }
\endeczcode

\eczcode{orth_spacetime_block}{Orthogonal Spacetime Block Code (OSTBC)}{~\NoCaseChange{\protect\cite{cite2095}}}
\codefieldsection{Description}
The codewords are \(T\times n\) matrices as defined for spacetime codes, with the additional condition that columns of the coding matrix are orthogonal or unitary. The parameter \(n\) is the number of transmit channels, and \(T\) is the number of time slots.
\codefieldsection{Protection}
If the matrix \(C-C'\), where \(C\) and \(C'\) are distinct codewords, has minimum rank \(b\), the code has diversity order \(bn_R\) \NoCaseChange{\protect\cite[{Sec. 28.2.1}]{cite2147}}, where \(n_R\) is the number of receivers. The maximum possible diversity order is \(nn_R\).
\codefieldsection{Rate}
The greatest rate which can be achieved is \(\frac{n_0+1}{2n_0}\), where either \(n=2n_0\) or \(n=2n_0-1\) for the orthogonal case \NoCaseChange{\protect\cite{cite2097}}.
\codefieldsection{Decoding}
\begin{eczvaluelist}
\item\relax Maximum-likelihood decoding can be achieved with only linear processing \NoCaseChange{\protect\cite{cite2148}}.
\end{eczvaluelist}
\codefieldsection{Parent}
\begin{eczvaluelist}
\item\relax
\flmRefsHyperref[eczindexfamilyrel]{code:spacetime_block}{Spacetime block code (STBC)} --- STBCs whose codewords are orthogonal or unitary matrices are OSTBCs.
\end{eczvaluelist}
\codefieldsection{Child}
\begin{eczvaluelist}
\item\relax
\flmRefsHyperref[eczindexfamilyrel]{code:alamouti}{Alamouti code}\end{eczvaluelist}
\codefieldsection{Cousin}
\begin{eczvaluelist}
\item\relax
\flmRefsHyperref[eczindexfamilyrel]{code:unitary}{Unitary code} --- Orthogonal spacetime block codes with complex coefficients for \(T=n\) are unitary codes.
\end{eczvaluelist}
\eczhbkcontributors{ Richard Barney, \eczhuVVA }
\endeczcode

\eczcode{product_matrix}{Product-matrix (PM) code}{~\NoCaseChange{\protect\cite{cite2149}}}
\codefieldsection{Description}
Code constructed using two explicit constructions, with each construction corresponding to one of the two extreme points of the storage-bandwidth trade-off curve \NoCaseChange{\protect\cite{cite2150}}.

For the MBR point, the parameters satisfy \(n-1 \ge d \ge k\), \(\alpha=d\beta\), and
\flmMathEnvironment{align}{}{
  M=\left(kd-\binom{k}{2}\right)\beta~.
}
For the MSR point, the parameters satisfy \(d \ge 2k-2\), \(\alpha=(d-k+1)\beta\), and \(M=k\alpha\).

PM codes are the first explicit constructions for all values of the system parameters \([n,k,d]\) at the MBR point, and for all parameters satisfying \([n,k,d \ge 2k-2]\) at the MSR point.
Both constructions are based on a common product-matrix framework, which makes them easy to implement.

\codefieldsection{Parent}
\begin{eczvaluelist}
\item\relax
\flmRefsHyperref[eczindexfamilyrel]{code:regenerating}{Regenerating code (RGC)}\end{eczvaluelist}
\codefieldsection{Cousins}
\begin{eczvaluelist}
\item\relax
\flmRefsHyperref[eczindexfamilyrel]{code:mbr}{Minimum-bandwidth regenerating (MBR) code} --- One of the two PM code constructions yields MBR codes for all \([n,k,d]\).
\item\relax
\flmRefsHyperref[eczindexfamilyrel]{code:msr}{Minimum-storage regenerating (MSR) code} --- One of the two PM code constructions yields MSR codes for all \([n,k,d \ge 2k-2]\).
\end{eczvaluelist}
\eczhbkcontributors{ Adway Patra, \eczhuVVA }
\endeczcode

\eczcode{rank_metric}{Rank-metric code}{~\NoCaseChange{\protect\cite{cite2134}}}
\codefieldsection{Alternative Names}
\begin{eczvaluelist}
\item\relax Delsarte rank-metric code
\end{eczvaluelist}
\eczhIndexCodeAliasName{rank_metric}{Delsarte rank-metric code}
\codefieldsection{Description}
Each codeword is a \textit{matrix} over \(\mathbb{F}_q\), with codewords forming a \(\mathbb{F}_q\)-linear subspace, and with the metric being the rank of the difference of matrices. The distance \(d\) is the minimum rank of all nonzero matrices in the code. Rank-metric codes on \(n\times m\) matrices are denoted as \([n\times m,k,d]_q\).

The number of codewords satisfies \(k \leq \max(n, m) M\), where \(M\) is the maximum rank of all matrices in the code. Codes that achieve this bound with equality are called \textit{Delsarte optimal anticodes}.

\codefieldsection{Protection}
Protects against errors with rank \(\leq \lfloor \frac{d-1}2 \rfloor\).

In particular, this includes criss-cross errors confined to a small number of rows and/or columns whenever the corresponding error matrix has sufficiently small rank \NoCaseChange{\protect\cite[{Sec. 5}]{cite189}}.

The complexity of decoding rank-metric codes is unknown but expected to be harder than that of binary linear codes \NoCaseChange{\protect\cite{cite2151}}.

Linear programming bounds have been derived \NoCaseChange{\protect\cite{cite2134,cite2152,cite2153}}.
See Ref. \NoCaseChange{\protect\cite{cite2137}} for a discussion of \flmRefsHyperref{ref113}{MacWilliams identities}.

\codefieldsection{Decoding}
\begin{eczvaluelist}
\item\relax Polynomial-reconstruction Berlekamp-Welch based decoder \NoCaseChange{\protect\cite{cite2154}}.
\item\relax Berlekamp-Massey based decoder \NoCaseChange{\protect\cite{cite2155}}.
\end{eczvaluelist}
\codefieldsection{Realizations}
\begin{eczvaluelist}
\item\relax Identity-Based Encryption \NoCaseChange{\protect\cite{cite322}}.
\item\relax Digital watermarking \NoCaseChange{\protect\cite{cite323}}.
\item\relax Network coding and streaming media broadcasting \NoCaseChange{\protect\cite{cite324}}.
\end{eczvaluelist}
\codefieldsection{Notes}
\begin{eczvaluelist}
\item\relax See Ref. \NoCaseChange{\protect\cite{cite946,cite2156,cite2157}\protect\cite[{Sec. 5}]{cite189}} for more details.
\end{eczvaluelist}
\codefieldsection{Parents}
\begin{eczvaluelist}
\item\relax
\flmRefsHyperref[eczindexfamilyrel]{code:sum_rank_metric}{Sum-rank-metric code} --- The sum-rank metric generalizes both the Hamming metric and the rank metric \NoCaseChange{\protect\cite{cite1259}}.
\item\relax
\flmRefsHyperref[eczindexfamilyrel]{code:2pt_homogeneous}{Two-point homogeneous-space code} --- Matrices of dimension \(m\times n\) over \(\mathbb{F}_q\) under the rank metric form a finite two-point homogeneous space \NoCaseChange{\protect\cite{cite2134,cite2152,cite2153}\protect\cite[{Table 2}]{cite985}}.
\end{eczvaluelist}
\codefieldsection{Children}
\begin{eczvaluelist}
\item\relax
\flmRefsHyperref[eczindexfamilyrel]{code:gabidulin}{Gabidulin code} --- Gabidulin codes over \(\mathbb{F}_{q^N}\), when expressed as matrices over \(\mathbb{F}_q\), are rank-metric codes (see Def. 14 in Ref. \NoCaseChange{\protect\cite{cite2137}}). The reverse is not always true, i.e., not every rank-metric code is Gabidulin \NoCaseChange{\protect\cite[{Rm. 16}]{cite2137}}.
\item\relax
\flmRefsHyperref[eczindexfamilyrel]{code:lrpc}{Low-rank parity-check (LRPC) code}\item\relax
\flmRefsHyperref[eczindexfamilyrel]{code:maximum_rank_distance}{Maximum-rank distance (MRD) code}\end{eczvaluelist}
\codefieldsection{Cousins}
\begin{eczvaluelist}
\item\relax
\flmRefsHyperref[eczindexfamilyrel]{code:subspace}{Subspace code} --- An \(m \times n\) rank-metric codeword \(A\) can be \textit{lifted} to a subspace codeword \((I | A)\) that generates an \(m\)-dimensional subspace \NoCaseChange{\protect\cite{cite293}\protect\cite[{Def. 14.5.21}]{cite202}}.
\item\relax
\flmRefsHyperref[eczindexfamilyrel]{code:finite_grassmann}{Constant-dimension code} --- Rank-metric codes can be lifted to make constant-dimension codes \NoCaseChange{\protect\cite{cite293,cite347}}; see review \NoCaseChange{\protect\cite{cite2128}}.
\item\relax
\flmRefsHyperref[eczindexfamilyrel]{code:quantum_gabidulin}{Quantum Gabidulin code} --- Quantum Gabidulin code and (classical) rank-metric code distances are based on ranks of the matrix representations of their corresponding errors.
\end{eczvaluelist}
\eczhbkcontributors{ Mazin Karjikar, Thomas Wrona, \eczhuVVA }
\endeczcode

\eczcode{regenerating}{Regenerating code (RGC)}{~\NoCaseChange{\protect\cite{cite2150}}}
\codefieldsection{Description}
An \([n,k,d,\alpha,\beta,M]_q\) Regenerating Code \(\mathcal{C}\) is an erasure-correcting code that
encodes \(M\) symbols from \(\mathbb{F}_q\) into an \(\alpha \times n\) matrix over \(\mathbb{F}_q\), with each column of the matrix
treated as a coordinate of a codeword.

The code has the following properties. Any \(k\) coordinates of the codeword can be used to recover the original \(M\) symbols.
Any erased coordinate, i.e., the \(\alpha\) symbols of that coordinate, can be recovered by contacting any \(d\) other coordinates and collecting \(\beta\) symbols from each one of them, where \(k \le d \le n-1\) and \(\beta \le \alpha\).

The connection of such codes to distributed storage is as follows. A file of size \(M\) symbols is encoded using the code, and the \(\alpha\) symbols of each coordinate are stored in geographically separated storage nodes.
Accessing any \(k\) such nodes gives back the original file.
Node failures are considered erasures in the codeword, and any such failed node can be regenerated by contacting \(d\) surviving nodes and downloading \(\beta\) symbols from each of them.

\codefieldsection{Protection}
Corrects up to \(n-k\) erasures on coordinates.
For standard erasure codes, like RS codes, total download bandwidth for recovery of a single node is \(k\alpha\) due to the MDS property.
For regenerating codes, it is \(d\beta\) which can be significantly less.
It was shown by network coding arguments that \(M \le \sum_{i=0}^{k-1}\min\{\alpha,(d-i)\beta\}\).
Depending on the relative values of \(\alpha\) and \(\beta\), a trade-off between storage per node and download bandwidth arises.
Two extreme points of this trade-off curve are the MBR (\(\alpha = d\beta\)) and MSR (\(\alpha = (d-k+1)\beta\)) codes.

\codefieldsection{Decoding}
\begin{eczvaluelist}
\item\relax If the recovered symbols are exactly equal to the erased symbols, the repair is called an \textit{exact repair}.
\item\relax If the recovered symbols are not exactly equal to the erased symbols but still preserve the code properties, the repair is called a \textit{functional repair}.
\end{eczvaluelist}
\codefieldsection{Parent}
\begin{eczvaluelist}
\item\relax
\flmRefsHyperref[eczindexfamilyrel]{code:array}{Array code}\end{eczvaluelist}
\codefieldsection{Children}
\begin{eczvaluelist}
\item\relax
\flmRefsHyperref[eczindexfamilyrel]{code:determinant}{Determinant code}\item\relax
\flmRefsHyperref[eczindexfamilyrel]{code:mbr}{Minimum-bandwidth regenerating (MBR) code} --- MBR codes are minimum-bandwidth extreme points in the storage-bandwidth trade-off curve and are characterized by \(\alpha = d\beta\).
\item\relax
\flmRefsHyperref[eczindexfamilyrel]{code:msr}{Minimum-storage regenerating (MSR) code} --- MSR codes are extreme points in the storage-bandwidth trade-off curve and are characterised by \(\alpha = (d-k+1)\beta\).
\item\relax
\flmRefsHyperref[eczindexfamilyrel]{code:product_matrix}{Product-matrix (PM) code}\end{eczvaluelist}
\codefieldsection{Cousin}
\begin{eczvaluelist}
\item\relax
\flmRefsHyperref[eczindexfamilyrel]{code:locally_recoverable}{Locally recoverable code (LRC)} --- RGCs and LRCs are related via the group repair with optimal access problem \NoCaseChange{\protect\cite{cite191}}.
\end{eczvaluelist}
\eczhbkcontributors{ Adway Patra, \eczhuVVA }
\endeczcode

\eczcode{rdp}{Row-Diagonal Parity (RDP) code}{~\NoCaseChange{\protect\cite{cite2158}}}
\codefieldsection{Description}
An MDS array code protecting against double erasures.

\codefieldsection{Parent}
\begin{eczvaluelist}
\item\relax
\flmRefsHyperref[eczindexfamilyrel]{code:mds_array}{MDS array code}\end{eczvaluelist}
\eczhbkcontributors{ \eczhuVVA }
\endeczcode

\eczcode{spacetime_block}{Spacetime block code (STBC)}{~\NoCaseChange{\protect\cite{cite2146,cite2159,cite2160,cite2161}}}
\codefieldsection{Description}
In a space-time block code, \(n\) transmitting antennas send symbols to \(m\) receiving antennas over \(T\) time slots.
These symbols can be arranged in a \(n \times T\) matrix where the rows correspond to the transmitting antennas, and the columns correspond to the time slots.
The codewords \(\{X\}\) making up the code are thus \(n \times T\) matrices.

\codefieldsection{Protection}
Provides protection against errors due to thermal noise and destructive interference arising from traversing an environment with scattering, reflection, and/or refraction.

Transmission occurs from \(n\) transmitting antennas to \(m\) receiving antennas over \(T\) time slots.
The typical noise model, a fading channel \NoCaseChange{\protect\cite{cite2162}}, multiplies an incoming codeword by an \(m \times n\) \textit{fading matrix} or \textit{damping matrix} \(H\) and adds random (typically Gaussian) noise using an \(m \times T\) matrix \(Z\),
\flmMathEnvironment{align}{}{
X\to HX+Z = Y~.
}
If \(H\) is known (unknown) to the receiver, then the receiver is called coherent (incoherent).

Decoding corresponds to choosing the codeword \(X\) that, when transformed under the channel, is closest to the corrupted output \(Y\) in the Frobenius norm,
\flmMathEnvironment{align}{}{
  \min_{X\in C}\|Y-HX\|^{2}~.
}

\codefieldsection{Rate}
If a codeword encodes \(k\) information symbols over \(T\) channel uses, then its rate is \(k/T\) symbols per channel use. A spacetime block code is said to be full rate when this equals the number \(n\) of transmit antennas, i.e., when \(k=nT\).
\codefieldsection{Realizations}
\begin{eczvaluelist}
\item\relax High data-rate wireless communication, e.g., WiMAX (IEEE 802.16m) \NoCaseChange{\protect\cite{cite342,cite343,cite344}}.
\end{eczvaluelist}
\codefieldsection{Parent}
\begin{eczvaluelist}
\item\relax
\flmRefsHyperref[eczindexfamilyrel]{code:spacetime}{Spacetime code (STC)} --- Spacetime codes also use spatial and temporal diversity, but do not necessarily use blocks as codewords.
\end{eczvaluelist}
\codefieldsection{Children}
\begin{eczvaluelist}
\item\relax
\flmRefsHyperref[eczindexfamilyrel]{code:linear_spacetime}{Linear STC}\item\relax
\flmRefsHyperref[eczindexfamilyrel]{code:orth_spacetime_block}{Orthogonal Spacetime Block Code (OSTBC)} --- STBCs whose codewords are orthogonal or unitary matrices are OSTBCs.
\item\relax
\flmRefsHyperref[eczindexfamilyrel]{code:spacetime_group}{Multi-channel group-orbit code}\end{eczvaluelist}
\eczhbkcontributors{ Richard Barney, \eczhuVVA }
\endeczcode

\eczcode{spacetime}{Spacetime code (STC)}{~\NoCaseChange{\protect\cite{cite2163}}}
\codefieldsection{Description}
Code designed for wireless transmission over multiple antennas and multiple time slots. A spacetime code uses a modulation scheme to encode a message into signals that are transmitted at different times from different antennas, thereby utilizing both spatial and temporal (i.e., \textit{spacetime}) degrees of freedom.

\codefieldsection{Rate}
Shannon capacity of various multiple-input multiple-output (MIMO) channels has been determined \NoCaseChange{\protect\cite{cite2164,cite2165,cite2166,cite2167}}. MIMO channel capacity when the channel is unknown to the sender and receiver \NoCaseChange{\protect\cite{cite2159,cite2168}} can be interpreted as a problem of placing points on the Grassmannian \NoCaseChange{\protect\cite{cite2169}}.
\codefieldsection{Notes}
\begin{eczvaluelist}
\item\relax See the chapter \NoCaseChange{\protect\cite{cite2170}} or Ref. \NoCaseChange{\protect\cite{cite2147}} for an introduction to spacetime coding.
\end{eczvaluelist}
\codefieldsection{Parent}
\begin{eczvaluelist}
\item\relax
\flmRefsHyperref[eczindexfamilyrel]{code:matrices_into_matrices}{Matrix-based code}\end{eczvaluelist}
\codefieldsection{Child}
\begin{eczvaluelist}
\item\relax
\flmRefsHyperref[eczindexfamilyrel]{code:spacetime_block}{Spacetime block code (STBC)} --- Spacetime codes also use spatial and temporal diversity, but do not necessarily use blocks as codewords.
\end{eczvaluelist}
\codefieldsection{Cousins}
\begin{eczvaluelist}
\item\relax
\flmRefsHyperref[eczindexfamilyrel]{code:grassmannian}{Grassmannian code} --- MIMO channel capacity when the channel is unknown to the sender and receiver \NoCaseChange{\protect\cite{cite2159,cite2168}} can be interpreted as a problem of placing points on the Grassmannian \NoCaseChange{\protect\cite{cite2169}}.
\item\relax
\flmRefsHyperref[eczindexfamilyrel]{code:stiefel}{Stiefel code} --- Stiefel codes can be thought of as spacetime codes \NoCaseChange{\protect\cite{cite2171}}
\item\relax
\flmRefsHyperref[eczindexfamilyrel]{code:hnss}{Hayden-Nezami-Salton-Sanders bosonic code} --- Hayden-Nezami-Salton-Sanders codes have been considered in the context of spacetime replication of quantum data \NoCaseChange{\protect\cite{cite512,cite2172}}, while STCs are designed to replicate classical data.
\item\relax
\flmRefsHyperref[eczindexfamilyrel]{code:cws}{Codeword stabilized (CWS) code} --- CWS codes have been considered in the context of spacetime replication of quantum data \NoCaseChange{\protect\cite{cite512,cite2172}}, while STCs are designed to replicate classical data.
\end{eczvaluelist}
\eczhbkcontributors{ \eczhuVVA }
\endeczcode

\eczcode{star}{Star code}{~\NoCaseChange{\protect\cite{cite2173}}}
\codefieldsection{Description}
An MDS array code protecting against triple erasures.

\codefieldsection{Parent}
\begin{eczvaluelist}
\item\relax
\flmRefsHyperref[eczindexfamilyrel]{code:mds_array}{MDS array code}\end{eczvaluelist}
\eczhbkcontributors{ \eczhuVVA }
\endeczcode

\eczcode{subspace}{Subspace code}{~\NoCaseChange{\protect\cite{cite292,cite2174,cite2175}}}
\codefieldsection{Description}
A code that is a set of subspaces of \(\mathbb{F}_q^n\). Codewords are the subspaces themselves, often represented by generator matrices in reduced-row echelon form, and distance is governed by various notions of subspace overlap.

\codefieldsection{Protection}
Subspace codes are quantified with respect to the subspace distance \NoCaseChange{\protect\cite{cite292}} or injection distance \NoCaseChange{\protect\cite{cite2174}}.

Generalizations of various bounds for ordinary \(q\)-ary codes have been developed for subspace codes; see \NoCaseChange{\protect\cite{cite347}}.

\codefieldsection{Decoding}
\begin{eczvaluelist}
\item\relax List decoding up to the Singleton bound \NoCaseChange{\protect\cite{cite882}}.
\end{eczvaluelist}
\codefieldsection{Realizations}
\begin{eczvaluelist}
\item\relax Packet-based transmission over linear networks \NoCaseChange{\protect\cite{cite292,cite293,cite347}}.
\end{eczvaluelist}
\codefieldsection{Notes}
\begin{eczvaluelist}
\item\relax Reviews of subspace codes \NoCaseChange{\protect\cite{cite2128,cite347}}.
\end{eczvaluelist}
\codefieldsection{Parent}
\begin{eczvaluelist}
\item\relax
\flmRefsHyperref[eczindexfamilyrel]{code:matrices_into_matrices}{Matrix-based code} --- Subspace codes are represented by generator matrices of subspaces of \(\mathbb{F}_q^n\).
\end{eczvaluelist}
\codefieldsection{Child}
\begin{eczvaluelist}
\item\relax
\flmRefsHyperref[eczindexfamilyrel]{code:finite_grassmann}{Constant-dimension code} --- Subspace codes of constant dimension reduce to constant-dimension codes.
\end{eczvaluelist}
\codefieldsection{Cousins}
\begin{eczvaluelist}
\item\relax
\flmRefsHyperref[eczindexfamilyrel]{code:projective}{Projective geometry code} --- Subspace codes are sets of subspaces of a projective space \(PG(n-1,q)\).
\item\relax
\flmRefsHyperref[eczindexfamilyrel]{code:gabidulin}{Gabidulin code} --- Gabidulin codes can be used to construct asymptotically good subspace codes \NoCaseChange{\protect\cite{cite167,cite292}}.
\item\relax
\flmRefsHyperref[eczindexfamilyrel]{code:rank_metric}{Rank-metric code} --- An \(m \times n\) rank-metric codeword \(A\) can be \textit{lifted} to a subspace codeword \((I | A)\) that generates an \(m\)-dimensional subspace \NoCaseChange{\protect\cite{cite293}\protect\cite[{Def. 14.5.21}]{cite202}}.
\item\relax
\flmRefsHyperref[eczindexfamilyrel]{code:poset}{Poset code} --- Poset-code and subspace-code distance metric families intersect only at the Hamming metric \NoCaseChange{\protect\cite{cite1010}}.
\end{eczvaluelist}
\eczhbkcontributors{ \eczhuVVA }
\endeczcode

\eczcode{subspace_design}{Subspace design}{~\NoCaseChange{\protect\cite{cite2176,cite881,cite986}}}
\codefieldsection{Alternative Names}
\begin{eczvaluelist}
\item\relax \(q\)-design
\item\relax Geometric design
\item\relax Design over finite fields
\item\relax Design over \(\mathbb{F}_q\)
\item\relax Design in the \(q\)-Johnson scheme
\item\relax Design in vector spaces
\end{eczvaluelist}
\eczhIndexCodeAliasName{subspace_design}{\(q\)-design}
\eczhIndexCodeAliasName{subspace_design}{Geometric design}
\eczhIndexCodeAliasName{subspace_design}{Design over finite fields}
\eczhIndexCodeAliasName{subspace_design}{Design over \(\mathbb{F}_q\)}
\eczhIndexCodeAliasName{subspace_design}{Design in the \(q\)-Johnson scheme}
\eczhIndexCodeAliasName{subspace_design}{Design in vector spaces}
\codefieldsection{Description}
A constant-dimension code that forms a design on the finite Grassmannian.
Subspace designs exist for all parameters in sufficiently large dimension that also satisfies divisibility constraints \NoCaseChange{\protect\cite{cite906,cite907}}.

An alternative but related definition is given in Refs. \NoCaseChange{\protect\cite{cite2177,cite882}}.

\codefieldsection{Notes}
\begin{eczvaluelist}
\item\relax See \NoCaseChange{\protect\cite{cite2178}} for a review and history of subspace designs.
\item\relax Popular summary of the existence of subspace designs in \flmHref{https://www.quantamagazine.org/mathematicians-find-hidden-structure-in-a-common-type-of-space-20230412}{Quanta Magazine}.
\end{eczvaluelist}
\codefieldsection{Parents}
\begin{eczvaluelist}
\item\relax
\flmRefsHyperref[eczindexfamilyrel]{code:finite_grassmann}{Constant-dimension code}\item\relax
\flmRefsHyperref[eczindexfamilyrel]{code:t-designs}{\(t\)-design} --- Subspace designs are designs on the finite-field Grassmannian (a.k.a. \(q\)-Johnson space or \(q\)-Johnson association scheme) \NoCaseChange{\protect\cite{cite912}\protect\cite[{Sec. 8.6}]{cite913}}.
\end{eczvaluelist}
\codefieldsection{Cousin}
\begin{eczvaluelist}
\item\relax
\flmRefsHyperref[eczindexfamilyrel]{code:combinatorial_design}{Combinatorial design} --- Combinatorial designs are designs in Johnson space, the space of all size-\(w\) subsets of a set with \(n\) elements. The \(q\)-Johnson spaces generalize this notion to subspaces and reduce to Johnson spaces at \(q=1\). In other words, combinatorial designs are designs over spaces of subsets, while subspace designs are designs over spaces of subspaces.
\end{eczvaluelist}
\eczhbkcontributors{ \eczhuVVA }
\endeczcode

\eczcode{sum_rank_metric}{Sum-rank-metric code}{~\NoCaseChange{\protect\cite{cite2179}}}
\codefieldsection{Description}
A code whose performance is evaluated in the sum-rank metric, which is a metric that generalizes both the Hamming metric and the rank metric.

\codefieldsection{Protection}
Linear programming bounds have been derived \NoCaseChange{\protect\cite{cite2180}}.

\codefieldsection{Parents}
\begin{eczvaluelist}
\item\relax
\flmRefsHyperref[eczindexfamilyrel]{code:matrices_into_matrices}{Matrix-based code}\item\relax
\flmRefsHyperref[eczindexfamilyrel]{code:distributed_storage}{Distributed-storage code} --- Sum-rank-metric codes are useful for distributed storage \NoCaseChange{\protect\cite{cite994}}.
\end{eczvaluelist}
\codefieldsection{Children}
\begin{eczvaluelist}
\item\relax
\flmRefsHyperref[eczindexfamilyrel]{code:maximum_sum_rank_distance}{Maximum-sum-rank distance (MSRD) code}\item\relax
\flmRefsHyperref[eczindexfamilyrel]{code:rank_metric}{Rank-metric code} --- The sum-rank metric generalizes both the Hamming metric and the rank metric \NoCaseChange{\protect\cite{cite1259}}.
\end{eczvaluelist}
\codefieldsection{Cousin}
\begin{eczvaluelist}
\item\relax
\flmRefsHyperref[eczindexfamilyrel]{code:bits_into_bits}{Binary code} --- The sum-rank metric generalizes both the Hamming metric and the rank metric \NoCaseChange{\protect\cite{cite1259}}.
\end{eczvaluelist}
\eczhbkcontributors{ \eczhuVVA }
\endeczcode

\eczcode{tensor}{Tensor-product code}{~\NoCaseChange{\protect\cite{cite2181,cite2182,cite966,cite2183}}}
\codefieldsection{Alternative Names}
\begin{eczvaluelist}
\item\relax Tensor code
\item\relax Kroneckerian code
\item\relax Product code
\end{eczvaluelist}
\eczhIndexCodeAliasName{tensor}{Tensor code}
\eczhIndexCodeAliasName{tensor}{Kroneckerian code}
\eczhIndexCodeAliasName{tensor}{Product code}
\codefieldsection{Description}
A matrix-based code \(C_A \otimes C_B\) constructed out of two linear binary or \(q\)-ary codes \(C_A,C_B\) in an outer-product construction. 
Its dual is sometimes called the \textit{check-product} code, denoted as \(C_{A}\boxplus C_{B}\).

Codewords are those matrices whose column vectors are in \(C_A=[n_A,k_A,d_A]\) and whose row vectors are in \(C_B=[n_B,k_B,d_B]\).
Codewords \(c\) of a tensor code satisfy the parity check equation \(H_A c H^{\text{T}}_B = 0\).

A check-product code forms the matrix subspace dual to its corresponding tensor-product code,
\flmMathEnvironment{align}{}{
  C_{A}\boxplus C_{B}=C_A \otimes \mathbb{F}_q^{n_B} + \mathbb{F}_q^{n_A} \otimes C_B~.\label{ref2184}
}
The parity-check matrix of this code is \(H_A \otimes H_B\), where \(H_{A,B}\) is the parity-check matrix of \(C_{A,B}\) \NoCaseChange{\protect\cite[{Lemma 3.3}]{cite2185}}.

\codefieldsection{Protection}
For linear codes \(C_A=[n_A,k_A,d_A]\) and \(C_B=[n_B,k_B,d_B]\), the resulting tensor code is \(C_A \otimes C_B=[n_A n_B,k_A k_B,d_A d_B]\).
Tensor codes can be useful for protecting against burst errors \NoCaseChange{\protect\cite{cite2186,cite2187}}.

Many (but not all \NoCaseChange{\protect\cite{cite2188}}) tensor codes are \textit{robustly testable} \NoCaseChange{\protect\cite{cite73,cite1483,cite1484}}, a property useful for constructing LTCs \NoCaseChange{\protect\cite{cite1331}}, including a family of \(c^3\)-LTCs \NoCaseChange{\protect\cite{cite88}}.
A property equivalent to robust testability is \(\kappa\)\textit{-product expansion} \NoCaseChange{\protect\cite{cite184}}.
For check-product codes, this property means that for every codeword \(c_1 + c_2 \in C_{A}\boxplus C_{B}\), split up according to \eqref{ref2184},
\flmMathEnvironment{align}{}{
  \kappa\left(\frac{\|c_{1}\|_{A}}{n_{A}}+\frac{\|c_{2}\|_{B}}{n_{B}}\right)\leq\frac{|c_{1}+c_{2}|}{n_{A}n_{B}}~,
}
where \(\|c_{1}\|_{A}\) (\(\|c_{2}\|_{B}\)) is the number of nonzero columns (rows) in \(c_1\) (\(c_2\)).

Check-product codes formed by two random linear codes are robustly testable \NoCaseChange{\protect\cite[{Thm. 1}]{cite2189}}, a property useful for constructing asymptotically good QLDPC codes \NoCaseChange{\protect\cite{cite2190,cite2189}} and proving distance bounds \NoCaseChange{\protect\cite{cite2191}}.

\codefieldsection{Rate}
Rate of the tensor-product code \(C_A \otimes C_B\) is a product of the rates of the codes \(C_A\) and \(C_B\).
\codefieldsection{Decoding}
\begin{eczvaluelist}
\item\relax The simple decoding algorithm (first decode all columns with \(C_A\), then all rows with \(C_B\)) corrects up to \((d_A d_B-1)/4 \) errors.
\item\relax Algorithms such as generalized minimum-distance decoding \NoCaseChange{\protect\cite{cite969}} or the min-sum algorithm can decode all errors of weight up to \((d_A d_B-1)/2\). Error location may be coupled with Viterbi decoding for every faulty sub-block \NoCaseChange{\protect\cite{cite348}}.
\end{eczvaluelist}
\codefieldsection{Realizations}
\begin{eczvaluelist}
\item\relax Construction can be used in magnetic recording by taking the tensor product of an RS code and a parity-check code \NoCaseChange{\protect\cite{cite348}}.
\end{eczvaluelist}
\codefieldsection{Notes}
\begin{eczvaluelist}
\item\relax See Refs. \NoCaseChange{\protect\cite{cite2192,cite993}\protect\cite[{Ch. 18}]{cite41}} for expositions.
\end{eczvaluelist}
\codefieldsection{Parent}
\begin{eczvaluelist}
\item\relax
\flmRefsHyperref[eczindexfamilyrel]{code:matrices_into_matrices}{Matrix-based code}\end{eczvaluelist}
\codefieldsection{Cousins}
\begin{eczvaluelist}
\item\relax
\flmRefsHyperref[eczindexfamilyrel]{code:parallel_concatenated}{Parallel concatenated code} --- Tensor-product codes can be viewed both as serial or parallel concatenated codes \NoCaseChange{\protect\cite{cite972}}.
\item\relax
\flmRefsHyperref[eczindexfamilyrel]{code:concatenated}{Concatenated code} --- Tensor-product codes can be viewed both as serial or parallel concatenated codes \NoCaseChange{\protect\cite{cite972}}.
\item\relax
\flmRefsHyperref[eczindexfamilyrel]{code:lr-cayley-complex}{Left-right Cayley complex code} --- Left-right Cayley complex codewords for a fixed graph vertex are codewords of a tensor code.
\item\relax
\flmRefsHyperref[eczindexfamilyrel]{code:ldpc}{Low-density parity-check (LDPC) code} --- Tensor products of random LDPC codes are robustly testable \NoCaseChange{\protect\cite{cite1483,cite1484}}.
\item\relax
\flmRefsHyperref[eczindexfamilyrel]{code:array}{Array code} --- Classical block array codes form a subclass of product codes, i.e., tensor-product codes \NoCaseChange{\protect\cite{cite189}}.
\item\relax
\flmRefsHyperref[eczindexfamilyrel]{code:reed_solomon}{Reed-Solomon (RS) code} --- Tensor codes constructed from RS codes are robustly testable \NoCaseChange{\protect\cite{cite2013}}.
\item\relax
\flmRefsHyperref[eczindexfamilyrel]{code:parity_check_tensor}{Parity-check tensor-product code} --- Tensor-product codewords (parity-check tensor-product parity-check matrices) are constructed via an outer product of the underlying codes (parity-check matrices).
\item\relax
\flmRefsHyperref[eczindexfamilyrel]{code:dhlv}{Dinur-Hsieh-Lin-Vidick (DHLV) code} --- Tensor codes are used in constructing quantum DHLV codes.
\item\relax
\flmRefsHyperref[eczindexfamilyrel]{code:check_product}{Quantum check-product code} --- Quantum check-product codes extend the concept of a check product, which yields the dual of a tensor code, to a product between a classical and a quantum code.
\item\relax
\flmRefsHyperref[eczindexfamilyrel]{code:classical_product}{Classical-product code} --- Tensor-product codes are utilized in classical-product code constructions.
\item\relax
\flmRefsHyperref[eczindexfamilyrel]{code:quantum_tanner}{Quantum Tanner code} --- Tensor codes are used in constructing quantum Tanner codes.
\item\relax
\flmRefsHyperref[eczindexfamilyrel]{code:quantum_tensor_product}{Quantum tensor-product code} --- Quantum tensor-product codes are quantum analogues of tensor-product codes.
\end{eczvaluelist}
\eczhbkcontributors{ Shashank Sule, \eczhuVVA }
\endeczcode

\eczcode{unitary_design}{Unitary \(t\)-design}{~\NoCaseChange{\protect\cite{cite887,cite888,cite889}}}
\codefieldsection{Description}
A unitary \(t\)-design is a subset of the unitary group that reproduces Haar averages of polynomials over the group up to degree \(t\) \NoCaseChange{\protect\cite[{Def. 1}]{cite889}}.
The design conditions are defined using the \(t\)th tensor product of the group's adjoint representation.

\codefieldsection{Protection}
Bounds on design size have been computed \NoCaseChange{\protect\cite{cite2193}}.

\codefieldsection{Notes}
\begin{eczvaluelist}
\item\relax Introductory material on unitary designs \NoCaseChange{\protect\cite{cite2194}}.
\end{eczvaluelist}
\codefieldsection{Parents}
\begin{eczvaluelist}
\item\relax
\flmRefsHyperref[eczindexfamilyrel]{code:unitary}{Unitary code} --- Unitary \(t\)-designs are designs on the unitary group \(U(N)\). Random \(n\)-qubit unitary circuits form approximate unitary designs on \(U(2^n)\) at a depth logarithmic in \(n\) \NoCaseChange{\protect\cite{cite890}}.
\item\relax
\flmRefsHyperref[eczindexfamilyrel]{code:t-designs}{\(t\)-design} --- Unitary \(t\)-designs are designs on the unitary group \(U(N)\).
\end{eczvaluelist}
\codefieldsection{Child}
\begin{eczvaluelist}
\item\relax
\flmRefsHyperref[eczindexfamilyrel]{code:clifford_group}{Clifford group} --- Stabilizer states on \(n\) qubits form 3-designs on complex projective spaces \(\mathbb{C}P^{2^n}\) \NoCaseChange{\protect\cite{cite937}}. The \flmRefsHyperref{ref409}{Clifford group} is a unitary 2-design \NoCaseChange{\protect\cite{cite938}} and a 3-design \NoCaseChange{\protect\cite{cite940,cite941}\protect\cite[{Thm. 1.6(B)}]{cite939}\protect\cite[{pg. 191}]{cite42}} on \(U(2^n)\). The \(\llbracket 2m,2m-2,2\rrbracket \) code when \(2m\) is a multiple of four obstructs the Clifford group from being a 4-design \NoCaseChange{\protect\cite{cite801}}.
\end{eczvaluelist}
\codefieldsection{Cousins}
\begin{eczvaluelist}
\item\relax
\flmRefsHyperref[eczindexfamilyrel]{code:kerdock}{Kerdock code} --- Kerdock codes correspond to cluster states, and the corresponding Clifford-group automorphisms of this set form a particular group \NoCaseChange{\protect\cite{cite934}} that is a unitary 2-design on \(U(2^n)\) \NoCaseChange{\protect\cite{cite935}}. As such, cluster states form complex projective on 2-designs \(\mathbb{C}P^{2^n}\). These are useful in matrix-vector multiplication \NoCaseChange{\protect\cite{cite936}}.
\item\relax
\flmRefsHyperref[eczindexfamilyrel]{code:hamiltonian}{Hamiltonian-based code} --- Evolving with a Hamiltonian that has constant locality does not yield a unitary 2-design, but increasing the locality slightly overcomes this and yields a design \NoCaseChange{\protect\cite{cite2195}}.
\item\relax
\flmRefsHyperref[eczindexfamilyrel]{code:local_haar_random}{Local Haar-random circuit qubit code} --- Local Haar-random circuits of polynomial depth form approximate unitary designs \NoCaseChange{\protect\cite{cite2196}}.
\item\relax
\flmRefsHyperref[eczindexfamilyrel]{code:qubit_stabilizer}{Qubit stabilizer code} --- Stabilizer states on \(n\) qubits form complex projective 3-designs, but not 4-designs, on \(\mathbb{C}P^{2^n-1}\) \NoCaseChange{\protect\cite{cite937}}. The \flmRefsHyperref{ref409}{Clifford group} is a unitary 2-design \NoCaseChange{\protect\cite{cite938}} and a 3-design \NoCaseChange{\protect\cite{cite940,cite941}\protect\cite[{Thm. 1.6(B)}]{cite939}\protect\cite[{pg. 191}]{cite42}} on \(U(2^n)\). The \(\llbracket 2m,2m-2,2\rrbracket \) code for \(2m\) being a multiple of four obstructs the Clifford group from being a 4-design \NoCaseChange{\protect\cite{cite801}}.
\item\relax
\flmRefsHyperref[eczindexfamilyrel]{code:surface}{Kitaev surface code} --- Unitary \(t\)-designs can be generated via coherent errors, syndrome extraction, and correction \NoCaseChange{\protect\cite{cite2197}}.
\item\relax
\flmRefsHyperref[eczindexfamilyrel]{code:qudits_into_qudits}{Modular-qudit code} --- The \flmRefsHyperref{ref2198}{prime-qudit Pauli group} is a unitary 1-design.
\item\relax
\flmRefsHyperref[eczindexfamilyrel]{code:qudit_stabilizer}{Modular-qudit stabilizer code} --- The prime-qudit Clifford group is a unitary 2-design on \(U(p^n)\) \NoCaseChange{\protect\cite{cite942}}.
\item\relax
\flmRefsHyperref[eczindexfamilyrel]{code:t_group}{Twisted \(1\)-group code} --- Twisted unitary \(t\)-groups \NoCaseChange{\protect\cite{cite2199}} generalize the idea of unitary \(t\)-groups \NoCaseChange{\protect\cite{cite939,cite2193,cite2200}}, which are subgroups of the unitary group that form unitary \(t\)-designs.
\end{eczvaluelist}
\eczhbkcontributors{ \eczhuVVA }
\endeczcode

\eczcode{unitary}{Unitary code}{~\NoCaseChange{\protect\cite{cite2193}}}
\codefieldsection{Description}
Encodes \(K\) states (codewords) into the unitary group \(U(N)\), with codewords represented by matrices.

\codefieldsection{Protection}
LP bounds on the unitary group have been established \NoCaseChange{\protect\cite{cite2201,cite2193}}.

\codefieldsection{Parents}
\begin{eczvaluelist}
\item\relax
\flmRefsHyperref[eczindexfamilyrel]{code:matrices_into_matrices}{Matrix-based code}\item\relax
\flmRefsHyperref[eczindexfamilyrel]{code:symmetric_space}{Symmetric-space code} --- The unitary group is a compact symmetric space \(G/H\) with \(G=U(N)\times U(N)\) and \(H = U(N)\) \NoCaseChange{\protect\cite[{Table 3}]{cite985}}.
\end{eczvaluelist}
\codefieldsection{Children}
\begin{eczvaluelist}
\item\relax
\flmRefsHyperref[eczindexfamilyrel]{code:alamouti}{Alamouti code} --- Codewords of the Alamouti code are two-dimensional unitary matrices.
\item\relax
\flmRefsHyperref[eczindexfamilyrel]{code:unitary_design}{Unitary \(t\)-design} --- Unitary \(t\)-designs are designs on the unitary group \(U(N)\). Random \(n\)-qubit unitary circuits form approximate unitary designs on \(U(2^n)\) at a depth logarithmic in \(n\) \NoCaseChange{\protect\cite{cite890}}.
\end{eczvaluelist}
\codefieldsection{Cousin}
\begin{eczvaluelist}
\item\relax
\flmRefsHyperref[eczindexfamilyrel]{code:orth_spacetime_block}{Orthogonal Spacetime Block Code (OSTBC)} --- Orthogonal spacetime block codes with complex coefficients for \(T=n\) are unitary codes.
\end{eczvaluelist}
\eczhbkcontributors{ \eczhuVVA }
\endeczcode

\eczcode{x_array}{X-code}{~\NoCaseChange{\protect\cite{cite2202}}}
\codefieldsection{Description}
An MDS array code with a simple geometrical construction that achieves optimal encoding and update complexity.

\codefieldsection{Parent}
\begin{eczvaluelist}
\item\relax
\flmRefsHyperref[eczindexfamilyrel]{code:mds_array}{MDS array code} --- X-codes are examples of MDS array codes \NoCaseChange{\protect\cite{cite192}}.
\end{eczvaluelist}
\eczhbkcontributors{ \eczhuVVA }
\endeczcode

\eczcode{ye_barg}{Ye-Barg code}{~\NoCaseChange{\protect\cite{cite190,cite191}}}
\codefieldsection{Description}
A member of an explicit family of MDS array codes with the optimal access property.
The constructions of Ye and Barg achieve optimal repair bandwidth for single-node repair and include optimal-access variants with nearly optimal sub-packetization \NoCaseChange{\protect\cite{cite190,cite191}}.

\codefieldsection{Parent}
\begin{eczvaluelist}
\item\relax
\flmRefsHyperref[eczindexfamilyrel]{code:mds_array}{MDS array code}\end{eczvaluelist}
\eczhbkcontributors{ \eczhuVVA }
\endeczcode

\eczcode{zigzag}{Zigzag code}{~\NoCaseChange{\protect\cite{cite192}}}
\codefieldsection{Description}
An MDS array code correcting two erasures with optimal rebuilding ratio; see Ref. \NoCaseChange{\protect\cite{cite192}} for definitions.

\codefieldsection{Parent}
\begin{eczvaluelist}
\item\relax
\flmRefsHyperref[eczindexfamilyrel]{code:mds_array}{MDS array code}\end{eczvaluelist}
\eczhbkcontributors{ \eczhuVVA }
\endeczcode

\onecolumngrid
\clearpage

\section{Analog Kingdom}

\begin{eczEpigraph}
\begin{quote}
\flmQuoteSetup{quote}%
So one of the fundamental problems in communication theory is determining the densest packing of balls in high-dimensional spaces. This geometrical way of representing signals, at the heart of Shannon's mathematical theory of communication, underlies the high-speed modems that we now take for granted. One of the most common coding schemes in use today works so well because the signals are represented as points in eight-dimensional space.
\flmQuoteAttributed{Neil J. A. Sloane, 1998}
\end{quote}
\end{eczEpigraph}

\twocolumngrid

\eczcode{lambda16}{\(\Lambda_{16}\) Barnes-Wall lattice}{~\NoCaseChange{\protect\cite{cite2203}}}
\eczhIndexCodeAliasName{lambda16}{Barnes-Wall lattice}
\codefieldsection{Description}
BW lattice in dimension \(16\).

\codefieldsection{Protection}
Exhibits the densest known lattice packing in 16 dimensions.

\codefieldsection{Parent}
\begin{eczvaluelist}
\item\relax
\flmRefsHyperref[eczindexfamilyrel]{code:barnes_wall}{Barnes-Wall (BW) lattice}\end{eczvaluelist}
\codefieldsection{Cousins}
\begin{eczvaluelist}
\item\relax
\flmRefsHyperref[eczindexfamilyrel]{code:leech}{\(\Lambda_{24}\) Leech lattice} --- The \(\Lambda_{16}\) Barnes-Wall lattice can be obtained from the Leech lattice by restricting to the fixed-point 16-dimensional subspace of an involution \NoCaseChange{\protect\cite[{Ch. 4, pg. 131}]{cite39}}.
\item\relax
\flmRefsHyperref[eczindexfamilyrel]{code:biorthogonal}{\([2^m,m+1,2^{m-1}]\) First-order RM code} --- Applying Construction B to the first-order RM\((1,4)\) code yields the \(\Lambda_{16}\) Barnes-Wall lattice \NoCaseChange{\protect\cite[{Ch. 4, pg. 130}]{cite39}\protect\cite[{Exam. 10.7.2}]{cite115}}.
\item\relax
\flmRefsHyperref[eczindexfamilyrel]{code:q-ary_repetition}{\(q\)-ary repetition code} --- The \(\Lambda_{16}\) Barnes-Wall lattice can be obtained from the \([4,1,4]\) repetition code over \(\mathbb{F}_9\) via Quebbemann's construction \NoCaseChange{\protect\cite[{Ch. 8, pg. 219}]{cite39}}.
\item\relax
\flmRefsHyperref[eczindexfamilyrel]{code:lambda16_shell}{\(\Lambda_{16}\) lattice-shell code} --- The \(\Lambda_{16}\) lattice-shell code is obtained from a shell of the \(\Lambda_{16}\) lattice.
\end{eczvaluelist}
\eczhbkcontributors{ \eczhuVVA }
\endeczcode

\eczcode{leech}{\(\Lambda_{24}\) Leech lattice}{~\NoCaseChange{\protect\cite{cite2204}}}
\eczhIndexCodeAliasName{leech}{Leech lattice}
\codefieldsection{Description}
Even unimodular lattice in 24 dimensions that exhibits optimal packing.
Its automorphism group is the Conway group Co\(_0\).

A generator matrix for the symplectic version \NoCaseChange{\protect\cite[{Appx. 2}]{cite2205}} is
\flmMathEnvironment{align}{}{
  \left(\begin{smallmatrix}
  24 & -10 & -10 & -10 & -10 & -10 & -10 & -10 & -10 & -10 & -10 & -10 & 3 & 6 & 6 & 6 & 6 & 6 & 6 & 6 & 6 & 6 & 6 & 6\\
  -10 & 6 & 4 & 4 & 4 & 4 & 4 & 4 & 4 & 4 & 4 & 4 & -1 & -2 & -2 & -3 & -2 & -2 & -2 & -3 & -3 & -3 & -2 & -3\\
  -10 & 4 & 6 & 4 & 4 & 4 & 4 & 4 & 4 & 4 & 4 & 4 & -1 & -2 & -3 & -2 & -2 & -2 & -3 & -3 & -3 & -2 & -3 & -2\\
  -10 & 4 & 4 & 6 & 4 & 4 & 4 & 4 & 4 & 4 & 4 & 4 & -1 & -3 & -2 & -2 & -2 & -3 & -3 & -3 & -2 & -3 & -2 & -2\\
  -10 & 4 & 4 & 4 & 6 & 4 & 4 & 4 & 4 & 4 & 4 & 4 & -1 & -2 & -2 & -2 & -3 & -3 & -3 & -2 & -3 & -2 & -2 & -3\\
  -10 & 4 & 4 & 4 & 4 & 6 & 4 & 4 & 4 & 4 & 4 & 4 & -1 & -2 & -2 & -3 & -3 & -3 & -2 & -3 & -2 & -2 & -3 & -2\\
  -10 & 4 & 4 & 4 & 4 & 4 & 6 & 4 & 4 & 4 & 4 & 4 & -1 & -2 & -3 & -3 & -3 & -2 & -3 & -2 & -2 & -3 & -2 & -2\\
  -10 & 4 & 4 & 4 & 4 & 4 & 4 & 6 & 4 & 4 & 4 & 4 & -1 & -3 & -3 & -3 & -2 & -3 & -2 & -2 & -3 & -2 & -2 & -2\\
  -10 & 4 & 4 & 4 & 4 & 4 & 4 & 4 & 6 & 4 & 4 & 4 & -1 & -3 & -3 & -2 & -3 & -2 & -2 & -3 & -2 & -2 & -2 & -3\\
  -10 & 4 & 4 & 4 & 4 & 4 & 4 & 4 & 4 & 6 & 4 & 4 & -1 & -3 & -2 & -3 & -2 & -2 & -3 & -2 & -2 & -2 & -3 & -3\\
  -10 & 4 & 4 & 4 & 4 & 4 & 4 & 4 & 4 & 4 & 6 & 4 & -1 & -2 & -3 & -2 & -2 & -3 & -2 & -2 & -2 & -3 & -3 & -3\\
  -10 & 4 & 4 & 4 & 4 & 4 & 4 & 4 & 4 & 4 & 4 & 6 & -1 & -3 & -2 & -2 & -3 & -2 & -2 & -2 & -3 & -3 & -3 & -2\\
  3 & -1 & -1 & -1 & -1 & -1 & -1 & -1 & -1 & -1 & -1 & -1 & 16 & 4 & 4 & 4 & 4 & 4 & 4 & 4 & 4 & 4 & 4 & 4\\
  6 & -2 & -2 & -3 & -2 & -2 & -2 & -3 & -3 & -3 & -2 & -3 & 4 & 4 & 2 & 2 & 2 & 2 & 2 & 2 & 2 & 2 & 2 & 2\\
  6 & -2 & -3 & -2 & -2 & -2 & -3 & -3 & -3 & -2 & -3 & -2 & 4 & 2 & 4 & 2 & 2 & 2 & 2 & 2 & 2 & 2 & 2 & 2\\
  6 & -3 & -2 & -2 & -2 & -3 & -3 & -3 & -2 & -3 & -2 & -2 & 4 & 2 & 2 & 4 & 2 & 2 & 2 & 2 & 2 & 2 & 2 & 2\\
  6 & -2 & -2 & -2 & -3 & -3 & -3 & -2 & -3 & -2 & -2 & -3 & 4 & 2 & 2 & 2 & 4 & 2 & 2 & 2 & 2 & 2 & 2 & 2\\
  6 & -2 & -2 & -3 & -3 & -3 & -2 & -3 & -2 & -2 & -3 & -2 & 4 & 2 & 2 & 2 & 2 & 4 & 2 & 2 & 2 & 2 & 2 & 2\\
  6 & -2 & -3 & -3 & -3 & -2 & -3 & -2 & -2 & -3 & -2 & -2 & 4 & 2 & 2 & 2 & 2 & 2 & 4 & 2 & 2 & 2 & 2 & 2\\
  6 & -3 & -3 & -3 & -2 & -3 & -2 & -2 & -3 & -2 & -2 & -2 & 4 & 2 & 2 & 2 & 2 & 2 & 2 & 4 & 2 & 2 & 2 & 2\\
  6 & -3 & -3 & -2 & -3 & -2 & -2 & -3 & -2 & -2 & -2 & -3 & 4 & 2 & 2 & 2 & 2 & 2 & 2 & 2 & 4 & 2 & 2 & 2\\
  6 & -3 & -2 & -3 & -2 & -2 & -3 & -2 & -2 & -2 & -3 & -3 & 4 & 2 & 2 & 2 & 2 & 2 & 2 & 2 & 2 & 4 & 2 & 2\\
  6 & -2 & -3 & -2 & -2 & -3 & -2 & -2 & -2 & -3 & -3 & -3 & 4 & 2 & 2 & 2 & 2 & 2 & 2 & 2 & 2 & 2 & 4 & 2\\
  6 & -3 & -2 & -2 & -3 & -2 & -2 & -2 & -3 & -3 & -3 & -2 & 4 & 2 & 2 & 2 & 2 & 2 & 2 & 2 & 2 & 2 & 2 & 4
  \end{smallmatrix}\right).
}

\codefieldsection{Protection}
The Leech lattice has a nominal coding gain of \(4\). It exhibits the densest packing \NoCaseChange{\protect\cite{cite2206}} and highest kissing number of 196560 in 24 dimensions.

\codefieldsection{Notes}
\begin{eczvaluelist}
\item\relax Popular summary of solution to the sphere-packing problem in \flmHref{https://www.quantamagazine.org/sphere-packing-solved-in-higher-dimensions-20160330/}{Quanta Magazine}.
\end{eczvaluelist}
\codefieldsection{Parents}
\begin{eczvaluelist}
\item\relax
\flmRefsHyperref[eczindexfamilyrel]{code:niemeier}{Niemeier lattice} --- The Leech lattice is the Niemeier lattice with minimal norm 4 \NoCaseChange{\protect\cite{cite2037}}. Every Niemeier lattice is a sublattice of the Leech lattice \NoCaseChange{\protect\cite{cite2207,cite2037}}. In the holy construction for the Niemeier lattice \(A_2^{12}\), the combinations for which the sum of all coefficients is zero form a copy of the Leech lattice \NoCaseChange{\protect\cite[{Ch. 24, pg. 510}]{cite39}}. The Leech lattice can be constructed from pseudo Golay codes via \flmTerm{term}{ref114}{}{Construction \(A_4\)} \NoCaseChange{\protect\cite{cite122,cite1198}}. The Leech lattice can be constructed from the extended quaternary Golay code via \flmTerm{term}{ref114}{}{Construction \(A_4\)} \NoCaseChange{\protect\cite[{3rd Ed., pg. xxxiii}]{cite39}} (see also \NoCaseChange{\protect\cite{cite1659,cite112,cite1198}}).
\item\relax
\flmRefsHyperref[eczindexfamilyrel]{code:univ_opt_analog}{Universally optimal sphere packing} --- The Leech lattice is universally optimal \NoCaseChange{\protect\cite{cite2208}}.
\end{eczvaluelist}
\codefieldsection{Cousins}
\begin{eczvaluelist}
\item\relax
\flmRefsHyperref[eczindexfamilyrel]{code:hexagonal}{\(A_2\) triangular lattice} --- The \(A_2\) hexagonal lattice can be specified as a section of the Leech lattice \NoCaseChange{\protect\cite[{Fig. 6.2}]{cite39}}.
\item\relax
\flmRefsHyperref[eczindexfamilyrel]{code:dthree}{\(D_3\) face-centered cubic (fcc) lattice} --- The \(D_3\) lattice can be specified as a section of the Leech lattice \NoCaseChange{\protect\cite[{Fig. 6.2}]{cite39}}. The Leech lattice can also be constructed via the Turyn construction and the holy construction using the octacode as the glue code; one of these constructions uses eight copies of the \(D_3\) fcc lattice \NoCaseChange{\protect\cite{cite2209,cite158}}.
\item\relax
\flmRefsHyperref[eczindexfamilyrel]{code:dfour}{\(D_4\) hyper-diamond lattice} --- The \(D_4\) lattice can be specified as a section of the Leech lattice \NoCaseChange{\protect\cite[{Fig. 6.2}]{cite39}}.
\item\relax
\flmRefsHyperref[eczindexfamilyrel]{code:esix}{\(E_6\) root lattice} --- The \(E_6\) lattice can be specified as a section of the Leech lattice \NoCaseChange{\protect\cite[{Fig. 6.2}]{cite39}}.
\item\relax
\flmRefsHyperref[eczindexfamilyrel]{code:eseven}{\(E_7\) root lattice} --- The \(E_7\) lattice can be specified as a section of the Leech lattice \NoCaseChange{\protect\cite[{Fig. 6.2}]{cite39}}.
\item\relax
\flmRefsHyperref[eczindexfamilyrel]{code:eeight}{\(E_8\) Gosset lattice} --- The \(E_8\) lattice can be specified as a section of the Leech lattice \NoCaseChange{\protect\cite[{Fig. 6.2}]{cite39}}. The Leech lattice admits a Turyn-type construction from two copies of the \(E_8\) lattice in the same space \NoCaseChange{\protect\cite[{Ch. 8, pg. 211}]{cite39}}.
\item\relax
\flmRefsHyperref[eczindexfamilyrel]{code:combinatorial_design}{Combinatorial design} --- The Leech lattice is completely determined by the Steiner system \(S(5,8,24)\) formed by the octads of the extended Golay code \NoCaseChange{\protect\cite[{Ch. 12, pg. 335, Thm. 6}]{cite39}}.
\item\relax
\flmRefsHyperref[eczindexfamilyrel]{code:extended_golay}{\([24, 12, 8]\) Extended Golay code} --- Two copies of the 24-dimensional packing obtained from the extended Golay code can be fitted together without overlap to form the Leech lattice \NoCaseChange{\protect\cite[{Ch. 5, pg. 145}]{cite39}}. Half of the lattice can be obtained using Construction \(B^{\star}\) \NoCaseChange{\protect\cite[{Exam. 10.7.3}]{cite115}}.
\item\relax
\flmRefsHyperref[eczindexfamilyrel]{code:ternary_golay}{\([11,6,5]_3\) Ternary Golay code} --- A 12-dimensional complex version of the Leech lattice can be obtained from the ternary Golay code \NoCaseChange{\protect\cite{cite1658,cite1659}\protect\cite[{pg. 200}]{cite39}}.
\item\relax
\flmRefsHyperref[eczindexfamilyrel]{code:quaternary_golay}{Extended quaternary Golay code} --- The Leech lattice can be constructed from the extended quaternary Golay code via \flmTerm{term}{ref114}{}{Construction \(A_4\)} \NoCaseChange{\protect\cite[{3rd Ed., pg. xxxiii}]{cite39}} (see also \NoCaseChange{\protect\cite{cite1659,cite112,cite1198}}).
\item\relax
\flmRefsHyperref[eczindexfamilyrel]{code:sharp_config}{Spherical sharp configuration} --- Several spherical sharp configurations are derived from the Leech lattice \NoCaseChange{\protect\cite{cite119}}.
\item\relax
\flmRefsHyperref[eczindexfamilyrel]{code:octacode}{Octacode} --- The Leech lattice can be constructed via the Turyn construction and the holy construction using the octacode as the glue code; one of these constructions uses eight copies of the \(D_3\) fcc lattice \NoCaseChange{\protect\cite{cite2209,cite158}}.
\item\relax
\flmRefsHyperref[eczindexfamilyrel]{code:modulation}{Modulation scheme} --- Codewords of the Leech lattice have been proposed to be used for a modulation scheme \NoCaseChange{\protect\cite{cite2210}}.
\item\relax
\flmRefsHyperref[eczindexfamilyrel]{code:pseudo_golay}{Pseudo Golay code} --- The Leech lattice can be constructed from pseudo Golay codes via \flmTerm{term}{ref114}{}{Construction \(A_4\)} \NoCaseChange{\protect\cite{cite122,cite1198}}.
\item\relax
\flmRefsHyperref[eczindexfamilyrel]{code:self_dual_over_z4}{Self-dual code over \(\mathbb{Z}_4\)} --- Each 4-frame of the Leech lattice corresponds to an extremal Type II self-dual code over \(\mathbb{Z}_4\) \NoCaseChange{\protect\cite{cite2211}}.
\item\relax
\flmRefsHyperref[eczindexfamilyrel]{code:lambda16}{\(\Lambda_{16}\) Barnes-Wall lattice} --- The \(\Lambda_{16}\) Barnes-Wall lattice can be obtained from the Leech lattice by restricting to the fixed-point 16-dimensional subspace of an involution \NoCaseChange{\protect\cite[{Ch. 4, pg. 131}]{cite39}}.
\item\relax
\flmRefsHyperref[eczindexfamilyrel]{code:coxeter_todd}{Coxeter-Todd \(K_{12}\) lattice} --- The Coxeter-Todd lattice can be realized as a subset of the Leech lattice \NoCaseChange{\protect\cite[{Ch. 4, pg. 128}]{cite39}}.
\item\relax
\flmRefsHyperref[eczindexfamilyrel]{code:higman-sims_graph}{Higman-Sims graph-adjacency code} --- The Higman-Sims graph occurs in the Leech lattice \NoCaseChange{\protect\cite{cite39}}.
\item\relax
\flmRefsHyperref[eczindexfamilyrel]{code:leech_shell}{\(\Lambda_{24}\) Leech lattice-shell code} --- The \(\Lambda_{24}\) lattice-shell code is obtained from a shell of the Leech lattice.
\end{eczvaluelist}
\eczhbkcontributors{ \eczhuVVA }
\endeczcode

\eczcode{hypercubic}{\(\mathbb{Z}^n\) hypercubic lattice}{}
\eczhIndexCodeAliasName{hypercubic}{hypercubic lattice}
\codefieldsection{Description}
Lattice-based code consisting of all integer vectors in \(n\) dimensions.
Its generator matrix is the \(n\)-dimensional identity matrix.
Its automorphism group consists of all coordinate permutations and sign changes.

\codefieldsection{Protection}
The \(\mathbb{Z}\) integer lattice solves the lattice quantization problem in one dimension with a second moment of \(G_1 = 1/12\).
The lattice has determinant 1, kissing number \(2n\), packing radius \(1/2\), covering radius \(\sqrt{n}/2\), and density \(V_{n}/2^{n}\) (with \(V_n\) the volume of the unit \(n\)-sphere).
See Ref. \NoCaseChange{\protect\cite{cite2212}} for further characterization.

\codefieldsection{Parents}
\begin{eczvaluelist}
\item\relax
\flmRefsHyperref[eczindexfamilyrel]{code:self_dual_lattice}{Unimodular lattice} --- The hypercubic lattice is odd and unimodular.
\item\relax
\flmRefsHyperref[eczindexfamilyrel]{code:root}{Root lattice}\end{eczvaluelist}
\codefieldsection{Cousins}
\begin{eczvaluelist}
\item\relax
\flmRefsHyperref[eczindexfamilyrel]{code:points_into_lattices}{Lattice} --- The generator matrix of a lattice-based code serves as a linear transformation that can be applied to the hypercubic lattice to obtain said code \NoCaseChange{\protect\cite[{Ch. 10}]{cite115}}.
\item\relax
\flmRefsHyperref[eczindexfamilyrel]{code:barnes_wall}{Barnes-Wall (BW) lattice} --- The hypercubic lattice for \(n=2\) is the \(m=0\) BW lattice.
\item\relax
\flmRefsHyperref[eczindexfamilyrel]{code:hypercube}{Hypercube code} --- Hypercube codewords form the minimal lattice shell code of the \(\mathbb{Z}^n\) hypercubic lattice when the lattice is shifted such that the center of a hypercube is at the origin.
\item\relax
\flmRefsHyperref[eczindexfamilyrel]{code:gkp}{Square-lattice GKP code} --- GKP codewords, when written in terms of coherent states, form a square lattice in phase space.
\item\relax
\flmRefsHyperref[eczindexfamilyrel]{code:488_color}{Square-octagon (4.8.8) color code} --- The 4.8.8 (square-octagon) tiling is obtained by applying a fattening procedure to the square lattice \NoCaseChange{\protect\cite{cite430}}.
\item\relax
\flmRefsHyperref[eczindexfamilyrel]{code:3d_bacon_shor}{3D Bacon-Shor code} --- 3D Bacon-Shor codes are defined on a hypercubic lattice.
\item\relax
\flmRefsHyperref[eczindexfamilyrel]{code:css_plaquette}{CSS-Plaquette code} --- CSS-Plaquette codes are defined on a hypercubic lattice.
\end{eczvaluelist}
\eczhbkcontributors{ \eczhuVVA }
\endeczcode

\eczcode{hexagonal}{\(A_2\) triangular lattice}{}
\codefieldsection{Alternative Names}
\begin{eczvaluelist}
\item\relax \(A_2\) hexagonal lattice
\end{eczvaluelist}
\eczhIndexCodeAliasName{hexagonal}{triangular lattice}
\eczhIndexCodeAliasName{hexagonal}{\(A_2\) hexagonal lattice}
\codefieldsection{Description}
Two-dimensional lattice that corresponds to the triangular tiling and that exhibits optimal packing, solving the packing, kissing, covering and quantization problems.
As a tiling, its dual (whose points lie at the centers of each triangle) is the honeycomb tiling.

Its generator matrix is
\flmMathEnvironment{align}{}{
  V=\begin{pmatrix}1 & 0\\
    -1/2 & \sqrt{3}/2
      \end{pmatrix}~.
}
All possible sublattices are characterized in Refs. \NoCaseChange{\protect\cite{cite2213,cite2214}} from the point of view of information transmission over the AWGN channel.

\codefieldsection{Protection}
The triangular lattice exhibits the densest packing with density \(\Delta = \pi/\sqrt{12} \approx 0.9069\) \NoCaseChange{\protect\cite[{Sec. 1.4}]{cite39}}, the highest kissing number of 6, and the thinnest covering with thickness \(\Theta = 2\pi/(3\sqrt{3})\approx 1.2092\) \NoCaseChange{\protect\cite{cite2215}} in two dimensions.
It solves the quantizer problem in two dimensions with \(G_2 = \frac{5}{36\sqrt{3}}\) \NoCaseChange{\protect\cite{cite2216,cite2217,cite2218,cite2219}}.
It also solves the Gaussian channel coding problem \NoCaseChange{\protect\cite{cite2217}}.

\codefieldsection{Realizations}
\begin{eczvaluelist}
\item\relax Wireless communication \NoCaseChange{\protect\cite{cite358,cite359}}.
\end{eczvaluelist}
\codefieldsection{Parents}
\begin{eczvaluelist}
\item\relax
\flmRefsHyperref[eczindexfamilyrel]{code:an}{\(A_n\) lattice}\item\relax
\flmRefsHyperref[eczindexfamilyrel]{code:an_dual}{\(A_n^{\perp}\) lattice} --- The \(A_2\) lattice is equivalent, up to rescaling, to its dual \(A_2^{\perp}\) \NoCaseChange{\protect\cite[{Ch. 1, pg. 13}]{cite39}}.
\end{eczvaluelist}
\codefieldsection{Cousins}
\begin{eczvaluelist}
\item\relax
\flmRefsHyperref[eczindexfamilyrel]{code:polygon}{Polygon code} --- The Voronoi cell of the triangular lattice is the hexagon.
\item\relax
\flmRefsHyperref[eczindexfamilyrel]{code:honeycomb}{Honeycomb tiling} --- The Voronoi cell of the triangular lattice (honeycomb tiling) is a hexagon (triangle). Triangular and hexagonal tilings are dual to each other as tilings, i.e., the vertices of one tiling lie at the centers of faces of the other. Points of the honeycomb tiling form two triangular lattices. The ruby tiling is a fattened honeycomb tiling interpolating between the honeycomb tiling and triangular lattice.
\item\relax
\flmRefsHyperref[eczindexfamilyrel]{code:leech}{\(\Lambda_{24}\) Leech lattice} --- The \(A_2\) hexagonal lattice can be specified as a section of the Leech lattice \NoCaseChange{\protect\cite[{Fig. 6.2}]{cite39}}.
\item\relax
\flmRefsHyperref[eczindexfamilyrel]{code:univ_opt_analog}{Universally optimal sphere packing} --- The triangular lattice is universally optimal among all lattices, but has not been proven to be optimal over all periodic packings \NoCaseChange{\protect\cite{cite2220}}.
\item\relax
\flmRefsHyperref[eczindexfamilyrel]{code:hexagonal_gkp}{Hexagonal GKP code} --- The hexagonal GKP code is based on the triangular lattice.
\end{eczvaluelist}
\eczhbkcontributors{ \eczhuVVA }
\endeczcode

\eczcode{an}{\(A_n\) lattice}{}
\eczhIndexCodeAliasName{an}{lattice}
\codefieldsection{Description}
Lattice-based \(n\)-dimensional code that can be simply defined in \(n+1\) dimensions as the set of integer vectors \(x\) lying in the hyperplane \(x_0+x_1+\cdots+x_{n} = 0\).

Its generator matrix can be represented by
\flmMathEnvironment{align}{}{
  \begin{pmatrix}-1 & \phantom{-}1 & 0 & 0 & \cdots & 0 & 0\\
  0 & -1 & \phantom{-}1 & 0 & \cdots & 0 & 0\\
  0 & 0 & -1 & \phantom{-}1 & \cdots & 0 & 0\\
  \vdots & \vdots & \vdots & \vdots & \ddots & \vdots & \vdots\\
  0 & 0 & 0 & 0 & \cdots & -1 & \phantom{-}1
  \end{pmatrix}~.
}

\codefieldsection{Protection}
The lattice has Gram determinant \(n+1\) (equivalently, fundamental volume \(\sqrt{n+1}\)), kissing number \(n(n+1)\), packing radius \(1/\sqrt{2}\), covering radius \(\sqrt{\frac{a\left(n+1-a\right)}{n+1}}\) (with \(a=\lfloor (n+1)/2 \rfloor\)), and center density \(V_n/(2^{n/2}\sqrt{n+1})\) (with \(V_n\) the volume of the unit \(n\)-sphere) \NoCaseChange{\protect\cite[{Chs. 4 and 21}]{cite39}}.

\codefieldsection{Parent}
\begin{eczvaluelist}
\item\relax
\flmRefsHyperref[eczindexfamilyrel]{code:root}{Root lattice}\end{eczvaluelist}
\codefieldsection{Children}
\begin{eczvaluelist}
\item\relax
\flmRefsHyperref[eczindexfamilyrel]{code:hexagonal}{\(A_2\) triangular lattice}\item\relax
\flmRefsHyperref[eczindexfamilyrel]{code:dthree}{\(D_3\) face-centered cubic (fcc) lattice} --- The \(D_3\) fcc lattice is equivalent to the \(A_3\) root lattice \NoCaseChange{\protect\cite[{Ch. 1, pg. 13}]{cite39}}.
\end{eczvaluelist}
\codefieldsection{Cousin}
\begin{eczvaluelist}
\item\relax
\flmRefsHyperref[eczindexfamilyrel]{code:an_dual}{\(A_n^{\perp}\) lattice} --- The \(A_n\) and \(A_n^{\perp}\) lattices are dual to each other.
\end{eczvaluelist}
\eczhbkcontributors{ \eczhuVVA }
\endeczcode

\eczcode{an_dual}{\(A_n^{\perp}\) lattice}{}
\eczhIndexCodeAliasName{an_dual}{lattice}
\codefieldsection{Description}
Lattice-based \(n\)-dimensional code whose codewords form the dual of the \(A_n\) lattice.

\codefieldsection{Protection}
Exhibits the thinnest covering in two dimensions and the thinnest lattice covering in dimensions three \NoCaseChange{\protect\cite{cite2221}}, four \NoCaseChange{\protect\cite{cite2222}}, and five \NoCaseChange{\protect\cite{cite2223,cite2224}}.

\codefieldsection{Parent}
\begin{eczvaluelist}
\item\relax
\flmRefsHyperref[eczindexfamilyrel]{code:points_into_lattices}{Lattice}\end{eczvaluelist}
\codefieldsection{Children}
\begin{eczvaluelist}
\item\relax
\flmRefsHyperref[eczindexfamilyrel]{code:bcc}{Body-centered cubic (bcc) lattice} --- The bcc lattice is the dual of the \(A_3=D_3\) fcc lattice.
\item\relax
\flmRefsHyperref[eczindexfamilyrel]{code:hexagonal}{\(A_2\) triangular lattice} --- The \(A_2\) lattice is equivalent, up to rescaling, to its dual \(A_2^{\perp}\) \NoCaseChange{\protect\cite[{Ch. 1, pg. 13}]{cite39}}.
\end{eczvaluelist}
\codefieldsection{Cousin}
\begin{eczvaluelist}
\item\relax
\flmRefsHyperref[eczindexfamilyrel]{code:an}{\(A_n\) lattice} --- The \(A_n\) and \(A_n^{\perp}\) lattices are dual to each other.
\end{eczvaluelist}
\eczhbkcontributors{ \eczhuVVA }
\endeczcode

\eczcode{bw32}{\(BW_{32}\) Barnes-Wall lattice}{~\NoCaseChange{\protect\cite{cite2203}}}
\eczhIndexCodeAliasName{bw32}{Barnes-Wall lattice}
\codefieldsection{Description}
BW lattice in dimension \(32\).

\codefieldsection{Parents}
\begin{eczvaluelist}
\item\relax
\flmRefsHyperref[eczindexfamilyrel]{code:barnes_wall}{Barnes-Wall (BW) lattice}\item\relax
\flmRefsHyperref[eczindexfamilyrel]{code:construction_a4}{Construction \(A_4\) lattice} --- The \(C_{m=5,r=1}\) code gives rise to the \(BW_{32}\) Barnes-Wall lattice via \flmTerm{term}{ref114}{}{Construction \(A_4\)} \NoCaseChange{\protect\cite{cite2225,cite112}}.
\end{eczvaluelist}
\codefieldsection{Cousins}
\begin{eczvaluelist}
\item\relax
\flmRefsHyperref[eczindexfamilyrel]{code:cmr}{\(C_{m,r}\) code} --- The \(C_{m=5,r=1}\) code gives rise to the \(BW_{32}\) Barnes-Wall lattice via \flmTerm{term}{ref114}{}{Construction \(A_4\)} \NoCaseChange{\protect\cite{cite2225,cite112}}.
\item\relax
\flmRefsHyperref[eczindexfamilyrel]{code:bw32_shell}{\(BW_{32}\) lattice-shell code} --- The \(BW_{32}\) lattice-shell code is obtained from a shell of the \(BW_{32}\) Barnes-Wall lattice.
\end{eczvaluelist}
\eczhbkcontributors{ \eczhuVVA }
\endeczcode

\eczcode{dthree}{\(D_3\) face-centered cubic (fcc) lattice}{}
\codefieldsection{Alternative Names}
\begin{eczvaluelist}
\item\relax Cannonball lattice
\end{eczvaluelist}
\eczhIndexCodeAliasName{dthree}{face-centered cubic (fcc) lattice}
\eczhIndexCodeAliasName{dthree}{Cannonball lattice}
\codefieldsection{Description}
Laminated three-dimensional lattice consisting of layers of triangular lattices.
\codefieldsection{Protection}
The \(D_3\) fcc lattice exhibits the densest packing and highest kissing number of 12 in three dimensions.
The \textit{Kepler conjecture} \NoCaseChange{\protect\cite{cite2226}} states that the \(D_3\) fcc lattice has the densest packing in 3D.

It was first proven that this lattice was the densest 3D lattice packing \NoCaseChange{\protect\cite{cite2227}}.
Determining the maximum density of any sphere packing in 3D was then reduced to a computationally tractable problem \NoCaseChange{\protect\cite{cite2217}}, which was solved \NoCaseChange{\protect\cite{cite2228}} and formalized in automated proof checking software \NoCaseChange{\protect\cite{cite2229}}.

\codefieldsection{Parents}
\begin{eczvaluelist}
\item\relax
\flmRefsHyperref[eczindexfamilyrel]{code:dn}{\(D_n\) checkerboard lattice} --- The \(D_3\) root lattice is equivalent to the \(A_3\) root lattice \NoCaseChange{\protect\cite[{Ch. 10}]{cite115}}.
\item\relax
\flmRefsHyperref[eczindexfamilyrel]{code:an}{\(A_n\) lattice} --- The \(D_3\) fcc lattice is equivalent to the \(A_3\) root lattice \NoCaseChange{\protect\cite[{Ch. 1, pg. 13}]{cite39}}.
\end{eczvaluelist}
\codefieldsection{Cousins}
\begin{eczvaluelist}
\item\relax
\flmRefsHyperref[eczindexfamilyrel]{code:rhombic_dodecahedron}{Rhombic dodecahedron code} --- The Voronoi cell of the \(D_3\) fcc lattice is a rhombic dodecahedron \NoCaseChange{\protect\cite[{Ch. 21, pg. 464}]{cite39}}.
\item\relax
\flmRefsHyperref[eczindexfamilyrel]{code:bcc}{Body-centered cubic (bcc) lattice} --- The bcc and fcc lattices are dual to each other.
\item\relax
\flmRefsHyperref[eczindexfamilyrel]{code:leech}{\(\Lambda_{24}\) Leech lattice} --- The \(D_3\) lattice can be specified as a section of the Leech lattice \NoCaseChange{\protect\cite[{Fig. 6.2}]{cite39}}. The Leech lattice can also be constructed via the Turyn construction and the holy construction using the octacode as the glue code; one of these constructions uses eight copies of the \(D_3\) fcc lattice \NoCaseChange{\protect\cite{cite2209,cite158}}.
\item\relax
\flmRefsHyperref[eczindexfamilyrel]{code:cubeoctahedron}{Cuboctahedron code} --- Cuboctahedron codewords form the minimal shell of the \(D_3\) face-centered cubic (fcc) lattice.
\end{eczvaluelist}
\eczhbkcontributors{ \eczhuVVA }
\endeczcode

\eczcode{dfour}{\(D_4\) hyper-diamond lattice}{}
\eczhIndexCodeAliasName{dfour}{hyper-diamond lattice}
\codefieldsection{Description}
BW lattice in dimension \(4\), which is the lattice corresponding to the \([4,3,2]\) SPC code via \flmTerm{term}{ref127}{}{Construction A}.
The lattice points form the \(\{3,3,4,3\}\) tessellation of 4-space \NoCaseChange{\protect\cite[{pg. 136}]{cite178}}.

\codefieldsection{Protection}
The \(D_4\) lattice has a density of \(\pi^2/16\approx 0.6169\) and nominal coding gain of \(\sqrt{2}\). It exhibits the densest lattice packing in four dimensions \NoCaseChange{\protect\cite{cite2230}}.

\codefieldsection{Parents}
\begin{eczvaluelist}
\item\relax
\flmRefsHyperref[eczindexfamilyrel]{code:dn}{\(D_n\) checkerboard lattice} --- The \(D_4\) lattice is equivalent, up to rescaling, to its dual \(D_4^{\perp}\) \NoCaseChange{\protect\cite[{Ch. 1, pg. 13}]{cite39}}.
\item\relax
\flmRefsHyperref[eczindexfamilyrel]{code:barnes_wall}{Barnes-Wall (BW) lattice}\end{eczvaluelist}
\codefieldsection{Cousins}
\begin{eczvaluelist}
\item\relax
\flmRefsHyperref[eczindexfamilyrel]{code:parity_check}{\([n,n-1,2]\) Single parity-check (SPC) code} --- The \(D_4\) lattice is constructed out of the \([4,3,2]\) SPC code via \flmTerm{term}{ref127}{}{Construction A} \NoCaseChange{\protect\cite[{pg. 138}]{cite39}}.
\item\relax
\flmRefsHyperref[eczindexfamilyrel]{code:repetition}{Repetition code} --- Construction \(A_c\) applied to the binary repetition code \(\{00,11\}\) over the Gaussian integers yields a Gaussian lattice whose corresponding real lattice is \(D_4\) \NoCaseChange{\protect\cite[{Ch. 7, pg. 202}]{cite39}}.
\item\relax
\flmRefsHyperref[eczindexfamilyrel]{code:24cell}{24-cell code} --- The Voronoi cell of the \(D_4\) lattice is a 24-cell \NoCaseChange{\protect\cite[{Ch. 21, pg. 464}]{cite39}}.
\item\relax
\flmRefsHyperref[eczindexfamilyrel]{code:leech}{\(\Lambda_{24}\) Leech lattice} --- The \(D_4\) lattice can be specified as a section of the Leech lattice \NoCaseChange{\protect\cite[{Fig. 6.2}]{cite39}}.
\item\relax
\flmRefsHyperref[eczindexfamilyrel]{code:dfour_shell}{\(D_4\) lattice-shell code} --- The \(D_4\) lattice-shell code is obtained from a shell of the \(D_4\) lattice.
\item\relax
\flmRefsHyperref[eczindexfamilyrel]{code:dfour_gkp}{\(D_4\) hyper-diamond GKP code} --- The \(D_4\) GKP code is built from the \(D_4\) lattice.
\item\relax
\flmRefsHyperref[eczindexfamilyrel]{code:4d_13_surface}{\((1,3)\) 4D toric code} --- The \((1,3)\) 4D toric code on a hyper-diamond lattice admits a transversal logical \(CCCZ\) gate \NoCaseChange{\protect\cite{cite479}}.
\end{eczvaluelist}
\eczhbkcontributors{ \eczhuVVA }
\endeczcode

\eczcode{dn}{\(D_n\) checkerboard lattice}{}
\eczhIndexCodeAliasName{dn}{checkerboard lattice}
\codefieldsection{Description}
Lattice consisting of all points whose coordinates add up to an even integer.

Its generator matrix can be represented by
\flmMathEnvironment{align}{}{
  \begin{pmatrix}-1 & -1 & 0 & 0 & \cdots & 0 & 0\\
  \phantom{-}1 & -1 & 0 & 0 & \cdots & 0 & 0\\
  0 & \phantom{-}1 & -1 & 0 & \cdots & 0 & 0\\
  0 & 0 & \phantom{-}1 & -1 & \cdots & 0 & 0\\
  \vdots & \vdots & \vdots & \vdots & \ddots & \vdots & \vdots\\
  0 & 0 & 0 & 0 & \cdots\phantom{-} & 1 & -1
  \end{pmatrix}~.
}

\codefieldsection{Protection}
Exhibits the densest lattice packing and highest known kissing number in four and five \NoCaseChange{\protect\cite{cite2231}} dimensions.

\codefieldsection{Parents}
\begin{eczvaluelist}
\item\relax
\flmRefsHyperref[eczindexfamilyrel]{code:root}{Root lattice}\item\relax
\flmRefsHyperref[eczindexfamilyrel]{code:construction_a}{Construction A code} --- \([n,n-1,2]\) SPC codes yield \(D_n\) checkerboard lattices via \flmTerm{term}{ref127}{}{Construction A} \NoCaseChange{\protect\cite[{Exam. 10.5.1}]{cite115}\protect\cite[{pg. 138}]{cite39}}.
\end{eczvaluelist}
\codefieldsection{Children}
\begin{eczvaluelist}
\item\relax
\flmRefsHyperref[eczindexfamilyrel]{code:dfour}{\(D_4\) hyper-diamond lattice} --- The \(D_4\) lattice is equivalent, up to rescaling, to its dual \(D_4^{\perp}\) \NoCaseChange{\protect\cite[{Ch. 1, pg. 13}]{cite39}}.
\item\relax
\flmRefsHyperref[eczindexfamilyrel]{code:dthree}{\(D_3\) face-centered cubic (fcc) lattice} --- The \(D_3\) root lattice is equivalent to the \(A_3\) root lattice \NoCaseChange{\protect\cite[{Ch. 10}]{cite115}}.
\end{eczvaluelist}
\codefieldsection{Cousin}
\begin{eczvaluelist}
\item\relax
\flmRefsHyperref[eczindexfamilyrel]{code:parity_check}{\([n,n-1,2]\) Single parity-check (SPC) code} --- \([n,n-1,2]\) SPC codes yield \(D_n\) checkerboard lattices via \flmTerm{term}{ref127}{}{Construction A} \NoCaseChange{\protect\cite[{Exam. 10.5.1}]{cite115}\protect\cite[{pg. 138}]{cite39}}.
\end{eczvaluelist}
\eczhbkcontributors{ \eczhuVVA }
\endeczcode

\eczcode{esix}{\(E_6\) root lattice}{}
\eczhIndexCodeAliasName{esix}{root lattice}
\codefieldsection{Description}
Exceptional root lattice in dimension \(6\).

A generating matrix for the lattice embedded in eight dimensions is \NoCaseChange{\protect\cite{cite39}}
\flmMathEnvironment{align}{}{
\begin{pmatrix}
0 & -1 & 1 & 0 & 0 & 0 & 0 & 0 \\
0 & 0 & -1 & 1 & 0 & 0 & 0 & 0 \\
0 & 0 & 0 & -1 & 1 & 0 & 0 & 0 \\
0 & 0 & 0 & 0 & -1 & 1 & 0 & 0 \\
0 & 0 & 0 & 0 & 0 & -1 & 1 & 0 \\
\frac{1}{2} & \frac{1}{2} & \frac{1}{2} & \frac{1}{2} & -\frac{1}{2} & -\frac{1}{2} & -\frac{1}{2} & -\frac{1}{2}
\end{pmatrix}~.
}

\codefieldsection{Protection}
The root \(E_6\) lattice exhibits the densest lattice packing \NoCaseChange{\protect\cite{cite2232,cite2233,cite2234,cite2235,cite2236}} and highest known kissing number in six dimensions.
\codefieldsection{Parent}
\begin{eczvaluelist}
\item\relax
\flmRefsHyperref[eczindexfamilyrel]{code:root}{Root lattice}\end{eczvaluelist}
\codefieldsection{Cousins}
\begin{eczvaluelist}
\item\relax
\flmRefsHyperref[eczindexfamilyrel]{code:q-ary_repetition}{\(q\)-ary repetition code} --- The \([3,1,3]_3\) ternary repetition code can be used to obtain the \(E_6\) root lattice \NoCaseChange{\protect\cite[{Exam. 10.5.4}]{cite115}\protect\cite[{Ch. 7, pg. 200}]{cite39}}.
\item\relax
\flmRefsHyperref[eczindexfamilyrel]{code:rect_hessian_polyhedron}{Rectified Hessian polyhedron code} --- The Voronoi cell of the \(E_6\) root lattice is the dual of the Gosset \(1_{22}\) polytope \NoCaseChange{\protect\cite[{Ch. 21, pg. 465}]{cite39}}.
\item\relax
\flmRefsHyperref[eczindexfamilyrel]{code:leech}{\(\Lambda_{24}\) Leech lattice} --- The \(E_6\) lattice can be specified as a section of the Leech lattice \NoCaseChange{\protect\cite[{Fig. 6.2}]{cite39}}.
\item\relax
\flmRefsHyperref[eczindexfamilyrel]{code:esix_shell}{\(E_6\) lattice-shell code} --- The \(E_6\) lattice-shell code is obtained from a shell of the \(E_6\) lattice.
\item\relax
\flmRefsHyperref[eczindexfamilyrel]{code:hessian_polyhedron}{Hessian polyhedron code} --- The 27 Hessian polyhedron codewords are intimately related to the \(E_6\) Lie group \NoCaseChange{\protect\cite{cite2237}}.
\end{eczvaluelist}
\eczhbkcontributors{ \eczhuVVA }
\endeczcode

\eczcode{eseven}{\(E_7\) root lattice}{}
\eczhIndexCodeAliasName{eseven}{root lattice}
\codefieldsection{Description}
Exceptional root lattice in dimension \(7\).

A generating matrix for the lattice embedded in eight dimensions is \NoCaseChange{\protect\cite{cite39}}
\flmMathEnvironment{align}{}{
\begin{pmatrix}
-1 & 1 & 0 & 0 & 0 & 0 & 0 & 0 \\
0 & -1 & 1 & 0 & 0 & 0 & 0 & 0 \\
0 & 0 & -1 & 1 & 0 & 0 & 0 & 0 \\
0 & 0 & 0 & -1 & 1 & 0 & 0 & 0 \\
0 & 0 & 0 & 0 & -1 & 1 & 0 & 0 \\
0 & 0 & 0 & 0 & 0 & -1 & 1 & 0 \\
\frac{1}{2} & \frac{1}{2} & \frac{1}{2} & \frac{1}{2} & -\frac{1}{2} & -\frac{1}{2} & -\frac{1}{2} & -\frac{1}{2}
\end{pmatrix}~.
}

The Voronoi cell of the lattice is the reciprocal of the Gosset \(2_{31}\) polytope \NoCaseChange{\protect\cite[{Ch. 21, pg. 465}]{cite39}}.

\codefieldsection{Protection}
The \(E_7\) root lattice exhibits the densest lattice packing \NoCaseChange{\protect\cite{cite2232,cite2233,cite2234,cite2235,cite2238}}.
\codefieldsection{Parents}
\begin{eczvaluelist}
\item\relax
\flmRefsHyperref[eczindexfamilyrel]{code:root}{Root lattice}\item\relax
\flmRefsHyperref[eczindexfamilyrel]{code:construction_a}{Construction A code} --- The \([7,3,4]\) simplex code yields the \(E_7\) root lattice via \flmTerm{term}{ref127}{}{Construction A} \NoCaseChange{\protect\cite{cite1204}\protect\cite[{Exam. 10.5.3}]{cite115}}.
\end{eczvaluelist}
\codefieldsection{Cousins}
\begin{eczvaluelist}
\item\relax
\flmRefsHyperref[eczindexfamilyrel]{code:hamming743}{\([7,4,3]\) Hamming code} --- The \([7,4,3]\) Hamming code yields the \(E_7^{\perp}\) lattice via \flmTerm{term}{ref127}{}{Construction A} \NoCaseChange{\protect\cite{cite1204}}.
\item\relax
\flmRefsHyperref[eczindexfamilyrel]{code:simplex734}{\([7,3,4]\) simplex code} --- The \([7,3,4]\) simplex code yields the \(E_7\) root lattice via \flmTerm{term}{ref127}{}{Construction A} \NoCaseChange{\protect\cite{cite1204}\protect\cite[{Exam. 10.5.3}]{cite115}\protect\cite[{pg. 138}]{cite39}}.
\item\relax
\flmRefsHyperref[eczindexfamilyrel]{code:leech}{\(\Lambda_{24}\) Leech lattice} --- The \(E_7\) lattice can be specified as a section of the Leech lattice \NoCaseChange{\protect\cite[{Fig. 6.2}]{cite39}}.
\item\relax
\flmRefsHyperref[eczindexfamilyrel]{code:eseven_shell}{\(E_7\) lattice-shell code} --- The \(E_7\) lattice-shell code is obtained from a shell of the \(E_7\) lattice.
\end{eczvaluelist}
\eczhbkcontributors{ \eczhuVVA }
\endeczcode

\eczcode{eeight}{\(E_8\) Gosset lattice}{~\NoCaseChange{\protect\cite{cite2239}}}
\eczhIndexCodeAliasName{eeight}{Gosset lattice}
\codefieldsection{Description}
Even unimodular BW lattice in dimension \(8\), consisting of the Cayley integral octonions rescaled by \(\sqrt{2}\).
The lattice corresponds to the \([8,4,4]\) Hamming code via \flmTerm{term}{ref127}{}{Construction A}.

A generator matrix is \NoCaseChange{\protect\cite{cite39}}
\flmMathEnvironment{align}{}{
  \begin{pmatrix}
  2 & 0 & 0 & 0 & 0 & 0 & 0 & 0 \\
  -1 & 1 & 0 & 0 & 0 & 0 & 0 & 0 \\
  0 & -1 & 1 & 0 & 0 & 0 & 0 & 0 \\
  0 & 0 & -1 & 1 & 0 & 0 & 0 & 0 \\
  0 & 0 & 0 & -1 & 1 & 0 & 0 & 0 \\
  0 & 0 & 0 & 0 & -1 & 1 & 0 & 0 \\
  0 & 0 & 0 & 0 & 0 & -1 & 1 & 0 \\
  \frac{1}{2} & \frac{1}{2} & \frac{1}{2} & \frac{1}{2} & \frac{1}{2} & \frac{1}{2} & \frac{1}{2} & \frac{1}{2}
  \end{pmatrix}.
}
A generator matrix for the symplectic version \NoCaseChange{\protect\cite[{Appx. 2}]{cite2205}} is
\flmMathEnvironment{align}{}{
  \begin{pmatrix}
  2 & 1 &  0 &  1 &  1 &  0 &  0 &  0 \\
  1 & 2 &  1 &  0 &  0 &  1 &  0 &  0 \\
  0 & 1 &  2 & -1 &  0 &  0 &  1 &  0 \\
  1 & 0 & -1 &  2 &  0 &  0 &  0 &  1 \\
  1 & 0 &  0 &  0 &  2 & -1 &  0 & -1 \\
  0 & 1 &  0 &  0 & -1 &  2 & -1 &  0 \\
  0 & 0 &  1 &  0 &  0 & -1 &  2 &  1 \\
  0 & 0 &  0 &  1 & -1 &  0 &  1 &  2
  \end{pmatrix}.
}

\codefieldsection{Protection}
The \(E_8\) lattice has a nominal coding gain of \(2\).
It exhibits the densest lattice packing \NoCaseChange{\protect\cite{cite2232,cite2233,cite2234,cite2235}}, the densest packing \NoCaseChange{\protect\cite{cite2206}}, and the highest kissing number of 240 \NoCaseChange{\protect\cite{cite2240}} in eight dimensions.

\codefieldsection{Notes}
\begin{eczvaluelist}
\item\relax Popular summary of solution to the sphere-packing problem in \flmHref{https://www.quantamagazine.org/sphere-packing-solved-in-higher-dimensions-20160330/}{Quanta Magazine}.
\end{eczvaluelist}
\codefieldsection{Parents}
\begin{eczvaluelist}
\item\relax
\flmRefsHyperref[eczindexfamilyrel]{code:self_dual_lattice}{Unimodular lattice} --- The \(E_8\) Gosset lattice is even and unimodular.
\item\relax
\flmRefsHyperref[eczindexfamilyrel]{code:root}{Root lattice}\item\relax
\flmRefsHyperref[eczindexfamilyrel]{code:barnes_wall}{Barnes-Wall (BW) lattice}\item\relax
\flmRefsHyperref[eczindexfamilyrel]{code:construction_a}{Construction A code} --- The \([8,4,4]\) extended Hamming code yields the \(E_8\) Gosset lattice via \flmTerm{term}{ref127}{}{Construction A} \NoCaseChange{\protect\cite[{Exam. 10.5.2}]{cite115}}.
\item\relax
\flmRefsHyperref[eczindexfamilyrel]{code:construction_a4}{Construction \(A_4\) lattice} --- The octacode yields the \(E_8\) Gosset lattice via \flmTerm{term}{ref114}{}{Construction \(A_4\)} \NoCaseChange{\protect\cite{cite2241,cite112}\protect\cite[{Exam. 12.5.13}]{cite126}}.
\item\relax
\flmRefsHyperref[eczindexfamilyrel]{code:univ_opt_analog}{Universally optimal sphere packing} --- The \(E_8\) Gosset lattice is universally optimal \NoCaseChange{\protect\cite{cite2208}}.
\end{eczvaluelist}
\codefieldsection{Cousins}
\begin{eczvaluelist}
\item\relax
\flmRefsHyperref[eczindexfamilyrel]{code:hamming844}{\([8,4,4]\) extended Hamming code} --- The \([8,4,4]\) extended Hamming code yields the \(E_8\) Gosset lattice via \flmTerm{term}{ref127}{}{Construction A} \NoCaseChange{\protect\cite[{Exam. 10.5.2}]{cite115}\protect\cite[{pg. 138}]{cite39}}.
\item\relax
\flmRefsHyperref[eczindexfamilyrel]{code:repetition}{Repetition code} --- The \([8,1,8]\) repetition code can be used to obtain the \(E_8\) Gosset lattice \NoCaseChange{\protect\cite[{Exam. 10.7.1}]{cite115}}.
\item\relax
\flmRefsHyperref[eczindexfamilyrel]{code:witting_polytope}{Witting polytope code} --- The Voronoi cell of the \(E_8\) Gosset lattice is the dual of the Gosset \(4_{21}\) polytope \NoCaseChange{\protect\cite[{Ch. 21, pg. 464}]{cite39}}.
\item\relax
\flmRefsHyperref[eczindexfamilyrel]{code:sharp_config}{Spherical sharp configuration} --- Several spherical sharp configurations are derived from the \(E_8\) Gosset lattice \NoCaseChange{\protect\cite{cite119}}.
\item\relax
\flmRefsHyperref[eczindexfamilyrel]{code:leech}{\(\Lambda_{24}\) Leech lattice} --- The \(E_8\) lattice can be specified as a section of the Leech lattice \NoCaseChange{\protect\cite[{Fig. 6.2}]{cite39}}. The Leech lattice admits a Turyn-type construction from two copies of the \(E_8\) lattice in the same space \NoCaseChange{\protect\cite[{Ch. 8, pg. 211}]{cite39}}.
\item\relax
\flmRefsHyperref[eczindexfamilyrel]{code:tetracode}{\([4,2,3]_3\) Tetracode} --- The \([4,2,3]_3\) tetracode can be used to obtain the \(E_8\) Gosset lattice \NoCaseChange{\protect\cite[{Exam. 10.5.5}]{cite115}\protect\cite[{Ch. 7, pg. 200}]{cite39}}.
\item\relax
\flmRefsHyperref[eczindexfamilyrel]{code:octacode}{Octacode} --- The octacode yields the \(E_8\) Gosset lattice via \flmTerm{term}{ref114}{}{Construction \(A_4\)} \NoCaseChange{\protect\cite{cite2241,cite112}\protect\cite[{Exam. 12.5.13}]{cite126}}.
\item\relax
\flmRefsHyperref[eczindexfamilyrel]{code:eeight_shell}{\(E_8\) Gosset lattice-shell code} --- The \(E_8\) lattice-shell code is obtained from a shell of the \(E_8\) lattice.
\item\relax
\flmRefsHyperref[eczindexfamilyrel]{code:lca_stabilizer}{Locally compact Abelian (LCA) stabilizer code} --- Integer symplectic matrices like the symplectic \(E_8\) generator matrix can be used to construct LCA stabilizer codes \NoCaseChange{\protect\cite{cite1428}}.
\item\relax
\flmRefsHyperref[eczindexfamilyrel]{code:majorana_hamming}{\(\llbracket 2^{m-1},2^{m-1}-m-1,4\rrbracket _{f}\) Hamming Majorana code} --- The logical subspace of the \(\llbracket 8,3,4\rrbracket _{f}\) Hamming Majorana code is a Cartan subspace of the \(E_8\) Lie algebra \NoCaseChange{\protect\cite{cite565}}.
\end{eczvaluelist}
\eczhbkcontributors{ \eczhuVVA }
\endeczcode

\eczcode{analog}{Analog code}{}
\codefieldsection{Alternative Names}
\begin{eczvaluelist}
\item\relax Code over \(\mathbb{R}\)
\end{eczvaluelist}
\eczhIndexCodeAliasName{analog}{Code over \(\mathbb{R}\)}

\codefieldsection{Kingdom root code}
for the \flmRefsHyperref{kingdom:analog}{Analog Kingdom}.
\codefieldsection{Description}
Encodes states (codewords) into continuous coordinates in the \(n\)-dimensional (real or complex) coordinate space (\(\mathbb{R}^n\) or \(\mathbb{C}^n\)).
Important subclasses include sphere packings, tilings, and modulation constellations.
The number of codewords may be infinite because the coordinate space is infinite, so various restricted versions have to be constructed in practice.

\codefieldsection{Parents}
\begin{eczvaluelist}
\item\relax
\flmRefsHyperref[eczindexfamilyrel]{code:block}{Block code}\item\relax
\flmRefsHyperref[eczindexfamilyrel]{code:group_classical}{Group-alphabet code} --- Analog code alphabets, such as \(\mathbb{R}^n\) or \(\mathbb{C}^n\), are additive groups.
\end{eczvaluelist}
\codefieldsection{Children}
\begin{eczvaluelist}
\item\relax
\flmRefsHyperref[eczindexfamilyrel]{code:integers_into_integers}{Integer-based code}\item\relax
\flmRefsHyperref[eczindexfamilyrel]{code:honeycomb}{Honeycomb tiling}\item\relax
\flmRefsHyperref[eczindexfamilyrel]{code:points_into_balls}{Bounded-energy code} --- Bounded-energy codes are analog codes constrained to lie on or inside a sphere.
\item\relax
\flmRefsHyperref[eczindexfamilyrel]{code:real_block}{Real block code} --- Real-number block codes encode continuous sets of real or complex numbers into a real or complex vector space.
\item\relax
\flmRefsHyperref[eczindexfamilyrel]{code:sphere_packing}{Sphere packing}\end{eczvaluelist}
\codefieldsection{Cousins}
\begin{eczvaluelist}
\item\relax
\flmRefsHyperref[eczindexfamilyrel]{code:bosonic_classical_into_quantum}{Bosonic c-q code} --- Any analog code can be embedded into a bosonic Hilbert space, and thus passed through a bosonic channel, by associating the reals with the configuration space of position states of bosonic modes.
\item\relax
\flmRefsHyperref[eczindexfamilyrel]{code:2pt_homogeneous}{Two-point homogeneous-space code} --- Euclidean space \(\mathbb{R}^n\) is a noncompact two-point homogeneous space and is, in fact, a noncompact three-point homogeneous space \NoCaseChange{\protect\cite[{Sec. 6.6.1.1}]{cite2242}}. One can also think of \(\mathbb{R}^n\) as a homogeneous space of the Euclidean group by the orthogonal group, \(E(n)/O(n)\) \NoCaseChange{\protect\cite[{Ch. XI}]{cite2243}}.
\item\relax
\flmRefsHyperref[eczindexfamilyrel]{code:lcc}{Locally correctable code (LCC)} --- LCCs can also be defined over the real or complex numbers, and there are no complex 2-query LCCs \NoCaseChange{\protect\cite{cite1070}}.
\item\relax
\flmRefsHyperref[eczindexfamilyrel]{code:oscillators}{Bosonic code} --- Bosonic codes are quantum counterparts of analog codes.
\end{eczvaluelist}
\eczhbkcontributors{ \eczhuVVA }
\endeczcode

\eczcode{analog_reed_solomon}{Analog RS code}{~\NoCaseChange{\protect\cite{cite2244,cite2245}}}
\codefieldsection{Description}
A block code over the real or complex numbers, called a real RS or complex RS (CRS) code respectively, whose encoding generalizes the RS-code construction over finite fields.
As in the finite-field case, codewords are obtained by evaluating low-degree polynomials at distinct real or complex evaluation points.

\codefieldsection{Decoding}
\begin{eczvaluelist}
\item\relax Syndrome repairing (SR) decoder \NoCaseChange{\protect\cite{cite2246}}.
\end{eczvaluelist}
\codefieldsection{Notes}
\begin{eczvaluelist}
\item\relax CRS codes are useful for compressed sensing \NoCaseChange{\protect\cite{cite2247,cite2248}}.
\item\relax CRS codes are potentially useful for Orthogonal Frequency-Division Multiplexing (OFDM) \NoCaseChange{\protect\cite{cite2249}}.
\end{eczvaluelist}
\codefieldsection{Parent}
\begin{eczvaluelist}
\item\relax
\flmRefsHyperref[eczindexfamilyrel]{code:real_block}{Real block code}\end{eczvaluelist}
\codefieldsection{Cousin}
\begin{eczvaluelist}
\item\relax
\flmRefsHyperref[eczindexfamilyrel]{code:reed_solomon}{Reed-Solomon (RS) code} --- Analog RS codes are versions of RS codes over the real and complex numbers.
\end{eczvaluelist}
\eczhbkcontributors{ \eczhuVVA }
\endeczcode

\eczcode{antipode}{Antipode sphere packing}{~\NoCaseChange{\protect\cite{cite2250}}}
\codefieldsection{Description}
Sphere packing constructed via the antipode construction.

\codefieldsection{Parent}
\begin{eczvaluelist}
\item\relax
\flmRefsHyperref[eczindexfamilyrel]{code:sphere_packing}{Sphere packing}\end{eczvaluelist}
\codefieldsection{Cousin}
\begin{eczvaluelist}
\item\relax
\flmRefsHyperref[eczindexfamilyrel]{code:anticode}{Anticode} --- The antipode and anticode constructions are morally similar \NoCaseChange{\protect\cite{cite39}}.
\end{eczvaluelist}
\eczhbkcontributors{ \eczhuVVA }
\endeczcode

\eczcode{barnes_wall}{Barnes-Wall (BW) lattice}{~\NoCaseChange{\protect\cite{cite2203,cite2251}}}
\codefieldsection{Description}
Member of a family of \(2^{m+1}\)-dimensional lattices, denoted as BW\(_{2^{m+1}}\), that are the densest lattices known.
Members include the integer square lattice \(\mathbb{Z}^2\), \(D_4\), the Gosset \(E_8\) lattice, and the \(\Lambda_{16}\) lattice, corresponding to \(m\in\{0,1,2,3\}\), respectively.

Its automorphism group is the \flmRefsHyperref{ref409}{real Clifford group} \NoCaseChange{\protect\cite{cite2103,cite918,cite2117}}.

\codefieldsection{Protection}
BW lattices in dimension \(2^{m+1}\) have a nominal coding gain of \(2^{m/2}\).
Their kissing number is \(K_{\text{min}} = \prod_{i=1}^{m+1} (2^i + 2)\).

\codefieldsection{Parent}
\begin{eczvaluelist}
\item\relax
\flmRefsHyperref[eczindexfamilyrel]{code:points_into_lattices}{Lattice}\end{eczvaluelist}
\codefieldsection{Children}
\begin{eczvaluelist}
\item\relax
\flmRefsHyperref[eczindexfamilyrel]{code:bw32}{\(BW_{32}\) Barnes-Wall lattice}\item\relax
\flmRefsHyperref[eczindexfamilyrel]{code:lambda16}{\(\Lambda_{16}\) Barnes-Wall lattice}\item\relax
\flmRefsHyperref[eczindexfamilyrel]{code:eeight}{\(E_8\) Gosset lattice}\item\relax
\flmRefsHyperref[eczindexfamilyrel]{code:dfour}{\(D_4\) hyper-diamond lattice}\end{eczvaluelist}
\codefieldsection{Cousins}
\begin{eczvaluelist}
\item\relax
\flmRefsHyperref[eczindexfamilyrel]{code:reed_muller}{Reed-Muller (RM) code} --- BW lattices are lattice analogues of RM codes in that both can be constructed recursively via a \(|u|u+v|\) construction \NoCaseChange{\protect\cite{cite1579,cite1580}}.

\item\relax
\flmRefsHyperref[eczindexfamilyrel]{code:qubit_stabilizer}{Qubit stabilizer code} --- Stabilizer states can be mapped into the first lattice shell of a BW lattice over a cyclotomic field, while the \flmRefsHyperref{ref409}{Clifford group} is related to the symmetry group of the lattice \NoCaseChange{\protect\cite{cite2117}}.
\item\relax
\flmRefsHyperref[eczindexfamilyrel]{code:clifford_group}{Clifford group} --- Stabilizer states can be mapped into the first lattice shell of a BW lattice over a cyclotomic field, while the \flmRefsHyperref{ref409}{Clifford group} is related to the symmetry group of the lattice \NoCaseChange{\protect\cite{cite2117}}.
\item\relax
\flmRefsHyperref[eczindexfamilyrel]{code:qudit_stabilizer}{Modular-qudit stabilizer code} --- Modular-qudit stabilizer states can be mapped into the first lattice shell of a BW lattice over a cyclotomic field, while the modular-qudit Clifford group is related to the symmetry group of the lattice \NoCaseChange{\protect\cite{cite2117}}.
\item\relax
\flmRefsHyperref[eczindexfamilyrel]{code:t-designs}{\(t\)-design} --- BW lattices support Grassmannian 6-designs \NoCaseChange{\protect\cite{cite918}}.
\item\relax
\flmRefsHyperref[eczindexfamilyrel]{code:grassmannian}{Grassmannian code} --- BW lattices support Grassmannian 6-designs \NoCaseChange{\protect\cite{cite918}}.
\item\relax
\flmRefsHyperref[eczindexfamilyrel]{code:hypercubic}{\(\mathbb{Z}^n\) hypercubic lattice} --- The hypercubic lattice for \(n=2\) is the \(m=0\) BW lattice.
\item\relax
\flmRefsHyperref[eczindexfamilyrel]{code:cmr}{\(C_{m,r}\) code} --- \(C_{m,r=1}\) codes give rise to certain BW lattices \NoCaseChange{\protect\cite{cite2225,cite112}}.
\item\relax
\flmRefsHyperref[eczindexfamilyrel]{code:sidelnikov}{Real-Clifford subgroup-orbit code} --- The automorphism group of BW lattices is a subgroup of index 2 of a \flmRefsHyperref{ref409}{real Clifford group} \NoCaseChange{\protect\cite{cite2109,cite2110}} (see \NoCaseChange{\protect\cite{cite2103,cite2117}} for an explanation).
\end{eczvaluelist}
\eczhbkcontributors{ \eczhuVVA }
\endeczcode

\eczcode{bcc}{Body-centered cubic (bcc) lattice}{}
\codefieldsection{Description}
Three-dimensional lattice consisting of all points \((x,y,z)\) whose integer components are either all even or all odd.

\codefieldsection{Protection}
The bcc lattice has density \(\Delta=\pi\sqrt{3}/8\approx 0.6802\).
It exhibits the thinnest lattice covering \NoCaseChange{\protect\cite{cite2221}} in three dimensions.
It solves the lattice quantizer problem in three dimensions with \(G_3 = \frac{19}{192\cdot 2^{1/3}}\approx 0.0785\) \NoCaseChange{\protect\cite{cite2252}}.

\codefieldsection{Parent}
\begin{eczvaluelist}
\item\relax
\flmRefsHyperref[eczindexfamilyrel]{code:an_dual}{\(A_n^{\perp}\) lattice} --- The bcc lattice is the dual of the \(A_3=D_3\) fcc lattice.
\end{eczvaluelist}
\codefieldsection{Cousins}
\begin{eczvaluelist}
\item\relax
\flmRefsHyperref[eczindexfamilyrel]{code:dthree}{\(D_3\) face-centered cubic (fcc) lattice} --- The bcc and fcc lattices are dual to each other.
\item\relax
\flmRefsHyperref[eczindexfamilyrel]{code:dual_lattice}{Dual lattice} --- The bcc and fcc lattices are dual to each other.
\item\relax
\flmRefsHyperref[eczindexfamilyrel]{code:rbh}{Raussendorf-Bravyi-Harrington (RBH) cluster-state code} --- The RBH code is defined on the bcc lattice.
\item\relax
\flmRefsHyperref[eczindexfamilyrel]{code:tetrahedral_color}{Tetrahedral color code} --- The tetrahedral color code is defined on a lattice of tetrahedra carved out of a suitably colored BCC lattice \NoCaseChange{\protect\cite{cite475}}.
\end{eczvaluelist}
\eczhbkcontributors{ \eczhuVVA }
\endeczcode

\eczcode{points_into_balls}{Bounded-energy code}{}
\codefieldsection{Alternative Names}
\begin{eczvaluelist}
\item\relax Spherical cluster
\end{eczvaluelist}
\eczhIndexCodeAliasName{points_into_balls}{Spherical cluster}
\codefieldsection{Description}
Code whose codewords are points lying on or inside a real or complex ball; the squared radius is called the \textit{energy}.

\codefieldsection{Parent}
\begin{eczvaluelist}
\item\relax
\flmRefsHyperref[eczindexfamilyrel]{code:analog}{Analog code} --- Bounded-energy codes are analog codes constrained to lie on or inside a sphere.
\end{eczvaluelist}
\codefieldsection{Children}
\begin{eczvaluelist}
\item\relax
\flmRefsHyperref[eczindexfamilyrel]{code:simplex_discrete}{Simplex integer-based code} --- Points of a discrete simplex \(\Delta_{q,N}\) lie in a \(q\)-ball of radius \(N\).
\item\relax
\flmRefsHyperref[eczindexfamilyrel]{code:modulation}{Modulation scheme}\item\relax
\flmRefsHyperref[eczindexfamilyrel]{code:points_into_spheres}{Constant-energy spherical code} --- Constant-energy codes are bounded-energy codes constrained to lie on a sphere.
\end{eczvaluelist}
\eczhbkcontributors{ \eczhuVVA }
\endeczcode

\eczcode{construction_a4}{Construction \(A_4\) lattice}{}
\codefieldsection{Alternative Names}
\begin{eczvaluelist}
\item\relax Mod-four lattice
\end{eczvaluelist}
\eczhIndexCodeAliasName{construction_a4}{Mod-four lattice}
\codefieldsection{Description}
A lattice that is constructed from a linear code over \(\mathbb{Z}_4\) using \flmTerm{term}{ref114}{}{Construction \(A_4\)}.

\begin{defterm}{Construction \(A_4\)}\label{ref114}
Construction \(A_4\) converts a linear code over \(\mathbb{Z}_4\) into a lattice \NoCaseChange{\protect\cite[{Sec. 12.5.3}]{cite126}}.
In the normalization used here, a code \(C\subseteq \mathbb{Z}_4^n\) yields the lattice
\flmMathEnvironment{align}{}{
  \Lambda(C)=\frac{1}{2}\{x\in\mathbb{Z}^n:x\bmod 4\in C\}~,
}
i.e., each codeword \(c\) determines all points \(x/2\) whose residue class modulo four is \(c\).
\end{defterm}

\codefieldsection{Parent}
\begin{eczvaluelist}
\item\relax
\flmRefsHyperref[eczindexfamilyrel]{code:points_into_lattices}{Lattice}\end{eczvaluelist}
\codefieldsection{Children}
\begin{eczvaluelist}
\item\relax
\flmRefsHyperref[eczindexfamilyrel]{code:bw32}{\(BW_{32}\) Barnes-Wall lattice} --- The \(C_{m=5,r=1}\) code gives rise to the \(BW_{32}\) Barnes-Wall lattice via \flmTerm{term}{ref114}{}{Construction \(A_4\)} \NoCaseChange{\protect\cite{cite2225,cite112}}.
\item\relax
\flmRefsHyperref[eczindexfamilyrel]{code:eeight}{\(E_8\) Gosset lattice} --- The octacode yields the \(E_8\) Gosset lattice via \flmTerm{term}{ref114}{}{Construction \(A_4\)} \NoCaseChange{\protect\cite{cite2241,cite112}\protect\cite[{Exam. 12.5.13}]{cite126}}.
\item\relax
\flmRefsHyperref[eczindexfamilyrel]{code:niemeier}{Niemeier lattice} --- Niemeier lattices can be constructed from quaternary codes over \(\mathbb{Z}_4\) via \flmTerm{term}{ref114}{}{Construction \(A_4\)} \NoCaseChange{\protect\cite{cite2253}}. These codes are the Harada-Kitazume codes \NoCaseChange{\protect\cite{cite2037}}.
\end{eczvaluelist}
\codefieldsection{Cousins}
\begin{eczvaluelist}
\item\relax
\flmRefsHyperref[eczindexfamilyrel]{code:quaternary_over_z4}{Linear code over \(\mathbb{Z}_4\)} --- Every linear code over \(\mathbb{Z}_4\) yields a lattice under \flmTerm{term}{ref114}{}{Construction \(A_4\)} \NoCaseChange{\protect\cite[{Sec. 12.5.3}]{cite126}}.
\item\relax
\flmRefsHyperref[eczindexfamilyrel]{code:self_dual_lattice}{Unimodular lattice} --- Type I (type II) codes over \(\mathbb{Z}_4\) yield type I (type II) lattices under \flmTerm{term}{ref114}{}{Construction \(A_4\)} \NoCaseChange{\protect\cite[{Thm. 12.5.12}]{cite126}}.
\end{eczvaluelist}
\eczhbkcontributors{ \eczhuVVA }
\endeczcode

\eczcode{construction_a}{Construction A code}{~\NoCaseChange{\protect\cite{cite2204}}}
\codefieldsection{Alternative Names}
\begin{eczvaluelist}
\item\relax Mod-two lattice
\end{eczvaluelist}
\eczhIndexCodeAliasName{construction_a}{Mod-two lattice}
\codefieldsection{Description}
Sphere packing constructed from a binary \((n,K)\) code using \flmTerm{term}{ref127}{}{Construction A}.

\begin{defterm}{Construction A}\label{ref127}
Construction A \NoCaseChange{\protect\cite{cite39}} converts a binary code into a sphere packing.
Each binary codeword \(c\) of the code is mapped to an infinite set of points \(x\) such that \(x = c\) modulo two.
The resulting lattice is the collection of all such sets,
\flmMathEnvironment{align}{}{
  \Lambda(C) = \{ x \in \mathbb{Z}^n \,\,\vert\,\, x \bmod 2 \in C \}
  = \bigcup_{c \in C} \left( c + 2 \mathbb{Z}^n \right)~.
}
For a nonlinear code, this is generally a periodic non-lattice sphere packing.
If the underlying code is linear, then the resulting set of points forms a lattice.
\end{defterm}

\codefieldsection{Parent}
\begin{eczvaluelist}
\item\relax
\flmRefsHyperref[eczindexfamilyrel]{code:sphere_packing}{Sphere packing}\end{eczvaluelist}
\codefieldsection{Children}
\begin{eczvaluelist}
\item\relax
\flmRefsHyperref[eczindexfamilyrel]{code:eeight}{\(E_8\) Gosset lattice} --- The \([8,4,4]\) extended Hamming code yields the \(E_8\) Gosset lattice via \flmTerm{term}{ref127}{}{Construction A} \NoCaseChange{\protect\cite[{Exam. 10.5.2}]{cite115}}.
\item\relax
\flmRefsHyperref[eczindexfamilyrel]{code:dn}{\(D_n\) checkerboard lattice} --- \([n,n-1,2]\) SPC codes yield \(D_n\) checkerboard lattices via \flmTerm{term}{ref127}{}{Construction A} \NoCaseChange{\protect\cite[{Exam. 10.5.1}]{cite115}\protect\cite[{pg. 138}]{cite39}}.
\item\relax
\flmRefsHyperref[eczindexfamilyrel]{code:eseven}{\(E_7\) root lattice} --- The \([7,3,4]\) simplex code yields the \(E_7\) root lattice via \flmTerm{term}{ref127}{}{Construction A} \NoCaseChange{\protect\cite{cite1204}\protect\cite[{Exam. 10.5.3}]{cite115}}.
\end{eczvaluelist}
\codefieldsection{Cousins}
\begin{eczvaluelist}
\item\relax
\flmRefsHyperref[eczindexfamilyrel]{code:bits_into_bits}{Binary code} --- Each binary code yields a sphere packing under \flmTerm{term}{ref127}{}{Construction A}.
\item\relax
\flmRefsHyperref[eczindexfamilyrel]{code:binary_linear}{Linear binary code} --- Every binary linear code yields a lattice under \flmTerm{term}{ref127}{}{Construction A}.
\item\relax
\flmRefsHyperref[eczindexfamilyrel]{code:best}{\((10,40,4)\) Best code} --- Using \flmTerm{term}{ref127}{}{Construction A}, the Best code yields \(P_{10c}\), a non-lattice sphere packing in 10 dimensions that is the densest known \NoCaseChange{\protect\cite{cite376}\protect\cite[{pg. 140}]{cite39}}.
\item\relax
\flmRefsHyperref[eczindexfamilyrel]{code:julin12}{Julin-Golay code} --- Using \flmTerm{term}{ref127}{}{Construction A}, the Julin-Golay codes yield non-lattice sphere-packings that hold records in 9 and 11 dimensions.
\end{eczvaluelist}
\eczhbkcontributors{ \eczhuVVA }
\endeczcode

\eczcode{coxeter_todd}{Coxeter-Todd \(K_{12}\) lattice}{~\NoCaseChange{\protect\cite{cite2254}}}
\codefieldsection{Description}
Even integral lattice in dimension \(12\) that gives the densest known lattice packing.
Its automorphism group was discovered by Mitchell \NoCaseChange{\protect\cite{cite176}}.
As a real lattice, \(K_{12}\) is equivalent, up to rescaling, to its dual \(K_{12}^{\perp}\) \NoCaseChange{\protect\cite[{Ch. 4, pg. 128}]{cite39}}.
For more details, see \NoCaseChange{\protect\cite{cite177}\protect\cite[{Sec. 4.9}]{cite39}}.

\codefieldsection{Protection}
The \(K_{12}\) lattice exhibits the densest known lattice packing in 12 dimensions.

\codefieldsection{Parent}
\begin{eczvaluelist}
\item\relax
\flmRefsHyperref[eczindexfamilyrel]{code:points_into_lattices}{Lattice}\end{eczvaluelist}
\codefieldsection{Cousins}
\begin{eczvaluelist}
\item\relax
\flmRefsHyperref[eczindexfamilyrel]{code:leech}{\(\Lambda_{24}\) Leech lattice} --- The Coxeter-Todd lattice can be realized as a subset of the Leech lattice \NoCaseChange{\protect\cite[{Ch. 4, pg. 128}]{cite39}}.
\item\relax
\flmRefsHyperref[eczindexfamilyrel]{code:hexacode}{\([6,3,4]_4\) Hexacode} --- The hexacode can be used to obtain the Coxeter-Todd \(K_{12}\) lattice \NoCaseChange{\protect\cite[{Exam. 10.5.6}]{cite115}\protect\cite[{Ch. 7, pg. 198}]{cite39}}.
\item\relax
\flmRefsHyperref[eczindexfamilyrel]{code:q-ary_repetition}{\(q\)-ary repetition code} --- Applying Construction \(B_c\) to the ternary repetition code of length \(6\) over the Eisenstein integers yields the Coxeter-Todd \(K_{12}\) lattice \NoCaseChange{\protect\cite[{Ch. 7, pg. 200}]{cite39}}.
\item\relax
\flmRefsHyperref[eczindexfamilyrel]{code:self_dual_lattice}{Unimodular lattice} --- The Coxeter-Todd lattice is unimodular when defined as a complex lattice over the Eisenstein integers \NoCaseChange{\protect\cite[{Ch. 4, pg. 128}]{cite39}}.
\end{eczvaluelist}
\eczhbkcontributors{ \eczhuVVA }
\endeczcode

\eczcode{dual_lattice}{Dual lattice}{}
\codefieldsection{Alternative Names}
\begin{eczvaluelist}
\item\relax Reciprocal lattice
\item\relax Polar lattice
\end{eczvaluelist}
\eczhIndexCodeAliasName{dual_lattice}{Reciprocal lattice}
\eczhIndexCodeAliasName{dual_lattice}{Polar lattice}
\codefieldsection{Description}
For any dimensional lattice \(L\), the dual lattice is the set of vectors whose inner products with the elements of \(L\) are integers.

More technically, the dual lattice is
\flmMathEnvironment{align}{}{
L^{\perp} = \{ y\in \mathbb{R}^{n} ~|~ x \cdot y \in \mathbb{Z} ~\forall~ x \in L\},
}
where the Euclidean inner product is used.

A lattice that is contained in its dual, \(L \subseteq L^\perp\), is called \textit{integral}.
The Gram matrix of such a lattice has integer entries, and its dual is contained in a suitably scaled version of itself, \(L^{\perp} \subseteq L/\det L\).
Integral lattices are classified into \textit{even} or \textit{odd}: an even lattice has every squared norm even, while an odd lattice has at least one vector of odd squared norm.

A lattice that is equal to its dual, \(L^\perp = L\), is called \textit{unimodular} or \textit{self-dual}.

\codefieldsection{Protection}
The Gram matrix of \(L^{\perp}\) is the inverse of that of \(L\).
The generator matrix of \(L^{\perp}\) is the transposed inverse of that of \(L\).

\codefieldsection{Parent}
\begin{eczvaluelist}
\item\relax
\flmRefsHyperref[eczindexfamilyrel]{code:points_into_lattices}{Lattice}\end{eczvaluelist}
\codefieldsection{Child}
\begin{eczvaluelist}
\item\relax
\flmRefsHyperref[eczindexfamilyrel]{code:self_dual_lattice}{Unimodular lattice}\end{eczvaluelist}
\codefieldsection{Cousins}
\begin{eczvaluelist}
\item\relax
\flmRefsHyperref[eczindexfamilyrel]{code:dual}{Dual linear code} --- Dual lattices are lattice analogues of dual codes.
\item\relax
\flmRefsHyperref[eczindexfamilyrel]{code:bcc}{Body-centered cubic (bcc) lattice} --- The bcc and fcc lattices are dual to each other.
\end{eczvaluelist}
\eczhbkcontributors{ \eczhuVVA }
\endeczcode

\eczcode{fsk}{Frequency-shift keying (FSK) modulation format}{}
\codefieldsection{Alternative Names}
\begin{eczvaluelist}
\item\relax Frequency-shift keying (FSK) modulation code
\item\relax Frequency-shift keying (FSK) modulation scheme
\item\relax Frequency-shift keying (FSK) signaling format
\end{eczvaluelist}
\eczhIndexCodeAliasName{fsk}{Frequency-shift keying (FSK) modulation code}
\eczhIndexCodeAliasName{fsk}{Frequency-shift keying (FSK) modulation scheme}
\eczhIndexCodeAliasName{fsk}{Frequency-shift keying (FSK) signaling format}
\codefieldsection{Description}
A \(q\)-ary frequency-shift keying (\(q\)-FSK) encodes one \(q\)-ary digit of information into signals with \(q\) different frequencies.
In its standard orthogonal form, each symbol is carried by one of \(q\) approximately orthogonal tones over a fixed symbol interval.

\codefieldsection{Parent}
\begin{eczvaluelist}
\item\relax
\flmRefsHyperref[eczindexfamilyrel]{code:modulation}{Modulation scheme}\end{eczvaluelist}
\codefieldsection{Cousin}
\begin{eczvaluelist}
\item\relax
\flmRefsHyperref[eczindexfamilyrel]{code:quantum_fsk}{Coherent FSK (CFSK) c-q modulation format} --- Coherent FSK c-q codes are classical-quantum analogues of FSK codes.
\end{eczvaluelist}
\eczhbkcontributors{ \eczhuVVA }
\endeczcode

\eczcode{honeycomb}{Honeycomb tiling}{}
\codefieldsection{Alternative Names}
\begin{eczvaluelist}
\item\relax Hexagonal tiling
\end{eczvaluelist}
\eczhIndexCodeAliasName{honeycomb}{Hexagonal tiling}
\codefieldsection{Description}
A two-dimensional point set whose points are vertices of hexagons.
It is not a lattice since its points do not form a group under addition.
As a tiling, its dual (whose points lie at the centers of each triangle) is the triangular tiling.
The \textit{ruby tiling} is a fattened honeycomb tiling interpolating between the honeycomb tiling and triangular lattice.

\codefieldsection{Parent}
\begin{eczvaluelist}
\item\relax
\flmRefsHyperref[eczindexfamilyrel]{code:analog}{Analog code}\end{eczvaluelist}
\codefieldsection{Cousins}
\begin{eczvaluelist}
\item\relax
\flmRefsHyperref[eczindexfamilyrel]{code:hexagonal}{\(A_2\) triangular lattice} --- The Voronoi cell of the triangular lattice (honeycomb tiling) is a hexagon (triangle). Triangular and hexagonal tilings are dual to each other as tilings, i.e., the vertices of one tiling lie at the centers of faces of the other. Points of the honeycomb tiling form two triangular lattices. The ruby tiling is a fattened honeycomb tiling interpolating between the honeycomb tiling and triangular lattice.
\item\relax
\flmRefsHyperref[eczindexfamilyrel]{code:polygon}{Polygon code} --- The faces of the honeycomb tiling are hexagons, while the faces of its dual triangular tiling are triangles.
\item\relax
\flmRefsHyperref[eczindexfamilyrel]{code:floquet_xyz_ruby}{Ruby Floquet code} --- The ruby Floquet code is defined on the ruby tiling.
\item\relax
\flmRefsHyperref[eczindexfamilyrel]{code:honeycomb_floquet}{Honeycomb Floquet code} --- The honeycomb Floquet code is defined on the honeycomb tiling.
\item\relax
\flmRefsHyperref[eczindexfamilyrel]{code:nonabelian_kitaev_honeycomb}{Non-Abelian Kitaev honeycomb code} --- The Kitaev honeycomb model is defined on the honeycomb tiling.
\item\relax
\flmRefsHyperref[eczindexfamilyrel]{code:4612_color}{Truncated trihexagonal (4.6.12) color code} --- The 4.6.12 (truncated trihexagonal or square-hexagon-dodecagon) tiling is obtained by applying a fattening procedure to the honeycomb tiling \NoCaseChange{\protect\cite{cite430}}.
\item\relax
\flmRefsHyperref[eczindexfamilyrel]{code:triangular_color}{Honeycomb (6.6.6) color code} --- The 6.6.6 color code is defined on the honeycomb tiling.
\item\relax
\flmRefsHyperref[eczindexfamilyrel]{code:kitaev_honeycomb}{Kitaev honeycomb code} --- The Kitaev honeycomb code is defined on the honeycomb tiling.
\end{eczvaluelist}
\eczhbkcontributors{ \eczhuVVA }
\endeczcode

\eczcode{integers_into_integers}{Integer-based code}{}
\codefieldsection{Alternative Names}
\begin{eczvaluelist}
\item\relax \(\mathbb{Z}\) code
\end{eczvaluelist}
\eczhIndexCodeAliasName{integers_into_integers}{\(\mathbb{Z}\) code}
\codefieldsection{Description}
Encodes \(K\) states (codewords) into \(n\) integer coordinates, i.e., into a subset of \(\mathbb{Z}^n\).
\codefieldsection{Protection}
A common distance between integer vectors is the \(\ell_1\) distance \NoCaseChange{\protect\cite{cite1675,cite2255}}.

\codefieldsection{Parent}
\begin{eczvaluelist}
\item\relax
\flmRefsHyperref[eczindexfamilyrel]{code:analog}{Analog code}\end{eczvaluelist}
\codefieldsection{Child}
\begin{eczvaluelist}
\item\relax
\flmRefsHyperref[eczindexfamilyrel]{code:simplex_discrete}{Simplex integer-based code}\end{eczvaluelist}
\codefieldsection{Cousin}
\begin{eczvaluelist}
\item\relax
\flmRefsHyperref[eczindexfamilyrel]{code:rotor}{Rotor code} --- Rotor codes are quantum counterparts of integer codes.
\end{eczvaluelist}
\eczhbkcontributors{ \eczhuVVA }
\endeczcode

\eczcode{points_into_lattices}{Lattice}{}
\codefieldsection{Description}
Encodes states (codewords) in coordinates of an \(n\)-dimensional lattice, i.e., a discrete set of points in Euclidean space \(\mathbb{R}^n\) that forms a group under vector addition when translated so that one point is at the origin. The number of codewords may be infinite because the ambient Euclidean space is unbounded, so various restricted versions have to be constructed in practice. Since lattices are closed under addition, lattice-based codes can be thought of as linear codes over the reals.

An \(n\)-dimensional lattice-based code can be defined using a generator matrix \(G\) of rank \(n\), where the rows of \(G\) are the lattice translation vectors \(g_i\).
Any lattice point \(x\) is a linear combination of translation vectors with integer coefficients \(c_i\), \(x = c_1 g_1 + c_2 g_2 + \cdots + c_n g_n\).
A lattice-based code can also be defined using the Gram matrix \(GG^T\).

The \textit{automorphism group} of a lattice is the group of all isometries that preserve the origin and map the lattice into itself.
This group is a subgroup of the orthogonal group \(O(n)\), which is the group of isometries in Euclidean space.
An orthogonal matrix \(U\) leaves the lattice invariant if there exists an integer matrix \(A\) of determinant \(\pm 1\) such that
\flmMathEnvironment{align}{}{
  AG=GU~.\label{ref2256}
}
The \textit{affine automorphism group} is the group obtained from adding lattice translations to the automorphism group.

Two lattices with generator matrices \(G,G^{\prime}\) are \textit{equivalent} if one is obtained from the other by the following combination of an orthogonal matrix \(U\) and a change of scale \(c\neq 0\),
\flmMathEnvironment{align}{}{
  G^{\prime}=cAGU~,
}
where \(A\) is an integer matrix of determinant \(\pm 1\).

\codefieldsection{Protection}
Lattices are characterized by the \textit{minimum (Euclidean) distance} \(d_{\text{min}}\) between two lattice points, the \textit{kissing number} \(K_{\text{min}}\) of nearest neighbors at each lattice point, and the \textit{volume} \(V=|\det G|\), which is the volume of the lattice's \textit{fundamental region} that can be used to tile all of \(\mathbb{R}^n\).

The minimum Euclidean distance is an analogue of the minimum distance of binary codes. Half of this distance is called the \textit{packing radius}.

The \textit{nominal coding gain} \(\gamma_{c}\) (a.k.a. \textit{Hermite parameter}) of an \(n\)-dimensional lattice is
\flmMathEnvironment{align}{}{
  \gamma_{c}=\left(d_{\text{min}}/\sqrt[n]{V}\right)^{2}~,
}
characterizing the ratio of the level of protection to the required spatial resources.
The \textit{density} of a lattice is the fraction of the total volume of space that is occupied by spheres of \textit{packing radius} \(\frac{1}{2}d_{\text{min}}\) centered at each point in the lattice,
\flmMathEnvironment{align}{}{
  \Delta=\frac{\text{volume of one sphere}}{V}~.
}

The covering radius of a lattice is defined similarly as above, but with the spheres' \textit{covering radius} now being the smallest one such that the spheres cover all space.
In general, finding the covering radius of a lattice is \(NP\)-hard \NoCaseChange{\protect\cite{cite2257}}.

The \textit{lattice quantizer problem} is to find a lattice whose \textit{fundamental Voronoi cell} \(\Pi\), the Voronoi cell at the origin, has the smallest possible normalized second moment,
\flmMathEnvironment{align}{}{
  G(\Pi)=\frac{\frac{1}{n}\int_{\Pi}x\cdot x\,\textnormal{d}x}{\text{Vol}(\Pi)^{1+2/n}}\,.
}
Higher-dimensional lattices yield quantizers with lower normalized second moments than the 1D integer lattice \NoCaseChange{\protect\cite{cite2258,cite2259}}.

The \textit{shortest vector problem} (SVP) asks for the shortest nonzero vector in a given lattice and is related to cryptographic protocols.
Solving SVP up to an error independent of lattice dimension is NP-complete \NoCaseChange{\protect\cite{cite2260,cite2261}}.
The Lenstra-Lenstra-Lovasz (LLL) algorithm solves SVP in polynomial time, but up to an error exponential in the dimension \NoCaseChange{\protect\cite{cite2262}}; see the book \NoCaseChange{\protect\cite{cite2263}}.

\codefieldsection{Rate}
Lattices with minimal-distance decoding achieve the capacity of the AWGN channel \NoCaseChange{\protect\cite{cite2264,cite2265,cite2266,cite2267}}.

\codefieldsection{Decoding}
\begin{eczvaluelist}
\item\relax Sphere decoder \NoCaseChange{\protect\cite{cite2268,cite2269}}.
\end{eczvaluelist}
\codefieldsection{Notes}
\begin{eczvaluelist}
\item\relax See books \NoCaseChange{\protect\cite{cite39,cite2270,cite2271}} for introductions and overviews of lattices.
\item\relax See LMFDB \NoCaseChange{\protect\cite{cite1816}} and Catalogue of Lattices \NoCaseChange{\protect\cite{cite2272}} for databases of lattices.
\item\relax Tables of bounds on kissing numbers \NoCaseChange{\protect\cite{cite2273}}. Popular summary of bounds on kissing numbers in 17-21 dimensions in \flmHref{https://www.quantamagazine.org/mathematicians-discover-new-way-for-spheres-to-kiss-20250115/}{Quanta Magazine}.
\item\relax See Refs. \NoCaseChange{\protect\cite{cite2274,cite2275,cite2276}} for various examples and implementations in Magma.
\end{eczvaluelist}
\codefieldsection{Parents}
\begin{eczvaluelist}
\item\relax
\flmRefsHyperref[eczindexfamilyrel]{code:sphere_packing}{Sphere packing} --- Lattices are sphere packings whose points form a group under addition.
\item\relax
\flmRefsHyperref[eczindexfamilyrel]{code:group_linear}{Linear code over \(G\)} --- Lattice-based codes are linear codes over \(G=\mathbb{R}\). Because any orthogonal matrix leaving the lattice invariant has a corresponding integer matrix (see lattice description), integer representations of groups can be used to obtain lattices \NoCaseChange{\protect\cite[{Ch. 3, Sec. 4.2}]{cite39}}.
\end{eczvaluelist}
\codefieldsection{Children}
\begin{eczvaluelist}
\item\relax
\flmRefsHyperref[eczindexfamilyrel]{code:construction_a4}{Construction \(A_4\) lattice}\item\relax
\flmRefsHyperref[eczindexfamilyrel]{code:an_dual}{\(A_n^{\perp}\) lattice}\item\relax
\flmRefsHyperref[eczindexfamilyrel]{code:barnes_wall}{Barnes-Wall (BW) lattice}\item\relax
\flmRefsHyperref[eczindexfamilyrel]{code:coxeter_todd}{Coxeter-Todd \(K_{12}\) lattice}\item\relax
\flmRefsHyperref[eczindexfamilyrel]{code:dual_lattice}{Dual lattice}\item\relax
\flmRefsHyperref[eczindexfamilyrel]{code:root}{Root lattice}\end{eczvaluelist}
\codefieldsection{Cousins}
\begin{eczvaluelist}
\item\relax
\flmRefsHyperref[eczindexfamilyrel]{code:2pt_homogeneous}{Two-point homogeneous-space code} --- The Levenshtein bound \NoCaseChange{\protect\cite{cite2277,cite2278,cite2088,cite171,cite914}} and Cohn-Elkies LP bound \NoCaseChange{\protect\cite{cite2279}} can be derived for sphere packings by thinking of \(\mathbb{R}^n\) as a homogeneous space of the Euclidean group by the orthogonal group, \(E(n)/O(n)\) \NoCaseChange{\protect\cite[{Ch. XI}]{cite2243}}.

\item\relax
\flmRefsHyperref[eczindexfamilyrel]{code:qam}{Quadrature-amplitude modulation (QAM) format} --- QAM encodings often consist of lattice constellations, i.e., finite sets of points scooped out of an infinite 2D lattice.
\item\relax
\flmRefsHyperref[eczindexfamilyrel]{code:hypercubic}{\(\mathbb{Z}^n\) hypercubic lattice} --- The generator matrix of a lattice-based code serves as a linear transformation that can be applied to the hypercubic lattice to obtain said code \NoCaseChange{\protect\cite[{Ch. 10}]{cite115}}.
\item\relax
\flmRefsHyperref[eczindexfamilyrel]{code:symmetric_space}{Symmetric-space code} --- Upper bounds on kissing numbers can be worked out by treating the sphere as a symmetric space \NoCaseChange{\protect\cite{cite2280}}.

\item\relax
\flmRefsHyperref[eczindexfamilyrel]{code:lexicographic}{Lexicographic code} --- Lexicographic codes are \(q\)-ary analogues of laminated lattices \NoCaseChange{\protect\cite{cite43}\protect\cite[{pg. 162}]{cite41}}.
\item\relax
\flmRefsHyperref[eczindexfamilyrel]{code:lattice_shell}{Lattice-shell code} --- Lattice-shell codes consist of lattice-point vectors, initially all of the same norm, normalized to one.
\item\relax
\flmRefsHyperref[eczindexfamilyrel]{code:quantum_lattice}{Quantum lattice code} --- Quantum lattice codes can be thought of as quantum analogues of lattices because they store information in quantum superpositions of points on a lattice in quantum phase space.
\end{eczvaluelist}
\eczhbkcontributors{ \eczhuVVA }
\endeczcode

\eczcode{modulation}{Modulation scheme}{}
\codefieldsection{Description}
A sphere packing mapped into a time-dependent electromagnetic signal \NoCaseChange{\protect\cite{cite193,cite194}}.
There is a close relation between abstract real-space encodings and modulation schemes, and certain simple sphere packings are often synonymous with their corresponding modulation schemes.

Linear modulation schemes encode points into amplitudes of basis waveforms.
\textit{Pulse-amplitude modulation (PAM)} associates each point with a real-valued amplitude of one quadrature of an electromagnetic waveform \NoCaseChange{\protect\cite[{Sec. 10.5}]{cite194}}.
\textit{Quadrature amplitude modulation (QAM)} associates each point in a two-dimensional constellation with a complex-valued two-quadrature amplitude of a band-limited signal \NoCaseChange{\protect\cite[{Ch. 16}]{cite194}}.

\codefieldsection{Notes}
\begin{eczvaluelist}
\item\relax See books \NoCaseChange{\protect\cite{cite2281,cite2282}} for an introduction to modulation schemes.
\end{eczvaluelist}
\codefieldsection{Parent}
\begin{eczvaluelist}
\item\relax
\flmRefsHyperref[eczindexfamilyrel]{code:points_into_balls}{Bounded-energy code}\end{eczvaluelist}
\codefieldsection{Children}
\begin{eczvaluelist}
\item\relax
\flmRefsHyperref[eczindexfamilyrel]{code:fsk}{Frequency-shift keying (FSK) modulation format}\item\relax
\flmRefsHyperref[eczindexfamilyrel]{code:qam}{Quadrature-amplitude modulation (QAM) format}\item\relax
\flmRefsHyperref[eczindexfamilyrel]{code:ppm}{Pulse-position modulation (PPM) format}\item\relax
\flmRefsHyperref[eczindexfamilyrel]{code:psk}{Phase-shift keying (PSK) modulation format} --- PSK is a modulation whose constellation consists of points arranged equidistantly on a circle.
\end{eczvaluelist}
\codefieldsection{Cousins}
\begin{eczvaluelist}
\item\relax
\flmRefsHyperref[eczindexfamilyrel]{code:leech}{\(\Lambda_{24}\) Leech lattice} --- Codewords of the Leech lattice have been proposed to be used for a modulation scheme \NoCaseChange{\protect\cite{cite2210}}.
\item\relax
\flmRefsHyperref[eczindexfamilyrel]{code:bosonic_classical_into_quantum}{Bosonic c-q code} --- Classical modulation schemes transmit classical signals over classical channels, while bosonic c-q modulation formats transmit quantum states over quantum channels and can use quantum-enhanced receivers.
\item\relax
\flmRefsHyperref[eczindexfamilyrel]{code:coherent_constellation}{Coherent-state constellation code} --- Coherent-state constellation codes are quantum counterparts of modulation schemes in that their codewords are superpositions of points in a constellation. Additionally, analog codes that achieve AWGN capacity \NoCaseChange{\protect\cite{cite2283}} can be used to develop capacity-achieving concatenations of coherent-state constellation codes with quantum polar codes \NoCaseChange{\protect\cite{cite930,cite931}}.
\end{eczvaluelist}
\eczhbkcontributors{ \eczhuVVA }
\endeczcode

\eczcode{niemeier}{Niemeier lattice}{~\NoCaseChange{\protect\cite{cite2284}}}
\codefieldsection{Description}
One of the 24 positive-definite even unimodular lattices of rank 24.
The 24 lattices are \(D_{24}\), \(D_{16}E_8\), \(E_8^3\), \(A_{24}\), \(D_{12}^2\), \(A_{17}E_7\), \(D_{10}E_7^2\), \(A_{15}D_9\), \(D_8^3\), \(A_{12}^2\), \(A_{11}D_7E_6\), \(E_6^4\), \(A_9^2D_6\), \(D_6^4\), \(A_8^3\), \(A_7^2D_5^2\), \(A_6^4\), \(A_5^4D_4\), \(D_4^6\), \(A_4^6\), \(A_3^8\), \(A_2^{12}\), \(A_1^{24}\), and \(\Lambda_{24}\) (the Leech lattice) \NoCaseChange{\protect\cite[{Table 16.1}]{cite39}}.

\codefieldsection{Parents}
\begin{eczvaluelist}
\item\relax
\flmRefsHyperref[eczindexfamilyrel]{code:self_dual_lattice}{Unimodular lattice} --- Niemeier lattices are even and unimodular.
\item\relax
\flmRefsHyperref[eczindexfamilyrel]{code:construction_a4}{Construction \(A_4\) lattice} --- Niemeier lattices can be constructed from quaternary codes over \(\mathbb{Z}_4\) via \flmTerm{term}{ref114}{}{Construction \(A_4\)} \NoCaseChange{\protect\cite{cite2253}}. These codes are the Harada-Kitazume codes \NoCaseChange{\protect\cite{cite2037}}.
\end{eczvaluelist}
\codefieldsection{Child}
\begin{eczvaluelist}
\item\relax
\flmRefsHyperref[eczindexfamilyrel]{code:leech}{\(\Lambda_{24}\) Leech lattice} --- The Leech lattice is the Niemeier lattice with minimal norm 4 \NoCaseChange{\protect\cite{cite2037}}. Every Niemeier lattice is a sublattice of the Leech lattice \NoCaseChange{\protect\cite{cite2207,cite2037}}. In the holy construction for the Niemeier lattice \(A_2^{12}\), the combinations for which the sum of all coefficients is zero form a copy of the Leech lattice \NoCaseChange{\protect\cite[{Ch. 24, pg. 510}]{cite39}}. The Leech lattice can be constructed from pseudo Golay codes via \flmTerm{term}{ref114}{}{Construction \(A_4\)} \NoCaseChange{\protect\cite{cite122,cite1198}}. The Leech lattice can be constructed from the extended quaternary Golay code via \flmTerm{term}{ref114}{}{Construction \(A_4\)} \NoCaseChange{\protect\cite[{3rd Ed., pg. xxxiii}]{cite39}} (see also \NoCaseChange{\protect\cite{cite1659,cite112,cite1198}}).
\end{eczvaluelist}
\codefieldsection{Cousins}
\begin{eczvaluelist}
\item\relax
\flmRefsHyperref[eczindexfamilyrel]{code:self_dual}{Self-dual linear code} --- The nine inequivalent \([24,12]\) doubly even self-dual codes \NoCaseChange{\protect\cite{cite2026}} yield certain Niemeier lattices via \flmTerm{term}{ref127}{}{Construction A} \NoCaseChange{\protect\cite{cite2037}}. Niemeier lattices can be constructed from ternary self-dual codes of length 24 \NoCaseChange{\protect\cite{cite2038}}.
\item\relax
\flmRefsHyperref[eczindexfamilyrel]{code:self_dual_over_zq}{Self-dual code over \(\mathbb{Z}_q\)} --- Extremal Type II self-dual codes of length 24 over \(\mathbb{Z}_6\) yield Niemeier lattices \NoCaseChange{\protect\cite{cite2285}}.
\item\relax
\flmRefsHyperref[eczindexfamilyrel]{code:octacode}{Octacode} --- The octacode is the glue code for the Niemeier lattice \(A_4^6\) \NoCaseChange{\protect\cite[{3rd Ed., pg. liv}]{cite39}}.
\item\relax
\flmRefsHyperref[eczindexfamilyrel]{code:tetracode}{\([4,2,3]_3\) Tetracode} --- The tetracode is the glue code for the Niemeier lattice \(E_6^4\) \NoCaseChange{\protect\cite[{Ch. 16, pg. 408}]{cite39}}.
\item\relax
\flmRefsHyperref[eczindexfamilyrel]{code:hexacode}{\([6,3,4]_4\) Hexacode} --- The hexacode is the glue code for the Niemeier lattice \(D_4^6\) \NoCaseChange{\protect\cite[{Ch. 16, pg. 408}]{cite39}}.
\item\relax
\flmRefsHyperref[eczindexfamilyrel]{code:ternary_golay}{\([11,6,5]_3\) Ternary Golay code} --- The extended ternary Golay code is the glue code for the Niemeier lattice \(A^{12}_2\) \NoCaseChange{\protect\cite[{Ch. 16, pg. 408}]{cite39}}.
\item\relax
\flmRefsHyperref[eczindexfamilyrel]{code:extended_golay}{\([24, 12, 8]\) Extended Golay code} --- The extended Golay code is the glue code for the Niemeier lattice \(A_1^{24}\) \NoCaseChange{\protect\cite[{Ch. 16, pg. 408}]{cite39}}.
\item\relax
\flmRefsHyperref[eczindexfamilyrel]{code:harada_kitazume}{Harada-Kitazume code} --- Niemeier lattices can be constructed from quaternary codes over \(\mathbb{Z}_4\) via \flmTerm{term}{ref114}{}{Construction \(A_4\)} \NoCaseChange{\protect\cite{cite2253}}. These codes are the Harada-Kitazume codes \NoCaseChange{\protect\cite{cite2037}}.
\end{eczvaluelist}
\eczhbkcontributors{ \eczhuVVA }
\endeczcode

\eczcode{pam}{Pulse-amplitude modulation (PAM) format}{}
\codefieldsection{Alternative Names}
\begin{eczvaluelist}
\item\relax Pulse-amplitude modulation (PAM) code
\item\relax Pulse-amplitude modulation (PAM) scheme
\item\relax Pulse-amplitude modulation (PAM) signaling format
\end{eczvaluelist}
\eczhIndexCodeAliasName{pam}{Pulse-amplitude modulation (PAM) code}
\eczhIndexCodeAliasName{pam}{Pulse-amplitude modulation (PAM) scheme}
\eczhIndexCodeAliasName{pam}{Pulse-amplitude modulation (PAM) signaling format}
\codefieldsection{Description}
Encodes a \(q\)-ary digit into a constellation of equally spaced points on the real line.
A standard \(q\)-PAM constellation can be written as \(\{(2i-q-1)\alpha\}_{i=1}^{q}\) for some real scaling factor \(\alpha\); for \(q=8\), this yields \(\{ \pm \alpha,\pm 3\alpha,\pm 5\alpha, \pm 7\alpha \}\).
The points in the constellation are typically associated with one quadrature of an electromagnetic signal.

\codefieldsection{Parent}
\begin{eczvaluelist}
\item\relax
\flmRefsHyperref[eczindexfamilyrel]{code:qam}{Quadrature-amplitude modulation (QAM) format} --- PAM codes can be thought of as QAM codes restricted to the real line. A \(q\times q\)-QAM code is informationally equivalent to two \(q\)-PAM codes.
\end{eczvaluelist}
\codefieldsection{Cousins}
\begin{eczvaluelist}
\item\relax
\flmRefsHyperref[eczindexfamilyrel]{code:hyperbolic}{Hyperbolic sphere packing} --- Hyperbolic PAM constellations may yield improved performance over Euclidean ones \NoCaseChange{\protect\cite{cite2286}}.
\item\relax
\flmRefsHyperref[eczindexfamilyrel]{code:bpsk}{Binary PSK (BPSK) modulation format} --- BPSK for real \(\alpha\) is the simplest non-trivial PAM encoding.
\end{eczvaluelist}
\eczhbkcontributors{ \eczhuVVA }
\endeczcode

\eczcode{qam}{Quadrature-amplitude modulation (QAM) format}{}
\codefieldsection{Alternative Names}
\begin{eczvaluelist}
\item\relax Quadrature-amplitude modulation (QAM) code
\item\relax Quadrature-amplitude modulation (QAM) scheme
\item\relax Quadrature-amplitude modulation (QAM) signaling format
\end{eczvaluelist}
\eczhIndexCodeAliasName{qam}{Quadrature-amplitude modulation (QAM) code}
\eczhIndexCodeAliasName{qam}{Quadrature-amplitude modulation (QAM) scheme}
\eczhIndexCodeAliasName{qam}{Quadrature-amplitude modulation (QAM) signaling format}
\codefieldsection{Description}
Encodes into a finite set of points in \(\mathbb{R}^{2}\), often treated as \(\mathbb{C}\).
Each point is associated with a complex amplitude of an electromagnetic signal, so information is encoded jointly in the in-phase and quadrature components \NoCaseChange{\protect\cite[{Ch. 16}]{cite194}}.

QAM schemes with \(q\) constellation points are often called \(q\)-QAM, and \(q\) is often chosen as a power of two to facilitate concatenation with a binary code.

\codefieldsection{Rate}
High-order QAM, especially with appropriate shaping and coding, can operate close to Shannon AWGN capacity at high signal-to-noise ratio \NoCaseChange{\protect\cite[{Fig. 11.8}]{cite2287}}.
\codefieldsection{Realizations}
\begin{eczvaluelist}
\item\relax Optical communication (e.g., Ref. \NoCaseChange{\protect\cite{cite310}}).
\item\relax Telephone-line modems: 1971 Codex 9600C and international standard V.29 used 16-QAM \NoCaseChange{\protect\cite{cite311}}.
\end{eczvaluelist}
\codefieldsection{Parent}
\begin{eczvaluelist}
\item\relax
\flmRefsHyperref[eczindexfamilyrel]{code:modulation}{Modulation scheme}\end{eczvaluelist}
\codefieldsection{Child}
\begin{eczvaluelist}
\item\relax
\flmRefsHyperref[eczindexfamilyrel]{code:pam}{Pulse-amplitude modulation (PAM) format} --- PAM codes can be thought of as QAM codes restricted to the real line. A \(q\times q\)-QAM code is informationally equivalent to two \(q\)-PAM codes.
\end{eczvaluelist}
\codefieldsection{Cousins}
\begin{eczvaluelist}
\item\relax
\flmRefsHyperref[eczindexfamilyrel]{code:points_into_lattices}{Lattice} --- QAM encodings often consist of lattice constellations, i.e., finite sets of points scooped out of an infinite 2D lattice.
\item\relax
\flmRefsHyperref[eczindexfamilyrel]{code:multimodegkp}{Gottesman-Kitaev-Preskill (GKP) code} --- Finite-energy GKP codes are quantum counterparts of lattice-based QAM codes in that both use a subset of points on a lattice.
\item\relax
\flmRefsHyperref[eczindexfamilyrel]{code:gray}{Gray code} --- 2D Gray codes are often concatenated with \(n=1\) lattice-based QAM codes so that the Hamming distance between the bitstrings encoded into the points is a discretized version of the Euclidean distance between the points.
\item\relax
\flmRefsHyperref[eczindexfamilyrel]{code:hyperbolic}{Hyperbolic sphere packing} --- Hyperbolic QAM constellations may yield improved performance over Euclidean ones \NoCaseChange{\protect\cite{cite2288}}.
\item\relax
\flmRefsHyperref[eczindexfamilyrel]{code:turbo}{Turbo code} --- Turbo codes concatenated with QAM codes offer a substantial coding gain \NoCaseChange{\protect\cite{cite2075}}.
\end{eczvaluelist}
\eczhbkcontributors{ \eczhuVVA }
\endeczcode

\eczcode{real_block}{Real block code}{~\NoCaseChange{\protect\cite{cite2244,cite2245}\protect\cite[{pg. 321}]{cite41}}}
\codefieldsection{Alternative Names}
\begin{eczvaluelist}
\item\relax Real-number block code
\item\relax Real code
\item\relax \(\mathbb{R}\) code
\end{eczvaluelist}
\eczhIndexCodeAliasName{real_block}{Real-number block code}
\eczhIndexCodeAliasName{real_block}{Real code}
\eczhIndexCodeAliasName{real_block}{\(\mathbb{R}\) code}
\codefieldsection{Description}
A block code encoding a \(k\)-dimensional vector of real or complex numbers into an \(n\)-dimensional real or complex vector space, typically by a linear or polynomial map.

\codefieldsection{Parent}
\begin{eczvaluelist}
\item\relax
\flmRefsHyperref[eczindexfamilyrel]{code:analog}{Analog code} --- Real-number block codes encode continuous sets of real or complex numbers into a real or complex vector space.
\end{eczvaluelist}
\codefieldsection{Child}
\begin{eczvaluelist}
\item\relax
\flmRefsHyperref[eczindexfamilyrel]{code:analog_reed_solomon}{Analog RS code}\end{eczvaluelist}
\codefieldsection{Cousin}
\begin{eczvaluelist}
\item\relax
\flmRefsHyperref[eczindexfamilyrel]{code:analog_stabilizer}{Analog stabilizer code} --- Analog stabilizer codes are quantum counterparts of real block codes.
\end{eczvaluelist}
\eczhbkcontributors{ \eczhuVVA }
\endeczcode

\eczcode{root}{Root lattice}{}
\codefieldsection{Description}
A lattice that is symmetric under a specific crystallographic reflection group; see \NoCaseChange{\protect\cite[{Table 4.1}]{cite39}} for the list of crystallographic reflection groups and their corresponding root lattices.
The root-lattice family consists of lattices \(A_n\), \(\mathbb{Z}^n\), or \(D_n\) for dimension \(n\), or \(E_{i}\) for \(i\in\{6,7,8\}\).
Their generator matrices can be taken to be the root matrices of the corresponding reflection groups.

\codefieldsection{Protection}
The densest lattice packings in dimensions \(3\) through \(8\) are root lattices \NoCaseChange{\protect\cite[{Table 1.1}]{cite39}}.

\codefieldsection{Parent}
\begin{eczvaluelist}
\item\relax
\flmRefsHyperref[eczindexfamilyrel]{code:points_into_lattices}{Lattice}\end{eczvaluelist}
\codefieldsection{Children}
\begin{eczvaluelist}
\item\relax
\flmRefsHyperref[eczindexfamilyrel]{code:eeight}{\(E_8\) Gosset lattice}\item\relax
\flmRefsHyperref[eczindexfamilyrel]{code:hypercubic}{\(\mathbb{Z}^n\) hypercubic lattice}\item\relax
\flmRefsHyperref[eczindexfamilyrel]{code:an}{\(A_n\) lattice}\item\relax
\flmRefsHyperref[eczindexfamilyrel]{code:dn}{\(D_n\) checkerboard lattice}\item\relax
\flmRefsHyperref[eczindexfamilyrel]{code:eseven}{\(E_7\) root lattice}\item\relax
\flmRefsHyperref[eczindexfamilyrel]{code:esix}{\(E_6\) root lattice}\end{eczvaluelist}
\eczhbkcontributors{ \eczhuVVA }
\endeczcode

\eczcode{simplex_discrete}{Simplex integer-based code}{~\NoCaseChange{\protect\cite{cite2289,cite1007,cite2290}}}
\codefieldsection{Alternative Names}
\begin{eczvaluelist}
\item\relax Multiset code
\end{eczvaluelist}
\eczhIndexCodeAliasName{simplex_discrete}{Multiset code}
\codefieldsection{Description}
An integer-based code over the \(\ell_1\) metric whose codewords are restricted to lie in a \flmRefsHyperref{ref655}{discrete simplex}.

\begin{defterm}{Discrete simplex}\label{ref2291}\label{ref655}
The discrete simplex \(\Delta_{q,N}\) of dimension \(q-1\) and discretization \(N\) is the set of length-\(q\) non-negative integer strings whose coordinates add up to \(N\),
\flmMathEnvironment{align}{}{
  \Delta_{q,N}=\left\{ \mathbf{n}\in\mathbb{Z}_{\geq 0}^{q}\,|\,\sum_j n_j =N\right\}~.
}
The number of points in a discrete simplex is
\flmMathEnvironment{align}{}{
  |\Delta_{q,N}|=\binom{N+q-1}{q-1}\,.
}
\end{defterm}

\codefieldsection{Protection}
There is a \flmRefsHyperref{ref85}{GV bound} for simplex integer-based codes over the \(\ell_1\) metric \NoCaseChange{\protect\cite{cite2292,cite2293,cite2294,cite2295}}.
Because all codewords have the same coordinate sum, the \(\ell_1\) distance between any two codewords is always even, so the natural error-correction radius is often stated in terms of half that distance.
Existence of certain codes is guaranteed by constructing Sidon sets per the Bose-Chowla theorem \NoCaseChange{\protect\cite{cite2296}\protect\cite[{Thm. 14}]{cite1007}}.

\codefieldsection{Parents}
\begin{eczvaluelist}
\item\relax
\flmRefsHyperref[eczindexfamilyrel]{code:integers_into_integers}{Integer-based code}\item\relax
\flmRefsHyperref[eczindexfamilyrel]{code:points_into_balls}{Bounded-energy code} --- Points of a discrete simplex \(\Delta_{q,N}\) lie in a \(q\)-ball of radius \(N\).
\end{eczvaluelist}
\codefieldsection{Cousins}
\begin{eczvaluelist}
\item\relax
\flmRefsHyperref[eczindexfamilyrel]{code:dna}{DNA storage code} --- Simplex integer-based codewords are intended to model multisets of DNA molecules \NoCaseChange{\protect\cite{cite1007}}. Points in a \flmRefsHyperref{ref655}{discrete simplex} are in one-to-one correspondence to multisets because their coordinates denote the multiplicity of each element in a given multiset \NoCaseChange{\protect\cite{cite1007}}.
\item\relax
\flmRefsHyperref[eczindexfamilyrel]{code:simplex_spherical}{Simplex spherical code} --- Codewords of simplex integer-based codes are restricted to lie in a \flmRefsHyperref{ref655}{discrete simplex}.
\item\relax
\flmRefsHyperref[eczindexfamilyrel]{code:simplex}{\([2^m-1,m,2^{m-1}]\) simplex code} --- Codewords of simplex integer-based codes are restricted to lie in a \flmRefsHyperref{ref655}{discrete simplex}.
\item\relax
\flmRefsHyperref[eczindexfamilyrel]{code:permutation_invariant}{Permutation-invariant (PI) code} --- Simplex integer-based codes can be partitioned into qudit PI codewords whose error-correction is guaranteed by the Tverberg theorem \NoCaseChange{\protect\cite[{Thm. VII.5}]{cite500}}.
\end{eczvaluelist}
\eczhbkcontributors{ \eczhuVVA }
\endeczcode

\eczcode{sphere_packing}{Sphere packing}{}
\codefieldsection{Description}
An analog code whose points can be thought of as forming centers of spheres that pack Euclidean space \(\mathbb{R}^n\).
Such packings can also be interpreted as complex sphere packings by mapping pairs of real coordinates to the complex plane.
Sphere packings provide ways of encoding digital or analog information into the frequency, amplitude, and phase of one or more analog waveforms for transmission through, e.g., an optical fiber or free space.
This is due to Kotelnikov's \NoCaseChange{\protect\cite{cite2297}} and Shannon's \NoCaseChange{\protect\cite{cite2298}} fundamental observation that a discretized electromagnetic signal of finite bandwidth and average power \(P\) can be represented as a vector in \(\mathbb{R}^n\) with squared norm \(nP\).
Questions of capacity of electromagnetic communication channels then translate to packing problems in \(\mathbb{R}^n\) \NoCaseChange{\protect\cite{cite39}}.

In the electromagnetic context, the information stored in the code is called the \textit{bitstream}, coordinates used for encoding are often called \textit{signal points} and form a \textit{constellation}, and \(\mathbb{R}^n\) is called the \textit{signal space}.

\codefieldsection{Protection}
Sphere packings can be used to transmit information using electromagnetic signals. The primary noise channel for such signals is the \textit{additive white Gaussian channel (AWGN)}, which adds a random Gaussian-distributed displacement with variance \(\sigma^2\) to each signal point.
Protection of a constellation thus depends on how far apart its points are in terms of the Euclidean distance.
As usual, there is a tradeoff between packing of space and level of protection.
For a given \(n,M,P\), the \textit{Gaussian channel coding problem} asks to find a set of \(M\) codewords of norm \(\leq nP\) that minimizes the error probability during transmission; see \NoCaseChange{\protect\cite[{Ch. 3}]{cite39}}.

The \textit{minimum distance} \(d\) of a constellation is the infimum over distances between any two of its points. The corresponding \textit{packing radius} is \(d/2\).
The \textit{density} \(\Delta\) is the fraction of the total volume of space that is occupied by spheres of packing radius centered at each point in the sphere packing.
Defining a density for infinite constellations can be done using a limit \NoCaseChange{\protect\cite[{pg. 349}]{cite115}}.

The Kabatiansky-Levenshtein bound \NoCaseChange{\protect\cite{cite2299}\protect\cite[{Ch. 9}]{cite39}} says that any sphere packing must satisfy \(\frac{1}{n}\log_{2}\Delta\lesssim-0.5990\) asymptotically with dimension \(n\).
This bound improved over the Rogers bound \NoCaseChange{\protect\cite{cite2300}} and the Sidelnikov bound \NoCaseChange{\protect\cite{cite2301}}. Afterwards came the Levenshtein bound \NoCaseChange{\protect\cite{cite2277,cite2278,cite2088,cite171,cite914}}, Cohn-Elkies LP bounds \NoCaseChange{\protect\cite{cite2279}}, and, most recently, an improvement of the Rogers bound \NoCaseChange{\protect\cite{cite2302}}.
For more details, see \NoCaseChange{\protect\cite[{Ch. 10.4}]{cite115}\protect\cite[{Ch. 1}]{cite39}}.
There exists an \(n\)-dimensional sphere packing whose density is at least \(c n^2 / 2^n\) for \(c\) being a universal constant \NoCaseChange{\protect\cite{cite2303}}.

The \textit{covering problem} asks how one can cover all of space by overlapping spheres in the most efficient way.
The \textit{thickness} \(\Theta\) (a.k.a. covering density or sparsity) of a covering is defined in the same way as the (packing) density, but with the spheres' \textit{covering radius} now being the smallest one such that the spheres cover all space.

There is an upper bound on the thickness of a covering in dimensions \(n>3\) \NoCaseChange{\protect\cite{cite2304}} as well as an asymptotic lower bound \NoCaseChange{\protect\cite{cite2305}},
\flmMathEnvironment{align}{}{
  \frac{n}{e\sqrt{e}}\lesssim\Theta\leq n\ln n+n\ln\ln n+5n~.
}

Sphere packings can be used as \textit{quantizers} or \textit{analog-to-digital converters}, which perform data compression by rounding a vector of real numbers to the point in the packing that is closest to the vector.
The set of all points closest to a point \(x\) is called the \textit{Voronoi cell of \(x\)}.
The \textit{quantizer problem} asks for a sphere packing that minimizes a dimensionless quantity proportional to the average mean-squared error per dimension; see \NoCaseChange{\protect\cite[{Sec. 3.2}]{cite39}}.
For dimension \(n\), this minimum is known as \(G_n\). It is known only for \(n=1,2\) and is attained by the integer and triangular lattices, respectively.

\codefieldsection{Rate}
The rate of a code consisting of \(M\) codewords that represent a signal of bandwidth \(W\) and duration \(T\) is defined as
\flmMathEnvironment{align}{}{
  R=\frac{1}{T}\log_{2}M\quad\quad\text{bits/s}~.
}

The Shannon capacity of the AWGN channel for a signal whose power is \(P\) is
\flmMathEnvironment{align}{}{
  C=W\log_{2}\left(1+\frac{P}{\sigma^{2}}\right)\,.
}
Random sphere packings achieve this capacity \NoCaseChange{\protect\cite{cite954}}; see the book \NoCaseChange{\protect\cite{cite39}} for more details.
Tradeoffs between various parameters have been analyzed \NoCaseChange{\protect\cite{cite2306}}. Deterministic sets of constellations from quadrature rules can also achieve capacity \NoCaseChange{\protect\cite{cite2283}}.

\codefieldsection{Decoding}
\begin{eczvaluelist}
\item\relax Each signal point is assigned its own Voronoi cell, and a received point is mapped back to the center of the Voronoi cell in which it is located upon reception.
\end{eczvaluelist}
\codefieldsection{Notes}
\begin{eczvaluelist}
\item\relax Database of sphere packings \NoCaseChange{\protect\cite{cite2307}}.
\item\relax See Refs. \NoCaseChange{\protect\cite{cite1032,cite1033}} for reviews of sphere packing.
\item\relax Popular summary of an improvement over the Rogers bound in \flmHref{https://www.quantamagazine.org/to-pack-spheres-tightly-mathematicians-throw-them-at-random-20240430/}{Quanta Magazine}.
\end{eczvaluelist}
\codefieldsection{Parent}
\begin{eczvaluelist}
\item\relax
\flmRefsHyperref[eczindexfamilyrel]{code:analog}{Analog code}\end{eczvaluelist}
\codefieldsection{Children}
\begin{eczvaluelist}
\item\relax
\flmRefsHyperref[eczindexfamilyrel]{code:antipode}{Antipode sphere packing}\item\relax
\flmRefsHyperref[eczindexfamilyrel]{code:construction_a}{Construction A code}\item\relax
\flmRefsHyperref[eczindexfamilyrel]{code:points_into_lattices}{Lattice} --- Lattices are sphere packings whose points form a group under addition.
\item\relax
\flmRefsHyperref[eczindexfamilyrel]{code:univ_opt_analog}{Universally optimal sphere packing}\end{eczvaluelist}
\codefieldsection{Cousins}
\begin{eczvaluelist}
\item\relax
\flmRefsHyperref[eczindexfamilyrel]{code:cft}{Conformal-field theory (CFT) code} --- The Cohn-Elkies linear programming bound can be recast as a conformal bootstrap problem, which is a way of utilizing symmetry to constrain correlation functions of conformal field theories \NoCaseChange{\protect\cite{cite2308,cite2309,cite2310}}.
\item\relax
\flmRefsHyperref[eczindexfamilyrel]{code:best}{\((10,40,4)\) Best code} --- Using \flmTerm{term}{ref127}{}{Construction A}, the Best code yields \(P_{10c}\), a non-lattice sphere packing in 10 dimensions that is the densest known \NoCaseChange{\protect\cite{cite376}\protect\cite[{pg. 140}]{cite39}}.
\item\relax
\flmRefsHyperref[eczindexfamilyrel]{code:julin12}{Julin-Golay code} --- Using \flmTerm{term}{ref127}{}{Construction A}, the Julin-Golay codes yield non-lattice sphere-packings that hold records in 9 and 11 dimensions.
\end{eczvaluelist}
\eczhbkcontributors{ \eczhuVVA }
\endeczcode

\eczcode{self_dual_lattice}{Unimodular lattice}{}
\codefieldsection{Alternative Names}
\begin{eczvaluelist}
\item\relax Self-dual lattice
\end{eczvaluelist}
\eczhIndexCodeAliasName{self_dual_lattice}{Self-dual lattice}
\codefieldsection{Description}
A lattice, scaled to be integral, that is equal to its dual, \(L^\perp = L\).
Unimodular lattices have \(\det L = \pm 1\).

\codefieldsection{Protection}
The minimum norm of a unimodular lattice satisfies
\flmMathEnvironment{align}{}{
  \mu\leq2\left\lfloor\frac{n}{24}\right\rfloor+2~,
}
where \(\lfloor \cdot \rfloor\) is the floor function.
This bound was originally established for even unimodular lattices \NoCaseChange{\protect\cite[{Cor. 21, Chap. 7}]{cite39}}, and later extended to all unimodular lattices, with the single exception \(n = 23\), using shadow theory \NoCaseChange{\protect\cite{cite2311}}.

\codefieldsection{Parent}
\begin{eczvaluelist}
\item\relax
\flmRefsHyperref[eczindexfamilyrel]{code:dual_lattice}{Dual lattice}\end{eczvaluelist}
\codefieldsection{Children}
\begin{eczvaluelist}
\item\relax
\flmRefsHyperref[eczindexfamilyrel]{code:eeight}{\(E_8\) Gosset lattice} --- The \(E_8\) Gosset lattice is even and unimodular.
\item\relax
\flmRefsHyperref[eczindexfamilyrel]{code:hypercubic}{\(\mathbb{Z}^n\) hypercubic lattice} --- The hypercubic lattice is odd and unimodular.
\item\relax
\flmRefsHyperref[eczindexfamilyrel]{code:niemeier}{Niemeier lattice} --- Niemeier lattices are even and unimodular.
\end{eczvaluelist}
\codefieldsection{Cousins}
\begin{eczvaluelist}
\item\relax
\flmRefsHyperref[eczindexfamilyrel]{code:self_dual}{Self-dual linear code} --- Unimodular lattices are lattice analogues of self-dual codes. There are several parallels between (doubly even) self-dual binary codes and (even) unimodular lattices \NoCaseChange{\protect\cite{cite39,cite42,cite2039}}. Even self-dual binary codes and even unimodular lattices define CFTs \NoCaseChange{\protect\cite{cite2033,cite2034,cite2035}}.
\item\relax
\flmRefsHyperref[eczindexfamilyrel]{code:self_dual_over_zq}{Self-dual code over \(\mathbb{Z}_q\)} --- There are parallels between self-dual codes over \(\mathbb{Z}_{2k}\) and even unimodular lattices \NoCaseChange{\protect\cite{cite1581,cite2312}}. Type I (type II) codes over \(\mathbb{Z}_4\) yield type I (type II) lattices under \flmTerm{term}{ref114}{}{Construction \(A_4\)} \NoCaseChange{\protect\cite[{Thm. 12.5.12}]{cite126}}.
\item\relax
\flmRefsHyperref[eczindexfamilyrel]{code:construction_a4}{Construction \(A_4\) lattice} --- Type I (type II) codes over \(\mathbb{Z}_4\) yield type I (type II) lattices under \flmTerm{term}{ref114}{}{Construction \(A_4\)} \NoCaseChange{\protect\cite[{Thm. 12.5.12}]{cite126}}.
\item\relax
\flmRefsHyperref[eczindexfamilyrel]{code:cft}{Conformal-field theory (CFT) code} --- Even self-dual binary codes and even unimodular lattices define CFTs \NoCaseChange{\protect\cite{cite2033,cite2034,cite2035}}.
\item\relax
\flmRefsHyperref[eczindexfamilyrel]{code:coxeter_todd}{Coxeter-Todd \(K_{12}\) lattice} --- The Coxeter-Todd lattice is unimodular when defined as a complex lattice over the Eisenstein integers \NoCaseChange{\protect\cite[{Ch. 4, pg. 128}]{cite39}}.
\item\relax
\flmRefsHyperref[eczindexfamilyrel]{code:self_dual_polytope}{Self-dual polytope code} --- Self-dual polytope codes are spherical analogues of self-dual lattices.
\item\relax
\flmRefsHyperref[eczindexfamilyrel]{code:spherical_design}{Spherical design} --- A union of \(t\) shells of self-dual lattices and their shadows form spherical \(t\)-designs \NoCaseChange{\protect\cite{cite2313}}.
\end{eczvaluelist}
\eczhbkcontributors{ \eczhuVVA }
\endeczcode

\eczcode{univ_opt_analog}{Universally optimal sphere packing}{~\NoCaseChange{\protect\cite{cite119}}}
\codefieldsection{Description}
A periodic sphere packing that (weakly) minimizes all completely monotonic potentials of square Euclidean distance among all periodic packings of the same density.

\codefieldsection{Parents}
\begin{eczvaluelist}
\item\relax
\flmRefsHyperref[eczindexfamilyrel]{code:sphere_packing}{Sphere packing}\item\relax
\flmRefsHyperref[eczindexfamilyrel]{code:univ_opt}{Universally optimal code}\end{eczvaluelist}
\codefieldsection{Children}
\begin{eczvaluelist}
\item\relax
\flmRefsHyperref[eczindexfamilyrel]{code:eeight}{\(E_8\) Gosset lattice} --- The \(E_8\) Gosset lattice is universally optimal \NoCaseChange{\protect\cite{cite2208}}.
\item\relax
\flmRefsHyperref[eczindexfamilyrel]{code:leech}{\(\Lambda_{24}\) Leech lattice} --- The Leech lattice is universally optimal \NoCaseChange{\protect\cite{cite2208}}.
\end{eczvaluelist}
\codefieldsection{Cousin}
\begin{eczvaluelist}
\item\relax
\flmRefsHyperref[eczindexfamilyrel]{code:hexagonal}{\(A_2\) triangular lattice} --- The triangular lattice is universally optimal among all lattices, but has not been proven to be optimal over all periodic packings \NoCaseChange{\protect\cite{cite2220}}.
\end{eczvaluelist}
\eczhbkcontributors{ Alexander Barg, \eczhuVVA }
\endeczcode

\onecolumngrid
\clearpage

\section{Spherical Kingdom}

\begin{eczEpigraph}
\begin{quote}
\flmQuoteSetup{quote}%
Life is better when well-rounded.
\flmQuoteAttributed{Maria James}
\end{quote}
\end{eczEpigraph}

\twocolumngrid

\eczcode{lambda16_shell}{\(\Lambda_{16}\) lattice-shell code}{}
\eczhIndexCodeAliasName{lambda16_shell}{lattice-shell code}
\codefieldsection{Description}
Spherical code whose codewords are points on the \(\Lambda_{16}\) Barnes-Wall lattice normalized to lie on the unit sphere.

The minimal shell of the lattice yields the \((16,4320,1)\) code.

\codefieldsection{Parents}
\begin{eczvaluelist}
\item\relax
\flmRefsHyperref[eczindexfamilyrel]{code:lattice_shell}{Lattice-shell code}\item\relax
\flmRefsHyperref[eczindexfamilyrel]{code:sidelnikov}{Real-Clifford subgroup-orbit code} --- The minimal \(\Lambda_{16}\) lattice-shell code is equivalent to the real Clifford subgroup-orbit code for \(n=16\).
\end{eczvaluelist}
\codefieldsection{Cousin}
\begin{eczvaluelist}
\item\relax
\flmRefsHyperref[eczindexfamilyrel]{code:lambda16}{\(\Lambda_{16}\) Barnes-Wall lattice} --- The \(\Lambda_{16}\) lattice-shell code is obtained from a shell of the \(\Lambda_{16}\) lattice.
\end{eczvaluelist}
\eczhbkcontributors{ \eczhuVVA }
\endeczcode

\eczcode{leech_shell}{\(\Lambda_{24}\) Leech lattice-shell code}{~\NoCaseChange{\protect\cite{cite2204}}}
\eczhIndexCodeAliasName{leech_shell}{Leech lattice-shell code}
\codefieldsection{Description}
Spherical code whose codewords are points on the \(\Lambda_{24}\) Leech lattice normalized to lie on the unit sphere.
The minimal shell of the lattice yields the \((24,196560,1)\) code, and recursively taking its kissing configurations yields the \((23,4600,1/3)\) and \((22,891,1/4)\) spherical codes \NoCaseChange{\protect\cite{cite385}}.
The \((24,196560,1)\) and \((23,4600,1/3)\) codes are optimal and unique for their parameters \NoCaseChange{\protect\cite{cite124}}.
The \((22,891,1/4)\) code is optimal and unique for its parameters \NoCaseChange{\protect\cite{cite2314,cite125}}.
Further recursive kissing configurations yield the \((21,336,1/5)\) and \((20,170,1/6)\) spherical codes \NoCaseChange{\protect\cite{cite124}}.

\codefieldsection{Protection}
The minimal-shell code yields an optimal solution to the kissing problem in 24D \NoCaseChange{\protect\cite{cite124}}.
This code saturates the Levenshtein bound \NoCaseChange{\protect\cite{cite2277,cite2278,cite2088,cite171,cite914,cite2315}\protect\cite[{pg. 337}]{cite39}} and is unique up to equivalence \NoCaseChange{\protect\cite{cite124}}.
Its kissing configuration also yields a unique \((23,4600,1/3)\) spherical code \NoCaseChange{\protect\cite{cite124}}.

\codefieldsection{Parent}
\begin{eczvaluelist}
\item\relax
\flmRefsHyperref[eczindexfamilyrel]{code:lattice_shell}{Lattice-shell code}\end{eczvaluelist}
\codefieldsection{Cousins}
\begin{eczvaluelist}
\item\relax
\flmRefsHyperref[eczindexfamilyrel]{code:leech}{\(\Lambda_{24}\) Leech lattice} --- The \(\Lambda_{24}\) lattice-shell code is obtained from a shell of the Leech lattice.
\item\relax
\flmRefsHyperref[eczindexfamilyrel]{code:sharp_config}{Spherical sharp configuration} --- The smallest-shell \((24,196560,1)\) code is a spherical sharp configuration \NoCaseChange{\protect\cite{cite2316,cite119}}. The \((23,4600,1/3)\) \NoCaseChange{\protect\cite{cite124,cite2314,cite125,cite119}\protect\cite[{Table 1}]{cite394}} and \((22,891,1/4)\) \NoCaseChange{\protect\cite{cite2314,cite125,cite119}\protect\cite[{Table 1}]{cite394}} spherical codes are also sharp configurations.
\item\relax
\flmRefsHyperref[eczindexfamilyrel]{code:spherical_design}{Spherical design} --- The smallest-shell \((24,196560,1)\) code is a tight and unique spherical 11-design \NoCaseChange{\protect\cite{cite124}\protect\cite[{Ch. 3}]{cite39}}. The \((23,4600,1/3)\) \NoCaseChange{\protect\cite{cite124,cite2314,cite125,cite119}\protect\cite[{Table 1}]{cite394}} and \((22,891,1/4)\) \NoCaseChange{\protect\cite{cite2314,cite125,cite119}\protect\cite[{Table 1}]{cite394}} spherical codes are also sharp configurations.
\item\relax
\flmRefsHyperref[eczindexfamilyrel]{code:mclaughlin}{McLaughlin spherical code} --- The \((23,552,1/5)\) McLaughlin code can be derived from a shell of the Leech lattice \NoCaseChange{\protect\cite{cite119,cite2317}}.
\end{eczvaluelist}
\eczhbkcontributors{ \eczhuVVA }
\endeczcode

\eczcode{231_polytope}{\(2_{31}\) polytope code}{}
\eczhIndexCodeAliasName{231_polytope}{polytope code}
\codefieldsection{Description}
An antipodal spherical \((7,126,1)\) code whose codewords are the vertices of the smallest shell of the \(E_7\) lattice \NoCaseChange{\protect\cite{cite232}}.

\codefieldsection{Notes}
\begin{eczvaluelist}
\item\relax See the corresponding Bendwavy database entry \NoCaseChange{\protect\cite{cite2318}}.
\end{eczvaluelist}
\codefieldsection{Parents}
\begin{eczvaluelist}
\item\relax
\flmRefsHyperref[eczindexfamilyrel]{code:polytope}{Polytope code}\item\relax
\flmRefsHyperref[eczindexfamilyrel]{code:eseven_shell}{\(E_7\) lattice-shell code} --- Codewords of the \(2_{31}\) polytope form the smallest shell of the \(E_7\) lattice \NoCaseChange{\protect\cite{cite232}}.
\item\relax
\flmRefsHyperref[eczindexfamilyrel]{code:spherical_design}{Spherical design} --- The 126 vertices of the \(2_{31}\) polytope form a spherical 5-design \NoCaseChange{\protect\cite{cite384}}.
\end{eczvaluelist}
\codefieldsection{Cousins}
\begin{eczvaluelist}
\item\relax
\flmRefsHyperref[eczindexfamilyrel]{code:delsarte_optimal}{Sharp configuration} --- The 63 antipodal pairs of vertices of the \(2_{31}\) polytope form a sharp configuration in \(\mathbb{R}P^6\) \NoCaseChange{\protect\cite{cite119}}.
\item\relax
\flmRefsHyperref[eczindexfamilyrel]{code:t-designs}{\(t\)-design} --- The 63 antipodal pairs of vertices of the \(2_{31}\) polytope form a 2-design in \(\mathbb{R}P^6\) \NoCaseChange{\protect\cite{cite119}}.
\item\relax
\flmRefsHyperref[eczindexfamilyrel]{code:real_projective}{Real projective space code} --- The 63 antipodal pairs of vertices of the \(2_{31}\) polytope form a sharp configuration and a 2-design in \(\mathbb{R}P^6\) \NoCaseChange{\protect\cite{cite119}}.
\end{eczvaluelist}
\eczhbkcontributors{ \eczhuVVA }
\endeczcode

\eczcode{241_polytope}{\(2_{41}\) polytope code}{}
\eczhIndexCodeAliasName{241_polytope}{polytope code}
\codefieldsection{Description}
An antipodal spherical \((8,2160,1/2)\) code whose codewords are the vertices of the second-smallest shell of the \(E_8\) lattice \NoCaseChange{\protect\cite{cite232}\protect\cite[{Table 10.3}]{cite115}}.

\codefieldsection{Notes}
\begin{eczvaluelist}
\item\relax See the corresponding Bendwavy database entry \NoCaseChange{\protect\cite{cite2319}}.
\end{eczvaluelist}
\codefieldsection{Parents}
\begin{eczvaluelist}
\item\relax
\flmRefsHyperref[eczindexfamilyrel]{code:polytope}{Polytope code}\item\relax
\flmRefsHyperref[eczindexfamilyrel]{code:eeight_shell}{\(E_8\) Gosset lattice-shell code} --- Codewords of the \(2_{41}\) polytope form the second-smallest shell of the \(E_8\) lattice \NoCaseChange{\protect\cite{cite232}}.
\item\relax
\flmRefsHyperref[eczindexfamilyrel]{code:spherical_design}{Spherical design} --- The \(2_{41}\) polytope code forms a spherical 7-design \NoCaseChange{\protect\cite{cite232}}.
\end{eczvaluelist}
\codefieldsection{Cousin}
\begin{eczvaluelist}
\item\relax
\flmRefsHyperref[eczindexfamilyrel]{code:witting_polytope}{Witting polytope code} --- Vertices of the \(2_{41}\) and \(4_{21}\) polytopes minimize each other's potential functions \NoCaseChange{\protect\cite{cite232}}.
\end{eczvaluelist}
\eczhbkcontributors{ Sergiy Borodachov, \eczhuVVA }
\endeczcode

\eczcode{hess_polytope}{\(3_{21}\) polytope code}{~\NoCaseChange{\protect\cite{cite2239}}}
\codefieldsection{Alternative Names}
\begin{eczvaluelist}
\item\relax Hess polytope code
\item\relax Hesse polytope code
\item\relax 7-ic semi-regular figure code
\end{eczvaluelist}
\eczhIndexCodeAliasName{hess_polytope}{polytope code}
\eczhIndexCodeAliasName{hess_polytope}{Hess polytope code}
\eczhIndexCodeAliasName{hess_polytope}{Hesse polytope code}
\eczhIndexCodeAliasName{hess_polytope}{7-ic semi-regular figure code}
\codefieldsection{Description}
Spherical \((7,56,1/3)\) code whose codewords are the vertices of the \(3_{21}\) polytope (a.k.a. the Hess polytope).
The vertices form the kissing configuration of the Witting polytope code.
The 1-skeleton of this polytope is the Gosset graph \NoCaseChange{\protect\cite{cite178}}.
The code is optimal and unique up to equivalence \NoCaseChange{\protect\cite{cite124,cite39,cite125}}.

Antipodal pairs of points of the \(3_{21}\) polytope code correspond to the 28 bitangent lines of a general quartic plane curve in the complex project plane \NoCaseChange{\protect\cite{cite117,cite118,cite119,cite120}}.
A representation of the codewords consists of all seven permutations of the eight vectors \((\pm 1,0,\pm 1,\pm 1,0,0,0)\).

\codefieldsection{Notes}
\begin{eczvaluelist}
\item\relax See the corresponding Bendwavy database entry \NoCaseChange{\protect\cite{cite2320}}.
\end{eczvaluelist}
\codefieldsection{Parents}
\begin{eczvaluelist}
\item\relax
\flmRefsHyperref[eczindexfamilyrel]{code:polytope}{Polytope code}\item\relax
\flmRefsHyperref[eczindexfamilyrel]{code:sharp_config}{Spherical sharp configuration} --- The \(3_{21}\) polytope code is a sharp configuration \NoCaseChange{\protect\cite{cite2321,cite119}}.
\end{eczvaluelist}
\codefieldsection{Cousins}
\begin{eczvaluelist}
\item\relax
\flmRefsHyperref[eczindexfamilyrel]{code:spherical_design}{Spherical design} --- The \(3_{21}\) polytope code forms a tight spherical 5-design \NoCaseChange{\protect\cite{cite385,cite124}\protect\cite[{Ch. 14}]{cite39}\protect\cite[{Table 1}]{cite119}}.
\item\relax
\flmRefsHyperref[eczindexfamilyrel]{code:eseven_shell}{\(E_7\) lattice-shell code} --- \(3_{21}\) polytope codewords form the minimal lattice-shell code of the \(E_7^{\perp}\) lattice \NoCaseChange{\protect\cite{cite2322}}.
\item\relax
\flmRefsHyperref[eczindexfamilyrel]{code:complex_projective}{Complex projective space code} --- Antipodal pairs of points of the \(3_{21}\) polytope code correspond to the 28 bitangent lines of a general quartic plane curve in the complex project plane \NoCaseChange{\protect\cite{cite117,cite118,cite119,cite120}}.
\item\relax
\flmRefsHyperref[eczindexfamilyrel]{code:real_projective}{Real projective space code} --- The 28 antipodal pairs of the \(3_{21}\) polytope code form 28 equiangular lines in \(\mathbb{R}^7\), achieving the absolute bound \NoCaseChange{\protect\cite{cite119}}. This is because the only inner product between distinct antipodal pairs is \(\pm 1/3\) \NoCaseChange{\protect\cite[{Table 1}]{cite119}}.
\item\relax
\flmRefsHyperref[eczindexfamilyrel]{code:witting_polytope}{Witting polytope code} --- \(3_{21}\) polytope codewords form the first recursive kissing configuration of the Witting polytope code \NoCaseChange{\protect\cite{cite124,cite119}\protect\cite[{Ch. 9, pg. 264}]{cite39}}. The Gosset graph is a subgraph of the graph formed by the vertices of the Witting polytope \NoCaseChange{\protect\cite[{Sec. 3.11}]{cite1385}}.
\item\relax
\flmRefsHyperref[eczindexfamilyrel]{code:hessian_polyhedron}{Hessian polyhedron code} --- The Hessian polyhedron code forms the next recursive kissing configuration after the \(3_{21}\) polytope code in the \(E_8\) lattice-shell/Witting polytope sequence \NoCaseChange{\protect\cite{cite124}}.
\end{eczvaluelist}
\eczhbkcontributors{ \eczhuVVA }
\endeczcode

\eczcode{bw32_shell}{\(BW_{32}\) lattice-shell code}{}
\eczhIndexCodeAliasName{bw32_shell}{lattice-shell code}
\codefieldsection{Description}
Spherical code whose codewords are points on the \(BW_{32}\) Barnes-Wall lattice normalized to lie on the unit sphere.

The minimal shell of the lattice yields the \((32,146880,1)\) code.

\codefieldsection{Parents}
\begin{eczvaluelist}
\item\relax
\flmRefsHyperref[eczindexfamilyrel]{code:lattice_shell}{Lattice-shell code}\item\relax
\flmRefsHyperref[eczindexfamilyrel]{code:sidelnikov}{Real-Clifford subgroup-orbit code} --- The minimal \(BW_{32}\) lattice-shell code is equivalent to the real Clifford subgroup-orbit code for \(n=32\).
\end{eczvaluelist}
\codefieldsection{Cousin}
\begin{eczvaluelist}
\item\relax
\flmRefsHyperref[eczindexfamilyrel]{code:bw32}{\(BW_{32}\) Barnes-Wall lattice} --- The \(BW_{32}\) lattice-shell code is obtained from a shell of the \(BW_{32}\) Barnes-Wall lattice.
\end{eczvaluelist}
\eczhbkcontributors{ \eczhuVVA }
\endeczcode

\eczcode{dfour_shell}{\(D_4\) lattice-shell code}{}
\eczhIndexCodeAliasName{dfour_shell}{lattice-shell code}
\codefieldsection{Description}
Spherical code whose codewords are points on the \(D_4\) lattice normalized to lie on the unit sphere.

\codefieldsection{Parent}
\begin{eczvaluelist}
\item\relax
\flmRefsHyperref[eczindexfamilyrel]{code:lattice_shell}{Lattice-shell code}\end{eczvaluelist}
\codefieldsection{Children}
\begin{eczvaluelist}
\item\relax
\flmRefsHyperref[eczindexfamilyrel]{code:24cell}{24-cell code} --- The 24-cell code is the minimal shell of the \(D_4\) lattice.
\item\relax
\flmRefsHyperref[eczindexfamilyrel]{code:disphenoidal288cell}{Disphenoidal 288-cell code} --- Disphenoidal 288-cell codewords are the union of two 24-point shells of the \(D_4\) lattice, with each shell making up the vertices of a 24-cell.
\end{eczvaluelist}
\codefieldsection{Cousins}
\begin{eczvaluelist}
\item\relax
\flmRefsHyperref[eczindexfamilyrel]{code:dfour}{\(D_4\) hyper-diamond lattice} --- The \(D_4\) lattice-shell code is obtained from a shell of the \(D_4\) lattice.
\item\relax
\flmRefsHyperref[eczindexfamilyrel]{code:spherical_design}{Spherical design} --- \(D_4\) \(2m\)-shell codes can form spherical designs \NoCaseChange{\protect\cite{cite2323}}.
\end{eczvaluelist}
\eczhbkcontributors{ \eczhuVVA }
\endeczcode

\eczcode{esix_shell}{\(E_6\) lattice-shell code}{}
\eczhIndexCodeAliasName{esix_shell}{lattice-shell code}
\codefieldsection{Description}
Spherical code obtained by taking vectors from a fixed shell of the \(E_6\) lattice and normalizing them to lie on the unit sphere.

The minimal shell of the lattice yields the \((6,72,1)\) code, whose codewords form the vertices of the \(1_{22}\) polytope and the rectified Hessian polyhedron.

\codefieldsection{Parent}
\begin{eczvaluelist}
\item\relax
\flmRefsHyperref[eczindexfamilyrel]{code:lattice_shell}{Lattice-shell code}\end{eczvaluelist}
\codefieldsection{Child}
\begin{eczvaluelist}
\item\relax
\flmRefsHyperref[eczindexfamilyrel]{code:rect_hessian_polyhedron}{Rectified Hessian polyhedron code} --- Rectified Hessian polyhedron codewords form the minimal shell of the \(E_6\) lattice.
\end{eczvaluelist}
\codefieldsection{Cousins}
\begin{eczvaluelist}
\item\relax
\flmRefsHyperref[eczindexfamilyrel]{code:esix}{\(E_6\) root lattice} --- The \(E_6\) lattice-shell code is obtained from a shell of the \(E_6\) lattice.
\item\relax
\flmRefsHyperref[eczindexfamilyrel]{code:delsarte_optimal}{Sharp configuration} --- The 36 antipodal pairs of the smallest \(E_6\) lattice shell form a sharp configuration in \(\mathbb{R}P^5\) \NoCaseChange{\protect\cite{cite119}}.
\item\relax
\flmRefsHyperref[eczindexfamilyrel]{code:t-designs}{\(t\)-design} --- The 36 antipodal pairs of the smallest \(E_6\) lattice shell form a 2-design in \(\mathbb{R}P^5\) \NoCaseChange{\protect\cite{cite119}}.
\item\relax
\flmRefsHyperref[eczindexfamilyrel]{code:real_projective}{Real projective space code} --- The 36 antipodal pairs of the smallest \(E_6\) lattice shell form a sharp configuration and a 2-design in \(\mathbb{R}P^5\) \NoCaseChange{\protect\cite{cite119}}.
\item\relax
\flmRefsHyperref[eczindexfamilyrel]{code:hessian_polyhedron}{Hessian polyhedron code} --- Double Hessian polyhedron codewords form the minimal lattice-shell code of the \(E_6^{\perp}\) lattice \NoCaseChange{\protect\cite{cite2322}}.
\end{eczvaluelist}
\eczhbkcontributors{ \eczhuVVA }
\endeczcode

\eczcode{eseven_shell}{\(E_7\) lattice-shell code}{}
\eczhIndexCodeAliasName{eseven_shell}{lattice-shell code}
\codefieldsection{Description}
Spherical code whose codewords are points on the \(E_7\) lattice normalized to lie on the unit sphere.

\codefieldsection{Parent}
\begin{eczvaluelist}
\item\relax
\flmRefsHyperref[eczindexfamilyrel]{code:lattice_shell}{Lattice-shell code}\end{eczvaluelist}
\codefieldsection{Child}
\begin{eczvaluelist}
\item\relax
\flmRefsHyperref[eczindexfamilyrel]{code:231_polytope}{\(2_{31}\) polytope code} --- Codewords of the \(2_{31}\) polytope form the smallest shell of the \(E_7\) lattice \NoCaseChange{\protect\cite{cite232}}.
\end{eczvaluelist}
\codefieldsection{Cousins}
\begin{eczvaluelist}
\item\relax
\flmRefsHyperref[eczindexfamilyrel]{code:eseven}{\(E_7\) root lattice} --- The \(E_7\) lattice-shell code is obtained from a shell of the \(E_7\) lattice.
\item\relax
\flmRefsHyperref[eczindexfamilyrel]{code:hess_polytope}{\(3_{21}\) polytope code} --- \(3_{21}\) polytope codewords form the minimal lattice-shell code of the \(E_7^{\perp}\) lattice \NoCaseChange{\protect\cite{cite2322}}.
\end{eczvaluelist}
\eczhbkcontributors{ \eczhuVVA }
\endeczcode

\eczcode{eeight_shell}{\(E_8\) Gosset lattice-shell code}{}
\eczhIndexCodeAliasName{eeight_shell}{Gosset lattice-shell code}
\codefieldsection{Description}
Spherical code whose codewords are points on the \(E_8\) Gosset lattice normalized to lie on the unit sphere.

The minimal shell of the lattice yields the \((8,240,1)\) code, whose codewords form the vertices of the \(4_{21}\) polytope and the Witting complex polytope.

\codefieldsection{Protection}
Smallest-shell code yields an optimal solution to the kissing problem in 8D \NoCaseChange{\protect\cite{cite2324,cite124}}.

\codefieldsection{Parent}
\begin{eczvaluelist}
\item\relax
\flmRefsHyperref[eczindexfamilyrel]{code:lattice_shell}{Lattice-shell code}\end{eczvaluelist}
\codefieldsection{Children}
\begin{eczvaluelist}
\item\relax
\flmRefsHyperref[eczindexfamilyrel]{code:241_polytope}{\(2_{41}\) polytope code} --- Codewords of the \(2_{41}\) polytope form the second-smallest shell of the \(E_8\) lattice \NoCaseChange{\protect\cite{cite232}}.
\item\relax
\flmRefsHyperref[eczindexfamilyrel]{code:witting_polytope}{Witting polytope code} --- The minimal shell of the lattice yields the \((8,240,1)\) code, whose codewords form the vertices of the \(4_{21}\) polytope.
\end{eczvaluelist}
\codefieldsection{Cousin}
\begin{eczvaluelist}
\item\relax
\flmRefsHyperref[eczindexfamilyrel]{code:eeight}{\(E_8\) Gosset lattice} --- The \(E_8\) lattice-shell code is obtained from a shell of the \(E_8\) lattice.
\end{eczvaluelist}
\eczhbkcontributors{ \eczhuVVA }
\endeczcode

\eczcode{120cell}{120-cell code}{~\NoCaseChange{\protect\cite{cite2325}}}
\codefieldsection{Alternative Names}
\begin{eczvaluelist}
\item\relax Hecatonicosachoron code
\item\relax Dodecaplex code
\item\relax Hyperdodecahedron code
\end{eczvaluelist}
\eczhIndexCodeAliasName{120cell}{Hecatonicosachoron code}
\eczhIndexCodeAliasName{120cell}{Dodecaplex code}
\eczhIndexCodeAliasName{120cell}{Hyperdodecahedron code}
\codefieldsection{Description}
Spherical \((4,600,(7-3\sqrt{5})/4)\) code whose codewords are the vertices of the 120-cell.
See \NoCaseChange{\protect\cite{cite178}\protect\cite[{Table 1}]{cite227}\protect\cite[{Table 3}]{cite228}} for explicit realizations of its 600 codewords.

\begin{flmFloat}{figure}{NumCap}\includegraphics[width=306.14173228346453bp,max width=\linewidth]{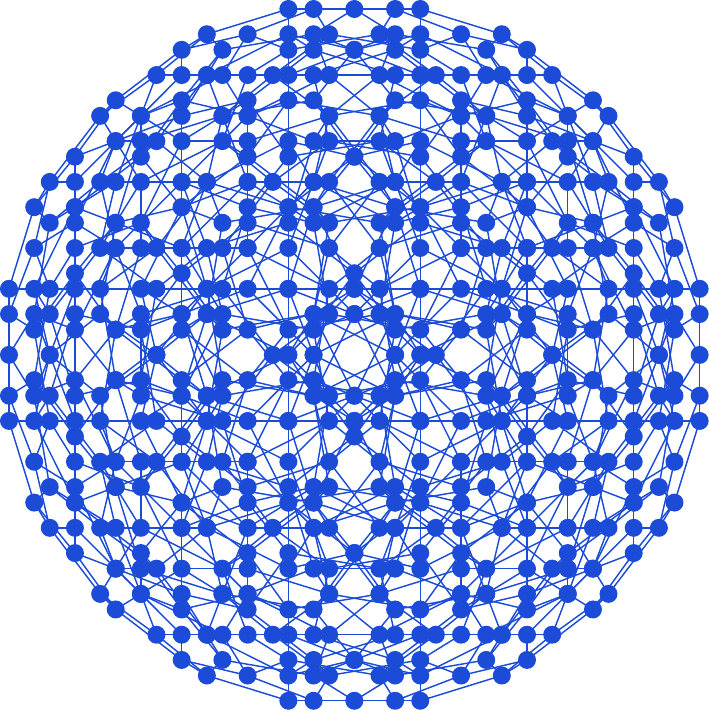}\caption{Projection of the coordinates of the 120-cell.}\label{ref2326}\end{flmFloat}

\codefieldsection{Notes}
\begin{eczvaluelist}
\item\relax The 120-cell code yields improved proofs of the Bell-Kochen-Specker (BKS) theorem \NoCaseChange{\protect\cite{cite227}}.
\item\relax See the corresponding Bendwavy database entry \NoCaseChange{\protect\cite{cite2327}}.
\end{eczvaluelist}
\codefieldsection{Parents}
\begin{eczvaluelist}
\item\relax
\flmRefsHyperref[eczindexfamilyrel]{code:polytope}{Polytope code}\item\relax
\flmRefsHyperref[eczindexfamilyrel]{code:spherical_design}{Spherical design} --- The code forms a spherical 11-design because its vertices can be divided into five 600-cells, each of which forms said design.
\end{eczvaluelist}
\codefieldsection{Cousins}
\begin{eczvaluelist}
\item\relax
\flmRefsHyperref[eczindexfamilyrel]{code:dual_polytope}{Dual polytope code} --- The 600-cell and 120-cell are dual to each other.
\item\relax
\flmRefsHyperref[eczindexfamilyrel]{code:24cell}{24-cell code} --- Vertices of a 120-cell can be split up into vertices of five 600-cells \NoCaseChange{\protect\cite{cite178,cite227}}, and vertices of a 600-cell can be split up into vertices of five 24-cells \NoCaseChange{\protect\cite{cite2328,cite178,cite229}}. Therefore, vertices of a 120-cell can be split up into vertices of 25 24-cells.
\item\relax
\flmRefsHyperref[eczindexfamilyrel]{code:600cell}{600-cell code} --- Vertices of a 120-cell can be split up into vertices of five 600-cells \NoCaseChange{\protect\cite{cite178,cite227}}. The 600-cell and 120-cell are dual to each other.
\end{eczvaluelist}
\eczhbkcontributors{ Shubham P. Jain, \eczhuVVA }
\endeczcode

\eczcode{24cell}{24-cell code}{~\NoCaseChange{\protect\cite{cite2325}}}
\codefieldsection{Alternative Names}
\begin{eczvaluelist}
\item\relax Icositetrachoron code
\item\relax Octaplex code
\item\relax Hyperdiamond code
\end{eczvaluelist}
\eczhIndexCodeAliasName{24cell}{Icositetrachoron code}
\eczhIndexCodeAliasName{24cell}{Octaplex code}
\eczhIndexCodeAliasName{24cell}{Hyperdiamond code}
\codefieldsection{Description}
Spherical \((4,24,1)\) code whose codewords are the vertices of the 24-cell.
Codewords form the minimal lattice-shell code of the \(D_4\) lattice.

A realization of the codewords consists of the 24 permutations of the four vectors \((0,0,\pm 1,\pm 1)\); see \NoCaseChange{\protect\cite[{Table 3}]{cite228}} for another realization.
A realization in terms of quaternion coordinates yields the 24 elements of the binary tetrahedral group \(2T\) \NoCaseChange{\protect\cite{cite230}}.
\begin{flmFloat}{figure}{NumCap}\includegraphics[width=306.14173228346453bp,max width=\linewidth]{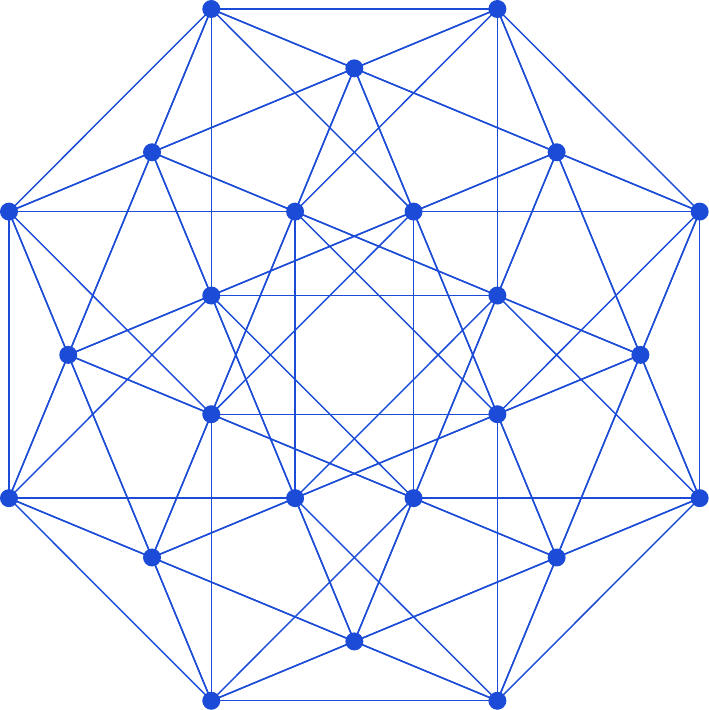}\caption{Projection of the coordinates of the 24-cell.}\label{ref2329}\end{flmFloat}

\codefieldsection{Protection}
Code yields an optimal solution to the kissing problem in 4D \NoCaseChange{\protect\cite{cite2330,cite2331}}.

\codefieldsection{Notes}
\begin{eczvaluelist}
\item\relax See \flmHref{https://www.gregegan.net/SCIENCE/24-cell/24-cell.html}{post} by G. Egan for more details.
\item\relax See the corresponding Bendwavy database entry \NoCaseChange{\protect\cite{cite2332}}.
\end{eczvaluelist}
\codefieldsection{Parents}
\begin{eczvaluelist}
\item\relax
\flmRefsHyperref[eczindexfamilyrel]{code:self_dual_polytope}{Self-dual polytope code} --- The 24-cell is self-dual.
\item\relax
\flmRefsHyperref[eczindexfamilyrel]{code:dfour_shell}{\(D_4\) lattice-shell code} --- The 24-cell code is the minimal shell of the \(D_4\) lattice.
\item\relax
\flmRefsHyperref[eczindexfamilyrel]{code:sidelnikov}{Real-Clifford subgroup-orbit code} --- The 24-cell code is equivalent to the real Clifford subgroup-orbit code for \(n=4\).
\end{eczvaluelist}
\codefieldsection{Cousins}
\begin{eczvaluelist}
\item\relax
\flmRefsHyperref[eczindexfamilyrel]{code:spherical_design}{Spherical design} --- The 24-cell code is a spherical 5-design \NoCaseChange{\protect\cite{cite377}}.
\item\relax
\flmRefsHyperref[eczindexfamilyrel]{code:group_classical}{Group-alphabet code} --- The 24-cell code has a quaternion-coordinate realization as the 24 elements of the binary tetrahedral group \(2T\), one of the three exceptional finite subgroups of \(SU(2)\) \NoCaseChange{\protect\cite{cite230}}.
\item\relax
\flmRefsHyperref[eczindexfamilyrel]{code:600cell}{600-cell code} --- Vertices of a 600-cell can be split up into vertices of five 24-cells \NoCaseChange{\protect\cite{cite2328,cite178,cite229}}.
\item\relax
\flmRefsHyperref[eczindexfamilyrel]{code:120cell}{120-cell code} --- Vertices of a 120-cell can be split up into vertices of five 600-cells \NoCaseChange{\protect\cite{cite178,cite227}}, and vertices of a 600-cell can be split up into vertices of five 24-cells \NoCaseChange{\protect\cite{cite2328,cite178,cite229}}. Therefore, vertices of a 120-cell can be split up into vertices of 25 24-cells.
\item\relax
\flmRefsHyperref[eczindexfamilyrel]{code:disphenoidal288cell}{Disphenoidal 288-cell code} --- Vertices of a disphenoidal 288-cell can be split up into vertices of a 24-cell and its dual 24-cell \NoCaseChange{\protect\cite[{Sec. 8.6}]{cite230}}.
\item\relax
\flmRefsHyperref[eczindexfamilyrel]{code:univ_opt_spherical}{Universally optimal spherical code} --- The 24-cell code is not universally optimal \NoCaseChange{\protect\cite{cite377}}, but comes quite close \NoCaseChange{\protect\cite[{Exam. 12.4.29}]{cite199}}.
\item\relax
\flmRefsHyperref[eczindexfamilyrel]{code:delsarte_optimal}{Sharp configuration} --- The 12 antipodal pairs of the 24-cell code form a sharp configuration and a 2-design in \(\mathbb{R}P^3\) \NoCaseChange{\protect\cite{cite119}}.
\item\relax
\flmRefsHyperref[eczindexfamilyrel]{code:t-designs}{\(t\)-design} --- The 12 antipodal pairs of the 24-cell code form a sharp configuration and a 2-design in \(\mathbb{R}P^3\) \NoCaseChange{\protect\cite{cite119}}.
\item\relax
\flmRefsHyperref[eczindexfamilyrel]{code:real_projective}{Real projective space code} --- The 12 antipodal pairs of the 24-cell code form a sharp configuration and a 2-design in \(\mathbb{R}P^3\) \NoCaseChange{\protect\cite{cite119}}. This is a special case of a family of real projective plane codes, constructed using Kerdock codes \NoCaseChange{\protect\cite{cite1411}} (cf. \NoCaseChange{\protect\cite{cite917}}).
\item\relax
\flmRefsHyperref[eczindexfamilyrel]{code:kerdock}{Kerdock code} --- The 12 antipodal pairs of the 24-cell code form a sharp configuration and a 2-design in \(\mathbb{R}P^3\) \NoCaseChange{\protect\cite{cite119}}. This is a special case of a family of real projective plane codes, constructed using Kerdock codes \NoCaseChange{\protect\cite{cite1411}} (cf. \NoCaseChange{\protect\cite{cite917}}).
\item\relax
\flmRefsHyperref[eczindexfamilyrel]{code:dfour}{\(D_4\) hyper-diamond lattice} --- The Voronoi cell of the \(D_4\) lattice is a 24-cell \NoCaseChange{\protect\cite[{Ch. 21, pg. 464}]{cite39}}.
\item\relax
\flmRefsHyperref[eczindexfamilyrel]{code:biorthogonal_spherical}{Biorthogonal spherical code} --- Vertices of a 24-cell can be split up into vertices of three 16-cells, which are biorthogonal spherical codes for \(n=4\) \NoCaseChange{\protect\cite{cite178}}. The vertices of a 24-cell are a union of the vertices of a tesseract and a 16-cell \NoCaseChange{\protect\cite[{Exam. 2.6}]{cite380}}.
\item\relax
\flmRefsHyperref[eczindexfamilyrel]{code:hypercube}{Hypercube code} --- The vertices of a 24-cell are a union of the vertices of a tesseract and a 16-cell \NoCaseChange{\protect\cite[{Exam. 2.6}]{cite380}}.
\item\relax
\flmRefsHyperref[eczindexfamilyrel]{code:2t_qutrit}{2T-qutrit code} --- The \(2T\)-qutrit code is constructed out of superpositions of coherent states whose amplitudes make up the binary tetrahedral group \(2T\), a.k.a. the 24-cell.
\item\relax
\flmRefsHyperref[eczindexfamilyrel]{code:quantum_sidelnikov}{Clifford subgroup-orbit QSC} --- Logical constellations of the Clifford subgroup-orbit code for \(r=1\) form vertices of 24-cells when mapped into the real sphere, while code constellations form vertices of a disphenoidal 288-cell.
\end{eczvaluelist}
\eczhbkcontributors{ Shubham P. Jain, \eczhuVVA }
\endeczcode

\eczcode{600cell}{600-cell code}{~\NoCaseChange{\protect\cite{cite2325}}}
\codefieldsection{Alternative Names}
\begin{eczvaluelist}
\item\relax Hexacosichoron code
\item\relax Tetraplex code
\item\relax Polytetrahedron code
\end{eczvaluelist}
\eczhIndexCodeAliasName{600cell}{Hexacosichoron code}
\eczhIndexCodeAliasName{600cell}{Tetraplex code}
\eczhIndexCodeAliasName{600cell}{Polytetrahedron code}
\codefieldsection{Description}
Spherical \((4,120,(3-\sqrt{5})/2)\) code whose codewords are the vertices of the 600-cell.
See \NoCaseChange{\protect\cite[{Table 1}]{cite229}\protect\cite[{Table 3}]{cite228}} for realizations of the 120 codewords.
A realization of the 600-cell can be given in terms of icosians, which are quaternion coordinates of the 120 elements of the binary icosahedral group \(2I \cong 2.A_5\) (a.k.a. the icosian group) \NoCaseChange{\protect\cite{cite230}\protect\cite[{Ch. 8, pg. 207}]{cite39}}.

\begin{flmFloat}{figure}{NumCap}\includegraphics[width=306.14173228346453bp,max width=\linewidth]{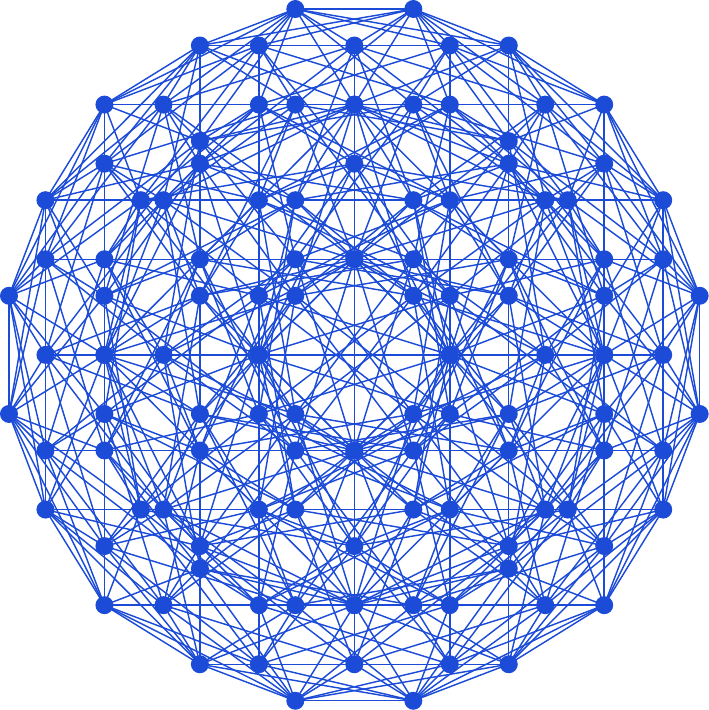}\caption{Projection of the coordinates of the \(600\)-cell.}\label{ref2333}\end{flmFloat}

\codefieldsection{Protection}
The 600-cell code is unique up to equivalence, which follows by saturating the Boroczky bound \NoCaseChange{\protect\cite{cite2334,cite378}}.

\codefieldsection{Notes}
\begin{eczvaluelist}
\item\relax The 600-cell code yields improved proofs of the Bell-Kochen-Specker (BKS) theorem \NoCaseChange{\protect\cite{cite229}}.
\item\relax See \flmHref{https://johncarlosbaez.wordpress.com/2017/12/16/the-600-cell/}{post} by J. Baez for more details.
\item\relax See the corresponding Bendwavy database entry \NoCaseChange{\protect\cite{cite2335}}.
\end{eczvaluelist}
\codefieldsection{Parents}
\begin{eczvaluelist}
\item\relax
\flmRefsHyperref[eczindexfamilyrel]{code:polytope}{Polytope code}\item\relax
\flmRefsHyperref[eczindexfamilyrel]{code:univ_opt_spherical}{Universally optimal spherical code} --- The 600-cell is universally optimal, but it is not a spherical sharp configuration \NoCaseChange{\protect\cite{cite119}\protect\cite[{Thm. 12.4.27}]{cite199}}.
\item\relax
\flmRefsHyperref[eczindexfamilyrel]{code:spherical_design}{Spherical design} --- The 600-cell code forms a spherical 11-design that is unique up to equivalence \NoCaseChange{\protect\cite{cite378}}.
\end{eczvaluelist}
\codefieldsection{Cousins}
\begin{eczvaluelist}
\item\relax
\flmRefsHyperref[eczindexfamilyrel]{code:group_classical}{Group-alphabet code} --- The 600-cell code has a quaternion-coordinate realization as the 120 elements of the binary icosahedral group \(2I \cong 2.A_5\), one of the three exceptional finite subgroups of \(SU(2)\) \NoCaseChange{\protect\cite{cite230}\protect\cite[{Ch. 8, pg. 207}]{cite39}}.
\item\relax
\flmRefsHyperref[eczindexfamilyrel]{code:dual_polytope}{Dual polytope code} --- The 600-cell and 120-cell are dual to each other.
\item\relax
\flmRefsHyperref[eczindexfamilyrel]{code:120cell}{120-cell code} --- Vertices of a 120-cell can be split up into vertices of five 600-cells \NoCaseChange{\protect\cite{cite178,cite227}}. The 600-cell and 120-cell are dual to each other.
\item\relax
\flmRefsHyperref[eczindexfamilyrel]{code:icosahedron}{Icosahedron code} --- A realization of the 600-cell can be done in terms of icosians, which are quaternion coordinates of the 120 elements of the binary icosahedral group \(2I \cong 2.A_5\) (a.k.a. the icosian group) \NoCaseChange{\protect\cite{cite230}\protect\cite[{Ch. 8, pg. 207}]{cite39}}.
\item\relax
\flmRefsHyperref[eczindexfamilyrel]{code:witting_polytope}{Witting polytope code} --- The 120 vertices of the 600-cell are the unit icosians, and these icosian units, together with their multiples by \((1-\sqrt{5})/2\), form the 240 minimal vectors of a version of the \(E_8\) lattice, i.e., the Witting polytope \NoCaseChange{\protect\cite[{Ch. 8, pg. 210}]{cite39}}.
\item\relax
\flmRefsHyperref[eczindexfamilyrel]{code:24cell}{24-cell code} --- Vertices of a 600-cell can be split up into vertices of five 24-cells \NoCaseChange{\protect\cite{cite2328,cite178,cite229}}.
\end{eczvaluelist}
\eczhbkcontributors{ Shubham P. Jain, \eczhuVVA }
\endeczcode

\eczcode{annealing_spherical}{Annealing-based spherical code}{~\NoCaseChange{\protect\cite{cite2336,cite2337,cite2338}}}
\codefieldsection{Description}
Code whose codewords are obtained from a simulated annealing or energy-repulsion numerical optimization procedure.

\codefieldsection{Parent}
\begin{eczvaluelist}
\item\relax
\flmRefsHyperref[eczindexfamilyrel]{code:spherical}{Spherical code}\end{eczvaluelist}
\eczhbkcontributors{ \eczhuVVA }
\endeczcode

\eczcode{antiprism}{Antiprism code}{}
\codefieldsection{Description}
Spherical \((3,2q)\) code for \(q \geq 2\) whose codewords are the vertices of a \(q\)-antiprism.

\codefieldsection{Protection}
The antiprism vertices consists of two \(q\)-gon vertices with the \(q\)-gons rotated by \(2\pi/q\) degrees.
The relative height and radii of the \(q\)-gons can be modulated while still staying on the sphere.
For the case when the two \(q\)-gons are such that the \(q=2,3\) cases reduce to the tetrahedron and octahedron, respectively, the antiprism is a spherical 3-design for \(q \geq 3\), and a \(2\)-design for \(q=2\) \NoCaseChange{\protect\cite{cite2339}}.
This can be seen as a consequence of \NoCaseChange{\protect\cite[{Lemma 6.11}]{cite232}}.

\codefieldsection{Notes}
\begin{eczvaluelist}
\item\relax See the corresponding Bendwavy database entry \NoCaseChange{\protect\cite{cite2340}}.
\end{eczvaluelist}
\codefieldsection{Parent}
\begin{eczvaluelist}
\item\relax
\flmRefsHyperref[eczindexfamilyrel]{code:polyhedron}{Polyhedron code}\end{eczvaluelist}
\codefieldsection{Child}
\begin{eczvaluelist}
\item\relax
\flmRefsHyperref[eczindexfamilyrel]{code:square_antiprism}{Square-antiprism code} --- The antiprism reduces to a square antiprism for \(q=4\).
\end{eczvaluelist}
\codefieldsection{Cousins}
\begin{eczvaluelist}
\item\relax
\flmRefsHyperref[eczindexfamilyrel]{code:spherical_design}{Spherical design} --- For the case when the two \(q\)-gons are such that the \(q=2,3\) cases reduce to the tetrahedron and octahedron, respectively, the antiprism is a spherical 3-design for \(q \geq 3\), and a \(2\)-design for \(q=2\) \NoCaseChange{\protect\cite{cite2339}}. This can be seen as a consequence of \NoCaseChange{\protect\cite[{Lemma 6.11}]{cite232}}.
\item\relax
\flmRefsHyperref[eczindexfamilyrel]{code:biorthogonal_spherical}{Biorthogonal spherical code} --- The antiprism reduces to the octahedron for \(q=3\).
\item\relax
\flmRefsHyperref[eczindexfamilyrel]{code:simplex_spherical}{Simplex spherical code} --- The antiprism reduces to the tetrahedron for \(q=2\).
\end{eczvaluelist}
\eczhbkcontributors{ \eczhuVVA }
\endeczcode

\eczcode{binary_antipodal}{Binary antipodal code}{}
\codefieldsection{Alternative Names}
\begin{eczvaluelist}
\item\relax Binary signal constellation
\end{eczvaluelist}
\eczhIndexCodeAliasName{binary_antipodal}{Binary signal constellation}
\codefieldsection{Description}
An \((n,K,4d/n)\) spherical code obtained from a binary \((n,K,d)\) code via the \flmRefsHyperref{ref38}{antipodal mapping}.

\begin{defterm}{Antipodal mapping}\label{ref2341}\label{ref38}
The antipodal mapping, also known as a \textit{Euclidean-space image} or \(Y_2\) construction), is a component-wise mapping from binary space into Euclidean space.
Each coordinate of a binary string is mapped into a sign, \(0\to +1\) and \(1 \to -1\) \NoCaseChange{\protect\cite[{Example 1.2.1}]{cite115}}.
\end{defterm}

\codefieldsection{Parent}
\begin{eczvaluelist}
\item\relax
\flmRefsHyperref[eczindexfamilyrel]{code:polyphase}{Polyphase code} --- The polyphase mapping for \(q=2\) reduces to the \flmRefsHyperref{ref38}{antipodal mapping}.
\end{eczvaluelist}
\codefieldsection{Children}
\begin{eczvaluelist}
\item\relax
\flmRefsHyperref[eczindexfamilyrel]{code:bpsk}{Binary PSK (BPSK) modulation format} --- A binary antipodal code can be thought of as a concatenation of a binary outer code with a BPSK inner code. A single-bit binary code yields a spherical \((n,2,4)\) spherical code under the \flmRefsHyperref{ref38}{antipodal mapping}, which is equivalent to the BPSK code for dimension \(n=2\).
\item\relax
\flmRefsHyperref[eczindexfamilyrel]{code:kerdock_spherical}{Kerdock spherical code}\end{eczvaluelist}
\codefieldsection{Cousins}
\begin{eczvaluelist}
\item\relax
\flmRefsHyperref[eczindexfamilyrel]{code:bits_into_bits}{Binary code} --- Binary antipodal codes are spherical codes obtained from binary codes via the \flmRefsHyperref{ref38}{antipodal mapping}.
\item\relax
\flmRefsHyperref[eczindexfamilyrel]{code:slepian_group}{Slepian group-orbit code} --- Any length-\(n\) binary linear code can be used to define a diagonal subgroup of \(n\)-dimensional rotation matrices with \(\pm 1\) on the diagonals via the \flmRefsHyperref{ref38}{antipodal mapping} \(0\to+1\) and \(1\to-1\). The orbit of this subgroup yields the corresponding Slepian group-orbit code; see \NoCaseChange{\protect\cite[{Thm. 8.5.2}]{cite115}}.
\item\relax
\flmRefsHyperref[eczindexfamilyrel]{code:biorthogonal_spherical}{Biorthogonal spherical code} --- Each first-order RM\((1,m)\) code maps to a \((2^m,2^{m+1})\) biorthogonal spherical code under the \flmRefsHyperref{ref38}{antipodal mapping} \NoCaseChange{\protect\cite{cite1181}\protect\cite[{Sec. 6.4}]{cite1165}\protect\cite[{pg. 19}]{cite115}}. In other words, first-order RM (biorthogonal spherical) codes form orthoplexes in Hamming (Euclidean) space.
\item\relax
\flmRefsHyperref[eczindexfamilyrel]{code:hypercube}{Hypercube code} --- Binary antipodal codes are subcodes of a hypercube code since the hypercube code corresponds to the Hamming \(n\)-cube (a.k.a. Boolean hypercube) embedded into the unit \(n\)-sphere.
\item\relax
\flmRefsHyperref[eczindexfamilyrel]{code:simplex_spherical}{Simplex spherical code} --- Binary simplex codes map to \((2^m,2^m+1)\) simplex spherical codes under the \flmRefsHyperref{ref38}{antipodal mapping} \NoCaseChange{\protect\cite[{Sec. 6.5.2}]{cite1165}\protect\cite[{pg. 18}]{cite115}}. In other words, simplex (simplex spherical) codes form simplices in Hamming (Euclidean) space.
\end{eczvaluelist}
\eczhbkcontributors{ \eczhuVVA }
\endeczcode

\eczcode{binary_balanced}{Binary balanced spherical code}{}
\codefieldsection{Description}
An \((n-1,K,\frac{nd}{nw-w^2})\) spherical code obtained from a constant-weight-\(w\) binary \((n,K,d)\) code via the component-wise binary balanced mapping.

The \textit{binary balanced mapping} (also known as \textit{the CW\(_2\) construction}) is defined as
\flmMathEnvironment{align}{}{
  \begin{split}
    0&\to\sqrt{\frac{w}{n\left(n-w\right)}}\\
    1&\to -\sqrt{\frac{n-w}{nw}}~.
  \end{split}
}
This construction can be extended to the general balanced binary construction CW\(_q\) for spherical code alphabets of size \(q\) \NoCaseChange{\protect\cite[{Sec. 6.6}]{cite115}}.

\codefieldsection{Notes}
\begin{eczvaluelist}
\item\relax See \NoCaseChange{\protect\cite[{Sec. 6.2}]{cite115}} for more details.
\end{eczvaluelist}
\codefieldsection{Parents}
\begin{eczvaluelist}
\item\relax
\flmRefsHyperref[eczindexfamilyrel]{code:spherical}{Spherical code}\item\relax
\flmRefsHyperref[eczindexfamilyrel]{code:concatenated}{Concatenated code} --- A binary balanced spherical code can be thought of as a concatenation of a constant-weight binary outer code with a shifted and scaled BPSK-like inner code.
\end{eczvaluelist}
\codefieldsection{Cousins}
\begin{eczvaluelist}
\item\relax
\flmRefsHyperref[eczindexfamilyrel]{code:constant_weight}{Constant-weight code} --- Binary balanced spherical codes are obtained from constant-weight binary codes.
\item\relax
\flmRefsHyperref[eczindexfamilyrel]{code:bpsk}{Binary PSK (BPSK) modulation format} --- A binary balanced spherical code can be thought of as a concatenation of a constant-weight binary outer code with a shifted and scaled BPSK-like inner code.
\end{eczvaluelist}
\eczhbkcontributors{ \eczhuVVA }
\endeczcode

\eczcode{bpsk}{Binary PSK (BPSK) modulation format}{~\NoCaseChange{\protect\cite{cite2342}}}
\codefieldsection{Alternative Names}
\begin{eczvaluelist}
\item\relax Binary PSK (BPSK) modulation code
\item\relax Binary PSK (BPSK) modulation scheme
\item\relax Binary PSK (BPSK) signaling format
\item\relax Binary antipodal modulation
\item\relax Phase-reversal keying (PRK)
\item\relax Antipodal signaling
\end{eczvaluelist}
\eczhIndexCodeAliasName{bpsk}{Binary PSK (BPSK) modulation code}
\eczhIndexCodeAliasName{bpsk}{Binary PSK (BPSK) modulation scheme}
\eczhIndexCodeAliasName{bpsk}{Binary PSK (BPSK) signaling format}
\eczhIndexCodeAliasName{bpsk}{Binary antipodal modulation}
\eczhIndexCodeAliasName{bpsk}{Phase-reversal keying (PRK)}
\eczhIndexCodeAliasName{bpsk}{Antipodal signaling}
\codefieldsection{Description}
Encodes one bit of information into a constellation of antipodal points \(\pm\alpha\) for complex \(\alpha\).
These points are typically associated with two phases of an electromagnetic signal.

\codefieldsection{Rate}
Achieves capacity of AWGN in the low signal-to-noise regime \NoCaseChange{\protect\cite{cite2283}} (see also \NoCaseChange{\protect\cite{cite2342}}). BPSK concatenated with classical-quantum polar codes approaches the Holevo capacity of the \flmRefsHyperref{ref498}{pure-loss} optical channel in the low-photon-number regime \NoCaseChange{\protect\cite{cite2343}}.
\codefieldsection{Realizations}
\begin{eczvaluelist}
\item\relax Telephone-line modems throughout 1950s and 1960s: Bell 103 and 202, as well as international standards V.21 \NoCaseChange{\protect\cite{cite237}} and V.23 \NoCaseChange{\protect\cite{cite238}}.
\end{eczvaluelist}
\codefieldsection{Parents}
\begin{eczvaluelist}
\item\relax
\flmRefsHyperref[eczindexfamilyrel]{code:psk}{Phase-shift keying (PSK) modulation format} --- BPSK is also known as 2-PSK.
\item\relax
\flmRefsHyperref[eczindexfamilyrel]{code:binary_antipodal}{Binary antipodal code} --- A binary antipodal code can be thought of as a concatenation of a binary outer code with a BPSK inner code. A single-bit binary code yields a spherical \((n,2,4)\) spherical code under the \flmRefsHyperref{ref38}{antipodal mapping}, which is equivalent to the BPSK code for dimension \(n=2\).
\end{eczvaluelist}
\codefieldsection{Cousins}
\begin{eczvaluelist}
\item\relax
\flmRefsHyperref[eczindexfamilyrel]{code:pam}{Pulse-amplitude modulation (PAM) format} --- BPSK for real \(\alpha\) is the simplest non-trivial PAM encoding.
\item\relax
\flmRefsHyperref[eczindexfamilyrel]{code:binary_linear}{Linear binary code} --- Concatenating binary linear codes with BPSK yields a standard way of digitizing the analog AWGN channel \NoCaseChange{\protect\cite[{Ch. 29}]{cite194}}.
\item\relax
\flmRefsHyperref[eczindexfamilyrel]{code:two-legged-cat}{Two-component cat code} --- BPSK (two-component cat) codes are used to transmit classical (quantum) information using (superpositions of) antipodal coherent states over classical (quantum) channels.
\item\relax
\flmRefsHyperref[eczindexfamilyrel]{code:polar_for_quantum}{Polar c-q code} --- BPSK concatenated with classical-quantum polar codes approaches the Holevo capacity of the \flmRefsHyperref{ref498}{pure-loss} optical channel in the low-photon-number regime \NoCaseChange{\protect\cite{cite2343}}.
\item\relax
\flmRefsHyperref[eczindexfamilyrel]{code:turbo}{Turbo code} --- Turbo codes can be concatenated with BPSK codes \NoCaseChange{\protect\cite{cite2076}}.
\item\relax
\flmRefsHyperref[eczindexfamilyrel]{code:binary_balanced}{Binary balanced spherical code} --- A binary balanced spherical code can be thought of as a concatenation of a constant-weight binary outer code with a shifted and scaled BPSK-like inner code.
\item\relax
\flmRefsHyperref[eczindexfamilyrel]{code:quantum_bpsk}{BPSK c-q modulation format} --- BPSK (BPSK c-q) codes are used to transmit classical information using antipodal coherent states over classical (quantum) channels.
\item\relax
\flmRefsHyperref[eczindexfamilyrel]{code:quantum_hadamard_bpsk}{Hadamard BPSK c-q modulation format} --- The Hadamard BPSK c-q code can be thought of as a concatenation of the Hadamard binary linear code with BPSK for the purposes of transmission of classical information over quantum channels.
\end{eczvaluelist}
\eczhbkcontributors{ \eczhuVVA }
\endeczcode

\eczcode{biorthogonal_spherical}{Biorthogonal spherical code}{}
\codefieldsection{Alternative Names}
\begin{eczvaluelist}
\item\relax Cross polytope code
\item\relax Hyperoctahedron code
\item\relax Orthoplex code
\item\relax Co-cube code
\end{eczvaluelist}
\eczhIndexCodeAliasName{biorthogonal_spherical}{Cross polytope code}
\eczhIndexCodeAliasName{biorthogonal_spherical}{Hyperoctahedron code}
\eczhIndexCodeAliasName{biorthogonal_spherical}{Orthoplex code}
\eczhIndexCodeAliasName{biorthogonal_spherical}{Co-cube code}
\codefieldsection{Description}
Spherical \((n,2n,2)\) code whose codewords are all permutations of the \(n\)-dimensional vectors \((0,0,\cdots,0,\pm1)\), up to normalization.
The code makes up the vertices of an \(n\)-orthoplex (a.k.a. hyperoctahedron or cross polytope).

For \(n=3\), biorthogonal spherical codewords make up the vertices of an octahedron.
For \(n=4\), codewords make up the vertices of a 16-cell, or, equivalently, the Möbius-Kantor complex polygon.
A quaternion realization of the vertices yields the quaternion group \(Q\).

The set of permutations of \((0,0,\cdots,0,1)\) forms an orthogonal set and yields the biorthogonal code when combined with the set of permutations of \((0,0,\cdots,0,-1)\).

\codefieldsection{Protection}
Biorthogonal spherical codes saturate the absolute bound for antipodal codes and the third Rankin bound \NoCaseChange{\protect\cite{cite115}}.
Biorthogonal codes are unique up to equivalence \NoCaseChange{\protect\cite[{pg. 19}]{cite115}}, which follows from saturating the Boroczky bound \NoCaseChange{\protect\cite{cite2334}}.
The octahedron is the optimal antipodal configuration of 6 points in 3D space \NoCaseChange{\protect\cite{cite2344}}.

\codefieldsection{Notes}
\begin{eczvaluelist}
\item\relax See the corresponding Bendwavy database entry \NoCaseChange{\protect\cite{cite2345}}.
\end{eczvaluelist}
\codefieldsection{Parents}
\begin{eczvaluelist}
\item\relax
\flmRefsHyperref[eczindexfamilyrel]{code:polytope}{Polytope code} --- Biorthogonal spherical codewords in 2 (3, 4, \(n\)) dimensions form the vertices of a square (octahedron, 16-cell, \(n\)-orthoplex).
\item\relax
\flmRefsHyperref[eczindexfamilyrel]{code:sharp_config}{Spherical sharp configuration}\item\relax
\flmRefsHyperref[eczindexfamilyrel]{code:lattice_shell}{Lattice-shell code} --- Biorthogonal codewords form the minimal shell of the \(\mathbb{Z}^n\) hypercubic lattice.
\item\relax
\flmRefsHyperref[eczindexfamilyrel]{code:permutation_spherical}{Permutation spherical code}\end{eczvaluelist}
\codefieldsection{Child}
\begin{eczvaluelist}
\item\relax
\flmRefsHyperref[eczindexfamilyrel]{code:qpsk}{Quadrature PSK (QPSK) modulation format} --- The QPSK is equivalent to the biorthogonal spherical code for \(n=2\).
\end{eczvaluelist}
\codefieldsection{Cousins}
\begin{eczvaluelist}
\item\relax
\flmRefsHyperref[eczindexfamilyrel]{code:spherical_design}{Spherical design} --- Biorthogonal spherical codes are the only tight spherical 3-designs \NoCaseChange{\protect\cite[{Tab. 9.3}]{cite115}}. A suitable weighted union of the vertices of a hypercube and an orthoplex forms a weighted spherical 5-design in dimensions \(\geq 3\) \NoCaseChange{\protect\cite[{Sec. 8.6, Ex. 5-2}]{cite379}\protect\cite[{Exam. 2.6}]{cite380}}.
\item\relax
\flmRefsHyperref[eczindexfamilyrel]{code:24cell}{24-cell code} --- Vertices of a 24-cell can be split up into vertices of three 16-cells, which are biorthogonal spherical codes for \(n=4\) \NoCaseChange{\protect\cite{cite178}}. The vertices of a 24-cell are a union of the vertices of a tesseract and a 16-cell \NoCaseChange{\protect\cite[{Exam. 2.6}]{cite380}}.
\item\relax
\flmRefsHyperref[eczindexfamilyrel]{code:dual_polytope}{Dual polytope code} --- Orthoplexes and hypercubes are dual to each other.
\item\relax
\flmRefsHyperref[eczindexfamilyrel]{code:hypercube}{Hypercube code} --- Orthoplexes and hypercubes are dual to each other. A suitable weighted union of the vertices of a hypercube and an orthoplex forms a weighted spherical 5-design in dimensions \(\geq 3\) \NoCaseChange{\protect\cite[{Sec. 8.6, Ex. 5-2}]{cite379}\protect\cite[{Exam. 2.6}]{cite380}}.
\item\relax
\flmRefsHyperref[eczindexfamilyrel]{code:binary_antipodal}{Binary antipodal code} --- Each first-order RM\((1,m)\) code maps to a \((2^m,2^{m+1})\) biorthogonal spherical code under the \flmRefsHyperref{ref38}{antipodal mapping} \NoCaseChange{\protect\cite{cite1181}\protect\cite[{Sec. 6.4}]{cite1165}\protect\cite[{pg. 19}]{cite115}}. In other words, first-order RM (biorthogonal spherical) codes form orthoplexes in Hamming (Euclidean) space.
\item\relax
\flmRefsHyperref[eczindexfamilyrel]{code:antiprism}{Antiprism code} --- The antiprism reduces to the octahedron for \(q=3\).
\item\relax
\flmRefsHyperref[eczindexfamilyrel]{code:biorthogonal}{\([2^m,m+1,2^{m-1}]\) First-order RM code} --- Each first-order RM code maps to a \((2^m,2^{m+1})\) biorthogonal spherical code under the \flmRefsHyperref{ref38}{antipodal mapping} \NoCaseChange{\protect\cite{cite1181}\protect\cite[{Sec. 6.4}]{cite1165}\protect\cite[{pg. 19}]{cite115}}. In other words, first-order RM (biorthogonal spherical) codes form orthoplexes in Hamming (Euclidean) space.
\item\relax
\flmRefsHyperref[eczindexfamilyrel]{code:ppm}{Pulse-position modulation (PPM) format} --- PPM codewords form a spherical code whose constellation consists of the standard basis vectors. Adjoining negatives yields the corresponding biorthogonal spherical code.
\item\relax
\flmRefsHyperref[eczindexfamilyrel]{code:rhombic_dodecahedron}{Rhombic dodecahedron code} --- The vertices of a rhombic dodecahedron are a union of the vertices of a cube and an octahedron.
\item\relax
\flmRefsHyperref[eczindexfamilyrel]{code:kerdock_spherical}{Kerdock spherical code} --- Kerdock spherical codes form spherical 3-designs because their codewords are unions of \(2^{2r-1}+1\) orthoplexes \NoCaseChange{\protect\cite{cite386}}.
\item\relax
\flmRefsHyperref[eczindexfamilyrel]{code:polyphase}{Polyphase code} --- Biorthogonal spherical codes for dimension \(n=p\) with \(p\) an odd prime admit a polyphase realization \NoCaseChange{\protect\cite[{Sec. 7.7}]{cite115}}.
\item\relax
\flmRefsHyperref[eczindexfamilyrel]{code:quantum_ppm}{Pulse-position (PPM) c-q modulation format} --- PPM c-q codewords are c-q spherical codes whose constellation consists of the standard basis vectors. Adjoining negatives yields the corresponding biorthogonal c-q spherical code.
\end{eczvaluelist}
\eczhbkcontributors{ \eczhuVVA }
\endeczcode

\eczcode{cgs_spherical}{Cameron-Goethals-Seidel (CGS) isotropic subspace code}{~\NoCaseChange{\protect\cite{cite2346}}}
\codefieldsection{Description}
Member of a \((q(q^2-q+1),(q+1)(q^3+1),2-2/q^2)\) family of spherical codes for any prime-power \(q\).
Constructed from generalized quadrangles, which in this case correspond to sets of totally isotropic points and lines in the projective space \(PG(5,q)\) \NoCaseChange{\protect\cite[{Exam. 9.4.5}]{cite115}}.
There exist multiple distinct spherical codes using this construction for \(q>3\) \NoCaseChange{\protect\cite{cite119,cite386}}.

\codefieldsection{Protection}
CGS isotropic subspace codes saturate the Levenshtein bound \NoCaseChange{\protect\cite[{pg. 64}]{cite115}}.
\codefieldsection{Parent}
\begin{eczvaluelist}
\item\relax
\flmRefsHyperref[eczindexfamilyrel]{code:sharp_config}{Spherical sharp configuration} --- CGS isotropic subspace codes are the only known spherical sharp configurations not derived from regular polytopes or lattices \NoCaseChange{\protect\cite{cite119}}.
\end{eczvaluelist}
\codefieldsection{Child}
\begin{eczvaluelist}
\item\relax
\flmRefsHyperref[eczindexfamilyrel]{code:hessian_polyhedron}{Hessian polyhedron code} --- The CGS isotropic subspace code for \(q=2\) reduces to the Hessian polytope.
\end{eczvaluelist}
\codefieldsection{Cousin}
\begin{eczvaluelist}
\item\relax
\flmRefsHyperref[eczindexfamilyrel]{code:projective}{Projective geometry code} --- CGS isotropic subspace codes are constructed from incidence matrices of \(PG(5,q)\) \NoCaseChange{\protect\cite[{Exam. 9.4.5}]{cite115}}.
\end{eczvaluelist}
\eczhbkcontributors{ \eczhuVVA }
\endeczcode

\eczcode{complex_hadamard}{Complex Hadamard spherical code}{~\NoCaseChange{\protect\cite{cite2347}}}
\codefieldsection{Description}
A spherical code obtained from particular complex Hadamard matrices \NoCaseChange{\protect\cite{cite2348}}.

\codefieldsection{Protection}
Covering radius can be bounded in terms of the code's spherical design strength \NoCaseChange{\protect\cite{cite2347}}.

\codefieldsection{Notes}
\begin{eczvaluelist}
\item\relax Database of complex Hadamard matrices \NoCaseChange{\protect\cite{cite2349}}.
\end{eczvaluelist}
\codefieldsection{Parent}
\begin{eczvaluelist}
\item\relax
\flmRefsHyperref[eczindexfamilyrel]{code:spherical}{Spherical code}\end{eczvaluelist}
\codefieldsection{Cousin}
\begin{eczvaluelist}
\item\relax
\flmRefsHyperref[eczindexfamilyrel]{code:stabilizer_over_gfqsq}{Hermitian Galois-qudit code} --- Complex Hadamard matrices can be used to build Hermitian \NoCaseChange{\protect\cite{cite2348}} and other \NoCaseChange{\protect\cite{cite2350}} Galois-qudit stabilizer codes.
\end{eczvaluelist}
\eczhbkcontributors{ \eczhuVVA }
\endeczcode

\eczcode{points_into_spheres}{Constant-energy spherical code}{}

\codefieldsection{Kingdom root code}
for the \flmRefsHyperref{kingdom:points_into_spheres}{Spherical Kingdom}.
\codefieldsection{Description}
Code whose codewords are points on a real or complex sphere whose radius squared is called the \textit{energy}.
Typically, only angular distances between points are relevant for code performance, so one can normalize codewords of a constant-energy code to obtain up a spherical code, i.e., a constant energy code with energy one.

\codefieldsection{Protection}
Constant-energy codes are sphere packings constrained to lie on a sphere, meaning that they can be used to transmit information through the AWGN channel.
For a given dimension \(n\), number of codewords \(M\), and average energy \(P\), the \textit{constant-energy Gaussian channel coding problem} asks to find a set of \(M\) codewords of energy \(nP\) that minimizes the error probability during transmission; see \NoCaseChange{\protect\cite[{Ch. 3}]{cite39}}.

\codefieldsection{Notes}
\begin{eczvaluelist}
\item\relax See \NoCaseChange{\protect\cite[{Ch. 7}]{cite115}} for more details.
\end{eczvaluelist}
\codefieldsection{Parents}
\begin{eczvaluelist}
\item\relax
\flmRefsHyperref[eczindexfamilyrel]{code:points_into_balls}{Bounded-energy code} --- Constant-energy codes are bounded-energy codes constrained to lie on a sphere.
\item\relax
\flmRefsHyperref[eczindexfamilyrel]{code:2pt_homogeneous}{Two-point homogeneous-space code} --- Real spheres are compact connected two-point homogeneous spaces with quotient \(SO(D+1)/SO(D)\) \NoCaseChange{\protect\cite[{Table 1}]{cite985}}. Complex spheres can be treated as real spheres of twice the dimension over \(\mathbb{R}\), with quotient \(SU(D+1)/SU(D)\). In fact, spheres are compact three-point homogeneous spaces \NoCaseChange{\protect\cite[{Sec. 6.6.1.1}]{cite2242}}.
\end{eczvaluelist}
\codefieldsection{Child}
\begin{eczvaluelist}
\item\relax
\flmRefsHyperref[eczindexfamilyrel]{code:spherical}{Spherical code}\end{eczvaluelist}
\codefieldsection{Cousins}
\begin{eczvaluelist}
\item\relax
\flmRefsHyperref[eczindexfamilyrel]{code:real_projective}{Real projective space code} --- Real projective space can be obtained from the sphere by identifying antipodal points, i.e., \(\mathbb{R}P^N = S^N/\mathbb{Z}_2\). As such, real projective space codes are in one-to-one correspondence with antipodal spherical codes, with each antipodal pair of spherical codewords corresponding to one line in projective space.
\item\relax
\flmRefsHyperref[eczindexfamilyrel]{code:qsc}{Quantum spherical code (QSC)} --- QSCs are quantum counterparts of spherical and constant-energy codes because they store information in quantum superpositions of points on a sphere in quantum phase space.
\end{eczvaluelist}
\eczhbkcontributors{ \eczhuVVA }
\endeczcode

\eczcode{cubeoctahedron}{Cuboctahedron code}{}
\codefieldsection{Alternative Names}
\begin{eczvaluelist}
\item\relax Rectified cube code
\item\relax Rectified octahedron code
\end{eczvaluelist}
\eczhIndexCodeAliasName{cubeoctahedron}{Rectified cube code}
\eczhIndexCodeAliasName{cubeoctahedron}{Rectified octahedron code}
\codefieldsection{Description}
Spherical \((3,12,1)\) code whose codewords are the vertices of the cuboctahedron.
Codewords form the minimal lattice-shell code of the \(D_3\) face-centered cubic (fcc) lattice.

\codefieldsection{Protection}
Code yields an optimal solution to the kissing problem in 3D \NoCaseChange{\protect\cite{cite2351}}.

\codefieldsection{Notes}
\begin{eczvaluelist}
\item\relax See the corresponding Bendwavy database entry \NoCaseChange{\protect\cite{cite2352}}.
\end{eczvaluelist}
\codefieldsection{Parents}
\begin{eczvaluelist}
\item\relax
\flmRefsHyperref[eczindexfamilyrel]{code:polyhedron}{Polyhedron code}\item\relax
\flmRefsHyperref[eczindexfamilyrel]{code:lattice_shell}{Lattice-shell code} --- Cuboctahedron codewords form the minimal shell of the \(D_3\) face-centered cubic (fcc) lattice.
\end{eczvaluelist}
\codefieldsection{Cousins}
\begin{eczvaluelist}
\item\relax
\flmRefsHyperref[eczindexfamilyrel]{code:dthree}{\(D_3\) face-centered cubic (fcc) lattice} --- Cuboctahedron codewords form the minimal shell of the \(D_3\) face-centered cubic (fcc) lattice.
\item\relax
\flmRefsHyperref[eczindexfamilyrel]{code:rhombic_dodecahedron}{Rhombic dodecahedron code} --- The rhombic dodecahedron and cuboctahedron are dual to each other \NoCaseChange{\protect\cite{cite2353}}.
\end{eczvaluelist}
\eczhbkcontributors{ \eczhuVVA }
\endeczcode

\eczcode{disphenoidal288cell}{Disphenoidal 288-cell code}{}
\codefieldsection{Description}
Spherical \((4,48,2-\sqrt{2})\) code \NoCaseChange{\protect\cite[{Exam. 1.2.6}]{cite115}} whose codewords are the vertices of the disphenoidal 288-cell.
Codewords are the union of two 24-point lattice shells of the \(D_4\) lattice.

The first shell consists of the 24 permutations of the four vectors \((0,0,\pm 1,\pm 1)\), and the second of the 16 vectors \((\pm 1,\pm 1,\pm 1,\pm 1)\) and the 8 permutations of the vectors \((0,0,0,\pm 2)\).
A realization in terms of quaternion coordinates yields the 48 elements of the binary octahedral group \(2O\) \NoCaseChange{\protect\cite[{Sec. 8.6}]{cite230}}.

\begin{flmFloat}{figure}{NumCap}\includegraphics[width=306.14173228346453bp,max width=\linewidth]{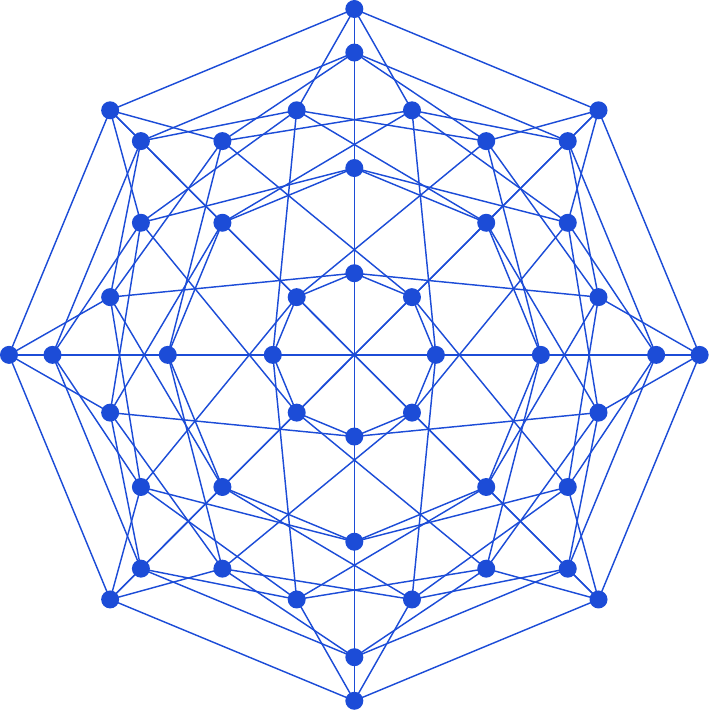}\caption{Projection of the coordinates of the disphenoidal 288-cell.}\label{ref2354}\end{flmFloat}

\codefieldsection{Parents}
\begin{eczvaluelist}
\item\relax
\flmRefsHyperref[eczindexfamilyrel]{code:dfour_shell}{\(D_4\) lattice-shell code} --- Disphenoidal 288-cell codewords are the union of two 24-point shells of the \(D_4\) lattice, with each shell making up the vertices of a 24-cell.
\item\relax
\flmRefsHyperref[eczindexfamilyrel]{code:spherical_design}{Spherical design} --- The disphenoidal 288-cell code forms a spherical 7-design \NoCaseChange{\protect\cite{cite381}}.
\end{eczvaluelist}
\codefieldsection{Cousins}
\begin{eczvaluelist}
\item\relax
\flmRefsHyperref[eczindexfamilyrel]{code:group_classical}{Group-alphabet code} --- The disphenoidal 288-cell code has a quaternion-coordinate realization as the 48 elements of the binary octahedral group \(2O\), one of the three exceptional finite subgroups of \(SU(2)\) \NoCaseChange{\protect\cite[{Sec. 8.6}]{cite230}}.
\item\relax
\flmRefsHyperref[eczindexfamilyrel]{code:sidelnikov}{Real-Clifford subgroup-orbit code} --- The disphenoidal 288-cell code is a group-orbit code with the group being the \flmRefsHyperref{ref409}{real Clifford group} in \(4\) dimensions.
\item\relax
\flmRefsHyperref[eczindexfamilyrel]{code:24cell}{24-cell code} --- Vertices of a disphenoidal 288-cell can be split up into vertices of a 24-cell and its dual 24-cell \NoCaseChange{\protect\cite[{Sec. 8.6}]{cite230}}.
\item\relax
\flmRefsHyperref[eczindexfamilyrel]{code:quantum_sidelnikov}{Clifford subgroup-orbit QSC} --- Logical constellations of the Clifford subgroup-orbit code for \(r=1\) form vertices of 24-cells when mapped into the real sphere, while code constellations form vertices of a disphenoidal 288-cell.
\end{eczvaluelist}
\eczhbkcontributors{ Shubham P. Jain, \eczhuVVA }
\endeczcode

\eczcode{dodecahedron}{Dodecahedron code}{}
\codefieldsection{Alternative Names}
\begin{eczvaluelist}
\item\relax Cosmohedron code
\end{eczvaluelist}
\eczhIndexCodeAliasName{dodecahedron}{Cosmohedron code}
\codefieldsection{Description}
Spherical \((3,20,2-2\sqrt{5}/3)\) code whose codewords are the vertices of the dodecahedron (alternatively, the centers of the faces of a icosahedron, the dodecahedron's dual polytope).

\codefieldsection{Notes}
\begin{eczvaluelist}
\item\relax See the corresponding Bendwavy database entry \NoCaseChange{\protect\cite{cite2355}}.
\end{eczvaluelist}
\codefieldsection{Parents}
\begin{eczvaluelist}
\item\relax
\flmRefsHyperref[eczindexfamilyrel]{code:polyhedron}{Polyhedron code}\item\relax
\flmRefsHyperref[eczindexfamilyrel]{code:spherical_design}{Spherical design} --- The dodecahedron code forms a spherical 5-design \NoCaseChange{\protect\cite{cite382}}.
\end{eczvaluelist}
\codefieldsection{Cousins}
\begin{eczvaluelist}
\item\relax
\flmRefsHyperref[eczindexfamilyrel]{code:dual_polytope}{Dual polytope code} --- The icosahedron and dodecahedron are dual to each other.
\item\relax
\flmRefsHyperref[eczindexfamilyrel]{code:icosahedron}{Icosahedron code} --- The icosahedron and dodecahedron are dual to each other.
\item\relax
\flmRefsHyperref[eczindexfamilyrel]{code:extended_golay}{\([24, 12, 8]\) Extended Golay code} --- The parity bits of the extended Golay code can be visualized to lie on the vertices of the icosahedron; see \NoCaseChange{\protect\cite{cite1196}} for more details. To construct the code, one can use the great dodecahedron to generate codewords by placing message bits on the faces and calculating the parity bits that live on the 12 vertices of the inner icosahedron.
\item\relax
\flmRefsHyperref[eczindexfamilyrel]{code:petersen}{\([15,6,5]\) Petersen cycle code} --- The Petersen graph can be thought of as a dodecahedron with antipodes identified \NoCaseChange{\protect\cite[{Appx. A.2.1}]{cite101}}.
\item\relax
\flmRefsHyperref[eczindexfamilyrel]{code:pentakis_dodecahedron}{Pentakis dodecahedron code} --- The pentakis dodecahedron is the convex hull of the icosahedron and dodecahedron.
\item\relax
\flmRefsHyperref[eczindexfamilyrel]{code:simplex_spherical}{Simplex spherical code} --- Vertices of a dodecahedron can be split up into vertices of five tetrahedra, which are simplex spherical codes for \(n=3\) \NoCaseChange{\protect\cite{cite178}}.
\item\relax
\flmRefsHyperref[eczindexfamilyrel]{code:quantum_dodecahedron}{\(\llbracket 16,4,3\rrbracket \) dodecahedral code} --- The \flmRefsHyperref{ref857}{encoder-respecting form} of the \(\llbracket 16,4,3\rrbracket \) dodecahedral code is the graph of vertices of a dodecahedron \NoCaseChange{\protect\cite{cite858}}.
\item\relax
\flmRefsHyperref[eczindexfamilyrel]{code:stellated_dodecahedron_css}{\(\llbracket 30,8,3\rrbracket \) Bring code} --- The qubits and stabilizer generators of the \(\llbracket 30,8,3\rrbracket \) Bring code lie on the vertices of the small stellated dodecahedron.
\end{eczvaluelist}
\eczhbkcontributors{ \eczhuVVA }
\endeczcode

\eczcode{dual_polytope}{Dual polytope code}{}
\codefieldsection{Alternative Names}
\begin{eczvaluelist}
\item\relax Reciprocal polytope code
\end{eczvaluelist}
\eczhIndexCodeAliasName{dual_polytope}{Reciprocal polytope code}
\codefieldsection{Description}
For a spherical code whose codewords are vertices of a convex polytope, the dual code consists of codewords corresponding to the facets of the original polytope, i.e., to the vertices of the polar dual polytope. For regular polytopes, these dual codewords can be represented by the normalized centers of the facets of the original polytope. The dual codewords make up the vertices of the polytope dual to the original polytope.

If the dual polytope is the same as the original polytope, the original polytope is said to be \textit{self-dual}. 
Self-dual polytopes have the same number of facets as vertices.

\codefieldsection{Parent}
\begin{eczvaluelist}
\item\relax
\flmRefsHyperref[eczindexfamilyrel]{code:polytope}{Polytope code}\end{eczvaluelist}
\codefieldsection{Child}
\begin{eczvaluelist}
\item\relax
\flmRefsHyperref[eczindexfamilyrel]{code:self_dual_polytope}{Self-dual polytope code}\end{eczvaluelist}
\codefieldsection{Cousins}
\begin{eczvaluelist}
\item\relax
\flmRefsHyperref[eczindexfamilyrel]{code:dodecahedron}{Dodecahedron code} --- The icosahedron and dodecahedron are dual to each other.
\item\relax
\flmRefsHyperref[eczindexfamilyrel]{code:icosahedron}{Icosahedron code} --- The icosahedron and dodecahedron are dual to each other.
\item\relax
\flmRefsHyperref[eczindexfamilyrel]{code:pentakis_dodecahedron}{Pentakis dodecahedron code} --- The pentakis dodecahedron and truncated icosahedron are dual to each other \NoCaseChange{\protect\cite[{pg. 55}]{cite2353}}.
\item\relax
\flmRefsHyperref[eczindexfamilyrel]{code:rhombic_dodecahedron}{Rhombic dodecahedron code} --- The rhombic dodecahedron and cuboctahedron are dual to each other \NoCaseChange{\protect\cite{cite2353}}.
\item\relax
\flmRefsHyperref[eczindexfamilyrel]{code:120cell}{120-cell code} --- The 600-cell and 120-cell are dual to each other.
\item\relax
\flmRefsHyperref[eczindexfamilyrel]{code:600cell}{600-cell code} --- The 600-cell and 120-cell are dual to each other.
\item\relax
\flmRefsHyperref[eczindexfamilyrel]{code:rect_hessian_polyhedron}{Rectified Hessian polyhedron code} --- The rectified and double Hessian polyhedra are dual to each other, analogous to the octahedron and cube.
\item\relax
\flmRefsHyperref[eczindexfamilyrel]{code:biorthogonal_spherical}{Biorthogonal spherical code} --- Orthoplexes and hypercubes are dual to each other.
\item\relax
\flmRefsHyperref[eczindexfamilyrel]{code:hypercube}{Hypercube code} --- Orthoplexes and hypercubes are dual to each other.
\end{eczvaluelist}
\eczhbkcontributors{ \eczhuVVA }
\endeczcode

\eczcode{hessian_polyhedron}{Hessian polyhedron code}{~\NoCaseChange{\protect\cite{cite2239,cite2356}}}
\codefieldsection{Alternative Names}
\begin{eczvaluelist}
\item\relax \(2_{21}\) polytope code
\item\relax Schläfli configuration
\end{eczvaluelist}
\eczhIndexCodeAliasName{hessian_polyhedron}{\(2_{21}\) polytope code}
\eczhIndexCodeAliasName{hessian_polyhedron}{Schläfli configuration}
\codefieldsection{Description}
Spherical \((6,27,3/2)\) code whose codewords are the vertices of the Hessian complex polyhedron and the \(2_{21}\) polytope.
Two copies of the code yield the \((6,54,1)\) \textit{double Hessian polyhedron} (a.k.a. diplo-Schläfli) code.
The code can be obtained from the Schläfli graph \NoCaseChange{\protect\cite[{Ch. 9}]{cite115}}.
The (antipodal pairs of) points of the (double) Hessian polyhedron correspond to the 27 lines on a smooth cubic surface in \(\mathbb{C}P^3\) \NoCaseChange{\protect\cite{cite116,cite117,cite118,cite119,cite120}}.

See \NoCaseChange{\protect\cite{cite2357}\protect\cite[{Exam. 1.2.5}]{cite115}\protect\cite[{pg. 119}]{cite231}} for a real (complex) realization of the 27 codewords.
\begin{flmFloat}{figure}{NumCap}\includegraphics[width=306.14173228346453bp,max width=\linewidth]{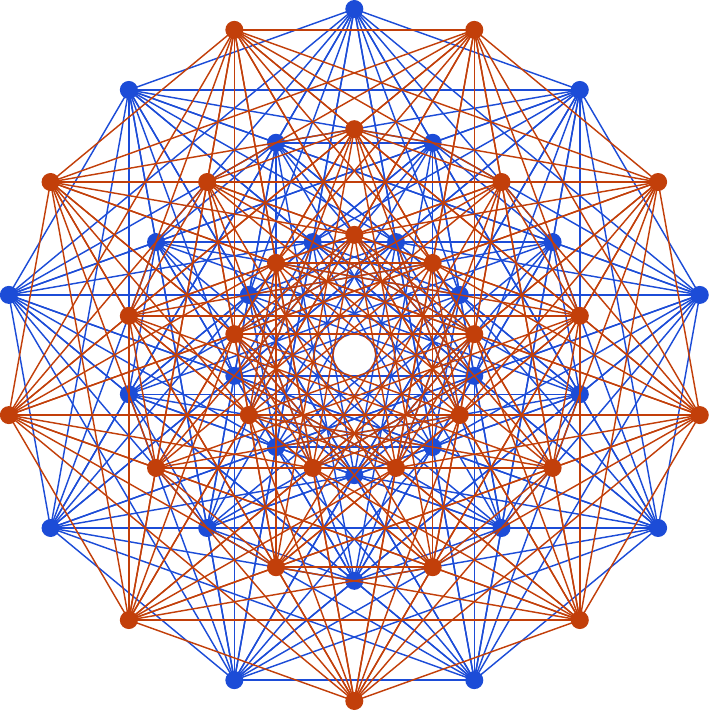}\caption{Projection of the coordinates of the Hessian and double Hessian polytopes.}\label{ref2358}\end{flmFloat}

\codefieldsection{Protection}
The Hessian polytope code has degree \(d=2\) and saturates the absolute bound \NoCaseChange{\protect\cite{cite115}}.
\codefieldsection{Realizations}
\begin{eczvaluelist}
\item\relax Quantum mechanical SIC-POVMs \NoCaseChange{\protect\cite{cite276}}.
\end{eczvaluelist}
\codefieldsection{Notes}
\begin{eczvaluelist}
\item\relax See the corresponding Bendwavy database entries for the Hessian complex polyhedron \NoCaseChange{\protect\cite{cite2359}} and its real-space embedding as the \(2_{21}\) polytope \NoCaseChange{\protect\cite{cite2360}}.
\end{eczvaluelist}
\codefieldsection{Parents}
\begin{eczvaluelist}
\item\relax
\flmRefsHyperref[eczindexfamilyrel]{code:self_dual_polytope}{Self-dual polytope code} --- The \(2_{21}\) polytope is self-dual \NoCaseChange{\protect\cite{cite116}}.
\item\relax
\flmRefsHyperref[eczindexfamilyrel]{code:cgs_spherical}{Cameron-Goethals-Seidel (CGS) isotropic subspace code} --- The CGS isotropic subspace code for \(q=2\) reduces to the Hessian polytope.
\end{eczvaluelist}
\codefieldsection{Cousins}
\begin{eczvaluelist}
\item\relax
\flmRefsHyperref[eczindexfamilyrel]{code:spherical_design}{Spherical design} --- The Hessian polytope code forms a tight spherical 4-design \NoCaseChange{\protect\cite[{Exam. 7.3}]{cite383}}. The double Hessian polytope code forms a spherical 5-design \NoCaseChange{\protect\cite{cite384}}.
\item\relax
\flmRefsHyperref[eczindexfamilyrel]{code:esix}{\(E_6\) root lattice} --- The 27 Hessian polyhedron codewords are intimately related to the \(E_6\) Lie group \NoCaseChange{\protect\cite{cite2237}}.
\item\relax
\flmRefsHyperref[eczindexfamilyrel]{code:esix_shell}{\(E_6\) lattice-shell code} --- Double Hessian polyhedron codewords form the minimal lattice-shell code of the \(E_6^{\perp}\) lattice \NoCaseChange{\protect\cite{cite2322}}.
\item\relax
\flmRefsHyperref[eczindexfamilyrel]{code:hess_polytope}{\(3_{21}\) polytope code} --- The Hessian polyhedron code forms the next recursive kissing configuration after the \(3_{21}\) polytope code in the \(E_8\) lattice-shell/Witting polytope sequence \NoCaseChange{\protect\cite{cite124}}.
\item\relax
\flmRefsHyperref[eczindexfamilyrel]{code:complex_projective}{Complex projective space code} --- The (antipodal pairs of) points of the (double) Hessian polyhedron correspond to the 27 lines on a smooth cubic surface in \(\mathbb{C}P^3\) \NoCaseChange{\protect\cite{cite117,cite118,cite119,cite120}}.
\item\relax
\flmRefsHyperref[eczindexfamilyrel]{code:witting_polytope}{Witting polytope code} --- The Hessian polyhedron code forms the next recursive kissing configuration after the \(3_{21}\) polytope code in the \(E_8\) lattice-shell/Witting polytope sequence \NoCaseChange{\protect\cite{cite124}}. The Schläfli graph is a subgraph of the graph formed by the vertices of the Witting polytope \NoCaseChange{\protect\cite[{Sec. 3.11}]{cite1385}}.
\item\relax
\flmRefsHyperref[eczindexfamilyrel]{code:rect_hessian_polyhedron}{Rectified Hessian polyhedron code} --- The Hessian and rectified Hessian polyhedra are analogues of the tetrahedron and octahedron in 3D complex space, while the double Hessian polyhedron is the analogue of a cube \NoCaseChange{\protect\cite[{pg. 127}]{cite231}}. The rectified and double Hessian polyhedra are dual to each other, just like the octahedron and cube. Moreover, the double Hessian consists of two Hessians, just like the cube can be constructed from two tetrahedra.
\item\relax
\flmRefsHyperref[eczindexfamilyrel]{code:hessian_qsc}{Hessian QSC} --- Each codeword of the Hessian QSC is a quantum superposition of vertices of a Hessian complex polyhedron.
\end{eczvaluelist}
\eczhbkcontributors{ Shubham P. Jain, \eczhuVVA }
\endeczcode

\eczcode{hypercube}{Hypercube code}{}
\codefieldsection{Alternative Names}
\begin{eczvaluelist}
\item\relax Measure polytope code
\end{eczvaluelist}
\eczhIndexCodeAliasName{hypercube}{Measure polytope code}
\codefieldsection{Description}
Spherical \((n,2^n,4/n)\) code whose codewords are vertices of an \(n\)-cube, i.e., all permutations and negations of the vector \((1,1,\cdots,1)\), up to normalization.

\codefieldsection{Protection}
The square (cube) is the optimal antipodal configuration of 4 (8) points in 2D (3D) space \NoCaseChange{\protect\cite{cite2344}}.

\codefieldsection{Notes}
\begin{eczvaluelist}
\item\relax See the corresponding Bendwavy database entry for the tesseract \NoCaseChange{\protect\cite{cite2361}}.
\end{eczvaluelist}
\codefieldsection{Parents}
\begin{eczvaluelist}
\item\relax
\flmRefsHyperref[eczindexfamilyrel]{code:polytope}{Polytope code} --- Hypercube codewords in 2 (3, 4, \(n\)) dimensions form the vertices of a square (cube, tesseract, \(n\)-cube).
\item\relax
\flmRefsHyperref[eczindexfamilyrel]{code:lattice_shell}{Lattice-shell code} --- Hypercube codewords form the minimal lattice shell code of the \(\mathbb{Z}^n\) hypercubic lattice when the lattice is shifted such that the center of a hypercube is at the origin.
\item\relax
\flmRefsHyperref[eczindexfamilyrel]{code:spherical_design}{Spherical design} --- Hypercube codes form spherical 3-designs. The weighted union of the vertices of a hypercube and an orthoplex form a weighted spherical 5-design in dimensions \(\geq 3\) \NoCaseChange{\protect\cite[{Sec. 8.6, Ex. 5-2}]{cite379}\protect\cite[{Exam. 2.6}]{cite380}}.
\item\relax
\flmRefsHyperref[eczindexfamilyrel]{code:polyphase}{Polyphase code}\end{eczvaluelist}
\codefieldsection{Cousins}
\begin{eczvaluelist}
\item\relax
\flmRefsHyperref[eczindexfamilyrel]{code:dual_polytope}{Dual polytope code} --- Orthoplexes and hypercubes are dual to each other.
\item\relax
\flmRefsHyperref[eczindexfamilyrel]{code:hypercubic}{\(\mathbb{Z}^n\) hypercubic lattice} --- Hypercube codewords form the minimal lattice shell code of the \(\mathbb{Z}^n\) hypercubic lattice when the lattice is shifted such that the center of a hypercube is at the origin.
\item\relax
\flmRefsHyperref[eczindexfamilyrel]{code:binary_antipodal}{Binary antipodal code} --- Binary antipodal codes are subcodes of a hypercube code since the hypercube code corresponds to the Hamming \(n\)-cube (a.k.a. Boolean hypercube) embedded into the unit \(n\)-sphere.
\item\relax
\flmRefsHyperref[eczindexfamilyrel]{code:bits_into_bits}{Binary code} --- Binary strings are elements of the Hamming \(n\)-cube (a.k.a. Boolean hypercube).
\item\relax
\flmRefsHyperref[eczindexfamilyrel]{code:24cell}{24-cell code} --- The vertices of a 24-cell are a union of the vertices of a tesseract and a 16-cell \NoCaseChange{\protect\cite[{Exam. 2.6}]{cite380}}.
\item\relax
\flmRefsHyperref[eczindexfamilyrel]{code:square_antiprism}{Square-antiprism code} --- The square antiprism can be obtained by stretching the cube and twisting the top of the cube by \(45\) degrees \NoCaseChange{\protect\cite[{pg. 72}]{cite115}}.
\item\relax
\flmRefsHyperref[eczindexfamilyrel]{code:rhombic_dodecahedron}{Rhombic dodecahedron code} --- The vertices of a rhombic dodecahedron are a union of the vertices of a cube and an octahedron.
\item\relax
\flmRefsHyperref[eczindexfamilyrel]{code:biorthogonal_spherical}{Biorthogonal spherical code} --- Orthoplexes and hypercubes are dual to each other. A suitable weighted union of the vertices of a hypercube and an orthoplex forms a weighted spherical 5-design in dimensions \(\geq 3\) \NoCaseChange{\protect\cite[{Sec. 8.6, Ex. 5-2}]{cite379}\protect\cite[{Exam. 2.6}]{cite380}}.
\item\relax
\flmRefsHyperref[eczindexfamilyrel]{code:hypercube_quantum}{\(\llbracket 2^D,D,2\rrbracket \) hypercube quantum code} --- \(\llbracket 2^D,D,2\rrbracket \) hypercube quantum code qubits are placed on vertices of a \(D\)-cube.
\item\relax
\flmRefsHyperref[eczindexfamilyrel]{code:kls}{Khesin-Lu-Shor code} --- The \flmRefsHyperref{ref857}{encoder-respecting form} of the \(\llbracket m 2^m / (m+1), 2^m / (m+1), d(m)\rrbracket \) Khesin-Lu-Shor code is the graph of a hypercube in \(m = 2^r - 1\) dimensions, and input nodes in the graph are codewords of the \([2^r-1,2^r-r-1,3]\) Hamming code \NoCaseChange{\protect\cite{cite858}}.
\item\relax
\flmRefsHyperref[eczindexfamilyrel]{code:stab_16_6_4}{\(\llbracket 16,6,4\rrbracket \) Tesseract color code} --- Stabilizer generators of both types of the tesseract color code are supported on each cube of a tesseract \NoCaseChange{\protect\cite{cite862,cite2362}}.
\item\relax
\flmRefsHyperref[eczindexfamilyrel]{code:cubic_surface}{\(\llbracket 8,3,2\rrbracket \) Surface code on a cube} --- The surface code on a cube, whose qubits lie on the vertices of a cube, is obtained by three-coloring the faces of a cube and placing \(X\), \(Y\), and \(Z\) stabilizer generators on each pair of faces of the same color.
\item\relax
\flmRefsHyperref[eczindexfamilyrel]{code:hemicubic}{Hemicubic code} --- Hemicubic codes are built from cellulations derived from hypercubes.
\end{eczvaluelist}
\eczhbkcontributors{ \eczhuVVA }
\endeczcode

\eczcode{icosahedron}{Icosahedron code}{}
\codefieldsection{Alternative Names}
\begin{eczvaluelist}
\item\relax Snub tetrahedron code
\end{eczvaluelist}
\eczhIndexCodeAliasName{icosahedron}{Snub tetrahedron code}
\codefieldsection{Description}
Spherical \((3,12,2-2/\sqrt{5})\) code whose codewords are the vertices of the icosahedron (alternatively, the centers of the faces of a dodecahedron, the icosahedron's dual polytope).

\codefieldsection{Protection}
Optimal configuration of 12 points in 3D space \NoCaseChange{\protect\cite[{pg. 76}]{cite115}}. Saturates the absolute bound for antipodal codes \NoCaseChange{\protect\cite[{pg. 314}]{cite115}}.
\codefieldsection{Notes}
\begin{eczvaluelist}
\item\relax See \flmHref{https://blogs.ams.org/visualinsight/2015/05/15/dodecahedron-with-5-tetrahedra}{post} by J. Baez for more details.
\item\relax See the corresponding Bendwavy database entry \NoCaseChange{\protect\cite{cite2363}}.
\end{eczvaluelist}
\codefieldsection{Parents}
\begin{eczvaluelist}
\item\relax
\flmRefsHyperref[eczindexfamilyrel]{code:polyhedron}{Polyhedron code}\item\relax
\flmRefsHyperref[eczindexfamilyrel]{code:sharp_config}{Spherical sharp configuration} --- The icosahedron is a sharp configuration \NoCaseChange{\protect\cite{cite2364,cite119}}.
\end{eczvaluelist}
\codefieldsection{Cousins}
\begin{eczvaluelist}
\item\relax
\flmRefsHyperref[eczindexfamilyrel]{code:spherical_design}{Spherical design} --- The icosahedron code forms a unique tight spherical 5-design \NoCaseChange{\protect\cite{cite385}\protect\cite[{Exam. 9.6.1}]{cite115}}.
\item\relax
\flmRefsHyperref[eczindexfamilyrel]{code:dual_polytope}{Dual polytope code} --- The icosahedron and dodecahedron are dual to each other.
\item\relax
\flmRefsHyperref[eczindexfamilyrel]{code:extended_golay}{\([24, 12, 8]\) Extended Golay code} --- The parity bits of the extended Golay code can be visualized to lie on the vertices of the icosahedron; see \NoCaseChange{\protect\cite{cite1196}} for more details. To construct the code, one can use the great dodecahedron to generate codewords by placing message bits on the faces and calculating the parity bits that live on the 12 vertices of the inner icosahedron.
\item\relax
\flmRefsHyperref[eczindexfamilyrel]{code:dodecahedron}{Dodecahedron code} --- The icosahedron and dodecahedron are dual to each other.
\item\relax
\flmRefsHyperref[eczindexfamilyrel]{code:pentakis_dodecahedron}{Pentakis dodecahedron code} --- The pentakis dodecahedron is the convex hull of the icosahedron and dodecahedron. The pentakis dodecahedron and truncated icosahedron are dual to each other \NoCaseChange{\protect\cite[{pg. 55}]{cite2353}}.
\item\relax
\flmRefsHyperref[eczindexfamilyrel]{code:600cell}{600-cell code} --- A realization of the 600-cell can be done in terms of icosians, which are quaternion coordinates of the 120 elements of the binary icosahedral group \(2I \cong 2.A_5\) (a.k.a. the icosian group) \NoCaseChange{\protect\cite{cite230}\protect\cite[{Ch. 8, pg. 207}]{cite39}}.
\item\relax
\flmRefsHyperref[eczindexfamilyrel]{code:quantum_icosahedron}{\(\llbracket 54,6,5\rrbracket \) five-covered icosahedral code} --- The \flmRefsHyperref{ref857}{encoder-respecting form} of the \(\llbracket 54,6,5\rrbracket \) five-covered icosahedral code is the graph of a five-cover of the icosahedron \NoCaseChange{\protect\cite{cite858}}.
\item\relax
\flmRefsHyperref[eczindexfamilyrel]{code:icosahedral_permutation_invariant}{\(\llparenthesis 7,2,3\rrparenthesis \) Pollatsek-Ruskai code} --- Binary icosahedral group \(2I\) gates can be realized transversally in the Pollatsek-Ruskai code \NoCaseChange{\protect\cite{cite647}}.
\end{eczvaluelist}
\eczhbkcontributors{ \eczhuVVA }
\endeczcode

\eczcode{kerdock_spherical}{Kerdock spherical code}{~\NoCaseChange{\protect\cite{cite2365,cite2366,cite1411}}}
\codefieldsection{Description}
Family of \((n=2^{2r},n^2,2-2/\sqrt{n})\) spherical codes for \(r \geq 2\), obtained from Kerdock codes via the \flmRefsHyperref{ref38}{antipodal mapping} \NoCaseChange{\protect\cite[{pg. 157}]{cite115}}.
These codes are optimal for their parameters for \(2\leq r\leq 5\), they are unique for \(r\in\{2,3\}\), and they form spherical 3-designs because their codewords are unions of \(2^{2r-1}+1\) orthoplexes \NoCaseChange{\protect\cite{cite386}}.
Their spherical images are also asymptotically optimal with respect to the Levenshtein and energy bounds discussed in Chapter 12 \NoCaseChange{\protect\cite[{Exam. 12.4.31}]{cite199}}.

\codefieldsection{Parents}
\begin{eczvaluelist}
\item\relax
\flmRefsHyperref[eczindexfamilyrel]{code:binary_antipodal}{Binary antipodal code}\item\relax
\flmRefsHyperref[eczindexfamilyrel]{code:spherical_design}{Spherical design} --- Kerdock spherical codes form spherical 3-designs because their codewords are unions of \(2^{2r-1}+1\) orthoplexes \NoCaseChange{\protect\cite{cite386}}.
\end{eczvaluelist}
\codefieldsection{Cousins}
\begin{eczvaluelist}
\item\relax
\flmRefsHyperref[eczindexfamilyrel]{code:kerdock}{Kerdock code} --- Kerdock spherical codes can be obtained from Kerdock codes using the \flmRefsHyperref{ref38}{antipodal mapping} \NoCaseChange{\protect\cite[{pg. 157}]{cite115}}.
\item\relax
\flmRefsHyperref[eczindexfamilyrel]{code:univ_opt_spherical}{Universally optimal spherical code} --- Kerdock spherical codes are almost universally optimal \NoCaseChange{\protect\cite{cite2367}}.
\item\relax
\flmRefsHyperref[eczindexfamilyrel]{code:biorthogonal_spherical}{Biorthogonal spherical code} --- Kerdock spherical codes form spherical 3-designs because their codewords are unions of \(2^{2r-1}+1\) orthoplexes \NoCaseChange{\protect\cite{cite386}}.
\end{eczvaluelist}
\eczhbkcontributors{ \eczhuVVA }
\endeczcode

\eczcode{laminated_spherical}{Laminated spherical code}{~\NoCaseChange{\protect\cite{cite2368}}}
\codefieldsection{Description}
Spherical code whose codewords are obtained from a recursive procedure that is similar to the procedure that creates laminated lattices.

\codefieldsection{Parent}
\begin{eczvaluelist}
\item\relax
\flmRefsHyperref[eczindexfamilyrel]{code:spherical}{Spherical code}\end{eczvaluelist}
\eczhbkcontributors{ \eczhuVVA }
\endeczcode

\eczcode{lattice_shell}{Lattice-shell code}{~\NoCaseChange{\protect\cite{cite2369,cite2370}}}
\codefieldsection{Description}
Spherical code whose codewords are scaled versions of points on a lattice.
A \(m\)-shell code consists of normalized lattice vectors \(x\) with squared norm \(\|x\|^2 = m\).
Each code is constructed by normalizing a set of lattice vectors in one or more \textit{shells}, i.e., sets of lattice points lying on a hypersphere.

\codefieldsection{Notes}
\begin{eczvaluelist}
\item\relax See \NoCaseChange{\protect\cite[{Ch. 10}]{cite115}} for tables of lattice-shell codes.
\end{eczvaluelist}
\codefieldsection{Parent}
\begin{eczvaluelist}
\item\relax
\flmRefsHyperref[eczindexfamilyrel]{code:spherical}{Spherical code}\end{eczvaluelist}
\codefieldsection{Children}
\begin{eczvaluelist}
\item\relax
\flmRefsHyperref[eczindexfamilyrel]{code:bw32_shell}{\(BW_{32}\) lattice-shell code}\item\relax
\flmRefsHyperref[eczindexfamilyrel]{code:dfour_shell}{\(D_4\) lattice-shell code}\item\relax
\flmRefsHyperref[eczindexfamilyrel]{code:eeight_shell}{\(E_8\) Gosset lattice-shell code}\item\relax
\flmRefsHyperref[eczindexfamilyrel]{code:eseven_shell}{\(E_7\) lattice-shell code}\item\relax
\flmRefsHyperref[eczindexfamilyrel]{code:esix_shell}{\(E_6\) lattice-shell code}\item\relax
\flmRefsHyperref[eczindexfamilyrel]{code:lambda16_shell}{\(\Lambda_{16}\) lattice-shell code}\item\relax
\flmRefsHyperref[eczindexfamilyrel]{code:leech_shell}{\(\Lambda_{24}\) Leech lattice-shell code}\item\relax
\flmRefsHyperref[eczindexfamilyrel]{code:cubeoctahedron}{Cuboctahedron code} --- Cuboctahedron codewords form the minimal shell of the \(D_3\) face-centered cubic (fcc) lattice.
\item\relax
\flmRefsHyperref[eczindexfamilyrel]{code:biorthogonal_spherical}{Biorthogonal spherical code} --- Biorthogonal codewords form the minimal shell of the \(\mathbb{Z}^n\) hypercubic lattice.
\item\relax
\flmRefsHyperref[eczindexfamilyrel]{code:hypercube}{Hypercube code} --- Hypercube codewords form the minimal lattice shell code of the \(\mathbb{Z}^n\) hypercubic lattice when the lattice is shifted such that the center of a hypercube is at the origin.
\end{eczvaluelist}
\codefieldsection{Cousins}
\begin{eczvaluelist}
\item\relax
\flmRefsHyperref[eczindexfamilyrel]{code:points_into_lattices}{Lattice} --- Lattice-shell codes consist of lattice-point vectors, initially all of the same norm, normalized to one.
\item\relax
\flmRefsHyperref[eczindexfamilyrel]{code:cyclic}{Cyclic code} --- Lattice-shell codewords are often permutations of a particular set of reference vectors, meaning that a cyclic permutation of a codeword yields another codeword.
\item\relax
\flmRefsHyperref[eczindexfamilyrel]{code:spherical_design}{Spherical design} --- Nonempty \(2m\)-shell codes of extremal even unimodular lattices in \(n\) dimensions form spherical \(t\)-designs with \(t=11\) (\(t=7\), \(t=3\)) if \(n \equiv 0\) (\(n \equiv 8\), \(n\equiv 16\)) modulo 24 \NoCaseChange{\protect\cite{cite2371,cite384}}. Shells of \(A_n\) and \(D_n\) lattices form infinite families of spherical 3-designs \NoCaseChange{\protect\cite[{Exam. 2.9}]{cite2372}}.
\end{eczvaluelist}
\eczhbkcontributors{ \eczhuVVA }
\endeczcode

\eczcode{mclaughlin}{McLaughlin spherical code}{~\NoCaseChange{\protect\cite{cite2373,cite2374}}}
\codefieldsection{Description}
The \((22,275,1/6)\) or \((23,552,1/5)\) code associated with the McLaughlin graph and the Leech lattice.
See Ref. \NoCaseChange{\protect\cite{cite387}} for explicit constructions of and relations between both codes.

\codefieldsection{Parent}
\begin{eczvaluelist}
\item\relax
\flmRefsHyperref[eczindexfamilyrel]{code:sharp_config}{Spherical sharp configuration} --- Both McLaughlin spherical codes are sharp configurations \NoCaseChange{\protect\cite{cite119,cite387}}. The \((22,275,1/6)\) code is a unique and tight spherical 4-design, while the \((23,552,1/5)\) code is a unique and tight spherical 5-design; see Ref. \NoCaseChange{\protect\cite[{Appx. A}]{cite119}}.
\end{eczvaluelist}
\codefieldsection{Cousins}
\begin{eczvaluelist}
\item\relax
\flmRefsHyperref[eczindexfamilyrel]{code:spherical_design}{Spherical design} --- Both McLaughlin spherical codes are sharp configurations \NoCaseChange{\protect\cite{cite119,cite387}}. The \((22,275,1/6)\) code is a unique and tight spherical 4-design, while the \((23,552,1/5)\) code is a unique and tight spherical 5-design; see Ref. \NoCaseChange{\protect\cite[{Appx. A}]{cite119}}.
\item\relax
\flmRefsHyperref[eczindexfamilyrel]{code:golay}{\([23, 12, 7]\) Golay code} --- The McLaughlin spherical code can be constructed from length-23 Golay codewords \NoCaseChange{\protect\cite{cite387}}.
\item\relax
\flmRefsHyperref[eczindexfamilyrel]{code:real_projective}{Real projective space code} --- The \((23,552,1/5)\) McLaughlin spherical code yields a set of \(276\) equiangular lines in 23 dimensions \NoCaseChange{\protect\cite{cite119,cite387}}.
\item\relax
\flmRefsHyperref[eczindexfamilyrel]{code:leech_shell}{\(\Lambda_{24}\) Leech lattice-shell code} --- The \((23,552,1/5)\) McLaughlin code can be derived from a shell of the Leech lattice \NoCaseChange{\protect\cite{cite119,cite2317}}.
\end{eczvaluelist}
\eczhbkcontributors{ \eczhuVVA }
\endeczcode

\eczcode{pentakis_dodecahedron}{Pentakis dodecahedron code}{}
\codefieldsection{Alternative Names}
\begin{eczvaluelist}
\item\relax Apiculated dodecahedron code
\item\relax Kisdodecahedron code
\end{eczvaluelist}
\eczhIndexCodeAliasName{pentakis_dodecahedron}{Apiculated dodecahedron code}
\eczhIndexCodeAliasName{pentakis_dodecahedron}{Kisdodecahedron code}
\codefieldsection{Description}
Spherical \((3,32,(9-\sqrt{5})/6)\) code whose codewords are the vertices of the pentakis dodecahedron, the convex hull of the icosahedron and dodecahedron.

\codefieldsection{Protection}
Optimal antipodal configuration of 32 points in 3D space \NoCaseChange{\protect\cite{cite2344}}.

\codefieldsection{Notes}
\begin{eczvaluelist}
\item\relax See the corresponding Bendwavy database entry \NoCaseChange{\protect\cite{cite2375}}.
\end{eczvaluelist}
\codefieldsection{Parents}
\begin{eczvaluelist}
\item\relax
\flmRefsHyperref[eczindexfamilyrel]{code:polyhedron}{Polyhedron code}\item\relax
\flmRefsHyperref[eczindexfamilyrel]{code:spherical_design}{Spherical design} --- Vertices of the pentakis dodecahedron form a weighted spherical 9-design \NoCaseChange{\protect\cite{cite388,cite389}\protect\cite[{Exam. 2.5}]{cite380}}.
\end{eczvaluelist}
\codefieldsection{Cousins}
\begin{eczvaluelist}
\item\relax
\flmRefsHyperref[eczindexfamilyrel]{code:dual_polytope}{Dual polytope code} --- The pentakis dodecahedron and truncated icosahedron are dual to each other \NoCaseChange{\protect\cite[{pg. 55}]{cite2353}}.
\item\relax
\flmRefsHyperref[eczindexfamilyrel]{code:icosahedron}{Icosahedron code} --- The pentakis dodecahedron is the convex hull of the icosahedron and dodecahedron. The pentakis dodecahedron and truncated icosahedron are dual to each other \NoCaseChange{\protect\cite[{pg. 55}]{cite2353}}.
\item\relax
\flmRefsHyperref[eczindexfamilyrel]{code:dodecahedron}{Dodecahedron code} --- The pentakis dodecahedron is the convex hull of the icosahedron and dodecahedron.
\end{eczvaluelist}
\eczhbkcontributors{ \eczhuVVA }
\endeczcode

\eczcode{permutation_spherical}{Permutation spherical code}{~\NoCaseChange{\protect\cite{cite2376,cite2377}}}
\codefieldsection{Description}
Slepian group-orbit code whose codewords are constructed from an arbitrary unit vector in two possible variants. Variant 1 consists of codewords that are permutations of the vector's coordinates, while Variant 2 consists of such permutations and all possible sign changes of the vector's components.

\codefieldsection{Rate}
Number of codewords cannot increase exponentially with dimension \(n\) \NoCaseChange{\protect\cite{cite2378}}
\codefieldsection{Decoding}
\begin{eczvaluelist}
\item\relax Efficient maximum-likelihood decoder determining the Voronoi region of an error word.
\end{eczvaluelist}
\codefieldsection{Notes}
\begin{eczvaluelist}
\item\relax See \NoCaseChange{\protect\cite[{Ch. 4}]{cite115}} for more details and tables of optimal codes.
\end{eczvaluelist}
\codefieldsection{Parent}
\begin{eczvaluelist}
\item\relax
\flmRefsHyperref[eczindexfamilyrel]{code:slepian_group}{Slepian group-orbit code} --- Permutations and sign changes can be implemented on vectors by orthogonal matrices, so permutation spherical codes are Slepian group-orbit codes.
\end{eczvaluelist}
\codefieldsection{Children}
\begin{eczvaluelist}
\item\relax
\flmRefsHyperref[eczindexfamilyrel]{code:biorthogonal_spherical}{Biorthogonal spherical code}\item\relax
\flmRefsHyperref[eczindexfamilyrel]{code:simplex_spherical}{Simplex spherical code}\end{eczvaluelist}
\eczhbkcontributors{ \eczhuVVA }
\endeczcode

\eczcode{petersen_spherical}{Petersen spherical code}{~\NoCaseChange{\protect\cite{cite390}}}
\codefieldsection{Description}
A \((4,10,1/6)\) spherical code whose codewords correspond to vertices of the Petersen graph.
Its Gram matrix is constructed by putting \(-2/3\) whenever two vertices are adjacent in the graph, and \(1/6\) otherwise.
The code is optimal for its parameters \NoCaseChange{\protect\cite{cite390}}.

\codefieldsection{Parent}
\begin{eczvaluelist}
\item\relax
\flmRefsHyperref[eczindexfamilyrel]{code:spherical_design}{Spherical design} --- The Petersen spherical code forms a spherical 2-design \NoCaseChange{\protect\cite{cite390}}.
\end{eczvaluelist}
\codefieldsection{Cousins}
\begin{eczvaluelist}
\item\relax
\flmRefsHyperref[eczindexfamilyrel]{code:petersen}{\([15,6,5]\) Petersen cycle code} --- The Petersen spherical code is a spherical realization associated with the Petersen cycle code.
\item\relax
\flmRefsHyperref[eczindexfamilyrel]{code:simplex_spherical}{Simplex spherical code} --- Codewords of the Petersen spherical code correspond to midpoints of the \(5\)-cell \NoCaseChange{\protect\cite{cite390}}.
\end{eczvaluelist}
\eczhbkcontributors{ \eczhuVVA }
\endeczcode

\eczcode{psk}{Phase-shift keying (PSK) modulation format}{}
\codefieldsection{Alternative Names}
\begin{eczvaluelist}
\item\relax Phase-shift keying (PSK) modulation code
\item\relax Phase-shift keying (PSK) modulation scheme
\item\relax Phase-shift keying (PSK) signaling format
\end{eczvaluelist}
\eczhIndexCodeAliasName{psk}{Phase-shift keying (PSK) modulation code}
\eczhIndexCodeAliasName{psk}{Phase-shift keying (PSK) modulation scheme}
\eczhIndexCodeAliasName{psk}{Phase-shift keying (PSK) signaling format}
\codefieldsection{Description}
A \(q\)-ary phase-shift keying (\(q\)-PSK) encodes one \(q\)-ary digit of information into a constellation of \(q\) points distributed equidistantly on a circle in \(\mathbb{C}\) or, equivalently, \(\mathbb{R}^2\).

For example, such a constellation could be
\flmMathEnvironment{align}{}{
  \{1,e^{i\frac{2\pi}{q}},\cdots,e^{i\frac{2\pi}{q}(q-1)}\}~.
}
Each point is typically associated with a complex amplitude of an electromagnetic signal, and information is encoded into the phase of that signal.

Concatenating PSK with \(q\)-ary codes yields a natural non-binary way of digitizing the analog AWGN channel \NoCaseChange{\protect\cite{cite2379,cite2380}}.

\codefieldsection{Rate}
As the constellation size \(q\) grows, \(q\)-PSK approaches the Shannon AWGN capacity of the corresponding two-dimensional signal set \NoCaseChange{\protect\cite[{Fig. 11.7}]{cite2287}}.
\codefieldsection{Realizations}
\begin{eczvaluelist}
\item\relax Telephone-line modems: 1967 Milgo 4400/48 and international standard V.27 used 8-PSK \NoCaseChange{\protect\cite{cite304}}.
\end{eczvaluelist}
\codefieldsection{Parents}
\begin{eczvaluelist}
\item\relax
\flmRefsHyperref[eczindexfamilyrel]{code:polygon}{Polygon code} --- The PSK constellation forms a \(q\)-gon.
\item\relax
\flmRefsHyperref[eczindexfamilyrel]{code:polyphase}{Polyphase code} --- A polyphase code can be thought of as a concatenation of a \(q\)-ary outer code with a PSK inner code. When the outer code is trivial, the construction reduces to a PSK code.
\item\relax
\flmRefsHyperref[eczindexfamilyrel]{code:modulation}{Modulation scheme} --- PSK is a modulation whose constellation consists of points arranged equidistantly on a circle.
\end{eczvaluelist}
\codefieldsection{Children}
\begin{eczvaluelist}
\item\relax
\flmRefsHyperref[eczindexfamilyrel]{code:bpsk}{Binary PSK (BPSK) modulation format} --- BPSK is also known as 2-PSK.
\item\relax
\flmRefsHyperref[eczindexfamilyrel]{code:qpsk}{Quadrature PSK (QPSK) modulation format} --- The QPSK is also known as 4-PSK.
\end{eczvaluelist}
\codefieldsection{Cousins}
\begin{eczvaluelist}
\item\relax
\flmRefsHyperref[eczindexfamilyrel]{code:gray}{Gray code} --- 1D Gray codes are often concatenated with PSKs so that the Hamming distance between the bitstrings encoded into the points is a discretized version of the Euclidean distance between the points.
\item\relax
\flmRefsHyperref[eczindexfamilyrel]{code:cat}{Cat code} --- PSK (cat) codes are used to transmit classical (quantum) information using (superpositions of) single-mode coherent states distributed on a circle over classical (quantum) channels.
\item\relax
\flmRefsHyperref[eczindexfamilyrel]{code:hyperbolic}{Hyperbolic sphere packing} --- Hyperbolic PSK constellations may yield improved performance over Euclidean ones \NoCaseChange{\protect\cite{cite2381}}.
\item\relax
\flmRefsHyperref[eczindexfamilyrel]{code:quantum_psk}{PSK c-q modulation format} --- PSK (PSK c-q) codes are used to transmit classical information using single-mode coherent states distributed on a circle over classical (quantum) channels.
\end{eczvaluelist}
\eczhbkcontributors{ \eczhuVVA }
\endeczcode

\eczcode{polygon}{Polygon code}{}
\codefieldsection{Description}
Spherical \((1,q,4\sin^2 \frac{\pi}{q})\) code for any \(q\geq2\) whose codewords are the vertices of a \(q\)-gon. Special cases include the line segment (\(q=2\)), triangle (\(q=3\)), square (\(q=4\)), pentagon (\(q=5\)), and hexagon (\(q=6\)).

\begin{flmFloat}{figure}{NumCap}\includegraphics[width=107.98652891338584bp,max width=\linewidth]{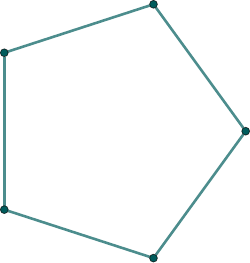}\caption{\(q\)-gon code for \(q=5\). Each codeword is a vertex of the \(5\)-gon.}\label{ref2382}\end{flmFloat}

\codefieldsection{Parents}
\begin{eczvaluelist}
\item\relax
\flmRefsHyperref[eczindexfamilyrel]{code:self_dual_polytope}{Self-dual polytope code} --- Polygons are self-dual.
\item\relax
\flmRefsHyperref[eczindexfamilyrel]{code:sharp_config}{Spherical sharp configuration}\end{eczvaluelist}
\codefieldsection{Child}
\begin{eczvaluelist}
\item\relax
\flmRefsHyperref[eczindexfamilyrel]{code:psk}{Phase-shift keying (PSK) modulation format} --- The PSK constellation forms a \(q\)-gon.
\end{eczvaluelist}
\codefieldsection{Cousins}
\begin{eczvaluelist}
\item\relax
\flmRefsHyperref[eczindexfamilyrel]{code:spherical_design}{Spherical design} --- A \(q\)-gon is a tight spherical \(q-1\) design.
\item\relax
\flmRefsHyperref[eczindexfamilyrel]{code:cat}{Cat code} --- The \(q(S+1)\)-component cat coherent-state constellation forms the vertices of a \(q(S+1)\)-gon.
\item\relax
\flmRefsHyperref[eczindexfamilyrel]{code:quantum_psk}{PSK c-q modulation format} --- The PSK coherent-state constellation forms the vertices of a \(q\)-gon.
\item\relax
\flmRefsHyperref[eczindexfamilyrel]{code:real_projective}{Real projective space code} --- For even \(q\), the \(q/2\) sets of antipodal pairs of a \(q\)-gon form a tight design on the projective line \(\mathbb{R}P^1\) \NoCaseChange{\protect\cite{cite917}}.
\item\relax
\flmRefsHyperref[eczindexfamilyrel]{code:t-designs}{\(t\)-design} --- For even \(q\), the \(q/2\) sets of antipodal pairs of a \(q\)-gon form a tight design on the projective line \(\mathbb{R}P^1\) \NoCaseChange{\protect\cite{cite917}}.
\item\relax
\flmRefsHyperref[eczindexfamilyrel]{code:honeycomb}{Honeycomb tiling} --- The faces of the honeycomb tiling are hexagons, while the faces of its dual triangular tiling are triangles.
\item\relax
\flmRefsHyperref[eczindexfamilyrel]{code:hexagonal}{\(A_2\) triangular lattice} --- The Voronoi cell of the triangular lattice is the hexagon.
\end{eczvaluelist}
\eczhbkcontributors{ Shubham P. Jain, \eczhuVVA }
\endeczcode

\eczcode{polyhedron}{Polyhedron code}{}
\codefieldsection{Description}
A polytope code in three dimensions, i.e., a spherical code whose codewords form vertices of a polyhedron.

\codefieldsection{Parent}
\begin{eczvaluelist}
\item\relax
\flmRefsHyperref[eczindexfamilyrel]{code:polytope}{Polytope code} --- Three-dimensional polytope codes are polyhedron codes.
\end{eczvaluelist}
\codefieldsection{Children}
\begin{eczvaluelist}
\item\relax
\flmRefsHyperref[eczindexfamilyrel]{code:antiprism}{Antiprism code}\item\relax
\flmRefsHyperref[eczindexfamilyrel]{code:cubeoctahedron}{Cuboctahedron code}\item\relax
\flmRefsHyperref[eczindexfamilyrel]{code:dodecahedron}{Dodecahedron code}\item\relax
\flmRefsHyperref[eczindexfamilyrel]{code:icosahedron}{Icosahedron code}\item\relax
\flmRefsHyperref[eczindexfamilyrel]{code:pentakis_dodecahedron}{Pentakis dodecahedron code}\item\relax
\flmRefsHyperref[eczindexfamilyrel]{code:rhombic_dodecahedron}{Rhombic dodecahedron code}\item\relax
\flmRefsHyperref[eczindexfamilyrel]{code:rhombicuboctahedron}{Rhombicuboctahedron code}\item\relax
\flmRefsHyperref[eczindexfamilyrel]{code:snub_cube}{Snub-cube code}\end{eczvaluelist}
\codefieldsection{Cousins}
\begin{eczvaluelist}
\item\relax
\flmRefsHyperref[eczindexfamilyrel]{code:ball_color}{Ball code} --- Polytopes dual to the hyperoctahedron, truncated octahedron, truncated cuboctahedron, and truncated icosidodecahedron are used to construct 3D ball codes.
\item\relax
\flmRefsHyperref[eczindexfamilyrel]{code:stellated_dodecahedron_css}{\(\llbracket 30,8,3\rrbracket \) Bring code} --- Bring code and related codes listed in \NoCaseChange{\protect\cite[{Table 1}]{cite2383}} arrange qubits and stabilizer generators on star polyhedra.
\end{eczvaluelist}
\eczhbkcontributors{ \eczhuVVA }
\endeczcode

\eczcode{polyphase}{Polyphase code}{~\NoCaseChange{\protect\cite{cite2384,cite2385,cite2386,cite2387,cite2388,cite2389,cite2390,cite2391,cite2392,cite2393,cite2394}}}
\codefieldsection{Description}
A spherical code obtained from a binary code, \(q\)-ary code, or \(q\)-ary code over \(\mathbb{Z}_q\) via a component-wise mapping of each \(q\)-ary digit to a \(q\)th root of unity in a generalization of the \flmRefsHyperref{ref38}{antipodal mapping}.

For example, for the case \(q=4\), one can map either the ring-based alphabet \(\mathbb{Z}_4 = \{0,1,2,3\}\) or the field-based alphabet \(\mathbb{F}_2^2 = \{00,01,10,11\}\) to the set \(\{1,\theta,\theta^2,\theta^3\}\) for some fourth root of unity \(\theta\).

\codefieldsection{Notes}
\begin{eczvaluelist}
\item\relax See \NoCaseChange{\protect\cite[{Ch. 7}]{cite115}} for more details.
\end{eczvaluelist}
\codefieldsection{Parents}
\begin{eczvaluelist}
\item\relax
\flmRefsHyperref[eczindexfamilyrel]{code:tlsc}{Torus-layer spherical code (TLSC)}\item\relax
\flmRefsHyperref[eczindexfamilyrel]{code:concatenated}{Concatenated code} --- A polyphase code can be thought of as a concatenation of a \(q\)-ary outer code with a PSK inner code.
\end{eczvaluelist}
\codefieldsection{Children}
\begin{eczvaluelist}
\item\relax
\flmRefsHyperref[eczindexfamilyrel]{code:psk}{Phase-shift keying (PSK) modulation format} --- A polyphase code can be thought of as a concatenation of a \(q\)-ary outer code with a PSK inner code. When the outer code is trivial, the construction reduces to a PSK code.
\item\relax
\flmRefsHyperref[eczindexfamilyrel]{code:hypercube}{Hypercube code}\item\relax
\flmRefsHyperref[eczindexfamilyrel]{code:binary_antipodal}{Binary antipodal code} --- The polyphase mapping for \(q=2\) reduces to the \flmRefsHyperref{ref38}{antipodal mapping}.
\end{eczvaluelist}
\codefieldsection{Cousins}
\begin{eczvaluelist}
\item\relax
\flmRefsHyperref[eczindexfamilyrel]{code:q-ary_digits_into_q-ary_digits}{\(q\)-ary code} --- Polyphase codes are spherical codes that can be obtained from \(q\)-ary codes.
\item\relax
\flmRefsHyperref[eczindexfamilyrel]{code:q-ary_over_zq}{\(q\)-ary code over \(\mathbb{Z}_q\)} --- Polyphase codes are spherical codes that can be obtained from \(q\)-ary codes over rings \(\mathbb{Z}_q\).
\item\relax
\flmRefsHyperref[eczindexfamilyrel]{code:simplex_spherical}{Simplex spherical code} --- Simplex spherical codes for dimension \(n=(p-1)/2\) with \(p\) an odd prime admit a polyphase realization \NoCaseChange{\protect\cite[{Sec. 7.7}]{cite115}}.
\item\relax
\flmRefsHyperref[eczindexfamilyrel]{code:biorthogonal_spherical}{Biorthogonal spherical code} --- Biorthogonal spherical codes for dimension \(n=p\) with \(p\) an odd prime admit a polyphase realization \NoCaseChange{\protect\cite[{Sec. 7.7}]{cite115}}.
\item\relax
\flmRefsHyperref[eczindexfamilyrel]{code:hexacode}{\([6,3,4]_4\) Hexacode} --- The hexacode is a complex spherical 3-design when embedded into the complex sphere via the polyphase mapping \NoCaseChange{\protect\cite{cite1671}}.
\end{eczvaluelist}
\eczhbkcontributors{ \eczhuVVA }
\endeczcode

\eczcode{polytope}{Polytope code}{}
\codefieldsection{Description}
Spherical code whose codewords are the vertices of a polytope, i.e., a geometrical figure bounded by lines, planes, and hyperplanes in either real \NoCaseChange{\protect\cite{cite178}} or complex \NoCaseChange{\protect\cite{cite231}} space.
A polytope in two (three, four) dimensions is called a polygon (polyhedron, polychoron).

\codefieldsection{Notes}
\begin{eczvaluelist}
\item\relax See \flmHref{https://polytope.miraheze.org/wiki/Main_Page}{Polytope Wiki} and \flmHref{https://bendwavy.org/klitzing/home.htm}{webpage by R. Klitzing} for lists of polytopes.
\end{eczvaluelist}
\codefieldsection{Parent}
\begin{eczvaluelist}
\item\relax
\flmRefsHyperref[eczindexfamilyrel]{code:slepian_group}{Slepian group-orbit code} --- Vertices of polytope codes typically form an orbit of the polytope's symmetry group.
\end{eczvaluelist}
\codefieldsection{Children}
\begin{eczvaluelist}
\item\relax
\flmRefsHyperref[eczindexfamilyrel]{code:polyhedron}{Polyhedron code} --- Three-dimensional polytope codes are polyhedron codes.
\item\relax
\flmRefsHyperref[eczindexfamilyrel]{code:120cell}{120-cell code}\item\relax
\flmRefsHyperref[eczindexfamilyrel]{code:600cell}{600-cell code}\item\relax
\flmRefsHyperref[eczindexfamilyrel]{code:rect_hessian_polyhedron}{Rectified Hessian polyhedron code}\item\relax
\flmRefsHyperref[eczindexfamilyrel]{code:231_polytope}{\(2_{31}\) polytope code}\item\relax
\flmRefsHyperref[eczindexfamilyrel]{code:hess_polytope}{\(3_{21}\) polytope code}\item\relax
\flmRefsHyperref[eczindexfamilyrel]{code:241_polytope}{\(2_{41}\) polytope code}\item\relax
\flmRefsHyperref[eczindexfamilyrel]{code:dual_polytope}{Dual polytope code}\item\relax
\flmRefsHyperref[eczindexfamilyrel]{code:biorthogonal_spherical}{Biorthogonal spherical code} --- Biorthogonal spherical codewords in 2 (3, 4, \(n\)) dimensions form the vertices of a square (octahedron, 16-cell, \(n\)-orthoplex).
\item\relax
\flmRefsHyperref[eczindexfamilyrel]{code:hypercube}{Hypercube code} --- Hypercube codewords in 2 (3, 4, \(n\)) dimensions form the vertices of a square (cube, tesseract, \(n\)-cube).
\end{eczvaluelist}
\codefieldsection{Cousin}
\begin{eczvaluelist}
\item\relax
\flmRefsHyperref[eczindexfamilyrel]{code:qsc}{Quantum spherical code (QSC)} --- QSCs can be constructed by using vertices of polytopes for logical constellations. The logical constellations form the vertices of the code constellation, a polytope compound.
\end{eczvaluelist}
\eczhbkcontributors{ \eczhuVVA }
\endeczcode

\eczcode{ppm}{Pulse-position modulation (PPM) format}{}
\codefieldsection{Alternative Names}
\begin{eczvaluelist}
\item\relax Pulse-position modulation (PPM) code
\item\relax Pulse-position modulation (PPM) scheme
\item\relax Pulse-position modulation (PPM) signaling format
\end{eczvaluelist}
\eczhIndexCodeAliasName{ppm}{Pulse-position modulation (PPM) code}
\eczhIndexCodeAliasName{ppm}{Pulse-position modulation (PPM) scheme}
\eczhIndexCodeAliasName{ppm}{Pulse-position modulation (PPM) signaling format}
\codefieldsection{Description}
A modulation code with \(q\) equal-energy signals in which each codeword has one pulse in one of \(q\) time slots and zeros elsewhere.

\codefieldsection{Realizations}
\begin{eczvaluelist}
\item\relax Greek hydraulic semaphore system \NoCaseChange{\protect\cite{cite306,cite307}}.
\item\relax Telegraph time-division multiplexing.
\item\relax Radio-control, fiber-optic communications, and deep-space communications.
\end{eczvaluelist}
\codefieldsection{Parents}
\begin{eczvaluelist}
\item\relax
\flmRefsHyperref[eczindexfamilyrel]{code:spherical}{Spherical code}\item\relax
\flmRefsHyperref[eczindexfamilyrel]{code:modulation}{Modulation scheme}\end{eczvaluelist}
\codefieldsection{Cousins}
\begin{eczvaluelist}
\item\relax
\flmRefsHyperref[eczindexfamilyrel]{code:biorthogonal_spherical}{Biorthogonal spherical code} --- PPM codewords form a spherical code whose constellation consists of the standard basis vectors. Adjoining negatives yields the corresponding biorthogonal spherical code.
\item\relax
\flmRefsHyperref[eczindexfamilyrel]{code:one_hot}{One-hot code} --- The PPM code is a continuous analogue of the one-hot code.
\item\relax
\flmRefsHyperref[eczindexfamilyrel]{code:quantum_ppm}{Pulse-position (PPM) c-q modulation format} --- PPM c-q codes are quantum analogues of PPM codes.
\end{eczvaluelist}
\eczhbkcontributors{ \eczhuVVA }
\endeczcode

\eczcode{qpsk}{Quadrature PSK (QPSK) modulation format}{~\NoCaseChange{\protect\cite{cite2395}}}
\codefieldsection{Alternative Names}
\begin{eczvaluelist}
\item\relax Quadrature PSK (QPSK) modulation code
\item\relax Quadrature PSK (QPSK) modulation scheme
\item\relax Quadrature PSK (QPSK) signaling format
\item\relax Quadriphase PSK modulation format
\item\relax 4-PSK modulation format
\item\relax 4-QAM modulation format
\end{eczvaluelist}
\eczhIndexCodeAliasName{qpsk}{Quadrature PSK (QPSK) modulation code}
\eczhIndexCodeAliasName{qpsk}{Quadrature PSK (QPSK) modulation scheme}
\eczhIndexCodeAliasName{qpsk}{Quadrature PSK (QPSK) signaling format}
\eczhIndexCodeAliasName{qpsk}{Quadriphase PSK modulation format}
\eczhIndexCodeAliasName{qpsk}{4-PSK modulation format}
\eczhIndexCodeAliasName{qpsk}{4-QAM modulation format}
\codefieldsection{Description}
A quaternary encoding into a constellation of four points distributed equidistantly on a circle.
For the case of \(\pi/4\)-QPSK, the constellation is \(\{e^{\pm i\frac{\pi}{4}},e^{\pm i\frac{3\pi}{4}}\}\).

\codefieldsection{Realizations}
\begin{eczvaluelist}
\item\relax Japanese and North American digital cellular and personal systems \NoCaseChange{\protect\cite{cite308}}.
\item\relax Telephone-line modems: 1962 Bell 201 and international standard V.24 \NoCaseChange{\protect\cite{cite309}}.
\end{eczvaluelist}
\codefieldsection{Parents}
\begin{eczvaluelist}
\item\relax
\flmRefsHyperref[eczindexfamilyrel]{code:psk}{Phase-shift keying (PSK) modulation format} --- The QPSK is also known as 4-PSK.
\item\relax
\flmRefsHyperref[eczindexfamilyrel]{code:biorthogonal_spherical}{Biorthogonal spherical code} --- The QPSK is equivalent to the biorthogonal spherical code for \(n=2\).
\end{eczvaluelist}
\codefieldsection{Cousin}
\begin{eczvaluelist}
\item\relax
\flmRefsHyperref[eczindexfamilyrel]{code:quaternary_over_z4}{Linear code over \(\mathbb{Z}_4\)} --- The Lee distance \(d_L\) between two digits \(a,b\in\mathbb{Z}_4\) governs how many cyclic shifts it takes to connect them, and is thus related to the Euclidean distance between \(i^a\) and \(i^b\) as \(2d_L(a,b) = \|i^a - i^b\|^2\).
Quaternary codes over the Lee metric are thus naturally mapped to QPSK codes.

\end{eczvaluelist}
\eczhbkcontributors{ \eczhuVVA }
\endeczcode

\eczcode{sidelnikov}{Real-Clifford subgroup-orbit code}{~\NoCaseChange{\protect\cite{cite2109,cite2110}}}
\codefieldsection{Description}
Slepian group-orbit code of dimension \(2^r\), approximate asymptotic size \(2.38 \cdot 2^{r(r+1)/2+1}\), and distance \(1\).
Code is constructed by applying elements of an index-two subgroup of the \flmRefsHyperref{ref409}{real Clifford group}, when taken as a subgroup of the orthogonal group \NoCaseChange{\protect\cite{cite2103}}, onto the vector \((1,0,0,\cdots,0)\).
This group is the automorphism group of the BW lattice, and the resulting codes coincide with the optimal spherical codes for dimensions \(\{4,8,16\}\).

Taking the orbit under the entire \flmRefsHyperref{ref409}{real Clifford group} yields spherical codes with twice the points and distance \(2-\sqrt{2}\).

\codefieldsection{Parents}
\begin{eczvaluelist}
\item\relax
\flmRefsHyperref[eczindexfamilyrel]{code:slepian_group}{Slepian group-orbit code}\item\relax
\flmRefsHyperref[eczindexfamilyrel]{code:spherical_design}{Spherical design} --- The orbit of any point under the real Clifford subgroup is a spherical 7-design \NoCaseChange{\protect\cite{cite391}}, and some are 11-designs \NoCaseChange{\protect\cite{cite392}}.
\end{eczvaluelist}
\codefieldsection{Children}
\begin{eczvaluelist}
\item\relax
\flmRefsHyperref[eczindexfamilyrel]{code:bw32_shell}{\(BW_{32}\) lattice-shell code} --- The minimal \(BW_{32}\) lattice-shell code is equivalent to the real Clifford subgroup-orbit code for \(n=32\).
\item\relax
\flmRefsHyperref[eczindexfamilyrel]{code:lambda16_shell}{\(\Lambda_{16}\) lattice-shell code} --- The minimal \(\Lambda_{16}\) lattice-shell code is equivalent to the real Clifford subgroup-orbit code for \(n=16\).
\item\relax
\flmRefsHyperref[eczindexfamilyrel]{code:24cell}{24-cell code} --- The 24-cell code is equivalent to the real Clifford subgroup-orbit code for \(n=4\).
\item\relax
\flmRefsHyperref[eczindexfamilyrel]{code:witting_polytope}{Witting polytope code} --- The Witting polytope code is equivalent to the real Clifford subgroup-orbit code for \(n=8\).
\end{eczvaluelist}
\codefieldsection{Cousins}
\begin{eczvaluelist}
\item\relax
\flmRefsHyperref[eczindexfamilyrel]{code:barnes_wall}{Barnes-Wall (BW) lattice} --- The automorphism group of BW lattices is a subgroup of index 2 of a \flmRefsHyperref{ref409}{real Clifford group} \NoCaseChange{\protect\cite{cite2109,cite2110}} (see \NoCaseChange{\protect\cite{cite2103,cite2117}} for an explanation).
\item\relax
\flmRefsHyperref[eczindexfamilyrel]{code:clifford_group}{Clifford group} --- The automorphism group of BW lattices is a subgroup of index 2 of a \flmRefsHyperref{ref409}{real Clifford group} \NoCaseChange{\protect\cite{cite2109,cite2110}} (see \NoCaseChange{\protect\cite{cite2103,cite2117}} for an explanation).
\item\relax
\flmRefsHyperref[eczindexfamilyrel]{code:disphenoidal288cell}{Disphenoidal 288-cell code} --- The disphenoidal 288-cell code is a group-orbit code with the group being the \flmRefsHyperref{ref409}{real Clifford group} in \(4\) dimensions.
\item\relax
\flmRefsHyperref[eczindexfamilyrel]{code:quantum_sidelnikov}{Clifford subgroup-orbit QSC} --- Clifford group-orbit QSCs are quantum counterparts of real Clifford subgroup-orbit codes.
\end{eczvaluelist}
\eczhbkcontributors{ \eczhuVVA }
\endeczcode

\eczcode{rect_hessian_polyhedron}{Rectified Hessian polyhedron code}{}
\codefieldsection{Alternative Names}
\begin{eczvaluelist}
\item\relax \(1_{22}\) polytope code
\end{eczvaluelist}
\eczhIndexCodeAliasName{rect_hessian_polyhedron}{\(1_{22}\) polytope code}
\codefieldsection{Description}
Spherical \((6,72,1)\) code whose codewords are the vertices of the rectified Hessian complex polyhedron and the \(1_{22}\) polytope.
Codewords form the minimal lattice-shell code of the \(E_6\) lattice, i.e., the 72 roots of \(E_6\) after normalization to the unit sphere.
See \NoCaseChange{\protect\cite[{pg. 127}]{cite231}\protect\cite[{pg. 126}]{cite39}} for realizations of the 72 codewords.

\codefieldsection{Notes}
\begin{eczvaluelist}
\item\relax See the corresponding Bendwavy database entries for the rectified Hessian complex polyhedron \NoCaseChange{\protect\cite{cite2396}} and its real-space embedding as the \(1_{22}\) polytope \NoCaseChange{\protect\cite{cite2397}}.
\end{eczvaluelist}
\codefieldsection{Parents}
\begin{eczvaluelist}
\item\relax
\flmRefsHyperref[eczindexfamilyrel]{code:polytope}{Polytope code}\item\relax
\flmRefsHyperref[eczindexfamilyrel]{code:esix_shell}{\(E_6\) lattice-shell code} --- Rectified Hessian polyhedron codewords form the minimal shell of the \(E_6\) lattice.
\item\relax
\flmRefsHyperref[eczindexfamilyrel]{code:spherical_design}{Spherical design} --- The rectified Hessian polyhedron code forms a spherical 5-design \NoCaseChange{\protect\cite{cite393}}.
\end{eczvaluelist}
\codefieldsection{Cousins}
\begin{eczvaluelist}
\item\relax
\flmRefsHyperref[eczindexfamilyrel]{code:hessian_polyhedron}{Hessian polyhedron code} --- The Hessian and rectified Hessian polyhedra are analogues of the tetrahedron and octahedron in 3D complex space, while the double Hessian polyhedron is the analogue of a cube \NoCaseChange{\protect\cite[{pg. 127}]{cite231}}. The rectified and double Hessian polyhedra are dual to each other, just like the octahedron and cube. Moreover, the double Hessian consists of two Hessians, just like the cube can be constructed from two tetrahedra.
\item\relax
\flmRefsHyperref[eczindexfamilyrel]{code:dual_polytope}{Dual polytope code} --- The rectified and double Hessian polyhedra are dual to each other, analogous to the octahedron and cube.
\item\relax
\flmRefsHyperref[eczindexfamilyrel]{code:esix}{\(E_6\) root lattice} --- The Voronoi cell of the \(E_6\) root lattice is the dual of the Gosset \(1_{22}\) polytope \NoCaseChange{\protect\cite[{Ch. 21, pg. 465}]{cite39}}.
\end{eczvaluelist}
\eczhbkcontributors{ \eczhuVVA }
\endeczcode

\eczcode{rhombic_dodecahedron}{Rhombic dodecahedron code}{}
\codefieldsection{Alternative Names}
\begin{eczvaluelist}
\item\relax Granatahedron code
\end{eczvaluelist}
\eczhIndexCodeAliasName{rhombic_dodecahedron}{Granatahedron code}
\codefieldsection{Description}
Spherical \((3,14,2-2/\sqrt{3})\) code whose codewords are the normalized vertices of the rhombic dodecahedron.
Equivalently, the codewords are the union of the vertices of a cube and an octahedron on the unit sphere.

\codefieldsection{Protection}
Optimal antipodal configuration of 14 points in 3D space \NoCaseChange{\protect\cite{cite2344}}.

\codefieldsection{Notes}
\begin{eczvaluelist}
\item\relax See the corresponding Bendwavy database entry \NoCaseChange{\protect\cite{cite2398}}.
\end{eczvaluelist}
\codefieldsection{Parent}
\begin{eczvaluelist}
\item\relax
\flmRefsHyperref[eczindexfamilyrel]{code:polyhedron}{Polyhedron code}\end{eczvaluelist}
\codefieldsection{Cousins}
\begin{eczvaluelist}
\item\relax
\flmRefsHyperref[eczindexfamilyrel]{code:dual_polytope}{Dual polytope code} --- The rhombic dodecahedron and cuboctahedron are dual to each other \NoCaseChange{\protect\cite{cite2353}}.
\item\relax
\flmRefsHyperref[eczindexfamilyrel]{code:cubeoctahedron}{Cuboctahedron code} --- The rhombic dodecahedron and cuboctahedron are dual to each other \NoCaseChange{\protect\cite{cite2353}}.
\item\relax
\flmRefsHyperref[eczindexfamilyrel]{code:hypercube}{Hypercube code} --- The vertices of a rhombic dodecahedron are a union of the vertices of a cube and an octahedron.
\item\relax
\flmRefsHyperref[eczindexfamilyrel]{code:biorthogonal_spherical}{Biorthogonal spherical code} --- The vertices of a rhombic dodecahedron are a union of the vertices of a cube and an octahedron.
\item\relax
\flmRefsHyperref[eczindexfamilyrel]{code:dthree}{\(D_3\) face-centered cubic (fcc) lattice} --- The Voronoi cell of the \(D_3\) fcc lattice is a rhombic dodecahedron \NoCaseChange{\protect\cite[{Ch. 21, pg. 464}]{cite39}}.
\item\relax
\flmRefsHyperref[eczindexfamilyrel]{code:rhombic_dodecahedron_surface}{\(\llbracket 14,3,3\rrbracket \) Rhombic dodecahedron surface code} --- The qubits of the \(\llbracket 14,3,3\rrbracket \) rhombic dodecahedron surface code lie on the vertices of the small rhombic dodecahedron.
\end{eczvaluelist}
\eczhbkcontributors{ \eczhuVVA }
\endeczcode

\eczcode{rhombicuboctahedron}{Rhombicuboctahedron code}{}
\codefieldsection{Alternative Names}
\begin{eczvaluelist}
\item\relax Small rhombicuboctahedron code
\item\relax Expanded octahedron code
\item\relax Expanded cube code
\end{eczvaluelist}
\eczhIndexCodeAliasName{rhombicuboctahedron}{Small rhombicuboctahedron code}
\eczhIndexCodeAliasName{rhombicuboctahedron}{Expanded octahedron code}
\eczhIndexCodeAliasName{rhombicuboctahedron}{Expanded cube code}
\codefieldsection{Description}
Spherical \((3,24,4/(5+2 \sqrt{2}\rrparenthesis \) code whose codewords are the vertices of the rhombicuboctahedron.

\codefieldsection{Protection}
Optimal antipodal configuration of 24 points in 3D space \NoCaseChange{\protect\cite{cite2344}}.

\codefieldsection{Notes}
\begin{eczvaluelist}
\item\relax See the corresponding Bendwavy database entry \NoCaseChange{\protect\cite{cite2399}}.
\end{eczvaluelist}
\codefieldsection{Parent}
\begin{eczvaluelist}
\item\relax
\flmRefsHyperref[eczindexfamilyrel]{code:polyhedron}{Polyhedron code}\end{eczvaluelist}
\eczhbkcontributors{ \eczhuVVA }
\endeczcode

\eczcode{self_dual_polytope}{Self-dual polytope code}{}
\codefieldsection{Alternative Names}
\begin{eczvaluelist}
\item\relax Self-reciprocal polytope code
\end{eczvaluelist}
\eczhIndexCodeAliasName{self_dual_polytope}{Self-reciprocal polytope code}
\codefieldsection{Description}
A spherical code whose codewords are the vertices of a self-dual polytope.

\codefieldsection{Parent}
\begin{eczvaluelist}
\item\relax
\flmRefsHyperref[eczindexfamilyrel]{code:dual_polytope}{Dual polytope code}\end{eczvaluelist}
\codefieldsection{Children}
\begin{eczvaluelist}
\item\relax
\flmRefsHyperref[eczindexfamilyrel]{code:polygon}{Polygon code} --- Polygons are self-dual.
\item\relax
\flmRefsHyperref[eczindexfamilyrel]{code:24cell}{24-cell code} --- The 24-cell is self-dual.
\item\relax
\flmRefsHyperref[eczindexfamilyrel]{code:hessian_polyhedron}{Hessian polyhedron code} --- The \(2_{21}\) polytope is self-dual \NoCaseChange{\protect\cite{cite116}}.
\item\relax
\flmRefsHyperref[eczindexfamilyrel]{code:witting_polytope}{Witting polytope code} --- The Witting polytope is self-dual as a complex polytope. The \(4_{21}\) polytope is not self-dual \NoCaseChange{\protect\cite{cite178}}.
\item\relax
\flmRefsHyperref[eczindexfamilyrel]{code:simplex_spherical}{Simplex spherical code} --- The simplex is self-dual.
\end{eczvaluelist}
\codefieldsection{Cousins}
\begin{eczvaluelist}
\item\relax
\flmRefsHyperref[eczindexfamilyrel]{code:self_dual_lattice}{Unimodular lattice} --- Self-dual polytope codes are spherical analogues of self-dual lattices.
\item\relax
\flmRefsHyperref[eczindexfamilyrel]{code:self_dual}{Self-dual linear code} --- Self-dual polytope codes are spherical analogues of self-dual linear codes.
\end{eczvaluelist}
\eczhbkcontributors{ \eczhuVVA }
\endeczcode

\eczcode{simplex_spherical}{Simplex spherical code}{}
\codefieldsection{Description}
Spherical \((n,n+1,2+2/n)\) code whose codewords are all permutations of the \(n+1\)-dimensional vector \((1,1,\cdots,1,-n)\), up to normalization, forming an \(n\)-simplex.
Codewords are all equidistant and their components add up to zero.
Simplex spherical codewords in 2 (3, 4) dimensions form the vertices of a triangle (tetrahedron, 5-cell).
In general, the code makes up the vertices of an \(n\)-simplex.
The union of a simplex and its antipodal simplex forms the vertices of a bi-simplex, which has \(2(n+1)\) vertices.

See \NoCaseChange{\protect\cite[{Sec. 7.7}]{cite115}} for a parameterization.

\begin{flmFloat}{figure}{NumCap}\includegraphics[width=107.98652891338584bp,max width=\linewidth]{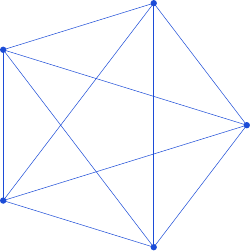}\caption{Projection of the coordinates of the 5-cell (4D simplex).}\label{ref2400}\end{flmFloat}

\codefieldsection{Protection}
Simplex spherical codes saturate the absolute bound, the Levenshtein bound and, for \(2 < \rho \leq 4\), the first two Rankin bounds \NoCaseChange{\protect\cite{cite115}}.
All simplex codes are unique up to equivalence \NoCaseChange{\protect\cite[{pg. 18}]{cite115}}, which follows from saturating the Boroczky bound \NoCaseChange{\protect\cite{cite2334}}.

\codefieldsection{Notes}
\begin{eczvaluelist}
\item\relax See the corresponding Bendwavy database entry for the 5-cell \NoCaseChange{\protect\cite{cite2401}}.
\end{eczvaluelist}
\codefieldsection{Parents}
\begin{eczvaluelist}
\item\relax
\flmRefsHyperref[eczindexfamilyrel]{code:self_dual_polytope}{Self-dual polytope code} --- The simplex is self-dual.
\item\relax
\flmRefsHyperref[eczindexfamilyrel]{code:sharp_config}{Spherical sharp configuration}\item\relax
\flmRefsHyperref[eczindexfamilyrel]{code:permutation_spherical}{Permutation spherical code}\end{eczvaluelist}
\codefieldsection{Cousins}
\begin{eczvaluelist}
\item\relax
\flmRefsHyperref[eczindexfamilyrel]{code:spherical_design}{Spherical design} --- Simplex spherical codes are the only tight spherical 2-designs \NoCaseChange{\protect\cite[{Tab. 9.3}]{cite115}}. The bi-simplex is a spherical 3-design since antipodal codes have zero averages over odd-degree polynomials.
\item\relax
\flmRefsHyperref[eczindexfamilyrel]{code:dodecahedron}{Dodecahedron code} --- Vertices of a dodecahedron can be split up into vertices of five tetrahedra, which are simplex spherical codes for \(n=3\) \NoCaseChange{\protect\cite{cite178}}.
\item\relax
\flmRefsHyperref[eczindexfamilyrel]{code:binary_antipodal}{Binary antipodal code} --- Binary simplex codes map to \((2^m,2^m+1)\) simplex spherical codes under the \flmRefsHyperref{ref38}{antipodal mapping} \NoCaseChange{\protect\cite[{Sec. 6.5.2}]{cite1165}\protect\cite[{pg. 18}]{cite115}}. In other words, simplex (simplex spherical) codes form simplices in Hamming (Euclidean) space.
\item\relax
\flmRefsHyperref[eczindexfamilyrel]{code:antiprism}{Antiprism code} --- The antiprism reduces to the tetrahedron for \(q=2\).
\item\relax
\flmRefsHyperref[eczindexfamilyrel]{code:simplex_discrete}{Simplex integer-based code} --- Codewords of simplex integer-based codes are restricted to lie in a \flmRefsHyperref{ref655}{discrete simplex}.
\item\relax
\flmRefsHyperref[eczindexfamilyrel]{code:simplex}{\([2^m-1,m,2^{m-1}]\) simplex code} --- Binary simplex codes map to \((2^m,2^m+1)\) simplex spherical codes under the \flmRefsHyperref{ref38}{antipodal mapping} \NoCaseChange{\protect\cite[{Sec. 6.5.2}]{cite1165}\protect\cite[{pg. 18}]{cite115}}. In other words, simplex (simplex spherical) codes form simplices in Hamming (Euclidean) space.
\item\relax
\flmRefsHyperref[eczindexfamilyrel]{code:polyphase}{Polyphase code} --- Simplex spherical codes for dimension \(n=(p-1)/2\) with \(p\) an odd prime admit a polyphase realization \NoCaseChange{\protect\cite[{Sec. 7.7}]{cite115}}.
\item\relax
\flmRefsHyperref[eczindexfamilyrel]{code:petersen_spherical}{Petersen spherical code} --- Codewords of the Petersen spherical code correspond to midpoints of the \(5\)-cell \NoCaseChange{\protect\cite{cite390}}.
\item\relax
\flmRefsHyperref[eczindexfamilyrel]{code:pauli_qsc}{Pauli tessellation QSC} --- Each codeword of the Pauli tessellation QSC is a quantum superposition of vertices of a tetrahedron with \(\pm 1\) coefficients.
\item\relax
\flmRefsHyperref[eczindexfamilyrel]{code:diagonal_clifford}{\(\llbracket 2^r-1,1,3\rrbracket \) simplex code} --- Each \(\llbracket 2^r-1,1,3\rrbracket \) simplex code is a color code whose qubits are placed on the vertices, edges, and faces of an \((r-1)\)-simplex \NoCaseChange{\protect\cite{cite475,cite832}}.
\end{eczvaluelist}
\eczhbkcontributors{ Shubham P. Jain, \eczhuVVA }
\endeczcode

\eczcode{slepian_group}{Slepian group-orbit code}{~\NoCaseChange{\protect\cite{cite2402,cite2403,cite2404}}}
\codefieldsection{Description}
Spherical code in \(n\) dimensions whose codewords correspond to points in an orbit of some \textit{initial vector} under a \textit{generating group} \(G\), which is a subgroup of the orthogonal group \(O(n)\) acting by Euclidean isometries.
Neither the vector nor the group are unique for a given code.

\codefieldsection{Protection}
Code properties depend on the relationship between the group and the initial vector, and the number of codewords is the number of cosets of an initial vector's symmetry subgroup in \(G\) per the orbit-stabilizer theorem.
See Refs. \NoCaseChange{\protect\cite{cite2405,cite2406,cite2407,cite2408}} for allowed code parameters.

\codefieldsection{Notes}
\begin{eczvaluelist}
\item\relax See \NoCaseChange{\protect\cite{cite2409}\protect\cite[{Ch. 8}]{cite115}} for more details and code tables.
\end{eczvaluelist}
\codefieldsection{Parents}
\begin{eczvaluelist}
\item\relax
\flmRefsHyperref[eczindexfamilyrel]{code:spherical}{Spherical code}\item\relax
\flmRefsHyperref[eczindexfamilyrel]{code:group_orbit}{Group-orbit code} --- Slepian group-orbit codes are group-orbit codes on spheres.
\end{eczvaluelist}
\codefieldsection{Children}
\begin{eczvaluelist}
\item\relax
\flmRefsHyperref[eczindexfamilyrel]{code:permutation_spherical}{Permutation spherical code} --- Permutations and sign changes can be implemented on vectors by orthogonal matrices, so permutation spherical codes are Slepian group-orbit codes.
\item\relax
\flmRefsHyperref[eczindexfamilyrel]{code:sidelnikov}{Real-Clifford subgroup-orbit code}\item\relax
\flmRefsHyperref[eczindexfamilyrel]{code:tlsc}{Torus-layer spherical code (TLSC)} --- Polyphase codewords can be implemented by acting on the all-ones initial vector by block-diagonal orthogonal matrices whose \(2\times 2\) rotation blocks encode the codeword components \NoCaseChange{\protect\cite[{Ch. 8}]{cite115}}. TLSC codes are generalizations of polyphase codes to other initial vectors and are examples of Abelian Slepian-group codes.
\item\relax
\flmRefsHyperref[eczindexfamilyrel]{code:polytope}{Polytope code} --- Vertices of polytope codes typically form an orbit of the polytope's symmetry group.
\end{eczvaluelist}
\codefieldsection{Cousins}
\begin{eczvaluelist}
\item\relax
\flmRefsHyperref[eczindexfamilyrel]{code:binary_linear}{Linear binary code} --- Any length-\(n\) binary linear code can be used to define a diagonal subgroup of \(n\)-dimensional rotation matrices with \(\pm 1\) on the diagonals via the \flmRefsHyperref{ref38}{antipodal mapping} \(0\to+1\) and \(1\to-1\). The orbit of this subgroup yields the corresponding Slepian group-orbit code; see \NoCaseChange{\protect\cite[{Thm. 8.5.2}]{cite115}}.
\item\relax
\flmRefsHyperref[eczindexfamilyrel]{code:binary_antipodal}{Binary antipodal code} --- Any length-\(n\) binary linear code can be used to define a diagonal subgroup of \(n\)-dimensional rotation matrices with \(\pm 1\) on the diagonals via the \flmRefsHyperref{ref38}{antipodal mapping} \(0\to+1\) and \(1\to-1\). The orbit of this subgroup yields the corresponding Slepian group-orbit code; see \NoCaseChange{\protect\cite[{Thm. 8.5.2}]{cite115}}.
\item\relax
\flmRefsHyperref[eczindexfamilyrel]{code:group_linear}{Linear code over \(G\)} --- Any finite-group code can be mapped to a Slepian group-orbit code by representing the group using orthogonal matrices \NoCaseChange{\protect\cite{cite2403}}.
\item\relax
\flmRefsHyperref[eczindexfamilyrel]{code:binary_group_orbit}{Binary group-orbit code} --- Binary group-orbit codes can be mapped into Slepian group-orbit codes via various mappings \NoCaseChange{\protect\cite[{Ch. 8}]{cite115}}.
\item\relax
\flmRefsHyperref[eczindexfamilyrel]{code:spherical_design}{Spherical design} --- Slepian group-orbit codes can form spherical designs for real \NoCaseChange{\protect\cite{cite2410,cite393}} or complex spheres \NoCaseChange{\protect\cite{cite2411}}. Polynomial invariants of a discrete subgroup \(G\) of the orthogonal group can be used to determine the real design strength of orbits of \(G\) \NoCaseChange{\protect\cite{cite2412}}. Let \(t+1\) be the degree of the lowest-degree \(G\)-invariant polynomial that is not a polynomial in the norm \(\left\Vert x\right\Vert^2\). Then, any orbit under \(G\) forms a Slepian group-orbit code that is also a spherical \(t\)-design.
\end{eczvaluelist}
\eczhbkcontributors{ \eczhuVVA }
\endeczcode

\eczcode{smith_spherical}{Smith \(40\)-point code}{~\NoCaseChange{\protect\cite{cite2413,cite2414}}}
\codefieldsection{Description}
A \((10,40,1/6)\) spherical code found by W. D. Smith \NoCaseChange{\protect\cite{cite2413}} and conjectured to be optimal in terms of minimizing potential energy functions \NoCaseChange{\protect\cite{cite2414}}.
A beautified set of coordinates can be found on the site \NoCaseChange{\protect\cite{cite2413}}.

\codefieldsection{Parent}
\begin{eczvaluelist}
\item\relax
\flmRefsHyperref[eczindexfamilyrel]{code:spherical}{Spherical code}\end{eczvaluelist}
\codefieldsection{Cousin}
\begin{eczvaluelist}
\item\relax
\flmRefsHyperref[eczindexfamilyrel]{code:sharp_config}{Spherical sharp configuration} --- The Smith spherical code is conjectured to be a global minimum of completely monotonic potential functions \NoCaseChange{\protect\cite{cite2414}}.
\end{eczvaluelist}
\eczhbkcontributors{ \eczhuVVA }
\endeczcode

\eczcode{snub_cube}{Snub-cube code}{}
\codefieldsection{Alternative Names}
\begin{eczvaluelist}
\item\relax Snub cuboctahedron code
\end{eczvaluelist}
\eczhIndexCodeAliasName{snub_cube}{Snub cuboctahedron code}
\codefieldsection{Description}
Spherical \((3,24,\frac{2(t^2+1)}{3t^2+2t+2})\) code whose codewords are the vertices of a snub cube, normalized to lie on the unit sphere.
Here, \(t \approx 1.839\) is the tribonacci constant, the real root of \(t^3-t^2-t-1=0\), and the minimum distance squared is approximately \(0.55384\).

\codefieldsection{Protection}
Optimal configuration of 24 points on \(S^2\) \NoCaseChange{\protect\cite[{pg. 78}]{cite115}}; see also the recent discussion in \NoCaseChange{\protect\cite[{Sec. 1.3, Conj. 1.4}]{cite386}}, which cites Robinson's geometric proof of optimality and asks for a semidefinite-programming proof.
\codefieldsection{Notes}
\begin{eczvaluelist}
\item\relax See the corresponding Bendwavy database entry \NoCaseChange{\protect\cite{cite2415}}.
\end{eczvaluelist}
\codefieldsection{Parent}
\begin{eczvaluelist}
\item\relax
\flmRefsHyperref[eczindexfamilyrel]{code:polyhedron}{Polyhedron code}\end{eczvaluelist}
\eczhbkcontributors{ \eczhuVVA }
\endeczcode

\eczcode{spherical}{Spherical code}{}
\codefieldsection{Description}
Code whose codewords are points on an \(n\)-dimensional sphere \(S^{n-1}\) with radius one.

A spherical code is typically denoted by parameters \((n,M,\rho)\), where \(n\) is the ambient dimension, \(M\) is the size or number of codewords, and \(\rho\) is the \textit{squared minimum distance}, i.e., the smallest squared Euclidean distance between pairs of distinct codewords,
\flmMathEnvironment{align}{}{
  \rho=\min\left\{ \left\Vert x-y\right\Vert ^{2}\,\text{s.t.}\,x,y\in C,\,\,x\neq y\right\}~.
}

A spherical code can be defined using the Gram matrix \(G = XX^T\), where the rows of \(X\) are the codeword vectors.
The Gram matrix is symmetric, positive semidefinite, and has all diagonal elements equal to one.
The code dimension is equal to the rank of \(G\), which can be less than the dimension of the codeword vectors.

Spherical codeword components are often taken from a discrete set of real values called an \textit{alphabet}.
For example, codewords of any length-\(n\) binary code can be mapped into spherical codewords with alphabet \(\{\pm 1/\sqrt{n} \}\) via the \flmRefsHyperref{ref38}{antipodal mapping} \(0\to +1\) and \(1 \to -1\) \NoCaseChange{\protect\cite[{Example 1.2.1}]{cite115}}.

\codefieldsection{Protection}
The Euclidean distance between two points is related to the dot product as
\flmMathEnvironment{align}{}{
  \left\Vert x-y\right\Vert^{2} = 2-2x \cdot y~,
}
where \(x\cdot y\) is the Euclidean inner product. As a result, the \textit{angular distance},
\flmMathEnvironment{align}{}{
  \theta=\arccos(x\cdot y) \in[0,\pi]~,
}
can be equivalently used to quantify code performance.

The size of an \((n,M,\rho)\) spherical code with \(d\) distances between distinct points satisfies the \textit{absolute bound}, \NoCaseChange{\protect\cite{cite385}}
\flmMathEnvironment{align}{}{
  M\leq{n+d-1 \choose d}+{n+d-2 \choose d-1}~.
}
The parameter \(d\) is sometimes called the \(degree\) of the code.
For antipodal codes, i.e., codes that are invariant under \(x\to-x\), the bound is
\flmMathEnvironment{align}{}{
  M\leq2{n+d-2 \choose d-1}~.
}

Denote \(A_n(\rho)\) to be the largest possible size of a spherical code with distance \(\rho\).
Spherical code parameters \((n,M,\rho)\) as well as \(A_n(\rho)\) satisfy the following three \textit{Rankin bounds} \NoCaseChange{\protect\cite{cite2416,cite2417,cite2418}}
\flmMathEnvironment{align}{}{
  \begin{split}
  \rho & \leq\frac{2M}{M-1}\\
  A_{n}\left(\rho\right) & \leq n+1\,\,\,\,\,\,\,\,\,\,\,\,\,\,\,\,\,\,2<\rho\leq4\\
  A_{n}\left(2\right) & \leq2n~.
  \end{split}
}
See \NoCaseChange{\protect\cite[{Ch. 1.2}]{cite39}} for other bounds on \(A_n(\rho)\).

Other bounds on spherical codes include the Fejes Toth bound \NoCaseChange{\protect\cite{cite2419}}, the Wyner bound \NoCaseChange{\protect\cite{cite2420}} and the apple-picking bound \NoCaseChange{\protect\cite{cite2337}}.

\codefieldsection{Rate}
The Kabatiansky-Levenshtein bound \NoCaseChange{\protect\cite{cite2299}\protect\cite[{Ch. 9}]{cite39}} is the best known upper bound on the rate of a spherical code with fixed Euclidean distance.
\codefieldsection{Realizations}
\begin{eczvaluelist}
\item\relax Spherical codes are relevant to modern Hopfield networks \NoCaseChange{\protect\cite{cite345,cite346}}
\end{eczvaluelist}
\codefieldsection{Notes}
\begin{eczvaluelist}
\item\relax See \NoCaseChange{\protect\cite{cite115,cite2413}} for more details and tables of optimal codes.
\item\relax See article \NoCaseChange{\protect\cite{cite2421}} for relations of spherical codes to other fields.
\end{eczvaluelist}
\codefieldsection{Parent}
\begin{eczvaluelist}
\item\relax
\flmRefsHyperref[eczindexfamilyrel]{code:points_into_spheres}{Constant-energy spherical code}\end{eczvaluelist}
\codefieldsection{Children}
\begin{eczvaluelist}
\item\relax
\flmRefsHyperref[eczindexfamilyrel]{code:slepian_group}{Slepian group-orbit code}\item\relax
\flmRefsHyperref[eczindexfamilyrel]{code:lattice_shell}{Lattice-shell code}\item\relax
\flmRefsHyperref[eczindexfamilyrel]{code:ppm}{Pulse-position modulation (PPM) format}\item\relax
\flmRefsHyperref[eczindexfamilyrel]{code:annealing_spherical}{Annealing-based spherical code}\item\relax
\flmRefsHyperref[eczindexfamilyrel]{code:laminated_spherical}{Laminated spherical code}\item\relax
\flmRefsHyperref[eczindexfamilyrel]{code:wrapped_spherical}{Wrapped spherical code}\item\relax
\flmRefsHyperref[eczindexfamilyrel]{code:binary_balanced}{Binary balanced spherical code}\item\relax
\flmRefsHyperref[eczindexfamilyrel]{code:complex_hadamard}{Complex Hadamard spherical code}\item\relax
\flmRefsHyperref[eczindexfamilyrel]{code:smith_spherical}{Smith \(40\)-point code}\item\relax
\flmRefsHyperref[eczindexfamilyrel]{code:univ_opt_spherical}{Universally optimal spherical code}\item\relax
\flmRefsHyperref[eczindexfamilyrel]{code:spherical_design}{Spherical design}\end{eczvaluelist}
\codefieldsection{Cousin}
\begin{eczvaluelist}
\item\relax
\flmRefsHyperref[eczindexfamilyrel]{code:qsc}{Quantum spherical code (QSC)} --- QSCs are quantum counterparts of spherical and constant-energy codes because they store information in quantum superpositions of points on a sphere in quantum phase space.
\end{eczvaluelist}
\eczhbkcontributors{ \eczhuVVA }
\endeczcode

\eczcode{spherical_design}{Spherical design}{~\NoCaseChange{\protect\cite{cite385}}}
\codefieldsection{Description}
Spherical code whose codewords are uniformly distributed in a way that is useful for determining averages of polynomials over the real sphere.
A spherical code is a spherical design of \textit{strength} \(t\), i.e., a \(t\)-design, if the average of any polynomial of degree up to \(t\) over its codewords is equal to the average over the entire sphere.
A \textit{weighed spherical design} is a generalization in which the average over codewords is non-uniform.

Spherical designs can also be defined for complex spheres, and there are ways to convert between the two \NoCaseChange{\protect\cite[{Lemma 3.6}]{cite383}}.

\codefieldsection{Protection}
The number of points \(|X|\) of an \(n\)-dimensional spherical design \(X\) is bounded by \NoCaseChange{\protect\cite{cite385}}
\flmMathEnvironment{align}{}{
  |X|\geq\begin{cases}
  {n+s-1 \choose n-1}+{n+s-2 \choose n-1} & t=2s\\
  2{n+s-1 \choose n-1} & t=2s+1
  \end{cases}~,
}
and designs saturating the above inequality are called \textit{tight}.

Spherical designs with asymptotically optimal cardinality exist for all \(n\) \NoCaseChange{\protect\cite{cite2422}}, proving the Korevaar-Meyers conjecture \NoCaseChange{\protect\cite{cite2423}}.

\codefieldsection{Notes}
\begin{eczvaluelist}
\item\relax See Refs. \NoCaseChange{\protect\cite{cite2372,cite115,cite2424,cite163,cite2425}\protect\cite[{pg. 89}]{cite39}} for reviews and examples on spherical designs.
\end{eczvaluelist}
\codefieldsection{Parents}
\begin{eczvaluelist}
\item\relax
\flmRefsHyperref[eczindexfamilyrel]{code:spherical}{Spherical code}\item\relax
\flmRefsHyperref[eczindexfamilyrel]{code:t-designs}{\(t\)-design} --- Spherical designs are designs on real or complex spheres.
\end{eczvaluelist}
\codefieldsection{Children}
\begin{eczvaluelist}
\item\relax
\flmRefsHyperref[eczindexfamilyrel]{code:sidelnikov}{Real-Clifford subgroup-orbit code} --- The orbit of any point under the real Clifford subgroup is a spherical 7-design \NoCaseChange{\protect\cite{cite391}}, and some are 11-designs \NoCaseChange{\protect\cite{cite392}}.
\item\relax
\flmRefsHyperref[eczindexfamilyrel]{code:dodecahedron}{Dodecahedron code} --- The dodecahedron code forms a spherical 5-design \NoCaseChange{\protect\cite{cite382}}.
\item\relax
\flmRefsHyperref[eczindexfamilyrel]{code:pentakis_dodecahedron}{Pentakis dodecahedron code} --- Vertices of the pentakis dodecahedron form a weighted spherical 9-design \NoCaseChange{\protect\cite{cite388,cite389}\protect\cite[{Exam. 2.5}]{cite380}}.
\item\relax
\flmRefsHyperref[eczindexfamilyrel]{code:120cell}{120-cell code} --- The code forms a spherical 11-design because its vertices can be divided into five 600-cells, each of which forms said design.
\item\relax
\flmRefsHyperref[eczindexfamilyrel]{code:600cell}{600-cell code} --- The 600-cell code forms a spherical 11-design that is unique up to equivalence \NoCaseChange{\protect\cite{cite378}}.
\item\relax
\flmRefsHyperref[eczindexfamilyrel]{code:disphenoidal288cell}{Disphenoidal 288-cell code} --- The disphenoidal 288-cell code forms a spherical 7-design \NoCaseChange{\protect\cite{cite381}}.
\item\relax
\flmRefsHyperref[eczindexfamilyrel]{code:rect_hessian_polyhedron}{Rectified Hessian polyhedron code} --- The rectified Hessian polyhedron code forms a spherical 5-design \NoCaseChange{\protect\cite{cite393}}.
\item\relax
\flmRefsHyperref[eczindexfamilyrel]{code:231_polytope}{\(2_{31}\) polytope code} --- The 126 vertices of the \(2_{31}\) polytope form a spherical 5-design \NoCaseChange{\protect\cite{cite384}}.
\item\relax
\flmRefsHyperref[eczindexfamilyrel]{code:241_polytope}{\(2_{41}\) polytope code} --- The \(2_{41}\) polytope code forms a spherical 7-design \NoCaseChange{\protect\cite{cite232}}.
\item\relax
\flmRefsHyperref[eczindexfamilyrel]{code:hypercube}{Hypercube code} --- Hypercube codes form spherical 3-designs. The weighted union of the vertices of a hypercube and an orthoplex form a weighted spherical 5-design in dimensions \(\geq 3\) \NoCaseChange{\protect\cite[{Sec. 8.6, Ex. 5-2}]{cite379}\protect\cite[{Exam. 2.6}]{cite380}}.
\item\relax
\flmRefsHyperref[eczindexfamilyrel]{code:kerdock_spherical}{Kerdock spherical code} --- Kerdock spherical codes form spherical 3-designs because their codewords are unions of \(2^{2r-1}+1\) orthoplexes \NoCaseChange{\protect\cite{cite386}}.
\item\relax
\flmRefsHyperref[eczindexfamilyrel]{code:sharp_config}{Spherical sharp configuration} --- Spherical sharp configurations are spherical designs of strength \(2m-1\) for some \(m\).
\item\relax
\flmRefsHyperref[eczindexfamilyrel]{code:petersen_spherical}{Petersen spherical code} --- The Petersen spherical code forms a spherical 2-design \NoCaseChange{\protect\cite{cite390}}.
\end{eczvaluelist}
\codefieldsection{Cousins}
\begin{eczvaluelist}
\item\relax
\flmRefsHyperref[eczindexfamilyrel]{code:slepian_group}{Slepian group-orbit code} --- Slepian group-orbit codes can form spherical designs for real \NoCaseChange{\protect\cite{cite2410,cite393}} or complex spheres \NoCaseChange{\protect\cite{cite2411}}. Polynomial invariants of a discrete subgroup \(G\) of the orthogonal group can be used to determine the real design strength of orbits of \(G\) \NoCaseChange{\protect\cite{cite2412}}. Let \(t+1\) be the degree of the lowest-degree \(G\)-invariant polynomial that is not a polynomial in the norm \(\left\Vert x\right\Vert^2\). Then, any orbit under \(G\) forms a Slepian group-orbit code that is also a spherical \(t\)-design.
\item\relax
\flmRefsHyperref[eczindexfamilyrel]{code:self_dual_lattice}{Unimodular lattice} --- A union of \(t\) shells of self-dual lattices and their shadows form spherical \(t\)-designs \NoCaseChange{\protect\cite{cite2313}}.
\item\relax
\flmRefsHyperref[eczindexfamilyrel]{code:combinatorial_design}{Combinatorial design} --- Spherical designs can be thought of as Euclidean analogues of combinatorial designs \NoCaseChange{\protect\cite{cite157}}.
\item\relax
\flmRefsHyperref[eczindexfamilyrel]{code:golay}{\([23, 12, 7]\) Golay code} --- The dual of the Golay code forms a spherical 3-design under the \flmRefsHyperref{ref38}{antipodal mapping} \NoCaseChange{\protect\cite[{Exam. 9.3}]{cite385}}.
\item\relax
\flmRefsHyperref[eczindexfamilyrel]{code:hexacode}{\([6,3,4]_4\) Hexacode} --- The hexacode is a complex spherical 3-design when embedded into the complex sphere via the polyphase mapping \NoCaseChange{\protect\cite{cite1671}}.
\item\relax
\flmRefsHyperref[eczindexfamilyrel]{code:dfour_shell}{\(D_4\) lattice-shell code} --- \(D_4\) \(2m\)-shell codes can form spherical designs \NoCaseChange{\protect\cite{cite2323}}.
\item\relax
\flmRefsHyperref[eczindexfamilyrel]{code:lattice_shell}{Lattice-shell code} --- Nonempty \(2m\)-shell codes of extremal even unimodular lattices in \(n\) dimensions form spherical \(t\)-designs with \(t=11\) (\(t=7\), \(t=3\)) if \(n \equiv 0\) (\(n \equiv 8\), \(n\equiv 16\)) modulo 24 \NoCaseChange{\protect\cite{cite2371,cite384}}. Shells of \(A_n\) and \(D_n\) lattices form infinite families of spherical 3-designs \NoCaseChange{\protect\cite[{Exam. 2.9}]{cite2372}}.
\item\relax
\flmRefsHyperref[eczindexfamilyrel]{code:leech_shell}{\(\Lambda_{24}\) Leech lattice-shell code} --- The smallest-shell \((24,196560,1)\) code is a tight and unique spherical 11-design \NoCaseChange{\protect\cite{cite124}\protect\cite[{Ch. 3}]{cite39}}. The \((23,4600,1/3)\) \NoCaseChange{\protect\cite{cite124,cite2314,cite125,cite119}\protect\cite[{Table 1}]{cite394}} and \((22,891,1/4)\) \NoCaseChange{\protect\cite{cite2314,cite125,cite119}\protect\cite[{Table 1}]{cite394}} spherical codes are also sharp configurations.
\item\relax
\flmRefsHyperref[eczindexfamilyrel]{code:polygon}{Polygon code} --- A \(q\)-gon is a tight spherical \(q-1\) design.
\item\relax
\flmRefsHyperref[eczindexfamilyrel]{code:antiprism}{Antiprism code} --- For the case when the two \(q\)-gons are such that the \(q=2,3\) cases reduce to the tetrahedron and octahedron, respectively, the antiprism is a spherical 3-design for \(q \geq 3\), and a \(2\)-design for \(q=2\) \NoCaseChange{\protect\cite{cite2339}}. This can be seen as a consequence of \NoCaseChange{\protect\cite[{Lemma 6.11}]{cite232}}.
\item\relax
\flmRefsHyperref[eczindexfamilyrel]{code:icosahedron}{Icosahedron code} --- The icosahedron code forms a unique tight spherical 5-design \NoCaseChange{\protect\cite{cite385}\protect\cite[{Exam. 9.6.1}]{cite115}}.
\item\relax
\flmRefsHyperref[eczindexfamilyrel]{code:24cell}{24-cell code} --- The 24-cell code is a spherical 5-design \NoCaseChange{\protect\cite{cite377}}.
\item\relax
\flmRefsHyperref[eczindexfamilyrel]{code:hessian_polyhedron}{Hessian polyhedron code} --- The Hessian polytope code forms a tight spherical 4-design \NoCaseChange{\protect\cite[{Exam. 7.3}]{cite383}}. The double Hessian polytope code forms a spherical 5-design \NoCaseChange{\protect\cite{cite384}}.
\item\relax
\flmRefsHyperref[eczindexfamilyrel]{code:hess_polytope}{\(3_{21}\) polytope code} --- The \(3_{21}\) polytope code forms a tight spherical 5-design \NoCaseChange{\protect\cite{cite385,cite124}\protect\cite[{Ch. 14}]{cite39}\protect\cite[{Table 1}]{cite119}}.
\item\relax
\flmRefsHyperref[eczindexfamilyrel]{code:witting_polytope}{Witting polytope code} --- The Witting polytope code forms a tight spherical 7-design \NoCaseChange{\protect\cite{cite124}\protect\cite[{Ch. 14}]{cite39}}.
\item\relax
\flmRefsHyperref[eczindexfamilyrel]{code:biorthogonal_spherical}{Biorthogonal spherical code} --- Biorthogonal spherical codes are the only tight spherical 3-designs \NoCaseChange{\protect\cite[{Tab. 9.3}]{cite115}}. A suitable weighted union of the vertices of a hypercube and an orthoplex forms a weighted spherical 5-design in dimensions \(\geq 3\) \NoCaseChange{\protect\cite[{Sec. 8.6, Ex. 5-2}]{cite379}\protect\cite[{Exam. 2.6}]{cite380}}.
\item\relax
\flmRefsHyperref[eczindexfamilyrel]{code:simplex_spherical}{Simplex spherical code} --- Simplex spherical codes are the only tight spherical 2-designs \NoCaseChange{\protect\cite[{Tab. 9.3}]{cite115}}. The bi-simplex is a spherical 3-design since antipodal codes have zero averages over odd-degree polynomials.
\item\relax
\flmRefsHyperref[eczindexfamilyrel]{code:mclaughlin}{McLaughlin spherical code} --- Both McLaughlin spherical codes are sharp configurations \NoCaseChange{\protect\cite{cite119,cite387}}. The \((22,275,1/6)\) code is a unique and tight spherical 4-design, while the \((23,552,1/5)\) code is a unique and tight spherical 5-design; see Ref. \NoCaseChange{\protect\cite[{Appx. A}]{cite119}}.
\end{eczvaluelist}
\eczhbkcontributors{ \eczhuVVA }
\endeczcode

\eczcode{sharp_config}{Spherical sharp configuration}{~\NoCaseChange{\protect\cite{cite2277,cite2278,cite2088,cite171,cite914,cite119}}}
\codefieldsection{Alternative Names}
\begin{eczvaluelist}
\item\relax Delsarte code on Euclidean spheres
\item\relax Spherical Delsarte code
\end{eczvaluelist}
\eczhIndexCodeAliasName{sharp_config}{Delsarte code on Euclidean spheres}
\eczhIndexCodeAliasName{sharp_config}{Spherical Delsarte code}
\codefieldsection{Description}
A spherical code that is a spherical design of strength \(2m-1\) for some \(m\) and that has \(m\) distances between distinct points.
All known spherical sharp configurations are either obtained from the Leech or \(E_8\) lattice, certain regular polytopes, or are CGS isotropic subspace spherical codes \NoCaseChange{\protect\cite[{Table 1}]{cite394}} (see also \NoCaseChange{\protect\cite[{Table 9.1}]{cite171}}).

\codefieldsection{Parents}
\begin{eczvaluelist}
\item\relax
\flmRefsHyperref[eczindexfamilyrel]{code:univ_opt_spherical}{Universally optimal spherical code} --- All sharp configurations are universally optimal \NoCaseChange{\protect\cite{cite119}}, but not all universally optimal spherical codes are sharp configurations. The one known exception is the 600-cell.
\item\relax
\flmRefsHyperref[eczindexfamilyrel]{code:delsarte_optimal}{Sharp configuration}\item\relax
\flmRefsHyperref[eczindexfamilyrel]{code:spherical_design}{Spherical design} --- Spherical sharp configurations are spherical designs of strength \(2m-1\) for some \(m\).
\end{eczvaluelist}
\codefieldsection{Children}
\begin{eczvaluelist}
\item\relax
\flmRefsHyperref[eczindexfamilyrel]{code:polygon}{Polygon code}\item\relax
\flmRefsHyperref[eczindexfamilyrel]{code:icosahedron}{Icosahedron code} --- The icosahedron is a sharp configuration \NoCaseChange{\protect\cite{cite2364,cite119}}.
\item\relax
\flmRefsHyperref[eczindexfamilyrel]{code:hess_polytope}{\(3_{21}\) polytope code} --- The \(3_{21}\) polytope code is a sharp configuration \NoCaseChange{\protect\cite{cite2321,cite119}}.
\item\relax
\flmRefsHyperref[eczindexfamilyrel]{code:witting_polytope}{Witting polytope code} --- The Witting polytope code is a sharp configuration \NoCaseChange{\protect\cite{cite2321,cite119}}.
\item\relax
\flmRefsHyperref[eczindexfamilyrel]{code:biorthogonal_spherical}{Biorthogonal spherical code}\item\relax
\flmRefsHyperref[eczindexfamilyrel]{code:simplex_spherical}{Simplex spherical code}\item\relax
\flmRefsHyperref[eczindexfamilyrel]{code:cgs_spherical}{Cameron-Goethals-Seidel (CGS) isotropic subspace code} --- CGS isotropic subspace codes are the only known spherical sharp configurations not derived from regular polytopes or lattices \NoCaseChange{\protect\cite{cite119}}.
\item\relax
\flmRefsHyperref[eczindexfamilyrel]{code:mclaughlin}{McLaughlin spherical code} --- Both McLaughlin spherical codes are sharp configurations \NoCaseChange{\protect\cite{cite119,cite387}}. The \((22,275,1/6)\) code is a unique and tight spherical 4-design, while the \((23,552,1/5)\) code is a unique and tight spherical 5-design; see Ref. \NoCaseChange{\protect\cite[{Appx. A}]{cite119}}.
\end{eczvaluelist}
\codefieldsection{Cousins}
\begin{eczvaluelist}
\item\relax
\flmRefsHyperref[eczindexfamilyrel]{code:eeight}{\(E_8\) Gosset lattice} --- Several spherical sharp configurations are derived from the \(E_8\) Gosset lattice \NoCaseChange{\protect\cite{cite119}}.
\item\relax
\flmRefsHyperref[eczindexfamilyrel]{code:leech}{\(\Lambda_{24}\) Leech lattice} --- Several spherical sharp configurations are derived from the Leech lattice \NoCaseChange{\protect\cite{cite119}}.
\item\relax
\flmRefsHyperref[eczindexfamilyrel]{code:leech_shell}{\(\Lambda_{24}\) Leech lattice-shell code} --- The smallest-shell \((24,196560,1)\) code is a spherical sharp configuration \NoCaseChange{\protect\cite{cite2316,cite119}}. The \((23,4600,1/3)\) \NoCaseChange{\protect\cite{cite124,cite2314,cite125,cite119}\protect\cite[{Table 1}]{cite394}} and \((22,891,1/4)\) \NoCaseChange{\protect\cite{cite2314,cite125,cite119}\protect\cite[{Table 1}]{cite394}} spherical codes are also sharp configurations.
\item\relax
\flmRefsHyperref[eczindexfamilyrel]{code:smith_spherical}{Smith \(40\)-point code} --- The Smith spherical code is conjectured to be a global minimum of completely monotonic potential functions \NoCaseChange{\protect\cite{cite2414}}.
\end{eczvaluelist}
\eczhbkcontributors{ Alexander Barg, \eczhuVVA }
\endeczcode

\eczcode{square_antiprism}{Square-antiprism code}{}
\codefieldsection{Description}
Spherical \((3,8,4(4-\sqrt{2})/7)\) code whose codewords are the vertices of the square antiprism \NoCaseChange{\protect\cite[{pg. 72}]{cite115}}.

\codefieldsection{Protection}
Optimal configuration of 8 points in 3D space \NoCaseChange{\protect\cite[{pg. 72}]{cite115}}.
\codefieldsection{Notes}
\begin{eczvaluelist}
\item\relax See the corresponding Bendwavy database entry \NoCaseChange{\protect\cite{cite2426}}.
\end{eczvaluelist}
\codefieldsection{Parent}
\begin{eczvaluelist}
\item\relax
\flmRefsHyperref[eczindexfamilyrel]{code:antiprism}{Antiprism code} --- The antiprism reduces to a square antiprism for \(q=4\).
\end{eczvaluelist}
\codefieldsection{Cousin}
\begin{eczvaluelist}
\item\relax
\flmRefsHyperref[eczindexfamilyrel]{code:hypercube}{Hypercube code} --- The square antiprism can be obtained by stretching the cube and twisting the top of the cube by \(45\) degrees \NoCaseChange{\protect\cite[{pg. 72}]{cite115}}.
\end{eczvaluelist}
\eczhbkcontributors{ \eczhuVVA }
\endeczcode

\eczcode{tlsc}{Torus-layer spherical code (TLSC)}{~\NoCaseChange{\protect\cite{cite2427}}}
\codefieldsection{Description}
Code whose codewords are elements of a foliation of the \(2n-1\)-dimensional hypersphere \(S^{2n-1}\) using flat tori \(S^1\times S^1 \times \cdots \times S^1\).
Related constructions include the spherical codes by Hopf foliations (SCHF) \NoCaseChange{\protect\cite{cite2428}}.

\codefieldsection{Decoding}
\begin{eczvaluelist}
\item\relax Efficiently decodable \NoCaseChange{\protect\cite{cite2427}}.
\end{eczvaluelist}
\codefieldsection{Notes}
\begin{eczvaluelist}
\item\relax See \NoCaseChange{\protect\cite[{Sec. 5.2}]{cite2429}} for an exposition.
\end{eczvaluelist}
\codefieldsection{Parent}
\begin{eczvaluelist}
\item\relax
\flmRefsHyperref[eczindexfamilyrel]{code:slepian_group}{Slepian group-orbit code} --- Polyphase codewords can be implemented by acting on the all-ones initial vector by block-diagonal orthogonal matrices whose \(2\times 2\) rotation blocks encode the codeword components \NoCaseChange{\protect\cite[{Ch. 8}]{cite115}}. TLSC codes are generalizations of polyphase codes to other initial vectors and are examples of Abelian Slepian-group codes.
\end{eczvaluelist}
\codefieldsection{Child}
\begin{eczvaluelist}
\item\relax
\flmRefsHyperref[eczindexfamilyrel]{code:polyphase}{Polyphase code}\end{eczvaluelist}
\eczhbkcontributors{ \eczhuVVA }
\endeczcode

\eczcode{univ_opt_spherical}{Universally optimal spherical code}{~\NoCaseChange{\protect\cite{cite2430,cite2091,cite2431,cite2421,cite119}}}
\codefieldsection{Description}
A spherical code that (weakly) minimizes all completely monotonic potentials on the sphere for its cardinality. See \NoCaseChange{\protect\cite{cite395}\protect\cite[{Sec. 12.4}]{cite199}} for further discussion.
In particular, all spherical sharp configurations and the 600-cell are universally optimal \NoCaseChange{\protect\cite[{Thm. 1.2}]{cite119}}.

\codefieldsection{Parents}
\begin{eczvaluelist}
\item\relax
\flmRefsHyperref[eczindexfamilyrel]{code:spherical}{Spherical code}\item\relax
\flmRefsHyperref[eczindexfamilyrel]{code:univ_opt}{Universally optimal code}\end{eczvaluelist}
\codefieldsection{Children}
\begin{eczvaluelist}
\item\relax
\flmRefsHyperref[eczindexfamilyrel]{code:600cell}{600-cell code} --- The 600-cell is universally optimal, but it is not a spherical sharp configuration \NoCaseChange{\protect\cite{cite119}\protect\cite[{Thm. 12.4.27}]{cite199}}.
\item\relax
\flmRefsHyperref[eczindexfamilyrel]{code:sharp_config}{Spherical sharp configuration} --- All sharp configurations are universally optimal \NoCaseChange{\protect\cite{cite119}}, but not all universally optimal spherical codes are sharp configurations. The one known exception is the 600-cell.
\end{eczvaluelist}
\codefieldsection{Cousins}
\begin{eczvaluelist}
\item\relax
\flmRefsHyperref[eczindexfamilyrel]{code:24cell}{24-cell code} --- The 24-cell code is not universally optimal \NoCaseChange{\protect\cite{cite377}}, but comes quite close \NoCaseChange{\protect\cite[{Exam. 12.4.29}]{cite199}}.
\item\relax
\flmRefsHyperref[eczindexfamilyrel]{code:kerdock_spherical}{Kerdock spherical code} --- Kerdock spherical codes are almost universally optimal \NoCaseChange{\protect\cite{cite2367}}.
\end{eczvaluelist}
\eczhbkcontributors{ Alexander Barg, \eczhuVVA }
\endeczcode

\eczcode{witting_polytope}{Witting polytope code}{}
\codefieldsection{Alternative Names}
\begin{eczvaluelist}
\item\relax \(4_{21}\) polytope code
\item\relax Gosset polytope code
\end{eczvaluelist}
\eczhIndexCodeAliasName{witting_polytope}{\(4_{21}\) polytope code}
\eczhIndexCodeAliasName{witting_polytope}{Gosset polytope code}
\codefieldsection{Description}
Spherical \((8,240,1)\) code whose codewords are the vertices of the Witting complex polytope, the \(4_{21}\) polytope, and the minimal lattice-shell code of the \(E_8\) lattice.
The code is optimal and unique up to equivalence \NoCaseChange{\protect\cite{cite124,cite39,cite125}}.
Antipodal pairs of points of the \(4_{21}\) polytope code correspond to the 120 tritangent planes of a canonical sextic curve in \(\mathbb{C}P^3\) \NoCaseChange{\protect\cite{cite117,cite118,cite119,cite120}}.

A representation of the codewords consists of the 112 vectors obtained from permutations of \((0,0,0,0,0,0,\pm 2,\pm 2)\) together with the 128 vectors \((\pm 1)^{\times 8}\) for which the number of minus signs is even. After normalization, these are precisely the 240 minimal vectors of the \(E_8\) lattice.
See \NoCaseChange{\protect\cite[{pg. 132}]{cite231}} for a complex representation.

Recursively taking its kissing configurations yields the \((7,56,1/3)\), \((6,27,1/4)\), \((5,16,1/5)\), \((4,10,1/6)\), and \((3,6,1/7)\) spherical codes \NoCaseChange{\protect\cite{cite124}}.

\begin{flmFloat}{figure}{NumCap}\includegraphics[width=306.14173228346453bp,max width=\linewidth]{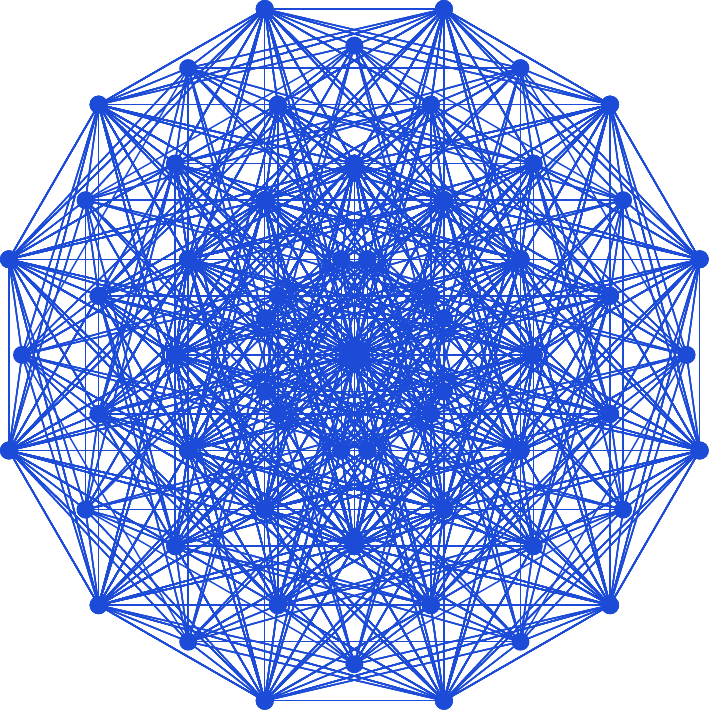}\caption{Projection of the coordinates of the Witting polytope.}\label{ref2432}\end{flmFloat}

\codefieldsection{Protection}
Code yields an optimal solution to the kissing problem in 8D \NoCaseChange{\protect\cite{cite2330,cite2331}}, saturates the Levenshtein bound \NoCaseChange{\protect\cite{cite2315}}, and is unique up to equivalence \NoCaseChange{\protect\cite{cite124}}.
It carries a 4-class association scheme, and the inner products with respect to any codeword are \(\pm 1\), \(\pm 1/2\), and \(0\) with multiplicities \(1\), \(56\), and \(126\), respectively \NoCaseChange{\protect\cite{cite124}}.

\codefieldsection{Notes}
\begin{eczvaluelist}
\item\relax The Witting polytope yields 40 states of a qudit of dimension 4 and a non-probabilistic version of Bell's theorem \NoCaseChange{\protect\cite{cite2433,cite2434,cite2435,cite2436}}.
\item\relax See the corresponding Bendwavy database entries for the Witting complex polytope \NoCaseChange{\protect\cite{cite2437}} and its real-space embedding as the \(4_{21}\) polytope \NoCaseChange{\protect\cite{cite2438}}.
\end{eczvaluelist}
\codefieldsection{Parents}
\begin{eczvaluelist}
\item\relax
\flmRefsHyperref[eczindexfamilyrel]{code:self_dual_polytope}{Self-dual polytope code} --- The Witting polytope is self-dual as a complex polytope. The \(4_{21}\) polytope is not self-dual \NoCaseChange{\protect\cite{cite178}}.
\item\relax
\flmRefsHyperref[eczindexfamilyrel]{code:eeight_shell}{\(E_8\) Gosset lattice-shell code} --- The minimal shell of the lattice yields the \((8,240,1)\) code, whose codewords form the vertices of the \(4_{21}\) polytope.
\item\relax
\flmRefsHyperref[eczindexfamilyrel]{code:sharp_config}{Spherical sharp configuration} --- The Witting polytope code is a sharp configuration \NoCaseChange{\protect\cite{cite2321,cite119}}.
\item\relax
\flmRefsHyperref[eczindexfamilyrel]{code:sidelnikov}{Real-Clifford subgroup-orbit code} --- The Witting polytope code is equivalent to the real Clifford subgroup-orbit code for \(n=8\).
\end{eczvaluelist}
\codefieldsection{Cousins}
\begin{eczvaluelist}
\item\relax
\flmRefsHyperref[eczindexfamilyrel]{code:spherical_design}{Spherical design} --- The Witting polytope code forms a tight spherical 7-design \NoCaseChange{\protect\cite{cite124}\protect\cite[{Ch. 14}]{cite39}}.
\item\relax
\flmRefsHyperref[eczindexfamilyrel]{code:complex_projective}{Complex projective space code} --- Antipodal pairs of points of the Witting polytope code correspond to the 120 tritangent planes of a canonical sextic curve in \(\mathbb{C}P^3\) \NoCaseChange{\protect\cite{cite117,cite118,cite119,cite120}}.
\item\relax
\flmRefsHyperref[eczindexfamilyrel]{code:delsarte_optimal}{Sharp configuration} --- The 120 antipodal pairs of the Witting polytope code form a sharp configuration in \(\mathbb{R}P^7\) \NoCaseChange{\protect\cite{cite119}}.
\item\relax
\flmRefsHyperref[eczindexfamilyrel]{code:t-designs}{\(t\)-design} --- Antipodal pairs of points of the Witting polytope code form a 3-design in \(\mathbb{R}P^7\) \NoCaseChange{\protect\cite{cite119}}.
\item\relax
\flmRefsHyperref[eczindexfamilyrel]{code:real_projective}{Real projective space code} --- The 120 antipodal pairs of the Witting polytope code form a sharp configuration and a 3-design in \(\mathbb{R}P^7\) \NoCaseChange{\protect\cite{cite119}}.
\item\relax
\flmRefsHyperref[eczindexfamilyrel]{code:eeight}{\(E_8\) Gosset lattice} --- The Voronoi cell of the \(E_8\) Gosset lattice is the dual of the Gosset \(4_{21}\) polytope \NoCaseChange{\protect\cite[{Ch. 21, pg. 464}]{cite39}}.
\item\relax
\flmRefsHyperref[eczindexfamilyrel]{code:600cell}{600-cell code} --- The 120 vertices of the 600-cell are the unit icosians, and these icosian units, together with their multiples by \((1-\sqrt{5})/2\), form the 240 minimal vectors of a version of the \(E_8\) lattice, i.e., the Witting polytope \NoCaseChange{\protect\cite[{Ch. 8, pg. 210}]{cite39}}.
\item\relax
\flmRefsHyperref[eczindexfamilyrel]{code:hessian_polyhedron}{Hessian polyhedron code} --- The Hessian polyhedron code forms the next recursive kissing configuration after the \(3_{21}\) polytope code in the \(E_8\) lattice-shell/Witting polytope sequence \NoCaseChange{\protect\cite{cite124}}. The Schläfli graph is a subgraph of the graph formed by the vertices of the Witting polytope \NoCaseChange{\protect\cite[{Sec. 3.11}]{cite1385}}.
\item\relax
\flmRefsHyperref[eczindexfamilyrel]{code:hess_polytope}{\(3_{21}\) polytope code} --- \(3_{21}\) polytope codewords form the first recursive kissing configuration of the Witting polytope code \NoCaseChange{\protect\cite{cite124,cite119}\protect\cite[{Ch. 9, pg. 264}]{cite39}}. The Gosset graph is a subgraph of the graph formed by the vertices of the Witting polytope \NoCaseChange{\protect\cite[{Sec. 3.11}]{cite1385}}.
\item\relax
\flmRefsHyperref[eczindexfamilyrel]{code:241_polytope}{\(2_{41}\) polytope code} --- Vertices of the \(2_{41}\) and \(4_{21}\) polytopes minimize each other's potential functions \NoCaseChange{\protect\cite{cite232}}.
\item\relax
\flmRefsHyperref[eczindexfamilyrel]{code:quantum_sidelnikov}{Clifford subgroup-orbit QSC} --- Logical constellations of the Clifford subgroup-orbit code for \(r=2\) form vertices of Witting polytopes.
\end{eczvaluelist}
\eczhbkcontributors{ Shubham P. Jain, \eczhuVVA }
\endeczcode

\eczcode{wrapped_spherical}{Wrapped spherical code}{~\NoCaseChange{\protect\cite{cite2439}}}
\codefieldsection{Description}
Spherical code in dimension \(n+1\) whose codewords are obtained from centers of spheres in a finite sphere packing of \(\mathbb{R}^{n}\) that is "wrapped" onto \(S^n\).

\codefieldsection{Rate}
Asymptotically maximal spherical coding density is obtained with the densest possible sphere packing.
\codefieldsection{Parent}
\begin{eczvaluelist}
\item\relax
\flmRefsHyperref[eczindexfamilyrel]{code:spherical}{Spherical code}\end{eczvaluelist}
\eczhbkcontributors{ \eczhuVVA }
\endeczcode

\onecolumngrid
\clearpage

\section{Ring Kingdom}

\begin{eczEpigraph}
\begin{quote}
\flmQuoteSetup{quote}%
One Ring to rule them all,\\
One Ring to find them,\\
One Ring to bring them all,\\
and in the darkness bind them.
\flmQuoteAttributed{J.R.R. Tolkien}
\end{quote}
\end{eczEpigraph}

\twocolumngrid

\eczcode{pentacode}{\((5,40,4)_{\mathbb{Z}_4}\) Pentacode}{~\NoCaseChange{\protect\cite{cite373}}}
\eczhIndexCodeAliasName{pentacode}{Pentacode}
\codefieldsection{Description}
Nonlinear \((5,40,4)_{\mathbb{Z}_4}\) code over \(\mathbb{Z}_4\) whose codewords are all cyclic permutations and negations of the strings \(01112\), \(03110\), \(21310\), and \(21132\).

\codefieldsection{Parents}
\begin{eczvaluelist}
\item\relax
\flmRefsHyperref[eczindexfamilyrel]{code:q-ary_over_zq}{\(q\)-ary code over \(\mathbb{Z}_q\)}\item\relax
\flmRefsHyperref[eczindexfamilyrel]{code:small_distance}{Small-distance block code}\end{eczvaluelist}
\codefieldsection{Cousins}
\begin{eczvaluelist}
\item\relax
\flmRefsHyperref[eczindexfamilyrel]{code:best}{\((10,40,4)\) Best code} --- Codewords of the Best code can be obtained by applying the Gray map to the pentacode \NoCaseChange{\protect\cite[{Sec. 2}]{cite373}}.
\item\relax
\flmRefsHyperref[eczindexfamilyrel]{code:q-ary_additive}{Additive \(q\)-ary code} --- A close relative of the pentacode is an additive quaternary code \NoCaseChange{\protect\cite{cite1713}}.
\end{eczvaluelist}
\eczhbkcontributors{ \eczhuVVA }
\endeczcode

\eczcode{self_dual_z6}{\([4,2,2]_{\mathbb{Z}_6}\) senary code}{~\NoCaseChange{\protect\cite{cite128}}}
\eczhIndexCodeAliasName{self_dual_z6}{senary code}
\codefieldsection{Description}
A self-dual code over \(\mathbb{Z}_6\) that is one of two such codes, up to permutations \NoCaseChange{\protect\cite{cite128}}.

A generator matrix for the code is
\flmMathEnvironment{align}{}{
  \begin{pmatrix}
  1 & 0 & 1 & 2 \\
  0 & 1 & 4 & 1
  \end{pmatrix}~.
}

\codefieldsection{Protection}
Distance \(d=2\) implies detection of any single-symbol error.

\codefieldsection{Parents}
\begin{eczvaluelist}
\item\relax
\flmRefsHyperref[eczindexfamilyrel]{code:self_dual_over_zq}{Self-dual code over \(\mathbb{Z}_q\)}\item\relax
\flmRefsHyperref[eczindexfamilyrel]{code:small_distance}{Small-distance block code}\end{eczvaluelist}
\eczhbkcontributors{ \eczhuVVA }
\endeczcode

\eczcode{cmr}{\(C_{m,r}\) code}{~\NoCaseChange{\protect\cite{cite1581}}}
\eczhIndexCodeAliasName{cmr}{code}
\codefieldsection{Description}
A member of a family of Type IV self-dual quaternary linear codes over \(\mathbb{Z}_4\) generated by \(\textnormal{RM}(r,m) + 2\textnormal{RM}(m-r-1,m)\) for \(3r \leq m-1\) \NoCaseChange{\protect\cite{cite121}}.

\codefieldsection{Parent}
\begin{eczvaluelist}
\item\relax
\flmRefsHyperref[eczindexfamilyrel]{code:self_dual_over_z4}{Self-dual code over \(\mathbb{Z}_4\)} --- The \(C_{m,r}\) code is a Type IV self-dual code over \(\mathbb{Z}_4\) \NoCaseChange{\protect\cite{cite121}}.
\end{eczvaluelist}
\codefieldsection{Cousins}
\begin{eczvaluelist}
\item\relax
\flmRefsHyperref[eczindexfamilyrel]{code:reed_muller}{Reed-Muller (RM) code} --- The \(C_{m,r}\) code is generated by \(\textnormal{RM}(r,m) + 2\textnormal{RM}(m-r-1,m)\) for \(3r \leq m-1\) \NoCaseChange{\protect\cite{cite121}}.
\item\relax
\flmRefsHyperref[eczindexfamilyrel]{code:barnes_wall}{Barnes-Wall (BW) lattice} --- \(C_{m,r=1}\) codes give rise to certain BW lattices \NoCaseChange{\protect\cite{cite2225,cite112}}.
\item\relax
\flmRefsHyperref[eczindexfamilyrel]{code:bw32}{\(BW_{32}\) Barnes-Wall lattice} --- The \(C_{m=5,r=1}\) code gives rise to the \(BW_{32}\) Barnes-Wall lattice via \flmTerm{term}{ref114}{}{Construction \(A_4\)} \NoCaseChange{\protect\cite{cite2225,cite112}}.
\item\relax
\flmRefsHyperref[eczindexfamilyrel]{code:klemm}{Klemm code} --- The Klemm code at \(m=8\) is the \(C_{m,r=0}\) code with parameters \([32,16,4]_{\mathbb{Z}_4}\) \NoCaseChange{\protect\cite{cite1581}}.
\end{eczvaluelist}
\eczhbkcontributors{ \eczhuVVA }
\endeczcode

\eczcode{q-ary_over_zq}{\(q\)-ary code over \(\mathbb{Z}_q\)}{}
\codefieldsection{Description}
A code encoding \(K\) states (codewords) in \(n\) coordinates over the ring \(\mathbb{Z}_q\) of integers modulo \(q\).
\codefieldsection{Protection}
In addition to the Hamming distance, codes over \(\mathbb{Z}_q\) are also defined over the Lee metric \NoCaseChange{\protect\cite{cite2440}}.
Linear programming bounds exist under this metric \NoCaseChange{\protect\cite{cite2441,cite2442}}.

\codefieldsection{Notes}
\begin{eczvaluelist}
\item\relax See books \NoCaseChange{\protect\cite{cite126,cite195}} for introductions.
\end{eczvaluelist}
\codefieldsection{Parents}
\begin{eczvaluelist}
\item\relax
\flmRefsHyperref[eczindexfamilyrel]{code:rings_into_rings}{Ring code}\item\relax
\flmRefsHyperref[eczindexfamilyrel]{code:symmetric_space}{Symmetric-space code} --- The space of \(q\)-ary codes over \(\mathbb{Z}_q\) under the Lee metric can be viewed as a finite symmetric space \(G/H\) with \(G = D_q \wr S_n\) \NoCaseChange{\protect\cite{cite2441,cite2442}\protect\cite[{Table 3}]{cite985}}.
\end{eczvaluelist}
\codefieldsection{Children}
\begin{eczvaluelist}
\item\relax
\flmRefsHyperref[eczindexfamilyrel]{code:bits_into_bits}{Binary code} --- A \(q\)-ary code over \(\mathbb{Z}_q\) reduces to a binary code at \(q=2\). Ternary computing may be more applicable than binary computing to cryptographic schemes \NoCaseChange{\protect\cite{cite1257,cite1258}}.
\item\relax
\flmRefsHyperref[eczindexfamilyrel]{code:ecoc}{Error-correcting output code (ECOC)}\item\relax
\flmRefsHyperref[eczindexfamilyrel]{code:pentacode}{\((5,40,4)_{\mathbb{Z}_4}\) Pentacode}\item\relax
\flmRefsHyperref[eczindexfamilyrel]{code:q-ary_linear_over_zq}{Linear code over \(\mathbb{Z}_q\)}\item\relax
\flmRefsHyperref[eczindexfamilyrel]{code:upc}{Universal Product Code (UPC)} --- The last digit of a UPC barcode is a base-10 check digit computed modulo 10 from an alternating weighted sum of the preceding digits \NoCaseChange{\protect\cite{cite962}}.
\end{eczvaluelist}
\codefieldsection{Cousins}
\begin{eczvaluelist}
\item\relax
\flmRefsHyperref[eczindexfamilyrel]{code:combinatorial_design}{Combinatorial design} --- Optimal constant-weight codes over \(\mathbb{Z}_q\) can be constructed \NoCaseChange{\protect\cite{cite131}} from a generalization of combinatorial designs to \(q\)-ary alphabets \NoCaseChange{\protect\cite{cite132,cite133}}.
\item\relax
\flmRefsHyperref[eczindexfamilyrel]{code:qudits_into_qudits}{Modular-qudit code} --- Modular-qudit codes are quantum counterparts of \(q\)-ary codes over \(\mathbb{Z}_q\).
\item\relax
\flmRefsHyperref[eczindexfamilyrel]{code:julin12}{Julin-Golay code} --- Julin codes can be obtained from simple nonlinear codes over \(\mathbb{Z}_4\) using the Gray map \NoCaseChange{\protect\cite{cite373}}.
\item\relax
\flmRefsHyperref[eczindexfamilyrel]{code:q-ary_constant_weight}{Constant-weight block code} --- Optimal constant-weight codes over \(\mathbb{Z}_q\) can be constructed \NoCaseChange{\protect\cite{cite131}} from a generalization of combinatorial designs to \(q\)-ary alphabets \NoCaseChange{\protect\cite{cite132,cite133}}.
\item\relax
\flmRefsHyperref[eczindexfamilyrel]{code:q-ary_digits_into_q-ary_digits}{\(q\)-ary code} --- \(q\)-ary codes for \(q=p\) prime are \(p\)-ary codes over \(\mathbb{Z}_p \cong \mathbb{F}_p\).
\item\relax
\flmRefsHyperref[eczindexfamilyrel]{code:polyphase}{Polyphase code} --- Polyphase codes are spherical codes that can be obtained from \(q\)-ary codes over rings \(\mathbb{Z}_q\).
\end{eczvaluelist}
\eczhbkcontributors{ \eczhuVVA }
\endeczcode

\eczcode{rings_linear}{\(R\)-linear code}{}
\codefieldsection{Description}
A code of length \(n\) over a ring \(R\) is \(R\)-linear if it is a submodule of \(R^n\).
Axiomatically, one can define such a code by assuming that the message set is a module over the alphabet and that encoding functions are module homomorphisms \NoCaseChange{\protect\cite{cite45}}.

For finite commutative Frobenius rings, the standard inner product, duality, and MacWilliams identities extend from the field case \NoCaseChange{\protect\cite[{Secs. 6.4 and 6.5}]{cite1145}}.
There is a standard form for codes over finite chain rings with maximal ideals \NoCaseChange{\protect\cite[{Thm. 2.12}]{cite2443}}.

\codefieldsection{Notes}
\begin{eczvaluelist}
\item\relax See Ref. \NoCaseChange{\protect\cite{cite2444}} for an introduction.
\item\relax See book \NoCaseChange{\protect\cite{cite2443}} for an introduction.
\end{eczvaluelist}
\codefieldsection{Parents}
\begin{eczvaluelist}
\item\relax
\flmRefsHyperref[eczindexfamilyrel]{code:rings_into_rings}{Ring code}\item\relax
\flmRefsHyperref[eczindexfamilyrel]{code:group_linear}{Linear code over \(G\)} --- \(R\)-linear codes are linear over \(G=R\) since rings and submodules are Abelian groups under addition.
\end{eczvaluelist}
\codefieldsection{Children}
\begin{eczvaluelist}
\item\relax
\flmRefsHyperref[eczindexfamilyrel]{code:q-ary_linear}{Linear \(q\)-ary code} --- Linear \(q\)-ary codes are \(\mathbb{F}_q\)-linear.
\item\relax
\flmRefsHyperref[eczindexfamilyrel]{code:dual_over_rings}{Dual linear code over \(R\)}\item\relax
\flmRefsHyperref[eczindexfamilyrel]{code:q-ary_linear_over_zq}{Linear code over \(\mathbb{Z}_q\)}\end{eczvaluelist}
\eczhbkcontributors{ \eczhuVVA }
\endeczcode

\eczcode{berlekamp}{Berlekamp code}{~\NoCaseChange{\protect\cite{cite2445}\protect\cite[{Ch. 9}]{cite974}}}
\codefieldsection{Description}
A linear \(p\)-ary code (for prime \(p\)) that has Lee distance 5 and whose construction resembles that of RS codes.
It is obtained by first constructing an RS-like parity-check matrix out of a certain \flmRefsHyperref{ref33}{field extension} of \(\mathbb{F}_p\) and then taking the \flmRefsHyperref{ref33}{subfield} subcode of the corresponding code; see \NoCaseChange{\protect\cite[{Ch. 10.6}]{cite195}}.

\codefieldsection{Parents}
\begin{eczvaluelist}
\item\relax
\flmRefsHyperref[eczindexfamilyrel]{code:q-ary_linear_over_zq}{Linear code over \(\mathbb{Z}_q\)}\item\relax
\flmRefsHyperref[eczindexfamilyrel]{code:q-ary_linear}{Linear \(q\)-ary code}\item\relax
\flmRefsHyperref[eczindexfamilyrel]{code:constacyclic}{Constacyclic code} --- Berlekamp codes are negacyclic \NoCaseChange{\protect\cite[{Ch. 9}]{cite974}}.
\end{eczvaluelist}
\codefieldsection{Cousins}
\begin{eczvaluelist}
\item\relax
\flmRefsHyperref[eczindexfamilyrel]{code:alternant}{Alternant code} --- Berlekamp codes reduce to narrow-sense alternant codes for \(p=2\) \NoCaseChange{\protect\cite[{Ch. 10.6}]{cite195}}.
\item\relax
\flmRefsHyperref[eczindexfamilyrel]{code:reed_solomon}{Reed-Solomon (RS) code} --- Berlekamp codes are obtained by first constructing an RS-like parity-check matrix out of a certain \flmRefsHyperref{ref33}{field extension} of \(\mathbb{F}_p\) and then taking the \flmRefsHyperref{ref33}{subfield} subcode of the corresponding code; see \NoCaseChange{\protect\cite[{Ch. 10.6}]{cite195}}.
\end{eczvaluelist}
\eczhbkcontributors{ \eczhuVVA }
\endeczcode

\eczcode{dual_over_z4}{Dual code over \(\mathbb{Z}_4\)}{}
\codefieldsection{Description}
For any linear code \(C\) over \(\mathbb{Z}_4\), the dual code is the set of quaternary strings that are orthogonal to the codewords of \(C\) under the standard inner product modulo \(4\).

\codefieldsection{Protection}
The dual of a code \(C=(n,4^{k_1} 2^{k_2})_{\mathbb{Z}_4}\) is \(C^{\perp} = (n,4^{n-k_1-k_2} 2^{k_2})_{\mathbb{Z}_4}\), whose generator matrix can be written in terms of the standard form of \(C\) \NoCaseChange{\protect\cite[{Sec. 6.2}]{cite1145}\protect\cite[{Prop. 1.2}]{cite123}}.

\codefieldsection{Parents}
\begin{eczvaluelist}
\item\relax
\flmRefsHyperref[eczindexfamilyrel]{code:quaternary_over_z4}{Linear code over \(\mathbb{Z}_4\)}\item\relax
\flmRefsHyperref[eczindexfamilyrel]{code:dual_over_zq}{Dual code over \(\mathbb{Z}_q\)}\end{eczvaluelist}
\codefieldsection{Child}
\begin{eczvaluelist}
\item\relax
\flmRefsHyperref[eczindexfamilyrel]{code:self_dual_over_z4}{Self-dual code over \(\mathbb{Z}_4\)}\end{eczvaluelist}
\codefieldsection{Cousin}
\begin{eczvaluelist}
\item\relax
\flmRefsHyperref[eczindexfamilyrel]{code:quaternary_reed_muller}{Quaternary RM (QRM) code} --- The dual of a QRM\((r,m)\) code is the QRM\((m-r-1,m)\) code \NoCaseChange{\protect\cite[{Thm. 19}]{cite158}}.
\end{eczvaluelist}
\eczhbkcontributors{ \eczhuVVA }
\endeczcode

\eczcode{dual_over_zq}{Dual code over \(\mathbb{Z}_q\)}{}
\codefieldsection{Description}
For any linear code \(C\) over \(\mathbb{Z}_q\), the dual code is the set of \(q\)-ary strings over \(\mathbb{Z}_q\) that are orthogonal to the codewords of \(C\) under the standard inner product modulo \(q\).

The dual code over \(\mathbb{Z}_q\) is
\flmMathEnvironment{align}{}{
  C^\perp = \{ y\in (\mathbb{Z}_q)^{n} ~|~ x \cdot y=0 \mod q \forall x\in C\}~.
}

\codefieldsection{Parents}
\begin{eczvaluelist}
\item\relax
\flmRefsHyperref[eczindexfamilyrel]{code:q-ary_linear_over_zq}{Linear code over \(\mathbb{Z}_q\)}\item\relax
\flmRefsHyperref[eczindexfamilyrel]{code:dual_over_rings}{Dual linear code over \(R\)}\end{eczvaluelist}
\codefieldsection{Children}
\begin{eczvaluelist}
\item\relax
\flmRefsHyperref[eczindexfamilyrel]{code:self_dual_over_zq}{Self-dual code over \(\mathbb{Z}_q\)}\item\relax
\flmRefsHyperref[eczindexfamilyrel]{code:dual_over_z4}{Dual code over \(\mathbb{Z}_4\)}\end{eczvaluelist}
\eczhbkcontributors{ \eczhuVVA }
\endeczcode

\eczcode{dual_over_rings}{Dual linear code over \(R\)}{}
\codefieldsection{Description}
For any linear code \(C\) over a ring \(R\), the dual code is the set of strings that are orthogonal to the codewords of \(C\) under some inner product.

\codefieldsection{Protection}
For linear codes over a finite commutative Frobenius ring \(R\), the dual code is linear and satisfies
\(|C||C^\perp|=|R|^n\) \NoCaseChange{\protect\cite[{Secs. 6.4.1 and 6.5}]{cite1145}}.

\codefieldsection{Parent}
\begin{eczvaluelist}
\item\relax
\flmRefsHyperref[eczindexfamilyrel]{code:rings_linear}{\(R\)-linear code}\end{eczvaluelist}
\codefieldsection{Children}
\begin{eczvaluelist}
\item\relax
\flmRefsHyperref[eczindexfamilyrel]{code:dual}{Dual linear code}\item\relax
\flmRefsHyperref[eczindexfamilyrel]{code:self_dual_over_rings}{Self-dual code over \(R\)}\item\relax
\flmRefsHyperref[eczindexfamilyrel]{code:dual_over_zq}{Dual code over \(\mathbb{Z}_q\)}\end{eczvaluelist}
\codefieldsection{Cousin}
\begin{eczvaluelist}
\item\relax
\flmRefsHyperref[eczindexfamilyrel]{code:dual_additive}{Dual additive code} --- Dual additive codes are additive analogues of dual linear codes over rings.
\end{eczvaluelist}
\eczhbkcontributors{ \eczhuVVA }
\endeczcode

\eczcode{quaternary_golay}{Extended quaternary Golay code}{~\NoCaseChange{\protect\cite{cite112}}}
\codefieldsection{Alternative Names}
\begin{eczvaluelist}
\item\relax \(\mathbb{Z}_4\) Golay code
\end{eczvaluelist}
\eczhIndexCodeAliasName{quaternary_golay}{\(\mathbb{Z}_4\) Golay code}
\codefieldsection{Description}
An extended quadratic residue quaternary linear \((24,4^{12},12)_{\mathbb{Z}_4}\) code that is a quaternary version of the Golay code.
The code has Lee distance 12, Hamming distance 8, and Euclidean distance 16 \NoCaseChange{\protect\cite{cite112}}. 
Under the \flmTerm{term}{ref81}{}{Gray map}, the code yields a formally self-dual binary \((48,2^{24},12)\) code whose distance distribution is the \flmRefsHyperref{ref113}{MacWilliams transform} of the distance distribution of its dual code \NoCaseChange{\protect\cite{cite112}}.
The lattice \(A(C)/2\) obtained from the code is the Leech lattice \NoCaseChange{\protect\cite{cite112}}.

\codefieldsection{Parent}
\begin{eczvaluelist}
\item\relax
\flmRefsHyperref[eczindexfamilyrel]{code:self_dual_over_z4}{Self-dual code over \(\mathbb{Z}_4\)} --- The extended quaternary Golay code is an extremal Type II self-dual code over \(\mathbb{Z}_4\) by virtue of its parameters \NoCaseChange{\protect\cite{cite2211}}.
\end{eczvaluelist}
\codefieldsection{Cousins}
\begin{eczvaluelist}
\item\relax
\flmRefsHyperref[eczindexfamilyrel]{code:extended_golay}{\([24, 12, 8]\) Extended Golay code} --- Codewords of the extended quaternary Golay code with entries 0 and 2 are of the form \(2c\), where \(c\) is a codeword of the extended Golay code. Its mod-two reduction (mapping \(0,1,2,3\) to \(0,1,0,1\)) also yields the extended Golay code \NoCaseChange{\protect\cite{cite112,cite1198}}. The quaternary Golay code can be constructed from the extended Golay code by Hensel lifting to \(\mathbb{Z}_4\) \NoCaseChange{\protect\cite{cite112,cite1199}\protect\cite[{3rd Ed., pg. xxxiii}]{cite39}}.
\item\relax
\flmRefsHyperref[eczindexfamilyrel]{code:combinatorial_design}{Combinatorial design} --- Supports of codewords of any fixed symmetrized type of the extended quaternary Golay code form a 5-design \NoCaseChange{\protect\cite{cite141,cite142,cite143}}.
\item\relax
\flmRefsHyperref[eczindexfamilyrel]{code:leech}{\(\Lambda_{24}\) Leech lattice} --- The Leech lattice can be constructed from the extended quaternary Golay code via \flmTerm{term}{ref114}{}{Construction \(A_4\)} \NoCaseChange{\protect\cite[{3rd Ed., pg. xxxiii}]{cite39}} (see also \NoCaseChange{\protect\cite{cite1659,cite112,cite1198}}).
\end{eczvaluelist}
\eczhbkcontributors{ \eczhuVVA }
\endeczcode

\eczcode{harada_kitazume}{Harada-Kitazume code}{~\NoCaseChange{\protect\cite{cite2037}}}
\codefieldsection{Description}
A member of a family of extremal Type II self-dual codes over \(\mathbb{Z}_4\) that yield all Niemeier lattices via \flmTerm{term}{ref114}{}{Construction \(A_4\)}.

\codefieldsection{Parent}
\begin{eczvaluelist}
\item\relax
\flmRefsHyperref[eczindexfamilyrel]{code:self_dual_over_z4}{Self-dual code over \(\mathbb{Z}_4\)} --- Harada-Kitazume codes are extremal Type II self-dual codes over \(\mathbb{Z}_4\) \NoCaseChange{\protect\cite{cite2037}}.
\end{eczvaluelist}
\codefieldsection{Cousins}
\begin{eczvaluelist}
\item\relax
\flmRefsHyperref[eczindexfamilyrel]{code:niemeier}{Niemeier lattice} --- Niemeier lattices can be constructed from quaternary codes over \(\mathbb{Z}_4\) via \flmTerm{term}{ref114}{}{Construction \(A_4\)} \NoCaseChange{\protect\cite{cite2253}}. These codes are the Harada-Kitazume codes \NoCaseChange{\protect\cite{cite2037}}.
\item\relax
\flmRefsHyperref[eczindexfamilyrel]{code:self_dual}{Self-dual linear code} --- Codewords consisting of 0 and 2 of nine Harada-Kitazume codes are of the form \(2c\), where \(c\) is a codeword of one of the nine corresponding \([24,12]\) doubly even self-dual codes \NoCaseChange{\protect\cite{cite2037}}.
\end{eczvaluelist}
\eczhbkcontributors{ \eczhuVVA }
\endeczcode

\eczcode{klemm}{Klemm code}{~\NoCaseChange{\protect\cite{cite2446}}}
\codefieldsection{Description}
A member of a family of self-dual linear \((4m,4^1 2^{4m-2})_{\mathbb{Z}_4}\) codes.
Its generator matrix consists of a sum of the generator matrix of the repetition code and twice the generator matrix of the SPC code \NoCaseChange{\protect\cite{cite121}}.

A generator matrix for this code is
\flmMathEnvironment{align}{}{
  \begin{pmatrix}
  1 & 1 & 1 & \cdots & 1 & 1\\
  0 & 2 & 0 & \cdots & 0 & 2\\
  0 & 0 & 2 & \cdots & 0 & 2\\
  0 & 0 & 0 & \ddots & \vdots & \vdots\\
  0 & 0 & 0 & \cdots & 2 & 2
  \end{pmatrix}\,.
}

\codefieldsection{Parent}
\begin{eczvaluelist}
\item\relax
\flmRefsHyperref[eczindexfamilyrel]{code:self_dual_over_z4}{Self-dual code over \(\mathbb{Z}_4\)} --- The Klemm code is a Type IV self-dual code over \(\mathbb{Z}_4\) \NoCaseChange{\protect\cite{cite121}}.
\end{eczvaluelist}
\codefieldsection{Cousins}
\begin{eczvaluelist}
\item\relax
\flmRefsHyperref[eczindexfamilyrel]{code:hamming844}{\([8,4,4]\) extended Hamming code} --- The binary image of the \(m=1\) Klemm code under the \flmTerm{term}{ref81}{}{Gray map} is the \([8,4,4]\) extended Hamming code \NoCaseChange{\protect\cite[{Exam. 3.2}]{cite123}}.
\item\relax
\flmRefsHyperref[eczindexfamilyrel]{code:repetition}{Repetition code} --- The generator matrix of the Klemm code consists of a sum of the generator matrix of the repetition code and twice the generator matrix of the SPC code \NoCaseChange{\protect\cite{cite121}}.
\item\relax
\flmRefsHyperref[eczindexfamilyrel]{code:parity_check}{\([n,n-1,2]\) Single parity-check (SPC) code} --- The generator matrix of the Klemm code consists of a sum of the generator matrix of the repetition code and twice the generator matrix of the SPC code \NoCaseChange{\protect\cite{cite121}}.
\item\relax
\flmRefsHyperref[eczindexfamilyrel]{code:cmr}{\(C_{m,r}\) code} --- The Klemm code at \(m=8\) is the \(C_{m,r=0}\) code with parameters \([32,16,4]_{\mathbb{Z}_4}\) \NoCaseChange{\protect\cite{cite1581}}.
\item\relax
\flmRefsHyperref[eczindexfamilyrel]{code:reed_muller}{Reed-Muller (RM) code} --- The Klemm code at \(m=4\) is generated by \(\textnormal{RM}(0,4) + 2\textnormal{RM}(3,4)\) \NoCaseChange{\protect\cite{cite1581}}.
\item\relax
\flmRefsHyperref[eczindexfamilyrel]{code:zrm}{ZRM code} --- The Klemm code at \(m=1\) is the ZRM\((1,2)\) code \NoCaseChange{\protect\cite[{Exam. 4.1}]{cite123}}.
\end{eczvaluelist}
\eczhbkcontributors{ \eczhuVVA }
\endeczcode

\eczcode{quaternary_over_z4}{Linear code over \(\mathbb{Z}_4\)}{}
\codefieldsection{Description}
A code that forms a subgroup of \(\mathbb{Z}_4^n\) under addition.
More technically, linear codes over \(\mathbb{Z}_4\) are submodules of \(\mathbb{Z}_4^n\).

Linear codes over \(\mathbb{Z}_4\) can be defined in terms of a \textit{generator matrix} \(G\), whose rows generate the \(k\)-dimensional codespace.
Given a message (i.e., logical string) \(x\), the corresponding encoded codeword is \(G^T x\). 

A code's generator matrix can be reduced via coordinate permutations to its \textit{standard form} 
\flmMathEnvironment{align}{}{
  \left(\begin{array}{ccc}
  I_{k_{1}} & A & B\\
  0 & 2I_{k_{2}} & 2C
  \end{array}\right)~,
}
where \(A,C\) are binary matrices, and where \(B\) is a quaternary matrix. 
Such a code encodes \(4^{k_1} 2^{k_2}\) codewords, consisting of \(k_1\) logical quaternary digits and \(k_2\) logical bits.
It is called an \((n,4^{k_1} 2^{k_2})_{\mathbb{Z}_4}\) linear code over \(\mathbb{Z}_4\) of \textit{type} \(4^{k_1} 2^{k_2}\).
The generator matrix forms a basis when there are no logical bits, i.e., the submodule is free if and only if \(k_2=0\) \NoCaseChange{\protect\cite{cite123}}.

Two linear codes over \(\mathbb{Z}_4\) are \textit{monomially equivalent} if the codewords of one code can be mapped into those of the other under a permutation and coordinate sign flips.
The \textit{monomial automorphism group} of a linear code over \(\mathbb{Z}_4\) is the largest subgroup of coordinate permutations and sign flips that maps the code onto itself.

\codefieldsection{Protection}
Quaternary codes over \(\mathbb{Z}_4\) come with a Hamming distance, which is the minimum number of nonzero coordinates of a nonzero codeword.
They also protect from cyclic shifts in each quaternary digit, which are quantified by the Lee metric.
The Lee weight of a digit is given by \(w_L(0)=0\), \(w_L(1)=1\), \(w_L(2)=2\), and \(w_L(3)=1\), and the Lee weight of a string is the sum of the Lee weights of its digits.
The \textit{Lee distance} of a code is the minimum Lee weight of a nonzero codeword \NoCaseChange{\protect\cite[{Secs. 6.2 and 6.3}]{cite1145}}.

The Lee distance \(d_L\) between two digits \(a,b\in\mathbb{Z}_4\) governs how many cyclic shifts it takes to connect them, and is thus related to the Euclidean distance between the vectors \(i^a\) and \(i^b\) as \(2d_L(a,b) = \|i^a - i^b\|^2\).
Quaternary codes over the Lee metric are thus naturally mapped to QPSK codes.

\flmRefsHyperref{ref113}{Weight enumerators} have also been extended to the Lee metric \NoCaseChange{\protect\cite{cite2446}}.

\codefieldsection{Notes}
\begin{eczvaluelist}
\item\relax Code \flmHref{http://quantumcodes.info/Z4/}{Database}, including quasi-cyclic and quasi-twisted codes \NoCaseChange{\protect\cite{cite2447}}.
\item\relax See books \NoCaseChange{\protect\cite{cite123,cite126,cite195}} for introductions.
\end{eczvaluelist}
\codefieldsection{Parent}
\begin{eczvaluelist}
\item\relax
\flmRefsHyperref[eczindexfamilyrel]{code:q-ary_linear_over_zq}{Linear code over \(\mathbb{Z}_q\)} --- Linear codes over \(\mathbb{Z}_4\) are linear \(q\)-ary codes over \(\mathbb{Z}_q\) for \(q=4\).
\end{eczvaluelist}
\codefieldsection{Children}
\begin{eczvaluelist}
\item\relax
\flmRefsHyperref[eczindexfamilyrel]{code:dual_over_z4}{Dual code over \(\mathbb{Z}_4\)}\item\relax
\flmRefsHyperref[eczindexfamilyrel]{code:quaternary_reed_muller}{Quaternary RM (QRM) code}\item\relax
\flmRefsHyperref[eczindexfamilyrel]{code:zrm}{ZRM code}\end{eczvaluelist}
\codefieldsection{Cousins}
\begin{eczvaluelist}
\item\relax
\flmRefsHyperref[eczindexfamilyrel]{code:binary_linear}{Linear binary code} --- A linear quaternary code over \(\mathbb{Z}_4\) of length \(n\), type \(4^{k_1}2^{k_2}\), and minimum Lee weight \(d\) maps under the \flmTerm{term}{ref81}{}{Gray map} to a binary code of length \(2n\), cardinality \(2^{2k_1+k_2}\), and minimum Hamming weight \(d\) \NoCaseChange{\protect\cite[{Sec. 6.3}]{cite1145}}.
\item\relax
\flmRefsHyperref[eczindexfamilyrel]{code:qpsk}{Quadrature PSK (QPSK) modulation format} --- The Lee distance \(d_L\) between two digits \(a,b\in\mathbb{Z}_4\) governs how many cyclic shifts it takes to connect them, and is thus related to the Euclidean distance between \(i^a\) and \(i^b\) as \(2d_L(a,b) = \|i^a - i^b\|^2\).
Quaternary codes over the Lee metric are thus naturally mapped to QPSK codes.

\item\relax
\flmRefsHyperref[eczindexfamilyrel]{code:construction_a4}{Construction \(A_4\) lattice} --- Every linear code over \(\mathbb{Z}_4\) yields a lattice under \flmTerm{term}{ref114}{}{Construction \(A_4\)} \NoCaseChange{\protect\cite[{Sec. 12.5.3}]{cite126}}.
\item\relax
\flmRefsHyperref[eczindexfamilyrel]{code:melas}{\([2^m -1, 2^m - 1 - 2m, 5]\) Melas code} --- The even-weight subcode of the Melas code can be lifted to a linear code over \(\mathbb{Z}_4\) \NoCaseChange{\protect\cite{cite105}}.
\item\relax
\flmRefsHyperref[eczindexfamilyrel]{code:hergert}{Hergert code} --- Hergert codes can be seen, via the \flmTerm{term}{ref81}{}{Gray map}, as linear codes over \(\mathbb{Z}_4\) \NoCaseChange{\protect\cite{cite158,cite123}}.
\item\relax
\flmRefsHyperref[eczindexfamilyrel]{code:gray}{Gray code} --- A linear code \(C\) over \(\mathbb{Z}_4\) can be mapped, via the \flmTerm{term}{ref81}{}{Gray map}, to a binary code. The binary code is linear if and only if doubling the component-wise product of any two codewords in \(C\) yields another codeword in \(C\) \NoCaseChange{\protect\cite[{Thm. 12.2.3}]{cite126}}. More specifically, a linear quaternary code over \(\mathbb{Z}_4\) of length \(n\), type \(4^{k_1}2^{k_2}\), and minimum Lee weight \(d\) maps under the \flmTerm{term}{ref81}{}{Gray map} to a binary code of length \(2n\), cardinality \(2^{2k_1+k_2}\), and minimum Hamming weight \(d\) \NoCaseChange{\protect\cite[{Sec. 6.3}]{cite1145}}.
\item\relax
\flmRefsHyperref[eczindexfamilyrel]{code:delsarte_goethals}{Delsarte-Goethals (DG) code} --- DG codes can be seen, via the \flmTerm{term}{ref81}{}{Gray map}, as extended linear cyclic codes over \(\mathbb{Z}_4\) \NoCaseChange{\protect\cite{cite158}}.
\item\relax
\flmRefsHyperref[eczindexfamilyrel]{code:reed_muller}{Reed-Muller (RM) code} --- Binary Reed-Muller codes are images of linear quaternary codes over \(\mathbb{Z}_4\) under the Gray map \NoCaseChange{\protect\cite[{Sec. 6.3}]{cite1145}}.
\end{eczvaluelist}
\eczhbkcontributors{ \eczhuVVA }
\endeczcode

\eczcode{q-ary_linear_over_zq}{Linear code over \(\mathbb{Z}_q\)}{}
\codefieldsection{Description}
A code encoding \(K\) states (codewords) in \(n\) coordinates over the ring \(\mathbb{Z}_q\) of integers modulo \(q\) that forms an Abelian subgroup of \(\mathbb{Z}_q^n\) under addition.
Since addition of \(m\) identical elements is equivalent to multiplying by \(m\), linear codes over \(\mathbb{Z}_q\) are automatically closed under scalar multiplication.
More technically, linear codes over \(\mathbb{Z}_q\) are submodules of \(\mathbb{Z}_q^n\).

Linear codes can be defined using generator matrices \NoCaseChange{\protect\cite{cite2448,cite2449,cite169}}.
There is a standard form of the generator matrix for even \(q\) and an upper-triangular permutation-equivalent standard form when \(q\) is a prime power \NoCaseChange{\protect\cite{cite1199,cite2312}}.  

\codefieldsection{Notes}
\begin{eczvaluelist}
\item\relax See book \NoCaseChange{\protect\cite{cite2443}} for an introduction.
\end{eczvaluelist}
\codefieldsection{Parents}
\begin{eczvaluelist}
\item\relax
\flmRefsHyperref[eczindexfamilyrel]{code:q-ary_over_zq}{\(q\)-ary code over \(\mathbb{Z}_q\)}\item\relax
\flmRefsHyperref[eczindexfamilyrel]{code:rings_linear}{\(R\)-linear code}\end{eczvaluelist}
\codefieldsection{Children}
\begin{eczvaluelist}
\item\relax
\flmRefsHyperref[eczindexfamilyrel]{code:binary_linear}{Linear binary code} --- Linear binary codes are linear \(q\)-ary codes over \(\mathbb{Z}_q\) for \(q=2\).
\item\relax
\flmRefsHyperref[eczindexfamilyrel]{code:ternary_golay}{\([11,6,5]_3\) Ternary Golay code}\item\relax
\flmRefsHyperref[eczindexfamilyrel]{code:hill_56_6_36}{\([56,6,36]_3\) Hill-cap code}\item\relax
\flmRefsHyperref[eczindexfamilyrel]{code:classical_fractal_liquid}{Classical fractal liquid code}\item\relax
\flmRefsHyperref[eczindexfamilyrel]{code:berlekamp}{Berlekamp code}\item\relax
\flmRefsHyperref[eczindexfamilyrel]{code:dual_over_zq}{Dual code over \(\mathbb{Z}_q\)}\item\relax
\flmRefsHyperref[eczindexfamilyrel]{code:quaternary_over_z4}{Linear code over \(\mathbb{Z}_4\)} --- Linear codes over \(\mathbb{Z}_4\) are linear \(q\)-ary codes over \(\mathbb{Z}_q\) for \(q=4\).
\end{eczvaluelist}
\codefieldsection{Cousins}
\begin{eczvaluelist}
\item\relax
\flmRefsHyperref[eczindexfamilyrel]{code:constantin_rao}{Constantin-Rao (CR) code} --- CR codes, and their special cases the VT codes, can be converted to ternary codes with nice structure via a \textit{binary-to-ternary} map \(00\to 0\), \(11\to 0\), \(01\to 1\), and \(10\to 2\) \NoCaseChange{\protect\cite{cite1186}}.
\item\relax
\flmRefsHyperref[eczindexfamilyrel]{code:q-ary_constant_weight}{Constant-weight block code} --- Constant-weight linear codes over \(\mathbb{Z}_q\) have been classified \NoCaseChange{\protect\cite{cite169}}.
\item\relax
\flmRefsHyperref[eczindexfamilyrel]{code:q-ary_bch}{Bose–Chaudhuri–Hocquenghem (BCH) code} --- BCH codes for \(q=p\) prime field can also work as codes over the Lee metric \NoCaseChange{\protect\cite{cite1740}}.
\item\relax
\flmRefsHyperref[eczindexfamilyrel]{code:q-ary_linear}{Linear \(q\)-ary code} --- \(q\)-ary linear codes for \(q=p\) prime are linear \(p\)-ary codes over \(\mathbb{Z}_p \cong \mathbb{F}_p\).
\item\relax
\flmRefsHyperref[eczindexfamilyrel]{code:qudit_css}{Modular-qudit CSS code} --- The modular-qudit CSS construction uses two related \(q\)-ary linear codes over \(\mathbb{Z}_q\), \(C_X\) and \(C_Z\).
\item\relax
\flmRefsHyperref[eczindexfamilyrel]{code:qudit_stabilizer}{Modular-qudit stabilizer code} --- Modular-qudit stabilizer codes are the closest quantum analogues of additive codes over \(\mathbb{Z}_q\) because addition in the ring corresponds to multiplication of stabilizers in the quantum case.
\end{eczvaluelist}
\eczhbkcontributors{ \eczhuVVA }
\endeczcode

\eczcode{octacode}{Octacode}{~\NoCaseChange{\protect\cite{cite2450,cite39,cite2023}}}
\codefieldsection{Description}
The unique self-dual linear \((8,4^4,6)_{\mathbb{Z}_4}\) code of Euclidean distance 8.
Its shortened version is called the \((7,4^3,6)_{\mathbb{Z}_4}\) \textit{heptacode}.

A generator matrix for this code is
\flmMathEnvironment{align}{}{
  \left(
  \begin{array}{cccccccc}
  1 & 1 & 2 & 1 & 3 & 0 & 0 & 0 \\
  1 & 0 & 1 & 2 & 1 & 3 & 0 & 0 \\
  1 & 0 & 0 & 1 & 2 & 1 & 3 & 0 \\
  1 & 0 & 0 & 0 & 1 & 2 & 1 & 3
  \end{array}
  \right)\,,
}
and a generator matrix for the heptacode is
\flmMathEnvironment{align}{}{
  \left(
  \begin{array}{ccccccc}
  1 & 1 & 3 & 2 & 1 & 0 & 0 \\
  0 & 1 & 1 & 3 & 2 & 1 & 0 \\
  0 & 0 & 1 & 1 & 3 & 2 & 1
  \end{array}
  \right)\,.
}

\codefieldsection{Parent}
\begin{eczvaluelist}
\item\relax
\flmRefsHyperref[eczindexfamilyrel]{code:self_dual_over_z4}{Self-dual code over \(\mathbb{Z}_4\)} --- The octacode is self-dual over \(\mathbb{Z}_4\).
\end{eczvaluelist}
\codefieldsection{Cousins}
\begin{eczvaluelist}
\item\relax
\flmRefsHyperref[eczindexfamilyrel]{code:cyclic}{Cyclic code} --- The heptacode is a cyclic code over \(\mathbb{Z}_4\) with generator polynomial \(x^3+3x^2+2x+3\) \NoCaseChange{\protect\cite{cite42}}.
\item\relax
\flmRefsHyperref[eczindexfamilyrel]{code:hamming743}{\([7,4,3]\) Hamming code} --- The heptacode can be obtained by Hensel-lifting the \([7,4,3]\) Hamming code to \(\mathbb{Z}_4\) \NoCaseChange{\protect\cite{cite1199,cite158}}.
\item\relax
\flmRefsHyperref[eczindexfamilyrel]{code:hamming844}{\([8,4,4]\) extended Hamming code} --- The mod-two reduction of the octacode is the \([8,4,4]\) extended Hamming code \NoCaseChange{\protect\cite{cite42}}. The octacode can be obtained by Hensel-lifting the \([8,4,4]\) extended Hamming code to \(\mathbb{Z}_4\) \NoCaseChange{\protect\cite{cite1199}}.
\item\relax
\flmRefsHyperref[eczindexfamilyrel]{code:eeight}{\(E_8\) Gosset lattice} --- The octacode yields the \(E_8\) Gosset lattice via \flmTerm{term}{ref114}{}{Construction \(A_4\)} \NoCaseChange{\protect\cite{cite2241,cite112}\protect\cite[{Exam. 12.5.13}]{cite126}}.
\item\relax
\flmRefsHyperref[eczindexfamilyrel]{code:projective}{Projective geometry code} --- Columns of the heptacode's (octacode's) generator matrix represent the seven (eight) points of a hyperoval (8-arc) in the projective Hjelmslev plane \(PHG(2,\mathbb{Z}_4)\) (\(PHG(3,\mathbb{Z}_4)\)) \NoCaseChange{\protect\cite{cite1974}\protect\cite[{Exam. 5}]{cite1147}}.
\item\relax
\flmRefsHyperref[eczindexfamilyrel]{code:simplex734}{\([7,3,4]\) simplex code} --- Codewords of the heptacode with entries 0 and 2 are of the form \(2c\), where \(c\) is a codeword of the \([7,3,4]\) simplex code \NoCaseChange{\protect\cite[{Exam. 5}]{cite1147}}.
\item\relax
\flmRefsHyperref[eczindexfamilyrel]{code:q-ary_quad_residue}{Quadratic-residue (QR) code} --- The octacode is equivalent to the length-eight quaternary extended quadratic-residue code over \(\mathbb{Z}_4\) \NoCaseChange{\protect\cite{cite112}}.
\item\relax
\flmRefsHyperref[eczindexfamilyrel]{code:leech}{\(\Lambda_{24}\) Leech lattice} --- The Leech lattice can be constructed via the Turyn construction and the holy construction using the octacode as the glue code; one of these constructions uses eight copies of the \(D_3\) fcc lattice \NoCaseChange{\protect\cite{cite2209,cite158}}.
\item\relax
\flmRefsHyperref[eczindexfamilyrel]{code:niemeier}{Niemeier lattice} --- The octacode is the glue code for the Niemeier lattice \(A_4^6\) \NoCaseChange{\protect\cite[{3rd Ed., pg. liv}]{cite39}}.
\item\relax
\flmRefsHyperref[eczindexfamilyrel]{code:nordstrom_robinson}{\((16,256,6)\) Nordstrom-Robinson (NR) code} --- The NR code is the image of the octacode under the \flmTerm{term}{ref81}{}{Gray map} \NoCaseChange{\protect\cite{cite1146,cite123}\protect\cite[{Sec. 6.3}]{cite1145}\protect\cite[{Thm. 12}]{cite158}}. The \((14, 64, 6)\) shortened NR code is the image of the heptacode under the \flmTerm{term}{ref81}{}{Gray map} \NoCaseChange{\protect\cite[{Exam. 5}]{cite1147}}.
\end{eczvaluelist}
\eczhbkcontributors{ \eczhuVVA }
\endeczcode

\eczcode{pseudo_golay}{Pseudo Golay code}{~\NoCaseChange{\protect\cite{cite2451,cite2452,cite122}}}
\codefieldsection{Description}
Any one of 13 quaternary extremal Type II self-dual linear codes over \(\mathbb{Z}_4\) of length 24 whose mod-two reduction (mapping \(0,1,2,3\) to \(0,1,0,1\)) is the Golay code \NoCaseChange{\protect\cite[{Thm. 11}]{cite122}}.
Each code has Lee distance 12, Hamming distance 8, and Euclidean distance 16 \NoCaseChange{\protect\cite[{Thm. 9}]{cite122}}. 

\codefieldsection{Parent}
\begin{eczvaluelist}
\item\relax
\flmRefsHyperref[eczindexfamilyrel]{code:self_dual_over_z4}{Self-dual code over \(\mathbb{Z}_4\)} --- Pseudo Golay codes are extremal Type II self-dual codes over \(\mathbb{Z}_4\) \NoCaseChange{\protect\cite[{Thm. 9}]{cite122}}.
\end{eczvaluelist}
\codefieldsection{Cousins}
\begin{eczvaluelist}
\item\relax
\flmRefsHyperref[eczindexfamilyrel]{code:extended_golay}{\([24, 12, 8]\) Extended Golay code} --- The mod-two reduction (mapping \(0,1,2,3\) to \(0,1,0,1\)) of all pseudo Golay codes yields the extended Golay code; see Ref. \NoCaseChange{\protect\cite{cite1198}}.
\item\relax
\flmRefsHyperref[eczindexfamilyrel]{code:combinatorial_design}{Combinatorial design} --- Supports of codewords of any fixed symmetrized type of pseudo Golay codes form a 5-design \NoCaseChange{\protect\cite{cite141,cite142,cite143}}.
\item\relax
\flmRefsHyperref[eczindexfamilyrel]{code:leech}{\(\Lambda_{24}\) Leech lattice} --- The Leech lattice can be constructed from pseudo Golay codes via \flmTerm{term}{ref114}{}{Construction \(A_4\)} \NoCaseChange{\protect\cite{cite122,cite1198}}.
\end{eczvaluelist}
\eczhbkcontributors{ \eczhuVVA }
\endeczcode

\eczcode{quaternary_reed_muller}{Quaternary RM (QRM) code}{~\NoCaseChange{\protect\cite{cite158}}}
\codefieldsection{Description}
A quaternary linear code over \(\mathbb{Z}_4\) whose binary mod-two reduction is an RM code.
This code family includes the quaternary codes whose Gray images are the binary Kerdock and Preparata codes.
The code is usually noted as QRM\((r,m)\), with its mod-two reduction yielding the RM code RM\((r,m)\) \NoCaseChange{\protect\cite[{Thm. 19}]{cite158}}.

\codefieldsection{Decoding}
\begin{eczvaluelist}
\item\relax QRM codes whose Gray images are Preparata codes can be decoded using a syndrome-calculation-based algorithm to correct all error patterns of Lee weight at most 2 and detect all (or, for some constructions, a subset of) error patterns of Lee weight 3 or 4 \NoCaseChange{\protect\cite{cite158}}.
\end{eczvaluelist}
\codefieldsection{Parent}
\begin{eczvaluelist}
\item\relax
\flmRefsHyperref[eczindexfamilyrel]{code:quaternary_over_z4}{Linear code over \(\mathbb{Z}_4\)}\end{eczvaluelist}
\codefieldsection{Cousins}
\begin{eczvaluelist}
\item\relax
\flmRefsHyperref[eczindexfamilyrel]{code:reed_muller}{Reed-Muller (RM) code} --- The mod-two reduction of the QRM\((r,m)\) code is the RM\((r,m)\) code \NoCaseChange{\protect\cite[{Thm. 19}]{cite158}}.
\item\relax
\flmRefsHyperref[eczindexfamilyrel]{code:dual_over_z4}{Dual code over \(\mathbb{Z}_4\)} --- The dual of a QRM\((r,m)\) code is the QRM\((m-r-1,m)\) code \NoCaseChange{\protect\cite[{Thm. 19}]{cite158}}.
\item\relax
\flmRefsHyperref[eczindexfamilyrel]{code:preparata}{Preparata code} --- The binary Preparata code is the Gray-map image of the quaternary code QRM\((m-2,m)\) \NoCaseChange{\protect\cite[{Thm. 19}]{cite158}}.
\item\relax
\flmRefsHyperref[eczindexfamilyrel]{code:kerdock}{Kerdock code} --- The binary Kerdock code is the Gray-map image of the quaternary code QRM\((1,m)\), an extended cyclic code over \(\mathbb{Z}_4\) \NoCaseChange{\protect\cite[{Thm. 19}]{cite158}} (see also Ref. \NoCaseChange{\protect\cite{cite1382}}).
\end{eczvaluelist}
\eczhbkcontributors{ \eczhuVVA }
\endeczcode

\eczcode{rings_into_rings}{Ring code}{}

\codefieldsection{Kingdom root code}
for the \flmRefsHyperref{kingdom:rings_into_rings}{Ring Kingdom}.
\codefieldsection{Description}
Encodes \(K\) states (codewords) in \(n\) coordinates over a finite ring \(R\).
\codefieldsection{Parents}
\begin{eczvaluelist}
\item\relax
\flmRefsHyperref[eczindexfamilyrel]{code:block}{Block code} --- Ring codes are block codes with \(\Sigma=R\).
\item\relax
\flmRefsHyperref[eczindexfamilyrel]{code:ecc_finite}{Finite-dimensional error-correcting code (ECC)}\item\relax
\flmRefsHyperref[eczindexfamilyrel]{code:group_classical}{Group-alphabet code} --- A ring \(R\) is an Abelian group under addition.
\end{eczvaluelist}
\codefieldsection{Children}
\begin{eczvaluelist}
\item\relax
\flmRefsHyperref[eczindexfamilyrel]{code:q-ary_digits_into_q-ary_digits}{\(q\)-ary code} --- Galois fields are rings. Codes over algebraic number fields have also been studied \NoCaseChange{\protect\cite{cite1686,cite1687,cite1688}}.
\item\relax
\flmRefsHyperref[eczindexfamilyrel]{code:q-ary_over_zq}{\(q\)-ary code over \(\mathbb{Z}_q\)}\item\relax
\flmRefsHyperref[eczindexfamilyrel]{code:rings_linear}{\(R\)-linear code}\end{eczvaluelist}
\eczhbkcontributors{ \eczhuVVA }
\endeczcode

\eczcode{self_dual_over_z4}{Self-dual code over \(\mathbb{Z}_4\)}{}
\codefieldsection{Description}
A linear code \(C\) over \(\mathbb{Z}_4\) that is equal to its dual, \(C^\perp = C\), where the dual is defined with respect to the standard inner product.
The code contains \(2^n\) codewords \NoCaseChange{\protect\cite[{Corr. 1.3}]{cite123}}.

\codefieldsection{Protection}
Extremal Type-II self-dual codes over \(\mathbb{Z}_4\) have been classified for \(n\leq 16\) \NoCaseChange{\protect\cite{cite2453,cite2454}}, and there are 4744 such codes at \(n=24\) \NoCaseChange{\protect\cite{cite2211}}.

\codefieldsection{Parents}
\begin{eczvaluelist}
\item\relax
\flmRefsHyperref[eczindexfamilyrel]{code:dual_over_z4}{Dual code over \(\mathbb{Z}_4\)}\item\relax
\flmRefsHyperref[eczindexfamilyrel]{code:self_dual_over_zq}{Self-dual code over \(\mathbb{Z}_q\)}\end{eczvaluelist}
\codefieldsection{Children}
\begin{eczvaluelist}
\item\relax
\flmRefsHyperref[eczindexfamilyrel]{code:cmr}{\(C_{m,r}\) code} --- The \(C_{m,r}\) code is a Type IV self-dual code over \(\mathbb{Z}_4\) \NoCaseChange{\protect\cite{cite121}}.
\item\relax
\flmRefsHyperref[eczindexfamilyrel]{code:harada_kitazume}{Harada-Kitazume code} --- Harada-Kitazume codes are extremal Type II self-dual codes over \(\mathbb{Z}_4\) \NoCaseChange{\protect\cite{cite2037}}.
\item\relax
\flmRefsHyperref[eczindexfamilyrel]{code:klemm}{Klemm code} --- The Klemm code is a Type IV self-dual code over \(\mathbb{Z}_4\) \NoCaseChange{\protect\cite{cite121}}.
\item\relax
\flmRefsHyperref[eczindexfamilyrel]{code:octacode}{Octacode} --- The octacode is self-dual over \(\mathbb{Z}_4\).
\item\relax
\flmRefsHyperref[eczindexfamilyrel]{code:pseudo_golay}{Pseudo Golay code} --- Pseudo Golay codes are extremal Type II self-dual codes over \(\mathbb{Z}_4\) \NoCaseChange{\protect\cite[{Thm. 9}]{cite122}}.
\item\relax
\flmRefsHyperref[eczindexfamilyrel]{code:quaternary_golay}{Extended quaternary Golay code} --- The extended quaternary Golay code is an extremal Type II self-dual code over \(\mathbb{Z}_4\) by virtue of its parameters \NoCaseChange{\protect\cite{cite2211}}.
\end{eczvaluelist}
\codefieldsection{Cousins}
\begin{eczvaluelist}
\item\relax
\flmRefsHyperref[eczindexfamilyrel]{code:gray}{Gray code} --- Under the \flmTerm{term}{ref81}{}{Gray map}, any self-dual code over \(\mathbb{Z}_4\) maps to a formally self-dual binary code \NoCaseChange{\protect\cite{cite112}}.
\item\relax
\flmRefsHyperref[eczindexfamilyrel]{code:self_dual}{Self-dual linear code} --- Under the \flmTerm{term}{ref81}{}{Gray map}, any self-dual code over \(\mathbb{Z}_4\) maps to a formally self-dual binary code \NoCaseChange{\protect\cite{cite112}}.
\item\relax
\flmRefsHyperref[eczindexfamilyrel]{code:leech}{\(\Lambda_{24}\) Leech lattice} --- Each 4-frame of the Leech lattice corresponds to an extremal Type II self-dual code over \(\mathbb{Z}_4\) \NoCaseChange{\protect\cite{cite2211}}.
\end{eczvaluelist}
\eczhbkcontributors{ \eczhuVVA }
\endeczcode

\eczcode{self_dual_over_zq}{Self-dual code over \(\mathbb{Z}_q\)}{}
\codefieldsection{Description}
A linear code \(C\) over \(\mathbb{Z}_q\) that is equal to its dual, \(C^\perp = C\), where the dual is defined with respect to the standard inner product.

\codefieldsection{Notes}
\begin{eczvaluelist}
\item\relax See books \NoCaseChange{\protect\cite{cite42}} for more on self-dual codes over \(\mathbb{Z}_q\).
\item\relax See \flmHref{https://www.math.is.tohoku.ac.jp/~munemasa/selfdualcodes.htm}{Database of self-dual codes} by M. Harada and A. Munemasa for a database of self-dual codes over \(\mathbb{Z}_{4}\), \(\mathbb{Z}_{6}\), \(\mathbb{Z}_{8}\), \(\mathbb{Z}_{9}\), and \(\mathbb{Z}_{10}\).
\end{eczvaluelist}
\codefieldsection{Parents}
\begin{eczvaluelist}
\item\relax
\flmRefsHyperref[eczindexfamilyrel]{code:dual_over_zq}{Dual code over \(\mathbb{Z}_q\)}\item\relax
\flmRefsHyperref[eczindexfamilyrel]{code:self_dual_over_rings}{Self-dual code over \(R\)}\end{eczvaluelist}
\codefieldsection{Children}
\begin{eczvaluelist}
\item\relax
\flmRefsHyperref[eczindexfamilyrel]{code:pless_symmetry}{\([2q+2,q+1]_3\) Pless symmetry code}\item\relax
\flmRefsHyperref[eczindexfamilyrel]{code:tetracode}{\([4,2,3]_3\) Tetracode}\item\relax
\flmRefsHyperref[eczindexfamilyrel]{code:self_dual_z6}{\([4,2,2]_{\mathbb{Z}_6}\) senary code}\item\relax
\flmRefsHyperref[eczindexfamilyrel]{code:self_dual_over_z4}{Self-dual code over \(\mathbb{Z}_4\)}\end{eczvaluelist}
\codefieldsection{Cousins}
\begin{eczvaluelist}
\item\relax
\flmRefsHyperref[eczindexfamilyrel]{code:niemeier}{Niemeier lattice} --- Extremal Type II self-dual codes of length 24 over \(\mathbb{Z}_6\) yield Niemeier lattices \NoCaseChange{\protect\cite{cite2285}}.
\item\relax
\flmRefsHyperref[eczindexfamilyrel]{code:self_dual_lattice}{Unimodular lattice} --- There are parallels between self-dual codes over \(\mathbb{Z}_{2k}\) and even unimodular lattices \NoCaseChange{\protect\cite{cite1581,cite2312}}. Type I (type II) codes over \(\mathbb{Z}_4\) yield type I (type II) lattices under \flmTerm{term}{ref114}{}{Construction \(A_4\)} \NoCaseChange{\protect\cite[{Thm. 12.5.12}]{cite126}}.
\item\relax
\flmRefsHyperref[eczindexfamilyrel]{code:ternary_golay}{\([11,6,5]_3\) Ternary Golay code} --- The extended ternary Golay code is self-dual \NoCaseChange{\protect\cite[{Rem. 4.2.6}]{cite40}}.
\end{eczvaluelist}
\eczhbkcontributors{ \eczhuVVA }
\endeczcode

\eczcode{self_dual_over_rings}{Self-dual code over \(R\)}{}
\codefieldsection{Description}
An additive linear code \(C\) over a ring \(R\) that is equal to its dual, \(C^\perp = C\), where the dual is defined with respect to some inner product.

For \(m=2^{s} p_{1}^{n_{1}} \cdots p_{r}^{n_{r}}\) with distinct odd primes \(p_i\), the group ring \(\mathbb{Z}_m G\) contains a self-dual group code if and only if all exponents \(n_i\) are even and either \(s\) or \(|G|\) is even \NoCaseChange{\protect\cite[{Thm. 16.12.6}]{cite196}}.

\codefieldsection{Parent}
\begin{eczvaluelist}
\item\relax
\flmRefsHyperref[eczindexfamilyrel]{code:dual_over_rings}{Dual linear code over \(R\)}\end{eczvaluelist}
\codefieldsection{Children}
\begin{eczvaluelist}
\item\relax
\flmRefsHyperref[eczindexfamilyrel]{code:self_dual}{Self-dual linear code} --- Self-dual linear codes are over fields, which are also rings.
\item\relax
\flmRefsHyperref[eczindexfamilyrel]{code:self_dual_over_zq}{Self-dual code over \(\mathbb{Z}_q\)}\end{eczvaluelist}
\eczhbkcontributors{ \eczhuVVA }
\endeczcode

\eczcode{upc}{Universal Product Code (UPC)}{~\NoCaseChange{\protect\cite{cite2455}}}
\codefieldsection{Description}
A checksum code used worldwide, in the form of a barcode, to identify products in stores.

\codefieldsection{Protection}
The last digit of a UPC barcode is a base-10 check digit computed modulo 10 from an alternating weighted sum of the preceding digits \NoCaseChange{\protect\cite{cite962}}.
\codefieldsection{Notes}
\begin{eczvaluelist}
\item\relax See Ref. \NoCaseChange{\protect\cite{cite962}} for a history.
\end{eczvaluelist}
\codefieldsection{Parents}
\begin{eczvaluelist}
\item\relax
\flmRefsHyperref[eczindexfamilyrel]{code:q-ary_over_zq}{\(q\)-ary code over \(\mathbb{Z}_q\)} --- The last digit of a UPC barcode is a base-10 check digit computed modulo 10 from an alternating weighted sum of the preceding digits \NoCaseChange{\protect\cite{cite962}}.
\item\relax
\flmRefsHyperref[eczindexfamilyrel]{code:checksum}{Checksum code} --- The last digit of a UPC barcode is a base-10 check digit computed modulo 10 from an alternating weighted sum of the preceding digits \NoCaseChange{\protect\cite{cite962}}.
\end{eczvaluelist}
\eczhbkcontributors{ \eczhuVVA }
\endeczcode

\eczcode{zrm}{ZRM code}{~\NoCaseChange{\protect\cite{cite158}}}
\codefieldsection{Description}
A quaternary linear code over \(\mathbb{Z}_4\) whose binary image under the \flmTerm{term}{ref81}{}{Gray map} is an RM code.
The code is usually denoted as ZRM\((r,m-1)\), with its image under the \flmTerm{term}{ref81}{}{Gray map} being the RM code RM\((r,m)\) \NoCaseChange{\protect\cite[{Thm. 7}]{cite158}}.
The code is generated by \(\textnormal{RM}(r-1,m-1) + 2\textnormal{RM}(r,m-1)\) \NoCaseChange{\protect\cite[{Thm. 7}]{cite158}}.

\codefieldsection{Parent}
\begin{eczvaluelist}
\item\relax
\flmRefsHyperref[eczindexfamilyrel]{code:quaternary_over_z4}{Linear code over \(\mathbb{Z}_4\)}\end{eczvaluelist}
\codefieldsection{Cousins}
\begin{eczvaluelist}
\item\relax
\flmRefsHyperref[eczindexfamilyrel]{code:reed_muller}{Reed-Muller (RM) code} --- The ZRM code is generated by \(\textnormal{RM}(r-1,m-1) + 2\textnormal{RM}(r,m-1)\) \NoCaseChange{\protect\cite[{Thm. 7}]{cite158}}. 
The image of the ZRM\((r,m-1)\) code under the \flmTerm{term}{ref81}{}{Gray map} is the RM\((r,m)\) code \NoCaseChange{\protect\cite[{Thm. 7}]{cite158}}.

\item\relax
\flmRefsHyperref[eczindexfamilyrel]{code:gray}{Gray code} --- The image of the ZRM\((r,m-1)\) code under the \flmTerm{term}{ref81}{}{Gray map} is the RM\((r,m)\) code \NoCaseChange{\protect\cite[{Thm. 7}]{cite158}}.
\item\relax
\flmRefsHyperref[eczindexfamilyrel]{code:combinatorial_design}{Combinatorial design} --- The weight-four codewords of the binary image of the dual of ZRM\((1,m)\) form a Steiner system that is identical to that formed by the weight-four codewords of an extended Hamming code \NoCaseChange{\protect\cite{cite158}}.
\item\relax
\flmRefsHyperref[eczindexfamilyrel]{code:extended_hamming}{\([2^m,2^m-m-1,4]\) Extended Hamming code} --- The weight-four codewords of the binary image of the dual of ZRM\((1,m)\) form a Steiner system that is identical to that formed by the weight-four codewords of an extended Hamming code \NoCaseChange{\protect\cite{cite158}}.
\item\relax
\flmRefsHyperref[eczindexfamilyrel]{code:preparata}{Preparata code} --- Each Preparata code is contained in a corresponding dual of ZRM\((1,m)\) \NoCaseChange{\protect\cite{cite158}}.
\item\relax
\flmRefsHyperref[eczindexfamilyrel]{code:kerdock}{Kerdock code} --- Each Kerdock code is contained in a corresponding ZRM\((2,m)\) code \NoCaseChange{\protect\cite{cite158}}.
\item\relax
\flmRefsHyperref[eczindexfamilyrel]{code:klemm}{Klemm code} --- The Klemm code at \(m=1\) is the ZRM\((1,2)\) code \NoCaseChange{\protect\cite[{Exam. 4.1}]{cite123}}.
\end{eczvaluelist}
\eczhbkcontributors{ \eczhuVVA }
\endeczcode

\onecolumngrid
\clearpage

\section{Group Kingdom}

\begin{eczEpigraph}
\begin{quote}
\flmQuoteSetup{quote}%
There is nothing you can do that can't be undone.
\flmQuoteAttributed{Anthony Zee}
\end{quote}
\end{eczEpigraph}

\twocolumngrid

\eczcode{binary-ternary}{Binary-ternary mixed code}{~\NoCaseChange{\protect\cite{cite360}}}
\codefieldsection{Description}
Encodes \(K\) states (codewords) in a string of \(n_1+n_2\) coordinates, with the first \(n_1\) coordinates being binary, and the last \(n_2\) coordinates being ternary.
\codefieldsection{Protection}
See Ref. \NoCaseChange{\protect\cite{cite2456,cite2457}} for bounds on binary-ternary mixed codes.

\codefieldsection{Notes}
\begin{eczvaluelist}
\item\relax Binary-ternary mixed codes have been used in football pools, in which \(n_2\) of the matches result in either a win, a loss, or a draw, while \(n_1\) of the matches are assumed to have only a win or a loss outcome \NoCaseChange{\protect\cite{cite360}}.
\end{eczvaluelist}
\codefieldsection{Parent}
\begin{eczvaluelist}
\item\relax
\flmRefsHyperref[eczindexfamilyrel]{code:mixed}{Mixed code}\end{eczvaluelist}
\codefieldsection{Cousin}
\begin{eczvaluelist}
\item\relax
\flmRefsHyperref[eczindexfamilyrel]{code:covering}{Covering code} --- See Ref. \NoCaseChange{\protect\cite{cite1762}} for bounds on binary-ternary mixed covering codes.
\end{eczvaluelist}
\eczhbkcontributors{ \eczhuVVA }
\endeczcode

\eczcode{binary_permutation}{Code in permutations}{~\NoCaseChange{\protect\cite{cite2376,cite2458,cite2459,cite2460,cite2461}}}
\codefieldsection{Alternative Names}
\begin{eczvaluelist}
\item\relax Permutation-based code
\end{eczvaluelist}
\eczhIndexCodeAliasName{binary_permutation}{Permutation-based code}
\codefieldsection{Description}
Encodes codewords into permutations of \(n\) objects.
Permutations can be mapped to inversion vectors \NoCaseChange{\protect\cite{cite2462}} over a mixed alphabet.

\codefieldsection{Protection}
Protects against errors in the Kendall tau distance on the space of permutations.
The Kendall distance between permutations \(\sigma\) and \(\pi\) is defined as the minimum number of adjacent transpositions required to change \(\sigma\) into \(\pi\).
Various bounds have been developed \NoCaseChange{\protect\cite{cite2463,cite2462}}, including LP bounds \NoCaseChange{\protect\cite{cite2464,cite2465}}.
The mapping to inversion vectors is not distance preserving, but the \(\ell_1\) distance between inversion vectors is a lower bound on the Kendall tau distance \NoCaseChange{\protect\cite{cite2462}}.

Other distances include the Ulam distance \NoCaseChange{\protect\cite{cite2466}}.

\codefieldsection{Rate}
Asymptotically good codes in the Ulam metric exist \NoCaseChange{\protect\cite{cite2467}}.
\codefieldsection{Notes}
\begin{eczvaluelist}
\item\relax Review of parallels between linear binary codes and permutation groups \NoCaseChange{\protect\cite{cite2468}}.
\end{eczvaluelist}
\codefieldsection{Parents}
\begin{eczvaluelist}
\item\relax
\flmRefsHyperref[eczindexfamilyrel]{code:group_classical}{Group-alphabet code} --- Codes in permutations are group-alphabet codes for the symmetric group \(G=S_n\).
\item\relax
\flmRefsHyperref[eczindexfamilyrel]{code:symmetric_space}{Symmetric-space code} --- The permutation group can be viewed as a finite symmetric space \(G/H\) with \(G = S_n \times S_n\) and \(H=S_n\) \NoCaseChange{\protect\cite{cite2464,cite2465}\protect\cite[{Table 3}]{cite985}}.
\end{eczvaluelist}
\codefieldsection{Child}
\begin{eczvaluelist}
\item\relax
\flmRefsHyperref[eczindexfamilyrel]{code:rank_modulation}{Rank-modulation code}\end{eczvaluelist}
\codefieldsection{Cousins}
\begin{eczvaluelist}
\item\relax
\flmRefsHyperref[eczindexfamilyrel]{code:convolutional}{Convolutional code} --- Convolutional codes in permutations have been constructed \NoCaseChange{\protect\cite{cite1758}}.
\item\relax
\flmRefsHyperref[eczindexfamilyrel]{code:mixed}{Mixed code} --- Permutation group elements can be mapped to inversion vectors \NoCaseChange{\protect\cite{cite2462}} over a mixed alphabet.
\item\relax
\flmRefsHyperref[eczindexfamilyrel]{code:t-designs}{\(t\)-design} --- The \(GA(n,\mathbb{F}_q)\) group is a permutation 2-design for general \(q\), and a 3-design for \(q=2\). This follows from the fact that the group acts transitively on ordered pairs of distinct points, and on ordered triples of distinct points for \(q=2\) \NoCaseChange{\protect\cite{cite919}}.
\end{eczvaluelist}
\eczhbkcontributors{ Jiaxin Huang, \eczhuVVA }
\endeczcode

\eczcode{dihedral}{Dihedral code}{~\NoCaseChange{\protect\cite{cite2469,cite2470}}}
\codefieldsection{Description}
A block code that encodes \(K\) states (codewords) into an alphabet whose coordinates are elements of the dihedral group.
\codefieldsection{Protection}
Transposition errors \NoCaseChange{\protect\cite{cite246,cite2471,cite2472}}.

\codefieldsection{Notes}
\begin{eczvaluelist}
\item\relax Dihedral codes may be relevant to computation over MAC \NoCaseChange{\protect\cite{cite2473}}.
\end{eczvaluelist}
\codefieldsection{Parent}
\begin{eczvaluelist}
\item\relax
\flmRefsHyperref[eczindexfamilyrel]{code:group_classical}{Group-alphabet code} --- Dihedral codes are group-alphabet codes for the dihedral group \(G=D_n\).
\end{eczvaluelist}
\eczhbkcontributors{ Alexander Barg, \eczhuVVA }
\endeczcode

\eczcode{group_classical}{Group-alphabet code}{}

\codefieldsection{Kingdom root code}
for the \flmRefsHyperref{kingdom:group_classical}{Group Kingdom}.
\codefieldsection{Description}
Encodes \(K\) states (codewords) using symbols drawn from a group \(G\), typically with the group operation inducing a natural notion of \flmRefsHyperref{ref20}{translation} or symmetry on the alphabet. The number of codewords may be infinite for infinite groups, so various restricted versions have to be constructed in practice.

\codefieldsection{Parent}
\begin{eczvaluelist}
\item\relax
\flmRefsHyperref[eczindexfamilyrel]{code:homogeneous_space_classical}{Homogeneous-space code} --- Homogeneous spaces \(G/H\) for trivial \(H\) reduce to group spaces. A group-\(G\) space can also be thought of as a multiplicity-free homogeneous space \((G\times G) / G\) \NoCaseChange{\protect\cite[{pg. 60}]{cite2474}}.
\end{eczvaluelist}
\codefieldsection{Children}
\begin{eczvaluelist}
\item\relax
\flmRefsHyperref[eczindexfamilyrel]{code:analog}{Analog code} --- Analog code alphabets, such as \(\mathbb{R}^n\) or \(\mathbb{C}^n\), are additive groups.
\item\relax
\flmRefsHyperref[eczindexfamilyrel]{code:dihedral}{Dihedral code} --- Dihedral codes are group-alphabet codes for the dihedral group \(G=D_n\).
\item\relax
\flmRefsHyperref[eczindexfamilyrel]{code:group_linear}{Linear code over \(G\)}\item\relax
\flmRefsHyperref[eczindexfamilyrel]{code:mixed}{Mixed code}\item\relax
\flmRefsHyperref[eczindexfamilyrel]{code:binary_permutation}{Code in permutations} --- Codes in permutations are group-alphabet codes for the symmetric group \(G=S_n\).
\item\relax
\flmRefsHyperref[eczindexfamilyrel]{code:matrices_into_matrices}{Matrix-based code} --- Matrix-based code alphabets are additive groups.
\item\relax
\flmRefsHyperref[eczindexfamilyrel]{code:rings_into_rings}{Ring code} --- A ring \(R\) is an Abelian group under addition.
\end{eczvaluelist}
\codefieldsection{Cousins}
\begin{eczvaluelist}
\item\relax
\flmRefsHyperref[eczindexfamilyrel]{code:group_quantum}{Group-based quantum code} --- Group-based quantum codes are quantum counterparts of group-alphabet codes.
\item\relax
\flmRefsHyperref[eczindexfamilyrel]{code:symmetric_space}{Symmetric-space code} --- Group spaces for Lie groups \(G\) are symmetric spaces \NoCaseChange{\protect\cite[{Table 6.1}]{cite2242}}.
\item\relax
\flmRefsHyperref[eczindexfamilyrel]{code:24cell}{24-cell code} --- The 24-cell code has a quaternion-coordinate realization as the 24 elements of the binary tetrahedral group \(2T\), one of the three exceptional finite subgroups of \(SU(2)\) \NoCaseChange{\protect\cite{cite230}}.
\item\relax
\flmRefsHyperref[eczindexfamilyrel]{code:600cell}{600-cell code} --- The 600-cell code has a quaternion-coordinate realization as the 120 elements of the binary icosahedral group \(2I \cong 2.A_5\), one of the three exceptional finite subgroups of \(SU(2)\) \NoCaseChange{\protect\cite{cite230}\protect\cite[{Ch. 8, pg. 207}]{cite39}}.
\item\relax
\flmRefsHyperref[eczindexfamilyrel]{code:disphenoidal288cell}{Disphenoidal 288-cell code} --- The disphenoidal 288-cell code has a quaternion-coordinate realization as the 48 elements of the binary octahedral group \(2O\), one of the three exceptional finite subgroups of \(SU(2)\) \NoCaseChange{\protect\cite[{Sec. 8.6}]{cite230}}.
\end{eczvaluelist}
\eczhbkcontributors{ \eczhuVVA }
\endeczcode

\eczcode{group_linear}{Linear code over \(G\)}{~\NoCaseChange{\protect\cite{cite2475,cite2476,cite2403}}}
\codefieldsection{Description}
Block code that encodes \(K\) states (codewords) in \(n\) coordinates over a group \(G\) such that the codewords form a subgroup of \(G^n\). In other words, the set of codewords is closed under the componentwise group operation. This notion generalizes linear codes over finite fields, but it does not require \(G\) to be a field or even Abelian.

For codes endowed with the Hamming metric on \(G^n\), the ambient isometry group contains coordinate permutations together with coordinatewise \flmRefsHyperref{ref20}{left translations} by elements of \(G\). The \textit{automorphism group} of a particular group code is the subgroup of those isometries that preserves the code, rather than the entire ambient group.

\codefieldsection{Rate}
Linear codes over non-Abelian \(G\) cannot have better parameters than those for Abelian groups \NoCaseChange{\protect\cite{cite2477}} and are asymptotically bad \NoCaseChange{\protect\cite{cite2478,cite2479}}.
\codefieldsection{Encoding}
\begin{eczvaluelist}
\item\relax Canonical encoder \NoCaseChange{\protect\cite{cite2480}}.
\end{eczvaluelist}
\codefieldsection{Parents}
\begin{eczvaluelist}
\item\relax
\flmRefsHyperref[eczindexfamilyrel]{code:group_classical}{Group-alphabet code}\item\relax
\flmRefsHyperref[eczindexfamilyrel]{code:group_orbit}{Group-orbit code} --- The set of codewords of a linear code over \(G\) can be thought of as an orbit of a particular codeword under the group formed by the code. However, group orbit codes do not have to be linear \NoCaseChange{\protect\cite[{Remark 8.4.3}]{cite115}}.
\item\relax
\flmRefsHyperref[eczindexfamilyrel]{code:block}{Block code} --- Linear codes over \(G\) are linear block codes with \(\Sigma=G\).
\end{eczvaluelist}
\codefieldsection{Children}
\begin{eczvaluelist}
\item\relax
\flmRefsHyperref[eczindexfamilyrel]{code:points_into_lattices}{Lattice} --- Lattice-based codes are linear codes over \(G=\mathbb{R}\). Because any orthogonal matrix leaving the lattice invariant has a corresponding integer matrix (see lattice description), integer representations of groups can be used to obtain lattices \NoCaseChange{\protect\cite[{Ch. 3, Sec. 4.2}]{cite39}}.
\item\relax
\flmRefsHyperref[eczindexfamilyrel]{code:q-ary_additive}{Additive \(q\)-ary code} --- Additive \(q\)-ary codes are linear over the additive group \(G=\mathbb{F}_q\). If \(q=p^m\), they are always \(\mathbb{F}_p\)-linear, but for \(m>1\) they need not be \(\mathbb{F}_q\)-linear.
\item\relax
\flmRefsHyperref[eczindexfamilyrel]{code:rings_linear}{\(R\)-linear code} --- \(R\)-linear codes are linear over \(G=R\) since rings and submodules are Abelian groups under addition.
\end{eczvaluelist}
\codefieldsection{Cousins}
\begin{eczvaluelist}
\item\relax
\flmRefsHyperref[eczindexfamilyrel]{code:group_gkp}{Group GKP code} --- Group GKP codes are quantum analogues of linear codes over groups.
\item\relax
\flmRefsHyperref[eczindexfamilyrel]{code:slepian_group}{Slepian group-orbit code} --- Any finite-group code can be mapped to a Slepian group-orbit code by representing the group using orthogonal matrices \NoCaseChange{\protect\cite{cite2403}}.
\end{eczvaluelist}
\eczhbkcontributors{ \eczhuVVA }
\endeczcode

\eczcode{mixed}{Mixed code}{}
\codefieldsection{Alternative Names}
\begin{eczvaluelist}
\item\relax Mixed-alphabet code
\item\relax Heterogeneous code
\end{eczvaluelist}
\eczhIndexCodeAliasName{mixed}{Mixed-alphabet code}
\eczhIndexCodeAliasName{mixed}{Heterogeneous code}
\codefieldsection{Description}
Encodes \(K\) states (codewords) in a string of two or more coordinates, each of which takes values in one of two or more possible groups.
\codefieldsection{Protection}
The Hamming, Singleton, and Plotkin bounds are straightforwardly extended to mixed alphabets \NoCaseChange{\protect\cite[{Thm. 5.1}]{cite2481}}.

\codefieldsection{Parent}
\begin{eczvaluelist}
\item\relax
\flmRefsHyperref[eczindexfamilyrel]{code:group_classical}{Group-alphabet code}\end{eczvaluelist}
\codefieldsection{Child}
\begin{eczvaluelist}
\item\relax
\flmRefsHyperref[eczindexfamilyrel]{code:binary-ternary}{Binary-ternary mixed code}\end{eczvaluelist}
\codefieldsection{Cousins}
\begin{eczvaluelist}
\item\relax
\flmRefsHyperref[eczindexfamilyrel]{code:perfect}{Perfect code} --- Perfect mixed codes with minimum distance \(3\) can be constructed from partitions of vector spaces \NoCaseChange{\protect\cite[{Thm. 3.3.13}]{cite70}}.
\item\relax
\flmRefsHyperref[eczindexfamilyrel]{code:orthogonal_array}{Orthogonal array (OA)} --- Orthogonal arrays generalized to mixed alphabets are called mixed-level orthogonal arrays \NoCaseChange{\protect\cite{cite210,cite211}} (see \NoCaseChange{\protect\cite[{Ch. 9}]{cite212}}). See Ref. \NoCaseChange{\protect\cite{cite213}} for bounds on mixed orthogonal arrays.
\item\relax
\flmRefsHyperref[eczindexfamilyrel]{code:combinatorial_design}{Combinatorial design} --- Combinatorial designs have been generalized to mixed alphabets \NoCaseChange{\protect\cite{cite148}}.
\item\relax
\flmRefsHyperref[eczindexfamilyrel]{code:binary_permutation}{Code in permutations} --- Permutation group elements can be mapped to inversion vectors \NoCaseChange{\protect\cite{cite2462}} over a mixed alphabet.
\item\relax
\flmRefsHyperref[eczindexfamilyrel]{code:hybrid_qudit_oscillator}{Mixed oscillator code} --- Mixed oscillator codes are examples of quantum analogues of mixed codes.
\end{eczvaluelist}
\eczhbkcontributors{ \eczhuVVA }
\endeczcode

\eczcode{rank_modulation}{Rank-modulation code}{~\NoCaseChange{\protect\cite{cite2482,cite2462,cite1907}}}
\codefieldsection{Description}
A permutation code designed for storing information in the relative rank order of a collection of cell levels rather than in their absolute values.
Many rank-modulation codes are obtained from \(q\)-ary linear codes, such as Lee-metric codes, RS codes \NoCaseChange{\protect\cite{cite1907}}, quadratic-residue codes, and most binary codes.

\codefieldsection{Rate}
Rank modulation codes with code distance of \flmRefsHyperref{ref65}{order} \(d=\Theta(n^{1+\epsilon})\) for \(\epsilon\in[0,1]\) achieve a rate of \(1-\epsilon\) \NoCaseChange{\protect\cite{cite2483}}.
\codefieldsection{Realizations}
\begin{eczvaluelist}
\item\relax Electronic devices where charges can either increase in an individual cell or decrease in a block of adjacent cells, e.g., flash memories \NoCaseChange{\protect\cite{cite325}}.
\end{eczvaluelist}
\codefieldsection{Parent}
\begin{eczvaluelist}
\item\relax
\flmRefsHyperref[eczindexfamilyrel]{code:binary_permutation}{Code in permutations}\end{eczvaluelist}
\codefieldsection{Cousins}
\begin{eczvaluelist}
\item\relax
\flmRefsHyperref[eczindexfamilyrel]{code:gray}{Gray code} --- The rank-modulation Gray code is an extension of the original binary Gray code to a code on the permutation group \NoCaseChange{\protect\cite{cite325}}.
\item\relax
\flmRefsHyperref[eczindexfamilyrel]{code:q-ary_linear}{Linear \(q\)-ary code} --- Almost all linear \(q\)-ary codes can be converted to rank-modulation codes \NoCaseChange{\protect\cite{cite1907}}.
\end{eczvaluelist}
\eczhbkcontributors{ Jiaxin Huang, \eczhuVVA }
\endeczcode

\onecolumngrid
\clearpage

\section{Homogeneous-space Kingdom}

\begin{eczEpigraph}
\begin{quote}
\flmQuoteSetup{quote}%
On a homogeneous space, you can't complain about moving to a bad neighborhood.
\flmQuoteAttributed{Rory Whybrow}
\end{quote}
\end{eczEpigraph}

\twocolumngrid

\eczcode{alternating_projection}{Alternating projection code}{~\NoCaseChange{\protect\cite{cite2484,cite2485}}}
\codefieldsection{Description}
A Grassmannian code constructed from the alternating projection method \NoCaseChange{\protect\cite{cite2484,cite2485}}.

\codefieldsection{Parent}
\begin{eczvaluelist}
\item\relax
\flmRefsHyperref[eczindexfamilyrel]{code:grassmannian}{Grassmannian code}\end{eczvaluelist}
\eczhbkcontributors{ \eczhuVVA }
\endeczcode

\eczcode{complex_projective}{Complex projective space code}{}
\codefieldsection{Alternative Names}
\begin{eczvaluelist}
\item\relax \(\mathbb{C}P^N\) code
\item\relax Packing in \(\mathbb{C}P^N\)
\end{eczvaluelist}
\eczhIndexCodeAliasName{complex_projective}{\(\mathbb{C}P^N\) code}
\eczhIndexCodeAliasName{complex_projective}{Packing in \(\mathbb{C}P^N\)}
\codefieldsection{Description}
Encodes \(K\) states (codewords) into a complex projective space \(\mathbb{C}P^N\), the space of lines in a complex vector space. The space for \(N=2\) is called the complex projective plane.

\codefieldsection{Notes}
\begin{eczvaluelist}
\item\relax Review and tables of packings in complex projective space \NoCaseChange{\protect\cite{cite2486}}.
\end{eczvaluelist}
\codefieldsection{Parents}
\begin{eczvaluelist}
\item\relax
\flmRefsHyperref[eczindexfamilyrel]{code:2pt_homogeneous}{Two-point homogeneous-space code} --- Compact two-point homogeneous spaces \(G/H\) reduce to complex projective spaces for \(G = SU(D+1)\) and \(H = U(D)\) \NoCaseChange{\protect\cite[{Ch. 9}]{cite39}\protect\cite[{Table 1}]{cite985}}.
\item\relax
\flmRefsHyperref[eczindexfamilyrel]{code:grassmannian}{Grassmannian code} --- Complex projective spaces \(\mathbb{C}P^N\) are complex Grassmannians \(G/H\) for \(G = U(N+1)\) and \(H = U(N)\times U(1)\).
\end{eczvaluelist}
\codefieldsection{Cousins}
\begin{eczvaluelist}
\item\relax
\flmRefsHyperref[eczindexfamilyrel]{code:qecc_finite}{Finite-dimensional quantum error-correcting code} --- Pure quantum states in an \((N+1)\)-dimensional Hilbert space are parameterized by points in the complex projective space \(\mathbb{C}P^N\). As such, (classical) complex projective codes can be associated with subsets of pure quantum states.
\item\relax
\flmRefsHyperref[eczindexfamilyrel]{code:t-designs}{\(t\)-design} --- Pure quantum states in an \((N+1)\)-dimensional Hilbert space are parameterized by points in the complex projective space \(\mathbb{C}P^N\). As such, complex projective designs are designs on the space of pure quantum states \NoCaseChange{\protect\cite{cite883,cite884,cite885}}. Symmetric informationally complete quantum measurements (SIC-POVMs) \NoCaseChange{\protect\cite{cite920,cite883}} and mutually unbiased bases (MUBs) \NoCaseChange{\protect\cite{cite921,cite922,cite923,cite924,cite925,cite926}} are important examples of such designs.
\item\relax
\flmRefsHyperref[eczindexfamilyrel]{code:kerdock}{Kerdock code} --- Kerdock codes correspond to cluster states, and the corresponding Clifford-group automorphisms of this set form a particular group \NoCaseChange{\protect\cite{cite934}} that is a unitary 2-design on \(U(2^n)\) \NoCaseChange{\protect\cite{cite935}}. As such, cluster states form complex projective on 2-designs \(\mathbb{C}P^{2^n}\). These are useful in matrix-vector multiplication \NoCaseChange{\protect\cite{cite936}}.
\item\relax
\flmRefsHyperref[eczindexfamilyrel]{code:clifford_group}{Clifford group} --- Stabilizer states on \(n\) qubits form 3-designs on complex projective spaces \(\mathbb{C}P^{2^n}\) \NoCaseChange{\protect\cite{cite937}}. The \flmRefsHyperref{ref409}{Clifford group} is a unitary 2-design \NoCaseChange{\protect\cite{cite938}} and a 3-design \NoCaseChange{\protect\cite{cite940,cite941}\protect\cite[{Thm. 1.6(B)}]{cite939}\protect\cite[{pg. 191}]{cite42}} on \(U(2^n)\). The \(\llbracket 2m,2m-2,2\rrbracket \) code when \(2m\) is a multiple of four obstructs the Clifford group from being a 4-design \NoCaseChange{\protect\cite{cite801}}.
\item\relax
\flmRefsHyperref[eczindexfamilyrel]{code:hessian_polyhedron}{Hessian polyhedron code} --- The (antipodal pairs of) points of the (double) Hessian polyhedron correspond to the 27 lines on a smooth cubic surface in \(\mathbb{C}P^3\) \NoCaseChange{\protect\cite{cite117,cite118,cite119,cite120}}.
\item\relax
\flmRefsHyperref[eczindexfamilyrel]{code:hess_polytope}{\(3_{21}\) polytope code} --- Antipodal pairs of points of the \(3_{21}\) polytope code correspond to the 28 bitangent lines of a general quartic plane curve in the complex project plane \NoCaseChange{\protect\cite{cite117,cite118,cite119,cite120}}.
\item\relax
\flmRefsHyperref[eczindexfamilyrel]{code:witting_polytope}{Witting polytope code} --- Antipodal pairs of points of the Witting polytope code correspond to the 120 tritangent planes of a canonical sextic curve in \(\mathbb{C}P^3\) \NoCaseChange{\protect\cite{cite117,cite118,cite119,cite120}}.
\item\relax
\flmRefsHyperref[eczindexfamilyrel]{code:cluster_state}{Cluster-state code} --- Kerdock codes correspond to cluster states, and the corresponding Clifford-group automorphisms of this set form a particular group \NoCaseChange{\protect\cite{cite934}} that is a unitary 2-design on \(U(2^n)\) \NoCaseChange{\protect\cite{cite935}}. As such, cluster states form complex projective 2-designs on \(\mathbb{C}P^{2^n-1}\). These are useful in matrix-vector multiplication \NoCaseChange{\protect\cite{cite936}}.
\item\relax
\flmRefsHyperref[eczindexfamilyrel]{code:qubit_stabilizer}{Qubit stabilizer code} --- Stabilizer states on \(n\) qubits form complex projective 3-designs, but not 4-designs, on \(\mathbb{C}P^{2^n-1}\) \NoCaseChange{\protect\cite{cite937}}. The \flmRefsHyperref{ref409}{Clifford group} is a unitary 2-design \NoCaseChange{\protect\cite{cite938}} and a 3-design \NoCaseChange{\protect\cite{cite940,cite941}\protect\cite[{Thm. 1.6(B)}]{cite939}\protect\cite[{pg. 191}]{cite42}} on \(U(2^n)\).
\item\relax
\flmRefsHyperref[eczindexfamilyrel]{code:qudit_stabilizer}{Modular-qudit stabilizer code} --- Stabilizer states on \(n\) prime-dimensional qudits form complex projective 2-designs on \(\mathbb{C}P^{p^n-1}\) \NoCaseChange{\protect\cite{cite937}}.
\item\relax
\flmRefsHyperref[eczindexfamilyrel]{code:galois_stabilizer}{Galois-qudit stabilizer code} --- Stabilizer states on \(n\) Galois qubits form 2-designs on complex projective spaces \(\mathbb{C}P^{p^{mn}}\) \NoCaseChange{\protect\cite{cite943}}.
\end{eczvaluelist}
\eczhbkcontributors{ \eczhuVVA }
\endeczcode

\eczcode{grassmannian}{Grassmannian code}{~\NoCaseChange{\protect\cite{cite2487,cite2488,cite2344,cite2489}}}
\codefieldsection{Description}
Encodes \(K\) states (codewords) into a compact Grassmannian, which includes the real, complex, or quaternionic Grassmannians.
Points in a real (complex) Grassmannian index fixed-dimension subspaces of real (complex) vector spaces.

\codefieldsection{Protection}
Optimal Grassmannian codes are robust to coordinate erasure \NoCaseChange{\protect\cite{cite272,cite2490}}.
Various code bounds have been formulated \NoCaseChange{\protect\cite{cite2491,cite2492,cite2493,cite2494,cite2489}}.

\codefieldsection{Realizations}
\begin{eczvaluelist}
\item\relax Multiple description coding \NoCaseChange{\protect\cite{cite272}}.
\item\relax Digital fingerprinting \NoCaseChange{\protect\cite{cite273}}.
\end{eczvaluelist}
\codefieldsection{Notes}
\begin{eczvaluelist}
\item\relax Tables of real Grassmannian codes \NoCaseChange{\protect\cite{cite2495}}
\end{eczvaluelist}
\codefieldsection{Parent}
\begin{eczvaluelist}
\item\relax
\flmRefsHyperref[eczindexfamilyrel]{code:symmetric_space}{Symmetric-space code} --- Grassmannians are symmetric spaces \(G/H\) for \(G = O(p+q)\) and \(H = O(p)\times O(q)\) in the case of real Grassmannians, \(G = U(p+q)\) and \(H = U(p)\times U(q)\) in the case of complex Grassmannians, and \(G = Sp(p+q)\) and \(H = Sp(p)\times Sp(q)\) in the case of quaternionic Grassmannians.
\end{eczvaluelist}
\codefieldsection{Children}
\begin{eczvaluelist}
\item\relax
\flmRefsHyperref[eczindexfamilyrel]{code:complex_projective}{Complex projective space code} --- Complex projective spaces \(\mathbb{C}P^N\) are complex Grassmannians \(G/H\) for \(G = U(N+1)\) and \(H = U(N)\times U(1)\).
\item\relax
\flmRefsHyperref[eczindexfamilyrel]{code:real_projective}{Real projective space code} --- Real projective spaces \(\mathbb{R}P^N\) are real Grassmannians \(G/H\) for \(G = O(N+1)\) and \(H = O(N)\times O(1)\).
\item\relax
\flmRefsHyperref[eczindexfamilyrel]{code:alternating_projection}{Alternating projection code}\end{eczvaluelist}
\codefieldsection{Cousins}
\begin{eczvaluelist}
\item\relax
\flmRefsHyperref[eczindexfamilyrel]{code:t-designs}{\(t\)-design} --- Designs have been formulated on Grassmannians \NoCaseChange{\protect\cite{cite927,cite928,cite918,cite911,cite929}}.
\item\relax
\flmRefsHyperref[eczindexfamilyrel]{code:barnes_wall}{Barnes-Wall (BW) lattice} --- BW lattices support Grassmannian 6-designs \NoCaseChange{\protect\cite{cite918}}.
\item\relax
\flmRefsHyperref[eczindexfamilyrel]{code:spacetime}{Spacetime code (STC)} --- MIMO channel capacity when the channel is unknown to the sender and receiver \NoCaseChange{\protect\cite{cite2159,cite2168}} can be interpreted as a problem of placing points on the Grassmannian \NoCaseChange{\protect\cite{cite2169}}.
\item\relax
\flmRefsHyperref[eczindexfamilyrel]{code:finite_grassmann}{Constant-dimension code} --- The finite-field Grassmannian is a finite analogue of the compact Grassmannians.
\end{eczvaluelist}
\eczhbkcontributors{ \eczhuVVA }
\endeczcode

\eczcode{homogeneous_space_classical}{Homogeneous-space code}{}
\codefieldsection{Alternative Names}
\begin{eczvaluelist}
\item\relax Coset-space code
\item\relax \(G/H\) code
\end{eczvaluelist}
\eczhIndexCodeAliasName{homogeneous_space_classical}{Coset-space code}
\eczhIndexCodeAliasName{homogeneous_space_classical}{\(G/H\) code}

\codefieldsection{Kingdom root code}
for the \flmRefsHyperref{kingdom:homogeneous_space_classical}{Homogeneous-space Kingdom}.
\codefieldsection{Description}
Encodes \(K\) states (codewords) into a homogeneous (a.k.a. coset) space \(G/H\), where \(G\) is a group and \(H\) is a subgroup of \(G\). The space is labeled by cosets of \(H\) in \(G\).
Notable groups include compact groups, locally compact Abelian groups, and finite groups. 

The two-sphere is a simple example of a homogeneous space with \(G/H = SO(3)/SO(2)\). Any point on the sphere can be obtained from any other point by a proper (i.e., \(SO(3)\)) rotation, which is equivalent to saying that \(SO(3)\) acts \textit{transitively}. One can then show that any point is invariant under the subgroup of rotations around the axis parallel to the point. This means that one can associate each point with a coset of \(SO(2)\) in \(SO(3)\).

\codefieldsection{Protection}
The space of normalizable functions on a homogeneous space, \(L^2(G/H)\), carries a representation of \(G\), which can further be decomposed into irreducible representations (irreps) and their multiplicities. 
Functions on \(G/H\) that transform as an irrep of \(G\) are called \(G\)\textit{-harmonics}. 

If the decomposition of a homogeneous space into \(G\)-irreps is \textit{multiplicity free}, there are no multiplicities, and the irreps and their internal indices are sufficient to completely label a basis for the space. 
In this case, the pair \((G,H)\) is called a Gelfand pair.
For example, the integer angular momentum \(J\) and its \(z\)-axis projection \(m\) are sufficient to completely label the spherical harmonics (i.e., form a good set of quantum numbers), which in turn can be used to expand any function on the two-sphere \(S^2 = SO(3)/SO(2)\). 
Therefore, \(( SO(3), SO(2) )\) is a Gelfand pair.

Functions that correspond to projections onto irreps are called \textit{zonal spherical functions} \NoCaseChange{\protect\cite{cite2496}}; these correspond to functions on the double coset space \(H \backslash G / H\) \NoCaseChange{\protect\cite[{Eq. (2.9)}]{cite2497}} and can be obtained by averaging harmonics over \(H\). 

The matrix algebra of zonal spherical functions is non-Abelian when there are multiplicities in the irrep decomposition since irrep projections can be tensored with arbitrary matrices on the irreps' multiplicity space and still commute with the group action, but not with each other.

For multiplicity-free spaces such as symmetric spaces, the zonal spherical functions form an Abelian algebra, and the behavior of such functions can be used to obtain bounds on code parameters such as the Levenshtein bound \NoCaseChange{\protect\cite{cite2277,cite2278,cite2088,cite171,cite914}}.

\codefieldsection{Notes}
\begin{eczvaluelist}
\item\relax See Refs. \NoCaseChange{\protect\cite{cite987,cite985}\protect\cite[{Ch. 9}]{cite39}} for reviews.
\end{eczvaluelist}
\codefieldsection{Parent}
\begin{eczvaluelist}
\item\relax
\flmRefsHyperref[eczindexfamilyrel]{code:ecc}{Error-correcting code (ECC)}\end{eczvaluelist}
\codefieldsection{Children}
\begin{eczvaluelist}
\item\relax
\flmRefsHyperref[eczindexfamilyrel]{code:group_classical}{Group-alphabet code} --- Homogeneous spaces \(G/H\) for trivial \(H\) reduce to group spaces. A group-\(G\) space can also be thought of as a multiplicity-free homogeneous space \((G\times G) / G\) \NoCaseChange{\protect\cite[{pg. 60}]{cite2474}}.
\item\relax
\flmRefsHyperref[eczindexfamilyrel]{code:stiefel}{Stiefel code} --- Homogeneous spaces \(G/H\) reduce to real Stiefel manifolds for \(G = O(n)\) and \(H = O(n-k)\), to complex Stiefel manifolds for \(G = U(n)\) and \(H = U(n-k)\), and to quaternionic Stiefel manifolds for \(G = Sp(n)\) and \(H = Sp(n-k)\).
\item\relax
\flmRefsHyperref[eczindexfamilyrel]{code:symmetric_space}{Symmetric-space code} --- Infinite symmetric spaces are homogeneous spaces with an appropriately defined inversion operation. Finite symmetric spaces are defined in coding theory as spaces admitting a generously transitive group action \NoCaseChange{\protect\cite[{Def. 4.5}]{cite987}\protect\cite[{Sec. 3.4}]{cite985}}. For multiplicity-free spaces such as symmetric spaces, the zonal spherical functions form an Abelian algebra, and the behavior of such functions can be used to obtain bounds on code parameters such as the Levenshtein bound \NoCaseChange{\protect\cite{cite2277,cite2278,cite2088,cite171,cite914}}.
\item\relax
\flmRefsHyperref[eczindexfamilyrel]{code:univ_opt}{Universally optimal code}\end{eczvaluelist}
\codefieldsection{Cousins}
\begin{eczvaluelist}
\item\relax
\flmRefsHyperref[eczindexfamilyrel]{code:flag_variety}{Flag-variety code} --- The flag variety is a finite homogeneous space \NoCaseChange{\protect\cite{cite28}}.
\item\relax
\flmRefsHyperref[eczindexfamilyrel]{code:homogeneous_space_quantum}{Homogeneous-space quantum code} --- Homogeneous-space quantum codes are quantum counterparts of homogeneous-space codes.
\end{eczvaluelist}
\eczhbkcontributors{ Alexander Barg, \eczhuVVA }
\endeczcode

\eczcode{hyperbolic}{Hyperbolic sphere packing}{~\NoCaseChange{\protect\cite{cite2498,cite2286,cite2499}}}
\codefieldsection{Description}
Encodes states (codewords) as points in hyperbolic space (in particular, the hyperbolic plane for dimension \(D=2\)).
Distances are measured using the hyperbolic geodesic metric, and finite codebooks correspond to hyperbolic constellations/sphere packings.

\codefieldsection{Protection}
Designed to communicate information over channels for which a Lorentzian metric is appropriate \NoCaseChange{\protect\cite{cite2500}}.
Linear programming (LP) bounds exist for hyperbolic spaces and for closed hyperbolic surfaces \NoCaseChange{\protect\cite{cite2501,cite2502,cite2503,cite2504,cite2505}}.

\codefieldsection{Parent}
\begin{eczvaluelist}
\item\relax
\flmRefsHyperref[eczindexfamilyrel]{code:2pt_homogeneous}{Two-point homogeneous-space code} --- Hyperbolic space in \(D\) dimensions is a symmetric space \(G/H\) for \(G = SO(D,1)\) the proper Lorentz group and \(H = O(D)\). The hyperbolic plane is the case \(D=2\). In fact, hyperbolic spaces are noncompact three-point homogeneous spaces \NoCaseChange{\protect\cite[{Sec. 6.6.1.1}]{cite2242}}.
\end{eczvaluelist}
\codefieldsection{Cousins}
\begin{eczvaluelist}
\item\relax
\flmRefsHyperref[eczindexfamilyrel]{code:pam}{Pulse-amplitude modulation (PAM) format} --- Hyperbolic PAM constellations may yield improved performance over Euclidean ones \NoCaseChange{\protect\cite{cite2286}}.
\item\relax
\flmRefsHyperref[eczindexfamilyrel]{code:psk}{Phase-shift keying (PSK) modulation format} --- Hyperbolic PSK constellations may yield improved performance over Euclidean ones \NoCaseChange{\protect\cite{cite2381}}.
\item\relax
\flmRefsHyperref[eczindexfamilyrel]{code:qam}{Quadrature-amplitude modulation (QAM) format} --- Hyperbolic QAM constellations may yield improved performance over Euclidean ones \NoCaseChange{\protect\cite{cite2288}}.
\item\relax
\flmRefsHyperref[eczindexfamilyrel]{code:tesselation}{Hyperbolic tessellation code} --- Hyperbolic tessellation codes are quantum counterparts of hyperbolic sphere packings because they store information in quantum superpositions of points on the hyperbolic plane.
\end{eczvaluelist}
\eczhbkcontributors{ \eczhuVVA }
\endeczcode

\eczcode{real_projective}{Real projective space code}{}
\codefieldsection{Alternative Names}
\begin{eczvaluelist}
\item\relax \(\mathbb{R}P^N\) code
\item\relax Packing in \(\mathbb{R}P^N\)
\end{eczvaluelist}
\eczhIndexCodeAliasName{real_projective}{\(\mathbb{R}P^N\) code}
\eczhIndexCodeAliasName{real_projective}{Packing in \(\mathbb{R}P^N\)}
\codefieldsection{Description}
Encodes \(K\) states (codewords) into a real projective space \(\mathbb{R}P^N\), the space of lines in real space. The space for \(N=2\) is called the projective plane.

\codefieldsection{Protection}
A common distance-related quantity on real projective space is the absolute inner product (a.k.a. coherence).
Equiangular lines are sets of lines which have the same inner product between each other; these lines are points in real projective space.
Sets of lines minimizing coherence are called \textit{Grassmannian frames} \NoCaseChange{\protect\cite{cite272}}, and equiangular tight frames are important examples of optimal line packings \NoCaseChange{\protect\cite{cite2506}}.
The Welch bound is a lower bound on the worst-case coherence of a code \NoCaseChange{\protect\cite{cite2507}}.

\codefieldsection{Parents}
\begin{eczvaluelist}
\item\relax
\flmRefsHyperref[eczindexfamilyrel]{code:2pt_homogeneous}{Two-point homogeneous-space code} --- Compact two-point homogeneous spaces \(G/H\) reduce to real projective spaces for \(G = SO(D+1)\) and \(H = O(D)\) \NoCaseChange{\protect\cite[{Ch. 9}]{cite39}\protect\cite[{Table 1}]{cite985}}.
\item\relax
\flmRefsHyperref[eczindexfamilyrel]{code:grassmannian}{Grassmannian code} --- Real projective spaces \(\mathbb{R}P^N\) are real Grassmannians \(G/H\) for \(G = O(N+1)\) and \(H = O(N)\times O(1)\).
\end{eczvaluelist}
\codefieldsection{Cousins}
\begin{eczvaluelist}
\item\relax
\flmRefsHyperref[eczindexfamilyrel]{code:points_into_spheres}{Constant-energy spherical code} --- Real projective space can be obtained from the sphere by identifying antipodal points, i.e., \(\mathbb{R}P^N = S^N/\mathbb{Z}_2\). As such, real projective space codes are in one-to-one correspondence with antipodal spherical codes, with each antipodal pair of spherical codewords corresponding to one line in projective space.
\item\relax
\flmRefsHyperref[eczindexfamilyrel]{code:kerdock}{Kerdock code} --- The 12 sets of antipodal pairs of the 24-cell code form a sharp configuration in the projective space \(\mathbb{R}P^3\) \NoCaseChange{\protect\cite{cite119}}. This is a special case of a family of real projective plane codes, constructed using Kerdock codes \NoCaseChange{\protect\cite{cite1411}} (cf. \NoCaseChange{\protect\cite{cite917}}).
\item\relax
\flmRefsHyperref[eczindexfamilyrel]{code:esix_shell}{\(E_6\) lattice-shell code} --- The 36 antipodal pairs of the smallest \(E_6\) lattice shell form a sharp configuration and a 2-design in \(\mathbb{R}P^5\) \NoCaseChange{\protect\cite{cite119}}.
\item\relax
\flmRefsHyperref[eczindexfamilyrel]{code:polygon}{Polygon code} --- For even \(q\), the \(q/2\) sets of antipodal pairs of a \(q\)-gon form a tight design on the projective line \(\mathbb{R}P^1\) \NoCaseChange{\protect\cite{cite917}}.
\item\relax
\flmRefsHyperref[eczindexfamilyrel]{code:24cell}{24-cell code} --- The 12 antipodal pairs of the 24-cell code form a sharp configuration and a 2-design in \(\mathbb{R}P^3\) \NoCaseChange{\protect\cite{cite119}}. This is a special case of a family of real projective plane codes, constructed using Kerdock codes \NoCaseChange{\protect\cite{cite1411}} (cf. \NoCaseChange{\protect\cite{cite917}}).
\item\relax
\flmRefsHyperref[eczindexfamilyrel]{code:231_polytope}{\(2_{31}\) polytope code} --- The 63 antipodal pairs of vertices of the \(2_{31}\) polytope form a sharp configuration and a 2-design in \(\mathbb{R}P^6\) \NoCaseChange{\protect\cite{cite119}}.
\item\relax
\flmRefsHyperref[eczindexfamilyrel]{code:hess_polytope}{\(3_{21}\) polytope code} --- The 28 antipodal pairs of the \(3_{21}\) polytope code form 28 equiangular lines in \(\mathbb{R}^7\), achieving the absolute bound \NoCaseChange{\protect\cite{cite119}}. This is because the only inner product between distinct antipodal pairs is \(\pm 1/3\) \NoCaseChange{\protect\cite[{Table 1}]{cite119}}.
\item\relax
\flmRefsHyperref[eczindexfamilyrel]{code:witting_polytope}{Witting polytope code} --- The 120 antipodal pairs of the Witting polytope code form a sharp configuration and a 3-design in \(\mathbb{R}P^7\) \NoCaseChange{\protect\cite{cite119}}.
\item\relax
\flmRefsHyperref[eczindexfamilyrel]{code:mclaughlin}{McLaughlin spherical code} --- The \((23,552,1/5)\) McLaughlin spherical code yields a set of \(276\) equiangular lines in 23 dimensions \NoCaseChange{\protect\cite{cite119,cite387}}.
\end{eczvaluelist}
\eczhbkcontributors{ \eczhuVVA }
\endeczcode

\eczcode{delsarte_optimal}{Sharp configuration}{~\NoCaseChange{\protect\cite{cite171,cite914,cite119}}}
\codefieldsection{Alternative Names}
\begin{eczvaluelist}
\item\relax Delsarte code
\end{eczvaluelist}
\eczhIndexCodeAliasName{delsarte_optimal}{Delsarte code}
\codefieldsection{Description}
A code \(W\) in a compact connected two-point homogeneous space with \(m=l(W)\) distinct distances such that either \(r(W) \geq 2m-1\), or \(r(W)=2m-2>0\) and \(W\) is diametrical \NoCaseChange{\protect\cite{cite171}}.

Sharp configurations attain the Levenshtein bound \NoCaseChange{\protect\cite{cite2277,cite2278,cite2088,cite171,cite914}}.
However, not all codes that attain the Levenshtein bound are sharp configurations.
See \NoCaseChange{\protect\cite[{Table 9.2}]{cite171}} for Levenshtein-bound achieving codes on various projective spaces.

\codefieldsection{Parents}
\begin{eczvaluelist}
\item\relax
\flmRefsHyperref[eczindexfamilyrel]{code:2pt_homogeneous}{Two-point homogeneous-space code}\item\relax
\flmRefsHyperref[eczindexfamilyrel]{code:univ_opt}{Universally optimal code} --- All sharp configurations are universally optimal \NoCaseChange{\protect\cite{cite119,cite173}}, but not all universally optimal codes are sharp configurations.
\item\relax
\flmRefsHyperref[eczindexfamilyrel]{code:t-designs}{\(t\)-design} --- Sharp configurations attain a universal bound expressed in terms of the minimal distance, the number of distances between codewords, and the strength of the design formed by the codewords.
\end{eczvaluelist}
\codefieldsection{Children}
\begin{eczvaluelist}
\item\relax
\flmRefsHyperref[eczindexfamilyrel]{code:delsarte_optimal_q-ary}{\(q\)-ary sharp configuration}\item\relax
\flmRefsHyperref[eczindexfamilyrel]{code:sharp_config}{Spherical sharp configuration}\end{eczvaluelist}
\codefieldsection{Cousins}
\begin{eczvaluelist}
\item\relax
\flmRefsHyperref[eczindexfamilyrel]{code:constant_weight}{Constant-weight code} --- See \NoCaseChange{\protect\cite[{Table 8.4}]{cite171}} for lists of constant-weight binary codes in Johnson spaces that are Delsarte codes, and hence maximum codes in that metric space.
\item\relax
\flmRefsHyperref[eczindexfamilyrel]{code:esix_shell}{\(E_6\) lattice-shell code} --- The 36 antipodal pairs of the smallest \(E_6\) lattice shell form a sharp configuration in \(\mathbb{R}P^5\) \NoCaseChange{\protect\cite{cite119}}.
\item\relax
\flmRefsHyperref[eczindexfamilyrel]{code:24cell}{24-cell code} --- The 12 antipodal pairs of the 24-cell code form a sharp configuration and a 2-design in \(\mathbb{R}P^3\) \NoCaseChange{\protect\cite{cite119}}.
\item\relax
\flmRefsHyperref[eczindexfamilyrel]{code:231_polytope}{\(2_{31}\) polytope code} --- The 63 antipodal pairs of vertices of the \(2_{31}\) polytope form a sharp configuration in \(\mathbb{R}P^6\) \NoCaseChange{\protect\cite{cite119}}.
\item\relax
\flmRefsHyperref[eczindexfamilyrel]{code:witting_polytope}{Witting polytope code} --- The 120 antipodal pairs of the Witting polytope code form a sharp configuration in \(\mathbb{R}P^7\) \NoCaseChange{\protect\cite{cite119}}.
\end{eczvaluelist}
\eczhbkcontributors{ Alexander Barg, \eczhuVVA }
\endeczcode

\eczcode{stiefel}{Stiefel code}{~\NoCaseChange{\protect\cite{cite2171}}}
\codefieldsection{Description}
Encodes \(K\) states (codewords) into a Stiefel manifold.
Points in a real (complex) Stiefel manifold are ordered orthonormal frames, equivalently \(n \times k\) real (complex) matrices \(X\) satisfying \(X^{T}X=I_k\) (\(X^{\dagger}X=I_k\)). The corresponding \(k\)-dimensional subspaces are instead parametrized by the Grassmannian.

\codefieldsection{Protection}
Bounds exist for codes on Stiefel manifolds \NoCaseChange{\protect\cite{cite2492,cite2494}}.

\codefieldsection{Parent}
\begin{eczvaluelist}
\item\relax
\flmRefsHyperref[eczindexfamilyrel]{code:homogeneous_space_classical}{Homogeneous-space code} --- Homogeneous spaces \(G/H\) reduce to real Stiefel manifolds for \(G = O(n)\) and \(H = O(n-k)\), to complex Stiefel manifolds for \(G = U(n)\) and \(H = U(n-k)\), and to quaternionic Stiefel manifolds for \(G = Sp(n)\) and \(H = Sp(n-k)\).
\end{eczvaluelist}
\codefieldsection{Cousins}
\begin{eczvaluelist}
\item\relax
\flmRefsHyperref[eczindexfamilyrel]{code:spacetime}{Spacetime code (STC)} --- Stiefel codes can be thought of as spacetime codes \NoCaseChange{\protect\cite{cite2171}}
\item\relax
\flmRefsHyperref[eczindexfamilyrel]{code:approximate_qecc}{Approximate quantum error-correcting code (AQECC)} --- Riemannian optimization techniques can be applied to design approximate QECCs since the set of unitary encoding maps \(U\) forms a Stiefel manifold \NoCaseChange{\protect\cite{cite2508}}.
\end{eczvaluelist}
\eczhbkcontributors{ \eczhuVVA }
\endeczcode

\eczcode{symmetric_space}{Symmetric-space code}{}
\codefieldsection{Description}
Encodes \(K\) states (codewords) into a symmetric space, which is a homogeneous space \(G/H\) with an additional symmetry property whose definition depends on whether the space is continuous or finite.

Continuous symmetric spaces are homogeneous spaces for which every point admits an involutive isometry fixing that point and reversing geodesics through it. The sphere is a basic example, giving the \(D\)-dimensional spherical symmetric space family \(SO(D+1)/SO(D)\). 
Cartan classified the compact symmetric spaces whose \(G\) are simple real Lie groups \NoCaseChange{\protect\cite{cite2509,cite2242}}.
These spaces include spheres, projective spaces, and Grassmannians.
Noncompact symmetric spaces include Euclidean and hyperbolic spaces.

Finite symmetric spaces are defined in coding theory as spaces admitting a \textit{generously transitive} group action, i.e., for all pairs of points \(x,y \in G/H\), there exists a \(g\in G\) such that \(g.x = y\) and \(g.y = x\) \NoCaseChange{\protect\cite[{Def. 4.5}]{cite987}\protect\cite[{Sec. 3.4}]{cite985}}.

\codefieldsection{Protection}
The decomposition of a symmetric space into \(G\)-irreps is multiplicity free.
Optimal codes have been formulated for quaternionic and octonionic projective spaces \NoCaseChange{\protect\cite{cite917,cite2510}}.

\codefieldsection{Notes}
\begin{eczvaluelist}
\item\relax See \NoCaseChange{\protect\cite{cite987,cite985}} for reviews.
\end{eczvaluelist}
\codefieldsection{Parent}
\begin{eczvaluelist}
\item\relax
\flmRefsHyperref[eczindexfamilyrel]{code:homogeneous_space_classical}{Homogeneous-space code} --- Infinite symmetric spaces are homogeneous spaces with an appropriately defined inversion operation. Finite symmetric spaces are defined in coding theory as spaces admitting a generously transitive group action \NoCaseChange{\protect\cite[{Def. 4.5}]{cite987}\protect\cite[{Sec. 3.4}]{cite985}}. For multiplicity-free spaces such as symmetric spaces, the zonal spherical functions form an Abelian algebra, and the behavior of such functions can be used to obtain bounds on code parameters such as the Levenshtein bound \NoCaseChange{\protect\cite{cite2277,cite2278,cite2088,cite171,cite914}}.
\end{eczvaluelist}
\codefieldsection{Children}
\begin{eczvaluelist}
\item\relax
\flmRefsHyperref[eczindexfamilyrel]{code:binary_permutation}{Code in permutations} --- The permutation group can be viewed as a finite symmetric space \(G/H\) with \(G = S_n \times S_n\) and \(H=S_n\) \NoCaseChange{\protect\cite{cite2464,cite2465}\protect\cite[{Table 3}]{cite985}}.
\item\relax
\flmRefsHyperref[eczindexfamilyrel]{code:2pt_homogeneous}{Two-point homogeneous-space code} --- A special class of symmetric spaces are the two-point homogeneous spaces (a.k.a. rank-one symmetric spaces \NoCaseChange{\protect\cite[{Table 6.1}]{cite2242}}), whose metric is \(G\)-invariant and for which any two pairs of points at the same distance can be mapped to each other by some \(g\in G\) \NoCaseChange{\protect\cite[{Def. 4.7}]{cite987}}.
\item\relax
\flmRefsHyperref[eczindexfamilyrel]{code:grassmannian}{Grassmannian code} --- Grassmannians are symmetric spaces \(G/H\) for \(G = O(p+q)\) and \(H = O(p)\times O(q)\) in the case of real Grassmannians, \(G = U(p+q)\) and \(H = U(p)\times U(q)\) in the case of complex Grassmannians, and \(G = Sp(p+q)\) and \(H = Sp(p)\times Sp(q)\) in the case of quaternionic Grassmannians.
\item\relax
\flmRefsHyperref[eczindexfamilyrel]{code:unitary}{Unitary code} --- The unitary group is a compact symmetric space \(G/H\) with \(G=U(N)\times U(N)\) and \(H = U(N)\) \NoCaseChange{\protect\cite[{Table 3}]{cite985}}.
\item\relax
\flmRefsHyperref[eczindexfamilyrel]{code:q-ary_constant_weight}{Constant-weight block code} --- The set of all weight-\(w\) \(q\)-ary strings of length \(n\) forms the \textit{nonbinary Johnson space} (a.k.a. \(q\)-ary Johnson space), a finite symmetric space \(G/H\) with \(G = S_{q-1} \wr S_n\) \NoCaseChange{\protect\cite{cite984}\protect\cite[{Sec. 8.8}]{cite913}\protect\cite[{Table 3}]{cite985}}. The number of such strings is \({n \choose w} (q-1)^w\). This reduces to the Johnson space for \(q=2\).
\item\relax
\flmRefsHyperref[eczindexfamilyrel]{code:poset}{Poset code} --- Ordered Hamming space can be viewed as a finite symmetric space \NoCaseChange{\protect\cite{cite214,cite215}\protect\cite[{Sec. 4.2.3}]{cite987}\protect\cite[{Table 3}]{cite985}}.
\item\relax
\flmRefsHyperref[eczindexfamilyrel]{code:q-ary_over_zq}{\(q\)-ary code over \(\mathbb{Z}_q\)} --- The space of \(q\)-ary codes over \(\mathbb{Z}_q\) under the Lee metric can be viewed as a finite symmetric space \(G/H\) with \(G = D_q \wr S_n\) \NoCaseChange{\protect\cite{cite2441,cite2442}\protect\cite[{Table 3}]{cite985}}.
\end{eczvaluelist}
\codefieldsection{Cousins}
\begin{eczvaluelist}
\item\relax
\flmRefsHyperref[eczindexfamilyrel]{code:points_into_lattices}{Lattice} --- Upper bounds on kissing numbers can be worked out by treating the sphere as a symmetric space \NoCaseChange{\protect\cite{cite2280}}.

\item\relax
\flmRefsHyperref[eczindexfamilyrel]{code:group_classical}{Group-alphabet code} --- Group spaces for Lie groups \(G\) are symmetric spaces \NoCaseChange{\protect\cite[{Table 6.1}]{cite2242}}.
\end{eczvaluelist}
\eczhbkcontributors{ \eczhuVVA }
\endeczcode

\eczcode{2pt_homogeneous}{Two-point homogeneous-space code}{~\NoCaseChange{\protect\cite{cite2299}}}
\codefieldsection{Alternative Names}
\begin{eczvaluelist}
\item\relax Rank-one symmetric space code
\end{eczvaluelist}
\eczhIndexCodeAliasName{2pt_homogeneous}{Rank-one symmetric space code}
\codefieldsection{Description}
Encodes \(K\) states (codewords) into a two-point homogeneous space \(G/H\), i.e., a homogeneous space with a \(G\)-invariant metric whose symmetry group acts transitively on pairs of points at any fixed distance.
In the compact connected case, these are precisely the rank-one symmetric spaces. Finite examples tag along as the corresponding combinatorial analogues, where pairs of points are indistinguishable up to the symmetry group once their mutual distance is fixed.

More technically, a two-point homogeneous space \(G/H\) is a homogeneous space with a \(G\)-invariant metric such that for any \(x,x',y,y'\in G/H\), there exists \(g\in G\) with \(gx=x'\) and \(gy=y'\) whenever \(d(x,y)=d(x',y')\) \NoCaseChange{\protect\cite[{Def. 4.7}]{cite987}}. For finite spaces, this is the metric-space analogue of distance transitivity on pairs, not ordinary two-transitivity on all ordered pairs \NoCaseChange{\protect\cite[{Table 2}]{cite985}}.

Two-point homogeneous spaces for compact connected \(G\) have been classified and include spheres, the real, complex, and quaternionic projective spaces, and the octonionic projective (a.k.a. Cayley) plane \NoCaseChange{\protect\cite{cite2511}\protect\cite[{Ch. 9}]{cite39}}. 
These compact examples have positive sectional curvature \NoCaseChange{\protect\cite{cite2512}\protect\cite[{Fig. 1}]{cite2513}}. 

Examples of two-point homogeneous spaces for finite \(G\) include Hamming space, Johnson space, and distance-transitive graphs \NoCaseChange{\protect\cite{cite1385}}.
Finite spaces have yet to be classified \NoCaseChange{\protect\cite{cite987}\protect\cite[{Ch. 9}]{cite39}\protect\cite[{Table 2}]{cite985}}.

Infinite two-point homogeneous spaces with constant curvature include spheres, Euclidean spaces, and hyperbolic spaces.
These spaces are three-point homogeneous (a.k.a. satisfy the mobility axiom), meaning that the \(G\) action maps any triangle to any other triangle with the same parameters \NoCaseChange{\protect\cite[{Sec. 6.6.1.1}]{cite2242}}.

\codefieldsection{Protection}
The zonal spherical functions of two-point homogeneous spaces depend only on the distance between points due to the pairwise distance-homogeneity of the \(G\) action. 
For the compact connected families, these spherical functions are Jacobi polynomials; in the spherical case, they reduce to Gegenbauer polynomials \NoCaseChange{\protect\cite[{Table 1}]{cite985}}.
This yields a general series of bounds on packings in \(G/H\) originating with the Kabatiansky-Levenshtein bound \NoCaseChange{\protect\cite{cite2299}\protect\cite[{Ch. 9}]{cite39}}; see Ref. \NoCaseChange{\protect\cite{cite914}} for a review. 

\codefieldsection{Notes}
\begin{eczvaluelist}
\item\relax See \NoCaseChange{\protect\cite{cite2152,cite987,cite985}\protect\cite[{Ch. 9}]{cite39}} for reviews.
\end{eczvaluelist}
\codefieldsection{Parent}
\begin{eczvaluelist}
\item\relax
\flmRefsHyperref[eczindexfamilyrel]{code:symmetric_space}{Symmetric-space code} --- A special class of symmetric spaces are the two-point homogeneous spaces (a.k.a. rank-one symmetric spaces \NoCaseChange{\protect\cite[{Table 6.1}]{cite2242}}), whose metric is \(G\)-invariant and for which any two pairs of points at the same distance can be mapped to each other by some \(g\in G\) \NoCaseChange{\protect\cite[{Def. 4.7}]{cite987}}.
\end{eczvaluelist}
\codefieldsection{Children}
\begin{eczvaluelist}
\item\relax
\flmRefsHyperref[eczindexfamilyrel]{code:hyperbolic}{Hyperbolic sphere packing} --- Hyperbolic space in \(D\) dimensions is a symmetric space \(G/H\) for \(G = SO(D,1)\) the proper Lorentz group and \(H = O(D)\). The hyperbolic plane is the case \(D=2\). In fact, hyperbolic spaces are noncompact three-point homogeneous spaces \NoCaseChange{\protect\cite[{Sec. 6.6.1.1}]{cite2242}}.
\item\relax
\flmRefsHyperref[eczindexfamilyrel]{code:complex_projective}{Complex projective space code} --- Compact two-point homogeneous spaces \(G/H\) reduce to complex projective spaces for \(G = SU(D+1)\) and \(H = U(D)\) \NoCaseChange{\protect\cite[{Ch. 9}]{cite39}\protect\cite[{Table 1}]{cite985}}.
\item\relax
\flmRefsHyperref[eczindexfamilyrel]{code:real_projective}{Real projective space code} --- Compact two-point homogeneous spaces \(G/H\) reduce to real projective spaces for \(G = SO(D+1)\) and \(H = O(D)\) \NoCaseChange{\protect\cite[{Ch. 9}]{cite39}\protect\cite[{Table 1}]{cite985}}.
\item\relax
\flmRefsHyperref[eczindexfamilyrel]{code:finite_grassmann}{Constant-dimension code} --- The finite-field Grassmannian (a.k.a. \(q\)-Johnson space) can be regarded as a finite two-point homogeneous space \(G/H\) where \(G = GL(n,\mathbb{F}_q)\) \NoCaseChange{\protect\cite{cite28}\protect\cite[{Sec. 4.2.1}]{cite987}\protect\cite[{Table 2}]{cite985}\protect\cite[{Ch. 9}]{cite39}\protect\cite[{Sec. 8.6}]{cite913}}.
\item\relax
\flmRefsHyperref[eczindexfamilyrel]{code:rank_metric}{Rank-metric code} --- Matrices of dimension \(m\times n\) over \(\mathbb{F}_q\) under the rank metric form a finite two-point homogeneous space \NoCaseChange{\protect\cite{cite2134,cite2152,cite2153}\protect\cite[{Table 2}]{cite985}}.
\item\relax
\flmRefsHyperref[eczindexfamilyrel]{code:delsarte_optimal}{Sharp configuration}\item\relax
\flmRefsHyperref[eczindexfamilyrel]{code:q-ary_digits_into_q-ary_digits}{\(q\)-ary code} --- Hamming space can be regarded as a finite two-point homogeneous space \(G/H\) where \(G = S_q \wr S_n\) is its isometry group \NoCaseChange{\protect\cite[{Sec. 5.3}]{cite987}\protect\cite[{Table 2}]{cite985}}.
\item\relax
\flmRefsHyperref[eczindexfamilyrel]{code:points_into_spheres}{Constant-energy spherical code} --- Real spheres are compact connected two-point homogeneous spaces with quotient \(SO(D+1)/SO(D)\) \NoCaseChange{\protect\cite[{Table 1}]{cite985}}. Complex spheres can be treated as real spheres of twice the dimension over \(\mathbb{R}\), with quotient \(SU(D+1)/SU(D)\). In fact, spheres are compact three-point homogeneous spaces \NoCaseChange{\protect\cite[{Sec. 6.6.1.1}]{cite2242}}.
\end{eczvaluelist}
\codefieldsection{Cousins}
\begin{eczvaluelist}
\item\relax
\flmRefsHyperref[eczindexfamilyrel]{code:ecc}{Error-correcting code (ECC)} --- ECCs and \(t\)-designs on two-point homogeneous spaces are intimately related via association schemes \NoCaseChange{\protect\cite{cite226,cite916}}.
\item\relax
\flmRefsHyperref[eczindexfamilyrel]{code:analog}{Analog code} --- Euclidean space \(\mathbb{R}^n\) is a noncompact two-point homogeneous space and is, in fact, a noncompact three-point homogeneous space \NoCaseChange{\protect\cite[{Sec. 6.6.1.1}]{cite2242}}. One can also think of \(\mathbb{R}^n\) as a homogeneous space of the Euclidean group by the orthogonal group, \(E(n)/O(n)\) \NoCaseChange{\protect\cite[{Ch. XI}]{cite2243}}.
\item\relax
\flmRefsHyperref[eczindexfamilyrel]{code:points_into_lattices}{Lattice} --- The Levenshtein bound \NoCaseChange{\protect\cite{cite2277,cite2278,cite2088,cite171,cite914}} and Cohn-Elkies LP bound \NoCaseChange{\protect\cite{cite2279}} can be derived for sphere packings by thinking of \(\mathbb{R}^n\) as a homogeneous space of the Euclidean group by the orthogonal group, \(E(n)/O(n)\) \NoCaseChange{\protect\cite[{Ch. XI}]{cite2243}}.

\item\relax
\flmRefsHyperref[eczindexfamilyrel]{code:constant_weight}{Constant-weight code} --- The set of all weight-\(w\) binary strings of length \(n\) forms the \textit{Johnson space} \(J(n,w)\), a finite two-point homogeneous space \(G/H\) with \(G = S_n\) and \(H = S_w \times S_{n-w}\) \NoCaseChange{\protect\cite{cite880,cite986,cite912,cite171}\protect\cite[{Sec. 4.2.1}]{cite987}\protect\cite[{Table 2}]{cite985}}.
\item\relax
\flmRefsHyperref[eczindexfamilyrel]{code:higman-sims_graph}{Higman-Sims graph-adjacency code} --- The Higman-Sims graph is distance-transitive, hence it is a finite two-point homogeneous space \NoCaseChange{\protect\cite{cite1385}}.
\item\relax
\flmRefsHyperref[eczindexfamilyrel]{code:hoffman-singleton_graph}{Hoffman-Singleton graph-adjacency code} --- The Hoffman-Singleton graph is distance-transitive, hence it is a finite two-point homogeneous space \NoCaseChange{\protect\cite{cite1385}}.
\item\relax
\flmRefsHyperref[eczindexfamilyrel]{code:t-designs}{\(t\)-design} --- Designs exist on compact connected two-point homogeneous spaces \NoCaseChange{\protect\cite{cite226,cite914,cite917}}. ECCs and \(t\)-designs on two-point homogeneous spaces are intimately related via association schemes \NoCaseChange{\protect\cite{cite226,cite916}}.
\end{eczvaluelist}
\eczhbkcontributors{ Alexander Barg, \eczhuVVA }
\endeczcode

\eczcode{univ_opt}{Universally optimal code}{~\NoCaseChange{\protect\cite{cite2299}}}
\codefieldsection{Description}
A code that minimizes, among all codes of the same cardinality, a large family of potential functions.

\codefieldsection{Notes}
\begin{eczvaluelist}
\item\relax See notes on universally optimal codes \NoCaseChange{\protect\cite{cite394}}.
\end{eczvaluelist}
\codefieldsection{Parent}
\begin{eczvaluelist}
\item\relax
\flmRefsHyperref[eczindexfamilyrel]{code:homogeneous_space_classical}{Homogeneous-space code}\end{eczvaluelist}
\codefieldsection{Children}
\begin{eczvaluelist}
\item\relax
\flmRefsHyperref[eczindexfamilyrel]{code:univ_opt_analog}{Universally optimal sphere packing}\item\relax
\flmRefsHyperref[eczindexfamilyrel]{code:delsarte_optimal}{Sharp configuration} --- All sharp configurations are universally optimal \NoCaseChange{\protect\cite{cite119,cite173}}, but not all universally optimal codes are sharp configurations.
\item\relax
\flmRefsHyperref[eczindexfamilyrel]{code:univ_opt_q-ary}{Universally optimal \(q\)-ary code}\item\relax
\flmRefsHyperref[eczindexfamilyrel]{code:univ_opt_spherical}{Universally optimal spherical code}\end{eczvaluelist}
\eczhbkcontributors{ Alexander Barg, \eczhuVVA }
\endeczcode

\part{Codes in the Quantum Domain}
\onecolumngrid

\begin{eczEpigraph}
\begin{quote}
\flmQuoteSetup{quote}%
Many of us are hopeful that quantum computers will become practical and useful computing devices some time during the 21st century. It is probably fair to say, though, that none of us can now envision exactly what the hardware of that machine of the future will be like; surely, it will be much different than the sort of hardware that experimental physicists are  investigating these days. But of one thing we can be quite confident—that a practical quantum computer will incorporate some type of error correction into its operation. Quantum computers are far more susceptible to making errors than conventional digital computers, and some method of controlling and correcting those errors will be needed to prevent a quantum computer from crashing.
\flmQuoteAttributed{John Preskill, 1997}
\end{quote}
\end{eczEpigraph}

\section{Property codes}

\twocolumngrid

\eczcode{g_covariant_erasure}{\(G\)-covariant erasure code}{~\NoCaseChange{\protect\cite{cite2514}}}
\codefieldsection{Description}
A \(G\)-covariant block code that serves as a proof-of-principle construction to demonstrate the existence of
\flmRefsHyperref{code:covariant}{\(G\)-covariant codes} where \(G\) is a finite
group, and the physical space is finite-dimensional.
This construction can be done for any erasure-correcting code.

Consider a finite group \(G\) acting on a finite set
\(A\) as a subgroup of the symmetric group on \(|A|\) elements, \(G \subset S_{|A|}\).
Let \(U_0: \mathsf{H}_{\text{logical}} \rightarrow \mathsf{H}_{\text{physical}}
= \mathsf{H}^{\otimes n}\) be any QECC, possibly non-covariant.  Define the covariant
encoder \(U \equiv U_0^{\otimes |A|}: \mathsf{H}_{\text{logical}}^{\otimes |A|}
\rightarrow \mathsf{H}_{\text{physical}}^{\otimes |A|}\) on \(|A|\).  Then, the group acts on codewords by index
permutation:
\flmMathEnvironment{align}{}{
V(g) | \phi_{a_1} \rangle | \phi_{a_2} \rangle \cdots | \phi_{a_{|A|}} \rangle = | \phi_{g^{-1} a_1} \rangle | \phi_{g^{-1} a_2} \rangle \cdots | \phi_{g^{-1} a_{|A|}} \rangle~,
}
where \(V(g)\) is the unitary representation of \(g \in G\) acting on the physical space. The action of \(V(g)\) is transversal with respect to the partition \(\mathsf{H}_{\text{physical}}^{\otimes |A|}\).

\codefieldsection{Protection}
Depends on the base encoding \(U_0\).

\codefieldsection{Parent}
\begin{eczvaluelist}
\item\relax
\flmRefsHyperref[eczindexfamilyrel]{code:covariant}{Covariant block quantum code} --- In a proof of principle demonstration, error-correcting codes that are finite-\(G\) covariant can be constructed from a base encoding \(U_0\).
\end{eczvaluelist}
\eczhbkcontributors{ Jack Davis, \eczhuVVA }
\endeczcode

\eczcode{nonabelian_covariant_erasure}{\(U(d)\)-covariant approximate erasure code}{~\NoCaseChange{\protect\cite{cite2515,cite2516}}}
\codefieldsection{Description}
Covariant code whose construction takes an arbitrary erasure-correcting code as input and yields an approximate QECC that is also covariant with respect to the unitary group.

\codefieldsection{Parents}
\begin{eczvaluelist}
\item\relax
\flmRefsHyperref[eczindexfamilyrel]{code:covariant}{Covariant block quantum code}\item\relax
\flmRefsHyperref[eczindexfamilyrel]{code:approximate_qecc}{Approximate quantum error-correcting code (AQECC)} --- Any finite-dimensional code covariant w.r.t. a continuous group has to be approximate because of the \flmRefsHyperref{ref721}{Eastin-Knill theorem}.
\end{eczvaluelist}
\codefieldsection{Cousin}
\begin{eczvaluelist}
\item\relax
\flmRefsHyperref[eczindexfamilyrel]{code:stab_5_1_3}{\(\llbracket 5,1,3\rrbracket \) Five-qubit perfect code} --- The five-qubit code can be used to construct an approximate code that is also covariant with respect to the unitary group.
\end{eczvaluelist}
\eczhbkcontributors{ \eczhuVVA }
\endeczcode

\eczcode{topological_abelian}{Abelian topological code}{}
\codefieldsection{Description}
Code whose codewords realize topological order associated with an Abelian anyon theory.
In 2D, this is equivalent to a unitary braided fusion category which is also an Abelian group under fusion \NoCaseChange{\protect\cite{cite574}}.
Unless otherwise noted, the phases discussed are bosonic.

\subsection{2D Abelian topological codes}

A theory is defined by an Abelian group \(A\) of anyon types whose multiplication relations define the fusion rules, and a set of exchange statistics \(\theta(a)\in U(1)\) obtained by exchanging two anyons of type \(a\in A\).
The exchange statistics in turn define braiding relations,
\flmMathEnvironment{align}{}{
  B(a,b) = \frac{\theta(ab)}{\theta(a)\theta(b)}~,
}
between all anyon pairs \(a,b\).

All 2D Abelian bosonic topological orders can be understood within the subsystem stabilizer formalism \NoCaseChange{\protect\cite{cite414}}.
As such, many of the operations one can perform on such codes have both a stabilizer and a topological-phase interpretation.
Stabilizer generators of 2D topological codes acting on 1D loops of qubits can be interpreted as one-form symmetries of the underlying phase realized by the code.
Identification of an anyon \(a\) with the vacuum is equivalent to adding string excitation operators corresponding to \(a\) to the stabilizer group and taking the center to get another stabilizer group.
Code states of this new stabilizer code correspond to a \flmRefsHyperref{ref410}{condensed phase} of the parent topological phase.
A related subsystem-code operation is to \flmRefsHyperref{ref666}{gauge out} anyon types by adjoining their short string operators to the gauge group; unlike condensation, the gauged-out anyons need not be bosons and are not identified with the vacuum \NoCaseChange{\protect\cite{cite414}}.
The remaining unidentified parent-phase anyons behave differently with respect to the new condensed-phase state.
Some become \textit{confined} while the remaining ones pick up new braiding relations.

A 2D Abelian anyon theory admits a gapped boundary iff it has a Lagrangian subgroup, i.e., a maximal subgroup of bosons with trivial mutual statistics whose order squares to that of the anyon group \NoCaseChange{\protect\cite{cite573,cite406,cite405}}.
It is conjectured that the logical dimension of an Abelian topological code on a torus is always a square \NoCaseChange{\protect\cite{cite573}}.

\subsection{3D Abelian topological codes}

While excitations are point-like in 2D topological codes, they can have higher dimension in 3D topological codes.
For example, there are three types of \(\mathbb{Z}_2\) Abelian topological orders in 3D: one with bosonic charge and loop excitations (BcBl) and two with fermionic charge excitations and bosonic (FcBl) or fermionic (FcFl) loop excitations, respectively \NoCaseChange{\protect\cite{cite579,cite455}}.
The first two correspond to the 3D surface code and 3D fermionic surface code, while the FcFl case is purported not to have a commuting projector Hamiltonian realization \NoCaseChange{\protect\cite{cite455}}.
There exists an invariant that distinguishes these \NoCaseChange{\protect\cite{cite455}}.

\codefieldsection{Rate}
\(\mathbb{Z}_q\) topological order cannot exist on any fractal geometry that is embeddable in two Euclidean dimensions \NoCaseChange{\protect\cite{cite2517}}.
\codefieldsection{Encoding}
\begin{eczvaluelist}
\item\relax Any local quantum circuit connecting ground states of topological orders with non-isomorphic Abelian groups must have depth that is at least linear in the diameter of the system \NoCaseChange{\protect\cite{cite2518}}.
\end{eczvaluelist}
\codefieldsection{Gates}
\begin{eczvaluelist}
\item\relax Clifford gates can be implemented by braiding defects; for qubit-based stabilizer codes realizing Abelian topological phases, see Refs. \NoCaseChange{\protect\cite{cite2519,cite2520}}. Most of such designs focus on the surface code \NoCaseChange{\protect\cite{cite442,cite2521,cite2522,cite2523,cite2524,cite2525}}.
\end{eczvaluelist}
\codefieldsection{Fault Tolerance}
\begin{eczvaluelist}
\item\relax Fault-tolerant logical operations can be interpreted as anyon \flmRefsHyperref{ref410}{condensation} events \NoCaseChange{\protect\cite{cite2526}}.
\item\relax Modular decoding, designed to overcome the backlog problem, is applicable to fault-tolerant protocols based on topological qubit stabilizer codes \NoCaseChange{\protect\cite{cite2527}}.
\end{eczvaluelist}
\codefieldsection{Parent}
\begin{eczvaluelist}
\item\relax
\flmRefsHyperref[eczindexfamilyrel]{code:topological}{Topological code}\end{eczvaluelist}
\codefieldsection{Children}
\begin{eczvaluelist}
\item\relax
\flmRefsHyperref[eczindexfamilyrel]{code:compactified_r}{Compactified \(\mathbb{R}\) gauge theory code} --- The compactified \(\mathbb{R}\) gauge theory code can be obtained from the analog surface code by \flmRefsHyperref{ref410}{condensing} certain anyons \NoCaseChange{\protect\cite{cite411}}. This results in a pinning of each mode to the space of periodic functions, which is the Hilbert space of a physical rotor, and can be thought of as compactification of the 2D \(\mathbb{R}\) gauge theory phase realized by the analog surface code.
\item\relax
\flmRefsHyperref[eczindexfamilyrel]{code:mbq}{Majorana box qubit} --- When treated as ground states of the code Hamiltonian, codewords of a single Kitaev chain realize \(\mathbb{Z}_2\) fermionic topological order.
\item\relax
\flmRefsHyperref[eczindexfamilyrel]{code:double_semion_string_net}{Double-semion string-net code} --- When treated as ground states of the code Hamiltonian, the double-semion string-net code states realize 2D double-semion topological order, a topological phase of matter that exists as the deconfined phase of the 2D twisted \(\mathbb{Z}_2\) gauge theory \NoCaseChange{\protect\cite{cite584}}.
\item\relax
\flmRefsHyperref[eczindexfamilyrel]{code:3d_color}{3D color code}\item\relax
\flmRefsHyperref[eczindexfamilyrel]{code:3d_fermionic_surface}{3D fermionic surface code} --- The 3D fermionic surface code realizes 3D \(\mathbb{Z}_2\) gauge theory with fermionic charge and bosonic loop excitations (FcBl), i.e., with an emergent fermion. The fermionic excitations endow the code with an anomalous two-form symmetry, which is argued to induce a non-trivial finite-temperature topological order \NoCaseChange{\protect\cite{cite2528}}.
\item\relax
\flmRefsHyperref[eczindexfamilyrel]{code:4d_13_surface}{\((1,3)\) 4D toric code} --- The \((1,3)\) 4D toric code realizes 4D \(\mathbb{Z}_2\) gauge theory with 1D \(Z\)-type and 3D \(X\)-type logical operators.
\item\relax
\flmRefsHyperref[eczindexfamilyrel]{code:4d_surface}{\((2,2)\) Loop toric code} --- The 4D loop toric code realizes 4D \(\mathbb{Z}_2\) gauge theory with only loop excitations \NoCaseChange{\protect\cite{cite2529}}.
\item\relax
\flmRefsHyperref[eczindexfamilyrel]{code:3d_kitaev_honeycomb}{3D Kitaev honeycomb code} --- One of the phases realized by the 3D Kitaev honeycomb Hamiltonian is that of the 3D fermionic surface code \NoCaseChange{\protect\cite{cite458}}.
\item\relax
\flmRefsHyperref[eczindexfamilyrel]{code:subsystem_three_fermion}{Three-fermion (3F) subsystem code} --- The 3F code is a 2D subsystem code characterized by 3F topological order \NoCaseChange{\protect\cite{cite414}}, which is chiral and modular.
\item\relax
\flmRefsHyperref[eczindexfamilyrel]{code:qudit_3d_surface}{Modular-qudit 3D surface code} --- The modular-qudit 3D surface code realizes 3D \(\mathbb{Z}_q\) gauge theory with bosonic charge and loop excitations (BcBl).
\item\relax
\flmRefsHyperref[eczindexfamilyrel]{code:tqd_abelian_stabilizer}{Abelian TQD stabilizer code} --- Every Abelian TQD code with Type-I and -II cocycles can be realized as a modular-qudit Pauli stabilizer code by starting from a stack of Abelian quantum double models (it suffices to take \(\prod_i \mathbb{Z}_{N_i^2}\) toric codes) and \flmRefsHyperref{ref410}{condensing} certain bosonic anyons \NoCaseChange{\protect\cite{cite405}}.

\item\relax
\flmRefsHyperref[eczindexfamilyrel]{code:qudit_znone}{\(\mathbb{Z}_q^{(1)}\) subsystem code} --- The \(\mathbb{Z}_q^{(1)}\) subsystem code is characterized by the \(\mathbb{Z}_q^{(1)}\) anyon theory \NoCaseChange{\protect\cite{cite638}}. The anyon theory has a single generator \(a \in \mathbb Z_N\) with \(\theta(a) =e^{\frac{2\pi i}{N}a^2}\). It is modular for odd prime \(q\) and non-modular otherwise.
\item\relax
\flmRefsHyperref[eczindexfamilyrel]{code:semion}{Chiral semion subsystem code} --- The semion code is a subsystem code characterized by the chiral semion topological phase.
\item\relax
\flmRefsHyperref[eczindexfamilyrel]{code:zthree_znine}{\(\mathbb{Z}_3\times\mathbb{Z}_9\)-fusion subsystem code} --- The \(\mathbb{Z}_3\times\mathbb{Z}_9\)-fusion subsystem code is characterized by a non-modular anyon theory with \(\mathbb{Z}_3\times\mathbb{Z}_9\) fusion rules.
\item\relax
\flmRefsHyperref[eczindexfamilyrel]{code:galois_color}{Galois-qudit color code} --- A Galois qudit for \(q=p^m\) can be decomposed into a Kronecker product of \(m\) modular qudits \NoCaseChange{\protect\cite{cite696,cite398,cite698,cite699,cite700}\protect\cite[{Sec. 5.3}]{cite697}}. Galois-qudit color codes yield Abelian quantum-double codes with Abelian-group topological order via this decomposition.
\item\relax
\flmRefsHyperref[eczindexfamilyrel]{code:galois_topological}{Galois-qudit surface code} --- A Galois qudit for \(q=p^m\) can be decomposed into a Kronecker product of \(m\) modular qudits \NoCaseChange{\protect\cite{cite696,cite398,cite698,cite699,cite700}\protect\cite[{Sec. 5.3}]{cite697}}. Galois-qudit surface codes yield Abelian quantum-double codes with \(\mathbb{F}_{p^m}\cong \mathbb{Z}_p^m\) topological order via this decomposition.
\end{eczvaluelist}
\codefieldsection{Cousins}
\begin{eczvaluelist}
\item\relax
\flmRefsHyperref[eczindexfamilyrel]{code:walker_wang}{Walker-Wang model code} --- Any Abelian anyon theory \(A\) can be realized at one of the surfaces of a 3D Walker-Wang model whose underlying theory is an Abelian TQD containing \(A\) as a subtheory \NoCaseChange{\protect\cite{cite471,cite472}\protect\cite[{Appx. H}]{cite414}}.
\item\relax
\flmRefsHyperref[eczindexfamilyrel]{code:3d_stabilizer}{3D lattice stabilizer code} --- Translation-invariant qubit 3D TQFT stabilizer models are conjectured to be equivalent, under a locality-preserving unitary, to multiple copies of the 3D surface code and/or the 3D fermionic surface code together with trivial ancillas \NoCaseChange{\protect\cite{cite456}}.
\item\relax
\flmRefsHyperref[eczindexfamilyrel]{code:gauss_law}{Gauss' law code} --- Gauge-group elements of a \(D\)-dimensional fermionic \(\mathbb{Z}_2\) gauge theory can arise from a single-error-correcting linear binary code \NoCaseChange{\protect\cite[{Thm. 1}]{cite78}}. There is a general correspondence between stabilizer codes and gauge theory, with the stabilizer group playing the role of the gauge group \NoCaseChange{\protect\cite{cite1365}}, and with the Gauss' law code being a specific example.
\item\relax
\flmRefsHyperref[eczindexfamilyrel]{code:plaquette_ising}{Plaquette Ising code} --- The 2D plaquette Ising model can be constructed by coupling layers of 1D \(\mathbb{Z}_2\) lattice gauge theory \NoCaseChange{\protect\cite{cite1283}}. A field-theoretic description of the 2D plaquette Ising model can be obtained by coupling layers of 1D gauge theory \NoCaseChange{\protect\cite{cite568}}.
\item\relax
\flmRefsHyperref[eczindexfamilyrel]{code:topological_classical}{Classical topological code} --- Some topological orders have classical analogues that can be used for error correction.
\item\relax
\flmRefsHyperref[eczindexfamilyrel]{code:tqd_abelian}{Abelian TQD code} --- Abelian TQDs with Type-I and -II cocycles account for all 2D Abelian topological orders that admit gapped boundaries \NoCaseChange{\protect\cite{cite573}}.
Conversely, every Abelian anyon theory is a subtheory of some TQD \NoCaseChange{\protect\cite[{Sec. 6.2}]{cite414}}. 
Any Abelian anyon theory \(A\) can be realized at one of the surfaces of a 3D Walker-Wang model whose underlying theory is an Abelian TQD containing \(A\) as a subtheory \NoCaseChange{\protect\cite{cite471,cite472}\protect\cite[{Appx. H}]{cite414}}.

\item\relax
\flmRefsHyperref[eczindexfamilyrel]{code:stabilizer}{Stabilizer code} --- There is a general correspondence between stabilizer codes and gauge theory, with the stabilizer group playing the role of the gauge group \NoCaseChange{\protect\cite{cite1365}}.
\item\relax
\flmRefsHyperref[eczindexfamilyrel]{code:translationally_invariant_subsystem}{Lattice subsystem code} --- All 2D Abelian bosonic topological orders can be realized as modular-qudit lattice subsystem codes by starting with a topological stabilizer code whose anyon theory contains the target Abelian anyon theory and then \flmRefsHyperref{ref666}{gauging out} complementary anyon types \NoCaseChange{\protect\cite{cite414}}.
One convenient choice for the parent stabilizer code is an Abelian TQD containing the target theory as a subtheory \NoCaseChange{\protect\cite[{Sec. 6.2}]{cite414}}.
The stabilizer generators of the new subsystem code may no longer be geometrically local.
Lattice subsystem stabilizer code Hamiltonians described by an Abelian anyon theory do not always realize the corresponding anyonic topological order in their ground-state subspace and may exhibit a rich phase diagram.
Non-Abelian topological orders are purported not to be realizable with Pauli stabilizer codes \NoCaseChange{\protect\cite{cite2530}}.

\item\relax
\flmRefsHyperref[eczindexfamilyrel]{code:analog_surface}{Analog surface code} --- The analog surface code realizes a straightforward extension of the modular-qudit surface code to infinite local dimension, \(q\to\infty\) \NoCaseChange{\protect\cite{cite2531}}.
The code realizes a phase of 2D \(\mathbb{R}\) gauge theory.
There are two types of anyons, \(e\) and \(m\), with each type being valued in a continuous domain as opposed to \(\mathbb{Z}_q\) for the qudit surface code.

\item\relax
\flmRefsHyperref[eczindexfamilyrel]{code:chern_simons_gkp}{\(U(1)_{2n} \times U(1)_{-2m}\) Chern-Simons GKP code} --- The \(U(1)_{2n} \times U(1)_{-2m}\) Chern-Simons GKP code realizes \(U(1)_{2n} \times U(1)_{-2m}\) Chern-Simons theory on bosonic modes. The code can be obtained from the analog surface code by \flmRefsHyperref{ref410}{condensing} certain anyons \NoCaseChange{\protect\cite{cite411}}.
\item\relax
\flmRefsHyperref[eczindexfamilyrel]{code:tiger_surface}{Tiger surface code} --- The tiger surface code is conjectured to realize phases of \(U(1)\) gauge theory.
\item\relax
\flmRefsHyperref[eczindexfamilyrel]{code:da}{Dynamical code} --- Useful measurement sequences of dynamical codes can be extracted from topological quantum field theory \NoCaseChange{\protect\cite{cite2532}}.
\item\relax
\flmRefsHyperref[eczindexfamilyrel]{code:invertible}{Chen-Hsin invertible-order code} --- Instances of the code in 4D realize the 3D \(\mathbb{Z}_2\) gauge theory with fermionic charge and either bosonic (FcBl) or fermionic (FcFl) loop excitations at their boundaries \NoCaseChange{\protect\cite{cite579,cite455}}; see Ref. \NoCaseChange{\protect\cite{cite580}} for a different lattice-model formulation of the FcBl boundary code.
\item\relax
\flmRefsHyperref[eczindexfamilyrel]{code:quantum_repetition}{Quantum repetition code} --- Product constructions built from the one-dimensional Ising/repetition code yield several topological phases \NoCaseChange{\protect\cite[{Fig. 8}]{cite1501}}.
\item\relax
\flmRefsHyperref[eczindexfamilyrel]{code:layer}{Layer code} --- The Layer code realizes 2D layers of \(\mathbb{Z}_2\) gauge theory coupled along defects.
\item\relax
\flmRefsHyperref[eczindexfamilyrel]{code:qcga}{Bivariate bicycle (BB) code} --- BB codes have been investigated in terms of their anyons and topological order \NoCaseChange{\protect\cite{cite2533}}.
\item\relax
\flmRefsHyperref[eczindexfamilyrel]{code:qubit_css}{Qubit CSS code} --- The \flmRefsHyperref{ref683}{mapping of qubit CSS codes to chain complexes} allows the application of structures from topology to error correction. Chain complexes describing some QLDPC codes \NoCaseChange{\protect\cite{cite484,cite485}}, and, more generally, CSS codes \NoCaseChange{\protect\cite{cite486}} can be "lifted" into higher-dimensional manifolds admitting some notion of geometric locality. Qubit CSS codes admit several dualities \NoCaseChange{\protect\cite{cite2534,cite469}}. In particular, a CSS code and two dual classical codes can be organized by the same 2-complex, and gauging \NoCaseChange{\protect\cite{cite462,cite463,cite233,cite464,cite465,cite466,cite467,cite468,cite469,cite470}} either classical code yields the same CSS code up to Hadamard \NoCaseChange{\protect\cite{cite469}}.
\item\relax
\flmRefsHyperref[eczindexfamilyrel]{code:qubit_stabilizer}{Qubit stabilizer code} --- Qubit stabilizer states can be interpreted as states that are preparable using the Euclidean path integral in 3D Chern-Simons theory, defined on manifolds that are toy models of AdS/CFT wormholes \NoCaseChange{\protect\cite{cite2535,cite2536}}.
\item\relax
\flmRefsHyperref[eczindexfamilyrel]{code:twist_defect_color}{Twist-defect color code} --- Twist-defect color codes realize \(\mathbb{Z}_2 \times \mathbb{Z}_2\) topological order with twist defects.
\item\relax
\flmRefsHyperref[eczindexfamilyrel]{code:twist_defect_surface}{Twist-defect surface code} --- Twist-defect surface codes realize \(\mathbb{Z}_2\) topological order with twist defects.
\item\relax
\flmRefsHyperref[eczindexfamilyrel]{code:three_fermion}{Three-fermion (3F) Walker-Wang model code} --- The gapped boundary of the 3F Walker-Wang model supports the 3F topological order \NoCaseChange{\protect\cite{cite477,cite478}}.
\end{eczvaluelist}
\eczhbkcontributors{ \eczhuVVA }
\endeczcode

\eczcode{approximate_oaecc}{Approximate operator-algebra QECC}{~\NoCaseChange{\protect\cite{cite2537,cite2538}}}
\codefieldsection{Description}
Code encoding quantum and/or classical information that approximately corrects against noise affecting operators forming an algebra.
\codefieldsection{Protection}
Given an algebra \(\mathcal{A}\), \(\mathcal{A}\) is \textit{\(\epsilon\)-correctable} under noise channel \(\mathcal{N}\) if there exists some quantum channel \(\mathcal{R}\) such that
\flmMathEnvironment{align}{}{
  ||(\mathcal{R}\circ\mathcal{N})-P_{\mathcal{A}}||_{\diamond}\leq\epsilon~,
}
where \(P_{\mathcal{A}}\) is the projector onto algebra \(\mathcal{A}\) and we use the diamond norm \(\diamond\) \NoCaseChange{\protect\cite{cite2539}}.

Let the minimal error for some algebra \(\mathcal{A}\) under noise channel \(\mathcal{N}\) be
\flmMathEnvironment{align}{}{
  \epsilon_{\mathcal{A}}=\min_{\mathcal{R}} ||\mathcal{R}\circ\mathcal{N}-P_{\mathcal{A}}||_{\diamond}~.
}
Let \(\delta_{\mathcal{A}}=||\mathcal{N}^C-\mathcal{N}^C\circ P_{\mathcal{A}'}||_{\diamond}\)
for the commutant \(\mathcal{A}'\) of algebra \(\mathcal{A}\) and the \flmRefsHyperref{ref2540}{complementary channel} \(\mathcal{N}^C\) of noise channel \(\mathcal{N}\). Then \NoCaseChange{\protect\cite{cite2537}},
\flmMathEnvironment{align}{}{
  \delta_{\mathcal{A}}^2/4\leq \epsilon_{\mathcal{A}}\leq 2\delta_{\mathcal{A}}^{1/2}~.
}

\subsection{Complementary channel formulation}

Given the projector \(\mathcal{P}_{\mathcal{A}}\) onto algebra \(\mathcal{A}\) and a noise channel \(\mathcal{N}\),
we can quantify approximate operator algebra error correction using worst-case entanglement fidelity \(F\) as
\flmMathEnvironment{align}{}{F(R\mathcal{N},\mathcal{P}_{\mathcal{A}})\geq 1-\epsilon}
for some small \(\epsilon\) and recovery channel \(R\).

This is equivalent to considering the \flmRefsHyperref{ref2540}{complementary channel} \(\mathcal{N}^C\)
with projector \(\mathcal{P}_{\mathcal{A}'}\) onto the commutant \(\mathcal{A}'\) of \(\mathcal{A}\) such that
\flmMathEnvironment{align}{}{
  F(\mathcal{N}^C,R'\mathcal{P}_{\mathcal{A}'})\geq 1-\epsilon
}
for that same value of \(\epsilon\) and some channel \(R'\).

This formulation gives a necessary and sufficient condition for approximate operator-algebra QECCs.
It implies the standard operator-algebra correctability conditions, \([A,V^{\dagger}E_i^{\dagger}E_jV]=0\) for all \(A\in\mathcal{A}\), in the exact limit \NoCaseChange{\protect\cite{cite2538}}.
In particular, the same ideas have been extended to locality-restricted settings, where recovery operations must respect spatial or subsystem constraints (see \NoCaseChange{\protect\cite{cite2541}} for the generalization).
Private algebras are correctable algebras for the complementary channel \NoCaseChange{\protect\cite{cite2542}}.

\codefieldsection{Parent}
\begin{eczvaluelist}
\item\relax
\flmRefsHyperref[eczindexfamilyrel]{code:quantum_into_quantum}{Quantum code}\end{eczvaluelist}
\codefieldsection{Child}
\begin{eczvaluelist}
\item\relax
\flmRefsHyperref[eczindexfamilyrel]{code:approximate_qecc}{Approximate quantum error-correcting code (AQECC)}\end{eczvaluelist}
\codefieldsection{Cousins}
\begin{eczvaluelist}
\item\relax
\flmRefsHyperref[eczindexfamilyrel]{code:oaecc}{Operator-algebra QECC (OAQECC)} --- Approximate OAQECCs correcting a noise channel exactly reduce to OAQECCs.
\item\relax
\flmRefsHyperref[eczindexfamilyrel]{code:holographic}{Holographic code} --- Properties of holographic codes are often quantified in the Heisenberg picture, i.e., in terms of operator algebras \NoCaseChange{\protect\cite{cite2543,cite2544,cite2545,cite2546}}.
\end{eczvaluelist}
\eczhbkcontributors{ Milan Tenn, \eczhuVVA }
\endeczcode

\eczcode{approximate_qecc}{Approximate quantum error-correcting code (AQECC)}{~\NoCaseChange{\protect\cite{cite859,cite2125,cite2547,cite2548,cite2537,cite2538,cite2549}}}
\codefieldsection{Description}
Encodes quantum information so that it is possible to approximately recover that information from noise up to an error bound in recovery.

Many families of approximate block quantum codes become exact in the \(n\to\infty\) limit (see children).
More generally, codes that become exact for some parameter values are called \textit{quasi-exact} \NoCaseChange{\protect\cite{cite790}}.

\codefieldsection{Protection}
Many of the state fidelity conditions that hold exactly for \flmRefsHyperref{code:qecc_finite}{(exact) QECCs} can be shown to hold up to some error \(\epsilon\) for approximate QECCs.
Approximate correction has been formulated for certain types of correlated noise \NoCaseChange{\protect\cite{cite2550}}.

\subsection{Input-output fidelity}

This is the primary notion of closeness between states before and after error correction. 
It is defined as
\flmMathEnvironment{align}{}{
  \langle\psi|\mathcal{N}(|\psi\rangle\langle\psi|)|\psi\rangle~,
}
where \(\mathcal{N}\) is the combination of encoding, noise, and recovery channels.
This quantity is difficult to compute, and there exist other quantities that are easier to work with and that also become 1 in the limit of perfect error correction.

\subsection{Entanglement fidelity}

Let \(f(\rho_1,\rho_2)\) be the fidelity between quantum states.
Let the entanglement fidelity between channels \(\mathcal{N}\) and \(\mathcal{M}\) be defined as
\flmMathEnvironment{align}{}{
  F_{\rho}(\mathcal{N},\mathcal{M})
  = f( (\mathcal{N}\otimes\mathrm{id})\ket{\psi}\bra{\psi}, (\mathcal{M}\otimes\mathrm{id})\ket{\psi}\bra{\psi} )~,
}
where \(\ket{\psi}\) is a purification of the mixed state \(\rho\).
The worst-case entanglement fidelity is then defined as
\flmMathEnvironment{align}{}{
  F(\mathcal{N},\mathcal{M})=\min_{\rho} F_{\rho}(\mathcal{N},\mathcal{M})~.
}

Now, based on the Bures distance and worst-case entanglement fidelity, we define
\flmMathEnvironment{align}{}{
  d(\mathcal{N},\mathcal{M})=\sqrt{1-F(\mathcal{N},\mathcal{M})}
}
as a measure of distance between quantum channels~\NoCaseChange{\protect\cite{cite2538}}.

Given some encoding map \(\mathcal{U}\) and some noise channel \(\mathcal{E}\),
the code described by \(\mathcal{U}\) is \textit{\(\epsilon\)-correctable} if there exists
some quantum channel \(\mathcal{D}\) such that
\flmMathEnvironment{align}{}{
  d(\mathcal{D}\mathcal{E}\mathcal{U}(\rho),\rho)\leq \epsilon
}
for all logical states \(\rho\)~\NoCaseChange{\protect\cite{cite2538}}.
When \(\epsilon=0\) we can derive the standard \flmTerm{term}{ref1043}{}{Knill-Laflamme conditions} \NoCaseChange{\protect\cite{cite2551}}.

Upper and lower bounds based on the average entanglement fidelity can be derived \NoCaseChange{\protect\cite[{Eq. (10)}]{cite2538}\protect\cite[{Eqs. (138-139)}]{cite2552}\protect\cite[{Eq. (139)}]{cite2553}}.
Riemannian optimization techniques can be applied to design approximate QECCs since the set of unitary encoding maps \(U\) forms a Stiefel manifold \NoCaseChange{\protect\cite{cite2508}}.

\subsection{Complementary channel formulation}

Given a noise channel \(\mathcal{E}\), there exists a recovery channel \(\mathcal{D}\) such that
\(F(\mathcal{D}\mathcal{E},\mathrm{id})=1-\epsilon\) iff there exists some \(\mathcal{D}'\) such that
for \flmRefsHyperref{ref2540}{complementary channel} \(\mathcal{E}^C\), \(F(\mathcal{E}^C,\mathcal{D}')=1-\epsilon\)
and \(\mathcal{D}'(\rho)=\rho_0\) for some fixed \(\rho_0\).
Note that \(F\) denotes worst case entanglement fidelity between channels.

We can generalize this by replacing \(\mathrm{id}\) with some channel \(M\). Given noise channel \(\mathcal{E}\),
there exists a recovery channel \(\mathcal{D}\) such that
\(F(\mathcal{D}\mathcal{E},\mathcal{M})=1-\epsilon\) iff there exists some \(\mathcal{D}'\) such that
for \flmRefsHyperref{ref2540}{complementary channel} \(\mathcal{E}^C\), \(F(\mathcal{E}^C,\mathcal{D}'\mathcal{M}^C)=1-\epsilon\) \NoCaseChange{\protect\cite{cite2538}}.
If \(\mathcal{M}^C\) is a projection, this dual formulation yields the near-optimal estimate
\flmMathEnvironment{align}{}{
  \frac{1}{2}d(\mathcal{E}^C,\mathcal{E}^C\mathcal{M}^C)
  \leq \min_{\mathcal{D}}d(\mathcal{D}\mathcal{E},\mathcal{M})
  \leq d(\mathcal{E}^C,\mathcal{E}^C\mathcal{M}^C)~,
}
reducing the recovery problem to estimating the information leaked to the environment \NoCaseChange{\protect\cite{cite2538}}.

\subsection{Approximate error-correction conditions}
Analogously to the \flmTerm{term}{ref1043}{}{Knill-Laflamme conditions} for (exact) QECCs, there exist various formulations of necessary and sufficient conditions for approximate error correction to determine if some code is \(\epsilon\)-correctable under a noise channel.

Necessary and sufficient conditions for approximate error correction can also be expressed in terms of
\flmRefsHyperref{ref2540}{complementary channels}.
Given some code defined by projector \(\Pi = U U^\dagger\), \(\Pi\) is \textit{\(\epsilon\)-correctable}
with respect to some noise channel \(\mathcal{E}\) if
\flmMathEnvironment{align}{}{
  \Pi E_i^{\dagger}E_j \Pi =\lambda_{ij}\Pi +\Pi B_{ij}\Pi ~,
}
where \(\Lambda(\rho)=\mathrm{Tr}(\rho)\sum_{ij} \lambda_{ij}|i\rangle\langle j|\)
is a density operator,
\flmMathEnvironment{align}{}{
  (\Lambda+B)(\rho)=\Lambda(\rho)+\sum_{ij}\mathrm{Tr}(\rho B_{ij})|i\rangle\langle j|
}
is the output state of the complementary noise channel \(\mathcal{E}^C = \Lambda+B\), and the Bures distance \(d(\Lambda+B,\Lambda)\le\epsilon\)~\NoCaseChange{\protect\cite{cite2538}}.
An alternative measure, the \textit{AQEC relative entropy}, measures the relative entropy between \(\Lambda + B\) and \(\Lambda\) \NoCaseChange{\protect\cite{cite2554}}.
The non-correctable contributions \(B_{ij}\) can be arranged in a signature vector that is amenable to numerical optimization in the space of Stiefel manifolds \NoCaseChange{\protect\cite{cite2555,cite528}}. 
The Frobenius norm of the matrix \(B_{ij}\) bounds the difference between the two \flmRefsHyperref{ref672}{quantum weight enumerators} \NoCaseChange{\protect\cite{cite2556}}.

In addition to the necessary and sufficient error correction conditions,
there exist sufficient conditions for AQECCs.
Given a noise channel \(\mathcal{U}(\rho)=\sum_{n} A_n \rho A_n^{\dagger}\) where \(\forall{n}\), \(A_n\) is a Kraus operator, and code projector \(\Pi \), express the following using polar decomposition, \(A_n \Pi =U_n \sqrt{\Pi A_n^{\dagger}A_n \Pi }\), and let \(p_n\) and \(p_n\lambda_n\) be the largest and smallest eigenvalues for \(\Pi A_n^{\dagger}A_n \Pi \).
Then, we are guaranteed that if
\flmMathEnvironment{align}{}{\Pi U_m^{\dagger}U_n \Pi =\delta_{mn} \Pi  \land p_n(1-\lambda_n)\le O( f(\epsilon) )}
we have a fidelity \(F \geq 1-O( f(\epsilon) )\) after recovery~\NoCaseChange{\protect\cite{cite859}}.

\subsection{Universal subspace AQECCs and alpha-bits}

Universal subspace approximate error correction is a type of approximate error correction that quantifies protection of information stored in (strict) subspaces of a logical space.
See also formulations of error correction for subsets that are not necessarily subspaces \NoCaseChange{\protect\cite{cite2557}}.

Given a subspace of a Hilbert space \(\mathsf{S}\) of dimension \(d\), noise channel \(\mathcal{E}\), and encoding \(\mathcal{U}\), we define the subspace as an \textit{\(\alpha\)-dit} with error \(\epsilon\) if, for all subspaces \(\tilde{\mathsf{S}}\) of dimension less than or equal to \(d^{\alpha}+1\),
there exists some channel \(\tilde{\mathcal{D}}\) such that
\flmMathEnvironment{align}{}{||(\tilde{\mathcal{D}}\circ \mathcal{E}\circ \mathcal{U})|\psi\rangle-|\psi\rangle||_1\leq \epsilon}
for all \(|\psi\rangle\in \tilde{\mathsf{S}}\)~\NoCaseChange{\protect\cite{cite2549}}.

Generalizing the notion of quantum information transmission and capacity of \flmRefsHyperref{code:qecc_finite}{(exact) QECCs}, one can achieve an \(\alpha\)-bit transmission rate \(r\) for quantum channel \(\mathcal{E}\) iff,
for sufficiently large \(d\) and \(n\), and for all \(\epsilon>0\), the channel \(\mathcal{E}^{\otimes n}\)
is able to transmit
\flmMathEnvironment{align}{}{\left\lceil \frac{n r}{\log(d)} \right\rceil\quad \textup{\(\alpha\)-dits}}
with total error \(\epsilon\) across those \(\alpha\)-dits.
The \(\alpha\)-bit capacity \(Q\) of \(\mathcal{E}\)
is defined as the supremum of achievable transmission rates~\NoCaseChange{\protect\cite{cite2549}}.

\subsection{Other metrics of approximate error correction}

\begin{enumerate}[(1)]\item \begin{defterm}{Code space complexity}\label{ref2558}\label{ref2559}
One can relate robustness of an approximate quantum code to the quantum \textit{circuit complexity} \NoCaseChange{\protect\cite{cite2560,cite2561,cite2562,cite2563}} of creating states in the codespace.
For a family of block codes, scaling as \flmRefsHyperref{ref65}{order} \(O(k/n)\) of a code parameter called the \textit{subsystem variance} characterizes the transition between code subspaces with low and high circuit complexity \NoCaseChange{\protect\cite{cite2564}}.
The Lovasz local lemma yields a trade-off between circuit complexity and local indistinguishability \NoCaseChange{\protect\cite{cite2565}}.
\end{defterm}
\item \textit{Integrity} measures how well a code state \(\psi\) and an orthogonal state can be distinguished in trace distance after noise and recovery are applied \NoCaseChange{\protect\cite{cite2566}}.
\end{enumerate}

\codefieldsection{Rate}
An extension of the \flmRefsHyperref{ref487}{BPT bound} to approximate codes is done in Ref. \NoCaseChange{\protect\cite{cite2567}}.
\codefieldsection{Encoding}
\begin{eczvaluelist}
\item\relax Given a decoder, an encoding that yields the optimal entanglement fidelity can be obtained by solving a semi-definite program \NoCaseChange{\protect\cite{cite2568,cite2547,cite2569,cite2570}}).
\item\relax Variational quantum circuit encoder \NoCaseChange{\protect\cite{cite2571}}.
\end{eczvaluelist}
\codefieldsection{Decoding}
\begin{eczvaluelist}
\item\relax Given an encoding and a noise channel, a decoder that yields the optimal entanglement fidelity can be obtained by solving a semi-definite program \NoCaseChange{\protect\cite{cite2568,cite2547,cite2572,cite2569,cite2570}}. This optimal decoder is robust to unexpected variations in the noise channel \NoCaseChange{\protect\cite{cite2573}}.
\item\relax The \textit{decoupling approach} a.k.a. the \textit{Uhlmann decoder} \NoCaseChange{\protect\cite{cite2574,cite2575,cite2576}}.
\item\relax Quantum machine-learning based decoders such as quantum convolutional neural networks \NoCaseChange{\protect\cite{cite2577}} and quantum autoencoders \NoCaseChange{\protect\cite{cite2578}}.
\item\relax The \textit{Leung recovery map} \NoCaseChange{\protect\cite{cite859}} for a noise channel whose Kraus operators \(E_j\) yield a diagonal QEC matrix, \(c_{ij}\propto\delta_{ij}\), has Kraus operators \(\Pi V_j^{\dagger}\), where \(\Pi\) is the codespace projection, and where \(V_j\) is the unitary from the polar decomposition of \(E_j \Pi\). This is the recovery used in the proof of the Knill-Laflamme conditions \NoCaseChange{\protect\cite[{Thm. 10.1}]{cite2579}}.
\item\relax A near-optimal recovery channel can be constructed from a saddle point in the complementary-channel optimization, achieving recovery error \(d(\widetilde{\mathcal{D}}\mathcal{E},\mathcal{M})\leq d(\mathcal{E}^C,\mathcal{E}^C\mathcal{M}^C)\) whenever \(\mathcal{M}^C\) is a projection \NoCaseChange{\protect\cite{cite2538}}.
\item\relax The \textit{Cafaro recovery map} \NoCaseChange{\protect\cite{cite2580}} can be obtained for noise Kraus operators if there exists a basis of error words with respect to which the uncorrectable piece in the Knill-Laflamme conditions is diagonal; see Ref. \NoCaseChange{\protect\cite{cite2581}}. The map recovers information perfectly for strictly correctable noise.
\item\relax The \textit{Petz recovery map} a.k.a. the \textit{transpose map} \NoCaseChange{\protect\cite{cite2582,cite2583,cite2584}}, a quantum channel determined by the codespace and noise channel, yields an infidelity of recovery that is at most twice away from the infidelity of the best possible recovery \NoCaseChange{\protect\cite{cite2585}}.  The fidelity can be expressed exactly as a function of the \flmTerm{term}{ref1043}{}{Knill-Laflamme conditions} \NoCaseChange{\protect\cite[{Thm. 1}]{cite2586}}, and it can be used to derive a generalization of the \flmTerm{term}{ref1043}{}{Knill-Laflamme conditions} for approximate QECCs \NoCaseChange{\protect\cite{cite2587,cite2588}}. Satisfaction of the \flmTerm{term}{ref1043}{}{Knill-Laflamme conditions} is sufficient but not necessary for the Petz recovery map to be the optimal recovery, and a necessary and sufficient condition has been derived \NoCaseChange{\protect\cite{cite2589}}. The infidelity of a modified Petz recovery map under erasure can be bounded using the conditional mutual information via the \textit{approximate Petz theorem} \NoCaseChange{\protect\cite{cite2590,cite2591,cite2567}}. In the case of topological codes, the Petz infidelity is related to the topological entanglement entropy \NoCaseChange{\protect\cite{cite2592}}. Modifications include the Petz-like decoder \NoCaseChange{\protect\cite{cite2593}}, the temporal Petz recovery map for dynamical codes \NoCaseChange{\protect\cite{cite2594}}, and a syndrome-based Petz recovery \NoCaseChange{\protect\cite{cite2595}}. The Petz map is related to quantum Bayes' rule \NoCaseChange{\protect\cite{cite2596}}.
\item\relax The Yoshida-Kitaev decoder for the Hayden-Preskill protocol \NoCaseChange{\protect\cite{cite2597}} can be extended to general QECCs \NoCaseChange{\protect\cite{cite2593}}.
\item\relax If parts of the \flmTerm{term}{ref1043}{}{Knill-Laflamme conditions} are violated, a deterministic recovery operation is not possible. However, a probabilistic recovery and a modified version of the conditions can still be constructed \NoCaseChange{\protect\cite{cite2598}}.
\end{eczvaluelist}
\codefieldsection{Notes}
\begin{eczvaluelist}
\item\relax See review \NoCaseChange{\protect\cite{cite2599}}.
\end{eczvaluelist}
\codefieldsection{Parent}
\begin{eczvaluelist}
\item\relax
\flmRefsHyperref[eczindexfamilyrel]{code:approximate_oaecc}{Approximate operator-algebra QECC}\end{eczvaluelist}
\codefieldsection{Children}
\begin{eczvaluelist}
\item\relax
\flmRefsHyperref[eczindexfamilyrel]{code:ampdamp}{Amplitude-damping (AD) code} --- Protection against \flmRefsHyperref{ref498}{AD} noise is typically approximate because the tensor product of Kraus operators with all \(\ell=0\) is typically corrected only up to some order in \(\gamma\) \NoCaseChange{\protect\cite{cite859,cite2600}}.
\item\relax
\flmRefsHyperref[eczindexfamilyrel]{code:rg_cat}{Renormalization group (RG) cat code} --- RG cat codes approximately protect against displacements that represent ultraviolet coherent operators.
\item\relax
\flmRefsHyperref[eczindexfamilyrel]{code:nonabelian_covariant_erasure}{\(U(d)\)-covariant approximate erasure code} --- Any finite-dimensional code covariant w.r.t. a continuous group has to be approximate because of the \flmRefsHyperref{ref721}{Eastin-Knill theorem}.
\item\relax
\flmRefsHyperref[eczindexfamilyrel]{code:w_state}{W-state code} --- The W-state code approximately protects against a single erasure while allowing for a universal transversal set of gates.
\item\relax
\flmRefsHyperref[eczindexfamilyrel]{code:quantum_random}{Random quantum code} --- Random codes typically correct errors on average.
\item\relax
\flmRefsHyperref[eczindexfamilyrel]{code:cft}{Conformal-field theory (CFT) code}\item\relax
\flmRefsHyperref[eczindexfamilyrel]{code:syk}{SYK code} --- SYK codes are approximately error correcting in that they satisfy certain error-correction conditions based on mutual information \NoCaseChange{\protect\cite{cite2601}}.
\item\relax
\flmRefsHyperref[eczindexfamilyrel]{code:circuit_to_hamiltonian}{Circuit-to-Hamiltonian approximate code}\item\relax
\flmRefsHyperref[eczindexfamilyrel]{code:eth}{Eigenstate thermalization hypothesis (ETH) code} --- ETH codes approximately protect against erasures in the thermodynamic limit. There is a link between ETH and approximate QEC, with fluctuations of the infinite-time average of certain observables expressible in terms of code error \NoCaseChange{\protect\cite{cite2602}}.
\item\relax
\flmRefsHyperref[eczindexfamilyrel]{code:reinforcement_learning}{Reinforcement-learning quantum code}\item\relax
\flmRefsHyperref[eczindexfamilyrel]{code:quantum_singleton}{Singleton-bound approaching AQECC}\item\relax
\flmRefsHyperref[eczindexfamilyrel]{code:mps}{Magnon code} --- Magnon codes approximately protect against erasures in the thermodynamic limit.
\item\relax
\flmRefsHyperref[eczindexfamilyrel]{code:vbs}{Valence-bond-solid (VBS) code} --- VBS codes approximately protect against erasures in the thermodynamic limit.
\item\relax
\flmRefsHyperref[eczindexfamilyrel]{code:landau_level}{Landau-level spin code} --- The Landau-level spin code approximately protects against rotational errors.
\end{eczvaluelist}
\codefieldsection{Cousins}
\begin{eczvaluelist}
\item\relax
\flmRefsHyperref[eczindexfamilyrel]{code:topological}{Topological code} --- In the case of topological codes, the Petz infidelity is related to the topological entanglement entropy \NoCaseChange{\protect\cite{cite2592}}.
\item\relax
\flmRefsHyperref[eczindexfamilyrel]{code:spt}{Symmetry-protected topological (SPT) code} --- Certain phases with continuous symmetries cannot be prepared using a constant-depth circuit, a consequence of the Lieb-Schult-Mattis theorem \NoCaseChange{\protect\cite{cite2603,cite2604,cite2605,cite2606}}. The theorem, in turn, can be linked to the circuit complexity of the underlying approximate error-correcting code \NoCaseChange{\protect\cite{cite2565}}.
\item\relax
\flmRefsHyperref[eczindexfamilyrel]{code:stiefel}{Stiefel code} --- Riemannian optimization techniques can be applied to design approximate QECCs since the set of unitary encoding maps \(U\) forms a Stiefel manifold \NoCaseChange{\protect\cite{cite2508}}.
\item\relax
\flmRefsHyperref[eczindexfamilyrel]{code:qlwc}{Quantum low-weight check (QLWC) code} --- A family of approximate non-stabilizer qubit QLWC codes with linear distance and rate has been constructed \NoCaseChange{\protect\cite{cite1633}} using unary codes that arise from the Feynman-Kitaev clock construction \NoCaseChange{\protect\cite{cite1634}}.
\item\relax
\flmRefsHyperref[eczindexfamilyrel]{code:numopt}{Numerically optimized bosonic code} --- Numerically optimized codes arising from optimization routines are often approximate QECCs.
\item\relax
\flmRefsHyperref[eczindexfamilyrel]{code:qudits_into_oscillators}{Qudit-into-oscillator code} --- Approximate QEC techniques of finding the entanglement fidelity can be adapted to bosonic codes with a finite-dimensional codespace \NoCaseChange{\protect\cite{cite496}}.
\item\relax
\flmRefsHyperref[eczindexfamilyrel]{code:gkp}{Square-lattice GKP code} --- Square-lattice GKP codes approximately protect against \flmRefsHyperref{ref498}{photon loss} \NoCaseChange{\protect\cite{cite2607,cite496,cite2608}}.
\item\relax
\flmRefsHyperref[eczindexfamilyrel]{code:covariant}{Covariant block quantum code} --- Normalizable constructions of infinite-dimensional \(G\)-covariant codes for continuous \(G\) are approximately error-correcting.
\item\relax
\flmRefsHyperref[eczindexfamilyrel]{code:holographic}{Holographic code} --- Universal subspace approximate error correction is used to model black holes \NoCaseChange{\protect\cite{cite2609}}.
\item\relax
\flmRefsHyperref[eczindexfamilyrel]{code:qecc}{Quantum error-correcting code (QECC)} --- QAECCs correcting a noise channel exactly reduce to QECCs.
\item\relax
\flmRefsHyperref[eczindexfamilyrel]{code:da}{Dynamical code} --- Approximate versions of dynamical codes have been formulated \NoCaseChange{\protect\cite{cite2594}}.
\item\relax
\flmRefsHyperref[eczindexfamilyrel]{code:quantum_secret_sharing}{Approximate secret-sharing code} --- Secret-sharing codes approximately correct errors on up to \(\lfloor (n-1)/2 \rfloor\) errors.
\end{eczvaluelist}
\eczhbkcontributors{ Milan Tenn, Manasi Shingane, \eczhuPhF, \eczhuVVA }
\endeczcode

\eczcode{asymmetric_qecc}{Asymmetric quantum code (AQC)}{~\NoCaseChange{\protect\cite{cite861,cite2610}}}
\codefieldsection{Alternative Names}
\begin{eczvaluelist}
\item\relax Noise-biased quantum code
\end{eczvaluelist}
\eczhIndexCodeAliasName{asymmetric_qecc}{Noise-biased quantum code}
\codefieldsection{Description}
Quantum systems can be roughly characterized by two types of noise, a bit-flip noise that maps canonical basis states into each other, and a phase-flip noise that induces relative phases between superpositions of such basis states.
A code cannot protect against both types of noise arbitrarily well, and there is a tradeoff between the two types of protection.
An AQC is one that performs much better against one type of noise than the other type.
Such codes typically have tunable distances against each noise type and include CSS codes, GKP codes, and QSCs.

\codefieldsection{Protection}
Noise channels for which one type of noise is more prominent than the other are called \textit{asymmetric-noise channels} or \textit{biased-noise channels}.
An example of a noise-biased channel is a Pauli channel of independent \(X\) and \(Z\)-type noise with \(p_X \gg p_Z\) or vice versa.

In the context of comparing weight as well as of determining distances for noise models biased toward \(X\)- or \(Z\)-type errors, an extended notation for asymmetric CSS block quantum codes is \(\llbracket n,k,(d_X,d_Z)\rrbracket \) or \(\llbracket n,k,d_X/d_Z\rrbracket \), where \(d_{X,Z}\) are the \(X\)- and \(Z\)-distances, respectively \NoCaseChange{\protect\cite{cite444}}.
An asymmetric Singleton bound and linear programming bounds for asymmetric CSS codes have been formulated \NoCaseChange{\protect\cite{cite1354}}, as well as asymmetric quantum GV bounds \NoCaseChange{\protect\cite{cite2611}} and Hamming and Singleton bounds for general asymmetric subsystem codes \NoCaseChange{\protect\cite{cite2612}}.
Asymmetric MDS codes have been characterized \NoCaseChange{\protect\cite{cite2612,cite2613}}.

\codefieldsection{Gates}
\begin{eczvaluelist}
\item\relax Taking into account noise bias can reduce resource overhead in magic-state distillation schemes \NoCaseChange{\protect\cite{cite2614}}.
\item\relax A CNOT gate continuously connected to the identity cannot be noise-bias-preserving in finite dimensions \NoCaseChange{\protect\cite{cite2615}\protect\cite[{Appx. A}]{cite2616}}.
\item\relax Qubit gates that preserve noise bias (after but not during each gate) correspond to permutations in the Pauli-\(X\) basis \NoCaseChange{\protect\cite{cite2617}}. There is a bias-preserving Hadamard test \NoCaseChange{\protect\cite{cite2617}} (see also Ref. \NoCaseChange{\protect\cite{cite2618}}).
\end{eczvaluelist}
\codefieldsection{Decoding}
\begin{eczvaluelist}
\item\relax Measurement-free error correction has been optimized for biased noise \NoCaseChange{\protect\cite{cite2619}}.
\end{eczvaluelist}
\codefieldsection{Fault Tolerance}
\begin{eczvaluelist}
\item\relax Fault-tolerant noise-bias-preserving computation scheme \NoCaseChange{\protect\cite{cite2615}}.
\item\relax Fault-tolerant circuits converting between asymmetric and symmetric subsystem codes \NoCaseChange{\protect\cite{cite2620,cite2621}}.
\end{eczvaluelist}
\codefieldsection{Threshold}
\begin{eczvaluelist}
\item\relax A lower bound on \flmRefsHyperref{ref515}{concatenated thresholds} with CSS codes under biased noise \NoCaseChange{\protect\cite{cite2622}}.
\end{eczvaluelist}
\codefieldsection{Notes}
\begin{eczvaluelist}
\item\relax See Ref. \NoCaseChange{\protect\cite{cite2623}} for a brief review of asymmetric quantum codes.
\end{eczvaluelist}
\codefieldsection{Parent}
\begin{eczvaluelist}
\item\relax
\flmRefsHyperref[eczindexfamilyrel]{code:qecc}{Quantum error-correcting code (QECC)}\end{eczvaluelist}
\codefieldsection{Cousins}
\begin{eczvaluelist}
\item\relax
\flmRefsHyperref[eczindexfamilyrel]{code:distance_balanced}{Distance-balanced code} --- Distance balancing is a procedure that can convert an asymmetric CSS code into a less asymmetric one.
\item\relax
\flmRefsHyperref[eczindexfamilyrel]{code:subsystem_surface}{Subsystem surface code} --- Subsystem surface codes perform well against biased circuit-level noise \NoCaseChange{\protect\cite{cite2624}}.
\item\relax
\flmRefsHyperref[eczindexfamilyrel]{code:clifford-deformed_surface}{Clifford-deformed surface code (CDSC)} --- Random Clifford deformation can improve performance of surface codes against biased noise \NoCaseChange{\protect\cite{cite2625,cite2626}}.
\item\relax
\flmRefsHyperref[eczindexfamilyrel]{code:xysurface}{XY surface code} --- XY surface codes perform well against biased noise \NoCaseChange{\protect\cite{cite2627}}.
\item\relax
\flmRefsHyperref[eczindexfamilyrel]{code:xyz_product}{XYZ product code} --- XYZ product codes can be used to protect against biased noise \NoCaseChange{\protect\cite{cite2628}}.
\item\relax
\flmRefsHyperref[eczindexfamilyrel]{code:xyz_color}{XYZ color code} --- XYZ color codes perform well against biased noise \NoCaseChange{\protect\cite{cite2629}}.
\item\relax
\flmRefsHyperref[eczindexfamilyrel]{code:twisted_xzzx}{Twisted XZZX toric code} --- Twisted XZZX codes perform well against biased noise \NoCaseChange{\protect\cite{cite2630,cite2631,cite2632}}; see also Ref. \NoCaseChange{\protect\cite{cite2633}}.
\item\relax
\flmRefsHyperref[eczindexfamilyrel]{code:xzzx}{XZZX surface code} --- The XZZX surface code can be foliated for a noise-bias preserving MBQC \NoCaseChange{\protect\cite{cite2634}} or FBQC \NoCaseChange{\protect\cite{cite2635}} protocol; see also \NoCaseChange{\protect\cite{cite2636}}.
\item\relax
\flmRefsHyperref[eczindexfamilyrel]{code:concatenated_steane}{Concatenated Steane code} --- Concatenating while taking into account noise bias can reduce resource overhead \NoCaseChange{\protect\cite{cite2621}}.
\item\relax
\flmRefsHyperref[eczindexfamilyrel]{code:quantum_mds}{Quantum maximum-distance-separable (MDS) code} --- An asymmetric Singleton bound and linear programming bounds for asymmetric CSS codes have been formulated  \NoCaseChange{\protect\cite{cite1354}}. Asymmetric MDS codes have been characterized \NoCaseChange{\protect\cite{cite2613}}.
\item\relax
\flmRefsHyperref[eczindexfamilyrel]{code:two_dimensional_hyperbolic_surface}{2D hyperbolic surface code} --- Asymmetric 2D hyperbolic surface codes have been constructed \NoCaseChange{\protect\cite{cite2637}}.
\item\relax
\flmRefsHyperref[eczindexfamilyrel]{code:surface}{Kitaev surface code} --- The surface code on the honeycomb tiling is an asymmetric CSS code \NoCaseChange{\protect\cite{cite2637}}.
\item\relax
\flmRefsHyperref[eczindexfamilyrel]{code:css}{Calderbank-Shor-Steane (CSS) stabilizer code} --- In the context of comparing weight as well as of determining distances for noise models biased toward \(X\)- or \(Z\)-type errors, an extended notation for asymmetric CSS block quantum codes is \(\llbracket n,k,(d_X,d_Z),w\rrbracket \) or \(\llbracket n,k,d_X/d_Z,w\rrbracket \).
\item\relax
\flmRefsHyperref[eczindexfamilyrel]{code:galois_css}{Galois-qudit CSS code} --- Most known Galois-qudit AQC families are derived from the asymmetric Galois-qudit CSS construction \NoCaseChange{\protect\cite[{Thm. 27.5.2}]{cite2024}}, and assuming the MDS conjecture, all possible parameters for \flmRefsHyperref{ref672}{pure} Galois-qudit CSS asymmetric MDS codes have been determined \NoCaseChange{\protect\cite[{Thm. 27.5.3}]{cite2024}}.
\item\relax
\flmRefsHyperref[eczindexfamilyrel]{code:quantum_reed_muller}{Quantum Reed-Muller (RM) code} --- Asymmetric quantum RM codes have been constructed \NoCaseChange{\protect\cite[{Lemma 4.1}]{cite1354}}.
\item\relax
\flmRefsHyperref[eczindexfamilyrel]{code:pg_ldpc}{Finite-geometry LDPC (FG-LDPC) code} --- FG-LDPC codes can be used to construct asymmetric CSS codes \NoCaseChange{\protect\cite{cite1355}\protect\cite[{Lemma 4.1}]{cite1354}}.
\item\relax
\flmRefsHyperref[eczindexfamilyrel]{code:quantum_hermitian_ag}{Quantum Hermitian AG code} --- One-point and two-point Hermitian codes can be used to construct asymmetric Galois-qudit CSS codes, and the two-point construction can improve on the corresponding one-point codes \NoCaseChange{\protect\cite{cite870}}.
\item\relax
\flmRefsHyperref[eczindexfamilyrel]{code:galois_polynomial}{Galois-qudit RS code} --- Asymmetric Galois-qudit RS codes have been constructed \NoCaseChange{\protect\cite{cite2638,cite2639}\protect\cite[{Sec. 17.3}]{cite872}}.
\item\relax
\flmRefsHyperref[eczindexfamilyrel]{code:bacon_shor}{Bacon-Shor code} --- Bacon-Shor code parameters against bit- and phase-noise can be optimized by changing the block geometry, yielding good performance against biased noise \NoCaseChange{\protect\cite{cite2640}}. A fault-tolerant teleportation-based computation scheme for asymmetric Bacon-Shor codes is effective against highly biased noise \NoCaseChange{\protect\cite{cite2641}}.
\item\relax
\flmRefsHyperref[eczindexfamilyrel]{code:gkp}{Square-lattice GKP code} --- GKP code parameters against position and momentum displacements can be tuned by the choice of lattice (e.g., square vs rectangular).
\item\relax
\flmRefsHyperref[eczindexfamilyrel]{code:binomial}{Binomial code} --- Binomial code parameters against loss/gain errors and dephasing can be tuned.
\item\relax
\flmRefsHyperref[eczindexfamilyrel]{code:qsc}{Quantum spherical code (QSC)} --- QSC code parameters against loss/gain errors and Gaussian rotations can be tuned.
\item\relax
\flmRefsHyperref[eczindexfamilyrel]{code:quantum_parity}{Quantum parity code (QPC)} --- QPC parameters against bit- and phase-noise can be tuned.
\item\relax
\flmRefsHyperref[eczindexfamilyrel]{code:eastab}{EA qubit stabilizer code} --- Entanglement can help decode asymmetric quantum codes \NoCaseChange{\protect\cite{cite2642}}.
\item\relax
\flmRefsHyperref[eczindexfamilyrel]{code:floquet}{Hastings-Haah Floquet code} --- Floquet codes can be adapted for asymmetric noise \NoCaseChange{\protect\cite{cite2643}}.
\item\relax
\flmRefsHyperref[eczindexfamilyrel]{code:quantum_cyclic}{Cyclic quantum code} --- Cyclic quantum codes can be adapted for asymmetric noise \NoCaseChange{\protect\cite{cite2644}}.
\item\relax
\flmRefsHyperref[eczindexfamilyrel]{code:xyz_hexagonal}{XYZ\(^2\) hexagonal stabilizer code} --- The XYZ\(^2\) hexagonal stabilizer code has high thresholds under biased noise \NoCaseChange{\protect\cite{cite2645}}.
\item\relax
\flmRefsHyperref[eczindexfamilyrel]{code:3d_stabilizer}{3D lattice stabilizer code} --- Applying Clifford deformations to various 3D stabilizer codes, including the 3D surface code, 3D color code, X-cube model code, and Sierpinski prism model code, yields a \(50\%\) code capacity threshold under infinitely biased Pauli noise \NoCaseChange{\protect\cite{cite2626}}.
\item\relax
\flmRefsHyperref[eczindexfamilyrel]{code:two-legged-cat}{Two-component cat code} --- Cat qubits provide an asymmetric-noise platform admitting bias-preserving \(X\), CNOT, and Toffoli gates \NoCaseChange{\protect\cite{cite2616,cite2646}}. A bias-preserving SWAP gate has also been proposed \NoCaseChange{\protect\cite{cite2647}}.
\item\relax
\flmRefsHyperref[eczindexfamilyrel]{code:quantum_lego}{Tensor-network code} --- Quantum Lego and more general tensor-network code optimization for biased noise can be done using reinforcement learning \NoCaseChange{\protect\cite{cite2648,cite2649}}.
\item\relax
\flmRefsHyperref[eczindexfamilyrel]{code:multisector_hypergraph}{Higher-dimensional homological product code} --- The \((a,b)\)-complex construction yields asymmetric CSS codes with \(d_X=\delta^a\) and \(d_Z=\delta^b\), allowing independent tuning of the two distances for channels with unequal bit-flip and phase-flip rates \NoCaseChange{\protect\cite{cite1613}}.
\item\relax
\flmRefsHyperref[eczindexfamilyrel]{code:qldpc}{Qubit QLDPC code} --- There are recipes to determine transversal gates for asymmetric qubit QLDPC codes \NoCaseChange{\protect\cite{cite771}}.
\item\relax
\flmRefsHyperref[eczindexfamilyrel]{code:compass_model}{Compass code} --- Families of random compass codes perform well against biased noise and spatially dependent (i.e., asymmetric) noise \NoCaseChange{\protect\cite{cite2650}}.
Clifford deformation can enhance the performance of compass codes against biased noise \NoCaseChange{\protect\cite{cite2651}}.

\item\relax
\flmRefsHyperref[eczindexfamilyrel]{code:galois_bch}{Galois-qudit BCH code} --- Asymmetric quantum BCH codes have been constructed \NoCaseChange{\protect\cite{cite2610,cite2652,cite2653}\protect\cite[{Lemma 4.4}]{cite1354}\protect\cite[{Sec. 17.3}]{cite872}}, including subsystem BCH codes \NoCaseChange{\protect\cite{cite2612}\protect\cite[{Sec. 9.3}]{cite872}}.
\item\relax
\flmRefsHyperref[eczindexfamilyrel]{code:abelian_lifted_product}{Abelian LP code} --- The Abelian LP construction has been adapted to accommodate noise bias, yielding bias-tailored LP codes \NoCaseChange{\protect\cite{cite2654}}.
\end{eczvaluelist}
\eczhbkcontributors{ \eczhuVVA }
\endeczcode

\eczcode{block_quantum}{Block quantum code}{}
\codefieldsection{Description}
An encoding of quantum information into a subspace of a \textit{multi-partite} (a.k.a. \textit{many-body}) quantum system, a physical space consisting of a tensor product of \(n > 1\) identical factors.
Each factor is referred to as a \textit{subsystem}, \textit{register}, \textit{site}, \textit{party}, or \textit{body}, depending on context.
The subsystems include qubits, modular qudits, Galois qudits, oscillators, groups, or categories.
For finite dimensional codes, the dimension of the underlying subsystem is denoted by \(q\) and is sometimes called the \textit{local dimension}.

While codewords \(c\) of block codes are elements of \(\Sigma^n\) for some alphabet \(\Sigma\), quantum states of block quantum codes are \(L^2\)-normalizable functions on \(\Sigma^n\).
Put differently, the configuration space of the canonical (a.k.a. computational) basis states \(|c\rangle\) of an \(n\)-body quantum system is the classical \(n\)-coordinate alphabet \(\Sigma^n \ni c\).

\codefieldsection{Protection}
A block quantum code over a finite alphabet \(\Sigma\) with \textit{distance} \(d\) detects errors acting on up to \(d-1\) subsystems, corrects erasure errors on up to \(d-1\) subsystems, or corrects errors acting on up to \(\lfloor (d-1)/2 \rfloor\) subsystems.
The subsystems that are erased are known to the receiver, and erasures of subsystems at unknown locations are called \textit{deletion errors} \NoCaseChange{\protect\cite{cite2655,cite2656,cite2657,cite2658}}.
More general forms of noise are caused by \textit{insertion errors} \NoCaseChange{\protect\cite{cite2655,cite2656,cite2657,cite2658}}, where subsystems are inserted into the block, and \textit{synchronization errors} (a.k.a. misalignment) \NoCaseChange{\protect\cite{cite1246}}, where the code block is misplaced in a larger block by one or more locations.
There are relations between deletion and insertion errors \NoCaseChange{\protect\cite{cite2659,cite2660}}.

The \textit{weight} of an operator on a tensor-product Hilbert space is the number of subsystems on which the operator acts non-trivially.
For example, an operator acting on two subsystems is called a weight-two operator or a two-body operator.

General noise models for block codes include \textit{stochastic noise}, in which every possible error is assigned a probability.
In the case of \textit{local stochastic noise}, the probability decreases rapidly (typically, exponentially) with the number of subsystems that an error acts on.
On the other hand, the \textit{adversarial noise} model consists of errors acting on at most a fixed number of subsystems.
Independent channels that are close to the identity can be approximated by adversarial \(t\)-subsystem error maps with an error that is exponentially small in \(t+1\), motivating the study of \(t\)-subsystem errors even when the physical noise is memoryless \NoCaseChange{\protect\cite[{Ch. 1}]{cite398}}.
Errors acting on subsystems in a geometrically local region are called \textit{burst errors} \NoCaseChange{\protect\cite{cite2661,cite2662}}.

\subsection{Bounds on code parameters}
Bounds on finite dimensional block code performance include the quantum Singleton bound, quantum Hamming bound, \flmRefsHyperref{ref1729}{quantum GV bound}, various quantum linear programming (LP) bounds \NoCaseChange{\protect\cite{cite2663,cite2664}} (see the book \NoCaseChange{\protect\cite{cite398}}), and other bounds \NoCaseChange{\protect\cite{cite2665,cite2666}}.
A code whose parameters attain the quantum Hamming bound (quantum Singleton bound) is called a perfect quantum code (a quantum MDS code).
We are often interested in how parameters of particular infinite block quantum code families scale with increasing block length \(n\), necessitating the use of \flmRefsHyperref{ref65}{asymptotic notation}.
A code family is called \textit{good} when its rate \(k/n\) and relative error-correction capability \(t/n\), equivalently relative distance up to constant factors, remain bounded away from zero as \(n\to\infty\) \NoCaseChange{\protect\cite[{Ch. 2}]{cite398}}.

\begin{defterm}{Quantum GV bound}\label{ref2667}\label{ref1729}
The \flmRefsHyperref{ref1729}{quantum GV bound} \NoCaseChange{\protect\cite{cite2668}} (see also Refs. \NoCaseChange{\protect\cite{cite2669,cite2670,cite2671,cite696,cite2672}}) for Galois qudits states that a \flmRefsHyperref{ref672}{pure} \(\llbracket n,k,d\rrbracket _q\) Galois-qudit stabilizer code exists if 
\flmMathEnvironment{align}{}{
  \frac{q^{n-k+2}-1}{q^{2}-1}>\sum_{j=1}^{d-1}(q^{2}-1)^{j-1}\binom{n}{j}~.
}
The bound gives rise to the \textit{asymptotic quantum GV bound} (i.e., quantum GV bound in the \(n\to\infty\) limit), expressed in terms of the maximum achievable rate \(R\) and relative distance \(\delta\),
\flmMathEnvironment{align}{}{
  R\geq 1-\delta\log_q(q+1) - h_{q}(\delta)~,
}
where \(h_q\) is the \flmRefsHyperref{ref85}{\(q\)-ary entropy function}.
\end{defterm}

\codefieldsection{Transversal and Permutation-Based Gates}
\begin{eczvaluelist}
\item\relax \begin{defterm}{Eastin-Knill theorem}\label{ref720}\label{ref721} \textit{Transversal gates} are logical gates on block codes that can be realized as tensor products of unitary operations acting on subsets of subsystems whose size is independent of \(n\). For subsets of size one, gates are sometimes called \textit{strongly transversal} if the single-subsystem unitaries are identical, and \textit{weakly transversal} otherwise. A universal gate set for a finite-dimensional block quantum code cannot be transversal for any code that detects single-block errors due to the Eastin-Knill theorem \NoCaseChange{\protect\cite{cite722}}. \end{defterm}
\item\relax A qudit code of length \(n\) with permutation automorphism subgroups \(N\triangleleft G\leq \mathrm{PAut}(Q)\) and simple non-Abelian quotient \(G/N\) must satisfy \(n\geq \mu(G/N)\) \NoCaseChange{\protect\cite[{Thm. S5}]{cite723}}.
\end{eczvaluelist}
\codefieldsection{Notes}
\begin{eczvaluelist}
\item\relax Tables of linear-programming upper bounds on general block quantum codes for various \(n\), \(k\), and \(q\), based on algorithms developed in Refs. \NoCaseChange{\protect\cite{cite2673,cite2674}}, are maintained by M. Grassl at this \flmHref{https://www.codetables.de/}{website}. A Magma implementation exists at this \flmHref{https://magma.maths.usyd.edu.au/magma/handbook/text/1976}{website}.
\item\relax States of block quantum codes can be classified in terms of the complexity of their underlying encoding circuit; see the Complexity Zoo Exhibit on Classes of Quantum States and Probability Distributions \NoCaseChange{\protect\cite{cite4}}.
\end{eczvaluelist}
\codefieldsection{Parent}
\begin{eczvaluelist}
\item\relax
\flmRefsHyperref[eczindexfamilyrel]{code:qecc}{Quantum error-correcting code (QECC)}\end{eczvaluelist}
\codefieldsection{Children}
\begin{eczvaluelist}
\item\relax
\flmRefsHyperref[eczindexfamilyrel]{code:oscillators}{Bosonic code} --- Bosonic codes are block quantum codes with \(\Sigma=\mathbb{R}\).
\item\relax
\flmRefsHyperref[eczindexfamilyrel]{code:covariant}{Covariant block quantum code} --- Covariant codes for \(n>1\) are block quantum codes.
\item\relax
\flmRefsHyperref[eczindexfamilyrel]{code:dynamic_gen}{Dynamically generated QECC}\item\relax
\flmRefsHyperref[eczindexfamilyrel]{code:quantum_locally_recoverable}{Quantum locally recoverable code (QLRC)}\item\relax
\flmRefsHyperref[eczindexfamilyrel]{code:quantum_mds}{Quantum maximum-distance-separable (MDS) code}\item\relax
\flmRefsHyperref[eczindexfamilyrel]{code:single_shot}{Single-shot code}\item\relax
\flmRefsHyperref[eczindexfamilyrel]{code:small_distance_quantum}{Small-distance block quantum code}\item\relax
\flmRefsHyperref[eczindexfamilyrel]{code:quantum_quasi_cyclic}{Quasi-cyclic quantum code}\item\relax
\flmRefsHyperref[eczindexfamilyrel]{code:quantum_lego}{Tensor-network code}\item\relax
\flmRefsHyperref[eczindexfamilyrel]{code:topological}{Topological code} --- Topological codes are block codes because an infinite family of tensor-product Hilbert spaces is required to formally define a phase of matter.
\item\relax
\flmRefsHyperref[eczindexfamilyrel]{code:qltc}{Quantum locally testable code (QLTC)}\item\relax
\flmRefsHyperref[eczindexfamilyrel]{code:qudits_into_qudits}{Modular-qudit code} --- Modular-qudit codes are block quantum codes with \(\Sigma=\mathbb{Z}_q\).
\item\relax
\flmRefsHyperref[eczindexfamilyrel]{code:galois_into_galois}{Galois-qudit code} --- Galois-qudit codes are block quantum codes with \(\Sigma=\mathbb{F}_q\).
\end{eczvaluelist}
\codefieldsection{Cousins}
\begin{eczvaluelist}
\item\relax
\flmRefsHyperref[eczindexfamilyrel]{code:single_subsystem}{Monolithic quantum code} --- Block quantum codes for \(n=1\) are monolithic codes.
\item\relax
\flmRefsHyperref[eczindexfamilyrel]{code:block}{Block code} --- Block quantum codes are quantum analogues of block codes.
\end{eczvaluelist}
\eczhbkcontributors{ \eczhuVVA }
\endeczcode

\eczcode{commuting_projector}{Commuting-projector Hamiltonian code}{}
\codefieldsection{Description}
Hamiltonian-based code whose Hamiltonian terms can be expressed as orthogonal projectors (i.e., Hermitian operators with eigenvalues 0 or 1) that commute with each other.

\codefieldsection{Protection}
Geometrically local commuting-projector code Hamiltonians on Euclidean manifolds are stable with respect to small perturbations when they satisfy the \flmRefsHyperref{ref2675}{TQO conditions}, meaning that a notion of a phase can be defined \NoCaseChange{\protect\cite{cite2676,cite2677,cite2678,cite2679}}.
This notion can be extended to semi-hyperbolic manifolds \NoCaseChange{\protect\cite{cite2680}} and non-geometrically local QLDPC codes exhibiting check soundness \NoCaseChange{\protect\cite{cite2681}} (see also \NoCaseChange{\protect\cite{cite2682}}).
Hamiltonians satisfying a Peierls condition are stable to off-diagonal perturbations \NoCaseChange{\protect\cite{cite2683}}.

2D topological order on qubit manifolds requires weight-four (four-body) commuting-projector Hamiltonian terms, i.e., it cannot be stabilized via weight-two (two-body) or weight-three (three-body) terms on nearly Euclidean geometries of qubits or qutrits \NoCaseChange{\protect\cite{cite2684,cite2685,cite2686}}.

Ground-state spaces of commuting-projector Hamiltonians with weight-two (two-body) terms cannot be used to suppress errors in adiabatic quantum computation \NoCaseChange{\protect\cite{cite2687}}, but this can be circumvented with excited-state subspaces \NoCaseChange{\protect\cite{cite2688}} or ground-state subspaces of subsystem code Hamiltonians, e.g., using BBS codes \NoCaseChange{\protect\cite{cite670,cite2689}}.
No eigenspace of a weight-two commuting-projector Hamiltonian can simultaneously have \(d > 2\) and dimension greater than 1 \NoCaseChange{\protect\cite{cite2688}}.

\codefieldsection{Parent}
\begin{eczvaluelist}
\item\relax
\flmRefsHyperref[eczindexfamilyrel]{code:hamiltonian}{Hamiltonian-based code} --- Geometrically local commuting-projector code Hamiltonians on Euclidean manifolds are stable with respect to small perturbations when they satisfy the \flmRefsHyperref{ref2675}{TQO conditions}, meaning that a notion of a phase can be defined \NoCaseChange{\protect\cite{cite2676,cite2677,cite2678,cite2679}}. This notion can be extended to semi-hyperbolic manifolds \NoCaseChange{\protect\cite{cite2680}} and non-geometrically local QLDPC codes exhibiting check soundness \NoCaseChange{\protect\cite{cite2681}} (see also \NoCaseChange{\protect\cite{cite2682}}). Hamiltonians admitting a Peierls condition are stable to off-diagonal perturbations \NoCaseChange{\protect\cite{cite2683}}.
\end{eczvaluelist}
\codefieldsection{Children}
\begin{eczvaluelist}
\item\relax
\flmRefsHyperref[eczindexfamilyrel]{code:cage_net}{Cage-net code} --- Cage-net codewords form ground-state subspaces of frustration-free commuting projector Hamiltonians.
\item\relax
\flmRefsHyperref[eczindexfamilyrel]{code:yetter_gauge_theory}{Two-gauge theory code} --- Two-gauge theory codewords span ground-state subspaces of frustration-free commuting-projector Hamiltonians.
\item\relax
\flmRefsHyperref[eczindexfamilyrel]{code:enriched_string_net}{Multi-fusion string-net code} --- Multi-fusion string-net codes form eigenspaces of frustration-free commuting projector Hamiltonians.
\item\relax
\flmRefsHyperref[eczindexfamilyrel]{code:enriched_walker_wang}{\(G\)-enriched Walker-Wang model code} --- \(G\)-enriched Walker-Wang model codewords form ground-state subspaces of frustration-free commuting projector Hamiltonians.
\item\relax
\flmRefsHyperref[eczindexfamilyrel]{code:stabilizer}{Stabilizer code} --- Codespace is the ground-state space of the \textit{code Hamiltonian}, which consists of an equal linear combination of stabilizer generators and which can be made into a frustration-free commuting-projector Hamiltonian.
\item\relax
\flmRefsHyperref[eczindexfamilyrel]{code:qltc}{Quantum locally testable code (QLTC)} --- Quantum LTC codespaces are ground-state spaces of \(u\)-local frustration-free commuting-projector Hamiltonians.
\end{eczvaluelist}
\codefieldsection{Cousins}
\begin{eczvaluelist}
\item\relax
\flmRefsHyperref[eczindexfamilyrel]{code:topological}{Topological code} --- Geometrically local commuting-projector code Hamiltonians on Euclidean manifolds are stable with respect to small perturbations when they satisfy the \flmRefsHyperref{ref2675}{TQO conditions}, meaning that a notion of a phase can be defined \NoCaseChange{\protect\cite{cite2676,cite2677,cite2678,cite2679}}. This notion can be extended to semi-hyperbolic manifolds \NoCaseChange{\protect\cite{cite2680}} and non-geometrically local QLDPC codes exhibiting check soundness \NoCaseChange{\protect\cite{cite2681}} (see also \NoCaseChange{\protect\cite{cite2682}}). Hamiltonians admitting a Peierls condition are stable to off-diagonal perturbations \NoCaseChange{\protect\cite{cite2683}}. 2D states admitting strict area-law entanglement necessarily have commuting-projector Hamiltonians \NoCaseChange{\protect\cite{cite2690}}. 2D topological order on qubit manifolds requires weight-four (four-body) commuting-projector Hamiltonian terms, i.e., it cannot be stabilized via weight-two (two-body) or weight-three (three-body) terms on nearly Euclidean geometries of qubits or qutrits \NoCaseChange{\protect\cite{cite2684,cite2685,cite2686}}.
\item\relax
\flmRefsHyperref[eczindexfamilyrel]{code:qubit_stabilizer}{Qubit stabilizer code} --- Qubit stabilizer codes are \textit{infectious}: if the ground-state subspace of an \(\ell\)-local commuting projector Hamiltonian contains a state close to a stabilizer state, then the entire ground-state space is close to a stabilizer code \NoCaseChange{\protect\cite{cite2691,cite2692}}.
\item\relax
\flmRefsHyperref[eczindexfamilyrel]{code:qubit_subsystem_stabilizer}{Subsystem qubit stabilizer code} --- Ground-state spaces of qubit commuting-projector Hamiltonians with weight-two (two-body) terms cannot be used to suppress errors in adiabatic quantum computation \NoCaseChange{\protect\cite{cite2687}}, but this can be circumvented with excited-state subspaces \NoCaseChange{\protect\cite{cite2688}} or ground-state subspaces of subsystem code Hamiltonians, e.g., using BBS codes \NoCaseChange{\protect\cite{cite670,cite2689}}.
\item\relax
\flmRefsHyperref[eczindexfamilyrel]{code:bravyi_bacon_shor}{Bravyi-Bacon-Shor (BBS) code} --- Ground-state spaces of qubit commuting-projector Hamiltonians with weight-two (two-body) terms cannot be used to suppress errors in adiabatic quantum computation \NoCaseChange{\protect\cite{cite2687}}, but this can be circumvented with excited-state subspaces \NoCaseChange{\protect\cite{cite2688}} or ground-state subspaces of subsystem code Hamiltonians, e.g., using BBS codes \NoCaseChange{\protect\cite{cite670,cite2689}}.
\item\relax
\flmRefsHyperref[eczindexfamilyrel]{code:binary_linear}{Linear binary code} --- Parity-check constraints defining a binary linear code can be encoded into a classical Ising model Hamiltonian, a commuting-projector model whose terms contain products of Pauli \(Z\) matrices participating in each parity check. Such Ising models are also frustration-free since the codewords satisfy all parity checks.
\item\relax
\flmRefsHyperref[eczindexfamilyrel]{code:classical_fractal_liquid}{Classical fractal liquid code} --- Classical fractal liquid codewords form the ground-state space of a class of exactly solvable spin-glass Ising models with three-body interactions.
\item\relax
\flmRefsHyperref[eczindexfamilyrel]{code:tqd_abelian}{Abelian TQD code} --- Many Abelian TQD code Hamiltonians were originally formulated as commuting-projector models \NoCaseChange{\protect\cite{cite2693}}.
\item\relax
\flmRefsHyperref[eczindexfamilyrel]{code:frustration_free}{Frustration-free Hamiltonian code} --- Frustration-free Hamiltonians can contain non-commuting projectors; an example is the AKLT model \NoCaseChange{\protect\cite{cite626}}. On the other hand, commuting-projector Hamiltonians can be frustrated; an example is the 1D classical Ising model on a circle for odd \(n\) with one two-body interaction having the opposite sign.
\item\relax
\flmRefsHyperref[eczindexfamilyrel]{code:double_semion_string_net}{Double-semion string-net code} --- A commuting-projector version of the double-semion string-net code can also be derived \NoCaseChange{\protect\cite{cite2694,cite2693}}.
\item\relax
\flmRefsHyperref[eczindexfamilyrel]{code:qldpc}{Qubit QLDPC code} --- Qubit QLDPC codes with check soundness, meaning that every weight-\(m\) stabilizer can be written as a product of \flmRefsHyperref{ref65}{order} \(O(m)\) stabilizer generators, are robust against few-body perturbations. This means that phases of matter can be defined from certain non-geometrically local QLDPC code Hamiltonians \NoCaseChange{\protect\cite{cite2681}}.
\end{eczvaluelist}
\eczhbkcontributors{ \eczhuVVA }
\endeczcode

\eczcode{quantum_concatenated}{Concatenated quantum code}{~\NoCaseChange{\protect\cite{cite2695}}}
\codefieldsection{Description}
A combination of two quantum codes, an inner code \(C_{\text{in}}\) and an outer code \(C_{\text{out}}\), where the physical subspace used for the inner code consists of the logical subspace of the outer code.
In other words, one first encodes in the inner code, and then encodes each of its physical registers in the outer code.
An inner \(C_{\text{in}} = \llparenthesis n_1,K,d_1\rrparenthesis _{q_1}\) and outer \(C_{\text{out}} = \llparenthesis n_2,q_1,d_2\rrparenthesis _{q_2}\) block quantum code yield an \(\llparenthesis n_1 n_2, K, d \geq d_1d_2\rrparenthesis _{q_2}\) concatenated block quantum code \NoCaseChange{\protect\cite{cite398}}.

More generally, one can concatenate in blocks: several physical registers of the inner code can be grouped together so that each such block carries one logical register of the outer code.
Encoding then proceeds by first applying the inner encoding to each block and then applying the outer encoding across the resulting blocks.

Concatenating an \(\llparenthesis n,q,d\rrparenthesis _q\) block quantum code can be done recursively, with the \(r\)\textit{th level} of concatenation yielding an \(\llparenthesis n^r,q,d^r\rrparenthesis _q\) code.

Other ways to combine quantum codes include pasting \NoCaseChange{\protect\cite{cite1694}}, and generalizations of concatenation exist \NoCaseChange{\protect\cite{cite2696,cite2697}}.

\codefieldsection{Encoding}
\begin{eczvaluelist}
\item\relax Standard encoding proceeds by first encoding into the inner code and then encoding each physical register of the inner code into the outer code.
\end{eczvaluelist}
\codefieldsection{Decoding}
\begin{eczvaluelist}
\item\relax Standard decoding proceeds in the reverse order: first decode the outer code blocks and then use the resulting data to decode the inner code.
\item\relax Maximum-likelihood decoding can be formulated as contraction of a tree tensor network, yielding exact decoders that improve on minimum-distance decoding for recursively concatenated block codes \NoCaseChange{\protect\cite{cite2698,cite400}}.
\end{eczvaluelist}
\codefieldsection{Fault Tolerance}
\begin{eczvaluelist}
\item\relax Recursive concatenation is the standard route to threshold-theorem constructions: if one level of a fault-tolerant simulation suppresses logical error below the physical error rate, then further concatenation suppresses it rapidly, yielding the family of protocols used in polylogarithmic-overhead threshold theorems \NoCaseChange{\protect\cite[{Chs. 10 and 14}]{cite398}}.
\end{eczvaluelist}
\codefieldsection{Notes}
\begin{eczvaluelist}
\item\relax See the book \NoCaseChange{\protect\cite{cite398}} for an introduction.
\end{eczvaluelist}
\codefieldsection{Parent}
\begin{eczvaluelist}
\item\relax
\flmRefsHyperref[eczindexfamilyrel]{code:quantum_lego}{Tensor-network code} --- Encoders for concatenated quantum codes correspond to tree tensor networks \NoCaseChange{\protect\cite{cite400}}.
\end{eczvaluelist}
\codefieldsection{Children}
\begin{eczvaluelist}
\item\relax
\flmRefsHyperref[eczindexfamilyrel]{code:group_quantum_parity}{Group-based QPC} --- A group-based QPC is a concatenation of a phase-flip group-based repetition code with a bit-flip group-based repetition code.
\item\relax
\flmRefsHyperref[eczindexfamilyrel]{code:oscillators_concatenated}{Concatenated bosonic code} --- A concatenated bosonic code is a bosonic code that can be thought of as a concatenation of a possibly non-bosonic inner code and a bosonic outer code.
\item\relax
\flmRefsHyperref[eczindexfamilyrel]{code:qubit_concatenated}{Concatenated qubit code}\end{eczvaluelist}
\codefieldsection{Cousins}
\begin{eczvaluelist}
\item\relax
\flmRefsHyperref[eczindexfamilyrel]{code:concatenated}{Concatenated code} --- Quantum codes can be concatenated with classical codes to yield good quantum codes \NoCaseChange{\protect\cite{cite971}}.
\item\relax
\flmRefsHyperref[eczindexfamilyrel]{code:concatenated_c-q}{Concatenated c-q code} --- Concatenated c-q codes are hybrid classical-into-quantum analogues of concatenated quantum codes.
\item\relax
\flmRefsHyperref[eczindexfamilyrel]{code:rotor_gkp}{Rotor GKP code} --- The rotor GKP code can be thought of as a concatenation of a homological rotor code and a modular-qudit GKP code \NoCaseChange{\protect\cite[{Fig. 3}]{cite2699}}.
\item\relax
\flmRefsHyperref[eczindexfamilyrel]{code:group_4_2_2}{\(\llbracket 4,2,2\rrbracket _{G}\) four group-qudit code} --- The \(|\overline{g_1=1,g_2}\rangle\) \(\llbracket 4,1,2\rrbracket _{G}\) subcode is the smallest group-based QPC, i.e., a concatenation of a bit-flip with a phase-flip group-based repetition code for that group.
\item\relax
\flmRefsHyperref[eczindexfamilyrel]{code:self_correct}{Self-correcting quantum code} --- A concatenated quantum code with self-simulating control elements based on work by Gacs \NoCaseChange{\protect\cite{cite1596,cite1597,cite1598,cite1599}} yields a self-correcting quantum memory in 2D \NoCaseChange{\protect\cite{cite2700}}.
\item\relax
\flmRefsHyperref[eczindexfamilyrel]{code:cws}{Codeword stabilized (CWS) code} --- CWS codes can be concatenated by applying generalized local complementation to their underlying graphs \NoCaseChange{\protect\cite{cite2701}}.
\item\relax
\flmRefsHyperref[eczindexfamilyrel]{code:qudit_cws}{Modular-qudit CWS code} --- Generalized concatenations of modular-qudit CWS codes yield a family of codes that have larger logical dimension than stabilizer codes and that asymptotically approach the modular-qudit Hamming bound \NoCaseChange{\protect\cite{cite2696}}.
\item\relax
\flmRefsHyperref[eczindexfamilyrel]{code:stab_9_1_3}{\(\llbracket 9,1,3\rrbracket _{\mathbb{Z}_q}\) modular-qudit code} --- The \(\llbracket 9,1,3\rrbracket _{\mathbb{Z}_q}\) modular-qudit code is a concatenation of a bit-flip with a phase-flip group repetition code for \(G=\mathbb{Z}_q\).
\item\relax
\flmRefsHyperref[eczindexfamilyrel]{code:ea_galois_stabilizer}{EA Galois-qudit stabilizer code} --- Concatenated EA Galois-qudit stabilizer codes have been studied \NoCaseChange{\protect\cite{cite2702,cite2703}}.
\item\relax
\flmRefsHyperref[eczindexfamilyrel]{code:galois_grs}{Galois-qudit GRS code} --- Concatenations of Galois-qudit GRS codes and random stabilizer codes can achieve the \flmRefsHyperref{ref1729}{quantum GV bound} \NoCaseChange{\protect\cite{cite2704}}.
\item\relax
\flmRefsHyperref[eczindexfamilyrel]{code:galois_polynomial}{Galois-qudit RS code} --- Recursive concatenations of quantum RS codes can be asymptotically good \NoCaseChange{\protect\cite{cite2705}}.
\end{eczvaluelist}
\eczhbkcontributors{ \eczhuVVA }
\endeczcode

\eczcode{constant_excitation}{Constant-excitation (CE) code}{~\NoCaseChange{\protect\cite{cite2706,cite2707,cite2708}}}
\codefieldsection{Description}
Code whose codewords lie in an eigenspace of fixed total energy or fixed total excitation number for the underlying quantum system.
For qubit codes, such a Hamiltonian is often the \textit{total spin Hamiltonian}, \(H=\sum_i Z_i\).
For spin-\(S\) codes, this generalizes to \(H=\sum_i J_z^{(i)}\), where \(J_z\) is the spin-\(S\) \(Z\)-operator.
For bosonic (and, similarly, for fermion) codes, such as Fock-state codes, codewords are often in an eigenspace with eigenvalue \(N>0\) of the \textit{total excitation} or \textit{energy Hamiltonian}, \(H=\sum_i \hat{n}_i\).

\codefieldsection{Protection}
CE codewords have to lie in the same excitation subspace in order to protect against changes in the total excitation number.

Fock-state CE codes lie in the constant-excitation Fock space and are in one-to-one correspondence with points on the \flmRefsHyperref{ref655}{discrete simplex}.
They are protected from identical \flmRefsHyperref{ref498}{AD} acting on all modes because the damping acts on all codewords identically \NoCaseChange{\protect\cite{cite859,cite2600}}.
The all-zero \flmRefsHyperref{ref498}{AD} Kraus operator acts identically on every state and so can be exactly correctable in the case of Fock-state CE codes.
For example, this operator's acting on a Fock state \(|\boldsymbol{m}\rangle\) depends only on the total occupation number \(|\boldsymbol{m}|=\sum_j m_j\) and not on the individual occupation numbers \(m_j\),
\flmMathEnvironment{align}{}{
  E_{0}^{\otimes n}|\boldsymbol{m}\rangle=\left(1-\gamma\right)^{|\boldsymbol{m}|/2}|\boldsymbol{m}\rangle~.
}
This effect extends to the damping portion, \(\left(1-\gamma\right)^{\hat{n}/2}\), of any \(\ell\neq 0\) \flmRefsHyperref{ref498}{AD} Kraus operators.

In similar fashion, qubit CE codes are protected from coherent noise in the form of transversal \(Z\)-rotations because such rotations act identically on all codewords \NoCaseChange{\protect\cite{cite2709,cite808}}.
In the case of CSS codes, all codes oblivious to such rotations are CE codes \NoCaseChange{\protect\cite{cite2709,cite808}}.
Stabilizer codes can be extended to codes that are protected against such coherent noise via an enlargement procedure \NoCaseChange{\protect\cite{cite808}}.

\codefieldsection{Rate}
Fock-state CE codes can be used in a protocol that achieves the two-way quantum capacity of the \flmRefsHyperref{ref498}{AD} Gaussian channel \NoCaseChange{\protect\cite{cite2710}}. For every \(K,t \geq 2\), there are explicitly constructible \(K\)-dimensional Fock-state CE codes with \(q=N=(K-1)t(t+1)\) modes, total excitation \(N\), and distance \(t+1\); there also exist families with logical dimension \(K = o(2^N)\) and distance of \flmRefsHyperref{ref65}{order} \(o(N/\log N)\) \NoCaseChange{\protect\cite{cite500}}.
\codefieldsection{Fault Tolerance}
\begin{eczvaluelist}
\item\relax Fault-tolerant QEC framework for CE CSS codes using modified Shor and Steane syndrome extraction, where weight-\(2w\) stabilizers are measured using \(w\)-CE cat states and zero-controlled NOT (\(\mathrm{C}_0 X\)) gates replace standard CNOT gates to preserve the constant-excitation structure \NoCaseChange{\protect\cite{cite524}}.
\end{eczvaluelist}
\codefieldsection{Parent}
\begin{eczvaluelist}
\item\relax
\flmRefsHyperref[eczindexfamilyrel]{code:hamiltonian}{Hamiltonian-based code} --- Constant-excitation codes are associated with a Hamiltonian governing the total excitations of the system.
\end{eczvaluelist}
\codefieldsection{Children}
\begin{eczvaluelist}
\item\relax
\flmRefsHyperref[eczindexfamilyrel]{code:chuang-leung-yamamoto}{Chuang-Leung-Yamamoto (CLY) code} --- Chuang-Leung-Yamamoto codewords are constructed out of Fock states with the same total excitation number.
\item\relax
\flmRefsHyperref[eczindexfamilyrel]{code:constant_excitation_permutation_invariant}{Ouyang-Chao constant-excitation PI code}\item\relax
\flmRefsHyperref[eczindexfamilyrel]{code:icosahedral_fock}{Icosahedral Fock-state code}\item\relax
\flmRefsHyperref[eczindexfamilyrel]{code:very-small-logical-qubit}{Very small logical qubit (VSLQ) code}\item\relax
\flmRefsHyperref[eczindexfamilyrel]{code:jump}{Jump code}\item\relax
\flmRefsHyperref[eczindexfamilyrel]{code:css_12_1_3}{\(\llbracket 12,1,3\rrbracket \) CE CSS code}\item\relax
\flmRefsHyperref[eczindexfamilyrel]{code:phantom_14_3_3}{\(\llbracket 14,3,3\rrbracket \) CE phantom code} --- This code is single-qubit Clifford equivalent to the \(\llbracket 14,3,3\rrbracket \) CE CSS code obtained by dual-rail concatenation of the \(\llbracket 7,3,2\rrbracket \) punctured hypercube code \NoCaseChange{\protect\cite{cite524}}.
\item\relax
\flmRefsHyperref[eczindexfamilyrel]{code:qubit_8_1_3}{\(\llparenthesis 8,2,3\rrparenthesis \) Plenio-Vedral-Knight CE code}\end{eczvaluelist}
\codefieldsection{Cousins}
\begin{eczvaluelist}
\item\relax
\flmRefsHyperref[eczindexfamilyrel]{code:ampdamp}{Amplitude-damping (AD) code} --- Fock-state and qubit CE codes exactly protect against the \flmRefsHyperref{ref498}{AD} Kraus operator \(E_{0}^{\otimes n}\) because it acts identically on all Fock (and qubit) states with the same excitation number \NoCaseChange{\protect\cite{cite859,cite2600}}.
\item\relax
\flmRefsHyperref[eczindexfamilyrel]{code:qubit_css}{Qubit CSS code} --- Qubit CE codes are protected from coherent noise in the form of transversal \(Z\)-rotations because such rotations act identically on all codewords \NoCaseChange{\protect\cite{cite2709,cite808}}.
In the case of qubit CSS codes, all codes oblivious to such rotations are CE codes \NoCaseChange{\protect\cite{cite2709,cite808}}.
Any \(\llbracket n,k,d\rrbracket \) CSS code can be made into an \(\llbracket mn,k,>d\rrbracket \) CE code \NoCaseChange{\protect\cite{cite2709}}.
Concatenating the dual-rail code with an inner \(\llbracket n,k,d\rrbracket \) qubit stabilizer code yields a degenerate \(\llbracket 2n,k,d\rrbracket \) constant-excitation stabilizer code that avoids coherent phase errors and is equivalent to a Pauli-rotated repetition-concatenated stabilizer code \NoCaseChange{\protect\cite{cite2711}}. CSS structure is preserved when the original code is CSS \NoCaseChange{\protect\cite{cite524}}.

\item\relax
\flmRefsHyperref[eczindexfamilyrel]{code:stab_5_1_3}{\(\llbracket 5,1,3\rrbracket \) Five-qubit perfect code} --- The five-qubit code can be concatenated with a particular decoherence-free subspace (DFS) \NoCaseChange{\protect\cite{cite2712,cite2713,cite2714,cite2715}} to yield a 20-qubit CE code \NoCaseChange{\protect\cite{cite2708,cite2716}}. Dual-rail concatenation of the five-qubit code yields a \(\llbracket 10,1,3\rrbracket \) CE stabilizer code \NoCaseChange{\protect\cite{cite524}}.
\item\relax
\flmRefsHyperref[eczindexfamilyrel]{code:qubit_stabilizer}{Qubit stabilizer code} --- Concatenating the dual-rail code with an inner \(\llbracket n,k,d\rrbracket \) qubit stabilizer code yields a degenerate \(\llbracket 2n,k,d\rrbracket \) constant-excitation stabilizer code that avoids coherent phase errors and is equivalent to a Pauli-rotated repetition-concatenated stabilizer code \NoCaseChange{\protect\cite{cite2711}}. CSS structure is preserved when the original code is CSS \NoCaseChange{\protect\cite{cite524}}.
\item\relax
\flmRefsHyperref[eczindexfamilyrel]{code:q-ary_constant_weight}{Constant-weight block code} --- Constant-weight codes are classical analogues of qubit constant-excitation codes.
\item\relax
\flmRefsHyperref[eczindexfamilyrel]{code:permutation_invariant}{Permutation-invariant (PI) code} --- Modular-qudit PI codes can be converted to constant-excitation Fock-state codes via the \flmRefsHyperref{ref499}{simplex mapping} \NoCaseChange{\protect\cite[{Prop. V.2}]{cite500}}. Any transversal gates are mapped to Gaussian gates on the Fock-state codes \NoCaseChange{\protect\cite{cite500}}.
\item\relax
\flmRefsHyperref[eczindexfamilyrel]{code:fermions}{Fermion code} --- Fermion codewords lying in a fixed fermion-number subspace have to lie in the same subspace in order to protect against changes in fermion number \NoCaseChange{\protect\cite{cite559}}.
\item\relax
\flmRefsHyperref[eczindexfamilyrel]{code:quantum_parity}{Quantum parity code (QPC)} --- QPCs for even \(m_1\) can be made into CE codes by a Pauli transformation (e.g., \(XIXI\cdots XI\)) applied to each block of \(m_1\) qubits.
\end{eczvaluelist}
\eczhbkcontributors{ Yinchen Liu, \eczhuVVA }
\endeczcode

\eczcode{covariant}{Covariant block quantum code}{~\NoCaseChange{\protect\cite{cite2514}}}
\codefieldsection{Alternative Names}
\begin{eczvaluelist}
\item\relax Equivariant block quantum code
\end{eczvaluelist}
\eczhIndexCodeAliasName{covariant}{Equivariant block quantum code}
\codefieldsection{Description}
A block code on \(n\) subsystems that admits a group \(G\) of transversal gates. The group has to be finite for finite-dimensional codes due to the \flmRefsHyperref{ref721}{Eastin-Knill theorem}.
Continuous-\(G\) covariant codes, necessarily infinite-dimensional, are relevant to error correction of quantum reference frames \NoCaseChange{\protect\cite{cite2514}} and error-corrected parameter estimation.

Denoting the code's encoding map as \(U\), covariance is equivalent to
\flmMathEnvironment{align}{}{
  \left(\bigotimes_{j=1}^{n}V_{j}\left(g\right)\right)U=UV_{L}\left(g\right)\quad\quad\forall g\in G\,,
}
where \(V_j(g)\) is a unitary representation of \(g\) acting on the \(j\) subsystem, and \(V_L\) is a unitary representation acting on the unencoded logical information.
In this way, covariant encoding maps are equivariant (i.e., commute) with group actions on the logical and physical spaces.

Almost always, the physical representation is defined to be the transversal one (with respect to some tensor-product decomposition), but can reduce to any representation when the code is a subspace of a larger space that is not expressed as a tensor product (\(n=1\)). More generally, a code is sometimes said to be \textit{time-covariant} if it admits a continuous-parameter \(U(1)\) family of gates, not necessarily transversal \NoCaseChange{\protect\cite{cite2717}}.

\codefieldsection{Protection}
Finite-dimensional codes correcting a single-subsystem erasure and admitting a continuous-parameter family of transversal gates (assuming \(n>1\)) cannot exist in finite
dimensions due to the \flmRefsHyperref{ref721}{Eastin-Knill theorem}. As a result, there is generally a tradeoff between covariance and error correction.

Exact error-correcting \(G\)-covariant codes can exist in infinite dimensions, but their codewords are non-normalizable, meaning that approximate constructions have to be considered that are only approximately error correcting.
On the other hand, there exist exact error-correcting codes in finite dimensions that are approximately covariant \NoCaseChange{\protect\cite{cite2718,cite2719}}.
Various bounds quantify the covariance-performance tradeoff \NoCaseChange{\protect\cite{cite2720,cite2721,cite2722,cite2718,cite2719,cite2723,cite2724,cite2565}}.
In particular, an approximate Eastin-Knill theorem implies that an \(SU(d_{L})\)-covariant code satisfies \(\epsilon_{\mathrm{worst}}\gtrsim [2n \max_{i}\ln d_{i}]^{-1}\), so for fixed \(n\) the local subsystem dimensions must grow exponentially in \(1/\epsilon_{\mathrm{worst}}\) to support a universal transversal gate set \NoCaseChange{\protect\cite[{Thm. 4}]{cite2720}}.

\codefieldsection{Transversal and Permutation-Based Gates}
\begin{eczvaluelist}
\item\relax \(G\)-covariant codes defined on a tensor product space consisting of \(n\) subsystems are equivalent to codes with a transversal gate set realizing \(G\).
\end{eczvaluelist}
\codefieldsection{Parent}
\begin{eczvaluelist}
\item\relax
\flmRefsHyperref[eczindexfamilyrel]{code:block_quantum}{Block quantum code} --- Covariant codes for \(n>1\) are block quantum codes.
\end{eczvaluelist}
\codefieldsection{Children}
\begin{eczvaluelist}
\item\relax
\flmRefsHyperref[eczindexfamilyrel]{code:group_4_2_2}{\(\llbracket 4,2,2\rrbracket _{G}\) four group-qudit code} --- The four group-qudit code is \((G\times G)\)-covariant, with transversal logical left and right multiplication gates \NoCaseChange{\protect\cite[{Sec. VIII.B}]{cite2720}}.
\item\relax
\flmRefsHyperref[eczindexfamilyrel]{code:g_covariant_erasure}{\(G\)-covariant erasure code} --- In a proof of principle demonstration, error-correcting codes that are finite-\(G\) covariant can be constructed from a base encoding \(U_0\).
\item\relax
\flmRefsHyperref[eczindexfamilyrel]{code:nonabelian_covariant_erasure}{\(U(d)\)-covariant approximate erasure code}\item\relax
\flmRefsHyperref[eczindexfamilyrel]{code:w_state}{W-state code} --- The W-state code approximately protects against a single erasure while allowing for a universal transversal set of gates.
\item\relax
\flmRefsHyperref[eczindexfamilyrel]{code:group_representation}{Group-representation code} --- Group-representation code projections are onto a single irrep of a subgroup of canonical or distinguished unitary operations on a Hilbert space. This makes them covariant w.r.t. that subgroup. More general covariant codes need not be projections onto a single irrep. Removing the restriction to distinguished operations and allowing all operations, every code projection on an \(N\)-dim Hilbert space can be expressed as a projection onto the irrep formed by the code-preserving subgroup of \(U(N)\). The same idea holds when \(N\) is taken to infinity. In other words, while all codes are covariant w.r.t. some group, group-representation codes are covariant w.r.t. a canonical or distinguished subgroup.
\end{eczvaluelist}
\codefieldsection{Cousins}
\begin{eczvaluelist}
\item\relax
\flmRefsHyperref[eczindexfamilyrel]{code:approximate_qecc}{Approximate quantum error-correcting code (AQECC)} --- Normalizable constructions of infinite-dimensional \(G\)-covariant codes for continuous \(G\) are approximately error-correcting.
\item\relax
\flmRefsHyperref[eczindexfamilyrel]{code:quantum_reed_muller}{Quantum Reed-Muller (RM) code} --- Quantum RM codes are approximately covariant and nearly saturate certain covariance-performance bounds \NoCaseChange{\protect\cite{cite2718}}.
\item\relax
\flmRefsHyperref[eczindexfamilyrel]{code:eth}{Eigenstate thermalization hypothesis (ETH) code} --- ETH codes consisting of \flmRefsHyperref{ref526}{Dicke states} are approximately \(U(1)\)-covariant and nearly saturate certain covariance-performance bounds \NoCaseChange{\protect\cite{cite2720,cite2718}}.
\item\relax
\flmRefsHyperref[eczindexfamilyrel]{code:quantum_random}{Random quantum code} --- Random \(U(1)\)-covariant \NoCaseChange{\protect\cite{cite2725}} and \(U(d)\)-covariant \NoCaseChange{\protect\cite{cite2720,cite2726}} approximate QECCs exist.
\item\relax
\flmRefsHyperref[eczindexfamilyrel]{code:group_gkp}{Group GKP code} --- Group-GKP codes corresponding to the \(G^{k_1} \subseteq G^{k_2} \subset G^{n}\) group construction admit \(X\)-type logical group-multiplication gates, and are thus covariant with respect to the induced \(G^{k_2}\)-action \NoCaseChange{\protect\cite{cite735}}.
\item\relax
\flmRefsHyperref[eczindexfamilyrel]{code:rotor_3_1_2}{\(\llbracket 3,1,2\rrbracket _{\mathbb{Z}}\) Three-rotor code} --- The three-rotor code is \(U(1)\)-covariant.
\item\relax
\flmRefsHyperref[eczindexfamilyrel]{code:rotor_5_1_3}{\(\llbracket 5,1,3\rrbracket _{\mathbb{Z}}\) Five-rotor code} --- The five-rotor code is \(U(1)\)-covariant.
\item\relax
\flmRefsHyperref[eczindexfamilyrel]{code:ame}{Perfect-tensor code} --- An \(SU(2)\)-invariant three-qudit perfect tensor exists \NoCaseChange{\protect\cite{cite2727}}, but invariant perfect tensors do not exist on four parties \NoCaseChange{\protect\cite{cite2728,cite2727,cite2729}}.
\item\relax
\flmRefsHyperref[eczindexfamilyrel]{code:metrological}{Metrological code} --- Any time-covariant QECC, i.e., a code admitting a continuous-parameter \(U(1)\) family of gates, is automatically a metrological code.
\item\relax
\flmRefsHyperref[eczindexfamilyrel]{code:vbs}{Valence-bond-solid (VBS) code} --- Two classes of (approximate) VBS codes have \(SU(q)\) transversal gates, i.e., are \(SU(q)\)-covariant \NoCaseChange{\protect\cite[{Tab. III}]{cite790}}.
\end{eczvaluelist}
\eczhbkcontributors{ Jack Davis, \eczhuVVA }
\endeczcode

\eczcode{quantum_cyclic}{Cyclic quantum code}{~\NoCaseChange{\protect\cite{cite2730}}}
\codefieldsection{Description}
A block quantum code such that cyclic permutations of the subsystems leave the codespace invariant. In other words, the automorphism group of the code contains the cyclic group \(\mathbb{Z}_n\).

An example \(\llbracket 17,9,3\rrbracket \) cyclic Hermitian qubit code
has stabilizer generated by cyclic shifts of the Pauli operators
\(XXYIXYZZYXIYXXIII\) and \(ZZXIZXYYXZIXZZIII\).
An example \(\llbracket 17,1,7\rrbracket \) cyclic Hermitian qubit code
has stabilizer generated by cyclic shifts of the Pauli operators
\(XYYIZZIYYXIIIIIII\) and \(ZXXIYYIXXZIIIIIII\).

\codefieldsection{Protection}
Cyclic symmetry guarantees that if a single subsystem is protected against some noise, then all other subsystems are also.
\codefieldsection{Encoding}
\begin{eczvaluelist}
\item\relax Linear feedback shift registers \NoCaseChange{\protect\cite{cite2731}}.
\end{eczvaluelist}
\codefieldsection{Decoding}
\begin{eczvaluelist}
\item\relax Linear feedback shift registers \NoCaseChange{\protect\cite{cite2731}}.
\item\relax Adapted from the Berlekamp decoding algorithm for classical BCH codes \NoCaseChange{\protect\cite{cite2730}}.
\end{eczvaluelist}
\codefieldsection{Notes}
\begin{eczvaluelist}
\item\relax Many examples have been found by computer algebra programs. Ref. \NoCaseChange{\protect\cite{cite2730}} gives examples of \(\llbracket 17,1,7\rrbracket \) and \(\llbracket 17,9,3\rrbracket \) quantum cyclic codes.
\end{eczvaluelist}
\codefieldsection{Parent}
\begin{eczvaluelist}
\item\relax
\flmRefsHyperref[eczindexfamilyrel]{code:quantum_quasi_cyclic}{Quasi-cyclic quantum code}\end{eczvaluelist}
\codefieldsection{Children}
\begin{eczvaluelist}
\item\relax
\flmRefsHyperref[eczindexfamilyrel]{code:rotor_5_1_3}{\(\llbracket 5,1,3\rrbracket _{\mathbb{Z}}\) Five-rotor code}\item\relax
\flmRefsHyperref[eczindexfamilyrel]{code:group_quantum_repetition}{Group-based quantum repetition code}\item\relax
\flmRefsHyperref[eczindexfamilyrel]{code:braunstein}{\(\llbracket 5,1,3\rrbracket _{\mathbb{R}}\) Braunstein five-mode code}\item\relax
\flmRefsHyperref[eczindexfamilyrel]{code:permutation_invariant}{Permutation-invariant (PI) code} --- The cyclic group of these codes is a subgroup of the \(S_n\) symmetric group used in permutation invariant codes.
\item\relax
\flmRefsHyperref[eczindexfamilyrel]{code:qubit_5_6_2}{\(\llparenthesis 5,6,2\rrparenthesis \) qubit code}\item\relax
\flmRefsHyperref[eczindexfamilyrel]{code:steane}{\(\llbracket 7,1,3\rrbracket \) Steane code} --- The Steane code is equivalent to a cyclic code via qubit permutations \NoCaseChange{\protect\cite[{Exam. 1}]{cite438}}.
\item\relax
\flmRefsHyperref[eczindexfamilyrel]{code:qubit_9_12_3}{\(\llparenthesis 9,12,3\rrparenthesis \) qubit code}\item\relax
\flmRefsHyperref[eczindexfamilyrel]{code:bipartite_cyclic_cluster}{Bipartite cyclic cluster (BCC) code} --- BCC codes are invariant under cyclic shifts by construction \NoCaseChange{\protect\cite{cite440}}.
\item\relax
\flmRefsHyperref[eczindexfamilyrel]{code:lacross}{La-cross code}\item\relax
\flmRefsHyperref[eczindexfamilyrel]{code:twisted_xzzx}{Twisted XZZX toric code}\item\relax
\flmRefsHyperref[eczindexfamilyrel]{code:qudit_5_1_3}{\(\llbracket 5,1,3\rrbracket _{\mathbb{Z}_q}\) modular-qudit code}\item\relax
\flmRefsHyperref[eczindexfamilyrel]{code:frobenius}{Frobenius code}\item\relax
\flmRefsHyperref[eczindexfamilyrel]{code:galois_5_1_3}{\(\llbracket 5,1,3\rrbracket _q\) Galois-qudit code}\end{eczvaluelist}
\codefieldsection{Cousins}
\begin{eczvaluelist}
\item\relax
\flmRefsHyperref[eczindexfamilyrel]{code:cyclic}{Cyclic code} --- Cyclic quantum codes are quantum analogues of cyclic codes.
\item\relax
\flmRefsHyperref[eczindexfamilyrel]{code:1d_stabilizer}{1D lattice stabilizer code} --- A 1D lattice stabilizer code with periodic boundary conditions is a quantum cyclic code.
\item\relax
\flmRefsHyperref[eczindexfamilyrel]{code:asymmetric_qecc}{Asymmetric quantum code (AQC)} --- Cyclic quantum codes can be adapted for asymmetric noise \NoCaseChange{\protect\cite{cite2644}}.
\end{eczvaluelist}
\eczhbkcontributors{ Simon Burton, Nolan Coble, \eczhuVVA }
\endeczcode

\eczcode{dynamic_gen}{Dynamically generated QECC}{~\NoCaseChange{\protect\cite{cite2732}}}
\codefieldsection{Description}
Block quantum code whose natural definition is in terms of a many-body scaling limit of a local dynamical process.
Such processes, which are often non-deterministic, update the code structure and can include random unitary evolution or non-commuting projective measurements.

\codefieldsection{Notes}
\begin{eczvaluelist}
\item\relax See \NoCaseChange{\protect\cite{cite2733}} for a pedagogical introduction to dynamical codes.
\end{eczvaluelist}
\codefieldsection{Parent}
\begin{eczvaluelist}
\item\relax
\flmRefsHyperref[eczindexfamilyrel]{code:block_quantum}{Block quantum code}\end{eczvaluelist}
\codefieldsection{Children}
\begin{eczvaluelist}
\item\relax
\flmRefsHyperref[eczindexfamilyrel]{code:random_circuit}{Random-circuit code}\item\relax
\flmRefsHyperref[eczindexfamilyrel]{code:spacetime_circuit}{Spacetime circuit code}\item\relax
\flmRefsHyperref[eczindexfamilyrel]{code:qudit_da}{Modular-qudit dynamical code} --- Dynamical code state initialization, logical gates, and error correction are done by a sequence of different (usually weight-two) stabilizer measurements.
\end{eczvaluelist}
\codefieldsection{Cousins}
\begin{eczvaluelist}
\item\relax
\flmRefsHyperref[eczindexfamilyrel]{code:general_qldpc}{QLDPC code} --- QLDPC codes can arise from a dynamical process \NoCaseChange{\protect\cite{cite2734}}.
\item\relax
\flmRefsHyperref[eczindexfamilyrel]{code:cluster_state}{Cluster-state code} --- MBQC is done using a measurement-based dynamical process.
\item\relax
\flmRefsHyperref[eczindexfamilyrel]{code:fusion}{Fusion-based quantum computing (FBQC) code} --- Building a fusion network is done using a measurement-based dynamical process.
\item\relax
\flmRefsHyperref[eczindexfamilyrel]{code:clifford-deformed_surface}{Clifford-deformed surface code (CDSC)} --- To create CDSCs, a dynamical process is applied on top of the surface code \NoCaseChange{\protect\cite{cite2625}}.
\end{eczvaluelist}
\eczhbkcontributors{ Michael Gullans, \eczhuVVA }
\endeczcode

\eczcode{eacq}{Entanglement-assisted (EA) hybrid QECC}{~\NoCaseChange{\protect\cite{cite2735,cite2736,cite671}}}
\codefieldsection{Description}
Code that encodes quantum and classical information and requires pre-shared
entanglement for transmission.

EA hybrid block quantum codes on \(n\) Galois qudits of dimension \(q\) are denoted by \(\llparenthesis n,k:c,d;e\rrparenthesis _q\), where \(k\) (\(c\)) is the number of encoded logical qudits (classical symbols), where \(d\) is the distance, and where \(e\) is the required number of pre-shared ebits.
Similarly, block codes on \(n\) modular qudits are denoted by \(\llparenthesis n,k:c,d;e\rrparenthesis _{\mathbb{Z}_q}\).

In alternative conventions (not used here), EA hybrid codes are called entanglement-assisted classical-quantum (EACQ) codes.
Here, we use the term classical-quantum for codes for transmitting classical information over quantum channels.

\codefieldsection{Protection}
If an EA hybrid code is viewed as transmitting \(C\) cbits,
\(Q\) qubits, and consuming \(E\) ebits, then the EA hybrid Singleton bound is
the set of triples \((C,Q,E)\) for which there exists a parameter
\(t\in[0,\log q]\) such that
\flmMathEnvironment{align}{}{
  C + 2Q & \leq (n-d+1)(\log q + t)~,\\
  Q - E & \leq (n-2d+2)t~,\\
  C + Q - E & \leq (n-d+1)\log q - (d-1)t~,
}
for \(q\)-ary physical systems \NoCaseChange{\protect\cite[{Thm. 8}]{cite2737}}.

\codefieldsection{Rate}
Trade-off between classical communication, quantum communication, and entanglement distribution has been examined \NoCaseChange{\protect\cite{cite2738,cite2739,cite2740}}; see also Ref. \NoCaseChange{\protect\cite{cite2741}}.
\codefieldsection{Notes}
\begin{eczvaluelist}
\item\relax Examples from the original paper include a \(\llbracket 9,1:3,3;0\rrbracket \) code obtained from the Shor code, a \(\llbracket 8,1:3,3;1\rrbracket \) code obtained from an \(\llbracket 8,1,3;1\rrbracket \) EAQECC, and a \(\llbracket 63,21:12,7;6\rrbracket \) code obtained from the \(\llbracket 63,21,9;6\rrbracket \) EAQECC built from a classical \([63,39,9]\) BCH code \NoCaseChange{\protect\cite{cite2735}}.
\item\relax Inside the EAOAQEC stabilizer framework, hybrid stabilizer codes are a proper subclass of the broader EA hybrid subspace codes because the EACQ transversal operators obey additional constraints not required in general \NoCaseChange{\protect\cite{cite856}}.
\end{eczvaluelist}
\codefieldsection{Parent}
\begin{eczvaluelist}
\item\relax
\flmRefsHyperref[eczindexfamilyrel]{code:eaoaecc}{Entanglement-assisted operator-algebra QECC (EAOA QECC)} --- An EAOA QECC that has no gauge structure (e.g., gauge qubits), that has a block structure that corresponds to a classical code, and that utilizes pre-shared entanglement is an EA hybrid QECC.
\end{eczvaluelist}
\codefieldsection{Cousins}
\begin{eczvaluelist}
\item\relax
\flmRefsHyperref[eczindexfamilyrel]{code:hybridqecc}{Hybrid QECC} --- EA hybrid codes utilize additional ancillary subsystems in a pre-shared entangled state, but reduce to hybrid QECCs when said subsystems are interpreted as noiseless physical subsystems.
\item\relax
\flmRefsHyperref[eczindexfamilyrel]{code:ea_classical_into_quantum}{Entanglement-assisted (EA) c-q code} --- EA c-q codes transmit only classical information with entanglement assistance, while EA hybrid QECCs transmit both classical and quantum information with entanglement assistance.
\item\relax
\flmRefsHyperref[eczindexfamilyrel]{code:eaqecc}{Entanglement-assisted (EA) QECC} --- An EA hybrid QECC storing no classical information reduces to an EA QECC. Conversely, any EA QECC can be converted into an EA hybrid QECC by using a portion of its logical subspace to store only classical information.
\item\relax
\flmRefsHyperref[eczindexfamilyrel]{code:eaoa_stabilizer}{EAOA qubit stabilizer code} --- The original EACQ formalism describes a proper subclass of EA hybrid subspace codes inside the EAOAQEC stabilizer framework; EACQ representability imposes extra constraints on the transversal operators beyond belonging to distinct normalizer cosets \NoCaseChange{\protect\cite{cite856}}.
\end{eczvaluelist}
\eczhbkcontributors{ \eczhuVVA }
\endeczcode

\eczcode{eaoecc}{Entanglement-assisted (EA) operator QECC}{~\NoCaseChange{\protect\cite{cite2742,cite2743}}}
\codefieldsection{Alternative Names}
\begin{eczvaluelist}
\item\relax EA subsystem QECC
\end{eczvaluelist}
\eczhIndexCodeAliasName{eaoecc}{EA subsystem QECC}
\codefieldsection{Description}
Subsystem QECC whose encoding and decoding utilize pre-shared entanglement between sender and receiver.

\codefieldsection{Parent}
\begin{eczvaluelist}
\item\relax
\flmRefsHyperref[eczindexfamilyrel]{code:eaoaecc}{Entanglement-assisted operator-algebra QECC (EAOA QECC)} --- An EAOA QECC that has gauge structure (e.g., gauge qubits), that has no block structure that corresponds to a classical code, and that utilizes pre-shared entanglement is an EAOQECC.
\end{eczvaluelist}
\codefieldsection{Cousins}
\begin{eczvaluelist}
\item\relax
\flmRefsHyperref[eczindexfamilyrel]{code:oecc}{Subsystem QECC} --- EAOQECCs utilize additional ancillary subsystems in a pre-shared entangled state, but reduce to subsystem QECCs when said subsystems are interpreted as noiseless physical subsystems.
\item\relax
\flmRefsHyperref[eczindexfamilyrel]{code:eaqecc}{Entanglement-assisted (EA) QECC} --- An EAOQECC reduces to an EA QECC when the gauge subsystem is trivial. Conversely, any EA QECC with a tensor-product logical subspace can be turned into an EAOQECC by treating a logical tensor factor as a gauge subsystem.
\end{eczvaluelist}
\eczhbkcontributors{ \eczhuVVA }
\endeczcode

\eczcode{eaqecc}{Entanglement-assisted (EA) QECC}{~\NoCaseChange{\protect\cite{cite2744,cite1429,cite1430}}}
\codefieldsection{Alternative Names}
\begin{eczvaluelist}
\item\relax Catalytic QECC
\end{eczvaluelist}
\eczhIndexCodeAliasName{eaqecc}{Catalytic QECC}
\codefieldsection{Description}
QECC whose encoding and decoding utilize pre-shared entanglement between sender and receiver.
\codefieldsection{Protection}
Pre-shared entanglement can be prepared in a way that is robust to noise \NoCaseChange{\protect\cite{cite2745}}.

\codefieldsection{Rate}
The EA quantum capacity is the highest rate of quantum information transmission through a quantum channel with arbitrarily small error rate and access to arbitrary amounts of entanglement \NoCaseChange{\protect\cite{cite2746}}.
The fault-tolerant EA capacity is the capacity for the more general case where the encoding and decoding maps are also assumed to undergo noise \NoCaseChange{\protect\cite{cite2747}}.

\codefieldsection{Notes}
\begin{eczvaluelist}
\item\relax See Ref. \NoCaseChange{\protect\cite{cite2748,cite2749}} for an introduction to EAQECCs.
\end{eczvaluelist}
\codefieldsection{Parent}
\begin{eczvaluelist}
\item\relax
\flmRefsHyperref[eczindexfamilyrel]{code:eaoaecc}{Entanglement-assisted operator-algebra QECC (EAOA QECC)} --- An EAOA QECC that has no gauge structure (e.g., gauge qubits), that has no block structure that corresponds to a classical code, and that utilizes pre-shared entanglement is an EA QECC.
\end{eczvaluelist}
\codefieldsection{Children}
\begin{eczvaluelist}
\item\relax
\flmRefsHyperref[eczindexfamilyrel]{code:ea_oscillators}{EA bosonic code}\item\relax
\flmRefsHyperref[eczindexfamilyrel]{code:ea_galois_into_galois}{EA Galois-qudit code}\end{eczvaluelist}
\codefieldsection{Cousins}
\begin{eczvaluelist}
\item\relax
\flmRefsHyperref[eczindexfamilyrel]{code:qecc}{Quantum error-correcting code (QECC)} --- EA QECCs utilize additional ancillary subsystems in a pre-shared entangled state, but reduce to QECCs when said subsystems are interpreted as noiseless physical subsystems.
\item\relax
\flmRefsHyperref[eczindexfamilyrel]{code:eacq}{Entanglement-assisted (EA) hybrid QECC} --- An EA hybrid QECC storing no classical information reduces to an EA QECC. Conversely, any EA QECC can be converted into an EA hybrid QECC by using a portion of its logical subspace to store only classical information.
\item\relax
\flmRefsHyperref[eczindexfamilyrel]{code:eaoecc}{Entanglement-assisted (EA) operator QECC} --- An EAOQECC reduces to an EA QECC when the gauge subsystem is trivial. Conversely, any EA QECC with a tensor-product logical subspace can be turned into an EAOQECC by treating a logical tensor factor as a gauge subsystem.
\item\relax
\flmRefsHyperref[eczindexfamilyrel]{code:ea_classical_into_quantum}{Entanglement-assisted (EA) c-q code} --- EA c-q codes transmit classical information with entanglement assistance, while EAQECCs transmit quantum information with entanglement assistance.
\item\relax
\flmRefsHyperref[eczindexfamilyrel]{code:metopt}{Error-corrected sensing code} --- Metrologically optimal codes can be thought of as being entanglement-assisted because they require error-free ancillas for optimal local parameter estimation, and the estimation procedure uses an entangling gate.
\item\relax
\flmRefsHyperref[eczindexfamilyrel]{code:ea_3_1_3-2}{\(\llbracket 3, 1, 3;2\rrbracket \) EA code} --- The \(\llbracket 3, 1, 3;2\rrbracket \) EA code is the first EA code.
\end{eczvaluelist}
\eczhbkcontributors{ \eczhuVVA }
\endeczcode

\eczcode{eaoaecc}{Entanglement-assisted operator-algebra QECC (EAOA QECC)}{~\NoCaseChange{\protect\cite{cite856}}}
\codefieldsection{Description}
A code family that encompasses ordinary (i.e., subspace) codes, subsystem codes, classical-quantum codes, hybrid codes, and their entanglement-assisted counterparts using an operator-algebraic framework.
In the EAOAQEC framework, the original EAQEC, EAOQEC, and EACQ formalisms appear as special cases, and the operator-algebra perspective also yields EA hybrid subspace and EA subsystem codes beyond those earlier settings \NoCaseChange{\protect\cite{cite856}}.
\codefieldsection{Protection}
For Pauli noise with noiseless receiver ebits, the EAOAQEC framework gives a unified error-correction criterion and an associated distance notion that specialize to the corresponding conditions for EAQEC, EAOQEC, and EACQ codes \NoCaseChange{\protect\cite{cite856}}.
\codefieldsection{Parent}
\begin{eczvaluelist}
\item\relax
\flmRefsHyperref[eczindexfamilyrel]{code:quantum_into_quantum}{Quantum code}\end{eczvaluelist}
\codefieldsection{Children}
\begin{eczvaluelist}
\item\relax
\flmRefsHyperref[eczindexfamilyrel]{code:ea_classical_into_quantum}{Entanglement-assisted (EA) c-q code} --- An EAOA QECC that has no gauge structure (e.g., gauge qubits), that has a block structure that corresponds to a classical code, that stores no quantum information, and that utilizes pre-shared entanglement is an EA c-q code.
\item\relax
\flmRefsHyperref[eczindexfamilyrel]{code:eacq}{Entanglement-assisted (EA) hybrid QECC} --- An EAOA QECC that has no gauge structure (e.g., gauge qubits), that has a block structure that corresponds to a classical code, and that utilizes pre-shared entanglement is an EA hybrid QECC.
\item\relax
\flmRefsHyperref[eczindexfamilyrel]{code:eaoecc}{Entanglement-assisted (EA) operator QECC} --- An EAOA QECC that has gauge structure (e.g., gauge qubits), that has no block structure that corresponds to a classical code, and that utilizes pre-shared entanglement is an EAOQECC.
\item\relax
\flmRefsHyperref[eczindexfamilyrel]{code:eaqecc}{Entanglement-assisted (EA) QECC} --- An EAOA QECC that has no gauge structure (e.g., gauge qubits), that has no block structure that corresponds to a classical code, and that utilizes pre-shared entanglement is an EA QECC.
\item\relax
\flmRefsHyperref[eczindexfamilyrel]{code:eaoa_qubits_into_qubits}{EAOA qubit code} --- An EAOA QECC defined over qubits is an EAOA qubit code.
\end{eczvaluelist}
\codefieldsection{Cousin}
\begin{eczvaluelist}
\item\relax
\flmRefsHyperref[eczindexfamilyrel]{code:oaecc}{Operator-algebra QECC (OAQECC)} --- EAOA QECCs use pre-shared entangled ancillary subsystems, while OAQECCs recover the same operator-algebraic structures when those ancillary subsystems are instead treated as noiseless physical subsystems.
\end{eczvaluelist}
\eczhbkcontributors{ \eczhuVVA }
\endeczcode

\eczcode{metopt}{Error-corrected sensing code}{~\NoCaseChange{\protect\cite{cite2750,cite2751}}}
\codefieldsection{Description}
Code that can be obtained via an optimization procedure that ensures correction against a set \(\cal{E}\) of errors as well as guaranteeing optimal precision in locally estimating a parameter using a noiseless ancilla. For tensor-product spaces consisting of \(n\) subsystems (e.g., qubits, modular qudits, or Galois qudits), the procedure can yield a code whose parameter estimation precision satisfies \textit{Heisenberg scaling}, i.e., scales quadratically with the number \(n\) of subsystems.

The conditions required for a code are that it corrects errors in the set \(\cal{E}\) and admits a continuous-parameter \(U(1)\) group of logical gates generated by some \textit{signal Hamiltonian} \(H\) (with the time of evolution by \(H\) the parameter that is to be estimated).
This means that \(H\) cannot itself be a detectable error, i.e., \(H\) cannot be expressed as a linear combination of the errors, a condition known as the \textit{Hamiltonian-not-in-Kraus-span} (HNKS) condition \NoCaseChange{\protect\cite{cite2752}} (alternatively, Hamiltonian-not-in-Lindblad-span, or HNLS, for Markovian noise \NoCaseChange{\protect\cite{cite2751}}; see also \NoCaseChange{\protect\cite{cite2753}}).
If these conditions are satisfied, a semidefinite-program based optimization procedure yields a metrologically optimal code.
The procedure has been generalized to more general groups, corresponding to multiparameter estimation \NoCaseChange{\protect\cite{cite2754}}.
If these conditions are not satisfied, Heisenberg scaling is not achievable, but metrologically optimal codes can still be obtained via another semidefinite-program based optimization procedure \NoCaseChange{\protect\cite{cite2755,cite2752}}.

Metrologically optimal QECCs require error-free ancillas for optimal local parameter estimation using an entangling gate.
In this sense, such codes can be thought of as being entanglement-assisted.
\textit{Ancilla-free} versions exist in the case when the noise commutes with the signal Hamiltonian \NoCaseChange{\protect\cite{cite2756,cite2757}}.

\codefieldsection{Decoding}
\begin{eczvaluelist}
\item\relax There is a condition under which autonomous QEC gives rise to metrology with optimal Heisenberg scaling \NoCaseChange{\protect\cite{cite2758}}.
\end{eczvaluelist}
\codefieldsection{Realizations}
\begin{eczvaluelist}
\item\relax A single physical qubit entangled with an NV spin was used to measure an incoming signal in a way that bit-flip errors on the qubit were correctable \NoCaseChange{\protect\cite{cite2759}}.
\end{eczvaluelist}
\codefieldsection{Parent}
\begin{eczvaluelist}
\item\relax
\flmRefsHyperref[eczindexfamilyrel]{code:qecc_finite}{Finite-dimensional quantum error-correcting code} --- Semidefinite-program optimization procedure for finding a metrologically optimal code holds for finite-dimensional spaces.
\end{eczvaluelist}
\codefieldsection{Child}
\begin{eczvaluelist}
\item\relax
\flmRefsHyperref[eczindexfamilyrel]{code:chebyshev}{Chebyshev code}\end{eczvaluelist}
\codefieldsection{Cousins}
\begin{eczvaluelist}
\item\relax
\flmRefsHyperref[eczindexfamilyrel]{code:eaqecc}{Entanglement-assisted (EA) QECC} --- Metrologically optimal codes can be thought of as being entanglement-assisted because they require error-free ancillas for optimal local parameter estimation, and the estimation procedure uses an entangling gate.
\item\relax
\flmRefsHyperref[eczindexfamilyrel]{code:hamiltonian}{Hamiltonian-based code} --- Metrologically optimal codes admit a \(U(1)\) set of gates generated by a signal Hamiltonian \(H\), meaning that there exists a basis of codewords that are eigenstates of the \(H\).
\item\relax
\flmRefsHyperref[eczindexfamilyrel]{code:qubit_css}{Qubit CSS code} --- Qubit CSS codes can be used for sensing whenever the HNLS condition is satisfied, with the quantum Fisher information related to the number of weight-two codewords of the dual code \NoCaseChange{\protect\cite{cite2760}}.
\item\relax
\flmRefsHyperref[eczindexfamilyrel]{code:metrological}{Metrological code} --- Error-corrected sensing codes are required to satisfy the \flmTerm{term}{ref1043}{}{Knill-Laflamme conditions}, while metrological codes need only satisfy the conditions partially.
\item\relax
\flmRefsHyperref[eczindexfamilyrel]{code:gnu_permutation_invariant}{GNU PI code} --- GNU codes can be used to sense signals within the PI subspace \NoCaseChange{\protect\cite{cite2761}}.
\end{eczvaluelist}
\eczhbkcontributors{ Esha Swaroop, Sisi Zhou, \eczhuVVA }
\endeczcode

\eczcode{qecc_finite}{Finite-dimensional quantum error-correcting code}{}
\codefieldsection{Description}
Encodes quantum information in a \(K\)-dimensional (\textit{logical}) subspace of an \(N\)-dimensional (\textit{physical}) Hilbert space such that it is possible to recover said information from errors. The logical subspace is spanned by a basis comprised of \textit{code basis states} or \textit{codewords}.
\codefieldsection{Protection}
Denoting Hilbert spaces by the letter \(\mathsf{H}\), a finite-dimensional quantum code \((U,\cal{E})\) is a partial isometry \(U:\mathsf{H}_{K}\to\mathsf{H}_{N}\) and a set of correctable errors \({\cal{E}}:\mathsf{H}_N\to\mathsf{H}_M\) with the following property: there exists a quantum operation \({\cal{D}}:\mathsf{H}_M\to \mathsf{H}_K\) such that for all \(E\in\cal{E}\) and states \(|\psi\rangle\in\mathsf{H}_{K}\),
\flmMathEnvironment{align}{}{
{\cal D}(EU|\psi\rangle\langle\psi|U^{\dagger}E^{\dagger})=c(E,|\psi\rangle)|\psi\rangle\langle\psi|}
for some constant \(c\) \NoCaseChange{\protect\cite{cite398}}. A code is said to \textit{protect against} or \textit{correct} the errors \(\mathcal{E}\).

\subsection{Knill-Laflamme error-correction conditions}

Equivalently, correction capability is determined by the quantum
error-correction conditions. A code that satisfies
these conditions approximately, i.e., up to some small quantifiable error, is
called an \flmRefsHyperref{code:approximate_qecc}{approximate code}.

\begin{defterm}{Knill-Laflamme conditions}\label{ref1043}
The Knill-Laflamme error-correction conditions \NoCaseChange{\protect\cite{cite2762,cite2763,cite2764}\protect\cite[{Thm. 10.1}]{cite2579}} are necessary and sufficient conditions for a code to successfully
correct a set of errors in a finite-dimensional Hilbert space.
A code (defined by a partial isometry \(U\)) with code space projector \(\Pi = U U^\dagger\)
can correct a set of errors \(\{ E_j \}\) if and only if
\flmMathEnvironment{align}{}{
  \Pi E_i^\dagger E_j \Pi = c_{ij}\, \Pi\qquad\text{for all \(i,j\),}
}
where the \textit{QEC matrix} elements \(c_{ij}\) are arbitrary complex numbers.  
The term \(\Pi E_i^\dagger E_j \Pi\) can be split into two types of conditions, the \textit{diagonal} (a.k.a. non-deformation or invariance) conditions \(\langle \psi| E_i^\dagger E_j | \psi\rangle\) for a codeword \(|\psi\rangle\), and the \textit{off-diagonal} (a.k.a. orthogonality or distinguishability) conditions \(\langle \psi| E_i^\dagger E_j | \phi\rangle\) for two orthogonal codewords \(|\psi\rangle\) and \(|\phi\rangle\).
By linearity of quantum error correction, if a code corrects a set of errors \(\mathcal{E}\), then it also corrects \(\operatorname{span}\mathcal{E}\).
\end{defterm}

For codewords whose basis states are chosen far enough apart (in some notion of distance) so that the off-diagonal conditions are automatically zero, the remaining diagonal conditions correspond to a system of non-linear constraints on the basis-expansion coefficients of the codewords.
In this setting, there exist finite QECCs with \(K = \lceil N/D\rceil/(D+1)\) that protect against an error set with \(D\) basis elements \NoCaseChange{\protect\cite[{Thm. 4}]{cite648}}, a consequence of the \textit{Tverberg theorem} \NoCaseChange{\protect\cite{cite2765,cite597,cite2766}}. For \(K = 2\) this theorem reduces to Radon's theorem \NoCaseChange{\protect\cite{cite596,cite597}}.

The Knill-Laflamme conditions can alternatively be expressed in terms of the \flmRefsHyperref{ref2540}{complementary channel}, in an entropic information-theoretic way via a data processing inequality \NoCaseChange{\protect\cite{cite2767,cite2768,cite2769,cite2770,cite2771}}, or can be interpreted thermodynamically \NoCaseChange{\protect\cite{cite2772}}.
They motivate higher-rank numerical ranges, which are generalizations of the numerical range of an operator \NoCaseChange{\protect\cite{cite2773,cite2774,cite2775}}.
They have been extended to sequences of multiple errors and rounds of correction \NoCaseChange{\protect\cite{cite2776}}.

\begin{defterm}{Degeneracy}\label{ref2777}\label{ref811}
A code is degenerate with respect to a noise model if different errors map code states to the same error subspace.
For a linearly independent error set \(\cal{E}\), degeneracy is equivalent to \(\text{rank}(c_{ij}) < |\cal{E}|\) \NoCaseChange{\protect\cite{cite398}}.
\end{defterm}

\subsection{Correctability of quantum channels}

From now on, we use \(\mathcal{E}\) to denote a noise channel constructed out of the set of errors \(E\) and let \(\mathcal{U}(\cdot)=U(\cdot)U^\dagger\) be the superoperator corresponding to the partial encoding isometry \(U\).
A noise channel is correctable if there exists a recovery channel \(\mathcal{D}\) such that
\flmMathEnvironment{align}{}{
  \mathcal{D}\mathcal{E}\mathcal{U}(\rho)=\rho
}
for all logical states \(\rho\).

The above is equivalent to the fidelity between \(\rho\) and \(\mathcal{D}\mathcal{E}\mathcal{U}(\rho)\) being one for any notion of distance between quantum states.
In particular, we can consider a scenario where we send only one part of an entangled state through a channel and determine whether the entanglement has been preserved during transmission.
Using the notion of entanglement fidelity, a quantum channel \(\mathcal{E}\) is exactly correctable iff there exists a quantum channel \(\mathcal{D}\) such that
\flmMathEnvironment{align}{}{
  (\mathcal{D}\mathcal{E}\mathcal{U}\otimes\mathrm{id})(\ket{\psi}\bra{\psi})=\ket{\psi}\bra{\psi}
}
for all states \(\rho\) and their corresponding purifications \(\ket{\psi}\) (i.e., states \(\ket{\psi}\) such that \(\text{Tr}_{2}(|\psi\rangle\langle\psi|)=\rho\)).

The above entanglement fidelity condition can be alternatively expressed using complementary channels.

\begin{defterm}{Complementary channel}\label{ref2778}\label{ref2540}
A complementary channel \(\mathcal{E}^C\) is obtained from a channel \(\mathcal{E}\) that acts on a system by interpreting the channel as coming from a unitary operation acting on a larger system-environment tensor-product space (i.e., performing an isometric extension) with the environment necessarily in a pure state, and then tracing out the system factor (instead of the second environmental factor) \NoCaseChange{\protect\cite[{Sec. 5.2.2}]{cite2779}}.
A noise channel \({\cal E}(\cdot)=\sum_{j}E_{j}(\cdot)E_{j}^{\dagger}\) admits a complementary channel of the form
\flmMathEnvironment{align}{}{
  {\cal E}^{C}(\cdot)=\sum_{j,k}\text{Tr}\{E_{j}(\cdot)E_{k}^{\dagger}\}|j\rangle\langle k|~.
}
\end{defterm}

A channel \(\mathcal{E}\) is correctable if  \(\mathcal{E}^C(\rho)=\rho_0\mathrm{Tr}(\rho)\) for some constant state \(\rho_0\), which is equivalent to the \flmTerm{term}{ref1043}{}{Knill-Laflamme conditions} \NoCaseChange{\protect\cite{cite2780,cite2538}}.
The logical and physical dimensions are related to the channel rank for non-degenerate codes via the quantum packing bound \NoCaseChange{\protect\cite{cite2666}}. 

Exact correctability can also be expressed using the \flmRefsHyperref{ref2781}{coherent information}.

\begin{defterm}{Coherent information}\label{ref2782}\label{ref2781}
Given a bipartite state \(\rho_{RQ}\), the \flmRefsHyperref{ref2781}{coherent information} in subsystem \(Q\) is
\flmMathEnvironment{align}{}{
  I_{c}(\rho_{RQ})=S(\rho_{Q})-S(\rho_{RQ})~.
}
For a channel \(\mathcal{E}:L\to Q\) and a pure input state \(\rho\) on \(R\otimes L\), the \flmRefsHyperref{ref2781}{coherent information} of the channel is
\flmMathEnvironment{align}{}{
  I_{c}(\mathcal{E},\rho)=S(\mathcal{E}(\rho_{L}\rrparenthesis -S\llparenthesis \mathrm{id}\otimes\mathcal{E})(\rho\rrparenthesis ~.
}
\flmRefsHyperref{ref2781}{Coherent information} cannot increase under further processing of the output, a statement known as the \textit{quantum data processing inequality} \NoCaseChange{\protect\cite{cite2767,cite2768,cite2769,cite2770,cite2771,cite398}}.
\end{defterm}

Exact correctability is equivalent to preservation of \flmRefsHyperref{ref2781}{coherent information}: a channel \(\mathcal{E}\) is exactly correctable on a code iff the \flmRefsHyperref{ref2781}{coherent information} after encoding and noise is the same as that of the logical input for every pure input state, and it is enough to check this on a maximally entangled state between the logical system and a reference \NoCaseChange{\protect\cite{cite2767,cite2769,cite398}}.

\codefieldsection{Rate}
The quantum channel capacity, i.e., the regularized \flmRefsHyperref{ref2781}{coherent information}, is the highest rate of quantum information transmission through a quantum channel with arbitrarily small error rate \NoCaseChange{\protect\cite{cite2783,cite2784,cite2785}}. 
In other words, the capacity formula implies that one can achieve a transmission rate
\(r\) over a quantum channel \(\mathcal{E}\) iff, for sufficiently large \(n\), \(m=\lfloor r n \rfloor\),
and for all \(\epsilon>0\),
\flmMathEnvironment{align}{}{
  \lVert\mathcal{D}\mathcal{E}\mathcal{U}-I^{\otimes m}\rVert_1\leq \epsilon
}
for some encoding channel \(\mathcal{U}\) and some recovery channel \(\mathcal{D}\).
The quantum capacity \(Q\) of \(\mathcal{E}\) is defined as the supremum over \(n\) of achievable transmission rates \NoCaseChange{\protect\cite{cite2786}}.
See \NoCaseChange{\protect\cite[{Ch. 24}]{cite535}} for definitions and a history.     

The fault-tolerant capacity is the capacity for the more general case where the encoding and decoding maps are also assumed to undergo noise \NoCaseChange{\protect\cite{cite1017}}.

Doeblin coefficients \NoCaseChange{\protect\cite{cite1029}} for quantum channels have been studied \NoCaseChange{\protect\cite{cite1030}}.

\codefieldsection{Decoding}
\begin{eczvaluelist}
\item\relax The operation \(\cal{D}\) in the definition of this code is called the decoder. However, the term \textit{decoder} can sometimes be used for the unencoder \(\cal{U}\) (i.e., the inverse of the encoder), which does not correct errors.
\item\relax There are several recovery maps which work for noise that is not exactly correctable; see \flmRefsHyperref{code:approximate_qecc}{AQECC} entry.
\item\relax QECCs are useful \NoCaseChange{\protect\cite{cite2787}} for the mean king's measurement problem \NoCaseChange{\protect\cite{cite2788}}.
\item\relax Protection can be implemented via \textit{autonomous QEC} (a.k.a. continuous QEC or continuous-time QEC) \NoCaseChange{\protect\cite{cite2789,cite2790,cite2791,cite2792,cite2793}} via, e.g., reservoir engineering \NoCaseChange{\protect\cite{cite2794}}; see review \NoCaseChange{\protect\cite{cite2795}}. There are analogues of the \flmTerm{term}{ref1043}{}{Knill-Laflamme conditions} for autonomous QEC \NoCaseChange{\protect\cite{cite2796,cite2797}}, and it has been adapted to non-Markovian noise \NoCaseChange{\protect\cite{cite2798}}. Information-theoretic bounds have been derived for open-loop control \NoCaseChange{\protect\cite{cite2799}}. Machine learning can be used to optimize autonomous QEC encoding and recovery \NoCaseChange{\protect\cite{cite2800}}.
\end{eczvaluelist}
\codefieldsection{Code Capacity Threshold}
\begin{eczvaluelist}
\item\relax \flmRefsHyperref{ref2781}{Coherent information} of the state under the action of a noise channel can be used to estimate the optimal threshold \NoCaseChange{\protect\cite{cite2801}}.
\end{eczvaluelist}
\codefieldsection{Parent}
\begin{eczvaluelist}
\item\relax
\flmRefsHyperref[eczindexfamilyrel]{code:qecc}{Quantum error-correcting code (QECC)} --- Finite-dimensional QECCs are a special case of quantum error-correcting codes, which can also include infinite-dimensional codes such as bosonic codes. The Knill-Laflamme conditions and the notion of correctability can be extended to infinite-dimensional codes.
\end{eczvaluelist}
\codefieldsection{Children}
\begin{eczvaluelist}
\item\relax
\flmRefsHyperref[eczindexfamilyrel]{code:metopt}{Error-corrected sensing code} --- Semidefinite-program optimization procedure for finding a metrologically optimal code holds for finite-dimensional spaces.
\item\relax
\flmRefsHyperref[eczindexfamilyrel]{code:quantum_mds}{Quantum maximum-distance-separable (MDS) code}\item\relax
\flmRefsHyperref[eczindexfamilyrel]{code:quantum_perfect}{Perfect quantum code}\item\relax
\flmRefsHyperref[eczindexfamilyrel]{code:single_shot}{Single-shot code}\item\relax
\flmRefsHyperref[eczindexfamilyrel]{code:qltc}{Quantum locally testable code (QLTC)}\item\relax
\flmRefsHyperref[eczindexfamilyrel]{code:qudits_into_qudits}{Modular-qudit code}\item\relax
\flmRefsHyperref[eczindexfamilyrel]{code:galois_into_galois}{Galois-qudit code}\item\relax
\flmRefsHyperref[eczindexfamilyrel]{code:spins_into_spins}{Spin code}\end{eczvaluelist}
\codefieldsection{Cousins}
\begin{eczvaluelist}
\item\relax
\flmRefsHyperref[eczindexfamilyrel]{code:ecc_finite}{Finite-dimensional error-correcting code (ECC)} --- Finite-dimensional QECCs are quantum analogues of finite-dimensional classical ECCs.
\item\relax
\flmRefsHyperref[eczindexfamilyrel]{code:complex_projective}{Complex projective space code} --- Pure quantum states in an \((N+1)\)-dimensional Hilbert space are parameterized by points in the complex projective space \(\mathbb{C}P^N\). As such, (classical) complex projective codes can be associated with subsets of pure quantum states.
\end{eczvaluelist}
\eczhbkcontributors{ Milan Tenn, \eczhuVVA }
\endeczcode

\eczcode{frustration_free}{Frustration-free Hamiltonian code}{}
\codefieldsection{Description}
Hamiltonian-based code whose Hamiltonian is frustration free, i.e., whose ground states minimize the energy of each term.

\codefieldsection{Protection}
Geometrically local frustration-free code Hamiltonians on Euclidean manifolds are stable with respect to sufficiently weak quasi-local perturbations when they satisfy \textit{local topological quantum order} (LTQO) together with the \textit{Local-Gap} condition; LTQO also implies an area law for the entanglement entropy of the ground-state subspace \NoCaseChange{\protect\cite{cite2802}}.
See also \NoCaseChange{\protect\cite{cite2803}}.

\codefieldsection{Encoding}
\begin{eczvaluelist}
\item\relax Lindbladian-based dissipative encoding can be constructed for a codespace that is the ground-state subspace of a frustration-free Hamiltonian \NoCaseChange{\protect\cite{cite2804,cite2805,cite2806,cite2807}}.
\end{eczvaluelist}
\codefieldsection{Parent}
\begin{eczvaluelist}
\item\relax
\flmRefsHyperref[eczindexfamilyrel]{code:hamiltonian}{Hamiltonian-based code}\end{eczvaluelist}
\codefieldsection{Children}
\begin{eczvaluelist}
\item\relax
\flmRefsHyperref[eczindexfamilyrel]{code:cage_net}{Cage-net code} --- Cage-net codewords form ground-state subspaces of frustration-free commuting projector Hamiltonians.
\item\relax
\flmRefsHyperref[eczindexfamilyrel]{code:yetter_gauge_theory}{Two-gauge theory code} --- Two-gauge theory codewords span ground-state subspaces of frustration-free commuting-projector Hamiltonians.
\item\relax
\flmRefsHyperref[eczindexfamilyrel]{code:enriched_string_net}{Multi-fusion string-net code} --- Multi-fusion string-net codes form eigenspaces of frustration-free commuting projector Hamiltonians.
\item\relax
\flmRefsHyperref[eczindexfamilyrel]{code:enriched_walker_wang}{\(G\)-enriched Walker-Wang model code} --- \(G\)-enriched Walker-Wang model codewords form ground-state subspaces of frustration-free commuting projector Hamiltonians.
\item\relax
\flmRefsHyperref[eczindexfamilyrel]{code:stabilizer}{Stabilizer code} --- Codespace is the ground-state space of the \textit{code Hamiltonian}, which consists of an equal linear combination of stabilizer generators and which can be made into a frustration-free commuting-projector Hamiltonian.
\item\relax
\flmRefsHyperref[eczindexfamilyrel]{code:qltc}{Quantum locally testable code (QLTC)} --- Quantum LTC codespaces are ground-state spaces of \(u\)-local frustration-free commuting-projector Hamiltonians.
\item\relax
\flmRefsHyperref[eczindexfamilyrel]{code:circuit_to_hamiltonian}{Circuit-to-Hamiltonian approximate code} --- Circuit-to-Hamiltonian approximate codes form the ground-state space of a frustration-free non-commuting projector Hamiltonian whose projectors are constant weight, but such that each physical qubit is acted on by \flmRefsHyperref{ref65}{order} \(O( \text{polylog}(n) )\) projectors.
\item\relax
\flmRefsHyperref[eczindexfamilyrel]{code:mps}{Magnon code} --- Magnon codewords are low-energy excited states of the frustration-free Heisenberg-XXX model Hamiltonian \NoCaseChange{\protect\cite{cite595}}.
\item\relax
\flmRefsHyperref[eczindexfamilyrel]{code:vbs}{Valence-bond-solid (VBS) code} --- VBS codewords are eigenstates of the frustration-free VBS Hamiltonian \NoCaseChange{\protect\cite{cite2808,cite790}}.
\end{eczvaluelist}
\codefieldsection{Cousins}
\begin{eczvaluelist}
\item\relax
\flmRefsHyperref[eczindexfamilyrel]{code:commuting_projector}{Commuting-projector Hamiltonian code} --- Frustration-free Hamiltonians can contain non-commuting projectors; an example is the AKLT model \NoCaseChange{\protect\cite{cite626}}. On the other hand, commuting-projector Hamiltonians can be frustrated; an example is the 1D classical Ising model on a circle for odd \(n\) with one two-body interaction having the opposite sign.
\item\relax
\flmRefsHyperref[eczindexfamilyrel]{code:topological}{Topological code} --- Geometrically local frustration-free code Hamiltonians on Euclidean manifolds are stable with respect to sufficiently weak quasi-local perturbations when they satisfy local topological quantum order together with the Local-Gap condition; LTQO also implies an area law for the entanglement entropy of the ground-state subspace \NoCaseChange{\protect\cite{cite2802}}. See also \NoCaseChange{\protect\cite{cite2803}}.
\item\relax
\flmRefsHyperref[eczindexfamilyrel]{code:binary_linear}{Linear binary code} --- Parity-check constraints defining a binary linear code can be encoded into a classical Ising model Hamiltonian, a commuting-projector model whose terms contain products of Pauli \(Z\) matrices participating in each parity check. Such Ising models are also frustration-free since the codewords satisfy all parity checks.
\item\relax
\flmRefsHyperref[eczindexfamilyrel]{code:eth}{Eigenstate thermalization hypothesis (ETH) code} --- ETH codewords are eigenstates of a local Hamiltonian whose eigenstates satisfy ETH, and many example codes are eigenstates of frustration-free Hamiltonians.
\item\relax
\flmRefsHyperref[eczindexfamilyrel]{code:movassagh_ouyang}{Movassagh-Ouyang Hamiltonian code} --- Movassagh-Ouyang codes reside in the ground space of a Hamiltonian. Justesen codes can be used to build a family of \(n\)-qubit Movassagh-Ouyang Hamiltonian spin codes encoding one logical qubit with linear distance. These codes form the ground-state subspace of a frustration-free geometrically local Hamiltonian \NoCaseChange{\protect\cite{cite1407}}.
\item\relax
\flmRefsHyperref[eczindexfamilyrel]{code:gnu_permutation_invariant}{GNU PI code} --- GNU codes lie within the ground state of ferromagnetic Heisenberg models without an external magnetic field \NoCaseChange{\protect\cite{cite2809}}.
\end{eczvaluelist}
\eczhbkcontributors{ \eczhuVVA }
\endeczcode

\eczcode{group_representation}{Group-representation code}{~\NoCaseChange{\protect\cite{cite646,cite647,cite2810}}}
\codefieldsection{Description}
Code whose projector is onto an irreducible representation of a subgroup \(G\) of a group of canonical or distinguished unitary operations, e.g., transversal gates in the case of block quantum codes, Gaussian operations in the case of bosonic codes, or \(SU(2)\) operations in the case of single-spin codes.

\codefieldsection{Protection}
Error correction ability is not guaranteed, but can be sought in the multiplicity space of the irrep in case there is more than one copy present.

\codefieldsection{Encoding}
\begin{eczvaluelist}
\item\relax General encoding map \NoCaseChange{\protect\cite[{Lemma 1}]{cite2810}}.
\end{eczvaluelist}
\codefieldsection{Gates}
\begin{eczvaluelist}
\item\relax By definition, a group \(G\) of gates can be realized on the code using the unitary operations used to define the code.
\end{eczvaluelist}
\codefieldsection{Parent}
\begin{eczvaluelist}
\item\relax
\flmRefsHyperref[eczindexfamilyrel]{code:covariant}{Covariant block quantum code} --- Group-representation code projections are onto a single irrep of a subgroup of canonical or distinguished unitary operations on a Hilbert space. This makes them covariant w.r.t. that subgroup. More general covariant codes need not be projections onto a single irrep. Removing the restriction to distinguished operations and allowing all operations, every code projection on an \(N\)-dim Hilbert space can be expressed as a projection onto the irrep formed by the code-preserving subgroup of \(U(N)\). The same idea holds when \(N\) is taken to infinity. In other words, while all codes are covariant w.r.t. some group, group-representation codes are covariant w.r.t. a canonical or distinguished subgroup.
\end{eczvaluelist}
\codefieldsection{Children}
\begin{eczvaluelist}
\item\relax
\flmRefsHyperref[eczindexfamilyrel]{code:tesselation}{Hyperbolic tessellation code} --- Hyperbolic tessellation-code projections are onto a copy of an irreducible representation of the proper triangle group associated with the tessellation, and the resulting logical gates are implemented geometrically by hyperbolic rotations \NoCaseChange{\protect\cite{cite2811}}.
\item\relax
\flmRefsHyperref[eczindexfamilyrel]{code:icosahedral_fock}{Icosahedral Fock-state code} --- Icosahedral Fock-state codes are group-representation codes with the \(G = 2I\) subgroup of Gaussian rotations \NoCaseChange{\protect\cite{cite646}}.
\item\relax
\flmRefsHyperref[eczindexfamilyrel]{code:one_hot_quantum}{One-hot quantum code} --- One-hot quantum codes are group-representation codes with the \(G = SU(q)\) subgroup of Gaussian rotations \NoCaseChange{\protect\cite{cite2810}}.
\item\relax
\flmRefsHyperref[eczindexfamilyrel]{code:clifford_qsc}{Clifford group-representation QSC} --- The Clifford group-representation QSC is a group-representation code with \(G\) being \flmRefsHyperref{ref409}{single-qubit Clifford group}, taken as the binary octahedral subgroup of the group \(SU(2)\) of Gaussian rotations.
\item\relax
\flmRefsHyperref[eczindexfamilyrel]{code:fourier_bosonic}{Bosonic quantum Fourier code} --- The bosonic quantum Fourier code is a group-representation code with \(G\) being the single-qubit Pauli group.
\item\relax
\flmRefsHyperref[eczindexfamilyrel]{code:pauli_qsc}{Pauli tessellation QSC} --- The Pauli tessellation QSC is a group-representation code with \(G\) being the single-qubit Pauli group.
\item\relax
\flmRefsHyperref[eczindexfamilyrel]{code:qutrit_pauli_gkp_subcode}{Qutrit-Pauli tessellation code} --- The qutrit-Pauli tessellation code is a group-representation code with \(G\) being the \flmRefsHyperref{ref2198}{single-qutrit Pauli group}.
\item\relax
\flmRefsHyperref[eczindexfamilyrel]{code:knill}{Knill code} --- Knill codes project onto a single irrep sector associated with a normal subgroup of the group formed by a \flmRefsHyperref{ref2812}{nice error basis} \NoCaseChange{\protect\cite[{Lemma 3.1}]{cite2813}}.
\item\relax
\flmRefsHyperref[eczindexfamilyrel]{code:t_group}{Twisted \(1\)-group code} --- Twisted \(1\)-group codes are group-representation codes with \(G\) being a twisted \(1\)-group.
\item\relax
\flmRefsHyperref[eczindexfamilyrel]{code:j_gross}{Clifford-group spin code} --- Clifford-group spin codes are group-representation codes with \(G\) being a subgroup of \(SU(2)\) \NoCaseChange{\protect\cite{cite646,cite2814}}.
\item\relax
\flmRefsHyperref[eczindexfamilyrel]{code:su3_spin}{\(SU(3)\) spin code} --- \(SU(3)\) spin codes are group-representation codes with \(G\) being a subgroup of \(SU(3)\) \NoCaseChange{\protect\cite{cite2815}}.
\end{eczvaluelist}
\codefieldsection{Cousins}
\begin{eczvaluelist}
\item\relax
\flmRefsHyperref[eczindexfamilyrel]{code:small_distance_quantum}{Small-distance block quantum code} --- See Ref. \NoCaseChange{\protect\cite{cite789}} for tables of distance-two codes with various families of transversal gates.
\item\relax
\flmRefsHyperref[eczindexfamilyrel]{code:two-mode_binomial}{Two-mode binomial code} --- An application of the group-representation encoding construction \NoCaseChange{\protect\cite[{Lemma 1}]{cite2810}} yields a family of two-mode codes that closely resemble the two-mode binomial codes \NoCaseChange{\protect\cite{cite2816}}.
\item\relax
\flmRefsHyperref[eczindexfamilyrel]{code:cat}{Cat code} --- Cat codes are not group representation codes with \(G\) being a cyclic group since their representation is reducible \NoCaseChange{\protect\cite{cite2810}}.
\item\relax
\flmRefsHyperref[eczindexfamilyrel]{code:qsc}{Quantum spherical code (QSC)} --- QSCs should be able to be formulated as group-representation codes whose group is that formed by the permutation representation of the code polytope symmetry group, but this representation may be reducible.
\item\relax
\flmRefsHyperref[eczindexfamilyrel]{code:stab_5_1_3}{\(\llbracket 5,1,3\rrbracket \) Five-qubit perfect code} --- The five-qubit code is a group-representation code with \(G\) being the \(2T\) subgroup of \(SU(2)\) \NoCaseChange{\protect\cite{cite2810}}.
\item\relax
\flmRefsHyperref[eczindexfamilyrel]{code:steane}{\(\llbracket 7,1,3\rrbracket \) Steane code} --- The Steane code is a group-representation code with \(G\) being the \(2O\) subgroup of \(SU(2)\) \NoCaseChange{\protect\cite{cite2810}}.
\end{eczvaluelist}
\eczhbkcontributors{ \eczhuVVA }
\endeczcode

\eczcode{hamiltonian}{Hamiltonian-based code}{}
\codefieldsection{Description}
Code whose codespace corresponds to a set of energy eigenstates of a quantum-mechanical Hamiltonian, i.e., a Hermitian operator whose expectation value measures the energy of its underlying physical system.
The codespace is typically a set of low-energy eigenstates or ground states, but can include subspaces of arbitrarily high energy.
Hamiltonians whose eigenstates are the canonical basis elements are called \textit{classical}; otherwise, a Hamiltonian is called \textit{quantum}.

A Hamiltonian whose ground states minimize the energy of each term is called a frustration-free Hamiltonian.
A Hamiltonian whose terms commute and can be written as orthogonal projectors (i.e., with eigenvalues zero or one) is called a commuting projector Hamiltonian.

For block quantum codes, the code Hamiltonian is typically written as a sum of terms, with each term acting on at most some number \(K\) of subsystems (i.e., being of weight at most \(K\)).
When \(K\) is independent of the total number of subsystems (e.g., \flmRefsHyperref{code:qldpc}{QLDPC codes}), the Hamiltonian is called a \(K\)\textit{-body} or \(K\)\textit{-local} Hamiltonian.
Otherwise, the Hamiltonian is called \textit{non-local}.
When the physical space is endowed with a geometry, the Hamiltonian is typically \textit{geometrically local}, consisting of operators acting on subsystems that occupy a region whose size is independent of the number of subsystems.

The notion of locality can be softened to include Hamiltonians whose terms each act non-trivially on all sites, but whose support on regions farther and farther from some designated central site decays super-polynomially with the radius from the center.
Such Hamiltonians are called \textit{quasi-local Hamiltonians} (a.k.a. almost local or approximately local).

\subsection{Quantum phases of matter}

Ground states of infinite families geometrically local block-code Hamiltonians on tessellations of Euclidean geometries are said to be in a particular \textit{phase of quantum matter}, i.e., a region in some parameter space of Hamiltonians "in which [ground] states possess properties that can be distinguished from those in other phases" \NoCaseChange{\protect\cite{cite2817}}.
For example, \flmRefsHyperref{code:topological}{topological code} Hamiltonians realize a particular phase called a topological phase.

Hamiltonians realizing different phases cannot be adiabatically deformed into one another without a closing of the energy gap between the ground and excited states.
Such adiabatic deformations naively would be generated by non-local Hamiltonians.
However, Hastings and Wen \NoCaseChange{\protect\cite{cite2818}} (see also \NoCaseChange{\protect\cite{cite2819,cite2820}}) showed that adiabatic evolution can in fact be generated by a quasi-local operator; such evolution is often called \textit{quasi-adiabatic evolution}, \textit{quasi-adiabatic continuation}, or \textit{spectral flow}.

Two states in the same phase can be deformed into one another by evolving, via quasi-adiabatic evolution, for a time independent of the system size \(n\) \NoCaseChange{\protect\cite[{Appendix}]{cite2821}}.
The unitary operation generated by a quasi-local Hamiltonian can be simulated by a quantum circuit, with the time of evolution determining the depth of the circuit.
Approximating such constant-time evolution with constant error using a quantum circuit can be done in constant depth for finite system size \NoCaseChange{\protect\cite{cite2822}}.
For infinite system size, Haah proved that every bounded local Hamiltonian evolution is a limit of a sequence of locality-preserving automorphisms (a.k.a. quantum cellular automata, or QCAs) \NoCaseChange{\protect\cite[{Thm. A.17}]{cite2823}}.

States in certain phases (e.g., \flmRefsHyperref{code:topological}{topological phases}) remain in said phases even after evolving for longer times (e.g., times that are logarithmic in \(n\)).
This means that circuits of non-constant depth may leave a state in the same phase, making the phase classification problem quantum computationally hard \NoCaseChange{\protect\cite{cite890}}.
More general phases of matter are also computationally hard to recognize, even for quantum computers \NoCaseChange{\protect\cite{cite2824}}.

A state \(|\psi\rangle\) in an \textit{invertible phase} \NoCaseChange{\protect\cite{cite2825}} can be deformed to a product state if it is first tensored with other states in related invertible phases that are realized on copies of the underlying lattice of \(|\psi\rangle\).
Invertible phases are sometimes referred to as \textit{short-range entangled phases} \NoCaseChange{\protect\cite{cite2826,cite2827}}, and their low-energy excitations are often characterized by invertible field theories \NoCaseChange{\protect\cite{cite2828}}.

\codefieldsection{Protection}
Ground states of many Hamiltonians can be easily written as tensor-network states or, in 1D, matrix product states (MPS).
A no-go theorem states that open-boundary MPS that form a degenerate ground-state space of a gapped local Hamiltonian yield codes with distance that is only constant in the number of qubits \(n\), so MPS excitation ansatze have to be used to achieve a distance scaling nontrivially with \(n\) \NoCaseChange{\protect\cite{cite595}} (see also Ref. \NoCaseChange{\protect\cite{cite1260}}).

\codefieldsection{Notes}
\begin{eczvaluelist}
\item\relax Reviews of various quantum phases of matter and many-body systems \NoCaseChange{\protect\cite{cite2829,cite2830,cite2831,cite2832,cite2833,cite2834,cite2835,cite2836}}.
\item\relax Book on rigorous results on stability of non-topological phases \NoCaseChange{\protect\cite{cite2837}}.
\end{eczvaluelist}
\codefieldsection{Parent}
\begin{eczvaluelist}
\item\relax
\flmRefsHyperref[eczindexfamilyrel]{code:qecc}{Quantum error-correcting code (QECC)}\end{eczvaluelist}
\codefieldsection{Children}
\begin{eczvaluelist}
\item\relax
\flmRefsHyperref[eczindexfamilyrel]{code:matrix_qm}{Matrix-model code} --- Matrix-model codewords for simple codes are eigenstates of a matrix-model Hamiltonian.
\item\relax
\flmRefsHyperref[eczindexfamilyrel]{code:topological}{Topological code} --- Codespace of a topological code is typically the ground-state or low-energy subspace of a geometrically local Hamiltonian admitting a topological phase.
Logical qubits can also be created via lattice defects or by appropriately scheduling measurements of gauge generators (see Floquet codes).
Geometrically local frustration-free code Hamiltonians on Euclidean manifolds are stable with respect to sufficiently weak quasi-local perturbations when they satisfy local topological quantum order together with the Local-Gap condition \NoCaseChange{\protect\cite{cite2802,cite2803}}.

\item\relax
\flmRefsHyperref[eczindexfamilyrel]{code:commuting_projector}{Commuting-projector Hamiltonian code} --- Geometrically local commuting-projector code Hamiltonians on Euclidean manifolds are stable with respect to small perturbations when they satisfy the \flmRefsHyperref{ref2675}{TQO conditions}, meaning that a notion of a phase can be defined \NoCaseChange{\protect\cite{cite2676,cite2677,cite2678,cite2679}}. This notion can be extended to semi-hyperbolic manifolds \NoCaseChange{\protect\cite{cite2680}} and non-geometrically local QLDPC codes exhibiting check soundness \NoCaseChange{\protect\cite{cite2681}} (see also \NoCaseChange{\protect\cite{cite2682}}). Hamiltonians admitting a Peierls condition are stable to off-diagonal perturbations \NoCaseChange{\protect\cite{cite2683}}.
\item\relax
\flmRefsHyperref[eczindexfamilyrel]{code:constant_excitation}{Constant-excitation (CE) code} --- Constant-excitation codes are associated with a Hamiltonian governing the total excitations of the system.
\item\relax
\flmRefsHyperref[eczindexfamilyrel]{code:frustration_free}{Frustration-free Hamiltonian code}\item\relax
\flmRefsHyperref[eczindexfamilyrel]{code:symmetry_protected_self_correct}{Symmetry-protected self-correcting quantum code}\item\relax
\flmRefsHyperref[eczindexfamilyrel]{code:cft}{Conformal-field theory (CFT) code} --- CFT codewords lie in the low-energy subspace of a conformal field theory (CFT), e.g., the quantum Ising model at its critical point.
\item\relax
\flmRefsHyperref[eczindexfamilyrel]{code:syk}{SYK code} --- The SYK code Hamiltonian is constructed out of non-commuting few-site terms, and every fermion participates in many interactions.
\item\relax
\flmRefsHyperref[eczindexfamilyrel]{code:eth}{Eigenstate thermalization hypothesis (ETH) code} --- ETH codewords are eigenstates of a local Hamiltonian whose eigenstates satisfy ETH, and many example codes are eigenstates of frustration-free Hamiltonians.
\item\relax
\flmRefsHyperref[eczindexfamilyrel]{code:movassagh_ouyang}{Movassagh-Ouyang Hamiltonian code} --- Movassagh-Ouyang codes reside in the ground space of a Hamiltonian. Justesen codes can be used to build a family of \(n\)-qubit Movassagh-Ouyang Hamiltonian spin codes encoding one logical qubit with linear distance. These codes form the ground-state subspace of a frustration-free geometrically local Hamiltonian \NoCaseChange{\protect\cite{cite1407}}.
\end{eczvaluelist}
\codefieldsection{Cousins}
\begin{eczvaluelist}
\item\relax
\flmRefsHyperref[eczindexfamilyrel]{code:unitary_design}{Unitary \(t\)-design} --- Evolving with a Hamiltonian that has constant locality does not yield a unitary 2-design, but increasing the locality slightly overcomes this and yields a design \NoCaseChange{\protect\cite{cite2195}}.
\item\relax
\flmRefsHyperref[eczindexfamilyrel]{code:ldpc}{Low-density parity-check (LDPC) code} --- There are relations between LDPC codes and statistical mechanical models of spin glasses \NoCaseChange{\protect\cite{cite1485,cite1467,cite1468,cite946}}.
\item\relax
\flmRefsHyperref[eczindexfamilyrel]{code:quantum_double}{Quantum-double code} --- Quantum double code Hamiltonians can be simulated, with the help of perturbation theory and the \(\llbracket 4,1,1,2\rrbracket \) subsystem code, by two-dimensional two-body Hamiltonians with non-commuting terms \NoCaseChange{\protect\cite{cite2838}}.
\item\relax
\flmRefsHyperref[eczindexfamilyrel]{code:general_qldpc}{QLDPC code} --- QLDPC code Hamiltonians can be simulated, with the help of perturbation theory, by two-dimensional Hamiltonians with non-commuting terms whose interactions scale with \(n\) \NoCaseChange{\protect\cite{cite2839}}.
\item\relax
\flmRefsHyperref[eczindexfamilyrel]{code:paircat}{Pair-cat code} --- Two-component pair-cat codewords form ground-state subspace of a multimode Kerr Hamiltonian.
\item\relax
\flmRefsHyperref[eczindexfamilyrel]{code:two-legged-cat}{Two-component cat code} --- The two-component cat code forms the ground-state subspace of a Kerr Hamiltonian \NoCaseChange{\protect\cite{cite2840}}.
\item\relax
\flmRefsHyperref[eczindexfamilyrel]{code:metopt}{Error-corrected sensing code} --- Metrologically optimal codes admit a \(U(1)\) set of gates generated by a signal Hamiltonian \(H\), meaning that there exists a basis of codewords that are eigenstates of the \(H\).
\item\relax
\flmRefsHyperref[eczindexfamilyrel]{code:holographic_tensor}{Holographic tensor-network code} --- Local Hamiltonians lying at the CFT boundary can be mapped into the AdS bulk using tools from Hamiltonian simulation theory \NoCaseChange{\protect\cite{cite2841}}.
\item\relax
\flmRefsHyperref[eczindexfamilyrel]{code:mbq}{Majorana box qubit} --- A Majorana box qubit forms a fixed-parity subspace of the ground-state subspace of one or more Kitaev Majorana chain Hamiltonians.
\item\relax
\flmRefsHyperref[eczindexfamilyrel]{code:tetron}{Tetron code} --- Embedding each physical qubit into two fermions via the tetron code is useful for exactly solving the Kitaev honeycomb model Hamiltonian \NoCaseChange{\protect\cite{cite537}} and other qubit Hamiltonians on certain graphs \NoCaseChange{\protect\cite{cite2842,cite2843}}. Majorana stabilizer groups can be converted into ordinary qubit stabilizer groups via the parton mapping, while their corresponding states are converted via the Gutzwiller projection \NoCaseChange{\protect\cite{cite2844}}.
\item\relax
\flmRefsHyperref[eczindexfamilyrel]{code:qubit_concatenated}{Concatenated qubit code} --- Concatenated stabilizer code Hamiltonians have been investigated \NoCaseChange{\protect\cite{cite2845}}.
\item\relax
\flmRefsHyperref[eczindexfamilyrel]{code:2d_color}{2D color code} --- 2D color code Hamiltonians can be simulated, with the help of perturbation theory, by two-dimensional weight-two (two-body) Hamiltonians with non-commuting terms \NoCaseChange{\protect\cite{cite2846}}.
\item\relax
\flmRefsHyperref[eczindexfamilyrel]{code:surface}{Kitaev surface code} --- While codewords of the surface code form ground states of the code's stabilizer Hamiltonian, they can also be ground states of other gapless Hamiltonians \NoCaseChange{\protect\cite{cite2847}}.
\item\relax
\flmRefsHyperref[eczindexfamilyrel]{code:3d_surface}{3D surface code} --- Stability of the 3D surface code against Hamiltonian perturbations was determined using a tensor-network representation \NoCaseChange{\protect\cite{cite2848}}. The phase diagram of the perturbed tensor network maps to that of a 3D Ising gauge theory.
\item\relax
\flmRefsHyperref[eczindexfamilyrel]{code:bacon_shor}{Bacon-Shor code} --- The 2D Bacon-Shor gauge-group Hamiltonian is the compass model \NoCaseChange{\protect\cite{cite656,cite657,cite658}}.
\end{eczvaluelist}
\eczhbkcontributors{ \eczhuVVA }
\endeczcode

\eczcode{holographic}{Holographic code}{~\NoCaseChange{\protect\cite{cite1667}}}
\codefieldsection{Description}
Block quantum code whose features serve to model aspects of the AdS/CFT holographic duality and, more generally, quantum gravity.

In the original exactly solvable toy models, a network of perfect tensors defines an isometric encoding map from bulk logical degrees of freedom to boundary physical degrees of freedom \NoCaseChange{\protect\cite{cite1667}}.
For connected boundary regions of negatively curved planar holographic states, the discrete Ryu-Takayanagi formula is satisfied exactly, and bulk operators admit multiple boundary reconstructions via tensor pushing and greedy-geodesic methods \NoCaseChange{\protect\cite{cite1667}}.

\codefieldsection{Notes}
\begin{eczvaluelist}
\item\relax Reviews of holographic codes \NoCaseChange{\protect\cite{cite2849,cite2850}}.
\item\relax The original paper also describes black-hole toy models by removing central tensors and interpreting the newly exposed bulk legs as black-hole microstate degrees of freedom \NoCaseChange{\protect\cite{cite1667}}.
\end{eczvaluelist}
\codefieldsection{Parent}
\begin{eczvaluelist}
\item\relax
\flmRefsHyperref[eczindexfamilyrel]{code:qecc}{Quantum error-correcting code (QECC)}\end{eczvaluelist}
\codefieldsection{Children}
\begin{eczvaluelist}
\item\relax
\flmRefsHyperref[eczindexfamilyrel]{code:matrix_qm}{Matrix-model code} --- Matrix-model codes are motivated by the AdS/CFT correspondence because it is manifest in continuous non-Abelian gauge theories with large gauge groups \NoCaseChange{\protect\cite{cite2851}}.
\item\relax
\flmRefsHyperref[eczindexfamilyrel]{code:rg_cat}{Renormalization group (RG) cat code} --- The RG cat code encoder has coarse-graining features reminiscent of holography \NoCaseChange{\protect\cite{cite2545}}.
\item\relax
\flmRefsHyperref[eczindexfamilyrel]{code:holographic_tensor}{Holographic tensor-network code} --- Holographic codes whose encoders are holographic tensor networks are holographic tensor-network codes.
\item\relax
\flmRefsHyperref[eczindexfamilyrel]{code:cft}{Conformal-field theory (CFT) code} --- CFT codewords lie in the low-energy subspace of a conformal field theory (CFT), e.g., the quantum Ising model at its critical point.
\item\relax
\flmRefsHyperref[eczindexfamilyrel]{code:kpt}{Kim-Preskill-Tang (KPT) code} --- The robustness of KPT codes does not rely on arguments from holographic duality, but such codes do aim to describe interiors of black holes.
\item\relax
\flmRefsHyperref[eczindexfamilyrel]{code:syk}{SYK code} --- In a holographic model \NoCaseChange{\protect\cite{cite2601}}, the large distance of these codes can be interpreted as being due to the emergence of a wormhole.
\end{eczvaluelist}
\codefieldsection{Cousins}
\begin{eczvaluelist}
\item\relax
\flmRefsHyperref[eczindexfamilyrel]{code:approximate_qecc}{Approximate quantum error-correcting code (AQECC)} --- Universal subspace approximate error correction is used to model black holes \NoCaseChange{\protect\cite{cite2609}}.
\item\relax
\flmRefsHyperref[eczindexfamilyrel]{code:approximate_oaecc}{Approximate operator-algebra QECC} --- Properties of holographic codes are often quantified in the Heisenberg picture, i.e., in terms of operator algebras \NoCaseChange{\protect\cite{cite2543,cite2544,cite2545,cite2546}}.
\item\relax
\flmRefsHyperref[eczindexfamilyrel]{code:2d_stabilizer}{2D lattice stabilizer code} --- 2D lattice stabilizer codes admit a bulk-boundary correspondence similar to that of holographic codes, namely, the boundary Hilbert space of the former cannot be realized via local degrees of freedom \NoCaseChange{\protect\cite{cite2852}}.
\item\relax
\flmRefsHyperref[eczindexfamilyrel]{code:translationally_invariant_stabilizer}{Lattice stabilizer code} --- Lattice stabilizer codes admit a bulk-boundary correspondence similar to that of holographic codes \NoCaseChange{\protect\cite{cite2853}}.
\item\relax
\flmRefsHyperref[eczindexfamilyrel]{code:multimodegkp}{Gottesman-Kitaev-Preskill (GKP) code} --- GKP codespaces exist in the CFT dual of a particular holographic framework \NoCaseChange{\protect\cite{cite2854,cite2855}}.
\item\relax
\flmRefsHyperref[eczindexfamilyrel]{code:qubit_stabilizer}{Qubit stabilizer code} --- Qubit stabilizer states can be interpreted as states that are preparable using the Euclidean path integral in 3D Chern-Simons theory, defined on manifolds that are toy models of AdS/CFT wormholes \NoCaseChange{\protect\cite{cite2535,cite2536}}.
\end{eczvaluelist}
\eczhbkcontributors{ Joel Rajakumar, \eczhuVVA }
\endeczcode

\eczcode{holographic_tensor}{Holographic tensor-network code}{~\NoCaseChange{\protect\cite{cite1667,cite2856,cite2857,cite2858}}}
\codefieldsection{Description}
Quantum Lego code whose encoding isometry forms a holographic tensor network, i.e., a tensor network associated with a tiling of hyperbolic space.
Physical qubits are associated with uncontracted tensor legs at the boundary of the tessellation, while logical qubits are associated with uncontracted legs in the bulk.
The number of layers emanating from the central point of the tiling is the \textit{radius} of the code.

The encoding map models radial time evolution for a fixed time slice in Anti de Sitter (AdS) space, mapping operators in the bulk of AdS, represented by logical qudits, onto operators on the boundary of the corresponding conformal field theory (CFT), represented by physical qudits.
See \NoCaseChange{\protect\cite[{Defn. 4.3}]{cite468}} for a technical formulation.

\codefieldsection{Protection}
Protects against erasure errors on the boundary.
Error-correction properties are often stated in the Heisenberg picture, i.e., in terms of which logical operators can be \textit{reconstructed} after erasures.
Specifically, bulk operators outside the entanglement wedges of the erased boundary operators can be reconstructed using the remaining boundary operators.
However, the protection can be nontrivial, and may only apply to a subalgebra of bulk operators \NoCaseChange{\protect\cite{cite2543,cite2544}}.

Typically, the encoding isometry \(U\) obeys the \textit{entanglement-wedge reconstruction condition}, which states that for any boundary region \(R\), any bulk operator \(O\) localized to the entanglement wedge of \(R\) must be implementable by some boundary operator \(O^{\prime}\) localized to \(R\). Formally, \(UO = O^{\prime}U\) and \([O^{\prime},UU^\dagger] = 0\). The entanglement wedge is the space enclosed within the Ryu–Takayanagi surface in the bulk (minimal surface) with boundary \(R\).

\codefieldsection{Encoding}
\begin{eczvaluelist}
\item\relax Quantum encoding maps are isometries, but non-isometric encodings are relevant to describing mappings into the interior of a black hole \NoCaseChange{\protect\cite{cite2859}} and de Sitter time evolution \NoCaseChange{\protect\cite{cite642}}. Trace-norm preserving encodings have also been studied \NoCaseChange{\protect\cite{cite2860}}.
\end{eczvaluelist}
\codefieldsection{Transversal and Permutation-Based Gates}
\begin{eczvaluelist}
\item\relax There exist holographic approximate codes with arbitrary transversal gate sets for any compact Lie group \NoCaseChange{\protect\cite{cite468}}. However, for sufficiently localized logical subsystems of holographic stabilizer codes, the set of transversally implementable logical operations is contained in the \flmRefsHyperref{ref409}{Clifford group} \NoCaseChange{\protect\cite{cite740}}.
\end{eczvaluelist}
\codefieldsection{Code Capacity Threshold}
\begin{eczvaluelist}
\item\relax The ideal holographic tensor-network code (perfect representation of AdS/CFT) should be able to protect a central bulk operator against erasures of half of the physical qubits on the boundary, in line with AdS-Rindler reconstruction \NoCaseChange{\protect\cite{cite1667}}.
\item\relax Holographic tensor-network codes are argued to have a \textit{algebraic threshold}, for which the error rate scales polynomially (as opposed to exponentially) in the thermodynamic limit \NoCaseChange{\protect\cite{cite2861}}. Such a threshold is governed by the underlying conformal field theory describing the boundary.
\end{eczvaluelist}
\codefieldsection{Notes}
\begin{eczvaluelist}
\item\relax There is a link between position verification and holography \NoCaseChange{\protect\cite{cite2862,cite2863}}.
\end{eczvaluelist}
\codefieldsection{Parents}
\begin{eczvaluelist}
\item\relax
\flmRefsHyperref[eczindexfamilyrel]{code:holographic}{Holographic code} --- Holographic codes whose encoders are holographic tensor networks are holographic tensor-network codes.
\item\relax
\flmRefsHyperref[eczindexfamilyrel]{code:quantum_lego}{Tensor-network code} --- Quantum Lego codes whose encoders are tensor networks discretizing hyperbolic space can be thought of as holographic codes. More generally, holographic tensor-network codes are types of quantum LEGO codes made from stabilizer codes where logical and physical legs are pre-assigned and logical legs are not contracted. In other words, logical legs resulting from the conversion of codes to tensors must remain logical in the final tensor network, and the same for physical. Contracting logical legs is another word for gluing two logical legs together.
\end{eczvaluelist}
\codefieldsection{Children}
\begin{eczvaluelist}
\item\relax
\flmRefsHyperref[eczindexfamilyrel]{code:happy}{Pastawski-Yoshida-Harlow-Preskill (HaPPY) code} --- The encoding of a HaPPY code is a holographic tensor network consisting of pentagon and hexagon \flmRefsHyperref{ref219}{perfect tensors}.
\item\relax
\flmRefsHyperref[eczindexfamilyrel]{code:holographic_5_1_2}{Surface-code-fragment (SCF) holographic code} --- The encoding of the heptagon holographic code is a holographic tensor network consisting of the encoding isometry for the \(\llbracket 5,1,2\rrbracket \) rotated surface code, which is a \flmRefsHyperref{code:block_perfect}{planar-perfect tensor}.
\item\relax
\flmRefsHyperref[eczindexfamilyrel]{code:holographic_6_1_3}{Six-qubit-tensor holographic code} --- The encoding of the six-qubit-tensor holographic code is a holographic tensor network consisting of the encoding isometry for the \(\llbracket 6,1,3\rrbracket \) six-qubit stabilizer code.
\item\relax
\flmRefsHyperref[eczindexfamilyrel]{code:holographic_hyperinvariant}{Hyperinvariant tensor-network (HTN) code} --- The encoding of an HTN code is a hyperinvariant tensor network.
\item\relax
\flmRefsHyperref[eczindexfamilyrel]{code:holographic_steane}{Heptagon holographic code} --- The encoding of the heptagon holographic code is a holographic tensor network consisting of the encoding isometry for the Steane code, which is a \flmRefsHyperref{code:block_perfect}{planar-perfect tensor}.
\item\relax
\flmRefsHyperref[eczindexfamilyrel]{code:holographic_subsystem}{Subsystem holographic code} --- The holographic hybrid code is constructed out of alternating isometries of the five-qubit and \(\llbracket 4,1,1,2\rrbracket \) Bacon-Shor codes.
\item\relax
\flmRefsHyperref[eczindexfamilyrel]{code:stab_3_1_2}{\(\llbracket 3,1,2\rrbracket _3\) Three-qutrit code} --- The three-qutrit code is a radius-one holographic tensor-network code and serves as a minimal model for holography \NoCaseChange{\protect\cite{cite2543,cite2864}}.
\end{eczvaluelist}
\codefieldsection{Cousins}
\begin{eczvaluelist}
\item\relax
\flmRefsHyperref[eczindexfamilyrel]{code:random_stabilizer}{Random stabilizer code} --- Random holographic tensor-network codes reproduce many aspects of holography \NoCaseChange{\protect\cite{cite2856,cite2857,cite2865}}.
\item\relax
\flmRefsHyperref[eczindexfamilyrel]{code:hamiltonian}{Hamiltonian-based code} --- Local Hamiltonians lying at the CFT boundary can be mapped into the AdS bulk using tools from Hamiltonian simulation theory \NoCaseChange{\protect\cite{cite2841}}.
\item\relax
\flmRefsHyperref[eczindexfamilyrel]{code:galois_grs}{Galois-qudit GRS code} --- Galois-qudit GRS codes can be used to construct holographic p-adic (i.e., tree-tensor-network) codes on Bruhat-Tits trees and buildings and on Drinfeld symmetric spaces \NoCaseChange{\protect\cite{cite2866,cite2867}}.
\item\relax
\flmRefsHyperref[eczindexfamilyrel]{code:qecc}{Quantum error-correcting code (QECC)} --- Quantum encoding maps are isometries, but non-isometric encodings are relevant to describing mappings into the interior of a black hole \NoCaseChange{\protect\cite{cite2859}} and de Sitter time evolution \NoCaseChange{\protect\cite{cite642}}. Trace-norm preserving encodings have also been studied \NoCaseChange{\protect\cite{cite2860}}.
\item\relax
\flmRefsHyperref[eczindexfamilyrel]{code:stab_15_1_3}{\(\llbracket 15,1,3\rrbracket \) quantum RM code} --- The \(\llbracket 15,1,3\rrbracket \) code serves as the local tensor in a holographic quantum-Lego code on a \(\{15,4\}\) tiling that supports a transversal non-Clifford operator \NoCaseChange{\protect\cite{cite2868}}.
\item\relax
\flmRefsHyperref[eczindexfamilyrel]{code:hyperbolic_surface}{Hyperbolic surface code} --- Both holographic tensor-network and hyperbolic surface codes utilize tessellations of hyperbolic surfaces. Encodings for the former are hyperbolically tiled tensor networks, while the latter is defined on hyperbolically tiled physical-qubit lattices.
\end{eczvaluelist}
\eczhbkcontributors{ Matthew Steinberg, \eczhuVVA }
\endeczcode

\eczcode{hybridqecc}{Hybrid QECC}{~\NoCaseChange{\protect\cite{cite2869,cite2870,cite2871,cite2735}}}
\codefieldsection{Description}
A quantum code which encodes both quantum and classical information.

Hybrid QECCs arise as the \(e=0\) subclass of the EACQ formalism, i.e., the classically enhanced quantum codes that do not require pre-shared entanglement \NoCaseChange{\protect\cite{cite2735}}.

In general, a different quantum code \(\mathsf{C}_j\) is associated with each classical
value \(j \in \{0, 1, \ldots, l-1\}\), and the Hilbert space decomposes as
\flmMathEnvironment{align}{}{
  \mathsf{H} = \bigoplus_{j=0}^{l-1} \mathsf{C}_j \oplus \mathsf{C}^{\perp}~,
}
where \(\mathsf{C}^{\perp}\) is the combined error space of all \(l\) codes.
The simplest example encodes a single qubit and a single classical bit (\(l = 2\)):
a different quantum code is associated with each value \(j \in \{0,1\}\), giving
\(\mathsf{H} = \mathsf{C}_0 \oplus \mathsf{C}_1 \oplus \mathsf{C}^{\perp}\).
The error-correction conditions require the \flmTerm{term}{ref1043}{}{Knill-Laflamme conditions} to hold within
each quantum code subspace \(\mathsf{C}_j\) \NoCaseChange{\protect\cite[{Eq. (3)}]{cite2872}}, and that error operators
map different code subspaces to mutually orthogonal subspaces \NoCaseChange{\protect\cite[{Eq. (4)}]{cite2872}}.

\codefieldsection{Rate}
The capacity of a hybrid quantum memory is determined by a convex region in the classical-quantum entropy plane \NoCaseChange{\protect\cite{cite2869}}. The quantum capacity for simultaneous transmission of classical and quantum information has been derived \NoCaseChange{\protect\cite{cite2870}}. The existence of a hybrid code protecting against a channel depends on certain matricial ranges \NoCaseChange{\protect\cite{cite2873}}.
\codefieldsection{Parent}
\begin{eczvaluelist}
\item\relax
\flmRefsHyperref[eczindexfamilyrel]{code:oaecc}{Operator-algebra QECC (OAQECC)} --- An OAQECC which has no gauge structure (e.g., gauge qubits) but has a block structure that corresponds to a classical code is a hybrid QECC.
\end{eczvaluelist}
\codefieldsection{Child}
\begin{eczvaluelist}
\item\relax
\flmRefsHyperref[eczindexfamilyrel]{code:hybrid_qubits_into_qubits}{Hybrid qubit code}\end{eczvaluelist}
\codefieldsection{Cousins}
\begin{eczvaluelist}
\item\relax
\flmRefsHyperref[eczindexfamilyrel]{code:qecc}{Quantum error-correcting code (QECC)} --- A hybrid QECC storing no classical information reduces to a QECC. Conversely, any QECC can be converted into a hybrid QECC by using a portion of its logical subspace to store only classical information.
\item\relax
\flmRefsHyperref[eczindexfamilyrel]{code:classical_into_quantum}{Classical-quantum (c-q) code} --- A hybrid QECC storing no quantum information reduces to a c-q code.
\item\relax
\flmRefsHyperref[eczindexfamilyrel]{code:eacq}{Entanglement-assisted (EA) hybrid QECC} --- EA hybrid codes utilize additional ancillary subsystems in a pre-shared entangled state, but reduce to hybrid QECCs when said subsystems are interpreted as noiseless physical subsystems.
\item\relax
\flmRefsHyperref[eczindexfamilyrel]{code:subsystem_quantum_parity}{Subsystem hypergraph product (SHP) code} --- Classical information can also be encoded in subsystem codes using their gauge qubits \NoCaseChange{\protect\cite{cite2874}}.
\end{eczvaluelist}
\eczhbkcontributors{ \eczhuVVA }
\endeczcode

\eczcode{knill}{Knill code}{~\NoCaseChange{\protect\cite{cite2813}}}
\codefieldsection{Alternative Names}
\begin{eczvaluelist}
\item\relax Clifford code
\end{eczvaluelist}
\eczhIndexCodeAliasName{knill}{Clifford code}
\codefieldsection{Description}
A group representation code whose projection is onto an irrep of a normal subgroup of the group \(G\) formed by a nice error basis.
More general definitions dropping the normality requirement also exist \NoCaseChange{\protect\cite{cite2875}}.
Knill codes yield stabilizer-like codes based on error bases that are non-Pauli but that nevertheless maintain many of the useful features of Pauli-type bases.

\begin{defterm}{Nice error basis}\label{ref2876}\label{ref2812}
  A nice error basis \NoCaseChange{\protect\cite{cite2877,cite2813,cite2878}} for a \(q\)-dimensional vector space is a set \(\{E_g~,~g\in G\}\) of unitary operators, where \(G\) is a (not necessarily Abelian) group of order \(q^2\) that is represented projectively. Namely,
  \flmMathEnvironment{align}{}{
    \text{tr}(E_{g})&=q\delta^{G}_{g,1}\\
    E_{g}E_{h}&=\omega_{g,h}E_{gh}
  }
for all group elements \(g,h\).
Above, \(\delta^{G}_{g,1}\) is the \flmRefsHyperref{ref20}{group Kronecker-delta function}.
This definition can naturally be extended to continuous groups.
\end{defterm}

The first example of an error basis based on a non-Abelian error group is due to S. Egner and consists of products of \(S\), Pauli, and Hadamard gates \NoCaseChange{\protect\cite{cite2813}}.
An example of a small non-stabilizer Knill code is presented in \NoCaseChange{\protect\cite[{Sec. 10.9}]{cite2879}}. 
Certain nice error bases have been classified and are related to the braid group \NoCaseChange{\protect\cite{cite2880}}.

\codefieldsection{Notes}
\begin{eczvaluelist}
\item\relax Catalogue of \flmRefsHyperref{ref2812}{nice error bases}, managed by A. Klappenecker and M. Rotteler, is available on \flmHref{https://people.engr.tamu.edu/andreas-klappenecker/ueb/ueb.html}{this website}.
\item\relax Many Knill codes are qubit stabilizer codes \NoCaseChange{\protect\cite{cite2881}}. A table of non-stabilizer Knill codes is available in Ref. \NoCaseChange{\protect\cite{cite2882}}. An infinite family is constructed in Ref. \NoCaseChange{\protect\cite{cite2883}}.
\end{eczvaluelist}
\codefieldsection{Parent}
\begin{eczvaluelist}
\item\relax
\flmRefsHyperref[eczindexfamilyrel]{code:group_representation}{Group-representation code} --- Knill codes project onto a single irrep sector associated with a normal subgroup of the group formed by a \flmRefsHyperref{ref2812}{nice error basis} \NoCaseChange{\protect\cite[{Lemma 3.1}]{cite2813}}.
\end{eczvaluelist}
\codefieldsection{Child}
\begin{eczvaluelist}
\item\relax
\flmRefsHyperref[eczindexfamilyrel]{code:stabilizer}{Stabilizer code} --- Stabilizer codes are Knill codes whose \flmRefsHyperref{ref2812}{nice error basis} is either the Pauli strings, modular-qudit Pauli strings, Galois-qudit Pauli strings, oscillator displacement operators, or rotor generalized Pauli strings.
\end{eczvaluelist}
\codefieldsection{Cousin}
\begin{eczvaluelist}
\item\relax
\flmRefsHyperref[eczindexfamilyrel]{code:oecc}{Subsystem QECC} --- Subsystem Knill codes can be formulated \NoCaseChange{\protect\cite{cite2884}}.
\end{eczvaluelist}
\eczhbkcontributors{ \eczhuVVA }
\endeczcode

\eczcode{metrological}{Metrological code}{~\NoCaseChange{\protect\cite{cite2717}}}
\codefieldsection{Description}
Two-dimensional subspace of a Hilbert space whose basis states satisfy only a part of the \flmTerm{term}{ref1043}{}{Knill-Laflamme conditions}. The satisfied part of the conditions ensures that the code can be used for local parameter estimation.

Letting \(\Pi = U U^\dagger\) be the codespace projector for encoding isometry \(U\) and projecting a pair of errors \(E_i,E_j\) from an error set \(\cal E\) into the two-dimensional codespace yields
\flmMathEnvironment{align}{}{
  \Pi E_{i}^{^{\dagger}}E_{j}\Pi=c_{ij}\,\Pi+x_{ij}\overline{X}+y_{ij}\overline{Y}+z_{ij}\overline{Z}
}
with error-matrix element \(c_{ij}\) and logical-error coefficients
\flmMathEnvironment{align}{}{
  \left\{ x,y,z\right\} _{ij}={\textstyle \frac{1}{2}}\text{Tr}\left(\{ \overline{X},\overline{Y},\overline{Z}\} E_{i}^{^{\dagger}}E_{j}\right)~.
}
If all three logical-error coefficients are zero, then the \flmTerm{term}{ref1043}{}{Knill-Laflamme conditions} are satisfied, and the code is a \flmRefsHyperref{code:qecc}{QECC}. If only one of the three coefficients is zero, then the code is the more general metrological code.

\codefieldsection{Protection}
Physical noise can cause logical errors along one of the three axes, i.e., either logical-\(X\), \(Y\), or \(Z\), depending on what basis is used. Codes protect against logical errors along the remaining two axes.

A metrological block quantum code has distance \(d\) if the above conditions are satisfied for an error set \(\cal E\) consisting of errors supported on \(d-1\) subsystems or fewer.

\codefieldsection{Parent}
\begin{eczvaluelist}
\item\relax
\flmRefsHyperref[eczindexfamilyrel]{code:quantum_into_quantum}{Quantum code}\end{eczvaluelist}
\codefieldsection{Cousins}
\begin{eczvaluelist}
\item\relax
\flmRefsHyperref[eczindexfamilyrel]{code:metopt}{Error-corrected sensing code} --- Error-corrected sensing codes are required to satisfy the \flmTerm{term}{ref1043}{}{Knill-Laflamme conditions}, while metrological codes need only satisfy the conditions partially.
\item\relax
\flmRefsHyperref[eczindexfamilyrel]{code:covariant}{Covariant block quantum code} --- Any time-covariant QECC, i.e., a code admitting a continuous-parameter \(U(1)\) family of gates, is automatically a metrological code.
\item\relax
\flmRefsHyperref[eczindexfamilyrel]{code:qubit_stabilizer}{Qubit stabilizer code} --- A joint \(+1\) and \(-1\) eigenstate of a set of stabilizers can form a metrological stabilizer code \NoCaseChange{\protect\cite{cite2717}}.
\item\relax
\flmRefsHyperref[eczindexfamilyrel]{code:qecc}{Quantum error-correcting code (QECC)} --- Metrological codes are logical-qubit codes that satisfy the \flmTerm{term}{ref1043}{}{Knill-Laflamme conditions} conditions only partially, and codes that satisfy them fully are QECCs.
\item\relax
\flmRefsHyperref[eczindexfamilyrel]{code:qetc}{Quantum error-transmuting code (QETC)} --- Metrological codes are also codes which satisfy a generalization of the \flmTerm{term}{ref1043}{}{Knill-Laflamme conditions}, albeit a different one.
\end{eczvaluelist}
\eczhbkcontributors{ \eczhuVVA }
\endeczcode

\eczcode{monitored_random_circuits}{Monitored random-circuit code}{~\NoCaseChange{\protect\cite{cite2885,cite2886,cite2887}}}
\codefieldsection{Description}
Error-correcting code arising from a monitored random circuit. Such a circuit is described by a series of intermittent random local projective Pauli measurements with random unitary time-evolution operators.

An important sub-family consists of \textit{Clifford monitored random circuits}, where unitaries are sampled from the \flmRefsHyperref{ref409}{Clifford group} \NoCaseChange{\protect\cite{cite2888}}.
When the rate of projective measurements is independently controlled by a probability parameter \(p\), there can exist two stable phases, one described by volume-law entanglement entropy and the other by area-law entanglement entropy.
The phases and their transition can be understood from the perspective of quantum error correction, information scrambling, and channel capacities \NoCaseChange{\protect\cite{cite2889,cite541}}.

In the volume-law or mixed phase (\( p < p_c \) for some critical probability \(p_c\)), the channel-capacity density remains nonzero on polynomial timescales and the purification time grows exponentially with system size \NoCaseChange{\protect\cite{cite541}}.
The monitored dynamics projects the system into a random error-correcting code, and for strong purification transitions this code can be capacity-achieving for the future unraveled evolution of the channel \NoCaseChange{\protect\cite{cite541}}.
In the area-law or pure phase (\( p > p_c \)), the channel-capacity density vanishes and the system purifies rapidly \NoCaseChange{\protect\cite{cite541}}.
With appropriately chosen evolution operators and measurements, the code is a stabilizer code whose parameters depend on time, \( \llbracket n,k(t),d(t)\rrbracket  \).
A similar notion applies to Haar random circuits with measurements \NoCaseChange{\protect\cite{cite2890}}.

\codefieldsection{Protection}
When \( p < p_c \), protects against monitored projective measurements by dynamically encoding information into a late-time code space. For one-dimensional stabilizer circuits, the average contiguous code length is efficiently computable and upper bounds the code distance; it is subextensive deep in the mixed phase and appears extensive near \( p_c \) \NoCaseChange{\protect\cite{cite541}}.
\codefieldsection{Rate}
Rate can be finite for \( p < p_c \) and vanishes for \( p > p_c \); in the 1+1-dimensional random Clifford model, the residual entropy density equals the channel-capacity density in the mixed phase \NoCaseChange{\protect\cite{cite541}}.
\codefieldsection{Encoding}
\begin{eczvaluelist}
\item\relax The dynamics of the monitored random circuit can be recast in the language of stabilizer codes \NoCaseChange{\protect\cite{cite541}}. The stabilizer group of the error-correcting code resulting from a monitored Clifford circuit either grows or shrinks with each time step, depending on which projective measurements were performed during the time step.
\item\relax For strong purification transitions, the monitored dynamics itself implements a single-copy capacity-achieving encoding for the future unraveled evolution of the channel \NoCaseChange{\protect\cite{cite541}}.
\end{eczvaluelist}
\codefieldsection{Decoding}
\begin{eczvaluelist}
\item\relax With access to the measurement record, recovery operations can reverse the future unraveled evolution with high fidelity; on the code space, the induced dynamics becomes effectively unitary in the thermodynamic limit \NoCaseChange{\protect\cite{cite541}}.
\end{eczvaluelist}
\codefieldsection{Threshold}
\begin{eczvaluelist}
\item\relax At the purification threshold \( p_c \), the channel-capacity density changes from finite to zero; above \( p_c \), the natural error-correction properties of the circuit can no longer protect an extensive amount of information \NoCaseChange{\protect\cite{cite541}}.
\item\relax These dynamically generated codes saturate the trade-off between the density of encoded information and the error-rate threshold \NoCaseChange{\protect\cite{cite541}}.
\end{eczvaluelist}
\codefieldsection{Realizations}
\begin{eczvaluelist}
\item\relax Measurement-induced quantum phases have been realized in a trapped-ion processor \NoCaseChange{\protect\cite{cite2891}}.
\end{eczvaluelist}
\codefieldsection{Notes}
\begin{eczvaluelist}
\item\relax Connections to information scrambling in black hole physics, as introduced in \NoCaseChange{\protect\cite[{Sec. 11}]{cite2889}}. In particular, monitored random circuits can be viewed as the Hayden-Preskill recovery problem \NoCaseChange{\protect\cite{cite2892}} running backwards in time. In this setting, the volume-law entanglement phase of the monitored circuit describes the phase when information can be recovered from an old black hole (i.e., a black hole that is maximally entangled with the early universe).
\item\relax Mapping monitored random circuits to statistical mechanics models can help estimate thresholds and code distances for these systems \NoCaseChange{\protect\cite{cite2893}}.
\end{eczvaluelist}
\codefieldsection{Parent}
\begin{eczvaluelist}
\item\relax
\flmRefsHyperref[eczindexfamilyrel]{code:random_circuit}{Random-circuit code} --- Monitored random circuits are random circuits where projective measurements are interspersed throughout the circuit and measurement results are recorded.
\end{eczvaluelist}
\codefieldsection{Cousins}
\begin{eczvaluelist}
\item\relax
\flmRefsHyperref[eczindexfamilyrel]{code:topological}{Topological code} --- Topological order can be generated in 2D monitored random circuits \NoCaseChange{\protect\cite{cite2894}}.
\item\relax
\flmRefsHyperref[eczindexfamilyrel]{code:random_stabilizer}{Random stabilizer code} --- An important sub-family of monitored random-circuit codes is the Clifford monitored random-circuit family, where unitaries are sampled from the \flmRefsHyperref{ref409}{Clifford group} \NoCaseChange{\protect\cite{cite2888}}.
\item\relax
\flmRefsHyperref[eczindexfamilyrel]{code:da}{Dynamical code} --- Both dynamical and monitored random circuit codes can have an instantaneous stabilizer group which evolves through unitary evolution and measurements. However, dynamical codewords are generated via a specific prescribed sequence of measurements, while random-circuit codes maintain a stabilizer group after any measurement. Dynamical codes have the additional capability of detecting errors induced during the measurement process; see Appx. A of Ref. \NoCaseChange{\protect\cite{cite536}}.
\item\relax
\flmRefsHyperref[eczindexfamilyrel]{code:crystalline_dynamic_gen}{Crystalline-circuit qubit code} --- Projective measurements can be included in crystalline-circuit codes in a spacetime translation-invariant fashion, turning such codes into \textit{monitored crystalline-circuit codes}. However, the unit cell of measurements must be large enough to avoid purification.
\end{eczvaluelist}
\eczhbkcontributors{ Elizabeth R. Bennewitz, \eczhuVVA }
\endeczcode

\eczcode{single_subsystem}{Monolithic quantum code}{}
\codefieldsection{Description}
A code constructed in a single quantum system, i.e., a physical space that is \textit{not} treated as a tensor product of \(n\) identical subsystems.
Examples include codes in a single qudit, spin, oscillator, or molecule.

\codefieldsection{Parent}
\begin{eczvaluelist}
\item\relax
\flmRefsHyperref[eczindexfamilyrel]{code:qecc}{Quantum error-correcting code (QECC)}\end{eczvaluelist}
\codefieldsection{Children}
\begin{eczvaluelist}
\item\relax
\flmRefsHyperref[eczindexfamilyrel]{code:molecular}{Molecular code}\item\relax
\flmRefsHyperref[eczindexfamilyrel]{code:rotor_gkp}{Rotor GKP code}\item\relax
\flmRefsHyperref[eczindexfamilyrel]{code:single-mode}{Single-mode bosonic code}\item\relax
\flmRefsHyperref[eczindexfamilyrel]{code:qudit_gkp}{Modular-qudit GKP code}\item\relax
\flmRefsHyperref[eczindexfamilyrel]{code:qudit_sign}{Modular-qudit shift-resistant code}\item\relax
\flmRefsHyperref[eczindexfamilyrel]{code:single_spin}{Single-spin code}\end{eczvaluelist}
\codefieldsection{Cousin}
\begin{eczvaluelist}
\item\relax
\flmRefsHyperref[eczindexfamilyrel]{code:block_quantum}{Block quantum code} --- Block quantum codes for \(n=1\) are monolithic codes.
\end{eczvaluelist}
\eczhbkcontributors{ \eczhuVVA }
\endeczcode

\eczcode{oaecc}{Operator-algebra QECC (OAQECC)}{~\NoCaseChange{\protect\cite{cite648,cite2895,cite2896,cite2897,cite2871,cite2898,cite2899,cite1040,cite2900,cite2545}}}
\codefieldsection{Description}
A code family that encompasses ordinary (i.e., subspace) codes, subsystem codes, classical-quantum codes, and hybrid codes using an operator-algebraic framework.

A simple example encompassing elements of all subfamilies encodes quantum information and a single classical bit into a direct sum of two subsystem codes.
A quantum subsystem code \(\mathsf{A}_j\otimes\mathsf{B}_j\), with \(\mathsf{A}_j\) the logical factor associated with the quantum information, and \(\mathsf{B}_j\) the gauge factor, is associated with each of the two values \(j\in\{1,2\}\) of the classical bit.
The corresponding decomposition of the Hilbert space \(\mathsf{H}\) is
\flmMathEnvironment{align}{}{
  \mathsf{H}=(\mathsf{A}_{1}\otimes\mathsf{B}_{1})\oplus(\mathsf{A}_{2}\otimes\mathsf{B}_{2})\oplus\mathsf{C}^{\perp}~,
}
where \(\mathsf{C}^\perp\) is the combined error space of both codes.
The above code reduces to a subsystem code when \(\mathsf{A}_{2}\otimes\mathsf{B}_{2}\) is trivial, reduces to a classical-quantum code when \(\mathsf{A}_{1,2}\) are both trivial, reduces to a hybrid code when \(\mathsf{B}_{1,2}\) are both trivial, and reduces to an ordinary (i.e., subspace) code when \(\mathsf{B}_1\) and \(\mathsf{A}_{2}\otimes\mathsf{B}_{2}\) are both trivial.

In general, an OAQECC is determined by a finite dimensional \(C^*\) algebra \(\mathcal{A}\) of operators on \(\mathsf{H}\).
This \textit{logical algebra} induces a decomposition of the Hilbert space as
\flmMathEnvironment{align}{}{\mathsf{H} = \bigoplus_\gamma \mathsf{A}_\gamma \otimes \mathsf{B}_\gamma,}
with respect to which \(\mathcal{A}\) takes the form
\flmMathEnvironment{align}{}{\mathcal{A} = \bigoplus_\gamma I_\gamma \otimes \mathcal{L}(\mathsf{B}_\gamma),}
where \(\mathcal{L}(\mathsf{B}_\gamma)\) denotes the full set of linear maps on \(\mathsf{B}_\gamma\).
The \(\mathsf{A}_{\gamma}\) factors can be used to store quantum information, \(\gamma\) indexes the block structure of the code, while \(\mathsf{B}_{\gamma}\) determine its gauge structure.
Together, the above forms the most general form of an information preserving structure \NoCaseChange{\protect\cite{cite648,cite2901,cite2902,cite2897,cite2903,cite2904}}.
Logical operators form the commutant of \(\mathcal{A}\) as a result of the double commutant (a.k.a. double centralizer) theorem \NoCaseChange{\protect\cite{cite2905}}.
\codefieldsection{Protection}
Given an error operation \(\mathcal{E}\), one says that \(\mathcal{A}\) is \textit{correctable} for \(\mathcal{E}\) if there exists a recovery operation \(\mathcal{R}\) such that
\flmMathEnvironment{align}{}{\Pi_{\mathcal{A}} (\mathcal{R} \circ \mathcal{E})^\dagger(X) \Pi_{\mathcal{A}} = X} for all \(X \in \mathcal{A}\), where \(\Pi_{\mathcal{A}}\) is the unit projection onto \(\mathcal{A}\).

Equivalently, \(\mathcal{A}\) is correctable for \(\mathcal{E}\) if there exists a recovery operation \(\mathcal{R}\) such that for any \(\gamma\) and density operators \(\rho_\gamma,\sigma_\gamma\) supported on \(\mathsf{A}_\gamma\) and \(\mathsf{B}_\gamma\), respectively, there exists a state \(\tau_\gamma\) supported on \(\mathsf{A}_\gamma\) such that
\flmMathEnvironment{align}{}{(\mathcal{R} \circ \mathcal{E})(\rho_\gamma \otimes \sigma_\gamma) = \tau_\gamma \otimes \sigma_\gamma.}

An algebraic condition for correctability can be given in terms of the Kraus operators \(E_j\) of \(\mathcal{E}\).
Indeed, \(\mathcal{A}\) is correctable for \(\mathcal{E}\) if \flmMathEnvironment{align}{}{\Pi_{\mathcal{A}} E_j^\dagger E_k \Pi_{\mathcal{A}} \in \mathcal{A}'}
for all \(j,k\), where \(\mathcal{A}'\) is the commutant of \(\mathcal{A}\).

Conversely, a \textit{private} algebra \(\mathcal{A}\) for a channel \(\mathcal{E}\) is one which is completely decohered by the channel \NoCaseChange{\protect\cite{cite2906}}.
In other words, no information about the algebra is retained after the action of the channel.
Tradeoffs between error correction and privacy have been studied \NoCaseChange{\protect\cite{cite2907}}.

\codefieldsection{Parent}
\begin{eczvaluelist}
\item\relax
\flmRefsHyperref[eczindexfamilyrel]{code:quantum_into_quantum}{Quantum code}\end{eczvaluelist}
\codefieldsection{Children}
\begin{eczvaluelist}
\item\relax
\flmRefsHyperref[eczindexfamilyrel]{code:classical_into_quantum}{Classical-quantum (c-q) code} --- An OAQECC that retains its block structure for storing classical information but stores no quantum information and has no gauge degrees of freedom (e.g., gauge qubits) is a c-q code.
\item\relax
\flmRefsHyperref[eczindexfamilyrel]{code:hybridqecc}{Hybrid QECC} --- An OAQECC which has no gauge structure (e.g., gauge qubits) but has a block structure that corresponds to a classical code is a hybrid QECC.
\item\relax
\flmRefsHyperref[eczindexfamilyrel]{code:oecc}{Subsystem QECC} --- An OAQECC which has gauge structure (e.g., gauge qubits) but no block structure is a subsystem QECC.
\item\relax
\flmRefsHyperref[eczindexfamilyrel]{code:qecc}{Quantum error-correcting code (QECC)} --- An OAQECC which has no gauge structure (e.g., gauge qubits) and no block structure is a QECC.
\item\relax
\flmRefsHyperref[eczindexfamilyrel]{code:oa_qubits_into_qubits}{OA qubit code} --- An OAQECC defined over qubits is an OA qubit code.
\end{eczvaluelist}
\codefieldsection{Cousins}
\begin{eczvaluelist}
\item\relax
\flmRefsHyperref[eczindexfamilyrel]{code:approximate_oaecc}{Approximate operator-algebra QECC} --- Approximate OAQECCs correcting a noise channel exactly reduce to OAQECCs.
\item\relax
\flmRefsHyperref[eczindexfamilyrel]{code:eaoaecc}{Entanglement-assisted operator-algebra QECC (EAOA QECC)} --- EAOA QECCs use pre-shared entangled ancillary subsystems, while OAQECCs recover the same operator-algebraic structures when those ancillary subsystems are instead treated as noiseless physical subsystems.
\end{eczvaluelist}
\eczhbkcontributors{ Michael Liaofan Liu, \eczhuVVA }
\endeczcode

\eczcode{quantum_perfect}{Perfect quantum code}{}
\codefieldsection{Description}
A type of block quantum code whose parameters satisfy the quantum Hamming bound with equality.

A \flmRefsHyperref{ref811}{non-degenerate} code constructed out of \(q\)-dimensional qudits and having parameters \(\llparenthesis n,K,2t+1\rrparenthesis \) is perfect if \(n\), \(K\), \(t\), and \(q\) are such that the quantum Hamming bound \NoCaseChange{\protect\cite{cite2669}},
\flmMathEnvironment{align}{}{
\sum_{j=0}^{t}(q^2-1)^{j}{n \choose j}\leq q^{n}/K
}
becomes an equality for such codes.
For example, for a qubit \(q=2\) code with one logical qubit (\(K=2\)) and \(t=1\), the bound becomes \(3n+1 \leq 2^{n-1}\).
The bound can be saturated only at certain \(n\).

For qubit codes with \(K=2^k\), one can work out an asymptotic Hamming bound in the large-\(n,k,t\) limit,
\flmMathEnvironment{align}{}{
\frac{k}{n}\leq 1-\frac{t}{n}\log_{2}3-h(t/n),
}
where \(h\) is the binary entropy function.

\flmRefsHyperref{ref811}{Degenerate} codes can in principle violate the quantum Hamming bound.
It was shown that qubit stabilizer codes correcting up to two errors \NoCaseChange{\protect\cite{cite736}}, qudit stabilizer codes up to distance two \NoCaseChange{\protect\cite{cite2908}}, qudit CSS codes of qudit dimension \(q\geq 5\) along with certain other codes \NoCaseChange{\protect\cite{cite2780}}, and qubit codes up to distance \(d\leq 127\) \NoCaseChange{\protect\cite{cite2909}} do not violate the bound.
A quantum Hamming-like bound exists for \flmRefsHyperref{ref811}{degenerate} qubit stabilizer codes \NoCaseChange{\protect\cite{cite2910}}.

\codefieldsection{Protection}
Perfect codes have been classified.
For qubits (\(q=2\)), the only nontrivial perfect codes are the stabilizer code family \(\llbracket (4^r-1)/3, (4^r-1)/3 - 2r, 3\rrbracket \) for \(r \geq 2\), obtained from Hamming codes over \(\mathbb{F}_4\) via the Hermitian construction \NoCaseChange{\protect\cite{cite1694,cite449}}. These codes are related to partial spreads in projective geometry \NoCaseChange{\protect\cite{cite1695}}.
For qudits, the corresponding family is the \(\llbracket \frac{q^{2r}-1}{q^{2}-1},\frac{q^{2r}-1}{q^{2}-1}-2r,3\rrbracket _q\) family of quantum twisted codes \NoCaseChange{\protect\cite{cite2911,cite2912}}.

\codefieldsection{Rate}
\(k/n\to 1\) asymptotically with \(n\).
\codefieldsection{Notes}
\begin{eczvaluelist}
\item\relax 
\end{eczvaluelist}
\codefieldsection{Parents}
\begin{eczvaluelist}
\item\relax
\flmRefsHyperref[eczindexfamilyrel]{code:small_distance_quantum}{Small-distance block quantum code} --- All non-trivial perfect codes have distance three.
\item\relax
\flmRefsHyperref[eczindexfamilyrel]{code:qecc_finite}{Finite-dimensional quantum error-correcting code}\end{eczvaluelist}
\codefieldsection{Child}
\begin{eczvaluelist}
\item\relax
\flmRefsHyperref[eczindexfamilyrel]{code:stab_5_1_3}{\(\llbracket 5,1,3\rrbracket \) Five-qubit perfect code} --- The five-qubit code is the smallest perfect code and is a member of the perfect qubit code family \(\llbracket (4^r-1)/3, (4^r-1)/3 - 2r, 3\rrbracket \) for \(r = 2\).
\end{eczvaluelist}
\codefieldsection{Cousins}
\begin{eczvaluelist}
\item\relax
\flmRefsHyperref[eczindexfamilyrel]{code:perfect}{Perfect code} --- A classical (quantum) perfect code saturates the classical (quantum) Hamming bound.
\item\relax
\flmRefsHyperref[eczindexfamilyrel]{code:stabilizer_over_gf4}{Hermitian qubit code} --- For qubits (\(q=2\)), the only nontrivial perfect codes are the stabilizer code family \(\llbracket (4^r-1)/3, (4^r-1)/3 - 2r, 3\rrbracket \) for \(r \geq 2\), obtained from Hamming codes over \(\mathbb{F}_4\) via the Hermitian construction \NoCaseChange{\protect\cite{cite1694,cite449}}. These codes are related to partial spreads in projective geometry \NoCaseChange{\protect\cite{cite1695}}.
\item\relax
\flmRefsHyperref[eczindexfamilyrel]{code:q-ary_hamming}{\(q\)-ary Hamming code} --- For qubits (\(q=2\)), the only nontrivial perfect codes are the stabilizer code family \(\llbracket (4^r-1)/3, (4^r-1)/3 - 2r, 3\rrbracket \) for \(r \geq 2\), obtained from Hamming codes over \(\mathbb{F}_4\) via the Hermitian construction \NoCaseChange{\protect\cite{cite1694,cite449}}. These codes are related to partial spreads in projective geometry \NoCaseChange{\protect\cite{cite1695}}.
\item\relax
\flmRefsHyperref[eczindexfamilyrel]{code:quantum_random}{Random quantum code} --- Haar random codes achieve the quantum Hamming bound \NoCaseChange{\protect\cite{cite2913}}.
\item\relax
\flmRefsHyperref[eczindexfamilyrel]{code:quantum_hamming}{\(\llbracket 2^r, 2^r-r-2, 3\rrbracket \) Gottesman code} --- \(\llbracket 2^r, 2^r-r-2, 3\rrbracket \) Gottesman codes saturate the asymptotic quantum Hamming bound.
\item\relax
\flmRefsHyperref[eczindexfamilyrel]{code:stab_15_7_3}{\(\llbracket 15, 7, 3\rrbracket \) quantum Hamming code} --- \(\llbracket 15, 7, 3\rrbracket \) quantum Hamming code is perfect as a CSS code, i.e., the number of its \(Z\)-type syndromes matches the number of \(X\)-type Pauli errors up to weight one \NoCaseChange{\protect\cite{cite791}}.
\item\relax
\flmRefsHyperref[eczindexfamilyrel]{code:data_syndrome}{Quantum data-syndrome (QDS) code} --- The quantum Hamming bound can be extended to QDS codes \NoCaseChange{\protect\cite{cite2914}}.
\item\relax
\flmRefsHyperref[eczindexfamilyrel]{code:qudit_cws}{Modular-qudit CWS code} --- Generalized concatenations of modular-qudit CWS codes yield a family of codes that have larger logical dimension than stabilizer codes and that asymptotically approach the modular-qudit Hamming bound \NoCaseChange{\protect\cite{cite2696}}.
\item\relax
\flmRefsHyperref[eczindexfamilyrel]{code:qudit_gkp}{Modular-qudit GKP code} --- The modular-qudit GKP code is not a block code, but it is perfect in the sense that each correctable error maps the logical space into a distinct error space.
\item\relax
\flmRefsHyperref[eczindexfamilyrel]{code:qudit_sign}{Modular-qudit shift-resistant code} --- The modular-qudit shift-resistant code is not a block code, but it is perfect in the sense that each correctable error maps the logical space into a distinct error space \NoCaseChange{\protect\cite{cite2915}}.
\item\relax
\flmRefsHyperref[eczindexfamilyrel]{code:quantum_twisted}{Quantum twisted code} --- The \(\llbracket \frac{q^{2r}-1}{q^{2}-1},q^{n-2r},3\rrbracket _q\) family of quantum twisted codes are the only perfect Galois-qudit codes \NoCaseChange{\protect\cite{cite2911,cite2912}}.
\end{eczvaluelist}
\eczhbkcontributors{ Mazin Karjikar, \eczhuVVA }
\endeczcode

\eczcode{ame}{Perfect-tensor code}{}
\codefieldsection{Alternative Names}
\begin{eczvaluelist}
\item\relax Absolutely maximally entangled (AME) code
\item\relax Maximally multipartite entangled state (MMES) code
\end{eczvaluelist}
\eczhIndexCodeAliasName{ame}{Absolutely maximally entangled (AME) code}
\eczhIndexCodeAliasName{ame}{Maximally multipartite entangled state (MMES) code}
\codefieldsection{Description}
Block quantum code encoding one subsystem into an odd number \(n\) subsystems whose encoding isometry is a perfect tensor.
This code stems from an AME\((n,q)\) \flmRefsHyperref{ref219}{AME state}, or equivalently, a \(\llparenthesis n+1,1,\lfloor (n+1)/2 \rfloor + 1\rrparenthesis \) code.

\begin{defterm}{Absolutely maximally entangled (AME) state}\label{ref2916}\label{ref219}
A state on \(n\) subsystems is \(d\)\textit{-uniform} \NoCaseChange{\protect\cite{cite2917,cite1670,cite220}} (a.k.a. \(d\)-undetermined \NoCaseChange{\protect\cite{cite2918}} or \(d\)-maximally mixed \NoCaseChange{\protect\cite{cite2919}}) if all reduced density matrices on up to \(d\) subsystems are maximally mixed.
A \(K\)-dimensional subspace of \((d-1)\)-uniform states of \(n\) subsystems is equivalent to a \flmRefsHyperref{ref672}{pure} \(\llparenthesis n,K,d\rrparenthesis \) block quantum code \NoCaseChange{\protect\cite{cite530,cite2920}}.
An AME state (a.k.a. maximally multi-partite entangled state or MMES \NoCaseChange{\protect\cite{cite2921,cite2922}}) is a \(\lfloor n/2 \rfloor\)-uniform state, corresponding to a \flmRefsHyperref{ref672}{pure} \(\llparenthesis n,1,\lfloor n/2 \rfloor + 1\rrparenthesis \) code.
The rank-\(n\) tensor formed by the encoding isometry of such codes is a \textit{perfect tensor} (a.k.a. multi-unitary tensor), meaning that it is proportional to an isometry for any bipartition of its indices into a set \(A\) and a complementary set \(A^{\perp}\) such that \(|A|\leq|A^{\perp}|\).
Absolutely maximal entanglement exists for non-normalizable states of continuous-variable (CV) systems, whose reduced density matrices are proportional to the infinite-dimensional identity matrix; such states are called CV AME or CV MMES \NoCaseChange{\protect\cite{cite2923,cite507}}.
Explicit Gaussian and non-Gaussian CV AME constructions are known \NoCaseChange{\protect\cite{cite507}}.
\end{defterm}

Stabilizer Galois-qudit perfect-tensor codes can be converted to \flmRefsHyperref{ref219}{AME states} via established shortening/lengthening procedures \NoCaseChange{\protect\cite{cite1653}\protect\cite[{Table 1}]{cite813}}.
For example, an \(\llbracket n,0,d\rrbracket \) AME state can be reinterpreted as an \(\llbracket n-1,1,d-1\rrbracket \) perfect-tensor code by designating one subsystem as the logical input leg of the encoding isometry \NoCaseChange{\protect\cite[{Sec. 3.5}]{cite736}}.
There exist infinite families of inequivalent \flmRefsHyperref{ref219}{AME states} \NoCaseChange{\protect\cite{cite2924}}.

\codefieldsection{Encoding}
\begin{eczvaluelist}
\item\relax Fault-tolerant \(d\)-uniform state preparation \NoCaseChange{\protect\cite{cite2925}}.
\item\relax Quantum circuits for non-stabilizer AME states \NoCaseChange{\protect\cite{cite2926}}.
\end{eczvaluelist}
\codefieldsection{Fault Tolerance}
\begin{eczvaluelist}
\item\relax Fault-tolerant \(d\)-uniform state preparation \NoCaseChange{\protect\cite{cite2925}}.
\end{eczvaluelist}
\codefieldsection{Notes}
\begin{eczvaluelist}
\item\relax See Ref. \NoCaseChange{\protect\cite{cite2927}} and corresponding \flmHref{https://tp.nt.uni-siegen.de/ame/ame.html}{Table of AME states}.
\item\relax \(d\)-uniform states are useful for masking quantum information \NoCaseChange{\protect\cite{cite2928}}.
\item\relax Quantum simulation of approximately \(d\)-uniform states is similar to that with random-state inputs in terms of Trotter error \NoCaseChange{\protect\cite{cite2929}}.
\item\relax See Ref. \NoCaseChange{\protect\cite{cite2930}} for a review.
\end{eczvaluelist}
\codefieldsection{Parent}
\begin{eczvaluelist}
\item\relax
\flmRefsHyperref[eczindexfamilyrel]{code:block_perfect}{Planar-perfect-tensor code} --- \flmRefsHyperref{code:block_perfect}{Planar-perfect tensors} are automatically \flmRefsHyperref{code:ame}{perfect tensors}.
\end{eczvaluelist}
\codefieldsection{Children}
\begin{eczvaluelist}
\item\relax
\flmRefsHyperref[eczindexfamilyrel]{code:rotor_3_1_2}{\(\llbracket 3,1,2\rrbracket _{\mathbb{Z}}\) Three-rotor code} --- Three-rotor codewords are CV AME states \NoCaseChange{\protect\cite{cite507}}.
\item\relax
\flmRefsHyperref[eczindexfamilyrel]{code:rotor_5_1_3}{\(\llbracket 5,1,3\rrbracket _{\mathbb{Z}}\) Five-rotor code} --- Five-rotor codewords are CV AME \NoCaseChange{\protect\cite{cite507}}.
\item\relax
\flmRefsHyperref[eczindexfamilyrel]{code:braunstein}{\(\llbracket 5,1,3\rrbracket _{\mathbb{R}}\) Braunstein five-mode code} --- Braunstein five-mode codewords are CV AME \NoCaseChange{\protect\cite{cite507}}.
\item\relax
\flmRefsHyperref[eczindexfamilyrel]{code:qudit_3_6_2}{\(\llparenthesis 3,6,2\rrparenthesis _{\mathbb{Z}_6}\) Euler code} --- The \(\llparenthesis 3,6,2\rrparenthesis _{\mathbb{Z}_6}\) Euler code is an example of a non-stabilizer perfect-tensor code \NoCaseChange{\protect\cite{cite2931}}.
\item\relax
\flmRefsHyperref[eczindexfamilyrel]{code:qudit_5_1_3}{\(\llbracket 5,1,3\rrbracket _{\mathbb{Z}_q}\) modular-qudit code} --- The \(\llbracket 5,1,3\rrbracket _{\mathbb{Z}_q}\) code is a perfect-tensor code because it stems from the \(\llbracket 6,0,4\rrbracket _{\mathbb{Z}_q}\) \flmRefsHyperref{ref219}{AME state} \NoCaseChange{\protect\cite[{Thm. 13}]{cite532}}.
\item\relax
\flmRefsHyperref[eczindexfamilyrel]{code:stab_3_1_2}{\(\llbracket 3,1,2\rrbracket _3\) Three-qutrit code} --- Three-qutrit codewords are AME, and the three-qutrit code stems from the \(\llbracket 4,0,3\rrbracket _3\) \flmRefsHyperref{ref219}{AME state} \NoCaseChange{\protect\cite{cite1925,cite151,cite2932}}.
\end{eczvaluelist}
\codefieldsection{Cousins}
\begin{eczvaluelist}
\item\relax
\flmRefsHyperref[eczindexfamilyrel]{code:quantum_mds}{Quantum maximum-distance-separable (MDS) code} --- \flmRefsHyperref{ref219}{AME states} for even \(n\) are examples of quantum MDS codes with no logical qubits \NoCaseChange{\protect\cite{cite1670,cite1926,cite2933}}.
A family of conjectured perfect-tensor codes is quantum MDS \NoCaseChange{\protect\cite{cite975}}.

\item\relax
\flmRefsHyperref[eczindexfamilyrel]{code:combinatorial_design}{Combinatorial design} --- Combinatorial designs and \(d\)-uniform quantum states are related \NoCaseChange{\protect\cite{cite151,cite152,cite153}}.
\item\relax
\flmRefsHyperref[eczindexfamilyrel]{code:orthogonal_array}{Orthogonal array (OA)} --- Orthogonal arrays and \(d\)-uniform quantum states are related \NoCaseChange{\protect\cite{cite220,cite152,cite221,cite222,cite223,cite224}}.
\item\relax
\flmRefsHyperref[eczindexfamilyrel]{code:mds}{Maximum distance separable (MDS) code} --- MDS codes can be used to obtain cluster states that are AME with minimal support \NoCaseChange{\protect\cite{cite1923,cite1924,cite1925,cite151,cite1926,cite1927}}.
\item\relax
\flmRefsHyperref[eczindexfamilyrel]{code:qudit_cluster_state}{Modular-qudit cluster-state code} --- MDS codes can be used to obtain cluster states that are AME with minimal support \NoCaseChange{\protect\cite{cite1923,cite1924,cite1925,cite151,cite1926,cite1927}}.
\item\relax
\flmRefsHyperref[eczindexfamilyrel]{code:galois_polynomial}{Galois-qudit RS code} --- \flmRefsHyperref{ref219}{AME states} for even \(n\) are examples of quantum MDS codes with no logical qubits \NoCaseChange{\protect\cite{cite1670,cite1926,cite2933}}. MDS RS codes can yield perfect tensors via the CSS and Hermitian constructions \NoCaseChange{\protect\cite{cite975}} (see also Refs. \NoCaseChange{\protect\cite{cite2866,cite2867}}).
\item\relax
\flmRefsHyperref[eczindexfamilyrel]{code:quantum_secret_sharing}{Approximate secret-sharing code} --- Perfect tensors are useful for quantum secret sharing and open-destination multi-party teleportation \NoCaseChange{\protect\cite{cite2934,cite1924,cite2935}}.
\item\relax
\flmRefsHyperref[eczindexfamilyrel]{code:qubit_stabilizer}{Qubit stabilizer code} --- The codespace of a qubit stabilizer code with \flmRefsHyperref{ref672}{pure distance} \(d_{\textnormal{pure}}\) is a \((d_{\textnormal{pure}}-1)\)-uniform space.
\item\relax
\flmRefsHyperref[eczindexfamilyrel]{code:reinforcement_learning}{Reinforcement-learning quantum code} --- Reinforcement learning \NoCaseChange{\protect\cite{cite2936}} and graph-based optimizers like PyTheus \NoCaseChange{\protect\cite{cite2937}} can be used to find AME states.
\item\relax
\flmRefsHyperref[eczindexfamilyrel]{code:cws}{Codeword stabilized (CWS) code} --- CWS codes can be constructed from \((d-1)\)-uniform states \NoCaseChange{\protect\cite{cite2938}}.
\item\relax
\flmRefsHyperref[eczindexfamilyrel]{code:covariant}{Covariant block quantum code} --- An \(SU(2)\)-invariant three-qudit perfect tensor exists \NoCaseChange{\protect\cite{cite2727}}, but invariant perfect tensors do not exist on four parties \NoCaseChange{\protect\cite{cite2728,cite2727,cite2729}}.
\item\relax
\flmRefsHyperref[eczindexfamilyrel]{code:rotor_cluster}{Rotor cluster-state code} --- Rotor AME cluster states exist for any number of modes \NoCaseChange{\protect\cite{cite507}}.
\item\relax
\flmRefsHyperref[eczindexfamilyrel]{code:analog_repetition}{Analog repetition code} --- Analog GHZ states are \(1\)-uniform for all \(n\) and CV AME for \(n=2,3\) \NoCaseChange{\protect\cite{cite2923,cite507}}.
\item\relax
\flmRefsHyperref[eczindexfamilyrel]{code:analog_stabilizer}{Analog stabilizer code} --- Analog stabilizer states are generically CV AME \NoCaseChange{\protect\cite{cite2923}}, and explicit constructions exist for any number of modes \NoCaseChange{\protect\cite{cite507}}. The codespace of an analog stabilizer code with \flmRefsHyperref{ref672}{pure distance} \(d_{\textnormal{pure}}\) is a \((d_{\textnormal{pure}}-1)\)-uniform space \NoCaseChange{\protect\cite{cite507}}. Normalizable finitely squeezed versions of infinitely squeezed Gaussian states are locally thermal, up to corrections in the squeezing parameter \NoCaseChange{\protect\cite{cite2921,cite2939}}.
\item\relax
\flmRefsHyperref[eczindexfamilyrel]{code:analog_surface}{Analog surface code} --- Analog surface-code states are \(3\)-uniform \NoCaseChange{\protect\cite{cite507}}.
\item\relax
\flmRefsHyperref[eczindexfamilyrel]{code:cv_cluster_state}{Analog cluster-state code} --- Analog cluster states are generically CV AME \NoCaseChange{\protect\cite{cite2923}}, and explicit constructions exist for any number of modes \NoCaseChange{\protect\cite{cite507}}.
\item\relax
\flmRefsHyperref[eczindexfamilyrel]{code:quantum_repetition}{Quantum repetition code} --- GHZ states are \(1\)-uniform for all \(n\) and AME for \(n=2,3\).
\item\relax
\flmRefsHyperref[eczindexfamilyrel]{code:stabilizer_over_gf4}{Hermitian qubit code} --- The sole codeword of some \(\llbracket n,0,d\rrbracket \) Hermitian codes is an \flmRefsHyperref{ref219}{AME state} \NoCaseChange{\protect\cite{cite2932}}.
\item\relax
\flmRefsHyperref[eczindexfamilyrel]{code:happy}{Pastawski-Yoshida-Harlow-Preskill (HaPPY) code} --- The encoding of a HaPPY code is a holographic tensor network consisting of pentagon and hexagon \flmRefsHyperref{ref219}{perfect tensors}.
\end{eczvaluelist}
\eczhbkcontributors{ \eczhuVVA }
\endeczcode

\eczcode{permutation_invariant}{Permutation-invariant (PI) code}{~\NoCaseChange{\protect\cite{cite2940}}}
\codefieldsection{Description}
Block quantum code such that any permutation of the subsystems leaves any codeword invariant.
In other words, the automorphism group of the code contains the symmetric group \(S_n\).

There is a notion of Wigner functions for PI subspaces \NoCaseChange{\protect\cite{cite2941}}. 

\subsection{Qudit Dicke states and the discrete simplex mapping}

For \(n\)-modular-qudit block codes with qudit dimension \(q\), an often used basis for the PI subspace consists of the qudit Dicke states. 

\begin{defterm}{Qudit Dicke states}\label{ref2942}\label{ref499}
A qudit Dicke state is an equal superposition of all qudit basis elements whose labels have the same composition,
\flmMathEnvironment{align}{}{
  |D_{\mathbf{c}}\rangle=\frac{1}{\sqrt{\binom{n}{\mathbf{c}}}}\sum_{\substack{\mathbf{n}\in\mathbb{Z}_{q}^{n}\\ C(\mathbf{n})=\mathbf{c} } }|\mathbf{n}\rangle\,,
}  
where \(\binom{n}{\mathbf{c}}\) is the multinomial coefficient.
Above, the \textit{composition} \(C\) of a qudit basis label \(\mathbf{n}\) tabulates the number of each type of element present in the label. For example, the label \(\mathbf{n}=(0313)\) has composition \(C(\mathbf{n})=(1102)\), whose coordinates denote the number of zeroes (one), number of ones (one), number of twos (zero), and number of threes (two) present in the label.  
The \(q=2\) case reduces to the \flmRefsHyperref{ref526}{Dicke-state} mapping.
\end{defterm}

Qudit Dicke states are in one-to-one correspondence with points on the \flmRefsHyperref{ref655}{discrete simplex} \(\Delta_{q,n}\), which houses the totally symmetric irrep of \(SU(q)\) \NoCaseChange{\protect\cite{cite500}}.
The same simplex labels define constant-excitation Fock-state codes and single-spin codes on the completely symmetric \(SU(q)\) irrep.
Applying the simplex mapping to a qudit PI code yields Fock-state and spin codes with the same distance \NoCaseChange{\protect\cite[{Prop. V.2, Prop. VI.2}]{cite500}}.
Wigner functions are also interconvertible between the Fock-state and single-spin spaces \NoCaseChange{\protect\cite{cite2943}}.

\codefieldsection{Protection}
Noise models can be categorized as those that cause the state to leave the maximally symmetric subspace and those that do not.
PI codes of distance \(d\) can protect against \(d-1\) deletion errors \NoCaseChange{\protect\cite{cite2655,cite2656,cite2657,cite2658}}, i.e., erasures of subsystems at unknown locations.

Other protection depends on the code family.
The GNU PI family (parameterized by \(t\)) protects against arbitrary weight \(t\) qubit errors and approximately corrects spontaneous decay errors \NoCaseChange{\protect\cite{cite2944,cite2945}}.
Other related codes protect against \flmRefsHyperref{ref498}{AD} \NoCaseChange{\protect\cite{cite2946}} while admitting a constant number of excitations.

\codefieldsection{Rate}
For every \(K,t \geq 2\), there are explicitly constructible PI codes with \(q=N=(K-1)t(t+1)\), length \(N\), and distance \(t+1\); there also exist families with logical dimension \(K = o(2^N)\) and distance of \flmRefsHyperref{ref65}{order} \(o(N/\log N)\) \NoCaseChange{\protect\cite{cite500}}.
\codefieldsection{Encoding}
\begin{eczvaluelist}
\item\relax State preparation of qudit Dicke states \NoCaseChange{\protect\cite{cite2947}}.
\end{eczvaluelist}
\codefieldsection{Transversal and Permutation-Based Gates}
\begin{eczvaluelist}
\item\relax Qudit Dicke states are in one-to-one correspondence with points on the \flmRefsHyperref{ref655}{discrete simplex} \(\Delta_{q,n}\), which houses the totally symmetric irrep of \(SU(q)\) \NoCaseChange{\protect\cite{cite500}}. Any transversal gates of the form \(U^{\otimes N}\), with \(U \in SU(q)\), will implement logical operations in a subgroup of \(SU(q)\).
\end{eczvaluelist}
\codefieldsection{Notes}
\begin{eczvaluelist}
\item\relax PI codes can be constructed using real polynomials for high-dimensional qudit spaces \NoCaseChange{\protect\cite{cite2948}}.
\item\relax Qubit and qudit PI codes obtained from numerical optimization routines are useful for entanglement distillation \NoCaseChange{\protect\cite[{Appx. B.1}]{cite2949}}.
\item\relax Qudit Dicke state preparation \NoCaseChange{\protect\cite{cite2950}}.
\end{eczvaluelist}
\codefieldsection{Parent}
\begin{eczvaluelist}
\item\relax
\flmRefsHyperref[eczindexfamilyrel]{code:quantum_cyclic}{Cyclic quantum code} --- The cyclic group of these codes is a subgroup of the \(S_n\) symmetric group used in permutation invariant codes.
\end{eczvaluelist}
\codefieldsection{Children}
\begin{eczvaluelist}
\item\relax
\flmRefsHyperref[eczindexfamilyrel]{code:constant_excitation_permutation_invariant}{Ouyang-Chao constant-excitation PI code}\item\relax
\flmRefsHyperref[eczindexfamilyrel]{code:very-small-logical-qubit}{Very small logical qubit (VSLQ) code}\item\relax
\flmRefsHyperref[eczindexfamilyrel]{code:w_state}{W-state code}\item\relax
\flmRefsHyperref[eczindexfamilyrel]{code:qubit_permutation_invariant}{PI qubit code}\item\relax
\flmRefsHyperref[eczindexfamilyrel]{code:three_qutrit_permutation_invariant}{\(\llparenthesis 3,2,2\rrparenthesis _3\) Three-qutrit single-deletion code}\item\relax
\flmRefsHyperref[eczindexfamilyrel]{code:t_group}{Twisted \(1\)-group code}\end{eczvaluelist}
\codefieldsection{Cousins}
\begin{eczvaluelist}
\item\relax
\flmRefsHyperref[eczindexfamilyrel]{code:simplex_discrete}{Simplex integer-based code} --- Simplex integer-based codes can be partitioned into qudit PI codewords whose error-correction is guaranteed by the Tverberg theorem \NoCaseChange{\protect\cite[{Thm. VII.5}]{cite500}}.
\item\relax
\flmRefsHyperref[eczindexfamilyrel]{code:constant_excitation}{Constant-excitation (CE) code} --- Modular-qudit PI codes can be converted to constant-excitation Fock-state codes via the \flmRefsHyperref{ref499}{simplex mapping} \NoCaseChange{\protect\cite[{Prop. V.2}]{cite500}}. Any transversal gates are mapped to Gaussian gates on the Fock-state codes \NoCaseChange{\protect\cite{cite500}}.
\item\relax
\flmRefsHyperref[eczindexfamilyrel]{code:fock_state}{Fock-state bosonic code} --- Modular-qudit PI codes can be converted to constant-excitation Fock-state codes via the \flmRefsHyperref{ref499}{simplex mapping} \NoCaseChange{\protect\cite[{Prop. V.2}]{cite500}}. Any transversal gates are mapped to Gaussian gates on the Fock-state codes \NoCaseChange{\protect\cite{cite500}}.
\item\relax
\flmRefsHyperref[eczindexfamilyrel]{code:single_spin}{Single-spin code} --- Modular-qudit PI codes can be converted to spin codes defined on the completely symmetric irrep of \(SU(q)\) via the \flmRefsHyperref{ref499}{simplex mapping} \NoCaseChange{\protect\cite[{Prop. VI.2}]{cite500}}. Any transversal gates are mapped to \(SU(q)\) gates on the spin codes \NoCaseChange{\protect\cite{cite500}}.
\item\relax
\flmRefsHyperref[eczindexfamilyrel]{code:insertion_deletion}{Editing code} --- PI codes of distance \(d\) can protect against \(d-1\) (quantum) deletion errors.
\end{eczvaluelist}
\eczhbkcontributors{ Benjamin Quiring, \eczhuVVA }
\endeczcode

\eczcode{block_perfect}{Planar-perfect-tensor code}{~\NoCaseChange{\protect\cite{cite2951,cite2952}}}
\codefieldsection{Alternative Names}
\begin{eczvaluelist}
\item\relax Block-perfect-tensor code
\item\relax Perfect-tangle code
\end{eczvaluelist}
\eczhIndexCodeAliasName{block_perfect}{Block-perfect-tensor code}
\eczhIndexCodeAliasName{block_perfect}{Perfect-tangle code}
\codefieldsection{Description}
Block quantum code whose encoding isometry is a block perfect tensor, i.e., a tensor which remains an isometry under partitions into two contiguous components in a fixed plane.
This code stems from a planar maximally entangled state \NoCaseChange{\protect\cite{cite2953}}.

\codefieldsection{Parent}
\begin{eczvaluelist}
\item\relax
\flmRefsHyperref[eczindexfamilyrel]{code:quantum_lego}{Tensor-network code}\end{eczvaluelist}
\codefieldsection{Children}
\begin{eczvaluelist}
\item\relax
\flmRefsHyperref[eczindexfamilyrel]{code:ame}{Perfect-tensor code} --- \flmRefsHyperref{code:block_perfect}{Planar-perfect tensors} are automatically \flmRefsHyperref{code:ame}{perfect tensors}.
\item\relax
\flmRefsHyperref[eczindexfamilyrel]{code:stab_5_1_2}{\(\llbracket 5,1,2\rrbracket \) rotated surface code} --- The \(\llbracket 5,1,2\rrbracket \) rotated surface code is the smallest SCF holographic code \NoCaseChange{\protect\cite{cite2954}}. The encoding of more general SCF holographic codes is a holographic tensor network consisting of the encoding isometry for the \(\llbracket 5,1,2\rrbracket \) rotated surface code, which is a \flmRefsHyperref{code:block_perfect}{planar-perfect tensor}.
\item\relax
\flmRefsHyperref[eczindexfamilyrel]{code:steane}{\(\llbracket 7,1,3\rrbracket \) Steane code} --- The Steane code is the smallest heptagon holographic code. The encoding of more general heptagon holographic codes is a holographic tensor network consisting of the encoding isometry for the Steane code, which is a \flmRefsHyperref{code:block_perfect}{planar-perfect tensor}.
\end{eczvaluelist}
\codefieldsection{Cousins}
\begin{eczvaluelist}
\item\relax
\flmRefsHyperref[eczindexfamilyrel]{code:category_quantum}{Category-based quantum code} --- Several modular fusion categories can be used to define \flmRefsHyperref{code:block_perfect}{planar-perfect tensor}s \NoCaseChange{\protect\cite{cite2951}}.
\item\relax
\flmRefsHyperref[eczindexfamilyrel]{code:holographic_5_1_2}{Surface-code-fragment (SCF) holographic code} --- The encoding of the heptagon holographic code is a holographic tensor network consisting of the encoding isometry for the \(\llbracket 5,1,2\rrbracket \) rotated surface code, which is a \flmRefsHyperref{code:block_perfect}{planar-perfect tensor}.
\item\relax
\flmRefsHyperref[eczindexfamilyrel]{code:holographic_steane}{Heptagon holographic code} --- The encoding of the heptagon holographic code is a holographic tensor network consisting of the encoding isometry for the Steane code, which is a \flmRefsHyperref{code:block_perfect}{planar-perfect tensor}.
\end{eczvaluelist}
\eczhbkcontributors{ \eczhuVVA }
\endeczcode

\eczcode{quantum_into_quantum}{Quantum code}{}
\codefieldsection{Description}
Code designed for transmission of quantum and/or classical information through a quantum channel for the purposes of robust storage, communication, or sensing. 
Transmission can be performed with side information or entanglement.

While codewords \(c\) of an ECC are elements of some alphabet \(\Sigma\), quantum codewords are \(L^2\)-normalizable complex functions on \(\Sigma\).
Put differently, the configuration space of the canonical (a.k.a. computational) basis states \(|c\rangle\) of a quantum system is the classical alphabet \(\Sigma\).
The table below lists the most common alphabets used in quantum codes, along with names of the corresponding systems.
  \begin{flmFloat}{table}{NumCap}\flmCellsBeginCenter
\long\def\flmTempTypesetThisTable#1{%
\begin{tblr}{#1,
  hspan=minimal,
  cell{1}{1}={}{c, font={\flmCellsHeaderFont}},
  cell{1}{2}={}{c, font={\flmCellsHeaderFont}},
  cell{2}{1}={}{c},
  cell{2}{2}={}{c},
  cell{3}{1}={}{c},
  cell{3}{2}={}{c},
  cell{4}{1}={}{c},
  cell{4}{2}={}{c},
  cell{5}{1}={}{c},
  cell{5}{2}={}{c},
  cell{6}{1}={}{c},
  cell{6}{2}={}{c},
  cell{7}{1}={}{c},
  cell{7}{2}={}{c},
  cell{8}{1}={}{c},
  cell{8}{2}={}{c},
  hline{2}={1}{.4pt,solid},
  hline{2}={2}{.4pt,solid}}%
\toprule
alphabet \(\Sigma\) & system \(L^2(\Sigma)\)\\

    \(\mathbb{Z}_{2}=\mathbb{F}_2\) & qubit
        \\

    \(\mathbb{F}_q\) & Galois qudit
        \\

    \(\mathbb{Z}_{q}\) & modular qudit
        \\

    \(\mathbb{R}\) & bosonic mode
        \\

    \(G\) & group-valued qudit
        \\

    \(G/H\) & coset-valued qudit
        \\

    \(\mathcal{C}\) & category-valued qudit
    \\
\bottomrule
\end{tblr}%
}%
\def\flmTmpMaxW{\dimexpr 0.96\linewidth\relax}%
\setbox0=\hbox{\flmTempTypesetThisTable{colspec={cc}}}%
\ifdim\wd0<\flmTmpMaxW\relax
  \leavevmode\box0 
\else
  \flmTempTypesetThisTable{width=\flmTmpMaxW,colspec={X[-1]X[-1]}}
\fi
\flmCellsEndCenter \caption{Table listing the most common alphabets (a.k.a. configuration spaces) used in quantum codes. Here, \(\mathbb{F}_q\) is a \flmRefsHyperref{ref33}{finite field}, \(G\) is a group, \(H\) is a subgroup of \(G\), and \(\mathcal{C}\) is a category.}\label{ref2955}\end{flmFloat}

\codefieldsection{Children}
\begin{eczvaluelist}
\item\relax
\flmRefsHyperref[eczindexfamilyrel]{code:approximate_oaecc}{Approximate operator-algebra QECC}\item\relax
\flmRefsHyperref[eczindexfamilyrel]{code:eaoaecc}{Entanglement-assisted operator-algebra QECC (EAOA QECC)}\item\relax
\flmRefsHyperref[eczindexfamilyrel]{code:oaecc}{Operator-algebra QECC (OAQECC)}\item\relax
\flmRefsHyperref[eczindexfamilyrel]{code:metrological}{Metrological code}\item\relax
\flmRefsHyperref[eczindexfamilyrel]{code:qetc}{Quantum error-transmuting code (QETC)}\end{eczvaluelist}
\eczhbkcontributors{ \eczhuVVA }
\endeczcode

\eczcode{qecc}{Quantum error-correcting code (QECC)}{}
\codefieldsection{Description}
Encodes quantum information in a (\textit{logical}) subspace of a
(\textit{physical}) Hilbert space such that it is possible to recover said
information from errors that act as linear maps on the physical space.
The state space of a QECC is contained in the space of complex \(L^2\)-normalizable functions of some configuration space, which usually corresponds to the alphabet of a classical code.

Since quantum information is encoded in quantum superpositions, an additional source of noise (not relevant to classical encodings) can affect the relative phase of such superpositions.
Quantum error-correcting codes have to protect against such \textit{phase-flip noise} while also protecting against conventional classical \textit{bit-flip} noise.
The better a code is at protecting against phase-flip noise, the worse it is at protecting against bit-flip noise, and vice versa, so there is a tradeoff between the two types of noise.

The logical subspace is spanned by a basis comprised of \textit{code basis states}
or \textit{codewords}. Codewords may not be normalizable if the physical
Hilbert space is infinite-dimensional, so approximate versions have to be constructed in
practice.

While all considered QECC states are complex functions, real or quaternionic function spaces can also be considered for QEC \NoCaseChange{\protect\cite{cite2956,cite2957}}. 

\codefieldsection{Protection}
Denoting Hilbert spaces by the letter \(\mathsf{H}\), a quantum code
\((U,\cal{E})\) is a partial isometry
\(U:\mathsf{H}_{\text{logical}}\to\mathsf{H}_{\text{physical}}\) with a set of
correctable errors \(\cal{E}\) with the following property: there exists a
quantum operation \(\cal{D}\) such that for all \(E\in\cal{E}\) and states
\(|\psi\rangle\in\mathsf{H}_{\text{logical}}\),
\flmMathEnvironment{align}{}{
  {\cal D} (EU|\psi\rangle\langle\psi|U^{\dagger}E^{\dagger})
  = c(E)|\psi\rangle\langle\psi|
}
for some constant \(c\) \NoCaseChange{\protect\cite{cite398}}.

Equivalently, correction capability is determined by the \flmTerm{term}{ref1043}{}{Knill-Laflamme conditions}, which may admit infinite terms due to non-normalizability of ideal code states in the case of codes with infinite-dimensional physical spaces. A code that satisfies these conditions approximately, i.e., up to some small quantifiable error, is called an \flmRefsHyperref{code:approximate_qecc}{approximate code}. These conditions can also be formulated in terms of a dual Heisenberg picture, where correctability is checked for some algebra of observables \NoCaseChange{\protect\cite{cite2958}}.

\begin{defterm}{Pseudo-threshold (a.k.a. break-even point)}\label{ref2959}\label{ref2960}
The ultimate goal of error correction is to make sure that the logical error rate is smaller than the underlying physical error rate.
For a noise model parameterized by a single physical error rate \(p\), the \textit{pseudo-threshold} or \textit{break-even point} is the smallest \(p\) at which the logical error rate after error correction is equal to \(p\).
\end{defterm}

\codefieldsection{Decoding}
\begin{eczvaluelist}
\item\relax The effect of an error is a mapping of the code subspace into another, potentially overlapping, subspace. To determine, or diagnose, the effect of the error in what is known as \textit{syndrome-based decoding}, one can measure one or more operators called \textit{check operators}, which resolve code and error spaces without collapsing the quantum information inside the spaces. The eigenvalues of check operators are called \textit{error syndromes}. One \textit{round} or \textit{cycle} of quantum error correction proceeds by extracting syndromes and performing correcting operations to map the error space containing the logical information back into the codespace. For some codes, correcting operations are not necessary because one can instead track which error space contains the logical information.
\end{eczvaluelist}
\codefieldsection{Notes}
\begin{eczvaluelist}
\item\relax See Refs. \NoCaseChange{\protect\cite{cite2961,cite2962,cite1634,cite2579,cite2963,cite2964,cite2965,cite2966,cite2967,cite2968,cite2969,cite2970,cite2971,cite2764,cite2972,cite1975,cite2973,cite2974,cite2733}} for overviews of quantum error correction.
\item\relax See Refs. \NoCaseChange{\protect\cite{cite2975,cite2976,cite398}} for books on quantum error correction.
\item\relax See video tutorials by \flmHref{https://www.youtube.com/watch?v=_ls3KczZL2c}{V. V. Albert}, \flmHref{https://www.youtube.com/watch?v=uD69GCYF9Zg}{S. M. Girvin}, \flmHref{https://www.youtube.com/watch?v=buIbd_aXAHw}{P. Shor}, \flmHref{https://www.youtube.com/watch?v=Je7sVJGKMgU}{B. Terhal}, and \flmHref{https://www.youtube.com/watch?v=mcwpe8iJ5uo}{J. Wright}.
\item\relax Quantum error correction was initially claimed not to be theoretically possible \NoCaseChange{\protect\cite{cite2977,cite2978}} and has been criticized since \NoCaseChange{\protect\cite{cite2979}}.
\item\relax Resource-theoretic interpretations of quantum error correction have been developed, including those that think of codes together with recovery operations as superchannels (a.k.a. quantum combs or bipartite operations) \NoCaseChange{\protect\cite{cite2980,cite2981,cite2982,cite2776,cite2983}}.
\item\relax QECC can be used as a mechanism for securing quantum computation \NoCaseChange{\protect\cite{cite2984}}.
\end{eczvaluelist}
\codefieldsection{Parent}
\begin{eczvaluelist}
\item\relax
\flmRefsHyperref[eczindexfamilyrel]{code:oaecc}{Operator-algebra QECC (OAQECC)} --- An OAQECC which has no gauge structure (e.g., gauge qubits) and no block structure is a QECC.
\end{eczvaluelist}
\codefieldsection{Children}
\begin{eczvaluelist}
\item\relax
\flmRefsHyperref[eczindexfamilyrel]{code:category_quantum}{Category-based quantum code}\item\relax
\flmRefsHyperref[eczindexfamilyrel]{code:homogeneous_space_quantum}{Homogeneous-space quantum code}\item\relax
\flmRefsHyperref[eczindexfamilyrel]{code:asymmetric_qecc}{Asymmetric quantum code (AQC)}\item\relax
\flmRefsHyperref[eczindexfamilyrel]{code:block_quantum}{Block quantum code}\item\relax
\flmRefsHyperref[eczindexfamilyrel]{code:holographic}{Holographic code}\item\relax
\flmRefsHyperref[eczindexfamilyrel]{code:hamiltonian}{Hamiltonian-based code}\item\relax
\flmRefsHyperref[eczindexfamilyrel]{code:qecc_finite}{Finite-dimensional quantum error-correcting code} --- Finite-dimensional QECCs are a special case of quantum error-correcting codes, which can also include infinite-dimensional codes such as bosonic codes. The Knill-Laflamme conditions and the notion of correctability can be extended to infinite-dimensional codes.
\item\relax
\flmRefsHyperref[eczindexfamilyrel]{code:single_subsystem}{Monolithic quantum code}\end{eczvaluelist}
\codefieldsection{Cousins}
\begin{eczvaluelist}
\item\relax
\flmRefsHyperref[eczindexfamilyrel]{code:approximate_qecc}{Approximate quantum error-correcting code (AQECC)} --- QAECCs correcting a noise channel exactly reduce to QECCs.
\item\relax
\flmRefsHyperref[eczindexfamilyrel]{code:ecc}{Error-correcting code (ECC)} --- Quantum information cannot be copied using a linear process \NoCaseChange{\protect\cite{cite1042}}, so one cannot send several copies of a quantum state through a channel as can be done for classical information. The \flmTerm{term}{ref1043}{}{Knill-Laflamme conditions} can similarly be formulated for classical codes \NoCaseChange{\protect\cite[{Sec. 3}]{cite1044}}, although they are not as widely used as those for quantum codes.
\item\relax
\flmRefsHyperref[eczindexfamilyrel]{code:metrological}{Metrological code} --- Metrological codes are logical-qubit codes that satisfy the \flmTerm{term}{ref1043}{}{Knill-Laflamme conditions} conditions only partially, and codes that satisfy them fully are QECCs.
\item\relax
\flmRefsHyperref[eczindexfamilyrel]{code:eaqecc}{Entanglement-assisted (EA) QECC} --- EA QECCs utilize additional ancillary subsystems in a pre-shared entangled state, but reduce to QECCs when said subsystems are interpreted as noiseless physical subsystems.
\item\relax
\flmRefsHyperref[eczindexfamilyrel]{code:hybridqecc}{Hybrid QECC} --- A hybrid QECC storing no classical information reduces to a QECC. Conversely, any QECC can be converted into a hybrid QECC by using a portion of its logical subspace to store only classical information.
\item\relax
\flmRefsHyperref[eczindexfamilyrel]{code:oecc}{Subsystem QECC} --- A subsystem QECC reduces to an ordinary (i.e., subspace) QECC when the gauge subsystem is trivial. Conversely, any QECC with a tensor-product logical subspace can be turned into a subsystem code by treating a logical tensor factor as a gauge subsystem.
\item\relax
\flmRefsHyperref[eczindexfamilyrel]{code:holographic_tensor}{Holographic tensor-network code} --- Quantum encoding maps are isometries, but non-isometric encodings are relevant to describing mappings into the interior of a black hole \NoCaseChange{\protect\cite{cite2859}} and de Sitter time evolution \NoCaseChange{\protect\cite{cite642}}. Trace-norm preserving encodings have also been studied \NoCaseChange{\protect\cite{cite2860}}.
\item\relax
\flmRefsHyperref[eczindexfamilyrel]{code:qetc}{Quantum error-transmuting code (QETC)} --- QETCs are quantum codes which satisfy a generalization of the \flmTerm{term}{ref1043}{}{Knill-Laflamme conditions}. QETCs for which the admissible logical error set consists solely of the identity are QECCs.
\item\relax
\flmRefsHyperref[eczindexfamilyrel]{code:shor_nine}{\(\llbracket 9,1,3\rrbracket \) Shor code} --- The Shor code is the first quantum error-correcting code.
\end{eczvaluelist}
\eczhbkcontributors{ \eczhuPhF, \eczhuVVA }
\endeczcode

\eczcode{qetc}{Quantum error-transmuting code (QETC)}{~\NoCaseChange{\protect\cite{cite2985}}}
\codefieldsection{Description}
Encodes quantum information in a (logical, \(k\)-qubit) subspace \(\mathsf{C}\) of a (physical, \(n\)-qubit) Hilbert space \(\mathsf{H}\) such that recovery is possible from a set of physical errors occurring \textit{up to} a pre-specified (smaller, but non-empty) admissible set of logical errors.
This is relevant to, e.g., simulation of noisy systems.
Most QETCs are stabilizer codes: \(\mathsf{C}\) is the subspace stabilised by an abelian subgroup \(\mathsf{S} \subset \mathcal{G}_n\) of the \flmRefsHyperref{ref663}{Pauli group} on \(n\) qubits.

\codefieldsection{Protection}
For a pair \(E=\{E_i\}\) and \(\overline{M} = \{\overline{m}_{\alpha}\}\) of physical and logical error sets, \(\mathsf{C}\) is a QETC if there exists a recovery operation \(\mathcal{R}\) such that for all physical noise channels \(\mathcal{E}\) with Kraus operators proportional to elements of \(\{E_i\}\), and density matrices \(\rho\) supported on \(\mathsf{C}\), we have:
\flmMathEnvironment{align}{}{
\mathcal{R} \circ \mathcal{E} (\rho) = \mathcal{M} (\rho).
}
Here \(\mathcal{M}\) is a logical noise channel whose Kraus operators are proportional to elements of the cosets \(\{\mathcal{A}(\overline{m}_{\alpha})\mathsf{S}\} \subset \mathcal{G}_n\), where
\flmMathEnvironment{align}{}{
        \mathcal{G}_k \xrightarrow[]{\mathcal{A}} \mathsf{N(S)}/\mathsf{S}
}
is a choice of isomorphism, \textit{i.e.} an identification of the logical \flmRefsHyperref{ref663}{Pauli group} with elements of the physical \flmRefsHyperref{ref663}{Pauli group}. \(\mathsf{N(S)}\) denotes the normaliser of \(\mathsf{S}\) in \(\mathcal{G}_n\).

Equivalently, this can be rephrased in terms of a generalization of the \flmTerm{term}{ref1043}{}{Knill-Laflamme conditions} for error-correction.
Label the cosets of \(\mathsf{N(S)}\) in \(\mathcal{G}_n\) by \(\mathfrak{n}\), and the physical errors in a given coset by \(E_{\mathfrak{n}} := \{E_{\mathfrak{n},i}\}\). Then \(\mathsf{C}\) is a QETC for a pair \(\{E, \overline{M}\}\) if and only if there exists an isomorphism \(\mathcal{A}: \overline{\mathcal{G}}_k \rightarrow \mathsf{N(S)}/\mathsf{S}\), such that for each \(\mathfrak{n}\) there exists a map:
\flmMathEnvironment{align}{}{
\pi_{\mathfrak{n}}: E_{\mathfrak{n}}  \rightarrow \mathcal{A}(\overline{M}),\qquad \pi_{\mathfrak{n}}(E_{\mathfrak{n},i}) := \overline{m}_{\pi_{\mathfrak{n}},i}
}
so that \(\forall\, E_{\mathfrak{n},i},\,E_{\mathfrak{n},j} \in E_{\mathfrak{n}}\):
\flmMathEnvironment{align}{}{
P E_{\mathfrak{n},i}^\dagger E_{\mathfrak{n},j} P = P{m}_{\pi_{\mathfrak{n}},i}^\dagger m_{\pi_{\mathfrak{n}},j}P.
}
Here, \(P\) is the projector onto \(\mathsf{C}\), and \( m_{\pi_{\mathfrak{n}},i}\) is a choice of representative of \(\overline{m}_{\pi_{\mathfrak{n}},i}\) in \(\mathsf{N(S)}\).

For a QETC, the \textit{effective} code distance \(d_{\text{eff}}\) is defined to be \(2w+1\), where \(w\) is the maximum weight such that the set of all errors with weight \(\leq w\) obey the generalised \flmTerm{term}{ref1043}{}{Knill-Laflamme conditions} above. The code \(\mathsf{C}\) is thus able to transmute the set of all Pauli errors of weight \(\leq w\) to the admissible error set \(\overline{M}\).

\codefieldsection{Parent}
\begin{eczvaluelist}
\item\relax
\flmRefsHyperref[eczindexfamilyrel]{code:quantum_into_quantum}{Quantum code}\end{eczvaluelist}
\codefieldsection{Children}
\begin{eczvaluelist}
\item\relax
\flmRefsHyperref[eczindexfamilyrel]{code:qetc_7_2}{\(\llbracket 7,2,2\rrbracket \) QETC}\item\relax
\flmRefsHyperref[eczindexfamilyrel]{code:bvc}{Ball-Verstraete-Cirac (BVC) code} --- The BVC code transmutes all single-qubit errors \NoCaseChange{\protect\cite{cite2985}}.
\item\relax
\flmRefsHyperref[eczindexfamilyrel]{code:derby_klassen}{Derby-Klassen (DK) code} --- The DK code transmutes all single-qubit errors \NoCaseChange{\protect\cite{cite2985}}.
\end{eczvaluelist}
\codefieldsection{Cousins}
\begin{eczvaluelist}
\item\relax
\flmRefsHyperref[eczindexfamilyrel]{code:qubit_stabilizer}{Qubit stabilizer code} --- Most QETCs are stabilizer codes: \(\mathsf{C}\) is the subspace stabilised by an abelian subgroup \(\mathsf{S} \subset \mathcal{G}_n\) of the \flmRefsHyperref{ref663}{Pauli group} on \(n\) qubits.
\item\relax
\flmRefsHyperref[eczindexfamilyrel]{code:qecc}{Quantum error-correcting code (QECC)} --- QETCs are quantum codes which satisfy a generalization of the \flmTerm{term}{ref1043}{}{Knill-Laflamme conditions}. QETCs for which the admissible logical error set consists solely of the identity are QECCs.
\item\relax
\flmRefsHyperref[eczindexfamilyrel]{code:oecc}{Subsystem QECC} --- Subsystem codes are QETCs whose admissible error group decomposes as \(M = I \otimes G\) within the logical and gauge tensor-product space \NoCaseChange{\protect\cite[{Sec. 4}]{cite2985}}.
\item\relax
\flmRefsHyperref[eczindexfamilyrel]{code:metrological}{Metrological code} --- Metrological codes are also codes which satisfy a generalization of the \flmTerm{term}{ref1043}{}{Knill-Laflamme conditions}, albeit a different one.
\end{eczvaluelist}
\eczhbkcontributors{ Toby S. Cubitt, Daniel Zhang, \eczhuVVA }
\endeczcode

\eczcode{quantum_locally_recoverable}{Quantum locally recoverable code (QLRC)}{~\NoCaseChange{\protect\cite{cite812}}}
\codefieldsection{Description}
A QLRC of locality \(r\) is a block quantum code whose code states can be recovered after a single erasure by accessing at most \(r-1\) other subsystems and applying a recovery map.

\codefieldsection{Protection}
A Singleton-like QLRC bound states that an \(\llparenthesis n,K,d\rrparenthesis _q\) QLRC of locality \(r\) and rate \(R = \frac{\log_q K}{n}\) must have relative distance \NoCaseChange{\protect\cite[{Thm. 35}]{cite812}}
\flmMathEnvironment{align}{}{
  \delta = \frac{d}{n} \leq \frac{1-R}{2} - \Omega\left(\frac{1}{r}\right)~,
}
implying that locality restricts the distance of the code.
Random QLRCs with qudit dimension \(q = 2^{O(r)}\) achieve a relative distance that is \flmRefsHyperref{ref65}{order} \(O(1/r)\) below the bound \NoCaseChange{\protect\cite[{Prop. 5}]{cite812}}.
Codes constructed with the help of AEL distance amplification \NoCaseChange{\protect\cite{cite493,cite494}} admit a gap of \flmRefsHyperref{ref65}{order} \(O(1/r^{1/4})\) \NoCaseChange{\protect\cite[{Prop. 6}]{cite812}}.
Folded quantum Tamo-Barg codes yield explicit QLRCs of arbitrary prime locality \(r\), rate at least \(R\), relative distance \(\delta \geq (1-R)/2 - O(1/\sqrt{r})\), and qudit dimension \(q = n^{O(r^2)}\) \NoCaseChange{\protect\cite[{Cor. 64}]{cite812}}.

QLRCs have been extended to codes with intersecting recovery sets, and a Singleton-like bound has been derived for such codes \NoCaseChange{\protect\cite{cite2986}}.

\codefieldsection{Decoding}
\begin{eczvaluelist}
\item\relax Codes constructed with the help of AEL distance amplification \NoCaseChange{\protect\cite{cite493,cite494}} admit efficient decoders \NoCaseChange{\protect\cite{cite812}}.
\end{eczvaluelist}
\codefieldsection{Parent}
\begin{eczvaluelist}
\item\relax
\flmRefsHyperref[eczindexfamilyrel]{code:block_quantum}{Block quantum code}\end{eczvaluelist}
\codefieldsection{Children}
\begin{eczvaluelist}
\item\relax
\flmRefsHyperref[eczindexfamilyrel]{code:qldpc}{Qubit QLDPC code} --- Qubit QLDPC codes are stabilizer QLRCs whose locality \(r \leq w\), where \(w\) is the maximum stabilizer-generator weight \NoCaseChange{\protect\cite{cite812}}.
\item\relax
\flmRefsHyperref[eczindexfamilyrel]{code:quantum_tamo_barg}{Quantum Tamo-Barg (QTB) code} --- Folded quantum Tamo-Barg codes yield explicit QLRCs of arbitrary prime locality \(r\), rate at least \(R\), relative distance \(\delta \geq (1-R)/2 - O(1/\sqrt{r})\), and qudit dimension \(q = n^{O(r^2)}\) \NoCaseChange{\protect\cite[{Cor. 64}]{cite812}}.
\end{eczvaluelist}
\codefieldsection{Cousins}
\begin{eczvaluelist}
\item\relax
\flmRefsHyperref[eczindexfamilyrel]{code:locally_recoverable}{Locally recoverable code (LRC)} --- QLRCs are quantum analogues of LRCs.
\item\relax
\flmRefsHyperref[eczindexfamilyrel]{code:galois_css}{Galois-qudit CSS code} --- A Galois-qudit CSS code is a QLRC of locality \(r\) if each qudit participates in at least one \(X\)-type and one \(Z\)-type stabilizer whose union of supports has weight \(\leq r\) \NoCaseChange{\protect\cite[{Corr. 34}]{cite812}}.
\item\relax
\flmRefsHyperref[eczindexfamilyrel]{code:quantum_random}{Random quantum code} --- Random QLRCs with qudit dimension \(q = 2^{O(r)}\) achieve a relative distance that is \flmRefsHyperref{ref65}{order} \(O(1/r)\) below the Singleton-like QLRC bound \NoCaseChange{\protect\cite[{Prop. 5}]{cite812}}.
\item\relax
\flmRefsHyperref[eczindexfamilyrel]{code:hypergraph_product}{Hypergraph product (HGP) code} --- A variant of the hypergraph product can be used to define QLRCs with intersecting recovery sets \NoCaseChange{\protect\cite{cite2986}}.
\item\relax
\flmRefsHyperref[eczindexfamilyrel]{code:ldc}{Locally decodable code (LDC)} --- There are quantum counterparts of LDCs, but they can be transformed into (classical) LDCs which can be decoded well on average \NoCaseChange{\protect\cite{cite1079}}.
\item\relax
\flmRefsHyperref[eczindexfamilyrel]{code:lcc}{Locally correctable code (LCC)} --- A quantum code cannot admit two disjoint local recovery sets for the same qudit unless that qudit is fixed, ruling out a natural quantum analogue of LCCs \NoCaseChange{\protect\cite[{Thm. 74}]{cite812}}.
\item\relax
\flmRefsHyperref[eczindexfamilyrel]{code:general_qldpc}{QLDPC code} --- Finite-dimensional block QLDPC stabilizer codes are QLRCs whose locality \(r \leq w\), where \(w\) is the maximum stabilizer-generator weight \NoCaseChange{\protect\cite{cite812}}.
\end{eczvaluelist}
\eczhbkcontributors{ \eczhuVVA }
\endeczcode

\eczcode{qltc}{Quantum locally testable code (QLTC)}{~\NoCaseChange{\protect\cite{cite2987}}}
\codefieldsection{Description}
A local commuting-projector Hamiltonian-based block quantum code which has a nonzero average-energy penalty for creating large errors. Informally, states that are far away from the codespace of a QLTC have to be excited states of a number of the code's local projectors that scales linearly with \(n\).

The average-energy penalty is quantified by the code's \textit{soundness} \(R\). Typically, one looks at how \(R\) scales with increasing code size for infinite families of codes, defining QLTC families as those for which the soundness is asymptotically constant. QLTC families that also have asymptotically constant distance, rate, and weight of local projectors are called \(c^3\)\textit{-QLTCs}; none have been found so far.

More technically, a QLTC is a code \(\mathsf{C}\) defined as the ground-state space of a commuting-projector Hamiltonian \(H\) consisting of a sum of \(r\) local projectors (where \(r\) typically scales linearly with \(n\)), each of which acts on exactly \(u\) qubits (for some constant \(u\)). Such a code is a \((u,R)\)-QLTC with soundness function \(R(\delta)\in[0,1]\) if
\flmMathEnvironment{align}{}{
  \forall \delta > 0,|\psi\rangle~:~\text{dist}(|\psi\rangle,C) \geq \delta n \Rightarrow \frac{1}{r}\langle\psi|H|\psi\rangle\geq R(\delta)~,\label{ref2988}
}
where \(\text{dist}(|\psi\rangle,\mathsf{C})\) is a particular distance function between the state \(|\psi\rangle\) and the codespace \(\mathsf{C}\) \NoCaseChange{\protect\cite[{Def. 13}]{cite2987}}. The locality parameter \(u\) is called the \textit{query complexity} of the code.

A qubit, modular-qudit, or Galois-qudit stabilizer code that is locally testable is called a \textit{stabilizer locally testable code (SLTC)}. In other words, the code admits a set of \(r\) \(u\)-local stabilizer generators \(S_i\) whose corresponding code Hamiltonian \(H=\frac{1}{2}\sum_{i=1}^r (I-S_i)\) satisfies the requirement of being a QLTC.

For example, the \(\llbracket n=2L^2,k=2,d=L\rrbracket \) toric code on an \(L\times L\) lattice is \textit{not} a QLTC because of the following argument. Let \(|\psi\rangle\) be a ground state that is excited by \(L/3\) Pauli strings, each of length \(L/2\). In order to fit on the lattice, such strings can, e.g., be horizontal and aligned next to each other in the vertical direction. The distance function \(\text{dist}(|\psi\rangle,\mathsf{C})\) is the weight of the smallest Pauli string that multiplies \(|\psi\rangle\) to yield a state in the codespace. In this case, that weight is the same as the weight of the perturbing string, i.e., \(L^2/6\), requiring \(\delta = 1/12\) to satisfy \eqref{ref2988}. There are \(2L/3\) violated Hamiltonian terms because each of the \(L/3\) strings violates only two stabilizer generators. However, there are \(r = 2(L^2-1)\) stabilizer generators, so the implication of \eqref{ref2988} is not satisfied for nonzero soundness as \(L\to\infty\) because \(\frac{1}{r}\langle\psi|H|\psi\rangle = \frac{2L/3}{2(L^2-1)}\to 0\).

\codefieldsection{Protection}
Distance balancing and weight reduction are useful for constructing QLTCs.
For CSS code families with \(w_X,w_Z,q_X,q_Z=O(\log n)\), Hastings' weight-reduction construction yields a QLDPC family whose soundness parameters \(\epsilon_X,\epsilon_Z\) deteriorate by at most polylogarithmic factors \NoCaseChange{\protect\cite[{Lemmas 6 and 7}]{cite2989}}.
Analogous scaling for a generalized distance-balancing scheme \NoCaseChange{\protect\cite{cite684}} is proven in \NoCaseChange{\protect\cite[{Thm. 1.1}]{cite2990}}.
Weight reduction can be used to construct codes of constant locality out of CSS QLTCs \NoCaseChange{\protect\cite[{Thm. 1.1}]{cite2991}}.

\emph{Soundness amplification} \NoCaseChange{\protect\cite[{Thm. 1.2}]{cite2991}} can be used to obtain a constant-soundness (i.e., \(R = \Omega(1)\)) QLTC family from a CSS family with a sub-constant value, with the former's locality being at most polynomial in \(1/R\).

AEL distance amplification \NoCaseChange{\protect\cite{cite493,cite494}} can be used to convert an \(\llbracket n^{\prime},k,d,w\rrbracket \) soundness-\(R\) CSS LTC family into one with the same dimension, linear distance, and block length differing by at most a constant factor, with \(w\) and \(R\) differing by at most a polynomial factor in \(w\) and \(n/d\) \NoCaseChange{\protect\cite[{Thm. 1.3}]{cite2991}}.

\codefieldsection{Notes}
\begin{eczvaluelist}
\item\relax It was shown in Ref. \NoCaseChange{\protect\cite{cite1104}} that existence of a QLTC with constant parameters would imply resolution of the \textit{No low-energy trivial states} (NLTS) conjecture \NoCaseChange{\protect\cite{cite2562}} (see also \NoCaseChange{\protect\cite{cite2992}}). QLTCs are believed to also be useful for solving the quantum PCP conjecture \NoCaseChange{\protect\cite{cite2993}}.
\end{eczvaluelist}
\codefieldsection{Parents}
\begin{eczvaluelist}
\item\relax
\flmRefsHyperref[eczindexfamilyrel]{code:block_quantum}{Block quantum code}\item\relax
\flmRefsHyperref[eczindexfamilyrel]{code:qecc_finite}{Finite-dimensional quantum error-correcting code}\item\relax
\flmRefsHyperref[eczindexfamilyrel]{code:commuting_projector}{Commuting-projector Hamiltonian code} --- Quantum LTC codespaces are ground-state spaces of \(u\)-local frustration-free commuting-projector Hamiltonians.
\item\relax
\flmRefsHyperref[eczindexfamilyrel]{code:frustration_free}{Frustration-free Hamiltonian code} --- Quantum LTC codespaces are ground-state spaces of \(u\)-local frustration-free commuting-projector Hamiltonians.
\end{eczvaluelist}
\codefieldsection{Child}
\begin{eczvaluelist}
\item\relax
\flmRefsHyperref[eczindexfamilyrel]{code:check_product}{Quantum check-product code} --- Quantum check-product constructions yield an SLTC code with constant soundness \(2\rho\) from a classical LTC code with soundness \(\rho\) \NoCaseChange{\protect\cite{cite2185}}. These form the first bona-fide QLTC family because they admit asymptotically constant soundness, but they are not practical because their distance is two.
\end{eczvaluelist}
\codefieldsection{Cousins}
\begin{eczvaluelist}
\item\relax
\flmRefsHyperref[eczindexfamilyrel]{code:general_qldpc}{QLDPC code} --- Stabilizer LTCs are QLDPC. More general QLTCs are not defined using Pauli strings, but the codespace is the ground-state subspace of a local Hamiltonian. In this sense, QLTCs are QLDPC codes.
\item\relax
\flmRefsHyperref[eczindexfamilyrel]{code:self_correct}{Self-correcting quantum code} --- The notion of an energy barrier in a self-correcting memory is intimately related to the soundness of a QLTC.
\item\relax
\flmRefsHyperref[eczindexfamilyrel]{code:qubit_css}{Qubit CSS code} --- A qubit CSS code defined by \(H_{Z}\) and \(H_{X}\) is locally testable with some soundness iff the constituent codes \(\ker H_{Z}\) and \(\ker H_{X}\) are locally testable with the same soundness \NoCaseChange{\protect\cite[{Fact 17}]{cite1104}}.
\item\relax
\flmRefsHyperref[eczindexfamilyrel]{code:distance_balanced}{Distance-balanced code} --- Distance balancing and weight reduction are useful for constructing QLTCs \NoCaseChange{\protect\cite{cite2989,cite2990,cite2991}}.
\item\relax
\flmRefsHyperref[eczindexfamilyrel]{code:ltc}{Locally testable code (LTC)} --- QLTCs are quantum analogues of LTCs.
\item\relax
\flmRefsHyperref[eczindexfamilyrel]{code:dlv}{Dinur-Lin-Vidick (DLV) code} --- DLV codes have linear dimension and inverse poly-logarithmic relative distance and soundness, assuming a conjecture about random linear maps \NoCaseChange{\protect\cite{cite2994}}. Applying distance amplification and soundness amplification yields asymptotically constant soundness, \flmRefsHyperref{ref65}{order} \(\Theta(n)\) distance, \flmRefsHyperref{ref65}{order} \(\Theta(n)\) dimension, but poly-logarithmic locality \NoCaseChange{\protect\cite[{Table 4}]{cite2991}}.
\item\relax
\flmRefsHyperref[eczindexfamilyrel]{code:hemicubic}{Hemicubic code} --- The hemicubic code family has asymptotically diminishing soundness that scales as \flmRefsHyperref{ref65}{order} \(\Omega(1/\log n)\), locality of stabilizer generators scaling as \flmRefsHyperref{ref65}{order} \(O(\log n)\), and distance of \flmRefsHyperref{ref65}{order} \(\Theta(\sqrt{n})\).
Soundness amplification and AEL distance amplification \NoCaseChange{\protect\cite{cite493,cite494}} can also yield improvements in various parameters \NoCaseChange{\protect\cite[{Table 3}]{cite2991}}.
Application of generalized distance balancing \NoCaseChange{\protect\cite{cite684}} to hemicubic codes using an asymptotically good classical code of length \(t\) yields \(O( 1/(\log(n) t^2) )\) soundness and \flmRefsHyperref{ref65}{order} \(\Theta(\sqrt{n}t)\) distance while maintaining locality scaling and at the expense of a dimension scaling as \flmRefsHyperref{ref65}{order} \(\Theta(t^2)\) \NoCaseChange{\protect\cite{cite2990}}.

\item\relax
\flmRefsHyperref[eczindexfamilyrel]{code:hypersphere_product}{Hypersphere product code} --- The hypersphere product code family has asymptotically diminishing soundness that scales as \flmRefsHyperref{ref65}{order} \(O(1/\log (n)^2)\), locality of stabilizer generators scaling as \flmRefsHyperref{ref65}{order} \(O(\log n/ \log\log n)\), and distance of \flmRefsHyperref{ref65}{order} \(\Theta(\sqrt{n})\). Applying Hastings' weight-reduction construction yields QLDPC families with distance \(\Theta^*(\sqrt{n})\) and inverse-polylogarithmic soundness \NoCaseChange{\protect\cite{cite2989}}. Application of generalized distance balancing \NoCaseChange{\protect\cite{cite684}} to hypersphere product codes using an asymptotically good classical code of length \(t\) yields \(O( 1/(\log(n)^2 t^2) )\) soundness and \flmRefsHyperref{ref65}{order} \(\Theta(\sqrt{n}t)\) distance while maintaining locality scaling and at the expense of a dimension scaling as \flmRefsHyperref{ref65}{order} \(\Theta(t^2)\) \NoCaseChange{\protect\cite{cite2990}}.
\end{eczvaluelist}
\eczhbkcontributors{ \eczhuVVA }
\endeczcode

\eczcode{quantum_mds}{Quantum maximum-distance-separable (MDS) code}{~\NoCaseChange{\protect\cite{cite2768,cite2769,cite532}}}
\codefieldsection{Description}
A type of block quantum code whose parameters satisfy the quantum Singleton bound with equality.

An \(\llparenthesis n,K,d\rrparenthesis \) code constructed out of \(q\)-dimensional qudits is a quantum MDS code if parameters \(n\), \(K\), \(d\), and \(q\) are such that they saturate the quantum Singleton bound \NoCaseChange{\protect\cite{cite2768,cite2769,cite532,cite2024}},
\flmMathEnvironment{align}{}{
K \leq q^{n-2(d-1)}
}
becomes an equality for such codes.
When \(K = q^k\) for some integer \(k\), the equality condition reduces to \(2(d-1) = n-k\).
Such codes are \flmRefsHyperref{ref672}{pure} \NoCaseChange{\protect\cite{cite532}}; see also \NoCaseChange{\protect\cite{cite2995}} mentioned in Ref. \NoCaseChange{\protect\cite{cite2920}}.
The length \(n\) of a quantum MDS code with distance \(d \geq 3\) is bounded by the qudit dimension, \(n \leq q^2 + d - 2\) \NoCaseChange{\protect\cite{cite2920}}.

\codefieldsection{Protection}
Given \(n\) and \(k\), MDS codes have the highest distance possible of all codes and so have the best possible error correction properties.
\codefieldsection{Notes}
\begin{eczvaluelist}
\item\relax See Ref. \NoCaseChange{\protect\cite{cite2996,cite2024}} for an overview of quantum MDS codes.
\item\relax Tables of quantum MDS codes \NoCaseChange{\protect\cite{cite2920}}.
\end{eczvaluelist}
\codefieldsection{Parents}
\begin{eczvaluelist}
\item\relax
\flmRefsHyperref[eczindexfamilyrel]{code:block_quantum}{Block quantum code}\item\relax
\flmRefsHyperref[eczindexfamilyrel]{code:qecc_finite}{Finite-dimensional quantum error-correcting code}\end{eczvaluelist}
\codefieldsection{Children}
\begin{eczvaluelist}
\item\relax
\flmRefsHyperref[eczindexfamilyrel]{code:iceberg}{\(\llbracket 2m,2m-2,2\rrbracket \) error-detecting code} --- The only nontrivial qubit MDS codes have parameters \(\llbracket 5,1,3\rrbracket \), \(\llbracket 6,0,4\rrbracket \), and \(\llbracket 2m,2m-2,2\rrbracket \) \NoCaseChange{\protect\cite[{Sec. 27.4}]{cite2024}}.
\item\relax
\flmRefsHyperref[eczindexfamilyrel]{code:stab_5_1_3}{\(\llbracket 5,1,3\rrbracket \) Five-qubit perfect code} --- The only nontrivial qubit MDS codes have parameters \(\llbracket 5,1,3\rrbracket \), \(\llbracket 6,0,4\rrbracket \), and \(\llbracket 2m,2m-2,2\rrbracket \) \NoCaseChange{\protect\cite[{Sec. 27.4}]{cite2024}}.
\item\relax
\flmRefsHyperref[eczindexfamilyrel]{code:stab_3_1_2}{\(\llbracket 3,1,2\rrbracket _3\) Three-qutrit code} --- The three-qutrit code is the smallest nontrivial quantum MDS code.
\item\relax
\flmRefsHyperref[eczindexfamilyrel]{code:stab_9_1_5}{\(\llbracket 9,1,5\rrbracket _3\) quantum Glynn code}\item\relax
\flmRefsHyperref[eczindexfamilyrel]{code:css_5_1_3}{\(\llbracket 5,1,3\rrbracket _4\) Galois-qudit CSS code} --- The \(\llbracket 5,1,3\rrbracket _4\) code saturates the quantum Singleton bound.
\item\relax
\flmRefsHyperref[eczindexfamilyrel]{code:galois_3_1_2}{\(\llbracket 3,1,2\rrbracket _4\) three-Galois-quartrit code} --- The \(\llbracket 3,1,2\rrbracket _4\) code saturates the quantum Singleton bound.
\item\relax
\flmRefsHyperref[eczindexfamilyrel]{code:galois_6_2_3}{\(\llbracket 6,2,3\rrbracket _{q}\) code}\item\relax
\flmRefsHyperref[eczindexfamilyrel]{code:galois_7_3_3}{\(\llbracket 7,3,3\rrbracket _{q}\) code}\end{eczvaluelist}
\codefieldsection{Cousins}
\begin{eczvaluelist}
\item\relax
\flmRefsHyperref[eczindexfamilyrel]{code:mds}{Maximum distance separable (MDS) code} --- Quantum MDS codes are quantum analogues of MDS codes.
\item\relax
\flmRefsHyperref[eczindexfamilyrel]{code:galois_reed_muller}{Galois-qudit quantum RM code} --- Quantum GRM codes yield quantum MDS families \(\llbracket q,q-2\nu-2,\nu+2\rrbracket _q\) for \(0 \leq \nu \leq (q-2)/2\), \(\llbracket q^2,q^2-2\nu-2,\nu+2\rrbracket _q\) for \(0 \leq \nu \leq q-2\), and punctured descendants \(\llbracket (\nu+1)q,(\nu+1)q-2\nu-2,\nu+2\rrbracket _q\) for \(0 \leq \nu \leq q-2\) \NoCaseChange{\protect\cite{cite828}}.
\item\relax
\flmRefsHyperref[eczindexfamilyrel]{code:q-ary_cyclic}{Cyclic linear \(q\)-ary code} --- Quantum MDS codes can be constructed from \(q\)-ary cyclic codes using the Hermitian construction \NoCaseChange{\protect\cite{cite979}}.
\item\relax
\flmRefsHyperref[eczindexfamilyrel]{code:stabilizer_over_gfqsq}{Hermitian Galois-qudit code} --- Many quantum MDS codes are constructed from Hermitian self-orthogonal codes over \(\mathbb{F}_{q^2}\) using the Hermitian construction \NoCaseChange{\protect\cite{cite975,cite976,cite977,cite978}}, in particular from cyclic \NoCaseChange{\protect\cite{cite979}}, constacyclic \NoCaseChange{\protect\cite{cite980,cite981,cite978}} and negacyclic \NoCaseChange{\protect\cite{cite982}} codes.
\item\relax
\flmRefsHyperref[eczindexfamilyrel]{code:constacyclic}{Constacyclic code} --- Many quantum MDS codes are constructed from Hermitian self-orthogonal codes over \(\mathbb{F}_{q^2}\) using the Hermitian construction \NoCaseChange{\protect\cite{cite975,cite976,cite977,cite978}}, in particular from cyclic \NoCaseChange{\protect\cite{cite979}}, constacyclic \NoCaseChange{\protect\cite{cite980,cite981,cite978}}, and negacyclic \NoCaseChange{\protect\cite{cite982}} codes.
\item\relax
\flmRefsHyperref[eczindexfamilyrel]{code:generalized_reed_solomon}{Generalized RS (GRS) code} --- Some quantum MDS codes are constructed from cyclic and constacyclic codes \NoCaseChange{\protect\cite{cite1653}} which are GRS codes \NoCaseChange{\protect\cite{cite1844,cite1845}}.
\item\relax
\flmRefsHyperref[eczindexfamilyrel]{code:skew-cyclic_galois_css}{Skew-cyclic CSS code} --- Some quantum MDS codes are constructed from cyclic and constacyclic codes using the Galois-qudit CSS construction \NoCaseChange{\protect\cite{cite815}}.
\item\relax
\flmRefsHyperref[eczindexfamilyrel]{code:good_qldpc}{Good QLDPC code} --- AEL distance amplification \NoCaseChange{\protect\cite{cite493,cite494}} can be used to construct asymptotically good QLDPC codes that approach the quantum Singleton bound \NoCaseChange{\protect\cite[{Corr. 5.3}]{cite495}}.
\item\relax
\flmRefsHyperref[eczindexfamilyrel]{code:asymmetric_qecc}{Asymmetric quantum code (AQC)} --- An asymmetric Singleton bound and linear programming bounds for asymmetric CSS codes have been formulated  \NoCaseChange{\protect\cite{cite1354}}. Asymmetric MDS codes have been characterized \NoCaseChange{\protect\cite{cite2613}}.
\item\relax
\flmRefsHyperref[eczindexfamilyrel]{code:ame}{Perfect-tensor code} --- \flmRefsHyperref{ref219}{AME states} for even \(n\) are examples of quantum MDS codes with no logical qubits \NoCaseChange{\protect\cite{cite1670,cite1926,cite2933}}.
A family of conjectured perfect-tensor codes is quantum MDS \NoCaseChange{\protect\cite{cite975}}.

\item\relax
\flmRefsHyperref[eczindexfamilyrel]{code:data_syndrome}{Quantum data-syndrome (QDS) code} --- The quantum Singleton bound can be extended to QDS codes \NoCaseChange{\protect\cite{cite2914}}.
\item\relax
\flmRefsHyperref[eczindexfamilyrel]{code:ea_mds}{EA MDS code} --- EA MDS codes are entanglement-assisted versions of quantum MDS codes.
\item\relax
\flmRefsHyperref[eczindexfamilyrel]{code:quantum_singleton}{Singleton-bound approaching AQECC} --- Singleton-bound approaching AQECCs asymptotically approach the quantum Singleton bound, rather than exactly saturating it at finite blocklength.
\item\relax
\flmRefsHyperref[eczindexfamilyrel]{code:galois_quad_residue}{Quantum quadratic-residue (QR) code} --- Almost all quantum QR codes for prime-dimensional qudits are quantum MDS \NoCaseChange{\protect\cite[{Corr. 11}]{cite532}}.
\item\relax
\flmRefsHyperref[eczindexfamilyrel]{code:galois_grs}{Galois-qudit GRS code} --- Some Galois-qudit GRS codes are quantum MDS \NoCaseChange{\protect\cite{cite823}}.
\item\relax
\flmRefsHyperref[eczindexfamilyrel]{code:galois_polynomial}{Galois-qudit RS code} --- A polynomial code is a quantum MDS code when \(n-k_1=k_1-k_2\).
\item\relax
\flmRefsHyperref[eczindexfamilyrel]{code:galois_subsystem_stabilizer}{Subsystem Galois-qudit stabilizer code} --- A subsystem Galois-qudit stabilizer code saturating the quantum Singleton bound must have a trivial gauge subsystem, i.e., there are no \(\llbracket n,n-2d+2,r>0,d\rrbracket _q\) codes \NoCaseChange{\protect\cite[{Thms. 19,20}]{cite1742}}. More generally, all \flmRefsHyperref{ref672}{pure} MDS subsystem stabilizer codes are derived from MDS stabilizer codes \NoCaseChange{\protect\cite{cite2997}}.
\end{eczvaluelist}
\eczhbkcontributors{ Qingfeng (Kee) Wang, \eczhuVVA }
\endeczcode

\eczcode{quantum_quasi_cyclic}{Quasi-cyclic quantum code}{~\NoCaseChange{\protect\cite{cite821}}}
\codefieldsection{Description}
A block code on \(n\) subsystems such that cyclic shifts of the subsystems by \(\ell\) positions leave the codespace invariant.
\codefieldsection{Parent}
\begin{eczvaluelist}
\item\relax
\flmRefsHyperref[eczindexfamilyrel]{code:block_quantum}{Block quantum code}\end{eczvaluelist}
\codefieldsection{Children}
\begin{eczvaluelist}
\item\relax
\flmRefsHyperref[eczindexfamilyrel]{code:quasi_cyclic_qldpc}{Quasi-cyclic QLDPC (QC-QLDPC) code}\item\relax
\flmRefsHyperref[eczindexfamilyrel]{code:quantum_cyclic}{Cyclic quantum code}\end{eczvaluelist}
\codefieldsection{Cousin}
\begin{eczvaluelist}
\item\relax
\flmRefsHyperref[eczindexfamilyrel]{code:quasi_cyclic}{Quasi-cyclic code} --- Quasi-cyclic quantum codes are quantum analogues of quasi-cyclic codes.
\end{eczvaluelist}
\eczhbkcontributors{ \eczhuVVA }
\endeczcode

\eczcode{quantum_random}{Random quantum code}{}
\codefieldsection{Description}
Quantum code whose construction is non-deterministic in some way, i.e., codes that utilize an element of randomness somewhere in their construction. Members of this class range from fully non-deterministic codes (e.g., random-circuit codes), to codes whose multi-step construction is deterministic with the exception of a single step (e.g., expander lifted-product codes).
\codefieldsection{Protection}
Certain random codes have nontrivial \flmRefsHyperref{ref2559}{codespace complexity} \NoCaseChange{\protect\cite{cite2564}}.

\codefieldsection{Rate}
Haar random codes achieve the quantum Hamming bound \NoCaseChange{\protect\cite{cite2913}}.
\codefieldsection{Parent}
\begin{eczvaluelist}
\item\relax
\flmRefsHyperref[eczindexfamilyrel]{code:approximate_qecc}{Approximate quantum error-correcting code (AQECC)} --- Random codes typically correct errors on average.
\end{eczvaluelist}
\codefieldsection{Child}
\begin{eczvaluelist}
\item\relax
\flmRefsHyperref[eczindexfamilyrel]{code:random_circuit}{Random-circuit code}\end{eczvaluelist}
\codefieldsection{Cousins}
\begin{eczvaluelist}
\item\relax
\flmRefsHyperref[eczindexfamilyrel]{code:random}{Random code} --- Random quantum codes are quantum analogues of random classical codes.
\item\relax
\flmRefsHyperref[eczindexfamilyrel]{code:quantum_perfect}{Perfect quantum code} --- Haar random codes achieve the quantum Hamming bound \NoCaseChange{\protect\cite{cite2913}}.
\item\relax
\flmRefsHyperref[eczindexfamilyrel]{code:bosonic_rotation}{Bosonic rotation code} --- Random bosonic rotation codes can outperform cat and binomial codes when loss rate is large relative to dephasing rate \NoCaseChange{\protect\cite{cite2998}}.
\item\relax
\flmRefsHyperref[eczindexfamilyrel]{code:ntru_gkp}{NTRU-GKP code} --- Several NTRU lattices come from randomized constructions, yielding constant-rate GKP code families whose largest decodable displacement length scales as \(O(\sqrt{n})\) with high probability.
\item\relax
\flmRefsHyperref[eczindexfamilyrel]{code:covariant}{Covariant block quantum code} --- Random \(U(1)\)-covariant \NoCaseChange{\protect\cite{cite2725}} and \(U(d)\)-covariant \NoCaseChange{\protect\cite{cite2720,cite2726}} approximate QECCs exist.
\item\relax
\flmRefsHyperref[eczindexfamilyrel]{code:quantum_locally_recoverable}{Quantum locally recoverable code (QLRC)} --- Random QLRCs with qudit dimension \(q = 2^{O(r)}\) achieve a relative distance that is \flmRefsHyperref{ref65}{order} \(O(1/r)\) below the Singleton-like QLRC bound \NoCaseChange{\protect\cite[{Prop. 5}]{cite812}}.
\end{eczvaluelist}
\eczhbkcontributors{ \eczhuVVA }
\endeczcode

\eczcode{random_circuit}{Random-circuit code}{~\NoCaseChange{\protect\cite{cite2999}}}
\codefieldsection{Description}
Code whose encoding is naturally constructed by randomly sampling from a large set of quantum circuits. Examples include short random Clifford circuits that define good quantum error-correcting codes \NoCaseChange{\protect\cite{cite540}} and monitored random circuits whose mixed phase dynamically generates error-protected subspaces with nonzero channel-capacity density on polynomial timescales \NoCaseChange{\protect\cite{cite541}}.
\codefieldsection{Protection}
A useful proxy and upper bound to the code distance \(d\) is the \textit{contiguous code distance}: the contiguous length (with periodic boundary conditions) of the shortest logical operator \NoCaseChange{\protect\cite{cite3000,cite541}}.
\codefieldsection{Notes}
\begin{eczvaluelist}
\item\relax See Refs. \NoCaseChange{\protect\cite{cite3001,cite3002}} for reviews on random-circuit codes.
\end{eczvaluelist}
\codefieldsection{Parents}
\begin{eczvaluelist}
\item\relax
\flmRefsHyperref[eczindexfamilyrel]{code:dynamic_gen}{Dynamically generated QECC}\item\relax
\flmRefsHyperref[eczindexfamilyrel]{code:quantum_random}{Random quantum code}\end{eczvaluelist}
\codefieldsection{Children}
\begin{eczvaluelist}
\item\relax
\flmRefsHyperref[eczindexfamilyrel]{code:monitored_random_circuits}{Monitored random-circuit code} --- Monitored random circuits are random circuits where projective measurements are interspersed throughout the circuit and measurement results are recorded.
\item\relax
\flmRefsHyperref[eczindexfamilyrel]{code:haar_random}{Haar-random qubit code}\item\relax
\flmRefsHyperref[eczindexfamilyrel]{code:local_haar_random}{Local Haar-random circuit qubit code}\end{eczvaluelist}
\codefieldsection{Cousins}
\begin{eczvaluelist}
\item\relax
\flmRefsHyperref[eczindexfamilyrel]{code:random_stabilizer}{Random stabilizer code} --- Random stabilizer codes can be constructed by sampling random Clifford circuits.
\item\relax
\flmRefsHyperref[eczindexfamilyrel]{code:crystalline_dynamic_gen}{Crystalline-circuit qubit code} --- Crystalline-circuit codes can be thought of as random-circuit codes with symmetries.
\end{eczvaluelist}
\eczhbkcontributors{ \eczhuVVA }
\endeczcode

\eczcode{self_complementary}{Self-complementary qubit code}{~\NoCaseChange{\protect\cite{cite1262,cite851}}}
\codefieldsection{Description}
A qubit code which admits a basis of codewords of the form \(|c\rangle+|\overline{c}\rangle\), where \(c\) is a bitstring and \(\overline{c}\) is its negation a.k.a. complement. 
Their codewords generalize the two-qubit Bell states and three-qubit GHZ states and are often called \textit{(qubit) cat states} or \textit{poor-man's GHZ states}.
Such codes were originally pointed out to perform well against \flmRefsHyperref{ref498}{AD} noise \NoCaseChange{\protect\cite{cite851}}.

\codefieldsection{Protection}
Self-complementary codes automatically protect against a single \(Z\) error and lie in the \(+1\)-eigenspace of the all-\(X\) Pauli string \NoCaseChange{\protect\cite{cite1262}}. 
They are at most distance-two since the minimal number of computational basis states in a logical state is two \NoCaseChange{\protect\cite[{Thm. 4, contrapositive}]{cite529}}.
Codes consisting of computational basis states whose bitstrings are sufficiently spaced apart correct at least one \flmRefsHyperref{ref498}{AD} error \NoCaseChange{\protect\cite[{Thm. 2.5}]{cite851}\protect\cite[{Thm. 2}]{cite1297}}.
Self-complementary stabilizer codes are qubit CSS codes with a single \(X\)-type generator given by the all-\(X\) string.

\codefieldsection{Parents}
\begin{eczvaluelist}
\item\relax
\flmRefsHyperref[eczindexfamilyrel]{code:small_distance_quantum}{Small-distance block quantum code} --- Self-complementary quantum codes are at most distance-two since the minimal number of computational basis states in a logical state is two \NoCaseChange{\protect\cite[{Thm. 4, contrapositive}]{cite529}}.
\item\relax
\flmRefsHyperref[eczindexfamilyrel]{code:ampdamp}{Amplitude-damping (AD) code} --- Self-complementary quantum codes consisting of computational basis states whose bitstrings are sufficiently spaced apart correct at least one \flmRefsHyperref{ref498}{AD} error \NoCaseChange{\protect\cite[{Thm. 2.5}]{cite851}\protect\cite[{Thm. 2}]{cite1297}}.
\end{eczvaluelist}
\codefieldsection{Children}
\begin{eczvaluelist}
\item\relax
\flmRefsHyperref[eczindexfamilyrel]{code:ampdamp_cws}{Amplitude-damping CWS code}\item\relax
\flmRefsHyperref[eczindexfamilyrel]{code:ssw}{Smolin-Smith-Wehner (SSW) code}\item\relax
\flmRefsHyperref[eczindexfamilyrel]{code:ampdamp_stabilizer}{\(\llbracket 2(m+1),m,2\rrbracket \) single-loss AD code} --- The \(\llbracket 2(m+1),m,2\rrbracket \) single-loss AD code is self-complementary, with the all-\(X\) stabilizer enforcing the complement-pair structure of the codewords \NoCaseChange{\protect\cite{cite1187}}.
\item\relax
\flmRefsHyperref[eczindexfamilyrel]{code:hypercube_quantum}{\(\llbracket 2^D,D,2\rrbracket \) hypercube quantum code} --- A basis of hypercube quantum codewords of the form \(|c\rangle+|\overline{c}\rangle\) can be obtained via the \flmRefsHyperref{code:qubit_css}{qubit CSS codeword construction} since their sole \(X\)-type stabilizer generator acts on all qubits.
\item\relax
\flmRefsHyperref[eczindexfamilyrel]{code:iceberg}{\(\llbracket 2m,2m-2,2\rrbracket \) error-detecting code}\end{eczvaluelist}
\codefieldsection{Cousins}
\begin{eczvaluelist}
\item\relax
\flmRefsHyperref[eczindexfamilyrel]{code:bits_into_bits}{Binary code} --- A binary code is called \textit{self-complementary} if, for each codeword \(c\), its negation \(\overline{c}\) is also a codeword \NoCaseChange{\protect\cite{cite1261}}. Any self-complementary \((n,K,d > 1)\) classical code yields an \(\llparenthesis n,K/2,2\rrparenthesis \) self-complementary quantum code whose quantum codewords are superpositions of the classical codewords and their complements \NoCaseChange{\protect\cite[{Lemma 1}]{cite1262}}. Self-complementary classical code parameters are governed by the Gray-Rankin bound \NoCaseChange{\protect\cite{cite1263}}.
\item\relax
\flmRefsHyperref[eczindexfamilyrel]{code:qubit_8_4_2}{\(\llparenthesis 8,16,2\rrparenthesis \) \(PG(3,2)\) code} --- The logical basis of the \(\llparenthesis 8,16,2\rrparenthesis \) \(PG(3,2)\) code contains a GHZ state and linear combinations of self-complementary states \NoCaseChange{\protect\cite{cite723}}.
\end{eczvaluelist}
\eczhbkcontributors{ \eczhuVVA }
\endeczcode

\eczcode{self_correct}{Self-correcting quantum code}{~\NoCaseChange{\protect\cite{cite480,cite481}}}
\codefieldsection{Alternative Names}
\begin{eczvaluelist}
\item\relax Self-correcting quantum memory
\item\relax Thermally stable encoding
\end{eczvaluelist}
\eczhIndexCodeAliasName{self_correct}{Self-correcting quantum memory}
\eczhIndexCodeAliasName{self_correct}{Thermally stable encoding}
\codefieldsection{Description}
A block quantum code that forms the ground-state subspace of an \(n\)-body geometrically local Hamiltonian whose logical information is recoverable for arbitrarily long times in the \(n\to\infty\) limit after interaction with a sufficiently cold thermal environment.
Typically, one also requires a decoder whose decoding time scales polynomially with \(n\) and a finite energy density.

The original criteria for a self-correcting quantum memory, informally known as the \textit{Caltech rules} \NoCaseChange{\protect\cite{cite676,cite3003}}, also required finite-spin Hamiltonians.
A concatenated quantum code with self-simulating control elements based on work by Gacs \NoCaseChange{\protect\cite{cite1596,cite1597,cite1598,cite1599}} yields a self-correcting quantum memory in 2D \NoCaseChange{\protect\cite{cite2700}}.

The effect of a Markovian thermal environment consists of a Lindbladian in Davies form admitting a Gibbs steady state at some temperature \(T\) \NoCaseChange{\protect\cite{cite1610}}.
To test whether a system is self-correcting, an initial codeword \(\rho(0)\) is evolved under the Davies Lindbladian and the code Hamiltonian (or, if we are to allow extra passive protection, the code Lindbladian) to the state \(\rho(t)\) at time \(t\), after which it is decoded via decoding map \(\cal{D}\).
The memory time \(\tau\) is defined to be
\flmMathEnvironment{align}{}{
  \tau=\sup\left\{ t>0\,|\left\Vert {\cal D}( \rho(t) )-\rho(0)\right\Vert _{1}<\epsilon\right\}
}
for some fixed \(\epsilon\).
For a self-correcting memory, there exists a critical temperature \(T_\star>0\) such that \(\tau\to\infty\) (typically, exponentially with \(n\)) as \(n\to\infty\) for any temperature \(T<T_{\star}\) and any codeword \(\rho(0)\).
A memory is \textit{partially self-correcting} if \(\tau\) scales polynomially with \(n\) up to some cutoff \(n_{max}\).
A self-correcting memory is typically associated with a (stable) phase of quantum matter.

\codefieldsection{Protection}
Self-correcting classical memories exist in two and higher dimensions, with the canonical example being the classical Ising model.
In that model, a classical bit is stored in the overall magnetization. The magnetization is thermally stable due to the fact that there is an \(n\)-dependent (i.e., \textit{macroscopic}) energy cost of flipping a contiguous region of physical bits \NoCaseChange{\protect\cite{cite3004,cite1610}}.
This cost scales with the surface area of the region, and the surface area is \(n\)-dependent for dimensions greater than one.

Self-correcting quantum memories are known in four and higher dimensions, and a concatenated construction with self-simulating control elements yields one in 2D \NoCaseChange{\protect\cite{cite2700}}. Existence in one dimension is impossible (see, e.g., Ref. \NoCaseChange{\protect\cite{cite3005}}), and existence in three dimensions remains an open question.
For similar reasons as the classical 2D Ising model is a self-correcting classical memory, the 4D loop toric code is a self-correcting quantum memory due to an \flmRefsHyperref{ref65}{order} \(O(n)\) energy cost of creating a logical error \NoCaseChange{\protect\cite{cite480,cite481}}.
On the other hand, the 2D surface code is not thermally stable \NoCaseChange{\protect\cite{cite3006,cite3007,cite3008,cite3009,cite3010}} because its string-like logical operators anti-commute with stabilizer generators supported only at their ends, and thus have a constant energy cost of creation.
There is a general upper bound on the relaxation rate of a qubit stabilizer or qubit subsystem stabilizer quantum memory interacting with a Markovian environment \NoCaseChange{\protect\cite{cite3011}}.

An \(n\)-dependent energy barrier to creating all logical errors is likely necessary for a thermally stable memory, having been shown as such for a large class of 2D topological phases \NoCaseChange{\protect\cite{cite3012,cite3013}} including Abelian \NoCaseChange{\protect\cite{cite3014}} and non-Abelian \NoCaseChange{\protect\cite{cite3015}} quantum doubles.
Two-dimensional stabilizer codes \NoCaseChange{\protect\cite{cite3000}} and encodings of frustration-free code Hamiltonians \NoCaseChange{\protect\cite{cite3016}} admit only constant-energy excitations, and so do not admit such a barrier.
No-go theorems for 3D models are much more restrictive \NoCaseChange{\protect\cite{cite3017}}, e.g., a 3D lattice stabilizer code with a locality-preserving non-Clifford gate cannot have a microscopic energy barrier \NoCaseChange{\protect\cite{cite3018}}.
2D stabilizer codes \NoCaseChange{\protect\cite{cite3000}} and encodings of frustration-free code Hamiltonians \NoCaseChange{\protect\cite{cite3016}} admit only constant-energy excitations, and so do not have an energy barrier.
More generally, translationally invariant CSS codes are not self-correcting at high temperature \NoCaseChange{\protect\cite{cite3019}}.
There exist several candidates for self-correction as well as several partially self-correcting memories (see cousins below).

The lifetime of a ground-state memory protected by a Hamiltonian alone can increase at most logarithmically with \(n\) under depolarizing noise \NoCaseChange{\protect\cite{cite3020,cite3021}}, and a clock Hamiltonian can saturate this bound \NoCaseChange{\protect\cite{cite3021}}. 

\codefieldsection{Notes}
\begin{eczvaluelist}
\item\relax Reviews of self-correcting memories \NoCaseChange{\protect\cite{cite3022,cite1610}}.
\end{eczvaluelist}
\codefieldsection{Parent}
\begin{eczvaluelist}
\item\relax
\flmRefsHyperref[eczindexfamilyrel]{code:symmetry_protected_self_correct}{Symmetry-protected self-correcting quantum code} --- Self-correcting quantum codes do no require a symmetry for protection, so in that sense they are protected by a trivial symmetry.
\end{eczvaluelist}
\codefieldsection{Cousins}
\begin{eczvaluelist}
\item\relax
\flmRefsHyperref[eczindexfamilyrel]{code:surface}{Kitaev surface code} --- The surface code is not thermally stable \NoCaseChange{\protect\cite{cite3006,cite3007,cite3008,cite3009,cite3010}} because its string-like logical operators anti-commute with stabilizer generators supported only at their ends, and thus have a constant energy cost of creation. Various candidates for self-correcting quantum memories have been constructed by coupling neighboring anyons in the code so as to prevent them from spreading \NoCaseChange{\protect\cite{cite3023,cite3024,cite3025,cite3026,cite3027,cite3028,cite3029,cite3030}}.
\item\relax
\flmRefsHyperref[eczindexfamilyrel]{code:2d_stabilizer}{2D lattice stabilizer code} --- 2D stabilizer codes \NoCaseChange{\protect\cite{cite3000}} and encodings of frustration-free code Hamiltonians \NoCaseChange{\protect\cite{cite3016}} admit only constant-energy excitations, and so do not have an energy barrier.
\item\relax
\flmRefsHyperref[eczindexfamilyrel]{code:quantum_double}{Quantum-double code} --- An \(n\)-dependent energy barrier to creating all logical errors is likely necessary for a thermally stable memory, having been shown as such for a large class of 2D topological phases \NoCaseChange{\protect\cite{cite3012,cite3013}} including Abelian \NoCaseChange{\protect\cite{cite3014}} and non-Abelian \NoCaseChange{\protect\cite{cite3015}} quantum doubles.
\item\relax
\flmRefsHyperref[eczindexfamilyrel]{code:3d_stabilizer}{3D lattice stabilizer code} --- 3D translationally-invariant qubit stabilizer code families with constant \(k\) support logical string operators and thus cannot be self-correcting \NoCaseChange{\protect\cite{cite3031}}. For non-constant \(k\), such families can support at most a logarithmic energy barrier \NoCaseChange{\protect\cite{cite3032}}.
\item\relax
\flmRefsHyperref[eczindexfamilyrel]{code:3d_surface}{3D surface code} --- The 3D welded surface code is partially self-correcting with a power-law energy barrier \NoCaseChange{\protect\cite{cite3033}}. The 3D toric code is a classical self-correcting memory, whose protected bit admits a membrane-like logical operator \NoCaseChange{\protect\cite{cite3003}}, but it is not a quantum self-correcting memory because the star terms thermalize \NoCaseChange{\protect\cite{cite3019}}.
\item\relax
\flmRefsHyperref[eczindexfamilyrel]{code:3d_subsystem_surface}{3D subsystem surface code} --- The 3D subsystem surface code is not a self-correcting quantum memory despite being a single-shot code \NoCaseChange{\protect\cite{cite3034}}.
\item\relax
\flmRefsHyperref[eczindexfamilyrel]{code:4d_surface}{\((2,2)\) Loop toric code} --- For similar reasons as the classical 2D Ising model is a self-correcting classical memory, the 4D loop toric code is a self-correcting quantum memory due to an \flmRefsHyperref{ref65}{order} \(O(n)\) energy cost of creating a logical error \NoCaseChange{\protect\cite{cite480,cite481}}.
\item\relax
\flmRefsHyperref[eczindexfamilyrel]{code:color}{Color code} --- The 6D color code is a self-correcting quantum memory and admits a fault-tolerant universal gate set in 7D \NoCaseChange{\protect\cite{cite3035}}.
\item\relax
\flmRefsHyperref[eczindexfamilyrel]{code:haah_cubic}{Haah cubic code (CC)} --- Cubic code 1 is partially self-correcting with a logarithmic energy barrier \NoCaseChange{\protect\cite{cite3036}}.
\item\relax
\flmRefsHyperref[eczindexfamilyrel]{code:translationally_invariant_stabilizer}{Lattice stabilizer code} --- Translationally invariant CSS codes are not self-correcting at high temperature \NoCaseChange{\protect\cite{cite3019}}.
\item\relax
\flmRefsHyperref[eczindexfamilyrel]{code:quantum_repetition}{Quantum repetition code} --- The bit-flip repetition code associated with the 2D classical Ising model is a self-correcting classical memory \NoCaseChange{\protect\cite[{Sec. V.A}]{cite1610}}.
\item\relax
\flmRefsHyperref[eczindexfamilyrel]{code:repetition}{Repetition code} --- The repetition code associated with the 2D classical Ising model is a self-correcting classical memory \NoCaseChange{\protect\cite{cite1609}\protect\cite[{Sec. V.A}]{cite1610}}.
\item\relax
\flmRefsHyperref[eczindexfamilyrel]{code:3d_bacon_shor}{3D Bacon-Shor code} --- 3D Bacon-Shor codes were conjectured to be self-correcting \NoCaseChange{\protect\cite{cite3037}}, but there remain issues to be resolved in order to validate this conjecture (see \NoCaseChange{\protect\cite[{Sec. IX.B}]{cite1610}}).
\item\relax
\flmRefsHyperref[eczindexfamilyrel]{code:expander}{Expander code} --- Constant-rate random (quantum) expander codes are self-correcting (quantum) memories, but have no thermodynamic phase transitions \NoCaseChange{\protect\cite{cite849}}.
\item\relax
\flmRefsHyperref[eczindexfamilyrel]{code:quantum_expander}{Quantum expander code} --- Constant-rate random (quantum) expander codes are self-correcting (quantum) memories, but have no thermodynamic phase transitions \NoCaseChange{\protect\cite{cite849}}.
\item\relax
\flmRefsHyperref[eczindexfamilyrel]{code:hypergraph_product}{Hypergraph product (HGP) code} --- There are bounds on the energy barrier of hypergraph product codes \NoCaseChange{\protect\cite{cite3038}}.
\item\relax
\flmRefsHyperref[eczindexfamilyrel]{code:general_qldpc}{QLDPC code} --- Linear confinement of QLDPC (LDPC) codes implies (classical) self-correction \NoCaseChange{\protect\cite{cite849}}.
\item\relax
\flmRefsHyperref[eczindexfamilyrel]{code:ldpc}{Low-density parity-check (LDPC) code} --- Linear confinement of QLDPC (LDPC) codes implies (classical) self-correction \NoCaseChange{\protect\cite{cite849}}.
\item\relax
\flmRefsHyperref[eczindexfamilyrel]{code:quantum_concatenated}{Concatenated quantum code} --- A concatenated quantum code with self-simulating control elements based on work by Gacs \NoCaseChange{\protect\cite{cite1596,cite1597,cite1598,cite1599}} yields a self-correcting quantum memory in 2D \NoCaseChange{\protect\cite{cite2700}}.
\item\relax
\flmRefsHyperref[eczindexfamilyrel]{code:matrix_qm}{Matrix-model code} --- Matrix-model codes are similar to self-correcting memories in the sense that memory time becomes infinite in the thermodynamic limit, but with corrections being polynomial in \(N\).
\item\relax
\flmRefsHyperref[eczindexfamilyrel]{code:cat_repetition}{Cat-repetition code} --- The cat-repetition code on a 2D mode lattice is a candidate for a memory that may be self-correcting, but only in the limit of infinite energy per mode \NoCaseChange{\protect\cite{cite3039}}.
\item\relax
\flmRefsHyperref[eczindexfamilyrel]{code:single_shot}{Single-shot code} --- The presence of an energy barrier (i.e., confinement) is sufficient for a code to be single shot, and is also conjectured to be necessary for a code to be a self-correcting memory. Linear confinement of QLDPC (LDPC) codes implies (classical) self-correction \NoCaseChange{\protect\cite{cite849}}.
\item\relax
\flmRefsHyperref[eczindexfamilyrel]{code:qltc}{Quantum locally testable code (QLTC)} --- The notion of an energy barrier in a self-correcting memory is intimately related to the soundness of a QLTC.
\item\relax
\flmRefsHyperref[eczindexfamilyrel]{code:cubic_theory}{Cubic theory code} --- A family of five-dimensional cubic theory codes with Abelian loop excitations and non-Abelian membrane excitations is argued to be self-correcting below a critical temperature via a Peierls argument \NoCaseChange{\protect\cite{cite576}}.
\item\relax
\flmRefsHyperref[eczindexfamilyrel]{code:layer}{Layer code} --- The energy barrier of excitations for layer codes constructed using asymptotically good QLDPC codes scales as \flmRefsHyperref{ref65}{order} \(\Theta(n^{1/3})\) \NoCaseChange{\protect\cite{cite3040}}. Layer codes are partially self-correcting quantum memories \NoCaseChange{\protect\cite{cite3041,cite3042}}. Layer codes constructed from random CSS codes have near-optimal scaling of code parameters and a polynomial energy barrier, exhibiting behavior consistent with partial self-correction \NoCaseChange{\protect\cite{cite3041}}.
\item\relax
\flmRefsHyperref[eczindexfamilyrel]{code:fractal_surface}{Fractal surface code} --- The classical codes underlying the fractal product code form classical self-correcting memories \NoCaseChange{\protect\cite{cite677,cite678,cite679}}.
\end{eczvaluelist}
\eczhbkcontributors{ Yi-Ting (Rick) Tu, \eczhuVVA }
\endeczcode

\eczcode{single_shot}{Single-shot code}{~\NoCaseChange{\protect\cite{cite480,cite838,cite675}}}
\codefieldsection{Description}
Block quantum qudit code whose error-syndrome weights increase linearly with the distance of the error state to the code space.

Measurement errors during decoding can yield the wrong syndrome outcome, which can cause error correction to fail even against correctable data errors.
A single-shot code is a block quantum code admitting a fault-tolerant error-correcting protocol that does not "fail too badly" when faced with noisy syndrome measurements.

The property typically implies that a sufficiently large number of error-correction rounds will keep both (sufficiently low-weight) data and measurement errors bounded, as opposed to yielding eventually uncorrectable residual errors.
The property is sufficient (but not necessary \NoCaseChange{\protect\cite{cite3043}}) to reduce the number of error-correction rounds required for fault-tolerant error correction.
The word "single" refers to the ability to decode well using syndrome data from only one measurement round, i.e., without using syndrome data from previous rounds.

In the loosest form of the single-shot property for qubit, modular-qudit, or Galois-qudit codes, given some data error \(e\), ideal data error syndrome \(s\), and measurement error \(m\), there exists an error-correction protocol that outputs a correction \(f\) such that the Hamming weight of the \textit{residual error} \(e-f\) is \textit{polynomial} in the weight of \(m\).
Note that the \textit{stabilizer-reduced weight} \NoCaseChange{\protect\cite{cite675}} of \(e\) is often used instead of the weight of \(e\), with the justification that many decoders are designed to obtain the minimum-weight error representative.

A related property is \textit{linear confinement}, which states that low-weight errors cause low-weight syndromes.
A code admits \((\gamma,\alpha)\) linear confinement if the (stabilizer-reduced) weight of the syndrome is proportional to the (stabilizer-reduced) weight of the data error (for data errors of weight less than \(\gamma\)) with proportionality constant \(\alpha\).
Linear confinement is sufficient for being single shot against local stochastic noise, and more general notions of confinement are sufficient for being single shot against adversarial noise \NoCaseChange{\protect\cite{cite844}}.

Under adversarial noise, good soundness of the measurement checks is a sufficient condition for the single-shot property, although it is not known to be necessary in general \NoCaseChange{\protect\cite{cite675}}.

\codefieldsection{Protection}
A \textit{single-shot distance} \(d_{\text{ss}}\) can be defined to quantify syndrome errors, and single-shot codes have \(d_{\text{ss}} = \infty\) \NoCaseChange{\protect\cite{cite675}}.

\codefieldsection{Threshold}
\begin{eczvaluelist}
\item\relax Residual errors do not become unwieldy after some system-size-independent number of cycles of faulty syndrome measurements, and a perfect decoder would be able to recover the information if the final residual error is correctable.
Consider acting on a state \(\rho\) with a noise channel \(\mathcal N\) with noise rate \(p\), followed by \(t\) rounds of faulty syndrome measurements \(\mathcal R\) with noise rate \(\eta\) and one perfect recovery (which can be substituted with destructive physical-qubit measurements in practice).
The failure probability of a single-shot code should decrease exponentially with the distance of the code,
\flmMathEnvironment{align}{}{
  p_{\text{fail}}&=1-F\left(\mathcal{R}[\mathcal{R}_{\eta}\mathcal{N}_{p}]^{t}(\rho),\rho\right)\\&=t\left(p/p_{\star}\right)^{d}~,
}
where \(F\) is a state fidelity, and where \(p_{\star}\) is called the \textit{sustainable threshold} \NoCaseChange{\protect\cite{cite844}}.
For any \(p\) below this threshold, some maximum measurement noise \(\eta_{\star}>0\) can be tolerated after sufficiently large \(t\).

The final ideal decoding step \(\mathcal{R}\) cannot be done non-destructively in practice due to noisy syndrome measurements, but information can still be recovered by measuring all logical qubits in the computational basis and correcting the outcomes.
If the code is single-shot, then such a procedure will output the correct logical information.

\end{eczvaluelist}
\codefieldsection{Parents}
\begin{eczvaluelist}
\item\relax
\flmRefsHyperref[eczindexfamilyrel]{code:block_quantum}{Block quantum code}\item\relax
\flmRefsHyperref[eczindexfamilyrel]{code:qecc_finite}{Finite-dimensional quantum error-correcting code}\end{eczvaluelist}
\codefieldsection{Children}
\begin{eczvaluelist}
\item\relax
\flmRefsHyperref[eczindexfamilyrel]{code:double_homological_product}{Campbell double homological product code} --- For a minimal input chain complex associated with a classical \([n,k,d]\) code, the Campbell double homological product code is a single-shot code with \(d_{\text{ss}}=\infty\), \((d,f)\)-soundness for \(f(x)=x^3/4\), and check redundancy bounded by \(<2\) \NoCaseChange{\protect\cite{cite675}}.
\item\relax
\flmRefsHyperref[eczindexfamilyrel]{code:quantum_expander}{Quantum expander code} --- Quantum expander codes are single-shot \NoCaseChange{\protect\cite{cite847}}.
\item\relax
\flmRefsHyperref[eczindexfamilyrel]{code:4d_surface}{\((2,2)\) Loop toric code} --- Single-shot QEC has been realized using the \(\llbracket 33,1,4\rrbracket \) loop toric code on the Quantinuum H2 device \NoCaseChange{\protect\cite{cite850}}.
\item\relax
\flmRefsHyperref[eczindexfamilyrel]{code:3d_subsystem_surface}{3D subsystem surface code} --- The 3D subsystem surface code is a single-shot code \NoCaseChange{\protect\cite{cite839,cite840}}; see Ref. \NoCaseChange{\protect\cite{cite841}} for an alternative formulation.
\end{eczvaluelist}
\codefieldsection{Cousins}
\begin{eczvaluelist}
\item\relax
\flmRefsHyperref[eczindexfamilyrel]{code:self_correct}{Self-correcting quantum code} --- The presence of an energy barrier (i.e., confinement) is sufficient for a code to be single shot, and is also conjectured to be necessary for a code to be a self-correcting memory. Linear confinement of QLDPC (LDPC) codes implies (classical) self-correction \NoCaseChange{\protect\cite{cite849}}.
\item\relax
\flmRefsHyperref[eczindexfamilyrel]{code:hypergraph_product}{Hypergraph product (HGP) code} --- Two-fold application of the hypergraph product to a pair of binary linear codes yields single-shot QLDPC codes that exploit redundancy in their stabilizer generators \NoCaseChange{\protect\cite{cite675}}.
\item\relax
\flmRefsHyperref[eczindexfamilyrel]{code:data_syndrome}{Quantum data-syndrome (QDS) code} --- QDS codes are closely related to single-shot codes because both use redundant syndrome information to suppress measurement errors in a single round of syndrome extraction \NoCaseChange{\protect\cite{cite675}}.
\item\relax
\flmRefsHyperref[eczindexfamilyrel]{code:homological_product}{Homological product code} --- It is conjectured that a particular class of codes called three-dimensional product codes is single shot \NoCaseChange{\protect\cite{cite844}}.
\item\relax
\flmRefsHyperref[eczindexfamilyrel]{code:quantum_tanner}{Quantum Tanner code} --- Certain quantum Tanner codes facilitate single-shot decoding \NoCaseChange{\protect\cite{cite846}}.
\item\relax
\flmRefsHyperref[eczindexfamilyrel]{code:qldpc}{Qubit QLDPC code} --- Qubit QLDPC codes satisfying linear confinement are single shot \NoCaseChange{\protect\cite{cite844}}. Any code that admits a local greedy decoder also satisfies linear confinement, and so is single shot \NoCaseChange{\protect\cite{cite848}}.
\item\relax
\flmRefsHyperref[eczindexfamilyrel]{code:qubit_stabilizer}{Qubit stabilizer code} --- Any stabilizer code can be single shot if sufficiently non-local high-weight stabilizer generators are used for syndrome measurements.  These can be obtained with a Gaussian elimination procedure \NoCaseChange{\protect\cite{cite675}}.
\item\relax
\flmRefsHyperref[eczindexfamilyrel]{code:hyperbolic_surface}{Hyperbolic surface code} --- A 4D hyperbolic surface code can be decoded with the Hastings decoder \NoCaseChange{\protect\cite{cite845}} in time \(O(n\log n)\) and with a logical error scaling inverse polynomially with \(n\).
\item\relax
\flmRefsHyperref[eczindexfamilyrel]{code:shyps}{Subsystem Hypergraph Product Simplex (SHYPS) code} --- SHYPS codes exhibit practical single-shot signatures, including logical error-rate stability under small-window sliding-window decoding and constant single-shot distance \(d_{\mathrm{ss}}=3\), which supports using one syndrome-extraction round between logical generators \NoCaseChange{\protect\cite{cite785}}.
\item\relax
\flmRefsHyperref[eczindexfamilyrel]{code:3d_subsystem_color}{3D subsystem color code} --- The 3D subsystem color code defined on the cube-truncated rhombic dodecahedral honeycomb, i.e., a tessellation of cubes and chamfered cubes (a.k.a. tetratruncated rhombic dodecahedra) \NoCaseChange{\protect\cite[{Fig. 1}]{cite832}}, is a single-shot code \NoCaseChange{\protect\cite{cite838,cite832}}.
\item\relax
\flmRefsHyperref[eczindexfamilyrel]{code:generalized_bicycle}{Generalized bicycle (GB) code} --- A qubit GB code \(\llbracket n,k,d\rrbracket _2\) has \(k\) non-trivial relations between the syndrome bits, which is expected to help with operation in a fault-tolerant regime (in the presence of syndrome measurement errors). See Ref. \NoCaseChange{\protect\cite{cite842}} for many examples of such codes. There is numerical evidence that a particular family is single shot \NoCaseChange{\protect\cite{cite843}}.
\end{eczvaluelist}
\eczhbkcontributors{ \eczhuVVA }
\endeczcode

\eczcode{small_distance_quantum}{Small-distance block quantum code}{}
\codefieldsection{Description}
A block quantum code on \(n\) subsystems that either detects or corrects errors on at most two subsystems, i.e., have distance \(\leq 5\).

See Refs. \NoCaseChange{\protect\cite{cite446,cite3044,cite3045}} for small-distance codes.

\codefieldsection{Parent}
\begin{eczvaluelist}
\item\relax
\flmRefsHyperref[eczindexfamilyrel]{code:block_quantum}{Block quantum code}\end{eczvaluelist}
\codefieldsection{Children}
\begin{eczvaluelist}
\item\relax
\flmRefsHyperref[eczindexfamilyrel]{code:current_mirror}{Kitaev current-mirror qubit code}\item\relax
\flmRefsHyperref[eczindexfamilyrel]{code:rotor_3_1_2}{\(\llbracket 3,1,2\rrbracket _{\mathbb{Z}}\) Three-rotor code}\item\relax
\flmRefsHyperref[eczindexfamilyrel]{code:zero_pi}{Zero-pi qubit code}\item\relax
\flmRefsHyperref[eczindexfamilyrel]{code:rotor_5_1_3}{\(\llbracket 5,1,3\rrbracket _{\mathbb{Z}}\) Five-rotor code}\item\relax
\flmRefsHyperref[eczindexfamilyrel]{code:group_10_1_4}{\(\llbracket 10,1,4\rrbracket _{G}\) tenfold code}\item\relax
\flmRefsHyperref[eczindexfamilyrel]{code:group_4_2_2}{\(\llbracket 4,2,2\rrbracket _{G}\) four group-qudit code}\item\relax
\flmRefsHyperref[eczindexfamilyrel]{code:braunstein}{\(\llbracket 5,1,3\rrbracket _{\mathbb{R}}\) Braunstein five-mode code}\item\relax
\flmRefsHyperref[eczindexfamilyrel]{code:lloyd_slotine}{\(\llbracket 9,1,3\rrbracket _{\mathbb{R}}\) Lloyd-Slotine code}\item\relax
\flmRefsHyperref[eczindexfamilyrel]{code:quantum_perfect}{Perfect quantum code} --- All non-trivial perfect codes have distance three.
\item\relax
\flmRefsHyperref[eczindexfamilyrel]{code:self_complementary}{Self-complementary qubit code} --- Self-complementary quantum codes are at most distance-two since the minimal number of computational basis states in a logical state is two \NoCaseChange{\protect\cite[{Thm. 4, contrapositive}]{cite529}}.
\item\relax
\flmRefsHyperref[eczindexfamilyrel]{code:rains}{\(\llparenthesis 2m+1,3 \times 2^{2m-3},2\rrparenthesis \) Rains code}\item\relax
\flmRefsHyperref[eczindexfamilyrel]{code:ruskai}{\(\llparenthesis 9,2,3\rrparenthesis \) Ruskai code}\item\relax
\flmRefsHyperref[eczindexfamilyrel]{code:unentangled_permutation_invariant}{\(\llparenthesis n,2,2\rrparenthesis \) Bravyi-Lee-Li-Yoshida PI code}\item\relax
\flmRefsHyperref[eczindexfamilyrel]{code:qubit_10_24_3}{\(\llparenthesis 10,24,3\rrparenthesis \) qubit code} --- The \(\llparenthesis 10,24,3\rrparenthesis \) qubit code can be combined to form an infinite family of distance-three qubit codes whose logical dimension is \(50\%\) larger than that of the optimal stabilizer code \NoCaseChange{\protect\cite{cite3046}}.
\item\relax
\flmRefsHyperref[eczindexfamilyrel]{code:ampdamp_numopt}{Numerically optimized four-qubit AD code}\item\relax
\flmRefsHyperref[eczindexfamilyrel]{code:qubit_6_2_3}{\(\llparenthesis 6,2,3\rrparenthesis \) transversal-\(\mathbb{Z}_{10}\) code}\item\relax
\flmRefsHyperref[eczindexfamilyrel]{code:qubit_8_1_3}{\(\llparenthesis 8,2,3\rrparenthesis \) Plenio-Vedral-Knight CE code}\item\relax
\flmRefsHyperref[eczindexfamilyrel]{code:qubit_8_4_2}{\(\llparenthesis 8,16,2\rrparenthesis \) \(PG(3,2)\) code}\item\relax
\flmRefsHyperref[eczindexfamilyrel]{code:qubit_9_12_3}{\(\llparenthesis 9,12,3\rrparenthesis \) qubit code} --- The \(\llparenthesis 9,12,3\rrparenthesis \) qubit code can be combined to form an infinite family of distance-three qubit codes whose logical dimension is \(50\%\) larger than that of the optimal stabilizer code \NoCaseChange{\protect\cite{cite3046}}.
\item\relax
\flmRefsHyperref[eczindexfamilyrel]{code:small_distance_qubit_stabilizer}{Small-distance qubit stabilizer code}\item\relax
\flmRefsHyperref[eczindexfamilyrel]{code:sslp}{Subset-Sum-Linear-Programming (SS-LP) code}\item\relax
\flmRefsHyperref[eczindexfamilyrel]{code:qudit_3_6_2}{\(\llparenthesis 3,6,2\rrparenthesis _{\mathbb{Z}_6}\) Euler code}\item\relax
\flmRefsHyperref[eczindexfamilyrel]{code:qudit_5_1_3}{\(\llbracket 5,1,3\rrbracket _{\mathbb{Z}_q}\) modular-qudit code}\item\relax
\flmRefsHyperref[eczindexfamilyrel]{code:qutrit_golay}{\(\llbracket 11,1,5\rrbracket _3\) qutrit Golay code}\item\relax
\flmRefsHyperref[eczindexfamilyrel]{code:stab_3_1_2}{\(\llbracket 3,1,2\rrbracket _3\) Three-qutrit code}\item\relax
\flmRefsHyperref[eczindexfamilyrel]{code:stab_9_1_3}{\(\llbracket 9,1,3\rrbracket _{\mathbb{Z}_q}\) modular-qudit code}\item\relax
\flmRefsHyperref[eczindexfamilyrel]{code:stab_9_1_5}{\(\llbracket 9,1,5\rrbracket _3\) quantum Glynn code}\item\relax
\flmRefsHyperref[eczindexfamilyrel]{code:three_qutrit_permutation_invariant}{\(\llparenthesis 3,2,2\rrparenthesis _3\) Three-qutrit single-deletion code}\item\relax
\flmRefsHyperref[eczindexfamilyrel]{code:qudit_hamming_css}{\(\llbracket 2^r-1, 2^r-2r-1, 3\rrbracket _p\) quantum Hamming code}\item\relax
\flmRefsHyperref[eczindexfamilyrel]{code:qutrit_small_triorthogonal}{\(\llbracket 9m-k,k,2\rrbracket _3\) triorthogonal code}\item\relax
\flmRefsHyperref[eczindexfamilyrel]{code:arvind}{\(\llparenthesis n,1+n(q-1),2\rrparenthesis _q\) union stabilizer code}\item\relax
\flmRefsHyperref[eczindexfamilyrel]{code:css_5_1_3}{\(\llbracket 5,1,3\rrbracket _4\) Galois-qudit CSS code}\item\relax
\flmRefsHyperref[eczindexfamilyrel]{code:galois_3_1_2}{\(\llbracket 3,1,2\rrbracket _4\) three-Galois-quartrit code}\item\relax
\flmRefsHyperref[eczindexfamilyrel]{code:galois_5_1_3}{\(\llbracket 5,1,3\rrbracket _q\) Galois-qudit code}\item\relax
\flmRefsHyperref[eczindexfamilyrel]{code:galois_6_2_3}{\(\llbracket 6,2,3\rrbracket _{q}\) code}\item\relax
\flmRefsHyperref[eczindexfamilyrel]{code:galois_7_3_3}{\(\llbracket 7,3,3\rrbracket _{q}\) code}\item\relax
\flmRefsHyperref[eczindexfamilyrel]{code:t_group}{Twisted \(1\)-group code} --- All twisted \(1\)-group codes have a distance \(d \geq 2\).
\end{eczvaluelist}
\codefieldsection{Cousins}
\begin{eczvaluelist}
\item\relax
\flmRefsHyperref[eczindexfamilyrel]{code:small_distance}{Small-distance block code} --- Small-distance block quantum codes are quantum analogues of small-distance block codes.
\item\relax
\flmRefsHyperref[eczindexfamilyrel]{code:group_representation}{Group-representation code} --- See Ref. \NoCaseChange{\protect\cite{cite789}} for tables of distance-two codes with various families of transversal gates.
\item\relax
\flmRefsHyperref[eczindexfamilyrel]{code:binary_dihedral_permutation_invariant}{Binary dihedral PI code} --- The first and second families of binary dihedral PI codes have distance three, and the third family has the member \(\llparenthesis 27,2,5\rrparenthesis \).
\item\relax
\flmRefsHyperref[eczindexfamilyrel]{code:quantum_plane_curve}{Quantum plane-curve code} --- The quantum plane-curve code for the Hermitian curve \(y^3 + y = x^4\) is a \(\llbracket 27,13,4\rrbracket _3\) qutrit code.
\end{eczvaluelist}
\eczhbkcontributors{ \eczhuVVA }
\endeczcode

\eczcode{oecc}{Subsystem QECC}{~\NoCaseChange{\protect\cite{cite3047,cite3048}}}
\codefieldsection{Alternative Names}
\begin{eczvaluelist}
\item\relax Operator QECC (OQECC)
\item\relax Gauge QECC
\end{eczvaluelist}
\eczhIndexCodeAliasName{oecc}{Operator QECC (OQECC)}
\eczhIndexCodeAliasName{oecc}{Gauge QECC}
\codefieldsection{Description}
A quantum code which encodes quantum information in a tensor factor of a subspace that is decomposed into a tensor product of subsystems.

A subsystem code encodes information in a subsystem \(\mathsf{A}\) of the code space \(\mathsf{C}\), which is part of the system Hilbert space \(\mathsf{H}\), as
\flmMathEnvironment{align}{}{
\mathsf{H}=\mathsf{C} \oplus \mathsf{C}^{\perp} = \mathsf{A} \otimes \mathsf{B} \oplus \mathsf{C}^{\perp}~.
}
Following an error, the encoded quantum information in subsystem \(\mathsf{A}\) can be recovered modulo an arbitrary error on the auxiliary or \textit{gauge} subsystem \(\mathsf{B}\). 
The gauge subsystem provides additional freedom to the error correction process: errors that act trivially on the information subsystem \(\mathsf{A}\) but nontrivially on \(\mathsf{B}\) need not be corrected.
The subsystem \(\mathsf{B}\) can encode \textit{gauge qubits} when its dimension is a power of two.
While all operator QECCs are also ordinary QECCs, the attachment of a gauge subsystem to a code allows for a wider variety of encoding procedures, fault-tolerant logical operations, and efficient error-correction protocols.

\codefieldsection{Protection}
The necessary and sufficient error-correction conditions are, for all errors \(E_a,E_b\) in an error set \(\cal{E}\) \NoCaseChange{\protect\cite{cite3049}}:
\flmMathEnvironment{align}{}{
\Pi E^{\dagger}_a E_b \Pi = I_{\mathsf{A}} \otimes g_{ab}^{\mathsf{B}}
}
where \(\Pi\) is a projector onto the codespace \(\mathsf{C}\), and \(g_{ab}^{\mathsf{B}}\) is an arbitrary operator on the gauge subsystem \(\mathsf{B}\).
This condition ensures that distinguishing and correcting errors based on their effect on subsystem \(\mathsf{A}\) is sufficient; errors that act identically on \(\mathsf{A}\) but differ on \(\mathsf{B}\) are considered equivalent.

These can be studied in the presence of continuous noise \NoCaseChange{\protect\cite{cite3050}}.

A \textit{unitarily recoverable subsystem} is a correctable subsystem whose logical information can be restored by a unitary operation, possibly into a different subsystem representation; thus, recovery is more relaxed than correction \NoCaseChange{\protect\cite{cite3051}}. In fact, every correctable subsystem is unitarily recoverable \NoCaseChange{\protect\cite[{Thm. 1}]{cite3051}}. For unital noise channels, \textit{unitarily correctable subsystems} are precisely the noiseless subsystems of \(\mathcal{E}^{\dagger}\circ\mathcal{E}\) \NoCaseChange{\protect\cite[{Thm. 2}]{cite3051}}; these are related to the multiplicative domain of the channel \NoCaseChange{\protect\cite{cite3052}} (see also \NoCaseChange{\protect\cite{cite3053}}).

No additional OQEC conditions are needed to tolerate imperfect initialization: under the standard subsystem-code conditions, the effective noise induced by population outside the code can only increase the fidelity with an ideally encoded state, and this robustness persists under encoded CPTP operations \NoCaseChange{\protect\cite[{Thms. 3,4}]{cite3054}}.

\codefieldsection{Encoding}
\begin{eczvaluelist}
\item\relax Subsystem QECCs are robust to initialization errors without modifying the standard OQEC conditions \NoCaseChange{\protect\cite{cite3054}}.
\end{eczvaluelist}
\codefieldsection{Decoding}
\begin{eczvaluelist}
\item\relax Petz recovery map provides a recovery operation that is near-optimal for certain subsystem codes \NoCaseChange{\protect\cite{cite2588}}.
\end{eczvaluelist}
\codefieldsection{Realizations}
\begin{eczvaluelist}
\item\relax A two-qubit unitarily recoverable subsystem code recovery has been realized in an optical system \NoCaseChange{\protect\cite{cite3055}}.
\end{eczvaluelist}
\codefieldsection{Notes}
\begin{eczvaluelist}
\item\relax See Refs. \NoCaseChange{\protect\cite{cite3056,cite3057,cite3058}} for an introduction to operator QEC.
\item\relax See \NoCaseChange{\protect\cite{cite2733}} for a pedagogical introduction to subsystem codes.
\end{eczvaluelist}
\codefieldsection{Parent}
\begin{eczvaluelist}
\item\relax
\flmRefsHyperref[eczindexfamilyrel]{code:oaecc}{Operator-algebra QECC (OAQECC)} --- An OAQECC which has gauge structure (e.g., gauge qubits) but no block structure is a subsystem QECC.
\end{eczvaluelist}
\codefieldsection{Child}
\begin{eczvaluelist}
\item\relax
\flmRefsHyperref[eczindexfamilyrel]{code:subsystem_group_quantum}{Subsystem group-based quantum code}\end{eczvaluelist}
\codefieldsection{Cousins}
\begin{eczvaluelist}
\item\relax
\flmRefsHyperref[eczindexfamilyrel]{code:qecc}{Quantum error-correcting code (QECC)} --- A subsystem QECC reduces to an ordinary (i.e., subspace) QECC when the gauge subsystem is trivial. Conversely, any QECC with a tensor-product logical subspace can be turned into a subsystem code by treating a logical tensor factor as a gauge subsystem.
\item\relax
\flmRefsHyperref[eczindexfamilyrel]{code:eaoecc}{Entanglement-assisted (EA) operator QECC} --- EAOQECCs utilize additional ancillary subsystems in a pre-shared entangled state, but reduce to subsystem QECCs when said subsystems are interpreted as noiseless physical subsystems.
\item\relax
\flmRefsHyperref[eczindexfamilyrel]{code:knill}{Knill code} --- Subsystem Knill codes can be formulated \NoCaseChange{\protect\cite{cite2884}}.
\item\relax
\flmRefsHyperref[eczindexfamilyrel]{code:qetc}{Quantum error-transmuting code (QETC)} --- Subsystem codes are QETCs whose admissible error group decomposes as \(M = I \otimes G\) within the logical and gauge tensor-product space \NoCaseChange{\protect\cite[{Sec. 4}]{cite2985}}.
\end{eczvaluelist}
\eczhbkcontributors{ Srilekha Gandhari, \eczhuVVA }
\endeczcode

\eczcode{symmetry_protected_self_correct}{Symmetry-protected self-correcting quantum code}{~\NoCaseChange{\protect\cite{cite3059}}}
\codefieldsection{Alternative Names}
\begin{eczvaluelist}
\item\relax Symmetry-protected self-correcting memory
\end{eczvaluelist}
\eczhIndexCodeAliasName{symmetry_protected_self_correct}{Symmetry-protected self-correcting memory}
\codefieldsection{Description}
A code which admits a restricted notion of thermal stability against symmetric perturbations, i.e., perturbations that commute with a set of operators forming a group \(G\) called the \textit{symmetry group}.

Given a symmetry group \(G\) and its unitary representation \(S\) on the \(n\)-site physical Hilbert space (in this case, a lattice), an operator \(O\) is \(G\)\textit{-symmetric} (a.k.a. respects the \(G\) symmetry) if \([S(g),O]=0\) for all \(g\in G\).
A symmetry-protected self-correcting memory is a ground-state encoding of an \(n\)-body \(G\)-symmetric geometrically local Hamiltonian whose logical information is recoverable for arbitrarily long times in the \(n\to\infty\) limit after a \(G\)-symmetric interaction with a thermal environment at sufficiently low temperature.

Tensor-product symmetries of the form \(S(g)=u(g)^{\otimes n}\), where \(u\) is a unitary representation of \(G\ni g\) on a site, cannot support symmetry-protected self-correction. One can instead use \textit{one-form symmetries}, i.e., symmetries generated by operators of the form
\flmMathEnvironment{align}{}{
  S_{\mathcal{M}}(g)=\bigotimes_{\text{sites}\in\mathcal{M}}u(g),
}
where \(\mathcal{M}\) runs over all closed codimension-one submanifolds of the lattice. Recent work further relaxed the requirement so that symmetries need only be enforced on the system's boundaries \NoCaseChange{\protect\cite{cite3060}}.

\codefieldsection{Protection}
The code is intended to be used as a self-correcting quantum memory when the symmetry is enforced, and protection is characterized by the scaling of the memory time \(\tau\) in the system size.

Another characterization of the protection property is the symmetric version of the energy barrier \(\Delta\), defined as follows.
For a given logical operator and a given decomposition into a product of local operators, we consider the maximal energy attained when implementing this logical operator stepwise with this decomposition. Then, \(\Delta\) is defined by minimizing this quantity over all logical operators and over those decompositions for which each local operator respects the symmetry. For some models \NoCaseChange{\protect\cite{cite3059}}, the linear growth of \(\Delta\) with system size \(n\) implies the exponential growth of \(\tau\) below a critical temperature.

\codefieldsection{Parent}
\begin{eczvaluelist}
\item\relax
\flmRefsHyperref[eczindexfamilyrel]{code:hamiltonian}{Hamiltonian-based code}\end{eczvaluelist}
\codefieldsection{Child}
\begin{eczvaluelist}
\item\relax
\flmRefsHyperref[eczindexfamilyrel]{code:self_correct}{Self-correcting quantum code} --- Self-correcting quantum codes do no require a symmetry for protection, so in that sense they are protected by a trivial symmetry.
\end{eczvaluelist}
\codefieldsection{Cousins}
\begin{eczvaluelist}
\item\relax
\flmRefsHyperref[eczindexfamilyrel]{code:spt}{Symmetry-protected topological (SPT) code} --- Symmetry-protected self-correcting quantum codes can realize symmetry-protected topological phases \NoCaseChange{\protect\cite{cite3059}}. Metastable subspaces of certain many-body Lindbladians can arise due to symmetry relations in the low-lying excitations \NoCaseChange{\protect\cite{cite3061}}.
\item\relax
\flmRefsHyperref[eczindexfamilyrel]{code:rbh}{Raussendorf-Bravyi-Harrington (RBH) cluster-state code} --- The RBH code can exhibit self-correction protected by a certain symmetry.
\item\relax
\flmRefsHyperref[eczindexfamilyrel]{code:3d_subsystem_color}{3D subsystem color code} --- A particular gauge-fixed version of a subsystem code on a 3D lattice yields a self-correcting memory protected by one-form symmetries \NoCaseChange{\protect\cite{cite466}\protect\cite[{Sec. IV D}]{cite3059}}.
The symmetric energy barrier grows linearly with the length of a side of the lattice. When the system is coupled locally to a thermal bath respecting the symmetry and below a critical temperature, the memory time grows exponentially with the side length.
The subsystem color code is not a self-correcting quantum memory if symmetry protection is removed \NoCaseChange{\protect\cite{cite3034}}.

\end{eczvaluelist}
\eczhbkcontributors{ Yi-Ting (Rick) Tu, \eczhuVVA }
\endeczcode

\eczcode{spt}{Symmetry-protected topological (SPT) code}{~\NoCaseChange{\protect\cite{cite3062,cite3063}}}
\codefieldsection{Description}
A code whose codewords form the ground-state or low-energy subspace of a code Hamiltonian realizing symmetry-protected topological (SPT) order.

\codefieldsection{Protection}
SPT codes typically do not offer protection against generic errors, but can protect against noise that respects the underlying symmetry.
One-form symmetries can be detected using tools from QEC \NoCaseChange{\protect\cite{cite3064}}.

\codefieldsection{Encoding}
\begin{eczvaluelist}
\item\relax Conjectured QCA encoder for SPTs defined by Stiefel-Whitney classes in arbitrary dimensions \NoCaseChange{\protect\cite{cite3065}}.
\end{eczvaluelist}
\codefieldsection{Gates}
\begin{eczvaluelist}
\item\relax There is a relation between the \flmTerm{term}{ref694}{}{Clifford hierarchy} and group cohomology \NoCaseChange{\protect\cite{cite3066}} (with the latter being useful for classifying SPTs).
\end{eczvaluelist}
\codefieldsection{Notes}
\begin{eczvaluelist}
\item\relax Review on generalized (i.e., non-tensor-product) symmetries \NoCaseChange{\protect\cite{cite3067}}.
\end{eczvaluelist}
\codefieldsection{Parent}
\begin{eczvaluelist}
\item\relax
\flmRefsHyperref[eczindexfamilyrel]{code:topological}{Topological code} --- SPT codes realize symmetry-protected topological phases.
\end{eczvaluelist}
\codefieldsection{Children}
\begin{eczvaluelist}
\item\relax
\flmRefsHyperref[eczindexfamilyrel]{code:kitaev_chain}{Kitaev chain code} --- The Kitaev chain is a 1D fermionic SPT (more precisely, a 1D topological superconductor) protected by fermion parity symmetry.
\item\relax
\flmRefsHyperref[eczindexfamilyrel]{code:three_fermion}{Three-fermion (3F) Walker-Wang model code} --- When treated as ground states of the code Hamiltonian, 3F Walker-Wang model code states realize a 3D time-reversal SPT order \NoCaseChange{\protect\cite{cite477}}. The 3F Walker-Wang QCA encoder \NoCaseChange{\protect\cite{cite3068,cite3069}} can be extended to SPTs in higher dimensions based on an exact bosonization duality \NoCaseChange{\protect\cite{cite3065}}.
\end{eczvaluelist}
\codefieldsection{Cousins}
\begin{eczvaluelist}
\item\relax
\flmRefsHyperref[eczindexfamilyrel]{code:bits_into_bits}{Binary code} --- SPT orders may be used for encoding classical information \NoCaseChange{\protect\cite{cite1260}}.
\item\relax
\flmRefsHyperref[eczindexfamilyrel]{code:surface}{Kitaev surface code} --- Gauging \NoCaseChange{\protect\cite{cite462,cite463,cite233,cite464,cite465,cite466,cite467,cite468,cite469,cite470}} the symmetry of a trivial 2D bosonic \(\mathbb{Z}_2\) Ising SPT yields the surface-code phase \NoCaseChange{\protect\cite[{Sec. IV}]{cite462}}.
\item\relax
\flmRefsHyperref[eczindexfamilyrel]{code:double_semion_string_net}{Double-semion string-net code} --- Gauging \NoCaseChange{\protect\cite{cite462,cite463,cite233,cite464,cite465,cite466,cite467,cite468,cite469,cite470}} the symmetry of a nontrivial 2D bosonic \(\mathbb{Z}_2\) Ising SPT yields the doubled-semion phase, with semionic \(\pi\)-flux excitations \NoCaseChange{\protect\cite[{Sec. IV}]{cite462}}.
\item\relax
\flmRefsHyperref[eczindexfamilyrel]{code:quantum_double}{Quantum-double code} --- The \(Q\) quantum double model can be obtained by gauging \NoCaseChange{\protect\cite{cite462,cite463,cite233,cite464,cite465,cite466,cite467,cite468,cite469,cite470}} symmetries of a Type I and Type III \(\mathbb{Z}_2^3\) SPT \NoCaseChange{\protect\cite{cite3070,cite3071}}.
\item\relax
\flmRefsHyperref[eczindexfamilyrel]{code:quantum_double_dihedral}{Dihedral \(G=D_m\) quantum-double code} --- The \(D_4\) quantum double model can be obtained by gauging \NoCaseChange{\protect\cite{cite462,cite463,cite233,cite464,cite465,cite466,cite467,cite468,cite469,cite470}} symmetries of a Type III \(\mathbb{Z}_2^3\) SPT \NoCaseChange{\protect\cite{cite575,cite3071}}.
\item\relax
\flmRefsHyperref[eczindexfamilyrel]{code:tqd}{Twisted quantum double (TQD) code} --- A TQD code Hamiltonian can be obtained by gauging \NoCaseChange{\protect\cite{cite462,cite463,cite233,cite464,cite465,cite466,cite467,cite468,cite469,cite470}} the symmetry of a particular 2D SPT model. The same group and cocycle data classifies both 2D SPTs and TQDs \NoCaseChange{\protect\cite{cite462,cite3072}}.
\item\relax
\flmRefsHyperref[eczindexfamilyrel]{code:translationally_invariant_stabilizer}{Lattice stabilizer code} --- Lattice CSS codes in \(D\) dimensions can be converted into SPT Hamiltonians in one less dimension \NoCaseChange{\protect\cite{cite466}}.
\item\relax
\flmRefsHyperref[eczindexfamilyrel]{code:gkp}{Square-lattice GKP code} --- The Segal-Bargmann representations of GKP states are the theta functions of the lowest Landau level on a torus \NoCaseChange{\protect\cite[{Sec. V}]{cite3073}\protect\cite[{Prop. 6.3}]{cite3074}} (see also Refs. \NoCaseChange{\protect\cite{cite3075,cite3076,cite3077}}).
\item\relax
\flmRefsHyperref[eczindexfamilyrel]{code:approximate_qecc}{Approximate quantum error-correcting code (AQECC)} --- Certain phases with continuous symmetries cannot be prepared using a constant-depth circuit, a consequence of the Lieb-Schult-Mattis theorem \NoCaseChange{\protect\cite{cite2603,cite2604,cite2605,cite2606}}. The theorem, in turn, can be linked to the circuit complexity of the underlying approximate error-correcting code \NoCaseChange{\protect\cite{cite2565}}.
\item\relax
\flmRefsHyperref[eczindexfamilyrel]{code:symmetry_protected_self_correct}{Symmetry-protected self-correcting quantum code} --- Symmetry-protected self-correcting quantum codes can realize symmetry-protected topological phases \NoCaseChange{\protect\cite{cite3059}}. Metastable subspaces of certain many-body Lindbladians can arise due to symmetry relations in the low-lying excitations \NoCaseChange{\protect\cite{cite3061}}.
\item\relax
\flmRefsHyperref[eczindexfamilyrel]{code:hexagonal_cz}{Hexagonal \(CZ\) code} --- The hexagonal \(CZ\) code can be obtained by gauging \NoCaseChange{\protect\cite{cite462,cite463,cite233,cite464,cite465,cite466,cite467,cite468,cite469,cite470}} the symmetry of a particular SPT \NoCaseChange{\protect\cite{cite575}}.
\item\relax
\flmRefsHyperref[eczindexfamilyrel]{code:invertible}{Chen-Hsin invertible-order code} --- Instances of the Chen-Hsin invertible-order code realize beyond-group-cohomology SPTs \NoCaseChange{\protect\cite{cite578}}.
\item\relax
\flmRefsHyperref[eczindexfamilyrel]{code:sierpinsky_fractal_liquid}{Sierpinski prism model code} --- Ungauging \NoCaseChange{\protect\cite{cite462,cite463,cite233,cite464,cite465,cite466,cite467,cite468,cite469,cite470}} yields explicit fracton-SPT constructions; in particular, the Yoshida first-order fractal spin-liquid code underlies a 2D fracton SPT protected by Sierpinski-triangle symmetries via a gapped domain wall \NoCaseChange{\protect\cite{cite466}}.
\item\relax
\flmRefsHyperref[eczindexfamilyrel]{code:two_foliated}{Two-foliated fracton code} --- Gauging a 3D paramagnet with planar subsystem symmetries in two directions yields the anisotropic lineon model; each symmetry charge becomes a lineon gauge charge, while certain pairs become planons \NoCaseChange{\protect\cite[{Sec. 4.1.2}]{cite467}}.
\item\relax
\flmRefsHyperref[eczindexfamilyrel]{code:cluster_state}{Cluster-state code} --- Cluster states defined on various lattices are representatives of SPT phases, and states realizing these phases can be resources for MBQC. 
In 1D, cluster states are examples of SPT phases with global symmetries \NoCaseChange{\protect\cite{cite3078,cite3079,cite3080,cite3081,cite3072}} and enable MBQC on a single qubit \NoCaseChange{\protect\cite{cite428,cite429}}. 
The square-lattice cluster state, which is the prototypical resource for universal MBQC \NoCaseChange{\protect\cite{cite428,cite429}}, and other 2D cluster states \NoCaseChange{\protect\cite{cite3082,cite3083,cite3084}} have SPT order protected by subsystem symmetries \NoCaseChange{\protect\cite{cite3085,cite3086,cite3082}}.
States like AKLT states and SPT fixed-point states can be efficiently converted into cluster states using local measurements and subsequently used as resources for MBQC \NoCaseChange{\protect\cite{cite3087,cite3079,cite3088,cite3089,cite3090,cite3091}}.
In 3D, cluster states belong to SPT phases protected by higher-form symmetries \NoCaseChange{\protect\cite{cite3092}} and enable universal fault-tolerant MBQC \NoCaseChange{\protect\cite{cite3093}}.
A cluster-like state, or a state that is in the same SPT phase as a cluster state, can be prepared in finite time \NoCaseChange{\protect\cite{cite3094}}. Cluster states can be created on various lattices \NoCaseChange{\protect\cite{cite3095}}.

\item\relax
\flmRefsHyperref[eczindexfamilyrel]{code:rbh}{Raussendorf-Bravyi-Harrington (RBH) cluster-state code} --- In 3D, cluster states belong to SPT phases protected by higher-form symmetries \NoCaseChange{\protect\cite{cite3092}} and enable universal fault-tolerant MBQC \NoCaseChange{\protect\cite{cite3093}}.
\item\relax
\flmRefsHyperref[eczindexfamilyrel]{code:square_lattice_cluster}{Square-lattice cluster-state code} --- The square-lattice cluster state, which is the prototypical resource for universal MBQC \NoCaseChange{\protect\cite{cite428,cite429}}, and other 2D cluster states \NoCaseChange{\protect\cite{cite3082,cite3083,cite3084}} have SPT order protected by subsystem symmetries \NoCaseChange{\protect\cite{cite3085,cite3086,cite3082}}.

\item\relax
\flmRefsHyperref[eczindexfamilyrel]{code:3d_color}{3D color code} --- Transversal action of \(T\) gates on color codes on general 3-manifolds realizes a \(CCZ\) gate on three logical qubits and is related to a topological invariant that is called the triple intersection number \NoCaseChange{\protect\cite{cite703}}. Transversal \(S\) gate on color codes on general 3-manifolds corresponds to a higher-form symmetry \NoCaseChange{\protect\cite{cite703}}.
\item\relax
\flmRefsHyperref[eczindexfamilyrel]{code:3d_subsystem_color}{3D subsystem color code} --- Ungauging \NoCaseChange{\protect\cite{cite462,cite463,cite233,cite464,cite465,cite466,cite467,cite468,cite469,cite470}} different stabilizer Hamiltonians of the 3D subsystem color code yields distinct SPT phases; in particular, one ungauges to three decoupled copies of the RBH model \NoCaseChange{\protect\cite{cite466}}.
\item\relax
\flmRefsHyperref[eczindexfamilyrel]{code:fracton}{Fracton stabilizer code} --- Certain 3D CSS fracton codes can be ungauged \NoCaseChange{\protect\cite{cite462,cite463,cite233,cite464,cite465,cite466,cite467,cite468,cite469,cite470}} into 2D fractal-like SPT Hamiltonians; the paper gives an explicit construction from the 3D fractal code \NoCaseChange{\protect\cite{cite466}}. In subsystem-symmetry gauging \NoCaseChange{\protect\cite{cite462,cite463,cite233,cite464,cite465,cite466,cite467,cite468,cite469,cite470}} constructions, symmetry charges transforming under planar symmetries in one, two, or three directions become planon, lineon, or fracton excitations, respectively \NoCaseChange{\protect\cite{cite467}}.
\item\relax
\flmRefsHyperref[eczindexfamilyrel]{code:qudit_cluster_state}{Modular-qudit cluster-state code} --- Qudit cluster states defined on 1D lattices are representatives of various SPT phases \NoCaseChange{\protect\cite{cite3096}}. 

\item\relax
\flmRefsHyperref[eczindexfamilyrel]{code:tqd_abelian_stabilizer}{Abelian TQD stabilizer code} --- Gauging \NoCaseChange{\protect\cite{cite462,cite463,cite233,cite464,cite465,cite466,cite467,cite468,cite469,cite470}} the \(1\)-form symmetries associated with gauge charges of Abelian TQD stabilizer codes yields Pauli stabilizer models of SPT phases classified by products of Type-I and Type-II cocycles \NoCaseChange{\protect\cite{cite405}}.
\item\relax
\flmRefsHyperref[eczindexfamilyrel]{code:mps}{Magnon code} --- Magnon codewords \NoCaseChange{\protect\cite{cite595}} are associated with 1D SPT orders \NoCaseChange{\protect\cite{cite3097,cite3098,cite3099,cite3100}}.
\item\relax
\flmRefsHyperref[eczindexfamilyrel]{code:vbs}{Valence-bond-solid (VBS) code} --- VBS codewords \NoCaseChange{\protect\cite{cite2808}} are associated with 1D SPT orders \NoCaseChange{\protect\cite{cite3097,cite3098,cite3099,cite3100}}.
\end{eczvaluelist}
\eczhbkcontributors{ \eczhuVVA }
\endeczcode

\eczcode{quantum_lego}{Tensor-network code}{~\NoCaseChange{\protect\cite{cite400,cite2858,cite2868,cite3101,cite3102}}}
\codefieldsection{Alternative Names}
\begin{eczvaluelist}
\item\relax Quantum Lego code
\end{eczvaluelist}
\eczhIndexCodeAliasName{quantum_lego}{Quantum Lego code}
\codefieldsection{Description}
Block quantum code constructed using a tensor-network-based graphical framework from atomic tensors a.k.a. \textit{quantum "Lego" blocks} \NoCaseChange{\protect\cite{cite2868}}, which can serve as encoding isometries for smaller quantum codes.
The class of codes constructed using the framework depends on the choice of atomic "Lego" blocks.

The individual "Lego" blocks and resulting quantum Lego codes can be stabilizer \NoCaseChange{\protect\cite{cite2858,cite3101}} or non-stabilizer \NoCaseChange{\protect\cite{cite2868,cite3102}}.
They need not be isometries, meaning that this class of codes generalizes \flmRefsHyperref{code:block_perfect}{planar-perfect tensor}-network codes.
However, both the logical and physical degrees of freedom must have the same local dimension.

For example, any stabilizer code can be built out of atomic blocks like the 2-site repetition code, single-site trivial stabilizer codes, and tensor products of the \(|0\rangle\) state.
Specifically, the \flmRefsHyperref{code:happy}{HaPPY} holographic code is a quantum Lego code whose atomic "Lego" block is the \flmRefsHyperref{code:stab_5_1_3}{five-qubit perfect} qubit code.

Many known codes can be created using this code's methods in order to further their understanding, including a 6-qubit implementation of the generalized Bacon-Shor code, the toric code, and the \(\llbracket 7,1,3\rrbracket \) Steane code.
Finite patches of the toric-code tensor network, equipped with boundary stopper tensors or repetition-code boundary tensors, reproduce surface-code patches and subsystem variants.
Local Hadamard modifications of alternating tensors yield XZZX surface-code variants, while re-interpreting alternating rows of surface-code physical legs as logical legs yields 2D Bacon-Shor codes and makes their relation to the quantum compass model explicit \NoCaseChange{\protect\cite{cite2868}}.
The same framework also yields flat-geometry perfect-tensor codes, holographic Reed-Muller codes with transversal non-Clifford gates, and a 3D subsystem code built from Steane tensors with localized cube-like stabilizers \NoCaseChange{\protect\cite{cite2868}}.
For example, a simple \( \llbracket 4,2,2\rrbracket  \) stabilizer code can be written as a rank 6 tensor.
Attaching two of these via gluing together one logical leg from each can produce a \(\llbracket 6,4,2\rrbracket \) stabilizer code.
Code optimization in this framework can be done using reinforcement learning \NoCaseChange{\protect\cite{cite2648,cite2649}}.

To construct a Lego code, the encoding map \(V\) for each code that is to be used in the construction is converted to a tensor by decomposing it using the formula
\flmMathEnvironment{align}{}{
V = \sum_{i_j} V_{i_1 \ldots i_{n+k}} | i_{k+1} \ldots i_{k+n} \rangle \langle i_1 \ldots i_k |~.
}
We then look at the codes graphically, treating each \(i_j\) as an edge dangling out of the tensor vertex \(V_{i_1 \ldots i_{n+k}}\). These edges are either connected to another tensor vertex's edges or left dangling. If the block codes are stabilizer, then each local tensor has unitary product stabilizers (UPS). The goal is to push each UPS through the tensor network until each dangling edge has only trivial support. Otherwise, a matching value is pushed through the edge and the process is repeated on the next tensor. If a UPS can be pushed through the whole network, then a UPS for the larger network has been found. The dangling legs (edges) and UPS of the whole network can then be converted to physical/logical elements and stabilizers/logical operators for a new quantum code.

\codefieldsection{Protection}
Stabilizer code distance can be calculated by tensor contraction \NoCaseChange{\protect\cite{cite3101}}, which can be optimized \NoCaseChange{\protect\cite{cite3103}}.
\flmRefsHyperref{ref672}{Quantum weight enumerators} are related to tensor-network structures like correlation tensor norms \NoCaseChange{\protect\cite{cite3104,cite3105}}.

\codefieldsection{Encoding}
\begin{eczvaluelist}
\item\relax Unitary-circuit encoding exists for a restricted class of tensor networks contractible via isometries \NoCaseChange{\protect\cite{cite2868}}.
\end{eczvaluelist}
\codefieldsection{Transversal and Permutation-Based Gates}
\begin{eczvaluelist}
\item\relax The quantum Lego framework yields an \(\llbracket 8,1,2\rrbracket \) stabilizer code that admits a transversal logical \(T\) gate that originates from that of a trivial (distance-one) \(\llbracket 7,1\rrbracket \) code. This code, in turn, is obtained from the \(\llbracket 15,1,3\rrbracket \) code \NoCaseChange{\protect\cite{cite786}}.
\end{eczvaluelist}
\codefieldsection{Decoding}
\begin{eczvaluelist}
\item\relax The decoder is created by creating a decoding quantum circuit with dangling legs replaced with input/output wires, and tensors converted to unitary gates. Maximum likelihood decoding can be used when the tensors are stabilizer codes.
\item\relax Tensor-network decoder when the tensor network is contractible via stabilizer isometries \NoCaseChange{\protect\cite{cite2858}}. Independent logical qubits can be decoded in parallel \NoCaseChange{\protect\cite{cite3106}}.
\item\relax Tensor-network-based decoder when the encoding unitary is known \NoCaseChange{\protect\cite{cite400}}.
\end{eczvaluelist}
\codefieldsection{Notes}
\begin{eczvaluelist}
\item\relax TensorNetworkCodes Julia software library \NoCaseChange{\protect\cite{cite3101}}.
\item\relax LEGO\(\_\)HQEC software tool \NoCaseChange{\protect\cite{cite3107}}.
\end{eczvaluelist}
\codefieldsection{Parent}
\begin{eczvaluelist}
\item\relax
\flmRefsHyperref[eczindexfamilyrel]{code:block_quantum}{Block quantum code}\end{eczvaluelist}
\codefieldsection{Children}
\begin{eczvaluelist}
\item\relax
\flmRefsHyperref[eczindexfamilyrel]{code:holographic_tensor}{Holographic tensor-network code} --- Quantum Lego codes whose encoders are tensor networks discretizing hyperbolic space can be thought of as holographic codes. More generally, holographic tensor-network codes are types of quantum LEGO codes made from stabilizer codes where logical and physical legs are pre-assigned and logical legs are not contracted. In other words, logical legs resulting from the conversion of codes to tensors must remain logical in the final tensor network, and the same for physical. Contracting logical legs is another word for gluing two logical legs together.
\item\relax
\flmRefsHyperref[eczindexfamilyrel]{code:block_perfect}{Planar-perfect-tensor code}\item\relax
\flmRefsHyperref[eczindexfamilyrel]{code:quantum_concatenated}{Concatenated quantum code} --- Encoders for concatenated quantum codes correspond to tree tensor networks \NoCaseChange{\protect\cite{cite400}}.
\item\relax
\flmRefsHyperref[eczindexfamilyrel]{code:qudit_stabilizer}{Modular-qudit stabilizer code} --- Modular-qudit stabilizer codes are quantum Lego codes built out of atomic blocks such as the 2-qudit repetition code, single-qudit trivial stabilizer codes, and tensor-products of the \(|0\rangle\) state \NoCaseChange{\protect\cite{cite3101}}.
\end{eczvaluelist}
\codefieldsection{Cousins}
\begin{eczvaluelist}
\item\relax
\flmRefsHyperref[eczindexfamilyrel]{code:steane}{\(\llbracket 7,1,3\rrbracket \) Steane code} --- The Steane code can be built from two \(\llbracket 4,2,2\rrbracket \) codes in the quantum Lego code framework \NoCaseChange{\protect\cite{cite2868}}.
\item\relax
\flmRefsHyperref[eczindexfamilyrel]{code:stab_6_4_2}{\(\llbracket 6,4,2\rrbracket \) error-detecting code} --- The \(\llbracket 6,4,2\rrbracket \) error-detecting code can be constructed out of two \(\llbracket 4,2,2\rrbracket \) codes in the quantum Lego code framework \NoCaseChange{\protect\cite{cite2868}}.
\item\relax
\flmRefsHyperref[eczindexfamilyrel]{code:stab_4_2_2}{\(\llbracket 4,2,2\rrbracket \) Four-qubit code} --- The Steane and \(\llbracket 6,4,2\rrbracket \) error-detecting codes can be built from two \(\llbracket 4,2,2\rrbracket \) codes in the quantum Lego code framework \NoCaseChange{\protect\cite{cite2868}}. The toric code can be constructed by arranging \(\llbracket 4,2,2\rrbracket \) tensors on a square lattice and recovering the star and plaquette operators by operator pushing \NoCaseChange{\protect\cite{cite2868}}.
\item\relax
\flmRefsHyperref[eczindexfamilyrel]{code:rotated_surface}{Rotated surface code} --- A tensor-network based modification of the rotated surface code improves performance against depolarizing noise by \(\approx 2\%\) \NoCaseChange{\protect\cite{cite3101}}.
\item\relax
\flmRefsHyperref[eczindexfamilyrel]{code:triangular_color}{Honeycomb (6.6.6) color code} --- Larger 6.6.6 color codes can be constructed by contracting legs of tensors of smaller codes \NoCaseChange{\protect\cite[{Fig. 5}]{cite3101}}.
\item\relax
\flmRefsHyperref[eczindexfamilyrel]{code:asymmetric_qecc}{Asymmetric quantum code (AQC)} --- Quantum Lego and more general tensor-network code optimization for biased noise can be done using reinforcement learning \NoCaseChange{\protect\cite{cite2648,cite2649}}.
\item\relax
\flmRefsHyperref[eczindexfamilyrel]{code:reinforcement_learning}{Reinforcement-learning quantum code} --- Quantum Lego and more general tensor-network code optimization for biased noise can be done using reinforcement learning \NoCaseChange{\protect\cite{cite2648,cite2649}}.
\item\relax
\flmRefsHyperref[eczindexfamilyrel]{code:surface}{Kitaev surface code} --- Planar surface codes arise from finite patches of the toric-code tensor network by contracting boundary legs with stopper tensors or repetition-code boundary tensors \NoCaseChange{\protect\cite{cite2868}}. The 2D Bacon-Shor code can also be obtained from a surface-code tensor network by reassigning every other row of dangling physical legs to logical legs; in this quantum-Lego picture, the gauge generators remain weight-two \(XX\) and \(ZZ\) operators and the construction makes explicit a connection to the quantum compass model \NoCaseChange{\protect\cite{cite2868}}.
\item\relax
\flmRefsHyperref[eczindexfamilyrel]{code:branching_mera}{Branching MERA code} --- Encoders for branching MERA codes are related to branching MERA tensor networks \NoCaseChange{\protect\cite{cite1534,cite400}}.
\item\relax
\flmRefsHyperref[eczindexfamilyrel]{code:quantum_polar}{Quantum polar code} --- Quantum polar encoding circuits can be viewed as branching-tree tensor networks \NoCaseChange{\protect\cite{cite400}}.
\item\relax
\flmRefsHyperref[eczindexfamilyrel]{code:xp_stabilizer}{XP stabilizer code} --- XP stabilizer codes can be understood through the Quantum Lego formalism \NoCaseChange{\protect\cite{cite786}}.
\item\relax
\flmRefsHyperref[eczindexfamilyrel]{code:stab_15_1_3}{\(\llbracket 15,1,3\rrbracket \) quantum RM code} --- The \(\llbracket 15,1,3\rrbracket \) code serves as the local tensor in a holographic quantum-Lego code on a \(\{15,4\}\) tiling that supports a transversal non-Clifford operator \NoCaseChange{\protect\cite{cite2868}}.
\item\relax
\flmRefsHyperref[eczindexfamilyrel]{code:stab_8_1_2}{\(\llbracket 8,1,2\rrbracket \) Shen-Wang-Cao code} --- The \(\llbracket 8,1,2\rrbracket \) code is an XP-regular code that can be obtained via the XP stabilizer formalism applied to the \(\llbracket 15,1,3\rrbracket \) Reed-Muller code \NoCaseChange{\protect\cite{cite786}}.
\item\relax
\flmRefsHyperref[eczindexfamilyrel]{code:quantum_convolutional}{Quantum convolutional code} --- Quantum convolutional encoding circuits can be viewed as matrix-product-state tensor networks \NoCaseChange{\protect\cite{cite400}}.
\item\relax
\flmRefsHyperref[eczindexfamilyrel]{code:quantum_parity}{Quantum parity code (QPC)} --- Encoders for a recursively concatenated QPCs are related to \textit{quantum trees} \NoCaseChange{\protect\cite{cite3108,cite3109}} and tree tensor networks \NoCaseChange{\protect\cite{cite400}}.
\item\relax
\flmRefsHyperref[eczindexfamilyrel]{code:toric}{Toric code} --- The toric code can be constructed by arranging \(\llbracket 4,2,2\rrbracket \) tensors on a square lattice and recovering the star and plaquette operators by operator pushing \NoCaseChange{\protect\cite{cite2868}}.
\item\relax
\flmRefsHyperref[eczindexfamilyrel]{code:triangle_surface}{Triangular surface code} --- Triangle surface codes can be reproduced by inserting a defect tensor and Hadamard-modified tensors into the surface-code tensor network \NoCaseChange{\protect\cite{cite2868}}.
\item\relax
\flmRefsHyperref[eczindexfamilyrel]{code:bacon_shor}{Bacon-Shor code} --- The 2D Bacon-Shor code can also be obtained from a surface-code tensor network by reassigning every other row of dangling physical legs to logical legs; in this quantum-Lego picture, the gauge generators remain weight-two \(XX\) and \(ZZ\) operators and the construction makes explicit a connection to the quantum compass model \NoCaseChange{\protect\cite{cite2868}}.
\end{eczvaluelist}
\eczhbkcontributors{ Thomas Wrona, \eczhuVVA }
\endeczcode

\eczcode{topological}{Topological code}{~\NoCaseChange{\protect\cite{cite423}}}
\codefieldsection{Description}
A code whose codewords form the ground-state or low-energy subspace of a (typically geometrically local) code Hamiltonian realizing a topological phase.
A topological phase may be \textit{bosonic} or \textit{fermionic}, i.e., constructed out of underlying subsystems whose operators commute or anti-commute with each other, respectively.
Unless otherwise noted, the phases discussed are bosonic.

\subsection{2D topological codes}

The physical Hilbert space of 2D topological codes consists of \(n\) subsystems which lie on edges, vertices, or faces of a tessellation of a 2D surface \(\Sigma^2\).
2D topological order requires weight-four (four-body) Hamiltonian terms, i.e., it cannot be stabilized via weight-two (two-body) or weight-three (three-body) terms on nearly Euclidean geometries of qubits or qutrits \NoCaseChange{\protect\cite{cite2684,cite2685,cite2686}}.

For subsystems with finite local dimension, topological phases are defined by their \textit{anyons} \NoCaseChange{\protect\cite{cite3110,cite3111,cite3112,cite3113}}, which are local bulk excitations of the code Hamiltonian defined on a lattice; see Refs. \NoCaseChange{\protect\cite{cite3114,cite3115,cite3116}} for more rigorous formulations.

Anyons are created in pairs by local operators, and two anyons lie in the same \textit{superselection sector} if a  local operator can convert one anyon into the other.
Each superselection sector is assumed to be labeled by one anyon type, and local operators cannot change superselection sectors.

Anyons can braid with themselves, with their \textit{exchange statistics} (a.k.a. \textit{topological spin}) defined by phases \(\theta(a)\in U(1)\) obtained by exchanging two anyons of each type \(a\).
They can also braid with each other, a process defined by \textit{braiding relations} \(B(a,b)\) for an anyon pair \(a,b\).
An anyon theory is called \textit{non-modular} or \textit{pre-modular} if there exists an anyon \(a\) that braids trivially with all anyons.

Anyons \(a\) and \(b\) can also fuse with each other, meaning that one considers both anyons as one anyon \(ab\) and decomposes \(ab\) into the anyons representing each superselection sector according to the anyons' \textit{fusion rules}.
For example, two anyons \(a,b\) may fuse to the trivial (i.e., vacuum) anyon \(1\), \(ab=1\), meaning that the composite excitation \(ab\) is indistinguishable from the case of no excitation.
Anyon fusion for many anyon theories is equivalent to a truncated Kronecker product of irreps of deformed Lie groups (i.e., "quantum groups"), which is a combination of the ordinary Kronecker product and a restriction to only decomposable representations \NoCaseChange{\protect\cite[{Secs. 4.4-4.5}]{cite3117}}.

The exchange statistics and fusion rules of anyons cannot be arbitrary and have to satisfy certain consistency relations.
Admissible exchange and fusion data are characterized by a \textit{unitary braided fusion category}.

Each anyon \(a\) has a \textit{quantum dimension} \(d_a\) associated with it.
The quantum dimensions add up to the \textit{total quantum dimension} \(D\),
\flmMathEnvironment{align}{}{
  \sum_{a}d_{a}^{2}=D^{2}~.
}
Quantum dimensions do not correspond to dimensions of vector spaces and may not be integer-valued.

An anyon theory that does not admit gapped boundaries (when put on a manifold with boundaries) is called \textit{chiral}; otherwise, it is \textit{non-chiral} or \textit{gapped}.
Chiral topological phases admit a nonzero value of the \textit{chiral central charge} \(c_{-}\).
A generalization \NoCaseChange{\protect\cite{cite423}} of the Gauss-Milgram sum rule for an anyon theory \(A\) admitting \(|A|\) anyon types,
\flmMathEnvironment{align}{}{
  \frac{1}{\sqrt{|A|}}\sum_{a\in A}d_{a}^{2}\theta_{a}=De^{i\frac{2\pi}{8}c_{-}}~,
}
relates the chiral central charge (modulo 8) to the exchange statistics and quantum dimensions.
Gapped anyon theories admit a Lagrangian subalgebra \NoCaseChange{\protect\cite{cite3118,cite3119,cite406}}.

The \textit{topological entanglement entropy} \NoCaseChange{\protect\cite{cite3120,cite3121}} of a 2D topological code is a measure of the entanglement between a region and its complement; this measure extracts the total quantum dimension \(D\) of the phase.
ZX calculus can be used to detect 2D topological order \NoCaseChange{\protect\cite{cite3122}}.
Other functions of code states extract the topological \(S\)-matrix \NoCaseChange{\protect\cite{cite3123,cite2518}} and the chiral central charge \(c_-\) \NoCaseChange{\protect\cite{cite3124}}.
No observable can distinguish topological order from product states \NoCaseChange{\protect\cite[{Appx. I}]{cite3125}}.

There is no 1D bosonic topological order at nonzero temperature \NoCaseChange{\protect\cite{cite3126,cite3127}}, and no commuting-projector model with nontrivial 2D topological order at nonzero temperature \NoCaseChange{\protect\cite{cite3128}} (see also Ref. \NoCaseChange{\protect\cite{cite3129}}). 
General thermal states are separable above a sufficiently high temperature \NoCaseChange{\protect\cite{cite3130}}.

\subsection{3D and higher-dimensional topological codes}

The physical Hilbert space of 3D topological codes consists of \(n\) subsystems which lie on edges, vertices, or faces of a tessellation of a 3D surface.
3D topological phases can have point-like and loop-like excitations, with the latter being created in pairs by 2D operators acting on subsystems supported on a plane or, more generally, a "membrane".

In the case when all point-like excitations satisfy bosonic braiding statistics, the topological phase can be realized by a Dijkgraaf-Witten gauge theory.
Such cases are thus characterized by the gauge theory's underlying data, a finite group and a cohomological cycle (i.e., cocycle) \NoCaseChange{\protect\cite{cite3131,cite3132}}.
This data is in one-to-one correspondence with pointed fusion two-categories \NoCaseChange{\protect\cite{cite3131}}.

Phases with fermionic point-like excitations are examples of \textit{beyond-group-cohomology phases} \NoCaseChange{\protect\cite{cite3133}}.
They have been classified \NoCaseChange{\protect\cite{cite3134}}, and some of them can be described by a two-gauge theory \NoCaseChange{\protect\cite{cite578}}.

The classification of 4D \NoCaseChange{\protect\cite{cite3135}} and higher-dimensional \NoCaseChange{\protect\cite{cite3136,cite3137,cite3138}} topological phases is ongoing.

\codefieldsection{Protection}
Geometrically local 2D commuting-projector topological code Hamiltonians satisfy the two \flmRefsHyperref{ref2675}{topological quantum order (TQO) conditions}, TQO-1 and TQO-2 \NoCaseChange{\protect\cite{cite2676,cite2677,cite2678,cite2679}}.
For geometrically local frustration-free Hamiltonians, LTQO generalizes the TQO conditions, and LTQO together with the Local-Gap condition yields stability of the spectral gap under sufficiently weak quasi-local perturbations \NoCaseChange{\protect\cite{cite2802,cite2803}}.

\begin{defterm}{TQO conditions}\label{ref3139}\label{ref2675}
The TQO-1 condition states that the distance of the ground-state-subspace code is macroscopic, i.e., grows as a positive power of the lattice size \NoCaseChange{\protect\cite{cite2677}}.
The TQO-2 condition relates the ground states of restrictions of the Hamiltonian to some geometrically local region to those of the full Hamiltonian.
Let \(\Pi_{N(X)}\) be the ground-state subspace projector of the Hamiltonian that includes all terms with at least some support on a geometrically local region \(X\), with \(N(X)\) consisting of the smallest region containing the support of all included terms.
TQO-2 states that any operator \(O_X\) that annihilates the codespace projector \(\Pi\) also has to annihilate the local projector \(\Pi_{N(X)}\),
\flmMathEnvironment{align}{}{
  O_{X}\Pi=0\quad\Rightarrow\quad O_{X}\Pi_{N(X)}=0~.
}
This condition implies that any operator supported solely on \(X\) cannot distinguish the global projector from the local one \NoCaseChange{\protect\cite{cite2676,cite2680}}.
\end{defterm}

A notion of topological order generalizing both the \flmRefsHyperref{ref3140}{cleaning lemma} and the \flmRefsHyperref{ref2675}{TQO conditions} is \textit{homogeneous topological order} \NoCaseChange{\protect\cite{cite3141}}.
Related topological order definitions include equivalence under course-graining (i.e., renormalization group) \NoCaseChange{\protect\cite{cite3142,cite3143}}.
See \NoCaseChange{\protect\cite[{Sec. 4}]{cite3141}} for a discussion.

Certain topological codes have nontrivial \flmRefsHyperref{ref2559}{codespace complexity} \NoCaseChange{\protect\cite{cite2564}}.

\codefieldsection{Rate}
The logical dimension \(K\) of 2D topological codes described by unitary modular fusion categories depends on the type of manifold \(\Sigma^2\) that is tessellated to form the many-body system.
For closed orientable manifolds \NoCaseChange{\protect\cite{cite3144,cite3145}},
\flmMathEnvironment{align}{}{
  K=\sum_{a\in A}\left(d_{a}/D\right)^{\chi(\Sigma^{2})}~,
}
and a generalization of the formula to the non-orientable case can be found in Ref. \NoCaseChange{\protect\cite{cite3146}}.

\codefieldsection{Encoding}
\begin{eczvaluelist}
\item\relax A depth of \flmRefsHyperref{ref65}{order} \(\Omega(L)\) is necessary for a unitary circuit to initialize in a 2D topologically ordered state using geometrically local gates on an \(L\times L\) lattice \NoCaseChange{\protect\cite{cite3147,cite3148}}, irrespective of whether the ground state admits Abelian or non-Abelian anyonic excitations.
However, only a finite-depth circuit and one round of measurements is required for non-Abelian topological orders with a Lagrangian subgroup \NoCaseChange{\protect\cite{cite3149}}.

\item\relax Algorithm that takes in reduced density matrices and outputs a circuit preparing the global state in polynomial time \NoCaseChange{\protect\cite{cite3150}}.
\end{eczvaluelist}
\codefieldsection{Gates}
\begin{eczvaluelist}
\item\relax Ising anyon braiding and fusion were studied in a phenomenological model that was the first to study error correction with non-Abelian anyons \NoCaseChange{\protect\cite{cite3151}}.
\item\relax Codes with non-Abelian anyons present two complications. First, their anyons are not always represented as tensor-product unitary operators and thus cannot be corrected by such operations. Second, fusing such anyons can yield more than one outcome, complicating decoding algorithms \NoCaseChange{\protect\cite{cite3152}}.
\end{eczvaluelist}
\codefieldsection{Threshold}
\begin{eczvaluelist}
\item\relax Topological-code families can be used to obtain a fault-tolerance threshold, although the numerical threshold value and overhead depend strongly on the syndrome-extraction and decoding protocol \NoCaseChange{\protect\cite[{Ch. 10}]{cite398}}.
\end{eczvaluelist}
\codefieldsection{Notes}
\begin{eczvaluelist}
\item\relax Ref. \NoCaseChange{\protect\cite{cite638,cite3153,cite2831,cite2825,cite3154,cite3155}\protect\cite[{Appx. F}]{cite537}} for introductions to topological phases.
\item\relax See \flmHref{https://anyonwiki.github.io/}{AnyonWiki} for lists of categories relevant to anyons.
\item\relax See \NoCaseChange{\protect\cite{cite2733}} for a pedagogical introduction to topological codes.
\end{eczvaluelist}
\codefieldsection{Parents}
\begin{eczvaluelist}
\item\relax
\flmRefsHyperref[eczindexfamilyrel]{code:block_quantum}{Block quantum code} --- Topological codes are block codes because an infinite family of tensor-product Hilbert spaces is required to formally define a phase of matter.
\item\relax
\flmRefsHyperref[eczindexfamilyrel]{code:hamiltonian}{Hamiltonian-based code} --- Codespace of a topological code is typically the ground-state or low-energy subspace of a geometrically local Hamiltonian admitting a topological phase.
Logical qubits can also be created via lattice defects or by appropriately scheduling measurements of gauge generators (see Floquet codes).
Geometrically local frustration-free code Hamiltonians on Euclidean manifolds are stable with respect to sufficiently weak quasi-local perturbations when they satisfy local topological quantum order together with the Local-Gap condition \NoCaseChange{\protect\cite{cite2802,cite2803}}.

\end{eczvaluelist}
\codefieldsection{Children}
\begin{eczvaluelist}
\item\relax
\flmRefsHyperref[eczindexfamilyrel]{code:yetter_gauge_theory}{Two-gauge theory code} --- Two-gauge theory codes realize lattice two-gauge theory for a finite two-group.
\item\relax
\flmRefsHyperref[eczindexfamilyrel]{code:enriched_string_net}{Multi-fusion string-net code} --- Enriched string-net codes realize 2D topological phases based on unitary multi-fusion categories.
\item\relax
\flmRefsHyperref[eczindexfamilyrel]{code:enriched_walker_wang}{\(G\)-enriched Walker-Wang model code} --- \(G\)-enriched Walker-Wang models realize 3D topological phases based on unitary \(G\)-crossed braided fusion categories.
\item\relax
\flmRefsHyperref[eczindexfamilyrel]{code:generalized_color}{Generalized 2D color code} --- A generalized color code for group \(G\) on the 4.8.8 lattice is equivalent to a \(G\) quantum double model and another \(G/[G,G]\) quantum double model defined using the Abelianization of \(G\).
\item\relax
\flmRefsHyperref[eczindexfamilyrel]{code:spt}{Symmetry-protected topological (SPT) code} --- SPT codes realize symmetry-protected topological phases.
\item\relax
\flmRefsHyperref[eczindexfamilyrel]{code:topological_abelian}{Abelian topological code}\item\relax
\flmRefsHyperref[eczindexfamilyrel]{code:nonabelian_kitaev_honeycomb}{Non-Abelian Kitaev honeycomb code} --- The Kitaev honeycomb model with a magnetic field is a qubit many-body system in the Ising-anyon phase, and the underlying code stores information in the fusion space of its non-Abelian anyonic excitations.
\end{eczvaluelist}
\codefieldsection{Cousins}
\begin{eczvaluelist}
\item\relax
\flmRefsHyperref[eczindexfamilyrel]{code:cluster_state}{Cluster-state code} --- There exist necessary and sufficient conditions for a family of cluster states to exhibit the TQO-1 property \NoCaseChange{\protect\cite{cite3156}}.
\item\relax
\flmRefsHyperref[eczindexfamilyrel]{code:string_net}{String-net code} --- String-net codes realize 2D topological phases based on unitary fusion categories. Any 2D many-body state satisfying the entanglement bootstrap axioms can be mapped into the ground-state subspace of a string-net model via a constant-depth unitary circuit \NoCaseChange{\protect\cite{cite3157}}. Different string-net models with Morita-equivalent input fusion categories and the same topological order are connected by constant-depth unitary circuits and therefore lie in the same phase \NoCaseChange{\protect\cite{cite3158,cite3159}}.
\item\relax
\flmRefsHyperref[eczindexfamilyrel]{code:translationally_invariant_stabilizer}{Lattice stabilizer code} --- Topological phases are not realizable using lattice stabilizer codes iff they have long-range magic \NoCaseChange{\protect\cite{cite2691}}.
\item\relax
\flmRefsHyperref[eczindexfamilyrel]{code:general_qldpc}{QLDPC code} --- Topological codes are not generally defined using Pauli strings or their qudit and bosonic generalizations. However, for appropriate tessellations, the codespace is the ground-state subspace of a geometrically local Hamiltonian. In this sense, topological codes are QLDPC codes. Geometrically local commuting-projector code Hamiltonians on Euclidean manifolds are stable with respect to small perturbations when they satisfy the \flmRefsHyperref{ref2675}{TQO conditions}, meaning that a notion of a phase can be defined \NoCaseChange{\protect\cite{cite2676,cite2677,cite2678,cite2679}}. This notion can be extended to semi-hyperbolic manifolds \NoCaseChange{\protect\cite{cite2680}} and non-geometrically local QLDPC codes exhibiting check soundness \NoCaseChange{\protect\cite{cite2681}} (see also \NoCaseChange{\protect\cite{cite2682}}).
\item\relax
\flmRefsHyperref[eczindexfamilyrel]{code:approximate_qecc}{Approximate quantum error-correcting code (AQECC)} --- In the case of topological codes, the Petz infidelity is related to the topological entanglement entropy \NoCaseChange{\protect\cite{cite2592}}.
\item\relax
\flmRefsHyperref[eczindexfamilyrel]{code:monitored_random_circuits}{Monitored random-circuit code} --- Topological order can be generated in 2D monitored random circuits \NoCaseChange{\protect\cite{cite2894}}.
\item\relax
\flmRefsHyperref[eczindexfamilyrel]{code:commuting_projector}{Commuting-projector Hamiltonian code} --- Geometrically local commuting-projector code Hamiltonians on Euclidean manifolds are stable with respect to small perturbations when they satisfy the \flmRefsHyperref{ref2675}{TQO conditions}, meaning that a notion of a phase can be defined \NoCaseChange{\protect\cite{cite2676,cite2677,cite2678,cite2679}}. This notion can be extended to semi-hyperbolic manifolds \NoCaseChange{\protect\cite{cite2680}} and non-geometrically local QLDPC codes exhibiting check soundness \NoCaseChange{\protect\cite{cite2681}} (see also \NoCaseChange{\protect\cite{cite2682}}). Hamiltonians admitting a Peierls condition are stable to off-diagonal perturbations \NoCaseChange{\protect\cite{cite2683}}. 2D states admitting strict area-law entanglement necessarily have commuting-projector Hamiltonians \NoCaseChange{\protect\cite{cite2690}}. 2D topological order on qubit manifolds requires weight-four (four-body) commuting-projector Hamiltonian terms, i.e., it cannot be stabilized via weight-two (two-body) or weight-three (three-body) terms on nearly Euclidean geometries of qubits or qutrits \NoCaseChange{\protect\cite{cite2684,cite2685,cite2686}}.
\item\relax
\flmRefsHyperref[eczindexfamilyrel]{code:frustration_free}{Frustration-free Hamiltonian code} --- Geometrically local frustration-free code Hamiltonians on Euclidean manifolds are stable with respect to sufficiently weak quasi-local perturbations when they satisfy local topological quantum order together with the Local-Gap condition; LTQO also implies an area law for the entanglement entropy of the ground-state subspace \NoCaseChange{\protect\cite{cite2802}}. See also \NoCaseChange{\protect\cite{cite2803}}.
\item\relax
\flmRefsHyperref[eczindexfamilyrel]{code:local_haar_random}{Local Haar-random circuit qubit code} --- Local Haar-random codewords, like topological codewords, are locally indistinguishable \NoCaseChange{\protect\cite{cite2196}}.
\item\relax
\flmRefsHyperref[eczindexfamilyrel]{code:eth}{Eigenstate thermalization hypothesis (ETH) code} --- ETH codewords, like topological codewords, are locally indistinguishable.
\item\relax
\flmRefsHyperref[eczindexfamilyrel]{code:fusion}{Fusion-based quantum computing (FBQC) code} --- Surface-code-based topological fault-tolerant protocols can be realized in FBQC, including topological features such as boundaries, defects, and twists, by modifying fusion measurements and, in some constructions, adding single-qubit measurements \NoCaseChange{\protect\cite{cite3160,cite3161}}.
\item\relax
\flmRefsHyperref[eczindexfamilyrel]{code:dhlv}{Dinur-Hsieh-Lin-Vidick (DHLV) code} --- DHLV codes are expected to realize topological quantum spin glass order \NoCaseChange{\protect\cite{cite3162}}.
\item\relax
\flmRefsHyperref[eczindexfamilyrel]{code:quantum_expander}{Quantum expander code} --- Quantum expander codes realize topological quantum spin glass order \NoCaseChange{\protect\cite{cite3162}}.
\item\relax
\flmRefsHyperref[eczindexfamilyrel]{code:fracton}{Fracton stabilizer code} --- Unlike topological phases, whose excitations can move in any direction, fracton phases are characterized by excitations whose movement is restricted. Fracton phases can be understood as topological defect networks, meaning that they can be described in the language of topological quantum field theory with defects \NoCaseChange{\protect\cite{cite3163,cite3164}}.
\item\relax
\flmRefsHyperref[eczindexfamilyrel]{code:expander_lifted_product}{Expander LP code} --- Expander lifted-product codes are expected to realize topological quantum spin glass order \NoCaseChange{\protect\cite{cite3162}}.
\end{eczvaluelist}
\eczhbkcontributors{ \eczhuVVA }
\endeczcode

\eczcode{w_state}{W-state code}{~\NoCaseChange{\protect\cite{cite2720}}}
\codefieldsection{Description}
Approximate block quantum code whose encoding resembles the structure of the
W state~\NoCaseChange{\protect\cite{cite527}}.
This code enables universal quantum computation with transversal gates.

The encoding is of a \(d_L\)-dimensional Hilbert space into \(n\) physical quantum systems, each associated with a Hilbert space
of dimension \(d_L+1\):
\flmMathEnvironment{align}{}{
  \ket\psi
  \to \frac{1}{\sqrt{n}}\bigl(\ket{\psi\perp\perp\ldots}
  + \ket{\perp\psi\perp\ldots} + \cdots
  + \ket{\perp\perp\ldots\psi}\bigr)\ ,
}
where on each physical system, \(\ket\perp\) denotes the \((d_L+1)\)-th basis state,
and \(\ket\psi\) is encoded using the first \(d_L\) basis states.

Indeed, to apply any logical unitary \(U\) it suffices to apply \(U\) on each physical system,
where the unitary is taken to act nontrivially only on the first \(d_L\) basis states
of each system.  Universal computation with transversal gates does not violate the
\flmRefsHyperref{ref721}{Eastin-Knill theorem} because this code is an approximate error-correcting
code~\NoCaseChange{\protect\cite{cite2514,cite2720}} rather than an exact error-correcting
code.
\codefieldsection{Protection}
The W state code is an approximate error-correcting code.  Intuitively, if a
subsystem is lost to the environment, the environment only gains access to
\(\ket\psi\) with probability of \flmRefsHyperref{ref65}{order} \(O(1/n)\). Under a single located erasure,
the worst-case entanglement infidelity of the W state code can be upper bounded as
\flmMathEnvironment{align}{}{
  \epsilon_{\mathrm{worst}} \leq \frac{\sqrt{2} + d_L}{\sqrt{n}}\ .
}

In contrast to the \flmRefsCref{code:eth}, the W state code does not saturate the scaling
\(1/n\) in worst-case entanglement infidelity which is known to be
optimal for covariant approximate error-correcting codes~\NoCaseChange{\protect\cite{cite2720}}.
\codefieldsection{Encoding}
\begin{eczvaluelist}
\item\relax There are complexity-theoretic bounds on \(W\)-state preparation \NoCaseChange{\protect\cite{cite2565}}.
\end{eczvaluelist}
\codefieldsection{Transversal and Permutation-Based Gates}
\begin{eczvaluelist}
\item\relax All logical gates can be implemented transversally. The logical unitary \(U_L\) can be performed with the physical unitary \(U_L\otimes U_L\otimes\cdots\otimes U_L\), where on the physical space \(U_L\) is taken to act trivially on \(\ket\perp\), i.e., \( U_L\ket\perp = \ket\perp\).
\end{eczvaluelist}
\codefieldsection{Parents}
\begin{eczvaluelist}
\item\relax
\flmRefsHyperref[eczindexfamilyrel]{code:covariant}{Covariant block quantum code} --- The W-state code approximately protects against a single erasure while allowing for a universal transversal set of gates.
\item\relax
\flmRefsHyperref[eczindexfamilyrel]{code:permutation_invariant}{Permutation-invariant (PI) code}\item\relax
\flmRefsHyperref[eczindexfamilyrel]{code:approximate_qecc}{Approximate quantum error-correcting code (AQECC)} --- The W-state code approximately protects against a single erasure while allowing for a universal transversal set of gates.
\end{eczvaluelist}
\codefieldsection{Cousin}
\begin{eczvaluelist}
\item\relax
\flmRefsHyperref[eczindexfamilyrel]{code:eth}{Eigenstate thermalization hypothesis (ETH) code} --- The W-state is not a unique ground state of any local Hamiltonian \NoCaseChange{\protect\cite{cite3165}}.
\end{eczvaluelist}
\eczhbkcontributors{ \eczhuPhF, \eczhuVVA }
\endeczcode

\onecolumngrid
\clearpage

\section{Qubit Kingdom}

\begin{eczEpigraph}
\begin{quote}
\flmQuoteSetup{quote}%
The view of quantum computers as able to “try all possible solutions in parallel,” and then instantly choose the correct one, is fundamentally mistaken.
\flmQuoteAttributed{Scott Aaronson}
\end{quote}
\end{eczEpigraph}

\twocolumngrid

\eczcode{qubit_10_24_3}{\(\llparenthesis 10,24,3\rrparenthesis \) qubit code}{~\NoCaseChange{\protect\cite{cite452}}}
\eczhIndexCodeAliasName{qubit_10_24_3}{qubit code}
\codefieldsection{Description}
Ten-qubit nonadditive CWS code saturating the linear-programming bound for one-error-correcting codes on ten qubits \NoCaseChange{\protect\cite{cite452}}.
It was constructed using the coding-clique graph-theoretic framework introduced in \NoCaseChange{\protect\cite{cite452}}, which unifies additive and nonadditive code constructions.

The \(\llparenthesis 10,24,3\rrparenthesis \) qubit code can be combined to form an infinite family of distance-three qubit codes whose logical dimension is \(50\%\) larger than that of the optimal stabilizer code \NoCaseChange{\protect\cite{cite3046}}.

\codefieldsection{Parents}
\begin{eczvaluelist}
\item\relax
\flmRefsHyperref[eczindexfamilyrel]{code:cws}{Codeword stabilized (CWS) code} --- The \(\llparenthesis 10,24,3\rrparenthesis \) qubit code is a CWS code \NoCaseChange{\protect\cite{cite3166}}.
\item\relax
\flmRefsHyperref[eczindexfamilyrel]{code:small_distance_quantum}{Small-distance block quantum code} --- The \(\llparenthesis 10,24,3\rrparenthesis \) qubit code can be combined to form an infinite family of distance-three qubit codes whose logical dimension is \(50\%\) larger than that of the optimal stabilizer code \NoCaseChange{\protect\cite{cite3046}}.
\end{eczvaluelist}
\eczhbkcontributors{ \eczhuVVA }
\endeczcode

\eczcode{quantum_goethals_preparata}{\(\llparenthesis 2^m,2^{2^m−5m+1},8\rrparenthesis \) Goethals-Preparata code}{~\NoCaseChange{\protect\cite{cite1369,cite1370}}}
\eczhIndexCodeAliasName{quantum_goethals_preparata}{Goethals-Preparata code}
\codefieldsection{Description}
Member of a family of nonadditive \(\llparenthesis 2^m,2^{2^m−5m+1},8\rrparenthesis \) CSS-like union stabilizer codes constructed using the classical Goethals and Preparata codes.

They can be viewed as union stabilizer codes constructed from the codespace of a \(\llbracket 2^m,2^m-7m+3,8\rrbracket \) code, itself obtained via \flmRefsCref{ref863}, together with the coset representatives used to obtain the Goethals and Preparata codes \NoCaseChange{\protect\cite[{Thm. 10.3}]{cite3167}}.

The Goethals and Preparata codes can each be used to obtain families of union stabilizer codes with distance 8 and 6, respectively \NoCaseChange{\protect\cite{cite1369}}.
A construction using the \(\mathbb{Z}_4\) versions of these codes and the \flmTerm{term}{ref81}{}{Gray map} yields qubit code families with similar parameters \NoCaseChange{\protect\cite{cite1371}}.

\codefieldsection{Parent}
\begin{eczvaluelist}
\item\relax
\flmRefsHyperref[eczindexfamilyrel]{code:non_stabilizer}{Union stabilizer (USt) code}\end{eczvaluelist}
\codefieldsection{Cousins}
\begin{eczvaluelist}
\item\relax
\flmRefsHyperref[eczindexfamilyrel]{code:goethals}{Goethals code} --- The \(\llparenthesis 2^m,2^{2^m−5m+1},8\rrparenthesis \) Goethals-Preparata code is constructed using the classical Goethals and Preparata codes \NoCaseChange{\protect\cite{cite1369,cite1370}}. A construction using the \(\mathbb{Z}_4\) versions of the Goethals and Preparata codes and the \flmTerm{term}{ref81}{}{Gray map} yields qubit code families with similar parameters \NoCaseChange{\protect\cite{cite1371}}.
\item\relax
\flmRefsHyperref[eczindexfamilyrel]{code:preparata}{Preparata code} --- The \(\llparenthesis 2^m,2^{2^m−5m+1},8\rrparenthesis \) Goethals-Preparata code is constructed using the classical Goethals and Preparata codes \NoCaseChange{\protect\cite{cite1369,cite1370}}. A construction using the \(\mathbb{Z}_4\) versions of the Goethals and Preparata codes and the \flmTerm{term}{ref81}{}{Gray map} yields qubit code families with similar parameters \NoCaseChange{\protect\cite{cite1371}}.
\item\relax
\flmRefsHyperref[eczindexfamilyrel]{code:gray}{Gray code} --- A construction using the \(\mathbb{Z}_4\) versions of the Goethals and Preparata codes and the \flmTerm{term}{ref81}{}{Gray map} yields qubit code families with similar parameters as the \(\llparenthesis 2^m,2^{2^m−5m+1},8\rrparenthesis \) Goethals-Preparata code \NoCaseChange{\protect\cite{cite1371}}.
\end{eczvaluelist}
\eczhbkcontributors{ \eczhuVVA }
\endeczcode

\eczcode{rains}{\(\llparenthesis 2m+1,3 \times 2^{2m-3},2\rrparenthesis \) Rains code}{~\NoCaseChange{\protect\cite{cite3168}}}
\eczhIndexCodeAliasName{rains}{Rains code}
\codefieldsection{Description}
Member of a family of \flmRefsHyperref{ref672}{pure} odd-length distance-two CWS codes with parameters \(\llparenthesis 2m+1,3 \times 2^{2m-3},2\rrparenthesis \) for all \(m \geq 2\), constructed recursively from the \(\llparenthesis 5,6,2\rrparenthesis \) code \NoCaseChange{\protect\cite{cite854,cite855}\protect\cite[{Lem. 5 and Thm. 4}]{cite446}\protect\cite[{Exam. 8}]{cite853}}.

\codefieldsection{Parents}
\begin{eczvaluelist}
\item\relax
\flmRefsHyperref[eczindexfamilyrel]{code:cws}{Codeword stabilized (CWS) code} --- The \(\llparenthesis 2m+1,3 \times 2^{2m-3},2\rrparenthesis \) qubit code family is a CWS family whose graph state is the union of the ring and Bell-pair graphs \NoCaseChange{\protect\cite{cite852,cite3166}}.
\item\relax
\flmRefsHyperref[eczindexfamilyrel]{code:small_distance_quantum}{Small-distance block quantum code}\end{eczvaluelist}
\codefieldsection{Child}
\begin{eczvaluelist}
\item\relax
\flmRefsHyperref[eczindexfamilyrel]{code:qubit_5_6_2}{\(\llparenthesis 5,6,2\rrparenthesis \) qubit code} --- The \(\llparenthesis 5,6,2\rrparenthesis \) code is the smallest nontrivial Rains code \NoCaseChange{\protect\cite{cite446}} (see also \NoCaseChange{\protect\cite{cite3044}\protect\cite[{Exam. 8}]{cite853}}).
\end{eczvaluelist}
\codefieldsection{Cousin}
\begin{eczvaluelist}
\item\relax
\flmRefsHyperref[eczindexfamilyrel]{code:ssw}{Smolin-Smith-Wehner (SSW) code} --- The SSW code outperforms the Rains codes in terms of code parameters at odd \(n > 11\) \NoCaseChange{\protect\cite{cite852,cite3166}}.
\end{eczvaluelist}
\eczhbkcontributors{ \eczhuVVA }
\endeczcode

\eczcode{four_qubit_permutation_invariant}{\(\llparenthesis 4,2,2\rrparenthesis \) Four-qubit single-deletion code}{~\NoCaseChange{\protect\cite{cite2655,cite2656}}}
\eczhIndexCodeAliasName{four_qubit_permutation_invariant}{Four-qubit single-deletion code}
\codefieldsection{Description}
Four-qubit PI code that is the smallest qubit code to correct one deletion error.

In terms of \flmRefsHyperref{ref526}{Dicke states}, a basis of logical codewords is
\flmMathEnvironment{align}{}{
\begin{split}
  |0_{L}\rangle&=\frac{1}{\sqrt{2}}\left(|D_{0}^{4}\rangle+|D_{4}^{4}\rangle\right)\\
  |1_{L}\rangle&=|D_{2}^{4}\rangle~.
\end{split}
}

\codefieldsection{Protection}
The smallest qubit PI code to correct one deletion error.

\codefieldsection{Parents}
\begin{eczvaluelist}
\item\relax
\flmRefsHyperref[eczindexfamilyrel]{code:gnu_permutation_invariant}{GNU PI code} --- The four-qubit single-deletion code is a GNU code for \(g=m=2\) \NoCaseChange{\protect\cite{cite2657}}.
\item\relax
\flmRefsHyperref[eczindexfamilyrel]{code:unentangled_permutation_invariant}{\(\llparenthesis n,2,2\rrparenthesis \) Bravyi-Lee-Li-Yoshida PI code} --- The Bravyi-Lee-Li-Yoshida code reduces to the four-qubit single-deletion code for \(n=4\).
\end{eczvaluelist}
\codefieldsection{Cousins}
\begin{eczvaluelist}
\item\relax
\flmRefsHyperref[eczindexfamilyrel]{code:binomial}{Binomial code} --- The four-qubit single-deletion code can be obtained from the "0-2-4" single-mode binomial code by substituting Fock states with \flmRefsHyperref{ref526}{Dicke states}.
\item\relax
\flmRefsHyperref[eczindexfamilyrel]{code:stab_4_2_2}{\(\llbracket 4,2,2\rrbracket \) Four-qubit code} --- Projecting the four-qubit code into the PI subspace yields the four-qubit single-deletion code. A basis of codewords for the four-qubit single-deletion code consists of the \(|\overline{00}\rangle\) and \(|\overline{01}\rangle+|\overline{10}\rangle+|\overline{11}\rangle\) states of the four-qubit code.
\item\relax
\flmRefsHyperref[eczindexfamilyrel]{code:combinatorial_permutation_invariant}{Combinatorial PI code} --- The combinatorial PI code \(Q_{1,1,1,-}\) is another example of a four-qubit code correcting a single deletion error \NoCaseChange{\protect\cite[{Sec. 5.1}]{cite3169}}.
\item\relax
\flmRefsHyperref[eczindexfamilyrel]{code:three_qutrit_permutation_invariant}{\(\llparenthesis 3,2,2\rrparenthesis _3\) Three-qutrit single-deletion code} --- The four-qubit (three-qutrit) single-deletion code is the smallest PI qubit (qutrit) code to correct one deletion error.
\end{eczvaluelist}
\eczhbkcontributors{ \eczhuVVA }
\endeczcode

\eczcode{qubit_5_6_2}{\(\llparenthesis 5,6,2\rrparenthesis \) qubit code}{~\NoCaseChange{\protect\cite{cite3168}}}
\eczhIndexCodeAliasName{qubit_5_6_2}{qubit code}
\codefieldsection{Description}
Five-qubit cyclic CWS code detecting a single-qubit error.
This code has a logical subspace whose dimension is larger than that of the \(\llbracket 5,2,2\rrbracket \) code, the best five-qubit stabilizer code with the same distance \NoCaseChange{\protect\cite{cite452}}.

Its codeword stabilizer consists of all cyclic shifts of \(ZXZII\).
A standard-form CWS presentation uses the five-cycle graph together with the classical codewords \(00000\), \(11010\), \(01101\), \(10110\), \(01011\), and \(10101\) \NoCaseChange{\protect\cite{cite852,cite452}}.
Its automorphism group is of size 3840 and given in Ref. \NoCaseChange{\protect\cite{cite3168}} (see also \NoCaseChange{\protect\cite[{Corr. 18}]{cite446}}).

\codefieldsection{Parents}
\begin{eczvaluelist}
\item\relax
\flmRefsHyperref[eczindexfamilyrel]{code:rains}{\(\llparenthesis 2m+1,3 \times 2^{2m-3},2\rrparenthesis \) Rains code} --- The \(\llparenthesis 5,6,2\rrparenthesis \) code is the smallest nontrivial Rains code \NoCaseChange{\protect\cite{cite446}} (see also \NoCaseChange{\protect\cite{cite3044}\protect\cite[{Exam. 8}]{cite853}}).
\item\relax
\flmRefsHyperref[eczindexfamilyrel]{code:arvind}{\(\llparenthesis n,1+n(q-1),2\rrparenthesis _q\) union stabilizer code} --- The \(\llparenthesis 5,6,2\rrparenthesis \) code is the \(\llparenthesis n,1+n(q-1),2\rrparenthesis _q\) union stabilizer code for \(n=5\) and \(q=2\) \NoCaseChange{\protect\cite{cite3170}}.
\item\relax
\flmRefsHyperref[eczindexfamilyrel]{code:quantum_cyclic}{Cyclic quantum code}\end{eczvaluelist}
\codefieldsection{Cousin}
\begin{eczvaluelist}
\item\relax
\flmRefsHyperref[eczindexfamilyrel]{code:stab_4_2_2}{\(\llbracket 4,2,2\rrbracket \) Four-qubit code} --- Tracing out any one qubit of the \(\llparenthesis 5,6,2\rrparenthesis \) code projector yields a \(\llparenthesis 4,4,2\rrparenthesis \) code; for this code, all five such partial traces are additive and therefore locally equivalent to the \(\llbracket 4,2,2\rrbracket \) code \NoCaseChange{\protect\cite[{Thm. 8 and Corr. 18}]{cite446}}.
\end{eczvaluelist}
\eczhbkcontributors{ \eczhuVVA }
\endeczcode

\eczcode{qubit_6_2_3}{\(\llparenthesis 6,2,3\rrparenthesis \) transversal-\(\mathbb{Z}_{10}\) code}{~\NoCaseChange{\protect\cite{cite528}}}
\eczhIndexCodeAliasName{qubit_6_2_3}{transversal-\(\mathbb{Z}_{10}\) code}
\codefieldsection{Description}
Six-qubit code that realizes gates from the group \(\mathbb{Z}_{10}\) transversally.
This is the smallest known distance-three code supporting a transversal gate outside of the \flmRefsHyperref{ref409}{Clifford group}.
See Ref. \NoCaseChange{\protect\cite{cite528}} for the codewords.

\codefieldsection{Transversal and Permutation-Based Gates}
\begin{eczvaluelist}
\item\relax The \(\mathbb{Z}_{10}\) group can be realized transversally \NoCaseChange{\protect\cite{cite528}}. This is the smallest known distance-three code supporting a transversal gate outside of the \flmRefsHyperref{ref409}{Clifford group}.
\end{eczvaluelist}
\codefieldsection{Parents}
\begin{eczvaluelist}
\item\relax
\flmRefsHyperref[eczindexfamilyrel]{code:qubits_into_qubits}{Qubit code}\item\relax
\flmRefsHyperref[eczindexfamilyrel]{code:small_distance_quantum}{Small-distance block quantum code}\end{eczvaluelist}
\eczhbkcontributors{ \eczhuVVA }
\endeczcode

\eczcode{icosahedral_permutation_invariant}{\(\llparenthesis 7,2,3\rrparenthesis \) Pollatsek-Ruskai code}{~\NoCaseChange{\protect\cite{cite2940,cite646,cite647}}}
\codefieldsection{Alternative Names}
\begin{eczvaluelist}
\item\relax \(\llparenthesis 7,2,3\rrparenthesis \) icosahedral code
\item\relax \(\llparenthesis 7,2,3\rrparenthesis \) Kubischta-Teixeira code
\end{eczvaluelist}
\eczhIndexCodeAliasName{icosahedral_permutation_invariant}{Pollatsek-Ruskai code}
\eczhIndexCodeAliasName{icosahedral_permutation_invariant}{\(\llparenthesis 7,2,3\rrparenthesis \) icosahedral code}
\eczhIndexCodeAliasName{icosahedral_permutation_invariant}{\(\llparenthesis 7,2,3\rrparenthesis \) Kubischta-Teixeira code}
\codefieldsection{Description}
Seven-qubit PI code that realizes gates from the binary icosahedral group transversally.
Can also be interpreted as a spin-\(7/2\) single-spin code.
The codespace projection is a projection onto an irrep of the binary icosahedral group \(2I\).
See Ref. \NoCaseChange{\protect\cite{cite528}} for other non-PI codes realizing \(2I\) gates transversally.

In terms of \flmRefsHyperref{ref526}{Dicke states}, the unnormalized logical states of one version \NoCaseChange{\protect\cite{cite647}} of this code are
\flmMathEnvironment{align}{}{
  \begin{split}
    |0_{L}\rangle&\propto \sqrt{15}|D_{0}^{7}\rangle+\sqrt{21}|D_{4}^{7}\rangle\\&\quad+\sqrt{7}\;|D_{2}^{7}\rangle-\sqrt{21}|D_{6}^{7}\rangle\,,\\|1_{L}\rangle&\propto X^{\otimes7}|0_{L}\rangle\,.
  \end{split}
}
Another version \NoCaseChange{\protect\cite{cite646}} of this code, converted into Dicke states, has unnormalized logical states
\flmMathEnvironment{align}{}{
  \begin{split}
    |0_{L}\rangle&\propto\sqrt{3}|D_{0}^{7}\rangle+\sqrt{7}|D_{5}^{7}\rangle\\
    |1_{L}\rangle&\propto\sqrt{7}|D_{2}^{7}\rangle-\sqrt{3}|D_{7}^{7}\rangle\,.
  \end{split}
}

\codefieldsection{Transversal and Permutation-Based Gates}
\begin{eczvaluelist}
\item\relax Binary icosahedral group \(2I\) gates can be realized transversally \NoCaseChange{\protect\cite{cite647}}. See Ref. \NoCaseChange{\protect\cite{cite528}} for other non-PI codes realizing \(2I\) gates transversally.
\end{eczvaluelist}
\codefieldsection{Parents}
\begin{eczvaluelist}
\item\relax
\flmRefsHyperref[eczindexfamilyrel]{code:combinatorial_permutation_invariant}{Combinatorial PI code} --- The Pollatsek-Ruskai code is equivalent to the \(Q_{2,1,2,-}\) combinatorial PI code \NoCaseChange{\protect\cite[{Sec. 5.2}]{cite3169}}. It is a seven-qubit PI code that realizes gates from the binary icosahedral group transversally.
\item\relax
\flmRefsHyperref[eczindexfamilyrel]{code:t_group}{Twisted \(1\)-group code} --- The \(\llparenthesis 7,2,3\rrparenthesis \) Pollatsek-Ruskai code admits a transversal representation of the twisted \(1\)-group \(2I\) \NoCaseChange{\protect\cite{cite2199}}.
\end{eczvaluelist}
\codefieldsection{Cousins}
\begin{eczvaluelist}
\item\relax
\flmRefsHyperref[eczindexfamilyrel]{code:steane}{\(\llbracket 7,1,3\rrbracket \) Steane code} --- The Pollatsek-Ruskai code can be continuously deformed to the Steane code \NoCaseChange{\protect\cite{cite2555}}.
\item\relax
\flmRefsHyperref[eczindexfamilyrel]{code:icosahedron}{Icosahedron code} --- Binary icosahedral group \(2I\) gates can be realized transversally in the Pollatsek-Ruskai code \NoCaseChange{\protect\cite{cite647}}.
\item\relax
\flmRefsHyperref[eczindexfamilyrel]{code:icosahedral_spin}{Icosahedral spin code} --- The \(\llparenthesis 7,2,3\rrparenthesis \) Pollatsek-Ruskai code maps to the icosahedral spin code via the \flmRefsHyperref{ref526}{Dicke state mapping} \NoCaseChange{\protect\cite{cite647}}.
\item\relax
\flmRefsHyperref[eczindexfamilyrel]{code:icosahedral_fock}{Icosahedral Fock-state code} --- The \(\llparenthesis 7,2,3\rrparenthesis \) Pollatsek-Ruskai code maps to the icosahedral Fock-state code via the \flmRefsHyperref{ref499}{simplex mapping} \NoCaseChange{\protect\cite{cite500}}.
\end{eczvaluelist}
\eczhbkcontributors{ \eczhuVVA }
\endeczcode

\eczcode{qubit_8_4_2}{\(\llparenthesis 8,16,2\rrparenthesis \) \(PG(3,2)\) code}{~\NoCaseChange{\protect\cite{cite723}}}
\eczhIndexCodeAliasName{qubit_8_4_2}{\(PG(3,2)\) code}
\codefieldsection{Description}
Eight-qubit code encoding four logical qubits whose logical basis consists of a GHZ state together with fifteen states built from the incidence geometry of the projective space \(PG(3,2)\).

\codefieldsection{Protection}
Distance two, and this is optimal because no \(\llparenthesis 8,16,3\rrparenthesis \) code exists \NoCaseChange{\protect\cite{cite723}}.
No Pauli stabilizer subsystem phantom code of type \(\llbracket 8,4,r,d\geq2\rrbracket \) exists, so this exceptional \(k=4\) phantom code is necessarily nonstabilizer \NoCaseChange{\protect\cite{cite723}}.

\codefieldsection{Transversal and Permutation-Based Gates}
\begin{eczvaluelist}
\item\relax Even physical-qubit permutations act as \(GL(4,\mathbb{F}_2)\) on the logical basis, and odd permutations extend the permutation automorphism group to the full symmetric group \(S_8\) \NoCaseChange{\protect\cite{cite723}}.
\item\relax \(T^{\otimes 8}\) is a transversal non-Clifford gate implementing \(2\ket{\overline{0}}\bra{\overline{0}}-I\) on the logical subspace \NoCaseChange{\protect\cite{cite723}}.
\item\relax A specific odd permutation implements a non-Clifford logical involution, and the full permutation automorphism group is \(S_8\) \NoCaseChange{\protect\cite{cite723}}.
\end{eczvaluelist}
\codefieldsection{Parents}
\begin{eczvaluelist}
\item\relax
\flmRefsHyperref[eczindexfamilyrel]{code:qubits_into_qubits}{Qubit code}\item\relax
\flmRefsHyperref[eczindexfamilyrel]{code:small_distance_quantum}{Small-distance block quantum code}\end{eczvaluelist}
\codefieldsection{Cousins}
\begin{eczvaluelist}
\item\relax
\flmRefsHyperref[eczindexfamilyrel]{code:phantom}{Phantom code} --- This is the exceptional nonstabilizer \(k=4\) qubit phantom code of minimal length eight that violates the generic bound \(n\geq 2^k-1\) \NoCaseChange{\protect\cite{cite723}}.
\item\relax
\flmRefsHyperref[eczindexfamilyrel]{code:self_complementary}{Self-complementary qubit code} --- The logical basis of the \(\llparenthesis 8,16,2\rrparenthesis \) \(PG(3,2)\) code contains a GHZ state and linear combinations of self-complementary states \NoCaseChange{\protect\cite{cite723}}.
\end{eczvaluelist}
\eczhbkcontributors{ \eczhuVVA }
\endeczcode

\eczcode{qubit_8_1_3}{\(\llparenthesis 8,2,3\rrparenthesis \) Plenio-Vedral-Knight CE code}{~\NoCaseChange{\protect\cite{cite2706}}}
\eczhIndexCodeAliasName{qubit_8_1_3}{Plenio-Vedral-Knight CE code}
\codefieldsection{Description}
An eight-qubit single-error-correcting code that is the first CE code.
Each logical state is a superposition of computational basis states with four excitations.

Admits codewords of the form
\flmMathEnvironment{align}{}{
\begin{split}
  |\overline{0}\rangle&=(|00001111\rangle+|11101000\rangle−|10010110\rangle−|01110001\rangle\\&+|11010100\rangle+|00110011\rangle+|01001101\rangle+|10101010\rangle)/\sqrt{8}\\
  |\overline{1}\rangle&=X^{\otimes8}|\overline{0}\rangle~.
\end{split}
}

\codefieldsection{Parents}
\begin{eczvaluelist}
\item\relax
\flmRefsHyperref[eczindexfamilyrel]{code:qubits_into_qubits}{Qubit code}\item\relax
\flmRefsHyperref[eczindexfamilyrel]{code:constant_excitation}{Constant-excitation (CE) code}\item\relax
\flmRefsHyperref[eczindexfamilyrel]{code:small_distance_quantum}{Small-distance block quantum code}\end{eczvaluelist}
\eczhbkcontributors{ \eczhuVVA }
\endeczcode

\eczcode{qubit_9_12_3}{\(\llparenthesis 9,12,3\rrparenthesis \) qubit code}{~\NoCaseChange{\protect\cite{cite3171}}}
\eczhIndexCodeAliasName{qubit_9_12_3}{qubit code}
\codefieldsection{Description}
Nine-qubit cyclic CWS code correcting a single-qubit error.
This code has a logical subspace whose dimension is larger than that of the \(\llbracket 9,3,3\rrbracket \) code, the best nine-qubit stabilizer code with the same distance \NoCaseChange{\protect\cite{cite449}}.

Its codeword stabilizer consists of all cyclic shifts of \(ZXZIIIIII\).
In the coding-clique framework, it is realized by the nine-vertex loop graph \(L_9\); within a graph search \NoCaseChange{\protect\cite{cite452}}, the realization \((L_9,12,3)\) is unique and there is no \((G,13,3)\) code on any nine-vertex graph.

The \(\llparenthesis 9,12,3\rrparenthesis \) qubit code can be combined to form an infinite family of distance-three qubit codes whose logical dimension is \(50\%\) larger than that of the optimal stabilizer code \NoCaseChange{\protect\cite{cite3046}}.

\codefieldsection{Decoding}
\begin{eczvaluelist}
\item\relax Fault-tolerant scheme that converts the required POVM into 10 binary measurements whose redundancy is guaranteed by a classical code \NoCaseChange{\protect\cite{cite3172}}.
\end{eczvaluelist}
\codefieldsection{Parents}
\begin{eczvaluelist}
\item\relax
\flmRefsHyperref[eczindexfamilyrel]{code:cws}{Codeword stabilized (CWS) code} --- The \(\llparenthesis 9,12,3\rrparenthesis \) qubit code is a cyclic CWS code \NoCaseChange{\protect\cite{cite852,cite3166}}.
\item\relax
\flmRefsHyperref[eczindexfamilyrel]{code:quantum_cyclic}{Cyclic quantum code}\item\relax
\flmRefsHyperref[eczindexfamilyrel]{code:small_distance_quantum}{Small-distance block quantum code} --- The \(\llparenthesis 9,12,3\rrparenthesis \) qubit code can be combined to form an infinite family of distance-three qubit codes whose logical dimension is \(50\%\) larger than that of the optimal stabilizer code \NoCaseChange{\protect\cite{cite3046}}.
\end{eczvaluelist}
\eczhbkcontributors{ \eczhuVVA }
\endeczcode

\eczcode{ruskai}{\(\llparenthesis 9,2,3\rrparenthesis \) Ruskai code}{~\NoCaseChange{\protect\cite{cite3173}}}
\eczhIndexCodeAliasName{ruskai}{Ruskai code}
\codefieldsection{Description}
Nine-qubit PI code that protects against single-qubit errors as well as two-qubit errors arising from exchange processes.

In terms of \flmRefsHyperref{ref526}{Dicke states}, the codewords are
\flmMathEnvironment{align}{}{
  \begin{split}
    |0_{L}\rangle&\propto|D_{0}^{9}\rangle+\sqrt{3}|D_{6}^{9}\rangle\\
    |1_{L}\rangle&\propto\sqrt{3}|D_{3}^{9}\rangle+|D_{9}^{9}\rangle~.
  \end{split}
}

\codefieldsection{Protection}
Protects against all single-qubit errors as well as two-qubit errors arising from exchange processes.

\codefieldsection{Parents}
\begin{eczvaluelist}
\item\relax
\flmRefsHyperref[eczindexfamilyrel]{code:gnu_permutation_invariant}{GNU PI code} --- The \(\llparenthesis 9,2,3\rrparenthesis \) Ruskai code is a GNU PI code \NoCaseChange{\protect\cite{cite2944}}.
\item\relax
\flmRefsHyperref[eczindexfamilyrel]{code:small_distance_quantum}{Small-distance block quantum code}\end{eczvaluelist}
\codefieldsection{Cousin}
\begin{eczvaluelist}
\item\relax
\flmRefsHyperref[eczindexfamilyrel]{code:shor_nine}{\(\llbracket 9,1,3\rrbracket \) Shor code} --- The \(\llparenthesis 9,2,3\rrparenthesis \) Ruskai code results from projecting the Shor code into the PI qubit subspace \NoCaseChange{\protect\cite{cite3173}}.
\end{eczvaluelist}
\eczhbkcontributors{ \eczhuVVA }
\endeczcode

\eczcode{unentangled_permutation_invariant}{\(\llparenthesis n,2,2\rrparenthesis \) Bravyi-Lee-Li-Yoshida PI code}{~\NoCaseChange{\protect\cite{cite529}}}
\eczhIndexCodeAliasName{unentangled_permutation_invariant}{Bravyi-Lee-Li-Yoshida PI code}
\codefieldsection{Description}
PI distance-two code on \(n\geq4\) qubits whose degree of entanglement vanishes asymptotically with \(n\) \NoCaseChange{\protect\cite[{Appx. D}]{cite529}} (cf. \NoCaseChange{\protect\cite{cite530}}).

In terms of \flmRefsHyperref{ref526}{Dicke states}, the codewords are
\flmMathEnvironment{align}{}{
  \begin{split}
    |0_{L}\rangle&=\sqrt{1-\frac{2}{n}}|D_{0}^{n}\rangle+\sqrt{\frac{2}{n}}|D_{n}^{n}\rangle\\
    |1_{L}\rangle&=|D_{2}^{n}\rangle~.
  \end{split}
}

\codefieldsection{Parents}
\begin{eczvaluelist}
\item\relax
\flmRefsHyperref[eczindexfamilyrel]{code:qubit_permutation_invariant}{PI qubit code}\item\relax
\flmRefsHyperref[eczindexfamilyrel]{code:movassagh_ouyang}{Movassagh-Ouyang Hamiltonian code} --- The \(\llparenthesis n,2,2\rrparenthesis \) PI code is a Movassagh-Ouyang Hamiltonian code constructed from a binary code consisting of all codewords of weight 0, 2, or \(n\) \NoCaseChange{\protect\cite[{Appx. D}]{cite529}}.
\item\relax
\flmRefsHyperref[eczindexfamilyrel]{code:small_distance_quantum}{Small-distance block quantum code}\end{eczvaluelist}
\codefieldsection{Child}
\begin{eczvaluelist}
\item\relax
\flmRefsHyperref[eczindexfamilyrel]{code:four_qubit_permutation_invariant}{\(\llparenthesis 4,2,2\rrparenthesis \) Four-qubit single-deletion code} --- The Bravyi-Lee-Li-Yoshida code reduces to the four-qubit single-deletion code for \(n=4\).
\end{eczvaluelist}
\codefieldsection{Cousin}
\begin{eczvaluelist}
\item\relax
\flmRefsHyperref[eczindexfamilyrel]{code:qubit_concatenated}{Concatenated qubit code} --- The Bravyi-Lee-Li-Yoshida PI code can be concatenated to yield codes that have higher distance and that admit codewords with vanishing entanglement \NoCaseChange{\protect\cite[{Appx. D}]{cite529}} (cf. \NoCaseChange{\protect\cite{cite530}}).
\end{eczvaluelist}
\eczhbkcontributors{ \eczhuVVA }
\endeczcode

\eczcode{4d_13_surface}{\((1,3)\) 4D toric code}{~\NoCaseChange{\protect\cite{cite479}}}
\eczhIndexCodeAliasName{4d_13_surface}{4D toric code}
\codefieldsection{Description}
A generalization of the Kitaev surface code defined on a 4D lattice.
The code is called a \((1,3)\) toric code because it admits 1D \(Z\)-type and 3D \(X\)-type logical operators.
In the hypercubic lattice version, qubits are placed on edges, each \(Z\)-type stabilizer generator is supported on cubes on the boundary of a hypercube, and \(X\)-type stabilizers are placed on the edges neighboring every vertex \NoCaseChange{\protect\cite{cite479}}.

\codefieldsection{Transversal and Permutation-Based Gates}
\begin{eczvaluelist}
\item\relax Logical \(CCCZ\) gate on a hyper-diamond lattice \NoCaseChange{\protect\cite{cite479}}.
\end{eczvaluelist}
\codefieldsection{Parents}
\begin{eczvaluelist}
\item\relax
\flmRefsHyperref[eczindexfamilyrel]{code:higher_dimensional_surface}{Homological code} --- The \((1,3)\) 4D toric code realizes 4D \(\mathbb{Z}_2\) gauge theory with 1D \(Z\)-type and 3D \(X\)-type logical operators.
\item\relax
\flmRefsHyperref[eczindexfamilyrel]{code:4d_stabilizer}{4D lattice stabilizer code}\item\relax
\flmRefsHyperref[eczindexfamilyrel]{code:topological_abelian}{Abelian topological code} --- The \((1,3)\) 4D toric code realizes 4D \(\mathbb{Z}_2\) gauge theory with 1D \(Z\)-type and 3D \(X\)-type logical operators.
\item\relax
\flmRefsHyperref[eczindexfamilyrel]{code:dijkgraaf_witten}{Dijkgraaf-Witten gauge theory code} --- An untwisted Dijkgraaf-Witten theory in 4D for the group \(G=\mathbb{Z}_2\) is a \((1,3)\) 4D toric code.
\end{eczvaluelist}
\codefieldsection{Cousins}
\begin{eczvaluelist}
\item\relax
\flmRefsHyperref[eczindexfamilyrel]{code:dfour}{\(D_4\) hyper-diamond lattice} --- The \((1,3)\) 4D toric code on a hyper-diamond lattice admits a transversal logical \(CCCZ\) gate \NoCaseChange{\protect\cite{cite479}}.
\item\relax
\flmRefsHyperref[eczindexfamilyrel]{code:multisector_hypergraph}{Higher-dimensional homological product code} --- The 4D \((1,3)\) planar (toric) code on a hypercubic lattice can be obtained from a particular choice of chain complex from a hypergraph product of four repetition codes \NoCaseChange{\protect\cite{cite1613}}.
\item\relax
\flmRefsHyperref[eczindexfamilyrel]{code:repetition}{Repetition code} --- The 4D \((1,3)\) planar (toric) code on a hypercubic lattice can be obtained from a particular choice of chain complex from a hypergraph product of four repetition codes \NoCaseChange{\protect\cite{cite1613}}.
\end{eczvaluelist}
\eczhbkcontributors{ Nathanan Tantivasadakarn, \eczhuVVA }
\endeczcode

\eczcode{4d_surface}{\((2,2)\) Loop toric code}{~\NoCaseChange{\protect\cite{cite480}}}
\codefieldsection{Alternative Names}
\begin{eczvaluelist}
\item\relax Kitaev tesseract code
\item\relax 4D surface code
\item\relax All-loop toric code
\item\relax \((2,2)\) 4D toric code
\end{eczvaluelist}
\eczhIndexCodeAliasName{4d_surface}{Loop toric code}
\eczhIndexCodeAliasName{4d_surface}{Kitaev tesseract code}
\eczhIndexCodeAliasName{4d_surface}{4D surface code}
\eczhIndexCodeAliasName{4d_surface}{All-loop toric code}
\eczhIndexCodeAliasName{4d_surface}{\((2,2)\) 4D toric code}
\codefieldsection{Description}
A generalization of the Kitaev surface code defined on a 4D lattice.
The code is called a \((2,2)\) toric code because it admits 2D membrane \(Z\)-type and \(X\)-type logical operators.
Both types of operators create 1D (i.e., loop) excitations at their edges.
The code serves as a self-correcting quantum memory \NoCaseChange{\protect\cite{cite480,cite481}}.

The open-boundary hypercubic realization is often called the \textit{tesseract code}.
It can be formulated using relative homology and encodes one logical qubit, in contrast to the six logical qubits of the periodic 4D toric code \NoCaseChange{\protect\cite{cite3174}}.

Qubits are placed on plaquettes, each \(Z\)-type stabilizer generator is supported on six plaquettes surrounding an edge, and \(X\)-type stabilizers are placed on the six plaquettes of every cube \NoCaseChange{\protect\cite{cite479}}.

\textit{Loop toric code} often either refers to the construction on
the 4D torus or is an alternative name for the general
construction.

The construction has been extended to modular qudits \NoCaseChange{\protect\cite{cite2529}}.

\codefieldsection{Protection}
Code parameters for an open hypercubic lattice of side-length \(L\) are \(\llbracket 6L^4 − 12L^3 + 10L^2 - 4L + 1, 1, L^2\rrbracket \) \NoCaseChange{\protect\cite{cite3174,cite850}}.
In the open-boundary/tesseract realization, all bulk stabilizer generators have weight six and each qubit participates in eight stabilizer checks \NoCaseChange{\protect\cite{cite3174}}.

\codefieldsection{Rate}
For the open-boundary tesseract family, \(k=1\) and \(n=6d^2-12d^{3/2}+10d-4\sqrt{d}+1\) because \(d=L^2\) and \(n=6L^4-12L^3+10L^2-4L+1\) \NoCaseChange{\protect\cite{cite3174}}.

\codefieldsection{Encoding}
\begin{eczvaluelist}
\item\relax Lindbladian-based dissipative encoding, for which codespace is steady-state space of a Lindbladian \NoCaseChange{\protect\cite{cite3175}}.
\end{eczvaluelist}
\codefieldsection{Transversal and Permutation-Based Gates}
\begin{eczvaluelist}
\item\relax Only logical \flmRefsHyperref{ref409}{Clifford gates} can be implemented transversally when defined on a hypercubic lattice \NoCaseChange{\protect\cite{cite479}}.
\end{eczvaluelist}
\codefieldsection{Gates}
\begin{eczvaluelist}
\item\relax Logical \(S\) gate using physical \(CS\) gates via the Pontryagin square \NoCaseChange{\protect\cite{cite2529}}.
\item\relax On a hypercubic lattice, electromagnetic duality implements a logical Hadamard gate \NoCaseChange{\protect\cite{cite576}}.
\item\relax On closed 4-manifolds, the gauged-SPT operator \(i^{\int \mathcal{P}(a)}\) realizes logical \(CZ\) gates on \(T^4\) and a logical \(S\) gate on \(\mathbb{CP}^2\) \NoCaseChange{\protect\cite{cite576}}.
\item\relax Single-shot lattice surgery for the 4D loop toric code can be formulated using the fault-complex formalism \NoCaseChange{\protect\cite{cite3176}}.
\end{eczvaluelist}
\codefieldsection{Decoding}
\begin{eczvaluelist}
\item\relax A local recovery procedure that dispenses with fast classical processing, and can even be formulated without explicit measurements, is possible when the code is realized in four or more spatial dimensions \NoCaseChange{\protect\cite{cite480}}.
\item\relax Single-shot repair-syndrome decoder for the open-boundary/tesseract code, followed by RG decoding on the repaired syndrome \NoCaseChange{\protect\cite{cite3174}}.
\item\relax Local automaton decoder \NoCaseChange{\protect\cite{cite3177}} based on Toom's rule for the classical 2D repetition code \NoCaseChange{\protect\cite{cite1592,cite1593,cite1594,cite1595}}.
\item\relax Local automaton decoder obtained from reinforcement learning \NoCaseChange{\protect\cite{cite1601}}.
\end{eczvaluelist}
\codefieldsection{Code Capacity Threshold}
\begin{eczvaluelist}
\item\relax Independent \(X,Z\) noise: \(2.117\%\) with Hastings decoder \NoCaseChange{\protect\cite{cite3177}} and \(7.3\%\) with RG decoder for the open-boundary 4D tesseract code \NoCaseChange{\protect\cite{cite3174}}. It is conjectured via a statistical-mechanical mapping that the optimal ML decoder yields a threshold of \(11.003\%\) \NoCaseChange{\protect\cite{cite3178}}.
\end{eczvaluelist}
\codefieldsection{Threshold}
\begin{eczvaluelist}
\item\relax Phenomenological noise model for the open-boundary 4D tesseract code: \(4.35\%\) with RG decoder \NoCaseChange{\protect\cite{cite3174}}, and \(4.3\%\) under improved BP-OSD decoder \NoCaseChange{\protect\cite{cite3179}}.
\item\relax Gate-based depolarizing noise: \(0.31\%\) with RG decoder for the open-boundary 4D tesseract code \NoCaseChange{\protect\cite{cite3174}}.
\item\relax \(1.59\%\) for independent \(X,Z\) noise and faulty syndrome measurements using the Hastings decoder \NoCaseChange{\protect\cite{cite3177}}.
\end{eczvaluelist}
\codefieldsection{Realizations}
\begin{eczvaluelist}
\item\relax Trapped ions: single-shot QEC realized using a \(\llbracket 33,1,4\rrbracket \) rotated version of the loop toric code on the Quantinuum H2 device \NoCaseChange{\protect\cite{cite850}}.
\end{eczvaluelist}
\codefieldsection{Parents}
\begin{eczvaluelist}
\item\relax
\flmRefsHyperref[eczindexfamilyrel]{code:higher_dimensional_surface}{Homological code} --- The 4D loop toric code realizes 4D \(\mathbb{Z}_2\) gauge theory with only loop excitations \NoCaseChange{\protect\cite{cite2529}}.
\item\relax
\flmRefsHyperref[eczindexfamilyrel]{code:4d_stabilizer}{4D lattice stabilizer code}\item\relax
\flmRefsHyperref[eczindexfamilyrel]{code:single_shot}{Single-shot code} --- Single-shot QEC has been realized using the \(\llbracket 33,1,4\rrbracket \) loop toric code on the Quantinuum H2 device \NoCaseChange{\protect\cite{cite850}}.
\item\relax
\flmRefsHyperref[eczindexfamilyrel]{code:topological_abelian}{Abelian topological code} --- The 4D loop toric code realizes 4D \(\mathbb{Z}_2\) gauge theory with only loop excitations \NoCaseChange{\protect\cite{cite2529}}.
\end{eczvaluelist}
\codefieldsection{Cousins}
\begin{eczvaluelist}
\item\relax
\flmRefsHyperref[eczindexfamilyrel]{code:multisector_hypergraph}{Higher-dimensional homological product code} --- The 4D loop planar (toric) code on a hypercubic lattice can be obtained from a particular choice of chain complex from a hypergraph product of four repetition codes \NoCaseChange{\protect\cite{cite1613}}.
\item\relax
\flmRefsHyperref[eczindexfamilyrel]{code:repetition}{Repetition code} --- The 4D loop planar (toric) code on a hypercubic lattice can be obtained from a particular choice of chain complex from a hypergraph product of four repetition codes \NoCaseChange{\protect\cite{cite1613}}.
\item\relax
\flmRefsHyperref[eczindexfamilyrel]{code:double_homological_product}{Campbell double homological product code} --- The 4D loop planar (toric) code on a hypercubic lattice can be obtained from a particular choice of chain complex from a hypergraph product of four repetition codes \NoCaseChange{\protect\cite{cite1613}}. As such, it is a particular Campbell double homological product code \NoCaseChange{\protect\cite[{table I}]{cite675}}.
\item\relax
\flmRefsHyperref[eczindexfamilyrel]{code:3d_surface}{3D surface code} --- Setting one linear size of the open-boundary tesseract construction to \(1\) yields the cubic/3D surface code \NoCaseChange{\protect\cite{cite3174}}.
\item\relax
\flmRefsHyperref[eczindexfamilyrel]{code:surface}{Kitaev surface code} --- Setting \(L_2=L_4=1\) in the open-boundary tesseract construction yields the planar surface code \NoCaseChange{\protect\cite{cite3174}}.
\item\relax
\flmRefsHyperref[eczindexfamilyrel]{code:self_correct}{Self-correcting quantum code} --- For similar reasons as the classical 2D Ising model is a self-correcting classical memory, the 4D loop toric code is a self-correcting quantum memory due to an \flmRefsHyperref{ref65}{order} \(O(n)\) energy cost of creating a logical error \NoCaseChange{\protect\cite{cite480,cite481}}.
\item\relax
\flmRefsHyperref[eczindexfamilyrel]{code:haah_cubic}{Haah cubic code (CC)} --- The energy of any partial implementation of CC1 is proportional to the boundary length, similar to the 4D toric code. This can potentially suppress the effects of thermal errors, but it is currently an open problem.
\end{eczvaluelist}
\eczhbkcontributors{ \eczhuVVA }
\endeczcode

\eczcode{stab_5_1_2_convolutional}{\((5,1,2)\)-convolutional code}{~\NoCaseChange{\protect\cite{cite3180}}}
\codefieldsection{Description}
Family of quantum convolutional codes that are 1D lattice generalizations of the five-qubit perfect code, with the former's lattice-translation symmetry being the extension of the latter's cyclic permutation symmetry.

Their stabilizer generators for semi-open boundary conditions are
\flmMathEnvironment{align}{}{
  \begin{array}{cccccccc}
  X & Z & I & I & I & I & I & \cdots\\
  Z & X & X & Z & I & I & I & \cdots\\
  I & Z & X & X & Z & I & I & \cdots\\
  I & I & Z & X & X & Z & I & \cdots\\
  \vdots & \vdots & \vdots & \vdots & \vdots & \vdots & \vdots & \ddots
  \end{array}~.
}
\codefieldsection{Parent}
\begin{eczvaluelist}
\item\relax
\flmRefsHyperref[eczindexfamilyrel]{code:quantum_convolutional}{Quantum convolutional code}\end{eczvaluelist}
\codefieldsection{Child}
\begin{eczvaluelist}
\item\relax
\flmRefsHyperref[eczindexfamilyrel]{code:stab_5_1_3}{\(\llbracket 5,1,3\rrbracket \) Five-qubit perfect code} --- The \((5,1,2)\)-convolutional code is a 1D lattice extension of the five-qubit perfect code, with the former's lattice-translation symmetry being the extension of the latter's cyclic permutation symmetry. The \((5,1,2)\)-convolutional code reduces to the five-qubit code for a five-qubit chain and periodic boundary conditions. See Ref. \NoCaseChange{\protect\cite{cite3181}} for the first few codes in a different extension of the five-qubit perfect code.
\end{eczvaluelist}
\codefieldsection{Cousin}
\begin{eczvaluelist}
\item\relax
\flmRefsHyperref[eczindexfamilyrel]{code:twisted_xzzx}{Twisted XZZX toric code} --- \((5,1,2)\)-convolutional codes (twisted XZZX toric codes) are 1D (2D) lattice extensions of the five-qubit perfect code.
\end{eczvaluelist}
\eczhbkcontributors{ \eczhuVVA }
\endeczcode

\eczcode{stab_10_1_2}{\(\llbracket 10,1,2\rrbracket \) Vasmer-Kubica code}{~\NoCaseChange{\protect\cite{cite687}}}
\eczhIndexCodeAliasName{stab_10_1_2}{Vasmer-Kubica code}
\codefieldsection{Description}
A stabilizer code obtained by morphing the \(\llbracket 15,1,3\rrbracket \) quantum Reed-Muller code on a subset whose child code is the \(\llbracket 8,3,2\rrbracket \) smallest interesting color code \NoCaseChange{\protect\cite{cite687}}.
It is the smallest known stabilizer code with a fault-tolerant logical \(T\) gate, implemented via physical \(T\), \(T^{\dagger}\), and \(CCZ\) gates \NoCaseChange{\protect\cite{cite687}}.

A stabilizer tableau for the code is \NoCaseChange{\protect\cite{cite3182}}
\flmMathEnvironment{align}{}{
\begin{array}{cccccccccc}
  X & X & X & X & I & I & I & X & I & I \\
  I & X & X & I & X & X & I & I & X & I \\
  I & I & X & X & I & X & X & I & I & X \\
  Z & Z & Z & Z & I & I & I & I & I & I \\
  I & Z & Z & I & Z & Z & I & I & I & I \\
  I & I & Z & Z & I & Z & Z & I & I & I \\
  I & I & Z & I & I & Z & I & Z & I & I \\
  I & I & Z & Z & I & I & I & I & Z & I \\
  I & Z & Z & I & I & I & I & I & I & Z
\end{array}~.
}
The minimum-weight logical-\(Z\) representatives are \(Z_1 Z_8\), \(Z_5 Z_9\), and \(Z_7 Z_{10}\), each of weight two, confirming \(d=2\).

\codefieldsection{Gates}
\begin{eczvaluelist}
\item\relax A logical \(\bar{T}\) gate is implemented by applying \(T^{\pm 1}\) to qubits \(1\)--\(7\) and \(CCZ\) to qubits \(8,9,10\) \NoCaseChange{\protect\cite{cite687}}.
\item\relax 10-to-1 magic-state distillation protocol that consumes seven \(|T\rangle\) states and one \(|CCZ\rangle\) state \NoCaseChange{\protect\cite{cite687}}.
\end{eczvaluelist}
\codefieldsection{Fault Tolerance}
\begin{eczvaluelist}
\item\relax A fault-tolerant universal gate set can be obtained via \flmRefsHyperref{ref410}{code switching} between the Steane code and the \(\llbracket 10,1,2\rrbracket \) code \NoCaseChange{\protect\cite{cite3182}}.
\end{eczvaluelist}
\codefieldsection{Realizations}
\begin{eczvaluelist}
\item\relax Trapped-ion devices: fault-tolerant universal gate set via \flmRefsHyperref{ref410}{code switching} between the Steane code and the \(\llbracket 10,1,2\rrbracket \) code on a device from the Monz group \NoCaseChange{\protect\cite{cite3182}}.
\end{eczvaluelist}
\codefieldsection{Parent}
\begin{eczvaluelist}
\item\relax
\flmRefsHyperref[eczindexfamilyrel]{code:morphed_diagonal_clifford}{\(\llbracket 2^r+r-1,1,2\rrbracket \) morphed simplex code} --- The \(\llbracket 10,1,2\rrbracket \) code is a specific instance of the \(\llbracket 2^r+r-1,1,2\rrbracket \) morphed simplex codes with \(r=3\) \NoCaseChange{\protect\cite[{Appx. C}]{cite687}}.
\end{eczvaluelist}
\codefieldsection{Cousins}
\begin{eczvaluelist}
\item\relax
\flmRefsHyperref[eczindexfamilyrel]{code:stab_15_1_3}{\(\llbracket 15,1,3\rrbracket \) quantum RM code} --- The \(\llbracket 10,1,2\rrbracket \) code is obtained by morphing the \(\llbracket 15,1,3\rrbracket \) code on a region whose child code is the \(\llbracket 8,3,2\rrbracket \) smallest interesting color code \NoCaseChange{\protect\cite{cite687}}.
\item\relax
\flmRefsHyperref[eczindexfamilyrel]{code:stab_8_3_2}{\(\llbracket 8,3,2\rrbracket \) Smallest interesting color code} --- The \(\llbracket 10,1,2\rrbracket \) code is obtained by morphing the \(\llbracket 15,1,3\rrbracket \) code on a region whose child code is the \(\llbracket 8,3,2\rrbracket \) smallest interesting color code \NoCaseChange{\protect\cite{cite687}}.
\item\relax
\flmRefsHyperref[eczindexfamilyrel]{code:steane}{\(\llbracket 7,1,3\rrbracket \) Steane code} --- A fault-tolerant universal gate set can be obtained via \flmRefsHyperref{ref410}{code switching} between the Steane code and the \(\llbracket 10,1,2\rrbracket \) code \NoCaseChange{\protect\cite{cite3182}}.
\end{eczvaluelist}
\eczhbkcontributors{ \eczhuVVA }
\endeczcode

\eczcode{eaoa_hamming}{\(\llbracket 10,1,3;1,3,4\rrbracket \) EAOA Hamming code}{~\NoCaseChange{\protect\cite{cite856}}}
\eczhIndexCodeAliasName{eaoa_hamming}{EAOA Hamming code}
\codefieldsection{Description}
An EAOA qubit stabilizer code constructed from the dual of a \([10,6,3]\) code obtained by shortening the classical \([15,11,3]\) Hamming code at five positions.
In the notation of the parent entry, the example of Ref. \NoCaseChange{\protect\cite{cite856}} is a \(\llbracket 10,1,3;1,3,4\rrbracket \) code: it encodes one logical qubit and four classical strings (equivalently, two classical bits), while retaining one gauge qubit and using three ebits.

\codefieldsection{Parent}
\begin{eczvaluelist}
\item\relax
\flmRefsHyperref[eczindexfamilyrel]{code:eaoa_stabilizer}{EAOA qubit stabilizer code}\end{eczvaluelist}
\codefieldsection{Cousins}
\begin{eczvaluelist}
\item\relax
\flmRefsHyperref[eczindexfamilyrel]{code:hamming}{\([2^r-1,2^r-r-1,3]\) Hamming code} --- The dual of a \([10,6,3]\) code obtained by shortening the \([15,11,3]\) Hamming code at five positions can be used to construct \(\llbracket 10,1,3;1,3,4\rrbracket \) EAOA Hamming code \NoCaseChange{\protect\cite{cite856}}.
\item\relax
\flmRefsHyperref[eczindexfamilyrel]{code:small_distance_qubit_stabilizer}{Small-distance qubit stabilizer code}\end{eczvaluelist}
\eczhbkcontributors{ \eczhuVVA }
\endeczcode

\eczcode{stab_10_2_3}{\(\llbracket 10,2,3\rrbracket \) binarized Galois-qudit code}{~\NoCaseChange{\protect\cite{cite514}}}
\eczhIndexCodeAliasName{stab_10_2_3}{binarized Galois-qudit code}
\codefieldsection{Description}
CSS code obtained by binarizing a \(\llbracket 5,1,3\rrbracket _4\) Galois-qudit CSS code in the self-dual normal basis \(\{\omega,\omega^2\}\).

A stabilizer tableau for the binarized code is \NoCaseChange{\protect\cite{cite514}}
\flmMathEnvironment{align}{}{
\begin{array}{cccccccccc}
  Z & I & Z & I & Z & I & Z & I & I & I \\
  I & Z & I & Z & I & Z & I & Z & I & I \\
  I & I & Z & I & Z & Z & I & Z & Z & I \\
  I & I & I & Z & Z & I & Z & Z & I & Z \\
  X & I & X & I & X & I & X & I & I & I \\
  I & X & I & X & I & X & I & X & I & I \\
  I & I & X & I & I & X & X & X & X & I \\
  I & I & I & X & X & X & X & I & I & X
\end{array}~.
}

\codefieldsection{Transversal and Permutation-Based Gates}
\begin{eczvaluelist}
\item\relax The code is permutation-equivalent to its Hadamard dual: transversal physical Hadamards followed by swapping the two qubits in each binarized \(\mathbb{F}_4\) coordinate preserve the codespace and implement a logical \(H^{\otimes 2}\) up to logical \(\mathrm{SWAP}\). This is the binarized version of the permutation-assisted Hadamard available from plain self-duality of the starting \(q=4\) Galois-qudit CSS code \NoCaseChange{\protect\cite{cite514}}.
\end{eczvaluelist}
\codefieldsection{Parents}
\begin{eczvaluelist}
\item\relax
\flmRefsHyperref[eczindexfamilyrel]{code:qubit_css}{Qubit CSS code}\item\relax
\flmRefsHyperref[eczindexfamilyrel]{code:small_distance_qubit_stabilizer}{Small-distance qubit stabilizer code}\end{eczvaluelist}
\codefieldsection{Cousins}
\begin{eczvaluelist}
\item\relax
\flmRefsHyperref[eczindexfamilyrel]{code:bc_phantom}{Binarized-and-concatenated (B\&C) phantom code} --- The binarized \(\llbracket 10,2,3\rrbracket \) code is not phantom, but concatenating each qubit pair with the \(\llbracket 4,2,2\rrbracket \) code yields a \(\llbracket 20,2,6\rrbracket \) B\&C phantom code admitting fold-diagonal logical \(SS\) gates \NoCaseChange{\protect\cite{cite514}}.
\item\relax
\flmRefsHyperref[eczindexfamilyrel]{code:stab_4_2_2}{\(\llbracket 4,2,2\rrbracket \) Four-qubit code} --- Concatenating each qubit pair of the binarized \(\llbracket 10,2,3\rrbracket \) code with the \(\llbracket 4,2,2\rrbracket \) code yields the \(\llbracket 20,2,6\rrbracket \) B\&C phantom code \NoCaseChange{\protect\cite{cite514,cite795}}.
\item\relax
\flmRefsHyperref[eczindexfamilyrel]{code:stab_20_2_6}{\(\llbracket 20,2,6\rrbracket \) B\&C phantom code} --- The \(\llbracket 20,2,6\rrbracket \) code is obtained by concatenating each qubit pair of the \(\llbracket 10,2,3\rrbracket \) binarized Galois-qudit code with the \(\llbracket 4,2,2\rrbracket \) code \NoCaseChange{\protect\cite{cite514}}.
\item\relax
\flmRefsHyperref[eczindexfamilyrel]{code:css_5_1_3}{\(\llbracket 5,1,3\rrbracket _4\) Galois-qudit CSS code} --- Binarizing the \(\llbracket 5,1,3\rrbracket _4\) code in the self-dual normal basis \(\{\omega,\omega^2\}\) yields a \(\llbracket 10,2,3\rrbracket \) qubit CSS code \NoCaseChange{\protect\cite{cite514}}.
\end{eczvaluelist}
\eczhbkcontributors{ \eczhuVVA }
\endeczcode

\eczcode{xzzx_10_2_3}{\(\llbracket 10,2,3\rrbracket \) rotated toric code}{~\NoCaseChange{\protect\cite{cite1316}\protect\cite[{Exam. 3}]{cite439}}}
\eczhIndexCodeAliasName{xzzx_10_2_3}{rotated toric code}
\codefieldsection{Description}
Rotated toric code that is the CSS form of the twisted XZZX toric code with parameters \(a=1\), \(b=3\) \NoCaseChange{\protect\cite{cite438,cite427}},
related to the XZZX form by Hadamard on the \(B\)-sublattice.
It is also the \flmRefsHyperref{ref436}{symplectic double} (a.k.a. genus-one double cover) of the \(\llbracket 5,1,3\rrbracket \) five-qubit perfect code \NoCaseChange{\protect\cite{cite439,cite435}}, the \flmRefsHyperref{ref436}{symplectic double} of the \(\llbracket 5,1,2\rrbracket \) rotated surface code \NoCaseChange{\protect\cite{cite435}}, and a BCC code \NoCaseChange{\protect\cite{cite440}}.

A stabilizer tableau for this code is
\flmMathEnvironment{align}{}{
\begin{array}{cccccccccc}
  Z & Z & I & Z & Z & I & I & I & I & I \\
  I & I & Z & Z & I & Z & Z & I & I & I \\
  I & I & I & I & Z & Z & I & Z & Z & I \\
  Z & I & I & I & I & I & Z & Z & I & Z \\
  X & I & X & X & I & I & I & I & I & X \\
  I & X & X & I & X & X & I & I & I & I \\
  I & I & I & X & X & I & X & X & I & I \\
  I & I & I & I & I & X & X & I & X & X
\end{array}~.
}

\codefieldsection{Protection}
Distance 3, correcting any single-qubit error.
Achieves the lower bound \(n \geq d^2+1 = 10\) for weight-four GB codes of odd distance \NoCaseChange{\protect\cite{cite3183}}.

\codefieldsection{Transversal and Permutation-Based Gates}
\begin{eczvaluelist}
\item\relax Transversal Hadamard-SWAP: qubit permutation \(m \mapsto -m\) (mod 10) paired with sublattice swap \(A \leftrightarrow B\), followed by Hadamard on all qubits \NoCaseChange{\protect\cite{cite440}}.
\item\relax Logical Hadamard without SWAP via \(m \mapsto 3m\) (mod 10) followed by Hadamard \NoCaseChange{\protect\cite{cite440}}.
\end{eczvaluelist}
\codefieldsection{Parents}
\begin{eczvaluelist}
\item\relax
\flmRefsHyperref[eczindexfamilyrel]{code:twisted_xzzx}{Twisted XZZX toric code} --- This is the \(d=3\) instance of the \(\llbracket d^2+1,2,d\rrbracket \) family of twisted XZZX toric codes (parameters \(a=1\), \(b=3\)), presented in its CSS form \NoCaseChange{\protect\cite{cite440}}.
\item\relax
\flmRefsHyperref[eczindexfamilyrel]{code:bipartite_cyclic_cluster}{Bipartite cyclic cluster (BCC) code} --- The \(\llbracket 10,2,3\rrbracket \) rotated toric code is a \(\llbracket d^2+1,2,d\rrbracket \) BCC code for \(d=3\) \NoCaseChange{\protect\cite{cite440}}. A non-CSS cyclic cluster code related to the \(\llbracket 10,2,3\rrbracket \) rotated toric code yields the \(\llbracket 5,1,3\rrbracket \) five-qubit perfect code for \(d=3\) \NoCaseChange{\protect\cite{cite440}}.
\item\relax
\flmRefsHyperref[eczindexfamilyrel]{code:small_distance_qubit_stabilizer}{Small-distance qubit stabilizer code}\end{eczvaluelist}
\codefieldsection{Cousins}
\begin{eczvaluelist}
\item\relax
\flmRefsHyperref[eczindexfamilyrel]{code:stab_5_1_3}{\(\llbracket 5,1,3\rrbracket \) Five-qubit perfect code} --- The \(\llbracket 10,2,3\rrbracket \) rotated toric code is the \flmRefsHyperref{ref436}{symplectic double} (a.k.a. genus-one double cover) of the five-qubit perfect code \NoCaseChange{\protect\cite{cite435}\protect\cite[{Exam. 3}]{cite439}}. A non-CSS cyclic cluster code related to the \(\llbracket 10,2,3\rrbracket \) rotated toric code yields the \(\llbracket 5,1,3\rrbracket \) five-qubit perfect code for \(d=3\) \NoCaseChange{\protect\cite{cite440}}.
\item\relax
\flmRefsHyperref[eczindexfamilyrel]{code:stab_5_1_2}{\(\llbracket 5,1,2\rrbracket \) rotated surface code} --- The \(\llbracket 10,2,3\rrbracket \) rotated toric code is the \flmRefsHyperref{ref436}{symplectic double} (a.k.a. genus-one double cover) of the \(\llbracket 5,1,2\rrbracket \) rotated surface code \NoCaseChange{\protect\cite{cite435}}.
\end{eczvaluelist}
\eczhbkcontributors{ \eczhuVVA }
\endeczcode

\eczcode{bb108}{\(\llbracket 108,8,10\rrbracket \) BB6 code}{~\NoCaseChange{\protect\cite{cite441}}}
\codefieldsection{Alternative Names}
\begin{eczvaluelist}
\item\relax \((9,6)\) BB6 code
\end{eczvaluelist}
\eczhIndexCodeAliasName{bb108}{BB6 code}
\eczhIndexCodeAliasName{bb108}{\((9,6)\) BB6 code}
\codefieldsection{Description}
A bivariate bicycle (BB) code with parameters \(\llbracket 108,8,10\rrbracket \) and weight-six stabilizer generators \NoCaseChange{\protect\cite{cite441}}.

One defining presentation uses \((\ell,m)=(9,6)\) with \(x^{\ell}=y^{m}=1\), and
\(A=x^3+y+y^2\), \(B=y^3+x+x^2\) in \(\mathbb{F}_2[x,y]/(x^{\ell}-1,y^{m}-1)\) \NoCaseChange{\protect\cite[{Table 3}]{cite441}}.

\codefieldsection{Rate}
Ancilla-added encoding rate is \(1/27\), using \(n_a=n=108\) ancilla qubits.
\codefieldsection{Parent}
\begin{eczvaluelist}
\item\relax
\flmRefsHyperref[eczindexfamilyrel]{code:qcga}{Bivariate bicycle (BB) code}\end{eczvaluelist}
\eczhbkcontributors{ \eczhuVVA }
\endeczcode

\eczcode{stab_11_1_5}{\(\llbracket 11,1,5\rrbracket \) quantum dodecacode}{~\NoCaseChange{\protect\cite{cite449}}}
\eczhIndexCodeAliasName{stab_11_1_5}{quantum dodecacode}
\codefieldsection{Description}
Eleven-qubit \flmRefsHyperref{ref672}{pure} stabilizer code that is the smallest qubit stabilizer code to correct two-qubit errors.
It can be obtained from the non-additive dodecacode by puncturing \NoCaseChange{\protect\cite[{Table IV}]{cite449}}.

A stabilizer tableau for the code is given by \NoCaseChange{\protect\cite{cite1647}\protect\cite[{Table 8.5}]{cite736}}
\flmMathEnvironment{align}{}{
\begin{array}{ccccccccccc}
  Z & Z & Z & Z & Z & Z & I & I & I & I & I \\
  X & X & X & X & X & X & I & I & I & I & I \\
  I & I & I & Z & X & Y & Y & Y & Y & X & Z \\
  I & I & I & X & Y & Z & Z & Z & Z & Y & X \\
  Z & Y & X & I & I & I & Z & Y & X & I & I \\
  X & Z & Y & I & I & I & X & Z & Y & I & I \\
  I & I & I & Z & Y & X & X & Y & Z & I & I \\
  I & I & I & X & Z & Y & Z & X & Y & I & I \\
  Z & X & Y & I & I & I & Z & Z & Z & X & Y \\
  Y & Z & X & I & I & I & Y & Y & Y & Z & X
\end{array}~.
}
The code has paired support despite not being Hermitian \NoCaseChange{\protect\cite{cite795}}.

\codefieldsection{Protection}
Smallest stabilizer code that protects against errors on any two qubits. Detects four-qubit errors.
\codefieldsection{Encoding}
\begin{eczvaluelist}
\item\relax Encoding circuit consisting of 32 gates constructed from reinforcement learning \NoCaseChange{\protect\cite{cite3184}}.
\end{eczvaluelist}
\codefieldsection{Parent}
\begin{eczvaluelist}
\item\relax
\flmRefsHyperref[eczindexfamilyrel]{code:small_distance_qubit_stabilizer}{Small-distance qubit stabilizer code}\end{eczvaluelist}
\codefieldsection{Cousin}
\begin{eczvaluelist}
\item\relax
\flmRefsHyperref[eczindexfamilyrel]{code:dodecacode}{\((12,4^6,6)_4\) Dodecacode} --- The dodecacode corresponds to a \(\llbracket 12,0,6\rrbracket \) quantum code in the \flmRefsHyperref{code:qubit_stabilizer}{\(\mathbb{F}_4\) representation}  \NoCaseChange{\protect\cite{cite449}}. The \(\llbracket 11,1,5\rrbracket \) quantum dodecacode code corresponds to the shortened dodecacode \NoCaseChange{\protect\cite{cite449,cite1647}}. A \flmRefsHyperref{ref672}{pure} \(\llbracket 10,1,4\rrbracket \) quantum code can be obtained from the doubly punctured dodecacode \NoCaseChange{\protect\cite{cite1647}}. These codes are not obtained from the Hermitian construction since none of the classical codes are linear.
\end{eczvaluelist}
\eczhbkcontributors{ \eczhuVVA }
\endeczcode

\eczcode{css_12_1_3}{\(\llbracket 12,1,3\rrbracket \) CE CSS code}{~\NoCaseChange{\protect\cite{cite524}}}
\eczhIndexCodeAliasName{css_12_1_3}{CE CSS code}
\codefieldsection{Description}
Twelve-qubit constant-excitation (CE) CSS code that encodes one logical qubit with distance three.
It is the smallest CE CSS code that corrects a single-qubit error \NoCaseChange{\protect\cite{cite524}}.
Codewords lie in a fixed Hamming-weight subspace, making the code immune to coherent noise in the form of transversal \(Z\)-rotations.

One stabilizer tableau for the code is, up to Pauli frame \NoCaseChange{\protect\cite{cite524}},
\flmMathEnvironment{align}{}{
\begin{array}{cccccccccccc}
  X & X & X & X & I & I & I & I & I & I & I & I \\
  I & I & X & X & X & X & I & I & I & I & I & I \\
  I & I & I & I & I & I & X & X & X & X & I & I \\
  I & I & I & I & I & I & I & I & X & X & X & X \\
  Z & I & Z & I & Z & I & Z & I & Z & I & Z & I \\
  Z & Z & I & I & I & I & I & I & I & I & I & I \\
  I & I & Z & Z & I & I & I & I & I & I & I & I \\
  I & I & I & I & Z & Z & I & I & I & I & I & I \\
  I & I & I & I & I & I & Z & Z & I & I & I & I \\
  I & I & I & I & I & I & I & I & Z & Z & I & I \\
  I & I & I & I & I & I & I & I & I & I & Z & Z
\end{array}~.
}

\codefieldsection{Protection}
Corrects any single-qubit error (distance \(d=3\)).
Protects from collective coherent noise in the form of transversal \(Z\)-rotations, since all codewords lie in the same Hamming-weight subspace \NoCaseChange{\protect\cite{cite2709,cite808}}.

\codefieldsection{Fault Tolerance}
\begin{eczvaluelist}
\item\relax Fault-tolerant syndrome extraction using modified Shor and Steane methods adapted for CE codes: weight-\(2w\) stabilizers are measured using \(w\)-CE cat states, and zero-controlled NOT (\(\mathrm{C}_0 X\)) gates replace standard CNOT gates to preserve the constant-excitation structure \NoCaseChange{\protect\cite{cite524}}.
\end{eczvaluelist}
\codefieldsection{Threshold}
\begin{eczvaluelist}
\item\relax Pseudo-threshold of \(\sim 9.28 \times 10^{-4}\) (circuit-level stochastic noise) and \(\sim 5.98 \times 10^{-4}\) (with coherent corrections, \(\gamma = 0.01\)) under collective coherent noise \NoCaseChange{\protect\cite{cite524}}.
\end{eczvaluelist}
\codefieldsection{Parents}
\begin{eczvaluelist}
\item\relax
\flmRefsHyperref[eczindexfamilyrel]{code:qubit_css}{Qubit CSS code}\item\relax
\flmRefsHyperref[eczindexfamilyrel]{code:constant_excitation}{Constant-excitation (CE) code}\item\relax
\flmRefsHyperref[eczindexfamilyrel]{code:qubit_concatenated}{Concatenated qubit code} --- This code is obtained by dual-rail concatenation of the \(\llbracket 6,1,2\rrbracket \) CSS code \NoCaseChange{\protect\cite[{ID 50}]{cite453}}.
\item\relax
\flmRefsHyperref[eczindexfamilyrel]{code:small_distance_qubit_stabilizer}{Small-distance qubit stabilizer code}\end{eczvaluelist}
\eczhbkcontributors{ \eczhuVVA }
\endeczcode

\eczcode{stab_12_2_2}{\(\llbracket 12,2,2\rrbracket \) CSS code}{~\NoCaseChange{\protect\cite{cite767}}}
\eczhIndexCodeAliasName{stab_12_2_2}{CSS code}
\codefieldsection{Description}
CSS code that admits a logical \(CS\) gate via application of physical \(T\) and \(T^{\dagger}\) gates.

A stabilizer tableau for the code is given by \NoCaseChange{\protect\cite[{Fig. 6}]{cite767}}
\flmMathEnvironment{align}{}{
\begin{array}{cccccccccccc}
  X & X & X & X & I & I & I & I & X & X & X & X \\
  I & I & I & I & X & X & X & X & X & X & X & X \\
  Z & Z & Z & Z & I & I & I & I & I & I & I & I \\
  Z & Z & I & I & Z & Z & I & I & I & I & I & I \\
  Z & I & Z & I & Z & I & Z & I & I & I & I & I \\
  I & Z & Z & I & Z & I & I & Z & I & I & I & I \\
  Z & I & I & I & Z & I & I & I & Z & I & I & I \\
  I & Z & I & I & Z & I & I & I & I & Z & I & I \\
  I & I & Z & I & Z & I & I & I & I & I & Z & I \\
  Z & Z & Z & I & Z & I & I & I & I & I & I & Z
\end{array}~.
}

\codefieldsection{Transversal and Permutation-Based Gates}
\begin{eczvaluelist}
\item\relax Logical \(CS_{01}\) gate via transversal application of \(T\) on qubits \(0,3,4,7,8,11\), \(T^3\) on qubits \(1,2,5,6\), and \(T^7 = T^{\dagger}\) on qubits \(9,10\) \NoCaseChange{\protect\cite[{Fig. 6}]{cite767}}.
\end{eczvaluelist}
\codefieldsection{Parents}
\begin{eczvaluelist}
\item\relax
\flmRefsHyperref[eczindexfamilyrel]{code:phantom}{Phantom code} --- Ref. \NoCaseChange{\protect\cite{cite514}} identifies a \(\llbracket 12,2,2\rrbracket \) CSS phantom code by exhaustive enumeration.
\item\relax
\flmRefsHyperref[eczindexfamilyrel]{code:small_distance_qubit_stabilizer}{Small-distance qubit stabilizer code}\end{eczvaluelist}
\codefieldsection{Cousin}
\begin{eczvaluelist}
\item\relax
\flmRefsHyperref[eczindexfamilyrel]{code:stab_8_3_2}{\(\llbracket 8,3,2\rrbracket \) Smallest interesting color code} --- The \(\llbracket 12,2,2\rrbracket \) CSS code can be obtained by joining two copies of the \(\llbracket 8,3,2\rrbracket \) code at a common face \NoCaseChange{\protect\cite{cite767}}.
\end{eczvaluelist}
\eczhbkcontributors{ \eczhuVVA }
\endeczcode

\eczcode{carbon}{\(\llbracket 12,2,4\rrbracket \) carbon code}{~\NoCaseChange{\protect\cite{cite525}}}
\codefieldsection{Alternative Names}
\begin{eczvaluelist}
\item\relax \(C_{12}\) code
\end{eczvaluelist}
\eczhIndexCodeAliasName{carbon}{carbon code}
\eczhIndexCodeAliasName{carbon}{\(C_{12}\) code}
\codefieldsection{Description}
Twelve-qubit CSS code based on Knill's \(C_4/C_6\) scheme \NoCaseChange{\protect\cite{cite525}}.
Using the concatenation convention of the Zoo, the carbon code can be viewed as a block concatenation with inner code \(\llbracket 4,2,2\rrbracket \) and outer code \(C_6\): three inner \(\llbracket 4,2,2\rrbracket \) blocks encode six intermediate qubits, which are then encoded into two logical qubits by the outer \(\llbracket 6,2,2\rrbracket \) code.

A stabilizer tableau for the code is given by \NoCaseChange{\protect\cite[{Table IV}]{cite525}}
\flmMathEnvironment{align}{}{
\begin{array}{cccccccccccc}
  X & X & X & X & I & I & I & I & I & I & I & I \\
  Z & Z & Z & Z & I & I & I & I & I & I & I & I \\
  I & I & I & I & X & X & X & X & I & I & I & I \\
  I & I & I & I & Z & Z & Z & Z & I & I & I & I \\
  I & I & I & I & I & I & I & I & X & X & X & X \\
  I & I & I & I & I & I & I & I & Z & Z & Z & Z \\
  X & X & I & I & I & X & I & X & X & I & I & X \\
  X & I & I & X & X & X & I & I & I & X & I & X \\
  Z & I & Z & I & I & I & Z & Z & Z & I & I & Z \\
  Z & I & I & Z & Z & I & Z & I & I & I & Z & Z
\end{array}~.
}

\codefieldsection{Encoding}
\begin{eczvaluelist}
\item\relax Simplified fault-tolerant state preparation circuits for the \(00\) and \(++\) states \NoCaseChange{\protect\cite{cite525}}.
\end{eczvaluelist}
\codefieldsection{Transversal and Permutation-Based Gates}
\begin{eczvaluelist}
\item\relax Two-block CNOT gates are transversal because the code is CSS.
\item\relax Automorphism groups of the underlying classical codes can yield transversal \flmRefsHyperref{ref409}{Clifford gates} when combined with qubit permutations \NoCaseChange{\protect\cite{cite763}}. In particular, logical Hadamard is realized by a transversal physical Hadamard followed by a qubit permutation, and a logical one-block CNOT is implemented by a qubit permutation \NoCaseChange{\protect\cite{cite448,cite525}}.
\end{eczvaluelist}
\codefieldsection{Decoding}
\begin{eczvaluelist}
\item\relax Syndrome extraction circuit based on Knill error correction (a.k.a. telecorrection \NoCaseChange{\protect\cite{cite3185}}), but using only one ancillary code block instead of two \NoCaseChange{\protect\cite[{Fig. 5}]{cite525}}.
\end{eczvaluelist}
\codefieldsection{Realizations}
\begin{eczvaluelist}
\item\relax Trapped-ion devices: Three rounds of error correction and post-selected fault-tolerant logical Bell-state preparation with logical error rates at least 5 times lower than physical rate on a quantum charge-coupled device (QCCD) \NoCaseChange{\protect\cite{cite3186}} by Microsoft and Quantinuum \NoCaseChange{\protect\cite{cite525}}.
\end{eczvaluelist}
\codefieldsection{Parents}
\begin{eczvaluelist}
\item\relax
\flmRefsHyperref[eczindexfamilyrel]{code:bc_phantom}{Binarized-and-concatenated (B\&C) phantom code} --- The carbon code is the B\&C phantom code obtained from the \(\llbracket 3,1,2\rrbracket _4\) Galois-qudit code \NoCaseChange{\protect\cite{cite514}}.
\item\relax
\flmRefsHyperref[eczindexfamilyrel]{code:small_distance_qubit_stabilizer}{Small-distance qubit stabilizer code}\end{eczvaluelist}
\codefieldsection{Cousins}
\begin{eczvaluelist}
\item\relax
\flmRefsHyperref[eczindexfamilyrel]{code:stab_4_2_2}{\(\llbracket 4,2,2\rrbracket \) Four-qubit code} --- The carbon code is a concatenation of the \(\llbracket 4,2,2\rrbracket \) code and the \(C_6\) code.
\item\relax
\flmRefsHyperref[eczindexfamilyrel]{code:stab_6_2_2}{\(\llbracket 6,2,2\rrbracket \) \(C_6\) code} --- The carbon code is a concatenation of the \(\llbracket 4,2,2\rrbracket \) code and the \(C_6\) code.
\item\relax
\flmRefsHyperref[eczindexfamilyrel]{code:galois_3_1_2}{\(\llbracket 3,1,2\rrbracket _4\) three-Galois-quartrit code} --- Binarizing this code and concatenating each qubit pair with the \(\llbracket 4,2,2\rrbracket \) code yields the \(\llbracket 12,2,4\rrbracket \) carbon code \NoCaseChange{\protect\cite{cite514}}.
\end{eczvaluelist}
\eczhbkcontributors{ Marcus P da Silva, Victory Omole, \eczhuVVA }
\endeczcode

\eczcode{quad_residue_13_1_5}{\(\llbracket 13,1,5\rrbracket \) quantum QR code}{~\NoCaseChange{\protect\cite[{pg. 11}]{cite818}}}
\eczhIndexCodeAliasName{quad_residue_13_1_5}{quantum QR code}
\codefieldsection{Description}
Thirteen-qubit cyclic Hermitian qubit code derived from a quaternary quadratic-residue code using the Hermitian construction \NoCaseChange{\protect\cite{cite819}\protect\cite[{pg. 11}]{cite818}}.
The code admits a stabilizer tableau whose rows are cyclic permutations of the Pauli string \(XXZZIZIIIZIZZ\) \NoCaseChange{\protect\cite[{ID 67067199e766ad364e845262}]{cite781}}.

\codefieldsection{Parents}
\begin{eczvaluelist}
\item\relax
\flmRefsHyperref[eczindexfamilyrel]{code:stabilizer_over_gf4}{Hermitian qubit code}\item\relax
\flmRefsHyperref[eczindexfamilyrel]{code:galois_quad_residue}{Quantum quadratic-residue (QR) code}\item\relax
\flmRefsHyperref[eczindexfamilyrel]{code:small_distance_qubit_stabilizer}{Small-distance qubit stabilizer code}\end{eczvaluelist}
\eczhbkcontributors{ \eczhuVVA }
\endeczcode

\eczcode{stab_13_1_5}{\(\llbracket 13,1,5\rrbracket \) twisted toric code}{~\NoCaseChange{\protect\cite{cite438}}}
\eczhIndexCodeAliasName{stab_13_1_5}{twisted toric code}
\codefieldsection{Description}
Thirteen-qubit twisted toric code whose stabilizer tableau consists of cyclic permutations of the \(XZZX\)-type Pauli string \(XIZZIXIIIIIII\).
The code can be thought of as a small twisted XZZX code \NoCaseChange{\protect\cite[{Exam. 11 and Fig. 3}]{cite438}}.

\codefieldsection{Transversal and Permutation-Based Gates}
\begin{eczvaluelist}
\item\relax No non-Pauli transversal gates.
\end{eczvaluelist}
\codefieldsection{Parents}
\begin{eczvaluelist}
\item\relax
\flmRefsHyperref[eczindexfamilyrel]{code:twisted_xzzx}{Twisted XZZX toric code} --- The \(\llbracket 13,1,5\rrbracket \) twisted toric code is a small twisted XZZX toric code \NoCaseChange{\protect\cite[{Exam. 11 and Fig. 3}]{cite438}}.
\item\relax
\flmRefsHyperref[eczindexfamilyrel]{code:frobenius}{Frobenius code}\item\relax
\flmRefsHyperref[eczindexfamilyrel]{code:stabilizer_over_gf4}{Hermitian qubit code} --- The \(\llbracket 13,1,5\rrbracket \) twisted toric code is Hermitian \NoCaseChange{\protect\cite[{ID 67067199e766ad364e845262}]{cite781}}.
\item\relax
\flmRefsHyperref[eczindexfamilyrel]{code:small_distance_qubit_stabilizer}{Small-distance qubit stabilizer code}\end{eczvaluelist}
\eczhbkcontributors{ Simon Burton, \eczhuVVA }
\endeczcode

\eczcode{phantom_14_3_3}{\(\llbracket 14,3,3\rrbracket \) CE phantom code}{~\NoCaseChange{\protect\cite{cite524,cite514}}}
\eczhIndexCodeAliasName{phantom_14_3_3}{CE phantom code}
\codefieldsection{Description}
CSS phantom code obtained by concatenating the \(\llbracket 7,3,(d_X=3,d_Z=2)\rrbracket \) punctured hypercube code with the two-qubit phase-flip repetition code.
The code is equivalent to the \(\llbracket 14,3,3\rrbracket \) constant-excitation (CE) CSS code obtained by applying dual-rail concatenation to the \(\llbracket 7,3,2\rrbracket \) punctured hypercube code, up to single-qubit Clifford gates, a physical-qubit permutation, and a Pauli frame \NoCaseChange{\protect\cite{cite524}}.

One stabilizer tableau for the code is
\flmMathEnvironment{align}{}{
\begin{smallmatrix}
  Z & Z & Z & Z & Z & Z & I & I & I & I & I & I & Z & Z \\
  I & I & Z & Z & I & I & Z & Z & I & I & Z & Z & Z & Z \\
  Z & Z & Z & Z & I & I & I & I & Z & Z & Z & Z & I & I \\
  X & X & I & I & I & I & I & I & I & I & I & I & I & I \\
  I & I & X & X & I & I & I & I & I & I & I & I & I & I \\
  I & I & I & I & X & X & I & I & I & I & I & I & I & I \\
  I & I & I & I & I & I & X & X & I & I & I & I & I & I \\
  I & I & I & I & I & I & I & I & X & X & I & I & I & I \\
  I & I & I & I & I & I & I & I & I & I & X & X & I & I \\
  I & I & I & I & I & I & I & I & I & I & I & I & X & X \\
  X & I & X & I & X & I & X & I & X & I & X & I & X & I
\end{smallmatrix}~.
}

\codefieldsection{Protection}
Corrects a single-qubit error. Its \(X\)- and \(Z\)-sector distances are \(d_X=3\) and \(d_Z=4\), respectively \NoCaseChange{\protect\cite{cite514}}.
\codefieldsection{Gates}
\begin{eczvaluelist}
\item\relax The code is phantom, so every ordered-pair in-block logical CNOT gate between its three logical qubits can be implemented by a physical-qubit permutation \NoCaseChange{\protect\cite{cite514}}.
\item\relax The Hadamard-dual code admits fold-diagonal logical \(S_iS_j\) and \(CZ_{ij}\) gates \NoCaseChange{\protect\cite{cite514}}.
\end{eczvaluelist}
\codefieldsection{Fault Tolerance}
\begin{eczvaluelist}
\item\relax In the locally Clifford-equivalent CE CSS frame, fault-tolerant syndrome extraction can use modified Shor and Steane methods adapted for CE codes: weight-\(2w\) stabilizers are measured using \(w\)-CE cat states, and zero-controlled NOT (\(\mathrm{C}_0 X\)) gates replace standard CNOT gates to preserve the constant-excitation structure \NoCaseChange{\protect\cite{cite524}}.
\end{eczvaluelist}
\codefieldsection{Parents}
\begin{eczvaluelist}
\item\relax
\flmRefsHyperref[eczindexfamilyrel]{code:phantom}{Phantom code} --- This \(\llbracket 14,3,3\rrbracket \) code is a CSS phantom code obtained from the punctured hypercube code and the two-qubit phase-flip repetition code \NoCaseChange{\protect\cite{cite514}}.
\item\relax
\flmRefsHyperref[eczindexfamilyrel]{code:constant_excitation}{Constant-excitation (CE) code} --- This code is single-qubit Clifford equivalent to the \(\llbracket 14,3,3\rrbracket \) CE CSS code obtained by dual-rail concatenation of the \(\llbracket 7,3,2\rrbracket \) punctured hypercube code \NoCaseChange{\protect\cite{cite524}}.
\item\relax
\flmRefsHyperref[eczindexfamilyrel]{code:qubit_concatenated}{Concatenated qubit code} --- This code is a concatenation of the \(\llbracket 7,3,2\rrbracket \) punctured hypercube code with the two-qubit phase-flip repetition code \NoCaseChange{\protect\cite{cite514}}.
\item\relax
\flmRefsHyperref[eczindexfamilyrel]{code:small_distance_qubit_stabilizer}{Small-distance qubit stabilizer code}\end{eczvaluelist}
\codefieldsection{Cousins}
\begin{eczvaluelist}
\item\relax
\flmRefsHyperref[eczindexfamilyrel]{code:xz_7_3_2}{\(\llbracket 7,3,2\rrbracket \) punctured hypercube code} --- Concatenating the \(\llbracket 7,3,(d_X=3,d_Z=2)\rrbracket \) punctured hypercube code with the two-qubit phase-flip repetition code yields this \(\llbracket 14,3,(d_X=3,d_Z=4)\rrbracket \) CSS phantom code \NoCaseChange{\protect\cite{cite514}}. Dual-rail concatenation of the same punctured hypercube code yields a single-qubit Clifford-equivalent CE CSS frame \NoCaseChange{\protect\cite{cite524}}.
\item\relax
\flmRefsHyperref[eczindexfamilyrel]{code:steane}{\(\llbracket 7,1,3\rrbracket \) Steane code} --- Dual-rail concatenation of the \(\llbracket 7,1,3\rrbracket \) Steane code yields a \(\llbracket 14,1,3\rrbracket \) CE CSS code, from which the locally Clifford-equivalent \(\llbracket 14,3,3\rrbracket \) CE CSS frame is obtained by removing two independent \(Z\)-type stabilizer generators \NoCaseChange{\protect\cite{cite524}}.
\item\relax
\flmRefsHyperref[eczindexfamilyrel]{code:quantum_repetition}{Quantum repetition code} --- The inner code in the construction is the two-qubit phase-flip repetition code \NoCaseChange{\protect\cite{cite514}}.
\end{eczvaluelist}
\eczhbkcontributors{ \eczhuVVA }
\endeczcode

\eczcode{rhombic_dodecahedron_surface}{\(\llbracket 14,3,3\rrbracket \) Rhombic dodecahedron surface code}{~\NoCaseChange{\protect\cite{cite3187}}}
\codefieldsection{Alternative Names}
\begin{eczvaluelist}
\item\relax Landahl jaunty code
\end{eczvaluelist}
\eczhIndexCodeAliasName{rhombic_dodecahedron_surface}{Rhombic dodecahedron surface code}
\eczhIndexCodeAliasName{rhombic_dodecahedron_surface}{Landahl jaunty code}
\codefieldsection{Description}
A \(\llbracket 14,3,3\rrbracket \) twist-defect surface code whose qubits lie on the vertices of a rhombic dodecahedron.
Its non-CSS nature is due to twist defects \NoCaseChange{\protect\cite{cite442}} stemming from the geometry of the polytope.
A local-Clifford-equivalent clean realization has only \(X\)- and \(Z\)-type operators on its four-valent vertices, and its symplectic double is a \(\llbracket 28,6,3\rrbracket \) genus-three code \NoCaseChange{\protect\cite{cite435}}.

A stabilizer tableau for the code is \NoCaseChange{\protect\cite[{Eq. (1)}]{cite3187}}
\flmMathEnvironment{align}{}{
\begin{smallmatrix}
  X & X & X & I & I & X & I & I & I & I & I & I & I & I \\
  I & I & X & X & I & I & X & X & I & I & I & I & I & I \\
  I & I & I & I & I & I & I & X & X & I & I & X & X & I \\
  X & I & I & I & I & I & I & I & I & X & I & I & X & X \\
  I & Y & Y & Y & Y & I & I & I & I & I & I & I & I & I \\
  I & I & Y & I & I & Y & Y & I & I & I & Y & I & I & I \\
  I & I & I & I & Y & I & I & I & Y & Y & I & I & Y & I \\
  I & I & I & I & I & I & I & I & I & I & Y & Y & Y & Y \\
  Z & Z & I & I & Z & I & I & I & I & Z & I & I & I & I \\
  I & I & I & Z & Z & I & I & Z & Z & I & I & I & I & I \\
  I & I & I & I & I & I & Z & Z & I & I & Z & Z & I & I
\end{smallmatrix}~.
}

\codefieldsection{Parents}
\begin{eczvaluelist}
\item\relax
\flmRefsHyperref[eczindexfamilyrel]{code:twist_defect_surface}{Twist-defect surface code} --- The rhombic dodecahedron surface code is a twist-defect surface code whose degree-three vertices can be interpreted as disclination twists \NoCaseChange{\protect\cite{cite3187}}.
\item\relax
\flmRefsHyperref[eczindexfamilyrel]{code:small_distance_qubit_stabilizer}{Small-distance qubit stabilizer code}\end{eczvaluelist}
\codefieldsection{Cousin}
\begin{eczvaluelist}
\item\relax
\flmRefsHyperref[eczindexfamilyrel]{code:rhombic_dodecahedron}{Rhombic dodecahedron code} --- The qubits of the \(\llbracket 14,3,3\rrbracket \) rhombic dodecahedron surface code lie on the vertices of the small rhombic dodecahedron.
\end{eczvaluelist}
\eczhbkcontributors{ \eczhuVVA }
\endeczcode

\eczcode{gross}{\(\llbracket 144,12,12\rrbracket \) gross code}{~\NoCaseChange{\protect\cite{cite441}}}
\codefieldsection{Alternative Names}
\begin{eczvaluelist}
\item\relax \((3,3)\) BB6 code
\end{eczvaluelist}
\eczhIndexCodeAliasName{gross}{gross code}
\eczhIndexCodeAliasName{gross}{\((3,3)\) BB6 code}
\codefieldsection{Description}
A BB code which requires less physical and ancilla qubits (for syndrome extraction) than the surface code with the same number of logical qubits and distance.
The gross code is equivalent to 8 copies of the surface code via a constant-depth Clifford circuit, and is an element of a larger family of 2D stabilizer codes \NoCaseChange{\protect\cite{cite443}}.
The name stems from the fact that a gross is a dozen dozen.

A different BB QLDPC code with the same parameters was introduced in \NoCaseChange{\protect\cite{cite3188}}.

\codefieldsection{Protection}
Admits a \flmRefsHyperref{ref2960}{pseudo-threshold} of \(\approx 0.7\%\) for the circuit-based noise model.
At physical error rate \(p=10^{-3}\), the code suppresses the logical error rate to about \(2\times 10^{-7}\), enough to preserve \(12\) logical qubits for nearly one million syndrome cycles using \(288\) total physical qubits \NoCaseChange{\protect\cite{cite441}}.

\codefieldsection{Rate}
An ancilla-added rate of \(1/24\). In contrast, the distance-13 surface code has ancilla-added rate \(1/338\).
\codefieldsection{Transversal and Permutation-Based Gates}
\begin{eczvaluelist}
\item\relax Logical Pauli operators and fold-transversal gates studied in Ref. \NoCaseChange{\protect\cite{cite719}}.
\end{eczvaluelist}
\codefieldsection{Gates}
\begin{eczvaluelist}
\item\relax \flmRefsHyperref{ref409}{Clifford gates} \NoCaseChange{\protect\cite{cite3189}}.
\end{eczvaluelist}
\codefieldsection{Decoding}
\begin{eczvaluelist}
\item\relax The GDG sliding-window decoder \NoCaseChange{\protect\cite{cite3190}}, with a realization achieving a worst-case decoding latency of 3ms per window.
\item\relax AC decoder is faster than ordinary BP-OSD with no reduction of fidelity \NoCaseChange{\protect\cite{cite3191}}.
\item\relax Transformer-based neural-network decoder \NoCaseChange{\protect\cite{cite3192}}.
\end{eczvaluelist}
\codefieldsection{Fault Tolerance}
\begin{eczvaluelist}
\item\relax Fault-tolerant modular quantum computing framework \NoCaseChange{\protect\cite{cite3193}}.
\end{eczvaluelist}
\codefieldsection{Realizations}
\begin{eczvaluelist}
\item\relax An FPGA implementation of the Relay-BP decoder \NoCaseChange{\protect\cite{cite3194}}.
\end{eczvaluelist}
\codefieldsection{Parent}
\begin{eczvaluelist}
\item\relax
\flmRefsHyperref[eczindexfamilyrel]{code:qcga}{Bivariate bicycle (BB) code}\end{eczvaluelist}
\codefieldsection{Cousin}
\begin{eczvaluelist}
\item\relax
\flmRefsHyperref[eczindexfamilyrel]{code:surface}{Kitaev surface code} --- The gross code requires less physical and ancilla qubits (for syndrome extraction) than the surface code with the same number of logical qubits and distance. The gross code is equivalent to 8 copies of the surface code via a constant-depth Clifford circuit, and is an element of a larger family of 2D stabilizer codes \NoCaseChange{\protect\cite{cite443}}. An architecture combining the surface and gross codes was proposed in \NoCaseChange{\protect\cite{cite3195}}.
\end{eczvaluelist}
\eczhbkcontributors{ \eczhuVVA }
\endeczcode

\eczcode{stab_15_7_3}{\(\llbracket 15, 7, 3\rrbracket \) quantum Hamming code}{~\NoCaseChange{\protect\cite{cite3196,cite861,cite3197}}}
\eczhIndexCodeAliasName{stab_15_7_3}{quantum Hamming code}
\codefieldsection{Description}
Self-dual quantum Hamming code that admits permutation-based CZ logical gates.
The code is constructed using the CSS construction from the \([15,11,3]\) Hamming code and its \([15,4,8]\) dual simplex code.

In the qubit order given by the nonzero binary four-tuples \(0001,0010,\ldots,1111\), one stabilizer tableau for the code is \NoCaseChange{\protect\cite[{ID 6705229219cca60cf657a8fd}]{cite781}}
\flmMathEnvironment{align}{}{
\begin{smallmatrix}
  I & I & I & I & I & I & I & Z & Z & Z & Z & Z & Z & Z & Z \\
  I & I & I & Z & Z & Z & Z & I & I & I & I & Z & Z & Z & Z \\
  I & Z & Z & I & I & Z & Z & I & I & Z & Z & I & I & Z & Z \\
  Z & I & Z & I & Z & I & Z & I & Z & I & Z & I & Z & I & Z \\
  I & I & I & I & I & I & I & X & X & X & X & X & X & X & X \\
  I & I & I & X & X & X & X & I & I & I & I & X & X & X & X \\
  I & X & X & I & I & X & X & I & I & X & X & I & I & X & X \\
  X & I & X & I & X & I & X & I & X & I & X & I & X & I & X
\end{smallmatrix}~.
}

\codefieldsection{Transversal and Permutation-Based Gates}
\begin{eczvaluelist}
\item\relax CNOT gate because it is a CSS code.
\item\relax In a suitable logical basis, single-qubit Clifford operations applied transversally yield the corresponding \flmRefsHyperref{ref409}{Clifford gates} on each of the seven logical qubits \NoCaseChange{\protect\cite{cite791}}.
\item\relax Automorphism groups of the underlying classical codes can yield transversal \flmRefsHyperref{ref409}{Clifford gates} when combined with qubit permutations \NoCaseChange{\protect\cite[{Sec. IV.A}]{cite763}}.
\item\relax Transversal interblock \(CCZ\) gate \NoCaseChange{\protect\cite{cite787}}.
\end{eczvaluelist}
\codefieldsection{Gates}
\begin{eczvaluelist}
\item\relax CZ gates can be performed using qubit permutations, and a \(CCZ\) gate can be performed using four ancilla qubits \NoCaseChange{\protect\cite{cite791}}.
\end{eczvaluelist}
\codefieldsection{Fault Tolerance}
\begin{eczvaluelist}
\item\relax \flmRefsHyperref{ref409}{Clifford gates} can be performed fault-tolerantly using two ancillary flag qubits, and a \(CCZ\) gate can be performed using four ancilla qubits \NoCaseChange{\protect\cite{cite791}}.
\end{eczvaluelist}
\codefieldsection{Parents}
\begin{eczvaluelist}
\item\relax
\flmRefsHyperref[eczindexfamilyrel]{code:quantum_hamming_css}{\(\llbracket 2^r-1, 2^r-2r-1, 3\rrbracket \) quantum Hamming code}\item\relax
\flmRefsHyperref[eczindexfamilyrel]{code:stabilizer_over_gf4}{Hermitian qubit code} --- The \(\llbracket 15,7,3\rrbracket \) is Hermitian \NoCaseChange{\protect\cite[{ID 6705229219cca60cf657a8fd}]{cite781}}.
\end{eczvaluelist}
\codefieldsection{Cousins}
\begin{eczvaluelist}
\item\relax
\flmRefsHyperref[eczindexfamilyrel]{code:quantum_perfect}{Perfect quantum code} --- \(\llbracket 15, 7, 3\rrbracket \) quantum Hamming code is perfect as a CSS code, i.e., the number of its \(Z\)-type syndromes matches the number of \(X\)-type Pauli errors up to weight one \NoCaseChange{\protect\cite{cite791}}.
\item\relax
\flmRefsHyperref[eczindexfamilyrel]{code:stab_15_1_3}{\(\llbracket 15,1,3\rrbracket \) quantum RM code} --- \flmRefsHyperref{ref666}{Gauging out} six of the seven logical qubits of the \(\llbracket 15,7,3\rrbracket \) code yields the \(\llbracket 15,1,3\rrbracket \) code \NoCaseChange{\protect\cite{cite726}}.
\item\relax
\flmRefsHyperref[eczindexfamilyrel]{code:stab_16_6_4}{\(\llbracket 16,6,4\rrbracket \) Tesseract color code} --- The \(\llbracket 15,7,3\rrbracket \) quantum Hamming code can be obtained by puncturing the tesseract color code \NoCaseChange{\protect\cite{cite862}}.
\end{eczvaluelist}
\eczhbkcontributors{ \eczhuVVA }
\endeczcode

\eczcode{stab_15_1_3}{\(\llbracket 15,1,3\rrbracket \) quantum RM code}{~\NoCaseChange{\protect\cite{cite792,cite3198,cite690}}}
\codefieldsection{Alternative Names}
\begin{eczvaluelist}
\item\relax Tetrahedral code
\end{eczvaluelist}
\eczhIndexCodeAliasName{stab_15_1_3}{quantum RM code}
\eczhIndexCodeAliasName{stab_15_1_3}{Tetrahedral code}
\codefieldsection{Description}
A \(\llbracket 15,1,3\rrbracket \) quantum Reed-Muller code that is most easily thought of as a tetrahedral 3D color code.
It can be constructed as a CSS code from the \([15,5,8]\) punctured Reed-Muller code and its even subcode, which explains its transversal \(T^\dagger\) gate \NoCaseChange{\protect\cite{cite398}}.

This code contains 15 qubits, represented by four vertices, four face centers, six edge centers, and one body center.
The tetrahedron is cellulated into four identical polyhedron cells by connecting the body center to all four face centers, where each face center is then connected by three adjacent edge centers.
Each colored cell corresponds to a weight-eight \(X\)-check, and each face corresponds to a weight-4 \(Z\)-check.
A logical \(Z\) is any weight-3 \(Z\)-string along an edge of the entire tetrahedron.
The logical \(X\) is any weight-7 \(X\)-face of the entire tetrahedron.

The code can also be thought of as a code on all corners of a tesseract except one \NoCaseChange{\protect\cite{cite3199}}.

In the qubit order given by the nonzero binary four-tuples \(0001,0010,\ldots,1111\), one stabilizer tableau for the code is \NoCaseChange{\protect\cite[{ID 6705228b19cca60cf657a8f6}]{cite781}}
\flmMathEnvironment{align}{}{
\begin{smallmatrix}
  Z & Z & I & Z & I & I & Z & I & I & I & I & I & I & I & I \\
  Z & Z & I & I & Z & Z & I & I & I & I & I & I & I & I & I \\
  Z & Z & I & I & I & I & I & Z & I & I & Z & I & I & I & I \\
  Z & Z & I & I & I & I & I & I & Z & Z & I & I & I & I & I \\
  Z & Z & I & I & I & I & I & I & I & I & I & Z & I & I & Z \\
  Z & Z & I & I & I & I & I & I & I & I & I & I & Z & Z & I \\
  Z & I & Z & Z & I & Z & I & I & I & I & I & I & I & I & I \\
  Z & I & Z & I & I & I & I & Z & I & Z & I & I & I & I & I \\
  Z & I & Z & I & I & I & I & I & I & I & I & Z & I & Z & I \\
  Z & I & I & Z & I & I & I & Z & I & I & I & I & Z & I & I \\
  I & I & I & I & I & I & I & X & X & X & X & X & X & X & X \\
  I & I & I & X & X & X & X & I & I & I & I & X & X & X & X \\
  I & X & X & I & I & X & X & I & I & X & X & I & I & X & X \\
  X & I & X & I & X & I & X & I & X & I & X & I & X & I & X
\end{smallmatrix}~.
}
The logical \(\ket{\overline{0}}\) and \(\ket{\overline{1}}\) states are superpositions of computational-basis words of weight \(0 \bmod 8\) and \(7 \bmod 8\).

\codefieldsection{Magic}
Magic-state yield parameter \( \gamma= \log_d (n/k)\approx 2.47\) \NoCaseChange{\protect\cite{cite101}\protect\cite[{Box 2}]{cite707}}.
\codefieldsection{Encoding}
\begin{eczvaluelist}
\item\relax Fault-tolerant logical zero and logical plus state preparation using reinforcement learning \NoCaseChange{\protect\cite{cite3200}}.
\end{eczvaluelist}
\codefieldsection{Transversal and Permutation-Based Gates}
\begin{eczvaluelist}
\item\relax A transversal logical \(T\) is implemented by applying a \(T^\dagger\) gate on every qubit \NoCaseChange{\protect\cite{cite792,cite793,cite707}}. This is the smallest qubit stabilizer code with a (strongly) transversal gate outside of the \flmRefsHyperref{ref409}{Clifford group} \NoCaseChange{\protect\cite{cite794}}.
\item\relax Transversal logical \(CS\) gates between code blocks, together with transversal \(T\) and CNOT, can be used to realize IQP-like sampling architectures on a hypercube connectivity graph \NoCaseChange{\protect\cite{cite759}}.
\end{eczvaluelist}
\codefieldsection{Gates}
\begin{eczvaluelist}
\item\relax Code is often used in magic-state distillation protocols because of its transversal \(T\) gate \NoCaseChange{\protect\cite{cite690}}. In the original 15-to-1 protocol, 15 noisy \(R_{\pi/8}\ket{+}\) magic states are used to implement the transversal \(T\) by compressed gate teleportation, the resulting encoded state is post-selected by 15-qubit error detection, and decoding yields an output magic state with error probability \(O(p^3)\) when the input states have independent error probability \(p\) \NoCaseChange{\protect\cite[{Sec. 13.5.1, Protocol 13.2}]{cite398}}.
\item\relax Non-transversal logical Hadamard \NoCaseChange{\protect\cite{cite3201}}.
\end{eczvaluelist}
\codefieldsection{Decoding}
\begin{eczvaluelist}
\item\relax Decoding has been studied for the BP, BP+OSD, and AutDEC decoders \NoCaseChange{\protect\cite{cite3202}}.
\item\relax An exact ML decoder based on the Walsh-Hadamard transform was developed for error correction and gauge fixing in the alternating Steane/\(\llbracket 15,1,3\rrbracket \) Clifford + \(T\) protocol \NoCaseChange{\protect\cite{cite731}}.
\end{eczvaluelist}
\codefieldsection{Fault Tolerance}
\begin{eczvaluelist}
\item\relax A fault-tolerant universal gate set can be done via \flmRefsHyperref{ref410}{code switching} between the Steane code and the \(\llbracket 15,1,3\rrbracket \) code \NoCaseChange{\protect\cite{cite731,cite787,cite793,cite3203,cite3204}}.
\item\relax Fault-tolerant logical zero and logical plus state preparation using reinforcement learning \NoCaseChange{\protect\cite{cite3200}}.
\end{eczvaluelist}
\codefieldsection{Realizations}
\begin{eczvaluelist}
\item\relax Trapped ions: \flmRefsHyperref{ref410}{code switching} between the Steane code and the \(\llbracket 15,1,3\rrbracket \) code as well as magic-state preparation and logical Bell measurements on the Steane code realized on the 28-qubit H2-1 device by Quantinuum \NoCaseChange{\protect\cite{cite3205}}.
\item\relax MBQC with the \(\llbracket 15,1,3\rrbracket \) code has been demonstrated in neutral atom arrays by the Lukin group \NoCaseChange{\protect\cite{cite3206}}.
\end{eczvaluelist}
\codefieldsection{Parents}
\begin{eczvaluelist}
\item\relax
\flmRefsHyperref[eczindexfamilyrel]{code:tetrahedral_color}{Tetrahedral color code} --- The \(\llbracket 15,1,3\rrbracket \) code is a tetrahedral color code defined on a single tetrahedron.
\item\relax
\flmRefsHyperref[eczindexfamilyrel]{code:diagonal_clifford}{\(\llbracket 2^r-1,1,3\rrbracket \) simplex code}\item\relax
\flmRefsHyperref[eczindexfamilyrel]{code:xs_stabilizer}{XS stabilizer code} --- The \(\llbracket 15,1,3\rrbracket \) code can be viewed as an XS stabilizer code \NoCaseChange{\protect\cite[{Exam. 6.4}]{cite798}}.
\item\relax
\flmRefsHyperref[eczindexfamilyrel]{code:quantum_triorthogonal}{Triorthogonal code} --- The \(\llbracket 15, 1, 3\rrbracket \) code is a triorthogonal code \NoCaseChange{\protect\cite{cite3207}}.
\end{eczvaluelist}
\codefieldsection{Cousins}
\begin{eczvaluelist}
\item\relax
\flmRefsHyperref[eczindexfamilyrel]{code:doubled_color}{Doubled color code} --- The \(\llbracket 15,1,3\rrbracket \) code can be viewed as a (gauge-fixed) doubled color code obtained from the Steane code via the doubling transformation \NoCaseChange{\protect\cite{cite731}}.
\item\relax
\flmRefsHyperref[eczindexfamilyrel]{code:steane}{\(\llbracket 7,1,3\rrbracket \) Steane code} --- The \(\llbracket 15,1,3\rrbracket \) code can be viewed as a (gauge-fixed) doubled color code obtained from the Steane code via the doubling transformation \NoCaseChange{\protect\cite{cite731}}. A fault-tolerant universal gate set can be done via \flmRefsHyperref{ref410}{code switching} between the Steane code and the \(\llbracket 15,1,3\rrbracket \) code \NoCaseChange{\protect\cite{cite787,cite793,cite3203,cite3204,cite3208}}. An \(\llbracket 105,1,3\rrbracket \) alternative concatenation of the \(\llbracket 15,1,3\rrbracket \) and Steane codes allows for a universal gate set consisting of gates that are close to transversal \NoCaseChange{\protect\cite{cite3209,cite775}}.
\item\relax
\flmRefsHyperref[eczindexfamilyrel]{code:concatenated_steane}{Concatenated Steane code} --- The \(\llbracket 105,1\rrbracket \) concatenation of the \(\llbracket 15,1,3\rrbracket \) and Steane codes allows for a universal gate set consisting of gates that are transversal w.r.t. to two different partitions \NoCaseChange{\protect\cite{cite3209,cite775}}.
\item\relax
\flmRefsHyperref[eczindexfamilyrel]{code:qubit_concatenated}{Concatenated qubit code} --- The concatenated \(\llbracket 15,1,3\rrbracket \) code has a \flmRefsHyperref{ref3210}{measurement threshold} less than one \NoCaseChange{\protect\cite{cite3211}}.
\item\relax
\flmRefsHyperref[eczindexfamilyrel]{code:quantum_lego}{Tensor-network code} --- The \(\llbracket 15,1,3\rrbracket \) code serves as the local tensor in a holographic quantum-Lego code on a \(\{15,4\}\) tiling that supports a transversal non-Clifford operator \NoCaseChange{\protect\cite{cite2868}}.
\item\relax
\flmRefsHyperref[eczindexfamilyrel]{code:holographic_tensor}{Holographic tensor-network code} --- The \(\llbracket 15,1,3\rrbracket \) code serves as the local tensor in a holographic quantum-Lego code on a \(\{15,4\}\) tiling that supports a transversal non-Clifford operator \NoCaseChange{\protect\cite{cite2868}}.
\item\relax
\flmRefsHyperref[eczindexfamilyrel]{code:cluster_state}{Cluster-state code} --- MBQC with the \(\llbracket 15,1,3\rrbracket \) code has been demonstrated in neutral atom arrays by the Lukin group \NoCaseChange{\protect\cite{cite3206}}.
\item\relax
\flmRefsHyperref[eczindexfamilyrel]{code:stab_10_1_2}{\(\llbracket 10,1,2\rrbracket \) Vasmer-Kubica code} --- The \(\llbracket 10,1,2\rrbracket \) code is obtained by morphing the \(\llbracket 15,1,3\rrbracket \) code on a region whose child code is the \(\llbracket 8,3,2\rrbracket \) smallest interesting color code \NoCaseChange{\protect\cite{cite687}}.
\item\relax
\flmRefsHyperref[eczindexfamilyrel]{code:stab_15_7_3}{\(\llbracket 15, 7, 3\rrbracket \) quantum Hamming code} --- \flmRefsHyperref{ref666}{Gauging out} six of the seven logical qubits of the \(\llbracket 15,7,3\rrbracket \) code yields the \(\llbracket 15,1,3\rrbracket \) code \NoCaseChange{\protect\cite{cite726}}.
\item\relax
\flmRefsHyperref[eczindexfamilyrel]{code:stab_8_1_2}{\(\llbracket 8,1,2\rrbracket \) Shen-Wang-Cao code} --- The \(\llbracket 8,1,2\rrbracket \) code is an XP-regular code that can be obtained via the XP stabilizer formalism applied to the \(\llbracket 15,1,3\rrbracket \) Reed-Muller code \NoCaseChange{\protect\cite{cite786}}.
\item\relax
\flmRefsHyperref[eczindexfamilyrel]{code:stab_8_3_2}{\(\llbracket 8,3,2\rrbracket \) Smallest interesting color code} --- The \(\llbracket 8,3,2\rrbracket \) code can be obtained from a subset of physical qubits of the \(\llbracket 15,1,3\rrbracket \) code \NoCaseChange{\protect\cite{cite687}}.
\item\relax
\flmRefsHyperref[eczindexfamilyrel]{code:sslp}{Subset-Sum-Linear-Programming (SS-LP) code} --- The \(\llparenthesis 7,2,3\rrparenthesis \) SS-LP code realizes the \(T\) gate transversally, but requires fewer qubits than the \(\llbracket 15,1,3\rrbracket \) quantum RM code.
\end{eczvaluelist}
\eczhbkcontributors{ Remmy Zen, Balint Pato, Qingfeng (Kee) Wang, \eczhuVVA }
\endeczcode

\eczcode{quantum_dodecahedron}{\(\llbracket 16,4,3\rrbracket \) dodecahedral code}{~\NoCaseChange{\protect\cite{cite858}}}
\eczhIndexCodeAliasName{quantum_dodecahedron}{dodecahedral code}
\codefieldsection{Description}
A \(\llbracket 16,4,3\rrbracket \) non-CSS qubit stabilizer code whose \flmRefsHyperref{ref857}{encoder-respecting form} is the graph of vertices of a dodecahedron \NoCaseChange{\protect\cite{cite858}}.

A stabilizer tableau for the code, obtained from the graph-to-tableau map applied to the dodecahedral graph, is \NoCaseChange{\protect\cite[{Fig. 19 and Eq. (10)}]{cite858}}
\flmMathEnvironment{align}{}{
\begin{smallmatrix}
  I & Z & X & Z & I & I & I & I & Z & I & I & I & I & I & I & I \\
  I & I & Z & X & X & Z & I & I & I & Z & I & Z & I & I & I & I \\
  I & I & I & I & Z & X & X & Z & I & I & I & I & Z & I & Z & I \\
  Z & I & I & I & I & I & Z & X & I & I & I & I & I & I & I & Z \\
  X & Z & Z & I & I & I & I & Z & X & I & Z & I & I & I & I & I \\
  Z & X & Z & Z & I & I & I & I & I & X & I & Z & I & I & I & I \\
  I & I & Z & X & I & I & I & I & Z & Z & X & I & Z & I & I & I \\
  I & I & I & I & Z & I & I & I & I & Z & I & X & I & Z & I & I \\
  I & I & I & I & I & Z & I & I & I & I & Z & I & X & I & Z & I \\
  I & I & I & I & Z & X & I & I & I & I & I & Z & Z & X & I & Z \\
  X & Z & I & I & I & I & Z & Z & I & I & I & I & Z & I & X & I \\
  Z & X & Z & I & I & I & I & Z & I & I & I & I & I & Z & I & X
\end{smallmatrix}~.
}

\codefieldsection{Protection}
The code saturates the paper's graph degree upper bound on distance, attaining distance 3 on a degree-3 graph \NoCaseChange{\protect\cite{cite858}}.
\codefieldsection{Rate}
The code has a rate of \(1/4\), higher than that of the five-qubit perfect code.
\codefieldsection{Encoding}
\begin{eczvaluelist}
\item\relax Encoding circuits of depth 5 are possible and optimal; a direct graph-based construction gives depth 6 before further optimization \NoCaseChange{\protect\cite{cite858}}.
\end{eczvaluelist}
\codefieldsection{Parent}
\begin{eczvaluelist}
\item\relax
\flmRefsHyperref[eczindexfamilyrel]{code:small_distance_qubit_stabilizer}{Small-distance qubit stabilizer code}\end{eczvaluelist}
\codefieldsection{Cousin}
\begin{eczvaluelist}
\item\relax
\flmRefsHyperref[eczindexfamilyrel]{code:dodecahedron}{Dodecahedron code} --- The \flmRefsHyperref{ref857}{encoder-respecting form} of the \(\llbracket 16,4,3\rrbracket \) dodecahedral code is the graph of vertices of a dodecahedron \NoCaseChange{\protect\cite{cite858}}.
\end{eczvaluelist}
\eczhbkcontributors{ \eczhuVVA }
\endeczcode

\eczcode{stab_16_6_4}{\(\llbracket 16,6,4\rrbracket \) Tesseract color code}{~\NoCaseChange{\protect\cite{cite3212,cite101,cite862,cite2362}}}
\codefieldsection{Alternative Names}
\begin{eczvaluelist}
\item\relax \(\llbracket 16,6,4\rrbracket \) hypercube code
\item\relax \(\llbracket 16,6,4\rrbracket \) 4D color code
\end{eczvaluelist}
\eczhIndexCodeAliasName{stab_16_6_4}{Tesseract color code}
\eczhIndexCodeAliasName{stab_16_6_4}{\(\llbracket 16,6,4\rrbracket \) hypercube code}
\eczhIndexCodeAliasName{stab_16_6_4}{\(\llbracket 16,6,4\rrbracket \) 4D color code}
\codefieldsection{Description}
A (hyperbolic self-dual CSS) 4D color code defined on a tesseract, with stabilizer generators of both types supported on each cube. 
A \(\llbracket 16,4,2,4\rrbracket \) tesseract subsystem code can be obtained from this code by using two logical qubits as gauge qubits \NoCaseChange{\protect\cite{cite483}}.

A stabilizer tableau for the code is \NoCaseChange{\protect\cite[{ID 67d2cf9965e067195651cfbe}]{cite781}}
\flmMathEnvironment{align}{}{
\begin{smallmatrix}
  X & X & X & X & X & X & X & X & X & X & X & X & X & X & X & X \\
  I & X & I & X & I & X & I & X & I & X & I & X & I & X & I & X \\
  I & I & X & X & I & I & X & X & I & I & X & X & I & I & X & X \\
  I & I & I & I & X & X & X & X & I & I & I & I & X & X & X & X \\
  I & I & I & I & I & I & I & I & X & X & X & X & X & X & X & X \\
  Z & Z & Z & Z & Z & Z & Z & Z & Z & Z & Z & Z & Z & Z & Z & Z \\
  I & Z & I & Z & I & Z & I & Z & I & Z & I & Z & I & Z & I & Z \\
  I & I & Z & Z & I & I & Z & Z & I & I & Z & Z & I & I & Z & Z \\
  I & I & I & I & Z & Z & Z & Z & I & I & I & I & Z & Z & Z & Z \\
  I & I & I & I & I & I & I & I & Z & Z & Z & Z & Z & Z & Z & Z
\end{smallmatrix}~.
}

\codefieldsection{Transversal and Permutation-Based Gates}
\begin{eczvaluelist}
\item\relax Global transversal \(S\) implements a logical circuit composed of \(CZ\) and \(Z\) gates \NoCaseChange{\protect\cite{cite753,cite724}}
\item\relax Transversal Hadamard can be chosen to swap three pairs of logical qubits, allowing even-weight Hadamard-product measurements in magic-state distillation protocols \NoCaseChange{\protect\cite[{Sec. III}]{cite101}}.
\end{eczvaluelist}
\codefieldsection{Gates}
\begin{eczvaluelist}
\item\relax Using this code as a hyperbolic inner code yields quartic magic-state distillation on six outputs; pipelining it with \(\llbracket 4,2,2\rrbracket \) inner-code checks lowers the non-Clifford cost from 390 to 246 noisy \(T\) gates \NoCaseChange{\protect\cite[{Sec. I.B.2}]{cite101}}.
\end{eczvaluelist}
\codefieldsection{Decoding}
\begin{eczvaluelist}
\item\relax Post-selected fault-tolerant syndrome extraction \NoCaseChange{\protect\cite{cite862,cite2362}}.
\end{eczvaluelist}
\codefieldsection{Fault Tolerance}
\begin{eczvaluelist}
\item\relax Post-selected fault-tolerant syndrome extraction \NoCaseChange{\protect\cite{cite862,cite2362}}.
\end{eczvaluelist}
\codefieldsection{Realizations}
\begin{eczvaluelist}
\item\relax Trapped-ion devices: logical graph and GHZ states of up to 12 logical qubits constructed using three copies of the \(\llbracket 16,4,2,4\rrbracket \) tesseract subsystem code, along with five rounds of post-selected fault-tolerant error correction in a device by Quantinuum \NoCaseChange{\protect\cite{cite483}}.
\item\relax Neutral atom arrays: deep circuits and 1D-cluster-state creation using 96 logical qubits and hundreds of logical teleportations by the Lukin group \NoCaseChange{\protect\cite{cite3206}}.
\end{eczvaluelist}
\codefieldsection{Parents}
\begin{eczvaluelist}
\item\relax
\flmRefsHyperref[eczindexfamilyrel]{code:color}{Color code} --- The tesseract color code is a 4D color code defined on a tesseract \NoCaseChange{\protect\cite{cite3212,cite101,cite862,cite2362}}.
\item\relax
\flmRefsHyperref[eczindexfamilyrel]{code:quantum_reed_muller}{Quantum Reed-Muller (RM) code} --- The tesseract color code is constructed from the \([16,5,8]\) first-order RM code via the CSS construction \NoCaseChange{\protect\cite{cite101,cite757}}.
\item\relax
\flmRefsHyperref[eczindexfamilyrel]{code:4d_stabilizer}{4D lattice stabilizer code} --- The tesseract color code is a 4D color code defined on a tesseract \NoCaseChange{\protect\cite{cite3212,cite101,cite862,cite2362}}.
\item\relax
\flmRefsHyperref[eczindexfamilyrel]{code:self_dual_css}{Self-dual CSS code}\item\relax
\flmRefsHyperref[eczindexfamilyrel]{code:small_distance_qubit_stabilizer}{Small-distance qubit stabilizer code}\end{eczvaluelist}
\codefieldsection{Cousins}
\begin{eczvaluelist}
\item\relax
\flmRefsHyperref[eczindexfamilyrel]{code:biorthogonal}{\([2^m,m+1,2^{m-1}]\) First-order RM code} --- The tesseract color code is constructed from the \([16,5,8]\) first-order RM code via the CSS construction \NoCaseChange{\protect\cite{cite101,cite757}}.
\item\relax
\flmRefsHyperref[eczindexfamilyrel]{code:stab_15_7_3}{\(\llbracket 15, 7, 3\rrbracket \) quantum Hamming code} --- The \(\llbracket 15,7,3\rrbracket \) quantum Hamming code can be obtained by puncturing the tesseract color code \NoCaseChange{\protect\cite{cite862}}.
\item\relax
\flmRefsHyperref[eczindexfamilyrel]{code:stab_8_3_2}{\(\llbracket 8,3,2\rrbracket \) Smallest interesting color code} --- Applying CNOT gates to the tesseract color code disentangles it into two \(\llbracket 8,3,2\rrbracket \) color codes \NoCaseChange{\protect\cite{cite483}}.
\item\relax
\flmRefsHyperref[eczindexfamilyrel]{code:stab_4_2_2}{\(\llbracket 4,2,2\rrbracket \) Four-qubit code} --- The \(\llbracket 16,4,2,4\rrbracket \) tesseract subsystem color code with particular gauge fixing can be obtained from four copies of the \(\llbracket 4,2,2\rrbracket \) code \NoCaseChange{\protect\cite{cite483}}.
\item\relax
\flmRefsHyperref[eczindexfamilyrel]{code:hypercube}{Hypercube code} --- Stabilizer generators of both types of the tesseract color code are supported on each cube of a tesseract \NoCaseChange{\protect\cite{cite862,cite2362}}.
\end{eczvaluelist}
\eczhbkcontributors{ Nolan Coble, \eczhuVVA }
\endeczcode

\eczcode{stab_17_1_5}{\(\llbracket 17,1,5\rrbracket \) 4.8.8 color code}{~\NoCaseChange{\protect\cite{cite710}}}
\eczhIndexCodeAliasName{stab_17_1_5}{4.8.8 color code}
\codefieldsection{Description}
Seventeen-qubit doubly even 2D color code that admits a transversal implementation of the logical Clifford group.
The smallest distance-five CSS code has \(n=17\) \NoCaseChange{\protect\cite{cite444}}.
It is also a normal self-dual CSS code whose transversal Hadamard acts logically, making it suitable as an inner code for fifth-order magic-state distillation \NoCaseChange{\protect\cite{cite101}}.

A stabilizer tableau for the code is \NoCaseChange{\protect\cite[{ID 672e4f2a4f33a38b84d76b7c}]{cite781}}
\flmMathEnvironment{align}{}{
\begin{smallmatrix}
  X & I & I & I & I & I & I & I & I & X & I & I & X & I & I & X & I \\
  I & X & I & I & I & I & I & I & I & X & I & I & X & I & I & I & X \\
  I & I & X & I & I & I & I & I & I & I & I & X & I & X & X & I & I \\
  I & I & I & X & I & I & I & I & X & X & I & I & X & I & I & I & I \\
  I & I & I & I & X & I & I & I & X & I & X & I & X & X & X & X & X \\
  I & I & I & I & I & X & I & I & I & I & X & X & I & X & I & I & I \\
  I & I & I & I & I & I & X & I & X & X & X & I & I & X & X & X & X \\
  I & I & I & I & I & I & I & X & I & I & X & X & I & I & X & I & I \\
  Z & I & I & I & I & I & I & I & I & Z & I & I & Z & I & I & Z & I \\
  I & Z & I & I & I & I & I & I & I & Z & I & I & Z & I & I & I & Z \\
  I & I & Z & I & I & I & I & I & I & I & I & Z & I & Z & Z & I & I \\
  I & I & I & Z & I & I & I & I & Z & Z & I & I & Z & I & I & I & I \\
  I & I & I & I & Z & I & I & I & Z & I & Z & I & Z & Z & Z & Z & Z \\
  I & I & I & I & I & Z & I & I & I & I & Z & Z & I & Z & I & I & I \\
  I & I & I & I & I & I & Z & I & Z & Z & Z & I & I & Z & Z & Z & Z \\
  I & I & I & I & I & I & I & Z & I & I & Z & Z & I & I & Z & I & I
\end{smallmatrix}~.
}

\codefieldsection{Encoding}
\begin{eczvaluelist}
\item\relax Logical zero state preparation using reinforcement learning \NoCaseChange{\protect\cite{cite3200}}.
\end{eczvaluelist}
\codefieldsection{Transversal and Permutation-Based Gates}
\begin{eczvaluelist}
\item\relax Transversal implementation of the logical Clifford group.
\end{eczvaluelist}
\codefieldsection{Gates}
\begin{eczvaluelist}
\item\relax As an inner code for magic-state distillation, it yields a 69-to-1 fifth-order routine, or a 49-to-1 pipelined routine when preceded by the Steane code \NoCaseChange{\protect\cite[{Sec. I.A.2}]{cite101}}.
\end{eczvaluelist}
\codefieldsection{Parents}
\begin{eczvaluelist}
\item\relax
\flmRefsHyperref[eczindexfamilyrel]{code:488_color}{Square-octagon (4.8.8) color code} --- The \(\llbracket 17,1,5\rrbracket \) color code can be defined on the 4.8.8 tiling \NoCaseChange{\protect\cite{cite731}}.
\item\relax
\flmRefsHyperref[eczindexfamilyrel]{code:quantum_divisible}{Quantum divisible code}\item\relax
\flmRefsHyperref[eczindexfamilyrel]{code:self_dual_css}{Self-dual CSS code}\item\relax
\flmRefsHyperref[eczindexfamilyrel]{code:small_distance_qubit_stabilizer}{Small-distance qubit stabilizer code} --- The smallest distance-five CSS code has \(n=17\) \NoCaseChange{\protect\cite{cite444}}.
\end{eczvaluelist}
\codefieldsection{Cousin}
\begin{eczvaluelist}
\item\relax
\flmRefsHyperref[eczindexfamilyrel]{code:stab_49_1_5}{\(\llbracket 49,1,5\rrbracket \) triorthogonal code} --- The \(\llbracket 49,1,5\rrbracket \) triorthogonal code can be viewed as a (gauge-fixed) doubled color code obtained from the \(\llbracket 17,1,5\rrbracket \) 4.8.8 color code via the doubling transformation \NoCaseChange{\protect\cite{cite731}}.
\end{eczvaluelist}
\eczhbkcontributors{ \eczhuVVA }
\endeczcode

\eczcode{ampdamp_stabilizer}{\(\llbracket 2(m+1),m,2\rrbracket \) single-loss AD code}{~\NoCaseChange{\protect\cite{cite1187}}}
\eczhIndexCodeAliasName{ampdamp_stabilizer}{single-loss AD code}
\codefieldsection{Description}
A member of a class of \(\llbracket 2(m+1),m,2\rrbracket \) CSS codes for \(m\geq 1\) that generalizes the \(\llbracket 4,1,2\rrbracket \) approximate amplitude-damping code of Ref. \NoCaseChange{\protect\cite{cite859}}.
Its \(Z\)-type generators are \(m+1\) pairwise products, with each qubit participating in only one check; the single \(X\)-type generator is the all-\(X\) string.

\codefieldsection{Protection}
The \(\llbracket 2(m+1),m\rrbracket \) family is designed to correct a single \flmRefsHyperref{ref498}{AD} error \NoCaseChange{\protect\cite{cite1187}}.

\codefieldsection{Parents}
\begin{eczvaluelist}
\item\relax
\flmRefsHyperref[eczindexfamilyrel]{code:qubit_css}{Qubit CSS code}\item\relax
\flmRefsHyperref[eczindexfamilyrel]{code:self_complementary}{Self-complementary qubit code} --- The \(\llbracket 2(m+1),m,2\rrbracket \) single-loss AD code is self-complementary, with the all-\(X\) stabilizer enforcing the complement-pair structure of the codewords \NoCaseChange{\protect\cite{cite1187}}.
\item\relax
\flmRefsHyperref[eczindexfamilyrel]{code:small_distance_qubit_stabilizer}{Small-distance qubit stabilizer code}\end{eczvaluelist}
\codefieldsection{Child}
\begin{eczvaluelist}
\item\relax
\flmRefsHyperref[eczindexfamilyrel]{code:css_4_1_2}{\(\llbracket 4,1,2\rrbracket \) Leung-Nielsen-Chuang-Yamamoto (LNCY) code} --- The \(\llbracket 4,1,2\rrbracket \) LNCY code (approximately) corrects a single \flmRefsHyperref{ref498}{AD} error \NoCaseChange{\protect\cite{cite859}} and is the smallest member of the amplitude-damping stabilizer family of Ref. \NoCaseChange{\protect\cite{cite1187}}.
\end{eczvaluelist}
\eczhbkcontributors{ \eczhuVVA }
\endeczcode

\eczcode{single_qubit_clifford}{\(\llbracket 2^{2r-1}-1,1,2^r-1\rrbracket \) quantum punctured RM code}{~\NoCaseChange{\protect\cite[{Ch. 7}]{cite2764}}}
\eczhIndexCodeAliasName{single_qubit_clifford}{quantum punctured RM code}
\codefieldsection{Description}
A quantum Reed-Muller code constructed from a punctured self-dual RM code and its even subcode for \(r \geq 2\).

\codefieldsection{Transversal and Permutation-Based Gates}
\begin{eczvaluelist}
\item\relax All single-qubit gates in the \flmRefsHyperref{ref409}{Clifford group}.
\end{eczvaluelist}
\codefieldsection{Parents}
\begin{eczvaluelist}
\item\relax
\flmRefsHyperref[eczindexfamilyrel]{code:quantum_reed_muller}{Quantum Reed-Muller (RM) code} --- The \(\llbracket 2^{2r-1}-1,1,2^r-1\rrbracket \) quantum punctured RM codes are special cases of the \(\llbracket \sum_{i=w+1}^m \binom{m}{i}, \sum_{i=0}^{w} \binom{m}{i}, \sum_{i=w+1}^{r+1} \binom{r+1}{i}\rrbracket \) family for \(m \to 2r-1\), \(w \to 0\), and \(r \to r-1\).
\item\relax
\flmRefsHyperref[eczindexfamilyrel]{code:self_dual_css}{Self-dual CSS code} --- Puncturing a self-dual RM code yields a classical punctured RM code whose dual is its even subcode; applying the CSS construction to the even subcode yields this self-dual CSS family \NoCaseChange{\protect\cite[{Sec. 7.15.3}]{cite2764}}.
\end{eczvaluelist}
\codefieldsection{Child}
\begin{eczvaluelist}
\item\relax
\flmRefsHyperref[eczindexfamilyrel]{code:steane}{\(\llbracket 7,1,3\rrbracket \) Steane code}\end{eczvaluelist}
\eczhbkcontributors{ Ian Teixeira, \eczhuVVA }
\endeczcode

\eczcode{majorana_hamming}{\(\llbracket 2^{m-1},2^{m-1}-m-1,4\rrbracket _{f}\) Hamming Majorana code}{~\NoCaseChange{\protect\cite{cite3212}}}
\eczhIndexCodeAliasName{majorana_hamming}{Hamming Majorana code}
\codefieldsection{Description}
A member of the \(\llbracket 2^{m-1},2^{m-1}-m-1,4\rrbracket _{f}\) family of Majorana stabilizer codes for \(m \geq 3\) constructed from a self-orthogonal first-order RM code (whose dual is the extended Hamming code).
A shortened \(\llbracket 2^{m-1}-1,2^{m-1}-m-2,3\rrbracket _{f}\) version can also be defined \NoCaseChange{\protect\cite[{Prop. 2.5.1}]{cite564}}.
The logical subspace of the \(\llbracket 8,3,4\rrbracket _{f}\) Hamming Majorana code is a Cartan subspace of the \(E_8\) Lie algebra \NoCaseChange{\protect\cite{cite565}}.

\codefieldsection{Parents}
\begin{eczvaluelist}
\item\relax
\flmRefsHyperref[eczindexfamilyrel]{code:majorana_reed_muller}{RM Majorana code} --- A Hamming Majorana code is constructed from a first-order RM code (whose dual is the extended Hamming code).
\item\relax
\flmRefsHyperref[eczindexfamilyrel]{code:small_distance_qubit_stabilizer}{Small-distance qubit stabilizer code}\end{eczvaluelist}
\codefieldsection{Cousins}
\begin{eczvaluelist}
\item\relax
\flmRefsHyperref[eczindexfamilyrel]{code:extended_hamming}{\([2^m,2^m-m-1,4]\) Extended Hamming code} --- A Hamming Majorana code is constructed from a first-order RM code (whose dual is the extended Hamming code).
\item\relax
\flmRefsHyperref[eczindexfamilyrel]{code:biorthogonal}{\([2^m,m+1,2^{m-1}]\) First-order RM code} --- A Hamming Majorana code is constructed from a first-order RM code (whose dual is the extended Hamming code).
\item\relax
\flmRefsHyperref[eczindexfamilyrel]{code:eeight}{\(E_8\) Gosset lattice} --- The logical subspace of the \(\llbracket 8,3,4\rrbracket _{f}\) Hamming Majorana code is a Cartan subspace of the \(E_8\) Lie algebra \NoCaseChange{\protect\cite{cite565}}.
\item\relax
\flmRefsHyperref[eczindexfamilyrel]{code:stab_4_2_2}{\(\llbracket 4,2,2\rrbracket \) Four-qubit code} --- The \(\llbracket 8,3,4\rrbracket _{f}\) Hamming Majorana code is a Majorana stabilizer code obtained by combining two four-qubit codes \NoCaseChange{\protect\cite{cite565}}.
\item\relax
\flmRefsHyperref[eczindexfamilyrel]{code:majorana_color}{Majorana color code} --- The \(\llbracket 8,3,4\rrbracket _{f}\) Hamming Majorana code can replace stacks of three tetrons in a 4.8.8 Majorana surface code to yield an order-3 Majorana color code with maximum stabilizer weight \(16\) \NoCaseChange{\protect\cite[{Table I}]{cite402}}.
\end{eczvaluelist}
\eczhbkcontributors{ \eczhuVVA }
\endeczcode

\eczcode{hypercube_quantum}{\(\llbracket 2^D,D,2\rrbracket \) hypercube quantum code}{~\NoCaseChange{\protect\cite{cite422,cite797}\protect\cite[{Exam. 3}]{cite687}}}
\codefieldsection{Alternative Names}
\begin{eczvaluelist}
\item\relax Hyperoctahedron code
\item\relax Hyperoctahedron color code
\end{eczvaluelist}
\eczhIndexCodeAliasName{hypercube_quantum}{hypercube quantum code}
\eczhIndexCodeAliasName{hypercube_quantum}{Hyperoctahedron code}
\eczhIndexCodeAliasName{hypercube_quantum}{Hyperoctahedron color code}
\codefieldsection{Description}
Member of a family of codes defined by placing qubits on a \(D\)-dimensional hypercube, \(Z\)-stabilizers on all two-dimensional faces, and an \(X\)-stabilizer on all vertices.
These codes realize gates at the \((D-1)\)-st level of the Clifford hierarchy.
The measured physical bit string can be post-processed into both a logical output string and stabilizer checks, enabling end-of-circuit error detection directly from classical samples \NoCaseChange{\protect\cite{cite759}}.
Puncturing the \(\llbracket 2^D,D,2\rrbracket \) hypercube quantum code yields the \(\llbracket 2^D-1,D,2\rrbracket \) punctured-hypercube family.

Higher-distance generalizations include a \(\llbracket 2^{2D},D,4\rrbracket \) hyperoctahedron family and a \(\llbracket 2^D(2^D+1),D,4\rrbracket \) family built from distance-two \(D\)-dimensional toric/surface-code blocks \NoCaseChange{\protect\cite{cite759}}.
Various other concatenations give families with increasing distance (see cousins).

In the color-code picture, they arise from hypercube-like lattices with no bulk qubits and opposite boundaries carrying the same color; after local Clifford disentangling, the transversal \(\widetilde{R_D}\) operator acts as a logical \(C^{D-1}Z\) gate on \(D\) decoupled distance-two toric/surface-code factors \NoCaseChange{\protect\cite{cite422}}.

\codefieldsection{Protection}
The code detects a single general error but has an \(X\)-distance \(d_X = 4\).
In encoded IQP sampling, this allows error detection without intermediate measurements by postselecting on the final stabilizer data extracted from measurement samples \NoCaseChange{\protect\cite{cite759}}.

\codefieldsection{Transversal and Permutation-Based Gates}
\begin{eczvaluelist}
\item\relax  \(CZ\), \(CCZ\), and generalized \(CZ\) gates at the \((D-1)\)-st level of the Clifford hierarchy \NoCaseChange{\protect\cite{cite797}\protect\cite[{Exam. 6.10}]{cite798}}. CNOT and SWAP gates can be realized by qubit permutations \NoCaseChange{\protect\cite{cite759}}.
\end{eczvaluelist}
\codefieldsection{Notes}
\begin{eczvaluelist}
\item\relax Degree-\(D\) instantaneous quantum polynomial (IQP) circuits \NoCaseChange{\protect\cite{cite3213}} can be realized on hypercube quantum codes in a hardware-efficient way; Ref. \NoCaseChange{\protect\cite{cite759}} proposes hypercube IQP (hIQP) circuits on a hypercube connectivity graph.
\item\relax For \(D=4\), Bell sampling on two copies of degree-\(4\) IQP circuits encoded in the \(\llbracket 16,4,2\rrbracket \) member is proposed as an efficiently classically verifiable quantum-advantage experiment \NoCaseChange{\protect\cite{cite759}}.
\end{eczvaluelist}
\codefieldsection{Parents}
\begin{eczvaluelist}
\item\relax
\flmRefsHyperref[eczindexfamilyrel]{code:ball_color}{Ball code} --- \(\llbracket 2^D,D,2\rrbracket \) hypercube quantum codes can be thought of as small ball codes constructed from hyperoctahedra \NoCaseChange{\protect\cite[{Exam. 3}]{cite687}}, or on lattices with no bulk qubits and cubic boundaries  \NoCaseChange{\protect\cite{cite422,cite797}}.
\item\relax
\flmRefsHyperref[eczindexfamilyrel]{code:quantum_reed_muller}{Quantum Reed-Muller (RM) code} --- \(\llbracket 2^D,D,2\rrbracket \) hypercube quantum codes are special cases of the \(\llbracket 2^m,{m \choose r}, 2^r\rrbracket \) quantum RM codes for \(m=D\) and \(r=1\) \NoCaseChange{\protect\cite{cite754,cite755,cite756,cite757}\protect\cite[{Exam. 8}]{cite724}}.
\item\relax
\flmRefsHyperref[eczindexfamilyrel]{code:phantom}{Phantom code} --- The \(\llbracket 2^D,D,2\rrbracket \) hypercube quantum codes are phantom codes: all ordered-pair in-block logical CNOT gates can be implemented by physical-qubit permutations \NoCaseChange{\protect\cite{cite514}}. The punctured hypercube family is unique among binary CSS phantom codes saturating \(n\geq 2^k-1\) for \(k=3\) and \(k\geq 5\) \NoCaseChange{\protect\cite[{Thm. 4}]{cite723}}.
\item\relax
\flmRefsHyperref[eczindexfamilyrel]{code:self_complementary}{Self-complementary qubit code} --- A basis of hypercube quantum codewords of the form \(|c\rangle+|\overline{c}\rangle\) can be obtained via the \flmRefsHyperref{code:qubit_css}{qubit CSS codeword construction} since their sole \(X\)-type stabilizer generator acts on all qubits.
\end{eczvaluelist}
\codefieldsection{Children}
\begin{eczvaluelist}
\item\relax
\flmRefsHyperref[eczindexfamilyrel]{code:stab_4_2_2}{\(\llbracket 4,2,2\rrbracket \) Four-qubit code} --- The \(\llbracket 4,2,2\rrbracket \) code is a hypercube code for \(D=2\).
\item\relax
\flmRefsHyperref[eczindexfamilyrel]{code:stab_8_3_2}{\(\llbracket 8,3,2\rrbracket \) Smallest interesting color code} --- The \(\llbracket 8,3,2\rrbracket \) code is a hypercube code for \(D=3\).
\end{eczvaluelist}
\codefieldsection{Cousins}
\begin{eczvaluelist}
\item\relax
\flmRefsHyperref[eczindexfamilyrel]{code:xp_stabilizer}{XP stabilizer code} --- The \(D\)th hypercube quantum code can be viewed as an XP stabilizer code with precision \(N = 2^D\) \NoCaseChange{\protect\cite[{Exam. 6.10}]{cite798}}.
\item\relax
\flmRefsHyperref[eczindexfamilyrel]{code:hypercube}{Hypercube code} --- \(\llbracket 2^D,D,2\rrbracket \) hypercube quantum code qubits are placed on vertices of a \(D\)-cube.
\item\relax
\flmRefsHyperref[eczindexfamilyrel]{code:quantum_repetition}{Quantum repetition code} --- The hypercube quantum code can be concatenated with a two-qubit quantum repetition code to yield a \(\llbracket 2^{D+1},D,4\rrbracket \) error-detecting code family \NoCaseChange{\protect\cite{cite759}}.
\item\relax
\flmRefsHyperref[eczindexfamilyrel]{code:higher_dimensional_surface}{Homological code} --- The hypercube quantum code can be concatenated with \(D\) distance-two \(D\)-dimensional toric/surface-code blocks to yield a \(\llbracket 2^D(2^D+1),D,4\rrbracket \) error-correcting family that admits a transversal implementation of the logical \(C^{D-1}Z\) gate \NoCaseChange{\protect\cite{cite759}}.
\item\relax
\flmRefsHyperref[eczindexfamilyrel]{code:qubit_concatenated}{Concatenated qubit code} --- The hypercube quantum code can be concatenated with a two-qubit quantum repetition code to yield a \(\llbracket 2^{D+1},D,4\rrbracket \) error-detecting code family \NoCaseChange{\protect\cite{cite759}}.
It can also be concatenated with \(D\) distance-two \(D\)-dimensional toric/surface-code blocks to yield a \(\llbracket 2^D(2^D+1),D,4\rrbracket \) error-correcting code family that admits a transversal implementation of the logical \(C^{D-1}Z\) gate \NoCaseChange{\protect\cite{cite759}}.

\item\relax
\flmRefsHyperref[eczindexfamilyrel]{code:morphed_diagonal_clifford}{\(\llbracket 2^r+r-1,1,2\rrbracket \) morphed simplex code} --- The \(\llbracket 2^r+r-1,1,2\rrbracket \) morphed simplex code is obtained by morphing the \(\llbracket 2^{r+1}-1,1,3\rrbracket \) simplex code on a region whose child code is a \(\llbracket 2^r,r,2\rrbracket \) hypercube code \NoCaseChange{\protect\cite[{Appx. C}]{cite687}}.
\end{eczvaluelist}
\eczhbkcontributors{ \eczhuVVA }
\endeczcode

\eczcode{quantum_hamming_css}{\(\llbracket 2^r-1, 2^r-2r-1, 3\rrbracket \) quantum Hamming code}{~\NoCaseChange{\protect\cite{cite861}}}
\eczhIndexCodeAliasName{quantum_hamming_css}{quantum Hamming code}
\codefieldsection{Description}
Member of a family of self-dual CSS codes constructed from \([2^r-1,2^r-r-1,3]=C_X=C_Z\) Hamming codes and their duals, the simplex codes.
The code's stabilizer generator matrix blocks \(H_{X}\) and \(H_{Z}\) are both the generator matrix for a simplex code.
The weight of each stabilizer generator is \(2^{r-1}\).

\codefieldsection{Protection}
Protects against any single qubit error.
\codefieldsection{Transversal and Permutation-Based Gates}
\begin{eczvaluelist}
\item\relax Pauli, Hadamard, and CNOT gates.
\end{eczvaluelist}
\codefieldsection{Decoding}
\begin{eczvaluelist}
\item\relax Efficient decoder \NoCaseChange{\protect\cite{cite3214}}.
\end{eczvaluelist}
\codefieldsection{Fault Tolerance}
\begin{eczvaluelist}
\item\relax Syndrome measurement can be done with two ancillary flag qubits \NoCaseChange{\protect\cite{cite3215}}.
\item\relax Concatenating a growing sequence of quantum Hamming codes yields fault-tolerant quantum computation with constant space overhead and quasi-polylogarithmic time overhead \NoCaseChange{\protect\cite{cite3214}}.
\item\relax Concatenating quantum Hamming codes on top of the \(\llbracket 4,2,2\rrbracket \) and \(C_6\) codes yields fault-tolerant quantum computation with constant space and quasi-polylogarithmic time overheads \NoCaseChange{\protect\cite{cite3216}}. In the optimized protocol of Ref. \NoCaseChange{\protect\cite{cite3216}}, a level-five \(C_4/C_6\) code underlies concatenated quantum Hamming codes \(\mathcal{Q}_5,\mathcal{Q}_6,\mathcal{Q}_7,\mathcal{Q}_7\), yielding a \(2.5\%\) threshold and space overheads \(162\) and \(373\) physical qubits per logical qubit at physical error rate \(0.1\%\) for logical CNOT error rates \(10^{-10}\) and \(10^{-24}\), respectively.
\item\relax A modified tower of interleaved quantum Hamming codes with reserved qubits and recursive hookless Pauli-product measurements yields fault-tolerant quantum computation on a 1D nearest-neighbor qubit line with asymptotic rate above \(5\%\), constant space overhead, quasi-polylogarithmic time overhead, and a threshold \NoCaseChange{\protect\cite{cite3217}}.
\end{eczvaluelist}
\codefieldsection{Threshold}
\begin{eczvaluelist}
\item\relax \flmRefsHyperref{ref3210}{Concatenated threshold} requiring constant-space and quasi-polylogarithmic time overhead \NoCaseChange{\protect\cite{cite3214}}.
\end{eczvaluelist}
\codefieldsection{Parents}
\begin{eczvaluelist}
\item\relax
\flmRefsHyperref[eczindexfamilyrel]{code:quantum_reed_muller}{Quantum Reed-Muller (RM) code} --- \(\llbracket 2^r-1, 2^r-2r-1, 3\rrbracket \) quantum Hamming codes are quantum RM codes because Hamming and simplex codes are both punctured RM codes.
\item\relax
\flmRefsHyperref[eczindexfamilyrel]{code:qudit_hamming_css}{\(\llbracket 2^r-1, 2^r-2r-1, 3\rrbracket _p\) quantum Hamming code} --- \(\llbracket 2^r-1, 2^r-2r-1, 3\rrbracket _p\) prime-qudit CSS codes for \(p=2\) reduce to \(\llbracket 2^r-1, 2^r-2r-1, 3\rrbracket \) quantum Hamming codes.
\item\relax
\flmRefsHyperref[eczindexfamilyrel]{code:self_dual_css}{Self-dual CSS code}\item\relax
\flmRefsHyperref[eczindexfamilyrel]{code:small_distance_qubit_stabilizer}{Small-distance qubit stabilizer code}\end{eczvaluelist}
\codefieldsection{Children}
\begin{eczvaluelist}
\item\relax
\flmRefsHyperref[eczindexfamilyrel]{code:stab_15_7_3}{\(\llbracket 15, 7, 3\rrbracket \) quantum Hamming code}\item\relax
\flmRefsHyperref[eczindexfamilyrel]{code:steane}{\(\llbracket 7,1,3\rrbracket \) Steane code}\end{eczvaluelist}
\codefieldsection{Cousins}
\begin{eczvaluelist}
\item\relax
\flmRefsHyperref[eczindexfamilyrel]{code:hamming}{\([2^r-1,2^r-r-1,3]\) Hamming code} --- Quantum Hamming codes result from applying the CSS construction to Hamming codes and their duals the simplex codes.
\item\relax
\flmRefsHyperref[eczindexfamilyrel]{code:simplex}{\([2^m-1,m,2^{m-1}]\) simplex code} --- Quantum Hamming codes result from applying the CSS construction to Hamming codes and their duals the simplex codes.
\item\relax
\flmRefsHyperref[eczindexfamilyrel]{code:qubit_concatenated}{Concatenated qubit code} --- Concatenating a growing sequence of quantum Hamming codes yields fault-tolerant quantum computation with constant space overhead and quasi-polylogarithmic time overhead \NoCaseChange{\protect\cite{cite3214}}.
Concatenating quantum Hamming codes on top of the \(\llbracket 4,2,2\rrbracket \) and \(C_6\) codes yields fault-tolerant quantum computation with constant space and quasi-polylogarithmic time overheads \NoCaseChange{\protect\cite{cite3216}}. In the optimized protocol of Ref. \NoCaseChange{\protect\cite{cite3216}}, a level-five \(C_4/C_6\) code underlies concatenated quantum Hamming codes \(\mathcal{Q}_5,\mathcal{Q}_6,\mathcal{Q}_7,\mathcal{Q}_7\), yielding a \(2.5\%\) threshold and space overheads \(162\) and \(373\) physical qubits per logical qubit at physical error rate \(0.1\%\) for logical CNOT error rates \(10^{-10}\) and \(10^{-24}\), respectively.
A modified tower of interleaved quantum Hamming codes with reserved qubits and recursive hookless Pauli-product measurements yields fault-tolerant quantum computation on a 1D nearest-neighbor qubit line with asymptotic rate above \(5\%\), constant space overhead, quasi-polylogarithmic time overhead, and a threshold \NoCaseChange{\protect\cite{cite3217}}.
Quantum Hamming codes can also be concatenated with surface codes \NoCaseChange{\protect\cite{cite3218}}.

\item\relax
\flmRefsHyperref[eczindexfamilyrel]{code:stab_4_2_2}{\(\llbracket 4,2,2\rrbracket \) Four-qubit code} --- Concatenating quantum Hamming codes on top of the \(\llbracket 4,2,2\rrbracket \) and \(C_6\) codes yields fault-tolerant quantum computation with constant space and quasi-polylogarithmic time overheads \NoCaseChange{\protect\cite{cite3216}}. In the optimized protocol of Ref. \NoCaseChange{\protect\cite{cite3216}}, a level-five \(C_4/C_6\) code underlies concatenated quantum Hamming codes \(\mathcal{Q}_5,\mathcal{Q}_6,\mathcal{Q}_7,\mathcal{Q}_7\), yielding a \(2.5\%\) threshold and space overheads \(162\) and \(373\) physical qubits per logical qubit at physical error rate \(0.1\%\) for logical CNOT error rates \(10^{-10}\) and \(10^{-24}\), respectively.
\item\relax
\flmRefsHyperref[eczindexfamilyrel]{code:stab_6_2_2}{\(\llbracket 6,2,2\rrbracket \) \(C_6\) code} --- Concatenating quantum Hamming codes on top of the \(\llbracket 4,2,2\rrbracket \) and \(C_6\) codes yields fault-tolerant quantum computation with constant space and quasi-polylogarithmic time overheads \NoCaseChange{\protect\cite{cite3216}}. In the optimized protocol of Ref. \NoCaseChange{\protect\cite{cite3216}}, a level-five \(C_4/C_6\) code underlies concatenated quantum Hamming codes \(\mathcal{Q}_5,\mathcal{Q}_6,\mathcal{Q}_7,\mathcal{Q}_7\), yielding a \(2.5\%\) threshold and space overheads \(162\) and \(373\) physical qubits per logical qubit at physical error rate \(0.1\%\) for logical CNOT error rates \(10^{-10}\) and \(10^{-24}\), respectively.
\item\relax
\flmRefsHyperref[eczindexfamilyrel]{code:surface}{Kitaev surface code} --- Quantum Hamming codes can be concatenated with surface codes \NoCaseChange{\protect\cite{cite3218}}. In a unified logical-CNOT comparison under circuit-level depolarizing noise, using the surface code as the underlying code gives a \(0.31\%\) threshold and requires space overhead \(4.5\times 10^3\) at physical error rate \(0.1\%\) to achieve logical CNOT error rate \(10^{-24}\), compared to \(3.7\times 10^2\) for the optimized \(C_4/C_6\)/Hamming construction \NoCaseChange{\protect\cite{cite3216}}.
\item\relax
\flmRefsHyperref[eczindexfamilyrel]{code:data_syndrome}{Quantum data-syndrome (QDS) code} --- Because every stabilizer generator has the same weight \(2^{r-1}\), quantum Hamming codes admit QDS extensions based on good binary syndrome-measurement codes \NoCaseChange{\protect\cite{cite2914}}.
\end{eczvaluelist}
\eczhbkcontributors{ Qingfeng (Kee) Wang, \eczhuVVA }
\endeczcode

\eczcode{diagonal_clifford}{\(\llbracket 2^r-1,1,3\rrbracket \) simplex code}{~\NoCaseChange{\protect\cite{cite3219,cite690,cite799}}}
\codefieldsection{Alternative Names}
\begin{eczvaluelist}
\item\relax \(\llbracket 2^r-1,1,3\rrbracket \) quantum RM code
\end{eczvaluelist}
\eczhIndexCodeAliasName{diagonal_clifford}{simplex code}
\eczhIndexCodeAliasName{diagonal_clifford}{\(\llbracket 2^r-1,1,3\rrbracket \) quantum RM code}
\codefieldsection{Description}
Member of a color-code family constructed from a punctured first-order RM\((1,m=r)\) code and its even subcode for \(r \geq 3\).
Each code transversally implements a diagonal gate at the \((r-1)\)st level of the \flmTerm{term}{ref694}{}{Clifford hierarchy} \NoCaseChange{\protect\cite{cite799,cite800}}.
Each code is a color code defined on a simplex in \(r-1\) dimensions \NoCaseChange{\protect\cite{cite475,cite832}}, where qubits are placed on the vertices, edges, and faces as well as on the simplex itself.

The family also admits an XP-stabilizer presentation at precision \(N = 2^{r-2}\) whose generators are symmetric in \(X\) and \(P\), and only \(2r\) such generators are needed to stabilize the codespace \NoCaseChange{\protect\cite[{Prop. 27}]{cite798}}.
Morphing the \(r\)-dimensional distance-three code in this family yields a \(\llbracket 2^r+r-1,1,2\rrbracket \) code with a fault-tolerant logical gate at the \((r-1)\)st level of the Clifford hierarchy \NoCaseChange{\protect\cite[{Appx. C}]{cite687}}.

\codefieldsection{Transversal and Permutation-Based Gates}
\begin{eczvaluelist}
\item\relax Each code transversally implements a diagonal gate at the \((r-1)\)st level of the \flmTerm{term}{ref694}{}{Clifford hierarchy} in the form of a \(Z\)-rotation by angle \(-\pi/2^{r-1}\) \NoCaseChange{\protect\cite{cite799,cite800}}. These are the smallest distance-three qubit stabilizer codes with such a (strongly) transversal gate \NoCaseChange{\protect\cite{cite794}}.
\end{eczvaluelist}
\codefieldsection{Fault Tolerance}
\begin{eczvaluelist}
\item\relax Fault-tolerant syndrome extraction circuits using flag qubits \NoCaseChange{\protect\cite{cite3220}}.
\end{eczvaluelist}
\codefieldsection{Parents}
\begin{eczvaluelist}
\item\relax
\flmRefsHyperref[eczindexfamilyrel]{code:quantum_reed_muller}{Quantum Reed-Muller (RM) code} --- \(\llbracket 2^r-1,1,3\rrbracket \) simplex codes are special cases of the \(\llbracket \sum_{i=w+1}^m \binom{m}{i}, \sum_{i=0}^{w} \binom{m}{i}, \sum_{i=w+1}^{r+1} \binom{r+1}{i}\rrbracket \) quantum RM codes for \(w=0\) and \(r=1\), with \(m\) equal to the present entry's parameter \(r\) \NoCaseChange{\protect\cite[{Thm. 1}]{cite701}}.
\item\relax
\flmRefsHyperref[eczindexfamilyrel]{code:color}{Color code} --- Each \(\llbracket 2^r-1,1,3\rrbracket \) simplex code is a color code defined on a simplex in \(r-1\) dimensions \NoCaseChange{\protect\cite{cite475,cite832}}.
\item\relax
\flmRefsHyperref[eczindexfamilyrel]{code:quantum_divisible}{Quantum divisible code} --- \(\llbracket 2^r-1,1,3\rrbracket \) simplex codes come from RM\((1,m=r)\) codes, which are \((r-1)\)-even \NoCaseChange{\protect\cite{cite1574,cite1575}}, and admit transversal gates at levels of the \flmTerm{term}{ref694}{}{Clifford hierarchy}. Building a tower of generalized divisible codes by starting with the Steane code yields the \(\llbracket 2^r-1,1,3\rrbracket \) simplex codes \NoCaseChange{\protect\cite[{Sec. VI.B}]{cite734}}.
\item\relax
\flmRefsHyperref[eczindexfamilyrel]{code:small_distance_qubit_stabilizer}{Small-distance qubit stabilizer code}\end{eczvaluelist}
\codefieldsection{Children}
\begin{eczvaluelist}
\item\relax
\flmRefsHyperref[eczindexfamilyrel]{code:stab_15_1_3}{\(\llbracket 15,1,3\rrbracket \) quantum RM code}\item\relax
\flmRefsHyperref[eczindexfamilyrel]{code:steane}{\(\llbracket 7,1,3\rrbracket \) Steane code}\end{eczvaluelist}
\codefieldsection{Cousins}
\begin{eczvaluelist}
\item\relax
\flmRefsHyperref[eczindexfamilyrel]{code:xp_stabilizer}{XP stabilizer code} --- Each \(\llbracket 2^r-1,1,3\rrbracket \) simplex code can be viewed as an XP stabilizer code with precision \(N = 2^{r-2}\) \NoCaseChange{\protect\cite[{Exam. 6.4}]{cite798}}.
\item\relax
\flmRefsHyperref[eczindexfamilyrel]{code:quantum_k-orthogonal}{\(k\)-orthogonal code} --- \(\llbracket 2^r-1,1,3\rrbracket \) simplex codes are \((r-1)\)-orthogonal \NoCaseChange{\protect\cite[{Lemma 2}]{cite794}}.
\item\relax
\flmRefsHyperref[eczindexfamilyrel]{code:biorthogonal}{\([2^m,m+1,2^{m-1}]\) First-order RM code} --- The \(\llbracket 2^r-1,1,3\rrbracket \) simplex code is constructed with a punctured first-order RM code and its even subcode.
\item\relax
\flmRefsHyperref[eczindexfamilyrel]{code:simplex_spherical}{Simplex spherical code} --- Each \(\llbracket 2^r-1,1,3\rrbracket \) simplex code is a color code whose qubits are placed on the vertices, edges, and faces of an \((r-1)\)-simplex \NoCaseChange{\protect\cite{cite475,cite832}}.
\item\relax
\flmRefsHyperref[eczindexfamilyrel]{code:binary_dihedral_permutation_invariant}{Binary dihedral PI code} --- The \(\llparenthesis 2^{r-1}+3,2,3\rrparenthesis \) family of binary dihedral PI codes realizes the same (strongly) transversal gates as the \(\llbracket 2^r-1,1,3\rrbracket \) quantum RM codes, but require fewer qubits in almost all cases.
\item\relax
\flmRefsHyperref[eczindexfamilyrel]{code:morphed_diagonal_clifford}{\(\llbracket 2^r+r-1,1,2\rrbracket \) morphed simplex code} --- The \(\llbracket 2^r+r-1,1,2\rrbracket \) morphed simplex code is obtained by morphing the \(\llbracket 2^{r+1}-1,1,3\rrbracket \) simplex code on a region whose child code is a \(\llbracket 2^r,r,2\rrbracket \) hypercube code \NoCaseChange{\protect\cite[{Appx. C}]{cite687}}.
\end{eczvaluelist}
\eczhbkcontributors{ \eczhuVVA }
\endeczcode

\eczcode{quantum_hamming}{\(\llbracket 2^r, 2^r-r-2, 3\rrbracket \) Gottesman code}{~\NoCaseChange{\protect\cite{cite1182,cite449}}}
\codefieldsection{Alternative Names}
\begin{eczvaluelist}
\item\relax \(\llbracket 2^r, 2^r-r-2, 3\rrbracket \) quantum Hamming code
\end{eczvaluelist}
\eczhIndexCodeAliasName{quantum_hamming}{Gottesman code}
\eczhIndexCodeAliasName{quantum_hamming}{\(\llbracket 2^r, 2^r-r-2, 3\rrbracket \) quantum Hamming code}
\codefieldsection{Description}
A family of \flmRefsHyperref{ref672}{pure} \NoCaseChange{\protect\cite{cite449}} non-CSS stabilizer codes of distance \(3\) that saturate the asymptotic quantum Hamming bound.

The family can be obtained from a modified CSS construction \NoCaseChange{\protect\cite{cite1182,cite861}} with a \([2^r,r+1,2^{r-1}] = C_2^{\perp}\) first-order RM code and a \([2^r,2^r-1,2] = C_1\) even-weight code \NoCaseChange{\protect\cite{cite1182,cite861}}.
The modification introduces signs between the codewords.

\codefieldsection{Protection}
Protects against any single qubit error.
\codefieldsection{Notes}
\begin{eczvaluelist}
\item\relax The code is useful for entanglement distillation \NoCaseChange{\protect\cite{cite3221}}.
\end{eczvaluelist}
\codefieldsection{Parent}
\begin{eczvaluelist}
\item\relax
\flmRefsHyperref[eczindexfamilyrel]{code:small_distance_qubit_stabilizer}{Small-distance qubit stabilizer code}\end{eczvaluelist}
\codefieldsection{Child}
\begin{eczvaluelist}
\item\relax
\flmRefsHyperref[eczindexfamilyrel]{code:stab_8_3_3}{\(\llbracket 8, 3, 3\rrbracket \) Eight-qubit Gottesman code}\end{eczvaluelist}
\codefieldsection{Cousins}
\begin{eczvaluelist}
\item\relax
\flmRefsHyperref[eczindexfamilyrel]{code:quantum_perfect}{Perfect quantum code} --- \(\llbracket 2^r, 2^r-r-2, 3\rrbracket \) Gottesman codes saturate the asymptotic quantum Hamming bound.
\item\relax
\flmRefsHyperref[eczindexfamilyrel]{code:hamming}{\([2^r-1,2^r-r-1,3]\) Hamming code} --- \(\llbracket 2^r, 2^r-r-2, 3\rrbracket \) Gottesman codes are analogues of Hamming codes in that they saturate the asymptotic Hamming bound.
\item\relax
\flmRefsHyperref[eczindexfamilyrel]{code:biorthogonal}{\([2^m,m+1,2^{m-1}]\) First-order RM code} --- Gottesman codes can be obtained from a modified CSS construction \NoCaseChange{\protect\cite{cite1182,cite861}} with a \([2^r,r+1,2^{r-1}] = C_2^{\perp}\) first-order RM code and a \([2^r,2^r-1,2] = C_1\) even-weight code \NoCaseChange{\protect\cite{cite1182,cite861}}.
\item\relax
\flmRefsHyperref[eczindexfamilyrel]{code:projective}{Projective geometry code} --- Gottesman codes are related to partial spreads in projective geometry \NoCaseChange{\protect\cite{cite1695}}.
\end{eczvaluelist}
\eczhbkcontributors{ Marianna Podzorova, \eczhuVVA }
\endeczcode

\eczcode{morphed_diagonal_clifford}{\(\llbracket 2^r+r-1,1,2\rrbracket \) morphed simplex code}{~\NoCaseChange{\protect\cite{cite687}}}
\codefieldsection{Alternative Names}
\begin{eczvaluelist}
\item\relax \(\llbracket 2^r+r-1,1,2\rrbracket \) morphed quantum RM code
\end{eczvaluelist}
\eczhIndexCodeAliasName{morphed_diagonal_clifford}{morphed simplex code}
\eczhIndexCodeAliasName{morphed_diagonal_clifford}{\(\llbracket 2^r+r-1,1,2\rrbracket \) morphed quantum RM code}
\codefieldsection{Description}
A member of a family of codes obtained by morphing the \(\llbracket 2^{r+1}-1,1,3\rrbracket \) simplex codes on a region whose child code is a \(\llbracket 2^r,r,2\rrbracket \) hypercube code \NoCaseChange{\protect\cite[{Appx. C}]{cite687}}.
The morphing process replaces a subset of qubits with their logical qubits, yielding a code with parameters \(\llbracket 2^r+r-1,1,2\rrbracket \) that inherits a diagonal gate at the \((r-1)\)st level of the Clifford hierarchy from the parent code.

\codefieldsection{Gates}
\begin{eczvaluelist}
\item\relax Each code implements a diagonal gate at the \((r-1)\)st level of the \flmTerm{term}{ref694}{}{Clifford hierarchy} using transversal operations and \(C^{r}Z\) gates \NoCaseChange{\protect\cite[{Appx. C}]{cite687}}.
\end{eczvaluelist}
\codefieldsection{Parents}
\begin{eczvaluelist}
\item\relax
\flmRefsHyperref[eczindexfamilyrel]{code:qubit_css}{Qubit CSS code}\item\relax
\flmRefsHyperref[eczindexfamilyrel]{code:small_distance_qubit_stabilizer}{Small-distance qubit stabilizer code}\end{eczvaluelist}
\codefieldsection{Children}
\begin{eczvaluelist}
\item\relax
\flmRefsHyperref[eczindexfamilyrel]{code:stab_10_1_2}{\(\llbracket 10,1,2\rrbracket \) Vasmer-Kubica code} --- The \(\llbracket 10,1,2\rrbracket \) code is a specific instance of the \(\llbracket 2^r+r-1,1,2\rrbracket \) morphed simplex codes with \(r=3\) \NoCaseChange{\protect\cite[{Appx. C}]{cite687}}.
\item\relax
\flmRefsHyperref[eczindexfamilyrel]{code:stab_5_1_2}{\(\llbracket 5,1,2\rrbracket \) rotated surface code} --- The \(\llbracket 5,1,2\rrbracket \) code is a specific instance of the \(\llbracket 2^r+r-1,1,2\rrbracket \) morphed simplex codes with \(r=2\) \NoCaseChange{\protect\cite[{Fig. 1}]{cite687}}.
\end{eczvaluelist}
\codefieldsection{Cousins}
\begin{eczvaluelist}
\item\relax
\flmRefsHyperref[eczindexfamilyrel]{code:diagonal_clifford}{\(\llbracket 2^r-1,1,3\rrbracket \) simplex code} --- The \(\llbracket 2^r+r-1,1,2\rrbracket \) morphed simplex code is obtained by morphing the \(\llbracket 2^{r+1}-1,1,3\rrbracket \) simplex code on a region whose child code is a \(\llbracket 2^r,r,2\rrbracket \) hypercube code \NoCaseChange{\protect\cite[{Appx. C}]{cite687}}.
\item\relax
\flmRefsHyperref[eczindexfamilyrel]{code:hypercube_quantum}{\(\llbracket 2^D,D,2\rrbracket \) hypercube quantum code} --- The \(\llbracket 2^r+r-1,1,2\rrbracket \) morphed simplex code is obtained by morphing the \(\llbracket 2^{r+1}-1,1,3\rrbracket \) simplex code on a region whose child code is a \(\llbracket 2^r,r,2\rrbracket \) hypercube code \NoCaseChange{\protect\cite[{Appx. C}]{cite687}}.
\end{eczvaluelist}
\eczhbkcontributors{ Michael Vasmer, \eczhuVVA }
\endeczcode

\eczcode{ring_cpc}{\(\llbracket 2^r+r, 2^r-r-2, 3\rrbracket \) Ring CPC code}{~\NoCaseChange{\protect\cite{cite860}}}
\eczhIndexCodeAliasName{ring_cpc}{Ring CPC code}
\codefieldsection{Description}
A family of \(\llbracket 2^r+r, 2^r-r-2, 3\rrbracket \) CPC codes for \(r \geq 3\) whose matrices are based on the shortened version of the \([2^r-1,2^r-r-1,3]\) Hamming code.
See \NoCaseChange{\protect\cite[{Thm. 4}]{cite860}} for their stabilizer generator matrix.

\codefieldsection{Parents}
\begin{eczvaluelist}
\item\relax
\flmRefsHyperref[eczindexfamilyrel]{code:cpc}{Coherent-parity-check (CPC) code}\item\relax
\flmRefsHyperref[eczindexfamilyrel]{code:small_distance_qubit_stabilizer}{Small-distance qubit stabilizer code}\end{eczvaluelist}
\codefieldsection{Cousin}
\begin{eczvaluelist}
\item\relax
\flmRefsHyperref[eczindexfamilyrel]{code:hamming}{\([2^r-1,2^r-r-1,3]\) Hamming code} --- The ring CPC code is obtained from the shortened Hamming code via the CPC construction \NoCaseChange{\protect\cite{cite860}}.
\end{eczvaluelist}
\eczhbkcontributors{ \eczhuVVA }
\endeczcode

\eczcode{stab_20_2_6}{\(\llbracket 20,2,6\rrbracket \) B\&C phantom code}{~\NoCaseChange{\protect\cite{cite514,cite795}}}
\eczhIndexCodeAliasName{stab_20_2_6}{B\&C phantom code}
\codefieldsection{Description}
Self-dual CSS code on 20 physical qubits encoding two logical qubits with distance 6.
The code is obtained by binarizing the \(\llbracket 5,1,3\rrbracket _4\) code in the self-dual normal basis \(\{\omega,\omega^2\}\) to the \(\llbracket 10,2,3\rrbracket \) binarized Galois-qudit code and then concatenating each qubit pair with the \(\llbracket 4,2,2\rrbracket \) code \NoCaseChange{\protect\cite{cite514}}.

Physical qubits are arranged in five blocks of four, where block \(j\) corresponds to the \(j\)-th \(\mathbb{F}_4\) qudit of the \(\llbracket 5,1,3\rrbracket _4\) code.
Stabilizer generators of the \(\llbracket 4,2,2\rrbracket \) inner code on each block are \(ZZZZ\) and \(XXXX\); the outer stabilizers are the lifted images of the eight stabilizer generators of the \(\llbracket 10,2,3\rrbracket \) code using the logical-operator map
\(X^\omega\mapsto XXII\), \(X^{\omega^2}\mapsto XIXI\), \(Z^\omega\mapsto IZIZ\), \(Z^{\omega^2}\mapsto IIZZ\) within each block.

A stabilizer tableau for the code is \NoCaseChange{\protect\cite{cite514}}
\flmMathEnvironment{align}{}{
\begin{smallmatrix}
  Z & Z & Z & Z & I & I & I & I & I & I & I & I & I & I & I & I & I & I & I & I \\
  I & I & I & I & Z & Z & Z & Z & I & I & I & I & I & I & I & I & I & I & I & I \\
  I & I & I & I & I & I & I & I & Z & Z & Z & Z & I & I & I & I & I & I & I & I \\
  I & I & I & I & I & I & I & I & I & I & I & I & Z & Z & Z & Z & I & I & I & I \\
  I & I & I & I & I & I & I & I & I & I & I & I & I & I & I & I & Z & Z & Z & Z \\
  I & Z & I & Z & I & Z & I & Z & I & Z & I & Z & I & Z & I & Z & I & I & I & I \\
  I & I & Z & Z & I & I & Z & Z & I & I & Z & Z & I & I & Z & Z & I & I & I & I \\
  I & I & I & I & I & Z & I & Z & I & Z & Z & I & I & I & Z & Z & I & Z & I & Z \\
  I & I & I & I & I & I & Z & Z & I & Z & I & Z & I & Z & Z & I & I & I & Z & Z \\
  X & X & X & X & I & I & I & I & I & I & I & I & I & I & I & I & I & I & I & I \\
  I & I & I & I & X & X & X & X & I & I & I & I & I & I & I & I & I & I & I & I \\
  I & I & I & I & I & I & I & I & X & X & X & X & I & I & I & I & I & I & I & I \\
  I & I & I & I & I & I & I & I & I & I & I & I & X & X & X & X & I & I & I & I \\
  I & I & I & I & I & I & I & I & I & I & I & I & I & I & I & I & X & X & X & X \\
  X & X & I & I & X & X & I & I & X & X & I & I & X & X & I & I & I & I & I & I \\
  X & I & X & I & X & I & X & I & X & I & X & I & X & I & X & I & I & I & I & I \\
  I & I & I & I & X & X & I & I & X & I & X & I & I & X & X & I & X & X & I & I \\
  I & I & I & I & X & I & X & I & I & X & X & I & X & X & I & I & X & I & X & I
\end{smallmatrix}~.
}
Rows 1--5 are the \(Z\)-type inner stabilizers of each \(\llbracket 4,2,2\rrbracket \) block; rows 6--9 are the lifted outer \(Z\)-type stabilizers; rows 10--14 are the \(X\)-type inner stabilizers; rows 15--18 are the lifted outer \(X\)-type stabilizers.
This code is equivalent, up to qubit permutations and single-qubit Clifford gates, to three \(\llbracket 20,2,6\rrbracket \) self-dual CSS codes in QECDB \NoCaseChange{\protect\cite[{ID 672e6e9c34d5954e0cb7d1b7}]{cite781}\protect\cite[{ID 67a388ef9d65c7b409826879}]{cite781}\protect\cite[{ID 67a3890a9d65c7b40982687b}]{cite781}}.

\codefieldsection{Protection}
Detects errors on up to 5 qubits and corrects errors on up to 2 qubits.

\codefieldsection{Encoding}
\begin{eczvaluelist}
\item\relax Non-fault-tolerant encoding circuits \NoCaseChange{\protect\cite{cite795}}.
\end{eczvaluelist}
\codefieldsection{Transversal and Permutation-Based Gates}
\begin{eczvaluelist}
\item\relax Logical CNOT gates in both directions between the two logical qubits are realized by qubit permutations within a code block \NoCaseChange{\protect\cite{cite514,cite795}}.
\item\relax Fold-diagonal logical \(SS\) gates are available from the self-duality of the outer \(\llbracket 5,1,3\rrbracket _4\) code and the inner \(\llbracket 4,2,2\rrbracket \) layer \NoCaseChange{\protect\cite{cite514}}.
\end{eczvaluelist}
\codefieldsection{Fault Tolerance}
\begin{eczvaluelist}
\item\relax Selective state filtering (post-selection): logical error rates of \(\approx 2\times 10^{-6}\) per round of Steane-style error correction at physical error rate \(p=10^{-3}\), without being fully fault-tolerant \NoCaseChange{\protect\cite{cite795}}.
\end{eczvaluelist}
\codefieldsection{Parents}
\begin{eczvaluelist}
\item\relax
\flmRefsHyperref[eczindexfamilyrel]{code:bc_phantom}{Binarized-and-concatenated (B\&C) phantom code} --- The \(\llbracket 20,2,6\rrbracket \) code is the B\&C phantom code obtained from the \(\llbracket 5,1,3\rrbracket _4\) Galois-qudit CSS code \NoCaseChange{\protect\cite{cite514,cite795}}.
\item\relax
\flmRefsHyperref[eczindexfamilyrel]{code:self_dual_css}{Self-dual CSS code} --- The \(\llbracket 20,2,6\rrbracket \) code is a self-dual CSS code obtained from the \(\llbracket 5,1,3\rrbracket \) code via the BLT mapping and concatenation with \(\llbracket 4,2,2\rrbracket \) \NoCaseChange{\protect\cite[{Corr. 2}]{cite795}\protect\cite[{Corr. 1}]{cite1432}}.
\end{eczvaluelist}
\codefieldsection{Cousins}
\begin{eczvaluelist}
\item\relax
\flmRefsHyperref[eczindexfamilyrel]{code:stab_5_1_3}{\(\llbracket 5,1,3\rrbracket \) Five-qubit perfect code} --- The \(\llbracket 20,2,6\rrbracket \) code is obtained from the \(\llbracket 5,1,3\rrbracket \) five-qubit code via the BLT mapping (Lemma 1) and concatenation with the \(\llbracket 4,2,2\rrbracket \) code (Corollary 2) \NoCaseChange{\protect\cite{cite795}\protect\cite[{Corr. 1}]{cite1432}}.
\item\relax
\flmRefsHyperref[eczindexfamilyrel]{code:stab_4_2_2}{\(\llbracket 4,2,2\rrbracket \) Four-qubit code} --- The \(\llbracket 20,2,6\rrbracket \) code is obtained by concatenating each qubit pair of the \(\llbracket 10,2,3\rrbracket \) binarized code with the \(\llbracket 4,2,2\rrbracket \) code \NoCaseChange{\protect\cite{cite514,cite795}}.
\item\relax
\flmRefsHyperref[eczindexfamilyrel]{code:stab_10_2_3}{\(\llbracket 10,2,3\rrbracket \) binarized Galois-qudit code} --- The \(\llbracket 20,2,6\rrbracket \) code is obtained by concatenating each qubit pair of the \(\llbracket 10,2,3\rrbracket \) binarized Galois-qudit code with the \(\llbracket 4,2,2\rrbracket \) code \NoCaseChange{\protect\cite{cite514}}.
\end{eczvaluelist}
\eczhbkcontributors{ \eczhuVVA }
\endeczcode

\eczcode{qubit_golay}{\(\llbracket 23, 1, 7\rrbracket \) Quantum Golay code}{~\NoCaseChange{\protect\cite{cite861}}}
\codefieldsection{Alternative Names}
\begin{eczvaluelist}
\item\relax Qubit Golay code
\end{eczvaluelist}
\eczhIndexCodeAliasName{qubit_golay}{Quantum Golay code}
\eczhIndexCodeAliasName{qubit_golay}{Qubit Golay code}
\codefieldsection{Description}
A \(\llbracket 23, 1, 7\rrbracket \) self-dual CSS code with eleven stabilizer generators of each type, and with each generator being weight eight.

The code's 11-by-23 stabilizer generator matrix blocks \(H_{X}\) and \(H_{Z}\) are both parity-check matrices of the classical Golay code.
Equivalently, it can be obtained from the \([24,12,8]\) extended Golay code by shortening on one bit to a self-orthogonal \([23,11,7]\) code \NoCaseChange{\protect\cite[{Appx. A.1.5}]{cite101}}.
It can be punctured twice to obtain a \(\llbracket 21,3,5\rrbracket \) code \NoCaseChange{\protect\cite{cite101}}.

The automorphism group of the code is \(M_{23}\) \NoCaseChange{\protect\cite{cite3222}}.

\codefieldsection{Protection}
Detects up to 6-qubit errors and corrects up to 3-qubit errors.
\codefieldsection{Magic}
Magic-state distillation scaling exponent \(\gamma=\log_2 23 \approx 4.52\)\NoCaseChange{\protect\cite{cite706}}.
\codefieldsection{Encoding}
\begin{eczvaluelist}
\item\relax Fault-tolerant depth-7 circuit consisting of 57 CNOT gates and preparing a logical-zero state \NoCaseChange{\protect\cite{cite796}}.
\item\relax Circuit with 56 entangling gates using reinforcement learning \NoCaseChange{\protect\cite{cite3223}}.
\end{eczvaluelist}
\codefieldsection{Transversal and Permutation-Based Gates}
\begin{eczvaluelist}
\item\relax \flmRefsHyperref{ref409}{Single-qubit Clifford group} by choosing \(\overline{U}=U^{\otimes 23}\) for every Clifford unitary \(U\) \NoCaseChange{\protect\cite{cite796}}.
\end{eczvaluelist}
\codefieldsection{Gates}
\begin{eczvaluelist}
\item\relax The Golay code can be used to perform magic-state distillation for the magic state defined as \(|T\rangle\langle T|=\frac{1}{2}(I+\frac{1}{\sqrt{3}}(X+Y+Z) )\), where \(|T\rangle\) is an eigenstate of the Clifford "facet" gate \(SH\) \NoCaseChange{\protect\cite{cite3224}}.
\item\relax Pipelining the \(\llbracket 23,1,7\rrbracket \) code after \(\llbracket 7,1,3\rrbracket \) and \(\llbracket 17,1,5\rrbracket \) inner-code stages yields a 95-to-1 seventh-order magic-state distillation protocol \NoCaseChange{\protect\cite[{Sec. V.B.3}]{cite101}}.
\end{eczvaluelist}
\codefieldsection{Decoding}
\begin{eczvaluelist}
\item\relax Meggitt decoding for the cyclic CSS structure was used in comparative logical-CNOT simulations \NoCaseChange{\protect\cite{cite3225}}.
\end{eczvaluelist}
\codefieldsection{Fault Tolerance}
\begin{eczvaluelist}
\item\relax Fault-tolerant depth-7 circuit consisting of 57 CNOT gates and preparing a logical-zero state \NoCaseChange{\protect\cite{cite796}}.
\end{eczvaluelist}
\codefieldsection{Threshold}
\begin{eczvaluelist}
\item\relax \(1.32\times 10^{-3}\)-per gate error rate for depolarizing noise upon recursive concatenation \NoCaseChange{\protect\cite{cite796}}, improving previous lower bounds \NoCaseChange{\protect\cite{cite518,cite3225,cite3226}}. A numerical study \NoCaseChange{\protect\cite{cite518}} found that the Golay code achieved the highest threshold among a dozen well-known codes at the time \NoCaseChange{\protect\cite{cite3225}}.
\end{eczvaluelist}
\codefieldsection{Notes}
\begin{eczvaluelist}
\item\relax See Ref. \NoCaseChange{\protect\cite{cite3227}} for more details.
\item\relax Two levels of concatenation of the qubit Golay code can tolerate high teleportation errors \NoCaseChange{\protect\cite{cite3228,cite3229}}.
\end{eczvaluelist}
\codefieldsection{Parents}
\begin{eczvaluelist}
\item\relax
\flmRefsHyperref[eczindexfamilyrel]{code:self_dual_css}{Self-dual CSS code}\item\relax
\flmRefsHyperref[eczindexfamilyrel]{code:galois_quad_residue}{Quantum quadratic-residue (QR) code} --- The Golay code is a qubit quantum QR code \NoCaseChange{\protect\cite{cite829,cite2914}}.
\item\relax
\flmRefsHyperref[eczindexfamilyrel]{code:data_syndrome}{Quantum data-syndrome (QDS) code} --- There exists a \(\llbracket 23,1,7:18\rrbracket \) QDS code based on the qubit Golay code, requiring 18 additional stabilizer measurements instead of 24 from the general cyclic construction \NoCaseChange{\protect\cite[{Ex. 15}]{cite2914}}.
\end{eczvaluelist}
\codefieldsection{Cousins}
\begin{eczvaluelist}
\item\relax
\flmRefsHyperref[eczindexfamilyrel]{code:golay}{\([23, 12, 7]\) Golay code} --- The qubit Golay code is a CSS code constructed with the Golay code.
\item\relax
\flmRefsHyperref[eczindexfamilyrel]{code:qutrit_golay}{\(\llbracket 11,1,5\rrbracket _3\) qutrit Golay code} --- The qubit Golay code is the qubit counterpart of the qutrit Golay code.
\item\relax
\flmRefsHyperref[eczindexfamilyrel]{code:quantum_triorthogonal}{Triorthogonal code} --- A \(\llbracket 95,1,7\rrbracket \) triorthogonal code with a transversal \(T\) gate can be obtained from the qubit Golay code via the doubling transformation \NoCaseChange{\protect\cite{cite3230}}.
\item\relax
\flmRefsHyperref[eczindexfamilyrel]{code:small_distance_qubit_stabilizer}{Small-distance qubit stabilizer code} --- The quantum Golay code can be punctured twice to obtain a \(\llbracket 21,3,5\rrbracket \) code.
\item\relax
\flmRefsHyperref[eczindexfamilyrel]{code:cat_concatenated}{Concatenated cat code} --- Two-component cat codes concatenated with Steane and Golay codes are estimated to be fault tolerant against \flmRefsHyperref{ref498}{photon loss} noise with rate \(\eta < 5\times 10^{-4}\) provided that \(\alpha > 1.2\) \NoCaseChange{\protect\cite{cite3231}}.
\end{eczvaluelist}
\eczhbkcontributors{ Yinchen Liu, \eczhuVVA }
\endeczcode

\eczcode{bb288}{\(\llbracket 288,12,18\rrbracket \) double-gross code}{~\NoCaseChange{\protect\cite{cite441}}}
\codefieldsection{Alternative Names}
\begin{eczvaluelist}
\item\relax \(\llbracket 288,12,18\rrbracket \) two-gross code
\item\relax \((12,12)\) BB6 code
\end{eczvaluelist}
\eczhIndexCodeAliasName{bb288}{double-gross code}
\eczhIndexCodeAliasName{bb288}{\(\llbracket 288,12,18\rrbracket \) two-gross code}
\eczhIndexCodeAliasName{bb288}{\((12,12)\) BB6 code}
\codefieldsection{Description}
A bivariate bicycle (BB) code with parameters \(\llbracket 288,12,18\rrbracket \) and weight-six stabilizer generators \NoCaseChange{\protect\cite{cite441}}.

One defining presentation uses \((\ell,m)=(12,12)\) with \(x^{\ell}=y^{m}=1\), and
\(A=x^3+y^2+y^7\), \(B=y^3+x+x^2\) in \(\mathbb{F}_2[x,y]/(x^{\ell}-1,y^{m}-1)\) \NoCaseChange{\protect\cite[{Table 3}]{cite441}}.

\codefieldsection{Rate}
Ancilla-added encoding rate is \(1/48\), using \(n_a=n=288\) ancilla qubits.
\codefieldsection{Parent}
\begin{eczvaluelist}
\item\relax
\flmRefsHyperref[eczindexfamilyrel]{code:qcga}{Bivariate bicycle (BB) code}\end{eczvaluelist}
\eczhbkcontributors{ \eczhuVVA }
\endeczcode

\eczcode{iceberg}{\(\llbracket 2m,2m-2,2\rrbracket \) error-detecting code}{~\NoCaseChange{\protect\cite{cite861,cite737,cite3232}}}
\codefieldsection{Alternative Names}
\begin{eczvaluelist}
\item\relax Iceberg code
\item\relax \(\llbracket 2m,2m-2,2\rrbracket \) quantum parity code
\end{eczvaluelist}
\eczhIndexCodeAliasName{iceberg}{error-detecting code}
\eczhIndexCodeAliasName{iceberg}{Iceberg code}
\eczhIndexCodeAliasName{iceberg}{\(\llbracket 2m,2m-2,2\rrbracket \) quantum parity code}
\codefieldsection{Description}
Self-complementary and self-dual CSS code for \(m\geq 2\) with generators \(\{XX\cdots X, ZZ\cdots Z\} \) acting on all \(2m\) physical qubits.
The code is constructed via the CSS construction from an SPC code and a repetition code \NoCaseChange{\protect\cite[{Sec. III}]{cite773}}.
This is the highest-rate distance-two code when an even number of qubits is used \NoCaseChange{\protect\cite{cite449}}.

Admits a basis such that each codeword is a superposition of a computational basis state labeled by an even-weight bitstring \(b\) and a state labeled by the negation of \(b\).
Its all-zero logical state is a conventional GHZ state.
Removing the \(Z\)-type generator expands the number of codewords to all combinations of bitstrings and their negations, yielding a code with \(k=2m-1\) \NoCaseChange{\protect\cite{cite3233}}.

All of its automorphisms lie in the \flmRefsHyperref{ref409}{Clifford group} \NoCaseChange{\protect\cite[{Thm. 13}]{cite446}}.

\codefieldsection{Protection}
Detects a single-qubit error.
\codefieldsection{Encoding}
\begin{eczvaluelist}
\item\relax Adaptive constant-depth circuit with geometrically local gates and measurements throughout \NoCaseChange{\protect\cite{cite3234,cite3235}}.
\end{eczvaluelist}
\codefieldsection{Transversal and Permutation-Based Gates}
\begin{eczvaluelist}
\item\relax Transveral CNOT gates can be performed by first teleporting qubits into different code blocks \NoCaseChange{\protect\cite[{Sec. VII}]{cite737}}.
\item\relax For \(2m\) being a multiple of four, this code houses a transversal representation of the single-qubit Clifford group \NoCaseChange{\protect\cite[{Sec. 3.1}]{cite801}}. Its \(n\)-block version houses a transversal representation of the \(n\)-qubit Clifford group \NoCaseChange{\protect\cite[{Secs. 3.1 and 3.4}]{cite801}}. The commutant of transversal representations of the Clifford group contains qubit permutations and projections onto the \(\llbracket 2m,2m-2,2\rrbracket \) error-detecting code \NoCaseChange{\protect\cite{cite802}}.
\end{eczvaluelist}
\codefieldsection{Gates}
\begin{eczvaluelist}
\item\relax Logical SWAP gates can be performed fault tolerantly using an ancilla qubit \NoCaseChange{\protect\cite[{Sec. VII}]{cite737}}.
\item\relax Universal set of gates, each of which is supported on two qubits \NoCaseChange{\protect\cite{cite3236}}.
\item\relax Fault-tolerant Clifford Trotter circuits that are linear in \(k\) using flag qubits via a solve-and-stitch algorithm and application of a logical identity circuit \NoCaseChange{\protect\cite{cite3237}}.
\end{eczvaluelist}
\codefieldsection{Decoding}
\begin{eczvaluelist}
\item\relax The \(\llbracket 2m,2m-2,2\rrbracket \) error-detecting code \NoCaseChange{\protect\cite{cite3238}} and its relative the code with single stabilizer \(XX\cdots X\) \NoCaseChange{\protect\cite{cite3239}} admit autonomous QEC against single \flmRefsHyperref{ref498}{AD} errors.
\end{eczvaluelist}
\codefieldsection{Fault Tolerance}
\begin{eczvaluelist}
\item\relax Logical SWAP gates can be performed fault tolerantly using an ancilla qubit \NoCaseChange{\protect\cite[{Sec. VII}]{cite737}}.
\item\relax Two-qubit fault-tolerant state preparation, error detection and projective measurements \NoCaseChange{\protect\cite{cite3215}} (see also \NoCaseChange{\protect\cite{cite3236}}).
\item\relax CNOT and Hadamard gates using only two extra qubits and four-qubit fault-tolerant \(CCZ\) gate \NoCaseChange{\protect\cite{cite791}}.
\item\relax Fault-tolerant Clifford Trotter circuits that are linear in \(k\) using flag qubits via a solve-and-stitch algorithm and application of a logical identity circuit \NoCaseChange{\protect\cite{cite3237}}.
\item\relax Weak fault tolerance: any single gate error can be detected by measuring stabilizers and utilizing extra ancillas \NoCaseChange{\protect\cite{cite3240}}.
\end{eczvaluelist}
\codefieldsection{Realizations}
\begin{eczvaluelist}
\item\relax Trapped-ion devices: the \(m=5\) code has been realized on a 12-qubit device by Quantinuum \NoCaseChange{\protect\cite{cite3236}}. The QAOA algorithm has been realized on the \(m=18\) code using 510 two-logical-qubit gates \NoCaseChange{\protect\cite{cite3241}}. State preparation, measurement, and a partially fault-tolerant simulation of the XY model demonstrated on a 98-qubit Quantinuum Helios device using the code and its concatenated version \NoCaseChange{\protect\cite{cite810}}.
\end{eczvaluelist}
\codefieldsection{Notes}
\begin{eczvaluelist}
\item\relax See description of the code in Ref. \NoCaseChange{\protect\cite{cite2764}}.
\item\relax The code solves \NoCaseChange{\protect\cite{cite3242}} the mean king's measurement problem \NoCaseChange{\protect\cite{cite2788}}.
\item\relax The code is useful for entanglement distillation \NoCaseChange{\protect\cite{cite3221}}.
\item\relax The code is used in a fault-tolerant implementation of the QAOA algorithm \NoCaseChange{\protect\cite{cite3243}}.
\item\relax The iceberg code can be used for robustly simulating \(SU(2)\) gauge theories \NoCaseChange{\protect\cite{cite1366}}.
\end{eczvaluelist}
\codefieldsection{Parents}
\begin{eczvaluelist}
\item\relax
\flmRefsHyperref[eczindexfamilyrel]{code:qmdpc}{Quantum multi-dimensional parity-check (QMDPC) code} --- The \(\llbracket 2m,2m-2,2\rrbracket \) error-detecting code is a 1D QMDPC.
\item\relax
\flmRefsHyperref[eczindexfamilyrel]{code:quantum_mds}{Quantum maximum-distance-separable (MDS) code} --- The only nontrivial qubit MDS codes have parameters \(\llbracket 5,1,3\rrbracket \), \(\llbracket 6,0,4\rrbracket \), and \(\llbracket 2m,2m-2,2\rrbracket \) \NoCaseChange{\protect\cite[{Sec. 27.4}]{cite2024}}.
\item\relax
\flmRefsHyperref[eczindexfamilyrel]{code:ball_color}{Ball code} --- The \(\llbracket 2m,2m-2,2\rrbracket \) error-detecting code is a ball color code \NoCaseChange{\protect\cite[{Sec. III.A}]{cite687}}.
\item\relax
\flmRefsHyperref[eczindexfamilyrel]{code:stabilizer_over_gf4}{Hermitian qubit code} --- The \(\llbracket 2m,2m-2,2\rrbracket \) error-detecting code is Hermitian \NoCaseChange{\protect\cite[{Table 6}]{cite454}}.
\item\relax
\flmRefsHyperref[eczindexfamilyrel]{code:self_dual_css}{Self-dual CSS code}\item\relax
\flmRefsHyperref[eczindexfamilyrel]{code:self_complementary}{Self-complementary qubit code}\end{eczvaluelist}
\codefieldsection{Children}
\begin{eczvaluelist}
\item\relax
\flmRefsHyperref[eczindexfamilyrel]{code:stab_4_2_2}{\(\llbracket 4,2,2\rrbracket \) Four-qubit code} --- The \(\llbracket 2m,2m-2,2\rrbracket \) error-detecting code for \(m=2\) reduces to the \(\llbracket 4,2,2\rrbracket \) code.
\item\relax
\flmRefsHyperref[eczindexfamilyrel]{code:stab_6_4_2}{\(\llbracket 6,4,2\rrbracket \) error-detecting code} --- The \(\llbracket 2m,2m-2,2\rrbracket \) error-detecting code for \(m=3\) reduces to the \(\llbracket 6,4,2\rrbracket \) error-detecting code.
\end{eczvaluelist}
\codefieldsection{Cousins}
\begin{eczvaluelist}
\item\relax
\flmRefsHyperref[eczindexfamilyrel]{code:parity_check}{\([n,n-1,2]\) Single parity-check (SPC) code} --- The \(\llbracket 2m,2m-2,2\rrbracket \) error-detecting code is constructed via the CSS construction from an SPC code and its dual repetition code \NoCaseChange{\protect\cite[{Sec. III}]{cite773}}.
\item\relax
\flmRefsHyperref[eczindexfamilyrel]{code:repetition}{Repetition code} --- The \(\llbracket 2m,2m-2,2\rrbracket \) error-detecting code is constructed via the CSS construction from an SPC code and its dual repetition code \NoCaseChange{\protect\cite[{Sec. III}]{cite773}}.
\item\relax
\flmRefsHyperref[eczindexfamilyrel]{code:4612_color}{Truncated trihexagonal (4.6.12) color code} --- The \(\llbracket 2m,2m-2,2\rrbracket \) error-detecting code for \(m=4\) is a color code defined on a single octagon of the 6.6.6 or 4.6.12 tilings.
\item\relax
\flmRefsHyperref[eczindexfamilyrel]{code:ampdamp}{Amplitude-damping (AD) code} --- The \(\llbracket 2m,2m-2,2\rrbracket \) error-detecting code \NoCaseChange{\protect\cite{cite3238}} and its relative the code with single stabilizer \(XX\cdots X\) \NoCaseChange{\protect\cite{cite3239}} admit autonomous QEC against single \flmRefsHyperref{ref498}{AD} errors.
\item\relax
\flmRefsHyperref[eczindexfamilyrel]{code:gauss_law}{Gauss' law code} --- The iceberg code can be used for robustly simulating \(SU(2)\) gauge theories \NoCaseChange{\protect\cite{cite1366}}.
\item\relax
\flmRefsHyperref[eczindexfamilyrel]{code:clifford_group}{Clifford group} --- Stabilizer states on \(n\) qubits form 3-designs on complex projective spaces \(\mathbb{C}P^{2^n}\) \NoCaseChange{\protect\cite{cite937}}. The \flmRefsHyperref{ref409}{Clifford group} is a unitary 2-design \NoCaseChange{\protect\cite{cite938}} and a 3-design \NoCaseChange{\protect\cite{cite940,cite941}\protect\cite[{Thm. 1.6(B)}]{cite939}\protect\cite[{pg. 191}]{cite42}} on \(U(2^n)\). The \(\llbracket 2m,2m-2,2\rrbracket \) code when \(2m\) is a multiple of four obstructs the Clifford group from being a 4-design \NoCaseChange{\protect\cite{cite801}}.
\item\relax
\flmRefsHyperref[eczindexfamilyrel]{code:group_4_2_2}{\(\llbracket 4,2,2\rrbracket _{G}\) four group-qudit code} --- The four group-qudit code can be extended to the \(\llbracket 2m,2m-2,2\rrbracket _{G}\) group-qudit code \NoCaseChange{\protect\cite[{Sec. VIII}]{cite2720}}. The latter reduces to the \(\llbracket 2m,2m-2,2\rrbracket \) error-detecting code for \(G=\mathbb{Z}_2\).
\item\relax
\flmRefsHyperref[eczindexfamilyrel]{code:jump}{Jump code} --- The subcode of the \(\llbracket 2m,2m-2,2\rrbracket \) error-detecting code consisting of codewords labeled by weight-\(m\) bitstrings is a \(\llparenthesis 2m,\frac{1}{2}{2m \choose m},1\rrparenthesis _{m}\) optimal jump code \NoCaseChange{\protect\cite{cite144}\protect\cite[{Corr. 9}]{cite145}}.
\item\relax
\flmRefsHyperref[eczindexfamilyrel]{code:hybrid_stabilizer}{Hybrid stabilizer code} --- The \(\llbracket 2m+1,2m+2:1,2\rrbracket \) hybrid stabilizer code \NoCaseChange{\protect\cite{cite671}} (extendable to modular qudits \NoCaseChange{\protect\cite{cite3244}}) is closely related to the \(\llbracket 2m,2m-2,2\rrbracket \) error-detecting code.
\item\relax
\flmRefsHyperref[eczindexfamilyrel]{code:qubit_stabilizer}{Qubit stabilizer code} --- The \(\llbracket 2m,2m-2,2\rrbracket \) code for \(2m\) being a multiple of four obstructs the Clifford group from being a 4-design \NoCaseChange{\protect\cite{cite801}}.
\item\relax
\flmRefsHyperref[eczindexfamilyrel]{code:trapezoid}{Trapezoid subsystem code} --- The trapezoid code family can be obtained from the \(\llbracket 2m,2m-2,2\rrbracket \) error-detecting code by using some logical qubits as gauge qubits and imposing a two-dimensional qubit geometry \NoCaseChange{\protect\cite{cite3245}}.
\item\relax
\flmRefsHyperref[eczindexfamilyrel]{code:galois_true_stabilizer}{True Galois-qudit stabilizer code} --- A naive extension of the iceberg code to Galois qudits keeps only two CSS-type generators, \(M_1(1)=X_{\alpha_1}\otimes\cdots\otimes X_{\alpha_n}\) and \(M_2(1)=Z_{\beta_1}\otimes\cdots\otimes Z_{\beta_n}\), with nonzero \(\alpha_i,\beta_i\in\mathbb{F}_q\) satisfying \(\sum_i \alpha_i\beta_i=0\). For prime-power dimensions with \(q=p^m\) and \(m>1\), this yields a Galois-qudit code of distance one that is generally not a true stabilizer code because the stabilizer is not closed under multiplication by arbitrary \(\gamma\in\mathbb{F}_q\). Adding all \(M_1(\gamma)\) and \(M_2(\gamma)\) to the stabilizer group recovers the corresponding true Galois-qudit CSS code of distance two \NoCaseChange{\protect\cite[{Sec. 8.2.2}]{cite398}}.
\end{eczvaluelist}
\eczhbkcontributors{ Connor Clayton, \eczhuVVA }
\endeczcode

\eczcode{ea_3_1_3-2}{\(\llbracket 3, 1, 3;2\rrbracket \) EA code}{~\NoCaseChange{\protect\cite{cite2744}}}
\eczhIndexCodeAliasName{ea_3_1_3-2}{EA code}
\codefieldsection{Description}
Distance-three EA stabilizer code encoding one logical qubit and using two ebits.
It is the smallest example of an EA code correcting an arbitrary single-qubit error.

\codefieldsection{Parent}
\begin{eczvaluelist}
\item\relax
\flmRefsHyperref[eczindexfamilyrel]{code:eastab}{EA qubit stabilizer code}\end{eczvaluelist}
\codefieldsection{Cousins}
\begin{eczvaluelist}
\item\relax
\flmRefsHyperref[eczindexfamilyrel]{code:eaqecc}{Entanglement-assisted (EA) QECC} --- The \(\llbracket 3, 1, 3;2\rrbracket \) EA code is the first EA code.
\item\relax
\flmRefsHyperref[eczindexfamilyrel]{code:stab_5_1_3}{\(\llbracket 5,1,3\rrbracket \) Five-qubit perfect code} --- The \(\llbracket 3, 1, 3;2\rrbracket \) EA code and the five-qubit code have the same stabilizers \NoCaseChange{\protect\cite{cite1430,cite2702}}.
\item\relax
\flmRefsHyperref[eczindexfamilyrel]{code:small_distance_qubit_stabilizer}{Small-distance qubit stabilizer code}\end{eczvaluelist}
\eczhbkcontributors{ \eczhuVVA }
\endeczcode

\eczcode{stellated_dodecahedron_css}{\(\llbracket 30,8,3\rrbracket \) Bring code}{~\NoCaseChange{\protect\cite{cite2383}}}
\codefieldsection{Alternative Names}
\begin{eczvaluelist}
\item\relax Small stellated dodecahedron code
\end{eczvaluelist}
\eczhIndexCodeAliasName{stellated_dodecahedron_css}{Bring code}
\eczhIndexCodeAliasName{stellated_dodecahedron_css}{Small stellated dodecahedron code}
\codefieldsection{Description}
A \(\llbracket 30,8,3\rrbracket \) hyperbolic surface code on a quotient of the \(\{5,5\}\) hyperbolic tiling called Bring's curve.
Its qubits and stabilizer generators lie on the vertices of the small stellated dodecahedron. It admits a set of weight-five stabilizer generators.

\codefieldsection{Transversal and Permutation-Based Gates}
\begin{eczvaluelist}
\item\relax \flmRefsHyperref{ref409}{Clifford group} of four of the eight logical qubits can be implemented by transversal gates combined with qubit permutations \NoCaseChange{\protect\cite{cite762}}.
\end{eczvaluelist}
\codefieldsection{Decoding}
\begin{eczvaluelist}
\item\relax Fault-tolerant parity-check schedules whose performance is similar to those of the surface-17 code, but with qubit overhead reduced by a factor of 2.6 \NoCaseChange{\protect\cite{cite3246}}.
\end{eczvaluelist}
\codefieldsection{Parents}
\begin{eczvaluelist}
\item\relax
\flmRefsHyperref[eczindexfamilyrel]{code:two_dimensional_hyperbolic_surface}{2D hyperbolic surface code}\item\relax
\flmRefsHyperref[eczindexfamilyrel]{code:small_distance_qubit_stabilizer}{Small-distance qubit stabilizer code}\end{eczvaluelist}
\codefieldsection{Cousins}
\begin{eczvaluelist}
\item\relax
\flmRefsHyperref[eczindexfamilyrel]{code:polyhedron}{Polyhedron code} --- Bring code and related codes listed in \NoCaseChange{\protect\cite[{Table 1}]{cite2383}} arrange qubits and stabilizer generators on star polyhedra.
\item\relax
\flmRefsHyperref[eczindexfamilyrel]{code:golay}{\([23, 12, 7]\) Golay code} --- The automorphism group of the parity-check matrix of the Golay code is the same as a certain automorphism group of the Bring code \NoCaseChange{\protect\cite{cite762}}.
\item\relax
\flmRefsHyperref[eczindexfamilyrel]{code:dodecahedron}{Dodecahedron code} --- The qubits and stabilizer generators of the \(\llbracket 30,8,3\rrbracket \) Bring code lie on the vertices of the small stellated dodecahedron.
\item\relax
\flmRefsHyperref[eczindexfamilyrel]{code:surface-17}{\(\llbracket 9,1,3\rrbracket \) Surface-17 code} --- Bring's code and the surface-17 code have been compared numerically \NoCaseChange{\protect\cite{cite2383}}.
\end{eczvaluelist}
\eczhbkcontributors{ Marc Serra-Peralta, \eczhuVVA }
\endeczcode

\eczcode{small_triorthogonal}{\(\llbracket 3k + 8, k, 2\rrbracket \) triorthogonal code}{~\NoCaseChange{\protect\cite[{Sec. VII}]{cite691}}}
\eczhIndexCodeAliasName{small_triorthogonal}{triorthogonal code}
\codefieldsection{Description}
Member of the \(\llbracket 3k + 8, k, 2\rrbracket \) family (for even \(k\)) of triorthogonal and quantum divisible codes that admit a transversal \(T\) gate and are relevant for magic-state distillation \NoCaseChange{\protect\cite{cite691}\protect\cite[{Sec. VI.C}]{cite734}}.

\codefieldsection{Magic}
The family yields the asymptotic exponent \(\gamma = \log_2 \frac{3k+8}{k} \to \log_2 3 \approx 1.6\) for sufficiently large \(k\) \NoCaseChange{\protect\cite[{Box 2}]{cite707}}; see \NoCaseChange{\protect\cite[{Table V}]{cite705}}.
\codefieldsection{Transversal and Permutation-Based Gates}
\begin{eczvaluelist}
\item\relax The code admits a transversal \(T\) gate \NoCaseChange{\protect\cite[{Lemma 2}]{cite691}}.
\end{eczvaluelist}
\codefieldsection{Parents}
\begin{eczvaluelist}
\item\relax
\flmRefsHyperref[eczindexfamilyrel]{code:quantum_triorthogonal}{Triorthogonal code}\item\relax
\flmRefsHyperref[eczindexfamilyrel]{code:quantum_divisible}{Quantum divisible code}\item\relax
\flmRefsHyperref[eczindexfamilyrel]{code:small_distance_qubit_stabilizer}{Small-distance qubit stabilizer code}\end{eczvaluelist}
\codefieldsection{Cousin}
\begin{eczvaluelist}
\item\relax
\flmRefsHyperref[eczindexfamilyrel]{code:quantum_h}{\(\llbracket k+4,k,2\rrbracket \) H code} --- The H code \(\llbracket k+4,k,2\rrbracket \) family yields the \(\llbracket 3k + 8, k, 2\rrbracket \) family of triorthogonal codes when level-lifted \NoCaseChange{\protect\cite[{Sec. VI.C}]{cite734}}.
\end{eczvaluelist}
\eczhbkcontributors{ Benjamin Quiring, \eczhuVVA }
\endeczcode

\eczcode{bacon_shor_4}{\(\llbracket 4,1,1,2\rrbracket \) Four-qubit subsystem code}{~\NoCaseChange{\protect\cite{cite3,cite3037}}}
\eczhIndexCodeAliasName{bacon_shor_4}{Four-qubit subsystem code}
\codefieldsection{Description}
Error-detecting four-qubit subsystem stabilizer code encoding one logical qubit and one gauge qubit.

The \(\llbracket 4,1,1,2\rrbracket \) code can be obtained by picking one of the logical qubits of the \(\llbracket 4,2,2\rrbracket \) four-qubit code to be a gauge qubit; e.g., see Ref. \NoCaseChange{\protect\cite{cite3247}}.
One particular gauge configuration has gauge group \(\mathsf{G}\) with gauge generators (excluding phases)
\flmMathEnvironment{align}{}{
\begin{array}{cccc}
  X & X & I & I \\
  I & I & X & X \\
  Z & I & Z & I \\
  I & Z & I & Z
\end{array}~.
}

\codefieldsection{Protection}
The code detects arbitrary single-qubit errors.
An equal-weight sum of its gauge generators has energy separation \(2(\sqrt{2}-1)\), so encoding into the corresponding ground subspace suppresses arbitrary single-qubit errors using non-commuting two-local Hamiltonian terms \NoCaseChange{\protect\cite{cite670}}.

\codefieldsection{Parent}
\begin{eczvaluelist}
\item\relax
\flmRefsHyperref[eczindexfamilyrel]{code:bacon_shor}{Bacon-Shor code} --- The four-qubit subsystem code is the shortest error-detecting Bacon-Shor code.
\end{eczvaluelist}
\codefieldsection{Cousins}
\begin{eczvaluelist}
\item\relax
\flmRefsHyperref[eczindexfamilyrel]{code:small_distance_qubit_stabilizer}{Small-distance qubit stabilizer code}\item\relax
\flmRefsHyperref[eczindexfamilyrel]{code:stab_4_2_2}{\(\llbracket 4,2,2\rrbracket \) Four-qubit code} --- The \(\llbracket 4,1,1,2\rrbracket \) code can be obtained by picking one of the logical qubits of the \(\llbracket 4,2,2\rrbracket \) four-qubit code to be a gauge qubit; e.g., see Ref. \NoCaseChange{\protect\cite{cite3247}}. One particular gauge configuration has gauge operators \(\{XXII,IIXX,ZIZI,IZIZ\}\).
\item\relax
\flmRefsHyperref[eczindexfamilyrel]{code:bravyi_bacon_shor_6}{\(\llbracket 6,2,3,2\rrbracket \) BBS code} --- Both the \(\llbracket 6,2,3,2\rrbracket \) BBS code and the four-qubit subsystem code can be used to suppress errors in adiabatic quantum computation \NoCaseChange{\protect\cite{cite670}}.
\item\relax
\flmRefsHyperref[eczindexfamilyrel]{code:quantum_double}{Quantum-double code} --- Quantum double code Hamiltonians can be simulated, with the help of perturbation theory and the four-qubit subsystem code, by two-dimensional two-body Hamiltonians with non-commuting terms \NoCaseChange{\protect\cite{cite2838}}.
\item\relax
\flmRefsHyperref[eczindexfamilyrel]{code:rbh}{Raussendorf-Bravyi-Harrington (RBH) cluster-state code} --- Concatenation of the RBH code with small codes such as the \(\llbracket 2,1\rrbracket \) repetition code, \(\llbracket 4,1,1,2\rrbracket \) subsystem code, or Steane code can improve thresholds \NoCaseChange{\protect\cite{cite3248}}.
\item\relax
\flmRefsHyperref[eczindexfamilyrel]{code:holographic_subsystem}{Subsystem holographic code} --- The holographic hybrid code is constructed out of alternating isometries of the five-qubit and \(\llbracket 4,1,1,2\rrbracket \) Bacon-Shor codes.
\end{eczvaluelist}
\eczhbkcontributors{ \eczhuVVA }
\endeczcode

\eczcode{css_4_1_2}{\(\llbracket 4,1,2\rrbracket \) Leung-Nielsen-Chuang-Yamamoto (LNCY) code}{~\NoCaseChange{\protect\cite{cite859}}}
\codefieldsection{Alternative Names}
\begin{eczvaluelist}
\item\relax \(\llbracket 4,1,2\rrbracket \) Leung code
\end{eczvaluelist}
\eczhIndexCodeAliasName{css_4_1_2}{Leung-Nielsen-Chuang-Yamamoto (LNCY) code}
\eczhIndexCodeAliasName{css_4_1_2}{\(\llbracket 4,1,2\rrbracket \) Leung code}
\codefieldsection{Description}
A four-qubit CSS stabilizer code that is the only qubit CSS code with such parameters.

A stabilizer tableau for the code is given by \NoCaseChange{\protect\cite[{ID 6}]{cite453}}
\flmMathEnvironment{align}{}{
\begin{array}{cccc}
  X & X & I & I \\
  I & I & X & X \\
  Z & Z & Z & Z
\end{array}~.
}
The code is depicted in \flmRefsCref{ref3249}.
\begin{flmFloat}{figure}{NumCap}\includegraphics[width=214.29921259842524bp,max width=\linewidth]{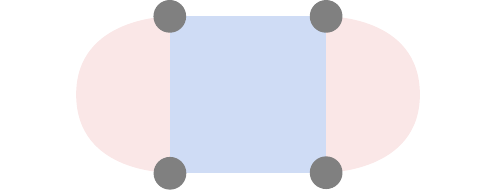}\caption{
  Stabilizer generators of the \(\llbracket 4,1,2\rrbracket \) LNCY code.
  The 4 data qubits (circles) are arranged on a \(2\times 2\) rotated surface code lattice with open boundaries.
  The bulk generator is a weight-four (four-body) \(Z\) operator and the two boundary generators are weight-two (two-body) \(X\) operators.
  Red regions correspond to \(X\) operators while blue regions correspond to \(Z\) operators.}\label{ref3249}\end{flmFloat}

The code admits the following basis of codewords,
\flmMathEnvironment{align}{}{
  \begin{split}
    |\overline{0}\rangle = (|0000\rangle + |1111\rangle)/\sqrt{2}~{\phantom{.}}\\
    |\overline{1}\rangle = (|0011\rangle + |1100\rangle)/\sqrt{2}~.
  \end{split}
}
It is realized as the \(\{|\overline{00}\rangle,|\overline{01}\rangle\}\) subcode of the \(\llbracket 4,2,2\rrbracket \) four-qubit code \NoCaseChange{\protect\cite{cite859}}, and the subcodes spanned by \(|\overline{00}\rangle\) and any other \(\llbracket 4,2,2\rrbracket \) codeword are equivalent to it.
Applying the Pauli string \(IXIX\) to its codewords yields an equivalent constant-excitation \(\llbracket 4,1,2\rrbracket \) code.

The code's \(\pm\)-basis codewords can be written as
\flmMathEnvironment{align}{}{
  |\overline{\pm}\rangle = \frac{1}{2}(|00\rangle \pm |11\rangle)^{\otimes 2}~.
}
This code can be thought of as a concatenation of a two-qubit bit-flip with a two-qubit phase-flip code.

\codefieldsection{Protection}
Detects a single-qubit error or single erasure as a distance-two code.
The code also approximately corrects a single \flmRefsHyperref{ref498}{AD} error, with recovery fidelity \(1-5\gamma^2+O(\gamma^3)\) \NoCaseChange{\protect\cite{cite859}}.
A \flmRefsHyperref{ref2540}{complementary-channel} analysis shows that the optimal worst-case entanglement-fidelity distance for this code under \flmRefsHyperref{ref498}{AD} noise is of order \(\gamma\), improving over the uncorrected order \(\sqrt{\gamma}\), and that no recovery can improve this asymptotic scaling \NoCaseChange{\protect\cite{cite2538}}.
The \(\{|\overline{01}\rangle,|\overline{11}\rangle\}\) \(\llbracket 4,1,2\rrbracket \) subcode \NoCaseChange{\protect\cite{cite3250}} also approximately corrects a single \flmRefsHyperref{ref498}{AD} error, and is a constant-excitation code.

\codefieldsection{Realizations}
\begin{eczvaluelist}
\item\relax Linear optical networks \NoCaseChange{\protect\cite{cite3251,cite3252}}.
\item\relax Superconducting-circuit devices \NoCaseChange{\protect\cite{cite3253,cite3254}}.
\item\relax Logical gates within one block \NoCaseChange{\protect\cite{cite3255}} and between two blocks \NoCaseChange{\protect\cite{cite3256}}, with the latter interpreted as lattice surgery between planar surface codes, were realized in superconducting circuits.
\item\relax Neutral atom arrays by Atom Computing ran the Bernstein-Vazirani algorithm on up to 28 logical qubits \NoCaseChange{\protect\cite{cite3257}}.
\item\relax Break-even performance has been demonstrated on a superconducting IBM device using the syndrome-based Petz recovery \NoCaseChange{\protect\cite{cite2595}}.
\item\relax Implementation of a universal logical gate set in superconducting circuits by Origin Quantum Computing \NoCaseChange{\protect\cite{cite3258}}.
\end{eczvaluelist}
\codefieldsection{Parents}
\begin{eczvaluelist}
\item\relax
\flmRefsHyperref[eczindexfamilyrel]{code:rotated_surface}{Rotated surface code} --- The \(\llbracket 4,1,2\rrbracket \) LNCY code is a small planar rotated surface code \NoCaseChange{\protect\cite{cite3256,cite3253,cite3254,cite3255}}.
\item\relax
\flmRefsHyperref[eczindexfamilyrel]{code:quantum_parity}{Quantum parity code (QPC)} --- The \(\llbracket 4,1,2\rrbracket \) LNCY code is the smallest QPC, i.e., a concatenation of a two-qubit bit-flip with a two-qubit phase-flip repetition code.
An \(\llbracket 8,1,2\rrbracket \) QPC correcting a single \flmRefsHyperref{ref498}{AD} error is equivalent to a concatenation of its constant-excitation version with the dual-rail code \NoCaseChange{\protect\cite{cite3250,cite3259,cite2711}}.

\item\relax
\flmRefsHyperref[eczindexfamilyrel]{code:ampdamp_stabilizer}{\(\llbracket 2(m+1),m,2\rrbracket \) single-loss AD code} --- The \(\llbracket 4,1,2\rrbracket \) LNCY code (approximately) corrects a single \flmRefsHyperref{ref498}{AD} error \NoCaseChange{\protect\cite{cite859}} and is the smallest member of the amplitude-damping stabilizer family of Ref. \NoCaseChange{\protect\cite{cite1187}}.
\item\relax
\flmRefsHyperref[eczindexfamilyrel]{code:jump}{Jump code} --- The \(\llbracket 4,1,2\rrbracket \) LNCY code \NoCaseChange{\protect\cite{cite3250}} is equivalent to a \(\llparenthesis 4,2,1\rrparenthesis _2\) jump code correcting a single \flmRefsHyperref{ref498}{AD} error.
A \(\llparenthesis 4,3,1\rrparenthesis _2\) jump code is a subcode of the \(\llbracket 4,2,2\rrbracket \) code and contains the \(\llbracket 4,1,2\rrbracket \) LNCY code as a subcode \NoCaseChange{\protect\cite{cite144}}.

\end{eczvaluelist}
\codefieldsection{Cousins}
\begin{eczvaluelist}
\item\relax
\flmRefsHyperref[eczindexfamilyrel]{code:stab_4_2_2}{\(\llbracket 4,2,2\rrbracket \) Four-qubit code} --- The \(\llbracket 4,1,2\rrbracket \) LNCY code is obtained as the \(\{|\overline{00}\rangle,|\overline{01}\rangle\}\) \(\llbracket 4,1,2\rrbracket \) subcode of the \(\llbracket 4,2,2\rrbracket \) four-qubit code \NoCaseChange{\protect\cite{cite859}}. A \(\llparenthesis 4,3,1\rrparenthesis _2\) jump code is a subcode of the \(\llbracket 4,2,2\rrbracket \) code and contains the \(\llbracket 4,1,2\rrbracket \) LNCY code as a subcode \NoCaseChange{\protect\cite{cite144}}.
\item\relax
\flmRefsHyperref[eczindexfamilyrel]{code:stab_4_1_2}{\(\llbracket 4,1,2\rrbracket \) twist-defect code} --- Adding \(XXII\) (\(XYZI\)) to the stabilizer group of the \(\llbracket 4,2,2\rrbracket \) code yields the \(\llbracket 4,1,2\rrbracket \) LNCY (twist-defect) code.
\item\relax
\flmRefsHyperref[eczindexfamilyrel]{code:binomial}{Binomial code} --- The \(\llbracket 4,1,2\rrbracket \) LNCY code reduces to the \(0,2,4\) binomial code when the basis labels in each codeword are written as in base-ten. Such a mapping can be generalized \NoCaseChange{\protect\cite{cite3260}}.
\item\relax
\flmRefsHyperref[eczindexfamilyrel]{code:heavy_hex}{Heavy-hexagon code} --- Magic states prepared using a \(\llbracket 4,1,2\rrbracket \) subcode can be injected into the heavy-hex code \NoCaseChange{\protect\cite{cite3261,cite3262}}.
The \(d=2\) heavy-hex code is closely related to the \(\llbracket 4,1,2\rrbracket \) LNCY code.

\item\relax
\flmRefsHyperref[eczindexfamilyrel]{code:qubit_concatenated}{Concatenated qubit code} --- The \(\llbracket 4,1,2\rrbracket \) LNCY code is the smallest QPC, i.e., a concatenation of a two-qubit bit-flip with a two-qubit phase-flip repetition code.
An \(\llbracket 8,1,2\rrbracket \) QPC correcting a single \flmRefsHyperref{ref498}{AD} error is equivalent to a concatenation of its constant-excitation version with the dual-rail code \NoCaseChange{\protect\cite{cite3250,cite3259,cite2711}}.
More generally, an \(\llbracket m^2,1,m\rrbracket \) QPC corrects \(m-1\) \flmRefsHyperref{ref498}{AD} errors \NoCaseChange{\protect\cite{cite3263}}.
Recursively concatenating a \(\llbracket 4,1,2\rrbracket \) LNCY subcode attains a threshold \NoCaseChange{\protect\cite{cite3264,cite3265}}.

\item\relax
\flmRefsHyperref[eczindexfamilyrel]{code:cluster_state}{Cluster-state code} --- A \(\llbracket 4,1,2\rrbracket \) LNCY code can be thought of as a cluster-state code \NoCaseChange{\protect\cite{cite3266}}.
\item\relax
\flmRefsHyperref[eczindexfamilyrel]{code:quantum_polar}{Quantum polar code} --- The \(\llbracket 4,1,2\rrbracket \) LNCY code is a small quantum polar encoding \NoCaseChange{\protect\cite{cite3267}}.
\item\relax
\flmRefsHyperref[eczindexfamilyrel]{code:ampdamp_numopt}{Numerically optimized four-qubit AD code} --- The numerically optimized four-qubit AD code can correct a single \flmRefsHyperref{ref498}{AD} error with higher entanglement fidelity than the \(\llbracket 4,1,2\rrbracket \) LNCY code \NoCaseChange{\protect\cite{cite859}}.
\end{eczvaluelist}
\eczhbkcontributors{ \eczhuVVA }
\endeczcode

\eczcode{stab_4_1_2}{\(\llbracket 4,1,2\rrbracket \) twist-defect code}{~\NoCaseChange{\protect\cite{cite803}}}
\eczhIndexCodeAliasName{stab_4_1_2}{twist-defect code}
\codefieldsection{Description}
A four-qubit non-CSS stabilizer code that can be interpreted as the smallest triangular color code with \(x\)-, \(y\)-, and \(z\)-type Pauli boundaries \NoCaseChange{\protect\cite[{Fig. 7}]{cite445}}, and equivalently as a small twist-defect surface code on a tetrahedron inscribed in a sphere \NoCaseChange{\protect\cite{cite435}}.
It is the only non-CSS qubit stabilizer code with parameters \(\llbracket 4,1,2\rrbracket \).
The code admits weight-three stabilizer generators and weight-two logical Pauli \(X,Y,Z\) operators.

A stabilizer tableau for the code is given by \NoCaseChange{\protect\cite[{ID 8}]{cite453}}
\flmMathEnvironment{align}{}{
\begin{array}{cccc}
  X & I & X & X \\
  Y & Y & I & Y \\
  Z & Z & Z & I
\end{array}~.
}

Stabilizer generators are shown in \flmRefsCref{ref3268}.
  \begin{flmFloat}{figure}{NumCap}\includegraphics[width=107.14960629921262bp,max width=\linewidth]{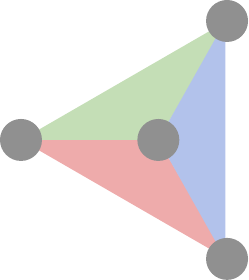}\caption{
    Stabilizer generators of the \(\llbracket 4,1,2\rrbracket \) twist-defect code, interpreted as a twist-defect color code. Each boundary type is shown in a different color (red, green, blue), corresponding to an \(X\)-type, \(Y\)-type, or \(Z\)-type Pauli supported on the vertices.
    }\label{ref3268}\end{flmFloat}

\codefieldsection{Protection}
Detects a single-qubit error or single erasure as a distance-two code.

\codefieldsection{Transversal and Permutation-Based Gates}
\begin{eczvaluelist}
\item\relax Weight-two transversal logical Pauli \(X,Y,Z\) operations \NoCaseChange{\protect\cite{cite803}}.
\end{eczvaluelist}
\codefieldsection{Gates}
\begin{eczvaluelist}
\item\relax A set of local Clifford operations and permutations (in the twist-defect realization, braiding the four genons) generates the full single-qubit Clifford group \NoCaseChange{\protect\cite{cite435}}.
\end{eczvaluelist}
\codefieldsection{Realizations}
\begin{eczvaluelist}
\item\relax Logical \flmRefsHyperref{ref409}{Clifford gates} were realized in a trapped-ion device by Quantinuum \NoCaseChange{\protect\cite{cite435}}.
\end{eczvaluelist}
\codefieldsection{Parents}
\begin{eczvaluelist}
\item\relax
\flmRefsHyperref[eczindexfamilyrel]{code:twist_defect_color}{Twist-defect color code} --- The \(\llbracket 4,1,2\rrbracket \) twist-defect code is the smallest triangular color code with \(x\)-, \(y\)-, and \(z\)-type Pauli boundaries, which make the code non-CSS \NoCaseChange{\protect\cite[{Fig. 7}]{cite445}}.
\item\relax
\flmRefsHyperref[eczindexfamilyrel]{code:twist_defect_surface}{Twist-defect surface code} --- The \(\llbracket 4,1,2\rrbracket \) twist-defect code is equivalent to a twist-defect surface code on a tetrahedron inscribed in a sphere \NoCaseChange{\protect\cite{cite435}} via a single-qubit \flmRefsHyperref{ref409}{Clifford circuit}.
\item\relax
\flmRefsHyperref[eczindexfamilyrel]{code:small_distance_qubit_stabilizer}{Small-distance qubit stabilizer code}\end{eczvaluelist}
\codefieldsection{Cousins}
\begin{eczvaluelist}
\item\relax
\flmRefsHyperref[eczindexfamilyrel]{code:stab_4_2_2}{\(\llbracket 4,2,2\rrbracket \) Four-qubit code} --- Adding \(XYZI\) to the stabilizer group of the \(\llbracket 4,2,2\rrbracket \) four-qubit code yields the \(\llbracket 4,1,2\rrbracket \) twist-defect subcode \NoCaseChange{\protect\cite{cite803}}.
\item\relax
\flmRefsHyperref[eczindexfamilyrel]{code:toric}{Toric code} --- The \flmRefsHyperref{ref436}{symplectic double} of the \(\llbracket 4,1,2\rrbracket \) twist-defect code is the \(\llbracket 8,2,2\rrbracket \) twisted toric code \NoCaseChange{\protect\cite{cite435}}.
\item\relax
\flmRefsHyperref[eczindexfamilyrel]{code:css_4_1_2}{\(\llbracket 4,1,2\rrbracket \) Leung-Nielsen-Chuang-Yamamoto (LNCY) code} --- Adding \(XXII\) (\(XYZI\)) to the stabilizer group of the \(\llbracket 4,2,2\rrbracket \) code yields the \(\llbracket 4,1,2\rrbracket \) LNCY (twist-defect) code.
\end{eczvaluelist}
\eczhbkcontributors{ \eczhuVVA }
\endeczcode

\eczcode{stab_4_2_2}{\(\llbracket 4,2,2\rrbracket \) Four-qubit code}{~\NoCaseChange{\protect\cite{cite3269,cite3270}}}
\codefieldsection{Alternative Names}
\begin{eczvaluelist}
\item\relax \(C_4\) code
\item\relax Little Shor code
\end{eczvaluelist}
\eczhIndexCodeAliasName{stab_4_2_2}{Four-qubit code}
\eczhIndexCodeAliasName{stab_4_2_2}{\(C_4\) code}
\eczhIndexCodeAliasName{stab_4_2_2}{Little Shor code}
\codefieldsection{Description}
A four-qubit hyperbolic self-dual CSS stabilizer code that is the smallest two-logical-qubit stabilizer code to detect a single-qubit error.
It is unique for its parameters \NoCaseChange{\protect\cite[{Thm. 8}]{cite446}}.

A stabilizer tableau for the code is given by \NoCaseChange{\protect\cite[{ID 9}]{cite453}}
\flmMathEnvironment{align}{}{
\begin{array}{cccc}
  Z & Z & Z & Z \\
  X & X & X & X
\end{array}~.
}
Its stabilizer generator matrix blocks, \(H_{X}=H_{Z}=(1,1,1,1)\), are both the parity-check matrix of the \([4,3,2]\) SPC code.
A basis of codewords is
\flmMathEnvironment{align}{}{
  \begin{split}
    |\overline{00}\rangle = (|0000\rangle + |1111\rangle)/\sqrt{2}~{\phantom{.}}\\
    |\overline{01}\rangle = (|0011\rangle + |1100\rangle)/\sqrt{2}~{\phantom{.}}\\
    |\overline{10}\rangle = (|0101\rangle + |1010\rangle)/\sqrt{2}~{\phantom{.}}\\
    |\overline{11}\rangle = (|0110\rangle + |1001\rangle)/\sqrt{2}~.
  \end{split}
}

\codefieldsection{Protection}
Detects a single-qubit error \NoCaseChange{\protect\cite{cite3269}} or single erasure \NoCaseChange{\protect\cite{cite3270}}.
It cannot correct arbitrary single-qubit errors because \( \lfloor \frac{d-1}{2} \rfloor =0 \).
An equivalent version of this code can suppress errors in adiabatic quantum computation by being used as an excited-state space of a particular Hamiltonian \NoCaseChange{\protect\cite{cite2688}}.

\codefieldsection{Magic}
Various magic-state distillation protocols exist for the \(\llbracket 4,2,2\rrbracket \) qubit code and the \(C_6\) code in what are known as Meier-Eastin-Knill (MEK) protocols \NoCaseChange{\protect\cite{cite708,cite101}}. In the inner/outer-code formulation of Ref. \NoCaseChange{\protect\cite{cite101}}, the \(\llbracket 4,2,2\rrbracket \) code is a hyperbolic self-dual inner code for quadratic distillation. For example, the magic-state yield parameter is \(\gamma = \log_2 5 \approx 2.322\) for a protocol using the \(\llbracket 10,2,2\rrbracket \) code \NoCaseChange{\protect\cite[{Box 2}]{cite707}}; see also \NoCaseChange{\protect\cite[{Table IV}]{cite705}}.
\codefieldsection{Transversal and Permutation-Based Gates}
\begin{eczvaluelist}
\item\relax A tensor product of Hadamard gates applies a Hadamard gate to both logical qubits, and a tensor product of \(S=\sqrt{Z}\) gates applies a \(CZ\) gate followed by a logical \(Z\) on both qubits \NoCaseChange{\protect\cite{cite804}} (see also \NoCaseChange{\protect\cite{cite805}}). A logical \(CZ\) gate is then realized by \(\sqrt{Z}\otimes\sqrt{Z}^{\dagger}\otimes\sqrt{Z}^{\dagger}\otimes\sqrt{Z}\). With a different logical basis, transversal Hadamard swaps the two logical qubits, enabling control-SWAP and \(H^{\otimes 2}\)-measurement routines for quadratic magic-state distillation \NoCaseChange{\protect\cite[{Sec. I.B.1}]{cite101}}.
\item\relax This code is the only four-qubit subspace to house a transversal representation of the single-qubit Clifford group but not of the single-qubit unitary group \NoCaseChange{\protect\cite[{Eq. (104)}]{cite801}}. Equivalently, its projector is the only extra generator of the fourth tensor-power Clifford commutant beyond qubit permutations \NoCaseChange{\protect\cite[{Thm. 1}]{cite801}}. Its \(n\)-block version is a \(\llbracket 4n,2n,2\rrbracket \) code, which houses a signed permutation representation of the \(n\)-qubit Clifford group \NoCaseChange{\protect\cite[{Sec. 3.1 and Appx. B}]{cite801}}.
\end{eczvaluelist}
\codefieldsection{Gates}
\begin{eczvaluelist}
\item\relax Some inter-block gates can be weight-two (two-body) with the help of perturbative gadgets, making it possible to suppress errors in adiabatic quantum computation \NoCaseChange{\protect\cite{cite2688}}.
\item\relax Logical \flmRefsHyperref{ref409}{Clifford circuits} for various qubit connectivities \NoCaseChange{\protect\cite{cite3271}}.
\end{eczvaluelist}
\codefieldsection{Decoding}
\begin{eczvaluelist}
\item\relax Erasure decoder \NoCaseChange{\protect\cite{cite3272}}.
\end{eczvaluelist}
\codefieldsection{Fault Tolerance}
\begin{eczvaluelist}
\item\relax Preparation of certain states, both magic and non-magic, along with transversal gates can be performed fault-tolerantly, but requires post-selection because the code cannot correct errors \NoCaseChange{\protect\cite{cite804}}. Magic states can be injected into surface and color codes since the code is a small instance of both \NoCaseChange{\protect\cite{cite3262}}.
\item\relax Knill's \(C_4/C_6\) architecture uses the \(\llbracket 4,2,2\rrbracket \) code at the first level and the \(C_6\) code at higher levels, together with error-correcting teleportation \NoCaseChange{\protect\cite{cite448}}. Later work refined the postselected-threshold analysis \NoCaseChange{\protect\cite{cite3273,cite3274}} (see also Ref. \NoCaseChange{\protect\cite{cite3275}}).
\item\relax Concatenating quantum Hamming codes on top of the \(\llbracket 4,2,2\rrbracket \) and \(C_6\) codes yields fault-tolerant quantum computation with constant space and quasi-polylogarithmic time overheads \NoCaseChange{\protect\cite{cite3216}}. In the optimized protocol of Ref. \NoCaseChange{\protect\cite{cite3216}}, a level-five \(C_4/C_6\) code underlies concatenated quantum Hamming codes \(\mathcal{Q}_5,\mathcal{Q}_6,\mathcal{Q}_7,\mathcal{Q}_7\), yielding a \(2.5\%\) threshold and space overheads \(162\) and \(373\) physical qubits per logical qubit at physical error rate \(0.1\%\) for logical CNOT error rates \(10^{-10}\) and \(10^{-24}\), respectively.
\item\relax Fault-tolerant implementation of the Deutsch-Jozsa algorithm \NoCaseChange{\protect\cite{cite3276}}.
\end{eczvaluelist}
\codefieldsection{Realizations}
\begin{eczvaluelist}
\item\relax See also \NoCaseChange{\protect\cite[{Tab. I}]{cite3277}} for more details on each experimental realization.
\item\relax Trapped-ion device by IonQ \NoCaseChange{\protect\cite{cite3278}}.
\item\relax Logical state preparation and flag-qubit error correction realized in superconducting-circuit devices by IBM \NoCaseChange{\protect\cite{cite3279,cite3280,cite3281,cite3261}}.
\item\relax The CZ magic state has been realized on an IBM heavy-hex superconducting circuit device \NoCaseChange{\protect\cite{cite3262}}.
\item\relax CPC gadgets for the \(\llbracket 4,2,2\rrbracket \) code have been implemented on the IBM 5Q superconducting device \NoCaseChange{\protect\cite{cite3282}}.
\item\relax An FPGA implementation of the collision clustering decoder \NoCaseChange{\protect\cite{cite3283}} realized on a Rigetti superconducting device \NoCaseChange{\protect\cite{cite3284}}.
\item\relax Neutral atom arrays: error detection, erasure correction, and post-selected fault-tolerant circuits demonstrated on 24 logical qubits on a 256-qubit device by Atom Computing, with each qubit encoded in the \(\llbracket 4,2,2\rrbracket \) code \NoCaseChange{\protect\cite{cite3257}}. 
Post-selected fault-tolerant realization of a benchmarking protocol \NoCaseChange{\protect\cite{cite804}}, preparation of the ground state of the single-impurity Anderson impurity model, and post-selected fault-tolerant logical Bell-state preparation demonstrated on one copy of the \(\llbracket 4,2,2\rrbracket \) code on a device by Infleqtion \NoCaseChange{\protect\cite{cite3277}}.
Error correction with mid-circuit erasure measurements and logical teleportation demonstrated by the Thompson group \NoCaseChange{\protect\cite{cite3285}}. Logical implementation of Shor's algorithm on a device by Infleqtion \NoCaseChange{\protect\cite{cite3286}}.

\item\relax The \(\llbracket 4,2,2\rrbracket \) code has been implemented on a star topology using superconducting devices and microwave cavities \NoCaseChange{\protect\cite{cite3287}}.
\item\relax Trapped-ion processor by AQT: modular logical-state teleportation between two four-qubit error-detecting code blocks without mid-circuit measurements \NoCaseChange{\protect\cite{cite3288}}.
\end{eczvaluelist}
\codefieldsection{Parents}
\begin{eczvaluelist}
\item\relax
\flmRefsHyperref[eczindexfamilyrel]{code:rotated_surface}{Rotated surface code} --- The \(\llbracket 4,2,2\rrbracket \) code is the smallest rotated toric code \NoCaseChange{\protect\cite{cite454}}.
\item\relax
\flmRefsHyperref[eczindexfamilyrel]{code:488_color}{Square-octagon (4.8.8) color code} --- The \(\llbracket 4,2,2\rrbracket \) code can be interpreted as a 2D color code on a square of the 4.8.8 tiling \NoCaseChange{\protect\cite{cite2526,cite3262}}. Removing \(X\) checks from blue octagons and \(Z\) checks from green octagons of the 4.8.8 color code yields a light 4.8.8 color code that is equivalent to concatenating the surface/toric code with the \(\llbracket 4,2,2\rrbracket \) code \NoCaseChange{\protect\cite{cite3289}}.
\item\relax
\flmRefsHyperref[eczindexfamilyrel]{code:triangular_color}{Honeycomb (6.6.6) color code} --- The \(\llbracket 4,2,2\rrbracket \) code can be interpreted as a 2D color code on a trapezoidal patch that makes up two-thirds of a hexagon of the 6.6.6 tiling \NoCaseChange{\protect\cite{cite2526,cite3262}}.
\item\relax
\flmRefsHyperref[eczindexfamilyrel]{code:4612_color}{Truncated trihexagonal (4.6.12) color code} --- The \(\llbracket 4,2,2\rrbracket \) code can be interpreted as a 2D color code on a square of the 4.6.12 tiling \NoCaseChange{\protect\cite{cite2526,cite3262}}. Concatenating the \(\llbracket 4,2,2\rrbracket \) code with two copies of the surface code on a hexagonal lattice yields the self-dual 4.6.12 color code \NoCaseChange{\protect\cite{cite3289}}.
\item\relax
\flmRefsHyperref[eczindexfamilyrel]{code:hypercube_quantum}{\(\llbracket 2^D,D,2\rrbracket \) hypercube quantum code} --- The \(\llbracket 4,2,2\rrbracket \) code is a hypercube code for \(D=2\).
\item\relax
\flmRefsHyperref[eczindexfamilyrel]{code:iceberg}{\(\llbracket 2m,2m-2,2\rrbracket \) error-detecting code} --- The \(\llbracket 2m,2m-2,2\rrbracket \) error-detecting code for \(m=2\) reduces to the \(\llbracket 4,2,2\rrbracket \) code.
\item\relax
\flmRefsHyperref[eczindexfamilyrel]{code:brickwork}{Brickwork \(XS\) stabilizer code} --- The \(\llbracket 4,2,2\rrbracket \) code can be interpreted as a brickwork code on a square of the overlapping rectangular tilings \NoCaseChange{\protect\cite{cite589}}.
\item\relax
\flmRefsHyperref[eczindexfamilyrel]{code:group_4_2_2}{\(\llbracket 4,2,2\rrbracket _{G}\) four group-qudit code} --- The four group-qudit code reduces to the four-qubit code for \(G=\mathbb{Z}_2\).
\end{eczvaluelist}
\codefieldsection{Cousins}
\begin{eczvaluelist}
\item\relax
\flmRefsHyperref[eczindexfamilyrel]{code:stab_5_1_3}{\(\llbracket 5,1,3\rrbracket \) Five-qubit perfect code} --- The \(\llbracket 4,2,2\rrbracket \) code can be derived from the five-qubit code using a protocol that converts an \(\llbracket n,k,d\rrbracket \) code into an \(\llbracket n-1, k+1, d-1\rrbracket \) code \NoCaseChange{\protect\cite[{Sec. 3.5}]{cite736}\protect\cite[{Fig. 3}]{cite2933}}.
\item\relax
\flmRefsHyperref[eczindexfamilyrel]{code:surface}{Kitaev surface code} --- Concatenating the \(\llbracket 4,2,2\rrbracket \) code with the surface code is equivalent to removing stabilizer generators from the 4.8.8 color code \NoCaseChange{\protect\cite{cite3289}}.
\item\relax
\flmRefsHyperref[eczindexfamilyrel]{code:toric}{Toric code} --- The toric code can be constructed by arranging \(\llbracket 4,2,2\rrbracket \) tensors on a square lattice and recovering the star and plaquette operators by operator pushing \NoCaseChange{\protect\cite{cite2868}}.
\item\relax
\flmRefsHyperref[eczindexfamilyrel]{code:qubit_concatenated}{Concatenated qubit code} --- Concatenations of \(\llbracket 4,2,2\rrbracket \) and \(C_6\) codes yield fault-tolerant quantum computation schemes \NoCaseChange{\protect\cite{cite448}} admitting a post-selected threshold \NoCaseChange{\protect\cite{cite3273,cite3274}} (see also Ref. \NoCaseChange{\protect\cite{cite3275}}).
Concatenating quantum Hamming codes on top of the \(\llbracket 4,2,2\rrbracket \) and \(C_6\) codes yields fault-tolerant quantum computation with constant space and quasi-polylogarithmic time overheads \NoCaseChange{\protect\cite{cite3216}}. In the optimized protocol of Ref. \NoCaseChange{\protect\cite{cite3216}}, a level-five \(C_4/C_6\) code underlies concatenated quantum Hamming codes \(\mathcal{Q}_5,\mathcal{Q}_6,\mathcal{Q}_7,\mathcal{Q}_7\), yielding a \(2.5\%\) threshold and space overheads \(162\) and \(373\) physical qubits per logical qubit at physical error rate \(0.1\%\) for logical CNOT error rates \(10^{-10}\) and \(10^{-24}\), respectively.
Concatenating the \(\llbracket 4,2,2\rrbracket \) code with the surface code is equivalent to removing stabilizer generators from the 4.8.8 color code \NoCaseChange{\protect\cite{cite3289}}.
The \(\llbracket 4,2,2\rrbracket \) code can be concatenated with two copies of the surface code to yield the 4.6.12 color code \NoCaseChange{\protect\cite{cite3289}}.

\item\relax
\flmRefsHyperref[eczindexfamilyrel]{code:stab_6_2_2}{\(\llbracket 6,2,2\rrbracket \) \(C_6\) code} --- Concatenations of \(\llbracket 4,2,2\rrbracket \) and \(C_6\) codes yield fault-tolerant quantum computation schemes \NoCaseChange{\protect\cite{cite448}} admitting a post-selected threshold \NoCaseChange{\protect\cite{cite3273,cite3274}} (see also Ref. \NoCaseChange{\protect\cite{cite3275}}) and the Meier-Eastin-Knill (MEK) magic-state distillation protocols \NoCaseChange{\protect\cite{cite708}}. Concatenating quantum Hamming codes on top of the \(\llbracket 4,2,2\rrbracket \) and \(C_6\) codes yields fault-tolerant quantum computation with constant space and quasi-polylogarithmic time overheads \NoCaseChange{\protect\cite{cite3216}}. In the optimized protocol of Ref. \NoCaseChange{\protect\cite{cite3216}}, a level-five \(C_4/C_6\) code underlies concatenated quantum Hamming codes \(\mathcal{Q}_5,\mathcal{Q}_6,\mathcal{Q}_7,\mathcal{Q}_7\), yielding a \(2.5\%\) threshold and space overheads \(162\) and \(373\) physical qubits per logical qubit at physical error rate \(0.1\%\) for logical CNOT error rates \(10^{-10}\) and \(10^{-24}\), respectively.
\item\relax
\flmRefsHyperref[eczindexfamilyrel]{code:steane}{\(\llbracket 7,1,3\rrbracket \) Steane code} --- The Steane code can be built from two \(\llbracket 4,2,2\rrbracket \) codes in the quantum Lego code framework \NoCaseChange{\protect\cite{cite2868}}. Ref. \NoCaseChange{\protect\cite{cite3216}} also introduces a \(C_4\)/Steane concatenated code, obtained by concatenating the \(\llbracket 4,2,2\rrbracket \) code with the Steane code, as an underlying code for further concatenation with quantum Hamming codes.
\item\relax
\flmRefsHyperref[eczindexfamilyrel]{code:stab_6_4_2}{\(\llbracket 6,4,2\rrbracket \) error-detecting code} --- The \(\llbracket 6,4,2\rrbracket \) error-detecting code can be constructed out of two \(\llbracket 4,2,2\rrbracket \) codes in the quantum Lego code framework \NoCaseChange{\protect\cite{cite2868}}.
\item\relax
\flmRefsHyperref[eczindexfamilyrel]{code:cpc}{Coherent-parity-check (CPC) code} --- CPC gadgets for the \(\llbracket 4,2,2\rrbracket \) code have been implemented on the IBM 5Q superconducting device \NoCaseChange{\protect\cite{cite3282}}.
\item\relax
\flmRefsHyperref[eczindexfamilyrel]{code:jump}{Jump code} --- A \(\llparenthesis 4,3,1\rrparenthesis _2\) jump code is a subcode of the \(\llbracket 4,2,2\rrbracket \) code and contains the \(\llbracket 4,1,2\rrbracket \) LNCY code as a subcode \NoCaseChange{\protect\cite{cite144}}.
\item\relax
\flmRefsHyperref[eczindexfamilyrel]{code:phantom}{Phantom code} --- The \(\llbracket 4,2,2\rrbracket \) code is the smallest phantom code: logical CNOT gates between its two logical qubits can be implemented by physical-qubit permutations \NoCaseChange{\protect\cite{cite514}}. Gluing copies of the \(\llbracket 4,2,2\rrbracket \) code with \(X\)-type stabilizers yields CSS phantom codes with parameters \(\llbracket 4m,2,(d_X=2,d_Z=2m)\rrbracket \), and puncturing one qubit from this construction yields \(\llbracket 4m-1,2,(d_X=2,d_Z=2m-1)\rrbracket \), for \(m\geq1\) \NoCaseChange{\protect\cite{cite514}}.
\item\relax
\flmRefsHyperref[eczindexfamilyrel]{code:parity_check}{\([n,n-1,2]\) Single parity-check (SPC) code} --- The \(\llbracket 4,2,2\rrbracket \) code is constructed from the \([4,3,2]\) SPC code via the CSS construction.
\item\relax
\flmRefsHyperref[eczindexfamilyrel]{code:dual_rail}{Dual-rail quantum code} --- An \(\llbracket 8,1,2\rrbracket \) QPC correcting a single \flmRefsHyperref{ref498}{AD} error is equivalent to a concatenation of the \(\{|\overline{01}\rangle,|\overline{11}\rangle\}\) (constant-excitation) subcode of the \(\llbracket 4,2,2\rrbracket \) code with the dual-rail code \NoCaseChange{\protect\cite{cite3250,cite3259,cite2711}}. More generally, an \(\llbracket m^2,1,m\rrbracket \) QPC corrects \(m-1\) \flmRefsHyperref{ref498}{AD} errors \NoCaseChange{\protect\cite{cite3263}}.
\item\relax
\flmRefsHyperref[eczindexfamilyrel]{code:gkp_concatenated}{Concatenated GKP code} --- Recursively concatenating the \(C_6\) and \(\llbracket 4,2,2\rrbracket \) codes with GKP codes achieves the hashing bound of the displacement channel \NoCaseChange{\protect\cite{cite3290}}.
\item\relax
\flmRefsHyperref[eczindexfamilyrel]{code:quantum_lego}{Tensor-network code} --- The Steane and \(\llbracket 6,4,2\rrbracket \) error-detecting codes can be built from two \(\llbracket 4,2,2\rrbracket \) codes in the quantum Lego code framework \NoCaseChange{\protect\cite{cite2868}}. The toric code can be constructed by arranging \(\llbracket 4,2,2\rrbracket \) tensors on a square lattice and recovering the star and plaquette operators by operator pushing \NoCaseChange{\protect\cite{cite2868}}.
\item\relax
\flmRefsHyperref[eczindexfamilyrel]{code:ladder}{Ladder Floquet code} --- The smallest example of the ladder Floquet code is a dynamical version of the \(\llbracket 4,2,2\rrbracket \) code \NoCaseChange{\protect\cite{cite3291,cite3247}}. The \(\llbracket 4,2,2\rrbracket \) code can be Floquetified in various ways \NoCaseChange{\protect\cite{cite3292,cite3293}}.
\item\relax
\flmRefsHyperref[eczindexfamilyrel]{code:majorana_hamming}{\(\llbracket 2^{m-1},2^{m-1}-m-1,4\rrbracket _{f}\) Hamming Majorana code} --- The \(\llbracket 8,3,4\rrbracket _{f}\) Hamming Majorana code is a Majorana stabilizer code obtained by combining two four-qubit codes \NoCaseChange{\protect\cite{cite565}}.
\item\relax
\flmRefsHyperref[eczindexfamilyrel]{code:hybrid_stabilizer}{Hybrid stabilizer code} --- The \(\llbracket 4,2,2\rrbracket \) codewords can be modified by signs to yield a \(\llbracket 4,1:1,2\rrbracket \) hybrid stabilizer code \NoCaseChange{\protect\cite{cite3294}}.
\item\relax
\flmRefsHyperref[eczindexfamilyrel]{code:stab_10_2_3}{\(\llbracket 10,2,3\rrbracket \) binarized Galois-qudit code} --- Concatenating each qubit pair of the binarized \(\llbracket 10,2,3\rrbracket \) code with the \(\llbracket 4,2,2\rrbracket \) code yields the \(\llbracket 20,2,6\rrbracket \) B\&C phantom code \NoCaseChange{\protect\cite{cite514,cite795}}.
\item\relax
\flmRefsHyperref[eczindexfamilyrel]{code:carbon}{\(\llbracket 12,2,4\rrbracket \) carbon code} --- The carbon code is a concatenation of the \(\llbracket 4,2,2\rrbracket \) code and the \(C_6\) code.
\item\relax
\flmRefsHyperref[eczindexfamilyrel]{code:stab_16_6_4}{\(\llbracket 16,6,4\rrbracket \) Tesseract color code} --- The \(\llbracket 16,4,2,4\rrbracket \) tesseract subsystem color code with particular gauge fixing can be obtained from four copies of the \(\llbracket 4,2,2\rrbracket \) code \NoCaseChange{\protect\cite{cite483}}.
\item\relax
\flmRefsHyperref[eczindexfamilyrel]{code:stab_20_2_6}{\(\llbracket 20,2,6\rrbracket \) B\&C phantom code} --- The \(\llbracket 20,2,6\rrbracket \) code is obtained by concatenating each qubit pair of the \(\llbracket 10,2,3\rrbracket \) binarized code with the \(\llbracket 4,2,2\rrbracket \) code \NoCaseChange{\protect\cite{cite514,cite795}}.
\item\relax
\flmRefsHyperref[eczindexfamilyrel]{code:css_4_1_2}{\(\llbracket 4,1,2\rrbracket \) Leung-Nielsen-Chuang-Yamamoto (LNCY) code} --- The \(\llbracket 4,1,2\rrbracket \) LNCY code is obtained as the \(\{|\overline{00}\rangle,|\overline{01}\rangle\}\) \(\llbracket 4,1,2\rrbracket \) subcode of the \(\llbracket 4,2,2\rrbracket \) four-qubit code \NoCaseChange{\protect\cite{cite859}}. A \(\llparenthesis 4,3,1\rrparenthesis _2\) jump code is a subcode of the \(\llbracket 4,2,2\rrbracket \) code and contains the \(\llbracket 4,1,2\rrbracket \) LNCY code as a subcode \NoCaseChange{\protect\cite{cite144}}.
\item\relax
\flmRefsHyperref[eczindexfamilyrel]{code:four_qubit_permutation_invariant}{\(\llparenthesis 4,2,2\rrparenthesis \) Four-qubit single-deletion code} --- Projecting the four-qubit code into the PI subspace yields the four-qubit single-deletion code. A basis of codewords for the four-qubit single-deletion code consists of the \(|\overline{00}\rangle\) and \(|\overline{01}\rangle+|\overline{10}\rangle+|\overline{11}\rangle\) states of the four-qubit code.
\item\relax
\flmRefsHyperref[eczindexfamilyrel]{code:stab_4_1_2}{\(\llbracket 4,1,2\rrbracket \) twist-defect code} --- Adding \(XYZI\) to the stabilizer group of the \(\llbracket 4,2,2\rrbracket \) four-qubit code yields the \(\llbracket 4,1,2\rrbracket \) twist-defect subcode \NoCaseChange{\protect\cite{cite803}}.
\item\relax
\flmRefsHyperref[eczindexfamilyrel]{code:qubit_5_6_2}{\(\llparenthesis 5,6,2\rrparenthesis \) qubit code} --- Tracing out any one qubit of the \(\llparenthesis 5,6,2\rrparenthesis \) code projector yields a \(\llparenthesis 4,4,2\rrparenthesis \) code; for this code, all five such partial traces are additive and therefore locally equivalent to the \(\llbracket 4,2,2\rrbracket \) code \NoCaseChange{\protect\cite[{Thm. 8 and Corr. 18}]{cite446}}.
\item\relax
\flmRefsHyperref[eczindexfamilyrel]{code:stab_5_1_2}{\(\llbracket 5,1,2\rrbracket \) rotated surface code} --- The \(\llbracket 5,1,2\rrbracket \) morphed Steane code is obtained by morphing the Steane code on a region whose child code is a \(\llbracket 4,2,2\rrbracket \) code \NoCaseChange{\protect\cite[{Fig. 1}]{cite687}}.
\item\relax
\flmRefsHyperref[eczindexfamilyrel]{code:stab_8_2_3}{\(\llbracket 8,2,3\rrbracket \) Hermitian code} --- Applying the BLT mapping to the \(\llbracket 8,2,3\rrbracket \) Hermitian code and concatenating each qubit pair with the \(\llbracket 4,2,2\rrbracket \) code yields a \(\llbracket 32,4,6\rrbracket \) self-dual CSS code \NoCaseChange{\protect\cite[{Corr. 2}]{cite795}}.
\item\relax
\flmRefsHyperref[eczindexfamilyrel]{code:bc_phantom}{Binarized-and-concatenated (B\&C) phantom code} --- The \(\llbracket 4,2,2\rrbracket \) code is used as the inner code for each binarized \(\mathbb{F}_4\)-qudit pair \NoCaseChange{\protect\cite{cite514}}.
\item\relax
\flmRefsHyperref[eczindexfamilyrel]{code:hypergraph_product}{Hypergraph product (HGP) code} --- There is a fault-tolerant universal computation scheme for hypergraph-product codes concatenated with the \(\llbracket 4,2,2\rrbracket \) code in which the full syndrome measurement on the lower hypergraph product code is performed only if an error is detected at the upper four-qubit code \NoCaseChange{\protect\cite{cite3295}}.
\item\relax
\flmRefsHyperref[eczindexfamilyrel]{code:quantum_hamming_css}{\(\llbracket 2^r-1, 2^r-2r-1, 3\rrbracket \) quantum Hamming code} --- Concatenating quantum Hamming codes on top of the \(\llbracket 4,2,2\rrbracket \) and \(C_6\) codes yields fault-tolerant quantum computation with constant space and quasi-polylogarithmic time overheads \NoCaseChange{\protect\cite{cite3216}}. In the optimized protocol of Ref. \NoCaseChange{\protect\cite{cite3216}}, a level-five \(C_4/C_6\) code underlies concatenated quantum Hamming codes \(\mathcal{Q}_5,\mathcal{Q}_6,\mathcal{Q}_7,\mathcal{Q}_7\), yielding a \(2.5\%\) threshold and space overheads \(162\) and \(373\) physical qubits per logical qubit at physical error rate \(0.1\%\) for logical CNOT error rates \(10^{-10}\) and \(10^{-24}\), respectively.
\item\relax
\flmRefsHyperref[eczindexfamilyrel]{code:self_dual_css}{Self-dual CSS code} --- Any \(\llbracket n,k,d\rrbracket \) qubit stabilizer code maps to a \(\llbracket 4n,2k,2d\rrbracket \) self-dual CSS code by applying the BLT mapping and concatenating each qubit pair with the \(\llbracket 4,2,2\rrbracket \) code \NoCaseChange{\protect\cite[{Corr. 2}]{cite795}\protect\cite[{Corr. 1}]{cite1432}}. The BLT mapping proceeds by first concatenating each qubit with the \flmRefsHyperref{code:tetron}{tetron code} to obtain an intermediate \(\llbracket 2n,k,2d\rrbracket _{f}\) Majorana stabilizer code.
\item\relax
\flmRefsHyperref[eczindexfamilyrel]{code:bacon_shor_4}{\(\llbracket 4,1,1,2\rrbracket \) Four-qubit subsystem code} --- The \(\llbracket 4,1,1,2\rrbracket \) code can be obtained by picking one of the logical qubits of the \(\llbracket 4,2,2\rrbracket \) four-qubit code to be a gauge qubit; e.g., see Ref. \NoCaseChange{\protect\cite{cite3247}}. One particular gauge configuration has gauge operators \(\{XXII,IIXX,ZIZI,IZIZ\}\).
\end{eczvaluelist}
\eczhbkcontributors{ Yinchen Liu, Antonio D. Córcoles, Qingfeng (Kee) Wang, \eczhuVVA }
\endeczcode

\eczcode{stab_47_1_11}{\(\llbracket 47,1,11\rrbracket \) quantum QR code}{~\NoCaseChange{\protect\cite{cite3225}}}
\eczhIndexCodeAliasName{stab_47_1_11}{quantum QR code}
\codefieldsection{Description}
A \(\llbracket 47,1,11\rrbracket \) self-dual CSS code constructed from a binary quadratic-residue code.
It is one of the shortest known qubit stabilizer codes of distance eleven that realizes the full Clifford group transversally \NoCaseChange{\protect\cite{cite760}}.

\codefieldsection{Protection}
Detects up to 10-qubit errors and corrects up to 5-qubit errors.
\codefieldsection{Transversal and Permutation-Based Gates}
\begin{eczvaluelist}
\item\relax All logical \flmRefsHyperref{ref409}{Clifford gates} are realized transversally \NoCaseChange{\protect\cite{cite760}}.
\end{eczvaluelist}
\codefieldsection{Decoding}
\begin{eczvaluelist}
\item\relax Algebraic decoder \NoCaseChange{\protect\cite{cite3225}}.
\end{eczvaluelist}
\codefieldsection{Parents}
\begin{eczvaluelist}
\item\relax
\flmRefsHyperref[eczindexfamilyrel]{code:self_dual_css}{Self-dual CSS code}\item\relax
\flmRefsHyperref[eczindexfamilyrel]{code:galois_quad_residue}{Quantum quadratic-residue (QR) code} --- The \(\llbracket 47,1,11\rrbracket \) code is a qubit quantum QR code \NoCaseChange{\protect\cite{cite3225,cite760}}.
\end{eczvaluelist}
\codefieldsection{Cousin}
\begin{eczvaluelist}
\item\relax
\flmRefsHyperref[eczindexfamilyrel]{code:self_dual_48_24_12}{\([48,24,12]\) self-dual code} --- Applying the puncture-and-CSS construction to the \([48,24,12]\) self-dual doubly even quadratic-residue code yields the \(\llbracket 47,1,11\rrbracket \) quantum QR code \NoCaseChange{\protect\cite{cite760}}.
\end{eczvaluelist}
\eczhbkcontributors{ \eczhuVVA }
\endeczcode

\eczcode{stab_49_1_5}{\(\llbracket 49,1,5\rrbracket \) triorthogonal code}{~\NoCaseChange{\protect\cite[{Appx. B}]{cite691}}}
\eczhIndexCodeAliasName{stab_49_1_5}{triorthogonal code}
\codefieldsection{Description}
Triorthogonal and quantum divisible code which is the smallest distance-five stabilizer code to admit a transversal \(T\) gate \NoCaseChange{\protect\cite{cite659}\protect\cite[{Appx. B}]{cite691}}.
It is one example of a level-three generalized divisible code obtainable from the doubling transformation \NoCaseChange{\protect\cite[{Sec. VI.D}]{cite734}}.
Its \(X\)-type stabilizers form a triply even linear binary code in the \flmRefsHyperref{ref817}{symplectic representation}.

\codefieldsection{Magic}
The code yields an exponent \(\gamma = \log 49 / \log 5 \approx 2.42\).
\codefieldsection{Transversal and Permutation-Based Gates}
\begin{eczvaluelist}
\item\relax The code admits a transversal \(T\) gate \NoCaseChange{\protect\cite[{Appx. B}]{cite691}}.
\end{eczvaluelist}
\codefieldsection{Parents}
\begin{eczvaluelist}
\item\relax
\flmRefsHyperref[eczindexfamilyrel]{code:quantum_triorthogonal}{Triorthogonal code}\item\relax
\flmRefsHyperref[eczindexfamilyrel]{code:quantum_divisible}{Quantum divisible code}\item\relax
\flmRefsHyperref[eczindexfamilyrel]{code:small_distance_qubit_stabilizer}{Small-distance qubit stabilizer code}\end{eczvaluelist}
\codefieldsection{Cousins}
\begin{eczvaluelist}
\item\relax
\flmRefsHyperref[eczindexfamilyrel]{code:doubled_color}{Doubled color code} --- The \(\llbracket 49,1,5\rrbracket \) triorthogonal code can be viewed as a (gauge-fixed) doubled color code obtained from the \(\llbracket 17,1,5\rrbracket \) 4.8.8 color code via the doubling transformation \NoCaseChange{\protect\cite{cite731}}.
\item\relax
\flmRefsHyperref[eczindexfamilyrel]{code:stab_17_1_5}{\(\llbracket 17,1,5\rrbracket \) 4.8.8 color code} --- The \(\llbracket 49,1,5\rrbracket \) triorthogonal code can be viewed as a (gauge-fixed) doubled color code obtained from the \(\llbracket 17,1,5\rrbracket \) 4.8.8 color code via the doubling transformation \NoCaseChange{\protect\cite{cite731}}.
\item\relax
\flmRefsHyperref[eczindexfamilyrel]{code:divisible}{Divisible code} --- The \(\llbracket 49,1,5\rrbracket \) triorthogonal code stabilizer generator matrix can be obtained from a triply even linear binary code \NoCaseChange{\protect\cite[{Appx. B}]{cite691}}.
\item\relax
\flmRefsHyperref[eczindexfamilyrel]{code:binary_dihedral_permutation_invariant}{Binary dihedral PI code} --- The \(\llparenthesis 27,2,5\rrparenthesis \) binary dihedral PI code realizes the \(T\) gate (strongly) transversally, but requires fewer qubits than the \(\llbracket 49,1,5\rrbracket \) triorthogonal code.
\end{eczvaluelist}
\eczhbkcontributors{ Benjamin Quiring, \eczhuVVA }
\endeczcode

\eczcode{stab_5_1_2}{\(\llbracket 5,1,2\rrbracket \) rotated surface code}{}
\codefieldsection{Alternative Names}
\begin{eczvaluelist}
\item\relax \(\llbracket 5,1,2\rrbracket \) morphed Steane code
\item\relax \(\llbracket 5,1,2\rrbracket \) rotated toric code
\item\relax \(\llbracket 5,1,2\rrbracket \) genon code
\item\relax \(\llbracket 5,1,2\rrbracket \) holographic code
\item\relax \(\llbracket 5,1,2\rrbracket \) planar-perfect code
\end{eczvaluelist}
\eczhIndexCodeAliasName{stab_5_1_2}{rotated surface code}
\eczhIndexCodeAliasName{stab_5_1_2}{\(\llbracket 5,1,2\rrbracket \) morphed Steane code}
\eczhIndexCodeAliasName{stab_5_1_2}{\(\llbracket 5,1,2\rrbracket \) rotated toric code}
\eczhIndexCodeAliasName{stab_5_1_2}{\(\llbracket 5,1,2\rrbracket \) genon code}
\eczhIndexCodeAliasName{stab_5_1_2}{\(\llbracket 5,1,2\rrbracket \) holographic code}
\eczhIndexCodeAliasName{stab_5_1_2}{\(\llbracket 5,1,2\rrbracket \) planar-perfect code}
\codefieldsection{Description}
A rotated surface code on one rung of a ladder, with one qubit on the rung, and four qubits surrounding it. This is the smallest code that implements a fault-tolerant logical \(S\) gate using a diagonal depth-one Clifford circuit \NoCaseChange{\protect\cite{cite447}}.

A stabilizer tableau for the code is given by \NoCaseChange{\protect\cite[{ID 18}]{cite453}}
\flmMathEnvironment{align}{}{
\begin{array}{ccccc}
  Z & Z & I & Z & I \\
  I & I & Z & Z & Z \\
  I & X & X & X & I \\
  X & I & I & X & X
\end{array}~.
}
The code is depicted in \flmRefsCref{ref3296}.

\begin{flmFloat}{figure}{NumCap}\includegraphics[width=107.14960629921262bp,max width=\linewidth]{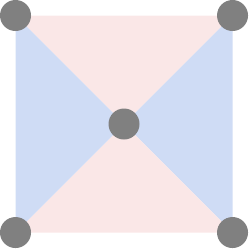}\caption{
  Stabilizer generators of the \(\llbracket 5,1,2\rrbracket \) rotated surface code.
  The 5 data qubits (circles) consist of 4 corner qubits and 1 center qubit.
  Each triangular region corresponds to a weight-three stabilizer generator.
  Red regions correspond to \(X\) operators while blue regions correspond to \(Z\) operators.}\label{ref3296}\end{flmFloat}

A non-CSS genon-code form of the same stabilizer group, local Clifford equivalent to the above via Hadamard on the four corner qubits, is \NoCaseChange{\protect\cite{cite435}}
\flmMathEnvironment{align}{}{
\begin{array}{ccccc}
  X & X & I & Z & I \\
  I & I & X & Z & X \\
  I & Z & Z & X & I \\
  Z & I & I & X & Z
\end{array}~.
}
The missing external stabilizer \(YYYIY\) (product of all four generators) forms the back face when the code is viewed as a genon code on a sphere.

\codefieldsection{Gates}
\begin{eczvaluelist}
\item\relax In the qubit order of the stabilizer tableau above, the physical action \(S_0 S_2 S_3^{\dagger} CZ_{1,4}\) implements the logical gate \(\bar{S}\) \NoCaseChange{\protect\cite[{Fig. 1}]{cite687}}.
\item\relax Fault-tolerant implementation of the \flmRefsHyperref{ref409}{single-qubit Clifford group} \NoCaseChange{\protect\cite[{Fig. 1}]{cite687}}.
\end{eczvaluelist}
\codefieldsection{Fault Tolerance}
\begin{eczvaluelist}
\item\relax Fault-tolerant implementation of the \flmRefsHyperref{ref409}{single-qubit Clifford group} \NoCaseChange{\protect\cite[{Fig. 1}]{cite687}}.
\end{eczvaluelist}
\codefieldsection{Parents}
\begin{eczvaluelist}
\item\relax
\flmRefsHyperref[eczindexfamilyrel]{code:rotated_surface}{Rotated surface code}\item\relax
\flmRefsHyperref[eczindexfamilyrel]{code:morphed_diagonal_clifford}{\(\llbracket 2^r+r-1,1,2\rrbracket \) morphed simplex code} --- The \(\llbracket 5,1,2\rrbracket \) code is a specific instance of the \(\llbracket 2^r+r-1,1,2\rrbracket \) morphed simplex codes with \(r=2\) \NoCaseChange{\protect\cite[{Fig. 1}]{cite687}}.
\item\relax
\flmRefsHyperref[eczindexfamilyrel]{code:holographic_5_1_2}{Surface-code-fragment (SCF) holographic code} --- The \(\llbracket 5,1,2\rrbracket \) rotated surface code is the smallest SCF holographic code \NoCaseChange{\protect\cite{cite2954}}. The encoding of more general SCF holographic codes is a holographic tensor network consisting of the encoding isometry for the \(\llbracket 5,1,2\rrbracket \) rotated surface code, which is a \flmRefsHyperref{code:block_perfect}{planar-perfect tensor}.
\item\relax
\flmRefsHyperref[eczindexfamilyrel]{code:block_perfect}{Planar-perfect-tensor code} --- The \(\llbracket 5,1,2\rrbracket \) rotated surface code is the smallest SCF holographic code \NoCaseChange{\protect\cite{cite2954}}. The encoding of more general SCF holographic codes is a holographic tensor network consisting of the encoding isometry for the \(\llbracket 5,1,2\rrbracket \) rotated surface code, which is a \flmRefsHyperref{code:block_perfect}{planar-perfect tensor}.
\end{eczvaluelist}
\codefieldsection{Cousins}
\begin{eczvaluelist}
\item\relax
\flmRefsHyperref[eczindexfamilyrel]{code:twist_defect_surface}{Twist-defect surface code} --- The \(\llbracket 5,1,2\rrbracket \) rotated surface code is a genon code on a sphere, with the missing external \(Y\)-type stabilizer forming the back of the sphere. More generally, any surface code with a single boundary component can be interpreted this way \NoCaseChange{\protect\cite{cite435}}.
\item\relax
\flmRefsHyperref[eczindexfamilyrel]{code:steane}{\(\llbracket 7,1,3\rrbracket \) Steane code} --- The \(\llbracket 5,1,2\rrbracket \) morphed Steane code is obtained by morphing the Steane code on a region whose child code is a \(\llbracket 4,2,2\rrbracket \) code \NoCaseChange{\protect\cite[{Fig. 1}]{cite687}}.
\item\relax
\flmRefsHyperref[eczindexfamilyrel]{code:stab_4_2_2}{\(\llbracket 4,2,2\rrbracket \) Four-qubit code} --- The \(\llbracket 5,1,2\rrbracket \) morphed Steane code is obtained by morphing the Steane code on a region whose child code is a \(\llbracket 4,2,2\rrbracket \) code \NoCaseChange{\protect\cite[{Fig. 1}]{cite687}}.
\item\relax
\flmRefsHyperref[eczindexfamilyrel]{code:xzzx_10_2_3}{\(\llbracket 10,2,3\rrbracket \) rotated toric code} --- The \(\llbracket 10,2,3\rrbracket \) rotated toric code is the \flmRefsHyperref{ref436}{symplectic double} (a.k.a. genus-one double cover) of the \(\llbracket 5,1,2\rrbracket \) rotated surface code \NoCaseChange{\protect\cite{cite435}}.
\end{eczvaluelist}
\eczhbkcontributors{ \eczhuVVA }
\endeczcode

\eczcode{stab_5_1_3}{\(\llbracket 5,1,3\rrbracket \) Five-qubit perfect code}{~\NoCaseChange{\protect\cite{cite3297,cite2763}}}
\codefieldsection{Alternative Names}
\begin{eczvaluelist}
\item\relax Laflamme code
\end{eczvaluelist}
\eczhIndexCodeAliasName{stab_5_1_3}{Five-qubit perfect code}
\eczhIndexCodeAliasName{stab_5_1_3}{Laflamme code}
\codefieldsection{Description}
Five-qubit cyclic stabilizer code that is the smallest qubit stabilizer code to correct a single-qubit error.

A stabilizer tableau for the code is given by \NoCaseChange{\protect\cite[{ID 21}]{cite453}}
\flmMathEnvironment{align}{}{
\begin{array}{ccccc}
  X & Z & Z & X & I \\
  I & X & Z & Z & X \\
  X & I & X & Z & Z \\
  Z & X & I & X & Z
\end{array}~.
}
A basis of codewords for the above stabilizer is \NoCaseChange{\protect\cite[{Ch. 3}]{cite398}}
\flmMathEnvironment{align}{}{
\begin{split}
|\overline{0}\rangle &= \tfrac{1}{4}(|00000\rangle + |10010\rangle + |01001\rangle - |11011\rangle \\
&\quad + |10100\rangle - |00110\rangle - |11101\rangle - |01111\rangle \\
&\quad + |01010\rangle - |11000\rangle - |00011\rangle - |10001\rangle \\
&\quad - |11110\rangle - |01100\rangle - |10111\rangle + |00101\rangle)\\
|\overline{1}\rangle &= \tfrac{1}{4}(|11111\rangle + |01101\rangle + |10110\rangle - |00100\rangle \\
&\quad + |01011\rangle - |11001\rangle - |00010\rangle - |10000\rangle \\
&\quad + |10101\rangle - |00111\rangle - |11100\rangle - |01110\rangle \\
&\quad - |00001\rangle - |10011\rangle - |01000\rangle + |11010\rangle)~.
\end{split}
}
Logical Pauli operators are \(\bar{X} = XXXXX\) and \(\bar{Z} = ZZZZZ\) \NoCaseChange{\protect\cite[{Table 3.4}]{cite398}}.
The code's automorphism group is the dihedral group of order 10 \NoCaseChange{\protect\cite{cite3222}}.
A graph-code realization of the code uses a pentagon graph with an additional central input node \NoCaseChange{\protect\cite{cite866}}.
The \flmRefsHyperref{ref857}{encoder-respecting form} can be taken to have this shape \NoCaseChange{\protect\cite{cite858}}.

It is the unique code for its parameters, up to equivalence \NoCaseChange{\protect\cite[{Corr. 10}]{cite446}}.
Any five-qubit \(2T\)-transversal stabilizer code with distance \(d>1\) must be the five-qubit code \NoCaseChange{\protect\cite{cite3298,cite3299}}.

This code is sometimes referred to as the DiVincenzo-Shor code after a paper that studied the code's syndrome extraction circuits \NoCaseChange{\protect\cite{cite3300}}.

\codefieldsection{Protection}
Smallest stabilizer code that protects against a single error on any one qubit. Detects two-qubit errors.
The five-qubit perfect code approximately corrects a single \flmRefsHyperref{ref498}{AD} error \NoCaseChange{\protect\cite{cite859}}.

\codefieldsection{Encoding}
\begin{eczvaluelist}
\item\relax Nine single- and two-qubit unitaries, six of which are CNOT gates \NoCaseChange{\protect\cite{cite3301}}.
\item\relax Four generalized control gates, four Hadamard, and one \(Z\) gate \NoCaseChange{\protect\cite[{Fig. 10.16}]{cite3302}}.
\item\relax Evolution under stabilizer Hamiltonian \NoCaseChange{\protect\cite{cite3303}}.
\item\relax Four CNOT and five CPHASE gates \NoCaseChange{\protect\cite{cite3304}}.
\item\relax Reinforcement-learning discovery of logical-state-preparation circuits \NoCaseChange{\protect\cite{cite3200}}.
\item\relax Fault-tolerant logical one and logical minus state preparation in all-to-all and 2D grid connectivity \NoCaseChange{\protect\cite{cite3200}}.
\end{eczvaluelist}
\codefieldsection{Transversal and Permutation-Based Gates}
\begin{eczvaluelist}
\item\relax A non-Pauli Hadamard-phase "facet" gate \(SH\) and three-qubit Clifford operation \(M_3\) \NoCaseChange{\protect\cite{cite737,cite736}}. These realize the \(2T\) binary tetrahedral subgroup of \(SU(2)\).
\item\relax The entire logical Clifford group can be realized using fold-transversal gates \NoCaseChange{\protect\cite{cite806,cite807,cite719}}.
\item\relax The code does not admit any \flmRefsHyperref{ref409}{non-Clifford} transversal gates \NoCaseChange{\protect\cite{cite446}}; in particular, see \NoCaseChange{\protect\cite{cite808}} for the case of collective \(Z\) rotations.
\item\relax Transversal gates can be interpreted as monodromies under a particular notion of parallel transport \NoCaseChange{\protect\cite[{Exam. 6.4.2}]{cite809}}.
\end{eczvaluelist}
\codefieldsection{Gates}
\begin{eczvaluelist}
\item\relax Magic-state distillation protocol \NoCaseChange{\protect\cite{cite690}}. One protocol distills the state \(\ket{R}=\cos\beta\ket{0}+e^{\mathrm{i}\pi/4}\sin\beta\ket{1}\), with \(\cos(2\beta)=1/\sqrt{3}\), using the fact that a transversal Clifford gate \(R\) is a gadget for the code; the protocol projects five noisy \(\ket{R}\) states onto the code space and suppresses the output error to \(O(p^2)\) for independent input error probability \(p\) \NoCaseChange{\protect\cite[{Sec. 13.5.3, Protocol 13.5}]{cite398}}.
\item\relax Pieceable fault-tolerant CZ, CNOT, and \(CCZ\) gates \NoCaseChange{\protect\cite{cite806}}.
\end{eczvaluelist}
\codefieldsection{Decoding}
\begin{eczvaluelist}
\item\relax Ideal transversal computational-basis measurement distinguishes logical basis states by the parity of the outcome string, but this is not a fault-tolerant measurement gadget because a single faulty measurement bit can flip the decoded logical outcome \NoCaseChange{\protect\cite[{Sec. 11.4}]{cite398}}.
\item\relax Fault-tolerant syndrome extraction circuits \NoCaseChange{\protect\cite{cite3300,cite3305}}.
\item\relax Syndrome extraction circuit optimized for a linear qubit architecture \NoCaseChange{\protect\cite{cite3306}}.
\item\relax Combined dynamical decoupling and error correction protocol on individually-controlled qubits with always-on Ising couplings \NoCaseChange{\protect\cite{cite3304}}.
\item\relax Syndrome extraction circuit using only CNOT-SWAP gates \NoCaseChange{\protect\cite{cite3307}}.
\item\relax Symmetric decoder correcting all weight-one Pauli errors. The resulting logical error channel after coherent noise has been explicitly derived \NoCaseChange{\protect\cite{cite3308}}.
\item\relax Inspired by the honeycomb Floquet code, various weight-two measurement schemes have been designed \NoCaseChange{\protect\cite{cite3309}}.
\end{eczvaluelist}
\codefieldsection{Fault Tolerance}
\begin{eczvaluelist}
\item\relax Pieceable fault-tolerant CZ, CNOT, and \(CCZ\) gates \NoCaseChange{\protect\cite{cite806}}.
\item\relax A fault-tolerant logical \(T\) gate can be obtained by encoding the five-qubit code's five physical qubits into the five logical qubits of a \(\llbracket 31,5,3\rrbracket \) outer quantum divisible CSS code preserved by transversal \(T^\dagger\); this layered construction can be viewed as a factorization of a \(\llbracket 31,1,3\rrbracket \) triorthogonal code and does not require magic-state distillation \NoCaseChange{\protect\cite{cite765}}.
\item\relax Syndrome measurement can be done with two ancillary flag qubits \NoCaseChange{\protect\cite{cite3215}}. The depth of syndrome extraction circuits can be lowered by using past syndrome values \NoCaseChange{\protect\cite{cite3310}}.
\item\relax Fault-tolerant logical one and logical minus state preparation in all-to-all and 2D grid connectivity \NoCaseChange{\protect\cite{cite3200}}.
\item\relax Inspired by the honeycomb Floquet code, various weight-two measurement schemes have been designed \NoCaseChange{\protect\cite{cite3309}}.
\end{eczvaluelist}
\codefieldsection{Realizations}
\begin{eczvaluelist}
\item\relax NMR: Implementation of perfect error correcting code on 5 spin subsystem of labeled crotonic acid for quantum network benchmarking \NoCaseChange{\protect\cite{cite3311}}. Single-qubit logical gates \NoCaseChange{\protect\cite{cite3312}}. Magic-state distillation using 7-qubit device \NoCaseChange{\protect\cite{cite3313}}.
\item\relax Superconducting qubits \NoCaseChange{\protect\cite{cite3314}}.
\item\relax Trapped-ion qubits: non-transversal CNOT gate between two logical qubits, including rounds of correction and fault-tolerant primitives such as flag qubits and pieceable fault tolerance, on a 12-qubit device by Quantinuum \NoCaseChange{\protect\cite{cite3315}}. Real-time magic-state distillation \NoCaseChange{\protect\cite{cite3316}}.
\item\relax Nitrogen-vacancy centers in diamond: fault-tolerant single-qubit Clifford operations using two ancillas \NoCaseChange{\protect\cite{cite3317}}. The fault-tolerant circuit yields better fidelity than the non-fault-tolerant circuit.
\end{eczvaluelist}
\codefieldsection{Parents}
\begin{eczvaluelist}
\item\relax
\flmRefsHyperref[eczindexfamilyrel]{code:twisted_xzzx}{Twisted XZZX toric code} --- Twisted XZZX codes are 2D lattice extensions of the five-qubit perfect code. The five-qubit code is a small twisted XZZX toric code \NoCaseChange{\protect\cite[{Exam. 11 and Fig. 3}]{cite438}\protect\cite[{Exam. 3}]{cite439}\protect\cite[{Fig. 1}]{cite427}}. Its genus-one double cover is a \(\llbracket 10,2,3\rrbracket \) toric code \NoCaseChange{\protect\cite{cite435}\protect\cite[{Exam. 3}]{cite439}}. The base code's transversal \(SH\) gate lifts to a logical \(CX \cdot SWAP\) gate on that double cover \NoCaseChange{\protect\cite{cite435}}.
\item\relax
\flmRefsHyperref[eczindexfamilyrel]{code:stab_5_1_2_convolutional}{\((5,1,2)\)-convolutional code} --- The \((5,1,2)\)-convolutional code is a 1D lattice extension of the five-qubit perfect code, with the former's lattice-translation symmetry being the extension of the latter's cyclic permutation symmetry. The \((5,1,2)\)-convolutional code reduces to the five-qubit code for a five-qubit chain and periodic boundary conditions. See Ref. \NoCaseChange{\protect\cite{cite3181}} for the first few codes in a different extension of the five-qubit perfect code.
\item\relax
\flmRefsHyperref[eczindexfamilyrel]{code:happy}{Pastawski-Yoshida-Harlow-Preskill (HaPPY) code} --- The five-qubit code is the smallest (i.e., radius-one) single-qubit HaPPY code. The five-qubit encoding isometry tiles various holographic codes because its corresponding encoding isometry tensor is a \flmRefsHyperref{ref219}{perfect tensor} \NoCaseChange{\protect\cite{cite1667}}.
\item\relax
\flmRefsHyperref[eczindexfamilyrel]{code:quantum_perfect}{Perfect quantum code} --- The five-qubit code is the smallest perfect code and is a member of the perfect qubit code family \(\llbracket (4^r-1)/3, (4^r-1)/3 - 2r, 3\rrbracket \) for \(r = 2\).
\item\relax
\flmRefsHyperref[eczindexfamilyrel]{code:stabilizer_over_gf4}{Hermitian qubit code} --- The five-qubit code is Hermitian \NoCaseChange{\protect\cite[{ID 21}]{cite453}}, and is derived from the \([5,3,3]_4\) shortened hexacode via the \flmRefsHyperref{code:stabilizer_over_gf4}{qubit Hermitian construction} \NoCaseChange{\protect\cite{cite1670}\protect\cite[{Exam. A}]{cite1666}}.
\item\relax
\flmRefsHyperref[eczindexfamilyrel]{code:quantum_mds}{Quantum maximum-distance-separable (MDS) code} --- The only nontrivial qubit MDS codes have parameters \(\llbracket 5,1,3\rrbracket \), \(\llbracket 6,0,4\rrbracket \), and \(\llbracket 2m,2m-2,2\rrbracket \) \NoCaseChange{\protect\cite[{Sec. 27.4}]{cite2024}}.
\item\relax
\flmRefsHyperref[eczindexfamilyrel]{code:frobenius}{Frobenius code} --- The \(\llbracket 5,1,3\rrbracket \) code is the smallest qubit Frobenius code \NoCaseChange{\protect\cite[{Table I}]{cite3318}}.
\item\relax
\flmRefsHyperref[eczindexfamilyrel]{code:qudit_5_1_3}{\(\llbracket 5,1,3\rrbracket _{\mathbb{Z}_q}\) modular-qudit code} --- The \(\llbracket 5,1,3\rrbracket _{\mathbb{Z}_q}\) modular-qudit code for \(q=2\) reduces to the five-qubit perfect code.
\item\relax
\flmRefsHyperref[eczindexfamilyrel]{code:galois_5_1_3}{\(\llbracket 5,1,3\rrbracket _q\) Galois-qudit code} --- The \(\llbracket 5,1,3\rrbracket _q\) Galois-qudit code for \(q=2\) reduces to the five-qubit perfect code.
\item\relax
\flmRefsHyperref[eczindexfamilyrel]{code:small_distance_qubit_stabilizer}{Small-distance qubit stabilizer code}\end{eczvaluelist}
\codefieldsection{Cousins}
\begin{eczvaluelist}
\item\relax
\flmRefsHyperref[eczindexfamilyrel]{code:group_representation}{Group-representation code} --- The five-qubit code is a group-representation code with \(G\) being the \(2T\) subgroup of \(SU(2)\) \NoCaseChange{\protect\cite{cite2810}}.
\item\relax
\flmRefsHyperref[eczindexfamilyrel]{code:majorana_stab}{Majorana stabilizer code} --- The five-qubit code Hamiltonian is local when expressed in terms of mutually commuting Majorana operators \NoCaseChange{\protect\cite{cite3319}}.
\item\relax
\flmRefsHyperref[eczindexfamilyrel]{code:qubits_into_qubits}{Qubit code} --- Every \(\llparenthesis 5,2,3\rrparenthesis \) qubit code is single-qubit-Clifford-equivalent equivalent to the five-qubit code \NoCaseChange{\protect\cite[{Corr. 10}]{cite446}}.
\item\relax
\flmRefsHyperref[eczindexfamilyrel]{code:qubit_concatenated}{Concatenated qubit code} --- The recursively concatenated five-qubit code has a \flmRefsHyperref{ref3210}{measurement threshold} of one \NoCaseChange{\protect\cite{cite3211}}. Code performance against general Pauli channels has been worked out \NoCaseChange{\protect\cite{cite3320,cite3321}}.
\item\relax
\flmRefsHyperref[eczindexfamilyrel]{code:cluster_state}{Cluster-state code} --- The five-qubit perfect code is equivalent via a single-qubit \flmRefsHyperref{ref409}{Clifford circuit} to a cluster-state code defined from a five-cycle (a.k.a. pentagon) graph and a classical repetition code \NoCaseChange{\protect\cite{cite852,cite3166,cite3322,cite868}\protect\cite[{Exam. 2}]{cite438}}.
\item\relax
\flmRefsHyperref[eczindexfamilyrel]{code:floquet}{Hastings-Haah Floquet code} --- Inspired by the honeycomb Floquet code, various weight-two measurement schemes have been designed for the five-qubit code \NoCaseChange{\protect\cite{cite3309}}.
\item\relax
\flmRefsHyperref[eczindexfamilyrel]{code:ampdamp}{Amplitude-damping (AD) code} --- The five-qubit perfect code approximately corrects a single \flmRefsHyperref{ref498}{AD} error \NoCaseChange{\protect\cite{cite859}}.
\item\relax
\flmRefsHyperref[eczindexfamilyrel]{code:quantum_divisible}{Quantum divisible code} --- A fault-tolerant logical \(T\) gate can be obtained by encoding the five-qubit code's five physical qubits into the five logical qubits of a \(\llbracket 31,5,3\rrbracket \) outer quantum divisible CSS code preserved by transversal \(T^\dagger\); this layered construction can be viewed as a factorization of a \(\llbracket 31,1,3\rrbracket \) triorthogonal code and does not require magic-state distillation \NoCaseChange{\protect\cite{cite765}}.
\item\relax
\flmRefsHyperref[eczindexfamilyrel]{code:quantum_triorthogonal}{Triorthogonal code} --- A fault-tolerant logical \(T\) gate can be obtained by encoding the five-qubit code's five physical qubits into the five logical qubits of a \(\llbracket 31,5,3\rrbracket \) outer quantum divisible CSS code preserved by transversal \(T^\dagger\); this layered construction can be viewed as a factorization of a \(\llbracket 31,1,3\rrbracket \) triorthogonal code and does not require magic-state distillation \NoCaseChange{\protect\cite{cite765}}.
\item\relax
\flmRefsHyperref[eczindexfamilyrel]{code:hexacode}{\([6,3,4]_4\) Hexacode} --- Applying the \flmRefsHyperref{code:stabilizer_over_gf4}{qubit Hermitian construction} to the hexacode yields the \(\llbracket 6,0,4\rrbracket \) quantum hexacode \NoCaseChange{\protect\cite{cite1670}}, and tracing out any one qubit of that code yields the \(\llbracket 5,1,3\rrbracket \) five-qubit code \NoCaseChange{\protect\cite[{Cor. 9 and Cor. 10}]{cite446}}.
\item\relax
\flmRefsHyperref[eczindexfamilyrel]{code:shortened_hexacode}{\([5,3,3]_4\) Shortened hexacode} --- The five-qubit code can be obtained either by applying the \flmRefsHyperref{code:stabilizer_over_gf4}{qubit Hermitian construction} to the shortened hexacode \NoCaseChange{\protect\cite[{Exam. A}]{cite1666}} or by tracing out a qubit of the \(\llbracket 6,0,4\rrbracket \) code \NoCaseChange{\protect\cite[{Appx. A}]{cite1667}}.
\item\relax
\flmRefsHyperref[eczindexfamilyrel]{code:group_10_1_4}{\(\llbracket 10,1,4\rrbracket _{G}\) tenfold code} --- The \(\llbracket 10,1,4\rrbracket _{G}\) Abelian group code for \(G=\mathbb{Z}_2\) is defined using a graph that is closely related to the \(\llbracket 5,1,3\rrbracket \) five-qubit code \NoCaseChange{\protect\cite{cite866}}. The former code can be obtained by converting the latter into a code that is oblivious to collective \(Z\)-type rotations \NoCaseChange{\protect\cite[{Exam. 6}]{cite808}}.
\item\relax
\flmRefsHyperref[eczindexfamilyrel]{code:gkp_concatenated}{Concatenated GKP code} --- GKP codes have been concatenated with the five-qubit code \NoCaseChange{\protect\cite{cite3323}}.
\item\relax
\flmRefsHyperref[eczindexfamilyrel]{code:nonabelian_covariant_erasure}{\(U(d)\)-covariant approximate erasure code} --- The five-qubit code can be used to construct an approximate code that is also covariant with respect to the unitary group.
\item\relax
\flmRefsHyperref[eczindexfamilyrel]{code:constant_excitation}{Constant-excitation (CE) code} --- The five-qubit code can be concatenated with a particular decoherence-free subspace (DFS) \NoCaseChange{\protect\cite{cite2712,cite2713,cite2714,cite2715}} to yield a 20-qubit CE code \NoCaseChange{\protect\cite{cite2708,cite2716}}. Dual-rail concatenation of the five-qubit code yields a \(\llbracket 10,1,3\rrbracket \) CE stabilizer code \NoCaseChange{\protect\cite{cite524}}.
\item\relax
\flmRefsHyperref[eczindexfamilyrel]{code:ea_3_1_3-2}{\(\llbracket 3, 1, 3;2\rrbracket \) EA code} --- The \(\llbracket 3, 1, 3;2\rrbracket \) EA code and the five-qubit code have the same stabilizers \NoCaseChange{\protect\cite{cite1430,cite2702}}.
\item\relax
\flmRefsHyperref[eczindexfamilyrel]{code:reinforcement_learning}{Reinforcement-learning quantum code} --- Various five-qubit codes, numerically obtained through variational techniques, can outperform the five-qubit perfect code against depolarizing noise \NoCaseChange{\protect\cite{cite3324}}.
\item\relax
\flmRefsHyperref[eczindexfamilyrel]{code:xzzx_10_2_3}{\(\llbracket 10,2,3\rrbracket \) rotated toric code} --- The \(\llbracket 10,2,3\rrbracket \) rotated toric code is the \flmRefsHyperref{ref436}{symplectic double} (a.k.a. genus-one double cover) of the five-qubit perfect code \NoCaseChange{\protect\cite{cite435}\protect\cite[{Exam. 3}]{cite439}}. A non-CSS cyclic cluster code related to the \(\llbracket 10,2,3\rrbracket \) rotated toric code yields the \(\llbracket 5,1,3\rrbracket \) five-qubit perfect code for \(d=3\) \NoCaseChange{\protect\cite{cite440}}.
\item\relax
\flmRefsHyperref[eczindexfamilyrel]{code:stab_20_2_6}{\(\llbracket 20,2,6\rrbracket \) B\&C phantom code} --- The \(\llbracket 20,2,6\rrbracket \) code is obtained from the \(\llbracket 5,1,3\rrbracket \) five-qubit code via the BLT mapping (Lemma 1) and concatenation with the \(\llbracket 4,2,2\rrbracket \) code (Corollary 2) \NoCaseChange{\protect\cite{cite795}\protect\cite[{Corr. 1}]{cite1432}}.
\item\relax
\flmRefsHyperref[eczindexfamilyrel]{code:stab_4_2_2}{\(\llbracket 4,2,2\rrbracket \) Four-qubit code} --- The \(\llbracket 4,2,2\rrbracket \) code can be derived from the five-qubit code using a protocol that converts an \(\llbracket n,k,d\rrbracket \) code into an \(\llbracket n-1, k+1, d-1\rrbracket \) code \NoCaseChange{\protect\cite[{Sec. 3.5}]{cite736}\protect\cite[{Fig. 3}]{cite2933}}.
\item\relax
\flmRefsHyperref[eczindexfamilyrel]{code:stab_6_1_3}{\(\llbracket 6,1,3\rrbracket \) Six-qubit stabilizer code} --- The \(\llbracket 6,1,3\rrbracket \) six-qubit code is one of two six-qubit distance-three codes that are unique up to equivalence \NoCaseChange{\protect\cite{cite449}\protect\cite[{ID 68}]{cite453}}, with the other code being decomposable and an extension of the five-qubit code \NoCaseChange{\protect\cite{cite451}\protect\cite[{ID 87}]{cite453}}.
\item\relax
\flmRefsHyperref[eczindexfamilyrel]{code:holographic_subsystem}{Subsystem holographic code} --- The holographic hybrid code is constructed out of alternating isometries of the five-qubit and \(\llbracket 4,1,1,2\rrbracket \) Bacon-Shor codes.
\item\relax
\flmRefsHyperref[eczindexfamilyrel]{code:css_5_1_3}{\(\llbracket 5,1,3\rrbracket _4\) Galois-qudit CSS code} --- The \(\llbracket 5,1,3\rrbracket _4\) Galois-qudit CSS code is the image of the \(\llbracket 5,1,3\rrbracket \) five-qubit code under the BLT mapping \NoCaseChange{\protect\cite[{Lemma 1}]{cite795}\protect\cite[{Lemma 1}]{cite1432}}.
\end{eczvaluelist}
\eczhbkcontributors{ Remmy Zen, Aleksander Kubica, Marianna Podzorova, Qingfeng (Kee) Wang, \eczhuVVA }
\endeczcode

\eczcode{quantum_icosahedron}{\(\llbracket 54,6,5\rrbracket \) five-covered icosahedral code}{~\NoCaseChange{\protect\cite{cite858}}}
\eczhIndexCodeAliasName{quantum_icosahedron}{five-covered icosahedral code}
\codefieldsection{Description}
A \(\llbracket 54,6,5\rrbracket \) qubit stabilizer code whose \flmRefsHyperref{ref857}{encoder-respecting form} is the graph of a five-cover of the icosahedron \NoCaseChange{\protect\cite{cite858}}.
The covering-space construction avoids the weight-three logical operators that occur for the bare icosahedral graph \NoCaseChange{\protect\cite{cite858}}.

\codefieldsection{Parent}
\begin{eczvaluelist}
\item\relax
\flmRefsHyperref[eczindexfamilyrel]{code:small_distance_qubit_stabilizer}{Small-distance qubit stabilizer code}\end{eczvaluelist}
\codefieldsection{Cousin}
\begin{eczvaluelist}
\item\relax
\flmRefsHyperref[eczindexfamilyrel]{code:icosahedron}{Icosahedron code} --- The \flmRefsHyperref{ref857}{encoder-respecting form} of the \(\llbracket 54,6,5\rrbracket \) five-covered icosahedral code is the graph of a five-cover of the icosahedron \NoCaseChange{\protect\cite{cite858}}.
\end{eczvaluelist}
\eczhbkcontributors{ \eczhuVVA }
\endeczcode

\eczcode{css_6_1_2}{\(\llbracket 6,1,2\rrbracket \) semi-self-dual CSS code}{~\NoCaseChange{\protect\cite{cite738}}}
\eczhIndexCodeAliasName{css_6_1_2}{semi-self-dual CSS code}
\codefieldsection{Description}
A six-qubit CSS stabilizer code that is an example of a semi-self-dual CSS code, i.e., a CSS code whose \(X\)-type stabilizers are contained in the \(Z\)-type stabilizers \NoCaseChange{\protect\cite{cite738}}.

A stabilizer tableau for the code is given by \NoCaseChange{\protect\cite[{ID 59}]{cite453}}
\flmMathEnvironment{align}{}{
\begin{array}{cccccc}
  Z & I & Z & I & I & Z \\
  I & Z & Z & I & Z & I \\
  I & I & I & Z & Z & Z \\
  X & I & X & X & X & I \\
  I & X & X & X & I & X
\end{array}~.
}
An equivalent self-dual semi-CSS presentation has generators \(XXXXII\), \(IIXXXX\), \(ZZZZII\), \(IIZZZZ\), and \(IYIYIY\) \NoCaseChange{\protect\cite{cite738}}.
A qubit permutation \(0\mapsto 0\), \(1\mapsto 2\), \(2\mapsto 4\), \(3\mapsto 5\), \(4\mapsto 1\), \(5\mapsto 3\), followed by the same one-qubit Clifford on every qubit sending \(X\mapsto Z\), \(Z\mapsto Y\), and \(Y\mapsto X\), converts the CSS generators to that semi-CSS presentation.
This makes the code a concrete example of the equivalence between the semi-self-dual CSS and self-dual semi-CSS cases in the \(U(\ell,R_8)\) family \NoCaseChange{\protect\cite{cite738}}.

\codefieldsection{Protection}
Detects any single-qubit error as a distance-two stabilizer code.

\codefieldsection{Transversal and Permutation-Based Gates}
\begin{eczvaluelist}
\item\relax The physical gate \(S^{\dagger}_{0}S^{\dagger}_{1}S^{\dagger}_{2}S_{3}S_{4}S_{5}\) implements the logical action \(\bar{S}\) \NoCaseChange{\protect\cite[{ID 59}]{cite453}}.
\item\relax In the equivalent self-dual semi-CSS presentation obtained by fixing one logical qubit of the \(\llbracket 6,2,2\rrbracket \) \(C_6\) code to \(|Y^{-}\rangle_L\), transversal gates support a fault-tolerant magic-state-preparation circuit \NoCaseChange{\protect\cite{cite738}\protect\cite[{ID 59}]{cite453}}.
\end{eczvaluelist}
\codefieldsection{Parents}
\begin{eczvaluelist}
\item\relax
\flmRefsHyperref[eczindexfamilyrel]{code:qubit_css}{Qubit CSS code}\item\relax
\flmRefsHyperref[eczindexfamilyrel]{code:small_distance_qubit_stabilizer}{Small-distance qubit stabilizer code}\end{eczvaluelist}
\codefieldsection{Cousin}
\begin{eczvaluelist}
\item\relax
\flmRefsHyperref[eczindexfamilyrel]{code:stab_6_2_2}{\(\llbracket 6,2,2\rrbracket \) \(C_6\) code} --- Fixing one logical qubit of the \(\llbracket 6,2,2\rrbracket \) \(C_6\) code to \(|Y^{-}\rangle_L\) yields this \(\llbracket 6,1,2\rrbracket \) code \NoCaseChange{\protect\cite{cite738}\protect\cite[{ID 59}]{cite453}}.
\end{eczvaluelist}
\eczhbkcontributors{ \eczhuVVA }
\endeczcode

\eczcode{majorana_6_1_3}{\(\llbracket 6,1,3\rrbracket _{f}\) Vijay-Fu Majorana code}{~\NoCaseChange{\protect\cite{cite566}}}
\eczhIndexCodeAliasName{majorana_6_1_3}{Vijay-Fu Majorana code}
\codefieldsection{Description}
A Majorana stabilizer code encoding a logical fermion into six physical fermions.
This code is the shortest code correcting single fermion-parity flips \NoCaseChange{\protect\cite{cite566}}.

\codefieldsection{Parents}
\begin{eczvaluelist}
\item\relax
\flmRefsHyperref[eczindexfamilyrel]{code:majorana_stab}{Majorana stabilizer code}\item\relax
\flmRefsHyperref[eczindexfamilyrel]{code:small_distance_qubit_stabilizer}{Small-distance qubit stabilizer code}\end{eczvaluelist}
\eczhbkcontributors{ \eczhuVVA }
\endeczcode

\eczcode{stab_6_1_3}{\(\llbracket 6,1,3\rrbracket \) Six-qubit stabilizer code}{~\NoCaseChange{\protect\cite{cite449,cite451}}}
\eczhIndexCodeAliasName{stab_6_1_3}{Six-qubit stabilizer code}
\codefieldsection{Description}
A degenerate, non-trivial \(\llbracket 6,1,3\rrbracket \) stabilizer code.
It is one of two six-qubit distance-three codes that are unique up to equivalence \NoCaseChange{\protect\cite{cite449}}, with the other code being decomposable and an extension of the five-qubit code \NoCaseChange{\protect\cite{cite451}\protect\cite[{ID 87}]{cite453}}.
The code admits fault-tolerant syndrome extraction using only one ancilla per stabilizer generator measurement.

A stabilizer tableau for the code is given by \NoCaseChange{\protect\cite[{ID 68}]{cite453}}
\flmMathEnvironment{align}{}{
\begin{array}{cccccc}
  X & I & I & I & I & Z \\
  I & X & X & Z & Z & I \\
  I & Y & I & Y & Z & Z \\
  I & X & Z & I & X & Z \\
  Z & Y & Z & Z & I & Y
\end{array}~.
}

Stabilizer generators and logical Pauli operators are presented in Refs. \NoCaseChange{\protect\cite{cite451,cite3325}}.
The code is equivalent to the graph code in Ref. \NoCaseChange{\protect\cite{cite3326}}.

\codefieldsection{Encoding}
\begin{eczvaluelist}
\item\relax CNOT and Hadamard gates \NoCaseChange{\protect\cite{cite451,cite3325}}.
\end{eczvaluelist}
\codefieldsection{Gates}
\begin{eczvaluelist}
\item\relax Logical CNOT gate \NoCaseChange{\protect\cite{cite451,cite3325}}.
\end{eczvaluelist}
\codefieldsection{Decoding}
\begin{eczvaluelist}
\item\relax Fault-tolerant syndrome extraction using a single ancilla \NoCaseChange{\protect\cite{cite3326}}.
\end{eczvaluelist}
\codefieldsection{Parents}
\begin{eczvaluelist}
\item\relax
\flmRefsHyperref[eczindexfamilyrel]{code:holographic_6_1_3}{Six-qubit-tensor holographic code} --- The \(\llbracket 6,1,3\rrbracket \) six-qubit stabilizer code is the smallest six-qubit-tensor holographic code. The encoding of more general SCF holographic codes is a holographic tensor network consisting of the encoding isometry for the \(\llbracket 6,1,3\rrbracket \) six-qubit stabilizer code.
\item\relax
\flmRefsHyperref[eczindexfamilyrel]{code:small_distance_qubit_stabilizer}{Small-distance qubit stabilizer code}\end{eczvaluelist}
\codefieldsection{Cousins}
\begin{eczvaluelist}
\item\relax
\flmRefsHyperref[eczindexfamilyrel]{code:qubit_subsystem_stabilizer}{Subsystem qubit stabilizer code} --- The \(\llbracket 6,1,3\rrbracket \) six-qubit code can be converted into a \(\llbracket 6,1,1,3\rrbracket \) subsystem code that saturates the subsystem Singleton bound while requiring only four stabilizer measurements during recovery \NoCaseChange{\protect\cite{cite451,cite3325}}.
\item\relax
\flmRefsHyperref[eczindexfamilyrel]{code:stab_5_1_3}{\(\llbracket 5,1,3\rrbracket \) Five-qubit perfect code} --- The \(\llbracket 6,1,3\rrbracket \) six-qubit code is one of two six-qubit distance-three codes that are unique up to equivalence \NoCaseChange{\protect\cite{cite449}\protect\cite[{ID 68}]{cite453}}, with the other code being decomposable and an extension of the five-qubit code \NoCaseChange{\protect\cite{cite451}\protect\cite[{ID 87}]{cite453}}.
\end{eczvaluelist}
\eczhbkcontributors{ Matthew Steinberg, \eczhuVVA }
\endeczcode

\eczcode{stab_6_2_2}{\(\llbracket 6,2,2\rrbracket \) \(C_6\) code}{~\NoCaseChange{\protect\cite{cite448}}}
\eczhIndexCodeAliasName{stab_6_2_2}{\(C_6\) code}
\codefieldsection{Description}
Error-detecting normal self-dual CSS code on three qubit pairs that encodes a logical qubit pair and detects any error acting on one pair \NoCaseChange{\protect\cite{cite448}}.
In Knill's \(C_4/C_6\) architecture, this code is used at the second and higher concatenation levels.

A stabilizer tableau for the code is given by \NoCaseChange{\protect\cite[{ID 126}]{cite453}}
\flmMathEnvironment{align}{}{
\begin{array}{cccccc}
  Z & Z & Z & Z & I & I \\
  Z & Z & I & I & Z & Z \\
  X & X & X & X & I & I \\
  X & X & I & I & X & X
\end{array}~.
}
Its logical operators are \(X_L = IIXXII\), \(Z_L = ZIIZZI\), \(X_S = IXIXXI\), and \(Z_S = IIIIZZ\) \NoCaseChange{\protect\cite{cite448}}.

\codefieldsection{Protection}
As a distance-two code, the \(C_6\) code detects any single-qubit error.
In the qubit-pair grouping used by Knill, it detects any error acting on one of the three pairs, and therefore can correct a pair error when its location is already known \NoCaseChange{\protect\cite{cite448}}.

\codefieldsection{Magic}
Various magic-state distillation protocols exist for the \(\llbracket 4,2,2\rrbracket \) qubit code and the \(C_6\) code in what are known as Meier-Eastin-Knill (MEK) protocols \NoCaseChange{\protect\cite{cite708}}. For example, the magic-state yield parameter is \(\gamma = \log_2 5 \approx 2.322\) for a protocol using the \(\llbracket 10,2,2\rrbracket \) code \NoCaseChange{\protect\cite[{Box 2}]{cite707}}; see also \NoCaseChange{\protect\cite[{Table IV}]{cite705}}.
\codefieldsection{Transversal and Permutation-Based Gates}
\begin{eczvaluelist}
\item\relax Transversal physical Hadamards preserve the codespace because the displayed \(X\)- and \(Z\)-check spaces coincide, making this a normal self-dual qubit CSS code \NoCaseChange{\protect\cite{cite810}}.
\end{eczvaluelist}
\codefieldsection{Gates}
\begin{eczvaluelist}
\item\relax Fault-tolerant magic-state preparation \NoCaseChange{\protect\cite{cite3327}}.
\end{eczvaluelist}
\codefieldsection{Fault Tolerance}
\begin{eczvaluelist}
\item\relax Knill's \(C_4/C_6\) architecture uses the \(\llbracket 4,2,2\rrbracket \) code at the first level and the \(C_6\) code at higher levels, together with error-correcting teleportation \NoCaseChange{\protect\cite{cite448}}. That paper gives evidence for postselected thresholds above \(0.03\) and extrapolates to \(0.06\), while the error-correcting architecture has evidence for a threshold above \(0.01\). Later work refined the postselected-threshold analysis \NoCaseChange{\protect\cite{cite3273,cite3274}} (see also Ref. \NoCaseChange{\protect\cite{cite3275}}).
\item\relax Concatenating quantum Hamming codes on top of the \(\llbracket 4,2,2\rrbracket \) and \(C_6\) codes yields fault-tolerant quantum computation with constant space and quasi-polylogarithmic time overheads \NoCaseChange{\protect\cite{cite3216}}. In the optimized protocol of Ref. \NoCaseChange{\protect\cite{cite3216}}, a level-five \(C_4/C_6\) code underlies concatenated quantum Hamming codes \(\mathcal{Q}_5,\mathcal{Q}_6,\mathcal{Q}_7,\mathcal{Q}_7\), yielding a \(2.5\%\) threshold and space overheads \(162\) and \(373\) physical qubits per logical qubit at physical error rate \(0.1\%\) for logical CNOT error rates \(10^{-10}\) and \(10^{-24}\), respectively.
\item\relax Fault-tolerant magic-state preparation \NoCaseChange{\protect\cite{cite3327}}.
\item\relax One of the code's logical qubits can be relaxed to a gauge qubit to yield a \(\llbracket 6,1,1,2\rrbracket \) subsystem qubit stabilizer code with a particular set of transversal gates. This code admits a fault-tolerant circuit relevant to magic-state preparation \NoCaseChange{\protect\cite{cite454}}.
\end{eczvaluelist}
\codefieldsection{Realizations}
\begin{eczvaluelist}
\item\relax Trapped ions: fault-tolerant magic-state preparation demonstrated on a 20-qubit H1-1 device by Quantinuum \NoCaseChange{\protect\cite{cite3327}}.
\end{eczvaluelist}
\codefieldsection{Parents}
\begin{eczvaluelist}
\item\relax
\flmRefsHyperref[eczindexfamilyrel]{code:2d_color}{2D color code} --- The \(C_6\) code is a color code on a ladder with three rungs and periodic boundary conditions, (a.k.a. a triangular prism with no top and bottom faces) \NoCaseChange{\protect\cite{cite2339}}. Purely \(Z\)- or \(X\)-type stabilizers lie on the three square faces of the ladder.
\item\relax
\flmRefsHyperref[eczindexfamilyrel]{code:quantum_h}{\(\llbracket k+4,k,2\rrbracket \) H code} --- The \(\llbracket k+4,k,2\rrbracket \) H code for \(k=2\) is the \(C_6\) code.
\item\relax
\flmRefsHyperref[eczindexfamilyrel]{code:goy}{\(\llbracket 6r,2r,2\rrbracket \) Ganti-Onunkwo-Young code} --- The Ganti-Onunkwo-Young code for \(r=1\) is the \(C_6\) code.
\item\relax
\flmRefsHyperref[eczindexfamilyrel]{code:kls}{Khesin-Lu-Shor code} --- The Khesin-Lu-Shor code for \(r=2\) and \(m=2^r - 1 = 3\) is the \(C_6\) code.
\item\relax
\flmRefsHyperref[eczindexfamilyrel]{code:stabilizer_over_gf4}{Hermitian qubit code} --- The \(C_6\) code is Hermitian \NoCaseChange{\protect\cite[{ID 126}]{cite453}\protect\cite[{Table 6}]{cite454}}.
\end{eczvaluelist}
\codefieldsection{Cousins}
\begin{eczvaluelist}
\item\relax
\flmRefsHyperref[eczindexfamilyrel]{code:css_6_1_2}{\(\llbracket 6,1,2\rrbracket \) semi-self-dual CSS code} --- Fixing one logical qubit of the \(\llbracket 6,2,2\rrbracket \) \(C_6\) code to \(|Y^{-}\rangle_L\) yields this \(\llbracket 6,1,2\rrbracket \) code \NoCaseChange{\protect\cite{cite738}\protect\cite[{ID 59}]{cite453}}.
\item\relax
\flmRefsHyperref[eczindexfamilyrel]{code:qubit_concatenated}{Concatenated qubit code} --- Concatenations of \(\llbracket 4,2,2\rrbracket \) and \(C_6\) codes yield fault-tolerant quantum computation schemes \NoCaseChange{\protect\cite{cite448}} admitting a post-selected threshold \NoCaseChange{\protect\cite{cite3273,cite3274}} (see also Ref. \NoCaseChange{\protect\cite{cite3275}}) and the Meier-Eastin-Knill (MEK) magic-state distillation protocols \NoCaseChange{\protect\cite{cite708}}. Concatenating quantum Hamming codes on top of the \(\llbracket 4,2,2\rrbracket \) and \(C_6\) codes yields fault-tolerant quantum computation with constant space and quasi-polylogarithmic time overheads \NoCaseChange{\protect\cite{cite3216}}. In the optimized protocol of Ref. \NoCaseChange{\protect\cite{cite3216}}, a level-five \(C_4/C_6\) code underlies concatenated quantum Hamming codes \(\mathcal{Q}_5,\mathcal{Q}_6,\mathcal{Q}_7,\mathcal{Q}_7\), yielding a \(2.5\%\) threshold and space overheads \(162\) and \(373\) physical qubits per logical qubit at physical error rate \(0.1\%\) for logical CNOT error rates \(10^{-10}\) and \(10^{-24}\), respectively.
\item\relax
\flmRefsHyperref[eczindexfamilyrel]{code:gkp_concatenated}{Concatenated GKP code} --- Recursively concatenating the \(C_6\) and \(\llbracket 4,2,2\rrbracket \) codes with GKP codes achieves the hashing bound of the displacement channel \NoCaseChange{\protect\cite{cite3290}}.
\item\relax
\flmRefsHyperref[eczindexfamilyrel]{code:ladder}{Ladder Floquet code} --- The ISG of the ladder code includes \(Z\)-type Pauli products around squares of the qubit ladder. These are also included in the checks of the \(C_6\) code.
\item\relax
\flmRefsHyperref[eczindexfamilyrel]{code:carbon}{\(\llbracket 12,2,4\rrbracket \) carbon code} --- The carbon code is a concatenation of the \(\llbracket 4,2,2\rrbracket \) code and the \(C_6\) code.
\item\relax
\flmRefsHyperref[eczindexfamilyrel]{code:stab_4_2_2}{\(\llbracket 4,2,2\rrbracket \) Four-qubit code} --- Concatenations of \(\llbracket 4,2,2\rrbracket \) and \(C_6\) codes yield fault-tolerant quantum computation schemes \NoCaseChange{\protect\cite{cite448}} admitting a post-selected threshold \NoCaseChange{\protect\cite{cite3273,cite3274}} (see also Ref. \NoCaseChange{\protect\cite{cite3275}}) and the Meier-Eastin-Knill (MEK) magic-state distillation protocols \NoCaseChange{\protect\cite{cite708}}. Concatenating quantum Hamming codes on top of the \(\llbracket 4,2,2\rrbracket \) and \(C_6\) codes yields fault-tolerant quantum computation with constant space and quasi-polylogarithmic time overheads \NoCaseChange{\protect\cite{cite3216}}. In the optimized protocol of Ref. \NoCaseChange{\protect\cite{cite3216}}, a level-five \(C_4/C_6\) code underlies concatenated quantum Hamming codes \(\mathcal{Q}_5,\mathcal{Q}_6,\mathcal{Q}_7,\mathcal{Q}_7\), yielding a \(2.5\%\) threshold and space overheads \(162\) and \(373\) physical qubits per logical qubit at physical error rate \(0.1\%\) for logical CNOT error rates \(10^{-10}\) and \(10^{-24}\), respectively.
\item\relax
\flmRefsHyperref[eczindexfamilyrel]{code:steane}{\(\llbracket 7,1,3\rrbracket \) Steane code} --- In Knill's \(C_4/C_6\) architecture, noisy \(\ket{\pi/8}\) states are injected using \(C_4/C_6\) logical Bell pairs and then purified by encoding them into the Steane code; Knill also proposed using the Steane code as a final concatenation level for the \(C_4/C_6\) scheme \NoCaseChange{\protect\cite{cite448}}.
\item\relax
\flmRefsHyperref[eczindexfamilyrel]{code:quantum_hamming_css}{\(\llbracket 2^r-1, 2^r-2r-1, 3\rrbracket \) quantum Hamming code} --- Concatenating quantum Hamming codes on top of the \(\llbracket 4,2,2\rrbracket \) and \(C_6\) codes yields fault-tolerant quantum computation with constant space and quasi-polylogarithmic time overheads \NoCaseChange{\protect\cite{cite3216}}. In the optimized protocol of Ref. \NoCaseChange{\protect\cite{cite3216}}, a level-five \(C_4/C_6\) code underlies concatenated quantum Hamming codes \(\mathcal{Q}_5,\mathcal{Q}_6,\mathcal{Q}_7,\mathcal{Q}_7\), yielding a \(2.5\%\) threshold and space overheads \(162\) and \(373\) physical qubits per logical qubit at physical error rate \(0.1\%\) for logical CNOT error rates \(10^{-10}\) and \(10^{-24}\), respectively.
\item\relax
\flmRefsHyperref[eczindexfamilyrel]{code:galois_3_1_2}{\(\llbracket 3,1,2\rrbracket _4\) three-Galois-quartrit code} --- Binarizing the \(\llbracket 3,1,2\rrbracket _4\) code in the self-dual normal basis \(\{\omega,\omega^2\}\) yields a \(\llbracket 6,2,2\rrbracket \) qubit CSS code equivalent to the \(C_6\) code after the qubit relabeling \((2\,3\,4\,5\,6)\) \NoCaseChange{\protect\cite{cite514}}.
\end{eczvaluelist}
\eczhbkcontributors{ \eczhuVVA }
\endeczcode

\eczcode{bravyi_bacon_shor_6}{\(\llbracket 6,2,3,2\rrbracket \) BBS code}{~\NoCaseChange{\protect\cite{cite3328}}}
\eczhIndexCodeAliasName{bravyi_bacon_shor_6}{BBS code}
\codefieldsection{Description}
Error-detecting six-qubit BBS subsystem code with parameters \(\llbracket 6,2,3,2\rrbracket \) that can suppress errors in adiabatic quantum computation \NoCaseChange{\protect\cite{cite670}}.

The code is defined by the \(3\times 3\) binary matrix
\flmMathEnvironment{align}{}{
A = \begin{pmatrix} 1 & 1 & 0 \\ 0 & 1 & 1 \\ 1 & 0 & 1 \end{pmatrix},
}
with physical qubits placed at the six nonzero entries of \(A\) and labeled \(1\)–\(6\) in row-major order \NoCaseChange{\protect\cite{cite3328}}.
The gauge generators (excluding phases) are
\flmMathEnvironment{align}{}{
\begin{array}{cccccc}
  X & X & I & I & I & I \\
  I & I & X & X & I & I \\
  I & I & I & I & X & X \\
  Z & I & I & I & Z & I \\
  I & Z & Z & I & I & I \\
  I & I & I & Z & I & Z
\end{array}~,
}
where each \(XX\) row corresponds to a same-row pair in \(A\) and each \(ZZ\) row to a same-column pair.

\codefieldsection{Protection}
This code detects arbitrary single-qubit errors and is used as an error-suppressing encoding with non-commuting two-local Hamiltonian terms built from gauge generators \NoCaseChange{\protect\cite{cite670}}.
For the symmetric choice of coefficients considered in \NoCaseChange{\protect\cite{cite670}}, the resulting two-local suppression Hamiltonian has energy separation \(4-2\sqrt{3}\).

\codefieldsection{Gates}
\begin{eczvaluelist}
\item\relax When both logical qubits are encoded in the same block, \(X_{L1}X_{L2}\) and \(Z_{L1}Z_{L2}\) are implementable using two-local physical interactions up to stabilizers \NoCaseChange{\protect\cite{cite670}}.
\end{eczvaluelist}
\codefieldsection{Parent}
\begin{eczvaluelist}
\item\relax
\flmRefsHyperref[eczindexfamilyrel]{code:trapezoid}{Trapezoid subsystem code} --- The even-logical-qubit trapezoid family at \(l=k=1\) reduces to the \(\llbracket 6,2,3,2\rrbracket \) BBS code.
\end{eczvaluelist}
\codefieldsection{Cousins}
\begin{eczvaluelist}
\item\relax
\flmRefsHyperref[eczindexfamilyrel]{code:small_distance_qubit_stabilizer}{Small-distance qubit stabilizer code}\item\relax
\flmRefsHyperref[eczindexfamilyrel]{code:bacon_shor_4}{\(\llbracket 4,1,1,2\rrbracket \) Four-qubit subsystem code} --- Both the \(\llbracket 6,2,3,2\rrbracket \) BBS code and the four-qubit subsystem code can be used to suppress errors in adiabatic quantum computation \NoCaseChange{\protect\cite{cite670}}.
\end{eczvaluelist}
\eczhbkcontributors{ \eczhuVVA }
\endeczcode

\eczcode{stab_6_4_2}{\(\llbracket 6,4,2\rrbracket \) error-detecting code}{~\NoCaseChange{\protect\cite{cite861,cite737,cite3329,cite3330}}}
\eczhIndexCodeAliasName{stab_6_4_2}{error-detecting code}
\codefieldsection{Description}
Self-complementary six-qubit code with rate \(2/3\) that is unique for its parameters, up to equivalence \NoCaseChange{\protect\cite[{Tab. III}]{cite449}}.
Concatenations of this code with itself yield the \(\llbracket 6^r,4^r,2^r\rrbracket \) level-\(r\) \textit{many-hypercube} code \NoCaseChange{\protect\cite{cite450}}.

A stabilizer tableau for the code is given by \NoCaseChange{\protect\cite[{ID 29}]{cite453}}
\flmMathEnvironment{align}{}{
\begin{array}{cccccc}
  Z & Z & Z & Z & Z & Z \\
  X & X & X & X & X & X
\end{array}~.
}
Stabilizer generators are shown in \flmRefsCref{ref3331}.
See \NoCaseChange{\protect\cite[{Appx. B}]{cite450}} for a set of logical Paulis.

  \begin{flmFloat}{figure}{NumCap}\includegraphics[width=180bp,max width=\linewidth]{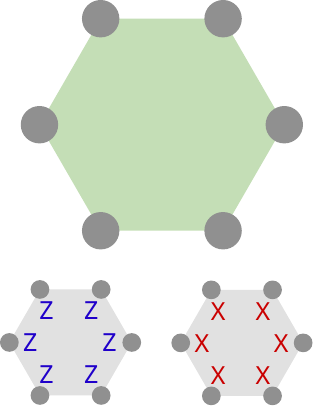}\caption{
    Stabilizer generators of the \(\llbracket 6,4,2\rrbracket \) error-detecting code.
    }\label{ref3331}\end{flmFloat}

\codefieldsection{Encoding}
\begin{eczvaluelist}
\item\relax See \NoCaseChange{\protect\cite[{Fig. 5}]{cite450}}.
\end{eczvaluelist}
\codefieldsection{Transversal and Permutation-Based Gates}
\begin{eczvaluelist}
\item\relax CNOT and Hadamard gates \NoCaseChange{\protect\cite[{Appx. B}]{cite450}}.
\item\relax A \(CZ\) gate implemented by transversal \(S\) and \(S^{\dagger}\) \NoCaseChange{\protect\cite{cite805}}; see also \NoCaseChange{\protect\cite{cite687}}.
\end{eczvaluelist}
\codefieldsection{Gates}
\begin{eczvaluelist}
\item\relax Universal \flmRefsHyperref{ref409}{Clifford gates} via the logical Clifford synthesis (LCS) algorithm \NoCaseChange{\protect\cite{cite3332}\protect\cite[{Sec. III}]{cite773}}.
\end{eczvaluelist}
\codefieldsection{Decoding}
\begin{eczvaluelist}
\item\relax Efficient decoder for the many-hypercube code \NoCaseChange{\protect\cite{cite450}}.
\end{eczvaluelist}
\codefieldsection{Realizations}
\begin{eczvaluelist}
\item\relax Trapped-ion devices: Bayesian quantum phase estimation on a device by Quantinuum \NoCaseChange{\protect\cite{cite3333}}.
\item\relax Neutral atom arrays: state initialization of the \(\llbracket 16,4,4\rrbracket \) doubly concatenated code (a.k.a., the level-two many-hypercube code) on a device by Infleqtion \NoCaseChange{\protect\cite{cite3286}}.
\end{eczvaluelist}
\codefieldsection{Parents}
\begin{eczvaluelist}
\item\relax
\flmRefsHyperref[eczindexfamilyrel]{code:iceberg}{\(\llbracket 2m,2m-2,2\rrbracket \) error-detecting code} --- The \(\llbracket 2m,2m-2,2\rrbracket \) error-detecting code for \(m=3\) reduces to the \(\llbracket 6,4,2\rrbracket \) error-detecting code.
\item\relax
\flmRefsHyperref[eczindexfamilyrel]{code:triangular_color}{Honeycomb (6.6.6) color code} --- The \(\llbracket 6,4,2\rrbracket \) error-detecting code is a color code defined on a single hexagon of the 6.6.6 or 4.6.12 tilings. The \(\llbracket 6,4,2\rrbracket \) code can be concatenated with the surface code to yield the 6.6.6 color code \NoCaseChange{\protect\cite[{Appx. A}]{cite3289}}.
\item\relax
\flmRefsHyperref[eczindexfamilyrel]{code:4612_color}{Truncated trihexagonal (4.6.12) color code} --- The \(\llbracket 6,4,2\rrbracket \) error-detecting code is a color code defined on a single hexagon of the 6.6.6 or 4.6.12 tilings.
\end{eczvaluelist}
\codefieldsection{Cousins}
\begin{eczvaluelist}
\item\relax
\flmRefsHyperref[eczindexfamilyrel]{code:qubit_concatenated}{Concatenated qubit code} --- Concatenations of this code with itself yield the level-\(r\) \(\llbracket 6^r,4^r,2^r\rrbracket \) many-hypercube code \NoCaseChange{\protect\cite{cite450}}. The \(\llbracket 6,4,2\rrbracket \) code can be concatenated with the surface code to yield the 6.6.6 color code \NoCaseChange{\protect\cite[{Appx. A}]{cite3289}}.
\item\relax
\flmRefsHyperref[eczindexfamilyrel]{code:quantum_lego}{Tensor-network code} --- The \(\llbracket 6,4,2\rrbracket \) error-detecting code can be constructed out of two \(\llbracket 4,2,2\rrbracket \) codes in the quantum Lego code framework \NoCaseChange{\protect\cite{cite2868}}.
\item\relax
\flmRefsHyperref[eczindexfamilyrel]{code:stab_4_2_2}{\(\llbracket 4,2,2\rrbracket \) Four-qubit code} --- The \(\llbracket 6,4,2\rrbracket \) error-detecting code can be constructed out of two \(\llbracket 4,2,2\rrbracket \) codes in the quantum Lego code framework \NoCaseChange{\protect\cite{cite2868}}.
\end{eczvaluelist}
\eczhbkcontributors{ \eczhuVVA }
\endeczcode

\eczcode{campbell_howard}{\(\llbracket 6k+2,3k,2\rrbracket \) Campbell-Howard code}{~\NoCaseChange{\protect\cite{cite754}}}
\eczhIndexCodeAliasName{campbell_howard}{Campbell-Howard code}
\codefieldsection{Description}
Family of \(\llbracket 6k+2,3k,2\rrbracket \) qubit stabilizer codes with quasi-transversal \(CCZ^{\otimes k}\) gates that are relevant to magic-state distillation.
In the synthillation framework, these distance-two codes realize batches of logical \(CCZ\) gates using physical \(T\) gates followed by a Clifford correction.

\codefieldsection{Protection}
Detects single-qubit errors.
\codefieldsection{Rate}
Encoding rate is \(3k/(6k+2)\), approaching \(1/2\) as \(k\to\infty\).
\codefieldsection{Magic}
A total of \(r\) rounds of magic-state distillation yields a magic-state yield parameter \(\gamma\to 1^{+}\) as \(k,r\rightarrow \infty\). This matches the Bravyi-Haah conjectured lower bound \(\gamma \geq 1\) for concatenated triorthogonal-matrix protocols \NoCaseChange{\protect\cite[{Sec. VI}]{cite691}}.
\codefieldsection{Transversal and Permutation-Based Gates}
\begin{eczvaluelist}
\item\relax Quasi-transversal \(CCZ^{\otimes k}\) gates \NoCaseChange{\protect\cite{cite754}}.
\end{eczvaluelist}
\codefieldsection{Parent}
\begin{eczvaluelist}
\item\relax
\flmRefsHyperref[eczindexfamilyrel]{code:small_distance_qubit_stabilizer}{Small-distance qubit stabilizer code} --- The family has distance \(2\).
\end{eczvaluelist}
\codefieldsection{Child}
\begin{eczvaluelist}
\item\relax
\flmRefsHyperref[eczindexfamilyrel]{code:stab_8_3_2}{\(\llbracket 8,3,2\rrbracket \) Smallest interesting color code} --- The \(\llbracket 8,3,2\rrbracket \) code is the \(k=1\) member of the \(\llbracket 6k+2,3k,2\rrbracket \) Campbell-Howard family with a quasi-transversal logical \(CCZ\) gate \NoCaseChange{\protect\cite{cite754}}.
\end{eczvaluelist}
\codefieldsection{Cousin}
\begin{eczvaluelist}
\item\relax
\flmRefsHyperref[eczindexfamilyrel]{code:quantum_triorthogonal}{Triorthogonal code} --- Campbell-Howard codes arise from the generalized-triorthogonal/quasitransversal \(G\)-matrix framework, which extends Bravyi-Haah triorthogonal matrices by allowing odd pair and triple overlaps entirely within the logical-row block \(K\) \NoCaseChange{\protect\cite[{Appx. D}]{cite754}}.
\end{eczvaluelist}
\eczhbkcontributors{ Earl T. Campbell, \eczhuVVA }
\endeczcode

\eczcode{goy}{\(\llbracket 6r,2r,2\rrbracket \) Ganti-Onunkwo-Young code}{~\NoCaseChange{\protect\cite{cite3334}}}
\eczhIndexCodeAliasName{goy}{Ganti-Onunkwo-Young code}
\codefieldsection{Description}
A member of the family of \(\llbracket 6r,2r,2\rrbracket \) CSS codes designed to suppress errors in adiabatic quantum computation.
All but two of its stabilizer generators are weight-two (two-body), and the remaining two are weight-\(4r\).

\codefieldsection{Gates}
\begin{eczvaluelist}
\item\relax The code supports weight-two logical operators and maintains a planar connectivity graph with only a small increase in graph degree.
\end{eczvaluelist}
\codefieldsection{Parents}
\begin{eczvaluelist}
\item\relax
\flmRefsHyperref[eczindexfamilyrel]{code:qubit_css}{Qubit CSS code}\item\relax
\flmRefsHyperref[eczindexfamilyrel]{code:small_distance_qubit_stabilizer}{Small-distance qubit stabilizer code}\end{eczvaluelist}
\codefieldsection{Child}
\begin{eczvaluelist}
\item\relax
\flmRefsHyperref[eczindexfamilyrel]{code:stab_6_2_2}{\(\llbracket 6,2,2\rrbracket \) \(C_6\) code} --- The Ganti-Onunkwo-Young code for \(r=1\) is the \(C_6\) code.
\end{eczvaluelist}
\codefieldsection{Cousin}
\begin{eczvaluelist}
\item\relax
\flmRefsHyperref[eczindexfamilyrel]{code:trapezoid}{Trapezoid subsystem code} --- The odd-\(m\) trapezoid family at \(l=k\) has parameters \(\llbracket 6k,2k,4k,2\rrbracket \) and reproduces the two-local subsystem construction used for universal Hamiltonian quantum computation in \NoCaseChange{\protect\cite{cite3335}}; this is a subsystem analogue of the \(\llbracket 6k,2k,2\rrbracket \) Ganti-Onunkwo-Young family.
\end{eczvaluelist}
\eczhbkcontributors{ \eczhuVVA }
\endeczcode

\eczcode{hybrid_7_1-1_3}{\(\llbracket 7, 1:1, 3\rrbracket \) hybrid stabilizer code}{~\NoCaseChange{\protect\cite{cite2872}}}
\eczhIndexCodeAliasName{hybrid_7_1-1_3}{hybrid stabilizer code}
\codefieldsection{Description}
A distance-three seven-qubit hybrid stabilizer code storing one qubit and one classical bit.
Admits a stabilizer generator set with a weight-two generator, which delineates the underlying classical code \NoCaseChange{\protect\cite[{Eq. (3)}]{cite671}}.
\codefieldsection{Parent}
\begin{eczvaluelist}
\item\relax
\flmRefsHyperref[eczindexfamilyrel]{code:hybrid_stabilizer}{Hybrid stabilizer code}\end{eczvaluelist}
\codefieldsection{Cousin}
\begin{eczvaluelist}
\item\relax
\flmRefsHyperref[eczindexfamilyrel]{code:small_distance_qubit_stabilizer}{Small-distance qubit stabilizer code}\end{eczvaluelist}
\eczhbkcontributors{ \eczhuVVA }
\endeczcode

\eczcode{bare_7_1_3}{\(\llbracket 7,1,3\rrbracket \) bare code}{~\NoCaseChange{\protect\cite{cite3336}}}
\eczhIndexCodeAliasName{bare_7_1_3}{bare code}
\codefieldsection{Description}
A \(\llbracket 7,1,3\rrbracket \) code that admits fault-tolerant syndrome extraction using only one ancilla per stabilizer generator measurement.

A stabilizer tableau for the code is given by \NoCaseChange{\protect\cite[{ID 108}]{cite453}}
\flmMathEnvironment{align}{}{
\begin{array}{ccccccc}
  X & I & I & I & Z & I & I \\
  I & X & I & I & Z & I & I \\
  I & I & X & I & I & Z & I \\
  I & I & I & X & I & I & Z \\
  I & I & Z & Z & I & Y & Y \\
  Z & Z & Z & I & X & X & Z
\end{array}~.
}
It is one of sixteen distinct indecomposable \(\llbracket 7,1,3\rrbracket \) codes \NoCaseChange{\protect\cite{cite452}}.

\codefieldsection{Decoding}
\begin{eczvaluelist}
\item\relax Fault-tolerant syndrome extraction using a single ancilla \NoCaseChange{\protect\cite{cite3336}}.
\end{eczvaluelist}
\codefieldsection{Parent}
\begin{eczvaluelist}
\item\relax
\flmRefsHyperref[eczindexfamilyrel]{code:small_distance_qubit_stabilizer}{Small-distance qubit stabilizer code}\end{eczvaluelist}
\codefieldsection{Cousin}
\begin{eczvaluelist}
\item\relax
\flmRefsHyperref[eczindexfamilyrel]{code:qetc_7_2}{\(\llbracket 7,2,2\rrbracket \) QETC} --- The stabilizer group of the \(\llbracket 7,2,2\rrbracket \) QETC, together with the logical-\(Z\) operator on the first logical qubit, generates the stabilizer group of a \(\llbracket 7,1,3\rrbracket \) code \NoCaseChange{\protect\cite{cite2985}} equivalent to the bare \(\llbracket 7,1,3\rrbracket \) code \NoCaseChange{\protect\cite{cite447}}.
\end{eczvaluelist}
\eczhbkcontributors{ \eczhuVVA }
\endeczcode

\eczcode{steane}{\(\llbracket 7,1,3\rrbracket \) Steane code}{~\NoCaseChange{\protect\cite{cite3337,cite3338}}}
\eczhIndexCodeAliasName{steane}{Steane code}
\codefieldsection{Description}
A \(\llbracket 7,1,3\rrbracket \) self-dual CSS code that is the smallest qubit CSS code to correct a single-qubit error \NoCaseChange{\protect\cite{cite451}}.
The code is constructed using the classical binary \([7,4,3]\) Hamming code for protecting against both \(X\) and \(Z\) errors.

A stabilizer tableau for the code is given by \NoCaseChange{\protect\cite[{ID 226}]{cite453}}
\flmMathEnvironment{align}{}{
\begin{array}{ccccccc}
  I & I & I & X & X & X & X \\
  I & X & X & I & I & X & X \\
  X & I & X & I & X & I & X \\
  I & I & I & Z & Z & Z & Z \\
  I & Z & Z & I & I & Z & Z \\
  Z & I & Z & I & Z & I & Z
\end{array}~.
}
The code's stabilizer generator matrix blocks \(H_{X}\) and \(H_{Z}\) are both the parity-check matrix of the \([7,4,3]\) Hamming code.
The checks can be thought of as lying on the three trapezoids of the following tiling of the triangle.

\begin{flmFloat}{figure}{NumCap}\includegraphics[width=203.175bp,max width=\linewidth]{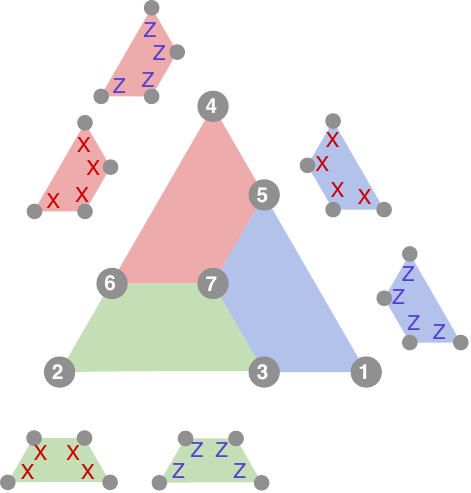}\caption{
  Stabilizer generators of the Steane code.
  }\label{ref3339}\end{flmFloat}
The Steane code can also be thought of as a code on all corners of a cube except one \NoCaseChange{\protect\cite{cite3199,cite3340}}, and the code's \flmRefsHyperref{ref857}{encoder-respecting form} is the graph of the full cube \NoCaseChange{\protect\cite{cite858}}.

A set of logical codewords is
\flmMathEnvironment{align}{}{
\begin{split}
  |\overline{0}\rangle&=\frac{1}{\sqrt{8}}\Big(|0000000\rangle+|1010101\rangle+|0110011\rangle+|1100110\rangle\\&\,\,\,\,\,\,\,\,+|0001111\rangle+|1011010\rangle+|0111100\rangle+|1101001\rangle\Big)\\|\overline{1}\rangle&=\frac{1}{\sqrt{8}}\Big(|1111111\rangle+|0101010\rangle+|1001100\rangle+|0011001\rangle\\&\,\,\,\,\,\,\,\,+|1110000\rangle+|0100101\rangle+|1000011\rangle+|0010110\rangle\Big)~.
\end{split}
}

The automorphism group of the code is \(PGL(3,2)\) \NoCaseChange{\protect\cite{cite3222}}.
It is one of sixteen distinct indecomposable \(\llbracket 7,1,3\rrbracket \) codes \NoCaseChange{\protect\cite{cite452}}.

\codefieldsection{Protection}
The Steane code is a distance 3 code. It detects errors on 2 qubits, corrects errors on 1 qubit.
\codefieldsection{Encoding}
\begin{eczvaluelist}
\item\relax Nine CNOT and four Hadamard gates \NoCaseChange{\protect\cite[{Fig. 10.14}]{cite3302}}.
\item\relax Evolution under stabilizer Hamiltonian \NoCaseChange{\protect\cite{cite3303}}.
\item\relax Fault-tolerant logical zero and logical plus state preparation on all-to-all and 2D grid qubit connectivity \NoCaseChange{\protect\cite{cite3200}}.
\item\relax Parity-check encoding with flag-bridge qubits on a square lattice connectivity \NoCaseChange{\protect\cite{cite3341}}.
\end{eczvaluelist}
\codefieldsection{Transversal and Permutation-Based Gates}
\begin{eczvaluelist}
\item\relax The \flmRefsHyperref{ref409}{single-qubit Clifford group} \NoCaseChange{\protect\cite{cite761,cite776}}. More generally, \(k\) copies of the Steane code form a \(\llbracket 7k,k,3\rrbracket \) code that admits a \(k\)-fold transversal implementation of the full \flmRefsHyperref{ref409}{Clifford group} on all \(k\) logical qubits, showing the tightness of a no-go theorem that requires at least \(k\)-fold transversal gadgets for the full Clifford group \NoCaseChange{\protect\cite{cite782}}.
\end{eczvaluelist}
\codefieldsection{Gates}
\begin{eczvaluelist}
\item\relax Fault-tolerant approximations of arbitrary single-qubit gates \NoCaseChange{\protect\cite{cite3342,cite3343}}.
\item\relax Non-fault-tolerant \(T\) gate \NoCaseChange{\protect\cite{cite3344}}.
\item\relax Fault-tolerant logical zero and magic state preparation \NoCaseChange{\protect\cite{cite3345}}. Magic-state preparation converts unbiased noise into biased noise \NoCaseChange{\protect\cite{cite3346}}.
\item\relax Because transversal Hadamard acts logically on the code, the Steane code serves as a normal self-dual inner code for magic-state distillation. One routine uses 14 noisy \(T\) gates and one noisy input magic state to produce one output with cubic error suppression, and it can be pipelined with the \(\llbracket 17,1,5\rrbracket \) code to obtain fifth-order suppression \NoCaseChange{\protect\cite[{Sec. I.A}]{cite101}}.
\item\relax Pieceable fault-tolerant \(CCZ\) gate \NoCaseChange{\protect\cite{cite806}}.
\end{eczvaluelist}
\codefieldsection{Decoding}
\begin{eczvaluelist}
\item\relax Shor error correction fidelity calculation \NoCaseChange{\protect\cite{cite3347,cite3348,cite3349}}.
\item\relax A \([15,3]\) syndrome-measurement code yields a QDS extension that uses the same 15 measurements as five-fold repetition of the three syndrome bits while achieving lower syndrome-decoding error \NoCaseChange{\protect\cite[{Fig. 1}]{cite2914}}.
\item\relax Fault-tolerant measurement-free error-correction cycle \NoCaseChange{\protect\cite{cite3350}}.
\end{eczvaluelist}
\codefieldsection{Fault Tolerance}
\begin{eczvaluelist}
\item\relax A fault-tolerant universal gate set can be done via \flmRefsHyperref{ref410}{code switching} between the Steane code and the \(\llbracket 15,1,3\rrbracket \) code \NoCaseChange{\protect\cite{cite731,cite787,cite793,cite3203,cite3204}}.
\item\relax A fault-tolerant universal gate set can be done via \flmRefsHyperref{ref410}{code switching} between the Steane code and the \(\llbracket 10,1,2\rrbracket \) code \NoCaseChange{\protect\cite{cite3182}}.
\item\relax A fault-tolerant logical \(T\) gate can be obtained by encoding the Steane code's seven physical qubits into the seven logical qubits of a \(\llbracket 63,7,3\rrbracket \) outer quantum divisible CSS code preserved by transversal \(T^\dagger\) \NoCaseChange{\protect\cite{cite765}}.
\item\relax Fault-tolerant logical zero and magic state preparation \NoCaseChange{\protect\cite{cite3345}}. Magic-state preparation converts unbiased noise into biased noise \NoCaseChange{\protect\cite{cite3346}}.
\item\relax Fault-tolerant logical zero and logical plus state preparation on all-to-all and 2D grid qubit connectivity \NoCaseChange{\protect\cite{cite3200}}.
\item\relax Pieceable fault-tolerant \(CCZ\) gate \NoCaseChange{\protect\cite{cite806}}.
\item\relax Syndrome measurement can be done with ancillary flag qubits \NoCaseChange{\protect\cite{cite3215}} or with no extra qubits \NoCaseChange{\protect\cite{cite3351}}. The depth of syndrome extraction circuits can be lowered by using past syndrome values \NoCaseChange{\protect\cite{cite3310}}.
\item\relax Computation of ground-state energy of the hydrogen molecule \NoCaseChange{\protect\cite{cite3352}}.
\item\relax Fault-tolerant measurement-free error-correction cycle \NoCaseChange{\protect\cite{cite3350}}.
\end{eczvaluelist}
\codefieldsection{Realizations}
\begin{eczvaluelist}
\item\relax Trapped-ion devices: seven-qubit device in Blatt group \NoCaseChange{\protect\cite{cite3353}}.
Ten-qubit QCCD device by Quantinuum \NoCaseChange{\protect\cite{cite3354}} realizing repeated syndrome extraction, real-time look-up-table decoding (yielding lower logical SPAM error rate than physical SPAM), and non-fault-tolerant magic-state distillation (see APS Physics Synopsis \NoCaseChange{\protect\cite{cite3355}}).
Fault-tolerant universal two-qubit gate set using T injection by Monz group \NoCaseChange{\protect\cite{cite3356}}.
Logical CNOT gate and Bell-pair creation between two logical qubits (yielding a logical fidelity higher than physical), including rounds of correction and fault-tolerant primitives such as flag qubits and pieceable fault tolerance, on a 20-qubit device by Quantinuum \NoCaseChange{\protect\cite{cite3315}}; logical fidelity interval of the combined preparation-CNOT-measurement procedure was higher than that of the unencoded physical qubits.
Multiple rounds of Steane error correction \NoCaseChange{\protect\cite{cite3357}}.
Fault-tolerant universal gate set via \flmRefsHyperref{ref410}{code switching} between the Steane code and the \(\llbracket 10,1,2\rrbracket \) code \NoCaseChange{\protect\cite{cite3182}}.
Post-selected fault-tolerant logical Bell-state preparation with logical error rates at least 10 times lower than physical rate on a device by Quantinuum \NoCaseChange{\protect\cite{cite525}}.
The quantum Fourier transform on three code blocks \NoCaseChange{\protect\cite{cite3358}}.
Fault-tolerant transversal and lattice-surgery state teleportation protocols as well as Knill error correction \NoCaseChange{\protect\cite{cite3359}}.
Rains shadow enumerators have been measured \NoCaseChange{\protect\cite{cite3360}}.
Inter-block CNOT gates have been characterized via cycle reconstruction \NoCaseChange{\protect\cite{cite3361}}.
\flmRefsHyperref{ref410}{Code switching} between the Steane code and the \(\llbracket 15,1,3\rrbracket \) code as well as magic-state preparation and logical Bell measurements on the Steane code realized on the 28-qubit H2-1 device by Quantinuum \NoCaseChange{\protect\cite{cite3205}}.
End-to-end fault-tolerant execution of QAOA and HHL circuits, including logical non-Clifford operations, with up to 12 logical qubits on Quantinuum systems \NoCaseChange{\protect\cite{cite3362}}.

\item\relax Neutral atom arrays: Lukin group. Ten logical qubits, transversal CNOT gate performed, logical ten-qubit GHZ state initialized with break-even fidelity, and fault-tolerant logical two-qubit GHZ state initialized \NoCaseChange{\protect\cite{cite3363}}. Deep-circuit protocols with dozens of logical qubits and hundreds of logical teleportations \NoCaseChange{\protect\cite{cite3206}}.
\end{eczvaluelist}
\codefieldsection{Notes}
\begin{eczvaluelist}
\item\relax Pedagogical explanation of QEC using the Steane code \NoCaseChange{\protect\cite{cite3364}}.
\item\relax The Steane code can be used for entanglement purification \NoCaseChange{\protect\cite{cite3365}}.
\end{eczvaluelist}
\codefieldsection{Parents}
\begin{eczvaluelist}
\item\relax
\flmRefsHyperref[eczindexfamilyrel]{code:triangular_color}{Honeycomb (6.6.6) color code} --- Steane code is a 2D color code defined on a seven-qubit patch of the 6.6.6 tiling.
\item\relax
\flmRefsHyperref[eczindexfamilyrel]{code:diagonal_clifford}{\(\llbracket 2^r-1,1,3\rrbracket \) simplex code}\item\relax
\flmRefsHyperref[eczindexfamilyrel]{code:quantum_hamming_css}{\(\llbracket 2^r-1, 2^r-2r-1, 3\rrbracket \) quantum Hamming code}\item\relax
\flmRefsHyperref[eczindexfamilyrel]{code:single_qubit_clifford}{\(\llbracket 2^{2r-1}-1,1,2^r-1\rrbracket \) quantum punctured RM code}\item\relax
\flmRefsHyperref[eczindexfamilyrel]{code:stabilizer_over_gf4}{Hermitian qubit code} --- The Steane code is Hermitian \NoCaseChange{\protect\cite[{ID 226}]{cite453}\protect\cite[{Table 6}]{cite454}}.
\item\relax
\flmRefsHyperref[eczindexfamilyrel]{code:quantum_cyclic}{Cyclic quantum code} --- The Steane code is equivalent to a cyclic code via qubit permutations \NoCaseChange{\protect\cite[{Exam. 1}]{cite438}}.
\item\relax
\flmRefsHyperref[eczindexfamilyrel]{code:galois_quad_residue}{Quantum quadratic-residue (QR) code} --- The Steane code is a qubit quantum QR code \NoCaseChange{\protect\cite{cite829,cite2914}}.
\item\relax
\flmRefsHyperref[eczindexfamilyrel]{code:data_syndrome}{Quantum data-syndrome (QDS) code} --- There exists a set of stabilizer generators for the Steane code that make it a QDS code; a \([15,3]\) syndrome-measurement code beats five-fold repeated syndrome extraction at the same measurement cost \NoCaseChange{\protect\cite{cite1970,cite2914}}.
\item\relax
\flmRefsHyperref[eczindexfamilyrel]{code:pg_qldpc}{Finite-geometry (FG) qubit QLDPC code} --- The Steane code is the \(m=1\) member of the \(\llbracket 2^{2m}+2^{m}+1,1,>2^{m}\rrbracket \) PG-QLDPC code family that is constructed from codes corresponding to lines and affine charts in \(PG(2,2^m)\) via the CSS construction \NoCaseChange{\protect\cite[{Def. 4.9}]{cite835}}.
\item\relax
\flmRefsHyperref[eczindexfamilyrel]{code:concatenated_steane}{Concatenated Steane code} --- The concatenated Steane code at level \(m=1\) is the Steane code.
\item\relax
\flmRefsHyperref[eczindexfamilyrel]{code:block_perfect}{Planar-perfect-tensor code} --- The Steane code is the smallest heptagon holographic code. The encoding of more general heptagon holographic codes is a holographic tensor network consisting of the encoding isometry for the Steane code, which is a \flmRefsHyperref{code:block_perfect}{planar-perfect tensor}.
\end{eczvaluelist}
\codefieldsection{Cousins}
\begin{eczvaluelist}
\item\relax
\flmRefsHyperref[eczindexfamilyrel]{code:group_representation}{Group-representation code} --- The Steane code is a group-representation code with \(G\) being the \(2O\) subgroup of \(SU(2)\) \NoCaseChange{\protect\cite{cite2810}}.
\item\relax
\flmRefsHyperref[eczindexfamilyrel]{code:cluster_state}{Cluster-state code} --- The Steane code is equivalent via a single-qubit Clifford unitary to a cluster-state code for a particular graph and classical code \NoCaseChange{\protect\cite[{Exam. 4}]{cite438}}. Four non-isomorphic graphs yield graph quantum codes that are equivalent to the Steane code under a single-qubit-\flmRefsHyperref{ref409}{Clifford circuit} \NoCaseChange{\protect\cite{cite867}}.
\item\relax
\flmRefsHyperref[eczindexfamilyrel]{code:eastab}{EA qubit stabilizer code} --- The Steane code is globally equivalent to a \(\llbracket 6,1,3;1\rrbracket \) EA CSS code, which the paper identifies as an example of the smallest one-ebit EA CSS code correcting an arbitrary single-qubit error on the sender's qubits \NoCaseChange{\protect\cite{cite451}}.
\item\relax
\flmRefsHyperref[eczindexfamilyrel]{code:stab_6_2_2}{\(\llbracket 6,2,2\rrbracket \) \(C_6\) code} --- In Knill's \(C_4/C_6\) architecture, noisy \(\ket{\pi/8}\) states are injected using \(C_4/C_6\) logical Bell pairs and then purified by encoding them into the Steane code; Knill also proposed using the Steane code as a final concatenation level for the \(C_4/C_6\) scheme \NoCaseChange{\protect\cite{cite448}}.
\item\relax
\flmRefsHyperref[eczindexfamilyrel]{code:quantum_divisible}{Quantum divisible code} --- A fault-tolerant logical \(T\) gate can be obtained by encoding the Steane code's seven physical qubits into the seven logical qubits of a \(\llbracket 63,7,3\rrbracket \) outer quantum divisible CSS code preserved by transversal \(T^\dagger\) \NoCaseChange{\protect\cite{cite765}}.
\item\relax
\flmRefsHyperref[eczindexfamilyrel]{code:hamming743}{\([7,4,3]\) Hamming code} --- The Steane code is constructed from the \([7,4,3]\) classical Hamming code via the CSS construction.
\item\relax
\flmRefsHyperref[eczindexfamilyrel]{code:dual_rail}{Dual-rail quantum code} --- The KLM protocol, one of the first protocols for fault-tolerant quantum computation, utilizes concatenations of the dual-rail code with a stabilizer code such as the Steane code \NoCaseChange{\protect\cite{cite3366,cite3367,cite3368}}.
\item\relax
\flmRefsHyperref[eczindexfamilyrel]{code:cat_concatenated}{Concatenated cat code} --- Two-component cat codes concatenated with Steane and Golay codes are estimated to be fault tolerant against \flmRefsHyperref{ref498}{photon loss} noise with rate \(\eta < 5\times 10^{-4}\) provided that \(\alpha > 1.2\) \NoCaseChange{\protect\cite{cite3231}}.
\item\relax
\flmRefsHyperref[eczindexfamilyrel]{code:quantum_lego}{Tensor-network code} --- The Steane code can be built from two \(\llbracket 4,2,2\rrbracket \) codes in the quantum Lego code framework \NoCaseChange{\protect\cite{cite2868}}.
\item\relax
\flmRefsHyperref[eczindexfamilyrel]{code:majorana_stab}{Majorana stabilizer code} --- Applying the CSS-to-Majorana map of \NoCaseChange{\protect\cite[{Lemma 2}]{cite1432}} to the \(\llbracket 7,1,3\rrbracket \) Steane code yields a seven-Majorana code encoding half a qubit; pairing two such odd-length copies gives a physical Majorana stabilizer code with odd logical operators \NoCaseChange{\protect\cite[{Sec. 8}]{cite1432}}.
\item\relax
\flmRefsHyperref[eczindexfamilyrel]{code:kls}{Khesin-Lu-Shor code} --- The \flmRefsHyperref{ref857}{encoder-respecting form} of both the Steane and Khesin-Lu-Shor codes is the graph of a hypercube \NoCaseChange{\protect\cite{cite858}}.
\item\relax
\flmRefsHyperref[eczindexfamilyrel]{code:stab_10_1_2}{\(\llbracket 10,1,2\rrbracket \) Vasmer-Kubica code} --- A fault-tolerant universal gate set can be obtained via \flmRefsHyperref{ref410}{code switching} between the Steane code and the \(\llbracket 10,1,2\rrbracket \) code \NoCaseChange{\protect\cite{cite3182}}.
\item\relax
\flmRefsHyperref[eczindexfamilyrel]{code:phantom_14_3_3}{\(\llbracket 14,3,3\rrbracket \) CE phantom code} --- Dual-rail concatenation of the \(\llbracket 7,1,3\rrbracket \) Steane code yields a \(\llbracket 14,1,3\rrbracket \) CE CSS code, from which the locally Clifford-equivalent \(\llbracket 14,3,3\rrbracket \) CE CSS frame is obtained by removing two independent \(Z\)-type stabilizer generators \NoCaseChange{\protect\cite{cite524}}.
\item\relax
\flmRefsHyperref[eczindexfamilyrel]{code:stab_15_1_3}{\(\llbracket 15,1,3\rrbracket \) quantum RM code} --- The \(\llbracket 15,1,3\rrbracket \) code can be viewed as a (gauge-fixed) doubled color code obtained from the Steane code via the doubling transformation \NoCaseChange{\protect\cite{cite731}}. A fault-tolerant universal gate set can be done via \flmRefsHyperref{ref410}{code switching} between the Steane code and the \(\llbracket 15,1,3\rrbracket \) code \NoCaseChange{\protect\cite{cite787,cite793,cite3203,cite3204,cite3208}}. An \(\llbracket 105,1,3\rrbracket \) alternative concatenation of the \(\llbracket 15,1,3\rrbracket \) and Steane codes allows for a universal gate set consisting of gates that are close to transversal \NoCaseChange{\protect\cite{cite3209,cite775}}.
\item\relax
\flmRefsHyperref[eczindexfamilyrel]{code:stab_4_2_2}{\(\llbracket 4,2,2\rrbracket \) Four-qubit code} --- The Steane code can be built from two \(\llbracket 4,2,2\rrbracket \) codes in the quantum Lego code framework \NoCaseChange{\protect\cite{cite2868}}. Ref. \NoCaseChange{\protect\cite{cite3216}} also introduces a \(C_4\)/Steane concatenated code, obtained by concatenating the \(\llbracket 4,2,2\rrbracket \) code with the Steane code, as an underlying code for further concatenation with quantum Hamming codes.
\item\relax
\flmRefsHyperref[eczindexfamilyrel]{code:stab_5_1_2}{\(\llbracket 5,1,2\rrbracket \) rotated surface code} --- The \(\llbracket 5,1,2\rrbracket \) morphed Steane code is obtained by morphing the Steane code on a region whose child code is a \(\llbracket 4,2,2\rrbracket \) code \NoCaseChange{\protect\cite[{Fig. 1}]{cite687}}.
\item\relax
\flmRefsHyperref[eczindexfamilyrel]{code:icosahedral_permutation_invariant}{\(\llparenthesis 7,2,3\rrparenthesis \) Pollatsek-Ruskai code} --- The Pollatsek-Ruskai code can be continuously deformed to the Steane code \NoCaseChange{\protect\cite{cite2555}}.
\item\relax
\flmRefsHyperref[eczindexfamilyrel]{code:rbh}{Raussendorf-Bravyi-Harrington (RBH) cluster-state code} --- Concatenation of the RBH code with small codes such as the \(\llbracket 2,1\rrbracket \) repetition code, \(\llbracket 4,1,1,2\rrbracket \) subsystem code, or Steane code can improve thresholds \NoCaseChange{\protect\cite{cite3248}}.
\end{eczvaluelist}
\eczhbkcontributors{ Remmy Zen, Eric Huang, Joseph T. Iosue, \eczhuVVA }
\endeczcode

\eczcode{twist_defect_7_1_3}{\(\llbracket 7,1,3\rrbracket \) twist-defect surface code}{~\NoCaseChange{\protect\cite{cite433}}}
\codefieldsection{Alternative Names}
\begin{eczvaluelist}
\item\relax \(\llbracket 7,1,3\rrbracket \) triangle code
\end{eczvaluelist}
\eczhIndexCodeAliasName{twist_defect_7_1_3}{twist-defect surface code}
\eczhIndexCodeAliasName{twist_defect_7_1_3}{\(\llbracket 7,1,3\rrbracket \) triangle code}
\codefieldsection{Description}
A \(\llbracket 7,1,3\rrbracket \) code (different from the Steane code) that is a small example of a twist-defect surface code.

A stabilizer tableau for the code is given by \NoCaseChange{\protect\cite[{ID 190}]{cite453}}
\flmMathEnvironment{align}{}{
\begin{array}{ccccccc}
  X & I & I & I & I & I & Z \\
  I & X & I & I & I & Z & I \\
  I & I & X & I & Z & I & I \\
  I & I & Z & X & X & Z & I \\
  I & Z & I & Y & I & Y & Z \\
  Z & I & I & Z & Z & I & X
\end{array}~.
}
It is one of sixteen distinct indecomposable \(\llbracket 7,1,3\rrbracket \) codes \NoCaseChange{\protect\cite{cite452}}.

\codefieldsection{Protection}
Fully fault-tolerant depolarizing-noise designs using \(13\) or \(15\) total qubits, including ancillas, have exREC pseudothresholds of order \(10^{-4}\) \NoCaseChange{\protect\cite{cite433}}.

\codefieldsection{Transversal and Permutation-Based Gates}
\begin{eczvaluelist}
\item\relax Admits certain transversal order-three single-qubit \flmRefsHyperref{ref409}{Clifford} gates (e.g., \(SH\)) \NoCaseChange{\protect\cite{cite433}}.
\end{eczvaluelist}
\codefieldsection{Gates}
\begin{eczvaluelist}
\item\relax Within the triangle-code architecture, supports the full logical \flmRefsHyperref{ref409}{Clifford group} using lattice surgery, 1-bit teleportation, and patch reorientation \NoCaseChange{\protect\cite{cite433}}.
\end{eczvaluelist}
\codefieldsection{Parents}
\begin{eczvaluelist}
\item\relax
\flmRefsHyperref[eczindexfamilyrel]{code:triangle_surface}{Triangular surface code}\item\relax
\flmRefsHyperref[eczindexfamilyrel]{code:small_distance_qubit_stabilizer}{Small-distance qubit stabilizer code}\end{eczvaluelist}
\eczhbkcontributors{ \eczhuVVA }
\endeczcode

\eczcode{xzzx_7_1_3}{\(\llbracket 7,1,3\rrbracket \) XZZX cyclic code}{~\NoCaseChange{\protect\cite{cite2630}}}
\eczhIndexCodeAliasName{xzzx_7_1_3}{XZZX cyclic code}
\codefieldsection{Description}
A \(\llbracket 7,1,3\rrbracket \) cyclic non-CSS code whose stabilizer generators are the seven cyclic shifts of the weight-four Pauli string \(XZIZXII\),
any six of which are independent.
The non-identity support of this generator is \(XZZX\) (at positions 0, 1, 3, 4), making this a member of the twisted XZZX code family.
It is one of sixteen distinct indecomposable \(\llbracket 7,1,3\rrbracket \) codes \NoCaseChange{\protect\cite{cite452}}.

A stabilizer tableau for the code is given by \NoCaseChange{\protect\cite[{ID 227}]{cite453}}
\flmMathEnvironment{align}{}{
\begin{array}{ccccccc}
  X & Z & I & Z & X & I & I \\
  I & X & Z & I & Z & X & I \\
  I & I & X & Z & I & Z & X \\
  X & I & I & X & Z & I & Z \\
  Z & X & I & I & X & Z & I \\
  I & Z & X & I & I & X & Z
\end{array}~.
}

\codefieldsection{Parents}
\begin{eczvaluelist}
\item\relax
\flmRefsHyperref[eczindexfamilyrel]{code:twisted_xzzx}{Twisted XZZX toric code} --- The \(\llbracket 7,1,3\rrbracket \) XZZX cyclic code is a cyclic non-CSS code whose generators are the cyclic shifts of the weight-four Pauli string \(XZIZXII\), which has \(XZZX\) support.
\item\relax
\flmRefsHyperref[eczindexfamilyrel]{code:small_distance_qubit_stabilizer}{Small-distance qubit stabilizer code}\end{eczvaluelist}
\eczhbkcontributors{ \eczhuVVA }
\endeczcode

\eczcode{hgp_7_2_2}{\(\llbracket 7,2,2\rrbracket \) HGP phantom code}{~\NoCaseChange{\protect\cite{cite514}}}
\eczhIndexCodeAliasName{hgp_7_2_2}{HGP phantom code}
\codefieldsection{Description}
Smallest member of a family of CSS phantom HGP codes obtained from the hypergraph product of a classical simplex code and a repetition code.

A stabilizer tableau for an equivalent code is given by \NoCaseChange{\protect\cite[{ID 521}]{cite453}}
\flmMathEnvironment{align}{}{
\begin{array}{ccccccc}
  Z & Z & I & I & I & I & Z \\
  I & I & Z & I & Z & I & Z \\
  I & I & I & Z & I & Z & Z \\
  I & X & X & X & I & I & X \\
  X & I & I & I & X & X & X
\end{array}~.
}

\codefieldsection{Parents}
\begin{eczvaluelist}
\item\relax
\flmRefsHyperref[eczindexfamilyrel]{code:phantom}{Phantom code} --- The \(\llbracket 7,2,2\rrbracket \) HGP code is the smallest member of a simplex/repetition HGP family of CSS phantom codes \NoCaseChange{\protect\cite{cite514}}.
\item\relax
\flmRefsHyperref[eczindexfamilyrel]{code:hypergraph_product}{Hypergraph product (HGP) code} --- This code is the hypergraph product of the \([3,2,2]\) simplex code and the \([2,1,2]\) repetition code \NoCaseChange{\protect\cite{cite514}}.
\item\relax
\flmRefsHyperref[eczindexfamilyrel]{code:small_distance_qubit_stabilizer}{Small-distance qubit stabilizer code}\end{eczvaluelist}
\codefieldsection{Cousins}
\begin{eczvaluelist}
\item\relax
\flmRefsHyperref[eczindexfamilyrel]{code:simplex}{\([2^m-1,m,2^{m-1}]\) simplex code} --- This code is constructed from the hypergraph product of the \([3,2,2]\) simplex code and the \([2,1,2]\) repetition code \NoCaseChange{\protect\cite{cite514}}.
\item\relax
\flmRefsHyperref[eczindexfamilyrel]{code:repetition}{Repetition code} --- This code is constructed from the hypergraph product of the \([3,2,2]\) simplex code and the \([2,1,2]\) repetition code \NoCaseChange{\protect\cite{cite514}}.
\end{eczvaluelist}
\eczhbkcontributors{ \eczhuVVA }
\endeczcode

\eczcode{qetc_7_2}{\(\llbracket 7,2,2\rrbracket \) QETC}{~\NoCaseChange{\protect\cite{cite2985}}}
\eczhIndexCodeAliasName{qetc_7_2}{QETC}
\codefieldsection{Description}
Seven-qubit \flmRefsHyperref{ref672}{pure} QETC that transmutes all single-qubit Pauli errors to logical phase errors.

A stabilizer tableau for the code is given by \NoCaseChange{\protect\cite[{Table 1}]{cite2985}}
\flmMathEnvironment{align}{}{
\begin{array}{ccccccc}
  X & X & Y & Y & Z & I & Z \\
  I & Z & X & Y & Y & X & Y \\
  I & I & I & I & I & Z & Z \\
  Z & Z & I & I & Z & I & Z \\
  Z & Z & Z & Z & I & I & I
\end{array}~.
}
The above stabilizer tableau is equivalent to \NoCaseChange{\protect\cite[{ID 646}]{cite453}} by applying \(H\) to qubits 1 and 2 and \(SH\) (with \(H\) applied first) to qubits 3 and 4, followed by the qubit relabeling \((1,2,3,4,5,6,7)\to(5,6,1,3,7,2,4)\).

\codefieldsection{Parents}
\begin{eczvaluelist}
\item\relax
\flmRefsHyperref[eczindexfamilyrel]{code:small_distance_qubit_stabilizer}{Small-distance qubit stabilizer code}\item\relax
\flmRefsHyperref[eczindexfamilyrel]{code:qetc}{Quantum error-transmuting code (QETC)}\end{eczvaluelist}
\codefieldsection{Cousin}
\begin{eczvaluelist}
\item\relax
\flmRefsHyperref[eczindexfamilyrel]{code:bare_7_1_3}{\(\llbracket 7,1,3\rrbracket \) bare code} --- The stabilizer group of the \(\llbracket 7,2,2\rrbracket \) QETC, together with the logical-\(Z\) operator on the first logical qubit, generates the stabilizer group of a \(\llbracket 7,1,3\rrbracket \) code \NoCaseChange{\protect\cite{cite2985}} equivalent to the bare \(\llbracket 7,1,3\rrbracket \) code \NoCaseChange{\protect\cite{cite447}}.
\end{eczvaluelist}
\eczhbkcontributors{ \eczhuVVA }
\endeczcode

\eczcode{xz_7_3_2}{\(\llbracket 7,3,2\rrbracket \) punctured hypercube code}{~\NoCaseChange{\protect\cite{cite736}}}
\eczhIndexCodeAliasName{xz_7_3_2}{punctured hypercube code}
\codefieldsection{Description}
A seven-qubit \flmRefsHyperref{ref672}{pure} CSS code that corrects a single \(X\) or a single \(Z\) error, but not a single \(Y\) error.
It is the punctured version of the \(\llbracket 8,3,2\rrbracket \) hypercube quantum code and is a phantom code \NoCaseChange{\protect\cite{cite514}}.

A CSS stabilizer tableau for the code is given by \NoCaseChange{\protect\cite[{Table 8.6}]{cite736}\protect\cite[{ID 475}]{cite453}}
\flmMathEnvironment{align}{}{
\begin{array}{ccccccc}
  Z & Z & Z & I & I & I & Z \\
  I & Z & I & Z & I & Z & Z \\
  Z & Z & I & I & Z & Z & I \\
  X & X & X & X & X & X & X
\end{array}~.
}

\codefieldsection{Protection}
Distance two. As the \(k=3\) punctured hypercube code, it saturates the general qubit phantom-code bound \(n\geq 2^k-1\) \NoCaseChange{\protect\cite{cite723}}.
\codefieldsection{Parents}
\begin{eczvaluelist}
\item\relax
\flmRefsHyperref[eczindexfamilyrel]{code:phantom}{Phantom code} --- This code is the punctured version of the \(\llbracket 8,3,2\rrbracket \) hypercube quantum code and is a phantom code \NoCaseChange{\protect\cite{cite514}}.
\item\relax
\flmRefsHyperref[eczindexfamilyrel]{code:quantum_reed_muller}{Quantum Reed-Muller (RM) code} --- The punctured hypercube family is a quantum Reed-Muller family built from shortened and punctured classical Reed-Muller codes \NoCaseChange{\protect\cite{cite723}}.
\item\relax
\flmRefsHyperref[eczindexfamilyrel]{code:small_distance_qubit_stabilizer}{Small-distance qubit stabilizer code}\end{eczvaluelist}
\codefieldsection{Cousins}
\begin{eczvaluelist}
\item\relax
\flmRefsHyperref[eczindexfamilyrel]{code:stab_8_3_2}{\(\llbracket 8,3,2\rrbracket \) Smallest interesting color code} --- The \(\llbracket 7,3,2\rrbracket \) code is obtained by puncturing one qubit from the \(\llbracket 8,3,2\rrbracket \) hypercube quantum code \NoCaseChange{\protect\cite{cite514}}.
\item\relax
\flmRefsHyperref[eczindexfamilyrel]{code:hamming743}{\([7,4,3]\) Hamming code} --- The \(\llbracket 7,3,2\rrbracket \) punctured hypercube code \(H_X\) check matrix is the parity-check matrix of the \([7,4,3]\) Hamming code, while its \(H_Z\) matrix is that of the SPC code.
\item\relax
\flmRefsHyperref[eczindexfamilyrel]{code:parity_check}{\([n,n-1,2]\) Single parity-check (SPC) code} --- The \(\llbracket 7,3,2\rrbracket \) punctured hypercube code \(H_X\) check matrix is the parity-check matrix of the \([7,4,3]\) Hamming code, while its \(H_Z\) matrix is that of the SPC code.
\item\relax
\flmRefsHyperref[eczindexfamilyrel]{code:phantom_14_3_3}{\(\llbracket 14,3,3\rrbracket \) CE phantom code} --- Concatenating the \(\llbracket 7,3,(d_X=3,d_Z=2)\rrbracket \) punctured hypercube code with the two-qubit phase-flip repetition code yields this \(\llbracket 14,3,(d_X=3,d_Z=4)\rrbracket \) CSS phantom code \NoCaseChange{\protect\cite{cite514}}. Dual-rail concatenation of the same punctured hypercube code yields a single-qubit Clifford-equivalent CE CSS frame \NoCaseChange{\protect\cite{cite524}}.
\end{eczvaluelist}
\eczhbkcontributors{ \eczhuVVA }
\endeczcode

\eczcode{bb72}{\(\llbracket 72,12,6\rrbracket \) BB6 code}{~\NoCaseChange{\protect\cite{cite441}}}
\codefieldsection{Alternative Names}
\begin{eczvaluelist}
\item\relax \((6,6)\) BB6 code
\end{eczvaluelist}
\eczhIndexCodeAliasName{bb72}{BB6 code}
\eczhIndexCodeAliasName{bb72}{\((6,6)\) BB6 code}
\codefieldsection{Description}
A bivariate bicycle (BB) code with parameters \(\llbracket 72,12,6\rrbracket \) and weight-six stabilizer generators \NoCaseChange{\protect\cite{cite441}}.

One defining presentation uses \((\ell,m)=(6,6)\) with \(x^{\ell}=y^{m}=1\), and
\(A=x^3+y+y^2\), \(B=y^3+x+x^2\) in \(\mathbb{F}_2[x,y]/(x^{\ell}-1,y^{m}-1)\) \NoCaseChange{\protect\cite[{Table 3}]{cite441}}.

\codefieldsection{Rate}
Ancilla-added encoding rate is \(1/12\), using \(n_a=n=72\) ancilla qubits.
\codefieldsection{Parent}
\begin{eczvaluelist}
\item\relax
\flmRefsHyperref[eczindexfamilyrel]{code:qcga}{Bivariate bicycle (BB) code}\end{eczvaluelist}
\eczhbkcontributors{ \eczhuVVA }
\endeczcode

\eczcode{hybrid_8_2-1_3}{\(\llbracket 8, 2:1, 3\rrbracket \) hybrid stabilizer code}{~\NoCaseChange{\protect\cite{cite2872}}}
\eczhIndexCodeAliasName{hybrid_8_2-1_3}{hybrid stabilizer code}
\codefieldsection{Description}
A code obtained from the \(\llbracket 8,3,3\rrbracket \) Gottesman code by using one of its logical qubits as a classical bit.
One can also use two logical qubits as classical bits, obtaining an \(\llbracket 8,1:2,3\rrbracket \) hybrid stabilizer code.
\codefieldsection{Parent}
\begin{eczvaluelist}
\item\relax
\flmRefsHyperref[eczindexfamilyrel]{code:hybrid_stabilizer}{Hybrid stabilizer code}\end{eczvaluelist}
\codefieldsection{Cousins}
\begin{eczvaluelist}
\item\relax
\flmRefsHyperref[eczindexfamilyrel]{code:small_distance_qubit_stabilizer}{Small-distance qubit stabilizer code}\item\relax
\flmRefsHyperref[eczindexfamilyrel]{code:stab_8_3_3}{\(\llbracket 8, 3, 3\rrbracket \) Eight-qubit Gottesman code} --- \(\llbracket 8, 2:1, 3\rrbracket \) hybrid stabilizer code is obtained from the \(\llbracket 8,3,3\rrbracket \) Gottesman code by using one of its logical qubits as a classical bit.
\end{eczvaluelist}
\eczhbkcontributors{ \eczhuVVA }
\endeczcode

\eczcode{stab_8_3_3}{\(\llbracket 8, 3, 3\rrbracket \) Eight-qubit Gottesman code}{~\NoCaseChange{\protect\cite{cite1182,cite3369,cite861,cite3370}}}
\eczhIndexCodeAliasName{stab_8_3_3}{Eight-qubit Gottesman code}
\codefieldsection{Description}
Eight-qubit \flmRefsHyperref{ref811}{non-degenerate} code that can be obtained from a modified CSS construction using the \([8,4,4]\) extended Hamming code and a \([8,7,2]\) even-weight code \NoCaseChange{\protect\cite{cite861}}.
The modification introduces signs between the codewords.

See \NoCaseChange{\protect\cite[{Table 3.3}]{cite736}} for its stabilizer generator matrix.
A stabilizer tableau for the code is given by \NoCaseChange{\protect\cite[{ID 6822}]{cite453}}
\flmMathEnvironment{align}{}{
\begin{array}{cccccccc}
  X & Y & Y & X & Z & I & I & Z \\
  Z & X & I & Z & X & I & Z & Z \\
  X & Z & Z & I & I & X & Z & Z \\
  Y & I & Y & Z & I & Z & X & Z \\
  Y & Z & X & I & Z & Z & I & Y
\end{array}~.
}
The code's automorphism group is \(\text{A}\Gamma\text{L}(1,8)\) \NoCaseChange{\protect\cite{cite3222}}.
It is unique for its parameters, up to equivalence \NoCaseChange{\protect\cite{cite452}\protect\cite[{pg. 386}]{cite42}}.

\codefieldsection{Transversal and Permutation-Based Gates}
\begin{eczvaluelist}
\item\relax Permutation-based gates \NoCaseChange{\protect\cite[{Sec. IV.D}]{cite763}}.
\item\relax No gates outside of the \flmRefsHyperref{ref663}{Pauli group} were found in Ref. \NoCaseChange{\protect\cite{cite805}}.
\end{eczvaluelist}
\codefieldsection{Gates}
\begin{eczvaluelist}
\item\relax Logical Trotter circuits can be implemented via symplectic transvections \NoCaseChange{\protect\cite{cite3371}}.
\end{eczvaluelist}
\codefieldsection{Parent}
\begin{eczvaluelist}
\item\relax
\flmRefsHyperref[eczindexfamilyrel]{code:quantum_hamming}{\(\llbracket 2^r, 2^r-r-2, 3\rrbracket \) Gottesman code}\end{eczvaluelist}
\codefieldsection{Cousins}
\begin{eczvaluelist}
\item\relax
\flmRefsHyperref[eczindexfamilyrel]{code:hamming844}{\([8,4,4]\) extended Hamming code} --- The \(\llbracket 8, 3, 3\rrbracket \) code is obtained via a modified CSS construction from the \([8,4,4]\) extended Hamming code.
\item\relax
\flmRefsHyperref[eczindexfamilyrel]{code:analog_stabilizer}{Analog stabilizer code} --- The eight-qubit Gottesman code has been extended to an analog stabilizer code \NoCaseChange{\protect\cite{cite3372}}.
\item\relax
\flmRefsHyperref[eczindexfamilyrel]{code:hybrid_8_2-1_3}{\(\llbracket 8, 2:1, 3\rrbracket \) hybrid stabilizer code} --- \(\llbracket 8, 2:1, 3\rrbracket \) hybrid stabilizer code is obtained from the \(\llbracket 8,3,3\rrbracket \) Gottesman code by using one of its logical qubits as a classical bit.
\end{eczvaluelist}
\eczhbkcontributors{ Feroz Ahmed Mian, \eczhuVVA }
\endeczcode

\eczcode{stab_8_1_2}{\(\llbracket 8,1,2\rrbracket \) Shen-Wang-Cao code}{~\NoCaseChange{\protect\cite{cite786}}}
\eczhIndexCodeAliasName{stab_8_1_2}{Shen-Wang-Cao code}
\codefieldsection{Description}
A stabilizer code that admits a logical \(T\) gate via application of physical \(T\) gates and a \(CZ\)-like gate.

A stabilizer tableau for the code is given by \NoCaseChange{\protect\cite[{ID 980}]{cite453}}.
\flmMathEnvironment{align}{}{
\begin{array}{cccccccc}
  X & X & I & I & X & X & I & I \\
  I & I & X & X & X & X & I & I \\
  I & I & I & I & I & I & X & X \\
  Z & I & I & Z & I & Z & Z & Z \\
  I & Z & I & Z & I & Z & Z & Z \\
  I & I & Z & Z & I & I & I & I \\
  I & I & I & I & Z & Z & I & I
\end{array}~.
}

\codefieldsection{Transversal and Permutation-Based Gates}
\begin{eczvaluelist}
\item\relax Logical \(T\) gate \(\bar{T}=T^{\otimes 6}\otimes K\) via physical \(T\) gates on the first six qubits and the two-qubit gate \(K\propto\cos(\pi/8)(I\otimes I)-i\sin(\pi/8)(Z\otimes Z)\propto\mathrm{diag}(1,e^{i\pi/4},e^{i\pi/4},1)\) on qubits 7 and 8 \NoCaseChange{\protect\cite{cite786}}.
\end{eczvaluelist}
\codefieldsection{Parents}
\begin{eczvaluelist}
\item\relax
\flmRefsHyperref[eczindexfamilyrel]{code:qubit_css}{Qubit CSS code}\item\relax
\flmRefsHyperref[eczindexfamilyrel]{code:small_distance_qubit_stabilizer}{Small-distance qubit stabilizer code}\end{eczvaluelist}
\codefieldsection{Cousins}
\begin{eczvaluelist}
\item\relax
\flmRefsHyperref[eczindexfamilyrel]{code:xp_stabilizer}{XP stabilizer code} --- The \(\llbracket 8,1,2\rrbracket \) code is an XP-regular code that can be obtained via the XP stabilizer formalism applied to the \(\llbracket 15,1,3\rrbracket \) Reed-Muller code \NoCaseChange{\protect\cite{cite786}}.
\item\relax
\flmRefsHyperref[eczindexfamilyrel]{code:stab_15_1_3}{\(\llbracket 15,1,3\rrbracket \) quantum RM code} --- The \(\llbracket 8,1,2\rrbracket \) code is an XP-regular code that can be obtained via the XP stabilizer formalism applied to the \(\llbracket 15,1,3\rrbracket \) Reed-Muller code \NoCaseChange{\protect\cite{cite786}}.
\item\relax
\flmRefsHyperref[eczindexfamilyrel]{code:quantum_lego}{Tensor-network code} --- The \(\llbracket 8,1,2\rrbracket \) code is an XP-regular code that can be obtained via the XP stabilizer formalism applied to the \(\llbracket 15,1,3\rrbracket \) Reed-Muller code \NoCaseChange{\protect\cite{cite786}}.
\end{eczvaluelist}
\eczhbkcontributors{ \eczhuVVA }
\endeczcode

\eczcode{stab_8_2_2}{\(\llbracket 8,2,2\rrbracket \) hyperbolic color code}{~\NoCaseChange{\protect\cite{cite702}}}
\eczhIndexCodeAliasName{stab_8_2_2}{hyperbolic color code}
\codefieldsection{Description}
An \(\llbracket 8,2,2\rrbracket \) hyperbolic color code defined on the projective plane. It is a self-dual CSS code.

A stabilizer tableau for the code is given by \NoCaseChange{\protect\cite[{ID 4926}]{cite453}}
\flmMathEnvironment{align}{}{
\begin{array}{cccccccc}
  Z & Z & Z & Z & I & I & I & I \\
  X & X & X & X & I & I & I & I \\
  Z & Z & I & I & Z & Z & I & I \\
  X & X & I & I & X & X & I & I \\
  Z & Z & I & I & I & I & Z & Z \\
  X & X & I & I & I & I & X & X
\end{array}~.
}

\codefieldsection{Transversal and Permutation-Based Gates}
\begin{eczvaluelist}
\item\relax Applying transversal \(S\) and \(S^{\dagger}\), \(\sqrt{X}\), and Hadamard gates yields various logical gates \NoCaseChange{\protect\cite{cite805}}. For instance, the physical gate \(S^{\dagger}_{2}S^{\dagger}_{3}S_{4}S_{5}S_{6}S_{7}CZ_{01}\) implements the logical action \(\bar{S}_{0} \bar{Z}_{1} \overline{CZ}_{01}\).
\end{eczvaluelist}
\codefieldsection{Parents}
\begin{eczvaluelist}
\item\relax
\flmRefsHyperref[eczindexfamilyrel]{code:hyperbolic_color}{Hyperbolic color code} --- The \(\llbracket 8,2,2\rrbracket \) hyperbolic color code is defined on the projective plane.
\item\relax
\flmRefsHyperref[eczindexfamilyrel]{code:stabilizer_over_gf4}{Hermitian qubit code} --- The \(\llbracket 8,2,2\rrbracket \) hyperbolic color code is Hermitian \NoCaseChange{\protect\cite[{ID 4926}]{cite453}}.
\item\relax
\flmRefsHyperref[eczindexfamilyrel]{code:self_dual_css}{Self-dual CSS code}\item\relax
\flmRefsHyperref[eczindexfamilyrel]{code:small_distance_qubit_stabilizer}{Small-distance qubit stabilizer code}\end{eczvaluelist}
\codefieldsection{Cousin}
\begin{eczvaluelist}
\item\relax
\flmRefsHyperref[eczindexfamilyrel]{code:shor_nine}{\(\llbracket 9,1,3\rrbracket \) Shor code} --- Like the Shor code, the \(\llbracket 8,2,2\rrbracket \) hyperbolic color code is a small code defined on the projective plane.
\end{eczvaluelist}
\eczhbkcontributors{ \eczhuVVA }
\endeczcode

\eczcode{stab_8_2_3}{\(\llbracket 8,2,3\rrbracket \) Hermitian code}{~\NoCaseChange{\protect\cite{cite862}}}
\eczhIndexCodeAliasName{stab_8_2_3}{Hermitian code}
\codefieldsection{Description}
A non-CSS Hermitian eight-qubit stabilizer code that is the only Hermitian code and that has the largest automorphism group among the 20 inequivalent \(\llbracket 8,2,3\rrbracket \) codes \NoCaseChange{\protect\cite{cite454}}.
The code has an exceptionally short fault-tolerant syndrome measurement sequence \NoCaseChange{\protect\cite{cite862,cite454}}.

A stabilizer tableau for the code is \NoCaseChange{\protect\cite{cite795}}
\flmMathEnvironment{align}{}{
\begin{array}{cccccccc}
  X & X & X & X & I & I & I & I \\
  Z & Z & Z & Z & I & I & I & I \\
  I & I & I & I & X & X & X & X \\
  I & I & I & I & Z & Z & Z & Z \\
  I & X & Y & Z & I & X & Y & Z \\
  I & Z & X & Y & I & Z & X & Y
\end{array}~,
}
which is equivalent to \NoCaseChange{\protect\cite[{ID 4947}]{cite453}}.

\codefieldsection{Decoding}
\begin{eczvaluelist}
\item\relax Short Shor-style syndrome extraction circuits can be used for syndrome-based decoding \NoCaseChange{\protect\cite{cite862}}.
\end{eczvaluelist}
\codefieldsection{Fault Tolerance}
\begin{eczvaluelist}
\item\relax Short Shor-style syndrome extraction circuits can be used for syndrome-based decoding \NoCaseChange{\protect\cite{cite862}}.
\end{eczvaluelist}
\codefieldsection{Parents}
\begin{eczvaluelist}
\item\relax
\flmRefsHyperref[eczindexfamilyrel]{code:stabilizer_over_gf4}{Hermitian qubit code}\item\relax
\flmRefsHyperref[eczindexfamilyrel]{code:small_distance_qubit_stabilizer}{Small-distance qubit stabilizer code}\end{eczvaluelist}
\codefieldsection{Cousins}
\begin{eczvaluelist}
\item\relax
\flmRefsHyperref[eczindexfamilyrel]{code:qubit_concatenated}{Concatenated qubit code} --- Applying the BLT mapping to the \(\llbracket 8,2,3\rrbracket \) Hermitian code and concatenating each qubit pair with the \(\llbracket 4,2,2\rrbracket \) code yields a \(\llbracket 32,4,6\rrbracket \) self-dual CSS code \NoCaseChange{\protect\cite[{Corr. 2}]{cite795}}.
\item\relax
\flmRefsHyperref[eczindexfamilyrel]{code:self_dual_css}{Self-dual CSS code} --- Applying the BLT mapping to the \(\llbracket 8,2,3\rrbracket \) Hermitian code and concatenating each qubit pair with the \(\llbracket 4,2,2\rrbracket \) code yields a \(\llbracket 32,4,6\rrbracket \) self-dual CSS code \NoCaseChange{\protect\cite[{Corr. 2}]{cite795}}.
\item\relax
\flmRefsHyperref[eczindexfamilyrel]{code:stab_4_2_2}{\(\llbracket 4,2,2\rrbracket \) Four-qubit code} --- Applying the BLT mapping to the \(\llbracket 8,2,3\rrbracket \) Hermitian code and concatenating each qubit pair with the \(\llbracket 4,2,2\rrbracket \) code yields a \(\llbracket 32,4,6\rrbracket \) self-dual CSS code \NoCaseChange{\protect\cite[{Corr. 2}]{cite795}}.
\end{eczvaluelist}
\eczhbkcontributors{ \eczhuVVA }
\endeczcode

\eczcode{stab_8_3_2}{\(\llbracket 8,3,2\rrbracket \) Smallest interesting color code}{~\NoCaseChange{\protect\cite{cite422,cite797}}}
\eczhIndexCodeAliasName{stab_8_3_2}{Smallest interesting color code}
\codefieldsection{Description}
Smallest 3D color code whose physical qubits lie on vertices of a cube and which admits a (weakly) transversal \(CCZ\) gate.

A stabilizer tableau for the code is given by \NoCaseChange{\protect\cite[{ID 4882}]{cite453}}
\flmMathEnvironment{align}{}{
\begin{array}{cccccccc}
  Z & Z & I & I & I & I & Z & Z \\
  Z & I & Z & Z & I & I & Z & I \\
  I & I & Z & I & Z & I & Z & Z \\
  Z & I & Z & I & I & Z & I & Z \\
  X & X & X & X & X & X & X & X
\end{array}~.
}

In encoded IQP sampling, the final measurement outcomes determine both the logical sample and stabilizer checks, enabling end-of-circuit error detection or postselected decoding directly from the classical samples \NoCaseChange{\protect\cite{cite759}}.

\codefieldsection{Transversal and Permutation-Based Gates}
\begin{eczvaluelist}
\item\relax \(CZ\) gates between any two logical qubits \NoCaseChange{\protect\cite{cite805}} and (weakly) transversal \(CCZ\) gate \NoCaseChange{\protect\cite{cite422,cite797,cite805}}.
\end{eczvaluelist}
\codefieldsection{Gates}
\begin{eczvaluelist}
\item\relax \(CCZ\) gate can be distilled in a fault-tolerant manner \NoCaseChange{\protect\cite{cite3373}}.
\item\relax Fault-tolerant and teleportation-free logical Hadamard \NoCaseChange{\protect\cite{cite3271}}.
\end{eczvaluelist}
\codefieldsection{Fault Tolerance}
\begin{eczvaluelist}
\item\relax \(CCZ\) gate can be distilled in a fault-tolerant manner \NoCaseChange{\protect\cite{cite3373}}.
\item\relax Fault-tolerant and teleportation-free logical Hadamard \NoCaseChange{\protect\cite{cite3271}}.
\item\relax Universal weakly fault-tolerant computation via code switching between this and another \(\llbracket 8,3,2\rrbracket \) CSS code in a postselected error-detecting regime \NoCaseChange{\protect\cite{cite3374}}.
\item\relax Fault-tolerant architecture \NoCaseChange{\protect\cite{cite3375}}.
\item\relax For hIQP sampling with decoding only in the final measurement round, error-detected \(\llbracket 8,3,2\rrbracket \) circuits outperform the \(\llbracket 16,3,4\rrbracket \) and \(\llbracket 15,1,3\rrbracket \) comparison circuits studied in Ref. \NoCaseChange{\protect\cite{cite759}} under its two-qubit-gate-noise model.
\end{eczvaluelist}
\codefieldsection{Realizations}
\begin{eczvaluelist}
\item\relax Trapped ions: one-qubit addition algorithm implemented fault-tolerantly on the Quantinuum H1-1 device \NoCaseChange{\protect\cite{cite3376}}. Trapped-ion processor by AQT: measurement-free universal fault-tolerant logical operations and a Grover-search demonstration \NoCaseChange{\protect\cite{cite3288}}.
\item\relax Superconducting circuits: fault-tolerant \(CCZ\) gate performed on IBM and IonQ devices \NoCaseChange{\protect\cite{cite3377}}.
\item\relax Neutral atom arrays: Lukin group \NoCaseChange{\protect\cite{cite3363}}. 48 logical qubits, 228 logical two-qubit gates, 48 logical \(CCZ\) gates, and error detection performed in 16 blocks. Circuit outcomes were sampled and cross-entropy (XEB) was calculated to verify quantumness. Logical entanglement entropy was measured \NoCaseChange{\protect\cite{cite3363}}.
\end{eczvaluelist}
\codefieldsection{Parents}
\begin{eczvaluelist}
\item\relax
\flmRefsHyperref[eczindexfamilyrel]{code:3d_color}{3D color code} --- The \(\llbracket 8,3,2\rrbracket \) code is the smallest non-trivial 3D color code.
\item\relax
\flmRefsHyperref[eczindexfamilyrel]{code:hypercube_quantum}{\(\llbracket 2^D,D,2\rrbracket \) hypercube quantum code} --- The \(\llbracket 8,3,2\rrbracket \) code is a hypercube code for \(D=3\).
\item\relax
\flmRefsHyperref[eczindexfamilyrel]{code:campbell_howard}{\(\llbracket 6k+2,3k,2\rrbracket \) Campbell-Howard code} --- The \(\llbracket 8,3,2\rrbracket \) code is the \(k=1\) member of the \(\llbracket 6k+2,3k,2\rrbracket \) Campbell-Howard family with a quasi-transversal logical \(CCZ\) gate \NoCaseChange{\protect\cite{cite754}}.
\end{eczvaluelist}
\codefieldsection{Cousins}
\begin{eczvaluelist}
\item\relax
\flmRefsHyperref[eczindexfamilyrel]{code:hamming844}{\([8,4,4]\) extended Hamming code} --- The \(\llbracket 8,3,2\rrbracket \) hypercube code \(H_X\) check matrix is the parity-check matrix of the \([8,4,4]\) extended Hamming code, while its \(H_Z\) matrix is that of the SPC code.
\item\relax
\flmRefsHyperref[eczindexfamilyrel]{code:parity_check}{\([n,n-1,2]\) Single parity-check (SPC) code} --- The \(\llbracket 8,3,2\rrbracket \) hypercube code \(H_X\) check matrix is the parity-check matrix of the \([8,4,4]\) extended Hamming code, while its \(H_Z\) matrix is that of the SPC code.
\item\relax
\flmRefsHyperref[eczindexfamilyrel]{code:xp_stabilizer}{XP stabilizer code} --- As the \(D=3\) member of the hypercube-code family, the \(\llbracket 8,3,2\rrbracket \) code can be viewed as an XP stabilizer code with precision \(N=8\) \NoCaseChange{\protect\cite[{Exam. 6.10}]{cite798}}.
\item\relax
\flmRefsHyperref[eczindexfamilyrel]{code:stab_15_1_3}{\(\llbracket 15,1,3\rrbracket \) quantum RM code} --- The \(\llbracket 8,3,2\rrbracket \) code can be obtained from a subset of physical qubits of the \(\llbracket 15,1,3\rrbracket \) code \NoCaseChange{\protect\cite{cite687}}.
\item\relax
\flmRefsHyperref[eczindexfamilyrel]{code:3d_surface}{3D surface code} --- Three cyclically rotated copies of the 3D surface/toric code admit a logical \(CCZ\) gate via transversal physical \(CCZ\) gates, and concatenating each such qubit triple with an \(\llbracket 8,3,2\rrbracket \) block yields a 3D toric/color family with parameters \(\llbracket 8n,3,2d\rrbracket \); its smallest member has parameters \(\llbracket 72,3,4\rrbracket \) \NoCaseChange{\protect\cite{cite759}}.
\item\relax
\flmRefsHyperref[eczindexfamilyrel]{code:qubit_concatenated}{Concatenated qubit code} --- Concatenating \(\llbracket 8,3,2\rrbracket \) blocks with triples of qubits drawn from three cyclically rotated 3D surface/toric codes yields a 3D toric/color family with parameters \(\llbracket 8n,3,2d\rrbracket \) and transversal logical \(CCZ\) implemented by physical \(T\) gates on the inner \(\llbracket 8,3,2\rrbracket \) blocks \NoCaseChange{\protect\cite{cite759}}.
\item\relax
\flmRefsHyperref[eczindexfamilyrel]{code:stab_10_1_2}{\(\llbracket 10,1,2\rrbracket \) Vasmer-Kubica code} --- The \(\llbracket 10,1,2\rrbracket \) code is obtained by morphing the \(\llbracket 15,1,3\rrbracket \) code on a region whose child code is the \(\llbracket 8,3,2\rrbracket \) smallest interesting color code \NoCaseChange{\protect\cite{cite687}}.
\item\relax
\flmRefsHyperref[eczindexfamilyrel]{code:stab_12_2_2}{\(\llbracket 12,2,2\rrbracket \) CSS code} --- The \(\llbracket 12,2,2\rrbracket \) CSS code can be obtained by joining two copies of the \(\llbracket 8,3,2\rrbracket \) code at a common face \NoCaseChange{\protect\cite{cite767}}.
\item\relax
\flmRefsHyperref[eczindexfamilyrel]{code:stab_16_6_4}{\(\llbracket 16,6,4\rrbracket \) Tesseract color code} --- Applying CNOT gates to the tesseract color code disentangles it into two \(\llbracket 8,3,2\rrbracket \) color codes \NoCaseChange{\protect\cite{cite483}}.
\item\relax
\flmRefsHyperref[eczindexfamilyrel]{code:xz_7_3_2}{\(\llbracket 7,3,2\rrbracket \) punctured hypercube code} --- The \(\llbracket 7,3,2\rrbracket \) code is obtained by puncturing one qubit from the \(\llbracket 8,3,2\rrbracket \) hypercube quantum code \NoCaseChange{\protect\cite{cite514}}.
\end{eczvaluelist}
\eczhbkcontributors{ \eczhuVVA }
\endeczcode

\eczcode{cubic_surface}{\(\llbracket 8,3,2\rrbracket \) Surface code on a cube}{~\NoCaseChange{\protect\cite{cite3187}}}
\codefieldsection{Alternative Names}
\begin{eczvaluelist}
\item\relax Landahl plucky code
\item\relax Cubic surface code
\end{eczvaluelist}
\eczhIndexCodeAliasName{cubic_surface}{Surface code on a cube}
\eczhIndexCodeAliasName{cubic_surface}{Landahl plucky code}
\eczhIndexCodeAliasName{cubic_surface}{Cubic surface code}
\codefieldsection{Description}
An \(\llbracket 8,3,2\rrbracket \) twist-defect surface code whose qubits lie on the vertices of a cube.
It is obtained by three-coloring the faces of a cube and placing \(X\), \(Y\), and \(Z\) stabilizer generators on each pair of faces of the same color.
Its non-CSS nature is due to twist defects \NoCaseChange{\protect\cite{cite442}} stemming from the geometry of the polytope.

A stabilizer tableau for the code is given by \NoCaseChange{\protect\cite[{ID 6851}]{cite453}}
\flmMathEnvironment{align}{}{
\begin{array}{cccccccc}
  I & Y & Y & I & Z & I & Z & I \\
  X & I & Z & Z & X & I & I & I \\
  I & X & I & X & Y & Y & I & I \\
  I & Z & I & I & I & Z & X & Z \\
  Z & I & I & Y & I & X & I & Y
\end{array}~.
}

\codefieldsection{Parents}
\begin{eczvaluelist}
\item\relax
\flmRefsHyperref[eczindexfamilyrel]{code:twist_defect_surface}{Twist-defect surface code} --- The surface code on a cube is a twist-defect surface code whose degree-three vertices can be interpreted as disclination twists \NoCaseChange{\protect\cite{cite3187}}.
\item\relax
\flmRefsHyperref[eczindexfamilyrel]{code:small_distance_qubit_stabilizer}{Small-distance qubit stabilizer code}\end{eczvaluelist}
\codefieldsection{Cousin}
\begin{eczvaluelist}
\item\relax
\flmRefsHyperref[eczindexfamilyrel]{code:hypercube}{Hypercube code} --- The surface code on a cube, whose qubits lie on the vertices of a cube, is obtained by three-coloring the faces of a cube and placing \(X\), \(Y\), and \(Z\) stabilizer generators on each pair of faces of the same color.
\end{eczvaluelist}
\eczhbkcontributors{ Andrew J. Landahl, Jim Harrington, \eczhuVVA }
\endeczcode

\eczcode{shor_nine}{\(\llbracket 9,1,3\rrbracket \) Shor code}{~\NoCaseChange{\protect\cite{cite3}}}
\eczhIndexCodeAliasName{shor_nine}{Shor code}
\codefieldsection{Description}
Nine-qubit \flmRefsHyperref{code:css}{CSS code} that is the first quantum error-correcting code \NoCaseChange{\protect\cite[{ID 8802}]{cite453}}.
Among indecomposable \(\llbracket 9,1,3\rrbracket \) CSS codes, the Shor code has the largest automorphism group \NoCaseChange{\protect\cite{cite454}}.

A set of logical codewords is
\flmMathEnvironment{align}{}{
\begin{split}
|\overline{0}\rangle&=\frac{1}{2\sqrt{2}}\left(|000\rangle+|111\rangle\right)^{\otimes3}\\
|\overline{1}\rangle&=\frac{1}{2\sqrt{2}}\left(|000\rangle-|111\rangle\right)^{\otimes3}~.
\end{split}
}
A stabilizer tableau for the code is
\flmMathEnvironment{align}{}{
\begin{array}{ccccccccc}
  Z & Z & I & I & I & I & I & I & I \\
  I & Z & Z & I & I & I & I & I & I \\
  I & I & I & Z & Z & I & I & I & I \\
  I & I & I & I & Z & Z & I & I & I \\
  I & I & I & I & I & I & Z & Z & I \\
  I & I & I & I & I & I & I & Z & Z \\
  X & X & X & X & X & X & I & I & I \\
  I & I & I & X & X & X & X & X & X
\end{array}~.
}
The \flmRefsHyperref{ref857}{encoder-respecting form} of the Shor code is a star-shaped tree graph \NoCaseChange{\protect\cite{cite858}}.
The code works by \flmRefsHyperref{code:qubit_concatenated}{concatenating} each qubit of a phase-flip \flmRefsHyperref{code:quantum_repetition}{repetition code} with a bit-flip \flmRefsHyperref{code:quantum_repetition}{repetition code}. Therefore, the code can correct both types of errors simultaneously.
The code is degenerate: for example, two \(Z\) errors in the same three-qubit block act identically on all codewords \NoCaseChange{\protect\cite[{Ch. 3}]{cite398}}.

\codefieldsection{Protection}
The code detects two-qubit errors or corrects an arbitrary single-qubit error. Since it corrects the single-qubit Pauli errors, linearity implies that it corrects arbitrary single-qubit errors and corresponding single-qubit error channels \NoCaseChange{\protect\cite[{Ch. 2}]{cite398}}. It also corrects two-qubit \flmRefsHyperref{ref498}{AD} errors \NoCaseChange{\protect\cite{cite3263}}.
\codefieldsection{Encoding}
\begin{eczvaluelist}
\item\relax Fault-tolerant logical zero and logical plus state preparation using reinforcement learning \NoCaseChange{\protect\cite{cite3200}}.
\item\relax Fault-tolerant measurement-free logical-zero state preparation \NoCaseChange{\protect\cite{cite3378}}.
\end{eczvaluelist}
\codefieldsection{Decoding}
\begin{eczvaluelist}
\item\relax Bit- and phase-flip circuits utilize CNOT and Hadamard gates \NoCaseChange{\protect\cite[{Fig. 10.6}]{cite3302}}.
\end{eczvaluelist}
\codefieldsection{Fault Tolerance}
\begin{eczvaluelist}
\item\relax Fault-tolerant logical zero and logical plus state preparation using reinforcement learning \NoCaseChange{\protect\cite{cite3200}}.
\item\relax Fault-tolerant measurement-free logical-zero state preparation \NoCaseChange{\protect\cite{cite3378}}.
\end{eczvaluelist}
\codefieldsection{Realizations}
\begin{eczvaluelist}
\item\relax Trapped-ion qubits: state preparation with 98.8(1)\% and 98.5(1)\% fidelity for state \(|\overline{0}\rangle\) and \(|\overline{1}\rangle\), respectively, by N. Linke group \NoCaseChange{\protect\cite{cite3379}}. Variants of the code to handle coherent noise studied and realized by K. Brown and C. Monroe groups \NoCaseChange{\protect\cite{cite3380}}.
\item\relax Optical systems: quantum teleportation of information implemented by J.-W. Pan group on a maximally entangled pair of one physical and one logical qubit with a fidelity of up to 78.6\% \NoCaseChange{\protect\cite{cite3381}}. All-photonic quantum repeater architecture tested on the same code \NoCaseChange{\protect\cite{cite3382}}.
\end{eczvaluelist}
\codefieldsection{Parents}
\begin{eczvaluelist}
\item\relax
\flmRefsHyperref[eczindexfamilyrel]{code:quantum_parity}{Quantum parity code (QPC)} --- The Shor code is part of the sub-family of \(\llbracket m^2,1,m\rrbracket \) QPCs.
\item\relax
\flmRefsHyperref[eczindexfamilyrel]{code:real_projective_plane}{Projective-plane surface code} --- The Shor code is one of the nine-qubit surface codes defined on the projective plane \NoCaseChange{\protect\cite[{Fig. 4}]{cite3383}\protect\cite[{Fig. 20}]{cite71}}.
\item\relax
\flmRefsHyperref[eczindexfamilyrel]{code:stab_9_1_3}{\(\llbracket 9,1,3\rrbracket _{\mathbb{Z}_q}\) modular-qudit code} --- The \(\llbracket 9,1,3\rrbracket _{\mathbb{Z}_q}\) modular-qudit code for \(q=2\) reduces to the \(\llbracket 9,1,3\rrbracket \) Shor code.
\item\relax
\flmRefsHyperref[eczindexfamilyrel]{code:small_distance_qubit_stabilizer}{Small-distance qubit stabilizer code}\end{eczvaluelist}
\codefieldsection{Cousins}
\begin{eczvaluelist}
\item\relax
\flmRefsHyperref[eczindexfamilyrel]{code:quantum_repetition}{Quantum repetition code} --- The Shor code is a concatenation of a three-qubit bit-flip with a three-qubit phase-flip repetition code.
\item\relax
\flmRefsHyperref[eczindexfamilyrel]{code:qubit_concatenated}{Concatenated qubit code} --- The Shor code is a concatenation of a three-qubit bit-flip with a three-qubit phase-flip repetition code.
\item\relax
\flmRefsHyperref[eczindexfamilyrel]{code:qecc}{Quantum error-correcting code (QECC)} --- The Shor code is the first quantum error-correcting code.
\item\relax
\flmRefsHyperref[eczindexfamilyrel]{code:cluster_state}{Cluster-state code} --- The Shor code admits a codeword that is the cluster state of a particular nine-vertex graph \NoCaseChange{\protect\cite{cite3322,cite868}}.
\item\relax
\flmRefsHyperref[eczindexfamilyrel]{code:lloyd_slotine}{\(\llbracket 9,1,3\rrbracket _{\mathbb{R}}\) Lloyd-Slotine code} --- The Lloyd-Slotine nine-mode code is a bosonic analogue of Shor's code.
\item\relax
\flmRefsHyperref[eczindexfamilyrel]{code:hybrid_stabilizer}{Hybrid stabilizer code} --- The Shor code can be modified into a degenerate \(\llbracket 9,1:3,3\rrbracket \) hybrid stabilizer code that still corrects arbitrary single-qubit errors \NoCaseChange{\protect\cite{cite2735}}.
\item\relax
\flmRefsHyperref[eczindexfamilyrel]{code:ruskai}{\(\llparenthesis 9,2,3\rrparenthesis \) Ruskai code} --- The \(\llparenthesis 9,2,3\rrparenthesis \) Ruskai code results from projecting the Shor code into the PI qubit subspace \NoCaseChange{\protect\cite{cite3173}}.
\item\relax
\flmRefsHyperref[eczindexfamilyrel]{code:stab_8_2_2}{\(\llbracket 8,2,2\rrbracket \) hyperbolic color code} --- Like the Shor code, the \(\llbracket 8,2,2\rrbracket \) hyperbolic color code is a small code defined on the projective plane.
\item\relax
\flmRefsHyperref[eczindexfamilyrel]{code:surface-17}{\(\llbracket 9,1,3\rrbracket \) Surface-17 code} --- Both Shor's code and surface-17 are \(\llbracket 9,1,3\rrbracket \) codes, but they are distinct (e.g., they have different \flmRefsHyperref{ref672}{quantum weight enumerators}).
\item\relax
\flmRefsHyperref[eczindexfamilyrel]{code:bacon_shor_9}{\(\llbracket 9,1,4,3\rrbracket \) Nine-qubit Bacon-Shor code} --- The Shor code is a CSS gauge fixing of the nine-qubit Bacon-Shor code \NoCaseChange{\protect\cite{cite3384,cite454}}.
\end{eczvaluelist}
\eczhbkcontributors{ Remmy Zen, Qingfeng (Kee) Wang, \eczhuVVA }
\endeczcode

\eczcode{surface-17}{\(\llbracket 9,1,3\rrbracket \) Surface-17 code}{~\NoCaseChange{\protect\cite{cite3385}}}
\codefieldsection{Alternative Names}
\begin{eczvaluelist}
\item\relax \(\llbracket 9,1,3\rrbracket \) rotated surface code
\end{eczvaluelist}
\eczhIndexCodeAliasName{surface-17}{Surface-17 code}
\eczhIndexCodeAliasName{surface-17}{\(\llbracket 9,1,3\rrbracket \) rotated surface code}
\codefieldsection{Description}
A \(\llbracket 9,1,3\rrbracket \) rotated surface code named for the sum of its 9 data qubits and 8 syndrome qubits.
It is one of the four inequivalent CSS gauge fixings of the nine-qubit Bacon-Shor code \NoCaseChange{\protect\cite{cite454}}.
It uses the smallest number of qubits to perform fault-tolerant error correction on a surface code with parallel syndrome extraction.

A stabilizer tableau for the code is given by \NoCaseChange{\protect\cite[{ID 8519}]{cite453}}.
\flmMathEnvironment{align}{}{
\begin{array}{ccccccccc}
  I & X & I & I & I & I & I & X & I \\
  I & I & I & X & X & I & I & I & I \\
  I & I & X & I & I & X & X & X & I \\
  X & I & I & X & I & X & I & I & X \\
  Z & I & I & I & I & I & I & I & Z \\
  I & I & Z & I & I & I & Z & I & I \\
  I & I & I & Z & Z & Z & Z & I & I \\
  I & Z & I & I & I & Z & I & Z & Z
\end{array}~.
}
The code is depicted in \flmRefsCref{ref3386}.

\begin{flmFloat}{figure}{NumCap}\includegraphics[width=107.98652891338584bp,max width=\linewidth]{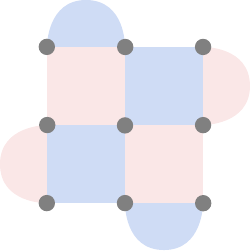}\caption{
  Stabilizer generators of the \(\llbracket 9,1,3\rrbracket \) surface-17 code.
  The 9 data qubits (circles) are arranged on a \(3\times 3\) rotated surface code lattice with open boundaries.
  The generators are weight-four (four-body) operators in the bulk and weight-two (two-body) operators on the boundaries.
  Red regions correspond to \(X\) operators while blue regions correspond to \(Z\) operators.}\label{ref3386}\end{flmFloat}

\codefieldsection{Protection}
Independent correction of single-qubit \(X\) and \(Z\) errors. Correction for some two-qubit \(X\) and \(Z\) errors.
Admits \flmRefsHyperref{ref2960}{pseudo-thresholds} of \(\approx 10^{-4}\) under depolarizing noise.

\codefieldsection{Encoding}
\begin{eczvaluelist}
\item\relax Measurement-free fault-tolerant logical zero state preparation in nearest-neighbor qubit connectivity \NoCaseChange{\protect\cite{cite3378}}.
\item\relax Fault-tolerant logical zero and logical plus state preparation in all-to-all connectivity, and fault-tolerant logical zero state preparation on 2D grids, with flag qubits \NoCaseChange{\protect\cite{cite3200}}.
\end{eczvaluelist}
\codefieldsection{Transversal and Permutation-Based Gates}
\begin{eczvaluelist}
\item\relax Pauli gates, CNOT gate, and \(H\) gate (with relabeling).
\end{eczvaluelist}
\codefieldsection{Decoding}
\begin{eczvaluelist}
\item\relax Lookup table \NoCaseChange{\protect\cite{cite3385}}.
\item\relax Syndrome extraction using Toffoli gates and qubit reset \NoCaseChange{\protect\cite{cite3387}}.
\end{eczvaluelist}
\codefieldsection{Fault Tolerance}
\begin{eczvaluelist}
\item\relax Measurement-free fault-tolerant logical zero state preparation in nearest-neighbor qubit connectivity \NoCaseChange{\protect\cite{cite3378}}.
\item\relax Fault-tolerant logical zero and logical plus state preparation in all-to-all connectivity, and fault-tolerant logical zero state preparation on 2D grids, with flag qubits \NoCaseChange{\protect\cite{cite3200}}.
\end{eczvaluelist}
\codefieldsection{Realizations}
\begin{eczvaluelist}
\item\relax Implemented at ETH Zurich by the Wallraff group
\NoCaseChange{\protect\cite{cite3388}} and on the Zuchongzhi 2.1 superconducting quantum processor \NoCaseChange{\protect\cite{cite3389}}.
Both experimental error rates are above the \flmRefsHyperref{ref2960}{pseudo-threshold} for this code relative to a single qubit; see Physics viewpoint for a summary \NoCaseChange{\protect\cite{cite3390}}.
Magic states have been created on the latter processor \NoCaseChange{\protect\cite{cite3391}}.
Lattice surgery on the surface-17 code has been realized by splitting the code into two repetition codes by the Wallraff group \NoCaseChange{\protect\cite{cite3392}}.
The device noise can be used to develop a decoder without relying on a theoretical noise model \NoCaseChange{\protect\cite{cite3393}}.

\end{eczvaluelist}
\codefieldsection{Notes}
\begin{eczvaluelist}
\item\relax Subject of various numerical studies examining the code under noise models and architectures specific to trapped ions \NoCaseChange{\protect\cite{cite3385,cite3394,cite3395}} and superconducting circuits \NoCaseChange{\protect\cite{cite3396,cite3397,cite3398}}.
\end{eczvaluelist}
\codefieldsection{Parents}
\begin{eczvaluelist}
\item\relax
\flmRefsHyperref[eczindexfamilyrel]{code:rotated_surface}{Rotated surface code}\item\relax
\flmRefsHyperref[eczindexfamilyrel]{code:small_distance_qubit_stabilizer}{Small-distance qubit stabilizer code}\end{eczvaluelist}
\codefieldsection{Cousins}
\begin{eczvaluelist}
\item\relax
\flmRefsHyperref[eczindexfamilyrel]{code:shor_nine}{\(\llbracket 9,1,3\rrbracket \) Shor code} --- Both Shor's code and surface-17 are \(\llbracket 9,1,3\rrbracket \) codes, but they are distinct (e.g., they have different \flmRefsHyperref{ref672}{quantum weight enumerators}).
\item\relax
\flmRefsHyperref[eczindexfamilyrel]{code:stellated_dodecahedron_css}{\(\llbracket 30,8,3\rrbracket \) Bring code} --- Bring's code and the surface-17 code have been compared numerically \NoCaseChange{\protect\cite{cite2383}}.
\item\relax
\flmRefsHyperref[eczindexfamilyrel]{code:bacon_shor_9}{\(\llbracket 9,1,4,3\rrbracket \) Nine-qubit Bacon-Shor code} --- The \(\llbracket 9,1,3\rrbracket \) rotated surface code is a CSS gauge fixing of the nine-qubit Bacon-Shor code \NoCaseChange{\protect\cite{cite454}}.
\end{eczvaluelist}
\eczhbkcontributors{ Remmy Zen, Kenneth R. Brown, \eczhuVVA }
\endeczcode

\eczcode{bacon_shor_9}{\(\llbracket 9,1,4,3\rrbracket \) Nine-qubit Bacon-Shor code}{~\NoCaseChange{\protect\cite{cite3,cite3037}}}
\eczhIndexCodeAliasName{bacon_shor_9}{Nine-qubit Bacon-Shor code}
\codefieldsection{Description}
Error-correcting nine-qubit subsystem stabilizer code encoding one logical qubit and four gauge qubits.
There are exactly four inequivalent CSS gauge fixings of the code, including the \flmRefsHyperref{code:shor_nine}{Shor code} and the \flmRefsHyperref{code:surface-17}{surface-17 code} \NoCaseChange{\protect\cite{cite454}}.

The stabilizer subgroup is generated by
\flmMathEnvironment{align}{}{
  \begin{array}{ccccccccc}
    X & X & X & X & X & X & I & I & I\\
    I & I & I & X & X & X & X & X & X\\
    Z & Z & I & Z & Z & I & Z & Z & I\\
    I & Z & Z & I & Z & Z & I & Z & Z
  \end{array}~,
}
while a convenient generating set for the gauge group is completed by the eight weight-two gauge operators
\flmMathEnvironment{align}{}{
  \begin{array}{ccccccccc}
  X & I & I & X & I & I & I & I & I\\
  I & X & I & I & X & I & I & I & I\\
  I & I & I & X & I & I & X & I & I\\
  I & I & I & I & X & I & I & X & I\\
  Z & Z & I & I & I & I & I & I & I\\
  I & I & I & Z & Z & I & I & I & I\\
  I & Z & Z & I & I & I & I & I & I\\
  I & I & I & I & Z & Z & I & I & I
  \end{array}~.
}
If the physical qubits are arranged in a three-by-three square, the \(Z\)-type (\(X\)-type) gauge operators are supported on qubits in the same row (column). 
The code reduces to the \flmRefsHyperref{code:shor_nine}{Shor code} for a particular gauge configuration.

\codefieldsection{Decoding}
\begin{eczvaluelist}
\item\relax Message passing for \(\llbracket 9,1,4,3\rrbracket \) Bacon-Shor code \NoCaseChange{\protect\cite{cite3399}}.
\end{eczvaluelist}
\codefieldsection{Code Capacity Threshold}
\begin{eczvaluelist}
\item\relax \(2.02 \times 10^{-5}\) \flmRefsHyperref{ref515}{concatenated threshold} for the recursively concatenated code \NoCaseChange{\protect\cite{cite3400}}.
\end{eczvaluelist}
\codefieldsection{Realizations}
\begin{eczvaluelist}
\item\relax Trapped-ion qubits: state preparation, logical measurement, and syndrome extraction (deferred to the end) demonstrated on a 13-qubit device by M. Cetina and C. Monroe groups \NoCaseChange{\protect\cite{cite3401}}.
\item\relax Neutral atom arrays: repeated error correction demonstrated on a device by Atom Computing \NoCaseChange{\protect\cite{cite3257}}.
\end{eczvaluelist}
\codefieldsection{Parent}
\begin{eczvaluelist}
\item\relax
\flmRefsHyperref[eczindexfamilyrel]{code:bacon_shor}{Bacon-Shor code} --- The nine-qubit Bacon-Shor code is the shortest error-correcting Bacon-Shor code.
\end{eczvaluelist}
\codefieldsection{Cousins}
\begin{eczvaluelist}
\item\relax
\flmRefsHyperref[eczindexfamilyrel]{code:shor_nine}{\(\llbracket 9,1,3\rrbracket \) Shor code} --- The Shor code is a CSS gauge fixing of the nine-qubit Bacon-Shor code \NoCaseChange{\protect\cite{cite3384,cite454}}.
\item\relax
\flmRefsHyperref[eczindexfamilyrel]{code:surface-17}{\(\llbracket 9,1,3\rrbracket \) Surface-17 code} --- The \(\llbracket 9,1,3\rrbracket \) rotated surface code is a CSS gauge fixing of the nine-qubit Bacon-Shor code \NoCaseChange{\protect\cite{cite454}}.
\item\relax
\flmRefsHyperref[eczindexfamilyrel]{code:small_distance_qubit_stabilizer}{Small-distance qubit stabilizer code}\end{eczvaluelist}
\eczhbkcontributors{ \eczhuVVA }
\endeczcode

\eczcode{stab_9_3_3}{\(\llbracket 9,3,3\rrbracket \) Quadric code}{~\NoCaseChange{\protect\cite{cite3402,cite1695}}}
\eczhIndexCodeAliasName{stab_9_3_3}{Quadric code}
\codefieldsection{Description}
Nine-qubit \flmRefsHyperref{ref672}{pure} Hermitian qubit code constructed from the almost MDS \([9,3,6]_4\) Hermitian self-orthogonal code.
It is the only \flmRefsHyperref{ref672}{pure} Hermitian code with its parameters and is the highest-distance qubit stabilizer code for its \(n\) and \(k\).

The code can be constructed from the elliptic quadric in \(PG(5, 2)\), or equivalently from the complement of the union of two disjoint hyperovals in \(PG(2, 4)\) \NoCaseChange{\protect\cite{cite1695}}.
A stabilizer tableau for the code is given by \NoCaseChange{\protect\cite[{ID 170235}]{cite453}}
\flmMathEnvironment{align}{}{
\begin{array}{ccccccccc}
  Y & X & X & Y & X & X & I & I & I \\
  Z & Z & X & I & I & I & X & Y & X \\
  Z & I & Z & Z & Z & I & Z & Y & I \\
  I & X & Z & Y & I & Z & I & Z & Z \\
  X & Z & Z & X & Z & Z & I & I & I \\
  X & I & X & X & X & I & X & Z & I
\end{array}~.
}

\codefieldsection{Parents}
\begin{eczvaluelist}
\item\relax
\flmRefsHyperref[eczindexfamilyrel]{code:stabilizer_over_gf4}{Hermitian qubit code} --- The \(\llbracket 9,3,3\rrbracket \) code is Hermitian \NoCaseChange{\protect\cite[{ID 170235}]{cite453}}.
\item\relax
\flmRefsHyperref[eczindexfamilyrel]{code:small_distance_qubit_stabilizer}{Small-distance qubit stabilizer code}\end{eczvaluelist}
\codefieldsection{Cousin}
\begin{eczvaluelist}
\item\relax
\flmRefsHyperref[eczindexfamilyrel]{code:projective}{Projective geometry code} --- The \(\llbracket 9,3,3\rrbracket \) quadric code can be constructed from the elliptic quadric in \(PG(5, 2)\) \NoCaseChange{\protect\cite{cite1695}}.
\end{eczvaluelist}
\eczhbkcontributors{ \eczhuVVA }
\endeczcode

\eczcode{bb90}{\(\llbracket 90,8,10\rrbracket \) BB6 code}{~\NoCaseChange{\protect\cite{cite441}}}
\codefieldsection{Alternative Names}
\begin{eczvaluelist}
\item\relax \((15,3)\) BB6 code
\end{eczvaluelist}
\eczhIndexCodeAliasName{bb90}{BB6 code}
\eczhIndexCodeAliasName{bb90}{\((15,3)\) BB6 code}
\codefieldsection{Description}
A bivariate bicycle (BB) code with parameters \(\llbracket 90,8,10\rrbracket \) and weight-six stabilizer generators \NoCaseChange{\protect\cite{cite441}}.

One defining presentation uses \((\ell,m)=(15,3)\) with \(x^{\ell}=y^{m}=1\), and
\(A=x^9+y+y^2\), \(B=1+x^2+x^7\) in \(\mathbb{F}_2[x,y]/(x^{\ell}-1,y^{m}-1)\) \NoCaseChange{\protect\cite[{Table 3}]{cite441}}.

\codefieldsection{Rate}
Ancilla-added encoding rate is \(2/45\approx 0.044\), using \(n_a=n=90\) ancilla qubits.
\codefieldsection{Parent}
\begin{eczvaluelist}
\item\relax
\flmRefsHyperref[eczindexfamilyrel]{code:qcga}{Bivariate bicycle (BB) code}\end{eczvaluelist}
\codefieldsection{Cousin}
\begin{eczvaluelist}
\item\relax
\flmRefsHyperref[eczindexfamilyrel]{code:balanced_product}{Balanced product (BP) code} --- The \(\llbracket 90,8,10\rrbracket \) BB code can be formulated as a balanced product of two cyclic codes \NoCaseChange{\protect\cite{cite3403}}.
\end{eczvaluelist}
\eczhbkcontributors{ \eczhuVVA }
\endeczcode

\eczcode{quantum_h}{\(\llbracket k+4,k,2\rrbracket \) H code}{~\NoCaseChange{\protect\cite{cite3404}}}
\eczhIndexCodeAliasName{quantum_h}{H code}
\codefieldsection{Description}
Family of \(\llbracket k+4,k,2\rrbracket \) self-dual CSS codes (for even \(k\)) with transversal Hadamard gates that are relevant to magic state distillation.
The four stabilizer generators are \(X_1X_2X_3X_4\), \(Z_1Z_2Z_3Z_4\), \(X_1X_2X_5X_6...X_{k+4}\), and \(Z_1Z_2Z_5Z_6...Z_{k+4}\).

\codefieldsection{Protection}
Detects weight-one Pauli errors. The \(r\)-level concatenated H code detects Pauli errors up to weight \(2^r-1\).
\codefieldsection{Rate}
The H codes are dense, i.e., the rate \(\frac{k}{k+4}\rightarrow 1\) as \(k \rightarrow \infty\). The distance is 2. However an \(r\)-level concatenation of H codes gives a distance of \(2^r\).
\codefieldsection{Magic}
A total of \(r\) rounds of magic-state distillation yields a magic-state yield parameter \(\gamma\to 1^{+}\) as \(k,r\rightarrow \infty\); see \NoCaseChange{\protect\cite[{Box 2}]{cite707}}. This matches the Bravyi-Haah conjectured lower bound \(\gamma \geq 1\) for concatenated triorthogonal-matrix protocols \NoCaseChange{\protect\cite[{Sec. VI}]{cite691}}.
\codefieldsection{Transversal and Permutation-Based Gates}
\begin{eczvaluelist}
\item\relax Hadamard and \(TXT^{\dagger}\) gates, with the latter Clifford-equivalent to Hadamard, and where \(T=\exp(i\pi(I-Z)/8)\) is the \(\pi/8\)-rotation gate.
\end{eczvaluelist}
\codefieldsection{Gates}
\begin{eczvaluelist}
\item\relax The H codes can be used for high-quality and high-efficiency magic-state distillation \NoCaseChange{\protect\cite{cite3404}}. Their associated multi-level magic-state protocols have an efficiency advantage over the 10-to-2 and 15-to-1 protocols for output error below \(10^{-7}\).
\end{eczvaluelist}
\codefieldsection{Parents}
\begin{eczvaluelist}
\item\relax
\flmRefsHyperref[eczindexfamilyrel]{code:generalized_quantum_divisible}{Generalized quantum divisible code} --- H codes are level-two generalized divisible codes \NoCaseChange{\protect\cite[{Sec. VI.C}]{cite734}}.
\item\relax
\flmRefsHyperref[eczindexfamilyrel]{code:self_dual_css}{Self-dual CSS code}\item\relax
\flmRefsHyperref[eczindexfamilyrel]{code:small_distance_qubit_stabilizer}{Small-distance qubit stabilizer code}\end{eczvaluelist}
\codefieldsection{Child}
\begin{eczvaluelist}
\item\relax
\flmRefsHyperref[eczindexfamilyrel]{code:stab_6_2_2}{\(\llbracket 6,2,2\rrbracket \) \(C_6\) code} --- The \(\llbracket k+4,k,2\rrbracket \) H code for \(k=2\) is the \(C_6\) code.
\end{eczvaluelist}
\codefieldsection{Cousin}
\begin{eczvaluelist}
\item\relax
\flmRefsHyperref[eczindexfamilyrel]{code:small_triorthogonal}{\(\llbracket 3k + 8, k, 2\rrbracket \) triorthogonal code} --- The H code \(\llbracket k+4,k,2\rrbracket \) family yields the \(\llbracket 3k + 8, k, 2\rrbracket \) family of triorthogonal codes when level-lifted \NoCaseChange{\protect\cite[{Sec. VI.C}]{cite734}}.
\end{eczvaluelist}
\eczhbkcontributors{ Xiao Xiao, \eczhuVVA }
\endeczcode

\eczcode{quantum_cap}{\(\llbracket n,n-2k,4\rrbracket \) Quantum cap code}{~\NoCaseChange{\protect\cite{cite1695,cite3405,cite62,cite3406,cite3407}}}
\eczhIndexCodeAliasName{quantum_cap}{Quantum cap code}
\codefieldsection{Description}
A distance-four \flmRefsHyperref{ref672}{pure} Hermitian qubit code constructed from a Hermitian self-orthogonal \([n,k]_4\) code associated with an \(n\)-cap in \(PG(k-1,4)\).

\codefieldsection{Rate}
Quantum cap codes can have a high rate and include codes with parameters \(\llbracket 6,0,4\rrbracket \) (from the hyperoval in \(PG(2,4)\)), \(\llbracket 12,4,4\rrbracket \) (from the union of two hyperovals in \(PG(3,4)\) on two planes meeting in an exterior line), \(\llbracket 40,30,4\rrbracket \) (from the 40-cap in \(AG(4,4)\)), \(\llbracket 41,31,4\rrbracket \) (from a 41-cap in \(PG(4,4)\)), \(\llbracket 126,114,4\rrbracket \) (from the Glynn 126-cap in \(PG(5,4)\)), \(\llbracket 756,740,4\rrbracket \) (from a 756-cap in \(PG(7,4)\)), and \(\llbracket 5040,5020,4\rrbracket \) (from a 5040-cap in \(PG(9,4)\)) \NoCaseChange{\protect\cite{cite1695}}, as well as \(\llbracket 12,2,4\rrbracket \), \(\llbracket 20,10,4\rrbracket \), or \(\llbracket 29,19,4\rrbracket \) \NoCaseChange{\protect\cite{cite3406}}.
\codefieldsection{Parents}
\begin{eczvaluelist}
\item\relax
\flmRefsHyperref[eczindexfamilyrel]{code:stabilizer_over_gf4}{Hermitian qubit code} --- A quantum cap code is a distance-four \flmRefsHyperref{ref672}{pure} Hermitian qubit code constructed by identifying its underlying Hermitian self-orthogonal \([n,k]_4\) code with a particular projective cap in \(PG(k-1,4)\).
\item\relax
\flmRefsHyperref[eczindexfamilyrel]{code:small_distance_qubit_stabilizer}{Small-distance qubit stabilizer code} --- Quantum cap codes can have a high rate and include codes with parameters \(\llbracket 6,0,4\rrbracket \), \(\llbracket 12,4,4\rrbracket \), \(\llbracket 40,30,4\rrbracket \), \(\llbracket 41,31,4\rrbracket \), \(\llbracket 126,114,4\rrbracket \), \(\llbracket 756,740,4\rrbracket \), and \(\llbracket 5040,5020,4\rrbracket \) \NoCaseChange{\protect\cite{cite1695}}, as well as \(\llbracket 12,2,4\rrbracket \), \(\llbracket 20,10,4\rrbracket \), or \(\llbracket 29,19,4\rrbracket \) \NoCaseChange{\protect\cite{cite3406}}.
\end{eczvaluelist}
\codefieldsection{Cousin}
\begin{eczvaluelist}
\item\relax
\flmRefsHyperref[eczindexfamilyrel]{code:projective}{Projective geometry code} --- A quantum cap code is a distance-four \flmRefsHyperref{ref672}{pure} Hermitian qubit code constructed by identifying its underlying Hermitian self-orthogonal \([n,k]_4\) code with a particular projective cap in \(PG(k-1,4)\).
\end{eczvaluelist}
\eczhbkcontributors{ \eczhuVVA }
\endeczcode

\eczcode{higher_dimensional_toric}{\(D\)-dimensional twisted toric code}{~\NoCaseChange{\protect\cite{cite3408,cite71,cite3409,cite3410}}}
\codefieldsection{Description}
Extension of the Kitaev toric code to higher-dimensional lattices with regular or shifted (a.k.a. twisted) boundary conditions.
Such boundary conditions yield qubit geometries that are tori \(\mathbb{R}^D/\Lambda\), where \(\Lambda\) is an arbitrary \(D\)-dimensional lattice.
Picking a hypercubic lattice yields the ordinary \(D\)-dimensional toric code.

It is conjectured that appropriate twisted boundary conditions yield multi-dimensional toric code families with sublinear distance scaling of \(n^{1-\epsilon}\) for any \(\epsilon>0\) and logarithmic-weight stabilizer generators \NoCaseChange{\protect\cite{cite3410,cite749}}.
At finite \(n\), twisting boundary conditions can reduce qubit overhead for a fixed distance \NoCaseChange{\protect\cite{cite749}}.

\codefieldsection{Protection}
In two dimensions, different choices for the periodic boundary conditions yield higher-rate codes with parameters \(\llbracket L^2+1,2,L\rrbracket \) for odd \(L\) \NoCaseChange{\protect\cite{cite71}}, and \(\llbracket L^2,2,L\rrbracket \) for even \(L\) \NoCaseChange{\protect\cite{cite3409}}.
Cyclic analogs of toric codes with parameters \(\llbracket t^2+(t+1)^2,1,2t+1\rrbracket \) are constructed in \NoCaseChange{\protect\cite{cite438}}.
Some higher-dimensional toric codes protect against burst errors \NoCaseChange{\protect\cite{cite3411}}.

\codefieldsection{Encoding}
\begin{eczvaluelist}
\item\relax Entangled logical states can be prepared by single-shot techniques \NoCaseChange{\protect\cite{cite749}} for twisted toric codes.
\end{eczvaluelist}
\codefieldsection{Gates}
\begin{eczvaluelist}
\item\relax Higher-dimensional toric codes can admit a cup product structure and can thus have logical gates in the \flmTerm{term}{ref694}{}{Clifford hierarchy} implemented by constant-depth \flmRefsHyperref{ref409}{Clifford circuits} \NoCaseChange{\protect\cite{cite1517}}.
\end{eczvaluelist}
\codefieldsection{Parents}
\begin{eczvaluelist}
\item\relax
\flmRefsHyperref[eczindexfamilyrel]{code:higher_dimensional_surface}{Homological code}\item\relax
\flmRefsHyperref[eczindexfamilyrel]{code:translationally_invariant_stabilizer}{Lattice stabilizer code}\end{eczvaluelist}
\codefieldsection{Child}
\begin{eczvaluelist}
\item\relax
\flmRefsHyperref[eczindexfamilyrel]{code:toric}{Toric code} --- The \(D\)-dimensional twisted toric code reduces to the toric code for \(D=2\) and a square lattice.
\end{eczvaluelist}
\codefieldsection{Cousins}
\begin{eczvaluelist}
\item\relax
\flmRefsHyperref[eczindexfamilyrel]{code:qldpc}{Qubit QLDPC code} --- It is conjectured that appropriate twisted boundary conditions yield multi-dimensional toric code families with sublinear distance scaling of \(N^{1-\epsilon}\) for any \(\epsilon>0\) and logarithmic-weight stabilizer generators \NoCaseChange{\protect\cite{cite3410}}. Assuming this conjecture, Hastings' weight-reduction construction yields QLDPC families with distance \(\Theta^*(N^{1-\epsilon})\) for any \(\epsilon>0\) \NoCaseChange{\protect\cite{cite2989}}.
\item\relax
\flmRefsHyperref[eczindexfamilyrel]{code:multisector_hypergraph}{Higher-dimensional homological product code} --- The non-twisted \(D\)-dimensional planar and toric codes on a hypercubic lattice can be obtained from a hypergraph product of \(D\) repetition codes \NoCaseChange{\protect\cite{cite1613}}.
\end{eczvaluelist}
\eczhbkcontributors{ Marcus P da Silva, \eczhuVVA }
\endeczcode

\eczcode{quantum_k-orthogonal}{\(k\)-orthogonal code}{~\NoCaseChange{\protect\cite{cite673,cite702,cite794}}}
\codefieldsection{Description}
Qubit stabilizer code whose \(X\)-type logicals and generators form a \(k\)-orthogonal matrix (defined below) in the \flmRefsHyperref{ref817}{symplectic representation}.
In other words, the overlap between any \(k\) \(X\)-type code-preserving Paulis (including the identity) is even.
The original definition is for qubit CSS codes \NoCaseChange{\protect\cite{cite673}}, but it can be extended to more general qubit stabilizer codes \NoCaseChange{\protect\cite[{Def. 1}]{cite794}}.
This entry is formulated for qubits, but an extension exists for modular qudits \NoCaseChange{\protect\cite{cite673}}.

A matrix is \(k\)-orthogonal \NoCaseChange{\protect\cite[{Def. 4}]{cite794}} if
\flmMathEnvironment{align}{}{
  |x^1|&\equiv 0 \mod 2 \\
  |x^1\cdot x^2|&\equiv 0 \mod 2 \\
  |x^1\cdot x^2\cdot x^3|&\equiv 0 \mod 2 \\
  &\vdots \\
  |x^1\cdot x^2\cdot x^3\cdot\ldots\cdot x^k|&\equiv 0 \mod 2
}
for all its rows \(x^j\), where the generalized dot-product notation means a sum of products of the respective coordinates of all vectors.

\codefieldsection{Parent}
\begin{eczvaluelist}
\item\relax
\flmRefsHyperref[eczindexfamilyrel]{code:qubit_stabilizer}{Qubit stabilizer code}\end{eczvaluelist}
\codefieldsection{Children}
\begin{eczvaluelist}
\item\relax
\flmRefsHyperref[eczindexfamilyrel]{code:quantum_pin}{Quantum pin code} --- Quantum pin codes are \(\ell\)-orthogonal, i.e., the overlap between any \(\ell\) stabilizers is even \NoCaseChange{\protect\cite{cite702}}.
\item\relax
\flmRefsHyperref[eczindexfamilyrel]{code:quantum_triorthogonal}{Triorthogonal code} --- \(k\)-orthogonal codes reduce to triorthogonal codes for \(k=3\).
\end{eczvaluelist}
\codefieldsection{Cousins}
\begin{eczvaluelist}
\item\relax
\flmRefsHyperref[eczindexfamilyrel]{code:qudit_color}{Modular-qudit lattice color code} --- The notion of \(k\)-orthogonality can be extended to modular-qudit codes and is known as \(k^{\star}\)-orthogonality \NoCaseChange{\protect\cite[{Def. 2}]{cite673}}. Modular-qudit lattice color codes defined on lattices in \(D\) spatial dimension whose \(X\)-type stabilizers are placed on cells of dimension \(\nu \leq D\) are \(k^{\star}\)-orthogonal for all \(k \leq \nu\) \NoCaseChange{\protect\cite[{Lemma 5}]{cite673}}.
\item\relax
\flmRefsHyperref[eczindexfamilyrel]{code:diagonal_clifford}{\(\llbracket 2^r-1,1,3\rrbracket \) simplex code} --- \(\llbracket 2^r-1,1,3\rrbracket \) simplex codes are \((r-1)\)-orthogonal \NoCaseChange{\protect\cite[{Lemma 2}]{cite794}}.
\end{eczvaluelist}
\eczhbkcontributors{ \eczhuVVA }
\endeczcode

\eczcode{2d_bosonization}{2D bosonization code}{~\NoCaseChange{\protect\cite{cite403}}}
\codefieldsection{Description}
A mapping between a 2D lattice quadratic Hamiltonian of Majorana modes and a 2D lattice of qubits.
The original exact 2D bosonization code \NoCaseChange{\protect\cite{cite403}} is a stabilizer code whose generators are products of plaquettes and stars of the surface code, with gauge constraints that project onto a toric-code-like subspace with emergent fermions \NoCaseChange{\protect\cite{cite403,cite404}}.
Finite-depth generalized local unitary Clifford circuits generate a family of equivalent local encodings with qubit-to-fermion ratio \(r = 1 + \frac{1}{2k}\) for any positive integer \(k\); the square-lattice compact encoding with \(r=1.5\) and the super-compact encoding with \(r=1.25\) are explicit examples \NoCaseChange{\protect\cite{cite404}}.

\codefieldsection{Protection}
The original code \NoCaseChange{\protect\cite{cite403}} can be converted via Clifford operations into codes whose distance runs up to \(7\) while preserving the code rate \NoCaseChange{\protect\cite{cite3412}}.

\codefieldsection{Encoding}
\begin{eczvaluelist}
\item\relax Tensor-network realization \NoCaseChange{\protect\cite{cite3413}}, extended to periodic boundary conditions and sectors of odd fermionic charge \NoCaseChange{\protect\cite{cite3414}}.
\end{eczvaluelist}
\codefieldsection{Parents}
\begin{eczvaluelist}
\item\relax
\flmRefsHyperref[eczindexfamilyrel]{code:bosonization}{Bosonization code}\item\relax
\flmRefsHyperref[eczindexfamilyrel]{code:2d_stabilizer}{2D lattice stabilizer code} --- The 2D bosonization code encodes fermionic modes into a 2D qubit stabilizer code.
\end{eczvaluelist}
\codefieldsection{Children}
\begin{eczvaluelist}
\item\relax
\flmRefsHyperref[eczindexfamilyrel]{code:bvc}{Ball-Verstraete-Cirac (BVC) code} --- The BVC code can be obtained from exact 2D bosonization by finite-depth generalized local unitaries after regrouping Majorana modes \NoCaseChange{\protect\cite{cite404}}.
\item\relax
\flmRefsHyperref[eczindexfamilyrel]{code:derby_klassen}{Derby-Klassen (DK) code} --- On the square lattice, the DK code is the \(r=1.5\) exact-bosonization construction after finite-depth generalized local unitaries and re-pairing of Majorana modes \NoCaseChange{\protect\cite{cite404}}.
\item\relax
\flmRefsHyperref[eczindexfamilyrel]{code:mlsc}{Majorana loop stabilizer code (MLSC)} --- The MLSC on a square lattice can be obtained from exact 2D bosonization by finite-depth generalized local unitaries \NoCaseChange{\protect\cite{cite404}}.
\item\relax
\flmRefsHyperref[eczindexfamilyrel]{code:super_compact}{Super-compact fermion-to-qubit code} --- The super-compact code is obtained from exact 2D bosonization by finite-depth generalized local unitaries \NoCaseChange{\protect\cite{cite404}}.
\end{eczvaluelist}
\codefieldsection{Cousins}
\begin{eczvaluelist}
\item\relax
\flmRefsHyperref[eczindexfamilyrel]{code:jw}{Jordan-Wigner transformation code} --- The exact 2D bosonization code can be converted by a linear-depth Clifford circuit into a Jordan-Wigner ordering path on the 2D lattice \NoCaseChange{\protect\cite[{Fig. 24}]{cite404}}.
\item\relax
\flmRefsHyperref[eczindexfamilyrel]{code:surface}{Kitaev surface code} --- The original 2D bosonization code \NoCaseChange{\protect\cite{cite403}} is a stabilizer code whose generators are products of plaquettes and stars of the surface code.
\item\relax
\flmRefsHyperref[eczindexfamilyrel]{code:nonabelian_kitaev_honeycomb}{Non-Abelian Kitaev honeycomb code} --- Embedding each physical qubit into two fermions via the tetron code allows the logical subspace of the Kitaev honeycomb model to be formulated as a joint eigenspace of certain Majorana operators \NoCaseChange{\protect\cite[{Sec. 4.1}]{cite3415}}, which admit braiding-based gates due to their non-Abelian statistics and which can be used for topological quantum computation.
When done in reverse, this embedding can be thought of as a 2D bosonization fermion-into-qubit encoding by converting to a relabeled square lattice and performing single-qubit rotations \NoCaseChange{\protect\cite{cite403}\protect\cite[{Sec. IV.B}]{cite404}}.

\item\relax
\flmRefsHyperref[eczindexfamilyrel]{code:kitaev_honeycomb}{Kitaev honeycomb code} --- Embedding each physical qubit into two fermions via the tetron code is useful for exactly solving the Kitaev honeycomb model Hamiltonian \NoCaseChange{\protect\cite{cite537}} and other qubit Hamiltonians on certain graphs \NoCaseChange{\protect\cite{cite2842,cite2843}}. When done in reverse, this embedding can be thought of as a 2D bosonization fermion-into-qubit encoding by converting to a relabeled square lattice and performing single-qubit rotations \NoCaseChange{\protect\cite{cite403}\protect\cite[{Sec. IV.B}]{cite404}}.
\end{eczvaluelist}
\eczhbkcontributors{ \eczhuVVA }
\endeczcode

\eczcode{2d_color}{2D color code}{~\NoCaseChange{\protect\cite{cite710,cite430}}}
\codefieldsection{Description}
Color code defined on a graph embedded in a two-dimensional surface.
Each face hosts two stabilizer generators, a Pauli-\(X\) and a Pauli-\(Z\) string acting on all the qubits of the face.

Most translation-invariant color codes are defined on trivalent planar graphs with three-colorable faces.
The three admissible uniform tilings are the 6.6.6 (honeycomb) tiling, the 4.8.8 (square octagon) tiling, and the 4.6.12 tiling \NoCaseChange{\protect\cite[{Fig. 1}]{cite432}}.
Non-uniform tilings include the [4.6.8, 6.8.8] and [4.6.8, 4.8.12] tilings \NoCaseChange{\protect\cite{cite402}}.
More general admissible tilings can be obtained via a fattening procedure \NoCaseChange{\protect\cite{cite430}}; see also a construction based on the more general quantum pin codes \NoCaseChange{\protect\cite{cite702}}.

Logical dimension is determined by the genus of the underlying surface (for closed surfaces) and the types of boundaries (for open surfaces).
There are six basic boundary types: three color boundaries corresponding to the three face colors \NoCaseChange{\protect\cite{cite475}} and three Pauli boundaries corresponding to the three Pauli labels \NoCaseChange{\protect\cite{cite445}}.

String operators are defined on paths along edges of the qubit lattice.
These paths can have branching points. Each path has two string operators, one corresponding to the \(X\) basis and one corresponding to the \(Z\) basis.
In correspondence with the coloring of the lattice faces, string operators also come in three colors.
A string of one color must end in a boundary of that same color.

\codefieldsection{Rate}
For general 2D manifolds, \(kd^2 \leq c(\log k)^2 n\) for some constant \(c\) in what can be thought of as an extension of the \flmRefsHyperref{ref487}{BPT bound} to codes on hyperbolic geometries \NoCaseChange{\protect\cite{cite837}}, meaning that color codes with finite rate can only achieve an asymptotic minimum distance that is logarithmic in \(n\).
\codefieldsection{Transversal and Permutation-Based Gates}
\begin{eczvaluelist}
\item\relax CNOT gate because the code is CSS.
\item\relax Hadamard gates for any qubit geometry which yields a self-dual CSS code.
\item\relax Certain triangular 2D color codes on suitably chosen lattices admit transversal implementations of the full Clifford group, including \(H\), \(S^\dagger\), and \(CNOT\), without selective addressing \NoCaseChange{\protect\cite{cite710}}.
\end{eczvaluelist}
\codefieldsection{Gates}
\begin{eczvaluelist}
\item\relax Magic-state distillation protocols \NoCaseChange{\protect\cite{cite3416}}.
\item\relax Non-clifford gates can be implemented via \flmRefsHyperref{ref410}{code switching} \NoCaseChange{\protect\cite{cite3416}}.
\end{eczvaluelist}
\codefieldsection{Decoding}
\begin{eczvaluelist}
\item\relax Projection decoder of \(O(n^4)\) complexity \NoCaseChange{\protect\cite{cite3417}}, modified to account for syndrome errors \NoCaseChange{\protect\cite{cite3418}}.
\item\relax Exact minimum-weight decoding is \(NP\)-hard for 2D color codes, including for solely Pauli-\(Z\) noise \NoCaseChange{\protect\cite{cite3419,cite3420}}.
\item\relax Chromöbius, an open-source implementation of the Möbius decoder, works for many 2D color codes \NoCaseChange{\protect\cite{cite3421}}.
\item\relax Concatenated MPWM decoder \NoCaseChange{\protect\cite{cite3422}}.
\item\relax Syndrome extraction circuits based on superdense coding and a middle-out strategy \NoCaseChange{\protect\cite{cite3421}}.
\end{eczvaluelist}
\codefieldsection{Parents}
\begin{eczvaluelist}
\item\relax
\flmRefsHyperref[eczindexfamilyrel]{code:color}{Color code}\item\relax
\flmRefsHyperref[eczindexfamilyrel]{code:generalized_color}{Generalized 2D color code} --- The generalized color code for \(G=\mathbb{Z}_2\) reduces to the 2D color code.
\item\relax
\flmRefsHyperref[eczindexfamilyrel]{code:twist_defect_color}{Twist-defect color code} --- Twist-defect color codes reduce to 2D color codes when there are no defects. See Ref. \NoCaseChange{\protect\cite{cite3423}} for an alternative non-CSS extension of 2D color codes.
\item\relax
\flmRefsHyperref[eczindexfamilyrel]{code:qudit_color}{Modular-qudit lattice color code} --- Modular-qudit 2D color codes reduce to 2D color codes for \(q=2\).
\item\relax
\flmRefsHyperref[eczindexfamilyrel]{code:galois_color}{Galois-qudit color code} --- Galois-qudit 2D color codes reduce to 2D color codes for \(q=2\).
\item\relax
\flmRefsHyperref[eczindexfamilyrel]{code:quantum_double_abelian}{Abelian quantum-double stabilizer code} --- When treated as ground states of the code Hamiltonian, states of the color code on a torus geometry realize \(\mathbb{Z}_2\times\mathbb{Z}_2\) topological order \NoCaseChange{\protect\cite{cite2846}}, equivalent to the phase realized by two copies of the toric code (i.e., the surface code on a torus) via a local constant-depth \flmRefsHyperref{ref409}{Clifford circuit} \NoCaseChange{\protect\cite{cite422}}.
This process can be viewed as an ungauging \NoCaseChange{\protect\cite{cite462,cite463,cite233,cite464,cite465,cite466,cite467,cite468,cite469,cite470}} of certain symmetries.

\end{eczvaluelist}
\codefieldsection{Children}
\begin{eczvaluelist}
\item\relax
\flmRefsHyperref[eczindexfamilyrel]{code:stab_6_2_2}{\(\llbracket 6,2,2\rrbracket \) \(C_6\) code} --- The \(C_6\) code is a color code on a ladder with three rungs and periodic boundary conditions, (a.k.a. a triangular prism with no top and bottom faces) \NoCaseChange{\protect\cite{cite2339}}. Purely \(Z\)- or \(X\)-type stabilizers lie on the three square faces of the ladder.
\item\relax
\flmRefsHyperref[eczindexfamilyrel]{code:4612_color}{Truncated trihexagonal (4.6.12) color code}\item\relax
\flmRefsHyperref[eczindexfamilyrel]{code:488_color}{Square-octagon (4.8.8) color code}\item\relax
\flmRefsHyperref[eczindexfamilyrel]{code:triangular_color}{Honeycomb (6.6.6) color code}\end{eczvaluelist}
\codefieldsection{Cousins}
\begin{eczvaluelist}
\item\relax
\flmRefsHyperref[eczindexfamilyrel]{code:surface}{Kitaev surface code} --- On closed surfaces, the 2D color code is equivalent to two decoupled copies of the 2D toric/surface code via a local constant-depth \flmRefsHyperref{ref409}{Clifford circuit} \NoCaseChange{\protect\cite{cite3424,cite422,cite3425}} and has the same topological entanglement entropy \NoCaseChange{\protect\cite{cite3426}}. For triangular patches with three differently colored boundaries, it is instead equivalent to a folded surface/toric code with two smooth and two rough boundaries \NoCaseChange{\protect\cite{cite422}}. The conversion process can be viewed as an ungauging \NoCaseChange{\protect\cite{cite462,cite463,cite233,cite464,cite465,cite466,cite467,cite468,cite469,cite470}} of certain symmetries. Conversely, the 2D color code can \flmRefsHyperref{ref410}{condense} to form the 2D surface code in nine different ways, i.e., by adding two-body hopping terms along one of its three triangular directions to the stabilizer group and then taking the center of the resulting nonabelian group \NoCaseChange{\protect\cite{cite2526}}. Both the surface and 2D color codes can be constructed from two distinct types of lattices, namely, 4-valent and 3-valent 3-colorable lattices, respectively \NoCaseChange{\protect\cite{cite3427}}.
\item\relax
\flmRefsHyperref[eczindexfamilyrel]{code:3d_color}{3D color code} --- Gauge fixing can be used to \flmRefsHyperref{ref410}{code switch} between 2D and 3D color codes, thereby yielding fault-tolerant computation with constant time overhead using only local quantum operations \NoCaseChange{\protect\cite{cite3428}}. There is a fault-tolerant measurement-free scheme for \flmRefsHyperref{ref410}{code switching} between 2D and 3D color codes \NoCaseChange{\protect\cite{cite3429}}.
\item\relax
\flmRefsHyperref[eczindexfamilyrel]{code:binary_linear}{Linear binary code} --- As CSS codes, variants of the 2D color code are constructed out of self-dual classical codes on cubic planar graphs \NoCaseChange{\protect\cite{cite1437}}.
\item\relax
\flmRefsHyperref[eczindexfamilyrel]{code:hamiltonian}{Hamiltonian-based code} --- 2D color code Hamiltonians can be simulated, with the help of perturbation theory, by two-dimensional weight-two (two-body) Hamiltonians with non-commuting terms \NoCaseChange{\protect\cite{cite2846}}.
\item\relax
\flmRefsHyperref[eczindexfamilyrel]{code:da_color_2d}{2D DA color code} --- At certain measurement rounds, the 2D DA color code realizes the instantaneous stabilizer group (ISG) of the 2D color code \NoCaseChange{\protect\cite[{Sec. III.A}]{cite2532}}.
\item\relax
\flmRefsHyperref[eczindexfamilyrel]{code:floquet_color}{Floquet color code} --- The parent topological phase of the Floquet color code is the \(\mathbb{Z}_2\times\mathbb{Z}_2\) 2D color-code phase.
\item\relax
\flmRefsHyperref[eczindexfamilyrel]{code:floquet_xyz_ruby}{Ruby Floquet code} --- Each ISG of the color-code schedule is FDLQC-equivalent to the 2D color code, and a parent stabilizer code is FDLQC-equivalent to two copies of the 2D color code \NoCaseChange{\protect\cite{cite533}}.
\item\relax
\flmRefsHyperref[eczindexfamilyrel]{code:floquet_3d_surface}{Floquet 3D surface code} --- A parent stabilizer code for the Floquet 3D surface code is FDQC-equivalent to a 3-foliated stack of 2D color codes \NoCaseChange{\protect\cite{cite533}}.
\item\relax
\flmRefsHyperref[eczindexfamilyrel]{code:floquet_xcube}{X-cube Floquet code} --- A parent stabilizer code for the rewinding X-cube Floquet code is FDQC-equivalent to a 3-foliated stack of 2D color codes \NoCaseChange{\protect\cite{cite533}}.
\item\relax
\flmRefsHyperref[eczindexfamilyrel]{code:majorana_color}{Majorana color code} --- The original Majorana color code is a fermionic analogue of a 2D color code in which one Majorana face operator doubles to matching \(X\)- and \(Z\)-type face checks, but the underlying cylinder graph need only be locally \(3\)-colorable and can support odd boundary logical operators \NoCaseChange{\protect\cite{cite1432}}. Later realizations stack Majorana surface-code layers and replace stacked building blocks with small Majorana fermion codes \NoCaseChange{\protect\cite{cite3212,cite3430,cite3431,cite402}}.
\item\relax
\flmRefsHyperref[eczindexfamilyrel]{code:derby_klassen}{Derby-Klassen (DK) code} --- The DK code on several tilings resembles the 2D color code with some vertex qubits removed \NoCaseChange{\protect\cite{cite3432}}.
\item\relax
\flmRefsHyperref[eczindexfamilyrel]{code:2d_subsystem_color}{2D subsystem color code} --- Gauge fixing relates 2D subsystem color codes to 2D color codes on the same lattice \NoCaseChange{\protect\cite{cite604,cite475}}; the original Union-Jack member is reproduced by the square-octagon-lattice construction of \NoCaseChange{\protect\cite{cite594}}.
\item\relax
\flmRefsHyperref[eczindexfamilyrel]{code:subsystem_three_fermion}{Three-fermion (3F) subsystem code} --- The 2D color code is equivalent to two decoupled copies of the 3F code in the sense that the same anyon theory describes the low-energy excitations of both codes \NoCaseChange{\protect\cite{cite3433}\protect\cite[{Appx. B}]{cite445}}.
\end{eczvaluelist}
\eczhbkcontributors{ Cella Kove, \eczhuVVA }
\endeczcode

\eczcode{da_color_2d}{2D DA color code}{~\NoCaseChange{\protect\cite{cite2532}}}
\codefieldsection{Description}
A 2D dynamical code constructed aperiodically that utilizes measurement sequences to encode logical information with automorphisms of the 2D color code.
The code is assembled from short measurement sequences that can realize all 72 automorphisms of the 2D color code.
On a stack of \(N\) triangular patches with a Pauli boundary, the code encodes \(N\) logical qubits.

The measurement sequence cycles the instantaneous stabilizer group (ISG) through different stabilizer groups, where certain measurement rounds realize the ISG of the 2D color code.
The parent topological phase underlying this dynamical code is realized by two copies of the 6.6.6 (honeycomb) color code \NoCaseChange{\protect\cite[{Sec. III.A}]{cite2532}}.

\codefieldsection{Gates}
\begin{eczvaluelist}
\item\relax On \(N\) layers of triangular patches (which encodes \(N\) logical qubits) with a \textit{Pauli boundary}, the measurement sequences can implement all Clifford logical gates via a sequence of two- and three-qubit Pauli measurements \NoCaseChange{\protect\cite[{Sec. IV.A}]{cite2532}}.
\end{eczvaluelist}
\codefieldsection{Parent}
\begin{eczvaluelist}
\item\relax
\flmRefsHyperref[eczindexfamilyrel]{code:da}{Dynamical code} --- The 2D DA color code is a dynamical code with an aperiodic measurement sequence realizing Clifford logical gates.
\end{eczvaluelist}
\codefieldsection{Cousins}
\begin{eczvaluelist}
\item\relax
\flmRefsHyperref[eczindexfamilyrel]{code:triangular_color}{Honeycomb (6.6.6) color code} --- The parent topological phase of the 2D DA color code is realized by two copies of the 6.6.6 color code \NoCaseChange{\protect\cite[{Sec. III.A}]{cite2532}}.
\item\relax
\flmRefsHyperref[eczindexfamilyrel]{code:2d_color}{2D color code} --- At certain measurement rounds, the 2D DA color code realizes the instantaneous stabilizer group (ISG) of the 2D color code \NoCaseChange{\protect\cite[{Sec. III.A}]{cite2532}}.
\item\relax
\flmRefsHyperref[eczindexfamilyrel]{code:surface}{Kitaev surface code} --- One of the instantaneous stabilizer groups of the 2D DA color code is that of stacks of surface codes \NoCaseChange{\protect\cite[{Sec. III.A}]{cite2532}}.
\end{eczvaluelist}
\eczhbkcontributors{ \eczhuVVA }
\endeczcode

\eczcode{two_dimensional_hyperbolic_surface}{2D hyperbolic surface code}{~\NoCaseChange{\protect\cite{cite3434,cite3435,cite3436}}}
\codefieldsection{Description}
Hyperbolic surface codes based on a tessellation of a closed 2D manifold with a hyperbolic geometry (i.e., non-Euclidean geometry, e.g., saddle surfaces when defined on a 2D plane).

For a tessellation involving regular polygons with \( r \) sides and \( s \) polygons meeting at each vertex, the number of logical qubits is given by \( k = (1-2/r - 2/s) n + 2 \).
Some possible tilings include \( \{r,s\}: \{7,3\}, \{5,4\} \).
The weights of the stabilizer generators depend on the tiling, with \(\{5,4\}\) having lower weight than \(\{7,3\}\).

A \textit{semi-hyperbolic surface code} \NoCaseChange{\protect\cite{cite3437}} is a code defined on a \(\{4,s\}\) tiling, but where each square is replaced with a square region of a 2D lattice.

\codefieldsection{Protection}
Protects against Pauli errors with distance \( d \propto \log(n) \). Code parameters are \( \llbracket n, (1-2/r - 2/s)  n + 2, O(\log n) \rrbracket  \)
\codefieldsection{Rate}
2D hyperbolic surface codes have an asymptotically constant encoding rate \( k/n \) with a distance scaling logarithmically with \( n\) when the surface is closed. The encoding rate depends on the tiling \( {r,s} \) and is given by \( k/n = (1-2/r - 2/s) + 2/n \), which approaches a constant value as the number of physical qubits grows. The weight of the stabilizers is \( r \) for \( Z \)-checks and \( s \) for \( X \)-checks. For open boundary conditions, the code reduces to constant distance.
\codefieldsection{Decoding}
\begin{eczvaluelist}
\item\relax Due to the symmetries of hyperbolic surface codes, optimal measurement schedules of the stabilizers can be found \NoCaseChange{\protect\cite{cite2624}}.
\item\relax Bounds on code capacity thresholds using ML decoding can be obtained by mapping the effect of noise on the code to a statistical mechanical model \NoCaseChange{\protect\cite{cite3438}}.
\item\relax Two flag-based decoders \NoCaseChange{\protect\cite{cite3439}}.
\end{eczvaluelist}
\codefieldsection{Code Capacity Threshold}
\begin{eczvaluelist}
\item\relax Bounds on code capacity thresholds using ML decoding can be obtained by mapping the effect of noise on the code to a statistical mechanical model \NoCaseChange{\protect\cite{cite3440}}.
\item\relax \(1.3\%\) for a phenomenological noise model for the \(\{4,5\}\)-hyperbolic surface code \NoCaseChange{\protect\cite{cite3437}}.
\end{eczvaluelist}
\codefieldsection{Threshold}
\begin{eczvaluelist}
\item\relax 1\(\%\) - 5\(\%\) for a \({5,4}\) tiling under minimum-weight decoding \NoCaseChange{\protect\cite{cite3441}}. For larger tilings, the lower bound on the distance decreases, suggesting the threshold will also decrease.
\end{eczvaluelist}
\codefieldsection{Notes}
\begin{eczvaluelist}
\item\relax See \NoCaseChange{\protect\cite[{Sec. III A}]{cite3442}} for a description of this code.
\item\relax Connection to percolation theory as shown in \NoCaseChange{\protect\cite{cite3443}}.
\item\relax A database of 2D hyperbolic surface codes is available in QECDB \NoCaseChange{\protect\cite{cite781}}, where the surface codes form the subset of non-self-dual codes.
\end{eczvaluelist}
\codefieldsection{Parent}
\begin{eczvaluelist}
\item\relax
\flmRefsHyperref[eczindexfamilyrel]{code:hyperbolic_surface}{Hyperbolic surface code}\end{eczvaluelist}
\codefieldsection{Child}
\begin{eczvaluelist}
\item\relax
\flmRefsHyperref[eczindexfamilyrel]{code:stellated_dodecahedron_css}{\(\llbracket 30,8,3\rrbracket \) Bring code}\end{eczvaluelist}
\codefieldsection{Cousins}
\begin{eczvaluelist}
\item\relax
\flmRefsHyperref[eczindexfamilyrel]{code:asymmetric_qecc}{Asymmetric quantum code (AQC)} --- Asymmetric 2D hyperbolic surface codes have been constructed \NoCaseChange{\protect\cite{cite2637}}.
\item\relax
\flmRefsHyperref[eczindexfamilyrel]{code:subsystem_hyperbolic_surface}{Subsystem hyperbolic surface code} --- Subsystem hyperbolic surface codes are subsystem versions of 2D hyperbolic surface codes.
\end{eczvaluelist}
\eczhbkcontributors{ Elizabeth R. Bennewitz, \eczhuVVA }
\endeczcode

\eczcode{2d_subsystem_color}{2D subsystem color code}{~\NoCaseChange{\protect\cite{cite604}}}
\codefieldsection{Alternative Names}
\begin{eczvaluelist}
\item\relax 2D gauge color code
\end{eczvaluelist}
\eczhIndexCodeAliasName{2d_subsystem_color}{2D gauge color code}
\codefieldsection{Description}
A subsystem version of the 2D color code.
The original topological subsystem-code example is defined on the Union Jack lattice \NoCaseChange{\protect\cite{cite604}}; the square-octagon-lattice hypergraph construction of \NoCaseChange{\protect\cite{cite594}} reproduces the same code from a complementary viewpoint.

\codefieldsection{Protection}
One family of subsystem codes has parameters \(\llbracket 3m,2g,2m+2g-2,d\rrbracket \), where \(m\) is the number of vertices of the original embedded two-colex, \(g\) is the genus of the surface embedding the two-colex, and the distance is bounded from below by the length of the smallest nontrivial homological cycle of the two-colex \(\Gamma\) \NoCaseChange{\protect\cite[{Construction B}]{cite660}\protect\cite[{Lemma 2}]{cite661}}.

\codefieldsection{Gates}
\begin{eczvaluelist}
\item\relax Braiding twist defects \NoCaseChange{\protect\cite{cite712}}.
\end{eczvaluelist}
\codefieldsection{Decoding}
\begin{eczvaluelist}
\item\relax For the Union-Jack/square-octagon member, decoding can be reduced to correcting \(Z\) errors using stabilizers of the topological color code \NoCaseChange{\protect\cite[{Sec. 4.1}]{cite594}}.
\end{eczvaluelist}
\codefieldsection{Code Capacity Threshold}
\begin{eczvaluelist}
\item\relax The threshold under ML decoding for depolarizing noise corresponds to the value of a critical point of a disordered spin model, calculated to be \(5.5(2)\%\) in Ref. \NoCaseChange{\protect\cite{cite3444}}.
\item\relax Erasure noise: \(50\%\) threshold error rate using the optimal erasure decoder \NoCaseChange{\protect\cite{cite3445}}, and \(9.7\%\) and \(44\%\) using gauge-fixing decoders \NoCaseChange{\protect\cite{cite3446,cite3447}}.
\end{eczvaluelist}
\codefieldsection{Parents}
\begin{eczvaluelist}
\item\relax
\flmRefsHyperref[eczindexfamilyrel]{code:subsystem_color}{Subsystem color code}\item\relax
\flmRefsHyperref[eczindexfamilyrel]{code:translationally_invariant_subsystem}{Lattice subsystem code}\end{eczvaluelist}
\codefieldsection{Child}
\begin{eczvaluelist}
\item\relax
\flmRefsHyperref[eczindexfamilyrel]{code:doubled_color}{Doubled color code}\end{eczvaluelist}
\codefieldsection{Cousin}
\begin{eczvaluelist}
\item\relax
\flmRefsHyperref[eczindexfamilyrel]{code:2d_color}{2D color code} --- Gauge fixing relates 2D subsystem color codes to 2D color codes on the same lattice \NoCaseChange{\protect\cite{cite604,cite475}}; the original Union-Jack member is reproduced by the square-octagon-lattice construction of \NoCaseChange{\protect\cite{cite594}}.
\end{eczvaluelist}
\eczhbkcontributors{ \eczhuVVA }
\endeczcode

\eczcode{3d_bacon_shor}{3D Bacon-Shor code}{~\NoCaseChange{\protect\cite{cite3037}}}
\codefieldsection{Description}
Generalization of the Bacon-Shor code to three dimensions that was conjectured to be a self-correcting memory.
It is defined on a cubic lattice and admits sheet-like stabilizer generators.

\codefieldsection{Protection}
On an \(L\times L\times L\) cubic lattice, the symmetric family has parameters \(\llbracket L^3,1,L\rrbracket \) \NoCaseChange{\protect\cite{cite3037}}.

\codefieldsection{Transversal and Permutation-Based Gates}
\begin{eczvaluelist}
\item\relax Logical \(CCZ\) gates on three code blocks of different orientations \NoCaseChange{\protect\cite{cite711}}.
\end{eczvaluelist}
\codefieldsection{Threshold}
\begin{eczvaluelist}
\item\relax The 3D Bacon-Shor code has two entanglement transitions \NoCaseChange{\protect\cite{cite3448}}.
\end{eczvaluelist}
\codefieldsection{Parents}
\begin{eczvaluelist}
\item\relax
\flmRefsHyperref[eczindexfamilyrel]{code:bravyi_bacon_shor}{Bravyi-Bacon-Shor (BBS) code}\item\relax
\flmRefsHyperref[eczindexfamilyrel]{code:translationally_invariant_subsystem}{Lattice subsystem code}\end{eczvaluelist}
\codefieldsection{Cousins}
\begin{eczvaluelist}
\item\relax
\flmRefsHyperref[eczindexfamilyrel]{code:hypercubic}{\(\mathbb{Z}^n\) hypercubic lattice} --- 3D Bacon-Shor codes are defined on a hypercubic lattice.
\item\relax
\flmRefsHyperref[eczindexfamilyrel]{code:self_correct}{Self-correcting quantum code} --- 3D Bacon-Shor codes were conjectured to be self-correcting \NoCaseChange{\protect\cite{cite3037}}, but there remain issues to be resolved in order to validate this conjecture (see \NoCaseChange{\protect\cite[{Sec. IX.B}]{cite1610}}).
\item\relax
\flmRefsHyperref[eczindexfamilyrel]{code:css_plaquette}{CSS-Plaquette code} --- 3D Bacon-Shor and CSS-Plaquette codes admit sheet-like and string-like stabilizer generators, respectively.
\end{eczvaluelist}
\eczhbkcontributors{ \eczhuVVA }
\endeczcode

\eczcode{3d_bosonization}{3D bosonization code}{~\NoCaseChange{\protect\cite{cite3449}}}
\codefieldsection{Description}
A mapping from a 3D lattice quadratic Hamiltonian of Majorana modes to a lattice of qubits which realizes a \(\mathbb{Z}_2\) gauge theory with a particular Gauss law.

\codefieldsection{Parents}
\begin{eczvaluelist}
\item\relax
\flmRefsHyperref[eczindexfamilyrel]{code:bosonization}{Bosonization code}\item\relax
\flmRefsHyperref[eczindexfamilyrel]{code:3d_stabilizer}{3D lattice stabilizer code} --- The 3D bosonization code encodes fermionic modes into a 3D qubit stabilizer code.
\end{eczvaluelist}
\codefieldsection{Cousins}
\begin{eczvaluelist}
\item\relax
\flmRefsHyperref[eczindexfamilyrel]{code:3d_fermionic_surface}{3D fermionic surface code} --- The 3D fermionic surface code is the result of applying the 3D bosonization mapping to a trivial fermionic theory \NoCaseChange{\protect\cite{cite3450}}. Twist defects in the 3D fermionic surface code take the form of Kitaev chains after the mapping \NoCaseChange{\protect\cite{cite3451,cite3450}}.
\item\relax
\flmRefsHyperref[eczindexfamilyrel]{code:three_fermion}{Three-fermion (3F) Walker-Wang model code} --- The 3F Walker-Wang QCA encoder \NoCaseChange{\protect\cite{cite3068,cite3069}} can be simplified using bosonization \NoCaseChange{\protect\cite{cite3452}}.
\end{eczvaluelist}
\eczhbkcontributors{ \eczhuVVA }
\endeczcode

\eczcode{3d_color}{3D color code}{~\NoCaseChange{\protect\cite{cite430}}}
\codefieldsection{Description}
Color code defined on a four-valent, four-colorable 3-colex in a 3-manifold.
In the original colex realization, qubits sit on vertices, \(X\)-type stabilizers are attached to 3-cells, and \(Z\)-type stabilizers are attached to faces \NoCaseChange{\protect\cite{cite430}}.

For a closed 3-manifold, the code encodes \(k=3h_1\) logical qubits, where \(h_1\) is the first Betti number \NoCaseChange{\protect\cite{cite430}}.
Logical operators can be represented by colored strings and colored membranes.
Excitations consist of point-like color charges at cell defects and loop-like color fluxes at face defects; winding a \(p\)-charge around a \(pq\)-flux produces a \(-1\) phase \NoCaseChange{\protect\cite{cite430}}.

There are 101 different types of boundaries for any uniform tiling \NoCaseChange{\protect\cite{cite3453}}; this was shown for the great rhombated cubic honeycomb (a.k.a. cantitruncated cubic honeycomb) uniform tiling, but is valid for general uniform tilings.

\codefieldsection{Protection}
On a closed 3-manifold with first Betti number \(h_1\), the 3D color code encodes \(k=3h_1\) logical qubits \NoCaseChange{\protect\cite{cite430}}.

\codefieldsection{Transversal and Permutation-Based Gates}
\begin{eczvaluelist}
\item\relax Transversal action of \(T\) gates on color codes on general 3-manifolds realizes a \(CCZ\) gate on three logical qubits and is related to a topological invariant that is called the triple intersection number; this gate is related to the fact that this code admits a cup product structure \NoCaseChange{\protect\cite{cite703}}.
\item\relax Transversal \(S\) gate on color codes on general 3-manifolds corresponds to a higher-form symmetry \NoCaseChange{\protect\cite{cite703}}.
\item\relax Universal transversal gates can be achieved using lattice surgery or code deformation \NoCaseChange{\protect\cite{cite712,cite713}}.
\item\relax Families of 3D color codes on quasi-hyperbolic, fibre-bundle, and Torelli mapping-torus 3-manifolds support collective logical \(CCZ\) gates via transversal \(T\) and individually addressable, parallelizable logical \(CZ\) gates via transversal \(S\) on codimension-1 submanifolds. Their rate-distance scalings are \(O(1/\log n)\) with \(d=O(\log n)\), \(O(1/\log^2 n)\) with \(d=\Omega(\log^2 n)\), and \(O(1)\) with distance scaling unknown, respectively \NoCaseChange{\protect\cite{cite703}}.
\end{eczvaluelist}
\codefieldsection{Gates}
\begin{eczvaluelist}
\item\relax Magic-state distillation protocols \NoCaseChange{\protect\cite{cite3416}}.
\item\relax Non-clifford gates can be implemented via \flmRefsHyperref{ref410}{code switching} \NoCaseChange{\protect\cite{cite3416}}.
\end{eczvaluelist}
\codefieldsection{Decoding}
\begin{eczvaluelist}
\item\relax Decoder that maps 3D color code to three copies of the 3D surface code \NoCaseChange{\protect\cite{cite3454}}.
\end{eczvaluelist}
\codefieldsection{Parents}
\begin{eczvaluelist}
\item\relax
\flmRefsHyperref[eczindexfamilyrel]{code:color}{Color code}\item\relax
\flmRefsHyperref[eczindexfamilyrel]{code:3d_stabilizer}{3D lattice stabilizer code}\item\relax
\flmRefsHyperref[eczindexfamilyrel]{code:qudit_color}{Modular-qudit lattice color code} --- Modular-qudit 3D color codes reduce to 3D color codes for \(q=2\).
\item\relax
\flmRefsHyperref[eczindexfamilyrel]{code:topological_abelian}{Abelian topological code}\end{eczvaluelist}
\codefieldsection{Children}
\begin{eczvaluelist}
\item\relax
\flmRefsHyperref[eczindexfamilyrel]{code:stab_8_3_2}{\(\llbracket 8,3,2\rrbracket \) Smallest interesting color code} --- The \(\llbracket 8,3,2\rrbracket \) code is the smallest non-trivial 3D color code.
\item\relax
\flmRefsHyperref[eczindexfamilyrel]{code:cubic_honeycomb_color}{Cubic honeycomb color code}\item\relax
\flmRefsHyperref[eczindexfamilyrel]{code:tetrahedral_color}{Tetrahedral color code}\end{eczvaluelist}
\codefieldsection{Cousins}
\begin{eczvaluelist}
\item\relax
\flmRefsHyperref[eczindexfamilyrel]{code:3d_surface}{3D surface code} --- On closed 3-manifolds, the 3D color code is equivalent to multiple decoupled copies of the 3D surface code via a local constant-depth \flmRefsHyperref{ref409}{Clifford circuit} \NoCaseChange{\protect\cite{cite3424,cite422,cite3425}}. This process can be viewed as an ungauging \NoCaseChange{\protect\cite{cite462,cite463,cite233,cite464,cite465,cite466,cite467,cite468,cite469,cite470}} of certain symmetries. This mapping can also be done via code concatenation \NoCaseChange{\protect\cite{cite715}}. In contrast to the 3D surface/toric code, the original colex Hamiltonian can be viewed as both a string-net condensate and a membrane-net condensate \NoCaseChange{\protect\cite{cite430}}.
\item\relax
\flmRefsHyperref[eczindexfamilyrel]{code:qubit_concatenated}{Concatenated qubit code} --- On closed 3-manifolds, the 3D color code is equivalent to multiple decoupled copies of the 3D surface code via a local constant-depth \flmRefsHyperref{ref409}{Clifford circuit} \NoCaseChange{\protect\cite{cite3424,cite422,cite3425}}. This process can be viewed as an ungauging \NoCaseChange{\protect\cite{cite462,cite463,cite233,cite464,cite465,cite466,cite467,cite468,cite469,cite470}} of certain symmetries. This mapping can also be done via code concatenation \NoCaseChange{\protect\cite{cite715}}.
\item\relax
\flmRefsHyperref[eczindexfamilyrel]{code:xs_stabilizer}{XS stabilizer code} --- The 3D color code on a particular lattice admits XS stabilizers; see \flmHref{https://www.youtube.com/watch?v=B8h5-ANc_-8}{talk by M. Kesselring at the 2020 FTQC conference}.
\item\relax
\flmRefsHyperref[eczindexfamilyrel]{code:spt}{Symmetry-protected topological (SPT) code} --- Transversal action of \(T\) gates on color codes on general 3-manifolds realizes a \(CCZ\) gate on three logical qubits and is related to a topological invariant that is called the triple intersection number \NoCaseChange{\protect\cite{cite703}}. Transversal \(S\) gate on color codes on general 3-manifolds corresponds to a higher-form symmetry \NoCaseChange{\protect\cite{cite703}}.
\item\relax
\flmRefsHyperref[eczindexfamilyrel]{code:da_color_3d}{3D DA color code} --- At certain measurement rounds, the 3D DA color code realizes the instantaneous stabilizer group (ISG) of the 3D color code \NoCaseChange{\protect\cite[{Sec. VI.A}]{cite2532}}.
\item\relax
\flmRefsHyperref[eczindexfamilyrel]{code:brickwork}{Brickwork \(XS\) stabilizer code} --- The brickwork \(XS\) stabilizer code can be obtained from a 3D color code \NoCaseChange{\protect\cite{cite589}}.
\item\relax
\flmRefsHyperref[eczindexfamilyrel]{code:haah_cubic}{Haah cubic code (CC)} --- The 3D color and cubic code families both include 3D codes that do not admit string-like operators.
\item\relax
\flmRefsHyperref[eczindexfamilyrel]{code:2d_color}{2D color code} --- Gauge fixing can be used to \flmRefsHyperref{ref410}{code switch} between 2D and 3D color codes, thereby yielding fault-tolerant computation with constant time overhead using only local quantum operations \NoCaseChange{\protect\cite{cite3428}}. There is a fault-tolerant measurement-free scheme for \flmRefsHyperref{ref410}{code switching} between 2D and 3D color codes \NoCaseChange{\protect\cite{cite3429}}.
\item\relax
\flmRefsHyperref[eczindexfamilyrel]{code:3d_subsystem_color}{3D subsystem color code} --- On a fixed 3D lattice, the 3D subsystem color code is gauge-related to the 3D color code; switching between the \((1,1)\) and \((1,2)\) members yields transversal \(CNOT\), \(H\), and \(R_3\) gates \NoCaseChange{\protect\cite{cite475}}.
\end{eczvaluelist}
\eczhbkcontributors{ Cella Kove, \eczhuVVA }
\endeczcode

\eczcode{da_color_3d}{3D DA color code}{~\NoCaseChange{\protect\cite{cite2532}}}
\codefieldsection{Description}
A 3D dynamical code constructed aperiodically that utilizes measurement sequences to encode logical information with automorphisms of the 3D color code.
The code represents the first step towards universal quantum computation with dynamical automorphism codes.

The measurement sequence cycles the instantaneous stabilizer group (ISG) through different stabilizer groups, where certain measurement rounds realize the ISG of the 3D color code.
The parent topological phase underlying this dynamical code is realized by three copies of the cubic honeycomb color code \NoCaseChange{\protect\cite[{Sec. VI.A}]{cite2532}}.

\codefieldsection{Gates}
\begin{eczvaluelist}
\item\relax A non-Clifford logical gate can be realized via adaptive two-qubit measurements \NoCaseChange{\protect\cite[{Sec. VII}]{cite2532}}.
\end{eczvaluelist}
\codefieldsection{Parent}
\begin{eczvaluelist}
\item\relax
\flmRefsHyperref[eczindexfamilyrel]{code:da}{Dynamical code} --- The 3D DA color code is a dynamical code with an aperiodic measurement sequence realizing a non-Clifford logical gate.
\end{eczvaluelist}
\codefieldsection{Cousins}
\begin{eczvaluelist}
\item\relax
\flmRefsHyperref[eczindexfamilyrel]{code:cubic_honeycomb_color}{Cubic honeycomb color code} --- The parent topological phase of the 3D DA color code is realized by three copies of the cubic honeycomb color code \NoCaseChange{\protect\cite[{Sec. VI.A}]{cite2532}}.
\item\relax
\flmRefsHyperref[eczindexfamilyrel]{code:3d_color}{3D color code} --- At certain measurement rounds, the 3D DA color code realizes the instantaneous stabilizer group (ISG) of the 3D color code \NoCaseChange{\protect\cite[{Sec. VI.A}]{cite2532}}.
\end{eczvaluelist}
\eczhbkcontributors{ \eczhuVVA }
\endeczcode

\eczcode{3d_fermionic_surface}{3D fermionic surface code}{~\NoCaseChange{\protect\cite{cite3455,cite628,cite3456,cite3449}}}
\codefieldsection{Alternative Names}
\begin{eczvaluelist}
\item\relax 3D toric code with emergent fermion
\item\relax Levin-Wen fermion model
\item\relax Fermionic-charge bosonic-loop (FcBl) surface code
\item\relax Twisted surface code
\end{eczvaluelist}
\eczhIndexCodeAliasName{3d_fermionic_surface}{3D toric code with emergent fermion}
\eczhIndexCodeAliasName{3d_fermionic_surface}{Levin-Wen fermion model}
\eczhIndexCodeAliasName{3d_fermionic_surface}{Fermionic-charge bosonic-loop (FcBl) surface code}
\eczhIndexCodeAliasName{3d_fermionic_surface}{Twisted surface code}
\codefieldsection{Description}
A non-CSS variant of the 3D Kitaev surface code that realizes \(\mathbb{Z}_2\) gauge theory with an emergent fermion, i.e., the fermionic-charge bosonic-loop (FcBl) phase \NoCaseChange{\protect\cite{cite455}}.
The model can be defined on a cubic lattice in several ways \NoCaseChange{\protect\cite[{Eq. (D45-46)}]{cite456}}.
Realizations on other lattices also exist \NoCaseChange{\protect\cite{cite457}}, and the phase of this code also exists in the 3D Kitaev honeycomb model \NoCaseChange{\protect\cite{cite458}}.

\textit{3D fermionic toric code} often either refers to the construction on the three-dimensional torus or is an alternative name for the general construction.
The construction on surfaces with boundaries is often called the
\textit{3D fermionic surface code}.
However, unlike the 3D surface code, an open (a.k.a. rough) boundary is not possible.
Twist defects in the form of Kitaev chains can be introduced as in the 2D surface code to store additional logicals \NoCaseChange{\protect\cite{cite3451,cite3450}}.

\codefieldsection{Transversal and Permutation-Based Gates}
\begin{eczvaluelist}
\item\relax \(CCZ\) and \(CS\) gates can be obtained for the fermionic 3D surface code on certain manifolds by circuits that can be interpreted as moving and spreading lattice realizations of Kitaev chain and \(p+ip\) defects \NoCaseChange{\protect\cite{cite714}}.
\end{eczvaluelist}
\codefieldsection{Parents}
\begin{eczvaluelist}
\item\relax
\flmRefsHyperref[eczindexfamilyrel]{code:qldpc}{Qubit QLDPC code}\item\relax
\flmRefsHyperref[eczindexfamilyrel]{code:3d_stabilizer}{3D lattice stabilizer code}\item\relax
\flmRefsHyperref[eczindexfamilyrel]{code:topological_abelian}{Abelian topological code} --- The 3D fermionic surface code realizes 3D \(\mathbb{Z}_2\) gauge theory with fermionic charge and bosonic loop excitations (FcBl), i.e., with an emergent fermion. The fermionic excitations endow the code with an anomalous two-form symmetry, which is argued to induce a non-trivial finite-temperature topological order \NoCaseChange{\protect\cite{cite2528}}.
\item\relax
\flmRefsHyperref[eczindexfamilyrel]{code:walker_wang}{Walker-Wang model code} --- The 3D fermionic surface code is a Walker-Wang model code with premodular input category \(\mathcal{C} = \text{sVec}\) consisting of a trivial anyon and a fermion.
\end{eczvaluelist}
\codefieldsection{Cousins}
\begin{eczvaluelist}
\item\relax
\flmRefsHyperref[eczindexfamilyrel]{code:3d_surface}{3D surface code} --- The 3D (fermionic) surface code is a CSS (non-CSS) code which realizes a \(\mathbb{Z}_2\) gauge theory in 3D (with an emergent fermion). Two copies of the 3D fermionic surface code are equivalent to a copy of the 3D surface code and a copy of the 3D fermionic surface code via anyon relabeling: the two incoming fermions, \(f_1\) and \(f_2\), can be re-organized into a boson \(f_1 f_2\) and fermion \(f_2\).
\item\relax
\flmRefsHyperref[eczindexfamilyrel]{code:kitaev_chain}{Kitaev chain code} --- The 3D fermionic surface code is the result of applying the 3D bosonization mapping to a trivial fermionic theory \NoCaseChange{\protect\cite{cite3450}}. Twist defects in the 3D fermionic surface code take the form of Kitaev chains after the mapping \NoCaseChange{\protect\cite{cite3451,cite3450}}.
\item\relax
\flmRefsHyperref[eczindexfamilyrel]{code:3d_bosonization}{3D bosonization code} --- The 3D fermionic surface code is the result of applying the 3D bosonization mapping to a trivial fermionic theory \NoCaseChange{\protect\cite{cite3450}}. Twist defects in the 3D fermionic surface code take the form of Kitaev chains after the mapping \NoCaseChange{\protect\cite{cite3451,cite3450}}.
\item\relax
\flmRefsHyperref[eczindexfamilyrel]{code:floquet_3d_fermionic_surface}{Floquet 3D fermionic surface code} --- Each ISG of the Floquet 3D fermionic surface code is FDLQC-equivalent to the 3D fermionic surface code \NoCaseChange{\protect\cite{cite533}}.
\item\relax
\flmRefsHyperref[eczindexfamilyrel]{code:3d_kitaev_honeycomb}{3D Kitaev honeycomb code} --- One of the phases realized by the 3D Kitaev honeycomb Hamiltonian is that of the 3D fermionic surface code \NoCaseChange{\protect\cite{cite458}}.
\end{eczvaluelist}
\eczhbkcontributors{ Nathanan Tantivasadakarn, \eczhuVVA }
\endeczcode

\eczcode{3d_kitaev_honeycomb}{3D Kitaev honeycomb code}{~\NoCaseChange{\protect\cite{cite458}}}
\codefieldsection{Description}
3D subsystem stabilizer code whose Hamiltonian is a 3D generalization of the Kitaev honeycomb model.
One of the phases realized by the 3D Kitaev honeycomb Hamiltonian is that of the 3D fermionic surface code \NoCaseChange{\protect\cite{cite458}}.

\codefieldsection{Parents}
\begin{eczvaluelist}
\item\relax
\flmRefsHyperref[eczindexfamilyrel]{code:qubit_subsystem_stabilizer}{Subsystem qubit stabilizer code}\item\relax
\flmRefsHyperref[eczindexfamilyrel]{code:translationally_invariant_subsystem}{Lattice subsystem code}\item\relax
\flmRefsHyperref[eczindexfamilyrel]{code:topological_abelian}{Abelian topological code} --- One of the phases realized by the 3D Kitaev honeycomb Hamiltonian is that of the 3D fermionic surface code \NoCaseChange{\protect\cite{cite458}}.
\end{eczvaluelist}
\codefieldsection{Cousins}
\begin{eczvaluelist}
\item\relax
\flmRefsHyperref[eczindexfamilyrel]{code:3d_fermionic_surface}{3D fermionic surface code} --- One of the phases realized by the 3D Kitaev honeycomb Hamiltonian is that of the 3D fermionic surface code \NoCaseChange{\protect\cite{cite458}}.
\item\relax
\flmRefsHyperref[eczindexfamilyrel]{code:kitaev_honeycomb}{Kitaev honeycomb code} --- The 3D Kitaev honeycomb model is a 3D generalization of the Kitaev honeycomb model.
\item\relax
\flmRefsHyperref[eczindexfamilyrel]{code:floquet_3d_fermionic_surface}{Floquet 3D fermionic surface code} --- The weight-two check operators of the Floquet 3D fermionic surface code are those of the 3D Kitaev honeycomb model \NoCaseChange{\protect\cite{cite458,cite533}}.
\end{eczvaluelist}
\eczhbkcontributors{ \eczhuVVA }
\endeczcode

\eczcode{3d_subsystem_color}{3D subsystem color code}{~\NoCaseChange{\protect\cite{cite475}}}
\codefieldsection{Alternative Names}
\begin{eczvaluelist}
\item\relax 3D gauge color code
\end{eczvaluelist}
\eczhIndexCodeAliasName{3d_subsystem_color}{3D gauge color code}
\codefieldsection{Description}
A subsystem version of the 3D color code defined on a 3-colex.

In the tetrahedral subsystem family introduced in \NoCaseChange{\protect\cite{cite475}}, qubits live on tetrahedra, boundary triangles, boundary edges, and boundary vertices of a colored tetrahedron, and the code encodes one logical qubit.
Gauge generators can be chosen to have weight four or six, while the corresponding stabilizer generators can be much larger.

\codefieldsection{Transversal and Permutation-Based Gates}
\begin{eczvaluelist}
\item\relax For the \((1,1)\) member, \(CNOT\) and Hadamard are transversal; gauge fixing to the \((1,2)\) code enables a transversal \(R_3\) gate, yielding a universal gate set without encoded ancillas \NoCaseChange{\protect\cite{cite475}}.
\end{eczvaluelist}
\codefieldsection{Fault Tolerance}
\begin{eczvaluelist}
\item\relax In the tetrahedral subsystem family, syndrome extraction can use 4- or 6-qubit gauge checks instead of directly measuring larger stabilizers, and gauge fixing uses only local quantum operations plus classical processing \NoCaseChange{\protect\cite{cite475}}.
\end{eczvaluelist}
\codefieldsection{Threshold}
\begin{eczvaluelist}
\item\relax Phenomenological noise: \(0.31\%\) under clustering decoder \NoCaseChange{\protect\cite{cite832}}.
\end{eczvaluelist}
\codefieldsection{Parents}
\begin{eczvaluelist}
\item\relax
\flmRefsHyperref[eczindexfamilyrel]{code:subsystem_color}{Subsystem color code}\item\relax
\flmRefsHyperref[eczindexfamilyrel]{code:translationally_invariant_subsystem}{Lattice subsystem code}\end{eczvaluelist}
\codefieldsection{Cousins}
\begin{eczvaluelist}
\item\relax
\flmRefsHyperref[eczindexfamilyrel]{code:3d_color}{3D color code} --- On a fixed 3D lattice, the 3D subsystem color code is gauge-related to the 3D color code; switching between the \((1,1)\) and \((1,2)\) members yields transversal \(CNOT\), \(H\), and \(R_3\) gates \NoCaseChange{\protect\cite{cite475}}.
\item\relax
\flmRefsHyperref[eczindexfamilyrel]{code:single_shot}{Single-shot code} --- The 3D subsystem color code defined on the cube-truncated rhombic dodecahedral honeycomb, i.e., a tessellation of cubes and chamfered cubes (a.k.a. tetratruncated rhombic dodecahedra) \NoCaseChange{\protect\cite[{Fig. 1}]{cite832}}, is a single-shot code \NoCaseChange{\protect\cite{cite838,cite832}}.
\item\relax
\flmRefsHyperref[eczindexfamilyrel]{code:symmetry_protected_self_correct}{Symmetry-protected self-correcting quantum code} --- A particular gauge-fixed version of a subsystem code on a 3D lattice yields a self-correcting memory protected by one-form symmetries \NoCaseChange{\protect\cite{cite466}\protect\cite[{Sec. IV D}]{cite3059}}.
The symmetric energy barrier grows linearly with the length of a side of the lattice. When the system is coupled locally to a thermal bath respecting the symmetry and below a critical temperature, the memory time grows exponentially with the side length.
The subsystem color code is not a self-correcting quantum memory if symmetry protection is removed \NoCaseChange{\protect\cite{cite3034}}.

\item\relax
\flmRefsHyperref[eczindexfamilyrel]{code:3d_surface}{3D surface code} --- The 3D subsystem color code can be ungauged \NoCaseChange{\protect\cite{cite462,cite463,cite233,cite464,cite465,cite466,cite467,cite468,cite469,cite470}} to obtain six copies of \(\mathbb{Z}_2\) gauge theory with one-form symmetries \NoCaseChange{\protect\cite{cite466}}.
\item\relax
\flmRefsHyperref[eczindexfamilyrel]{code:spt}{Symmetry-protected topological (SPT) code} --- Ungauging \NoCaseChange{\protect\cite{cite462,cite463,cite233,cite464,cite465,cite466,cite467,cite468,cite469,cite470}} different stabilizer Hamiltonians of the 3D subsystem color code yields distinct SPT phases; in particular, one ungauges to three decoupled copies of the RBH model \NoCaseChange{\protect\cite{cite466}}.
\item\relax
\flmRefsHyperref[eczindexfamilyrel]{code:rbh}{Raussendorf-Bravyi-Harrington (RBH) cluster-state code} --- Different stabilizer Hamiltonians of the 3D subsystem color code correspond to distinct SPT phases; one ungauges to three decoupled copies of the RBH model \NoCaseChange{\protect\cite{cite466}}.
The RBH code for a certain boundary Hamiltonian is dual to the 3D subsystem color code \NoCaseChange{\protect\cite[{Sec. IV.C.1}]{cite3059}}.

\end{eczvaluelist}
\eczhbkcontributors{ \eczhuVVA }
\endeczcode

\eczcode{3d_subsystem_surface}{3D subsystem surface code}{~\NoCaseChange{\protect\cite{cite839}}}
\codefieldsection{Alternative Names}
\begin{eczvaluelist}
\item\relax 3D subsystem toric code
\end{eczvaluelist}
\eczhIndexCodeAliasName{3d_subsystem_surface}{3D subsystem toric code}
\codefieldsection{Description}
Subsystem generalization of the surface code on a 3D cubic lattice with gauge-group generators of weight at most three.

\codefieldsection{Parents}
\begin{eczvaluelist}
\item\relax
\flmRefsHyperref[eczindexfamilyrel]{code:subsystem_higher_dimensional_surface}{Subsystem homological code}\item\relax
\flmRefsHyperref[eczindexfamilyrel]{code:translationally_invariant_subsystem}{Lattice subsystem code}\item\relax
\flmRefsHyperref[eczindexfamilyrel]{code:single_shot}{Single-shot code} --- The 3D subsystem surface code is a single-shot code \NoCaseChange{\protect\cite{cite839,cite840}}; see Ref. \NoCaseChange{\protect\cite{cite841}} for an alternative formulation.
\end{eczvaluelist}
\codefieldsection{Cousins}
\begin{eczvaluelist}
\item\relax
\flmRefsHyperref[eczindexfamilyrel]{code:3d_surface}{3D surface code} --- The 3D subsystem surface code is a subsystem version of the 3D surface code.
\item\relax
\flmRefsHyperref[eczindexfamilyrel]{code:quantum_double_abelian}{Abelian quantum-double stabilizer code} --- The 3D subsystem surface code Hamiltonian phase diagram exhibits \(\mathbb{Z}_2\) topological order \NoCaseChange{\protect\cite{cite3034}}.
\item\relax
\flmRefsHyperref[eczindexfamilyrel]{code:self_correct}{Self-correcting quantum code} --- The 3D subsystem surface code is not a self-correcting quantum memory despite being a single-shot code \NoCaseChange{\protect\cite{cite3034}}.
\item\relax
\flmRefsHyperref[eczindexfamilyrel]{code:floquet_3d_surface}{Floquet 3D surface code} --- A planar Floquet 3D surface code stacked with two planar 3D subsystem surface codes prepares an instantaneous state equivalent to a 3D surface code stacked with two checkerboard model codes, enabling a logical \(CCZ\) gate \NoCaseChange{\protect\cite{cite533}}.
\end{eczvaluelist}
\eczhbkcontributors{ \eczhuVVA }
\endeczcode

\eczcode{3d_surface}{3D surface code}{~\NoCaseChange{\protect\cite{cite480,cite3457}}}
\codefieldsection{Alternative Names}
\begin{eczvaluelist}
\item\relax 3D toric code
\item\relax 3D cubic code
\item\relax Bosonic-charge bosonic-loop (BcBl) surface code
\end{eczvaluelist}
\eczhIndexCodeAliasName{3d_surface}{3D toric code}
\eczhIndexCodeAliasName{3d_surface}{3D cubic code}
\eczhIndexCodeAliasName{3d_surface}{Bosonic-charge bosonic-loop (BcBl) surface code}
\codefieldsection{Description}
A generalization of the Kitaev surface code defined on a 3D cubic lattice.
Qubits are placed on edges, \(Z\)-type stabilizer generators are placed on square plaquettes oriented in all three directions, and \(X\)-type stabilizers are placed on the six edges neighboring every vertex \NoCaseChange{\protect\cite{cite459}}.

\textit{3D toric code} often either refers to the construction on
the three-dimensional torus or is an alternative name for the general
construction.
The construction on surfaces with boundaries is often called the
\textit{3D planar code}.
In the open-boundary hypercubic family of \NoCaseChange{\protect\cite{cite3174}}, setting one linear dimension of the tesseract construction to \(1\) yields a single-qubit \textit{cubic code} that interpolates between the planar surface code and the 4D tesseract code.
There exists a rotated version of the 3D surface code, the \textit{3D rotated surface code}, akin to the (2D) rotated surface code \NoCaseChange{\protect\cite{cite2626}}.
The \textit{welded surface code} \NoCaseChange{\protect\cite{cite3033}} consists of several 3D surface codes stitched together in a way that the distance scales faster than the linear size of the system.
In the rectified picture, three 3D surface codes can be supported on the same rectified cubic lattice, and the corresponding cubic-lattice realization is a gauge choice of the 3D Bacon-Shor code \NoCaseChange{\protect\cite{cite715}}.

Related models \NoCaseChange{\protect\cite{cite430,cite3458}} on lattices with certain colorability are equivalent to several copies of the 3D surface code \NoCaseChange{\protect\cite{cite456}}.

\codefieldsection{Protection}
The planar 3D surface code family on a cubic lattice of length \(L\) has parameters \(\llbracket 2L(L-1)^2+L^3,1,d_X=L^2,d_Z=L\rrbracket \), while the 3D toric code has parameters \(\llbracket 3L^3,3,d_X=L^2,d_Z=L\rrbracket \).
Rectangular open-boundary 3D versions furnish single-qubit \(\llbracket 71,1,6\rrbracket \), \(\llbracket 177,1,9\rrbracket \), \(\llbracket 331,1,12\rrbracket \), and \(\llbracket 616,1,16\rrbracket \) codes, illustrating an \(n\propto 3d^2\) tradeoff between the 2D surface and 4D tesseract families \NoCaseChange{\protect\cite{cite3174}}.

Stability against Hamiltonian perturbations was determined using a tensor-network representation \NoCaseChange{\protect\cite{cite2848}}. The phase diagram of the perturbed tensor network maps to that of a 3D Ising gauge theory.

\codefieldsection{Transversal and Permutation-Based Gates}
\begin{eczvaluelist}
\item\relax For a stack of three 3D surface codes on the same rectified cubic lattice, pairwise logical \(CZ\) and triple logical \(CCZ\) gates are transversal \NoCaseChange{\protect\cite{cite715}}.
\end{eczvaluelist}
\codefieldsection{Gates}
\begin{eczvaluelist}
\item\relax There is a CZ gate for the 3D toric code on a Klein bottle \(\times S^1\) \NoCaseChange{\protect\cite{cite3459}}.
\item\relax Lattice surgery \NoCaseChange{\protect\cite{cite715}}.
\item\relax Single-shot lattice surgery for the 3D toric/surface code can be formulated using the fault-complex formalism \NoCaseChange{\protect\cite{cite3176}}.
\item\relax 3D and Hybrid 2D-3D surface code computation using lattice surgery and without magic-state distillation \NoCaseChange{\protect\cite{cite715}}.
\item\relax Fault-tolerant Hadamard gate using teleportation and error correction \NoCaseChange{\protect\cite{cite715}}.
\item\relax Three distance-\(d\) 3D surface/toric codes with open boundaries and cyclically permuted lattice axes admit a logical \(CCZ\) gate via transversal physical \(CCZ\) gates; concatenating each supporting qubit triple with an \(\llbracket 8,3,2\rrbracket \) block yields a \(\llbracket 8n,3,2d\rrbracket \) 3D toric/color family whose smallest member has parameters \(\llbracket 72,3,4\rrbracket \) \NoCaseChange{\protect\cite{cite759}}.
\item\relax Various inter-code \(CZ\) and \(CCZ\) gates implemented via constant-depth circuits on stacks or coupled collections of 3D surface/toric codes \NoCaseChange{\protect\cite{cite422,cite715,cite3460,cite2517,cite703,cite3461,cite3453}}, with \(CZ\) gates formulated in terms of the slant product \NoCaseChange{\protect\cite{cite3066,cite3450}} or cup product \NoCaseChange{\protect\cite{cite1517}} structures.
\end{eczvaluelist}
\codefieldsection{Decoding}
\begin{eczvaluelist}
\item\relax Flip decoder and its modification p-flip \NoCaseChange{\protect\cite{cite3462}}.
\item\relax Tensor-network decoder \NoCaseChange{\protect\cite{cite3463}}.
\item\relax Efficient MWPM decoder for 3D toric and 3D welded surface codes handling string-like syndromes only \NoCaseChange{\protect\cite{cite3464}}.
\item\relax Generalization of linear-time ML erasure decoder \NoCaseChange{\protect\cite{cite3465}} to 3D surface codes \NoCaseChange{\protect\cite{cite3464}}.
\item\relax Equivariant machine learning decoder \NoCaseChange{\protect\cite{cite3466}}.
\end{eczvaluelist}
\codefieldsection{Fault Tolerance}
\begin{eczvaluelist}
\item\relax Fault-tolerant Hadamard gate using teleportation and error correction \NoCaseChange{\protect\cite{cite715}}.
\end{eczvaluelist}
\codefieldsection{Code Capacity Threshold}
\begin{eczvaluelist}
\item\relax Independent \(X,Z\) noise: \(12\%\) for bit-flip and \(3\%\) for phase-flip channels with MWPM decoder for 3D toric code \NoCaseChange{\protect\cite{cite3464}}, and \(17.2\%\) for the surface-like logical operator together with \(3.3\%\) for the line-like logical operator of the 3D cubic code under RG decoding \NoCaseChange{\protect\cite{cite3174}}.
\item\relax Erasure noise: \(24.8\%\) with generalization of linear-time ML erasure decoder \NoCaseChange{\protect\cite{cite3465}} to 3D surface codes \NoCaseChange{\protect\cite{cite3464}}. No threshold was observed for the 3D welded surface code \NoCaseChange{\protect\cite{cite3464}}.
\end{eczvaluelist}
\codefieldsection{Threshold}
\begin{eczvaluelist}
\item\relax Phenomenological noise model for the 3D toric code: \(2.90(2)\%\) under BP-OSD decoder \NoCaseChange{\protect\cite{cite844}}, \(7.1\%\) under improved BP-OSD \NoCaseChange{\protect\cite{cite3179}}, and \(2.6\%\) under flip decoder \NoCaseChange{\protect\cite{cite3462}}. For the line-like logical operator of the 3D cubic code, RG decoding yields \(7.3\%\) \NoCaseChange{\protect\cite{cite3174}}. For 3D surface code: \(3.08(4)\%\) under flip decoder \NoCaseChange{\protect\cite{cite844}}. Optimal thresholds of \(11\%\) under \(X\)-type and \(2\%\) under \(Z\)-type noise derived in Ref. \NoCaseChange{\protect\cite{cite3467}}.
\end{eczvaluelist}
\codefieldsection{Parents}
\begin{eczvaluelist}
\item\relax
\flmRefsHyperref[eczindexfamilyrel]{code:higher_dimensional_surface}{Homological code}\item\relax
\flmRefsHyperref[eczindexfamilyrel]{code:qudit_3d_surface}{Modular-qudit 3D surface code} --- The qudit 3D surface code reduces to the 3D surface code for \(q=2\). The 3D surface code realizes 3D \(\mathbb{Z}_2\) gauge theory with bosonic charge and loop excitations (BcBl). The welded surface code does not satisfy homogeneous topological order \NoCaseChange{\protect\cite{cite3141}}.
\end{eczvaluelist}
\codefieldsection{Cousins}
\begin{eczvaluelist}
\item\relax
\flmRefsHyperref[eczindexfamilyrel]{code:bacon_shor}{Bacon-Shor code} --- In the rectified-cubic construction, the resulting cubic-lattice 3D surface codes are particular gauge choices of the 3D Bacon-Shor code \NoCaseChange{\protect\cite{cite715}}.
\item\relax
\flmRefsHyperref[eczindexfamilyrel]{code:chamon}{Chamon model code} --- The 3D planar and toric code on a cubic lattice can be obtained from a hypergraph product of three repetition codes \NoCaseChange{\protect\cite{cite1613}\protect\cite[{Exam. A.1}]{cite1612}}. The Chamon code is an XYZ product of three repetition codes \NoCaseChange{\protect\cite[{Sec. 3.4}]{cite645}}.
\item\relax
\flmRefsHyperref[eczindexfamilyrel]{code:rotated_surface}{Rotated surface code} --- There exists a rotated version of the 3D surface code, akin to the (2D) rotated surface code \NoCaseChange{\protect\cite{cite2626}}.
\item\relax
\flmRefsHyperref[eczindexfamilyrel]{code:hamiltonian}{Hamiltonian-based code} --- Stability of the 3D surface code against Hamiltonian perturbations was determined using a tensor-network representation \NoCaseChange{\protect\cite{cite2848}}. The phase diagram of the perturbed tensor network maps to that of a 3D Ising gauge theory.
\item\relax
\flmRefsHyperref[eczindexfamilyrel]{code:multisector_hypergraph}{Higher-dimensional homological product code} --- The 3D planar and toric code on a cubic lattice can be obtained from a hypergraph product of three repetition codes \NoCaseChange{\protect\cite{cite1613}\protect\cite[{Exam. A.1}]{cite1612}}.
\item\relax
\flmRefsHyperref[eczindexfamilyrel]{code:repetition}{Repetition code} --- The 3D planar and toric code on a cubic lattice can be obtained from a hypergraph product of three repetition codes \NoCaseChange{\protect\cite{cite1613}\protect\cite[{Exam. A.1}]{cite1612}}.
\item\relax
\flmRefsHyperref[eczindexfamilyrel]{code:self_correct}{Self-correcting quantum code} --- The 3D welded surface code is partially self-correcting with a power-law energy barrier \NoCaseChange{\protect\cite{cite3033}}. The 3D toric code is a classical self-correcting memory, whose protected bit admits a membrane-like logical operator \NoCaseChange{\protect\cite{cite3003}}, but it is not a quantum self-correcting memory because the star terms thermalize \NoCaseChange{\protect\cite{cite3019}}.
\item\relax
\flmRefsHyperref[eczindexfamilyrel]{code:floquet_3d_surface}{Floquet 3D surface code} --- The G-round ISG is FDLQC-equivalent to the 3D surface code, while the other rounds are FDLQC-equivalent to two copies of the 3D surface code up to non-local stabilizers \NoCaseChange{\protect\cite{cite533}}.
\item\relax
\flmRefsHyperref[eczindexfamilyrel]{code:floquet_xcube}{X-cube Floquet code} --- The B- and first R-round ISGs of the rewinding schedule are FDLQC-equivalent to the product of the X-cube model, a 3D surface code, and a 3-foliated stack of 2D surface codes up to non-local stabilizers \NoCaseChange{\protect\cite{cite533}}.
\item\relax
\flmRefsHyperref[eczindexfamilyrel]{code:stab_8_3_2}{\(\llbracket 8,3,2\rrbracket \) Smallest interesting color code} --- Three cyclically rotated copies of the 3D surface/toric code admit a logical \(CCZ\) gate via transversal physical \(CCZ\) gates, and concatenating each such qubit triple with an \(\llbracket 8,3,2\rrbracket \) block yields a 3D toric/color family with parameters \(\llbracket 8n,3,2d\rrbracket \); its smallest member has parameters \(\llbracket 72,3,4\rrbracket \) \NoCaseChange{\protect\cite{cite759}}.
\item\relax
\flmRefsHyperref[eczindexfamilyrel]{code:sierpinsky_fractal_liquid}{Sierpinski prism model code} --- The Sierpinski prism model code admits a topological defect network construction out of 3D surface codes on triangular prisms \NoCaseChange{\protect\cite{cite3163,cite3164}}.
\item\relax
\flmRefsHyperref[eczindexfamilyrel]{code:haah_cubic}{Haah cubic code (CC)} --- The Haah B-code admits a topological defect network construction out of two copies of the 3D surface code \NoCaseChange{\protect\cite{cite3163}}.
\item\relax
\flmRefsHyperref[eczindexfamilyrel]{code:xcube}{X-cube model code} --- The X-cube model admits a topological defect network construction out of 3D surface codes \NoCaseChange{\protect\cite{cite3163}}.
\item\relax
\flmRefsHyperref[eczindexfamilyrel]{code:3d_color}{3D color code} --- On closed 3-manifolds, the 3D color code is equivalent to multiple decoupled copies of the 3D surface code via a local constant-depth \flmRefsHyperref{ref409}{Clifford circuit} \NoCaseChange{\protect\cite{cite3424,cite422,cite3425}}. This process can be viewed as an ungauging \NoCaseChange{\protect\cite{cite462,cite463,cite233,cite464,cite465,cite466,cite467,cite468,cite469,cite470}} of certain symmetries. This mapping can also be done via code concatenation \NoCaseChange{\protect\cite{cite715}}. In contrast to the 3D surface/toric code, the original colex Hamiltonian can be viewed as both a string-net condensate and a membrane-net condensate \NoCaseChange{\protect\cite{cite430}}.
\item\relax
\flmRefsHyperref[eczindexfamilyrel]{code:tetrahedral_color}{Tetrahedral color code} --- A tetrahedral 3D color code with four differently colored boundaries is equivalent, via a local \flmRefsHyperref{ref409}{Clifford circuit}, to three 3D surface codes attached along one boundary, with condensation of a composite electric charge on that attached boundary \NoCaseChange{\protect\cite{cite422}}.
\item\relax
\flmRefsHyperref[eczindexfamilyrel]{code:klein_bottle}{Klein-bottle surface code} --- There is a CZ gate for the 3D toric code on a Klein bottle \(\times S^1\) \NoCaseChange{\protect\cite{cite3459}}.
\item\relax
\flmRefsHyperref[eczindexfamilyrel]{code:3d_fermionic_surface}{3D fermionic surface code} --- The 3D (fermionic) surface code is a CSS (non-CSS) code which realizes a \(\mathbb{Z}_2\) gauge theory in 3D (with an emergent fermion). Two copies of the 3D fermionic surface code are equivalent to a copy of the 3D surface code and a copy of the 3D fermionic surface code via anyon relabeling: the two incoming fermions, \(f_1\) and \(f_2\), can be re-organized into a boson \(f_1 f_2\) and fermion \(f_2\).
\item\relax
\flmRefsHyperref[eczindexfamilyrel]{code:4d_surface}{\((2,2)\) Loop toric code} --- Setting one linear size of the open-boundary tesseract construction to \(1\) yields the cubic/3D surface code \NoCaseChange{\protect\cite{cite3174}}.
\item\relax
\flmRefsHyperref[eczindexfamilyrel]{code:fractal_surface}{Fractal surface code} --- Fractal surface codes are obtained by removing qubits from the 3D surface code on a cubic lattice.
\item\relax
\flmRefsHyperref[eczindexfamilyrel]{code:3d_subsystem_color}{3D subsystem color code} --- The 3D subsystem color code can be ungauged \NoCaseChange{\protect\cite{cite462,cite463,cite233,cite464,cite465,cite466,cite467,cite468,cite469,cite470}} to obtain six copies of \(\mathbb{Z}_2\) gauge theory with one-form symmetries \NoCaseChange{\protect\cite{cite466}}.
\item\relax
\flmRefsHyperref[eczindexfamilyrel]{code:3d_subsystem_surface}{3D subsystem surface code} --- The 3D subsystem surface code is a subsystem version of the 3D surface code.
\end{eczvaluelist}
\eczhbkcontributors{ Nathanan Tantivasadakarn, Aleksander Kubica, \eczhuVVA }
\endeczcode

\eczcode{ampdamp_cws}{Amplitude-damping CWS code}{~\NoCaseChange{\protect\cite{cite1297}}}
\codefieldsection{Description}
Self-complementary CWS code that is designed to detect and correct \flmRefsHyperref{ref498}{AD} errors.

Ref. \NoCaseChange{\protect\cite{cite1297}} constructs such codes by starting from classical self-complementary single-error-correcting codes for the binary asymmetric channel and lifting them to quantum codewords of the form \(|u\rangle+|\overline{u}\rangle\).
In particular, the paper uses nonlinear Constantin-Rao and related Varshamov-Tenengol'ts codes, and also a binary-to-ternary mapping to obtain improved families for short block lengths.

\codefieldsection{Encoding}
\begin{eczvaluelist}
\item\relax Because the resulting codes are of CWS type, encoding circuits can be obtained from the underlying CWS construction \NoCaseChange{\protect\cite{cite1297}}.
\end{eczvaluelist}
\codefieldsection{Decoding}
\begin{eczvaluelist}
\item\relax Because the resulting codes are of CWS type, decoding circuits can be obtained from the underlying CWS construction \NoCaseChange{\protect\cite{cite1297}}.
\end{eczvaluelist}
\codefieldsection{Parents}
\begin{eczvaluelist}
\item\relax
\flmRefsHyperref[eczindexfamilyrel]{code:cws}{Codeword stabilized (CWS) code}\item\relax
\flmRefsHyperref[eczindexfamilyrel]{code:self_complementary}{Self-complementary qubit code}\end{eczvaluelist}
\codefieldsection{Cousin}
\begin{eczvaluelist}
\item\relax
\flmRefsHyperref[eczindexfamilyrel]{code:constantin_rao}{Constantin-Rao (CR) code} --- Amplitude-damping CWS codes can be obtained from CR codes \NoCaseChange{\protect\cite{cite1297}}.
\end{eczvaluelist}
\eczhbkcontributors{ \eczhuVVA }
\endeczcode

\eczcode{anisotropic_z2_laplacian}{Anisotropic \(\mathbb{Z}_2\) Laplacian model code}{~\NoCaseChange{\protect\cite{cite460,cite461}}}
\codefieldsection{Description}
A graph-based analogue of a Type-I fracton phase with lineon-like restricted mobility \NoCaseChange{\protect\cite{cite460,cite461}}.

On generic sparse graphs, the Laplacian seed is not rank deficient enough to generate fracton order \NoCaseChange{\protect\cite{cite1350}}.
Instead, the model exhibits partially confined point-like excitations in a phase more akin to nonlocal topological order.

\codefieldsection{Parents}
\begin{eczvaluelist}
\item\relax
\flmRefsHyperref[eczindexfamilyrel]{code:hypergraph_product}{Hypergraph product (HGP) code} --- The anisotropic \(\mathbb{Z}_2\) Laplacian model is the hypergraph product of a cyclic repetition code and a Laplacian code \NoCaseChange{\protect\cite{cite1350}}.
\item\relax
\flmRefsHyperref[eczindexfamilyrel]{code:fracton}{Fracton stabilizer code} --- The anisotropic \(\mathbb{Z}_2\) Laplacian model code is a graph-based analogue of a Type-I fracton phase with lineon-like restricted mobility.
\end{eczvaluelist}
\codefieldsection{Cousins}
\begin{eczvaluelist}
\item\relax
\flmRefsHyperref[eczindexfamilyrel]{code:repetition}{Repetition code} --- The anisotropic \(\mathbb{Z}_2\) Laplacian model is the hypergraph product of a cyclic repetition code and a Laplacian code \NoCaseChange{\protect\cite{cite1350}}.
\item\relax
\flmRefsHyperref[eczindexfamilyrel]{code:laplacian}{Laplacian code} --- The anisotropic \(\mathbb{Z}_2\) Laplacian model is the hypergraph product of a cyclic repetition code and a Laplacian code \NoCaseChange{\protect\cite{cite1350}}.
\end{eczvaluelist}
\eczhbkcontributors{ \eczhuVVA }
\endeczcode

\eczcode{aqm}{Auxiliary qubit mapping (AQM) code}{~\NoCaseChange{\protect\cite{cite3468,cite3469}}}
\codefieldsection{Description}
A concatenation of the JW transformation code with a qubit stabilizer code.

\codefieldsection{Parents}
\begin{eczvaluelist}
\item\relax
\flmRefsHyperref[eczindexfamilyrel]{code:fermions_into_qubits}{Fermion-into-qubit code}\item\relax
\flmRefsHyperref[eczindexfamilyrel]{code:qubit_concatenated}{Concatenated qubit code}\end{eczvaluelist}
\codefieldsection{Cousin}
\begin{eczvaluelist}
\item\relax
\flmRefsHyperref[eczindexfamilyrel]{code:jw}{Jordan-Wigner transformation code} --- The AQM fermion-into-qubit code reduces to the JW transformation code when the outer code is trivial.
\end{eczvaluelist}
\eczhbkcontributors{ \eczhuVVA }
\endeczcode

\eczcode{bacon_shor}{Bacon-Shor code}{~\NoCaseChange{\protect\cite{cite3,cite3470,cite3037}}}
\codefieldsection{Description}
Subsystem CSS code defined on an \(m_1 \times m_2\) lattice of qubits that generalizes the \(\llbracket 9,1,3\rrbracket \) (subspace) Shor code.
It is said to be \textit{symmetric} when \(m_1=m_2\), and \textit{asymmetric} otherwise.

The \(X\)-type and \(Z\)-type stabilizers are defined as \(X\) and \(Z\) operators acting on all qubits on adjacent columns and rows, respectively. Let \(O_{i,j}\) denote an operator acting on the qubit at a position \((i,j)\) on the lattice, with \(i\in\{0,1,\ldots ,m_1-1\}\) and \(j\in\{0,1,\ldots,m_2-1\}\). The code's stabilizer group is
\flmMathEnvironment{align}{}{
\mathsf{S}=\langle X_{i,*}X_{i+1,*},Z_{*,j}Z_{*,j+1}\rangle~,
}
with generators expressed as products of nearest-neighbor two-qubit gauge operators,
\flmMathEnvironment{align}{}{
\begin{split}
X_{i,*}X_{i+1,*}= \bigotimes_{k=0}^{m_2-1} X_{i,k}X_{i+1,k} \\
Z_{*,j}Z_{*,j+1}=\bigotimes_{k=0}^{m_1-1} Z_{k,j}Z_{k,j+1}~.
\end{split}
}
Syndrome extraction can be done by measuring these gauge operators, which act on fewer qubits and are local.

A Floquet version of the Bacon-Shor code admits a period-four measurement sequence that utilizes its gauge degrees of freedom as defects evolving across measurement rounds. This \textit{Floquet-Bacon-Shor} code saturates the \flmRefsHyperref{ref492}{subsystem BT bound}.
Applying a period-four measurement schedule to the original Bacon-Shor code yields a numerical threshold under circuit-level noise \NoCaseChange{\protect\cite{cite3471}}.

\codefieldsection{Protection}
The \(\llbracket m_1 m_2,1,min(m_1,m_2)\rrbracket \) variant has distance \(d=min(m_1,m_2)\).
In a symmetric 3-dimensional case (defined on a cubic lattice) with \(L^3\) qubits, the code has the parameters \(\llbracket L^3,1,L\rrbracket \).
Bacon-Shor code parameters can be optimized by changing the block geometry, yielding good performance against biased noise \NoCaseChange{\protect\cite{cite2640}}.

\codefieldsection{Rate}
A non-LDPC family of Bacon-Shor codes achieves a distance of \flmRefsHyperref{ref65}{order} \(\Omega(n^{1-\epsilon})\) with sparse gauge operators.
\codefieldsection{Transversal and Permutation-Based Gates}
\begin{eczvaluelist}
\item\relax Logical Hadamard is transversal in symmetric Bacon-Shor codes up to a qubit permutation \NoCaseChange{\protect\cite{cite716}} and can be implemented with teleportation \NoCaseChange{\protect\cite{cite717}}.
\item\relax Bacon-Shor codes on an \(m \times m^k\) lattice admit transversal \(k\)-qubit-controlled \(Z\) gates \NoCaseChange{\protect\cite{cite711}}.
\end{eczvaluelist}
\codefieldsection{Gates}
\begin{eczvaluelist}
\item\relax Pieceably fault-tolerant circuits can be employed to construct non-transversal gates effectively \NoCaseChange{\protect\cite{cite3472}}.
\item\relax Subsystem lattice surgery \NoCaseChange{\protect\cite{cite3473}}.
\item\relax Measurement-free deformation protocol realizing the \(CCZ\) gate \NoCaseChange{\protect\cite{cite3474}}.
\end{eczvaluelist}
\codefieldsection{Decoding}
\begin{eczvaluelist}
\item\relax Majority-voting decoder \NoCaseChange{\protect\cite{cite3225}}.
\item\relax Steane error correction can outperform Shor error correction for this code \NoCaseChange{\protect\cite{cite3475}}.
\item\relax Utilizing the mapping of the effect of the noise to a statistical mechanical model \NoCaseChange{\protect\cite{cite480,cite3476}} yields several copies of the 1D Ising model \NoCaseChange{\protect\cite[{Sec. V.B}]{cite604}}.
\item\relax While check operators are few-body, stabilizer weights scale with the number of qubits, and stabilizer expectation values are obtained by taking products of gauge-operator expectation values. It is thus not clear how to extract stabilizer values in a fault-tolerant manner \NoCaseChange{\protect\cite{cite3477,cite536}}.
\item\relax Autonomous QEC \NoCaseChange{\protect\cite{cite3478}}.
\item\relax Applying a period-four measurement schedule to the original Bacon-Shor code yields a numerical threshold under circuit-level noise \NoCaseChange{\protect\cite{cite3471}}.
\end{eczvaluelist}
\codefieldsection{Fault Tolerance}
\begin{eczvaluelist}
\item\relax Fault-tolerant teleportation-based computation scheme for asymmetric Bacon-Shor codes that is effective against highly biased noise \NoCaseChange{\protect\cite{cite2641}}.
\item\relax Pieceably fault-tolerant circuits can be employed to construct non-transversal gates effectively \NoCaseChange{\protect\cite{cite3472}}.
\end{eczvaluelist}
\codefieldsection{Code Capacity Threshold}
\begin{eczvaluelist}
\item\relax The number of check operators scales sublinearly with system size, so the Bacon-Shor codes alone do not exhibit a topological threshold in the \(m_1,m_2 \to \infty\) limit \NoCaseChange{\protect\cite{cite3479}}. However, a threshold can be obtained from concatenated Bacon-Shor codes that are further restricted to planar geometries, whose recovery circuit is a subset of a circuit used by a larger bona-fide Bacon-Shor code \NoCaseChange{\protect\cite{cite3480}}. This threshold differs from a \flmRefsHyperref{ref515}{concatenated threshold} in that there are no long-range connectivity requirements.
\item\relax Lower bounds for the \flmRefsHyperref{ref515}{concatenated threshold} of various small Bacon-Shor codes are tabulated in \NoCaseChange{\protect\cite[{Table I}]{cite716}}.
\end{eczvaluelist}
\codefieldsection{Threshold}
\begin{eczvaluelist}
\item\relax The Bacon-Shor code has a \flmRefsHyperref{ref3210}{measurement threshold} of zero \NoCaseChange{\protect\cite{cite3211}}.
\item\relax Applying a period-four measurement schedule to the original Bacon-Shor code yields a numerical threshold under circuit-level noise \NoCaseChange{\protect\cite{cite3471}}.
\end{eczvaluelist}
\codefieldsection{Realizations}
\begin{eczvaluelist}
\item\relax Superconducting qubits: The Floquet-Bacon-Shor code has been realized on a 3-by-3 lattice \NoCaseChange{\protect\cite{cite3481}}.
\end{eczvaluelist}
\codefieldsection{Notes}
\begin{eczvaluelist}
\item\relax See \NoCaseChange{\protect\cite[{Sec. III.C1}]{cite2967}} for an exposition.
\end{eczvaluelist}
\codefieldsection{Parents}
\begin{eczvaluelist}
\item\relax
\flmRefsHyperref[eczindexfamilyrel]{code:bravyi_bacon_shor}{Bravyi-Bacon-Shor (BBS) code}\item\relax
\flmRefsHyperref[eczindexfamilyrel]{code:subsystem_quantum_parity}{Subsystem hypergraph product (SHP) code}\item\relax
\flmRefsHyperref[eczindexfamilyrel]{code:compass_model}{Compass code} --- A compass code on a fully non-colored lattice reduces to the Bacon-Shor code.
\end{eczvaluelist}
\codefieldsection{Children}
\begin{eczvaluelist}
\item\relax
\flmRefsHyperref[eczindexfamilyrel]{code:bacon_shor_4}{\(\llbracket 4,1,1,2\rrbracket \) Four-qubit subsystem code} --- The four-qubit subsystem code is the shortest error-detecting Bacon-Shor code.
\item\relax
\flmRefsHyperref[eczindexfamilyrel]{code:bacon_shor_9}{\(\llbracket 9,1,4,3\rrbracket \) Nine-qubit Bacon-Shor code} --- The nine-qubit Bacon-Shor code is the shortest error-correcting Bacon-Shor code.
\end{eczvaluelist}
\codefieldsection{Cousins}
\begin{eczvaluelist}
\item\relax
\flmRefsHyperref[eczindexfamilyrel]{code:hamiltonian}{Hamiltonian-based code} --- The 2D Bacon-Shor gauge-group Hamiltonian is the compass model \NoCaseChange{\protect\cite{cite656,cite657,cite658}}.
\item\relax
\flmRefsHyperref[eczindexfamilyrel]{code:floquet}{Hastings-Haah Floquet code} --- A Floquet version of the Bacon-Shor code admits a period-four measurement sequence that utilizes its gauge degrees of freedom as defects evolving across measurement rounds. This \textit{Floquet-Bacon-Shor} code saturates the \flmRefsHyperref{ref492}{subsystem BT bound}. Applying a period-four measurement schedule to the original Bacon-Shor code yields a numerical threshold under circuit-level noise \NoCaseChange{\protect\cite{cite3471}}.
\item\relax
\flmRefsHyperref[eczindexfamilyrel]{code:quantum_lego}{Tensor-network code} --- The 2D Bacon-Shor code can also be obtained from a surface-code tensor network by reassigning every other row of dangling physical legs to logical legs; in this quantum-Lego picture, the gauge generators remain weight-two \(XX\) and \(ZZ\) operators and the construction makes explicit a connection to the quantum compass model \NoCaseChange{\protect\cite{cite2868}}.
\item\relax
\flmRefsHyperref[eczindexfamilyrel]{code:surface}{Kitaev surface code} --- The 2D Bacon-Shor code can also be obtained from a surface-code tensor network by reassigning every other row of dangling physical legs to logical legs; in this quantum-Lego picture, the gauge generators remain weight-two \(XX\) and \(ZZ\) operators and the construction makes explicit a connection to the quantum compass model \NoCaseChange{\protect\cite{cite2868}}.
\item\relax
\flmRefsHyperref[eczindexfamilyrel]{code:plaquette_ising}{Plaquette Ising code} --- Ungauging \NoCaseChange{\protect\cite{cite462,cite463,cite233,cite464,cite465,cite466,cite467,cite468,cite469,cite470}} the \(Z\)-type symmetries of the 2D Bacon-Shor gauge Hamiltonian yields the Xu-Moore model, with emergent horizontal and vertical rigid \(X\)-type symmetries \NoCaseChange{\protect\cite{cite466}}.
\item\relax
\flmRefsHyperref[eczindexfamilyrel]{code:asymmetric_qecc}{Asymmetric quantum code (AQC)} --- Bacon-Shor code parameters against bit- and phase-noise can be optimized by changing the block geometry, yielding good performance against biased noise \NoCaseChange{\protect\cite{cite2640}}. A fault-tolerant teleportation-based computation scheme for asymmetric Bacon-Shor codes is effective against highly biased noise \NoCaseChange{\protect\cite{cite2641}}.
\item\relax
\flmRefsHyperref[eczindexfamilyrel]{code:majorana_subsystem}{Majorana subsystem stabilizer code} --- Bacon-Shor codes can be fermionized into fermionic subsystem codes with two-body terms \NoCaseChange{\protect\cite{cite3482}}.
\item\relax
\flmRefsHyperref[eczindexfamilyrel]{code:hybrid_bacon_shor}{OA Bacon-Shor code} --- Hybrid Bacon-Shor codes are obtained from Bacon-Shor subsystem codes either by gauge fixing gauge qubits into classical registers \NoCaseChange{\protect\cite{cite2874}} or by promoting a nontrivial subset of normalizer cosets to classical sectors \NoCaseChange{\protect\cite{cite3483}}.
\item\relax
\flmRefsHyperref[eczindexfamilyrel]{code:gnu_permutation_invariant}{GNU PI code} --- GNU codes of length \((2t+1)^2\) result from projecting Bacon-Shor codes into the PI qubit subspace \NoCaseChange{\protect\cite{cite2944}}.
\item\relax
\flmRefsHyperref[eczindexfamilyrel]{code:quantum_parity}{Quantum parity code (QPC)} --- Bacon-Shor codes reduce to QPCs when all \(X\)-type gauge generators are fixed \NoCaseChange{\protect\cite[{pg. 6}]{cite2650}}.
\item\relax
\flmRefsHyperref[eczindexfamilyrel]{code:3d_surface}{3D surface code} --- In the rectified-cubic construction, the resulting cubic-lattice 3D surface codes are particular gauge choices of the 3D Bacon-Shor code \NoCaseChange{\protect\cite{cite715}}.
\item\relax
\flmRefsHyperref[eczindexfamilyrel]{code:heavy_hex}{Heavy-hexagon code} --- Bacon-Shor stabilizers are used to measure the X-type stabilizers of the code.
\end{eczvaluelist}
\eczhbkcontributors{ Mazin Karjikar, Srilekha Gandhari, \eczhuVVA }
\endeczcode

\eczcode{ball_color}{Ball code}{~\NoCaseChange{\protect\cite{cite687}}}
\codefieldsection{Description}
A distance-two color code defined on a colorable \(D\)-ball, equivalently on a \(D\)-colex with boundary \NoCaseChange{\protect\cite[{Appx. A}]{cite687}}.
In the morphing construction of Ref. \NoCaseChange{\protect\cite{cite687}}, ball codes arise as the child codes associated with the morphed ball-like regions.
This family includes hypercube codes (defined on balls constructed from hyperoctahedra) and 3D ball codes (defined on duals of certain Archimedean solids).

\codefieldsection{Protection}
A \(D\)-dimensional ball code defined on a ball-like region \(B_v\) has parameters \(\llbracket |B_v^D|,|B_v^1|-D,2\rrbracket \) \NoCaseChange{\protect\cite[{Lemma 2}]{cite687}}.
The hypercube code family has parameters \(\llbracket 2^D,D,2\rrbracket \) \NoCaseChange{\protect\cite[{Exam. 3}]{cite687}}.
3D ball codes on duals of the truncated octahedron, truncated cuboctahedron, and truncated icosidodecahedron have parameters \(\llbracket 24,11,2\rrbracket \), \(\llbracket 48,23,2\rrbracket \), and \(\llbracket 120,59,2\rrbracket \), respectively \NoCaseChange{\protect\cite[{Exam. 4}]{cite687}}.

\codefieldsection{Magic}
The 3D ball codes on duals of the truncated octahedron, truncated cuboctahedron, and truncated icosidodecahedron have \(\gamma\) close to one \NoCaseChange{\protect\cite{cite687}}.
\codefieldsection{Transversal and Permutation-Based Gates}
\begin{eczvaluelist}
\item\relax The 3D ball codes on duals of the truncated octahedron, truncated cuboctahedron, and truncated icosidodecahedron admit logical \(CCZ\)-type gates implemented by physical \(T\) and \(T^\dagger\) gates \NoCaseChange{\protect\cite[{Exam. 4, Thm. 5}]{cite687}}.
\end{eczvaluelist}
\codefieldsection{Parents}
\begin{eczvaluelist}
\item\relax
\flmRefsHyperref[eczindexfamilyrel]{code:color}{Color code} --- Ball codes are color codes defined on a \(D\)-dimensional colex \NoCaseChange{\protect\cite[{Appx. A}]{cite687}}.
\item\relax
\flmRefsHyperref[eczindexfamilyrel]{code:small_distance_qubit_stabilizer}{Small-distance qubit stabilizer code}\end{eczvaluelist}
\codefieldsection{Children}
\begin{eczvaluelist}
\item\relax
\flmRefsHyperref[eczindexfamilyrel]{code:hypercube_quantum}{\(\llbracket 2^D,D,2\rrbracket \) hypercube quantum code} --- \(\llbracket 2^D,D,2\rrbracket \) hypercube quantum codes can be thought of as small ball codes constructed from hyperoctahedra \NoCaseChange{\protect\cite[{Exam. 3}]{cite687}}, or on lattices with no bulk qubits and cubic boundaries  \NoCaseChange{\protect\cite{cite422,cite797}}.
\item\relax
\flmRefsHyperref[eczindexfamilyrel]{code:iceberg}{\(\llbracket 2m,2m-2,2\rrbracket \) error-detecting code} --- The \(\llbracket 2m,2m-2,2\rrbracket \) error-detecting code is a ball color code \NoCaseChange{\protect\cite[{Sec. III.A}]{cite687}}.
\end{eczvaluelist}
\codefieldsection{Cousin}
\begin{eczvaluelist}
\item\relax
\flmRefsHyperref[eczindexfamilyrel]{code:polyhedron}{Polyhedron code} --- Polytopes dual to the hyperoctahedron, truncated octahedron, truncated cuboctahedron, and truncated icosidodecahedron are used to construct 3D ball codes.
\end{eczvaluelist}
\eczhbkcontributors{ \eczhuVVA }
\endeczcode

\eczcode{bvc}{Ball-Verstraete-Cirac (BVC) code}{~\NoCaseChange{\protect\cite{cite3484,cite3485}}}
\codefieldsection{Alternative Names}
\begin{eczvaluelist}
\item\relax Verstraete-Cirac code
\item\relax Auxiliary fermion code
\end{eczvaluelist}
\eczhIndexCodeAliasName{bvc}{Verstraete-Cirac code}
\eczhIndexCodeAliasName{bvc}{Auxiliary fermion code}
\codefieldsection{Description}
A 2D fermion-into-qubit encoding that builds upon the JW transformation by eliminating the weight-\(O(n)\) non-local \(Z\)-type string at the expense of introducing an auxiliary qubit per site and local gauge constraints.
See \NoCaseChange{\protect\cite[{Sec. IV.B}]{cite404}} for details.

\codefieldsection{Protection}
Some single-qubit errors are detectable, with the rest inducing low-weight fermionic dephasing noise \NoCaseChange{\protect\cite{cite3486}}.

\codefieldsection{Parents}
\begin{eczvaluelist}
\item\relax
\flmRefsHyperref[eczindexfamilyrel]{code:2d_bosonization}{2D bosonization code} --- The BVC code can be obtained from exact 2D bosonization by finite-depth generalized local unitaries after regrouping Majorana modes \NoCaseChange{\protect\cite{cite404}}.
\item\relax
\flmRefsHyperref[eczindexfamilyrel]{code:qetc}{Quantum error-transmuting code (QETC)} --- The BVC code transmutes all single-qubit errors \NoCaseChange{\protect\cite{cite2985}}.
\end{eczvaluelist}
\codefieldsection{Cousin}
\begin{eczvaluelist}
\item\relax
\flmRefsHyperref[eczindexfamilyrel]{code:rotated_surface}{Rotated surface code} --- An appropriately chosen stabilizer generator set for the BVC code contains the stabilizers of the rotated surface code \NoCaseChange{\protect\cite{cite3432}}.
\end{eczvaluelist}
\eczhbkcontributors{ \eczhuVVA }
\endeczcode

\eczcode{bb5}{BB5 code}{~\NoCaseChange{\protect\cite{cite407}}}
\codefieldsection{Description}
A BB code with weight-five stabilizer generators
(contrasting with the weight-six checks of standard BB codes), designed and
benchmarked for long chains of trapped ions \NoCaseChange{\protect\cite{cite407}}.

Two highlighted instances are \(\llbracket 30,4,5\rrbracket \) and \(\llbracket 48,4,7\rrbracket \), which improve
the best-known BB6 distances at the same \(\llbracket n,k\rrbracket \): respectively
\(\llbracket 30,4,4\rrbracket \) and \(\llbracket 48,4,6\rrbracket \) \NoCaseChange{\protect\cite{cite407}}.

\codefieldsection{Fault Tolerance}
\begin{eczvaluelist}
\item\relax Circuit-level simulations in \NoCaseChange{\protect\cite{cite407}} use BP-OSD for BB5 and BB6 instances under an ion-chain noise model.
\item\relax For physical error rate \(10^{-3}\), the \(\llbracket 48,4,7\rrbracket \) BB5 instance achieves logical error rate per syndrome round and per logical qubit \(\approx 5\times 10^{-5}\), about \(4\times\) lower than the best BB6 baseline considered in \NoCaseChange{\protect\cite{cite407}}. In that comparison, it also matches the logical error rate of a distance-7 surface code while using about \(4\times\) fewer physical qubits per logical qubit.
\end{eczvaluelist}
\codefieldsection{Parent}
\begin{eczvaluelist}
\item\relax
\flmRefsHyperref[eczindexfamilyrel]{code:qcga}{Bivariate bicycle (BB) code}\end{eczvaluelist}
\eczhbkcontributors{ \eczhuVVA }
\endeczcode

\eczcode{bicycle}{Bicycle code}{~\NoCaseChange{\protect\cite{cite682}}}
\codefieldsection{Description}
A CSS code whose stabilizer generator matrix blocks are \(H_{X}=H_{Z}=(A|A^T)\), where \(A\) is a circulant matrix.
The fact that \(A\) commutes with its transpose ensures that the CSS condition is satisfied.
Bicycle codes are the first QLDPC codes.

A notable example is an \(\llbracket 2^n,2^{(n+1)/2},2^{(n-1)/2}\rrbracket \) code constructed from the repetition code and the Cayley graph of \(\mathbb{Z}_2^n\) \NoCaseChange{\protect\cite{cite3487}}.

\codefieldsection{Parents}
\begin{eczvaluelist}
\item\relax
\flmRefsHyperref[eczindexfamilyrel]{code:qubit_generalized_homological_product_css}{Generalized homological-product qubit CSS code}\item\relax
\flmRefsHyperref[eczindexfamilyrel]{code:generalized_bicycle}{Generalized bicycle (GB) code} --- A GB code whose circulants satisfy \(B = A^T\) reduces to a bicycle code.

\end{eczvaluelist}
\codefieldsection{Cousin}
\begin{eczvaluelist}
\item\relax
\flmRefsHyperref[eczindexfamilyrel]{code:qldpc}{Qubit QLDPC code} --- Bicycle codes are the first QLDPC codes \NoCaseChange{\protect\cite{cite682}}.
\end{eczvaluelist}
\eczhbkcontributors{ \eczhuVVA }
\endeczcode

\eczcode{bc_phantom}{Binarized-and-concatenated (B\&C) phantom code}{~\NoCaseChange{\protect\cite{cite514}}}
\codefieldsection{Description}
Member of a family of \(k=2\) CSS phantom codes obtained from a \(q=4\) Galois-qudit CSS code by binarizing each \(\mathbb{F}_4\) qudit into two qubits and then concatenating each qubit pair with the \(\llbracket 4,2,2\rrbracket \) code \NoCaseChange{\protect\cite{cite514}}.

The construction starts from a \(\llbracket n,1,d\rrbracket _4\) CSS code satisfying a condition that realizes the Frobenius transform \(\gamma\mapsto\gamma^2\) by a coordinate permutation; together with field multiplication \(\gamma\mapsto\alpha\gamma\), this supplies all \(\mathrm{GL}(2,\mathbb{F}_2)\) transformations on the two binary components of the logical \(\mathbb{F}_4\) qudit after concatenation.
Binarization uses the self-dual normal basis \(\{\omega,\omega^2\}\) of \(\mathbb{F}_4\).
The subsequent \(\llbracket 4,2,2\rrbracket \) layer maps a single \(\mathbb{F}_4\)-qudit Pauli to four-qubit Paulis, e.g.,
\(X^\omega\mapsto XXII\), \(X^{\omega^2}\mapsto XIXI\), \(X^1\mapsto IXXI\), and similarly \(Z^{\omega^2}\mapsto IIZZ\), \(Z^\omega\mapsto IZIZ\), \(Z^1\mapsto IZZI\) \NoCaseChange{\protect\cite{cite514}}.

\codefieldsection{Protection}
A starting \(\llbracket n,1,d\rrbracket _4\) CSS code yields a qubit CSS code with parameters \(\llbracket 4n,2,\geq 2d\rrbracket \) after binarization and concatenation with the \(\llbracket 4,2,2\rrbracket \) code \NoCaseChange{\protect\cite{cite514}}.
The inner \(\llbracket 4,2,2\rrbracket \) blocks contribute many weight-four stabilizers, while completing the stabilizer group may require generators whose weight is at least the code distance.

\codefieldsection{Transversal and Permutation-Based Gates}
\begin{eczvaluelist}
\item\relax Ordinary CSS self-duality of the starting \(q=4\) Galois-qudit CSS code yields a permutation-assisted logical Hadamard on the resulting qubit code. Hermitian self-duality, meaning that the \(Z\)-type check space is the Frobenius conjugate of the \(X\)-type check space, additionally yields a logical \(CZ\) gate \NoCaseChange{\protect\cite{cite514}}.
\end{eczvaluelist}
\codefieldsection{Parents}
\begin{eczvaluelist}
\item\relax
\flmRefsHyperref[eczindexfamilyrel]{code:phantom}{Phantom code}\item\relax
\flmRefsHyperref[eczindexfamilyrel]{code:qubit_concatenated}{Concatenated qubit code} --- Each qubit pair obtained by binarizing one \(\mathbb{F}_4\) qudit is concatenated with the \(\llbracket 4,2,2\rrbracket \) code \NoCaseChange{\protect\cite{cite514}}.
\end{eczvaluelist}
\codefieldsection{Children}
\begin{eczvaluelist}
\item\relax
\flmRefsHyperref[eczindexfamilyrel]{code:carbon}{\(\llbracket 12,2,4\rrbracket \) carbon code} --- The carbon code is the B\&C phantom code obtained from the \(\llbracket 3,1,2\rrbracket _4\) Galois-qudit code \NoCaseChange{\protect\cite{cite514}}.
\item\relax
\flmRefsHyperref[eczindexfamilyrel]{code:stab_20_2_6}{\(\llbracket 20,2,6\rrbracket \) B\&C phantom code} --- The \(\llbracket 20,2,6\rrbracket \) code is the B\&C phantom code obtained from the \(\llbracket 5,1,3\rrbracket _4\) Galois-qudit CSS code \NoCaseChange{\protect\cite{cite514,cite795}}.
\end{eczvaluelist}
\codefieldsection{Cousins}
\begin{eczvaluelist}
\item\relax
\flmRefsHyperref[eczindexfamilyrel]{code:galois_quad_residue}{Quantum quadratic-residue (QR) code} --- A concrete B\&C phantom family starts from CSS quantum QR codes over \(\mathbb{F}_4\) \NoCaseChange{\protect\cite{cite514}}.
\item\relax
\flmRefsHyperref[eczindexfamilyrel]{code:galois_3_1_2}{\(\llbracket 3,1,2\rrbracket _4\) three-Galois-quartrit code} --- Binarizing the \(\llbracket 3,1,2\rrbracket _4\) Galois-qudit CSS code and concatenating each qubit pair with the \(\llbracket 4,2,2\rrbracket \) code yields the \(\llbracket 12,2,4\rrbracket \) carbon code \NoCaseChange{\protect\cite{cite514}}.
\item\relax
\flmRefsHyperref[eczindexfamilyrel]{code:stab_4_2_2}{\(\llbracket 4,2,2\rrbracket \) Four-qubit code} --- The \(\llbracket 4,2,2\rrbracket \) code is used as the inner code for each binarized \(\mathbb{F}_4\)-qudit pair \NoCaseChange{\protect\cite{cite514}}.
\item\relax
\flmRefsHyperref[eczindexfamilyrel]{code:stab_10_2_3}{\(\llbracket 10,2,3\rrbracket \) binarized Galois-qudit code} --- The binarized \(\llbracket 10,2,3\rrbracket \) code is not phantom, but concatenating each qubit pair with the \(\llbracket 4,2,2\rrbracket \) code yields a \(\llbracket 20,2,6\rrbracket \) B\&C phantom code admitting fold-diagonal logical \(SS\) gates \NoCaseChange{\protect\cite{cite514}}.
\end{eczvaluelist}
\eczhbkcontributors{ \eczhuVVA }
\endeczcode

\eczcode{binary_dihedral_permutation_invariant}{Binary dihedral PI code}{~\NoCaseChange{\protect\cite{cite3488}}}
\codefieldsection{Description}
Multi-qubit PI code designed to realize gates from the binary dihedral group transversally.
Can also be interpreted as a single-spin code.
The codespace projection is a projection onto an irrep of the \textit{binary dihedral group} \( \mathsf{BD}_{2N} = \langle\omega I, X, P\rangle \) of order \(8N\), where \( \omega \) is a \( 2N \)th root of unity, and \( P = \text{diag} ( 1, \omega^2) \).

The construction includes three families and a handful of particular codes.
The first family has parameters \(\llparenthesis 2m+3,2,3\rrparenthesis \) for \(m\) not a power of two, realizing binary dihedral transversal gates that are not possible to realize in any qubit stabilizer code \NoCaseChange{\protect\cite[{Prop. 1}]{cite3488}}.
The second family is the case of \(m\) being a power of two, corresponding to \(\llparenthesis 2^{m-1}+3,2,3\rrparenthesis \) codes, each realizing a member of the \flmTerm{term}{ref694}{}{Clifford hierarchy} transversally.
The third family consists of \(\llparenthesis n,2,d\rrparenthesis \) codes with \(n = \frac{1}{4}(3d^2+6d-7+2(d\text{ mod }8) )\), realizing \(S\) and \(T\) gates transversally.
The handful of codes have distance 5 (7, 9, 11, 13) and encode in 27 (49, 73, 107, 147) qubits, all realizing transversal \(T\) gates.

\codefieldsection{Transversal and Permutation-Based Gates}
\begin{eczvaluelist}
\item\relax Binary dihedral group gates can be realized transversally, which include subgroups of any level of the \flmTerm{term}{ref694}{}{Clifford hierarchy} and subgroups which cannot be realized by any qubit stabilizer code.
\end{eczvaluelist}
\codefieldsection{Parent}
\begin{eczvaluelist}
\item\relax
\flmRefsHyperref[eczindexfamilyrel]{code:qubit_permutation_invariant}{PI qubit code}\end{eczvaluelist}
\codefieldsection{Cousins}
\begin{eczvaluelist}
\item\relax
\flmRefsHyperref[eczindexfamilyrel]{code:small_distance_quantum}{Small-distance block quantum code} --- The first and second families of binary dihedral PI codes have distance three, and the third family has the member \(\llparenthesis 27,2,5\rrparenthesis \).
\item\relax
\flmRefsHyperref[eczindexfamilyrel]{code:combinatorial_permutation_invariant}{Combinatorial PI code} --- The \(Q_{3,1,2m-4,+}\) and \(Q_{3,1,2^m-4,+}\) combinatorial PI codes reduce to the \(\llparenthesis 2m+3,2,3\rrparenthesis \) and \(\llparenthesis 2^{m-1}+3,2,3\rrparenthesis \) binary dihedral PI codes, respectively \NoCaseChange{\protect\cite[{Prop. 5.6}]{cite3169}} (see also \NoCaseChange{\protect\cite{cite727}}).
\item\relax
\flmRefsHyperref[eczindexfamilyrel]{code:xp_stabilizer}{XP stabilizer code} --- Binary dihedral permutation invariant codewords form error spaces of XP stabilizer codes.
\item\relax
\flmRefsHyperref[eczindexfamilyrel]{code:diagonal_clifford}{\(\llbracket 2^r-1,1,3\rrbracket \) simplex code} --- The \(\llparenthesis 2^{r-1}+3,2,3\rrparenthesis \) family of binary dihedral PI codes realizes the same (strongly) transversal gates as the \(\llbracket 2^r-1,1,3\rrbracket \) quantum RM codes, but require fewer qubits in almost all cases.
\item\relax
\flmRefsHyperref[eczindexfamilyrel]{code:stab_49_1_5}{\(\llbracket 49,1,5\rrbracket \) triorthogonal code} --- The \(\llparenthesis 27,2,5\rrparenthesis \) binary dihedral PI code realizes the \(T\) gate (strongly) transversally, but requires fewer qubits than the \(\llbracket 49,1,5\rrbracket \) triorthogonal code.
\item\relax
\flmRefsHyperref[eczindexfamilyrel]{code:quantum_triorthogonal}{Triorthogonal code} --- There exist binary dihedral PI codes that have distance 5 (7, 9, 11, 13) and encode in 27 (49, 73, 107, 147) qubits, all realizing transversal \(T\) gates.
\item\relax
\flmRefsHyperref[eczindexfamilyrel]{code:j_gross}{Clifford-group spin code} --- Binary dihedral PI codes can be interpreted as Clifford single-spin codes via the \flmRefsHyperref{ref526}{Dicke-state mapping}.
\item\relax
\flmRefsHyperref[eczindexfamilyrel]{code:sslp}{Subset-Sum-Linear-Programming (SS-LP) code} --- SS-LP codes are optimized to admit diagonal gates transversally and include \(\llparenthesis 7,2,3\rrparenthesis \) codes that realize the \(\mathsf{BD}_{16}\) and \(\mathsf{BD}_{32}\) groups transversally, yielding \(T\) and \(\sqrt{T}\) gates, respectively. Larger codes include an \(\llparenthesis 8,2,3\rrparenthesis \) code that transversally realizes \(\mathsf{BD}_{64}\).
\end{eczvaluelist}
\eczhbkcontributors{ \eczhuVVA }
\endeczcode

\eczcode{qcga}{Bivariate bicycle (BB) code}{~\NoCaseChange{\protect\cite{cite441}}}
\codefieldsection{Description}
One of several Abelian 2BGA codes which admit time-optimal syndrome measurement circuits that can be implemented in a two-layer
architecture, a generalization of the square-lattice architecture
optimal for the surface codes.
Codes can be classified by the weight of their checks, e.g., by BB\(w\) where \(w\) is the check weight.

The qubit connectivity graph is not quite a 2D grid and is instead decomposable into two planar subgraphs of degree three; there exists an optimized layout minimizing Euclidean communication distance for check operators \NoCaseChange{\protect\cite{cite3489}}.
There are \(n\) \(X\) and \(Z\) check operators, with each one of weight six.

See Refs. \NoCaseChange{\protect\cite{cite3490,cite3491}} for examples of self-dual BB codes.
Several variants and generalizations exist \NoCaseChange{\protect\cite{cite3492,cite3493}}.
There exist qudit BB codes that achieve \(kd^2 / n = 20\) \NoCaseChange{\protect\cite{cite3494}}.

\codefieldsection{Protection}
Admits an \(0.8\%\) \flmRefsHyperref{ref2960}{pseudo-threshold} for circuit-level noise under BP-OSD decoder \NoCaseChange{\protect\cite{cite441}} (cf. \NoCaseChange{\protect\cite{cite3495}}).

\codefieldsection{Rate}
When ancilla qubit overhead is included, the encoding rate surpasses that of the surface code. A general \(\llbracket n,k,d\rrbracket \) bivariate bicycle code requires \(n\) ancilla qubits for encoding, meaning that its \textit{ancilla-added encoding rate} is \(k/2n\).
\codefieldsection{Transversal and Permutation-Based Gates}
\begin{eczvaluelist}
\item\relax Logical Pauli operators and fold-transversal gates studied in Ref. \NoCaseChange{\protect\cite{cite718,cite719}}.
\end{eczvaluelist}
\codefieldsection{Gates}
\begin{eczvaluelist}
\item\relax Certain bivariate bicycle codes admit a cup product structure and can thus have logical gates in the \flmTerm{term}{ref694}{}{Clifford hierarchy} implemented by constant-depth Clifford circuits \NoCaseChange{\protect\cite{cite1517}}.
\end{eczvaluelist}
\codefieldsection{Decoding}
\begin{eczvaluelist}
\item\relax Syndrome extraction circuit requires seven layers of CNOT gates regardless of code length. BP-OSD decoder \NoCaseChange{\protect\cite{cite1247}} has been extended \NoCaseChange{\protect\cite{cite441}} to account for measurement errors (i.e., the circuit-based noise model \NoCaseChange{\protect\cite{cite3495}}).
\item\relax The depth-7 syndrome extraction schedules studied in Ref. \NoCaseChange{\protect\cite{cite441}} are not generally \flmRefsHyperref{ref3496}{distance-preserving}; for the \(\llbracket 144,12,12\rrbracket \) code, all \(936\) depth-7 variants obtained by reordering the CNOT layers satisfy \(d_{\mathrm{circ}}\leq 10 < d\).
\item\relax Some long-range check operators can be measured less frequently than others \NoCaseChange{\protect\cite{cite3497}}.
\item\relax Syndrome extraction circuits called \textit{morphing circuits} \NoCaseChange{\protect\cite{cite3188}}, generalizing circuits for the color code \NoCaseChange{\protect\cite{cite3421}}.
\item\relax Decoding under circuit-level noise has been studied for the BP, BP+OSD, and AutDEC decoders \NoCaseChange{\protect\cite{cite3202}}.
\item\relax Transformer-based neural-network decoder \NoCaseChange{\protect\cite{cite3192}}.
\item\relax Matching decoder \NoCaseChange{\protect\cite{cite3498}}.
\end{eczvaluelist}
\codefieldsection{Fault Tolerance}
\begin{eczvaluelist}
\item\relax Fault-tolerant state initialization using lattice surgery techniques \NoCaseChange{\protect\cite{cite3499,cite848}} and an ancillary surface code \NoCaseChange{\protect\cite{cite441}}.
\end{eczvaluelist}
\codefieldsection{Realizations}
\begin{eczvaluelist}
\item\relax Superconducting circuits: syndrome extraction has been implemented for the \(\llbracket 18,4,4\rrbracket \) BB code on a 32-qubit Kunlun device by the Wang, Song, and Deng groups \NoCaseChange{\protect\cite{cite3500}}. The same is also shown for an \(\llbracket 18,6,3\rrbracket \) code obtained by removing two check operators from the former code \NoCaseChange{\protect\cite{cite3500}}.
\end{eczvaluelist}
\codefieldsection{Notes}
\begin{eczvaluelist}
\item\relax A database of bivariate bicycle codes is available in QECDB \NoCaseChange{\protect\cite{cite781}}.
\end{eczvaluelist}
\codefieldsection{Parents}
\begin{eczvaluelist}
\item\relax
\flmRefsHyperref[eczindexfamilyrel]{code:qubit_generalized_homological_product_css}{Generalized homological-product qubit CSS code}\item\relax
\flmRefsHyperref[eczindexfamilyrel]{code:2bga}{Two-block group-algebra (2BGA) codes} --- Bivariate bicycle codes are Abelian 2BGA codes over groups of the form \(\mathbb{Z}_{r} \times \mathbb{Z}_{s}\).
\item\relax
\flmRefsHyperref[eczindexfamilyrel]{code:abelian_lifted_product}{Abelian LP code} --- Bivariate bicycle codes are Abelian LP codes over groups of the form \(\mathbb{Z}_{r} \times \mathbb{Z}_{s}\).
\item\relax
\flmRefsHyperref[eczindexfamilyrel]{code:2d_stabilizer}{2D lattice stabilizer code} --- Bivariate bicycle codes are defined on 2D lattices with periodic boundary conditions, and versions with open boundary conditions have been investigated \NoCaseChange{\protect\cite{cite443,cite3501}}. Bivariate bicycle codes are on par with the surface code in terms of threshold, but admit a much higher ancilla-added encoding rate at the expense of having non-geometrically local weight-six check operators. BB codes have been investigated in terms of their anyons and topological order \NoCaseChange{\protect\cite{cite2533}}.
\end{eczvaluelist}
\codefieldsection{Children}
\begin{eczvaluelist}
\item\relax
\flmRefsHyperref[eczindexfamilyrel]{code:bb108}{\(\llbracket 108,8,10\rrbracket \) BB6 code}\item\relax
\flmRefsHyperref[eczindexfamilyrel]{code:bb288}{\(\llbracket 288,12,18\rrbracket \) double-gross code}\item\relax
\flmRefsHyperref[eczindexfamilyrel]{code:bb5}{BB5 code}\item\relax
\flmRefsHyperref[eczindexfamilyrel]{code:bb72}{\(\llbracket 72,12,6\rrbracket \) BB6 code}\item\relax
\flmRefsHyperref[eczindexfamilyrel]{code:bb90}{\(\llbracket 90,8,10\rrbracket \) BB6 code}\item\relax
\flmRefsHyperref[eczindexfamilyrel]{code:gross}{\(\llbracket 144,12,12\rrbracket \) gross code}\end{eczvaluelist}
\codefieldsection{Cousins}
\begin{eczvaluelist}
\item\relax
\flmRefsHyperref[eczindexfamilyrel]{code:triangular_color}{Honeycomb (6.6.6) color code} --- Certain bivariate bicycle codes are equivalent to a family of 6.6.6 color codes \NoCaseChange{\protect\cite{cite3501}}.
\item\relax
\flmRefsHyperref[eczindexfamilyrel]{code:topological_abelian}{Abelian topological code} --- BB codes have been investigated in terms of their anyons and topological order \NoCaseChange{\protect\cite{cite2533}}.
\item\relax
\flmRefsHyperref[eczindexfamilyrel]{code:generalized_bicycle}{Generalized bicycle (GB) code} --- GB codes (BB codes) are 2BGA codes over the cyclic group \(\mathbb{Z}_{\ell}\) (Abelian group \(\mathbb{Z}_{r} \times \mathbb{Z}_{s}\)). The two codes are the same when \(r\) and \(s\) are relatively prime due to the isomorphism \(\mathbb{Z}_{r} \times \mathbb{Z}_{s} \cong \mathbb{Z}_{\ell = rs}\).
\end{eczvaluelist}
\eczhbkcontributors{ Mark Webster, Leonid Pryadko, \eczhuVVA }
\endeczcode

\eczcode{bosonization}{Bosonization code}{~\NoCaseChange{\protect\cite{cite3502,cite3503,cite3504,cite3505}}}
\codefieldsection{Description}
A mapping that maps a \(D\)-dimensional lattice quadratic Hamiltonian of Majorana modes into a lattice of qubits.
The resulting qubit code can realize various topological phases, depending on the initial Majorana-mode Hamiltonian and its symmetries.

A general mapping for quadratic Hamiltonians was constructed in Ref. \NoCaseChange{\protect\cite{cite3502}}, while others considered higher-order products of Majorana modes that correspond to symmetry constraints \NoCaseChange{\protect\cite{cite3504,cite3503}}.

\codefieldsection{Parents}
\begin{eczvaluelist}
\item\relax
\flmRefsHyperref[eczindexfamilyrel]{code:fermions_into_qubits}{Fermion-into-qubit code}\item\relax
\flmRefsHyperref[eczindexfamilyrel]{code:qldpc}{Qubit QLDPC code} --- The \(D\)-dimensional bosonization code encodes fermionic modes into a \(D\)-dimensional qubit stabilizer code.
\item\relax
\flmRefsHyperref[eczindexfamilyrel]{code:translationally_invariant_stabilizer}{Lattice stabilizer code} --- The \(D\)-dimensional bosonization code encodes fermionic modes into a \(D\)-dimensional qubit stabilizer code.
\end{eczvaluelist}
\codefieldsection{Children}
\begin{eczvaluelist}
\item\relax
\flmRefsHyperref[eczindexfamilyrel]{code:2d_bosonization}{2D bosonization code}\item\relax
\flmRefsHyperref[eczindexfamilyrel]{code:3d_bosonization}{3D bosonization code}\end{eczvaluelist}
\codefieldsection{Cousins}
\begin{eczvaluelist}
\item\relax
\flmRefsHyperref[eczindexfamilyrel]{code:haah_cubic}{Haah cubic code (CC)} --- Bosonization can be used to realize a Haah cubic code with an emergent fermion from a Majorana stabilizer code \NoCaseChange{\protect\cite{cite3503}}. This code is shown to be distinct from the original code \NoCaseChange{\protect\cite{cite3506}}.
\item\relax
\flmRefsHyperref[eczindexfamilyrel]{code:fibonacci_fractal_liquid}{Fibonacci fractal spin-liquid code} --- Bosonization can be used to realize a Fibonacci fractal spin-liquid code with an emergent fermion from a Majorana stabilizer code \NoCaseChange{\protect\cite{cite3503}}. This code is shown to be distinct from the original code \NoCaseChange{\protect\cite{cite3506}}.
\item\relax
\flmRefsHyperref[eczindexfamilyrel]{code:xcube}{X-cube model code} --- Bosonization can be used to realize an X-cube model code with an emergent fermion from a Majorana stabilizer code \NoCaseChange{\protect\cite{cite3503}}, but this model has the same stabilizer group as the original X-cube model \NoCaseChange{\protect\cite{cite3504}}.
\item\relax
\flmRefsHyperref[eczindexfamilyrel]{code:three_fermion}{Three-fermion (3F) Walker-Wang model code} --- The 3F Walker-Wang QCA encoder \NoCaseChange{\protect\cite{cite3068,cite3069}} can be extended to SPTs in higher dimensions based on an exact bosonization duality \NoCaseChange{\protect\cite{cite3065}}.
\end{eczvaluelist}
\eczhbkcontributors{ Nathanan Tantivasadakarn, \eczhuVVA }
\endeczcode

\eczcode{branching_mera}{Branching MERA code}{~\NoCaseChange{\protect\cite{cite1534,cite400,cite1535}}}
\codefieldsection{Description}
Qubit stabilizer code whose encoding circuit corresponds to a branching MERA \NoCaseChange{\protect\cite{cite543}} tensor network.
These codes generalize quantum polar codes by reinstating the disentanglers omitted in the branching-tree tensor-network construction \NoCaseChange{\protect\cite{cite400}}.

\codefieldsection{Protection}
Numerical evidence indicates that channel polarization rapidly suppresses channels that are bad in both quadratures, allowing good performance without entanglement assistance \NoCaseChange{\protect\cite{cite400}}.
\codefieldsection{Rate}
Numerics on depolarizing and erasure channels indicate low block-error rates at rates approaching coherent information, with better finite-size behavior than quantum polar codes \NoCaseChange{\protect\cite{cite400}}.
\codefieldsection{Encoding}
\begin{eczvaluelist}
\item\relax Encoding uses a reversed branching-MERA CNOT circuit, with non-data inputs frozen to either \(|0\rangle\) or \(|+\rangle\) according to channel selection \NoCaseChange{\protect\cite{cite400}}.
\end{eczvaluelist}
\codefieldsection{Decoding}
\begin{eczvaluelist}
\item\relax Decoding can be formulated as tensor-network contraction and supports successive-cancellation-style decoding inherited from branching-MERA constructions \NoCaseChange{\protect\cite{cite1534,cite400}}.
\item\relax A symmetric decoder that jointly uses \(x\)- and \(z\)-error information remains efficiently contractible and improves finite-size performance over standard quantum polar decoding \NoCaseChange{\protect\cite{cite400}}.
\end{eczvaluelist}
\codefieldsection{Parent}
\begin{eczvaluelist}
\item\relax
\flmRefsHyperref[eczindexfamilyrel]{code:eastab}{EA qubit stabilizer code}\end{eczvaluelist}
\codefieldsection{Child}
\begin{eczvaluelist}
\item\relax
\flmRefsHyperref[eczindexfamilyrel]{code:quantum_polar}{Quantum polar code} --- Branching MERA codes generalize quantum polar codes by restoring the branching-MERA disentanglers while retaining efficient tensor-network decoding \NoCaseChange{\protect\cite{cite400}}.
\end{eczvaluelist}
\codefieldsection{Cousins}
\begin{eczvaluelist}
\item\relax
\flmRefsHyperref[eczindexfamilyrel]{code:polar}{Polar code} --- Classical versions of branching MERA codes can be thought of as extensions of polar codes \NoCaseChange{\protect\cite{cite1534,cite1535}}.
\item\relax
\flmRefsHyperref[eczindexfamilyrel]{code:quantum_lego}{Tensor-network code} --- Encoders for branching MERA codes are related to branching MERA tensor networks \NoCaseChange{\protect\cite{cite1534,cite400}}.
\end{eczvaluelist}
\eczhbkcontributors{ \eczhuVVA }
\endeczcode

\eczcode{bravyi_bacon_shor}{Bravyi-Bacon-Shor (BBS) code}{~\NoCaseChange{\protect\cite{cite3328}}}
\codefieldsection{Alternative Names}
\begin{eczvaluelist}
\item\relax Generalized Bacon-Shor code
\end{eczvaluelist}
\eczhIndexCodeAliasName{bravyi_bacon_shor}{Generalized Bacon-Shor code}
\codefieldsection{Description}
A CSS subsystem stabilizer code generalizing Bacon-Shor codes to a larger set of qubit geometries.
Defined through a binary matrix \(A\) such that physical qubits live on sites \((i,j)\) whenever \(A_{i,j}=1\).
The gauge group is generated by 2-qubit operators, including \(XX\) interactions between any two qubits sharing a column in \(A\), and \(ZZ\) interactions between any two qubits sharing a row.

\codefieldsection{Protection}
The code parameters are \(n=\sum_{i,j}A_{i,j}\), \(k=\text{rank}(A)\), and \(d\) is the minimum of the distances of the linear binary codes generated by the rows and columns of \(A\) \NoCaseChange{\protect\cite{cite3328}}.

\codefieldsection{Rate}
A class of BBS codes \NoCaseChange{\protect\cite{cite3507}} saturate the subsystem bound \(kd = O(n)\) \NoCaseChange{\protect\cite{cite3328}}.
\codefieldsection{Parent}
\begin{eczvaluelist}
\item\relax
\flmRefsHyperref[eczindexfamilyrel]{code:qubit_subsystem_css}{Subsystem qubit CSS code}\end{eczvaluelist}
\codefieldsection{Children}
\begin{eczvaluelist}
\item\relax
\flmRefsHyperref[eczindexfamilyrel]{code:bacon_shor}{Bacon-Shor code}\item\relax
\flmRefsHyperref[eczindexfamilyrel]{code:trapezoid}{Trapezoid subsystem code}\item\relax
\flmRefsHyperref[eczindexfamilyrel]{code:3d_bacon_shor}{3D Bacon-Shor code}\end{eczvaluelist}
\codefieldsection{Cousins}
\begin{eczvaluelist}
\item\relax
\flmRefsHyperref[eczindexfamilyrel]{code:subsystem_quantum_parity}{Subsystem hypergraph product (SHP) code} --- The BBS code construction can utilize different classical codes in different rows and columns of \(A\), while the subsystem construction does not; see \NoCaseChange{\protect\cite[{pg. 4}]{cite3328}}.
Subsystem hypergraph product and BBS codes have been numerically compared \NoCaseChange{\protect\cite{cite665}}.

\item\relax
\flmRefsHyperref[eczindexfamilyrel]{code:commuting_projector}{Commuting-projector Hamiltonian code} --- Ground-state spaces of qubit commuting-projector Hamiltonians with weight-two (two-body) terms cannot be used to suppress errors in adiabatic quantum computation \NoCaseChange{\protect\cite{cite2687}}, but this can be circumvented with excited-state subspaces \NoCaseChange{\protect\cite{cite2688}} or ground-state subspaces of subsystem code Hamiltonians, e.g., using BBS codes \NoCaseChange{\protect\cite{cite670,cite2689}}.
\end{eczvaluelist}
\eczhbkcontributors{ Srilekha Gandhari, \eczhuVVA }
\endeczcode

\eczcode{bksf}{Bravyi-Kitaev superfast (BKSF) code}{~\NoCaseChange{\protect\cite{cite557}}}
\codefieldsection{Alternative Names}
\begin{eczvaluelist}
\item\relax Loop-stabilized fermion simulation (LSFS) code
\end{eczvaluelist}
\eczhIndexCodeAliasName{bksf}{Loop-stabilized fermion simulation (LSFS) code}
\codefieldsection{Description}
A single-error-detecting fermion-into-qubit encoding defined on a 2D qubit lattice whose stabilizers are associated with loops in the lattice.
For the square-lattice edge ordering used in Ref. \NoCaseChange{\protect\cite{cite404}}, the BKSF logical operators coincide with exact 2D bosonization on the dual lattice after relabeling \(X\) and \(Y\).
The code can be generalized to a single error-correcting code (i.e., with distance three) on graphs of degree \(\geq 6\) \NoCaseChange{\protect\cite{cite408}}.

\codefieldsection{Protection}
The code can detect single-qubit errors \NoCaseChange{\protect\cite{cite3508}}.
A generalized BKSF code has distance 3 on a graph with degree \(\geq 6\), while original BKSF code cannot correct single-qubit errors on graphs of degree \(< 6\) \NoCaseChange{\protect\cite{cite408}}.

\codefieldsection{Parent}
\begin{eczvaluelist}
\item\relax
\flmRefsHyperref[eczindexfamilyrel]{code:mlsc}{Majorana loop stabilizer code (MLSC)} --- The BKSF code can be thought of as a particular MLSC \NoCaseChange{\protect\cite{cite3508}}.
\end{eczvaluelist}
\eczhbkcontributors{ \eczhuVVA }
\endeczcode

\eczcode{bkt}{Bravyi-Kitaev transformation (BKT) code}{~\NoCaseChange{\protect\cite{cite557}}}
\codefieldsection{Description}
A fermion-into-qubit encoding that maps Majorana operators into Pauli strings of weight \(\lceil \log_2(n+1) \rceil\).
The code can be reformulated in terms of Fenwick trees \NoCaseChange{\protect\cite{cite554}}, and the Pauli-string weight can be further optimized to yield the \textit{segmented Bravyi-Kitaev (SBK) transformation code} \NoCaseChange{\protect\cite{cite555}} (see also Ref. \NoCaseChange{\protect\cite{cite556}}).

\codefieldsection{Notes}
\begin{eczvaluelist}
\item\relax Review on the BKT \NoCaseChange{\protect\cite{cite3509}}.
\end{eczvaluelist}
\codefieldsection{Parent}
\begin{eczvaluelist}
\item\relax
\flmRefsHyperref[eczindexfamilyrel]{code:fermions_into_qubits}{Fermion-into-qubit code}\end{eczvaluelist}
\codefieldsection{Cousins}
\begin{eczvaluelist}
\item\relax
\flmRefsHyperref[eczindexfamilyrel]{code:jw}{Jordan-Wigner transformation code} --- The weight of a Majorana operator in the BKT (JW transformation) code scales logarithmically (linearly) with \(n\), with the former demonstrating an exponential improvement \NoCaseChange{\protect\cite{cite3510}}.
\item\relax
\flmRefsHyperref[eczindexfamilyrel]{code:ternary_tree_fermion}{Ternary-tree fermion-into-qubit code} --- The ternary-tree fermion-into-qubit code improves over the BKT code by a factor of \(\approx 1.58\) in the weight of encoded fermionic operators \NoCaseChange{\protect\cite{cite3510}}.
\end{eczvaluelist}
\eczhbkcontributors{ \eczhuVVA }
\endeczcode

\eczcode{brickwork}{Brickwork \(XS\) stabilizer code}{~\NoCaseChange{\protect\cite{cite589}}}
\codefieldsection{Description}
An \(XS\) stabilizer code that realizes the topological order of the Type-III \(G=\mathbb{Z}^3_2\) TQD model \NoCaseChange{\protect\cite{cite575,cite576}}, which is the same topological order as the \(G=D_4\) quantum double \NoCaseChange{\protect\cite{cite577}}.
Its qubits are placed on a 2D square lattice, and the stabilizers are defined using two overlapping rectangular tilings.

\codefieldsection{Decoding}
\begin{eczvaluelist}
\item\relax Just-in-time decoder \NoCaseChange{\protect\cite{cite589}}.
\end{eczvaluelist}
\codefieldsection{Parents}
\begin{eczvaluelist}
\item\relax
\flmRefsHyperref[eczindexfamilyrel]{code:xs_stabilizer}{XS stabilizer code} --- The brickwork \(XS\) stabilizer code is an \(XS\) stabilizer code \NoCaseChange{\protect\cite{cite589}}.
\item\relax
\flmRefsHyperref[eczindexfamilyrel]{code:tqd_abelian}{Abelian TQD code} --- The ground-state subspace of the brickwork \(XS\) stabilizer code realizes the topological order of the Type-III \(G=\mathbb{Z}^3_2\) TQD model \NoCaseChange{\protect\cite{cite575,cite576}}, which is the same topological order as the \(G=D_4\) quantum double \NoCaseChange{\protect\cite{cite577}}.
\end{eczvaluelist}
\codefieldsection{Child}
\begin{eczvaluelist}
\item\relax
\flmRefsHyperref[eczindexfamilyrel]{code:stab_4_2_2}{\(\llbracket 4,2,2\rrbracket \) Four-qubit code} --- The \(\llbracket 4,2,2\rrbracket \) code can be interpreted as a brickwork code on a square of the overlapping rectangular tilings \NoCaseChange{\protect\cite{cite589}}.
\end{eczvaluelist}
\codefieldsection{Cousins}
\begin{eczvaluelist}
\item\relax
\flmRefsHyperref[eczindexfamilyrel]{code:quantum_double_dihedral}{Dihedral \(G=D_m\) quantum-double code} --- The ground-state subspace of the brickwork \(XS\) stabilizer code realizes the topological order of the non-Abelian Type-III \(G=\mathbb{Z}^3_2\) TQD model \NoCaseChange{\protect\cite{cite575,cite576}}, which is the same topological order as the ordinary \(G=D_4\) quantum double \NoCaseChange{\protect\cite{cite577}}.
\item\relax
\flmRefsHyperref[eczindexfamilyrel]{code:3d_color}{3D color code} --- The brickwork \(XS\) stabilizer code can be obtained from a 3D color code \NoCaseChange{\protect\cite{cite589}}.
\item\relax
\flmRefsHyperref[eczindexfamilyrel]{code:hexagonal_cz}{Hexagonal \(CZ\) code} --- The brickwork \(XS\) stabilizer code and the hexagonal \(CZ\) code realize the same topological phase and are equivalent via a local unitary \NoCaseChange{\protect\cite{cite589,cite3511}}.
\end{eczvaluelist}
\eczhbkcontributors{ Benjamin J. Brown, \eczhuVVA }
\endeczcode

\eczcode{nonlocal_lowdepth}{Brown-Fawzi Clifford-circuit code}{~\NoCaseChange{\protect\cite{cite540}}}
\codefieldsection{Description}
An \(\llbracket n,k\rrbracket \) stabilizer code whose encoder is a random \flmRefsHyperref{ref409}{Clifford circuit} of depth of \flmRefsHyperref{ref65}{order} \(O(\log^3 n)\).

An \(n\)-qubit quantum encoding circuit with \(O(n^2 \log n)\) random two-qubit \flmRefsHyperref{ref409}{Clifford gates} applied to pairs of randomly chosen qubits yields a code with distance \(d\) with probability \(1 - \Omega(1/n^8)\), provided that \flmMathEnvironment{equation}{}{
  \frac{k}{n} < 1 - \frac{d}{n}\log_2 3 - h\left(\frac{d}{n}\right)~,
}
where \(h\) is the entropy function.
Since two gates acting on disjoint qubits can be executed simultaneously, the depth of a circuit of this size is typically of \flmRefsHyperref{ref65}{order} \(O(\log^3 n)\).

\codefieldsection{Rate}
The achievable distance of these codes is asymptotically the same as a code whose encoder is a random (not necessarily log-depth) general Clifford unitary \NoCaseChange{\protect\cite{cite540}}.
\codefieldsection{Encoding}
\begin{eczvaluelist}
\item\relax Random \(\log^3\)-depth \flmRefsHyperref{ref409}{Clifford circuit}.
\end{eczvaluelist}
\codefieldsection{Parents}
\begin{eczvaluelist}
\item\relax
\flmRefsHyperref[eczindexfamilyrel]{code:qubit_stabilizer}{Qubit stabilizer code}\item\relax
\flmRefsHyperref[eczindexfamilyrel]{code:random_stabilizer}{Random stabilizer code}\end{eczvaluelist}
\codefieldsection{Cousin}
\begin{eczvaluelist}
\item\relax
\flmRefsHyperref[eczindexfamilyrel]{code:circuit_to_hamiltonian}{Circuit-to-Hamiltonian approximate code} --- Circuit-to-Hamiltonian approximate codes are constructed by converting the encoding circuit of a Brown-Fawzi random Clifford-circuit code into a Hamiltonian using the spacetime circuit-to-Hamiltonian construction \NoCaseChange{\protect\cite{cite3512,cite3513}}.
\end{eczvaluelist}
\eczhbkcontributors{ Srilekha Gandhari, \eczhuVVA }
\endeczcode

\eczcode{hermitian_qldpc}{Camara-Ollivier-Tillich code}{~\NoCaseChange{\protect\cite{cite3514}}}
\codefieldsection{Description}
A Hermitian qubit QLDPC code whose stabilizer generator matrix is constructed using two nested subgroups of \(\mathbb{F}_4^n\).

Examples include a \((4,8)\)-regular 8736-qubit code and a \((4,8)\)-regular 3600-qubit code, both of rate one half.

\codefieldsection{Decoding}
\begin{eczvaluelist}
\item\relax Iterative error estimation based on the MIN-SUM and SUM-PRODUCT algorithms \NoCaseChange{\protect\cite{cite3514}}.
\end{eczvaluelist}
\codefieldsection{Parent}
\begin{eczvaluelist}
\item\relax
\flmRefsHyperref[eczindexfamilyrel]{code:stabilizer_over_gf4}{Hermitian qubit code}\end{eczvaluelist}
\eczhbkcontributors{ \eczhuVVA }
\endeczcode

\eczcode{double_homological_product}{Campbell double homological product code}{~\NoCaseChange{\protect\cite{cite675}}}
\codefieldsection{Description}
A multi-dimensional HGP code derived from two applications of the hypergraph product to a classical code, resulting in a length-\(4\) chain complex.
The construction method allows for the use of two different classical codes as inputs, with Ref. \NoCaseChange{\protect\cite{cite675}} assuming identical input codes for simplicity.

Explicitly, the resulting chain complex is
\flmMathEnvironment{align}{}{
A_{-2}\xrightarrow{\partial_{-2}}A_{-1}\xrightarrow{\partial_{-1}}A_{0}\xrightarrow{\partial_{0}}A_{1}\xrightarrow{\partial_{1}}A_{2}\,.
}
The boundary maps \(\partial_j\) are constructed using tensor products of the original boundary maps, ensuring the chain condition \(\partial_{j+1} \partial_j = 0\).
The additional parts of the chain complex yield metachecks to detect measurement errors, enabling single-shot error correction.

\codefieldsection{Protection}
For the minimal length-one chain complex associated with a classical \([n, k, d]\) code, the double homological product yields a quantum code with parameters \(\llbracket n^4 + 4n^2(n-k)^2 + (n-k)^4, k^4, \geq d\rrbracket \) and single-shot distance \(d_{\text{ss}}=\infty\) \NoCaseChange{\protect\cite{cite675}}.

In this minimal-input setting, the Campbell double homological product code is \((d, f)\)-sound with \(f(x) = x^3/4\), meaning that small syndromes can be explained by errors whose weight grows at most cubically in the syndrome weight \NoCaseChange{\protect\cite{cite675}}. Its check redundancy is bounded by \(<2\), and the construction preserves LDPC properties if the original classical-code family is LDPC.

\codefieldsection{Rate}
If the input classical-code family has asymptotically constant rate, then the resulting Campbell double homological product family also has asymptotically constant rate because both \(k_Q\) and \(n_Q\) scale as \(n^4\) \NoCaseChange{\protect\cite{cite675}}.
\codefieldsection{Decoding}
\begin{eczvaluelist}
\item\relax The minimum-weight decoder optimizes the recovery operation \( E_{\text{rec}} \) to minimize the residual error \( E_{\text{rec}} \cdot E \) given a noisy syndrome \( s = \sigma(E) + u \). The decoder's performance is intrinsically tied to the code's soundness: when the code is \((t, f)\)-sound, the minimum-weight decoder guarantees that the residual error's min-weight scales as \( f(2|u|) \) for measurement errors \( |u| < t/2 \) \NoCaseChange{\protect\cite{cite675}}. This property is particularly robust in double homological product codes, where soundness follows a cubic scaling (\( f(x) \sim x^3 \)).
\item\relax A meta-check-based decoder operates through a two-stage process: first, it identifies a minimal correction \( s_{\text{rec}} \) to the syndrome \( s \) such that the repaired syndrome \( s + s_{\text{rec}} \) satisfies all metachecks (\( H(s + s_{\text{rec}}) = 0 \)). Second, it computes a minimal-weight physical error \( E_{\text{rec}} \) consistent with the repaired syndrome. This approach uniquely tolerates up to \( \lfloor (d_{\textit{ss}} - 1)/2 \rfloor \) measurement errors in a single round (where \(d_{\textit{ss}}\) is the single-shot distance), eliminating the need for repeated syndrome measurements.
\end{eczvaluelist}
\codefieldsection{Parents}
\begin{eczvaluelist}
\item\relax
\flmRefsHyperref[eczindexfamilyrel]{code:multisector_hypergraph}{Higher-dimensional homological product code}\item\relax
\flmRefsHyperref[eczindexfamilyrel]{code:single_shot}{Single-shot code} --- For a minimal input chain complex associated with a classical \([n,k,d]\) code, the Campbell double homological product code is a single-shot code with \(d_{\text{ss}}=\infty\), \((d,f)\)-soundness for \(f(x)=x^3/4\), and check redundancy bounded by \(<2\) \NoCaseChange{\protect\cite{cite675}}.
\end{eczvaluelist}
\codefieldsection{Cousin}
\begin{eczvaluelist}
\item\relax
\flmRefsHyperref[eczindexfamilyrel]{code:4d_surface}{\((2,2)\) Loop toric code} --- The 4D loop planar (toric) code on a hypercubic lattice can be obtained from a particular choice of chain complex from a hypergraph product of four repetition codes \NoCaseChange{\protect\cite{cite1613}}. As such, it is a particular Campbell double homological product code \NoCaseChange{\protect\cite[{table I}]{cite675}}.
\end{eczvaluelist}
\eczhbkcontributors{ Feroz Ahmed Mian, \eczhuVVA }
\endeczcode

\eczcode{capped_color}{Capped color code (CCC)}{~\NoCaseChange{\protect\cite{cite3515}}}
\codefieldsection{Description}
A non-geometrically local subsystem color code consisting of two layers of 2D color code stacked together and topped (or capped) by a single qubit.
Gauge fixing yields two types of codes, capped color codes in H or T form.
Layers of 2D color codes can also be stacked together in a recursive construction, yielding \textit{recursive capped color codes} (RCCCs).

\codefieldsection{Transversal and Permutation-Based Gates}
\begin{eczvaluelist}
\item\relax Capped color codes in H (T) form admit a transversal Hadamard (T) gate.
\end{eczvaluelist}
\codefieldsection{Fault Tolerance}
\begin{eczvaluelist}
\item\relax Fault-tolerant syndrome extraction and error correction for capped color codes in H form \NoCaseChange{\protect\cite{cite3515}}.
\item\relax Fault-tolerant T gate implementation \NoCaseChange{\protect\cite{cite3515}}.
\end{eczvaluelist}
\codefieldsection{Parent}
\begin{eczvaluelist}
\item\relax
\flmRefsHyperref[eczindexfamilyrel]{code:subsystem_color}{Subsystem color code}\end{eczvaluelist}
\eczhbkcontributors{ \eczhuVVA }
\endeczcode

\eczcode{chamon}{Chamon model code}{~\NoCaseChange{\protect\cite{cite3516,cite3517}}}
\codefieldsection{Alternative Names}
\begin{eczvaluelist}
\item\relax Chamon-Bravyi-Leemhuis-Terhal (CBLT) code
\end{eczvaluelist}
\eczhIndexCodeAliasName{chamon}{Chamon-Bravyi-Leemhuis-Terhal (CBLT) code}
\codefieldsection{Description}
A foliated type-I fracton non-CSS code defined on a cubic lattice using one weight-eight stabilizer generator acting on the eight vertices of each cube in the lattice \NoCaseChange{\protect\cite[{Eq. (D38)}]{cite456}}.

In the realization as an XYZ product of three repetition codes \NoCaseChange{\protect\cite[{Sec. 3.4}]{cite645}}, qubits live on the vertices and faces of a cubic lattice, each stabilizer generator has weight six, and the natural logical operators are membrane-like rather than string-like \NoCaseChange{\protect\cite{cite645}}.

Variants include a CSS model that is expected to have the same excitation structure \NoCaseChange{\protect\cite{cite233}} and a modified Chamon code based on the XYZ product code construction \NoCaseChange{\protect\cite{cite645}}.

\codefieldsection{Protection}
Flexible string operators confined to planes orthogonal to \([1,1,1]^T\) imply \(d = O(\sqrt{N})\) for the stabilizer version \NoCaseChange{\protect\cite{cite645}}.
\codefieldsection{Rate}
For the stabilizer version on an \(n_1 \times n_2 \times n_3\) lattice, the number of logical qubits is \(k = 4 \gcd(n_1,n_2,n_3)\) \NoCaseChange{\protect\cite{cite645}}.
\codefieldsection{Decoding}
\begin{eczvaluelist}
\item\relax Repetition-based decoder, based on the three underlying repetition codes and improved by pre-treatment with a probabilistic greedy local algorithm \NoCaseChange{\protect\cite{cite3518}}.
\end{eczvaluelist}
\codefieldsection{Code Capacity Threshold}
\begin{eczvaluelist}
\item\relax Depolarizing noise: \(4.92\%\) with repetition-based decoder \NoCaseChange{\protect\cite{cite3518}}.
\end{eczvaluelist}
\codefieldsection{Parents}
\begin{eczvaluelist}
\item\relax
\flmRefsHyperref[eczindexfamilyrel]{code:xyz_product}{XYZ product code} --- The Chamon model code can be obtained from an XYZ product of three repetition codes \NoCaseChange{\protect\cite{cite1611}}, in a construction different from the 3D surface code; see \NoCaseChange{\protect\cite[{Sec. 3.4}]{cite645}}.
\item\relax
\flmRefsHyperref[eczindexfamilyrel]{code:fracton}{Fracton stabilizer code} --- The Chamon model is a 4-foliated type-I fracton code \NoCaseChange{\protect\cite{cite3519}} and is the first example of a fracton phase \NoCaseChange{\protect\cite{cite456}}.
\end{eczvaluelist}
\codefieldsection{Cousins}
\begin{eczvaluelist}
\item\relax
\flmRefsHyperref[eczindexfamilyrel]{code:repetition}{Repetition code} --- The Chamon model code can be obtained from an XYZ product of three repetition codes \NoCaseChange{\protect\cite{cite1611}}, in a construction different from the 3D surface code; see \NoCaseChange{\protect\cite[{Sec. 3.4}]{cite645}}.
\item\relax
\flmRefsHyperref[eczindexfamilyrel]{code:3d_surface}{3D surface code} --- The 3D planar and toric code on a cubic lattice can be obtained from a hypergraph product of three repetition codes \NoCaseChange{\protect\cite{cite1613}\protect\cite[{Exam. A.1}]{cite1612}}. The Chamon code is an XYZ product of three repetition codes \NoCaseChange{\protect\cite[{Sec. 3.4}]{cite645}}.
\item\relax
\flmRefsHyperref[eczindexfamilyrel]{code:xzzx}{XZZX surface code} --- The Chamon model code can be obtained from an XYZ product of three repetition codes \NoCaseChange{\protect\cite{cite1611}}; see \NoCaseChange{\protect\cite[{Sec. 3.4}]{cite645}}. Using only two repetition codes in the analogous 2D construction yields the XZZX code, making it a 2D analogue of the Chamon code \NoCaseChange{\protect\cite[{Sec. 2}]{cite645}}.
\end{eczvaluelist}
\eczhbkcontributors{ Zongyuan Wang, \eczhuVVA }
\endeczcode

\eczcode{checkerboard}{Checkerboard model code}{~\NoCaseChange{\protect\cite{cite3520}}}
\codefieldsection{Description}
A foliated type-I fracton code defined on a cubic lattice that admits weight-eight  \(X\)- and \(Z\)-type stabilizer generators on the eight vertices of each cube in the lattice.
A tetrahedral Ising model can be used to obtain the checkerboard model by gauging \NoCaseChange{\protect\cite{cite462,cite463,cite233,cite464,cite465,cite466,cite467,cite468,cite469,cite470}} its subsystem symmetry \NoCaseChange{\protect\cite{cite233}}.
In that construction, the checkerboard model is self-dual under exchange of \(X\)- and \(Z\)-type stabilizers, and its composites include dimension-1 and dimension-2 excitations, i.e., lineons and planons in later terminology, with anyonic mutual and self-statistics \NoCaseChange{\protect\cite{cite233}}.

Variants include the twisted checkerboard model \NoCaseChange{\protect\cite{cite3521}}.

\codefieldsection{Decoding}
\begin{eczvaluelist}
\item\relax Parallelized matching decoder \NoCaseChange{\protect\cite{cite3522}}.
\end{eczvaluelist}
\codefieldsection{Code Capacity Threshold}
\begin{eczvaluelist}
\item\relax Independent \(X,Z\) noise: \(\approx 7.5\%\), higher than 3D surface code and color code \NoCaseChange{\protect\cite{cite3523}}.
\end{eczvaluelist}
\codefieldsection{Parents}
\begin{eczvaluelist}
\item\relax
\flmRefsHyperref[eczindexfamilyrel]{code:qubit_css}{Qubit CSS code}\item\relax
\flmRefsHyperref[eczindexfamilyrel]{code:qldpc}{Qubit QLDPC code}\item\relax
\flmRefsHyperref[eczindexfamilyrel]{code:fracton}{Fracton stabilizer code} --- The checkerboard model is equivalent to two copies of the X-cube model via a local constant-depth unitary \NoCaseChange{\protect\cite{cite3524}}. Hence, it is a foliated type-I fracton code.
\item\relax
\flmRefsHyperref[eczindexfamilyrel]{code:lifted_product}{Lifted-product (LP) code} --- The checkerboard model code can be formulated directly as an LP code \NoCaseChange{\protect\cite{cite1350}}.
\end{eczvaluelist}
\codefieldsection{Cousins}
\begin{eczvaluelist}
\item\relax
\flmRefsHyperref[eczindexfamilyrel]{code:xcube}{X-cube model code} --- The checkerboard model is equivalent to two copies of the X-cube model via a local constant-depth unitary \NoCaseChange{\protect\cite{cite3524}}.
\item\relax
\flmRefsHyperref[eczindexfamilyrel]{code:floquet_3d_surface}{Floquet 3D surface code} --- A planar Floquet 3D surface code stacked with two planar 3D subsystem surface codes prepares an instantaneous state equivalent to a 3D surface code stacked with two checkerboard model codes, enabling a logical \(CCZ\) gate \NoCaseChange{\protect\cite{cite533}}.
\item\relax
\flmRefsHyperref[eczindexfamilyrel]{code:floquet_fracton}{Fracton Floquet code} --- The ISG of the Fracton Floquet code can be that of the X-cube model code or the checkerboard model code.
\end{eczvaluelist}
\eczhbkcontributors{ \eczhuVVA }
\endeczcode

\eczcode{invertible}{Chen-Hsin invertible-order code}{~\NoCaseChange{\protect\cite{cite578}}}
\codefieldsection{Description}
A geometrically local commuting-projector code that realizes beyond-group-cohomology invertible topological phases in arbitrary dimensions.
Its code Hamiltonian terms include Pauli-\(Z\) operators and products of Pauli-\(X\) operators and \(CZ\) gates \NoCaseChange{\protect\cite[{Eq. (3.25)}]{cite578}}.
Instances of the code in 4D realize the 3D \(\mathbb{Z}_2\) gauge theory with fermionic charge and either bosonic (FcBl) or fermionic (FcFl) loop excitations at their boundaries \NoCaseChange{\protect\cite{cite579,cite455}}; see Ref. \NoCaseChange{\protect\cite{cite580}} for a different lattice-model formulation of the FcBl boundary code.

\codefieldsection{Encoding}
\begin{eczvaluelist}
\item\relax QCA encoder \NoCaseChange{\protect\cite{cite578,cite3065}}.
\end{eczvaluelist}
\codefieldsection{Parents}
\begin{eczvaluelist}
\item\relax
\flmRefsHyperref[eczindexfamilyrel]{code:clifford_hierarchy}{Clifford-hierarchy stabilizer code} --- Chen-Hsin invertible-order code Hamiltonian terms include Pauli and \(CZ\) operators, making them Clifford stabilizer codes.
\item\relax
\flmRefsHyperref[eczindexfamilyrel]{code:yetter_gauge_theory}{Two-gauge theory code} --- Chen-Hsin invertible-order codes realize beyond-group-cohomology invertible topological phases of order two and four in arbitrary dimensions. These phases are described by invertible two-gauge theories \NoCaseChange{\protect\cite[{pg. 11}]{cite578}}.
\end{eczvaluelist}
\codefieldsection{Cousins}
\begin{eczvaluelist}
\item\relax
\flmRefsHyperref[eczindexfamilyrel]{code:topological_abelian}{Abelian topological code} --- Instances of the code in 4D realize the 3D \(\mathbb{Z}_2\) gauge theory with fermionic charge and either bosonic (FcBl) or fermionic (FcFl) loop excitations at their boundaries \NoCaseChange{\protect\cite{cite579,cite455}}; see Ref. \NoCaseChange{\protect\cite{cite580}} for a different lattice-model formulation of the FcBl boundary code.
\item\relax
\flmRefsHyperref[eczindexfamilyrel]{code:spt}{Symmetry-protected topological (SPT) code} --- Instances of the Chen-Hsin invertible-order code realize beyond-group-cohomology SPTs \NoCaseChange{\protect\cite{cite578}}.
\item\relax
\flmRefsHyperref[eczindexfamilyrel]{code:xp_stabilizer}{XP stabilizer code} --- The Chen-Hsin invertible-order code can be embedded into a larger codespace such that all diagonal logical operators are represented by XP operators \NoCaseChange{\protect\cite[{Sec. 4.3}]{cite767}}.
\end{eczvaluelist}
\eczhbkcontributors{ \eczhuVVA }
\endeczcode

\eczcode{circuit_to_hamiltonian}{Circuit-to-Hamiltonian approximate code}{~\NoCaseChange{\protect\cite{cite581}}}
\codefieldsection{Description}
Approximate qubit block code that forms the ground-state space of a frustration-free Hamiltonian with non-commuting terms.
Its distance and logical-qubit number are both of \flmRefsHyperref{ref65}{order} \(\Omega(n/\log^5 n)\) \NoCaseChange{\protect\cite[{Thm. 3.1}]{cite581}}.
The code is an approximate non-stabilizer QLWC code since the Hamiltonian consists of non-commuting 9-local non-Pauli projectors, with each qubit acted on by \flmRefsHyperref{ref65}{order} \(O( \text{polylog}(n) )\) projectors.

The code is constructed by converting the encoding circuit of a Brown-Fawzi random Clifford-circuit code into a Hamiltonian using the spacetime circuit-to-Hamiltonian construction \NoCaseChange{\protect\cite{cite3512,cite3513}} (a generalization of the Feynman-Kitaev clock construction \NoCaseChange{\protect\cite{cite1634}}).
The ground-state subspace of this Hamiltonian is the \(\epsilon\)-approximate code with infidelity of recovery \(\epsilon = O( 1/\text{polylog}(n) )\).

Using Markov-chain techniques, the gap of the Hamiltonian can be proven to be of \flmRefsHyperref{ref65}{order} \(\Omega(D^{-2}n^{-3.09}\log^{-6} n)\) for an \(n\)-qubit input circuit of depth \(D\).

\codefieldsection{Protection}
Circuit-to-Hamiltonian approximate codes have nontrivial \flmRefsHyperref{ref2559}{codespace complexity} \NoCaseChange{\protect\cite{cite2564}}.

\codefieldsection{Encoding}
\begin{eczvaluelist}
\item\relax There exists a circuit of size polynomial in \(n\) whose terms act on at most \(\log (n)+2\) qubits \NoCaseChange{\protect\cite[{Thm. 3.3}]{cite581}}.
\end{eczvaluelist}
\codefieldsection{Decoding}
\begin{eczvaluelist}
\item\relax Local detection of Pauli errors can be done using circuits of depth of \flmRefsHyperref{ref65}{order} \(O( \text{polylog}(n) )\) based on exact decoders for the Brown-Fawzi code \NoCaseChange{\protect\cite[{Lemma 3.2}]{cite581}}.
\end{eczvaluelist}
\codefieldsection{Parents}
\begin{eczvaluelist}
\item\relax
\flmRefsHyperref[eczindexfamilyrel]{code:qubits_into_qubits}{Qubit code}\item\relax
\flmRefsHyperref[eczindexfamilyrel]{code:approximate_qecc}{Approximate quantum error-correcting code (AQECC)}\item\relax
\flmRefsHyperref[eczindexfamilyrel]{code:frustration_free}{Frustration-free Hamiltonian code} --- Circuit-to-Hamiltonian approximate codes form the ground-state space of a frustration-free non-commuting projector Hamiltonian whose projectors are constant weight, but such that each physical qubit is acted on by \flmRefsHyperref{ref65}{order} \(O( \text{polylog}(n) )\) projectors.
\end{eczvaluelist}
\codefieldsection{Cousins}
\begin{eczvaluelist}
\item\relax
\flmRefsHyperref[eczindexfamilyrel]{code:qlwc}{Quantum low-weight check (QLWC) code} --- The circuit-to-Hamiltonian code construction yields approximate codes whose distance and logical-qubit number are both of \flmRefsHyperref{ref65}{order} \(\Omega(n/\log^5 n)\) \NoCaseChange{\protect\cite[{Thm. 3.1}]{cite581}}.
These codes are approximate non-stabilizer QLWC codes since the Hamiltonian consists of non-commuting 9-local non-Pauli projectors, with each qubit acted on by \flmRefsHyperref{ref65}{order} \(O( \text{polylog}(n) )\) projectors.
\item\relax
\flmRefsHyperref[eczindexfamilyrel]{code:nonlocal_lowdepth}{Brown-Fawzi Clifford-circuit code} --- Circuit-to-Hamiltonian approximate codes are constructed by converting the encoding circuit of a Brown-Fawzi random Clifford-circuit code into a Hamiltonian using the spacetime circuit-to-Hamiltonian construction \NoCaseChange{\protect\cite{cite3512,cite3513}}.
\end{eczvaluelist}
\eczhbkcontributors{ \eczhuVVA }
\endeczcode

\eczcode{classical_product}{Classical-product code}{~\NoCaseChange{\protect\cite{cite3525,cite204,cite833}}}
\codefieldsection{Description}
A QLDPC qubit CSS code constructed by separately constructing the \(X\) and \(Z\) check matrices using product constructions from classical codes. A particular \(\llbracket 512,174,8\rrbracket \) code performed well \NoCaseChange{\protect\cite{cite204}} against erasure and depolarizing noise when compared to other notable CSS codes, such as the asymptotically good quantum Tanner codes. These codes have been generalized to the \textit{intersecting subset code} family \NoCaseChange{\protect\cite{cite833}}.

For example, letting \(H_i^x\) and \(H_i^z\) be the \(X\)- and \(Z\)-check matrices of CSS codes \(C_i\) with \(i\in\{1,2,3,4\}\), the 2-fold symmetric classical product code is given by
\flmMathEnvironment{align}{}{
H_{\otimes}^x &:=\left(\begin{array}{c}
H_1^x \otimes H_2^x \otimes I \otimes I \\
I \otimes I \otimes H_3^x \otimes H_4^x
\end{array}\right) \\
H_{\otimes}^z &:=\left(\begin{array}{c}
H_1^z \otimes I \otimes H_3^z \otimes I \\
I \otimes H_2^z \otimes I \otimes H_4^z
\end{array}\right)~.
}
\codefieldsection{Parents}
\begin{eczvaluelist}
\item\relax
\flmRefsHyperref[eczindexfamilyrel]{code:qubit_css}{Qubit CSS code}\item\relax
\flmRefsHyperref[eczindexfamilyrel]{code:qldpc}{Qubit QLDPC code}\end{eczvaluelist}
\codefieldsection{Cousins}
\begin{eczvaluelist}
\item\relax
\flmRefsHyperref[eczindexfamilyrel]{code:quantum_tanner}{Quantum Tanner code} --- A \(\llbracket 512,174,8\rrbracket \) classical-product code performed well \NoCaseChange{\protect\cite{cite204}} against erasure and depolarizing noise when compared to a member of an asymptotically good quantum Tanner code family.
\item\relax
\flmRefsHyperref[eczindexfamilyrel]{code:parity_check}{\([n,n-1,2]\) Single parity-check (SPC) code} --- SPC codes are used as component codes in classical-product code constructions.
\item\relax
\flmRefsHyperref[eczindexfamilyrel]{code:tensor}{Tensor-product code} --- Tensor-product codes are utilized in classical-product code constructions.
\end{eczvaluelist}
\eczhbkcontributors{ Hengyun (Harry) Zhou, \eczhuVVA }
\endeczcode

\eczcode{clifford-deformed_surface}{Clifford-deformed surface code (CDSC)}{~\NoCaseChange{\protect\cite{cite2625}}}
\codefieldsection{Description}
A generally non-CSS derivative of the surface code defined by applying a translationally invariant constant-depth \flmRefsHyperref{ref409}{Clifford circuit} to the original (CSS) surface code.
Unlike the surface code, CDSCs include codes whose thresholds and subthreshold performance are enhanced under noise biased towards dephasing.
Examples of CDSCs include the XY code, XZZX code, and random CDSCs.

\codefieldsection{Protection}
As a stabilizer code, \(\llbracket n=O(d^2), k=O(1), d\rrbracket \).
\codefieldsection{Fault Tolerance}
\begin{eczvaluelist}
\item\relax In order to leverage the benefits of CDSCs into practical universal computation, we have to implement syndrome measurement circuits and fault-tolerant logical gates in a bias-preserving way.
\end{eczvaluelist}
\codefieldsection{Code Capacity Threshold}
\begin{eczvaluelist}
\item\relax Depolarizing noise: the threshold under ML decoding corresponds to the value of a critical point of the weight-two (two-body) two-dimensional random-bond Ising model (RBIM) on the Nishimori line \NoCaseChange{\protect\cite{cite3526,cite480,cite2625}}. Utilizing this statistical mechanical mapping yields a phase diagram for a CDSC.
\item\relax A class of random CDSCs, parametrized by the probabilities \(\Pi_{XZ},~ \Pi_{YZ}\) of \(X\leftrightarrow Z\) and \(Y\leftrightarrow Z\) Pauli permutations, respectively, has \(50\%\) code capacity threshold at infinite \(Z\) bias. Certain translation-invariant CDSCs such as the XY code and the XZZX code also have \(50\%\) code capacity threshold at infinite \(Z\) bias.
\item\relax XZZX code and the \((0.5,\Pi_{YZ})\) random CDSCs have a \(50\%\) code capacity threshold for noise infinitely biased towards either Pauli-\(X\), \(Y\), or \(Z\) errors.
\end{eczvaluelist}
\codefieldsection{Parents}
\begin{eczvaluelist}
\item\relax
\flmRefsHyperref[eczindexfamilyrel]{code:qldpc}{Qubit QLDPC code}\item\relax
\flmRefsHyperref[eczindexfamilyrel]{code:quantum_double_abelian}{Abelian quantum-double stabilizer code} --- When treated as ground states of the code Hamiltonian, surface codewords realize \(\mathbb{Z}_2\) topological order, a topological phase of matter that also exists in \(\mathbb{Z}_2\) lattice gauge theory \NoCaseChange{\protect\cite{cite3527}}. Local Clifford deformation preserves this topological order.
\end{eczvaluelist}
\codefieldsection{Child}
\begin{eczvaluelist}
\item\relax
\flmRefsHyperref[eczindexfamilyrel]{code:surface}{Kitaev surface code} --- CDSC codes are deformations of the surface code via constant-depth \flmRefsHyperref{ref409}{Clifford circuits} that may not be CSS.
\end{eczvaluelist}
\codefieldsection{Cousins}
\begin{eczvaluelist}
\item\relax
\flmRefsHyperref[eczindexfamilyrel]{code:dynamic_gen}{Dynamically generated QECC} --- To create CDSCs, a dynamical process is applied on top of the surface code \NoCaseChange{\protect\cite{cite2625}}.
\item\relax
\flmRefsHyperref[eczindexfamilyrel]{code:random_stabilizer}{Random stabilizer code} --- Many useful CDSCs are constructed using random \flmRefsHyperref{ref409}{Clifford circuits}.
\item\relax
\flmRefsHyperref[eczindexfamilyrel]{code:3d_stabilizer}{3D lattice stabilizer code} --- Applying Clifford deformations to various 3D stabilizer codes, including the 3D surface code, 3D color code, X-cube model code, and Sierpinski prism model code, yields a \(50\%\) code capacity threshold under infinitely biased Pauli noise \NoCaseChange{\protect\cite{cite2626}}.
\item\relax
\flmRefsHyperref[eczindexfamilyrel]{code:asymmetric_qecc}{Asymmetric quantum code (AQC)} --- Random Clifford deformation can improve performance of surface codes against biased noise \NoCaseChange{\protect\cite{cite2625,cite2626}}.
\item\relax
\flmRefsHyperref[eczindexfamilyrel]{code:compass_model}{Compass code} --- Clifford deformation can enhance the performance of compass codes against biased noise \NoCaseChange{\protect\cite{cite2651}}.
\end{eczvaluelist}
\eczhbkcontributors{ Aleksander Kubica, Liang Jiang, Steven T. Flammia, Michael Gullans, Arpit Dua, \eczhuVVA }
\endeczcode

\eczcode{clifford_hierarchy}{Clifford-hierarchy stabilizer code}{~\NoCaseChange{\protect\cite{cite725}}}
\codefieldsection{Description}
A qubit code whose codespace is a joint eigenspace of a subset of operators in the \flmRefsHyperref{ref2118}{Clifford hierarchy}.
The stabilizing set, which need not be a group, contains Pauli strings and operators at any level \(m\) of the \flmRefsHyperref{ref2118}{Clifford hierarchy}, generalizing qubit stabilizer codes (\(m=1\)) and \textit{Clifford stabilizer codes} (\(m=2\)).

Clifford-hierarchy codes in \(D\) spatial dimensions include \((D+1)\)-dimensional Dijkgraaf-Witten gauge theories with non-Abelian topological order \NoCaseChange{\protect\cite{cite725}}.
A \(D\)-dimensional code can be constructed from a twisted \(\mathbb{Z}_2^{D+1}\) gauge theory with Dijkgraaf-Witten twist \((-1)^{\int a_1 \cup a_2 \cup \cdots \cup a_{D+1}}\), where the stabilizers include gates at the \(D\)th level of the Clifford hierarchy in addition to Pauli \(X\) operators.

\codefieldsection{Transversal and Permutation-Based Gates}
\begin{eczvaluelist}
\item\relax A transversal logical \(\text{diag}(1, e^{i2\pi/2^D})\) gate at the \(D\)th level of the Clifford hierarchy in \((D-1)\) spatial dimensions \NoCaseChange{\protect\cite{cite725}}.
\end{eczvaluelist}
\codefieldsection{Parent}
\begin{eczvaluelist}
\item\relax
\flmRefsHyperref[eczindexfamilyrel]{code:qubits_into_qubits}{Qubit code}\end{eczvaluelist}
\codefieldsection{Children}
\begin{eczvaluelist}
\item\relax
\flmRefsHyperref[eczindexfamilyrel]{code:cubic_theory}{Cubic theory code} --- Cubic theory codes are joint eigenspaces of commuting non-Pauli stabilizers built from Pauli \(X\), Pauli \(Z\), and \(CZ\) operators, placing them at the second level of the Clifford hierarchy.
\item\relax
\flmRefsHyperref[eczindexfamilyrel]{code:invertible}{Chen-Hsin invertible-order code} --- Chen-Hsin invertible-order code Hamiltonian terms include Pauli and \(CZ\) operators, making them Clifford stabilizer codes.
\item\relax
\flmRefsHyperref[eczindexfamilyrel]{code:xp_stabilizer}{XP stabilizer code} --- XP stabilizer codes are joint eigenspaces of operators in the binary dihedral group, a subgroup consisting of Pauli strings and elements of a level of the \flmRefsHyperref{ref2118}{Clifford hierarchy}.
\end{eczvaluelist}
\codefieldsection{Cousins}
\begin{eczvaluelist}
\item\relax
\flmRefsHyperref[eczindexfamilyrel]{code:clifford_group}{Clifford group} --- Clifford-hierarchy codes are joint eigenspaces of subsets of the \flmRefsHyperref{ref2118}{Clifford hierarchy}, whose second level is the \flmRefsHyperref{ref409}{Clifford group}.
\item\relax
\flmRefsHyperref[eczindexfamilyrel]{code:yetter_gauge_theory}{Two-gauge theory code} --- Clifford-hierarchy codes in \(D\) spatial dimensions include \((D+1)\)-dimensional Dijkgraaf-Witten gauge theories with non-Abelian topological order \NoCaseChange{\protect\cite{cite725}}.
A \(D\)-dimensional code can be constructed from a twisted \(\mathbb{Z}_2^{D+1}\) gauge theory with Dijkgraaf-Witten twist \((-1)^{\int a_1 \cup a_2 \cup \cdots \cup a_{D+1}}\), where the stabilizers include gates at the \(D\)th level of the Clifford hierarchy in addition to Pauli \(X\) operators.

\end{eczvaluelist}
\eczhbkcontributors{ \eczhuVVA }
\endeczcode

\eczcode{cluster_state}{Cluster-state code}{~\NoCaseChange{\protect\cite{cite3528}}}
\codefieldsection{Alternative Names}
\begin{eczvaluelist}
\item\relax Graph-state code
\end{eczvaluelist}
\eczhIndexCodeAliasName{cluster_state}{Graph-state code}
\codefieldsection{Description}
A code based on a cluster state (a.k.a. graph state) and often used in measurement-based quantum computation (MBQC) \NoCaseChange{\protect\cite{cite428,cite429}} (a.k.a. one-way quantum processing), which substitutes the temporal dimension necessary for decoding a conventional code with a spatial dimension.
This is done by encoding the computation into the features of the cluster state's graph.

Cluster states are stabilizer states defined on a graph.
There is one stabilizer generator \(S_j\) per graph vertex \(j\) of the form
\flmMathEnvironment{align}{}{
  S_j = X_{j} \prod_{k\in N(j)} Z_k~,
}
where the neighborhood \(N(j)\) is the set of vertices which share an edge with \(j\).

A type of cluster-state code can be built from a cluster state by applying the CWS construction using a linear binary code, in which codewords are obtained by applying \(Z\)-type operators defined by the code to the cluster state; see, e.g., Ref. \NoCaseChange{\protect\cite{cite3529}}. \flmRefsHyperref{ref672}{Pure} CWS codes can be constructed from uniform graphs \NoCaseChange{\protect\cite{cite2938}}.

By contrast, MBQC schemes utilize single cluster states.
An MBQC scheme can be constructed out of any qubit CSS code (via \textit{foliation} \NoCaseChange{\protect\cite{cite3530}}; see also Ref. \NoCaseChange{\protect\cite{cite3531}}) or qubit stabilizer code \NoCaseChange{\protect\cite{cite3532}}.
In the fault-complex formalism, foliation of a CSS code is expressed as a homological product of the code's chain complex with a repetition-code complex \NoCaseChange{\protect\cite{cite3176}}.
The original one-way MBQC scheme uses a two-dimensional cluster state \NoCaseChange{\protect\cite{cite428,cite429}}.
Fault-tolerant MBQC based on topological error correction uses the three-dimensional RBH cluster state on the bcc lattice \NoCaseChange{\protect\cite{cite3533,cite3534}}.

Since they are stabilizer states, two cluster states are considered local-Clifford equivalent if they can be mapped into each other under a tensor product of arbitrary \flmRefsHyperref{ref409}{single-qubit Clifford} operations.
For graph states, any such equivalence can be generated by a finite sequence of local complementations (LCs), where local complementation about a vertex complements the induced subgraph on its neighborhood \NoCaseChange{\protect\cite{cite3535,cite3536}}.
More generally, any two local-unitary equivalent cluster states can be mapped into each other by operations at some level of the hierarchy \NoCaseChange{\protect\cite{cite3537}}, meaning that general tensor-product unitaries are not needed.
Cluster states on \(\leq 19\) qubits are LC equivalent if and only if they are local-unitary equivalent \NoCaseChange{\protect\cite{cite3537}}, and there exists a cluster state on \(27\) qubits that is local-unitary but not LC equivalent \NoCaseChange{\protect\cite{cite3538,cite3539,cite3540}}.
Local complementation has been generalized to graphically characterize local unitary equivalence under local Clifford operations and, more generally, at other levels of the \flmTerm{term}{ref694}{}{Clifford hierarchy} \NoCaseChange{\protect\cite{cite3537}}.

\codefieldsection{Protection}
Protection is related to the stabilizer code underlying the cluster state.
There is no physical error correction, and decoding output is simply used to update the Pauli frame.

There exist necessary and sufficient conditions for a family of cluster states to exhibit \flmRefsHyperref{ref2675}{TQO-1} \NoCaseChange{\protect\cite{cite3156}}.
\flmRefsHyperref{ref672}{Quantum weight enumerators} of cluster state codes are known as sector weights \NoCaseChange{\protect\cite{cite3541,cite3542,cite3543,cite3544}}.

Potential and limits of MBQC using probabilistic gates has been studied \NoCaseChange{\protect\cite{cite3545}}.
Cluster states can be optimized to be robust against qubit erasure \NoCaseChange{\protect\cite{cite3546}}.

\codefieldsection{Encoding}
\begin{eczvaluelist}
\item\relax Initialization of all qubits in the \(|+\rangle\) state and action of \(CZ\) gates along the edges of the graph.
\item\relax Optimal-depth preparation based on graph coloring \NoCaseChange{\protect\cite{cite3547}}.
\item\relax A cluster-like state, or a state that is in the same SPT phase as a cluster state, can be prepared in finite time \NoCaseChange{\protect\cite{cite3094}}.
\item\relax ZX calculus based encoder representation \NoCaseChange{\protect\cite{cite3529}}.
\end{eczvaluelist}
\codefieldsection{Gates}
\begin{eczvaluelist}
\item\relax In the original one-way MBQC scheme on a square-lattice cluster state, \(Z\)-basis measurements remove qubits from the resource, \(X\)-basis measurements teleport quantum information along a wire, adaptive equatorial-basis measurements on a five-qubit chain implement arbitrary single-qubit \(SU(2)\) rotations, and a four-qubit pattern implements CNOT. Random measurement outcomes generate Pauli byproduct operators that are tracked classically and can modify later measurement bases \NoCaseChange{\protect\cite{cite429}}.
\item\relax The computation is encoded in pre-determined fashion via topological features of the cluster state's graph, such as boundaries, defects, or twists. Such features can be created using \(Z\)-type measurements, which effectively cut a qubit off from the cluster state. Non-Clifford gates are performed by inserting \flmRefsHyperref{ref409}{non-Clifford} states into particular \textit{singular} qubits. More generally, any gate protocol of a qubit stabilizer code yields an MBQC protocol \NoCaseChange{\protect\cite{cite3532}}. To perform the computation, subsets qubits are measured, e.g., along one two-dimensional slice of a 3D lattice for each time step. This effectively teleports the logical information into the remaining unmeasured portion of the cluster state. The computation terminates after all qubits are measured.
The entire cluster state does not need to be created at the start of the computation. Instead, the portion of the cluster state in the extra dimension can be initialized as the computation progresses.
\item\relax Single-qubit Clifford operations mapping one cluster state to another can be realized as local complementations acting on the underlying graph \NoCaseChange{\protect\cite{cite3536}}. In one-way computation, Clifford gates can be parallelized \NoCaseChange{\protect\cite{cite3548}}.
\end{eczvaluelist}
\codefieldsection{Decoding}
\begin{eczvaluelist}
\item\relax MBQC syndrome extraction is performed by multiplying certain single-qubit \(X\)-type measurements, which yield syndrome values.
\end{eczvaluelist}
\codefieldsection{Fault Tolerance}
\begin{eczvaluelist}
\item\relax Photonic architecture \NoCaseChange{\protect\cite{cite3549}}.
\item\relax Generalized foliation procedures exist for noise-bias preserving MBQC \NoCaseChange{\protect\cite{cite2634}}.
\item\relax Fault complexes provide a representation of foliated cluster-state protocols that makes fault distances, logical correlations, and related fault-tolerance properties computable from the underlying CSS code \NoCaseChange{\protect\cite{cite3176}}.
\end{eczvaluelist}
\codefieldsection{Code Capacity Threshold}
\begin{eczvaluelist}
\item\relax Independent \(X,Z\) noise: \(p_X = 2.9\%\) under MWPM decoding \NoCaseChange{\protect\cite{cite3533}}. The threshold under ML decoding corresponds to the value of a critical point of the 3D random-plaquette \(\mathbb{Z}_2\) gauge theory (3D-RPGM) via the statistical mechanical mapping \NoCaseChange{\protect\cite{cite480}}, calculated to be \(3.3 \%\) \NoCaseChange{\protect\cite{cite3550}} (see also \NoCaseChange{\protect\cite{cite3551}}).
\end{eczvaluelist}
\codefieldsection{Threshold}
\begin{eczvaluelist}
\item\relax There is a simple proof of a threshold for MBQC \NoCaseChange{\protect\cite{cite3552}}.
\end{eczvaluelist}
\codefieldsection{Realizations}
\begin{eczvaluelist}
\item\relax Polarizations of photons: quantum computation \NoCaseChange{\protect\cite{cite3553,cite3554}} and single-qubit error correction on an 8-qubit cluster state \NoCaseChange{\protect\cite{cite3555}}.
\end{eczvaluelist}
\codefieldsection{Notes}
\begin{eczvaluelist}
\item\relax See Refs. \NoCaseChange{\protect\cite{cite3556,cite3557}} for a review of cluster states and their applications.
\item\relax Cluster states are useful for entanglement purification \NoCaseChange{\protect\cite{cite3558}}.
\item\relax The original one-way MBQC paper also noted that universal computation remains possible on irregular occupied 2D clusters above the percolation threshold, and that finite clusters can be reused by re-entangling successive computation segments \NoCaseChange{\protect\cite{cite429}}.
\item\relax Graph-State Compass (GSC) Python software library for manipulating local Clifford equivalence classes of cluster states \NoCaseChange{\protect\cite{cite3540,cite3559,cite3560}}.
\end{eczvaluelist}
\codefieldsection{Parents}
\begin{eczvaluelist}
\item\relax
\flmRefsHyperref[eczindexfamilyrel]{code:qubit_stabilizer}{Qubit stabilizer code} --- Cluster-state codes are particular qubit stabilizer codes. Any qubit stabilizer code is equivalent to a graph quantum code via a single-qubit \flmRefsHyperref{ref409}{Clifford circuit} \NoCaseChange{\protect\cite{cite3561}} (see also \NoCaseChange{\protect\cite{cite3536,cite867}}). As a corollary, any qubit stabilizer state is equivalent to a cluster state under a single-qubit \flmRefsHyperref{ref409}{Clifford circuit} \NoCaseChange{\protect\cite{cite3536}\protect\cite[{Appx. A}]{cite3562}}. There are algorithms that determine whether two stabilizer states are equivalent that work by checking whether their corresponding cluster states are equivalent \NoCaseChange{\protect\cite{cite3563,cite3564,cite3565}}. Any fault-tolerant scheme based on qubit stabilizer codes can be mapped into a cluster-state based MBQC protocol \NoCaseChange{\protect\cite{cite3532}}.
\item\relax
\flmRefsHyperref[eczindexfamilyrel]{code:cws}{Codeword stabilized (CWS) code} --- A type of cluster-state code can be built from a cluster state by applying the CWS construction using a linear binary code, in which codewords are obtained by applying \(Z\)-type operators defined by the code to the cluster state; see, e.g., Ref. \NoCaseChange{\protect\cite{cite3529}}. \flmRefsHyperref{ref672}{Pure} CWS codes can be constructed from uniform graphs \NoCaseChange{\protect\cite{cite2938}}.
\item\relax
\flmRefsHyperref[eczindexfamilyrel]{code:qudit_cluster_state}{Modular-qudit cluster-state code} --- Modular-qudit cluster-state codes reduce to cluster-state codes for \(q=2\).
\end{eczvaluelist}
\codefieldsection{Children}
\begin{eczvaluelist}
\item\relax
\flmRefsHyperref[eczindexfamilyrel]{code:rbh}{Raussendorf-Bravyi-Harrington (RBH) cluster-state code}\item\relax
\flmRefsHyperref[eczindexfamilyrel]{code:square_lattice_cluster}{Square-lattice cluster-state code}\item\relax
\flmRefsHyperref[eczindexfamilyrel]{code:tree_cluster}{Tree cluster-state code}\end{eczvaluelist}
\codefieldsection{Cousins}
\begin{eczvaluelist}
\item\relax
\flmRefsHyperref[eczindexfamilyrel]{code:fusion}{Fusion-based quantum computing (FBQC) code} --- FBQC and MBQC are both computational models in which computation is done by measuring resource states (which are qubit stabilizer states). The difference between the two is in how the states are constructed. FBQC is based exclusively on two-qubit measurements tailored to photonic platforms. These measurements require a foliation with more qubits but one which can be built by fusing smaller modules.
\item\relax
\flmRefsHyperref[eczindexfamilyrel]{code:qubit_css}{Qubit CSS code} --- A resource cluster state can be constructed out of any qubit CSS code via foliation. Conversely, CSS codes can be constructed out of cluster states \NoCaseChange{\protect\cite{cite3530}}. 
In the fault-complex formalism, foliation of a CSS code is expressed as a homological product of the code's chain complex with a repetition-code complex \NoCaseChange{\protect\cite{cite3176}}.

\item\relax
\flmRefsHyperref[eczindexfamilyrel]{code:homological_product}{Homological product code} --- In the fault-complex formalism, foliation of a CSS code is expressed as a homological product of the code's chain complex with a repetition-code complex \NoCaseChange{\protect\cite{cite3176}}.

\item\relax
\flmRefsHyperref[eczindexfamilyrel]{code:quantum_repetition}{Quantum repetition code} --- GHZ states can be used as resource states for MBQC protocols \NoCaseChange{\protect\cite{cite3566,cite3567,cite3568}}.
In the fault-complex formalism, foliation of a CSS code is expressed as a homological product of the code's chain complex with a repetition-code complex \NoCaseChange{\protect\cite{cite3176}}.

\item\relax
\flmRefsHyperref[eczindexfamilyrel]{code:xp_stabilizer}{XP stabilizer code} --- XP stabilizer states are in one-to-one correspondence with weighted hypergraph states \NoCaseChange{\protect\cite{cite798,cite768}}, which generalize both weighted graph states \NoCaseChange{\protect\cite{cite3569,cite3556,cite3570}} and hypergraph states \NoCaseChange{\protect\cite{cite3571,cite3572,cite3573}}. The latter can also be utilized in MBQC schemes \NoCaseChange{\protect\cite{cite3574,cite3575}}.
\item\relax
\flmRefsHyperref[eczindexfamilyrel]{code:dynamic_gen}{Dynamically generated QECC} --- MBQC is done using a measurement-based dynamical process.
\item\relax
\flmRefsHyperref[eczindexfamilyrel]{code:spt}{Symmetry-protected topological (SPT) code} --- Cluster states defined on various lattices are representatives of SPT phases, and states realizing these phases can be resources for MBQC. 
In 1D, cluster states are examples of SPT phases with global symmetries \NoCaseChange{\protect\cite{cite3078,cite3079,cite3080,cite3081,cite3072}} and enable MBQC on a single qubit \NoCaseChange{\protect\cite{cite428,cite429}}. 
The square-lattice cluster state, which is the prototypical resource for universal MBQC \NoCaseChange{\protect\cite{cite428,cite429}}, and other 2D cluster states \NoCaseChange{\protect\cite{cite3082,cite3083,cite3084}} have SPT order protected by subsystem symmetries \NoCaseChange{\protect\cite{cite3085,cite3086,cite3082}}.
States like AKLT states and SPT fixed-point states can be efficiently converted into cluster states using local measurements and subsequently used as resources for MBQC \NoCaseChange{\protect\cite{cite3087,cite3079,cite3088,cite3089,cite3090,cite3091}}.
In 3D, cluster states belong to SPT phases protected by higher-form symmetries \NoCaseChange{\protect\cite{cite3092}} and enable universal fault-tolerant MBQC \NoCaseChange{\protect\cite{cite3093}}.
A cluster-like state, or a state that is in the same SPT phase as a cluster state, can be prepared in finite time \NoCaseChange{\protect\cite{cite3094}}. Cluster states can be created on various lattices \NoCaseChange{\protect\cite{cite3095}}.

\item\relax
\flmRefsHyperref[eczindexfamilyrel]{code:translationally_invariant_stabilizer}{Lattice stabilizer code} --- Cluster states defined on various lattices are representatives of SPT phases, and states realizing these phases can be resources for MBQC. 
In 1D, cluster states are examples of SPT phases with global symmetries \NoCaseChange{\protect\cite{cite3078,cite3079,cite3080,cite3081,cite3072}} and enable MBQC on a single qubit \NoCaseChange{\protect\cite{cite428,cite429}}. 
The square-lattice cluster state, which is the prototypical resource for universal MBQC \NoCaseChange{\protect\cite{cite428,cite429}}, and other 2D cluster states \NoCaseChange{\protect\cite{cite3082,cite3083,cite3084}} have SPT order protected by subsystem symmetries \NoCaseChange{\protect\cite{cite3085,cite3086,cite3082}}.
States like AKLT states and SPT fixed-point states can be efficiently converted into cluster states using local measurements and subsequently used as resources for MBQC \NoCaseChange{\protect\cite{cite3087,cite3079,cite3088,cite3089,cite3090,cite3091}}.
In 3D, cluster states belong to SPT phases protected by higher-form symmetries \NoCaseChange{\protect\cite{cite3092}} and enable universal fault-tolerant MBQC \NoCaseChange{\protect\cite{cite3093}}.
A cluster-like state, or a state that is in the same SPT phase as a cluster state, can be prepared in finite time \NoCaseChange{\protect\cite{cite3094}}. Cluster states can be created on various lattices \NoCaseChange{\protect\cite{cite3095}}.

\item\relax
\flmRefsHyperref[eczindexfamilyrel]{code:t-designs}{\(t\)-design} --- Kerdock codes correspond to cluster states, and the corresponding Clifford-group automorphisms of this set form a particular group \NoCaseChange{\protect\cite{cite934}} that is a unitary 2-design on \(U(2^n)\) \NoCaseChange{\protect\cite{cite935}}. As such, cluster states form complex projective 2-designs on \(\mathbb{C}P^{2^n-1}\). These are useful in matrix-vector multiplication \NoCaseChange{\protect\cite{cite936}}.
\item\relax
\flmRefsHyperref[eczindexfamilyrel]{code:complex_projective}{Complex projective space code} --- Kerdock codes correspond to cluster states, and the corresponding Clifford-group automorphisms of this set form a particular group \NoCaseChange{\protect\cite{cite934}} that is a unitary 2-design on \(U(2^n)\) \NoCaseChange{\protect\cite{cite935}}. As such, cluster states form complex projective 2-designs on \(\mathbb{C}P^{2^n-1}\). These are useful in matrix-vector multiplication \NoCaseChange{\protect\cite{cite936}}.
\item\relax
\flmRefsHyperref[eczindexfamilyrel]{code:surface}{Kitaev surface code} --- Foliating the surface code yields a cluster state on the Lieb lattice \NoCaseChange{\protect\cite{cite3576,cite469}}. See also the discussion of the embedding on \NoCaseChange{\protect\cite[{pg. 8}]{cite423}}.
\item\relax
\flmRefsHyperref[eczindexfamilyrel]{code:kerdock}{Kerdock code} --- Kerdock codes correspond to cluster states, and the corresponding Clifford-group automorphisms of this set form a particular group \NoCaseChange{\protect\cite{cite934}} that is a unitary 2-design on \(U(2^n)\) \NoCaseChange{\protect\cite{cite935}}. As such, cluster states form complex projective on 2-designs \(\mathbb{C}P^{2^n}\). These are useful in matrix-vector multiplication \NoCaseChange{\protect\cite{cite936}}.
\item\relax
\flmRefsHyperref[eczindexfamilyrel]{code:clifford_group}{Clifford group} --- Kerdock codes correspond to cluster states, and the corresponding Clifford-group automorphisms of this set form a particular group \NoCaseChange{\protect\cite{cite934}} that is a unitary 2-design on \(U(2^n)\) \NoCaseChange{\protect\cite{cite935}}. As such, cluster states form complex projective 2-designs on \(\mathbb{C}P^{2^n}\). These are useful in matrix-vector multiplication \NoCaseChange{\protect\cite{cite936}}.
\item\relax
\flmRefsHyperref[eczindexfamilyrel]{code:dual_rail}{Dual-rail quantum code} --- The KLM protocol can be combined with cluster states in various ways to yield MBQC protocols \NoCaseChange{\protect\cite{cite3577,cite3578,cite3579}}; see review \NoCaseChange{\protect\cite{cite3580}}.
\item\relax
\flmRefsHyperref[eczindexfamilyrel]{code:gkp-cluster-state}{GKP CV-cluster-state code} --- GKP CV-cluster-state codes reduce to cluster-state codes concatenated with single-mode GKP codes \NoCaseChange{\protect\cite{cite415}} when all physical modes are initialized in GKP states.
\item\relax
\flmRefsHyperref[eczindexfamilyrel]{code:gkp_concatenated}{Concatenated GKP code} --- GKP codes have been concatenated with cluster-state codes \NoCaseChange{\protect\cite{cite415}}.
\item\relax
\flmRefsHyperref[eczindexfamilyrel]{code:topological}{Topological code} --- There exist necessary and sufficient conditions for a family of cluster states to exhibit the TQO-1 property \NoCaseChange{\protect\cite{cite3156}}.
\item\relax
\flmRefsHyperref[eczindexfamilyrel]{code:stab_15_1_3}{\(\llbracket 15,1,3\rrbracket \) quantum RM code} --- MBQC with the \(\llbracket 15,1,3\rrbracket \) code has been demonstrated in neutral atom arrays by the Lukin group \NoCaseChange{\protect\cite{cite3206}}.
\item\relax
\flmRefsHyperref[eczindexfamilyrel]{code:css_4_1_2}{\(\llbracket 4,1,2\rrbracket \) Leung-Nielsen-Chuang-Yamamoto (LNCY) code} --- A \(\llbracket 4,1,2\rrbracket \) LNCY code can be thought of as a cluster-state code \NoCaseChange{\protect\cite{cite3266}}.
\item\relax
\flmRefsHyperref[eczindexfamilyrel]{code:stab_5_1_3}{\(\llbracket 5,1,3\rrbracket \) Five-qubit perfect code} --- The five-qubit perfect code is equivalent via a single-qubit \flmRefsHyperref{ref409}{Clifford circuit} to a cluster-state code defined from a five-cycle (a.k.a. pentagon) graph and a classical repetition code \NoCaseChange{\protect\cite{cite852,cite3166,cite3322,cite868}\protect\cite[{Exam. 2}]{cite438}}.
\item\relax
\flmRefsHyperref[eczindexfamilyrel]{code:steane}{\(\llbracket 7,1,3\rrbracket \) Steane code} --- The Steane code is equivalent via a single-qubit Clifford unitary to a cluster-state code for a particular graph and classical code \NoCaseChange{\protect\cite[{Exam. 4}]{cite438}}. Four non-isomorphic graphs yield graph quantum codes that are equivalent to the Steane code under a single-qubit-\flmRefsHyperref{ref409}{Clifford circuit} \NoCaseChange{\protect\cite{cite867}}.
\item\relax
\flmRefsHyperref[eczindexfamilyrel]{code:shor_nine}{\(\llbracket 9,1,3\rrbracket \) Shor code} --- The Shor code admits a codeword that is the cluster state of a particular nine-vertex graph \NoCaseChange{\protect\cite{cite3322,cite868}}.
\item\relax
\flmRefsHyperref[eczindexfamilyrel]{code:haah_cubic}{Haah cubic code (CC)} --- A short-range entangled cluster-state model with fractal \(X\)-type symmetries on both sublattices can be built from the cubic-code gauging data. Gauging one sublattice yields, up to a local circuit, either the cubic code or its ungauged fractal-symmetry Ising model, while gauging both sublattices returns the cluster model up to local swaps and Hadamards \NoCaseChange{\protect\cite[{Secs. III.F, IV.C}]{cite464}}.
\item\relax
\flmRefsHyperref[eczindexfamilyrel]{code:bipartite_cyclic_cluster}{Bipartite cyclic cluster (BCC) code} --- BCC codes are obtained by applying Hadamard on the \(B\)-sublattice of a bipartite cluster state, converting CZ-gate preparation into CNOT-gate preparation \NoCaseChange{\protect\cite{cite440}}.
\item\relax
\flmRefsHyperref[eczindexfamilyrel]{code:concatenated_steane}{Concatenated Steane code} --- The cluster state corresponding to the concatenated Steane code has been worked out \NoCaseChange{\protect\cite{cite3535}}.
\item\relax
\flmRefsHyperref[eczindexfamilyrel]{code:xzzx}{XZZX surface code} --- XZZX surface code can be foliated for a noise-bias preserving MBQC \NoCaseChange{\protect\cite{cite2634}} or FBQC \NoCaseChange{\protect\cite{cite2635}} protocol; see also \NoCaseChange{\protect\cite{cite2636}}.
\end{eczvaluelist}
\eczhbkcontributors{ David T. Stephen, Yaron Jarach, \eczhuVVA }
\endeczcode

\eczcode{cws}{Codeword stabilized (CWS) code}{~\NoCaseChange{\protect\cite{cite853,cite852}}}
\codefieldsection{Description}
A code defined using a cluster state and a set of \(Z\)-type Pauli strings defined by a binary classical code.

The CWS construction takes in \( \mathcal{Q} = (\mathcal{G},\mathcal{C}) \), where \(\mathcal{G}\) is a graph, and where \(\mathcal{C}\) is an \((n,K,d)\) binary code.
From the graph, we form the unique cluster state \( |\mathcal{G} \rangle \).
From the binary code, we form Pauli \(Z\)-type operators \( W_i = Z^{c_{i,1}} \otimes \cdots \otimes Z^{c_{i,n}} \), where \(c_{i,j} \) is the \(j\)-th coordinate of the \(i\)-th classical codeword.
The CWS codewords are then \( | i \rangle =  W_i | \mathcal{G} \rangle \).

The above definition corresponds to the \textit{standard form} of CWS codes.
Since any stabilizer state is equivalent to a cluster state under a single-qubit \flmRefsHyperref{ref409}{Clifford circuit} \NoCaseChange{\protect\cite{cite3536}\protect\cite[{Appx. A}]{cite3562}}, any code whose underlying state is a non-cluster stabilizer state can be rewritten in standard CWS form \NoCaseChange{\protect\cite{cite852}}.

The term CWS was coined in Ref. \NoCaseChange{\protect\cite{cite852}}, and their approach is equivalent to another approach \NoCaseChange{\protect\cite{cite853}} based on Boolean functions (see Ref. \NoCaseChange{\protect\cite{cite3581}}).
In an alternative convention (not used here), CWS codes are defined from an underlying stabilizer state that is not necessarily a cluster state.

\codefieldsection{Protection}
In standard form, error detection reduces to detecting the induced binary error patterns \(C_{S}(E)\) with the underlying classical code, together with the requirement that any error satisfying \(C_{S}(E)=0\) commute with every word operator \NoCaseChange{\protect\cite{cite852}}.
The code distance of \(\mathcal{Q} = ( \mathcal{G},\mathcal{C}) \) is upper bounded by the distance of the classical code \(\mathcal{C} \).
A CWS code is degenerate if and only if it is \flmRefsHyperref{ref672}{impure} \NoCaseChange{\protect\cite{cite3582}}.
The \flmRefsHyperref{ref672}{pure distance} is upper bounded by \(\delta + 1\), where \(\delta\) is the minimum degree of \(\mathcal{G}\) \NoCaseChange{\protect\cite{cite3583,cite3584}}.
For additive CWS codes realizable from a fixed graph \(\mathcal{G}\), Ref. \NoCaseChange{\protect\cite{cite438}} derives upper bounds on the distance and proves a Gilbert-Varshamov existence bound matching the standard pure-stabilizer bound up to the graph-state distance \(d'(\mathcal{G})\).

\codefieldsection{Encoding}
\begin{eczvaluelist}
\item\relax If the classical code \( \mathcal{C} \) has an encoder of complexity \(f(n)\), then the CWS code \( \mathcal{Q} = (\mathcal{G},\mathcal{C}) \) has an encoder of complexity \(\max( n^2, f(n) )\), obtained by preparing the graph state and applying the classical encoder \NoCaseChange{\protect\cite{cite852}}.
\item\relax Sequential encoder related to MBQC \NoCaseChange{\protect\cite{cite2938}}.
\end{eczvaluelist}
\codefieldsection{Decoding}
\begin{eczvaluelist}
\item\relax There is no known \textit{efficient} algorithm to decode \textit{non-additive} (non-stabilizer) CWS codes.
\item\relax Clustered bounded-distance decoder \NoCaseChange{\protect\cite{cite3585,cite3586,cite3587}}.
\item\relax Structured error recovery \NoCaseChange{\protect\cite{cite3582}}, which reduces to syndrome-based recovery for additive (i.e., stabilizer) CWS codes.
\end{eczvaluelist}
\codefieldsection{Notes}
\begin{eczvaluelist}
\item\relax See Ref. \NoCaseChange{\protect\cite{cite3167}} for an overview of CWS codes.
\end{eczvaluelist}
\codefieldsection{Parents}
\begin{eczvaluelist}
\item\relax
\flmRefsHyperref[eczindexfamilyrel]{code:non_stabilizer}{Union stabilizer (USt) code} --- Any CWS code can be written as a USt whose (\(K=1\)) stabilizer code is the cluster state and whose coset representatives are constructed from the binary classical code. Conversely, USt codes are equivalent to CWS codes via a single-qubit \flmRefsHyperref{ref409}{Clifford circuit} as follows \NoCaseChange{\protect\cite{cite3585,cite3587}\protect\cite[{Sec. 10.4}]{cite3167}}. The set of coset representatives of any USt can be extended to a larger set iterating over the underlying stabilizer code such that all codewords can be obtained from a single stabilizer state. Then, one can apply a single-qubit Clifford transformation to map said stabilizer state into a cluster state.
\item\relax
\flmRefsHyperref[eczindexfamilyrel]{code:qudit_cws}{Modular-qudit CWS code} --- Modular-qudit CWS codes reduce to CWS codes for \(q=2\).
\item\relax
\flmRefsHyperref[eczindexfamilyrel]{code:galois_cws}{Galois-qudit CWS code} --- Galois-qudit CWS codes reduce to CWS codes for \(q=2\).
\end{eczvaluelist}
\codefieldsection{Children}
\begin{eczvaluelist}
\item\relax
\flmRefsHyperref[eczindexfamilyrel]{code:ampdamp_cws}{Amplitude-damping CWS code}\item\relax
\flmRefsHyperref[eczindexfamilyrel]{code:rains}{\(\llparenthesis 2m+1,3 \times 2^{2m-3},2\rrparenthesis \) Rains code} --- The \(\llparenthesis 2m+1,3 \times 2^{2m-3},2\rrparenthesis \) qubit code family is a CWS family whose graph state is the union of the ring and Bell-pair graphs \NoCaseChange{\protect\cite{cite852,cite3166}}.
\item\relax
\flmRefsHyperref[eczindexfamilyrel]{code:ssw}{Smolin-Smith-Wehner (SSW) code} --- SSW codes can be formulated as CWS codes \NoCaseChange{\protect\cite{cite852,cite3166}}.
\item\relax
\flmRefsHyperref[eczindexfamilyrel]{code:qubit_10_24_3}{\(\llparenthesis 10,24,3\rrparenthesis \) qubit code} --- The \(\llparenthesis 10,24,3\rrparenthesis \) qubit code is a CWS code \NoCaseChange{\protect\cite{cite3166}}.
\item\relax
\flmRefsHyperref[eczindexfamilyrel]{code:qubit_9_12_3}{\(\llparenthesis 9,12,3\rrparenthesis \) qubit code} --- The \(\llparenthesis 9,12,3\rrparenthesis \) qubit code is a cyclic CWS code \NoCaseChange{\protect\cite{cite852,cite3166}}.
\item\relax
\flmRefsHyperref[eczindexfamilyrel]{code:cluster_state}{Cluster-state code} --- A type of cluster-state code can be built from a cluster state by applying the CWS construction using a linear binary code, in which codewords are obtained by applying \(Z\)-type operators defined by the code to the cluster state; see, e.g., Ref. \NoCaseChange{\protect\cite{cite3529}}. \flmRefsHyperref{ref672}{Pure} CWS codes can be constructed from uniform graphs \NoCaseChange{\protect\cite{cite2938}}.
\end{eczvaluelist}
\codefieldsection{Cousins}
\begin{eczvaluelist}
\item\relax
\flmRefsHyperref[eczindexfamilyrel]{code:movassagh_ouyang}{Movassagh-Ouyang Hamiltonian code} --- The Movassagh-Ouyang codes overlap the CWS codes but neither family is contained in the other \NoCaseChange{\protect\cite{cite1407}}.
\item\relax
\flmRefsHyperref[eczindexfamilyrel]{code:spacetime}{Spacetime code (STC)} --- CWS codes have been considered in the context of spacetime replication of quantum data \NoCaseChange{\protect\cite{cite512,cite2172}}, while STCs are designed to replicate classical data.
\item\relax
\flmRefsHyperref[eczindexfamilyrel]{code:quantum_concatenated}{Concatenated quantum code} --- CWS codes can be concatenated by applying generalized local complementation to their underlying graphs \NoCaseChange{\protect\cite{cite2701}}.
\item\relax
\flmRefsHyperref[eczindexfamilyrel]{code:ea_qubits_into_qubits}{EA qubit code} --- EA CWS codes have been formulated \NoCaseChange{\protect\cite{cite3588}}.
\item\relax
\flmRefsHyperref[eczindexfamilyrel]{code:ame}{Perfect-tensor code} --- CWS codes can be constructed from \((d-1)\)-uniform states \NoCaseChange{\protect\cite{cite2938}}.
\item\relax
\flmRefsHyperref[eczindexfamilyrel]{code:xp_stabilizer}{XP stabilizer code} --- The orbit representatives of XP codes play a similar role to the word operators of CWS codes, and non-XP-regular codes have a similar structure \NoCaseChange{\protect\cite{cite798}}.
\item\relax
\flmRefsHyperref[eczindexfamilyrel]{code:qubit_stabilizer}{Qubit stabilizer code} --- CWS codes whose underlying classical code is a linear binary code are qubit stabilizer codes containing a cluster-state codeword \NoCaseChange{\protect\cite{cite852,cite3266}}.
Conversely, stabilizer codes admit a CWS standard form \NoCaseChange{\protect\cite{cite852}}; at the state level, the underlying stabilizer state can be mapped to a cluster state by a single-qubit \flmRefsHyperref{ref409}{Clifford circuit} \NoCaseChange{\protect\cite{cite3536}\protect\cite[{Appx. A}]{cite3562}}.

\end{eczvaluelist}
\eczhbkcontributors{ Eric Kubischta, \eczhuVVA }
\endeczcode

\eczcode{cpc}{Coherent-parity-check (CPC) code}{~\NoCaseChange{\protect\cite{cite860,cite3282,cite3589}}}
\codefieldsection{Description}
A qubit stabilizer code for which two binary linear codes are used to directly construct encoding and decoding circuits against \(X\)- and \(Z\)-type errors, respectively, via ZX calculus \NoCaseChange{\protect\cite{cite3590,cite3591}}.
CPC codes can be obtained from numerical search \NoCaseChange{\protect\cite{cite3282}}.

\codefieldsection{Parent}
\begin{eczvaluelist}
\item\relax
\flmRefsHyperref[eczindexfamilyrel]{code:qubit_stabilizer}{Qubit stabilizer code} --- CPC codes are a type of stabilizer code. A teleported version of the CPC construction, the Clifford noise reduction (CliNR) scheme, can reduce noise in \flmRefsHyperref{ref409}{Clifford circuits} with Pauli measurements with at most a three-fold overhead in the number of qubits and gates \NoCaseChange{\protect\cite{cite3592,cite3593}}. There is a simple formula for the probability that a \flmRefsHyperref{ref409}{Clifford circuit} contains a logical error \NoCaseChange{\protect\cite{cite3589}}.
\end{eczvaluelist}
\codefieldsection{Children}
\begin{eczvaluelist}
\item\relax
\flmRefsHyperref[eczindexfamilyrel]{code:ring_cpc}{\(\llbracket 2^r+r, 2^r-r-2, 3\rrbracket \) Ring CPC code}\item\relax
\flmRefsHyperref[eczindexfamilyrel]{code:qubit_css}{Qubit CSS code} --- CSS codes are a subset of CPC codes \NoCaseChange{\protect\cite{cite860}}, with the latter not requiring the two classical codes to be related.
\end{eczvaluelist}
\codefieldsection{Cousins}
\begin{eczvaluelist}
\item\relax
\flmRefsHyperref[eczindexfamilyrel]{code:binary_linear}{Linear binary code} --- The CPC Construction uses two binary linear codes.
\item\relax
\flmRefsHyperref[eczindexfamilyrel]{code:hamming}{\([2^r-1,2^r-r-1,3]\) Hamming code} --- \textit{Tripartite CPC codes} are constructed from Hamming codes via the CPC construction \NoCaseChange{\protect\cite[{Thm. 4}]{cite860}}.
\item\relax
\flmRefsHyperref[eczindexfamilyrel]{code:stab_4_2_2}{\(\llbracket 4,2,2\rrbracket \) Four-qubit code} --- CPC gadgets for the \(\llbracket 4,2,2\rrbracket \) code have been implemented on the IBM 5Q superconducting device \NoCaseChange{\protect\cite{cite3282}}.
\end{eczvaluelist}
\eczhbkcontributors{ \eczhuVVA }
\endeczcode

\eczcode{color}{Color code}{~\NoCaseChange{\protect\cite{cite710,cite430}}}
\codefieldsection{Description}
Member of a family of qubit CSS codes defined on particular \(D\)-dimensional graphs.

In the colex realization introduced in \NoCaseChange{\protect\cite{cite430}}, qubits are placed on vertices of a \(D\)-colex, and for any integers \(p,q\in\{1,\dots,D-1\}\) with \(p+q=D\), \(Z\)-type stabilizers are attached to \((p+1)\)-cells while \(X\)-type stabilizers are attached to \((q+1)\)-cells. On a closed \(D\)-manifold, the resulting commuting-projector Hamiltonian encodes \(k=\binom{D}{p} h_p\) logical qubits, where \(h_p\) is the \(p\)th Betti number \NoCaseChange{\protect\cite{cite430}}.

One family is defined on a \(D\)-dimensional graph which satisfies two properties: (1) the graph is a homogeneous simplicial \(D\)-complex obtained as a triangulation of the interior of a \(D\)-simplex, and (2) the graph is \(D+1\)-colorable.
Qubits are placed on the \(D\)-simplices and generators are supported on suitable simplices \NoCaseChange{\protect\cite{cite3594,cite726,cite3416}}.
Admissible graphs can be obtained via a fattening procedure \NoCaseChange{\protect\cite{cite430}}.
See also a construction based on the more general quantum pin codes \NoCaseChange{\protect\cite{cite702}}.

\codefieldsection{Protection}
As with the surface code, the code distance depends on the specific kind of lattice used to define the code. More precisely, the distance depends on the homology of logical string operators \NoCaseChange{\protect\cite{cite3594}}.

\codefieldsection{Transversal and Permutation-Based Gates}
\begin{eczvaluelist}
\item\relax Some color codes on \(D\)-dimensional lattices can transversally implement a gate at the \(D\)th level of the \flmTerm{term}{ref694}{}{Clifford hierarchy} in the form of a \(Z\)-rotation by angle \(\pi/2^{D-1}\) \NoCaseChange{\protect\cite[{Fig. 3}]{cite726}}.
\end{eczvaluelist}
\codefieldsection{Decoding}
\begin{eczvaluelist}
\item\relax In contrast to the surface code, the color code can suffer from unremovable \flmRefsHyperref{ref3496}{hook errors} due to the specifics of its syndrome extraction circuits. Fault-tolerant decoders thus have to utilize additional flag qubits.
\end{eczvaluelist}
\codefieldsection{Fault Tolerance}
\begin{eczvaluelist}
\item\relax The 6D color code is a self-correcting quantum memory and admits fault-tolerant universal gate set in 7D \NoCaseChange{\protect\cite{cite3035}}.
\end{eczvaluelist}
\codefieldsection{Notes}
\begin{eczvaluelist}
\item\relax See Ref. \NoCaseChange{\protect\cite{cite3594}} for an overview of color codes.
\end{eczvaluelist}
\codefieldsection{Parents}
\begin{eczvaluelist}
\item\relax
\flmRefsHyperref[eczindexfamilyrel]{code:qldpc}{Qubit QLDPC code}\item\relax
\flmRefsHyperref[eczindexfamilyrel]{code:quantum_pin}{Quantum pin code} --- Color codes are special cases of quantum pin codes \NoCaseChange{\protect\cite[{Sec. II.E}]{cite702}}
\end{eczvaluelist}
\codefieldsection{Children}
\begin{eczvaluelist}
\item\relax
\flmRefsHyperref[eczindexfamilyrel]{code:stab_16_6_4}{\(\llbracket 16,6,4\rrbracket \) Tesseract color code} --- The tesseract color code is a 4D color code defined on a tesseract \NoCaseChange{\protect\cite{cite3212,cite101,cite862,cite2362}}.
\item\relax
\flmRefsHyperref[eczindexfamilyrel]{code:diagonal_clifford}{\(\llbracket 2^r-1,1,3\rrbracket \) simplex code} --- Each \(\llbracket 2^r-1,1,3\rrbracket \) simplex code is a color code defined on a simplex in \(r-1\) dimensions \NoCaseChange{\protect\cite{cite475,cite832}}.
\item\relax
\flmRefsHyperref[eczindexfamilyrel]{code:2d_color}{2D color code}\item\relax
\flmRefsHyperref[eczindexfamilyrel]{code:3d_color}{3D color code}\item\relax
\flmRefsHyperref[eczindexfamilyrel]{code:ball_color}{Ball code} --- Ball codes are color codes defined on a \(D\)-dimensional colex \NoCaseChange{\protect\cite[{Appx. A}]{cite687}}.
\item\relax
\flmRefsHyperref[eczindexfamilyrel]{code:hyperbolic_color}{Hyperbolic color code}\item\relax
\flmRefsHyperref[eczindexfamilyrel]{code:quasi_hyperbolic_color}{Quasi-hyperbolic color code}\end{eczvaluelist}
\codefieldsection{Cousins}
\begin{eczvaluelist}
\item\relax
\flmRefsHyperref[eczindexfamilyrel]{code:self_dual_css}{Self-dual CSS code} --- Color codes often have self-dual \(X\)- and \(Z\)-type bulk stabilizer structure, but boundary choices can prevent the full code from being self-dual. Thus, only color-code geometries for which transversal Hadamard is a logical operation are self-dual CSS codes.
\item\relax
\flmRefsHyperref[eczindexfamilyrel]{code:higher_dimensional_surface}{Homological code} --- For the common realization with point-like electric excitations, the color code on a \(D\)-dimensional closed manifold is equivalent to \(D\) decoupled copies of the \(D\)-dimensional toric/surface code via a local constant-depth \flmRefsHyperref{ref409}{Clifford circuit} \NoCaseChange{\protect\cite{cite3424,cite422,cite3425}} (see also \NoCaseChange{\protect\cite[{Exam. 4}]{cite739}}).
On a \(D\)-simplex-like lattice with \(D+1\) differently colored boundaries, the corresponding toric-code copies are attached along a common \((D-1)\)-dimensional boundary rather than fully decoupled \NoCaseChange{\protect\cite{cite422}}.
The reverse of this process can be viewed as gauging \NoCaseChange{\protect\cite{cite462,cite463,cite233,cite464,cite465,cite466,cite467,cite468,cite469,cite470}} certain symmetries.
Morphing subsets of colorable \(D\)-balls produces hybrid color-toric codes that interpolate between the color code and \(D\) copies of the toric code (up to ancillas when all balls of one color are morphed), while inheriting the parent color code's fault-tolerant gates \NoCaseChange{\protect\cite{cite687}}.
Several hybrid color-surface codes exist \NoCaseChange{\protect\cite{cite687,cite3595}}.

\item\relax
\flmRefsHyperref[eczindexfamilyrel]{code:self_correct}{Self-correcting quantum code} --- The 6D color code is a self-correcting quantum memory and admits a fault-tolerant universal gate set in 7D \NoCaseChange{\protect\cite{cite3035}}.
\item\relax
\flmRefsHyperref[eczindexfamilyrel]{code:cubic_theory}{Cubic theory code} --- The cubic theory in \(D\) spacetime dimensions can be obtained by twisted compactification of a generalized color code in \(D+1\) spacetime dimensions; in particular, the five-dimensional cubic theory arises from a twisted compactification of the 6D color code \NoCaseChange{\protect\cite{cite576}}.
\item\relax
\flmRefsHyperref[eczindexfamilyrel]{code:subsystem_hypergraph}{Sarvepalli-Brown subsystem code} --- Sarvepalli-Brown subsystem codes can be derived from color codes \NoCaseChange{\protect\cite[{Thm. 3}]{cite660}}.
\item\relax
\flmRefsHyperref[eczindexfamilyrel]{code:subsystem_color}{Subsystem color code} --- Gauge fixing relates subsystem color codes to conventional color codes defined on the same lattice \NoCaseChange{\protect\cite{cite475}}.
\end{eczvaluelist}
\eczhbkcontributors{ Balint Pato, Xiaozhen Fu, \eczhuVVA }
\endeczcode

\eczcode{combinatorial_permutation_invariant}{Combinatorial PI code}{~\NoCaseChange{\protect\cite{cite3169}}}
\codefieldsection{Alternative Names}
\begin{eczvaluelist}
\item\relax AAB code
\end{eczvaluelist}
\eczhIndexCodeAliasName{combinatorial_permutation_invariant}{AAB code}
\codefieldsection{Description}
A member of a family of PI quantum codes whose correction properties are derived from solving a family of combinatorial identities.
The code encodes one logical qubit in superpositions of \flmRefsHyperref{ref526}{Dicke states} whose coefficients are square roots of ratios of binomial coefficients.

\codefieldsection{Protection}
A code \(Q_{g,m,\delta,\epsilon}\) is defined for nonnegative integers \(g\), \(m\), and \(\delta\) as well as a sign \(\epsilon\) \NoCaseChange{\protect\cite[{Construction 5.1}]{cite3169}}.
The number of qubits is \(n = 2g+m+\delta+1\).
The code corrects errors on up to \(t\) qubits for \(m\geq t\), \(\delta\geq 2t\), and either \((g\geq 2t,\epsilon=-)\) or \((g\geq 2t+1,\epsilon=+)\).
Under the same conditions, the code also corrects all patterns of \(2t\) deletions \NoCaseChange{\protect\cite{cite3169}}.

\codefieldsection{Transversal and Permutation-Based Gates}
\begin{eczvaluelist}
\item\relax A class of combinatorial PI codes called \((b,g,1)\)-codes has been identified that admits logical gates in the diagonal \flmTerm{term}{ref694}{}{Clifford hierarchy} from transversal \(Z\)-axis rotations \NoCaseChange{\protect\cite{cite727}}.
\end{eczvaluelist}
\codefieldsection{Notes}
\begin{eczvaluelist}
\item\relax See Quantum News and Views article \NoCaseChange{\protect\cite{cite3596}}.
\end{eczvaluelist}
\codefieldsection{Parent}
\begin{eczvaluelist}
\item\relax
\flmRefsHyperref[eczindexfamilyrel]{code:qubit_permutation_invariant}{PI qubit code}\end{eczvaluelist}
\codefieldsection{Child}
\begin{eczvaluelist}
\item\relax
\flmRefsHyperref[eczindexfamilyrel]{code:icosahedral_permutation_invariant}{\(\llparenthesis 7,2,3\rrparenthesis \) Pollatsek-Ruskai code} --- The Pollatsek-Ruskai code is equivalent to the \(Q_{2,1,2,-}\) combinatorial PI code \NoCaseChange{\protect\cite[{Sec. 5.2}]{cite3169}}. It is a seven-qubit PI code that realizes gates from the binary icosahedral group transversally.
\end{eczvaluelist}
\codefieldsection{Cousins}
\begin{eczvaluelist}
\item\relax
\flmRefsHyperref[eczindexfamilyrel]{code:binary_dihedral_permutation_invariant}{Binary dihedral PI code} --- The \(Q_{3,1,2m-4,+}\) and \(Q_{3,1,2^m-4,+}\) combinatorial PI codes reduce to the \(\llparenthesis 2m+3,2,3\rrparenthesis \) and \(\llparenthesis 2^{m-1}+3,2,3\rrparenthesis \) binary dihedral PI codes, respectively \NoCaseChange{\protect\cite[{Prop. 5.6}]{cite3169}} (see also \NoCaseChange{\protect\cite{cite727}}).
\item\relax
\flmRefsHyperref[eczindexfamilyrel]{code:gnu_permutation_invariant}{GNU PI code} --- Combinatorial PI codes \(Q_{g,(m-1)/2,g-1,+}\) are GNU codes for odd \(m\) \NoCaseChange{\protect\cite[{Prop. 5.4}]{cite3169}}.
\item\relax
\flmRefsHyperref[eczindexfamilyrel]{code:four_qubit_permutation_invariant}{\(\llparenthesis 4,2,2\rrparenthesis \) Four-qubit single-deletion code} --- The combinatorial PI code \(Q_{1,1,1,-}\) is another example of a four-qubit code correcting a single deletion error \NoCaseChange{\protect\cite[{Sec. 5.1}]{cite3169}}.
\end{eczvaluelist}
\eczhbkcontributors{ \eczhuVVA }
\endeczcode

\eczcode{compass_model}{Compass code}{~\NoCaseChange{\protect\cite{cite2650}}}
\codefieldsection{Description}
Subspace or subsystem CSS code defined by gauge-fixing the Bacon-Shor code, i.e., the code whose gauge group consists of terms in the compass model Hamiltonian \NoCaseChange{\protect\cite{cite656,cite657,cite658}} on a square lattice.
Families of random codes perform well against biased noise and spatially dependent (i.e., asymmetric) noise.

The gauge fixing proceeds by denoting plaquettes by \(X\) or \(Z\) type using two colors, and fixing or \textit{cutting} the corresponding \(X\) or \(Z\)-type gauge generators at the respective plaquettes.
A fully colored lattice yields a subspace code, but allowing for non-colored plaquettes yields a subsystem code.
A fully non-colored lattice reduces to the Bacon-Shor code.

The \textit{surface-density} compass code family is obtained by randomly cutting \(X\)-type stabilizers at only plaquettes of one color in a checkerboard coloring; it interpolates between Bacon-Shor codes and rotated surface codes.
The \textit{Shor-density} compass code family is obtained by randomly cutting \(X\)-type stabilizers at any plaquette; it interpolates between Bacon-Shor codes and QPCs.

\codefieldsection{Protection}
Provides some protection against coherent noise \NoCaseChange{\protect\cite{cite3597}}.

\codefieldsection{Decoding}
\begin{eczvaluelist}
\item\relax Asymmetrically-weighted variant of the union-find decoder \NoCaseChange{\protect\cite{cite2650}}.
\end{eczvaluelist}
\codefieldsection{Code Capacity Threshold}
\begin{eczvaluelist}
\item\relax See \NoCaseChange{\protect\cite[{Sec. IV}]{cite2650}} for tables of code-capacity thresholds against spatially dependent and biased noise.
\end{eczvaluelist}
\codefieldsection{Parents}
\begin{eczvaluelist}
\item\relax
\flmRefsHyperref[eczindexfamilyrel]{code:qubit_subsystem_css}{Subsystem qubit CSS code}\item\relax
\flmRefsHyperref[eczindexfamilyrel]{code:translationally_invariant_subsystem}{Lattice subsystem code}\end{eczvaluelist}
\codefieldsection{Children}
\begin{eczvaluelist}
\item\relax
\flmRefsHyperref[eczindexfamilyrel]{code:bacon_shor}{Bacon-Shor code} --- A compass code on a fully non-colored lattice reduces to the Bacon-Shor code.
\item\relax
\flmRefsHyperref[eczindexfamilyrel]{code:heavy_hex}{Heavy-hexagon code} --- The heavy-hex code is a compass code on a heavy-hexagonal lattice, combining weight-two \(XX\) and \(ZZ\) gauge operators that are partially gauge-fixed to yield surface-code \(Z\)-type stabilizers and Bacon-Shor \(X\)-type stabilizers \NoCaseChange{\protect\cite{cite3598}}.
\end{eczvaluelist}
\codefieldsection{Cousins}
\begin{eczvaluelist}
\item\relax
\flmRefsHyperref[eczindexfamilyrel]{code:rotated_surface}{Rotated surface code} --- The surface-density compass code family interpolates between Bacon-Shor codes and rotated surface codes.
\item\relax
\flmRefsHyperref[eczindexfamilyrel]{code:quantum_parity}{Quantum parity code (QPC)} --- The Shor-density compass code family interpolates between Bacon-Shor codes and QPCs.
\item\relax
\flmRefsHyperref[eczindexfamilyrel]{code:random_stabilizer}{Random stabilizer code} --- Compass code families are constructed by randomly assigning stabilizers to plaquettes of a square lattice.
\item\relax
\flmRefsHyperref[eczindexfamilyrel]{code:clifford-deformed_surface}{Clifford-deformed surface code (CDSC)} --- Clifford deformation can enhance the performance of compass codes against biased noise \NoCaseChange{\protect\cite{cite2651}}.
\item\relax
\flmRefsHyperref[eczindexfamilyrel]{code:asymmetric_qecc}{Asymmetric quantum code (AQC)} --- Families of random compass codes perform well against biased noise and spatially dependent (i.e., asymmetric) noise \NoCaseChange{\protect\cite{cite2650}}.
Clifford deformation can enhance the performance of compass codes against biased noise \NoCaseChange{\protect\cite{cite2651}}.

\end{eczvaluelist}
\eczhbkcontributors{ \eczhuVVA }
\endeczcode

\eczcode{qubit_concatenated}{Concatenated qubit code}{}
\codefieldsection{Description}
A concatenated code whose outer code is a qubit code. In other words, a qubit code that can be thought of as a concatenation of an inner qubit code and an outer qubit code.
An inner \(C_{\text{in}} = \llparenthesis n_1,K,d_1\rrparenthesis \) and outer \(C_{\text{out}} = \llparenthesis n_2,2,d_2\rrparenthesis \) qubit code yield an \(\llparenthesis n_1 n_2, K, d \geq d_1d_2\rrparenthesis \) concatenated qubit code.

Concatenating an \(\llparenthesis n,2,d\rrparenthesis \) qubit code can be done recursively, with the \(r\)\textit{th level} of concatenation yielding an \(\llparenthesis n^r,2,d^r\rrparenthesis \) code.

\codefieldsection{Protection}
Any distance-three recursively concatenated code protects against an open set of errors \NoCaseChange{\protect\cite{cite3599}}.
Concatenating stabilizer codes can help protect against catastrophic errors such as cosmic rays \NoCaseChange{\protect\cite{cite3600}}.

\codefieldsection{Decoding}
\begin{eczvaluelist}
\item\relax Adaptive syndrome extraction for a concatenation of a small error-detecting code and a high-rate, high-distance QLDPC code \NoCaseChange{\protect\cite{cite3295}}.
\item\relax The effective channel for a concatenation of codes is the composition of the codes' effective channels \NoCaseChange{\protect\cite{cite3321}}.
\item\relax Message passing algorithm for concatenated codes can be equivalent to ML decoding \NoCaseChange{\protect\cite{cite2698}}.
\end{eczvaluelist}
\codefieldsection{Fault Tolerance}
\begin{eczvaluelist}
\item\relax Fault-tolerant message passing between devices \NoCaseChange{\protect\cite{cite3601}}.
\item\relax Blocklet concatenation uses concatenation and transversal gates in a way that is tailored to FBQC platforms \NoCaseChange{\protect\cite{cite3602}}.
\end{eczvaluelist}
\codefieldsection{Threshold}
\begin{eczvaluelist}
\item\relax The first methods to achieve a \flmRefsHyperref{ref515}{concatenated threshold} against local stochastic noise use concatenated qubit stabilizer codes \NoCaseChange{\protect\cite{cite516,cite3364,cite3603,cite826,cite519,cite3604,cite3605,cite520}}; see the book \NoCaseChange{\protect\cite{cite398}}.

\end{eczvaluelist}
\codefieldsection{Parents}
\begin{eczvaluelist}
\item\relax
\flmRefsHyperref[eczindexfamilyrel]{code:qubits_into_qubits}{Qubit code}\item\relax
\flmRefsHyperref[eczindexfamilyrel]{code:quantum_concatenated}{Concatenated quantum code}\end{eczvaluelist}
\codefieldsection{Children}
\begin{eczvaluelist}
\item\relax
\flmRefsHyperref[eczindexfamilyrel]{code:css_12_1_3}{\(\llbracket 12,1,3\rrbracket \) CE CSS code} --- This code is obtained by dual-rail concatenation of the \(\llbracket 6,1,2\rrbracket \) CSS code \NoCaseChange{\protect\cite[{ID 50}]{cite453}}.
\item\relax
\flmRefsHyperref[eczindexfamilyrel]{code:phantom_14_3_3}{\(\llbracket 14,3,3\rrbracket \) CE phantom code} --- This code is a concatenation of the \(\llbracket 7,3,2\rrbracket \) punctured hypercube code with the two-qubit phase-flip repetition code \NoCaseChange{\protect\cite{cite514}}.
\item\relax
\flmRefsHyperref[eczindexfamilyrel]{code:quantum_turbo}{Quantum turbo code}\item\relax
\flmRefsHyperref[eczindexfamilyrel]{code:bc_phantom}{Binarized-and-concatenated (B\&C) phantom code} --- Each qubit pair obtained by binarizing one \(\mathbb{F}_4\) qudit is concatenated with the \(\llbracket 4,2,2\rrbracket \) code \NoCaseChange{\protect\cite{cite514}}.
\item\relax
\flmRefsHyperref[eczindexfamilyrel]{code:aqm}{Auxiliary qubit mapping (AQM) code}\item\relax
\flmRefsHyperref[eczindexfamilyrel]{code:concatenated_steane}{Concatenated Steane code} --- The combination of the concatenated Steane code and QLDPC codes with non-vanishing rate yields fault-tolerant quantum computation with constant space and polylogarithmic time overheads, even when classical computation time is taken into account \NoCaseChange{\protect\cite{cite3606}}.
\item\relax
\flmRefsHyperref[eczindexfamilyrel]{code:hierarchical}{Hierarchical code} --- Using the concatenation convention of the Zoo, hierarchical codes are concatenations of constant-rate QLDPC (inner) codes with rotated surface (outer) codes. The cited paper \NoCaseChange{\protect\cite{cite3607}} uses the opposite inner/outer terminology. The block length of the outer code is picked to grow logarithmically with the block length of the inner code.
\item\relax
\flmRefsHyperref[eczindexfamilyrel]{code:yoked_surface}{Yoked surface code} --- Using the concatenation convention of the Zoo, a yoked surface code is a concatenation of a QMDPC code (inner code) with a rotated surface code (outer code). The cited paper \NoCaseChange{\protect\cite{cite523}} uses the opposite inner/outer terminology.
\item\relax
\flmRefsHyperref[eczindexfamilyrel]{code:xyz_hexagonal}{XYZ\(^2\) hexagonal stabilizer code} --- The XYZ\(^2\) hexagonal stabilizer code can be viewed as a concatenation of the \(YZZY\) surface code with one of the possible \(\llbracket 2,1\rrbracket \) repetition codes, with the case of the bit-flip repetition code yielding a concatenation of the surface code with the dual-rail code \NoCaseChange{\protect\cite{cite2645}}.
\end{eczvaluelist}
\codefieldsection{Cousins}
\begin{eczvaluelist}
\item\relax
\flmRefsHyperref[eczindexfamilyrel]{code:hamiltonian}{Hamiltonian-based code} --- Concatenated stabilizer code Hamiltonians have been investigated \NoCaseChange{\protect\cite{cite2845}}.
\item\relax
\flmRefsHyperref[eczindexfamilyrel]{code:fusion}{Fusion-based quantum computing (FBQC) code} --- Blocklet concatenation uses concatenation and transversal gates in a way that is tailored to FBQC platforms \NoCaseChange{\protect\cite{cite3602}}.
\item\relax
\flmRefsHyperref[eczindexfamilyrel]{code:gauss_law}{Gauss' law code} --- The Gauss' law code can be concatenated to form a stabilizer code for fault-tolerant quantum simulation of the underlying gauge theory \NoCaseChange{\protect\cite{cite79,cite78}}.
\item\relax
\flmRefsHyperref[eczindexfamilyrel]{code:ampdamp}{Amplitude-damping (AD) code} --- Using the dual-rail code as an outer code with an inner \(\llbracket n,k,d\rrbracket \) qubit code yields an \(\llbracket 2n,k\rrbracket \) code correcting \(d-1\) qubit \flmRefsHyperref{ref498}{AD} errors \NoCaseChange{\protect\cite{cite3263}}.
\item\relax
\flmRefsHyperref[eczindexfamilyrel]{code:eastab}{EA qubit stabilizer code} --- There exist concatenated EA qubit stabilizer codes that saturate the EA quantum Griesmer and Plotkin bounds \NoCaseChange{\protect\cite{cite3608}}.
\item\relax
\flmRefsHyperref[eczindexfamilyrel]{code:unentangled_permutation_invariant}{\(\llparenthesis n,2,2\rrparenthesis \) Bravyi-Lee-Li-Yoshida PI code} --- The Bravyi-Lee-Li-Yoshida PI code can be concatenated to yield codes that have higher distance and that admit codewords with vanishing entanglement \NoCaseChange{\protect\cite[{Appx. D}]{cite529}} (cf. \NoCaseChange{\protect\cite{cite530}}).
\item\relax
\flmRefsHyperref[eczindexfamilyrel]{code:hypercube_quantum}{\(\llbracket 2^D,D,2\rrbracket \) hypercube quantum code} --- The hypercube quantum code can be concatenated with a two-qubit quantum repetition code to yield a \(\llbracket 2^{D+1},D,4\rrbracket \) error-detecting code family \NoCaseChange{\protect\cite{cite759}}.
It can also be concatenated with \(D\) distance-two \(D\)-dimensional toric/surface-code blocks to yield a \(\llbracket 2^D(2^D+1),D,4\rrbracket \) error-correcting code family that admits a transversal implementation of the logical \(C^{D-1}Z\) gate \NoCaseChange{\protect\cite{cite759}}.

\item\relax
\flmRefsHyperref[eczindexfamilyrel]{code:stab_15_1_3}{\(\llbracket 15,1,3\rrbracket \) quantum RM code} --- The concatenated \(\llbracket 15,1,3\rrbracket \) code has a \flmRefsHyperref{ref3210}{measurement threshold} less than one \NoCaseChange{\protect\cite{cite3211}}.
\item\relax
\flmRefsHyperref[eczindexfamilyrel]{code:css_4_1_2}{\(\llbracket 4,1,2\rrbracket \) Leung-Nielsen-Chuang-Yamamoto (LNCY) code} --- The \(\llbracket 4,1,2\rrbracket \) LNCY code is the smallest QPC, i.e., a concatenation of a two-qubit bit-flip with a two-qubit phase-flip repetition code.
An \(\llbracket 8,1,2\rrbracket \) QPC correcting a single \flmRefsHyperref{ref498}{AD} error is equivalent to a concatenation of its constant-excitation version with the dual-rail code \NoCaseChange{\protect\cite{cite3250,cite3259,cite2711}}.
More generally, an \(\llbracket m^2,1,m\rrbracket \) QPC corrects \(m-1\) \flmRefsHyperref{ref498}{AD} errors \NoCaseChange{\protect\cite{cite3263}}.
Recursively concatenating a \(\llbracket 4,1,2\rrbracket \) LNCY subcode attains a threshold \NoCaseChange{\protect\cite{cite3264,cite3265}}.

\item\relax
\flmRefsHyperref[eczindexfamilyrel]{code:stab_4_2_2}{\(\llbracket 4,2,2\rrbracket \) Four-qubit code} --- Concatenations of \(\llbracket 4,2,2\rrbracket \) and \(C_6\) codes yield fault-tolerant quantum computation schemes \NoCaseChange{\protect\cite{cite448}} admitting a post-selected threshold \NoCaseChange{\protect\cite{cite3273,cite3274}} (see also Ref. \NoCaseChange{\protect\cite{cite3275}}).
Concatenating quantum Hamming codes on top of the \(\llbracket 4,2,2\rrbracket \) and \(C_6\) codes yields fault-tolerant quantum computation with constant space and quasi-polylogarithmic time overheads \NoCaseChange{\protect\cite{cite3216}}. In the optimized protocol of Ref. \NoCaseChange{\protect\cite{cite3216}}, a level-five \(C_4/C_6\) code underlies concatenated quantum Hamming codes \(\mathcal{Q}_5,\mathcal{Q}_6,\mathcal{Q}_7,\mathcal{Q}_7\), yielding a \(2.5\%\) threshold and space overheads \(162\) and \(373\) physical qubits per logical qubit at physical error rate \(0.1\%\) for logical CNOT error rates \(10^{-10}\) and \(10^{-24}\), respectively.
Concatenating the \(\llbracket 4,2,2\rrbracket \) code with the surface code is equivalent to removing stabilizer generators from the 4.8.8 color code \NoCaseChange{\protect\cite{cite3289}}.
The \(\llbracket 4,2,2\rrbracket \) code can be concatenated with two copies of the surface code to yield the 4.6.12 color code \NoCaseChange{\protect\cite{cite3289}}.

\item\relax
\flmRefsHyperref[eczindexfamilyrel]{code:stab_5_1_3}{\(\llbracket 5,1,3\rrbracket \) Five-qubit perfect code} --- The recursively concatenated five-qubit code has a \flmRefsHyperref{ref3210}{measurement threshold} of one \NoCaseChange{\protect\cite{cite3211}}. Code performance against general Pauli channels has been worked out \NoCaseChange{\protect\cite{cite3320,cite3321}}.
\item\relax
\flmRefsHyperref[eczindexfamilyrel]{code:stab_6_2_2}{\(\llbracket 6,2,2\rrbracket \) \(C_6\) code} --- Concatenations of \(\llbracket 4,2,2\rrbracket \) and \(C_6\) codes yield fault-tolerant quantum computation schemes \NoCaseChange{\protect\cite{cite448}} admitting a post-selected threshold \NoCaseChange{\protect\cite{cite3273,cite3274}} (see also Ref. \NoCaseChange{\protect\cite{cite3275}}) and the Meier-Eastin-Knill (MEK) magic-state distillation protocols \NoCaseChange{\protect\cite{cite708}}. Concatenating quantum Hamming codes on top of the \(\llbracket 4,2,2\rrbracket \) and \(C_6\) codes yields fault-tolerant quantum computation with constant space and quasi-polylogarithmic time overheads \NoCaseChange{\protect\cite{cite3216}}. In the optimized protocol of Ref. \NoCaseChange{\protect\cite{cite3216}}, a level-five \(C_4/C_6\) code underlies concatenated quantum Hamming codes \(\mathcal{Q}_5,\mathcal{Q}_6,\mathcal{Q}_7,\mathcal{Q}_7\), yielding a \(2.5\%\) threshold and space overheads \(162\) and \(373\) physical qubits per logical qubit at physical error rate \(0.1\%\) for logical CNOT error rates \(10^{-10}\) and \(10^{-24}\), respectively.
\item\relax
\flmRefsHyperref[eczindexfamilyrel]{code:stab_6_4_2}{\(\llbracket 6,4,2\rrbracket \) error-detecting code} --- Concatenations of this code with itself yield the level-\(r\) \(\llbracket 6^r,4^r,2^r\rrbracket \) many-hypercube code \NoCaseChange{\protect\cite{cite450}}. The \(\llbracket 6,4,2\rrbracket \) code can be concatenated with the surface code to yield the 6.6.6 color code \NoCaseChange{\protect\cite[{Appx. A}]{cite3289}}.
\item\relax
\flmRefsHyperref[eczindexfamilyrel]{code:stab_8_2_3}{\(\llbracket 8,2,3\rrbracket \) Hermitian code} --- Applying the BLT mapping to the \(\llbracket 8,2,3\rrbracket \) Hermitian code and concatenating each qubit pair with the \(\llbracket 4,2,2\rrbracket \) code yields a \(\llbracket 32,4,6\rrbracket \) self-dual CSS code \NoCaseChange{\protect\cite[{Corr. 2}]{cite795}}.
\item\relax
\flmRefsHyperref[eczindexfamilyrel]{code:stab_8_3_2}{\(\llbracket 8,3,2\rrbracket \) Smallest interesting color code} --- Concatenating \(\llbracket 8,3,2\rrbracket \) blocks with triples of qubits drawn from three cyclically rotated 3D surface/toric codes yields a 3D toric/color family with parameters \(\llbracket 8n,3,2d\rrbracket \) and transversal logical \(CCZ\) implemented by physical \(T\) gates on the inner \(\llbracket 8,3,2\rrbracket \) blocks \NoCaseChange{\protect\cite{cite759}}.
\item\relax
\flmRefsHyperref[eczindexfamilyrel]{code:shor_nine}{\(\llbracket 9,1,3\rrbracket \) Shor code} --- The Shor code is a concatenation of a three-qubit bit-flip with a three-qubit phase-flip repetition code.
\item\relax
\flmRefsHyperref[eczindexfamilyrel]{code:phantom}{Phantom code} --- Concatenating a phantom outer code with a one-logical-qubit inner quantum code preserves phantomness.
\item\relax
\flmRefsHyperref[eczindexfamilyrel]{code:layer}{Layer code} --- Each pair of surface-code squares in a layer code is fused (or quasi-concatenated) with perpendicular surface-code squares via lattice surgery.
\item\relax
\flmRefsHyperref[eczindexfamilyrel]{code:quantum_divisible}{Quantum divisible code} --- A fault-tolerant \(T\) gate on the five-qubit or Steane code can be obtained by concatenating with particular quantum divisible codes \NoCaseChange{\protect\cite{cite765}}.
\item\relax
\flmRefsHyperref[eczindexfamilyrel]{code:rbh}{Raussendorf-Bravyi-Harrington (RBH) cluster-state code} --- Concatenation of the RBH code with small codes such as the \(\llbracket 2,1\rrbracket \) repetition code, \(\llbracket 4,1,1,2\rrbracket \) subsystem code, or Steane code can improve thresholds \NoCaseChange{\protect\cite{cite3248}}.
\item\relax
\flmRefsHyperref[eczindexfamilyrel]{code:hypergraph_product}{Hypergraph product (HGP) code} --- There is a fault-tolerant universal computation scheme for hypergraph-product codes concatenated with the \(\llbracket 4,2,2\rrbracket \) code in which the full syndrome measurement on the lower hypergraph product code is performed only if an error is detected at the upper four-qubit code \NoCaseChange{\protect\cite{cite3295}}.
\item\relax
\flmRefsHyperref[eczindexfamilyrel]{code:quantum_hamming_css}{\(\llbracket 2^r-1, 2^r-2r-1, 3\rrbracket \) quantum Hamming code} --- Concatenating a growing sequence of quantum Hamming codes yields fault-tolerant quantum computation with constant space overhead and quasi-polylogarithmic time overhead \NoCaseChange{\protect\cite{cite3214}}.
Concatenating quantum Hamming codes on top of the \(\llbracket 4,2,2\rrbracket \) and \(C_6\) codes yields fault-tolerant quantum computation with constant space and quasi-polylogarithmic time overheads \NoCaseChange{\protect\cite{cite3216}}. In the optimized protocol of Ref. \NoCaseChange{\protect\cite{cite3216}}, a level-five \(C_4/C_6\) code underlies concatenated quantum Hamming codes \(\mathcal{Q}_5,\mathcal{Q}_6,\mathcal{Q}_7,\mathcal{Q}_7\), yielding a \(2.5\%\) threshold and space overheads \(162\) and \(373\) physical qubits per logical qubit at physical error rate \(0.1\%\) for logical CNOT error rates \(10^{-10}\) and \(10^{-24}\), respectively.
A modified tower of interleaved quantum Hamming codes with reserved qubits and recursive hookless Pauli-product measurements yields fault-tolerant quantum computation on a 1D nearest-neighbor qubit line with asymptotic rate above \(5\%\), constant space overhead, quasi-polylogarithmic time overhead, and a threshold \NoCaseChange{\protect\cite{cite3217}}.
Quantum Hamming codes can also be concatenated with surface codes \NoCaseChange{\protect\cite{cite3218}}.

\item\relax
\flmRefsHyperref[eczindexfamilyrel]{code:3d_color}{3D color code} --- On closed 3-manifolds, the 3D color code is equivalent to multiple decoupled copies of the 3D surface code via a local constant-depth \flmRefsHyperref{ref409}{Clifford circuit} \NoCaseChange{\protect\cite{cite3424,cite422,cite3425}}. This process can be viewed as an ungauging \NoCaseChange{\protect\cite{cite462,cite463,cite233,cite464,cite465,cite466,cite467,cite468,cite469,cite470}} of certain symmetries. This mapping can also be done via code concatenation \NoCaseChange{\protect\cite{cite715}}.
\item\relax
\flmRefsHyperref[eczindexfamilyrel]{code:subsystem_product}{Subsystem homological product code} --- Concatenated CSS stabilizer codes are gauge-fixed SP codes \NoCaseChange{\protect\cite[{Thm. 4}]{cite664}}.
\end{eczvaluelist}
\eczhbkcontributors{ \eczhuVVA }
\endeczcode

\eczcode{concatenated_steane}{Concatenated Steane code}{~\NoCaseChange{\protect\cite{cite516,cite517}}}
\codefieldsection{Description}
A member of the family of \(\llbracket 7^m,1,3^m\rrbracket \) CSS codes, each of which is a recursive level-\(m\) concatenation of the Steane code.
This family is one of the first to admit a \flmRefsHyperref{ref515}{concatenated threshold} \NoCaseChange{\protect\cite{cite516,cite517,cite518,cite519,cite520}}.

\codefieldsection{Protection}
Code performance against general Pauli channels has been worked out \NoCaseChange{\protect\cite{cite3320,cite3321}}.
\codefieldsection{Decoding}
\begin{eczvaluelist}
\item\relax A simple message-passing decoder from level 1 to level 2 corrects all weight-four errors for the \(\llbracket 49,1,9\rrbracket \) code and was used in the comparative threshold study of Ref. \NoCaseChange{\protect\cite{cite3225}}.
\item\relax There exist fault-tolerant syndrome extraction protocols for the concatenated Steane code \NoCaseChange{\protect\cite{cite3609}}.
\item\relax Randomized compiling helps reduce logical error rate for some noise models \NoCaseChange{\protect\cite{cite3610}}.
\end{eczvaluelist}
\codefieldsection{Fault Tolerance}
\begin{eczvaluelist}
\item\relax Fault-tolerant computation can be done on nearest-neighbor arrays \NoCaseChange{\protect\cite{cite3611}}.
\item\relax There exist fault-tolerant syndrome extraction protocols for the concatenated Steane code \NoCaseChange{\protect\cite{cite3609}}.
\item\relax The combination of the concatenated Steane code and QLDPC codes with non-vanishing rate yields fault-tolerant quantum computation with constant space and polylogarithmic time overheads, even when classical computation time is taken into account \NoCaseChange{\protect\cite{cite3606}}.
\end{eczvaluelist}
\codefieldsection{Code Capacity Threshold}
\begin{eczvaluelist}
\item\relax This family is one of the first to admit a \flmRefsHyperref{ref515}{concatenated threshold} \NoCaseChange{\protect\cite{cite516,cite517,cite518,cite519,cite3604,cite3605,cite520}}; see the book \NoCaseChange{\protect\cite{cite398}}.
\end{eczvaluelist}
\codefieldsection{Threshold}
\begin{eczvaluelist}
\item\relax Between \(1.78\%\) and \(11.5\%\) with faulty photon detectors when combined with the dual-rail code at the first concatenation step in a variant of the KLM protocol \NoCaseChange{\protect\cite{cite3367,cite3368}}.
\item\relax For the adversarial-stochastic exRec analysis of the concatenated 7-qubit protocol, a crude bound gives \(p_T > 3.6\times 10^{-6}\), while circuit optimization together with careful counting of malignant sets improves this to \(p_T \geq 2.7\times 10^{-5}\) \NoCaseChange{\protect\cite[{Sec. 14.7.4}]{cite398}}.
\item\relax When used as the underlying code of a Steane/Hamming concatenation in a unified logical-CNOT comparison under circuit-level depolarizing noise, the threshold is \(0.030\%\); at physical error rate \(0.1\%\), this underlying code cannot suppress the logical error rate to \(10^{-24}\), while at \(0.01\%\) it requires space overhead \(6.1\times 10^3\) \NoCaseChange{\protect\cite{cite3216}}.
\item\relax The recursively concatenated Steane code has a \flmRefsHyperref{ref3210}{measurement threshold} of one \NoCaseChange{\protect\cite{cite3211}}.
\end{eczvaluelist}
\codefieldsection{Parents}
\begin{eczvaluelist}
\item\relax
\flmRefsHyperref[eczindexfamilyrel]{code:holographic_steane}{Heptagon holographic code} --- A recursively concatenated Steane code is a heptagon holographic code on a tree tensor network.
\item\relax
\flmRefsHyperref[eczindexfamilyrel]{code:qubit_concatenated}{Concatenated qubit code} --- The combination of the concatenated Steane code and QLDPC codes with non-vanishing rate yields fault-tolerant quantum computation with constant space and polylogarithmic time overheads, even when classical computation time is taken into account \NoCaseChange{\protect\cite{cite3606}}.
\end{eczvaluelist}
\codefieldsection{Child}
\begin{eczvaluelist}
\item\relax
\flmRefsHyperref[eczindexfamilyrel]{code:steane}{\(\llbracket 7,1,3\rrbracket \) Steane code} --- The concatenated Steane code at level \(m=1\) is the Steane code.
\end{eczvaluelist}
\codefieldsection{Cousins}
\begin{eczvaluelist}
\item\relax
\flmRefsHyperref[eczindexfamilyrel]{code:qldpc}{Qubit QLDPC code} --- The combination of the concatenated Steane code and QLDPC codes with non-vanishing rate yields fault-tolerant quantum computation with constant space and polylogarithmic time overheads, even when classical computation time is taken into account \NoCaseChange{\protect\cite{cite3606}}.
\item\relax
\flmRefsHyperref[eczindexfamilyrel]{code:cluster_state}{Cluster-state code} --- The cluster state corresponding to the concatenated Steane code has been worked out \NoCaseChange{\protect\cite{cite3535}}.
\item\relax
\flmRefsHyperref[eczindexfamilyrel]{code:asymmetric_qecc}{Asymmetric quantum code (AQC)} --- Concatenating while taking into account noise bias can reduce resource overhead \NoCaseChange{\protect\cite{cite2621}}.
\item\relax
\flmRefsHyperref[eczindexfamilyrel]{code:stab_15_1_3}{\(\llbracket 15,1,3\rrbracket \) quantum RM code} --- The \(\llbracket 105,1\rrbracket \) concatenation of the \(\llbracket 15,1,3\rrbracket \) and Steane codes allows for a universal gate set consisting of gates that are transversal w.r.t. to two different partitions \NoCaseChange{\protect\cite{cite3209,cite775}}.
\end{eczvaluelist}
\eczhbkcontributors{ \eczhuVVA }
\endeczcode

\eczcode{cft}{Conformal-field theory (CFT) code}{~\NoCaseChange{\protect\cite{cite582,cite2564,cite583}}}
\codefieldsection{Description}
Approximate code whose codewords lie in the low-energy subspace of a conformal field theory, e.g., the quantum Ising model at its critical point \NoCaseChange{\protect\cite{cite582,cite583}}.
Its encoding is argued to perform source coding (i.e., compression) as well as channel coding (i.e., error correction) \NoCaseChange{\protect\cite{cite582}}.

\codefieldsection{Protection}
Code performance is quantified by a lower bound on the entanglement fidelity in terms of the conditional mutual information \NoCaseChange{\protect\cite[{Eq. (9)}]{cite582}}; see also \NoCaseChange{\protect\cite[{Appx. A}]{cite2608}}.
Certain CFT codes have indefinite \flmRefsHyperref{ref2559}{codespace complexity}, and their protection depends on the minimum scaling dimension of the underlying CFT \NoCaseChange{\protect\cite{cite2564}}.
The coherent information of a combined noise and recovery channel can be perturbatively expanded \NoCaseChange{\protect\cite{cite583}}.

\codefieldsection{Code Capacity Threshold}
\begin{eczvaluelist}
\item\relax Threshold under dephasing depends on the structure of the conformal field theory, with the 1D critical Ising model admitting a finite threshold against certain dephasing noise \NoCaseChange{\protect\cite{cite583}}.
\end{eczvaluelist}
\codefieldsection{Parents}
\begin{eczvaluelist}
\item\relax
\flmRefsHyperref[eczindexfamilyrel]{code:qubits_into_qubits}{Qubit code}\item\relax
\flmRefsHyperref[eczindexfamilyrel]{code:hamiltonian}{Hamiltonian-based code} --- CFT codewords lie in the low-energy subspace of a conformal field theory (CFT), e.g., the quantum Ising model at its critical point.
\item\relax
\flmRefsHyperref[eczindexfamilyrel]{code:approximate_qecc}{Approximate quantum error-correcting code (AQECC)}\item\relax
\flmRefsHyperref[eczindexfamilyrel]{code:holographic}{Holographic code} --- CFT codewords lie in the low-energy subspace of a conformal field theory (CFT), e.g., the quantum Ising model at its critical point.
\end{eczvaluelist}
\codefieldsection{Cousins}
\begin{eczvaluelist}
\item\relax
\flmRefsHyperref[eczindexfamilyrel]{code:self_dual_lattice}{Unimodular lattice} --- Even self-dual binary codes and even unimodular lattices define CFTs \NoCaseChange{\protect\cite{cite2033,cite2034,cite2035}}.
\item\relax
\flmRefsHyperref[eczindexfamilyrel]{code:sphere_packing}{Sphere packing} --- The Cohn-Elkies linear programming bound can be recast as a conformal bootstrap problem, which is a way of utilizing symmetry to constrain correlation functions of conformal field theories \NoCaseChange{\protect\cite{cite2308,cite2309,cite2310}}.
\item\relax
\flmRefsHyperref[eczindexfamilyrel]{code:self_dual}{Self-dual linear code} --- Even self-dual binary codes and even unimodular lattices define CFTs \NoCaseChange{\protect\cite{cite2033,cite2034,cite2035}}. Self-dual ternary codes define superconformal field theories (SCFTs) \NoCaseChange{\protect\cite{cite2036}}.
\item\relax
\flmRefsHyperref[eczindexfamilyrel]{code:qubit_css}{Qubit CSS code} --- There is a holographic relation between qubit CSS codes describing CFTs and qubit stabilizer codes describing path integrals over certain topologies \NoCaseChange{\protect\cite{cite3612}}.
\end{eczvaluelist}
\eczhbkcontributors{ \eczhuVVA }
\endeczcode

\eczcode{crystalline_dynamic_gen}{Crystalline-circuit qubit code}{~\NoCaseChange{\protect\cite{cite3613}}}
\codefieldsection{Description}
Code dynamically generated by constant-depth unitary \flmRefsHyperref{ref409}{Clifford circuits} defined on a lattice with some crystalline symmetry. A notable example is the circuit defined on a rotated square lattice with vertices corresponding to iSWAP gates and edges decorated by \(R_X[\pi/2]\), a single-qubit rotation by \(\pi/2\) around the \(X\)-axis. This circuit is invariant under space-time translations by a unit cell \((T, a)\) and all transformations of the square lattice point group \(D_4\).

The input state to the circuit is taken to be a product stabilizer state with finite entropy density. If the input is translation-invariant, then this periodicity is preserved by the circuit at all later times, so the code is a quantum quasi-cyclic code with unit cell \(a\). The initial product state recurs after a time \(\tau(n)\) that is linear in \(n\) for \(n=a 2^k\), but is thought to be exponential for generic \(n\).

\codefieldsection{Protection}
The code protects against Pauli errors. The circuit composed of iSWAP and \(R_X[\pi/2]\) gates on the square lattice is a "good scrambler" with non-fractal operator spreading and thus behaves like a random circuit in that regard, motivating the use of contiguous code distance as a proxy for code distance.

For the \(D_4\) example above, the unit cell \(a=2\), and the initial product group is chosen to have code rate \(1/2\). The parameters of the code are \(\llbracket n, n/2, d(t)\rrbracket \), and the contiguous code distance \NoCaseChange{\protect\cite{cite3000}} grows linearly before saturating at \(O(n)\).

Selecting the code defined by the stabilizer group at the time when the contiguous distance is maximized and subjecting it to random erasures, an optimal threshold of \(1/4\) is achieved for a subset of system sizes \NoCaseChange{\protect\cite{cite3465,cite3614}}.
The subthreshold scaling is competitive with random codes, which obey the random matrix theory ansatz \NoCaseChange{\protect\cite{cite3614}}.

\codefieldsection{Encoding}
\begin{eczvaluelist}
\item\relax Clifford quantum cellular automaton (CQCA) that preserves crystalline symmetry.
\end{eczvaluelist}
\codefieldsection{Parents}
\begin{eczvaluelist}
\item\relax
\flmRefsHyperref[eczindexfamilyrel]{code:qldpc}{Qubit QLDPC code}\item\relax
\flmRefsHyperref[eczindexfamilyrel]{code:random_stabilizer}{Random stabilizer code}\item\relax
\flmRefsHyperref[eczindexfamilyrel]{code:translationally_invariant_stabilizer}{Lattice stabilizer code}\end{eczvaluelist}
\codefieldsection{Cousins}
\begin{eczvaluelist}
\item\relax
\flmRefsHyperref[eczindexfamilyrel]{code:random_circuit}{Random-circuit code} --- Crystalline-circuit codes can be thought of as random-circuit codes with symmetries.
\item\relax
\flmRefsHyperref[eczindexfamilyrel]{code:monitored_random_circuits}{Monitored random-circuit code} --- Projective measurements can be included in crystalline-circuit codes in a spacetime translation-invariant fashion, turning such codes into \textit{monitored crystalline-circuit codes}. However, the unit cell of measurements must be large enough to avoid purification.
\end{eczvaluelist}
\eczhbkcontributors{ Grace M. Sommers, \eczhuVVA }
\endeczcode

\eczcode{css_plaquette}{CSS-Plaquette code}{~\NoCaseChange{\protect\cite{cite3448}}}
\codefieldsection{Description}
Generalization of the Bacon-Shor code to three dimensions, defined on a cubic lattice and admitting string-like stabilizer generators.

\codefieldsection{Threshold}
\begin{eczvaluelist}
\item\relax The CSS-Plaquette code has one entanglement transition \NoCaseChange{\protect\cite{cite3448}}.
\end{eczvaluelist}
\codefieldsection{Parents}
\begin{eczvaluelist}
\item\relax
\flmRefsHyperref[eczindexfamilyrel]{code:qubit_subsystem_css}{Subsystem qubit CSS code}\item\relax
\flmRefsHyperref[eczindexfamilyrel]{code:translationally_invariant_subsystem}{Lattice subsystem code}\end{eczvaluelist}
\codefieldsection{Cousins}
\begin{eczvaluelist}
\item\relax
\flmRefsHyperref[eczindexfamilyrel]{code:3d_bacon_shor}{3D Bacon-Shor code} --- 3D Bacon-Shor and CSS-Plaquette codes admit sheet-like and string-like stabilizer generators, respectively.
\item\relax
\flmRefsHyperref[eczindexfamilyrel]{code:hypercubic}{\(\mathbb{Z}^n\) hypercubic lattice} --- CSS-Plaquette codes are defined on a hypercubic lattice.
\end{eczvaluelist}
\eczhbkcontributors{ \eczhuVVA }
\endeczcode

\eczcode{css-t}{CSS-T code}{~\NoCaseChange{\protect\cite{cite3615}}}
\codefieldsection{Description}
A CSS code for which a physical transversal \(T\) gate is either the identity (up to a global phase) or a logical gate.
CSS-T codes are constructed from a pair of linear binary codes via the CSS construction, with the pair satisfying certain conditions \NoCaseChange{\protect\cite{cite1315}}.

\codefieldsection{Rate}
Asymptotically good CSS-T codes exist \NoCaseChange{\protect\cite{cite3616}}.
\codefieldsection{Transversal and Permutation-Based Gates}
\begin{eczvaluelist}
\item\relax A physical transversal \(T\) gate is either the identity (up to a global phase) or a logical gate \NoCaseChange{\protect\cite{cite724}}.
\end{eczvaluelist}
\codefieldsection{Notes}
\begin{eczvaluelist}
\item\relax A database of CSS-T codes is available in QECDB \NoCaseChange{\protect\cite{cite781}}.
\end{eczvaluelist}
\codefieldsection{Parent}
\begin{eczvaluelist}
\item\relax
\flmRefsHyperref[eczindexfamilyrel]{code:qubit_css}{Qubit CSS code}\end{eczvaluelist}
\codefieldsection{Cousins}
\begin{eczvaluelist}
\item\relax
\flmRefsHyperref[eczindexfamilyrel]{code:binary_linear}{Linear binary code} --- CSS-T codes are constructed from a pair of linear binary codes via the CSS construction, with the pair satisfying certain conditions \NoCaseChange{\protect\cite{cite1315}}.
\item\relax
\flmRefsHyperref[eczindexfamilyrel]{code:binary_cyclic}{Cyclic linear binary code} --- Binary cyclic and extended cyclic codes can be used to construct CSS-T codes via the CSS construction \NoCaseChange{\protect\cite{cite1315}}.
\item\relax
\flmRefsHyperref[eczindexfamilyrel]{code:quantum_reed_muller}{Quantum Reed-Muller (RM) code} --- Certain quantum RM codes are CSS-T codes \NoCaseChange{\protect\cite{cite3617,cite3618,cite1315}}.
\item\relax
\flmRefsHyperref[eczindexfamilyrel]{code:quantum_triorthogonal}{Triorthogonal code} --- Triorthogonal and CSS-T codes overlap, but neither is a subset of the other \NoCaseChange{\protect\cite{cite760}}. CSS-T codes reduce to triorthogonal codes when the logical action of the physical transversal \(T\) gate is a logical \(T\) gate on all encoded qubits. Triorthogonality is necessary for physical transversal \(T\) gates on each qubit to realize the identity logical gate \NoCaseChange{\protect\cite[{Thm. 12}]{cite724}}. The \(X\)-type stabilizer generator matrix for a CSS-T code is always triorthogonal \NoCaseChange{\protect\cite[{Corr. 5}]{cite1315}}.
\end{eczvaluelist}
\eczhbkcontributors{ Narayanan Rengaswamy, \eczhuVVA }
\endeczcode

\eczcode{cubic_honeycomb_color}{Cubic honeycomb color code}{~\NoCaseChange{\protect\cite{cite430}}}
\codefieldsection{Description}
3D color code defined on a four-colorable bitruncated cubic honeycomb uniform tiling.

\codefieldsection{Transversal and Permutation-Based Gates}
\begin{eczvaluelist}
\item\relax A code family on a truncated cube with particular boundary conditions admits a transversal control-\(S\) gate via physical \(T\) and \(T^{\dagger}\) gates \NoCaseChange{\protect\cite{cite728}}.
\end{eczvaluelist}
\codefieldsection{Parent}
\begin{eczvaluelist}
\item\relax
\flmRefsHyperref[eczindexfamilyrel]{code:3d_color}{3D color code}\end{eczvaluelist}
\codefieldsection{Cousin}
\begin{eczvaluelist}
\item\relax
\flmRefsHyperref[eczindexfamilyrel]{code:da_color_3d}{3D DA color code} --- The parent topological phase of the 3D DA color code is realized by three copies of the cubic honeycomb color code \NoCaseChange{\protect\cite[{Sec. VI.A}]{cite2532}}.
\end{eczvaluelist}
\eczhbkcontributors{ \eczhuVVA }
\endeczcode

\eczcode{cubic_theory}{Cubic theory code}{~\NoCaseChange{\protect\cite{cite576}}}
\codefieldsection{Alternative Names}
\begin{eczvaluelist}
\item\relax Magic stabilizer code
\end{eczvaluelist}
\eczhIndexCodeAliasName{cubic_theory}{Magic stabilizer code}
\codefieldsection{Description}
A geometrically local commuting-projector code family defined on triangulations in arbitrary spatial dimensions.
Its Hamiltonian contains Pauli-\(Z\) flux terms and non-Pauli Gauss-law terms built from products of Pauli-\(X\) operators and \(CZ\) gates.
These commuting non-Pauli stabilizers realize higher-form \(\mathbb{Z}_2^3\) gauge theories with Abelian electric excitations and non-Abelian magnetic excitations.

For \(l=m=n=2\) in \(D=6\) spacetime dimensions, the code is a candidate non-Abelian self-correcting quantum memory with Abelian loop excitations and non-Abelian membrane excitations \NoCaseChange{\protect\cite{cite576}}.

The construction is based on first constructing a model for an SPT phase \NoCaseChange{\protect\cite{cite3100,cite3619}}, gauging its symmetries \NoCaseChange{\protect\cite{cite462,cite463,cite233,cite464,cite465,cite466,cite467,cite468,cite469,cite470}}, and making all terms commute outside of the ground-state subspace by projecting them to zero flux.

\codefieldsection{Protection}
On suitable triangulations of the Wu 5-manifold, a family of five-dimensional cubic theory codes has parameters \(\llbracket O(n),3,O(n^{2/5})\rrbracket \) \NoCaseChange{\protect\cite{cite576}}.

\codefieldsection{Gates}
\begin{eczvaluelist}
\item\relax In five spatial dimensions, a constant-depth circuit built from physical \(CCZ\) and SWAP gates implements the logical gate \(\overline{\mathrm{SWAP}}_{1,2}\overline{\mathrm{CCZ}}_{1,2,3}\) on the three logical qubits supported by the Wu-manifold family \NoCaseChange{\protect\cite{cite576}}.
\end{eczvaluelist}
\codefieldsection{Decoding}
\begin{eczvaluelist}
\item\relax Probabilistic local cellular-automaton decoder \NoCaseChange{\protect\cite{cite576}}.
\end{eczvaluelist}
\codefieldsection{Fault Tolerance}
\begin{eczvaluelist}
\item\relax The Wu-manifold family combines the logical non-Clifford gate with code parameters \(\llbracket O(n),3,O(n^{2/5})\rrbracket \), giving \(O(d^{5/2})\) space-time overhead for this topological scheme \NoCaseChange{\protect\cite{cite576}}.
\end{eczvaluelist}
\codefieldsection{Parents}
\begin{eczvaluelist}
\item\relax
\flmRefsHyperref[eczindexfamilyrel]{code:clifford_hierarchy}{Clifford-hierarchy stabilizer code} --- Cubic theory codes are joint eigenspaces of commuting non-Pauli stabilizers built from Pauli \(X\), Pauli \(Z\), and \(CZ\) operators, placing them at the second level of the Clifford hierarchy.
\item\relax
\flmRefsHyperref[eczindexfamilyrel]{code:yetter_gauge_theory}{Two-gauge theory code} --- Cubic theory codes realize higher-form \(\mathbb{Z}_2^3\) gauge theories with non-Abelian excitations in arbitrary dimensions.
\end{eczvaluelist}
\codefieldsection{Child}
\begin{eczvaluelist}
\item\relax
\flmRefsHyperref[eczindexfamilyrel]{code:hexagonal_cz}{Hexagonal \(CZ\) code} --- The \(2+1\)D cubic theory with \(l=m=n=1\) realizes the same topological order as the Type-III \(\mathbb{Z}_2^3\) twisted quantum double / \(G=D_4\) quantum double, and the hexagonal \(CZ\) code is a hexagonal-lattice realization of this phase \NoCaseChange{\protect\cite{cite575,cite576}}.
\end{eczvaluelist}
\codefieldsection{Cousins}
\begin{eczvaluelist}
\item\relax
\flmRefsHyperref[eczindexfamilyrel]{code:self_correct}{Self-correcting quantum code} --- A family of five-dimensional cubic theory codes with Abelian loop excitations and non-Abelian membrane excitations is argued to be self-correcting below a critical temperature via a Peierls argument \NoCaseChange{\protect\cite{cite576}}.
\item\relax
\flmRefsHyperref[eczindexfamilyrel]{code:color}{Color code} --- The cubic theory in \(D\) spacetime dimensions can be obtained by twisted compactification of a generalized color code in \(D+1\) spacetime dimensions; in particular, the five-dimensional cubic theory arises from a twisted compactification of the 6D color code \NoCaseChange{\protect\cite{cite576}}.
\item\relax
\flmRefsHyperref[eczindexfamilyrel]{code:quantum_double_dihedral}{Dihedral \(G=D_m\) quantum-double code} --- For \(D=3\) with \(l=m=n=1\), the cubic theory is equivalent to the \(G=D_4\) quantum double, i.e. the non-Abelian Type-III \(\mathbb{Z}_2^3\) twisted quantum double \NoCaseChange{\protect\cite{cite576}}.
\item\relax
\flmRefsHyperref[eczindexfamilyrel]{code:xp_stabilizer}{XP stabilizer code} --- The cubic theory code can be embedded into a larger codespace such that all diagonal logical operators are represented by XP operators \NoCaseChange{\protect\cite[{Sec. 4.3}]{cite767}}.
\end{eczvaluelist}
\eczhbkcontributors{ Guanyu Zhu, \eczhuVVA }
\endeczcode

\eczcode{cyclic_hgp}{Cyclic Hypergraph Product Code}{~\NoCaseChange{\protect\cite{cite3620}}}
\codefieldsection{Description}
The hypergraph product code constructed using two low-weight circulant matrices. The code family \(\mathrm{C2}\) is the product of a cyclic LDPC code with itself, and the family \(\mathrm{CxR}\) is the product of the cyclic code with a repetition code. 

The construction of \(\mathrm{C2}\) uses a single generating polynomial \(\sum a_ix^i\) for both factors, while the construction of \(\mathrm{CxR}\) uses the generating polynomial \(\sum a_ix^i\) along with the polynomial \(1+x\), where \(a_i\in\{0,1\}\).

\codefieldsection{Decoding}
\begin{eczvaluelist}
\item\relax BP-OSD decoder
\end{eczvaluelist}
\codefieldsection{Parent}
\begin{eczvaluelist}
\item\relax
\flmRefsHyperref[eczindexfamilyrel]{code:hypergraph_product}{Hypergraph product (HGP) code} --- A cyclic hypergraph product code is a hypergraph product code constructed using two circulant matrices.
\end{eczvaluelist}
\codefieldsection{Children}
\begin{eczvaluelist}
\item\relax
\flmRefsHyperref[eczindexfamilyrel]{code:lresc}{Long-range enhanced surface code (LRESC)} --- LRESCs are constructed using a hypergraph product of a concatenated LDPC-repetition code with itself.
\item\relax
\flmRefsHyperref[eczindexfamilyrel]{code:toric}{Toric code} --- The toric code can be obtained from a hypergraph product of two repetition codes \NoCaseChange{\protect\cite[{Exam. 6}]{cite1316}}. Other hypergraph products of two repetition codes yield the related \(\llbracket 2d^2-2d+1,1,d\rrbracket \) CSS code family \NoCaseChange{\protect\cite[{Exam. 5}]{cite1316}}.
\end{eczvaluelist}
\codefieldsection{Cousins}
\begin{eczvaluelist}
\item\relax
\flmRefsHyperref[eczindexfamilyrel]{code:qc_ldpc}{Quasi-cyclic LDPC (QC-LDPC) code} --- A classical cyclic LDPC code with parameters \([n,k,d]\) yields a \(\mathrm{C2}\) code with parameters \(\llbracket 2n^2,2k^2,d\rrbracket \) and a \(\mathrm{CxR}\) code with parameters \(\llbracket 2nd,2k,d\rrbracket \).
\item\relax
\flmRefsHyperref[eczindexfamilyrel]{code:lacross}{La-cross code} --- The La-cross code is a reduced block length, full-rank cyclic HGP code with generator polynomials of the form \(1+x+x^k\)
\item\relax
\flmRefsHyperref[eczindexfamilyrel]{code:generalized_bicycle}{Generalized bicycle (GB) code} --- Cyclic HGP codes and GB codes both use circulant matrices as building blocks.
\end{eczvaluelist}
\eczhbkcontributors{ Arda Aydin, \eczhuVVA }
\endeczcode

\eczcode{derby_klassen}{Derby-Klassen (DK) code}{~\NoCaseChange{\protect\cite{cite412,cite3621}}}
\codefieldsection{Alternative Names}
\begin{eczvaluelist}
\item\relax Compact encoding
\end{eczvaluelist}
\eczhIndexCodeAliasName{derby_klassen}{Compact encoding}
\codefieldsection{Description}
A fermion-into-qubit code defined on regular tilings with maximum degree 4 whose stabilizers are associated with loops in the tiling.
The code outperforms several other encodings in terms of encoding rate \NoCaseChange{\protect\cite[{Table I}]{cite412}}.
It has been extended for models with several modes per site \NoCaseChange{\protect\cite{cite413}}.

\codefieldsection{Protection}
Some single-qubit errors are detectable, with the rest inducing low-weight fermionic dephasing noise \NoCaseChange{\protect\cite{cite3486}}.

\codefieldsection{Parents}
\begin{eczvaluelist}
\item\relax
\flmRefsHyperref[eczindexfamilyrel]{code:2d_bosonization}{2D bosonization code} --- On the square lattice, the DK code is the \(r=1.5\) exact-bosonization construction after finite-depth generalized local unitaries and re-pairing of Majorana modes \NoCaseChange{\protect\cite{cite404}}.
\item\relax
\flmRefsHyperref[eczindexfamilyrel]{code:qetc}{Quantum error-transmuting code (QETC)} --- The DK code transmutes all single-qubit errors \NoCaseChange{\protect\cite{cite2985}}.
\end{eczvaluelist}
\codefieldsection{Cousins}
\begin{eczvaluelist}
\item\relax
\flmRefsHyperref[eczindexfamilyrel]{code:xzzx}{XZZX surface code} --- The DK code encodes fermions into excitations of the Wen plaquette model \NoCaseChange{\protect\cite{cite3622}}.
\item\relax
\flmRefsHyperref[eczindexfamilyrel]{code:2d_color}{2D color code} --- The DK code on several tilings resembles the 2D color code with some vertex qubits removed \NoCaseChange{\protect\cite{cite3432}}.
\item\relax
\flmRefsHyperref[eczindexfamilyrel]{code:super_compact}{Super-compact fermion-to-qubit code} --- On the square lattice, the super-compact code is obtained by further transforming the \(r=1.5\) compact/DK construction to a qubit-to-fermion ratio of \(1.25\) \NoCaseChange{\protect\cite{cite404}}.
\end{eczvaluelist}
\eczhbkcontributors{ \eczhuVVA }
\endeczcode

\eczcode{dhlv}{Dinur-Hsieh-Lin-Vidick (DHLV) code}{~\NoCaseChange{\protect\cite{cite2190}}}
\codefieldsection{Description}
A family of asymptotically good QLDPC codes which are related to expander LP codes in that the roles of the check operators and physical qubits are exchanged.
\codefieldsection{Rate}
Asymptotically good QLDPC codes.
\codefieldsection{Decoding}
\begin{eczvaluelist}
\item\relax Linear-time decoder utilizing the small set-flip decoder \NoCaseChange{\protect\cite{cite3623}} for \(Z\) errors and a reconstruction procedure for \(X\) errors \NoCaseChange{\protect\cite{cite2190}}.
\end{eczvaluelist}
\codefieldsection{Parents}
\begin{eczvaluelist}
\item\relax
\flmRefsHyperref[eczindexfamilyrel]{code:qubit_generalized_homological_product_css}{Generalized homological-product qubit CSS code}\item\relax
\flmRefsHyperref[eczindexfamilyrel]{code:lifted_product}{Lifted-product (LP) code} --- DHLV codes are LP codes \NoCaseChange{\protect\cite[{Footnote 7}]{cite1101}}.
\end{eczvaluelist}
\codefieldsection{Cousins}
\begin{eczvaluelist}
\item\relax
\flmRefsHyperref[eczindexfamilyrel]{code:good_qldpc}{Good QLDPC code} --- DHLV code construction yields asymptotically good QLDPC codes.
\item\relax
\flmRefsHyperref[eczindexfamilyrel]{code:regular_binary_tanner}{Regular binary Tanner code} --- Regular binary Tanner codes are used in constructing quantum DHLV codes.
\item\relax
\flmRefsHyperref[eczindexfamilyrel]{code:tensor}{Tensor-product code} --- Tensor codes are used in constructing quantum DHLV codes.
\item\relax
\flmRefsHyperref[eczindexfamilyrel]{code:balanced_product}{Balanced product (BP) code} --- DHLV codes can be obtained from a balanced product of two expander codes \NoCaseChange{\protect\cite{cite1101}}.
\item\relax
\flmRefsHyperref[eczindexfamilyrel]{code:topological}{Topological code} --- DHLV codes are expected to realize topological quantum spin glass order \NoCaseChange{\protect\cite{cite3162}}.
\end{eczvaluelist}
\eczhbkcontributors{ \eczhuVVA }
\endeczcode

\eczcode{dlv}{Dinur-Lin-Vidick (DLV) code}{~\NoCaseChange{\protect\cite{cite2994}}}
\codefieldsection{Description}
Member of a family of codes constructed using cubical chain complexes, which are \(t\)-order extensions of the complexes underlying expander codes (\(t=1\)) and expander lifted-product codes (\(t=2\)).

For \(t=4\), assuming a conjecture about random linear maps, there exists a quantum locally testable family with linear dimension and inverse poly-logarithmic relative distance and soundness.
Applying weight reduction yields \flmRefsHyperref{ref65}{order} \(\Omega(1/\text{polylog}n)\) soundness, \(\Omega(n/\text{polylog}n)\) distance and dimension, and constant locality \NoCaseChange{\protect\cite[{Table 4}]{cite2991}}.
Applying distance amplification and soundness amplification yields asymptotically constant soundness, \flmRefsHyperref{ref65}{order} \(\Theta(n)\) distance, \flmRefsHyperref{ref65}{order} \(\Theta(n)\) dimension, but poly-logarithmic locality \NoCaseChange{\protect\cite[{Table 4}]{cite2991}}.

\codefieldsection{Parent}
\begin{eczvaluelist}
\item\relax
\flmRefsHyperref[eczindexfamilyrel]{code:qubit_generalized_homological_product_css}{Generalized homological-product qubit CSS code} --- DLV codes are codes constructed using a cubical chain complex, which is a \(t\)-order extension of the chain complexes underlying quantum generalized homological product CSS codes.
\end{eczvaluelist}
\codefieldsection{Cousin}
\begin{eczvaluelist}
\item\relax
\flmRefsHyperref[eczindexfamilyrel]{code:qltc}{Quantum locally testable code (QLTC)} --- DLV codes have linear dimension and inverse poly-logarithmic relative distance and soundness, assuming a conjecture about random linear maps \NoCaseChange{\protect\cite{cite2994}}. Applying distance amplification and soundness amplification yields asymptotically constant soundness, \flmRefsHyperref{ref65}{order} \(\Theta(n)\) distance, \flmRefsHyperref{ref65}{order} \(\Theta(n)\) dimension, but poly-logarithmic locality \NoCaseChange{\protect\cite[{Table 4}]{cite2991}}.
\end{eczvaluelist}
\eczhbkcontributors{ \eczhuVVA }
\endeczcode

\eczcode{double_semion_string_net}{Double-semion string-net code}{~\NoCaseChange{\protect\cite{cite3624,cite462,cite473,cite3625}}}
\codefieldsection{Description}
An \(XS\) stabilizer code that realizes the 2D double semion topological phase.
The model can be extended to other spatial dimensions \NoCaseChange{\protect\cite{cite588}}.

\codefieldsection{Parents}
\begin{eczvaluelist}
\item\relax
\flmRefsHyperref[eczindexfamilyrel]{code:xs_stabilizer}{XS stabilizer code} --- The double-semion string-net code is an \(XS\) stabilizer code \NoCaseChange{\protect\cite[{Fig. 1}]{cite3625}}.
\item\relax
\flmRefsHyperref[eczindexfamilyrel]{code:tqd_abelian}{Abelian TQD code} --- When treated as ground states of the code Hamiltonian, the double-semion string-net code states realize 2D double-semion topological order, a topological phase of matter that exists as the deconfined phase of the 2D twisted \(\mathbb{Z}_2\) gauge theory \NoCaseChange{\protect\cite{cite584}}.
\item\relax
\flmRefsHyperref[eczindexfamilyrel]{code:topological_abelian}{Abelian topological code} --- When treated as ground states of the code Hamiltonian, the double-semion string-net code states realize 2D double-semion topological order, a topological phase of matter that exists as the deconfined phase of the 2D twisted \(\mathbb{Z}_2\) gauge theory \NoCaseChange{\protect\cite{cite584}}.
\end{eczvaluelist}
\codefieldsection{Cousins}
\begin{eczvaluelist}
\item\relax
\flmRefsHyperref[eczindexfamilyrel]{code:string_net}{String-net code} --- The string-net model code for the category \(\text{Vec}^{\omega}\mathbb{Z}_2\) for a nontrivial cocycle is the double semion string-net code.
\item\relax
\flmRefsHyperref[eczindexfamilyrel]{code:double_semion}{Double-semion stabilizer code} --- The double-semion stabilizer code and the double-semion string-net code both realize the double semion topological phase, but the former is a modular-qudit Pauli stabilizer code while the latter is an \(XS\) stabilizer code. Their ground-state subspaces are connected by a finite-depth circuit with ancillas \NoCaseChange{\protect\cite{cite405}}. A commuting-projector version of the double-semion string-net code can also be derived \NoCaseChange{\protect\cite{cite2694,cite2693}}.
\item\relax
\flmRefsHyperref[eczindexfamilyrel]{code:surface}{Kitaev surface code} --- There is a logical basis for both the toric and double-semion string-net codes where each codeword is a superposition of states corresponding to all noncontractible loops of a particular homotopy type. The superposition is equal for the toric code, whereas an odd number of loops appear with a \(-1\) coefficient for the double semion.
\item\relax
\flmRefsHyperref[eczindexfamilyrel]{code:commuting_projector}{Commuting-projector Hamiltonian code} --- A commuting-projector version of the double-semion string-net code can also be derived \NoCaseChange{\protect\cite{cite2694,cite2693}}.
\item\relax
\flmRefsHyperref[eczindexfamilyrel]{code:spt}{Symmetry-protected topological (SPT) code} --- Gauging \NoCaseChange{\protect\cite{cite462,cite463,cite233,cite464,cite465,cite466,cite467,cite468,cite469,cite470}} the symmetry of a nontrivial 2D bosonic \(\mathbb{Z}_2\) Ising SPT yields the doubled-semion phase, with semionic \(\pi\)-flux excitations \NoCaseChange{\protect\cite[{Sec. IV}]{cite462}}.
\end{eczvaluelist}
\eczhbkcontributors{ \eczhuVVA }
\endeczcode

\eczcode{doubled_color}{Doubled color code}{~\NoCaseChange{\protect\cite{cite731,cite3626,cite3627}}}
\codefieldsection{Description}
Family of \(\llbracket 2t^3+8t^2+6t-1,1,2t+1\rrbracket \) subsystem color codes (with \(t\geq 1\)), constructed using a generalization of the doubling transformation \NoCaseChange{\protect\cite{cite659}}, that admit a Clifford + \(T\) transversal gate set using gauge fixing.

The family is embedded into a 2D honeycomb lattice with two qubits per site; the \(C\)-code has spatially local face-supported gauge generators, and the \(T\)-code is reached by local gauge fixing using edge-supported stabilizer measurements \NoCaseChange{\protect\cite{cite731}}.

\codefieldsection{Transversal and Permutation-Based Gates}
\begin{eczvaluelist}
\item\relax Doubled color codes are triply even, so they yield a transversal \(T\) gate \NoCaseChange{\protect\cite{cite731}}. Using gauge fixing, the codes admit a Clifford + \(T\) transversal gate set.
\end{eczvaluelist}
\codefieldsection{Decoding}
\begin{eczvaluelist}
\item\relax ML decoder that can utilize a history of syndromes, based on the Walsh-Hadamard transform \NoCaseChange{\protect\cite{cite731}}.
\end{eczvaluelist}
\codefieldsection{Parent}
\begin{eczvaluelist}
\item\relax
\flmRefsHyperref[eczindexfamilyrel]{code:2d_subsystem_color}{2D subsystem color code}\end{eczvaluelist}
\codefieldsection{Cousins}
\begin{eczvaluelist}
\item\relax
\flmRefsHyperref[eczindexfamilyrel]{code:quantum_divisible}{Quantum divisible code} --- Doubled color codes are subsystem codes constructed using a generalization of the doubling transformation \NoCaseChange{\protect\cite{cite659}} that combines doubly even linear binary codes to make triply even codes.
The doubling transformation is a special case of level lifting (from two to three) \NoCaseChange{\protect\cite[{Sec. VI.D}]{cite734}}.

\item\relax
\flmRefsHyperref[eczindexfamilyrel]{code:stab_15_1_3}{\(\llbracket 15,1,3\rrbracket \) quantum RM code} --- The \(\llbracket 15,1,3\rrbracket \) code can be viewed as a (gauge-fixed) doubled color code obtained from the Steane code via the doubling transformation \NoCaseChange{\protect\cite{cite731}}.
\item\relax
\flmRefsHyperref[eczindexfamilyrel]{code:stab_49_1_5}{\(\llbracket 49,1,5\rrbracket \) triorthogonal code} --- The \(\llbracket 49,1,5\rrbracket \) triorthogonal code can be viewed as a (gauge-fixed) doubled color code obtained from the \(\llbracket 17,1,5\rrbracket \) 4.8.8 color code via the doubling transformation \NoCaseChange{\protect\cite{cite731}}.
\end{eczvaluelist}
\eczhbkcontributors{ \eczhuVVA }
\endeczcode

\eczcode{da}{Dynamical code}{~\NoCaseChange{\protect\cite{cite536,cite2532}}}
\codefieldsection{Alternative Names}
\begin{eczvaluelist}
\item\relax Dynamical automorphism (DA) code
\item\relax Floquet code
\end{eczvaluelist}
\eczhIndexCodeAliasName{da}{Dynamical automorphism (DA) code}
\eczhIndexCodeAliasName{da}{Floquet code}
\codefieldsection{Description}
Dynamically generated stabilizer-based code whose (not necessarily periodic) sequence of few-body measurements implements state initialization, logical gates and error detection.

After each measurement in the sequence, the codespace is a joint \(+1\) eigenspace of an \textit{instantaneous stabilizer group (ISG)}, i.e., a particular stabilizer group corresponding to the measurement.
The ISG specifies the state of the system as a Pauli stabilizer state at a particular round of measurement, and it evolves into a (potentially) different ISG via \flmRefsHyperref{ref410}{code switching} using the group \(\mathsf{F}\) of check operators measured in the next step in the sequence.

As opposed to subsystem codes, only specific measurement sequences maintain the codespace, and not all sequences implement error detection.
Aperiodic measurement sequences provide a way to implement logical gates \NoCaseChange{\protect\cite{cite2532}}.

For dynamical codes based on topological phases, the phase associated with each ISG of the code can be obtained from a single \textit{parent topological phase} associated with the dynamical code \NoCaseChange{\protect\cite{cite2526}} via \flmRefsHyperref{ref410}{anyon condensation}.
In this way, measurements cycle logical quantum information between the various condensed phases.

\codefieldsection{Protection}
\subsection{Classification of stabilizers by masking}
There exists an efficient classical algorithm that tracks information learned by syndrome extraction at each step \NoCaseChange{\protect\cite{cite3628}}.
The algorithm performs the following classification of stabilizers into unmasked, temporarily masked, and permanently masked stabilizers.

An unmasked stabilizer is a stabilizer whose outcome can be obtained by measurements.
In general, it is not obvious to determine if a stabilizer can be unmasked as its eigenvalue may only be revealed indirectly as a product of several measurements.
A temporarily masked stabilizer is a stabilizer whose syndrome cannot be obtained by the given sequence but could possibly be obtained with future measurements.
A permanently masked stabilizer is a stabilizer whose outcome is irreversibly lost by the given sequence.

For a masked stabilizer code with a set \(U\) of \(l\) masked stabilizers, its \textit{masked distance} is given by:
\flmMathEnvironment{equation}{}{
    d_{\mathrm{u}} = \min\:\text{wt}\{ \mathsf{N}(U)\backslash \mathsf{G}\}~.
}
Above, \(\mathsf{G}\) is a gauge group defined from the algorithm that depends partly on the freedom in the choice of \flmRefsHyperref{ref3629}{destabilizers} for the temporarily masked stabilizers, and partly on the measurement sequence which fixes the \flmRefsHyperref{ref3629}{destabilizers} for the permanently masked stabilizers.

\codefieldsection{Encoding}
\begin{eczvaluelist}
\item\relax A dynamical code with \(r\) stabilizer generators can be initialized by a measurement sequence in at most \(r\) cycles \NoCaseChange{\protect\cite{cite3628}}.
\end{eczvaluelist}
\codefieldsection{Gates}
\begin{eczvaluelist}
\item\relax Code bounds for gates in the \flmTerm{term}{ref694}{}{Clifford hierarchy} (similar to the \flmRefsHyperref{ref3630}{BK bounds}) can be formulated for QLDPC codes that are embedded in a \(D\)-dimensional lattice but that admit some long-range connectivity \NoCaseChange{\protect\cite{cite3628}}.
\end{eczvaluelist}
\codefieldsection{Decoding}
\begin{eczvaluelist}
\item\relax Temporal Petz recovery map \NoCaseChange{\protect\cite{cite2594}}.
\end{eczvaluelist}
\codefieldsection{Parents}
\begin{eczvaluelist}
\item\relax
\flmRefsHyperref[eczindexfamilyrel]{code:qubits_into_qubits}{Qubit code}\item\relax
\flmRefsHyperref[eczindexfamilyrel]{code:qudit_da}{Modular-qudit dynamical code}\end{eczvaluelist}
\codefieldsection{Children}
\begin{eczvaluelist}
\item\relax
\flmRefsHyperref[eczindexfamilyrel]{code:da_color_2d}{2D DA color code} --- The 2D DA color code is a dynamical code with an aperiodic measurement sequence realizing Clifford logical gates.
\item\relax
\flmRefsHyperref[eczindexfamilyrel]{code:da_color_3d}{3D DA color code} --- The 3D DA color code is a dynamical code with an aperiodic measurement sequence realizing a non-Clifford logical gate.
\item\relax
\flmRefsHyperref[eczindexfamilyrel]{code:floquet}{Hastings-Haah Floquet code} --- Periodic Floquet codes are dynamical codes with periodic measurement sequences.
\end{eczvaluelist}
\codefieldsection{Cousins}
\begin{eczvaluelist}
\item\relax
\flmRefsHyperref[eczindexfamilyrel]{code:topological_abelian}{Abelian topological code} --- Useful measurement sequences of dynamical codes can be extracted from topological quantum field theory \NoCaseChange{\protect\cite{cite2532}}.
\item\relax
\flmRefsHyperref[eczindexfamilyrel]{code:approximate_qecc}{Approximate quantum error-correcting code (AQECC)} --- Approximate versions of dynamical codes have been formulated \NoCaseChange{\protect\cite{cite2594}}.
\item\relax
\flmRefsHyperref[eczindexfamilyrel]{code:qubit_subsystem_stabilizer}{Subsystem qubit stabilizer code} --- A dynamical code can be viewed as a subsystem qubit stabilizer code, albeit one with fewer logical qubits.
\item\relax
\flmRefsHyperref[eczindexfamilyrel]{code:monitored_random_circuits}{Monitored random-circuit code} --- Both dynamical and monitored random circuit codes can have an instantaneous stabilizer group which evolves through unitary evolution and measurements. However, dynamical codewords are generated via a specific prescribed sequence of measurements, while random-circuit codes maintain a stabilizer group after any measurement. Dynamical codes have the additional capability of detecting errors induced during the measurement process; see Appx. A of Ref. \NoCaseChange{\protect\cite{cite536}}.
\item\relax
\flmRefsHyperref[eczindexfamilyrel]{code:majorana_stab}{Majorana stabilizer code} --- Dynamical codes are viable candidates for storage in Majorana-qubit devices \NoCaseChange{\protect\cite{cite3631}}.
\item\relax
\flmRefsHyperref[eczindexfamilyrel]{code:qldpc}{Qubit QLDPC code} --- Using ZX calculus, an \(\llbracket n,k,d\rrbracket \) qubit stabilizer code admitting stabilizer generators of weight no more than \(m\) can be \textit{Floquetified} into an \(\llbracket n+\lceil m/2 \rceil+\ell,k,d^{\prime}\rrbracket \) dynamical code with single- and two-qubit operations, where \(\ell \leq \log_{2} m\) and \(d^{\prime} \geq d\) \NoCaseChange{\protect\cite{cite3293}} (see also Ref. \NoCaseChange{\protect\cite{cite3292}}). 
A more general locality-preserving \textit{spacetime concatenation} procedure yields a dynamical code out of any qubit stabilizer code by structuring measurement gadgets using low-weight measurements while ensuring the preservation of logical information \NoCaseChange{\protect\cite{cite3632}}. 
In particular, spacetime concatenation reformulates the notion of a dynamical code associated with a stabilizer code in terms of code concatenation for every qubit (spatial concatenation) and measurements between these codes (temporal concatenation), leading to a temporal evolution of the stabilizer state \NoCaseChange{\protect\cite{cite3632}}. 
A matrix rank condition on the bond operators connecting the gadgets, called the Bond-Kernel-Rank Condition, and a strict locality preservation condition (SLPC), along with preservation of the operator algebra of the stabilizer code under the gadget action, preserves fault-tolerance and the spacetime distance of the code \NoCaseChange{\protect\cite{cite3632}}.

\end{eczvaluelist}
\eczhbkcontributors{ Arpit Dua, Xiaozhen Fu, Shankar N. Balasubramanian, Nathanan Tantivasadakarn, \eczhuVVA }
\endeczcode

\eczcode{ea_design_qldpc}{EA combinatorial-design QLDPC code}{~\NoCaseChange{\protect\cite{cite138}}}
\codefieldsection{Description}
One of several EA QLDPC code families constructed from combinatorial designs.

\codefieldsection{Parent}
\begin{eczvaluelist}
\item\relax
\flmRefsHyperref[eczindexfamilyrel]{code:ea_qldpc}{EA QLDPC code}\end{eczvaluelist}
\codefieldsection{Cousin}
\begin{eczvaluelist}
\item\relax
\flmRefsHyperref[eczindexfamilyrel]{code:combinatorial_design}{Combinatorial design} --- Combinatorial designs can be used to construct EA QLDPC codes \NoCaseChange{\protect\cite{cite138}}.
\end{eczvaluelist}
\eczhbkcontributors{ \eczhuVVA }
\endeczcode

\eczcode{ea_pg_qldpc}{EA FG-QLDPC code}{~\NoCaseChange{\protect\cite{cite544}}}
\codefieldsection{Description}
One of several EA QLDPC code families constructed from finite-geometry LDPC (FG-LDPC) codes.
The construction includes families whose entanglement-consumption rate \(c/n\) decreases with block length \(n\) \NoCaseChange{\protect\cite{cite544}}.
Two such FG-based families require only one ebit (\(c=1\)) independent of code length \NoCaseChange{\protect\cite{cite544}}.

\codefieldsection{Parent}
\begin{eczvaluelist}
\item\relax
\flmRefsHyperref[eczindexfamilyrel]{code:ea_qldpc}{EA QLDPC code}\end{eczvaluelist}
\codefieldsection{Cousins}
\begin{eczvaluelist}
\item\relax
\flmRefsHyperref[eczindexfamilyrel]{code:pg_ldpc}{Finite-geometry LDPC (FG-LDPC) code} --- EA FG-QLDPC codes are entanglement-assisted quantum analogues of finite-geometry LDPC codes.
\item\relax
\flmRefsHyperref[eczindexfamilyrel]{code:pg_qldpc}{Finite-geometry (FG) qubit QLDPC code} --- EA FG-QLDPC codes are entanglement-assisted versions of FG qubit QLDPC codes.
\end{eczvaluelist}
\eczhbkcontributors{ \eczhuVVA }
\endeczcode

\eczcode{ea_qc_qldpc}{EA QC-QLDPC code}{~\NoCaseChange{\protect\cite{cite546}}}
\codefieldsection{Description}
One of several EA QLDPC code families constructed from classical QC-LDPC codes with girth at least six.
The entanglement assistance removes the dual-containing constraint in the CSS construction, avoiding many 4-cycles while retaining SPA decoding \NoCaseChange{\protect\cite{cite546}}.

\codefieldsection{Decoding}
\begin{eczvaluelist}
\item\relax Sum-product algorithm (SPA) decoder \NoCaseChange{\protect\cite{cite544}}.
\end{eczvaluelist}
\codefieldsection{Parent}
\begin{eczvaluelist}
\item\relax
\flmRefsHyperref[eczindexfamilyrel]{code:ea_qldpc}{EA QLDPC code}\end{eczvaluelist}
\codefieldsection{Cousins}
\begin{eczvaluelist}
\item\relax
\flmRefsHyperref[eczindexfamilyrel]{code:qc_ldpc}{Quasi-cyclic LDPC (QC-LDPC) code} --- EA QC-QLDPC codes are entanglement-assisted quantum analogues of QC-LDPC codes.
\item\relax
\flmRefsHyperref[eczindexfamilyrel]{code:quasi_cyclic_qldpc}{Quasi-cyclic QLDPC (QC-QLDPC) code} --- EA QC-QLDPC codes are entanglement-assisted versions of QC-QLDPC codes.
\end{eczvaluelist}
\eczhbkcontributors{ \eczhuVVA }
\endeczcode

\eczcode{ea_qldpc}{EA QLDPC code}{}
\codefieldsection{Description}
EA qubit stabilizer code for which the number of sites participating in each stabilizer generator and the number of stabilizer generators that each site participates in are both bounded by a constant \(w\) as \(n\to\infty\).

\codefieldsection{Encoding}
\begin{eczvaluelist}
\item\relax Encoder adapted for an all-optical implementation \NoCaseChange{\protect\cite{cite3633}}.
\end{eczvaluelist}
\codefieldsection{Decoding}
\begin{eczvaluelist}
\item\relax Decoder adapted for an all-optical implementation \NoCaseChange{\protect\cite{cite3633}}.
\end{eczvaluelist}
\codefieldsection{Notes}
\begin{eczvaluelist}
\item\relax Review of EA QLDPC codes provided in Ref. \NoCaseChange{\protect\cite{cite3634}}.
\end{eczvaluelist}
\codefieldsection{Parent}
\begin{eczvaluelist}
\item\relax
\flmRefsHyperref[eczindexfamilyrel]{code:eastab}{EA qubit stabilizer code}\end{eczvaluelist}
\codefieldsection{Children}
\begin{eczvaluelist}
\item\relax
\flmRefsHyperref[eczindexfamilyrel]{code:ea_design_qldpc}{EA combinatorial-design QLDPC code}\item\relax
\flmRefsHyperref[eczindexfamilyrel]{code:ea_pg_qldpc}{EA FG-QLDPC code}\item\relax
\flmRefsHyperref[eczindexfamilyrel]{code:ea_qc_qldpc}{EA QC-QLDPC code}\end{eczvaluelist}
\codefieldsection{Cousins}
\begin{eczvaluelist}
\item\relax
\flmRefsHyperref[eczindexfamilyrel]{code:qldpc}{Qubit QLDPC code} --- EA QLDPC codes utilize additional ancillary qubits in a pre-shared entangled state, but reduce to qubit QLDPC codes when said qubits are interpreted as noiseless physical qubits.
\item\relax
\flmRefsHyperref[eczindexfamilyrel]{code:ldpc}{Low-density parity-check (LDPC) code} --- There exist necessary and sufficient conditions for an EA QLDPC code consuming \(e=1\) ebit that is obtainable from a pair of LDPC codes \NoCaseChange{\protect\cite{cite1488}}.
\end{eczvaluelist}
\eczhbkcontributors{ \eczhuVVA }
\endeczcode

\eczcode{ea_quantum_convolutional}{EA quantum convolutional code}{~\NoCaseChange{\protect\cite{cite547,cite548,cite549}}}
\codefieldsection{Description}
A quantum convolutional code designed to utilize pre-shared entanglement between sender and receiver \NoCaseChange{\protect\cite{cite547,cite548,cite549}}.
Entanglement assistance removes the self-orthogonality constraint that ordinary quantum convolutional codes inherit from the stabilizer formalism, allowing arbitrary classical convolutional codes to be imported into quantum ones \NoCaseChange{\protect\cite{cite549}}.
In some constructions, the additional ebits also reduce the memory requirements of the encoding circuit \NoCaseChange{\protect\cite{cite550}}.
\codefieldsection{Encoding}
\begin{eczvaluelist}
\item\relax Importing arbitrary classical quaternary convolutional codes into entanglement-assisted quantum convolutional encoders \NoCaseChange{\protect\cite{cite547,cite549}}.
\end{eczvaluelist}
\codefieldsection{Parent}
\begin{eczvaluelist}
\item\relax
\flmRefsHyperref[eczindexfamilyrel]{code:eastab}{EA qubit stabilizer code}\end{eczvaluelist}
\codefieldsection{Cousins}
\begin{eczvaluelist}
\item\relax
\flmRefsHyperref[eczindexfamilyrel]{code:convolutional}{Convolutional code} --- EA quantum convolutional codes are entanglement-assisted quantum analogues of convolutional codes.
\item\relax
\flmRefsHyperref[eczindexfamilyrel]{code:quantum_convolutional}{Quantum convolutional code} --- EA quantum convolutional codes are entanglement-assisted versions of quantum convolutional codes.
\end{eczvaluelist}
\eczhbkcontributors{ \eczhuVVA }
\endeczcode

\eczcode{ea_turbo}{EA quantum turbo code}{~\NoCaseChange{\protect\cite{cite3635,cite3636}}}
\codefieldsection{Description}
A quantum turbo code which uses pre-shared entanglement.
This allows its encoder to be both recursive and non-catastrophic.
\codefieldsection{Rate}
Maximal-entanglement EA quantum turbo codes come close to achieving the EA hashing bound \NoCaseChange{\protect\cite{cite3636}}; see \NoCaseChange{\protect\cite[{Footnote 2}]{cite3637}}.
\codefieldsection{Encoding}
\begin{eczvaluelist}
\item\relax As opposed to quantum turbo codes with no pre-shared entanglement, EA encoders can be both recursive and non-catastrophic \NoCaseChange{\protect\cite{cite3636}}.
\end{eczvaluelist}
\codefieldsection{Parent}
\begin{eczvaluelist}
\item\relax
\flmRefsHyperref[eczindexfamilyrel]{code:eastab}{EA qubit stabilizer code}\end{eczvaluelist}
\codefieldsection{Cousins}
\begin{eczvaluelist}
\item\relax
\flmRefsHyperref[eczindexfamilyrel]{code:turbo}{Turbo code} --- EA quantum turbo codes are entanglement-assisted quantum analogues of turbo codes.
\item\relax
\flmRefsHyperref[eczindexfamilyrel]{code:quantum_turbo}{Quantum turbo code} --- EA quantum turbo codes are entanglement-assisted versions of quantum turbo codes.
\item\relax
\flmRefsHyperref[eczindexfamilyrel]{code:maximal_entanglement_galois_stabilizer}{Maximal-entanglement EA Galois-qudit stabilizer code} --- Maximal-entanglement EA quantum turbo codes come close to achieving the EA hashing bound \NoCaseChange{\protect\cite{cite3636}}; see \NoCaseChange{\protect\cite[{Footnote 2}]{cite3637}}.
\end{eczvaluelist}
\eczhbkcontributors{ \eczhuVVA }
\endeczcode

\eczcode{ea_qubits_into_qubits}{EA qubit code}{}
\codefieldsection{Description}
Qubit code designed to utilize pre-shared entanglement between sender and receiver.

\codefieldsection{Protection}
The \flmRefsHyperref{ref1729}{quantum GV bound} and Plotkin bound have been extended to EA qubit codes \NoCaseChange{\protect\cite{cite3638}}.

\codefieldsection{Rate}
There are EA versions of classical and quantum capacities \NoCaseChange{\protect\cite{cite3639}}, and the ratio of the entanglement-assisted and unassisted classical capacities of a channel is bounded by a function of the input channel's dimension \NoCaseChange{\protect\cite{cite3640}}. EA hashing bounds on the minimum entanglement required to achieve the entanglement-assisted channel capacity are derived \NoCaseChange{\protect\cite{cite2744}}.
\codefieldsection{Encoding}
\begin{eczvaluelist}
\item\relax Encoding algorithm \NoCaseChange{\protect\cite{cite548}}.
\end{eczvaluelist}
\codefieldsection{Decoding}
\begin{eczvaluelist}
\item\relax Decoding algorithm \NoCaseChange{\protect\cite{cite548}}.
\end{eczvaluelist}
\codefieldsection{Parents}
\begin{eczvaluelist}
\item\relax
\flmRefsHyperref[eczindexfamilyrel]{code:eaoa_qubits_into_qubits}{EAOA qubit code} --- An EAOA qubit code with no gauge or block structure is an EA qubit code.
\item\relax
\flmRefsHyperref[eczindexfamilyrel]{code:ea_galois_into_galois}{EA Galois-qudit code} --- EA Galois-qudit codes reduce to EA qubit codes for \(q=2\).
\end{eczvaluelist}
\codefieldsection{Child}
\begin{eczvaluelist}
\item\relax
\flmRefsHyperref[eczindexfamilyrel]{code:eastab}{EA qubit stabilizer code}\end{eczvaluelist}
\codefieldsection{Cousins}
\begin{eczvaluelist}
\item\relax
\flmRefsHyperref[eczindexfamilyrel]{code:qubits_into_qubits}{Qubit code} --- EA qubit codes utilize additional ancillary qubits in a pre-shared entangled state, but reduce to ordinary qubit codes when said qubits are interpreted as noiseless physical qubits.
\item\relax
\flmRefsHyperref[eczindexfamilyrel]{code:cws}{Codeword stabilized (CWS) code} --- EA CWS codes have been formulated \NoCaseChange{\protect\cite{cite3588}}.
\end{eczvaluelist}
\eczhbkcontributors{ \eczhuVVA }
\endeczcode

\eczcode{eastab}{EA qubit stabilizer code}{~\NoCaseChange{\protect\cite{cite1429,cite1430}}}
\codefieldsection{Description}
A code constructed using a variation of the stabilizer formalism designed to utilize pre-shared entanglement between sender and receiver.
A code is typically denoted as \(\llbracket n,k;e\rrbracket \) or \(\llbracket n,k,d;e\rrbracket \), where \(d\) is the distance of the EA code and \(e\) is the number of required pre-shared maximally entangled Bell states (ebits).
While other entangled states can be used, there is always a choice of generators such that Bell states suffice while still using the fewest ebits.

The dual of an EA qubit stabilizer code is also an EA qubit stabilizer code whose logical qubits and ebits are interchanged, \(k\leftrightarrow e\) \NoCaseChange{\protect\cite{cite3638}}.

An \(\llbracket n,k+e;e\rrbracket \) EA stabilizer code can be constructed from an ordinary \(\llbracket n,k\rrbracket \) stabilizer code with check matrix \(H=(A|B)\), where the required number of ebits is \(e = \text{rank}(AB^T+BA^T)/2\) \NoCaseChange{\protect\cite{cite3641}}.
Alternatively, given a linear \([n,k,d]_{q^2}\) code \(C\) with parity check matrix \(H\), the Hermitian dual \(C^{\perp_H}\) stabilizes an EA-QEC with parameters \(\llbracket n,2k-n+c,d;c\rrbracket _q\), where \(c=\text{rank}(HH^\dagger)\) is the number of required ebits \NoCaseChange{\protect\cite[{Thm. 27.6.1}]{cite2024}}.

\codefieldsection{Protection}
Ancillary shared entanglement is assumed to be perfect, but this assumption can be relaxed \NoCaseChange{\protect\cite{cite3642}}.
There are quantum Griesmer \NoCaseChange{\protect\cite{cite3643}} and Plotkin \NoCaseChange{\protect\cite{cite3644}} bounds for EA qubit stabilizer codes.

\codefieldsection{Rate}
Asymptotically good EA qubit stabilizer codes exist \NoCaseChange{\protect\cite{cite1431}}.
\codefieldsection{Encoding}
\begin{eczvaluelist}
\item\relax Encoders and decoders of a minimal EA qubit stabilizer code should be realizable using hyper-entangled states \NoCaseChange{\protect\cite{cite3645}}.
\item\relax Fault-tolerant encoders utilizing pre-shared entanglement \NoCaseChange{\protect\cite{cite3646}}.
\end{eczvaluelist}
\codefieldsection{Decoding}
\begin{eczvaluelist}
\item\relax Encoders and decoders of a minimal EA qubit stabilizer code should be realizable using hyper-entangled states \NoCaseChange{\protect\cite{cite3645}}.
\end{eczvaluelist}
\codefieldsection{Notes}
\begin{eczvaluelist}
\item\relax Tables of bounds and examples of EA qubit (and EA qutrit) stabilizer codes for various \(n\) and \(k\), based on algorithms developed in Refs. \NoCaseChange{\protect\cite{cite3647,cite3648}}, are maintained by M. Grassl at this \flmHref{https://www.codetables.de/}{website}.
\item\relax See Ref. \NoCaseChange{\protect\cite{cite3648}} for code tables and bounds on performance.
\item\relax See Ref. \NoCaseChange{\protect\cite{cite2763}} for related notions.
\end{eczvaluelist}
\codefieldsection{Parents}
\begin{eczvaluelist}
\item\relax
\flmRefsHyperref[eczindexfamilyrel]{code:ea_qubits_into_qubits}{EA qubit code}\item\relax
\flmRefsHyperref[eczindexfamilyrel]{code:eaoa_stabilizer}{EAOA qubit stabilizer code} --- An EAOA qubit stabilizer code with no gauge or hybrid structure is an EA qubit stabilizer code. Conversely, any \(\llbracket n,q+c,d_1;e\rrbracket \) EA qubit stabilizer code can be converted into an \(\llbracket n,q:c,d_2;e\rrbracket \) EA hybrid stabilizer code by repurposing \(c\) logical qubits as classical bits, and any \(\llbracket n,q:c,d_2;e\rrbracket \) EA hybrid code can be converted into an \(\llbracket n,q,d_3;e\rrbracket \) EA qubit stabilizer code by absorbing the classical degrees of freedom back into the quantum code, with \(d_1 \leq d_2 \leq d_3\) \NoCaseChange{\protect\cite[{Thm. 6}]{cite2735}}.
\item\relax
\flmRefsHyperref[eczindexfamilyrel]{code:ea_galois_stabilizer}{EA Galois-qudit stabilizer code} --- EA Galois-qudit stabilizer codes reduce to EA qubit stabilizer codes for \(q=2\).
\end{eczvaluelist}
\codefieldsection{Children}
\begin{eczvaluelist}
\item\relax
\flmRefsHyperref[eczindexfamilyrel]{code:branching_mera}{Branching MERA code}\item\relax
\flmRefsHyperref[eczindexfamilyrel]{code:ea_3_1_3-2}{\(\llbracket 3, 1, 3;2\rrbracket \) EA code}\item\relax
\flmRefsHyperref[eczindexfamilyrel]{code:ea_qldpc}{EA QLDPC code}\item\relax
\flmRefsHyperref[eczindexfamilyrel]{code:ea_quantum_convolutional}{EA quantum convolutional code}\item\relax
\flmRefsHyperref[eczindexfamilyrel]{code:ea_turbo}{EA quantum turbo code}\end{eczvaluelist}
\codefieldsection{Cousins}
\begin{eczvaluelist}
\item\relax
\flmRefsHyperref[eczindexfamilyrel]{code:qubit_stabilizer}{Qubit stabilizer code} --- EA qubit stabilizer codes utilize additional ancillary qubits in a pre-shared entangled state, but reduce to qubit stabilizer codes when said qubits are interpreted as noiseless physical qubits. Qubit stabilizer codes can be used to obtain shortened EA qubit stabilizer codes \NoCaseChange{\protect\cite{cite3649}}.
\item\relax
\flmRefsHyperref[eczindexfamilyrel]{code:binary_linear}{Linear binary code} --- Any linear binary code can be used to construct an EA qubit stabilizer code \NoCaseChange{\protect\cite{cite1429,cite1430,cite1431}}.
\item\relax
\flmRefsHyperref[eczindexfamilyrel]{code:q-ary_linear}{Linear \(q\)-ary code} --- Any quaternary linear code can be used to construct an EA qubit stabilizer code \NoCaseChange{\protect\cite{cite1430}}.
\item\relax
\flmRefsHyperref[eczindexfamilyrel]{code:qubit_css}{Qubit CSS code} --- As opposed to CSS codes, EA qubit stabilizer codes can be constructed from any linear binary code.
\item\relax
\flmRefsHyperref[eczindexfamilyrel]{code:hybrid_qudit_oscillator}{Mixed oscillator code} --- Encoders and decoders of a minimal EA qubit stabilizer code should be realizable using hyper-entangled states \NoCaseChange{\protect\cite{cite3645}}.
\item\relax
\flmRefsHyperref[eczindexfamilyrel]{code:qubit_concatenated}{Concatenated qubit code} --- There exist concatenated EA qubit stabilizer codes that saturate the EA quantum Griesmer and Plotkin bounds \NoCaseChange{\protect\cite{cite3608}}.
\item\relax
\flmRefsHyperref[eczindexfamilyrel]{code:ea_mds}{EA MDS code} --- There exist concatenated EA qubit stabilizer codes that saturate the EA quantum Singleton bound \NoCaseChange{\protect\cite{cite3608}}.
\item\relax
\flmRefsHyperref[eczindexfamilyrel]{code:q-ary_additive}{Additive \(q\)-ary code} --- There is a relation between quaternary additive codes and EA qubit stabilizer codes \NoCaseChange{\protect\cite{cite1714}}.
\item\relax
\flmRefsHyperref[eczindexfamilyrel]{code:asymmetric_qecc}{Asymmetric quantum code (AQC)} --- Entanglement can help decode asymmetric quantum codes \NoCaseChange{\protect\cite{cite2642}}.
\item\relax
\flmRefsHyperref[eczindexfamilyrel]{code:steane}{\(\llbracket 7,1,3\rrbracket \) Steane code} --- The Steane code is globally equivalent to a \(\llbracket 6,1,3;1\rrbracket \) EA CSS code, which the paper identifies as an example of the smallest one-ebit EA CSS code correcting an arbitrary single-qubit error on the sender's qubits \NoCaseChange{\protect\cite{cite451}}.
\item\relax
\flmRefsHyperref[eczindexfamilyrel]{code:purity_testing}{Purity-testing stabilizer code} --- Purity-testing stabilizer codes are relevant to testing the purity of an entangled Bell state shared by two parties \NoCaseChange{\protect\cite{cite3650}}.
\item\relax
\flmRefsHyperref[eczindexfamilyrel]{code:quantum_reed_muller}{Quantum Reed-Muller (RM) code} --- EA versions of quantum RM codes and their quantum tensor-product variants can be constructed \NoCaseChange{\protect\cite{cite3651}}.
\end{eczvaluelist}
\eczhbkcontributors{ Lane G. Gunderman, \eczhuVVA }
\endeczcode

\eczcode{eaoa_qubits_into_qubits}{EAOA qubit code}{~\NoCaseChange{\protect\cite{cite856}}}

\codefieldsection{Kingdom root code}
for the \flmRefsHyperref{kingdom:qubits_into_qubits}{Qubit Kingdom}.
\codefieldsection{Description}
Entanglement-assisted qubit code in the operator-algebra framework.
This family encompasses ordinary entanglement-assisted subspace qubit codes, entanglement-assisted subsystem qubit codes, entanglement-assisted hybrid qubit codes, and their operator-algebra generalizations.
\codefieldsection{Parent}
\begin{eczvaluelist}
\item\relax
\flmRefsHyperref[eczindexfamilyrel]{code:eaoaecc}{Entanglement-assisted operator-algebra QECC (EAOA QECC)} --- An EAOA QECC defined over qubits is an EAOA qubit code.
\end{eczvaluelist}
\codefieldsection{Children}
\begin{eczvaluelist}
\item\relax
\flmRefsHyperref[eczindexfamilyrel]{code:ea_qubits_into_qubits}{EA qubit code} --- An EAOA qubit code with no gauge or block structure is an EA qubit code.
\item\relax
\flmRefsHyperref[eczindexfamilyrel]{code:eaoa_stabilizer}{EAOA qubit stabilizer code} --- EAOA qubit stabilizer codes are EAOA qubit codes described within a generalized stabilizer formalism.
\end{eczvaluelist}
\codefieldsection{Cousin}
\begin{eczvaluelist}
\item\relax
\flmRefsHyperref[eczindexfamilyrel]{code:oa_qubits_into_qubits}{OA qubit code} --- EAOA qubit codes utilize additional ancillary qubits in a pre-shared entangled state, but reduce to ordinary OA qubit codes when said qubits are interpreted as noiseless physical qubits.
\end{eczvaluelist}
\eczhbkcontributors{ \eczhuVVA }
\endeczcode

\eczcode{eaoa_stabilizer}{EAOA qubit stabilizer code}{~\NoCaseChange{\protect\cite{cite856}}}
\codefieldsection{Description}
Entanglement-assisted qubit stabilizer code in the operator-algebra framework. In the generalized stabilizer formalism of \NoCaseChange{\protect\cite{cite856}}, such a code is specified on an extended qubit space by Pauli data consisting of stabilizer, gauge, logical, and sector-labeling operators, and is viewed on the original system as using noiseless ebits shared with a receiver.

EAOA qubit stabilize codes are denoted by \(\llbracket n,k; r,e,c_b\rrbracket \) or \(\llbracket n,k,d; r,e,c_b\rrbracket \), where \(n\) is the number of transmitted physical qubits, \(k\) is the number of logical qubits, \(r\) is the number of gauge qubits, \(e\) is the number of ebits, and \(c_b\) is the number of classical strings encoded. When the hybrid component encodes \(c\) classical bits, one has \(c_b=2^c\).
This family encompasses ordinary entanglement-assisted qubit stabilizer codes, entanglement-assisted subsystem stabilizer codes, entanglement-assisted hybrid stabilizer codes, and operator-algebra generalizations described within that stabilizer formalism.
The framework also exhibits EA hybrid subspace and EA subsystem stabilizer codes that lie outside the earlier EACQ and EAOQECC formalisms \NoCaseChange{\protect\cite{cite856}}.

There exist four constructions of EAOA qubit stabilizer codes with distance lower bounds:
gauge fixing \(\llbracket n,k,d;r,e,c_b\rrbracket  \to \llbracket n,k,d';r-y,e,\leq 2^y c_b\rrbracket \),
clean qubits \(\llbracket n,k,d;r,0,c_b\rrbracket  \to \llbracket n-e,k,d';r,e,c_b\rrbracket \),
entanglement-assisted gauge fixing \(\llbracket n,k,d;r,0,c_b\rrbracket  \to \llbracket n,k,d';r-e,e,c_b\rrbracket \), and
general gauge fixing \(\llbracket n,k,d;r,e,c_b\rrbracket  \to \llbracket n,k,d';r-y_I-y_S,e+y_S,\leq 2^{y_I}c_b\rrbracket \)
\NoCaseChange{\protect\cite{cite856}}.
\codefieldsection{Protection}
For stabilizer-described qubit subclasses, the EAOAQEC framework yields explicit Pauli error-correction conditions for errors acting on the sender's qubits under the usual assumption that the receiver's halves of the ebits are noiseless \NoCaseChange{\protect\cite{cite856}}.
Its dressed distance is the minimum weight over logical-centralizer operators outside the isotropic-plus-gauge subgroup together with inter-sector cosets, reducing to the usual distance notions for EAQEC, EAOQECC, EACQ, OAQEC, and related stabilizer code families in the appropriate limits \NoCaseChange{\protect\cite[{Def. 2 and Rem. 1}]{cite856}}.
\codefieldsection{Parent}
\begin{eczvaluelist}
\item\relax
\flmRefsHyperref[eczindexfamilyrel]{code:eaoa_qubits_into_qubits}{EAOA qubit code} --- EAOA qubit stabilizer codes are EAOA qubit codes described within a generalized stabilizer formalism.
\end{eczvaluelist}
\codefieldsection{Children}
\begin{eczvaluelist}
\item\relax
\flmRefsHyperref[eczindexfamilyrel]{code:eastab}{EA qubit stabilizer code} --- An EAOA qubit stabilizer code with no gauge or hybrid structure is an EA qubit stabilizer code. Conversely, any \(\llbracket n,q+c,d_1;e\rrbracket \) EA qubit stabilizer code can be converted into an \(\llbracket n,q:c,d_2;e\rrbracket \) EA hybrid stabilizer code by repurposing \(c\) logical qubits as classical bits, and any \(\llbracket n,q:c,d_2;e\rrbracket \) EA hybrid code can be converted into an \(\llbracket n,q,d_3;e\rrbracket \) EA qubit stabilizer code by absorbing the classical degrees of freedom back into the quantum code, with \(d_1 \leq d_2 \leq d_3\) \NoCaseChange{\protect\cite[{Thm. 6}]{cite2735}}.
\item\relax
\flmRefsHyperref[eczindexfamilyrel]{code:eaoa_hamming}{\(\llbracket 10,1,3;1,3,4\rrbracket \) EAOA Hamming code}\end{eczvaluelist}
\codefieldsection{Cousins}
\begin{eczvaluelist}
\item\relax
\flmRefsHyperref[eczindexfamilyrel]{code:qubit_stabilizer_oaqecc}{Operator-algebra (OA) qubit stabilizer code} --- EAOA qubit stabilizer codes utilize additional ancillary subsystems in a pre-shared entangled state, but reduce to OA qubit stabilizer codes when said subsystems are interpreted as noiseless physical subsystems.
\item\relax
\flmRefsHyperref[eczindexfamilyrel]{code:hybrid_stabilizer}{Hybrid stabilizer code} --- EA hybrid qubit stabilizer codes utilize additional ancillary subsystems in a pre-shared entangled state, but reduce to hybrid qubit stabilizer codes when said subsystems are interpreted as noiseless physical subsystems. In the original EA hybrid stabilizer formalism, an \(\llbracket n,q:c,d;e\rrbracket \) EA hybrid stabilizer code is specified by a pair \((\mathcal{S}_Q,\mathcal{S}_C)\) of quantum and classical stabilizer groups, and in the equivalent symplectic formalism by a quantum parity-check matrix together with a classical parity-check matrix \NoCaseChange{\protect\cite[{Thms. 1-4}]{cite2735}}. Inside the EAOAQEC stabilizer framework, hybrid stabilizer codes are a proper subclass of the broader EA hybrid subspace codes because the EACQ transversal operators obey additional constraints not required in general \NoCaseChange{\protect\cite{cite856}}.
\item\relax
\flmRefsHyperref[eczindexfamilyrel]{code:eacq}{Entanglement-assisted (EA) hybrid QECC} --- The original EACQ formalism describes a proper subclass of EA hybrid subspace codes inside the EAOAQEC stabilizer framework; EACQ representability imposes extra constraints on the transversal operators beyond belonging to distinct normalizer cosets \NoCaseChange{\protect\cite{cite856}}.
\item\relax
\flmRefsHyperref[eczindexfamilyrel]{code:subsystem_color}{Subsystem color code} --- The 15-qubit subsystem color code yields several EAOA qubit stabilizer constructions, including \(\llbracket 13,1,3;6,2,3\rrbracket \), \(\llbracket 15,1,3;5,1,2\rrbracket \), and \(\llbracket 15,1,3;4,1,4\rrbracket \) examples obtained via clean-qubits and entanglement-assisted gauge-fixing constructions \NoCaseChange{\protect\cite{cite856}}.
\end{eczvaluelist}
\eczhbkcontributors{ \eczhuVVA }
\endeczcode

\eczcode{eth}{Eigenstate thermalization hypothesis (ETH) code}{~\NoCaseChange{\protect\cite{cite3652}}}
\codefieldsection{Alternative Names}
\begin{eczvaluelist}
\item\relax Thermodynamic code
\end{eczvaluelist}
\eczhIndexCodeAliasName{eth}{Thermodynamic code}
\codefieldsection{Description}
An \(n\)-qubit approximate code whose codespace is formed by eigenstates of a translationally-invariant quantum many-body system which satisfies the Eigenstate Thermalization Hypothesis (ETH).
ETH ensures that codewords cannot be locally distinguished in the thermodynamic limit.
Relevant many-body systems include 1D non-interacting spin chains or frustration-free systems such as Motzkin chains and Heisenberg models.

ETH requires that for ordered energy eigenstates \(|E_l\rangle\) and any local observable \(O\),
\flmMathEnvironment{align}{}{
|\langle E_l|O|E_l\rangle-\langle E_{l+1}|O|E_{l+1}\rangle|\leq\exp(-cn)
}
for a constant \(c\).
This implies that energy eigenstates around some energy \(\bar E\) are approximately locally indistinguishable from one another, as their reduced density matrices on any subsystem are both approximately thermal at energy \(\bar E\).
In this way, global information is protected from local measurements by the environment as \(n\to\infty\).

\codefieldsection{Protection}
Approximately protects against erasure errors at known locations. Translation invariance alone is sufficient for good approximate error-correcting properties in a many-body spectrum, including in integrable models \NoCaseChange{\protect\cite{cite3652}}. The ETH code generated from the spectrum of the translation-invariant 1D Heisenberg spin chain \NoCaseChange{\protect\cite{cite3652}} has recovery infidelity against the erasure of a constant number of sites scaling as \(\epsilon_\text{worst}=O(1/n)\) \NoCaseChange{\protect\cite{cite2720}}.

The ETH code defined on a Heisenberg spin chain has unbounded \flmRefsHyperref{ref2559}{codespace complexity} \NoCaseChange{\protect\cite{cite2564}}.

\codefieldsection{Decoding}
\begin{eczvaluelist}
\item\relax An explicit universal recovery channel for the ETH code is given in \NoCaseChange{\protect\cite{cite3653}}.
\end{eczvaluelist}
\codefieldsection{Parents}
\begin{eczvaluelist}
\item\relax
\flmRefsHyperref[eczindexfamilyrel]{code:qubits_into_qubits}{Qubit code} --- ETH codewords are eigenstates of a local Hamiltonian whose eigenstates satisfy ETH.
\item\relax
\flmRefsHyperref[eczindexfamilyrel]{code:hamiltonian}{Hamiltonian-based code} --- ETH codewords are eigenstates of a local Hamiltonian whose eigenstates satisfy ETH, and many example codes are eigenstates of frustration-free Hamiltonians.
\item\relax
\flmRefsHyperref[eczindexfamilyrel]{code:approximate_qecc}{Approximate quantum error-correcting code (AQECC)} --- ETH codes approximately protect against erasures in the thermodynamic limit. There is a link between ETH and approximate QEC, with fluctuations of the infinite-time average of certain observables expressible in terms of code error \NoCaseChange{\protect\cite{cite2602}}.
\end{eczvaluelist}
\codefieldsection{Cousins}
\begin{eczvaluelist}
\item\relax
\flmRefsHyperref[eczindexfamilyrel]{code:topological}{Topological code} --- ETH codewords, like topological codewords, are locally indistinguishable.
\item\relax
\flmRefsHyperref[eczindexfamilyrel]{code:qubit_permutation_invariant}{PI qubit code} --- Several instances of ETH codes contain PI qubit codewords.
\item\relax
\flmRefsHyperref[eczindexfamilyrel]{code:spins_into_spins}{Spin code} --- Relevant many-body systems housing ETH codes include 1D non-interacting spin chains or frustration-free systems such as Motzkin chains and Heisenberg models.
\item\relax
\flmRefsHyperref[eczindexfamilyrel]{code:frustration_free}{Frustration-free Hamiltonian code} --- ETH codewords are eigenstates of a local Hamiltonian whose eigenstates satisfy ETH, and many example codes are eigenstates of frustration-free Hamiltonians.
\item\relax
\flmRefsHyperref[eczindexfamilyrel]{code:covariant}{Covariant block quantum code} --- ETH codes consisting of \flmRefsHyperref{ref526}{Dicke states} are approximately \(U(1)\)-covariant and nearly saturate certain covariance-performance bounds \NoCaseChange{\protect\cite{cite2720,cite2718}}.
\item\relax
\flmRefsHyperref[eczindexfamilyrel]{code:w_state}{W-state code} --- The W-state is not a unique ground state of any local Hamiltonian \NoCaseChange{\protect\cite{cite3165}}.
\item\relax
\flmRefsHyperref[eczindexfamilyrel]{code:mps}{Magnon code} --- Magnon codes have been shown to protect against non-geometrically local noise, while ETH codes protect only against erasures on geometrically local patches.
\end{eczvaluelist}
\eczhbkcontributors{ Chris Fechisin, \eczhuVVA }
\endeczcode

\eczcode{fermions}{Fermion code}{~\NoCaseChange{\protect\cite{cite557}}}
\codefieldsection{Description}
Finite-dimensional quantum error-correcting code encoding a logical qudit or fermionic Hilbert space into a physical Fock space of fermionic modes.
Codes are typically described using Majorana operators, which are linear combinations of fermionic creation and annihilation operators \NoCaseChange{\protect\cite{cite557}}.
Majorana operators may either be considered individually or paired in various ways into creation and annihilation operators to yield fermionic modes.
They form a Clifford algebra and can be interpreted as Ising anyons in certain contexts.

Admissible codewords include fermionic states, a subset of which is the Gaussian fermionic states \NoCaseChange{\protect\cite{cite3654,cite3655,cite3656,cite3657,cite3658}}.
Gaussian fermionic states are analogues of (non-displaced) Gaussian bosonic states; they are labeled by points in a Grassmannian and are sometimes called fermionic coherent states \NoCaseChange{\protect\cite{cite3659}}. 
Fermionic analogues of ordinary (bosonic) coherent states are the fermionic coherent states labeled by Grassmann numbers \NoCaseChange{\protect\cite{cite3660,cite3661}}.
A Wigner function formalism has been developed for fermionic states \NoCaseChange{\protect\cite{cite3662}}.

\codefieldsection{Protection}
Majorana analogues of common qubit noise channels have been developed \NoCaseChange{\protect\cite{cite3663}}.

\codefieldsection{Encoding}
\begin{eczvaluelist}
\item\relax A fermionic code using fermion Fock states as codewords cannot protect against occupation-number errors (i.e., dephasing) and does not admit fermionic logical operators \NoCaseChange{\protect\cite{cite559,cite3664}}.
\end{eczvaluelist}
\codefieldsection{Gates}
\begin{eczvaluelist}
\item\relax Clifford operations on fermionic codes, shown \NoCaseChange{\protect\cite{cite3654}} to be equivalent to match gates \NoCaseChange{\protect\cite{cite3665}}, can be formulated using \textit{Fermionic Linear Optics}, a classically simulable model of computation \NoCaseChange{\protect\cite{cite3654,cite3655,cite3656,cite3666,cite3657,cite3658}}. The structure of the Majorana Clifford group has been studied \NoCaseChange{\protect\cite{cite3667}}.
\item\relax Non-Clifford gates can be done using gate teleportation, in which a gate can be obtained from a particular magic state (a.k.a. resource state) \NoCaseChange{\protect\cite{cite3666,cite3668,cite3669,cite3670,cite3671}}.
\item\relax General gates include qubit-like \(S\), \(T\), and \(CZ\) gates acting on either logical qubit or logical fermionic encodings. Fermionic gates include braiding gates which correspond to exchanging Majorana modes. Hybrid gates include \(CZ_{qf}\) gates between a logical qubit and a logical fermion. The braiding, \(CZ_{f}\), \(CZ_{qf}\), Hadamard, \(S\), and \(T\) gates are universal \NoCaseChange{\protect\cite{cite559}}.
\item\relax Logical-fermion circuits constructed out of certain transversal gates do not admit a lower \(T\) gate count than logical-qubit circuits \NoCaseChange{\protect\cite{cite559}}.
\item\relax Using fermion codes with logical fermion encodings and the fermionic fast Fourier transform \NoCaseChange{\protect\cite{cite3672}} can yield exponential improvements in circuit depth over fermion-into-qubit encodings \NoCaseChange{\protect\cite{cite559}}.
\end{eczvaluelist}
\codefieldsection{Notes}
\begin{eczvaluelist}
\item\relax See Ref. \NoCaseChange{\protect\cite{cite3673}} for an introduction to Majorana-based qubits.
\end{eczvaluelist}
\codefieldsection{Parent}
\begin{eczvaluelist}
\item\relax
\flmRefsHyperref[eczindexfamilyrel]{code:qubits_into_qubits}{Qubit code} --- The Majorana operator algebra is isomorphic to the qubit Pauli-operator algebra via various fermion-into-qubit encodings.
\end{eczvaluelist}
\codefieldsection{Children}
\begin{eczvaluelist}
\item\relax
\flmRefsHyperref[eczindexfamilyrel]{code:syk}{SYK code}\item\relax
\flmRefsHyperref[eczindexfamilyrel]{code:majorana_stab}{Majorana stabilizer code}\end{eczvaluelist}
\codefieldsection{Cousins}
\begin{eczvaluelist}
\item\relax
\flmRefsHyperref[eczindexfamilyrel]{code:oscillators}{Bosonic code} --- Bosonic (fermionic) codes are associated with bosonic (fermionic) degrees of freedom.
\item\relax
\flmRefsHyperref[eczindexfamilyrel]{code:fermions_into_qubits}{Fermion-into-qubit code} --- Fermion (fermion-into-qubit) codes encode logical information into a physical space of fermionic modes (qubits).
The Majorana operator algebra is isomorphic to the qubit Pauli-operator algebra via various fermion-into-qubit encodings.
Using fermion codes with logical fermion encodings and the fermionic fast Fourier transform \NoCaseChange{\protect\cite{cite3672}} can yield exponential improvements in circuit depth over fermion-into-qubit encodings \NoCaseChange{\protect\cite{cite559}}.

\item\relax
\flmRefsHyperref[eczindexfamilyrel]{code:constant_excitation}{Constant-excitation (CE) code} --- Fermion codewords lying in a fixed fermion-number subspace have to lie in the same subspace in order to protect against changes in fermion number \NoCaseChange{\protect\cite{cite559}}.
\end{eczvaluelist}
\eczhbkcontributors{ Michael Gullans, Alexander Schuckert, \eczhuVVA }
\endeczcode

\eczcode{fermions_into_qubits}{Fermion-into-qubit code}{}
\codefieldsection{Description}
Qubit stabilizer code encoding a logical fermionic Hilbert space into a physical space of \(n\) qubits.
Such codes are primarily intended for simulating fermionic systems on quantum computers, and some of them have error-detecting, correcting, and transmuting properties.

The first fermion-into-qubit code is the Jordan-Wigner transformation code, a trivial \(\llbracket n,n\rrbracket \) stabilizer code encoding Majorana operators into Pauli strings of weight \(O(n)\).
This is necessary to ensure that Majorana operators satisfy the proper anti-commutation relations.

Subsequent encodings consisted of stabilizer codes with \(k < n\), ensuring anti-commutation through the long-range entanglement of the codestates.
This makes it possible to reduce the Pauli weight of a Majorana operator by multiplying by a stabilizer.
For example, for stabilizer constraints associated with loops of a 2D lattice, applying loop constraints to high-weight fermionic string-like operators yields operators of lower weight.

See \NoCaseChange{\protect\cite[{Table I}]{cite412}\protect\cite[{Table I}]{cite404}} for comparisons of various fermion-into-qubit codes.
On the square lattice, explicit local 2D encodings with qubit-fermion ratios \(2\), \(1.5\), and \(1.25\) are exhibited in Ref. \NoCaseChange{\protect\cite{cite404}}.
In addition to the children of this entry, various custom encodings exist \NoCaseChange{\protect\cite{cite3622,cite3674,cite3675,cite3676,cite3677,cite3678}} that can be tailored to the quantum simulation problem of interest.

\codefieldsection{Encoding}
\begin{eczvaluelist}
\item\relax Any two locality-preserving fermion-into-qubit mappings in two spatial dimensions can be related by a finite-depth generalized local unitary transformation \NoCaseChange{\protect\cite{cite404}}.
\end{eczvaluelist}
\codefieldsection{Parent}
\begin{eczvaluelist}
\item\relax
\flmRefsHyperref[eczindexfamilyrel]{code:qubit_stabilizer}{Qubit stabilizer code} --- Fermion-into-qubit codes are qubit stabilizer codes that encode a logical fermionic Hilbert space into a physical space of \(n\) qubits.
\end{eczvaluelist}
\codefieldsection{Children}
\begin{eczvaluelist}
\item\relax
\flmRefsHyperref[eczindexfamilyrel]{code:aqm}{Auxiliary qubit mapping (AQM) code}\item\relax
\flmRefsHyperref[eczindexfamilyrel]{code:bkt}{Bravyi-Kitaev transformation (BKT) code}\item\relax
\flmRefsHyperref[eczindexfamilyrel]{code:bosonization}{Bosonization code}\item\relax
\flmRefsHyperref[eczindexfamilyrel]{code:jw}{Jordan-Wigner transformation code}\item\relax
\flmRefsHyperref[eczindexfamilyrel]{code:ternary_tree_fermion}{Ternary-tree fermion-into-qubit code}\end{eczvaluelist}
\codefieldsection{Cousins}
\begin{eczvaluelist}
\item\relax
\flmRefsHyperref[eczindexfamilyrel]{code:twist_defect_surface}{Twist-defect surface code} --- Treating a twist-defect surface codespace as a logical fermion encoding yields a fermion-into-qubit code \NoCaseChange{\protect\cite{cite3679}}.
\item\relax
\flmRefsHyperref[eczindexfamilyrel]{code:gauss_law}{Gauss' law code} --- Gauge-group elements of a \(D\)-dimensional fermionic \(\mathbb{Z}_2\) gauge theory can arise from a single-error-correcting linear binary code \NoCaseChange{\protect\cite[{Thm. 1}]{cite78}}.
\item\relax
\flmRefsHyperref[eczindexfamilyrel]{code:fermions}{Fermion code} --- Fermion (fermion-into-qubit) codes encode logical information into a physical space of fermionic modes (qubits).
The Majorana operator algebra is isomorphic to the qubit Pauli-operator algebra via various fermion-into-qubit encodings.
Using fermion codes with logical fermion encodings and the fermionic fast Fourier transform \NoCaseChange{\protect\cite{cite3672}} can yield exponential improvements in circuit depth over fermion-into-qubit encodings \NoCaseChange{\protect\cite{cite559}}.

\end{eczvaluelist}
\eczhbkcontributors{ Yijia Xu, \eczhuVVA }
\endeczcode

\eczcode{fiber_bundle}{Fiber-bundle code}{~\NoCaseChange{\protect\cite{cite3477}}}
\codefieldsection{Alternative Names}
\begin{eczvaluelist}
\item\relax Twisted product code
\end{eczvaluelist}
\eczhIndexCodeAliasName{fiber_bundle}{Twisted product code}
\codefieldsection{Description}
A CSS code constructed by combining one code as the base and another as the fiber of a fiber bundle.
In particular, taking a random LDPC code as the base and a cyclic repetition code as the fiber yields, after distance balancing, a QLDPC code with distance of \flmRefsHyperref{ref65}{order} \(\Omega( n^{3/5}/\text{polylog}(n) )\) and rate of \flmRefsHyperref{ref65}{order} \(\Omega( n^{-2/5}/\text{polylog}(n) )\).

\codefieldsection{Rate}
Rate \(k/n = \Omega( n^{-2/5}/\text{polylog}(n) )\), distance \(d=\Omega( n^{3/5}/\text{polylog}(n) )\). This is the first QLDPC code to achieve a distance scaling better than \(\sqrt{n}~\text{polylog}(n)\).
\codefieldsection{Decoding}
\begin{eczvaluelist}
\item\relax Greedy algorithm can be used to efficiently decode \(X\) errors, but no known efficient decoding of \(Z\) errors yet \NoCaseChange{\protect\cite{cite3477}}.
\end{eczvaluelist}
\codefieldsection{Parents}
\begin{eczvaluelist}
\item\relax
\flmRefsHyperref[eczindexfamilyrel]{code:qubit_generalized_homological_product_css}{Generalized homological-product qubit CSS code}\item\relax
\flmRefsHyperref[eczindexfamilyrel]{code:balanced_product}{Balanced product (BP) code} --- Fiber-bundle codes can be formulated in terms of a balanced product \NoCaseChange{\protect\cite{cite434}}.
\end{eczvaluelist}
\codefieldsection{Child}
\begin{eczvaluelist}
\item\relax
\flmRefsHyperref[eczindexfamilyrel]{code:homological_product}{Homological product code} --- A fiber-bundle code can be viewed as a homological product code with a twisted product.
\end{eczvaluelist}
\codefieldsection{Cousins}
\begin{eczvaluelist}
\item\relax
\flmRefsHyperref[eczindexfamilyrel]{code:lifted_product}{Lifted-product (LP) code} --- The specific fiber-bundle QLDPC code achieving a distance scaling better than \(\sqrt{n}~\text{polylog}(n)\) can also be formulated directly as an LP code (see published version of Ref.\NoCaseChange{\protect\cite{cite3477}}).
Lifted products of a length-one with a length-\(m\) chain complex can be thought of as fiber-bundle codes \NoCaseChange{\protect\cite{cite434}}.

\item\relax
\flmRefsHyperref[eczindexfamilyrel]{code:distance_balanced}{Distance-balanced code} --- Fiber-bundle code constructions use distance balancing and weight reduction to increase distance.
\item\relax
\flmRefsHyperref[eczindexfamilyrel]{code:random_stabilizer}{Random stabilizer code} --- Taking a random LDPC code as the base and a cyclic repetition code as the fiber yields, after distance balancing, a QLDPC code with distance of \flmRefsHyperref{ref65}{order} \(\Omega( n^{3/5}\text{polylog}(n) )\) and rate of \flmRefsHyperref{ref65}{order} \(\Omega( n^{-2/5}\text{polylog}(n) )\).
\end{eczvaluelist}
\eczhbkcontributors{ Jon Nelson, \eczhuVVA }
\endeczcode

\eczcode{fibonacci_fractal_liquid}{Fibonacci fractal spin-liquid code}{~\NoCaseChange{\protect\cite{cite1348}}}
\codefieldsection{Alternative Names}
\begin{eczvaluelist}
\item\relax Fibonacci prism model code
\end{eczvaluelist}
\eczhIndexCodeAliasName{fibonacci_fractal_liquid}{Fibonacci prism model code}
\codefieldsection{Description}
A fractal type-I fracton CSS code defined on a cubic lattice \NoCaseChange{\protect\cite[{Eq. (D23)}]{cite456}}.

\codefieldsection{Parents}
\begin{eczvaluelist}
\item\relax
\flmRefsHyperref[eczindexfamilyrel]{code:hypergraph_product}{Hypergraph product (HGP) code} --- The Fibonacci fractal spin-liquid code is a hypergraph product of the repetition code and the Fibonacci code \NoCaseChange{\protect\cite{cite1348}}, and can be formulated directly as a BP code \NoCaseChange{\protect\cite{cite1350}}.
\item\relax
\flmRefsHyperref[eczindexfamilyrel]{code:fracton}{Fracton stabilizer code} --- The Fibonacci fractal spin-liquid code is a fractal type-I fracton code \NoCaseChange{\protect\cite{cite456}}.
\end{eczvaluelist}
\codefieldsection{Cousins}
\begin{eczvaluelist}
\item\relax
\flmRefsHyperref[eczindexfamilyrel]{code:fibonacci_model}{Fibonacci code} --- The Fibonacci fractal spin-liquid code is a hypergraph product of the repetition code and the Fibonacci code \NoCaseChange{\protect\cite{cite1348}}, and can be formulated directly as a BP code \NoCaseChange{\protect\cite{cite1350}}.
\item\relax
\flmRefsHyperref[eczindexfamilyrel]{code:repetition}{Repetition code} --- The Fibonacci fractal spin-liquid code is a hypergraph product of the repetition code and the Fibonacci code \NoCaseChange{\protect\cite{cite1348}}, and can be formulated directly as a BP code \NoCaseChange{\protect\cite{cite1350}}.
\item\relax
\flmRefsHyperref[eczindexfamilyrel]{code:balanced_product}{Balanced product (BP) code} --- The Fibonacci fractal spin-liquid code is a hypergraph product of the repetition code and the Fibonacci code \NoCaseChange{\protect\cite{cite1348}}, and can be formulated directly as a BP code \NoCaseChange{\protect\cite{cite1350}}.
\item\relax
\flmRefsHyperref[eczindexfamilyrel]{code:bosonization}{Bosonization code} --- Bosonization can be used to realize a Fibonacci fractal spin-liquid code with an emergent fermion from a Majorana stabilizer code \NoCaseChange{\protect\cite{cite3503}}. This code is shown to be distinct from the original code \NoCaseChange{\protect\cite{cite3506}}.
\end{eczvaluelist}
\eczhbkcontributors{ \eczhuVVA }
\endeczcode

\eczcode{pg_qldpc}{Finite-geometry (FG) qubit QLDPC code}{~\NoCaseChange{\protect\cite{cite1356,cite834}\protect\cite[{Ch. 14}]{cite872}\protect\cite[{Sec. 4.1}]{cite835}}}
\codefieldsection{Description}
CSS code constructed from linear binary codes whose parity-check or generator matrices are incidence matrices of points, hyperplanes, or other structures in finite geometries.
These codes can be interpreted as quantum versions of FG-LDPC codes, but some of them \NoCaseChange{\protect\cite{cite834,cite835}} are not strictly QLDPC.

\codefieldsection{Parents}
\begin{eczvaluelist}
\item\relax
\flmRefsHyperref[eczindexfamilyrel]{code:qubit_css}{Qubit CSS code}\item\relax
\flmRefsHyperref[eczindexfamilyrel]{code:qldpc}{Qubit QLDPC code}\end{eczvaluelist}
\codefieldsection{Child}
\begin{eczvaluelist}
\item\relax
\flmRefsHyperref[eczindexfamilyrel]{code:steane}{\(\llbracket 7,1,3\rrbracket \) Steane code} --- The Steane code is the \(m=1\) member of the \(\llbracket 2^{2m}+2^{m}+1,1,>2^{m}\rrbracket \) PG-QLDPC code family that is constructed from codes corresponding to lines and affine charts in \(PG(2,2^m)\) via the CSS construction \NoCaseChange{\protect\cite[{Def. 4.9}]{cite835}}.
\end{eczvaluelist}
\codefieldsection{Cousins}
\begin{eczvaluelist}
\item\relax
\flmRefsHyperref[eczindexfamilyrel]{code:pg_ldpc}{Finite-geometry LDPC (FG-LDPC) code} --- Quantum versions of PG-LDPC and EG-LDPC codes can be constructed via the CSS construction \NoCaseChange{\protect\cite{cite1356,cite834}}.
\item\relax
\flmRefsHyperref[eczindexfamilyrel]{code:projective}{Projective geometry code} --- PG-QLDPC codes are constructed from linear binary codes whose parity-check or generator matrices are incidence matrices of structures in finite geometries.
\item\relax
\flmRefsHyperref[eczindexfamilyrel]{code:multisector_hypergraph}{Higher-dimensional homological product code} --- Multi-dimensional homological products of PG-QLDPC codes yield families whose stabilizer-generator weights scale logarithmically with \(n\) \NoCaseChange{\protect\cite[{Cor. 2.23}]{cite835}\protect\cite[{Sec. 4.1}]{cite835}}.
\item\relax
\flmRefsHyperref[eczindexfamilyrel]{code:ea_pg_qldpc}{EA FG-QLDPC code} --- EA FG-QLDPC codes are entanglement-assisted versions of FG qubit QLDPC codes.
\end{eczvaluelist}
\eczhbkcontributors{ \eczhuVVA }
\endeczcode

\eczcode{floquet_3d_fermionic_surface}{Floquet 3D fermionic surface code}{~\NoCaseChange{\protect\cite{cite533}}}
\codefieldsection{Description}
A 3D Floquet code on a trivalent lattice whose weight-two checks are the \(XX\), \(YY\), and \(ZZ\) edge terms of the 3D Kitaev honeycomb model \NoCaseChange{\protect\cite{cite458,cite533}}.

A rewinding sixteen-round schedule yields ISGs that are FDLQC-equivalent to the 3D fermionic surface code.
The rewinding avoids measuring all non-contractible-loop logical operators, so on periodic boundaries the Floquet code preserves a single logical qubit even though the static 3D fermionic surface code has three \NoCaseChange{\protect\cite{cite533}}.

\codefieldsection{Rate}
With periodic boundary conditions, the rewinding schedule preserves a single logical qubit because two logical operators are inferred during the cycle \NoCaseChange{\protect\cite{cite533}}.
\codefieldsection{Parent}
\begin{eczvaluelist}
\item\relax
\flmRefsHyperref[eczindexfamilyrel]{code:floquet}{Hastings-Haah Floquet code}\end{eczvaluelist}
\codefieldsection{Cousins}
\begin{eczvaluelist}
\item\relax
\flmRefsHyperref[eczindexfamilyrel]{code:3d_fermionic_surface}{3D fermionic surface code} --- Each ISG of the Floquet 3D fermionic surface code is FDLQC-equivalent to the 3D fermionic surface code \NoCaseChange{\protect\cite{cite533}}.
\item\relax
\flmRefsHyperref[eczindexfamilyrel]{code:3d_kitaev_honeycomb}{3D Kitaev honeycomb code} --- The weight-two check operators of the Floquet 3D fermionic surface code are those of the 3D Kitaev honeycomb model \NoCaseChange{\protect\cite{cite458,cite533}}.
\end{eczvaluelist}
\eczhbkcontributors{ Arpit Dua, \eczhuVVA }
\endeczcode

\eczcode{floquet_3d_surface}{Floquet 3D surface code}{~\NoCaseChange{\protect\cite{cite533}}}
\codefieldsection{Description}
A 3D Floquet code on a truncated cubic honeycomb with pairs of physical qubits on vertices.
It is constructed from three stacks of square-octagon Floquet toric codes, coupled by interlayer \(YY\) measurements in a coupled-layer construction \NoCaseChange{\protect\cite{cite534,cite533}}.

The rewinding schedule \(GBRBGR\) yields ISGs that are FDLQC-equivalent either to the 3D surface code or to two copies of the 3D surface code up to non-local stabilizers.
A parent stabilizer code for this Floquet code is FDQC-equivalent to a 3-foliated stack of 2D color codes \NoCaseChange{\protect\cite{cite533}}.

\codefieldsection{Rate}
On a three-torus of linear size \(L\), the Floquet code preserves three logical qubits \NoCaseChange{\protect\cite{cite533}}.
\codefieldsection{Gates}
\begin{eczvaluelist}
\item\relax A planar Floquet 3D surface code stacked with two planar 3D subsystem surface codes prepares an instantaneous state equivalent to a 3D surface code stacked with two checkerboard model codes, enabling a logical \(CCZ\) gate \NoCaseChange{\protect\cite{cite533}}.
\end{eczvaluelist}
\codefieldsection{Parent}
\begin{eczvaluelist}
\item\relax
\flmRefsHyperref[eczindexfamilyrel]{code:floquet}{Hastings-Haah Floquet code}\end{eczvaluelist}
\codefieldsection{Cousins}
\begin{eczvaluelist}
\item\relax
\flmRefsHyperref[eczindexfamilyrel]{code:3d_surface}{3D surface code} --- The G-round ISG is FDLQC-equivalent to the 3D surface code, while the other rounds are FDLQC-equivalent to two copies of the 3D surface code up to non-local stabilizers \NoCaseChange{\protect\cite{cite533}}.
\item\relax
\flmRefsHyperref[eczindexfamilyrel]{code:2d_color}{2D color code} --- A parent stabilizer code for the Floquet 3D surface code is FDQC-equivalent to a 3-foliated stack of 2D color codes \NoCaseChange{\protect\cite{cite533}}.
\item\relax
\flmRefsHyperref[eczindexfamilyrel]{code:3d_subsystem_surface}{3D subsystem surface code} --- A planar Floquet 3D surface code stacked with two planar 3D subsystem surface codes prepares an instantaneous state equivalent to a 3D surface code stacked with two checkerboard model codes, enabling a logical \(CCZ\) gate \NoCaseChange{\protect\cite{cite533}}.
\item\relax
\flmRefsHyperref[eczindexfamilyrel]{code:checkerboard}{Checkerboard model code} --- A planar Floquet 3D surface code stacked with two planar 3D subsystem surface codes prepares an instantaneous state equivalent to a 3D surface code stacked with two checkerboard model codes, enabling a logical \(CCZ\) gate \NoCaseChange{\protect\cite{cite533}}.
\end{eczvaluelist}
\eczhbkcontributors{ Arpit Dua, \eczhuVVA }
\endeczcode

\eczcode{floquet_color}{Floquet color code}{~\NoCaseChange{\protect\cite{cite3680,cite538,cite2526}}}
\codefieldsection{Alternative Names}
\begin{eczvaluelist}
\item\relax CSS Floquet toric code
\item\relax \(\mathbb{Z}_2\) Floquet code
\item\relax CSS honeycomb code
\end{eczvaluelist}
\eczhIndexCodeAliasName{floquet_color}{CSS Floquet toric code}
\eczhIndexCodeAliasName{floquet_color}{\(\mathbb{Z}_2\) Floquet code}
\eczhIndexCodeAliasName{floquet_color}{CSS honeycomb code}
\codefieldsection{Description}
2D Floquet code on a trivalent 2D lattice whose parent topological phase is the \(\mathbb{Z}_2\times\mathbb{Z}_2\) 2D color-code phase and whose measurements cycle logical quantum information between the nine \(\mathbb{Z}_2\) surface-code \flmRefsHyperref{ref410}{condensed phases} of the parent phase.
The code's ISG is the stabilizer group of one of the nine surface codes.

This older use of the term \emph{Floquet color code} refers to the CSS/honeycomb construction of Refs. \NoCaseChange{\protect\cite{cite538,cite2526}}, and is distinct from the ruby-lattice Floquet color code of Ref. \NoCaseChange{\protect\cite{cite533}}.

\codefieldsection{Decoding}
\begin{eczvaluelist}
\item\relax Period-six measurement sequence utilizing two-qubit measurements \NoCaseChange{\protect\cite{cite538}}.
\end{eczvaluelist}
\codefieldsection{Fault Tolerance}
\begin{eczvaluelist}
\item\relax Fault-tolerant measurement-based computation can be realized using the foliated Floquet color code \NoCaseChange{\protect\cite{cite3681}}.
\end{eczvaluelist}
\codefieldsection{Realizations}
\begin{eczvaluelist}
\item\relax Plaquette stabilizer measurement realized on the IBM Falcon superconducting-qubit device \NoCaseChange{\protect\cite{cite3682}}
\end{eczvaluelist}
\codefieldsection{Parent}
\begin{eczvaluelist}
\item\relax
\flmRefsHyperref[eczindexfamilyrel]{code:floquet}{Hastings-Haah Floquet code}\end{eczvaluelist}
\codefieldsection{Cousins}
\begin{eczvaluelist}
\item\relax
\flmRefsHyperref[eczindexfamilyrel]{code:2d_color}{2D color code} --- The parent topological phase of the Floquet color code is the \(\mathbb{Z}_2\times\mathbb{Z}_2\) 2D color-code phase.
\item\relax
\flmRefsHyperref[eczindexfamilyrel]{code:surface}{Kitaev surface code} --- The ISG of the Floquet color code is the stabilizer group of one of nine realizations of the \(\mathbb{Z}_2\) 2D surface code.
\item\relax
\flmRefsHyperref[eczindexfamilyrel]{code:floquet_fracton}{Fracton Floquet code} --- The fracton Floquet code is obtained via a 3D generalization of the construction used in the Floquet color code \NoCaseChange{\protect\cite{cite538}}.
\end{eczvaluelist}
\eczhbkcontributors{ Nathanan Tantivasadakarn, \eczhuVVA }
\endeczcode

\eczcode{fcc_fracton}{Four Color Cube (FCC) fracton model code}{~\NoCaseChange{\protect\cite{cite534}}}
\codefieldsection{Description}
A fracton code obtained from four coupled X-cube models using p-membrane condensation.
A modular-qudit generalization has been proposed \NoCaseChange{\protect\cite{cite474}}.

\codefieldsection{Rate}
The logical space on a cubic lattice of length \(L\) with periodic boundary conditions is \(32L - 24\) qubits.
\codefieldsection{Parents}
\begin{eczvaluelist}
\item\relax
\flmRefsHyperref[eczindexfamilyrel]{code:qubit_css}{Qubit CSS code}\item\relax
\flmRefsHyperref[eczindexfamilyrel]{code:qldpc}{Qubit QLDPC code}\item\relax
\flmRefsHyperref[eczindexfamilyrel]{code:fracton}{Fracton stabilizer code}\end{eczvaluelist}
\codefieldsection{Cousin}
\begin{eczvaluelist}
\item\relax
\flmRefsHyperref[eczindexfamilyrel]{code:xcube}{X-cube model code} --- The FCC fracton model code is obtained from four coupled X-cube models using p-membrane condensation. \NoCaseChange{\protect\cite{cite534}}.
\end{eczvaluelist}
\eczhbkcontributors{ \eczhuVVA }
\endeczcode

\eczcode{fractal_surface}{Fractal surface code}{~\NoCaseChange{\protect\cite{cite676,cite2517,cite3683}}}
\codefieldsection{Description}
Kitaev surface code on a fractal geometry, which is obtained by removing qubits from the surface code on a cubic lattice.
A related construction, the \textit{fractal product code}, is a hypergraph product of two classical codes defined on a Sierpinski carpet graph \NoCaseChange{\protect\cite{cite676}}. 
The underlying classical codes form classical self-correcting memories \NoCaseChange{\protect\cite{cite677,cite678,cite679}}.

\codefieldsection{Decoding}
\begin{eczvaluelist}
\item\relax Sweep local automaton decoder \NoCaseChange{\protect\cite{cite3683}}.
\end{eczvaluelist}
\codefieldsection{Parent}
\begin{eczvaluelist}
\item\relax
\flmRefsHyperref[eczindexfamilyrel]{code:higher_dimensional_surface}{Homological code} --- Fractal surface codes are obtained by removing qubits from the 3D surface code on a cubic lattice.
\end{eczvaluelist}
\codefieldsection{Cousins}
\begin{eczvaluelist}
\item\relax
\flmRefsHyperref[eczindexfamilyrel]{code:3d_surface}{3D surface code} --- Fractal surface codes are obtained by removing qubits from the 3D surface code on a cubic lattice.
\item\relax
\flmRefsHyperref[eczindexfamilyrel]{code:hypergraph_product}{Hypergraph product (HGP) code} --- The related fractal product code is a hypergraph product of two classical codes defined on a Sierpinski carpet graph \NoCaseChange{\protect\cite{cite676}}.
\item\relax
\flmRefsHyperref[eczindexfamilyrel]{code:binary_linear}{Linear binary code} --- The fractal product code is a hypergraph product of two classical codes defined on a Sierpinski carpet graph \NoCaseChange{\protect\cite{cite676}}.
\item\relax
\flmRefsHyperref[eczindexfamilyrel]{code:self_correct}{Self-correcting quantum code} --- The classical codes underlying the fractal product code form classical self-correcting memories \NoCaseChange{\protect\cite{cite677,cite678,cite679}}.
\end{eczvaluelist}
\eczhbkcontributors{ \eczhuVVA }
\endeczcode

\eczcode{floquet_fracton}{Fracton Floquet code}{~\NoCaseChange{\protect\cite{cite538}}}
\codefieldsection{Description}
3D Floquet code whose qubits are placed on vertices of a truncated cubic honeycomb.
Its weight-two check operators are placed on edges of each truncated cube, while weight-three check operators are placed on each triangle.
Its ISG can be that of the X-cube model code or the checkerboard model code.
On a three-torus of size \(L_x \times L_y \times L_z\), the code consists of \(n= 48L_xL_yL_z\) physical qubits and encodes \(k= 2(L_x+L_y+L_z)-6\) logical qubits.

\codefieldsection{Decoding}
\begin{eczvaluelist}
\item\relax Period-six measurement sequence utilizing two- and three-qubit measurements \NoCaseChange{\protect\cite{cite538}}.
\end{eczvaluelist}
\codefieldsection{Parent}
\begin{eczvaluelist}
\item\relax
\flmRefsHyperref[eczindexfamilyrel]{code:floquet}{Hastings-Haah Floquet code}\end{eczvaluelist}
\codefieldsection{Cousins}
\begin{eczvaluelist}
\item\relax
\flmRefsHyperref[eczindexfamilyrel]{code:xcube}{X-cube model code} --- The ISG of the Fracton Floquet code can be that of the X-cube model code or the checkerboard model code.
\item\relax
\flmRefsHyperref[eczindexfamilyrel]{code:checkerboard}{Checkerboard model code} --- The ISG of the Fracton Floquet code can be that of the X-cube model code or the checkerboard model code.
\item\relax
\flmRefsHyperref[eczindexfamilyrel]{code:floquet_color}{Floquet color code} --- The fracton Floquet code is obtained via a 3D generalization of the construction used in the Floquet color code \NoCaseChange{\protect\cite{cite538}}.
\end{eczvaluelist}
\eczhbkcontributors{ Nathanan Tantivasadakarn, \eczhuVVA }
\endeczcode

\eczcode{freedman_meyer_luo}{Freedman-Meyer-Luo code}{~\NoCaseChange{\protect\cite{cite3684}}}
\codefieldsection{Description}
Hyperbolic surface code constructed using cellulation of a Riemannian Manifold \(M\) exhibiting systolic freedom \NoCaseChange{\protect\cite{cite680}}. Codes derived from such manifolds can achieve distances scaling better than \(\sqrt{n}\), something that is impossible using closed 2D surfaces or 2D surfaces with boundaries \NoCaseChange{\protect\cite{cite681}}. Improved codes are obtained by studying a weak family of Riemann metrics on closed 4-dimensional manifolds \(S^2\otimes S^2\) with the \(\mathbb{Z}_2\)-homology. 

\codefieldsection{Protection}
4D manifolds with weak systolic freedom yield \(\llbracket n,2,\Omega(\sqrt{n \sqrt{\log n}})\rrbracket \) surface codes.
\codefieldsection{Rate}
Codes held a 20-year record the best lower bound on asymptotic scaling of the minimum code distance, \(d=\Omega(\sqrt{n \sqrt{\log n}})\), broken by Ramanujan tensor-product codes.
\codefieldsection{Notes}
\begin{eczvaluelist}
\item\relax See thesis by Fetaya for pedagogical exposition \NoCaseChange{\protect\cite{cite3685}}.
\end{eczvaluelist}
\codefieldsection{Parent}
\begin{eczvaluelist}
\item\relax
\flmRefsHyperref[eczindexfamilyrel]{code:hyperbolic_surface}{Hyperbolic surface code}\end{eczvaluelist}
\codefieldsection{Cousin}
\begin{eczvaluelist}
\item\relax
\flmRefsHyperref[eczindexfamilyrel]{code:ramanujan_tensor_product}{High-dimensional expander (HDX) code} --- Ramanujan codes broke 20-year record on minimum code distance set by Freedman-Meyer-Luo codes.
\end{eczvaluelist}
\eczhbkcontributors{ Xinyuan Zheng, \eczhuVVA }
\endeczcode

\eczcode{fusion}{Fusion-based quantum computing (FBQC) code}{~\NoCaseChange{\protect\cite{cite3686}}}
\codefieldsection{Description}
Code whose codewords are resource states used in an FBQC scheme.

FBQC is a fault-tolerant model of quantum computation built from small constant-sized entangled resource states together with destructive entangling measurements called \textit{fusions} \NoCaseChange{\protect\cite{cite3686}}.
Resource states are stabilizer states and can be described, up to local Clifford transformations, by graph states \NoCaseChange{\protect\cite{cite3686}}.
Unlike standard MBQC, which first prepares a large cluster state and then computes using single-qubit measurements, FBQC integrates entanglement generation, syndrome extraction, and logical processing into the fusion measurements themselves.
This makes FBQC particularly natural for photonic platforms, where Bell-type fusion measurements are native operations.

\codefieldsection{Protection}
Protects against erasure, Pauli errors, \flmRefsHyperref{ref498}{photon loss}, fusion failure from non-determinism, and faults in resource-state preparation. Redundancy in fusion outcomes is captured by the check-operator group. Fusion measurement outcomes form a syndrome that can be decoded to infer the logical Pauli frame, rather than by applying physical recovery operations \NoCaseChange{\protect\cite{cite3686}}.
\codefieldsection{Encoding}
\begin{eczvaluelist}
\item\relax Resource-state generators, which produce small constant-sized stabilizer states, together with Bell-fusion measurements.
\end{eczvaluelist}
\codefieldsection{Gates}
\begin{eczvaluelist}
\item\relax \flmRefsHyperref{ref409}{Clifford gates} are performed by introducing and manipulating topological features such as boundaries, defects, or twists through modified fusion bases and, in some constructions, single-qubit measurements. Logical gates can also be performed by code deformation.
Non-Clifford gates are performed by magic-state injection.
\end{eczvaluelist}
\codefieldsection{Decoding}
\begin{eczvaluelist}
\item\relax Surface-code-based FBQC schemes often admit a syndrome-graph description, allowing the use of decoders such as minimum-weight matching and union-find \NoCaseChange{\protect\cite{cite3686,cite3160}}.
\end{eczvaluelist}
\codefieldsection{Fault Tolerance}
\begin{eczvaluelist}
\item\relax Fusion networks can be constructed so that the surviving stabilizers and check operators realize topological surface-code fault tolerance \NoCaseChange{\protect\cite{cite3686}}.
\item\relax More generally, any three-dimensional cell complex in which each edge has four incident faces defines a surface-code fusion complex, yielding a large family of FBQC fault-tolerant protocols \NoCaseChange{\protect\cite{cite3161}}.
\end{eczvaluelist}
\codefieldsection{Threshold}
\begin{eczvaluelist}
\item\relax Under the hardware-agnostic fusion error model, pedagogical FBQC schemes have reported thresholds of \(11.98\%\) against erasure in each fusion measurement and \(1.07\%\) against Pauli error \NoCaseChange{\protect\cite{cite3686}}.
\item\relax For a linear-optical ballistic scheme, reported thresholds include \(43.2\%\) against fusion failure and \(10.4\%\) \flmRefsHyperref{ref498}{photon loss} per fusion \NoCaseChange{\protect\cite{cite3686}}.
\item\relax For surface-code logical blocks compiled to FBQC, the threshold for fault-tolerant logical gates was found to agree, within numerical uncertainty, with the bulk memory threshold \NoCaseChange{\protect\cite{cite3160}}.
\end{eczvaluelist}
\codefieldsection{Parent}
\begin{eczvaluelist}
\item\relax
\flmRefsHyperref[eczindexfamilyrel]{code:qubit_stabilizer}{Qubit stabilizer code} --- The resource states in FBQC are small stabilizer states, and the surviving stabilizers after fusion determine the encoded output state (conditioned on measurement outcomes).
\end{eczvaluelist}
\codefieldsection{Cousins}
\begin{eczvaluelist}
\item\relax
\flmRefsHyperref[eczindexfamilyrel]{code:topological}{Topological code} --- Surface-code-based topological fault-tolerant protocols can be realized in FBQC, including topological features such as boundaries, defects, and twists, by modifying fusion measurements and, in some constructions, adding single-qubit measurements \NoCaseChange{\protect\cite{cite3160,cite3161}}.
\item\relax
\flmRefsHyperref[eczindexfamilyrel]{code:dual_rail}{Dual-rail quantum code} --- FBQC resource states are concatenated with dual-rail codes to increase loss detection.
\item\relax
\flmRefsHyperref[eczindexfamilyrel]{code:dynamic_gen}{Dynamically generated QECC} --- Building a fusion network is done using a measurement-based dynamical process.
\item\relax
\flmRefsHyperref[eczindexfamilyrel]{code:cat_concatenated}{Concatenated cat code} --- The four-component cat code can be concatenated with the XZZX code to yield a fusion-based computation scheme on a 2D lattice \NoCaseChange{\protect\cite{cite3687}}.
\item\relax
\flmRefsHyperref[eczindexfamilyrel]{code:gkp}{Square-lattice GKP code} --- GKP states can be used to perform computation in a fusion-based encoding \NoCaseChange{\protect\cite{cite3688}}.
\item\relax
\flmRefsHyperref[eczindexfamilyrel]{code:qubit_concatenated}{Concatenated qubit code} --- Blocklet concatenation uses concatenation and transversal gates in a way that is tailored to FBQC platforms \NoCaseChange{\protect\cite{cite3602}}.
\item\relax
\flmRefsHyperref[eczindexfamilyrel]{code:cluster_state}{Cluster-state code} --- FBQC and MBQC are both computational models in which computation is done by measuring resource states (which are qubit stabilizer states). The difference between the two is in how the states are constructed. FBQC is based exclusively on two-qubit measurements tailored to photonic platforms. These measurements require a foliation with more qubits but one which can be built by fusing smaller modules.
\end{eczvaluelist}
\eczhbkcontributors{ Yaron Jarach, Dhruv Devulapalli, \eczhuVVA }
\endeczcode

\eczcode{five_squares}{Generalized five-squares code}{~\NoCaseChange{\protect\cite{cite594,cite660,cite661}}}
\codefieldsection{Description}
Member of a family of subsystem codes that are generalizations \NoCaseChange{\protect\cite{cite660,cite661}} of a code defined on a three-valent hypergraph associated with the five-squares lattice \NoCaseChange{\protect\cite{cite594}}.
The original five-squares code is a 2D topological subsystem code with local two-qubit gauge generators; on a torus, it encodes two logical qubits \NoCaseChange{\protect\cite{cite594}}.

\codefieldsection{Decoding}
\begin{eczvaluelist}
\item\relax For the original five-squares code, preprocessing maps decoding onto two copies of the toric code, after which one can use minimum-weight matching or renormalization-group decoding \NoCaseChange{\protect\cite{cite594}}.
\item\relax Generalized five-squares codes can also be decoded via a mapping to two copies of the surface code \NoCaseChange{\protect\cite{cite661}}.
\end{eczvaluelist}
\codefieldsection{Code Capacity Threshold}
\begin{eczvaluelist}
\item\relax For depolarizing noise, the original five-squares code has a threshold around \(1.5\%\) under the simple decoder and around \(2\%\) under the improved decoder \NoCaseChange{\protect\cite{cite594}}.
\end{eczvaluelist}
\codefieldsection{Parents}
\begin{eczvaluelist}
\item\relax
\flmRefsHyperref[eczindexfamilyrel]{code:subsystem_hypergraph}{Sarvepalli-Brown subsystem code} --- Generalized five-squares codes are special cases of Sarvepalli-Brown subsystem codes \NoCaseChange{\protect\cite[{Sec. II.B}]{cite661}}.
\item\relax
\flmRefsHyperref[eczindexfamilyrel]{code:translationally_invariant_subsystem}{Lattice subsystem code}\end{eczvaluelist}
\codefieldsection{Cousin}
\begin{eczvaluelist}
\item\relax
\flmRefsHyperref[eczindexfamilyrel]{code:toric}{Toric code} --- For the original five-squares code, preprocessing maps decoding onto two copies of the toric code \NoCaseChange{\protect\cite{cite594}}.
\end{eczvaluelist}
\eczhbkcontributors{ \eczhuVVA }
\endeczcode

\eczcode{qubit_generalized_homological_product_css}{Generalized homological-product qubit CSS code}{}
\codefieldsection{Description}
A qubit CSS code whose properties are determined from an underlying chain complex via the \flmRefsHyperref{ref683}{qubit CSS-to-homology correspondence}. This complex often consists of some type of product of other chain complexes.

\codefieldsection{Parents}
\begin{eczvaluelist}
\item\relax
\flmRefsHyperref[eczindexfamilyrel]{code:qubit_css}{Qubit CSS code}\item\relax
\flmRefsHyperref[eczindexfamilyrel]{code:qldpc}{Qubit QLDPC code} --- Homological products are a primary tool for generating qubit QLDPC codes with favorable parameters. Typically, whenever the input codes are binary LDPC or qubit QLDPC, the resulting code will be qubit QLDPC with non geometrically local stabilizer generators.
\item\relax
\flmRefsHyperref[eczindexfamilyrel]{code:generalized_homological_product_css}{Generalized homological-product CSS code}\end{eczvaluelist}
\codefieldsection{Children}
\begin{eczvaluelist}
\item\relax
\flmRefsHyperref[eczindexfamilyrel]{code:fiber_bundle}{Fiber-bundle code}\item\relax
\flmRefsHyperref[eczindexfamilyrel]{code:lossless_expander}{Lossless expander balanced-product code}\item\relax
\flmRefsHyperref[eczindexfamilyrel]{code:qcga}{Bivariate bicycle (BB) code}\item\relax
\flmRefsHyperref[eczindexfamilyrel]{code:bicycle}{Bicycle code}\item\relax
\flmRefsHyperref[eczindexfamilyrel]{code:dhlv}{Dinur-Hsieh-Lin-Vidick (DHLV) code}\item\relax
\flmRefsHyperref[eczindexfamilyrel]{code:lcs}{Lift-connected surface (LCS) code}\item\relax
\flmRefsHyperref[eczindexfamilyrel]{code:multisector_hypergraph}{Higher-dimensional homological product code}\item\relax
\flmRefsHyperref[eczindexfamilyrel]{code:dlv}{Dinur-Lin-Vidick (DLV) code} --- DLV codes are codes constructed using a cubical chain complex, which is a \(t\)-order extension of the chain complexes underlying quantum generalized homological product CSS codes.
\item\relax
\flmRefsHyperref[eczindexfamilyrel]{code:generalized_quantum_tanner}{Generalized quantum Tanner code}\item\relax
\flmRefsHyperref[eczindexfamilyrel]{code:higher_dimensional_surface}{Homological code} --- The generalized surface code is constructed from chain complexes arising from cell complexes of the underlying manifold. Such complexes are not necessarily products of two non-trivial complexes, but the manifolds are picked so that their homology ensures favorable code properties.
\end{eczvaluelist}
\codefieldsection{Cousin}
\begin{eczvaluelist}
\item\relax
\flmRefsHyperref[eczindexfamilyrel]{code:quantum_pin}{Quantum pin code} --- One can construct quantum pin codes from any chain complex \NoCaseChange{\protect\cite[{Sec. II.F}]{cite702}}.
\end{eczvaluelist}
\eczhbkcontributors{ \eczhuVVA }
\endeczcode

\eczcode{generalized_quantum_divisible}{Generalized quantum divisible code}{~\NoCaseChange{\protect\cite{cite734}}}
\codefieldsection{Description}
A level-\(\nu\) generalized quantum divisible code is a CSS code whose \(X\)-type stabilizers, in the \flmRefsHyperref{ref817}{symplectic representation}, have zero norm and form a \((\nu,t)\)-null matrix (defined below) with respect to some odd-integer vector \(t\) \NoCaseChange{\protect\cite[{Def. V.1}]{cite734}}.
Such codes admit gates at the \(\nu\)th level of the \flmTerm{term}{ref694}{}{Clifford hierarchy}.
Such codes can also be level-lifted \NoCaseChange{\protect\cite[{Theorem V.6}]{cite734}}, \(\nu\to\nu+1\), which recursively yields towers of generalized divisible codes from a particular ground code.

Given an odd-integer coefficient length-\(n\) vector \(t\), two vectors \(v,w\) are \((\nu,t)\)\textit{-orthogonal} if
\flmMathEnvironment{align}{}{
	\sum_i v_i t_i w_i \equiv 0 \mod 2^{\nu-1}~.
}
A matrix whose rows make up such vectors is called \((\nu,t)\)-orthogonal.

\codefieldsection{Transversal and Permutation-Based Gates}
\begin{eczvaluelist}
\item\relax A level-\(\nu\) generalized quantum divisible code admits a diagonal transversal gate at the \(\nu\)th level of the \flmTerm{term}{ref694}{}{Clifford hierarchy} \NoCaseChange{\protect\cite[{Lemma V.3}]{cite734}}.
\end{eczvaluelist}
\codefieldsection{Parent}
\begin{eczvaluelist}
\item\relax
\flmRefsHyperref[eczindexfamilyrel]{code:qubit_css}{Qubit CSS code} --- Generalized quantum divisible codes are CSS codes. Any self-dual CSS code yields a level-three generalized quantum divisible code when level-lifted \NoCaseChange{\protect\cite[{Thm. V.6}]{cite734}}.
\end{eczvaluelist}
\codefieldsection{Children}
\begin{eczvaluelist}
\item\relax
\flmRefsHyperref[eczindexfamilyrel]{code:quantum_divisible}{Quantum divisible code} --- Generalized level-\(\nu\) quantum divisible codes reduce to quantum level-\(\nu\) divisible codes when \(t\) is a vector with \(\pm 1\) entries.
The classical code formed by their \(X\)-type stabilizer generator matrix is \(\nu\)-even \NoCaseChange{\protect\cite[{pg. 10}]{cite734}}.
Both types of codes realize transversal gates outside of the \flmRefsHyperref{ref409}{Clifford group}.

\item\relax
\flmRefsHyperref[eczindexfamilyrel]{code:quantum_h}{\(\llbracket k+4,k,2\rrbracket \) H code} --- H codes are level-two generalized divisible codes \NoCaseChange{\protect\cite[{Sec. VI.C}]{cite734}}.
\end{eczvaluelist}
\codefieldsection{Cousins}
\begin{eczvaluelist}
\item\relax
\flmRefsHyperref[eczindexfamilyrel]{code:quantum_triorthogonal}{Triorthogonal code} --- Triorthogonal codes are stabilizer codes, while generalized quantum divisible codes are CSS codes. Every level-three generalized divisible code is a triorthogonal code, but whether the converse is true or false is not known \NoCaseChange{\protect\cite[{Sec. VI.C}]{cite734}}.
\item\relax
\flmRefsHyperref[eczindexfamilyrel]{code:random_stabilizer}{Random stabilizer code} --- Random CSS codes \NoCaseChange{\protect\cite{cite3196}} can be used to construct families of \(\llbracket O(d^{\nu−1}), \Omega(d), d\rrbracket \) level-\(\nu\) generalized quantum divisible codes \NoCaseChange{\protect\cite[{Sec. VI.A}]{cite734}}.
\item\relax
\flmRefsHyperref[eczindexfamilyrel]{code:self_dual_css}{Self-dual CSS code} --- Any self-dual CSS code yields a level-three generalized quantum divisible code when level-lifted \NoCaseChange{\protect\cite[{Thm. V.6}]{cite734}}.
\end{eczvaluelist}
\eczhbkcontributors{ Connor Clayton, \eczhuVVA }
\endeczcode

\eczcode{generalized_quantum_tanner}{Generalized quantum Tanner code}{~\NoCaseChange{\protect\cite{cite686}}}
\codefieldsection{Description}
An extension of quantum Tanner codes to codes constructed from two commuting regular graphs with the same vertex set.
This allows for code construction using finite sets and Schreier graphs, yielding a broader family of square complexes.

\codefieldsection{Parent}
\begin{eczvaluelist}
\item\relax
\flmRefsHyperref[eczindexfamilyrel]{code:qubit_generalized_homological_product_css}{Generalized homological-product qubit CSS code}\end{eczvaluelist}
\codefieldsection{Child}
\begin{eczvaluelist}
\item\relax
\flmRefsHyperref[eczindexfamilyrel]{code:quantum_tanner}{Quantum Tanner code} --- Generalized quantum Tanner codes constructed out of bipartite double covers of Cayley graphs reduce to quantum Tanner codes \NoCaseChange{\protect\cite{cite686}}.
\end{eczvaluelist}
\eczhbkcontributors{ \eczhuVVA }
\endeczcode

\eczcode{generalized_shor}{Generalized Shor code}{~\NoCaseChange{\protect\cite{cite3037,cite1433}}}
\codefieldsection{Description}
Qubit CSS code constructed by concatenating two classical codes in a way that generalizes the Shor and quantum parity codes.

This \(\llbracket n_1n_2,k_1k_2,\min(d_1,d_2)\rrbracket \) code can be defined \NoCaseChange{\protect\cite{cite1448}} via the CSS construction applied to two binary linear codes, \(C_X\) and \(C_Z\), satisfying \(C_X^{\perp}\subset C_Z\).
These codes are in turn constructed from two more binary linear codes, \(C_1 = [n_1, k_1, d_1]\) and \(C_2 = [n_2, k_2, d_2]\), with parity-check matrices \(H_1\) and \(H_2\) and generator matrices \(G_1\) and \(G_2\), respectively.
The parity-check matrices of \(C_X\) and \(C_Z\) are then
\flmMathEnvironment{align}{}{
\begin{split}
H_X &= H_1 \otimes I_{n_2}\\
H_Z &= G_1 \otimes H_2~.
\end{split}
}

Based on the above construction, the Hilbert space on \(n_1n_2\) qubits can be decomposed as a direct sum of tensor products of Hilbert spaces of lower dimensions, as outlined in \NoCaseChange{\protect\cite{cite3037}}.

\codefieldsection{Protection}
Has distance \(d=\min(d_1,d_2)\).
\codefieldsection{Decoding}
\begin{eczvaluelist}
\item\relax Efficient decoder \NoCaseChange{\protect\cite{cite3689}}.
\end{eczvaluelist}
\codefieldsection{Realizations}
\begin{eczvaluelist}
\item\relax The \(\llbracket m^2,1,m\rrbracket \) codes for \(m\leq 7\) have been realized in trapped-ion quantum devices \NoCaseChange{\protect\cite{cite3379}}.
\end{eczvaluelist}
\codefieldsection{Notes}
\begin{eczvaluelist}
\item\relax Non-deterministic linear-optical encoding \NoCaseChange{\protect\cite{cite3259}} whose success probability \(P_{E}\) is determined by the efficiency \(\eta\) of the photonic encoding circuit. A threshold \(\eta > 0.82 \) exists for the efficiency, above which \(P_{E}\to 1\) as \(m_1\to\infty\) for a particular \(m_2\).
\item\relax Studied in the context of error-corrected quantum repeaters \NoCaseChange{\protect\cite{cite3690}}.
\end{eczvaluelist}
\codefieldsection{Parent}
\begin{eczvaluelist}
\item\relax
\flmRefsHyperref[eczindexfamilyrel]{code:qubit_css}{Qubit CSS code}\end{eczvaluelist}
\codefieldsection{Child}
\begin{eczvaluelist}
\item\relax
\flmRefsHyperref[eczindexfamilyrel]{code:quantum_parity}{Quantum parity code (QPC)}\end{eczvaluelist}
\codefieldsection{Cousin}
\begin{eczvaluelist}
\item\relax
\flmRefsHyperref[eczindexfamilyrel]{code:subsystem_quantum_parity}{Subsystem hypergraph product (SHP) code} --- In a \(\llbracket n_1n_2, k_1k_2, min(d_1, d_2)\rrbracket \) generalized Shor code, error correction is achieved by measuring \((n_1−k_1)n_2+k_1(n_2−k_2)\) stabilizer generators \NoCaseChange{\protect\cite{cite3384}}. The SHP code achieves the same degree of correctability, but requires only \((n_1−k_1)k_2+k_1(n_2−k_2)\) stabilizer measurements.
\end{eczvaluelist}
\eczhbkcontributors{ \eczhuVVA }
\endeczcode

\eczcode{gnu_permutation_invariant}{GNU PI code}{~\NoCaseChange{\protect\cite{cite2944,cite2945}}}
\codefieldsection{Description}
PI code whose codewords can be expressed as superpositions of \flmRefsHyperref{ref526}{Dicke states} with coefficients are square-roots of the binomial distribution.

In terms of \flmRefsHyperref{ref526}{Dicke states}, logical codewords for codes encoding a single qubit \NoCaseChange{\protect\cite{cite2944}} are
\flmMathEnvironment{align}{}{
|\overline{\pm}\rangle = \sum_{\ell=0}^{m} \frac{(\pm 1)^\ell}{\sqrt{2^m}} \sqrt{m \choose \ell} |D^n_{g \ell}\rangle~.
}
Here, \(n\) is the number of particles used for encoding \(1\) qubit, and \(g, m \leq n\) are arbitrary positive integers.
Codes with higher logical dimension are developed in Ref. \NoCaseChange{\protect\cite{cite2945}}.
Each Dicke state in the code can be \textit{shifted} by adding a shift \(s\) to both \(n\) and \(g\).

\codefieldsection{Protection}
Depends on the family. One family which is completely symmetrized versions of Bacon-Shor codes (parameterized by \(t\)) protects against arbitrary weight-\(t\) spin errors. Additionally, codes with large enough length \((t+1)(3t+1)+t\) can approximately correct \(t\) spontaneous decay errors.
\codefieldsection{Decoding}
\begin{eczvaluelist}
\item\relax For a family of shifted gnu codes, decoding can be done using projection, probability amplitude rebalancing, and gate teleportation in time \(O(n^2)\) \NoCaseChange{\protect\cite{cite2657}}.
\item\relax Syndrome extraction protocol for insertion errors \NoCaseChange{\protect\cite{cite3691}}.
\end{eczvaluelist}
\codefieldsection{Notes}
\begin{eczvaluelist}
\item\relax The degree of entanglement in (non-concatenated) GNU codes scales at most logarithmically in their distance \NoCaseChange{\protect\cite[{Appx. D}]{cite529}}.
\end{eczvaluelist}
\codefieldsection{Parent}
\begin{eczvaluelist}
\item\relax
\flmRefsHyperref[eczindexfamilyrel]{code:qudit_gnu_permutation_invariant}{Qudit GNU PI code} --- Qudit GNU codes encoding logical qubits reduce to GNU codes.
\end{eczvaluelist}
\codefieldsection{Children}
\begin{eczvaluelist}
\item\relax
\flmRefsHyperref[eczindexfamilyrel]{code:ruskai}{\(\llparenthesis 9,2,3\rrparenthesis \) Ruskai code} --- The \(\llparenthesis 9,2,3\rrparenthesis \) Ruskai code is a GNU PI code \NoCaseChange{\protect\cite{cite2944}}.
\item\relax
\flmRefsHyperref[eczindexfamilyrel]{code:quantum_repetition}{Quantum repetition code} --- GNU codewords for \(g=1\) reduce to the phase-flip repetition code.
\item\relax
\flmRefsHyperref[eczindexfamilyrel]{code:four_qubit_permutation_invariant}{\(\llparenthesis 4,2,2\rrparenthesis \) Four-qubit single-deletion code} --- The four-qubit single-deletion code is a GNU code for \(g=m=2\) \NoCaseChange{\protect\cite{cite2657}}.
\end{eczvaluelist}
\codefieldsection{Cousins}
\begin{eczvaluelist}
\item\relax
\flmRefsHyperref[eczindexfamilyrel]{code:combinatorial_permutation_invariant}{Combinatorial PI code} --- Combinatorial PI codes \(Q_{g,(m-1)/2,g-1,+}\) are GNU codes for odd \(m\) \NoCaseChange{\protect\cite[{Prop. 5.4}]{cite3169}}.
\item\relax
\flmRefsHyperref[eczindexfamilyrel]{code:bacon_shor}{Bacon-Shor code} --- GNU codes of length \((2t+1)^2\) result from projecting Bacon-Shor codes into the PI qubit subspace \NoCaseChange{\protect\cite{cite2944}}.
\item\relax
\flmRefsHyperref[eczindexfamilyrel]{code:frustration_free}{Frustration-free Hamiltonian code} --- GNU codes lie within the ground state of ferromagnetic Heisenberg models without an external magnetic field \NoCaseChange{\protect\cite{cite2809}}.
\item\relax
\flmRefsHyperref[eczindexfamilyrel]{code:binomial}{Binomial code} --- Binomial codes and GNU codes related via the Holstein-Primakoff mapping \NoCaseChange{\protect\cite{cite651,cite652,cite653}}. A qudit generalization of GNU codes can be obtained from qudit binomial codes \NoCaseChange{\protect\cite[{Appx. C}]{cite496}}.
\item\relax
\flmRefsHyperref[eczindexfamilyrel]{code:metopt}{Error-corrected sensing code} --- GNU codes can be used to sense signals within the PI subspace \NoCaseChange{\protect\cite{cite2761}}.
\item\relax
\flmRefsHyperref[eczindexfamilyrel]{code:ae}{Æ code} --- Many well-performing Æ codes can be mapped into GNU codes via the \flmRefsHyperref{ref526}{Dicke state mapping}.
\end{eczvaluelist}
\eczhbkcontributors{ Benjamin Quiring, \eczhuVVA }
\endeczcode

\eczcode{golden}{Golden code}{~\NoCaseChange{\protect\cite{cite3692}}}
\codefieldsection{Description}
Variant of the Guth-Lubotzky hyperbolic surface code that uses regular tessellations of 4-dimensional hyperbolic space.

\codefieldsection{Protection}
Nonvanishing rate and asymptotic distance lower bounded by \(n^0.1\).
\codefieldsection{Rate}
Nonvanishing rate and asymptotic distance lower bounded by \(n^0.1\). However, the smallest number of physical qubits in this family is 234,000.
\codefieldsection{Parent}
\begin{eczvaluelist}
\item\relax
\flmRefsHyperref[eczindexfamilyrel]{code:four_dimensional_hyperbolic}{Guth-Lubotzky code}\end{eczvaluelist}
\eczhbkcontributors{ Micah Shaw, \eczhuVVA }
\endeczcode

\eczcode{four_dimensional_hyperbolic}{Guth-Lubotzky code}{~\NoCaseChange{\protect\cite{cite3693}}}
\codefieldsection{Description}
Homological linear-rate code based on cellulations of certain 4D hyperbolic manifolds with particular homology and systolic properties.

Guth and Lubotzky \NoCaseChange{\protect\cite{cite3693}} show that there exists \(\epsilon\), a 4D hyperbolic manifold \(M\), and a sequence of manifolds \(M_i\) such that
each \(M_i\) is a finite sheeted covering of \(M\), and the 4D volumes of the manifolds \(\text{Vol}_4(M_i)\) of the sequence tend to infinity.
Also, the dimension of the second homology and size of systoles are bounded by \(H_2(M_i, \mathbb{Z}_2) \geq \frac{\text{Vol}_4(M_i)}{100}\) and \(\text{Sys}_2(M_i) \geq \text{Vol}_4(M_i)^\epsilon\), respectively.

Then given any cellulation of \(M\), it can naturally be extended to cellulations for each of the manifolds \(M_i\) and used to define CSS codes via the homological construction by choosing the size three chain complex consisting of the \(3,2\) and \(1\)-cells of the cellulations.

For dense cellulations (i.e. large \(n\)) the number of physical qubits for these codes will scale with the volume of the manifolds.
Therefore, bounds on the dimension of the second homology and size of systoles are achieved in terms of \(n\) for large \(n\).

\codefieldsection{Protection}
Protection stems from the relationship between properties of manifolds and CSS codes derived from their cellulation. The number of physical \(k\) qubits and distance \(d\) of the code will scale as \flmRefsHyperref{ref65}{order} \(\Omega(n)\) and \(\Omega(n^\epsilon)\), respectively. A later explicit construction yields codes with \(d \geq c n^{0.2}\), while the same work notes the upper bound \(d = O(n^{0.3})\) for the Guth-Lubotzky construction \NoCaseChange{\protect\cite{cite3694}}.
\codefieldsection{Rate}
An explicit construction based on Coxeter groups yields a lower bound of \(13/72\) on the asymptotic rate \NoCaseChange{\protect\cite{cite3695}}.
\codefieldsection{Threshold}
\begin{eczvaluelist}
\item\relax Phenomenological noise: data is consistent with a threshold of about \(4\%\) using BP-OSD or cellular-automaton decoders \NoCaseChange{\protect\cite{cite3695}}.
\end{eczvaluelist}
\codefieldsection{Parent}
\begin{eczvaluelist}
\item\relax
\flmRefsHyperref[eczindexfamilyrel]{code:hyperbolic_surface}{Hyperbolic surface code}\end{eczvaluelist}
\codefieldsection{Child}
\begin{eczvaluelist}
\item\relax
\flmRefsHyperref[eczindexfamilyrel]{code:golden}{Golden code}\end{eczvaluelist}
\eczhbkcontributors{ Seyed Sajjad Nezhadi, \eczhuVVA }
\endeczcode

\eczcode{haah_cubic}{Haah cubic code (CC)}{~\NoCaseChange{\protect\cite{cite3032}}}
\codefieldsection{Description}
A 3D lattice stabilizer code on a length-\(L\) cubic lattice with one or two qubits per site.
Admits two types of stabilizer generators with support on each cube of the lattice.
In the non-CSS case, these two are related by spatial inversion.
For CSS codes, we require that the product of all corner operators is the identity.
We lastly require that there are no non-trivial string operators, meaning that single-site operators are a phase, and any period one logical operator \(l \in \mathsf{S}^{\perp}\) is just a phase.

Haah showed in his original construction that there is exactly one non-CSS code of this form, and 17 CSS codes \NoCaseChange{\protect\cite{cite3032}}.
The non-CSS code is labeled code 0, and the rest are numbered from 1 - 17.
Codes CC1-CC4, CC7, CC8, and CC10 do not have string logical operators \NoCaseChange{\protect\cite{cite3032,cite456}}.
The original cubic code in this family can be obtained by gauging \NoCaseChange{\protect\cite{cite462,cite463,cite233,cite464,cite465,cite466,cite467,cite468,cite469,cite470}} a fractal-symmetry Ising model \NoCaseChange{\protect\cite[{Sec. IV.B}]{cite464}}; related fracton gauging constructions also appear in \NoCaseChange{\protect\cite{cite233,cite464}}.

Under renormalization group flow \NoCaseChange{\protect\cite{cite3696}}, cubic code 1 fragments into itself and the \textit{Haah B-code} (a.k.a. \textit{CC1B}), which has four qubits per unit cell \NoCaseChange{\protect\cite{cite3697}\protect\cite[{Eq. (D2)}]{cite456}}.
In this context, cubic code 1 is sometimes called the \textit{Haah A-code} or \textit{CC1A}.
Cubic codes 11-17 fragment into combinations of themselves, their corresponding B-codes, and stacks of surface codes \NoCaseChange{\protect\cite[{Table 1}]{cite3697}}.

The Haah A-code can be written in a similar form as the Sierpinski prism model code \NoCaseChange{\protect\cite{cite3164}}.
The Haah B-code admits a topological defect network construction out of two copies of the 3D surface code \NoCaseChange{\protect\cite{cite3163}}.

Encodings using geometries with boundaries as well as lattice defects have been studied \NoCaseChange{\protect\cite{cite3698}}.
CC1A and CC1B have been generalized to manifolds more general than 3D lattices \NoCaseChange{\protect\cite{cite3699,cite3700}}.

\codefieldsection{Protection}
Cubic codes protect against simultaneous independent Pauli errors on different sites (not qubits, since there can be 2 qubits per site). Codes CC0-CC4 are known to have distance \(d \ge L\), meaning they can achieve macroscopic code distance as \(L\to\infty\).
\codefieldsection{Decoding}
\begin{eczvaluelist}
\item\relax Hard-decisions RG decoder \NoCaseChange{\protect\cite{cite3036}}.
\item\relax BP-OSD decoder \NoCaseChange{\protect\cite{cite1247}}.
\end{eczvaluelist}
\codefieldsection{Threshold}
\begin{eczvaluelist}
\item\relax The encoding rate depends on the code implemented, but code CC0 has been shown to have \(k \ge L\) on a periodic finite cubic lattice of side length \(L\). In general, we expect the number of logical qubits to scale as \(k = \Omega(L)\).
\end{eczvaluelist}
\codefieldsection{Parents}
\begin{eczvaluelist}
\item\relax
\flmRefsHyperref[eczindexfamilyrel]{code:qldpc}{Qubit QLDPC code}\item\relax
\flmRefsHyperref[eczindexfamilyrel]{code:qudit_cubic}{Qudit cubic code}\end{eczvaluelist}
\codefieldsection{Cousins}
\begin{eczvaluelist}
\item\relax
\flmRefsHyperref[eczindexfamilyrel]{code:surface}{Kitaev surface code} --- Under renormalization group flow \NoCaseChange{\protect\cite{cite3696}}, cubic codes 11-17 fragment into combinations of themselves, their corresponding B-codes, and stacks of surface codes \NoCaseChange{\protect\cite[{Table 1}]{cite3697}}.
\item\relax
\flmRefsHyperref[eczindexfamilyrel]{code:3d_surface}{3D surface code} --- The Haah B-code admits a topological defect network construction out of two copies of the 3D surface code \NoCaseChange{\protect\cite{cite3163}}.
\item\relax
\flmRefsHyperref[eczindexfamilyrel]{code:3d_color}{3D color code} --- The 3D color and cubic code families both include 3D codes that do not admit string-like operators.
\item\relax
\flmRefsHyperref[eczindexfamilyrel]{code:4d_surface}{\((2,2)\) Loop toric code} --- The energy of any partial implementation of CC1 is proportional to the boundary length, similar to the 4D toric code. This can potentially suppress the effects of thermal errors, but it is currently an open problem.
\item\relax
\flmRefsHyperref[eczindexfamilyrel]{code:generalized_bicycle}{Generalized bicycle (GB) code} --- A GB code for the group \(G=\mathbb{Z}_{L}^{\times 3}\) is a cubic code \NoCaseChange{\protect\cite[{Sec. III.A}]{cite674}}.
\item\relax
\flmRefsHyperref[eczindexfamilyrel]{code:cluster_state}{Cluster-state code} --- A short-range entangled cluster-state model with fractal \(X\)-type symmetries on both sublattices can be built from the cubic-code gauging data. Gauging one sublattice yields, up to a local circuit, either the cubic code or its ungauged fractal-symmetry Ising model, while gauging both sublattices returns the cluster model up to local swaps and Hadamards \NoCaseChange{\protect\cite[{Secs. III.F, IV.C}]{cite464}}.
\item\relax
\flmRefsHyperref[eczindexfamilyrel]{code:lifted_product}{Lifted-product (LP) code} --- A lifted-product code constructed with coefficients in the ring \(R=\mathbb{F}_2[x,y,z]/(x^L-1,y^L-1,z^L-1)\) is a cubic code \NoCaseChange{\protect\cite[{Appx. B}]{cite184}}.
\item\relax
\flmRefsHyperref[eczindexfamilyrel]{code:sierpinsky_fractal_liquid}{Sierpinski prism model code} --- The Haah A-code can be written in a similar form as the Sierpinski prism model code \NoCaseChange{\protect\cite{cite3164}}.
\item\relax
\flmRefsHyperref[eczindexfamilyrel]{code:fibonacci_model}{Fibonacci code} --- The Fibonacci code is designed to mimic the fractal properties of (quantum) Haah cubic code so that studying the former can help us toward the development of an efficient algorithm for the latter \NoCaseChange{\protect\cite{cite1349}}.
\item\relax
\flmRefsHyperref[eczindexfamilyrel]{code:self_correct}{Self-correcting quantum code} --- Cubic code 1 is partially self-correcting with a logarithmic energy barrier \NoCaseChange{\protect\cite{cite3036}}.
\item\relax
\flmRefsHyperref[eczindexfamilyrel]{code:bosonization}{Bosonization code} --- Bosonization can be used to realize a Haah cubic code with an emergent fermion from a Majorana stabilizer code \NoCaseChange{\protect\cite{cite3503}}. This code is shown to be distinct from the original code \NoCaseChange{\protect\cite{cite3506}}.
\end{eczvaluelist}
\eczhbkcontributors{ Siddharth Taneja, \eczhuVVA }
\endeczcode

\eczcode{haar_random}{Haar-random qubit code}{~\NoCaseChange{\protect\cite{cite2784,cite2574,cite2785,cite3701}}}
\codefieldsection{Description}
Haar-random codewords are generated in a process involving averaging over unitary operations distributed according to the Haar measure. Haar-random codes are used to prove statements about the capacity of a quantum channel to transmit quantum information \NoCaseChange{\protect\cite{cite535}}, but encoding and decoding in such \(n\)-qubit codes quickly becomes impractical as \(n\to\infty\).

There are different approaches to create Haar-random codewords. In the construction of Ref. \NoCaseChange{\protect\cite{cite2574}}, codewords are produced by performing a unitarily covariant projective measurement on a \textit{typical} subspace of a tensor-power state. Reference \NoCaseChange{\protect\cite{cite2574}} showed that coherent-information rates are achievable by encoding in such Haar-random codes. In particular, Haar-random codes achieve asymptotically vanishing decoding error in the \(n\to\infty\) limit by proving that the encoded information becomes decoupled from the environment. This is a necessary and sufficient condition for successful decoding since measurements of the environment should never reveal the encoded information \NoCaseChange{\protect\cite{cite2767}}.

Intuitively, coupling with the environment can be decreased by projecting the system onto a random codespace. The more qubits that are randomly discarded, the more the codespace is decoupled from the environment. One may ask what is the least amount of qubits that can be discarded, i.e. the largest remaining codespace, that still achieves decoupling. It can be shown through the decoupling inequality \NoCaseChange{\protect\cite{cite3702}} that the largest possible dimension of the random codespace that achieves arbitrarily large decoupling is exponential in the coherent information of the channel. Therefore, there exist codes that can transmit information at rates governed by coherent information. Furthermore, these codes can be constructed with high probability by performing a Haar-random isometry embedding \(k\) logical qubits into an \(n\)-qubit physical space. Such an isometry can be produced by QR decomposition of a Gaussian random matrix \NoCaseChange{\protect\cite{cite3703}}.

\codefieldsection{Rate}
Haar-random qubit codes attain the regularized coherent information of certain noise channels in the limit of large \(n\) \NoCaseChange{\protect\cite{cite3704}}.
\codefieldsection{Threshold}
\begin{eczvaluelist}
\item\relax Haar-random qubit codes have a \flmRefsHyperref{ref3210}{measurement threshold} of one \NoCaseChange{\protect\cite{cite3211}}.
\end{eczvaluelist}
\codefieldsection{Parents}
\begin{eczvaluelist}
\item\relax
\flmRefsHyperref[eczindexfamilyrel]{code:qubits_into_qubits}{Qubit code}\item\relax
\flmRefsHyperref[eczindexfamilyrel]{code:random_circuit}{Random-circuit code}\end{eczvaluelist}
\codefieldsection{Cousins}
\begin{eczvaluelist}
\item\relax
\flmRefsHyperref[eczindexfamilyrel]{code:local_haar_random}{Local Haar-random circuit qubit code} --- Approximating the random projections through \(t\)-designs is necessary in order to make the Haar-random qubit protocol practical. Replacing with random \flmRefsHyperref{ref409}{Clifford gates} is especially convenient since the \flmRefsHyperref{ref409}{Clifford group} forms a unitary 2-design and produces stabilizer codes.
\item\relax
\flmRefsHyperref[eczindexfamilyrel]{code:clifford_group}{Clifford group} --- Approximating the random projections through \(t\)-designs is necessary in order to make the Haar-random qubit protocol practical. Replacing with random \flmRefsHyperref{ref409}{Clifford gates} is especially convenient since the \flmRefsHyperref{ref409}{Clifford group} forms a unitary 2-design and produces stabilizer codes.
\end{eczvaluelist}
\eczhbkcontributors{ Jon Nelson, \eczhuVVA }
\endeczcode

\eczcode{floquet}{Hastings-Haah Floquet code}{~\NoCaseChange{\protect\cite{cite536}}}
\codefieldsection{Alternative Names}
\begin{eczvaluelist}
\item\relax Periodic Floquet code
\end{eczvaluelist}
\eczhIndexCodeAliasName{floquet}{Periodic Floquet code}
\codefieldsection{Description}
Dynamical code whose sequence of check-operator measurements is periodic.
The original Hastings-Haah construction introduced periodic measurement schedules that dynamically generate logical qubits even when the underlying subsystem code has fewer or no logical qubits \NoCaseChange{\protect\cite{cite536}}.
Its basic examples are the 2D honeycomb Floquet code and the 1D ladder Floquet code.

\codefieldsection{Protection}
In the original Hastings-Haah examples, periodic measurements protect against single-qubit Pauli faults and measurement faults: the honeycomb Floquet code on a torus stores two logical qubits with distance proportional to linear size, while the ladder Floquet code is an error-detecting toy model \NoCaseChange{\protect\cite{cite536}}. Spacetime errors for periodic Floquet codes have been studied in Ref. \NoCaseChange{\protect\cite{cite3705}}.
\codefieldsection{Fault Tolerance}
\begin{eczvaluelist}
\item\relax Periodic Floquet codes on tri-colorable lattices can be made fault-tolerant in the presence of dead qubits \NoCaseChange{\protect\cite{cite3706,cite3707}}.
\end{eczvaluelist}
\codefieldsection{Parent}
\begin{eczvaluelist}
\item\relax
\flmRefsHyperref[eczindexfamilyrel]{code:da}{Dynamical code} --- Periodic Floquet codes are dynamical codes with periodic measurement sequences.
\end{eczvaluelist}
\codefieldsection{Children}
\begin{eczvaluelist}
\item\relax
\flmRefsHyperref[eczindexfamilyrel]{code:floquet_color}{Floquet color code}\item\relax
\flmRefsHyperref[eczindexfamilyrel]{code:floquet_xyz_ruby}{Ruby Floquet code}\item\relax
\flmRefsHyperref[eczindexfamilyrel]{code:honeycomb_floquet}{Honeycomb Floquet code} --- The honeycomb Floquet code is the first 2D Floquet code.
\item\relax
\flmRefsHyperref[eczindexfamilyrel]{code:floquet_3d_fermionic_surface}{Floquet 3D fermionic surface code}\item\relax
\flmRefsHyperref[eczindexfamilyrel]{code:floquet_3d_surface}{Floquet 3D surface code}\item\relax
\flmRefsHyperref[eczindexfamilyrel]{code:floquet_fracton}{Fracton Floquet code}\item\relax
\flmRefsHyperref[eczindexfamilyrel]{code:floquet_xcube}{X-cube Floquet code}\item\relax
\flmRefsHyperref[eczindexfamilyrel]{code:hyperbolic_floquet}{Hyperbolic Floquet code}\item\relax
\flmRefsHyperref[eczindexfamilyrel]{code:ladder}{Ladder Floquet code} --- The ladder Floquet code is the first 1D Floquet code.
\end{eczvaluelist}
\codefieldsection{Cousins}
\begin{eczvaluelist}
\item\relax
\flmRefsHyperref[eczindexfamilyrel]{code:asymmetric_qecc}{Asymmetric quantum code (AQC)} --- Floquet codes can be adapted for asymmetric noise \NoCaseChange{\protect\cite{cite2643}}.
\item\relax
\flmRefsHyperref[eczindexfamilyrel]{code:stab_5_1_3}{\(\llbracket 5,1,3\rrbracket \) Five-qubit perfect code} --- Inspired by the honeycomb Floquet code, various weight-two measurement schemes have been designed for the five-qubit code \NoCaseChange{\protect\cite{cite3309}}.
\item\relax
\flmRefsHyperref[eczindexfamilyrel]{code:bacon_shor}{Bacon-Shor code} --- A Floquet version of the Bacon-Shor code admits a period-four measurement sequence that utilizes its gauge degrees of freedom as defects evolving across measurement rounds. This \textit{Floquet-Bacon-Shor} code saturates the \flmRefsHyperref{ref492}{subsystem BT bound}. Applying a period-four measurement schedule to the original Bacon-Shor code yields a numerical threshold under circuit-level noise \NoCaseChange{\protect\cite{cite3471}}.
\end{eczvaluelist}
\eczhbkcontributors{ \eczhuVVA }
\endeczcode

\eczcode{heavy_hex}{Heavy-hexagon code}{~\NoCaseChange{\protect\cite{cite3598}}}
\codefieldsection{Description}
Subsystem stabilizer code on the heavy-hexagonal point set that combines Bacon-Shor and surface-code stabilizers.
Encodes one logical qubit into \(n=(5d^2-2d-1)/2\) physical qubits with distance \(d\).
The heavy-hexagonal point set allows for low degree (at most 3) connectivity between all the data and ancilla qubits, which is suitable for fixed-frequency transmon qubits subject to frequency collision errors.
The code can be split into a surface and a Bacon-Shor code, with the idling qubits of one code serving as the physical qubits of the other \NoCaseChange{\protect\cite{cite662}}.

Data qubits and ancillas of the code are placed on a heavy-hexagonal point set, i.e., the vertices and edges of a tiling of hexagons. A subset of the ancilla qubits are flag qubits used for detecting high-weight errors arising from fewer faults. The code stabilizers for detecting \(X\)-type errors are measured by measuring weight-two \(Z\)-type gauge operators whose product produces stabilizers of the surface code. \(X\)-type stabilizers are column operators corresponding to stabilizers of the Bacon-Shor code, which are measured by taking products of weight-four and weight-two \(X\)-type gauge operators.

\codefieldsection{Protection}
Protects against Pauli noise. The code has no threshold for \(Z\)-type Pauli errors since they are detected by Bacon-Shor-type stabilizers.
\codefieldsection{Rate}
\(1/n\) for a distance-\(d\) heavy-hexagon code on \(n = (5d^2-2d-1)/2\) qubits.
\codefieldsection{Encoding}
\begin{eczvaluelist}
\item\relax For a logical-zero state, prepare all data qubits in the physical-zero state and then measure the \(X\)-type Bacon-Shor stabilizers. For logical-plus state, prepare all data qubits in the physical-plus state and then measure \(Z\)-type surface code stabilizers.
\item\relax Stabilizer measurement encoding circuits have a constant depth of 10 time steps (excluding ancilla state preparation and measurement).
\end{eczvaluelist}
\codefieldsection{Transversal and Permutation-Based Gates}
\begin{eczvaluelist}
\item\relax CNOT gates are transversal for this code. However, for most architectures, all logical gates would be implemented using lattice surgery methods.
\end{eczvaluelist}
\codefieldsection{Gates}
\begin{eczvaluelist}
\item\relax Universal gate set achieved with magic state injection and lattice surgery.
\item\relax Magic-state injection with and without flag qubits \NoCaseChange{\protect\cite{cite3708}}.
\end{eczvaluelist}
\codefieldsection{Decoding}
\begin{eczvaluelist}
\item\relax Any graph-based decoder can be used, such as MWPM and Union Find. However, edge weights must be dynamically renormalized using flag-qubit measurement outcomes after each syndrome measurement round.
\item\relax Machine-learning \NoCaseChange{\protect\cite{cite3709}} and neural-network \NoCaseChange{\protect\cite{cite3710}} decoders.
\end{eczvaluelist}
\codefieldsection{Fault Tolerance}
\begin{eczvaluelist}
\item\relax All logical gates can be fault-tolerantly implemented using lattice surgery and magic state injection.
\item\relax Stabilizer measurements are measured fault-tolerantly using one-flag circuits since some single-fault events can result in weight-two data qubit errors which are parallel to the code's logical operators. Hence, using information from the flag-qubit measurements is crucial to fault-tolerantly measure the code stabilizers.
\end{eczvaluelist}
\codefieldsection{Threshold}
\begin{eczvaluelist}
\item\relax \(0.45\%\) for \(X\) errors under a full circuit-level depolarizing noise model (obtained from Monte Carlo simulations).
\item\relax \(Z\)-errors have no threshold given the \(X\)-type Bacon-Shor stabilizers.
\end{eczvaluelist}
\codefieldsection{Realizations}
\begin{eczvaluelist}
\item\relax Superconducting qubits: Logical state preparation and flag-qubit error correction realized in superconducting-circuit devices (specifically, fixed-frequency transmon qubit architectures) by IBM for \(d=2\) \NoCaseChange{\protect\cite{cite3279,cite3261}} and \(d=3\) \NoCaseChange{\protect\cite{cite3711}}. Simultaneous syndrome extraction and logical Bell-state preparation for both the embedded surface and Bacon-Shor codes of distance \(\leq 4\) on an IBM 133-qubit device \NoCaseChange{\protect\cite{cite662}}. Embedded rotated surface code magic-state injection implemented on IBM fez device \NoCaseChange{\protect\cite{cite3712}}.
\end{eczvaluelist}
\codefieldsection{Parent}
\begin{eczvaluelist}
\item\relax
\flmRefsHyperref[eczindexfamilyrel]{code:compass_model}{Compass code} --- The heavy-hex code is a compass code on a heavy-hexagonal lattice, combining weight-two \(XX\) and \(ZZ\) gauge operators that are partially gauge-fixed to yield surface-code \(Z\)-type stabilizers and Bacon-Shor \(X\)-type stabilizers \NoCaseChange{\protect\cite{cite3598}}.
\end{eczvaluelist}
\codefieldsection{Cousins}
\begin{eczvaluelist}
\item\relax
\flmRefsHyperref[eczindexfamilyrel]{code:surface}{Kitaev surface code} --- Surface code stabilizers are used to measure the Z-type stabilizers of the code. There are various ways to embed the surface code into the heavy-hex lattice \NoCaseChange{\protect\cite{cite3713}}.
\item\relax
\flmRefsHyperref[eczindexfamilyrel]{code:bacon_shor}{Bacon-Shor code} --- Bacon-Shor stabilizers are used to measure the X-type stabilizers of the code.
\item\relax
\flmRefsHyperref[eczindexfamilyrel]{code:css_4_1_2}{\(\llbracket 4,1,2\rrbracket \) Leung-Nielsen-Chuang-Yamamoto (LNCY) code} --- Magic states prepared using a \(\llbracket 4,1,2\rrbracket \) subcode can be injected into the heavy-hex code \NoCaseChange{\protect\cite{cite3261,cite3262}}.
The \(d=2\) heavy-hex code is closely related to the \(\llbracket 4,1,2\rrbracket \) LNCY code.

\item\relax
\flmRefsHyperref[eczindexfamilyrel]{code:rotated_surface}{Rotated surface code} --- A rotated surface code can be mapped onto a heavy square lattice, resulting in a code similar to the heavy-hexagon code \NoCaseChange{\protect\cite{cite3598}}.
\item\relax
\flmRefsHyperref[eczindexfamilyrel]{code:xysurface}{XY surface code} --- XY surface code can be adapted for a heavy-hexagonal point set \NoCaseChange{\protect\cite{cite3714}}.
\item\relax
\flmRefsHyperref[eczindexfamilyrel]{code:xzzx}{XZZX surface code} --- XZZX surface code can be adapted for a heavy-hexagonal point set \NoCaseChange{\protect\cite{cite3714}}.
\end{eczvaluelist}
\eczhbkcontributors{ Christopher Chamberland, \eczhuVVA }
\endeczcode

\eczcode{hemicubic}{Hemicubic code}{~\NoCaseChange{\protect\cite{cite1269}}}
\codefieldsection{Description}
Homological code constructed out of cubes in high dimensions.
The hemicubic code family has asymptotically diminishing soundness that scales as \flmRefsHyperref{ref65}{order} \(\Omega(1/\log n)\), locality of stabilizer generators scaling as \flmRefsHyperref{ref65}{order} \(O(\log n)\), and distance of \flmRefsHyperref{ref65}{order} \(\Theta(\sqrt{n})\).

\codefieldsection{Decoding}
\begin{eczvaluelist}
\item\relax Polynomial-time decoding algorithm that corrects arbitrary errors of size up to the minimum distance multiplied by polylogarithmic factors \NoCaseChange{\protect\cite{cite1269}}. This was the first polynomial-time decoding algorithm for quantum locally testable codes.
\end{eczvaluelist}
\codefieldsection{Parent}
\begin{eczvaluelist}
\item\relax
\flmRefsHyperref[eczindexfamilyrel]{code:higher_dimensional_surface}{Homological code}\end{eczvaluelist}
\codefieldsection{Cousins}
\begin{eczvaluelist}
\item\relax
\flmRefsHyperref[eczindexfamilyrel]{code:qltc}{Quantum locally testable code (QLTC)} --- The hemicubic code family has asymptotically diminishing soundness that scales as \flmRefsHyperref{ref65}{order} \(\Omega(1/\log n)\), locality of stabilizer generators scaling as \flmRefsHyperref{ref65}{order} \(O(\log n)\), and distance of \flmRefsHyperref{ref65}{order} \(\Theta(\sqrt{n})\).
Soundness amplification and AEL distance amplification \NoCaseChange{\protect\cite{cite493,cite494}} can also yield improvements in various parameters \NoCaseChange{\protect\cite[{Table 3}]{cite2991}}.
Application of generalized distance balancing \NoCaseChange{\protect\cite{cite684}} to hemicubic codes using an asymptotically good classical code of length \(t\) yields \(O( 1/(\log(n) t^2) )\) soundness and \flmRefsHyperref{ref65}{order} \(\Theta(\sqrt{n}t)\) distance while maintaining locality scaling and at the expense of a dimension scaling as \flmRefsHyperref{ref65}{order} \(\Theta(t^2)\) \NoCaseChange{\protect\cite{cite2990}}.

\item\relax
\flmRefsHyperref[eczindexfamilyrel]{code:distance_balanced}{Distance-balanced code} --- Application of generalized distance balancing \NoCaseChange{\protect\cite{cite684}} to hemicubic codes using an asymptotically good classical code of length \(t\) yields \(O( 1/(\log(n) t^2) )\) soundness and \flmRefsHyperref{ref65}{order} \(\Theta(\sqrt{n}t)\) distance while maintaining locality scaling and at the expense of a dimension scaling as \flmRefsHyperref{ref65}{order} \(\Theta(t^2)\) \NoCaseChange{\protect\cite{cite2990}}.
\item\relax
\flmRefsHyperref[eczindexfamilyrel]{code:hypercube}{Hypercube code} --- Hemicubic codes are built from cellulations derived from hypercubes.
\end{eczvaluelist}
\eczhbkcontributors{ \eczhuVVA }
\endeczcode

\eczcode{holographic_steane}{Heptagon holographic code}{~\NoCaseChange{\protect\cite{cite2952}}}
\codefieldsection{Alternative Names}
\begin{eczvaluelist}
\item\relax Holographic Steane code
\end{eczvaluelist}
\eczhIndexCodeAliasName{holographic_steane}{Holographic Steane code}
\codefieldsection{Description}
Holographic tensor-network code constructed out of a network of encoding isometries of the Steane code.
Depending on how the isometry tensors are contracted, there is a zero-rate and a finite-rate code family.

\codefieldsection{Decoding}
\begin{eczvaluelist}
\item\relax Optimal erasure decoder \NoCaseChange{\protect\cite{cite2952}}.
\end{eczvaluelist}
\codefieldsection{Code Capacity Threshold}
\begin{eczvaluelist}
\item\relax \(~33\%\) under erasures using optimal erasure decoder for the finite-rate family, and \(50\%\) for the zero-rate family \NoCaseChange{\protect\cite{cite2952}}.
\item\relax Depolarizing noise: \(9.4\%\) using tensor-network decoder, and \(\approx 7\%\) using integer optimization decoder \NoCaseChange{\protect\cite{cite3106}}.
\item\relax \(18.985\%\) against depolarizing noise for zero-rate code under tensor-network decoder \NoCaseChange{\protect\cite{cite3715}}.
\end{eczvaluelist}
\codefieldsection{Parents}
\begin{eczvaluelist}
\item\relax
\flmRefsHyperref[eczindexfamilyrel]{code:qubit_css}{Qubit CSS code}\item\relax
\flmRefsHyperref[eczindexfamilyrel]{code:holographic_tensor}{Holographic tensor-network code} --- The encoding of the heptagon holographic code is a holographic tensor network consisting of the encoding isometry for the Steane code, which is a \flmRefsHyperref{code:block_perfect}{planar-perfect tensor}.
\end{eczvaluelist}
\codefieldsection{Child}
\begin{eczvaluelist}
\item\relax
\flmRefsHyperref[eczindexfamilyrel]{code:concatenated_steane}{Concatenated Steane code} --- A recursively concatenated Steane code is a heptagon holographic code on a tree tensor network.
\end{eczvaluelist}
\codefieldsection{Cousin}
\begin{eczvaluelist}
\item\relax
\flmRefsHyperref[eczindexfamilyrel]{code:block_perfect}{Planar-perfect-tensor code} --- The encoding of the heptagon holographic code is a holographic tensor network consisting of the encoding isometry for the Steane code, which is a \flmRefsHyperref{code:block_perfect}{planar-perfect tensor}.
\end{eczvaluelist}
\eczhbkcontributors{ \eczhuVVA }
\endeczcode

\eczcode{stabilizer_over_gf4}{Hermitian qubit code}{~\NoCaseChange{\protect\cite{cite449}}}
\codefieldsection{Alternative Names}
\begin{eczvaluelist}
\item\relax Calderbank-Rains-Shor-Sloane (CRSS) code
\item\relax \(\mathbb{F}_4\)-linear qubit stabilizer code
\item\relax \(M_{3}\) code
\end{eczvaluelist}
\eczhIndexCodeAliasName{stabilizer_over_gf4}{Calderbank-Rains-Shor-Sloane (CRSS) code}
\eczhIndexCodeAliasName{stabilizer_over_gf4}{\(\mathbb{F}_4\)-linear qubit stabilizer code}
\eczhIndexCodeAliasName{stabilizer_over_gf4}{\(M_{3}\) code}
\codefieldsection{Description}
A qubit stabilizer code constructed from a Hermitian self-orthogonal linear quaternary code using the Hermitian construction.

Hermitian codes are in one-to-one correspondence with Hermitian self-orthogonal additive codes via the \flmRefsHyperref{ref1778}{\(\mathbb{F}_4\) representation}.
Quaternary linear codes are Hermitian self-orthogonal (self-dual) iff they are trace-Hermitian self-orthogonal (self-dual) additive \NoCaseChange{\protect\cite[{Thm. 27.4.1}]{cite2024}\protect\cite[{Thm. 9.10.3}]{cite126}}.
In other words, if the underlying quaternary code is linear, then the \flmRefsHyperref{ref33}{field trace} can be removed from the definition of inner product.

An additive self-orthogonal code \(C \subseteq \mathbb{F}_4^n\) of size \(2^r\) yields an \(\llbracket n,n-r\rrbracket \) qubit stabilizer code \NoCaseChange{\protect\cite{cite3716}\protect\cite[{Thm. 2}]{cite449}}.
When \(C\) is \(\mathbb{F}_4\)-linear of parameters \([n,k]_4\), trace self-orthogonality is equivalent to Hermitian self-orthogonality, so \(|C|=4^k\) and the construction specializes to a Hermitian \(\llbracket n,n-2k\rrbracket \) qubit stabilizer code.
In the standard-form analysis of the linear case, the associated additive code has type \(4^{k}2^{0}\), i.e., the \(k_1=0\) case of Ref. \NoCaseChange{\protect\cite[{Sec. 4.4}]{cite2664}}; equivalently, the parameters satisfy \(k \equiv n\) mod 2 \NoCaseChange{\protect\cite{cite454}}.
The stabilizer generator matrix is of the form
\flmMathEnvironment{align}{}{
H=\begin{pmatrix}H\\
\alpha H
\end{pmatrix}~,
}
where \(H\) is the parity-check matrix of the classical code.

Every Hermitian qubit stabilizer code has stabilizer generators with \textit{paired support} \NoCaseChange{\protect\cite[{Def. 4}]{cite795}}: Pauli operators \(P\) and \(Q\) have paired support if \(P\), \(Q\), and \(P \cdot Q\) all have the same support, and a set of generators has paired support if it can be partitioned into such pairs. The converse does not hold in general.

All code automorphisms lie in the \flmRefsHyperref{ref409}{Clifford group} \NoCaseChange{\protect\cite[{Corr. 16}]{cite446}}.

\codefieldsection{Protection}
For an additive self-orthogonal code \(C \subseteq \mathbb{F}_4^n\), the resulting qubit stabilizer code has distance
\flmMathEnvironment{align}{}{
d=\min\{\operatorname{wt}(v):v \in C^{\perp}\setminus C\}~,
}
where \(\perp\) denotes duality under the trace inner product \NoCaseChange{\protect\cite{cite3716}\protect\cite[{Thm. 2}]{cite449}}.
The distance \(d_C\) of the classical code is the Hermitian code's \flmRefsHyperref{ref672}{pure distance}, and it is equal to the code distance for a \flmRefsHyperref{ref811}{non-degenerate} code \NoCaseChange{\protect\cite{cite983}}.
There is an equivalence between \flmRefsHyperref{ref672}{pure} \(\llbracket n,n-2k\rrbracket \) Hermitian qubit codes and certain sets of points in projective space \(PG(k-1,4)\) \NoCaseChange{\protect\cite[{Thm. 2.8}]{cite1695}}.

\codefieldsection{Transversal and Permutation-Based Gates}
\begin{eczvaluelist}
\item\relax All code automorphisms lie in the \flmRefsHyperref{ref409}{Clifford group} \NoCaseChange{\protect\cite[{Corr. 16}]{cite446}}, so transversal physical gates implement only Clifford logical gates.
\item\relax Transversal \(SH\) and \(HS\) "facet" gates (a.k.a. \(M_3\) gates) which cyclically permute Paulis as \(X \to Y\), \(Y \to Z\), and \(Z \to X\) \NoCaseChange{\protect\cite[{Sec. 8.2}]{cite736}}.
\item\relax The three-block transversal gate mapping each physical \(X \to XYZ\) and each \(Z \to ZXY\) implements a logical gate \NoCaseChange{\protect\cite{cite737}\protect\cite[{Exam. 2}]{cite532}}.
\item\relax A qubit stabilizer code is Hermitian if and only if a transversal \(R\) gate leaves the stabilizer group invariant \NoCaseChange{\protect\cite{cite454}}.
\item\relax Hermitian qubit codes admit the group \(U(\ell,\mathbb{F}_4)\) of diagonal transversal gates on \(\ell\) codeblocks \NoCaseChange{\protect\cite{cite738}}.
\end{eczvaluelist}
\codefieldsection{Gates}
\begin{eczvaluelist}
\item\relax Signed weight enumerators \NoCaseChange{\protect\cite{cite3717}} determine performance of magic \(T\)-state distillation protocols \NoCaseChange{\protect\cite{cite2040}}.
\end{eczvaluelist}
\codefieldsection{Fault Tolerance}
\begin{eczvaluelist}
\item\relax Characterizing fault-tolerant multi-qubit gates under the \flmRefsHyperref{ref1778}{\(\mathbb{F}_4\) representation} may involve characterizing all global automorphisms of some number of copies of a code that preserve the symplectic inner product \NoCaseChange{\protect\cite[{pg. 9}]{cite532}}.
\end{eczvaluelist}
\codefieldsection{Notes}
\begin{eczvaluelist}
\item\relax Hermitian \(\llbracket n,0,d\rrbracket \) codes, corresponding to a self-dual \(\mathbb{F}_4\) representation, are always \flmRefsHyperref{ref672}{pure} \NoCaseChange{\protect\cite[{Def. 2.4}]{cite1695}}. See Refs. \NoCaseChange{\protect\cite{cite3718,cite1645}} for tables. Bounds on self-dual \(\llbracket n,0,d\rrbracket \) Hermitian codes based on graphs have been derived \NoCaseChange{\protect\cite{cite1373,cite1976}}.
\item\relax Qubit Hermitian codes for \(n < 10\) have been classified \NoCaseChange{\protect\cite{cite454}}.
\end{eczvaluelist}
\codefieldsection{Parents}
\begin{eczvaluelist}
\item\relax
\flmRefsHyperref[eczindexfamilyrel]{code:qubit_stabilizer}{Qubit stabilizer code}\item\relax
\flmRefsHyperref[eczindexfamilyrel]{code:stabilizer_over_gfqsq}{Hermitian Galois-qudit code}\end{eczvaluelist}
\codefieldsection{Children}
\begin{eczvaluelist}
\item\relax
\flmRefsHyperref[eczindexfamilyrel]{code:iceberg}{\(\llbracket 2m,2m-2,2\rrbracket \) error-detecting code} --- The \(\llbracket 2m,2m-2,2\rrbracket \) error-detecting code is Hermitian \NoCaseChange{\protect\cite[{Table 6}]{cite454}}.
\item\relax
\flmRefsHyperref[eczindexfamilyrel]{code:quantum_cap}{\(\llbracket n,n-2k,4\rrbracket \) Quantum cap code} --- A quantum cap code is a distance-four \flmRefsHyperref{ref672}{pure} Hermitian qubit code constructed by identifying its underlying Hermitian self-orthogonal \([n,k]_4\) code with a particular projective cap in \(PG(k-1,4)\).
\item\relax
\flmRefsHyperref[eczindexfamilyrel]{code:quad_residue_13_1_5}{\(\llbracket 13,1,5\rrbracket \) quantum QR code}\item\relax
\flmRefsHyperref[eczindexfamilyrel]{code:stab_13_1_5}{\(\llbracket 13,1,5\rrbracket \) twisted toric code} --- The \(\llbracket 13,1,5\rrbracket \) twisted toric code is Hermitian \NoCaseChange{\protect\cite[{ID 67067199e766ad364e845262}]{cite781}}.
\item\relax
\flmRefsHyperref[eczindexfamilyrel]{code:stab_15_7_3}{\(\llbracket 15, 7, 3\rrbracket \) quantum Hamming code} --- The \(\llbracket 15,7,3\rrbracket \) is Hermitian \NoCaseChange{\protect\cite[{ID 6705229219cca60cf657a8fd}]{cite781}}.
\item\relax
\flmRefsHyperref[eczindexfamilyrel]{code:stab_5_1_3}{\(\llbracket 5,1,3\rrbracket \) Five-qubit perfect code} --- The five-qubit code is Hermitian \NoCaseChange{\protect\cite[{ID 21}]{cite453}}, and is derived from the \([5,3,3]_4\) shortened hexacode via the \flmRefsHyperref{code:stabilizer_over_gf4}{qubit Hermitian construction} \NoCaseChange{\protect\cite{cite1670}\protect\cite[{Exam. A}]{cite1666}}.
\item\relax
\flmRefsHyperref[eczindexfamilyrel]{code:stab_6_2_2}{\(\llbracket 6,2,2\rrbracket \) \(C_6\) code} --- The \(C_6\) code is Hermitian \NoCaseChange{\protect\cite[{ID 126}]{cite453}\protect\cite[{Table 6}]{cite454}}.
\item\relax
\flmRefsHyperref[eczindexfamilyrel]{code:steane}{\(\llbracket 7,1,3\rrbracket \) Steane code} --- The Steane code is Hermitian \NoCaseChange{\protect\cite[{ID 226}]{cite453}\protect\cite[{Table 6}]{cite454}}.
\item\relax
\flmRefsHyperref[eczindexfamilyrel]{code:stab_8_2_2}{\(\llbracket 8,2,2\rrbracket \) hyperbolic color code} --- The \(\llbracket 8,2,2\rrbracket \) hyperbolic color code is Hermitian \NoCaseChange{\protect\cite[{ID 4926}]{cite453}}.
\item\relax
\flmRefsHyperref[eczindexfamilyrel]{code:stab_8_2_3}{\(\llbracket 8,2,3\rrbracket \) Hermitian code}\item\relax
\flmRefsHyperref[eczindexfamilyrel]{code:stab_9_3_3}{\(\llbracket 9,3,3\rrbracket \) Quadric code} --- The \(\llbracket 9,3,3\rrbracket \) code is Hermitian \NoCaseChange{\protect\cite[{ID 170235}]{cite453}}.
\item\relax
\flmRefsHyperref[eczindexfamilyrel]{code:hermitian_qldpc}{Camara-Ollivier-Tillich code}\end{eczvaluelist}
\codefieldsection{Cousins}
\begin{eczvaluelist}
\item\relax
\flmRefsHyperref[eczindexfamilyrel]{code:dual}{Dual linear code} --- Hermitian qubit codes are constructed from Hermitian self-orthogonal linear codes over \(\mathbb{F}_4\) via the \flmRefsHyperref{ref1778}{\(\mathbb{F}_4\) representation}.
\item\relax
\flmRefsHyperref[eczindexfamilyrel]{code:constacyclic}{Constacyclic code} --- Duadic constacyclic codes yield many examples of Hermitian qubit codes \NoCaseChange{\protect\cite{cite983}}.
\item\relax
\flmRefsHyperref[eczindexfamilyrel]{code:graph}{Graph-adjacency code} --- Bounds on self-dual \(\llbracket n,0,d\rrbracket \) Hermitian codes based on graphs have been derived \NoCaseChange{\protect\cite{cite1373}}.
\item\relax
\flmRefsHyperref[eczindexfamilyrel]{code:ame}{Perfect-tensor code} --- The sole codeword of some \(\llbracket n,0,d\rrbracket \) Hermitian codes is an \flmRefsHyperref{ref219}{AME state} \NoCaseChange{\protect\cite{cite2932}}.
\item\relax
\flmRefsHyperref[eczindexfamilyrel]{code:self_dual}{Self-dual linear code} --- Hermitian qubit codes are constructed from Hermitian self-orthogonal linear codes over \(\mathbb{F}_4\) via the \flmRefsHyperref{ref1778}{\(\mathbb{F}_4\) representation}. This relation yields bounds on self-dual codes over \(\mathbb{F}_4\) \NoCaseChange{\protect\cite{cite2040}}.
\item\relax
\flmRefsHyperref[eczindexfamilyrel]{code:qubit_css}{Qubit CSS code} --- A Hermitian qubit code that can be put into CSS form via single-qubit Clifford operations remains Hermitian \NoCaseChange{\protect\cite{cite454}}.
\item\relax
\flmRefsHyperref[eczindexfamilyrel]{code:projective}{Projective geometry code} --- There is an equivalence between \flmRefsHyperref{ref672}{pure} \(\llbracket n,n-2k\rrbracket \) Hermitian qubit codes and certain sets of points in projective space \(PG(k-1,4)\) \NoCaseChange{\protect\cite[{Thm. 2.8}]{cite1695}}.
\item\relax
\flmRefsHyperref[eczindexfamilyrel]{code:divisible}{Divisible code} --- The commutation requirement for a Hermitian stabilizer code implies that its underlying Hermitian self-orthogonal linear code over \(\mathbb{F}_4\) is even; the converse is also true \NoCaseChange{\protect\cite[{Thm. 1.4.10}]{cite126}}.
\item\relax
\flmRefsHyperref[eczindexfamilyrel]{code:quantum_perfect}{Perfect quantum code} --- For qubits (\(q=2\)), the only nontrivial perfect codes are the stabilizer code family \(\llbracket (4^r-1)/3, (4^r-1)/3 - 2r, 3\rrbracket \) for \(r \geq 2\), obtained from Hamming codes over \(\mathbb{F}_4\) via the Hermitian construction \NoCaseChange{\protect\cite{cite1694,cite449}}. These codes are related to partial spreads in projective geometry \NoCaseChange{\protect\cite{cite1695}}.
\item\relax
\flmRefsHyperref[eczindexfamilyrel]{code:quantum_bch}{Qubit BCH code} --- Hermitian self-orthogonal quaternary BCH codes are used to construct a subset of qubit BCH codes via the Hermitian construction.
\item\relax
\flmRefsHyperref[eczindexfamilyrel]{code:frobenius}{Frobenius code} --- Frobenius Hermitian codes have been completely classified; no such codes exist when \(t\) is odd \NoCaseChange{\protect\cite{cite3318}}.
\end{eczvaluelist}
\eczhbkcontributors{ Simon Burton, Marianna Podzorova, \eczhuVVA }
\endeczcode

\eczcode{hexagonal_cz}{Hexagonal \(CZ\) code}{~\NoCaseChange{\protect\cite{cite575}}}
\codefieldsection{Description}
A hexagonal-lattice realization of the \(2+1\)D \(l=m=n=1\) cubic theory / Type-III \(\mathbb{Z}_2^3\) twisted quantum double phase.
Its stabilizers are products of Pauli-\(Z\) operators and \(CZ\) gates \NoCaseChange{\protect\cite[{Fig. 6}]{cite575}\protect\cite[{Fig. 3}]{cite589}}.
The ground-state subspace of the hexagonal \(CZ\) code realizes the topological order of the Type-III \(G=\mathbb{Z}^3_2\) Abelian TQD model \NoCaseChange{\protect\cite{cite575,cite576}}, which is the same topological order as the \(G=D_4\) non-Abelian quantum double \NoCaseChange{\protect\cite{cite577}}.
The stabilizers include \(CZ\) operators acting on hexagonal loops, but a reduced version exists where only two \(CZ\) gates act on each loop \NoCaseChange{\protect\cite{cite589}}.

\codefieldsection{Gates}
\begin{eczvaluelist}
\item\relax The hexagonal \(CZ\) code can be obtained from two surface codes by gauging \NoCaseChange{\protect\cite{cite462,cite463,cite233,cite464,cite465,cite466,cite467,cite468,cite469,cite470}} their logical \(CZ\) gate \NoCaseChange{\protect\cite{cite589}}. Gates on the two surface codes in the third level of the Clifford hierarchy, such as \(CZ\) gates, can be realized fault-tolerantly by performing this procedure and reversing it \NoCaseChange{\protect\cite{cite572,cite589}}.
\item\relax There is a constant-depth circuit implementing a transversal logical \(T\) gate via an emergent automorphism symmetry of the underlying \(\mathbb{D}_4\) topological order \NoCaseChange{\protect\cite{cite725}}.
\end{eczvaluelist}
\codefieldsection{Fault Tolerance}
\begin{eczvaluelist}
\item\relax The hexagonal \(CZ\) code can be obtained from two surface codes by gauging \NoCaseChange{\protect\cite{cite462,cite463,cite233,cite464,cite465,cite466,cite467,cite468,cite469,cite470}} their logical \(CZ\) gate \NoCaseChange{\protect\cite{cite589}}. Gates on the two surface codes in the third level of the Clifford hierarchy, such as \(CZ\) gates, can be realized fault-tolerantly by performing this procedure and reversing it \NoCaseChange{\protect\cite{cite572,cite589}}.
\end{eczvaluelist}
\codefieldsection{Realizations}
\begin{eczvaluelist}
\item\relax Signatures of the phase detected in a 27-qubit trapped-ion device by Quantinuum \NoCaseChange{\protect\cite{cite3719}}. Preparation of ground states and braiding of anyons has also been performed.
\end{eczvaluelist}
\codefieldsection{Notes}
\begin{eczvaluelist}
\item\relax Popular summary of realization of non-Abelian topological order in \flmHref{https://www.quantamagazine.org/physicists-create-elusive-particles-that-remember-their-pasts-20230509/}{Quanta Magazine}.
\end{eczvaluelist}
\codefieldsection{Parents}
\begin{eczvaluelist}
\item\relax
\flmRefsHyperref[eczindexfamilyrel]{code:cubic_theory}{Cubic theory code} --- The \(2+1\)D cubic theory with \(l=m=n=1\) realizes the same topological order as the Type-III \(\mathbb{Z}_2^3\) twisted quantum double / \(G=D_4\) quantum double, and the hexagonal \(CZ\) code is a hexagonal-lattice realization of this phase \NoCaseChange{\protect\cite{cite575,cite576}}.
\item\relax
\flmRefsHyperref[eczindexfamilyrel]{code:tqd}{Twisted quantum double (TQD) code} --- The ground-state subspace of the hexagonal \(CZ\) code realizes the topological order of the Type-III \(G=\mathbb{Z}^3_2\) Abelian TQD model \NoCaseChange{\protect\cite{cite575,cite576}}, which is the same topological order as the \(G=D_4\) quantum double \NoCaseChange{\protect\cite{cite577}}. There is a constant-depth circuit implementing a transversal logical \(T\) gate via an emergent automorphism symmetry of the underlying \(\mathbb{D}_4\) topological order \NoCaseChange{\protect\cite{cite725}}.
\end{eczvaluelist}
\codefieldsection{Cousins}
\begin{eczvaluelist}
\item\relax
\flmRefsHyperref[eczindexfamilyrel]{code:quantum_double_dihedral}{Dihedral \(G=D_m\) quantum-double code} --- The ground-state subspace of the hexagonal \(CZ\) code realizes the topological order of the Type-III \(G=\mathbb{Z}^3_2\) Abelian TQD model \NoCaseChange{\protect\cite{cite575,cite576}}, which is the same topological order as the \(G=D_4\) quantum double \NoCaseChange{\protect\cite{cite577}}. There is a constant-depth circuit implementing a transversal logical \(T\) gate via an emergent automorphism symmetry of the underlying \(\mathbb{D}_4\) topological order \NoCaseChange{\protect\cite{cite725}}.
\item\relax
\flmRefsHyperref[eczindexfamilyrel]{code:surface}{Kitaev surface code} --- The hexagonal \(CZ\) code can be obtained from two surface codes by gauging \NoCaseChange{\protect\cite{cite462,cite463,cite233,cite464,cite465,cite466,cite467,cite468,cite469,cite470}} their logical \(CZ\) gate \NoCaseChange{\protect\cite{cite589}}. Gates on the two surface codes in the third level of the Clifford hierarchy, such as \(CZ\) gates, can be realized fault-tolerantly by performing this procedure and reversing it \NoCaseChange{\protect\cite{cite572,cite589}}.
\item\relax
\flmRefsHyperref[eczindexfamilyrel]{code:spt}{Symmetry-protected topological (SPT) code} --- The hexagonal \(CZ\) code can be obtained by gauging \NoCaseChange{\protect\cite{cite462,cite463,cite233,cite464,cite465,cite466,cite467,cite468,cite469,cite470}} the symmetry of a particular SPT \NoCaseChange{\protect\cite{cite575}}.
\item\relax
\flmRefsHyperref[eczindexfamilyrel]{code:brickwork}{Brickwork \(XS\) stabilizer code} --- The brickwork \(XS\) stabilizer code and the hexagonal \(CZ\) code realize the same topological phase and are equivalent via a local unitary \NoCaseChange{\protect\cite{cite589,cite3511}}.
\end{eczvaluelist}
\eczhbkcontributors{ Benjamin J. Brown, \eczhuVVA }
\endeczcode

\eczcode{hierarchical}{Hierarchical code}{~\NoCaseChange{\protect\cite{cite3607}}}
\codefieldsection{Description}
Member of a family of \(\llbracket n,k,d\rrbracket \) qubit stabilizer codes resulting from a concatenation of a constant-rate \flmRefsHyperref{code:qldpc}{QLDPC code} with a \flmRefsHyperref{code:rotated_surface}{rotated surface code}.
Concatenation allows for syndrome extraction to be performed on a 2D geometry while maintaining a threshold at the expense of a logarithmically vanishing rate.
The growing syndrome extraction circuit depth allows known bounds in the literature to be weakened \NoCaseChange{\protect\cite{cite521,cite522}}.
\codefieldsection{Rate}
Rate \flmRefsHyperref{ref65}{scales} as \(\Omega(1/\log(n)^2)\).
\codefieldsection{Decoding}
\begin{eczvaluelist}
\item\relax Decoding is performed as in a standard \flmRefsHyperref{code:qubit_concatenated}{concatenated code} using decoders for the inner and outer codes. The syndrome extraction circuit depth for the outer code is optimized using a permutation routing algorithm \NoCaseChange{\protect\cite{cite3720}}. The bilayer architecture allows for logical entangling gates between logical surface-code patches.
\item\relax Soft output decoding \NoCaseChange{\protect\cite{cite3721}}.
\end{eczvaluelist}
\codefieldsection{Fault Tolerance}
\begin{eczvaluelist}
\item\relax 2D geometrically local syndrome extraction circuits of depth of \flmRefsHyperref{ref65}{order} \(O(\sqrt{n}/R)\) that utilize Clifford and SWAP gates of range \(R\) and that require \flmRefsHyperref{ref65}{order} \(O(n)\) data and ancilla qubits. Such parameters (including a range of one) are possible while maintaining a threshold because of the concatenation step. This reduces the noise that would otherwise accumulate within a growing-depth syndrome extraction circuit. A key idea is that constant-depth syndrome extraction is not a necessary condition for fault-tolerance.
\end{eczvaluelist}
\codefieldsection{Threshold}
\begin{eczvaluelist}
\item\relax Threshold exists for the locally decaying error model; see \NoCaseChange{\protect\cite[{Thm. 1.3}]{cite3607}}. However, the logical error rate below threshold falls super-polynomially (as opposed to exponentially) with the code distance. The code family possesses a threshold equal to that of surface codes given by tuning the inner code size for any fixed physical error rate.
\end{eczvaluelist}
\codefieldsection{Parents}
\begin{eczvaluelist}
\item\relax
\flmRefsHyperref[eczindexfamilyrel]{code:qldpc}{Qubit QLDPC code}\item\relax
\flmRefsHyperref[eczindexfamilyrel]{code:qubit_concatenated}{Concatenated qubit code} --- Using the concatenation convention of the Zoo, hierarchical codes are concatenations of constant-rate QLDPC (inner) codes with rotated surface (outer) codes. The cited paper \NoCaseChange{\protect\cite{cite3607}} uses the opposite inner/outer terminology. The block length of the outer code is picked to grow logarithmically with the block length of the inner code.
\end{eczvaluelist}
\codefieldsection{Cousin}
\begin{eczvaluelist}
\item\relax
\flmRefsHyperref[eczindexfamilyrel]{code:rotated_surface}{Rotated surface code} --- Hierarchical codes are concatenations of constant-rate QLDPC codes with rotated surface codes.
\end{eczvaluelist}
\eczhbkcontributors{ Christopher A. Pattison, \eczhuVVA }
\endeczcode

\eczcode{ramanujan_tensor_product}{High-dimensional expander (HDX) code}{~\NoCaseChange{\protect\cite{cite3722,cite684}}}
\codefieldsection{Description}
CSS code obtained by applying the generalized distance-balancing/product construction of Ref. \NoCaseChange{\protect\cite{cite684}} to a Ramanujan-complex quantum code and an asymptotically good classical LDPC code.

Ramanujan quantum codes are defined using LSV Ramanujan complexes, which are simplicial complexes that generalize Ramanujan graphs \NoCaseChange{\protect\cite{cite75,cite76}}.
The auxiliary classical code is viewed as a 1-dimensional chain complex, and the output code is defined on the co-complex of the product of the two co-complexes.
Using a 2D LSV complex yields a QLDPC family with \(K=\Omega(\sqrt{n/\log n})\) and \(D=\Omega(\sqrt{n \log n})\), while using a 3D LSV complex yields \(K=\Omega(\sqrt{n}/\log n)\) and \(D=\Omega(\sqrt{n}\log n)\).

\codefieldsection{Protection}
The unbalanced component code from a 2D LSV complex can have \(d_X=\Omega(\log n)\) and \(d_Z=\Omega(n)\), while one from a 3D LSV complex can have \(d_X=\Omega(\log^2 n)\) and \(d_Z=\Omega(n)\) \NoCaseChange{\protect\cite{cite684}}. After distance balancing, the resulting HDX family has minimum distance \(D=\Omega(\sqrt{n \log n})\) in the 2D case and \(D=\Omega(\sqrt{n}\log n)\) in the 3D case.
\codefieldsection{Rate}
For 2D LSV complexes, the rate is of \flmRefsHyperref{ref65}{order} \(\Omega(1/\sqrt{n \log n})\), with minimum distance \(D=\Omega(\sqrt{n \log n})\). For 3D LSV complexes, the rate is \(\Omega( 1/(\sqrt{n}\log n) )\), with minimum distance \(D=\Omega(\sqrt{n}\log n)\).
\codefieldsection{Decoding}
\begin{eczvaluelist}
\item\relax For HDX codes built from 2D LSV complexes, \(X\)-error decoding reduces to polynomial-time cycle-code decoding on the 1-skeleton together with decoding of the auxiliary classical LDPC code \NoCaseChange{\protect\cite{cite684}}.
\item\relax For the 2D construction, \(Z\)-errors of linear weight admit a local decoder based on coboundary expansion; replacing the component complex by the 2-skeleton of a 3D LSV complex preserves 2D-type asymptotic parameters while giving linear-time \(Z\)-decoding with unit-weight local corrections \NoCaseChange{\protect\cite{cite684}}.
\end{eczvaluelist}
\codefieldsection{Notes}
\begin{eczvaluelist}
\item\relax Codes were first to break a 20-year record set by the \flmRefsHyperref{code:freedman_meyer_luo}{Freedman-Meyer-Luo code} for the lower bound on scaling of the minimum distance \NoCaseChange{\protect\cite{cite3442}}.
\end{eczvaluelist}
\codefieldsection{Parents}
\begin{eczvaluelist}
\item\relax
\flmRefsHyperref[eczindexfamilyrel]{code:homological_product}{Homological product code} --- Ramanujan codes result from a tensor product of a classical-code and a quantum-code chain complex.
\item\relax
\flmRefsHyperref[eczindexfamilyrel]{code:iterated_ramanujan}{Tensor-product HDX code} --- Ramanujan codes result from a tensor product of a classical-code and a quantum-code chain complex.
\end{eczvaluelist}
\codefieldsection{Cousins}
\begin{eczvaluelist}
\item\relax
\flmRefsHyperref[eczindexfamilyrel]{code:distance_balanced}{Distance-balanced code} --- Ramanujan tensor-product constructions use distance balancing to increase distance.
\item\relax
\flmRefsHyperref[eczindexfamilyrel]{code:hypergraph_product}{Hypergraph product (HGP) code} --- Ramanujan codes utilize the hypergraph product with a twist, which is an automorphism on one of the complexes in the tensor product, in order to increase distance \NoCaseChange{\protect\cite{cite3442}}.
\item\relax
\flmRefsHyperref[eczindexfamilyrel]{code:freedman_meyer_luo}{Freedman-Meyer-Luo code} --- Ramanujan codes broke 20-year record on minimum code distance set by Freedman-Meyer-Luo codes.
\end{eczvaluelist}
\eczhbkcontributors{ Xiaozhen Fu, \eczhuVVA }
\endeczcode

\eczcode{multisector_hypergraph}{Higher-dimensional homological product code}{~\NoCaseChange{\protect\cite{cite835,cite2989,cite675,cite1613}}}
\codefieldsection{Alternative Names}
\begin{eczvaluelist}
\item\relax Higher-dimensional tensor product code
\item\relax Multi-sector homological product code
\item\relax Multi-sector tensor product code
\item\relax Iterative homological product code
\item\relax Iterative tensor product code
\end{eczvaluelist}
\eczhIndexCodeAliasName{multisector_hypergraph}{Higher-dimensional tensor product code}
\eczhIndexCodeAliasName{multisector_hypergraph}{Multi-sector homological product code}
\eczhIndexCodeAliasName{multisector_hypergraph}{Multi-sector tensor product code}
\eczhIndexCodeAliasName{multisector_hypergraph}{Iterative homological product code}
\eczhIndexCodeAliasName{multisector_hypergraph}{Iterative tensor product code}
\codefieldsection{Description}
A qubit CSS code formulated using a tensor product of two or more chain complexes, each of length one or greater.
The number of chain complexes participating in the product is the \textit{dimension} of the code.
When all chain complexes are length-one, meaning that they correspond to classical codes, the code is called a \textit{higher-dimensional HGP code} (a.k.a. multi-sector HGP code or iterative HGP code).

The stabilizer generator matrices of an \(m\)-dimensional homological product code are the boundary and co-boundary operators of a 2-dimensional chain complex contained within an \(m\)-complex that is recursively constructed by taking the tensor product of an \((m-1)\)-complex and a 1-complex.
There is freedom in choosing which 2-dimensional chain complex to pick for the code.

\codefieldsection{Protection}
The Künneth formula gives code properties in terms of those of the underlying chain complexes.
Hypergraph products of multiple classical codes were considered first \NoCaseChange{\protect\cite{cite835}}.
Homological products of CSS-code and classical chain complexes are used as a building block in Hastings' weight-reduction construction \NoCaseChange{\protect\cite{cite2989}}, while later work studied explicit product-code families built from length-two chain complexes \NoCaseChange{\protect\cite{cite675}} (which correspond to qubit CSS codes per the \flmRefsHyperref{ref683}{qubit CSS-to-homology correspondence}).

Ref. \NoCaseChange{\protect\cite{cite1613}} gives explicit parameter formulas for the tensor product \(\mathcal{A}\times\mathcal{K}(P)\) of an \(m\)-complex \(\mathcal{A}\) with the 1-complex \(\mathcal{K}(P)\) built from an \(r\times c\) binary matrix \(P\).
Let \(u=\mathrm{rank}(P)\) and \(\delta\) be the minimum distance of the binary code with parity-check matrix \(P\).
Writing \(n_j\), \(k_j\), \(d_j\) for the block length, logical dimension, and homological distance of the \(j\)-th sector of \(\mathcal{A}\), and primes for the corresponding parameters of the product, the Künneth theorem gives
\(n_j'=n_{j-1}c+n_j r\) and \(k_j'=k_{j-1}(c-u)+k_j(r-u)\).
The main result of \NoCaseChange{\protect\cite[{Thm. 1}]{cite1613}} establishes tight bounds on the homological distance:
\(d_j'=\min(d_j,d_{j-1}\delta)\) when \(r>u\) (P does not have full row rank), and \(d_j'=d_{j-1}\delta\) when \(r=u\) (P has full row rank).

\codefieldsection{Rate}
A notable special case uses a single full-row-rank \(r\times c\) seed matrix \(P\) (so \(r=u\), \(\kappa=c-r\) logical bits per block, classical distance \(\delta\), row/column weights bounded by \((\omega,\upsilon)\)).
Taking \(a\) copies of \(\mathcal{K}(P)\) and \(b\) copies of the dual 1-complex \(\tilde{\mathcal{K}}=\mathcal{K}(P^T)\) yields the \((a+b)\)-complex \(\mathcal{K}^{(a,b)}=\mathcal{K}^{\times a}\times \tilde{\mathcal{K}}^{\times b}\), which has a single non-trivial homology sector \(H_a\) with
\flmMathEnvironment{align}{}{
  n_a&=\sum_{i=0}^{a} c^{2i} r^{a+b-2i} {a\choose i}{b\choose i} < (r+c)^{a+b}~,\\
  k_a&=\kappa^{a+b}~,
}
yielding a quantum CSS code with \(d_X=\delta^a\), \(d_Z=\delta^b\), and stabilizer-generator weights bounded by \((a+b)\max(\omega,\upsilon)\) \NoCaseChange{\protect\cite{cite1613}}.
The parameters \(a\) and \(b\) can be chosen independently to trade off \(X\)- and \(Z\)-distance, enabling asymmetric codes optimized for channels with unequal bit-flip and phase-flip rates.
For asymptotically good LDPC seed-code families, these higher-dimensional HGP families have finite rate and \(d_X d_Z\) scaling linearly with block length.

\codefieldsection{Transversal and Permutation-Based Gates}
\begin{eczvaluelist}
\item\relax Transversal gates for multi-dimensional HGP codes of all dimensions lie in the Clifford group \NoCaseChange{\protect\cite{cite739}}.
\end{eczvaluelist}
\codefieldsection{Gates}
\begin{eczvaluelist}
\item\relax Gates in the \flmTerm{term}{ref694}{}{Clifford hierarchy} can be implemented via constant-depth circuits \NoCaseChange{\protect\cite{cite739}}.
\item\relax Parallel Pauli product measurements via homomorphic CNOT gates for 3- and 4-dimensional HGP codes \NoCaseChange{\protect\cite{cite1555}}.
\end{eczvaluelist}
\codefieldsection{Fault Tolerance}
\begin{eczvaluelist}
\item\relax For \((a,b)\) constructions with \(a,b>1\), the check matrices \(G_X=K_a\) and \(G_Z=K_{a+1}^T\) satisfy many linear relations coming from neighboring boundary maps; these redundant checks can be used to handle syndrome-measurement errors in repeated-measurement schemes \NoCaseChange{\protect\cite{cite1613}}.
\end{eczvaluelist}
\codefieldsection{Parent}
\begin{eczvaluelist}
\item\relax
\flmRefsHyperref[eczindexfamilyrel]{code:qubit_generalized_homological_product_css}{Generalized homological-product qubit CSS code}\end{eczvaluelist}
\codefieldsection{Children}
\begin{eczvaluelist}
\item\relax
\flmRefsHyperref[eczindexfamilyrel]{code:double_homological_product}{Campbell double homological product code}\item\relax
\flmRefsHyperref[eczindexfamilyrel]{code:iterated_ramanujan}{Tensor-product HDX code} --- Tensor-product HDX codes result from iterated homological products of length-two chain complexes (i.e., quantum codes) based on Ramanujan complexes \NoCaseChange{\protect\cite{cite3442}}.
\item\relax
\flmRefsHyperref[eczindexfamilyrel]{code:homological_product}{Homological product code} --- Multi-dimensional homological products of two length-two chain complexes reduce to homological product codes.
\end{eczvaluelist}
\codefieldsection{Cousins}
\begin{eczvaluelist}
\item\relax
\flmRefsHyperref[eczindexfamilyrel]{code:binary_linear}{Linear binary code} --- \(D\)-dimensional HGP codes are constructed using a hypergraph product of \(D\) linear binary codes.
\item\relax
\flmRefsHyperref[eczindexfamilyrel]{code:binary_cyclic}{Cyclic linear binary code} --- Higher-dimensional homological-product codes can be constructed out of CSS codes that in turn stem from cyclic codes \NoCaseChange{\protect\cite[{Sec. 4.2}]{cite835}}.
\item\relax
\flmRefsHyperref[eczindexfamilyrel]{code:quantum_reed_muller}{Quantum Reed-Muller (RM) code} --- Higher-dimensional homological-product codes can be constructed out of quantum RM codes \NoCaseChange{\protect\cite[{Sec. 4.3}]{cite835}}.
\item\relax
\flmRefsHyperref[eczindexfamilyrel]{code:qc_ldpc}{Quasi-cyclic LDPC (QC-LDPC) code} --- Higher-dimensional hypergraph-product codes can be constructed out of QC-LDPC codes \NoCaseChange{\protect\cite[{Table III}]{cite1555}}.
\item\relax
\flmRefsHyperref[eczindexfamilyrel]{code:asymmetric_qecc}{Asymmetric quantum code (AQC)} --- The \((a,b)\)-complex construction yields asymmetric CSS codes with \(d_X=\delta^a\) and \(d_Z=\delta^b\), allowing independent tuning of the two distances for channels with unequal bit-flip and phase-flip rates \NoCaseChange{\protect\cite{cite1613}}.
\item\relax
\flmRefsHyperref[eczindexfamilyrel]{code:quantum_expander}{Quantum expander code} --- Quantum expander codes have been generalized to hypergraph products of 3 or more expander codes \NoCaseChange{\protect\cite{cite3723}}.
\item\relax
\flmRefsHyperref[eczindexfamilyrel]{code:xyz_product}{XYZ product code} --- The XYZ product code is a non-CSS three-fold variant of the hypergraph product built from three classical linear binary codes \NoCaseChange{\protect\cite{cite645}}.
\item\relax
\flmRefsHyperref[eczindexfamilyrel]{code:pg_qldpc}{Finite-geometry (FG) qubit QLDPC code} --- Multi-dimensional homological products of PG-QLDPC codes yield families whose stabilizer-generator weights scale logarithmically with \(n\) \NoCaseChange{\protect\cite[{Cor. 2.23}]{cite835}\protect\cite[{Sec. 4.1}]{cite835}}.
\item\relax
\flmRefsHyperref[eczindexfamilyrel]{code:3d_surface}{3D surface code} --- The 3D planar and toric code on a cubic lattice can be obtained from a hypergraph product of three repetition codes \NoCaseChange{\protect\cite{cite1613}\protect\cite[{Exam. A.1}]{cite1612}}.
\item\relax
\flmRefsHyperref[eczindexfamilyrel]{code:4d_13_surface}{\((1,3)\) 4D toric code} --- The 4D \((1,3)\) planar (toric) code on a hypercubic lattice can be obtained from a particular choice of chain complex from a hypergraph product of four repetition codes \NoCaseChange{\protect\cite{cite1613}}.
\item\relax
\flmRefsHyperref[eczindexfamilyrel]{code:4d_surface}{\((2,2)\) Loop toric code} --- The 4D loop planar (toric) code on a hypercubic lattice can be obtained from a particular choice of chain complex from a hypergraph product of four repetition codes \NoCaseChange{\protect\cite{cite1613}}.
\item\relax
\flmRefsHyperref[eczindexfamilyrel]{code:higher_dimensional_toric}{\(D\)-dimensional twisted toric code} --- The non-twisted \(D\)-dimensional planar and toric codes on a hypercubic lattice can be obtained from a hypergraph product of \(D\) repetition codes \NoCaseChange{\protect\cite{cite1613}}.
\item\relax
\flmRefsHyperref[eczindexfamilyrel]{code:subsystem_product}{Subsystem homological product code} --- SP codes are projected higher-dimensional HGP codes \NoCaseChange{\protect\cite{cite664}}.
\end{eczvaluelist}
\eczhbkcontributors{ Nathanan Tantivasadakarn, Feroz Ahmed Mian, \eczhuVVA }
\endeczcode

\eczcode{higher_dimensional_surface}{Homological code}{~\NoCaseChange{\protect\cite{cite480,cite3684,cite1304,cite426}}}
\codefieldsection{Alternative Names}
\begin{eczvaluelist}
\item\relax Generalized surface code
\end{eczvaluelist}
\eczhIndexCodeAliasName{higher_dimensional_surface}{Generalized surface code}
\codefieldsection{Description}
A CSS extension of the Kitaev surface code to arbitrary manifolds.
The version on a Euclidean manifold of some fixed dimension is called the \(D\)\textit{-dimensional "surface"} or \(D\)\textit{-dimensional toric} code.

Given a cellulation of a manifold, qubits are put on \(p\)-dimensional faces, \(X\)-type stabilizers
are associated with \((p-1)\)-faces, while \(Z\)-type stabilizers are associated with \((p+1)\)-faces.
Here, \(p\) ranges between \(1\) and \(D-1\).

Lattice surface codes in \(D\) spatial dimensions can be partially classified by the dimension of their stabilizer generators (and corresponding excitations).
There are \((p,q)\) \textit{surface codes} for \(q = D-p\) \NoCaseChange{\protect\cite{cite575}}.
Applying this construction to the dual lattice of a \((p,q)\) surface code yields a \((q,p)\) surface code.
All lattice surface codes have bosonic \(e\) and \(m\) excitations of dimension \(p-1\) and \(q-1\), respectively. Their logical operators are also of dimension \(p\) and \(q\), respectively.

In 2D, there is only the \((1,1)\) surface code, which is equivalent to the Kitaev surface code and which admits point-like \(e\) and \(m\) excitations.
In 3D, there are the \((1,2)\) and \((2,1)\) 3D surface codes, which are equivalent by Hadamard gates. Both admit point-like excitations of one type and loop-like excitations of the other.
In 4D, there are the \((1,3)\), \((2,2)\), and \((3,1)\) 4D surface codes, with the first and last being equivalent by Hadamard gates. The \((1,3)\) code admits point-like \(e\) excitations and 2D membrane \(m\) excitations, while the \((2,2)\) loop toric code admits loop-like \(e\) and \(m\) excitations.

Open-boundary hypercubic realizations give a family of \((d_1,d_2)\)-surface codes whose logical \(\overline{X}\) and \(\overline{Z}\) operators have dimensions \(d_1\) and \(d_2\), respectively; the ordinary surface code, the 3D cubic code, and the 4D tesseract code are examples \NoCaseChange{\protect\cite{cite3174}}.

\codefieldsection{Protection}
The 2D members of the family obey the Bravyi-Terhal no-go theorem: geometrically local stabilizer generators allow distance at most \(O(L)\) and only an \(O(1)\) energy barrier, ruling out self-correction in two dimensions \NoCaseChange{\protect\cite{cite3000}}.
By contrast, the 4D \((2,2)\) loop surface code serves as a self-correcting quantum memory, while surface codes in higher dimensions can have distances not possible in lower dimensions.

\codefieldsection{Rate}
Rate depends on the underlying cellulation and manifold \NoCaseChange{\protect\cite{cite480,cite426}}.
For general 2D
manifolds, \(kd^2\leq c(\log k)^2 n\) for some constant \(c\)
\NoCaseChange{\protect\cite{cite837}}, meaning that (1) 2D surface codes with bounded
geometry have distance scaling at most as \(O(\sqrt{n})\)
\NoCaseChange{\protect\cite{cite1419,cite681}}, and (2) surface codes with
finite rate can only achieve an asymptotic minimum distance that is
logarithmic in \(n\).
Higher-dimensional manifolds yield distances scaling more favorably.
Loewner's theorem
provides an upper bound for any bounded-geometry surface code
\NoCaseChange{\protect\cite{cite3684}}.

\codefieldsection{Transversal and Permutation-Based Gates}
\begin{eczvaluelist}
\item\relax Locality preserving operations can be determined for stacks of homological codes in any dimension \NoCaseChange{\protect\cite{cite741}}.
\end{eczvaluelist}
\codefieldsection{Decoding}
\begin{eczvaluelist}
\item\relax Local automaton decoders based on Toom's rule and its generalization, the sweep rule \NoCaseChange{\protect\cite{cite3724,cite3725,cite3416}}.
\item\relax Improved BP-OSD decoder \NoCaseChange{\protect\cite{cite3179}}.
\item\relax Renormalization group (RG) decoder \NoCaseChange{\protect\cite{cite3174}}.
\end{eczvaluelist}
\codefieldsection{Code Capacity Threshold}
\begin{eczvaluelist}
\item\relax \(>0\%\) threshold with sweep decoder for lattice surface codes in various dimensions \NoCaseChange{\protect\cite{cite3416}}.
\end{eczvaluelist}
\codefieldsection{Notes}
\begin{eczvaluelist}
\item\relax 2D and 3D surface code \flmHref{https://gui.quantumcodes.io/}{visualization
tool}.

\item\relax \NoCaseChange{\protect\cite{cite3726}} on the role of homology in constructing surface codes by D. Browne.
\end{eczvaluelist}
\codefieldsection{Parent}
\begin{eczvaluelist}
\item\relax
\flmRefsHyperref[eczindexfamilyrel]{code:qubit_generalized_homological_product_css}{Generalized homological-product qubit CSS code} --- The generalized surface code is constructed from chain complexes arising from cell complexes of the underlying manifold. Such complexes are not necessarily products of two non-trivial complexes, but the manifolds are picked so that their homology ensures favorable code properties.
\end{eczvaluelist}
\codefieldsection{Children}
\begin{eczvaluelist}
\item\relax
\flmRefsHyperref[eczindexfamilyrel]{code:surface}{Kitaev surface code} --- The surface-code CSS stabilizer generator prescription is extendable to higher-dimensional manifolds.
\item\relax
\flmRefsHyperref[eczindexfamilyrel]{code:3d_surface}{3D surface code}\item\relax
\flmRefsHyperref[eczindexfamilyrel]{code:4d_13_surface}{\((1,3)\) 4D toric code} --- The \((1,3)\) 4D toric code realizes 4D \(\mathbb{Z}_2\) gauge theory with 1D \(Z\)-type and 3D \(X\)-type logical operators.
\item\relax
\flmRefsHyperref[eczindexfamilyrel]{code:4d_surface}{\((2,2)\) Loop toric code} --- The 4D loop toric code realizes 4D \(\mathbb{Z}_2\) gauge theory with only loop excitations \NoCaseChange{\protect\cite{cite2529}}.
\item\relax
\flmRefsHyperref[eczindexfamilyrel]{code:fractal_surface}{Fractal surface code} --- Fractal surface codes are obtained by removing qubits from the 3D surface code on a cubic lattice.
\item\relax
\flmRefsHyperref[eczindexfamilyrel]{code:hemicubic}{Hemicubic code}\item\relax
\flmRefsHyperref[eczindexfamilyrel]{code:higher_dimensional_toric}{\(D\)-dimensional twisted toric code}\item\relax
\flmRefsHyperref[eczindexfamilyrel]{code:hypersphere_product}{Hypersphere product code}\item\relax
\flmRefsHyperref[eczindexfamilyrel]{code:hyperbolic_surface}{Hyperbolic surface code}\end{eczvaluelist}
\codefieldsection{Cousins}
\begin{eczvaluelist}
\item\relax
\flmRefsHyperref[eczindexfamilyrel]{code:translationally_invariant_stabilizer}{Lattice stabilizer code} --- Lattice surface codes in \(D\) spatial dimensions can be partially classified by the dimension of their stabilizer generators (and corresponding excitations).
There are \((p,q)\) \textit{surface codes} for \(p+q=D\) realized by \(Z\)-type stabilizer generators of dimension \(p\) and \(X\)-type stabilizer generators of dimension \(q\).
The two corresponding types of excitations are of dimension \(p-1\) and \(q-1\), respectively.

\item\relax
\flmRefsHyperref[eczindexfamilyrel]{code:homological_classical}{Cycle code} --- Cycle codes feature in generalizations of the surface code \NoCaseChange{\protect\cite{cite1304}}.
\item\relax
\flmRefsHyperref[eczindexfamilyrel]{code:hypercube_quantum}{\(\llbracket 2^D,D,2\rrbracket \) hypercube quantum code} --- The hypercube quantum code can be concatenated with \(D\) distance-two \(D\)-dimensional toric/surface-code blocks to yield a \(\llbracket 2^D(2^D+1),D,4\rrbracket \) error-correcting family that admits a transversal implementation of the logical \(C^{D-1}Z\) gate \NoCaseChange{\protect\cite{cite759}}.
\item\relax
\flmRefsHyperref[eczindexfamilyrel]{code:color}{Color code} --- For the common realization with point-like electric excitations, the color code on a \(D\)-dimensional closed manifold is equivalent to \(D\) decoupled copies of the \(D\)-dimensional toric/surface code via a local constant-depth \flmRefsHyperref{ref409}{Clifford circuit} \NoCaseChange{\protect\cite{cite3424,cite422,cite3425}} (see also \NoCaseChange{\protect\cite[{Exam. 4}]{cite739}}).
On a \(D\)-simplex-like lattice with \(D+1\) differently colored boundaries, the corresponding toric-code copies are attached along a common \((D-1)\)-dimensional boundary rather than fully decoupled \NoCaseChange{\protect\cite{cite422}}.
The reverse of this process can be viewed as gauging \NoCaseChange{\protect\cite{cite462,cite463,cite233,cite464,cite465,cite466,cite467,cite468,cite469,cite470}} certain symmetries.
Morphing subsets of colorable \(D\)-balls produces hybrid color-toric codes that interpolate between the color code and \(D\) copies of the toric code (up to ancillas when all balls of one color are morphed), while inheriting the parent color code's fault-tolerant gates \NoCaseChange{\protect\cite{cite687}}.
Several hybrid color-surface codes exist \NoCaseChange{\protect\cite{cite687,cite3595}}.

\item\relax
\flmRefsHyperref[eczindexfamilyrel]{code:quasi_hyperbolic_color}{Quasi-hyperbolic color code} --- Quasi-hyperbolic color codes are related to quasi-hyperbolic surface codes via a constant-depth \flmRefsHyperref{ref409}{Clifford circuit} \NoCaseChange{\protect\cite{cite703}}.
\item\relax
\flmRefsHyperref[eczindexfamilyrel]{code:subsystem_higher_dimensional_surface}{Subsystem homological code} --- Subsystem homological codes are subsystem versions of homological codes, with gauge-group generators of lower weight than the corresponding surface-code stabilizers.
\end{eczvaluelist}
\eczhbkcontributors{ \eczhuVVA }
\endeczcode

\eczcode{homological_product}{Homological product code}{~\NoCaseChange{\protect\cite{cite2562,cite835,cite2989}}}
\codefieldsection{Alternative Names}
\begin{eczvaluelist}
\item\relax Tensor product code
\end{eczvaluelist}
\eczhIndexCodeAliasName{homological_product}{Tensor product code}
\codefieldsection{Description}
CSS code formulated using the tensor product of two chain complexes of length one or greater (see \flmRefsCref{ref683}).

Homological products and ordinary tensor products of chain complexes differ in a way that depends on whether the underlying code is defined by a general or a length-three chain complex \NoCaseChange{\protect\cite[{Sec. 3.4.3}]{cite835}}.

\codefieldsection{Protection}
Given two codes \(\llbracket n_i, k_i, d_i, w_i\rrbracket \) for \(i\in\{1,2\}\), where \(w_i\) denotes the maximum hamming weight of all rows and columns of \(\partial_i\), the homological product code has parameter \(\llbracket n=n_1 n_2, k=k_1 k_2, d\leq d_1 d_2, w\leq w_1+w_2\rrbracket \).
From this formula, and the fact that a randomly selected boundary operator \(\partial\) yields a CSS code that is good with high probability, we see that the product code has \(k=\Theta(n)\) and \(w=O(\sqrt{n})\) with high probability.
The main result in Ref. \NoCaseChange{\protect\cite{cite3727}} is to show that the product code has linear distance with high probability as well.
To sum up, it is shown that we have a family of \(\llbracket n,k=c_1 n, d=c_2 n, w=c_3 \sqrt{n}\rrbracket \) codes given small enough \(c_1,c_2,c_3\).

\codefieldsection{Gates}
\begin{eczvaluelist}
\item\relax Universal set of gates can be obtained by fault-tolerantly mapping between different encoded representations of a given logical state \NoCaseChange{\protect\cite{cite3728}}.
\item\relax Parallel Pauli product measurements via homomorphic CNOT gates \NoCaseChange{\protect\cite{cite1555}}.
\end{eczvaluelist}
\codefieldsection{Decoding}
\begin{eczvaluelist}
\item\relax Union-find decoder \NoCaseChange{\protect\cite{cite3729}}.
\item\relax BP-OSD-like post-processing \NoCaseChange{\protect\cite{cite1247}}.
\end{eczvaluelist}
\codefieldsection{Fault Tolerance}
\begin{eczvaluelist}
\item\relax Universal set of gates can be obtained by fault-tolerantly mapping between different encoded representations of a given logical state \NoCaseChange{\protect\cite{cite3728}}.
\end{eczvaluelist}
\codefieldsection{Parents}
\begin{eczvaluelist}
\item\relax
\flmRefsHyperref[eczindexfamilyrel]{code:multisector_hypergraph}{Higher-dimensional homological product code} --- Multi-dimensional homological products of two length-two chain complexes reduce to homological product codes.
\item\relax
\flmRefsHyperref[eczindexfamilyrel]{code:fiber_bundle}{Fiber-bundle code} --- A fiber-bundle code can be viewed as a homological product code with a twisted product.
\end{eczvaluelist}
\codefieldsection{Children}
\begin{eczvaluelist}
\item\relax
\flmRefsHyperref[eczindexfamilyrel]{code:hypergraph_product}{Hypergraph product (HGP) code} --- A homological-product code of length-one chain complexes reduces to an HGP code, which is also a special case of multi-dimensional homological products of two length-one chain complexes.
\item\relax
\flmRefsHyperref[eczindexfamilyrel]{code:ramanujan_tensor_product}{High-dimensional expander (HDX) code} --- Ramanujan codes result from a tensor product of a classical-code and a quantum-code chain complex.
\item\relax
\flmRefsHyperref[eczindexfamilyrel]{code:square_homological_product}{Square homological product code} --- Square homological product codes are homological product codes whose boundary operators are square matrices \NoCaseChange{\protect\cite[{Sec. 3.4}]{cite835}}.
\end{eczvaluelist}
\codefieldsection{Cousins}
\begin{eczvaluelist}
\item\relax
\flmRefsHyperref[eczindexfamilyrel]{code:random_stabilizer}{Random stabilizer code} --- Random homological codes are asymptotically good with high probability \NoCaseChange{\protect\cite[{Thm. 1}]{cite2562}}.
\item\relax
\flmRefsHyperref[eczindexfamilyrel]{code:single_shot}{Single-shot code} --- It is conjectured that a particular class of codes called three-dimensional product codes is single shot \NoCaseChange{\protect\cite{cite844}}.
\item\relax
\flmRefsHyperref[eczindexfamilyrel]{code:cluster_state}{Cluster-state code} --- In the fault-complex formalism, foliation of a CSS code is expressed as a homological product of the code's chain complex with a repetition-code complex \NoCaseChange{\protect\cite{cite3176}}.

\item\relax
\flmRefsHyperref[eczindexfamilyrel]{code:subsystem_product}{Subsystem homological product code} --- SP codes reduce to homological product codes when there are no gauge qubits \NoCaseChange{\protect\cite{cite664}}.
\item\relax
\flmRefsHyperref[eczindexfamilyrel]{code:distance_balanced}{Distance-balanced code} --- Distance balancing relies on taking a homological product of chain complexes corresponding to a classical and a quantum code.
\end{eczvaluelist}
\eczhbkcontributors{ Xinyuan Zheng, \eczhuVVA }
\endeczcode

\eczcode{triangular_color}{Honeycomb (6.6.6) color code}{~\NoCaseChange{\protect\cite{cite710}}}
\codefieldsection{Description}
2D color code defined on a (typically triangular) patch of the 6.6.6 (honeycomb) tiling.
The usual triangular patch has three differently colored boundaries, encodes one logical qubit, and is local-Clifford equivalent to a folded surface/toric code with two smooth and two rough boundaries \NoCaseChange{\protect\cite{cite422}}.

Stabilizer generators are shown in \flmRefsCref{ref3730}.
  \begin{flmFloat}{figure}{NumCap}\includegraphics[width=315bp,max width=\linewidth]{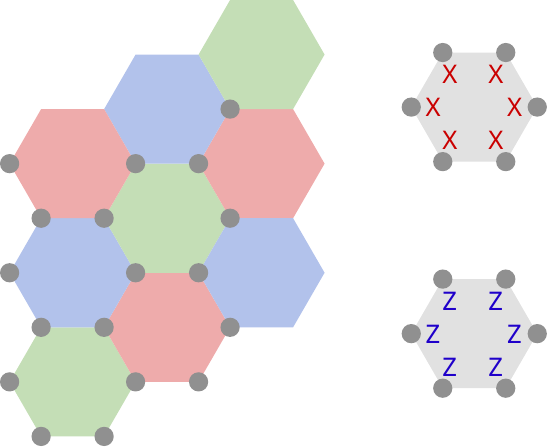}\caption{
    Stabilizer generators of the 6.6.6 color code.
    }\label{ref3730}\end{flmFloat}

\codefieldsection{Protection}
There is a \(\llbracket (3d^2+1)/4, 1, d\rrbracket \) code family \NoCaseChange{\protect\cite[{Fig. 2}]{cite432}} and a \(\llbracket (3d-1)^2/4, 1, d\rrbracket \) code family \NoCaseChange{\protect\cite{cite3731}}.

\codefieldsection{Transversal and Permutation-Based Gates}
\begin{eczvaluelist}
\item\relax CNOT gate because the code is CSS.
\item\relax Hadamard gates for any qubit geometry which yields a self-dual CSS code.
\end{eczvaluelist}
\codefieldsection{Gates}
\begin{eczvaluelist}
\item\relax Lattice surgery scheme for a hybrid 6.6.6-4.8.8 layout yields lower resource overhead when compared to analogous surface code scheme \NoCaseChange{\protect\cite{cite3732}}.
\item\relax Low-overhead magic-state distillation circuit using flag qubits \NoCaseChange{\protect\cite{cite3733}} or lattice surgery \NoCaseChange{\protect\cite{cite3734}}.
\end{eczvaluelist}
\codefieldsection{Decoding}
\begin{eczvaluelist}
\item\relax Distance-three measurement schedule based on detector error models \NoCaseChange{\protect\cite{cite3735}}.
\item\relax Message-passing decoder \NoCaseChange{\protect\cite{cite3736}}.
\item\relax Adaptation of the restriction decoder \NoCaseChange{\protect\cite{cite3731}}.
\item\relax Neural-network decoder \NoCaseChange{\protect\cite{cite3737}}.
\item\relax Möbius matching decoder gives low logical failure rate \NoCaseChange{\protect\cite{cite3738}} and has an open-source implementation called Chromöbius \NoCaseChange{\protect\cite{cite3421}}.
\item\relax AMBP4, a quaternary version \NoCaseChange{\protect\cite{cite3739}} of the MBP decoder \NoCaseChange{\protect\cite{cite3740}}.
\item\relax MaxSAT-based decoder \NoCaseChange{\protect\cite{cite3741}}.
\item\relax Height-bound decision-tree decoder (DTD) \NoCaseChange{\protect\cite{cite3742}}.
\item\relax Most likely error (MLE) decoder \NoCaseChange{\protect\cite{cite3743}}.
\item\relax Neural network decoder \NoCaseChange{\protect\cite{cite3743}}.
\end{eczvaluelist}
\codefieldsection{Fault Tolerance}
\begin{eczvaluelist}
\item\relax Fault-tolerant syndrome extraction circuits using flag qubits \NoCaseChange{\protect\cite{cite3220,cite3731}}.
\end{eczvaluelist}
\codefieldsection{Code Capacity Threshold}
\begin{eczvaluelist}
\item\relax Independent \(X,Z\) noise: \(p_X = 7.8\%\) under message-passing decoder \NoCaseChange{\protect\cite{cite3736}}, \(8.7\%\) under projection decoder \NoCaseChange{\protect\cite{cite3417}}, \(\geq 6\%\) under rescaling decoder \NoCaseChange{\protect\cite{cite3744}}, \(9.0\%\) under Möbius matching decoder \NoCaseChange{\protect\cite{cite3738}}, \(10.1\%\) under MaxSAT-based decoder \NoCaseChange{\protect\cite{cite3741}}, and \(8.2\%\) under concatenated MWPM decoder \NoCaseChange{\protect\cite{cite3422}}. The threshold under ML decoding corresponds to the value of a critical point of the two-dimensional three-body random-bond Ising model (RBIM) on the Nishimori line \NoCaseChange{\protect\cite{cite3526,cite3745}}, calculated to be \(10.9(2)\%\) in Ref. \NoCaseChange{\protect\cite{cite3745}} and \(10.97(1)\%\) in Ref. \NoCaseChange{\protect\cite{cite3746}}.
\item\relax Depolarizing channel: \(12.6\%\) under the restriction decoder \NoCaseChange{\protect\cite{cite3731}} and the projection decoder \NoCaseChange{\protect\cite{cite3417}}, and \(\approx 14.5\%\) under AMBP4 decoding \NoCaseChange{\protect\cite[{Fig. 12}]{cite3739}}.
\end{eczvaluelist}
\codefieldsection{Threshold}
\begin{eczvaluelist}
\item\relax The threshold under ML decoding with measurement errors corresponds to the value of a critical point of a three-dimensional disordered Ising model, estimated to be \(4.8(2)\%\) \NoCaseChange{\protect\cite{cite3747}}.
\item\relax Circuit-level noise: \(0.2\%\) using two flag qubits per stabilizer generator and the restriction decoder \NoCaseChange{\protect\cite{cite3731}}, and \(0.46\%\) under concatenated MWPM decoder \NoCaseChange{\protect\cite{cite3422}}.
\item\relax A \flmRefsHyperref{ref3210}{measurement threshold} of one \NoCaseChange{\protect\cite{cite3211}}.
\end{eczvaluelist}
\codefieldsection{Realizations}
\begin{eczvaluelist}
\item\relax Superconducting qubits: transversal \flmRefsHyperref{ref409}{Clifford gates}, randomized logical benchmarking, and magic-state injection demonstrated on distance-three and five 6.6.6 color codes on the Willow device by Google Quantum AI \NoCaseChange{\protect\cite{cite3743}}. 
Logical state teleportation using lattice surgery performed between two distance-three color codes.
Magic-state cultivation was demonstrated on a device by Google Quantum AI by code switching between a distance-three 6.6.6 color code and distance-five \(XZZX\) surface code and decoding with the Tesseract decoder \NoCaseChange{\protect\cite{cite3748}}.

\end{eczvaluelist}
\codefieldsection{Parents}
\begin{eczvaluelist}
\item\relax
\flmRefsHyperref[eczindexfamilyrel]{code:2d_color}{2D color code}\item\relax
\flmRefsHyperref[eczindexfamilyrel]{code:lifted_product}{Lifted-product (LP) code} --- The 6.6.6 color code can be formulated directly as an LP code \NoCaseChange{\protect\cite{cite1350}}.
\end{eczvaluelist}
\codefieldsection{Children}
\begin{eczvaluelist}
\item\relax
\flmRefsHyperref[eczindexfamilyrel]{code:stab_4_2_2}{\(\llbracket 4,2,2\rrbracket \) Four-qubit code} --- The \(\llbracket 4,2,2\rrbracket \) code can be interpreted as a 2D color code on a trapezoidal patch that makes up two-thirds of a hexagon of the 6.6.6 tiling \NoCaseChange{\protect\cite{cite2526,cite3262}}.
\item\relax
\flmRefsHyperref[eczindexfamilyrel]{code:stab_6_4_2}{\(\llbracket 6,4,2\rrbracket \) error-detecting code} --- The \(\llbracket 6,4,2\rrbracket \) error-detecting code is a color code defined on a single hexagon of the 6.6.6 or 4.6.12 tilings. The \(\llbracket 6,4,2\rrbracket \) code can be concatenated with the surface code to yield the 6.6.6 color code \NoCaseChange{\protect\cite[{Appx. A}]{cite3289}}.
\item\relax
\flmRefsHyperref[eczindexfamilyrel]{code:steane}{\(\llbracket 7,1,3\rrbracket \) Steane code} --- Steane code is a 2D color code defined on a seven-qubit patch of the 6.6.6 tiling.
\end{eczvaluelist}
\codefieldsection{Cousins}
\begin{eczvaluelist}
\item\relax
\flmRefsHyperref[eczindexfamilyrel]{code:honeycomb}{Honeycomb tiling} --- The 6.6.6 color code is defined on the honeycomb tiling.
\item\relax
\flmRefsHyperref[eczindexfamilyrel]{code:gkp_concatenated}{Concatenated GKP code} --- GKP codes have been concatenated with the 6.6.6 color code \NoCaseChange{\protect\cite{cite3323}}.
\item\relax
\flmRefsHyperref[eczindexfamilyrel]{code:quantum_lego}{Tensor-network code} --- Larger 6.6.6 color codes can be constructed by contracting legs of tensors of smaller codes \NoCaseChange{\protect\cite[{Fig. 5}]{cite3101}}.
\item\relax
\flmRefsHyperref[eczindexfamilyrel]{code:da_color_2d}{2D DA color code} --- The parent topological phase of the 2D DA color code is realized by two copies of the 6.6.6 color code \NoCaseChange{\protect\cite[{Sec. III.A}]{cite2532}}.
\item\relax
\flmRefsHyperref[eczindexfamilyrel]{code:floquet_xyz_ruby}{Ruby Floquet code} --- One third of the time during the XYZ ruby measurement schedule, the ISG is that of the 6.6.6 color code concatenated with a three-qubit repetition code.
\item\relax
\flmRefsHyperref[eczindexfamilyrel]{code:qcga}{Bivariate bicycle (BB) code} --- Certain bivariate bicycle codes are equivalent to a family of 6.6.6 color codes \NoCaseChange{\protect\cite{cite3501}}.
\item\relax
\flmRefsHyperref[eczindexfamilyrel]{code:488_color}{Square-octagon (4.8.8) color code} --- Lattice surgery scheme for a hybrid 6.6.6-4.8.8 layout yields lower resource overhead when compared to analogous surface code scheme \NoCaseChange{\protect\cite{cite3732}}.
\item\relax
\flmRefsHyperref[eczindexfamilyrel]{code:xyz_color}{XYZ color code} --- The XYZ color code is obtained from the 6.6.6 color code by applying single-qubit Clifford rotations on a subset of qubits such that the \(X\)- and \(Z\)-type generators are mapped to \(XZXZXZ\) and \(ZYZYZY\), respectively.
\end{eczvaluelist}
\eczhbkcontributors{ Eric Huang, Cella Kove, \eczhuVVA }
\endeczcode

\eczcode{honeycomb_floquet}{Honeycomb Floquet code}{~\NoCaseChange{\protect\cite{cite536}}}
\codefieldsection{Description}
2D Floquet code based on the Kitaev honeycomb model \NoCaseChange{\protect\cite{cite537}} whose logical qubits are generated through a particular sequence of measurements.
A CSS version of the code has been proposed which loosens the restriction of which sequences to use \NoCaseChange{\protect\cite{cite538}}.
The code has also been generalized to arbitrary non-chiral, Abelian topological order \NoCaseChange{\protect\cite{cite539}}.

The code is defined on a honeycomb tiling with a physical qubit located at each vertex. Edges are labeled \(x\), \(y\), and \(z\), such that one edge of each label meets at every vertex. Check operators are defined as \(XX\) acting on any two qubits joined by an \(x\) edge, and similarly for \(y\) and \(z\). The honeycomb tiling is 3-colorable, so the hexagons may be labeled 0, 1, 2 such that no two neighboring hexagons have the same label.

The code-generating measurement pattern consists of measuring the check operators located on all of the \(r\)-labeled edges in round \(r\) mod 3. The code space is the \(+1\) eigenspace of the instantaneous stabilizer group (ISG). The ISG specifies the state of the system as a Pauli stabilizer state at a particular round of measurement, and it evolves into a (potentially) different ISG depending on the check operators measured.
When the same check operators are instead regarded as a static subsystem code, there are no logical qubits; the logical qubits are created dynamically by the measurement ordering \NoCaseChange{\protect\cite{cite536}}.

\codefieldsection{Protection}
Protective features similar to the surface code: on a torus geometry, the code protects two logical qubits with a code distance proportional to the linear size of the torus. After each complete measurement round in the steady state, a depth-one local Clifford circuit maps the ISG to that of a toric code on a hexagonal superlattice, making the logical content explicit \NoCaseChange{\protect\cite{cite536}}. Properties of the code with open boundaries are discussed in Refs. \NoCaseChange{\protect\cite{cite3749,cite3750}}, and various other generalizations have been proposed \NoCaseChange{\protect\cite{cite3751}}.
\codefieldsection{Encoding}
\begin{eczvaluelist}
\item\relax Initialization can be performed by preparing each pair of qubits on an edge in some particular state independently specified by the effective-one-qubit operators (two-qubit Pauli strings centered on an edge) and then beginning the check measurement sequence. This is analogous to projecting a state into the code space by measuring stabilizers.
\end{eczvaluelist}
\codefieldsection{Gates}
\begin{eczvaluelist}
\item\relax There are two types of logical operators, \textit{inner} and \textit{outer}. An inner logical operator is the product of check operators on a homologically nontrivial cycle. They belong to the stabilizer group as a subsystem code. Outer logical operators have an interpretation in terms of magnetic and electric operators of an embedded toric/surface code, and they do not belong to the stabilizer group of the associated subsystem code.
\item\relax Fermionic string excitations can be condensed along 1D paths, yielding twist defects \NoCaseChange{\protect\cite{cite3752}}. Such excitations are created by taking products of plaquette check operators (that are present in all ISGs) over some region, multiplying them to yield a Pauli string on the boundary of said region, and cutting this string. Information is processed by braiding and fusing defects, which are located at the boundaries of the strings.
\item\relax Certain gates \NoCaseChange{\protect\cite{cite3751}} can be performed by considering adiabatic paths in the space of Hamiltonians \NoCaseChange{\protect\cite{cite3753,cite3754,cite3751}}, yielding an instance of holonomic quantum computation \NoCaseChange{\protect\cite{cite3755}}.
Fault-tolerant gates should be interpretable as monodromies under a particular notion of parallel transport \NoCaseChange{\protect\cite{cite809}}.

\end{eczvaluelist}
\codefieldsection{Decoding}
\begin{eczvaluelist}
\item\relax The ISG has a static subgroup for all time steps \(r\geq 3\) – that is, a subgroup which remains a subgroup of the ISG for all future times – given by so-called \textit{plaquette stabilizers}. These are stabilizers consisting of products of check operators around homologically trivial paths. The syndrome bits correspond to the eigenvalues of the plaquette stabilizers. Because of the structure of the check operators, only one-third of all plaquettes are measured each round. The syndrome bits must therefore be represented by a lattice in spacetime, to reflect when and where the outcome was obtained.
\end{eczvaluelist}
\codefieldsection{Fault Tolerance}
\begin{eczvaluelist}
\item\relax One can run a fault-tolerant decoding algorithm by (1) bipartitioning the syndrome lattice into two graphs which are congruent to the Cayley graph of the free Abelian group with three generators (up to boundary conditions) and (2) performing a matching algorithm to deduce errors.
\end{eczvaluelist}
\codefieldsection{Threshold}
\begin{eczvaluelist}
\item\relax \(0.2\%-0.3\%\) in a controlled-not circuit model with a correlated minimum-weight perfect-matching decoder \NoCaseChange{\protect\cite{cite3756}}.
\item\relax \(1.5\%<p<2.0\%\) in a circuit model with native weight-two measurements and a correlated minimum-weight perfect-matching decoder \NoCaseChange{\protect\cite{cite3756}}. Here, \(p\) is the collective error rate of the two-body measurement gate, including both measurement and correlated data depolarization error processes.
\item\relax Against circuit-level noise: within \(0.2\% − 0.3\%\) for SD6 (standard depolarizing 6-step cycle), \(0.1\% − 0.15\%\) for SI1000 (superconducting-inspired 1000 ns cycle), and \(1.5\% − 2.0\%\) for EM3 (entangling-measurement 3-step cycle) \NoCaseChange{\protect\cite{cite3757,cite3631}}.
\end{eczvaluelist}
\codefieldsection{Realizations}
\begin{eczvaluelist}
\item\relax Plaquette stabilizer measurement realized on the IBM Falcon superconducting-qubit device \NoCaseChange{\protect\cite{cite3682}}
\end{eczvaluelist}
\codefieldsection{Parents}
\begin{eczvaluelist}
\item\relax
\flmRefsHyperref[eczindexfamilyrel]{code:floquet}{Hastings-Haah Floquet code} --- The honeycomb Floquet code is the first 2D Floquet code.
\item\relax
\flmRefsHyperref[eczindexfamilyrel]{code:qudit_honeycomb}{Modular-qudit honeycomb Floquet code} --- The modular-qudit honeycomb Floquet code reduces to the Hastings-Haah Floquet code for \(q=2\).
\end{eczvaluelist}
\codefieldsection{Cousins}
\begin{eczvaluelist}
\item\relax
\flmRefsHyperref[eczindexfamilyrel]{code:qudit_znone}{\(\mathbb{Z}_q^{(1)}\) subsystem code} --- The dynamically generated logical qubit of the honeycomb Floquet code is generated by appropriately scheduling measurements of the gauge generators of the \(\mathbb{Z}_{q=2}^{(1)}\) subsystem stabilizer code corresponding to the Kitaev honeycomb model. However, since this subsystem code has zero logical qubits, the instantaneous stabilizer codes of the honeycomb code cannot be interpreted as gauge-fixed versions of this subsystem code.
\item\relax
\flmRefsHyperref[eczindexfamilyrel]{code:surface}{Kitaev surface code} --- Measurement of each check operator of the honeycomb Floquet code involves two qubits and projects the state of the two qubits to a two-dimensional subspace, which we regard as an effective qubit. 
These effective qubits form a surface code on an enlarged honeycomb tiling \NoCaseChange{\protect\cite[{Fig. 2}]{cite536}}.
Electric and magnetic operators on the embedded surface code correspond to outer logical operators of the Floquet code.
In fact, outer logical operators transition back and forth from magnetic to electric surface code operators under the measurement dynamics.
Inspired by the honeycomb Floquet code, various weight-two measurement schemes have been designed \NoCaseChange{\protect\cite{cite3758,cite3759,cite3760}}, with the scheme in Ref. \NoCaseChange{\protect\cite{cite3759}} being a special case of DWR.
Numerical comparisons have been performed \NoCaseChange{\protect\cite{cite3761}}.

\item\relax
\flmRefsHyperref[eczindexfamilyrel]{code:twist_defect_surface}{Twist-defect surface code} --- Fermionic string excitations of the honeycomb Floquet code can be condensed along 1D paths, yielding twist defects \NoCaseChange{\protect\cite{cite3752}}.
\item\relax
\flmRefsHyperref[eczindexfamilyrel]{code:subsystem_color}{Subsystem color code} --- Both honeycomb and subsystem color codes are generated via periodic sequences of measurements. However, any measurement sequence can be performed on the color code without destroying the logical qubits, while honeycomb codes can be maintained only with specific sequences. Honeycomb codes require a shorter measurement cycle and use fewer qubits at the given code distance \NoCaseChange{\protect\cite{cite536}}.
\item\relax
\flmRefsHyperref[eczindexfamilyrel]{code:majorana_stab}{Majorana stabilizer code} --- The Honeycomb code admits a convenient representation in terms of Majorana fermions. This leads to a possible physical realization of the code in terms of tetrons \NoCaseChange{\protect\cite{cite401}}, where each physical qubit is composed of four Majorana modes.
\item\relax
\flmRefsHyperref[eczindexfamilyrel]{code:qldpc}{Qubit QLDPC code} --- The Floquet check operators are weight-two, and each qubit participates in one check each round.
\item\relax
\flmRefsHyperref[eczindexfamilyrel]{code:kitaev_honeycomb}{Kitaev honeycomb code} --- The Kitaev honeycomb model Hamiltonian is a sum of checks of the honeycomb Floquet code \NoCaseChange{\protect\cite{cite536}}.
\item\relax
\flmRefsHyperref[eczindexfamilyrel]{code:honeycomb}{Honeycomb tiling} --- The honeycomb Floquet code is defined on the honeycomb tiling.
\item\relax
\flmRefsHyperref[eczindexfamilyrel]{code:real_projective_plane}{Projective-plane surface code} --- Implementing the honeycomb Floquet code on a non-orientable cross-cap geometry allows for a logical-\(HZ\) gate to be implemented via a measurement schedule \NoCaseChange{\protect\cite{cite3762}}.
\end{eczvaluelist}
\eczhbkcontributors{ Chris Fechisin, \eczhuVVA }
\endeczcode

\eczcode{hh_fracton}{Hsieh-Halasz (HH) code}{~\NoCaseChange{\protect\cite{cite3763}}}
\codefieldsection{Description}
Member of one of two families of fracton codes, named HH-I and HH-II, defined on a cubic lattice with two qubits per site.
HH-I (HH-II) is a CSS (non-CSS) stabilizer code family, with the former identified as a foliated type-I fracton code that is decomposable into two separate lattice models \NoCaseChange{\protect\cite{cite456}}.
The sorting analysis of Ref. \NoCaseChange{\protect\cite{cite456}} leaves HH-II inconclusive, consistent with either a fractal type-I or a type-II fracton phase.

\codefieldsection{Parents}
\begin{eczvaluelist}
\item\relax
\flmRefsHyperref[eczindexfamilyrel]{code:qldpc}{Qubit QLDPC code}\item\relax
\flmRefsHyperref[eczindexfamilyrel]{code:fracton}{Fracton stabilizer code} --- Both HH-I and HH-II are fracton codes; HH-I is identified as foliated type-I, while HH-II remains inconclusive between fractal type-I and type-II in the sorting analysis of Ref. \NoCaseChange{\protect\cite{cite456}}.
\end{eczvaluelist}
\eczhbkcontributors{ \eczhuVVA }
\endeczcode

\eczcode{hhb_fracton}{Hsieh-Halasz-Balents (HHB) code}{~\NoCaseChange{\protect\cite{cite3764}}}
\codefieldsection{Description}
Member of one of two families of fracton codes, named HHB model A and B, defined on a cubic lattice with two qubits per site.
Both are expected to be foliated type-I fracton codes \NoCaseChange{\protect\cite[{Eqs. (D42-D43)}]{cite456}}.

\codefieldsection{Parents}
\begin{eczvaluelist}
\item\relax
\flmRefsHyperref[eczindexfamilyrel]{code:qldpc}{Qubit QLDPC code}\item\relax
\flmRefsHyperref[eczindexfamilyrel]{code:fracton}{Fracton stabilizer code} --- Both HHB models are expected to be foliated type-I fracton codes \NoCaseChange{\protect\cite[{Eqs. (D42-D43)}]{cite456}}.
\end{eczvaluelist}
\eczhbkcontributors{ \eczhuVVA }
\endeczcode

\eczcode{hurwitz_surface}{Hurwitz surface code}{~\NoCaseChange{\protect\cite{cite3765}}}
\codefieldsection{Description}
Homological code constructed on triangulations of Hurwitz surfaces.

\codefieldsection{Rate}
Constant rate and vanishing distance \NoCaseChange{\protect\cite{cite3765}}.
\codefieldsection{Parent}
\begin{eczvaluelist}
\item\relax
\flmRefsHyperref[eczindexfamilyrel]{code:hyperbolic_surface}{Hyperbolic surface code}\end{eczvaluelist}
\eczhbkcontributors{ \eczhuVVA }
\endeczcode

\eczcode{hybrid_convolutional}{Hybrid convolutional code}{~\NoCaseChange{\protect\cite{cite3766}}}
\codefieldsection{Description}
A quantum convolutional code which protects both quantum and classical information.
\codefieldsection{Parent}
\begin{eczvaluelist}
\item\relax
\flmRefsHyperref[eczindexfamilyrel]{code:hybrid_stabilizer}{Hybrid stabilizer code}\end{eczvaluelist}
\codefieldsection{Cousins}
\begin{eczvaluelist}
\item\relax
\flmRefsHyperref[eczindexfamilyrel]{code:convolutional}{Convolutional code} --- Hybrid convolutional codes are hybrid c-q analogues of convolutional codes.
\item\relax
\flmRefsHyperref[eczindexfamilyrel]{code:quantum_convolutional}{Quantum convolutional code} --- Hybrid convolutional codes are hybrid analogues of quantum convolutional codes.
\end{eczvaluelist}
\eczhbkcontributors{ \eczhuVVA }
\endeczcode

\eczcode{hybrid_qubits_into_qubits}{Hybrid qubit code}{~\NoCaseChange{\protect\cite{cite2735,cite2872}}}
\codefieldsection{Description}
A qubit code which stores both quantum and classical information.
Usually denoted as \(\llparenthesis n,K:M\rrparenthesis \) or \(\llparenthesis n,K:M,d\rrparenthesis \), where \(K\) is the dimension of the underlying quantum code, \(M\) is the size of the classical code, and \(d\) is the distance.

\codefieldsection{Protection}
Any qubit code can be converted into a hybrid qubit code by using some of its logical qubits to store only classical information \NoCaseChange{\protect\cite{cite2735}}.
An \(\llparenthesis n,K:M\rrparenthesis \) hybrid qubit code can detect more errors than an \(\llparenthesis n,KM\rrparenthesis \) qubit code \NoCaseChange{\protect\cite{cite3767}}.
A hybrid Hamming bound has been constructed \NoCaseChange{\protect\cite{cite3294}}.

\flmRefsHyperref{ref672}{Quantum weight enumerators}, quantum MacWilliams identities, and linear programming bounds have been extended to hybrid qubit codes \NoCaseChange{\protect\cite{cite2872,cite3767,cite671}}.

\codefieldsection{Parents}
\begin{eczvaluelist}
\item\relax
\flmRefsHyperref[eczindexfamilyrel]{code:oa_qubits_into_qubits}{OA qubit code} --- An OA qubit code that has no gauge structure (e.g., gauge qubits) but has a block structure that corresponds to a classical code is a hybrid qubit code.
\item\relax
\flmRefsHyperref[eczindexfamilyrel]{code:hybridqecc}{Hybrid QECC}\end{eczvaluelist}
\codefieldsection{Child}
\begin{eczvaluelist}
\item\relax
\flmRefsHyperref[eczindexfamilyrel]{code:hybrid_stabilizer}{Hybrid stabilizer code} --- An \(\llbracket n,k:c,d\rrbracket \) hybrid stabilizer code is an \(\llparenthesis n,2^k:2^c,d\rrparenthesis \) hybrid qubit code.
\end{eczvaluelist}
\codefieldsection{Cousins}
\begin{eczvaluelist}
\item\relax
\flmRefsHyperref[eczindexfamilyrel]{code:qubits_into_qubits}{Qubit code} --- A hybrid qubit code storing no classical information reduces to a qubit code. Conversely, any qubit code can be converted into a hybrid qubit code by using some of its logical qubits to store only classical information \NoCaseChange{\protect\cite{cite2735}}.
\item\relax
\flmRefsHyperref[eczindexfamilyrel]{code:qubit_classical_into_quantum}{Qubit c-q code} --- A hybrid qubit code storing no quantum information reduces to a qubit c-q code.
\end{eczvaluelist}
\eczhbkcontributors{ \eczhuVVA }
\endeczcode

\eczcode{hybrid_stabilizer}{Hybrid stabilizer code}{~\NoCaseChange{\protect\cite{cite2735,cite2872}}}
\codefieldsection{Description}
A qubit stabilizer code which stores both quantum and classical information.
Usually denoted as \(\llbracket n,k:c\rrbracket \) or \(\llbracket n,k:c,d\rrbracket \), where \(k\) (\(c\)) is the number of encoded qubits (classical bits), and where \(d\) is the distance.

The algebraic structure of a hybrid stabilizer code is the same as that of a USt code whose cosets are indexed by a linear binary code:
both codes utilize codewords of an inner \(\llbracket n,k\rrbracket \) qubit stabilizer code \(\mathsf{C}\) and its cosets \(t \mathsf{C}\), where the \(2^c\) Pauli strings \(t\) correspond to the outer \([n,c]\) linear binary code.
However, the hybrid stabilizer code does not utilize superpositions of codewords of \(t \mathsf{C}\) and \(t^{\prime} \mathsf{C}\) for \(t \neq t^{\prime}\) since the different coset blocks correspond to classical codewords.

\codefieldsection{Parents}
\begin{eczvaluelist}
\item\relax
\flmRefsHyperref[eczindexfamilyrel]{code:hybrid_qubits_into_qubits}{Hybrid qubit code} --- An \(\llbracket n,k:c,d\rrbracket \) hybrid stabilizer code is an \(\llparenthesis n,2^k:2^c,d\rrparenthesis \) hybrid qubit code.
\item\relax
\flmRefsHyperref[eczindexfamilyrel]{code:qubit_stabilizer_oaqecc}{Operator-algebra (OA) qubit stabilizer code} --- An OA stabilizer code which has no gauge qubits but has a block structure that corresponds to a linear binary code is a hybrid stabilizer code.
\end{eczvaluelist}
\codefieldsection{Children}
\begin{eczvaluelist}
\item\relax
\flmRefsHyperref[eczindexfamilyrel]{code:hybrid_7_1-1_3}{\(\llbracket 7, 1:1, 3\rrbracket \) hybrid stabilizer code}\item\relax
\flmRefsHyperref[eczindexfamilyrel]{code:hybrid_8_2-1_3}{\(\llbracket 8, 2:1, 3\rrbracket \) hybrid stabilizer code}\item\relax
\flmRefsHyperref[eczindexfamilyrel]{code:hybrid_convolutional}{Hybrid convolutional code}\end{eczvaluelist}
\codefieldsection{Cousins}
\begin{eczvaluelist}
\item\relax
\flmRefsHyperref[eczindexfamilyrel]{code:qubit_stabilizer}{Qubit stabilizer code} --- A hybrid stabilizer code storing no classical information reduces to a qubit stabilizer code.
Conversely, any qubit stabilizer code can be converted into a hybrid stabilizer code by using some of its qubits to store only classical information \NoCaseChange{\protect\cite{cite2735}}.

\item\relax
\flmRefsHyperref[eczindexfamilyrel]{code:non_stabilizer}{Union stabilizer (USt) code} --- The algebraic structure of a hybrid stabilizer code is the same as that of a USt code whose cosets are indexed by a linear binary code \NoCaseChange{\protect\cite{cite2735}}.
\item\relax
\flmRefsHyperref[eczindexfamilyrel]{code:shor_nine}{\(\llbracket 9,1,3\rrbracket \) Shor code} --- The Shor code can be modified into a degenerate \(\llbracket 9,1:3,3\rrbracket \) hybrid stabilizer code that still corrects arbitrary single-qubit errors \NoCaseChange{\protect\cite{cite2735}}.
\item\relax
\flmRefsHyperref[eczindexfamilyrel]{code:iceberg}{\(\llbracket 2m,2m-2,2\rrbracket \) error-detecting code} --- The \(\llbracket 2m+1,2m+2:1,2\rrbracket \) hybrid stabilizer code \NoCaseChange{\protect\cite{cite671}} (extendable to modular qudits \NoCaseChange{\protect\cite{cite3244}}) is closely related to the \(\llbracket 2m,2m-2,2\rrbracket \) error-detecting code.
\item\relax
\flmRefsHyperref[eczindexfamilyrel]{code:stab_4_2_2}{\(\llbracket 4,2,2\rrbracket \) Four-qubit code} --- The \(\llbracket 4,2,2\rrbracket \) codewords can be modified by signs to yield a \(\llbracket 4,1:1,2\rrbracket \) hybrid stabilizer code \NoCaseChange{\protect\cite{cite3294}}.
\item\relax
\flmRefsHyperref[eczindexfamilyrel]{code:qubit_subsystem_stabilizer}{Subsystem qubit stabilizer code} --- Hybrid stabilizer codes can be constructed from subsystem qubit stabilizer codes by using the gauge qubits of the latter to store classical information \NoCaseChange{\protect\cite[{Thm. 4}]{cite2874}}.
\item\relax
\flmRefsHyperref[eczindexfamilyrel]{code:subsystem_quantum_parity}{Subsystem hypergraph product (SHP) code} --- Hybrid stabilizer codes can be constructed from SHP codes by using the gauge qubits of the latter to store classical information \NoCaseChange{\protect\cite[{Sec. 4}]{cite2874}}.
\item\relax
\flmRefsHyperref[eczindexfamilyrel]{code:eaoa_stabilizer}{EAOA qubit stabilizer code} --- EA hybrid qubit stabilizer codes utilize additional ancillary subsystems in a pre-shared entangled state, but reduce to hybrid qubit stabilizer codes when said subsystems are interpreted as noiseless physical subsystems. In the original EA hybrid stabilizer formalism, an \(\llbracket n,q:c,d;e\rrbracket \) EA hybrid stabilizer code is specified by a pair \((\mathcal{S}_Q,\mathcal{S}_C)\) of quantum and classical stabilizer groups, and in the equivalent symplectic formalism by a quantum parity-check matrix together with a classical parity-check matrix \NoCaseChange{\protect\cite[{Thms. 1-4}]{cite2735}}. Inside the EAOAQEC stabilizer framework, hybrid stabilizer codes are a proper subclass of the broader EA hybrid subspace codes because the EACQ transversal operators obey additional constraints not required in general \NoCaseChange{\protect\cite{cite856}}.
\end{eczvaluelist}
\eczhbkcontributors{ \eczhuVVA }
\endeczcode

\eczcode{hyperbolic_color}{Hyperbolic color code}{~\NoCaseChange{\protect\cite{cite837,cite836,cite702}}}
\codefieldsection{Description}
An extension of the color code construction to hyperbolic manifolds.
As opposed to there being only three types of uniform three-valent and three-colorable lattice tilings in the 2D Euclidean plane, there is an infinite number of admissible hyperbolic tilings in the 2D hyperbolic plane \NoCaseChange{\protect\cite{cite836}}.
Certain double covers of hyperbolic tilings also yield admissible tilings \NoCaseChange{\protect\cite{cite837}}.
Other admissible hyperbolic tilings can be obtained via a fattening procedure \NoCaseChange{\protect\cite{cite430}}; see also a construction based on the more general quantum pin codes \NoCaseChange{\protect\cite{cite702}}.

\codefieldsection{Protection}
The use of hyperbolic surfaces allows one to circumvent bounds on code parameters (such as the \flmTerm{term}{ref3768}{}{BPT bound}) that are valid for lattice geometries.
Hyperbolic color codes can have high rate but tend to have small distance.
For example, a \(\{4g,4g\}\) tiling with periodic boundary conditions (i.e., a \(g\)-torus) yields a \(\llbracket 4g+8,4g,4\rrbracket \) code family \NoCaseChange{\protect\cite{cite836}}.
More examples, such as the \(\llbracket 160,20,8\rrbracket \) code on the 4.10.10 tiling, are provided in \NoCaseChange{\protect\cite[{Sec. V.A}]{cite702}}.

\codefieldsection{Rate}
In the double-cover construction \NoCaseChange{\protect\cite{cite837}}, an \(\{\ell,m\}\) input tiling yields a code family with an asymptotic rate of \(1 - 2/\ell - 2/m\).
\codefieldsection{Decoding}
\begin{eczvaluelist}
\item\relax Two flag-based decoders \NoCaseChange{\protect\cite{cite3439}}.
\end{eczvaluelist}
\codefieldsection{Notes}
\begin{eczvaluelist}
\item\relax A database of 2D hyperbolic color codes is available in QECDB \NoCaseChange{\protect\cite{cite781}}, where the color codes form the subset of self-dual codes.
\end{eczvaluelist}
\codefieldsection{Parent}
\begin{eczvaluelist}
\item\relax
\flmRefsHyperref[eczindexfamilyrel]{code:color}{Color code}\end{eczvaluelist}
\codefieldsection{Child}
\begin{eczvaluelist}
\item\relax
\flmRefsHyperref[eczindexfamilyrel]{code:stab_8_2_2}{\(\llbracket 8,2,2\rrbracket \) hyperbolic color code} --- The \(\llbracket 8,2,2\rrbracket \) hyperbolic color code is defined on the projective plane.
\end{eczvaluelist}
\codefieldsection{Cousins}
\begin{eczvaluelist}
\item\relax
\flmRefsHyperref[eczindexfamilyrel]{code:hyperbolic_surface}{Hyperbolic surface code} --- Hyperbolic color codes and hyperbolic surface codes are both defined on hyperbolic tilings.
\item\relax
\flmRefsHyperref[eczindexfamilyrel]{code:small_distance_qubit_stabilizer}{Small-distance qubit stabilizer code} --- Many hyperbolic color codes have distance \(\leq 5\).
\end{eczvaluelist}
\eczhbkcontributors{ Guanyu Zhu, \eczhuVVA }
\endeczcode

\eczcode{hyperbolic_floquet}{Hyperbolic Floquet code}{~\NoCaseChange{\protect\cite{cite3750,cite3769,cite3770}}}
\codefieldsection{Description}
Floquet code whose check-operators correspond to edges of a hyperbolic lattice of degree 3.
\codefieldsection{Protection}
Code distance is at most \(O(\log n)\) due to the hyperbolic qubit geometry \NoCaseChange{\protect\cite{cite3750}}, but semi-hyperbolic lattices yield \(O(\sqrt{n})\) distance \NoCaseChange{\protect\cite{cite3769}}.

A useful concept is the \textit{embedded distance} \NoCaseChange{\protect\cite{cite3769}}, which is the distance of the stabilizer code lying inside the subspace defined by measurement outcomes of the weight-two parity checks of the code.

\codefieldsection{Rate}
Finite encoding rate whose value depends on the hyperbolic lattice. The asymptotic rate is 1/8 for a lattice of octagons \NoCaseChange{\protect\cite{cite3770}}.
\codefieldsection{Decoding}
\begin{eczvaluelist}
\item\relax Syndrome structure allows for MWPM decoding.
\end{eczvaluelist}
\codefieldsection{Threshold}
\begin{eczvaluelist}
\item\relax \(0.1\%\) under standard circuit-level depolarizing noise \NoCaseChange{\protect\cite{cite3769}}.
\item\relax \(0.1\%\) under phenomenological error model including depolarizing and measurement errors for the octagonal codes \NoCaseChange{\protect\cite{cite3770}}.
\end{eczvaluelist}
\codefieldsection{Notes}
\begin{eczvaluelist}
\item\relax The code may be suitable for distributed storage \NoCaseChange{\protect\cite{cite3771}}.
\end{eczvaluelist}
\codefieldsection{Parent}
\begin{eczvaluelist}
\item\relax
\flmRefsHyperref[eczindexfamilyrel]{code:floquet}{Hastings-Haah Floquet code}\end{eczvaluelist}
\eczhbkcontributors{ Ali Fahimniya, \eczhuVVA }
\endeczcode

\eczcode{hyperbolic_surface}{Hyperbolic surface code}{}
\codefieldsection{Description}
An extension of the Kitaev surface code construction to hyperbolic manifolds.
Given a cellulation of a hyperbolic manifold of arbitrary dimension, qubits are put on \(i\)-dimensional faces, \(X\)-type stabilizers are associated with \((i-1)\)-faces, while \(Z\)-type stabilizers are associated with \(i+1\)-faces.

\codefieldsection{Protection}
Constructions (see code children below) have yielded distances scaling favorably with the number of qubits. The use of hyperbolic surfaces allows one to circumvent bounds on code parameters (such as the \flmTerm{term}{ref3768}{}{BPT bound}) that are valid for lattice geometries.
\codefieldsection{Gates}
\begin{eczvaluelist}
\item\relax \((1,D-1)\) surface codes on hyperbolic geometries admit a fault-tolerant implementation of \(C^D Z\) gates \NoCaseChange{\protect\cite{cite3461}}.
\item\relax Higher-dimensional hyperbolic surface codes can admit a cup product structure and can thus have logical gates in the \flmTerm{term}{ref694}{}{Clifford hierarchy} implemented by constant-depth \flmRefsHyperref{ref409}{Clifford circuits} \NoCaseChange{\protect\cite{cite1517}}.
\end{eczvaluelist}
\codefieldsection{Decoding}
\begin{eczvaluelist}
\item\relax Hastings decoder \NoCaseChange{\protect\cite{cite845}}.
\end{eczvaluelist}
\codefieldsection{Parent}
\begin{eczvaluelist}
\item\relax
\flmRefsHyperref[eczindexfamilyrel]{code:higher_dimensional_surface}{Homological code}\end{eczvaluelist}
\codefieldsection{Children}
\begin{eczvaluelist}
\item\relax
\flmRefsHyperref[eczindexfamilyrel]{code:two_dimensional_hyperbolic_surface}{2D hyperbolic surface code}\item\relax
\flmRefsHyperref[eczindexfamilyrel]{code:four_dimensional_hyperbolic}{Guth-Lubotzky code}\item\relax
\flmRefsHyperref[eczindexfamilyrel]{code:freedman_meyer_luo}{Freedman-Meyer-Luo code}\item\relax
\flmRefsHyperref[eczindexfamilyrel]{code:hurwitz_surface}{Hurwitz surface code}\end{eczvaluelist}
\codefieldsection{Cousins}
\begin{eczvaluelist}
\item\relax
\flmRefsHyperref[eczindexfamilyrel]{code:holographic_tensor}{Holographic tensor-network code} --- Both holographic tensor-network and hyperbolic surface codes utilize tessellations of hyperbolic surfaces. Encodings for the former are hyperbolically tiled tensor networks, while the latter is defined on hyperbolically tiled physical-qubit lattices.
\item\relax
\flmRefsHyperref[eczindexfamilyrel]{code:single_shot}{Single-shot code} --- A 4D hyperbolic surface code can be decoded with the Hastings decoder \NoCaseChange{\protect\cite{cite845}} in time \(O(n\log n)\) and with a logical error scaling inverse polynomially with \(n\).
\item\relax
\flmRefsHyperref[eczindexfamilyrel]{code:hyperbolic_color}{Hyperbolic color code} --- Hyperbolic color codes and hyperbolic surface codes are both defined on hyperbolic tilings.
\end{eczvaluelist}
\eczhbkcontributors{ \eczhuVVA }
\endeczcode

\eczcode{hypergraph_product}{Hypergraph product (HGP) code}{~\NoCaseChange{\protect\cite{cite1448,cite1316}}}
\codefieldsection{Alternative Names}
\begin{eczvaluelist}
\item\relax Quantum hypergraph (QHG) code
\item\relax Tillich-Zemor product code
\item\relax HP code
\end{eczvaluelist}
\eczhIndexCodeAliasName{hypergraph_product}{Quantum hypergraph (QHG) code}
\eczhIndexCodeAliasName{hypergraph_product}{Tillich-Zemor product code}
\eczhIndexCodeAliasName{hypergraph_product}{HP code}
\codefieldsection{Description}
A member of a family of CSS codes whose stabilizer generator matrix is obtained from a hypergraph product of two classical linear binary codes.

More technically, the \(X\)- and \(Z\)-type stabilizer generator matrices of a hypergraph product code are, respectively, the boundary and coboundary operators of the 2-complex obtained from the tensor product of a chain complex and cochain complex corresponding to two classical linear binary \textit{seed} codes.
Let the two seed codes be \(C_i\) for \(i\in\{1,2\}\) with parameters \([n_i, k_i, d_i]\), defined as the kernel of \(r_i \times n_i\) check matrices \(H_i\) of rank \(n_i - k_i\).
The hypergraph product yields two classical codes \(C_{X,Z}\) with parity-check matrices
\flmMathEnvironment{align}{}{
  H_{X}&=\begin{pmatrix}H_{1}\otimes I_{n_{2}} & \,\,I_{r_{1}}\otimes H_{2}^{T}\end{pmatrix}\\
  H_{Z}&=\begin{pmatrix}I_{n_{1}}\otimes H_{2} & \,\,H_{1}^{T}\otimes I_{r_{2}}\end{pmatrix}~,
}
where \(I_m\) is the \(m\)-dimensional identity matrix.
These two codes then yield a hypergraph product code via the CSS construction.
The case when the two seed codes are equal, \(C_1=C_2\), is called a \textit{square hypergraph product code}.
If, in addition, \(\text{im} H = \text{im} H^T\), the hypergraph product code is called a \textit{symmetric hypergraph product code} \NoCaseChange{\protect\cite{cite742}}.

In terms of the \flmRefsCref{ref683}, the hypergraph product can be viewed as a homological product of two length-one chain complexes. The resulting code corresponds to the length-two chain complex that is called the \textit{total chain complex} of the product of the input complexes (see \NoCaseChange{\protect\cite[{Sec. II.C}]{cite3772}}).

\codefieldsection{Protection}
If \([n_i, k_i, d_i]\) and \([r_i, k_i^T, d_i^T]\) are the parameters of the codes \(\mathrm{ker}H_i\) and \(\mathrm{ker}H_i^T\), respectively, taking \(d_i^T=\infty\) when \(k_i^T=0\), then the hypergraph product has parameters \(\llbracket n_1 n_2 + r_1 r_2, k_1 k_2 + k_1^T k_2^T, \min(d_1, d_2, d_1^T, d_2^T)\rrbracket \).

An algebraic reformulation of HGP codes, together with rate-improved square, symmetric, and two-tile variants, was given in Ref. \NoCaseChange{\protect\cite{cite1316}}.
Using square seed parity-check matrices yields \(\llbracket 2n_1 n_2,2k_1 k_2,\min(d_1,d_2)\rrbracket \), symmetric seeds yield \(\llbracket n_1 n_2,k_1 k_2,\min(d_1,d_2)\rrbracket \), and two-tile cyclic constructions yield \(\llbracket n_1^2,2k_1^2,d_1\rrbracket \); these variants improve the rate of the original Tillich-Zemor family by factors up to four at small block length \NoCaseChange{\protect\cite{cite1316}}.

\codefieldsection{Encoding}
\begin{eczvaluelist}
\item\relax Fault-tolerant state preparation via dimension jump \NoCaseChange{\protect\cite{cite3773}}.
\end{eczvaluelist}
\codefieldsection{Transversal and Permutation-Based Gates}
\begin{eczvaluelist}
\item\relax Hadamard (up to logical SWAP gates) and control-\(Z\) on all logical qubits \NoCaseChange{\protect\cite{cite742}}.
\item\relax Patch-transversal gates inherited from the automorphism group of the underlying classical codes \NoCaseChange{\protect\cite[{Appx. D}]{cite743}}.
\item\relax Orientation-preserving constant-depth circuits can only implement gates in the Clifford group \NoCaseChange{\protect\cite{cite739}}.
\item\relax Permutation-based gates (automorphism gadgets) can be inherited from the underlying classical codes in the hypergraph construction \NoCaseChange{\protect\cite{cite744}}.
\end{eczvaluelist}
\codefieldsection{Gates}
\begin{eczvaluelist}
\item\relax Code deformation techniques yield \flmRefsHyperref{ref409}{Clifford gates} \NoCaseChange{\protect\cite{cite3774}}.
\item\relax Pieceable fault-tolerant circuits, transversal gates, and magic-state injection yield a universal gate set for symmetric hypergraph product codes \NoCaseChange{\protect\cite{cite742}}.
\item\relax Targeted logical gates \NoCaseChange{\protect\cite{cite3775}}.
\item\relax Logical gates via Dehn twists for hypergraph products of cyclic codes \NoCaseChange{\protect\cite{cite3403}}.
\end{eczvaluelist}
\codefieldsection{Decoding}
\begin{eczvaluelist}
\item\relax Single-ancilla syndrome extraction circuits do not admit \flmRefsHyperref{ref3496}{hook errors} \NoCaseChange{\protect\cite{cite3776}}.
\item\relax ReShape decoder that uses minimum weight decoders for the classical codes used in the hypergraph construction \NoCaseChange{\protect\cite{cite3777}}.
\item\relax 2D geometrically local syndrome extraction circuits with depth \flmRefsHyperref{ref65}{order} \(O(\sqrt{n})\) using \flmRefsHyperref{ref65}{order} \(O(n)\) ancilla qubits \NoCaseChange{\protect\cite{cite521}}.
\item\relax BP-OSD decoder \NoCaseChange{\protect\cite{cite1247}} and an improved BP-OSD decoder \NoCaseChange{\protect\cite{cite3179}}.
\item\relax Erasure correction can be implemented approximately with \(O(n^2)\) operations with quantum generalizations \NoCaseChange{\protect\cite{cite3778}} of the peeling and pruned peeling decoders \NoCaseChange{\protect\cite{cite99}}, with a probabilistic version running in \(O(n^{1.5})\) operations. Other nearly optimal erasure decoders exist \NoCaseChange{\protect\cite{cite3779,cite3780}}. Initial hypergraph product codes can be further optimized against the erasure channel using reinforcement learning \NoCaseChange{\protect\cite{cite3781}}.
\item\relax Syndrome measurements are \flmRefsHyperref{ref3496}{distance-preserving} because syndrome extraction circuits can be designed to avoid \flmRefsHyperref{ref3496}{hook errors} \NoCaseChange{\protect\cite{cite3782}}.
\item\relax Generalization \NoCaseChange{\protect\cite{cite3783}} of Viderman's algorithm for expander codes \NoCaseChange{\protect\cite{cite1338}}.
\item\relax Linear time iterative decoder \NoCaseChange{\protect\cite{cite3784}}.
\end{eczvaluelist}
\codefieldsection{Fault Tolerance}
\begin{eczvaluelist}
\item\relax Pieceable fault-tolerant circuits, transversal gates, and magic-state injection yield a universal gate set for symmetric hypergraph product codes \NoCaseChange{\protect\cite{cite742}}.
\item\relax Single-ancilla syndrome extraction circuits do not admit \flmRefsHyperref{ref3496}{hook errors} \NoCaseChange{\protect\cite{cite3776}}.
\item\relax There is a fault-tolerant universal computation scheme for hypergraph-product codes concatenated with the \(\llbracket 4,2,2\rrbracket \) code in which the full syndrome measurement on the lower hypergraph product code is performed only if an error is detected at the upper four-qubit code \NoCaseChange{\protect\cite{cite3295}}.
\end{eczvaluelist}
\codefieldsection{Code Capacity Threshold}
\begin{eczvaluelist}
\item\relax Some thresholds were determined in Ref. \NoCaseChange{\protect\cite{cite3441}}.
\item\relax Bounds on code capacity thresholds using ML decoding can be obtained by mapping the effect of noise on the code to a statistical mechanical model \NoCaseChange{\protect\cite{cite3438}}. For example, a threshold of \(7\%\) was obtained under independent \(X\) and \(Z\) noise for codes obtained from random \((3,4)\)-regular Gallager codes.
\end{eczvaluelist}
\codefieldsection{Threshold}
\begin{eczvaluelist}
\item\relax Circuit-level noise: \(0.1\%\) with all-to-all connected syndrome extraction circuits \NoCaseChange{\protect\cite{cite521}} and DiVincenzo-Aliferis syndrome extraction circuits \NoCaseChange{\protect\cite{cite3785}} combined with non-local gates \NoCaseChange{\protect\cite{cite3786}}. No threshold observed above physical noise rates at or above \(10^{-6}\) using 2D geometrically local syndrome extraction circuits.
\end{eczvaluelist}
\codefieldsection{Notes}
\begin{eczvaluelist}
\item\relax A database of hypergraph-product codes is available in QECDB \NoCaseChange{\protect\cite{cite781}}.
\end{eczvaluelist}
\codefieldsection{Parents}
\begin{eczvaluelist}
\item\relax
\flmRefsHyperref[eczindexfamilyrel]{code:homological_product}{Homological product code} --- A homological-product code of length-one chain complexes reduces to an HGP code, which is also a special case of multi-dimensional homological products of two length-one chain complexes.
\item\relax
\flmRefsHyperref[eczindexfamilyrel]{code:sc_qldpc}{Quantum spatially coupled (SC-QLDPC) code} --- Hypergraph-product stabilizer generator matrices can be used as sub-matrices to define a 2D SC-QLDPC code \NoCaseChange{\protect\cite{cite644}}.
\item\relax
\flmRefsHyperref[eczindexfamilyrel]{code:galois_hypergraph_product}{Galois-qudit HGP code} --- Hypergraph product codes are Galois-qudit hypergraph-product codes for qudit dimension \(q=2\).
\end{eczvaluelist}
\codefieldsection{Children}
\begin{eczvaluelist}
\item\relax
\flmRefsHyperref[eczindexfamilyrel]{code:hgp_7_2_2}{\(\llbracket 7,2,2\rrbracket \) HGP phantom code} --- This code is the hypergraph product of the \([3,2,2]\) simplex code and the \([2,1,2]\) repetition code \NoCaseChange{\protect\cite{cite514}}.
\item\relax
\flmRefsHyperref[eczindexfamilyrel]{code:anisotropic_z2_laplacian}{Anisotropic \(\mathbb{Z}_2\) Laplacian model code} --- The anisotropic \(\mathbb{Z}_2\) Laplacian model is the hypergraph product of a cyclic repetition code and a Laplacian code \NoCaseChange{\protect\cite{cite1350}}.
\item\relax
\flmRefsHyperref[eczindexfamilyrel]{code:fibonacci_fractal_liquid}{Fibonacci fractal spin-liquid code} --- The Fibonacci fractal spin-liquid code is a hypergraph product of the repetition code and the Fibonacci code \NoCaseChange{\protect\cite{cite1348}}, and can be formulated directly as a BP code \NoCaseChange{\protect\cite{cite1350}}.
\item\relax
\flmRefsHyperref[eczindexfamilyrel]{code:sierpinsky_fractal_liquid}{Sierpinski prism model code} --- The Sierpinski prism model code is a hypergraph product of the repetition code and the Newman-Moore code \NoCaseChange{\protect\cite{cite1501,cite1350}}.
\item\relax
\flmRefsHyperref[eczindexfamilyrel]{code:two_foliated}{Two-foliated fracton code} --- The two-foliated fracton code is a hypergraph product of the repetition code and the plaquette Ising code on a square lattice with periodic boundary conditions \NoCaseChange{\protect\cite{cite1517}}.
\item\relax
\flmRefsHyperref[eczindexfamilyrel]{code:cyclic_hgp}{Cyclic Hypergraph Product Code} --- A cyclic hypergraph product code is a hypergraph product code constructed using two circulant matrices.
\item\relax
\flmRefsHyperref[eczindexfamilyrel]{code:lacross}{La-cross code} --- La-cross codes are constructed using a hypergraph product of a cyclic LDPC code with itself.
\item\relax
\flmRefsHyperref[eczindexfamilyrel]{code:quantum_expander}{Quantum expander code}\item\relax
\flmRefsHyperref[eczindexfamilyrel]{code:tillichzemor}{Tillich-Zémor code}\end{eczvaluelist}
\codefieldsection{Cousins}
\begin{eczvaluelist}
\item\relax
\flmRefsHyperref[eczindexfamilyrel]{code:ltc}{Locally testable code (LTC)} --- Applying the hypergraph product to an LTC yields a code which provides an explicit example of \textit{No Low-Error Trivial States (NLETS)} \NoCaseChange{\protect\cite{cite1104}}.
\item\relax
\flmRefsHyperref[eczindexfamilyrel]{code:xyz_product}{XYZ product code} --- Hypergraph (XYZ) product codes are constructed out of hypergraph products of two (three) classical linear codes.
\item\relax
\flmRefsHyperref[eczindexfamilyrel]{code:reinforcement_learning}{Reinforcement-learning quantum code} --- Using reinforcement learning, hypergraph product codes can be further optimized against the erasure channel \NoCaseChange{\protect\cite{cite3781}} and can be weight reduced while maintaining distance \NoCaseChange{\protect\cite{cite3787}}.
\item\relax
\flmRefsHyperref[eczindexfamilyrel]{code:stab_4_2_2}{\(\llbracket 4,2,2\rrbracket \) Four-qubit code} --- There is a fault-tolerant universal computation scheme for hypergraph-product codes concatenated with the \(\llbracket 4,2,2\rrbracket \) code in which the full syndrome measurement on the lower hypergraph product code is performed only if an error is detected at the upper four-qubit code \NoCaseChange{\protect\cite{cite3295}}.
\item\relax
\flmRefsHyperref[eczindexfamilyrel]{code:qubit_concatenated}{Concatenated qubit code} --- There is a fault-tolerant universal computation scheme for hypergraph-product codes concatenated with the \(\llbracket 4,2,2\rrbracket \) code in which the full syndrome measurement on the lower hypergraph product code is performed only if an error is detected at the upper four-qubit code \NoCaseChange{\protect\cite{cite3295}}.
\item\relax
\flmRefsHyperref[eczindexfamilyrel]{code:pinwheel}{Pinwheel code} --- The hypergraph product of a pinwheel code with a cyclic repetition code yields a local Type-I fracton model in three dimensions, while the hypergraph product of two pinwheel codes yields a local Type-II fracton model in four dimensions \NoCaseChange{\protect\cite{cite1350}}.
\item\relax
\flmRefsHyperref[eczindexfamilyrel]{code:compactified_r}{Compactified \(\mathbb{R}\) gauge theory code} --- The compactified \(\mathbb{R}\) gauge theory code is constructed from a hypergraph product of two repetition codes over the integers.
\item\relax
\flmRefsHyperref[eczindexfamilyrel]{code:tiger_surface}{Tiger surface code} --- The tiger surface code is constructed from a hypergraph product of two repetition codes over the integers.
\item\relax
\flmRefsHyperref[eczindexfamilyrel]{code:quantum_locally_recoverable}{Quantum locally recoverable code (QLRC)} --- A variant of the hypergraph product can be used to define QLRCs with intersecting recovery sets \NoCaseChange{\protect\cite{cite2986}}.
\item\relax
\flmRefsHyperref[eczindexfamilyrel]{code:single_shot}{Single-shot code} --- Two-fold application of the hypergraph product to a pair of binary linear codes yields single-shot QLDPC codes that exploit redundancy in their stabilizer generators \NoCaseChange{\protect\cite{cite675}}.
\item\relax
\flmRefsHyperref[eczindexfamilyrel]{code:self_correct}{Self-correcting quantum code} --- There are bounds on the energy barrier of hypergraph product codes \NoCaseChange{\protect\cite{cite3038}}.
\item\relax
\flmRefsHyperref[eczindexfamilyrel]{code:phantom}{Phantom code} --- Some hypergraph-product constructions, such as products of a classical simplex code and a repetition code, yield phantom codes, while the smallest examples have lower rates than the phantom quantum RM constructions \NoCaseChange{\protect\cite{cite514}}.
\item\relax
\flmRefsHyperref[eczindexfamilyrel]{code:quantum_rainbow}{Quantum rainbow code} --- Hypergraph products of color codes yield quantum rainbow codes with growing distance and transversal gates in the \flmTerm{term}{ref694}{}{Clifford hierarchy}. In particular, utilizing this construction for quasi-hyperbolic color codes yields an \(\llbracket n,O(n),O(\log n)\rrbracket \) triorthogonal code family satisfying the necessary conditions for the magic-state yield parameter \(\gamma\) to become arbitrarily small \NoCaseChange{\protect\cite{cite704}}.
\item\relax
\flmRefsHyperref[eczindexfamilyrel]{code:ramanujan_tensor_product}{High-dimensional expander (HDX) code} --- Ramanujan codes utilize the hypergraph product with a twist, which is an automorphism on one of the complexes in the tensor product, in order to increase distance \NoCaseChange{\protect\cite{cite3442}}.
\item\relax
\flmRefsHyperref[eczindexfamilyrel]{code:rotated_surface}{Rotated surface code} --- Periodic checkerboard or rotated-toric codes on the same lattice can be obtained from hypergraph products of two cyclic linear binary codes with palindromic check polynomials \NoCaseChange{\protect\cite[{Sec. IV.D}]{cite1316}}.
\item\relax
\flmRefsHyperref[eczindexfamilyrel]{code:surface}{Kitaev surface code} --- The planar surface code on a square lattice can be obtained from a hypergraph product of two repetition codes with appropriate boundary checks.
\item\relax
\flmRefsHyperref[eczindexfamilyrel]{code:fractal_surface}{Fractal surface code} --- The related fractal product code is a hypergraph product of two classical codes defined on a Sierpinski carpet graph \NoCaseChange{\protect\cite{cite676}}.
\item\relax
\flmRefsHyperref[eczindexfamilyrel]{code:subsystem_quantum_parity}{Subsystem hypergraph product (SHP) code} --- Two SHP codes can be gauge-fixed to yield an HGP code \NoCaseChange{\protect\cite[{Sec. III}]{cite665}}. The SHP and HGP code constructions yield the same dimension and minimum distance, but the former does not yield QLDPC codes; see \NoCaseChange{\protect\cite[{pg. 18}]{cite1448}}.
\item\relax
\flmRefsHyperref[eczindexfamilyrel]{code:generalized_bicycle}{Generalized bicycle (GB) code} --- An arbitrary qubit GB code of length \(2\ell\) is equivalent to a rotated HGP code with periodicity vectors \(\vec{L}_{1}\) and \(\vec{L}_{2}\) such that \(\lvert \vec{L}_{1}\times\vec{L}_{2}\rvert=\ell\) \NoCaseChange{\protect\cite{cite3183}}.
\end{eczvaluelist}
\eczhbkcontributors{ Shi Jie Samuel Tan, Christopher A. Pattison, Joschka Roffe, \eczhuVVA }
\endeczcode

\eczcode{holographic_hyperinvariant}{Hyperinvariant tensor-network (HTN) code}{~\NoCaseChange{\protect\cite{cite3788}}}
\codefieldsection{Alternative Names}
\begin{eczvaluelist}
\item\relax Evenbly code
\end{eczvaluelist}
\eczhIndexCodeAliasName{holographic_hyperinvariant}{Evenbly code}
\codefieldsection{Description}
Holographic tensor-network code constructed out of a hyperinvariant tensor network \NoCaseChange{\protect\cite{cite639}}, i.e., a MERA-like network admitting a hyperbolic geometry.
The network is defined using two layers A and B, with constituent tensors satisfying isometry conditions (a.k.a. multitensor constraints).

This code produces boundary correlation functions that align with those expected from conformal field theory (CFT) boundary states.
HTN codes exhibit state-dependent breakdown of complementary recovery, consistent with quantum gravity corrections in AdS/CFT.

\codefieldsection{Code Capacity Threshold}
\begin{eczvaluelist}
\item\relax \(19.1\%\) under depolarizing noise and \(50\%\) under erasure noise for a \(\{5,4\}\) tiling \NoCaseChange{\protect\cite{cite3789}}.
\item\relax \(40\%\) under erasure noise for constant-rate version of the code \NoCaseChange{\protect\cite{cite3789}}.
\end{eczvaluelist}
\codefieldsection{Parents}
\begin{eczvaluelist}
\item\relax
\flmRefsHyperref[eczindexfamilyrel]{code:qubit_stabilizer}{Qubit stabilizer code}\item\relax
\flmRefsHyperref[eczindexfamilyrel]{code:holographic_tensor}{Holographic tensor-network code} --- The encoding of an HTN code is a hyperinvariant tensor network.
\end{eczvaluelist}
\codefieldsection{Cousin}
\begin{eczvaluelist}
\item\relax
\flmRefsHyperref[eczindexfamilyrel]{code:group_4_2_2}{\(\llbracket 4,2,2\rrbracket _{G}\) four group-qudit code} --- The explicit 4-ququart encoding tensor \(A'\) used in the HTN code is a \(\llbracket 4,1,2\rrbracket _{\mathbb{Z}_4}\) subcode of the \(\llbracket 4,2,2\rrbracket _{\mathbb{Z}_4}\) four group-qudit code \NoCaseChange{\protect\cite[{Sec. IID}]{cite3788}}.
\end{eczvaluelist}
\eczhbkcontributors{ \eczhuVVA }
\endeczcode

\eczcode{hypersphere_product}{Hypersphere product code}{~\NoCaseChange{\protect\cite{cite3410}}}
\codefieldsection{Description}
Homological code based on products of hyperspheres.
The hypersphere product code family has asymptotically diminishing soundness that scales as \flmRefsHyperref{ref65}{order} \(O(1/\log (n)^2)\), locality of stabilizer generators scaling as \flmRefsHyperref{ref65}{order} \(O(\log n/ \log\log n)\), and distance of \flmRefsHyperref{ref65}{order} \(\Theta(\sqrt{n})\).

\codefieldsection{Parent}
\begin{eczvaluelist}
\item\relax
\flmRefsHyperref[eczindexfamilyrel]{code:higher_dimensional_surface}{Homological code}\end{eczvaluelist}
\codefieldsection{Cousins}
\begin{eczvaluelist}
\item\relax
\flmRefsHyperref[eczindexfamilyrel]{code:qltc}{Quantum locally testable code (QLTC)} --- The hypersphere product code family has asymptotically diminishing soundness that scales as \flmRefsHyperref{ref65}{order} \(O(1/\log (n)^2)\), locality of stabilizer generators scaling as \flmRefsHyperref{ref65}{order} \(O(\log n/ \log\log n)\), and distance of \flmRefsHyperref{ref65}{order} \(\Theta(\sqrt{n})\). Applying Hastings' weight-reduction construction yields QLDPC families with distance \(\Theta^*(\sqrt{n})\) and inverse-polylogarithmic soundness \NoCaseChange{\protect\cite{cite2989}}. Application of generalized distance balancing \NoCaseChange{\protect\cite{cite684}} to hypersphere product codes using an asymptotically good classical code of length \(t\) yields \(O( 1/(\log(n)^2 t^2) )\) soundness and \flmRefsHyperref{ref65}{order} \(\Theta(\sqrt{n}t)\) distance while maintaining locality scaling and at the expense of a dimension scaling as \flmRefsHyperref{ref65}{order} \(\Theta(t^2)\) \NoCaseChange{\protect\cite{cite2990}}.
\item\relax
\flmRefsHyperref[eczindexfamilyrel]{code:distance_balanced}{Distance-balanced code} --- Application of generalized distance balancing \NoCaseChange{\protect\cite{cite684}} to hypersphere product codes using an asymptotically good classical code of length \(t\) yields \(O( 1/(\log(n)^2 t^2) )\) soundness and \flmRefsHyperref{ref65}{order} \(\Theta(\sqrt{n}t)\) distance while maintaining locality scaling and at the expense of a dimension scaling as \flmRefsHyperref{ref65}{order} \(\Theta(t^2)\) \NoCaseChange{\protect\cite{cite2990}}.
\end{eczvaluelist}
\eczhbkcontributors{ \eczhuVVA }
\endeczcode

\eczcode{jw}{Jordan-Wigner transformation code}{~\NoCaseChange{\protect\cite{cite3790,cite558,cite3791}}}
\codefieldsection{Description}
A mapping between qubit Pauli strings and Majorana operators that can be thought of as a trivial \(\llbracket n,n\rrbracket \) code.
The mapping is best described as converting a chain of \(n\) qubits into a chain of \(2n\) Majorana modes (i.e., \(n\) fermionic modes).
It maps Majorana operators into Pauli strings of weight \(O(n)\).

The Majorana modes \(\{\gamma_j\}\) are defined from Pauli strings as follows,
\flmMathEnvironment{align}{}{
\begin{split}
  \gamma_{0}&=Z\\
 -\gamma_{1}&=Y\\
  \gamma_{2}&=X\otimes Z\\
 -\gamma_{3}&=X\otimes Y\\
  \gamma_{4}&=X\otimes X\otimes Z\\
 -\gamma_{5}&=X\otimes X\otimes Y\\&\vdots
\end{split}
}
The \(X\)-type Pauli strings ensure that the resulting Majorana operators satisfy the appropriate anti-commutation relations, namely, \(\{\gamma_i,\gamma_j\} = 2\delta_{ij}\).

\codefieldsection{Encoding}
\begin{eczvaluelist}
\item\relax Circuit of depth linear in the number of qubits \(n\). The depth can be reduced for particle-preserving systems \NoCaseChange{\protect\cite{cite3792}} and in other contexts \NoCaseChange{\protect\cite{cite3793}}.
\end{eczvaluelist}
\codefieldsection{Parent}
\begin{eczvaluelist}
\item\relax
\flmRefsHyperref[eczindexfamilyrel]{code:fermions_into_qubits}{Fermion-into-qubit code}\end{eczvaluelist}
\codefieldsection{Cousins}
\begin{eczvaluelist}
\item\relax
\flmRefsHyperref[eczindexfamilyrel]{code:majorana_stab}{Majorana stabilizer code} --- A Majorana stabilizer code is a stabilizer code whose stabilizers are composed of Majorana fermion operators, which are in turn realizable using Pauli strings via the Jordan-Wigner mapping.

\item\relax
\flmRefsHyperref[eczindexfamilyrel]{code:kitaev_chain}{Kitaev chain code} --- The Kitaev chain code can be thought of as the Majorana stabilizer analogue of the quantum repetition code \NoCaseChange{\protect\cite{cite559}} and is related to that code via the Jordan-Wigner transformation \NoCaseChange{\protect\cite{cite3794}}.
\item\relax
\flmRefsHyperref[eczindexfamilyrel]{code:aqm}{Auxiliary qubit mapping (AQM) code} --- The AQM fermion-into-qubit code reduces to the JW transformation code when the outer code is trivial.
\item\relax
\flmRefsHyperref[eczindexfamilyrel]{code:bkt}{Bravyi-Kitaev transformation (BKT) code} --- The weight of a Majorana operator in the BKT (JW transformation) code scales logarithmically (linearly) with \(n\), with the former demonstrating an exponential improvement \NoCaseChange{\protect\cite{cite3510}}.
\item\relax
\flmRefsHyperref[eczindexfamilyrel]{code:2d_bosonization}{2D bosonization code} --- The exact 2D bosonization code can be converted by a linear-depth Clifford circuit into a Jordan-Wigner ordering path on the 2D lattice \NoCaseChange{\protect\cite[{Fig. 24}]{cite404}}.
\end{eczvaluelist}
\eczhbkcontributors{ \eczhuVVA }
\endeczcode

\eczcode{jump}{Jump code}{~\NoCaseChange{\protect\cite{cite3250,cite144,cite145}}}
\codefieldsection{Description}
A CE code designed to detect and correct \flmRefsHyperref{ref498}{AD} errors.
An \(\llparenthesis n,K\rrparenthesis \) jump code is denoted as \(\llparenthesis n,K,t\rrparenthesis _w\) (which conflicts with modular-qudit notation), where \(t\) is the maximum number of qubits that can be corrected after each one has undergone a jump error \(|0\rangle\langle 1|\), and where each codeword is a uniform superposition of qubit basis states with Hamming weight \(w\).

\codefieldsection{Protection}
Various code bounds, including an upper bound on \(K\) given the other parameters, are provided in Ref. \NoCaseChange{\protect\cite{cite145}}.
For example, one can apply bit flips to all qubits of an \(\llparenthesis n,K,t\rrparenthesis _w\) jump code to obtain an \(\llparenthesis n,K,t\rrparenthesis _{n-w}\) jump code.

\codefieldsection{Rate}
An infinite family of jump codes asymptotically attains an upper bound on \(K\) \NoCaseChange{\protect\cite[{Thm. 27}]{cite145}}.
\codefieldsection{Gates}
\begin{eczvaluelist}
\item\relax Two-qubit entangling gate \NoCaseChange{\protect\cite{cite3795}}.
\end{eczvaluelist}
\codefieldsection{Notes}
\begin{eczvaluelist}
\item\relax Survey of jump codes \NoCaseChange{\protect\cite{cite3796}}.
\end{eczvaluelist}
\codefieldsection{Parents}
\begin{eczvaluelist}
\item\relax
\flmRefsHyperref[eczindexfamilyrel]{code:qubits_into_qubits}{Qubit code}\item\relax
\flmRefsHyperref[eczindexfamilyrel]{code:ampdamp}{Amplitude-damping (AD) code} --- Jump codes are designed to protect against qubit \flmRefsHyperref{ref498}{AD} noise.
\item\relax
\flmRefsHyperref[eczindexfamilyrel]{code:constant_excitation}{Constant-excitation (CE) code}\end{eczvaluelist}
\codefieldsection{Child}
\begin{eczvaluelist}
\item\relax
\flmRefsHyperref[eczindexfamilyrel]{code:css_4_1_2}{\(\llbracket 4,1,2\rrbracket \) Leung-Nielsen-Chuang-Yamamoto (LNCY) code} --- The \(\llbracket 4,1,2\rrbracket \) LNCY code \NoCaseChange{\protect\cite{cite3250}} is equivalent to a \(\llparenthesis 4,2,1\rrparenthesis _2\) jump code correcting a single \flmRefsHyperref{ref498}{AD} error.
A \(\llparenthesis 4,3,1\rrparenthesis _2\) jump code is a subcode of the \(\llbracket 4,2,2\rrbracket \) code and contains the \(\llbracket 4,1,2\rrbracket \) LNCY code as a subcode \NoCaseChange{\protect\cite{cite144}}.

\end{eczvaluelist}
\codefieldsection{Cousins}
\begin{eczvaluelist}
\item\relax
\flmRefsHyperref[eczindexfamilyrel]{code:chuang-leung-yamamoto}{Chuang-Leung-Yamamoto (CLY) code} --- Jump codes can be thought of as qubit analogues of uniform CLY codes.
\item\relax
\flmRefsHyperref[eczindexfamilyrel]{code:iceberg}{\(\llbracket 2m,2m-2,2\rrbracket \) error-detecting code} --- The subcode of the \(\llbracket 2m,2m-2,2\rrbracket \) error-detecting code consisting of codewords labeled by weight-\(m\) bitstrings is a \(\llparenthesis 2m,\frac{1}{2}{2m \choose m},1\rrparenthesis _{m}\) optimal jump code \NoCaseChange{\protect\cite{cite144}\protect\cite[{Corr. 9}]{cite145}}.
\item\relax
\flmRefsHyperref[eczindexfamilyrel]{code:combinatorial_design}{Combinatorial design} --- Certain types of combinatorial designs can be used to obtain jump codes \NoCaseChange{\protect\cite{cite144,cite145,cite146}}.
\item\relax
\flmRefsHyperref[eczindexfamilyrel]{code:self_dual}{Self-dual linear code} --- Iso-dual codes can be used to construct jump codes \NoCaseChange{\protect\cite{cite145}}.
\item\relax
\flmRefsHyperref[eczindexfamilyrel]{code:stab_4_2_2}{\(\llbracket 4,2,2\rrbracket \) Four-qubit code} --- A \(\llparenthesis 4,3,1\rrparenthesis _2\) jump code is a subcode of the \(\llbracket 4,2,2\rrbracket \) code and contains the \(\llbracket 4,1,2\rrbracket \) LNCY code as a subcode \NoCaseChange{\protect\cite{cite144}}.
\end{eczvaluelist}
\eczhbkcontributors{ \eczhuVVA }
\endeczcode

\eczcode{kls}{Khesin-Lu-Shor code}{~\NoCaseChange{\protect\cite{cite858}}}
\codefieldsection{Description}
A family of \(\llbracket m 2^m / (m+1), 2^m / (m+1), d(m)\rrbracket \) qubit CSS codes derived from the Hamming code, where \(m = 2^r - 1\).
Their \flmRefsHyperref{ref857}{encoder-respecting form} is the graph of a hypercube in \(m\) dimensions, and input nodes in the graph are codewords of the \([2^r-1,2^r-r-1,3]\) Hamming code \NoCaseChange{\protect\cite{cite858}}.

\codefieldsection{Protection}
The code distance satisfies \(\lfloor (m-1)/2 \rfloor \leq d(m) \leq m\) and is conjectured to be \(m\) for \(m \geq 7\) \NoCaseChange{\protect\cite{cite858}}.
\codefieldsection{Decoding}
\begin{eczvaluelist}
\item\relax A greedy graph decoder on the hypercube representation corrects at least \(\lfloor (m-1)/4 \rfloor - 1\) Pauli errors \NoCaseChange{\protect\cite{cite858}}.
\end{eczvaluelist}
\codefieldsection{Parent}
\begin{eczvaluelist}
\item\relax
\flmRefsHyperref[eczindexfamilyrel]{code:qubit_css}{Qubit CSS code}\end{eczvaluelist}
\codefieldsection{Child}
\begin{eczvaluelist}
\item\relax
\flmRefsHyperref[eczindexfamilyrel]{code:stab_6_2_2}{\(\llbracket 6,2,2\rrbracket \) \(C_6\) code} --- The Khesin-Lu-Shor code for \(r=2\) and \(m=2^r - 1 = 3\) is the \(C_6\) code.
\end{eczvaluelist}
\codefieldsection{Cousins}
\begin{eczvaluelist}
\item\relax
\flmRefsHyperref[eczindexfamilyrel]{code:hamming}{\([2^r-1,2^r-r-1,3]\) Hamming code} --- The \flmRefsHyperref{ref857}{encoder-respecting form} of the \(\llbracket m 2^m / (m+1), 2^m / (m+1), d(m)\rrbracket \) Khesin-Lu-Shor code is the graph of a hypercube in \(m = 2^r - 1\) dimensions, and input nodes in the graph are codewords of the \([2^r-1,2^r-r-1,3]\) Hamming code \NoCaseChange{\protect\cite{cite858}}.
\item\relax
\flmRefsHyperref[eczindexfamilyrel]{code:hypercube}{Hypercube code} --- The \flmRefsHyperref{ref857}{encoder-respecting form} of the \(\llbracket m 2^m / (m+1), 2^m / (m+1), d(m)\rrbracket \) Khesin-Lu-Shor code is the graph of a hypercube in \(m = 2^r - 1\) dimensions, and input nodes in the graph are codewords of the \([2^r-1,2^r-r-1,3]\) Hamming code \NoCaseChange{\protect\cite{cite858}}.
\item\relax
\flmRefsHyperref[eczindexfamilyrel]{code:steane}{\(\llbracket 7,1,3\rrbracket \) Steane code} --- The \flmRefsHyperref{ref857}{encoder-respecting form} of both the Steane and Khesin-Lu-Shor codes is the graph of a hypercube \NoCaseChange{\protect\cite{cite858}}.
\end{eczvaluelist}
\eczhbkcontributors{ Andrey Boris Khesin, \eczhuVVA }
\endeczcode

\eczcode{kpt}{Kim-Preskill-Tang (KPT) code}{~\NoCaseChange{\protect\cite{cite640}}}
\codefieldsection{Description}
An approximate quantum error-correcting code that protects the encoded interior of a black hole from computationally bounded exterior observers.
Under the assumption that the Hawking radiation emitted by an old black hole is pseudorandom, there exists a subspace of the radiation system that encodes the black hole interior, entangled with the late outgoing Hawking quanta.
The logical operators of this code, called ghost operators in \NoCaseChange{\protect\cite{cite640}}, commute with efficient operations acting on the radiation, protecting the interior up to corrections exponentially small in the black hole's entropy.
The construction is state dependent: the encoding depends on the state that collapsed to form the black hole \NoCaseChange{\protect\cite{cite640}}.

This code has been tested in various models of gravity \NoCaseChange{\protect\cite{cite640,cite3797}}.

\codefieldsection{Protection}
Protection relies on the pseudorandomness of the radiation and on restricting noise to efficient quantum computations acting on the radiation alone.
Such operations commute with the interior logical algebra up to errors exponentially small in the remaining black-hole entropy \NoCaseChange{\protect\cite{cite640}}.

\codefieldsection{Parents}
\begin{eczvaluelist}
\item\relax
\flmRefsHyperref[eczindexfamilyrel]{code:qubits_into_qubits}{Qubit code}\item\relax
\flmRefsHyperref[eczindexfamilyrel]{code:holographic}{Holographic code} --- The robustness of KPT codes does not rely on arguments from holographic duality, but such codes do aim to describe interiors of black holes.
\end{eczvaluelist}
\codefieldsection{Cousin}
\begin{eczvaluelist}
\item\relax
\flmRefsHyperref[eczindexfamilyrel]{code:syk}{SYK code} --- The Brownian SYK model can be used to demonstrate the complexity-based error-correction of KPT codes \NoCaseChange{\protect\cite{cite640}}.
\end{eczvaluelist}
\eczhbkcontributors{ \eczhuVVA }
\endeczcode

\eczcode{kitaev_chain}{Kitaev chain code}{~\NoCaseChange{\protect\cite{cite558}}}
\codefieldsection{Alternative Names}
\begin{eczvaluelist}
\item\relax Majorana repetition code
\end{eczvaluelist}
\eczhIndexCodeAliasName{kitaev_chain}{Majorana repetition code}
\codefieldsection{Description}
A Majorana stabilizer code obtained from the ground-state subspace of the Kitaev Majorana chain in its fermionic topological phase \NoCaseChange{\protect\cite{cite558}}. Its codespace is stabilized by nearest-neighbor Majorana bilinears, while two unpaired edge Majoranas furnish one logical fermionic mode. Under parity-preserving noise, it behaves as the Majorana analogue of the repetition code \NoCaseChange{\protect\cite{cite559}}.

At the fixed-point limit, the code Hamiltonian is proportional to \(-\sum_{j=1}^{n-1} i \gamma_{2j}\gamma_{2j+1}\), so the codespace is the common \(+1\) eigenspace of the stabilizers \(S_j=i\gamma_{2j}\gamma_{2j+1}\). The two unpaired edge operators \(\gamma_{1}\) and \(\gamma_{2n}\) are the \textit{Majorana zero modes (MZMs)} or \textit{Majorana edge modes (MEMs)}; they commute with all stabilizers and define a logical fermionic mode \(f_{\mathrm{L}}=(\gamma_{1}+i\gamma_{2n})/2\). Via the Jordan-Wigner transformation, the model maps to the 1D quantum Ising chain in its symmetry-breaking phase.
The code can be thought of as the Majorana stabilizer analogue of the quantum repetition code: parity-preserving dephasing operators \(Z_j=i\gamma_{2j-1}\gamma_{2j}\) play the role of repetition-code bit flips, while the logical Majorana operators have odd weight and therefore encode a logical fermion \NoCaseChange{\protect\cite{cite559}}.

The two basis states of a single chain have opposite fermionic parity.
Therefore, a single chain does not by itself furnish a fixed-parity qubit encoding; coherent superpositions between the two basis states are not directly accessible in an isolated fermionic system with parity superselection.
One can combine two such code blocks to form a Majorana box qubit, which is the fixed-parity subspace of the combined codespace.
Odd numbers of code blocks also contain fixed-parity logical subspaces in their codespace.

\codefieldsection{Protection}
In the fixed-point limit, local parity-preserving bilinears such as \(Z_j=i\gamma_{2j-1}\gamma_{2j}\) anticommute with nearby stabilizers, so the chain behaves as a repetition code against dephasing or phase errors \NoCaseChange{\protect\cite{cite559}}. For a finite chain, the splitting between the two code states is exponentially small in the separation between the edge modes \NoCaseChange{\protect\cite{cite558}}.
As a Majorana stabilizer code, however, its distance is \(1\) because a single MZM is already a logical operator. The code therefore does not protect against single-Majorana, parity-violating errors such as quasiparticle poisoning.
Disorder may help with protection \NoCaseChange{\protect\cite{cite3798}}.

\codefieldsection{Gates}
\begin{eczvaluelist}
\item\relax Braiding, \(S\), and \(T\) phase gates, fermion \(CZ_f\), and mixed qubit-fermion \(CZ_{qf}\) gates are described for logical fermions encoded in this repetition code \NoCaseChange{\protect\cite{cite559}}.
\end{eczvaluelist}
\codefieldsection{Decoding}
\begin{eczvaluelist}
\item\relax Local automaton decoder based on a self-dual cellular automaton \NoCaseChange{\protect\cite{cite3799}}.
\item\relax Syndrome extraction of the stabilizers \(S_j=i\gamma_{2j}\gamma_{2j+1}\) can be performed by interfacing with a qubit ancilla and mixed qubit-fermion gates \NoCaseChange{\protect\cite{cite559}}.
\end{eczvaluelist}
\codefieldsection{Realizations}
\begin{eczvaluelist}
\item\relax Photonic systems: braiding of topological Majorana modes has been simulated in a device that has a different notion of locality than a bona fide fermionic system \NoCaseChange{\protect\cite{cite3800}}.
\item\relax Superconducting circuits: preparation \NoCaseChange{\protect\cite{cite3801}}, braiding \NoCaseChange{\protect\cite{cite3802}}, and detection of Majorana edge modes \NoCaseChange{\protect\cite{cite3802,cite3803}} have been simulated in devices that have a different notion of locality than a bona fide fermionic system.
\end{eczvaluelist}
\codefieldsection{Notes}
\begin{eczvaluelist}
\item\relax See notes \NoCaseChange{\protect\cite{cite3794}} for a description of this code.
\end{eczvaluelist}
\codefieldsection{Parents}
\begin{eczvaluelist}
\item\relax
\flmRefsHyperref[eczindexfamilyrel]{code:majorana_stab}{Majorana stabilizer code}\item\relax
\flmRefsHyperref[eczindexfamilyrel]{code:spt}{Symmetry-protected topological (SPT) code} --- The Kitaev chain is a 1D fermionic SPT (more precisely, a 1D topological superconductor) protected by fermion parity symmetry.
\item\relax
\flmRefsHyperref[eczindexfamilyrel]{code:small_distance_qubit_stabilizer}{Small-distance qubit stabilizer code}\end{eczvaluelist}
\codefieldsection{Cousins}
\begin{eczvaluelist}
\item\relax
\flmRefsHyperref[eczindexfamilyrel]{code:mbq}{Majorana box qubit} --- Majorana box qubit codes are defined to be positive-parity logical subspaces of two or more Kitaev-chain code blocks. The parameter \(n\) in the MBQ code definition corresponds to the number of Kitaev chains used in the construction, and not the total number of physical Majorana modes of the chains.
\item\relax
\flmRefsHyperref[eczindexfamilyrel]{code:quantum_repetition}{Quantum repetition code} --- The Kitaev chain code can be thought of as the Majorana stabilizer analogue of the quantum repetition code \NoCaseChange{\protect\cite{cite559}} and is related to that code via the Jordan-Wigner transformation \NoCaseChange{\protect\cite{cite3794}}.
\item\relax
\flmRefsHyperref[eczindexfamilyrel]{code:jw}{Jordan-Wigner transformation code} --- The Kitaev chain code can be thought of as the Majorana stabilizer analogue of the quantum repetition code \NoCaseChange{\protect\cite{cite559}} and is related to that code via the Jordan-Wigner transformation \NoCaseChange{\protect\cite{cite3794}}.
\item\relax
\flmRefsHyperref[eczindexfamilyrel]{code:3d_fermionic_surface}{3D fermionic surface code} --- The 3D fermionic surface code is the result of applying the 3D bosonization mapping to a trivial fermionic theory \NoCaseChange{\protect\cite{cite3450}}. Twist defects in the 3D fermionic surface code take the form of Kitaev chains after the mapping \NoCaseChange{\protect\cite{cite3451,cite3450}}.
\end{eczvaluelist}
\eczhbkcontributors{ \eczhuVVA }
\endeczcode

\eczcode{kitaev_honeycomb}{Kitaev honeycomb code}{~\NoCaseChange{\protect\cite{cite537,cite594,cite414}}}
\codefieldsection{Description}
Subsystem qubit stabilizer code underlying the Kitaev honeycomb model \NoCaseChange{\protect\cite{cite537,cite594}}.
Its gauge generators are the two-qubit \(XX\), \(YY\), and \(ZZ\) link operators on the three edge types of the honeycomb lattice \NoCaseChange{\protect\cite[{Sec. 3.2}]{cite594}}.
Its stabilizer group is generated by loop operators, and syndrome extraction can be reduced to ordered measurements of the two-qubit link operators \NoCaseChange{\protect\cite[{Sec. 3.2}]{cite594}}.
This is the \(q=2\) instance of the \(\mathbb{Z}_q^{(1)}\) subsystem code and does not encode any logical qubits \NoCaseChange{\protect\cite{cite594}\protect\cite[{Sec. 7.3}]{cite414}}.

The original Kitaev honeycomb spin model is exactly solvable by mapping spins to Majorana fermions in a static \(\mathbb{Z}_2\) gauge field, yielding three gapped \(A\) phases and one gapless \(B\) phase \NoCaseChange{\protect\cite{cite537}}.
Its ground state lies in the vortex-free sector, and the gapped \(A\) phases realize Abelian \(\mathbb{Z}_2\) topological order \NoCaseChange{\protect\cite{cite537}}.

\codefieldsection{Encoding}
\begin{eczvaluelist}
\item\relax The geometric entanglement measure of a ground state of the Kitaev honeycomb model and any state with anomalous one-form symmetry scales as \flmRefsHyperref{ref65}{order} \(\Omega(n)\) \NoCaseChange{\protect\cite{cite3804}}.
\end{eczvaluelist}
\codefieldsection{Realizations}
\begin{eczvaluelist}
\item\relax Neutral atom arrays: realized on a 72 qubit device with 32 ancillas by the Lukin group, where a fermion-into-qubit mapping was used to recast this model in terms of simulated fermionic degrees of freedom and simulate other fermionic Hamiltonians \NoCaseChange{\protect\cite{cite3805}}.
\item\relax Superconducting qubits: driven version of the Kitaev honeycomb model \NoCaseChange{\protect\cite{cite3806}} realized by the Pollmann group on the Sycamore and Willow devices by Google Quantum AI \NoCaseChange{\protect\cite{cite3807}}.
\end{eczvaluelist}
\codefieldsection{Parents}
\begin{eczvaluelist}
\item\relax
\flmRefsHyperref[eczindexfamilyrel]{code:qubit_subsystem_stabilizer}{Subsystem qubit stabilizer code}\item\relax
\flmRefsHyperref[eczindexfamilyrel]{code:qudit_znone}{\(\mathbb{Z}_q^{(1)}\) subsystem code} --- The Kitaev honeycomb code is the \(q=2\) instance of the \(\mathbb{Z}_q^{(1)}\) subsystem code \NoCaseChange{\protect\cite[{Sec. 7.3}]{cite414}}.
\end{eczvaluelist}
\codefieldsection{Cousins}
\begin{eczvaluelist}
\item\relax
\flmRefsHyperref[eczindexfamilyrel]{code:tetron}{Tetron code} --- Embedding each physical qubit into two fermions via the tetron code is useful for exactly solving the Kitaev honeycomb model Hamiltonian \NoCaseChange{\protect\cite{cite537}} and other qubit Hamiltonians on certain graphs \NoCaseChange{\protect\cite{cite2842,cite2843}}.
\item\relax
\flmRefsHyperref[eczindexfamilyrel]{code:2d_bosonization}{2D bosonization code} --- Embedding each physical qubit into two fermions via the tetron code is useful for exactly solving the Kitaev honeycomb model Hamiltonian \NoCaseChange{\protect\cite{cite537}} and other qubit Hamiltonians on certain graphs \NoCaseChange{\protect\cite{cite2842,cite2843}}. When done in reverse, this embedding can be thought of as a 2D bosonization fermion-into-qubit encoding by converting to a relabeled square lattice and performing single-qubit rotations \NoCaseChange{\protect\cite{cite403}\protect\cite[{Sec. IV.B}]{cite404}}.
\item\relax
\flmRefsHyperref[eczindexfamilyrel]{code:surface}{Kitaev surface code} --- The Kitaev honeycomb code can be obtained from the square-lattice surface code by \flmRefsHyperref{ref666}{gauging out} the anyon \(em\) \NoCaseChange{\protect\cite[{Sec. 7.3}]{cite414}}. During this process, the square lattice is effectively expanded to a honeycomb tiling \NoCaseChange{\protect\cite[{Fig. 12}]{cite414}}.
\item\relax
\flmRefsHyperref[eczindexfamilyrel]{code:honeycomb}{Honeycomb tiling} --- The Kitaev honeycomb code is defined on the honeycomb tiling.
\item\relax
\flmRefsHyperref[eczindexfamilyrel]{code:honeycomb_floquet}{Honeycomb Floquet code} --- The Kitaev honeycomb model Hamiltonian is a sum of checks of the honeycomb Floquet code \NoCaseChange{\protect\cite{cite536}}.
\item\relax
\flmRefsHyperref[eczindexfamilyrel]{code:nonabelian_kitaev_honeycomb}{Non-Abelian Kitaev honeycomb code} --- The gauge-group generators of the Kitaev honeycomb code are terms of the Kitaev honeycomb model Hamiltonian. Adding a magnetic field to this Hamiltonian for particular parameter values yields the non-Abelian Ising-anyon phase, whose anyons encode the logical information of the non-Abelian Kitaev honeycomb code \NoCaseChange{\protect\cite{cite537}}.
\item\relax
\flmRefsHyperref[eczindexfamilyrel]{code:matching}{Matching code} --- Matching codes were inspired by the \(\mathbb{Z}_2\) topological order phase of the Kitaev honeycomb model \NoCaseChange{\protect\cite{cite537}}.
\item\relax
\flmRefsHyperref[eczindexfamilyrel]{code:3d_kitaev_honeycomb}{3D Kitaev honeycomb code} --- The 3D Kitaev honeycomb model is a 3D generalization of the Kitaev honeycomb model.
\end{eczvaluelist}
\eczhbkcontributors{ \eczhuVVA }
\endeczcode

\eczcode{surface}{Kitaev surface code}{~\NoCaseChange{\protect\cite{cite2125,cite3808,cite423,cite3809}}
}
\codefieldsection{Description}
A family of Abelian topological \flmRefsHyperref{code:css}{CSS stabilizer} codes
whose generators are few-body \(X\)-type and \(Z\)-type Pauli strings
associated to the stars and plaquettes, respectively, of a cellulation of a
two-dimensional surface (with a qubit located at each edge of the
cellulation).
Codewords correspond to ground states of the surface code Hamiltonian, and error operators create or annihilate pairs of anyonic charges or vortices.

The construction on a torus is called the toric code, while the construction on a planar patch with boundaries is called the \textit{planar code} \NoCaseChange{\protect\cite{cite3383,cite3809}}.
Boundary segments come in two types, \textit{open} (a.k.a. rough or primal) and \textit{closed} (a.k.a. smooth or dual); logical operators on patches with boundary are naturally described by relative homology classes \NoCaseChange{\protect\cite{cite3809}}.
In the homological formulation, a cellulation of a closed surface \(M\) encodes \(k=2-\chi(M)\) qubits, while a cellulation of a surface with boundary encodes \(k=1-\chi(M)\) qubits; cutting handles of closed surfaces yields planar patches with holes \NoCaseChange{\protect\cite{cite71}}.
A \textit{mixed boundary} consists of an interleaving of the two boundary types \NoCaseChange{\protect\cite{cite426}}.

\codefieldsection{Protection}
The original planar code on a square-lattice patch with different boundary conditions on the vertical and horizontal edges is a \(\llbracket L^2+(L-1)^2,1,L\rrbracket \) CSS code \NoCaseChange{\protect\cite{cite3383}}.
On a closed orientable surface of genus \(g\), the codespace has dimension \(2^{2g}\), i.e., the code encodes \(k=2g\) logical qubits; such higher-genus surfaces have been investigated \NoCaseChange{\protect\cite{cite3809,cite3810}}.
Planar patches with holes or mixed boundaries can encode multiple logical qubits, and mixed-boundary layouts can reduce planar overhead by about a factor of three compared with punctured square-lattice constructions \NoCaseChange{\protect\cite{cite426}}.
The surface code has also been shown to be resilient to burst errors \NoCaseChange{\protect\cite{cite3811}}.

\subsection{Topological order and gauge-theory analogy}
When treated as ground states of the code Hamiltonian, the code states realize \(\mathbb{Z}_2\) topological order, a topological phase of matter that also exists in \(\mathbb{Z}_2\) lattice gauge theory \NoCaseChange{\protect\cite{cite3527}}.
For sufficiently weak local perturbations on closed surfaces, the splitting of this topological ground-state degeneracy is exponentially small in the shortest linear lattice size \NoCaseChange{\protect\cite{cite423}}.
This order does not persist at nonzero temperature \NoCaseChange{\protect\cite{cite3128,cite3812}}.

Pauli noise operators can be organized into anyonic strings of the gauge theory, which cause excitations of the ground-state subspace.
The inability of local errors to distinguish the codewords translates to the "topologically protected" degeneracy of the ground state, rigorously formulated by the \flmRefsHyperref{ref2675}{TQO-1 condition}.
The joint \(+1\)-eigenspace of the \(Z\)-type Paulis corresponds to the subspace that conserves \(\mathbb{Z}_2\) flux, while the joint \(+1\)-eigenspace of \(X\)-type operators corresponds to the subspace that preserves \(\mathbb{Z}_2\) gauge symmetry (a one-form symmetry).
Logical Pauli operators correspond to non-contractible Wilson loops in the case of closed boundaries, and to paths connecting different types of boundaries in the case of open boundaries.

Behavior under Hamiltonian \(X\)-type and \(Z\)-type perturbations is related to an anisotropic 3D gauge Higgs model \NoCaseChange{\protect\cite{cite3813,cite3814,cite3815,cite3816,cite3817}}.
In order to corrupt logical states, any local noise must bring the code state out of the topological order \NoCaseChange{\protect\cite{cite3812}}.  

Alternatively, there is a general correspondence between stabilizer codes and gauge theory, with the stabilizer group playing the role of the gauge group \NoCaseChange{\protect\cite{cite1365}}.
In this interpretation, both the \(X\) and \(Z\) stabilizers are gauge group elements.

\codefieldsection{Rate}
Both the planar and toric codes saturate the \flmRefsHyperref{ref487}{BPT bound}, which states that \(k d^2 = O(L^2)\) for codes on a 2D lattice of length \(O(L)\).

\codefieldsection{Encoding}
\begin{eczvaluelist}
\item\relax A depth-\(L^2\) circuit that grows the code out of a small patch on an \(L\times L\) square lattice using CNOT gates (i.e., "local moves") \NoCaseChange{\protect\cite{cite480,cite3818}}.
\item\relax Teleportation-based state injection into the planar code \NoCaseChange{\protect\cite{cite3819}}.
\item\relax Graph-state based adaptive circuit \NoCaseChange{\protect\cite{cite3093,cite3820}}.
\item\relax For an \(L\times L\) lattice, deterministic state preparation can be done with a geometrically local unitary \(O(L)\)-depth circuit \NoCaseChange{\protect\cite{cite3821,cite3822}} or an \(O(\log{L})\)-depth unitary circuit with non-local two-qubit gates \NoCaseChange{\protect\cite{cite3818,cite3823,cite3824}} (matching lower bounds \NoCaseChange{\protect\cite{cite3147,cite3148,cite3825}}). The geometric entanglement measure of a ground state of the surface code scales as \flmRefsHyperref{ref65}{order} \(\Omega(L^2)\) \NoCaseChange{\protect\cite{cite3804}}.
\item\relax Stabilizer measurement-based circuit of linear depth \NoCaseChange{\protect\cite{cite480,cite3826}}.
\item\relax Any geometrically local unitary circuit on a lattice \(\Lambda\) that prepares a state whose energy density with respect to the surface code Hamiltonian is \(\epsilon\) must have depth of \flmRefsHyperref{ref65}{order} \(\Omega( \min(\sqrt{|\Lambda|}, 1/\epsilon^{\frac{1-\alpha}{2}}) )\) for any \(\alpha>0\) \NoCaseChange{\protect\cite{cite3827}}.
\item\relax Single-shot state preparation \NoCaseChange{\protect\cite{cite3828}}, with MWPM decoding for such schemes \NoCaseChange{\protect\cite{cite3829}}.
\item\relax Various techniques to generate lattices useful for particular architectures \NoCaseChange{\protect\cite{cite3830}} or removing lattice defects \NoCaseChange{\protect\cite{cite3831,cite3832}} exist.
\item\relax Fault-tolerant constant-depth encoder and unencoder using measurements \NoCaseChange{\protect\cite{cite1436}}.
\end{eczvaluelist}
\codefieldsection{Transversal and Permutation-Based Gates}
\begin{eczvaluelist}
\item\relax Folded surface codes, which are local-Clifford equivalent to triangular color codes with three differently colored boundaries, admit transversal \flmRefsHyperref{ref409}{Clifford gates} \NoCaseChange{\protect\cite{cite422,cite745}}.
\item\relax Fold-transversal initialization of the \(|Y\rangle\) logical state \NoCaseChange{\protect\cite{cite745,cite746,cite747,cite748}}.

\end{eczvaluelist}
\codefieldsection{Gates}
\begin{eczvaluelist}
\item\relax \flmRefsHyperref{ref409}{Clifford gates} can be implemented via lattice surgery
\NoCaseChange{\protect\cite{cite3833,cite3834,cite3835,cite3836}}. Gauging logical operators directly generalizes lattice surgery and recovers conventional surface-code lattice surgery for suitable graph choices \NoCaseChange{\protect\cite{cite470}}.

\item\relax Logical Hadamard gate \NoCaseChange{\protect\cite{cite3837}}.

\item\relax Non-Clifford gates can be implemented using magic-state distillation
\NoCaseChange{\protect\cite{cite3838}}, Dehn twists \NoCaseChange{\protect\cite{cite3437,cite3839}}, or
just-in-time decoding \NoCaseChange{\protect\cite{cite3840,cite3841,cite589}}.

\item\relax Non-stabilizer surface-code states can be prepared by augmenting the code with a quantum double model \NoCaseChange{\protect\cite{cite3842,cite3843}}.

\item\relax ZX calculus \NoCaseChange{\protect\cite{cite3590,cite3591}} can be used to reduce the complexity of surface-code lattice surgery diagrams \NoCaseChange{\protect\cite{cite3844}} and
to reduce \(T\)-gate counts in magic-state distillation protocols \NoCaseChange{\protect\cite{cite3845,cite3846}}.

\item\relax Transversal injection method to prepare non-stabilizer states \NoCaseChange{\protect\cite{cite3847}}.

\item\relax Logical CZ gate from physical CZ gates \NoCaseChange{\protect\cite{cite575,cite3848,cite3450}}, related to the fact that the code admits a cup product structure \NoCaseChange{\protect\cite{cite1517}}.
\item\relax Certain gates can be performed adiabatically \NoCaseChange{\protect\cite{cite3753,cite3754,cite3849}}, yielding an instance of holonomic quantum computation \NoCaseChange{\protect\cite{cite3755}}.
Fault-tolerant gates should be interpretable as monodromies under a particular notion of parallel transport \NoCaseChange{\protect\cite{cite809}}.

\item\relax A combination of fold-transversal gates, Dehn twists and single-shot logical Pauli measurements generates the logical \flmRefsHyperref{ref409}{Clifford group} \NoCaseChange{\protect\cite{cite3850}}.

\item\relax Magic-state cultivation that avoids grafting \NoCaseChange{\protect\cite{cite3851}}.

\end{eczvaluelist}
\codefieldsection{Decoding}
\begin{eczvaluelist}
\item\relax Using data from multiple syndrome measurements prior to decoding allows for correcting syndrome measurement errors. The surface code requires \flmRefsHyperref{ref65}{order} \(O(d)\) extraction rounds in order to gain a reliable estimate. Syndrome measurements are \flmRefsHyperref{ref3496}{distance-preserving} because syndrome extraction circuits can be designed to avoid \flmRefsHyperref{ref3496}{hook errors} \NoCaseChange{\protect\cite{cite480}}.
\item\relax Syndrome extraction circuits consist of CNOT gates and ancillary measurements since this is a stabilizer code \NoCaseChange{\protect\cite{cite2522}}. Measurement schedules can be optimized using spacetime circuit codes to yield what is known as the \textit{3CX surface code} \NoCaseChange{\protect\cite{cite3852}}. Schedules can also be optimized via ZX calculus \NoCaseChange{\protect\cite{cite3590,cite3591}}. Inspired by the honeycomb Floquet code, various weight-two measurement schemes have been designed \NoCaseChange{\protect\cite{cite3758,cite3759,cite3760}}, with the scheme in Ref. \NoCaseChange{\protect\cite{cite3759}} being a special case of DWR.
\item\relax Fault-tolerant syndrome extraction circuits using three-qubit gates \NoCaseChange{\protect\cite{cite3853,cite3854}}.
\item\relax Expanding diamonds decoder correcting errors of some maximum fractal dimension \NoCaseChange{\protect\cite{cite3855}}. The sub-threshold failure probability scales as \((p/p_{\text{th}})^{d^\beta}\), where \(p_{\text{th}}\) is the threshold and \(\beta = \log_3 2\).
\item\relax Minimum weight perfect-matching (MWPM) \NoCaseChange{\protect\cite{cite480,cite3856,cite3857,cite3858}} (based on work by Edmonds on finding a matching in a graph \NoCaseChange{\protect\cite{cite3859,cite3860}}), which takes time up to polynomial in \(n\) for the surface code. For the case of the surface code, minimum-weight decoding reduces to MWPM \NoCaseChange{\protect\cite{cite480,cite3859,cite3861}}. MWPM solves the MPE decoding problem exactly for independent \(X\) and \(Z\) noise. Minimum-weight decoding is \(NP\)-hard for more general Pauli noise and for transversal-CNOT decoding with Pauli-\(Z\) and measurement bit-flip errors \NoCaseChange{\protect\cite{cite3862,cite3420}}. PyMatching is a Python software library for implementing MWPM \NoCaseChange{\protect\cite{cite3863}}.
\item\relax The Bravyi-Suchara-Vargo (BSV) tensor network decoder \NoCaseChange{\protect\cite{cite3864}} exactly solves the ML decoding problem under independent \(X,Z\) noise for the surface code and has complexity of \flmRefsHyperref{ref65}{order} \(O(n^2)\); the decoder provides an efficient tensor-network contraction for the partition function resulting from the statistical mechanical mapping, which is known to be solvable for an Ising model on a planar graph \NoCaseChange{\protect\cite{cite3865}}. ML decoding \NoCaseChange{\protect\cite{cite480}} is \(\#P\)-hard in general for the surface code \NoCaseChange{\protect\cite{cite3862}}.
\item\relax Union-find decoder \NoCaseChange{\protect\cite{cite3866}} uses the \textit{union-find data structure} \NoCaseChange{\protect\cite{cite3867,cite3868,cite3869}}, solving the MPE decoding problem exactly for low-weight errors under depolarizing noise. A subsequent modification utilizes the continuous signal obtained in the physical implementation of the stabilizer measurement (as opposed to discretizing the signal into a syndrome bit) \NoCaseChange{\protect\cite{cite3870}}. Belief union find is a combination of belief-propagation and union-find \NoCaseChange{\protect\cite{cite3871}}. Strictly local (as opposed to partially local) union find \NoCaseChange{\protect\cite{cite3872}} has a worst-case runtime of \flmRefsHyperref{ref65}{order} \(O(d^3)\) in the distance \(d\).
\item\relax Modified MWPM decoders: topological code Autotune \NoCaseChange{\protect\cite{cite3873}}; pipeline MWPM (accounting for correlations between events) \NoCaseChange{\protect\cite{cite3874,cite3875}}; modification tailored to asymmetric noise \NoCaseChange{\protect\cite{cite3876}}; parity blossom MWPM and fusion blossom MWPM \NoCaseChange{\protect\cite{cite3877}}, a modification utilizing the continuous signal obtained in the physical implementation of the stabilizer measurement (as opposed to discretizing the signal into a syndrome bit) \NoCaseChange{\protect\cite{cite3870}}; belief perfect matching (a combination of belief-propagation and MWPM) \NoCaseChange{\protect\cite{cite3871}}; spanning tree matching (STM) and rapid-fire (RFire) decoders \NoCaseChange{\protect\cite{cite3878}}; ordered decoding based on MWPM \NoCaseChange{\protect\cite{cite3879}}; Micro Blossom adapted for a parallelized architecture \NoCaseChange{\protect\cite{cite3880}}; logical observable MWPM and a windowed version \NoCaseChange{\protect\cite{cite3881}}. Combining, or \textit{harmonizing}, various decoders can improve performance \NoCaseChange{\protect\cite{cite3882}}. One such example is the Libra decoder \NoCaseChange{\protect\cite{cite3883}}, a combination of MWPM decoders and matching synthesis.
\item\relax Renormalization group (RG) \NoCaseChange{\protect\cite{cite3884,cite3885,cite3886}}; see Ref. \NoCaseChange{\protect\cite{cite3887}} for the planar surface code.
\item\relax Linear-time ML erasure decoder \NoCaseChange{\protect\cite{cite3465}}.
\item\relax Linear-time decoder for general noise, including coherent noise and correlated noise \NoCaseChange{\protect\cite{cite3888}}.
\item\relax Markov-chain Monte Carlo \NoCaseChange{\protect\cite{cite3889}}.
\item\relax Cellular automaton decoders \NoCaseChange{\protect\cite{cite3890,cite3027,cite3028}}; see also \NoCaseChange{\protect\cite{cite3029}}.
\item\relax Neural network \NoCaseChange{\protect\cite{cite3891,cite3892,cite3893,cite3894,cite3895,cite3896,cite3897}}, reinforcement learning \NoCaseChange{\protect\cite{cite3898,cite3899,cite3900,cite3901}}, and transformer-based \NoCaseChange{\protect\cite{cite3902,cite3903}} decoders like the AlphaQubit series \NoCaseChange{\protect\cite{cite3904,cite3905}}.
\item\relax Lightweight low-latency look-up table (LILLIPUT) decoder for small surface codes \NoCaseChange{\protect\cite{cite3906}}.
\item\relax Decoders can be augmented with a pre-decoder \NoCaseChange{\protect\cite{cite3907,cite3908}}, which can allow for some processing to be done inside the cryogenic environment of the quantum system \NoCaseChange{\protect\cite{cite3909}}.
\item\relax Sliding-window \NoCaseChange{\protect\cite{cite3910,cite3911}}, parallel-window \NoCaseChange{\protect\cite{cite3910}}, and predictive-window \NoCaseChange{\protect\cite{cite3912}} parallelizable decoders, designed to overcome the backlog problem, can be combined with many inner decoders, such as MWPM or union-find.
\item\relax Modifications of BP: generalized belief propagation (GBP) \NoCaseChange{\protect\cite{cite3913}}, based on a classical version \NoCaseChange{\protect\cite{cite3914}}; AMBP4, a quaternary version \NoCaseChange{\protect\cite{cite3739}} of the MBP decoder \NoCaseChange{\protect\cite{cite3740}} of complexity \(O(n\log\log n)\); blockBP, a combination of BP and tensor-network decoders \NoCaseChange{\protect\cite{cite3915}}; machine-learning inspired modifications \NoCaseChange{\protect\cite{cite3916}}. See Ref. \NoCaseChange{\protect\cite{cite3917}} for a review of BP decoders. The min-sum decoder, a simple variant of BP, cannot be used to attain the benefits of codes with distance greater than 9 \NoCaseChange{\protect\cite{cite3918}}.
\item\relax A color-code decoder can be used for the surface code \NoCaseChange{\protect\cite{cite3919}}.
\item\relax Progressive-Proximity Bit-Flipping (PPBF) decoder \NoCaseChange{\protect\cite{cite3920}}.
\item\relax Collision clustering decoder \NoCaseChange{\protect\cite{cite3283}}.
\item\relax Quasi-local Lindbladian decoder based on the approximate Petz theorem \NoCaseChange{\protect\cite{cite3921}}.
\item\relax Exclusive decoder family incorporating post-selection on decoding instances deemed not too difficult \NoCaseChange{\protect\cite{cite3922}}.
\item\relax Quantum version of the Tsirelson local automaton decoder \NoCaseChange{\protect\cite{cite3923}}.
\item\relax Bubble clustering decoder \NoCaseChange{\protect\cite{cite3924}}.
\item\relax Union-Intersection Union-Find (UIUF) decoder \NoCaseChange{\protect\cite{cite3925}}.
\end{eczvaluelist}
\codefieldsection{Fault Tolerance}
\begin{eczvaluelist}
\item\relax Transversal (\flmRefsHyperref{ref409}{non-Clifford}) \(CCZ\) gate by bringing 2D surface codes together and using just-in-time decoding \NoCaseChange{\protect\cite{cite3840,cite3841}}. Gate can be simulated by taking 2D slices out of 3D surface codes \NoCaseChange{\protect\cite{cite3926}}.
\item\relax Flag fault-tolerant syndrome extraction \NoCaseChange{\protect\cite{cite3220}}.
\item\relax Homomorphic measurement protocols for arbitrary surface codes \NoCaseChange{\protect\cite{cite3927}}.
\item\relax Non-geometrically local connectivity can reduce overhead cost \NoCaseChange{\protect\cite{cite3928}}.
\item\relax Magic-state distillation protocols \NoCaseChange{\protect\cite{cite2522,cite3929,cite3930,cite3931}} leading up to magic-state cultivation \NoCaseChange{\protect\cite{cite3932}}.
\item\relax Framework of fault tolerance utilizing ZX calculus \NoCaseChange{\protect\cite{cite3590,cite3591}} that is applicable to MBQC, FBQC, and conventional computation versions of the surface code \NoCaseChange{\protect\cite{cite3933}}.
\item\relax Syndrome extraction circuits consisting of CNOT gates and ancillary measurements \NoCaseChange{\protect\cite{cite2522}}. Measurement schedules can be optimized using spacetime circuit codes to yield what is known as the \textit{3CX surface code} \NoCaseChange{\protect\cite{cite3852}}. Schedules can also be optimized via ZX calculus \NoCaseChange{\protect\cite{cite3590,cite3591}}. Inspired by the honeycomb Floquet code, various weight-two measurement schemes have been designed \NoCaseChange{\protect\cite{cite3758,cite3759,cite3760}}, with the scheme in Ref. \NoCaseChange{\protect\cite{cite3759}} being a special case of DWR.
\item\relax LUCI framework for syndrome extraction circuits \NoCaseChange{\protect\cite{cite3934,cite3935}}.
\item\relax Fault-tolerant constant-depth encoder and unencoder using measurements \NoCaseChange{\protect\cite{cite1436}}.
\end{eczvaluelist}
\codefieldsection{Threshold}
\begin{eczvaluelist}
\item\relax Circuit-level noise: \(1.8\%\) under correlated CNOT-gate errors and single-qubit depolarizing noise \NoCaseChange{\protect\cite{cite3936}} with optimal decoder \NoCaseChange{\protect\cite{cite3937}}, and \(0.35\%\) under independent \(X,Z\) noise with optimal decoder \NoCaseChange{\protect\cite{cite3937}}. Also, \(0.57\%\) for depolarizing noise on data and syndrome qubits as well as initialization, gate, and measurement errors under MWPM decoding \NoCaseChange{\protect\cite{cite2522}}. For this model, a logical qubit with a \(10^{-14}\) logical error rate requires between \(10^3\) to \(10^4\) physical qubits and a target gate fidelity above \(99.9\%\). Later work gave a rigorous threshold proof for arbitrarily reliable computation under local stochastic circuit noise with physical error rate \(p < 7.4\times 10^{-4}\) \NoCaseChange{\protect\cite{cite3938}}. Thresholds of \(0.5-2.9\%\) have been observed for various noise models \NoCaseChange{\protect\cite{cite3939,cite3534,cite3495,cite3940,cite3941,cite3942,cite3937}}. A threshold of \(0.41\%\) when concatenated with the \(\llbracket 4,2,2\rrbracket \) code \NoCaseChange{\protect\cite{cite3289}}. The union-find decoder has a finite threshold under circuit-level local stochastic noise \NoCaseChange{\protect\cite{cite3943}}.
\item\relax Phenomenological noise: \(3.3\%\) for square tiling \NoCaseChange{\protect\cite{cite3550}}, and \(2.93(2)\%\) using several rounds of syndrome measurement \NoCaseChange{\protect\cite{cite3939}}.
\item\relax Fabrication errors \NoCaseChange{\protect\cite{cite3944}}.
\item\relax Quasistatic phase damping and readout noise: \(2.85\%\) \NoCaseChange{\protect\cite{cite3945}}.
\item\relax When used as the underlying code of a surface/Hamming concatenation and benchmarked by a logical CNOT implemented via lattice surgery under circuit-level depolarizing noise, the threshold is \(0.31\%\), and achieving logical CNOT error rate \(10^{-24}\) at physical error rate \(0.1\%\) requires space overhead \(4.5\times 10^3\) \NoCaseChange{\protect\cite{cite3216}}.
\item\relax Thresholds for various measurement schedules, including that of the 3CX surface code, have been obtained \NoCaseChange{\protect\cite{cite3946}}.
\end{eczvaluelist}
\codefieldsection{Realizations}
\begin{eczvaluelist}
\item\relax Signatures of the corresponding topological phase of matter detected in superconducting circuits \NoCaseChange{\protect\cite{cite3947}} and two-dimensional neutral atom arrays \NoCaseChange{\protect\cite{cite3948}}.

\item\relax Measurement schedules associated with the 3CX surface code realized in superconducting qubits on the Willow device by Google Quantum AI \NoCaseChange{\protect\cite{cite3949}}.
\end{eczvaluelist}
\codefieldsection{Notes}
\begin{eczvaluelist}
\item\relax Introduction to computation with the surface code \NoCaseChange{\protect\cite{cite3950,cite3951}}.
\item\relax Tutorials from error-correction perspective by
\flmHref{https://boulderschool.yale.edu/2023/boulder-school-2023-lecture-notes}{A. Kubica} and \flmHref{https://boulderschool.yale.edu/2018/boulder-school-2018-lecture-notes}{J. Haah}
and condensed-matter perspective by
\flmHref{https://boulderschool.yale.edu/2016/boulder-school-2016-lecture-notes}{M. Levin
and C. Nayak}.

\item\relax Review of surface code decoders \NoCaseChange{\protect\cite{cite3952}}.
\item\relax Hardware requirements for implementing surface code QEC can be reduced by utilizing structure in the time slices of the QEC circuits \NoCaseChange{\protect\cite{cite3852}}. Various optimization and calibration routines exist \NoCaseChange{\protect\cite{cite3953}}.

\item\relax A database of surface codes is available in QECDB \NoCaseChange{\protect\cite{cite781}}.
\end{eczvaluelist}
\codefieldsection{Parents}
\begin{eczvaluelist}
\item\relax
\flmRefsHyperref[eczindexfamilyrel]{code:higher_dimensional_surface}{Homological code} --- The surface-code CSS stabilizer generator prescription is extendable to higher-dimensional manifolds.
\item\relax
\flmRefsHyperref[eczindexfamilyrel]{code:twist_defect_surface}{Twist-defect surface code} --- Twist-defect surface codes reduce to surface codes when there are no defects.
\item\relax
\flmRefsHyperref[eczindexfamilyrel]{code:clifford-deformed_surface}{Clifford-deformed surface code (CDSC)} --- CDSC codes are deformations of the surface code via constant-depth \flmRefsHyperref{ref409}{Clifford circuits} that may not be CSS.
\item\relax
\flmRefsHyperref[eczindexfamilyrel]{code:lcs}{Lift-connected surface (LCS) code} --- LCS codes consist of sparsely interconnected stacks of surface codes.
\item\relax
\flmRefsHyperref[eczindexfamilyrel]{code:qudit_surface}{Modular-qudit surface code} --- The modular-qudit surface code for \(q=2\) reduces to the surface code.
\item\relax
\flmRefsHyperref[eczindexfamilyrel]{code:galois_topological}{Galois-qudit surface code} --- The Galois-qudit surface code for \(q=2\) reduces to the surface code.
\end{eczvaluelist}
\codefieldsection{Children}
\begin{eczvaluelist}
\item\relax
\flmRefsHyperref[eczindexfamilyrel]{code:klein_bottle}{Klein-bottle surface code} --- The Klein bottle surface code is the surface code on a Klein bottle.
\item\relax
\flmRefsHyperref[eczindexfamilyrel]{code:real_projective_plane}{Projective-plane surface code} --- The projective-plane surface code is the surface code on \(\mathbb{R}P^2\).
\item\relax
\flmRefsHyperref[eczindexfamilyrel]{code:rotated_surface}{Rotated surface code} --- The lattice of the rotated surface code can be obtained by taking the medial graph of the surface code lattice (treated as a graph) and applying a procedure to construct the check operators \NoCaseChange{\protect\cite{cite425,cite426}\protect\cite[{Fig. 8}]{cite427}}. Applying the quantum Tanner transformation to the surface code yields the rotated surface code \NoCaseChange{\protect\cite{cite3954,cite3955}}. The rotated surface code presents certain savings over the original surface code \NoCaseChange{\protect\cite{cite3956}}.
\item\relax
\flmRefsHyperref[eczindexfamilyrel]{code:toric}{Toric code} --- The toric code is the surface code on a 2D torus.
\item\relax
\flmRefsHyperref[eczindexfamilyrel]{code:xysurface}{XY surface code} --- The XY surface code is obtained from the surface code by applying \(H\sqrt{Z}H\) to all qubits, thereby exchanging \(Z\leftrightarrow Y\). While it is equivalent to a CSS surface code with the same distance, but other properties like noise-bias performance can differ significantly.
\end{eczvaluelist}
\codefieldsection{Cousins}
\begin{eczvaluelist}
\item\relax
\flmRefsHyperref[eczindexfamilyrel]{code:layer}{Layer code} --- Layer codes are combinations of constant-rate QLDPC codes with surface codes built using lattice surgery.
\item\relax
\flmRefsHyperref[eczindexfamilyrel]{code:lresc}{Long-range enhanced surface code (LRESC)} --- LRESCs reduce to planar surface codes when a trivial LDPC code is used in the hypergraph product.
\item\relax
\flmRefsHyperref[eczindexfamilyrel]{code:lacross}{La-cross code} --- La-cross codes with periodic (open) boundary conditions reduce to the toric (planar surface) code at \(k=1\).
\item\relax
\flmRefsHyperref[eczindexfamilyrel]{code:quantum_double}{Quantum-double code} --- On closed surfaces, a quantum-double model with \(G=\mathbb{Z}_2\) reduces to the surface code; on a torus, this is the toric code. Quantum doubles with open boundary conditions also reduce to surface codes on open surfaces \NoCaseChange{\protect\cite{cite3957,cite3958,cite3959,cite3960,cite730}}. Non-stabilizer surface-code states can be prepared by augmenting the surface code with a quantum double model \NoCaseChange{\protect\cite{cite3842,cite3843,cite3961}}.
\item\relax
\flmRefsHyperref[eczindexfamilyrel]{code:hamiltonian}{Hamiltonian-based code} --- While codewords of the surface code form ground states of the code's stabilizer Hamiltonian, they can also be ground states of other gapless Hamiltonians \NoCaseChange{\protect\cite{cite2847}}.
\item\relax
\flmRefsHyperref[eczindexfamilyrel]{code:unitary_design}{Unitary \(t\)-design} --- Unitary \(t\)-designs can be generated via coherent errors, syndrome extraction, and correction \NoCaseChange{\protect\cite{cite2197}}.
\item\relax
\flmRefsHyperref[eczindexfamilyrel]{code:hypergraph_product}{Hypergraph product (HGP) code} --- The planar surface code on a square lattice can be obtained from a hypergraph product of two repetition codes with appropriate boundary checks.
\item\relax
\flmRefsHyperref[eczindexfamilyrel]{code:repetition}{Repetition code} --- The planar surface code on a square lattice can be obtained from a hypergraph product of two repetition codes with appropriate boundary checks.
\item\relax
\flmRefsHyperref[eczindexfamilyrel]{code:2d_stabilizer}{2D lattice stabilizer code} --- Translation-invariant 2D qubit lattice stabilizer codes are equivalent to several copies of the Kitaev surface code via a local constant-depth qudit Clifford circuit \NoCaseChange{\protect\cite{cite603,cite3962,cite3963}}.
\item\relax
\flmRefsHyperref[eczindexfamilyrel]{code:asymmetric_qecc}{Asymmetric quantum code (AQC)} --- The surface code on the honeycomb tiling is an asymmetric CSS code \NoCaseChange{\protect\cite{cite2637}}.
\item\relax
\flmRefsHyperref[eczindexfamilyrel]{code:quantum_lego}{Tensor-network code} --- Planar surface codes arise from finite patches of the toric-code tensor network by contracting boundary legs with stopper tensors or repetition-code boundary tensors \NoCaseChange{\protect\cite{cite2868}}. The 2D Bacon-Shor code can also be obtained from a surface-code tensor network by reassigning every other row of dangling physical legs to logical legs; in this quantum-Lego picture, the gauge generators remain weight-two \(XX\) and \(ZZ\) operators and the construction makes explicit a connection to the quantum compass model \NoCaseChange{\protect\cite{cite2868}}.
\item\relax
\flmRefsHyperref[eczindexfamilyrel]{code:spt}{Symmetry-protected topological (SPT) code} --- Gauging \NoCaseChange{\protect\cite{cite462,cite463,cite233,cite464,cite465,cite466,cite467,cite468,cite469,cite470}} the symmetry of a trivial 2D bosonic \(\mathbb{Z}_2\) Ising SPT yields the surface-code phase \NoCaseChange{\protect\cite[{Sec. IV}]{cite462}}.
\item\relax
\flmRefsHyperref[eczindexfamilyrel]{code:self_correct}{Self-correcting quantum code} --- The surface code is not thermally stable \NoCaseChange{\protect\cite{cite3006,cite3007,cite3008,cite3009,cite3010}} because its string-like logical operators anti-commute with stabilizer generators supported only at their ends, and thus have a constant energy cost of creation. Various candidates for self-correcting quantum memories have been constructed by coupling neighboring anyons in the code so as to prevent them from spreading \NoCaseChange{\protect\cite{cite3023,cite3024,cite3025,cite3026,cite3027,cite3028,cite3029,cite3030}}.
\item\relax
\flmRefsHyperref[eczindexfamilyrel]{code:da_color_2d}{2D DA color code} --- One of the instantaneous stabilizer groups of the 2D DA color code is that of stacks of surface codes \NoCaseChange{\protect\cite[{Sec. III.A}]{cite2532}}.
\item\relax
\flmRefsHyperref[eczindexfamilyrel]{code:floquet_color}{Floquet color code} --- The ISG of the Floquet color code is the stabilizer group of one of nine realizations of the \(\mathbb{Z}_2\) 2D surface code.
\item\relax
\flmRefsHyperref[eczindexfamilyrel]{code:honeycomb_floquet}{Honeycomb Floquet code} --- Measurement of each check operator of the honeycomb Floquet code involves two qubits and projects the state of the two qubits to a two-dimensional subspace, which we regard as an effective qubit. 
These effective qubits form a surface code on an enlarged honeycomb tiling \NoCaseChange{\protect\cite[{Fig. 2}]{cite536}}.
Electric and magnetic operators on the embedded surface code correspond to outer logical operators of the Floquet code.
In fact, outer logical operators transition back and forth from magnetic to electric surface code operators under the measurement dynamics.
Inspired by the honeycomb Floquet code, various weight-two measurement schemes have been designed \NoCaseChange{\protect\cite{cite3758,cite3759,cite3760}}, with the scheme in Ref. \NoCaseChange{\protect\cite{cite3759}} being a special case of DWR.
Numerical comparisons have been performed \NoCaseChange{\protect\cite{cite3761}}.

\item\relax
\flmRefsHyperref[eczindexfamilyrel]{code:floquet_xcube}{X-cube Floquet code} --- The rewinding schedule has rounds whose ISGs include a 3-foliated stack of 2D surface codes as an FDLQC-equivalent factor \NoCaseChange{\protect\cite{cite533}}.
\item\relax
\flmRefsHyperref[eczindexfamilyrel]{code:spacetime_circuit}{Spacetime circuit code} --- Stabilizer generators of a spacetime code are called \textit{detectors} in Refs. \NoCaseChange{\protect\cite{cite3964,cite667}}.
\item\relax
\flmRefsHyperref[eczindexfamilyrel]{code:majorana_surface}{Majorana surface code} --- Majorana surface codes map non-uniquely to bosonic surface codes: replacing each tetron in a 4.8.8 code by a qubit yields the square-lattice surface code, while 6.6.6 and 4.6.12 codes map to rotated-square and Kagome-lattice surface-code realizations, respectively \NoCaseChange{\protect\cite{cite402}}.
\item\relax
\flmRefsHyperref[eczindexfamilyrel]{code:hexagonal_cz}{Hexagonal \(CZ\) code} --- The hexagonal \(CZ\) code can be obtained from two surface codes by gauging \NoCaseChange{\protect\cite{cite462,cite463,cite233,cite464,cite465,cite466,cite467,cite468,cite469,cite470}} their logical \(CZ\) gate \NoCaseChange{\protect\cite{cite589}}. Gates on the two surface codes in the third level of the Clifford hierarchy, such as \(CZ\) gates, can be realized fault-tolerantly by performing this procedure and reversing it \NoCaseChange{\protect\cite{cite572,cite589}}.
\item\relax
\flmRefsHyperref[eczindexfamilyrel]{code:double_semion_string_net}{Double-semion string-net code} --- There is a logical basis for both the toric and double-semion string-net codes where each codeword is a superposition of states corresponding to all noncontractible loops of a particular homotopy type. The superposition is equal for the toric code, whereas an odd number of loops appear with a \(-1\) coefficient for the double semion.
\item\relax
\flmRefsHyperref[eczindexfamilyrel]{code:reinforcement_learning}{Reinforcement-learning quantum code} --- Reinforcement learners can be used to optimize the geometry of the surface code to be more suited to a noise channel \NoCaseChange{\protect\cite{cite3965}}.
\item\relax
\flmRefsHyperref[eczindexfamilyrel]{code:stab_4_2_2}{\(\llbracket 4,2,2\rrbracket \) Four-qubit code} --- Concatenating the \(\llbracket 4,2,2\rrbracket \) code with the surface code is equivalent to removing stabilizer generators from the 4.8.8 color code \NoCaseChange{\protect\cite{cite3289}}.
\item\relax
\flmRefsHyperref[eczindexfamilyrel]{code:2d_bosonization}{2D bosonization code} --- The original 2D bosonization code \NoCaseChange{\protect\cite{cite403}} is a stabilizer code whose generators are products of plaquettes and stars of the surface code.
\item\relax
\flmRefsHyperref[eczindexfamilyrel]{code:haah_cubic}{Haah cubic code (CC)} --- Under renormalization group flow \NoCaseChange{\protect\cite{cite3696}}, cubic codes 11-17 fragment into combinations of themselves, their corresponding B-codes, and stacks of surface codes \NoCaseChange{\protect\cite[{Table 1}]{cite3697}}.
\item\relax
\flmRefsHyperref[eczindexfamilyrel]{code:xcube}{X-cube model code} --- The X-cube model can be constructed by coupling layers of the surface code \NoCaseChange{\protect\cite{cite534,cite3164}}.
\item\relax
\flmRefsHyperref[eczindexfamilyrel]{code:cluster_state}{Cluster-state code} --- Foliating the surface code yields a cluster state on the Lieb lattice \NoCaseChange{\protect\cite{cite3576,cite469}}. See also the discussion of the embedding on \NoCaseChange{\protect\cite[{pg. 8}]{cite423}}.
\item\relax
\flmRefsHyperref[eczindexfamilyrel]{code:rbh}{Raussendorf-Bravyi-Harrington (RBH) cluster-state code} --- The RBH state encodes the temporal gate operations on the surface code into a third spatial dimension \NoCaseChange{\protect\cite{cite3533,cite3534}}. In addition, one possible 2D boundary of the RBH code is effectively a 2D toric code.
\item\relax
\flmRefsHyperref[eczindexfamilyrel]{code:gross}{\(\llbracket 144,12,12\rrbracket \) gross code} --- The gross code requires less physical and ancilla qubits (for syndrome extraction) than the surface code with the same number of logical qubits and distance. The gross code is equivalent to 8 copies of the surface code via a constant-depth Clifford circuit, and is an element of a larger family of 2D stabilizer codes \NoCaseChange{\protect\cite{cite443}}. An architecture combining the surface and gross codes was proposed in \NoCaseChange{\protect\cite{cite3195}}.
\item\relax
\flmRefsHyperref[eczindexfamilyrel]{code:quantum_hamming_css}{\(\llbracket 2^r-1, 2^r-2r-1, 3\rrbracket \) quantum Hamming code} --- Quantum Hamming codes can be concatenated with surface codes \NoCaseChange{\protect\cite{cite3218}}. In a unified logical-CNOT comparison under circuit-level depolarizing noise, using the surface code as the underlying code gives a \(0.31\%\) threshold and requires space overhead \(4.5\times 10^3\) at physical error rate \(0.1\%\) to achieve logical CNOT error rate \(10^{-24}\), compared to \(3.7\times 10^2\) for the optimized \(C_4/C_6\)/Hamming construction \NoCaseChange{\protect\cite{cite3216}}.
\item\relax
\flmRefsHyperref[eczindexfamilyrel]{code:2d_color}{2D color code} --- On closed surfaces, the 2D color code is equivalent to two decoupled copies of the 2D toric/surface code via a local constant-depth \flmRefsHyperref{ref409}{Clifford circuit} \NoCaseChange{\protect\cite{cite3424,cite422,cite3425}} and has the same topological entanglement entropy \NoCaseChange{\protect\cite{cite3426}}. For triangular patches with three differently colored boundaries, it is instead equivalent to a folded surface/toric code with two smooth and two rough boundaries \NoCaseChange{\protect\cite{cite422}}. The conversion process can be viewed as an ungauging \NoCaseChange{\protect\cite{cite462,cite463,cite233,cite464,cite465,cite466,cite467,cite468,cite469,cite470}} of certain symmetries. Conversely, the 2D color code can \flmRefsHyperref{ref410}{condense} to form the 2D surface code in nine different ways, i.e., by adding two-body hopping terms along one of its three triangular directions to the stabilizer group and then taking the center of the resulting nonabelian group \NoCaseChange{\protect\cite{cite2526}}. Both the surface and 2D color codes can be constructed from two distinct types of lattices, namely, 4-valent and 3-valent 3-colorable lattices, respectively \NoCaseChange{\protect\cite{cite3427}}.
\item\relax
\flmRefsHyperref[eczindexfamilyrel]{code:4d_surface}{\((2,2)\) Loop toric code} --- Setting \(L_2=L_4=1\) in the open-boundary tesseract construction yields the planar surface code \NoCaseChange{\protect\cite{cite3174}}.
\item\relax
\flmRefsHyperref[eczindexfamilyrel]{code:xzzx}{XZZX surface code} --- The XZZX surface code on a square lattice with non-twisted periodic boundary conditions is obtained from a surface code by applying Hadamard gates on a subset of qubits such that \(XXXX\) and \(ZZZZ\) generators are both mapped to \(XZXZ\). While this code is equivalent to a CSS surface code with the same distance, other properties like noise-bias performance can differ significantly. Twisted XZZX surface codes are generally not equivalent to CSS surface codes via a single-qubit Clifford circuit and permutation.
\item\relax
\flmRefsHyperref[eczindexfamilyrel]{code:bacon_shor}{Bacon-Shor code} --- The 2D Bacon-Shor code can also be obtained from a surface-code tensor network by reassigning every other row of dangling physical legs to logical legs; in this quantum-Lego picture, the gauge generators remain weight-two \(XX\) and \(ZZ\) operators and the construction makes explicit a connection to the quantum compass model \NoCaseChange{\protect\cite{cite2868}}.
\item\relax
\flmRefsHyperref[eczindexfamilyrel]{code:kitaev_honeycomb}{Kitaev honeycomb code} --- The Kitaev honeycomb code can be obtained from the square-lattice surface code by \flmRefsHyperref{ref666}{gauging out} the anyon \(em\) \NoCaseChange{\protect\cite[{Sec. 7.3}]{cite414}}. During this process, the square lattice is effectively expanded to a honeycomb tiling \NoCaseChange{\protect\cite[{Fig. 12}]{cite414}}.
\item\relax
\flmRefsHyperref[eczindexfamilyrel]{code:heavy_hex}{Heavy-hexagon code} --- Surface code stabilizers are used to measure the Z-type stabilizers of the code. There are various ways to embed the surface code into the heavy-hex lattice \NoCaseChange{\protect\cite{cite3713}}.
\item\relax
\flmRefsHyperref[eczindexfamilyrel]{code:subsystem_three_fermion}{Three-fermion (3F) subsystem code} --- One version of the 3F subsystem code can be obtained from two copies of the square-lattice surface code by \flmRefsHyperref{ref666}{gauging out} the anyons \(e_1m_1e_2\) and \(e_2m_2\) \NoCaseChange{\protect\cite[{Sec. 7.4}]{cite414}}.
\item\relax
\flmRefsHyperref[eczindexfamilyrel]{code:subsystem_surface}{Subsystem surface code} --- Subsystem surface codes are subsystem versions of surface codes.
\item\relax
\flmRefsHyperref[eczindexfamilyrel]{code:fracton}{Fracton stabilizer code} --- Foliated type-I fracton phase codes can be grown by \textit{foliation}, i.e., stacking copies of the 2D surface code; see \NoCaseChange{\protect\cite[{Eq. (D32)}]{cite456}}.
\item\relax
\flmRefsHyperref[eczindexfamilyrel]{code:generalized_bicycle}{Generalized bicycle (GB) code} --- Any non-trivial qubit GB code of row weight four and distance \(d\geq 3\) is equivalent to a square-lattice surface code \NoCaseChange{\protect\cite{cite3183}}.
\end{eczvaluelist}
\eczhbkcontributors{ Shouzhen (Bailey) Gu, Marcus P da Silva, Tony Lau, Hassan Shapourian, Michael Vasmer, \eczhuPhF, \eczhuVVA }
\endeczcode

\eczcode{klein_bottle}{Klein-bottle surface code}{~\NoCaseChange{\protect\cite{cite3966}}}
\codefieldsection{Description}
A family of Kitaev surface codes on the non-orientable Klein bottle.
\codefieldsection{Parent}
\begin{eczvaluelist}
\item\relax
\flmRefsHyperref[eczindexfamilyrel]{code:surface}{Kitaev surface code} --- The Klein bottle surface code is the surface code on a Klein bottle.
\end{eczvaluelist}
\codefieldsection{Cousin}
\begin{eczvaluelist}
\item\relax
\flmRefsHyperref[eczindexfamilyrel]{code:3d_surface}{3D surface code} --- There is a CZ gate for the 3D toric code on a Klein bottle \(\times S^1\) \NoCaseChange{\protect\cite{cite3459}}.
\end{eczvaluelist}
\eczhbkcontributors{ \eczhuVVA }
\endeczcode

\eczcode{lacross}{La-cross code}{~\NoCaseChange{\protect\cite{cite3967}}}
\codefieldsection{Description}
Code constructed using a hypergraph product of two copies of a cyclic LDPC code.
The construction uses cyclic LDPC codes with \flmRefsHyperref{ref67}{generating polynomials} \(1+x+x^k\) for some \(k\).
Using a length-\(n\) seed code yields an \(\llbracket 2n^2,2k^2\rrbracket \) family for periodic boundary conditions and an \(\llbracket (n-k)^2+n^2,k^2\rrbracket \) family for open boundary conditions.
\codefieldsection{Parents}
\begin{eczvaluelist}
\item\relax
\flmRefsHyperref[eczindexfamilyrel]{code:hypergraph_product}{Hypergraph product (HGP) code} --- La-cross codes are constructed using a hypergraph product of a cyclic LDPC code with itself.
\item\relax
\flmRefsHyperref[eczindexfamilyrel]{code:quantum_cyclic}{Cyclic quantum code}\end{eczvaluelist}
\codefieldsection{Cousins}
\begin{eczvaluelist}
\item\relax
\flmRefsHyperref[eczindexfamilyrel]{code:qc_ldpc}{Quasi-cyclic LDPC (QC-LDPC) code} --- La-cross codes are constructed using a hypergraph product of a cyclic LDPC code with itself.
\item\relax
\flmRefsHyperref[eczindexfamilyrel]{code:cyclic_hgp}{Cyclic Hypergraph Product Code} --- The La-cross code is a reduced block length, full-rank cyclic HGP code with generator polynomials of the form \(1+x+x^k\)
\item\relax
\flmRefsHyperref[eczindexfamilyrel]{code:lresc}{Long-range enhanced surface code (LRESC)} --- La-cross codes yield LRESCs for \(k=2\). La-cross codes have a number of long-range stabilizers that scales linearly with code size, while the number of LRESC long-range stabilizers can be tuned to scale between the square-root of the size and linearly in the size.
\item\relax
\flmRefsHyperref[eczindexfamilyrel]{code:surface}{Kitaev surface code} --- La-cross codes with periodic (open) boundary conditions reduce to the toric (planar surface) code at \(k=1\).
\end{eczvaluelist}
\eczhbkcontributors{ \eczhuVVA }
\endeczcode

\eczcode{ladder}{Ladder Floquet code}{~\NoCaseChange{\protect\cite{cite536}}}
\codefieldsection{Description}
1D Floquet code defined on a ladder qubit geometry, with one qubit per vertex.
The check operators consist of \(ZZ\) checks on each rung and alternating \(XX\) and \(YY\) checks on the legs.
The period-four measurement schedule measures \(ZZ\), \(XX\), \(ZZ\), and \(YY\) in rounds \(0,1,2,3\) mod \(4\), respectively, dynamically generating one logical qubit \NoCaseChange{\protect\cite{cite536}}.

\codefieldsection{Protection}
For \(r\geq 4\), the ISG is generated by plaquette stabilizers on the ladder squares together with the most recently measured checks. The code has effective distance \(2\): it detects all single-qubit Pauli errors, and chains of check errors produce syndrome defects only at their endpoints, analogously to a repetition code \NoCaseChange{\protect\cite{cite536}}.
\codefieldsection{Decoding}
\begin{eczvaluelist}
\item\relax Syndrome bits are the plaquette-stabilizer eigenvalues, recorded every time they can be inferred from consecutive rounds of measurements \NoCaseChange{\protect\cite{cite536}}.
\item\relax Period-four measurement sequence utilizing two-qubit measurements \NoCaseChange{\protect\cite{cite536}}.
\end{eczvaluelist}
\codefieldsection{Parent}
\begin{eczvaluelist}
\item\relax
\flmRefsHyperref[eczindexfamilyrel]{code:floquet}{Hastings-Haah Floquet code} --- The ladder Floquet code is the first 1D Floquet code.
\end{eczvaluelist}
\codefieldsection{Cousins}
\begin{eczvaluelist}
\item\relax
\flmRefsHyperref[eczindexfamilyrel]{code:stab_4_2_2}{\(\llbracket 4,2,2\rrbracket \) Four-qubit code} --- The smallest example of the ladder Floquet code is a dynamical version of the \(\llbracket 4,2,2\rrbracket \) code \NoCaseChange{\protect\cite{cite3291,cite3247}}. The \(\llbracket 4,2,2\rrbracket \) code can be Floquetified in various ways \NoCaseChange{\protect\cite{cite3292,cite3293}}.
\item\relax
\flmRefsHyperref[eczindexfamilyrel]{code:stab_6_2_2}{\(\llbracket 6,2,2\rrbracket \) \(C_6\) code} --- The ISG of the ladder code includes \(Z\)-type Pauli products around squares of the qubit ladder. These are also included in the checks of the \(C_6\) code.
\end{eczvaluelist}
\eczhbkcontributors{ \eczhuVVA }
\endeczcode

\eczcode{layer}{Layer code}{~\NoCaseChange{\protect\cite{cite3040}}}
\codefieldsection{Description}
Member of a family of qubit QLDPC CSS codes with stabilizer generator weights \(\leq 6\) that are obtained by coupling layers of 2D surface codes according to the Tanner graph of a QLDPC code (or a more general qubit stabilizer code).
Geometric locality is maintained because, instead of being concatenated, each pair of parallel surface-code squares is fused (or quasi-concatenated) with perpendicular surface-code squares via lattice surgery.

\codefieldsection{Rate}
Layer codes achieve the 3D \flmRefsHyperref{ref487}{BPT bound}, with parameters  \(\llbracket n,\Theta(n^{1/3}),\Theta(n^{1/3})\rrbracket \), when asymptotically good QLDPC codes are used in the construction.
\codefieldsection{Decoding}
\begin{eczvaluelist}
\item\relax Decoders against stochastic and adversarial noise \NoCaseChange{\protect\cite{cite3041}}.
\end{eczvaluelist}
\codefieldsection{Parents}
\begin{eczvaluelist}
\item\relax
\flmRefsHyperref[eczindexfamilyrel]{code:qubit_css}{Qubit CSS code}\item\relax
\flmRefsHyperref[eczindexfamilyrel]{code:qldpc}{Qubit QLDPC code} --- Layer codes are constructed by coupling layers of 2D surface codes according to the Tanner graph of a QLDPC code.
\end{eczvaluelist}
\codefieldsection{Cousins}
\begin{eczvaluelist}
\item\relax
\flmRefsHyperref[eczindexfamilyrel]{code:topological_abelian}{Abelian topological code} --- The Layer code realizes 2D layers of \(\mathbb{Z}_2\) gauge theory coupled along defects.
\item\relax
\flmRefsHyperref[eczindexfamilyrel]{code:fracton}{Fracton stabilizer code} --- Layer codes are non-translation invariant 3D lattice stabilizer codes that can be viewed as fracton topological defect networks \NoCaseChange{\protect\cite{cite3040}}.
\item\relax
\flmRefsHyperref[eczindexfamilyrel]{code:good_qldpc}{Good QLDPC code} --- Layer codes achieve the 3D \flmRefsHyperref{ref487}{BPT bound}, with parameters  \(\llbracket n,\Theta(n^{1/3}),\Theta(n^{1/3})\rrbracket \), when asymptotically good QLDPC codes are used in the construction.
\item\relax
\flmRefsHyperref[eczindexfamilyrel]{code:qubit_concatenated}{Concatenated qubit code} --- Each pair of surface-code squares in a layer code is fused (or quasi-concatenated) with perpendicular surface-code squares via lattice surgery.
\item\relax
\flmRefsHyperref[eczindexfamilyrel]{code:self_correct}{Self-correcting quantum code} --- The energy barrier of excitations for layer codes constructed using asymptotically good QLDPC codes scales as \flmRefsHyperref{ref65}{order} \(\Theta(n^{1/3})\) \NoCaseChange{\protect\cite{cite3040}}. Layer codes are partially self-correcting quantum memories \NoCaseChange{\protect\cite{cite3041,cite3042}}. Layer codes constructed from random CSS codes have near-optimal scaling of code parameters and a polynomial energy barrier, exhibiting behavior consistent with partial self-correction \NoCaseChange{\protect\cite{cite3041}}.
\item\relax
\flmRefsHyperref[eczindexfamilyrel]{code:surface}{Kitaev surface code} --- Layer codes are combinations of constant-rate QLDPC codes with surface codes built using lattice surgery.
\end{eczvaluelist}
\eczhbkcontributors{ \eczhuVVA }
\endeczcode

\eczcode{lcs}{Lift-connected surface (LCS) code}{~\NoCaseChange{\protect\cite{cite3968}}}
\codefieldsection{Description}
Member of one of several families of lifted-product codes that consist of sparsely interconnected stacks of surface codes.

\codefieldsection{Protection}
Code distance is linear with \(n\) up to some maximum number of qubits.

\codefieldsection{Code Capacity Threshold}
\begin{eczvaluelist}
\item\relax \(6.7\%\) and \(7.7\%\) under bit-flip noise and BP+OSD decoding for two families of LCS codes.
\end{eczvaluelist}
\codefieldsection{Threshold}
\begin{eczvaluelist}
\item\relax \(2.9\%\) and \(3.2\%\) under phenomenological noise and BP+OSD decoding for two families of LCS codes.
\end{eczvaluelist}
\codefieldsection{Parents}
\begin{eczvaluelist}
\item\relax
\flmRefsHyperref[eczindexfamilyrel]{code:qubit_generalized_homological_product_css}{Generalized homological-product qubit CSS code}\item\relax
\flmRefsHyperref[eczindexfamilyrel]{code:lifted_product}{Lifted-product (LP) code}\end{eczvaluelist}
\codefieldsection{Child}
\begin{eczvaluelist}
\item\relax
\flmRefsHyperref[eczindexfamilyrel]{code:surface}{Kitaev surface code} --- LCS codes consist of sparsely interconnected stacks of surface codes.
\end{eczvaluelist}
\eczhbkcontributors{ \eczhuVVA }
\endeczcode

\eczcode{local_haar_random}{Local Haar-random circuit qubit code}{~\NoCaseChange{\protect\cite{cite2196}}}
\codefieldsection{Description}
An \(n\)-qubit code whose codewords are a pair of approximately locally indistinguishable states produced by starting with any two orthogonal \(n\)-qubit states and acting with a random unitary circuit of depth polynomial in \(n\).
Two states are \textit{locally indistinguishable} if they cannot be distinguished by local measurements. A single layer of the encoding circuit is composed of about \(n/2\) two-qubit nearest-neighbor gates run in parallel, with each gate drawn randomly from the Haar distribution on two-qubit unitaries.

The above circuit elements act on nearest-neighbor qubits arranged in a line, i.e., a 1D geometry (\(D=1\)); codes for higher-dimensional geometries require \(O(n^{1/D})\)-depth circuits \NoCaseChange{\protect\cite{cite2196}}. Follow-up work \NoCaseChange{\protect\cite{cite3614}} showed that, at the erasure threshold, 1D random circuits require \(O(\sqrt{n})\) depth, whereas dimensions \(D \geq 2\) retain \(O(\log n)\)-depth scaling.
This result has in turn been extended to other types of Pauli noise \NoCaseChange{\protect\cite{cite3969}}, while the previous result applies to erasure noise.

\codefieldsection{Protection}
In a 1D geometry, the local Haar-random circuit qubit code approximately detects any error with support on a segment of length \(\leq n/4\), with deviations exponentially suppressed in \(n\). 
There is a phase transition in error correction power vs error rate \(p\), with a critical depth of \flmRefsHyperref{ref65}{order} \(O(1/p)\) \NoCaseChange{\protect\cite{cite3970}}.

\codefieldsection{Encoding}
\begin{eczvaluelist}
\item\relax Random local circuit of depth proportional to \(n^{\alpha}\), with \(\alpha\) depending on system geometry.
\end{eczvaluelist}
\codefieldsection{Parents}
\begin{eczvaluelist}
\item\relax
\flmRefsHyperref[eczindexfamilyrel]{code:qubits_into_qubits}{Qubit code}\item\relax
\flmRefsHyperref[eczindexfamilyrel]{code:random_circuit}{Random-circuit code}\end{eczvaluelist}
\codefieldsection{Cousins}
\begin{eczvaluelist}
\item\relax
\flmRefsHyperref[eczindexfamilyrel]{code:topological}{Topological code} --- Local Haar-random codewords, like topological codewords, are locally indistinguishable \NoCaseChange{\protect\cite{cite2196}}.
\item\relax
\flmRefsHyperref[eczindexfamilyrel]{code:unitary_design}{Unitary \(t\)-design} --- Local Haar-random circuits of polynomial depth form approximate unitary designs \NoCaseChange{\protect\cite{cite2196}}.
\item\relax
\flmRefsHyperref[eczindexfamilyrel]{code:1d_stabilizer}{1D lattice stabilizer code} --- In a 1D geometry, the local Haar-random circuit qubit code approximately detects any error with support on a segment of length \(\leq n/4\), with deviations exponentially suppressed in \(n\). There is a phase transition in error correction power vs error rate \(p\), with a critical depth of \flmRefsHyperref{ref65}{order} \(O(1/p)\) \NoCaseChange{\protect\cite{cite3970}}.
\item\relax
\flmRefsHyperref[eczindexfamilyrel]{code:haar_random}{Haar-random qubit code} --- Approximating the random projections through \(t\)-designs is necessary in order to make the Haar-random qubit protocol practical. Replacing with random \flmRefsHyperref{ref409}{Clifford gates} is especially convenient since the \flmRefsHyperref{ref409}{Clifford group} forms a unitary 2-design and produces stabilizer codes.
\end{eczvaluelist}
\eczhbkcontributors{ Jonathan Kunjummen, \eczhuVVA }
\endeczcode

\eczcode{approximate_log_depth}{Log-depth geometrically local Clifford-circuit code}{~\NoCaseChange{\protect\cite{cite3971}}}
\codefieldsection{Description}
A random \(\llbracket n,k\rrbracket \) stabilizer code whose encoder is a random \flmRefsHyperref{ref409}{Clifford circuit} of depth of \flmRefsHyperref{ref65}{order} \(O(\log n)\) on a 1D Euclidean geometry.

\codefieldsection{Rate}
Log-depth \flmRefsHyperref{ref409}{Clifford circuits} on a 1D geometry yield approximate codes whose encoding rate achieves the hashing bound for Pauli noise and the channel capacity for erasure errors \NoCaseChange{\protect\cite{cite3972,cite3973}}.
\codefieldsection{Encoding}
\begin{eczvaluelist}
\item\relax Random \(\log\)-depth \flmRefsHyperref{ref409}{Clifford circuit} on a 1D Euclidean geometry.
\end{eczvaluelist}
\codefieldsection{Decoding}
\begin{eczvaluelist}
\item\relax Minimum-weight decoding using tropical tensor networks \NoCaseChange{\protect\cite{cite3971}}.
\end{eczvaluelist}
\codefieldsection{Fault Tolerance}
\begin{eczvaluelist}
\item\relax Fault-tolerant state preparation \NoCaseChange{\protect\cite{cite3971}}.
\end{eczvaluelist}
\codefieldsection{Parents}
\begin{eczvaluelist}
\item\relax
\flmRefsHyperref[eczindexfamilyrel]{code:qubit_stabilizer}{Qubit stabilizer code}\item\relax
\flmRefsHyperref[eczindexfamilyrel]{code:random_stabilizer}{Random stabilizer code} --- Log-depth \flmRefsHyperref{ref409}{Clifford circuits} on a 1D geometry yield approximate codes whose encoding rate achieves the hashing bound for Pauli noise and the channel capacity for erasure errors \NoCaseChange{\protect\cite{cite3972,cite3973}}.
\end{eczvaluelist}
\codefieldsection{Cousin}
\begin{eczvaluelist}
\item\relax
\flmRefsHyperref[eczindexfamilyrel]{code:1d_stabilizer}{1D lattice stabilizer code} --- Log-depth \flmRefsHyperref{ref409}{Clifford circuits} on a 1D geometry yield approximate codes whose encoding rate achieves the hashing bound for Pauli noise and the channel capacity for erasure errors \NoCaseChange{\protect\cite{cite3972,cite3973}}.
\end{eczvaluelist}
\eczhbkcontributors{ \eczhuVVA }
\endeczcode

\eczcode{lresc}{Long-range enhanced surface code (LRESC)}{~\NoCaseChange{\protect\cite{cite743}}}
\codefieldsection{Description}
Code constructed using a hypergraph product of two copies of a concatenated LDPC-repetition seed code.
This family interpolates between surface codes and hypergraph codes since the hypergraph product of two repetition codes yields the planar surface code.
The construction uses small \([3,2,2]\) and \([6,2,4]\) LDPC codes concatenated with \([4,1,4]\) and \([2,1,2]\) repetition codes, respectively.
An example using a \([5,2,3]\) code is also presented.
\codefieldsection{Gates}
\begin{eczvaluelist}
\item\relax Patch-transversal gates for suitable seed codes \NoCaseChange{\protect\cite{cite743}}.
\end{eczvaluelist}
\codefieldsection{Realizations}
\begin{eczvaluelist}
\item\relax Preparation of GHZ state of four logical qubits with beyond break-even fidelity in a \(\llbracket 25,4,3\rrbracket \) LRESC \NoCaseChange{\protect\cite{cite3974}}.
\end{eczvaluelist}
\codefieldsection{Parent}
\begin{eczvaluelist}
\item\relax
\flmRefsHyperref[eczindexfamilyrel]{code:cyclic_hgp}{Cyclic Hypergraph Product Code} --- LRESCs are constructed using a hypergraph product of a concatenated LDPC-repetition code with itself.
\end{eczvaluelist}
\codefieldsection{Cousins}
\begin{eczvaluelist}
\item\relax
\flmRefsHyperref[eczindexfamilyrel]{code:lacross}{La-cross code} --- La-cross codes yield LRESCs for \(k=2\). La-cross codes have a number of long-range stabilizers that scales linearly with code size, while the number of LRESC long-range stabilizers can be tuned to scale between the square-root of the size and linearly in the size.
\item\relax
\flmRefsHyperref[eczindexfamilyrel]{code:ldpc}{Low-density parity-check (LDPC) code} --- LRESCs are constructed using a hypergraph product of two copies of a concatenated LDPC-repetition seed code.
\item\relax
\flmRefsHyperref[eczindexfamilyrel]{code:repetition}{Repetition code} --- LRESCs are constructed using a hypergraph product of two copies of a concatenated LDPC-repetition seed code.
\item\relax
\flmRefsHyperref[eczindexfamilyrel]{code:concatenated}{Concatenated code} --- LRESCs are constructed using a hypergraph product of two copies of a concatenated LDPC-repetition seed code.
\item\relax
\flmRefsHyperref[eczindexfamilyrel]{code:surface}{Kitaev surface code} --- LRESCs reduce to planar surface codes when a trivial LDPC code is used in the hypergraph product.
\end{eczvaluelist}
\eczhbkcontributors{ \eczhuVVA }
\endeczcode

\eczcode{lossless_expander}{Lossless expander balanced-product code}{~\NoCaseChange{\protect\cite{cite186,cite188,cite187}}}
\codefieldsection{Description}
QLDPC code constructed by taking the balanced product of lossless expander graphs.
Using one part of a quantum-code chain complex constructed with one-sided loss expanders \NoCaseChange{\protect\cite{cite185}} yields a \(c^3\)-LTC \NoCaseChange{\protect\cite{cite186}}.
Using two-sided expanders \NoCaseChange{\protect\cite{cite187}} yields an asymptotically good QLDPC code family \NoCaseChange{\protect\cite{cite188}}.

\codefieldsection{Rate}
Asymptotically good QLDPC codes \NoCaseChange{\protect\cite{cite188,cite187}}.
\codefieldsection{Parents}
\begin{eczvaluelist}
\item\relax
\flmRefsHyperref[eczindexfamilyrel]{code:qubit_generalized_homological_product_css}{Generalized homological-product qubit CSS code}\item\relax
\flmRefsHyperref[eczindexfamilyrel]{code:balanced_product}{Balanced product (BP) code}\end{eczvaluelist}
\codefieldsection{Cousins}
\begin{eczvaluelist}
\item\relax
\flmRefsHyperref[eczindexfamilyrel]{code:ltc}{Locally testable code (LTC)} --- Using one part of a quantum-code chain complex constructed with one-sided loss expanders yields a \(c^3\)-LTC \NoCaseChange{\protect\cite{cite186}}.
\item\relax
\flmRefsHyperref[eczindexfamilyrel]{code:good_qldpc}{Good QLDPC code} --- Taking a balanced product of two-sided expanders \NoCaseChange{\protect\cite{cite187}} yields an asymptotically good QLDPC code family \NoCaseChange{\protect\cite{cite188}}.
\end{eczvaluelist}
\eczhbkcontributors{ \eczhuVVA }
\endeczcode

\eczcode{mbq}{Majorana box qubit}{~\NoCaseChange{\protect\cite{cite558,cite3975,cite401}}}
\codefieldsection{Description}
A family of Majorana stabilizer codes obtained by fixing the total fermion parity of \(n\) fermionic modes, equivalently \(2n\) Majorana zero modes, within the ground-state subspace of \(n\) Kitaev Majorana chain Hamiltonians.
The resulting positive-parity subspace encodes \(n-1\) logical qubits and has Majorana distance \(2\).

The \(\llbracket 2,1,2\rrbracket _{f}\) member is called the tetron Majorana code, while an \(\llbracket 3,2,2\rrbracket _{f}\) extension using three Kitaev chains and housing two logical qubits of the same parity is called the \textit{hexon Majorana code}.
Similarly, the \(\llbracket 4,3,2\rrbracket _{f}\) \textit{octon}, \(\llbracket 5,4,2\rrbracket _{f}\) \textit{decon}, and \(\llbracket 6,5,2\rrbracket _{f}\) \textit{dodecon} are codes defined by the positive-parity subspace of \(4\), \(5\), and \(6\) fermionic modes, respectively \NoCaseChange{\protect\cite{cite402}}.

\codefieldsection{Protection}
Fixing total fermion parity detects single-Majorana events as parity violations if that parity is measured.
However, these distance-two box-qubit codes do not correct errors: two-Majorana operators act within the fixed-parity sector and can implement logical Pauli errors without changing the stabilizer outcome, while treating the parity constraint as a Hamiltonian term turns single-Majorana events into leakage errors \NoCaseChange{\protect\cite{cite402}}.

\codefieldsection{Gates}
\begin{eczvaluelist}
\item\relax Braiding and fusion of MZMs, which act as Ising anyons \NoCaseChange{\protect\cite{cite3976,cite3977}}.
\end{eczvaluelist}
\codefieldsection{Decoding}
\begin{eczvaluelist}
\item\relax Qubit readout can be done by charge sensing \NoCaseChange{\protect\cite{cite3977,cite3975,cite401,cite3978}}.
\end{eczvaluelist}
\codefieldsection{Fault Tolerance}
\begin{eczvaluelist}
\item\relax Fault-tolerant computation scheme \NoCaseChange{\protect\cite{cite3979}}.
\end{eczvaluelist}
\codefieldsection{Parents}
\begin{eczvaluelist}
\item\relax
\flmRefsHyperref[eczindexfamilyrel]{code:majorana_stab}{Majorana stabilizer code}\item\relax
\flmRefsHyperref[eczindexfamilyrel]{code:small_distance_qubit_stabilizer}{Small-distance qubit stabilizer code}\item\relax
\flmRefsHyperref[eczindexfamilyrel]{code:topological_abelian}{Abelian topological code} --- When treated as ground states of the code Hamiltonian, codewords of a single Kitaev chain realize \(\mathbb{Z}_2\) fermionic topological order.
\item\relax
\flmRefsHyperref[eczindexfamilyrel]{code:qldpc}{Qubit QLDPC code} --- The Majorana box qubit is a 1D qubit stabilizer code with respect to the Majorana operator basis.
\item\relax
\flmRefsHyperref[eczindexfamilyrel]{code:1d_stabilizer}{1D lattice stabilizer code} --- The Majorana box qubit is a 1D qubit stabilizer code with respect to the Majorana operator basis.
\end{eczvaluelist}
\codefieldsection{Child}
\begin{eczvaluelist}
\item\relax
\flmRefsHyperref[eczindexfamilyrel]{code:tetron}{Tetron code} --- The Majorana box qubit for \(n=2\) is the tetron code.
\end{eczvaluelist}
\codefieldsection{Cousins}
\begin{eczvaluelist}
\item\relax
\flmRefsHyperref[eczindexfamilyrel]{code:hamiltonian}{Hamiltonian-based code} --- A Majorana box qubit forms a fixed-parity subspace of the ground-state subspace of one or more Kitaev Majorana chain Hamiltonians.
\item\relax
\flmRefsHyperref[eczindexfamilyrel]{code:majorana_surface}{Majorana surface code} --- The 4.8.8, 6.6.6, and 4.6.12 Majorana surface-code families realize logical tetrons and hexons as fault-tolerant versions of these small Majorana blocks, using tetrons, hexons, or dodecons as parity-fixed building blocks \NoCaseChange{\protect\cite{cite402}}.
\item\relax
\flmRefsHyperref[eczindexfamilyrel]{code:majorana_color}{Majorana color code} --- Majorana color codes are obtained by stacking Majorana surface-code layers and replacing stacked building blocks by small Majorana fermion codes such as hexons, octons, and a \(\llbracket 10,4,4\rrbracket _{f}\) decon-based code \NoCaseChange{\protect\cite[{Sec. V}]{cite402}}.
\item\relax
\flmRefsHyperref[eczindexfamilyrel]{code:kitaev_chain}{Kitaev chain code} --- Majorana box qubit codes are defined to be positive-parity logical subspaces of two or more Kitaev-chain code blocks. The parameter \(n\) in the MBQ code definition corresponds to the number of Kitaev chains used in the construction, and not the total number of physical Majorana modes of the chains.
\end{eczvaluelist}
\eczhbkcontributors{ \eczhuVVA }
\endeczcode

\eczcode{majorana_checkerboard}{Majorana checkerboard code}{~\NoCaseChange{\protect\cite{cite3520}}}
\codefieldsection{Alternative Names}
\begin{eczvaluelist}
\item\relax Majorana cubic model code
\end{eczvaluelist}
\eczhIndexCodeAliasName{majorana_checkerboard}{Majorana cubic model code}
\codefieldsection{Description}
A Majorana analogue of the X-cube model defined on a cubic lattice.
The code admits weight-eight Majorana stabilizer generators on the eight vertices of each cube of a checkerboard sublattice.

\codefieldsection{Parents}
\begin{eczvaluelist}
\item\relax
\flmRefsHyperref[eczindexfamilyrel]{code:majorana_stab}{Majorana stabilizer code}\item\relax
\flmRefsHyperref[eczindexfamilyrel]{code:qldpc}{Qubit QLDPC code} --- The Majorana checkerboard code is a 3D qubit stabilizer code with respect to the Majorana operator basis.
\item\relax
\flmRefsHyperref[eczindexfamilyrel]{code:fracton}{Fracton stabilizer code} --- The Majorana checkerboard code is a foliated type-I fracton code \NoCaseChange{\protect\cite{cite3980}}.
\end{eczvaluelist}
\codefieldsection{Cousin}
\begin{eczvaluelist}
\item\relax
\flmRefsHyperref[eczindexfamilyrel]{code:xcube}{X-cube model code} --- The Majorana checkerboard code is equivalent via a constant-depth unitary to a semionic version of the X-cube model and some decoupled fermionic modes \NoCaseChange{\protect\cite{cite3980}}.
\end{eczvaluelist}
\eczhbkcontributors{ \eczhuVVA }
\endeczcode

\eczcode{majorana_color}{Majorana color code}{~\NoCaseChange{\protect\cite{cite1432,cite3212,cite3430,cite3431,cite402}}}
\codefieldsection{Description}
A fermionic analogue of a 2D color code.

In the original construction \NoCaseChange{\protect\cite{cite1432}}, Majorana modes occupy the vertices of a trivalent graph embedded on a cylinder, each face operator is the product of the Majoranas around an even-length face, and the graph is only locally \(3\)-colorable.
The code encodes one qubit whose two odd logical operators live on the opposite cylinder boundaries, while an even logical operator is supported on a string connecting the boundaries.

Later hardware-oriented Majorana color-code constructions realize related codes by concatenating Majorana surface codes with small Majorana fermion codes \NoCaseChange{\protect\cite{cite3212,cite3430,cite3431,cite402}}.
Equivalently, they are multiple Majorana-surface-code layers whose stacked building blocks are replaced by an outer \(\llbracket n_f,k,d_m\rrbracket _{f}\) Majorana code; hexon, octon, \(\llbracket 8,3,4\rrbracket _{f}\), and \(\llbracket 10,4,4\rrbracket _{f}\) outer codes give concrete order-2, -3, and -4 examples \NoCaseChange{\protect\cite{cite402}}.

\codefieldsection{Protection}
For the original cylindrical family with circumference \(R\) and length \(L\), the code distance scales as \(d=\Omega( \min(R,L) )\), while the minimum diameter of an even logical operator obeys \(l_{\rm even}=\Omega(L)\) \NoCaseChange{\protect\cite[{Sec. 7}]{cite1432}}.
The code therefore interpolates between Kitaev-chain-like protection by fermion-parity superselection (\(R=O(1)\), \(L\gg 1\)) and distance-based protection when both linear dimensions are macroscopic.

\codefieldsection{Rate}
Concatenating a 4.8.8 Majorana surface code with an outer \(\llbracket n_f,k,d_m\rrbracket _{f}\) code yields a fermionic mode overhead of \(\frac{2n_f}{k d_m^2} d^2\) per logical qubit of distance \(d\) \NoCaseChange{\protect\cite{cite402}}.

\codefieldsection{Gates}
\begin{eczvaluelist}
\item\relax Color-to-surface-code lattice surgery \NoCaseChange{\protect\cite{cite3431}}.
\item\relax Logical tetrons and hexons can be encoded in Majorana color codes and manipulated by ordinary, twist-based, or surface-to-color-code lattice surgery \NoCaseChange{\protect\cite{cite3431,cite402}}.
\end{eczvaluelist}
\codefieldsection{Fault Tolerance}
\begin{eczvaluelist}
\item\relax Ordinary and twist-based lattice surgery can be made fault tolerant, and surface-to-color-code surgery reduces ancilla-measurement weight for the 16-Majorana-stabilizer families \NoCaseChange{\protect\cite{cite402}} (see also \NoCaseChange{\protect\cite{cite3981}}).
\end{eczvaluelist}
\codefieldsection{Parents}
\begin{eczvaluelist}
\item\relax
\flmRefsHyperref[eczindexfamilyrel]{code:majorana_stab}{Majorana stabilizer code}\item\relax
\flmRefsHyperref[eczindexfamilyrel]{code:qldpc}{Qubit QLDPC code} --- The Majorana color code is a 2D qubit stabilizer code with respect to the Majorana operator basis.
\item\relax
\flmRefsHyperref[eczindexfamilyrel]{code:2d_stabilizer}{2D lattice stabilizer code} --- The Majorana color code is a 2D qubit stabilizer code with respect to the Majorana operator basis.
\end{eczvaluelist}
\codefieldsection{Cousins}
\begin{eczvaluelist}
\item\relax
\flmRefsHyperref[eczindexfamilyrel]{code:2d_color}{2D color code} --- The original Majorana color code is a fermionic analogue of a 2D color code in which one Majorana face operator doubles to matching \(X\)- and \(Z\)-type face checks, but the underlying cylinder graph need only be locally \(3\)-colorable and can support odd boundary logical operators \NoCaseChange{\protect\cite{cite1432}}. Later realizations stack Majorana surface-code layers and replace stacked building blocks with small Majorana fermion codes \NoCaseChange{\protect\cite{cite3212,cite3430,cite3431,cite402}}.
\item\relax
\flmRefsHyperref[eczindexfamilyrel]{code:majorana_surface}{Majorana surface code} --- The original Majorana color code is a fermionic analogue of a 2D color code in which one Majorana face operator doubles to matching \(X\)- and \(Z\)-type face checks, but the underlying cylinder graph need only be locally \(3\)-colorable and can support odd boundary logical operators \NoCaseChange{\protect\cite{cite1432}}. Later realizations stack Majorana surface-code layers and replace stacked building blocks with small Majorana fermion codes \NoCaseChange{\protect\cite{cite3212,cite3430,cite3431,cite402}}.
\item\relax
\flmRefsHyperref[eczindexfamilyrel]{code:majorana_subsystem}{Majorana subsystem stabilizer code} --- A particular self-dual stabilizer Hamiltonian within the 3D subsystem color code admits a Majorana variant whose boundaries support 2D Majorana color codes \NoCaseChange{\protect\cite{cite466}}.
\item\relax
\flmRefsHyperref[eczindexfamilyrel]{code:mbq}{Majorana box qubit} --- Majorana color codes are obtained by stacking Majorana surface-code layers and replacing stacked building blocks by small Majorana fermion codes such as hexons, octons, and a \(\llbracket 10,4,4\rrbracket _{f}\) decon-based code \NoCaseChange{\protect\cite[{Sec. V}]{cite402}}.
\item\relax
\flmRefsHyperref[eczindexfamilyrel]{code:majorana_hamming}{\(\llbracket 2^{m-1},2^{m-1}-m-1,4\rrbracket _{f}\) Hamming Majorana code} --- The \(\llbracket 8,3,4\rrbracket _{f}\) Hamming Majorana code can replace stacks of three tetrons in a 4.8.8 Majorana surface code to yield an order-3 Majorana color code with maximum stabilizer weight \(16\) \NoCaseChange{\protect\cite[{Table I}]{cite402}}.
\end{eczvaluelist}
\eczhbkcontributors{ \eczhuVVA }
\endeczcode

\eczcode{mlsc}{Majorana loop stabilizer code (MLSC)}{~\NoCaseChange{\protect\cite{cite3508}}}
\codefieldsection{Description}
A single-error-correcting fermion-into-qubit encoding defined on a 2D qubit lattice whose stabilizers are associated with loops in the lattice.

\codefieldsection{Protection}
The code can correct single-qubit errors \NoCaseChange{\protect\cite{cite3508}}.

\codefieldsection{Parent}
\begin{eczvaluelist}
\item\relax
\flmRefsHyperref[eczindexfamilyrel]{code:2d_bosonization}{2D bosonization code} --- The MLSC on a square lattice can be obtained from exact 2D bosonization by finite-depth generalized local unitaries \NoCaseChange{\protect\cite{cite404}}.
\end{eczvaluelist}
\codefieldsection{Child}
\begin{eczvaluelist}
\item\relax
\flmRefsHyperref[eczindexfamilyrel]{code:bksf}{Bravyi-Kitaev superfast (BKSF) code} --- The BKSF code can be thought of as a particular MLSC \NoCaseChange{\protect\cite{cite3508}}.
\end{eczvaluelist}
\eczhbkcontributors{ \eczhuVVA }
\endeczcode

\eczcode{majorana_stab}{Majorana stabilizer code}{~\NoCaseChange{\protect\cite{cite1432}}}
\codefieldsection{Description}
A stabilizer code whose stabilizers are products of an even number of Majorana fermion operators, analogous to Pauli strings for a traditional stabilizer code and referred to as \textit{Majorana stabilizers}.
The codespace is the mutual \(+1\) eigenspace of all Majorana stabilizers.

Codes can be denoted as \(\llbracket n,k,d\rrbracket _{f}\) \NoCaseChange{\protect\cite{cite566}}, where \(n\) is the number of fermionic modes (equivalently, \(2n\) Majorana modes).
Two copies of an \(n\)-Majorana mode code may be combined to form a single \(n\)-fermion code by using one copy for the real parts of each fermion, and the other copy for the imaginary parts \NoCaseChange{\protect\cite{cite559}}.
Codes that admit a logical operator of even (odd) weight are called even (odd) Majorana codes \NoCaseChange{\protect\cite{cite750}}.
Even Majorana codes encode logical qubits, and odd Majorana codes have at least one logical Majorana fermion.
For odd codes, an additional protection parameter is the minimum diameter \(l_{\rm even}\) of an even logical operator \NoCaseChange{\protect\cite{cite1432}}.

In some cases, Majorana-based stabilizer codes are designed to protect against fermionic noise \NoCaseChange{\protect\cite{cite558}} and are thus useful for physical platforms based on fermions.
In other cases, Majorana-based frameworks are helpful for understanding conventional qubit stabilizer codes designed for qubit-based platforms.

\codefieldsection{Protection}
Detects products of Majorana operators with weight up to \(d-1\).
Physically, protects against dephasing errors caused by coupling of fermion density to the environment and bit-flip errors caused by quasiparticle poisoning processes.
For odd Majorana codes, the physically relevant logical protection is also governed by the minimum diameter \(l_{\rm even}\) of an even logical operator \NoCaseChange{\protect\cite{cite1432}}.

Code bounds have been developed for small codes \NoCaseChange{\protect\cite{cite3212}}.
LP bounds for Majorana codes have been developed based on the identification of Majorana operators with the Clifford algebra \NoCaseChange{\protect\cite{cite564}}.

\codefieldsection{Encoding}
\begin{eczvaluelist}
\item\relax Unitary encoding using fermionic Clifford operations \NoCaseChange{\protect\cite{cite3982}}.
\end{eczvaluelist}
\codefieldsection{Transversal and Permutation-Based Gates}
\begin{eczvaluelist}
\item\relax Transversal Clifford operations are discussed in Ref. \NoCaseChange{\protect\cite{cite750}}.
\end{eczvaluelist}
\codefieldsection{Gates}
\begin{eczvaluelist}
\item\relax Some gates can be implemented through braiding of the computational anyons. Circuit-based gates can be converted into braid patterns via quantum compiling algorithms \NoCaseChange{\protect\cite{cite3983}}.
\end{eczvaluelist}
\codefieldsection{Parents}
\begin{eczvaluelist}
\item\relax
\flmRefsHyperref[eczindexfamilyrel]{code:fermions}{Fermion code}\item\relax
\flmRefsHyperref[eczindexfamilyrel]{code:qubit_stabilizer}{Qubit stabilizer code} --- A Majorana stabilizer code is a stabilizer code whose stabilizers are composed of Majorana fermion operators, which are in turn realizable using Pauli strings via the Jordan-Wigner mapping.
Any \(\llbracket n,k,d\rrbracket \) stabilizer code can be mapped into a \(\llbracket 2n,k,2d\rrbracket _{f}\) Majorana stabilizer code by concatenating with the tetron code \NoCaseChange{\protect\cite{cite537}\protect\cite[{Lemma 1}]{cite1432}}.
Embedding each physical qubit into two fermions via the tetron code is useful for exactly solving the Kitaev honeycomb model Hamiltonian \NoCaseChange{\protect\cite{cite537}} and other qubit Hamiltonians on certain graphs \NoCaseChange{\protect\cite{cite2842,cite2843}}.
Majorana stabilizer groups can be converted into ordinary qubit stabilizer groups via the parton mapping, while their corresponding states are converted via the Gutzwiller projection \NoCaseChange{\protect\cite{cite2844}}.

\end{eczvaluelist}
\codefieldsection{Children}
\begin{eczvaluelist}
\item\relax
\flmRefsHyperref[eczindexfamilyrel]{code:majorana_color}{Majorana color code}\item\relax
\flmRefsHyperref[eczindexfamilyrel]{code:majorana_surface}{Majorana surface code}\item\relax
\flmRefsHyperref[eczindexfamilyrel]{code:majorana_6_1_3}{\(\llbracket 6,1,3\rrbracket _{f}\) Vijay-Fu Majorana code}\item\relax
\flmRefsHyperref[eczindexfamilyrel]{code:majorana_checkerboard}{Majorana checkerboard code}\item\relax
\flmRefsHyperref[eczindexfamilyrel]{code:kitaev_chain}{Kitaev chain code}\item\relax
\flmRefsHyperref[eczindexfamilyrel]{code:mbq}{Majorana box qubit}\item\relax
\flmRefsHyperref[eczindexfamilyrel]{code:majorana_reed_muller}{RM Majorana code}\end{eczvaluelist}
\codefieldsection{Cousins}
\begin{eczvaluelist}
\item\relax
\flmRefsHyperref[eczindexfamilyrel]{code:dual}{Dual linear code} --- Classical self-orthogonal codes can be used to construct Majorana stabilizer codes \NoCaseChange{\protect\cite{cite566,cite1783,cite1784}}. The direct relationship between the two codes follows from expressing the Majorana strings as binary vectors – akin to the \flmRefsHyperref{ref817}{symplectic representation} – and observing that the binary stabilizer matrix \(S\) for such a Majorana stabilizer code satisfies \(S\cdot S^T=0\) because it has commuting stabilizers, which is precisely the condition \(G\cdot G^T=0\) on the generator matrix \(G\) of a self-orthogonal classical code. A self-orthogonal classical code \(C\) with parameters \([2N,k,d]\) yields a Majorana stabilizer code with parameters \(\llbracket N,N-k,d^\perp\rrbracket _f\), where \(d^\perp\) is the code distance of the dual code \(C^\perp\).
\item\relax
\flmRefsHyperref[eczindexfamilyrel]{code:qubit_css}{Qubit CSS code} --- Every \(\llbracket n,k,d\rrbracket _f\) Majorana stabilizer code is associated with a \(\llbracket 2n,2k,d\rrbracket \) qubit CSS code whose \(X\)- and \(Z\)-check supports coincide \NoCaseChange{\protect\cite[{Lemma 2}]{cite1432}}. An odd-length self-dual CSS code can be converted into a complex-fermion code by replacing qubit \(Z\)-type and \(X\)-type operators with \(\gamma\)-type and \(\tilde{\gamma}\)-type Majorana operators, respectively \NoCaseChange{\protect\cite{cite559}}.
\item\relax
\flmRefsHyperref[eczindexfamilyrel]{code:steane}{\(\llbracket 7,1,3\rrbracket \) Steane code} --- Applying the CSS-to-Majorana map of \NoCaseChange{\protect\cite[{Lemma 2}]{cite1432}} to the \(\llbracket 7,1,3\rrbracket \) Steane code yields a seven-Majorana code encoding half a qubit; pairing two such odd-length copies gives a physical Majorana stabilizer code with odd logical operators \NoCaseChange{\protect\cite[{Sec. 8}]{cite1432}}.
\item\relax
\flmRefsHyperref[eczindexfamilyrel]{code:binary_linear}{Linear binary code} --- When constructing a Majorana stabilizer code from a self-orthogonal classical code with an odd number of bits and generator matrix \(G\), a more complex procedure must be applied to ensure that the fermion code has an even number of Majorana zero modes, and thus a physical Hilbert space \NoCaseChange{\protect\cite{cite1432,cite566}}. Rather than taking \(G\) to be the stabilizer matrix as in the even case, we take \(G\oplus G\). This is a concatenation of classical codes as in the CSS construction and it yields a mapping \([2n-1,k,d]\rightarrow \llbracket 2n-1,2n-1-k,d^\perp\rrbracket _f\). This procedure may be further generalized by concatenating two different self-orthogonal classical codes with an odd number of bits, as is often done in the CSS construction.
\item\relax
\flmRefsHyperref[eczindexfamilyrel]{code:binary_cyclic}{Cyclic linear binary code} --- Cyclic binary linear codes can be used to construct translation-invariant Majorana stabilizer codes, provided that they are also self-orthogonal \NoCaseChange{\protect\cite{cite566}}.
\item\relax
\flmRefsHyperref[eczindexfamilyrel]{code:stabilizer}{Stabilizer code} --- Majorana stabilizer codes are useful for Majorana-based architectures, where the degrees of freedom are electrons, and the notion of locality is different than all other code kingdoms.
\item\relax
\flmRefsHyperref[eczindexfamilyrel]{code:qudit_stabilizer}{Modular-qudit stabilizer code} --- Majorana stabilizer codes can be extended to modular qudits, yielding parafermion stabilizer codes \NoCaseChange{\protect\cite{cite3984}}.
\item\relax
\flmRefsHyperref[eczindexfamilyrel]{code:jw}{Jordan-Wigner transformation code} --- A Majorana stabilizer code is a stabilizer code whose stabilizers are composed of Majorana fermion operators, which are in turn realizable using Pauli strings via the Jordan-Wigner mapping.

\item\relax
\flmRefsHyperref[eczindexfamilyrel]{code:honeycomb_floquet}{Honeycomb Floquet code} --- The Honeycomb code admits a convenient representation in terms of Majorana fermions. This leads to a possible physical realization of the code in terms of tetrons \NoCaseChange{\protect\cite{cite401}}, where each physical qubit is composed of four Majorana modes.
\item\relax
\flmRefsHyperref[eczindexfamilyrel]{code:da}{Dynamical code} --- Dynamical codes are viable candidates for storage in Majorana-qubit devices \NoCaseChange{\protect\cite{cite3631}}.
\item\relax
\flmRefsHyperref[eczindexfamilyrel]{code:majorana_subsystem}{Majorana subsystem stabilizer code} --- Subsystem qubit stabilizer codes have been formulated in terms of Majorana operators \NoCaseChange{\protect\cite{cite3482}}.
\item\relax
\flmRefsHyperref[eczindexfamilyrel]{code:stab_5_1_3}{\(\llbracket 5,1,3\rrbracket \) Five-qubit perfect code} --- The five-qubit code Hamiltonian is local when expressed in terms of mutually commuting Majorana operators \NoCaseChange{\protect\cite{cite3319}}.
\item\relax
\flmRefsHyperref[eczindexfamilyrel]{code:tfim}{Transverse-field Ising model (TFIM) code} --- The TFIM code stabilizers can be expressed in terms of Majorana operators.
\item\relax
\flmRefsHyperref[eczindexfamilyrel]{code:quantum_parity}{Quantum parity code (QPC)} --- QPCs for \(m_1=m_2\) can be conveniently expressed in terms of mutually commuting Majorana operators \NoCaseChange{\protect\cite{cite557}}.
\item\relax
\flmRefsHyperref[eczindexfamilyrel]{code:happy}{Pastawski-Yoshida-Harlow-Preskill (HaPPY) code} --- The pentagon HaPPY code Hamiltonian can be expressed in terms of mutually commuting weight-two (two-body) Majorana operators \NoCaseChange{\protect\cite{cite3985}}.
\item\relax
\flmRefsHyperref[eczindexfamilyrel]{code:self_dual_css}{Self-dual CSS code} --- An odd-length self-dual CSS code can be converted into a complex-fermion code by replacing qubit \(Z\)-type and \(X\)-type operators with \(\gamma\)-type and \(\tilde{\gamma}\)-type Majorana operators, respectively \NoCaseChange{\protect\cite{cite559}}.
\end{eczvaluelist}
\eczhbkcontributors{ Michael Gullans, Alexander Schuckert, Chris Fechisin, \eczhuVVA }
\endeczcode

\eczcode{majorana_subsystem}{Majorana subsystem stabilizer code}{~\NoCaseChange{\protect\cite{cite3482}}}
\codefieldsection{Description}
A Majorana subsystem code with some of its logical qubits denoted as \textit{gauge} qubits and not used for storage of logical information.

\codefieldsection{Parent}
\begin{eczvaluelist}
\item\relax
\flmRefsHyperref[eczindexfamilyrel]{code:qubit_subsystem_stabilizer}{Subsystem qubit stabilizer code} --- Subsystem qubit stabilizer codes have been formulated in terms of Majorana operators \NoCaseChange{\protect\cite{cite3482}}.
\end{eczvaluelist}
\codefieldsection{Cousins}
\begin{eczvaluelist}
\item\relax
\flmRefsHyperref[eczindexfamilyrel]{code:majorana_stab}{Majorana stabilizer code} --- Subsystem qubit stabilizer codes have been formulated in terms of Majorana operators \NoCaseChange{\protect\cite{cite3482}}.
\item\relax
\flmRefsHyperref[eczindexfamilyrel]{code:translationally_invariant_subsystem}{Lattice subsystem code} --- Translationally invariant subsystem codes have been formulated in terms of Majorana operators \NoCaseChange{\protect\cite{cite3482}}.
\item\relax
\flmRefsHyperref[eczindexfamilyrel]{code:bacon_shor}{Bacon-Shor code} --- Bacon-Shor codes can be fermionized into fermionic subsystem codes with two-body terms \NoCaseChange{\protect\cite{cite3482}}.
\item\relax
\flmRefsHyperref[eczindexfamilyrel]{code:subsystem_color}{Subsystem color code} --- A particular self-dual stabilizer Hamiltonian within the 3D subsystem color code admits a Majorana variant whose boundaries support 2D Majorana color codes \NoCaseChange{\protect\cite{cite466}}.
\item\relax
\flmRefsHyperref[eczindexfamilyrel]{code:majorana_color}{Majorana color code} --- A particular self-dual stabilizer Hamiltonian within the 3D subsystem color code admits a Majorana variant whose boundaries support 2D Majorana color codes \NoCaseChange{\protect\cite{cite466}}.
\item\relax
\flmRefsHyperref[eczindexfamilyrel]{code:qubit_stabilizer}{Qubit stabilizer code} --- The B\(\mapsto\)F mapping yields Majorana subsystem codes from qubit stabilizer codes such that their gauge groups contain tetrons \NoCaseChange{\protect\cite{cite3987}\protect\cite[{Sec. IV}]{cite3986}}. The output Majorana subsystem codes can correct odd-weight errors.
\item\relax
\flmRefsHyperref[eczindexfamilyrel]{code:tetron}{Tetron code} --- The B\(\mapsto\)F mapping yields Majorana subsystem codes from qubit stabilizer codes such that their gauge groups contain tetrons \NoCaseChange{\protect\cite{cite3987}\protect\cite[{Sec. IV}]{cite3986}}. The output Majorana subsystem codes can correct odd-weight errors.
\end{eczvaluelist}
\eczhbkcontributors{ \eczhuVVA }
\endeczcode

\eczcode{majorana_surface}{Majorana surface code}{~\NoCaseChange{\protect\cite{cite3988,cite3989}}}
\codefieldsection{Description}
Fermionic analogue of the surface code defined on a three-colorable 2D tiling whose face operators are non-overlapping even-Majorana stabilizers.
Open patches with four or six alternating colored boundaries encode logical tetrons or hexons.
The uniform 4.8.8, 6.6.6, and 4.6.12 tilings yield families with tetron, hexon, or dodecon building blocks and with twist-based lattice surgery supporting minimal-overhead logical Clifford gates \NoCaseChange{\protect\cite{cite402}}.

\codefieldsection{Protection}
Under quasiparticle-poisoning noise, single Majorana operators flip the syndromes of adjacent stabilizers and can be decoded with surface-code methods.
If the shortest logical operator has Majorana weight \(d_m\), then the code corrects up to \(d_m/2-1\) Majorana errors; the corresponding qubit distance is naturally labeled by \(d=d_m/2\) because Pauli errors on tetron/hexon hardware involve pairs of Majoranas \NoCaseChange{\protect\cite{cite402}}.
Implementations that treat one color of stabilizers as parity-fixing constraints reduce measured stabilizer weight, at the cost that parity-violating single-Majorana events become leakage unless that color is also measured \NoCaseChange{\protect\cite{cite402}}.

\codefieldsection{Rate}
For code distance \(d=d_m/2\), the 4.8.8, 6.6.6, and 4.6.12 families require \(4d^2\), \(6d^2+\mathcal{O}(d)\), and \(12d^2+\mathcal{O}(d)\) Majoranas per logical qubit, respectively \NoCaseChange{\protect\cite[{Table I}]{cite402}}.
Their maximum stabilizer weights for fault-tolerant lattice surgery can be reduced to \(8\), \(6\), and \(6\) Majoranas, respectively, compared to \(10\) Majoranas for bosonic twist-based surface-code surgery \NoCaseChange{\protect\cite{cite402}}.

\codefieldsection{Gates}
\begin{eczvaluelist}
\item\relax All logical Clifford gates, including CNOT, can be implemented with zero time overhead by classically tracking them and measuring Pauli products via ordinary and twist-based lattice surgery \NoCaseChange{\protect\cite{cite402}}.
\item\relax Surface-code state injection of noisy magic states into logical tetrons \NoCaseChange{\protect\cite[{Sec. IV.C}]{cite402}}.
\end{eczvaluelist}
\codefieldsection{Fault Tolerance}
\begin{eczvaluelist}
\item\relax Repeated syndrome rounds make ordinary and twist-based lattice surgery fault tolerant against both data and measurement errors \NoCaseChange{\protect\cite{cite402}} (see also \NoCaseChange{\protect\cite{cite3981}}).
\end{eczvaluelist}
\codefieldsection{Parents}
\begin{eczvaluelist}
\item\relax
\flmRefsHyperref[eczindexfamilyrel]{code:majorana_stab}{Majorana stabilizer code}\item\relax
\flmRefsHyperref[eczindexfamilyrel]{code:qldpc}{Qubit QLDPC code} --- The Majorana surface code is a 2D qubit stabilizer code with respect to the Majorana operator basis.
\item\relax
\flmRefsHyperref[eczindexfamilyrel]{code:2d_stabilizer}{2D lattice stabilizer code} --- The Majorana surface code is a 2D qubit stabilizer code with respect to the Majorana operator basis.
\end{eczvaluelist}
\codefieldsection{Cousins}
\begin{eczvaluelist}
\item\relax
\flmRefsHyperref[eczindexfamilyrel]{code:surface}{Kitaev surface code} --- Majorana surface codes map non-uniquely to bosonic surface codes: replacing each tetron in a 4.8.8 code by a qubit yields the square-lattice surface code, while 6.6.6 and 4.6.12 codes map to rotated-square and Kagome-lattice surface-code realizations, respectively \NoCaseChange{\protect\cite{cite402}}.
\item\relax
\flmRefsHyperref[eczindexfamilyrel]{code:majorana_color}{Majorana color code} --- The original Majorana color code is a fermionic analogue of a 2D color code in which one Majorana face operator doubles to matching \(X\)- and \(Z\)-type face checks, but the underlying cylinder graph need only be locally \(3\)-colorable and can support odd boundary logical operators \NoCaseChange{\protect\cite{cite1432}}. Later realizations stack Majorana surface-code layers and replace stacked building blocks with small Majorana fermion codes \NoCaseChange{\protect\cite{cite3212,cite3430,cite3431,cite402}}.
\item\relax
\flmRefsHyperref[eczindexfamilyrel]{code:mbq}{Majorana box qubit} --- The 4.8.8, 6.6.6, and 4.6.12 Majorana surface-code families realize logical tetrons and hexons as fault-tolerant versions of these small Majorana blocks, using tetrons, hexons, or dodecons as parity-fixed building blocks \NoCaseChange{\protect\cite{cite402}}.
\item\relax
\flmRefsHyperref[eczindexfamilyrel]{code:tetron}{Tetron code} --- Four-boundary Majorana surface-code patches are logical tetrons, i.e., higher-distance versions of the tetron code \NoCaseChange{\protect\cite{cite402}}.
\end{eczvaluelist}
\eczhbkcontributors{ \eczhuVVA }
\endeczcode

\eczcode{matching}{Matching code}{~\NoCaseChange{\protect\cite{cite3990}}}
\codefieldsection{Description}
Member of a class of qubit stabilizer codes based on the Abelian phase of the Kitaev honeycomb model.
\codefieldsection{Realizations}
\begin{eczvaluelist}
\item\relax Braiding of defects has been demonstrated for a five-qubit version of the code \NoCaseChange{\protect\cite{cite3991}}.
\end{eczvaluelist}
\codefieldsection{Parents}
\begin{eczvaluelist}
\item\relax
\flmRefsHyperref[eczindexfamilyrel]{code:qldpc}{Qubit QLDPC code}\item\relax
\flmRefsHyperref[eczindexfamilyrel]{code:quantum_double_abelian}{Abelian quantum-double stabilizer code} --- Matching codes were inspired by the \(\mathbb{Z}_2\) topological order phase of the Kitaev honeycomb model \NoCaseChange{\protect\cite{cite537}}.
\end{eczvaluelist}
\codefieldsection{Child}
\begin{eczvaluelist}
\item\relax
\flmRefsHyperref[eczindexfamilyrel]{code:xyz_hexagonal}{XYZ\(^2\) hexagonal stabilizer code}\end{eczvaluelist}
\codefieldsection{Cousin}
\begin{eczvaluelist}
\item\relax
\flmRefsHyperref[eczindexfamilyrel]{code:kitaev_honeycomb}{Kitaev honeycomb code} --- Matching codes were inspired by the \(\mathbb{Z}_2\) topological order phase of the Kitaev honeycomb model \NoCaseChange{\protect\cite{cite537}}.
\end{eczvaluelist}
\eczhbkcontributors{ \eczhuVVA }
\endeczcode

\eczcode{movassagh_ouyang}{Movassagh-Ouyang Hamiltonian code}{~\NoCaseChange{\protect\cite{cite1407}}}
\codefieldsection{Description}
This is a family of codes derived via an algorithm that takes as input \textit{any} binary classical code and outputs a quantum code (note that this framework can be extended to \(q\)-ary codes).  
The algorithm is probabilistic but succeeds almost surely if the classical code is random. 
An explicit code construction does exist for linear distance codes encoding one logical qubit using Radon's theorem \NoCaseChange{\protect\cite{cite596,cite597}}. 
For finite rate codes, there is no rigorous proof that the construction algorithm succeeds, and approximate constructions are described instead.

This family strictly generalizes CSS codes (because CSS codes come only from linear or self-orthogonal classical codes). These codes can be shown to be realized as a subspace of the ground space of a (geometrically) local Hamiltonian.

\codefieldsection{Protection}
Let \(C \subset \{0,1,\dots,q-1\}^n\) be a classical code with distance \(d_x\). Let \(d_z\) satisfy \(q^n > 2 V_q(d_z-1) -1\), where \(V_q(r)\) is the volume of the \(q\)-ary Hamming ball of radius \(r\). Then the algorithm produces a quantum code with distance \(d = \min(d_x,d_z)\). Asymptotically, the distance scales linearly with \(n\).
\codefieldsection{Rate}
The rate depends on the classical code, but distance can scale linearly with \(n\).
\codefieldsection{Parents}
\begin{eczvaluelist}
\item\relax
\flmRefsHyperref[eczindexfamilyrel]{code:qubits_into_qubits}{Qubit code}\item\relax
\flmRefsHyperref[eczindexfamilyrel]{code:hamiltonian}{Hamiltonian-based code} --- Movassagh-Ouyang codes reside in the ground space of a Hamiltonian. Justesen codes can be used to build a family of \(n\)-qubit Movassagh-Ouyang Hamiltonian spin codes encoding one logical qubit with linear distance. These codes form the ground-state subspace of a frustration-free geometrically local Hamiltonian \NoCaseChange{\protect\cite{cite1407}}.
\end{eczvaluelist}
\codefieldsection{Child}
\begin{eczvaluelist}
\item\relax
\flmRefsHyperref[eczindexfamilyrel]{code:unentangled_permutation_invariant}{\(\llparenthesis n,2,2\rrparenthesis \) Bravyi-Lee-Li-Yoshida PI code} --- The \(\llparenthesis n,2,2\rrparenthesis \) PI code is a Movassagh-Ouyang Hamiltonian code constructed from a binary code consisting of all codewords of weight 0, 2, or \(n\) \NoCaseChange{\protect\cite[{Appx. D}]{cite529}}.
\end{eczvaluelist}
\codefieldsection{Cousins}
\begin{eczvaluelist}
\item\relax
\flmRefsHyperref[eczindexfamilyrel]{code:qubit_stabilizer}{Qubit stabilizer code} --- Many, but not all, Movassagh-Ouyang codes are stabilizer codes.
\item\relax
\flmRefsHyperref[eczindexfamilyrel]{code:bits_into_bits}{Binary code} --- Movassagh-Ouyang codes are constructed from classical binary codes.
\item\relax
\flmRefsHyperref[eczindexfamilyrel]{code:justesen}{Justesen code} --- Justesen codes can be used to build a family of \(n\)-qubit Movassagh-Ouyang Hamiltonian spin codes encoding one logical qubit with linear distance. These codes form the ground-state subspace of a frustration-free geometrically local Hamiltonian \NoCaseChange{\protect\cite{cite1407}}.
\item\relax
\flmRefsHyperref[eczindexfamilyrel]{code:spins_into_spins}{Spin code} --- Justesen codes can be used to build a family of \(n\)-qubit Movassagh-Ouyang Hamiltonian spin codes encoding one logical qubit with linear distance. These codes form the ground-state subspace of a frustration-free geometrically local Hamiltonian \NoCaseChange{\protect\cite{cite1407}}.
\item\relax
\flmRefsHyperref[eczindexfamilyrel]{code:frustration_free}{Frustration-free Hamiltonian code} --- Movassagh-Ouyang codes reside in the ground space of a Hamiltonian. Justesen codes can be used to build a family of \(n\)-qubit Movassagh-Ouyang Hamiltonian spin codes encoding one logical qubit with linear distance. These codes form the ground-state subspace of a frustration-free geometrically local Hamiltonian \NoCaseChange{\protect\cite{cite1407}}.
\item\relax
\flmRefsHyperref[eczindexfamilyrel]{code:cws}{Codeword stabilized (CWS) code} --- The Movassagh-Ouyang codes overlap the CWS codes but neither family is contained in the other \NoCaseChange{\protect\cite{cite1407}}.
\item\relax
\flmRefsHyperref[eczindexfamilyrel]{code:qubit_css}{Qubit CSS code} --- Qubit CSS codes encoding one logical qubit are a subset of Movassagh-Ouyang codes.
\end{eczvaluelist}
\eczhbkcontributors{ Eric Kubischta, \eczhuVVA }
\endeczcode

\eczcode{nonabelian_kitaev_honeycomb}{Non-Abelian Kitaev honeycomb code}{~\NoCaseChange{\protect\cite{cite537}}}
\codefieldsection{Description}
Code whose logical subspace in the gapped non-Abelian phase of the Kitaev honeycomb model with a magnetic field is labeled by different fusion outcomes of Ising anyons \NoCaseChange{\protect\cite{cite537}}.

The original Kitaev honeycomb spin model is exactly solvable by mapping spins to Majorana fermions in a static \(\mathbb{Z}_2\) gauge field (equivalently, embedding each physical qubit into two fermions via the tetron code \NoCaseChange{\protect\cite[{Sec. 4.1}]{cite3415}}), yielding three gapped \(A\) phases and one gapless \(B\) phase \NoCaseChange{\protect\cite{cite537}}.
A magnetic field opens a gap in phase \(B\) of the underlying Kitaev honeycomb code and yields the non-Abelian Ising-anyon phase \NoCaseChange{\protect\cite{cite537}} (a.k.a. \(p+ip\) superconducting phase \NoCaseChange{\protect\cite{cite3992}}).
In the honeycomb model with magnetic field, the spectral Chern number is \(\nu=\pm 1\) depending on the field direction; more generally, gapped free-fermion phases with \(\mathbb{Z}_2\) vortices are classified by a spectral Chern number \(\nu\), and their anyonic properties depend on \(\nu \bmod 16\) \NoCaseChange{\protect\cite{cite537}}.

Ising anyons also exist in other phases, such as the fractional quantum Hall phase \NoCaseChange{\protect\cite{cite3993}}.

\codefieldsection{Encoding}
\begin{eczvaluelist}
\item\relax Anyon initialization via quantum control \NoCaseChange{\protect\cite{cite3994}}.
\end{eczvaluelist}
\codefieldsection{Gates}
\begin{eczvaluelist}
\item\relax \flmRefsHyperref{ref409}{Clifford gates} can be performed by braiding Majorana operators and Pauli measurements can be performed by measuring certain Majorana operators \NoCaseChange{\protect\cite{cite3993,cite3415}}.
\item\relax CPHASE gate or a \(\pi/8\) rotation with the help of ancilla states completes a universal gate set \NoCaseChange{\protect\cite{cite3993,cite3415}}.
\end{eczvaluelist}
\codefieldsection{Fault Tolerance}
\begin{eczvaluelist}
\item\relax One can distill ancilla states to arbitrary precision for sufficiently small noise rates and assuming perfect Clifford operations \NoCaseChange{\protect\cite{cite3993}}.
\end{eczvaluelist}
\codefieldsection{Parents}
\begin{eczvaluelist}
\item\relax
\flmRefsHyperref[eczindexfamilyrel]{code:qubits_into_qubits}{Qubit code} --- The Kitaev honeycomb model with a magnetic field is a qubit many-body system in the Ising-anyon phase, and the underlying code stores information in the fusion space of its non-Abelian anyonic excitations.
\item\relax
\flmRefsHyperref[eczindexfamilyrel]{code:topological}{Topological code} --- The Kitaev honeycomb model with a magnetic field is a qubit many-body system in the Ising-anyon phase, and the underlying code stores information in the fusion space of its non-Abelian anyonic excitations.
\end{eczvaluelist}
\codefieldsection{Cousins}
\begin{eczvaluelist}
\item\relax
\flmRefsHyperref[eczindexfamilyrel]{code:kitaev_honeycomb}{Kitaev honeycomb code} --- The gauge-group generators of the Kitaev honeycomb code are terms of the Kitaev honeycomb model Hamiltonian. Adding a magnetic field to this Hamiltonian for particular parameter values yields the non-Abelian Ising-anyon phase, whose anyons encode the logical information of the non-Abelian Kitaev honeycomb code \NoCaseChange{\protect\cite{cite537}}.
\item\relax
\flmRefsHyperref[eczindexfamilyrel]{code:tetron}{Tetron code} --- Embedding each physical qubit into two fermions via the tetron code allows the logical subspace of the Kitaev honeycomb model to be formulated as a joint eigenspace of certain Majorana operators \NoCaseChange{\protect\cite[{Sec. 4.1}]{cite3415}}, which admit braiding-based gates due to their non-Abelian statistics and which can be used for topological quantum computation.
When done in reverse, this embedding can be thought of as a 2D bosonization fermion-into-qubit encoding by converting to a relabeled square lattice and performing single-qubit rotations \NoCaseChange{\protect\cite{cite403}\protect\cite[{Sec. IV.B}]{cite404}}.

\item\relax
\flmRefsHyperref[eczindexfamilyrel]{code:2d_bosonization}{2D bosonization code} --- Embedding each physical qubit into two fermions via the tetron code allows the logical subspace of the Kitaev honeycomb model to be formulated as a joint eigenspace of certain Majorana operators \NoCaseChange{\protect\cite[{Sec. 4.1}]{cite3415}}, which admit braiding-based gates due to their non-Abelian statistics and which can be used for topological quantum computation.
When done in reverse, this embedding can be thought of as a 2D bosonization fermion-into-qubit encoding by converting to a relabeled square lattice and performing single-qubit rotations \NoCaseChange{\protect\cite{cite403}\protect\cite[{Sec. IV.B}]{cite404}}.

\item\relax
\flmRefsHyperref[eczindexfamilyrel]{code:honeycomb}{Honeycomb tiling} --- The Kitaev honeycomb model is defined on the honeycomb tiling.
\end{eczvaluelist}
\eczhbkcontributors{ \eczhuVVA }
\endeczcode

\eczcode{ampdamp_numopt}{Numerically optimized four-qubit AD code}{~\NoCaseChange{\protect\cite{cite3995,cite3996}}}
\codefieldsection{Description}
One of several four-qubit codes that can (approximately) correct a single \flmRefsHyperref{ref498}{AD} error with higher fidelity than the \(\llbracket 4,1,2\rrbracket \) subcodes of the \(\llbracket 4,2,2\rrbracket \) code.

A code obtained by a biconvex optimization of the entanglement fidelity admits a codeword basis of \NoCaseChange{\protect\cite{cite3995}}
\flmMathEnvironment{align}{}{
\begin{split}
|\overline{0}\rangle&=\sqrt{1-\frac{1}{2(1-\gamma)^{2}}}|0000\rangle+\frac{1}{\sqrt{2}(1-\gamma)}|1111\rangle\\
|\overline{1}\rangle&=\frac{1}{2}(|0011\rangle+|0101\rangle-|1010\rangle+|1100\rangle)
\end{split}
}
for \flmRefsHyperref{ref498}{AD} error rate \(\gamma\).
Another code, obtained from a machine-learning optimization \NoCaseChange{\protect\cite{cite3996}}, admits a codeword basis of
\flmMathEnvironment{align}{}{
\begin{split}
|\bar{0}\rangle&=\sqrt{\frac{1}{1+(1-\gamma)^{-4}}}\left(|0000\rangle+(1-\gamma)^{-2}|1111\rangle\right)\\
|\bar{1}\rangle&=\sqrt{\frac{1}{2}}\left(|0011\rangle+|1100\rangle\right).
\end{split}
}

\codefieldsection{Encoding}
\begin{eczvaluelist}
\item\relax Analytical encoding channel \NoCaseChange{\protect\cite{cite3995}}.
\end{eczvaluelist}
\codefieldsection{Decoding}
\begin{eczvaluelist}
\item\relax Analytical recovery channel \NoCaseChange{\protect\cite{cite3995}}.
\end{eczvaluelist}
\codefieldsection{Parents}
\begin{eczvaluelist}
\item\relax
\flmRefsHyperref[eczindexfamilyrel]{code:qubits_into_qubits}{Qubit code}\item\relax
\flmRefsHyperref[eczindexfamilyrel]{code:ampdamp}{Amplitude-damping (AD) code}\item\relax
\flmRefsHyperref[eczindexfamilyrel]{code:numopt}{Numerically optimized bosonic code} --- Numerically optimized four-qubit AD codes can be obtained from a biconvex optimization of the entanglement fidelity \NoCaseChange{\protect\cite{cite3995}}.
\item\relax
\flmRefsHyperref[eczindexfamilyrel]{code:small_distance_quantum}{Small-distance block quantum code}\end{eczvaluelist}
\codefieldsection{Cousins}
\begin{eczvaluelist}
\item\relax
\flmRefsHyperref[eczindexfamilyrel]{code:css_4_1_2}{\(\llbracket 4,1,2\rrbracket \) Leung-Nielsen-Chuang-Yamamoto (LNCY) code} --- The numerically optimized four-qubit AD code can correct a single \flmRefsHyperref{ref498}{AD} error with higher entanglement fidelity than the \(\llbracket 4,1,2\rrbracket \) LNCY code \NoCaseChange{\protect\cite{cite859}}.
\item\relax
\flmRefsHyperref[eczindexfamilyrel]{code:reinforcement_learning}{Reinforcement-learning quantum code} --- Numerically optimized four-qubit AD codes can be obtained from a machine-learning optimization \NoCaseChange{\protect\cite{cite3996}}.
\end{eczvaluelist}
\eczhbkcontributors{ \eczhuVVA }
\endeczcode

\eczcode{hybrid_bacon_shor}{OA Bacon-Shor code}{~\NoCaseChange{\protect\cite{cite2874,cite3483}}}
\codefieldsection{Description}
Family of OA qubit stabilizer codes derived from Bacon-Shor subsystem codes by using their extra gauge structure to store classical information.

In the earlier gauge-fixing construction, one fixes commuting gauge operators and uses the corresponding gauge qubits to encode classical messages, yielding hybrid stabilizer codes with separate quantum and classical distances \NoCaseChange{\protect\cite{cite2874}}.
For the symmetric \(n\times n\) Bacon-Shor code, this gives the family \(\llbracket n^2,1:(n-1)^2,n:2\rrbracket _2\), including the \(\llbracket 9,1:4,3:2\rrbracket _2\) code.

In the later operator-algebra construction, one instead chooses a nontrivial subset \(\mathcal{T}_0\) of coset representatives of the normalizer of the Bacon-Shor stabilizer group inside the Pauli group, so that the classical information is carried by distinct normalizer-coset sectors \NoCaseChange{\protect\cite{cite3483}}.
For the symmetric \(\ell\times\ell\) Bacon-Shor code, choosing \(\mathcal{T}_0\) generated by \(\prod_{i=1}^{\lfloor \ell / 2\rfloor} X_{(2i,1)}\) and \(\prod_{j=1}^{\lfloor \ell / 2\rfloor} Z_{(1,2j-1)}\) yields a \(\llbracket \ell^2,1:2,\lceil (\ell-1)/2 \rceil\rrbracket \) OA Bacon-Shor code, whose smallest nontrivial error-correcting member is the \(\llbracket 16,1:2,2\rrbracket \) code.
\codefieldsection{Protection}
In the gauge-fixing construction, the underlying Bacon-Shor subsystem code retains its quantum distance while the classical distance is determined by the chosen gauge fixing; in particular, the symmetric \(n\times n\) family above has parameters \(\llbracket n^2,1:(n-1)^2,n:2\rrbracket _2\) \NoCaseChange{\protect\cite{cite2874}}.

In the operator-algebra construction, for the canonical stabilizer generators of the symmetric \(\ell\times\ell\) Bacon-Shor code, each single-qubit Pauli error anti-commutes with at most two \(X\)-type and two \(Z\)-type stabilizer generators.
Therefore, choosing \(X\)-type and \(Z\)-type coset representatives from classical linear codes \(C_X\) and \(C_Z\) with parameters \([\ell-1,k_X,d_X]\) and \([\ell-1,k_Z,d_Z]\), respectively, yields a hybrid Bacon-Shor code with \(|\mathcal{T}_0|=2^{k_X+k_Z}\) sectors and distance at least
\flmMathEnvironment{align}{}{
  \min\left(\ell,\left\lceil d_X/2 \right\rceil,\left\lceil d_Z/2 \right\rceil\right)~.
}
Choosing both \(C_X\) and \(C_Z\) to be the \([\ell-1,1,\ell-1]\) repetition code saturates this construction and yields the \(\llbracket \ell^2,1:2,\lceil (\ell-1)/2 \rceil\rrbracket \) family.
For \(\ell=8\), taking both \(C_X\) and \(C_Z\) to be the \([7,4,3]\) Hamming code yields a \(\llbracket 64,1:8,2\rrbracket \) OA Bacon-Shor code \NoCaseChange{\protect\cite{cite3483}}.
\codefieldsection{Parent}
\begin{eczvaluelist}
\item\relax
\flmRefsHyperref[eczindexfamilyrel]{code:qubit_stabilizer_oaqecc}{Operator-algebra (OA) qubit stabilizer code}\end{eczvaluelist}
\codefieldsection{Cousin}
\begin{eczvaluelist}
\item\relax
\flmRefsHyperref[eczindexfamilyrel]{code:bacon_shor}{Bacon-Shor code} --- Hybrid Bacon-Shor codes are obtained from Bacon-Shor subsystem codes either by gauge fixing gauge qubits into classical registers \NoCaseChange{\protect\cite{cite2874}} or by promoting a nontrivial subset of normalizer cosets to classical sectors \NoCaseChange{\protect\cite{cite3483}}.
\end{eczvaluelist}
\eczhbkcontributors{ \eczhuVVA }
\endeczcode

\eczcode{oa_qubits_into_qubits}{OA qubit code}{}
\codefieldsection{Alternative Names}
\begin{eczvaluelist}
\item\relax Hybrid subsystem qubit code
\end{eczvaluelist}
\eczhIndexCodeAliasName{oa_qubits_into_qubits}{Hybrid subsystem qubit code}

\codefieldsection{Kingdom root code}
for the \flmRefsHyperref{kingdom:qubits_into_qubits}{Qubit Kingdom}.
\codefieldsection{Description}
An OAQECC family that encompasses ordinary (i.e., subspace) qubit codes, subsystem qubit codes, and hybrid qubit codes using an operator-algebraic framework.

A simple example encompassing elements of all three subfamilies encodes a single logical qubit and a single classical bit into a direct sum of two subsystem qubit codes.
A quantum subsystem code \(\mathsf{A}_j\otimes\mathsf{B}_j\), with \(\mathsf{A}_j\) the logical qubit factor, and \(\mathsf{B}_j\) the gauge qubit factor, is associated with each of the two classical-bit values, labeled by \(j\in\{1,2\}\).
The corresponding decomposition of the Hilbert space \(\mathsf{H}\) is
\flmMathEnvironment{align}{}{
  \mathsf{H}=(\mathsf{A}_{1}\otimes\mathsf{B}_{1})\oplus(\mathsf{A}_{2}\otimes\mathsf{B}_{2})\oplus\mathsf{C}^{\perp}~,
}
where \(\mathsf{C}^\perp\) is the combined error space of both codes.
The above code reduces to a subsystem code when \(\mathsf{A}_{2}\otimes\mathsf{B}_{2}\) is trivial, reduces to a hybrid code when \(\mathsf{B}_{1,2}\) are both trivial, and reduces to an ordinary (i.e., subspace) qubit code when \(\mathsf{B}_{1}\) and \(\mathsf{A}_{2}\otimes\mathsf{B}_{2}\) are both trivial.
\codefieldsection{Parent}
\begin{eczvaluelist}
\item\relax
\flmRefsHyperref[eczindexfamilyrel]{code:oaecc}{Operator-algebra QECC (OAQECC)} --- An OAQECC defined over qubits is an OA qubit code.
\end{eczvaluelist}
\codefieldsection{Children}
\begin{eczvaluelist}
\item\relax
\flmRefsHyperref[eczindexfamilyrel]{code:hybrid_qubits_into_qubits}{Hybrid qubit code} --- An OA qubit code that has no gauge structure (e.g., gauge qubits) but has a block structure that corresponds to a classical code is a hybrid qubit code.
\item\relax
\flmRefsHyperref[eczindexfamilyrel]{code:qubit_stabilizer_oaqecc}{Operator-algebra (OA) qubit stabilizer code}\item\relax
\flmRefsHyperref[eczindexfamilyrel]{code:qubits_into_qubits}{Qubit code} --- An OA qubit code which has no gauge qubits and no block structure is a qubit code.
\item\relax
\flmRefsHyperref[eczindexfamilyrel]{code:subsystem_qubits_into_qubits}{Subsystem qubit code} --- An OA qubit code which has gauge structure (e.g., gauge qubits) but no block structure is a subsystem qubit code.
\end{eczvaluelist}
\codefieldsection{Cousin}
\begin{eczvaluelist}
\item\relax
\flmRefsHyperref[eczindexfamilyrel]{code:eaoa_qubits_into_qubits}{EAOA qubit code} --- EAOA qubit codes utilize additional ancillary qubits in a pre-shared entangled state, but reduce to ordinary OA qubit codes when said qubits are interpreted as noiseless physical qubits.
\end{eczvaluelist}
\eczhbkcontributors{ \eczhuVVA }
\endeczcode

\eczcode{qubit_stabilizer_oaqecc}{Operator-algebra (OA) qubit stabilizer code}{~\NoCaseChange{\protect\cite{cite3483}}}
\codefieldsection{Alternative Names}
\begin{eczvaluelist}
\item\relax Hybrid subsystem qubit stabilizer code
\end{eczvaluelist}
\eczhIndexCodeAliasName{qubit_stabilizer_oaqecc}{Hybrid subsystem qubit stabilizer code}
\codefieldsection{Description}
An OAQECC in which the commutant \(\mathcal{A}'\) of the logical algebra \(\mathcal{A}\) arises as the \flmRefsHyperref{ref205}{group algebra} of a subgroup \(\mathsf{G}\) of the \(n\)-qubit \flmRefsHyperref{ref663}{Pauli group} \(\mathsf{P}_n\).

The stabilizer \(\mathsf{S}\) is the center of \(\mathsf{G}\) modulo factors of \(i I\).
The quotient \(\mathsf{P}_n / \mathsf{N(S)}\), where \(\mathsf{N(S)}\) is the normalizer of \(\mathsf{S}\), is in bijective correspondence with the factors of the logical algebra \(\mathcal{A}\).
\codefieldsection{Protection}
Specialized conditions for the correctability of \(\mathcal{A}\) with respect to an error operation \(\mathcal{E}\) with operation elements \(\{E_j\}_j\) can be given in group theoretic terms.
Indeed, \(\mathcal{A}\) is correctable for \(\mathcal{E}\) if, for all \(j,k\),
\flmMathEnvironment{align}{}{
E_{j}^{\dagger}E_{k}&\notin(\mathsf{N(S)}-\mathsf{G})\cup\\&
\,\,\,\,\,\,\,\,\,\,\,\,\,\cup\left(\bigcup_{\tau,\sigma:\tau\mathsf{N(S)}\neq\sigma\mathsf{N(S)}}\tau\mathsf{N(S)}\sigma\right)~.
}
\codefieldsection{Parent}
\begin{eczvaluelist}
\item\relax
\flmRefsHyperref[eczindexfamilyrel]{code:oa_qubits_into_qubits}{OA qubit code}\end{eczvaluelist}
\codefieldsection{Children}
\begin{eczvaluelist}
\item\relax
\flmRefsHyperref[eczindexfamilyrel]{code:hybrid_stabilizer}{Hybrid stabilizer code} --- An OA stabilizer code which has no gauge qubits but has a block structure that corresponds to a linear binary code is a hybrid stabilizer code.
\item\relax
\flmRefsHyperref[eczindexfamilyrel]{code:hybrid_bacon_shor}{OA Bacon-Shor code}\item\relax
\flmRefsHyperref[eczindexfamilyrel]{code:qubit_stabilizer}{Qubit stabilizer code} --- An OA qubit stabilizer code storing no classical information and admitting no gauge qubits is a qubit stabilizer code.
\item\relax
\flmRefsHyperref[eczindexfamilyrel]{code:qubit_subsystem_stabilizer}{Subsystem qubit stabilizer code} --- An OA qubit stabilizer code storing no classical information but retaining gauge qubits for its quantum code is a subsystem qubit stabilizer code.
\end{eczvaluelist}
\codefieldsection{Cousins}
\begin{eczvaluelist}
\item\relax
\flmRefsHyperref[eczindexfamilyrel]{code:eaoa_stabilizer}{EAOA qubit stabilizer code} --- EAOA qubit stabilizer codes utilize additional ancillary subsystems in a pre-shared entangled state, but reduce to OA qubit stabilizer codes when said subsystems are interpreted as noiseless physical subsystems.
\item\relax
\flmRefsHyperref[eczindexfamilyrel]{code:quantum_synchronizable}{Quantum synchronizable code} --- Quantum synchronizable versions of qubit subsystem codes, hybrid codes, and OA qubit stabilizer codes have been constructed \NoCaseChange{\protect\cite{cite3997}}.
\end{eczvaluelist}
\eczhbkcontributors{ Michael Liaofan Liu, \eczhuVVA }
\endeczcode

\eczcode{happy}{Pastawski-Yoshida-Harlow-Preskill (HaPPY) code}{~\NoCaseChange{\protect\cite{cite1667}}}
\codefieldsection{Alternative Names}
\begin{eczvaluelist}
\item\relax Perfect holographic code
\end{eczvaluelist}
\eczhIndexCodeAliasName{happy}{Perfect holographic code}
\codefieldsection{Description}
Holographic code constructed from six-leg five-qubit \flmRefsHyperref{ref219}{perfect tensors} placed on hyperbolic pentagon and hexagon tilings.
The code serves as a minimal model for several aspects of the AdS/CFT holographic duality \NoCaseChange{\protect\cite{cite641}} and potentially a dS/CFT duality \NoCaseChange{\protect\cite{cite642}}.

It has been generalized to higher dimensions \NoCaseChange{\protect\cite{cite3998}} and to include gauge-like degrees of freedom on the links of the tensor network \NoCaseChange{\protect\cite{cite3999,cite468}}.
In the lifted version, the bulk symmetry of the HaPPY code can be interpreted as arising from restricting a bulk gauge-like theory to a fixed-flux sector \NoCaseChange{\protect\cite{cite468}}.
All boundary global symmetries must be dual to bulk gauge symmetries, and vice versa \NoCaseChange{\protect\cite{cite4000}}.

The construction below is described for qubits, but the underlying five-leg perfect tensor also has modular-qudit and oscillator extensions, and a rotor version can be stacked into an approximately error-correcting \(U(1)\)-covariant holographic code \NoCaseChange{\protect\cite{cite2720}}.
Encoding is accomplished using a tensor network of five-qubit encoding isometries, which are six-legged \flmRefsHyperref{ref219}{perfect tensors} (with five legs corresponding to the physical qubits and one for the encoded logical qubit).

To construct the encoding, one first uniformly tiles the hyperbolic AdS/CFT disc using pentagons and hexagons.
Then, one places a 6-legged five-qubit encoding tensor at each hexagon and pentagon, contracting legs between neighboring shapes and leaving one leg uncontracted at each pentagon.
This construction forms an encoding isometry from the uncontracted legs in the bulk to the uncontracted legs at the boundary.

The \textit{single-qubit HaPPY code} has a central pentagon encoding one bulk operator and hexagons tiling all other layers.
The \textit{pentagon-hexagon HaPPY code} has alternating layers of pentagons and hexagons in the tiling.
The \textit{pentagon HaPPY code} (a.k.a. the hyperbolic pentagon code, or HyPeC) consists of a purely pentagonal tiling.

\codefieldsection{Protection}
Protects against erasure errors and Pauli errors on the boundary qubits.
\codefieldsection{Rate}
The pentagon HaPPY code has an asymptotic rate \(\frac{1}{\sqrt{5}} \approx 0.447\). The pentagon-hexagon HaPPY code has a rate of \(0.299\) if the last layer is a pentagon layer and a rate of \(0.088\) if the last layer is a hexagon layer.
\codefieldsection{Encoding}
\begin{eczvaluelist}
\item\relax Heisenberg-picture encoding is done through \textit{tensor pushing}. Each bulk operator (logical) is pushed to an operator supported on a portion of the boundary region (physical). Pushing all the bulk operators through results in reconstruction of the boundary.
\item\relax ZX calculus based encoder for the pentagon HaPPY code \NoCaseChange{\protect\cite{cite3529}}.
\end{eczvaluelist}
\codefieldsection{Transversal and Permutation-Based Gates}
\begin{eczvaluelist}
\item\relax Any transversal gate of the five-qubit code is a transversal gate of the HaPPY code since the HaPPY code is constructed from five-qubit encoding tensors, which are covariant under such gates.
\item\relax For locality-preserving physical gates on the boundary, the set of transversally implementable logical operations in the bulk is strictly contained in the \flmRefsHyperref{ref409}{Clifford group} \NoCaseChange{\protect\cite{cite740}}.
\end{eczvaluelist}
\codefieldsection{Decoding}
\begin{eczvaluelist}
\item\relax Hierarchical recovery model \NoCaseChange{\protect\cite{cite1667}}.
\item\relax The greedy algorithm reconstructs bulk operators by iteratively absorbing tensors for which at least half of the legs are already included; the resulting greedy geodesic gives an explicit boundary reconstruction region \NoCaseChange{\protect\cite{cite1667}}.
\end{eczvaluelist}
\codefieldsection{Code Capacity Threshold}
\begin{eczvaluelist}
\item\relax \(26\%\) for boundary erasure errors on the pentagon-hexagon HaPPY code under the greedy decoder \NoCaseChange{\protect\cite{cite1667}}.
\item\relax Lower bound of \(1/12 \approx 8.3\%\) for boundary erasure errors on the single-qubit HaPPY code under hierarchical recovery \NoCaseChange{\protect\cite{cite1667}}. Numerical evidence indicates the threshold may be closer to \(50\%\).
\item\relax There is no threshold for the pentagon HaPPY code as a constant number of errors (four) can make bulk recovery impossible \NoCaseChange{\protect\cite{cite1667}}.
\item\relax \(16.3\%\) for boundary Pauli errors on the single-qubit HaPPY code with 3 layers using integer optimization decoder \NoCaseChange{\protect\cite{cite2954}}.
\item\relax \(50\%\) against biased Pauli noise for single-qubit HaPPY code under tensor-network decoder \NoCaseChange{\protect\cite{cite3715}}.
\end{eczvaluelist}
\codefieldsection{Threshold}
\begin{eczvaluelist}
\item\relax A single-qubit HaPPY code has a \flmRefsHyperref{ref3210}{measurement threshold} of one \NoCaseChange{\protect\cite{cite4001}}.
\end{eczvaluelist}
\codefieldsection{Notes}
\begin{eczvaluelist}
\item\relax Reference \NoCaseChange{\protect\cite{cite642}} discusses the HaPPY code for an AdS\_3 space and its relation to a dS\_2 \textit{braneworld} with a conformal boundary.
\end{eczvaluelist}
\codefieldsection{Parents}
\begin{eczvaluelist}
\item\relax
\flmRefsHyperref[eczindexfamilyrel]{code:qubit_stabilizer}{Qubit stabilizer code} --- The HaPPY code is a stabilizer code because it is defined by a contracted network of stabilizer tensors; see \NoCaseChange{\protect\cite[{Thm. 6}]{cite1667}}.
\item\relax
\flmRefsHyperref[eczindexfamilyrel]{code:holographic_tensor}{Holographic tensor-network code} --- The encoding of a HaPPY code is a holographic tensor network consisting of pentagon and hexagon \flmRefsHyperref{ref219}{perfect tensors}.
\end{eczvaluelist}
\codefieldsection{Child}
\begin{eczvaluelist}
\item\relax
\flmRefsHyperref[eczindexfamilyrel]{code:stab_5_1_3}{\(\llbracket 5,1,3\rrbracket \) Five-qubit perfect code} --- The five-qubit code is the smallest (i.e., radius-one) single-qubit HaPPY code. The five-qubit encoding isometry tiles various holographic codes because its corresponding encoding isometry tensor is a \flmRefsHyperref{ref219}{perfect tensor} \NoCaseChange{\protect\cite{cite1667}}.
\end{eczvaluelist}
\codefieldsection{Cousins}
\begin{eczvaluelist}
\item\relax
\flmRefsHyperref[eczindexfamilyrel]{code:majorana_stab}{Majorana stabilizer code} --- The pentagon HaPPY code Hamiltonian can be expressed in terms of mutually commuting weight-two (two-body) Majorana operators \NoCaseChange{\protect\cite{cite3985}}.
\item\relax
\flmRefsHyperref[eczindexfamilyrel]{code:ame}{Perfect-tensor code} --- The encoding of a HaPPY code is a holographic tensor network consisting of pentagon and hexagon \flmRefsHyperref{ref219}{perfect tensors}.
\end{eczvaluelist}
\eczhbkcontributors{ Joel Rajakumar, \eczhuVVA }
\endeczcode

\eczcode{phantom}{Phantom code}{~\NoCaseChange{\protect\cite{cite514}}}
\codefieldsection{Description}
Qubit CSS code for which, in some logical basis, every ordered-pair logical \(\overline{\mathrm{CNOT}}_{ab}\) gate between logical qubits in the same code block can be implemented by a physical-qubit permutation \NoCaseChange{\protect\cite{cite514}}.
The definition has been extended to non-CSS and non-qubit codes \NoCaseChange{\protect\cite{cite723}}.

\codefieldsection{Protection}
CSS phantom codes obey a Hamming-type constraint: if \(d=d_{\mu}\) for \(\mu\in\{X,Z\}\), then \(\eta(2^k-1)\leq B(n,d)\), where \(\eta\) counts weight-\(d\) logical operators in a fixed \(\mu\)-type logical equivalence class and \(B(n,d)\leq {n \choose d}\) is the maximum size of a binary length-\(n\) code whose pairwise sums have weight at least \(d\) \NoCaseChange{\protect\cite{cite514}}.
Any qubit phantom code of distance \(d\geq 2\) encoding \(k\geq 2\) logical qubits with \(k\neq 4\) obeys \(n\geq 2^k-1\), equivalently \(k\leq \log_2(n+1)\); this parameter bound also holds for non-CSS phantom codes and for qubit subspace or subsystem phantom-LU codes \NoCaseChange{\protect\cite{cite723}}.

\codefieldsection{Transversal and Permutation-Based Gates}
\begin{eczvaluelist}
\item\relax For CSS phantom codes, interblock \(\overline{\mathrm{CNOT}}\) gates are transversal. Combining transversal interblock CNOTs with in-block permutation CNOTs implements any logical CNOT circuit on \(2^a\) phantom-code blocks in physical depth at most \(4(2^a-1)\), up to a residual logical-qubit permutation; for unidirectional CNOT circuits, the bound is \(2(2^a-1)\) while preserving logical-qubit order \NoCaseChange{\protect\cite{cite514}}.
\item\relax A stabilizer code supporting a logical gate by qubit permutations cannot admit any strictly transversal logical gate that does not commute with that permutation-implemented logical gate, ruling out strictly transversal implementations of several gates on phantom codes \NoCaseChange{\protect\cite{cite514}}.
\item\relax Additional logical Clifford and non-Clifford gates can arise from code automorphisms combining local Cliffords and qubit permutations, from fold-diagonal gates using patterned one- and two-qubit diagonal interactions, and from non-uniform diagonal single-qubit rotations \NoCaseChange{\protect\cite{cite514}}.
\end{eczvaluelist}
\codefieldsection{Gates}
\begin{eczvaluelist}
\item\relax Certain phantom quantum RM codes admit the full logical Clifford group via fold-\(\overline{S}_i\overline{S}_j\) gates and teleported Hadamards, and admit a distance-two magic-gate scheme by temporarily projecting into hypercube-code subspaces \NoCaseChange{\protect\cite{cite514}}.
\end{eczvaluelist}
\codefieldsection{Decoding}
\begin{eczvaluelist}
\item\relax Spatiotemporal sliding-window correlated list and most-likely-error decoders for Steane-style error correction \NoCaseChange{\protect\cite{cite514}}.
\end{eczvaluelist}
\codefieldsection{Fault Tolerance}
\begin{eczvaluelist}
\item\relax Preselection-based fault-tolerant state preparation and Steane-style error correction for non-LDPC phantom quantum RM codes \NoCaseChange{\protect\cite{cite514}}.
\end{eczvaluelist}
\codefieldsection{Notes}
\begin{eczvaluelist}
\item\relax Ref. \NoCaseChange{\protect\cite{cite514}} exhaustively enumerates all \(2.71\times10^{10}\) inequivalent CSS codes with \(n\leq14\), identifying \(1.39\times10^5\) CSS phantom codes, and uses SAT-based search to find further examples up to \(n=21\).
\end{eczvaluelist}
\codefieldsection{Parent}
\begin{eczvaluelist}
\item\relax
\flmRefsHyperref[eczindexfamilyrel]{code:qubit_css}{Qubit CSS code}\end{eczvaluelist}
\codefieldsection{Children}
\begin{eczvaluelist}
\item\relax
\flmRefsHyperref[eczindexfamilyrel]{code:hypercube_quantum}{\(\llbracket 2^D,D,2\rrbracket \) hypercube quantum code} --- The \(\llbracket 2^D,D,2\rrbracket \) hypercube quantum codes are phantom codes: all ordered-pair in-block logical CNOT gates can be implemented by physical-qubit permutations \NoCaseChange{\protect\cite{cite514}}. The punctured hypercube family is unique among binary CSS phantom codes saturating \(n\geq 2^k-1\) for \(k=3\) and \(k\geq 5\) \NoCaseChange{\protect\cite[{Thm. 4}]{cite723}}.
\item\relax
\flmRefsHyperref[eczindexfamilyrel]{code:stab_12_2_2}{\(\llbracket 12,2,2\rrbracket \) CSS code} --- Ref. \NoCaseChange{\protect\cite{cite514}} identifies a \(\llbracket 12,2,2\rrbracket \) CSS phantom code by exhaustive enumeration.
\item\relax
\flmRefsHyperref[eczindexfamilyrel]{code:phantom_14_3_3}{\(\llbracket 14,3,3\rrbracket \) CE phantom code} --- This \(\llbracket 14,3,3\rrbracket \) code is a CSS phantom code obtained from the punctured hypercube code and the two-qubit phase-flip repetition code \NoCaseChange{\protect\cite{cite514}}.
\item\relax
\flmRefsHyperref[eczindexfamilyrel]{code:hgp_7_2_2}{\(\llbracket 7,2,2\rrbracket \) HGP phantom code} --- The \(\llbracket 7,2,2\rrbracket \) HGP code is the smallest member of a simplex/repetition HGP family of CSS phantom codes \NoCaseChange{\protect\cite{cite514}}.
\item\relax
\flmRefsHyperref[eczindexfamilyrel]{code:xz_7_3_2}{\(\llbracket 7,3,2\rrbracket \) punctured hypercube code} --- This code is the punctured version of the \(\llbracket 8,3,2\rrbracket \) hypercube quantum code and is a phantom code \NoCaseChange{\protect\cite{cite514}}.
\item\relax
\flmRefsHyperref[eczindexfamilyrel]{code:bc_phantom}{Binarized-and-concatenated (B\&C) phantom code}\end{eczvaluelist}
\codefieldsection{Cousins}
\begin{eczvaluelist}
\item\relax
\flmRefsHyperref[eczindexfamilyrel]{code:quantum_reed_muller}{Quantum Reed-Muller (RM) code} --- Some quantum RM codes are phantom after selected logical qubits of a parent quantum RM code are fixed to \(\ket{\overline{0}}\) or \(\ket{\overline{+}}\), promoting the corresponding logical operators to stabilizers \NoCaseChange{\protect\cite{cite514}}.
\item\relax
\flmRefsHyperref[eczindexfamilyrel]{code:qubit_concatenated}{Concatenated qubit code} --- Concatenating a phantom outer code with a one-logical-qubit inner quantum code preserves phantomness.
\item\relax
\flmRefsHyperref[eczindexfamilyrel]{code:hypergraph_product}{Hypergraph product (HGP) code} --- Some hypergraph-product constructions, such as products of a classical simplex code and a repetition code, yield phantom codes, while the smallest examples have lower rates than the phantom quantum RM constructions \NoCaseChange{\protect\cite{cite514}}.
\item\relax
\flmRefsHyperref[eczindexfamilyrel]{code:stab_4_2_2}{\(\llbracket 4,2,2\rrbracket \) Four-qubit code} --- The \(\llbracket 4,2,2\rrbracket \) code is the smallest phantom code: logical CNOT gates between its two logical qubits can be implemented by physical-qubit permutations \NoCaseChange{\protect\cite{cite514}}. Gluing copies of the \(\llbracket 4,2,2\rrbracket \) code with \(X\)-type stabilizers yields CSS phantom codes with parameters \(\llbracket 4m,2,(d_X=2,d_Z=2m)\rrbracket \), and puncturing one qubit from this construction yields \(\llbracket 4m-1,2,(d_X=2,d_Z=2m-1)\rrbracket \), for \(m\geq1\) \NoCaseChange{\protect\cite{cite514}}.
\item\relax
\flmRefsHyperref[eczindexfamilyrel]{code:qubit_8_4_2}{\(\llparenthesis 8,16,2\rrparenthesis \) \(PG(3,2)\) code} --- This is the exceptional nonstabilizer \(k=4\) qubit phantom code of minimal length eight that violates the generic bound \(n\geq 2^k-1\) \NoCaseChange{\protect\cite{cite723}}.
\end{eczvaluelist}
\eczhbkcontributors{ \eczhuVVA }
\endeczcode

\eczcode{qubit_permutation_invariant}{PI qubit code}{}
\codefieldsection{Description}
Block quantum code defined on two-dimensional subsystems such that any permutation of the subsystems leaves any codeword invariant.

\begin{defterm}{Dicke states}\label{ref4002}\label{ref526}
For \(n\)-qubit block codes, an often used basis for the \(n+1\)-dimensional PI subspace consists of the Dicke states \(|D^n_w\rangle\) -- normalized PI states of \(w\) excitations, i.e., a normalized sum over all binary-string basis elements with \(w\) ones and \(n - w\) zeroes.
For example, the single-excitation Dicke state, known as a \(W\) \textit{state}, on three qubits is
\flmMathEnvironment{align}{}{
  |D_{1}^{3}\rangle=\frac{1}{\sqrt{3}}\left(|001\rangle+|010\rangle+|100\rangle\right)~.
}
The \(n+1\)-dimensional PI space can be thought of as a standalone spin-\(n/2\) quantum system, yielding a way to convert between PI qubit codes and \(SU(2)\) spin codes.
A single-spin code for the \(SU(2)\) group correcting spherical tensors can be mapped into a PI qubit code with an analogous distance \NoCaseChange{\protect\cite{cite2814}\protect\cite[{Thm. 1}]{cite3488}}.
\end{defterm}

\codefieldsection{Protection}
Permutation invariant qubit codes of distance \(d\) can protect against \(d-1\) deletion errors \NoCaseChange{\protect\cite{cite2655,cite2656}}, i.e., erasures of subsystems at unknown locations.
There are also simplified conditions on insertion errors \NoCaseChange{\protect\cite{cite4003}}.

\codefieldsection{Encoding}
\begin{eczvaluelist}
\item\relax With quantum harmonic oscillators (superconducting charge qubits in an ultrastrong coupling regime) in \(O(N)\) as in \NoCaseChange{\protect\cite{cite4004}}. Can be done in \(O(N^2)\) steps using quantum circuits \NoCaseChange{\protect\cite{cite4005}}, or using geometric phase gates in \(O(N)\) \NoCaseChange{\protect\cite{cite4006}}.
\item\relax Finite-depth quantum circuits with LOCC for Dicke states \NoCaseChange{\protect\cite{cite4007,cite4008,cite4009}}.
\item\relax Preparation of sparse Dicke states using the combinatorial number system \NoCaseChange{\protect\cite{cite2011}}.
\end{eczvaluelist}
\codefieldsection{Gates}
\begin{eczvaluelist}
\item\relax There is a measurement-free code-switching protocol between a qubit stabilizer code and a PI qubit code \NoCaseChange{\protect\cite{cite727}}.
\end{eczvaluelist}
\codefieldsection{Decoding}
\begin{eczvaluelist}
\item\relax Schur-Weyl-transform based decoder \NoCaseChange{\protect\cite{cite2761}}. Here, one first measures nested total angular momenta, i.e., that of the first qubit, the first and second, followed by the first, second, and third, etc. Then, for codes with spacing, one measures the projection of the angular momentum modulo the spacing. Recovery can be performed by applying geometric phase gates \NoCaseChange{\protect\cite{cite4010}} or the quantum Schur transform. This decoder has been generalized to work with insertion errors \NoCaseChange{\protect\cite{cite3691}}.
\end{eczvaluelist}
\codefieldsection{Parents}
\begin{eczvaluelist}
\item\relax
\flmRefsHyperref[eczindexfamilyrel]{code:qubits_into_qubits}{Qubit code}\item\relax
\flmRefsHyperref[eczindexfamilyrel]{code:permutation_invariant}{Permutation-invariant (PI) code}\end{eczvaluelist}
\codefieldsection{Children}
\begin{eczvaluelist}
\item\relax
\flmRefsHyperref[eczindexfamilyrel]{code:ampdamp_post_selected}{Post-selected PI code}\item\relax
\flmRefsHyperref[eczindexfamilyrel]{code:binary_dihedral_permutation_invariant}{Binary dihedral PI code}\item\relax
\flmRefsHyperref[eczindexfamilyrel]{code:combinatorial_permutation_invariant}{Combinatorial PI code}\item\relax
\flmRefsHyperref[eczindexfamilyrel]{code:qudit_gnu_permutation_invariant}{Qudit GNU PI code}\item\relax
\flmRefsHyperref[eczindexfamilyrel]{code:unentangled_permutation_invariant}{\(\llparenthesis n,2,2\rrparenthesis \) Bravyi-Lee-Li-Yoshida PI code}\end{eczvaluelist}
\codefieldsection{Cousins}
\begin{eczvaluelist}
\item\relax
\flmRefsHyperref[eczindexfamilyrel]{code:qubit_stabilizer}{Qubit stabilizer code} --- There is a measurement-free code-switching protocol between a qubit stabilizer code and a PI qubit code \NoCaseChange{\protect\cite{cite727}}.
\item\relax
\flmRefsHyperref[eczindexfamilyrel]{code:eth}{Eigenstate thermalization hypothesis (ETH) code} --- Several instances of ETH codes contain PI qubit codewords.
\item\relax
\flmRefsHyperref[eczindexfamilyrel]{code:j_gross}{Clifford-group spin code} --- Clifford codes for spins housing representations of \(SU(2)\) yield PI qubit codes with non-trivial distance when the single spin-\(n/2\) is treated as the permutationally invariant subspace of \(n\) qubits via the \flmRefsHyperref{ref526}{Dicke-state mapping}. The subgroup of gates of a Clifford-group spin code is implemented transversally via this mapping \NoCaseChange{\protect\cite{cite2814}}.
\item\relax
\flmRefsHyperref[eczindexfamilyrel]{code:single_spin}{Single-spin code} --- Single-spin codes are subspaces of a single large \(SU(2)\) spin, which can be either standalone or correspond to the PI subspace of a set of spins via the \flmRefsHyperref{ref526}{Dicke state mapping}.
\end{eczvaluelist}
\eczhbkcontributors{ \eczhuVVA }
\endeczcode

\eczcode{ampdamp_post_selected}{Post-selected PI code}{~\NoCaseChange{\protect\cite{cite2598}}}
\codefieldsection{Description}
PI qubit code whose recovery succeeds at protecting against \flmRefsHyperref{ref498}{AD} errors with a success probability less than one.

The simplest code admits a codeword basis of
\flmMathEnvironment{align}{}{
\begin{split}
|\overline{0}\rangle&=\frac{1}{\sqrt{3}}\left(|100\rangle+|010\rangle+|001\rangle\right)\\
|\overline{1}\rangle&=|111\rangle~.
\end{split}
}
The code violates the diagonal part of the \flmTerm{term}{ref1043}{}{Knill-Laflamme conditions}.
Nevertheless, the code admits a probabilistic recovery that protects against single losses and yields an infidelity of \flmRefsHyperref{ref65}{order} \(O(\gamma^2)\) in the noise rate \(\gamma\).
The failure probability of the recovery is of the same \flmRefsHyperref{ref65}{order} as the probability of the single loss errors, i.e., \(O(\gamma)\).

\codefieldsection{Realizations}
\begin{eczvaluelist}
\item\relax Superconducting circuits: IBM quantum hardware \NoCaseChange{\protect\cite{cite4011}}.
\end{eczvaluelist}
\codefieldsection{Parents}
\begin{eczvaluelist}
\item\relax
\flmRefsHyperref[eczindexfamilyrel]{code:qubit_permutation_invariant}{PI qubit code}\item\relax
\flmRefsHyperref[eczindexfamilyrel]{code:ampdamp}{Amplitude-damping (AD) code}\end{eczvaluelist}
\eczhbkcontributors{ Sourav Dutta, Aditya Jain, Prabha Mandayam, \eczhuVVA }
\endeczcode

\eczcode{real_projective_plane}{Projective-plane surface code}{~\NoCaseChange{\protect\cite{cite3383,cite71}}}
\codefieldsection{Description}
A family of Kitaev surface codes on the non-orientable 2-dimensional compact manifold \(\mathbb{R}P^2\) (in contrast to a genus-\(g\) surface).
Whereas genus-\(g\) surface codes require \(2g\) logical qubits, qubit codes on \(\mathbb{R}P^2\) are made from a single logical qubit.
\codefieldsection{Protection}
If \(\mathcal{C}\) is a cellulation of \(\mathbb{R}P^2\), then the bit-flip distance \(d_X\) is the shortest cycle in \(\mathcal{C}\), and the phase-flip distance \(d_Z\) is the shortest cycle in the dual cellulation \(\mathcal{C}^*\).
\codefieldsection{Rate}
The rate is \(1/n\), where \(n\) is the number of edges of the particular cellulation.
\codefieldsection{Gates}
\begin{eczvaluelist}
\item\relax Fault-tolerant Hadamard gate via constant-depth \flmRefsHyperref{ref409}{Clifford circuit} \NoCaseChange{\protect\cite{cite3762}}.
\item\relax Complete logical gate set for a stack of projective-plane surface codes \NoCaseChange{\protect\cite{cite3762}}.
\end{eczvaluelist}
\codefieldsection{Fault Tolerance}
\begin{eczvaluelist}
\item\relax Fault-tolerant Hadamard gate via constant-depth \flmRefsHyperref{ref409}{Clifford circuit} \NoCaseChange{\protect\cite{cite3762}}.
\end{eczvaluelist}
\codefieldsection{Parent}
\begin{eczvaluelist}
\item\relax
\flmRefsHyperref[eczindexfamilyrel]{code:surface}{Kitaev surface code} --- The projective-plane surface code is the surface code on \(\mathbb{R}P^2\).
\end{eczvaluelist}
\codefieldsection{Child}
\begin{eczvaluelist}
\item\relax
\flmRefsHyperref[eczindexfamilyrel]{code:shor_nine}{\(\llbracket 9,1,3\rrbracket \) Shor code} --- The Shor code is one of the nine-qubit surface codes defined on the projective plane \NoCaseChange{\protect\cite[{Fig. 4}]{cite3383}\protect\cite[{Fig. 20}]{cite71}}.
\end{eczvaluelist}
\codefieldsection{Cousins}
\begin{eczvaluelist}
\item\relax
\flmRefsHyperref[eczindexfamilyrel]{code:honeycomb_floquet}{Honeycomb Floquet code} --- Implementing the honeycomb Floquet code on a non-orientable cross-cap geometry allows for a logical-\(HZ\) gate to be implemented via a measurement schedule \NoCaseChange{\protect\cite{cite3762}}.
\item\relax
\flmRefsHyperref[eczindexfamilyrel]{code:stab_9_1_3}{\(\llbracket 9,1,3\rrbracket _{\mathbb{Z}_q}\) modular-qudit code} --- The qudit Shor code is a small qudit surface code on a Möbius strip with smooth boundary, which is obtained from removing a face of the tessellation of the projective plane \(\mathbb{R}P^2\) \NoCaseChange{\protect\cite[{Fig. 4}]{cite3383}}.
\end{eczvaluelist}
\eczhbkcontributors{ Eric Kubischta, \eczhuVVA }
\endeczcode

\eczcode{purity_testing}{Purity-testing stabilizer code}{~\NoCaseChange{\protect\cite{cite3650}}}
\codefieldsection{Description}
A qubit stabilizer code that is constructed from a normal rational curve and that is relevant to testing the purity of an entangled Bell state shared by two parties \NoCaseChange{\protect\cite{cite3650}}.
Purity-testing stabilizer codes have been generalized to come from more general non-projective codes \NoCaseChange{\protect\cite{cite4012}}.

\codefieldsection{Parent}
\begin{eczvaluelist}
\item\relax
\flmRefsHyperref[eczindexfamilyrel]{code:qubit_stabilizer}{Qubit stabilizer code}\end{eczvaluelist}
\codefieldsection{Cousins}
\begin{eczvaluelist}
\item\relax
\flmRefsHyperref[eczindexfamilyrel]{code:projective}{Projective geometry code} --- Purity-testing stabilizer codes are constructed from normal rational curves.
\item\relax
\flmRefsHyperref[eczindexfamilyrel]{code:eastab}{EA qubit stabilizer code} --- Purity-testing stabilizer codes are relevant to testing the purity of an entangled Bell state shared by two parties \NoCaseChange{\protect\cite{cite3650}}.
\item\relax
\flmRefsHyperref[eczindexfamilyrel]{code:quantum_secret_sharing}{Approximate secret-sharing code} --- The purity-testing protocol of Ref. \NoCaseChange{\protect\cite{cite3650}} can be improved using approximate codes similar to the approximate secret-sharing codes \NoCaseChange{\protect\cite{cite4013}}.
\end{eczvaluelist}
\eczhbkcontributors{ \eczhuVVA }
\endeczcode

\eczcode{check_product}{Quantum check-product code}{~\NoCaseChange{\protect\cite{cite2185}}}
\codefieldsection{Description}
CSS code constructed from an extension of the check product (between two classical codes) to a product between a classical and a quantum code.

\codefieldsection{Parents}
\begin{eczvaluelist}
\item\relax
\flmRefsHyperref[eczindexfamilyrel]{code:qubit_css}{Qubit CSS code}\item\relax
\flmRefsHyperref[eczindexfamilyrel]{code:qltc}{Quantum locally testable code (QLTC)} --- Quantum check-product constructions yield an SLTC code with constant soundness \(2\rho\) from a classical LTC code with soundness \(\rho\) \NoCaseChange{\protect\cite{cite2185}}. These form the first bona-fide QLTC family because they admit asymptotically constant soundness, but they are not practical because their distance is two.
\end{eczvaluelist}
\codefieldsection{Cousins}
\begin{eczvaluelist}
\item\relax
\flmRefsHyperref[eczindexfamilyrel]{code:tensor}{Tensor-product code} --- Quantum check-product codes extend the concept of a check product, which yields the dual of a tensor code, to a product between a classical and a quantum code.
\item\relax
\flmRefsHyperref[eczindexfamilyrel]{code:quantum_tensor_product}{Quantum tensor-product code} --- Quantum check-product codes extend the concept of a check product, which yields the dual of a tensor code, to a product between a classical and a quantum code.
\item\relax
\flmRefsHyperref[eczindexfamilyrel]{code:distance_balanced}{Distance-balanced code} --- Quantum check-product code constructions use distance balancing to increase distance \NoCaseChange{\protect\cite{cite2185}}.
\end{eczvaluelist}
\eczhbkcontributors{ \eczhuVVA }
\endeczcode

\eczcode{quantum_convolutional}{Quantum convolutional code}{~\NoCaseChange{\protect\cite{cite4014,cite4015,cite3180,cite4016}}}
\codefieldsection{Description}
1D translationally invariant qubit stabilizer code whose stabilizer group can be partitioned into constant-size subsets of constant support and of constant overlap between neighboring sets.
Initially formulated as a quantum analogue of convolutional codes, which were designed to protect a continuous and never-ending stream of information.
Precise formulations sometimes begin with a finite-dimensional lattice, with the intent to take the thermodynamic limit; logical dimension can be infinite as well.

Quantum convolutional codes, like their classical counterparts, can also be understood in terms of frames. Let each encoding frame take in \(n\) qubits, carry \(m\) qubits of information between frames, and act on them with \(n-k\) Pauli generators. Each generator, countably infinite in length, must commute with each \(n\) register shift of itself, but need not commute with the other generators \NoCaseChange{\protect\cite{cite4017}}. The \(m\) qubits of information carried between each frame are also stabilized by additional memory Pauli operators. It is known that the minimal value for \(m\) is given by \(\text{dim}(M)-\frac{1}{2}\text{rank}(M)\), with \(M\) being the matrix containing the required commutation relations of the memory qubits \NoCaseChange{\protect\cite{cite4018,cite4019,cite3641}}. These operators can be efficiently determined \NoCaseChange{\protect\cite{cite4020}}.
\codefieldsection{Encoding}
\begin{eczvaluelist}
\item\relax Encoding is efficient and uses only \flmRefsHyperref{ref409}{Clifford gates}. Some encoders yield \textit{catastrophic} errors, i.e., errors that require a circuit of infinite depth to correct \NoCaseChange{\protect\cite[{Def. 4.1}]{cite4016}}.
\item\relax Pearl-necklace encoding \NoCaseChange{\protect\cite{cite3180,cite4016,cite4021,cite4022}}.
\item\relax Quantum shift register encoding \NoCaseChange{\protect\cite{cite4023}}.
\item\relax Encoding circuits can be viewed as matrix-product-state tensor networks \NoCaseChange{\protect\cite{cite400}}.
\end{eczvaluelist}
\codefieldsection{Decoding}
\begin{eczvaluelist}
\item\relax Quantum Viterbi decoder \NoCaseChange{\protect\cite{cite4015,cite3180,cite4016}}.
\item\relax ML decoder \NoCaseChange{\protect\cite{cite3180}}.
\end{eczvaluelist}
\codefieldsection{Notes}
\begin{eczvaluelist}
\item\relax See Refs. \NoCaseChange{\protect\cite{cite3180,cite1666,cite4021,cite4024}} for explicit and simple examples.
\item\relax See Ref. \NoCaseChange{\protect\cite{cite4025}} and the book \NoCaseChange{\protect\cite{cite398}} for an introduction to quantum convolutional codes.
\item\relax Quantum convolutional codes are briefly reviewed in \NoCaseChange{\protect\cite[{Sec. 27.6}]{cite2024}}.
\end{eczvaluelist}
\codefieldsection{Parents}
\begin{eczvaluelist}
\item\relax
\flmRefsHyperref[eczindexfamilyrel]{code:qldpc}{Qubit QLDPC code}\item\relax
\flmRefsHyperref[eczindexfamilyrel]{code:1d_stabilizer}{1D lattice stabilizer code} --- Quantum convolutional codes are lattice stabilizer codes on a semi-infinite or infinite lattice in one dimension \NoCaseChange{\protect\cite{cite4026}}. Some notions may be extendable to non-stabilizer codes \NoCaseChange{\protect\cite[{Sec. 4}]{cite4016}}.
Any prime-qudit code can be converted using a constant-depth \flmRefsHyperref{ref409}{Clifford circuit} to several copies of the 1D repetition code along with some trivial codes \NoCaseChange{\protect\cite{cite3963}}.

\end{eczvaluelist}
\codefieldsection{Children}
\begin{eczvaluelist}
\item\relax
\flmRefsHyperref[eczindexfamilyrel]{code:quantum_irregular_convolutional}{Quantum irregular convolutional code (QIRCC)}\item\relax
\flmRefsHyperref[eczindexfamilyrel]{code:quantum_turbo}{Quantum turbo code}\item\relax
\flmRefsHyperref[eczindexfamilyrel]{code:stab_5_1_2_convolutional}{\((5,1,2)\)-convolutional code}\end{eczvaluelist}
\codefieldsection{Cousins}
\begin{eczvaluelist}
\item\relax
\flmRefsHyperref[eczindexfamilyrel]{code:quantum_lego}{Tensor-network code} --- Quantum convolutional encoding circuits can be viewed as matrix-product-state tensor networks \NoCaseChange{\protect\cite{cite400}}.
\item\relax
\flmRefsHyperref[eczindexfamilyrel]{code:generalized_reed_solomon}{Generalized RS (GRS) code} --- GRS codes can be used to construct quantum convolutional codes \NoCaseChange{\protect\cite[{Ch. 12}]{cite872}}.
\item\relax
\flmRefsHyperref[eczindexfamilyrel]{code:generalized_reed_muller}{Generalized RM (GRM) code} --- GRM codes can be used to construct quantum convolutional codes \NoCaseChange{\protect\cite{cite1838,cite1839}\protect\cite[{Sec. 12.4}]{cite872}}.
\item\relax
\flmRefsHyperref[eczindexfamilyrel]{code:convolutional}{Convolutional code} --- Quantum convolutional codes are quantum analogues of convolutional codes.
\item\relax
\flmRefsHyperref[eczindexfamilyrel]{code:ea_quantum_convolutional}{EA quantum convolutional code} --- EA quantum convolutional codes are entanglement-assisted versions of quantum convolutional codes.
\item\relax
\flmRefsHyperref[eczindexfamilyrel]{code:hybrid_convolutional}{Hybrid convolutional code} --- Hybrid convolutional codes are hybrid analogues of quantum convolutional codes.
\item\relax
\flmRefsHyperref[eczindexfamilyrel]{code:data_syndrome}{Quantum data-syndrome (QDS) code} --- The QDS code framework has been extended to quantum convolutional codes \NoCaseChange{\protect\cite{cite4027}}.
\item\relax
\flmRefsHyperref[eczindexfamilyrel]{code:quantum_reed_muller}{Quantum Reed-Muller (RM) code} --- Quantum convolutional codes can be derived from quantum RM codes \NoCaseChange{\protect\cite{cite4028}}.
\end{eczvaluelist}
\eczhbkcontributors{ Lane G. Gunderman, \eczhuVVA }
\endeczcode

\eczcode{data_syndrome}{Quantum data-syndrome (QDS) code}{~\NoCaseChange{\protect\cite{cite1970,cite4029,cite4030,cite4031,cite2914}}}
\codefieldsection{Description}
Stabilizer code designed to correct both data qubit errors and syndrome measurement errors simultaneously due to extra redundancy in its stabilizer generators.

The redundancy can be added to any \(\llbracket n,n-m\rrbracket \) qubit stabilizer code by expanding its stabilizer generator matrix \(H\) as
\flmMathEnvironment{align}{}{
  H_{DS}=\begin{pmatrix}H & I_{m} & 0\\
  0 & A^{T} & I_{r}
  \end{pmatrix}~,
}
where the redundancy is provided by the underlying \([m+r,m]\) \textit{syndrome measurement code} with generator matrix \(G= (I_m|A)\) \NoCaseChange{\protect\cite{cite2914}}.

\codefieldsection{Protection}
Protects against both physical qubit and syndrome measurement errors.
An \(\llbracket n,k,d:r\rrbracket \) QDS code corrects any combination of \(t_{\mathrm{D}}\) data-qubit errors and \(t_{\mathrm{S}}\) syndrome-bit errors whenever \(t_{\mathrm{D}}+t_{\mathrm{S}}<d/2\), and its distance cannot exceed that of the underlying stabilizer code \NoCaseChange{\protect\cite{cite2914}}.

Random QDS codes with \(r\leq n-k\) can attain the stabilizer Gilbert-Varshamov bound \NoCaseChange{\protect\cite[{Thm. 12}]{cite2914}}.
Quantum Singleton bounds, quantum Hamming bounds, and \flmRefsHyperref{ref672}{quantum MacWilliams identities} can be extended to QDS codes.
Single-error-correcting QDS codes stemming from \flmRefsHyperref{ref672}{impure} stabilizer codes must satisfy a variant of the quantum Hamming bound \NoCaseChange{\protect\cite{cite4032}}.

\codefieldsection{Gates}
\begin{eczvaluelist}
\item\relax Fault-tolerant flag-based non-transversal logical gates \NoCaseChange{\protect\cite{cite3201}}.
\end{eczvaluelist}
\codefieldsection{Decoding}
\begin{eczvaluelist}
\item\relax Syndrome errors are decoded using redundant stabilizer measurements.
\item\relax Syndrome-measurement codes can outperform repeated syndrome extraction; for the Steane code, a \([15,3]\) syndrome-measurement code uses the same 15 measurements as five-fold repetition of three syndrome bits while achieving lower syndrome-decoding error \NoCaseChange{\protect\cite[{Fig. 1}]{cite2914}}.
\end{eczvaluelist}
\codefieldsection{Fault Tolerance}
\begin{eczvaluelist}
\item\relax Shor error correction can be recast as a QDS code whose underlying matrix \(A\) is the identity matrix \(I_m\) repeated \(\ell\) times \NoCaseChange{\protect\cite{cite2914}}.
\item\relax Fault-tolerant flag-based non-transversal logical gates \NoCaseChange{\protect\cite{cite3201}}.
\end{eczvaluelist}
\codefieldsection{Notes}
\begin{eczvaluelist}
\item\relax QDS codes can be used to estimate physical Pauli noise up to their \flmRefsHyperref{ref672}{pure distance} \NoCaseChange{\protect\cite{cite4033}}, and logical Pauli noise for any correctable physical noise \NoCaseChange{\protect\cite{cite4034}}.
\end{eczvaluelist}
\codefieldsection{Parent}
\begin{eczvaluelist}
\item\relax
\flmRefsHyperref[eczindexfamilyrel]{code:qubit_stabilizer}{Qubit stabilizer code} --- QDS codes are stabilizer codes whose stabilizer generators encode extra redundancy (via a linear binary code) so as to protect from syndrome measurement errors.
\end{eczvaluelist}
\codefieldsection{Children}
\begin{eczvaluelist}
\item\relax
\flmRefsHyperref[eczindexfamilyrel]{code:steane}{\(\llbracket 7,1,3\rrbracket \) Steane code} --- There exists a set of stabilizer generators for the Steane code that make it a QDS code; a \([15,3]\) syndrome-measurement code beats five-fold repeated syndrome extraction at the same measurement cost \NoCaseChange{\protect\cite{cite1970,cite2914}}.
\item\relax
\flmRefsHyperref[eczindexfamilyrel]{code:qubit_golay}{\(\llbracket 23, 1, 7\rrbracket \) Quantum Golay code} --- There exists a \(\llbracket 23,1,7:18\rrbracket \) QDS code based on the qubit Golay code, requiring 18 additional stabilizer measurements instead of 24 from the general cyclic construction \NoCaseChange{\protect\cite[{Ex. 15}]{cite2914}}.
\end{eczvaluelist}
\codefieldsection{Cousins}
\begin{eczvaluelist}
\item\relax
\flmRefsHyperref[eczindexfamilyrel]{code:quantum_hamming_css}{\(\llbracket 2^r-1, 2^r-2r-1, 3\rrbracket \) quantum Hamming code} --- Because every stabilizer generator has the same weight \(2^{r-1}\), quantum Hamming codes admit QDS extensions based on good binary syndrome-measurement codes \NoCaseChange{\protect\cite{cite2914}}.
\item\relax
\flmRefsHyperref[eczindexfamilyrel]{code:quantum_mds}{Quantum maximum-distance-separable (MDS) code} --- The quantum Singleton bound can be extended to QDS codes \NoCaseChange{\protect\cite{cite2914}}.
\item\relax
\flmRefsHyperref[eczindexfamilyrel]{code:quantum_perfect}{Perfect quantum code} --- The quantum Hamming bound can be extended to QDS codes \NoCaseChange{\protect\cite{cite2914}}.
\item\relax
\flmRefsHyperref[eczindexfamilyrel]{code:binary_linear}{Linear binary code} --- The QDS code construction employs a particular binary linear code to provide protection against syndrome measurement errors.
\item\relax
\flmRefsHyperref[eczindexfamilyrel]{code:quantum_convolutional}{Quantum convolutional code} --- The QDS code framework has been extended to quantum convolutional codes \NoCaseChange{\protect\cite{cite4027}}.
\item\relax
\flmRefsHyperref[eczindexfamilyrel]{code:single_shot}{Single-shot code} --- QDS codes are closely related to single-shot codes because both use redundant syndrome information to suppress measurement errors in a single round of syndrome extraction \NoCaseChange{\protect\cite{cite675}}.
\item\relax
\flmRefsHyperref[eczindexfamilyrel]{code:narrow_sense_q-ary_bch}{Primitive narrow-sense BCH code} --- Primitive narrow-sense BCH codes can be used as the syndrome measurement codes of a QDS code \NoCaseChange{\protect\cite{cite1969}}. This construction requires fewer measurements than a previous general construction \NoCaseChange{\protect\cite{cite1970}}.
\item\relax
\flmRefsHyperref[eczindexfamilyrel]{code:qubit_subsystem_stabilizer}{Subsystem qubit stabilizer code} --- The DS construction can be extended to subsystem qubit stabilizer codes \NoCaseChange{\protect\cite{cite4032}}.
\item\relax
\flmRefsHyperref[eczindexfamilyrel]{code:galois_quad_residue}{Quantum quadratic-residue (QR) code} --- CSS QDS codes can be constructed from dual-containing cyclic codes without reducing distance; for \(p=8j-1\), quantum QR codes yield \(\llbracket p,1,d:r\rrbracket \) QDS codes with \(r\leq p+1\) \NoCaseChange{\protect\cite[{Thms. 13,14}]{cite2914}}.
\end{eczvaluelist}
\eczhbkcontributors{ \eczhuVVA }
\endeczcode

\eczcode{quantum_divisible}{Quantum divisible code}{~\NoCaseChange{\protect\cite{cite4035,cite756}}}
\codefieldsection{Description}
A level-\(\nu\) quantum divisible code is a CSS code whose \(X\)-type stabilizers form a \(\nu\)-even linear binary code in the \flmRefsHyperref{ref817}{symplectic representation} and which admits a transversal gate at the \(\nu\)th level of the \flmTerm{term}{ref694}{}{Clifford hierarchy}.
A CSS code is \textit{doubly even} (\textit{triply even}) if all \(X\)-type stabilizers have weight divisible by four (eight), i.e., if they form a doubly even (triply even) linear binary code.

The definition can be generalized to \textit{weakly} \(\nu\)-\textit{divisible} (see, e.g., Ref. \NoCaseChange{\protect\cite{cite760}}), which means that there exist some disjoint qubit subsets \(M^{\pm}\) such that 
\flmMathEnvironment{align}{}{
  | x \cap M^{+} | - | x \cap M^{-} | \equiv 0 \mod \nu
}
for all rows \(x\) of the code's \(X\)-type stabilizer generator matrix.
CSS codes satisfying the above with \(\nu = 2\) (\(\nu = 4\), \(\nu = 8\)) are called \textit{weak even} (\textit{weak doubly even}, \textit{weak triply even}).
This generalization reduces to the original definition when \(M^{+}\) is the full set of qubits, and \(M^{-}\) the empty set.

An alternative definition \NoCaseChange{\protect\cite{cite764,cite765}}, not used here, is a CSS code defined from two linear binary codes \(C_{1,2}\) such that it is quantum divisible with \(\nu > 1\), and all weights in each coset of \(C_2\) in \(C_1\) are congruent to \(\nu\).
For example \NoCaseChange{\protect\cite{cite765}}, if \(C_2\) is the first-order RM\((1,m)\) code, and \(C_1/ C_2\) consists of quadratic forms with a bounded rank, then \(\llbracket n = 2^m − 1, 1 \leq k \leq 1 + \sum_{i=1}^{m-4}(m − i), d = 3\rrbracket \) is a family of such codes.

\codefieldsection{Transversal and Permutation-Based Gates}
\begin{eczvaluelist}
\item\relax A self-dual weakly doubly even \(\llbracket n,1,d\rrbracket \) CSS code admits a partitioned transversal physical \(S\) gate that realizes \(\overline{S}^m\), where \(m=|M^+|-|M^-| \pmod 4\); for odd \(m\), together with transversal Hadamard and CNOT, this yields the full logical Clifford group transversally \NoCaseChange{\protect\cite{cite731}\protect\cite[{Lemma 4}]{cite760}}.
\item\relax A weakly triply even \(\llbracket n,1,d\rrbracket \) CSS code with a strongly transversal logical \(X\) gate admits a partitioned transversal physical \(T\) gate that realizes \(\overline{T}^m\), where \(m=|M^+|-|M^-| \pmod 8\) \NoCaseChange{\protect\cite{cite731}\protect\cite[{Lemma 3}]{cite760}}.
\item\relax If the \(X\)-type stabilizers of a CSS code form an \(\nu\)-even classical code, and if all \(X\)-type logicals are \((\nu-1)\)-even, then the code admits a diagonal transversal gate in the \(\nu\)th level of the \flmTerm{term}{ref694}{}{Clifford hierarchy} \NoCaseChange{\protect\cite[{Prop. 8}]{cite702}}.
\end{eczvaluelist}
\codefieldsection{Gates}
\begin{eczvaluelist}
\item\relax The \(\llbracket 2^m − 1, 1 \leq k \leq 1 + \sum_{i=1}^{m-4}(m − i), 3\rrbracket \) quantum divisible code family can serve as outer codes of either the five-qubit \(\llbracket 5,1,3\rrbracket \) or Steane \(\llbracket 7,1,3\rrbracket \) code to realize a \(T\) gate on the inner code \NoCaseChange{\protect\cite{cite765}}.
For example, when \(m=5\) (\(m=6\)), the resulting \(\llbracket 31,5,3\rrbracket \) (\(\llbracket 63,7,3\rrbracket \)) code yields the \(T\) gate on the inner five-qubit (Steane) code.
The induced logical gate on the \(k\) logical qubits is, up to global phase, \(\exp{(\frac{i \pi}{8} Z^{\otimes k})}\), which decomposes into a \(T\) gate on every logical qubit, controlled-Phase\(^\dagger\) on every pair, and \(CCZ\) on every triple \NoCaseChange{\protect\cite{cite765}}.

\end{eczvaluelist}
\codefieldsection{Fault Tolerance}
\begin{eczvaluelist}
\item\relax The \(T\) gate realized by concatenating members of the \(\llbracket 2^m − 1, 1 \leq k \leq 1 + \sum_{i=1}^{m-4}(m − i), 3\rrbracket \) quantum divisible code family with either the five-qubit \(\llbracket 5,1,3\rrbracket \) or Steane \(\llbracket 7,1,3\rrbracket \) code is fault-tolerant and does not require magic-state distillation \NoCaseChange{\protect\cite{cite765}}.
The gate is performed on the inner five-qubit/Steane code and does require encoding and decoding algorithms to pass between the inner and outer codes.

\end{eczvaluelist}
\codefieldsection{Realizations}
\begin{eczvaluelist}
\item\relax Triply even codes can yield secure multi-party quantum computation \NoCaseChange{\protect\cite{cite4036}}.
\end{eczvaluelist}
\codefieldsection{Parent}
\begin{eczvaluelist}
\item\relax
\flmRefsHyperref[eczindexfamilyrel]{code:generalized_quantum_divisible}{Generalized quantum divisible code} --- Generalized level-\(\nu\) quantum divisible codes reduce to quantum level-\(\nu\) divisible codes when \(t\) is a vector with \(\pm 1\) entries.
The classical code formed by their \(X\)-type stabilizer generator matrix is \(\nu\)-even \NoCaseChange{\protect\cite[{pg. 10}]{cite734}}.
Both types of codes realize transversal gates outside of the \flmRefsHyperref{ref409}{Clifford group}.

\end{eczvaluelist}
\codefieldsection{Children}
\begin{eczvaluelist}
\item\relax
\flmRefsHyperref[eczindexfamilyrel]{code:stab_17_1_5}{\(\llbracket 17,1,5\rrbracket \) 4.8.8 color code}\item\relax
\flmRefsHyperref[eczindexfamilyrel]{code:small_triorthogonal}{\(\llbracket 3k + 8, k, 2\rrbracket \) triorthogonal code}\item\relax
\flmRefsHyperref[eczindexfamilyrel]{code:stab_49_1_5}{\(\llbracket 49,1,5\rrbracket \) triorthogonal code}\item\relax
\flmRefsHyperref[eczindexfamilyrel]{code:diagonal_clifford}{\(\llbracket 2^r-1,1,3\rrbracket \) simplex code} --- \(\llbracket 2^r-1,1,3\rrbracket \) simplex codes come from RM\((1,m=r)\) codes, which are \((r-1)\)-even \NoCaseChange{\protect\cite{cite1574,cite1575}}, and admit transversal gates at levels of the \flmTerm{term}{ref694}{}{Clifford hierarchy}. Building a tower of generalized divisible codes by starting with the Steane code yields the \(\llbracket 2^r-1,1,3\rrbracket \) simplex codes \NoCaseChange{\protect\cite[{Sec. VI.B}]{cite734}}.
\end{eczvaluelist}
\codefieldsection{Cousins}
\begin{eczvaluelist}
\item\relax
\flmRefsHyperref[eczindexfamilyrel]{code:divisible}{Divisible code} --- The \(X\)-type stabilizers of a level-\(\nu\) quantum divisible code form a \(\nu\)-even linear binary code.
\item\relax
\flmRefsHyperref[eczindexfamilyrel]{code:biorthogonal}{\([2^m,m+1,2^{m-1}]\) First-order RM code} --- Quantum divisible codes can be constructed out of first-order RM\((1,m)\) codes \NoCaseChange{\protect\cite{cite765}}.
\item\relax
\flmRefsHyperref[eczindexfamilyrel]{code:quantum_triorthogonal}{Triorthogonal code} --- The \(\llbracket 31,5,3\rrbracket \) member together with the five-qubit code can be viewed as a factorization of a \(\llbracket 31,1,3\rrbracket \) triorthogonal code \NoCaseChange{\protect\cite{cite765}}.
\item\relax
\flmRefsHyperref[eczindexfamilyrel]{code:qubit_concatenated}{Concatenated qubit code} --- A fault-tolerant \(T\) gate on the five-qubit or Steane code can be obtained by concatenating with particular quantum divisible codes \NoCaseChange{\protect\cite{cite765}}.
\item\relax
\flmRefsHyperref[eczindexfamilyrel]{code:quasi_cyclic}{Quasi-cyclic code} --- Certain double circulant codes can be used to construct doubly even \(\llbracket 55,1,11\rrbracket \) and \(\llbracket 87,1,15\rrbracket \) codes \NoCaseChange{\protect\cite{cite1123}}.
\item\relax
\flmRefsHyperref[eczindexfamilyrel]{code:stab_5_1_3}{\(\llbracket 5,1,3\rrbracket \) Five-qubit perfect code} --- A fault-tolerant logical \(T\) gate can be obtained by encoding the five-qubit code's five physical qubits into the five logical qubits of a \(\llbracket 31,5,3\rrbracket \) outer quantum divisible CSS code preserved by transversal \(T^\dagger\); this layered construction can be viewed as a factorization of a \(\llbracket 31,1,3\rrbracket \) triorthogonal code and does not require magic-state distillation \NoCaseChange{\protect\cite{cite765}}.
\item\relax
\flmRefsHyperref[eczindexfamilyrel]{code:steane}{\(\llbracket 7,1,3\rrbracket \) Steane code} --- A fault-tolerant logical \(T\) gate can be obtained by encoding the Steane code's seven physical qubits into the seven logical qubits of a \(\llbracket 63,7,3\rrbracket \) outer quantum divisible CSS code preserved by transversal \(T^\dagger\) \NoCaseChange{\protect\cite{cite765}}.
\item\relax
\flmRefsHyperref[eczindexfamilyrel]{code:quantum_reed_muller}{Quantum Reed-Muller (RM) code} --- Fault-tolerant universal computation can be achieved via \flmRefsHyperref{ref410}{code switching} between the \(\llbracket 127,1,15\rrbracket \) self-dual doubly even punctured quantum RM code and the \(\llbracket 127,1,7\rrbracket \) triply even punctured quantum RM code \NoCaseChange{\protect\cite{cite757}}.
\item\relax
\flmRefsHyperref[eczindexfamilyrel]{code:self_dual_css}{Self-dual CSS code} --- A self-dual weakly doubly even \(\llbracket n,1,d\rrbracket \) CSS code admits a partitioned transversal physical \(S\) gate that realizes \(\overline{S}^m\), where \(m=|M^+|-|M^-| \pmod 4\); for odd \(m\), together with transversal Hadamard and CNOT, this yields the full logical Clifford group transversally \NoCaseChange{\protect\cite{cite731}\protect\cite[{Lemma 4}]{cite760}}.
\item\relax
\flmRefsHyperref[eczindexfamilyrel]{code:doubled_color}{Doubled color code} --- Doubled color codes are subsystem codes constructed using a generalization of the doubling transformation \NoCaseChange{\protect\cite{cite659}} that combines doubly even linear binary codes to make triply even codes.
The doubling transformation is a special case of level lifting (from two to three) \NoCaseChange{\protect\cite[{Sec. VI.D}]{cite734}}.

\item\relax
\flmRefsHyperref[eczindexfamilyrel]{code:galois_quad_residue}{Quantum quadratic-residue (QR) code} --- Qubit quantum QR codes are doubly even and admit transversal implementations of the \flmRefsHyperref{ref409}{single-qubit Clifford group} \NoCaseChange{\protect\cite{cite760}}. They yield a family of high-distance triorthogonal and weak triply even codes via the doubling transformation \NoCaseChange{\protect\cite{cite760}}; such codes admit transversal implementations of the \(T\) gate.
\end{eczvaluelist}
\eczhbkcontributors{ Jingzhen Hu, \eczhuVVA }
\endeczcode

\eczcode{quantum_expander}{Quantum expander code}{~\NoCaseChange{\protect\cite{cite4037}}}
\codefieldsection{Alternative Names}
\begin{eczvaluelist}
\item\relax Quantum Sipser-Spielman code
\item\relax Expander HGP code
\end{eczvaluelist}
\eczhIndexCodeAliasName{quantum_expander}{Quantum Sipser-Spielman code}
\eczhIndexCodeAliasName{quantum_expander}{Expander HGP code}
\codefieldsection{Description}
CSS code constructed from a hypergraph product of bipartite expander graphs \NoCaseChange{\protect\cite{cite74}} with bounded left and right vertex degrees. For every bipartite graph there is an associated matrix (the parity check matrix) with columns indexed by the left vertices, rows indexed by the right vertices, and 1 entries whenever a left and right vertex are connected. This matrix can serve as the parity check matrix of a classical code. Two bipartite expander graphs can be used to construct a quantum CSS code (the quantum expander code) via the hypergraph product of their parity check matrices.

\codefieldsection{Protection}
The code family has distance scaling as \flmRefsHyperref{ref65}{order} \(\Omega(n^{1/2})\), and the small-set-flip decoder corrects a constant fraction of that many adversarial Pauli errors \NoCaseChange{\protect\cite{cite4037}}.
\codefieldsection{Rate}
\(\llbracket n,k=\Theta(n),d=O(\sqrt{n})\rrbracket \) code with asymptotically constant rate.
\codefieldsection{Encoding}
\begin{eczvaluelist}
\item\relax Single-shot state preparation with constant space-time overhead \NoCaseChange{\protect\cite{cite3723}}.
\end{eczvaluelist}
\codefieldsection{Gates}
\begin{eczvaluelist}
\item\relax Dimensional jump protocols between various quantum expander codes \NoCaseChange{\protect\cite{cite3723}}.
\end{eczvaluelist}
\codefieldsection{Decoding}
\begin{eczvaluelist}
\item\relax Small set-flip linear-time decoder, which corrects \flmRefsHyperref{ref65}{order} \(\Omega(n^{1/2})\) adversarial errors \NoCaseChange{\protect\cite{cite4037}}. The decoder has been generalized to hypergraph products of 3 or more expander codes \NoCaseChange{\protect\cite{cite3723}}.
\item\relax Log-time decoder \NoCaseChange{\protect\cite{cite847}}.
\item\relax Constant-time decoder \NoCaseChange{\protect\cite{cite4038}}.
\item\relax 2D geometrically local syndrome extraction circuits acting on a patch of \(N\) physical qubits must have depth of \flmRefsHyperref{ref65}{order} \(\Omega(n/\sqrt{N})\) or greater. More generally, there is a tradeoff between the depth \(D\) and width \(W\) of a syndrome extraction circuit, namely, \(D \geq n/\sqrt{W}\) \NoCaseChange{\protect\cite{cite521}}.
\end{eczvaluelist}
\codefieldsection{Fault Tolerance}
\begin{eczvaluelist}
\item\relax Fault-tolerance with constant overhead can be achieved \NoCaseChange{\protect\cite{cite847}}.
\end{eczvaluelist}
\codefieldsection{Threshold}
\begin{eczvaluelist}
\item\relax Locally stochastic noise: \(2.7 \cdot 10^{-16}\) \NoCaseChange{\protect\cite{cite4039}}.
\item\relax Dimensional jump protocols between various quantum expander codes have a threshold under local stochastic noise \NoCaseChange{\protect\cite{cite3723}}.
\end{eczvaluelist}
\codefieldsection{Parents}
\begin{eczvaluelist}
\item\relax
\flmRefsHyperref[eczindexfamilyrel]{code:hypergraph_product}{Hypergraph product (HGP) code}\item\relax
\flmRefsHyperref[eczindexfamilyrel]{code:galois_expander}{Galois-qudit expander code}\item\relax
\flmRefsHyperref[eczindexfamilyrel]{code:single_shot}{Single-shot code} --- Quantum expander codes are single-shot \NoCaseChange{\protect\cite{cite847}}.
\end{eczvaluelist}
\codefieldsection{Cousins}
\begin{eczvaluelist}
\item\relax
\flmRefsHyperref[eczindexfamilyrel]{code:expander}{Expander code} --- Quantum expander codes are quantum analogues of expander codes.
\item\relax
\flmRefsHyperref[eczindexfamilyrel]{code:topological}{Topological code} --- Quantum expander codes realize topological quantum spin glass order \NoCaseChange{\protect\cite{cite3162}}.
\item\relax
\flmRefsHyperref[eczindexfamilyrel]{code:multisector_hypergraph}{Higher-dimensional homological product code} --- Quantum expander codes have been generalized to hypergraph products of 3 or more expander codes \NoCaseChange{\protect\cite{cite3723}}.
\item\relax
\flmRefsHyperref[eczindexfamilyrel]{code:self_correct}{Self-correcting quantum code} --- Constant-rate random (quantum) expander codes are self-correcting (quantum) memories, but have no thermodynamic phase transitions \NoCaseChange{\protect\cite{cite849}}.
\end{eczvaluelist}
\eczhbkcontributors{ Nolan Coble, \eczhuVVA }
\endeczcode

\eczcode{quantum_irregular_convolutional}{Quantum irregular convolutional code (QIRCC)}{~\NoCaseChange{\protect\cite{cite4040}}}
\codefieldsection{Description}
Quantum convolutional code whose stabilizer group consists of different constant-size subsets. 
\codefieldsection{Parent}
\begin{eczvaluelist}
\item\relax
\flmRefsHyperref[eczindexfamilyrel]{code:quantum_convolutional}{Quantum convolutional code}\end{eczvaluelist}
\codefieldsection{Cousin}
\begin{eczvaluelist}
\item\relax
\flmRefsHyperref[eczindexfamilyrel]{code:irregular_convolutional}{Irregular convolutional code (IRCC)} --- Quantum irregular convolutional codes are quantum analogues of irregular convolutional codes.
\end{eczvaluelist}
\eczhbkcontributors{ \eczhuVVA }
\endeczcode

\eczcode{qmdpc}{Quantum multi-dimensional parity-check (QMDPC) code}{~\NoCaseChange{\protect\cite{cite523}}}
\codefieldsection{Description}
High-rate low-distance CSS code whose qubits lie on a \(D\)-dimensional rectangle, with \(X\)-type stabilizer generators defined on each \(D-1\)-dimensional rectangle.
The \(Z\)-type stabilizer generators are defined via permutations in order to commute with the \(X\)-type generators.

For example, the \(D=2\) square geometry corresponds to a \(\llbracket n^2,n^2-4n+2,4\rrbracket \) code, with \(X\)-type stabilizer generators defined on rows and columns.

\codefieldsection{Protection}
The general construction for a \(D\)-dimensional rectangle with sides \(n_i\) yields a \(\llbracket \prod_{i=1}^{D}n_{i},2\prod_{i=1}^{D}(n_{i}-1)-\prod_{i=1}^{D}n_{i},2^{D}\rrbracket \) code family.
\codefieldsection{Parent}
\begin{eczvaluelist}
\item\relax
\flmRefsHyperref[eczindexfamilyrel]{code:qubit_css}{Qubit CSS code}\end{eczvaluelist}
\codefieldsection{Child}
\begin{eczvaluelist}
\item\relax
\flmRefsHyperref[eczindexfamilyrel]{code:iceberg}{\(\llbracket 2m,2m-2,2\rrbracket \) error-detecting code} --- The \(\llbracket 2m,2m-2,2\rrbracket \) error-detecting code is a 1D QMDPC.
\end{eczvaluelist}
\codefieldsection{Cousins}
\begin{eczvaluelist}
\item\relax
\flmRefsHyperref[eczindexfamilyrel]{code:yoked_surface}{Yoked surface code} --- Yoked surface codes are concatenations of QMDPC codes with rotated surface codes.
\item\relax
\flmRefsHyperref[eczindexfamilyrel]{code:small_distance_qubit_stabilizer}{Small-distance qubit stabilizer code} --- QMDPC codes for dimensions \(D \leq 2\) are examples of small distance qubit stabilizer codes.
\end{eczvaluelist}
\eczhbkcontributors{ \eczhuVVA }
\endeczcode

\eczcode{quantum_parity}{Quantum parity code (QPC)}{~\NoCaseChange{\protect\cite{cite3,cite4041,cite3259}}}
\codefieldsection{Alternative Names}
\begin{eczvaluelist}
\item\relax Subspace Shor code
\end{eczvaluelist}
\eczhIndexCodeAliasName{quantum_parity}{Subspace Shor code}
\codefieldsection{Description}
A \(\llbracket m_1 m_2,1,\min(m_1,m_2)\rrbracket \) CSS code family obtained from concatenating an \(m_1\)-qubit bit-flip repetition code with an \(m_2\)-qubit phase-flip repetition code.

A set of logical codewords is
\flmMathEnvironment{align}{}{
\begin{split}
|\overline{0}\rangle&=\frac{1}{2^{m_2/2}}\left(|0\rangle^{\otimes m_1}+|1\rangle^{\otimes m_1}\right)^{\otimes m_2}\\
|\overline{1}\rangle&=\frac{1}{2^{m_2/2}}\left(|0\rangle^{\otimes m_1}-|1\rangle^{\otimes m_1}\right)^{\otimes m_2}~.
\end{split}
}

\codefieldsection{Protection}
Has distance \(d=\min(m_1,m_2)\).
\codefieldsection{Encoding}
\begin{eczvaluelist}
\item\relax Encoders for a recursively concatenated QPCs are related to \textit{quantum trees} \NoCaseChange{\protect\cite{cite3108,cite3109,cite4042}} and tree tensor networks \NoCaseChange{\protect\cite{cite400}}.
\item\relax Linear-optical encoding \NoCaseChange{\protect\cite{cite4043}}.
\end{eczvaluelist}
\codefieldsection{Decoding}
\begin{eczvaluelist}
\item\relax Teleportation-based QEC \NoCaseChange{\protect\cite{cite3690}}.
\end{eczvaluelist}
\codefieldsection{Threshold}
\begin{eczvaluelist}
\item\relax All optical scheme using QPCs concatenated with either Steane or Golay codes \NoCaseChange{\protect\cite{cite4044}}.
\end{eczvaluelist}
\codefieldsection{Realizations}
\begin{eczvaluelist}
\item\relax The \(\llbracket m^2,1,m\rrbracket \) codes for \(m\leq 7\) have been realized in trapped-ion quantum devices \NoCaseChange{\protect\cite{cite3379}}.
\item\relax QPCs have been discussed independently in the context of superconducting circuits \NoCaseChange{\protect\cite[{Eq. (1)}]{cite4045}\protect\cite[{Eqs. (8-10)}]{cite4046}}, and aspects of such designs have been realized in experiments \NoCaseChange{\protect\cite{cite4047}}.
\end{eczvaluelist}
\codefieldsection{Notes}
\begin{eczvaluelist}
\item\relax Non-deterministic linear-optical encoding \NoCaseChange{\protect\cite{cite3259}} whose success probability \(P_{E}\) is determined by the efficiency \(\eta\) of the photonic encoding circuit. A threshold \(\eta > 0.82 \) exists for the efficiency, above which \(P_{E}\to 1\) as \(m_1\to\infty\) given particular \(m_2\).
\end{eczvaluelist}
\codefieldsection{Parents}
\begin{eczvaluelist}
\item\relax
\flmRefsHyperref[eczindexfamilyrel]{code:generalized_shor}{Generalized Shor code}\item\relax
\flmRefsHyperref[eczindexfamilyrel]{code:group_quantum_parity}{Group-based QPC} --- A \(\llbracket m_1 m_2,1,\min(m_1,m_2)\rrbracket _G\) group-based QPC reduces to a QPC for \(G=\mathbb{Z}_2\).
\end{eczvaluelist}
\codefieldsection{Children}
\begin{eczvaluelist}
\item\relax
\flmRefsHyperref[eczindexfamilyrel]{code:quantum_repetition}{Quantum repetition code} --- A \(\llbracket m_1 m_2,1,\min(m_1,m_2)\rrbracket \) QPC reduces to a repetition code when \(m_1\) or \(m_2\) is one.
\item\relax
\flmRefsHyperref[eczindexfamilyrel]{code:css_4_1_2}{\(\llbracket 4,1,2\rrbracket \) Leung-Nielsen-Chuang-Yamamoto (LNCY) code} --- The \(\llbracket 4,1,2\rrbracket \) LNCY code is the smallest QPC, i.e., a concatenation of a two-qubit bit-flip with a two-qubit phase-flip repetition code.
An \(\llbracket 8,1,2\rrbracket \) QPC correcting a single \flmRefsHyperref{ref498}{AD} error is equivalent to a concatenation of its constant-excitation version with the dual-rail code \NoCaseChange{\protect\cite{cite3250,cite3259,cite2711}}.

\item\relax
\flmRefsHyperref[eczindexfamilyrel]{code:shor_nine}{\(\llbracket 9,1,3\rrbracket \) Shor code} --- The Shor code is part of the sub-family of \(\llbracket m^2,1,m\rrbracket \) QPCs.
\end{eczvaluelist}
\codefieldsection{Cousins}
\begin{eczvaluelist}
\item\relax
\flmRefsHyperref[eczindexfamilyrel]{code:quantum_lego}{Tensor-network code} --- Encoders for a recursively concatenated QPCs are related to \textit{quantum trees} \NoCaseChange{\protect\cite{cite3108,cite3109}} and tree tensor networks \NoCaseChange{\protect\cite{cite400}}.
\item\relax
\flmRefsHyperref[eczindexfamilyrel]{code:bacon_shor}{Bacon-Shor code} --- Bacon-Shor codes reduce to QPCs when all \(X\)-type gauge generators are fixed \NoCaseChange{\protect\cite[{pg. 6}]{cite2650}}.
\item\relax
\flmRefsHyperref[eczindexfamilyrel]{code:majorana_stab}{Majorana stabilizer code} --- QPCs for \(m_1=m_2\) can be conveniently expressed in terms of mutually commuting Majorana operators \NoCaseChange{\protect\cite{cite557}}.
\item\relax
\flmRefsHyperref[eczindexfamilyrel]{code:constant_excitation}{Constant-excitation (CE) code} --- QPCs for even \(m_1\) can be made into CE codes by a Pauli transformation (e.g., \(XIXI\cdots XI\)) applied to each block of \(m_1\) qubits.
\item\relax
\flmRefsHyperref[eczindexfamilyrel]{code:ampdamp}{Amplitude-damping (AD) code} --- An \(\llbracket 8,1,2\rrbracket \) QPC correcting a single \flmRefsHyperref{ref498}{AD} error is equivalent to a concatenation of the \(\{|\overline{01}\rangle,|\overline{11}\rangle\}\) (constant-excitation) subcode of the \(\llbracket 4,2,2\rrbracket \) code with the dual-rail code \NoCaseChange{\protect\cite{cite3250,cite3259,cite2711}}. More generally, an \(\llbracket m^2,1,m\rrbracket \) QPC corrects \(m-1\) \flmRefsHyperref{ref498}{AD} errors \NoCaseChange{\protect\cite{cite3263}}.
\item\relax
\flmRefsHyperref[eczindexfamilyrel]{code:rbh}{Raussendorf-Bravyi-Harrington (RBH) cluster-state code} --- QPCs can be concatenated with RBH codes \NoCaseChange{\protect\cite{cite4048}}.
\item\relax
\flmRefsHyperref[eczindexfamilyrel]{code:dual_rail}{Dual-rail quantum code} --- An \(\llbracket 8,1,2\rrbracket \) QPC correcting a single \flmRefsHyperref{ref498}{AD} error is equivalent to a concatenation of the \(\{|\overline{01}\rangle,|\overline{11}\rangle\}\) (constant-excitation) subcode of the \(\llbracket 4,2,2\rrbracket \) code with the dual-rail code \NoCaseChange{\protect\cite{cite3250,cite3259,cite2711}}. More generally, an \(\llbracket m^2,1,m\rrbracket \) QPC corrects \(m-1\) \flmRefsHyperref{ref498}{AD} errors \NoCaseChange{\protect\cite{cite3263}}.
\item\relax
\flmRefsHyperref[eczindexfamilyrel]{code:two-mode_binomial}{Two-mode binomial code} --- Two-mode binomial codes can be concatenated with repetition codes to yield bosonic analogues of QPCs \NoCaseChange{\protect\cite{cite4049}}.
\item\relax
\flmRefsHyperref[eczindexfamilyrel]{code:gkp_concatenated}{Concatenated GKP code} --- GKP codes have been concatenated with QPCs \NoCaseChange{\protect\cite{cite4050}}.
\item\relax
\flmRefsHyperref[eczindexfamilyrel]{code:asymmetric_qecc}{Asymmetric quantum code (AQC)} --- QPC parameters against bit- and phase-noise can be tuned.
\item\relax
\flmRefsHyperref[eczindexfamilyrel]{code:compass_model}{Compass code} --- The Shor-density compass code family interpolates between Bacon-Shor codes and QPCs.
\end{eczvaluelist}
\eczhbkcontributors{ Xinyuan Zheng, \eczhuVVA }
\endeczcode

\eczcode{quantum_pin}{Quantum pin code}{~\NoCaseChange{\protect\cite{cite702}}}
\codefieldsection{Description}
Member of a family of CSS codes that encompasses both quantum RM and color codes and that is defined using intersections of pinned sets.

\codefieldsection{Magic}
A family of punctured pin codes admits \(\gamma \approx 1.6\) \NoCaseChange{\protect\cite[{Table VII}]{cite702}}.
\codefieldsection{Parents}
\begin{eczvaluelist}
\item\relax
\flmRefsHyperref[eczindexfamilyrel]{code:quantum_rainbow}{Quantum rainbow code} --- Quantum pin codes are a special case of quantum rainbow codes \NoCaseChange{\protect\cite{cite704}}.
\item\relax
\flmRefsHyperref[eczindexfamilyrel]{code:quantum_k-orthogonal}{\(k\)-orthogonal code} --- Quantum pin codes are \(\ell\)-orthogonal, i.e., the overlap between any \(\ell\) stabilizers is even \NoCaseChange{\protect\cite{cite702}}.
\end{eczvaluelist}
\codefieldsection{Children}
\begin{eczvaluelist}
\item\relax
\flmRefsHyperref[eczindexfamilyrel]{code:quantum_reed_muller}{Quantum Reed-Muller (RM) code} --- Quantum RM codes are special cases of quantum pin codes \NoCaseChange{\protect\cite[{Sec. II.D}]{cite702}}.
\item\relax
\flmRefsHyperref[eczindexfamilyrel]{code:color}{Color code} --- Color codes are special cases of quantum pin codes \NoCaseChange{\protect\cite[{Sec. II.E}]{cite702}}
\end{eczvaluelist}
\codefieldsection{Cousins}
\begin{eczvaluelist}
\item\relax
\flmRefsHyperref[eczindexfamilyrel]{code:subsystem_color}{Subsystem color code} --- Quantum pin codes have a subsystem version that can be viewed as a generalization of subsystem color codes \NoCaseChange{\protect\cite{cite702}}.
\item\relax
\flmRefsHyperref[eczindexfamilyrel]{code:qubit_generalized_homological_product_css}{Generalized homological-product qubit CSS code} --- One can construct quantum pin codes from any chain complex \NoCaseChange{\protect\cite[{Sec. II.F}]{cite702}}.
\end{eczvaluelist}
\eczhbkcontributors{ \eczhuVVA }
\endeczcode

\eczcode{quantum_polar}{Quantum polar code}{~\NoCaseChange{\protect\cite{cite4051}}}
\codefieldsection{Description}
Entanglement-assisted CSS code utilized in a quantum polar coding scheme producing entangled pairs of qubits between sender and receiver. In such a scheme, the amplitude and phase information of a quantum state is handled in complementary fashion \NoCaseChange{\protect\cite{cite551}} using an encoding based on classical polar codes. Variants of the initial scheme have been developed for degradable channels \NoCaseChange{\protect\cite{cite552}} and extended to arbitrary channels \NoCaseChange{\protect\cite{cite553}}.

The scheme requires some a priori quantum side information in the general case, making the associated code entanglement-assisted \NoCaseChange{\protect\cite{cite4051}}. 
They require assistance only to determine positions to store information which optimally protect against both bit and phase noise. Without this assistance, they are just CSS codes constructed out of polar codes.
The requirement of having quantum side information vanishes when the sum of the amplitude channel fidelity and the phase channel fidelity is not greater than 1. 
It is shown to vanish for the case of degradable noise channels \NoCaseChange{\protect\cite{cite553}}. 
A more complicated quantum polar-coding scheme that does not require pre-shared entanglement has also been derived \NoCaseChange{\protect\cite{cite4052}}.

\codefieldsection{Protection}
Protects against Pauli noise and erasures.
\codefieldsection{Rate}
The rate approaches the symmetric coherent information of arbitrary quantum channels \NoCaseChange{\protect\cite{cite552}}.
\codefieldsection{Encoding}
\begin{eczvaluelist}
\item\relax Encoding circuits can be viewed as branching-tree tensor networks \NoCaseChange{\protect\cite{cite400}}.
\end{eczvaluelist}
\codefieldsection{Decoding}
\begin{eczvaluelist}
\item\relax Arikan-style successive-cancellation decoding can be recast as a tensor-network contraction \NoCaseChange{\protect\cite{cite400}}.
\item\relax Quantum successive-cancellation list decoder (SCL-E) for quantum polar codes that do not need entanglement assistance \NoCaseChange{\protect\cite{cite4053}}.
\item\relax Numerics for tensor-network-based decoders on depolarizing and erasure channels indicate that good performance is possible without entanglement assistance, with channel polarization rapidly suppressing channels that are bad in both quadratures \NoCaseChange{\protect\cite{cite400}}.
\end{eczvaluelist}
\codefieldsection{Fault Tolerance}
\begin{eczvaluelist}
\item\relax State preparation of a single logical qubit \NoCaseChange{\protect\cite{cite4054}}.
\end{eczvaluelist}
\codefieldsection{Parent}
\begin{eczvaluelist}
\item\relax
\flmRefsHyperref[eczindexfamilyrel]{code:branching_mera}{Branching MERA code} --- Branching MERA codes generalize quantum polar codes by restoring the branching-MERA disentanglers while retaining efficient tensor-network decoding \NoCaseChange{\protect\cite{cite400}}.
\end{eczvaluelist}
\codefieldsection{Cousins}
\begin{eczvaluelist}
\item\relax
\flmRefsHyperref[eczindexfamilyrel]{code:qubit_css}{Qubit CSS code} --- Quantum polar codes are CSS codes used in an entanglement generation scheme that generally requires entanglement assistance. They require assistance only to determine positions to store information which optimally protect against both bit and phase noise. Without this assistance, they are just CSS codes constructed out of polar codes. A variant of quantum polar codes exists that does not require entanglement assistance \NoCaseChange{\protect\cite{cite4052}}.
\item\relax
\flmRefsHyperref[eczindexfamilyrel]{code:polar}{Polar code} --- Without entanglement assistance, quantum polar codes are CSS codes constructed out of polar codes.
\item\relax
\flmRefsHyperref[eczindexfamilyrel]{code:quantum_lego}{Tensor-network code} --- Quantum polar encoding circuits can be viewed as branching-tree tensor networks \NoCaseChange{\protect\cite{cite400}}.
\item\relax
\flmRefsHyperref[eczindexfamilyrel]{code:gkp_concatenated}{Concatenated GKP code} --- Concatenation of GKP codes with quantum polar codes achieves a rate against the displacement channel \NoCaseChange{\protect\cite{cite4055}}.
\item\relax
\flmRefsHyperref[eczindexfamilyrel]{code:css_4_1_2}{\(\llbracket 4,1,2\rrbracket \) Leung-Nielsen-Chuang-Yamamoto (LNCY) code} --- The \(\llbracket 4,1,2\rrbracket \) LNCY code is a small quantum polar encoding \NoCaseChange{\protect\cite{cite3267}}.
\item\relax
\flmRefsHyperref[eczindexfamilyrel]{code:quantum_reed_muller}{Quantum Reed-Muller (RM) code} --- There are codes interpolating between quantum RM and quantum polar codes \NoCaseChange{\protect\cite{cite4056}}.
\end{eczvaluelist}
\eczhbkcontributors{ Richard Barney, \eczhuVVA }
\endeczcode

\eczcode{quantum_rainbow}{Quantum rainbow code}{~\NoCaseChange{\protect\cite{cite704}}}
\codefieldsection{Description}
A CSS code whose qubits are associated with vertices of a simplex graph with \(m+1\) colors.

\codefieldsection{Magic}
Hypergraph products of color codes yield quantum rainbow codes with growing distance and transversal gates in the \flmTerm{term}{ref694}{}{Clifford hierarchy}. In particular, utilizing this construction for quasi-hyperbolic color codes \NoCaseChange{\protect\cite{cite703}} yields an \(\llbracket n,O(n),O(\log n)\rrbracket \) triorthogonal code family satisfying the necessary conditions for the magic-state yield parameter \(\gamma\) to become arbitrarily small \NoCaseChange{\protect\cite{cite704}}.
\codefieldsection{Transversal and Permutation-Based Gates}
\begin{eczvaluelist}
\item\relax Hypergraph products of color codes yield quantum rainbow codes with growing distance and transversal gates in the \flmTerm{term}{ref694}{}{Clifford hierarchy} \NoCaseChange{\protect\cite{cite704}}.
\end{eczvaluelist}
\codefieldsection{Parent}
\begin{eczvaluelist}
\item\relax
\flmRefsHyperref[eczindexfamilyrel]{code:qubit_css}{Qubit CSS code}\end{eczvaluelist}
\codefieldsection{Child}
\begin{eczvaluelist}
\item\relax
\flmRefsHyperref[eczindexfamilyrel]{code:quantum_pin}{Quantum pin code} --- Quantum pin codes are a special case of quantum rainbow codes \NoCaseChange{\protect\cite{cite704}}.
\end{eczvaluelist}
\codefieldsection{Cousins}
\begin{eczvaluelist}
\item\relax
\flmRefsHyperref[eczindexfamilyrel]{code:hypergraph_product}{Hypergraph product (HGP) code} --- Hypergraph products of color codes yield quantum rainbow codes with growing distance and transversal gates in the \flmTerm{term}{ref694}{}{Clifford hierarchy}. In particular, utilizing this construction for quasi-hyperbolic color codes yields an \(\llbracket n,O(n),O(\log n)\rrbracket \) triorthogonal code family satisfying the necessary conditions for the magic-state yield parameter \(\gamma\) to become arbitrarily small \NoCaseChange{\protect\cite{cite704}}.
\item\relax
\flmRefsHyperref[eczindexfamilyrel]{code:quantum_triorthogonal}{Triorthogonal code} --- Hypergraph products of color codes yield quantum rainbow codes with growing distance and transversal gates in the \flmTerm{term}{ref694}{}{Clifford hierarchy}. In particular, utilizing this construction for quasi-hyperbolic color codes \NoCaseChange{\protect\cite{cite703}} yields an \(\llbracket n,O(n),O(\log n)\rrbracket \) triorthogonal code family satisfying the necessary conditions for the magic-state yield parameter \(\gamma\) to become arbitrarily small \NoCaseChange{\protect\cite{cite704}}.
\item\relax
\flmRefsHyperref[eczindexfamilyrel]{code:quasi_hyperbolic_color}{Quasi-hyperbolic color code} --- Hypergraph products of color codes yield quantum rainbow codes with growing distance and transversal gates in the \flmTerm{term}{ref694}{}{Clifford hierarchy}. In particular, utilizing this construction for quasi-hyperbolic color codes yields an \(\llbracket n,O(n),O(\log n)\rrbracket \) triorthogonal code family satisfying the necessary conditions for the magic-state yield parameter \(\gamma\) to become arbitrarily small \NoCaseChange{\protect\cite{cite704}}.
\end{eczvaluelist}
\eczhbkcontributors{ \eczhuVVA }
\endeczcode

\eczcode{quantum_reed_muller}{Quantum Reed-Muller (RM) code}{~\NoCaseChange{\protect\cite{cite3198,cite4057}}}
\codefieldsection{Description}
A CSS code formed from a classical RM code or its punctured/shortened versions.
Such codes often admit transversal logical gates in the \flmTerm{term}{ref694}{}{Clifford hierarchy}.

Ordinary, punctured, or shortened RM codes can be used to construct quantum RM codes.
For example, the original construction \NoCaseChange{\protect\cite{cite3198}} uses a general RM\((r,m)\) code for the \(X\)-type stabilizers, and an RM\((r-1,m)\) code for the \(Z\)-type stabilizers.

Non-CSS codes can be derived from such codes by modifying the \(X\)-type stabilizers \NoCaseChange{\protect\cite{cite3198}}.

\codefieldsection{Protection}
Detects errors on \(d-1\) qubits, corrects errors on \(\left\lfloor (d-1)/2 \right\rfloor\) qubits.
\codefieldsection{Rate}
Dimension is \(k = 2^r - {r \choose t} + 2 \sum_{i=0}^{t-1} {r \choose i}\). CSS codes formed from RM codes achieve channel capacity on erasure channels \NoCaseChange{\protect\cite{cite4058}}.
\codefieldsection{Magic}
The family constructed out of shortened RM codes with parameters \(\llbracket \sum_{i=w+1}^m \binom{m}{i}, \sum_{i=0}^{w} \binom{m}{i}, \sum_{i=w+1}^{r+1} \binom{r+1}{i}\rrbracket \) for integers \(m > 2r\) and \(r > w \geq 0\) yields protocols with an exponent of \(\gamma < 0.678\), with the fewest-resource protocol with \(\gamma < 1\) requiring a code with parameters \(\{r,w,m\} = \{19,14,3r+1\}\) such that \(n \approx 2^{58}\) qubits \NoCaseChange{\protect\cite[{Corr. 1}]{cite701}}. This refutes a conjecture that no protocol could achieve \(\gamma < 1\) \NoCaseChange{\protect\cite{cite691}}.
\codefieldsection{Transversal and Permutation-Based Gates}
\begin{eczvaluelist}
\item\relax Stabilizer generators can be defined as Pauli strings acting on subsets of qubits corresponding to subcubes of the Hamming \(n\)-cube (a.k.a. Boolean hypercube) \NoCaseChange{\protect\cite{cite753}}. Transversal \(Z\)-rotations by angles \(\pi/2^k\) acting on subcubes can implement logical multi-controlled-\(Z\) gates \NoCaseChange{\protect\cite{cite753}}.
\item\relax The \(\llbracket 2^m,{m \choose r}, 2^{\min(r,m-r)}\rrbracket \) family, where \(r\) divides \(m\), admits diagonal gates in the form of \(Z\)-rotations by angle \(\pi/2^{m/r}\) \NoCaseChange{\protect\cite{cite754,cite755,cite756}\protect\cite[{Exam. 8 and Thm. 19}]{cite724}}. Of these, the self-dual sub-family for \(m=2r\) admits logical Clifford group gates via permutations, transversal gates, and fold-transversal gates with the help of ancillas \NoCaseChange{\protect\cite{cite757}}; the ancilla requirement was later removed \NoCaseChange{\protect\cite{cite758}}.
\item\relax The \(\llbracket 64,15,4\rrbracket \) member is highlighted in Ref. \NoCaseChange{\protect\cite{cite759}} as a candidate for transversal IQP sampling because a transversal \(T\) gate implements a logical circuit consisting of 15 \(CCZ\) gates.
\item\relax The family constructed out of shortened RM codes with parameters \(\llbracket \sum_{i=w+1}^m \binom{m}{i}, \sum_{i=0}^{w} \binom{m}{i}, \sum_{i=w+1}^{r+1} \binom{r+1}{i}\rrbracket \) for integers \(m > 2r\) and \(r > w \geq 0\) admits a transversal gate at the \(\nu\)th level in the hierarchy whenever \(m > \nu r\) \NoCaseChange{\protect\cite[{Thm. 1}]{cite701}}.
\end{eczvaluelist}
\codefieldsection{Fault Tolerance}
\begin{eczvaluelist}
\item\relax Gate switching protocol for universal computation \NoCaseChange{\protect\cite{cite793}}.
\item\relax Fault-tolerant universal computation can be achieved via \flmRefsHyperref{ref410}{code switching} between the \(\llbracket 127,1,15\rrbracket \) self-dual doubly even punctured quantum RM code and the \(\llbracket 127,1,7\rrbracket \) triply even punctured quantum RM code \NoCaseChange{\protect\cite{cite757}}.
\end{eczvaluelist}
\codefieldsection{Parents}
\begin{eczvaluelist}
\item\relax
\flmRefsHyperref[eczindexfamilyrel]{code:quantum_pin}{Quantum pin code} --- Quantum RM codes are special cases of quantum pin codes \NoCaseChange{\protect\cite[{Sec. II.D}]{cite702}}.
\item\relax
\flmRefsHyperref[eczindexfamilyrel]{code:qudit_reed_muller}{Prime-qudit RM code} --- Prime-qudit RM codes reduce to quantum RM codes when \(q=p=2\).
\end{eczvaluelist}
\codefieldsection{Children}
\begin{eczvaluelist}
\item\relax
\flmRefsHyperref[eczindexfamilyrel]{code:hypercube_quantum}{\(\llbracket 2^D,D,2\rrbracket \) hypercube quantum code} --- \(\llbracket 2^D,D,2\rrbracket \) hypercube quantum codes are special cases of the \(\llbracket 2^m,{m \choose r}, 2^r\rrbracket \) quantum RM codes for \(m=D\) and \(r=1\) \NoCaseChange{\protect\cite{cite754,cite755,cite756,cite757}\protect\cite[{Exam. 8}]{cite724}}.
\item\relax
\flmRefsHyperref[eczindexfamilyrel]{code:stab_16_6_4}{\(\llbracket 16,6,4\rrbracket \) Tesseract color code} --- The tesseract color code is constructed from the \([16,5,8]\) first-order RM code via the CSS construction \NoCaseChange{\protect\cite{cite101,cite757}}.
\item\relax
\flmRefsHyperref[eczindexfamilyrel]{code:xz_7_3_2}{\(\llbracket 7,3,2\rrbracket \) punctured hypercube code} --- The punctured hypercube family is a quantum Reed-Muller family built from shortened and punctured classical Reed-Muller codes \NoCaseChange{\protect\cite{cite723}}.
\item\relax
\flmRefsHyperref[eczindexfamilyrel]{code:diagonal_clifford}{\(\llbracket 2^r-1,1,3\rrbracket \) simplex code} --- \(\llbracket 2^r-1,1,3\rrbracket \) simplex codes are special cases of the \(\llbracket \sum_{i=w+1}^m \binom{m}{i}, \sum_{i=0}^{w} \binom{m}{i}, \sum_{i=w+1}^{r+1} \binom{r+1}{i}\rrbracket \) quantum RM codes for \(w=0\) and \(r=1\), with \(m\) equal to the present entry's parameter \(r\) \NoCaseChange{\protect\cite[{Thm. 1}]{cite701}}.
\item\relax
\flmRefsHyperref[eczindexfamilyrel]{code:quantum_hamming_css}{\(\llbracket 2^r-1, 2^r-2r-1, 3\rrbracket \) quantum Hamming code} --- \(\llbracket 2^r-1, 2^r-2r-1, 3\rrbracket \) quantum Hamming codes are quantum RM codes because Hamming and simplex codes are both punctured RM codes.
\item\relax
\flmRefsHyperref[eczindexfamilyrel]{code:single_qubit_clifford}{\(\llbracket 2^{2r-1}-1,1,2^r-1\rrbracket \) quantum punctured RM code} --- The \(\llbracket 2^{2r-1}-1,1,2^r-1\rrbracket \) quantum punctured RM codes are special cases of the \(\llbracket \sum_{i=w+1}^m \binom{m}{i}, \sum_{i=0}^{w} \binom{m}{i}, \sum_{i=w+1}^{r+1} \binom{r+1}{i}\rrbracket \) family for \(m \to 2r-1\), \(w \to 0\), and \(r \to r-1\).
\end{eczvaluelist}
\codefieldsection{Cousins}
\begin{eczvaluelist}
\item\relax
\flmRefsHyperref[eczindexfamilyrel]{code:reed_muller}{Reed-Muller (RM) code} --- Quantum RM codes are constructed from RM codes via the CSS construction. There is a relation between RM code performance against correlated generalizations of multiple-access channels (MACs) and quantum RM code performance against Pauli channels \NoCaseChange{\protect\cite{cite1565}}.
\item\relax
\flmRefsHyperref[eczindexfamilyrel]{code:quantum_convolutional}{Quantum convolutional code} --- Quantum convolutional codes can be derived from quantum RM codes \NoCaseChange{\protect\cite{cite4028}}.
\item\relax
\flmRefsHyperref[eczindexfamilyrel]{code:eastab}{EA qubit stabilizer code} --- EA versions of quantum RM codes and their quantum tensor-product variants can be constructed \NoCaseChange{\protect\cite{cite3651}}.
\item\relax
\flmRefsHyperref[eczindexfamilyrel]{code:quantum_tensor_product}{Quantum tensor-product code} --- EA versions of quantum RM codes and their quantum tensor-product variants can be constructed \NoCaseChange{\protect\cite{cite3651}}.
\item\relax
\flmRefsHyperref[eczindexfamilyrel]{code:quantum_divisible}{Quantum divisible code} --- Fault-tolerant universal computation can be achieved via \flmRefsHyperref{ref410}{code switching} between the \(\llbracket 127,1,15\rrbracket \) self-dual doubly even punctured quantum RM code and the \(\llbracket 127,1,7\rrbracket \) triply even punctured quantum RM code \NoCaseChange{\protect\cite{cite757}}.
\item\relax
\flmRefsHyperref[eczindexfamilyrel]{code:quantum_polar}{Quantum polar code} --- There are codes interpolating between quantum RM and quantum polar codes \NoCaseChange{\protect\cite{cite4056}}.
\item\relax
\flmRefsHyperref[eczindexfamilyrel]{code:asymmetric_qecc}{Asymmetric quantum code (AQC)} --- Asymmetric quantum RM codes have been constructed \NoCaseChange{\protect\cite[{Lemma 4.1}]{cite1354}}.
\item\relax
\flmRefsHyperref[eczindexfamilyrel]{code:covariant}{Covariant block quantum code} --- Quantum RM codes are approximately covariant and nearly saturate certain covariance-performance bounds \NoCaseChange{\protect\cite{cite2718}}.
\item\relax
\flmRefsHyperref[eczindexfamilyrel]{code:majorana_reed_muller}{RM Majorana code} --- RM Majorana (quantum RM) codes are designed for fermionic (qubit) noise.
\item\relax
\flmRefsHyperref[eczindexfamilyrel]{code:phantom}{Phantom code} --- Some quantum RM codes are phantom after selected logical qubits of a parent quantum RM code are fixed to \(\ket{\overline{0}}\) or \(\ket{\overline{+}}\), promoting the corresponding logical operators to stabilizers \NoCaseChange{\protect\cite{cite514}}.
\item\relax
\flmRefsHyperref[eczindexfamilyrel]{code:css-t}{CSS-T code} --- Certain quantum RM codes are CSS-T codes \NoCaseChange{\protect\cite{cite3617,cite3618,cite1315}}.
\item\relax
\flmRefsHyperref[eczindexfamilyrel]{code:quantum_triorthogonal}{Triorthogonal code} --- Classification of triorthogonal codes yields a connection to RM code polynomials \NoCaseChange{\protect\cite{cite3207}}.
\item\relax
\flmRefsHyperref[eczindexfamilyrel]{code:multisector_hypergraph}{Higher-dimensional homological product code} --- Higher-dimensional homological-product codes can be constructed out of quantum RM codes \NoCaseChange{\protect\cite[{Sec. 4.3}]{cite835}}.
\item\relax
\flmRefsHyperref[eczindexfamilyrel]{code:self_dual_css}{Self-dual CSS code} --- The \(\llbracket 2^m,{m \choose r}, 2^{\min(r,m-r)}\rrbracket \) quantum RM family contains a self-dual sub-family for \(m=2r\), which admits logical Clifford group gates via permutations, transversal gates, and fold-transversal gates \NoCaseChange{\protect\cite{cite757,cite758}}.
\end{eczvaluelist}
\eczhbkcontributors{ Benjamin Quiring, \eczhuVVA }
\endeczcode

\eczcode{quantum_repetition}{Quantum repetition code}{~\NoCaseChange{\protect\cite{cite4059}}}
\codefieldsection{Description}
Encodes \(1\) qubit into \(n\) qubits according to \(|0\rangle\to|\phi_0\rangle^{\otimes n}\) and \(|1\rangle\to|\phi_1\rangle^{\otimes n}\). The code is called a \textit{bit-flip} code when \(|\phi_i\rangle = |i\rangle\), and a \textit{phase-flip} code when \(|\phi_0\rangle = |+\rangle\) and \(|\phi_1\rangle = |-\rangle\).
This repetition-style encoding does not clone an arbitrary quantum state; instead, it extends the copying of computational-basis states linearly to entangled codewords  \NoCaseChange{\protect\cite[{Ch. 2}]{cite398}}.

The \(\pm\)-basis codewords for the bit-flip code are \textit{GHZ states} \NoCaseChange{\protect\cite{cite4060,cite4061,cite4062}} (a.k.a. qubit cat states) \(|0\rangle^{\otimes n}\pm|1\rangle^{\otimes n}\). These are ground states of the 1D \textit{classical Ising model} Hamiltonian \(H=\sum_{i} Z_{i}Z_{i+1}\).

The \(\pm\)-basis codewords for the phase-flip code are expanded in the computational basis as
\flmMathEnvironment{align}{}{
  \begin{split}
    |\overline{+}\rangle =\frac{1}{\sqrt{2^{n-1}}}\sum_{\sum_{i}v_{i}=0}|v_{1},\cdots,v_{n}\rangle~{\phantom{,}}\\
    |\overline{-}\rangle =\frac{1}{\sqrt{2^{n-1}}}\sum_{\sum_{i}v_{i}=1}|v_{1},\cdots,v_{n}\rangle~,
  \end{split}
}
showing that the phase-flip code stores information in the total parity of the qubits.
For example, an early code realized in devices is the 2-qubit phase-flip code \NoCaseChange{\protect\cite{cite4041}}, which encodes a logical qubit into Bell states \(|00\rangle+|11\rangle\) and \(|01\rangle+|10\rangle\).

\codefieldsection{Protection}
Bit-flip code corrects bit-flip errors \(X\) on \(\left\lfloor (n-1)/2\right\rfloor\) qubits and does not detect any phase-flip errors \(Z\).
Phase-flip code corrects phase-flip errors \(Z\) on \(\left\lfloor (n-1)/2\right\rfloor\) qubits and does not detect any bit-flip errors \(X\).

Because they protect against only one type of noise, both codes can be thought of as a classical \([n,1,n]\) repetition code embedded in a quantum system.
Nevertheless, the phase-flip code can offer some degree of protection in particular physical systems based on superconducting circuits \NoCaseChange{\protect\cite{cite4063,cite4064}}.

\codefieldsection{Encoding}
\begin{eczvaluelist}
\item\relax Non-deterministic encoders for various specific states of the 2-qubit phase-flip code \NoCaseChange{\protect\cite{cite4065}}.
\item\relax Fault-tolerant GHZ-state preparation with small qubit registers \NoCaseChange{\protect\cite{cite4066}}.
\item\relax Unitary circuit of depth logarithmic in \(n\) \NoCaseChange{\protect\cite{cite4067}}. Any circuit has to have range \(n\) because Ghz states are locally indistinguishable \NoCaseChange{\protect\cite{cite4068}}.
\item\relax Adaptive constant-depth circuit with geometrically local gates and measurements throughout \NoCaseChange{\protect\cite{cite3234,cite3235}}.
\item\relax Lindbladian-based dissipative encoding and autonomous QEC passively protecting against bit flips \NoCaseChange{\protect\cite{cite2791,cite4069}}.
\item\relax Error-corrected GHZ state distillation for Steane error correction \NoCaseChange{\protect\cite{cite4070}}.
\item\relax Approximate Hamiltonian-based encoding of GHZ states \NoCaseChange{\protect\cite{cite4071}}.
\end{eczvaluelist}
\codefieldsection{Gates}
\begin{eczvaluelist}
\item\relax Toffoli magic-state preparation protocol \NoCaseChange{\protect\cite{cite4072}}.
\end{eczvaluelist}
\codefieldsection{Decoding}
\begin{eczvaluelist}
\item\relax Fault-tolerant syndrome detection \NoCaseChange{\protect\cite{cite4073}}.
\item\relax Autonomous QEC for the 3-qubit bit-flip code \NoCaseChange{\protect\cite{cite4074}}.
\item\relax Machine learning algorithm to implement autonomous QEC for the three-qubit quantum repetition code \NoCaseChange{\protect\cite{cite4075}}.
\item\relax Quantum cellular automata for majority voting and two-line voting \NoCaseChange{\protect\cite{cite4076}}.
\item\relax Quantum version of the Tsirelson local automaton decoder \NoCaseChange{\protect\cite{cite3923}}.
\item\relax Planar decoder designed to work under circuit-level noise \NoCaseChange{\protect\cite{cite4077}}.
\item\relax Single-rule and shearing-rule local automaton decoders \NoCaseChange{\protect\cite{cite4078}}.
\end{eczvaluelist}
\codefieldsection{Fault Tolerance}
\begin{eczvaluelist}
\item\relax Fault-tolerant syndrome detection \NoCaseChange{\protect\cite{cite4073}}.
\item\relax An \(m\)-qubit GHZ state, i.e., qubit cat state, can serve as an ancilla for fault-tolerant measurement of a weight-\(m\) Pauli operator by coupling each ancilla qubit to one data qubit, applying Hadamards to the ancilla, and reading out the parity of the measurement outcomes \NoCaseChange{\protect\cite{cite398}}.
\item\relax Toffoli magic-state preparation protocol \NoCaseChange{\protect\cite{cite4072}}.
\end{eczvaluelist}
\codefieldsection{Code Capacity Threshold}
\begin{eczvaluelist}
\item\relax Independent \(X\) noise: \(50\%\) with RG decoder for quantum repetition code arranged on a 1D or 2D lattice \NoCaseChange{\protect\cite{cite3174}}.
\item\relax A nonzero threshold exists under the single-rule local automaton decoders \NoCaseChange{\protect\cite{cite4078}}.
\end{eczvaluelist}
\codefieldsection{Threshold}
\begin{eczvaluelist}
\item\relax Phenomenological noise: \(11\%\) and \(17.2\%\) with RG decoder for quantum repetition code arranged on a 1D and 2D lattice, respectively \NoCaseChange{\protect\cite{cite3174}}.
\item\relax Threshold under real-time geometrically local decoder based on introducing an ancillary buffer and confining spacetime interactions between anyons  \NoCaseChange{\protect\cite{cite3030}}.
\end{eczvaluelist}
\codefieldsection{Realizations}
\begin{eczvaluelist}
\item\relax NMR: 2-qubit phase-flip code \NoCaseChange{\protect\cite{cite4079,cite4080}}; 3-qubit bit-flip code \NoCaseChange{\protect\cite{cite4081}}; 3-qubit phase-flip code \NoCaseChange{\protect\cite{cite4082,cite4083}}, with up to two rounds of error correction in liquid-state NMR \NoCaseChange{\protect\cite{cite4084}}. Such codes were used to characterize noise \NoCaseChange{\protect\cite{cite4085}}.
\item\relax Linear optics: 2-qubit phase-flip code \NoCaseChange{\protect\cite{cite4086}}.
\item\relax Trapped ions: 3-qubit bit-flip code by Wineland group \NoCaseChange{\protect\cite{cite4087}}, and 3-qubit phase-flip algorithm implemented in 3 cycles on high fidelity gate operations \NoCaseChange{\protect\cite{cite4088}}.
Both phase- and bit-flip codes for 31 qubits and their stabilizer measurements have been realized by Quantinuum \NoCaseChange{\protect\cite{cite3186}}.
Multiple rounds of Steane error correction \NoCaseChange{\protect\cite{cite3357}}.

\item\relax Superconducting circuits: 3-qubit phase-flip and bit-flip code by Schoelkopf group \NoCaseChange{\protect\cite{cite4089,cite4090}}; 3-qubit bit-flip code \NoCaseChange{\protect\cite{cite4091}}; 3-qubit phase-flip code up to 3 cycles of error correction \NoCaseChange{\protect\cite{cite4092}}; IBM 15-qubit device \NoCaseChange{\protect\cite{cite4093}}; IBM Rochester device using 43-qubit code \NoCaseChange{\protect\cite{cite4094}}; Google system performing up to 8 error-correction cycles on 5 and 9 qubits \NoCaseChange{\protect\cite{cite4095}}; Google Quantum AI Sycamore utilizing up to 11 physical qubits and running 50 correction rounds \NoCaseChange{\protect\cite{cite3254}}; Google Quantum AI Sycamore utilizing up to 25 qubits for comparison of logical error scaling with a quantum code \NoCaseChange{\protect\cite{cite4096}} (see also \NoCaseChange{\protect\cite{cite4097}}). 
Google Quantum AI follow-up experiment on codes up to (classical) distance 29, demonstrating exponential suppression to an error floor of \(10^{-10}\) \NoCaseChange{\protect\cite{cite4098}}. 
Ising-model Nishimori phase transition realized for GHZ states on 54 qubits on a 127 qubit IBM device \NoCaseChange{\protect\cite{cite4099}}. 
GHZ state on 75 qubits made on an IBM device \NoCaseChange{\protect\cite{cite4100}}. 
Implementation of planar decoder for codes with distances between 3 and 11 on 72-qubit superconducting device \NoCaseChange{\protect\cite{cite4077}}.
Lattice surgery on the surface-17 code has been realized by splitting the code into two repetition codes by the Wallraff group \NoCaseChange{\protect\cite{cite3392}}.
2-qubit phase-flip code error detection realized with one ancilla by the Simakov group \NoCaseChange{\protect\cite{cite4101}}. GHZ state of 120 qubits realized on IBM device \NoCaseChange{\protect\cite{cite4102}}.

\item\relax Autonomous QEC protocols have been implemented on a 3-qubit superconducting qubit device \NoCaseChange{\protect\cite{cite4103}}.
\item\relax Semiconductor spin-qubit devices: 3-qubit devices at RIKEN \NoCaseChange{\protect\cite{cite4104}} and Delft \NoCaseChange{\protect\cite{cite4105}}.
\item\relax Nitrogen-vacancy centers in diamond: 3-qubit phase-flip code \NoCaseChange{\protect\cite{cite4106,cite4107}} (see also Ref. \NoCaseChange{\protect\cite{cite4108}}).
\item\relax Repetition codes are used in quantum annealing protocols \NoCaseChange{\protect\cite{cite803,cite4109,cite4110}}.
\item\relax Neutral atom arrays: 41 rounds of syndrome extraction and heralded logical Bell state preparation \NoCaseChange{\protect\cite{cite4111}}.
\end{eczvaluelist}
\codefieldsection{Notes}
\begin{eczvaluelist}
\item\relax Repetition codes can be used to benchmark device performance \NoCaseChange{\protect\cite{cite4112}}.
\end{eczvaluelist}
\codefieldsection{Parents}
\begin{eczvaluelist}
\item\relax
\flmRefsHyperref[eczindexfamilyrel]{code:quantum_parity}{Quantum parity code (QPC)} --- A \(\llbracket m_1 m_2,1,\min(m_1,m_2)\rrbracket \) QPC reduces to a repetition code when \(m_1\) or \(m_2\) is one.
\item\relax
\flmRefsHyperref[eczindexfamilyrel]{code:1d_stabilizer}{1D lattice stabilizer code} --- The codespace of the quantum repetition code is the ground-state space of a frustration-free 1D classical Ising model with nearest-neighbor interactions.
\item\relax
\flmRefsHyperref[eczindexfamilyrel]{code:qldpc}{Qubit QLDPC code} --- The codespace of the quantum repetition code is the ground-state space of a frustration-free 1D classical Ising model with nearest-neighbor interactions.
\item\relax
\flmRefsHyperref[eczindexfamilyrel]{code:small_distance_qubit_stabilizer}{Small-distance qubit stabilizer code}\item\relax
\flmRefsHyperref[eczindexfamilyrel]{code:gnu_permutation_invariant}{GNU PI code} --- GNU codewords for \(g=1\) reduce to the phase-flip repetition code.
\item\relax
\flmRefsHyperref[eczindexfamilyrel]{code:group_quantum_repetition}{Group-based quantum repetition code} --- Group-based quantum repetition codes reduce to quantum repetition codes for \(G = \mathbb{Z}_2\).
\end{eczvaluelist}
\codefieldsection{Child}
\begin{eczvaluelist}
\item\relax
\flmRefsHyperref[eczindexfamilyrel]{code:tetron}{Tetron code} --- The tetron code is a special case of the quantum repetition code with \(n=2\).
\end{eczvaluelist}
\codefieldsection{Cousins}
\begin{eczvaluelist}
\item\relax
\flmRefsHyperref[eczindexfamilyrel]{code:fracton}{Fracton stabilizer code} --- Product constructions built from the one-dimensional Ising/repetition code yield several fracton phases \NoCaseChange{\protect\cite[{Fig. 8}]{cite1501}}.
\item\relax
\flmRefsHyperref[eczindexfamilyrel]{code:topological_abelian}{Abelian topological code} --- Product constructions built from the one-dimensional Ising/repetition code yield several topological phases \NoCaseChange{\protect\cite[{Fig. 8}]{cite1501}}.
\item\relax
\flmRefsHyperref[eczindexfamilyrel]{code:ame}{Perfect-tensor code} --- GHZ states are \(1\)-uniform for all \(n\) and AME for \(n=2,3\).
\item\relax
\flmRefsHyperref[eczindexfamilyrel]{code:repetition}{Repetition code} --- A quantum repetition code can be thought of as a classical \([n,1,n]\) repetition code embedded in a quantum system.
\item\relax
\flmRefsHyperref[eczindexfamilyrel]{code:dual_rail}{Dual-rail quantum code} --- The dual-rail code is an error space of the quantum repetition code for \(n=2\) and is stabilized by \(-ZZ\).
\item\relax
\flmRefsHyperref[eczindexfamilyrel]{code:very-small-logical-qubit}{Very small logical qubit (VSLQ) code} --- Parts of the VSLQ codewords resemble the two-qubit phase-flip repetition code, though the code cannot correct phase errors. Unlike the phase-flip code, the VSLQ code can correct for single \flmRefsHyperref{ref498}{photon loss} because it uses the second excited state in the construction, which remains distinct from the vacuum even after \flmRefsHyperref{ref498}{photon loss}.
\item\relax
\flmRefsHyperref[eczindexfamilyrel]{code:cat_repetition}{Cat-repetition code} --- The cat-repetition code is obtained by encoding each qubit of a quantum repetition code into a two-component cat code in its cat-state basis \NoCaseChange{\protect\cite{cite2616,cite2646,cite2647,cite4072,cite4113}}.
\item\relax
\flmRefsHyperref[eczindexfamilyrel]{code:two-legged-cat}{Two-component cat code} --- Two-component cat and quantum repetition codes can be thought of as classical codes because they protect against only one type of noise. Two-component cat codes (quantum repetition) codes suppress cavity dephasing (bit-flip) noise exponentially with \(|\alpha|^2\) (\(n\)). The stability offered by cat codes has been linked to several favorable properties of phases of matter associated with the repetition-code Hamiltonian \NoCaseChange{\protect\cite{cite4114,cite4115}}.
\item\relax
\flmRefsHyperref[eczindexfamilyrel]{code:dfour_gkp}{\(D_4\) hyper-diamond GKP code} --- The \(D_4\) hyper-diamond GKP code can be seen as a concatenation of a rotated square-lattice GKP code with a repetition code \NoCaseChange{\protect\cite{cite482}}. This is related to the fact that the four-bit repetition code yields the \(D_4\) hyper-diamond lattice via \flmTerm{term}{ref127}{}{Construction A}.
\item\relax
\flmRefsHyperref[eczindexfamilyrel]{code:gkp_concatenated}{Concatenated GKP code} --- Concatenating a three-qubit quantum repetition code with GKP codes can correct some two-bit-flip errors \NoCaseChange{\protect\cite{cite3290}} (see also \NoCaseChange{\protect\cite{cite4116}}).
\item\relax
\flmRefsHyperref[eczindexfamilyrel]{code:coherent_state_repetition}{Coherent-state repetition code} --- Two-component cat codes in the coherent-state basis have been concatenated with quantum repetition codes \NoCaseChange{\protect\cite{cite4117,cite4118}}.
\item\relax
\flmRefsHyperref[eczindexfamilyrel]{code:self_correct}{Self-correcting quantum code} --- The bit-flip repetition code associated with the 2D classical Ising model is a self-correcting classical memory \NoCaseChange{\protect\cite[{Sec. V.A}]{cite1610}}.
\item\relax
\flmRefsHyperref[eczindexfamilyrel]{code:floquet_xyz_ruby}{Ruby Floquet code} --- One third of the time during the XYZ ruby measurement schedule, the ISG is that of the 6.6.6 color code concatenated with a three-qubit repetition code. One round of the color-code schedule is exactly the 2D color code concatenated with a three-qubit repetition code \NoCaseChange{\protect\cite{cite533}}.
\item\relax
\flmRefsHyperref[eczindexfamilyrel]{code:floquet_xcube}{X-cube Floquet code} --- The G-round of the rewinding X-cube Floquet code is exactly the canonical X-cube model concatenated with four-qubit repetition codes on composite green edges \NoCaseChange{\protect\cite{cite533}}.
\item\relax
\flmRefsHyperref[eczindexfamilyrel]{code:kitaev_chain}{Kitaev chain code} --- The Kitaev chain code can be thought of as the Majorana stabilizer analogue of the quantum repetition code \NoCaseChange{\protect\cite{cite559}} and is related to that code via the Jordan-Wigner transformation \NoCaseChange{\protect\cite{cite3794}}.
\item\relax
\flmRefsHyperref[eczindexfamilyrel]{code:hypercube_quantum}{\(\llbracket 2^D,D,2\rrbracket \) hypercube quantum code} --- The hypercube quantum code can be concatenated with a two-qubit quantum repetition code to yield a \(\llbracket 2^{D+1},D,4\rrbracket \) error-detecting code family \NoCaseChange{\protect\cite{cite759}}.
\item\relax
\flmRefsHyperref[eczindexfamilyrel]{code:phantom_14_3_3}{\(\llbracket 14,3,3\rrbracket \) CE phantom code} --- The inner code in the construction is the two-qubit phase-flip repetition code \NoCaseChange{\protect\cite{cite514}}.
\item\relax
\flmRefsHyperref[eczindexfamilyrel]{code:shor_nine}{\(\llbracket 9,1,3\rrbracket \) Shor code} --- The Shor code is a concatenation of a three-qubit bit-flip with a three-qubit phase-flip repetition code.
\item\relax
\flmRefsHyperref[eczindexfamilyrel]{code:tfim}{Transverse-field Ising model (TFIM) code} --- When written in the computational basis, the phase-flip and TFIM codewords are superpositions of qubit states of fixed total parity. The superposition is equal for the phase-flip code, whereas some states appear with a \(-1\) coefficient for the TFIM code. However, the TFIM code can be encoded in constant depth.
\item\relax
\flmRefsHyperref[eczindexfamilyrel]{code:cluster_state}{Cluster-state code} --- GHZ states can be used as resource states for MBQC protocols \NoCaseChange{\protect\cite{cite3566,cite3567,cite3568}}.
In the fault-complex formalism, foliation of a CSS code is expressed as a homological product of the code's chain complex with a repetition-code complex \NoCaseChange{\protect\cite{cite3176}}.

\item\relax
\flmRefsHyperref[eczindexfamilyrel]{code:qudit_sign}{Modular-qudit shift-resistant code} --- Both the quantum repetition and modular-qudit shift-resistant codes protect against only one type of noise.
\end{eczvaluelist}
\eczhbkcontributors{ Mazin Karjikar, \eczhuVVA }
\endeczcode

\eczcode{sc_qldpc}{Quantum spatially coupled (SC-QLDPC) code}{~\NoCaseChange{\protect\cite{cite643,cite4119,cite4120,cite644}}}
\codefieldsection{Description}
QLDPC code whose stabilizer generator matrix resembles the parity-check matrix of SC-LDPC codes.
There exist CSS \NoCaseChange{\protect\cite{cite643}} and stabilizer constructions \NoCaseChange{\protect\cite{cite644}}.
In either case, the stabilizer generator matrix is constructed by "spatially" coupling sub-matrix blocks in chain-like fashion (or, more generally, in grid-like fashion) to yield a band matrix.
The sub-matrix blocks have to satisfy certain conditions amongst themselves so that the resulting band matrix is a stabilizer generator matrix.
Matrices corresponding to translationally invariant chains are called \textit{time-invariant}, and otherwise are called \textit{time-varying}.

A finite-length chain is then capped by imposing either open boundary conditions (yielding \textit{non-tail-biting} SC-QLDPC codes) or periodic boundary conditions (yielding \textit{tail-biting} SC-QLDPC codes).
Both constructions \NoCaseChange{\protect\cite{cite643,cite644}} are tail-biting.

In the stabilizer construction \NoCaseChange{\protect\cite{cite644}}, the structure of the band matrix allows codes to be concisely defined in terms of \textit{characteristic polynomials}, whose coefficients are the sub-matrix blocks and which resemble the \flmRefsHyperref{ref4121}{Pauli-to-polynomial mapping} associated with translationally invariant stabilizer codes.
Some CSS code constructions can be used to define sub-matrix blocks, yielding spatially coupled (i.e., translationally invariant) extensions of such codes.

For example, the \(3\times 3\) toric code can be expressed as an SC-QLDPC code with stabilizer generator matrix given in \flmRefsCref{ref4122}.
\begin{flmFloat}{figure}{NumCap}\includegraphics[width=298.080009bp,max width=\linewidth]{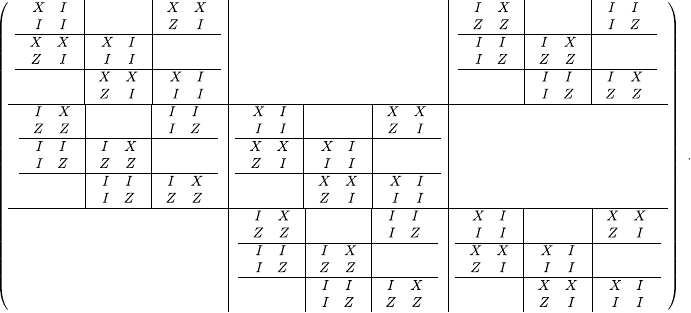}\caption{Stabilizer generator matrix of the \(3\times 3\) toric code, expressed
      as an SC-QLDPC code.}\label{ref4122}\end{flmFloat}

\codefieldsection{Parents}
\begin{eczvaluelist}
\item\relax
\flmRefsHyperref[eczindexfamilyrel]{code:qldpc}{Qubit QLDPC code}\item\relax
\flmRefsHyperref[eczindexfamilyrel]{code:translationally_invariant_stabilizer}{Lattice stabilizer code} --- Stabilizer generator matrices of SC-QLDPC codes on infinite-length chains or grids define a class of lattice stabilizer codes.
\end{eczvaluelist}
\codefieldsection{Children}
\begin{eczvaluelist}
\item\relax
\flmRefsHyperref[eczindexfamilyrel]{code:hypergraph_product}{Hypergraph product (HGP) code} --- Hypergraph-product stabilizer generator matrices can be used as sub-matrices to define a 2D SC-QLDPC code \NoCaseChange{\protect\cite{cite644}}.
\item\relax
\flmRefsHyperref[eczindexfamilyrel]{code:xyz_product}{XYZ product code} --- XYZ product stabilizer generator matrices can be used as sub-matrices to define a 2D SC-QLDPC code \NoCaseChange{\protect\cite{cite644}}.
\end{eczvaluelist}
\codefieldsection{Cousins}
\begin{eczvaluelist}
\item\relax
\flmRefsHyperref[eczindexfamilyrel]{code:sc_ldpc}{Spatially coupled LDPC (SC-LDPC) code} --- SC-QLDPC code stabilizer-generator matrices have similar block form as the parity-check matrices of SC-LDPC codes.
\item\relax
\flmRefsHyperref[eczindexfamilyrel]{code:quasi_cyclic}{Quasi-cyclic code} --- Quasi-cyclic binary code parity-check matrices can be used as sub-matrices to define a 1D SC-QLDPC code \NoCaseChange{\protect\cite{cite643}}.
\item\relax
\flmRefsHyperref[eczindexfamilyrel]{code:generalized_bicycle}{Generalized bicycle (GB) code} --- Qubit GB codes can be categorized as 1D SC-QLDPC codes, see \NoCaseChange{\protect\cite[{Remark 7}]{cite644}}.
\end{eczvaluelist}
\eczhbkcontributors{ \eczhuVVA }
\endeczcode

\eczcode{quantum_synchronizable}{Quantum synchronizable code}{~\NoCaseChange{\protect\cite{cite1246}}}
\codefieldsection{Description}
A qubit stabilizer code designed to protect against synchronization errors (a.k.a. misalignment), which are errors that misalign the code block in a larger block by one or more locations. 

In such a setting, the qubits are arranged in a line and embedded into a larger block that represents a stream of information coming into the receiver.
The receiver may not properly identify the first qubit in the desired code block, leading to a misalignment of the block.
A quantum synchronizable code is denoted by \((a_l,a_r)-\llbracket n+a_l+a_r,k\rrbracket \), correcting misalignment by up to \(a_l\) (\(a_r\)) qubits to the left (right).

The initial construction of quantum synchronizable codes was based on the CSS construction \NoCaseChange{\protect\cite{cite1246}}, but later work extended it to non-CSS stabilizer codes \NoCaseChange{\protect\cite{cite3997}}.
\codefieldsection{Protection}
In the original CSS construction based on cyclic codes \(C \subset D\) with parameters \([n,k_1,d_1]\) and \([n,k_2,d_2]\), one obtains an \((a_l,a_r)-\llbracket n+a_l+a_r,2k_1-n\rrbracket \) code whenever \(a_l+a_r<k_2-k_1\), correcting at least \(\lfloor (d_1-1)/2\rfloor\) phase errors and \(\lfloor (d_2-1)/2\rfloor\) bit errors \NoCaseChange{\protect\cite[{Thm. 1}]{cite1246}}.

\codefieldsection{Parent}
\begin{eczvaluelist}
\item\relax
\flmRefsHyperref[eczindexfamilyrel]{code:qubit_stabilizer}{Qubit stabilizer code}\end{eczvaluelist}
\codefieldsection{Cousins}
\begin{eczvaluelist}
\item\relax
\flmRefsHyperref[eczindexfamilyrel]{code:binary_cyclic}{Cyclic linear binary code} --- The original construction of quantum synchronizable codes is based on pairs of binary cyclic codes satisfying \(C^\perp \subseteq C \subset D\) \NoCaseChange{\protect\cite{cite1246}}.
\item\relax
\flmRefsHyperref[eczindexfamilyrel]{code:bch}{Binary BCH code} --- BCH codes can be used to construct quantum synchronizable codes via the CSS construction \NoCaseChange{\protect\cite{cite1246}}.
\item\relax
\flmRefsHyperref[eczindexfamilyrel]{code:binary_quad_residue}{Binary quadratic-residue (QR) code} --- Binary QR codes can be used to construct quantum synchronizable codes via the CSS construction \NoCaseChange{\protect\cite{cite1279}}.
\item\relax
\flmRefsHyperref[eczindexfamilyrel]{code:qubit_stabilizer_oaqecc}{Operator-algebra (OA) qubit stabilizer code} --- Quantum synchronizable versions of qubit subsystem codes, hybrid codes, and OA qubit stabilizer codes have been constructed \NoCaseChange{\protect\cite{cite3997}}.
\end{eczvaluelist}
\eczhbkcontributors{ \eczhuVVA }
\endeczcode

\eczcode{quantum_tanner}{Quantum Tanner code}{~\NoCaseChange{\protect\cite{cite1414}}}
\codefieldsection{Description}
Member of a family of QLDPC codes based on two compatible classical Tanner codes defined on a two-dimensional Cayley complex, a complex constructed from Cayley graphs of groups.
For certain choices of codes and complex, the resulting codes have asymptotically good parameters.
See Ref. \NoCaseChange{\protect\cite{cite685}} for explicit instances based on dihedral groups.
This construction has been generalized to Schreier graphs \NoCaseChange{\protect\cite{cite686}}.

The underlying geometric complex of the code is a left-right Cayley complex \(\operatorname{Cay}_2(A,G,B)\), where \(G\) is a finite group and \(A=A^{-1}\), \(B=B^{-1}\) are two symmetric generating sets satisfying the total no-conjugacy condition: \(ag\ne gb\) for any \(g\in G\), \(a\in A\), and \(b\in B\).
The vertices of the complex are group elements, i.e., \(V=G\). If necessary, the double cover of the graph should be taken so that the graph is bipartite, \(V=V_0\sqcup V_1\).

There are two types of edges in the resulting complex,
  \flmMathEnvironment{align}{}{
  \begin{split}
    E_A &= \{(g,ag): g\in G, a\in A\}\\
    E_B &= \{(g,gb): g\in G, b\in B\}~.
  \end{split}
  }
The faces are squares defined by quadruples,
  \flmMathEnvironment{align}{}{
    Q = \{(g,ag,gb,agb): g\in G, a\in A, b\in B\}~.
  }
Two additional graphs can be obtained from \(\operatorname{Cay}_2(A,G,B)\) by taking the diagonals of the squares as edges, \(\mathcal G_0^\square = (V_0, Q)\) and \(\mathcal G_1^\square = (V_1, Q)\).

In the quantum Tanner construction, qubits are placed on the squares of the left-right Cayley complex.
Two classical codes \(C_A\) and \(C_B\) of blocklengths \(|A|\) and \(|B|\), respectively, are chosen, yielding local codes \(C_0 = C_A\otimes C_B\) and \(C_1 = C_A^\perp\otimes C_B^\perp\).
The quantum Tanner code is a CSS code defined by the classical Tanner codes \(C_Z = T(\mathcal G_0^\square, C_0^\perp)\) and \(C_X = T(\mathcal G_1^\square, C_1^\perp)\).
The figure below depicts an example of a stabilizer generator.

\begin{flmFloat}{figure}{NumCap}\includegraphics[width=657.7628231004622bp,max width=\linewidth]{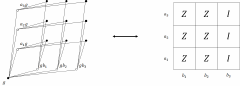}\caption{An example of a \(Z\) generator on a \(V_0\) local view when \(C_A = \operatorname{span}\{111\}\) and \(C_B=\operatorname{span}\{110, 011\}\). The faces incident to a \(V_0\) vertex are in bijection with the set \(A\times B\), and a codeword of \(C_0 = C_A\otimes C_B\) can be described using this set.}\label{ref4123}\end{flmFloat}

To achieve asymptotically good parameters, fixed classical local codes are chosen so that their dual tensor codes are sufficiently robust, and the left-right Cayley complexes are chosen to be sufficiently expanding. The family is defined using a family of groups \(G\) of increasing size but constant-size generating sets \(A\), \(B\).

\codefieldsection{Protection}
For correctly chosen complexes and local codes, the distance scales as \(d=\Theta(n)\). Minimum distance bound obtained using robustness of dual \flmRefsHyperref{code:tensor}{tensor-product} codes \NoCaseChange{\protect\cite{cite2191}}.
\codefieldsection{Rate}
Asymptotically good QLDPC codes. When \(C_A\) and \(C_B\) are chosen to have rates not equal to a half, the number of encoded qubits scales as \(k=\Theta(n)\).
\codefieldsection{Decoding}
\begin{eczvaluelist}
\item\relax Linear-time potential-based decoder similar to the small-set-flip decoder for quantum expander codes \NoCaseChange{\protect\cite{cite3623}}.
\item\relax Linear-time decoder \NoCaseChange{\protect\cite{cite4124}}.
\item\relax Logarithmic-time mismatch decomposition decoder \NoCaseChange{\protect\cite{cite2191}}.
\end{eczvaluelist}
\codefieldsection{Code Capacity Threshold}
\begin{eczvaluelist}
\item\relax Independent \(X,Z\) noise: lower bound under potential-based decoder \NoCaseChange{\protect\cite[{Corr. 15}]{cite3623}}.
\end{eczvaluelist}
\codefieldsection{Realizations}
\begin{eczvaluelist}
\item\relax Used to obtain explicit lower bounds in the sum-of-squares game \NoCaseChange{\protect\cite{cite4125}}.
\item\relax States that, on average, achieve small violations of check operators for quantum Tanner codes require a circuit of non-constant depth to make. They are used in the proof \NoCaseChange{\protect\cite{cite4126}} of the NLTS conjecture \NoCaseChange{\protect\cite{cite2562}}.
\end{eczvaluelist}
\codefieldsection{Notes}
\begin{eczvaluelist}
\item\relax For details, see talk by \flmHref{https://www.youtube.com/watch?v=5GO3BtJuo3I}{A. Leverrier}.
\end{eczvaluelist}
\codefieldsection{Parent}
\begin{eczvaluelist}
\item\relax
\flmRefsHyperref[eczindexfamilyrel]{code:generalized_quantum_tanner}{Generalized quantum Tanner code} --- Generalized quantum Tanner codes constructed out of bipartite double covers of Cayley graphs reduce to quantum Tanner codes \NoCaseChange{\protect\cite{cite686}}.
\end{eczvaluelist}
\codefieldsection{Child}
\begin{eczvaluelist}
\item\relax
\flmRefsHyperref[eczindexfamilyrel]{code:rotated_surface}{Rotated surface code} --- Applying the quantum Tanner transformation to the surface code yields the rotated surface code \NoCaseChange{\protect\cite{cite3954,cite3955}}.
\end{eczvaluelist}
\codefieldsection{Cousins}
\begin{eczvaluelist}
\item\relax
\flmRefsHyperref[eczindexfamilyrel]{code:single_shot}{Single-shot code} --- Certain quantum Tanner codes facilitate single-shot decoding \NoCaseChange{\protect\cite{cite846}}.
\item\relax
\flmRefsHyperref[eczindexfamilyrel]{code:good_qldpc}{Good QLDPC code} --- Quantum Tanner code construction yields asymptotically good QLDPC codes.
\item\relax
\flmRefsHyperref[eczindexfamilyrel]{code:regular_binary_tanner}{Regular binary Tanner code} --- Regular binary Tanner codes are used in constructing quantum Tanner codes.
\item\relax
\flmRefsHyperref[eczindexfamilyrel]{code:tensor}{Tensor-product code} --- Tensor codes are used in constructing quantum Tanner codes.
\item\relax
\flmRefsHyperref[eczindexfamilyrel]{code:expander_lifted_product}{Expander LP code} --- Quantum Tanner codes are an attempt to construct asymptotically good QLDPC codes that are similar to but simpler than expander lifted-product codes; see Ref. \NoCaseChange{\protect\cite{cite4124}} for connection between the codes.
\item\relax
\flmRefsHyperref[eczindexfamilyrel]{code:lr-cayley-complex}{Left-right Cayley complex code} --- Applying the CSS construction to two left-right Cayley complex codes yields quantum Tanner codes, and one can simultaneously prove a linear distance for the quantum code and local testability for one of its constituent classical codes \NoCaseChange{\protect\cite{cite1414}}.
\item\relax
\flmRefsHyperref[eczindexfamilyrel]{code:classical_product}{Classical-product code} --- A \(\llbracket 512,174,8\rrbracket \) classical-product code performed well \NoCaseChange{\protect\cite{cite204}} against erasure and depolarizing noise when compared to a member of an asymptotically good quantum Tanner code family.
\end{eczvaluelist}
\eczhbkcontributors{ Shouzhen (Bailey) Gu, \eczhuVVA }
\endeczcode

\eczcode{quantum_tensor_product}{Quantum tensor-product code}{~\NoCaseChange{\protect\cite{cite4024,cite1131}}}
\codefieldsection{Description}
CSS code constructed from a tensor code. In some cases, only one of the classical codes forming the tensor code needs to be self-orthogonal. 

\codefieldsection{Protection}
If one of the classical codes forming the tensor code protects against burst errors, the resulting quantum code does also \NoCaseChange{\protect\cite{cite1131}}.

\codefieldsection{Parent}
\begin{eczvaluelist}
\item\relax
\flmRefsHyperref[eczindexfamilyrel]{code:qubit_css}{Qubit CSS code}\end{eczvaluelist}
\codefieldsection{Cousins}
\begin{eczvaluelist}
\item\relax
\flmRefsHyperref[eczindexfamilyrel]{code:tensor}{Tensor-product code} --- Quantum tensor-product codes are quantum analogues of tensor-product codes.
\item\relax
\flmRefsHyperref[eczindexfamilyrel]{code:reversible}{Reversible code} --- Reversible cyclic codes can be used to construct quantum tensor-product codes \NoCaseChange{\protect\cite{cite1131}}.
\item\relax
\flmRefsHyperref[eczindexfamilyrel]{code:q-ary_cyclic}{Cyclic linear \(q\)-ary code} --- Reversible cyclic codes can be used to construct quantum tensor-product codes \NoCaseChange{\protect\cite{cite1131}}.
\item\relax
\flmRefsHyperref[eczindexfamilyrel]{code:mds}{Maximum distance separable (MDS) code} --- MDS codes can be used to construct quantum tensor-product codes \NoCaseChange{\protect\cite{cite1131}}.
\item\relax
\flmRefsHyperref[eczindexfamilyrel]{code:galois_polynomial}{Galois-qudit RS code} --- Product codes constructed from a self-orthogonal and an arbitrary RS code yield an RS code \NoCaseChange{\protect\cite{cite4024}}.
\item\relax
\flmRefsHyperref[eczindexfamilyrel]{code:check_product}{Quantum check-product code} --- Quantum check-product codes extend the concept of a check product, which yields the dual of a tensor code, to a product between a classical and a quantum code.
\item\relax
\flmRefsHyperref[eczindexfamilyrel]{code:quantum_reed_muller}{Quantum Reed-Muller (RM) code} --- EA versions of quantum RM codes and their quantum tensor-product variants can be constructed \NoCaseChange{\protect\cite{cite3651}}.
\end{eczvaluelist}
\eczhbkcontributors{ \eczhuVVA }
\endeczcode

\eczcode{quantum_turbo}{Quantum turbo code}{~\NoCaseChange{\protect\cite{cite4127,cite399}}}
\codefieldsection{Description}
A quantum version of the turbo code, obtained from an interleaved serial quantum concatenation \NoCaseChange{\protect\cite[{Def. 30}]{cite399}} of quantum convolutional codes.
The interleaver induces long-range entanglement and can increase the minimum distance relative to the constituent convolutional codes \NoCaseChange{\protect\cite{cite400}}.
\codefieldsection{Encoding}
\begin{eczvaluelist}
\item\relax Encoders of codes with polynomial distance yield catastrophic errors, but codes with bounded distance admit non-catastrophic encoders.
\end{eczvaluelist}
\codefieldsection{Decoding}
\begin{eczvaluelist}
\item\relax Turbo decoder \NoCaseChange{\protect\cite[{Sec. V}]{cite399}}.
\item\relax Modified decoder yields improvement over the memoryless depolarizing channel \NoCaseChange{\protect\cite{cite3636}}.
\item\relax Iterative decoding is analogous to a mean-field treatment of two matrix-product-state chains coupled by random non-local interactions \NoCaseChange{\protect\cite{cite400}}.
\item\relax EXIT charts \NoCaseChange{\protect\cite{cite4128}}.
\end{eczvaluelist}
\codefieldsection{Parents}
\begin{eczvaluelist}
\item\relax
\flmRefsHyperref[eczindexfamilyrel]{code:quantum_convolutional}{Quantum convolutional code}\item\relax
\flmRefsHyperref[eczindexfamilyrel]{code:qubit_concatenated}{Concatenated qubit code}\end{eczvaluelist}
\codefieldsection{Cousins}
\begin{eczvaluelist}
\item\relax
\flmRefsHyperref[eczindexfamilyrel]{code:turbo}{Turbo code} --- Quantum turbo codes are quantum analogues of turbo codes.
\item\relax
\flmRefsHyperref[eczindexfamilyrel]{code:ea_turbo}{EA quantum turbo code} --- EA quantum turbo codes are entanglement-assisted versions of quantum turbo codes.
\end{eczvaluelist}
\eczhbkcontributors{ \eczhuVVA }
\endeczcode

\eczcode{quasi_hyperbolic_color}{Quasi-hyperbolic color code}{~\NoCaseChange{\protect\cite{cite703}}}
\codefieldsection{Description}
An extension of the color code construction to quasi-hyperbolic 3-manifolds, e.g., a product of a 2D hyperbolic surface and a circle.

\codefieldsection{Protection}
There exists a quasi-hyperbolic family with rate of \flmRefsHyperref{ref65}{order} \(O(1/\log n)\) and distance of \flmRefsHyperref{ref65}{order} \(O(\log n)\) \NoCaseChange{\protect\cite{cite703}}.
These codes support collective logical \(CCZ\) gates via transversal \(T\) and exponentially many individually addressable, parallelizable logical \(CZ\) gates via transversal \(S\) on codimension-1 submanifolds \NoCaseChange{\protect\cite{cite703}}.

\codefieldsection{Rate}
A Torelli mapping-torus construction yields a code with constant rate, although its distance scaling is currently unknown \NoCaseChange{\protect\cite{cite703}}.
\codefieldsection{Fault Tolerance}
\begin{eczvaluelist}
\item\relax The quasi-hyperbolic family supports collective logical \(CCZ\) gates via transversal \(T\) and exponentially many individually addressable, parallelizable logical \(CZ\) gates via transversal \(S\) on codimension-1 submanifolds \NoCaseChange{\protect\cite{cite703}}.
\end{eczvaluelist}
\codefieldsection{Parent}
\begin{eczvaluelist}
\item\relax
\flmRefsHyperref[eczindexfamilyrel]{code:color}{Color code}\end{eczvaluelist}
\codefieldsection{Cousins}
\begin{eczvaluelist}
\item\relax
\flmRefsHyperref[eczindexfamilyrel]{code:higher_dimensional_surface}{Homological code} --- Quasi-hyperbolic color codes are related to quasi-hyperbolic surface codes via a constant-depth \flmRefsHyperref{ref409}{Clifford circuit} \NoCaseChange{\protect\cite{cite703}}.
\item\relax
\flmRefsHyperref[eczindexfamilyrel]{code:quantum_rainbow}{Quantum rainbow code} --- Hypergraph products of color codes yield quantum rainbow codes with growing distance and transversal gates in the \flmTerm{term}{ref694}{}{Clifford hierarchy}. In particular, utilizing this construction for quasi-hyperbolic color codes yields an \(\llbracket n,O(n),O(\log n)\rrbracket \) triorthogonal code family satisfying the necessary conditions for the magic-state yield parameter \(\gamma\) to become arbitrarily small \NoCaseChange{\protect\cite{cite704}}.
\end{eczvaluelist}
\eczhbkcontributors{ Guanyu Zhu, \eczhuVVA }
\endeczcode

\eczcode{quantum_bch}{Qubit BCH code}{~\NoCaseChange{\protect\cite{cite861,cite3270,cite449,cite4129,cite4130}}}
\codefieldsection{Description}
Qubit stabilizer code constructed from a self-orthogonal binary BCH code via the CSS construction, from a Hermitian self-orthogonal quaternary BCH code via the Hermitian construction, or by taking a Euclidean self-orthogonal BCH code over \(\mathbb{F}_{2^m}\), converting it to a binary code, and applying the CSS construction.

\codefieldsection{Gates}
\begin{eczvaluelist}
\item\relax Magic-state distillation protocols \NoCaseChange{\protect\cite{cite705}}.
\end{eczvaluelist}
\codefieldsection{Threshold}
\begin{eczvaluelist}
\item\relax Semi-analytical estimates of \flmRefsHyperref{ref515}{concatenated thresholds} were given in \NoCaseChange{\protect\cite{cite518}}. In the broader comparative study of Ref. \NoCaseChange{\protect\cite{cite3225}}, BCH codes larger than \(\llbracket 47,1,11\rrbracket \) were not simulated because encoded-CNOT ex-Recs became impractical; although their large \(t/n\) suggests potentially good thresholds, the resulting overhead was argued to limit their usefulness as bottom codes.
\end{eczvaluelist}
\codefieldsection{Notes}
\begin{eczvaluelist}
\item\relax Qubit BCH codes for small \(n\) are tabulated in Ref. \NoCaseChange{\protect\cite{cite4130}}.
\end{eczvaluelist}
\codefieldsection{Parents}
\begin{eczvaluelist}
\item\relax
\flmRefsHyperref[eczindexfamilyrel]{code:qubit_stabilizer}{Qubit stabilizer code}\item\relax
\flmRefsHyperref[eczindexfamilyrel]{code:galois_bch}{Galois-qudit BCH code} --- Galois-qudit BCH codes for \(q=2\) reduce to qubit BCH codes.
\end{eczvaluelist}
\codefieldsection{Cousins}
\begin{eczvaluelist}
\item\relax
\flmRefsHyperref[eczindexfamilyrel]{code:bch}{Binary BCH code} --- Binary BCH codes are used to construct a subset of qubit BCH codes via the CSS construction.
\item\relax
\flmRefsHyperref[eczindexfamilyrel]{code:q-ary_bch}{Bose–Chaudhuri–Hocquenghem (BCH) code} --- BCH codes are used to construct qubit BCH codes via the CSS construction or the Hermitian construction.
\item\relax
\flmRefsHyperref[eczindexfamilyrel]{code:qubit_css}{Qubit CSS code} --- Some qubit BCH codes are CSS.
\item\relax
\flmRefsHyperref[eczindexfamilyrel]{code:stabilizer_over_gf4}{Hermitian qubit code} --- Hermitian self-orthogonal quaternary BCH codes are used to construct a subset of qubit BCH codes via the Hermitian construction.
\end{eczvaluelist}
\eczhbkcontributors{ \eczhuVVA }
\endeczcode

\eczcode{qubits_into_qubits}{Qubit code}{}
\codefieldsection{Alternative Names}
\begin{eczvaluelist}
\item\relax Qubit subspace code
\end{eczvaluelist}
\eczhIndexCodeAliasName{qubits_into_qubits}{Qubit subspace code}

\codefieldsection{Kingdom root code}
for the \flmRefsHyperref{kingdom:qubits_into_qubits}{Qubit Kingdom}.
\codefieldsection{Description}
Encodes \(K\)-dimensional Hilbert space into a \(2^n\)-dimensional (i.e., \(n\)-qubit) Hilbert space.
Usually denoted as \(\llparenthesis n,K\rrparenthesis \) or \(\llparenthesis n,K,d\rrparenthesis \), where \(d\) is the code's distance.

The qubit codes are \textit{equivalent} if the codespace of one code can be mapped into that of the other under a tensor product of single-qubit unitary operations and a qubit permutation.
Equivalent qubit codes have the same logical dimension and distance, and their decoding problems have the same computational complexity up to the corresponding relabeling of errors \NoCaseChange{\protect\cite[{Ch. 2}]{cite398}}.

\codefieldsection{Protection}
An \(\llparenthesis n,K,d\rrparenthesis \) code with distance \(d\) detects errors acting on up to \(d-1\) qubits, corrects erasure errors on up to \(d-1\) qubits \NoCaseChange{\protect\cite{cite3270}}, or corrects errors acting on up to \(\lfloor (d-1)/2 \rfloor\) qubits.
Combinations of errors and erasures can also be corrected \NoCaseChange{\protect\cite{cite4131}}.
The number of correctable errors is often called the \textit{decoding radius}, and it is upper bounded by half of the code distance.
As a result, qubit codes cannot tolerate adversarial errors on more than \((1-R)/4\) registers, where \(R = \log_2 K/n\) is the code rate.

\subsection{Pauli-string error basis}\label{ref663}

A convenient and often considered error set is the \textit{Pauli error} or \textit{Pauli string} basis.

\begin{defterm}{Pauli strings}\label{ref4132}
For a single qubit, this set consists of products of powers of the Pauli matrices
\flmMathEnvironment{align}{}{
  X=\begin{pmatrix}0 & 1\\
  1 & 0
  \end{pmatrix}~,
}
\flmMathEnvironment{align}{}{
  Y=iXZ=\begin{pmatrix}0 & -i\\
  i & 0
  \end{pmatrix}~,
}
\flmMathEnvironment{align}{}{
  Z=\begin{pmatrix}1 & 0\\
  0 & -1
  \end{pmatrix}~.
}
The operator \(X\) is a bit flip, \(Z\) is a phase flip, and \(Y\) is a combined bit and phase flip up to an overall physically irrelevant phase \NoCaseChange{\protect\cite[{Ch. 1}]{cite398}}.
For multiple qubits, error set elements are tensor products of elements of the single-qubit error set.
Tensor products of \(X\) (\(Z\)) Paulis acting on different qubits are called \(X\)\textit{-type} (\(Z\)\textit{-type}) Pauli strings.
Combining the \(X\)-type and \(Z\)-type strings with \(i\) forms a group called the \textit{Pauli group} on \(n\) qubits, while combining them with \(-1\) forms the \textit{real Pauli group}.
\end{defterm}

The Pauli error set is a unitary and Hermitian basis for linear operators on the multi-qubit Hilbert space that is orthonormal under the Hilbert-Schmidt inner product; it is a prototypical \flmRefsHyperref{ref2812}{nice error basis}.
The distance associated with this set is often the minimum weight of a Pauli string that implements a nontrivial logical operation in the code.

\subsection{Noise channels}

A quantum channel that admits a set of Pauli strings as its Kraus operators is called a \textit{Pauli channel}, and such channels are typically more tractable than the more general, non-Pauli channels.
Relevant Pauli channels include dephasing noise and depolarizing noise (a.k.a. Werner-Holevo channel \NoCaseChange{\protect\cite{cite4133}}).
A single-qubit dephasing channel has Kraus operators proportional to \(I\) and \(Z\), while the single-qubit depolarizing channel has equal probabilities for the \(X\), \(Y\), and \(Z\) errors; both are Pauli channels \NoCaseChange{\protect\cite[{Ch. 1}]{cite398}}.
One can extract a binary memoryless symmetric channel from a Pauli channel that is a classical counterpart to the Pauli channel \NoCaseChange{\protect\cite{cite4134}}.

Relevant non-Pauli channels are \flmRefsHyperref{ref498}{AD} noise, erasure (which maps all qubit states into a third state \(|e\rangle\) outside of the qubit Hilbert space), and biased erasure \NoCaseChange{\protect\cite{cite4135}} (in which case only the \(|1\rangle\) qubit state is mapped to \(|e\rangle\)).
Erasure channels are easier to correct than general channels because one can, in principle, measure whether the state left the qubit Hilbert space and thereby learn which qubits were erased without measuring the stored qubit state \NoCaseChange{\protect\cite[{Ch. 1}]{cite398}}.
Noise can be correlated in space or in time, with the latter being an example of a non-Markovian phenomenon \NoCaseChange{\protect\cite{cite4136,cite4137}\protect\cite[{Ch. 1}]{cite398}}.

\subsection{Quantum weight enumerators and pure distance}\label{ref672}

\begin{defterm}{Quantum weight enumerator}\label{ref4138}
Determining protection and bounds on code parameters can also be done using the code's Shor-Laflamme \textit{quantum weight enumerator} \NoCaseChange{\protect\cite{cite4139}} (cf. \flmRefsHyperref{ref113}{weight enumerators})
  \flmMathEnvironment{align}{}{
  \begin{split}
    A(x)&=\sum_{j=0}^{n}A_{j}x^{j}\\
    A_{j}&=\frac{1}{K^{2}}\sum_{\text{wt-}j\text{ Paulis }P}\left|\text{tr}(P\Pi)\right|^{2}~,
  \end{split}
  }
where \(K=\mathrm{tr}(\Pi)\) is the dimension of the code subspace, \(\Pi\) is the code projection, and where the sum is over the Pauli group modulo the subgroup of phases (hence, the dagger below is necessary in case the coset representative is not Hermitian).

The dual quantum weight enumerator is
  \flmMathEnvironment{align}{}{
  \begin{split}
    B(x)&=\sum_{j=0}^{n}B_{j}x^{j}\\
    B_{j}&=\frac{1}{K}\sum_{\text{wt-}j\text{ Paulis }P}\text{tr}(P\Pi P^{\dagger}\Pi)~,
  \end{split}
  }
and the two satisfy the \textit{quantum MacWilliams identity} \NoCaseChange{\protect\cite{cite4139}}; see \NoCaseChange{\protect\cite[{Ch. 7}]{cite398}}.
Their coefficients satisfy \(A_0=B_0=1\) and \(B_j\geq A_j\geq 0\) for all \(j\) \NoCaseChange{\protect\cite{cite4139,cite398}}.
This identity gives rise to quantum linear programming (LP) bounds \NoCaseChange{\protect\cite{cite2663,cite2664}}; see the book \NoCaseChange{\protect\cite{cite398}}.
Weight enumerators give rise to an analogue of Poisson summation for qubit and, more generally, modular-qudit stabilizer codes \NoCaseChange{\protect\cite{cite4140}}.
\end{defterm}

\begin{defterm}{Pure distance}\label{ref4141}
The distance \(d\) of a qubit code is the smallest integer \(0<j=d\) at which the \flmRefsHyperref{ref672}{quantum weight enumerator} is not equal to its dual, \(A_j \neq B_j\) \NoCaseChange{\protect\cite{cite4142}}.
A code is called \textit{pure} if \(A_j = 0\) for all \(0 < j < d\); otherwise, the code is called \textit{impure}.
The \textit{pure distance} \NoCaseChange{\protect\cite{cite1970,cite4033}} (a.k.a. diagonal distance \NoCaseChange{\protect\cite{cite3583,cite3584}}) \(d_{\textnormal{pure}}\) is the smallest integer \(1 < j=d_{\textnormal{pure}}\) at which \(A_j > 0\).
Codes for which \(d_{\textnormal{pure}} < d\) are impure, otherwise they are pure.
For impure codes, there exists a Pauli error of weight less than the \(d\) that has a nonzero expectation value with respect to a code state.

Degenerate qubit codes are impure, but impure codes may not be degenerate \NoCaseChange{\protect\cite{cite449,cite398}}.
There are subtleties with defining \flmRefsHyperref{ref811}{degeneracy} for non-stabilizer qubit codes with even distance \NoCaseChange{\protect\cite{cite398}}.
\end{defterm}

Other types of quantum weight enumerators are the Rains unitary enumerators \NoCaseChange{\protect\cite{cite4143}} and the \textit{Rains shadow enumerators} \NoCaseChange{\protect\cite{cite2663}} (see also \NoCaseChange{\protect\cite{cite1647}}), and \textit{signed weight enumerators} taking into account the sign of the expectation value of a Pauli string \NoCaseChange{\protect\cite{cite3717}}.
For qubit codes, the shadow enumerator coefficients are nonnegative and are determined by the Shor-Laflamme enumerator \NoCaseChange{\protect\cite{cite2663,cite398}} via
\flmMathEnvironment{align}{}{
  Sh(x)=\frac{K}{2^n}(1+3x)^n A\left(\frac{x-1}{1+3x}\right)~.
}
Rains shadow enumerators are related to Bell sampling \NoCaseChange{\protect\cite{cite3360}}.
These notions can be generalized to qudit codes and other error bases \NoCaseChange{\protect\cite{cite4144,cite4145,cite4146,cite3102}}.
There are techniques to compute them for general codes \NoCaseChange{\protect\cite{cite3102}}.
Semidefinite programming (SDP) hierarchies and a quantum Delsarte bound have been developed for qubit codes, with rational infeasibility certificates later yielding rigorous non-existence proofs and improved upper bounds for small code parameters \NoCaseChange{\protect\cite{cite4147,cite4148}}.

\codefieldsection{Rate}
Exact two-way assisted capacities have been obtained for the erasure and dephasing channels \NoCaseChange{\protect\cite{cite4149}}. There are many bounds on the quantum capacity of the depolarizing channel (e.g., \NoCaseChange{\protect\cite{cite4150}}); see review \NoCaseChange{\protect\cite{cite4151}}. The optimal asymptotic error exponent of entanglement distillation is given by the reverse relative entropy of entanglement, a single-letter quantity \NoCaseChange{\protect\cite{cite4152}}.
\codefieldsection{Transversal and Permutation-Based Gates}
\begin{eczvaluelist}
\item\relax A qubit code is \(U\)-\textit{quasi-transversal} if it can realize the logical gate \(U\) in the third level of the \flmTerm{term}{ref694}{}{Clifford hierarchy} using the physical gate \(C T^{\otimes n}\), where \(C\) is some Clifford gate \NoCaseChange{\protect\cite[{Def. 4}]{cite754}}.
\item\relax If a qubit code \(Q\) of length \(n\) has compact subgroups \(N\triangleleft G\leq \mathrm{Aut}(Q)\) such that \(G/N\) is finite, non-Abelian, simple, and not \(A_5\), then \(n\) is at least the minimal permutation degree \(\mu(G/N)\) \NoCaseChange{\protect\cite[{Thm. 1}]{cite723}}.
\end{eczvaluelist}
\codefieldsection{Gates}
\begin{eczvaluelist}
\item\relax Computing with \flmRefsHyperref{ref409}{Clifford gates}, Pauli measurements, and classical feedforward acting on stabilizer states only can be efficiently simulated on a classical computer by tracking stabilizer and logical generators, according to the \textit{Gottesman-Knill theorem} \NoCaseChange{\protect\cite{cite2115,cite2116}}.
There is a canonical form for \flmRefsHyperref{ref409}{Clifford circuits} \NoCaseChange{\protect\cite{cite2119,cite2120}} and many algorithms for simulating them \NoCaseChange{\protect\cite{cite2121,cite2122,cite2123}}.
Universal quantum computing can be achieved using \flmRefsHyperref{ref409}{Clifford gates} and a single type of \flmRefsHyperref{ref409}{non-Clifford} gate, such as the \(T\) gate \NoCaseChange{\protect\cite{cite2124}}.
More generally, the \textit{Solovay-Kitaev} theorem \NoCaseChange{\protect\cite{cite2125,cite1634}} states that any subset of gates that generates a dense subgroup of the full \(n\)-qubit gate group can be used to construct any gate to arbitrary accuracy (see \NoCaseChange{\protect\cite{cite2126}\protect\cite[{Appx. 3}]{cite2127}}). The task of approximating a desired gate by \flmRefsHyperref{ref409}{Clifford gates} and a fixed set of \flmRefsHyperref{ref409}{non-Clifford} gates is called \textit{gate compilation} or \textit{circuit synthesis}.

\item\relax Non-Clifford gates are typically more difficult to implement than \flmRefsHyperref{ref409}{Clifford gates} and so are treated as a resource. Gate errors in circuit synthesis can sometimes add up destructively \NoCaseChange{\protect\cite{cite4153}}. There is a threshold against depolarizing noise for any single-qubit gate that determines if the gate enables universal quantum computation \NoCaseChange{\protect\cite{cite4154,cite4155}}.
\item\relax The most studied set of universal gates is generated by the Clifford+\(T\) gate set. Exactly optimizing \(T\)-gate count in circuit synthesis is \(NP\)-hard \NoCaseChange{\protect\cite{cite4156,cite4157}}. Gate compilation can be done using various heuristic procedures \NoCaseChange{\protect\cite{cite4158,cite4159,cite1578,cite4160,cite4161,cite4162,cite4163,cite4164,cite4165,cite4166}}, e.g., \textit{ZX calculus} (a.k.a. Penrose spin calculus) \NoCaseChange{\protect\cite{cite4167,cite4168,cite4169,cite4170}}, reinforcement learning \NoCaseChange{\protect\cite{cite4171,cite4172,cite4173,cite4174,cite4175,cite4176}}, genetic algorithms \NoCaseChange{\protect\cite{cite4177}}, or Hermitian lattices \NoCaseChange{\protect\cite{cite4178}}.
There is an optimal asymptotic scaling of the number of T gates needed to prepare an arbitrary state \NoCaseChange{\protect\cite{cite4179,cite4180}}.

\item\relax Other gate sets for generating universal gates are Clifford + \(\sqrt{T}\) \NoCaseChange{\protect\cite{cite4181}}, Toffoli and Hadamard \NoCaseChange{\protect\cite{cite4182,cite4183}}, cosine-sine \NoCaseChange{\protect\cite{cite4184}}, and icosahedral super-golden gates \NoCaseChange{\protect\cite{cite4185,cite4186,cite4187,cite4188}}. The \(n\)-qubit Toffoli gates can be exactly realized using at least \(nT\) gates \NoCaseChange{\protect\cite{cite4189}}, but this can be relaxed at the expense of some errors \NoCaseChange{\protect\cite{cite4190}}.
\item\relax \begin{defterm}{Clifford hierarchy}\label{ref694}\label{ref2118} The Clifford hierarchy \NoCaseChange{\protect\cite{cite3219,cite800,cite4191,cite4192,cite4193}} is a tower of gate sets which includes Pauli and \flmRefsHyperref{ref409}{Clifford gates} at its first two levels, and \flmRefsHyperref{ref409}{non-Clifford} gates at higher levels. The \(k\)th level is defined recursively by \flmMathEnvironment{align}{}{ C_k = \{ U | U P U^{\dagger} \in C_{k-1} \}~, } where \(P\) is any Pauli matrix, where \(C_1\) is the \flmRefsHyperref{ref663}{Pauli group}, and where \(C_2\) is the \flmRefsHyperref{ref409}{Clifford group}. Gates for one qubit have been classified \NoCaseChange{\protect\cite{cite4194}}. \end{defterm}
\item\relax Arbitrary \(n\)-qubit circuits can be implemented fault-tolerantly in a 3D architecture using \(O(n^{3/2}\log^3 n)\) qubits, and in a 2D architecture using only \(O(n^2 \log^3 n)\) qubits \NoCaseChange{\protect\cite{cite4195}}.
\item\relax Fault-tolerant gates can be done for any code supporting a transversal implementation of Pauli gates using generalized gate teleportation \NoCaseChange{\protect\cite{cite4196}}.
\end{eczvaluelist}
\codefieldsection{Decoding}
\begin{eczvaluelist}
\item\relax Syndrome measurements are assumed to be perfect in the \textit{code-capacity model}. Incorporating faulty syndrome measurements can be done using the \textit{phenomenological noise model}, which simulates errors during syndrome extraction by flipping some of the bits of the measured syndrome bitstring. In the more involved \textit{circuit-level noise model}, every component of the syndrome extraction circuit can be faulty.
\item\relax The decoder determining the most likely error given a noise channel is called the \textit{maximum probability error} (MPE) decoder. For few-qubit codes (\(n\) is small), MPE decoding can be based on creating a lookup table. For infinite code families, the size of such a table scales exponentially with \(n\), so approximate decoding algorithms scaling polynomially with \(n\) have to be used.
\item\relax \begin{defterm}{Effective distance and hook errors}\label{ref4197}\label{ref3496} Decoders are characterized by an effective distance (a.k.a. \textit{circuit-level distance} or fault distance), the minimum number of faulty operations during syndrome measurement that is required to make an undetectable error. A code is \textit{distance-preserving} if it admits a decoder whose circuit-level distance is equal to the code distance. A particularly dangerous class of syndrome measurement circuit faults are \textit{hook errors}, which are ancilla faults that cause more than one data-qubit error \NoCaseChange{\protect\cite{cite480}}. Hook errors occur at specific places in a syndrome extraction circuit and can sometimes be removed by re-ordering the gates of the circuit. If not, the use of \textit{flag qubits} (see \NoCaseChange{\protect\cite{cite398}}) to detect hook errors may be necessary to yield fault-tolerant decoders.  \end{defterm}
\end{eczvaluelist}
\codefieldsection{Fault Tolerance}
\begin{eczvaluelist}
\item\relax There are lower bounds on the overhead of fault-tolerant QEC in terms of the capacity of the noise channel \NoCaseChange{\protect\cite{cite4198}}. A more stringent bound applies to geometrically local QEC due to the fact that locality constrains the growth of the entanglement that is needed for protection \NoCaseChange{\protect\cite{cite522}}.
\item\relax Arbitrary \(n\)-qubit circuits can be implemented fault-tolerantly in a 3D architecture using \(O(n^{3/2}\log^3 n)\) qubits, and in a 2D architecture using only \(O(n^2 \log^3 n)\) qubits \NoCaseChange{\protect\cite{cite4195}}.
\item\relax Fault-tolerant gates can be done for any code supporting a transversal implementation of Pauli gates using generalized gate teleportation \NoCaseChange{\protect\cite{cite4196}}.
\end{eczvaluelist}
\codefieldsection{Threshold}
\begin{eczvaluelist}
\item\relax \begin{defterm}{Computational threshold}\label{ref4199}\label{ref515}
A fault-tolerant computational threshold is the maximum noise rate in a particular single-parameter noise model below which any logical computation of size \(M\) can be executed on a physical-qubit architecture to arbitrary accuracy and with an overhead of \flmRefsHyperref{ref65}{order} \(O(M\text{polylog}M)\).
The first methods to achieve a computational threshold use recursively concatenated stabilizer code families \NoCaseChange{\protect\cite{cite516,cite3364,cite3603,cite826,cite519,cite3604,cite3605,cite520}}; such a threshold is called a \textit{concatenated threshold}.
Initially proven under local stochastic noise, the concatenated threshold theorem also holds for various types of non-Markovian noise \NoCaseChange{\protect\cite{cite4200,cite3604,cite3605,cite4201}} and leakage errors \NoCaseChange{\protect\cite{cite4202}}.
This theorem can be rephrased in terms of Bernoulli site percolation \NoCaseChange{\protect\cite{cite4203}}.
The resulting concatenated code is highly \flmRefsHyperref{ref811}{degenerate}, with all but an exponentially small fraction of generators having small weights. 
Circuit and measurement designs have to take care of the few stabilizer generators with large weights in order to be fault tolerant, but measurement duration may not pose a threat to scalability \NoCaseChange{\protect\cite{cite3785}}.
While generic concatenated methods yield a computational threshold with overhead \(O(M\text{polylog}M)\), concatenations using quantum Hamming codes can additionally attain constant space overhead with quasi-polylogarithmic time overhead \NoCaseChange{\protect\cite{cite3214,cite3216}}, and concatenations of the Steane code and certain QLDPC codes further improve this time overhead to polylogarithmic while keeping constant space overhead \NoCaseChange{\protect\cite{cite3606}}.
Subsequently, thresholds were determined for infinite families of lattice stabilizer codes, starting with the toric code \NoCaseChange{\protect\cite{cite480}}; such a threshold is colloquially called a \textit{topological threshold}.
When different classes of circuit locations have different error rates, the single-number threshold generalizes to a \textit{threshold surface} in the space of error-rate vectors. A one-level crossing where a particular logical location becomes more reliable is then not a full-fledged threshold for the entire circuit; deciding whether a protocol really improves may require following the coupled recurrence relations through additional concatenation levels \NoCaseChange{\protect\cite[{Sec. 14.7.6}]{cite398}}.
Fault-tolerant computations with no notion of locality can be made local on a 2D or 3D geometry with minimal overhead \NoCaseChange{\protect\cite{cite4195}}.
\end{defterm}

\item\relax There is an upper bound on the threshold under local update recovery that is derived via quantum optimal transport \NoCaseChange{\protect\cite{cite4204}} (see also Ref. \NoCaseChange{\protect\cite{cite4205}}).
\item\relax There is a threshold against depolarizing noise for any single-qubit gate that determines if the gate enables universal quantum computation \NoCaseChange{\protect\cite{cite4154,cite4155}}.
\item\relax \begin{defterm}{Measurement threshold}\label{ref4206}\label{ref3210} One can derive conditions quantifying how many random single-qubit measurements can be made without destroying the logical information \NoCaseChange{\protect\cite{cite3211}}. The measurement threshold is the maximum total probability that a single qubit is measured in a random \(X\), \(Y\), or \(Z\) basis at which the logical information is still recoverable. The measurement threshold is at least as large as the erasure threshold \NoCaseChange{\protect\cite[{Thm. 4}]{cite3211}}. \end{defterm}
\item\relax There is a dynamical phase transition between a bounded-error and an unbounded-error phase for a model of qubits weakly coupled to a refrigerator \NoCaseChange{\protect\cite{cite4207}}.
\end{eczvaluelist}
\codefieldsection{Notes}
\begin{eczvaluelist}
\item\relax There is a relation between one-way entanglement distillation protocols and QECCs \NoCaseChange{\protect\cite{cite2763}}.
\item\relax Qubit error correction is required for unconditionally secure quantum key distribution \NoCaseChange{\protect\cite{cite4208}}.
\item\relax See \flmHref{https://github.com/qiskit-community/qiskit-qec}{Qiskit QEC framework} for realizing protocols on IBM machines.
\item\relax Any logical state \(\psi\) of an \(\llparenthesis n,2^k,d\rrparenthesis \) qubit code obeys \(E_h(\psi) \geq \left(d/2^h-1\right)H^{-1}(k/n)\), giving a distance- and rate-dependent lower bound on geometric entanglement \NoCaseChange{\protect\cite[{Thm. 3(i)}]{cite529}}.
\item\relax Expanding any logical state of a distance-\(d\) qubit code in any computational basis requires at least \(2^{d-1}\) basis states \NoCaseChange{\protect\cite[{Thm. 4}]{cite529}}.
\end{eczvaluelist}
\codefieldsection{Parents}
\begin{eczvaluelist}
\item\relax
\flmRefsHyperref[eczindexfamilyrel]{code:oa_qubits_into_qubits}{OA qubit code} --- An OA qubit code which has no gauge qubits and no block structure is a qubit code.
\item\relax
\flmRefsHyperref[eczindexfamilyrel]{code:qudits_into_qudits}{Modular-qudit code} --- Modular-qudit quantum codes for \(q=2\) correspond to qubit codes. Modular-qudit codes \NoCaseChange{\protect\cite{cite4209}}, circuits \NoCaseChange{\protect\cite{cite4210}}, and magic-state distillation schemes \NoCaseChange{\protect\cite{cite692,cite752}} can have advantages over their qubit counterparts. Modular qudits are useful for simulating gauge theories \NoCaseChange{\protect\cite{cite4211,cite4212}}. There are several ways to embed one or more qubits into a single modular qudit, yielding efficient qubit gate decompositions \NoCaseChange{\protect\cite{cite4213}}.
\item\relax
\flmRefsHyperref[eczindexfamilyrel]{code:galois_into_galois}{Galois-qudit code} --- Galois-qudit quantum codes for \(q=2\) correspond to qubit codes.
\item\relax
\flmRefsHyperref[eczindexfamilyrel]{code:spins_into_spins}{Spin code} --- Spin codes with spin \(\ell=1/2\) correspond to qubit codes since the single-qubit Pauli matrices generate the Lie algebra of \(SU(2)\).
\end{eczvaluelist}
\codefieldsection{Children}
\begin{eczvaluelist}
\item\relax
\flmRefsHyperref[eczindexfamilyrel]{code:da}{Dynamical code}\item\relax
\flmRefsHyperref[eczindexfamilyrel]{code:haar_random}{Haar-random qubit code}\item\relax
\flmRefsHyperref[eczindexfamilyrel]{code:local_haar_random}{Local Haar-random circuit qubit code}\item\relax
\flmRefsHyperref[eczindexfamilyrel]{code:cft}{Conformal-field theory (CFT) code}\item\relax
\flmRefsHyperref[eczindexfamilyrel]{code:kpt}{Kim-Preskill-Tang (KPT) code}\item\relax
\flmRefsHyperref[eczindexfamilyrel]{code:fermions}{Fermion code} --- The Majorana operator algebra is isomorphic to the qubit Pauli-operator algebra via various fermion-into-qubit encodings.
\item\relax
\flmRefsHyperref[eczindexfamilyrel]{code:nonabelian_kitaev_honeycomb}{Non-Abelian Kitaev honeycomb code} --- The Kitaev honeycomb model with a magnetic field is a qubit many-body system in the Ising-anyon phase, and the underlying code stores information in the fusion space of its non-Abelian anyonic excitations.
\item\relax
\flmRefsHyperref[eczindexfamilyrel]{code:circuit_to_hamiltonian}{Circuit-to-Hamiltonian approximate code}\item\relax
\flmRefsHyperref[eczindexfamilyrel]{code:clifford_hierarchy}{Clifford-hierarchy stabilizer code}\item\relax
\flmRefsHyperref[eczindexfamilyrel]{code:eth}{Eigenstate thermalization hypothesis (ETH) code} --- ETH codewords are eigenstates of a local Hamiltonian whose eigenstates satisfy ETH.
\item\relax
\flmRefsHyperref[eczindexfamilyrel]{code:jump}{Jump code}\item\relax
\flmRefsHyperref[eczindexfamilyrel]{code:movassagh_ouyang}{Movassagh-Ouyang Hamiltonian code}\item\relax
\flmRefsHyperref[eczindexfamilyrel]{code:non_stabilizer}{Union stabilizer (USt) code}\item\relax
\flmRefsHyperref[eczindexfamilyrel]{code:qubit_permutation_invariant}{PI qubit code}\item\relax
\flmRefsHyperref[eczindexfamilyrel]{code:qubit_concatenated}{Concatenated qubit code}\item\relax
\flmRefsHyperref[eczindexfamilyrel]{code:reinforcement_learning}{Reinforcement-learning quantum code}\item\relax
\flmRefsHyperref[eczindexfamilyrel]{code:ampdamp_numopt}{Numerically optimized four-qubit AD code}\item\relax
\flmRefsHyperref[eczindexfamilyrel]{code:qubit_6_2_3}{\(\llparenthesis 6,2,3\rrparenthesis \) transversal-\(\mathbb{Z}_{10}\) code}\item\relax
\flmRefsHyperref[eczindexfamilyrel]{code:qubit_8_1_3}{\(\llparenthesis 8,2,3\rrparenthesis \) Plenio-Vedral-Knight CE code}\item\relax
\flmRefsHyperref[eczindexfamilyrel]{code:qubit_8_4_2}{\(\llparenthesis 8,16,2\rrparenthesis \) \(PG(3,2)\) code}\item\relax
\flmRefsHyperref[eczindexfamilyrel]{code:sslp}{Subset-Sum-Linear-Programming (SS-LP) code}\end{eczvaluelist}
\codefieldsection{Cousins}
\begin{eczvaluelist}
\item\relax
\flmRefsHyperref[eczindexfamilyrel]{code:clifford_group}{Clifford group} --- Computing with \flmRefsHyperref{ref409}{Clifford gates}, Pauli measurements, and classical feedforward acting on stabilizer states only can be efficiently simulated on a classical computer by tracking stabilizer and logical generators, according to the \textit{Gottesman-Knill theorem} \NoCaseChange{\protect\cite{cite2115,cite2116}}.
There is a canonical form for \flmRefsHyperref{ref409}{Clifford circuits} \NoCaseChange{\protect\cite{cite2119,cite2120}} and many algorithms for simulating them \NoCaseChange{\protect\cite{cite2121,cite2122,cite2123}}.
Universal quantum computing can be achieved using \flmRefsHyperref{ref409}{Clifford gates} and a single type of \flmRefsHyperref{ref409}{non-Clifford} gate, such as the \(T\) gate \NoCaseChange{\protect\cite{cite2124}}.
More generally, the \textit{Solovay-Kitaev} theorem \NoCaseChange{\protect\cite{cite2125,cite1634}} states that any subset of gates that generates a dense subgroup of the full \(n\)-qubit gate group can be used to construct any gate to arbitrary accuracy (see \NoCaseChange{\protect\cite{cite2126}\protect\cite[{Appx. 3}]{cite2127}}). The task of approximating a desired gate by \flmRefsHyperref{ref409}{Clifford gates} and a fixed set of \flmRefsHyperref{ref409}{non-Clifford} gates is called \textit{gate compilation} or \textit{circuit synthesis}.

\item\relax
\flmRefsHyperref[eczindexfamilyrel]{code:qubit_classical_into_quantum}{Qubit c-q code} --- Qubit c-q codes are qubit codes designed to transmit classical information.
\item\relax
\flmRefsHyperref[eczindexfamilyrel]{code:bits_into_bits}{Binary code} --- Qubit codes are quantum counterparts of binary codes.
\item\relax
\flmRefsHyperref[eczindexfamilyrel]{code:gray}{Gray code} --- Gray codes are useful for optimizing qubit unitary circuits \NoCaseChange{\protect\cite{cite1381}} and for encoding qudits in multiple qubits \NoCaseChange{\protect\cite{cite506}}.
\item\relax
\flmRefsHyperref[eczindexfamilyrel]{code:reed_muller}{Reed-Muller (RM) code} --- Optimizing T gates in a qubit circuit that uses CNOT and T gates is equivalent to decoding a particular RM code \NoCaseChange{\protect\cite{cite1578}}.
\item\relax
\flmRefsHyperref[eczindexfamilyrel]{code:ea_qubits_into_qubits}{EA qubit code} --- EA qubit codes utilize additional ancillary qubits in a pre-shared entangled state, but reduce to ordinary qubit codes when said qubits are interpreted as noiseless physical qubits.
\item\relax
\flmRefsHyperref[eczindexfamilyrel]{code:hybrid_qubits_into_qubits}{Hybrid qubit code} --- A hybrid qubit code storing no classical information reduces to a qubit code. Conversely, any qubit code can be converted into a hybrid qubit code by using some of its logical qubits to store only classical information \NoCaseChange{\protect\cite{cite2735}}.
\item\relax
\flmRefsHyperref[eczindexfamilyrel]{code:stab_5_1_3}{\(\llbracket 5,1,3\rrbracket \) Five-qubit perfect code} --- Every \(\llparenthesis 5,2,3\rrparenthesis \) qubit code is single-qubit-Clifford-equivalent equivalent to the five-qubit code \NoCaseChange{\protect\cite[{Corr. 10}]{cite446}}.
\item\relax
\flmRefsHyperref[eczindexfamilyrel]{code:subsystem_qubits_into_qubits}{Subsystem qubit code} --- Subsystem qubit codes reduce to (subspace) qubit codes when there is no gauge subsystem.
\end{eczvaluelist}
\eczhbkcontributors{ Sam Gunn, \eczhuVVA }
\endeczcode

\eczcode{qubit_css}{Qubit CSS code}{~\NoCaseChange{\protect\cite{cite3196,cite3337,cite3338}}}
\codefieldsection{Alternative Names}
\begin{eczvaluelist}
\item\relax Qubit Euclidean code
\end{eczvaluelist}
\eczhIndexCodeAliasName{qubit_css}{Qubit Euclidean code}
\codefieldsection{Description}
An \(\llbracket n,k,d\rrbracket \) stabilizer code admitting a set of stabilizer generators that are either \(Z\)-type or \(X\)-type Pauli strings.
Codes can be defined from two classical codes and/or chain complexes over \(\mathbb{Z}_2\) per the \flmRefsHyperref{ref683}{qubit CSS-to-homology correspondence} below.

The stabilizer generator matrix is of the form
\flmMathEnvironment{align}{}{
H=\begin{pmatrix}0 & H_{Z}\\
H_{X} & 0
\end{pmatrix}\label{ref4214}
}
such that the rows of the two blocks must be orthogonal
\flmMathEnvironment{align}{}{
H_X H_Z^T=0~.\label{ref4215}
}
The above condition guarantees that the \(X\)-stabilizer generators, defined in the \flmRefsHyperref{ref817}{symplectic representation} as rows of \(H_X\), commute with the \(Z\)-stabilizer generators associated with \(H_Z\).
A qubit stabilizer code is a qubit CSS code if and only if \(\text{rank}H_X + \text{rank} H_Z = n-k\) \NoCaseChange{\protect\cite[{Lemma 7.4}]{cite454}}.

Encoding is based on two related \flmRefsHyperref{code:binary_linear}{binary linear codes}, an \([n,k_X,\delta_X]\) code \(C_X\) and \([n,k_Z,\delta_Z]\) code \(C_Z\), satisfying \(C_X^\perp \subseteq C_Z\).
The resulting CSS code has \(k=k_X+k_Z-n\) logical qubits.
The \(H_X\) (\(H_Z\)) block of \(H\) \eqref{ref4214} is the parity-check matrix of the code \(C_Z\) (\(C_X\)).
The requirement \(C_X^\perp \subseteq C_Z\) guarantees \eqref{ref4215} and also implies  \(C_Z^\perp \subseteq C_X \).
Basis states for the code are, for coset representatives \(\gamma \in C_X/C_Z^\perp\),
\flmMathEnvironment{align}{}{
|\gamma + C_Z^\perp \rangle = \frac{1}{\sqrt{|C_Z^\perp|}} \sum_{\eta \in C_Z^\perp} |\gamma + \eta\rangle.
}
After a Hadamard transform on every qubit, the same code can be described using superpositions over cosets of \(C_X^\perp\) in \(C_Z\), exchanging the roles of bit-flip and phase-flip protection.

Inequivalent CSS codes up to \(n=14\) qubits have been classified \NoCaseChange{\protect\cite{cite514}}.

\subsection{CSS-to-homology correspondence}

\begin{defterm}{Qubit CSS-to-homology correspondence}\label{ref4216}\label{ref683}
CSS codes and their properties can be formulated in terms of homology theory, yielding a powerful correspondence between codes and chain complexes, the primary homological structures.
There exists a many-to-one mapping from size three chain complexes to CSS codes \NoCaseChange{\protect\cite{cite2125,cite71,cite3727,cite4217,cite3966}} that allows one to extract code properties from topological features of the complexes.
Codes constructed in this manner are sometimes called \textit{homological CSS codes}, but they are equivalent to CSS codes.
This mapping of codes to manifolds allows the application of structures from topology to error correction, yielding \flmRefsHyperref{code:generalized_homological_product}{various QLDPC codes} with favorable properties.
\end{defterm}

A \textit{chain complex} of size three is given by binary vector spaces \(A_2\), \(A_1\), \(A_0\) and binary matrices \(\partial_{i=1,2}\) (called \textit{boundary operators}) from \(A_i\) to \(A_{i-1}\) that satisfy \(\partial_1 \partial_2 = 0\). Such a complex is typically denoted as
\flmMathEnvironment{align}{}{
A_2 \xrightarrow{\partial_2} A_1 \xrightarrow{\partial_1} A_0~.\label{ref4218}
}
One constructs a CSS code by associating a physical qubit to every basis element of \(A_1\), and defining parity-check matrices \(H_X=\partial_1\) and \(H_Z=\partial_2^T\). That way, the spaces \(A_0\) and \(A_2\) can be associated with \(X\)-type and \(Z\)-type Pauli operators, respectively, and boundary operators determine the Paulis making up the stabilizer generators. The requirement \(\partial_1 \partial_2 = 0\) guarantees that the \(X\)-stabilizer generators associated with \(H_X\) commute with the \(Z\)-stabilizer generators associated with \(H_Z\).
The number of encoded logical qubits is equal to the dimension of the first \(\mathbb{Z}_2\)-homology of the chain complex, \(H_1(\partial, \mathbb{Z}_2) = \frac{\text{Ker}(\partial_1)}{\text{Im}(\partial_2)}\).
See \NoCaseChange{\protect\cite[{Table 3.2}]{cite3951}} for a Rosetta stone comparing statistical mechanical models, CSS codes, and chain complexes.

Usually, the chain complex \eqref{ref4218} used in the construction comes from the chain complex associated with a cellulation of a manifold. When the manifold is a two-dimensional surface, its entire chain is used.
Higher-dimensional manifolds allow for longer chain complexes, and one can use the three largest non-trivial vector spaces in its chain.

CSS codes saturate a type of \textit{error correction uncertainty relation} \NoCaseChange{\protect\cite[{Thm. 3}]{cite3337}}, which is a special case of an entropic uncertainty relation between a pair of bases \NoCaseChange{\protect\cite{cite4219,cite4220,cite4221}}.
The code state \(\sum_{c\in C_{Z}}|c\rangle\) can be expressed in terms of either basis states labeled by the code \(C_{Z}\) or its dual, satisfying, with equality, the relation
\flmMathEnvironment{align}{}{
  |C_{Z}||C_{Z}^{\perp}| \geq 2^{n}\,.
}

\codefieldsection{Protection}
The quantity \(\min\{\delta_X,\delta_Z\}\) is the CSS code's \flmRefsHyperref{ref672}{pure distance} \NoCaseChange{\protect\cite{cite204}}, and it is equal to the code distance for a \flmRefsHyperref{ref811}{non-degenerate} code.
To find the code distance of a \flmRefsHyperref{ref811}{degenerate} CSS code, we have to first remove the codewords of the smaller codes as those codewords correspond to stabilizer generators instead of logical operators.
The general formulae are
\flmMathEnvironment{align}{}{
d_{X}&=\min\{ w_H(c) | c \in C_X \setminus C_Z^\perp \} \geq \delta_X \\
d_{Z}&=\min\{ w_H(c) | c \in C_Z \setminus C_X^\perp \} \geq \delta_Z \\
d&=\min\{d_X,d_Z\}~,
}
where \(w_H\) is the Hamming weight of a codeword.
In the homology correspondence, the code distance is equal to the minimum of the combinatorial (\(d-1\))-systole of the cellulated \(d\)-dimensional manifold and its dual.

A CSS code has \textit{stabilizer weight} \(w\) if the highest weight of any stabilizer generator is \(w\), i.e., any row of \(H_X\) and \(H_Z\) has weight at most \(w\).
\textit{Strong CSS codes} are codes for which there exists a set of \(X\) and \(Z\) stabilizer generators of equal weight.
In the context of comparing weight as well as of determining distances for noise models biased toward \(X\)- or \(Z\)-type errors, an extended notation for asymmetric qubit CSS codes is \(\llbracket n,k,(d_X,d_Z),w\rrbracket \) or \(\llbracket n,k,d_X/d_Z,w\rrbracket \).

\begin{defterm}{Steane enlargement}\label{ref4222}\label{ref863}
An \(\llbracket n,2k-n,d\rrbracket \) CSS code can be converted to a \(\llbracket n,k+k^{\prime}−n,\min(d,\left\lceil 3d^{\prime}/2\right\rceil )\rrbracket \) code for particular \(k^{\prime}\) and \(d^{\prime}\) via the Steane enlargement construction \NoCaseChange{\protect\cite{cite4129,cite2024}}.
\end{defterm}

\codefieldsection{Rate}
For a depolarizing channel with probability \(p\), CSS codes allowing for arbitrarily accurate recovery exist with asymptotic rate \(1-2h(p)\), where \(h\) is the binary entropy function \NoCaseChange{\protect\cite{cite480,cite4120}}.
\codefieldsection{Encoding}
\begin{eczvaluelist}
\item\relax Steane Latin-rectangle encoder \NoCaseChange{\protect\cite{cite4223,cite4224}}.
\item\relax Stabilizer measurement \NoCaseChange{\protect\cite{cite3826}}.
\item\relax Clusterization, i.e., measurement of a particular cluster state \NoCaseChange{\protect\cite{cite3530}}.
\item\relax Entanglement purification \NoCaseChange{\protect\cite{cite4225}}.
\item\relax Reinforcement-learning discovery of logical-state-preparation circuits \NoCaseChange{\protect\cite{cite3200}}.
\item\relax There is a correspondence between qubit CSS codes and phase-free ZX calculus diagrams \NoCaseChange{\protect\cite{cite4226}}. ZX calculus provides a canonical form for the encoding circuit \NoCaseChange{\protect\cite{cite4227}}.
\item\relax Automated fault-tolerant encoding circuit synthesis \NoCaseChange{\protect\cite{cite4228}}.
\end{eczvaluelist}
\codefieldsection{Transversal and Permutation-Based Gates}
\begin{eczvaluelist}
\item\relax Transversal CNOT gates preserve the logical subspace, up to \(X\)-type Paulis, iff a qubit stabilizer code is CSS \NoCaseChange{\protect\cite{cite761,cite398}}. The Paulis are necessary for when the code is stabilized by stabilizers with a minus in front of them, e.g., \(-XXXX\) and \(ZZZZ\).
\item\relax \textit{Fold-transversal} \NoCaseChange{\protect\cite{cite422,cite745,cite762}} \flmRefsHyperref{ref409}{Clifford gates} are transversal gates combined with qubit permutations. Some of these can be obtained from automorphism groups of the underlying classical codes \NoCaseChange{\protect\cite[{Thms. 2-3}]{cite763}}.
\item\relax Necessary and sufficient conditions for diagonal physical gates on a CSS code to induce logical gates in the \flmTerm{term}{ref694}{}{Clifford hierarchy} have been formulated \NoCaseChange{\protect\cite{cite766}\protect\cite[{Thm. 9}]{cite764}\protect\cite[{Thm. 5}]{cite765}}. There are routines that can determine what diagonal gates in the \flmTerm{term}{ref694}{}{Clifford hierarchy} are realized by a code \NoCaseChange{\protect\cite{cite767,cite768}}.
\item\relax CSS code families with asymptotic rate \(> 1/3\) and distance at \(\geq 3\) do not admit logical qubit permutations from physical permutations \NoCaseChange{\protect\cite{cite769}}.
\item\relax Diagonal transversal \flmRefsHyperref{ref409}{Clifford gates} on \(\ell\) codeblocks of a CSS code form \(GL(\ell,\mathbb{F}_2)\) for non-self-dual CSS codes, \(U(\ell,R_8)\) for \textit{semi-self-dual CSS codes} (i.e., CSS codes whose \(X\)-type stabilizers are contained in the \(Z\)-type stabilizers), and \(Sp(2\ell,\mathbb{F}_2)\) for self-dual CSS codes \NoCaseChange{\protect\cite{cite738}}.
\item\relax Diagonal transversal gate groups can be defined using a set of equations \NoCaseChange{\protect\cite{cite770}}.
\end{eczvaluelist}
\codefieldsection{Gates}
\begin{eczvaluelist}
\item\relax LDPC CSS code symmetries called \(XZ\)-dualities allow for fold-transversal gates, i.e., transversal gates followed by qubit permutations \NoCaseChange{\protect\cite{cite762}}.
\item\relax Generalized lattice surgery \NoCaseChange{\protect\cite{cite4229}}.
\item\relax Cohomology invariants give rise to logical gates implemented by constant-depth \flmRefsHyperref{ref409}{Clifford circuits} for codes admitting a cup product structure \NoCaseChange{\protect\cite{cite703,cite4230,cite1517,cite4231}}. For example, a diagonal \textit{copy-cup} gate in the \(m\)th level of the \flmTerm{term}{ref694}{}{Clifford hierarchy} can be implemented on a code admitting an \(m\)-fold cup product \NoCaseChange{\protect\cite{cite1517}}.
\item\relax Fault-tolerant CNOT gate using generalized lattice surgery \NoCaseChange{\protect\cite{cite4232}}.
\end{eczvaluelist}
\codefieldsection{Decoding}
\begin{eczvaluelist}
\item\relax CSS syndrome decoding splits into two classical decoding problems: measuring the \(Z\)-type stabilizers yields the classical syndrome for bit-flip errors with respect to \(C_X\), while measuring the \(X\)-type stabilizers yields the classical syndrome for phase errors with respect to \(C_Z\). In the Hadamard basis, phase-error decoding is ordinary classical syndrome decoding.
\item\relax Coherent decoders allow for measurement-free error correction \NoCaseChange{\protect\cite{cite4233}}. One method is table/multi-control decoding \NoCaseChange{\protect\cite{cite4234}}, which scales exponentially with the number of ancillas used in syndrome measurement. A fault-tolerant measurement-free scheme for low-distance CSS codes is formulated in Ref. \NoCaseChange{\protect\cite{cite3350}}. Another method, the Ising-based decoder, utilizes the mapping of the effect of the noise to a statistical mechanical model \NoCaseChange{\protect\cite{cite480,cite3476}} such that the decoding problem maps to preparation of the ground state of an Ising model. See \NoCaseChange{\protect\cite[{Table 3.2}]{cite3951}} for a Rosetta stone comparing statistical mechanical models, CSS codes, and chain complexes. Models for bit- and phase-flip noise can be dual to one another \NoCaseChange{\protect\cite{cite4235}}.
\item\relax Transformer-based decoder \NoCaseChange{\protect\cite{cite4236}}.
\item\relax MaxSAT decoder \NoCaseChange{\protect\cite{cite4237}}.
\end{eczvaluelist}
\codefieldsection{Fault Tolerance}
\begin{eczvaluelist}
\item\relax Steane error correction \NoCaseChange{\protect\cite{cite4238}}, where fault-tolerance is ensured by preparing ancillary encoded states and extracting syndromes via \(CNOT\) gates.
\item\relax Encoded \(\ket{0}\) and \(\ket{+}\) ancillas for Steane error correction can be prepared by hierarchical Steane-style verification; without code-specific optimizations, a two-level procedure uses at least \((t+1)^2\) noisy ancillas for a distance-\(2t+1\) CSS code \NoCaseChange{\protect\cite[{Secs. 13.1.2-13.1.3}]{cite398}}.
\item\relax Steane's method also yields non-destructive logical Pauli measurement for CSS codes by coupling the data block transversally to encoded \(\ket{0}\) or \(\ket{+}\) ancillas and classically decoding the ancilla measurement results \NoCaseChange{\protect\cite{cite398}}.
\item\relax Transversal computational-basis measurement followed by classical decoding is a fault-tolerant gadget for logical measurement of all encoded qubits \NoCaseChange{\protect\cite[{Sec. 11.4}]{cite398}}.
\item\relax Fault-tolerant error correction and logical measurements using flag qubits for distance-three cyclic CSS codes \NoCaseChange{\protect\cite{cite4239}}. Parallel syndrome extraction for distance-three codes can be done fault-tolerantly using one flag qubit \NoCaseChange{\protect\cite{cite4240}}. \flmRefsHyperref{ref3496}{Distance-preserving} flag fault-tolerant error correction can be done using lookup tables for small codes \NoCaseChange{\protect\cite{cite4241}}.
\item\relax Homomorphic gadgets fault-tolerant measurement unify Steane and Shor error correction \NoCaseChange{\protect\cite{cite3927}}.
\item\relax A fault-tolerant error-correction protocol using \(O(d\log d)\) syndrome measurements can be applied to any CSS code with distance \(d \geq \Omega(n^{\alpha})\) for any \(\alpha > 0\) \NoCaseChange{\protect\cite{cite3043}}.
\item\relax Fault-tolerant measurement-free scheme for low-distance CSS codes \NoCaseChange{\protect\cite{cite3350}}.
\item\relax Automated fault-tolerant encoding circuit synthesis \NoCaseChange{\protect\cite{cite4228}}.
\item\relax Fault-tolerant homological measurement of logical Pauli operators \NoCaseChange{\protect\cite{cite4242}}.
\item\relax Fault-tolerant CNOT gate using generalized lattice surgery \NoCaseChange{\protect\cite{cite4232}}.
\end{eczvaluelist}
\codefieldsection{Code Capacity Threshold}
\begin{eczvaluelist}
\item\relax Bounds on code capacity thresholds for various noise models exist in terms of stabilizer generator weights \NoCaseChange{\protect\cite{cite3441,cite4243}}.
\end{eczvaluelist}
\codefieldsection{Realizations}
\begin{eczvaluelist}
\item\relax Fully homomorphic encryption \NoCaseChange{\protect\cite{cite4244}}.
\item\relax Cryptographic applications stemming from the monogamy of entanglement of CSS code and error words \NoCaseChange{\protect\cite{cite4245}}.
\end{eczvaluelist}
\codefieldsection{Notes}
\begin{eczvaluelist}
\item\relax See Refs. \NoCaseChange{\protect\cite{cite861,cite2579,cite2764,cite398,cite2024}} for simple examples of CSS codes.
\item\relax Introduction to \flmRefsCref{ref683} by \flmHref{https://www.youtube.com/watch?v=SeLpWg_8qlc}{M. Hastings}; see also Refs. \NoCaseChange{\protect\cite{cite4246,cite3951,cite3726}}.
\item\relax Entanglement purification protocols with qubit CSS codes are related to quantum key distribution (QKD) \NoCaseChange{\protect\cite{cite4247}}.
\item\relax Qubit CSS codes can be used in quantum repeaters \NoCaseChange{\protect\cite{cite4248}}.
\end{eczvaluelist}
\codefieldsection{Parents}
\begin{eczvaluelist}
\item\relax
\flmRefsHyperref[eczindexfamilyrel]{code:cpc}{Coherent-parity-check (CPC) code} --- CSS codes are a subset of CPC codes \NoCaseChange{\protect\cite{cite860}}, with the latter not requiring the two classical codes to be related.
\item\relax
\flmRefsHyperref[eczindexfamilyrel]{code:qudit_css}{Modular-qudit CSS code} --- Modular-qudit CSS codes for \(q=2\) are qubit CSS codes.
\item\relax
\flmRefsHyperref[eczindexfamilyrel]{code:galois_css}{Galois-qudit CSS code} --- Galois-qudit CSS codes for \(q=2\) are qubit CSS codes.
\end{eczvaluelist}
\codefieldsection{Children}
\begin{eczvaluelist}
\item\relax
\flmRefsHyperref[eczindexfamilyrel]{code:ampdamp_stabilizer}{\(\llbracket 2(m+1),m,2\rrbracket \) single-loss AD code}\item\relax
\flmRefsHyperref[eczindexfamilyrel]{code:goy}{\(\llbracket 6r,2r,2\rrbracket \) Ganti-Onunkwo-Young code}\item\relax
\flmRefsHyperref[eczindexfamilyrel]{code:kls}{Khesin-Lu-Shor code}\item\relax
\flmRefsHyperref[eczindexfamilyrel]{code:qmdpc}{Quantum multi-dimensional parity-check (QMDPC) code}\item\relax
\flmRefsHyperref[eczindexfamilyrel]{code:stab_10_2_3}{\(\llbracket 10,2,3\rrbracket \) binarized Galois-qudit code}\item\relax
\flmRefsHyperref[eczindexfamilyrel]{code:css_12_1_3}{\(\llbracket 12,1,3\rrbracket \) CE CSS code}\item\relax
\flmRefsHyperref[eczindexfamilyrel]{code:css_6_1_2}{\(\llbracket 6,1,2\rrbracket \) semi-self-dual CSS code}\item\relax
\flmRefsHyperref[eczindexfamilyrel]{code:stab_8_1_2}{\(\llbracket 8,1,2\rrbracket \) Shen-Wang-Cao code}\item\relax
\flmRefsHyperref[eczindexfamilyrel]{code:generalized_shor}{Generalized Shor code}\item\relax
\flmRefsHyperref[eczindexfamilyrel]{code:phantom}{Phantom code}\item\relax
\flmRefsHyperref[eczindexfamilyrel]{code:checkerboard}{Checkerboard model code}\item\relax
\flmRefsHyperref[eczindexfamilyrel]{code:fcc_fracton}{Four Color Cube (FCC) fracton model code}\item\relax
\flmRefsHyperref[eczindexfamilyrel]{code:layer}{Layer code}\item\relax
\flmRefsHyperref[eczindexfamilyrel]{code:xcube}{X-cube model code}\item\relax
\flmRefsHyperref[eczindexfamilyrel]{code:holographic_5_1_2}{Surface-code-fragment (SCF) holographic code}\item\relax
\flmRefsHyperref[eczindexfamilyrel]{code:holographic_steane}{Heptagon holographic code}\item\relax
\flmRefsHyperref[eczindexfamilyrel]{code:generalized_quantum_divisible}{Generalized quantum divisible code} --- Generalized quantum divisible codes are CSS codes. Any self-dual CSS code yields a level-three generalized quantum divisible code when level-lifted \NoCaseChange{\protect\cite[{Thm. V.6}]{cite734}}.
\item\relax
\flmRefsHyperref[eczindexfamilyrel]{code:css-t}{CSS-T code}\item\relax
\flmRefsHyperref[eczindexfamilyrel]{code:quantum_rainbow}{Quantum rainbow code}\item\relax
\flmRefsHyperref[eczindexfamilyrel]{code:yoked_surface}{Yoked surface code}\item\relax
\flmRefsHyperref[eczindexfamilyrel]{code:check_product}{Quantum check-product code}\item\relax
\flmRefsHyperref[eczindexfamilyrel]{code:classical_product}{Classical-product code}\item\relax
\flmRefsHyperref[eczindexfamilyrel]{code:pg_qldpc}{Finite-geometry (FG) qubit QLDPC code}\item\relax
\flmRefsHyperref[eczindexfamilyrel]{code:qubit_generalized_homological_product_css}{Generalized homological-product qubit CSS code}\item\relax
\flmRefsHyperref[eczindexfamilyrel]{code:quantum_tensor_product}{Quantum tensor-product code}\item\relax
\flmRefsHyperref[eczindexfamilyrel]{code:morphed_diagonal_clifford}{\(\llbracket 2^r+r-1,1,2\rrbracket \) morphed simplex code}\item\relax
\flmRefsHyperref[eczindexfamilyrel]{code:self_dual_css}{Self-dual CSS code} --- Self-dual CSS codes are qubit CSS codes whose stabilizer group is preserved by transversal Hadamard.
\end{eczvaluelist}
\codefieldsection{Cousins}
\begin{eczvaluelist}
\item\relax
\flmRefsHyperref[eczindexfamilyrel]{code:qubit_stabilizer}{Qubit stabilizer code} --- Qubit CSS codes are qubit stabilizer codes whose stabilizer groups admit a generating set of pure-\(X\) and pure-\(Z\) Pauli strings. 
Transversal CNOT gates preserve the logical subspace iff a qubit stabilizer code is CSS \NoCaseChange{\protect\cite{cite761,cite398}}.
Any \(\llbracket n,k,d\rrbracket \) stabilizer code can be mapped onto a \(\llbracket 2n,2k,\geq d\rrbracket \) \flmRefsHyperref{code:two_block_quantum}{two-block CSS code} via \flmRefsHyperref{ref436}{symplectic doubling}, which preserves geometric locality of a code up to a constant factor.
For any non-CSS qubit stabilizer code \(\mathsf{C}\), there exists a CSS code \(\mathsf{C}^{\prime}\) such that \(\mathsf{C} = DQ\mathsf{C}^{\prime}\), where \(D\) is a diagonal Clifford operator, and where \(Q\) is an element of an XP stabilizer group \NoCaseChange{\protect\cite[{Prop. B.3.1}]{cite768}}.
There is a holographic relation between qubit CSS codes describing CFTs and qubit stabilizer codes describing path integrals over certain topologies \NoCaseChange{\protect\cite{cite3612}}.

\item\relax
\flmRefsHyperref[eczindexfamilyrel]{code:movassagh_ouyang}{Movassagh-Ouyang Hamiltonian code} --- Qubit CSS codes encoding one logical qubit are a subset of Movassagh-Ouyang codes.
\item\relax
\flmRefsHyperref[eczindexfamilyrel]{code:two_block_quantum}{Two-block CSS code} --- Any \(\llbracket n,k,d\rrbracket \) stabilizer code can be mapped onto a \(\llbracket 2n,2k,\geq d\rrbracket \) \flmRefsHyperref{code:two_block_quantum}{two-block CSS code} via \flmRefsHyperref{ref436}{symplectic doubling}, which preserves geometric locality of a code up to a constant factor.
\item\relax
\flmRefsHyperref[eczindexfamilyrel]{code:binary_linear}{Linear binary code} --- The CSS construction uses two related binary linear codes, \(C_X\) and \(C_Z\).
\item\relax
\flmRefsHyperref[eczindexfamilyrel]{code:alternant}{Alternant code} --- Alternant codes used in the CSS construction yield quantum codes that asymptotically achieve the \flmRefsHyperref{ref1729}{quantum GV bound} \NoCaseChange{\protect\cite{cite1730}}.
\item\relax
\flmRefsHyperref[eczindexfamilyrel]{code:random_stabilizer}{Random stabilizer code} --- Random CSS codes asymptotically achieve linear distance with high probability, achieving the \flmRefsHyperref{ref1729}{quantum GV bound} \NoCaseChange{\protect\cite{cite3196}}.
\item\relax
\flmRefsHyperref[eczindexfamilyrel]{code:binary_quantum_goppa}{Binary quantum Goppa code} --- Quantum Goppa codes can exceed the \flmRefsHyperref{ref1729}{quantum GV bound} \NoCaseChange{\protect\cite{cite4249}}.
\item\relax
\flmRefsHyperref[eczindexfamilyrel]{code:qubit_subsystem_stabilizer}{Subsystem qubit stabilizer code} --- Qubit CSS "seed" codes can be used to produce subsystem qubit stabilizer codes \NoCaseChange{\protect\cite{cite4250}}.
\item\relax
\flmRefsHyperref[eczindexfamilyrel]{code:translationally_invariant_stabilizer}{Lattice stabilizer code} --- The \flmRefsHyperref{ref683}{mapping of qubit CSS codes to chain complexes} allows the application of structures from topology to error correction. Chain complexes describing some QLDPC codes \NoCaseChange{\protect\cite{cite484,cite485}}, and, more generally, CSS codes \NoCaseChange{\protect\cite{cite486}} can be "lifted" into higher-dimensional manifolds admitting some notion of geometric locality. Qubit CSS codes admit several dualities \NoCaseChange{\protect\cite{cite2534,cite469}}. In particular, a CSS code and two dual classical codes can be organized by the same 2-complex, and gauging \NoCaseChange{\protect\cite{cite462,cite463,cite233,cite464,cite465,cite466,cite467,cite468,cite469,cite470}} either classical code yields the same CSS code up to Hadamard \NoCaseChange{\protect\cite{cite469}}.
\item\relax
\flmRefsHyperref[eczindexfamilyrel]{code:topological_abelian}{Abelian topological code} --- The \flmRefsHyperref{ref683}{mapping of qubit CSS codes to chain complexes} allows the application of structures from topology to error correction. Chain complexes describing some QLDPC codes \NoCaseChange{\protect\cite{cite484,cite485}}, and, more generally, CSS codes \NoCaseChange{\protect\cite{cite486}} can be "lifted" into higher-dimensional manifolds admitting some notion of geometric locality. Qubit CSS codes admit several dualities \NoCaseChange{\protect\cite{cite2534,cite469}}. In particular, a CSS code and two dual classical codes can be organized by the same 2-complex, and gauging \NoCaseChange{\protect\cite{cite462,cite463,cite233,cite464,cite465,cite466,cite467,cite468,cite469,cite470}} either classical code yields the same CSS code up to Hadamard \NoCaseChange{\protect\cite{cite469}}.
\item\relax
\flmRefsHyperref[eczindexfamilyrel]{code:cft}{Conformal-field theory (CFT) code} --- There is a holographic relation between qubit CSS codes describing CFTs and qubit stabilizer codes describing path integrals over certain topologies \NoCaseChange{\protect\cite{cite3612}}.
\item\relax
\flmRefsHyperref[eczindexfamilyrel]{code:homological_classical}{Cycle code} --- Cycle codes, including the Petersen cycle and Hoffman-Singleton cycle codes, feature in magic-state distillation protocols \NoCaseChange{\protect\cite[{Appx. A.2.1}]{cite101}\protect\cite[{Sec. VII.A}]{cite705}}.
\item\relax
\flmRefsHyperref[eczindexfamilyrel]{code:coxeter}{Coxeter code} --- Coxeter codes can be used to make qubit CSS codes \NoCaseChange{\protect\cite{cite1298}}.
\item\relax
\flmRefsHyperref[eczindexfamilyrel]{code:ampdamp}{Amplitude-damping (AD) code} --- An \(\llbracket n,k,d_Z=t+1,d_X=2t+1\rrbracket \) qubit CSS code protects against \(t\) \flmRefsHyperref{ref498}{AD} errors \NoCaseChange{\protect\cite{cite3263}\protect\cite[{Sec. 8.7}]{cite736}}.
\item\relax
\flmRefsHyperref[eczindexfamilyrel]{code:cat_concatenated}{Concatenated cat code} --- Stabilizers of CSS codes concatenated with two-component cat codes in their coherent-state basis come directly from the CSS codes via the mapping \(X \to (-1)^{\hat n}\) and \(Z \to \hat a\) \NoCaseChange{\protect\cite{cite382}}. Stabilizers of CSS codes concatenated with two-component cat codes in their cat-state basis come directly from the CSS codes via the mapping \(Z \to (-1)^{\hat n}\) and \(X \to \hat a\). In both cases, one type of noise is handled actively via syndrome extraction and correction, while the other type is handled passively via stabilizing dissipation.
\item\relax
\flmRefsHyperref[eczindexfamilyrel]{code:metopt}{Error-corrected sensing code} --- Qubit CSS codes can be used for sensing whenever the HNLS condition is satisfied, with the quantum Fisher information related to the number of weight-two codewords of the dual code \NoCaseChange{\protect\cite{cite2760}}.
\item\relax
\flmRefsHyperref[eczindexfamilyrel]{code:constant_excitation}{Constant-excitation (CE) code} --- Qubit CE codes are protected from coherent noise in the form of transversal \(Z\)-rotations because such rotations act identically on all codewords \NoCaseChange{\protect\cite{cite2709,cite808}}.
In the case of qubit CSS codes, all codes oblivious to such rotations are CE codes \NoCaseChange{\protect\cite{cite2709,cite808}}.
Any \(\llbracket n,k,d\rrbracket \) CSS code can be made into an \(\llbracket mn,k,>d\rrbracket \) CE code \NoCaseChange{\protect\cite{cite2709}}.
Concatenating the dual-rail code with an inner \(\llbracket n,k,d\rrbracket \) qubit stabilizer code yields a degenerate \(\llbracket 2n,k,d\rrbracket \) constant-excitation stabilizer code that avoids coherent phase errors and is equivalent to a Pauli-rotated repetition-concatenated stabilizer code \NoCaseChange{\protect\cite{cite2711}}. CSS structure is preserved when the original code is CSS \NoCaseChange{\protect\cite{cite524}}.

\item\relax
\flmRefsHyperref[eczindexfamilyrel]{code:qltc}{Quantum locally testable code (QLTC)} --- A qubit CSS code defined by \(H_{Z}\) and \(H_{X}\) is locally testable with some soundness iff the constituent codes \(\ker H_{Z}\) and \(\ker H_{X}\) are locally testable with the same soundness \NoCaseChange{\protect\cite[{Fact 17}]{cite1104}}.
\item\relax
\flmRefsHyperref[eczindexfamilyrel]{code:eastab}{EA qubit stabilizer code} --- As opposed to CSS codes, EA qubit stabilizer codes can be constructed from any linear binary code.
\item\relax
\flmRefsHyperref[eczindexfamilyrel]{code:quantum_polar}{Quantum polar code} --- Quantum polar codes are CSS codes used in an entanglement generation scheme that generally requires entanglement assistance. They require assistance only to determine positions to store information which optimally protect against both bit and phase noise. Without this assistance, they are just CSS codes constructed out of polar codes. A variant of quantum polar codes exists that does not require entanglement assistance \NoCaseChange{\protect\cite{cite4052}}.
\item\relax
\flmRefsHyperref[eczindexfamilyrel]{code:majorana_stab}{Majorana stabilizer code} --- Every \(\llbracket n,k,d\rrbracket _f\) Majorana stabilizer code is associated with a \(\llbracket 2n,2k,d\rrbracket \) qubit CSS code whose \(X\)- and \(Z\)-check supports coincide \NoCaseChange{\protect\cite[{Lemma 2}]{cite1432}}. An odd-length self-dual CSS code can be converted into a complex-fermion code by replacing qubit \(Z\)-type and \(X\)-type operators with \(\gamma\)-type and \(\tilde{\gamma}\)-type Majorana operators, respectively \NoCaseChange{\protect\cite{cite559}}.
\item\relax
\flmRefsHyperref[eczindexfamilyrel]{code:xp_stabilizer}{XP stabilizer code} --- Each XP-regular code can be mapped to a CSS code with the same diagonal logical operators and similar non-diagonal logical operators \NoCaseChange{\protect\cite{cite798}}.
\item\relax
\flmRefsHyperref[eczindexfamilyrel]{code:non_stabilizer}{Union stabilizer (USt) code} --- An \(\llbracket n,2k-n,d\rrbracket \) CSS code can be converted to a \(\llbracket n,k+k^{\prime}−n,\min(d,\left\lceil 3d^{\prime}/2\right\rceil )\rrbracket \) code for particular \(k^{\prime}\) and \(d^{\prime}\) via \flmRefsCref{ref863}. This code can be treated as a union stabilizer code \NoCaseChange{\protect\cite{cite1369}}.
\item\relax
\flmRefsHyperref[eczindexfamilyrel]{code:stabilizer_over_gf4}{Hermitian qubit code} --- A Hermitian qubit code that can be put into CSS form via single-qubit Clifford operations remains Hermitian \NoCaseChange{\protect\cite{cite454}}.
\item\relax
\flmRefsHyperref[eczindexfamilyrel]{code:cluster_state}{Cluster-state code} --- A resource cluster state can be constructed out of any qubit CSS code via foliation. Conversely, CSS codes can be constructed out of cluster states \NoCaseChange{\protect\cite{cite3530}}. 
In the fault-complex formalism, foliation of a CSS code is expressed as a homological product of the code's chain complex with a repetition-code complex \NoCaseChange{\protect\cite{cite3176}}.

\item\relax
\flmRefsHyperref[eczindexfamilyrel]{code:quantum_bch}{Qubit BCH code} --- Some qubit BCH codes are CSS.
\item\relax
\flmRefsHyperref[eczindexfamilyrel]{code:qubit_subsystem_css}{Subsystem qubit CSS code} --- Subsystem qubit CSS codes reduce to (subspace) CSS qubit codes when there is no gauge subsystem.
\end{eczvaluelist}
\eczhbkcontributors{ Michael Gullans, Christopher A. Pattison, Balint Pato, Armin Gerami, Leonid Pryadko, Seyed Sajjad Nezhadi, \eczhuVVA }
\endeczcode

\eczcode{qldpc}{Qubit QLDPC code}{}
\codefieldsection{Alternative Names}
\begin{eczvaluelist}
\item\relax Sparse qubit stabilizer code
\end{eczvaluelist}
\eczhIndexCodeAliasName{qldpc}{Sparse qubit stabilizer code}
\codefieldsection{Description}
Member of a family of \(\llbracket n,k,d\rrbracket \) qubit stabilizer codes for which the number of sites participating in each stabilizer generator and the number of stabilizer generators that each site participates in are both bounded by a constant \(w\) as \(n\to\infty\).
The code can be denoted by \(\llbracket n,k,d,w\rrbracket \).
Sometimes, the two parameters are explicitly stated: each site of an \((l,w)\)\textit{-regular qubit QLDPC code} is acted on by \(\leq l\) generators of weight \(\leq w\).

Qubit QLDPC codes can correct many stochastic errors far beyond the distance, which may not scale as favorably.
Together with more accurate, faster, and easier-to-parallelize measurements than those of general stabilizer codes, this property makes QLDPC codes interesting in practice.

A \textit{geometrically local qubit stabilizer code} is a qubit QLDPC code where the sites involved in any syndrome value are contained in a fixed volume that does not scale with \(n\).
As opposed to general stabilizer codes, syndrome extraction of the constant-weight check operators of a QLDPC code can be done using a constant-depth circuit.

\codefieldsection{Protection}
Detects errors on \(d-1\) sites, corrects errors on \(\left\lfloor (d-1)/2 \right\rfloor\) sites.
Code distance may not be a reliable marker of code performance.

Since qubit QLDPC codes are stabilizer QLRCs whose locality \(r \leq w\), their relative distance is bounded by \NoCaseChange{\protect\cite[{Thm. 35}]{cite812}}
\flmMathEnvironment{align}{}{
  \delta = \frac{d}{n} \leq \frac{1}{2} - \Omega\left(\frac{1}{r}\right)~.
}

\codefieldsection{Rate}
Asymptotic scaling of \(k\) and \(d\) with \(n\) depends heavily on the code construction.
Bounds generalizing the \flmRefsHyperref{ref487}{BPT bound} to qubit QLDPC codes depend on the separation profile of the code's underlying connectivity graph \NoCaseChange{\protect\cite{cite4251,cite4252}}.
A constant relative minimum distance can be achieved only for graphs that contain expanders \NoCaseChange{\protect\cite{cite4251}}.
Conversely, a code with parameters \(k\) and \(d\) requires a graph with \flmRefsHyperref{ref65}{order} \(\Omega(d)\) edges of length of \flmRefsHyperref{ref65}{order} \(\Omega(d/n^{1/D})\) \NoCaseChange{\protect\cite{cite4253}}.
Random qubit QLDPC codes found by solving certain constraint satisfaction problems (CSPs) practically achieve the capacity of the erasure channel \NoCaseChange{\protect\cite{cite4254}}.

Qubit QLDPC codes cannot attain the capacity of the erasure channel \NoCaseChange{\protect\cite{cite3443}}, but this capacity can be attained by code families with weight \(w = O(\text{polylog}n)\) \NoCaseChange{\protect\cite{cite4255}}.
There are bounds on their performance against erasure noise \NoCaseChange{\protect\cite{cite3443}}.

\codefieldsection{Encoding}
\begin{eczvaluelist}
\item\relax Fault-tolerant encoders utilizing pre-shared entanglement for qubit QLDPC codes \NoCaseChange{\protect\cite{cite3646}}.
\item\relax Any logical state of an \(s\)-sparse qubit QLDPC code with \(d > s^4 2^{5h}\) has depth-\(h\) geometric entanglement \(\Omega(d)\), equivalently exponentially small overlap with any depth-\(h\) topologically trivial state, even allowing arbitrary ancillas \NoCaseChange{\protect\cite[{Thm. 1}]{cite529}}.
\item\relax For a qubit QLDPC code with parameters \((n,k,d)\), any logical state has depth-\(h\) geometric entanglement \(\Omega\!\left(n H^{-1}(k/n)/2^{4h}\right)\); in particular, constant-rate families require \(\Omega(n/2^{4h})\) entanglement \NoCaseChange{\protect\cite[{Thm. 3(ii)}]{cite529}}. Here, \(H^{-1}\) is the inverse of the binary entropy function \(H(p) = -p\log_2 p - (1-p)\log_2(1-p)\).
\item\relax Fault-tolerant state preparation can be done in an overhead that is constant with the number of qubits \(n\) \NoCaseChange{\protect\cite{cite4256}}.
\end{eczvaluelist}
\codefieldsection{Transversal and Permutation-Based Gates}
\begin{eczvaluelist}
\item\relax There are recipes to determine transversal gates for asymmetric qubit QLDPC codes \NoCaseChange{\protect\cite{cite771}}.
\end{eczvaluelist}
\codefieldsection{Gates}
\begin{eczvaluelist}
\item\relax Fault-tolerant logical measurements by gauging logical operators have worst-case qubit overhead \(O(W\log^{2}W)\) for a weight-\(W\) Pauli and improve earlier QLDPC measurement schemes \NoCaseChange{\protect\cite{cite470}}. This can be used for a generalization of lattice surgery for CSS QLDPC codes \NoCaseChange{\protect\cite{cite4257}}. There are conditions on when fault-tolerant surgery can be performed with constant-time overhead \NoCaseChange{\protect\cite{cite4258}}.
\item\relax Repetition-code adapter for logical Pauli measurements and logical CNOT gates via Dehn twists \NoCaseChange{\protect\cite{cite4259}}.
\item\relax Fault-tolerant logical measurements based on an extractor system and allowing for universal computation \NoCaseChange{\protect\cite{cite4260}}.
\item\relax Fault-tolerant batched gadgets for CSS QLDPC codes with constant spacetime overhead \NoCaseChange{\protect\cite{cite3491}}.
\end{eczvaluelist}
\codefieldsection{Decoding}
\begin{eczvaluelist}
\item\relax Iterative error estimation based on the MIN-SUM and SUM-PRODUCT algorithms \NoCaseChange{\protect\cite{cite3514}}.
\item\relax Quantum belief propagation (BP) decoder \NoCaseChange{\protect\cite{cite4261,cite4262,cite4263}} is a quantum version of the classical BP decoder, but performance suffers due to degeneracy \NoCaseChange{\protect\cite{cite4264}}. Various post-processing algorithms have been proposed (see below and also Refs. \NoCaseChange{\protect\cite{cite4265,cite4266}}).
\item\relax BP-OSD decoder, scaling as \(O(n^3)\), adds a post-processing step based on ordered statistics decoding (OSD) to the belief propagation (BP) decoder \NoCaseChange{\protect\cite{cite1247}}.
\item\relax Neural network BP decoders \NoCaseChange{\protect\cite{cite4267,cite4268}} and GNN decoders \NoCaseChange{\protect\cite{cite4269,cite4270}} for qubit codes.
\item\relax Partially and fully decoupled BP decoders, which use the decoupling representation, yield improvements against depolarizing noise \NoCaseChange{\protect\cite{cite4271}}.
\item\relax Message-passing decoder utilizing stabilizer inactivation (MP-SI a.k.a. BP-SI) for CSS-type QLDPC qubit codes \NoCaseChange{\protect\cite{cite4272}}.
\item\relax BP localized statistics decoding (BP-LSD) that exploits error clustering \NoCaseChange{\protect\cite{cite4273}}.
\item\relax Syndrome-based linear programming (SB-LP) algorithm can be applied as a post-processing step after syndrome-based min-sum (SM-MS) decoding \NoCaseChange{\protect\cite{cite4274}}.
\item\relax BP guided decimation (BPGD) decoder \NoCaseChange{\protect\cite{cite4275}}.
\item\relax SymBreak decoder, which adaptively modifies the decoding graph to break the degeneracy of the BP decoder \NoCaseChange{\protect\cite{cite4276}}.
\item\relax Ambiguity clustering (AC) decoder, in which measurement data is divided into clusters and decoded independently \NoCaseChange{\protect\cite{cite3191}}.
\item\relax 2D geometrically local syndrome extraction circuits with bounded depth using \flmRefsHyperref{ref65}{order} \(O(n^2)\) ancilla qubits \NoCaseChange{\protect\cite{cite521}}. For CSS codes, syndrome extraction can be implemented in constant depth \NoCaseChange{\protect\cite{cite4277}}.
\item\relax Soft (i.e., analog) syndrome iterative BP for CSS-type QLDPC codes, utilizing the continuous signal obtained in the physical implementation of the stabilizer measurement (as opposed to discretizing the signal into a syndrome bit) \NoCaseChange{\protect\cite{cite4278}}.
\item\relax The MWPM decoder for surface codes may be generalizable to QLDPC codes \NoCaseChange{\protect\cite{cite4279}}.
\item\relax Extensions of the union-find decoder for qubit QLDPC codes \NoCaseChange{\protect\cite{cite4280,cite4281,cite4282}}.
\item\relax Sliding-window decoding \NoCaseChange{\protect\cite{cite4283}}.
\item\relax Closed-branch decoder \NoCaseChange{\protect\cite{cite4284}}.
\item\relax BP with guided decimation guessing (GDG) sliding-window decoder for CSS qubit codes \NoCaseChange{\protect\cite{cite3190}}.
\item\relax Performing \(d\) syndrome extraction rounds obtains an \flmRefsHyperref{ref3496}{effective distance} of \(d\) for a qubit QLDPC code \NoCaseChange{\protect\cite{cite4285}}.
\item\relax BP plus ordered Tanner forest (BP+OTF) almost-linear time decoder \NoCaseChange{\protect\cite{cite4286}}.
\item\relax Cluster decoder \NoCaseChange{\protect\cite{cite3780}}.
\item\relax BP approximate degenerate OSD (BP+ADOSD) decoder \NoCaseChange{\protect\cite{cite4287}}.
\item\relax Decision tree decoders (DTDs), one that provably finds the minimum-weight correction, and one that is heuristic \NoCaseChange{\protect\cite{cite3742}}.
\item\relax AutDEC decoder for codes with large automorphism groups \NoCaseChange{\protect\cite{cite3202}}.
\item\relax Tesseract ML decoder \NoCaseChange{\protect\cite{cite4288}}.
\item\relax Relay-BP decoder \NoCaseChange{\protect\cite{cite4289}}.
\item\relax HyperBlossom \NoCaseChange{\protect\cite{cite4290}}.
\item\relax Post-selection strategies for clustering based decoders \NoCaseChange{\protect\cite{cite4291}}.
\item\relax Decoder switching between soft-output and higher-accuracy decoders \NoCaseChange{\protect\cite{cite4292}}.
\item\relax Graph augmentation and rewiring for interference (GARI) framework for circuit-level noise \NoCaseChange{\protect\cite{cite4293}}.
\end{eczvaluelist}
\codefieldsection{Fault Tolerance}
\begin{eczvaluelist}
\item\relax Lattice surgery techniques with ancilla qubits \NoCaseChange{\protect\cite{cite3499,cite848,cite3189}}. In one such technique, one first performs a logical measurement by \flmRefsHyperref{ref410}{code switching} into a code whose stabilizer group includes the original stabilizers together with the logical Paulis that are to be measured. Then, one can reduce the weight of the output code using \flmRefsHyperref{ref491}{weight reduction}.
\item\relax Fault-tolerance with constant overhead can be performed on certain qubit QLDPC codes \NoCaseChange{\protect\cite{cite4285}}, e.g., quantum expander codes \NoCaseChange{\protect\cite{cite847}}.
\item\relax Error-corrected GHZ state distillation for Steane error correction \NoCaseChange{\protect\cite{cite4070}}.
\item\relax Fault-tolerant logical measurements by gauging logical operators have worst-case qubit overhead \(O(W\log^{2}W)\) for a weight-\(W\) Pauli and improve earlier QLDPC measurement schemes \NoCaseChange{\protect\cite{cite470}}. This can be used for a generalization of lattice surgery for CSS QLDPC codes \NoCaseChange{\protect\cite{cite4257}}. There are conditions on when fault-tolerant surgery can be performed with constant-time overhead \NoCaseChange{\protect\cite{cite4258}}.
\item\relax Fault-tolerant logical measurements based on an extractor system and allowing for universal computation \NoCaseChange{\protect\cite{cite4260}}.
\item\relax Fault-tolerant batched gadgets for CSS QLDPC codes with constant spacetime overhead \NoCaseChange{\protect\cite{cite3491}}.
\item\relax High-rate surgery, which yields parallelizable logical Pauli-product measurements \NoCaseChange{\protect\cite{cite4294}}.
\item\relax Fault-tolerant state preparation can be done in an overhead that is constant with the number of qubits \(n\) \NoCaseChange{\protect\cite{cite4256}}.
\end{eczvaluelist}
\codefieldsection{Code Capacity Threshold}
\begin{eczvaluelist}
\item\relax Bounds on code capacity thresholds using ML decoding can be obtained by mapping the effect of noise on the code to a statistical mechanical model \NoCaseChange{\protect\cite{cite480,cite3441,cite4295}}. In particular, any family of qubit QLDPC codes with superlogarithmic distance achieves a threshold \NoCaseChange{\protect\cite{cite3441}}.
\item\relax Bounds on code capacity thresholds for various noise models exist in terms of stabilizer generator weights \NoCaseChange{\protect\cite{cite4243}}.
\end{eczvaluelist}
\codefieldsection{Threshold}
\begin{eczvaluelist}
\item\relax Qubit QLDPC codes with a constant encoding rate can reduce the overhead of fault-tolerant quantum computation to be constant \NoCaseChange{\protect\cite{cite4285}}.
\end{eczvaluelist}
\codefieldsection{Notes}
\begin{eczvaluelist}
\item\relax Links to code tables of notable QLDPC codes \NoCaseChange{\protect\cite{cite3442}}.
\item\relax Collection of QLDPC qubit codes based on hyperbolic tilings in the QEC-Pages software library \NoCaseChange{\protect\cite{cite4296}}.
\item\relax High-rate QLDPC codes can be used for Bell-pair distillation \NoCaseChange{\protect\cite{cite4297}}.
\item\relax Qldpc code circUIT Simulator (QUITS) Python software library for simulating QLDPC code circuits \NoCaseChange{\protect\cite{cite4298,cite4299}}.
\item\relax See \NoCaseChange{\protect\cite{cite2733}} for a pedagogical introduction to QLDPC codes.
\end{eczvaluelist}
\codefieldsection{Parents}
\begin{eczvaluelist}
\item\relax
\flmRefsHyperref[eczindexfamilyrel]{code:qubit_stabilizer}{Qubit stabilizer code}\item\relax
\flmRefsHyperref[eczindexfamilyrel]{code:quantum_locally_recoverable}{Quantum locally recoverable code (QLRC)} --- Qubit QLDPC codes are stabilizer QLRCs whose locality \(r \leq w\), where \(w\) is the maximum stabilizer-generator weight \NoCaseChange{\protect\cite{cite812}}.
\item\relax
\flmRefsHyperref[eczindexfamilyrel]{code:general_qldpc}{QLDPC code}\end{eczvaluelist}
\codefieldsection{Children}
\begin{eczvaluelist}
\item\relax
\flmRefsHyperref[eczindexfamilyrel]{code:crystalline_dynamic_gen}{Crystalline-circuit qubit code}\item\relax
\flmRefsHyperref[eczindexfamilyrel]{code:spacetime_circuit}{Spacetime circuit code} --- Spacetime circuit codes are useful for constructing fault-tolerant syndrome extraction circuits for qubit QLDPC codes. General spacetime circuit codes can be sparsified to yield QLDPC spacetime circuit codes \NoCaseChange{\protect\cite{cite667}}.
\item\relax
\flmRefsHyperref[eczindexfamilyrel]{code:majorana_color}{Majorana color code} --- The Majorana color code is a 2D qubit stabilizer code with respect to the Majorana operator basis.
\item\relax
\flmRefsHyperref[eczindexfamilyrel]{code:majorana_surface}{Majorana surface code} --- The Majorana surface code is a 2D qubit stabilizer code with respect to the Majorana operator basis.
\item\relax
\flmRefsHyperref[eczindexfamilyrel]{code:majorana_checkerboard}{Majorana checkerboard code} --- The Majorana checkerboard code is a 3D qubit stabilizer code with respect to the Majorana operator basis.
\item\relax
\flmRefsHyperref[eczindexfamilyrel]{code:mbq}{Majorana box qubit} --- The Majorana box qubit is a 1D qubit stabilizer code with respect to the Majorana operator basis.
\item\relax
\flmRefsHyperref[eczindexfamilyrel]{code:quantum_repetition}{Quantum repetition code} --- The codespace of the quantum repetition code is the ground-state space of a frustration-free 1D classical Ising model with nearest-neighbor interactions.
\item\relax
\flmRefsHyperref[eczindexfamilyrel]{code:tfim}{Transverse-field Ising model (TFIM) code}\item\relax
\flmRefsHyperref[eczindexfamilyrel]{code:quantum_convolutional}{Quantum convolutional code}\item\relax
\flmRefsHyperref[eczindexfamilyrel]{code:bosonization}{Bosonization code} --- The \(D\)-dimensional bosonization code encodes fermionic modes into a \(D\)-dimensional qubit stabilizer code.
\item\relax
\flmRefsHyperref[eczindexfamilyrel]{code:checkerboard}{Checkerboard model code}\item\relax
\flmRefsHyperref[eczindexfamilyrel]{code:fcc_fracton}{Four Color Cube (FCC) fracton model code}\item\relax
\flmRefsHyperref[eczindexfamilyrel]{code:haah_cubic}{Haah cubic code (CC)}\item\relax
\flmRefsHyperref[eczindexfamilyrel]{code:hh_fracton}{Hsieh-Halasz (HH) code}\item\relax
\flmRefsHyperref[eczindexfamilyrel]{code:hhb_fracton}{Hsieh-Halasz-Balents (HHB) code}\item\relax
\flmRefsHyperref[eczindexfamilyrel]{code:layer}{Layer code} --- Layer codes are constructed by coupling layers of 2D surface codes according to the Tanner graph of a QLDPC code.
\item\relax
\flmRefsHyperref[eczindexfamilyrel]{code:xcube}{X-cube model code}\item\relax
\flmRefsHyperref[eczindexfamilyrel]{code:rbh}{Raussendorf-Bravyi-Harrington (RBH) cluster-state code}\item\relax
\flmRefsHyperref[eczindexfamilyrel]{code:square_lattice_cluster}{Square-lattice cluster-state code}\item\relax
\flmRefsHyperref[eczindexfamilyrel]{code:hierarchical}{Hierarchical code}\item\relax
\flmRefsHyperref[eczindexfamilyrel]{code:classical_product}{Classical-product code}\item\relax
\flmRefsHyperref[eczindexfamilyrel]{code:pg_qldpc}{Finite-geometry (FG) qubit QLDPC code}\item\relax
\flmRefsHyperref[eczindexfamilyrel]{code:qubit_generalized_homological_product_css}{Generalized homological-product qubit CSS code} --- Homological products are a primary tool for generating qubit QLDPC codes with favorable parameters. Typically, whenever the input codes are binary LDPC or qubit QLDPC, the resulting code will be qubit QLDPC with non geometrically local stabilizer generators.
\item\relax
\flmRefsHyperref[eczindexfamilyrel]{code:sc_qldpc}{Quantum spatially coupled (SC-QLDPC) code}\item\relax
\flmRefsHyperref[eczindexfamilyrel]{code:color}{Color code}\item\relax
\flmRefsHyperref[eczindexfamilyrel]{code:twist_defect_color}{Twist-defect color code}\item\relax
\flmRefsHyperref[eczindexfamilyrel]{code:matching}{Matching code}\item\relax
\flmRefsHyperref[eczindexfamilyrel]{code:3d_fermionic_surface}{3D fermionic surface code}\item\relax
\flmRefsHyperref[eczindexfamilyrel]{code:clifford-deformed_surface}{Clifford-deformed surface code (CDSC)}\item\relax
\flmRefsHyperref[eczindexfamilyrel]{code:twist_defect_surface}{Twist-defect surface code}\item\relax
\flmRefsHyperref[eczindexfamilyrel]{code:three_fermion}{Three-fermion (3F) Walker-Wang model code}\end{eczvaluelist}
\codefieldsection{Cousins}
\begin{eczvaluelist}
\item\relax
\flmRefsHyperref[eczindexfamilyrel]{code:ldpc}{Low-density parity-check (LDPC) code} --- Qubit QLDPC codes are quantum analogues of binary LDPC codes.
\item\relax
\flmRefsHyperref[eczindexfamilyrel]{code:commuting_projector}{Commuting-projector Hamiltonian code} --- Qubit QLDPC codes with check soundness, meaning that every weight-\(m\) stabilizer can be written as a product of \flmRefsHyperref{ref65}{order} \(O(m)\) stabilizer generators, are robust against few-body perturbations. This means that phases of matter can be defined from certain non-geometrically local QLDPC code Hamiltonians \NoCaseChange{\protect\cite{cite2681}}.
\item\relax
\flmRefsHyperref[eczindexfamilyrel]{code:single_shot}{Single-shot code} --- Qubit QLDPC codes satisfying linear confinement are single shot \NoCaseChange{\protect\cite{cite844}}. Any code that admits a local greedy decoder also satisfies linear confinement, and so is single shot \NoCaseChange{\protect\cite{cite848}}.
\item\relax
\flmRefsHyperref[eczindexfamilyrel]{code:ldgm}{Low-density generator-matrix (LDGM) code} --- LDGM codes can yield CSS \NoCaseChange{\protect\cite{cite1442,cite1443,cite1444,cite1445}} and non-CSS \NoCaseChange{\protect\cite{cite1446,cite1447}} qubit QLDPC codes. Some of the LDGM-based CSS codes have \(n\)-independent minimum distance and no code capacity threshold \NoCaseChange{\protect\cite[{Sec. 4.2}]{cite1448}}.
\item\relax
\flmRefsHyperref[eczindexfamilyrel]{code:random_stabilizer}{Random stabilizer code} --- Random qubit QLDPC codes found by solving certain constraint satisfaction problems (CSPs) practically achieve the capacity of the erasure channel \NoCaseChange{\protect\cite{cite4254}}.
\item\relax
\flmRefsHyperref[eczindexfamilyrel]{code:algebraic_ldpc}{Algebraic LDPC code} --- Algebraic LDPC codes made from Latin squares can be used to make qubit QLDPC codes \NoCaseChange{\protect\cite[{Ch. 15}]{cite872}}.
\item\relax
\flmRefsHyperref[eczindexfamilyrel]{code:asymmetric_qecc}{Asymmetric quantum code (AQC)} --- There are recipes to determine transversal gates for asymmetric qubit QLDPC codes \NoCaseChange{\protect\cite{cite771}}.
\item\relax
\flmRefsHyperref[eczindexfamilyrel]{code:2d_stabilizer}{2D lattice stabilizer code} --- Chain complexes describing qubit QLDPC codes can be converted to 2D lattice stabilizer codes \NoCaseChange{\protect\cite{cite489}}.
\item\relax
\flmRefsHyperref[eczindexfamilyrel]{code:translationally_invariant_stabilizer}{Lattice stabilizer code} --- Chain complexes describing some QLDPC codes \NoCaseChange{\protect\cite{cite484,cite485}}, and, more generally, CSS codes \NoCaseChange{\protect\cite{cite486}} can be 'lifted' into higher-dimensional manifolds admitting some notion of geometric locality. In addition, chain complexes describing QLDPC codes can be converted to 2D lattice stabilizer codes \NoCaseChange{\protect\cite{cite489}}.
\item\relax
\flmRefsHyperref[eczindexfamilyrel]{code:sparse_subsystem}{QLDPC subsystem code} --- Any qubit QLDPC code with stabilizer-generator weights \(w_i\) can be mapped constructively to a sparse subsystem qubit code with the same number of logical qubits and distance, using \(n=O(\sum_i w_i)\) physical qubits and constant-weight gauge generators \NoCaseChange{\protect\cite{cite668}}.
\item\relax
\flmRefsHyperref[eczindexfamilyrel]{code:honeycomb_floquet}{Honeycomb Floquet code} --- The Floquet check operators are weight-two, and each qubit participates in one check each round.
\item\relax
\flmRefsHyperref[eczindexfamilyrel]{code:da}{Dynamical code} --- Using ZX calculus, an \(\llbracket n,k,d\rrbracket \) qubit stabilizer code admitting stabilizer generators of weight no more than \(m\) can be \textit{Floquetified} into an \(\llbracket n+\lceil m/2 \rceil+\ell,k,d^{\prime}\rrbracket \) dynamical code with single- and two-qubit operations, where \(\ell \leq \log_{2} m\) and \(d^{\prime} \geq d\) \NoCaseChange{\protect\cite{cite3293}} (see also Ref. \NoCaseChange{\protect\cite{cite3292}}). 
A more general locality-preserving \textit{spacetime concatenation} procedure yields a dynamical code out of any qubit stabilizer code by structuring measurement gadgets using low-weight measurements while ensuring the preservation of logical information \NoCaseChange{\protect\cite{cite3632}}. 
In particular, spacetime concatenation reformulates the notion of a dynamical code associated with a stabilizer code in terms of code concatenation for every qubit (spatial concatenation) and measurements between these codes (temporal concatenation), leading to a temporal evolution of the stabilizer state \NoCaseChange{\protect\cite{cite3632}}. 
A matrix rank condition on the bond operators connecting the gadgets, called the Bond-Kernel-Rank Condition, and a strict locality preservation condition (SLPC), along with preservation of the operator algebra of the stabilizer code under the gadget action, preserves fault-tolerance and the spacetime distance of the code \NoCaseChange{\protect\cite{cite3632}}.

\item\relax
\flmRefsHyperref[eczindexfamilyrel]{code:ea_qldpc}{EA QLDPC code} --- EA QLDPC codes utilize additional ancillary qubits in a pre-shared entangled state, but reduce to qubit QLDPC codes when said qubits are interpreted as noiseless physical qubits.
\item\relax
\flmRefsHyperref[eczindexfamilyrel]{code:bicycle}{Bicycle code} --- Bicycle codes are the first QLDPC codes \NoCaseChange{\protect\cite{cite682}}.
\item\relax
\flmRefsHyperref[eczindexfamilyrel]{code:concatenated_steane}{Concatenated Steane code} --- The combination of the concatenated Steane code and QLDPC codes with non-vanishing rate yields fault-tolerant quantum computation with constant space and polylogarithmic time overheads, even when classical computation time is taken into account \NoCaseChange{\protect\cite{cite3606}}.
\item\relax
\flmRefsHyperref[eczindexfamilyrel]{code:higher_dimensional_toric}{\(D\)-dimensional twisted toric code} --- It is conjectured that appropriate twisted boundary conditions yield multi-dimensional toric code families with sublinear distance scaling of \(N^{1-\epsilon}\) for any \(\epsilon>0\) and logarithmic-weight stabilizer generators \NoCaseChange{\protect\cite{cite3410}}. Assuming this conjecture, Hastings' weight-reduction construction yields QLDPC families with distance \(\Theta^*(N^{1-\epsilon})\) for any \(\epsilon>0\) \NoCaseChange{\protect\cite{cite2989}}.
\end{eczvaluelist}
\eczhbkcontributors{ Austin Yubo He, Eugene Tang, Xiaozhen Fu, \eczhuVVA }
\endeczcode

\eczcode{qubit_stabilizer}{Qubit stabilizer code}{~\NoCaseChange{\protect\cite{cite3369,cite736}}}
\codefieldsection{Alternative Names}
\begin{eczvaluelist}
\item\relax Binary stabilizer code
\item\relax Pauli stabilizer code
\item\relax Symplectic code
\item\relax Additive quantum code
\item\relax Additive CWS code
\item\relax Clifford code
\end{eczvaluelist}
\eczhIndexCodeAliasName{qubit_stabilizer}{Binary stabilizer code}
\eczhIndexCodeAliasName{qubit_stabilizer}{Pauli stabilizer code}
\eczhIndexCodeAliasName{qubit_stabilizer}{Symplectic code}
\eczhIndexCodeAliasName{qubit_stabilizer}{Additive quantum code}
\eczhIndexCodeAliasName{qubit_stabilizer}{Additive CWS code}
\eczhIndexCodeAliasName{qubit_stabilizer}{Clifford code}
\codefieldsection{Description}
An \(\llparenthesis n,2^k,d\rrparenthesis \) qubit stabilizer code is denoted as \(\llbracket n,k\rrbracket \) or \(\llbracket n,k,d\rrbracket \), where \(k\) is the number of logical qubits (code dimension), and where \(d\) is the code's distance.
The logical subspace is the joint eigenspace of commuting Pauli operators forming the code's stabilizer group \(\mathsf{S}\).
Traditionally, the logical subspace is the joint \(+1\) eigenspace of a set of \(2^{n-k}\) commuting Pauli operators which do not contain \(-I\).
The distance is the minimum weight of a Pauli string that implements a nontrivial logical operation in the code.

Qubit stabilizer codes form the joint \(+1\)-eigenspace of a stabilizer group \(\mathsf{S}\), i.e., a group of commuting Paulis that does not contain \(-I\) and is generated by \(r=n-k\) generators.
The table below summarizes the relevant groups and their sizes for a qubit stabilizer code.
  \begin{flmFloat}{table}{NumCap}\flmCellsBeginCenter
\long\def\flmTempTypesetThisTable#1{%
\begin{tblr}{#1,
  hspan=minimal,
  cell{1}{1}={}{c, font={\flmCellsHeaderFont}},
  cell{1}{2}={}{c, font={\flmCellsHeaderFont}},
  cell{1}{3}={}{c, font={\flmCellsHeaderFont}},
  cell{2}{1}={}{c},
  cell{2}{2}={}{c},
  cell{2}{3}={}{c},
  cell{3}{1}={}{c},
  cell{3}{2}={}{c},
  cell{3}{3}={}{c},
  cell{4}{1}={}{c},
  cell{4}{2}={}{c},
  cell{4}{3}={}{c},
  hline{2}={1}{.4pt,solid},
  hline{2}={2}{.4pt,solid},
  hline{2}={3}{.4pt,solid}}%
\toprule
purpose & symbol & size\\

    stabilizer group & \(\mathsf{S}\) & \(2^{n-k}\)
        \\

    code-preserving Paulis & \(\mathsf{N}(\mathsf{S})\) & \(4\cdot 2^{n+k}\)
        \\

    logical Paulis & \(\mathsf{N}(\mathsf{S})/\mathsf{S}\) & \(4^{k}\)
    \\
\bottomrule
\end{tblr}%
}%
\def\flmTmpMaxW{\dimexpr 0.96\linewidth\relax}%
\setbox0=\hbox{\flmTempTypesetThisTable{colspec={ccc}}}%
\ifdim\wd0<\flmTmpMaxW\relax
  \leavevmode\box0 
\else
  \flmTempTypesetThisTable{width=\flmTmpMaxW,colspec={X[-1]X[-1]X[-1]}}
\fi
\flmCellsEndCenter \caption{Groups relevant to qubit stabilizer codes. The normalizer \(\mathsf{N}(\mathsf{S})\) (technically, the centralizer, but these are equivalent for this case) is the group formed by all elements of the \(n\)-qubit Pauli group that commute with all elements in \(\mathsf{S}\). The normalizer is defined so as to include \(i\) and its powers as elements, while the stabilizer group is not.}\label{ref4300}\end{flmFloat}

Two qubit stabilizer codes are \textit{equivalent} if the codespace of one code can be mapped into that of the other under a tensor product of elements of the \flmRefsHyperref{ref409}{single-qubit Clifford group} and a qubit permutation (see Refs. \NoCaseChange{\protect\cite{cite449,cite454}}).
Equivalence under single-qubit Clifford operations is not the same as equivalence under a tensor product of arbitrary single-qubit unitary operations \NoCaseChange{\protect\cite{cite3538}} (see also Ref. \NoCaseChange{\protect\cite{cite4301}} about this false LC-LU conjecture).
Any qubit stabilizer code is equivalent to a graph quantum code via a single-qubit \flmRefsHyperref{ref409}{Clifford circuit} \NoCaseChange{\protect\cite{cite3561}} (see also \NoCaseChange{\protect\cite{cite867}}).
At the state level, every qubit stabilizer state is equivalent to a graph state under a single-qubit \flmRefsHyperref{ref409}{Clifford circuit}, and local-Clifford equivalence between graph states is generated by local complementations of the underlying graph \NoCaseChange{\protect\cite{cite3536}}.
In the binary symplectic representation, local-Clifford equivalence of stabilizer states can be tested by solving a system of \(n^2\) linear equations in the \(4n\) entries of the local Clifford operator \NoCaseChange{\protect\cite{cite3536}} (see also \NoCaseChange{\protect\cite{cite3563,cite3564,cite3565}}).

A qubit stabilizer code is \textit{decomposable} if there exists a permutation that maps the stabilizer group into a tensor product of two stabilizer groups acting on disjoint sets of qubits.
Otherwise, the code is indecomposable.
For example, a \(2n\)-qubit code consisting of two copies of an \(n\)-qubit indecomposable code is decomposable. 

\subsection{Code representations}

Qubit stabilizer states are quadratic-phase states on affine subspaces of \(\mathbb{Z}_2^n\): every stabilizer state can be written, up to global phase, as
\flmMathEnvironment{align}{}{
|\psi\rangle = 2^{-k/2}\sum_{x\in A} i^{q(x)}(-1)^{\ell(x)} |x\rangle
}
for some affine subspace \(A\subseteq \mathbb{Z}_2^n\), linear function \(\ell\), and \(\mathbb{Z}_4\)-valued quadratic function \(q\) \NoCaseChange{\protect\cite{cite2104,cite4302,cite2106}}.
There are efficient ways to compute stabilizer inner products and other functions \NoCaseChange{\protect\cite{cite4303}}.
For any logical state of a distance-\(d\) stabilizer code, the squared overlap with any \(n\)-qubit product state is at most \(2^{1-d}\) \NoCaseChange{\protect\cite[{Thm. 2}]{cite529}}.

Instead of being represented by a basis of codewords, stabilizer codes can be concisely defined and represented by a presentation of the generators of the stabilizer group.  
A set of generators is not unique, and various stabilizer codes admit generators with certain locality properties.

Stabilizer generators can be arranged as rows of a matrix, forming a \textit{stabilizer tableau}. 
For example, a stabilizer tableau of the two-qubit Bell state \(|00\rangle + |11\rangle\) is
\flmMathEnvironment{align}{}{
\begin{array}{cc}
Z & Z\\
X & X
\end{array}~.
}
Another stabilizer tableau can be defined by letting \(X \to Y\). 

Pauli strings, and the corresponding stabilizer tableaus, can be represented in other ways.

\begin{defterm}{Symplectic representation}\label{ref4304}\label{ref817}
In the symplectic representation, the single-qubit identity, \(X\), \(Y\), and \(Z\) Pauli matrices are represented using two bits as \((0|0)\), \((1|0)\), \((1|1)\), and \((0|1)\), respectively.
In other words, the single-qubit Pauli string \(X^a Z^b\) is converted to the vector \(a|b\).
The multi-qubit version follows naturally.
\end{defterm}

Each stabilizer code can be represented by a \((n-k) \times 2n\) \textit{check matrix} (a.k.a. \textit{stabilizer generator matrix}) \(H=(A|B)\), where each row \((a|b)\) is the \flmRefsHyperref{ref817}{symplectic representation} of an element from a set of generating elements of the stabilizer group.
The check matrix can be brought into standard form (a.k.a. canonical form) via Gaussian elimination \NoCaseChange{\protect\cite{cite2579,cite2120}}.

A pair of \(n\)-qubit stabilizers with \flmRefsHyperref{ref817}{symplectic representation} \((a|b)\) and \((a^{\prime}|b^{\prime})\) commute iff their symplectic inner product is zero,
\flmMathEnvironment{align}{}{
  a \cdot b^{\prime} + a^{\prime}\cdot b = \sum_{j=1}^{n} a_j b^{\prime}_j + a^{\prime}_j b_j = 0~.
}
The row space of the check matrix forms a symplectic self-orthogonal binary linear code of length \(2n\).

Another correspondence between qubit Pauli matrices and elements of the \flmRefsHyperref{ref33}{quaternary Galois field} \(\mathbb{F}_4\) yields the one-to-one correspondence between qubit stabilizer codes and trace-Hermitian self-orthogonal additive quaternary codes.

\begin{defterm}{\(\mathbb{F}_4\) representation}\label{ref4305}\label{ref1778}
An \(n\)-qubit Pauli stabilizer can be represented as a length-\(n\) quaternary vector using the one-to-one correspondence between the four Pauli matrices \(\{I,X,Y,Z\}\) and the four elements \(\{0,1,\omega^2,\omega\}\) of the \flmRefsHyperref{ref33}{quaternary Galois field} \(\mathbb{F}_4\).
\end{defterm}

The sets of \(\mathbb{F}_4\)-represented vectors for all generators yield a trace-Hermitian self-orthogonal additive quaternary code.
In other words, an additive self-orthogonal code \(C \subseteq \mathbb{F}_4^n\) of size \(2^r\) yields an \(\llbracket n,n-r\rrbracket \) qubit stabilizer code \NoCaseChange{\protect\cite{cite3716}\protect\cite[{Thm. 2}]{cite449}}.
This classical code corresponds to the stabilizer group \(\mathsf{S}\) while its trace-Hermitian dual corresponds to the normalizer \(\mathsf{N(S)}\).
In the case of stabilizer states, the correspondence is between such states and trace-Hermitian self-dual quaternary codes; such codes, and therefore such states, have been classified up to equivalence for \(n \leq 12\) \NoCaseChange{\protect\cite{cite3718,cite1645}}.
There is a complete set of invariants characterizing stabilizer states up to equivalence \NoCaseChange{\protect\cite{cite4306,cite4307}}.

ZX calculus is complete, sound, and universal for qubit stabilizer codes \NoCaseChange{\protect\cite{cite4308}}.
Any stabilizer code can be represented by a \textit{ZX canonical form} (ZXCF) \NoCaseChange{\protect\cite{cite858}}, and there exist two other representations \NoCaseChange{\protect\cite{cite4309,cite858}} that utilize ZX calculus.
\begin{defterm}{Encoder-respecting form}\label{ref4310}\label{ref857}
In an \textit{encoder-respecting form}, each qubit stabilizer code \NoCaseChange{\protect\cite{cite858}} (see also Ref. \NoCaseChange{\protect\cite{cite4311}}) is represented by a semi-bipartite graph with \(k\) input and \(n\) output nodes in which the \(k\) input nodes are not connected to each other.
Conversion from stabilizer tableaus to graphs can be done in time of \flmRefsHyperref{ref65}{order} \(O(n^3)\), while the inverse map from graphs to stabilizer codes can be computed in \(O(n^2)\), both using ZX calculus \NoCaseChange{\protect\cite{cite858}}.
Properties of the underlying graph are related to properties of the code; for example, bipartite encoder-respecting graphs yield CSS codes, and graph degree controls bounds on code distance, stabilizer weight, and encoding-circuit depth \NoCaseChange{\protect\cite{cite858}}.
\end{defterm}

Alternative representations include the \textit{decoupling representation}, in which Pauli strings are represented as vectors over \(\mathbb{F}_2\) using three bits \NoCaseChange{\protect\cite{cite4271}}.

\codefieldsection{Protection}
Detects errors on up to \(d-1\) qubits, and corrects erasure errors on up to \(d-1\) qubits.
There are algorithms to calculate the minimum distance \NoCaseChange{\protect\cite{cite3101,cite4312,cite4313,cite4314}}.
Computing the distance exactly or approximately is generally \(NP\)-complete, and is \(NP\)-hard for \flmRefsHyperref{ref811}{non-degenerate} codes \NoCaseChange{\protect\cite{cite4315}}.
Distance approximation, minimum-weight stabilizer-generator selection, and decoding are approximately optimal strategies for various quantum lights-out (QLO) games that can be played on the codes' \flmRefsHyperref{ref857}{encoder-respecting form} \NoCaseChange{\protect\cite{cite858}}.

There is the following analogue of the \flmTerm{term}{ref1043}{}{Knill-Laflamme conditions} for qubit stabilizer codes.
Define the normalizer \(\mathsf{N(S)}\) of \(\mathsf{S}\) to be the set of all Pauli operators that commute with all \(S\in\mathsf{S}\).
A stabilizer code can correct a Pauli error set \({\mathcal{E}}\) if and only if \(E^\dagger F \notin \mathsf{N(S)}\setminus \mathsf{S}\) for all \(E,F \in {\mathcal{E}}\).

There are subtleties with defining \flmRefsHyperref{ref811}{degeneracy} for non-stabilizer qubit codes with even distance \NoCaseChange{\protect\cite{cite398}}, but they are resolved for stabilizer codes.
A stabilizer code is \flmRefsHyperref{ref811}{degenerate} with respect to \(\mathcal{E}\) if and only if \(E^\dagger F \in \mathsf{S}\) for some Pauli strings \(E,F \in \mathcal{E}\).
As a distance-\(d\) code, a stabilizer code is degenerate if it admits a non-identity stabilizer whose weight is lower than the distance \NoCaseChange{\protect\cite{cite398}}.
Since that stabilizer is in the normalizer, a stabilizer code is \flmRefsHyperref{ref811}{degenerate} if and only if it is \flmRefsHyperref{ref672}{impure}.
The \flmRefsHyperref{ref672}{pure distance} of a qubit stabilizer code is the minimum weight of a non-identity stabilizer, and a qubit stabilizer code is \flmRefsHyperref{ref672}{pure} if the weight of the lowest-weight stabilizer is the code distance. 
There is a toolkit for studying \flmRefsHyperref{ref811}{degenerate} codes for Pauli channels \NoCaseChange{\protect\cite{cite4316}}.
The \flmRefsHyperref{ref1729}{quantum GV bound} can be extended into a three-way tradeoff between distance, rate, and stabilizer weight of a qubit stabilizer code \NoCaseChange{\protect\cite{cite858}}.  

\begin{defterm}{Cleaning lemma}\label{ref4317}\label{ref3140}
If all logical operators act trivially on some subset of qubits in a stabilizer code, then any logical Pauli operator can be represented on the complementary qubit subset via a stabilizer.
More technically, given any subset \(M\) of qubits that is correctable (under erasure), any logical Pauli operator \(P\) can be \textit{cleaned off} of \(M\) using a stabilizer \(S\) such that \(PS\) is supported on \(M^{\perp}\).
More generally, for any \(M\), we have \(g(M)+g(M^{\perp}) = 2k\), where \(g(M)\) is the number of logical-\(X\) and logical-\(Z\) Pauli operators supported fully on \(M\) (up to stabilizers).
The Cleaning Lemma was originally proven \NoCaseChange{\protect\cite{cite3000}}, where an analogous result is stated for subsystem codes; see also Ref. \NoCaseChange{\protect\cite{cite4318}}.
\end{defterm}

Qubit stabilizer codes can be interleaved to protect against burst errors \NoCaseChange{\protect\cite{cite2661}}.
Entropic conditions have been formulated for random projective measurement noise \NoCaseChange{\protect\cite{cite3211}}.
An effective logical channel can be derived for qubit stabilizer codes under Pauli noise \NoCaseChange{\protect\cite{cite4319,cite4320}}.

\codefieldsection{Rate}
The \textit{ancilla-added rate} is defined as the ratio of logical qubits to the total number of physical qubits including ancillas, \(k/(n+c)\), where \(c\) is the number of ancilla qubits. The \textit{hashing bound} states that there is a qubit stabilizer code achieving a rate \(R = 1 - H(\mathbf{p})\) for a Pauli noise channel with Pauli error probabilities \(\mathbf{p}=(p_I,p_X,p_Y,p_Z)\), where \(H(\mathbf{p})\) is the entropy of the argument \NoCaseChange{\protect\cite[{Thm. 23.6.2}]{cite2779}}. Finite block length bounds and a refinement of the hashing bound have been developed \NoCaseChange{\protect\cite{cite2120}}.
\codefieldsection{Encoding}
\begin{eczvaluelist}
\item\relax \flmRefsHyperref{ref409}{Clifford circuits}, i.e., those consisting of CNOT, Hadamard, and certain phase gates \NoCaseChange{\protect\cite{cite3370}}, using an algorithm \NoCaseChange{\protect\cite{cite2123}} based on the Gottesman-Knill theorem \NoCaseChange{\protect\cite{cite2121}} or using ZX calculus \NoCaseChange{\protect\cite{cite3590,cite3591}}. 
Generic constructions use \(O(n^2)\) Clifford gates and can be adjusted to realize a chosen set of logical Pauli operators \NoCaseChange{\protect\cite{cite398}}.
Encoding circuits can be done in linear depth, up to logarithmic corrections, for most practical use cases \NoCaseChange{\protect\cite{cite4321}}. 
\begin{defterm}{Destabilizers}\label{ref4322}\label{ref3629}
A Clifford encoding circuit maps the first \(k\) qubits to the logical qubits of the code, and the Pauli \(Z\) operators of the last \(r = n-k\) qubits are mapped into a set of stabilizer generators.
The set of Pauli \(X\) operators of the first \(r\) qubits that are mapped into a set of generators for the destabilizer group \NoCaseChange{\protect\cite{cite2121,cite4323}}.
Each such generator anticommutes with only one stabilizer generator while commuting with the rest of the stabilizer generators.
\end{defterm}

\item\relax For a code given in \flmRefsHyperref{ref857}{encoder-respecting form} with maximum graph degree \(\Delta\), there is an efficient construction of an encoding circuit of depth at most \(2\Delta+3\) \NoCaseChange{\protect\cite{cite858}}.
\item\relax Circuits obtained by first constructing the CWS form of the code \NoCaseChange{\protect\cite{cite4324,cite852}}. These consist of \(n\) Hadamard gates, a classical encoder which takes at most \(n\) CX gates for a single-qubit encoding code, and at most \(n(n-1)/2\) CZ gates to create the needed graph state.
\item\relax Lindbladian-based dissipative encoding \NoCaseChange{\protect\cite{cite4325,cite4326}}, for which the codespace is the steady-state space of a Lindbladian. This does not give a speedup, in terms of scaling with \(n\), over circuit-based encoders \NoCaseChange{\protect\cite{cite4327}}. A finite-time non-Lindbladian dissipative encoder may always be constructed \NoCaseChange{\protect\cite{cite4328}}.
\item\relax Algorithm for fault-tolerant magic-state initialization \NoCaseChange{\protect\cite{cite4329}}.
\item\relax Measurement-based purification protocol yielding arbitrary logical states from initial thermal states \NoCaseChange{\protect\cite{cite4330}}.
\item\relax State injection protocols that generalize lattice surgery for the surface code \NoCaseChange{\protect\cite{cite749}}.
\item\relax The overlap with any state prepared by a depth-\(h\) Clifford circuit with a product state is at most \(2^{1-d/2^h}\) \NoCaseChange{\protect\cite{cite529}}.
\end{eczvaluelist}
\codefieldsection{Transversal and Permutation-Based Gates}
\begin{eczvaluelist}
\item\relax Logical Pauli transformations \NoCaseChange{\protect\cite{cite736,cite772,cite773}}.
\item\relax The four-block transversal gate mapping each \(X \to IXXX\) and each \(Z \to IZZZ\) implements the same logical gate on all qubits \NoCaseChange{\protect\cite{cite736}\protect\cite[{Exam. 1}]{cite532}}. More generally, in the \(t\)-block setting, the commutant of the \(t\)-fold action of the \(n\)-qubit Clifford group contains invertible operators realizing the stochastic orthogonal group \(O_t(2)\), yielding inter-block symmetries in tensor powers of stabilizer constructions \NoCaseChange{\protect\cite{cite774}}.
\item\relax Transversal logical gates are in a finite level of the \flmTerm{term}{ref694}{}{Clifford hierarchy}, which is shown using stabilizer \textit{disjointness} \NoCaseChange{\protect\cite{cite775}} (see also \NoCaseChange{\protect\cite{cite776,cite777}}).
Transversal gates for \(n\in\{1,2\}\) are semi-Clifford \NoCaseChange{\protect\cite{cite778}}.

\item\relax No stabilizer code can implement a classical universal gate set transversally \NoCaseChange{\protect\cite{cite779}}.
\item\relax Fold-transversal gates have been extended from qubit CSS codes to qubit stabilizer codes, and there is an algorithm to determine them from the stabilizer group \NoCaseChange{\protect\cite{cite719}}.
\item\relax Computation can be sped up substantially for codes that admit transversal measurements of logical \(X\) and \(Z\) \NoCaseChange{\protect\cite{cite780}}.
\item\relax Diagonal transversal \flmRefsHyperref{ref409}{Clifford gates} on multiple code blocks must form one of six families of matrix groups: \(O(\ell,\mathbb{F}_2)\) generically, \(U(\ell,\mathbb{F}_4)\) for Hermitian qubit codes, \(GL(\ell,\mathbb{F}_2)\) for non-self-dual CSS codes, \(O( \ell, \mathbb{F}_2[x]/(x^2) )\) for self-dual non-CSS codes, \(U(\ell,R_8)\) for intermediate semi-self-dual CSS or self-dual semi-CSS cases, and \(Sp(2\ell,\mathbb{F}_2)\) for self-dual CSS codes \NoCaseChange{\protect\cite{cite738}}. There are two \(\llbracket 8,1,3\rrbracket \) self-dual non-CSS codes; see QECDB \NoCaseChange{\protect\cite{cite781}}.
\item\relax An entangling diagonal transversal two-qubit \flmRefsHyperref{ref409}{Clifford gate} exists only for codes equivalent to CSS or self-dual codes; canonical representatives are the inter-block CNOT in the CSS case and a \(Y\)-controlled-\(Y\) gate, locally Clifford equivalent to \(CZ\), in the self-dual case \NoCaseChange{\protect\cite{cite738}}.
\item\relax Implementing the full Clifford group on \(k\) logical qubits requires \(k\)-fold transversal gates; in particular, stabilizer codes cannot admit a transversal (1-fold) implementation of the full \flmRefsHyperref{ref409}{Clifford group} on more than one logical qubit, a fold-transversal (2-fold) implementation on more than two logical qubits, or a code-automorphism implementation on more than one logical qubit \NoCaseChange{\protect\cite{cite782}}. These bounds are tight: \(k\) copies of the \(\llbracket 7,1,3\rrbracket \) Steane code form a \(\llbracket 7k,k,3\rrbracket \) code admitting a \(k\)-fold transversal \flmRefsHyperref{ref409}{Clifford group} \NoCaseChange{\protect\cite{cite782}}.
\end{eczvaluelist}
\codefieldsection{Gates}
\begin{eczvaluelist}
\item\relax Logical \flmRefsHyperref{ref409}{Clifford gates} can be performed by physical \flmRefsHyperref{ref409}{Clifford circuits} that permute logical Pauli operators \NoCaseChange{\protect\cite{cite773}}.
\item\relax With pieceable fault-tolerance, any \flmRefsHyperref{ref811}{non-degenerate} stabilizer code with a complete set of fault-tolerant single-qubit gates in the \flmRefsHyperref{ref409}{Clifford group} has a universal set of non-transversal fault-tolerant gates \NoCaseChange{\protect\cite{cite806}}.
\item\relax Non-Clifford gates can be done using \textit{gate teleportation}, in which a gate can be obtained from a particular \textit{magic state} (a.k.a. resource state) \NoCaseChange{\protect\cite{cite3219,cite717}}. Such protocols can be made fault tolerant with the help of magic-state distillation \NoCaseChange{\protect\cite{cite690}} as well as state-injection and purification schemes based on concatenated error-detecting codes \NoCaseChange{\protect\cite{cite448}}. There exist various performance metrics for magic-state distillation \NoCaseChange{\protect\cite{cite4331,cite4332,cite4333,cite4189}} focusing mostly on distilling \(T\) gates. This Clifford+T gate set is self-testable \NoCaseChange{\protect\cite{cite4334}}. Gate errors in magic-state distillation protocols can sometimes add up destructively \NoCaseChange{\protect\cite{cite4153}}. The Hadamard gate cannot be obtained from a magic state \NoCaseChange{\protect\cite{cite4335}}. A magic state arising from a generalized controlled \(Z\) gate is a type of hypergraph state \NoCaseChange{\protect\cite{cite3571,cite3572,cite3573}} (see \NoCaseChange{\protect\cite{cite703,cite759}}). The Toffoli gate can be distilled from a particular two-qubit state \NoCaseChange{\protect\cite{cite4336}}. Magic-state protocol fidelity is upper bounded by the fidelity of protocols that have undergone stabilizer reduction, and there exist non-distillable states outside of the stabilizer octahedron \NoCaseChange{\protect\cite{cite4337,cite4338}}.
\item\relax Certain operations can be implemented in a fault-tolerant version \NoCaseChange{\protect\cite{cite4339,cite4340,cite4341,cite4342}} of holonomic quantum computation \NoCaseChange{\protect\cite{cite3755}}.
\item\relax Magic-state distillation and circuit compilation based on the SWAP test \NoCaseChange{\protect\cite{cite4343}}.
\item\relax Logical Clifford synthesis (LCS) taking in a code and a logical Clifford operation and producing a circuit acting on the physical qubits \NoCaseChange{\protect\cite{cite3332}}.
\item\relax Clifford stabilizer circuits can be compiled using stabilizer tableau manipulation \NoCaseChange{\protect\cite{cite4344}}.
\item\relax A teleported version of the CPC construction, the Clifford noise reduction (CliNR) scheme, can reduce noise in \flmRefsHyperref{ref409}{Clifford circuits} with Pauli measurements with at most a three-fold overhead in the number of qubits and gates \NoCaseChange{\protect\cite{cite3592,cite3593}}. There is a simple formula for the probability that a \flmRefsHyperref{ref409}{Clifford circuit} contains a logical error \NoCaseChange{\protect\cite{cite3589}}.
\item\relax Logical Trotter circuits for any stabilizer code can be implemented via symplectic transvections \NoCaseChange{\protect\cite{cite3371}}.
\item\relax Hardware-tailored logical \flmRefsHyperref{ref409}{Clifford circuits} \NoCaseChange{\protect\cite{cite3271}}.
\item\relax The \flmRefsHyperref{ref3630}{Bravyi-Koenig bound} can be generalized to arbitrary qubit stabilizer codes \NoCaseChange{\protect\cite{cite739}}.
\item\relax Fault-tolerant flag-based non-transversal logical gates \NoCaseChange{\protect\cite{cite3201}}.
\end{eczvaluelist}
\codefieldsection{Decoding}
\begin{eczvaluelist}
\item\relax The size of the circuit extracting the syndrome depends on the weight of its corresponding stabilizer generator. Syndrome extraction circuits can be simulated efficiently using dedicated software (e.g., STIM \NoCaseChange{\protect\cite{cite3964}}) and there are many general schemes for generating them \NoCaseChange{\protect\cite{cite4345}} (see also \NoCaseChange{\protect\cite{cite3172}}). Noise can be characterized without interrupting syndrome extraction \NoCaseChange{\protect\cite{cite4346}}. Decoding of qubit stabilizer codes is an approximately optimal strategy for various quantum lights-out (QLO) games that can be played on the codes' \flmRefsHyperref{ref857}{encoder-respecting form} \NoCaseChange{\protect\cite{cite858}}.
\item\relax Instead of applying the Pauli recovery from a fault-tolerant syndrome-extraction gadget, one can usually track it classically as a \textit{Pauli frame}; Clifford operations update the frame by conjugation, while measurement-based non-Clifford gadgets must be interpreted relative to the current frame \NoCaseChange{\protect\cite{cite398}}.
\item\relax A greedy graph decoder for \flmRefsHyperref{ref857}{encoder-respecting forms} corrects all recoverable errors for sufficiently sparse graphs, including graphs with girth at least 13 and some lower-girth families with additional spacing constraints \NoCaseChange{\protect\cite{cite858}}.
\item\relax DiVincenzo-Aliferis syndrome extraction circuits \NoCaseChange{\protect\cite{cite3785}}.
\item\relax Greedy syndrome measurement schedule \NoCaseChange{\protect\cite{cite3439}}.
\item\relax Dynamical weight reduction (DWR) scheme in which measurements of smaller-weight Paulis yield the outcome of a larger-weight Pauli via the use of ZX calculus and ancillary qubits \NoCaseChange{\protect\cite{cite4347}}.
\item\relax Ancilla modes can be used for syndrome extraction instead of ancilla qubits \NoCaseChange{\protect\cite{cite4348}}, and using two-component cat codes \NoCaseChange{\protect\cite{cite4349}} yields fault-tolerant syndrome extraction circuits.
\item\relax Autonomous QEC protocol \NoCaseChange{\protect\cite{cite4350}}.
\item\relax MPE decoding, i.e., the process of finding the most likely error, is \(NP\)-complete in the worst case \NoCaseChange{\protect\cite{cite4351,cite4352}}. If the noise model is such that the most likely error is the lowest-weight error, then ML decoding is called \textit{minimum-weight} decoding. Maximum-likelihood (ML) decoding (a.k.a.\ degenerate maximum-likelihood decoding), i.e., the process of finding the most likely error class (up to degeneracy of errors), is \(\#P\)-complete in the worst case \NoCaseChange{\protect\cite{cite4353}}. Decoding a random qubit stabilizer code (i.e., the learning stabilizers with noise, or LSN, problem \NoCaseChange{\protect\cite{cite4354}}) is at least as hard as decoding a random classical code at constant rate \NoCaseChange{\protect\cite{cite4355}}.
\item\relax Incorporating faulty syndrome measurements can be done by performing spacetime decoding, i.e., using data from past rounds for decoding syndromes in any given round. If a decoder does not process syndrome data sufficiently quickly, it can lead to the \textit{backlog problem} \NoCaseChange{\protect\cite{cite2967}}, slowing down the computation.
\item\relax Splitting decoders \NoCaseChange{\protect\cite{cite4356}}.
\item\relax Trellis decoder, which builds a compact representation of the algebraic structure of the normalizer \(\mathsf{N(S)}\) \NoCaseChange{\protect\cite{cite4127}}.
\item\relax Quantum extension of GRAND decoder \NoCaseChange{\protect\cite{cite4357}}.
\item\relax Deep neural-network probabilistic decoder \NoCaseChange{\protect\cite{cite4358}}.
\item\relax Generalized belief propagation (GBP) \NoCaseChange{\protect\cite{cite3913}} based on a classical version \NoCaseChange{\protect\cite{cite3914}}.
\item\relax Integer optimization decoder \NoCaseChange{\protect\cite{cite2954}}.
\item\relax Autonomous Lindbladian based decoders for codes encoding a single logical qubit \NoCaseChange{\protect\cite{cite4359}}.
\item\relax For codes encoding a single logical qubit, logical information can be extracted by single-qubit operations and classical communication \NoCaseChange{\protect\cite{cite4360}}.
\item\relax Correlated decoding can improve performance of Clifford and \flmRefsHyperref{ref409}{non-Clifford} entangling gates \NoCaseChange{\protect\cite{cite4361}}.
\item\relax Detector graphs \NoCaseChange{\protect\cite{cite3964,cite4362}} and detector error models \NoCaseChange{\protect\cite{cite3735}} can be used to design syndrome extraction circuits and logical measurements.
\item\relax Fault-tolerant constant-depth unencoder transforming logical states into physical states using single-qubit measurements \NoCaseChange{\protect\cite{cite1436}}.
\item\relax Degenerate erasure decoder showing near ML decoding for various codes \NoCaseChange{\protect\cite{cite4363}}.
\item\relax Tensor-network decoder for non-Markovian noise \NoCaseChange{\protect\cite{cite4364}}.
\item\relax Delayed-erasure conversion decoder \NoCaseChange{\protect\cite{cite4365}}.
\item\relax Trickle-down dissipative correction \NoCaseChange{\protect\cite{cite4366}}.
\end{eczvaluelist}
\codefieldsection{Fault Tolerance}
\begin{eczvaluelist}
\item\relax Shor error correction \NoCaseChange{\protect\cite{cite761,cite3300}} (see also Steane's ancilla factory \NoCaseChange{\protect\cite{cite1123}}), in which fault tolerance against syndrome extraction errors is ensured by simply repeating syndrome measurements. A modification uses adaptive measurements \NoCaseChange{\protect\cite{cite4367}}.
\item\relax Gates in the \flmTerm{term}{ref694}{}{Clifford hierarchy} can be done using \textit{gate teleportation}, in which a gate can be obtained from a particular \textit{magic state} \NoCaseChange{\protect\cite{cite3219,cite717}}. Such protocols can be made fault tolerant with the help of magic-state distillation \NoCaseChange{\protect\cite{cite690}}. See review on magic-state distillation \NoCaseChange{\protect\cite{cite4368}}.
\item\relax Logical Bell measurements can be done transversally, and thus fault tolerantly, by performing bitwise Bell measurements for each pair of qubits (with each member of the pair taken from one of the two code blocks) and processing the result.
\item\relax With pieceable fault-tolerance, any \flmRefsHyperref{ref811}{non-degenerate} stabilizer code with a complete set of fault-tolerant single-qubit gates in the \flmRefsHyperref{ref409}{Clifford group} has a universal set of non-transversal fault-tolerant gates \NoCaseChange{\protect\cite{cite806}}.
\item\relax Generalization of Steane error correction for stabilizer codes \NoCaseChange{\protect\cite[{Sec. 3.6}]{cite3472}}.
\item\relax Fault-tolerant error correction scheme by Knill (a.k.a. telecorrection \NoCaseChange{\protect\cite{cite3185}}), which is based on teleportation \NoCaseChange{\protect\cite{cite448,cite4369}}. A variant of it has been termed the Fibonacci scheme \NoCaseChange{\protect\cite{cite4370}}.
\item\relax For a non-CSS stabilizer code, the logical Bell-state ancilla needed for Knill error correction can be prepared by treating two code blocks as one larger stabilizer state and verifying the combined stabilizer via the Shor-style preparation procedure \NoCaseChange{\protect\cite[{Sec. 13.1.3}]{cite398}}.
\item\relax Fault-tolerant error correction using flag qubits for codes satisfying certain conditions \NoCaseChange{\protect\cite{cite3220}}.
\item\relax GHZ state distillation for Steane error correction \NoCaseChange{\protect\cite{cite4371}}.
\item\relax Syndrome extraction using flag qubits and classical codes \NoCaseChange{\protect\cite{cite4372}}.
\item\relax Fault-tolerant constant-depth unencoder transforming logical states into physical states using single-qubit measurements \NoCaseChange{\protect\cite{cite1436}}.
\item\relax Post-selection based algorithm preparing magic state corresponding to arbitrary rotations \NoCaseChange{\protect\cite{cite4373}}.
\item\relax \flmRefsHyperref{ref410}{Code switching} can be done using only transversal gates for qubit stabilizer codes \NoCaseChange{\protect\cite{cite4374}}.
\item\relax Flag-Proxy Networks (FPNs) \NoCaseChange{\protect\cite{cite3439}}.
\item\relax A logical Pauli can be \flmRefsHyperref{ref666}{gauged out} to yield a fault-tolerant measurement with worst-case qubit overhead \(O(W\log^{2}W)\) for a weight-\(W\) Pauli \NoCaseChange{\protect\cite{cite470}}.
\item\relax Automated fault-tolerant circuit synthesis using boolean satisfiability \NoCaseChange{\protect\cite{cite4375}}.
\item\relax Algorithm for fault-tolerant magic-state initialization \NoCaseChange{\protect\cite{cite4329}}.
\item\relax Fault tolerance of qubit stabilizer codes has been formalized \NoCaseChange{\protect\cite{cite4376}}.
\item\relax Fault-tolerant flag-based non-transversal logical gates \NoCaseChange{\protect\cite{cite3201}}.
\end{eczvaluelist}
\codefieldsection{Code Capacity Threshold}
\begin{eczvaluelist}
\item\relax Bounds on code capacity thresholds using ML decoding can be obtained by mapping the effect of noise on the code to a statistical mechanical model \NoCaseChange{\protect\cite{cite480,cite3441,cite4295,cite4377}}. The AQEC relative entropy is related to the resulting threshold \NoCaseChange{\protect\cite{cite2554}}.
\end{eczvaluelist}
\codefieldsection{Threshold}
\begin{eczvaluelist}
\item\relax Certain operations can be implemented in a fault-tolerant version \NoCaseChange{\protect\cite{cite4339,cite4341}} of holonomic quantum computation \NoCaseChange{\protect\cite{cite3755}}.
\end{eczvaluelist}
\codefieldsection{Notes}
\begin{eczvaluelist}
\item\relax Introductions to stabilizer codes can be found in \NoCaseChange{\protect\cite{cite736,cite4378,cite2764,cite4379,cite4380,cite2024}}.
\item\relax Tables of bounds and examples of stabilizer codes for various \(n\) and \(k\), based on algorithms developed in Ref. \NoCaseChange{\protect\cite{cite2673}}, are maintained by M. Grassl at this \flmHref{https://www.codetables.de/}{website}. A Magma implementation exists at this \flmHref{https://magma.maths.usyd.edu.au/magma/handbook/text/1976}{website}.
\item\relax See Quantum Codes qubit stabilizer database, maintained by N. Aydin, P. Liu, and B. Yoshino \NoCaseChange{\protect\cite{cite4381,cite4382}}, at this \flmHref{http://quantumcodes.info/}{website}.
\item\relax See the QEC database \NoCaseChange{\protect\cite{cite781}}.
\item\relax Entanglement purification protocols with qubit stabilizer codes are related to quantum key distribution (QKD) \NoCaseChange{\protect\cite{cite4383}}. There is a correspondence between stabilizer codes and bilocal Clifford entanglement distillation circuits \NoCaseChange{\protect\cite{cite4384}}. Purification protocols using two-way classical channels can exceed the quantum Hamming and quantum Singleton bounds \NoCaseChange{\protect\cite{cite4385}}.
\item\relax Qubit stabilizer codes can be used to estimate physical Pauli noise up to their \flmRefsHyperref{ref672}{pure distance} \NoCaseChange{\protect\cite{cite4033}}, and logical Pauli noise for any correctable physical noise \NoCaseChange{\protect\cite{cite4034}}.
\item\relax The stabilizer formalism has been gamified \NoCaseChange{\protect\cite{cite4386}}.
\item\relax There is a relation between magic states and the onset of contextuality \NoCaseChange{\protect\cite{cite4387}}.
\item\relax Codes can be found via genetic algorithms \NoCaseChange{\protect\cite{cite4314}}.
\item\relax Stim Python software library for simulating and analyzing qubit stabilizer circuits \NoCaseChange{\protect\cite{cite3964,cite4388}}.
\item\relax PanQEC Python software library for simulating and visualizing qubit stabilizer codes \NoCaseChange{\protect\cite{cite2626,cite4389}}.
\item\relax QuantumSavory Julia software library that includes stabilizer formalism tools, symbolic manipulation, and a tool for drawing quantum circuits \NoCaseChange{\protect\cite{cite4390,cite4391}}.
\end{eczvaluelist}
\codefieldsection{Parents}
\begin{eczvaluelist}
\item\relax
\flmRefsHyperref[eczindexfamilyrel]{code:non_stabilizer}{Union stabilizer (USt) code} --- A stabilizer code with stabilizer group \(\mathsf{S}\) can be thought of as a USt with only the identity coset representative.
Conversely, if the set of coset representatives of a USt form a linear binary code, then they can be absorbed into a stabilizer group that defines the USt.

\item\relax
\flmRefsHyperref[eczindexfamilyrel]{code:xp_stabilizer}{XP stabilizer code} --- XP stabilizer codes reduce to qubit stabilizer codes for \(N=2\).
\item\relax
\flmRefsHyperref[eczindexfamilyrel]{code:qubit_stabilizer_oaqecc}{Operator-algebra (OA) qubit stabilizer code} --- An OA qubit stabilizer code storing no classical information and admitting no gauge qubits is a qubit stabilizer code.
\item\relax
\flmRefsHyperref[eczindexfamilyrel]{code:qudit_stabilizer}{Modular-qudit stabilizer code} --- Modular-qudit stabilizer codes for \(q=2\) correspond to qubit stabilizer codes.
Modular-qudit stabilizer codes for prime-dimensional qudits \(q=p\) inherit most of the features of qubit stabilizer codes, including encoding an integer number of qudits and a \flmRefsHyperref{ref2198}{modular-qudit Pauli group} with a unique number of generators.
Conversely, qubit codes can be extended to modular-qudit codes by decorating appropriate generators with powers.
For example, \(\llbracket 4,2,2\rrbracket \) qubit code generators can be adjusted to \(ZZZZ\) and \(XX^{-1} XX^{-1}\).
A systematic procedure extending a qubit code to prime-qudit codes involves putting its generator matrix into local-dimension-invariant (LDI) form  \NoCaseChange{\protect\cite{cite4392}}.
Various bounds exist on the distance of the resulting codes \NoCaseChange{\protect\cite{cite820,cite4393}}.

\item\relax
\flmRefsHyperref[eczindexfamilyrel]{code:galois_true_stabilizer}{True Galois-qudit stabilizer code} --- True Galois-qudit stabilizer codes for \(q=2\) correspond to qubit stabilizer codes.
\end{eczvaluelist}
\codefieldsection{Children}
\begin{eczvaluelist}
\item\relax
\flmRefsHyperref[eczindexfamilyrel]{code:approximate_log_depth}{Log-depth geometrically local Clifford-circuit code}\item\relax
\flmRefsHyperref[eczindexfamilyrel]{code:nonlocal_lowdepth}{Brown-Fawzi Clifford-circuit code}\item\relax
\flmRefsHyperref[eczindexfamilyrel]{code:majorana_stab}{Majorana stabilizer code} --- A Majorana stabilizer code is a stabilizer code whose stabilizers are composed of Majorana fermion operators, which are in turn realizable using Pauli strings via the Jordan-Wigner mapping.
Any \(\llbracket n,k,d\rrbracket \) stabilizer code can be mapped into a \(\llbracket 2n,k,2d\rrbracket _{f}\) Majorana stabilizer code by concatenating with the tetron code \NoCaseChange{\protect\cite{cite537}\protect\cite[{Lemma 1}]{cite1432}}.
Embedding each physical qubit into two fermions via the tetron code is useful for exactly solving the Kitaev honeycomb model Hamiltonian \NoCaseChange{\protect\cite{cite537}} and other qubit Hamiltonians on certain graphs \NoCaseChange{\protect\cite{cite2842,cite2843}}.
Majorana stabilizer groups can be converted into ordinary qubit stabilizer groups via the parton mapping, while their corresponding states are converted via the Gutzwiller projection \NoCaseChange{\protect\cite{cite2844}}.

\item\relax
\flmRefsHyperref[eczindexfamilyrel]{code:small_distance_qubit_stabilizer}{Small-distance qubit stabilizer code} --- Criteria for the existence of single error-correcting qubit stabilizer codes have been developed \NoCaseChange{\protect\cite{cite4394}}. Qubit stabilizer codes for \(n < 10\) have been classified \NoCaseChange{\protect\cite{cite454}}, all of which have small distance. Qubit stabilizer codes have been partially enumerated up to twelve qubits \NoCaseChange{\protect\cite{cite4395}}. There are two \(\llbracket 8,1,3\rrbracket \) self-dual non-CSS codes; see QECDB \NoCaseChange{\protect\cite{cite781}}.
\item\relax
\flmRefsHyperref[eczindexfamilyrel]{code:cpc}{Coherent-parity-check (CPC) code} --- CPC codes are a type of stabilizer code. A teleported version of the CPC construction, the Clifford noise reduction (CliNR) scheme, can reduce noise in \flmRefsHyperref{ref409}{Clifford circuits} with Pauli measurements with at most a three-fold overhead in the number of qubits and gates \NoCaseChange{\protect\cite{cite3592,cite3593}}. There is a simple formula for the probability that a \flmRefsHyperref{ref409}{Clifford circuit} contains a logical error \NoCaseChange{\protect\cite{cite3589}}.
\item\relax
\flmRefsHyperref[eczindexfamilyrel]{code:data_syndrome}{Quantum data-syndrome (QDS) code} --- QDS codes are stabilizer codes whose stabilizer generators encode extra redundancy (via a linear binary code) so as to protect from syndrome measurement errors.
\item\relax
\flmRefsHyperref[eczindexfamilyrel]{code:fermions_into_qubits}{Fermion-into-qubit code} --- Fermion-into-qubit codes are qubit stabilizer codes that encode a logical fermionic Hilbert space into a physical space of \(n\) qubits.
\item\relax
\flmRefsHyperref[eczindexfamilyrel]{code:stabilizer_over_gf4}{Hermitian qubit code}\item\relax
\flmRefsHyperref[eczindexfamilyrel]{code:happy}{Pastawski-Yoshida-Harlow-Preskill (HaPPY) code} --- The HaPPY code is a stabilizer code because it is defined by a contracted network of stabilizer tensors; see \NoCaseChange{\protect\cite[{Thm. 6}]{cite1667}}.
\item\relax
\flmRefsHyperref[eczindexfamilyrel]{code:holographic_6_1_3}{Six-qubit-tensor holographic code}\item\relax
\flmRefsHyperref[eczindexfamilyrel]{code:holographic_hyperinvariant}{Hyperinvariant tensor-network (HTN) code}\item\relax
\flmRefsHyperref[eczindexfamilyrel]{code:quantum_k-orthogonal}{\(k\)-orthogonal code}\item\relax
\flmRefsHyperref[eczindexfamilyrel]{code:cluster_state}{Cluster-state code} --- Cluster-state codes are particular qubit stabilizer codes. Any qubit stabilizer code is equivalent to a graph quantum code via a single-qubit \flmRefsHyperref{ref409}{Clifford circuit} \NoCaseChange{\protect\cite{cite3561}} (see also \NoCaseChange{\protect\cite{cite3536,cite867}}). As a corollary, any qubit stabilizer state is equivalent to a cluster state under a single-qubit \flmRefsHyperref{ref409}{Clifford circuit} \NoCaseChange{\protect\cite{cite3536}\protect\cite[{Appx. A}]{cite3562}}. There are algorithms that determine whether two stabilizer states are equivalent that work by checking whether their corresponding cluster states are equivalent \NoCaseChange{\protect\cite{cite3563,cite3564,cite3565}}. Any fault-tolerant scheme based on qubit stabilizer codes can be mapped into a cluster-state based MBQC protocol \NoCaseChange{\protect\cite{cite3532}}.
\item\relax
\flmRefsHyperref[eczindexfamilyrel]{code:fusion}{Fusion-based quantum computing (FBQC) code} --- The resource states in FBQC are small stabilizer states, and the surviving stabilizers after fusion determine the encoded output state (conditioned on measurement outcomes).
\item\relax
\flmRefsHyperref[eczindexfamilyrel]{code:purity_testing}{Purity-testing stabilizer code}\item\relax
\flmRefsHyperref[eczindexfamilyrel]{code:qldpc}{Qubit QLDPC code}\item\relax
\flmRefsHyperref[eczindexfamilyrel]{code:quantum_bch}{Qubit BCH code}\item\relax
\flmRefsHyperref[eczindexfamilyrel]{code:quantum_synchronizable}{Quantum synchronizable code}\end{eczvaluelist}
\codefieldsection{Cousins}
\begin{eczvaluelist}
\item\relax
\flmRefsHyperref[eczindexfamilyrel]{code:cws}{Codeword stabilized (CWS) code} --- CWS codes whose underlying classical code is a linear binary code are qubit stabilizer codes containing a cluster-state codeword \NoCaseChange{\protect\cite{cite852,cite3266}}.
Conversely, stabilizer codes admit a CWS standard form \NoCaseChange{\protect\cite{cite852}}; at the state level, the underlying stabilizer state can be mapped to a cluster state by a single-qubit \flmRefsHyperref{ref409}{Clifford circuit} \NoCaseChange{\protect\cite{cite3536}\protect\cite[{Appx. A}]{cite3562}}.

\item\relax
\flmRefsHyperref[eczindexfamilyrel]{code:binary_linear}{Linear binary code} --- Qubit stabilizer codes are the closest quantum analogues of binary linear codes because addition modulo two corresponds to multiplication of stabilizers in the quantum case. 
Any binary linear code can be thought of as a qubit stabilizer code with \(Z\)-type stabilizer generators \NoCaseChange{\protect\cite{cite1434}\protect\cite[{Table I}]{cite1433}}. 
The stabilizer generators are extracted from rows of the parity-check matrix, while logical \(X\) Paulis correspond to rows of the generator matrix.
States close to the equal superposition of all bitstrings within Hamming distance \(b\) of a binary linear code can be prepared efficiently \NoCaseChange{\protect\cite{cite1435}}.
Binary linear codes can be used for error-corrected entanglement distillation protocols \NoCaseChange{\protect\cite{cite1436}}.

\item\relax
\flmRefsHyperref[eczindexfamilyrel]{code:dual}{Dual linear code} --- Qubit stabilizer codes are in one-to-one correspondence with symplectic self-orthogonal binary linear codes of length \(2n\) via the \flmRefsHyperref{ref817}{symplectic representation}.
\item\relax
\flmRefsHyperref[eczindexfamilyrel]{code:dual_additive}{Dual additive code} --- Qubit stabilizer codes are in one-to-one correspondence with trace-Hermitian self-orthogonal additive quaternary codes of length \(n\) via the \flmRefsHyperref{ref1778}{\(\mathbb{F}_4\) representation}.
\item\relax
\flmRefsHyperref[eczindexfamilyrel]{code:single_shot}{Single-shot code} --- Any stabilizer code can be single shot if sufficiently non-local high-weight stabilizer generators are used for syndrome measurements.  These can be obtained with a Gaussian elimination procedure \NoCaseChange{\protect\cite{cite675}}.
\item\relax
\flmRefsHyperref[eczindexfamilyrel]{code:t-designs}{\(t\)-design} --- Stabilizer states on \(n\) qubits form complex projective 3-designs, but not 4-designs, on \(\mathbb{C}P^{2^n-1}\) \NoCaseChange{\protect\cite{cite937}}. The \flmRefsHyperref{ref409}{Clifford group} is a unitary 2-design \NoCaseChange{\protect\cite{cite938}} and a 3-design \NoCaseChange{\protect\cite{cite940,cite941}\protect\cite[{Thm. 1.6(B)}]{cite939}\protect\cite[{pg. 191}]{cite42}} on \(U(2^n)\). The \(\llbracket 2m,2m-2,2\rrbracket \) code for \(2m\) being a multiple of four obstructs the Clifford group from being a 4-design \NoCaseChange{\protect\cite{cite801}}.
\item\relax
\flmRefsHyperref[eczindexfamilyrel]{code:unitary_design}{Unitary \(t\)-design} --- Stabilizer states on \(n\) qubits form complex projective 3-designs, but not 4-designs, on \(\mathbb{C}P^{2^n-1}\) \NoCaseChange{\protect\cite{cite937}}. The \flmRefsHyperref{ref409}{Clifford group} is a unitary 2-design \NoCaseChange{\protect\cite{cite938}} and a 3-design \NoCaseChange{\protect\cite{cite940,cite941}\protect\cite[{Thm. 1.6(B)}]{cite939}\protect\cite[{pg. 191}]{cite42}} on \(U(2^n)\). The \(\llbracket 2m,2m-2,2\rrbracket \) code for \(2m\) being a multiple of four obstructs the Clifford group from being a 4-design \NoCaseChange{\protect\cite{cite801}}.
\item\relax
\flmRefsHyperref[eczindexfamilyrel]{code:complex_projective}{Complex projective space code} --- Stabilizer states on \(n\) qubits form complex projective 3-designs, but not 4-designs, on \(\mathbb{C}P^{2^n-1}\) \NoCaseChange{\protect\cite{cite937}}. The \flmRefsHyperref{ref409}{Clifford group} is a unitary 2-design \NoCaseChange{\protect\cite{cite938}} and a 3-design \NoCaseChange{\protect\cite{cite940,cite941}\protect\cite[{Thm. 1.6(B)}]{cite939}\protect\cite[{pg. 191}]{cite42}} on \(U(2^n)\).
\item\relax
\flmRefsHyperref[eczindexfamilyrel]{code:iceberg}{\(\llbracket 2m,2m-2,2\rrbracket \) error-detecting code} --- The \(\llbracket 2m,2m-2,2\rrbracket \) code for \(2m\) being a multiple of four obstructs the Clifford group from being a 4-design \NoCaseChange{\protect\cite{cite801}}.
\item\relax
\flmRefsHyperref[eczindexfamilyrel]{code:ampdamp}{Amplitude-damping (AD) code} --- Concatenating the dual-rail code with an inner \(\llbracket n,k,d\rrbracket \) qubit stabilizer code yields a degenerate \(\llbracket 2n,k,d\rrbracket \) constant-excitation stabilizer code that corrects \(d-1\) \flmRefsHyperref{ref498}{AD} errors \NoCaseChange{\protect\cite{cite3263,cite2711}}.
\item\relax
\flmRefsHyperref[eczindexfamilyrel]{code:two-legged-cat}{Two-component cat code} --- Ancilla modes can be used for syndrome extraction instead of ancilla qubits \NoCaseChange{\protect\cite{cite4348}}, and using two-component cat codes \NoCaseChange{\protect\cite{cite4349}} yields fault-tolerant syndrome extraction circuits.
\item\relax
\flmRefsHyperref[eczindexfamilyrel]{code:projective}{Projective geometry code} --- \(\llbracket n,k,d\rrbracket \) qubit stabilizer codes with no weight-one stabilizers are equivalent to particular "quantum" sets of lines in projective space \(PG(n-k-1,2)\) \NoCaseChange{\protect\cite{cite1976,cite1977}\protect\cite[{Thm. 3.7}]{cite1975}}. This equivalence is stated in the case of \flmRefsHyperref{ref672}{pure} qubit stabilizer codes with distance two or greater in \NoCaseChange{\protect\cite[{Thm. 2.6}]{cite1695}}.
\item\relax
\flmRefsHyperref[eczindexfamilyrel]{code:holographic}{Holographic code} --- Qubit stabilizer states can be interpreted as states that are preparable using the Euclidean path integral in 3D Chern-Simons theory, defined on manifolds that are toy models of AdS/CFT wormholes \NoCaseChange{\protect\cite{cite2535,cite2536}}.
\item\relax
\flmRefsHyperref[eczindexfamilyrel]{code:topological_abelian}{Abelian topological code} --- Qubit stabilizer states can be interpreted as states that are preparable using the Euclidean path integral in 3D Chern-Simons theory, defined on manifolds that are toy models of AdS/CFT wormholes \NoCaseChange{\protect\cite{cite2535,cite2536}}.
\item\relax
\flmRefsHyperref[eczindexfamilyrel]{code:galois_css}{Galois-qudit CSS code} --- Any \(\llbracket n,k,d\rrbracket \) qubit stabilizer code can be mapped to an \(\llbracket n,k,d\rrbracket _4\) Galois-qudit CSS code via the BLT mapping \NoCaseChange{\protect\cite[{Lemma 1}]{cite795}\protect\cite[{Lemma 1}]{cite1432}}: each stabilizer \(P = \bigotimes_j P_j\) generates an \(XX\)-type stabilizer via \(\mathcal{D}_X\) (\(I\mapsto II,\,X\mapsto XI,\,Z\mapsto IX,\,Y\mapsto XX\)) and a \(ZZ\)-type stabilizer via \(\mathcal{D}_Z\) (\(I\mapsto II,\,X\mapsto IZ,\,Z\mapsto ZI,\,Y\mapsto ZZ\)).
\item\relax
\flmRefsHyperref[eczindexfamilyrel]{code:barnes_wall}{Barnes-Wall (BW) lattice} --- Stabilizer states can be mapped into the first lattice shell of a BW lattice over a cyclotomic field, while the \flmRefsHyperref{ref409}{Clifford group} is related to the symmetry group of the lattice \NoCaseChange{\protect\cite{cite2117}}.
\item\relax
\flmRefsHyperref[eczindexfamilyrel]{code:clifford_group}{Clifford group} --- Computing with \flmRefsHyperref{ref409}{Clifford gates}, Pauli measurements, and classical feedforward acting on stabilizer states only can be efficiently simulated on a classical computer by tracking stabilizer and logical generators, according to the \textit{Gottesman-Knill theorem} \NoCaseChange{\protect\cite{cite2115,cite2116}}. Stabilizer states can be mapped into the first lattice shell of a BW lattice over a cyclotomic field, while the \flmRefsHyperref{ref409}{Clifford group} is related to the symmetry group of the lattice \NoCaseChange{\protect\cite{cite2117}}.
\item\relax
\flmRefsHyperref[eczindexfamilyrel]{code:ame}{Perfect-tensor code} --- The codespace of a qubit stabilizer code with \flmRefsHyperref{ref672}{pure distance} \(d_{\textnormal{pure}}\) is a \((d_{\textnormal{pure}}-1)\)-uniform space.
\item\relax
\flmRefsHyperref[eczindexfamilyrel]{code:commuting_projector}{Commuting-projector Hamiltonian code} --- Qubit stabilizer codes are \textit{infectious}: if the ground-state subspace of an \(\ell\)-local commuting projector Hamiltonian contains a state close to a stabilizer state, then the entire ground-state space is close to a stabilizer code \NoCaseChange{\protect\cite{cite2691,cite2692}}.
\item\relax
\flmRefsHyperref[eczindexfamilyrel]{code:constant_excitation}{Constant-excitation (CE) code} --- Concatenating the dual-rail code with an inner \(\llbracket n,k,d\rrbracket \) qubit stabilizer code yields a degenerate \(\llbracket 2n,k,d\rrbracket \) constant-excitation stabilizer code that avoids coherent phase errors and is equivalent to a Pauli-rotated repetition-concatenated stabilizer code \NoCaseChange{\protect\cite{cite2711}}. CSS structure is preserved when the original code is CSS \NoCaseChange{\protect\cite{cite524}}.
\item\relax
\flmRefsHyperref[eczindexfamilyrel]{code:metrological}{Metrological code} --- A joint \(+1\) and \(-1\) eigenstate of a set of stabilizers can form a metrological stabilizer code \NoCaseChange{\protect\cite{cite2717}}.
\item\relax
\flmRefsHyperref[eczindexfamilyrel]{code:qetc}{Quantum error-transmuting code (QETC)} --- Most QETCs are stabilizer codes: \(\mathsf{C}\) is the subspace stabilised by an abelian subgroup \(\mathsf{S} \subset \mathcal{G}_n\) of the \flmRefsHyperref{ref663}{Pauli group} on \(n\) qubits.
\item\relax
\flmRefsHyperref[eczindexfamilyrel]{code:eastab}{EA qubit stabilizer code} --- EA qubit stabilizer codes utilize additional ancillary qubits in a pre-shared entangled state, but reduce to qubit stabilizer codes when said qubits are interpreted as noiseless physical qubits. Qubit stabilizer codes can be used to obtain shortened EA qubit stabilizer codes \NoCaseChange{\protect\cite{cite3649}}.
\item\relax
\flmRefsHyperref[eczindexfamilyrel]{code:majorana_subsystem}{Majorana subsystem stabilizer code} --- The B\(\mapsto\)F mapping yields Majorana subsystem codes from qubit stabilizer codes such that their gauge groups contain tetrons \NoCaseChange{\protect\cite{cite3987}\protect\cite[{Sec. IV}]{cite3986}}. The output Majorana subsystem codes can correct odd-weight errors.
\item\relax
\flmRefsHyperref[eczindexfamilyrel]{code:tetron}{Tetron code} --- Any \(\llbracket n,k,d\rrbracket \) stabilizer code can be mapped into a \(\llbracket 2n,k,2d\rrbracket _{f}\) Majorana stabilizer code by concatenating with the tetron code \NoCaseChange{\protect\cite{cite537}\protect\cite[{Lemma 1}]{cite1432}}.
Embedding each physical qubit into two fermions via the tetron code is useful for exactly solving the Kitaev honeycomb model Hamiltonian \NoCaseChange{\protect\cite{cite537}} and other qubit Hamiltonians on certain graphs \NoCaseChange{\protect\cite{cite2842,cite2843}}. Majorana stabilizer groups can be converted into ordinary qubit stabilizer groups via the parton mapping, while their corresponding states are converted via the Gutzwiller projection \NoCaseChange{\protect\cite{cite2844}}.

\item\relax
\flmRefsHyperref[eczindexfamilyrel]{code:movassagh_ouyang}{Movassagh-Ouyang Hamiltonian code} --- Many, but not all, Movassagh-Ouyang codes are stabilizer codes.
\item\relax
\flmRefsHyperref[eczindexfamilyrel]{code:hybrid_stabilizer}{Hybrid stabilizer code} --- A hybrid stabilizer code storing no classical information reduces to a qubit stabilizer code.
Conversely, any qubit stabilizer code can be converted into a hybrid stabilizer code by using some of its qubits to store only classical information \NoCaseChange{\protect\cite{cite2735}}.

\item\relax
\flmRefsHyperref[eczindexfamilyrel]{code:qubit_permutation_invariant}{PI qubit code} --- There is a measurement-free code-switching protocol between a qubit stabilizer code and a PI qubit code \NoCaseChange{\protect\cite{cite727}}.
\item\relax
\flmRefsHyperref[eczindexfamilyrel]{code:qubit_css}{Qubit CSS code} --- Qubit CSS codes are qubit stabilizer codes whose stabilizer groups admit a generating set of pure-\(X\) and pure-\(Z\) Pauli strings. 
Transversal CNOT gates preserve the logical subspace iff a qubit stabilizer code is CSS \NoCaseChange{\protect\cite{cite761,cite398}}.
Any \(\llbracket n,k,d\rrbracket \) stabilizer code can be mapped onto a \(\llbracket 2n,2k,\geq d\rrbracket \) \flmRefsHyperref{code:two_block_quantum}{two-block CSS code} via \flmRefsHyperref{ref436}{symplectic doubling}, which preserves geometric locality of a code up to a constant factor.
For any non-CSS qubit stabilizer code \(\mathsf{C}\), there exists a CSS code \(\mathsf{C}^{\prime}\) such that \(\mathsf{C} = DQ\mathsf{C}^{\prime}\), where \(D\) is a diagonal Clifford operator, and where \(Q\) is an element of an XP stabilizer group \NoCaseChange{\protect\cite[{Prop. B.3.1}]{cite768}}.
There is a holographic relation between qubit CSS codes describing CFTs and qubit stabilizer codes describing path integrals over certain topologies \NoCaseChange{\protect\cite{cite3612}}.

\item\relax
\flmRefsHyperref[eczindexfamilyrel]{code:self_dual_css}{Self-dual CSS code} --- Any \(\llbracket n,k,d\rrbracket \) qubit stabilizer code maps to a \(\llbracket 4n,2k,2d\rrbracket \) self-dual CSS code by applying the BLT mapping and concatenating each qubit pair with the \(\llbracket 4,2,2\rrbracket \) code \NoCaseChange{\protect\cite[{Corr. 2}]{cite795}\protect\cite[{Corr. 1}]{cite1432}}. The BLT mapping proceeds by first concatenating each qubit with the \flmRefsHyperref{code:tetron}{tetron code} to obtain an intermediate \(\llbracket 2n,k,2d\rrbracket _{f}\) Majorana stabilizer code.
\item\relax
\flmRefsHyperref[eczindexfamilyrel]{code:qubit_subsystem_stabilizer}{Subsystem qubit stabilizer code} --- Subsystem qubit stabilizer codes reduce to qubit stabilizer codes when there are no gauge qubits.
An \(\llbracket n,k,d\rrbracket \) qubit stabilizer code can be converted into an \flmRefsHyperref{ref65}{order} \(\llbracket O(\ell \delta n),k,\Omega(d/w)\rrbracket \) subsystem qubit stabilizer code with weight-three gauge operators via the wire-code mapping \NoCaseChange{\protect\cite{cite490}}, which uses \flmRefsHyperref{ref491}{weight reduction}. 
Here, \(w\) and \(\delta\) are the weight and degree of the input code's Tanner graph, while \(\ell\) is the length of the longest edge of a particular embedding of that graph.

\end{eczvaluelist}
\eczhbkcontributors{ Qingfeng (Kee) Wang, Lane G. Gunderman, Leonid Pryadko, Daniel Gottesman, \eczhuVVA }
\endeczcode

\eczcode{qudit_gnu_permutation_invariant}{Qudit GNU PI code}{~\NoCaseChange{\protect\cite{cite2948}}}
\codefieldsection{Description}
Extension of the GNU PI codes to those encoding logical qudits into physical qubits.
Codewords can be expressed as superpositions of \flmRefsHyperref{ref526}{Dicke states} with coefficients given by square roots of polynomial coefficients, with the case of binomial coefficients reducing to the GNU PI codes.

\codefieldsection{Parents}
\begin{eczvaluelist}
\item\relax
\flmRefsHyperref[eczindexfamilyrel]{code:qubit_permutation_invariant}{PI qubit code}\item\relax
\flmRefsHyperref[eczindexfamilyrel]{code:ampdamp}{Amplitude-damping (AD) code} --- Qudit GNU PI codes protect against \flmRefsHyperref{ref498}{AD} errors.
\end{eczvaluelist}
\codefieldsection{Child}
\begin{eczvaluelist}
\item\relax
\flmRefsHyperref[eczindexfamilyrel]{code:gnu_permutation_invariant}{GNU PI code} --- Qudit GNU codes encoding logical qubits reduce to GNU codes.
\end{eczvaluelist}
\eczhbkcontributors{ \eczhuVVA }
\endeczcode

\eczcode{rbh}{Raussendorf-Bravyi-Harrington (RBH) cluster-state code}{~\NoCaseChange{\protect\cite{cite4396,cite3533,cite3534}}}
\codefieldsection{Alternative Names}
\begin{eczvaluelist}
\item\relax Raussendorf-Harrington-Goyal (RHG) cluster-state code
\end{eczvaluelist}
\eczhIndexCodeAliasName{rbh}{Raussendorf-Harrington-Goyal (RHG) cluster-state code}
\codefieldsection{Description}
A three-dimensional cluster-state code defined on the bcc lattice (i.e., a cubic lattice with qubits on edges and faces).

The MBQC version of the code is defined as the unique ground state of a certain code Hamiltonian. This state is the resource state used in the first MBQC scheme \NoCaseChange{\protect\cite{cite3533,cite3534}}.
It encodes the temporal gate operations on the surface code into a third spatial dimension.

Addition of certain boundary Hamiltonians yields a degenerate ground-state space that serves as an example of a symmetry-protected self-correcting memory \NoCaseChange{\protect\cite{cite3059}}.

\codefieldsection{Protection}
Exhibits symmetry-protected self-correction \NoCaseChange{\protect\cite{cite3059}}. The energy barrier for symmetry-preserving excitations outside of the code space grows linearly with the lattice width. When the system is coupled locally to a thermal bath respecting the symmetry and below a critical temperature, the memory time grows exponentially with the lattice width.

\codefieldsection{Gates}
\begin{eczvaluelist}
\item\relax The computation is encoded in a pre-determined fashion via topological features of the lattice, such as boundaries, defects, or twists. For example, qubits may be encoded in 2D defects along slices of the surface code, and \flmRefsHyperref{ref409}{Clifford gates} are encoded by spatially braiding the defects along the 3rd dimension. Non-Clifford gates are performed by inserting \flmRefsHyperref{ref409}{non-Clifford} states into particular \textit{singular} qubits.
To perform the computation, qubits along the extra dimension are measured, e.g., along one two-dimensional slice per time step. This effectively teleports the logical information into the remaining unmeasured portion of the cluster state.
\end{eczvaluelist}
\codefieldsection{Decoding}
\begin{eczvaluelist}
\item\relax MBQC syndrome extraction consists of single-qubit measurements and classical post-processing. The six \(X\)-measurements of qubits on the faces of a cube of the bcc lattice multiply to the product of the six cluster-state stabilizers whose vertices are on the faces of the cube. Such measurements, if done on a 2D slice, also yield \(Z\)-type syndromes on the next slice.
\item\relax Minimum weight perfect-matching (MWPM) \NoCaseChange{\protect\cite{cite480,cite3858}} (based on work by Edmonds on finding a matching in a graph \NoCaseChange{\protect\cite{cite3859,cite3860}}).
\end{eczvaluelist}
\codefieldsection{Threshold}
\begin{eczvaluelist}
\item\relax For the topological 3D cluster-state scheme of Raussendorf-Harrington-Goyal, the reported threshold is \(1.4\%\) for local depolarizing noise and \(0.11\%\) per location in a circuit-level model with preparation, gate, storage, and measurement errors \NoCaseChange{\protect\cite{cite3533}}.
\item\relax Various thresholds for optical quantum computing schemes with RBH codes \NoCaseChange{\protect\cite{cite4397,cite4398}}.
\item\relax \(0.75\%\) for preparation, gate, storage, and measurement errors \NoCaseChange{\protect\cite{cite3534}}.
\item\relax \(24.9\%\) under erasure noise \NoCaseChange{\protect\cite{cite4399}}.
\item\relax Concatenation of the RBH code with small codes such as the \(\llbracket 2,1\rrbracket \) repetition code, \(\llbracket 4,1,1,2\rrbracket \) subsystem code, or Steane code can improve thresholds \NoCaseChange{\protect\cite{cite3248}}.
\end{eczvaluelist}
\codefieldsection{Notes}
\begin{eczvaluelist}
\item\relax Introduction to MBQC protocols with the RBH state \NoCaseChange{\protect\cite{cite3951}}.
\item\relax Blind quantum computation is possible with the RBH state \NoCaseChange{\protect\cite{cite4400}}.
\end{eczvaluelist}
\codefieldsection{Parents}
\begin{eczvaluelist}
\item\relax
\flmRefsHyperref[eczindexfamilyrel]{code:cluster_state}{Cluster-state code}\item\relax
\flmRefsHyperref[eczindexfamilyrel]{code:qldpc}{Qubit QLDPC code}\item\relax
\flmRefsHyperref[eczindexfamilyrel]{code:3d_stabilizer}{3D lattice stabilizer code}\item\relax
\flmRefsHyperref[eczindexfamilyrel]{code:walker_wang}{Walker-Wang model code} --- The Walker-Wang model code reduces to the RBH cluster-state code when the input category \(\mathcal{C}\) is that of the surface code \NoCaseChange{\protect\cite[{Sec. V.A}]{cite478}}.
\end{eczvaluelist}
\codefieldsection{Cousins}
\begin{eczvaluelist}
\item\relax
\flmRefsHyperref[eczindexfamilyrel]{code:symmetry_protected_self_correct}{Symmetry-protected self-correcting quantum code} --- The RBH code can exhibit self-correction protected by a certain symmetry.
\item\relax
\flmRefsHyperref[eczindexfamilyrel]{code:spt}{Symmetry-protected topological (SPT) code} --- In 3D, cluster states belong to SPT phases protected by higher-form symmetries \NoCaseChange{\protect\cite{cite3092}} and enable universal fault-tolerant MBQC \NoCaseChange{\protect\cite{cite3093}}.
\item\relax
\flmRefsHyperref[eczindexfamilyrel]{code:surface}{Kitaev surface code} --- The RBH state encodes the temporal gate operations on the surface code into a third spatial dimension \NoCaseChange{\protect\cite{cite3533,cite3534}}. In addition, one possible 2D boundary of the RBH code is effectively a 2D toric code.
\item\relax
\flmRefsHyperref[eczindexfamilyrel]{code:bcc}{Body-centered cubic (bcc) lattice} --- The RBH code is defined on the bcc lattice.
\item\relax
\flmRefsHyperref[eczindexfamilyrel]{code:qubit_concatenated}{Concatenated qubit code} --- Concatenation of the RBH code with small codes such as the \(\llbracket 2,1\rrbracket \) repetition code, \(\llbracket 4,1,1,2\rrbracket \) subsystem code, or Steane code can improve thresholds \NoCaseChange{\protect\cite{cite3248}}.
\item\relax
\flmRefsHyperref[eczindexfamilyrel]{code:bacon_shor_4}{\(\llbracket 4,1,1,2\rrbracket \) Four-qubit subsystem code} --- Concatenation of the RBH code with small codes such as the \(\llbracket 2,1\rrbracket \) repetition code, \(\llbracket 4,1,1,2\rrbracket \) subsystem code, or Steane code can improve thresholds \NoCaseChange{\protect\cite{cite3248}}.
\item\relax
\flmRefsHyperref[eczindexfamilyrel]{code:steane}{\(\llbracket 7,1,3\rrbracket \) Steane code} --- Concatenation of the RBH code with small codes such as the \(\llbracket 2,1\rrbracket \) repetition code, \(\llbracket 4,1,1,2\rrbracket \) subsystem code, or Steane code can improve thresholds \NoCaseChange{\protect\cite{cite3248}}.
\item\relax
\flmRefsHyperref[eczindexfamilyrel]{code:hybrid_cat}{Hybrid cat code} --- Hybrid cat codes can be concatenated with RBH codes \NoCaseChange{\protect\cite{cite4401}}.
\item\relax
\flmRefsHyperref[eczindexfamilyrel]{code:quantum_parity}{Quantum parity code (QPC)} --- QPCs can be concatenated with RBH codes \NoCaseChange{\protect\cite{cite4048}}.
\item\relax
\flmRefsHyperref[eczindexfamilyrel]{code:3d_subsystem_color}{3D subsystem color code} --- Different stabilizer Hamiltonians of the 3D subsystem color code correspond to distinct SPT phases; one ungauges to three decoupled copies of the RBH model \NoCaseChange{\protect\cite{cite466}}.
The RBH code for a certain boundary Hamiltonian is dual to the 3D subsystem color code \NoCaseChange{\protect\cite[{Sec. IV.C.1}]{cite3059}}.

\end{eczvaluelist}
\eczhbkcontributors{ Yi-Ting (Rick) Tu, \eczhuVVA }
\endeczcode

\eczcode{reinforcement_learning}{Reinforcement-learning quantum code}{~\NoCaseChange{\protect\cite{cite4402,cite2936,cite3965}}}
\codefieldsection{Description}
An approximate qubit code obtained from a numerical optimization involving a reinforcement learning agent.

\codefieldsection{Protection}
Depends on the parameter being optimized.

\codefieldsection{Rate}
Neural network codes can be obtained by optimizing the coherent information \NoCaseChange{\protect\cite{cite2936}}.
\codefieldsection{Encoding}
\begin{eczvaluelist}
\item\relax Both codes and encoding circuits can be obtained via a reinforcement learning agent \NoCaseChange{\protect\cite{cite3184}}.
\end{eczvaluelist}
\codefieldsection{Notes}
\begin{eczvaluelist}
\item\relax See a review on the use of artificial intelligence in quantum error correction \NoCaseChange{\protect\cite{cite4403}}.
\end{eczvaluelist}
\codefieldsection{Parents}
\begin{eczvaluelist}
\item\relax
\flmRefsHyperref[eczindexfamilyrel]{code:qubits_into_qubits}{Qubit code}\item\relax
\flmRefsHyperref[eczindexfamilyrel]{code:approximate_qecc}{Approximate quantum error-correcting code (AQECC)}\end{eczvaluelist}
\codefieldsection{Cousins}
\begin{eczvaluelist}
\item\relax
\flmRefsHyperref[eczindexfamilyrel]{code:surface}{Kitaev surface code} --- Reinforcement learners can be used to optimize the geometry of the surface code to be more suited to a noise channel \NoCaseChange{\protect\cite{cite3965}}.
\item\relax
\flmRefsHyperref[eczindexfamilyrel]{code:numopt}{Numerically optimized bosonic code} --- Numerically optimized bosonic codes can be obtained via reinforcement learning \NoCaseChange{\protect\cite{cite4404,cite4405}}.
\item\relax
\flmRefsHyperref[eczindexfamilyrel]{code:small_distance_qubit_stabilizer}{Small-distance qubit stabilizer code} --- 13 inequivalent \(\llbracket 9,3,3\rrbracket \) codes, along with others, have been found via reinforcement learning \NoCaseChange{\protect\cite{cite3184}}.
\item\relax
\flmRefsHyperref[eczindexfamilyrel]{code:stab_5_1_3}{\(\llbracket 5,1,3\rrbracket \) Five-qubit perfect code} --- Various five-qubit codes, numerically obtained through variational techniques, can outperform the five-qubit perfect code against depolarizing noise \NoCaseChange{\protect\cite{cite3324}}.
\item\relax
\flmRefsHyperref[eczindexfamilyrel]{code:quantum_lego}{Tensor-network code} --- Quantum Lego and more general tensor-network code optimization for biased noise can be done using reinforcement learning \NoCaseChange{\protect\cite{cite2648,cite2649}}.
\item\relax
\flmRefsHyperref[eczindexfamilyrel]{code:ame}{Perfect-tensor code} --- Reinforcement learning \NoCaseChange{\protect\cite{cite2936}} and graph-based optimizers like PyTheus \NoCaseChange{\protect\cite{cite2937}} can be used to find AME states.
\item\relax
\flmRefsHyperref[eczindexfamilyrel]{code:ampdamp_numopt}{Numerically optimized four-qubit AD code} --- Numerically optimized four-qubit AD codes can be obtained from a machine-learning optimization \NoCaseChange{\protect\cite{cite3996}}.
\item\relax
\flmRefsHyperref[eczindexfamilyrel]{code:hypergraph_product}{Hypergraph product (HGP) code} --- Using reinforcement learning, hypergraph product codes can be further optimized against the erasure channel \NoCaseChange{\protect\cite{cite3781}} and can be weight reduced while maintaining distance \NoCaseChange{\protect\cite{cite3787}}.
\end{eczvaluelist}
\eczhbkcontributors{ \eczhuVVA }
\endeczcode

\eczcode{majorana_reed_muller}{RM Majorana code}{~\NoCaseChange{\protect\cite{cite566,cite1783}}}
\codefieldsection{Description}
A Majorana stabilizer code constructed from a self-orthogonal RM code.
These codes have the additional property that the global fermion parity is fixed in the codespace. 
Logical measurements are reduced to parity measurements of some subset of Majorana fermions in the code.

\codefieldsection{Protection}
Code parameters are \(\llbracket 2^{m-1},2^{m-1} - \sum_{j=0}^{r} {m \choose j},2^{r+1}\rrbracket _{f}\), which are optimal per Majorana LP bounds for \(r = 0\) and \(m \geq 1\), and \(r=1\) and \(m=4,5\) \NoCaseChange{\protect\cite[{Sec. 3.4}]{cite564}}.

\codefieldsection{Parent}
\begin{eczvaluelist}
\item\relax
\flmRefsHyperref[eczindexfamilyrel]{code:majorana_stab}{Majorana stabilizer code}\end{eczvaluelist}
\codefieldsection{Child}
\begin{eczvaluelist}
\item\relax
\flmRefsHyperref[eczindexfamilyrel]{code:majorana_hamming}{\(\llbracket 2^{m-1},2^{m-1}-m-1,4\rrbracket _{f}\) Hamming Majorana code} --- A Hamming Majorana code is constructed from a first-order RM code (whose dual is the extended Hamming code).
\end{eczvaluelist}
\codefieldsection{Cousins}
\begin{eczvaluelist}
\item\relax
\flmRefsHyperref[eczindexfamilyrel]{code:reed_muller}{Reed-Muller (RM) code} --- RM Majorana codes are constructed from self-orthogonal RM codes.
\item\relax
\flmRefsHyperref[eczindexfamilyrel]{code:quantum_reed_muller}{Quantum Reed-Muller (RM) code} --- RM Majorana (quantum RM) codes are designed for fermionic (qubit) noise.
\end{eczvaluelist}
\eczhbkcontributors{ Rui Okada, \eczhuVVA }
\endeczcode

\eczcode{rotated_surface}{Rotated surface code}{~\NoCaseChange{\protect\cite{cite425,cite3427,cite1316,cite3385}}}
\codefieldsection{Alternative Names}
\begin{eczvaluelist}
\item\relax Checkerboard code
\item\relax Medial surface code
\item\relax Rectified surface code
\end{eczvaluelist}
\eczhIndexCodeAliasName{rotated_surface}{Checkerboard code}
\eczhIndexCodeAliasName{rotated_surface}{Medial surface code}
\eczhIndexCodeAliasName{rotated_surface}{Rectified surface code}
\codefieldsection{Description}
Variant of the surface code obtained by taking the medial graph of the surface code lattice (treated as a graph) and applying a procedure to construct the check operators \NoCaseChange{\protect\cite{cite425,cite426}\protect\cite[{Fig. 8}]{cite427}}.
On a square lattice, this amounts to a rotation by 45 degrees such that qubits are on vertices, and both \(X\)- and \(Z\)-type check operators occupy plaquettes in an alternating checkerboard pattern.

Stabilizer generators for this code on a square lattice are shown in \flmRefsCref{ref4406}.
\begin{flmFloat}{figure}{NumCap}\includegraphics[width=175.49999319685043bp,max width=\linewidth]{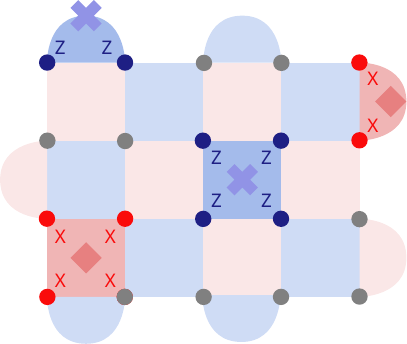}\caption{
  Stabilizer generators of a 2D rotated surface code with open boundaries.
  The generators are weight-four (four-body) operators on the corners of squares in the bulk and weight-two (two-body) operators on the boundaries.
  Red regions correspond to \(X\) operators while blue regions correspond to \(Z\) operators.}\label{ref4406}\end{flmFloat}

\codefieldsection{Protection}
The \(\llbracket L^2,1,L\rrbracket \) planar rotated surface code variant \NoCaseChange{\protect\cite{cite425}} includes the \(\llbracket 9,1,3\rrbracket \) \flmRefsHyperref{code:surface-17}{surface-17} code, named as such because 8 ancilla qubits are used for check operator measurements alongside the 9 physical qubits.
The \(\llbracket L^2,2,L\rrbracket \) periodic variant is a rotated toric or checkerboard code, whose smallest example is the \(\llbracket 4,2,2\rrbracket \) code.
Non-bipartite rotated toric codes with odd distance include the family \(\llbracket t^2+(t+1)^2,1,2t+1\rrbracket \) \NoCaseChange{\protect\cite[{Exam. 4}]{cite1316}}.

\codefieldsection{Encoding}
\begin{eczvaluelist}
\item\relax Unitary encoder based on code conversion between rotated and regular surface codes \NoCaseChange{\protect\cite{cite4407}}.
\end{eczvaluelist}
\codefieldsection{Transversal and Permutation-Based Gates}
\begin{eczvaluelist}
\item\relax Fold-transversal \(S\) gate \NoCaseChange{\protect\cite{cite783,cite748}}.
\end{eczvaluelist}
\codefieldsection{Gates}
\begin{eczvaluelist}
\item\relax Injection of the \(|Y\rangle\) state \NoCaseChange{\protect\cite{cite4408}}.
\end{eczvaluelist}
\codefieldsection{Decoding}
\begin{eczvaluelist}
\item\relax Only certain syndrome extraction schedules are \flmRefsHyperref{ref3496}{distance-preserving} \NoCaseChange{\protect\cite{cite3385}}.
\item\relax Local neural-network using 3D convolutions, combined with a separate global decoder \NoCaseChange{\protect\cite{cite3893}}.
\item\relax Iterative CNOT decoder \NoCaseChange{\protect\cite{cite4409}}.
\item\relax Fault-tolerant BP (FTBP) decoder \NoCaseChange{\protect\cite{cite4410}}.
\end{eczvaluelist}
\codefieldsection{Fault Tolerance}
\begin{eczvaluelist}
\item\relax A particular choice of CNOT gates during syndrome extraction is required to avoid \flmRefsHyperref{ref3496}{hook errors} and be fault-tolerant to syndrome qubit errors \NoCaseChange{\protect\cite{cite480,cite2522,cite3385}}.
\end{eczvaluelist}
\codefieldsection{Threshold}
\begin{eczvaluelist}
\item\relax Thresholds for various amounts of erasure, Pauli, correlated, and measurement noise are known \NoCaseChange{\protect\cite{cite4411,cite4412}}.
\end{eczvaluelist}
\codefieldsection{Realizations}
\begin{eczvaluelist}
\item\relax Teleportation transition of distance-seven rotated surface-code states on a 125-qubit superconducting processor \NoCaseChange{\protect\cite{cite4413}}.
\end{eczvaluelist}
\codefieldsection{Parents}
\begin{eczvaluelist}
\item\relax
\flmRefsHyperref[eczindexfamilyrel]{code:surface}{Kitaev surface code} --- The lattice of the rotated surface code can be obtained by taking the medial graph of the surface code lattice (treated as a graph) and applying a procedure to construct the check operators \NoCaseChange{\protect\cite{cite425,cite426}\protect\cite[{Fig. 8}]{cite427}}. Applying the quantum Tanner transformation to the surface code yields the rotated surface code \NoCaseChange{\protect\cite{cite3954,cite3955}}. The rotated surface code presents certain savings over the original surface code \NoCaseChange{\protect\cite{cite3956}}.
\item\relax
\flmRefsHyperref[eczindexfamilyrel]{code:quantum_tanner}{Quantum Tanner code} --- Applying the quantum Tanner transformation to the surface code yields the rotated surface code \NoCaseChange{\protect\cite{cite3954,cite3955}}.
\end{eczvaluelist}
\codefieldsection{Children}
\begin{eczvaluelist}
\item\relax
\flmRefsHyperref[eczindexfamilyrel]{code:css_4_1_2}{\(\llbracket 4,1,2\rrbracket \) Leung-Nielsen-Chuang-Yamamoto (LNCY) code} --- The \(\llbracket 4,1,2\rrbracket \) LNCY code is a small planar rotated surface code \NoCaseChange{\protect\cite{cite3256,cite3253,cite3254,cite3255}}.
\item\relax
\flmRefsHyperref[eczindexfamilyrel]{code:stab_4_2_2}{\(\llbracket 4,2,2\rrbracket \) Four-qubit code} --- The \(\llbracket 4,2,2\rrbracket \) code is the smallest rotated toric code \NoCaseChange{\protect\cite{cite454}}.
\item\relax
\flmRefsHyperref[eczindexfamilyrel]{code:stab_5_1_2}{\(\llbracket 5,1,2\rrbracket \) rotated surface code}\item\relax
\flmRefsHyperref[eczindexfamilyrel]{code:surface-17}{\(\llbracket 9,1,3\rrbracket \) Surface-17 code}\end{eczvaluelist}
\codefieldsection{Cousins}
\begin{eczvaluelist}
\item\relax
\flmRefsHyperref[eczindexfamilyrel]{code:hypergraph_product}{Hypergraph product (HGP) code} --- Periodic checkerboard or rotated-toric codes on the same lattice can be obtained from hypergraph products of two cyclic linear binary codes with palindromic check polynomials \NoCaseChange{\protect\cite[{Sec. IV.D}]{cite1316}}.
\item\relax
\flmRefsHyperref[eczindexfamilyrel]{code:binary_cyclic}{Cyclic linear binary code} --- Periodic checkerboard or rotated-toric codes on the same lattice can be obtained from hypergraph products of two cyclic linear binary codes with palindromic check polynomials \NoCaseChange{\protect\cite[{Sec. IV.D}]{cite1316}}.
\item\relax
\flmRefsHyperref[eczindexfamilyrel]{code:heavy_hex}{Heavy-hexagon code} --- A rotated surface code can be mapped onto a heavy square lattice, resulting in a code similar to the heavy-hexagon code \NoCaseChange{\protect\cite{cite3598}}.
\item\relax
\flmRefsHyperref[eczindexfamilyrel]{code:hierarchical}{Hierarchical code} --- Hierarchical codes are concatenations of constant-rate QLDPC codes with rotated surface codes.
\item\relax
\flmRefsHyperref[eczindexfamilyrel]{code:yoked_surface}{Yoked surface code} --- Yoked surface codes are concatenations of QMDPC codes with rotated surface codes.
\item\relax
\flmRefsHyperref[eczindexfamilyrel]{code:plaquette_ising}{Plaquette Ising code} --- The 2D plaquette Ising model can be thought of as the rotated surface code whose \(X\)-type stabilizer generators have been converted to \(Z\)-type stabilizer generators.
\item\relax
\flmRefsHyperref[eczindexfamilyrel]{code:cat_concatenated}{Concatenated cat code} --- Cat codes have been concatenated with rotated surface codes \NoCaseChange{\protect\cite{cite4072}}.
\item\relax
\flmRefsHyperref[eczindexfamilyrel]{code:gkp_surface_concatenated}{GKP-surface code} --- GKP codes have been concatenated with rotated surface codes \NoCaseChange{\protect\cite{cite417,cite418,cite419,cite420}}.
\item\relax
\flmRefsHyperref[eczindexfamilyrel]{code:quantum_lego}{Tensor-network code} --- A tensor-network based modification of the rotated surface code improves performance against depolarizing noise by \(\approx 2\%\) \NoCaseChange{\protect\cite{cite3101}}.
\item\relax
\flmRefsHyperref[eczindexfamilyrel]{code:bvc}{Ball-Verstraete-Cirac (BVC) code} --- An appropriately chosen stabilizer generator set for the BVC code contains the stabilizers of the rotated surface code \NoCaseChange{\protect\cite{cite3432}}.
\item\relax
\flmRefsHyperref[eczindexfamilyrel]{code:toric}{Toric code} --- Rotating the square lattice by \(\pi/4\) and choosing periodicity vectors on the rotated checkerboard lattice yields periodic checkerboard or rotated-toric variants with the same \(\llbracket L^2,2,L\rrbracket \) scaling, as well as non-bipartite odd-distance families with parameters \(\llbracket t^2+(t+1)^2,1,2t+1\rrbracket \) \NoCaseChange{\protect\cite[{Sec. III}]{cite1316}}.
\item\relax
\flmRefsHyperref[eczindexfamilyrel]{code:3d_surface}{3D surface code} --- There exists a rotated version of the 3D surface code, akin to the (2D) rotated surface code \NoCaseChange{\protect\cite{cite2626}}.
\item\relax
\flmRefsHyperref[eczindexfamilyrel]{code:xzzx}{XZZX surface code} --- The XZZX code is obtained from the rotated surface code by applying Hadamard gates on a subset of qubits such that \(XXXX\) and \(ZZZZ\) generators are both mapped to \(XZXZ\). Both rotated and XZZX codes offer improved performance over the original surface code for biased noise \NoCaseChange{\protect\cite{cite4414}}.
\item\relax
\flmRefsHyperref[eczindexfamilyrel]{code:compass_model}{Compass code} --- The surface-density compass code family interpolates between Bacon-Shor codes and rotated surface codes.
\item\relax
\flmRefsHyperref[eczindexfamilyrel]{code:subsystem_rotated_surface}{Subsystem rotated surface code} --- Subsystem rotated surface codes are subsystem versions of rotated surface codes.
\end{eczvaluelist}
\eczhbkcontributors{ Eric Huang, Marcus P da Silva, \eczhuVVA }
\endeczcode

\eczcode{floquet_xyz_ruby}{Ruby Floquet code}{~\NoCaseChange{\protect\cite{cite533,cite542}}}
\codefieldsection{Alternative Names}
\begin{eczvaluelist}
\item\relax Floquet color code
\item\relax Ruby Floquet color code
\end{eczvaluelist}
\eczhIndexCodeAliasName{floquet_xyz_ruby}{Floquet color code}
\eczhIndexCodeAliasName{floquet_xyz_ruby}{Ruby Floquet color code}
\codefieldsection{Description}
2D Floquet code whose qubits are placed on vertices of a ruby tiling, with weight-two Pauli check operators on \(x\)-, \(y\)-, and \(z\)-labeled edges \NoCaseChange{\protect\cite{cite542}}.
The code admits two different measurement schedules, the XYZ ruby schedule and the color-code schedule.

One third of the time during the XYZ ruby measurement schedule, its ISG is that of the 6.6.6 color code concatenated with a three-qubit repetition code.
Together, all ISGs generate the gauge group of the 3F subsystem code.

A different three-round color-code schedule \NoCaseChange{\protect\cite{cite533}} admits ISGs are FDLQC-equivalent to the 2D color code; in one round, the ISG is exactly the color code concatenated with a three-qubit repetition code.
The color-code schedule has a \(\mathbb{Z}_3\) automorphism of the dynamically generated logical operators, while the rewinding schedule \(012102\) and another six-round schedule both trivialize that automorphism.
A parent stabilizer code for this schedule is FDLQC-equivalent to two copies of the 2D color code \NoCaseChange{\protect\cite{cite533}}.

\codefieldsection{Rate}
On a torus, the color-code schedule encodes four logical qubits, two of which are dynamically generated relative to the underlying subsystem code \NoCaseChange{\protect\cite{cite533}}.
\codefieldsection{Gates}
\begin{eczvaluelist}
\item\relax In the round whose ISG is the 2D color code concatenated with a three-qubit repetition code, the usual transversal logical Clifford gates of the color code remain available \NoCaseChange{\protect\cite{cite533}}.
\end{eczvaluelist}
\codefieldsection{Fault Tolerance}
\begin{eczvaluelist}
\item\relax Pairs of consecutive ISGs of the color-code schedule are locally reversible, and the paper argues that this suggests a non-zero fault-tolerant threshold \NoCaseChange{\protect\cite{cite533}}.
\end{eczvaluelist}
\codefieldsection{Threshold}
\begin{eczvaluelist}
\item\relax Circuit-level noise: \(\approx 0.18\%\) using BP-OSD decoder \NoCaseChange{\protect\cite{cite542}}.
\end{eczvaluelist}
\codefieldsection{Parent}
\begin{eczvaluelist}
\item\relax
\flmRefsHyperref[eczindexfamilyrel]{code:floquet}{Hastings-Haah Floquet code}\end{eczvaluelist}
\codefieldsection{Cousins}
\begin{eczvaluelist}
\item\relax
\flmRefsHyperref[eczindexfamilyrel]{code:triangular_color}{Honeycomb (6.6.6) color code} --- One third of the time during the XYZ ruby measurement schedule, the ISG is that of the 6.6.6 color code concatenated with a three-qubit repetition code.
\item\relax
\flmRefsHyperref[eczindexfamilyrel]{code:2d_color}{2D color code} --- Each ISG of the color-code schedule is FDLQC-equivalent to the 2D color code, and a parent stabilizer code is FDLQC-equivalent to two copies of the 2D color code \NoCaseChange{\protect\cite{cite533}}.
\item\relax
\flmRefsHyperref[eczindexfamilyrel]{code:quantum_repetition}{Quantum repetition code} --- One third of the time during the XYZ ruby measurement schedule, the ISG is that of the 6.6.6 color code concatenated with a three-qubit repetition code. One round of the color-code schedule is exactly the 2D color code concatenated with a three-qubit repetition code \NoCaseChange{\protect\cite{cite533}}.
\item\relax
\flmRefsHyperref[eczindexfamilyrel]{code:honeycomb}{Honeycomb tiling} --- The ruby Floquet code is defined on the ruby tiling.
\item\relax
\flmRefsHyperref[eczindexfamilyrel]{code:subsystem_three_fermion}{Three-fermion (3F) subsystem code} --- Together, all ISGs of the ruby Floquet code generate the gauge group of the 3F subsystem code.
\end{eczvaluelist}
\eczhbkcontributors{ Julio Carlos Magdalena De La Fuente, \eczhuVVA }
\endeczcode

\eczcode{subsystem_hypergraph}{Sarvepalli-Brown subsystem code}{~\NoCaseChange{\protect\cite{cite660}}}
\codefieldsection{Description}
Member of a family of non-CSS subsystem codes constructed from hypergraphs that satisfy certain assumptions \NoCaseChange{\protect\cite[{Construction C}]{cite660}}.

\codefieldsection{Parent}
\begin{eczvaluelist}
\item\relax
\flmRefsHyperref[eczindexfamilyrel]{code:qubit_subsystem_stabilizer}{Subsystem qubit stabilizer code}\end{eczvaluelist}
\codefieldsection{Child}
\begin{eczvaluelist}
\item\relax
\flmRefsHyperref[eczindexfamilyrel]{code:five_squares}{Generalized five-squares code} --- Generalized five-squares codes are special cases of Sarvepalli-Brown subsystem codes \NoCaseChange{\protect\cite[{Sec. II.B}]{cite661}}.
\end{eczvaluelist}
\codefieldsection{Cousin}
\begin{eczvaluelist}
\item\relax
\flmRefsHyperref[eczindexfamilyrel]{code:color}{Color code} --- Sarvepalli-Brown subsystem codes can be derived from color codes \NoCaseChange{\protect\cite[{Thm. 3}]{cite660}}.
\end{eczvaluelist}
\eczhbkcontributors{ \eczhuVVA }
\endeczcode

\eczcode{self_dual_css}{Self-dual CSS code}{}
\codefieldsection{Alternative Names}
\begin{eczvaluelist}
\item\relax Weakly self-dual CSS code
\item\relax Symmetric CSS code
\item\relax Self-orthogonal CSS code
\item\relax Homogeneous CSS code
\end{eczvaluelist}
\eczhIndexCodeAliasName{self_dual_css}{Weakly self-dual CSS code}
\eczhIndexCodeAliasName{self_dual_css}{Symmetric CSS code}
\eczhIndexCodeAliasName{self_dual_css}{Self-orthogonal CSS code}
\eczhIndexCodeAliasName{self_dual_css}{Homogeneous CSS code}
\codefieldsection{Description}
A qubit CSS code for which transversal Hadamard is a logical operation.
Equivalently, the stabilizer group is preserved by exchanging \(X\)-type and \(Z\)-type Pauli operators.

Specializing the CSS construction to the case when \(C_Z=[n,k]\) is dual-containing and \(C_X=C_Z\) yields an \(\llbracket n,2k-n\rrbracket \) self-dual qubit CSS code (a.k.a. weakly self-dual, symmetric \NoCaseChange{\protect\cite{cite788}}, self-orthogonal, or homogeneous \NoCaseChange{\protect\cite{cite4415}} qubit CSS code).
Its \(X\)-type and \(Z\)-type stabilizers are identically supported, and transversal Hadamard preserves the stabilizer group.

Self-dual CSS codes split into \textit{normal} and \textit{hyperbolic} classes depending on whether transversal Hadamard acts as logical Hadamards or as pairwise logical swaps in a suitable logical basis \NoCaseChange{\protect\cite[{Sec. III}]{cite101}}.
Hyperbolic codes necessarily encode an even number of logical qubits and have even blocklength and distance \NoCaseChange{\protect\cite[{Sec. III}]{cite101}}.

\codefieldsection{Magic}
Normal and hyperbolic self-dual CSS codes yield magic-state distillation protocols with asymptotically constant space overhead and yield parameter \(\gamma \to 1^{+}\) \NoCaseChange{\protect\cite{cite705}\protect\cite[{Thms. 4.1 and 4.2}]{cite101}}.
\codefieldsection{Transversal and Permutation-Based Gates}
\begin{eczvaluelist}
\item\relax Self-dual CSS codes admit a transversal Hadamard gate. There are criteria for when such codes realize logical gates from tensor products of \(S\) and \(S^{\dagger}\) gates \NoCaseChange{\protect\cite{cite784}}.
\item\relax Diagonal transversal \flmRefsHyperref{ref409}{Clifford gates} on \(\ell\) codeblocks of a CSS code form \(Sp(2\ell,\mathbb{F}_2)\) for self-dual CSS codes \NoCaseChange{\protect\cite{cite738}}.
\item\relax A self-dual weakly doubly even \(\llbracket n,1,d\rrbracket \) CSS code admits a partitioned transversal physical \(S\) gate that realizes \(\overline{S}^m\), where \(m=|M^+|-|M^-| \pmod 4\); for odd \(m\), together with transversal Hadamard and CNOT, this yields the full logical Clifford group transversally \NoCaseChange{\protect\cite{cite731}\protect\cite[{Lemma 4}]{cite760}}.
\end{eczvaluelist}
\codefieldsection{Fault Tolerance}
\begin{eczvaluelist}
\item\relax Any self-dual CSS code with bounded-weight stabilizer generators admits flag fault-tolerant syndrome extraction \NoCaseChange{\protect\cite{cite3220}}.
\item\relax Triorthogonal codes realizing logical \(T\) gates using only physical \(T\) gates can be paired up with self-dual CSS codes to yield a transversal CNOT gate and universal fault-tolerant gates using Steane error correction \NoCaseChange{\protect\cite{cite788}}.
\end{eczvaluelist}
\codefieldsection{Parent}
\begin{eczvaluelist}
\item\relax
\flmRefsHyperref[eczindexfamilyrel]{code:qubit_css}{Qubit CSS code} --- Self-dual CSS codes are qubit CSS codes whose stabilizer group is preserved by transversal Hadamard.
\end{eczvaluelist}
\codefieldsection{Children}
\begin{eczvaluelist}
\item\relax
\flmRefsHyperref[eczindexfamilyrel]{code:iceberg}{\(\llbracket 2m,2m-2,2\rrbracket \) error-detecting code}\item\relax
\flmRefsHyperref[eczindexfamilyrel]{code:stab_16_6_4}{\(\llbracket 16,6,4\rrbracket \) Tesseract color code}\item\relax
\flmRefsHyperref[eczindexfamilyrel]{code:stab_17_1_5}{\(\llbracket 17,1,5\rrbracket \) 4.8.8 color code}\item\relax
\flmRefsHyperref[eczindexfamilyrel]{code:stab_18_2_5}{\(\llbracket 18,2,5\rrbracket \) BCC code} --- The stabilizer group of the \(\llbracket 18,2,5\rrbracket \) BCC code is invariant under transversal Hadamard \NoCaseChange{\protect\cite{cite440}}.
\item\relax
\flmRefsHyperref[eczindexfamilyrel]{code:stab_20_2_6}{\(\llbracket 20,2,6\rrbracket \) B\&C phantom code} --- The \(\llbracket 20,2,6\rrbracket \) code is a self-dual CSS code obtained from the \(\llbracket 5,1,3\rrbracket \) code via the BLT mapping and concatenation with \(\llbracket 4,2,2\rrbracket \) \NoCaseChange{\protect\cite[{Corr. 2}]{cite795}\protect\cite[{Corr. 1}]{cite1432}}.
\item\relax
\flmRefsHyperref[eczindexfamilyrel]{code:stab_8_2_2}{\(\llbracket 8,2,2\rrbracket \) hyperbolic color code}\item\relax
\flmRefsHyperref[eczindexfamilyrel]{code:stab_47_1_11}{\(\llbracket 47,1,11\rrbracket \) quantum QR code}\item\relax
\flmRefsHyperref[eczindexfamilyrel]{code:quantum_h}{\(\llbracket k+4,k,2\rrbracket \) H code}\item\relax
\flmRefsHyperref[eczindexfamilyrel]{code:qubit_golay}{\(\llbracket 23, 1, 7\rrbracket \) Quantum Golay code}\item\relax
\flmRefsHyperref[eczindexfamilyrel]{code:quantum_hamming_css}{\(\llbracket 2^r-1, 2^r-2r-1, 3\rrbracket \) quantum Hamming code}\item\relax
\flmRefsHyperref[eczindexfamilyrel]{code:single_qubit_clifford}{\(\llbracket 2^{2r-1}-1,1,2^r-1\rrbracket \) quantum punctured RM code} --- Puncturing a self-dual RM code yields a classical punctured RM code whose dual is its even subcode; applying the CSS construction to the even subcode yields this self-dual CSS family \NoCaseChange{\protect\cite[{Sec. 7.15.3}]{cite2764}}.
\end{eczvaluelist}
\codefieldsection{Cousins}
\begin{eczvaluelist}
\item\relax
\flmRefsHyperref[eczindexfamilyrel]{code:dual}{Dual linear code} --- Self-dual CSS codes arise from dual-containing (equivalently, self-orthogonal a.k.a. weakly self-dual) binary linear codes.
\item\relax
\flmRefsHyperref[eczindexfamilyrel]{code:qubit_stabilizer}{Qubit stabilizer code} --- Any \(\llbracket n,k,d\rrbracket \) qubit stabilizer code maps to a \(\llbracket 4n,2k,2d\rrbracket \) self-dual CSS code by applying the BLT mapping and concatenating each qubit pair with the \(\llbracket 4,2,2\rrbracket \) code \NoCaseChange{\protect\cite[{Corr. 2}]{cite795}\protect\cite[{Corr. 1}]{cite1432}}. The BLT mapping proceeds by first concatenating each qubit with the \flmRefsHyperref{code:tetron}{tetron code} to obtain an intermediate \(\llbracket 2n,k,2d\rrbracket _{f}\) Majorana stabilizer code.
\item\relax
\flmRefsHyperref[eczindexfamilyrel]{code:tetron}{Tetron code} --- Any \(\llbracket n,k,d\rrbracket \) qubit stabilizer code maps to a \(\llbracket 4n,2k,2d\rrbracket \) self-dual CSS code by applying the BLT mapping and concatenating each qubit pair with the \(\llbracket 4,2,2\rrbracket \) code \NoCaseChange{\protect\cite[{Corr. 2}]{cite795}\protect\cite[{Corr. 1}]{cite1432}}. The BLT mapping proceeds by first concatenating each qubit with the \flmRefsHyperref{code:tetron}{tetron code} to obtain an intermediate \(\llbracket 2n,k,2d\rrbracket _{f}\) Majorana stabilizer code.
\item\relax
\flmRefsHyperref[eczindexfamilyrel]{code:stab_4_2_2}{\(\llbracket 4,2,2\rrbracket \) Four-qubit code} --- Any \(\llbracket n,k,d\rrbracket \) qubit stabilizer code maps to a \(\llbracket 4n,2k,2d\rrbracket \) self-dual CSS code by applying the BLT mapping and concatenating each qubit pair with the \(\llbracket 4,2,2\rrbracket \) code \NoCaseChange{\protect\cite[{Corr. 2}]{cite795}\protect\cite[{Corr. 1}]{cite1432}}. The BLT mapping proceeds by first concatenating each qubit with the \flmRefsHyperref{code:tetron}{tetron code} to obtain an intermediate \(\llbracket 2n,k,2d\rrbracket _{f}\) Majorana stabilizer code.
\item\relax
\flmRefsHyperref[eczindexfamilyrel]{code:quantum_reed_muller}{Quantum Reed-Muller (RM) code} --- The \(\llbracket 2^m,{m \choose r}, 2^{\min(r,m-r)}\rrbracket \) quantum RM family contains a self-dual sub-family for \(m=2r\), which admits logical Clifford group gates via permutations, transversal gates, and fold-transversal gates \NoCaseChange{\protect\cite{cite757,cite758}}.
\item\relax
\flmRefsHyperref[eczindexfamilyrel]{code:quantum_divisible}{Quantum divisible code} --- A self-dual weakly doubly even \(\llbracket n,1,d\rrbracket \) CSS code admits a partitioned transversal physical \(S\) gate that realizes \(\overline{S}^m\), where \(m=|M^+|-|M^-| \pmod 4\); for odd \(m\), together with transversal Hadamard and CNOT, this yields the full logical Clifford group transversally \NoCaseChange{\protect\cite{cite731}\protect\cite[{Lemma 4}]{cite760}}.
\item\relax
\flmRefsHyperref[eczindexfamilyrel]{code:generalized_quantum_divisible}{Generalized quantum divisible code} --- Any self-dual CSS code yields a level-three generalized quantum divisible code when level-lifted \NoCaseChange{\protect\cite[{Thm. V.6}]{cite734}}.
\item\relax
\flmRefsHyperref[eczindexfamilyrel]{code:quantum_triorthogonal}{Triorthogonal code} --- Triorthogonal codes realizing logical \(T\) gates using only physical \(T\) gates can be paired up with self-dual CSS codes to yield a transversal CNOT gate and universal fault-tolerant gates using Steane error correction \NoCaseChange{\protect\cite{cite788}}.
\item\relax
\flmRefsHyperref[eczindexfamilyrel]{code:majorana_stab}{Majorana stabilizer code} --- An odd-length self-dual CSS code can be converted into a complex-fermion code by replacing qubit \(Z\)-type and \(X\)-type operators with \(\gamma\)-type and \(\tilde{\gamma}\)-type Majorana operators, respectively \NoCaseChange{\protect\cite{cite559}}.
\item\relax
\flmRefsHyperref[eczindexfamilyrel]{code:stab_8_2_3}{\(\llbracket 8,2,3\rrbracket \) Hermitian code} --- Applying the BLT mapping to the \(\llbracket 8,2,3\rrbracket \) Hermitian code and concatenating each qubit pair with the \(\llbracket 4,2,2\rrbracket \) code yields a \(\llbracket 32,4,6\rrbracket \) self-dual CSS code \NoCaseChange{\protect\cite[{Corr. 2}]{cite795}}.
\item\relax
\flmRefsHyperref[eczindexfamilyrel]{code:color}{Color code} --- Color codes often have self-dual \(X\)- and \(Z\)-type bulk stabilizer structure, but boundary choices can prevent the full code from being self-dual. Thus, only color-code geometries for which transversal Hadamard is a logical operation are self-dual CSS codes.
\end{eczvaluelist}
\eczhbkcontributors{ \eczhuVVA }
\endeczcode

\eczcode{sierpinsky_fractal_liquid}{Sierpinski prism model code}{~\NoCaseChange{\protect\cite{cite4416,cite1348}}}
\codefieldsection{Alternative Names}
\begin{eczvaluelist}
\item\relax Sierpinski fractal spin-liquid (SFSL) code
\item\relax Yoshida first-order fractal spin-liquid code
\end{eczvaluelist}
\eczhIndexCodeAliasName{sierpinsky_fractal_liquid}{Sierpinski fractal spin-liquid (SFSL) code}
\eczhIndexCodeAliasName{sierpinsky_fractal_liquid}{Yoshida first-order fractal spin-liquid code}
\codefieldsection{Description}
A fractal type-I fracton CSS code defined on a cubic lattice \NoCaseChange{\protect\cite[{Eq. (D22)}]{cite456}}.
The code admits an excitation-moving operator shaped like a Sierpinski triangle \NoCaseChange{\protect\cite[{Fig. 2}]{cite456}}.

\codefieldsection{Parents}
\begin{eczvaluelist}
\item\relax
\flmRefsHyperref[eczindexfamilyrel]{code:hypergraph_product}{Hypergraph product (HGP) code} --- The Sierpinski prism model code is a hypergraph product of the repetition code and the Newman-Moore code \NoCaseChange{\protect\cite{cite1501,cite1350}}.
\item\relax
\flmRefsHyperref[eczindexfamilyrel]{code:fracton}{Fracton stabilizer code} --- The Sierpinski prism model code is a fractal type-I fracton code \NoCaseChange{\protect\cite{cite456}}.
\end{eczvaluelist}
\codefieldsection{Cousins}
\begin{eczvaluelist}
\item\relax
\flmRefsHyperref[eczindexfamilyrel]{code:newman_moore}{Newman-Moore code} --- The Sierpinski prism model code is a hypergraph product of the repetition code and the Newman-Moore code \NoCaseChange{\protect\cite{cite1501,cite1350}}.
\item\relax
\flmRefsHyperref[eczindexfamilyrel]{code:repetition}{Repetition code} --- The Sierpinski prism model code is a hypergraph product of the repetition code and the Newman-Moore code \NoCaseChange{\protect\cite{cite1501,cite1350}}.
\item\relax
\flmRefsHyperref[eczindexfamilyrel]{code:spt}{Symmetry-protected topological (SPT) code} --- Ungauging \NoCaseChange{\protect\cite{cite462,cite463,cite233,cite464,cite465,cite466,cite467,cite468,cite469,cite470}} yields explicit fracton-SPT constructions; in particular, the Yoshida first-order fractal spin-liquid code underlies a 2D fracton SPT protected by Sierpinski-triangle symmetries via a gapped domain wall \NoCaseChange{\protect\cite{cite466}}.
\item\relax
\flmRefsHyperref[eczindexfamilyrel]{code:3d_surface}{3D surface code} --- The Sierpinski prism model code admits a topological defect network construction out of 3D surface codes on triangular prisms \NoCaseChange{\protect\cite{cite3163,cite3164}}.
\item\relax
\flmRefsHyperref[eczindexfamilyrel]{code:haah_cubic}{Haah cubic code (CC)} --- The Haah A-code can be written in a similar form as the Sierpinski prism model code \NoCaseChange{\protect\cite{cite3164}}.
\end{eczvaluelist}
\eczhbkcontributors{ Nathanan Tantivasadakarn, \eczhuVVA }
\endeczcode

\eczcode{holographic_6_1_3}{Six-qubit-tensor holographic code}{~\NoCaseChange{\protect\cite{cite2858}}}
\codefieldsection{Description}
Holographic tensor-network code constructed out of a network of encoding isometries of the \(\llbracket 6,1,3\rrbracket \) six-qubit stabilizer code.
The structure of the isometry is similar to that of the heptagon holographic code since both isometries are rank-six tensors, but the isometry in this case is neither a \flmRefsHyperref{ref219}{perfect tensor} nor a \flmRefsHyperref{code:block_perfect}{planar-perfect tensor}.

\codefieldsection{Rate}
Zero-rate version of the code surpasses the hashing bound for certain Pauli noise \NoCaseChange{\protect\cite{cite3715}}.
\codefieldsection{Code Capacity Threshold}
\begin{eczvaluelist}
\item\relax \(18.8\%\) under depolarizing noise using tensor-network decoder \NoCaseChange{\protect\cite{cite2858}}.
\end{eczvaluelist}
\codefieldsection{Parents}
\begin{eczvaluelist}
\item\relax
\flmRefsHyperref[eczindexfamilyrel]{code:qubit_stabilizer}{Qubit stabilizer code}\item\relax
\flmRefsHyperref[eczindexfamilyrel]{code:holographic_tensor}{Holographic tensor-network code} --- The encoding of the six-qubit-tensor holographic code is a holographic tensor network consisting of the encoding isometry for the \(\llbracket 6,1,3\rrbracket \) six-qubit stabilizer code.
\end{eczvaluelist}
\codefieldsection{Child}
\begin{eczvaluelist}
\item\relax
\flmRefsHyperref[eczindexfamilyrel]{code:stab_6_1_3}{\(\llbracket 6,1,3\rrbracket \) Six-qubit stabilizer code} --- The \(\llbracket 6,1,3\rrbracket \) six-qubit stabilizer code is the smallest six-qubit-tensor holographic code. The encoding of more general SCF holographic codes is a holographic tensor network consisting of the encoding isometry for the \(\llbracket 6,1,3\rrbracket \) six-qubit stabilizer code.
\end{eczvaluelist}
\eczhbkcontributors{ \eczhuVVA }
\endeczcode

\eczcode{small_distance_qubit_stabilizer}{Small-distance qubit stabilizer code}{}
\codefieldsection{Description}
A qubit stabilizer code that either detects or corrects errors on at most two subsystems, i.e., has distance \(\leq 5\).

Criteria for the existence of single error-correcting qubit stabilizer codes have been developed \NoCaseChange{\protect\cite{cite4394}}. Qubit stabilizer codes for \(n < 10\) have been classified \NoCaseChange{\protect\cite{cite454}}, all of which have small distance. Qubit stabilizer codes have been partially enumerated up to twelve qubits \NoCaseChange{\protect\cite{cite4395}}.
Ref. \NoCaseChange{\protect\cite{cite514}} enumerates all \(2.71\times10^{10}\) inequivalent CSS codes with \(n\leq14\).

\codefieldsection{Parents}
\begin{eczvaluelist}
\item\relax
\flmRefsHyperref[eczindexfamilyrel]{code:qubit_stabilizer}{Qubit stabilizer code} --- Criteria for the existence of single error-correcting qubit stabilizer codes have been developed \NoCaseChange{\protect\cite{cite4394}}. Qubit stabilizer codes for \(n < 10\) have been classified \NoCaseChange{\protect\cite{cite454}}, all of which have small distance. Qubit stabilizer codes have been partially enumerated up to twelve qubits \NoCaseChange{\protect\cite{cite4395}}. There are two \(\llbracket 8,1,3\rrbracket \) self-dual non-CSS codes; see QECDB \NoCaseChange{\protect\cite{cite781}}.
\item\relax
\flmRefsHyperref[eczindexfamilyrel]{code:small_distance_quantum}{Small-distance block quantum code}\end{eczvaluelist}
\codefieldsection{Children}
\begin{eczvaluelist}
\item\relax
\flmRefsHyperref[eczindexfamilyrel]{code:majorana_6_1_3}{\(\llbracket 6,1,3\rrbracket _{f}\) Vijay-Fu Majorana code}\item\relax
\flmRefsHyperref[eczindexfamilyrel]{code:kitaev_chain}{Kitaev chain code}\item\relax
\flmRefsHyperref[eczindexfamilyrel]{code:mbq}{Majorana box qubit}\item\relax
\flmRefsHyperref[eczindexfamilyrel]{code:majorana_hamming}{\(\llbracket 2^{m-1},2^{m-1}-m-1,4\rrbracket _{f}\) Hamming Majorana code}\item\relax
\flmRefsHyperref[eczindexfamilyrel]{code:ampdamp_stabilizer}{\(\llbracket 2(m+1),m,2\rrbracket \) single-loss AD code}\item\relax
\flmRefsHyperref[eczindexfamilyrel]{code:goy}{\(\llbracket 6r,2r,2\rrbracket \) Ganti-Onunkwo-Young code}\item\relax
\flmRefsHyperref[eczindexfamilyrel]{code:quantum_cap}{\(\llbracket n,n-2k,4\rrbracket \) Quantum cap code} --- Quantum cap codes can have a high rate and include codes with parameters \(\llbracket 6,0,4\rrbracket \), \(\llbracket 12,4,4\rrbracket \), \(\llbracket 40,30,4\rrbracket \), \(\llbracket 41,31,4\rrbracket \), \(\llbracket 126,114,4\rrbracket \), \(\llbracket 756,740,4\rrbracket \), and \(\llbracket 5040,5020,4\rrbracket \) \NoCaseChange{\protect\cite{cite1695}}, as well as \(\llbracket 12,2,4\rrbracket \), \(\llbracket 20,10,4\rrbracket \), or \(\llbracket 29,19,4\rrbracket \) \NoCaseChange{\protect\cite{cite3406}}.
\item\relax
\flmRefsHyperref[eczindexfamilyrel]{code:quantum_hamming}{\(\llbracket 2^r, 2^r-r-2, 3\rrbracket \) Gottesman code}\item\relax
\flmRefsHyperref[eczindexfamilyrel]{code:quantum_icosahedron}{\(\llbracket 54,6,5\rrbracket \) five-covered icosahedral code}\item\relax
\flmRefsHyperref[eczindexfamilyrel]{code:quantum_repetition}{Quantum repetition code}\item\relax
\flmRefsHyperref[eczindexfamilyrel]{code:ring_cpc}{\(\llbracket 2^r+r, 2^r-r-2, 3\rrbracket \) Ring CPC code}\item\relax
\flmRefsHyperref[eczindexfamilyrel]{code:stab_10_2_3}{\(\llbracket 10,2,3\rrbracket \) binarized Galois-qudit code}\item\relax
\flmRefsHyperref[eczindexfamilyrel]{code:xzzx_10_2_3}{\(\llbracket 10,2,3\rrbracket \) rotated toric code}\item\relax
\flmRefsHyperref[eczindexfamilyrel]{code:stab_11_1_5}{\(\llbracket 11,1,5\rrbracket \) quantum dodecacode}\item\relax
\flmRefsHyperref[eczindexfamilyrel]{code:carbon}{\(\llbracket 12,2,4\rrbracket \) carbon code}\item\relax
\flmRefsHyperref[eczindexfamilyrel]{code:css_12_1_3}{\(\llbracket 12,1,3\rrbracket \) CE CSS code}\item\relax
\flmRefsHyperref[eczindexfamilyrel]{code:stab_12_2_2}{\(\llbracket 12,2,2\rrbracket \) CSS code}\item\relax
\flmRefsHyperref[eczindexfamilyrel]{code:quad_residue_13_1_5}{\(\llbracket 13,1,5\rrbracket \) quantum QR code}\item\relax
\flmRefsHyperref[eczindexfamilyrel]{code:stab_13_1_5}{\(\llbracket 13,1,5\rrbracket \) twisted toric code}\item\relax
\flmRefsHyperref[eczindexfamilyrel]{code:phantom_14_3_3}{\(\llbracket 14,3,3\rrbracket \) CE phantom code}\item\relax
\flmRefsHyperref[eczindexfamilyrel]{code:rhombic_dodecahedron_surface}{\(\llbracket 14,3,3\rrbracket \) Rhombic dodecahedron surface code}\item\relax
\flmRefsHyperref[eczindexfamilyrel]{code:quantum_dodecahedron}{\(\llbracket 16,4,3\rrbracket \) dodecahedral code}\item\relax
\flmRefsHyperref[eczindexfamilyrel]{code:stab_16_6_4}{\(\llbracket 16,6,4\rrbracket \) Tesseract color code}\item\relax
\flmRefsHyperref[eczindexfamilyrel]{code:stab_17_1_5}{\(\llbracket 17,1,5\rrbracket \) 4.8.8 color code} --- The smallest distance-five CSS code has \(n=17\) \NoCaseChange{\protect\cite{cite444}}.
\item\relax
\flmRefsHyperref[eczindexfamilyrel]{code:stab_18_2_5}{\(\llbracket 18,2,5\rrbracket \) BCC code}\item\relax
\flmRefsHyperref[eczindexfamilyrel]{code:stab_4_1_2}{\(\llbracket 4,1,2\rrbracket \) twist-defect code}\item\relax
\flmRefsHyperref[eczindexfamilyrel]{code:stab_5_1_3}{\(\llbracket 5,1,3\rrbracket \) Five-qubit perfect code}\item\relax
\flmRefsHyperref[eczindexfamilyrel]{code:css_6_1_2}{\(\llbracket 6,1,2\rrbracket \) semi-self-dual CSS code}\item\relax
\flmRefsHyperref[eczindexfamilyrel]{code:stab_6_1_3}{\(\llbracket 6,1,3\rrbracket \) Six-qubit stabilizer code}\item\relax
\flmRefsHyperref[eczindexfamilyrel]{code:bare_7_1_3}{\(\llbracket 7,1,3\rrbracket \) bare code}\item\relax
\flmRefsHyperref[eczindexfamilyrel]{code:hgp_7_2_2}{\(\llbracket 7,2,2\rrbracket \) HGP phantom code}\item\relax
\flmRefsHyperref[eczindexfamilyrel]{code:qetc_7_2}{\(\llbracket 7,2,2\rrbracket \) QETC}\item\relax
\flmRefsHyperref[eczindexfamilyrel]{code:twist_defect_7_1_3}{\(\llbracket 7,1,3\rrbracket \) twist-defect surface code}\item\relax
\flmRefsHyperref[eczindexfamilyrel]{code:xz_7_3_2}{\(\llbracket 7,3,2\rrbracket \) punctured hypercube code}\item\relax
\flmRefsHyperref[eczindexfamilyrel]{code:xzzx_7_1_3}{\(\llbracket 7,1,3\rrbracket \) XZZX cyclic code}\item\relax
\flmRefsHyperref[eczindexfamilyrel]{code:cubic_surface}{\(\llbracket 8,3,2\rrbracket \) Surface code on a cube}\item\relax
\flmRefsHyperref[eczindexfamilyrel]{code:stab_8_1_2}{\(\llbracket 8,1,2\rrbracket \) Shen-Wang-Cao code}\item\relax
\flmRefsHyperref[eczindexfamilyrel]{code:stab_8_2_2}{\(\llbracket 8,2,2\rrbracket \) hyperbolic color code}\item\relax
\flmRefsHyperref[eczindexfamilyrel]{code:stab_8_2_3}{\(\llbracket 8,2,3\rrbracket \) Hermitian code}\item\relax
\flmRefsHyperref[eczindexfamilyrel]{code:shor_nine}{\(\llbracket 9,1,3\rrbracket \) Shor code}\item\relax
\flmRefsHyperref[eczindexfamilyrel]{code:stab_9_3_3}{\(\llbracket 9,3,3\rrbracket \) Quadric code}\item\relax
\flmRefsHyperref[eczindexfamilyrel]{code:surface-17}{\(\llbracket 9,1,3\rrbracket \) Surface-17 code}\item\relax
\flmRefsHyperref[eczindexfamilyrel]{code:tfim}{Transverse-field Ising model (TFIM) code}\item\relax
\flmRefsHyperref[eczindexfamilyrel]{code:campbell_howard}{\(\llbracket 6k+2,3k,2\rrbracket \) Campbell-Howard code} --- The family has distance \(2\).
\item\relax
\flmRefsHyperref[eczindexfamilyrel]{code:quantum_h}{\(\llbracket k+4,k,2\rrbracket \) H code}\item\relax
\flmRefsHyperref[eczindexfamilyrel]{code:small_triorthogonal}{\(\llbracket 3k + 8, k, 2\rrbracket \) triorthogonal code}\item\relax
\flmRefsHyperref[eczindexfamilyrel]{code:stab_49_1_5}{\(\llbracket 49,1,5\rrbracket \) triorthogonal code}\item\relax
\flmRefsHyperref[eczindexfamilyrel]{code:diagonal_clifford}{\(\llbracket 2^r-1,1,3\rrbracket \) simplex code}\item\relax
\flmRefsHyperref[eczindexfamilyrel]{code:morphed_diagonal_clifford}{\(\llbracket 2^r+r-1,1,2\rrbracket \) morphed simplex code}\item\relax
\flmRefsHyperref[eczindexfamilyrel]{code:quantum_hamming_css}{\(\llbracket 2^r-1, 2^r-2r-1, 3\rrbracket \) quantum Hamming code}\item\relax
\flmRefsHyperref[eczindexfamilyrel]{code:ball_color}{Ball code}\item\relax
\flmRefsHyperref[eczindexfamilyrel]{code:stellated_dodecahedron_css}{\(\llbracket 30,8,3\rrbracket \) Bring code}\end{eczvaluelist}
\codefieldsection{Cousins}
\begin{eczvaluelist}
\item\relax
\flmRefsHyperref[eczindexfamilyrel]{code:ea_3_1_3-2}{\(\llbracket 3, 1, 3;2\rrbracket \) EA code}\item\relax
\flmRefsHyperref[eczindexfamilyrel]{code:eaoa_hamming}{\(\llbracket 10,1,3;1,3,4\rrbracket \) EAOA Hamming code}\item\relax
\flmRefsHyperref[eczindexfamilyrel]{code:hybrid_7_1-1_3}{\(\llbracket 7, 1:1, 3\rrbracket \) hybrid stabilizer code}\item\relax
\flmRefsHyperref[eczindexfamilyrel]{code:hybrid_8_2-1_3}{\(\llbracket 8, 2:1, 3\rrbracket \) hybrid stabilizer code}\item\relax
\flmRefsHyperref[eczindexfamilyrel]{code:reinforcement_learning}{Reinforcement-learning quantum code} --- 13 inequivalent \(\llbracket 9,3,3\rrbracket \) codes, along with others, have been found via reinforcement learning \NoCaseChange{\protect\cite{cite3184}}.
\item\relax
\flmRefsHyperref[eczindexfamilyrel]{code:qmdpc}{Quantum multi-dimensional parity-check (QMDPC) code} --- QMDPC codes for dimensions \(D \leq 2\) are examples of small distance qubit stabilizer codes.
\item\relax
\flmRefsHyperref[eczindexfamilyrel]{code:qubit_golay}{\(\llbracket 23, 1, 7\rrbracket \) Quantum Golay code} --- The quantum Golay code can be punctured twice to obtain a \(\llbracket 21,3,5\rrbracket \) code.
\item\relax
\flmRefsHyperref[eczindexfamilyrel]{code:hyperbolic_color}{Hyperbolic color code} --- Many hyperbolic color codes have distance \(\leq 5\).
\item\relax
\flmRefsHyperref[eczindexfamilyrel]{code:bacon_shor_4}{\(\llbracket 4,1,1,2\rrbracket \) Four-qubit subsystem code}\item\relax
\flmRefsHyperref[eczindexfamilyrel]{code:bacon_shor_9}{\(\llbracket 9,1,4,3\rrbracket \) Nine-qubit Bacon-Shor code}\item\relax
\flmRefsHyperref[eczindexfamilyrel]{code:bravyi_bacon_shor_6}{\(\llbracket 6,2,3,2\rrbracket \) BBS code}\item\relax
\flmRefsHyperref[eczindexfamilyrel]{code:trapezoid}{Trapezoid subsystem code}\end{eczvaluelist}
\eczhbkcontributors{ \eczhuVVA }
\endeczcode

\eczcode{ssw}{Smolin-Smith-Wehner (SSW) code}{~\NoCaseChange{\protect\cite{cite1262,cite854}}}
\codefieldsection{Description}
A family of \(\llparenthesis n=4k+2l+3,M_{k,l},2\rrparenthesis \) self-complementary CWS codes, where \(M_{k,l} \approx 2^{n-2}(1-\sqrt{2/(\pi(n-1\rrparenthesis })\).
For \(n \geq 11\), these codes have a logical subspace whose dimension is larger than that of the largest stabilizer code for the same \(n\) and \(d\).
Ref. \NoCaseChange{\protect\cite{cite452}} augments a star-graph-based subfamily \(\llparenthesis 4n+1,M_n,2\rrparenthesis \) by one additional graph-state basis word, yielding \(\llparenthesis 4n+1,M_n+1,2\rrparenthesis \) codes with \(M_n=2^{4n-1}-\frac{1}{2}\binom{4n}{2n}\).
In the CWS description of Ref. \NoCaseChange{\protect\cite{cite852}}, the underlying stabilizer state is locally Clifford-equivalent to a GHZ state and its standard-form graph is a star graph.

\codefieldsection{Realizations}
\begin{eczvaluelist}
\item\relax The \(\llparenthesis 5,5,2\rrparenthesis \) SSW code has been realized in an NMR device \NoCaseChange{\protect\cite{cite4417}}.
\end{eczvaluelist}
\codefieldsection{Parents}
\begin{eczvaluelist}
\item\relax
\flmRefsHyperref[eczindexfamilyrel]{code:cws}{Codeword stabilized (CWS) code} --- SSW codes can be formulated as CWS codes \NoCaseChange{\protect\cite{cite852,cite3166}}.
\item\relax
\flmRefsHyperref[eczindexfamilyrel]{code:self_complementary}{Self-complementary qubit code}\end{eczvaluelist}
\codefieldsection{Cousin}
\begin{eczvaluelist}
\item\relax
\flmRefsHyperref[eczindexfamilyrel]{code:rains}{\(\llparenthesis 2m+1,3 \times 2^{2m-3},2\rrparenthesis \) Rains code} --- The SSW code outperforms the Rains codes in terms of code parameters at odd \(n > 11\) \NoCaseChange{\protect\cite{cite852,cite3166}}.
\end{eczvaluelist}
\eczhbkcontributors{ \eczhuVVA }
\endeczcode

\eczcode{spacetime_circuit}{Spacetime circuit code}{~\NoCaseChange{\protect\cite{cite668,cite4418,cite667}}}
\codefieldsection{Description}
Qubit stabilizer code constructed from a \flmRefsHyperref{ref409}{Clifford circuit}, i.e., a circuit made up of \flmRefsHyperref{ref409}{Clifford gates} and Pauli measurements, in order to detect and correct circuit faults.
The code utilizes redundancy in the measurement outcomes of a circuit to correct circuit faults, which correspond to Pauli errors of the code.

The structure of the \flmRefsHyperref{ref409}{Clifford circuit} yields correlations between the circuit's possible measurement outcomes.
The set of outcomes can be made into a classical binary linear code called the \textit{outcome code} \NoCaseChange{\protect\cite[{Corr. 2}]{cite667}}.
The spacetime circuit code is defined such that its error syndromes can be backpropagated to obtain the parity checks of the outcome code.
In other words, both codes have the same set of parity check outcomes.

More technically, given an \([m,k]\) outcome code associated with an \(n\)-qubit circuit of depth \(\Delta\) with \(m\) measurements and \(2^k\) outcomes, the corresponding spacetime circuit code is an \(\llbracket  n (\Delta + 1), n (\Delta + 1) - (m - k) \rrbracket \) code \NoCaseChange{\protect\cite[{Thm. 2}]{cite667}}.

The spacetime circuit code is the stabilizer-code counterpart of earlier subsystem constructions \NoCaseChange{\protect\cite{cite668,cite4418}}, which dealt with restricted families of \flmRefsHyperref{ref409}{Clifford circuits}.
A more general construction includes circuits with intermediate and multi-qubit measurements \NoCaseChange{\protect\cite{cite667}}.

Many features of the spacetime circuit formalism can be understood through ZX calculus \NoCaseChange{\protect\cite{cite542}}.
Two circuits are \textit{fault-equivalent} if all undetectable faults on one circuit have a corresponding fault on the other \NoCaseChange{\protect\cite{cite4419}}.

\codefieldsection{Decoding}
\begin{eczvaluelist}
\item\relax A most-likely error decoder for the spacetime code can be converted into a most-likely fault decoder for the underlying circuit \NoCaseChange{\protect\cite{cite667}}.
\item\relax Efficient decoders can be constructed for some circuits, especially when the resulting outcome and spacetime codes are LDPC \NoCaseChange{\protect\cite{cite667}}.
\end{eczvaluelist}
\codefieldsection{Fault Tolerance}
\begin{eczvaluelist}
\item\relax The outcome-code distance is a lower bound on the \flmRefsHyperref{ref3496}{circuit-level distance}, and the bound need not be tight because some undetectable fault configurations are logically trivial. The circuit-level distance corresponds to the minimum-weight outcome codeword that is not in the kernel of the logical effect matrix \NoCaseChange{\protect\cite{cite4420}}.
\end{eczvaluelist}
\codefieldsection{Notes}
\begin{eczvaluelist}
\item\relax See \NoCaseChange{\protect\cite[{Sec. 2}]{cite3742}} for a brief overview of spacetime circuits.
\end{eczvaluelist}
\codefieldsection{Parents}
\begin{eczvaluelist}
\item\relax
\flmRefsHyperref[eczindexfamilyrel]{code:qldpc}{Qubit QLDPC code} --- Spacetime circuit codes are useful for constructing fault-tolerant syndrome extraction circuits for qubit QLDPC codes. General spacetime circuit codes can be sparsified to yield QLDPC spacetime circuit codes \NoCaseChange{\protect\cite{cite667}}.
\item\relax
\flmRefsHyperref[eczindexfamilyrel]{code:dynamic_gen}{Dynamically generated QECC}\end{eczvaluelist}
\codefieldsection{Cousins}
\begin{eczvaluelist}
\item\relax
\flmRefsHyperref[eczindexfamilyrel]{code:binary_linear}{Linear binary code} --- The set of measurement outcomes of a \flmRefsHyperref{ref409}{Clifford circuit} can be made into a classical binary linear code.
Error syndromes of the spacetime circuit code can be used to obtain the parity checks of the outcome code.

\item\relax
\flmRefsHyperref[eczindexfamilyrel]{code:surface}{Kitaev surface code} --- Stabilizer generators of a spacetime code are called \textit{detectors} in Refs. \NoCaseChange{\protect\cite{cite3964,cite667}}.
\item\relax
\flmRefsHyperref[eczindexfamilyrel]{code:qubit_subsystem_stabilizer}{Subsystem qubit stabilizer code} --- Spacetime circuit codes can be upgraded to subsystem codes by \flmRefsHyperref{ref666}{gauging out} a subgroup of the logical \flmRefsHyperref{ref663}{Pauli group} which causes trivial faults in the corresponding \flmRefsHyperref{ref409}{Clifford circuit}.
\item\relax
\flmRefsHyperref[eczindexfamilyrel]{code:ldpc}{Low-density parity-check (LDPC) code} --- There is an equivalence between \flmRefsHyperref{ref409}{Clifford circuits} and LDPC codes with bit-check symmetry \NoCaseChange{\protect\cite{cite1487}}.
\item\relax
\flmRefsHyperref[eczindexfamilyrel]{code:subsystem_spacetime_circuit}{Subsystem spacetime circuit code} --- Spacetime circuit codes can yield subsystem spacetime circuit codes by \flmRefsHyperref{ref666}{gauging out} a subgroup of the logical \flmRefsHyperref{ref663}{Pauli group} which causes trivial faults in the corresponding \flmRefsHyperref{ref409}{Clifford circuit}. This construction is used to show the existence of geometrically local subsystem codes that nearly saturate the \flmRefsHyperref{ref492}{subsystem BT bound} \NoCaseChange{\protect\cite{cite668}}.
\end{eczvaluelist}
\eczhbkcontributors{ Marcus P da Silva, David Aasen, \eczhuVVA }
\endeczcode

\eczcode{square_homological_product}{Square homological product code}{~\NoCaseChange{\protect\cite{cite3727}}}
\codefieldsection{Alternative Names}
\begin{eczvaluelist}
\item\relax Single-sector homological code
\item\relax Bravyi-Hastings homological code
\item\relax Square tensor product code
\end{eczvaluelist}
\eczhIndexCodeAliasName{square_homological_product}{Single-sector homological code}
\eczhIndexCodeAliasName{square_homological_product}{Bravyi-Hastings homological code}
\eczhIndexCodeAliasName{square_homological_product}{Square tensor product code}
\codefieldsection{Description}
Homological product code whose underlying quantum-code boundary operators are square matrices (see \flmRefsCref{ref683}).

Each base code is associated with the chain complex \( C_i \longrightarrow C_i\longrightarrow C_i\) such that the boundary operator (a.k.a. parity-check matrix) satisfies \(H_i^{2}=0\) \NoCaseChange{\protect\cite[{Def. 3.8}]{cite835}}. 
The parity-check matrix of the resulting product code is 
\flmMathEnvironment{align}{}{
  H_1 \otimes I_2 + I_1 \otimes H_2~,
}
where \(I_i\) is the identity on the check space of code \(i\).
The logical dimension \(k = k_1 k_2\).

\codefieldsection{Protection}
Square homological-product codes admit different properties than those with rectangular boundary operators \NoCaseChange{\protect\cite[{Sec. 3.4}]{cite835}}.

\codefieldsection{Parent}
\begin{eczvaluelist}
\item\relax
\flmRefsHyperref[eczindexfamilyrel]{code:homological_product}{Homological product code} --- Square homological product codes are homological product codes whose boundary operators are square matrices \NoCaseChange{\protect\cite[{Sec. 3.4}]{cite835}}.
\end{eczvaluelist}
\eczhbkcontributors{ \eczhuVVA }
\endeczcode

\eczcode{square_lattice_cluster}{Square-lattice cluster-state code}{~\NoCaseChange{\protect\cite{cite3528,cite428,cite429}}}
\codefieldsection{Description}
A code based on the cluster state on a square lattice that was used in the first proposal for MBQC \NoCaseChange{\protect\cite{cite428,cite429}}.
In the one-way model, the pre-entangled square-lattice cluster is a universal resource, and the computation is carried out entirely by adaptive single-qubit measurements.

\codefieldsection{Protection}
Random measurement outcomes induce Pauli byproduct operators that are tracked classically and propagated to later measurements or the final readout \NoCaseChange{\protect\cite{cite429}}.
For computations longer than the available lattice extent, the original paper proposed splitting the computation into consecutive segments and stabilizing each segment using standard error-correction techniques \NoCaseChange{\protect\cite{cite429}}.

\codefieldsection{Encoding}
\begin{eczvaluelist}
\item\relax Initialization of each qubit in the \(|+\rangle\) state followed by nearest-neighbor Ising-type entangling evolution, equivalently controlled-phase gates on the edges of the square lattice, prepares the resource state \NoCaseChange{\protect\cite{cite429}}.
\end{eczvaluelist}
\codefieldsection{Gates}
\begin{eczvaluelist}
\item\relax Measurements in the \(Z\) basis remove qubits from the lattice to carve out the computation network. \(X\)-basis measurements propagate quantum information along a wire, adaptive equatorial-basis measurements on a five-qubit chain implement arbitrary single-qubit \(SU(2)\) rotations, and a four-qubit pattern implements CNOT between neighboring wires. Later measurement bases can depend on earlier outcomes because of the tracked Pauli byproducts \NoCaseChange{\protect\cite{cite429}}.
\item\relax Universal MBQC remains possible on irregular occupied sublattices above the percolation threshold because wires and gates can be bent and stretched without changing circuit topology \NoCaseChange{\protect\cite{cite429}}.
\end{eczvaluelist}
\codefieldsection{Realizations}
\begin{eczvaluelist}
\item\relax Encoding on 72 qubits of the Zuchongzhi 3.1 quantum processor \NoCaseChange{\protect\cite{cite4421}}.
\end{eczvaluelist}
\codefieldsection{Notes}
\begin{eczvaluelist}
\item\relax The original proposal discussed implementations using neutral atoms in optical lattices with controlled collisions and capacitively coupled quantum dots \NoCaseChange{\protect\cite{cite429}}.
\end{eczvaluelist}
\codefieldsection{Parents}
\begin{eczvaluelist}
\item\relax
\flmRefsHyperref[eczindexfamilyrel]{code:cluster_state}{Cluster-state code}\item\relax
\flmRefsHyperref[eczindexfamilyrel]{code:qldpc}{Qubit QLDPC code}\item\relax
\flmRefsHyperref[eczindexfamilyrel]{code:2d_stabilizer}{2D lattice stabilizer code}\end{eczvaluelist}
\codefieldsection{Cousin}
\begin{eczvaluelist}
\item\relax
\flmRefsHyperref[eczindexfamilyrel]{code:spt}{Symmetry-protected topological (SPT) code} --- The square-lattice cluster state, which is the prototypical resource for universal MBQC \NoCaseChange{\protect\cite{cite428,cite429}}, and other 2D cluster states \NoCaseChange{\protect\cite{cite3082,cite3083,cite3084}} have SPT order protected by subsystem symmetries \NoCaseChange{\protect\cite{cite3085,cite3086,cite3082}}.

\end{eczvaluelist}
\eczhbkcontributors{ \eczhuVVA }
\endeczcode

\eczcode{488_color}{Square-octagon (4.8.8) color code}{~\NoCaseChange{\protect\cite{cite710}}}
\codefieldsection{Description}
2D color code defined on a patch of the 4.8.8 (square-octagon) tiling, which itself is obtained by applying a fattening procedure to the square lattice \NoCaseChange{\protect\cite{cite430}}.
An equivalent description uses the Tetrakis square tiling (a.k.a. the Union Jack lattice), which is dual to the 4.8.8 lattice \NoCaseChange{\protect\cite{cite431}}.
Among the three semiregular triangular 2D color-code families, the 4.8.8 family uses the fewest physical qubits for a given distance and is the only one of the three with transversal implementations of the full Clifford group \NoCaseChange{\protect\cite{cite432}}.

Stabilizer generators are shown in \flmRefsCref{ref4422}.
  \begin{flmFloat}{figure}{NumCap}\includegraphics[width=315bp,max width=\linewidth]{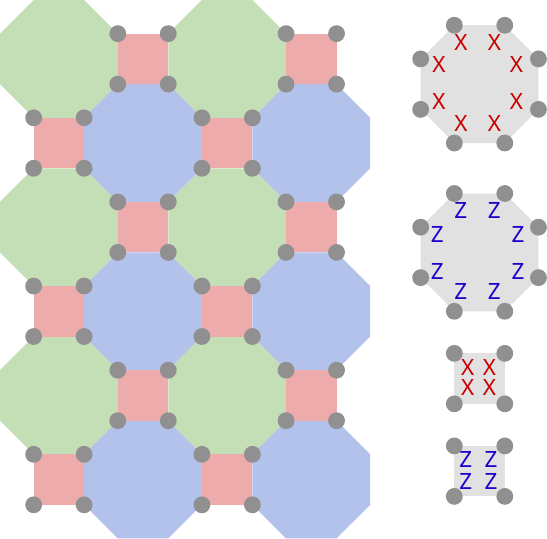}\caption{
    Stabilizer generators of the 4.8.8 color code.
    }\label{ref4422}\end{flmFloat}

Different boundaries affect the logical dimension \NoCaseChange{\protect\cite{cite4423}}.

\codefieldsection{Protection}
There is a \(\llbracket (d^2-1)/2+d, 1, d\rrbracket \) code family for any odd distance \(d\) \NoCaseChange{\protect\cite[{Fig. 2}]{cite432}}.

\codefieldsection{Transversal and Permutation-Based Gates}
\begin{eczvaluelist}
\item\relax CNOT gate because the code is CSS.
\item\relax Hadamard gates for any qubit geometry which yields a self-dual CSS code.
\item\relax Transversal \(S\) gate \NoCaseChange{\protect\cite{cite710,cite432}}.
\item\relax Transversal logical \flmRefsHyperref{ref409}{Clifford gates} in the Union Jack formulation \NoCaseChange{\protect\cite{cite431}}.
\item\relax Single-qubit Clifford and CNOT gates between qubits encoded in holes in the lattice can be implemented via braiding \NoCaseChange{\protect\cite{cite713}}.
\end{eczvaluelist}
\codefieldsection{Gates}
\begin{eczvaluelist}
\item\relax Color-code lattice surgery \NoCaseChange{\protect\cite{cite4424}}.
\item\relax Lattice surgery scheme for a hybrid 6.6.6-4.8.8 layout yields lower resource overhead when compared to analogous surface code scheme \NoCaseChange{\protect\cite{cite3732}}.
\end{eczvaluelist}
\codefieldsection{Decoding}
\begin{eczvaluelist}
\item\relax Fault-tolerant syndrome extraction circuits \NoCaseChange{\protect\cite{cite432}}.
\item\relax Matching decoder \NoCaseChange{\protect\cite{cite4425,cite4424}}.
\item\relax Integer-program (IP) decoder \NoCaseChange{\protect\cite{cite432}}.
\item\relax Two-copy surface-code decoder \NoCaseChange{\protect\cite{cite4426}}.
\end{eczvaluelist}
\codefieldsection{Fault Tolerance}
\begin{eczvaluelist}
\item\relax Color-code lattice surgery \NoCaseChange{\protect\cite{cite4424}}.
\item\relax Fault-tolerant syndrome extraction circuits \NoCaseChange{\protect\cite{cite432}}.
\end{eczvaluelist}
\codefieldsection{Code Capacity Threshold}
\begin{eczvaluelist}
\item\relax Independent \(X,Z\) noise: \(p_X = 10.56(1)\%\) under IP decoder \NoCaseChange{\protect\cite{cite432}}, \(8.87\%\) under matching decoder \NoCaseChange{\protect\cite{cite4425}}, \(7.60(2)\%\) under projection decoder \NoCaseChange{\protect\cite{cite3418}}, and \(8.7\%\) under two-copy surface-code decoder \NoCaseChange{\protect\cite{cite4426}} (see \NoCaseChange{\protect\cite[{Table I}]{cite432}}). The threshold under ML decoding corresponds to the value of a critical point of a two-dimensional three-body random-bond Ising model (RBIM) on the Nishimori line \NoCaseChange{\protect\cite{cite3526,cite3745}}, calculated to be \(10.9(2)\%\) in Ref. \NoCaseChange{\protect\cite{cite3745}} (and in the Union Jack formulation in Ref. \NoCaseChange{\protect\cite{cite431}}) and \(10.925(5)\%\) in Ref. \NoCaseChange{\protect\cite{cite3746}}.
\end{eczvaluelist}
\codefieldsection{Threshold}
\begin{eczvaluelist}
\item\relax Phenomenological noise: \(3.05(4)\%\) under IP decoder \NoCaseChange{\protect\cite[{Table I}]{cite432}} and \(2.08(1)\%\) under projection decoder \NoCaseChange{\protect\cite{cite3418}}.
\item\relax Circuit-level noise: \(0.082(3)\%\) under IP decoder, \(0.143(1)\%\) under projection decoder \NoCaseChange{\protect\cite{cite3418}}, \(0.143\%\) under matching decoder \NoCaseChange{\protect\cite{cite4424}}, and an analytic lower bound of \(\approx 0.1\%\) \NoCaseChange{\protect\cite{cite4425}} (see \NoCaseChange{\protect\cite[{Table I}]{cite432}}).
\end{eczvaluelist}
\codefieldsection{Realizations}
\begin{eczvaluelist}
\item\relax Neutral atom arrays: logical magic-state distillation using distance-three and five 4.8.8 color codes, observing an improvement in logical fidelity on a device by Quera \NoCaseChange{\protect\cite{cite4427}}.
\end{eczvaluelist}
\codefieldsection{Parent}
\begin{eczvaluelist}
\item\relax
\flmRefsHyperref[eczindexfamilyrel]{code:2d_color}{2D color code}\end{eczvaluelist}
\codefieldsection{Children}
\begin{eczvaluelist}
\item\relax
\flmRefsHyperref[eczindexfamilyrel]{code:stab_17_1_5}{\(\llbracket 17,1,5\rrbracket \) 4.8.8 color code} --- The \(\llbracket 17,1,5\rrbracket \) color code can be defined on the 4.8.8 tiling \NoCaseChange{\protect\cite{cite731}}.
\item\relax
\flmRefsHyperref[eczindexfamilyrel]{code:stab_4_2_2}{\(\llbracket 4,2,2\rrbracket \) Four-qubit code} --- The \(\llbracket 4,2,2\rrbracket \) code can be interpreted as a 2D color code on a square of the 4.8.8 tiling \NoCaseChange{\protect\cite{cite2526,cite3262}}. Removing \(X\) checks from blue octagons and \(Z\) checks from green octagons of the 4.8.8 color code yields a light 4.8.8 color code that is equivalent to concatenating the surface/toric code with the \(\llbracket 4,2,2\rrbracket \) code \NoCaseChange{\protect\cite{cite3289}}.
\end{eczvaluelist}
\codefieldsection{Cousins}
\begin{eczvaluelist}
\item\relax
\flmRefsHyperref[eczindexfamilyrel]{code:triangular_color}{Honeycomb (6.6.6) color code} --- Lattice surgery scheme for a hybrid 6.6.6-4.8.8 layout yields lower resource overhead when compared to analogous surface code scheme \NoCaseChange{\protect\cite{cite3732}}.
\item\relax
\flmRefsHyperref[eczindexfamilyrel]{code:hypercubic}{\(\mathbb{Z}^n\) hypercubic lattice} --- The 4.8.8 (square-octagon) tiling is obtained by applying a fattening procedure to the square lattice \NoCaseChange{\protect\cite{cite430}}.
\item\relax
\flmRefsHyperref[eczindexfamilyrel]{code:gkp_concatenated}{Concatenated GKP code} --- GKP codes have been concatenated with 4.8.8 color codes \NoCaseChange{\protect\cite{cite4428}}.
\end{eczvaluelist}
\eczhbkcontributors{ Eric Huang, \eczhuVVA }
\endeczcode

\eczcode{stellated_color}{Stellated color code}{~\NoCaseChange{\protect\cite{cite445}}}
\codefieldsection{Description}
A non-CSS color-code family on a lattice patch with a single central puncture that hosts a twist defect connected to the boundary by a domain wall.

The family is parameterized by a rotational symmetry order \(s\); for odd \(s\), the code encodes \(k=s-1\) logical qubits, while for even \(s\), it encodes \(k=s-2\) logical qubits \NoCaseChange{\protect\cite{cite445}}.

\codefieldsection{Rate}
Code families yield the following values of the constant \(c\) in the \flmRefsHyperref{ref487}{BPT bound}, \(k d^2 \leq c n\). On the 4.8.8 lattice, stellated color codes have \(c=4-\frac{4}{s}\) for odd \(s\) and \(c=4-\frac{8}{s}\) for even \(s\), approaching \(4\) as \(s\) grows. On the 6.6.6 lattice, they have \(c=\frac{8}{3}-\frac{8}{3s}\) for odd \(s\) and \(c=\frac{8}{3}-\frac{16}{3s}\) for even \(s\), approaching \(\frac{8}{3}\) \NoCaseChange{\protect\cite{cite445}}.
\codefieldsection{Parent}
\begin{eczvaluelist}
\item\relax
\flmRefsHyperref[eczindexfamilyrel]{code:twist_defect_color}{Twist-defect color code}\end{eczvaluelist}
\codefieldsection{Cousins}
\begin{eczvaluelist}
\item\relax
\flmRefsHyperref[eczindexfamilyrel]{code:triangle_surface}{Triangular surface code} --- Stellated color codes are color-code analogues of triangle surface codes in that both encode logical information in lattices with a single twist defect. Instances of the former can be obtained by fattening \NoCaseChange{\protect\cite{cite430}} the vertices of the latter \NoCaseChange{\protect\cite{cite445}}.
\item\relax
\flmRefsHyperref[eczindexfamilyrel]{code:stellated_surface}{Stellated surface code} --- Stellated color codes are color-code analogues of stellated surface codes; the surface-code family has the same rotational parameter \(s\), but half the asymptotic \(c\)-value of the 4.8.8 stellated color-code family \NoCaseChange{\protect\cite{cite445}}.
\end{eczvaluelist}
\eczhbkcontributors{ \eczhuVVA }
\endeczcode

\eczcode{stellated_surface}{Stellated surface code}{~\NoCaseChange{\protect\cite{cite445}}}
\codefieldsection{Description}
A twist-defect surface-code family parameterized by a rotational symmetry order \(s\), with a central toric-code twist connected to the boundary by a domain wall.
The \(s=3\) member is the triangular surface code \NoCaseChange{\protect\cite{cite433}}.

In the plaquette construction of Ref. \NoCaseChange{\protect\cite[{Appx. D}]{cite445}}, qubits lie on vertices, each plaquette hosts one stabilizer, and plaquettes along the central domain wall act in mixed Pauli bases.

\codefieldsection{Rate}
Stellated surface codes have \(c=2-\frac{2}{s}\) for odd \(s\) and \(c=2-\frac{4}{s}\) for even \(s\), both approaching \(2\) as \(s\) grows \NoCaseChange{\protect\cite[{Appx. D}]{cite445}}.
\codefieldsection{Parent}
\begin{eczvaluelist}
\item\relax
\flmRefsHyperref[eczindexfamilyrel]{code:twist_defect_surface}{Twist-defect surface code}\end{eczvaluelist}
\codefieldsection{Child}
\begin{eczvaluelist}
\item\relax
\flmRefsHyperref[eczindexfamilyrel]{code:triangle_surface}{Triangular surface code} --- The triangular surface code is the \(s=3\) member of the stellated surface-code family \NoCaseChange{\protect\cite[{Appx. D}]{cite445}}.
\end{eczvaluelist}
\codefieldsection{Cousin}
\begin{eczvaluelist}
\item\relax
\flmRefsHyperref[eczindexfamilyrel]{code:stellated_color}{Stellated color code} --- Stellated color codes are color-code analogues of stellated surface codes; the surface-code family has the same rotational parameter \(s\), but half the asymptotic \(c\)-value of the 4.8.8 stellated color-code family \NoCaseChange{\protect\cite{cite445}}.
\end{eczvaluelist}
\eczhbkcontributors{ \eczhuVVA }
\endeczcode

\eczcode{sslp}{Subset-Sum-Linear-Programming (SS-LP) code}{~\NoCaseChange{\protect\cite{cite528}}}
\codefieldsection{Description}
Qubit block quantum code that encodes a logical qubit and that is constructed using the Subset-Sum-Linear-Programming (SS-LP) numerical construction.
SS-LP codes are optimized to admit diagonal gates transversally and include \(\llparenthesis 7,2,3\rrparenthesis \) codes that realize the \(\mathsf{BD}_{16}\) and \(\mathsf{BD}_{32}\) groups transversally, yielding \(T\) and \(\sqrt{T}\) gates, respectively. Larger codes include an \(\llparenthesis 8,2,3\rrparenthesis \) code that transversally realizes \(\mathsf{BD}_{64}\).

\codefieldsection{Transversal and Permutation-Based Gates}
\begin{eczvaluelist}
\item\relax SS-LP codes are optimized to admit diagonal gates transversally and include \(\llparenthesis 7,2,3\rrparenthesis \) codes that realize the \(\mathsf{BD}_{16}\) and \(\mathsf{BD}_{32}\) groups transversally, yielding \(T\) and \(\sqrt{T}\) gates, respectively. Larger codes include an \(\llparenthesis 8,2,3\rrparenthesis \) code that transversally realizes \(\mathsf{BD}_{64}\).
\end{eczvaluelist}
\codefieldsection{Parents}
\begin{eczvaluelist}
\item\relax
\flmRefsHyperref[eczindexfamilyrel]{code:qubits_into_qubits}{Qubit code}\item\relax
\flmRefsHyperref[eczindexfamilyrel]{code:small_distance_quantum}{Small-distance block quantum code}\end{eczvaluelist}
\codefieldsection{Cousins}
\begin{eczvaluelist}
\item\relax
\flmRefsHyperref[eczindexfamilyrel]{code:stab_15_1_3}{\(\llbracket 15,1,3\rrbracket \) quantum RM code} --- The \(\llparenthesis 7,2,3\rrparenthesis \) SS-LP code realizes the \(T\) gate transversally, but requires fewer qubits than the \(\llbracket 15,1,3\rrbracket \) quantum RM code.
\item\relax
\flmRefsHyperref[eczindexfamilyrel]{code:binary_dihedral_permutation_invariant}{Binary dihedral PI code} --- SS-LP codes are optimized to admit diagonal gates transversally and include \(\llparenthesis 7,2,3\rrparenthesis \) codes that realize the \(\mathsf{BD}_{16}\) and \(\mathsf{BD}_{32}\) groups transversally, yielding \(T\) and \(\sqrt{T}\) gates, respectively. Larger codes include an \(\llparenthesis 8,2,3\rrparenthesis \) code that transversally realizes \(\mathsf{BD}_{64}\).
\end{eczvaluelist}
\eczhbkcontributors{ \eczhuVVA }
\endeczcode

\eczcode{subsystem_color}{Subsystem color code}{~\NoCaseChange{\protect\cite{cite604,cite475}}}
\codefieldsection{Alternative Names}
\begin{eczvaluelist}
\item\relax Gauge color code
\end{eczvaluelist}
\eczhIndexCodeAliasName{subsystem_color}{Gauge color code}
\codefieldsection{Description}
A subsystem version of the color code.

Subsystem color codes form a \((d,e)\) family on punctured \(D\)-colexes, or equivalently on suitably colored simplicial \(D\)-balls, encoding one logical qubit and interpolating between conventional and subsystem color codes via gauge fixing \NoCaseChange{\protect\cite{cite475}}.
Examples include 2D subsystem color codes obtained by expanding the vertices of a two-colex embedded in a surface of genus \(g\), where each vertex is split into a triangle and each edge into a pair of edges.

The stabilizer group may contain generators of unbounded weight, distinguishing these codes from stabilizer codes with bounded-weight generators for which some logical qubits were re-assigned to be gauge qubits.

Gauge fixing between subsystem color codes defined on the same lattice can be implemented using local measurements and classical processing analogous to error correction \NoCaseChange{\protect\cite{cite475}}.

\codefieldsection{Transversal and Permutation-Based Gates}
\begin{eczvaluelist}
\item\relax For a \(D\)-dimensional \((d,e)\) gauge color code, \(CNOT\) is transversal, Hadamard is transversal when \(d=e\), and \(R_n=\operatorname{diag}(1,e^{2\pi i/2^n})\) is transversal whenever \(D \geq n(D-e)\) \NoCaseChange{\protect\cite{cite475}}.
\end{eczvaluelist}
\codefieldsection{Decoding}
\begin{eczvaluelist}
\item\relax Clustering decoder \NoCaseChange{\protect\cite{cite832}}.
\item\relax Erasure decoder \NoCaseChange{\protect\cite{cite3447}}.
\item\relax Gauge-fixing decoders \NoCaseChange{\protect\cite{cite3446,cite3447}}.
\end{eczvaluelist}
\codefieldsection{Parents}
\begin{eczvaluelist}
\item\relax
\flmRefsHyperref[eczindexfamilyrel]{code:qubit_subsystem_css}{Subsystem qubit CSS code}\item\relax
\flmRefsHyperref[eczindexfamilyrel]{code:qudit_subsystem_color}{Modular-qudit subsystem color code} --- Modular-qudit subsystem color codes reduce to subsystem color codes for \(q=2\).
\end{eczvaluelist}
\codefieldsection{Children}
\begin{eczvaluelist}
\item\relax
\flmRefsHyperref[eczindexfamilyrel]{code:2d_subsystem_color}{2D subsystem color code}\item\relax
\flmRefsHyperref[eczindexfamilyrel]{code:3d_subsystem_color}{3D subsystem color code}\item\relax
\flmRefsHyperref[eczindexfamilyrel]{code:capped_color}{Capped color code (CCC)}\end{eczvaluelist}
\codefieldsection{Cousins}
\begin{eczvaluelist}
\item\relax
\flmRefsHyperref[eczindexfamilyrel]{code:color}{Color code} --- Gauge fixing relates subsystem color codes to conventional color codes defined on the same lattice \NoCaseChange{\protect\cite{cite475}}.
\item\relax
\flmRefsHyperref[eczindexfamilyrel]{code:eaoa_stabilizer}{EAOA qubit stabilizer code} --- The 15-qubit subsystem color code yields several EAOA qubit stabilizer constructions, including \(\llbracket 13,1,3;6,2,3\rrbracket \), \(\llbracket 15,1,3;5,1,2\rrbracket \), and \(\llbracket 15,1,3;4,1,4\rrbracket \) examples obtained via clean-qubits and entanglement-assisted gauge-fixing constructions \NoCaseChange{\protect\cite{cite856}}.
\item\relax
\flmRefsHyperref[eczindexfamilyrel]{code:honeycomb_floquet}{Honeycomb Floquet code} --- Both honeycomb and subsystem color codes are generated via periodic sequences of measurements. However, any measurement sequence can be performed on the color code without destroying the logical qubits, while honeycomb codes can be maintained only with specific sequences. Honeycomb codes require a shorter measurement cycle and use fewer qubits at the given code distance \NoCaseChange{\protect\cite{cite536}}.
\item\relax
\flmRefsHyperref[eczindexfamilyrel]{code:majorana_subsystem}{Majorana subsystem stabilizer code} --- A particular self-dual stabilizer Hamiltonian within the 3D subsystem color code admits a Majorana variant whose boundaries support 2D Majorana color codes \NoCaseChange{\protect\cite{cite466}}.
\item\relax
\flmRefsHyperref[eczindexfamilyrel]{code:quantum_pin}{Quantum pin code} --- Quantum pin codes have a subsystem version that can be viewed as a generalization of subsystem color codes \NoCaseChange{\protect\cite{cite702}}.
\end{eczvaluelist}
\eczhbkcontributors{ Yi-Ting (Rick) Tu, \eczhuVVA }
\endeczcode

\eczcode{holographic_subsystem}{Subsystem holographic code}{~\NoCaseChange{\protect\cite{cite4429}}}
\codefieldsection{Alternative Names}
\begin{eczvaluelist}
\item\relax Holographic hybrid code
\end{eczvaluelist}
\eczhIndexCodeAliasName{holographic_subsystem}{Holographic hybrid code}
\codefieldsection{Description}
A holographic tensor-network code constructed out of alternating isometries of the five-qubit and \(\llbracket 4,1,1,2\rrbracket \) Bacon-Shor codes.

\codefieldsection{Parents}
\begin{eczvaluelist}
\item\relax
\flmRefsHyperref[eczindexfamilyrel]{code:qubit_subsystem_stabilizer}{Subsystem qubit stabilizer code} --- The holographic hybrid code is constructed out of alternating isometries of the five-qubit and \(\llbracket 4,1,1,2\rrbracket \) Bacon-Shor codes.
\item\relax
\flmRefsHyperref[eczindexfamilyrel]{code:holographic_tensor}{Holographic tensor-network code} --- The holographic hybrid code is constructed out of alternating isometries of the five-qubit and \(\llbracket 4,1,1,2\rrbracket \) Bacon-Shor codes.
\end{eczvaluelist}
\codefieldsection{Cousins}
\begin{eczvaluelist}
\item\relax
\flmRefsHyperref[eczindexfamilyrel]{code:stab_5_1_3}{\(\llbracket 5,1,3\rrbracket \) Five-qubit perfect code} --- The holographic hybrid code is constructed out of alternating isometries of the five-qubit and \(\llbracket 4,1,1,2\rrbracket \) Bacon-Shor codes.
\item\relax
\flmRefsHyperref[eczindexfamilyrel]{code:bacon_shor_4}{\(\llbracket 4,1,1,2\rrbracket \) Four-qubit subsystem code} --- The holographic hybrid code is constructed out of alternating isometries of the five-qubit and \(\llbracket 4,1,1,2\rrbracket \) Bacon-Shor codes.
\end{eczvaluelist}
\eczhbkcontributors{ \eczhuVVA }
\endeczcode

\eczcode{subsystem_higher_dimensional_surface}{Subsystem homological code}{}
\codefieldsection{Alternative Names}
\begin{eczvaluelist}
\item\relax Subsystem generalized surface code
\end{eczvaluelist}
\eczhIndexCodeAliasName{subsystem_higher_dimensional_surface}{Subsystem generalized surface code}
\codefieldsection{Description}
A subsystem CSS code that is a subsystem version of the homological code, defined on cellulations of manifolds in arbitrary dimensions.
Gauge-group generators are of lower weight than the stabilizers of the corresponding surface code, enabling fault-tolerant syndrome extraction with simpler circuits.
The stabilizer group may contain generators of unbounded weight, distinguishing these codes from stabilizer codes with bounded-weight generators for which some logical qubits were re-assigned to be gauge qubits.

\codefieldsection{Parents}
\begin{eczvaluelist}
\item\relax
\flmRefsHyperref[eczindexfamilyrel]{code:qubit_subsystem_css}{Subsystem qubit CSS code}\item\relax
\flmRefsHyperref[eczindexfamilyrel]{code:sparse_subsystem}{QLDPC subsystem code}\end{eczvaluelist}
\codefieldsection{Children}
\begin{eczvaluelist}
\item\relax
\flmRefsHyperref[eczindexfamilyrel]{code:3d_subsystem_surface}{3D subsystem surface code}\item\relax
\flmRefsHyperref[eczindexfamilyrel]{code:subsystem_hyperbolic_surface}{Subsystem hyperbolic surface code}\item\relax
\flmRefsHyperref[eczindexfamilyrel]{code:subsystem_rotated_surface}{Subsystem rotated surface code}\item\relax
\flmRefsHyperref[eczindexfamilyrel]{code:subsystem_surface}{Subsystem surface code}\end{eczvaluelist}
\codefieldsection{Cousin}
\begin{eczvaluelist}
\item\relax
\flmRefsHyperref[eczindexfamilyrel]{code:higher_dimensional_surface}{Homological code} --- Subsystem homological codes are subsystem versions of homological codes, with gauge-group generators of lower weight than the corresponding surface-code stabilizers.
\end{eczvaluelist}
\eczhbkcontributors{ \eczhuVVA }
\endeczcode

\eczcode{subsystem_product}{Subsystem homological product code}{~\NoCaseChange{\protect\cite{cite664}}}
\codefieldsection{Description}
A CSS subsystem code constructed from a product of two (subspace) CSS codes.
The case for qubits is formulated below, but these codes have also been extended to Galois qudits \NoCaseChange{\protect\cite{cite664}}.

Denote the two CSS codes' parity-check matrix blocks as \(H_X^i, H_Z^i\) for \(i \in \{A, B\}\).
SP codes can be constructed by the following gauge generating matrices
\flmMathEnvironment{align}{}{
\begin{split}
  \label{sub:gauge}
    G_X=\left(\begin{array}{c}H_X^A \otimes I \\ I \otimes
              H_X^B \end{array}\right)
    G_Z=\left(\begin{array}{c}H_Z^A \otimes I \\ I \otimes
              H_Z^B \end{array}\right)~,
\end{split}
}
where \(I\) is the identity matrix with size chosen to match the dimensions.

A stabilizer generator matrix can be written in terms of
the codes' generating matrices, \(L_X^i, L_Z^i\) for \(i \in \{A, B\}\):
\flmMathEnvironment{align}{}{
\begin{split}
  \label{sub:stabilizer}
    H_X=\left(\begin{array}{c}H_X^A \otimes H_X^B \\
              H_X^A \otimes L_X^B \\
              L_X^A \otimes H_X^B \end{array}\right),
    H_Z=\left(\begin{array}{c}H_Z^A \otimes H_Z^B \\
              H_Z^A \otimes L_Z^B \\
              L_Z^A \otimes H_Z^B \end{array}\right)~.
\end{split}
}
The null space of \(G\) excluding \(H\) gives logical generating matrices in canonical pairs
\flmMathEnvironment{align}{}{
\begin{split}
  L_{X}&=\left(L_{X}^{A}\otimes L_{X}^{B}\right)\\
  L_{Z}&=\left(L_{Z}^{A}\otimes L_{Z}^{B}\right)~,
\end{split}
}
which satisfy \(L_{X}L_{Z}^{T}=I\).

\codefieldsection{Protection}
If the CSS codes have parameters \(\llbracket n_i,k_i,d_{i},d_{i}\rrbracket \) and sparsity \(\{r_i,c_i\}\), for \(i=A,B\) respectively,
the SP code has parameters \(\llbracket n_An_B,k_Ak_B,d\leq d_Ad_B\rrbracket \) and sparsity \(\{\max (r_A,r_B), c_A+c_B\}\).
Note the distance relation holds for both \(X\) and \(Z\), hence we omit the \(X/Z\) subscript.

\codefieldsection{Parent}
\begin{eczvaluelist}
\item\relax
\flmRefsHyperref[eczindexfamilyrel]{code:qubit_subsystem_css}{Subsystem qubit CSS code}\end{eczvaluelist}
\codefieldsection{Child}
\begin{eczvaluelist}
\item\relax
\flmRefsHyperref[eczindexfamilyrel]{code:subsystem_quantum_parity}{Subsystem hypergraph product (SHP) code} --- SP codes reduce to SHP codes when constructed from two classical codes instead of quantum CSS codes \NoCaseChange{\protect\cite{cite664}}.
\end{eczvaluelist}
\codefieldsection{Cousins}
\begin{eczvaluelist}
\item\relax
\flmRefsHyperref[eczindexfamilyrel]{code:homological_product}{Homological product code} --- SP codes reduce to homological product codes when there are no gauge qubits \NoCaseChange{\protect\cite{cite664}}.
\item\relax
\flmRefsHyperref[eczindexfamilyrel]{code:multisector_hypergraph}{Higher-dimensional homological product code} --- SP codes are projected higher-dimensional HGP codes \NoCaseChange{\protect\cite{cite664}}.
\item\relax
\flmRefsHyperref[eczindexfamilyrel]{code:qubit_concatenated}{Concatenated qubit code} --- Concatenated CSS stabilizer codes are gauge-fixed SP codes \NoCaseChange{\protect\cite[{Thm. 4}]{cite664}}.
\end{eczvaluelist}
\eczhbkcontributors{ Weilei Zeng, \eczhuVVA }
\endeczcode

\eczcode{subsystem_hyperbolic_surface}{Subsystem hyperbolic surface code}{~\NoCaseChange{\protect\cite{cite2624}}}
\codefieldsection{Description}
Subsystem generalization of the surface code on a 2D hyperbolic tessellation with gauge-group generators of weight at most three.
An \(\{r,4\}\) hyperbolic tessellation with \(E\) edges yields a \(\llbracket 3E/2,(1/2-2/r)E+2,(1-2/r)E,d\rrbracket \) subsystem code.

\codefieldsection{Protection}
Distance \(d\) is bounded between \(d_X/2\) and \(d_X\), where \(d_X\) is the \(X\)-distance of the subspace hyperbolic surface code derived from the same tessellation \NoCaseChange{\protect\cite[{Appx. F}]{cite2624}}.

\codefieldsection{Parent}
\begin{eczvaluelist}
\item\relax
\flmRefsHyperref[eczindexfamilyrel]{code:subsystem_higher_dimensional_surface}{Subsystem homological code}\end{eczvaluelist}
\codefieldsection{Cousin}
\begin{eczvaluelist}
\item\relax
\flmRefsHyperref[eczindexfamilyrel]{code:two_dimensional_hyperbolic_surface}{2D hyperbolic surface code} --- Subsystem hyperbolic surface codes are subsystem versions of 2D hyperbolic surface codes.
\end{eczvaluelist}
\eczhbkcontributors{ \eczhuVVA }
\endeczcode

\eczcode{subsystem_quantum_parity}{Subsystem hypergraph product (SHP) code}{~\NoCaseChange{\protect\cite{cite1433,cite665}}}
\codefieldsection{Alternative Names}
\begin{eczvaluelist}
\item\relax Subsystem generalized Shor code
\item\relax Bacon-Casaccino subsystem code
\end{eczvaluelist}
\eczhIndexCodeAliasName{subsystem_quantum_parity}{Subsystem generalized Shor code}
\eczhIndexCodeAliasName{subsystem_quantum_parity}{Bacon-Casaccino subsystem code}
\codefieldsection{Description}
A CSS subsystem version of the generalized Shor code that has the same parameters as the subspace version, but requires fewer stabilizer measurements, resulting in a simpler error recovery routine.
The code can also be thought of as a subsystem version of an HGP code because two such codes reduce to an HGP code upon gauge fixing \NoCaseChange{\protect\cite[{Sec. III}]{cite665}}.
The code can be obtained from a generalized Shor code by removing certain stabilizers that do not affect the code distance.

The \(X\)- and \(Z\)-type gauge generators of this CSS \(\llbracket n_1n_2,k_1k_2,\min(d_1,d_2)\rrbracket \) code correspond to rows of the following two respective matrices,
\flmMathEnvironment{align}{}{
\begin{split}
G_{X}&=H_{1}\otimes I_{n_{2}}\\
G_{Z}&=I_{n_{1}}\otimes H_{2}~,
\end{split}
}
where \(H_{1,2}\) are the parity-check matrices of two binary linear codes, \(C_1 = [n_1, k_1, d_1]\) and \(C_2 = [n_2, k_2, d_2]\)
\NoCaseChange{\protect\cite{cite665}}.

\codefieldsection{Decoding}
\begin{eczvaluelist}
\item\relax Efficient decoder \NoCaseChange{\protect\cite{cite3689}}.
\end{eczvaluelist}
\codefieldsection{Parents}
\begin{eczvaluelist}
\item\relax
\flmRefsHyperref[eczindexfamilyrel]{code:subsystem_lifted_product}{Subsystem lifted-product (SLP) code} --- SLP codes reduce to SHP codes when the lift is trivial.
\item\relax
\flmRefsHyperref[eczindexfamilyrel]{code:subsystem_product}{Subsystem homological product code} --- SP codes reduce to SHP codes when constructed from two classical codes instead of quantum CSS codes \NoCaseChange{\protect\cite{cite664}}.
\end{eczvaluelist}
\codefieldsection{Children}
\begin{eczvaluelist}
\item\relax
\flmRefsHyperref[eczindexfamilyrel]{code:bacon_shor}{Bacon-Shor code}\item\relax
\flmRefsHyperref[eczindexfamilyrel]{code:shyps}{Subsystem Hypergraph Product Simplex (SHYPS) code}\end{eczvaluelist}
\codefieldsection{Cousins}
\begin{eczvaluelist}
\item\relax
\flmRefsHyperref[eczindexfamilyrel]{code:hypergraph_product}{Hypergraph product (HGP) code} --- Two SHP codes can be gauge-fixed to yield an HGP code \NoCaseChange{\protect\cite[{Sec. III}]{cite665}}. The SHP and HGP code constructions yield the same dimension and minimum distance, but the former does not yield QLDPC codes; see \NoCaseChange{\protect\cite[{pg. 18}]{cite1448}}.
\item\relax
\flmRefsHyperref[eczindexfamilyrel]{code:generalized_shor}{Generalized Shor code} --- In a \(\llbracket n_1n_2, k_1k_2, min(d_1, d_2)\rrbracket \) generalized Shor code, error correction is achieved by measuring \((n_1−k_1)n_2+k_1(n_2−k_2)\) stabilizer generators \NoCaseChange{\protect\cite{cite3384}}. The SHP code achieves the same degree of correctability, but requires only \((n_1−k_1)k_2+k_1(n_2−k_2)\) stabilizer measurements.
\item\relax
\flmRefsHyperref[eczindexfamilyrel]{code:hybridqecc}{Hybrid QECC} --- Classical information can also be encoded in subsystem codes using their gauge qubits \NoCaseChange{\protect\cite{cite2874}}.
\item\relax
\flmRefsHyperref[eczindexfamilyrel]{code:hybrid_stabilizer}{Hybrid stabilizer code} --- Hybrid stabilizer codes can be constructed from SHP codes by using the gauge qubits of the latter to store classical information \NoCaseChange{\protect\cite[{Sec. 4}]{cite2874}}.
\item\relax
\flmRefsHyperref[eczindexfamilyrel]{code:bravyi_bacon_shor}{Bravyi-Bacon-Shor (BBS) code} --- The BBS code construction can utilize different classical codes in different rows and columns of \(A\), while the subsystem construction does not; see \NoCaseChange{\protect\cite[{pg. 4}]{cite3328}}.
Subsystem hypergraph product and BBS codes have been numerically compared \NoCaseChange{\protect\cite{cite665}}.

\end{eczvaluelist}
\eczhbkcontributors{ Sarah Meng Li, \eczhuVVA }
\endeczcode

\eczcode{shyps}{Subsystem Hypergraph Product Simplex (SHYPS) code}{~\NoCaseChange{\protect\cite{cite785}}}
\codefieldsection{Description}
Family of quantum LDPC codes obtained by combining the \flmRefsHyperref{code:subsystem_quantum_parity}{subsystem hypergraph product code} construction with classical \flmRefsHyperref{code:simplex}{simplex codes}. 
The results are CSS subsystem codes with weight-three gauge generators and code parameters \(\llbracket n=(2^r − 1)^2, k=r^2, d=2^{r-1}\rrbracket \) for \(r \geq 3\).

Due to their symmetric structure, SHYPS codes inherit the large automorphism group of the underlying classical simplex codes. 
More precisely, \(| Aut( SHYPS(r) ) | \geq |GL(r,\mathbb{F}_2)|^2\), which is exponential in the number of logical qubits. 
This large automorphism group can be leveraged to obtain a depth-one fault-tolerant implementation for a large set of logical Clifford operators. 
This set of depth-one Clifford generators is sufficiently large to allow for efficient compilation, i.e., any \(m\)-qubit Clifford operator can be executed in depth of \flmRefsHyperref{ref65}{order} \(O(m)\). 
SHYPS codes exhibit practical single-shot features, so only \flmRefsHyperref{ref65}{order} \(O(m)\) rounds of syndrome extraction are required to fault-tolerantly execute any logical \(m\)-qubit \flmRefsHyperref{ref409}{Clifford circuit}.

\codefieldsection{Protection}
Memory simulations of the \(\llbracket 49, 9, 4\rrbracket \) and \(\llbracket 225, 16, 8\rrbracket \) SHYPS codes under circuit-level noise, using a sliding-window BPLSD decoder, yield pseudo-thresholds of approximately \(0.32\%\) and \(0.35\%\), respectively \NoCaseChange{\protect\cite{cite785}}.
Depth-126 logical \flmRefsHyperref{ref409}{Clifford-circuit} simulations on two blocks of the \(\llbracket 49, 9, 4\rrbracket \) SHYPS code were also performed in \NoCaseChange{\protect\cite{cite785}}.

\codefieldsection{Rate}
The exact encoding rate is \(k/n = r^2/(2^r-1)^2\), i.e., \flmRefsHyperref{ref65}{asymptotically} as \(\Theta( (\log n)^2/n )\) \NoCaseChange{\protect\cite{cite785}}.
\codefieldsection{Transversal and Permutation-Based Gates}
\begin{eczvaluelist}
\item\relax Cross-block transversal CNOT gates \(\prod_{i=1}^n CNOT_{i, n+(\sigma_1\otimes\sigma_2)(i)}\) for \(\sigma_1, \sigma_2\) automorphisms of the classical simplex code. These operators implement a generating set of cross-block logical CNOT gates.
\item\relax In-block phase-type fold-transversal gates with \(Sp(2k,\mathbb{F}_2)\) representation \(\begin{pmatrix} I & (\sigma \otimes \sigma^T)\tau \\ 0 & 1\end{pmatrix}\) for \(\sigma\) an automorphism of the classical simplex code, and \(\tau\) a self-inverse permutation which acts like \(\tau (e_i \otimes e_j) = e_j \otimes e_i\) for the canonical basis \(\{e_i\}\) of \(\mathbb{F}_2^{\sqrt{n}}\). These operators implement a generating set of logical in-block diagonal gates \NoCaseChange{\protect\cite{cite785}}.
\item\relax Cross-block phase-type fold-transversal gates \(\prod_{i=1}^n CZ_{i, n+(\sigma_1\otimes\sigma_2)\tau(i)}\) for \(\sigma_1, \sigma_2\) automorphisms of the classical simplex code and \(\tau\) as above. These operators implement a generating set of logical cross-block diagonal gates \NoCaseChange{\protect\cite{cite785}}.
\item\relax Fold-transversal Hadamard gate \(H^{\otimes n} \tau \), with \(\tau\) as above. Implements logical Hadamard-SWAP operator \(H^{\otimes k} \tau_k \), with \(\tau_k\) defined analogously to \(\tau\) \NoCaseChange{\protect\cite{cite785}}.
\end{eczvaluelist}
\codefieldsection{Gates}
\begin{eczvaluelist}
\item\relax Arbitrary \(m\)-qubit logical \flmRefsHyperref{ref409}{Clifford gates} can be implemented in \(4m( 1+o(1) )\) logical cycles, each consisting of a depth-one physical Clifford layer followed by a depth-six syndrome-extraction round \NoCaseChange{\protect\cite{cite785}}.
\item\relax Worst-case logical Clifford operation on \(b\) blocks can be implemented fault-tolerantly in depth roughly \(4bk\) using at most \(b\) auxiliary code blocks \NoCaseChange{\protect\cite{cite785}}.
\end{eczvaluelist}
\codefieldsection{Decoding}
\begin{eczvaluelist}
\item\relax BPLSD decoder \NoCaseChange{\protect\cite{cite785}}.
\end{eczvaluelist}
\codefieldsection{Fault Tolerance}
\begin{eczvaluelist}
\item\relax Logical Clifford operation on \(b\) blocks can be implemented fault-tolerantly in depth \(4br^2( 1+o(1) )\) while remaining compatible with one syndrome-extraction round between logical generators \NoCaseChange{\protect\cite{cite785}}.
\end{eczvaluelist}
\codefieldsection{Parent}
\begin{eczvaluelist}
\item\relax
\flmRefsHyperref[eczindexfamilyrel]{code:subsystem_quantum_parity}{Subsystem hypergraph product (SHP) code}\end{eczvaluelist}
\codefieldsection{Cousins}
\begin{eczvaluelist}
\item\relax
\flmRefsHyperref[eczindexfamilyrel]{code:simplex}{\([2^m-1,m,2^{m-1}]\) simplex code} --- SHYPS code gauge generator matrices are constructed from hypergraph products of simplex codes \NoCaseChange{\protect\cite{cite785}}.
\item\relax
\flmRefsHyperref[eczindexfamilyrel]{code:single_shot}{Single-shot code} --- SHYPS codes exhibit practical single-shot signatures, including logical error-rate stability under small-window sliding-window decoding and constant single-shot distance \(d_{\mathrm{ss}}=3\), which supports using one syndrome-extraction round between logical generators \NoCaseChange{\protect\cite{cite785}}.
\end{eczvaluelist}
\eczhbkcontributors{ Alexis Schotte, \eczhuVVA }
\endeczcode

\eczcode{subsystem_lifted_product}{Subsystem lifted-product (SLP) code}{~\NoCaseChange{\protect\cite{cite4250}}}
\codefieldsection{Description}
Member of a family of subsystem CSS codes constructed from a subsystem hypergraph product of a \flmRefsHyperref{ref47}{lifted} binary linear code.

\codefieldsection{Parent}
\begin{eczvaluelist}
\item\relax
\flmRefsHyperref[eczindexfamilyrel]{code:qubit_subsystem_css}{Subsystem qubit CSS code}\end{eczvaluelist}
\codefieldsection{Child}
\begin{eczvaluelist}
\item\relax
\flmRefsHyperref[eczindexfamilyrel]{code:subsystem_quantum_parity}{Subsystem hypergraph product (SHP) code} --- SLP codes reduce to SHP codes when the lift is trivial.
\end{eczvaluelist}
\codefieldsection{Cousins}
\begin{eczvaluelist}
\item\relax
\flmRefsHyperref[eczindexfamilyrel]{code:lifted_product}{Lifted-product (LP) code} --- SLP codes reduce to (subspace) LP codes when there is no gauge subsystem.
\item\relax
\flmRefsHyperref[eczindexfamilyrel]{code:binary_linear}{Linear binary code} --- SLP codes are constructed from a subsystem hypergraph product of a \flmRefsHyperref{ref47}{lifted} binary linear code.
\end{eczvaluelist}
\eczhbkcontributors{ \eczhuVVA }
\endeczcode

\eczcode{subsystem_qubits_into_qubits}{Subsystem qubit code}{}
\codefieldsection{Alternative Names}
\begin{eczvaluelist}
\item\relax Gauge qubit code
\end{eczvaluelist}
\eczhIndexCodeAliasName{subsystem_qubits_into_qubits}{Gauge qubit code}

\codefieldsection{Kingdom root code}
for the \flmRefsHyperref{kingdom:qubits_into_qubits}{Qubit Kingdom}.
\codefieldsection{Description}
Subsystem QECC encoding into a \(2^n\)-dimensional (i.e., \(n\)-qubit) Hilbert space.

\codefieldsection{Transversal and Permutation-Based Gates}
\begin{eczvaluelist}
\item\relax If a subsystem qubit code \(Q\) of length \(n\) has compact subgroups \(N\triangleleft G\leq \mathrm{Aut}(Q)\) such that \(G/N\) is finite, non-Abelian, simple, and not \(A_5\), then \(n\) is at least the minimal permutation degree \(\mu(G/N)\) \NoCaseChange{\protect\cite[{Thm. 1}]{cite723}}.
\end{eczvaluelist}
\codefieldsection{Parents}
\begin{eczvaluelist}
\item\relax
\flmRefsHyperref[eczindexfamilyrel]{code:oa_qubits_into_qubits}{OA qubit code} --- An OA qubit code which has gauge structure (e.g., gauge qubits) but no block structure is a subsystem qubit code.
\item\relax
\flmRefsHyperref[eczindexfamilyrel]{code:subsystem_qudits_into_qudits}{Subsystem modular-qudit code} --- Subsystem modular-qudit codes reduce to subsystem qubit codes for qudit dimension \(q=2\).
\item\relax
\flmRefsHyperref[eczindexfamilyrel]{code:subsystem_galois_into_galois}{Subsystem Galois-qudit code} --- Subsystem Galois-qudit quantum codes for \(q=2\) correspond to subsystem qubit codes.
\end{eczvaluelist}
\codefieldsection{Child}
\begin{eczvaluelist}
\item\relax
\flmRefsHyperref[eczindexfamilyrel]{code:qubit_subsystem_stabilizer}{Subsystem qubit stabilizer code}\end{eczvaluelist}
\codefieldsection{Cousin}
\begin{eczvaluelist}
\item\relax
\flmRefsHyperref[eczindexfamilyrel]{code:qubits_into_qubits}{Qubit code} --- Subsystem qubit codes reduce to (subspace) qubit codes when there is no gauge subsystem.
\end{eczvaluelist}
\eczhbkcontributors{ \eczhuVVA }
\endeczcode

\eczcode{qubit_subsystem_css}{Subsystem qubit CSS code}{~\NoCaseChange{\protect\cite{cite2884,cite1742,cite4430}}}
\codefieldsection{Description}
Subsystem qubit stabilizer code which admits a set of gauge-group generators which consist of either all-\(Z\) or all-\(X\) Pauli strings.
This ensures that the code's stabilizer group is also CSS.

The gauge-group generators can be expressed as a matrix using the symplectic representation. This matrix is of the form
\flmMathEnvironment{align}{}{
G=\begin{pmatrix}0 & G_{Z}\\
G_{X} & 0
\end{pmatrix}~.\label{ref4431}
}
The two matrix blocks, \(G_{Z}\) and \(G_X\), correspond to the parity-check matrices of two \flmRefsHyperref{code:binary_linear}{binary linear codes}, an \([n,k_X,d_X]\) code \(C_X\) and an \([n,k_Z,d_Z]\) code \(C_Z\), respectively.
Code parameters can be expressed in terms of only data associated with these two classical codes \NoCaseChange{\protect\cite{cite1742,cite2884}}.
Explicit basis-state constructions are given in Ref. \NoCaseChange{\protect\cite{cite4432}}.

\codefieldsection{Protection}
For any code whose gauge group is generated by \(XX\) and \(ZZ\), the weight of an \(X\)-type (\(Z\)-type) single-qubit bare-logical operator is lower-bounded by the number of \(Z\)-type (\(X\)-type) bare-logical operators acting on its
supporting logical qubits \NoCaseChange{\protect\cite{cite3335,cite4433}}.

\codefieldsection{Decoding}
\begin{eczvaluelist}
\item\relax Steane-type decoder utilizing data from the underlying classical codes \NoCaseChange{\protect\cite{cite4432}}.
\end{eczvaluelist}
\codefieldsection{Parents}
\begin{eczvaluelist}
\item\relax
\flmRefsHyperref[eczindexfamilyrel]{code:qubit_subsystem_stabilizer}{Subsystem qubit stabilizer code} --- Subsystem qubit CSS codes are subsystem qubit stabilizer codes whose gauge groups admit a generating set of pure-\(X\) and pure-\(Z\) Pauli strings. Any \(\llbracket n,k,r,d\rrbracket \) subsystem qubit stabilizer code can be mapped onto a \(\llbracket 2n,2k,2r,\geq d\rrbracket \) subsystem CSS code via \flmRefsHyperref{ref436}{symplectic doubling}, which preserves geometric locality of a code up to a constant factor. Every subsystem qubit stabilizer code can be constructed from two nested subsystem CSS codes satisfying certain constraints \NoCaseChange{\protect\cite{cite4432}}.
\item\relax
\flmRefsHyperref[eczindexfamilyrel]{code:qudit_subsystem_css}{Subsystem modular-qudit CSS code} --- Subsystem modular-qudit CSS codes reduce to subsystem qubit CSS codes for \(q=2\).
\item\relax
\flmRefsHyperref[eczindexfamilyrel]{code:galois_subsystem_css}{Subsystem Galois-qudit CSS code} --- Subsystem Galois-qudit CSS codes reduce to subsystem qubit CSS codes for \(q=2\).
\end{eczvaluelist}
\codefieldsection{Children}
\begin{eczvaluelist}
\item\relax
\flmRefsHyperref[eczindexfamilyrel]{code:bravyi_bacon_shor}{Bravyi-Bacon-Shor (BBS) code}\item\relax
\flmRefsHyperref[eczindexfamilyrel]{code:compass_model}{Compass code}\item\relax
\flmRefsHyperref[eczindexfamilyrel]{code:subsystem_lifted_product}{Subsystem lifted-product (SLP) code}\item\relax
\flmRefsHyperref[eczindexfamilyrel]{code:subsystem_product}{Subsystem homological product code}\item\relax
\flmRefsHyperref[eczindexfamilyrel]{code:subsystem_color}{Subsystem color code}\item\relax
\flmRefsHyperref[eczindexfamilyrel]{code:css_plaquette}{CSS-Plaquette code}\item\relax
\flmRefsHyperref[eczindexfamilyrel]{code:subsystem_higher_dimensional_surface}{Subsystem homological code}\end{eczvaluelist}
\codefieldsection{Cousin}
\begin{eczvaluelist}
\item\relax
\flmRefsHyperref[eczindexfamilyrel]{code:qubit_css}{Qubit CSS code} --- Subsystem qubit CSS codes reduce to (subspace) CSS qubit codes when there is no gauge subsystem.
\end{eczvaluelist}
\eczhbkcontributors{ \eczhuVVA }
\endeczcode

\eczcode{qubit_subsystem_stabilizer}{Subsystem qubit stabilizer code}{~\NoCaseChange{\protect\cite{cite3384}}}
\codefieldsection{Alternative Names}
\begin{eczvaluelist}
\item\relax Gauge qubit stabilizer code
\end{eczvaluelist}
\eczhIndexCodeAliasName{qubit_subsystem_stabilizer}{Gauge qubit stabilizer code}
\codefieldsection{Description}
A stabilizer code with some of its logical qubits denoted as \textit{gauge} qubits and not used for storage of logical information.
Note that this doesn't lead to new codes but does lead to new error correction and fault tolerance procedures.
Subsystem codes are denoted by \(\llbracket n,k,g,d\rrbracket \), similar to stabilizer codes, but with an extra parameter \(g\) denoting the number of gauge qubits.

Subsystem qubit stabilizer codes are defined by a gauge group \(\mathsf{G}\), together with a stabilizer subgroup \(\mathsf{S}\) of the center of \(\mathsf{G}\).
The table below summarizes the relevant groups and their sizes for a subsystem qubit stabilizer code.
  \begin{flmFloat}{table}{NumCap}\flmCellsBeginCenter
\long\def\flmTempTypesetThisTable#1{%
\begin{tblr}{#1,
  hspan=minimal,
  cell{1}{1}={}{c, font={\flmCellsHeaderFont}},
  cell{1}{2}={}{c, font={\flmCellsHeaderFont}},
  cell{1}{3}={}{c, font={\flmCellsHeaderFont}},
  cell{2}{1}={}{c},
  cell{2}{2}={}{c},
  cell{2}{3}={}{c},
  cell{3}{1}={}{c},
  cell{3}{2}={}{c},
  cell{3}{3}={}{c},
  cell{4}{1}={}{c},
  cell{4}{2}={}{c},
  cell{4}{3}={}{c},
  cell{5}{1}={}{c},
  cell{5}{2}={}{c},
  cell{5}{3}={}{c},
  cell{6}{1}={}{c},
  cell{6}{2}={}{c},
  cell{6}{3}={}{c},
  cell{7}{1}={}{c},
  cell{7}{2}={}{c},
  cell{7}{3}={}{c},
  cell{8}{1}={}{c},
  cell{8}{2}={}{c},
  cell{8}{3}={}{c},
  hline{2}={1}{.4pt,solid},
  hline{2}={2}{.4pt,solid},
  hline{2}={3}{.4pt,solid}}%
\toprule
purpose & symbol & size\\

    gauge group & \(\mathsf{G}\) & \(4\cdot 2^{n-k+g}\)
        \\

    stabilizer group & \(\mathsf{S}\) & \(2^{n-k-g}\)
        \\

    code-preserving Paulis & \(\mathsf{N}(\mathsf{S})\) & \(4\cdot 2^{n+k+g}\)
        \\

    logical Paulis & \(\mathsf{N}(\mathsf{S})/\mathsf{G}\) & \(4^{k}\)
        \\

    gauge Paulis & \(\frac{\mathsf{N}(\mathsf{S})}{\mathsf{N}(\mathsf{S})/\mathsf{G}\times\left\langle i,\mathsf{S}\right\rangle }=\mathsf{G}/\left\langle i,\mathsf{S}\right\rangle\) & \(4^{g}\)
        \\

    gauge-preserving Paulis & \(\mathsf{N}(\mathsf{G})\) & \(4\cdot 2^{n+k-g}\)
        \\

    bare logicals & \(\mathsf{N}(\mathsf{G})/\left\langle i,\mathsf{S}\right\rangle\) & \(4^{k}\)
    \\
\bottomrule
\end{tblr}%
}%
\def\flmTmpMaxW{\dimexpr 0.96\linewidth\relax}%
\setbox0=\hbox{\flmTempTypesetThisTable{colspec={ccc}}}%
\ifdim\wd0<\flmTmpMaxW\relax
  \leavevmode\box0 
\else
  \flmTempTypesetThisTable{width=\flmTmpMaxW,colspec={X[-1]X[-1]X[-1]}}
\fi
\flmCellsEndCenter \caption{Groups relevant to subsystem qubit stabilizer codes. The centralizer \(\mathsf{N}(\mathsf{S})\) is the group formed by all elements of the \(n\)-qubit Pauli group that commute with all elements in \(\mathsf{S}\). 
    The notation \(\mathsf{N}\) is used because the centralizer of a stabilizer group \(\mathsf{S}\) is the same as the group's normalizer, the group formed by all elements of the \(n\)-qubit Pauli group that conjugate \(\mathsf{S}\) to itself.
    This equivalence is no longer the case for non-stabilizer Pauli groups, e.g., the normalizer of the Pauli group is the group itself, while its centralizer is just \(\langle i \rangle\).
    The gauge group and centralizer are defined so as to include \(i\) and its powers as elements, while the stabilizer group is not.}\label{ref4434}\end{flmFloat}

To create these codes proceed as follows.
Choose \(2n\) operators \(\{ \tilde{X}_j,\tilde{Z}_j\}_{j=1}^n\) from \(\mathsf{P}_n\), the \flmRefsHyperref{ref663}{Pauli group} on \(n\) qubits, such that they obey the same commutation relations as the regular \(n\)-qubit Pauli generators \( \{X_j,Z_j\}_{j=1}^n \) (the subscript on these latter operators indicates the single qubit the Pauli matrix acts on).
The tilde operators might act on more than one physical (or \textit{bare}) qubit but they behave as if they acted only on a single qubit.
WLOG we can choose a stabilizer group as \( \mathsf{S} = \langle \tilde{Z}_1,\dots, \tilde{Z}_s \rangle \). It follows that the normalizer of \(\mathsf{S} \) is \( \mathsf{N}(\mathsf{S}) = \langle i, \tilde{Z}_1,\dots, \tilde{Z}_n, \tilde{X}_{s+1},\dots, \tilde{X}_n \rangle \).
We now choose a gauge group as \( \mathsf{G} = \langle i, \tilde{Z}_1,\dots, \tilde{Z}_s, \tilde{X}_{s+1}, \tilde{Z}_{s+1}, \dots, \tilde{X}_{s+g}, \tilde{Z}_{s+g} \rangle \) with \( s + g \leq n \).
The logical group is chosen as \( \mathsf{L} = \mathsf{N}(\mathsf{S})/\mathsf{G} \simeq \langle \tilde{X}_{s+g+1},\tilde{Z}_{s+g+1}, \dots, \tilde{X}_n,\tilde{Z}_n \rangle \).
Now the codespace \( C \) is as usual the \(+1\) eigenspace of the stabilizer \( \mathsf{S} \).
But the gauge and logical groups have further decomposed this space into \( C = A \otimes B \simeq (\mathbb{C}^2)^{\otimes k} \otimes (\mathbb{C}^2)^{\otimes g} \).
Thus the Hilbert space is partitioned into 3 sets; \(k\) logical qubits, \(g\) gauge qubits, and \(s\) error-syndrome qubits, with \(s+g+k=n\).

\codefieldsection{Protection}
Detects errors on \(d-1\) qubits, corrects errors on \(\left\lfloor (d-1)/2 \right\rfloor\) qubits.

There is the following analogue of the \flmTerm{term}{ref1043}{}{Knill-Laflamme conditions} for subsystem qubit stabilizer codes.
A set of errors \( \{ E_a \} \) is correctable iff \( E_aE_b \not\in \mathsf{N}(\mathsf{S}) \setminus \mathsf{G} \) for all pairs \(a,b\). The distance of the code is the minimal weight of operators in \( \mathsf{N}(\mathsf{S}) \setminus \mathsf{G}\).

\flmRefsHyperref{ref672}{Pure} \(\llbracket n,k,g,d\rrbracket \) subsystem stabilizer codes codes satisfy \(k+g\leq n-2d+2\), and there are no \(\llbracket n,n-2d+2,g>0,d\rrbracket \) codes \NoCaseChange{\protect\cite[{Cor. 12; Thm. 20}]{cite1742}}.

Entropic conditions have been formulated for random projective measurement noise \NoCaseChange{\protect\cite{cite3211}}.

\codefieldsection{Encoding}
\begin{eczvaluelist}
\item\relax A subsystem codeword can be encoded with the \flmRefsHyperref{ref409}{Clifford circuits} of the stabilizer code corresponding to treating all gauge qubits as logical qubits. One can use the standard-form or the conjugation method \NoCaseChange{\protect\cite{cite4435,cite4144}}.
\end{eczvaluelist}
\codefieldsection{Gates}
\begin{eczvaluelist}
\item\relax Logical \flmRefsHyperref{ref409}{Clifford gates} can be implemented fault-tolerantly for subsystem codes of distance at least three \NoCaseChange{\protect\cite{cite3204}}.
\end{eczvaluelist}
\codefieldsection{Fault Tolerance}
\begin{eczvaluelist}
\item\relax Logical \flmRefsHyperref{ref409}{Clifford gates} can be implemented fault-tolerantly for subsystem codes of distance at least three \NoCaseChange{\protect\cite{cite3204}}.
\end{eczvaluelist}
\codefieldsection{Code Capacity Threshold}
\begin{eczvaluelist}
\item\relax For correlated Pauli noise, bounds can be obtained by mapping the effect of noise on the code to a statistical mechanical model \NoCaseChange{\protect\cite{cite4377}}.
\end{eczvaluelist}
\codefieldsection{Notes}
\begin{eczvaluelist}
\item\relax See Ref. \NoCaseChange{\protect\cite{cite4436}} for algorithms and lists of possible tilings of particular subsystem codes.
\item\relax Subsystem qubit stabilizer codes can be used to estimate logical Pauli noise for any correctable physical noise \NoCaseChange{\protect\cite{cite4034}}.
\item\relax Subsystem qubit stabilizer codes with two or more gauge qubits admit quantum contextuality \NoCaseChange{\protect\cite{cite4437}}.
\end{eczvaluelist}
\codefieldsection{Parents}
\begin{eczvaluelist}
\item\relax
\flmRefsHyperref[eczindexfamilyrel]{code:subsystem_qubits_into_qubits}{Subsystem qubit code}\item\relax
\flmRefsHyperref[eczindexfamilyrel]{code:qubit_stabilizer_oaqecc}{Operator-algebra (OA) qubit stabilizer code} --- An OA qubit stabilizer code storing no classical information but retaining gauge qubits for its quantum code is a subsystem qubit stabilizer code.
\item\relax
\flmRefsHyperref[eczindexfamilyrel]{code:qudit_subsystem_stabilizer}{Subsystem modular-qudit stabilizer code} --- Subsystem modular-qudit stabilizer codes reduce to subsystem qubit stabilizer codes for qudit dimension \(q=2\).
\item\relax
\flmRefsHyperref[eczindexfamilyrel]{code:galois_subsystem_stabilizer}{Subsystem Galois-qudit stabilizer code} --- Subsystem Galois-qudit stabilizer codes reduce to subsystem qubit stabilizer codes for qudit dimension \(q=2\).
\end{eczvaluelist}
\codefieldsection{Children}
\begin{eczvaluelist}
\item\relax
\flmRefsHyperref[eczindexfamilyrel]{code:majorana_subsystem}{Majorana subsystem stabilizer code} --- Subsystem qubit stabilizer codes have been formulated in terms of Majorana operators \NoCaseChange{\protect\cite{cite3482}}.
\item\relax
\flmRefsHyperref[eczindexfamilyrel]{code:holographic_subsystem}{Subsystem holographic code} --- The holographic hybrid code is constructed out of alternating isometries of the five-qubit and \(\llbracket 4,1,1,2\rrbracket \) Bacon-Shor codes.
\item\relax
\flmRefsHyperref[eczindexfamilyrel]{code:subsystem_hypergraph}{Sarvepalli-Brown subsystem code}\item\relax
\flmRefsHyperref[eczindexfamilyrel]{code:qubit_subsystem_css}{Subsystem qubit CSS code} --- Subsystem qubit CSS codes are subsystem qubit stabilizer codes whose gauge groups admit a generating set of pure-\(X\) and pure-\(Z\) Pauli strings. Any \(\llbracket n,k,r,d\rrbracket \) subsystem qubit stabilizer code can be mapped onto a \(\llbracket 2n,2k,2r,\geq d\rrbracket \) subsystem CSS code via \flmRefsHyperref{ref436}{symplectic doubling}, which preserves geometric locality of a code up to a constant factor. Every subsystem qubit stabilizer code can be constructed from two nested subsystem CSS codes satisfying certain constraints \NoCaseChange{\protect\cite{cite4432}}.
\item\relax
\flmRefsHyperref[eczindexfamilyrel]{code:subsystem_spacetime_circuit}{Subsystem spacetime circuit code}\item\relax
\flmRefsHyperref[eczindexfamilyrel]{code:3d_kitaev_honeycomb}{3D Kitaev honeycomb code}\item\relax
\flmRefsHyperref[eczindexfamilyrel]{code:kitaev_honeycomb}{Kitaev honeycomb code}\item\relax
\flmRefsHyperref[eczindexfamilyrel]{code:subsystem_three_fermion}{Three-fermion (3F) subsystem code}\end{eczvaluelist}
\codefieldsection{Cousins}
\begin{eczvaluelist}
\item\relax
\flmRefsHyperref[eczindexfamilyrel]{code:qubit_stabilizer}{Qubit stabilizer code} --- Subsystem qubit stabilizer codes reduce to qubit stabilizer codes when there are no gauge qubits.
An \(\llbracket n,k,d\rrbracket \) qubit stabilizer code can be converted into an \flmRefsHyperref{ref65}{order} \(\llbracket O(\ell \delta n),k,\Omega(d/w)\rrbracket \) subsystem qubit stabilizer code with weight-three gauge operators via the wire-code mapping \NoCaseChange{\protect\cite{cite490}}, which uses \flmRefsHyperref{ref491}{weight reduction}. 
Here, \(w\) and \(\delta\) are the weight and degree of the input code's Tanner graph, while \(\ell\) is the length of the longest edge of a particular embedding of that graph.

\item\relax
\flmRefsHyperref[eczindexfamilyrel]{code:q-ary_bch}{Bose–Chaudhuri–Hocquenghem (BCH) code} --- BCH codes yield subsystem stabilizer codes via the subsystem extension of the Hermitian construction to subsystem codes \NoCaseChange{\protect\cite[{Exam. 1}]{cite1742}}.
\item\relax
\flmRefsHyperref[eczindexfamilyrel]{code:reed_solomon}{Reed-Solomon (RS) code} --- Primitive RS codes yield subsystem stabilizer codes via the subsystem extension of the Hermitian construction to subsystem codes \NoCaseChange{\protect\cite[{Exam. 3}]{cite1742}}.
\item\relax
\flmRefsHyperref[eczindexfamilyrel]{code:commuting_projector}{Commuting-projector Hamiltonian code} --- Ground-state spaces of qubit commuting-projector Hamiltonians with weight-two (two-body) terms cannot be used to suppress errors in adiabatic quantum computation \NoCaseChange{\protect\cite{cite2687}}, but this can be circumvented with excited-state subspaces \NoCaseChange{\protect\cite{cite2688}} or ground-state subspaces of subsystem code Hamiltonians, e.g., using BBS codes \NoCaseChange{\protect\cite{cite670,cite2689}}.
\item\relax
\flmRefsHyperref[eczindexfamilyrel]{code:da}{Dynamical code} --- A dynamical code can be viewed as a subsystem qubit stabilizer code, albeit one with fewer logical qubits.
\item\relax
\flmRefsHyperref[eczindexfamilyrel]{code:spacetime_circuit}{Spacetime circuit code} --- Spacetime circuit codes can be upgraded to subsystem codes by \flmRefsHyperref{ref666}{gauging out} a subgroup of the logical \flmRefsHyperref{ref663}{Pauli group} which causes trivial faults in the corresponding \flmRefsHyperref{ref409}{Clifford circuit}.
\item\relax
\flmRefsHyperref[eczindexfamilyrel]{code:hybrid_stabilizer}{Hybrid stabilizer code} --- Hybrid stabilizer codes can be constructed from subsystem qubit stabilizer codes by using the gauge qubits of the latter to store classical information \NoCaseChange{\protect\cite[{Thm. 4}]{cite2874}}.
\item\relax
\flmRefsHyperref[eczindexfamilyrel]{code:stab_6_1_3}{\(\llbracket 6,1,3\rrbracket \) Six-qubit stabilizer code} --- The \(\llbracket 6,1,3\rrbracket \) six-qubit code can be converted into a \(\llbracket 6,1,1,3\rrbracket \) subsystem code that saturates the subsystem Singleton bound while requiring only four stabilizer measurements during recovery \NoCaseChange{\protect\cite{cite451,cite3325}}.
\item\relax
\flmRefsHyperref[eczindexfamilyrel]{code:data_syndrome}{Quantum data-syndrome (QDS) code} --- The DS construction can be extended to subsystem qubit stabilizer codes \NoCaseChange{\protect\cite{cite4032}}.
\item\relax
\flmRefsHyperref[eczindexfamilyrel]{code:qubit_css}{Qubit CSS code} --- Qubit CSS "seed" codes can be used to produce subsystem qubit stabilizer codes \NoCaseChange{\protect\cite{cite4250}}.
\item\relax
\flmRefsHyperref[eczindexfamilyrel]{code:balanced_product}{Balanced product (BP) code} --- The original explicit balanced-product family is first constructed as a horizontal subsystem balanced-product code built from expander codes and cyclic repetition codes \NoCaseChange{\protect\cite{cite434}}.
\item\relax
\flmRefsHyperref[eczindexfamilyrel]{code:distance_balanced}{Distance-balanced code} --- The weight reduction procedure of Ref. \NoCaseChange{\protect\cite{cite4438}} has been extended to subsystem qubit stabilizer codes \NoCaseChange{\protect\cite{cite490}}.
\end{eczvaluelist}
\eczhbkcontributors{ Eric Kubischta, \eczhuPhF, \eczhuVVA }
\endeczcode

\eczcode{subsystem_rotated_surface}{Subsystem rotated surface code}{~\NoCaseChange{\protect\cite{cite3479}}}
\codefieldsection{Description}
Subsystem version of the rotated surface code.

For example, a \(\llbracket 3d^2-2d,1,d\rrbracket \) family admits weight-six \(X,Z\)-type stabilizer generators on hexagonal plaquettes and weight-three gauge generators on bow-tie shaped regions on the dual lattice.

\codefieldsection{Parents}
\begin{eczvaluelist}
\item\relax
\flmRefsHyperref[eczindexfamilyrel]{code:subsystem_higher_dimensional_surface}{Subsystem homological code}\item\relax
\flmRefsHyperref[eczindexfamilyrel]{code:translationally_invariant_subsystem}{Lattice subsystem code}\end{eczvaluelist}
\codefieldsection{Cousin}
\begin{eczvaluelist}
\item\relax
\flmRefsHyperref[eczindexfamilyrel]{code:rotated_surface}{Rotated surface code} --- Subsystem rotated surface codes are subsystem versions of rotated surface codes.
\end{eczvaluelist}
\eczhbkcontributors{ \eczhuVVA }
\endeczcode

\eczcode{subsystem_spacetime_circuit}{Subsystem spacetime circuit code}{~\NoCaseChange{\protect\cite{cite668,cite4418}}}
\codefieldsection{Description}
Subsystem stabilizer code obtained from a spacetime circuit code by \flmRefsHyperref{ref666}{gauging out} logical operators that correspond to circuit faults with trivial effect \NoCaseChange{\protect\cite[{Sec. 5.4}]{cite667}}.
In the original circuit-to-code construction, each circuit element is replaced by low-weight gauge generators enforcing its input-output relations, yielding subsystem codes from restricted Clifford postselection circuits \NoCaseChange{\protect\cite{cite668}}.

An \(\llbracket n,k,d\rrbracket \) stabilizer code can be mapped into a sparse subsystem code with the same \(k\) and \(d\) as follows.
One can take the fault-tolerant syndrome extraction circuit associated with the stabilizer code, construct its spacetime circuit code, and then \flmRefsHyperref{ref666}{gauge out} qubits corresponding to trivial faults.
The subsystem code can be made geometrically local at the cost of more ancilla qubits \NoCaseChange{\protect\cite{cite668}}.

\codefieldsection{Protection}
When derived from a fault-tolerant error-detecting circuit for an \(\llbracket n,k,d\rrbracket \) stabilizer code, the resulting subsystem spacetime circuit code has parameters \(\llbracket O(|V|),k,d\rrbracket \), so the base-code distance is preserved \NoCaseChange{\protect\cite{cite668}}.

\codefieldsection{Rate}
The spacetime circuit code construction is used to show the existence of spatially local subsystem codes that nearly saturate the \flmRefsHyperref{ref492}{subsystem BT bound} \NoCaseChange{\protect\cite{cite668}}.
\codefieldsection{Fault Tolerance}
\begin{eczvaluelist}
\item\relax Fault-tolerant measurement gadget that is a modification based on the DiVincenzo-Shor cat-state method \NoCaseChange{\protect\cite{cite761,cite3300}}.
\item\relax The original construction uses \(|+\rangle\)-state vertex qubits together with expander-graph parity checks to obtain \(O(w)\)-size postselection gadgets for measuring weight-\(w\) stabilizers, with gates acting on at most ten wires \NoCaseChange{\protect\cite{cite668}}.
\end{eczvaluelist}
\codefieldsection{Parents}
\begin{eczvaluelist}
\item\relax
\flmRefsHyperref[eczindexfamilyrel]{code:qubit_subsystem_stabilizer}{Subsystem qubit stabilizer code}\item\relax
\flmRefsHyperref[eczindexfamilyrel]{code:sparse_subsystem}{QLDPC subsystem code}\end{eczvaluelist}
\codefieldsection{Cousin}
\begin{eczvaluelist}
\item\relax
\flmRefsHyperref[eczindexfamilyrel]{code:spacetime_circuit}{Spacetime circuit code} --- Spacetime circuit codes can yield subsystem spacetime circuit codes by \flmRefsHyperref{ref666}{gauging out} a subgroup of the logical \flmRefsHyperref{ref663}{Pauli group} which causes trivial faults in the corresponding \flmRefsHyperref{ref409}{Clifford circuit}. This construction is used to show the existence of geometrically local subsystem codes that nearly saturate the \flmRefsHyperref{ref492}{subsystem BT bound} \NoCaseChange{\protect\cite{cite668}}.
\end{eczvaluelist}
\eczhbkcontributors{ Xiaozhen Fu, \eczhuVVA }
\endeczcode

\eczcode{subsystem_surface}{Subsystem surface code}{~\NoCaseChange{\protect\cite{cite669}}}
\codefieldsection{Description}
Subsystem version of the surface code defined on a square lattice with qubits placed at every vertex and center of every edge.
Its gauge checks are weight-three triangle operators of type \(XXX\) and \(ZZZ\) \NoCaseChange{\protect\cite{cite669}}.

For example, a \(\llbracket 3L^2,2,L\rrbracket \) family has weight-six \(X,Z\)-type stabilizers supported on two of the four triangles of each plaquette.

\codefieldsection{Fault Tolerance}
\begin{eczvaluelist}
\item\relax Gauge fixing and changing the order in which check operators are measured yields a fault-tolerant decoder \NoCaseChange{\protect\cite{cite2624}}.
\end{eczvaluelist}
\codefieldsection{Code Capacity Threshold}
\begin{eczvaluelist}
\item\relax Independent \(X,Z\) noise: the threshold under ML decoding corresponds to the value of a critical point of the two-dimensional hexagonal-lattice random-bond Ising model (RBIM) on the Nishimori line \NoCaseChange{\protect\cite{cite3526,cite669}}, calculated to be around \(7\%\) in Ref. \NoCaseChange{\protect\cite{cite4439}}.
\end{eczvaluelist}
\codefieldsection{Threshold}
\begin{eczvaluelist}
\item\relax \(0.81\%\) threshold for circuit-level depolarizing noise under a variant of MWPM and using gauge-fixing and specific measurement schedules \NoCaseChange{\protect\cite{cite2624}}, improving the \(0.67\%\) threshold for standard measurement schedules \NoCaseChange{\protect\cite{cite669}}.
\item\relax \(2.22\%\) threshold for circuit-level infinitely biased noise under a variant of MWPM and using gauge-fixing and specific measurement schedules \NoCaseChange{\protect\cite{cite2624}}, improving the \(0.52\%\) threshold with standard measurement schedules.
\end{eczvaluelist}
\codefieldsection{Notes}
\begin{eczvaluelist}
\item\relax See \NoCaseChange{\protect\cite[{Sec. III.C3}]{cite2967}} for an exposition.
\end{eczvaluelist}
\codefieldsection{Parents}
\begin{eczvaluelist}
\item\relax
\flmRefsHyperref[eczindexfamilyrel]{code:subsystem_higher_dimensional_surface}{Subsystem homological code}\item\relax
\flmRefsHyperref[eczindexfamilyrel]{code:translationally_invariant_subsystem}{Lattice subsystem code}\end{eczvaluelist}
\codefieldsection{Cousins}
\begin{eczvaluelist}
\item\relax
\flmRefsHyperref[eczindexfamilyrel]{code:surface}{Kitaev surface code} --- Subsystem surface codes are subsystem versions of surface codes.
\item\relax
\flmRefsHyperref[eczindexfamilyrel]{code:asymmetric_qecc}{Asymmetric quantum code (AQC)} --- Subsystem surface codes perform well against biased circuit-level noise \NoCaseChange{\protect\cite{cite2624}}.
\end{eczvaluelist}
\eczhbkcontributors{ \eczhuVVA }
\endeczcode

\eczcode{super_compact}{Super-compact fermion-to-qubit code}{~\NoCaseChange{\protect\cite{cite404}}}
\codefieldsection{Alternative Names}
\begin{eczvaluelist}
\item\relax Super-compact encoding
\end{eczvaluelist}
\eczhIndexCodeAliasName{super_compact}{Super-compact encoding}
\codefieldsection{Description}
A 2D fermion-into-qubit encoding on the square lattice obtained from exact 2D bosonization by a finite-depth generalized local unitary Clifford circuit, followed by re-pairing of Majorana modes and a slight lattice deformation.
The code uses \(1.25\) qubits per fermion, improving on the square-lattice compact encoding with ratio \(r=1.5\).
Its fermion-parity, hopping, and stabilizer operators have weights \(1\)-\(2\), \(2\)-\(6\), and \(12\), respectively \NoCaseChange{\protect\cite[{Table I}]{cite404}}.

\codefieldsection{Parent}
\begin{eczvaluelist}
\item\relax
\flmRefsHyperref[eczindexfamilyrel]{code:2d_bosonization}{2D bosonization code} --- The super-compact code is obtained from exact 2D bosonization by finite-depth generalized local unitaries \NoCaseChange{\protect\cite{cite404}}.
\end{eczvaluelist}
\codefieldsection{Cousin}
\begin{eczvaluelist}
\item\relax
\flmRefsHyperref[eczindexfamilyrel]{code:derby_klassen}{Derby-Klassen (DK) code} --- On the square lattice, the super-compact code is obtained by further transforming the \(r=1.5\) compact/DK construction to a qubit-to-fermion ratio of \(1.25\) \NoCaseChange{\protect\cite{cite404}}.
\end{eczvaluelist}
\eczhbkcontributors{ \eczhuVVA }
\endeczcode

\eczcode{holographic_5_1_2}{Surface-code-fragment (SCF) holographic code}{~\NoCaseChange{\protect\cite{cite2954}}}
\codefieldsection{Description}
Holographic tensor-network code constructed out of a network of encoding isometries of the \(\llbracket 5,1,2\rrbracket \) rotated surface code.
The structure of the isometry is similar to that of the HaPPY code since both isometries are rank-six tensors.
In the case of the SCF holographic code, the isometry is only a \flmRefsHyperref{code:block_perfect}{planar-perfect tensor} (as opposed to a \flmRefsHyperref{ref219}{perfect tensor}).

\codefieldsection{Rate}
Zero-rate version of the code surpasses the hashing bound under certain Pauli noise \NoCaseChange{\protect\cite{cite3715}}.
\codefieldsection{Code Capacity Threshold}
\begin{eczvaluelist}
\item\relax \(7.1\%\) and \(8.2\%\) for even- and odd-radii reduced-rate codes, respectively, under depolarizing noise using the integer-optimization decoder \NoCaseChange{\protect\cite{cite2954}}.
\end{eczvaluelist}
\codefieldsection{Parents}
\begin{eczvaluelist}
\item\relax
\flmRefsHyperref[eczindexfamilyrel]{code:qubit_css}{Qubit CSS code}\item\relax
\flmRefsHyperref[eczindexfamilyrel]{code:holographic_tensor}{Holographic tensor-network code} --- The encoding of the heptagon holographic code is a holographic tensor network consisting of the encoding isometry for the \(\llbracket 5,1,2\rrbracket \) rotated surface code, which is a \flmRefsHyperref{code:block_perfect}{planar-perfect tensor}.
\end{eczvaluelist}
\codefieldsection{Child}
\begin{eczvaluelist}
\item\relax
\flmRefsHyperref[eczindexfamilyrel]{code:stab_5_1_2}{\(\llbracket 5,1,2\rrbracket \) rotated surface code} --- The \(\llbracket 5,1,2\rrbracket \) rotated surface code is the smallest SCF holographic code \NoCaseChange{\protect\cite{cite2954}}. The encoding of more general SCF holographic codes is a holographic tensor network consisting of the encoding isometry for the \(\llbracket 5,1,2\rrbracket \) rotated surface code, which is a \flmRefsHyperref{code:block_perfect}{planar-perfect tensor}.
\end{eczvaluelist}
\codefieldsection{Cousin}
\begin{eczvaluelist}
\item\relax
\flmRefsHyperref[eczindexfamilyrel]{code:block_perfect}{Planar-perfect-tensor code} --- The encoding of the heptagon holographic code is a holographic tensor network consisting of the encoding isometry for the \(\llbracket 5,1,2\rrbracket \) rotated surface code, which is a \flmRefsHyperref{code:block_perfect}{planar-perfect tensor}.
\end{eczvaluelist}
\eczhbkcontributors{ \eczhuVVA }
\endeczcode

\eczcode{syk}{SYK code}{~\NoCaseChange{\protect\cite{cite4440,cite2601}}}
\codefieldsection{Description}
Approximate \(n\)-fermionic code whose codewords are low-energy states of the Sachdev-Ye-Kitaev (SYK) Hamiltonian \NoCaseChange{\protect\cite{cite560,cite561}} or other low-rank SYK models \NoCaseChange{\protect\cite{cite562,cite563}}.

\codefieldsection{Rate}
SYK codes can have a constant rate and distance scaling as \(n^c\) for some power \(c\) \NoCaseChange{\protect\cite{cite2601}}.
\codefieldsection{Threshold}
\begin{eczvaluelist}
\item\relax The coherent information of noise channels that either break or conserve fermion parity has been calculated for the SYK thermofield double state \NoCaseChange{\protect\cite{cite4441}}.
\end{eczvaluelist}
\codefieldsection{Parents}
\begin{eczvaluelist}
\item\relax
\flmRefsHyperref[eczindexfamilyrel]{code:fermions}{Fermion code}\item\relax
\flmRefsHyperref[eczindexfamilyrel]{code:approximate_qecc}{Approximate quantum error-correcting code (AQECC)} --- SYK codes are approximately error correcting in that they satisfy certain error-correction conditions based on mutual information \NoCaseChange{\protect\cite{cite2601}}.
\item\relax
\flmRefsHyperref[eczindexfamilyrel]{code:hamiltonian}{Hamiltonian-based code} --- The SYK code Hamiltonian is constructed out of non-commuting few-site terms, and every fermion participates in many interactions.
\item\relax
\flmRefsHyperref[eczindexfamilyrel]{code:holographic}{Holographic code} --- In a holographic model \NoCaseChange{\protect\cite{cite2601}}, the large distance of these codes can be interpreted as being due to the emergence of a wormhole.
\end{eczvaluelist}
\codefieldsection{Cousin}
\begin{eczvaluelist}
\item\relax
\flmRefsHyperref[eczindexfamilyrel]{code:kpt}{Kim-Preskill-Tang (KPT) code} --- The Brownian SYK model can be used to demonstrate the complexity-based error-correction of KPT codes \NoCaseChange{\protect\cite{cite640}}.
\end{eczvaluelist}
\eczhbkcontributors{ \eczhuVVA }
\endeczcode

\eczcode{iterated_ramanujan}{Tensor-product HDX code}{~\NoCaseChange{\protect\cite{cite4442}}}
\codefieldsection{Description}
A code constructed in a similar way as the HDX code, but utilizing iterated homological products of \textit{multiple} Ramanujan complexes and then applying distance balancing. For any fixed tensor-power parameter \(c\), these yield explicit QLDPC codes with distance scaling as \(\sqrt{n}\log^{c} n\), improving on the original HDX construction by replacing a single logarithmic enhancement with arbitrarily high fixed polylogarithmic enhancement. The utility of such tensor products comes from the fact that one of the Ramanujan complexes is a \textit{collective cosystolic expander} as opposed to just a cosystolic expander.
\codefieldsection{Protection}
Construction yields explicit QLDPC codes with distance \(\sqrt{n}\log^c n\) for any fixed \(c\), using the \(c\)-fold tensor product of Ramanujan complexes followed by distance balancing \NoCaseChange{\protect\cite{cite4442}}.
\codefieldsection{Parent}
\begin{eczvaluelist}
\item\relax
\flmRefsHyperref[eczindexfamilyrel]{code:multisector_hypergraph}{Higher-dimensional homological product code} --- Tensor-product HDX codes result from iterated homological products of length-two chain complexes (i.e., quantum codes) based on Ramanujan complexes \NoCaseChange{\protect\cite{cite3442}}.
\end{eczvaluelist}
\codefieldsection{Child}
\begin{eczvaluelist}
\item\relax
\flmRefsHyperref[eczindexfamilyrel]{code:ramanujan_tensor_product}{High-dimensional expander (HDX) code} --- Ramanujan codes result from a tensor product of a classical-code and a quantum-code chain complex.
\end{eczvaluelist}
\eczhbkcontributors{ \eczhuVVA }
\endeczcode

\eczcode{ternary_tree_fermion}{Ternary-tree fermion-into-qubit code}{~\NoCaseChange{\protect\cite{cite3510}}}
\codefieldsection{Description}
A fermion-into-qubit encoding defined on ternary trees that maps Majorana operators into Pauli strings of weight \(\lceil \log_3 (2n+1) \rceil\).

\codefieldsection{Gates}
\begin{eczvaluelist}
\item\relax Fermion permutations on \(N\) modes can be done with a circuit of depth \flmRefsHyperref{ref65}{order} \(O(\log^2 N)\) \NoCaseChange{\protect\cite{cite4443}}.
\end{eczvaluelist}
\codefieldsection{Parent}
\begin{eczvaluelist}
\item\relax
\flmRefsHyperref[eczindexfamilyrel]{code:fermions_into_qubits}{Fermion-into-qubit code}\end{eczvaluelist}
\codefieldsection{Cousin}
\begin{eczvaluelist}
\item\relax
\flmRefsHyperref[eczindexfamilyrel]{code:bkt}{Bravyi-Kitaev transformation (BKT) code} --- The ternary-tree fermion-into-qubit code improves over the BKT code by a factor of \(\approx 1.58\) in the weight of encoded fermionic operators \NoCaseChange{\protect\cite{cite3510}}.
\end{eczvaluelist}
\eczhbkcontributors{ \eczhuVVA }
\endeczcode

\eczcode{tetrahedral_color}{Tetrahedral color code}{~\NoCaseChange{\protect\cite{cite475,cite726}}}
\codefieldsection{Description}
A 3D color code defined on a colored tetrahedron cut from a suitably colored BCC lattice \NoCaseChange{\protect\cite{cite475}}.
Qubits are placed on tetrahedra, on the triangles covering the tetrahedron faces, on the edges along the tetrahedron edges, and on the tetrahedron vertices.
The code has both string-like and sheet-like logical operators \NoCaseChange{\protect\cite{cite476}}.

\codefieldsection{Rate}
The tetrahedral family with linear size parameter \(L\) has \(n=1+4L+6L^2+4L^3\) physical qubits and encodes one logical qubit \NoCaseChange{\protect\cite{cite475}}.
\codefieldsection{Transversal and Permutation-Based Gates}
\begin{eczvaluelist}
\item\relax A \(\llbracket 5d^3-12d^2+16,3,d\rrbracket \) close relative of this code admits a logical \(CCZ\) gate via single-qubit rotations; for this family, stabilizers remain asymptotically constant-weight and can be gauge-reduced to weight at most six \NoCaseChange{\protect\cite{cite759}}.
\end{eczvaluelist}
\codefieldsection{Fault Tolerance}
\begin{eczvaluelist}
\item\relax Fault-tolerant quantum computation designed for a 2D architecture \NoCaseChange{\protect\cite{cite4444}}.
\end{eczvaluelist}
\codefieldsection{Threshold}
\begin{eczvaluelist}
\item\relax \(0.46\%\) with clustering decoder \NoCaseChange{\protect\cite{cite476}}.
\item\relax \(1.9\%\) for 1D string-like logical operators and \(27.6\%\) for 2D sheet-like operators for 3D codes with noise models using optimal decoding and perfect measurements \NoCaseChange{\protect\cite{cite476}}.
\end{eczvaluelist}
\codefieldsection{Parent}
\begin{eczvaluelist}
\item\relax
\flmRefsHyperref[eczindexfamilyrel]{code:3d_color}{3D color code}\end{eczvaluelist}
\codefieldsection{Child}
\begin{eczvaluelist}
\item\relax
\flmRefsHyperref[eczindexfamilyrel]{code:stab_15_1_3}{\(\llbracket 15,1,3\rrbracket \) quantum RM code} --- The \(\llbracket 15,1,3\rrbracket \) code is a tetrahedral color code defined on a single tetrahedron.
\end{eczvaluelist}
\codefieldsection{Cousins}
\begin{eczvaluelist}
\item\relax
\flmRefsHyperref[eczindexfamilyrel]{code:bcc}{Body-centered cubic (bcc) lattice} --- The tetrahedral color code is defined on a lattice of tetrahedra carved out of a suitably colored BCC lattice \NoCaseChange{\protect\cite{cite475}}.
\item\relax
\flmRefsHyperref[eczindexfamilyrel]{code:3d_surface}{3D surface code} --- A tetrahedral 3D color code with four differently colored boundaries is equivalent, via a local \flmRefsHyperref{ref409}{Clifford circuit}, to three 3D surface codes attached along one boundary, with condensation of a composite electric charge on that attached boundary \NoCaseChange{\protect\cite{cite422}}.
\end{eczvaluelist}
\eczhbkcontributors{ \eczhuVVA }
\endeczcode

\eczcode{tetron}{Tetron code}{~\NoCaseChange{\protect\cite{cite4445,cite537,cite1432,cite401}}}
\codefieldsection{Alternative Names}
\begin{eczvaluelist}
\item\relax Kitaev-Wen Majorana mapping
\item\relax Kitaev honeycomb mapping
\item\relax Bravyi-Leemhuis-Terhal (BLT) Majorana mapping
\item\relax Majorana representation
\item\relax Parton construction
\end{eczvaluelist}
\eczhIndexCodeAliasName{tetron}{Kitaev-Wen Majorana mapping}
\eczhIndexCodeAliasName{tetron}{Kitaev honeycomb mapping}
\eczhIndexCodeAliasName{tetron}{Bravyi-Leemhuis-Terhal (BLT) Majorana mapping}
\eczhIndexCodeAliasName{tetron}{Majorana representation}
\eczhIndexCodeAliasName{tetron}{Parton construction}
\codefieldsection{Description}
A \(\llbracket 2,1,2\rrbracket _{f}\) Majorana box qubit encoding a logical qubit into four Majorana modes, equivalently into the fixed-total-parity sector of two physical fermionic modes.
Four Majorana zero modes are the smallest aggregate that supports a qubit in a fixed fermion-parity sector \NoCaseChange{\protect\cite{cite401}}.
This code can be concatenated with various qubit codes such as surface codes and color codes.
Four-boundary Majorana surface-code patches are logical tetrons, i.e., higher-distance analogues of this physical tetron block \NoCaseChange{\protect\cite{cite402}}.

Embedding each physical qubit into two fermions via the tetron code is useful for exactly solving the Kitaev honeycomb model Hamiltonian \NoCaseChange{\protect\cite{cite537}} and other qubit Hamiltonians on certain graphs \NoCaseChange{\protect\cite{cite2842,cite2843}}.
It has been used throughout condensed matter physics under the name of the Majorana representation \NoCaseChange{\protect\cite{cite4446,cite4447}} or parton construction \NoCaseChange{\protect\cite{cite4448}}, allowing for a mean-field treatment of many models that are otherwise not amenable.
Majorana stabilizer groups can be converted into ordinary qubit stabilizer groups via the parton mapping, while their corresponding states are converted via the Gutzwiller projection \NoCaseChange{\protect\cite{cite2844}}.

\codefieldsection{Parents}
\begin{eczvaluelist}
\item\relax
\flmRefsHyperref[eczindexfamilyrel]{code:mbq}{Majorana box qubit} --- The Majorana box qubit for \(n=2\) is the tetron code.
\item\relax
\flmRefsHyperref[eczindexfamilyrel]{code:quantum_repetition}{Quantum repetition code} --- The tetron code is a special case of the quantum repetition code with \(n=2\).
\end{eczvaluelist}
\codefieldsection{Cousins}
\begin{eczvaluelist}
\item\relax
\flmRefsHyperref[eczindexfamilyrel]{code:hamiltonian}{Hamiltonian-based code} --- Embedding each physical qubit into two fermions via the tetron code is useful for exactly solving the Kitaev honeycomb model Hamiltonian \NoCaseChange{\protect\cite{cite537}} and other qubit Hamiltonians on certain graphs \NoCaseChange{\protect\cite{cite2842,cite2843}}. Majorana stabilizer groups can be converted into ordinary qubit stabilizer groups via the parton mapping, while their corresponding states are converted via the Gutzwiller projection \NoCaseChange{\protect\cite{cite2844}}.
\item\relax
\flmRefsHyperref[eczindexfamilyrel]{code:majorana_surface}{Majorana surface code} --- Four-boundary Majorana surface-code patches are logical tetrons, i.e., higher-distance versions of the tetron code \NoCaseChange{\protect\cite{cite402}}.
\item\relax
\flmRefsHyperref[eczindexfamilyrel]{code:qubit_stabilizer}{Qubit stabilizer code} --- Any \(\llbracket n,k,d\rrbracket \) stabilizer code can be mapped into a \(\llbracket 2n,k,2d\rrbracket _{f}\) Majorana stabilizer code by concatenating with the tetron code \NoCaseChange{\protect\cite{cite537}\protect\cite[{Lemma 1}]{cite1432}}.
Embedding each physical qubit into two fermions via the tetron code is useful for exactly solving the Kitaev honeycomb model Hamiltonian \NoCaseChange{\protect\cite{cite537}} and other qubit Hamiltonians on certain graphs \NoCaseChange{\protect\cite{cite2842,cite2843}}. Majorana stabilizer groups can be converted into ordinary qubit stabilizer groups via the parton mapping, while their corresponding states are converted via the Gutzwiller projection \NoCaseChange{\protect\cite{cite2844}}.

\item\relax
\flmRefsHyperref[eczindexfamilyrel]{code:majorana_subsystem}{Majorana subsystem stabilizer code} --- The B\(\mapsto\)F mapping yields Majorana subsystem codes from qubit stabilizer codes such that their gauge groups contain tetrons \NoCaseChange{\protect\cite{cite3987}\protect\cite[{Sec. IV}]{cite3986}}. The output Majorana subsystem codes can correct odd-weight errors.
\item\relax
\flmRefsHyperref[eczindexfamilyrel]{code:nonabelian_kitaev_honeycomb}{Non-Abelian Kitaev honeycomb code} --- Embedding each physical qubit into two fermions via the tetron code allows the logical subspace of the Kitaev honeycomb model to be formulated as a joint eigenspace of certain Majorana operators \NoCaseChange{\protect\cite[{Sec. 4.1}]{cite3415}}, which admit braiding-based gates due to their non-Abelian statistics and which can be used for topological quantum computation.
When done in reverse, this embedding can be thought of as a 2D bosonization fermion-into-qubit encoding by converting to a relabeled square lattice and performing single-qubit rotations \NoCaseChange{\protect\cite{cite403}\protect\cite[{Sec. IV.B}]{cite404}}.

\item\relax
\flmRefsHyperref[eczindexfamilyrel]{code:self_dual_css}{Self-dual CSS code} --- Any \(\llbracket n,k,d\rrbracket \) qubit stabilizer code maps to a \(\llbracket 4n,2k,2d\rrbracket \) self-dual CSS code by applying the BLT mapping and concatenating each qubit pair with the \(\llbracket 4,2,2\rrbracket \) code \NoCaseChange{\protect\cite[{Corr. 2}]{cite795}\protect\cite[{Corr. 1}]{cite1432}}. The BLT mapping proceeds by first concatenating each qubit with the \flmRefsHyperref{code:tetron}{tetron code} to obtain an intermediate \(\llbracket 2n,k,2d\rrbracket _{f}\) Majorana stabilizer code.
\item\relax
\flmRefsHyperref[eczindexfamilyrel]{code:toric}{Toric code} --- Coherent physical errors in the toric code are expected to become incoherent logical errors under syndrome measurement; see corroborating numerical studies performed by embedding each physical qubit into two fermions via the tetron code \NoCaseChange{\protect\cite{cite4449}} as well as deriving analytical bounds \NoCaseChange{\protect\cite{cite4450}}.
\item\relax
\flmRefsHyperref[eczindexfamilyrel]{code:kitaev_honeycomb}{Kitaev honeycomb code} --- Embedding each physical qubit into two fermions via the tetron code is useful for exactly solving the Kitaev honeycomb model Hamiltonian \NoCaseChange{\protect\cite{cite537}} and other qubit Hamiltonians on certain graphs \NoCaseChange{\protect\cite{cite2842,cite2843}}.
\end{eczvaluelist}
\eczhbkcontributors{ \eczhuVVA }
\endeczcode

\eczcode{subsystem_three_fermion}{Three-fermion (3F) subsystem code}{~\NoCaseChange{\protect\cite{cite602,cite414}}}
\codefieldsection{Description}
2D subsystem stabilizer code whose low-energy excitations realize the three-fermion anyon theory \NoCaseChange{\protect\cite{cite601,cite602,cite603}}.
One version uses two qubits at each site \NoCaseChange{\protect\cite{cite414}}, while other manifestations utilize a single qubit per site and only weight-two (two-body) interactions \NoCaseChange{\protect\cite{cite602,cite604}}.
All are expected to be equivalent to each other via a local constant-depth \flmRefsHyperref{ref409}{Clifford circuit}.

\codefieldsection{Parents}
\begin{eczvaluelist}
\item\relax
\flmRefsHyperref[eczindexfamilyrel]{code:qubit_subsystem_stabilizer}{Subsystem qubit stabilizer code}\item\relax
\flmRefsHyperref[eczindexfamilyrel]{code:translationally_invariant_subsystem}{Lattice subsystem code}\item\relax
\flmRefsHyperref[eczindexfamilyrel]{code:topological_abelian}{Abelian topological code} --- The 3F code is a 2D subsystem code characterized by 3F topological order \NoCaseChange{\protect\cite{cite414}}, which is chiral and modular.
\end{eczvaluelist}
\codefieldsection{Cousins}
\begin{eczvaluelist}
\item\relax
\flmRefsHyperref[eczindexfamilyrel]{code:surface}{Kitaev surface code} --- One version of the 3F subsystem code can be obtained from two copies of the square-lattice surface code by \flmRefsHyperref{ref666}{gauging out} the anyons \(e_1m_1e_2\) and \(e_2m_2\) \NoCaseChange{\protect\cite[{Sec. 7.4}]{cite414}}.
\item\relax
\flmRefsHyperref[eczindexfamilyrel]{code:three_fermion}{Three-fermion (3F) Walker-Wang model code} --- The (three-dimensional) 3F Walker-Wang model cluster-like state encodes the temporal gate operations on the (two-dimensional) 3F subsystem code into a third spatial dimension \NoCaseChange{\protect\cite{cite478}}.
\item\relax
\flmRefsHyperref[eczindexfamilyrel]{code:2d_color}{2D color code} --- The 2D color code is equivalent to two decoupled copies of the 3F code in the sense that the same anyon theory describes the low-energy excitations of both codes \NoCaseChange{\protect\cite{cite3433}\protect\cite[{Appx. B}]{cite445}}.
\item\relax
\flmRefsHyperref[eczindexfamilyrel]{code:floquet_xyz_ruby}{Ruby Floquet code} --- Together, all ISGs of the ruby Floquet code generate the gauge group of the 3F subsystem code.
\end{eczvaluelist}
\eczhbkcontributors{ Nathanan Tantivasadakarn, \eczhuVVA }
\endeczcode

\eczcode{three_fermion}{Three-fermion (3F) Walker-Wang model code}{~\NoCaseChange{\protect\cite{cite477,cite478}}}
\codefieldsection{Description}
A 3D lattice stabilizer code whose bulk realizes a 3D time-reversal SPT order \NoCaseChange{\protect\cite{cite477}} and whose gapped boundary supports the 2D three-fermion (3F) topological order.
The code can be used as a resource state for fault-tolerant MBQC \NoCaseChange{\protect\cite{cite478}}.

\codefieldsection{Encoding}
\begin{eczvaluelist}
\item\relax 3F QCA encoder \NoCaseChange{\protect\cite{cite3068,cite3069}}, which can be simplified using bosonization \NoCaseChange{\protect\cite{cite3452}} and can be extended to SPTs in higher dimensions based on an exact bosonization duality \NoCaseChange{\protect\cite{cite3065}}.
\end{eczvaluelist}
\codefieldsection{Gates}
\begin{eczvaluelist}
\item\relax \flmRefsHyperref{ref409}{Clifford gates} can be performed by braiding and fusing symmetry defects in the MBQC model.
\end{eczvaluelist}
\codefieldsection{Fault Tolerance}
\begin{eczvaluelist}
\item\relax Fault-tolerant MBQC protocol by encoding in, braiding, and fusing symmetry defects.
\end{eczvaluelist}
\codefieldsection{Parents}
\begin{eczvaluelist}
\item\relax
\flmRefsHyperref[eczindexfamilyrel]{code:qldpc}{Qubit QLDPC code}\item\relax
\flmRefsHyperref[eczindexfamilyrel]{code:3d_stabilizer}{3D lattice stabilizer code}\item\relax
\flmRefsHyperref[eczindexfamilyrel]{code:walker_wang}{Walker-Wang model code} --- The Walker-Wang model code reduces to the 3F model code when the input category \(\mathcal{C}=3F\) \NoCaseChange{\protect\cite{cite478}}. When treated as ground states of the code Hamiltonian, 3F Walker-Wang model code states realize a 3D time-reversal SPT order \NoCaseChange{\protect\cite{cite477}}, while the gapped boundary supports the 3F anyon theory.
\item\relax
\flmRefsHyperref[eczindexfamilyrel]{code:spt}{Symmetry-protected topological (SPT) code} --- When treated as ground states of the code Hamiltonian, 3F Walker-Wang model code states realize a 3D time-reversal SPT order \NoCaseChange{\protect\cite{cite477}}. The 3F Walker-Wang QCA encoder \NoCaseChange{\protect\cite{cite3068,cite3069}} can be extended to SPTs in higher dimensions based on an exact bosonization duality \NoCaseChange{\protect\cite{cite3065}}.
\end{eczvaluelist}
\codefieldsection{Cousins}
\begin{eczvaluelist}
\item\relax
\flmRefsHyperref[eczindexfamilyrel]{code:3d_bosonization}{3D bosonization code} --- The 3F Walker-Wang QCA encoder \NoCaseChange{\protect\cite{cite3068,cite3069}} can be simplified using bosonization \NoCaseChange{\protect\cite{cite3452}}.
\item\relax
\flmRefsHyperref[eczindexfamilyrel]{code:bosonization}{Bosonization code} --- The 3F Walker-Wang QCA encoder \NoCaseChange{\protect\cite{cite3068,cite3069}} can be extended to SPTs in higher dimensions based on an exact bosonization duality \NoCaseChange{\protect\cite{cite3065}}.
\item\relax
\flmRefsHyperref[eczindexfamilyrel]{code:topological_abelian}{Abelian topological code} --- The gapped boundary of the 3F Walker-Wang model supports the 3F topological order \NoCaseChange{\protect\cite{cite477,cite478}}.
\item\relax
\flmRefsHyperref[eczindexfamilyrel]{code:subsystem_three_fermion}{Three-fermion (3F) subsystem code} --- The (three-dimensional) 3F Walker-Wang model cluster-like state encodes the temporal gate operations on the (two-dimensional) 3F subsystem code into a third spatial dimension \NoCaseChange{\protect\cite{cite478}}.
\end{eczvaluelist}
\eczhbkcontributors{ \eczhuVVA }
\endeczcode

\eczcode{tillichzemor}{Tillich-Zémor code}{~\NoCaseChange{\protect\cite{cite4451}}}
\codefieldsection{Alternative Names}
\begin{eczvaluelist}
\item\relax Quantum \((n, m, r)\)-structured LDPC code
\end{eczvaluelist}
\eczhIndexCodeAliasName{tillichzemor}{Quantum \((n, m, r)\)-structured LDPC code}
\codefieldsection{Description}
A family of \(\llbracket n^2 + m^2, (n - \text{rank}([C \mid M]) )^2 + (m - \text{rank}([C \mid M]^\top) )^2, d\rrbracket \) quantum LDPC codes constructed via the hypergraph product of two classical \((n, m, r)\)-structured LDPC seed codes

A code's parity-check matrix \(H = [C \mid M]\) consists of an \(m \times m\) circulant core \(C\) with column weight \(2\) (enabling linear-time encoding) and an \(m \times (n-m)\) matrix \(M\) with column weight \(r \geq 3\) and no zero rows.
The resulting code has block length \(N = n^2 + m^2\) and code dimension \(K = (n - \text{rank}([C \mid M]) )^2 + (m - \text{rank}([C \mid M]^\top) )^2\). 

\codefieldsection{Protection}
The structured construction inherits the classical seed code's sub-linear distance scaling of \flmRefsHyperref{ref65}{order} \(O(n^{\frac{r-2}{r-1}+\epsilon})\) while preserving efficient encoding; see Ref. \NoCaseChange{\protect\cite{cite4452}} for distance upper bounds.

\codefieldsection{Parent}
\begin{eczvaluelist}
\item\relax
\flmRefsHyperref[eczindexfamilyrel]{code:hypergraph_product}{Hypergraph product (HGP) code}\end{eczvaluelist}
\eczhbkcontributors{ Feroz Ahmed Mian, \eczhuVVA }
\endeczcode

\eczcode{toric}{Toric code}{~\NoCaseChange{\protect\cite{cite423,cite2125,cite3808}}
}
\codefieldsection{Description}
Version of the Kitaev surface code on a square lattice with periodic boundary conditions, encoding two logical qubits.
Being the first manifestation of the surface code, "toric code" is often an alternative name for the general construction.
\textit{Twisted toric code} \NoCaseChange{\protect\cite[{Fig. 8}]{cite434}} refers to the construction on a torus with twisted (a.k.a. shifted) boundary conditions.
In the original Hamiltonian construction, open Pauli-\(Z\) and Pauli-\(X\) strings create pairs of electric charges and magnetic vortices, and braiding one type around the other yields the nontrivial Abelian anyonic phase \NoCaseChange{\protect\cite{cite423}}.

The stabilizers of the toric code are generated
by star operators \(A_v\) and plaquette operators \(B_p\).
Each star operator is a product of four Pauli-\(X\) operators on the edges adjacent to a vertex   \(v\) of the lattice; each plaquette operator is a product of four Pauli-\(Z\) operators applied to the edges adjacent to a face, or plaquette, \(p\) of the
lattice (\flmRefsCref{ref4453}).

\begin{flmFloat}{figure}{NumCap}\includegraphics[width=116.10000000000001bp,max width=\linewidth]{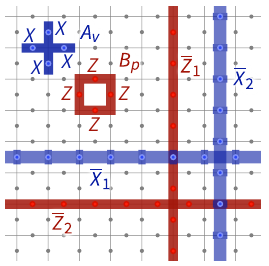}\caption{Stabilizer generators and logical operators of the toric code.
    The star operators \(A_v\) and the plaquette operators
    \(B_p\) generate the stabilizer group.  The logical
    operators are strings that wrap around the torus.}\label{ref4453}\end{flmFloat}

We denote by
\(\overline{X}_i,\overline{Z}_i\) the logical Pauli-\(X\) and Pauli-\(Z\)
operator of the \(i\)-th logical qubit (with \(i\in\{1,2\}\)).  They are represented by strings of Pauli-\(X\) operators or Pauli-\(Z\) operators that wrap around the torus, as shown in \flmRefsCref{ref4453}.

\codefieldsection{Protection}
Toric code on an \(L\times L\) torus is a \(\llbracket 2L^2,2,L\rrbracket \) CSS code.
The number of error patterns can be used to bound the ground-state energy of a \(\pm J\) Ising model \NoCaseChange{\protect\cite{cite4454}}.
Coherent physical errors in the toric code are expected to become incoherent logical errors under syndrome measurement; see corroborating numerical studies performed by embedding each physical qubit into two fermions via the tetron code \NoCaseChange{\protect\cite{cite4449}} as well as deriving analytical bounds \NoCaseChange{\protect\cite{cite4450}}.
More generally, there is a tensor-network routine that calculates the effective logical channel \NoCaseChange{\protect\cite{cite4455}}

\codefieldsection{Encoding}
\begin{eczvaluelist}
\item\relax Lindbladian-based dissipative encoding for the toric code \NoCaseChange{\protect\cite{cite4326}} that does not give a speedup relative to circuit-based encoders \NoCaseChange{\protect\cite{cite4327}}.
\end{eczvaluelist}
\codefieldsection{Transversal and Permutation-Based Gates}
\begin{eczvaluelist}
\item\relax Transversal logical Pauli gates correspond to Pauli strings on non-trivial loops of the torus.
\end{eczvaluelist}
\codefieldsection{Gates}
\begin{eczvaluelist}
\item\relax Logical \(CX\) gate for the \(\llbracket 12,2,3\rrbracket \) twisted toric code \NoCaseChange{\protect\cite{cite3271}}.
\end{eczvaluelist}
\codefieldsection{Code Capacity Threshold}
\begin{eczvaluelist}
\item\relax Independent \(X,Z\) noise: \(p_X = 10.31\%\) under MWPM decoding \NoCaseChange{\protect\cite{cite3939}} (see also Ref. \NoCaseChange{\protect\cite{cite3864}}), \(9.9\%\) under BP-OSD decoding \NoCaseChange{\protect\cite{cite1481}}, and \(8.9\%\) under GBP decoding \NoCaseChange{\protect\cite{cite3913}}. 
The threshold under ML decoding corresponds to the value of a critical point of a two-dimensional random-bond Ising model (RBIM) on the Nishimori line \NoCaseChange{\protect\cite{cite3526,cite480}}, calculated to be \(10.94 \pm 0.02\%\) in Ref. \NoCaseChange{\protect\cite{cite4456}}, \(10.93(2)\%\) in Ref. \NoCaseChange{\protect\cite{cite4457}}, \(10.9187\%\) in Ref. \NoCaseChange{\protect\cite{cite3940}}, \(10.917(3)\%\) in Ref. \NoCaseChange{\protect\cite{cite4458}}, \(10.939(6)\%\) in Ref. \NoCaseChange{\protect\cite{cite4459}}, and estimated to be between \(10.9\%\) and \(11\%\) in Ref. \NoCaseChange{\protect\cite{cite3864}}.
The model for the case of the toric code has been thoroughly investigated \NoCaseChange{\protect\cite{cite4460,cite4461}}.
The Bravyi-Suchara-Vargo (BSV) tensor network decoder \NoCaseChange{\protect\cite{cite3864}} exactly solves the ML decoding problem under independent \(X,Z\) noise for the surface code and has complexity of \flmRefsHyperref{ref65}{order} \(O(n^2)\); the decoder provides an efficient tensor-network contraction for the partition function resulting from the statistical mechanical mapping, which is known to be solvable for an Ising model on a planar graph \NoCaseChange{\protect\cite{cite3865}}.
ML decoding \NoCaseChange{\protect\cite{cite480}} is \(\#P\)-hard in general for the surface code \NoCaseChange{\protect\cite{cite3862}}.
Above values are for one type of noise only, and the ML threshold for combined \(X\) and \(Z\) noise is \(2p_X - p_X^2 \approx 20.6\%\) \NoCaseChange{\protect\cite[{Table 1}]{cite3913}}. 
Thresholds for various lattices have been obtained in Refs. \NoCaseChange{\protect\cite{cite2633,cite4462}}.

\item\relax Depolarizing noise: between \(17\%\) and \(18.5\%\) under BSV tensor-network decoding \NoCaseChange{\protect\cite{cite3864}}, \(14\%\) under GBP decoding \NoCaseChange{\protect\cite{cite3913}}, \(16.5\%\) under recursive MWPM \NoCaseChange{\protect\cite{cite4463}}, between \(16\%\) and \(17.5\%\) under AMBP4 (depending on whether surface or toric code is considered) \NoCaseChange{\protect\cite{cite3740}}, and between \(15\%\) and \(16\%\) under RG \NoCaseChange{\protect\cite{cite3884}}, Markov-chain \NoCaseChange{\protect\cite{cite3889}}, or MWPM \NoCaseChange{\protect\cite{cite3936}} decoding. The threshold under ML decoding corresponds to the value of a critical point of the disordered eight-vertex Ising model, calculated to be \(18.9(3)\%\) \NoCaseChange{\protect\cite{cite4464}} (see also APS Physics viewpoint \NoCaseChange{\protect\cite{cite4465}}).
\item\relax Erasure noise: \(50\%\) for square tiling \NoCaseChange{\protect\cite{cite4466,cite4467}}. There is an inverse relationship between coordination number of the syndrome graph, with the threshold corresponding to a percolation transition \NoCaseChange{\protect\cite{cite4468}}.
\item\relax \flmRefsHyperref{ref498}{AD} noise: \(39\%\) \NoCaseChange{\protect\cite{cite4469}}.
\item\relax Correlated noise: the threshold under ML decoding corresponds to the value of a critical point of a particular random-bond Ising model (RBIM) \NoCaseChange{\protect\cite{cite4470,cite4471}}. A threshold of \(10.04(6)\%\) under mildly correlated bit-flip noise is obtained in Ref. \NoCaseChange{\protect\cite{cite4377}}.
\item\relax The toric code has a \flmRefsHyperref{ref3210}{measurement threshold} of one \NoCaseChange{\protect\cite{cite3211}}.
\item\relax Coherent noise: the threshold under ML decoding corresponds to the value of a critical point of a particular random-bond Ising model (RBIM) called the complex-coupled Ashkin-Teller model \NoCaseChange{\protect\cite{cite4472,cite4473}}. Another statistical mechanical mapping has been studied for \(X\)-type noise channels interpolating between coherent and incoherent noise \NoCaseChange{\protect\cite{cite4474}}.
\item\relax Threshold of \(1.5\%\) under real-time geometrically local decoder based on introducing an ancillary buffer and confining spacetime interactions between anyons  \NoCaseChange{\protect\cite{cite3030}}.
\end{eczvaluelist}
\codefieldsection{Threshold}
\begin{eczvaluelist}
\item\relax The threshold under ML decoding with measurement errors corresponds to the value of a critical point of a three-dimensional random plaquette model \NoCaseChange{\protect\cite{cite480,cite3939}}.
\item\relax \(0.133\%\) for independent \(X,Z\) noise and faulty syndrome measurements using a local automaton decoder \NoCaseChange{\protect\cite{cite3177}}.
\item\relax Toric-code thresholds for post-selected QEC can be studied with statistical mechanical models \NoCaseChange{\protect\cite{cite4475}}.
\end{eczvaluelist}
\codefieldsection{Realizations}
\begin{eczvaluelist}
\item\relax Neutral atom arrays: One cycle of syndrome readout on 19-qubit planar and 24-qubit toric codes \NoCaseChange{\protect\cite{cite4476}}.

\end{eczvaluelist}
\codefieldsection{Parents}
\begin{eczvaluelist}
\item\relax
\flmRefsHyperref[eczindexfamilyrel]{code:surface}{Kitaev surface code} --- The toric code is the surface code on a 2D torus.
\item\relax
\flmRefsHyperref[eczindexfamilyrel]{code:higher_dimensional_toric}{\(D\)-dimensional twisted toric code} --- The \(D\)-dimensional twisted toric code reduces to the toric code for \(D=2\) and a square lattice.
\item\relax
\flmRefsHyperref[eczindexfamilyrel]{code:cyclic_hgp}{Cyclic Hypergraph Product Code} --- The toric code can be obtained from a hypergraph product of two repetition codes \NoCaseChange{\protect\cite[{Exam. 6}]{cite1316}}. Other hypergraph products of two repetition codes yield the related \(\llbracket 2d^2-2d+1,1,d\rrbracket \) CSS code family \NoCaseChange{\protect\cite[{Exam. 5}]{cite1316}}.
\end{eczvaluelist}
\codefieldsection{Cousins}
\begin{eczvaluelist}
\item\relax
\flmRefsHyperref[eczindexfamilyrel]{code:string_net}{String-net code} --- The toric code is the Turaev-Viro/Levin-Wen string-net code for the \(\mathbb{Z}_2\) input category; equivalently, the construction of Ref. \NoCaseChange{\protect\cite{cite599}} on a genus-one handlebody yields the toric code.
\item\relax
\flmRefsHyperref[eczindexfamilyrel]{code:lifted_product}{Lifted-product (LP) code} --- A lifted-product code for the ring \(R=\mathbb{F}_2[x,y]/(x^L-1,y^L-1)\) is the toric code \NoCaseChange{\protect\cite[{Appx. B}]{cite184}}.
\item\relax
\flmRefsHyperref[eczindexfamilyrel]{code:balanced_product}{Balanced product (BP) code} --- Twisted toric codes can be obtained from balanced products of cyclic graphs over a cyclic group \NoCaseChange{\protect\cite[{Fig. 8}]{cite434}}.
\item\relax
\flmRefsHyperref[eczindexfamilyrel]{code:repetition}{Repetition code} --- The toric code can be obtained from a hypergraph product of two repetition codes \NoCaseChange{\protect\cite[{Exam. 6}]{cite1316}}. Other hypergraph products of two repetition codes yield the related \(\llbracket 2d^2-2d+1,1,d\rrbracket \) CSS code family \NoCaseChange{\protect\cite[{Exam. 5}]{cite1316}}.
\item\relax
\flmRefsHyperref[eczindexfamilyrel]{code:tetron}{Tetron code} --- Coherent physical errors in the toric code are expected to become incoherent logical errors under syndrome measurement; see corroborating numerical studies performed by embedding each physical qubit into two fermions via the tetron code \NoCaseChange{\protect\cite{cite4449}} as well as deriving analytical bounds \NoCaseChange{\protect\cite{cite4450}}.
\item\relax
\flmRefsHyperref[eczindexfamilyrel]{code:rotated_surface}{Rotated surface code} --- Rotating the square lattice by \(\pi/4\) and choosing periodicity vectors on the rotated checkerboard lattice yields periodic checkerboard or rotated-toric variants with the same \(\llbracket L^2,2,L\rrbracket \) scaling, as well as non-bipartite odd-distance families with parameters \(\llbracket t^2+(t+1)^2,1,2t+1\rrbracket \) \NoCaseChange{\protect\cite[{Sec. III}]{cite1316}}.
\item\relax
\flmRefsHyperref[eczindexfamilyrel]{code:quantum_lego}{Tensor-network code} --- The toric code can be constructed by arranging \(\llbracket 4,2,2\rrbracket \) tensors on a square lattice and recovering the star and plaquette operators by operator pushing \NoCaseChange{\protect\cite{cite2868}}.
\item\relax
\flmRefsHyperref[eczindexfamilyrel]{code:toric_classical}{Hansen toric code} --- The toric code is not to be confused with the CSS code constructed from a polynomial evaluation code on a toric variety \NoCaseChange{\protect\cite{cite1880}}.
\item\relax
\flmRefsHyperref[eczindexfamilyrel]{code:gkp_surface_concatenated}{GKP-surface code} --- GKP codes have been concatenated with toric codes \NoCaseChange{\protect\cite{cite416}}.
\item\relax
\flmRefsHyperref[eczindexfamilyrel]{code:stab_4_1_2}{\(\llbracket 4,1,2\rrbracket \) twist-defect code} --- The \flmRefsHyperref{ref436}{symplectic double} of the \(\llbracket 4,1,2\rrbracket \) twist-defect code is the \(\llbracket 8,2,2\rrbracket \) twisted toric code \NoCaseChange{\protect\cite{cite435}}.
\item\relax
\flmRefsHyperref[eczindexfamilyrel]{code:stab_4_2_2}{\(\llbracket 4,2,2\rrbracket \) Four-qubit code} --- The toric code can be constructed by arranging \(\llbracket 4,2,2\rrbracket \) tensors on a square lattice and recovering the star and plaquette operators by operator pushing \NoCaseChange{\protect\cite{cite2868}}.
\item\relax
\flmRefsHyperref[eczindexfamilyrel]{code:five_squares}{Generalized five-squares code} --- For the original five-squares code, preprocessing maps decoding onto two copies of the toric code \NoCaseChange{\protect\cite{cite594}}.
\end{eczvaluelist}
\eczhbkcontributors{ \eczhuVVA }
\endeczcode

\eczcode{tfim}{Transverse-field Ising model (TFIM) code}{~\NoCaseChange{\protect\cite{cite4477}}}
\codefieldsection{Description}
A 1D translationally invariant stabilizer code whose encoding is a constant-depth circuit of nearest-neighbor gates on alternating even and odd bonds that consist of transverse-field Ising Hamiltonian interactions. The code allows for perfect state transfer of arbitrary distance using local operations and classical communications (LOCC).
\codefieldsection{Protection}
Code distance is 1 for open boundary conditions similar to a repetition code, and 3 for periodic boundary conditions with an encoding circuit depth of 4.
\codefieldsection{Encoding}
\begin{eczvaluelist}
\item\relax 1D geometrically local constant-depth brickwork circuit of nearest-neighbor gates on alternating even and odd bonds. Gates are generated by interaction terms of the transverse-field Ising Hamiltonian.
\end{eczvaluelist}
\codefieldsection{Parents}
\begin{eczvaluelist}
\item\relax
\flmRefsHyperref[eczindexfamilyrel]{code:small_distance_qubit_stabilizer}{Small-distance qubit stabilizer code}\item\relax
\flmRefsHyperref[eczindexfamilyrel]{code:qldpc}{Qubit QLDPC code}\item\relax
\flmRefsHyperref[eczindexfamilyrel]{code:1d_stabilizer}{1D lattice stabilizer code}\end{eczvaluelist}
\codefieldsection{Cousins}
\begin{eczvaluelist}
\item\relax
\flmRefsHyperref[eczindexfamilyrel]{code:majorana_stab}{Majorana stabilizer code} --- The TFIM code stabilizers can be expressed in terms of Majorana operators.
\item\relax
\flmRefsHyperref[eczindexfamilyrel]{code:quantum_repetition}{Quantum repetition code} --- When written in the computational basis, the phase-flip and TFIM codewords are superpositions of qubit states of fixed total parity. The superposition is equal for the phase-flip code, whereas some states appear with a \(-1\) coefficient for the TFIM code. However, the TFIM code can be encoded in constant depth.
\end{eczvaluelist}
\eczhbkcontributors{ \eczhuVVA }
\endeczcode

\eczcode{trapezoid}{Trapezoid subsystem code}{~\NoCaseChange{\protect\cite{cite3335,cite3245}}}
\codefieldsection{Description}
A member of a family of BBS codes with weight-two (two-body) gauge generators designed to suppress errors in adiabatic quantum computation.

The family consists of odd-\(m\) codes with \(m=2k+1\) and parameters \(\llbracket 4k+2l,2k,2k+2l-2,2\rrbracket \), together with even-\(m\) codes with \(m=2k\) and parameters \(\llbracket 4k+2l-2,2k-1,2k+2l-3,2\rrbracket \), where \(1 \leq l \leq k\) \NoCaseChange{\protect\cite{cite3245}}.

\codefieldsection{Protection}
These are distance-two subsystem codes, so they detect arbitrary single-qubit errors. In the energy-penalty setting of Hamiltonian quantum computation, the \(l=1\) subfamily maximizes the code rate and has the largest penalty gap within the trapezoid family \NoCaseChange{\protect\cite{cite3245}}.

\codefieldsection{Gates}
\begin{eczvaluelist}
\item\relax Single-qubit dressed logical operators are two-local, and products of two dressed logical operators of the same Pauli type can also be implemented using two-local physical interactions up to gauge operators \NoCaseChange{\protect\cite{cite3335,cite3245}}.
\end{eczvaluelist}
\codefieldsection{Parents}
\begin{eczvaluelist}
\item\relax
\flmRefsHyperref[eczindexfamilyrel]{code:bravyi_bacon_shor}{Bravyi-Bacon-Shor (BBS) code}\item\relax
\flmRefsHyperref[eczindexfamilyrel]{code:translationally_invariant_subsystem}{Lattice subsystem code}\end{eczvaluelist}
\codefieldsection{Child}
\begin{eczvaluelist}
\item\relax
\flmRefsHyperref[eczindexfamilyrel]{code:bravyi_bacon_shor_6}{\(\llbracket 6,2,3,2\rrbracket \) BBS code} --- The even-logical-qubit trapezoid family at \(l=k=1\) reduces to the \(\llbracket 6,2,3,2\rrbracket \) BBS code.
\end{eczvaluelist}
\codefieldsection{Cousins}
\begin{eczvaluelist}
\item\relax
\flmRefsHyperref[eczindexfamilyrel]{code:iceberg}{\(\llbracket 2m,2m-2,2\rrbracket \) error-detecting code} --- The trapezoid code family can be obtained from the \(\llbracket 2m,2m-2,2\rrbracket \) error-detecting code by using some logical qubits as gauge qubits and imposing a two-dimensional qubit geometry \NoCaseChange{\protect\cite{cite3245}}.
\item\relax
\flmRefsHyperref[eczindexfamilyrel]{code:goy}{\(\llbracket 6r,2r,2\rrbracket \) Ganti-Onunkwo-Young code} --- The odd-\(m\) trapezoid family at \(l=k\) has parameters \(\llbracket 6k,2k,4k,2\rrbracket \) and reproduces the two-local subsystem construction used for universal Hamiltonian quantum computation in \NoCaseChange{\protect\cite{cite3335}}; this is a subsystem analogue of the \(\llbracket 6k,2k,2\rrbracket \) Ganti-Onunkwo-Young family.
\item\relax
\flmRefsHyperref[eczindexfamilyrel]{code:small_distance_qubit_stabilizer}{Small-distance qubit stabilizer code}\end{eczvaluelist}
\eczhbkcontributors{ \eczhuVVA }
\endeczcode

\eczcode{tree_cluster}{Tree cluster-state code}{~\NoCaseChange{\protect\cite{cite4478,cite4479,cite4480,cite4481,cite4482}}}
\codefieldsection{Description}
Code obtained from a cluster state on a tree graph (e.g., a star graph \NoCaseChange{\protect\cite{cite4479,cite4480}}) that has been proposed in the context of quantum repeater and MBQC architectures.

\codefieldsection{Protection}
Some tree cluster-state codes have shown good performance over the depolarizing channel \NoCaseChange{\protect\cite{cite3562}}.

\codefieldsection{Gates}
\begin{eczvaluelist}
\item\relax Cluster states constructed from star clusters can be used to perform universal MBQC with probabilistic two-qubit gates \NoCaseChange{\protect\cite{cite4480}}.
\item\relax Three-tree cluster states can be fused into larger tree and hexagonal cluster states using postselected Bell measurements \NoCaseChange{\protect\cite{cite415}}.
\end{eczvaluelist}
\codefieldsection{Parent}
\begin{eczvaluelist}
\item\relax
\flmRefsHyperref[eczindexfamilyrel]{code:cluster_state}{Cluster-state code}\end{eczvaluelist}
\eczhbkcontributors{ \eczhuVVA }
\endeczcode

\eczcode{triangle_surface}{Triangular surface code}{~\NoCaseChange{\protect\cite{cite433}}}
\codefieldsection{Alternative Names}
\begin{eczvaluelist}
\item\relax Triangle surface code
\end{eczvaluelist}
\eczhIndexCodeAliasName{triangle_surface}{Triangle surface code}
\codefieldsection{Description}
A member of a twist-defect surface code family with a single central twist whose planar layout fits within a triangle.
Triangle codes can be viewed as three conjoined surface-code patches projected into two dimensions, with weight-four plaquette stabilizers and weight-two edge stabilizers \NoCaseChange{\protect\cite{cite433}}.
Symmetric distance-\(d\) triangle codes use \(3d^2/4+1/4\) data qubits, i.e., about \(25\%\) fewer than the rotated surface code for a given odd distance.
Logical \(\overline{X}\), \(\overline{Y}\), and \(\overline{Z}\) operators can be supported on the three sides of the triangle, enabling initialization and measurement in any Pauli basis.

The size of the triangular patches and which patch encodes data versus acts as ancillas for gates depends on the initialization and measurement procedures.
See Ref. \NoCaseChange{\protect\cite{cite433}} for tables and figures.

\codefieldsection{Rate}
Symmetric distance-\(d\) triangle codes use \(3d^2/4+1/4\) data qubits per logical qubit. Including ancillas needed for planar Clifford computation, the CC, BC, and CS architectures use \(3d^2+O(d)\), \(9d^2/4+O(d)\), and \(6d^2/4+O(d)\) physical qubits per logical qubit, respectively \NoCaseChange{\protect\cite[{Table 2}]{cite433}}.
\codefieldsection{Encoding}
\begin{eczvaluelist}
\item\relax Code conversion (CC) initialization and measurement method, in which the surface code is used to hold data between gates in patches.
\item\relax Basis-state conversion (BC) initialization and measurement method in which one initializes and then measures a logical Pauli eigenstate. To do this, triangle ancilla qubits are required outside of the triangle patches that hold the data. That is, the ancilla patches must be empty of data and be adjacent to the side that contains the logical Pauli that needs to be measured or initialized.
\item\relax The CAT states (CS) initialization and measurement method uses a row of \(d\) ancilla qubits along some edge of a triangle code with distance \(d\) to create and verify a GHZ state that is used to measure the logical operator along the same edge. Creating this GHZ state takes \(O(d)\) time steps. To reliably measure the logical state, the GHZ state must be measured \(O(d)\) times, resulting in \(O(d^2)\) time for logical measurement. Initialization is a similar procedure that requires \(O(d^2)\) time for logical-operator measurements that occur \(O(d)\) times as well as \(O(d^2)\) time to project the code onto a logical state \NoCaseChange{\protect\cite{cite433}}.
\end{eczvaluelist}
\codefieldsection{Transversal and Permutation-Based Gates}
\begin{eczvaluelist}
\item\relax Triangle codes admit transversal order-three single-qubit gates in the \flmRefsHyperref{ref409}{Clifford group}, e.g., \(\bar{SH}\) \NoCaseChange{\protect\cite{cite433}}.
\end{eczvaluelist}
\codefieldsection{Gates}
\begin{eczvaluelist}
\item\relax Triangle codes admit a distillation-free implementation of the full Clifford group using lattice surgery, 1-bit teleportation, and patch reorientation \NoCaseChange{\protect\cite{cite433}}.
\item\relax Performing single-qubit gates in the \flmRefsHyperref{ref409}{Clifford group} using the CC procedure requires surface code patches to be embedded in triangle patches. This procedure requires \(O(d)\) Clifford gate times for \(H,S,CNOT\) \NoCaseChange{\protect\cite{cite433}}.
\item\relax The BC procedure requires \(O(1)\) time to perform \(H,S\) gates and \(O(d)\) time to perform \(CNOT\). Sometimes reorientation of the sides is required, and that takes \(O(d)\) to perform \NoCaseChange{\protect\cite{cite433}}.
\end{eczvaluelist}
\codefieldsection{Decoding}
\begin{eczvaluelist}
\item\relax The decoding uses a single decoding graph since the triangle code is not a CSS code. Nodes of the graph are located at each stabilizer (center of the triangle graph) and have red or blue edges, where red associates with \(X\) errors and blue with \(Z\) errors. To take into account any errors from measuring the error syndrome, a three-dimensional stack of decoding graphs is laid on top of the code with vertical edges connecting qubits between layers \NoCaseChange{\protect\cite{cite433}}.
\end{eczvaluelist}
\codefieldsection{Fault Tolerance}
\begin{eczvaluelist}
\item\relax The symmetry of triangle codes allows for fault-tolerant measurement and encoding in any Pauli basis \NoCaseChange{\protect\cite{cite433}}.
\item\relax A non-fault-tolerant circuit initializes the triangle code. To guarantee fault-tolerance, post-selection is performed on trivial measurements of the syndrome and of the logical Pauli, depending on the basis of the logical states \NoCaseChange{\protect\cite{cite433}}.
\item\relax Making syndrome extraction fault tolerant requires a specific ordering of syndrome measurements so as to avoid \flmRefsHyperref{ref3496}{hook errors} \NoCaseChange{\protect\cite{cite433}}.
\end{eczvaluelist}
\codefieldsection{Code Capacity Threshold}
\begin{eczvaluelist}
\item\relax \(10\%\) under either bit-flip or bit-phase noise for ideal syndrome measurements. The decoder used is a decoding graph with the same weight assigned to each edge, and Dijkstra's algorithm is used to compute the total weight of any path \NoCaseChange{\protect\cite{cite433}}.
\end{eczvaluelist}
\codefieldsection{Threshold}
\begin{eczvaluelist}
\item\relax \(3.2\%\) bit-flip error-correction threshold for noisy syndrome measurements and \(2.6\%\) for bit-phase flip noise. The decoder used is a decoding graph as described above \NoCaseChange{\protect\cite{cite433}}.
\item\relax In general, the triangular surface code has a threshold of similar magnitude to the toric code for uncorrelated \(X\) and \(Z\) errors. For correlated errors, the triangle code has a lower threshold by a factor of about \(36\) \NoCaseChange{\protect\cite{cite433}}.
\end{eczvaluelist}
\codefieldsection{Parent}
\begin{eczvaluelist}
\item\relax
\flmRefsHyperref[eczindexfamilyrel]{code:stellated_surface}{Stellated surface code} --- The triangular surface code is the \(s=3\) member of the stellated surface-code family \NoCaseChange{\protect\cite[{Appx. D}]{cite445}}.
\end{eczvaluelist}
\codefieldsection{Child}
\begin{eczvaluelist}
\item\relax
\flmRefsHyperref[eczindexfamilyrel]{code:twist_defect_7_1_3}{\(\llbracket 7,1,3\rrbracket \) twist-defect surface code}\end{eczvaluelist}
\codefieldsection{Cousins}
\begin{eczvaluelist}
\item\relax
\flmRefsHyperref[eczindexfamilyrel]{code:quantum_lego}{Tensor-network code} --- Triangle surface codes can be reproduced by inserting a defect tensor and Hadamard-modified tensors into the surface-code tensor network \NoCaseChange{\protect\cite{cite2868}}.
\item\relax
\flmRefsHyperref[eczindexfamilyrel]{code:stellated_color}{Stellated color code} --- Stellated color codes are color-code analogues of triangle surface codes in that both encode logical information in lattices with a single twist defect. Instances of the former can be obtained by fattening \NoCaseChange{\protect\cite{cite430}} the vertices of the latter \NoCaseChange{\protect\cite{cite445}}.
\end{eczvaluelist}
\eczhbkcontributors{ Raley Roberts, \eczhuVVA }
\endeczcode

\eczcode{quantum_triorthogonal}{Triorthogonal code}{~\NoCaseChange{\protect\cite[{Secs. III-IV}]{cite691}}}
\codefieldsection{Description}
Qubit CSS code whose \(X\)-type logicals and stabilizer generators form a triorthogonal matrix (defined below) in the \flmRefsHyperref{ref817}{symplectic representation}.

An \(m \times n\) binary matrix is triorthogonal if its rows \(r_1, \ldots, r_m\) satisfy \(|r_i \cdot r_j| = 0\) and \(|r_i \cdot r_j \cdot r_k| = 0\) modulo \(2\), with binary addition and multiplication.
The triorthogonal CSS code associated with the matrix is constructed by mapping nonzero entries in even-weight rows to \(X\)-type stabilizer generators, odd-weight rows to \(X\)-type logical operators, and \(Z\) operators for each row in the orthogonal complement.

Generalized versions of triorthogonality allow odd pair and triple overlaps inside the logical-row block \(K\) while keeping every overlap involving an \(S\)-row even, which yields quasitransversal logical gates beyond a single logical \(T\) gate \NoCaseChange{\protect\cite[{Appx. D}]{cite754}\protect\cite[{Sec. 2.3}]{cite756}}.

\codefieldsection{Protection}
Weight \(t\) Pauli errors, where \(t\) depends on the family. For example, Ref. \NoCaseChange{\protect\cite[{Sec. VII}]{cite691}} provides a family of distance \(2\) codes.
\codefieldsection{Encoding}
\begin{eczvaluelist}
\item\relax Encoder for magic states for the code constructed in \NoCaseChange{\protect\cite{cite691}}.
\end{eczvaluelist}
\codefieldsection{Transversal and Permutation-Based Gates}
\begin{eczvaluelist}
\item\relax Transversal action of \(T\) gates on all qubits, followed by a particular pattern of \(CZ\) and \(S\) gates, will realize a logical \(T\) gate \NoCaseChange{\protect\cite[{Lemma 2}]{cite691}\protect\cite[{Rem. 3}]{cite760}}. When an additional condition on logical-\(X\) operators is satisfied, the \(CZ\) and \(S\) gates are no longer necessary \NoCaseChange{\protect\cite[{Thm. 14}]{cite724}}.
\item\relax Triorthogonality is necessary but not sufficient for physical transversal \(T\) gates on each qubit to realize the identity logical gate \NoCaseChange{\protect\cite[{Thm. 12}]{cite724}}.
\item\relax Certain codes realize controlled-controlled-\(Z\) gates \NoCaseChange{\protect\cite{cite787}}, realized via physical \(CCZ\) gates on three code blocks.
\item\relax Triorthogonal codes realizing logical \(T\) gates using only physical \(T\) gates can be paired up with self-dual CSS codes to yield a transversal CNOT gate \NoCaseChange{\protect\cite{cite788}}.
\end{eczvaluelist}
\codefieldsection{Decoding}
\begin{eczvaluelist}
\item\relax Steane error correction \NoCaseChange{\protect\cite{cite788}}.
\end{eczvaluelist}
\codefieldsection{Fault Tolerance}
\begin{eczvaluelist}
\item\relax Universal fault-tolerant gates can be performed without magic-state distillation \NoCaseChange{\protect\cite{cite787,cite3204}}.
\item\relax Universal fault-tolerant gates can be achieved by pairing with a self-dual CSS code and using Steane error correction \NoCaseChange{\protect\cite{cite788}}.
\end{eczvaluelist}
\codefieldsection{Notes}
\begin{eczvaluelist}
\item\relax Reference \NoCaseChange{\protect\cite{cite3207}} presents a classification of triorthogonal codes up to \(n + k \leq 38\) by associating each triorthogonal code with a RM code polynomial.
\item\relax A database of triorthogonal codes is available in QECDB \NoCaseChange{\protect\cite{cite781}}.
\end{eczvaluelist}
\codefieldsection{Parents}
\begin{eczvaluelist}
\item\relax
\flmRefsHyperref[eczindexfamilyrel]{code:quantum_k-orthogonal}{\(k\)-orthogonal code} --- \(k\)-orthogonal codes reduce to triorthogonal codes for \(k=3\).
\item\relax
\flmRefsHyperref[eczindexfamilyrel]{code:qudit_triorthogonal}{Prime-qudit triorthogonal code} --- Prime-qudit triorthogonal codes reduce to triorthogonal codes when \(p=2\).
\end{eczvaluelist}
\codefieldsection{Children}
\begin{eczvaluelist}
\item\relax
\flmRefsHyperref[eczindexfamilyrel]{code:stab_15_1_3}{\(\llbracket 15,1,3\rrbracket \) quantum RM code} --- The \(\llbracket 15, 1, 3\rrbracket \) code is a triorthogonal code \NoCaseChange{\protect\cite{cite3207}}.
\item\relax
\flmRefsHyperref[eczindexfamilyrel]{code:small_triorthogonal}{\(\llbracket 3k + 8, k, 2\rrbracket \) triorthogonal code}\item\relax
\flmRefsHyperref[eczindexfamilyrel]{code:stab_49_1_5}{\(\llbracket 49,1,5\rrbracket \) triorthogonal code}\end{eczvaluelist}
\codefieldsection{Cousins}
\begin{eczvaluelist}
\item\relax
\flmRefsHyperref[eczindexfamilyrel]{code:quantum_reed_muller}{Quantum Reed-Muller (RM) code} --- Classification of triorthogonal codes yields a connection to RM code polynomials \NoCaseChange{\protect\cite{cite3207}}.
\item\relax
\flmRefsHyperref[eczindexfamilyrel]{code:self_dual}{Self-dual linear code} --- Self-dual binary codes can be used to construct triorthogonal codes \NoCaseChange{\protect\cite{cite2041}}.
\item\relax
\flmRefsHyperref[eczindexfamilyrel]{code:css-t}{CSS-T code} --- Triorthogonal and CSS-T codes overlap, but neither is a subset of the other \NoCaseChange{\protect\cite{cite760}}. CSS-T codes reduce to triorthogonal codes when the logical action of the physical transversal \(T\) gate is a logical \(T\) gate on all encoded qubits. Triorthogonality is necessary for physical transversal \(T\) gates on each qubit to realize the identity logical gate \NoCaseChange{\protect\cite[{Thm. 12}]{cite724}}. The \(X\)-type stabilizer generator matrix for a CSS-T code is always triorthogonal \NoCaseChange{\protect\cite[{Corr. 5}]{cite1315}}.
\item\relax
\flmRefsHyperref[eczindexfamilyrel]{code:binary_dihedral_permutation_invariant}{Binary dihedral PI code} --- There exist binary dihedral PI codes that have distance 5 (7, 9, 11, 13) and encode in 27 (49, 73, 107, 147) qubits, all realizing transversal \(T\) gates.
\item\relax
\flmRefsHyperref[eczindexfamilyrel]{code:stab_5_1_3}{\(\llbracket 5,1,3\rrbracket \) Five-qubit perfect code} --- A fault-tolerant logical \(T\) gate can be obtained by encoding the five-qubit code's five physical qubits into the five logical qubits of a \(\llbracket 31,5,3\rrbracket \) outer quantum divisible CSS code preserved by transversal \(T^\dagger\); this layered construction can be viewed as a factorization of a \(\llbracket 31,1,3\rrbracket \) triorthogonal code and does not require magic-state distillation \NoCaseChange{\protect\cite{cite765}}.
\item\relax
\flmRefsHyperref[eczindexfamilyrel]{code:campbell_howard}{\(\llbracket 6k+2,3k,2\rrbracket \) Campbell-Howard code} --- Campbell-Howard codes arise from the generalized-triorthogonal/quasitransversal \(G\)-matrix framework, which extends Bravyi-Haah triorthogonal matrices by allowing odd pair and triple overlaps entirely within the logical-row block \(K\) \NoCaseChange{\protect\cite[{Appx. D}]{cite754}}.
\item\relax
\flmRefsHyperref[eczindexfamilyrel]{code:generalized_quantum_divisible}{Generalized quantum divisible code} --- Triorthogonal codes are stabilizer codes, while generalized quantum divisible codes are CSS codes. Every level-three generalized divisible code is a triorthogonal code, but whether the converse is true or false is not known \NoCaseChange{\protect\cite[{Sec. VI.C}]{cite734}}.
\item\relax
\flmRefsHyperref[eczindexfamilyrel]{code:quantum_divisible}{Quantum divisible code} --- The \(\llbracket 31,5,3\rrbracket \) member together with the five-qubit code can be viewed as a factorization of a \(\llbracket 31,1,3\rrbracket \) triorthogonal code \NoCaseChange{\protect\cite{cite765}}.
\item\relax
\flmRefsHyperref[eczindexfamilyrel]{code:quantum_rainbow}{Quantum rainbow code} --- Hypergraph products of color codes yield quantum rainbow codes with growing distance and transversal gates in the \flmTerm{term}{ref694}{}{Clifford hierarchy}. In particular, utilizing this construction for quasi-hyperbolic color codes \NoCaseChange{\protect\cite{cite703}} yields an \(\llbracket n,O(n),O(\log n)\rrbracket \) triorthogonal code family satisfying the necessary conditions for the magic-state yield parameter \(\gamma\) to become arbitrarily small \NoCaseChange{\protect\cite{cite704}}.
\item\relax
\flmRefsHyperref[eczindexfamilyrel]{code:qubit_golay}{\(\llbracket 23, 1, 7\rrbracket \) Quantum Golay code} --- A \(\llbracket 95,1,7\rrbracket \) triorthogonal code with a transversal \(T\) gate can be obtained from the qubit Golay code via the doubling transformation \NoCaseChange{\protect\cite{cite3230}}.
\item\relax
\flmRefsHyperref[eczindexfamilyrel]{code:self_dual_css}{Self-dual CSS code} --- Triorthogonal codes realizing logical \(T\) gates using only physical \(T\) gates can be paired up with self-dual CSS codes to yield a transversal CNOT gate and universal fault-tolerant gates using Steane error correction \NoCaseChange{\protect\cite{cite788}}.
\item\relax
\flmRefsHyperref[eczindexfamilyrel]{code:galois_quad_residue}{Quantum quadratic-residue (QR) code} --- Qubit quantum QR codes are doubly even and admit transversal implementations of the \flmRefsHyperref{ref409}{single-qubit Clifford group} \NoCaseChange{\protect\cite{cite760}}. They yield a family of high-distance triorthogonal and weak triply even codes via the doubling transformation \NoCaseChange{\protect\cite{cite760}}; such codes admit transversal implementations of the \(T\) gate.
\item\relax
\flmRefsHyperref[eczindexfamilyrel]{code:quantum_ag}{Quantum AG code} --- By defining a generalization of triorthogonal matrices to Galois qudits of dimension \(q=2^m\), one can construct an asymptotically good family of quantum AG codes that admits a diagonal transversal gate at the third level of the \flmTerm{term}{ref694}{}{Clifford hierarchy} and attains a zero magic-state yield parameter, \(\gamma = 0\) \NoCaseChange{\protect\cite{cite695}}. This code can be treated as a qubit code by decomposing each Galois qudit into a Kronecker product of \(m\) qubits; see \NoCaseChange{\protect\cite{cite696,cite398,cite698,cite699,cite700}\protect\cite[{Sec. 5.3}]{cite697}}. Two other asymptotically good families \NoCaseChange{\protect\cite{cite699,cite698}} admit a transversal \(CCZ\) gate (a different diagonal gate at the third level of the \flmTerm{term}{ref694}{}{Clifford hierarchy}) and achieve \(\gamma \to 0\) with constant alphabet size.
\end{eczvaluelist}
\eczhbkcontributors{ Shubham P. Jain, Narayanan Rengaswamy, Benjamin Quiring, \eczhuVVA }
\endeczcode

\eczcode{4612_color}{Truncated trihexagonal (4.6.12) color code}{~\NoCaseChange{\protect\cite{cite432}}}
\codefieldsection{Description}
2D color code defined on a (typically triangular) patch of the 4.6.12 (truncated trihexagonal or square-hexagon-dodecagon) tiling.

Stabilizer generators are shown in \flmRefsCref{ref4483}.
  \begin{flmFloat}{figure}{NumCap}\includegraphics[width=216bp,max width=\linewidth]{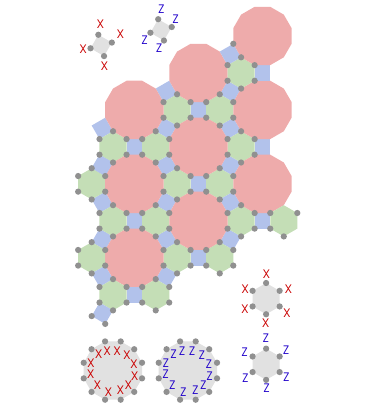}\caption{
    Stabilizer generators of the 4.6.12 color code.
    }\label{ref4483}\end{flmFloat}

\codefieldsection{Protection}
For triangular patches, there is a \(\llbracket (3d^2+5)/2-3d, 1, d\rrbracket \) code family \NoCaseChange{\protect\cite[{Fig. 2}]{cite432}}.

\codefieldsection{Transversal and Permutation-Based Gates}
\begin{eczvaluelist}
\item\relax CNOT gate because the code is CSS.
\item\relax Hadamard gates for any qubit geometry which yields a self-dual CSS code.
\end{eczvaluelist}
\codefieldsection{Parent}
\begin{eczvaluelist}
\item\relax
\flmRefsHyperref[eczindexfamilyrel]{code:2d_color}{2D color code}\end{eczvaluelist}
\codefieldsection{Children}
\begin{eczvaluelist}
\item\relax
\flmRefsHyperref[eczindexfamilyrel]{code:stab_4_2_2}{\(\llbracket 4,2,2\rrbracket \) Four-qubit code} --- The \(\llbracket 4,2,2\rrbracket \) code can be interpreted as a 2D color code on a square of the 4.6.12 tiling \NoCaseChange{\protect\cite{cite2526,cite3262}}. Concatenating the \(\llbracket 4,2,2\rrbracket \) code with two copies of the surface code on a hexagonal lattice yields the self-dual 4.6.12 color code \NoCaseChange{\protect\cite{cite3289}}.
\item\relax
\flmRefsHyperref[eczindexfamilyrel]{code:stab_6_4_2}{\(\llbracket 6,4,2\rrbracket \) error-detecting code} --- The \(\llbracket 6,4,2\rrbracket \) error-detecting code is a color code defined on a single hexagon of the 6.6.6 or 4.6.12 tilings.
\end{eczvaluelist}
\codefieldsection{Cousins}
\begin{eczvaluelist}
\item\relax
\flmRefsHyperref[eczindexfamilyrel]{code:honeycomb}{Honeycomb tiling} --- The 4.6.12 (truncated trihexagonal or square-hexagon-dodecagon) tiling is obtained by applying a fattening procedure to the honeycomb tiling \NoCaseChange{\protect\cite{cite430}}.
\item\relax
\flmRefsHyperref[eczindexfamilyrel]{code:iceberg}{\(\llbracket 2m,2m-2,2\rrbracket \) error-detecting code} --- The \(\llbracket 2m,2m-2,2\rrbracket \) error-detecting code for \(m=4\) is a color code defined on a single octagon of the 6.6.6 or 4.6.12 tilings.
\end{eczvaluelist}
\eczhbkcontributors{ Eric Huang, \eczhuVVA }
\endeczcode

\eczcode{twist_defect_color}{Twist-defect color code}{~\NoCaseChange{\protect\cite{cite442,cite712,cite4484}}}
\codefieldsection{Alternative Names}
\begin{eczvaluelist}
\item\relax Color code with a twist
\end{eczvaluelist}
\eczhIndexCodeAliasName{twist_defect_color}{Color code with a twist}
\codefieldsection{Description}
A non-CSS extension of the 2D color code whose non-CSS stabilizer generators are associated with twist defects of the associated lattice.
These twists terminate domain walls that permute color labels, Pauli labels, or interchange the two.

For lattices with dislocations and rotational disclinations, twist-defect stabilizer generators are placed at the location of the dislocations.
Logical dimension is determined by the genus of the underlying surface (for closed surfaces), types of boundaries (for open surfaces), and any twist defects present.

\codefieldsection{Protection}
Code properties depend on the number and size of the twist defects.
There are 72 types of twist defects in the 2D color code, organized by the \(S_{3}\wr\mathbb{Z}_{2}\) symmetry of the anyons \NoCaseChange{\protect\cite{cite445}}.

\codefieldsection{Gates}
\begin{eczvaluelist}
\item\relax \flmRefsHyperref{ref409}{Clifford gates} can be implemented via twist-based lattice surgery \NoCaseChange{\protect\cite{cite3834}} or braiding twist defects
\NoCaseChange{\protect\cite{cite4485}}.

\item\relax Domino twists \NoCaseChange{\protect\cite{cite4486}}.
\end{eczvaluelist}
\codefieldsection{Parents}
\begin{eczvaluelist}
\item\relax
\flmRefsHyperref[eczindexfamilyrel]{code:qldpc}{Qubit QLDPC code}\item\relax
\flmRefsHyperref[eczindexfamilyrel]{code:2d_stabilizer}{2D lattice stabilizer code}\end{eczvaluelist}
\codefieldsection{Children}
\begin{eczvaluelist}
\item\relax
\flmRefsHyperref[eczindexfamilyrel]{code:stab_4_1_2}{\(\llbracket 4,1,2\rrbracket \) twist-defect code} --- The \(\llbracket 4,1,2\rrbracket \) twist-defect code is the smallest triangular color code with \(x\)-, \(y\)-, and \(z\)-type Pauli boundaries, which make the code non-CSS \NoCaseChange{\protect\cite[{Fig. 7}]{cite445}}.
\item\relax
\flmRefsHyperref[eczindexfamilyrel]{code:2d_color}{2D color code} --- Twist-defect color codes reduce to 2D color codes when there are no defects. See Ref. \NoCaseChange{\protect\cite{cite3423}} for an alternative non-CSS extension of 2D color codes.
\item\relax
\flmRefsHyperref[eczindexfamilyrel]{code:stellated_color}{Stellated color code}\item\relax
\flmRefsHyperref[eczindexfamilyrel]{code:xyz_color}{XYZ color code}\end{eczvaluelist}
\codefieldsection{Cousins}
\begin{eczvaluelist}
\item\relax
\flmRefsHyperref[eczindexfamilyrel]{code:topological_abelian}{Abelian topological code} --- Twist-defect color codes realize \(\mathbb{Z}_2 \times \mathbb{Z}_2\) topological order with twist defects.
\item\relax
\flmRefsHyperref[eczindexfamilyrel]{code:twist_defect_surface}{Twist-defect surface code} --- Twist-defect color codes and twist-defect surface codes both encode using twist defects in topological stabilizer codes.
\end{eczvaluelist}
\eczhbkcontributors{ \eczhuVVA }
\endeczcode

\eczcode{twist_defect_surface}{Twist-defect surface code}{~\NoCaseChange{\protect\cite{cite442,cite4487,cite433,cite445,cite427,cite3160,cite435}\protect\cite[{Sec. 12, pt. 2}]{cite537}}
}
\codefieldsection{Alternative Names}
\begin{eczvaluelist}
\item\relax Surface code with a twist
\item\relax Genon surface code
\end{eczvaluelist}
\eczhIndexCodeAliasName{twist_defect_surface}{Surface code with a twist}
\eczhIndexCodeAliasName{twist_defect_surface}{Genon surface code}
\codefieldsection{Description}
A non-CSS extension of the 2D surface-code construction whose non-CSS stabilizer generators are associated with twist defects of the associated lattice.
A related construction \NoCaseChange{\protect\cite{cite435}} doubles the number of qubits in the lattice via \flmRefsHyperref{ref436}{symplectic doubling}.

For lattices with dislocations and rotational disclinations, twist-defect stabilizer generators are placed at the location of the dislocations to yield a stabilizer code whose logical dimension depends on the defects.
Logical dimension is determined by the genus of the underlying surface (for closed surfaces), types of boundaries (for open surfaces), and any twist defects present.

A simple example is a surface code on a lattice with a single lattice dislocation which hosts a weight-five non-CSS twist-defect stabilizer generator \NoCaseChange{\protect\cite[{Fig. 2}]{cite442}}.
More generally, given a graph embedded in a 2D manifold, qubits are placed on vertices, stabilizers on faces, and twist defects are associated to odd-degree vertices.

\codefieldsection{Protection}
Code properties depend on the number and size of the twist defects.

\codefieldsection{Rate}
Twist-defect surface codes have negative curvature around their defects, and thus circumvent the \flmRefsHyperref{ref487}{BPT bound} for codes on Euclidean lattices.
\codefieldsection{Gates}
\begin{eczvaluelist}
\item\relax \flmRefsHyperref{ref409}{Clifford gates} can be implemented via twist-based lattice surgery \NoCaseChange{\protect\cite{cite4488}} or braiding twist defects
\NoCaseChange{\protect\cite{cite4489,cite442,cite2521,cite2522,cite2523,cite2524,cite2525}}.

\item\relax State injection protocols yield arbitrary logical rotations \NoCaseChange{\protect\cite{cite4487}}.
\item\relax Symplectic doubles of codes yield fault-tolerant \flmRefsHyperref{ref409}{Clifford gates} performed via Dehn twists \NoCaseChange{\protect\cite{cite435}}.
\end{eczvaluelist}
\codefieldsection{Fault Tolerance}
\begin{eczvaluelist}
\item\relax Fault-tolerant measurement of defects \NoCaseChange{\protect\cite{cite4487}}.
\item\relax Twisted double covers of codes yield fault-tolerant \flmRefsHyperref{ref409}{Clifford gates} performed via Dehn twists \NoCaseChange{\protect\cite{cite435}}.
\end{eczvaluelist}
\codefieldsection{Realizations}
\begin{eczvaluelist}
\item\relax Ground state of the toric code has been implemented with and without twists, and the non-Abelian braiding behavior of the twists, which realize Ising anyons, has been demonstrated \NoCaseChange{\protect\cite{cite4490}}.

\item\relax Logical \flmRefsHyperref{ref409}{Clifford gates} arising from a \(\llbracket 4,1,2\rrbracket \) twist-defect surface-code protocol, together with lifted gates on its \(\llbracket 8,2,2\rrbracket \) and \(\llbracket 10,2,3\rrbracket \) double covers, were realized on a trapped-ion device by Quantinuum \NoCaseChange{\protect\cite{cite435}}.
\end{eczvaluelist}
\codefieldsection{Parents}
\begin{eczvaluelist}
\item\relax
\flmRefsHyperref[eczindexfamilyrel]{code:qldpc}{Qubit QLDPC code}\item\relax
\flmRefsHyperref[eczindexfamilyrel]{code:2d_stabilizer}{2D lattice stabilizer code}\end{eczvaluelist}
\codefieldsection{Children}
\begin{eczvaluelist}
\item\relax
\flmRefsHyperref[eczindexfamilyrel]{code:rhombic_dodecahedron_surface}{\(\llbracket 14,3,3\rrbracket \) Rhombic dodecahedron surface code} --- The rhombic dodecahedron surface code is a twist-defect surface code whose degree-three vertices can be interpreted as disclination twists \NoCaseChange{\protect\cite{cite3187}}.
\item\relax
\flmRefsHyperref[eczindexfamilyrel]{code:stab_4_1_2}{\(\llbracket 4,1,2\rrbracket \) twist-defect code} --- The \(\llbracket 4,1,2\rrbracket \) twist-defect code is equivalent to a twist-defect surface code on a tetrahedron inscribed in a sphere \NoCaseChange{\protect\cite{cite435}} via a single-qubit \flmRefsHyperref{ref409}{Clifford circuit}.
\item\relax
\flmRefsHyperref[eczindexfamilyrel]{code:cubic_surface}{\(\llbracket 8,3,2\rrbracket \) Surface code on a cube} --- The surface code on a cube is a twist-defect surface code whose degree-three vertices can be interpreted as disclination twists \NoCaseChange{\protect\cite{cite3187}}.
\item\relax
\flmRefsHyperref[eczindexfamilyrel]{code:surface}{Kitaev surface code} --- Twist-defect surface codes reduce to surface codes when there are no defects.
\item\relax
\flmRefsHyperref[eczindexfamilyrel]{code:stellated_surface}{Stellated surface code}\item\relax
\flmRefsHyperref[eczindexfamilyrel]{code:xzzx}{XZZX surface code} --- XZZX toric and planar codes can be treated in the general twist-defect surface code formalism \NoCaseChange{\protect\cite{cite427}}.
\end{eczvaluelist}
\codefieldsection{Cousins}
\begin{eczvaluelist}
\item\relax
\flmRefsHyperref[eczindexfamilyrel]{code:topological_abelian}{Abelian topological code} --- Twist-defect surface codes realize \(\mathbb{Z}_2\) topological order with twist defects.
\item\relax
\flmRefsHyperref[eczindexfamilyrel]{code:qudit_surface}{Modular-qudit surface code} --- Twist-defect surface codes have been extended to prime-dimensional qudits \NoCaseChange{\protect\cite{cite4491}}.
\item\relax
\flmRefsHyperref[eczindexfamilyrel]{code:honeycomb_floquet}{Honeycomb Floquet code} --- Fermionic string excitations of the honeycomb Floquet code can be condensed along 1D paths, yielding twist defects \NoCaseChange{\protect\cite{cite3752}}.
\item\relax
\flmRefsHyperref[eczindexfamilyrel]{code:stab_5_1_2}{\(\llbracket 5,1,2\rrbracket \) rotated surface code} --- The \(\llbracket 5,1,2\rrbracket \) rotated surface code is a genon code on a sphere, with the missing external \(Y\)-type stabilizer forming the back of the sphere. More generally, any surface code with a single boundary component can be interpreted this way \NoCaseChange{\protect\cite{cite435}}.
\item\relax
\flmRefsHyperref[eczindexfamilyrel]{code:fermions_into_qubits}{Fermion-into-qubit code} --- Treating a twist-defect surface codespace as a logical fermion encoding yields a fermion-into-qubit code \NoCaseChange{\protect\cite{cite3679}}.
\item\relax
\flmRefsHyperref[eczindexfamilyrel]{code:twist_defect_color}{Twist-defect color code} --- Twist-defect color codes and twist-defect surface codes both encode using twist defects in topological stabilizer codes.
\end{eczvaluelist}
\eczhbkcontributors{ \eczhuVVA }
\endeczcode

\eczcode{twisted_xzzx}{Twisted XZZX toric code}{~\NoCaseChange{\protect\cite{cite438}}}
\codefieldsection{Alternative Names}
\begin{eczvaluelist}
\item\relax XZZX cyclic code
\item\relax Cyclic toric code
\item\relax Generalized toric code (GTC)
\item\relax Genus-one genon code
\end{eczvaluelist}
\eczhIndexCodeAliasName{twisted_xzzx}{XZZX cyclic code}
\eczhIndexCodeAliasName{twisted_xzzx}{Cyclic toric code}
\eczhIndexCodeAliasName{twisted_xzzx}{Generalized toric code (GTC)}
\eczhIndexCodeAliasName{twisted_xzzx}{Genus-one genon code}
\codefieldsection{Description}
A cyclic code that can be thought of as the XZZX toric code with shifted (a.k.a twisted) boundary conditions.
Admits a set of stabilizer generators that are equivalent to cyclic shifts of a particular weight-four \(XZZX\) Pauli string.

Codes encode either one or two logical qubits, depending on qubit geometry, and perform well against biased noise \NoCaseChange{\protect\cite{cite2632}}.
See Ref. \NoCaseChange{\protect\cite{cite435}} for a table of some of these for small instances, where they are called genus-one genon codes.

\codefieldsection{Protection}
A family of \(\llbracket a^2+b^2,k,d\rrbracket \) cyclic codes exists for all \(b > a \geq 1\) such that \(\text{gcd}(a,b)=1\) \NoCaseChange{\protect\cite[{Thm. 3.9}]{cite427}}.
Here, \(k=1\) (\(k=2\)) and \(d=a+b\) (\(d=\max(a,b)\)) for odd \(n\) (even \(n\)).
The subfamily \(\llbracket d^2+1,2,d\rrbracket \) (i.e., \(a=1,b=d\) for odd \(d\)) includes \(\llbracket 10,2,3\rrbracket \), \(\llbracket 26,2,5\rrbracket \), \(\llbracket 50,2,7\rrbracket \), \(\ldots\) \NoCaseChange{\protect\cite[{Exam. 1}]{cite1316}}.
The subfamily \(\llbracket t^2+(t+1)^2,1,2t+1\rrbracket \) (i.e., \(a=t,b=t+1\)) includes \(\llbracket 5,1,3\rrbracket \), \(\llbracket 13,1,5\rrbracket \), \(\llbracket 25,1,7\rrbracket \), \(\ldots\) \NoCaseChange{\protect\cite[{Exam. 4}]{cite1316}}.
Small instances are tabulated in Ref. \NoCaseChange{\protect\cite{cite435}} as genus-one genon codes.
Other types of distances have been considered for this code \NoCaseChange{\protect\cite{cite2632}}.

\codefieldsection{Decoding}
\begin{eczvaluelist}
\item\relax Fault-tolerant syndrome extraction circuits using flag qubits \NoCaseChange{\protect\cite{cite2632}}.
\item\relax AMBP4, a quaternary version \NoCaseChange{\protect\cite{cite3739}} of the MBP decoder \NoCaseChange{\protect\cite{cite3740}}.
\item\relax Fault-tolerant BP (FTBP) decoder \NoCaseChange{\protect\cite{cite4410}}.
\end{eczvaluelist}
\codefieldsection{Fault Tolerance}
\begin{eczvaluelist}
\item\relax Fault-tolerant syndrome extraction circuits using flag qubits \NoCaseChange{\protect\cite{cite2632}}.
\end{eczvaluelist}
\codefieldsection{Code Capacity Threshold}
\begin{eczvaluelist}
\item\relax Depolarizing noise: \(17.5\%\) under AMBP4 decoding for the \(\llbracket (m^2+1)/2,1,m\rrbracket \) family \NoCaseChange{\protect\cite[{Fig. 10}]{cite3739}}.
\item\relax Biased noise: between \(20\%\) and \(45\%\) at noise bias ranging from 1 to 10 under MWPM \NoCaseChange{\protect\cite[{Fig. 5}]{cite2632}}.
\end{eczvaluelist}
\codefieldsection{Threshold}
\begin{eczvaluelist}
\item\relax Phenomenological noise: between \(3\%\) and \(10\%\) at noise bias ranging from 1 to 4 under MWPM \NoCaseChange{\protect\cite[{Fig. 5}]{cite2632}}.
\end{eczvaluelist}
\codefieldsection{Parents}
\begin{eczvaluelist}
\item\relax
\flmRefsHyperref[eczindexfamilyrel]{code:xzzx}{XZZX surface code} --- Imposing twisted (a.k.a. shifted) boundary conditions on the toric XZZX code yields the twisted XZZX code \NoCaseChange{\protect\cite[{Exam. 11 and Fig. 3}]{cite438}\protect\cite[{Fig. 6}]{cite427}}.
\item\relax
\flmRefsHyperref[eczindexfamilyrel]{code:quantum_cyclic}{Cyclic quantum code}\end{eczvaluelist}
\codefieldsection{Children}
\begin{eczvaluelist}
\item\relax
\flmRefsHyperref[eczindexfamilyrel]{code:xzzx_10_2_3}{\(\llbracket 10,2,3\rrbracket \) rotated toric code} --- This is the \(d=3\) instance of the \(\llbracket d^2+1,2,d\rrbracket \) family of twisted XZZX toric codes (parameters \(a=1\), \(b=3\)), presented in its CSS form \NoCaseChange{\protect\cite{cite440}}.
\item\relax
\flmRefsHyperref[eczindexfamilyrel]{code:stab_13_1_5}{\(\llbracket 13,1,5\rrbracket \) twisted toric code} --- The \(\llbracket 13,1,5\rrbracket \) twisted toric code is a small twisted XZZX toric code \NoCaseChange{\protect\cite[{Exam. 11 and Fig. 3}]{cite438}}.
\item\relax
\flmRefsHyperref[eczindexfamilyrel]{code:stab_5_1_3}{\(\llbracket 5,1,3\rrbracket \) Five-qubit perfect code} --- Twisted XZZX codes are 2D lattice extensions of the five-qubit perfect code. The five-qubit code is a small twisted XZZX toric code \NoCaseChange{\protect\cite[{Exam. 11 and Fig. 3}]{cite438}\protect\cite[{Exam. 3}]{cite439}\protect\cite[{Fig. 1}]{cite427}}. Its genus-one double cover is a \(\llbracket 10,2,3\rrbracket \) toric code \NoCaseChange{\protect\cite{cite435}\protect\cite[{Exam. 3}]{cite439}}. The base code's transversal \(SH\) gate lifts to a logical \(CX \cdot SWAP\) gate on that double cover \NoCaseChange{\protect\cite{cite435}}.
\item\relax
\flmRefsHyperref[eczindexfamilyrel]{code:xzzx_7_1_3}{\(\llbracket 7,1,3\rrbracket \) XZZX cyclic code} --- The \(\llbracket 7,1,3\rrbracket \) XZZX cyclic code is a cyclic non-CSS code whose generators are the cyclic shifts of the weight-four Pauli string \(XZIZXII\), which has \(XZZX\) support.
\end{eczvaluelist}
\codefieldsection{Cousins}
\begin{eczvaluelist}
\item\relax
\flmRefsHyperref[eczindexfamilyrel]{code:stab_5_1_2_convolutional}{\((5,1,2)\)-convolutional code} --- \((5,1,2)\)-convolutional codes (twisted XZZX toric codes) are 1D (2D) lattice extensions of the five-qubit perfect code.
\item\relax
\flmRefsHyperref[eczindexfamilyrel]{code:asymmetric_qecc}{Asymmetric quantum code (AQC)} --- Twisted XZZX codes perform well against biased noise \NoCaseChange{\protect\cite{cite2630,cite2631,cite2632}}; see also Ref. \NoCaseChange{\protect\cite{cite2633}}.
\item\relax
\flmRefsHyperref[eczindexfamilyrel]{code:bipartite_cyclic_cluster}{Bipartite cyclic cluster (BCC) code} --- The \(\llbracket d^2+1,2,d\rrbracket \) twisted XZZX toric codes (parameters \(a=1,b=d\)) are Clifford-equivalent to BCC codes for odd \(d\) \NoCaseChange{\protect\cite{cite440}}.
\end{eczvaluelist}
\eczhbkcontributors{ \eczhuVVA }
\endeczcode

\eczcode{two_foliated}{Two-foliated fracton code}{~\NoCaseChange{\protect\cite{cite4492,cite467}}}
\codefieldsection{Alternative Names}
\begin{eczvaluelist}
\item\relax Anisotropic lineon code
\end{eczvaluelist}
\eczhIndexCodeAliasName{two_foliated}{Anisotropic lineon code}
\codefieldsection{Description}
A type-I fracton code obtained by gauging \NoCaseChange{\protect\cite{cite462,cite463,cite233,cite464,cite465,cite466,cite467,cite468,cite469,cite470}} a 3D paramagnet with planar subsystem symmetries in two directions.
In that construction, the gauge charges are lineons and the flux excitations are also lineons moving in the same direction, yielding the anisotropic lineon model \NoCaseChange{\protect\cite[{Sec. 4.1.2}]{cite467}}.

\codefieldsection{Gates}
\begin{eczvaluelist}
\item\relax The code admits a cup product structure and a logical CZ gate from physical CZ gates \NoCaseChange{\protect\cite{cite1517}}.
\end{eczvaluelist}
\codefieldsection{Parents}
\begin{eczvaluelist}
\item\relax
\flmRefsHyperref[eczindexfamilyrel]{code:hypergraph_product}{Hypergraph product (HGP) code} --- The two-foliated fracton code is a hypergraph product of the repetition code and the plaquette Ising code on a square lattice with periodic boundary conditions \NoCaseChange{\protect\cite{cite1517}}.
\item\relax
\flmRefsHyperref[eczindexfamilyrel]{code:fracton}{Fracton stabilizer code} --- The two-foliated fracton code is a foliated type-I fracton code.
\end{eczvaluelist}
\codefieldsection{Cousins}
\begin{eczvaluelist}
\item\relax
\flmRefsHyperref[eczindexfamilyrel]{code:plaquette_ising}{Plaquette Ising code} --- The two-foliated fracton code is a hypergraph product of the repetition code and the plaquette Ising code on a square lattice with periodic boundary conditions \NoCaseChange{\protect\cite{cite1517}}.
\item\relax
\flmRefsHyperref[eczindexfamilyrel]{code:repetition}{Repetition code} --- The two-foliated fracton code is a hypergraph product of the repetition code and the plaquette Ising code on a square lattice with periodic boundary conditions \NoCaseChange{\protect\cite{cite1517}}.
\item\relax
\flmRefsHyperref[eczindexfamilyrel]{code:spt}{Symmetry-protected topological (SPT) code} --- Gauging a 3D paramagnet with planar subsystem symmetries in two directions yields the anisotropic lineon model; each symmetry charge becomes a lineon gauge charge, while certain pairs become planons \NoCaseChange{\protect\cite[{Sec. 4.1.2}]{cite467}}.
\end{eczvaluelist}
\eczhbkcontributors{ \eczhuVVA }
\endeczcode

\eczcode{non_stabilizer}{Union stabilizer (USt) code}{~\NoCaseChange{\protect\cite{cite3168,cite4493,cite4494,cite3170,cite1369}}}
\codefieldsection{Alternative Names}
\begin{eczvaluelist}
\item\relax Non-stabilizer code
\item\relax Quotient space quantum code (QSQC)
\end{eczvaluelist}
\eczhIndexCodeAliasName{non_stabilizer}{Non-stabilizer code}
\eczhIndexCodeAliasName{non_stabilizer}{Quotient space quantum code (QSQC)}
\codefieldsection{Description}
A qubit code whose codespace consists of a direct sum of a qubit stabilizer codespace and one or more of that stabilizer code's error spaces.

Given a subset \(T\) of coset representatives of \(\mathsf{N}(\mathsf{S})/\mathsf{S}\) of a stabilizer code \(\llbracket n,k\rrbracket \) with codespace \(\mathsf{C}\) and stabilizer group \(\mathsf{S}\), one can construct the USt with codespace \NoCaseChange{\protect\cite[{Def. 10.1}]{cite3167}}
\flmMathEnvironment{align}{}{
  \mathsf{C}_{\text{USt}}=\bigoplus_{t\in T}t\mathsf{C}~.
}
The parameters of the USt are \(\llparenthesis n,2^k |T|\rrparenthesis \), where \(|T|\) is the number of chosen coset representatives.
A USt is \textit{CSS-like} when the underlying stabilizer code is CSS and the coset representatives are chosen from the two classical codes underlying the CSS code.

Union stabilizer codes constructed in Ref. \NoCaseChange{\protect\cite{cite3170}} include the \(\llparenthesis 33, 155, 3\rrparenthesis \) and \(\llparenthesis 15, 8, 3\rrparenthesis \) codes.

\codefieldsection{Protection}
The distance does not exceed that of the original code \(\mathsf{C}\) unless that codespace is 1D \NoCaseChange{\protect\cite{cite3582}}.
Distance bounds are calculated in Refs. \NoCaseChange{\protect\cite{cite3167,cite4495}} using various formulations.
Since USt codes are equivalent to CWS codes via a single-qubit Clifford circuit, a USt code is degenerate if and only if it is \flmRefsHyperref{ref672}{impure} \NoCaseChange{\protect\cite{cite3582}}.

\codefieldsection{Decoding}
\begin{eczvaluelist}
\item\relax Error-detection algorithm \NoCaseChange{\protect\cite{cite3585,cite3586,cite3587}}.
\end{eczvaluelist}
\codefieldsection{Notes}
\begin{eczvaluelist}
\item\relax See Ref. \NoCaseChange{\protect\cite{cite3167}} for an overview of union stabilizer codes.
\end{eczvaluelist}
\codefieldsection{Parents}
\begin{eczvaluelist}
\item\relax
\flmRefsHyperref[eczindexfamilyrel]{code:qubits_into_qubits}{Qubit code}\item\relax
\flmRefsHyperref[eczindexfamilyrel]{code:qudit_non_stabilizer}{Modular-qudit USt code} --- Modular-qudit union stabilizer codes reduce to union stabilizer codes for \(q=2\).
\item\relax
\flmRefsHyperref[eczindexfamilyrel]{code:galois_non_stabilizer}{Galois-qudit USt code} --- Galois-qudit union stabilizer codes reduce to union stabilizer codes for \(q=2\).
\end{eczvaluelist}
\codefieldsection{Children}
\begin{eczvaluelist}
\item\relax
\flmRefsHyperref[eczindexfamilyrel]{code:cws}{Codeword stabilized (CWS) code} --- Any CWS code can be written as a USt whose (\(K=1\)) stabilizer code is the cluster state and whose coset representatives are constructed from the binary classical code. Conversely, USt codes are equivalent to CWS codes via a single-qubit \flmRefsHyperref{ref409}{Clifford circuit} as follows \NoCaseChange{\protect\cite{cite3585,cite3587}\protect\cite[{Sec. 10.4}]{cite3167}}. The set of coset representatives of any USt can be extended to a larger set iterating over the underlying stabilizer code such that all codewords can be obtained from a single stabilizer state. Then, one can apply a single-qubit Clifford transformation to map said stabilizer state into a cluster state.
\item\relax
\flmRefsHyperref[eczindexfamilyrel]{code:quantum_goethals_preparata}{\(\llparenthesis 2^m,2^{2^m−5m+1},8\rrparenthesis \) Goethals-Preparata code}\item\relax
\flmRefsHyperref[eczindexfamilyrel]{code:qubit_stabilizer}{Qubit stabilizer code} --- A stabilizer code with stabilizer group \(\mathsf{S}\) can be thought of as a USt with only the identity coset representative.
Conversely, if the set of coset representatives of a USt form a linear binary code, then they can be absorbed into a stabilizer group that defines the USt.

\end{eczvaluelist}
\codefieldsection{Cousins}
\begin{eczvaluelist}
\item\relax
\flmRefsHyperref[eczindexfamilyrel]{code:qubit_css}{Qubit CSS code} --- An \(\llbracket n,2k-n,d\rrbracket \) CSS code can be converted to a \(\llbracket n,k+k^{\prime}−n,\min(d,\left\lceil 3d^{\prime}/2\right\rceil )\rrbracket \) code for particular \(k^{\prime}\) and \(d^{\prime}\) via \flmRefsCref{ref863}. This code can be treated as a union stabilizer code \NoCaseChange{\protect\cite{cite1369}}.
\item\relax
\flmRefsHyperref[eczindexfamilyrel]{code:hybrid_stabilizer}{Hybrid stabilizer code} --- The algebraic structure of a hybrid stabilizer code is the same as that of a USt code whose cosets are indexed by a linear binary code \NoCaseChange{\protect\cite{cite2735}}.
\end{eczvaluelist}
\eczhbkcontributors{ \eczhuVVA }
\endeczcode

\eczcode{floquet_xcube}{X-cube Floquet code}{~\NoCaseChange{\protect\cite{cite4496}}}
\codefieldsection{Description}
A 3D Floquet code on the truncated cubic honeycomb, built from coupled layers of square-octagon Floquet toric codes.

Its original measurement schedule yields ISGs that are stacks of 2D surface codes or are FDLQC-equivalent to the X-cube model code \NoCaseChange{\protect\cite{cite4496}}.
A rewinding period-six schedule \(GBRBGR\) yields only fracton ISGs: the G-round is the canonical X-cube model concatenated with four-qubit repetition codes on composite green edges, the B- and first R-rounds are FDLQC-equivalent to the product of the X-cube model, a 3D surface code, and a 3-foliated stack of 2D surface codes up to non-local stabilizers, and the second R-round is FDLQC-equivalent to the X-cube model together with a 3-foliated stack of 2D surface codes \NoCaseChange{\protect\cite{cite533}}.
A parent stabilizer code for the rewinding schedule is FDQC-equivalent to a 3-foliated stack of 2D color codes \NoCaseChange{\protect\cite{cite533}}.

\codefieldsection{Rate}
The original code has subextensive logical dimension \NoCaseChange{\protect\cite{cite4496}}; for even \(L\), the rewinding schedule preserves \(6L-3\) logical qubits on an \(L\times L\times L\) three-torus \NoCaseChange{\protect\cite{cite533}}.
\codefieldsection{Decoding}
\begin{eczvaluelist}
\item\relax Period-six measurement sequence utilizing two-qubit measurements \NoCaseChange{\protect\cite{cite4496}}.
\item\relax A rewinding period-six schedule \(GBRBGR\) yields ISGs FDLQC-equivalent to the X-cube model together with a 3D surface-code factor and/or a 3-foliated stack of 2D surface codes \NoCaseChange{\protect\cite{cite533}}.
\end{eczvaluelist}
\codefieldsection{Code Capacity Threshold}
\begin{eczvaluelist}
\item\relax It is argued that this code has a threshold in Ref. \NoCaseChange{\protect\cite{cite4496}}.
\end{eczvaluelist}
\codefieldsection{Parent}
\begin{eczvaluelist}
\item\relax
\flmRefsHyperref[eczindexfamilyrel]{code:floquet}{Hastings-Haah Floquet code}\end{eczvaluelist}
\codefieldsection{Cousins}
\begin{eczvaluelist}
\item\relax
\flmRefsHyperref[eczindexfamilyrel]{code:xcube}{X-cube model code} --- The G-round of the rewinding X-cube Floquet code is the canonical X-cube model concatenated with four-qubit repetition codes, while the other rounds remain FDLQC-equivalent to the X-cube model together with additional topological factors \NoCaseChange{\protect\cite{cite533}}.
\item\relax
\flmRefsHyperref[eczindexfamilyrel]{code:surface}{Kitaev surface code} --- The rewinding schedule has rounds whose ISGs include a 3-foliated stack of 2D surface codes as an FDLQC-equivalent factor \NoCaseChange{\protect\cite{cite533}}.
\item\relax
\flmRefsHyperref[eczindexfamilyrel]{code:3d_surface}{3D surface code} --- The B- and first R-round ISGs of the rewinding schedule are FDLQC-equivalent to the product of the X-cube model, a 3D surface code, and a 3-foliated stack of 2D surface codes up to non-local stabilizers \NoCaseChange{\protect\cite{cite533}}.
\item\relax
\flmRefsHyperref[eczindexfamilyrel]{code:quantum_repetition}{Quantum repetition code} --- The G-round of the rewinding X-cube Floquet code is exactly the canonical X-cube model concatenated with four-qubit repetition codes on composite green edges \NoCaseChange{\protect\cite{cite533}}.
\item\relax
\flmRefsHyperref[eczindexfamilyrel]{code:2d_color}{2D color code} --- A parent stabilizer code for the rewinding X-cube Floquet code is FDQC-equivalent to a 3-foliated stack of 2D color codes \NoCaseChange{\protect\cite{cite533}}.
\end{eczvaluelist}
\eczhbkcontributors{ \eczhuVVA }
\endeczcode

\eczcode{xcube}{X-cube model code}{~\NoCaseChange{\protect\cite{cite233}}}
\codefieldsection{Description}
A foliated type-I fracton CSS code on a cubic lattice with qubits on edges, cube stabilizers, and three cross-shaped vertex stabilizers for each vertex \NoCaseChange{\protect\cite{cite233}}.
It supports a subextensive number of logical qubits.

In the generalized lattice-gauge-theory construction of Ref. \NoCaseChange{\protect\cite{cite233}}, the X-cube model is the quantum dual of the 3D plaquette Ising model in a transverse field.
Its fundamental excitations include immobile fractons created at the corners of membrane operators and dimension-1 quasiparticles, i.e., lineons in later terminology, created at the ends of straight Wilson lines \NoCaseChange{\protect\cite{cite233}}.
On an \(L_x\times L_y\times L_z\) three-torus, the ground-state degeneracy is \(2^{2L_x+2L_y+2L_z-3}\), reflecting the model's subextensive encoding rate \NoCaseChange{\protect\cite{cite233}}.

Variants include several generalized X-cube models \NoCaseChange{\protect\cite{cite1501}}.
A non-stabilizer commuting-projector code constructed by stacking layers of the double-semion string-net model, called the semionic X-cube model \NoCaseChange{\protect\cite{cite534}}, is equivalent to the X-cube model \NoCaseChange{\protect\cite{cite4492}} (see also Refs. \NoCaseChange{\protect\cite{cite3980,cite4497}}).

\codefieldsection{Decoding}
\begin{eczvaluelist}
\item\relax Parallelized matching decoder \NoCaseChange{\protect\cite{cite3522}}.
\end{eczvaluelist}
\codefieldsection{Code Capacity Threshold}
\begin{eczvaluelist}
\item\relax Independent \(X,Z\) noise: minimum threshold \(\approx 7.5\%\), higher than those reported for the 3D surface code and color code \NoCaseChange{\protect\cite{cite3523}}.
\end{eczvaluelist}
\codefieldsection{Parents}
\begin{eczvaluelist}
\item\relax
\flmRefsHyperref[eczindexfamilyrel]{code:qubit_css}{Qubit CSS code}\item\relax
\flmRefsHyperref[eczindexfamilyrel]{code:qldpc}{Qubit QLDPC code}\item\relax
\flmRefsHyperref[eczindexfamilyrel]{code:qudit_xcube}{Qudit X-cube model code} --- The qudit X-cube model code reduces to the X-cube model code for \(q=2\). The X-cube model is a foliated type-I fracton code \NoCaseChange{\protect\cite{cite4498,cite456}}.
\end{eczvaluelist}
\codefieldsection{Cousins}
\begin{eczvaluelist}
\item\relax
\flmRefsHyperref[eczindexfamilyrel]{code:quantum_inspired}{Quantum-inspired classical block code} --- According to Ref. \NoCaseChange{\protect\cite{cite1349}}, a classical analogue of the X-cube model is the eight-vertex model \NoCaseChange{\protect\cite{cite1989,cite1990,cite1991}}.
\item\relax
\flmRefsHyperref[eczindexfamilyrel]{code:double_semion}{Double-semion stabilizer code} --- A non-stabilizer commuting-projector code constructed by stacking layers of the double-semion string-net model, called the semionic X-cube model \NoCaseChange{\protect\cite{cite534}}, is equivalent to the X-cube model \NoCaseChange{\protect\cite{cite4492}} (see also Refs. \NoCaseChange{\protect\cite{cite3980,cite4497}}).
\item\relax
\flmRefsHyperref[eczindexfamilyrel]{code:string_net}{String-net code} --- A non-stabilizer commuting-projector code constructed by stacking layers of the double-semion string-net model, called the semionic X-cube model \NoCaseChange{\protect\cite{cite534}}, is equivalent to the X-cube model \NoCaseChange{\protect\cite{cite4492}} (see also Refs. \NoCaseChange{\protect\cite{cite3980,cite4497}}).
\item\relax
\flmRefsHyperref[eczindexfamilyrel]{code:surface}{Kitaev surface code} --- The X-cube model can be constructed by coupling layers of the surface code \NoCaseChange{\protect\cite{cite534,cite3164}}.
\item\relax
\flmRefsHyperref[eczindexfamilyrel]{code:3d_surface}{3D surface code} --- The X-cube model admits a topological defect network construction out of 3D surface codes \NoCaseChange{\protect\cite{cite3163}}.
\item\relax
\flmRefsHyperref[eczindexfamilyrel]{code:plaquette_ising}{Plaquette Ising code} --- The 3D plaquette Ising model can be used to obtain the X-cube model by gauging \NoCaseChange{\protect\cite{cite462,cite463,cite233,cite464,cite465,cite466,cite467,cite468,cite469,cite470}} its subsystem symmetry \NoCaseChange{\protect\cite{cite233}}.
\item\relax
\flmRefsHyperref[eczindexfamilyrel]{code:floquet_fracton}{Fracton Floquet code} --- The ISG of the Fracton Floquet code can be that of the X-cube model code or the checkerboard model code.
\item\relax
\flmRefsHyperref[eczindexfamilyrel]{code:floquet_xcube}{X-cube Floquet code} --- The G-round of the rewinding X-cube Floquet code is the canonical X-cube model concatenated with four-qubit repetition codes, while the other rounds remain FDLQC-equivalent to the X-cube model together with additional topological factors \NoCaseChange{\protect\cite{cite533}}.
\item\relax
\flmRefsHyperref[eczindexfamilyrel]{code:majorana_checkerboard}{Majorana checkerboard code} --- The Majorana checkerboard code is equivalent via a constant-depth unitary to a semionic version of the X-cube model and some decoupled fermionic modes \NoCaseChange{\protect\cite{cite3980}}.
\item\relax
\flmRefsHyperref[eczindexfamilyrel]{code:bosonization}{Bosonization code} --- Bosonization can be used to realize an X-cube model code with an emergent fermion from a Majorana stabilizer code \NoCaseChange{\protect\cite{cite3503}}, but this model has the same stabilizer group as the original X-cube model \NoCaseChange{\protect\cite{cite3504}}.
\item\relax
\flmRefsHyperref[eczindexfamilyrel]{code:checkerboard}{Checkerboard model code} --- The checkerboard model is equivalent to two copies of the X-cube model via a local constant-depth unitary \NoCaseChange{\protect\cite{cite3524}}.
\item\relax
\flmRefsHyperref[eczindexfamilyrel]{code:fcc_fracton}{Four Color Cube (FCC) fracton model code} --- The FCC fracton model code is obtained from four coupled X-cube models using p-membrane condensation. \NoCaseChange{\protect\cite{cite534}}.
\end{eczvaluelist}
\eczhbkcontributors{ Ke Liu (刘科 子竞), \eczhuVVA }
\endeczcode

\eczcode{xp_stabilizer}{XP stabilizer code}{~\NoCaseChange{\protect\cite{cite798}}}
\codefieldsection{Alternative Names}
\begin{eczvaluelist}
\item\relax Weighted hypergraph code
\end{eczvaluelist}
\eczhIndexCodeAliasName{xp_stabilizer}{Weighted hypergraph code}
\codefieldsection{Description}
The XP Stabilizer formalism is a generalization of the XS and Pauli stabilizer formalisms, with stabilizer generators taken from the group \( \mathsf{BD}_{2N}^{\otimes n} = \langle\omega I, X, P\rangle^{\otimes n} \), which is the tensor product of the binary dihedral group of order \(8N\).
Here, \(N\) is called the \textit{precision}, \( \omega \) is a \( 2N \)th root of unity, and \( P = \text{diag} ( 1, \omega^2) \).
The codespace is a \(+1\) eigenspace of a set of XP stabilizer generators, which need not commute to define a valid codespace.

XP stabilizer states are in one-to-one correspondence with weighted hypergraph states \NoCaseChange{\protect\cite{cite798,cite768}}, which generalize both weighted graph states \NoCaseChange{\protect\cite{cite3569,cite3556,cite3570}} and hypergraph states \NoCaseChange{\protect\cite{cite3571,cite3572,cite3573}}.
XP stabilizer codes are classified into XP-regular and XP-non-regular, where the former admits logical dimension \(K=2^k\) (for some integer \(k\)) and can be mapped to a CSS code with the same diagonal logical operators and similar non-diagonal logical operators, while the latter can have arbitrary dimension, are non-additive, and resemble CWS codes \NoCaseChange{\protect\cite{cite798}}.
Measurement of diagonal Pauli operators can be classically simulated efficiently on any XP code, but estimating outcome probabilities for general precision-\(4\) XP operators is NP-complete and measurement can take one outside the XP formalism \NoCaseChange{\protect\cite{cite798}}.

\codefieldsection{Encoding}
\begin{eczvaluelist}
\item\relax Initialization of all qubits in the \(|+\rangle\) state and action of generalized controlled \(Z\) gates on multi-edges of the underlying hypergraph \NoCaseChange{\protect\cite{cite798,cite768}}.
\end{eczvaluelist}
\codefieldsection{Parent}
\begin{eczvaluelist}
\item\relax
\flmRefsHyperref[eczindexfamilyrel]{code:clifford_hierarchy}{Clifford-hierarchy stabilizer code} --- XP stabilizer codes are joint eigenspaces of operators in the binary dihedral group, a subgroup consisting of Pauli strings and elements of a level of the \flmRefsHyperref{ref2118}{Clifford hierarchy}.
\end{eczvaluelist}
\codefieldsection{Children}
\begin{eczvaluelist}
\item\relax
\flmRefsHyperref[eczindexfamilyrel]{code:xs_stabilizer}{XS stabilizer code} --- The XP stabilizer formalism reduces to the XS formalism at \(N=4\).
\item\relax
\flmRefsHyperref[eczindexfamilyrel]{code:qubit_stabilizer}{Qubit stabilizer code} --- XP stabilizer codes reduce to qubit stabilizer codes for \(N=2\).
\end{eczvaluelist}
\codefieldsection{Cousins}
\begin{eczvaluelist}
\item\relax
\flmRefsHyperref[eczindexfamilyrel]{code:qubit_css}{Qubit CSS code} --- Each XP-regular code can be mapped to a CSS code with the same diagonal logical operators and similar non-diagonal logical operators \NoCaseChange{\protect\cite{cite798}}.
\item\relax
\flmRefsHyperref[eczindexfamilyrel]{code:cws}{Codeword stabilized (CWS) code} --- The orbit representatives of XP codes play a similar role to the word operators of CWS codes, and non-XP-regular codes have a similar structure \NoCaseChange{\protect\cite{cite798}}.
\item\relax
\flmRefsHyperref[eczindexfamilyrel]{code:quantum_lego}{Tensor-network code} --- XP stabilizer codes can be understood through the Quantum Lego formalism \NoCaseChange{\protect\cite{cite786}}.
\item\relax
\flmRefsHyperref[eczindexfamilyrel]{code:cubic_theory}{Cubic theory code} --- The cubic theory code can be embedded into a larger codespace such that all diagonal logical operators are represented by XP operators \NoCaseChange{\protect\cite[{Sec. 4.3}]{cite767}}.
\item\relax
\flmRefsHyperref[eczindexfamilyrel]{code:invertible}{Chen-Hsin invertible-order code} --- The Chen-Hsin invertible-order code can be embedded into a larger codespace such that all diagonal logical operators are represented by XP operators \NoCaseChange{\protect\cite[{Sec. 4.3}]{cite767}}.
\item\relax
\flmRefsHyperref[eczindexfamilyrel]{code:binary_dihedral_permutation_invariant}{Binary dihedral PI code} --- Binary dihedral permutation invariant codewords form error spaces of XP stabilizer codes.
\item\relax
\flmRefsHyperref[eczindexfamilyrel]{code:hypercube_quantum}{\(\llbracket 2^D,D,2\rrbracket \) hypercube quantum code} --- The \(D\)th hypercube quantum code can be viewed as an XP stabilizer code with precision \(N = 2^D\) \NoCaseChange{\protect\cite[{Exam. 6.10}]{cite798}}.
\item\relax
\flmRefsHyperref[eczindexfamilyrel]{code:stab_8_1_2}{\(\llbracket 8,1,2\rrbracket \) Shen-Wang-Cao code} --- The \(\llbracket 8,1,2\rrbracket \) code is an XP-regular code that can be obtained via the XP stabilizer formalism applied to the \(\llbracket 15,1,3\rrbracket \) Reed-Muller code \NoCaseChange{\protect\cite{cite786}}.
\item\relax
\flmRefsHyperref[eczindexfamilyrel]{code:stab_8_3_2}{\(\llbracket 8,3,2\rrbracket \) Smallest interesting color code} --- As the \(D=3\) member of the hypercube-code family, the \(\llbracket 8,3,2\rrbracket \) code can be viewed as an XP stabilizer code with precision \(N=8\) \NoCaseChange{\protect\cite[{Exam. 6.10}]{cite798}}.
\item\relax
\flmRefsHyperref[eczindexfamilyrel]{code:cluster_state}{Cluster-state code} --- XP stabilizer states are in one-to-one correspondence with weighted hypergraph states \NoCaseChange{\protect\cite{cite798,cite768}}, which generalize both weighted graph states \NoCaseChange{\protect\cite{cite3569,cite3556,cite3570}} and hypergraph states \NoCaseChange{\protect\cite{cite3571,cite3572,cite3573}}. The latter can also be utilized in MBQC schemes \NoCaseChange{\protect\cite{cite3574,cite3575}}.
\item\relax
\flmRefsHyperref[eczindexfamilyrel]{code:diagonal_clifford}{\(\llbracket 2^r-1,1,3\rrbracket \) simplex code} --- Each \(\llbracket 2^r-1,1,3\rrbracket \) simplex code can be viewed as an XP stabilizer code with precision \(N = 2^{r-2}\) \NoCaseChange{\protect\cite[{Exam. 6.4}]{cite798}}.
\end{eczvaluelist}
\eczhbkcontributors{ Muhammad Junaid Aftab, \eczhuVVA }
\endeczcode

\eczcode{xs_stabilizer}{XS stabilizer code}{~\NoCaseChange{\protect\cite{cite3625}}}
\codefieldsection{Description}
A type of stabilizer code where stabilizer generators are elements of the group \( \{\alpha I, X, \sqrt{Z}\}^{\otimes n} \), with \( S = \sqrt{Z} = \text{diag} (1, i)\). The codespace is a joint \(+1\) eigenspace of a set of stabilizer generators, which need not commute to define a valid codespace.

The phases in the qubit-basis expansion of an XS stabilizer state are polynomials of degree three or below \NoCaseChange{\protect\cite{cite3625}}.

\codefieldsection{Parent}
\begin{eczvaluelist}
\item\relax
\flmRefsHyperref[eczindexfamilyrel]{code:xp_stabilizer}{XP stabilizer code} --- The XP stabilizer formalism reduces to the XS formalism at \(N=4\).
\end{eczvaluelist}
\codefieldsection{Children}
\begin{eczvaluelist}
\item\relax
\flmRefsHyperref[eczindexfamilyrel]{code:brickwork}{Brickwork \(XS\) stabilizer code} --- The brickwork \(XS\) stabilizer code is an \(XS\) stabilizer code \NoCaseChange{\protect\cite{cite589}}.
\item\relax
\flmRefsHyperref[eczindexfamilyrel]{code:double_semion_string_net}{Double-semion string-net code} --- The double-semion string-net code is an \(XS\) stabilizer code \NoCaseChange{\protect\cite[{Fig. 1}]{cite3625}}.
\item\relax
\flmRefsHyperref[eczindexfamilyrel]{code:stab_15_1_3}{\(\llbracket 15,1,3\rrbracket \) quantum RM code} --- The \(\llbracket 15,1,3\rrbracket \) code can be viewed as an XS stabilizer code \NoCaseChange{\protect\cite[{Exam. 6.4}]{cite798}}.
\end{eczvaluelist}
\codefieldsection{Cousins}
\begin{eczvaluelist}
\item\relax
\flmRefsHyperref[eczindexfamilyrel]{code:tqd_abelian}{Abelian TQD code} --- Abelian TQD models for the groups \(\mathbb{Z}_2^k\) can be realized as XS stabilizer codes \NoCaseChange{\protect\cite{cite3625}}. Upon gauging some symmetries \NoCaseChange{\protect\cite{cite462,cite463,cite233,cite464,cite465,cite466,cite467,cite468,cite469,cite470}}, a Type-III \(\mathbb{Z}_2^3\) TQD realizes the same topological order as the \(G=D_4\) quantum double model \NoCaseChange{\protect\cite{cite577,cite575}}.
\item\relax
\flmRefsHyperref[eczindexfamilyrel]{code:3d_color}{3D color code} --- The 3D color code on a particular lattice admits XS stabilizers; see \flmHref{https://www.youtube.com/watch?v=B8h5-ANc_-8}{talk by M. Kesselring at the 2020 FTQC conference}.
\end{eczvaluelist}
\eczhbkcontributors{ \eczhuVVA }
\endeczcode

\eczcode{xysurface}{XY surface code}{~\NoCaseChange{\protect\cite{cite2627}}}
\codefieldsection{Alternative Names}
\begin{eczvaluelist}
\item\relax Tailored surface code (TSC)
\end{eczvaluelist}
\eczhIndexCodeAliasName{xysurface}{Tailored surface code (TSC)}
\codefieldsection{Description}
A variant of the surface code whose generators are \(XXXX\)  and \(YYYY\), obtained by mapping \(Z \to Y\) in the surface code.
While this code is equivalent to a CSS surface code with the same distance, other properties like noise-bias performance can differ significantly.

\codefieldsection{Protection}
As a stabilizer code, \(\llbracket n=O(d^2), k=O(1), d\rrbracket \).
\codefieldsection{Code Capacity Threshold}
\begin{eczvaluelist}
\item\relax \(50\%\) at infinite \(Z\) bias with maximum-likelihood decoder \NoCaseChange{\protect\cite{cite4499}}.
\item\relax \(18.7\%\) for standard depolarizing noise with maximum-likelihood decoder \NoCaseChange{\protect\cite{cite4499}}.
\end{eczvaluelist}
\codefieldsection{Threshold}
\begin{eczvaluelist}
\item\relax \(6.32(3)\%\) for infinite \(Z\) bias, and thresholds of \(\approx 5\%\) for \(Z\) bias around \(\eta = 100\) using a variant of the minimum-weight perfect matching decoder \NoCaseChange{\protect\cite{cite4500}}.
\end{eczvaluelist}
\codefieldsection{Parent}
\begin{eczvaluelist}
\item\relax
\flmRefsHyperref[eczindexfamilyrel]{code:surface}{Kitaev surface code} --- The XY surface code is obtained from the surface code by applying \(H\sqrt{Z}H\) to all qubits, thereby exchanging \(Z\leftrightarrow Y\). While it is equivalent to a CSS surface code with the same distance, but other properties like noise-bias performance can differ significantly.
\end{eczvaluelist}
\codefieldsection{Cousins}
\begin{eczvaluelist}
\item\relax
\flmRefsHyperref[eczindexfamilyrel]{code:heavy_hex}{Heavy-hexagon code} --- XY surface code can be adapted for a heavy-hexagonal point set \NoCaseChange{\protect\cite{cite3714}}.
\item\relax
\flmRefsHyperref[eczindexfamilyrel]{code:asymmetric_qecc}{Asymmetric quantum code (AQC)} --- XY surface codes perform well against biased noise \NoCaseChange{\protect\cite{cite2627}}.
\end{eczvaluelist}
\eczhbkcontributors{ Arpit Dua, \eczhuVVA }
\endeczcode

\eczcode{xyz_color}{XYZ color code}{~\NoCaseChange{\protect\cite{cite2629}}}
\codefieldsection{Description}
A variant of the 6.6.6 color code whose generators are \(XZXZXZ\) and \(ZYZYZY\) Pauli strings associated to each hexagonal in the hexagonal (6.6.6) tiling. 
A further variation called the \textit{domain wall color code} admits generators of the form \(XXXZZZ\) and \(ZZZXXX\) \NoCaseChange{\protect\cite{cite437}}.
While such codes are equivalent to CSS color codes with the same distance, other properties like noise-bias performance can differ significantly.

\codefieldsection{Decoding}
\begin{eczvaluelist}
\item\relax Efficient ML decoder at infinite bias \NoCaseChange{\protect\cite{cite2629}}.
\item\relax Cellular-automaton decoder \NoCaseChange{\protect\cite{cite2629}}.
\end{eczvaluelist}
\codefieldsection{Code Capacity Threshold}
\begin{eczvaluelist}
\item\relax \(50\%\) threshold for noise infinitely biased towards \(X\) or \(Y\) or \(Z\) errors using cellular-automaton decoder \NoCaseChange{\protect\cite{cite2629}}.
\item\relax Independent \(X,Y\) noise: threshold value of the sum of both noise probabilities is between \(9\%\) and \(14\%\), depending on the noise bias \NoCaseChange{\protect\cite{cite2629}}.
\end{eczvaluelist}
\codefieldsection{Parent}
\begin{eczvaluelist}
\item\relax
\flmRefsHyperref[eczindexfamilyrel]{code:twist_defect_color}{Twist-defect color code}\end{eczvaluelist}
\codefieldsection{Cousins}
\begin{eczvaluelist}
\item\relax
\flmRefsHyperref[eczindexfamilyrel]{code:triangular_color}{Honeycomb (6.6.6) color code} --- The XYZ color code is obtained from the 6.6.6 color code by applying single-qubit Clifford rotations on a subset of qubits such that the \(X\)- and \(Z\)-type generators are mapped to \(XZXZXZ\) and \(ZYZYZY\), respectively.
\item\relax
\flmRefsHyperref[eczindexfamilyrel]{code:xzzx}{XZZX surface code} --- The XZZX surface (XYZ color) is a non-CSS analogue of the rotated surface (6.6.6 color) code such that the two codes are related by single-qubit Clifford rotations.
\item\relax
\flmRefsHyperref[eczindexfamilyrel]{code:fracton}{Fracton stabilizer code} --- The XYZ color code resembles a Type-II fracton code in the limit of infinite noise bias \NoCaseChange{\protect\cite{cite2629}}.
\item\relax
\flmRefsHyperref[eczindexfamilyrel]{code:asymmetric_qecc}{Asymmetric quantum code (AQC)} --- XYZ color codes perform well against biased noise \NoCaseChange{\protect\cite{cite2629}}.
\end{eczvaluelist}
\eczhbkcontributors{ \eczhuVVA }
\endeczcode

\eczcode{xyz_product}{XYZ product code}{~\NoCaseChange{\protect\cite{cite1611,cite645}}}
\codefieldsection{Description}
A non-CSS QLDPC code obtained from a three-fold variant of the hypergraph product applied to three classical binary codes with parity-check matrices \(H_1,H_2,H_3\).  Unlike CSS three-fold hypergraph product codes, the third input code acts through Pauli-\(Y\) checks \NoCaseChange{\protect\cite{cite645}}.
Under mild assumptions, the code dimension is determined by a tensor Sylvester equation over \(\mathbb{F}_2\), and the minimum-distance problem reduces up to constant factors to how closely a related inhomogeneous tensor Sylvester equation can be satisfied \NoCaseChange{\protect\cite{cite645}}.  When the underlying classical codes are repetition codes, the construction yields the \flmRefsHyperref{code:chamon}{Chamon model code}.

\codefieldsection{Protection}
The natural logical operators are membrane-like two-dimensional objects rather than strings \NoCaseChange{\protect\cite{cite645}}.  For cyclic XYZ product codes, fractal operators yield large-weight errors with constant-weight syndromes, ruling out local testability \NoCaseChange{\protect\cite{cite645}}.

\codefieldsection{Rate}
The logical dimension depends on properties of the input classical codes, specifically similarity invariants of the matrices \(H_i H_i^T\). It is conjectured that specific instances of XYZ product codes have a constant encoding rate and a minimum distance of \(d \in \Theta(n^{2/3})\) \NoCaseChange{\protect\cite{cite645}}.

\codefieldsection{Parents}
\begin{eczvaluelist}
\item\relax
\flmRefsHyperref[eczindexfamilyrel]{code:sc_qldpc}{Quantum spatially coupled (SC-QLDPC) code} --- XYZ product stabilizer generator matrices can be used as sub-matrices to define a 2D SC-QLDPC code \NoCaseChange{\protect\cite{cite644}}.
\item\relax
\flmRefsHyperref[eczindexfamilyrel]{code:generalized_homological_product}{Generalized homological-product code} --- The XYZ product code is a non-CSS three-fold variant of the hypergraph product built from three classical linear binary codes \NoCaseChange{\protect\cite{cite645}}.
\end{eczvaluelist}
\codefieldsection{Child}
\begin{eczvaluelist}
\item\relax
\flmRefsHyperref[eczindexfamilyrel]{code:chamon}{Chamon model code} --- The Chamon model code can be obtained from an XYZ product of three repetition codes \NoCaseChange{\protect\cite{cite1611}}, in a construction different from the 3D surface code; see \NoCaseChange{\protect\cite[{Sec. 3.4}]{cite645}}.
\end{eczvaluelist}
\codefieldsection{Cousins}
\begin{eczvaluelist}
\item\relax
\flmRefsHyperref[eczindexfamilyrel]{code:binary_linear}{Linear binary code} --- The XYZ product code is a non-CSS three-fold variant of the hypergraph product built from three classical linear binary codes \NoCaseChange{\protect\cite{cite645}}.
\item\relax
\flmRefsHyperref[eczindexfamilyrel]{code:multisector_hypergraph}{Higher-dimensional homological product code} --- The XYZ product code is a non-CSS three-fold variant of the hypergraph product built from three classical linear binary codes \NoCaseChange{\protect\cite{cite645}}.
\item\relax
\flmRefsHyperref[eczindexfamilyrel]{code:asymmetric_qecc}{Asymmetric quantum code (AQC)} --- XYZ product codes can be used to protect against biased noise \NoCaseChange{\protect\cite{cite2628}}.
\item\relax
\flmRefsHyperref[eczindexfamilyrel]{code:hypergraph_product}{Hypergraph product (HGP) code} --- Hypergraph (XYZ) product codes are constructed out of hypergraph products of two (three) classical linear codes.
\end{eczvaluelist}
\eczhbkcontributors{ Finnegan Voichick, \eczhuVVA }
\endeczcode

\eczcode{xyz_hexagonal}{XYZ\(^2\) hexagonal stabilizer code}{~\NoCaseChange{\protect\cite{cite4501,cite4502}}}
\codefieldsection{Description}
An instance of the matching code based on the Kitaev honeycomb model. It is described on a honeycomb tiling with \(XYZXYZ\) stabilizers on each hexagonal plaquette. Each vertical pair of qubits has an \(XX\), \(YY\), or \(ZZ\) link stabilizer depending on the orientation of the plaquette stabilizers.
\codefieldsection{Protection}
As a stabilizer code with boundaries, protects a single qubit with parameters \(\llbracket 2 d^2, 1, d\rrbracket \). Isolated \(X\), \(Y\), and \(Z\) errors lead to unidirectional pairs of plaquette defects along the three directions of the honeycomb tiling.
\codefieldsection{Decoding}
\begin{eczvaluelist}
\item\relax Maximum-likelihood decoding using the EWD decoder \NoCaseChange{\protect\cite{cite4503}}.
\item\relax Sequential decoder \NoCaseChange{\protect\cite{cite2645}}.
\end{eczvaluelist}
\codefieldsection{Code Capacity Threshold}
\begin{eczvaluelist}
\item\relax \(50\%\) for pure \(Z\), \(Y\), or \(Z\) noise under maximum-likelihood decoding.
\item\relax Threshold matches that of the \(XZZX\) code for various bias levels of \(X\), \(Y\), or \(Z\) biased noise  under maximum-likelihood decoding.
\item\relax \(\approx 18\%\) for depolarizing noise under maximum-likelihood decoding.
\item\relax \(18.3\%\) under biased noise \NoCaseChange{\protect\cite{cite2645}}.
\end{eczvaluelist}
\codefieldsection{Parents}
\begin{eczvaluelist}
\item\relax
\flmRefsHyperref[eczindexfamilyrel]{code:matching}{Matching code}\item\relax
\flmRefsHyperref[eczindexfamilyrel]{code:qubit_concatenated}{Concatenated qubit code} --- The XYZ\(^2\) hexagonal stabilizer code can be viewed as a concatenation of the \(YZZY\) surface code with one of the possible \(\llbracket 2,1\rrbracket \) repetition codes, with the case of the bit-flip repetition code yielding a concatenation of the surface code with the dual-rail code \NoCaseChange{\protect\cite{cite2645}}.
\end{eczvaluelist}
\codefieldsection{Cousins}
\begin{eczvaluelist}
\item\relax
\flmRefsHyperref[eczindexfamilyrel]{code:xzzx}{XZZX surface code} --- The XYZ\(^2\) hexagonal stabilizer code can be viewed as a concatenation of the \(YZZY\) surface code with one of the possible \(\llbracket 2,1\rrbracket \) repetition codes, with the case of the bit-flip repetition code yielding a concatenation of the surface code with the dual-rail code \NoCaseChange{\protect\cite{cite2645}}.
\item\relax
\flmRefsHyperref[eczindexfamilyrel]{code:dual_rail}{Dual-rail quantum code} --- The XYZ\(^2\) hexagonal stabilizer code can be viewed as a concatenation of the \(YZZY\) surface code with one of the possible \(\llbracket 2,1\rrbracket \) repetition codes, with the case of the bit-flip repetition code yielding a concatenation of the surface code with the dual-rail code \NoCaseChange{\protect\cite{cite2645}}.
\item\relax
\flmRefsHyperref[eczindexfamilyrel]{code:asymmetric_qecc}{Asymmetric quantum code (AQC)} --- The XYZ\(^2\) hexagonal stabilizer code has high thresholds under biased noise \NoCaseChange{\protect\cite{cite2645}}.
\end{eczvaluelist}
\eczhbkcontributors{ Basudha Srivastava, \eczhuVVA }
\endeczcode

\eczcode{xzzx}{XZZX surface code}{~\NoCaseChange{\protect\cite{cite4445,cite438,cite4504,cite2631}}}
\codefieldsection{Alternative Names}
\begin{eczvaluelist}
\item\relax Wen plaquette model
\end{eczvaluelist}
\eczhIndexCodeAliasName{xzzx}{Wen plaquette model}
\codefieldsection{Description}
A variant of the rotated surface code whose generators are \(XZZX\) Pauli strings associated, clockwise, to the vertices of each face of a two-dimensional lattice (with a qubit located at each vertex of the tessellation).

\textit{XZZX toric code} often either refers to the construction on the two-dimensional torus or is an alternative name for the general construction.
\textit{Twisted XZZX toric code} refers to the construction on a torus with twisted (a.k.a. shifted) boundary conditions; these need not be equivalent to twisted toric codes because they can be non-CSS.
The construction on surfaces with boundaries is often called the
\textit{XZZX planar code}.
On a closed lattice, the Wen plaquette realization of the XZZX toric code has the same \(\mathbb{Z}_2\) topological order as the toric code, and translation by one lattice unit exchanges the \(e\) and \(m\) anyons \NoCaseChange{\protect\cite[{Appx. C}]{cite465}}.

Stabilizer generators for this code are shown in \flmRefsCref{ref4505}.
\begin{flmFloat}{figure}{NumCap}\includegraphics[width=278.5716bp,max width=\linewidth]{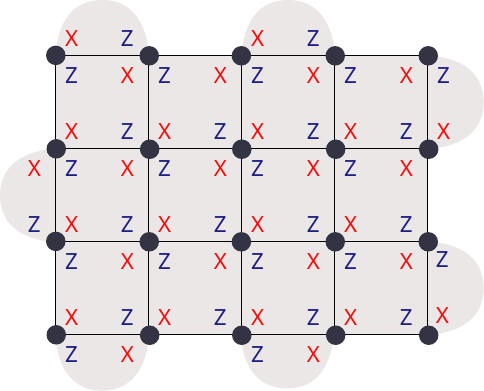}\caption{
  Stabilizer generators of an XZZX planar code with open boundaries.
  The generators are \(XZZX\) operators on the corners of squares in the bulk and \(XZ\) operators on the boundaries.
  }\label{ref4505}\end{flmFloat}

\codefieldsection{Protection}
As a stabilizer code, \(\llbracket n=O(d^2), k=O(1), d\rrbracket \).
\codefieldsection{Decoding}
\begin{eczvaluelist}
\item\relax MWPM decoder, which can be used for \(X\) and \(Z\) noise. For \(Y\) noise, a variant of the matching decoder could be used like it is used for the XY code in Ref. \NoCaseChange{\protect\cite{cite4500}}. Decoding complexity scales as \flmRefsHyperref{ref65}{order} \(O(n^3)\) because the code is non-CSS \NoCaseChange{\protect\cite{cite3739}\protect\cite[{Supplement}]{cite4500}}.
\end{eczvaluelist}
\codefieldsection{Code Capacity Threshold}
\begin{eczvaluelist}
\item\relax For large but finite \(X\)- or \(Z\)-biased noise, the code's thresholds exceed the zero-rate hashing bound.  The difference of the threshold from the hashing bound exceeds \(2.9\%\) at a \(Z\) or \(X\) bias of 300.
\item\relax \(50\%\) threshold for noise infinitely biased towards \(X\) or \(Y\) or \(Z\) errors using a maximum-likelihood decoder.
\item\relax Depolarizing noise: \(18.7(1)\%\) under tensor-network decoder \NoCaseChange{\protect\cite{cite2627}} and \(17.5\%\) under AMBP4 \NoCaseChange{\protect\cite{cite3740}}.
\end{eczvaluelist}
\codefieldsection{Threshold}
\begin{eczvaluelist}
\item\relax \(\approx 4.5\%\) using minimum-weight perfect matching decoder for depolarizing noise (bias \(\eta=0.5\)); \(\approx 10\%\) for infinite \(Z\) bias.
\item\relax \(4.15\%\) when \(98\%\) of depolarizing errors are converted into erasure errors with union-find decoder on a planar code, vs. \(0.937\%\) for pure depolarizing noise. The dominant source of noise in neutral atom arrays is spontaneous decay into detectable energy levels outside of the computational subspace. Since that decay occurs in a Rydberg level that is accessible from only one of the hyperfine states used for storage, the resulting channel is biased erasure \NoCaseChange{\protect\cite{cite4506}}.
\item\relax \(0.817\%\) and \(0.940\%\) with minimum-weight perfect matching and belief-matching decoder, respectively, for biased circuit-level noise \NoCaseChange{\protect\cite{cite3871}}.
\end{eczvaluelist}
\codefieldsection{Realizations}
\begin{eczvaluelist}
\item\relax Superconducting circuits: Distance-five 25-qubit code implemented on a superconducting quantum processor by Google Quantum AI \NoCaseChange{\protect\cite{cite4096}}.
This code outperformed the average of several instances of the smaller distance-three nine-qubit \(XZZX\) variant of the \flmRefsHyperref{code:surface-17}{surface-17} code realized on the same device, both in terms of logical error probability over 25 cycles and in terms of logical error per cycle.
This increase in error-correcting capabilities while using more physical qubits supports the notion of an error threshold.
Braiding of defects has been demonstrated for the distance-five code \NoCaseChange{\protect\cite{cite4507}}. Leakage errors have been handled in a separate work in a distance-three code \NoCaseChange{\protect\cite{cite4097}}.
Google Quantum AI follow-up experiment realizing distance-5 and distance-7 codes with 100 rounds of correction using the Libra and transformer-based decoders. The logical error rate is suppressed by a factor of \(\approx 2\), demonstrating beyond-break-even error correction with a block quantum code \NoCaseChange{\protect\cite{cite4098}}.
Magic-state cultivation was demonstrated on a device by Google Quantum AI by code switching between a distance-three 6.6.6 color code and distance-five \(XZZX\) surface code and decoding with the Tesseract decoder \NoCaseChange{\protect\cite{cite3748}}.

\item\relax Neutral atom arrays: Lukin group. Transversal CNOT gates performed on distance \(3\), \(5\), and \(7\) codes \NoCaseChange{\protect\cite{cite3363}}. Below-threshold performance on distance \(3\) and \(5\) codes with multiple rounds of syndrome extraction and error correction \NoCaseChange{\protect\cite{cite3206}}.
\end{eczvaluelist}
\codefieldsection{Notes}
\begin{eczvaluelist}
\item\relax A single \(X\) or \(Z\) error gives rise to two nearby defects, which can be viewed as endpoints of a string. That way, multiple \(Z\) errors can be decomposed into a combination of diagonal strings.
\item\relax Originally formulated as an example of \(\mathbb{Z}_2\) topological order in the Wen plaquette model \NoCaseChange{\protect\cite{cite4445}}.
\item\relax Popular summary of the Google Quantum AI above-threshold result in \flmHref{https://www.quantamagazine.org/quantum-computers-cross-critical-error-threshold-20241209/}{Quanta Magazine}.
\end{eczvaluelist}
\codefieldsection{Parent}
\begin{eczvaluelist}
\item\relax
\flmRefsHyperref[eczindexfamilyrel]{code:twist_defect_surface}{Twist-defect surface code} --- XZZX toric and planar codes can be treated in the general twist-defect surface code formalism \NoCaseChange{\protect\cite{cite427}}.
\end{eczvaluelist}
\codefieldsection{Child}
\begin{eczvaluelist}
\item\relax
\flmRefsHyperref[eczindexfamilyrel]{code:twisted_xzzx}{Twisted XZZX toric code} --- Imposing twisted (a.k.a. shifted) boundary conditions on the toric XZZX code yields the twisted XZZX code \NoCaseChange{\protect\cite[{Exam. 11 and Fig. 3}]{cite438}\protect\cite[{Fig. 6}]{cite427}}.
\end{eczvaluelist}
\codefieldsection{Cousins}
\begin{eczvaluelist}
\item\relax
\flmRefsHyperref[eczindexfamilyrel]{code:quantum_double_abelian}{Abelian quantum-double stabilizer code} --- The XZZX surface code is an example of \(\mathbb{Z}_2\) topological order as manifest in the Wen plaquette model \NoCaseChange{\protect\cite{cite4445}}.
\item\relax
\flmRefsHyperref[eczindexfamilyrel]{code:rotated_surface}{Rotated surface code} --- The XZZX code is obtained from the rotated surface code by applying Hadamard gates on a subset of qubits such that \(XXXX\) and \(ZZZZ\) generators are both mapped to \(XZXZ\). Both rotated and XZZX codes offer improved performance over the original surface code for biased noise \NoCaseChange{\protect\cite{cite4414}}.
\item\relax
\flmRefsHyperref[eczindexfamilyrel]{code:chamon}{Chamon model code} --- The Chamon model code can be obtained from an XYZ product of three repetition codes \NoCaseChange{\protect\cite{cite1611}}; see \NoCaseChange{\protect\cite[{Sec. 3.4}]{cite645}}. Using only two repetition codes in the analogous 2D construction yields the XZZX code, making it a 2D analogue of the Chamon code \NoCaseChange{\protect\cite[{Sec. 2}]{cite645}}.
\item\relax
\flmRefsHyperref[eczindexfamilyrel]{code:repetition}{Repetition code} --- The Chamon model code can be obtained from an XYZ product of three repetition codes \NoCaseChange{\protect\cite{cite1611}}; see \NoCaseChange{\protect\cite[{Sec. 3.4}]{cite645}}. Using only two repetition codes in the analogous 2D construction yields the XZZX code, making it a 2D analogue of the Chamon code \NoCaseChange{\protect\cite[{Sec. 2}]{cite645}}.
\item\relax
\flmRefsHyperref[eczindexfamilyrel]{code:fracton}{Fracton stabilizer code} --- Subsystem symmetries play a role in finite-bias decoders for both XZZX and fracton codes \NoCaseChange{\protect\cite{cite3522}}. The XZZX surface code resembles a Type-I fracton code with lineons in the limit of infinite noise bias \NoCaseChange{\protect\cite{cite2629}}.
\item\relax
\flmRefsHyperref[eczindexfamilyrel]{code:heavy_hex}{Heavy-hexagon code} --- XZZX surface code can be adapted for a heavy-hexagonal point set \NoCaseChange{\protect\cite{cite3714}}.
\item\relax
\flmRefsHyperref[eczindexfamilyrel]{code:cluster_state}{Cluster-state code} --- XZZX surface code can be foliated for a noise-bias preserving MBQC \NoCaseChange{\protect\cite{cite2634}} or FBQC \NoCaseChange{\protect\cite{cite2635}} protocol; see also \NoCaseChange{\protect\cite{cite2636}}.
\item\relax
\flmRefsHyperref[eczindexfamilyrel]{code:surface}{Kitaev surface code} --- The XZZX surface code on a square lattice with non-twisted periodic boundary conditions is obtained from a surface code by applying Hadamard gates on a subset of qubits such that \(XXXX\) and \(ZZZZ\) generators are both mapped to \(XZXZ\). While this code is equivalent to a CSS surface code with the same distance, other properties like noise-bias performance can differ significantly. Twisted XZZX surface codes are generally not equivalent to CSS surface codes via a single-qubit Clifford circuit and permutation.
\item\relax
\flmRefsHyperref[eczindexfamilyrel]{code:cat_concatenated}{Concatenated cat code} --- The four-component cat code can be concatenated with the XZZX code to yield a fusion-based computation scheme on a 2D lattice \NoCaseChange{\protect\cite{cite3687}}.
\item\relax
\flmRefsHyperref[eczindexfamilyrel]{code:gkp_surface_concatenated}{GKP-surface code} --- GKP codes have been concatenated with XZZX surface codes \NoCaseChange{\protect\cite{cite421}}.
\item\relax
\flmRefsHyperref[eczindexfamilyrel]{code:asymmetric_qecc}{Asymmetric quantum code (AQC)} --- The XZZX surface code can be foliated for a noise-bias preserving MBQC \NoCaseChange{\protect\cite{cite2634}} or FBQC \NoCaseChange{\protect\cite{cite2635}} protocol; see also \NoCaseChange{\protect\cite{cite2636}}.
\item\relax
\flmRefsHyperref[eczindexfamilyrel]{code:derby_klassen}{Derby-Klassen (DK) code} --- The DK code encodes fermions into excitations of the Wen plaquette model \NoCaseChange{\protect\cite{cite3622}}.
\item\relax
\flmRefsHyperref[eczindexfamilyrel]{code:xyz_color}{XYZ color code} --- The XZZX surface (XYZ color) is a non-CSS analogue of the rotated surface (6.6.6 color) code such that the two codes are related by single-qubit Clifford rotations.
\item\relax
\flmRefsHyperref[eczindexfamilyrel]{code:xyz_hexagonal}{XYZ\(^2\) hexagonal stabilizer code} --- The XYZ\(^2\) hexagonal stabilizer code can be viewed as a concatenation of the \(YZZY\) surface code with one of the possible \(\llbracket 2,1\rrbracket \) repetition codes, with the case of the bit-flip repetition code yielding a concatenation of the surface code with the dual-rail code \NoCaseChange{\protect\cite{cite2645}}.
\end{eczvaluelist}
\eczhbkcontributors{ Eric Huang, Arpit Dua, Marianna Podzorova, \eczhuVVA }
\endeczcode

\eczcode{yoked_surface}{Yoked surface code}{~\NoCaseChange{\protect\cite{cite523}}}
\codefieldsection{Description}
Member of a family of \(\llbracket n,k,d\rrbracket \) qubit CSS codes resulting from a concatenation of a \flmRefsHyperref{code:qmdpc}{QMDPC code} with a \flmRefsHyperref{code:rotated_surface}{rotated surface code}.
Concatenation does not impose additional connectivity constraints and can triple the number of logical qubits per physical qubit when compared to the original surface code.
Concatenation with 1D (2D) QMDPC yields codes with twice (four times) the distance.
Using the concatenation convention of the Zoo, the stabilizer generators of the inner QMDPC code are referred to as \textit{yokes} in this context; the cited paper \NoCaseChange{\protect\cite{cite523}} uses the opposite inner/outer terminology.
\codefieldsection{Decoding}
\begin{eczvaluelist}
\item\relax Soft information from the outer surface codes can be utilized via a message passing algorithm \NoCaseChange{\protect\cite{cite2698}}.
\item\relax Yokes can be measured using lattice surgery \NoCaseChange{\protect\cite{cite523}}.
\end{eczvaluelist}
\codefieldsection{Parents}
\begin{eczvaluelist}
\item\relax
\flmRefsHyperref[eczindexfamilyrel]{code:qubit_css}{Qubit CSS code}\item\relax
\flmRefsHyperref[eczindexfamilyrel]{code:qubit_concatenated}{Concatenated qubit code} --- Using the concatenation convention of the Zoo, a yoked surface code is a concatenation of a QMDPC code (inner code) with a rotated surface code (outer code). The cited paper \NoCaseChange{\protect\cite{cite523}} uses the opposite inner/outer terminology.
\end{eczvaluelist}
\codefieldsection{Cousins}
\begin{eczvaluelist}
\item\relax
\flmRefsHyperref[eczindexfamilyrel]{code:qmdpc}{Quantum multi-dimensional parity-check (QMDPC) code} --- Yoked surface codes are concatenations of QMDPC codes with rotated surface codes.
\item\relax
\flmRefsHyperref[eczindexfamilyrel]{code:rotated_surface}{Rotated surface code} --- Yoked surface codes are concatenations of QMDPC codes with rotated surface codes.
\end{eczvaluelist}
\eczhbkcontributors{ \eczhuVVA }
\endeczcode

\onecolumngrid
\clearpage

\section{Modular-qudit Kingdom}

\begin{eczEpigraph}
\begin{quote}
\flmQuoteSetup{quote}%
A lady of 80 named Gertie\\
Had a boyfriend of 60 named Bertie.\\
She told him emphatically\\
That viewed mathematically\\
By modulo 50, she's 30.
\flmQuoteAttributed{Patrick Vennebush}
\end{quote}
\end{eczEpigraph}

\twocolumngrid

\eczcode{three_qutrit_permutation_invariant}{\(\llparenthesis 3,2,2\rrparenthesis _3\) Three-qutrit single-deletion code}{~\NoCaseChange{\protect\cite{cite500}}}
\eczhIndexCodeAliasName{three_qutrit_permutation_invariant}{Three-qutrit single-deletion code}
\codefieldsection{Description}
Three-qutrit PI code that is the smallest qutrit PI code to correct one deletion error.

The code admits the following logical codewords:
\flmMathEnvironment{align}{}{
  |\overline{0}\rangle &\propto|000\rangle+|111\rangle+|222\rangle\\
  |\overline{1}\rangle &\propto|012\rangle+|021\rangle+|102\rangle+|120\rangle+|201\rangle+|210\rangle~.
}

\codefieldsection{Protection}
The smallest qutrit PI code to correct one deletion error.

\codefieldsection{Parents}
\begin{eczvaluelist}
\item\relax
\flmRefsHyperref[eczindexfamilyrel]{code:qudits_into_qudits}{Modular-qudit code}\item\relax
\flmRefsHyperref[eczindexfamilyrel]{code:permutation_invariant}{Permutation-invariant (PI) code}\item\relax
\flmRefsHyperref[eczindexfamilyrel]{code:small_distance_quantum}{Small-distance block quantum code}\end{eczvaluelist}
\codefieldsection{Cousins}
\begin{eczvaluelist}
\item\relax
\flmRefsHyperref[eczindexfamilyrel]{code:four_qubit_permutation_invariant}{\(\llparenthesis 4,2,2\rrparenthesis \) Four-qubit single-deletion code} --- The four-qubit (three-qutrit) single-deletion code is the smallest PI qubit (qutrit) code to correct one deletion error.
\item\relax
\flmRefsHyperref[eczindexfamilyrel]{code:wasilewski-banaszek}{Wasilewski-Banaszek code} --- The three-qutrit single-deletion code maps to the Wasilewski-Banaszek code via the \flmRefsHyperref{ref499}{simplex mapping} \NoCaseChange{\protect\cite{cite500}}.
\item\relax
\flmRefsHyperref[eczindexfamilyrel]{code:stab_3_1_2}{\(\llbracket 3,1,2\rrbracket _3\) Three-qutrit code} --- Projecting the three-qutrit code into the PI qutrit subspace yields the three-qutrit single-deletion code \NoCaseChange{\protect\cite{cite500}}.
\end{eczvaluelist}
\eczhbkcontributors{ \eczhuVVA }
\endeczcode

\eczcode{qudit_3_6_2}{\(\llparenthesis 3,6,2\rrparenthesis _{\mathbb{Z}_6}\) Euler code}{~\NoCaseChange{\protect\cite{cite2931}}}
\eczhIndexCodeAliasName{qudit_3_6_2}{Euler code}
\codefieldsection{Description}
Three-qudit error-detecting code with logical dimension \(K=6\) that is obtained from a particular \flmRefsHyperref{ref219}{AME state} that serves as a solution of a quantum analogue of the classical problem of 36 officers of Euler.
The code is obtained from a \(\llparenthesis 4,1,3\rrparenthesis _{\mathbb{Z}_6}\) code.

\codefieldsection{Encoding}
\begin{eczvaluelist}
\item\relax Quantum circuit encoding the state into a qubit system \NoCaseChange{\protect\cite{cite2926}}.
\end{eczvaluelist}
\codefieldsection{Notes}
\begin{eczvaluelist}
\item\relax Popular summary in \flmHref{https://www.quantamagazine.org/eulers-243-year-old-impossible-puzzle-gets-a-quantum-solution-20220110/}{Quanta Magazine}.
\end{eczvaluelist}
\codefieldsection{Parents}
\begin{eczvaluelist}
\item\relax
\flmRefsHyperref[eczindexfamilyrel]{code:qudits_into_qudits}{Modular-qudit code}\item\relax
\flmRefsHyperref[eczindexfamilyrel]{code:ame}{Perfect-tensor code} --- The \(\llparenthesis 3,6,2\rrparenthesis _{\mathbb{Z}_6}\) Euler code is an example of a non-stabilizer perfect-tensor code \NoCaseChange{\protect\cite{cite2931}}.
\item\relax
\flmRefsHyperref[eczindexfamilyrel]{code:small_distance_quantum}{Small-distance block quantum code}\end{eczvaluelist}
\eczhbkcontributors{ Karol Życzkowski, \eczhuVVA }
\endeczcode

\eczcode{qutrit_golay}{\(\llbracket 11,1,5\rrbracket _3\) qutrit Golay code}{~\NoCaseChange{\protect\cite{cite706}}}
\eczhIndexCodeAliasName{qutrit_golay}{qutrit Golay code}
\codefieldsection{Description}
An \(\llbracket 11,1,5\rrbracket _3\) code constructed from the ternary Golay code via the CSS construction.
The code's stabilizer generator matrix blocks \(H_{X}\) and \(H_{Z}\) are both the generator matrix of the ternary Golay code.

\codefieldsection{Magic}
Magic-state distillation scaling exponent \(\gamma=\log_3(1728\times 11) \approx 8.97\), where the \(1728\) factor comes from the fact that one round of distillation succeeds with probability \(\approx 1/1728\) \NoCaseChange{\protect\cite{cite706}}.
\codefieldsection{Transversal and Permutation-Based Gates}
\begin{eczvaluelist}
\item\relax All single-qutrit encoded Clifford gates \NoCaseChange{\protect\cite{cite706}}.
\end{eczvaluelist}
\codefieldsection{Gates}
\begin{eczvaluelist}
\item\relax Magic-state distillation of the strange state \(|S\rangle=\frac{1}{\sqrt{2}}(|1\rangle-|2\rangle)\) and the Norell state \(|N\rangle=\frac{1}{\sqrt{2}}(|1\rangle+|2\rangle)\), with the former achieving a cubic error suppression \NoCaseChange{\protect\cite{cite706}}.
\end{eczvaluelist}
\codefieldsection{Parents}
\begin{eczvaluelist}
\item\relax
\flmRefsHyperref[eczindexfamilyrel]{code:qudit_css}{Modular-qudit CSS code}\item\relax
\flmRefsHyperref[eczindexfamilyrel]{code:galois_quad_residue}{Quantum quadratic-residue (QR) code} --- The qutrit Golay code is a qutrit quantum QR code since the ternary Golay code is a QR code.
\item\relax
\flmRefsHyperref[eczindexfamilyrel]{code:small_distance_quantum}{Small-distance block quantum code}\end{eczvaluelist}
\codefieldsection{Cousins}
\begin{eczvaluelist}
\item\relax
\flmRefsHyperref[eczindexfamilyrel]{code:ternary_golay}{\([11,6,5]_3\) Ternary Golay code} --- The qutrit Golay code is a CSS code constructed from the ternary Golay code.
\item\relax
\flmRefsHyperref[eczindexfamilyrel]{code:qudit_cluster_state}{Modular-qudit cluster-state code} --- The qutrit Golay code can be realized as a modular-qudit cluster-state code \NoCaseChange{\protect\cite[{Fig. 2}]{cite706}}.
\item\relax
\flmRefsHyperref[eczindexfamilyrel]{code:qubit_golay}{\(\llbracket 23, 1, 7\rrbracket \) Quantum Golay code} --- The qubit Golay code is the qubit counterpart of the qutrit Golay code.
\end{eczvaluelist}
\eczhbkcontributors{ Yinchen Liu, \eczhuVVA }
\endeczcode

\eczcode{qudit_hamming_css}{\(\llbracket 2^r-1, 2^r-2r-1, 3\rrbracket _p\) quantum Hamming code}{~\NoCaseChange{\protect\cite{cite820}}}
\eczhIndexCodeAliasName{qudit_hamming_css}{quantum Hamming code}
\codefieldsection{Description}
A family of CSS codes extending \flmRefsHyperref{code:quantum_hamming_css}{quantum Hamming codes} to prime qudits of dimension \(p\) by expressing the qubit code stabilizers in local-dimension-invariant (LDI) form \NoCaseChange{\protect\cite{cite820}}.
\codefieldsection{Parents}
\begin{eczvaluelist}
\item\relax
\flmRefsHyperref[eczindexfamilyrel]{code:qudit_reed_muller}{Prime-qudit RM code} --- The \(\llbracket 2^r-1, 2^r-2r-1, 3\rrbracket _p\) quantum Hamming code family extends the qubit quantum Hamming family to prime qudits using local-dimension-invariant representations \NoCaseChange{\protect\cite{cite820}}.
\item\relax
\flmRefsHyperref[eczindexfamilyrel]{code:small_distance_quantum}{Small-distance block quantum code}\end{eczvaluelist}
\codefieldsection{Child}
\begin{eczvaluelist}
\item\relax
\flmRefsHyperref[eczindexfamilyrel]{code:quantum_hamming_css}{\(\llbracket 2^r-1, 2^r-2r-1, 3\rrbracket \) quantum Hamming code} --- \(\llbracket 2^r-1, 2^r-2r-1, 3\rrbracket _p\) prime-qudit CSS codes for \(p=2\) reduce to \(\llbracket 2^r-1, 2^r-2r-1, 3\rrbracket \) quantum Hamming codes.
\end{eczvaluelist}
\eczhbkcontributors{ Lane G. Gunderman, \eczhuVVA }
\endeczcode

\eczcode{stab_3_1_2}{\(\llbracket 3,1,2\rrbracket _3\) Three-qutrit code}{~\NoCaseChange{\protect\cite{cite4508}}}
\eczhIndexCodeAliasName{stab_3_1_2}{Three-qutrit code}
\codefieldsection{Description}
A \(\llbracket 3,1,2\rrbracket _3\) prime-qudit CSS code that is the smallest qutrit stabilizer code to detect a single-qutrit error.
It has stabilizer generators \(ZZZ\) and \(XXX\). The code defines a quantum secret-sharing scheme and serves as a minimal model for the AdS/CFT holographic duality. It is also the smallest non-trivial instance of a quantum maximum distance separable code (QMDS), saturating the quantum Singleton bound.

The codewords are
\flmMathEnvironment{align}{}{
  \begin{split}
    | \overline{0} \rangle &= \frac{1}{\sqrt{3}} (| 000 \rangle + | 111 \rangle + | 222 \rangle) \\
    | \overline{1} \rangle &= \frac{1}{\sqrt{3}} (| 012 \rangle + | 120 \rangle + | 201 \rangle) \\
    | \overline{2} \rangle &= \frac{1}{\sqrt{3}} (| 021 \rangle + | 102 \rangle + | 210 \rangle)~.
  \end{split}
}
The elements in the superposition of each logical codeword are related to each other via cyclic permutations.

\codefieldsection{Protection}
Detects single qutrit errors and protects against a single-qutrit erasure. It is the smallest single-erasure correcting qudit code for \(q>2\), and there does not exist a three-qubit code with analogous properties.

The code is an example of a \(\llparenthesis 2,3\rrparenthesis \) threshold scheme where a secret (the quantum information) is split into \(n=3\) shares and can be reconstructed from any \(k=2\) of them.

The key property of this code is that the reduced density matrix of any single qutrit is maximally mixed, meaning no information can be extracted from that qutrit. Therefore, a single qutrit tells you nothing about the encoded message, but access to any pair of qutrits reveals the secret.

\codefieldsection{Encoding}
\begin{eczvaluelist}
\item\relax In addition to thinking about the encoding of states, it is also interesting to look at the transformation of operators from the physical space into the logical space. Due to the unique structure and recovery protocol of the three-qutrit code, the representation of a logical operator \( \overline{O} \) is not unique. Instead, \( \overline{O} \) can be constructed from unitary matrices with support on only two out of the three qutrits. Therefore, the logical operator has valid representations supported on different pairs of qutrits. This operator construction is directly analogous to the construction of operators in the bulk (at the center) of the AdS\(_3\)-Rindler reconstruction. The three-qutrit code can then be used to describe how these local bulk operators are protected against localized boundary errors \NoCaseChange{\protect\cite{cite2543}}.
\end{eczvaluelist}
\codefieldsection{Decoding}
\begin{eczvaluelist}
\item\relax The quantum information (the secret) can be recovered from a unitary transformation acting on only two qutrits, \( U_{ij} \otimes I \), where \(U_{ij}\) acts on qutrits \(i,j\) and \(I\) is the identity on the remaining qutrit. By the cyclic structure of the codewords, this unitary transformation performs a permutation that recovers the information and stores it in one of the two qutrits involved in recovery.
\end{eczvaluelist}
\codefieldsection{Notes}
\begin{eczvaluelist}
\item\relax Connections to AdS/CFT from the perspective of how arbitrary operators are encoded into the logical space. This encoding is analogous and helps explain why operators acting on the bulk are protected against localized boundary errors \NoCaseChange{\protect\cite{cite2543}}.
\end{eczvaluelist}
\codefieldsection{Parents}
\begin{eczvaluelist}
\item\relax
\flmRefsHyperref[eczindexfamilyrel]{code:polynomial}{Prime-qudit RS code} --- The three-qutrit code is the smallest member of a family of \(\llbracket 2m-1,1,m\rrbracket _{p}\) prime-qudit quantum RS codes for \(p=3\) and \(m=2\) \NoCaseChange{\protect\cite{cite4508}}.
\item\relax
\flmRefsHyperref[eczindexfamilyrel]{code:holographic_tensor}{Holographic tensor-network code} --- The three-qutrit code is a radius-one holographic tensor-network code and serves as a minimal model for holography \NoCaseChange{\protect\cite{cite2543,cite2864}}.
\item\relax
\flmRefsHyperref[eczindexfamilyrel]{code:ame}{Perfect-tensor code} --- Three-qutrit codewords are AME, and the three-qutrit code stems from the \(\llbracket 4,0,3\rrbracket _3\) \flmRefsHyperref{ref219}{AME state} \NoCaseChange{\protect\cite{cite1925,cite151,cite2932}}.
\item\relax
\flmRefsHyperref[eczindexfamilyrel]{code:quantum_mds}{Quantum maximum-distance-separable (MDS) code} --- The three-qutrit code is the smallest nontrivial quantum MDS code.
\item\relax
\flmRefsHyperref[eczindexfamilyrel]{code:small_distance_quantum}{Small-distance block quantum code}\end{eczvaluelist}
\codefieldsection{Cousins}
\begin{eczvaluelist}
\item\relax
\flmRefsHyperref[eczindexfamilyrel]{code:quantum_secret_sharing}{Approximate secret-sharing code} --- The three-qutrit code defines a minimal secret-sharing scheme \NoCaseChange{\protect\cite{cite4508}} that is substantially generalized by approximate secret-sharing codes.
\item\relax
\flmRefsHyperref[eczindexfamilyrel]{code:three_qutrit_permutation_invariant}{\(\llparenthesis 3,2,2\rrparenthesis _3\) Three-qutrit single-deletion code} --- Projecting the three-qutrit code into the PI qutrit subspace yields the three-qutrit single-deletion code \NoCaseChange{\protect\cite{cite500}}.
\item\relax
\flmRefsHyperref[eczindexfamilyrel]{code:rotor_3_1_2}{\(\llbracket 3,1,2\rrbracket _{\mathbb{Z}}\) Three-rotor code} --- The three-rotor code is a rotor analogue of the three-qutrit code.
\end{eczvaluelist}
\eczhbkcontributors{ Felix Huber, Elizabeth R. Bennewitz, \eczhuVVA }
\endeczcode

\eczcode{qudit_5_1_3}{\(\llbracket 5,1,3\rrbracket _{\mathbb{Z}_q}\) modular-qudit code}{~\NoCaseChange{\protect\cite{cite531,cite532}}}
\eczhIndexCodeAliasName{qudit_5_1_3}{modular-qudit code}
\codefieldsection{Description}
Modular-qudit stabilizer code that generalizes the five-qubit perfect code using properties of the multiplicative group \(\mathbb{Z}_q\) \NoCaseChange{\protect\cite{cite531}}; see also \NoCaseChange{\protect\cite[{Thm. 13}]{cite532}}. It has four stabilizer generators consisting of \(X Z Z^\dagger X^\dagger I\) and its cyclic permutations.

The components of the encoding isometry in the computational basis (with \(a\) being the logical qudit index) are \NoCaseChange{\protect\cite[{Sec. VI.B}]{cite2720}}
\flmMathEnvironment{align}{}{
    T_{aklmnp}=\delta_{a,k+l+m+n+p}^{\mathbb{Z}_{q}}\frac{1}{q^{2}}\omega^{kl+lm+mn+np+pk}~,
}
where \(\omega\) is a primitive \(q\)th root of unity, and where \(\delta^{\mathbb{Z}_{q}}\) is the \flmRefsHyperref{ref20}{\(\mathbb{Z}_q\) Kronecker-delta function}.

\codefieldsection{Protection}
Protects against a single error on any one qudit. Detects two-qudit errors.
\codefieldsection{Encoding}
\begin{eczvaluelist}
\item\relax Generalized CNOT, Toffoli, and quantum Fourier transform gates.
\item\relax Encoders for prime-dimensional qudits \NoCaseChange{\protect\cite{cite4209,cite4509}}.
\end{eczvaluelist}
\codefieldsection{Gates}
\begin{eczvaluelist}
\item\relax Magic-state distillation for the \(q=3\) case \NoCaseChange{\protect\cite{cite4510}}.
\end{eczvaluelist}
\codefieldsection{Decoding}
\begin{eczvaluelist}
\item\relax Decoder for prime-dimensional qudits \NoCaseChange{\protect\cite{cite4209}}.
\end{eczvaluelist}
\codefieldsection{Parents}
\begin{eczvaluelist}
\item\relax
\flmRefsHyperref[eczindexfamilyrel]{code:qudit_stabilizer}{Modular-qudit stabilizer code}\item\relax
\flmRefsHyperref[eczindexfamilyrel]{code:ame}{Perfect-tensor code} --- The \(\llbracket 5,1,3\rrbracket _{\mathbb{Z}_q}\) code is a perfect-tensor code because it stems from the \(\llbracket 6,0,4\rrbracket _{\mathbb{Z}_q}\) \flmRefsHyperref{ref219}{AME state} \NoCaseChange{\protect\cite[{Thm. 13}]{cite532}}.
\item\relax
\flmRefsHyperref[eczindexfamilyrel]{code:quantum_cyclic}{Cyclic quantum code}\item\relax
\flmRefsHyperref[eczindexfamilyrel]{code:small_distance_quantum}{Small-distance block quantum code}\end{eczvaluelist}
\codefieldsection{Child}
\begin{eczvaluelist}
\item\relax
\flmRefsHyperref[eczindexfamilyrel]{code:stab_5_1_3}{\(\llbracket 5,1,3\rrbracket \) Five-qubit perfect code} --- The \(\llbracket 5,1,3\rrbracket _{\mathbb{Z}_q}\) modular-qudit code for \(q=2\) reduces to the five-qubit perfect code.
\end{eczvaluelist}
\codefieldsection{Cousins}
\begin{eczvaluelist}
\item\relax
\flmRefsHyperref[eczindexfamilyrel]{code:graph_quantum}{Graph quantum code} --- The \(\llbracket 5,1,3\rrbracket _{\mathbb{Z}_q}\) code admits a graph-quantum-code realization for \(G=\mathbb{Z}_q\) \NoCaseChange{\protect\cite{cite866}}.
\item\relax
\flmRefsHyperref[eczindexfamilyrel]{code:rotor_5_1_3}{\(\llbracket 5,1,3\rrbracket _{\mathbb{Z}}\) Five-rotor code} --- The five-rotor code is a rotor analogue of the five-qudit code.
\item\relax
\flmRefsHyperref[eczindexfamilyrel]{code:group_10_1_4}{\(\llbracket 10,1,4\rrbracket _{G}\) tenfold code} --- The \(\llbracket 10,1,4\rrbracket _{G}\) Abelian group code for \(G=\mathbb{Z}_q\) is defined using a graph that is closely related to the \(\llbracket 5,1,3\rrbracket _{\mathbb{Z}_q}\) modular-qudit code \NoCaseChange{\protect\cite{cite866}}.
\item\relax
\flmRefsHyperref[eczindexfamilyrel]{code:braunstein}{\(\llbracket 5,1,3\rrbracket _{\mathbb{R}}\) Braunstein five-mode code} --- The Braunstein five-mode code is a bosonic analogue of the five-qudit code.
\end{eczvaluelist}
\eczhbkcontributors{ Sarah Meng Li, \eczhuVVA }
\endeczcode

\eczcode{stab_9_1_3}{\(\llbracket 9,1,3\rrbracket _{\mathbb{Z}_q}\) modular-qudit code}{~\NoCaseChange{\protect\cite{cite4511}}}
\eczhIndexCodeAliasName{stab_9_1_3}{modular-qudit code}
\codefieldsection{Description}
Modular-qudit CSS code that generalizes the \(\llbracket 9,1,3\rrbracket \) Shor code to \(q\)-level systems.

\codefieldsection{Protection}
Protects against any quantum error arising from any one of the nine quantum registers.
\codefieldsection{Encoding}
\begin{eczvaluelist}
\item\relax Generalized CNOT, Toffoli, and quantum Fourier transform gates.
\end{eczvaluelist}
\codefieldsection{Parents}
\begin{eczvaluelist}
\item\relax
\flmRefsHyperref[eczindexfamilyrel]{code:qudit_css}{Modular-qudit CSS code}\item\relax
\flmRefsHyperref[eczindexfamilyrel]{code:group_quantum_parity}{Group-based QPC} --- The \(\llbracket 9,1,3\rrbracket _{G}\) group-based QPC reduces to the \(\llbracket 9,1,3\rrbracket _{\mathbb{Z}_q}\) modular-qudit code for \(G=\mathbb{Z}_q\).
\item\relax
\flmRefsHyperref[eczindexfamilyrel]{code:small_distance_quantum}{Small-distance block quantum code}\end{eczvaluelist}
\codefieldsection{Child}
\begin{eczvaluelist}
\item\relax
\flmRefsHyperref[eczindexfamilyrel]{code:shor_nine}{\(\llbracket 9,1,3\rrbracket \) Shor code} --- The \(\llbracket 9,1,3\rrbracket _{\mathbb{Z}_q}\) modular-qudit code for \(q=2\) reduces to the \(\llbracket 9,1,3\rrbracket \) Shor code.
\end{eczvaluelist}
\codefieldsection{Cousins}
\begin{eczvaluelist}
\item\relax
\flmRefsHyperref[eczindexfamilyrel]{code:real_projective_plane}{Projective-plane surface code} --- The qudit Shor code is a small qudit surface code on a Möbius strip with smooth boundary, which is obtained from removing a face of the tessellation of the projective plane \(\mathbb{R}P^2\) \NoCaseChange{\protect\cite[{Fig. 4}]{cite3383}}.
\item\relax
\flmRefsHyperref[eczindexfamilyrel]{code:group_quantum_repetition}{Group-based quantum repetition code} --- The \(\llbracket 9,1,3\rrbracket _{\mathbb{Z}_q}\) modular-qudit code is a concatenation of a bit-flip with a phase-flip group repetition code for \(G=\mathbb{Z}_q\).
\item\relax
\flmRefsHyperref[eczindexfamilyrel]{code:quantum_concatenated}{Concatenated quantum code} --- The \(\llbracket 9,1,3\rrbracket _{\mathbb{Z}_q}\) modular-qudit code is a concatenation of a bit-flip with a phase-flip group repetition code for \(G=\mathbb{Z}_q\).
\end{eczvaluelist}
\eczhbkcontributors{ Sarah Meng Li, \eczhuVVA }
\endeczcode

\eczcode{stab_9_1_5}{\(\llbracket 9,1,5\rrbracket _3\) quantum Glynn code}{~\NoCaseChange{\protect\cite{cite1653}}}
\eczhIndexCodeAliasName{stab_9_1_5}{quantum Glynn code}
\codefieldsection{Description}
Nine-qutrit \flmRefsHyperref{ref672}{pure} Hermitian code that is the smallest qutrit stabilizer code to correct two-qutrit errors.

See \NoCaseChange{\protect\cite[{Exam. 7}]{cite1655}} for its stabilizer generator matrix.

\codefieldsection{Protection}
Smallest stabilizer code that protects against errors on any two qutrits. Detects four-qutrit errors.
\codefieldsection{Parents}
\begin{eczvaluelist}
\item\relax
\flmRefsHyperref[eczindexfamilyrel]{code:qudit_stabilizer}{Modular-qudit stabilizer code}\item\relax
\flmRefsHyperref[eczindexfamilyrel]{code:stabilizer_over_gfqsq}{Hermitian Galois-qudit code}\item\relax
\flmRefsHyperref[eczindexfamilyrel]{code:quantum_mds}{Quantum maximum-distance-separable (MDS) code}\item\relax
\flmRefsHyperref[eczindexfamilyrel]{code:small_distance_quantum}{Small-distance block quantum code}\end{eczvaluelist}
\codefieldsection{Cousin}
\begin{eczvaluelist}
\item\relax
\flmRefsHyperref[eczindexfamilyrel]{code:glynn}{\([10,5,6]_9\) Glynn code} --- Applying the \flmRefsHyperref{code:stabilizer_over_gfqsq}{Hermitian construction} to the Glynn code yields a \(\llbracket 10,0,6\rrbracket _3\) state \NoCaseChange{\protect\cite{cite1653,cite1654}}. The \(\llbracket 9,1,5\rrbracket _3\) quantum Glynn code can be obtained by applying the \flmRefsHyperref{code:stabilizer_over_gfqsq}{Hermitian construction} to the shortened Glynn code \NoCaseChange{\protect\cite[{Corr. 4}]{cite1653}} (cf. \NoCaseChange{\protect\cite[{Exam. 7}]{cite1655}}).
\end{eczvaluelist}
\eczhbkcontributors{ \eczhuVVA }
\endeczcode

\eczcode{qutrit_small_triorthogonal}{\(\llbracket 9m-k,k,2\rrbracket _3\) triorthogonal code}{~\NoCaseChange{\protect\cite{cite709}}}
\eczhIndexCodeAliasName{qutrit_small_triorthogonal}{triorthogonal code}
\codefieldsection{Description}
Member of the \(\llbracket 9m-k,k,2\rrbracket _3\) family of triorthogonal qutrit codes (for \(k\leq 3m-2\)) that admit a transversal diagonal gate in the third level of the \flmRefsHyperref{ref751}{qudit Clifford hierarchy} and that are relevant for magic-state distillation.

\codefieldsection{Magic}
For \(k = 3m-2\), the family yields the magic-state yield parameter \(\gamma = \log_2 (2+\frac{6}{3m-2}) \to 1\) as \(m\to\infty\) \NoCaseChange{\protect\cite{cite709}}.
\codefieldsection{Parents}
\begin{eczvaluelist}
\item\relax
\flmRefsHyperref[eczindexfamilyrel]{code:qudit_triorthogonal}{Prime-qudit triorthogonal code}\item\relax
\flmRefsHyperref[eczindexfamilyrel]{code:small_distance_quantum}{Small-distance block quantum code}\end{eczvaluelist}
\eczhbkcontributors{ \eczhuVVA }
\endeczcode

\eczcode{zthree_znine}{\(\mathbb{Z}_3\times\mathbb{Z}_9\)-fusion subsystem code}{~\NoCaseChange{\protect\cite{cite414}}}
\codefieldsection{Description}
Modular-qudit 2D subsystem stabilizer code whose low-energy excitations realize a non-modular anyon theory with \(\mathbb{Z}_3\times\mathbb{Z}_9\) fusion rules.
Encodes two qutrits when put on a torus.

\codefieldsection{Parents}
\begin{eczvaluelist}
\item\relax
\flmRefsHyperref[eczindexfamilyrel]{code:qudit_subsystem_stabilizer}{Subsystem modular-qudit stabilizer code}\item\relax
\flmRefsHyperref[eczindexfamilyrel]{code:translationally_invariant_subsystem}{Lattice subsystem code}\item\relax
\flmRefsHyperref[eczindexfamilyrel]{code:topological_abelian}{Abelian topological code} --- The \(\mathbb{Z}_3\times\mathbb{Z}_9\)-fusion subsystem code is characterized by a non-modular anyon theory with \(\mathbb{Z}_3\times\mathbb{Z}_9\) fusion rules.
\end{eczvaluelist}
\codefieldsection{Cousin}
\begin{eczvaluelist}
\item\relax
\flmRefsHyperref[eczindexfamilyrel]{code:quantum_double_abelian}{Abelian quantum-double stabilizer code} --- The \(\mathbb{Z}_3\times\mathbb{Z}_9\)-fusion subsystem code can be obtained from a stack of \(q=3\) and \(q=9\) square-lattice qudit surface codes by \flmRefsHyperref{ref666}{gauging out} the anyons \(m_1^{-1}e_2^3\) and \(m_2^{-1}\) \NoCaseChange{\protect\cite[{Sec. 7.5}]{cite414}}.
\end{eczvaluelist}
\eczhbkcontributors{ Nathanan Tantivasadakarn, \eczhuVVA }
\endeczcode

\eczcode{qudit_znone}{\(\mathbb{Z}_q^{(1)}\) subsystem code}{~\NoCaseChange{\protect\cite{cite637,cite414}}}
\eczhIndexCodeAliasName{qudit_znone}{subsystem code}
\codefieldsection{Description}
Modular-qudit subsystem code, based on the Kitaev honeycomb model \NoCaseChange{\protect\cite{cite537}} and its generalization \NoCaseChange{\protect\cite{cite637}}, that is characterized by the \(\mathbb{Z}_q^{(1)}\) anyon theory \NoCaseChange{\protect\cite{cite638}}, which is modular for odd prime \(q\) and non-modular otherwise. Encodes a single \(q\)-dimensional qudit when put on a torus for odd \(q\), and a \(q/2\)-dimensional qudit for even \(q\). This code can be constructed using geometrically local gauge generators, but does not admit geometrically local stabilizer generators. For \(q=2\), the code reduces to the subsystem code underlying the Kitaev honeycomb model code as well as the honeycomb Floquet code.
\codefieldsection{Parents}
\begin{eczvaluelist}
\item\relax
\flmRefsHyperref[eczindexfamilyrel]{code:qudit_subsystem_stabilizer}{Subsystem modular-qudit stabilizer code}\item\relax
\flmRefsHyperref[eczindexfamilyrel]{code:translationally_invariant_subsystem}{Lattice subsystem code}\item\relax
\flmRefsHyperref[eczindexfamilyrel]{code:topological_abelian}{Abelian topological code} --- The \(\mathbb{Z}_q^{(1)}\) subsystem code is characterized by the \(\mathbb{Z}_q^{(1)}\) anyon theory \NoCaseChange{\protect\cite{cite638}}. The anyon theory has a single generator \(a \in \mathbb Z_N\) with \(\theta(a) =e^{\frac{2\pi i}{N}a^2}\). It is modular for odd prime \(q\) and non-modular otherwise.
\end{eczvaluelist}
\codefieldsection{Child}
\begin{eczvaluelist}
\item\relax
\flmRefsHyperref[eczindexfamilyrel]{code:kitaev_honeycomb}{Kitaev honeycomb code} --- The Kitaev honeycomb code is the \(q=2\) instance of the \(\mathbb{Z}_q^{(1)}\) subsystem code \NoCaseChange{\protect\cite[{Sec. 7.3}]{cite414}}.
\end{eczvaluelist}
\codefieldsection{Cousins}
\begin{eczvaluelist}
\item\relax
\flmRefsHyperref[eczindexfamilyrel]{code:qudit_surface}{Modular-qudit surface code} --- The \(\mathbb{Z}_q^{(1)}\) subsystem code can be obtained from the \(\mathbb{Z}_q\) square-lattice surface code by \flmRefsHyperref{ref666}{gauging out} the anyon \(e^{-1} m\) and applying transversal Clifford gates \NoCaseChange{\protect\cite[{Sec. 7.3}]{cite414}}. During this process, the square lattice is effectively expanded to a honeycomb tiling \NoCaseChange{\protect\cite[{Fig. 12}]{cite414}}.
\item\relax
\flmRefsHyperref[eczindexfamilyrel]{code:double_semion}{Double-semion stabilizer code} --- The anyonic exchange statistics of \(\mathbb{Z}_4^{(1)}\) subsystem code resemble those of the double semion code, but its fusion rules realize the \(\mathbb{Z}_4\) group.
\item\relax
\flmRefsHyperref[eczindexfamilyrel]{code:honeycomb_floquet}{Honeycomb Floquet code} --- The dynamically generated logical qubit of the honeycomb Floquet code is generated by appropriately scheduling measurements of the gauge generators of the \(\mathbb{Z}_{q=2}^{(1)}\) subsystem stabilizer code corresponding to the Kitaev honeycomb model. However, since this subsystem code has zero logical qubits, the instantaneous stabilizer codes of the honeycomb code cannot be interpreted as gauge-fixed versions of this subsystem code.
\item\relax
\flmRefsHyperref[eczindexfamilyrel]{code:semion}{Chiral semion subsystem code} --- The semion code can be obtained from the \(\mathbb{Z}_4^{(1)}\) subsystem code by \flmRefsHyperref{ref410}{condensing} the anyon \(s^2\) \NoCaseChange{\protect\cite[{Fig. 15}]{cite414}}.
\end{eczvaluelist}
\eczhbkcontributors{ Nathanan Tantivasadakarn, \eczhuVVA }
\endeczcode

\eczcode{quantum_double_abelian}{Abelian quantum-double stabilizer code}{~\NoCaseChange{\protect\cite{cite423}}}
\codefieldsection{Description}
Modular-qudit stabilizer code whose codewords realize 2D modular gapped Abelian topological order with trivial cocycle.
The corresponding anyon theory is defined by an Abelian group.
The \(G=\mathbb{Z}_2\) instance on a torus is the toric code, and cyclic-group instances reduce to modular-qudit surface codes.
All such codes can be realized by a stack of modular-qudit surface codes because all finite Abelian groups are direct products of cyclic groups.

There exists an invariant that can be computed to uniquely characterize the anyons of a state in an Abelian quantum-double topological phase \NoCaseChange{\protect\cite{cite2518}}.

\codefieldsection{Protection}
Error-correcting properties established in Ref. \NoCaseChange{\protect\cite{cite3116}} using operator algebra theory.
Correcting the maximum number of correctable errors is \(NP\)-complete \NoCaseChange{\protect\cite{cite4512}}.

\codefieldsection{Encoding}
\begin{eczvaluelist}
\item\relax Any geometrically local unitary circuit connecting two quantum double models whose groups are not isomorphic must have depth at least linear in \(n\) \NoCaseChange{\protect\cite{cite2518}}.
\end{eczvaluelist}
\codefieldsection{Decoding}
\begin{eczvaluelist}
\item\relax Efficient decoder correcting below the code distance \NoCaseChange{\protect\cite{cite4512}}.
\end{eczvaluelist}
\codefieldsection{Parents}
\begin{eczvaluelist}
\item\relax
\flmRefsHyperref[eczindexfamilyrel]{code:tqd_abelian_stabilizer}{Abelian TQD stabilizer code} --- The anyon theory corresponding to Abelian quantum double codes is defined by an Abelian group and trivial cocycle. Stacks of Abelian quantum double models are the starting point for constructing all Abelian TQD stabilizer codes by condensing bosons; for \(G=\prod_i \mathbb{Z}_{N_i}\), it suffices to use \(\prod_i \mathbb{Z}_{N_i^2}\) quantum doubles \NoCaseChange{\protect\cite{cite405}}. Conversely, every Abelian anyon theory is a subtheory of some Abelian TQD \NoCaseChange{\protect\cite[{Sec. 6.2}]{cite414}}. Upon gauging some symmetries \NoCaseChange{\protect\cite{cite462,cite463,cite233,cite464,cite465,cite466,cite467,cite468,cite469,cite470}}, Type-I and II \(\mathbb{Z}_2^3\) TQDs realize the same topological order as certain Abelian quantum double models \NoCaseChange{\protect\cite{cite577,cite575}}.
\item\relax
\flmRefsHyperref[eczindexfamilyrel]{code:quantum_double}{Quantum-double code} --- The anyon theory corresponding to (Abelian) quantum double codes is defined by an (Abelian) group.
\end{eczvaluelist}
\codefieldsection{Children}
\begin{eczvaluelist}
\item\relax
\flmRefsHyperref[eczindexfamilyrel]{code:2d_color}{2D color code} --- When treated as ground states of the code Hamiltonian, states of the color code on a torus geometry realize \(\mathbb{Z}_2\times\mathbb{Z}_2\) topological order \NoCaseChange{\protect\cite{cite2846}}, equivalent to the phase realized by two copies of the toric code (i.e., the surface code on a torus) via a local constant-depth \flmRefsHyperref{ref409}{Clifford circuit} \NoCaseChange{\protect\cite{cite422}}.
This process can be viewed as an ungauging \NoCaseChange{\protect\cite{cite462,cite463,cite233,cite464,cite465,cite466,cite467,cite468,cite469,cite470}} of certain symmetries.

\item\relax
\flmRefsHyperref[eczindexfamilyrel]{code:matching}{Matching code} --- Matching codes were inspired by the \(\mathbb{Z}_2\) topological order phase of the Kitaev honeycomb model \NoCaseChange{\protect\cite{cite537}}.
\item\relax
\flmRefsHyperref[eczindexfamilyrel]{code:clifford-deformed_surface}{Clifford-deformed surface code (CDSC)} --- When treated as ground states of the code Hamiltonian, surface codewords realize \(\mathbb{Z}_2\) topological order, a topological phase of matter that also exists in \(\mathbb{Z}_2\) lattice gauge theory \NoCaseChange{\protect\cite{cite3527}}. Local Clifford deformation preserves this topological order.
\item\relax
\flmRefsHyperref[eczindexfamilyrel]{code:qudit_surface}{Modular-qudit surface code} --- Modular-qudit surface code Hamiltonians admit topological phases associated with \(\mathbb{Z}_q\) topological order \NoCaseChange{\protect\cite{cite424}}.
\end{eczvaluelist}
\codefieldsection{Cousins}
\begin{eczvaluelist}
\item\relax
\flmRefsHyperref[eczindexfamilyrel]{code:2d_stabilizer}{2D lattice stabilizer code} --- Translation-invariant 2D prime-qudit lattice stabilizer codes are equivalent to several copies of the prime-qudit surface code and a trivial code via a local constant-depth qudit Clifford circuit \NoCaseChange{\protect\cite{cite4513}}.
\item\relax
\flmRefsHyperref[eczindexfamilyrel]{code:xzzx}{XZZX surface code} --- The XZZX surface code is an example of \(\mathbb{Z}_2\) topological order as manifest in the Wen plaquette model \NoCaseChange{\protect\cite{cite4445}}.
\item\relax
\flmRefsHyperref[eczindexfamilyrel]{code:3d_subsystem_surface}{3D subsystem surface code} --- The 3D subsystem surface code Hamiltonian phase diagram exhibits \(\mathbb{Z}_2\) topological order \NoCaseChange{\protect\cite{cite3034}}.
\item\relax
\flmRefsHyperref[eczindexfamilyrel]{code:zthree_znine}{\(\mathbb{Z}_3\times\mathbb{Z}_9\)-fusion subsystem code} --- The \(\mathbb{Z}_3\times\mathbb{Z}_9\)-fusion subsystem code can be obtained from a stack of \(q=3\) and \(q=9\) square-lattice qudit surface codes by \flmRefsHyperref{ref666}{gauging out} the anyons \(m_1^{-1}e_2^3\) and \(m_2^{-1}\) \NoCaseChange{\protect\cite[{Sec. 7.5}]{cite414}}.
\item\relax
\flmRefsHyperref[eczindexfamilyrel]{code:galois_color}{Galois-qudit color code} --- A Galois qudit for \(q=p^m\) can be decomposed into a Kronecker product of \(m\) modular qudits \NoCaseChange{\protect\cite{cite696,cite398,cite698,cite699,cite700}\protect\cite[{Sec. 5.3}]{cite697}}. Galois-qudit color codes yield Abelian quantum-double codes with Abelian-group topological order via this decomposition.
\item\relax
\flmRefsHyperref[eczindexfamilyrel]{code:galois_topological}{Galois-qudit surface code} --- A Galois qudit for \(q=p^m\) can be decomposed into a Kronecker product of \(m\) modular qudits \NoCaseChange{\protect\cite{cite696,cite398,cite698,cite699,cite700}\protect\cite[{Sec. 5.3}]{cite697}}. Galois-qudit surface codes yield Abelian quantum-double codes with \(\mathbb{F}_{p^m}\cong \mathbb{Z}_p^m\) topological order via this decomposition.
\end{eczvaluelist}
\eczhbkcontributors{ \eczhuVVA }
\endeczcode

\eczcode{tqd_abelian_stabilizer}{Abelian TQD stabilizer code}{~\NoCaseChange{\protect\cite{cite405}}}
\codefieldsection{Description}
Modular-qudit stabilizer code whose codewords realize a 2D Abelian twisted-quantum-double topological order on composite-dimensional qudits.
For every finite Abelian group \(G=\prod_i \mathbb{Z}_{N_i}\) and every product of Type-I and Type-II cocycles, there is a Pauli stabilizer Hamiltonian realizing the corresponding Abelian TQD \NoCaseChange{\protect\cite{cite405}}.
Equivalently, these codes exhaust the 2D Abelian topological orders that admit gapped boundaries \NoCaseChange{\protect\cite{cite405,cite406}}.

\codefieldsection{Rate}
On a torus, the ground-state Hilbert-space dimension is \(|G|^2\) for underlying group \(G=\prod_i \mathbb{Z}_{N_i}\) \NoCaseChange{\protect\cite{cite405}}.
\codefieldsection{Parents}
\begin{eczvaluelist}
\item\relax
\flmRefsHyperref[eczindexfamilyrel]{code:qudit_stabilizer}{Modular-qudit stabilizer code}\item\relax
\flmRefsHyperref[eczindexfamilyrel]{code:2d_stabilizer}{2D lattice stabilizer code} --- For every finite Abelian group \(G=\prod_i \mathbb{Z}_{N_i}\) and every product of Type-I and Type-II cocycles, there is a 2D modular-qudit Pauli stabilizer Hamiltonian on composite-dimensional qudits realizing the corresponding Abelian TQD \NoCaseChange{\protect\cite{cite405}}.
\item\relax
\flmRefsHyperref[eczindexfamilyrel]{code:tqd_abelian}{Abelian TQD code} --- Every Abelian TQD code with Type-I and -II cocycles can be realized as a modular-qudit Pauli stabilizer code by starting from a stack of Abelian quantum double models (it suffices to take \(\prod_i \mathbb{Z}_{N_i^2}\) toric codes) and \flmRefsHyperref{ref410}{condensing} certain bosonic anyons \NoCaseChange{\protect\cite{cite405}}.
\item\relax
\flmRefsHyperref[eczindexfamilyrel]{code:topological_abelian}{Abelian topological code} --- Every Abelian TQD code with Type-I and -II cocycles can be realized as a modular-qudit Pauli stabilizer code by starting from a stack of Abelian quantum double models (it suffices to take \(\prod_i \mathbb{Z}_{N_i^2}\) toric codes) and \flmRefsHyperref{ref410}{condensing} certain bosonic anyons \NoCaseChange{\protect\cite{cite405}}.

\end{eczvaluelist}
\codefieldsection{Children}
\begin{eczvaluelist}
\item\relax
\flmRefsHyperref[eczindexfamilyrel]{code:double_semion}{Double-semion stabilizer code} --- When treated as ground states of the code Hamiltonian, the double-semion stabilizer code states realize 2D double-semion topological order, i.e., the Abelian TQD for \(G=\mathbb{Z}_2\) with nontrivial Type-I cocycle, a topological phase that also exists as the deconfined phase of the 2D twisted \(\mathbb{Z}_2\) gauge theory \NoCaseChange{\protect\cite{cite584,cite405}}.
\item\relax
\flmRefsHyperref[eczindexfamilyrel]{code:quantum_double_abelian}{Abelian quantum-double stabilizer code} --- The anyon theory corresponding to Abelian quantum double codes is defined by an Abelian group and trivial cocycle. Stacks of Abelian quantum double models are the starting point for constructing all Abelian TQD stabilizer codes by condensing bosons; for \(G=\prod_i \mathbb{Z}_{N_i}\), it suffices to use \(\prod_i \mathbb{Z}_{N_i^2}\) quantum doubles \NoCaseChange{\protect\cite{cite405}}. Conversely, every Abelian anyon theory is a subtheory of some Abelian TQD \NoCaseChange{\protect\cite[{Sec. 6.2}]{cite414}}. Upon gauging some symmetries \NoCaseChange{\protect\cite{cite462,cite463,cite233,cite464,cite465,cite466,cite467,cite468,cite469,cite470}}, Type-I and II \(\mathbb{Z}_2^3\) TQDs realize the same topological order as certain Abelian quantum double models \NoCaseChange{\protect\cite{cite577,cite575}}.
\end{eczvaluelist}
\codefieldsection{Cousin}
\begin{eczvaluelist}
\item\relax
\flmRefsHyperref[eczindexfamilyrel]{code:spt}{Symmetry-protected topological (SPT) code} --- Gauging \NoCaseChange{\protect\cite{cite462,cite463,cite233,cite464,cite465,cite466,cite467,cite468,cite469,cite470}} the \(1\)-form symmetries associated with gauge charges of Abelian TQD stabilizer codes yields Pauli stabilizer models of SPT phases classified by products of Type-I and Type-II cocycles \NoCaseChange{\protect\cite{cite405}}.
\end{eczvaluelist}
\eczhbkcontributors{ \eczhuVVA }
\endeczcode

\eczcode{semion}{Chiral semion subsystem code}{~\NoCaseChange{\protect\cite{cite414}}}
\codefieldsection{Description}
Modular-qudit subsystem stabilizer code with qudit dimension \(q=4\) that is characterized by the chiral semion topological phase.
The code admits a set of geometrically local stabilizer generators on a torus.

\codefieldsection{Parents}
\begin{eczvaluelist}
\item\relax
\flmRefsHyperref[eczindexfamilyrel]{code:qudit_subsystem_stabilizer}{Subsystem modular-qudit stabilizer code}\item\relax
\flmRefsHyperref[eczindexfamilyrel]{code:translationally_invariant_subsystem}{Lattice subsystem code}\item\relax
\flmRefsHyperref[eczindexfamilyrel]{code:topological_abelian}{Abelian topological code} --- The semion code is a subsystem code characterized by the chiral semion topological phase.
\end{eczvaluelist}
\codefieldsection{Cousins}
\begin{eczvaluelist}
\item\relax
\flmRefsHyperref[eczindexfamilyrel]{code:double_semion}{Double-semion stabilizer code} --- The semion code can be obtained from the double-semion stabilizer code by \flmRefsHyperref{ref666}{gauging out} the anyon \(\bar{s}\) \NoCaseChange{\protect\cite[{Fig. 15}]{cite414}}.
\item\relax
\flmRefsHyperref[eczindexfamilyrel]{code:qudit_znone}{\(\mathbb{Z}_q^{(1)}\) subsystem code} --- The semion code can be obtained from the \(\mathbb{Z}_4^{(1)}\) subsystem code by \flmRefsHyperref{ref410}{condensing} the anyon \(s^2\) \NoCaseChange{\protect\cite[{Fig. 15}]{cite414}}.
\item\relax
\flmRefsHyperref[eczindexfamilyrel]{code:3d_semion}{Chiral semion Walker-Wang model code} --- A unitary QCA encoder applied to product state realizes the 3D chiral semion Walker-Wang model code, which in turn admits 2D chiral semion topological order if truncated at one of its surfaces \NoCaseChange{\protect\cite{cite471,cite472}}.
\end{eczvaluelist}
\eczhbkcontributors{ Nathanan Tantivasadakarn, \eczhuVVA }
\endeczcode

\eczcode{3d_semion}{Chiral semion Walker-Wang model code}{~\NoCaseChange{\protect\cite{cite472}}}
\codefieldsection{Description}
A 3D lattice modular-qudit stabilizer code with qudit dimension \(q=4\) whose low-energy excitations on boundaries realize the chiral semion topological order.
The model admits 2D chiral semion topological order at one of its surfaces \NoCaseChange{\protect\cite{cite471,cite472}}.
The corresponding phase can also be realized via a non-stabilizer Hamiltonian \NoCaseChange{\protect\cite{cite473}}.

\codefieldsection{Encoding}
\begin{eczvaluelist}
\item\relax A unitary QCA encoder applied to product state realizes the 3D chiral semion Walker-Wang model code, which in turn admits 2D chiral semion topological order if truncated at one of its surfaces \NoCaseChange{\protect\cite{cite471,cite472}}.
\end{eczvaluelist}
\codefieldsection{Parents}
\begin{eczvaluelist}
\item\relax
\flmRefsHyperref[eczindexfamilyrel]{code:qudit_stabilizer}{Modular-qudit stabilizer code}\item\relax
\flmRefsHyperref[eczindexfamilyrel]{code:3d_stabilizer}{3D lattice stabilizer code}\item\relax
\flmRefsHyperref[eczindexfamilyrel]{code:walker_wang}{Walker-Wang model code} --- The Walker-Wang model code reduces to the chiral semion model code when the input category is \(\mathcal{C}=\mathbb{Z}_{2}^{(1/2)}\), or alternatively \(\mathcal{C}=\mathbb{Z}_{4}^{(1)}\) after condensing a \(\mathbb{Z}_{2}\)-transparent boson.
\item\relax
\flmRefsHyperref[eczindexfamilyrel]{code:tqt}{Twisted quantum triple (TQT) code} --- When treated as ground states of the code Hamiltonian, the code states realize 3D double-semion topological order, a topological phase of matter that exists as the deconfined phase of the 3D twisted \(\mathbb{Z}_2\) gauge theory \NoCaseChange{\protect\cite{cite584}}.
\end{eczvaluelist}
\codefieldsection{Cousin}
\begin{eczvaluelist}
\item\relax
\flmRefsHyperref[eczindexfamilyrel]{code:semion}{Chiral semion subsystem code} --- A unitary QCA encoder applied to product state realizes the 3D chiral semion Walker-Wang model code, which in turn admits 2D chiral semion topological order if truncated at one of its surfaces \NoCaseChange{\protect\cite{cite471,cite472}}.
\end{eczvaluelist}
\eczhbkcontributors{ Nathanan Tantivasadakarn, \eczhuVVA }
\endeczcode

\eczcode{double_semion}{Double-semion stabilizer code}{~\NoCaseChange{\protect\cite{cite3624,cite405}}}
\codefieldsection{Alternative Names}
\begin{eczvaluelist}
\item\relax Doubled semion model code
\end{eczvaluelist}
\eczhIndexCodeAliasName{double_semion}{Doubled semion model code}
\codefieldsection{Description}
A 2D lattice modular-qudit stabilizer code with qudit dimension \(q=4\) that realizes the 2D double semion topological phase.
The code can be obtained from a \(\mathbb{Z}_4\) toric-code ground state by \flmRefsHyperref{ref410}{condensing} the emergent boson \(e^2 m^2\); in the stabilizer construction this condensation is implemented by two-body measurements \NoCaseChange{\protect\cite{cite405,cite414}}.
Its ground-state subspace can be mapped to that of the double-semion string-net model by a finite-depth quantum circuit with ancillas \NoCaseChange{\protect\cite{cite405}}.

This stabilizer code family is inequivalent to a CSS code via a constant-depth Clifford circuit \NoCaseChange{\protect\cite[{Thm. 1.1}]{cite4514}}.
Similarly, the double semion model has a sign problem \NoCaseChange{\protect\cite{cite4514,cite4515}} that cannot be eliminated via such a circuit.
However, the sign problem can be eliminated via a non-local circuit \NoCaseChange{\protect\cite{cite4516}}.

\codefieldsection{Parent}
\begin{eczvaluelist}
\item\relax
\flmRefsHyperref[eczindexfamilyrel]{code:tqd_abelian_stabilizer}{Abelian TQD stabilizer code} --- When treated as ground states of the code Hamiltonian, the double-semion stabilizer code states realize 2D double-semion topological order, i.e., the Abelian TQD for \(G=\mathbb{Z}_2\) with nontrivial Type-I cocycle, a topological phase that also exists as the deconfined phase of the 2D twisted \(\mathbb{Z}_2\) gauge theory \NoCaseChange{\protect\cite{cite584,cite405}}.
\end{eczvaluelist}
\codefieldsection{Cousins}
\begin{eczvaluelist}
\item\relax
\flmRefsHyperref[eczindexfamilyrel]{code:qudit_surface}{Modular-qudit surface code} --- The exchange statistics of the anyon for the double-semion code coincides with a subset of anyons in the \(\mathbb{Z}_4\) surface code, but the fusion rules are different. The double-semion code can be obtained from the \(\mathbb{Z}_4\) surface code by \flmRefsHyperref{ref410}{condensing} the anyon \(e^2 m^2\) \NoCaseChange{\protect\cite{cite414}} or by gauging \NoCaseChange{\protect\cite{cite462,cite463,cite233,cite464,cite465,cite466,cite467,cite468,cite469,cite470}} the one-form symmetry associated with said anyon \NoCaseChange{\protect\cite[{Footnote 20}]{cite414}}.
\item\relax
\flmRefsHyperref[eczindexfamilyrel]{code:double_semion_string_net}{Double-semion string-net code} --- The double-semion stabilizer code and the double-semion string-net code both realize the double semion topological phase, but the former is a modular-qudit Pauli stabilizer code while the latter is an \(XS\) stabilizer code. Their ground-state subspaces are connected by a finite-depth circuit with ancillas \NoCaseChange{\protect\cite{cite405}}. A commuting-projector version of the double-semion string-net code can also be derived \NoCaseChange{\protect\cite{cite2694,cite2693}}.
\item\relax
\flmRefsHyperref[eczindexfamilyrel]{code:xcube}{X-cube model code} --- A non-stabilizer commuting-projector code constructed by stacking layers of the double-semion string-net model, called the semionic X-cube model \NoCaseChange{\protect\cite{cite534}}, is equivalent to the X-cube model \NoCaseChange{\protect\cite{cite4492}} (see also Refs. \NoCaseChange{\protect\cite{cite3980,cite4497}}).
\item\relax
\flmRefsHyperref[eczindexfamilyrel]{code:qudit_znone}{\(\mathbb{Z}_q^{(1)}\) subsystem code} --- The anyonic exchange statistics of \(\mathbb{Z}_4^{(1)}\) subsystem code resemble those of the double semion code, but its fusion rules realize the \(\mathbb{Z}_4\) group.
\item\relax
\flmRefsHyperref[eczindexfamilyrel]{code:semion}{Chiral semion subsystem code} --- The semion code can be obtained from the double-semion stabilizer code by \flmRefsHyperref{ref666}{gauging out} the anyon \(\bar{s}\) \NoCaseChange{\protect\cite[{Fig. 15}]{cite414}}.
\end{eczvaluelist}
\eczhbkcontributors{ Nathanan Tantivasadakarn, \eczhuVVA }
\endeczcode

\eczcode{fracton}{Fracton stabilizer code}{~\NoCaseChange{\protect\cite{cite3032}}}
\codefieldsection{Description}
A 3D modular-qudit stabilizer code whose codewords make up the ground-state space of a Hamiltonian in a fracton phase.
Unlike topological phases, whose excitations can move in any direction, fracton phases are characterized by excitations whose movement is restricted.

Qubit fracton stabilizer codes are commonly grouped into the following three sub-types \NoCaseChange{\protect\cite{cite456}}:
\begin{enumerate}[(1)]\item \textit{Foliated type-I fracton phase}: Excitations are mobile in less than 3 dimensions, but codes can be grown by \textit{foliation}, i.e., stacking copies of the 2D surface code and applying a constant-depth circuit \NoCaseChange{\protect\cite{cite4517}}.
\item \textit{Fractal type-I fracton phase}: Excitations are mobile in less than 3 dimensions, and codes are not foliated.
\item \textit{Type-II fracton phase}: Excitations are not mobile in any dimension and there are no string operators.
\end{enumerate}

Fracton phases can be understood as topological defect networks, meaning that they can be described in the language of topological quantum field theory with defects \NoCaseChange{\protect\cite{cite3163,cite3164}}.

\codefieldsection{Parents}
\begin{eczvaluelist}
\item\relax
\flmRefsHyperref[eczindexfamilyrel]{code:qudit_stabilizer}{Modular-qudit stabilizer code}\item\relax
\flmRefsHyperref[eczindexfamilyrel]{code:3d_stabilizer}{3D lattice stabilizer code}\end{eczvaluelist}
\codefieldsection{Children}
\begin{eczvaluelist}
\item\relax
\flmRefsHyperref[eczindexfamilyrel]{code:majorana_checkerboard}{Majorana checkerboard code} --- The Majorana checkerboard code is a foliated type-I fracton code \NoCaseChange{\protect\cite{cite3980}}.
\item\relax
\flmRefsHyperref[eczindexfamilyrel]{code:anisotropic_z2_laplacian}{Anisotropic \(\mathbb{Z}_2\) Laplacian model code} --- The anisotropic \(\mathbb{Z}_2\) Laplacian model code is a graph-based analogue of a Type-I fracton phase with lineon-like restricted mobility.
\item\relax
\flmRefsHyperref[eczindexfamilyrel]{code:chamon}{Chamon model code} --- The Chamon model is a 4-foliated type-I fracton code \NoCaseChange{\protect\cite{cite3519}} and is the first example of a fracton phase \NoCaseChange{\protect\cite{cite456}}.
\item\relax
\flmRefsHyperref[eczindexfamilyrel]{code:checkerboard}{Checkerboard model code} --- The checkerboard model is equivalent to two copies of the X-cube model via a local constant-depth unitary \NoCaseChange{\protect\cite{cite3524}}. Hence, it is a foliated type-I fracton code.
\item\relax
\flmRefsHyperref[eczindexfamilyrel]{code:fcc_fracton}{Four Color Cube (FCC) fracton model code}\item\relax
\flmRefsHyperref[eczindexfamilyrel]{code:fibonacci_fractal_liquid}{Fibonacci fractal spin-liquid code} --- The Fibonacci fractal spin-liquid code is a fractal type-I fracton code \NoCaseChange{\protect\cite{cite456}}.
\item\relax
\flmRefsHyperref[eczindexfamilyrel]{code:sierpinsky_fractal_liquid}{Sierpinski prism model code} --- The Sierpinski prism model code is a fractal type-I fracton code \NoCaseChange{\protect\cite{cite456}}.
\item\relax
\flmRefsHyperref[eczindexfamilyrel]{code:hh_fracton}{Hsieh-Halasz (HH) code} --- Both HH-I and HH-II are fracton codes; HH-I is identified as foliated type-I, while HH-II remains inconclusive between fractal type-I and type-II in the sorting analysis of Ref. \NoCaseChange{\protect\cite{cite456}}.
\item\relax
\flmRefsHyperref[eczindexfamilyrel]{code:hhb_fracton}{Hsieh-Halasz-Balents (HHB) code} --- Both HHB models are expected to be foliated type-I fracton codes \NoCaseChange{\protect\cite[{Eqs. (D42-D43)}]{cite456}}.
\item\relax
\flmRefsHyperref[eczindexfamilyrel]{code:two_foliated}{Two-foliated fracton code} --- The two-foliated fracton code is a foliated type-I fracton code.
\item\relax
\flmRefsHyperref[eczindexfamilyrel]{code:fractal_liquid}{Type-II fractal spin-liquid code} --- The type-II fractal spin-liquid code is a type-II fracton code \NoCaseChange{\protect\cite{cite1348}}.
\item\relax
\flmRefsHyperref[eczindexfamilyrel]{code:qudit_cubic}{Qudit cubic code} --- Haah cubic \NoCaseChange{\protect\cite{cite3032}} codes 1-4, 7, 8, and 10 do not have string logical operators and are the first examples of Type-II fracton phases. The remaining cubic codes are fractal Type-I fracton codes \NoCaseChange{\protect\cite{cite456,cite4518}}. The qutrit models in \NoCaseChange{\protect\cite[{Eqs. (D11-D12)}]{cite456}} are likely Type-II, with no string operators found numerically up to width 20, while the \(q=5\) qudit model in \NoCaseChange{\protect\cite[{Eq. (D13)}]{cite456}} satisfies a proven no-string condition and is Type-II.
\item\relax
\flmRefsHyperref[eczindexfamilyrel]{code:qudit_xcube}{Qudit X-cube model code}\end{eczvaluelist}
\codefieldsection{Cousins}
\begin{eczvaluelist}
\item\relax
\flmRefsHyperref[eczindexfamilyrel]{code:topological}{Topological code} --- Unlike topological phases, whose excitations can move in any direction, fracton phases are characterized by excitations whose movement is restricted. Fracton phases can be understood as topological defect networks, meaning that they can be described in the language of topological quantum field theory with defects \NoCaseChange{\protect\cite{cite3163,cite3164}}.
\item\relax
\flmRefsHyperref[eczindexfamilyrel]{code:surface}{Kitaev surface code} --- Foliated type-I fracton phase codes can be grown by \textit{foliation}, i.e., stacking copies of the 2D surface code; see \NoCaseChange{\protect\cite[{Eq. (D32)}]{cite456}}.
\item\relax
\flmRefsHyperref[eczindexfamilyrel]{code:spt}{Symmetry-protected topological (SPT) code} --- Certain 3D CSS fracton codes can be ungauged \NoCaseChange{\protect\cite{cite462,cite463,cite233,cite464,cite465,cite466,cite467,cite468,cite469,cite470}} into 2D fractal-like SPT Hamiltonians; the paper gives an explicit construction from the 3D fractal code \NoCaseChange{\protect\cite{cite466}}. In subsystem-symmetry gauging \NoCaseChange{\protect\cite{cite462,cite463,cite233,cite464,cite465,cite466,cite467,cite468,cite469,cite470}} constructions, symmetry charges transforming under planar symmetries in one, two, or three directions become planon, lineon, or fracton excitations, respectively \NoCaseChange{\protect\cite{cite467}}.
\item\relax
\flmRefsHyperref[eczindexfamilyrel]{code:pinwheel}{Pinwheel code} --- The hypergraph product of a pinwheel code with a cyclic repetition code yields a local Type-I fracton model in three dimensions, while the hypergraph product of two pinwheel codes yields a local Type-II fracton model in four dimensions \NoCaseChange{\protect\cite{cite1350}}.
\item\relax
\flmRefsHyperref[eczindexfamilyrel]{code:cage_net}{Cage-net code} --- The cage-net construction can be used to realize various fracton phases, stabilizer and otherwise.
\item\relax
\flmRefsHyperref[eczindexfamilyrel]{code:groupoid_surface}{Groupoid toric code} --- Some groupoid toric code models admit fracton-like features such as extensive ground-state degeneracy and excitations with restricted mobility.
\item\relax
\flmRefsHyperref[eczindexfamilyrel]{code:quantum_repetition}{Quantum repetition code} --- Product constructions built from the one-dimensional Ising/repetition code yield several fracton phases \NoCaseChange{\protect\cite[{Fig. 8}]{cite1501}}.
\item\relax
\flmRefsHyperref[eczindexfamilyrel]{code:layer}{Layer code} --- Layer codes are non-translation invariant 3D lattice stabilizer codes that can be viewed as fracton topological defect networks \NoCaseChange{\protect\cite{cite3040}}.
\item\relax
\flmRefsHyperref[eczindexfamilyrel]{code:xyz_color}{XYZ color code} --- The XYZ color code resembles a Type-II fracton code in the limit of infinite noise bias \NoCaseChange{\protect\cite{cite2629}}.
\item\relax
\flmRefsHyperref[eczindexfamilyrel]{code:xzzx}{XZZX surface code} --- Subsystem symmetries play a role in finite-bias decoders for both XZZX and fracton codes \NoCaseChange{\protect\cite{cite3522}}. The XZZX surface code resembles a Type-I fracton code with lineons in the limit of infinite noise bias \NoCaseChange{\protect\cite{cite2629}}.
\end{eczvaluelist}
\eczhbkcontributors{ \eczhuVVA }
\endeczcode

\eczcode{frobenius}{Frobenius code}{~\NoCaseChange{\protect\cite{cite3318}}}
\codefieldsection{Description}
A cyclic prime-qudit stabilizer code whose length \(n\) divides \(p^t + 1\) for some positive integer \(t\).

\codefieldsection{Decoding}
\begin{eczvaluelist}
\item\relax Adapted from the Berlekamp decoding algorithm for classical BCH codes. There exists a polynomial-time quantum algorithm to correct errors of weight at most \(\tau\), where \(\delta=2\tau+1\) is the BCH distance of the code \NoCaseChange{\protect\cite{cite3318}}. 
\end{eczvaluelist}
\codefieldsection{Notes}
\begin{eczvaluelist}
\item\relax Frobenius Hermitian codes have been completely classified; no such codes exist when \(t\) is odd \NoCaseChange{\protect\cite{cite3318}}.
\end{eczvaluelist}
\codefieldsection{Parents}
\begin{eczvaluelist}
\item\relax
\flmRefsHyperref[eczindexfamilyrel]{code:qudit_stabilizer}{Modular-qudit stabilizer code}\item\relax
\flmRefsHyperref[eczindexfamilyrel]{code:quantum_cyclic}{Cyclic quantum code}\end{eczvaluelist}
\codefieldsection{Children}
\begin{eczvaluelist}
\item\relax
\flmRefsHyperref[eczindexfamilyrel]{code:stab_13_1_5}{\(\llbracket 13,1,5\rrbracket \) twisted toric code}\item\relax
\flmRefsHyperref[eczindexfamilyrel]{code:stab_5_1_3}{\(\llbracket 5,1,3\rrbracket \) Five-qubit perfect code} --- The \(\llbracket 5,1,3\rrbracket \) code is the smallest qubit Frobenius code \NoCaseChange{\protect\cite[{Table I}]{cite3318}}.
\end{eczvaluelist}
\codefieldsection{Cousin}
\begin{eczvaluelist}
\item\relax
\flmRefsHyperref[eczindexfamilyrel]{code:stabilizer_over_gf4}{Hermitian qubit code} --- Frobenius Hermitian codes have been completely classified; no such codes exist when \(t\) is odd \NoCaseChange{\protect\cite{cite3318}}.
\end{eczvaluelist}
\eczhbkcontributors{ En-Jui Kuo, Nolan Coble, \eczhuVVA }
\endeczcode

\eczcode{qudit_3d_surface}{Modular-qudit 3D surface code}{~\NoCaseChange{\protect\cite{cite459,cite4519}}}
\codefieldsection{Description}
A generalization of the 3D surface code to modular qudits.
Qudits are placed on edges, \(Z\)-type stabilizer generators are placed on square plaquettes oriented in all three directions, and \(X\)-type stabilizers are placed on the six edges neighboring every vertex \NoCaseChange{\protect\cite{cite459}}.

\codefieldsection{Parents}
\begin{eczvaluelist}
\item\relax
\flmRefsHyperref[eczindexfamilyrel]{code:qudit_css}{Modular-qudit CSS code}\item\relax
\flmRefsHyperref[eczindexfamilyrel]{code:quantum_triple}{Quantum-triple code} --- A quantum triple model for the group \(G=\mathbb{Z}_q\) is a modular-qudit 3D surface code.
\item\relax
\flmRefsHyperref[eczindexfamilyrel]{code:generalized_homological_product_css}{Generalized homological-product CSS code}\item\relax
\flmRefsHyperref[eczindexfamilyrel]{code:3d_stabilizer}{3D lattice stabilizer code}\item\relax
\flmRefsHyperref[eczindexfamilyrel]{code:topological_abelian}{Abelian topological code} --- The modular-qudit 3D surface code realizes 3D \(\mathbb{Z}_q\) gauge theory with bosonic charge and loop excitations (BcBl).
\end{eczvaluelist}
\codefieldsection{Child}
\begin{eczvaluelist}
\item\relax
\flmRefsHyperref[eczindexfamilyrel]{code:3d_surface}{3D surface code} --- The qudit 3D surface code reduces to the 3D surface code for \(q=2\). The 3D surface code realizes 3D \(\mathbb{Z}_2\) gauge theory with bosonic charge and loop excitations (BcBl). The welded surface code does not satisfy homogeneous topological order \NoCaseChange{\protect\cite{cite3141}}.
\end{eczvaluelist}
\eczhbkcontributors{ \eczhuVVA }
\endeczcode

\eczcode{qudit_cluster_state}{Modular-qudit cluster-state code}{~\NoCaseChange{\protect\cite{cite4520}}}
\codefieldsection{Alternative Names}
\begin{eczvaluelist}
\item\relax Modular-qudit graph-state code
\end{eczvaluelist}
\eczhIndexCodeAliasName{qudit_cluster_state}{Modular-qudit graph-state code}
\codefieldsection{Description}
A code based on a modular-qudit cluster state.

Modular-qudit cluster states are modular-qudit stabilizer states defined on a graph.
There is one modular-qudit stabilizer generator \(S_v\) per graph vertex \(v\) of the form \NoCaseChange{\protect\cite[{Eq. (1)}]{cite4520}}
\flmMathEnvironment{align}{}{
  S_v = X^{\dagger}_{v} \prod_{w\in N(v)} Z_w~,
}
where the neighborhood \(N(v)\) is the set of vertices which share an edge with \(v\).

\codefieldsection{Encoding}
\begin{eczvaluelist}
\item\relax Operators forming the information group can be used to track how logical information is encoded \NoCaseChange{\protect\cite{cite4521}}.
\end{eczvaluelist}
\codefieldsection{Gates}
\begin{eczvaluelist}
\item\relax 1D modular-qudit cluster states \NoCaseChange{\protect\cite{cite4520,cite4522}} are resources for universal MBQC.
\end{eczvaluelist}
\codefieldsection{Realizations}
\begin{eczvaluelist}
\item\relax Quantum computation with cluster states has been realized using photons in the time and frequency domains \NoCaseChange{\protect\cite{cite4523}}.
\end{eczvaluelist}
\codefieldsection{Parents}
\begin{eczvaluelist}
\item\relax
\flmRefsHyperref[eczindexfamilyrel]{code:qudit_stabilizer}{Modular-qudit stabilizer code} --- Modular-qudit cluster-state codes are particular modular-qudit stabilizer codes. Any modular-qudit stabilizer code is equivalent to a graph quantum code for \(G=\mathbb{Z}_q\) via a single-modular-qudit Clifford circuit \NoCaseChange{\protect\cite{cite3561}} (see also \NoCaseChange{\protect\cite{cite867}}).
\item\relax
\flmRefsHyperref[eczindexfamilyrel]{code:qudit_cws}{Modular-qudit CWS code} --- A type of modular-qudit cluster-state code can be built from a modular-qudit cluster state by applying the modular-qudit CWS construction using a linear \(q\)-ary code, in which codewords are obtained by applying modular-qudit \(Z\)-type operators defined by the code to the modular-qudit cluster state; see, e.g., Ref. \NoCaseChange{\protect\cite{cite3529}}.
\item\relax
\flmRefsHyperref[eczindexfamilyrel]{code:graph_quantum}{Graph quantum code} --- Graph quantum codes for \(G=\mathbb{Z}_q\) reduce to modular-qudit cluster-state codes.
\item\relax
\flmRefsHyperref[eczindexfamilyrel]{code:hopf_cluster_state}{Hopf-algebra cluster-state code} --- Hopf-algebra cluster-state codes reduce to modular-qudit cluster-state codes when the Hopf algebra reduces to the group \(\mathbb{Z}_q\).
\end{eczvaluelist}
\codefieldsection{Child}
\begin{eczvaluelist}
\item\relax
\flmRefsHyperref[eczindexfamilyrel]{code:cluster_state}{Cluster-state code} --- Modular-qudit cluster-state codes reduce to cluster-state codes for \(q=2\).
\end{eczvaluelist}
\codefieldsection{Cousins}
\begin{eczvaluelist}
\item\relax
\flmRefsHyperref[eczindexfamilyrel]{code:spt}{Symmetry-protected topological (SPT) code} --- Qudit cluster states defined on 1D lattices are representatives of various SPT phases \NoCaseChange{\protect\cite{cite3096}}. 

\item\relax
\flmRefsHyperref[eczindexfamilyrel]{code:1d_stabilizer}{1D lattice stabilizer code} --- Qudit cluster states defined on 1D lattices are representatives of various 1D SPT phases \NoCaseChange{\protect\cite{cite3096}}. 

\item\relax
\flmRefsHyperref[eczindexfamilyrel]{code:ame}{Perfect-tensor code} --- MDS codes can be used to obtain cluster states that are AME with minimal support \NoCaseChange{\protect\cite{cite1923,cite1924,cite1925,cite151,cite1926,cite1927}}.
\item\relax
\flmRefsHyperref[eczindexfamilyrel]{code:qutrit_golay}{\(\llbracket 11,1,5\rrbracket _3\) qutrit Golay code} --- The qutrit Golay code can be realized as a modular-qudit cluster-state code \NoCaseChange{\protect\cite[{Fig. 2}]{cite706}}.
\end{eczvaluelist}
\eczhbkcontributors{ \eczhuVVA }
\endeczcode

\eczcode{qudits_into_qudits}{Modular-qudit code}{}
\codefieldsection{Alternative Names}
\begin{eczvaluelist}
\item\relax \(\mathbb{Z}_q\)-qudit code
\item\relax Modular-qudit subspace code
\end{eczvaluelist}
\eczhIndexCodeAliasName{qudits_into_qudits}{\(\mathbb{Z}_q\)-qudit code}
\eczhIndexCodeAliasName{qudits_into_qudits}{Modular-qudit subspace code}

\codefieldsection{Kingdom root code}
for the \flmRefsHyperref{kingdom:qudits_into_qudits}{Modular-qudit Kingdom}.
\codefieldsection{Description}
Encodes a \(K\)-dimensional Hilbert space into a \(q^n\)-dimensional (\(n\)-qudit) Hilbert space, with canonical qudit states \(|k\rangle\) labeled by elements \(k\) of the group \(\mathbb{Z}_q\) of integers \textit{modulo} \(q\).
Usually denoted as \(\llparenthesis n,K\rrparenthesis _{\mathbb{Z}_q}\) or \(\llparenthesis n,K,d\rrparenthesis _{\mathbb{Z}_q}\), whenever the code's distance \(d\) is defined, and with \(q=p\) when the dimension is prime.

There exists an analogue of the Wigner function for modular qudits \NoCaseChange{\protect\cite{cite4524,cite4525,cite4526}}.

\codefieldsection{Protection}
An \(\llparenthesis n,K,d\rrparenthesis _{\mathbb{Z}_q}\) code with distance \(d\) detects errors acting on up to \(d-1\) modular qudits, corrects erasure errors on up to \(d-1\) modular qudits, or corrects errors acting on up to \(\lfloor (d-1)/2 \rfloor\) modular qudits.

\subsection{Modular-qudit Pauli-string error basis}

A convenient and often considered error set is the modular-qudit analogue \NoCaseChange{\protect\cite{cite4527,cite4528}} of the Pauli string basis for \flmRefsHyperref{code:qubits_into_qubits}{qubit} codes.

\begin{defterm}{Modular-qudit Pauli strings}\label{ref4529}\label{ref2198}
For a single qudit, this set consists of products of powers of the modular-qudit Pauli matrices \(X\) and \(Z\), which act on computational basis states \(|k\rangle\) for \(k\in\mathbb{Z}_q\) as
\flmMathEnvironment{align}{}{
  X\left|k\right\rangle =\left|k+1\right\rangle \,\,\text{ and }\,\,Z\left|k\right\rangle =e^{i\frac{2\pi}{q}k}\left|k\right\rangle ~,
}
with addition performed modulo \(q\).
For multiple qudits, error set elements are tensor products of elements of the single-qudit error set.
Tensor products of \(X\) (\(Z\)) modular-qudit Paulis acting on different qudits are called \(X\)\textit{-type} (\(Z\)\textit{-type}) modular-qudit Pauli strings.
Combining the \(X\)-type and \(Z\)-type strings with a primitive \(2q\)th root of unity forms a group called the \textit{modular-qudit Pauli group} \NoCaseChange{\protect\cite{cite2111}}.
\end{defterm}

For prime \(q\), this Pauli structure agrees with the usual prime-qudit Pauli group, but for prime-power \(q=p^m\) with \(m>1\) it differs from the \flmRefsHyperref{code:galois_into_galois}{Galois-qudit} choice based on field structure. 
Modular-qudit Pauli matrices \NoCaseChange{\protect\cite{cite4530,cite4531}} are also known as Weyl operators \NoCaseChange{\protect\cite{cite4532}}, Sylvester-t'Hooft generators \NoCaseChange{\protect\cite{cite4533,cite4534}}, shift and boost operators \NoCaseChange{\protect\cite{cite4535}}, or clock and shift matrices \NoCaseChange{\protect\cite{cite4536}}; they are special cases of Manin's quantum plane \NoCaseChange{\protect\cite{cite4537,cite4538,cite4539,cite4540}}.

The Pauli error set is a unitary basis for linear operators on the multi-qudit Hilbert space that is orthonormal under the Hilbert-Schmidt inner product; it is a \flmRefsHyperref{ref2812}{nice error basis}. The distance associated with this set is often the minimum weight of a qudit Pauli string that implements a nontrivial logical operation in the code.

\codefieldsection{Rate}
Non-stabilizer states yield higher quantum capacity of the discrete beamsplitter channel \NoCaseChange{\protect\cite{cite4541}}.
\codefieldsection{Gates}
\begin{eczvaluelist}
\item\relax The normalizer of the \flmRefsHyperref{ref2198}{modular-qudit Pauli group} is the \textit{modular-qudit Clifford group} \NoCaseChange{\protect\cite{cite4527,cite2111,cite4302,cite4522,cite71,cite4542,cite4543,cite4544,cite4545}}.
There is a standard form for modular-qudit Clifford-group operators \NoCaseChange{\protect\cite[{Lemma 4}]{cite2111}}, and any modular-qudit Clifford gate can be constructed from phase-shift and quantum Fourier transform gates \NoCaseChange{\protect\cite{cite4546}}.
For prime qudit dimension \(p\), the discrete Fourier transform, quadratic phase gate, and SUM gate generate the Clifford group \NoCaseChange{\protect\cite{cite398}}.
Clifford circuits on prime qudits, together with Pauli measurements and classical feedforward, admit efficient classical simulation by tracking stabilizer and logical generators; when measuring a Pauli outside the stabilizer normalizer, the outcome is uniformly distributed in \(\mathbb{Z}_p\) \NoCaseChange{\protect\cite{cite398}}.
Universal computing can be achieved using qudit Clifford gates and a single type of non-Clifford gate, such as the \(T\) gate \NoCaseChange{\protect\cite{cite673}}.
Non-Clifford gates are typically more difficult to implement than Clifford gates and so are treated as a resource.
There is a normal form for Clifford+\(T\) operators for qutrits \NoCaseChange{\protect\cite{cite4547}} and, more generally, odd prime qudits \NoCaseChange{\protect\cite{cite4548}}.
Optimizing non-Clifford-gate count can be done using various procedures; see Refs. \NoCaseChange{\protect\cite{cite4549,cite4550,cite4551,cite4552,cite4553}} for qutrit codes.
There are simulation algorithms for modular-qudit Clifford-dominated circuits \NoCaseChange{\protect\cite{cite4554}}.

\item\relax \begin{defterm}{Qudit Clifford hierarchy}\label{ref4555}\label{ref751} The modular-qudit Clifford hierarchy \NoCaseChange{\protect\cite{cite3219,cite4556,cite3018,cite800}} is a tower of gate sets which includes modular-qudit Pauli and modular-qudit Clifford gates at its first two levels, and non-Clifford qudit gates at higher levels. The \(k\)th level is defined recursively by \flmMathEnvironment{align}{}{ C_k = \{ U | U P U^{\dagger} \in C_{k-1} \}~, } where \(P\) is any modular-qudit Pauli matrix, and \(C_1\) is the \flmRefsHyperref{ref2198}{modular-qudit Pauli group}. Gates for one prime-dimensional qudit have been classified \NoCaseChange{\protect\cite{cite4194}}. \end{defterm}
\end{eczvaluelist}
\codefieldsection{Decoding}
\begin{eczvaluelist}
\item\relax For few-qudit codes (\(n\) is small), decoding can be based on a lookup table. For infinite code families, the size of such a table scales exponentially with \(n\), so approximate decoding algorithms scaling polynomially with \(n\) have to be used. The decoder determining the most likely error given a noise channel is called the \textit{maximum-likelihood} (ML) decoder.
\end{eczvaluelist}
\codefieldsection{Notes}
\begin{eczvaluelist}
\item\relax Review of qudit quantum computation \NoCaseChange{\protect\cite{cite4545}}.
\item\relax For odd prime qudit dimension, the discrete-Wigner-function analysis of \NoCaseChange{\protect\cite{cite4155}} determines depolarizing-noise thresholds beyond which non-stabilizer states and non-Clifford gates become Clifford-simulable, and identifies maximally robust such resources.
\item\relax Weight distribution of a code depends on the average entanglement of codewords \NoCaseChange{\protect\cite{cite1670,cite4557}}.
\item\relax Qudit Cirq library \NoCaseChange{\protect\cite{cite4558}}.
\item\relax See \NoCaseChange{\protect\cite[{Ch. 8}]{cite4559}} for a side-by-side introduction to modular and Galois qudits.
\end{eczvaluelist}
\codefieldsection{Parents}
\begin{eczvaluelist}
\item\relax
\flmRefsHyperref[eczindexfamilyrel]{code:block_quantum}{Block quantum code} --- Modular-qudit codes are block quantum codes with \(\Sigma=\mathbb{Z}_q\).
\item\relax
\flmRefsHyperref[eczindexfamilyrel]{code:qecc_finite}{Finite-dimensional quantum error-correcting code}\item\relax
\flmRefsHyperref[eczindexfamilyrel]{code:group_quantum}{Group-based quantum code} --- Group quantum codes whose physical spaces are constructed using modular-integer groups \(\mathbb{Z}_q\) are modular-qudit codes.
\end{eczvaluelist}
\codefieldsection{Children}
\begin{eczvaluelist}
\item\relax
\flmRefsHyperref[eczindexfamilyrel]{code:qubits_into_qubits}{Qubit code} --- Modular-qudit quantum codes for \(q=2\) correspond to qubit codes. Modular-qudit codes \NoCaseChange{\protect\cite{cite4209}}, circuits \NoCaseChange{\protect\cite{cite4210}}, and magic-state distillation schemes \NoCaseChange{\protect\cite{cite692,cite752}} can have advantages over their qubit counterparts. Modular qudits are useful for simulating gauge theories \NoCaseChange{\protect\cite{cite4211,cite4212}}. There are several ways to embed one or more qubits into a single modular qudit, yielding efficient qubit gate decompositions \NoCaseChange{\protect\cite{cite4213}}.
\item\relax
\flmRefsHyperref[eczindexfamilyrel]{code:qudit_da}{Modular-qudit dynamical code}\item\relax
\flmRefsHyperref[eczindexfamilyrel]{code:qudit_3_6_2}{\(\llparenthesis 3,6,2\rrparenthesis _{\mathbb{Z}_6}\) Euler code}\item\relax
\flmRefsHyperref[eczindexfamilyrel]{code:qudit_non_stabilizer}{Modular-qudit USt code}\item\relax
\flmRefsHyperref[eczindexfamilyrel]{code:three_qutrit_permutation_invariant}{\(\llparenthesis 3,2,2\rrparenthesis _3\) Three-qutrit single-deletion code}\end{eczvaluelist}
\codefieldsection{Cousins}
\begin{eczvaluelist}
\item\relax
\flmRefsHyperref[eczindexfamilyrel]{code:unitary_design}{Unitary \(t\)-design} --- The \flmRefsHyperref{ref2198}{prime-qudit Pauli group} is a unitary 1-design.
\item\relax
\flmRefsHyperref[eczindexfamilyrel]{code:q-ary_over_zq}{\(q\)-ary code over \(\mathbb{Z}_q\)} --- Modular-qudit codes are quantum counterparts of \(q\)-ary codes over \(\mathbb{Z}_q\).
\item\relax
\flmRefsHyperref[eczindexfamilyrel]{code:bosonic_q-ary_expansion}{Bosonic \(q\)-ary expansion} --- The bosonic \(q\)-ary expansion allows one to map between prime-dimensional qudit states and a Fock subspace of a single mode.
\item\relax
\flmRefsHyperref[eczindexfamilyrel]{code:subsystem_qudits_into_qudits}{Subsystem modular-qudit code} --- Subsystem modular-qudit codes reduce to (subspace) modular-qudit codes when there is no gauge subsystem.
\item\relax
\flmRefsHyperref[eczindexfamilyrel]{code:galois_into_galois}{Galois-qudit code} --- A Galois qudit for \(q=p^m\) can be decomposed into a Kronecker product of \(m\) modular qudits; see \NoCaseChange{\protect\cite{cite696,cite398,cite698,cite699,cite700}\protect\cite[{Sec. 5.3}]{cite697}}.
The two coincide when \(q\) is prime, and reduce to qubits when \(q=2\).
However, Pauli matrices for the two types of qudits are defined differently.
See \NoCaseChange{\protect\cite[{Ch. 8}]{cite4559}} for a side-by-side introduction to modular and Galois qudits.

\end{eczvaluelist}
\eczhbkcontributors{ \eczhuVVA }
\endeczcode

\eczcode{qudit_css}{Modular-qudit CSS code}{~\NoCaseChange{\protect\cite{cite3196,cite3337,cite3338,cite532,cite71}}}
\codefieldsection{Description}
An \(\llparenthesis n,K,d\rrparenthesis _q\) modular-qudit stabilizer code admitting a set of stabilizer generators that
are either \(Z\)-type or \(X\)-type Pauli strings.
Codes can be defined from two classical codes and/or chain complexes over the ring \(\mathbb{Z}_q\) via an extension of \flmRefsHyperref{ref683}{qubit CSS-to-homology correspondence} to modular qudits \NoCaseChange{\protect\cite{cite71}}.
The homology group of the logical operators has a torsion component because the chain complexes are defined over a ring, which yields codes whose logical dimension is not a power of \(q\).

The stabilizer generator matrix, taking values from \(\mathbb{Z}_q\), is of the form
\flmMathEnvironment{align}{}{
H=\begin{pmatrix}0 & H_{Z}\\
H_{X} & 0
\end{pmatrix}\label{ref4560}
}
such that the rows of the two blocks must be orthogonal
\flmMathEnvironment{align}{}{
H_X H_Z^T=0~.\label{ref4561}
}
The above condition guarantees that the \(X\)-stabilizer generators, defined in the \flmRefsHyperref{ref4562}{modular symplectic representation} as rows of \(H_X\), commute with the \(Z\)-stabilizer generators associated with \(H_Z\).

For composite \(q\), such codes need not encode an integer number of qudits, but there are general structure theorems \NoCaseChange{\protect\cite[{Thm. 3.3}]{cite4563}\protect\cite[{Thm. 8.1.1}]{cite4564}}, with the latter relating to homology.
An \((n,K_X,d_X)_{\mathbb{Z}_q}\) linear code \(C_X\) and an \((n,K_Z,d_Z)_{\mathbb{Z}_q}\) linear code \(C_Z\), satisfying \(C_X^\perp \subseteq C_Z\), yield an \(\llparenthesis n,K_X K_Z / q^n\rrparenthesis _{\mathbb{Z}_q}\) modular-qudit CSS code with distance \(d\geq\min\{d_X,d_Z\}\).
Specializing to the case when \(C_Z=(n,K,d)_{\mathbb{Z}_q}\) is dual-containing yields an \(\llparenthesis n,K^2 / q^n\rrparenthesis _{\mathbb{Z}_q}\) \textit{self-dual modular-qudit CSS code} with \(C_X = C_Z\) \NoCaseChange{\protect\cite[{Corr. 3.4}]{cite4563}}.

For prime \(q=p\), the logical dimension returns to being a power of \(q\): encoding is based on two related \flmRefsHyperref{code:q-ary_linear}{\(p\)-ary linear codes}, an \([n,k_X,d_X]_p \) code \(C_X\) and \([n,k_Z,d_Z]_p \) code \(C_Z\),
satisfying \(C_X^\perp \subseteq C_Z\). The resulting CSS code has \(k=k_X+k_Z-n\) logical qubits and distance \(d\geq\min\{d_X,d_Z\}\).
Specializing to the case when \(C_Z=[n,k,d]_p\) is dual-containing yields an \(\llbracket n,2k-n,\geq d_Z\rrbracket _p\) \textit{self-dual prime-qudit CSS code} with \(C_X = C_Z\).
The \(H_X\) (\(H_Z\)) block of \(H\) \eqref{ref4560} is the parity-check matrix of the code \(C_X\) (\(C_Z\)). 
The requirement \(C_X^\perp \subseteq C_Z\) guarantees \eqref{ref4561}.
Basis states for the code are, for coset representatives \(\gamma \in C_X/C_Z^\perp\),
\flmMathEnvironment{align}{}{
|\gamma + C_Z^\perp \rangle = \frac{1}{\sqrt{|C_Z^\perp|}} \sum_{\eta \in C_Z^\perp} |\gamma + \eta\rangle.
}

\codefieldsection{Transversal and Permutation-Based Gates}
\begin{eczvaluelist}
\item\relax Modular-qudit generalizations of CNOT gates are transversal interblock gates for all modular-qudit CSS codes.
\item\relax Scalar multiplication physical gates, which map \(|x\rangle \mapsto |ax\rangle\) for \(a \in \mathbb{Z}_q^\times\), the mutiplicative group modulo \(q\).
\end{eczvaluelist}
\codefieldsection{Parents}
\begin{eczvaluelist}
\item\relax
\flmRefsHyperref[eczindexfamilyrel]{code:qudit_stabilizer}{Modular-qudit stabilizer code} --- Modular-qudit CSS codes are modular-qudit stabilizer codes whose stabilizer groups admit a generating set of pure-\(X\) and pure-\(Z\) Pauli strings. 
Any \(\llbracket n,k,d\rrbracket _{\mathbb{Z}_q}\) stabilizer code can be mapped onto a \(\llbracket 2n,2k,\geq d\rrbracket _{\mathbb{Z}_q}\) \flmRefsHyperref{code:two_block_quantum}{two-block CSS code} code via \flmRefsHyperref{ref436}{symplectic doubling}, which preserves geometric locality of a code up to a constant factor.

\item\relax
\flmRefsHyperref[eczindexfamilyrel]{code:css}{Calderbank-Shor-Steane (CSS) stabilizer code}\end{eczvaluelist}
\codefieldsection{Children}
\begin{eczvaluelist}
\item\relax
\flmRefsHyperref[eczindexfamilyrel]{code:qubit_css}{Qubit CSS code} --- Modular-qudit CSS codes for \(q=2\) are qubit CSS codes.
\item\relax
\flmRefsHyperref[eczindexfamilyrel]{code:qutrit_golay}{\(\llbracket 11,1,5\rrbracket _3\) qutrit Golay code}\item\relax
\flmRefsHyperref[eczindexfamilyrel]{code:stab_9_1_3}{\(\llbracket 9,1,3\rrbracket _{\mathbb{Z}_q}\) modular-qudit code}\item\relax
\flmRefsHyperref[eczindexfamilyrel]{code:polynomial}{Prime-qudit RS code}\item\relax
\flmRefsHyperref[eczindexfamilyrel]{code:qudit_reed_muller}{Prime-qudit RM code}\item\relax
\flmRefsHyperref[eczindexfamilyrel]{code:fractal_liquid}{Type-II fractal spin-liquid code}\item\relax
\flmRefsHyperref[eczindexfamilyrel]{code:qudit_xcube}{Qudit X-cube model code}\item\relax
\flmRefsHyperref[eczindexfamilyrel]{code:qudit_triorthogonal}{Prime-qudit triorthogonal code}\item\relax
\flmRefsHyperref[eczindexfamilyrel]{code:qudit_gkp}{Modular-qudit GKP code}\item\relax
\flmRefsHyperref[eczindexfamilyrel]{code:qudit_sign}{Modular-qudit shift-resistant code}\item\relax
\flmRefsHyperref[eczindexfamilyrel]{code:qudit_3d_surface}{Modular-qudit 3D surface code}\item\relax
\flmRefsHyperref[eczindexfamilyrel]{code:qudit_color}{Modular-qudit lattice color code}\item\relax
\flmRefsHyperref[eczindexfamilyrel]{code:qudit_surface}{Modular-qudit surface code}\end{eczvaluelist}
\codefieldsection{Cousins}
\begin{eczvaluelist}
\item\relax
\flmRefsHyperref[eczindexfamilyrel]{code:two_block_quantum}{Two-block CSS code} --- Any \(\llbracket n,k,d\rrbracket _{\mathbb{Z}_q}\) stabilizer code can be mapped onto a \(\llbracket 2n,2k,\geq d\rrbracket _{\mathbb{Z}_q}\) \flmRefsHyperref{code:two_block_quantum}{two-block CSS code} code via \flmRefsHyperref{ref436}{symplectic doubling}, which preserves geometric locality of a code up to a constant factor.
\item\relax
\flmRefsHyperref[eczindexfamilyrel]{code:q-ary_linear_over_zq}{Linear code over \(\mathbb{Z}_q\)} --- The modular-qudit CSS construction uses two related \(q\)-ary linear codes over \(\mathbb{Z}_q\), \(C_X\) and \(C_Z\).
\item\relax
\flmRefsHyperref[eczindexfamilyrel]{code:qudit_subsystem_css}{Subsystem modular-qudit CSS code} --- Subsystem modular-qudit CSS codes reduce to (subspace) modular-qudit CSS codes when there is no gauge subsystem.
\end{eczvaluelist}
\eczhbkcontributors{ Leonid Pryadko, \eczhuVVA }
\endeczcode

\eczcode{qudit_cws}{Modular-qudit CWS code}{~\NoCaseChange{\protect\cite{cite3583,cite3044,cite4565}}}
\codefieldsection{Description}
A CWS code for modular qudits, defined using a modular-qudit cluster state and a set of modular-qudit \(Z\)-type Pauli strings defined by a \(q\)-ary classical code over \(\mathbb{Z}_q\).

The modular-qudit CWS construction takes in \( \mathcal{Q} = (\mathcal{G},\mathcal{C}) \), where \(\mathcal{G}\) is a graph, and where \(\mathcal{C}\) is an \((n,K,d)_{\mathbb{Z}_q}\) \(q\)-ary code over \(\mathbb{Z}_q\).
From the graph, we form the modular-qudit cluster state \( |\mathcal{G} \rangle \).
From the \(q\)-ary code, we form modular-qudit Pauli \(Z\)-type operators \( W_i = Z^{c_{i,1}} \otimes \cdots \otimes Z^{c_{i,n}} \), where \(c_{i,j} \) is the \(j\)-th coordinate of the \(i\)-th classical codeword.
The codewords are then \( | i \rangle =  W_i | \mathcal{G} \rangle \).

In an alternative convention (not used here), CWS codes are defined from an underlying modular-qudit stabilizer state that is not necessarily a cluster state.

\codefieldsection{Notes}
\begin{eczvaluelist}
\item\relax See Ref. \NoCaseChange{\protect\cite{cite3583}} for qudit CWS code tables.
\end{eczvaluelist}
\codefieldsection{Parent}
\begin{eczvaluelist}
\item\relax
\flmRefsHyperref[eczindexfamilyrel]{code:qudit_non_stabilizer}{Modular-qudit USt code} --- Any modular-qudit CWS code can be written as a modular-qudit USt whose (\(K=1\)) stabilizer code is the modular-qudit cluster state and whose coset representatives are constructed from the \(q\)-ary classical code over \(\mathbb{Z}_q\). Prime-dimensional modular-qudit CWS codes have a unique representation as USt codes \NoCaseChange{\protect\cite{cite4566}}. Conversely, modular-qudit USt codes are equivalent to modular-qudit CWS codes via a single-modular-qudit Clifford circuit as follows \NoCaseChange{\protect\cite{cite3585,cite3587}\protect\cite[{Sec. 10.4}]{cite3167}}. The set of coset representatives of any modular-qudit USt can be extended to a larger set iterating over the underlying stabilizer code such that all codewords can be obtained from a single stabilizer state. Then, one can apply a single-qudit Clifford transformation to map said modular-qudit stabilizer state into a modular-qudit cluster state.
\end{eczvaluelist}
\codefieldsection{Children}
\begin{eczvaluelist}
\item\relax
\flmRefsHyperref[eczindexfamilyrel]{code:cws}{Codeword stabilized (CWS) code} --- Modular-qudit CWS codes reduce to CWS codes for \(q=2\).
\item\relax
\flmRefsHyperref[eczindexfamilyrel]{code:qudit_cluster_state}{Modular-qudit cluster-state code} --- A type of modular-qudit cluster-state code can be built from a modular-qudit cluster state by applying the modular-qudit CWS construction using a linear \(q\)-ary code, in which codewords are obtained by applying modular-qudit \(Z\)-type operators defined by the code to the modular-qudit cluster state; see, e.g., Ref. \NoCaseChange{\protect\cite{cite3529}}.
\end{eczvaluelist}
\codefieldsection{Cousins}
\begin{eczvaluelist}
\item\relax
\flmRefsHyperref[eczindexfamilyrel]{code:quantum_perfect}{Perfect quantum code} --- Generalized concatenations of modular-qudit CWS codes yield a family of codes that have larger logical dimension than stabilizer codes and that asymptotically approach the modular-qudit Hamming bound \NoCaseChange{\protect\cite{cite2696}}.
\item\relax
\flmRefsHyperref[eczindexfamilyrel]{code:quantum_concatenated}{Concatenated quantum code} --- Generalized concatenations of modular-qudit CWS codes yield a family of codes that have larger logical dimension than stabilizer codes and that asymptotically approach the modular-qudit Hamming bound \NoCaseChange{\protect\cite{cite2696}}.
\item\relax
\flmRefsHyperref[eczindexfamilyrel]{code:qudit_stabilizer}{Modular-qudit stabilizer code} --- Modular-qudit CWS codes whose underlying classical code is a linear \(q\)-ary code over \(\mathbb{Z}_q\) are modular-qudit stabilizer codes containing a cluster-state codeword; see \NoCaseChange{\protect\cite[{Corr. 4-5}]{cite4567}}, which defines CWS codes as admitting an underlying stabilizer state that is not necessarily a cluster state.
\end{eczvaluelist}
\eczhbkcontributors{ \eczhuVVA }
\endeczcode

\eczcode{qudit_da}{Modular-qudit dynamical code}{}
\codefieldsection{Alternative Names}
\begin{eczvaluelist}
\item\relax Modular-qudit DA code
\item\relax Modular-qudit aperiodic Floquet code
\end{eczvaluelist}
\eczhIndexCodeAliasName{qudit_da}{Modular-qudit DA code}
\eczhIndexCodeAliasName{qudit_da}{Modular-qudit aperiodic Floquet code}
\codefieldsection{Description}
Dynamically generated stabilizer-based modular-qudit code whose (not necessarily periodic) sequence of few-body measurements implements state initialization, logical gates and error detection.

\codefieldsection{Parents}
\begin{eczvaluelist}
\item\relax
\flmRefsHyperref[eczindexfamilyrel]{code:qudits_into_qudits}{Modular-qudit code}\item\relax
\flmRefsHyperref[eczindexfamilyrel]{code:dynamic_gen}{Dynamically generated QECC} --- Dynamical code state initialization, logical gates, and error correction are done by a sequence of different (usually weight-two) stabilizer measurements.
\end{eczvaluelist}
\codefieldsection{Children}
\begin{eczvaluelist}
\item\relax
\flmRefsHyperref[eczindexfamilyrel]{code:da}{Dynamical code}\item\relax
\flmRefsHyperref[eczindexfamilyrel]{code:qudit_honeycomb}{Modular-qudit honeycomb Floquet code}\end{eczvaluelist}
\codefieldsection{Cousins}
\begin{eczvaluelist}
\item\relax
\flmRefsHyperref[eczindexfamilyrel]{code:random_stabilizer}{Random stabilizer code} --- Dynamical codes admit instantaneous stabilizer groups, and dynamical code state initialization, logical gates, and error correction are done by a sequence of different (usually weight-two) stabilizer measurements.
\item\relax
\flmRefsHyperref[eczindexfamilyrel]{code:translationally_invariant_stabilizer}{Lattice stabilizer code} --- Dynamical codes are typically defined on 2D and 3D lattices, but they are not conventional stabilizer codes in that they use \flmRefsHyperref{ref410}{code switching} for error correction and gates.
\end{eczvaluelist}
\eczhbkcontributors{ \eczhuVVA }
\endeczcode

\eczcode{qudit_gkp}{Modular-qudit GKP code}{~\NoCaseChange{\protect\cite[{Sec. II}]{cite513}}}
\codefieldsection{Alternative Names}
\begin{eczvaluelist}
\item\relax Shift-resistant code
\item\relax Pre-GKP code
\end{eczvaluelist}
\eczhIndexCodeAliasName{qudit_gkp}{Shift-resistant code}
\eczhIndexCodeAliasName{qudit_gkp}{Pre-GKP code}
\codefieldsection{Description}
Modular-qudit analogue of the GKP code.
Encodes a qudit into a larger qudit and protects against Pauli shifts up to some maximum value.

The simplest example requires an 18-dimensional qudit and admits stabilizer generators \(Z^6\) and \(X^6\).
A set of logical codewords is
\flmMathEnvironment{align}{}{
\begin{split}
|\overline{0}\rangle&=\frac{1}{\sqrt{3}}\left(|0\rangle+|6\rangle+|12\rangle\right)\\|\overline{1}\rangle&=\frac{1}{\sqrt{3}}\left(|3\rangle+|9\rangle+|15\rangle\right)~, \end{split}
}
and logical operators are \(Z^3\) and \(X^3\).

More generally, for qudit dimension \(q = r_1 r_2 K\) for some positive integers \(r_1\), \(r_2\), and logical dimension \(K\), the stabilizer generators are \(Z^{r_1 K}\) and \(X^{r_2 K}\).

\codefieldsection{Protection}
The above simple code corrects any Pauli string \(X^{a}Z^{b}\) with \(|a|,|b|\leq 1\).
A general code protects against any shift errors for which \(|a| < r_1/2\) and \(|b| < r_2/2\).

\codefieldsection{Gates}
\begin{eczvaluelist}
\item\relax Not all logical Clifford gates can be realized using \(q\)-dimensional modular-qudit Clifford gates for certain values of \(r_1,r_2\) \NoCaseChange{\protect\cite{cite4546}}.
\end{eczvaluelist}
\codefieldsection{Parents}
\begin{eczvaluelist}
\item\relax
\flmRefsHyperref[eczindexfamilyrel]{code:qudit_css}{Modular-qudit CSS code}\item\relax
\flmRefsHyperref[eczindexfamilyrel]{code:single_subsystem}{Monolithic quantum code}\end{eczvaluelist}
\codefieldsection{Cousins}
\begin{eczvaluelist}
\item\relax
\flmRefsHyperref[eczindexfamilyrel]{code:gkp}{Square-lattice GKP code} --- The square-lattice GKP code can be obtained from the modular-qudit code by taking the physical qudit dimension to be infinite \NoCaseChange{\protect\cite[{Sec. II}]{cite513}}.
\item\relax
\flmRefsHyperref[eczindexfamilyrel]{code:quantum_perfect}{Perfect quantum code} --- The modular-qudit GKP code is not a block code, but it is perfect in the sense that each correctable error maps the logical space into a distinct error space.
\item\relax
\flmRefsHyperref[eczindexfamilyrel]{code:rotor_gkp}{Rotor GKP code} --- The rotor GKP code can be thought of as a concatenation of a homological rotor code and a modular-qudit GKP code \NoCaseChange{\protect\cite[{Fig. 3}]{cite2699}}.
\item\relax
\flmRefsHyperref[eczindexfamilyrel]{code:qudit_sign}{Modular-qudit shift-resistant code} --- The modular-qudit shift-resistant code requires a smaller physical qudit dimension but protects against only one type of error \NoCaseChange{\protect\cite{cite2915}}.
\end{eczvaluelist}
\eczhbkcontributors{ \eczhuVVA }
\endeczcode

\eczcode{qudit_honeycomb}{Modular-qudit honeycomb Floquet code}{~\NoCaseChange{\protect\cite{cite4568}}}
\codefieldsection{Description}
A modular-qudit extension of the honeycomb Floquet code.

\codefieldsection{Parent}
\begin{eczvaluelist}
\item\relax
\flmRefsHyperref[eczindexfamilyrel]{code:qudit_da}{Modular-qudit dynamical code}\end{eczvaluelist}
\codefieldsection{Child}
\begin{eczvaluelist}
\item\relax
\flmRefsHyperref[eczindexfamilyrel]{code:honeycomb_floquet}{Honeycomb Floquet code} --- The modular-qudit honeycomb Floquet code reduces to the Hastings-Haah Floquet code for \(q=2\).
\end{eczvaluelist}
\eczhbkcontributors{ \eczhuVVA }
\endeczcode

\eczcode{qudit_color}{Modular-qudit lattice color code}{~\NoCaseChange{\protect\cite{cite673}}}
\codefieldsection{Description}
Extension of the color code to lattices of modular qudits.
Codes are defined analogously to qubit color codes on suitable lattices of any spatial dimension, but a directionality is required in order to make the modular-qudit stabilizers commute.
This can be done by puncturing a hyperspherical lattice \NoCaseChange{\protect\cite{cite475}} or constructing a star-bipartition; see \NoCaseChange{\protect\cite[{Sec. III}]{cite673}}.
Logical dimension is determined by the genus of the underlying surface (for closed surfaces), types of boundaries (for open surfaces), and/or any twist defects present.

\codefieldsection{Transversal and Permutation-Based Gates}
\begin{eczvaluelist}
\item\relax Some modular-qudit lattice color codes on \(D\)-dimensional lattices can transversally implement a gate at the \((D-1)\)st level of the \flmRefsHyperref{ref751}{qudit Clifford hierarchy} \NoCaseChange{\protect\cite{cite673}}.
\end{eczvaluelist}
\codefieldsection{Gates}
\begin{eczvaluelist}
\item\relax Modular-qudit lattice color codes whose \(X\)-type stabilizers are placed on cells of dimension \(\nu\) support transversal gates in the \(\nu\)th level of the \flmRefsHyperref{ref751}{qudit Clifford hierarchy} as long as \(\nu! \neq 0\) modulo the qudit dimension \NoCaseChange{\protect\cite[{Thm. 1}]{cite673}}. These codes saturate the \flmRefsHyperref{ref3630}{Bravyi-Koenig bound}. In particular, 3D modular-qudit color codes admit a transversal modular-qudit \(T\) gate.
\item\relax For odd prime qudit dimension, charge-and-color-permuting twist defects can be used to implement generalized \flmRefsHyperref{ref409}{Clifford gates} \NoCaseChange{\protect\cite{cite4569}}.
\end{eczvaluelist}
\codefieldsection{Decoding}
\begin{eczvaluelist}
\item\relax Generalized Color Clustering (GCC) decoder \NoCaseChange{\protect\cite{cite4570}}.
\end{eczvaluelist}
\codefieldsection{Parents}
\begin{eczvaluelist}
\item\relax
\flmRefsHyperref[eczindexfamilyrel]{code:qudit_css}{Modular-qudit CSS code}\item\relax
\flmRefsHyperref[eczindexfamilyrel]{code:translationally_invariant_stabilizer}{Lattice stabilizer code} --- Modular-qudit lattice color codes are defined analogous to qubit color codes on suitable lattices of any spatial dimension, but a directionality is required in order to make the modular-qudit stabilizers commute \NoCaseChange{\protect\cite[{Sec. III}]{cite673}}.
\item\relax
\flmRefsHyperref[eczindexfamilyrel]{code:generalized_homological_product_css}{Generalized homological-product CSS code}\end{eczvaluelist}
\codefieldsection{Children}
\begin{eczvaluelist}
\item\relax
\flmRefsHyperref[eczindexfamilyrel]{code:2d_color}{2D color code} --- Modular-qudit 2D color codes reduce to 2D color codes for \(q=2\).
\item\relax
\flmRefsHyperref[eczindexfamilyrel]{code:3d_color}{3D color code} --- Modular-qudit 3D color codes reduce to 3D color codes for \(q=2\).
\end{eczvaluelist}
\codefieldsection{Cousins}
\begin{eczvaluelist}
\item\relax
\flmRefsHyperref[eczindexfamilyrel]{code:generalized_color}{Generalized 2D color code} --- The generalized color code for \(G=\mathbb{Z}_q\) reduces to the 2D modular-qudit color code.
\item\relax
\flmRefsHyperref[eczindexfamilyrel]{code:quantum_k-orthogonal}{\(k\)-orthogonal code} --- The notion of \(k\)-orthogonality can be extended to modular-qudit codes and is known as \(k^{\star}\)-orthogonality \NoCaseChange{\protect\cite[{Def. 2}]{cite673}}. Modular-qudit lattice color codes defined on lattices in \(D\) spatial dimension whose \(X\)-type stabilizers are placed on cells of dimension \(\nu \leq D\) are \(k^{\star}\)-orthogonal for all \(k \leq \nu\) \NoCaseChange{\protect\cite[{Lemma 5}]{cite673}}.
\item\relax
\flmRefsHyperref[eczindexfamilyrel]{code:qudit_subsystem_color}{Modular-qudit subsystem color code} --- Gauge fixing a modular-qudit subsystem color code yields a modular-qudit color code, allowing a logical Hadamard gate and universal computation \NoCaseChange{\protect\cite[{Sec. VI}]{cite673}}.
\end{eczvaluelist}
\eczhbkcontributors{ \eczhuVVA }
\endeczcode

\eczcode{qudit_sign}{Modular-qudit shift-resistant code}{~\NoCaseChange{\protect\cite{cite2915}}}
\codefieldsection{Description}
Monolithic code encoding a qubit into a single modular qudit and protecting against either \(Z\)-type or \(X\)-type modular-qudit Pauli shifts.

The simplest example requires a 6-dimensional qudit.
The bit-flip version admits codewords \(|0\rangle\) and \(|3\rangle\) and corrects a single \(X\)-type shift.
The phase-flip version admits codewords 
\flmMathEnvironment{align}{}{
\begin{split}
|\overline{0}\rangle&=\frac{1}{\sqrt{6}}\sum_{j=0}^{5}|j\rangle\\|\overline{1}\rangle&=\frac{1}{\sqrt{6}}\sum_{j=0}^{5}(-1)^{j}|j\rangle\,.
\end{split}
}
Both codes are modular-qudit CSS codes with stabilizer generators \(Z^2\) and \(X^2\), respectively.

\codefieldsection{Parents}
\begin{eczvaluelist}
\item\relax
\flmRefsHyperref[eczindexfamilyrel]{code:qudit_css}{Modular-qudit CSS code}\item\relax
\flmRefsHyperref[eczindexfamilyrel]{code:single_subsystem}{Monolithic quantum code}\end{eczvaluelist}
\codefieldsection{Cousins}
\begin{eczvaluelist}
\item\relax
\flmRefsHyperref[eczindexfamilyrel]{code:qudit_gkp}{Modular-qudit GKP code} --- The modular-qudit shift-resistant code requires a smaller physical qudit dimension but protects against only one type of error \NoCaseChange{\protect\cite{cite2915}}.
\item\relax
\flmRefsHyperref[eczindexfamilyrel]{code:quantum_perfect}{Perfect quantum code} --- The modular-qudit shift-resistant code is not a block code, but it is perfect in the sense that each correctable error maps the logical space into a distinct error space \NoCaseChange{\protect\cite{cite2915}}.
\item\relax
\flmRefsHyperref[eczindexfamilyrel]{code:quantum_repetition}{Quantum repetition code} --- Both the quantum repetition and modular-qudit shift-resistant codes protect against only one type of noise.
\end{eczvaluelist}
\eczhbkcontributors{ \eczhuVVA }
\endeczcode

\eczcode{qudit_stabilizer}{Modular-qudit stabilizer code}{~\NoCaseChange{\protect\cite{cite736,cite4571}}}
\codefieldsection{Description}
An \(\llparenthesis n,K,d\rrparenthesis _q\) modular-qudit code whose logical subspace is the joint eigenspace of commuting qudit Pauli operators forming the code's stabilizer group \(\mathsf{S}\) \NoCaseChange{\protect\cite[{Sec. 3.6}]{cite736}}.
Traditionally, the logical subspace is the joint \(+1\) eigenspace, and the stabilizer group does not contain \(e^{i \phi} I\) for any \(\phi \neq 0\).
The distance \(d\) is the minimum weight of a qudit Pauli string that implements a nontrivial logical operation in the code.

A modular-qudit stabilizer code encoding an integer number of qudits (\(K=q^k\)) is denoted as \(\llbracket n,k\rrbracket _{\mathbb{Z}_q}\) or \(\llbracket n,k,d\rrbracket _{\mathbb{Z}_q}\).
For composite \(q\), such codes need not encode an integer number of qudits, with \(K=q^n/|\mathsf{S}|\) \NoCaseChange{\protect\cite{cite4543}\protect\cite[{Sec. 3.6}]{cite736}}.
This is because \(|{\mathsf{S}}|\) need not be a power of \(q\), as group generators may have different orders. 
As a result, \(\llbracket n,k,d\rrbracket \) notation is often used with non-integer \(k=\log_q K\), and the code dimension can be inferred from the prime decomposition of \(q\) \NoCaseChange{\protect\cite{cite4572}}.
\textit{Prime-qudit} stabilizer codes, where \(q=p\) for some prime \(p\), do not suffer from this issue and encode \(k\) logical qudits, with \(K=p^k\).

\begin{defterm}{Modular symplectic representation}\label{ref4573}\label{ref4562}
The single modular-qudit Pauli string \(X_{a} Z_{b}\) for \(a,b\in \mathbb{Z}_q\) is converted to the vector \((a|b)\in \mathbb{Z}_q^2\).
The multi modular-qudit version follows naturally.
\end{defterm}

Each code can be represented by a \textit{check matrix} (a.k.a. \textit{stabilizer generator matrix}) \(H=(A|B)\), where each row \((a|b)\) is the \flmRefsHyperref{ref4562}{modular symplectic representation} of a stabilizer generator. The check matrix can be brought into standard form via Gaussian elimination \NoCaseChange{\protect\cite{cite4543}}.

Modular-qudit stabilizer states can be expressed in terms of linear and quadratic functions over \(\mathbb{Z}_q^n\) \NoCaseChange{\protect\cite{cite4302}}.
They correspond to the set of states with positive Wigner functions \NoCaseChange{\protect\cite{cite4535,cite4574}} (see \NoCaseChange{\protect\cite[{Thm. 8.4}]{cite774}} for a robust version of Hudson's theorem for odd prime-dimensional qudits).
Stabilizer states saturate various uncertainty relations \NoCaseChange{\protect\cite{cite4575}}.
General modular-qudit stabilizer codes can equivalently \NoCaseChange{\protect\cite{cite3561}} be defined using graphs, yielding an analytical form for the codewords \NoCaseChange{\protect\cite{cite866}}.

There is a \flmRefsHyperref{ref1729}{quantum GV bound} for modular-qudit stabilizer codes \NoCaseChange{\protect\cite{cite4576}}.

\codefieldsection{Protection}
Detects errors on up to \(d-1\) qudits, and corrects erasure errors on up to \(d-1\) qudits. More generally, define the normalizer \(\mathsf{N(S)}\) of \(\mathsf{S}\) to be the set of all Pauli operators that commute with all \(S\in\mathsf{S}\). A stabilizer code can correct a Pauli error set \({\mathcal{E}}\) if and only if \(E^\dagger F \notin \mathsf{N(S)}\setminus \mathsf{S}\) for all \(E,F \in {\mathcal{E}}\).
\codefieldsection{Magic}
The \textit{magic-state yield parameter} \(\gamma = \log_d(n/k)\) quantifies the overhead cost of magic-state distillation per the original protocol \NoCaseChange{\protect\cite{cite690,cite691}}.
\codefieldsection{Encoding}
\begin{eczvaluelist}
\item\relax Encoder circuits for prime-qudit stabilizer codes \NoCaseChange{\protect\cite{cite4509}}.
\end{eczvaluelist}
\codefieldsection{Transversal and Permutation-Based Gates}
\begin{eczvaluelist}
\item\relax All qudit stabilizer codes realize modular qudit Pauli transformations transversally.
\end{eczvaluelist}
\codefieldsection{Gates}
\begin{eczvaluelist}
\item\relax Gates in the \flmRefsHyperref{ref751}{qudit Clifford hierarchy} can be done using \textit{qudit gate teleportation}, in which a gate can be obtained from a particular \textit{qudit magic state}. Magic states that are eigenstates of qudit Clifford operators have been classified for prime qudit dimensions 3 and 5 \NoCaseChange{\protect\cite{cite4577}}.
\end{eczvaluelist}
\codefieldsection{Decoding}
\begin{eczvaluelist}
\item\relax Trellis decoder for prime-dimensional qudits, which builds a compact representation of the algebraic structure of the normalizer \(\mathsf{N(S)}\) \NoCaseChange{\protect\cite{cite4578}}.
\end{eczvaluelist}
\codefieldsection{Notes}
\begin{eczvaluelist}
\item\relax Distance upper bounds for Galois-qudit stabilizer codes for various \(n\) and \(k\), based on algorithms developed in Refs. \NoCaseChange{\protect\cite{cite2673,cite2674}} and maintained by M. Grassl at this \flmHref{https://www.codetables.de/}{website}, hold for general modular-qudit codes because they are based on linear programming.
\item\relax A standardized definition of the qudit stabilizer group is developed in \NoCaseChange{\protect\cite{cite4543}}.
\item\relax The number of modular-qudit stabilizer codes was determined in Refs. \NoCaseChange{\protect\cite{cite4535,cite4579}}.
\end{eczvaluelist}
\codefieldsection{Parents}
\begin{eczvaluelist}
\item\relax
\flmRefsHyperref[eczindexfamilyrel]{code:qudit_non_stabilizer}{Modular-qudit USt code} --- A modular-qudit stabilizer code with stabilizer group \(\mathsf{S}\) can be thought of as a modular-qudit USt with only the identity coset representative. Conversely, if \(K = q^k\), and if the set of coset representatives of a modular-qudit USt form a \(q\)-ary linear code over \(\mathbb{Z}_q\), then they can be absorbed into a modular-qudit stabilizer group that defines the USt.
\item\relax
\flmRefsHyperref[eczindexfamilyrel]{code:stabilizer}{Stabilizer code}\item\relax
\flmRefsHyperref[eczindexfamilyrel]{code:quantum_lego}{Tensor-network code} --- Modular-qudit stabilizer codes are quantum Lego codes built out of atomic blocks such as the 2-qudit repetition code, single-qudit trivial stabilizer codes, and tensor-products of the \(|0\rangle\) state \NoCaseChange{\protect\cite{cite3101}}.
\end{eczvaluelist}
\codefieldsection{Children}
\begin{eczvaluelist}
\item\relax
\flmRefsHyperref[eczindexfamilyrel]{code:qubit_stabilizer}{Qubit stabilizer code} --- Modular-qudit stabilizer codes for \(q=2\) correspond to qubit stabilizer codes.
Modular-qudit stabilizer codes for prime-dimensional qudits \(q=p\) inherit most of the features of qubit stabilizer codes, including encoding an integer number of qudits and a \flmRefsHyperref{ref2198}{modular-qudit Pauli group} with a unique number of generators.
Conversely, qubit codes can be extended to modular-qudit codes by decorating appropriate generators with powers.
For example, \(\llbracket 4,2,2\rrbracket \) qubit code generators can be adjusted to \(ZZZZ\) and \(XX^{-1} XX^{-1}\).
A systematic procedure extending a qubit code to prime-qudit codes involves putting its generator matrix into local-dimension-invariant (LDI) form  \NoCaseChange{\protect\cite{cite4392}}.
Various bounds exist on the distance of the resulting codes \NoCaseChange{\protect\cite{cite820,cite4393}}.

\item\relax
\flmRefsHyperref[eczindexfamilyrel]{code:qudit_5_1_3}{\(\llbracket 5,1,3\rrbracket _{\mathbb{Z}_q}\) modular-qudit code}\item\relax
\flmRefsHyperref[eczindexfamilyrel]{code:stab_9_1_5}{\(\llbracket 9,1,5\rrbracket _3\) quantum Glynn code}\item\relax
\flmRefsHyperref[eczindexfamilyrel]{code:fracton}{Fracton stabilizer code}\item\relax
\flmRefsHyperref[eczindexfamilyrel]{code:frobenius}{Frobenius code}\item\relax
\flmRefsHyperref[eczindexfamilyrel]{code:qudit_cluster_state}{Modular-qudit cluster-state code} --- Modular-qudit cluster-state codes are particular modular-qudit stabilizer codes. Any modular-qudit stabilizer code is equivalent to a graph quantum code for \(G=\mathbb{Z}_q\) via a single-modular-qudit Clifford circuit \NoCaseChange{\protect\cite{cite3561}} (see also \NoCaseChange{\protect\cite{cite867}}).
\item\relax
\flmRefsHyperref[eczindexfamilyrel]{code:qudit_css}{Modular-qudit CSS code} --- Modular-qudit CSS codes are modular-qudit stabilizer codes whose stabilizer groups admit a generating set of pure-\(X\) and pure-\(Z\) Pauli strings. 
Any \(\llbracket n,k,d\rrbracket _{\mathbb{Z}_q}\) stabilizer code can be mapped onto a \(\llbracket 2n,2k,\geq d\rrbracket _{\mathbb{Z}_q}\) \flmRefsHyperref{code:two_block_quantum}{two-block CSS code} code via \flmRefsHyperref{ref436}{symplectic doubling}, which preserves geometric locality of a code up to a constant factor.

\item\relax
\flmRefsHyperref[eczindexfamilyrel]{code:3d_semion}{Chiral semion Walker-Wang model code}\item\relax
\flmRefsHyperref[eczindexfamilyrel]{code:tqd_abelian_stabilizer}{Abelian TQD stabilizer code}\end{eczvaluelist}
\codefieldsection{Cousins}
\begin{eczvaluelist}
\item\relax
\flmRefsHyperref[eczindexfamilyrel]{code:qudit_cws}{Modular-qudit CWS code} --- Modular-qudit CWS codes whose underlying classical code is a linear \(q\)-ary code over \(\mathbb{Z}_q\) are modular-qudit stabilizer codes containing a cluster-state codeword; see \NoCaseChange{\protect\cite[{Corr. 4-5}]{cite4567}}, which defines CWS codes as admitting an underlying stabilizer state that is not necessarily a cluster state.
\item\relax
\flmRefsHyperref[eczindexfamilyrel]{code:q-ary_linear_over_zq}{Linear code over \(\mathbb{Z}_q\)} --- Modular-qudit stabilizer codes are the closest quantum analogues of additive codes over \(\mathbb{Z}_q\) because addition in the ring corresponds to multiplication of stabilizers in the quantum case.
\item\relax
\flmRefsHyperref[eczindexfamilyrel]{code:t-designs}{\(t\)-design} --- Stabilizer states on \(n\) prime-dimensional qudits form complex projective 2-designs on \(\mathbb{C}P^{p^n-1}\), and they form 3-designs if and only if \(p=2\) \NoCaseChange{\protect\cite{cite937}}. The prime-qudit Clifford group is a unitary 2-design on \(U(p^n)\) \NoCaseChange{\protect\cite{cite942}}.
\item\relax
\flmRefsHyperref[eczindexfamilyrel]{code:unitary_design}{Unitary \(t\)-design} --- The prime-qudit Clifford group is a unitary 2-design on \(U(p^n)\) \NoCaseChange{\protect\cite{cite942}}.
\item\relax
\flmRefsHyperref[eczindexfamilyrel]{code:complex_projective}{Complex projective space code} --- Stabilizer states on \(n\) prime-dimensional qudits form complex projective 2-designs on \(\mathbb{C}P^{p^n-1}\) \NoCaseChange{\protect\cite{cite937}}.
\item\relax
\flmRefsHyperref[eczindexfamilyrel]{code:barnes_wall}{Barnes-Wall (BW) lattice} --- Modular-qudit stabilizer states can be mapped into the first lattice shell of a BW lattice over a cyclotomic field, while the modular-qudit Clifford group is related to the symmetry group of the lattice \NoCaseChange{\protect\cite{cite2117}}.
\item\relax
\flmRefsHyperref[eczindexfamilyrel]{code:rotor_stabilizer}{Rotor stabilizer code} --- By combining the paper's bounded-phase-space and integer-local-dimension constructions, prime-qudit stabilizer codes can be algebraically imported into rotor-code settings \NoCaseChange{\protect\cite[{Sec. 3.2}]{cite4580}}.
\item\relax
\flmRefsHyperref[eczindexfamilyrel]{code:analog_stabilizer}{Analog stabilizer code} --- Prime-qudit stabilizer codes can be transformed into analog stabilizer codes on the same number of modes and logical modes, with distance at least as large as that of the original code \NoCaseChange{\protect\cite[{Thm. 12}]{cite4580}}.
\item\relax
\flmRefsHyperref[eczindexfamilyrel]{code:majorana_stab}{Majorana stabilizer code} --- Majorana stabilizer codes can be extended to modular qudits, yielding parafermion stabilizer codes \NoCaseChange{\protect\cite{cite3984}}.
\item\relax
\flmRefsHyperref[eczindexfamilyrel]{code:qudit_subsystem_stabilizer}{Subsystem modular-qudit stabilizer code} --- Subsystem modular-qudit stabilizer codes reduce to modular-qudit stabilizer codes when there are no gauge qudits.
\item\relax
\flmRefsHyperref[eczindexfamilyrel]{code:galois_stabilizer}{Galois-qudit stabilizer code} --- Recalling that \(q=p^m\), Galois-qudit stabilizer codes can also be treated as prime-qudit stabilizer codes on \(mn\) qudits, giving \(k=nm-r\) \NoCaseChange{\protect\cite{cite696}}. The case \(m=1\) reduces to conventional prime-qudit stabilizer codes on \(n\) qudits. A modular-qudit stabilizer code with composite dimension \(q\) contains a subcode that is isomorphic to a \(p\)-dimensional prime-qudit stabilizer code for every prime factor \(p\) of \(q\), and the distance of the full stabilizer code is bounded by the distance of this subcode \NoCaseChange{\protect\cite{cite4581}}.
\end{eczvaluelist}
\eczhbkcontributors{ Leonid Pryadko, Qingfeng (Kee) Wang, \eczhuVVA }
\endeczcode

\eczcode{qudit_subsystem_color}{Modular-qudit subsystem color code}{~\NoCaseChange{\protect\cite{cite673}}}
\codefieldsection{Alternative Names}
\begin{eczvaluelist}
\item\relax Modular-qudit gauge color code
\end{eczvaluelist}
\eczhIndexCodeAliasName{qudit_subsystem_color}{Modular-qudit gauge color code}
\codefieldsection{Description}
An extension of subsystem color codes to modular qudits.
Codes are defined analogously to qubit subsystem color codes, but a directionality is required in order to make the modular-qudit stabilizers commute \NoCaseChange{\protect\cite[{Sec. VII}]{cite673}}.

\codefieldsection{Gates}
\begin{eczvaluelist}
\item\relax A logical Hadamard gate can be achieved by gauge fixing, which yields universal computation \NoCaseChange{\protect\cite[{Sec. VI}]{cite673}}.
\end{eczvaluelist}
\codefieldsection{Parents}
\begin{eczvaluelist}
\item\relax
\flmRefsHyperref[eczindexfamilyrel]{code:qudit_subsystem_css}{Subsystem modular-qudit CSS code}\item\relax
\flmRefsHyperref[eczindexfamilyrel]{code:sparse_subsystem}{QLDPC subsystem code}\end{eczvaluelist}
\codefieldsection{Child}
\begin{eczvaluelist}
\item\relax
\flmRefsHyperref[eczindexfamilyrel]{code:subsystem_color}{Subsystem color code} --- Modular-qudit subsystem color codes reduce to subsystem color codes for \(q=2\).
\end{eczvaluelist}
\codefieldsection{Cousins}
\begin{eczvaluelist}
\item\relax
\flmRefsHyperref[eczindexfamilyrel]{code:qudit_color}{Modular-qudit lattice color code} --- Gauge fixing a modular-qudit subsystem color code yields a modular-qudit color code, allowing a logical Hadamard gate and universal computation \NoCaseChange{\protect\cite[{Sec. VI}]{cite673}}.
\item\relax
\flmRefsHyperref[eczindexfamilyrel]{code:translationally_invariant_subsystem}{Lattice subsystem code} --- Modular-qudit subsystem lattice color codes are defined analogously to qubit subsystem lattice color codes on suitable lattices of any spatial dimension, but a directionality is required in order to make the modular-qudit stabilizers commute \NoCaseChange{\protect\cite[{Sec. VII}]{cite673}}.
\end{eczvaluelist}
\eczhbkcontributors{ \eczhuVVA }
\endeczcode

\eczcode{qudit_surface}{Modular-qudit surface code}{~\NoCaseChange{\protect\cite{cite423,cite71,cite424,cite4582}}}
\codefieldsection{Alternative Names}
\begin{eczvaluelist}
\item\relax \(\mathbb{Z}_q\) surface code
\end{eczvaluelist}
\eczhIndexCodeAliasName{qudit_surface}{\(\mathbb{Z}_q\) surface code}
\codefieldsection{Description}
Extension of the surface code to prime-dimensional \NoCaseChange{\protect\cite{cite423,cite424}} and more general modular qudits.
Stabilizer generators are few-body \(X\)-type and \(Z\)-type Pauli strings associated to the stars and plaquettes, respectively, of a tessellation of a two-dimensional surface.
Since qudits have more than one \(X\) and \(Z\)-type operator, various sets of stabilizer generators can be defined.

Ground-state degeneracy and the associated phase depend on the qudit dimension and the stabilizer generators.
In particular, the torus ground-state degeneracy can depend on the lattice periods, and the model can realize either topological order or symmetry-protected topological order \NoCaseChange{\protect\cite{cite4582,cite4583}}.

More generally, the \(\mathbb{Z}_q\) toric-code Hamiltonian admits rotor limits related to compact \(U(1)\) lattice gauge theory and a continuum coupled-oscillator limit whose ground states coincide with those of the CV surface code \NoCaseChange{\protect\cite{cite2531}}.

\codefieldsection{Protection}
The code distance is the minimum weight of a non-trivial homology or cohomology class, i.e., the length of the shortest non-contractible cycle in the cellulation or in its dual \NoCaseChange{\protect\cite{cite71}}.
Planar discs with holes yield local modular-qudit surface-code families whose logical operators are described by relative homology classes \NoCaseChange{\protect\cite{cite71}}.

\codefieldsection{Gates}
\begin{eczvaluelist}
\item\relax A magic-state preparation routine for the \(\mathbb{Z}_4\) surface code traverses through the \(D_4\) quantum double model \NoCaseChange{\protect\cite{cite4584}}.
\item\relax Gates can be implemented through topological operations corresponding to elements of the mapping class group, which is generated by Dehn-twists along non-contractible cycles \NoCaseChange{\protect\cite{cite3839}}.
\end{eczvaluelist}
\codefieldsection{Decoding}
\begin{eczvaluelist}
\item\relax Renormalization group decoder \NoCaseChange{\protect\cite{cite4585,cite3886}}.
\end{eczvaluelist}
\codefieldsection{Realizations}
\begin{eczvaluelist}
\item\relax State preparation, anyon creation, anyon fusion, and transfer of entanglement between anyons and defects in a 24-qutrit trapped-ion device by Quantinuum \NoCaseChange{\protect\cite{cite4586}}.
\end{eczvaluelist}
\codefieldsection{Notes}
\begin{eczvaluelist}
\item\relax The simplest \flmHref{https://web.archive.org/web/20161223121819/http://citizensciencegames.com/games/decodoku/}{Decodoku game} is based on the qudit surface code with \( q=10\). See related \flmHref{https://github.com/quantumjim/qec_lectures}{Qiskit tutorial}.
\end{eczvaluelist}
\codefieldsection{Parents}
\begin{eczvaluelist}
\item\relax
\flmRefsHyperref[eczindexfamilyrel]{code:qudit_css}{Modular-qudit CSS code}\item\relax
\flmRefsHyperref[eczindexfamilyrel]{code:quantum_double_abelian}{Abelian quantum-double stabilizer code} --- Modular-qudit surface code Hamiltonians admit topological phases associated with \(\mathbb{Z}_q\) topological order \NoCaseChange{\protect\cite{cite424}}.
\item\relax
\flmRefsHyperref[eczindexfamilyrel]{code:generalized_homological_product_css}{Generalized homological-product CSS code}\end{eczvaluelist}
\codefieldsection{Child}
\begin{eczvaluelist}
\item\relax
\flmRefsHyperref[eczindexfamilyrel]{code:surface}{Kitaev surface code} --- The modular-qudit surface code for \(q=2\) reduces to the surface code.
\end{eczvaluelist}
\codefieldsection{Cousins}
\begin{eczvaluelist}
\item\relax
\flmRefsHyperref[eczindexfamilyrel]{code:quantum_double_dihedral}{Dihedral \(G=D_m\) quantum-double code} --- The \(D_3\) quantum double model can be obtained by gauging \NoCaseChange{\protect\cite{cite462,cite463,cite233,cite464,cite465,cite466,cite467,cite468,cite469,cite470}} the charge conjugation symmetry of the \(\mathbb{Z}_3\) surface code \NoCaseChange{\protect\cite{cite3071}}. A magic-state preparation routine for the \(\mathbb{Z}_4\) surface code traverses through the \(D_4\) quantum double model \NoCaseChange{\protect\cite{cite4584}}.
\item\relax
\flmRefsHyperref[eczindexfamilyrel]{code:hopf_quantum_double}{Hopf-algebra quantum-double code} --- The modular-qudit surface code can be generalized to a Hopf-algebra quantum-double code whose ground states remain the same but whose excitations are based on quasitriangular semisimple Hopf algebras of \(\mathbb{Z}_q\) \NoCaseChange{\protect\cite{cite4587}}.
\item\relax
\flmRefsHyperref[eczindexfamilyrel]{code:compactified_r}{Compactified \(\mathbb{R}\) gauge theory code} --- The compactified \(\mathbb{R}\) gauge theory code can be thought of as a realization of the \(q\to\infty\) \(U(1)\) rotor limit \NoCaseChange{\protect\cite{cite2531}} of the qudit surface code as a bosonic stabilizer code.
\item\relax
\flmRefsHyperref[eczindexfamilyrel]{code:analog_surface}{Analog surface code} --- The analog surface code can be thought of as a realization of the \(q\to\infty\) \(\mathbb{R}\) oscillator limit \NoCaseChange{\protect\cite{cite2531}} of the qudit surface code as a bosonic stabilizer code.
\item\relax
\flmRefsHyperref[eczindexfamilyrel]{code:tiger_surface}{Tiger surface code} --- The tiger surface code can be thought of as a realization of the \(q\to\infty\) \(U(1)\) rotor limit \NoCaseChange{\protect\cite{cite2531}} of the qudit surface code as a tiger code.
\item\relax
\flmRefsHyperref[eczindexfamilyrel]{code:twist_defect_surface}{Twist-defect surface code} --- Twist-defect surface codes have been extended to prime-dimensional qudits \NoCaseChange{\protect\cite{cite4491}}.
\item\relax
\flmRefsHyperref[eczindexfamilyrel]{code:qudit_xcube}{Qudit X-cube model code} --- A field-theoretic description of the qudit X-cube model can be obtained by coupling layers of 2D \(\mathbb{Z}_q\) gauge theory \NoCaseChange{\protect\cite{cite568}}.
\item\relax
\flmRefsHyperref[eczindexfamilyrel]{code:double_semion}{Double-semion stabilizer code} --- The exchange statistics of the anyon for the double-semion code coincides with a subset of anyons in the \(\mathbb{Z}_4\) surface code, but the fusion rules are different. The double-semion code can be obtained from the \(\mathbb{Z}_4\) surface code by \flmRefsHyperref{ref410}{condensing} the anyon \(e^2 m^2\) \NoCaseChange{\protect\cite{cite414}} or by gauging \NoCaseChange{\protect\cite{cite462,cite463,cite233,cite464,cite465,cite466,cite467,cite468,cite469,cite470}} the one-form symmetry associated with said anyon \NoCaseChange{\protect\cite[{Footnote 20}]{cite414}}.
\item\relax
\flmRefsHyperref[eczindexfamilyrel]{code:qudit_znone}{\(\mathbb{Z}_q^{(1)}\) subsystem code} --- The \(\mathbb{Z}_q^{(1)}\) subsystem code can be obtained from the \(\mathbb{Z}_q\) square-lattice surface code by \flmRefsHyperref{ref666}{gauging out} the anyon \(e^{-1} m\) and applying transversal Clifford gates \NoCaseChange{\protect\cite[{Sec. 7.3}]{cite414}}. During this process, the square lattice is effectively expanded to a honeycomb tiling \NoCaseChange{\protect\cite[{Fig. 12}]{cite414}}.
\end{eczvaluelist}
\eczhbkcontributors{ Ian Teixeira, \eczhuVVA }
\endeczcode

\eczcode{qudit_non_stabilizer}{Modular-qudit USt code}{~\NoCaseChange{\protect\cite{cite3583,cite3044}}}
\codefieldsection{Description}
A modular-qudit code whose codespace consists of a direct sum of a modular-qudit stabilizer codespace and one or more of that stabilizer code's error spaces.

Given a subset \(T\) of coset representatives of \(\mathsf{N}(\mathsf{S})/\mathsf{S}\) of a modular-qudit stabilizer code \(\llparenthesis n,K\rrparenthesis \) with codespace \(\mathsf{C}\) and stabilizer group \(\mathsf{S}\), one can construct the modular-qudit USt with codespace
\flmMathEnvironment{align}{}{
  \mathsf{C}_{\text{USt}}=\bigoplus_{t\in T}t\mathsf{C}~.
}
The parameters of the USt are \(\llparenthesis n,K|T|,d\rrparenthesis \), where \(|T|\) is the number of chosen coset representatives.
A modular-qudit USt is \textit{CSS-like} when the underlying stabilizer code is CSS.

\codefieldsection{Parent}
\begin{eczvaluelist}
\item\relax
\flmRefsHyperref[eczindexfamilyrel]{code:qudits_into_qudits}{Modular-qudit code}\end{eczvaluelist}
\codefieldsection{Children}
\begin{eczvaluelist}
\item\relax
\flmRefsHyperref[eczindexfamilyrel]{code:non_stabilizer}{Union stabilizer (USt) code} --- Modular-qudit union stabilizer codes reduce to union stabilizer codes for \(q=2\).
\item\relax
\flmRefsHyperref[eczindexfamilyrel]{code:qudit_cws}{Modular-qudit CWS code} --- Any modular-qudit CWS code can be written as a modular-qudit USt whose (\(K=1\)) stabilizer code is the modular-qudit cluster state and whose coset representatives are constructed from the \(q\)-ary classical code over \(\mathbb{Z}_q\). Prime-dimensional modular-qudit CWS codes have a unique representation as USt codes \NoCaseChange{\protect\cite{cite4566}}. Conversely, modular-qudit USt codes are equivalent to modular-qudit CWS codes via a single-modular-qudit Clifford circuit as follows \NoCaseChange{\protect\cite{cite3585,cite3587}\protect\cite[{Sec. 10.4}]{cite3167}}. The set of coset representatives of any modular-qudit USt can be extended to a larger set iterating over the underlying stabilizer code such that all codewords can be obtained from a single stabilizer state. Then, one can apply a single-qudit Clifford transformation to map said modular-qudit stabilizer state into a modular-qudit cluster state.
\item\relax
\flmRefsHyperref[eczindexfamilyrel]{code:qudit_stabilizer}{Modular-qudit stabilizer code} --- A modular-qudit stabilizer code with stabilizer group \(\mathsf{S}\) can be thought of as a modular-qudit USt with only the identity coset representative. Conversely, if \(K = q^k\), and if the set of coset representatives of a modular-qudit USt form a \(q\)-ary linear code over \(\mathbb{Z}_q\), then they can be absorbed into a modular-qudit stabilizer group that defines the USt.
\end{eczvaluelist}
\eczhbkcontributors{ \eczhuVVA }
\endeczcode

\eczcode{qudit_reed_muller}{Prime-qudit RM code}{~\NoCaseChange{\protect\cite{cite828,cite692,cite752}}}
\codefieldsection{Description}
Modular-qudit stabilizer code constructed from GRM codes or their duals via the modular-qudit CSS construction.

For prime local dimension \(q\), CSS constructions from \(\mathrm{GRM}_q(\nu_1,m) \subseteq \mathrm{GRM}_q(\nu_2,m)\) yield \flmRefsHyperref{ref672}{pure} \(\llbracket q^m,k(\nu_2)-k(\nu_1),\min\{d(\nu_1^\perp),d(\nu_2)\}\rrbracket _q\) codes \NoCaseChange{\protect\cite{cite828}}.
The special case \(m=1\) gives quantum MDS codes \(\llbracket q,q-2\nu-2,\nu+2\rrbracket _q\) for \(0 \leq \nu \leq (q-2)/2\) \NoCaseChange{\protect\cite{cite828}}.
An odd-prime-qudit CSS code family constructed from first-order punctured GRM codes transversally implements a diagonal gate at any level of the \flmRefsHyperref{ref751}{qudit Clifford hierarchy} \NoCaseChange{\protect\cite{cite692,cite752}}.

\codefieldsection{Rate}
For the CSS family from \(\mathrm{GRM}_q(\nu_1,m) \subseteq \mathrm{GRM}_q(\nu_2,m)\), the rate is \(( k(\nu_2)-k(\nu_1) )/q^m\) \NoCaseChange{\protect\cite{cite828}}.
\codefieldsection{Magic}
An odd-prime-qudit CSS code family constructed from first-order punctured GRM codes can be used for qudit magic-state distillation; see \NoCaseChange{\protect\cite[{Table I}]{cite692}} for yields.
\codefieldsection{Transversal and Permutation-Based Gates}
\begin{eczvaluelist}
\item\relax An odd-prime-qudit CSS code family constructed from first-order punctured GRM codes transversally implements a diagonal gate at any level of the \flmRefsHyperref{ref751}{qudit Clifford hierarchy} \NoCaseChange{\protect\cite{cite692,cite752}}.
\end{eczvaluelist}
\codefieldsection{Parents}
\begin{eczvaluelist}
\item\relax
\flmRefsHyperref[eczindexfamilyrel]{code:qudit_css}{Modular-qudit CSS code}\item\relax
\flmRefsHyperref[eczindexfamilyrel]{code:galois_reed_muller}{Galois-qudit quantum RM code} --- Galois-qudit RM codes reduce to prime-qudit RM codes when \(q\) is prime.
\end{eczvaluelist}
\codefieldsection{Children}
\begin{eczvaluelist}
\item\relax
\flmRefsHyperref[eczindexfamilyrel]{code:quantum_reed_muller}{Quantum Reed-Muller (RM) code} --- Prime-qudit RM codes reduce to quantum RM codes when \(q=p=2\).
\item\relax
\flmRefsHyperref[eczindexfamilyrel]{code:qudit_hamming_css}{\(\llbracket 2^r-1, 2^r-2r-1, 3\rrbracket _p\) quantum Hamming code} --- The \(\llbracket 2^r-1, 2^r-2r-1, 3\rrbracket _p\) quantum Hamming code family extends the qubit quantum Hamming family to prime qudits using local-dimension-invariant representations \NoCaseChange{\protect\cite{cite820}}.
\end{eczvaluelist}
\eczhbkcontributors{ \eczhuVVA }
\endeczcode

\eczcode{polynomial}{Prime-qudit RS code}{~\NoCaseChange{\protect\cite{cite4588}}}
\codefieldsection{Alternative Names}
\begin{eczvaluelist}
\item\relax Prime-qudit polynomial code (QPyC)
\end{eczvaluelist}
\eczhIndexCodeAliasName{polynomial}{Prime-qudit polynomial code (QPyC)}
\codefieldsection{Description}
Prime-qudit CSS code constructed using two RS codes.

The original construction \NoCaseChange{\protect\cite{cite4588}} was for a qubit code (\(p=2\)) by using a basis for a larger Galois field over \(\mathbb{F}_2\), yielding an \(\llbracket kN,k(N-2K),K+1\rrbracket \) qubit code from a \([N,K,\delta]_{2^k}\) RS code with \(N=2^k-1\) and \(K=N-\delta+1\).

An alternative construction \NoCaseChange{\protect\cite{cite398}} yields an \(\llbracket n,k,d\rrbracket _{p>n}\) prime-qudit CSS code with \(d=\min(n-g,g+2-k)\) that is constructed using two RS codes over \(\mathbb{F}_p=\mathbb{Z}_p\).
Let \(\{\alpha_1,\cdots,\alpha_n\}\) be \(n\) distinct nonzero elements of \(\mathbb{Z}_p\), and let \(g\) be a number satisfying \(0\leq k \leq g < n\). Then, define degree-\(g\) polynomials
\flmMathEnvironment{align}{}{
  f_{\mu\cup c}\left(x\right)=\mu_{0}+\mu_{1}x+\cdots+\mu_{k-1}x^{k-1}+c_{k}x^{k}+\cdots+c_{g}x^{g}\,,
}
where the first \(k\) coefficients are indexed by the coefficient vector \(\mu\in\mathbb{Z}_p^{ k}\), and the remaining coefficients are indexed by the vector \(c\in\mathbb{Z}_p^{ (g+1-k)}\).
Logical states, labeled by \(\mu\), are superpositions of canonical basis states whose \(i\)th entry is \(f_{\mu\cup c}\) evaluated at \(\alpha_i\), summed over all possible vectors \(c\),
\flmMathEnvironment{align}{}{
  |\overline{\mu}\rangle=\sum_{c\in\mathbb{Z}_{p}^{(g+1-k)}}|f_{\mu\cup c}(\alpha_{1}),f_{\mu\cup c}(\alpha_{2}),\cdots,f_{\mu\cup c}(\alpha_{n})\rangle.
}

\codefieldsection{Magic}
Triorthogonal \(p\)-dimensional prime-qudit RS codes achieve a magic-state yield parameter \(\gamma = O(1/\log p)\) \NoCaseChange{\protect\cite{cite693}}.
\codefieldsection{Parents}
\begin{eczvaluelist}
\item\relax
\flmRefsHyperref[eczindexfamilyrel]{code:qudit_css}{Modular-qudit CSS code}\item\relax
\flmRefsHyperref[eczindexfamilyrel]{code:galois_polynomial}{Galois-qudit RS code} --- Galois-qudit RS codes for prime-dimensional qudits are prime-qudit RS codes.
\end{eczvaluelist}
\codefieldsection{Child}
\begin{eczvaluelist}
\item\relax
\flmRefsHyperref[eczindexfamilyrel]{code:stab_3_1_2}{\(\llbracket 3,1,2\rrbracket _3\) Three-qutrit code} --- The three-qutrit code is the smallest member of a family of \(\llbracket 2m-1,1,m\rrbracket _{p}\) prime-qudit quantum RS codes for \(p=3\) and \(m=2\) \NoCaseChange{\protect\cite{cite4508}}.
\end{eczvaluelist}
\codefieldsection{Cousin}
\begin{eczvaluelist}
\item\relax
\flmRefsHyperref[eczindexfamilyrel]{code:qudit_triorthogonal}{Prime-qudit triorthogonal code} --- Triorthogonal \(p\)-dimensional prime-qudit RS codes achieve a magic-state yield parameter \(\gamma = O(1/\log p)\) \NoCaseChange{\protect\cite{cite693}}.
\end{eczvaluelist}
\eczhbkcontributors{ Qingfeng (Kee) Wang, Manasi Shingane, \eczhuVVA }
\endeczcode

\eczcode{qudit_triorthogonal}{Prime-qudit triorthogonal code}{~\NoCaseChange{\protect\cite{cite693}}}
\codefieldsection{Description}
An \(m \times n\) matrix over \(\mathbb{F}_p=\mathbb{Z}_p\) is triorthogonal if its rows \(r_1, \ldots, r_m\) satisfy \(|r_i \cdot r_j| = 0\) and \(|r_i \cdot r_j \cdot r_k| = 0\) modulo \(p\), where addition and multiplication are done on \(\mathbb{F}_p\).
The triorthogonal prime-qudit CSS code associated with the matrix is constructed by mapping nonzero entries in self-orthogonal rows to \(X\) operators, and \(Z\) operators for each row in the orthogonal complement \NoCaseChange{\protect\cite{cite693,cite709}}.

\codefieldsection{Transversal and Permutation-Based Gates}
\begin{eczvaluelist}
\item\relax Admits a transversal gate from the third level of the \flmRefsHyperref{ref751}{qudit Clifford hierarchy} \NoCaseChange{\protect\cite{cite693}}.
\end{eczvaluelist}
\codefieldsection{Parents}
\begin{eczvaluelist}
\item\relax
\flmRefsHyperref[eczindexfamilyrel]{code:qudit_css}{Modular-qudit CSS code}\item\relax
\flmRefsHyperref[eczindexfamilyrel]{code:galois_css}{Galois-qudit CSS code}\end{eczvaluelist}
\codefieldsection{Children}
\begin{eczvaluelist}
\item\relax
\flmRefsHyperref[eczindexfamilyrel]{code:quantum_triorthogonal}{Triorthogonal code} --- Prime-qudit triorthogonal codes reduce to triorthogonal codes when \(p=2\).
\item\relax
\flmRefsHyperref[eczindexfamilyrel]{code:qutrit_small_triorthogonal}{\(\llbracket 9m-k,k,2\rrbracket _3\) triorthogonal code}\end{eczvaluelist}
\codefieldsection{Cousin}
\begin{eczvaluelist}
\item\relax
\flmRefsHyperref[eczindexfamilyrel]{code:polynomial}{Prime-qudit RS code} --- Triorthogonal \(p\)-dimensional prime-qudit RS codes achieve a magic-state yield parameter \(\gamma = O(1/\log p)\) \NoCaseChange{\protect\cite{cite693}}.
\end{eczvaluelist}
\eczhbkcontributors{ \eczhuVVA }
\endeczcode

\eczcode{qudit_cubic}{Qudit cubic code}{~\NoCaseChange{\protect\cite{cite4589,cite4590,cite2531,cite4519}}}
\codefieldsection{Description}
Generalization of the Haah cubic code to modular qudits.

\codefieldsection{Protection}
Performance over the erasure and depolarizing channels was studied in Ref. \NoCaseChange{\protect\cite{cite4591}}.

\codefieldsection{Parent}
\begin{eczvaluelist}
\item\relax
\flmRefsHyperref[eczindexfamilyrel]{code:fracton}{Fracton stabilizer code} --- Haah cubic \NoCaseChange{\protect\cite{cite3032}} codes 1-4, 7, 8, and 10 do not have string logical operators and are the first examples of Type-II fracton phases. The remaining cubic codes are fractal Type-I fracton codes \NoCaseChange{\protect\cite{cite456,cite4518}}. The qutrit models in \NoCaseChange{\protect\cite[{Eqs. (D11-D12)}]{cite456}} are likely Type-II, with no string operators found numerically up to width 20, while the \(q=5\) qudit model in \NoCaseChange{\protect\cite[{Eq. (D13)}]{cite456}} satisfies a proven no-string condition and is Type-II.
\end{eczvaluelist}
\codefieldsection{Child}
\begin{eczvaluelist}
\item\relax
\flmRefsHyperref[eczindexfamilyrel]{code:haah_cubic}{Haah cubic code (CC)}\end{eczvaluelist}
\codefieldsection{Cousins}
\begin{eczvaluelist}
\item\relax
\flmRefsHyperref[eczindexfamilyrel]{code:homological_rotor}{Homological rotor code} --- The qudit cubic code can be generalized to rotors \NoCaseChange{\protect\cite{cite4590,cite2531}}.
\item\relax
\flmRefsHyperref[eczindexfamilyrel]{code:analog_stabilizer}{Analog stabilizer code} --- The qudit cubic code can be generalized to oscillators \NoCaseChange{\protect\cite{cite2531}}.
\end{eczvaluelist}
\eczhbkcontributors{ \eczhuVVA }
\endeczcode

\eczcode{qudit_xcube}{Qudit X-cube model code}{~\NoCaseChange{\protect\cite{cite4519}}}
\codefieldsection{Description}
Generalization of the X-cube model code to modular qudits.

\codefieldsection{Parents}
\begin{eczvaluelist}
\item\relax
\flmRefsHyperref[eczindexfamilyrel]{code:qudit_css}{Modular-qudit CSS code}\item\relax
\flmRefsHyperref[eczindexfamilyrel]{code:fracton}{Fracton stabilizer code}\item\relax
\flmRefsHyperref[eczindexfamilyrel]{code:cage_net}{Cage-net code} --- A field-theoretic description of the qudit X-cube model can be obtained by coupling layers of 2D \(\mathbb{Z}_q\) gauge theory \NoCaseChange{\protect\cite{cite568}}. For three orthogonal foliations with \(\mathbb{Z}_q\) layers, the string-membrane-net model is equivalent to the \(\mathbb{Z}_q\) X-cube model \NoCaseChange{\protect\cite{cite569}}. String-membrane-net models are phase-equivalent to cage-net models under generalized local unitaries \NoCaseChange{\protect\cite{cite568}}.
\end{eczvaluelist}
\codefieldsection{Child}
\begin{eczvaluelist}
\item\relax
\flmRefsHyperref[eczindexfamilyrel]{code:xcube}{X-cube model code} --- The qudit X-cube model code reduces to the X-cube model code for \(q=2\). The X-cube model is a foliated type-I fracton code \NoCaseChange{\protect\cite{cite4498,cite456}}.
\end{eczvaluelist}
\codefieldsection{Cousin}
\begin{eczvaluelist}
\item\relax
\flmRefsHyperref[eczindexfamilyrel]{code:qudit_surface}{Modular-qudit surface code} --- A field-theoretic description of the qudit X-cube model can be obtained by coupling layers of 2D \(\mathbb{Z}_q\) gauge theory \NoCaseChange{\protect\cite{cite568}}.
\end{eczvaluelist}
\eczhbkcontributors{ \eczhuVVA }
\endeczcode

\eczcode{subsystem_qudits_into_qudits}{Subsystem modular-qudit code}{}
\codefieldsection{Alternative Names}
\begin{eczvaluelist}
\item\relax Gauge modular-qudit code
\end{eczvaluelist}
\eczhIndexCodeAliasName{subsystem_qudits_into_qudits}{Gauge modular-qudit code}

\codefieldsection{Kingdom root code}
for the \flmRefsHyperref{kingdom:qudits_into_qudits}{Modular-qudit Kingdom}.
\codefieldsection{Description}
Subsystem QECC encoding into a \(q^n\)-dimensional Hilbert space consisting of \(n\) modular qudits.

\codefieldsection{Parent}
\begin{eczvaluelist}
\item\relax
\flmRefsHyperref[eczindexfamilyrel]{code:subsystem_group_quantum}{Subsystem group-based quantum code} --- Subsystem group quantum codes whose physical spaces are constructed using modular-integer groups \(\mathbb{Z}_q\) are subsystem modular-qudit codes.
\end{eczvaluelist}
\codefieldsection{Children}
\begin{eczvaluelist}
\item\relax
\flmRefsHyperref[eczindexfamilyrel]{code:subsystem_qubits_into_qubits}{Subsystem qubit code} --- Subsystem modular-qudit codes reduce to subsystem qubit codes for qudit dimension \(q=2\).
\item\relax
\flmRefsHyperref[eczindexfamilyrel]{code:qudit_subsystem_stabilizer}{Subsystem modular-qudit stabilizer code}\end{eczvaluelist}
\codefieldsection{Cousin}
\begin{eczvaluelist}
\item\relax
\flmRefsHyperref[eczindexfamilyrel]{code:qudits_into_qudits}{Modular-qudit code} --- Subsystem modular-qudit codes reduce to (subspace) modular-qudit codes when there is no gauge subsystem.
\end{eczvaluelist}
\eczhbkcontributors{ \eczhuVVA }
\endeczcode

\eczcode{qudit_subsystem_css}{Subsystem modular-qudit CSS code}{}
\codefieldsection{Description}
Modular-qudit subsystem stabilizer code which admits a set of gauge-group generators which consist of either all-\(Z\) or all-\(X\) modular-qudit Pauli strings.
This ensures that the code's stabilizer group is also CSS.

The gauge group generators can be expressed as a matrix using the symplectic reprensetation. This matrix is of the form
\flmMathEnvironment{align}{}{
H=\begin{pmatrix}0 & H_{Z}\\
H_{X} & 0
\end{pmatrix}~.\label{ref4592}
}
The two matrix blocks, \(H_{Z}\) and \(H_X\), correspond to the parity-check matrices of two \flmRefsHyperref{code:q-ary_linear}{\(q\)-ary linear codes}, an \([n,k_X,d_X]_q\) code \(C_X\) and \([n,k_Z,d_Z]_q\) code \(C_Z\), respectively.
For prime-dimensional qudits, code parameters and code basis states have been expressed in terms of only data associated with these two classical codes \NoCaseChange{\protect\cite{cite1742,cite4432}}.

\begin{defterm}{Symplectic doubling}\label{ref4593}\label{ref436}
Any \(\llbracket n,k,r,d\rrbracket _{\mathbb{Z}_q}\) subsystem stabilizer code can be mapped onto a \(\llbracket 2n,2k,2r,\geq d\rrbracket _{\mathbb{Z}_q}\) subsystem CSS code, with the mapping preserving geometric locality of a code up to a constant factor \NoCaseChange{\protect\cite{cite4432}} (see also \NoCaseChange{\protect\cite{cite1432}\protect\cite[{Thm. 1}]{cite439}}).
In the \flmRefsHyperref{ref4562}{modular symplectic representation}, the gauge-group generator matrix of the former is mapped into that of latter as follows,
\flmMathEnvironment{align}{}{
  \begin{pmatrix}G_{X} & G_{Z}\end{pmatrix}
  \to
  \begin{pmatrix}
  0     &     0 & G_{Z} & -G_{X}\\
  G_{X} & G_{Z} &     0 &      0
  \end{pmatrix}~,
}
where the first two columns of the latter matrix correspond to the \(X\)-type part of the gauge-group generator matrix of the output subsystem CSS code.
In the case of a stabilizer code, the stabilizer generator matrix is mapped instead to yield a \flmRefsHyperref{code:two_block_quantum}{two-block CSS code} (see \NoCaseChange{\protect\cite[{Thm. 1}]{cite439}} for the case of qubit stabilizer codes).
For geometrically local 2D stabilizer codes with twist defects, this mapping yields a twisted double cover of the underlying qudit geometry \NoCaseChange{\protect\cite{cite435}}.
\end{defterm}

\codefieldsection{Decoding}
\begin{eczvaluelist}
\item\relax Steane-type decoder utilizing data from the underlying classical codes \NoCaseChange{\protect\cite{cite4432}}.
\end{eczvaluelist}
\codefieldsection{Parents}
\begin{eczvaluelist}
\item\relax
\flmRefsHyperref[eczindexfamilyrel]{code:qudit_subsystem_stabilizer}{Subsystem modular-qudit stabilizer code} --- Subsystem modular-qudit CSS codes are subsystem modular-qudit stabilizer codes whose gauge groups admit a generating set of pure-\(X\) and pure-\(Z\) Pauli strings. Any \(\llbracket n,k,r,d\rrbracket _{\mathbb{Z}_q}\) subsystem stabilizer code can be mapped onto a \(\llbracket 2n,2k,2r,\geq d\rrbracket _{\mathbb{Z}_q}\) subsystem CSS code via \flmRefsHyperref{ref436}{symplectic doubling}, which preserves geometric locality of a code up to a constant factor. Every subsystem prime-qudit stabilizer code can be constructed from two nested subsystem prime-qudit CSS codes satisfying certain constraints \NoCaseChange{\protect\cite{cite4432}}.
\item\relax
\flmRefsHyperref[eczindexfamilyrel]{code:subsystem_css}{Subsystem CSS code}\end{eczvaluelist}
\codefieldsection{Children}
\begin{eczvaluelist}
\item\relax
\flmRefsHyperref[eczindexfamilyrel]{code:qubit_subsystem_css}{Subsystem qubit CSS code} --- Subsystem modular-qudit CSS codes reduce to subsystem qubit CSS codes for \(q=2\).
\item\relax
\flmRefsHyperref[eczindexfamilyrel]{code:qudit_subsystem_color}{Modular-qudit subsystem color code}\end{eczvaluelist}
\codefieldsection{Cousin}
\begin{eczvaluelist}
\item\relax
\flmRefsHyperref[eczindexfamilyrel]{code:qudit_css}{Modular-qudit CSS code} --- Subsystem modular-qudit CSS codes reduce to (subspace) modular-qudit CSS codes when there is no gauge subsystem.
\end{eczvaluelist}
\eczhbkcontributors{ \eczhuVVA }
\endeczcode

\eczcode{qudit_subsystem_stabilizer}{Subsystem modular-qudit stabilizer code}{}
\codefieldsection{Alternative Names}
\begin{eczvaluelist}
\item\relax Gauge modular-qudit stabilizer code
\end{eczvaluelist}
\eczhIndexCodeAliasName{qudit_subsystem_stabilizer}{Gauge modular-qudit stabilizer code}
\codefieldsection{Description}
Modular-qudit generalization of a subsystem qubit stabilizer code.
Can be obtained by taking a modular-qudit stabilizer code and assigning some of its logical qudits to be gauge qudits.
For composite qudit dimensions, such codes need not encode an integer number of qudits.

Subsystem stabilizer codes are defined by a gauge group \(\mathsf{G}\) and a stabilizer group \(\mathsf{S}\), both subgroups of the \(n\)-\flmRefsHyperref{ref2198}{modular-qudit Pauli group} \(\mathsf{P}_n\) that satisfy \(\mathsf{Z}(\mathsf{G})=\mathsf{S}\), where \(\mathsf{Z}\) denotes taking the center of a group.

A code can be constructed by starting with either group.
Given an \(\mathsf{S}\), one can pick any \(\mathsf{G}\) satisfying \(\mathsf{S}\subseteq\mathsf{G}\subseteq\mathsf{N(S)}\), where \(\mathsf{N(S)}\) is the normalizer of the stabilizer group within \(\mathsf{P}_n\).
Alternatively, given a \(\mathsf{G}\), one defines \(\mathsf{S}\) to be the center of the gauge group.

The logical Pauli group is \(\mathsf{N}(\mathsf{G})/\mathsf{S}\).
As such, the case when \(\mathsf{G}=\mathsf{S}\) reduces to an ordinary stabilizer code, while the case \(\mathsf{G}=\mathsf{N(S)}\) reduces to a trivial code.

One can gauge fix \NoCaseChange{\protect\cite{cite3384}} an Abelian subgroup of the gauge group by adding it to the stabilizer group.
\begin{defterm}{Gauge fixing}\label{ref4594}\label{ref4595}
Gauge fixing is a map between subsystem codes that is done using an Abelian subgroup \(\mathsf{F}\subseteq\mathsf{G}\),
\flmMathEnvironment{align}{}{
\begin{split}
  \mathsf{S}&\to\left\langle \mathsf{S},\mathsf{F}\right\rangle \\
  \mathsf{G}&\to\mathsf{N}_{\mathsf{G}}\left(\mathsf{F}\right)~,
\end{split}
}
where \(\mathsf{N}_{\mathsf{G}}\left(\mathsf{F}\right)\) is the normalizer of \(\mathsf{F}\) within \(\mathsf{G}\).
\end{defterm}
Gauge fixing can be used to switch between different stabilizer codes that yield different gauge sets in a process known as \textit{gauge switching}.
Gauge fixing also encompasses lattice surgery and code deformation \NoCaseChange{\protect\cite{cite4596}}.

One can also gauge out a subgroup \(\mathsf{F}\) of the \flmRefsHyperref{ref2198}{modular-qudit Pauli group} by adding it to the gauge group.
\begin{defterm}{Gauging out}\label{ref4597}\label{ref666}
Gauging out is a map between subsystem codes that is done using a subgroup \(\mathsf{F}\subseteq\mathsf{P}_n\),
\flmMathEnvironment{align}{}{
\begin{split}
  \mathsf{S}&\to\mathsf{Z}\left(\left\langle \mathsf{G},\mathsf{F}\right\rangle \right)\\
  \mathsf{G}&\to\left\langle \mathsf{G},\mathsf{F}\right\rangle ~.
\end{split}
}
The stabilizer group of the output subsystem code is a subgroup of that of the input code, \(\mathsf{Z}\left(\left\langle \mathsf{G},\mathsf{F}\right\rangle \right)\subseteq\mathsf{Z}\left(\mathsf{G}\right)\).
When \(\mathsf{F}\) is a subgroup of the logical Pauli group, this is also called \textit{gauging}.
If \(\mathsf{F}\) is itself a Pauli group of \(m\) logical qudits of the original subsystem code, then gauging out those qudits is equivalent to treating them as gauge qubits.
Gauging out should not be confused with \textit{gauging} (or ungauging) symmetries \NoCaseChange{\protect\cite{cite462,cite465,cite466,cite467}}, a different process rooted in gauge theory which can be done to stabilizer or subsystem codes and which can change \(n\).
\end{defterm}

\codefieldsection{Decoding}
\begin{eczvaluelist}
\item\relax Syndrome measurements are obtained by first measuring gauge operators of the code and taking their products, which give the stabilizer measurement outcomes. The order in which gauge operators are measured is important since they do not commute. There is a necessary and sufficient condition for inferring the stabilizer syndrome from the measurements of the gauge generators \NoCaseChange{\protect\cite[{Appendix}]{cite594}}.
\item\relax Decoder for certain geometrically local subsystem codes from hypergraphs \NoCaseChange{\protect\cite{cite661}}.
\end{eczvaluelist}
\codefieldsection{Parents}
\begin{eczvaluelist}
\item\relax
\flmRefsHyperref[eczindexfamilyrel]{code:subsystem_qudits_into_qudits}{Subsystem modular-qudit code}\item\relax
\flmRefsHyperref[eczindexfamilyrel]{code:subsystem_stabilizer}{Subsystem stabilizer code}\end{eczvaluelist}
\codefieldsection{Children}
\begin{eczvaluelist}
\item\relax
\flmRefsHyperref[eczindexfamilyrel]{code:qubit_subsystem_stabilizer}{Subsystem qubit stabilizer code} --- Subsystem modular-qudit stabilizer codes reduce to subsystem qubit stabilizer codes for qudit dimension \(q=2\).
\item\relax
\flmRefsHyperref[eczindexfamilyrel]{code:qudit_subsystem_css}{Subsystem modular-qudit CSS code} --- Subsystem modular-qudit CSS codes are subsystem modular-qudit stabilizer codes whose gauge groups admit a generating set of pure-\(X\) and pure-\(Z\) Pauli strings. Any \(\llbracket n,k,r,d\rrbracket _{\mathbb{Z}_q}\) subsystem stabilizer code can be mapped onto a \(\llbracket 2n,2k,2r,\geq d\rrbracket _{\mathbb{Z}_q}\) subsystem CSS code via \flmRefsHyperref{ref436}{symplectic doubling}, which preserves geometric locality of a code up to a constant factor. Every subsystem prime-qudit stabilizer code can be constructed from two nested subsystem prime-qudit CSS codes satisfying certain constraints \NoCaseChange{\protect\cite{cite4432}}.
\item\relax
\flmRefsHyperref[eczindexfamilyrel]{code:qudit_znone}{\(\mathbb{Z}_q^{(1)}\) subsystem code}\item\relax
\flmRefsHyperref[eczindexfamilyrel]{code:semion}{Chiral semion subsystem code}\item\relax
\flmRefsHyperref[eczindexfamilyrel]{code:zthree_znine}{\(\mathbb{Z}_3\times\mathbb{Z}_9\)-fusion subsystem code}\end{eczvaluelist}
\codefieldsection{Cousin}
\begin{eczvaluelist}
\item\relax
\flmRefsHyperref[eczindexfamilyrel]{code:qudit_stabilizer}{Modular-qudit stabilizer code} --- Subsystem modular-qudit stabilizer codes reduce to modular-qudit stabilizer codes when there are no gauge qudits.
\end{eczvaluelist}
\eczhbkcontributors{ \eczhuVVA }
\endeczcode

\eczcode{fractal_liquid}{Type-II fractal spin-liquid code}{~\NoCaseChange{\protect\cite{cite1348}}}
\codefieldsection{Description}
A type-II fracton prime-qudit CSS code defined on a cubic lattice \NoCaseChange{\protect\cite[{Eqs. (D9-D10)}]{cite456}}.

\codefieldsection{Parents}
\begin{eczvaluelist}
\item\relax
\flmRefsHyperref[eczindexfamilyrel]{code:qudit_css}{Modular-qudit CSS code}\item\relax
\flmRefsHyperref[eczindexfamilyrel]{code:fracton}{Fracton stabilizer code} --- The type-II fractal spin-liquid code is a type-II fracton code \NoCaseChange{\protect\cite{cite1348}}.
\end{eczvaluelist}
\eczhbkcontributors{ \eczhuVVA }
\endeczcode

\onecolumngrid
\clearpage

\section{Galois-qudit Kingdom}

\begin{eczEpigraph}
\begin{quote}
\flmQuoteSetup{quote}%
At twenty, solved what none could see—\\
Equations' hidden symmetry.\\
A duel at dawn, then silence fell:\\
Groups lived on where the genius fell.
\flmQuoteAttributed{Generated with Claude AI (Anthropic)}
\end{quote}
\end{eczEpigraph}

\twocolumngrid

\eczcode{arvind}{\(\llparenthesis n,1+n(q-1),2\rrparenthesis _q\) union stabilizer code}{~\NoCaseChange{\protect\cite{cite3170}}}
\eczhIndexCodeAliasName{arvind}{union stabilizer code}
\codefieldsection{Description}
Member of a family of \(\llparenthesis n,1+n(q-1),2\rrparenthesis _q\) Galois-qudit union stabilizer codes for odd \(n\).

\codefieldsection{Protection}
Distance \(d=2\) implies detection of any single-qudit error.

\codefieldsection{Parents}
\begin{eczvaluelist}
\item\relax
\flmRefsHyperref[eczindexfamilyrel]{code:galois_non_stabilizer}{Galois-qudit USt code}\item\relax
\flmRefsHyperref[eczindexfamilyrel]{code:small_distance_quantum}{Small-distance block quantum code}\end{eczvaluelist}
\codefieldsection{Child}
\begin{eczvaluelist}
\item\relax
\flmRefsHyperref[eczindexfamilyrel]{code:qubit_5_6_2}{\(\llparenthesis 5,6,2\rrparenthesis \) qubit code} --- The \(\llparenthesis 5,6,2\rrparenthesis \) code is the \(\llparenthesis n,1+n(q-1),2\rrparenthesis _q\) union stabilizer code for \(n=5\) and \(q=2\) \NoCaseChange{\protect\cite{cite3170}}.
\end{eczvaluelist}
\eczhbkcontributors{ \eczhuVVA }
\endeczcode

\eczcode{stab_18_2_5}{\(\llbracket 18,2,5\rrbracket \) BCC code}{~\NoCaseChange{\protect\cite{cite440}}}
\eczhIndexCodeAliasName{stab_18_2_5}{BCC code}
\codefieldsection{Description}
BCC code on 18 qubits encoding 2 logical qubits with distance 5, found by computer search \NoCaseChange{\protect\cite{cite440}}.

The code is defined by \(\mathcal{S}=\{5,11,15,17\}\) in the \flmRefsHyperref{code:bipartite_cyclic_cluster}{BCC code} framework,
with even qubits \((0,2,\ldots,16)\) forming the \(A\)-sublattice and odd qubits \((1,3,\ldots,17)\) forming the \(B\)-sublattice.
The subset \(\{5,11,17\}\subset\mathcal{S}\) is closed under adding 6 (mod 18),
so the conjugated stabilizer \(UX_0X_6U^\dagger\) has large cancellations, yielding weight-4 stabilizers
\(X_{-3}X_0X_3X_6\) and its cyclic shifts by even integers (and likewise with \(X\to Z\)).

A stabilizer tableau for the code is
\flmMathEnvironment{align}{}{
\begin{smallmatrix}
  X & X & X & I & I & X & I & X & I & I & I & X & I & X & I & X & I & I \\
  I & I & X & X & X & I & I & X & I & X & I & I & I & X & I & X & I & X \\
  I & X & I & I & X & X & X & I & I & X & I & X & I & I & I & X & I & X \\
  I & X & I & X & I & I & X & X & X & I & I & X & I & X & I & I & I & X \\
  I & X & I & X & I & X & I & I & X & X & X & I & I & X & I & X & I & I \\
  I & I & I & X & I & X & I & X & I & I & X & X & X & I & I & X & I & X \\
  I & X & I & I & I & X & I & X & I & X & I & I & X & X & X & I & I & X \\
  I & X & I & X & I & I & I & X & I & X & I & X & I & I & X & X & X & I \\
  I & Z & Z & Z & I & I & Z & I & Z & I & Z & I & I & I & Z & I & Z & I \\
  Z & I & I & Z & Z & Z & I & I & Z & I & Z & I & Z & I & I & I & Z & I \\
  Z & I & Z & I & I & Z & Z & Z & I & I & Z & I & Z & I & Z & I & I & I \\
  I & I & Z & I & Z & I & I & Z & Z & Z & I & I & Z & I & Z & I & Z & I \\
  Z & I & I & I & Z & I & Z & I & I & Z & Z & Z & I & I & Z & I & Z & I \\
  Z & I & Z & I & I & I & Z & I & Z & I & I & Z & Z & Z & I & I & Z & I \\
  Z & I & Z & I & Z & I & I & I & Z & I & Z & I & I & Z & Z & Z & I & I \\
  I & I & Z & I & Z & I & Z & I & I & I & Z & I & Z & I & I & Z & Z & Z
\end{smallmatrix}~.
}

\codefieldsection{Transversal and Permutation-Based Gates}
\begin{eczvaluelist}
\item\relax Transversal Hadamard-SWAP logical gate (inherited from BCC code structure) \NoCaseChange{\protect\cite{cite440}}.
\item\relax Logical Hadamard without SWAP: the code is self-dual, so transversal physical Hadamard implements logical \(H^{\otimes 2}\) \NoCaseChange{\protect\cite{cite440}}.
\end{eczvaluelist}
\codefieldsection{Decoding}
\begin{eczvaluelist}
\item\relax An ancilla syndrome extraction scheme with 18 shared ancilla qubits (one per data qubit position across two code blocks) detects all single bit-flip errors; a further reduction to 9 entangled ancilla qubits is numerically shown to preserve fault tolerance to distance 5 \NoCaseChange{\protect\cite{cite440}}.
\end{eczvaluelist}
\codefieldsection{Fault Tolerance}
\begin{eczvaluelist}
\item\relax An ancilla syndrome extraction scheme with 18 shared ancilla qubits (one per data qubit position across two code blocks) detects all single bit-flip errors; a further reduction to 9 entangled ancilla qubits is numerically shown to preserve fault tolerance to distance 5 \NoCaseChange{\protect\cite{cite440}}.
\end{eczvaluelist}
\codefieldsection{Parents}
\begin{eczvaluelist}
\item\relax
\flmRefsHyperref[eczindexfamilyrel]{code:bipartite_cyclic_cluster}{Bipartite cyclic cluster (BCC) code} --- The \(\llbracket 18,2,5\rrbracket \) BCC code is the smallest BCC code with parameters \(\llbracket n,2,5\rrbracket \) \NoCaseChange{\protect\cite{cite440}}.
\item\relax
\flmRefsHyperref[eczindexfamilyrel]{code:self_dual_css}{Self-dual CSS code} --- The stabilizer group of the \(\llbracket 18,2,5\rrbracket \) BCC code is invariant under transversal Hadamard \NoCaseChange{\protect\cite{cite440}}.
\item\relax
\flmRefsHyperref[eczindexfamilyrel]{code:small_distance_qubit_stabilizer}{Small-distance qubit stabilizer code}\end{eczvaluelist}
\eczhbkcontributors{ \eczhuVVA }
\endeczcode

\eczcode{galois_3_1_2}{\(\llbracket 3,1,2\rrbracket _4\) three-Galois-quartrit code}{~\NoCaseChange{\protect\cite{cite4598,cite514}}}
\eczhIndexCodeAliasName{galois_3_1_2}{three-Galois-quartrit code}
\codefieldsection{Description}
Three-Galois-qudit CSS code over \(\mathbb{F}_4=\{0,1,\omega,\omega^2\}\) that encodes one logical Galois qudit and detects a single-qudit error.

Its \(X\)- and \(Z\)-type stabilizer check matrices are both
\flmMathEnvironment{align}{}{
  H_X=H_Z=\begin{pmatrix}1&\omega&\omega^2\end{pmatrix}~.
}
Since the code is a true stabilizer code, multiplication of this row by \(\omega\) or \(\omega^2\) also yields stabilizers \NoCaseChange{\protect\cite{cite514}}.

\codefieldsection{Protection}
Detects a single Galois-qudit error. It is a quantum MDS code, saturating the quantum Singleton bound.
\codefieldsection{Parents}
\begin{eczvaluelist}
\item\relax
\flmRefsHyperref[eczindexfamilyrel]{code:galois_polynomial}{Galois-qudit RS code} --- The \(\llbracket 3,1,2\rrbracket _4\) code is constructed from the shortened RS\(_4\) code \NoCaseChange{\protect\cite{cite514}}.
\item\relax
\flmRefsHyperref[eczindexfamilyrel]{code:galois_quad_residue}{Quantum quadratic-residue (QR) code} --- The \(\llbracket 3,1,2\rrbracket _4\) code is constructed from the shortened RS\(_4\) code \NoCaseChange{\protect\cite{cite514}}.
\item\relax
\flmRefsHyperref[eczindexfamilyrel]{code:quantum_mds}{Quantum maximum-distance-separable (MDS) code} --- The \(\llbracket 3,1,2\rrbracket _4\) code saturates the quantum Singleton bound.
\item\relax
\flmRefsHyperref[eczindexfamilyrel]{code:small_distance_quantum}{Small-distance block quantum code}\end{eczvaluelist}
\codefieldsection{Cousins}
\begin{eczvaluelist}
\item\relax
\flmRefsHyperref[eczindexfamilyrel]{code:reed_solomon_4}{\([4,2,3]_4\) RS\(_4\) code} --- The \(\llbracket 3,1,2\rrbracket _4\) code is constructed from the shortened RS\(_4\) code \NoCaseChange{\protect\cite{cite514}}.
\item\relax
\flmRefsHyperref[eczindexfamilyrel]{code:stab_6_2_2}{\(\llbracket 6,2,2\rrbracket \) \(C_6\) code} --- Binarizing the \(\llbracket 3,1,2\rrbracket _4\) code in the self-dual normal basis \(\{\omega,\omega^2\}\) yields a \(\llbracket 6,2,2\rrbracket \) qubit CSS code equivalent to the \(C_6\) code after the qubit relabeling \((2\,3\,4\,5\,6)\) \NoCaseChange{\protect\cite{cite514}}.
\item\relax
\flmRefsHyperref[eczindexfamilyrel]{code:carbon}{\(\llbracket 12,2,4\rrbracket \) carbon code} --- Binarizing this code and concatenating each qubit pair with the \(\llbracket 4,2,2\rrbracket \) code yields the \(\llbracket 12,2,4\rrbracket \) carbon code \NoCaseChange{\protect\cite{cite514}}.
\item\relax
\flmRefsHyperref[eczindexfamilyrel]{code:bc_phantom}{Binarized-and-concatenated (B\&C) phantom code} --- Binarizing the \(\llbracket 3,1,2\rrbracket _4\) Galois-qudit CSS code and concatenating each qubit pair with the \(\llbracket 4,2,2\rrbracket \) code yields the \(\llbracket 12,2,4\rrbracket \) carbon code \NoCaseChange{\protect\cite{cite514}}.
\end{eczvaluelist}
\eczhbkcontributors{ \eczhuVVA }
\endeczcode

\eczcode{css_5_1_3}{\(\llbracket 5,1,3\rrbracket _4\) Galois-qudit CSS code}{~\NoCaseChange{\protect\cite{cite514}}}
\eczhIndexCodeAliasName{css_5_1_3}{Galois-qudit CSS code}
\codefieldsection{Description}
Five-Galois-qudit CSS code over \(\mathbb{F}_4=\{0,1,\omega,\omega^2\}\) that encodes one logical Galois qudit and corrects a single-qudit error.

Its \(X\)- and \(Z\)-type stabilizer check matrices are
\flmMathEnvironment{align}{}{
  H_X&=\begin{pmatrix}1&1&1&1&0\\0&1&\omega&\omega^2&1\end{pmatrix},\\
  H_Z&=\begin{pmatrix}1&1&1&1&0\\0&1&\omega^2&\omega&1\end{pmatrix}~.
}
Since the code is a true stabilizer code, multiplication of these rows by \(\omega\) or \(\omega^2\) also yields stabilizers \NoCaseChange{\protect\cite{cite514}}.

\codefieldsection{Parents}
\begin{eczvaluelist}
\item\relax
\flmRefsHyperref[eczindexfamilyrel]{code:galois_quad_residue}{Quantum quadratic-residue (QR) code} --- The \(\llbracket 5,1,3\rrbracket _4\) code is obtained from the shortened hexacode \NoCaseChange{\protect\cite{cite514}}.
\item\relax
\flmRefsHyperref[eczindexfamilyrel]{code:quantum_mds}{Quantum maximum-distance-separable (MDS) code} --- The \(\llbracket 5,1,3\rrbracket _4\) code saturates the quantum Singleton bound.
\item\relax
\flmRefsHyperref[eczindexfamilyrel]{code:small_distance_quantum}{Small-distance block quantum code}\end{eczvaluelist}
\codefieldsection{Cousins}
\begin{eczvaluelist}
\item\relax
\flmRefsHyperref[eczindexfamilyrel]{code:stab_5_1_3}{\(\llbracket 5,1,3\rrbracket \) Five-qubit perfect code} --- The \(\llbracket 5,1,3\rrbracket _4\) Galois-qudit CSS code is the image of the \(\llbracket 5,1,3\rrbracket \) five-qubit code under the BLT mapping \NoCaseChange{\protect\cite[{Lemma 1}]{cite795}\protect\cite[{Lemma 1}]{cite1432}}.
\item\relax
\flmRefsHyperref[eczindexfamilyrel]{code:shortened_hexacode}{\([5,3,3]_4\) Shortened hexacode} --- The \(\llbracket 5,1,3\rrbracket _4\) code is obtained from the shortened hexacode \NoCaseChange{\protect\cite{cite514}}.
\item\relax
\flmRefsHyperref[eczindexfamilyrel]{code:stab_10_2_3}{\(\llbracket 10,2,3\rrbracket \) binarized Galois-qudit code} --- Binarizing the \(\llbracket 5,1,3\rrbracket _4\) code in the self-dual normal basis \(\{\omega,\omega^2\}\) yields a \(\llbracket 10,2,3\rrbracket \) qubit CSS code \NoCaseChange{\protect\cite{cite514}}.
\end{eczvaluelist}
\eczhbkcontributors{ \eczhuVVA }
\endeczcode

\eczcode{galois_5_1_3}{\(\llbracket 5,1,3\rrbracket _q\) Galois-qudit code}{~\NoCaseChange{\protect\cite{cite398}}}
\eczhIndexCodeAliasName{galois_5_1_3}{Galois-qudit code}
\codefieldsection{Description}
True stabilizer code that generalizes the five-qubit perfect code to Galois qudits of prime-power dimension \(q=p^m\). It has \(4m\) stabilizer generators expressed as \(X_{\gamma} Z_{\gamma} Z_{-\gamma} X_{-\gamma} I\) and its cyclic permutations, with \(\gamma\) iterating over basis elements of \(\mathbb{F}_q\) over \(\mathbb{F}_p\).

\codefieldsection{Notes}
\begin{eczvaluelist}
\item\relax This code is described in a QEC2014 talk by \flmHref{https://www.qec14.ethz.ch/slides/DanielGottesman.pdf}{Gottesman}.
\end{eczvaluelist}
\codefieldsection{Parents}
\begin{eczvaluelist}
\item\relax
\flmRefsHyperref[eczindexfamilyrel]{code:galois_true_stabilizer}{True Galois-qudit stabilizer code}\item\relax
\flmRefsHyperref[eczindexfamilyrel]{code:quantum_cyclic}{Cyclic quantum code}\item\relax
\flmRefsHyperref[eczindexfamilyrel]{code:small_distance_quantum}{Small-distance block quantum code}\end{eczvaluelist}
\codefieldsection{Child}
\begin{eczvaluelist}
\item\relax
\flmRefsHyperref[eczindexfamilyrel]{code:stab_5_1_3}{\(\llbracket 5,1,3\rrbracket \) Five-qubit perfect code} --- The \(\llbracket 5,1,3\rrbracket _q\) Galois-qudit code for \(q=2\) reduces to the five-qubit perfect code.
\end{eczvaluelist}
\codefieldsection{Cousins}
\begin{eczvaluelist}
\item\relax
\flmRefsHyperref[eczindexfamilyrel]{code:graph_quantum}{Graph quantum code} --- The \(\llbracket 5,1,3\rrbracket _q\) code admits a graph-quantum-code realization for the group \(G=\mathbb{F}_q\) \NoCaseChange{\protect\cite{cite866}}.
\item\relax
\flmRefsHyperref[eczindexfamilyrel]{code:group_10_1_4}{\(\llbracket 10,1,4\rrbracket _{G}\) tenfold code} --- The \(\llbracket 10,1,4\rrbracket _{G}\) Abelian group code for \(G=\mathbb{F}_q\) is defined using a graph that is closely related to the \(\llbracket 5,1,3\rrbracket _{q}\) Galois-qudit code \NoCaseChange{\protect\cite{cite866}}.
\end{eczvaluelist}
\eczhbkcontributors{ Sarah Meng Li, \eczhuVVA }
\endeczcode

\eczcode{galois_6_2_3}{\(\llbracket 6,2,3\rrbracket _{q}\) code}{~\NoCaseChange{\protect\cite{cite830,cite813}}}
\eczhIndexCodeAliasName{galois_6_2_3}{code}
\codefieldsection{Description}
Six-qudit MDS error-correcting code defined for Galois-qudit dimension \(q=3\) \NoCaseChange{\protect\cite{cite830}}, \(q=2^2\) \NoCaseChange{\protect\cite{cite831}}, and \(q \geq 5\) \NoCaseChange{\protect\cite{cite830}\protect\cite[{Exam. 33}]{cite813}}.
This code cannot exist for qubits (\(q=2\)).

\codefieldsection{Encoding}
\begin{eczvaluelist}
\item\relax Three different encoding circuits for \(q=3\) \NoCaseChange{\protect\cite{cite4599}}.
\end{eczvaluelist}
\codefieldsection{Parents}
\begin{eczvaluelist}
\item\relax
\flmRefsHyperref[eczindexfamilyrel]{code:galois_true_stabilizer}{True Galois-qudit stabilizer code} --- The code is a non-CSS stabilizer code in general \NoCaseChange{\protect\cite{cite831}}.
\item\relax
\flmRefsHyperref[eczindexfamilyrel]{code:quantum_mds}{Quantum maximum-distance-separable (MDS) code}\item\relax
\flmRefsHyperref[eczindexfamilyrel]{code:small_distance_quantum}{Small-distance block quantum code}\end{eczvaluelist}
\codefieldsection{Cousin}
\begin{eczvaluelist}
\item\relax
\flmRefsHyperref[eczindexfamilyrel]{code:graph_quantum}{Graph quantum code} --- The \(\llbracket 6,2,3\rrbracket _{q}\) code family contains examples of graph quantum codes \NoCaseChange{\protect\cite{cite830}}.
\end{eczvaluelist}
\eczhbkcontributors{ \eczhuVVA }
\endeczcode

\eczcode{galois_7_3_3}{\(\llbracket 7,3,3\rrbracket _{q}\) code}{~\NoCaseChange{\protect\cite{cite830,cite813}}}
\eczhIndexCodeAliasName{galois_7_3_3}{code}
\codefieldsection{Description}
Seven-qudit MDS error-detecting code defined for Galois-qudit dimension \(q=3\) \NoCaseChange{\protect\cite{cite830}} and \(q \geq 7\) \NoCaseChange{\protect\cite{cite830}\protect\cite[{Exam. 33}]{cite813}}.
This code cannot exist for qubits (\(q=2\)).

\codefieldsection{Parents}
\begin{eczvaluelist}
\item\relax
\flmRefsHyperref[eczindexfamilyrel]{code:galois_true_stabilizer}{True Galois-qudit stabilizer code}\item\relax
\flmRefsHyperref[eczindexfamilyrel]{code:quantum_mds}{Quantum maximum-distance-separable (MDS) code}\item\relax
\flmRefsHyperref[eczindexfamilyrel]{code:small_distance_quantum}{Small-distance block quantum code}\end{eczvaluelist}
\codefieldsection{Cousin}
\begin{eczvaluelist}
\item\relax
\flmRefsHyperref[eczindexfamilyrel]{code:graph_quantum}{Graph quantum code} --- The \(\llbracket 7,3,3\rrbracket _{q}\) code family contains examples of graph quantum codes \NoCaseChange{\protect\cite{cite830}}.
\end{eczvaluelist}
\eczhbkcontributors{ \eczhuVVA }
\endeczcode

\eczcode{abelian_lifted_product}{Abelian LP code}{~\NoCaseChange{\protect\cite{cite1247,cite674}}}
\codefieldsection{Description}
A lifted-product code whose lift group \(G\) is Abelian.
The case of \(G\) being a cyclic group is a GB code (a.k.a. a quasi-cyclic LP code) \NoCaseChange{\protect\cite[{Sec. III.E}]{cite674}}.
A particular family with \(G=\mathbb{Z}_{\ell}\) yields codes with parameters \(\llbracket n,k=\Theta(\log n),d=\Theta(n/\log n)\rrbracket \) \NoCaseChange{\protect\cite{cite674}}.

The Abelian LP construction has been adapted to accommodate noise bias, yielding \textit{bias-tailored LP codes} \NoCaseChange{\protect\cite{cite2654}}.
See Refs. \NoCaseChange{\protect\cite{cite1247,cite674,cite1556,cite848}} for other explicit examples.

\codefieldsection{Rate}
For cyclic groups \(G=\mathbb{Z}_{\ell}\) with \(\ell=\Theta(n/\log n)\), quasi-cyclic expander LP codes yield families with parameters \(\llbracket n,k=\Theta(\log n),d=\Theta(n/\log n)\rrbracket \) \NoCaseChange{\protect\cite{cite674}}. Related balanced-product reformulations and other explicit Abelian LP constructions appear in \NoCaseChange{\protect\cite{cite434,cite1557}}.
\codefieldsection{Decoding}
\begin{eczvaluelist}
\item\relax Ensemble BP decoder for codes without short cycles of length 4 \NoCaseChange{\protect\cite{cite1357}}.
\item\relax Efficient decoder correcting \flmRefsHyperref{ref65}{order} \(\Theta(n/\log n)\) errors \NoCaseChange{\protect\cite{cite4600}}.
\end{eczvaluelist}
\codefieldsection{Parent}
\begin{eczvaluelist}
\item\relax
\flmRefsHyperref[eczindexfamilyrel]{code:lifted_product}{Lifted-product (LP) code}\end{eczvaluelist}
\codefieldsection{Children}
\begin{eczvaluelist}
\item\relax
\flmRefsHyperref[eczindexfamilyrel]{code:qcga}{Bivariate bicycle (BB) code} --- Bivariate bicycle codes are Abelian LP codes over groups of the form \(\mathbb{Z}_{r} \times \mathbb{Z}_{s}\).
\item\relax
\flmRefsHyperref[eczindexfamilyrel]{code:generalized_bicycle}{Generalized bicycle (GB) code} --- A code GB\((a,b)\) with circulants of size \(\ell\) is a special case of a lifted-product code LP\((A,B)\) code over the Abelian \flmRefsHyperref{ref205}{group algebra} \(\mathbb{F}_q[\mathbb{Z}_{\ell}]\) associated with a cyclic group, with \(1\times 1\) matrices \(A=a(x)\), \(B=b(x)\) given by the corresponding polynomials.
Quasi-cyclic LP codes, i.e., LP codes constructed from cyclic groups, are equivalent to GB codes \NoCaseChange{\protect\cite[{Sec. III.E}]{cite674}}.

\end{eczvaluelist}
\codefieldsection{Cousins}
\begin{eczvaluelist}
\item\relax
\flmRefsHyperref[eczindexfamilyrel]{code:qc_ldpc}{Quasi-cyclic LDPC (QC-LDPC) code} --- QC-LDPC codes can be \flmRefsHyperref{ref47}{lifted} to yield various Abelian LP codes \NoCaseChange{\protect\cite{cite1556,cite1357,cite843}}. Conversely, the Abelian LP construction yields notable families of QC-LDPC codes \NoCaseChange{\protect\cite{cite1557}}.
\item\relax
\flmRefsHyperref[eczindexfamilyrel]{code:pg_ldpc}{Finite-geometry LDPC (FG-LDPC) code} --- FG-LDPC codes can be used to construct Abelian LP codes \NoCaseChange{\protect\cite{cite1357}}.
\item\relax
\flmRefsHyperref[eczindexfamilyrel]{code:expander_lifted_product}{Expander LP code} --- For cyclic groups \(G=\mathbb{Z}_{\ell}\) with \(\ell=\Theta(n/\log n)\), quasi-cyclic expander LP codes yield families with parameters \(\llbracket n,k=\Theta(\log n),d=\Theta(n/\log n)\rrbracket \) \NoCaseChange{\protect\cite{cite674}}. Related balanced-product reformulations and other explicit Abelian LP constructions appear in \NoCaseChange{\protect\cite{cite434,cite1557}}.
\item\relax
\flmRefsHyperref[eczindexfamilyrel]{code:asymmetric_qecc}{Asymmetric quantum code (AQC)} --- The Abelian LP construction has been adapted to accommodate noise bias, yielding bias-tailored LP codes \NoCaseChange{\protect\cite{cite2654}}.
\item\relax
\flmRefsHyperref[eczindexfamilyrel]{code:gkp_concatenated}{Concatenated GKP code} --- GKP codes have been concatenated with Abelian LP codes \NoCaseChange{\protect\cite{cite1556}} that are in turn based on QC-LDPC codes \NoCaseChange{\protect\cite{cite1550}}. Concatenating Abelian LP codes with GKP codes can surpass the CSS Hamming bound \NoCaseChange{\protect\cite{cite1556}}.
\item\relax
\flmRefsHyperref[eczindexfamilyrel]{code:2bga}{Two-block group-algebra (2BGA) codes} --- Abelian 2BGA codes are LP\((a,b)\) codes, constructed from a pair of one-by-one matrices \(a,b\in \mathbb{F}_q[G]\) in a \flmRefsHyperref{ref205}{group algebra} of an Abelian group \(G\).
\end{eczvaluelist}
\eczhbkcontributors{ \eczhuVVA }
\endeczcode

\eczcode{quantum_secret_sharing}{Approximate secret-sharing code}{~\NoCaseChange{\protect\cite{cite2548}}}
\codefieldsection{Description}
A family of \( \llbracket n,k,d\rrbracket _q \) CSS codes approximately correcting errors on up to \(\lfloor (n-1)/2 \rfloor\) Galois qudits, i.e., with approximate distance approaching the no-cloning bound \(n/2\). Constructed using a \flmRefsHyperref{ref811}{non-degenerate} CSS code, such as a polynomial quantum code, and a classical authentication scheme. The code can be viewed as a \(t\)-error-tolerant secret sharing scheme. Since the code yields a small logical subspace using large registers that contain both classical and quantum information, it is not useful for practical error correction problems, but instead demonstrates the power of approximate quantum error correction.
\codefieldsection{Protection}
Corrects up to \(\lfloor (n-1)/2 \rfloor\) errors with fidelity exponentially close to 1.
\codefieldsection{Encoding}
\begin{eczvaluelist}
\item\relax Uses a quantum authentication scheme, which is a keyed system in which a valid state has high fidelity, and a classical secret-sharing scheme.
\end{eczvaluelist}
\codefieldsection{Decoding}
\begin{eczvaluelist}
\item\relax Decoding is analogous to reconstruction in a secret sharing scheme and is done in polynomial time. The only required operations are verification of quantum authentication, which is a pair of polynomial-time quantum algorithms that check if the fidelity of the received state is close to 1, and erasure correction for a stabilizer code, which involves solving a system of linear equations.
\end{eczvaluelist}
\codefieldsection{Parent}
\begin{eczvaluelist}
\item\relax
\flmRefsHyperref[eczindexfamilyrel]{code:galois_css}{Galois-qudit CSS code} --- The code required to construct this code must be a \flmRefsHyperref{ref811}{non-degenerate} Galois-qudit CSS code.
\end{eczvaluelist}
\codefieldsection{Cousins}
\begin{eczvaluelist}
\item\relax
\flmRefsHyperref[eczindexfamilyrel]{code:approximate_qecc}{Approximate quantum error-correcting code (AQECC)} --- Secret-sharing codes approximately correct errors on up to \(\lfloor (n-1)/2 \rfloor\) errors.
\item\relax
\flmRefsHyperref[eczindexfamilyrel]{code:galois_polynomial}{Galois-qudit RS code} --- Polynomial codes can be used for a specific construction of this code.
\item\relax
\flmRefsHyperref[eczindexfamilyrel]{code:reed_solomon}{Reed-Solomon (RS) code} --- The classical information in this code is encoded using an RS code.
\item\relax
\flmRefsHyperref[eczindexfamilyrel]{code:purity_testing}{Purity-testing stabilizer code} --- The purity-testing protocol of Ref. \NoCaseChange{\protect\cite{cite3650}} can be improved using approximate codes similar to the approximate secret-sharing codes \NoCaseChange{\protect\cite{cite4013}}.
\item\relax
\flmRefsHyperref[eczindexfamilyrel]{code:ame}{Perfect-tensor code} --- Perfect tensors are useful for quantum secret sharing and open-destination multi-party teleportation \NoCaseChange{\protect\cite{cite2934,cite1924,cite2935}}.
\item\relax
\flmRefsHyperref[eczindexfamilyrel]{code:stab_3_1_2}{\(\llbracket 3,1,2\rrbracket _3\) Three-qutrit code} --- The three-qutrit code defines a minimal secret-sharing scheme \NoCaseChange{\protect\cite{cite4508}} that is substantially generalized by approximate secret-sharing codes.
\item\relax
\flmRefsHyperref[eczindexfamilyrel]{code:quantum_singleton}{Singleton-bound approaching AQECC} --- Quantum secret-sharing codes have asymptotically decaying rate and require qudit dimension to increase exponentially with \(n\), while Singleton-bound approaching AQECCs have constant rate and qudit dimension.
\end{eczvaluelist}
\eczhbkcontributors{ Manasi Shingane, \eczhuVVA }
\endeczcode

\eczcode{balanced_product}{Balanced product (BP) code}{~\NoCaseChange{\protect\cite{cite434}}}
\codefieldsection{Description}
Family of CSS quantum codes obtained from two classical-code chain complexes that share a common group symmetry.
The balanced product can be understood as taking the usual tensor or hypergraph product and then quotienting by the shared symmetry action.
This can reduce the overall number of physical qubits \(n\) while, in favorable cases, preserving the number of encoded qubits and the code distance, thereby improving the encoding rate \(k/n\) and normalized distance \(d/n\) compared to the underlying tensor or hypergraph product.

For trivial group action, the construction reduces to a hypergraph product code.
For cyclic groups, it overlaps with fiber-bundle and lifted-product constructions \NoCaseChange{\protect\cite{cite434}}.

\codefieldsection{Rate}
The original explicit balanced-product family is first constructed as a horizontal subsystem code with \(k \in \Theta(n^{2/3})\), \(d_X \in \Omega(n^{1/3})\), and \(d_Z \in \Theta(n)\); after distance balancing, it yields an LDPC family with \(k \in \Theta(n^{4/5})\) and \(d \in \Omega(n^{3/5})\) \NoCaseChange{\protect\cite{cite434}}. For balanced products of two good classical LDPC codes over groups of order \(\Theta(n)\), the original paper proves constant encoding rate and conjectures linear distance \NoCaseChange{\protect\cite{cite434}}.
\codefieldsection{Gates}
\begin{eczvaluelist}
\item\relax Logical gates via Dehn twists for balanced products of cyclic codes \NoCaseChange{\protect\cite{cite3403}}.
\end{eczvaluelist}
\codefieldsection{Decoding}
\begin{eczvaluelist}
\item\relax BP-OSD decoder \NoCaseChange{\protect\cite{cite1247}}.
\end{eczvaluelist}
\codefieldsection{Parents}
\begin{eczvaluelist}
\item\relax
\flmRefsHyperref[eczindexfamilyrel]{code:galois_css}{Galois-qudit CSS code}\item\relax
\flmRefsHyperref[eczindexfamilyrel]{code:generalized_homological_product_css}{Generalized homological-product CSS code} --- Balanced product codes result from a tensor product of two classical-code chain complexes, followed by a factoring out of certain symmetries.
\end{eczvaluelist}
\codefieldsection{Children}
\begin{eczvaluelist}
\item\relax
\flmRefsHyperref[eczindexfamilyrel]{code:fiber_bundle}{Fiber-bundle code} --- Fiber-bundle codes can be formulated in terms of a balanced product \NoCaseChange{\protect\cite{cite434}}.
\item\relax
\flmRefsHyperref[eczindexfamilyrel]{code:lossless_expander}{Lossless expander balanced-product code}\item\relax
\flmRefsHyperref[eczindexfamilyrel]{code:lifted_product}{Lifted-product (LP) code} --- Coarsely speaking, a lifted product is a balanced product where the group \(G\) acts freely. In principle, a lifted product can be defined for rings that are more general than \flmRefsHyperref{ref205}{group algebras} \( \mathbb{F}_q G \).
\end{eczvaluelist}
\codefieldsection{Cousins}
\begin{eczvaluelist}
\item\relax
\flmRefsHyperref[eczindexfamilyrel]{code:qubit_subsystem_stabilizer}{Subsystem qubit stabilizer code} --- The original explicit balanced-product family is first constructed as a horizontal subsystem balanced-product code built from expander codes and cyclic repetition codes \NoCaseChange{\protect\cite{cite434}}.
\item\relax
\flmRefsHyperref[eczindexfamilyrel]{code:distance_balanced}{Distance-balanced code} --- Applying distance balancing to the explicit subsystem balanced-product family of Ref. \NoCaseChange{\protect\cite{cite434}} yields an LDPC code family with \(k \in \Theta(n^{4/5})\) and \(d \in \Omega(n^{3/5})\).
\item\relax
\flmRefsHyperref[eczindexfamilyrel]{code:lr-cayley-complex}{Left-right Cayley complex code} --- Left-right Cayley complexes can be obtained via a balanced product of \(G\)-graphs \NoCaseChange{\protect\cite{cite88}}.
\item\relax
\flmRefsHyperref[eczindexfamilyrel]{code:fibonacci_fractal_liquid}{Fibonacci fractal spin-liquid code} --- The Fibonacci fractal spin-liquid code is a hypergraph product of the repetition code and the Fibonacci code \NoCaseChange{\protect\cite{cite1348}}, and can be formulated directly as a BP code \NoCaseChange{\protect\cite{cite1350}}.
\item\relax
\flmRefsHyperref[eczindexfamilyrel]{code:bb90}{\(\llbracket 90,8,10\rrbracket \) BB6 code} --- The \(\llbracket 90,8,10\rrbracket \) BB code can be formulated as a balanced product of two cyclic codes \NoCaseChange{\protect\cite{cite3403}}.
\item\relax
\flmRefsHyperref[eczindexfamilyrel]{code:dhlv}{Dinur-Hsieh-Lin-Vidick (DHLV) code} --- DHLV codes can be obtained from a balanced product of two expander codes \NoCaseChange{\protect\cite{cite1101}}.
\item\relax
\flmRefsHyperref[eczindexfamilyrel]{code:toric}{Toric code} --- Twisted toric codes can be obtained from balanced products of cyclic graphs over a cyclic group \NoCaseChange{\protect\cite[{Fig. 8}]{cite434}}.
\item\relax
\flmRefsHyperref[eczindexfamilyrel]{code:galois_expander}{Galois-qudit expander code} --- Balanced products of the RS-based expander-code complexes in \NoCaseChange{\protect\cite{cite689}} yield \([n,k\geq n^{1-\epsilon},d\geq n/\operatorname{poly}(\log n)]_q\) LTCs exhibiting the multiplication property.
\item\relax
\flmRefsHyperref[eczindexfamilyrel]{code:2bga}{Two-block group-algebra (2BGA) codes} --- 2BGA codes can be formulated directly as balanced product codes \NoCaseChange{\protect\cite[{Rem. C.1}]{cite718}}.
\end{eczvaluelist}
\eczhbkcontributors{ Finnegan Voichick, Nikolas Breuckmann, \eczhuVVA }
\endeczcode

\eczcode{binary_quantum_goppa}{Binary quantum Goppa code}{~\NoCaseChange{\protect\cite{cite4601,cite697,cite4249}}}
\codefieldsection{Description}
A quantum AG code obtained from algebraic-geometric Goppa codes via the Galois-qudit CSS construction.

For \(q=2^m\), let \(F/\mathbb{F}_q\) be an algebraic function field of one variable with an automorphism \(\sigma\) of order two fixing \(\mathbb{F}_q\), and let \(P_1,\ldots,P_n\) be pairwise distinct degree-one places such that \(\sigma P_i \neq P_j\) for all \(i,j\).
If \(\eta\) is a differential with \(v_{P_i}(\eta)=v_{\sigma P_i}(\eta)=-1\), \(\operatorname{res}_{P_i}(\eta)=1\), and \(\operatorname{res}_{\sigma P_i}(\eta)=-1\), and if \(G\) is a \(\sigma\)-invariant divisor satisfying \(v_{P_i}(G)=v_{\sigma P_i}(G)=0\), then
\flmMathEnvironment{align}{}{
  C(G)=\{(f(P_1),\ldots,f(P_n),f(\sigma P_1),\ldots, f(\sigma P_n) )\,|\,f\in\mathcal{L}(G)\}\subseteq \mathbb{F}_q^{2n}
}
obeys \(C(G)^{\perp_s}=C(H)\), where \(H=(P_1+\cdots+P_n+\sigma P_1+\cdots+\sigma P_n)-G+(\eta)\) \NoCaseChange{\protect\cite[{Ch. 5}]{cite697}}.
Whenever \(G \geq H\), this yields an \(\llbracket n,k,d\rrbracket _q\) quantum stabilizer code with
\flmMathEnvironment{align}{}{
  k=\dim G-\dim(G-P_1-\cdots-P_n-\sigma P_1-\cdots-\sigma P_n)-n
}
and \(d \geq n-\lfloor \deg G/2 \rfloor\) \NoCaseChange{\protect\cite{cite4601,cite4249}\protect\cite[{Ch. 5}]{cite697}}.

\codefieldsection{Protection}
For a code with distance \(d\), detects errors on up to \(d-1\) qudits and corrects errors on up to \(\lfloor (d-1)/2 \rfloor\) qudits. In Matsumoto's construction, \(d \geq n-\lfloor \deg G/2 \rfloor\).
\codefieldsection{Rate}
For every \(m \geq 2\), there are binary quantum Goppa-code families with \(\liminf k_i/n_i \geq 1-\frac{2}{2^m-1}-4m\delta\) and \(\liminf d_i/n_i \geq \delta\) \NoCaseChange{\protect\cite{cite4601,cite697}}.
\codefieldsection{Encoding}
\begin{eczvaluelist}
\item\relax Encoding defined in Ref. \NoCaseChange{\protect\cite{cite4602}} uses a technique from Ref. \NoCaseChange{\protect\cite{cite3716}} to encode quantum stabilizer codes.
\end{eczvaluelist}
\codefieldsection{Decoding}
\begin{eczvaluelist}
\item\relax Farran's decoder can be used for the underlying AG construction; under the bound \(2\,\mathrm{wt}(e)+1 \leq n-\lfloor \deg G/2 \rfloor\), the relevant minimum-weight error can be recovered in \(O(n^{2.81})\) time \NoCaseChange{\protect\cite{cite4603,cite697}}.
\end{eczvaluelist}
\codefieldsection{Parents}
\begin{eczvaluelist}
\item\relax
\flmRefsHyperref[eczindexfamilyrel]{code:galois_css}{Galois-qudit CSS code}\item\relax
\flmRefsHyperref[eczindexfamilyrel]{code:quantum_ag}{Quantum AG code}\end{eczvaluelist}
\codefieldsection{Cousins}
\begin{eczvaluelist}
\item\relax
\flmRefsHyperref[eczindexfamilyrel]{code:goppa}{Goppa code} --- Classical Goppa codes over various algebraic curves are used to construct quantum Goppa codes.
\item\relax
\flmRefsHyperref[eczindexfamilyrel]{code:qubit_css}{Qubit CSS code} --- Quantum Goppa codes can exceed the \flmRefsHyperref{ref1729}{quantum GV bound} \NoCaseChange{\protect\cite{cite4249}}.
\end{eczvaluelist}
\eczhbkcontributors{ Adam Wills, Manasi Shingane, \eczhuVVA }
\endeczcode

\eczcode{bipartite_cyclic_cluster}{Bipartite cyclic cluster (BCC) code}{~\NoCaseChange{\protect\cite{cite440}}}
\codefieldsection{Description}
Cyclic CSS code constructed from a bipartite cluster state with cyclic invariance,
emphasizing simplicity of state preparation over simplicity of stabilizers.

A BCC code encodes \(k\) logical qubits into \(n = k n_0\) physical qubits.
Qubits are labeled by pairs \((m,j)\) with \(m\) defined modulo \(n_0\) and \(1 \leq j \leq k\),
partitioned into sets \(A\) (indices \(j \in \mathcal{A}\)) and \(B\) (indices \(j \notin \mathcal{A}\)).
The code is defined by a bipartite graph \(G\) on the qubits, with edges only between \(A\) and \(B\),
that is invariant under cyclic shifts \((m,j) \to (m+1,j)\).
Code states are prepared by initializing \(A\)-qubits in \(|\pm\rangle\) and \(B\)-qubits in \(|0/1\rangle\),
then applying CNOT gates (source in \(A\), target in \(B\)) along the edges of \(G\).
The \(2^k\) resulting basis states confirm that \(k\) logical qubits are encoded.
After a Hadamard on each \(B\)-qubit, this gives a CSS code.

The \(X\)-type stabilizers arise from conjugating \(X_{m,j} X_{m+1,j}\) (for \(j \in \mathcal{A}\)) by the CNOT circuit,
and the \(Z\)-type stabilizers from \(Z_{m,j} Z_{m+1,j}\) (for \(j \in \mathcal{B}\)).

For \(k=2\), the graph \(G\) is parameterized by a set \(\mathcal{S}\) of odd integers modulo \(n = 2n_0\):
even qubit \(m\) connects to odd qubit \(m'\) iff \((m'-m) \bmod n \in \mathcal{S}\).

A related non-CSS construction, the \textit{cyclic cluster code}, drops the bipartiteness requirement on \(G\).
For \(d=3\), this construction applied to the \(\llbracket 10,2,3\rrbracket \) BCC code yields the \(\llbracket 5,1,3\rrbracket \) five-qubit perfect code \NoCaseChange{\protect\cite{cite440}}.

\codefieldsection{Protection}
The distance satisfies \(d \leq |\mathcal{S}| + 1\),
since \(U^\dagger X_{m,j} U\) for \(j \in \mathcal{A}\) gives a logical operator of weight \(|\mathcal{S}|+1\) \NoCaseChange{\protect\cite{cite440}}.
A weight-four GB code of odd distance \(d\) satisfies \(n \geq 1 + d^2\) \NoCaseChange{\protect\cite{cite3183}},
so the \(\llbracket d^2+1,2,d\rrbracket \) BCC codes are optimal among weight-four BCC codes.

\codefieldsection{Encoding}
\begin{eczvaluelist}
\item\relax Initialize \(A\)-qubits in \(|+\rangle\) and \(B\)-qubits in \(|0\rangle\), then apply CNOT gates along edges of \(G\), followed by Hadamard on each \(B\)-qubit.
\end{eczvaluelist}
\codefieldsection{Transversal and Permutation-Based Gates}
\begin{eczvaluelist}
\item\relax For \(k=2\), all BCC codes admit a transversal Hadamard-SWAP logical gate via qubit permutation \((m,1) \mapsto (-m,2)\) followed by Hadamard on all qubits \NoCaseChange{\protect\cite{cite440}}.
\item\relax The \(\llbracket d^2+1,2,d\rrbracket \) BCC codes also admit a logical Hadamard without SWAP via \(m \mapsto md \pmod{n}\) followed by Hadamard \NoCaseChange{\protect\cite{cite440}}.
\end{eczvaluelist}
\codefieldsection{Fault Tolerance}
\begin{eczvaluelist}
\item\relax For distance-3 BCC codes (\(|\mathcal{S}|=2\)), a single error produces at most weight-1 logical error, so fault tolerance is preserved \NoCaseChange{\protect\cite{cite440}}.
\end{eczvaluelist}
\codefieldsection{Parents}
\begin{eczvaluelist}
\item\relax
\flmRefsHyperref[eczindexfamilyrel]{code:generalized_bicycle}{Generalized bicycle (GB) code} --- BCC codes are GB codes with weight-two circulant \(A\) (polynomial \(a(x)=1+x\)) \NoCaseChange{\protect\cite{cite440}}.
\item\relax
\flmRefsHyperref[eczindexfamilyrel]{code:quantum_cyclic}{Cyclic quantum code} --- BCC codes are invariant under cyclic shifts by construction \NoCaseChange{\protect\cite{cite440}}.
\end{eczvaluelist}
\codefieldsection{Children}
\begin{eczvaluelist}
\item\relax
\flmRefsHyperref[eczindexfamilyrel]{code:xzzx_10_2_3}{\(\llbracket 10,2,3\rrbracket \) rotated toric code} --- The \(\llbracket 10,2,3\rrbracket \) rotated toric code is a \(\llbracket d^2+1,2,d\rrbracket \) BCC code for \(d=3\) \NoCaseChange{\protect\cite{cite440}}. A non-CSS cyclic cluster code related to the \(\llbracket 10,2,3\rrbracket \) rotated toric code yields the \(\llbracket 5,1,3\rrbracket \) five-qubit perfect code for \(d=3\) \NoCaseChange{\protect\cite{cite440}}.
\item\relax
\flmRefsHyperref[eczindexfamilyrel]{code:stab_18_2_5}{\(\llbracket 18,2,5\rrbracket \) BCC code} --- The \(\llbracket 18,2,5\rrbracket \) BCC code is the smallest BCC code with parameters \(\llbracket n,2,5\rrbracket \) \NoCaseChange{\protect\cite{cite440}}.
\end{eczvaluelist}
\codefieldsection{Cousins}
\begin{eczvaluelist}
\item\relax
\flmRefsHyperref[eczindexfamilyrel]{code:cluster_state}{Cluster-state code} --- BCC codes are obtained by applying Hadamard on the \(B\)-sublattice of a bipartite cluster state, converting CZ-gate preparation into CNOT-gate preparation \NoCaseChange{\protect\cite{cite440}}.
\item\relax
\flmRefsHyperref[eczindexfamilyrel]{code:twisted_xzzx}{Twisted XZZX toric code} --- The \(\llbracket d^2+1,2,d\rrbracket \) twisted XZZX toric codes (parameters \(a=1,b=d\)) are Clifford-equivalent to BCC codes for odd \(d\) \NoCaseChange{\protect\cite{cite440}}.
\end{eczvaluelist}
\eczhbkcontributors{ \eczhuVVA }
\endeczcode

\eczcode{distance_balanced}{Distance-balanced code}{~\NoCaseChange{\protect\cite{cite2989,cite4604,cite684}}}
\codefieldsection{Description}
Galois-qudit CSS code obtained from a CSS code by increasing the smaller of the \(X\)- and \(Z\)-distances using a homological-product-based balancing step or one of its generalizations.
The initial code is said to be \textit{unbalanced}, i.e., tailored to noise biased toward either bit- or phase-flip errors, and the procedure can result in a code that treats both types of errors on a more equal footing.

In the original construction \NoCaseChange{\protect\cite[{Sec. 4}]{cite2989}}, if \(C\) is a QLDPC CSS code then applying the balancing step with parameter \(l\) yields \(\tilde K=K\), \(\tilde d_X=l d_X\), \(\tilde d_Z=d_Z\), and \(\tilde N=O(Nl)\), so choosing \(l\approx d_Z/d_X\) balances the two distances.
In the generalized construction \NoCaseChange{\protect\cite[{Thm. 4.2}]{cite684}}, combining a component quantum code \(\mathcal{Q}\) with a classical code \(C\) yields a new code with \(K=k(\mathcal{Q})k(C)\), \(D_X=d_X(\mathcal{Q})d(C)\), and \(D_Z=d_Z(\mathcal{Q})\), so choosing \(d(C)\approx d_Z/d_X\) balances the two distances.
The original distance-balancing procedure \NoCaseChange{\protect\cite{cite2989}}, later generalized in this way, can yield QLDPC codes \NoCaseChange{\protect\cite[{Thm. 1}]{cite2989}}.

\begin{defterm}{Weight reduction}\label{ref4605}\label{ref491}
Various procedures performing \textit{weight reduction} \NoCaseChange{\protect\cite{cite2989,cite4604,cite4438}} take in a stabilizer code and output a longer code with bounded stabilizer-generator weight.
Hastings' original construction \NoCaseChange{\protect\cite{cite2989}} makes a qubit CSS code QLDPC while preserving the number of logical qubits and keeping the block length polynomial in the original one.
The weight reduction procedure of Ref. \NoCaseChange{\protect\cite{cite4438}} has been extended to subsystem qubit stabilizer codes \NoCaseChange{\protect\cite{cite490}}.
\end{defterm}

\codefieldsection{Decoding}
\begin{eczvaluelist}
\item\relax If the auxiliary classical LDPC code corrects all error patterns of weight \(<\alpha |A|\), then the resulting product code has a polynomial-time decoder for \(X\)-errors of weight \(< \alpha |A| d_X/2\), where \(d_X\) is the \(X\)-distance of the component quantum code \NoCaseChange{\protect\cite{cite684}}.
\item\relax If the component 2D complex has a polynomial-time decoder for \(Z\)-errors of weight \(< w\), then the resulting distance-balanced code also has a polynomial-time decoder for \(Z\)-errors of weight \(< w\) \NoCaseChange{\protect\cite{cite684}}.
\item\relax The effective distance of single-ancilla syndrome extraction QLDPC code circuits can be preserved under weight reduction \NoCaseChange{\protect\cite{cite3776}}. The distance balancing technique of Ref. \NoCaseChange{\protect\cite{cite684}} preserves the \flmRefsHyperref{ref3496}{effective distance} of single-ancilla syndrome extraction circuits \NoCaseChange{\protect\cite{cite3776}}.
\end{eczvaluelist}
\codefieldsection{Fault Tolerance}
\begin{eczvaluelist}
\item\relax Single-ancilla syndrome extraction circuits that, for the most part, preserve the \flmRefsHyperref{ref3496}{effective distance} of weight-reduced qLDPC codes \NoCaseChange{\protect\cite{cite3776}}. The distance balancing technique of Ref. \NoCaseChange{\protect\cite{cite684}} preserves \flmRefsHyperref{ref3496}{effective distance} \NoCaseChange{\protect\cite{cite3776}}.
\end{eczvaluelist}
\codefieldsection{Parents}
\begin{eczvaluelist}
\item\relax
\flmRefsHyperref[eczindexfamilyrel]{code:galois_css}{Galois-qudit CSS code}\item\relax
\flmRefsHyperref[eczindexfamilyrel]{code:generalized_homological_product_css}{Generalized homological-product CSS code}\end{eczvaluelist}
\codefieldsection{Cousins}
\begin{eczvaluelist}
\item\relax
\flmRefsHyperref[eczindexfamilyrel]{code:homological_product}{Homological product code} --- Distance balancing relies on taking a homological product of chain complexes corresponding to a classical and a quantum code.
\item\relax
\flmRefsHyperref[eczindexfamilyrel]{code:qubit_subsystem_stabilizer}{Subsystem qubit stabilizer code} --- The weight reduction procedure of Ref. \NoCaseChange{\protect\cite{cite4438}} has been extended to subsystem qubit stabilizer codes \NoCaseChange{\protect\cite{cite490}}.
\item\relax
\flmRefsHyperref[eczindexfamilyrel]{code:gkp-cluster-state}{GKP CV-cluster-state code} --- \flmRefsHyperref{ref491}{Weight reduction} has been studied in the context of GKP CV-cluster-state codes \NoCaseChange{\protect\cite{cite4438}}.
\item\relax
\flmRefsHyperref[eczindexfamilyrel]{code:general_qldpc}{QLDPC code} --- Lattice surgery techniques for QLDPC codes \NoCaseChange{\protect\cite{cite3499,cite848}} utilize \flmRefsHyperref{ref491}{weight reduction}. Single-ancilla syndrome extraction circuits that, for the most part, preserve the \flmRefsHyperref{ref3496}{effective distance} of weight-reduced qLDPC codes \NoCaseChange{\protect\cite{cite3776}}.
\item\relax
\flmRefsHyperref[eczindexfamilyrel]{code:asymmetric_qecc}{Asymmetric quantum code (AQC)} --- Distance balancing is a procedure that can convert an asymmetric CSS code into a less asymmetric one.
\item\relax
\flmRefsHyperref[eczindexfamilyrel]{code:qltc}{Quantum locally testable code (QLTC)} --- Distance balancing and weight reduction are useful for constructing QLTCs \NoCaseChange{\protect\cite{cite2989,cite2990,cite2991}}.
\item\relax
\flmRefsHyperref[eczindexfamilyrel]{code:fiber_bundle}{Fiber-bundle code} --- Fiber-bundle code constructions use distance balancing and weight reduction to increase distance.
\item\relax
\flmRefsHyperref[eczindexfamilyrel]{code:ramanujan_tensor_product}{High-dimensional expander (HDX) code} --- Ramanujan tensor-product constructions use distance balancing to increase distance.
\item\relax
\flmRefsHyperref[eczindexfamilyrel]{code:check_product}{Quantum check-product code} --- Quantum check-product code constructions use distance balancing to increase distance \NoCaseChange{\protect\cite{cite2185}}.
\item\relax
\flmRefsHyperref[eczindexfamilyrel]{code:hemicubic}{Hemicubic code} --- Application of generalized distance balancing \NoCaseChange{\protect\cite{cite684}} to hemicubic codes using an asymptotically good classical code of length \(t\) yields \(O( 1/(\log(n) t^2) )\) soundness and \flmRefsHyperref{ref65}{order} \(\Theta(\sqrt{n}t)\) distance while maintaining locality scaling and at the expense of a dimension scaling as \flmRefsHyperref{ref65}{order} \(\Theta(t^2)\) \NoCaseChange{\protect\cite{cite2990}}.
\item\relax
\flmRefsHyperref[eczindexfamilyrel]{code:hypersphere_product}{Hypersphere product code} --- Application of generalized distance balancing \NoCaseChange{\protect\cite{cite684}} to hypersphere product codes using an asymptotically good classical code of length \(t\) yields \(O( 1/(\log(n)^2 t^2) )\) soundness and \flmRefsHyperref{ref65}{order} \(\Theta(\sqrt{n}t)\) distance while maintaining locality scaling and at the expense of a dimension scaling as \flmRefsHyperref{ref65}{order} \(\Theta(t^2)\) \NoCaseChange{\protect\cite{cite2990}}.
\item\relax
\flmRefsHyperref[eczindexfamilyrel]{code:balanced_product}{Balanced product (BP) code} --- Applying distance balancing to the explicit subsystem balanced-product family of Ref. \NoCaseChange{\protect\cite{cite434}} yields an LDPC code family with \(k \in \Theta(n^{4/5})\) and \(d \in \Omega(n^{3/5})\).
\end{eczvaluelist}
\eczhbkcontributors{ \eczhuVVA }
\endeczcode

\eczcode{ea_galois_into_galois}{EA Galois-qudit code}{}

\codefieldsection{Kingdom root code}
for the \flmRefsHyperref{kingdom:galois_into_galois}{Galois-qudit Kingdom}.
\codefieldsection{Description}
Galois-qudit code designed to utilize pre-shared entanglement between sender and receiver.

\codefieldsection{Parent}
\begin{eczvaluelist}
\item\relax
\flmRefsHyperref[eczindexfamilyrel]{code:eaqecc}{Entanglement-assisted (EA) QECC}\end{eczvaluelist}
\codefieldsection{Children}
\begin{eczvaluelist}
\item\relax
\flmRefsHyperref[eczindexfamilyrel]{code:ea_qubits_into_qubits}{EA qubit code} --- EA Galois-qudit codes reduce to EA qubit codes for \(q=2\).
\item\relax
\flmRefsHyperref[eczindexfamilyrel]{code:ea_galois_stabilizer}{EA Galois-qudit stabilizer code}\item\relax
\flmRefsHyperref[eczindexfamilyrel]{code:ea_mds}{EA MDS code}\end{eczvaluelist}
\codefieldsection{Cousin}
\begin{eczvaluelist}
\item\relax
\flmRefsHyperref[eczindexfamilyrel]{code:galois_into_galois}{Galois-qudit code} --- EA Galois-qudit codes utilize additional ancillary Galois qudits in a pre-shared entangled state, but reduce to ordinary Galois-qudit codes when said qudits are interpreted as noiseless physical qudits.
\end{eczvaluelist}
\eczhbkcontributors{ \eczhuVVA }
\endeczcode

\eczcode{ea_galois_stabilizer}{EA Galois-qudit stabilizer code}{~\NoCaseChange{\protect\cite{cite4606}}}
\codefieldsection{Description}
A Galois-qudit stabilizer code constructed using a variation of the stabilizer formalism designed to utilize pre-shared entanglement between sender and receiver.
A code is typically denoted as \(\llbracket n,k;e\rrbracket _q\) or \(\llbracket n,k,d;e\rrbracket _q\), where \(d\) is the distance of the EA code and \(e\) is the number of required pre-shared maximally entangled Galois-qudit states.

\codefieldsection{Decoding}
\begin{eczvaluelist}
\item\relax Syndrome extraction and computation based on classical additive codes \NoCaseChange{\protect\cite{cite4607}}.
\end{eczvaluelist}
\codefieldsection{Parent}
\begin{eczvaluelist}
\item\relax
\flmRefsHyperref[eczindexfamilyrel]{code:ea_galois_into_galois}{EA Galois-qudit code}\end{eczvaluelist}
\codefieldsection{Children}
\begin{eczvaluelist}
\item\relax
\flmRefsHyperref[eczindexfamilyrel]{code:eastab}{EA qubit stabilizer code} --- EA Galois-qudit stabilizer codes reduce to EA qubit stabilizer codes for \(q=2\).
\item\relax
\flmRefsHyperref[eczindexfamilyrel]{code:ea_quantum_lcd}{EA quantum LCD code}\item\relax
\flmRefsHyperref[eczindexfamilyrel]{code:maximal_entanglement_galois_stabilizer}{Maximal-entanglement EA Galois-qudit stabilizer code}\end{eczvaluelist}
\codefieldsection{Cousins}
\begin{eczvaluelist}
\item\relax
\flmRefsHyperref[eczindexfamilyrel]{code:galois_stabilizer}{Galois-qudit stabilizer code} --- EA Galois-qudit stabilizer codes utilize additional ancillary Galois-qudits in a pre-shared entangled state, but reduce to Galois-qudit stabilizer codes when said qudits are interpreted as noiseless physical qudits. Pure Galois-qudit codes can be used to make EA Galois-qudit stabilizer codes \NoCaseChange{\protect\cite{cite4606}\protect\cite[{Thm. 10}]{cite545}}.
\item\relax
\flmRefsHyperref[eczindexfamilyrel]{code:galois_grs}{Galois-qudit GRS code} --- Galois-qudit GRS codes can be used to construct EA Galois-qudit stabilizer codes \NoCaseChange{\protect\cite{cite1911,cite4608}}.
\item\relax
\flmRefsHyperref[eczindexfamilyrel]{code:quantum_concatenated}{Concatenated quantum code} --- Concatenated EA Galois-qudit stabilizer codes have been studied \NoCaseChange{\protect\cite{cite2702,cite2703}}.
\item\relax
\flmRefsHyperref[eczindexfamilyrel]{code:ag}{Algebraic-geometry (AG) code} --- Certain AG codes can be used to construct EA Galois-qudit stabilizer codes \NoCaseChange{\protect\cite{cite1723}}.
\end{eczvaluelist}
\eczhbkcontributors{ \eczhuVVA }
\endeczcode

\eczcode{ea_mds}{EA MDS code}{~\NoCaseChange{\protect\cite{cite1430,cite4609,cite545}}}
\codefieldsection{Description}
EA Galois-qudit code whose parameters make the EAQECC Singleton bound \NoCaseChange{\protect\cite[{Thm. 6}]{cite545}} become an equality.

The original EAQECC Singleton bound \NoCaseChange{\protect\cite{cite1429,cite1430}} was shown to be erroneous \NoCaseChange{\protect\cite{cite3647}} and corrected in Ref. \NoCaseChange{\protect\cite{cite545}}.

\codefieldsection{Parent}
\begin{eczvaluelist}
\item\relax
\flmRefsHyperref[eczindexfamilyrel]{code:ea_galois_into_galois}{EA Galois-qudit code}\end{eczvaluelist}
\codefieldsection{Cousins}
\begin{eczvaluelist}
\item\relax
\flmRefsHyperref[eczindexfamilyrel]{code:quantum_mds}{Quantum maximum-distance-separable (MDS) code} --- EA MDS codes are entanglement-assisted versions of quantum MDS codes.
\item\relax
\flmRefsHyperref[eczindexfamilyrel]{code:mds}{Maximum distance separable (MDS) code} --- MDS codes give rise to families of EA Galois-qudit codes that saturate the original (erroneous) EAQECC Singleton bound \NoCaseChange{\protect\cite{cite1928}}.
\item\relax
\flmRefsHyperref[eczindexfamilyrel]{code:eastab}{EA qubit stabilizer code} --- There exist concatenated EA qubit stabilizer codes that saturate the EA quantum Singleton bound \NoCaseChange{\protect\cite{cite3608}}.
\end{eczvaluelist}
\eczhbkcontributors{ \eczhuVVA }
\endeczcode

\eczcode{ea_quantum_lcd}{EA quantum LCD code}{~\NoCaseChange{\protect\cite{cite1911}}}
\codefieldsection{Description}
An EA Galois-qudit stabilizer code constructed from an LCD code.
This family includes the first asymptotically good EA Galois-qudit codes.

\codefieldsection{Rate}
Asymptotically good maximal-entanglement EA Galois-qudit stabilizer codes can be constructed from LCD codes \NoCaseChange{\protect\cite{cite1911}}.
\codefieldsection{Parent}
\begin{eczvaluelist}
\item\relax
\flmRefsHyperref[eczindexfamilyrel]{code:ea_galois_stabilizer}{EA Galois-qudit stabilizer code}\end{eczvaluelist}
\codefieldsection{Cousins}
\begin{eczvaluelist}
\item\relax
\flmRefsHyperref[eczindexfamilyrel]{code:lcd}{Linear code with complementary dual (LCD)} --- Asymptotically good maximal-entanglement EA Galois-qudit stabilizer codes can be constructed from LCD codes \NoCaseChange{\protect\cite{cite1911}}.
\item\relax
\flmRefsHyperref[eczindexfamilyrel]{code:maximal_entanglement_galois_stabilizer}{Maximal-entanglement EA Galois-qudit stabilizer code} --- Asymptotically good maximal-entanglement EA Galois-qudit stabilizer codes can be constructed from LCD codes \NoCaseChange{\protect\cite{cite1911}}.
\end{eczvaluelist}
\eczhbkcontributors{ \eczhuVVA }
\endeczcode

\eczcode{expander_lifted_product}{Expander LP code}{~\NoCaseChange{\protect\cite{cite184}}}
\codefieldsection{Description}
Family of \(G\)-lifted product codes constructed using two classical \flmRefsHyperref{code:expander}{expander codes}, equivalently two regular \flmRefsHyperref{code:tanner}{Tanner codes} defined on the same expander graph \NoCaseChange{\protect\cite{cite74}}. For certain parameters, this construction yields the first asymptotically good QLDPC codes. Classical codes resulting from the same lifted-product complexes are one of the first two families of \(c^3\)-LTCs \NoCaseChange{\protect\cite{cite184}}.

An expander lifted-product code family is constructed as follows. First, take the Cayley graph of a finite group \(G\).
Second, take the double cover of the graph, resulting in a graph that satisfies the requirements of participating in a \(G\)-lifted product (i.e., the resulting graph is a free \({\mathbb{F}}_q G\)-module). Third, create two \flmRefsHyperref{code:tanner}{Tanner codes} on that graph, in which parity-check supports are defined by the graph and the local constraints are specified by two short classical codes (chosen randomly in the original proof). Fourth, take the \(G\)-lifted product of those two Tanner codes.

The small classical codes used in the construction of good QLDPC codes are required to have a certain product-expansion property \NoCaseChange{\protect\cite[{Lemma 10 in}]{cite184}}; it is proven that random codes satisfy said property in the thermodynamic limit.

\codefieldsection{Protection}
Code performance strongly depends on \(G\). Certain non-Abelian groups yield asymptotically good QLDPC codes with parameters \(\llbracket n,k=\Theta(n),d=\Theta(n)\rrbracket \) \NoCaseChange{\protect\cite{cite184}}. For cyclic Abelian groups \(G=\mathbb{Z}_{\ell}\) with \(\ell=\Theta(n/\log n)\), quasi-cyclic expander LP codes yield families with parameters \(\llbracket n,k=\Theta(\log n),d=\Theta(n/\log n)\rrbracket \) \NoCaseChange{\protect\cite{cite674}}.
\codefieldsection{Rate}
Expander lifted-product codes for non-Abelian groups include the first examples \NoCaseChange{\protect\cite{cite184}} of (asymptotically) \textit{good QLDPC codes}, i.e., codes with asymptotically constant rate and linear distance. For cyclic Abelian groups \(G=\mathbb{Z}_{\ell}\) with \(\ell=\Theta(n/\log n)\), quasi-cyclic expander LP codes yield families with parameters \(\llbracket n,k=\Theta(\log n),d=\Theta(n/\log n)\rrbracket \) \NoCaseChange{\protect\cite{cite674}}. Related balanced-product reformulations and other explicit Abelian LP constructions appear in \NoCaseChange{\protect\cite{cite434,cite1557}}.
\codefieldsection{Gates}
\begin{eczvaluelist}
\item\relax Certain qubit expander LP codes can admit a cup product structure and can thus have logical gates in the \flmTerm{term}{ref694}{}{Clifford hierarchy} implemented by constant-depth Clifford circuits \NoCaseChange{\protect\cite{cite1517}}.
\end{eczvaluelist}
\codefieldsection{Decoding}
\begin{eczvaluelist}
\item\relax Linear-time decoder \NoCaseChange{\protect\cite{cite4124}}.
\item\relax Logarithmic-time subroutine \NoCaseChange{\protect\cite{cite2191}}.
\end{eczvaluelist}
\codefieldsection{Notes}
\begin{eczvaluelist}
\item\relax Construction outlined in talk by \flmHref{https://www.youtube.com/watch?v=k7LuOiOBYyQ}{R. O'Donnell}.
\item\relax Popular summary in \flmHref{https://www.quantamagazine.org/qubits-can-be-as-safe-as-bits-researchers-show-20220106}{Quanta Magazine}.
\end{eczvaluelist}
\codefieldsection{Parent}
\begin{eczvaluelist}
\item\relax
\flmRefsHyperref[eczindexfamilyrel]{code:lifted_product}{Lifted-product (LP) code}\end{eczvaluelist}
\codefieldsection{Cousins}
\begin{eczvaluelist}
\item\relax
\flmRefsHyperref[eczindexfamilyrel]{code:good_qldpc}{Good QLDPC code} --- Lifted products of certain classical Tanner codes are the first asymptotically good QLDPC codes.
\item\relax
\flmRefsHyperref[eczindexfamilyrel]{code:q-ary_ltc}{\(q\)-ary linear LTC} --- Classical codes resulting from the expander lifted-product construction are one of the first two families of \(c^3\)-LTCs \NoCaseChange{\protect\cite{cite184}}.
\item\relax
\flmRefsHyperref[eczindexfamilyrel]{code:expander}{Expander code} --- Expander LP codes are lifted products of expander codes with different local codes \NoCaseChange{\protect\cite{cite184}}.
\item\relax
\flmRefsHyperref[eczindexfamilyrel]{code:random}{Random code} --- Expander lifted-product codes are quantum CSS codes that utilize short classical codes in their construction which need to satisfy some properties \NoCaseChange{\protect\cite[{Lemma 10}]{cite184}}. It is shown that such codes exist, but they are not explicitly constructed. Such codes can be obtained by repeated random sampling or by performing a search of all codes of desired length. Nevertheless, since the length of the desired short codes does not scale with \(n\), this construction is effectively explicit.
\item\relax
\flmRefsHyperref[eczindexfamilyrel]{code:topological}{Topological code} --- Expander lifted-product codes are expected to realize topological quantum spin glass order \NoCaseChange{\protect\cite{cite3162}}.
\item\relax
\flmRefsHyperref[eczindexfamilyrel]{code:quantum_tanner}{Quantum Tanner code} --- Quantum Tanner codes are an attempt to construct asymptotically good QLDPC codes that are similar to but simpler than expander lifted-product codes; see Ref. \NoCaseChange{\protect\cite{cite4124}} for connection between the codes.
\item\relax
\flmRefsHyperref[eczindexfamilyrel]{code:abelian_lifted_product}{Abelian LP code} --- For cyclic groups \(G=\mathbb{Z}_{\ell}\) with \(\ell=\Theta(n/\log n)\), quasi-cyclic expander LP codes yield families with parameters \(\llbracket n,k=\Theta(\log n),d=\Theta(n/\log n)\rrbracket \) \NoCaseChange{\protect\cite{cite674}}. Related balanced-product reformulations and other explicit Abelian LP constructions appear in \NoCaseChange{\protect\cite{cite434,cite1557}}.
\end{eczvaluelist}
\eczhbkcontributors{ Pavel Panteleev, \eczhuVVA }
\endeczcode

\eczcode{galois_fqrs}{Folded quantum RS (FQRS) code}{~\NoCaseChange{\protect\cite{cite495}}}
\codefieldsection{Description}
CSS code on \(q^m\)-dimensional Galois-qudits that is constructed from folded RS (FRS) codes (i.e., an RS code whose coordinates have been grouped together) via the Galois-qudit CSS construction.
This code is used to construct Singleton-bound approaching approximate quantum codes.

More technically, an \(m\)-folded quantum RS code is a member of the \(\llbracket n/m, R \cdot n/m, d/m\rrbracket _{q^m}\) CSS code family for any \(0<R<1\).
See \NoCaseChange{\protect\cite[{Defn. 3.8}]{cite495}} for an expression of the codewords.
A folded quantum generalized RS (GRS) code can be defined in similar fashion from GRS codes \NoCaseChange{\protect\cite[{Sec. 3}]{cite495}}.

\codefieldsection{Protection}
For every \(\gamma>0\) and \(0<R<1\), there are folded quantum RS code families of rate \(R\) with local dimension \(q=n^{O(1/\gamma^2)}\) that are \(\llparenthesis 1-R-\gamma)/2,n^{O(1/\gamma)})\)-quantum list-decodable \NoCaseChange{\protect\cite[{Thm. 3.9}]{cite495}}.

\codefieldsection{Decoding}
\begin{eczvaluelist}
\item\relax Efficiently \(\llparenthesis 1-R-\gamma)/2,n^{O(1/\gamma)})\)-quantum list-decodable codes exist for every fixed \(\gamma>0\) and rate \(0<R<1\) \NoCaseChange{\protect\cite[{Thm. 3.9}]{cite495}}.
\end{eczvaluelist}
\codefieldsection{Parent}
\begin{eczvaluelist}
\item\relax
\flmRefsHyperref[eczindexfamilyrel]{code:galois_css}{Galois-qudit CSS code} --- Folding a quantum polynomial code on \(q\)-dimensional Galois qudits yields an FQRS code on \(q^m\)-dimensional Galois qudits.
\end{eczvaluelist}
\codefieldsection{Cousins}
\begin{eczvaluelist}
\item\relax
\flmRefsHyperref[eczindexfamilyrel]{code:folded_reed_solomon}{Folded RS (FRS) code} --- Folded quantum RS codes are quantum analogues of folded RS codes.
\item\relax
\flmRefsHyperref[eczindexfamilyrel]{code:generalized_reed_solomon}{Generalized RS (GRS) code} --- A folded quantum generalized RS (GRS) code can be constructed in similar fashion from GRS codes as FQRS codes are constructed from FRS codes \NoCaseChange{\protect\cite[{Sec. 3}]{cite495}}.
\item\relax
\flmRefsHyperref[eczindexfamilyrel]{code:quantum_singleton}{Singleton-bound approaching AQECC} --- Singleton-bound approaching AQECCs are built using folded quantum Reed-Solomon (FQRS) codes \NoCaseChange{\protect\cite{cite495}}.
\item\relax
\flmRefsHyperref[eczindexfamilyrel]{code:galois_polynomial}{Galois-qudit RS code} --- A FQRS code with no extra grouping (\(m=1\)) reduces to a Galois-qudit RS code that is CSS.
\end{eczvaluelist}
\eczhbkcontributors{ Sam Gunn, \eczhuVVA }
\endeczcode

\eczcode{galois_bch}{Galois-qudit BCH code}{~\NoCaseChange{\protect\cite{cite4610,cite1838,cite4611,cite4612,cite4144,cite4613,cite865,cite4614,cite4615}\protect\cite[{Ch. 4}]{cite872}}}
\codefieldsection{Description}
True Galois-qudit stabilizer code constructed from BCH codes via either the Hermitian construction or the Galois-qudit CSS construction.
Parameters can be improved by applying \flmRefsCref{ref863} \NoCaseChange{\protect\cite{cite864}}, e.g., as in Ref. \NoCaseChange{\protect\cite{cite865}}.

\codefieldsection{Notes}
\begin{eczvaluelist}
\item\relax See Ref. \NoCaseChange{\protect\cite{cite2996}} for an overview of quantum BCH codes.
\end{eczvaluelist}
\codefieldsection{Parent}
\begin{eczvaluelist}
\item\relax
\flmRefsHyperref[eczindexfamilyrel]{code:galois_true_stabilizer}{True Galois-qudit stabilizer code} --- Galois-qudit BCH codes can be constructed via the CSS construction or the Hermitian construction.
\end{eczvaluelist}
\codefieldsection{Child}
\begin{eczvaluelist}
\item\relax
\flmRefsHyperref[eczindexfamilyrel]{code:quantum_bch}{Qubit BCH code} --- Galois-qudit BCH codes for \(q=2\) reduce to qubit BCH codes.
\end{eczvaluelist}
\codefieldsection{Cousins}
\begin{eczvaluelist}
\item\relax
\flmRefsHyperref[eczindexfamilyrel]{code:q-ary_bch}{Bose–Chaudhuri–Hocquenghem (BCH) code} --- Galois-qudit BCH codes are quantum analogues of q-ary BCH codes.
\item\relax
\flmRefsHyperref[eczindexfamilyrel]{code:galois_css}{Galois-qudit CSS code} --- Galois-qudit BCH codes can be constructed via the CSS construction or the Hermitian construction.
\item\relax
\flmRefsHyperref[eczindexfamilyrel]{code:stabilizer_over_gfqsq}{Hermitian Galois-qudit code} --- Galois-qudit BCH codes can be constructed via the CSS construction or the Hermitian construction.
\item\relax
\flmRefsHyperref[eczindexfamilyrel]{code:asymmetric_qecc}{Asymmetric quantum code (AQC)} --- Asymmetric quantum BCH codes have been constructed \NoCaseChange{\protect\cite{cite2610,cite2652,cite2653}\protect\cite[{Lemma 4.4}]{cite1354}\protect\cite[{Sec. 17.3}]{cite872}}, including subsystem BCH codes \NoCaseChange{\protect\cite{cite2612}\protect\cite[{Sec. 9.3}]{cite872}}.
\item\relax
\flmRefsHyperref[eczindexfamilyrel]{code:galois_subsystem_stabilizer}{Subsystem Galois-qudit stabilizer code} --- Asymmetric quantum BCH codes have been constructed \NoCaseChange{\protect\cite{cite2610,cite2652,cite2653}\protect\cite[{Lemma 4.4}]{cite1354}\protect\cite[{Sec. 17.3}]{cite872}}, including subsystem BCH codes \NoCaseChange{\protect\cite{cite2612}\protect\cite[{Sec. 9.3}]{cite872}}.
\item\relax
\flmRefsHyperref[eczindexfamilyrel]{code:general_qldpc}{QLDPC code} --- Some Galois-qudit BCH codes are QLDPC \NoCaseChange{\protect\cite{cite4616}\protect\cite[{Ch. 16}]{cite872}}.
\item\relax
\flmRefsHyperref[eczindexfamilyrel]{code:quasi_cyclic_qldpc}{Quasi-cyclic QLDPC (QC-QLDPC) code} --- Some Galois-qudit BCH codes are QC-QLDPC \NoCaseChange{\protect\cite[{Ch. 16}]{cite872}}.
\end{eczvaluelist}
\eczhbkcontributors{ \eczhuVVA }
\endeczcode

\eczcode{galois_into_galois}{Galois-qudit code}{}
\codefieldsection{Alternative Names}
\begin{eczvaluelist}
\item\relax \(\mathbb{F}_q\)-qudit code
\item\relax \(\mathbb{F}_q\)-qudit code
\item\relax Galois-qudit subspace code
\end{eczvaluelist}
\eczhIndexCodeAliasName{galois_into_galois}{\(\mathbb{F}_q\)-qudit code}
\eczhIndexCodeAliasName{galois_into_galois}{\(\mathbb{F}_q\)-qudit code}
\eczhIndexCodeAliasName{galois_into_galois}{Galois-qudit subspace code}

\codefieldsection{Kingdom root code}
for the \flmRefsHyperref{kingdom:galois_into_galois}{Galois-qudit Kingdom}.
\codefieldsection{Description}
Encodes \(K\)-dimensional Hilbert space into a \(q^n\)-dimensional (\(n\)-qudit) Hilbert space, with canonical qudit states \(|k\rangle\) labeled by elements \(k\) of the \textit{Galois field} \(\mathbb{F}_q\) and with \(q\) being a power of a prime \(p\).

Codes can be denoted as \(\llparenthesis n,K\rrparenthesis _q\) or \(\llparenthesis n,K,d\rrparenthesis _q\), whenever the code's distance \(d\) is defined.
This notation differentiates between Galois-qudit and \(\llparenthesis n,K,d\rrparenthesis _{\mathbb{Z}_q}\) modular-qudit codes, although the same notation is usually used for both.

There exists an analogue of the Wigner function for Galois qudits \NoCaseChange{\protect\cite{cite4524,cite4525}}.

\codefieldsection{Protection}
An \(\llparenthesis n,K,d\rrparenthesis _q\) code with distance \(d\) detects errors acting on up to \(d-1\) Galois qudits, corrects erasure errors on up to \(d-1\) Galois qudits, or corrects errors acting on up to \(\lfloor (d-1)/2 \rfloor\) Galois qudits.

\subsection{Galois-qudit Pauli-string error basis}
A convenient and often considered error set is the Galois-qudit analogue of the Pauli string set for \flmRefsHyperref{code:qubits_into_qubits}{qubit} codes.

\begin{defterm}{Galois-qudit Pauli strings}\label{ref4617}\label{ref4618}
For a single Galois qudit, this set consists of products of \(X\)-type and \(Z\)-type operators labeled by elements \(\beta \in \mathbb{F}_q\), which act on computational basis states \(|\gamma\rangle\) for \(\gamma\in \mathbb{F}_q\) as
\flmMathEnvironment{align}{}{
  X_{\beta}\left|\gamma\right\rangle =\left|\gamma+\beta\right\rangle \,\,\text{ and }\,\,Z_{\beta}\left|\gamma\right\rangle =e^{i\frac{2\pi}{p}\text{tr}(\beta\gamma)}\left|\gamma\right\rangle~,
}
where \(\text{tr}\) is the \flmRefsHyperref{ref33}{field trace}.
For multiple Galois qudits, error set elements are tensor products of elements of the single-qudit error set.
Tensor products of \(X\) (\(Z\)) Galois-qudit Paulis acting on different qudits are called \(X\)\textit{-type} (\(Z\)\textit{-type}) Galois-qudit Pauli strings.
Combining the \(X\)-type and \(Z\)-type strings with a \(p\)th root of unity forms a group called the \textit{Galois-qudit Pauli group} on \(n\) Galois qudits.
\end{defterm}

The Galois-qudit Pauli error set is a unitary basis for linear operators on the multi-qudit Hilbert space that is orthonormal under the Hilbert-Schmidt inner product; it is a \flmRefsHyperref{ref2812}{nice error basis}. The distance associated with this set is often the minimum weight of a Galois qudit Pauli string that implements a nontrivial logical operation in the code.

\codefieldsection{Gates}
\begin{eczvaluelist}
\item\relax The normalizer of the \flmRefsHyperref{ref4618}{Galois-qudit Pauli group}  is the Galois-qudit Clifford group \NoCaseChange{\protect\cite{cite4619}\protect\cite[{Ch. 8}]{cite4559}}.
\end{eczvaluelist}
\codefieldsection{Decoding}
\begin{eczvaluelist}
\item\relax For few-qudit codes (\(n\) is small), decoding can be based on a lookup table. For infinite code families, the size of such a table scales exponentially with \(n\), so approximate decoding algorithms scaling polynomially with \(n\) have to be used. The decoder determining the most likely error given a noise channel is called the \textit{maximum-likelihood} (ML) decoder.
\item\relax RL-on-Greedy decoder based on reinforcement learning \NoCaseChange{\protect\cite{cite4620}}.
\end{eczvaluelist}
\codefieldsection{Notes}
\begin{eczvaluelist}
\item\relax Introduction to Galois qudits by \flmHref{https://www.qec14.ethz.ch/slides/DanielGottesman.pdf}{Gottesman}.
\item\relax CodingTheory Julia software library \NoCaseChange{\protect\cite{cite1906}}.
\item\relax See \NoCaseChange{\protect\cite[{Ch. 8}]{cite4559}} for a side-by-side introduction to modular and Galois qudits.
\end{eczvaluelist}
\codefieldsection{Parents}
\begin{eczvaluelist}
\item\relax
\flmRefsHyperref[eczindexfamilyrel]{code:block_quantum}{Block quantum code} --- Galois-qudit codes are block quantum codes with \(\Sigma=\mathbb{F}_q\).
\item\relax
\flmRefsHyperref[eczindexfamilyrel]{code:qecc_finite}{Finite-dimensional quantum error-correcting code}\item\relax
\flmRefsHyperref[eczindexfamilyrel]{code:group_quantum}{Group-based quantum code} --- A Galois qudit for \(q=p^m\) can be decomposed into a Kronecker product of \(m\) modular qudits \NoCaseChange{\protect\cite{cite696}}; see \NoCaseChange{\protect\cite[{Sec. 5.3}]{cite697}}.
Interpreted this way, Galois-qudit codes are group quantum codes whose physical spaces are constructed using Galois fields \(\mathbb{F}_q\) as groups. More general versions of such qudits can be valued in a Galois ring \NoCaseChange{\protect\cite{cite4621}}, over which there also exists a Fourier transform \NoCaseChange{\protect\cite{cite4622}}.

\end{eczvaluelist}
\codefieldsection{Children}
\begin{eczvaluelist}
\item\relax
\flmRefsHyperref[eczindexfamilyrel]{code:qubits_into_qubits}{Qubit code} --- Galois-qudit quantum codes for \(q=2\) correspond to qubit codes.
\item\relax
\flmRefsHyperref[eczindexfamilyrel]{code:galois_non_stabilizer}{Galois-qudit USt code}\end{eczvaluelist}
\codefieldsection{Cousins}
\begin{eczvaluelist}
\item\relax
\flmRefsHyperref[eczindexfamilyrel]{code:qudits_into_qudits}{Modular-qudit code} --- A Galois qudit for \(q=p^m\) can be decomposed into a Kronecker product of \(m\) modular qudits; see \NoCaseChange{\protect\cite{cite696,cite398,cite698,cite699,cite700}\protect\cite[{Sec. 5.3}]{cite697}}.
The two coincide when \(q\) is prime, and reduce to qubits when \(q=2\).
However, Pauli matrices for the two types of qudits are defined differently.
See \NoCaseChange{\protect\cite[{Ch. 8}]{cite4559}} for a side-by-side introduction to modular and Galois qudits.

\item\relax
\flmRefsHyperref[eczindexfamilyrel]{code:q-ary_digits_into_q-ary_digits}{\(q\)-ary code} --- Galois-qudit codes are quantum counterparts of \(q\)-ary codes.
\item\relax
\flmRefsHyperref[eczindexfamilyrel]{code:ea_galois_into_galois}{EA Galois-qudit code} --- EA Galois-qudit codes utilize additional ancillary Galois qudits in a pre-shared entangled state, but reduce to ordinary Galois-qudit codes when said qudits are interpreted as noiseless physical qudits.
\item\relax
\flmRefsHyperref[eczindexfamilyrel]{code:subsystem_galois_into_galois}{Subsystem Galois-qudit code} --- Subsystem Galois-qudit codes reduce to (subspace) Galois-qudit codes when there is no gauge subsystem.
\end{eczvaluelist}
\eczhbkcontributors{ \eczhuVVA }
\endeczcode

\eczcode{galois_color}{Galois-qudit color code}{~\NoCaseChange{\protect\cite{cite4623}}}
\codefieldsection{Alternative Names}
\begin{eczvaluelist}
\item\relax \(\mathbb{F}_q\)-qudit color code
\end{eczvaluelist}
\eczhIndexCodeAliasName{galois_color}{\(\mathbb{F}_q\)-qudit color code}
\codefieldsection{Description}
Extension of the color code to 2D lattices of Galois qudits.

\codefieldsection{Parents}
\begin{eczvaluelist}
\item\relax
\flmRefsHyperref[eczindexfamilyrel]{code:galois_css}{Galois-qudit CSS code}\item\relax
\flmRefsHyperref[eczindexfamilyrel]{code:2d_stabilizer}{2D lattice stabilizer code}\item\relax
\flmRefsHyperref[eczindexfamilyrel]{code:generalized_homological_product_css}{Generalized homological-product CSS code}\item\relax
\flmRefsHyperref[eczindexfamilyrel]{code:topological_abelian}{Abelian topological code} --- A Galois qudit for \(q=p^m\) can be decomposed into a Kronecker product of \(m\) modular qudits \NoCaseChange{\protect\cite{cite696,cite398,cite698,cite699,cite700}\protect\cite[{Sec. 5.3}]{cite697}}. Galois-qudit color codes yield Abelian quantum-double codes with Abelian-group topological order via this decomposition.
\item\relax
\flmRefsHyperref[eczindexfamilyrel]{code:quantum_double}{Quantum-double code} --- A Galois qudit for \(q=p^m\) can be decomposed into a Kronecker product of \(m\) modular qudits \NoCaseChange{\protect\cite{cite696,cite398,cite698,cite699,cite700}\protect\cite[{Sec. 5.3}]{cite697}}. Galois-qudit color codes yield Abelian quantum-double codes with Abelian-group topological order via this decomposition.
\end{eczvaluelist}
\codefieldsection{Child}
\begin{eczvaluelist}
\item\relax
\flmRefsHyperref[eczindexfamilyrel]{code:2d_color}{2D color code} --- Galois-qudit 2D color codes reduce to 2D color codes for \(q=2\).
\end{eczvaluelist}
\codefieldsection{Cousin}
\begin{eczvaluelist}
\item\relax
\flmRefsHyperref[eczindexfamilyrel]{code:quantum_double_abelian}{Abelian quantum-double stabilizer code} --- A Galois qudit for \(q=p^m\) can be decomposed into a Kronecker product of \(m\) modular qudits \NoCaseChange{\protect\cite{cite696,cite398,cite698,cite699,cite700}\protect\cite[{Sec. 5.3}]{cite697}}. Galois-qudit color codes yield Abelian quantum-double codes with Abelian-group topological order via this decomposition.
\end{eczvaluelist}
\eczhbkcontributors{ \eczhuVVA }
\endeczcode

\eczcode{galois_css}{Galois-qudit CSS code}{~\NoCaseChange{\protect\cite{cite3196,cite3337,cite3338,cite2813,cite4624,cite975,cite4625}}}
\codefieldsection{Alternative Names}
\begin{eczvaluelist}
\item\relax Euclidean code
\end{eczvaluelist}
\eczhIndexCodeAliasName{galois_css}{Euclidean code}
\codefieldsection{Description}
An \(\llbracket n,k,d\rrbracket _q \) Galois-qudit true stabilizer code admitting a set of stabilizer generators that
are either \(Z\)-type or \(X\)-type Galois-qudit Pauli strings.
Codes can be defined from chain complexes over \(\mathbb{F}_q\) via an extension of \flmRefsHyperref{ref683}{qubit CSS-to-homology correspondence} to Galois qudits.

The stabilizer generator matrix \NoCaseChange{\protect\cite[{Def. 2}]{cite4619}}, taking values from \(\mathbb{F}_q\), is of the form
\flmMathEnvironment{align}{}{
H=\begin{pmatrix}0 & H_{Z}\\
H_{X} & 0
\end{pmatrix}\label{ref4626}
}
such that the rows of the two blocks must be orthogonal
\flmMathEnvironment{align}{}{
H_X H_Z^T=0~.\label{ref4627}
}
The above condition guarantees that the \(X\)-stabilizer generators, defined in the \flmRefsHyperref{ref873}{Galois symplectic representation} as rows of \(H_X\), commute with the \(Z\)-stabilizer generators associated with \(H_Z\).

Encoding is based on two related \flmRefsHyperref{code:q-ary_linear}{\(q\)-ary linear codes},
an \([n,k_X,d_X]_q \) code \(C_X\) and \([n,k_Z,d_Z]_q \) code \(C_Z\),
satisfying \(C_X^\perp \subseteq C_Z\) \NoCaseChange{\protect\cite[{Thm. 27.4.2}]{cite2024}}.
The resulting CSS code has \(k=k_X+k_Z-n\) logical Galois qudits and distance \(d\geq\min\{d_X,d_Z\}\).
The \(H_X\) (\(H_Z\)) block of \(H\) \eqref{ref4626} is the parity-check matrix of the code \(C_X\) (\(C_Z\)). The requirement \(C_X^\perp \subseteq C_Z\) guarantees \eqref{ref4627}.
Specializing to the case when \(C_Z=[n,k,d]_q\) is dual-containing yields a \(\llbracket n,2k-n,\geq d_Z\rrbracket _q\) \textit{self-dual Galois-qudit CSS code} with \(C_X = C_Z\).
When the field is a quadratic extension, such as \(\mathbb{F}_4/\mathbb{F}_2\), a Galois-qudit CSS code is \textit{Hermitian self-dual} when its \(Z\)-type check code is the Frobenius conjugate of its \(X\)-type check code, \(C_Z=\overline{C_X}\).
Basis states for the code are, for coset representatives \(\gamma \in C_X/C_Z^\perp\),
\flmMathEnvironment{align}{}{
|\gamma + C_Z^\perp \rangle = \frac{1}{\sqrt{|C_Z^\perp|}} \sum_{\eta \in C_Z^\perp} |\gamma + \eta\rangle.
}

Galois-qudit CSS codes can also be understood in terms of graphs via the \textit{reflexive stabilizer} framework, which also allows one to define a code for a given set of Pauli errors \NoCaseChange{\protect\cite{cite4628}}.

\codefieldsection{Protection}
Detects errors on \(d-1\) qudits, corrects errors on \(\left\lfloor (d-1)/2 \right\rfloor\) qudits.
A quantum version of the Griesmer bound has been derived for Galois-qudit CSS codes \NoCaseChange{\protect\cite{cite1864}}.

An \(\llbracket n,k,d\rrbracket _q\) CSS code can be propagated to an \(\llbracket n-2,k,d-1\rrbracket _q\) code \NoCaseChange{\protect\cite{cite4629}}.

\codefieldsection{Encoding}
\begin{eczvaluelist}
\item\relax Fault-tolerant encoding \NoCaseChange{\protect\cite{cite4630}}.
\end{eczvaluelist}
\codefieldsection{Parents}
\begin{eczvaluelist}
\item\relax
\flmRefsHyperref[eczindexfamilyrel]{code:galois_true_stabilizer}{True Galois-qudit stabilizer code} --- The Galois-qudit CSS construction yields a true stabilizer code \NoCaseChange{\protect\cite[{Sec. 8.2.2}]{cite398}}.
\item\relax
\flmRefsHyperref[eczindexfamilyrel]{code:css}{Calderbank-Shor-Steane (CSS) stabilizer code}\end{eczvaluelist}
\codefieldsection{Children}
\begin{eczvaluelist}
\item\relax
\flmRefsHyperref[eczindexfamilyrel]{code:qubit_css}{Qubit CSS code} --- Galois-qudit CSS codes for \(q=2\) are qubit CSS codes.
\item\relax
\flmRefsHyperref[eczindexfamilyrel]{code:qudit_triorthogonal}{Prime-qudit triorthogonal code}\item\relax
\flmRefsHyperref[eczindexfamilyrel]{code:quantum_secret_sharing}{Approximate secret-sharing code} --- The code required to construct this code must be a \flmRefsHyperref{ref811}{non-degenerate} Galois-qudit CSS code.
\item\relax
\flmRefsHyperref[eczindexfamilyrel]{code:quantum_singleton}{Singleton-bound approaching AQECC}\item\relax
\flmRefsHyperref[eczindexfamilyrel]{code:skew-cyclic_galois_css}{Skew-cyclic CSS code}\item\relax
\flmRefsHyperref[eczindexfamilyrel]{code:two_block_quantum}{Two-block CSS code}\item\relax
\flmRefsHyperref[eczindexfamilyrel]{code:galois_quad_residue}{Quantum quadratic-residue (QR) code}\item\relax
\flmRefsHyperref[eczindexfamilyrel]{code:binary_quantum_goppa}{Binary quantum Goppa code}\item\relax
\flmRefsHyperref[eczindexfamilyrel]{code:quantum_tamo_barg}{Quantum Tamo-Barg (QTB) code}\item\relax
\flmRefsHyperref[eczindexfamilyrel]{code:galois_fqrs}{Folded quantum RS (FQRS) code} --- Folding a quantum polynomial code on \(q\)-dimensional Galois qudits yields an FQRS code on \(q^m\)-dimensional Galois qudits.
\item\relax
\flmRefsHyperref[eczindexfamilyrel]{code:balanced_product}{Balanced product (BP) code}\item\relax
\flmRefsHyperref[eczindexfamilyrel]{code:distance_balanced}{Distance-balanced code}\item\relax
\flmRefsHyperref[eczindexfamilyrel]{code:galois_color}{Galois-qudit color code}\item\relax
\flmRefsHyperref[eczindexfamilyrel]{code:galois_topological}{Galois-qudit surface code}\end{eczvaluelist}
\codefieldsection{Cousins}
\begin{eczvaluelist}
\item\relax
\flmRefsHyperref[eczindexfamilyrel]{code:q-ary_linear}{Linear \(q\)-ary code} --- The Galois-qudit CSS construction uses two related \(q\)-ary linear codes, \(C_X\) and \(C_Z\).
\item\relax
\flmRefsHyperref[eczindexfamilyrel]{code:q-ary_cyclic}{Cyclic linear \(q\)-ary code} --- Galois CSS codes can be constructed using self-orthogonal \(q\)-ary cyclic codes \NoCaseChange{\protect\cite{cite1769}}.
\item\relax
\flmRefsHyperref[eczindexfamilyrel]{code:griesmer}{Griesmer code} --- A quantum version of the Griesmer bound has been derived for Galois-qudit CSS codes \NoCaseChange{\protect\cite{cite1864}} and Galois-qudit stabilizer codes \NoCaseChange{\protect\cite{cite1865}}.
\item\relax
\flmRefsHyperref[eczindexfamilyrel]{code:asymmetric_qecc}{Asymmetric quantum code (AQC)} --- Most known Galois-qudit AQC families are derived from the asymmetric Galois-qudit CSS construction \NoCaseChange{\protect\cite[{Thm. 27.5.2}]{cite2024}}, and assuming the MDS conjecture, all possible parameters for \flmRefsHyperref{ref672}{pure} Galois-qudit CSS asymmetric MDS codes have been determined \NoCaseChange{\protect\cite[{Thm. 27.5.3}]{cite2024}}.
\item\relax
\flmRefsHyperref[eczindexfamilyrel]{code:quantum_locally_recoverable}{Quantum locally recoverable code (QLRC)} --- A Galois-qudit CSS code is a QLRC of locality \(r\) if each qudit participates in at least one \(X\)-type and one \(Z\)-type stabilizer whose union of supports has weight \(\leq r\) \NoCaseChange{\protect\cite[{Corr. 34}]{cite812}}.
\item\relax
\flmRefsHyperref[eczindexfamilyrel]{code:qubit_stabilizer}{Qubit stabilizer code} --- Any \(\llbracket n,k,d\rrbracket \) qubit stabilizer code can be mapped to an \(\llbracket n,k,d\rrbracket _4\) Galois-qudit CSS code via the BLT mapping \NoCaseChange{\protect\cite[{Lemma 1}]{cite795}\protect\cite[{Lemma 1}]{cite1432}}: each stabilizer \(P = \bigotimes_j P_j\) generates an \(XX\)-type stabilizer via \(\mathcal{D}_X\) (\(I\mapsto II,\,X\mapsto XI,\,Z\mapsto IX,\,Y\mapsto XX\)) and a \(ZZ\)-type stabilizer via \(\mathcal{D}_Z\) (\(I\mapsto II,\,X\mapsto IZ,\,Z\mapsto ZI,\,Y\mapsto ZZ\)).
\item\relax
\flmRefsHyperref[eczindexfamilyrel]{code:galois_bch}{Galois-qudit BCH code} --- Galois-qudit BCH codes can be constructed via the CSS construction or the Hermitian construction.
\item\relax
\flmRefsHyperref[eczindexfamilyrel]{code:galois_duadic}{Quantum duadic code} --- Quantum duadic codes can be constructed via the CSS construction or the Hermitian construction.
\item\relax
\flmRefsHyperref[eczindexfamilyrel]{code:galois_reed_muller}{Galois-qudit quantum RM code} --- Galois-qudit RM codes admit a CSS subfamily built from nested GRM codes \( \mathrm{GRM}_q(v_1,m) \subseteq \mathrm{GRM}_q(v_2,m) \) \NoCaseChange{\protect\cite{cite828}}.
\item\relax
\flmRefsHyperref[eczindexfamilyrel]{code:galois_subsystem_css}{Subsystem Galois-qudit CSS code} --- Subsystem Galois-qudit CSS codes reduce to (subspace) Galois-qudit CSS codes when there is no gauge subsystem.
\end{eczvaluelist}
\eczhbkcontributors{ Leonid Pryadko, Daniel Gottesman, Qingfeng (Kee) Wang, \eczhuVVA }
\endeczcode

\eczcode{galois_cws}{Galois-qudit CWS code}{}
\codefieldsection{Description}
A CWS code for Galois qudits, defined using a Galois-qudit cluster state and a set of Galois-qudit \(Z\)-type Pauli strings defined by a \(q\)-ary classical code.

This entry has not yet been developed explicitly in the literature, so the formulation below is a conjectural Galois-qudit adaptation of the modular-qudit CWS constructions in Refs. \NoCaseChange{\protect\cite{cite3583,cite3044}}.
The Galois-qudit CWS construction takes in \( \mathcal{Q} = (\mathcal{G},\mathcal{C}) \), where \(\mathcal{G}\) is a graph, and where \(\mathcal{C}\) is an \((n,K,d)_q\) \(q\)-ary code.
From the graph, we form the Galois-qudit cluster state \( |\mathcal{G} \rangle \).
From the \(q\)-ary code, we form Galois-qudit Pauli \(Z\)-type operators \( W_i = Z_{c_{i,1}} \otimes \cdots \otimes Z_{c_{i,n}} \), where \(c_{i,j} \) is the \(j\)-th coordinate of the \(i\)-th classical codeword.
The codewords are then \( | i \rangle =  W_i | \mathcal{G} \rangle \).

\codefieldsection{Parent}
\begin{eczvaluelist}
\item\relax
\flmRefsHyperref[eczindexfamilyrel]{code:galois_non_stabilizer}{Galois-qudit USt code} --- Any Galois-qudit CWS code can be written as a USt whose (\(K=1\)) stabilizer code is the Galois-qudit cluster state and whose coset representatives are constructed from the \(q\)-ary classical code.
\end{eczvaluelist}
\codefieldsection{Child}
\begin{eczvaluelist}
\item\relax
\flmRefsHyperref[eczindexfamilyrel]{code:cws}{Codeword stabilized (CWS) code} --- Galois-qudit CWS codes reduce to CWS codes for \(q=2\).
\end{eczvaluelist}
\codefieldsection{Cousins}
\begin{eczvaluelist}
\item\relax
\flmRefsHyperref[eczindexfamilyrel]{code:graph_quantum}{Graph quantum code} --- A type of Galois-qudit cluster-state code can be built from a Galois-qudit cluster state by applying the conjectural Galois-qudit CWS construction using a linear \(q\)-ary code, in which codewords are obtained by applying Galois-qudit \(Z\)-type operators defined by the code to the Galois-qudit cluster state.
\item\relax
\flmRefsHyperref[eczindexfamilyrel]{code:galois_stabilizer}{Galois-qudit stabilizer code} --- Galois-qudit CWS codes whose underlying classical code is a linear \(q\)-ary code are Galois-qudit stabilizer codes containing a cluster-state codeword.
\end{eczvaluelist}
\eczhbkcontributors{ \eczhuVVA }
\endeczcode

\eczcode{galois_expander}{Galois-qudit expander code}{~\NoCaseChange{\protect\cite{cite689}}}
\codefieldsection{Alternative Names}
\begin{eczvaluelist}
\item\relax Galois-qudit Sipser-Spielman code
\end{eczvaluelist}
\eczhIndexCodeAliasName{galois_expander}{Galois-qudit Sipser-Spielman code}
\codefieldsection{Description}
Galois-qudit CSS code obtained from tensor products of chain complexes associated with an explicit family of expander codes with Reed-Solomon local checks.

In the explicit construction of \NoCaseChange{\protect\cite{cite689}}, these expander-code complexes contain planted GRM codewords, yielding a \textit{multiplication property} that allows QLDPC Galois-qudit quantum expander codes with transversal \(C^{r-1} Z\) gates while achieving \(D\geq N^{1/r}/\operatorname{poly}(\log N)\) and \(w\leq\operatorname{poly}(\log N)\).

\codefieldsection{Magic}
For every integer \(r\geq 2\) and every \(\epsilon>0\), the construction yields \(\llbracket N,K\geq N^{1-\epsilon},D\geq N^{1/r}/\operatorname{poly}(\log N)\rrbracket _q\) QLDPC Galois-qudit quantum expander codes with transversal \(C^{r-1} Z\) gates and stabilizer weight \(w\leq\operatorname{poly}(\log N)\) \NoCaseChange{\protect\cite{cite689}}. This construction allows for arbitrarily small magic-state yield parameter \(\gamma\).
\codefieldsection{Transversal and Permutation-Based Gates}
\begin{eczvaluelist}
\item\relax For every integer \(r\geq 2\), there are QLDPC Galois-qudit quantum expander codes with transversal \(C^{r-1} Z\) gates and parameters \(\llbracket N,K\geq N^{1-\epsilon},D\geq N^{1/r}/\operatorname{poly}(\log N)\rrbracket _q\) \NoCaseChange{\protect\cite{cite689}}. By decomposing each Galois qudit into a Kronecker product of qubits, this yields an explicit qubit CSS QLDPC family with parameters \(\llbracket N,K\geq N^{1-\epsilon},D\geq N^{1/r}/\operatorname{poly}(\log N)\rrbracket \), stabilizer weight \(\operatorname{poly}(\log N)\), and transversal \(C^{r-1}Z\) gates acting on \(N^{1-\epsilon}\) disjoint logical \(r\)-tuples.
\end{eczvaluelist}
\codefieldsection{Parent}
\begin{eczvaluelist}
\item\relax
\flmRefsHyperref[eczindexfamilyrel]{code:galois_hypergraph_product}{Galois-qudit HGP code}\end{eczvaluelist}
\codefieldsection{Child}
\begin{eczvaluelist}
\item\relax
\flmRefsHyperref[eczindexfamilyrel]{code:quantum_expander}{Quantum expander code}\end{eczvaluelist}
\codefieldsection{Cousins}
\begin{eczvaluelist}
\item\relax
\flmRefsHyperref[eczindexfamilyrel]{code:reed_solomon}{Reed-Solomon (RS) code} --- The explicit expander-code construction with Reed-Solomon local checks in \NoCaseChange{\protect\cite{cite689}} yields \(\llbracket N,K\geq N^{1-\epsilon},D\geq N^{1/r}/\operatorname{poly}(\log N)\rrbracket _q\) QLDPC Galois-qudit quantum expander codes with transversal \(C^{r-1} Z\) gates. Balanced products of the same RS-based complexes also yield \([n,k\geq n^{1-\epsilon},d\geq n/\operatorname{poly}(\log n)]_q\) LTCs exhibiting the multiplication property.
\item\relax
\flmRefsHyperref[eczindexfamilyrel]{code:expander}{Expander code} --- The explicit expander-code construction of \NoCaseChange{\protect\cite{cite689}} yields \(\llbracket N,K\geq N^{1-\epsilon},D\geq N^{1/r}/\operatorname{poly}(\log N)\rrbracket _q\) QLDPC Galois-qudit quantum expander codes with transversal \(C^{r-1} Z\) gates. Balanced products of the same expander-code complexes also yield \([n,k\geq n^{1-\epsilon},d\geq n/\operatorname{poly}(\log n)]_q\) LTCs exhibiting the multiplication property.
\item\relax
\flmRefsHyperref[eczindexfamilyrel]{code:generalized_reed_muller}{Generalized RM (GRM) code} --- The explicit expander-code construction of \NoCaseChange{\protect\cite{cite689}} contains planted GRM codewords.
\item\relax
\flmRefsHyperref[eczindexfamilyrel]{code:balanced_product}{Balanced product (BP) code} --- Balanced products of the RS-based expander-code complexes in \NoCaseChange{\protect\cite{cite689}} yield \([n,k\geq n^{1-\epsilon},d\geq n/\operatorname{poly}(\log n)]_q\) LTCs exhibiting the multiplication property.
\item\relax
\flmRefsHyperref[eczindexfamilyrel]{code:q-ary_ltc}{\(q\)-ary linear LTC} --- Balanced products of the RS-based expander-code complexes in \NoCaseChange{\protect\cite{cite689}} yield \([n,k\geq n^{1-\epsilon},d\geq n/\operatorname{poly}(\log n)]_q\) LTCs exhibiting the multiplication property.
\end{eczvaluelist}
\eczhbkcontributors{ \eczhuVVA }
\endeczcode

\eczcode{galois_grs}{Galois-qudit GRS code}{~\NoCaseChange{\protect\cite{cite826,cite823}}}
\codefieldsection{Description}
A true \(q\)-Galois-qudit stabilizer code constructed from GRS codes via either the Hermitian construction \NoCaseChange{\protect\cite{cite823,cite824,cite825}} or the Galois-qudit CSS construction \NoCaseChange{\protect\cite{cite826,cite827}}.

\codefieldsection{Rate}
Concatenations of quantum GRS codes and random stabilizer codes can achieve the \flmRefsHyperref{ref1729}{quantum GV bound} \NoCaseChange{\protect\cite{cite2704}}.
\codefieldsection{Parent}
\begin{eczvaluelist}
\item\relax
\flmRefsHyperref[eczindexfamilyrel]{code:quantum_ag}{Quantum AG code} --- Galois-qudit GRS codes can be constructed via the CSS construction or the Hermitian construction from GRS codes, which are evaluation AG codes.
\end{eczvaluelist}
\codefieldsection{Child}
\begin{eczvaluelist}
\item\relax
\flmRefsHyperref[eczindexfamilyrel]{code:galois_polynomial}{Galois-qudit RS code}\end{eczvaluelist}
\codefieldsection{Cousins}
\begin{eczvaluelist}
\item\relax
\flmRefsHyperref[eczindexfamilyrel]{code:generalized_reed_solomon}{Generalized RS (GRS) code} --- Galois-qudit GRS codes are quantum analogues of generalized RS codes.
\item\relax
\flmRefsHyperref[eczindexfamilyrel]{code:quantum_mds}{Quantum maximum-distance-separable (MDS) code} --- Some Galois-qudit GRS codes are quantum MDS \NoCaseChange{\protect\cite{cite823}}.
\item\relax
\flmRefsHyperref[eczindexfamilyrel]{code:stabilizer_over_gfqsq}{Hermitian Galois-qudit code} --- Galois-qudit GRS codes can be constructed via the CSS construction or the Hermitian construction.
\item\relax
\flmRefsHyperref[eczindexfamilyrel]{code:quantum_concatenated}{Concatenated quantum code} --- Concatenations of Galois-qudit GRS codes and random stabilizer codes can achieve the \flmRefsHyperref{ref1729}{quantum GV bound} \NoCaseChange{\protect\cite{cite2704}}.
\item\relax
\flmRefsHyperref[eczindexfamilyrel]{code:random_stabilizer}{Random stabilizer code} --- Concatenations of Galois-qudit GRS codes and random stabilizer codes can achieve the \flmRefsHyperref{ref1729}{quantum GV bound} \NoCaseChange{\protect\cite{cite2704}}.
\item\relax
\flmRefsHyperref[eczindexfamilyrel]{code:holographic_tensor}{Holographic tensor-network code} --- Galois-qudit GRS codes can be used to construct holographic p-adic (i.e., tree-tensor-network) codes on Bruhat-Tits trees and buildings and on Drinfeld symmetric spaces \NoCaseChange{\protect\cite{cite2866,cite2867}}.
\item\relax
\flmRefsHyperref[eczindexfamilyrel]{code:ea_galois_stabilizer}{EA Galois-qudit stabilizer code} --- Galois-qudit GRS codes can be used to construct EA Galois-qudit stabilizer codes \NoCaseChange{\protect\cite{cite1911,cite4608}}.
\end{eczvaluelist}
\eczhbkcontributors{ \eczhuVVA }
\endeczcode

\eczcode{galois_hypergraph_product}{Galois-qudit HGP code}{}
\codefieldsection{Alternative Names}
\begin{eczvaluelist}
\item\relax Galois-qudit quantum hypergraph (QHG) code
\item\relax Galois-qudit Tillich-Zemor product code
\end{eczvaluelist}
\eczhIndexCodeAliasName{galois_hypergraph_product}{Galois-qudit quantum hypergraph (QHG) code}
\eczhIndexCodeAliasName{galois_hypergraph_product}{Galois-qudit Tillich-Zemor product code}
\codefieldsection{Description}
A member of a family of Galois-qudit CSS codes whose stabilizer generator matrix is obtained from a hypergraph product of two classical linear \(q\)-ary codes.

\codefieldsection{Parent}
\begin{eczvaluelist}
\item\relax
\flmRefsHyperref[eczindexfamilyrel]{code:lifted_product}{Lifted-product (LP) code} --- Lifted-product codes for trivial lift are Galois-qudit hypergraph-product codes.
\end{eczvaluelist}
\codefieldsection{Children}
\begin{eczvaluelist}
\item\relax
\flmRefsHyperref[eczindexfamilyrel]{code:hypergraph_product}{Hypergraph product (HGP) code} --- Hypergraph product codes are Galois-qudit hypergraph-product codes for qudit dimension \(q=2\).
\item\relax
\flmRefsHyperref[eczindexfamilyrel]{code:galois_expander}{Galois-qudit expander code}\end{eczvaluelist}
\codefieldsection{Cousins}
\begin{eczvaluelist}
\item\relax
\flmRefsHyperref[eczindexfamilyrel]{code:q-ary_linear}{Linear \(q\)-ary code} --- Galois-qudit HGP codes are constructed out of two classical linear \(q\)-ary codes.
\item\relax
\flmRefsHyperref[eczindexfamilyrel]{code:2bga}{Two-block group-algebra (2BGA) codes} --- A 2BGA code \(LP(a,b)\) is constructible as a hypergraph-product code when the support subgroups generated by \(a\) and \(b\) are disjoint. In that case, the commuting matrices simultaneously acquire hypergraph-product Kronecker-product form, and the code can be obtained from a pair of classical group-algebra codes \NoCaseChange{\protect\cite[{Statements 8 and 12}]{cite842}}.

\end{eczvaluelist}
\eczhbkcontributors{ \eczhuVVA }
\endeczcode

\eczcode{galois_reed_muller}{Galois-qudit quantum RM code}{~\NoCaseChange{\protect\cite{cite828}}}
\codefieldsection{Description}
True Galois-qudit stabilizer code constructed from generalized Reed-Muller (GRM) codes via the Galois-qudit Hermitian construction, the Galois-qudit CSS construction, or directly from their parity-check matrices \NoCaseChange{\protect\cite{cite828}\protect\cite[{Sec. 4.2}]{cite829}}.

The CSS construction yields the code
\flmMathEnvironment{align}{}{
  \llbracket q^m,k(v_2)-k(v_1),\min\{d(v_1^{\perp}),d(v_2)\}\rrbracket _q
}
constructed from the generalized Reed-Muller codes RM\(_q(v_1,m)\) and RM\(_q(v_2,m)\), with \(0\leq v_1 \leq v_2 \leq m(q-1)-1\) \NoCaseChange{\protect\cite{cite828}}.
The parameters are
\flmMathEnvironment{align}{}{
  k(v) &= \sum_{j=0}^{m}(-1)^{j}\dbinom{m}{j}\dbinom{m+v-jq}{v-jq} \\
  d(v) &= (R+1)q^{Q}~,
}
where \(m(q-1)-v=(q-1)Q+R\) so that \(0\leq R\leq q-1\).
Here \(0\leq v_1,v_2 \leq m(q-1)-1\), \(q\) is a prime power, and \(m\) is a positive integer.

Using the code GRM\(_{q^2}(v,m)\) for \(0\leq v \leq m(q-1)-1\), the Hermitian construction yields the \flmRefsHyperref{ref672}{pure} quantum code
\flmMathEnvironment{align}{}{
  \llbracket q^{2m},q^{2m}-2k(v),d(v^{\perp})\rrbracket _q
}
\NoCaseChange{\protect\cite{cite828}}, where
\flmMathEnvironment{align}{}{
  k(v) &= \sum_{j=0}^{m}(-1)^{j}\dbinom{m}{j}\dbinom{m+v-jq^2}{v-jq^2} \\
  d(v^{\perp}) &= (R+1)q^{2Q}~,
}
with \(v+1 = (q^2 - 1)Q + R\).

For a CSS code constructed from classical codes \(C_1\) and \(C_2\), the punctured code is defined as
\flmMathEnvironment{align}{}{
  P(C) = \{(a_ib_i)_{i=1}^{n} \mid a \in C_1, b \in C_2^{\perp}\}^{\perp}~.
}
Quantum RM codes can be punctured to any length \(r\), provided
\flmMathEnvironment{align}{}{
  P(C) = \mathcal{R}_q(v_2-v_1,m)
}
has a codeword of this weight.
Likewise, the Hermitian puncture code contains \(\mathcal{R}_{q^2}(\mu,m)^\perp|_{\mathbb{F}_q}\) for \((q+1)\nu \leq \mu \leq m(q^2-1)-1\), yielding punctured descendants with distance at least that of the parent code \NoCaseChange{\protect\cite{cite828}}.

\codefieldsection{Protection}
The CSS family is \flmRefsHyperref{ref672}{pure} with distance \(\min\{d(v_1^\perp),d(v_2)\}\), while the Hermitian family is \flmRefsHyperref{ref672}{pure} with distance \(d(v^\perp)\); punctured descendants retain at least the parent distance \NoCaseChange{\protect\cite{cite828}}.
QRM\(_{d}(m)\) quantum codes are \(\mathcal{M}_{d}^{m}\) distillation codes of distance \(D=2\). We define a \(\mathcal{M}_{d}^{m}\) distillation code as any \(n\) Galois-qudit stabilizer code \(C\) having the following properties: (a) All \(M \in \mathcal{M}_{d}^{m}\) are transversal so that \(M^{\otimes n}C(M^{\otimes n})^{\dagger} = M_{L}^{\dagger}CM_{L}\), (b) the code has distance \(D \geq 2\), and (c) the code has logical pauli operators \(X_{L} = X[\mathbf{1}]\) and \(Z_{L} = Z[(d-1)\mathbf{1}]\). Here, \(\mathbf{1}\) is a shorthand for the vector \((1,1, \dots, 1)\).

\codefieldsection{Rate}
The CSS family has rate \( (k(v_2)-k(v_1) )/q^m\), while the Hermitian family has rate \(1-2k(v)/q^{2m}\) \NoCaseChange{\protect\cite{cite828}}.
\codefieldsection{Parent}
\begin{eczvaluelist}
\item\relax
\flmRefsHyperref[eczindexfamilyrel]{code:galois_true_stabilizer}{True Galois-qudit stabilizer code} --- Galois-qudit RM codes can be constructed via the Galois-qudit Hermitian construction, the Galois-qudit CSS construction, or directly from their parity-check matrices \NoCaseChange{\protect\cite{cite828}\protect\cite[{Sec. 4.2}]{cite829}}.
\end{eczvaluelist}
\codefieldsection{Child}
\begin{eczvaluelist}
\item\relax
\flmRefsHyperref[eczindexfamilyrel]{code:qudit_reed_muller}{Prime-qudit RM code} --- Galois-qudit RM codes reduce to prime-qudit RM codes when \(q\) is prime.
\end{eczvaluelist}
\codefieldsection{Cousins}
\begin{eczvaluelist}
\item\relax
\flmRefsHyperref[eczindexfamilyrel]{code:generalized_reed_muller}{Generalized RM (GRM) code} --- Generalized RM codes can be used to construct Galois-qudit RM codes via the Galois-qudit Hermitian construction, the Galois-qudit CSS construction, or directly from their parity-check matrices \NoCaseChange{\protect\cite{cite828}\protect\cite[{Sec. 4.2}]{cite829}}.
\item\relax
\flmRefsHyperref[eczindexfamilyrel]{code:projective_reed_muller}{Projective RM (PRM) code} --- Projective RM codes can be used to construct Galois-qudit RM codes \NoCaseChange{\protect\cite{cite1839}\protect\cite[{Sec. 6.2}]{cite872}}.
\item\relax
\flmRefsHyperref[eczindexfamilyrel]{code:galois_css}{Galois-qudit CSS code} --- Galois-qudit RM codes admit a CSS subfamily built from nested GRM codes \( \mathrm{GRM}_q(v_1,m) \subseteq \mathrm{GRM}_q(v_2,m) \) \NoCaseChange{\protect\cite{cite828}}.
\item\relax
\flmRefsHyperref[eczindexfamilyrel]{code:stabilizer_over_gfqsq}{Hermitian Galois-qudit code} --- Galois-qudit RM codes admit a Hermitian subfamily built from \(\mathrm{GRM}_{q^2}(v,m)\) codes contained in their Hermitian duals \NoCaseChange{\protect\cite{cite828}}.
\item\relax
\flmRefsHyperref[eczindexfamilyrel]{code:quantum_mds}{Quantum maximum-distance-separable (MDS) code} --- Quantum GRM codes yield quantum MDS families \(\llbracket q,q-2\nu-2,\nu+2\rrbracket _q\) for \(0 \leq \nu \leq (q-2)/2\), \(\llbracket q^2,q^2-2\nu-2,\nu+2\rrbracket _q\) for \(0 \leq \nu \leq q-2\), and punctured descendants \(\llbracket (\nu+1)q,(\nu+1)q-2\nu-2,\nu+2\rrbracket _q\) for \(0 \leq \nu \leq q-2\) \NoCaseChange{\protect\cite{cite828}}.
\end{eczvaluelist}
\eczhbkcontributors{ Shuubham Ojha, \eczhuVVA }
\endeczcode

\eczcode{galois_polynomial}{Galois-qudit RS code}{~\NoCaseChange{\protect\cite{cite826}}}
\codefieldsection{Alternative Names}
\begin{eczvaluelist}
\item\relax Galois-qudit polynomial code (QPyC)
\end{eczvaluelist}
\eczhIndexCodeAliasName{galois_polynomial}{Galois-qudit polynomial code (QPyC)}
\codefieldsection{Description}
A Galois-qudit CSS code family (with \(q>n\)) constructed using two RS codes over \(\mathbb{F}_q\).

Let \(C_1\) be a \([n,k_1,d_1]_q\) RS code and \(C_2^\perp\) be a \([n,k_2,d_2]_q\) RS code, modified such that \(C_2^\perp \subseteq C_1\) and \(0\le k_2 \le k_1 \le n\).
Then, a polynomial code is a \flmRefsHyperref{ref811}{non-degenerate} \(\llbracket n,k_2,d\rrbracket _q\) Galois-qudit CSS code with \(d=\min(n-k_1+1,k_1-k_2+1)\).
The polynomial code is the span of the basis codewords over \(\mathbb{F}_q\),
\flmMathEnvironment{align}{}{
|\overline{\beta_0,\cdots,\beta_{k_2-1}}\rangle
=
\sum_{(\beta_{k_2},\cdots,\beta_{k_1-1})\in \mathbb{F}_q}
\bigotimes_{i=1}^{n}
\left| \sum_{j=0}^{k_1-1} \beta_j \alpha_i^j \right\rangle,
}
where \((\alpha_1, \cdots, \alpha_n)\) are \(n\) distinct points chosen for code \(C_1\) from \(\mathbb{F}_q\setminus \{0\}\).

\codefieldsection{Protection}
Galois-qudit RS codes can be adapted for insertion and deletion noise \NoCaseChange{\protect\cite{cite4631}}.

\codefieldsection{Magic}
Punctured RS codes can be used for magic-state distillation with a spacetime overhead of \((\log \frac{1}{\epsilon})^{\gamma}\), with the magic-state scaling exponent \(\gamma \to 0\) with decreasing \(\epsilon\) \NoCaseChange{\protect\cite{cite688}}.
\codefieldsection{Transversal and Permutation-Based Gates}
\begin{eczvaluelist}
\item\relax There exists an \flmRefsHyperref{ref65}{order} \(\llbracket n,\Theta(n),\Theta(n)\rrbracket _{n^2}\) punctured RS code family that admits transversal \(CCZ\) gates for any three logical qubits \NoCaseChange{\protect\cite{cite733}}. This code can be treated as a qubit code by decomposing each Galois qudit into a Kronecker product of several qubits; see \NoCaseChange{\protect\cite{cite696,cite398,cite698,cite699,cite700}\protect\cite[{Sec. 5.3}]{cite697}}. This yields a qubit code family that is asymptotically good up to poly-logarithmic factors \NoCaseChange{\protect\cite{cite733}}.
\item\relax Quantum RS codes can be concatenated with quantum multiplication-friendly codes to yield CSS codes over a constant prime-power alphabet \(q\) admitting a transversal \(CCZ_q\) gate (and, for \(q \geq 5\), also a transversal \(U_q\) gate) with \flmRefsHyperref{ref65}{nearly linear} code dimension and distance, \(k,d = \Omega(n/2^{O(\log^* n)})\), where \(\log^* n\) is the iterated logarithm (the number of times \(\log\) must be applied to \(n\) before the result is \(\leq 1\)) \NoCaseChange{\protect\cite{cite699}}.
\end{eczvaluelist}
\codefieldsection{Fault Tolerance}
\begin{eczvaluelist}
\item\relax The original prime-field polynomial-code construction admits a fault-tolerant universal gate set beyond the standard transversal gates for qudit CSS codes \NoCaseChange{\protect\cite{cite826}}. Generalized NOT, SUM, SWAP, scalar multiplication, and phase gates are implemented transversally. A logical Fourier transform is implemented by weighted physical Fourier transforms followed by \flmRefsHyperref{ref410}{code switching}: a length-\(m\) degree-\(d\) polynomial code is mapped to the corresponding Fourier-dual polynomial code of degree \(m-d-1\). A transversal product gate \(|a\rangle|b\rangle|c\rangle \mapsto |a\rangle|b\rangle|c+ab\rangle\) maps into a degree-\(2d\) polynomial code, after which degree reduction switches back to the original degree.
\item\relax Aharonov and Ben-Or performed the code switching needed for their Fourier and product gates using concatenation and degree reduction \NoCaseChange{\protect\cite{cite826}}. The same switching between polynomial codes of different degree can also be implemented by teleportation using encoded Bell states; this uses the transversal SUM gate between different-degree quantum RS codes, with the lower-degree code serving as the control.
\item\relax Quantum RS codes yield fault-tolerant quantum computation with constant space and optimal time overheads \NoCaseChange{\protect\cite{cite688}}, improving over previous schemes \NoCaseChange{\protect\cite{cite3214,cite3216}}.
\end{eczvaluelist}
\codefieldsection{Parent}
\begin{eczvaluelist}
\item\relax
\flmRefsHyperref[eczindexfamilyrel]{code:galois_grs}{Galois-qudit GRS code}\end{eczvaluelist}
\codefieldsection{Children}
\begin{eczvaluelist}
\item\relax
\flmRefsHyperref[eczindexfamilyrel]{code:polynomial}{Prime-qudit RS code} --- Galois-qudit RS codes for prime-dimensional qudits are prime-qudit RS codes.
\item\relax
\flmRefsHyperref[eczindexfamilyrel]{code:galois_3_1_2}{\(\llbracket 3,1,2\rrbracket _4\) three-Galois-quartrit code} --- The \(\llbracket 3,1,2\rrbracket _4\) code is constructed from the shortened RS\(_4\) code \NoCaseChange{\protect\cite{cite514}}.
\end{eczvaluelist}
\codefieldsection{Cousins}
\begin{eczvaluelist}
\item\relax
\flmRefsHyperref[eczindexfamilyrel]{code:galois_fqrs}{Folded quantum RS (FQRS) code} --- A FQRS code with no extra grouping (\(m=1\)) reduces to a Galois-qudit RS code that is CSS.
\item\relax
\flmRefsHyperref[eczindexfamilyrel]{code:reed_solomon}{Reed-Solomon (RS) code} --- Galois-qudit RS codes are CSS codes constructed from RS codes.
\item\relax
\flmRefsHyperref[eczindexfamilyrel]{code:quantum_mds}{Quantum maximum-distance-separable (MDS) code} --- A polynomial code is a quantum MDS code when \(n-k_1=k_1-k_2\).
\item\relax
\flmRefsHyperref[eczindexfamilyrel]{code:quantum_concatenated}{Concatenated quantum code} --- Recursive concatenations of quantum RS codes can be asymptotically good \NoCaseChange{\protect\cite{cite2705}}.
\item\relax
\flmRefsHyperref[eczindexfamilyrel]{code:asymmetric_qecc}{Asymmetric quantum code (AQC)} --- Asymmetric Galois-qudit RS codes have been constructed \NoCaseChange{\protect\cite{cite2638,cite2639}\protect\cite[{Sec. 17.3}]{cite872}}.
\item\relax
\flmRefsHyperref[eczindexfamilyrel]{code:ame}{Perfect-tensor code} --- \flmRefsHyperref{ref219}{AME states} for even \(n\) are examples of quantum MDS codes with no logical qubits \NoCaseChange{\protect\cite{cite1670,cite1926,cite2933}}. MDS RS codes can yield perfect tensors via the CSS and Hermitian constructions \NoCaseChange{\protect\cite{cite975}} (see also Refs. \NoCaseChange{\protect\cite{cite2866,cite2867}}).
\item\relax
\flmRefsHyperref[eczindexfamilyrel]{code:quantum_tensor_product}{Quantum tensor-product code} --- Product codes constructed from a self-orthogonal and an arbitrary RS code yield an RS code \NoCaseChange{\protect\cite{cite4024}}.
\item\relax
\flmRefsHyperref[eczindexfamilyrel]{code:quantum_secret_sharing}{Approximate secret-sharing code} --- Polynomial codes can be used for a specific construction of this code.
\end{eczvaluelist}
\eczhbkcontributors{ Qingfeng (Kee) Wang, \eczhuVVA }
\endeczcode

\eczcode{galois_stabilizer}{Galois-qudit stabilizer code}{~\NoCaseChange{\protect\cite{cite696}}}
\codefieldsection{Description}
An \(\llparenthesis n,K,d\rrparenthesis _q\) Galois-qudit code whose logical subspace is the joint eigenspace of commuting Galois-qudit Pauli operators forming the code's stabilizer group \(\mathsf{S}\). Traditionally, the logical subspace is the joint \(+1\) eigenspace, and the stabilizer group does not contain \(e^{i \phi} I\) for any \(\phi \neq 0\). The distance \(d\) is the minimum weight of a Galois-qudit Pauli string that implements a nontrivial logical operation in the code.

A Galois-qudit stabilizer code encoding an integer number of qudits (\(K=q^k\)) is denoted as \(\llbracket n,k\rrbracket _q\) or \(\llbracket n,k,d\rrbracket _q\).
This notation differentiates between Galois-qudit and modular-qudit \(\llbracket n,k,d\rrbracket _{\mathbb{Z}_q}\) stabilizer codes, although the same notation is usually used for both.
Galois-qudit stabilizer codes need not encode an integer number of qudits, with \(K=q^{n-\frac{r}{m}}\), where \(r\) is the number of generators of the stabilizer group, and \(q=p^m\) given prime \(p\) for all Galois qudits.
As a result, \(\llbracket n,k,d\rrbracket \) notation is often used with non-integer \(k=\log_q K\).

\begin{defterm}{Galois symplectic representation}\label{ref4632}\label{ref873}
The single Galois-qudit Pauli string \(X_{a} Z_{b}\) for \(a,b\in \mathbb{F}_q\) is converted to the vector \((a|b)\in \mathbb{F}_q^2\).
The multi Galois-qudit version follows naturally.
\end{defterm}

A pair of Galois-qudit stabilizers on \(n\) Galois qudits with \flmRefsHyperref{ref873}{Galois symplectic representation} vectors \((a|b)\) and \((a^{\prime}|b^{\prime})\) commute iff their \textit{trace symplectic inner product} is zero,
\flmMathEnvironment{align}{}{
\text{tr}(a \cdot b^{\prime} - a^{\prime}\cdot b) = \sum_{j=1}^{n} \text{tr}(a_j b^{\prime}_j - a^{\prime}_j b_j) = 0~.
}
\flmRefsHyperref{ref873}{Galois symplectic representations} of stabilizer group elements form a trace-symplectic self-orthogonal linear code over \(\mathbb{F}_q^{2n}\).
The trace-symplectic inner product reduces to the \textit{symplectic inner product} when the \flmRefsHyperref{ref33}{field trace} is removed, and a symplectic self-orthogonal set of vectors is automatically trace-symplectic self-orthogonal.

Another correspondence between Galois-qudit Pauli matrices and elements of the Galois field \(\mathbb{F}_{q^2}\) yields the one-to-one correspondence between Galois-qudit stabilizer codes and trace-alternating self-orthogonal additive codes over \(\mathbb{F}_{q^2}\) \NoCaseChange{\protect\cite[{Thm. 15}]{cite813}}.

\begin{defterm}{\(\mathbb{F}_{q^2}\) representation}\label{ref4633}\label{ref1779}
An \(n\)-qudit Galois-qudit Pauli stabilizer can be represented as a length-\(n\) vector over \(\mathbb{F}_{q^2}\) using the one-to-one correspondence between the \(q^2\) Galois-qudit Pauli matrices and elements of \(\mathbb{F}_{q^2}\).
Given a \flmRefsHyperref{ref33}{basis} \((\beta,\beta^q)\) for \(\mathbb{F}_{q^2}\) over \(\mathbb{F}_q\), the vector \((a|b)\in \mathbb{F}_q^2\) (representing a Galois-qudit Pauli string in the \flmRefsHyperref{ref873}{Galois symplectic representation}) is in one-to-one correspondence with element \(a \beta + b \beta^q \in \mathbb{F}_{q^2}\) \NoCaseChange{\protect\cite[{Lem. 14}]{cite813}\protect\cite[{Thm. 27.4.1}]{cite2024}}.
\end{defterm}

The sets of \(\mathbb{F}_{q^2}\)-represented vectors for all generators yield a trace-alternating self-orthogonal additive code over \(\mathbb{F}_{q^2}\).

Recalling that \(q=p^m\), Galois-qudit stabilizer codes can also be treated as prime-qudit stabilizer codes on \(mn\) qudits, giving \(k=nm-r\) \NoCaseChange{\protect\cite{cite696}}.
In principle, Galois-qudit stabilizer states can be expressed in terms of linear and quadratic functions over \(\mathbb{Z}_p^{mn}\) \NoCaseChange{\protect\cite{cite4302}}.
Such states correspond to the set of states with positive Wigner functions \NoCaseChange{\protect\cite{cite4535,cite4574}}.

Galois-qudit stabilizer codes can equivalently \NoCaseChange{\protect\cite{cite3561}} (see also \NoCaseChange{\protect\cite{cite867}}) be defined using graphs, yielding an analytical form for the codewords \NoCaseChange{\protect\cite{cite866}}.

\codefieldsection{Protection}
Detects errors on up to \(d-1\) qudits, and corrects erasure errors on up to \(d-1\) qudits. Corrects errors on \(\left\lfloor (d-1)/2 \right\rfloor\) qudits.
There are algorithms to calculate the minimum distance \NoCaseChange{\protect\cite{cite4312}}.
There are established shortening/lengthening procedures for \flmRefsHyperref{ref672}{pure} Galois-qudit stabilizer codes \NoCaseChange{\protect\cite{cite1653}\protect\cite[{Table 1}]{cite813}}.

\codefieldsection{Encoding}
\begin{eczvaluelist}
\item\relax Encoder with \(O(n^2)\) gates can be determined in classical runtime of \flmRefsHyperref{ref65}{order} \(O(n^3)\) \NoCaseChange{\protect\cite{cite4619}}.
\end{eczvaluelist}
\codefieldsection{Gates}
\begin{eczvaluelist}
\item\relax As opposed to modular qudits for composite \(q\), Galois qudits inherit most of the properties of the prime-qudit Clifford group due to the correspondence between a \(q=p^m\) Galois qudit and \(m\) prime qudits of dimension \(p\) \NoCaseChange{\protect\cite{cite696}}.
\end{eczvaluelist}
\codefieldsection{Decoding}
\begin{eczvaluelist}
\item\relax Syndrome extraction and computation based on classical additive codes \NoCaseChange{\protect\cite{cite4607}}.
\end{eczvaluelist}
\codefieldsection{Notes}
\begin{eczvaluelist}
\item\relax Tables of bounds and examples of Galois-qudit stabilizer codes for various \(n\) and \(k\), based on algorithms developed in Refs. \NoCaseChange{\protect\cite{cite2673,cite2674}}, are maintained by M. Grassl at this \flmHref{https://www.codetables.de/}{website}. A Magma implementation exists at this \flmHref{https://magma.maths.usyd.edu.au/magma/handbook/text/1976}{website}. A modular-qudit stabilizer code with composite dimension \(q\) contains a subcode that is isomorphic to a \(p\)-dimensional prime-qudit stabilizer code for every prime factor \(p\) of \(q\), and the distance of the full stabilizer code is upper bound by the distance of this subcode \NoCaseChange{\protect\cite{cite4581}}.
\item\relax The number of Galois-qudit stabilizer codes was determined in Ref. \NoCaseChange{\protect\cite{cite4535}}.
\item\relax See Quantum Codes qudit stabilizer database, maintained by N. Aydin, P. Liu, and B. Yoshino \NoCaseChange{\protect\cite{cite4381,cite4382}}, at this \flmHref{http://quantumcodes.info/}{website}.
\item\relax Review of nonbinary stabilizer codes \NoCaseChange{\protect\cite{cite1839}}.
\end{eczvaluelist}
\codefieldsection{Parents}
\begin{eczvaluelist}
\item\relax
\flmRefsHyperref[eczindexfamilyrel]{code:galois_non_stabilizer}{Galois-qudit USt code} --- A Galois-qudit stabilizer code with stabilizer group \(\mathsf{S}\) can be thought of as a Galois-qudit USt with only the identity coset representative. Conversely, if \(K = q^k\), and if the set of coset representatives of a Galois-qudit USt form a \(q\)-ary linear code, then they can be absorbed into a Galois-qudit stabilizer group that defines the USt.
\item\relax
\flmRefsHyperref[eczindexfamilyrel]{code:stabilizer}{Stabilizer code}\end{eczvaluelist}
\codefieldsection{Child}
\begin{eczvaluelist}
\item\relax
\flmRefsHyperref[eczindexfamilyrel]{code:galois_true_stabilizer}{True Galois-qudit stabilizer code}\end{eczvaluelist}
\codefieldsection{Cousins}
\begin{eczvaluelist}
\item\relax
\flmRefsHyperref[eczindexfamilyrel]{code:galois_cws}{Galois-qudit CWS code} --- Galois-qudit CWS codes whose underlying classical code is a linear \(q\)-ary code are Galois-qudit stabilizer codes containing a cluster-state codeword.
\item\relax
\flmRefsHyperref[eczindexfamilyrel]{code:qudit_stabilizer}{Modular-qudit stabilizer code} --- Recalling that \(q=p^m\), Galois-qudit stabilizer codes can also be treated as prime-qudit stabilizer codes on \(mn\) qudits, giving \(k=nm-r\) \NoCaseChange{\protect\cite{cite696}}. The case \(m=1\) reduces to conventional prime-qudit stabilizer codes on \(n\) qudits. A modular-qudit stabilizer code with composite dimension \(q\) contains a subcode that is isomorphic to a \(p\)-dimensional prime-qudit stabilizer code for every prime factor \(p\) of \(q\), and the distance of the full stabilizer code is bounded by the distance of this subcode \NoCaseChange{\protect\cite{cite4581}}.
\item\relax
\flmRefsHyperref[eczindexfamilyrel]{code:q-ary_additive}{Additive \(q\)-ary code} --- Galois-qudit stabilizer codes are the closest quantum analogues of additive codes over \(\mathbb{F}_q\) because addition in the field corresponds to multiplication of stabilizers in the quantum case.
\item\relax
\flmRefsHyperref[eczindexfamilyrel]{code:dual_additive}{Dual additive code} --- Galois-qudit stabilizer codes are in one-to-one correspondence with trace-symplectic self-orthogonal additive codes of length \(2n\) over \(\mathbb{F}_q\) via the \flmRefsHyperref{ref873}{Galois symplectic representation} \NoCaseChange{\protect\cite{cite696}}. They are also in one-to-one correspondence with trace-alternating self-orthogonal additive codes of length \(n\) over \(\mathbb{F}_{q^2}\) via the \flmRefsHyperref{ref1779}{\(\mathbb{F}_{q^2}\) representation}.
\item\relax
\flmRefsHyperref[eczindexfamilyrel]{code:t-designs}{\(t\)-design} --- Stabilizer states on \(n\) Galois qubits form 2-designs on complex projective spaces \(\mathbb{C}P^{p^{mn}}\) \NoCaseChange{\protect\cite{cite943}}.
\item\relax
\flmRefsHyperref[eczindexfamilyrel]{code:complex_projective}{Complex projective space code} --- Stabilizer states on \(n\) Galois qubits form 2-designs on complex projective spaces \(\mathbb{C}P^{p^{mn}}\) \NoCaseChange{\protect\cite{cite943}}.
\item\relax
\flmRefsHyperref[eczindexfamilyrel]{code:rotor_stabilizer}{Rotor stabilizer code} --- Galois-qudit stabilizer codes can be imported into integral-domain settings, and rotor codes and their parameters can be obtained through a synthesis of these cases \NoCaseChange{\protect\cite[{Thm. 13; Sec. 3.2}]{cite4580}}.
\item\relax
\flmRefsHyperref[eczindexfamilyrel]{code:graph_quantum}{Graph quantum code} --- Graph quantum codes for \(G=\mathbb{F}_q\) are a subset of Galois-qudit stabilizer codes \NoCaseChange{\protect\cite{cite3561}}. Any Galois-qudit stabilizer code is equivalent to a graph quantum code for \(G=\mathbb{F}_q\) via a single-Galois-qudit Clifford circuit \NoCaseChange{\protect\cite{cite3561}} (see also \NoCaseChange{\protect\cite{cite3536,cite867}}).
\item\relax
\flmRefsHyperref[eczindexfamilyrel]{code:analog_stabilizer}{Analog stabilizer code} --- Galois-qudit stabilizer codes can be transformed into analog stabilizer codes; if the original code has \(r\) linearly independent generators in the symplectic representation, the resulting analog code has parameters \(\llbracket n,n-r,d^{\prime}\rrbracket _{\mathbb{R}}\) with \(d^{\prime}\geq d\) \NoCaseChange{\protect\cite[{Thm. 13}]{cite4580}}.
\item\relax
\flmRefsHyperref[eczindexfamilyrel]{code:ea_galois_stabilizer}{EA Galois-qudit stabilizer code} --- EA Galois-qudit stabilizer codes utilize additional ancillary Galois-qudits in a pre-shared entangled state, but reduce to Galois-qudit stabilizer codes when said qudits are interpreted as noiseless physical qudits. Pure Galois-qudit codes can be used to make EA Galois-qudit stabilizer codes \NoCaseChange{\protect\cite{cite4606}\protect\cite[{Thm. 10}]{cite545}}.
\item\relax
\flmRefsHyperref[eczindexfamilyrel]{code:galois_subsystem_stabilizer}{Subsystem Galois-qudit stabilizer code} --- Subsystem Galois-qudit stabilizer codes reduce to Galois-qudit stabilizer codes when there are no gauge qudits.
\end{eczvaluelist}
\eczhbkcontributors{ Markus Grassl, Leonid Pryadko, Qingfeng (Kee) Wang, \eczhuVVA }
\endeczcode

\eczcode{galois_topological}{Galois-qudit surface code}{~\NoCaseChange{\protect\cite{cite424,cite4634}}}
\codefieldsection{Alternative Names}
\begin{eczvaluelist}
\item\relax \(\mathbb{F}_q\)-qudit surface code
\end{eczvaluelist}
\eczhIndexCodeAliasName{galois_topological}{\(\mathbb{F}_q\)-qudit surface code}
\codefieldsection{Description}
Extension of the surface code to 2D lattices of Galois qudits.

\codefieldsection{Parents}
\begin{eczvaluelist}
\item\relax
\flmRefsHyperref[eczindexfamilyrel]{code:galois_css}{Galois-qudit CSS code}\item\relax
\flmRefsHyperref[eczindexfamilyrel]{code:2d_stabilizer}{2D lattice stabilizer code}\item\relax
\flmRefsHyperref[eczindexfamilyrel]{code:generalized_homological_product_css}{Generalized homological-product CSS code}\item\relax
\flmRefsHyperref[eczindexfamilyrel]{code:topological_abelian}{Abelian topological code} --- A Galois qudit for \(q=p^m\) can be decomposed into a Kronecker product of \(m\) modular qudits \NoCaseChange{\protect\cite{cite696,cite398,cite698,cite699,cite700}\protect\cite[{Sec. 5.3}]{cite697}}. Galois-qudit surface codes yield Abelian quantum-double codes with \(\mathbb{F}_{p^m}\cong \mathbb{Z}_p^m\) topological order via this decomposition.
\item\relax
\flmRefsHyperref[eczindexfamilyrel]{code:quantum_double}{Quantum-double code} --- A Galois qudit for \(q=p^m\) can be decomposed into a Kronecker product of \(m\) modular qudits \NoCaseChange{\protect\cite{cite696,cite398,cite698,cite699,cite700}\protect\cite[{Sec. 5.3}]{cite697}}. Galois-qudit surface codes yield Abelian quantum-double codes with \(\mathbb{F}_{p^m}\cong \mathbb{Z}_p^m\) topological order via this decomposition.
\end{eczvaluelist}
\codefieldsection{Child}
\begin{eczvaluelist}
\item\relax
\flmRefsHyperref[eczindexfamilyrel]{code:surface}{Kitaev surface code} --- The Galois-qudit surface code for \(q=2\) reduces to the surface code.
\end{eczvaluelist}
\codefieldsection{Cousins}
\begin{eczvaluelist}
\item\relax
\flmRefsHyperref[eczindexfamilyrel]{code:quantum_double_abelian}{Abelian quantum-double stabilizer code} --- A Galois qudit for \(q=p^m\) can be decomposed into a Kronecker product of \(m\) modular qudits \NoCaseChange{\protect\cite{cite696,cite398,cite698,cite699,cite700}\protect\cite[{Sec. 5.3}]{cite697}}. Galois-qudit surface codes yield Abelian quantum-double codes with \(\mathbb{F}_{p^m}\cong \mathbb{Z}_p^m\) topological order via this decomposition.
\item\relax
\flmRefsHyperref[eczindexfamilyrel]{code:2bga}{Two-block group-algebra (2BGA) codes} --- Any non-trivial 2BGA code with total row weight \(W\leq 4\) is equivalent to a direct sum of rotated Galois-qudit surface codes \NoCaseChange{\protect\cite[{Sec. IV.F}]{cite842}}.

\end{eczvaluelist}
\eczhbkcontributors{ \eczhuVVA }
\endeczcode

\eczcode{galois_non_stabilizer}{Galois-qudit USt code}{~\NoCaseChange{\protect\cite{cite3168,cite4493,cite4494,cite3170,cite1369,cite4635}}}
\codefieldsection{Alternative Names}
\begin{eczvaluelist}
\item\relax Galois-qudit non-stabilizer code
\end{eczvaluelist}
\eczhIndexCodeAliasName{galois_non_stabilizer}{Galois-qudit non-stabilizer code}
\codefieldsection{Description}
A Galois-qudit code whose codespace consists of a direct sum of a Galois-qudit stabilizer codespace and one or more of that stabilizer code's error spaces.

Given a subset \(T\) of coset representatives of \(\mathsf{N}(\mathsf{S})/\mathsf{S}\) of a Galois-qudit stabilizer code \(\llparenthesis n,K\rrparenthesis \) with codespace \(\mathsf{C}\) and stabilizer group \(\mathsf{S}\), one can construct the Galois-qudit USt with codespace
\flmMathEnvironment{align}{}{
  \mathsf{C}_{\text{USt}}=\bigoplus_{t\in T}t\mathsf{C}~.
}
The parameters of the USt are \(\llparenthesis n,K|T|\rrparenthesis \), where \(|T|\) is the number of chosen coset representatives.
A Galois-qudit USt is \textit{CSS-like} when the underlying stabilizer code is CSS and the coset representatives are chosen from the two classical codes underlying the CSS code.

Union stabilizer codes generalize stabilizer codes by modifying the original stabilizer code projection with elements of a subset \(\mathsf{B}\subset\mathsf{S}\) called the \textit{Fourier description} \NoCaseChange{\protect\cite[{Thm. 2.7}]{cite3170}}.
When \(\mathsf{B}\) is a subgroup of \(\mathsf{S}\), then the code reduces to an ordinary stabilizer code.

The \(\llparenthesis n, \lceil\frac{q^n}{n(q^2-1)}\rceil,2\rrparenthesis _q\) family of Galois-qudit non-stabilizer codes is constructed in Ref. \NoCaseChange{\protect\cite{cite3170}}.

\codefieldsection{Parent}
\begin{eczvaluelist}
\item\relax
\flmRefsHyperref[eczindexfamilyrel]{code:galois_into_galois}{Galois-qudit code}\end{eczvaluelist}
\codefieldsection{Children}
\begin{eczvaluelist}
\item\relax
\flmRefsHyperref[eczindexfamilyrel]{code:non_stabilizer}{Union stabilizer (USt) code} --- Galois-qudit union stabilizer codes reduce to union stabilizer codes for \(q=2\).
\item\relax
\flmRefsHyperref[eczindexfamilyrel]{code:arvind}{\(\llparenthesis n,1+n(q-1),2\rrparenthesis _q\) union stabilizer code}\item\relax
\flmRefsHyperref[eczindexfamilyrel]{code:galois_cws}{Galois-qudit CWS code} --- Any Galois-qudit CWS code can be written as a USt whose (\(K=1\)) stabilizer code is the Galois-qudit cluster state and whose coset representatives are constructed from the \(q\)-ary classical code.
\item\relax
\flmRefsHyperref[eczindexfamilyrel]{code:galois_stabilizer}{Galois-qudit stabilizer code} --- A Galois-qudit stabilizer code with stabilizer group \(\mathsf{S}\) can be thought of as a Galois-qudit USt with only the identity coset representative. Conversely, if \(K = q^k\), and if the set of coset representatives of a Galois-qudit USt form a \(q\)-ary linear code, then they can be absorbed into a Galois-qudit stabilizer group that defines the USt.
\end{eczvaluelist}
\codefieldsection{Cousin}
\begin{eczvaluelist}
\item\relax
\flmRefsHyperref[eczindexfamilyrel]{code:projective}{Projective geometry code} --- Galois-qudit USt codes can be obtained from lines in projective space \NoCaseChange{\protect\cite{cite1977,cite1978}}.
\end{eczvaluelist}
\eczhbkcontributors{ Jiaxin Huang, \eczhuVVA }
\endeczcode

\eczcode{generalized_bicycle}{Generalized bicycle (GB) code}{~\NoCaseChange{\protect\cite{cite439,cite3183}}}
\codefieldsection{Alternative Names}
\begin{eczvaluelist}
\item\relax Hyperbicycle code
\item\relax Quasi-cyclic LP code
\end{eczvaluelist}
\eczhIndexCodeAliasName{generalized_bicycle}{Hyperbicycle code}
\eczhIndexCodeAliasName{generalized_bicycle}{Quasi-cyclic LP code}
\codefieldsection{Description}
A quasi-cyclic Galois-qudit CSS code constructed using a generalized version of the bicycle ansatz \NoCaseChange{\protect\cite{cite682}} from a pair of equivalent index-two quasi-cyclic linear codes.
Equivalently, the codes can be constructed via the lifted-product construction for \(G\) being a cyclic group \NoCaseChange{\protect\cite[{Sec. III.E}]{cite674}}.

Various instances of qubit GB codes are constructed in Ref. \NoCaseChange{\protect\cite{cite3183}} (for \(k=2\)) and in Refs. \NoCaseChange{\protect\cite{cite842,cite4636,cite843}}.

The stabilizer generator matrix of a \(\llbracket  n=2\ell,k,d\rrbracket _q\) GB\((a,b)\) code, constructed from polynomials \(a(x)\) and \(b(x)\), can be refined to the form
\flmMathEnvironment{align}{}{
  H_{X}=(A|B), H_{Z}^{T}=\begin{pmatrix}B\\-A\end{pmatrix}~,
}
where \(A=a(P)\) and \(B=b(P)\) are  \(\ell\times\ell\) circulant matrices, and \(P\) is the permutation matrix of a one-step length-\(\ell\) cyclic shift.
This refinement is a special case of \flmRefsHyperref{ref436}{symplectic doubling}.

With any GB\((a,b)\) code, there is an associated \(q\)-ary cyclic classical code \(C_{h(x)}^{\perp}=C_{g(x)}\) of length \(\ell\), with the check and generating polynomials
\flmMathEnvironment{align}{}{
  h(x)=\text{gcd}(a(x),b(x),x^{\ell}-1) \quad\text{and}\quad g(x)=\frac{x^{\ell}-1}{h(x)}~,
}
respectively.
The number of qudits encoded in such a GB code is \(k=2\deg h(x)\), twice the dimension of the underlying classical code \NoCaseChange{\protect\cite{cite1247}}.
The syndrome spaces of \(H_X\) and \(H_Z\) are cyclic codes generated by \(h(x)\) and its reciprocal polynomial \(h^{*}(x)\), respectively, so choosing \(h(x)\) from a cyclic code with distance \(d_s\) protects the syndrome bits by classical cyclic codes of distance \(d_s\) \NoCaseChange{\protect\cite{cite1247}}.

Two codes GB\((a,b)\) and GB\((a',b')\) of the same size \(n=2\ell\) are equivalent if one of the following conditions is satisfied \NoCaseChange{\protect\cite{cite3183}}:
\begin{enumerate}[(1)]\item \(a'(x)=a(x^{m})\) mod \(x^{\ell}-1\), \(b'(x)=b(x^{m})\) mod \(x^{\ell}-1\) for some \(m\) mutually prime with \(\ell\), gcd\((m,\ell)=1\);
\item \(a'(x)=b(x), b'(x)=a(x)\);
\item \(a'(x)\) and \(b'(x)\) are the reciprocal polynomials of \(a(x)\) and \(b(x)\), respectively;
\item \(a'(x)=\delta a(x), b'(x)=b(x)\), for some \(0\neq\delta\in \mathbb{F}_q\);
\item \(a'(x)=f(x)a(x), b'(x)=f(x)b(x)\), for some polynomial \(f(x)\in \mathbb{F}_q[x]\) such that gcd\((f,x^{\ell}-1)=1\).
\end{enumerate}

The following modified construction yields \textit{asymmetric bicycle (AB) codes}: \(\mathcal{Q}'=\text{CSS}(H'_{X},H_{Z})\) has \(H'_{X}=(A_{1}|B_{1}),A_{1}=a_{1}(P),B_{1}=b_{1}(P)\), where \(a_{1}=\frac{a(x)}{\text{gcd}(a,b)},b_{1}=\frac{b(x)}{\text{gcd}(a,b)}\).
The distance of a GB code is equal to the distance of its associated AB code \NoCaseChange{\protect\cite{cite3183}}.

\codefieldsection{Protection}
Given the parameters \([n_{0}=2\ell,k_{0},d_{0}]_q\) of the classical linear quasi-cyclic code QC\((a,b)\), the quantum CSS code GB\((a,b)\) has parameters \(\llbracket  2\ell,2k_{0}-2\ell,d\rrbracket _q\)  where \(d\geq d_{0}\).

Consider a quasi-cyclic QC\((a,b)\) in the special case \(a(x)=f(x)h(x),b(x)=h(x)\), where for some polynomial \(r(x)\), \(\text{gcd}(a(x),b(x),x^{\ell}-1)=p(x)\) is a factor of the generating polynomial, \(g(x)=p(x)q(x)\). Then the distance of the QC code satisfies the following two bounds:
\begin{enumerate}[(a)]\item If \(r(x)=0\), \(d_{0}\geq\text{min}\{d[q],1+d[p]\}\);
\item Otherwise, if \(\text{gcd}(r(x),x^{\ell}-1)=1\), then \(d_{0}\geq\text{min}\{2d[q],d[p]/\text{wgt}(r)\}\).
\end{enumerate}
Here, \(h(x)=\text{gcd}(a(x),b(x),x^{\ell}-1)\) and \(g(x)=\frac{x^{\ell}-1}{h(x)}\), and \(d[q]\) is the distance of the linear cyclic code generated by \(q(x)\) \NoCaseChange{\protect\cite{cite3183}}.

Let \(x^{\ell}-1=g(x)h(x)\) with \(g(x)\in \mathbb{F}_q[x]\) irreducible, and
\flmMathEnvironment{align}{}{
  d_{GV}=\text{max }d:\sum_{s=1}^{d-1}(q-1)^{s}\left[\begin{pmatrix}2\ell\\
  s \end{pmatrix}-\begin{pmatrix}\ell\\ s
  \end{pmatrix}\right]\,.\label{ref4637}
}
Then, there exists \(f(x)\in \mathbb{F}_q[x]\) such that the length-\(2\ell\) quasi-cyclic code QC\((hf,h)\) has distance \(d\geq\text{min}(d[g],d_{GV})\), where \(d[g]\) is the distance of the cyclic code generated by \(g(x)\) \NoCaseChange{\protect\cite{cite3183}}.

\subsection{GB codes with linear distance}
Let \(\ell\) be such that \(\text{ord}_{\ell}(q)=\ell-1\), where \(\text{ord}_{\ell}(q)\) is the multiplicative order function of \(q\) modulo \(\ell\).
This ensures that \(x^{\ell}-1\) has only two irreducible factors in \(\mathbb{F}_q[x]\), \(h(x)=1-x\), and \(g(x)=1+x+\cdots+x^{\ell-1}\).
Then, there is a GB code with parameters \(\llbracket  2\ell,2,d\geq d_{GV}\rrbracket _q\) \NoCaseChange{\protect\cite{cite3183}}.

An incommensurate GB code with row weight \(w\) and parameters \(\llbracket  n=2\ell,k,d\rrbracket _p\) is equivalent to a CSS code local in \(D\leq w-1\) dimensions (\(D\leq w-2\) if \(\ell\) is prime). Its parameters satisfy the inequalities \(d\leq\mathcal{O}(n^{1-1/D})\) and \(kd^{2/(D-1)}\leq\mathcal{O}(n)\) \NoCaseChange{\protect\cite{cite3183}}.

A weight-four GB code of an odd distance \(d=2r+1\) must have length \(n\geq 1+d^{2}\).
For an even distance \(d=2r\), the length \(n\geq d^{2}\) \NoCaseChange{\protect\cite{cite3183}}.

\codefieldsection{Rate}
GB codes can achieve an asymptotic rate of \(1/4\) \NoCaseChange{\protect\cite{cite3183}}.
For an odd prime \(\ell\), let a prime \(p\) be a quadratic-residue modulo \(\ell\), i.e. \(p=m^{2}\text{mod}\ell\) for some integer \(m\).
Then, \(x^{\ell}-1\) has only three irreducible factors in \(\mathbb{F}_p[x]\), and there is a quadratic-residue cyclic code \([\ell,(\ell+1)/2, d]_p\) with \(d\geq\sqrt{\ell}\) and an irreducible generator polynomial.
Using the GV distance \(d_{GV}\), a prime-field GB code with parameters \(\llbracket  2\ell,(\ell-1)/2,d\geq \ell^{1/2}\rrbracket _p\) exists.
There exist GB codes that achieve the Hashing bound \NoCaseChange{\protect\cite{cite4638}}.

\codefieldsection{Decoding}
\begin{eczvaluelist}
\item\relax BP-OSD decoder \NoCaseChange{\protect\cite{cite1247}}.
\end{eczvaluelist}
\codefieldsection{Code Capacity Threshold}
\begin{eczvaluelist}
\item\relax Depolarizing noise: \(15\%\) for a family of 6-limited \(\llbracket 2^{m+1}-2,2m\rrbracket \) GB codes with BP-OSD decoder \NoCaseChange{\protect\cite[{Appx. C}]{cite1247}}.
\end{eczvaluelist}
\codefieldsection{Parents}
\begin{eczvaluelist}
\item\relax
\flmRefsHyperref[eczindexfamilyrel]{code:2bga}{Two-block group-algebra (2BGA) codes} --- A code GB\((a,b)\) with circulants of size \(\ell\) is a 2BGA code over the cyclic group \(\mathbb{Z}_{\ell}\).
More precisely, for the cyclic group \(\mathbb{Z}_{\ell}\equiv \langle x|x^\ell=1\rangle \), any element \(a\) of the \flmRefsHyperref{ref205}{group algebra} \(\mathbb{F}_q[\mathbb{Z}_{\ell}]\) can be seen as a polynomial \(a(x)\in \mathbb{F}_q[x]\) over the group generator \(x\), where the polynomial degree \(\deg a(x)<\ell\).
The 2BGA code LP\((a,b)\) is then just a generalized bicycle code GB\([a(x),b(x)]\) constructed from the polynomials \(a(x)\) and \(b(x)\) corresponding to \(a,b\in \mathbb{F}_q[\mathbb{Z}_{\ell}]\).

\item\relax
\flmRefsHyperref[eczindexfamilyrel]{code:abelian_lifted_product}{Abelian LP code} --- A code GB\((a,b)\) with circulants of size \(\ell\) is a special case of a lifted-product code LP\((A,B)\) code over the Abelian \flmRefsHyperref{ref205}{group algebra} \(\mathbb{F}_q[\mathbb{Z}_{\ell}]\) associated with a cyclic group, with \(1\times 1\) matrices \(A=a(x)\), \(B=b(x)\) given by the corresponding polynomials.
Quasi-cyclic LP codes, i.e., LP codes constructed from cyclic groups, are equivalent to GB codes \NoCaseChange{\protect\cite[{Sec. III.E}]{cite674}}.

\item\relax
\flmRefsHyperref[eczindexfamilyrel]{code:translationally_invariant_stabilizer}{Lattice stabilizer code} --- Incommensurate GB codes of row weight \(w\) are equivalent to CSS codes local in \(D \leq w-1\) Euclidean dimensions, or in \(D \leq w-2\) dimensions when \(\ell\) is prime \NoCaseChange{\protect\cite[{Statement 13}]{cite3183}\protect\cite[{Sec. II.A}]{cite4639}}.
\end{eczvaluelist}
\codefieldsection{Children}
\begin{eczvaluelist}
\item\relax
\flmRefsHyperref[eczindexfamilyrel]{code:bicycle}{Bicycle code} --- A GB code whose circulants satisfy \(B = A^T\) reduces to a bicycle code.

\item\relax
\flmRefsHyperref[eczindexfamilyrel]{code:bipartite_cyclic_cluster}{Bipartite cyclic cluster (BCC) code} --- BCC codes are GB codes with weight-two circulant \(A\) (polynomial \(a(x)=1+x\)) \NoCaseChange{\protect\cite{cite440}}.
\end{eczvaluelist}
\codefieldsection{Cousins}
\begin{eczvaluelist}
\item\relax
\flmRefsHyperref[eczindexfamilyrel]{code:sc_qldpc}{Quantum spatially coupled (SC-QLDPC) code} --- Qubit GB codes can be categorized as 1D SC-QLDPC codes, see \NoCaseChange{\protect\cite[{Remark 7}]{cite644}}.
\item\relax
\flmRefsHyperref[eczindexfamilyrel]{code:general_qldpc}{QLDPC code} --- Stabilizer generators of the code GB\((a,b)\) have weights given by the sum of weights of polynomials \(a(x)\) and \(b(x)\).
The GB code ansatz is convenient for designing QLDPC codes and several extensions exist \NoCaseChange{\protect\cite{cite4640}}.

\item\relax
\flmRefsHyperref[eczindexfamilyrel]{code:single_shot}{Single-shot code} --- A qubit GB code \(\llbracket n,k,d\rrbracket _2\) has \(k\) non-trivial relations between the syndrome bits, which is expected to help with operation in a fault-tolerant regime (in the presence of syndrome measurement errors). See Ref. \NoCaseChange{\protect\cite{cite842}} for many examples of such codes. There is numerical evidence that a particular family is single shot \NoCaseChange{\protect\cite{cite843}}.
\item\relax
\flmRefsHyperref[eczindexfamilyrel]{code:hypergraph_product}{Hypergraph product (HGP) code} --- An arbitrary qubit GB code of length \(2\ell\) is equivalent to a rotated HGP code with periodicity vectors \(\vec{L}_{1}\) and \(\vec{L}_{2}\) such that \(\lvert \vec{L}_{1}\times\vec{L}_{2}\rvert=\ell\) \NoCaseChange{\protect\cite{cite3183}}.
\item\relax
\flmRefsHyperref[eczindexfamilyrel]{code:surface}{Kitaev surface code} --- Any non-trivial qubit GB code of row weight four and distance \(d\geq 3\) is equivalent to a square-lattice surface code \NoCaseChange{\protect\cite{cite3183}}.
\item\relax
\flmRefsHyperref[eczindexfamilyrel]{code:bch}{Binary BCH code} --- There exist examples of GB codes whose syndromes are protected by small BCH codes \NoCaseChange{\protect\cite{cite1247}}.
\item\relax
\flmRefsHyperref[eczindexfamilyrel]{code:haah_cubic}{Haah cubic code (CC)} --- A GB code for the group \(G=\mathbb{Z}_{L}^{\times 3}\) is a cubic code \NoCaseChange{\protect\cite[{Sec. III.A}]{cite674}}.
\item\relax
\flmRefsHyperref[eczindexfamilyrel]{code:qcga}{Bivariate bicycle (BB) code} --- GB codes (BB codes) are 2BGA codes over the cyclic group \(\mathbb{Z}_{\ell}\) (Abelian group \(\mathbb{Z}_{r} \times \mathbb{Z}_{s}\)). The two codes are the same when \(r\) and \(s\) are relatively prime due to the isomorphism \(\mathbb{Z}_{r} \times \mathbb{Z}_{s} \cong \mathbb{Z}_{\ell = rs}\).
\item\relax
\flmRefsHyperref[eczindexfamilyrel]{code:cyclic_hgp}{Cyclic Hypergraph Product Code} --- Cyclic HGP codes and GB codes both use circulant matrices as building blocks.
\end{eczvaluelist}
\eczhbkcontributors{ Leonid Pryadko, Renyu Wang, \eczhuVVA }
\endeczcode

\eczcode{stabilizer_over_gfqsq}{Hermitian Galois-qudit code}{~\NoCaseChange{\protect\cite{cite2912,cite696}\protect\cite[{Corr. 5}]{cite532}}}
\codefieldsection{Alternative Names}
\begin{eczvaluelist}
\item\relax \(\mathbb{F}_{q^2}\)-linear stabilizer code
\end{eczvaluelist}
\eczhIndexCodeAliasName{stabilizer_over_gfqsq}{\(\mathbb{F}_{q^2}\)-linear stabilizer code}
\codefieldsection{Description}
An \(\llbracket n,k,d\rrbracket _q\) true Galois-qudit stabilizer code constructed from a Hermitian self-orthogonal linear code over \(\mathbb{F}_{q^2}\) using the one-to-one correspondence between the Galois-qudit Pauli matrices and elements of the Galois field \(\mathbb{F}_{q^2}\).

Galois-qudit stabilizer codes are in one-to-one correspondence with trace-alternating self-orthogonal additive codes of length \(n\) over \(\mathbb{F}_{q^2}\) via the \flmRefsHyperref{ref1779}{\(\mathbb{F}_{q^2}\) representation}.
Hermitian self-orthogonal linear codes over \(\mathbb{F}_{q^2}\) are automatically trace-alternating self-orthogonal, and applying this mapping to such codes yields Hermitian codes \NoCaseChange{\protect\cite[{Corr. 19}]{cite813}}.

More generally, a trace-alternating self-orthogonal additive code \(C \subseteq \mathbb{F}_{q^2}^n\) of size \(q^r\) yields an \(\llbracket n,n-r\rrbracket _q\) true stabilizer code \NoCaseChange{\protect\cite[{Corr. 1}]{cite696}}.
When \(C\) is Hermitian self-orthogonal and \(\mathbb{F}_{q^2}\)-linear of parameters \([n,k]_{q^2}\), the construction specializes to a Hermitian \(\llbracket n,n-2k\rrbracket _q\) true stabilizer code; this is called the \textit{Hermitian Galois-qudit construction} \NoCaseChange{\protect\cite[{Corr. 19}]{cite813}}.
The Hermitian construction was first proven via the \flmRefsHyperref{ref873}{Galois symplectic representation} (showing self-orthogonality under the trace-symplectic inner product; see Ref. \NoCaseChange{\protect\cite{cite696}}, Corr. 1).
There is an isomorphism between the \flmRefsHyperref{ref873}{Galois-symplectic} and \flmRefsHyperref{ref1779}{\(\mathbb{F}_{q^2}\) representations} \NoCaseChange{\protect\cite[{Thm. 27.4.1}]{cite2024}}.

It has also been extended to \(q^{2m}\)-ary Hermitian self-orthogonal linear codes \NoCaseChange{\protect\cite{cite4641}}, and similar constructions were formulated in Ref. \NoCaseChange{\protect\cite{cite4642}}.
\textit{Quantum construction X} \NoCaseChange{\protect\cite[{Thm. 27.4.4}]{cite2024}}, related to (classical) Construction X and Construction XX, allows for the use of nearly self-orthogonal codes \NoCaseChange{\protect\cite{cite4643,cite4644,cite2674}}; see Ref. \NoCaseChange{\protect\cite{cite4645}} for a review.

\codefieldsection{Protection}
For a trace-alternating self-orthogonal additive code \(C \subseteq \mathbb{F}_{q^2}^n\), the resulting true stabilizer code has distance
\flmMathEnvironment{align}{}{
d=\min\{\operatorname{wt}(v):v \in C^{\perp}\setminus C\}~,
}
where \(\perp\) denotes duality under the trace-alternating inner product \NoCaseChange{\protect\cite[{Corr. 1}]{cite696}}.
If \(C\) is Hermitian self-orthogonal and \(\mathbb{F}_{q^2}\)-linear, then
\flmMathEnvironment{align}{}{
d=\min\{\operatorname{wt}(v):v \in C^{\perp_h}\setminus C\}~,
}
so the code distance is bounded below by the Hermitian dual distance of \(C\) \NoCaseChange{\protect\cite[{Corr. 19}]{cite813}}.

\codefieldsection{Parent}
\begin{eczvaluelist}
\item\relax
\flmRefsHyperref[eczindexfamilyrel]{code:galois_true_stabilizer}{True Galois-qudit stabilizer code} --- Hermitian codes are true stabilizer codes because they are based on Hermitian self-orthogonal linear (as opposed to additive) codes over \(\mathbb{F}_{q^2}\).
\end{eczvaluelist}
\codefieldsection{Children}
\begin{eczvaluelist}
\item\relax
\flmRefsHyperref[eczindexfamilyrel]{code:stabilizer_over_gf4}{Hermitian qubit code}\item\relax
\flmRefsHyperref[eczindexfamilyrel]{code:stab_9_1_5}{\(\llbracket 9,1,5\rrbracket _3\) quantum Glynn code}\item\relax
\flmRefsHyperref[eczindexfamilyrel]{code:quantum_twisted}{Quantum twisted code}\item\relax
\flmRefsHyperref[eczindexfamilyrel]{code:quantum_plane_curve}{Quantum plane-curve code}\end{eczvaluelist}
\codefieldsection{Cousins}
\begin{eczvaluelist}
\item\relax
\flmRefsHyperref[eczindexfamilyrel]{code:dual}{Dual linear code} --- Hermitian codes are constructed from Hermitian self-orthogonal linear codes over \(\mathbb{F}_{q^2}\) via the \flmRefsHyperref{ref1779}{\(\mathbb{F}_{q^2}\) representation}.
\item\relax
\flmRefsHyperref[eczindexfamilyrel]{code:matrix_product}{Matrix-product code} --- Hermitian self-orthogonal matrix-product codes over \(\mathbb{F}_{q^2}\) can be used to construct quantum codes via the Hermitian construction \NoCaseChange{\protect\cite{cite1913,cite1914}}.
\item\relax
\flmRefsHyperref[eczindexfamilyrel]{code:galois_subsystem_stabilizer}{Subsystem Galois-qudit stabilizer code} --- The Hermitian construction has been extended to subsystem Galois-qudit stabilizer codes \NoCaseChange{\protect\cite{cite1742}}.
\item\relax
\flmRefsHyperref[eczindexfamilyrel]{code:complex_hadamard}{Complex Hadamard spherical code} --- Complex Hadamard matrices can be used to build Hermitian \NoCaseChange{\protect\cite{cite2348}} and other \NoCaseChange{\protect\cite{cite2350}} Galois-qudit stabilizer codes.
\item\relax
\flmRefsHyperref[eczindexfamilyrel]{code:gkp_concatenated}{Concatenated GKP code} --- Concatenations of square-lattice GKP codes with Hermitian Galois-qudit codes achieve the capacity for all loss rates \NoCaseChange{\protect\cite{cite4055}}.
\item\relax
\flmRefsHyperref[eczindexfamilyrel]{code:quantum_mds}{Quantum maximum-distance-separable (MDS) code} --- Many quantum MDS codes are constructed from Hermitian self-orthogonal codes over \(\mathbb{F}_{q^2}\) using the Hermitian construction \NoCaseChange{\protect\cite{cite975,cite976,cite977,cite978}}, in particular from cyclic \NoCaseChange{\protect\cite{cite979}}, constacyclic \NoCaseChange{\protect\cite{cite980,cite981,cite978}} and negacyclic \NoCaseChange{\protect\cite{cite982}} codes.
\item\relax
\flmRefsHyperref[eczindexfamilyrel]{code:galois_bch}{Galois-qudit BCH code} --- Galois-qudit BCH codes can be constructed via the CSS construction or the Hermitian construction.
\item\relax
\flmRefsHyperref[eczindexfamilyrel]{code:galois_duadic}{Quantum duadic code} --- Quantum duadic codes can be constructed via the CSS construction or the Hermitian construction.
\item\relax
\flmRefsHyperref[eczindexfamilyrel]{code:galois_reed_muller}{Galois-qudit quantum RM code} --- Galois-qudit RM codes admit a Hermitian subfamily built from \(\mathrm{GRM}_{q^2}(v,m)\) codes contained in their Hermitian duals \NoCaseChange{\protect\cite{cite828}}.
\item\relax
\flmRefsHyperref[eczindexfamilyrel]{code:galois_grs}{Galois-qudit GRS code} --- Galois-qudit GRS codes can be constructed via the CSS construction or the Hermitian construction.
\end{eczvaluelist}
\eczhbkcontributors{ \eczhuVVA }
\endeczcode

\eczcode{lifted_product}{Lifted-product (LP) code}{~\NoCaseChange{\protect\cite{cite1247,cite674,cite184}}}
\codefieldsection{Alternative Names}
\begin{eczvaluelist}
\item\relax Panteleev-Kalachev (PK) code
\end{eczvaluelist}
\eczhIndexCodeAliasName{lifted_product}{Panteleev-Kalachev (PK) code}
\codefieldsection{Description}
Galois-qudit code that utilizes the notion of a lifted product in its construction. Lifted products of certain classical Tanner codes are the first (asymptotically) \textit{good QLDPC codes}.

A code can be defined by \(LP(A,B)\), where \(A\) and \(B\) are a pair of matrices with elements from a \flmRefsHyperref{ref205}{group algebra}.
Heuristically, the code is constructed as a hypergraph product code over the \flmRefsHyperref{ref205}{group algebra}, with each entry subsequently extended into a matrix.

More technically, a \textit{lifted product over} a ring \(R\) is a product of two chain complexes whose chains are free modules over \(R\).
An interesting case is when \(R=\mathbb{F}_q [G]\), the \flmRefsHyperref{ref205}{group-\(G\) algebra} over the finite field \({\mathbb{F}}_q = \mathbb{F}_q\); in this case, the product can be called a \(G\)-\textit{lifted product}.
Just like its further generalization the balanced product, a lifted product code generalizes a hypergraph product code in that a reduction of symmetry is exploited to decrease the number of physical qubits required.
The first version of this construction appeared as a family of generalized hypergraph product codes that contains hypergraph product codes in the case where one of the two input parity-check matrices is square \NoCaseChange{\protect\cite{cite1247}}.

The key operation behind the \(G\)-lifted product is the \(G\)-\textit{lift}, a \flmRefsHyperref{ref205}{group-algebraic} version of the \flmRefsHyperref{ref47}{lifting} procedure of protograph LDPC codes.
A combination of the lift and the usual hypergraph product yields lifted-product codes.
The two operations commute: one can first take the usual hypergraph product of two chain complexes, and then lift the resulting product complex; equivalently, one can take the hypergraph product of the two lifted complexes.

\codefieldsection{Protection}
Code performance strongly depends on the group \(G\) used in the product \NoCaseChange{\protect\cite{cite674}}.
\codefieldsection{Rate}
There is no known simple way to compute the logical dimension \(k\) in the general case \NoCaseChange{\protect\cite{cite674}}.
\codefieldsection{Gates}
\begin{eczvaluelist}
\item\relax Transversal dimension jump, a code switching protocol between two LP codes \NoCaseChange{\protect\cite{cite4646}}.
\end{eczvaluelist}
\codefieldsection{Decoding}
\begin{eczvaluelist}
\item\relax Linear time iterative decoder \NoCaseChange{\protect\cite{cite3784}}.
\end{eczvaluelist}
\codefieldsection{Notes}
\begin{eczvaluelist}
\item\relax Formerly known as \textit{generalized hypergraph product codes} \NoCaseChange{\protect\cite{cite1247}}, and later renamed to lifted-product codes \NoCaseChange{\protect\cite{cite674,cite3442}}.
\end{eczvaluelist}
\codefieldsection{Parent}
\begin{eczvaluelist}
\item\relax
\flmRefsHyperref[eczindexfamilyrel]{code:balanced_product}{Balanced product (BP) code} --- Coarsely speaking, a lifted product is a balanced product where the group \(G\) acts freely. In principle, a lifted product can be defined for rings that are more general than \flmRefsHyperref{ref205}{group algebras} \( \mathbb{F}_q G \).
\end{eczvaluelist}
\codefieldsection{Children}
\begin{eczvaluelist}
\item\relax
\flmRefsHyperref[eczindexfamilyrel]{code:checkerboard}{Checkerboard model code} --- The checkerboard model code can be formulated directly as an LP code \NoCaseChange{\protect\cite{cite1350}}.
\item\relax
\flmRefsHyperref[eczindexfamilyrel]{code:dhlv}{Dinur-Hsieh-Lin-Vidick (DHLV) code} --- DHLV codes are LP codes \NoCaseChange{\protect\cite[{Footnote 7}]{cite1101}}.
\item\relax
\flmRefsHyperref[eczindexfamilyrel]{code:lcs}{Lift-connected surface (LCS) code}\item\relax
\flmRefsHyperref[eczindexfamilyrel]{code:triangular_color}{Honeycomb (6.6.6) color code} --- The 6.6.6 color code can be formulated directly as an LP code \NoCaseChange{\protect\cite{cite1350}}.
\item\relax
\flmRefsHyperref[eczindexfamilyrel]{code:galois_hypergraph_product}{Galois-qudit HGP code} --- Lifted-product codes for trivial lift are Galois-qudit hypergraph-product codes.
\item\relax
\flmRefsHyperref[eczindexfamilyrel]{code:abelian_lifted_product}{Abelian LP code}\item\relax
\flmRefsHyperref[eczindexfamilyrel]{code:expander_lifted_product}{Expander LP code}\item\relax
\flmRefsHyperref[eczindexfamilyrel]{code:2bga}{Two-block group-algebra (2BGA) codes} --- 2BGA codes are LP\((a,b)\) codes, constructed from a pair of one-by-one matrices \(a,b\in \mathbb{F}_q[G]\) in a \flmRefsHyperref{ref205}{group algebra}.
\end{eczvaluelist}
\codefieldsection{Cousins}
\begin{eczvaluelist}
\item\relax
\flmRefsHyperref[eczindexfamilyrel]{code:haah_cubic}{Haah cubic code (CC)} --- A lifted-product code constructed with coefficients in the ring \(R=\mathbb{F}_2[x,y,z]/(x^L-1,y^L-1,z^L-1)\) is a cubic code \NoCaseChange{\protect\cite[{Appx. B}]{cite184}}.
\item\relax
\flmRefsHyperref[eczindexfamilyrel]{code:fiber_bundle}{Fiber-bundle code} --- The specific fiber-bundle QLDPC code achieving a distance scaling better than \(\sqrt{n}~\text{polylog}(n)\) can also be formulated directly as an LP code (see published version of Ref.\NoCaseChange{\protect\cite{cite3477}}).
Lifted products of a length-one with a length-\(m\) chain complex can be thought of as fiber-bundle codes \NoCaseChange{\protect\cite{cite434}}.

\item\relax
\flmRefsHyperref[eczindexfamilyrel]{code:toric}{Toric code} --- A lifted-product code for the ring \(R=\mathbb{F}_2[x,y]/(x^L-1,y^L-1)\) is the toric code \NoCaseChange{\protect\cite[{Appx. B}]{cite184}}.
\item\relax
\flmRefsHyperref[eczindexfamilyrel]{code:subsystem_lifted_product}{Subsystem lifted-product (SLP) code} --- SLP codes reduce to (subspace) LP codes when there is no gauge subsystem.
\item\relax
\flmRefsHyperref[eczindexfamilyrel]{code:two_block_quantum}{Two-block CSS code} --- LP codes can be constructed using non-square matrices and taking a hypergraph product over a group algebra, while two-block CSS codes are constructed directly using square matrices.
\end{eczvaluelist}
\eczhbkcontributors{ Finnegan Voichick, Pavel Panteleev, \eczhuVVA }
\endeczcode

\eczcode{maximal_entanglement_galois_stabilizer}{Maximal-entanglement EA Galois-qudit stabilizer code}{~\NoCaseChange{\protect\cite{cite3637,cite1431,cite1911,cite4647}}}
\codefieldsection{Description}
An \(\llbracket n,k,d;e\rrbracket _q\) EA Galois-qudit stabilizer code for which \(e = n-k\), i.e., the number of required pre-shared maximally entangled Galois-qudit pairs saturates the defining maximal-entanglement condition.

\codefieldsection{Rate}
Maximal entanglement is required to achieve the EA hashing bound for the depolarizing channel using the father protocol from Refs. \NoCaseChange{\protect\cite{cite4648,cite4649}}; see \NoCaseChange{\protect\cite[{Footnote 2}]{cite3637}}.
\codefieldsection{Parent}
\begin{eczvaluelist}
\item\relax
\flmRefsHyperref[eczindexfamilyrel]{code:ea_galois_stabilizer}{EA Galois-qudit stabilizer code}\end{eczvaluelist}
\codefieldsection{Cousins}
\begin{eczvaluelist}
\item\relax
\flmRefsHyperref[eczindexfamilyrel]{code:lcd}{Linear code with complementary dual (LCD)} --- Asymptotically good maximal-entanglement EA Galois-qudit stabilizer codes can be constructed from LCD codes \NoCaseChange{\protect\cite{cite1911}}.
\item\relax
\flmRefsHyperref[eczindexfamilyrel]{code:ea_turbo}{EA quantum turbo code} --- Maximal-entanglement EA quantum turbo codes come close to achieving the EA hashing bound \NoCaseChange{\protect\cite{cite3636}}; see \NoCaseChange{\protect\cite[{Footnote 2}]{cite3637}}.
\item\relax
\flmRefsHyperref[eczindexfamilyrel]{code:ea_quantum_lcd}{EA quantum LCD code} --- Asymptotically good maximal-entanglement EA Galois-qudit stabilizer codes can be constructed from LCD codes \NoCaseChange{\protect\cite{cite1911}}.
\end{eczvaluelist}
\eczhbkcontributors{ \eczhuVVA }
\endeczcode

\eczcode{quantum_ag}{Quantum AG code}{~\NoCaseChange{\protect\cite{cite4650,cite4602}}}
\codefieldsection{Description}
True Galois-qudit stabilizer code constructed from evaluation AG codes via the Galois-qudit Hermitian construction or the Galois-qudit CSS construction.

\codefieldsection{Rate}
Quantum AG codes can be asymptotically good \NoCaseChange{\protect\cite{cite4650,cite4602}}. There exist three such families \NoCaseChange{\protect\cite{cite695,cite699,cite698}} that admit a diagonal transversal gate at the third level of the \flmTerm{term}{ref694}{}{Clifford hierarchy}.
\codefieldsection{Magic}
By defining a generalization of triorthogonal matrices to Galois qudits of dimension \(q=2^m\), one can construct an asymptotically good family of quantum AG codes that admits a diagonal transversal gate at the third level of the \flmTerm{term}{ref694}{}{Clifford hierarchy} and attains a zero magic-state yield parameter, \(\gamma = 0\) \NoCaseChange{\protect\cite{cite695}}. This code can be treated as a qubit code by decomposing each Galois qudit into a Kronecker product of \(m\) qubits; see \NoCaseChange{\protect\cite{cite696,cite398,cite698,cite699,cite700}\protect\cite[{Sec. 5.3}]{cite697}}. Two other asymptotically good families \NoCaseChange{\protect\cite{cite699,cite698}} admit a transversal \(CCZ\) gate (a different diagonal gate at the third level of the \flmTerm{term}{ref694}{}{Clifford hierarchy}) and achieve \(\gamma \to 0\) with constant alphabet size.
\codefieldsection{Encoding}
\begin{eczvaluelist}
\item\relax Encoding defined in Ref. \NoCaseChange{\protect\cite{cite4602}} uses a technique from Ref. \NoCaseChange{\protect\cite{cite3716}} to encode quantum stabilizer codes.
\end{eczvaluelist}
\codefieldsection{Transversal and Permutation-Based Gates}
\begin{eczvaluelist}
\item\relax There exist three asymptotically good code families \NoCaseChange{\protect\cite{cite695,cite699,cite698}} that admit a diagonal transversal gate at the third level of the \flmTerm{term}{ref694}{}{Clifford hierarchy}.
\item\relax By decomposing each Galois qudit into a Kronecker product of qubits, the family of Ref. \NoCaseChange{\protect\cite{cite698}} yields an explicit asymptotically good qubit CSS code family with parameters \(\llbracket N,K=\Theta(N),D=\Theta(N)\rrbracket \) on which \(\overline{CCZ}^{\otimes K}\) is realized by a transversal application of physical \(CCZ\) gates on a constant fraction of qubits.
\item\relax There exists an asymptotically good code family that admits three-Galois-qudit non-Clifford gates for any three logical Galois qudits \NoCaseChange{\protect\cite{cite733}}.
\end{eczvaluelist}
\codefieldsection{Parent}
\begin{eczvaluelist}
\item\relax
\flmRefsHyperref[eczindexfamilyrel]{code:galois_true_stabilizer}{True Galois-qudit stabilizer code} --- Quantum AG codes can be constructed via the Galois-qudit CSS construction or the Galois-qudit Hermitian construction.
\end{eczvaluelist}
\codefieldsection{Children}
\begin{eczvaluelist}
\item\relax
\flmRefsHyperref[eczindexfamilyrel]{code:binary_quantum_goppa}{Binary quantum Goppa code}\item\relax
\flmRefsHyperref[eczindexfamilyrel]{code:quantum_hermitian_ag}{Quantum Hermitian AG code}\item\relax
\flmRefsHyperref[eczindexfamilyrel]{code:quantum_plane_curve}{Quantum plane-curve code}\item\relax
\flmRefsHyperref[eczindexfamilyrel]{code:galois_grs}{Galois-qudit GRS code} --- Galois-qudit GRS codes can be constructed via the CSS construction or the Hermitian construction from GRS codes, which are evaluation AG codes.
\end{eczvaluelist}
\codefieldsection{Cousins}
\begin{eczvaluelist}
\item\relax
\flmRefsHyperref[eczindexfamilyrel]{code:evaluation}{Evaluation AG code} --- Quantum AG codes are quantum analogues of evaluation AG codes.
\item\relax
\flmRefsHyperref[eczindexfamilyrel]{code:quantum_triorthogonal}{Triorthogonal code} --- By defining a generalization of triorthogonal matrices to Galois qudits of dimension \(q=2^m\), one can construct an asymptotically good family of quantum AG codes that admits a diagonal transversal gate at the third level of the \flmTerm{term}{ref694}{}{Clifford hierarchy} and attains a zero magic-state yield parameter, \(\gamma = 0\) \NoCaseChange{\protect\cite{cite695}}. This code can be treated as a qubit code by decomposing each Galois qudit into a Kronecker product of \(m\) qubits; see \NoCaseChange{\protect\cite{cite696,cite398,cite698,cite699,cite700}\protect\cite[{Sec. 5.3}]{cite697}}. Two other asymptotically good families \NoCaseChange{\protect\cite{cite699,cite698}} admit a transversal \(CCZ\) gate (a different diagonal gate at the third level of the \flmTerm{term}{ref694}{}{Clifford hierarchy}) and achieve \(\gamma \to 0\) with constant alphabet size.
\item\relax
\flmRefsHyperref[eczindexfamilyrel]{code:shimura}{Tsfasman-Vladut-Zink (TVZ) code} --- The AG codes used in an asymptotically good construction of quantum AG codes with non-Clifford transversal gates \NoCaseChange{\protect\cite{cite699}} are those of the TVZ codes.
\item\relax
\flmRefsHyperref[eczindexfamilyrel]{code:elliptic}{Elliptic code} --- Elliptic codes can be used to construct quantum AG codes \NoCaseChange{\protect\cite{cite871}}.
\end{eczvaluelist}
\eczhbkcontributors{ \eczhuVVA }
\endeczcode

\eczcode{galois_duadic}{Quantum duadic code}{~\NoCaseChange{\protect\cite{cite4651,cite4652,cite4653,cite819}\protect\cite[{Ch. 5}]{cite872}}}
\codefieldsection{Description}
True Galois-qudit stabilizer code constructed from \(q\)-ary duadic codes via the Hermitian construction or the Galois-qudit CSS construction.
Large subclasses of quantum duadic codes are degenerate \NoCaseChange{\protect\cite[{Sec. 5.5}]{cite872}}.

Quantum construction X yields an infinite family of quantum duadic codes with favorable parameters \NoCaseChange{\protect\cite{cite819}}.

\codefieldsection{Parent}
\begin{eczvaluelist}
\item\relax
\flmRefsHyperref[eczindexfamilyrel]{code:galois_true_stabilizer}{True Galois-qudit stabilizer code} --- Quantum duadic codes can be constructed via the CSS construction or the Hermitian construction.
\end{eczvaluelist}
\codefieldsection{Child}
\begin{eczvaluelist}
\item\relax
\flmRefsHyperref[eczindexfamilyrel]{code:galois_quad_residue}{Quantum quadratic-residue (QR) code} --- Quantum QR codes are quantum duadic codes since QR codes are duadic codes.
\end{eczvaluelist}
\codefieldsection{Cousins}
\begin{eczvaluelist}
\item\relax
\flmRefsHyperref[eczindexfamilyrel]{code:q-ary_duadic}{\(q\)-ary duadic code} --- Quantum duadic codes are quantum analogues of \(q\)-ary duadic codes.
\item\relax
\flmRefsHyperref[eczindexfamilyrel]{code:galois_css}{Galois-qudit CSS code} --- Quantum duadic codes can be constructed via the CSS construction or the Hermitian construction.
\item\relax
\flmRefsHyperref[eczindexfamilyrel]{code:stabilizer_over_gfqsq}{Hermitian Galois-qudit code} --- Quantum duadic codes can be constructed via the CSS construction or the Hermitian construction.
\end{eczvaluelist}
\eczhbkcontributors{ \eczhuVVA }
\endeczcode

\eczcode{quantum_gabidulin}{Quantum Gabidulin code}{~\NoCaseChange{\protect\cite{cite2138}}}
\codefieldsection{Description}
A Galois-qudit stabilizer code over \(n\) Galois qudits of dimension \(q = 2^n \) that is useful in protecting against faults in qubit Clifford circuits acting on stacked quantum memories.
This code can be treated as a code on an \(n\times n\) qubit stacked memory by decomposing each Galois qudit into a Kronecker product of \(n\) qubits; see \NoCaseChange{\protect\cite{cite696,cite398,cite698,cite699,cite700}\protect\cite[{Sec. 5.3}]{cite697}}.

A quantum Gabidulin code is defined using two Gabidulin codes with associated parameters \(r,s\), respectively, such that \(r+s = n\) \NoCaseChange{\protect\cite{cite2138}}.

\codefieldsection{Protection}
The code distance is the minimum rank distance --- the rank of the field element of the lowest-rank undetectable Galois-qudit Pauli error, with the rank calculated by writing the element as an \(n\times n\) binary matrix.
The code is useful in protecting against faults in \(n\)-qubit Clifford circuits with \(n\) layers, which preserve the minimum rank distance.

\codefieldsection{Parent}
\begin{eczvaluelist}
\item\relax
\flmRefsHyperref[eczindexfamilyrel]{code:galois_true_stabilizer}{True Galois-qudit stabilizer code}\end{eczvaluelist}
\codefieldsection{Cousins}
\begin{eczvaluelist}
\item\relax
\flmRefsHyperref[eczindexfamilyrel]{code:gabidulin}{Gabidulin code} --- A quantum Gabidulin code is defined using two Gabidulin codes with associated parameters \(r,s\), respectively, such that \(r+s = n\) \NoCaseChange{\protect\cite{cite2138}}.
\item\relax
\flmRefsHyperref[eczindexfamilyrel]{code:rank_metric}{Rank-metric code} --- Quantum Gabidulin code and (classical) rank-metric code distances are based on ranks of the matrix representations of their corresponding errors.
\end{eczvaluelist}
\eczhbkcontributors{ \eczhuVVA }
\endeczcode

\eczcode{quantum_hermitian_ag}{Quantum Hermitian AG code}{~\NoCaseChange{\protect\cite{cite869,cite870}}}
\codefieldsection{Description}
Quantum AG code constructed from Hermitian AG codes via the Galois-qudit Hermitian construction or the Galois-qudit CSS construction.
The underlying classical codes can be constructed from one-point \NoCaseChange{\protect\cite{cite869}} or two-point \NoCaseChange{\protect\cite{cite870}} Hermitian codes on Hermitian curves (see also Ref. \NoCaseChange{\protect\cite{cite871}}).
In parameter ranges where two-point Hermitian codes improve on one-point codes, the resulting quantum codes can also have improved parameters \NoCaseChange{\protect\cite{cite870}}.

\codefieldsection{Parent}
\begin{eczvaluelist}
\item\relax
\flmRefsHyperref[eczindexfamilyrel]{code:quantum_ag}{Quantum AG code}\end{eczvaluelist}
\codefieldsection{Cousins}
\begin{eczvaluelist}
\item\relax
\flmRefsHyperref[eczindexfamilyrel]{code:hermitian}{Hermitian code} --- Quantum Hermitian AG codes are quantum analogues of Hermitian codes.
\item\relax
\flmRefsHyperref[eczindexfamilyrel]{code:asymmetric_qecc}{Asymmetric quantum code (AQC)} --- One-point and two-point Hermitian codes can be used to construct asymmetric Galois-qudit CSS codes, and the two-point construction can improve on the corresponding one-point codes \NoCaseChange{\protect\cite{cite870}}.
\end{eczvaluelist}
\eczhbkcontributors{ \eczhuVVA }
\endeczcode

\eczcode{quantum_plane_curve}{Quantum plane-curve code}{~\NoCaseChange{\protect\cite{cite4620}}}
\codefieldsection{Description}
Quantum AG code constructed from plane-curve codes via the Galois-qudit Hermitian construction.
Code parameters are \(\llbracket q^3,q^3+q^2-3q-2r,r+2q-q^2\rrbracket _q\), where \(r\) is an integer satisfying \(q^2 - 2 \leq r \leq q^2 + q - 3\), and where the underlying plane curve is \(y^q + y = x^{q-1}\).

\codefieldsection{Parents}
\begin{eczvaluelist}
\item\relax
\flmRefsHyperref[eczindexfamilyrel]{code:stabilizer_over_gfqsq}{Hermitian Galois-qudit code}\item\relax
\flmRefsHyperref[eczindexfamilyrel]{code:quantum_ag}{Quantum AG code}\end{eczvaluelist}
\codefieldsection{Cousins}
\begin{eczvaluelist}
\item\relax
\flmRefsHyperref[eczindexfamilyrel]{code:plane_curve}{Plane-curve evaluation code} --- Quantum plane-curve codes are quantum analogues of plane-curve evaluation codes.
\item\relax
\flmRefsHyperref[eczindexfamilyrel]{code:small_distance_quantum}{Small-distance block quantum code} --- The quantum plane-curve code for the Hermitian curve \(y^3 + y = x^4\) is a \(\llbracket 27,13,4\rrbracket _3\) qutrit code.
\end{eczvaluelist}
\eczhbkcontributors{ \eczhuVVA }
\endeczcode

\eczcode{galois_quad_residue}{Quantum quadratic-residue (QR) code}{~\NoCaseChange{\protect\cite{cite532,cite813,cite829}}}
\codefieldsection{Description}
Galois-qudit \(\llbracket n,1\rrbracket _q\) \flmRefsHyperref{ref672}{pure} self-dual Galois-qudit CSS code constructed from a dual-containing QR code via the Galois-qudit CSS construction.
For \(q\) not divisible by \(n\), its distance satisfies \(d^2-d+1 \geq n\) when \(n \equiv 3\) modulo 4 \NoCaseChange{\protect\cite[{Thm. 40}]{cite813}} and \(d \geq \sqrt{n}\) when \(n\equiv 1\) modulo 4 \NoCaseChange{\protect\cite[{Thm. 41}]{cite813}}.

\codefieldsection{Protection}
For qubit quantum QR codes obtained from extended binary QR codes, explicit examples satisfy \(n \leq d^2-d+1\) and \(d \leq 4\lfloor (n+1)/24 \rfloor + 3\) \NoCaseChange{\protect\cite{cite760}}.

\codefieldsection{Transversal and Permutation-Based Gates}
\begin{eczvaluelist}
\item\relax Qubit quantum QR codes admit transversal implementations of the \flmRefsHyperref{ref409}{single-qubit Clifford group} \NoCaseChange{\protect\cite{cite760}}. They yield a family of high-distance triorthogonal codes \NoCaseChange{\protect\cite{cite760}} via the doubling transformation \NoCaseChange{\protect\cite{cite731}}; such codes admit transversal implementations of the \(T\) gate.
\end{eczvaluelist}
\codefieldsection{Parents}
\begin{eczvaluelist}
\item\relax
\flmRefsHyperref[eczindexfamilyrel]{code:galois_css}{Galois-qudit CSS code}\item\relax
\flmRefsHyperref[eczindexfamilyrel]{code:galois_duadic}{Quantum duadic code} --- Quantum QR codes are quantum duadic codes since QR codes are duadic codes.
\end{eczvaluelist}
\codefieldsection{Children}
\begin{eczvaluelist}
\item\relax
\flmRefsHyperref[eczindexfamilyrel]{code:quad_residue_13_1_5}{\(\llbracket 13,1,5\rrbracket \) quantum QR code}\item\relax
\flmRefsHyperref[eczindexfamilyrel]{code:steane}{\(\llbracket 7,1,3\rrbracket \) Steane code} --- The Steane code is a qubit quantum QR code \NoCaseChange{\protect\cite{cite829,cite2914}}.
\item\relax
\flmRefsHyperref[eczindexfamilyrel]{code:stab_47_1_11}{\(\llbracket 47,1,11\rrbracket \) quantum QR code} --- The \(\llbracket 47,1,11\rrbracket \) code is a qubit quantum QR code \NoCaseChange{\protect\cite{cite3225,cite760}}.
\item\relax
\flmRefsHyperref[eczindexfamilyrel]{code:qubit_golay}{\(\llbracket 23, 1, 7\rrbracket \) Quantum Golay code} --- The Golay code is a qubit quantum QR code \NoCaseChange{\protect\cite{cite829,cite2914}}.
\item\relax
\flmRefsHyperref[eczindexfamilyrel]{code:qutrit_golay}{\(\llbracket 11,1,5\rrbracket _3\) qutrit Golay code} --- The qutrit Golay code is a qutrit quantum QR code since the ternary Golay code is a QR code.
\item\relax
\flmRefsHyperref[eczindexfamilyrel]{code:css_5_1_3}{\(\llbracket 5,1,3\rrbracket _4\) Galois-qudit CSS code} --- The \(\llbracket 5,1,3\rrbracket _4\) code is obtained from the shortened hexacode \NoCaseChange{\protect\cite{cite514}}.
\item\relax
\flmRefsHyperref[eczindexfamilyrel]{code:galois_3_1_2}{\(\llbracket 3,1,2\rrbracket _4\) three-Galois-quartrit code} --- The \(\llbracket 3,1,2\rrbracket _4\) code is constructed from the shortened RS\(_4\) code \NoCaseChange{\protect\cite{cite514}}.
\end{eczvaluelist}
\codefieldsection{Cousins}
\begin{eczvaluelist}
\item\relax
\flmRefsHyperref[eczindexfamilyrel]{code:q-ary_quad_residue}{Quadratic-residue (QR) code} --- Quantum quadratic-residue codes are quantum analogues of \(q\)-ary quadratic-residue codes.
\item\relax
\flmRefsHyperref[eczindexfamilyrel]{code:quantum_mds}{Quantum maximum-distance-separable (MDS) code} --- Almost all quantum QR codes for prime-dimensional qudits are quantum MDS \NoCaseChange{\protect\cite[{Corr. 11}]{cite532}}.
\item\relax
\flmRefsHyperref[eczindexfamilyrel]{code:quantum_triorthogonal}{Triorthogonal code} --- Qubit quantum QR codes are doubly even and admit transversal implementations of the \flmRefsHyperref{ref409}{single-qubit Clifford group} \NoCaseChange{\protect\cite{cite760}}. They yield a family of high-distance triorthogonal and weak triply even codes via the doubling transformation \NoCaseChange{\protect\cite{cite760}}; such codes admit transversal implementations of the \(T\) gate.
\item\relax
\flmRefsHyperref[eczindexfamilyrel]{code:quantum_divisible}{Quantum divisible code} --- Qubit quantum QR codes are doubly even and admit transversal implementations of the \flmRefsHyperref{ref409}{single-qubit Clifford group} \NoCaseChange{\protect\cite{cite760}}. They yield a family of high-distance triorthogonal and weak triply even codes via the doubling transformation \NoCaseChange{\protect\cite{cite760}}; such codes admit transversal implementations of the \(T\) gate.
\item\relax
\flmRefsHyperref[eczindexfamilyrel]{code:bc_phantom}{Binarized-and-concatenated (B\&C) phantom code} --- A concrete B\&C phantom family starts from CSS quantum QR codes over \(\mathbb{F}_4\) \NoCaseChange{\protect\cite{cite514}}.
\item\relax
\flmRefsHyperref[eczindexfamilyrel]{code:data_syndrome}{Quantum data-syndrome (QDS) code} --- CSS QDS codes can be constructed from dual-containing cyclic codes without reducing distance; for \(p=8j-1\), quantum QR codes yield \(\llbracket p,1,d:r\rrbracket \) QDS codes with \(r\leq p+1\) \NoCaseChange{\protect\cite[{Thms. 13,14}]{cite2914}}.
\end{eczvaluelist}
\eczhbkcontributors{ \eczhuVVA }
\endeczcode

\eczcode{quantum_tamo_barg}{Quantum Tamo-Barg (QTB) code}{~\NoCaseChange{\protect\cite{cite812}}}
\codefieldsection{Description}
A member of a family of Galois-qudit CSS codes whose underlying classical codes consist of Tamo-Barg codes together with specific low-weight codewords.
Folded versions of QTB codes, or \textit{FQTB codes}, defined on qudits whose dimension depends on \(n\), yield explicit examples of QLRCs of arbitrary locality \(r\) \NoCaseChange{\protect\cite[{Cor. 64}]{cite812}}.

\codefieldsection{Protection}
A family of QTBs can be defined for every prime \(r\), rate \(R\in(0,1)\), and qudit dimension \(q = n+1\) such that their relative distance is \(\delta \geq 1 - \sqrt{(1+R)/2} - O(1/r)\) \NoCaseChange{\protect\cite[{Thm. 62}]{cite812}}.

\textit{Folding} these codes by combining qudits into larger qudits yields FQTB codes with relative distance \(\delta \geq (1-R)/2 - O(1/\sqrt{r})\) \NoCaseChange{\protect\cite[{Cor. 64}]{cite812}} and qudit dimension \(q = n^{O(r^2)}\).
This relative distance is of \flmRefsHyperref{ref65}{order} \(O(1/\sqrt{r})\) below the Singleton-like QLRC bound.

\codefieldsection{Decoding}
\begin{eczvaluelist}
\item\relax Polynomially efficient decoder for QTB codes against errors acting on a number of subsystems that can go up to half of the distance bound proved for the family \NoCaseChange{\protect\cite[{Thm. 69}]{cite812}}. The decoder is based on decoding RS codes, and its runtime is independent of the locality \(r\).
\item\relax Polynomially efficient decoder for FQTB codes against errors acting on a number of subsystems that can go up to half of the distance bound proved for the family \NoCaseChange{\protect\cite[{Cor. 72}]{cite812}}. The runtime depends on the locality \(r\).
\end{eczvaluelist}
\codefieldsection{Parents}
\begin{eczvaluelist}
\item\relax
\flmRefsHyperref[eczindexfamilyrel]{code:galois_css}{Galois-qudit CSS code}\item\relax
\flmRefsHyperref[eczindexfamilyrel]{code:quantum_locally_recoverable}{Quantum locally recoverable code (QLRC)} --- Folded quantum Tamo-Barg codes yield explicit QLRCs of arbitrary prime locality \(r\), rate at least \(R\), relative distance \(\delta \geq (1-R)/2 - O(1/\sqrt{r})\), and qudit dimension \(q = n^{O(r^2)}\) \NoCaseChange{\protect\cite[{Cor. 64}]{cite812}}.
\end{eczvaluelist}
\codefieldsection{Cousin}
\begin{eczvaluelist}
\item\relax
\flmRefsHyperref[eczindexfamilyrel]{code:tamo_barg}{Tamo-Barg code} --- QTB codes are CSS codes constructed from Tamo-Barg codes.
\end{eczvaluelist}
\eczhbkcontributors{ \eczhuVVA }
\endeczcode

\eczcode{quantum_twisted}{Quantum twisted code}{~\NoCaseChange{\protect\cite{cite2912}}}
\codefieldsection{Description}
Hermitian stabilizer code constructed from twisted BCH codes.

\codefieldsection{Notes}
\begin{eczvaluelist}
\item\relax Tables of quantum twisted codes are available at \flmHref{https://web.archive.org/web/19991106022224/http://www.mathi.uni-heidelberg.de/~yves/Matritzen/QTBCH/QTBCHIndex.html}{Yves Edel's home page}.
\end{eczvaluelist}
\codefieldsection{Parent}
\begin{eczvaluelist}
\item\relax
\flmRefsHyperref[eczindexfamilyrel]{code:stabilizer_over_gfqsq}{Hermitian Galois-qudit code}\end{eczvaluelist}
\codefieldsection{Cousins}
\begin{eczvaluelist}
\item\relax
\flmRefsHyperref[eczindexfamilyrel]{code:quantum_perfect}{Perfect quantum code} --- The \(\llbracket \frac{q^{2r}-1}{q^{2}-1},q^{n-2r},3\rrbracket _q\) family of quantum twisted codes are the only perfect Galois-qudit codes \NoCaseChange{\protect\cite{cite2911,cite2912}}.
\item\relax
\flmRefsHyperref[eczindexfamilyrel]{code:twisted_bch}{Twisted BCH code} --- Quantum twisted codes are quantum analogues of twisted BCH codes.
\end{eczvaluelist}
\eczhbkcontributors{ \eczhuVVA }
\endeczcode

\eczcode{quantum_singleton}{Singleton-bound approaching AQECC}{~\NoCaseChange{\protect\cite{cite495}}}
\codefieldsection{Description}
A member of an approximate quantum code family of rate \(R\) that can tolerate adversarial errors nearly saturating the quantum Singleton bound of \((1-R)/2\).
The formulation of such codes relies on a notion of \textit{quantum list decoding} \NoCaseChange{\protect\cite{cite814,cite495}}.

One construction first builds constant-alphabet quantum list-decodable CSS codes from folded quantum Reed-Solomon outer codes, random CSS inner codes, and quantum Alon-Edmonds-Luby distance amplification/alphabet reduction, and then compiles them into AQECCs using purity-testing codes and robust secret sharing \NoCaseChange{\protect\cite{cite495}}.
The resulting codes are efficiently encodable and decodable, and descriptions of the codes can be sampled by an efficient randomized algorithm with \(2^{-\Omega(n)}\) failure probability.

\codefieldsection{Protection}
For any \(\gamma>0\) and rate \(0<R<1\), these approximate quantum \(\llbracket n,R \cdot n\rrbracket _q\) codes have constant Galois-qudit dimension \(q=2^{O(1/\gamma^5)}\) and correct errors acting on \((1-R-\gamma) \cdot n/2\) registers, up to a recovery error of \(2^{-\Omega(n)}\) \NoCaseChange{\protect\cite{cite495}}.
\codefieldsection{Rate}
For any target rate \(R\in(0,1)\), codes can tolerate adversarial errors on nearly a \((1-R)/2\) fraction of registers while keeping constant alphabet size \NoCaseChange{\protect\cite{cite495}}.
\codefieldsection{Encoding}
\begin{eczvaluelist}
\item\relax Efficient encoding.
\end{eczvaluelist}
\codefieldsection{Decoding}
\begin{eczvaluelist}
\item\relax Efficient decoder based on quantum list decoding together with purity-testing and robust-secret-sharing post-processing \NoCaseChange{\protect\cite{cite495}}.
\end{eczvaluelist}
\codefieldsection{Parents}
\begin{eczvaluelist}
\item\relax
\flmRefsHyperref[eczindexfamilyrel]{code:galois_css}{Galois-qudit CSS code}\item\relax
\flmRefsHyperref[eczindexfamilyrel]{code:approximate_qecc}{Approximate quantum error-correcting code (AQECC)}\end{eczvaluelist}
\codefieldsection{Cousins}
\begin{eczvaluelist}
\item\relax
\flmRefsHyperref[eczindexfamilyrel]{code:quantum_mds}{Quantum maximum-distance-separable (MDS) code} --- Singleton-bound approaching AQECCs asymptotically approach the quantum Singleton bound, rather than exactly saturating it at finite blocklength.
\item\relax
\flmRefsHyperref[eczindexfamilyrel]{code:galois_fqrs}{Folded quantum RS (FQRS) code} --- Singleton-bound approaching AQECCs are built using folded quantum Reed-Solomon (FQRS) codes \NoCaseChange{\protect\cite{cite495}}.
\item\relax
\flmRefsHyperref[eczindexfamilyrel]{code:quantum_secret_sharing}{Approximate secret-sharing code} --- Quantum secret-sharing codes have asymptotically decaying rate and require qudit dimension to increase exponentially with \(n\), while Singleton-bound approaching AQECCs have constant rate and qudit dimension.
\item\relax
\flmRefsHyperref[eczindexfamilyrel]{code:good_qldpc}{Good QLDPC code} --- AEL distance amplification \NoCaseChange{\protect\cite{cite493,cite494}} can be used to construct constant-alphabet QLDPC CSS codes of any target rate \(R\) and relative distance \((1-R-\gamma)/2\) that are decodable in linear time up to half that distance \NoCaseChange{\protect\cite[{Corr. 5.3}]{cite495}}. The AEL distance-amplification framework also yields constant-alphabet approximate quantum codes that decode nearly up to the quantum Singleton bound \NoCaseChange{\protect\cite{cite495}}.
\end{eczvaluelist}
\eczhbkcontributors{ Sam Gunn, \eczhuVVA }
\endeczcode

\eczcode{skew-cyclic_galois_css}{Skew-cyclic CSS code}{~\NoCaseChange{\protect\cite{cite4654,cite815}}}
\codefieldsection{Description}
A member of a family of Galois-qudit CSS codes constructed from skew-cyclic classical codes over rings \NoCaseChange{\protect\cite[{Thm. 5.8}]{cite815}}.
See related study \NoCaseChange{\protect\cite{cite816}} that uses cyclic codes over rings. 

\codefieldsection{Parent}
\begin{eczvaluelist}
\item\relax
\flmRefsHyperref[eczindexfamilyrel]{code:galois_css}{Galois-qudit CSS code}\end{eczvaluelist}
\codefieldsection{Cousins}
\begin{eczvaluelist}
\item\relax
\flmRefsHyperref[eczindexfamilyrel]{code:skew_cyclic}{Skew-cyclic code} --- Skew-cyclic CSS codes are constructed from classical skew-cyclic codes over rings.
\item\relax
\flmRefsHyperref[eczindexfamilyrel]{code:quantum_mds}{Quantum maximum-distance-separable (MDS) code} --- Some quantum MDS codes are constructed from cyclic and constacyclic codes using the Galois-qudit CSS construction \NoCaseChange{\protect\cite{cite815}}.
\end{eczvaluelist}
\eczhbkcontributors{ Nolan Coble, \eczhuVVA }
\endeczcode

\eczcode{subsystem_galois_into_galois}{Subsystem Galois-qudit code}{}
\codefieldsection{Alternative Names}
\begin{eczvaluelist}
\item\relax Gauge Galois-qudit code
\end{eczvaluelist}
\eczhIndexCodeAliasName{subsystem_galois_into_galois}{Gauge Galois-qudit code}

\codefieldsection{Kingdom root code}
for the \flmRefsHyperref{kingdom:galois_into_galois}{Galois-qudit Kingdom}.
\codefieldsection{Description}
Subsystem QECC encoding into a \(q^n\)-dimensional Hilbert space consisting of \(n\) Galois qudits.

\codefieldsection{Parent}
\begin{eczvaluelist}
\item\relax
\flmRefsHyperref[eczindexfamilyrel]{code:subsystem_group_quantum}{Subsystem group-based quantum code} --- A Galois qudit for \(q=p^m\) can be decomposed into a Kronecker product of \(m\) modular qudits \NoCaseChange{\protect\cite{cite696}}; see \NoCaseChange{\protect\cite[{Sec. 5.3}]{cite697}}.
Interpreted this way, subsystem Galois-qudit codes are subsystem group quantum codes whose physical spaces are constructed using Galois fields \(\mathbb{F}_q\) as groups. More general versions of such qudits can be valued in a Galois ring \NoCaseChange{\protect\cite{cite4621}}, over which there also exists a Fourier transform \NoCaseChange{\protect\cite{cite4622}}.

\end{eczvaluelist}
\codefieldsection{Children}
\begin{eczvaluelist}
\item\relax
\flmRefsHyperref[eczindexfamilyrel]{code:subsystem_qubits_into_qubits}{Subsystem qubit code} --- Subsystem Galois-qudit quantum codes for \(q=2\) correspond to subsystem qubit codes.
\item\relax
\flmRefsHyperref[eczindexfamilyrel]{code:galois_subsystem_stabilizer}{Subsystem Galois-qudit stabilizer code}\end{eczvaluelist}
\codefieldsection{Cousin}
\begin{eczvaluelist}
\item\relax
\flmRefsHyperref[eczindexfamilyrel]{code:galois_into_galois}{Galois-qudit code} --- Subsystem Galois-qudit codes reduce to (subspace) Galois-qudit codes when there is no gauge subsystem.
\end{eczvaluelist}
\eczhbkcontributors{ \eczhuVVA }
\endeczcode

\eczcode{galois_subsystem_css}{Subsystem Galois-qudit CSS code}{~\NoCaseChange{\protect\cite{cite1742,cite4430}}}
\codefieldsection{Alternative Names}
\begin{eczvaluelist}
\item\relax Euclidean construction subsystem code
\end{eczvaluelist}
\eczhIndexCodeAliasName{galois_subsystem_css}{Euclidean construction subsystem code}
\codefieldsection{Description}
Galois-qudit subsystem stabilizer code which admits a set of gauge-group generators which consist of either all-\(Z\) or all-\(X\) Galois-qudit Pauli strings.

These codes can be constructed from classical codes via a subsystem generalization of the CSS construction or the Hermitian construction \NoCaseChange{\protect\cite{cite2884,cite1742}}.

\codefieldsection{Parents}
\begin{eczvaluelist}
\item\relax
\flmRefsHyperref[eczindexfamilyrel]{code:galois_subsystem_stabilizer}{Subsystem Galois-qudit stabilizer code}\item\relax
\flmRefsHyperref[eczindexfamilyrel]{code:subsystem_css}{Subsystem CSS code}\end{eczvaluelist}
\codefieldsection{Child}
\begin{eczvaluelist}
\item\relax
\flmRefsHyperref[eczindexfamilyrel]{code:qubit_subsystem_css}{Subsystem qubit CSS code} --- Subsystem Galois-qudit CSS codes reduce to subsystem qubit CSS codes for \(q=2\).
\end{eczvaluelist}
\codefieldsection{Cousin}
\begin{eczvaluelist}
\item\relax
\flmRefsHyperref[eczindexfamilyrel]{code:galois_css}{Galois-qudit CSS code} --- Subsystem Galois-qudit CSS codes reduce to (subspace) Galois-qudit CSS codes when there is no gauge subsystem.
\end{eczvaluelist}
\eczhbkcontributors{ \eczhuVVA }
\endeczcode

\eczcode{galois_subsystem_stabilizer}{Subsystem Galois-qudit stabilizer code}{~\NoCaseChange{\protect\cite{cite2884}}}
\codefieldsection{Alternative Names}
\begin{eczvaluelist}
\item\relax Gauge Galois-qudit stabilizer code
\end{eczvaluelist}
\eczhIndexCodeAliasName{galois_subsystem_stabilizer}{Gauge Galois-qudit stabilizer code}
\codefieldsection{Description}
Galois-qudit generalization of a subsystem qubit stabilizer code.
Can be obtained by taking a Galois-qudit stabilizer code and assigning some of its logical qudits to be gauge qudits.

\codefieldsection{Protection}
There are Gilbert-Varshamov-type lower bounds, linear-programming upper bounds, and \flmRefsHyperref{ref672}{pure-code} Singleton and Hamming bounds for subsystem Galois-qudit stabilizer codes \NoCaseChange{\protect\cite{cite1742,cite4655}}.
A subsystem Galois-qudit stabilizer code saturating the quantum Singleton bound must have a trivial gauge subsystem, i.e., there are no \(\llbracket n,n-2d+2,r>0,d\rrbracket _q\) codes \NoCaseChange{\protect\cite[{Thms. 19,20}]{cite1742}}.

\codefieldsection{Parents}
\begin{eczvaluelist}
\item\relax
\flmRefsHyperref[eczindexfamilyrel]{code:subsystem_galois_into_galois}{Subsystem Galois-qudit code}\item\relax
\flmRefsHyperref[eczindexfamilyrel]{code:subsystem_stabilizer}{Subsystem stabilizer code}\end{eczvaluelist}
\codefieldsection{Children}
\begin{eczvaluelist}
\item\relax
\flmRefsHyperref[eczindexfamilyrel]{code:qubit_subsystem_stabilizer}{Subsystem qubit stabilizer code} --- Subsystem Galois-qudit stabilizer codes reduce to subsystem qubit stabilizer codes for qudit dimension \(q=2\).
\item\relax
\flmRefsHyperref[eczindexfamilyrel]{code:galois_subsystem_css}{Subsystem Galois-qudit CSS code}\end{eczvaluelist}
\codefieldsection{Cousins}
\begin{eczvaluelist}
\item\relax
\flmRefsHyperref[eczindexfamilyrel]{code:galois_stabilizer}{Galois-qudit stabilizer code} --- Subsystem Galois-qudit stabilizer codes reduce to Galois-qudit stabilizer codes when there are no gauge qudits.
\item\relax
\flmRefsHyperref[eczindexfamilyrel]{code:quantum_mds}{Quantum maximum-distance-separable (MDS) code} --- A subsystem Galois-qudit stabilizer code saturating the quantum Singleton bound must have a trivial gauge subsystem, i.e., there are no \(\llbracket n,n-2d+2,r>0,d\rrbracket _q\) codes \NoCaseChange{\protect\cite[{Thms. 19,20}]{cite1742}}. More generally, all \flmRefsHyperref{ref672}{pure} MDS subsystem stabilizer codes are derived from MDS stabilizer codes \NoCaseChange{\protect\cite{cite2997}}.
\item\relax
\flmRefsHyperref[eczindexfamilyrel]{code:galois_bch}{Galois-qudit BCH code} --- Asymmetric quantum BCH codes have been constructed \NoCaseChange{\protect\cite{cite2610,cite2652,cite2653}\protect\cite[{Lemma 4.4}]{cite1354}\protect\cite[{Sec. 17.3}]{cite872}}, including subsystem BCH codes \NoCaseChange{\protect\cite{cite2612}\protect\cite[{Sec. 9.3}]{cite872}}.
\item\relax
\flmRefsHyperref[eczindexfamilyrel]{code:stabilizer_over_gfqsq}{Hermitian Galois-qudit code} --- The Hermitian construction has been extended to subsystem Galois-qudit stabilizer codes \NoCaseChange{\protect\cite{cite1742}}.
\end{eczvaluelist}
\eczhbkcontributors{ \eczhuVVA }
\endeczcode

\eczcode{galois_true_stabilizer}{True Galois-qudit stabilizer code}{~\NoCaseChange{\protect\cite{cite696,cite398}}}
\codefieldsection{Alternative Names}
\begin{eczvaluelist}
\item\relax Linear stabilizer code
\end{eczvaluelist}
\eczhIndexCodeAliasName{galois_true_stabilizer}{Linear stabilizer code}
\codefieldsection{Description}
A \(\llbracket n,k,d\rrbracket _q\) stabilizer code whose stabilizer's \flmRefsHyperref{ref873}{Galois symplectic representation} forms a linear subspace. In other words, the set of \(q\)-ary vectors representing the stabilizer group is closed under both addition and multiplication by elements of \(\mathbb{F}_q\). In contrast, Galois-qudit stabilizer codes admit sets of vectors that are closed under addition only.

The number of generators \(r\) for a true stabilizer code is a multiple of \(m\) (recall that \(q=p^m\) for Galois qudits). As a result, the number \(k=n-r/m\) of logical qudits is an integer.

Each code can be represented by a stabilizer generator matrix \(H=(A|B)\), where each row \((a|b)\) is the \flmRefsHyperref{ref873}{Galois symplectic representation} of a stabilizer generator.

\codefieldsection{Protection}
Detects errors on up to \(d-1\) qudits, and corrects erasure errors on up to \(d-1\) qudits.
\codefieldsection{Notes}
\begin{eczvaluelist}
\item\relax See Ref. \NoCaseChange{\protect\cite{cite2996,cite2024}} for introductions to various stabilizer code constructions.
\end{eczvaluelist}
\codefieldsection{Parent}
\begin{eczvaluelist}
\item\relax
\flmRefsHyperref[eczindexfamilyrel]{code:galois_stabilizer}{Galois-qudit stabilizer code}\end{eczvaluelist}
\codefieldsection{Children}
\begin{eczvaluelist}
\item\relax
\flmRefsHyperref[eczindexfamilyrel]{code:qubit_stabilizer}{Qubit stabilizer code} --- True Galois-qudit stabilizer codes for \(q=2\) correspond to qubit stabilizer codes.
\item\relax
\flmRefsHyperref[eczindexfamilyrel]{code:galois_5_1_3}{\(\llbracket 5,1,3\rrbracket _q\) Galois-qudit code}\item\relax
\flmRefsHyperref[eczindexfamilyrel]{code:galois_6_2_3}{\(\llbracket 6,2,3\rrbracket _{q}\) code} --- The code is a non-CSS stabilizer code in general \NoCaseChange{\protect\cite{cite831}}.
\item\relax
\flmRefsHyperref[eczindexfamilyrel]{code:galois_7_3_3}{\(\llbracket 7,3,3\rrbracket _{q}\) code}\item\relax
\flmRefsHyperref[eczindexfamilyrel]{code:galois_bch}{Galois-qudit BCH code} --- Galois-qudit BCH codes can be constructed via the CSS construction or the Hermitian construction.
\item\relax
\flmRefsHyperref[eczindexfamilyrel]{code:galois_css}{Galois-qudit CSS code} --- The Galois-qudit CSS construction yields a true stabilizer code \NoCaseChange{\protect\cite[{Sec. 8.2.2}]{cite398}}.
\item\relax
\flmRefsHyperref[eczindexfamilyrel]{code:galois_duadic}{Quantum duadic code} --- Quantum duadic codes can be constructed via the CSS construction or the Hermitian construction.
\item\relax
\flmRefsHyperref[eczindexfamilyrel]{code:quantum_ag}{Quantum AG code} --- Quantum AG codes can be constructed via the Galois-qudit CSS construction or the Galois-qudit Hermitian construction.
\item\relax
\flmRefsHyperref[eczindexfamilyrel]{code:galois_reed_muller}{Galois-qudit quantum RM code} --- Galois-qudit RM codes can be constructed via the Galois-qudit Hermitian construction, the Galois-qudit CSS construction, or directly from their parity-check matrices \NoCaseChange{\protect\cite{cite828}\protect\cite[{Sec. 4.2}]{cite829}}.
\item\relax
\flmRefsHyperref[eczindexfamilyrel]{code:quantum_gabidulin}{Quantum Gabidulin code}\item\relax
\flmRefsHyperref[eczindexfamilyrel]{code:stabilizer_over_gfqsq}{Hermitian Galois-qudit code} --- Hermitian codes are true stabilizer codes because they are based on Hermitian self-orthogonal linear (as opposed to additive) codes over \(\mathbb{F}_{q^2}\).
\end{eczvaluelist}
\codefieldsection{Cousins}
\begin{eczvaluelist}
\item\relax
\flmRefsHyperref[eczindexfamilyrel]{code:q-ary_linear}{Linear \(q\)-ary code} --- A true Galois-qudit stabilizer code is the closest quantum analogue of a linear code over \(\mathbb{F}_q\) because the \(q\)-ary vectors corresponding to the \flmRefsHyperref{ref873}{Galois symplectic representation} of the stabilizers form a linear subspace.
\item\relax
\flmRefsHyperref[eczindexfamilyrel]{code:iceberg}{\(\llbracket 2m,2m-2,2\rrbracket \) error-detecting code} --- A naive extension of the iceberg code to Galois qudits keeps only two CSS-type generators, \(M_1(1)=X_{\alpha_1}\otimes\cdots\otimes X_{\alpha_n}\) and \(M_2(1)=Z_{\beta_1}\otimes\cdots\otimes Z_{\beta_n}\), with nonzero \(\alpha_i,\beta_i\in\mathbb{F}_q\) satisfying \(\sum_i \alpha_i\beta_i=0\). For prime-power dimensions with \(q=p^m\) and \(m>1\), this yields a Galois-qudit code of distance one that is generally not a true stabilizer code because the stabilizer is not closed under multiplication by arbitrary \(\gamma\in\mathbb{F}_q\). Adding all \(M_1(\gamma)\) and \(M_2(\gamma)\) to the stabilizer group recovers the corresponding true Galois-qudit CSS code of distance two \NoCaseChange{\protect\cite[{Sec. 8.2.2}]{cite398}}.
\end{eczvaluelist}
\eczhbkcontributors{ Daniel Gottesman, \eczhuVVA }
\endeczcode

\eczcode{two_block_quantum}{Two-block CSS code}{~\NoCaseChange{\protect\cite{cite439}}}
\codefieldsection{Alternative Names}
\begin{eczvaluelist}
\item\relax Two-sublattice code
\item\relax Two-square-block code
\end{eczvaluelist}
\eczhIndexCodeAliasName{two_block_quantum}{Two-sublattice code}
\eczhIndexCodeAliasName{two_block_quantum}{Two-square-block code}
\codefieldsection{Description}
Galois-qudit CSS code whose stabilizer generator matrices \(H_X=(A_1,B_1)\) and \(H_Z=(B^T_2,-A^T_2)\), are constructed from four matrices satisfying \(A_1 B_2 - B_1 A_2 = 0\).
In the case the two pairs are equal, we have \(H_X=(A,B)\) and \(H_Z=(B^T,-A^T)\), constructed from a pair of square commuting matrices \(A\) and \(B\).

Generalized constructions utilizing more than two blocks have also been considered \NoCaseChange{\protect\cite{cite4656}}.

\codefieldsection{Protection}
Code parameters are generally unknown, although they can be formally expressed in terms of ranks of some matrices related to \(A\) and \(B\).
The corresponding expressions, as well as some upper and lower bounds on parameters are given in \NoCaseChange{\protect\cite{cite842}}.

\codefieldsection{Parent}
\begin{eczvaluelist}
\item\relax
\flmRefsHyperref[eczindexfamilyrel]{code:galois_css}{Galois-qudit CSS code}\end{eczvaluelist}
\codefieldsection{Child}
\begin{eczvaluelist}
\item\relax
\flmRefsHyperref[eczindexfamilyrel]{code:2bga}{Two-block group-algebra (2BGA) codes} --- 2BGA codes are two-block quantum codes whose commuting matrices are constructed with the help of a \flmRefsHyperref{ref205}{group algebra}.
\end{eczvaluelist}
\codefieldsection{Cousins}
\begin{eczvaluelist}
\item\relax
\flmRefsHyperref[eczindexfamilyrel]{code:general_qldpc}{QLDPC code} --- When matrices \(A\) and \(B\) have row and column weights bounded by \(W\), a two-block CSS code is a quantum LDPC code with stabilizer generators bounded by \(2W\).
\item\relax
\flmRefsHyperref[eczindexfamilyrel]{code:lifted_product}{Lifted-product (LP) code} --- LP codes can be constructed using non-square matrices and taking a hypergraph product over a group algebra, while two-block CSS codes are constructed directly using square matrices.
\item\relax
\flmRefsHyperref[eczindexfamilyrel]{code:qubit_css}{Qubit CSS code} --- Any \(\llbracket n,k,d\rrbracket \) stabilizer code can be mapped onto a \(\llbracket 2n,2k,\geq d\rrbracket \) \flmRefsHyperref{code:two_block_quantum}{two-block CSS code} via \flmRefsHyperref{ref436}{symplectic doubling}, which preserves geometric locality of a code up to a constant factor.
\item\relax
\flmRefsHyperref[eczindexfamilyrel]{code:qudit_css}{Modular-qudit CSS code} --- Any \(\llbracket n,k,d\rrbracket _{\mathbb{Z}_q}\) stabilizer code can be mapped onto a \(\llbracket 2n,2k,\geq d\rrbracket _{\mathbb{Z}_q}\) \flmRefsHyperref{code:two_block_quantum}{two-block CSS code} code via \flmRefsHyperref{ref436}{symplectic doubling}, which preserves geometric locality of a code up to a constant factor.
\end{eczvaluelist}
\eczhbkcontributors{ Leonid Pryadko, \eczhuVVA }
\endeczcode

\eczcode{2bga}{Two-block group-algebra (2BGA) codes}{~\NoCaseChange{\protect\cite{cite4657,cite4658,cite842}}}
\codefieldsection{Alternative Names}
\begin{eczvaluelist}
\item\relax Non-Abelian GB code
\item\relax LR code
\end{eczvaluelist}
\eczhIndexCodeAliasName{2bga}{Non-Abelian GB code}
\eczhIndexCodeAliasName{2bga}{LR code}
\codefieldsection{Description}
2BGA codes are the one-by-one, or smallest, \flmRefsHyperref{code:lifted_product}{LP codes}:
\(LP(a,b)\) is defined by a pair of \flmRefsHyperref{ref205}{group algebra} elements
\(a,b\in \mathbb{F}_q[G]\), where \(G\) is a finite group.
If \(|G|=\ell\), then the code has length \(n=2\ell\).

A cyclic group \(G\) yields a \flmRefsHyperref{code:generalized_bicycle}{GB code}, while an Abelian group yields an \textit{Abelian 2BGA code} \NoCaseChange{\protect\cite{cite4659}}, i.e., a quasi-Abelian analog of an index-two quasi-cyclic code.
In the special case where the support subgroups generated by \(a\) and \(b\) are disjoint, the code reduces to a square-matrix \flmRefsHyperref{code:galois_hypergraph_product}{Galois-qudit hypergraph-product code} built from a pair of classical group-algebra codes \NoCaseChange{\protect\cite{cite842}}.

A \(Z\)-type logical subspace of \(LP(a,b)\) can be represented by pairs
\((u,v)\in \mathbb{F}_q[G]\times \mathbb{F}_q[G]\) satisfying
\flmMathEnvironment{align}{}{
  au+vb=0,
}
with any two pairs \((u,v)\) and \((u',v')\) such that \(u'=u+w b\)
and \(v'=v-aw\) considered equivalent.
The order in the products is relevant when the group is non-Abelian.

For example, consider the
alternating group \(G=A_4=T\), also known as the rotation group of a
regular tetrahedron,
\flmMathEnvironment{align}{}{
  T=\langle x,y|x^3=(yx)^3=y^2=1\rangle,\quad |T|=12,
} and the
binary algebra \(\mathbb{F}_2[T]\). Select \(a=1+x+y+x^{-1}yx\)
and \(b=1+x+y+yx\) to get an \emph{essentially non-Abelian} 2BGA code
LP\([a,b]\) with parameters \(\llbracket 24,5,3\rrbracket \) \NoCaseChange{\protect\cite{cite842}}.

\codefieldsection{Protection}
Some upper and lower bounds on parameters and many examples of 2BGA codes are given in Ref. \NoCaseChange{\protect\cite{cite842}}.
When the support groups generated by \(a\) and \(b\) intersect trivially, the code is equivalent to a square-matrix Galois-qudit hypergraph-product code and its distance reduces to \(d=\min(d_A^{\perp},d_B^{\perp})\); classical finite-rate group-algebra codes therefore yield finite-rate 2BGA families with \(d=\mathcal{O}(n^{1/2})\) \NoCaseChange{\protect\cite{cite842}}.
The code dimension \(k\) for Abelian 2BGA codes is always even \NoCaseChange{\protect\cite{cite4659}}.

\codefieldsection{Rate}
The 2BGA construction gives some of the best short codes with small stabilizer weights.

Exhaustive enumeration in \NoCaseChange{\protect\cite{cite842}} covered inequivalent connected 2BGA codes with row weight
\(W\le 8\), up to length \(n\le 100\) for Abelian groups and \(n\le 200\) for non-Abelian groups.
Among the resulting codes with \(kd\ge n\) are GB codes with parameters
\(\llbracket 70,8,10\rrbracket \) and \(\llbracket 72,10,9\rrbracket \), Abelian 2BGA codes over groups such as
\(\mathbb{Z}_m\times \mathbb{Z}_2\) (index-four quasi-cyclic codes) with parameters
\(\llbracket 48,8,6\rrbracket \) and \(\llbracket 56,8,7\rrbracket \), and non-Abelian codes with parameters
\(\llbracket 64,8,8\rrbracket \), \(\llbracket 82,10,9\rrbracket \), \(\llbracket 96,10,12\rrbracket \), and \(\llbracket 96,12,10\rrbracket \).
All of these examples have stabilizer generators of weight \(W=8\).
\codefieldsection{Transversal and Permutation-Based Gates}
\begin{eczvaluelist}
\item\relax Logical Pauli operators and fold-transversal gates have been studied \NoCaseChange{\protect\cite{cite718,cite719}}.
\end{eczvaluelist}
\codefieldsection{Gates}
\begin{eczvaluelist}
\item\relax Certain qubit 2BGA codes can admit a cup product structure and can thus have logical gates in the \flmTerm{term}{ref694}{}{Clifford hierarchy} implemented by constant-depth Clifford circuits \NoCaseChange{\protect\cite{cite1517}}.
\end{eczvaluelist}
\codefieldsection{Notes}
\begin{eczvaluelist}
\item\relax The idea of generating stabilizers according to a Cayley graph of a group was independently proposed by Hastings \NoCaseChange{\protect\cite{cite4657}}; see Ref. \NoCaseChange{\protect\cite{cite3699}}, which contains an error in Prop. 3.2 \NoCaseChange{\protect\cite{cite4660}}.
\item\relax A database of 2BGA codes is available in QECDB \NoCaseChange{\protect\cite{cite781}}.
\end{eczvaluelist}
\codefieldsection{Parents}
\begin{eczvaluelist}
\item\relax
\flmRefsHyperref[eczindexfamilyrel]{code:lifted_product}{Lifted-product (LP) code} --- 2BGA codes are LP\((a,b)\) codes, constructed from a pair of one-by-one matrices \(a,b\in \mathbb{F}_q[G]\) in a \flmRefsHyperref{ref205}{group algebra}.
\item\relax
\flmRefsHyperref[eczindexfamilyrel]{code:two_block_quantum}{Two-block CSS code} --- 2BGA codes are two-block quantum codes whose commuting matrices are constructed with the help of a \flmRefsHyperref{ref205}{group algebra}.
\end{eczvaluelist}
\codefieldsection{Children}
\begin{eczvaluelist}
\item\relax
\flmRefsHyperref[eczindexfamilyrel]{code:qcga}{Bivariate bicycle (BB) code} --- Bivariate bicycle codes are Abelian 2BGA codes over groups of the form \(\mathbb{Z}_{r} \times \mathbb{Z}_{s}\).
\item\relax
\flmRefsHyperref[eczindexfamilyrel]{code:generalized_bicycle}{Generalized bicycle (GB) code} --- A code GB\((a,b)\) with circulants of size \(\ell\) is a 2BGA code over the cyclic group \(\mathbb{Z}_{\ell}\).
More precisely, for the cyclic group \(\mathbb{Z}_{\ell}\equiv \langle x|x^\ell=1\rangle \), any element \(a\) of the \flmRefsHyperref{ref205}{group algebra} \(\mathbb{F}_q[\mathbb{Z}_{\ell}]\) can be seen as a polynomial \(a(x)\in \mathbb{F}_q[x]\) over the group generator \(x\), where the polynomial degree \(\deg a(x)<\ell\).
The 2BGA code LP\((a,b)\) is then just a generalized bicycle code GB\([a(x),b(x)]\) constructed from the polynomials \(a(x)\) and \(b(x)\) corresponding to \(a,b\in \mathbb{F}_q[\mathbb{Z}_{\ell}]\).

\end{eczvaluelist}
\codefieldsection{Cousins}
\begin{eczvaluelist}
\item\relax
\flmRefsHyperref[eczindexfamilyrel]{code:balanced_product}{Balanced product (BP) code} --- 2BGA codes can be formulated directly as balanced product codes \NoCaseChange{\protect\cite[{Rem. C.1}]{cite718}}.
\item\relax
\flmRefsHyperref[eczindexfamilyrel]{code:general_qldpc}{QLDPC code} --- Given \flmRefsHyperref{ref205}{group algebra} elements \(a,b\in \mathbb{F}_q[G]\) with weights \(W_a\) and \(W_b\) (i.e., number of nonzero terms in the expansion), the 2BGA code LP\((a,b)\) has stabilizer
generators of uniform weight \(W_a+W_b\).

\item\relax
\flmRefsHyperref[eczindexfamilyrel]{code:group}{Group-algebra code} --- A 2BGA code \(LP(a,b)\) is constructible as a hypergraph-product code when the support subgroups generated by \(a\) and \(b\) are disjoint. In that case, the commuting matrices simultaneously acquire hypergraph-product Kronecker-product form, and the code can be obtained from a pair of classical group-algebra codes \NoCaseChange{\protect\cite[{Statements 8 and 12}]{cite842}}.

\item\relax
\flmRefsHyperref[eczindexfamilyrel]{code:galois_hypergraph_product}{Galois-qudit HGP code} --- A 2BGA code \(LP(a,b)\) is constructible as a hypergraph-product code when the support subgroups generated by \(a\) and \(b\) are disjoint. In that case, the commuting matrices simultaneously acquire hypergraph-product Kronecker-product form, and the code can be obtained from a pair of classical group-algebra codes \NoCaseChange{\protect\cite[{Statements 8 and 12}]{cite842}}.

\item\relax
\flmRefsHyperref[eczindexfamilyrel]{code:galois_topological}{Galois-qudit surface code} --- Any non-trivial 2BGA code with total row weight \(W\leq 4\) is equivalent to a direct sum of rotated Galois-qudit surface codes \NoCaseChange{\protect\cite[{Sec. IV.F}]{cite842}}.

\item\relax
\flmRefsHyperref[eczindexfamilyrel]{code:abelian_lifted_product}{Abelian LP code} --- Abelian 2BGA codes are LP\((a,b)\) codes, constructed from a pair of one-by-one matrices \(a,b\in \mathbb{F}_q[G]\) in a \flmRefsHyperref{ref205}{group algebra} of an Abelian group \(G\).
\item\relax
\flmRefsHyperref[eczindexfamilyrel]{code:apm_ldpc}{Affine-permutation-matrix LDPC (APM-LDPC) code} --- APM-LDPC codes can be used to construct non-Abelian 2BGA codes based on the affine permutation group \NoCaseChange{\protect\cite{cite1214}}.
\end{eczvaluelist}
\eczhbkcontributors{ Leonid Pryadko, \eczhuVVA }
\endeczcode

\onecolumngrid
\clearpage

\section{Bosonic Kingdom}

\begin{eczEpigraph}
\begin{quote}
\flmQuoteSetup{quote}%
Heisenberg gets pulled over for speeding.\\
The cop asks Heisenberg "Do you know how fast you were going?"\\
Heisenberg replies, "No, but I know exactly where I am!"\\
The officer looks at him confused and says "you were going 108 miles per hour!"\\
Heisenberg throws his arms up and cries, "Great! Now I'm lost!"
\end{quote}
\end{eczEpigraph}

\twocolumngrid

\eczcode{braunstein}{\(\llbracket 5,1,3\rrbracket _{\mathbb{R}}\) Braunstein five-mode code}{~\NoCaseChange{\protect\cite{cite4661}}}
\codefieldsection{Alternative Names}
\begin{eczvaluelist}
\item\relax Five-wavepacket code
\end{eczvaluelist}
\eczhIndexCodeAliasName{braunstein}{Braunstein five-mode code}
\eczhIndexCodeAliasName{braunstein}{Five-wavepacket code}
\codefieldsection{Description}
An analog stabilizer version of the five-qubit perfect code, encoding one mode into five and correcting arbitrary errors on any one mode.

\codefieldsection{Encoding}
\begin{eczvaluelist}
\item\relax Seven beam splitters \NoCaseChange{\protect\cite{cite4662}}.
\end{eczvaluelist}
\codefieldsection{Decoding}
\begin{eczvaluelist}
\item\relax Error correction can be done using linear-optical elements and feedback \NoCaseChange{\protect\cite{cite4663}}.
\end{eczvaluelist}
\codefieldsection{Parents}
\begin{eczvaluelist}
\item\relax
\flmRefsHyperref[eczindexfamilyrel]{code:analog_stabilizer}{Analog stabilizer code}\item\relax
\flmRefsHyperref[eczindexfamilyrel]{code:quantum_cyclic}{Cyclic quantum code}\item\relax
\flmRefsHyperref[eczindexfamilyrel]{code:ame}{Perfect-tensor code} --- Braunstein five-mode codewords are CV AME \NoCaseChange{\protect\cite{cite507}}.
\item\relax
\flmRefsHyperref[eczindexfamilyrel]{code:small_distance_quantum}{Small-distance block quantum code}\end{eczvaluelist}
\codefieldsection{Cousin}
\begin{eczvaluelist}
\item\relax
\flmRefsHyperref[eczindexfamilyrel]{code:qudit_5_1_3}{\(\llbracket 5,1,3\rrbracket _{\mathbb{Z}_q}\) modular-qudit code} --- The Braunstein five-mode code is a bosonic analogue of the five-qudit code.
\end{eczvaluelist}
\eczhbkcontributors{ \eczhuVVA }
\endeczcode

\eczcode{lloyd_slotine}{\(\llbracket 9,1,3\rrbracket _{\mathbb{R}}\) Lloyd-Slotine code}{~\NoCaseChange{\protect\cite{cite4664,cite4661}}}
\codefieldsection{Alternative Names}
\begin{eczvaluelist}
\item\relax Nine-wavepacket code
\end{eczvaluelist}
\eczhIndexCodeAliasName{lloyd_slotine}{Lloyd-Slotine code}
\eczhIndexCodeAliasName{lloyd_slotine}{Nine-wavepacket code}
\codefieldsection{Description}
An analog stabilizer version of Shor's nine-qubit code, encoding one mode into nine and correcting arbitrary errors on any one mode.

The nullifiers for this code are
\flmMathEnvironment{align}{}{
\begin{split}
&\hat{x}_1 - \hat{x}_2~, \hat{x}_2 - \hat{x}_3~, \hat{x}_4 - \hat{x}_5~, \hat{x}_5 - \hat{x}_6~, \hat{x}_7 - \hat{x}_8~, \hat{x}_8 - \hat{x}_9~,\\
&(\hat{p}_1 + \hat{p}_2 + \hat{p}_3) - (\hat{p}_4 + \hat{p}_5 + \hat{p}_6)~,\\
&(\hat{p}_4 + \hat{p}_5 + \hat{p}_6) - (\hat{p}_7 +\hat{p}_8 + \hat{p}_9)~.
\end{split}
}
Logical mode operators are generated by
\flmMathEnvironment{align}{}{
\begin{split}
\bar q &= \hat{q}_1 + \hat{q}_4 + \hat{q}_7~, \\
\bar p &= \hat{p}_1 + \hat{p}_2 + \hat{p}_3~.
\end{split}
}

\codefieldsection{Encoding}
\begin{eczvaluelist}
\item\relax Twelve beam splitters \NoCaseChange{\protect\cite{cite4663}}.
\end{eczvaluelist}
\codefieldsection{Decoding}
\begin{eczvaluelist}
\item\relax Syndromes are real-valued, and decoding is done by a continuous version of majority voting (a.k.a. triple modular redundancy).
\end{eczvaluelist}
\codefieldsection{Realizations}
\begin{eczvaluelist}
\item\relax Optical network by the Furusawa group \NoCaseChange{\protect\cite{cite4665}}.
\end{eczvaluelist}
\codefieldsection{Parents}
\begin{eczvaluelist}
\item\relax
\flmRefsHyperref[eczindexfamilyrel]{code:analog_stabilizer}{Analog stabilizer code}\item\relax
\flmRefsHyperref[eczindexfamilyrel]{code:oscillator_css}{Bosonic CSS code}\item\relax
\flmRefsHyperref[eczindexfamilyrel]{code:group_quantum_parity}{Group-based QPC} --- The \(\llbracket 9,1,3\rrbracket _{G}\) group-based QPC reduces to the \(\llbracket 9,1,3\rrbracket _{\mathbb{R}}\) Lloyd-Slotine code for \(G=\mathbb{R}\).
\item\relax
\flmRefsHyperref[eczindexfamilyrel]{code:small_distance_quantum}{Small-distance block quantum code}\end{eczvaluelist}
\codefieldsection{Cousin}
\begin{eczvaluelist}
\item\relax
\flmRefsHyperref[eczindexfamilyrel]{code:shor_nine}{\(\llbracket 9,1,3\rrbracket \) Shor code} --- The Lloyd-Slotine nine-mode code is a bosonic analogue of Shor's code.
\end{eczvaluelist}
\eczhbkcontributors{ \eczhuVVA }
\endeczcode

\eczcode{chi2}{\(\chi^{(2)}\) code}{~\NoCaseChange{\protect\cite{cite4666}}}
\eczhIndexCodeAliasName{chi2}{code}
\codefieldsection{Description}
A \(3n\)-mode bosonic Fock-state code that requires only linear optics and the \(\chi^{(2)}\) optical nonlinear interaction for encoding, decoding, and logical gates.
Codewords lie in Fock-state subspaces that are invariant under Hermitian combinations of the \(\chi^{(2)}\) nonlinearities \(abc^\dagger\) and \(i abc^\dagger\), where \(a\), \(b\), and \(c\) are lowering operators acting on one of the \(n\) triples of modes on which the codes are defined.
Codewords are also \(+1\) eigenstates of stabilizer-like \textit{symmetry operators}, and photon parities are error syndromes.

\codefieldsection{Protection}
Codes protect against loss, gain, and dephasing errors conditional on the knowledge of the total number of photons lost.
\codefieldsection{Encoding}
\begin{eczvaluelist}
\item\relax Linear optics and \(\chi^{(2)}\) interactions.
\end{eczvaluelist}
\codefieldsection{Gates}
\begin{eczvaluelist}
\item\relax Linear optics and \(\chi^{(2)}\) interactions yield a universal set of gates.
\end{eczvaluelist}
\codefieldsection{Decoding}
\begin{eczvaluelist}
\item\relax Linear optics and \(\chi^{(2)}\) interactions.
\end{eczvaluelist}
\codefieldsection{Parent}
\begin{eczvaluelist}
\item\relax
\flmRefsHyperref[eczindexfamilyrel]{code:fock_state}{Fock-state bosonic code}\end{eczvaluelist}
\codefieldsection{Cousins}
\begin{eczvaluelist}
\item\relax
\flmRefsHyperref[eczindexfamilyrel]{code:tiger}{Tiger code} --- A three-mode tiger code with \(G=(2,2,-2)\), \(H=\left(\begin{smallmatrix}0&1&1\\1&0&1\end{smallmatrix}\right)\), and equal syndrome parameters has the same Fock-state support as one of the \(\chi^{(2)}\) codes \NoCaseChange{\protect\cite{cite4667}}.
\item\relax
\flmRefsHyperref[eczindexfamilyrel]{code:two-mode_binomial}{Two-mode binomial code} --- Two-mode binomial codes \NoCaseChange{\protect\cite[{Eqs. (90-91)}]{cite4666}} are closely related to three-mode \(\chi^2\) binomial codes \NoCaseChange{\protect\cite[{Eqs. (61-62)}]{cite4666}}.
\end{eczvaluelist}
\eczhbkcontributors{ \eczhuVVA }
\endeczcode

\eczcode{dfour_gkp}{\(D_4\) hyper-diamond GKP code}{~\NoCaseChange{\protect\cite{cite482}}}
\eczhIndexCodeAliasName{dfour_gkp}{hyper-diamond GKP code}
\codefieldsection{Description}
Two-mode GKP qubit-into-oscillator code based on the \(D_4\) hyper-diamond lattice \NoCaseChange{\protect\cite{cite482}}.

\codefieldsection{Gates}
\begin{eczvaluelist}
\item\relax Logical Clifford operations are given by passive Gaussian unitaries. Non-Clifford gates can be done through Kerr-type interactions.
\end{eczvaluelist}
\codefieldsection{Parents}
\begin{eczvaluelist}
\item\relax
\flmRefsHyperref[eczindexfamilyrel]{code:gkp_concatenated}{Concatenated GKP code} --- The \(D_4\) hyper-diamond GKP code can be seen as a concatenation of a rotated square-lattice GKP code with a repetition code \NoCaseChange{\protect\cite{cite482}}. This is related to the fact that the four-bit repetition code yields the \(D_4\) hyper-diamond lattice via \flmTerm{term}{ref127}{}{Construction A}.
\item\relax
\flmRefsHyperref[eczindexfamilyrel]{code:4d_stabilizer}{4D lattice stabilizer code}\item\relax
\flmRefsHyperref[eczindexfamilyrel]{code:qudits_into_oscillators}{Qudit-into-oscillator code}\end{eczvaluelist}
\codefieldsection{Cousins}
\begin{eczvaluelist}
\item\relax
\flmRefsHyperref[eczindexfamilyrel]{code:dfour}{\(D_4\) hyper-diamond lattice} --- The \(D_4\) GKP code is built from the \(D_4\) lattice.
\item\relax
\flmRefsHyperref[eczindexfamilyrel]{code:quantum_repetition}{Quantum repetition code} --- The \(D_4\) hyper-diamond GKP code can be seen as a concatenation of a rotated square-lattice GKP code with a repetition code \NoCaseChange{\protect\cite{cite482}}. This is related to the fact that the four-bit repetition code yields the \(D_4\) hyper-diamond lattice via \flmTerm{term}{ref127}{}{Construction A}.
\item\relax
\flmRefsHyperref[eczindexfamilyrel]{code:gkp-stabilizer}{Oscillator-into-oscillator GKP code} --- \(D_4\) hyper-diamond GKP codes may be optimal for oscillator-into-oscillator GKP codes utilizing two ancilla modes \NoCaseChange{\protect\cite{cite4668}}.
\end{eczvaluelist}
\eczhbkcontributors{ \eczhuVVA }
\endeczcode

\eczcode{chern_simons_gkp}{\(U(1)_{2n} \times U(1)_{-2m}\) Chern-Simons GKP code}{~\NoCaseChange{\protect\cite{cite411}}}
\eczhIndexCodeAliasName{chern_simons_gkp}{Chern-Simons GKP code}
\codefieldsection{Description}
A non-CSS multimode GKP code defined on a 2D mode lattice that encodes a qudit logical space and whose excitations are characterized by the \(U(1)_{2n} \times U(1)_{-2m}\) Chern-Simons theory.
The code can be obtained from the analog surface code by \flmRefsHyperref{ref410}{condensing} certain anyons \NoCaseChange{\protect\cite{cite411}}. 

The \(U(1)_{2} \times U(1)_{-4}\) case admits a topological order that is Witt non-trivial, i.e., that does not admit a gapped boundary. This order is not chiral (i.e., chiral central charge is zero) and does not admit a bosonic anyon.

\codefieldsection{Parents}
\begin{eczvaluelist}
\item\relax
\flmRefsHyperref[eczindexfamilyrel]{code:multimodegkp}{Gottesman-Kitaev-Preskill (GKP) code}\item\relax
\flmRefsHyperref[eczindexfamilyrel]{code:2d_stabilizer}{2D lattice stabilizer code}\end{eczvaluelist}
\codefieldsection{Cousins}
\begin{eczvaluelist}
\item\relax
\flmRefsHyperref[eczindexfamilyrel]{code:topological_abelian}{Abelian topological code} --- The \(U(1)_{2n} \times U(1)_{-2m}\) Chern-Simons GKP code realizes \(U(1)_{2n} \times U(1)_{-2m}\) Chern-Simons theory on bosonic modes. The code can be obtained from the analog surface code by \flmRefsHyperref{ref410}{condensing} certain anyons \NoCaseChange{\protect\cite{cite411}}.
\item\relax
\flmRefsHyperref[eczindexfamilyrel]{code:analog_surface}{Analog surface code} --- The \(U(1)_{2n} \times U(1)_{-2m}\) Chern-Simons GKP code can be obtained from the analog surface code by \flmRefsHyperref{ref410}{condensing} charge-flux composite anyons \NoCaseChange{\protect\cite{cite411}}.
\end{eczvaluelist}
\eczhbkcontributors{ \eczhuVVA }
\endeczcode

\eczcode{2t_qutrit}{2T-qutrit code}{~\NoCaseChange{\protect\cite{cite4669}}}
\codefieldsection{Description}
Two-mode qutrit code constructed out of superpositions of coherent states whose amplitudes make up the binary tetrahedral group \(2T\), a.k.a. the 24-cell.

The codespace is a particular three-dimensional subspace of the 24-dimensional two-mode coherent-state subspace,
\flmMathEnvironment{align}{}{
  \mathrm{Span}( \{|\sqrt{2} e^{i (2k+1) \pi/4} \alpha\rangle |0\rangle,  |0\rangle |\sqrt{2} e^{i (2k+1) \pi/4} \alpha\rangle, |e^{i k\pi/2} \alpha\rangle |e^{i \ell \pi/2} \alpha\rangle \: : \: 0\leq  k, \ell \leq 3\})
}
for any \(\alpha \geq 0\).
A basis can be constructed whose elements are uniform superpositions of coherent states whose amplitudes make up cosets of the quaternion subgroup \(Q\) in \(2T\).
\begin{flmFloat}{figure}{NumCap}\includegraphics[width=324bp,max width=\linewidth]{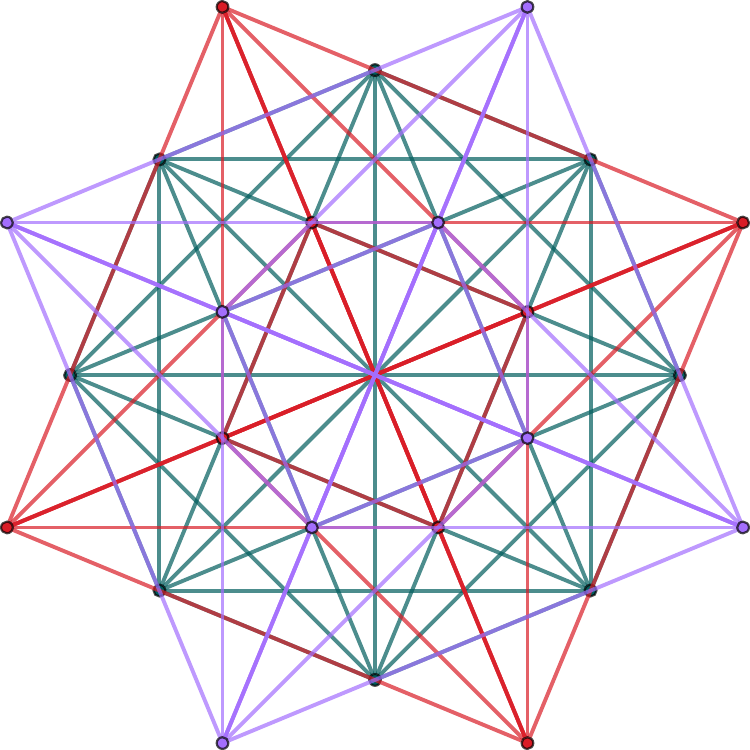}\caption{Projection of the \( 4\{3\}4 \) polytope with logical constellations marked in different colours.}\label{ref4670}\end{flmFloat}

\codefieldsection{Gates}
\begin{eczvaluelist}
\item\relax Logical phase-flip can be implemented using an excitation-preserving Gaussian transformation. Degree-four polynomial in the lowering operators of the two modes serves as a non-unitary logical bit-flip. Rotations of either mode by \(\pi/4\) are logical gates that swap two logical codewords.
\end{eczvaluelist}
\codefieldsection{Parent}
\begin{eczvaluelist}
\item\relax
\flmRefsHyperref[eczindexfamilyrel]{code:qsc}{Quantum spherical code (QSC)} --- The \(2T\)-qutrit is a QSC on the two-dimensional complex sphere whose code constellation is the \(4\{3\}4\) complex polytope.
\end{eczvaluelist}
\codefieldsection{Cousins}
\begin{eczvaluelist}
\item\relax
\flmRefsHyperref[eczindexfamilyrel]{code:two-mode_binomial}{Two-mode binomial code} --- The \(2T\)-qutrit code reduces to the two-mode "0-2-4" binomial code as \(\alpha\to 0\).
\item\relax
\flmRefsHyperref[eczindexfamilyrel]{code:24cell}{24-cell code} --- The \(2T\)-qutrit code is constructed out of superpositions of coherent states whose amplitudes make up the binary tetrahedral group \(2T\), a.k.a. the 24-cell.
\end{eczvaluelist}
\eczhbkcontributors{ Shubham P. Jain, \eczhuVVA }
\endeczcode

\eczcode{ampdamp}{Amplitude-damping (AD) code}{~\NoCaseChange{\protect\cite{cite859,cite2600}}}
\codefieldsection{Description}
Block quantum code on either qubits or bosonic modes that is designed to detect and correct qubit or bosonic \flmRefsHyperref{ref498}{AD} errors, respectively.

\codefieldsection{Protection}
\begin{defterm}{Amplitude damping noise}\label{ref4671}\label{ref498}
The amplitude damping (AD) channel is a bosonic channel that models loss of particles in a bosonic mode (a.k.a. photon loss, pure loss, or fiber attenuation).
Its Kraus operators are proportional to powers of a mode's annihilation operator \(a\), with the power signifying the number of particles lost during the error,
\flmMathEnvironment{align}{}{
  E_{\ell}=\left(\frac{\gamma}{1-\gamma}\right)^{\ell/2}\frac{a^{\ell}}{\sqrt{\ell!}}\left(1-\gamma\right)^{\hat{n}/2}\,,
}
where \(\gamma\in[0,1)\) is the noise rate \NoCaseChange{\protect\cite{cite4672,cite496}}.
For multiple modes, error set elements are tensor products of elements of the single-mode error set. 
The fixed point of this channel for any truncation of Fock space is unique \NoCaseChange{\protect\cite{cite4673}}.

Restricting the channel to the first two Fock states \(\{|0\rangle,|1\rangle\}\) yields the non-Pauli qubit AD channel, which requires protecting against the loss error \(E_1\propto X+iY\) (instead of \(X\) and \(Y\) Pauli errors individually). Both channels are called AD since the context makes clear which one is being referred to.
In this restriction, the qubit AD channel has Kraus operators \(A_0=\ket{0}\bra{0}+\sqrt{1-p}\ket{1}\bra{1}\) and \(A_1=\sqrt{p}\ket{0}\bra{1}\); \(A_1\) is the decay event and \(A_0\) is the no-jump operator whose action still changes the relative amplitudes \NoCaseChange{\protect\cite[{Ch. 1}]{cite398}}.
Other extensions to qudits are also known \NoCaseChange{\protect\cite{cite4674}}.
\end{defterm}

Protection against \flmRefsHyperref{ref498}{AD} noise is typically approximate because the tensor product of Kraus operators with all \(\ell=0\) is typically corrected only up to some order in \(\gamma\) \NoCaseChange{\protect\cite{cite859,cite2600}}.
For example, a qubit code that corrects a single \flmRefsHyperref{ref498}{AD} error is one for which all tensor products \(E_{\ell_1}\otimes\cdots\otimes E_{\ell_n}\) with \(\ell_1+\cdots + \ell_n \leq 1\) are correctable (per the \flmTerm{term}{ref1043}{}{Knill-Laflamme conditions}) up to \flmRefsHyperref{ref65}{order} \(O(\gamma^2)\).

Certain codes also have intrinsic protection against \flmRefsHyperref{ref498}{AD}, such as constant-excitation codes (CE), QSCs, or self-complementary codes.
Amplitude damping can be thought of as a quantum analogue to asymmetric noise \NoCaseChange{\protect\cite{cite1186}}.

\codefieldsection{Rate}
The quantum capacity of the \flmRefsHyperref{ref498}{AD} channel is \(\max\{0, \log \frac{1-\gamma}{\gamma}\} \) \NoCaseChange{\protect\cite{cite4675}}. Quantum capacities of the qubit \flmRefsHyperref{ref498}{AD} channel are also determined \NoCaseChange{\protect\cite{cite4676,cite4677}}, including of channels with memory \NoCaseChange{\protect\cite{cite4678,cite4679}}. Capacities of qudit extensions have also been studied \NoCaseChange{\protect\cite{cite4674}}.
\codefieldsection{Parents}
\begin{eczvaluelist}
\item\relax
\flmRefsHyperref[eczindexfamilyrel]{code:oscillators}{Bosonic code} --- Restricting the AD channel to the first two Fock states \(\{|0\rangle,|1\rangle\}\) yields the non-Pauli qubit AD channel, which requires protecting against the loss error \(E_1\propto X+iY\) (instead of \(X\) and \(Y\) Pauli errors individually). Qubit AD codes are thus a special case of bosonic AD codes.
\item\relax
\flmRefsHyperref[eczindexfamilyrel]{code:approximate_qecc}{Approximate quantum error-correcting code (AQECC)} --- Protection against \flmRefsHyperref{ref498}{AD} noise is typically approximate because the tensor product of Kraus operators with all \(\ell=0\) is typically corrected only up to some order in \(\gamma\) \NoCaseChange{\protect\cite{cite859,cite2600}}.
\end{eczvaluelist}
\codefieldsection{Children}
\begin{eczvaluelist}
\item\relax
\flmRefsHyperref[eczindexfamilyrel]{code:fock_state}{Fock-state bosonic code} --- Fock-state codes are designed to protect against bosonic \flmRefsHyperref{ref498}{AD} noise.
\item\relax
\flmRefsHyperref[eczindexfamilyrel]{code:squeezed_cat}{Squeezed cat code} --- Squeezing of coherent states allows for approximate protection against a single \flmRefsHyperref{ref498}{photon loss}.
\item\relax
\flmRefsHyperref[eczindexfamilyrel]{code:qsc}{Quantum spherical code (QSC)} --- QSC codewords are superpositions of coherent states with the same energy, but coherent states are not eigenstates of the energy Hamiltonian. The \flmRefsHyperref{ref498}{AD} Kraus operator \(E_{0}^{\otimes n}\) acts identically on each coherent state by shrinking the radius of the QSC's sphere.
\item\relax
\flmRefsHyperref[eczindexfamilyrel]{code:squeezed_fock_state}{Squeezed Fock-state code} --- The squeezed Fock-state code approximately protects against loss and dephasing errors, becoming exact in the \(r\to\infty\) limit.
\item\relax
\flmRefsHyperref[eczindexfamilyrel]{code:jump}{Jump code} --- Jump codes are designed to protect against qubit \flmRefsHyperref{ref498}{AD} noise.
\item\relax
\flmRefsHyperref[eczindexfamilyrel]{code:self_complementary}{Self-complementary qubit code} --- Self-complementary quantum codes consisting of computational basis states whose bitstrings are sufficiently spaced apart correct at least one \flmRefsHyperref{ref498}{AD} error \NoCaseChange{\protect\cite[{Thm. 2.5}]{cite851}\protect\cite[{Thm. 2}]{cite1297}}.
\item\relax
\flmRefsHyperref[eczindexfamilyrel]{code:ampdamp_post_selected}{Post-selected PI code}\item\relax
\flmRefsHyperref[eczindexfamilyrel]{code:qudit_gnu_permutation_invariant}{Qudit GNU PI code} --- Qudit GNU PI codes protect against \flmRefsHyperref{ref498}{AD} errors.
\item\relax
\flmRefsHyperref[eczindexfamilyrel]{code:ampdamp_numopt}{Numerically optimized four-qubit AD code}\end{eczvaluelist}
\codefieldsection{Cousins}
\begin{eczvaluelist}
\item\relax
\flmRefsHyperref[eczindexfamilyrel]{code:qubit_css}{Qubit CSS code} --- An \(\llbracket n,k,d_Z=t+1,d_X=2t+1\rrbracket \) qubit CSS code protects against \(t\) \flmRefsHyperref{ref498}{AD} errors \NoCaseChange{\protect\cite{cite3263}\protect\cite[{Sec. 8.7}]{cite736}}.
\item\relax
\flmRefsHyperref[eczindexfamilyrel]{code:qubit_concatenated}{Concatenated qubit code} --- Using the dual-rail code as an outer code with an inner \(\llbracket n,k,d\rrbracket \) qubit code yields an \(\llbracket 2n,k\rrbracket \) code correcting \(d-1\) qubit \flmRefsHyperref{ref498}{AD} errors \NoCaseChange{\protect\cite{cite3263}}.
\item\relax
\flmRefsHyperref[eczindexfamilyrel]{code:hamming}{\([2^r-1,2^r-r-1,3]\) Hamming code} --- Ref. \NoCaseChange{\protect\cite{cite1187}} presents a \(\llbracket 7,3\rrbracket \) qubit stabilizer code for a single \flmRefsHyperref{ref498}{AD} error based on the classical \([7,4,3]\) Hamming code.
\item\relax
\flmRefsHyperref[eczindexfamilyrel]{code:dual_rail}{Dual-rail quantum code} --- Dual-rail qubits can be used to convert leakage and \flmRefsHyperref{ref498}{AD} noise into erasure noise \NoCaseChange{\protect\cite{cite1187,cite4680}}. Concatenating the dual-rail code with an \(\llbracket n,k,d\rrbracket \) stabilizer code yields an \(\llbracket 2n,k,d\rrbracket \) constant-excitation code \NoCaseChange{\protect\cite{cite2711}} that protects against \(d-1\) \flmRefsHyperref{ref498}{AD} errors \NoCaseChange{\protect\cite{cite3263}}.
\item\relax
\flmRefsHyperref[eczindexfamilyrel]{code:constant_excitation}{Constant-excitation (CE) code} --- Fock-state and qubit CE codes exactly protect against the \flmRefsHyperref{ref498}{AD} Kraus operator \(E_{0}^{\otimes n}\) because it acts identically on all Fock (and qubit) states with the same excitation number \NoCaseChange{\protect\cite{cite859,cite2600}}.
\item\relax
\flmRefsHyperref[eczindexfamilyrel]{code:iceberg}{\(\llbracket 2m,2m-2,2\rrbracket \) error-detecting code} --- The \(\llbracket 2m,2m-2,2\rrbracket \) error-detecting code \NoCaseChange{\protect\cite{cite3238}} and its relative the code with single stabilizer \(XX\cdots X\) \NoCaseChange{\protect\cite{cite3239}} admit autonomous QEC against single \flmRefsHyperref{ref498}{AD} errors.
\item\relax
\flmRefsHyperref[eczindexfamilyrel]{code:stab_5_1_3}{\(\llbracket 5,1,3\rrbracket \) Five-qubit perfect code} --- The five-qubit perfect code approximately corrects a single \flmRefsHyperref{ref498}{AD} error \NoCaseChange{\protect\cite{cite859}}.
\item\relax
\flmRefsHyperref[eczindexfamilyrel]{code:quantum_parity}{Quantum parity code (QPC)} --- An \(\llbracket 8,1,2\rrbracket \) QPC correcting a single \flmRefsHyperref{ref498}{AD} error is equivalent to a concatenation of the \(\{|\overline{01}\rangle,|\overline{11}\rangle\}\) (constant-excitation) subcode of the \(\llbracket 4,2,2\rrbracket \) code with the dual-rail code \NoCaseChange{\protect\cite{cite3250,cite3259,cite2711}}. More generally, an \(\llbracket m^2,1,m\rrbracket \) QPC corrects \(m-1\) \flmRefsHyperref{ref498}{AD} errors \NoCaseChange{\protect\cite{cite3263}}.
\item\relax
\flmRefsHyperref[eczindexfamilyrel]{code:qubit_stabilizer}{Qubit stabilizer code} --- Concatenating the dual-rail code with an inner \(\llbracket n,k,d\rrbracket \) qubit stabilizer code yields a degenerate \(\llbracket 2n,k,d\rrbracket \) constant-excitation stabilizer code that corrects \(d-1\) \flmRefsHyperref{ref498}{AD} errors \NoCaseChange{\protect\cite{cite3263,cite2711}}.
\end{eczvaluelist}
\eczhbkcontributors{ \eczhuVVA }
\endeczcode

\eczcode{cv_cluster_state}{Analog cluster-state code}{~\NoCaseChange{\protect\cite{cite4681,cite4682,cite4683}}}
\codefieldsection{Alternative Names}
\begin{eczvaluelist}
\item\relax CV cluster-state code
\item\relax CV graph-state code
\item\relax Bosonic cluster-state code
\end{eczvaluelist}
\eczhIndexCodeAliasName{cv_cluster_state}{CV cluster-state code}
\eczhIndexCodeAliasName{cv_cluster_state}{CV graph-state code}
\eczhIndexCodeAliasName{cv_cluster_state}{Bosonic cluster-state code}
\codefieldsection{Description}
A code based on a continuous-variable (CV), or analog, cluster state.
Such a state can be used to perform MBQC of logical modes, which substitutes the temporal dimension necessary for decoding a conventional code with a spatial dimension.
The exact analog cluster state is non-normalizable, so approximate constructions have to be considered.

Analog cluster states are analog stabilizer states defined on a graph.
There is one nullifier \(\hat{\eta}_j\) per graph vertex \(j\) of the form
\flmMathEnvironment{align}{}{
  \hat{\eta}_j = \hat{p}_{j} - \sum_{k\in N(j)} V_{jk} \hat{x}_k~,
}
where the neighborhood \(N(j)\) is the set of vertices which share an edge with \(j\), and where \(V_{jk}\) is a weighted (real-valued) adjacency matrix of a graph \NoCaseChange{\protect\cite{cite4684}}.

Analog cluster states, like cluster states, can be defined on various geometries.
Analog cluster states defined on a 1D array of modes are called quantum wires \NoCaseChange{\protect\cite{cite4685,cite4686}}, not to be confused with the Kitaev quantum wire, a fermion code.
Analog cluster states defined on a 1D ladder are sometimes called dual-rail, not to be confused with the dual-rail code.

\codefieldsection{Protection}
Protection is related to the analog stabilizer code underlying the analog cluster state.

\codefieldsection{Encoding}
\begin{eczvaluelist}
\item\relax Initialization of all modes in momentum eigenstates and action of gates of the form \(\exp(iV_{jk}\hat{x}_{j}\hat{x}_{k})\). The normalizable version substitutes momentum eigenstates with finitely squeezed states.
\item\relax Squeezers and beam-splitters \NoCaseChange{\protect\cite{cite4687}}.
\end{eczvaluelist}
\codefieldsection{Gates}
\begin{eczvaluelist}
\item\relax Combination of linear-optical gates and homodyne measurements on subsets of vertices \NoCaseChange{\protect\cite{cite4682,cite4683}}.
\item\relax Gaussian operations can be realized as operations acting on graphs underlying a cluster state. They can be done in any order, demonstrating parallelism \NoCaseChange{\protect\cite{cite4682,cite4683}}.
\item\relax Magic-state distillation is required for universal computation \NoCaseChange{\protect\cite{cite4682,cite4683}}.
\end{eczvaluelist}
\codefieldsection{Realizations}
\begin{eczvaluelist}
\item\relax Analog cluster states on a number of modes ranging from tens to millions \NoCaseChange{\protect\cite{cite4688,cite4689,cite4690}} have been synthesized in photonic degrees of freedom. A \(12\times N\) mode cluster state, where \(N\) is the number of clock cycles of the experiment, has been realized in a photonic device by Xanadu \NoCaseChange{\protect\cite{cite4691}}.
\item\relax Required primitives for Gaussian gates have been realized \NoCaseChange{\protect\cite{cite4692}}.
\end{eczvaluelist}
\codefieldsection{Notes}
\begin{eczvaluelist}
\item\relax See Ref. \NoCaseChange{\protect\cite{cite3557}} for a review of analog cluster states and their applications.
\end{eczvaluelist}
\codefieldsection{Parents}
\begin{eczvaluelist}
\item\relax
\flmRefsHyperref[eczindexfamilyrel]{code:analog_stabilizer}{Analog stabilizer code} --- Analog cluster-state codes are particular analog stabilizer codes. Any analog stabilizer state is equivalent to an analog cluster state under a single-mode Gaussian circuit \NoCaseChange{\protect\cite{cite507}}. Relaxing the real weighted adjacency matrix of an analog cluster state to be complex yields a description of a general analog (i.e., Gaussian) stabilizer state \NoCaseChange{\protect\cite{cite4693}}. Pure Gaussian states, which are normalizable approximate versions of analog stabilizer states, are not equivalent to finitely squeezed analog cluster states via Gaussian local unitaries \NoCaseChange{\protect\cite{cite4684}}.
\item\relax
\flmRefsHyperref[eczindexfamilyrel]{code:graph_quantum}{Graph quantum code} --- Graph quantum codes for \(G=\mathbb{R}\) reduce to analog cluster-state codes.
\end{eczvaluelist}
\codefieldsection{Cousins}
\begin{eczvaluelist}
\item\relax
\flmRefsHyperref[eczindexfamilyrel]{code:ame}{Perfect-tensor code} --- Analog cluster states are generically CV AME \NoCaseChange{\protect\cite{cite2923}}, and explicit constructions exist for any number of modes \NoCaseChange{\protect\cite{cite507}}.
\item\relax
\flmRefsHyperref[eczindexfamilyrel]{code:1d_stabilizer}{1D lattice stabilizer code} --- Analog cluster states defined on a 1D array of modes are called quantum wires \NoCaseChange{\protect\cite{cite4685,cite4686}}, not to be confused with the Kitaev quantum wire, a fermion code. Analog cluster states defined on a 1D ladder are sometimes called dual-rail, not to be confused with the dual-rail code.
\item\relax
\flmRefsHyperref[eczindexfamilyrel]{code:gkp-cluster-state}{GKP CV-cluster-state code} --- GKP CV-cluster-state codes reduce to analog-cluster-state codes when all physical modes are initialized in momentum states.
\end{eczvaluelist}
\eczhbkcontributors{ \eczhuVVA }
\endeczcode

\eczcode{analog_repetition}{Analog repetition code}{}
\codefieldsection{Alternative Names}
\begin{eczvaluelist}
\item\relax Gaussian repetition code
\end{eczvaluelist}
\eczhIndexCodeAliasName{analog_repetition}{Gaussian repetition code}
\codefieldsection{Description}
An \(\llbracket n,1\rrbracket _{\mathbb{R}}\) analog stabilizer version of the quantum repetition code, encoding the position states of one mode into an odd number \(n\) of modes.

There are two variants, a bit- and a phase-flip code, whose encoding for \(n=3\) is
\flmMathEnvironment{align}{}{
  |\overline{x}_{\text{bit}}\rangle&\rightarrow|x,x,x\rangle\\
  |\overline{x}_{\text{phase}}\rangle&\rightarrow \int dx_{1}dx_{2}dx_{3}\delta(x_{1}+x_{2}+x_{3}-x)|x_{1},x_{2},x_{3}\rangle~.
}

Nullifiers for the bit-flip analog repetition code are differences \(\hat{x}_{j+1} - \hat{x}_{j}\).
Bit-flip codewords can be superposed to yield the logical momentum basis of \textit{analog GHZ states}
\flmMathEnvironment{align}{}{
  |\overline{p}\rangle=\int dx e^{ipx}|x\rangle^{\otimes n}~,
}
a bosonic version of GHZ states.
At \(p=0\), the above is an analog stabilizer state nullified by the bit-flip nullifiers and the total momentum operator \(\hat{p}_1+\hat{p}_2+\cdots+\hat{p}_n\) \NoCaseChange{\protect\cite{cite4694}}.
For \(n=2\), this state is known as an \textit{EPR pair} \NoCaseChange{\protect\cite{cite4695}}, an infinitely squeezed version of the two-mode squeezed (TMS) a.k.a. twin-beam state. 

\codefieldsection{Realizations}
\begin{eczvaluelist}
\item\relax Quantum teleportation \NoCaseChange{\protect\cite{cite4696}}, secret sharing \NoCaseChange{\protect\cite{cite4697}}, and superdense coding \NoCaseChange{\protect\cite{cite4698}} protocols have been realized with analog GHZ states for \(n=2,3\).
\end{eczvaluelist}
\codefieldsection{Notes}
\begin{eczvaluelist}
\item\relax Analog GHZ states are useful for quantum teleportation \NoCaseChange{\protect\cite{cite4699}}.
\end{eczvaluelist}
\codefieldsection{Parents}
\begin{eczvaluelist}
\item\relax
\flmRefsHyperref[eczindexfamilyrel]{code:analog_stabilizer}{Analog stabilizer code}\item\relax
\flmRefsHyperref[eczindexfamilyrel]{code:oscillator_css}{Bosonic CSS code}\item\relax
\flmRefsHyperref[eczindexfamilyrel]{code:1d_stabilizer}{1D lattice stabilizer code}\item\relax
\flmRefsHyperref[eczindexfamilyrel]{code:group_quantum_repetition}{Group-based quantum repetition code} --- Group-based quantum repetition codes reduce to analog repetition codes for \(G = \mathbb{R}\).
\end{eczvaluelist}
\codefieldsection{Cousins}
\begin{eczvaluelist}
\item\relax
\flmRefsHyperref[eczindexfamilyrel]{code:niset_andersen_cerf}{Niset-Andersen-Cerf code} --- EPR pairs are used in an encoding of the Niset-Andersen-Cerf code \NoCaseChange{\protect\cite{cite4700}}.
\item\relax
\flmRefsHyperref[eczindexfamilyrel]{code:ame}{Perfect-tensor code} --- Analog GHZ states are \(1\)-uniform for all \(n\) and CV AME for \(n=2,3\) \NoCaseChange{\protect\cite{cite2923,cite507}}.
\end{eczvaluelist}
\eczhbkcontributors{ \eczhuVVA }
\endeczcode

\eczcode{analog_stabilizer}{Analog stabilizer code}{~\NoCaseChange{\protect\cite{cite3372,cite507}}}
\codefieldsection{Alternative Names}
\begin{eczvaluelist}
\item\relax Gaussian stabilizer code
\item\relax Linear stabilizer code
\item\relax Symplectic stabilizer code
\item\relax Wavepacket code
\item\relax Infinitely squeezed state code
\end{eczvaluelist}
\eczhIndexCodeAliasName{analog_stabilizer}{Gaussian stabilizer code}
\eczhIndexCodeAliasName{analog_stabilizer}{Linear stabilizer code}
\eczhIndexCodeAliasName{analog_stabilizer}{Symplectic stabilizer code}
\eczhIndexCodeAliasName{analog_stabilizer}{Wavepacket code}
\eczhIndexCodeAliasName{analog_stabilizer}{Infinitely squeezed state code}
\codefieldsection{Description}
An oscillator-into-oscillator stabilizer code encoding logical oscillator modes into \(n\) physical modes. If the code is defined by \(r\) independent nullifiers, then it is denoted by \(\llbracket n,n-r\rrbracket _{\mathbb{R}}\) and encodes \(k=n-r\) logical modes \NoCaseChange{\protect\cite{cite507}}.
Any analog stabilizer state can be thought of as a pure Gaussian state that has been infinitely squeezed on all modes \NoCaseChange{\protect\cite{cite507}}.

Analog stabilizer codes admit a continuous stabilizer group of displacements. This group can equivalently be defined in terms of its Lie algebra. The codespace is equivalently the common \(0\)-eigenvalue eigenspace of the Lie algebra generators, which are mutually commuting linear combinations of oscillator position and momentum operators called \textit{nullifiers} \NoCaseChange{\protect\cite{cite4683}} or \textit{annihilators}.
An analog stabilizer code admitting a set of nullifiers such that each nullifier consists of either position or momentum operators is called an \textit{analog CSS code}.

\codefieldsection{Protection}
Protects against erasures of, or any errors on, at most \(d-1\) modes.
If an error operator does not commute with a nullifier, then that error is detectable. 
There are conditions on the encoding circuit which guarantee that the code can correct errors \NoCaseChange{\protect\cite{cite4662}}.
The code can be further optimized to increase resolution between syndrome spaces for certain noise \NoCaseChange{\protect\cite{cite4662}}.

Protection of logical modes against small displacements or other errors acting in every physical mode cannot be done using only Gaussian resources \NoCaseChange{\protect\cite{cite4701}}. For Gaussian displacement noise, linear oscillator encodings can at best squeeze the logical noise between conjugate quadratures rather than suppressing it in both simultaneously \NoCaseChange{\protect\cite[{Sec. VI}]{cite416}} (see also \NoCaseChange{\protect\cite{cite4702,cite4703}}). 
There are no such restrictions for non-Gaussian noise \NoCaseChange{\protect\cite{cite4704,cite4662}}.

\codefieldsection{Encoding}
\begin{eczvaluelist}
\item\relax Gaussian circuit applied to \(k\) modes storing logical information and \(n-k\) modes initialized in a position state.
\end{eczvaluelist}
\codefieldsection{Decoding}
\begin{eczvaluelist}
\item\relax Homodyne measurement of nullifiers yields real-valued syndromes, and recovery can be performed by displacements conditional on the syndromes.
\end{eczvaluelist}
\codefieldsection{Realizations}
\begin{eczvaluelist}
\item\relax One-sided device-independent QKD \NoCaseChange{\protect\cite{cite4705}}.
\end{eczvaluelist}
\codefieldsection{Parents}
\begin{eczvaluelist}
\item\relax
\flmRefsHyperref[eczindexfamilyrel]{code:oscillator_stabilizer}{Bosonic stabilizer code} --- Analog stabilizer codes are bosonic stabilizer codes with a continuous stabilizer group, corresponding to linear constraints on positions and momenta.
\item\relax
\flmRefsHyperref[eczindexfamilyrel]{code:oscillators_into_oscillators}{Oscillator-into-oscillator code}\end{eczvaluelist}
\codefieldsection{Children}
\begin{eczvaluelist}
\item\relax
\flmRefsHyperref[eczindexfamilyrel]{code:analog_repetition}{Analog repetition code}\item\relax
\flmRefsHyperref[eczindexfamilyrel]{code:analog_surface}{Analog surface code}\item\relax
\flmRefsHyperref[eczindexfamilyrel]{code:braunstein}{\(\llbracket 5,1,3\rrbracket _{\mathbb{R}}\) Braunstein five-mode code}\item\relax
\flmRefsHyperref[eczindexfamilyrel]{code:cv_cluster_state}{Analog cluster-state code} --- Analog cluster-state codes are particular analog stabilizer codes. Any analog stabilizer state is equivalent to an analog cluster state under a single-mode Gaussian circuit \NoCaseChange{\protect\cite{cite507}}. Relaxing the real weighted adjacency matrix of an analog cluster state to be complex yields a description of a general analog (i.e., Gaussian) stabilizer state \NoCaseChange{\protect\cite{cite4693}}. Pure Gaussian states, which are normalizable approximate versions of analog stabilizer states, are not equivalent to finitely squeezed analog cluster states via Gaussian local unitaries \NoCaseChange{\protect\cite{cite4684}}.
\item\relax
\flmRefsHyperref[eczindexfamilyrel]{code:hnss}{Hayden-Nezami-Salton-Sanders bosonic code}\item\relax
\flmRefsHyperref[eczindexfamilyrel]{code:lloyd_slotine}{\(\llbracket 9,1,3\rrbracket _{\mathbb{R}}\) Lloyd-Slotine code}\end{eczvaluelist}
\codefieldsection{Cousins}
\begin{eczvaluelist}
\item\relax
\flmRefsHyperref[eczindexfamilyrel]{code:gkp-stabilizer}{Oscillator-into-oscillator GKP code} --- Analog stabilizer codes protect logical modes against arbitrarily large displacements on a few modes, while oscillator-into-oscillator GKP codes protect an infinite-dimensional logical space against sufficiently small displacements in any number of modes. Encoding in analog-stabilizer (oscillator-into-oscillator GKP) codes can be done by a Gaussian operation acting on a tensor product of an arbitrary state in the first mode and position states (GKP states) on the remaining modes. For Gaussian displacement noise, linear oscillator encodings only squeeze the logical noise between conjugate quadratures \NoCaseChange{\protect\cite[{Sec. VI}]{cite416}}, and protection of logical modes against small displacements cannot be done using only Gaussian resources \NoCaseChange{\protect\cite{cite4702,cite4701}}, so oscillator-into-oscillator GKP codes can be thought of as analog stabilizer encodings utilizing non-Gaussian GKP resource states.
\item\relax
\flmRefsHyperref[eczindexfamilyrel]{code:qudit_stabilizer}{Modular-qudit stabilizer code} --- Prime-qudit stabilizer codes can be transformed into analog stabilizer codes on the same number of modes and logical modes, with distance at least as large as that of the original code \NoCaseChange{\protect\cite[{Thm. 12}]{cite4580}}.
\item\relax
\flmRefsHyperref[eczindexfamilyrel]{code:galois_stabilizer}{Galois-qudit stabilizer code} --- Galois-qudit stabilizer codes can be transformed into analog stabilizer codes; if the original code has \(r\) linearly independent generators in the symplectic representation, the resulting analog code has parameters \(\llbracket n,n-r,d^{\prime}\rrbracket _{\mathbb{R}}\) with \(d^{\prime}\geq d\) \NoCaseChange{\protect\cite[{Thm. 13}]{cite4580}}.
\item\relax
\flmRefsHyperref[eczindexfamilyrel]{code:stab_8_3_3}{\(\llbracket 8, 3, 3\rrbracket \) Eight-qubit Gottesman code} --- The eight-qubit Gottesman code has been extended to an analog stabilizer code \NoCaseChange{\protect\cite{cite3372}}.
\item\relax
\flmRefsHyperref[eczindexfamilyrel]{code:ame}{Perfect-tensor code} --- Analog stabilizer states are generically CV AME \NoCaseChange{\protect\cite{cite2923}}, and explicit constructions exist for any number of modes \NoCaseChange{\protect\cite{cite507}}. The codespace of an analog stabilizer code with \flmRefsHyperref{ref672}{pure distance} \(d_{\textnormal{pure}}\) is a \((d_{\textnormal{pure}}-1)\)-uniform space \NoCaseChange{\protect\cite{cite507}}. Normalizable finitely squeezed versions of infinitely squeezed Gaussian states are locally thermal, up to corrections in the squeezing parameter \NoCaseChange{\protect\cite{cite2921,cite2939}}.
\item\relax
\flmRefsHyperref[eczindexfamilyrel]{code:real_block}{Real block code} --- Analog stabilizer codes are quantum counterparts of real block codes.
\item\relax
\flmRefsHyperref[eczindexfamilyrel]{code:ea_analog_stabilizer}{EA analog stabilizer code} --- EA analog stabilizer codes utilize additional ancillary modes in a pre-shared entangled state, but reduce to ordinary analog stabilizer codes when said modes are interpreted as noiseless physical modes.
\item\relax
\flmRefsHyperref[eczindexfamilyrel]{code:qudit_cubic}{Qudit cubic code} --- The qudit cubic code can be generalized to oscillators \NoCaseChange{\protect\cite{cite2531}}.
\end{eczvaluelist}
\eczhbkcontributors{ \eczhuVVA }
\endeczcode

\eczcode{analog_surface}{Analog surface code}{~\NoCaseChange{\protect\cite{cite4706}}}
\codefieldsection{Alternative Names}
\begin{eczvaluelist}
\item\relax \(\mathbb{R}\) gauge theory code
\item\relax Continuous-variable (CV) surface code
\end{eczvaluelist}
\eczhIndexCodeAliasName{analog_surface}{\(\mathbb{R}\) gauge theory code}
\eczhIndexCodeAliasName{analog_surface}{Continuous-variable (CV) surface code}
\codefieldsection{Description}
An analog CSS version of the Kitaev surface code realizing a phase of 2D \(\mathbb{R}\) gauge theory.

\codefieldsection{Notes}
\begin{eczvaluelist}
\item\relax See \NoCaseChange{\protect\cite[{Sec. III.C2}]{cite2967}} for an exposition.
\end{eczvaluelist}
\codefieldsection{Parents}
\begin{eczvaluelist}
\item\relax
\flmRefsHyperref[eczindexfamilyrel]{code:analog_stabilizer}{Analog stabilizer code}\item\relax
\flmRefsHyperref[eczindexfamilyrel]{code:oscillator_css}{Bosonic CSS code}\item\relax
\flmRefsHyperref[eczindexfamilyrel]{code:2d_stabilizer}{2D lattice stabilizer code}\end{eczvaluelist}
\codefieldsection{Cousins}
\begin{eczvaluelist}
\item\relax
\flmRefsHyperref[eczindexfamilyrel]{code:ame}{Perfect-tensor code} --- Analog surface-code states are \(3\)-uniform \NoCaseChange{\protect\cite{cite507}}.
\item\relax
\flmRefsHyperref[eczindexfamilyrel]{code:qudit_surface}{Modular-qudit surface code} --- The analog surface code can be thought of as a realization of the \(q\to\infty\) \(\mathbb{R}\) oscillator limit \NoCaseChange{\protect\cite{cite2531}} of the qudit surface code as a bosonic stabilizer code.
\item\relax
\flmRefsHyperref[eczindexfamilyrel]{code:topological_abelian}{Abelian topological code} --- The analog surface code realizes a straightforward extension of the modular-qudit surface code to infinite local dimension, \(q\to\infty\) \NoCaseChange{\protect\cite{cite2531}}.
The code realizes a phase of 2D \(\mathbb{R}\) gauge theory.
There are two types of anyons, \(e\) and \(m\), with each type being valued in a continuous domain as opposed to \(\mathbb{Z}_q\) for the qudit surface code.

\item\relax
\flmRefsHyperref[eczindexfamilyrel]{code:compactified_r}{Compactified \(\mathbb{R}\) gauge theory code} --- The compactified \(\mathbb{R}\) gauge theory code can be obtained from the analog surface code by \flmRefsHyperref{ref410}{condensing} certain anyons \NoCaseChange{\protect\cite{cite411}}. This results in a pinning of each mode to the space of periodic functions, which is the Hilbert space of a physical rotor, and can be thought of as compactification of the 2D \(\mathbb{R}\) gauge theory phase realized by the analog surface code.
\item\relax
\flmRefsHyperref[eczindexfamilyrel]{code:chern_simons_gkp}{\(U(1)_{2n} \times U(1)_{-2m}\) Chern-Simons GKP code} --- The \(U(1)_{2n} \times U(1)_{-2m}\) Chern-Simons GKP code can be obtained from the analog surface code by \flmRefsHyperref{ref410}{condensing} charge-flux composite anyons \NoCaseChange{\protect\cite{cite411}}.
\item\relax
\flmRefsHyperref[eczindexfamilyrel]{code:gkp_surface_concatenated}{GKP-surface code} --- \flmRefsHyperref{ref410}{Condensing} pure fluxes and charges in the analog surface code yields toric-GKP codes \NoCaseChange{\protect\cite{cite411}}.
\end{eczvaluelist}
\eczhbkcontributors{ \eczhuVVA }
\endeczcode

\eczcode{binomial}{Binomial code}{~\NoCaseChange{\protect\cite{cite4707}}}
\codefieldsection{Description}
Bosonic rotation codes designed to approximately protect against errors consisting of powers of raising and lowering operators up to some maximum power. Binomial codes can be thought of as spin-coherent states embedded into an oscillator \NoCaseChange{\protect\cite{cite496}}.

A simple example of a binomial code is the \textit{"0-2-4"} qubit code with codewords
\flmMathEnvironment{align}{}{
\begin{split}
  |\overline{0}\rangle&=\frac{1}{\sqrt{2}}\left(|0\rangle+|4\rangle\right)\\
  |\overline{1}\rangle&=|2\rangle~,
\end{split}
}
constructed out of binomial states \NoCaseChange{\protect\cite{cite4708}}.

General \(q\)-dimensional qudit \((N, S)\) binomial codeword states are \(\{|\overline{i}\rangle\mid i\in \mathbb Z_q \}\), where
  \flmMathEnvironment{align}{}{
    |\overline{i}\rangle = \frac{1}{\sqrt{q^N}} \sum_{\substack{p=0\\p\equiv i \pmod{q}}}^{(q-1)(N+1)} \sqrt{\binom{N+1}{p}_q} \ket{p(S+1)}.
  }
  The set \(\{\ket{i}\mid i \in \mathbb{N}\}\) is the set of Fock states. Also, \(\binom{N+1}{p}_q\) are extended binomial coefficients, or polynomial coefficients, defined recursively as
  \flmMathEnvironment{align}{}{
    \binom{n}{m}_1 \equiv 1,\quad \binom{n}{m}_q \equiv \sum_{k=0}^n \binom{n}{k}\binom{k}{m-k}_{q-1}.
  }
  The extended binomial coefficients \( \binom{n}{m}_q \) are also the coefficients of \( x^m \) in the polynomial \( (1 + x + \cdots + x^{q-1})^n \).

\codefieldsection{Protection}
An \((N, S)\) binomial code protects against \(L\) boson losses, \(G\) boson gains, and dephasing up to \(\hat{n}^{D}\), where \(S=L+G\) and \(N = \mathrm{max}(L,G,2D)\).  Binomial codes approximately protect against continuous-time \flmRefsHyperref{ref498}{AD}, boson loss and gain, and dephasing.
\codefieldsection{Encoding}
\begin{eczvaluelist}
\item\relax State preparation using spin-boson interactions \NoCaseChange{\protect\cite{cite4709}}.
\end{eczvaluelist}
\codefieldsection{Gates}
\begin{eczvaluelist}
\item\relax Error-detecting \(CCZ\) and \(cSWAP\) gates for "0-2-4" code using three-level ancilla \NoCaseChange{\protect\cite{cite4710}}.
\item\relax Single logical-qubit rotations \NoCaseChange{\protect\cite{cite4711}}.
\item\relax Amplitude-mixing error-transparent gates \NoCaseChange{\protect\cite{cite4712}}.
\end{eczvaluelist}
\codefieldsection{Decoding}
\begin{eczvaluelist}
\item\relax Photon loss and dephasing errors can be detected by measuring the phase-space rotation \(\exp\left(2\pi\mathrm{i} \hat{n} / (S+1)\right)\) and the check operator \((J_x/J)^2\) in the spin-coherent state language, where \(J\) is the total angular momentum and \(J_x\) is the angular momentum in the \(x\) direction \NoCaseChange{\protect\cite{cite496}}. This type of error correction fails for errors that are products of \flmRefsHyperref{ref498}{photon loss/gain} and dephasing errors. However, for certain \((N,S)\) instances of the binomial code, detection of these types of errors can be done.
\item\relax Recovery can be done via projective measurements and unitary operations in a version of the Cafaro recovery map \NoCaseChange{\protect\cite{cite4707,cite496}}.
\item\relax Fault-tolerant scheme that converts the required POVM into binary measurements whose redundancy is guaranteed by a classical code \NoCaseChange{\protect\cite{cite3172}}.
\end{eczvaluelist}
\codefieldsection{Realizations}
\begin{eczvaluelist}
\item\relax Microwave cavities coupled to superconducting circuits: state transfer between a binomial codeword to another system \NoCaseChange{\protect\cite{cite4713}}, error-correction protocol nearly reaching break-even \NoCaseChange{\protect\cite{cite4714}}, a teleported CNOT gate \NoCaseChange{\protect\cite{cite4715}}, and fault-tolerant logical operations utilizing three-level ancillas \NoCaseChange{\protect\cite{cite4716}}. A realization of the "0-2-4" encoding is the first to go beyond break-even error-correction and yields a logical lifetime that exceeds the cavity lifetime by \(16\%\) \NoCaseChange{\protect\cite{cite4717}} (see also \NoCaseChange{\protect\cite{cite4718}}). See Ref. \NoCaseChange{\protect\cite{cite4719}} for another experiment. The 0-2-4 binomial code has been used to store entangled states \NoCaseChange{\protect\cite{cite4720}}.
\item\relax Motional degree of freedom of a trapped ion: binomial state preparation for \(S=2\) realized by Tan group \NoCaseChange{\protect\cite{cite4721}}.
\end{eczvaluelist}
\codefieldsection{Notes}
\begin{eczvaluelist}
\item\relax The mean occupation number, or average Fock-state number in maximally-mixed state of the code, is \((N+1)(S+1)(q-1)/2 \), where \(q\) is the qudit dimension.
\end{eczvaluelist}
\codefieldsection{Parent}
\begin{eczvaluelist}
\item\relax
\flmRefsHyperref[eczindexfamilyrel]{code:bosonic_rotation}{Bosonic rotation code} --- One can verify by direct calculation that the logical states are eigenstates of the discrete rotation operator. One has freedom in the exact form of the primitive state to choose; see \NoCaseChange{\protect\cite[{App. B.2}]{cite4722}}.
\end{eczvaluelist}
\codefieldsection{Cousins}
\begin{eczvaluelist}
\item\relax
\flmRefsHyperref[eczindexfamilyrel]{code:cat}{Cat code} --- For a fixed \(S\), binomial codes with \(N \to \infty\) coincide with cat codes as \(\alpha \to \infty\) \NoCaseChange{\protect\cite{cite4707}}.
\item\relax
\flmRefsHyperref[eczindexfamilyrel]{code:number_phase}{Number-phase code} --- In the limit as \(N,S \to \infty\), phase measurement in the binomial code has vanishing variance, just like in a number-phase code \NoCaseChange{\protect\cite{cite4722}}.
\item\relax
\flmRefsHyperref[eczindexfamilyrel]{code:chebyshev}{Chebyshev code} --- Chebyshev codes resemble binomial codes, and a class of binomial codes have similar error-correcting properties \NoCaseChange{\protect\cite{cite2756}}.
\item\relax
\flmRefsHyperref[eczindexfamilyrel]{code:two-mode_binomial}{Two-mode binomial code} --- Two-mode binomial codes are two-mode analogues of binomial codes.
\item\relax
\flmRefsHyperref[eczindexfamilyrel]{code:asymmetric_qecc}{Asymmetric quantum code (AQC)} --- Binomial code parameters against loss/gain errors and dephasing can be tuned.
\item\relax
\flmRefsHyperref[eczindexfamilyrel]{code:gnu_permutation_invariant}{GNU PI code} --- Binomial codes and GNU codes related via the Holstein-Primakoff mapping \NoCaseChange{\protect\cite{cite651,cite652,cite653}}. A qudit generalization of GNU codes can be obtained from qudit binomial codes \NoCaseChange{\protect\cite[{Appx. C}]{cite496}}.
\item\relax
\flmRefsHyperref[eczindexfamilyrel]{code:css_4_1_2}{\(\llbracket 4,1,2\rrbracket \) Leung-Nielsen-Chuang-Yamamoto (LNCY) code} --- The \(\llbracket 4,1,2\rrbracket \) LNCY code reduces to the \(0,2,4\) binomial code when the basis labels in each codeword are written as in base-ten. Such a mapping can be generalized \NoCaseChange{\protect\cite{cite3260}}.
\item\relax
\flmRefsHyperref[eczindexfamilyrel]{code:four_qubit_permutation_invariant}{\(\llparenthesis 4,2,2\rrparenthesis \) Four-qubit single-deletion code} --- The four-qubit single-deletion code can be obtained from the "0-2-4" single-mode binomial code by substituting Fock states with \flmRefsHyperref{ref526}{Dicke states}.
\item\relax
\flmRefsHyperref[eczindexfamilyrel]{code:ae}{Æ code} --- Many well-performing Æ codes can be mapped into shifted versions of binomial codes via the Holstein-Primakoff mapping.
\end{eczvaluelist}
\eczhbkcontributors{ Thomas Wrona, Joseph T. Iosue, \eczhuVVA }
\endeczcode

\eczcode{bosonic_q-ary_expansion}{Bosonic \(q\)-ary expansion}{~\NoCaseChange{\protect\cite{cite4723}}}
\codefieldsection{Description}
A one-to-one mapping between basis states on \(n\) prime-dimensional qudits (of dimension \(q=p\)) and the subspace of the first \(p^n\) single-mode Fock states.
While this mapping offers a way to map qudits into a single mode, noise models for the two code families induce different notions of locality and thus qualitatively different physical interpretations \NoCaseChange{\protect\cite{cite497}}.

The mapping allows one to think of the Fock subspace as a tensor product of \(n\) \(p\)-dimensional qudits by performing a \(p\)-ary expansion of the Fock-state labels and treating each digit as an index of a qudit basis state.
The Fock state integer label \(N\geq 0\) is expanded in the \(p\)-ary expansion as
\flmMathEnvironment{align}{}{
    N=\sum_{\mu=0}^{\infty}b_{\mu}p^{\mu}~,
}
with each \(p\)-ary digit \(b_{\mu}\in\mathbb{Z}_p\) corresponding to the basis-state label of qudit \(\mu\).

In the binary case, the first qubit's \(Z\)-operator is the parity operator \(Z_0=(-1)^{\hat{n}}\), while the second qubit's \(Z\)-operator is the two-photon parity \(Z_1=(-1)^{\frac{1}{2}\hat{n}(\hat{n}-1)}\) \NoCaseChange{\protect\cite{cite4724,cite4725}}. 
These satisfy \(Z_{1}aZ_{1}=aZ_{0}\).

Pauli operators for the constituent qudits can be expressed in terms of bosonic raising and lowering operators.
The modular-qudit Pauli-\(Z\) operator for qudit \(\mu\) is the Fock-space rotation
\flmMathEnvironment{align}{}{
  Z_{\mu}=\exp\left[i\frac{2\pi}{p}{\hat{n} \choose p^{\mu}}\right]~,
}
where \(\hat n\) is the mode's occupation number operator.
This can be proven by Lucas's theorem.

The Pauli-\(X\) operator is expressed as
\flmMathEnvironment{align}{}{
  X_{\mu}=\frac{1-P_{\mu}^{(p-1)}}{\sqrt{\left(\hat{n}+p^{\mu}\right)_{p^{\mu}}}}a^{p^{\mu}}+\frac{P_{\mu}^{(p-1)}}{\sqrt{\left(\hat{n}\right)_{p^{\mu}(p-1)}}}a^{\dagger p^{\mu}\left(p-1\right)}~,
}
where \(\left(a\right)_{b}\) is the falling factorial, and where the qudit projector is
\flmMathEnvironment{align}{}{
  P_{\mu}^{(k)}=\frac{1}{p}\sum_{l\in\mathbb{Z}_{p}}Z_{\mu}^{l}e^{-i\frac{2\pi}{p}kl}~.
}

\codefieldsection{Parents}
\begin{eczvaluelist}
\item\relax
\flmRefsHyperref[eczindexfamilyrel]{code:fock_state}{Fock-state bosonic code} --- The bosonic \(q\)-ary expansion allows one to map between prime-dimensional qudit states and a Fock subspace of a single mode.
\item\relax
\flmRefsHyperref[eczindexfamilyrel]{code:single-mode}{Single-mode bosonic code}\end{eczvaluelist}
\codefieldsection{Cousin}
\begin{eczvaluelist}
\item\relax
\flmRefsHyperref[eczindexfamilyrel]{code:qudits_into_qudits}{Modular-qudit code} --- The bosonic \(q\)-ary expansion allows one to map between prime-dimensional qudit states and a Fock subspace of a single mode.
\end{eczvaluelist}
\eczhbkcontributors{ \eczhuVVA }
\endeczcode

\eczcode{oscillators}{Bosonic code}{}
\codefieldsection{Alternative Names}
\begin{eczvaluelist}
\item\relax Continuous-variable (CV) quantum code
\item\relax Oscillator code
\item\relax Quantum modulation scheme
\end{eczvaluelist}
\eczhIndexCodeAliasName{oscillators}{Continuous-variable (CV) quantum code}
\eczhIndexCodeAliasName{oscillators}{Oscillator code}
\eczhIndexCodeAliasName{oscillators}{Quantum modulation scheme}

\codefieldsection{Kingdom root code}
for the \flmRefsHyperref{kingdom:oscillators}{Bosonic Kingdom}.
\codefieldsection{Description}
Encodes logical Hilbert space, finite- or infinite-dimensional, into a physical Hilbert space that contains at least one \textit{oscillator} (a.k.a. \textit{bosonic mode} or \textit{qumode}).

States of a single oscillator correspond to \(L^2\)-normalizable functions on \(\mathbb{R}\) that have finite energy, finite variance, and finite values of all other moments (where the energy operator is defined to be the harmonic oscillator Hamiltonian); such functions form \textit{Schwartz space} (a.k.a. nuclear space \NoCaseChange{\protect\cite{cite4726}}), a subspace of Hilbert space \NoCaseChange{\protect\cite{cite4727}}.
Ideal codewords may not be normalizable because the space is infinite-dimensional, so approximate versions have to be constructed in practice.

States can be represented by a series via a basis expansion, such as that in the countable basis of Fock states \(|n\rangle\) with \(n\geq 0\) for a single oscillator.
Alternatively, states can be represented as functions over the reals by expanding in a continuous "basis" (more technically, set of tempered distributions in the space dual to Schwartz space), such as the position "basis" \(|y\rangle\) with \(y\in\mathbb{R}\) or the momentum "basis" \(|p\rangle\) with \(p\in\mathbb{R}\).
A third option is to use coherent states \(|\alpha\rangle\) with \(\alpha\in\mathbb{C}\), which are eigenstates of the annihilation operator, which correspond to classical electromagnetic signals, and which resolve the identity \NoCaseChange{\protect\cite{cite4728,cite4729,cite4730,cite3661}}.
States can further be represented as functions over the joint position-momentum phase space in the Wigner function formalism \NoCaseChange{\protect\cite{cite4731,cite4732}}.
GKP states have negative Wigner functions, but the alternative Zak-Gross Wigner function represents them positively \NoCaseChange{\protect\cite{cite4733}}.

An important subset of states is formed by the \textit{Gaussian states}, which are in one-to-one correspondence with a (displacement) vector and covariance matrix \NoCaseChange{\protect\cite{cite4734,cite4735,cite4736,cite4737,cite4738,cite4739,cite3657,cite4740,cite3658}}.
Pure Gaussian states correspond to the pure states with positive Wigner functions, a result known as Hudson's theorem \NoCaseChange{\protect\cite{cite4741,cite4742}}.
Pure Gaussian states can be obtained from the \textit{vacuum Fock state} \(|n=0\rangle\) via a Gaussian unitary transformation (defined below). 
Any coherent state can be obtained from the vacuum Fock state, itself a coherent state, by a displacement.
There is a de Finetti theorem for Gaussian states \NoCaseChange{\protect\cite{cite4743}}.

\codefieldsection{Protection}
\subsection{Displacement error basis}

An error set relevant to \flmRefsHyperref{code:oscillator_stabilizer}{bosonic stabilizer} codes is the set of \textit{displacement operators} (a.k.a. Weyl operators \NoCaseChange{\protect\cite{cite4532}}), a bosonic analogue of the Pauli string basis for \flmRefsHyperref{code:qubits_into_qubits}{qubit} codes.

\begin{defterm}{Displacement operators}\label{ref4744}\label{ref4745}
For a single mode, its elements are products of exponentials of the mode's position and momentum operators, acting on the mode's position states \(|y\rangle\) for \(y\in\mathbb{R}\) as
\flmMathEnvironment{align}{}{
  e^{-iq\hat{p}}\left|y\right\rangle =\left|y+q\right\rangle \,\,\text{ and }\,\,e^{iq\hat{x}}\left|y\right\rangle =e^{iq y}\left|y\right\rangle ~,
}
where \(q\in\mathbb{R}\).
The former is also called a translation, while the latter is called a modulation in signal processing.
For multiple modes, error set elements are tensor products of elements of the single-oscillator error set, characterized by the vector of coefficients \(\xi\in\mathbb{R}^{2n}\).
\end{defterm}

The displacement error set is a unitary basis for bounded operators on the \(n\)-mode Hilbert space that is Dirac-orthonormal under the Hilbert-Schmidt inner product.
Expanding a bounded operator in terms of displacements is called the \textit{Fourier-Weyl transform} (a.k.a.  Fourier-Weyl relation) \NoCaseChange{\protect\cite{cite4747}\protect\cite[{Eq. (4.11)}]{cite4746}}.
For the expansion of Gaussian unitary operations in terms of displacements, see \NoCaseChange{\protect\cite[{Eq. (7.62)}]{cite4748}}.

There are two definitions of code distance associated with displacements.
The definition inherited from qubit codes is the minimum weight of a displacement operator (i.e., number of nonzero entries in \(\xi\)) that implements a nontrivial logical operation in the code. The second definition is the minimum Euclidean distance (i.e., \(\ell^2\)-norm of \(\xi\)) such that the corresponding displacement implements a nontrivial logical operation in the code.
\flmRefsHyperref{ref672}{Quantum weight enumerators} and the Cohn-Elkies bound have been extended to the case of displacement noise \NoCaseChange{\protect\cite{cite4749}}.

\subsection{Loss and gain operators}

An error set relevant to \flmRefsHyperref{code:fock_state}{Fock-state bosonic} codes is the set of loss operators associated with the \flmRefsHyperref{ref498}{AD} channel, a common form of physical noise in bosonic systems. 
For a single mode, loss operators are proportional to powers of the mode's annihilation operator \(a=(\hat{x}+i\hat{p})/\sqrt{2}\), where \(\hat x\) (\(\hat p\)) is the mode's position (momentum) operator, and with the power signifying the number of particles lost during the error. 
For multiple modes, error set elements are tensor products of elements of the single-mode error set. 
Quantum Hamming bounds have been extended to the case of loss noise \NoCaseChange{\protect\cite{cite4666}} 

\subsection{Number-phase operators}

A related error set is the set of powers of the \textit{Susskind–Glogower phase operator} \(\frac{1}{\sqrt{a a^\dagger}} a\) and its adjoint \NoCaseChange{\protect\cite{cite4750,cite4751,cite4752}} along with Fock-space rotations generated by the occupation number operator \(a^\dagger a\).
These can also be obtained from qudit Pauli matrices through a limiting procedure \NoCaseChange{\protect\cite{cite4752}} and allow one to expand trace-class operators despite not forming an orthonormal set \NoCaseChange{\protect\cite{cite4727}}. These operators correspond to the \textit{number-phase interpretation}, a polar-like decomposition of a single mode, complementing the cartesian-like decomposition in terms of position and momentum displacements.
This decomposition can be called a number-phase rotor, which differs from the ordinary \(U(1)\) rotor in the absence of states of negative angular momentum.  
Mathematically, the restriction of an ordinary rotor to states of non-negative momentum is a projection onto Hardy space, and the phase operator is an example of a Toeplitz operator on that space \NoCaseChange{\protect\cite{cite4753}}.

\subsection{Noise channels}

\textit{Gaussian channels} are quantum channels that map Gaussian states to Gaussian states \NoCaseChange{\protect\cite{cite4754,cite4755,cite4756,cite4757,cite4758,cite4759,cite4760}}; their Kraus representation is calculated in Ref. \NoCaseChange{\protect\cite{cite4761}}.
These include the \flmRefsHyperref{ref498}{AD} channel and the displacement noise channel.
The algebraic structure of the Lie algebra of Gaussian channels is the same as that of the super-Poincare algebra in three-dimensional spacetime \NoCaseChange{\protect\cite{cite4762}}.

An important non-Gaussian noise channel is the dephasing noise channel, which applies a random rotation in phase space about the origin.

\codefieldsection{Rate}
The quantum capacity of the \flmRefsHyperref{ref498}{AD} channel \NoCaseChange{\protect\cite{cite4675}} and the dephasing noise channel \NoCaseChange{\protect\cite{cite4763}} are both known.
The capacity of the displacement noise channel, the quantum analogue of AWGN, has been bounded using GKP codes \NoCaseChange{\protect\cite{cite4764,cite2608}}.
Exact two-way assisted capacities have been obtained for the \flmRefsHyperref{ref498}{AD} channels and quantum limited amplifiers in what is known as the PLOB bound \NoCaseChange{\protect\cite{cite4149}}.
Bounds exist on the two-way quantum and secret-key capacities for some prominent Gaussian channels \NoCaseChange{\protect\cite{cite4765,cite4766,cite4767,cite4768,cite4769,cite4770,cite4771,cite4772,cite4773,cite4774}} and non-Markovian channels, i.e., channels with memory effects \NoCaseChange{\protect\cite{cite4775}}. 
The continuous-variable erasure channel has a known quantum capacity \NoCaseChange{\protect\cite{cite4776,cite4777}}.
Non-Gaussian channel capacities can be bounded for single \NoCaseChange{\protect\cite{cite4778}} and multiple \NoCaseChange{\protect\cite[{Lemma 14}]{cite4668}} modes.
Non-asymptotic bounds exist for memoryless channels \NoCaseChange{\protect\cite{cite4779}} as well as for those with memory effects \NoCaseChange{\protect\cite{cite4780}}.
The optimal asymptotic error exponent of entanglement distillation is given by the reverse relative entropy of entanglement, a single-letter quantity \NoCaseChange{\protect\cite{cite4781}}.

\codefieldsection{Gates}
\begin{eczvaluelist}
\item\relax Displacement operations form a group called the Heisenberg-Weyl group, the oscillator analogue to the \flmRefsHyperref{ref663}{Pauli group}. Analogues of (non-Pauli) Clifford-group transformations are the \textit{Gaussian unitary transformations} (a.k.a. symplectic, Bogoliubov-Valatin, or linear canonical transformations) \NoCaseChange{\protect\cite{cite3660,cite4782,cite4739,cite4783}}, which are unitaries generated by quadratic polynomials in positions and momenta. The Gaussian unitary transformation group permutes displacement operators amongst themselves, and, up to any phases, is equivalent to the symplectic group \(Sp(2n,\mathbb{R})\). Every Gaussian unitary can be decomposed into single-mode squeezing gates sandwiched by passive linear-optical transformations in what is known as the Bloch-Messiah (a.k.a. Euler) decomposition \NoCaseChange{\protect\cite{cite4784,cite4746}}.
\item\relax Computing using Gaussian states and Gaussian unitaries only can be efficiently simulated on a classical computer \NoCaseChange{\protect\cite{cite4785,cite4786,cite4787}}, and there are efficient algorithms to do so \NoCaseChange{\protect\cite{cite4736,cite4788}}. This remains true even if superpositions of Gaussian states are considered \NoCaseChange{\protect\cite{cite4789,cite4790}}, but is no longer the case when the number of modes scales exponentially \NoCaseChange{\protect\cite{cite4791}}.
\item\relax A gate generated by a cubic or higher-degree polynomial is required to make a universal gate set on the oscillator (an infinite-dimensional version of the Solovay-Kitaev theorem) \NoCaseChange{\protect\cite{cite4792,cite4735,cite4793}}. 
The cubic phase gate \NoCaseChange{\protect\cite{cite513}} is a common gate used in tandem with Gaussian gates for universality, and Fock-state matrix elements of the cubic and quartic phase gates have been derived \NoCaseChange{\protect\cite{cite4794}} (see also Refs. \NoCaseChange{\protect\cite{cite4795,cite4796}}). 
Arbitrary-degree polynomial gates are well defined, but cubic or higher versions of squeezing admit subtle mathematical properties \NoCaseChange{\protect\cite{cite4797,cite4798,cite4799,cite4800}}. 
Unitaries generated by polynomials of position and momentum can exactly realize any finite-dimensional unitary evolution, and any physical bosonic unitary evolution can be approximated by a finite-dimensional unitary evolution \NoCaseChange{\protect\cite{cite4801,cite4793,cite4802}}. 
Certain non-quadratic Hamiltonians yield infinite-energy states in finite time \NoCaseChange{\protect\cite{cite4803}}.
See Ref. \NoCaseChange{\protect\cite{cite4804}} for bosonic computational complexity classes. 
The stellar rank \NoCaseChange{\protect\cite{cite4805}} and symplectic rank \NoCaseChange{\protect\cite{cite4806}} quantify the degree of non-Gaussianity of bosonic states. 
Functional MPS can be used to simulate evolution of non-Gaussian states \NoCaseChange{\protect\cite{cite4807}}.
Universal bosonic quantum computations can be simulated in exponential time on a classical computer \NoCaseChange{\protect\cite{cite4808}}.

\item\relax There is a bosonic analogue of the \flmTerm{term}{ref694}{}{Clifford hierarchy} \NoCaseChange{\protect\cite{cite4809,cite4810}}.
\item\relax Controllability of bosonic states has been proven when the normalizable state space is restricted to Schwartz space \NoCaseChange{\protect\cite{cite4811}} and using polynomials in position and momentum \NoCaseChange{\protect\cite{cite4793}}.
\item\relax Measurements can be performed by homodyne, heterodyne, and generalized homodyne measurements \NoCaseChange{\protect\cite{cite4812}}.
\item\relax The number-phase interpretation allows for the mapping of rotor Clifford gates into the oscillator, some of which become non-unitary (e.g., conditional occupation number addition) \NoCaseChange{\protect\cite{cite2699}}.
\item\relax ZX calculus has been extended to bosonic codes for both Gaussian operators \NoCaseChange{\protect\cite{cite4813}}, Fock-state based operators \NoCaseChange{\protect\cite{cite4814}}, and passive linear-optical transformations \NoCaseChange{\protect\cite{cite4815}}. An earlier graphical calculus exists for Gaussian pure states \NoCaseChange{\protect\cite{cite4693}}.
\item\relax Circuits can be decomposed into a series of primitives such as quantum lattice gates, which are exponentials of cosines and sines of position and momentum \NoCaseChange{\protect\cite{cite4816}}.
\item\relax Number-phase teleportation can be done using a two-mode squeezed state \NoCaseChange{\protect\cite{cite4817}}.
\end{eczvaluelist}
\codefieldsection{Notes}
\begin{eczvaluelist}
\item\relax For an introduction to continuous-variable quantum systems, see reviews \NoCaseChange{\protect\cite{cite4818,cite4819,cite4820,cite4821,cite4822,cite497,cite4727}} and books \NoCaseChange{\protect\cite{cite4823,cite4746,cite4824}}.
\item\relax See \NoCaseChange{\protect\cite{cite2733}} for a pedagogical introduction to bosonic codes.
\item\relax See video tutorial by \flmHref{https://www.youtube.com/watch?v=zQQI3Ov6xyw}{V. V. Albert}.
\end{eczvaluelist}
\codefieldsection{Parents}
\begin{eczvaluelist}
\item\relax
\flmRefsHyperref[eczindexfamilyrel]{code:block_quantum}{Block quantum code} --- Bosonic codes are block quantum codes with \(\Sigma=\mathbb{R}\).
\item\relax
\flmRefsHyperref[eczindexfamilyrel]{code:hybrid_qudit_oscillator}{Mixed oscillator code} --- Mixed oscillator codes defined only on oscillators reduce to oscillator codes.
\end{eczvaluelist}
\codefieldsection{Children}
\begin{eczvaluelist}
\item\relax
\flmRefsHyperref[eczindexfamilyrel]{code:ampdamp}{Amplitude-damping (AD) code} --- Restricting the AD channel to the first two Fock states \(\{|0\rangle,|1\rangle\}\) yields the non-Pauli qubit AD channel, which requires protecting against the loss error \(E_1\propto X+iY\) (instead of \(X\) and \(Y\) Pauli errors individually). Qubit AD codes are thus a special case of bosonic AD codes.
\item\relax
\flmRefsHyperref[eczindexfamilyrel]{code:coherent_constellation}{Coherent-state constellation code}\item\relax
\flmRefsHyperref[eczindexfamilyrel]{code:numopt}{Numerically optimized bosonic code}\item\relax
\flmRefsHyperref[eczindexfamilyrel]{code:oscillators_concatenated}{Concatenated bosonic code}\item\relax
\flmRefsHyperref[eczindexfamilyrel]{code:oscillators_into_oscillators}{Oscillator-into-oscillator code} --- Oscillator-into-oscillator codes are bosonic codes with an infinite-dimensional logical subspace.
\item\relax
\flmRefsHyperref[eczindexfamilyrel]{code:qudits_into_oscillators}{Qudit-into-oscillator code} --- Qudit-into-oscillator codes are bosonic codes with a finite-dimensional logical subspace.
\item\relax
\flmRefsHyperref[eczindexfamilyrel]{code:oscillator_stabilizer}{Bosonic stabilizer code}\item\relax
\flmRefsHyperref[eczindexfamilyrel]{code:tiger}{Tiger code} --- Tiger codewords are superpositions of coherent states with the same energy, but coherent states are not eigenstates of the energy Hamiltonian. The \flmRefsHyperref{ref498}{AD} Kraus operator \(E_{0}^{\otimes n}\) acts identically on each coherent state by shrinking the radius of the QSC's sphere.
\item\relax
\flmRefsHyperref[eczindexfamilyrel]{code:homological_number-phase}{Homological number-phase code} --- Homological number-phase codes are bosonic codes encoding logical qudits and/or logical rotors.
\item\relax
\flmRefsHyperref[eczindexfamilyrel]{code:penrose}{Penrose tiling code} --- Penrose tiling codes encode information into Penrose tilings, which are non-periodic tilings of \(\mathbb{R}^n\).
\item\relax
\flmRefsHyperref[eczindexfamilyrel]{code:qutrit_pauli_gkp_subcode}{Qutrit-Pauli tessellation code}\end{eczvaluelist}
\codefieldsection{Cousins}
\begin{eczvaluelist}
\item\relax
\flmRefsHyperref[eczindexfamilyrel]{code:analog}{Analog code} --- Bosonic codes are quantum counterparts of analog codes.
\item\relax
\flmRefsHyperref[eczindexfamilyrel]{code:t-designs}{\(t\)-design} --- Gaussian states, under a particular measure, do not form rigged two-designs \NoCaseChange{\protect\cite{cite932}}.
\item\relax
\flmRefsHyperref[eczindexfamilyrel]{code:single_spin}{Single-spin code} --- Bosonic states are typically represented with the assumption that a common phase reference exists, and the superselection rule compliant (SSRC) framework yields expressions without this assumption \NoCaseChange{\protect\cite{cite4825,cite4826,cite4827,cite4828,cite4829,cite4830,cite4831}}. In this framework, single-mode states can be treated as two-mode states in a fixed subspace of total occupation number \(N\) in the limit \(N \to \infty\). Passive Gaussian operations acting on the fixed-photon subspace of two modes realize \(U(2)\) transformations in the Jordan-Schwinger boson mapping \NoCaseChange{\protect\cite{cite4832,cite4833,cite4834,cite653}}.
\item\relax
\flmRefsHyperref[eczindexfamilyrel]{code:bosonic_classical_into_quantum}{Bosonic c-q code} --- Bosonic c-q codes are bosonic codes designed to transmit classical information.
\item\relax
\flmRefsHyperref[eczindexfamilyrel]{code:ea_oscillators}{EA bosonic code} --- EA bosonic codes utilize additional ancillary modes in a pre-shared entangled state, but reduce to ordinary bosonic codes when said modes are interpreted as noiseless physical modes.
\item\relax
\flmRefsHyperref[eczindexfamilyrel]{code:fermions}{Fermion code} --- Bosonic (fermionic) codes are associated with bosonic (fermionic) degrees of freedom.
\end{eczvaluelist}
\eczhbkcontributors{ \eczhuVVA }
\endeczcode

\eczcode{oscillator_css}{Bosonic CSS code}{}
\codefieldsection{Alternative Names}
\begin{eczvaluelist}
\item\relax CV CSS code
\item\relax Oscillator CSS code
\end{eczvaluelist}
\eczhIndexCodeAliasName{oscillator_css}{CV CSS code}
\eczhIndexCodeAliasName{oscillator_css}{Oscillator CSS code}
\codefieldsection{Description}
Bosonic stabilizer code admitting a set of stabilizer generators that are either position or momentum displacements.

\codefieldsection{Parents}
\begin{eczvaluelist}
\item\relax
\flmRefsHyperref[eczindexfamilyrel]{code:oscillator_stabilizer}{Bosonic stabilizer code}\item\relax
\flmRefsHyperref[eczindexfamilyrel]{code:css}{Calderbank-Shor-Steane (CSS) stabilizer code}\end{eczvaluelist}
\codefieldsection{Children}
\begin{eczvaluelist}
\item\relax
\flmRefsHyperref[eczindexfamilyrel]{code:homological_cv}{Integer-homology bosonic CSS code} --- Integer-homology bosonic CSS codes are constructed from chain complexes over the integers and realize homological rotor codes out of continuous displacement stabilizer groups. The stabilizer group is continuous, but contains discrete components in the form of the single-mode GKP stabilizers.
\item\relax
\flmRefsHyperref[eczindexfamilyrel]{code:analog_repetition}{Analog repetition code}\item\relax
\flmRefsHyperref[eczindexfamilyrel]{code:analog_surface}{Analog surface code}\item\relax
\flmRefsHyperref[eczindexfamilyrel]{code:hnss}{Hayden-Nezami-Salton-Sanders bosonic code}\item\relax
\flmRefsHyperref[eczindexfamilyrel]{code:lloyd_slotine}{\(\llbracket 9,1,3\rrbracket _{\mathbb{R}}\) Lloyd-Slotine code}\item\relax
\flmRefsHyperref[eczindexfamilyrel]{code:gkp}{Square-lattice GKP code}\end{eczvaluelist}
\eczhbkcontributors{ \eczhuVVA }
\endeczcode

\eczcode{fourier_bosonic}{Bosonic quantum Fourier code}{~\NoCaseChange{\protect\cite{cite4835}}}
\codefieldsection{Description}
Two-mode non-uniform QSC encoding two logical qubits whose projection is onto a copy of an irreducible representation of the \flmRefsHyperref{ref663}{single-qubit Pauli group}.
This code is an extension of the single-logical-qubit code in \NoCaseChange{\protect\cite[{Eq. (10)}]{cite2810}}, storing an extra logical qubit in the multiplicity space of the Pauli group.

The code admits the following basis of codewords for complex \(\alpha > 0\), up to normalization:
\flmMathEnvironment{align}{}{
\begin{split}
|\overline{0,0}\rangle&\propto\left(\left|\alpha\right\rangle -\left|-\alpha\right\rangle \right)\left(\left|i\alpha\right\rangle +\left|-i\alpha\right\rangle \right)\\
|\overline{0,1}\rangle&\propto\left(\left|i\alpha\right\rangle -\left|-i\alpha\right\rangle \right)\left(\left|\alpha\right\rangle +\left|-\alpha\right\rangle \right)\\
|\overline{1,0}\rangle&\propto\left(\left|i\alpha\right\rangle +\left|-i\alpha\right\rangle \right)\left(\left|\alpha\right\rangle -\left|-\alpha\right\rangle \right)\\
|\overline{1,1}\rangle&\propto\left(\left|\alpha\right\rangle +\left|-\alpha\right\rangle \right)\left(\left|i\alpha\right\rangle -\left|-i\alpha\right\rangle \right)
\end{split}
}

\codefieldsection{Gates}
\begin{eczvaluelist}
\item\relax The \flmRefsHyperref{ref663}{single-qubit Pauli group} can be realized via Gaussian rotations \NoCaseChange{\protect\cite{cite4835}}.
\item\relax Kerr interactions yield some Clifford gates. There is a Hadamard gate, up to a global rotation \NoCaseChange{\protect\cite{cite4835}}.
\item\relax A logical \(ZZ\)-gate can be performed using squeezing operators and quantum Zeno effect \NoCaseChange{\protect\cite{cite4835}}.
\end{eczvaluelist}
\codefieldsection{Decoding}
\begin{eczvaluelist}
\item\relax The code is stabilized by the two-mode parity operator and annihilated by the operators \(\hat{a}_1^4 - \alpha^4\) and \(\hat{a}_1^2 \hat{a}_2^2 + \alpha^4\) \NoCaseChange{\protect\cite{cite4835}}.
\end{eczvaluelist}
\codefieldsection{Parents}
\begin{eczvaluelist}
\item\relax
\flmRefsHyperref[eczindexfamilyrel]{code:qsc}{Quantum spherical code (QSC)} --- The bosonic quantum Fourier code has non-uniform \(\pm 1\) coefficients.
\item\relax
\flmRefsHyperref[eczindexfamilyrel]{code:group_representation}{Group-representation code} --- The bosonic quantum Fourier code is a group-representation code with \(G\) being the single-qubit Pauli group.
\end{eczvaluelist}
\codefieldsection{Cousin}
\begin{eczvaluelist}
\item\relax
\flmRefsHyperref[eczindexfamilyrel]{code:pauli_qsc}{Pauli tessellation QSC} --- The bosonic quantum Fourier code and the Pauli group-representation QSC are both group-representation codes with \(G\) being the single-qubit Pauli group.
\end{eczvaluelist}
\eczhbkcontributors{ \eczhuVVA }
\endeczcode

\eczcode{bosonic_rotation}{Bosonic rotation code}{~\NoCaseChange{\protect\cite{cite4722}}}
\codefieldsection{Alternative Names}
\begin{eczvaluelist}
\item\relax Rotationally symmetric bosonic (RSB) code
\end{eczvaluelist}
\eczhIndexCodeAliasName{bosonic_rotation}{Rotationally symmetric bosonic (RSB) code}
\codefieldsection{Description}
A single-mode Fock-state bosonic code whose codespace is preserved by a phase-space rotation by a multiple of \(2\pi/N\) for some \(N\). The rotation symmetry ensures that encoded states have support only on every \(N^{\textrm{th}}\) Fock state. For example, single-mode Fock-state codes for \(N=2\) encoding a qubit admit basis states that are, respectively, supported on Fock state sets \(\{|0\rangle,|4\rangle,|8\rangle,\cdots\}\) and \(\{|2\rangle,|6\rangle,|10\rangle,\cdots\}\).

Codewords can be uniquely specified by choosing a \emph{primitive} state \(|\Theta\rangle\). To ensure valid (orthogonal and nonzero) codewords, \(|\Theta\rangle\) must satisfy the following requirement: for each \(j \in \mathbb{Z}_q\), \(|\Theta\rangle\) must have support on at least one Fock state \(|(k_j q+j)N\rangle\) for some \(k_j \in \mathbb{N}_0\). A set of logical codewords is then obtained by projecting \(|\Theta\rangle\) onto the \(q\) eigenspaces of the discrete rotation operator, equivalently by summing the \(qN\) rotated copies of \(|\Theta\rangle\) with appropriate phases.

\codefieldsection{Protection}
Losses or gains less than \(N\) are detectable. Dephasing rotations \(\exp(\mathrm{i}\theta \hat{n})\) can be detected whenever \(\theta\) is roughly less than \(\pi/N\). To get precise bounds on \(\theta\), one needs to analyze the particular bosonic rotation code.
\codefieldsection{Encoding}
\begin{eczvaluelist}
\item\relax The optimal way to prepare codewords depends on the exact rotation code in question \NoCaseChange{\protect\cite{cite4722}}.
\end{eczvaluelist}
\codefieldsection{Gates}
\begin{eczvaluelist}
\item\relax The logical Pauli-\(Z\) gate can be the discrete rotation operator \(\mathrm{e}^{\mathrm{i} \pi \hat n /N}\), and the logical Pauli-\(X\) gate can be the Susskind–Glogower phase operator \(\sum_{n=0}^\infty |n\rangle\bra{n+N}\).
\item\relax For qubit codes, a logical phase gate is \(S = \mathrm{e}^{\pi \mathrm{i} \hat n^2 / 2N^2}\).
\item\relax The \(T = \mathrm{diag}(1,\exp(\mathrm{i}\pi/4\rrparenthesis \) gate can be done via gate teleportation and a resource state \(\vert 0_N\rangle + \exp(\mathrm{i}\pi/4) \vert 1_N \rangle\).
\item\relax A controlled-rotation gate between an order \(N\) rotation code and an order \(M\) rotation code is \(\mathrm{CROT}_{NM} = \mathrm{e}^{(2\pi\mathrm{i} / qNM) \hat n \otimes \hat n}\).
\end{eczvaluelist}
\codefieldsection{Decoding}
\begin{eczvaluelist}
\item\relax For qubit rotation codes, one can distinguish the computational-basis codewords destructively by performing a Fock-state number measurement. If a Fock state \(|n\rangle\) is measured, then one rounds to the nearest multiple of \(N\) and infers the logical value from the parity of that multiple \NoCaseChange{\protect\cite{cite4722}}.
\item\relax One can distinguish states in the dual basis by performing phase estimation on \(\mathrm{e}^{\mathrm{i} \theta \hat n}\). One then rounds the resulting \(\theta\) to the nearest number \(2\pi j / qN\) in order to determine which dual basis state \(j \in \mathbb Z_q\) it came from \NoCaseChange{\protect\cite{cite4722}}.
\item\relax Autonomous QEC for \(S=1\) codes \NoCaseChange{\protect\cite{cite4836}}.
\item\relax Decoder \NoCaseChange{\protect\cite{cite4722}} based on measuring in the phase-state basis and using Knill error correction (a.k.a. telecorrection \NoCaseChange{\protect\cite{cite3185}}), which is based on teleportation \NoCaseChange{\protect\cite{cite448,cite4369}}.
\item\relax Performance under non-Markovian noise has been investigated \NoCaseChange{\protect\cite{cite4837}}.
\end{eczvaluelist}
\codefieldsection{Fault Tolerance}
\begin{eczvaluelist}
\item\relax Decoder based on measuring in the phase-state basis and using Knill error correction \NoCaseChange{\protect\cite{cite4722}} is fault-tolerant under circuit-level noise \NoCaseChange{\protect\cite{cite4838}}.
\end{eczvaluelist}
\codefieldsection{Parents}
\begin{eczvaluelist}
\item\relax
\flmRefsHyperref[eczindexfamilyrel]{code:fock_state}{Fock-state bosonic code} --- Single-mode Fock-state codes are typically rotationally invariant.
\item\relax
\flmRefsHyperref[eczindexfamilyrel]{code:single-mode}{Single-mode bosonic code}\end{eczvaluelist}
\codefieldsection{Children}
\begin{eczvaluelist}
\item\relax
\flmRefsHyperref[eczindexfamilyrel]{code:binomial}{Binomial code} --- One can verify by direct calculation that the logical states are eigenstates of the discrete rotation operator. One has freedom in the exact form of the primitive state to choose; see \NoCaseChange{\protect\cite[{App. B.2}]{cite4722}}.
\item\relax
\flmRefsHyperref[eczindexfamilyrel]{code:number_phase}{Number-phase code} --- Number-phase codes are bosonic rotation codes whose primitive state is a Pegg-Barnett phase state \NoCaseChange{\protect\cite{cite501}}.
\item\relax
\flmRefsHyperref[eczindexfamilyrel]{code:squeezed_vacuum}{Squeezed-vacuum code} --- Squeezed-vacuum codes are qubit \flmRefsHyperref{code:bosonic_rotation}{rotation-symmetric bosonic codes} with \(m\)-fold rotational symmetry in phase space, constructed from \(m\) primitive squeezed vacuum states arranged at evenly-spaced angles.
\item\relax
\flmRefsHyperref[eczindexfamilyrel]{code:cat}{Cat code} --- The cat code is a bosonic rotation code whose primitive state is the coherent state \(|\alpha\rangle\) \NoCaseChange{\protect\cite{cite4722}}.
\end{eczvaluelist}
\codefieldsection{Cousin}
\begin{eczvaluelist}
\item\relax
\flmRefsHyperref[eczindexfamilyrel]{code:quantum_random}{Random quantum code} --- Random bosonic rotation codes can outperform cat and binomial codes when loss rate is large relative to dephasing rate \NoCaseChange{\protect\cite{cite2998}}.
\end{eczvaluelist}
\eczhbkcontributors{ Joseph T. Iosue, \eczhuVVA }
\endeczcode

\eczcode{oscillator_stabilizer}{Bosonic stabilizer code}{}
\codefieldsection{Alternative Names}
\begin{eczvaluelist}
\item\relax CV stabilizer code
\item\relax Oscillator stabilizer code
\end{eczvaluelist}
\eczhIndexCodeAliasName{oscillator_stabilizer}{CV stabilizer code}
\eczhIndexCodeAliasName{oscillator_stabilizer}{Oscillator stabilizer code}
\codefieldsection{Description}
Bosonic code whose codespace is defined as the common \(+1\) eigenspace of a group of mutually commuting displacement operators.
Displacements form the stabilizers of the code, and have continuous eigenvalues, in contrast with the discrete set of eigenvalues of qubit stabilizers.
As a result, exact codewords are non-normalizable, so approximate constructions have to be considered.
Stabilizer groups are any locally compact Abelian subgroups of \(\mathbb{R}^n\), can themselves contain discrete or continuous subgroups, and can admit logical qudit and/or oscillator logical subspaces.

Stabilizer codewords encoding a finite-dimensional codespace admit a discrete infinite stabilizer group and encode quantum information in a lattice.
Such \flmRefsHyperref{code:qudits_into_oscillators}{qudit-into-oscillator} stabilizer codes are \flmRefsHyperref{code:gkp}{GKP} and \flmRefsHyperref{code:multimodegkp}{multimode GKP} codes.

Stabilizer codewords encoding a logical oscillator (i.e., CV quantum information) admit either a discrete or a continuous stabilizer group.
The former, called oscillator-into-oscillator GKP codes, are obtained from multimode GKP codes by removing stabilizer generators for some of the modes.
The latter encode information in hyperplanes and can be defined in terms of the continuous group's Lie algebra, i.e., as the common \(0\)-eigenvalue eigenspace of mutually commuting linear combinations of oscillator position and momentum operators called \textit{nullifiers} \NoCaseChange{\protect\cite{cite4683}} or \textit{annihilators}. An \flmRefsHyperref{code:oscillators_into_oscillators}{oscillator-into-oscillator} stabilizer code encoding \(k\) logical modes into \(n\) physical modes is denoted as \(\llbracket n,k,d\rrbracket _{\mathbb{R}}\), where \(d\) is the code's distance.

\codefieldsection{Protection}
Protective properties can be delineated in terms of the nullifiers or displacements, and the most natural noise model for such codes is displacement noise. If an error operator does not commute with a stabilizer group element, then that error is detectable. Oscillator-into-oscillator stabilizer codes protect against erasures of a subset of modes, while GKP codes protect against sufficiently small displacements in any number of modes.
\codefieldsection{Gates}
\begin{eczvaluelist}
\item\relax General gates can be done using the bosonic analogue of gate teleportation \NoCaseChange{\protect\cite{cite4809}}.
\end{eczvaluelist}
\codefieldsection{Parents}
\begin{eczvaluelist}
\item\relax
\flmRefsHyperref[eczindexfamilyrel]{code:oscillators}{Bosonic code}\item\relax
\flmRefsHyperref[eczindexfamilyrel]{code:stabilizer}{Stabilizer code}\end{eczvaluelist}
\codefieldsection{Children}
\begin{eczvaluelist}
\item\relax
\flmRefsHyperref[eczindexfamilyrel]{code:analog_stabilizer}{Analog stabilizer code} --- Analog stabilizer codes are bosonic stabilizer codes with a continuous stabilizer group, corresponding to linear constraints on positions and momenta.
\item\relax
\flmRefsHyperref[eczindexfamilyrel]{code:quantum_lattice}{Quantum lattice code} --- Quantum lattice codes are bosonic stabilizer codes with a countably infinite stabilizer group, corresponding to modular constraints on positions and momenta.
\item\relax
\flmRefsHyperref[eczindexfamilyrel]{code:oscillator_css}{Bosonic CSS code}\end{eczvaluelist}
\codefieldsection{Cousins}
\begin{eczvaluelist}
\item\relax
\flmRefsHyperref[eczindexfamilyrel]{code:number_phase}{Number-phase code} --- Number-phase codewords span the joint right eigenspace of the \(N\)th power of the Susskind-Glogower phase operator and the bosonic rotation operator \NoCaseChange{\protect\cite{cite4722}}. These operators no longer form a group since the phase operator is not unitary.
\item\relax
\flmRefsHyperref[eczindexfamilyrel]{code:homological_number-phase}{Homological number-phase code} --- Homological number-phase codewords span the joint right eigenspace of powers of the non-unitary Susskind–Glogower phase operators and unitary bosonic rotation operators.
\end{eczvaluelist}
\eczhbkcontributors{ \eczhuVVA }
\endeczcode

\eczcode{cat}{Cat code}{~\NoCaseChange{\protect\cite{cite4839}}}
\codefieldsection{Alternative Names}
\begin{eczvaluelist}
\item\relax Superposition of coherent states (SCS)
\end{eczvaluelist}
\eczhIndexCodeAliasName{cat}{Superposition of coherent states (SCS)}
\codefieldsection{Description}
Rotation-symmetric bosonic Fock-state code encoding a \(q\)-dimensional qudit into one oscillator which utilizes a constellation of \(q(S+1)\) coherent states distributed equidistantly around a circle in phase space of radius \(\alpha\).

Codewords for a qubit code (\(q=2\)) consist of a coherent state \(|\alpha\rangle\) projected onto a subspace of Fock state number modulo \(2(S+1)\). The logical state \(|\overline{0}\rangle\) is in the \(\{|0\rangle , |2(S+1)\rangle , |4(S+1)\rangle \cdots \}\) Fock-state subspace, while \(|\overline{1}\rangle\) is in the \(\{|(S+1)\rangle, |3(S+1)\rangle , |5(S+1)\rangle , |7(S+1)\rangle \cdots \}\) subspace.
These projected coherent states make up generalized cat states \NoCaseChange{\protect\cite{cite4840,cite4841}}. 

\codefieldsection{Protection}
Due to the spacing between sets of Fock states, the distance between two distinct logical states is \(d=S+1\). Hence, this code is able to detect up to \(S\) \flmRefsHyperref{ref498}{photon-loss} errors.
\codefieldsection{Encoding}
\begin{eczvaluelist}
\item\relax Lindbladian-based dissipative encoding and autonomous QEC utilizing multi-photon generalization of two-photon absorption \NoCaseChange{\protect\cite{cite4842,cite4843,cite4844,cite4845}}. Encoding passively protects against modal dephasing, suppressing dephasing noise exponentially with \(|\alpha|^2\) \NoCaseChange{\protect\cite{cite4846}}.
\item\relax Approximate cat states can be prepared using Gaussian operations and photon detectors \NoCaseChange{\protect\cite{cite4847}}.
\end{eczvaluelist}
\codefieldsection{Gates}
\begin{eczvaluelist}
\item\relax Holonomic gates utilizing the Berry phase of coherent states are universal \NoCaseChange{\protect\cite{cite4848}}.
\item\relax Universal gates for the \(S=1\) code can be performed using squeezing operators and quantum Zeno effect and a rotation based on the Kerr nonlinearity \NoCaseChange{\protect\cite{cite4846}}.
\item\relax Error-detecting \(CCZ\) and \(cSWAP\) gates for four-component cat code using three-level ancilla \NoCaseChange{\protect\cite{cite4710}}.
\item\relax Universal set of error-corrected operations tolerating a single \flmRefsHyperref{ref498}{photon loss} and an arbitrary ancilla fault \NoCaseChange{\protect\cite{cite4849}}.
\end{eczvaluelist}
\codefieldsection{Decoding}
\begin{eczvaluelist}
\item\relax Measuring the Fock-state number modulo \(S+1\) can be used to determine if \flmRefsHyperref{ref498}{photon-loss} or excitation errors occurred. For \(S=1\), this is the occupation number parity \NoCaseChange{\protect\cite{cite4722}}.
\end{eczvaluelist}
\codefieldsection{Fault Tolerance}
\begin{eczvaluelist}
\item\relax Universal set of error-corrected operations tolerating a single \flmRefsHyperref{ref498}{photon loss} and an arbitrary ancilla fault \NoCaseChange{\protect\cite{cite4849}}.
\item\relax Linear-optical noise suppression and mitigation scheme \NoCaseChange{\protect\cite{cite4850}}.
\end{eczvaluelist}
\codefieldsection{Realizations}
\begin{eczvaluelist}
\item\relax Parity-syndrome measurement tested \NoCaseChange{\protect\cite{cite4851}} and implemented for the four-component (\(S=1\)) cat code \NoCaseChange{\protect\cite{cite4852}} in a microwave cavity coupled to a superconducting circuit. The latter work \NoCaseChange{\protect\cite{cite4852}} is the first to reach break-even error-correction, where the lifetime of a logical qubit is on par with the cavity lifetime, despite protection against dephasing not being implemented. A fault-tolerant version of parity measurement has also been realized \NoCaseChange{\protect\cite{cite4853}}.
\end{eczvaluelist}
\codefieldsection{Parents}
\begin{eczvaluelist}
\item\relax
\flmRefsHyperref[eczindexfamilyrel]{code:bosonic_rotation}{Bosonic rotation code} --- The cat code is a bosonic rotation code whose primitive state is the coherent state \(|\alpha\rangle\) \NoCaseChange{\protect\cite{cite4722}}.
\item\relax
\flmRefsHyperref[eczindexfamilyrel]{code:cat_repetition}{Cat-repetition code} --- The cat-repetition code for \(n=1\) reduces to the cat code.
\end{eczvaluelist}
\codefieldsection{Child}
\begin{eczvaluelist}
\item\relax
\flmRefsHyperref[eczindexfamilyrel]{code:two-legged-cat}{Two-component cat code} --- The cat code reduces to its two-component version for \(S=0\).
\end{eczvaluelist}
\codefieldsection{Cousins}
\begin{eczvaluelist}
\item\relax
\flmRefsHyperref[eczindexfamilyrel]{code:number_phase}{Number-phase code} --- In the limit as \(N,S \to \infty\), phase measurement in the cat code has vanishing variance, just like in a number-phase code \NoCaseChange{\protect\cite{cite4722}}. Conversely, a cat code can be thought of as an appropriately regularized number-phase code.
\item\relax
\flmRefsHyperref[eczindexfamilyrel]{code:group_representation}{Group-representation code} --- Cat codes are not group representation codes with \(G\) being a cyclic group since their representation is reducible \NoCaseChange{\protect\cite{cite2810}}.
\item\relax
\flmRefsHyperref[eczindexfamilyrel]{code:psk}{Phase-shift keying (PSK) modulation format} --- PSK (cat) codes are used to transmit classical (quantum) information using (superpositions of) single-mode coherent states distributed on a circle over classical (quantum) channels.
\item\relax
\flmRefsHyperref[eczindexfamilyrel]{code:polygon}{Polygon code} --- The \(q(S+1)\)-component cat coherent-state constellation forms the vertices of a \(q(S+1)\)-gon.
\item\relax
\flmRefsHyperref[eczindexfamilyrel]{code:quantum_psk}{PSK c-q modulation format} --- PSK c-q (cat) codes are used to transmit classical (quantum) information using (superpositions of) single-mode coherent states distributed on a circle over quantum channels.
\item\relax
\flmRefsHyperref[eczindexfamilyrel]{code:hybrid_cat}{Hybrid cat code} --- Hybrid cat codewords consist of a bosonic mode in either coherent or cat states.
\item\relax
\flmRefsHyperref[eczindexfamilyrel]{code:paircat}{Pair-cat code} --- Cat (pair-cat) codewords are superpositions of coherent (pair-coherent) states. Many cat-code protocols have analogues for the two-mode pair-cat codes.
\item\relax
\flmRefsHyperref[eczindexfamilyrel]{code:binomial}{Binomial code} --- For a fixed \(S\), binomial codes with \(N \to \infty\) coincide with cat codes as \(\alpha \to \infty\) \NoCaseChange{\protect\cite{cite4707}}.
\end{eczvaluelist}
\eczhbkcontributors{ Alexander Grimm, Joseph T. Iosue, Yijia Xu, \eczhuVVA }
\endeczcode

\eczcode{cat_repetition}{Cat-repetition code}{~\NoCaseChange{\protect\cite{cite4854,cite2616,cite2646}}}
\codefieldsection{Description}
A concatenated qubit-into-\(n\)-mode code obtained by encoding each qubit of a quantum repetition code into a two-component cat code in its cat-state basis.

A basis of codewords for the two-component case is
\flmMathEnvironment{align}{}{
  |\overline{\pm}\rangle\propto\left(\left|\alpha\right\rangle \pm\left|-\alpha\right\rangle \right)^{\otimes n}
}
for any complex \(\alpha\).

\codefieldsection{Protection}
The code can detect arbitrary losses in up to \(n/2\) modes.  
The cat-repetition code on a 2D mode lattice is a candidate for a memory that may be self-correcting, but only in the limit of infinite energy per mode \NoCaseChange{\protect\cite{cite3039}}.

\codefieldsection{Gates}
\begin{eczvaluelist}
\item\relax Fault-tolerant logical \(X\), CNOT, and Toffoli gates and a logical Hadamard synthesized from state preparation and measurement in the dual basis, without magic-state preparation or distillation \NoCaseChange{\protect\cite{cite2616}}.
\item\relax Using a physical bias-preserving CX between cat qubits, the logical \(\overline{\mathrm{CX}}\) gadget can be implemented transversally between repetition-code blocks; magic-state preparation using transversal \(ZZ(\theta)\) gates was also analyzed \NoCaseChange{\protect\cite{cite2646}}.
\end{eczvaluelist}
\codefieldsection{Decoding}
\begin{eczvaluelist}
\item\relax A measurement-code decoder for the repetition layer yields a \(\overline{\mathrm{CX}}\)-gadget threshold of about \(6\times 10^{-3}\) for \(n=5\); reaching a comparable threshold with naive repeated-syndrome decoding requires \(n=11\) and \(r=5\) \NoCaseChange{\protect\cite{cite2646}}.
\end{eczvaluelist}
\codefieldsection{Fault Tolerance}
\begin{eczvaluelist}
\item\relax Fault-tolerant logical \(X\), CNOT, and Toffoli gates and a logical Hadamard synthesized from state preparation and measurement in the dual basis, without magic-state preparation or distillation \NoCaseChange{\protect\cite{cite2616}}.
\end{eczvaluelist}
\codefieldsection{Threshold}
\begin{eczvaluelist}
\item\relax For dephasing bias \(\eta=10^4\), the cat-based logical \(\overline{\mathrm{CX}}\) gadget has threshold \(7.5\times 10^{-3}\), compared with \(3.55\times 10^{-3}\) for an earlier scheme; at \(\varepsilon=2.5\times 10^{-3}\), its circuit volume is about five times smaller \NoCaseChange{\protect\cite{cite2646}}.
\end{eczvaluelist}
\codefieldsection{Realizations}
\begin{eczvaluelist}
\item\relax Superconducting circuit devices: a repetition code out of two-component cat qubits has been realized for distances 3 and 5 \NoCaseChange{\protect\cite{cite4855}}.
\end{eczvaluelist}
\codefieldsection{Parent}
\begin{eczvaluelist}
\item\relax
\flmRefsHyperref[eczindexfamilyrel]{code:cat_concatenated}{Concatenated cat code} --- The cat-repetition code is a concatenation whose outer code is the cat code in its cat-state basis.
\end{eczvaluelist}
\codefieldsection{Child}
\begin{eczvaluelist}
\item\relax
\flmRefsHyperref[eczindexfamilyrel]{code:cat}{Cat code} --- The cat-repetition code for \(n=1\) reduces to the cat code.
\end{eczvaluelist}
\codefieldsection{Cousins}
\begin{eczvaluelist}
\item\relax
\flmRefsHyperref[eczindexfamilyrel]{code:quantum_repetition}{Quantum repetition code} --- The cat-repetition code is obtained by encoding each qubit of a quantum repetition code into a two-component cat code in its cat-state basis \NoCaseChange{\protect\cite{cite2616,cite2646,cite2647,cite4072,cite4113}}.
\item\relax
\flmRefsHyperref[eczindexfamilyrel]{code:self_correct}{Self-correcting quantum code} --- The cat-repetition code on a 2D mode lattice is a candidate for a memory that may be self-correcting, but only in the limit of infinite energy per mode \NoCaseChange{\protect\cite{cite3039}}.
\item\relax
\flmRefsHyperref[eczindexfamilyrel]{code:coherent_state_repetition}{Coherent-state repetition code} --- The cat (coherent-state) repetition code is a concatenation whose outer code is the (two-component) cat code in its cat (coherent-state) basis. For the two-component case, both reduce to the two-component cat code at \(n=1\).
\end{eczvaluelist}
\eczhbkcontributors{ \eczhuVVA }
\endeczcode

\eczcode{chebyshev}{Chebyshev code}{~\NoCaseChange{\protect\cite{cite2756}}}
\codefieldsection{Description}
Single-mode bosonic Fock-state code that can be used for error-corrected sensing of a signal Hamiltonian \({\hat n}^s\), where \({\hat n}\) is the occupation number operator. 

Codewords for the \(s\)th-order Chebyshev code are
\flmMathEnvironment{align}{}{
\begin{split}
\ket{\overline 0} &=\sum_{k \text{~even}}^{[0,s]} \tilde{c}_k \Ket{\left\lfloor M\sin^2\left( k\pi/{2s}\right) \right\rfloor},\\
\ket{\overline 1} &= \sum_{k \text{~odd}}^{[0,s]} \tilde{c}_k \Ket{\left\lfloor M\sin^2 \left(k\pi/{2s}\right) \right\rfloor},
\end{split}
}
where \(\tilde{c}_k>0\) can be obtained by solving a system of \flmRefsHyperref{ref65}{order} \(O(s^2)\) linear equations, and where \(\lfloor x \rfloor\) is the floor function. The code approaches optimality for sensing the signal Hamiltonian as \(M\) increases.

\codefieldsection{Protection}
The \(s\)th-order code corrects errors from the set \(\{I,a,a^{\dagger},{\hat n},{\hat n}^2,\cdots,{\hat n}^{s-1}\}\).
\codefieldsection{Parents}
\begin{eczvaluelist}
\item\relax
\flmRefsHyperref[eczindexfamilyrel]{code:single-mode}{Single-mode bosonic code}\item\relax
\flmRefsHyperref[eczindexfamilyrel]{code:metopt}{Error-corrected sensing code}\end{eczvaluelist}
\codefieldsection{Cousin}
\begin{eczvaluelist}
\item\relax
\flmRefsHyperref[eczindexfamilyrel]{code:binomial}{Binomial code} --- Chebyshev codes resemble binomial codes, and a class of binomial codes have similar error-correcting properties \NoCaseChange{\protect\cite{cite2756}}.
\end{eczvaluelist}
\eczhbkcontributors{ \eczhuVVA }
\endeczcode

\eczcode{chuang-leung-yamamoto}{Chuang-Leung-Yamamoto (CLY) code}{~\NoCaseChange{\protect\cite{cite2600}}}
\codefieldsection{Description}
Bosonic Fock-state code that encodes \(k\) qubits into \(n\) oscillators, with each oscillator restricted to having at most \(N\) excitations. Codewords are superpositions of oscillator Fock states which have exactly \(N\) total excitations, and are either uniform (i.e., balanced) superpositions or unbalanced superpositions.

Codes can be denoted as \(\llbracket N,n,2^k,d\rrbracket \), which conflicts with \flmRefsHyperref{code:stabilizer}{stabilizer code} notation.

\codefieldsection{Protection}
Protects against \flmRefsHyperref{ref498}{AD} for up to \(t = d-1\) excitation losses. Defining the \textit{spacing} between two Fock states \(|u_1\cdots u_n\rangle\) and \(|v_1\cdots v_n\rangle\),
\flmMathEnvironment{align}{}{
\text{Spacing}(u,v) = \frac{1}{2}\sum_{i=1}^n |u_i - v_i|,
}
the code distance \(d\) can be defined as the minimal spacing between Fock states making up the codewords.

\codefieldsection{Rate}
Code rate is \(\frac{k}{n \log_2(N+1)}\). To correct the loss of up to \(t\) excitations with \(K+1\) codewords, a code exists with scaling \(N \sim t^3 K/2\).
\codefieldsection{Encoding}
\begin{eczvaluelist}
\item\relax Photon Fock state input into a network of beamsplitters, phase shifters, and Kerr media. These operations all preserve total photon number. Beamsplitters and phase shifters take annihilation operators to linear combinations of annihilation operators, and the transformation matrix is unitary. The operations corresponding to Kerr nonlinear media are diagonal in the Fock basis, but they implement phases that in general depend nonlinearly on the number of photons in each mode. State preparation may require ancillary modes and be conditioned on photon-number measurement results.
\end{eczvaluelist}
\codefieldsection{Decoding}
\begin{eczvaluelist}
\item\relax Destructive decoding with a photon number measurement on each mode.
\item\relax State can be decoded with a network of beamsplitters, phase shifters, and Kerr media.
\end{eczvaluelist}
\codefieldsection{Parents}
\begin{eczvaluelist}
\item\relax
\flmRefsHyperref[eczindexfamilyrel]{code:fock_state}{Fock-state bosonic code} --- Chuang-Leung-Yamamoto codes are multi-mode Fock-state codes.
\item\relax
\flmRefsHyperref[eczindexfamilyrel]{code:constant_excitation}{Constant-excitation (CE) code} --- Chuang-Leung-Yamamoto codewords are constructed out of Fock states with the same total excitation number.
\end{eczvaluelist}
\codefieldsection{Children}
\begin{eczvaluelist}
\item\relax
\flmRefsHyperref[eczindexfamilyrel]{code:one_hot_quantum}{One-hot quantum code}\item\relax
\flmRefsHyperref[eczindexfamilyrel]{code:two-mode_binomial}{Two-mode binomial code}\end{eczvaluelist}
\codefieldsection{Cousin}
\begin{eczvaluelist}
\item\relax
\flmRefsHyperref[eczindexfamilyrel]{code:jump}{Jump code} --- Jump codes can be thought of as qubit analogues of uniform CLY codes.
\end{eczvaluelist}
\eczhbkcontributors{ Dhruv Devulapalli, Jonathan Kunjummen, \eczhuVVA }
\endeczcode

\eczcode{clifford_qsc}{Clifford group-representation QSC}{~\NoCaseChange{\protect\cite{cite2810}}}
\codefieldsection{Description}
Non-uniform QSC whose projection is onto a copy of an irreducible representation of the \flmRefsHyperref{ref409}{single-qubit Clifford group}, taken as the binary octahedral subgroup of the group \(SU(2)\) of Gaussian rotations.
Its codewords consist of non-uniform superpositions of 40 coherent states drawn from a 48-element Clifford-group orbit.

\codefieldsection{Gates}
\begin{eczvaluelist}
\item\relax The \flmRefsHyperref{ref409}{single-qubit Clifford group} can be realized via Gaussian rotations. The \(T\) and \(CZ\) gates can be realized using quartic Kerr operations \NoCaseChange{\protect\cite{cite2810}}.
\end{eczvaluelist}
\codefieldsection{Parents}
\begin{eczvaluelist}
\item\relax
\flmRefsHyperref[eczindexfamilyrel]{code:qsc}{Quantum spherical code (QSC)} --- The Clifford group-representation QSC has non-uniform coefficients.
\item\relax
\flmRefsHyperref[eczindexfamilyrel]{code:group_representation}{Group-representation code} --- The Clifford group-representation QSC is a group-representation code with \(G\) being \flmRefsHyperref{ref409}{single-qubit Clifford group}, taken as the binary octahedral subgroup of the group \(SU(2)\) of Gaussian rotations.
\end{eczvaluelist}
\eczhbkcontributors{ \eczhuVVA }
\endeczcode

\eczcode{quantum_sidelnikov}{Clifford subgroup-orbit QSC}{~\NoCaseChange{\protect\cite{cite382}}}
\codefieldsection{Alternative Names}
\begin{eczvaluelist}
\item\relax Sidelnikov QSC
\end{eczvaluelist}
\eczhIndexCodeAliasName{quantum_sidelnikov}{Sidelnikov QSC}
\codefieldsection{Description}
A \(\llparenthesis 2^r,2,2-\sqrt{2},8\rrparenthesis \) QSC for \(r \geq 1\) constructed using the real Clifford subgroup-orbit code.
Logical constellations are constructed by applying elements of an index-two subgroup of the \flmRefsHyperref{ref409}{real Clifford group}, when taken as a subgroup of the orthogonal group \NoCaseChange{\protect\cite{cite2103}} to \(2\) different vectors on the complex sphere.
The code is known as the \textit{Witting code} for \(r=2\) because its two logical constellations form vertices of Witting polytopes.

\codefieldsection{Parent}
\begin{eczvaluelist}
\item\relax
\flmRefsHyperref[eczindexfamilyrel]{code:qsc}{Quantum spherical code (QSC)}\end{eczvaluelist}
\codefieldsection{Cousins}
\begin{eczvaluelist}
\item\relax
\flmRefsHyperref[eczindexfamilyrel]{code:sidelnikov}{Real-Clifford subgroup-orbit code} --- Clifford group-orbit QSCs are quantum counterparts of real Clifford subgroup-orbit codes.
\item\relax
\flmRefsHyperref[eczindexfamilyrel]{code:witting_polytope}{Witting polytope code} --- Logical constellations of the Clifford subgroup-orbit code for \(r=2\) form vertices of Witting polytopes.
\item\relax
\flmRefsHyperref[eczindexfamilyrel]{code:24cell}{24-cell code} --- Logical constellations of the Clifford subgroup-orbit code for \(r=1\) form vertices of 24-cells when mapped into the real sphere, while code constellations form vertices of a disphenoidal 288-cell.
\item\relax
\flmRefsHyperref[eczindexfamilyrel]{code:disphenoidal288cell}{Disphenoidal 288-cell code} --- Logical constellations of the Clifford subgroup-orbit code for \(r=1\) form vertices of 24-cells when mapped into the real sphere, while code constellations form vertices of a disphenoidal 288-cell.
\end{eczvaluelist}
\eczhbkcontributors{ Shubham P. Jain, \eczhuVVA }
\endeczcode

\eczcode{coherent_constellation}{Coherent-state constellation code}{}
\codefieldsection{Description}
Qudit-into-oscillator code whose codewords can succinctly be expressed as superpositions of a countable set of coherent states that is called a \textit{constellation}. Some useful constellations form a group (see \flmRefsHyperref{code:gkp}{gkp}, \flmRefsHyperref{code:cat}{cat} or \flmRefsHyperref{code:2t_qutrit}{\(2T\)-qutrit} codes) while others make up a Gaussian quadrature rule \NoCaseChange{\protect\cite{cite930,cite931}}.

\codefieldsection{Rate}
Coherent-state constellation codes consisting of points from a Gaussian quadrature rule can be concatenated with quantum polar codes to achieve the Gaussian coherent information of the thermal noise channel \NoCaseChange{\protect\cite{cite930,cite931}}.
\codefieldsection{Parent}
\begin{eczvaluelist}
\item\relax
\flmRefsHyperref[eczindexfamilyrel]{code:oscillators}{Bosonic code}\end{eczvaluelist}
\codefieldsection{Children}
\begin{eczvaluelist}
\item\relax
\flmRefsHyperref[eczindexfamilyrel]{code:rg_cat}{Renormalization group (RG) cat code}\item\relax
\flmRefsHyperref[eczindexfamilyrel]{code:qsc}{Quantum spherical code (QSC)} --- Coherent-state QSCs are coherent-state constellation codes constrained to lie on a sphere.
\item\relax
\flmRefsHyperref[eczindexfamilyrel]{code:quantum_lattice}{Quantum lattice code} --- Quantum lattice codewords can be written as superpositions of coherent states lying on a lattice in phase space \NoCaseChange{\protect\cite{cite513,cite496}}.
\end{eczvaluelist}
\codefieldsection{Cousins}
\begin{eczvaluelist}
\item\relax
\flmRefsHyperref[eczindexfamilyrel]{code:oscillators_concatenated}{Concatenated bosonic code} --- Coherent-state constellation codes consisting of points from a Gaussian quadrature rule can be concatenated with quantum polar codes to achieve the Gaussian coherent information of the thermal noise channel \NoCaseChange{\protect\cite{cite930,cite931}}.
\item\relax
\flmRefsHyperref[eczindexfamilyrel]{code:t-designs}{\(t\)-design} --- Coherent-state constellation codes consisting of points from a Gaussian quadrature rule can be concatenated with quantum polar codes to achieve the Gaussian coherent information of the thermal noise channel \NoCaseChange{\protect\cite{cite930,cite931}}.
\item\relax
\flmRefsHyperref[eczindexfamilyrel]{code:modulation}{Modulation scheme} --- Coherent-state constellation codes are quantum counterparts of modulation schemes in that their codewords are superpositions of points in a constellation. Additionally, analog codes that achieve AWGN capacity \NoCaseChange{\protect\cite{cite2283}} can be used to develop capacity-achieving concatenations of coherent-state constellation codes with quantum polar codes \NoCaseChange{\protect\cite{cite930,cite931}}.
\item\relax
\flmRefsHyperref[eczindexfamilyrel]{code:coherent_state_c-q}{Coherent-state c-q modulation format} --- Coherent-state c-q codes encode classical alphabets into constellations of coherent states, while coherent-state constellation codes encode quantum information into superpositions of coherent states.
\item\relax
\flmRefsHyperref[eczindexfamilyrel]{code:tiger}{Tiger code} --- Tiger codewords consist of continuous and compact coherent-state constellations \NoCaseChange{\protect\cite{cite4667}}.
\end{eczvaluelist}
\eczhbkcontributors{ Armin Gerami, \eczhuVVA }
\endeczcode

\eczcode{coherent_state_repetition}{Coherent-state repetition code}{~\NoCaseChange{\protect\cite{cite4117,cite4118}}}
\codefieldsection{Description}
A concatenated qubit-into-\(n\)-mode code (for odd \(n\)) whose inner code is a quantum repetition code and whose outer code is the two-component cat code in its coherent-state basis. 

A basis of codewords is
\flmMathEnvironment{align}{}{
  |\overline{\pm}\rangle\propto\left|\pm\alpha\right\rangle ^{\otimes n}
}
for \(|\alpha| > 0\).

\codefieldsection{Protection}
For odd \(n\), the code has \(d_X=1\) and minimum Euclidean distance \(d_Z = 4n\) \NoCaseChange{\protect\cite{cite4667}}, so it does not protect against losses but suppresses dephasing exponentially in \(n|\alpha|^2\).

\codefieldsection{Encoding}
\begin{eczvaluelist}
\item\relax Lindbladian-based dissipative encoding with dissipators \(\hat{a}_j \hat{a}_k - \alpha^2\) for neighboring modes \(j,k\) on a 1D line as well as local dissipators \(\hat{a}_j^2 - \alpha^2\). Encoding passively protects against cavity dephasing.
\end{eczvaluelist}
\codefieldsection{Parents}
\begin{eczvaluelist}
\item\relax
\flmRefsHyperref[eczindexfamilyrel]{code:cat_concatenated}{Concatenated cat code} --- The coherent-state repetition code is a concatenation whose outer code is the cat code in its coherent-state basis.
\item\relax
\flmRefsHyperref[eczindexfamilyrel]{code:tiger}{Tiger code} --- For odd \(n\), the coherent-state repetition code is a tiger code whose matrix \(G\) is the cyclic repetition generator matrix over the integers and whose matrix \(H\) is zero \NoCaseChange{\protect\cite{cite4667}}. For even \(n\), or after removing the last row to impose open boundaries, the construction yields a logical-rotor variant instead of a logical qubit.
\end{eczvaluelist}
\codefieldsection{Child}
\begin{eczvaluelist}
\item\relax
\flmRefsHyperref[eczindexfamilyrel]{code:two-legged-cat}{Two-component cat code} --- The coherent-state repetition code for \(n=1\) reduces to the two-component cat code.
\end{eczvaluelist}
\codefieldsection{Cousins}
\begin{eczvaluelist}
\item\relax
\flmRefsHyperref[eczindexfamilyrel]{code:cat_repetition}{Cat-repetition code} --- The cat (coherent-state) repetition code is a concatenation whose outer code is the (two-component) cat code in its cat (coherent-state) basis. For the two-component case, both reduce to the two-component cat code at \(n=1\).
\item\relax
\flmRefsHyperref[eczindexfamilyrel]{code:quantum_repetition}{Quantum repetition code} --- Two-component cat codes in the coherent-state basis have been concatenated with quantum repetition codes \NoCaseChange{\protect\cite{cite4117,cite4118}}.
\end{eczvaluelist}
\eczhbkcontributors{ \eczhuVVA }
\endeczcode

\eczcode{compactified_r}{Compactified \(\mathbb{R}\) gauge theory code}{~\NoCaseChange{\protect\cite{cite411}}}
\codefieldsection{Description}
An integer-homology bosonic CSS code realizing 2D \(U(1)\) gauge theory on bosonic modes.
The code can be obtained from the analog surface code by \flmRefsHyperref{ref410}{condensing} certain anyons \NoCaseChange{\protect\cite{cite411}}. 
This results in a pinning of each mode to the space of periodic functions, which is the Hilbert space of a physical rotor, and can be thought of as compactification of the 2D \(\mathbb{R}\) gauge theory phase realized by the analog surface code.

\codefieldsection{Parents}
\begin{eczvaluelist}
\item\relax
\flmRefsHyperref[eczindexfamilyrel]{code:homological_cv}{Integer-homology bosonic CSS code} --- The compactified \(\mathbb{R}\) gauge theory code realizes \(U(1)\) gauge theory on bosonic modes.
\item\relax
\flmRefsHyperref[eczindexfamilyrel]{code:2d_stabilizer}{2D lattice stabilizer code}\item\relax
\flmRefsHyperref[eczindexfamilyrel]{code:topological_abelian}{Abelian topological code} --- The compactified \(\mathbb{R}\) gauge theory code can be obtained from the analog surface code by \flmRefsHyperref{ref410}{condensing} certain anyons \NoCaseChange{\protect\cite{cite411}}. This results in a pinning of each mode to the space of periodic functions, which is the Hilbert space of a physical rotor, and can be thought of as compactification of the 2D \(\mathbb{R}\) gauge theory phase realized by the analog surface code.
\end{eczvaluelist}
\codefieldsection{Cousins}
\begin{eczvaluelist}
\item\relax
\flmRefsHyperref[eczindexfamilyrel]{code:analog_surface}{Analog surface code} --- The compactified \(\mathbb{R}\) gauge theory code can be obtained from the analog surface code by \flmRefsHyperref{ref410}{condensing} certain anyons \NoCaseChange{\protect\cite{cite411}}. This results in a pinning of each mode to the space of periodic functions, which is the Hilbert space of a physical rotor, and can be thought of as compactification of the 2D \(\mathbb{R}\) gauge theory phase realized by the analog surface code.
\item\relax
\flmRefsHyperref[eczindexfamilyrel]{code:qudit_surface}{Modular-qudit surface code} --- The compactified \(\mathbb{R}\) gauge theory code can be thought of as a realization of the \(q\to\infty\) \(U(1)\) rotor limit \NoCaseChange{\protect\cite{cite2531}} of the qudit surface code as a bosonic stabilizer code.
\item\relax
\flmRefsHyperref[eczindexfamilyrel]{code:repetition}{Repetition code} --- The compactified \(\mathbb{R}\) gauge theory code is constructed from a hypergraph product of two repetition codes over the integers.
\item\relax
\flmRefsHyperref[eczindexfamilyrel]{code:hypergraph_product}{Hypergraph product (HGP) code} --- The compactified \(\mathbb{R}\) gauge theory code is constructed from a hypergraph product of two repetition codes over the integers.
\item\relax
\flmRefsHyperref[eczindexfamilyrel]{code:tiger_surface}{Tiger surface code} --- Both the compactified \(\mathbb{R}\) gauge theory and tiger surface code are constructed from a hypergraph product of two repetition codes over the integers.
\end{eczvaluelist}
\eczhbkcontributors{ \eczhuVVA }
\endeczcode

\eczcode{oscillators_concatenated}{Concatenated bosonic code}{}
\codefieldsection{Description}
A concatenated code whose outer code is a bosonic code. In other words, a bosonic code that can be thought of as a concatenation of a possibly non-bosonic inner code and a bosonic outer code.

\codefieldsection{Decoding}
\begin{eczvaluelist}
\item\relax Decoder exploiting analog information from the outer code for bosonic codes concatenated with qubit QLDPC codes \NoCaseChange{\protect\cite{cite1612}}.
\end{eczvaluelist}
\codefieldsection{Parents}
\begin{eczvaluelist}
\item\relax
\flmRefsHyperref[eczindexfamilyrel]{code:oscillators}{Bosonic code}\item\relax
\flmRefsHyperref[eczindexfamilyrel]{code:quantum_concatenated}{Concatenated quantum code} --- A concatenated bosonic code is a bosonic code that can be thought of as a concatenation of a possibly non-bosonic inner code and a bosonic outer code.
\end{eczvaluelist}
\codefieldsection{Children}
\begin{eczvaluelist}
\item\relax
\flmRefsHyperref[eczindexfamilyrel]{code:cat_concatenated}{Concatenated cat code}\item\relax
\flmRefsHyperref[eczindexfamilyrel]{code:gkp_concatenated}{Concatenated GKP code}\end{eczvaluelist}
\codefieldsection{Cousins}
\begin{eczvaluelist}
\item\relax
\flmRefsHyperref[eczindexfamilyrel]{code:hybrid_cat}{Hybrid cat code} --- Hybrid cat codes can be concatenated with RBH codes \NoCaseChange{\protect\cite{cite4401}}.
\item\relax
\flmRefsHyperref[eczindexfamilyrel]{code:coherent_constellation}{Coherent-state constellation code} --- Coherent-state constellation codes consisting of points from a Gaussian quadrature rule can be concatenated with quantum polar codes to achieve the Gaussian coherent information of the thermal noise channel \NoCaseChange{\protect\cite{cite930,cite931}}.
\item\relax
\flmRefsHyperref[eczindexfamilyrel]{code:dual_rail}{Dual-rail quantum code} --- The KLM protocol, one of the first protocols for fault-tolerant quantum computation, utilizes concatenations of the dual-rail code with a stabilizer code such as the Steane code \NoCaseChange{\protect\cite{cite3366,cite3367,cite3368}}. Concatenating the dual-rail code with an \(\llbracket n,k,d\rrbracket \) stabilizer code yields an \(\llbracket 2n,k,d\rrbracket \) constant-excitation code \NoCaseChange{\protect\cite{cite2711}} that protects against \(d-1\) \flmRefsHyperref{ref498}{AD} errors \NoCaseChange{\protect\cite{cite3263}}. Using the concatenation convention of the Zoo, concatenating the inner dual-rail code with an outer single-mode bosonic code yields several gates that are independent of the outer code \NoCaseChange{\protect\cite{cite4856}}.
\item\relax
\flmRefsHyperref[eczindexfamilyrel]{code:two-mode_binomial}{Two-mode binomial code} --- Two-mode binomial codes can be concatenated with repetition codes to yield bosonic analogues of QPCs \NoCaseChange{\protect\cite{cite4049}}.
\end{eczvaluelist}
\eczhbkcontributors{ \eczhuVVA }
\endeczcode

\eczcode{cat_concatenated}{Concatenated cat code}{}
\codefieldsection{Description}
A concatenated code obtained by encoding the physical qubits of an inner qubit code into cat-code states.
Most examples concatenate a qubit stabilizer code with the two-component cat code in its cat-state basis.

\codefieldsection{Protection}
The cat code can exponentially suppress one effective Pauli error channel with the size of its coherent states, so an inner qubit code such as a quantum repetition code can exploit a large noise bias while still ensuring good performance \NoCaseChange{\protect\cite{cite2616,cite2646,cite2647,cite4113}}.

\codefieldsection{Parents}
\begin{eczvaluelist}
\item\relax
\flmRefsHyperref[eczindexfamilyrel]{code:qsc}{Quantum spherical code (QSC)}\item\relax
\flmRefsHyperref[eczindexfamilyrel]{code:oscillators_concatenated}{Concatenated bosonic code}\end{eczvaluelist}
\codefieldsection{Children}
\begin{eczvaluelist}
\item\relax
\flmRefsHyperref[eczindexfamilyrel]{code:cat_repetition}{Cat-repetition code} --- The cat-repetition code is a concatenation whose outer code is the cat code in its cat-state basis.
\item\relax
\flmRefsHyperref[eczindexfamilyrel]{code:coherent_state_repetition}{Coherent-state repetition code} --- The coherent-state repetition code is a concatenation whose outer code is the cat code in its coherent-state basis.
\end{eczvaluelist}
\codefieldsection{Cousins}
\begin{eczvaluelist}
\item\relax
\flmRefsHyperref[eczindexfamilyrel]{code:qubit_css}{Qubit CSS code} --- Stabilizers of CSS codes concatenated with two-component cat codes in their coherent-state basis come directly from the CSS codes via the mapping \(X \to (-1)^{\hat n}\) and \(Z \to \hat a\) \NoCaseChange{\protect\cite{cite382}}. Stabilizers of CSS codes concatenated with two-component cat codes in their cat-state basis come directly from the CSS codes via the mapping \(Z \to (-1)^{\hat n}\) and \(X \to \hat a\). In both cases, one type of noise is handled actively via syndrome extraction and correction, while the other type is handled passively via stabilizing dissipation.
\item\relax
\flmRefsHyperref[eczindexfamilyrel]{code:rotated_surface}{Rotated surface code} --- Cat codes have been concatenated with rotated surface codes \NoCaseChange{\protect\cite{cite4072}}.
\item\relax
\flmRefsHyperref[eczindexfamilyrel]{code:ldpc}{Low-density parity-check (LDPC) code} --- Cat codes have been concatenated with LDPC codes (treated as qubit stabilizer codes) \NoCaseChange{\protect\cite{cite1486}}.
\item\relax
\flmRefsHyperref[eczindexfamilyrel]{code:lhz}{Lechner-Hauke-Zoller (LHZ) code} --- LHZ parity-codes have been concatenated with cat codes \NoCaseChange{\protect\cite{cite4857}}.
\item\relax
\flmRefsHyperref[eczindexfamilyrel]{code:steane}{\(\llbracket 7,1,3\rrbracket \) Steane code} --- Two-component cat codes concatenated with Steane and Golay codes are estimated to be fault tolerant against \flmRefsHyperref{ref498}{photon loss} noise with rate \(\eta < 5\times 10^{-4}\) provided that \(\alpha > 1.2\) \NoCaseChange{\protect\cite{cite3231}}.
\item\relax
\flmRefsHyperref[eczindexfamilyrel]{code:qubit_golay}{\(\llbracket 23, 1, 7\rrbracket \) Quantum Golay code} --- Two-component cat codes concatenated with Steane and Golay codes are estimated to be fault tolerant against \flmRefsHyperref{ref498}{photon loss} noise with rate \(\eta < 5\times 10^{-4}\) provided that \(\alpha > 1.2\) \NoCaseChange{\protect\cite{cite3231}}.
\item\relax
\flmRefsHyperref[eczindexfamilyrel]{code:xzzx}{XZZX surface code} --- The four-component cat code can be concatenated with the XZZX code to yield a fusion-based computation scheme on a 2D lattice \NoCaseChange{\protect\cite{cite3687}}.
\item\relax
\flmRefsHyperref[eczindexfamilyrel]{code:fusion}{Fusion-based quantum computing (FBQC) code} --- The four-component cat code can be concatenated with the XZZX code to yield a fusion-based computation scheme on a 2D lattice \NoCaseChange{\protect\cite{cite3687}}.
\end{eczvaluelist}
\eczhbkcontributors{ \eczhuVVA }
\endeczcode

\eczcode{gkp_concatenated}{Concatenated GKP code}{~\NoCaseChange{\protect\cite{cite3290}}}
\codefieldsection{Description}
A concatenated code whose outer code is a GKP code. In other words, a bosonic code that can be thought of as a concatenation of an arbitrary inner code and another bosonic outer code. Most examples encode physical qubits of an inner stabilizer code into the square-lattice GKP code.

\codefieldsection{Protection}
The analog syndrome information of the outer GKP code can improve protection of the inner code. As an example, concatenating a three-qubit quantum repetition code with GKP codes can correct some two-bit-flip errors \NoCaseChange{\protect\cite{cite3290}}.

\codefieldsection{Rate}
Recursively concatenating the \(C_6\) and \(\llbracket 4,2,2\rrbracket \) codes with GKP codes achieves the hashing bound of the displacement channel \NoCaseChange{\protect\cite{cite3290}}. 
Concatenating Abelian LP codes with GKP codes can surpass the CSS Hamming bound \NoCaseChange{\protect\cite{cite1556}}. 
Particular families of GKP codes achieve the capacity of \flmRefsHyperref{ref498}{AD} and amplification channels for some loss rates \NoCaseChange{\protect\cite{cite4858}}. 
Concatenations of square-lattice GKP codes with Hermitian Galois-qudit codes achieve the capacity for all loss rates \NoCaseChange{\protect\cite{cite4055}}. 
Concatenation of GKP codes with quantum polar codes achieves a rate against the displacement channel \NoCaseChange{\protect\cite{cite4055}}.

\codefieldsection{Gates}
\begin{eczvaluelist}
\item\relax Linear-optical computation \NoCaseChange{\protect\cite{cite4859}}.
\end{eczvaluelist}
\codefieldsection{Decoding}
\begin{eczvaluelist}
\item\relax Circuit-level soft information decoder \NoCaseChange{\protect\cite{cite4860}}.
\end{eczvaluelist}
\codefieldsection{Code Capacity Threshold}
\begin{eczvaluelist}
\item\relax \(0.599\) threshold displacement standard deviation for GKP-repetition code \NoCaseChange{\protect\cite{cite4116}}.
\item\relax \(0.59\) threshold displacement standard deviation for GKP-color code \NoCaseChange{\protect\cite{cite4428}}.
\item\relax A \flmRefsHyperref{ref515}{concatenated threshold} with GKP codes on the lowest level exists for general Markovian noise \NoCaseChange{\protect\cite{cite4861}}.
\item\relax There is an upper bound on the threshold under local update recovery that is derived via quantum optimal transport \NoCaseChange{\protect\cite{cite4205}}.
\end{eczvaluelist}
\codefieldsection{Notes}
\begin{eczvaluelist}
\item\relax Bosonic Pauli+ model is a numerical simulation tool for concatenated GKP codes \NoCaseChange{\protect\cite{cite4862}}.
\end{eczvaluelist}
\codefieldsection{Parents}
\begin{eczvaluelist}
\item\relax
\flmRefsHyperref[eczindexfamilyrel]{code:multimodegkp}{Gottesman-Kitaev-Preskill (GKP) code}\item\relax
\flmRefsHyperref[eczindexfamilyrel]{code:oscillators_concatenated}{Concatenated bosonic code}\end{eczvaluelist}
\codefieldsection{Children}
\begin{eczvaluelist}
\item\relax
\flmRefsHyperref[eczindexfamilyrel]{code:dfour_gkp}{\(D_4\) hyper-diamond GKP code} --- The \(D_4\) hyper-diamond GKP code can be seen as a concatenation of a rotated square-lattice GKP code with a repetition code \NoCaseChange{\protect\cite{cite482}}. This is related to the fact that the four-bit repetition code yields the \(D_4\) hyper-diamond lattice via \flmTerm{term}{ref127}{}{Construction A}.
\item\relax
\flmRefsHyperref[eczindexfamilyrel]{code:gkp_surface_concatenated}{GKP-surface code}\end{eczvaluelist}
\codefieldsection{Cousins}
\begin{eczvaluelist}
\item\relax
\flmRefsHyperref[eczindexfamilyrel]{code:cluster_state}{Cluster-state code} --- GKP codes have been concatenated with cluster-state codes \NoCaseChange{\protect\cite{cite415}}.
\item\relax
\flmRefsHyperref[eczindexfamilyrel]{code:quantum_repetition}{Quantum repetition code} --- Concatenating a three-qubit quantum repetition code with GKP codes can correct some two-bit-flip errors \NoCaseChange{\protect\cite{cite3290}} (see also \NoCaseChange{\protect\cite{cite4116}}).
\item\relax
\flmRefsHyperref[eczindexfamilyrel]{code:stab_4_2_2}{\(\llbracket 4,2,2\rrbracket \) Four-qubit code} --- Recursively concatenating the \(C_6\) and \(\llbracket 4,2,2\rrbracket \) codes with GKP codes achieves the hashing bound of the displacement channel \NoCaseChange{\protect\cite{cite3290}}.
\item\relax
\flmRefsHyperref[eczindexfamilyrel]{code:stab_6_2_2}{\(\llbracket 6,2,2\rrbracket \) \(C_6\) code} --- Recursively concatenating the \(C_6\) and \(\llbracket 4,2,2\rrbracket \) codes with GKP codes achieves the hashing bound of the displacement channel \NoCaseChange{\protect\cite{cite3290}}.
\item\relax
\flmRefsHyperref[eczindexfamilyrel]{code:abelian_lifted_product}{Abelian LP code} --- GKP codes have been concatenated with Abelian LP codes \NoCaseChange{\protect\cite{cite1556}} that are in turn based on QC-LDPC codes \NoCaseChange{\protect\cite{cite1550}}. Concatenating Abelian LP codes with GKP codes can surpass the CSS Hamming bound \NoCaseChange{\protect\cite{cite1556}}.
\item\relax
\flmRefsHyperref[eczindexfamilyrel]{code:quantum_parity}{Quantum parity code (QPC)} --- GKP codes have been concatenated with QPCs \NoCaseChange{\protect\cite{cite4050}}.
\item\relax
\flmRefsHyperref[eczindexfamilyrel]{code:488_color}{Square-octagon (4.8.8) color code} --- GKP codes have been concatenated with 4.8.8 color codes \NoCaseChange{\protect\cite{cite4428}}.
\item\relax
\flmRefsHyperref[eczindexfamilyrel]{code:triangular_color}{Honeycomb (6.6.6) color code} --- GKP codes have been concatenated with the 6.6.6 color code \NoCaseChange{\protect\cite{cite3323}}.
\item\relax
\flmRefsHyperref[eczindexfamilyrel]{code:stab_5_1_3}{\(\llbracket 5,1,3\rrbracket \) Five-qubit perfect code} --- GKP codes have been concatenated with the five-qubit code \NoCaseChange{\protect\cite{cite3323}}.
\item\relax
\flmRefsHyperref[eczindexfamilyrel]{code:quantum_polar}{Quantum polar code} --- Concatenation of GKP codes with quantum polar codes achieves a rate against the displacement channel \NoCaseChange{\protect\cite{cite4055}}.
\item\relax
\flmRefsHyperref[eczindexfamilyrel]{code:stabilizer_over_gfqsq}{Hermitian Galois-qudit code} --- Concatenations of square-lattice GKP codes with Hermitian Galois-qudit codes achieve the capacity for all loss rates \NoCaseChange{\protect\cite{cite4055}}.
\item\relax
\flmRefsHyperref[eczindexfamilyrel]{code:tesselation}{Hyperbolic tessellation code} --- The qubit-Pauli tessellation GKP code \NoCaseChange{\protect\cite{cite2811}} is the Euclidean \(\{2,4,4\}\) member of the curvature-dependent tessellation-code framework. It is a two-mode code in which each Cartesian direction is a single-mode qubit GKP code, making the full code a 2-to-1 concatenated qubit encoding. The logical \flmRefsHyperref{ref663}{single-qubit Pauli group} is implemented geometrically by one \(\pi\) rotation and two \(\pi/2\) rotations on the Euclidean tessellation \NoCaseChange{\protect\cite{cite2811}}.
\item\relax
\flmRefsHyperref[eczindexfamilyrel]{code:gkp-cluster-state}{GKP CV-cluster-state code} --- GKP CV-cluster-state codes reduce to cluster-state codes concatenated with single-mode GKP codes \NoCaseChange{\protect\cite{cite415}} when all physical modes are initialized in GKP states. Analog QEC with GKP codes concatenated with surface-code and 3D-cluster-state fault-tolerant schemes can reach a threshold displacement standard deviation of about \(0.607\) for ideal syndrome measurements and reduce the squeezing required for topologically protected MBQC to \(9.8\) dB \NoCaseChange{\protect\cite{cite415}}.
\item\relax
\flmRefsHyperref[eczindexfamilyrel]{code:gkp-stabilizer}{Oscillator-into-oscillator GKP code} --- Oscillator-into-oscillator GKP codes concatenated with qubit-into-oscillator GKP codes can outperform more conventional concatenations of qubit-into-oscillator GKP codes with qubit stabilizer codes \NoCaseChange{\protect\cite{cite4863}}.
\end{eczvaluelist}
\eczhbkcontributors{ \eczhuVVA }
\endeczcode

\eczcode{dual_rail}{Dual-rail quantum code}{~\NoCaseChange{\protect\cite{cite4864,cite4865,cite4866}}}
\codefieldsection{Description}
Two-mode bosonic code encoding a logical qubit in Fock states with one excitation.
The logical-zero state is represented by \(|10\rangle\), while the logical-one state is represented by \(|01\rangle\).
This encoding is often realized in temporal or spatial modes, corresponding to a \textit{time-bin} or \textit{frequency-bin} encoding.
Two different types of photon polarization can also be used.

This code is a DFS \NoCaseChange{\protect\cite{cite2712,cite2713,cite2714,cite2715}} with respect to phase errors \NoCaseChange{\protect\cite{cite4867}}.

\codefieldsection{Protection}
This is an error-detecting code against one \flmRefsHyperref{ref498}{photon loss} event; it is often used in photonic quantum devices because of its ease of realization. A single loss event can be detected because, after the loss occurs, the output state \(|00\rangle\) is orthogonal to the codespace. Recovery is not possible, so a successful run of a quantum circuit is conditioned on not losing a photon during the circuit.

Photon loss from a dual-rail or polarization encoding maps the qubit outside its two-dimensional codespace into a vacuum state, so it is naturally modeled as an erasure rather than as a qubit amplitude-damping channel \NoCaseChange{\protect\cite[{Ch. 1}]{cite398}}.

For Deutsch''s  problem specifically, this code protects against errors resulting in states that have the correct photon number, but in the wrong modes \NoCaseChange{\protect\cite{cite4864}}.

\codefieldsection{Encoding}
\begin{eczvaluelist}
\item\relax Optimal control pulses \NoCaseChange{\protect\cite{cite4868}}
\end{eczvaluelist}
\codefieldsection{Gates}
\begin{eczvaluelist}
\item\relax General gates are performed using two-body Hamiltonian rotations \NoCaseChange{\protect\cite{cite4867}}.
\item\relax Bosonic gates include beamsplitters \NoCaseChange{\protect\cite{cite4869}} and Kerr nonlinearities. In particular, a cross Kerr rotation at angle \(\pi\) induces a \(CZ\) gate. Universal quantum computing can be achieved using the KLM protocol \NoCaseChange{\protect\cite{cite3366}} with only linear optical elements and photon detectors.
\item\relax Photon-number-conserving universal quantum logic can be implemented using continuous-time quantum walks on dual-rail transmon arrays \NoCaseChange{\protect\cite{cite4870}}.
\item\relax Dynamical-decoupling protocols \NoCaseChange{\protect\cite{cite4867,cite4871}}.
\item\relax A probabilistic CZ gate via a non-linear sign-shift gate, which transforms the Fock states \( \alpha|0\rangle+\beta|1\rangle+\gamma|2\rangle\) into \(\alpha|0\rangle+\beta|1\rangle-\gamma|2\rangle \), followed by measurement \NoCaseChange{\protect\cite{cite4872}}.
\item\relax Error-detecting \(CCZ\) and \(cSWAP\) gates using three-level ancilla \NoCaseChange{\protect\cite{cite4710}}.
\item\relax Cavity-assisted bias-preserving CNOT gate \NoCaseChange{\protect\cite{cite4873}}.
\end{eczvaluelist}
\codefieldsection{Fault Tolerance}
\begin{eczvaluelist}
\item\relax Dual-rail qubits can be used to convert leakage and \flmRefsHyperref{ref498}{AD} noise into erasure noise \NoCaseChange{\protect\cite{cite1187,cite4680}}.
\end{eczvaluelist}
\codefieldsection{Realizations}
\begin{eczvaluelist}
\item\relax The dual-rail code is ubiquitous in linear-optical quantum devices and is behind the KLM protocol, one of the first proposals for fault-tolerant computation. See reviews \NoCaseChange{\protect\cite{cite4874,cite4875,cite3580}} for more details.
\item\relax Superconducting circuit devices: Gates have been demonstrated in the Schoelkopf group at Yale University \NoCaseChange{\protect\cite{cite4869}}. Error detection has been demonstrated in 3D cavities in the Devoret group at Yale University \NoCaseChange{\protect\cite{cite4868}} and Amazon Web Services \NoCaseChange{\protect\cite{cite4876}} using transmon qubits, following earlier theoretical proposals \NoCaseChange{\protect\cite{cite4877,cite4680}}. GHZ and Bell states as well as universal gates have been implemented on a four-qubit dual-rail code by the Yu group \NoCaseChange{\protect\cite{cite3361}}. Logical readout in 3D cavities has been demonstrated by Quantum Circuits Inc. \NoCaseChange{\protect\cite{cite4878}}. Cavity-assisted bias-preserving CNOT gate has been demonstrated \NoCaseChange{\protect\cite{cite4873}}.
\item\relax Photonic platforms: state preparation and measurement fidelity of \(99.98\%\) in the C telecom band by PsiQuantum \NoCaseChange{\protect\cite{cite4879}}.
\end{eczvaluelist}
\codefieldsection{Parents}
\begin{eczvaluelist}
\item\relax
\flmRefsHyperref[eczindexfamilyrel]{code:one_hot_quantum}{One-hot quantum code}\item\relax
\flmRefsHyperref[eczindexfamilyrel]{code:two-mode_binomial}{Two-mode binomial code} --- The two-mode binomial code for \(S=N=0\) reduces to the dual-rail code.
\end{eczvaluelist}
\codefieldsection{Cousins}
\begin{eczvaluelist}
\item\relax
\flmRefsHyperref[eczindexfamilyrel]{code:oscillators_concatenated}{Concatenated bosonic code} --- The KLM protocol, one of the first protocols for fault-tolerant quantum computation, utilizes concatenations of the dual-rail code with a stabilizer code such as the Steane code \NoCaseChange{\protect\cite{cite3366,cite3367,cite3368}}. Concatenating the dual-rail code with an \(\llbracket n,k,d\rrbracket \) stabilizer code yields an \(\llbracket 2n,k,d\rrbracket \) constant-excitation code \NoCaseChange{\protect\cite{cite2711}} that protects against \(d-1\) \flmRefsHyperref{ref498}{AD} errors \NoCaseChange{\protect\cite{cite3263}}. Using the concatenation convention of the Zoo, concatenating the inner dual-rail code with an outer single-mode bosonic code yields several gates that are independent of the outer code \NoCaseChange{\protect\cite{cite4856}}.
\item\relax
\flmRefsHyperref[eczindexfamilyrel]{code:steane}{\(\llbracket 7,1,3\rrbracket \) Steane code} --- The KLM protocol, one of the first protocols for fault-tolerant quantum computation, utilizes concatenations of the dual-rail code with a stabilizer code such as the Steane code \NoCaseChange{\protect\cite{cite3366,cite3367,cite3368}}.
\item\relax
\flmRefsHyperref[eczindexfamilyrel]{code:single-mode}{Single-mode bosonic code} --- Using the concatenation convention of the Zoo, concatenating the inner dual-rail code with an outer single-mode bosonic code yields several gates that are independent of the outer code \NoCaseChange{\protect\cite{cite4856}}.
\item\relax
\flmRefsHyperref[eczindexfamilyrel]{code:ampdamp}{Amplitude-damping (AD) code} --- Dual-rail qubits can be used to convert leakage and \flmRefsHyperref{ref498}{AD} noise into erasure noise \NoCaseChange{\protect\cite{cite1187,cite4680}}. Concatenating the dual-rail code with an \(\llbracket n,k,d\rrbracket \) stabilizer code yields an \(\llbracket 2n,k,d\rrbracket \) constant-excitation code \NoCaseChange{\protect\cite{cite2711}} that protects against \(d-1\) \flmRefsHyperref{ref498}{AD} errors \NoCaseChange{\protect\cite{cite3263}}.
\item\relax
\flmRefsHyperref[eczindexfamilyrel]{code:quantum_parity}{Quantum parity code (QPC)} --- An \(\llbracket 8,1,2\rrbracket \) QPC correcting a single \flmRefsHyperref{ref498}{AD} error is equivalent to a concatenation of the \(\{|\overline{01}\rangle,|\overline{11}\rangle\}\) (constant-excitation) subcode of the \(\llbracket 4,2,2\rrbracket \) code with the dual-rail code \NoCaseChange{\protect\cite{cite3250,cite3259,cite2711}}. More generally, an \(\llbracket m^2,1,m\rrbracket \) QPC corrects \(m-1\) \flmRefsHyperref{ref498}{AD} errors \NoCaseChange{\protect\cite{cite3263}}.
\item\relax
\flmRefsHyperref[eczindexfamilyrel]{code:stab_4_2_2}{\(\llbracket 4,2,2\rrbracket \) Four-qubit code} --- An \(\llbracket 8,1,2\rrbracket \) QPC correcting a single \flmRefsHyperref{ref498}{AD} error is equivalent to a concatenation of the \(\{|\overline{01}\rangle,|\overline{11}\rangle\}\) (constant-excitation) subcode of the \(\llbracket 4,2,2\rrbracket \) code with the dual-rail code \NoCaseChange{\protect\cite{cite3250,cite3259,cite2711}}. More generally, an \(\llbracket m^2,1,m\rrbracket \) QPC corrects \(m-1\) \flmRefsHyperref{ref498}{AD} errors \NoCaseChange{\protect\cite{cite3263}}.
\item\relax
\flmRefsHyperref[eczindexfamilyrel]{code:cluster_state}{Cluster-state code} --- The KLM protocol can be combined with cluster states in various ways to yield MBQC protocols \NoCaseChange{\protect\cite{cite3577,cite3578,cite3579}}; see review \NoCaseChange{\protect\cite{cite3580}}.
\item\relax
\flmRefsHyperref[eczindexfamilyrel]{code:quantum_repetition}{Quantum repetition code} --- The dual-rail code is an error space of the quantum repetition code for \(n=2\) and is stabilized by \(-ZZ\).
\item\relax
\flmRefsHyperref[eczindexfamilyrel]{code:fusion}{Fusion-based quantum computing (FBQC) code} --- FBQC resource states are concatenated with dual-rail codes to increase loss detection.
\item\relax
\flmRefsHyperref[eczindexfamilyrel]{code:xyz_hexagonal}{XYZ\(^2\) hexagonal stabilizer code} --- The XYZ\(^2\) hexagonal stabilizer code can be viewed as a concatenation of the \(YZZY\) surface code with one of the possible \(\llbracket 2,1\rrbracket \) repetition codes, with the case of the bit-flip repetition code yielding a concatenation of the surface code with the dual-rail code \NoCaseChange{\protect\cite{cite2645}}.
\end{eczvaluelist}
\eczhbkcontributors{ Yinchen Liu, Esha Swaroop, Dhruv Devulapalli, Aniket Maiti, \eczhuVVA }
\endeczcode

\eczcode{ea_analog_stabilizer}{EA analog stabilizer code}{~\NoCaseChange{\protect\cite{cite4880}}}
\codefieldsection{Description}
Constructed using a variation of the analog stabilizer formalism designed to utilize pre-shared entanglement between sender and receiver.

\codefieldsection{Protection}
Optimal code parameters have been determined \NoCaseChange{\protect\cite{cite3641}}.

\codefieldsection{Parent}
\begin{eczvaluelist}
\item\relax
\flmRefsHyperref[eczindexfamilyrel]{code:ea_oscillators}{EA bosonic code}\end{eczvaluelist}
\codefieldsection{Cousin}
\begin{eczvaluelist}
\item\relax
\flmRefsHyperref[eczindexfamilyrel]{code:analog_stabilizer}{Analog stabilizer code} --- EA analog stabilizer codes utilize additional ancillary modes in a pre-shared entangled state, but reduce to ordinary analog stabilizer codes when said modes are interpreted as noiseless physical modes.
\end{eczvaluelist}
\eczhbkcontributors{ \eczhuVVA }
\endeczcode

\eczcode{ea_oscillators}{EA bosonic code}{}

\codefieldsection{Kingdom root code}
for the \flmRefsHyperref{kingdom:oscillators}{Bosonic Kingdom}.
\codefieldsection{Description}
Bosonic code designed to utilize pre-shared entanglement between sender and receiver.

\codefieldsection{Parent}
\begin{eczvaluelist}
\item\relax
\flmRefsHyperref[eczindexfamilyrel]{code:eaqecc}{Entanglement-assisted (EA) QECC}\end{eczvaluelist}
\codefieldsection{Child}
\begin{eczvaluelist}
\item\relax
\flmRefsHyperref[eczindexfamilyrel]{code:ea_analog_stabilizer}{EA analog stabilizer code}\end{eczvaluelist}
\codefieldsection{Cousin}
\begin{eczvaluelist}
\item\relax
\flmRefsHyperref[eczindexfamilyrel]{code:oscillators}{Bosonic code} --- EA bosonic codes utilize additional ancillary modes in a pre-shared entangled state, but reduce to ordinary bosonic codes when said modes are interpreted as noiseless physical modes.
\end{eczvaluelist}
\eczhbkcontributors{ \eczhuVVA }
\endeczcode

\eczcode{fock_state}{Fock-state bosonic code}{}
\codefieldsection{Description}
Qudit-into-oscillator code whose protection against \flmRefsHyperref{ref498}{AD} noise (i.e., photon loss) stems from the use of disjoint sets of Fock states for the construction of each code basis state. The simplest example is the dual-rail code, which has codewords consisting of single Fock states \(|10\rangle\) and \(|01\rangle\). This code can detect a single loss error since a loss operator in either mode maps one of the codewords to a different Fock state \(|00\rangle\). More involved codewords consist of several well-separated Fock states such that multiple loss events can be detected and corrected.
\codefieldsection{Protection}
Code distance \(d\) is the minimum distance (assuming some metric) between any two labels of Fock states corresponding to different code basis states. For a single mode, \(d\) is the minimum absolute value of the difference between any two Fock-state labels; such codes can detect up to \(d-1\) loss events. Multimode distances can be defined analogously; see, e.g., \flmRefsHyperref{code:chuang-leung-yamamoto}{Chuang-Leung-Yamamoto codes}. There are tradeoffs in how well a Fock-state code protects against loss/gain errors and dephasing noise \NoCaseChange{\protect\cite{cite4881}}.
\codefieldsection{Rate}
For every \(K,t \geq 2\), there are explicitly constructible \(K\)-dimensional Fock-state codes with \(q=N=(K-1)t(t+1)\) modes, total excitation \(N\), and bosonic distance \(t+1\); there also exist families with logical dimension \(K = o(2^N)\) and distance of \flmRefsHyperref{ref65}{order} \(o(N/\log N)\) \NoCaseChange{\protect\cite{cite500}}.
\codefieldsection{Parents}
\begin{eczvaluelist}
\item\relax
\flmRefsHyperref[eczindexfamilyrel]{code:qudits_into_oscillators}{Qudit-into-oscillator code}\item\relax
\flmRefsHyperref[eczindexfamilyrel]{code:ampdamp}{Amplitude-damping (AD) code} --- Fock-state codes are designed to protect against bosonic \flmRefsHyperref{ref498}{AD} noise.
\end{eczvaluelist}
\codefieldsection{Children}
\begin{eczvaluelist}
\item\relax
\flmRefsHyperref[eczindexfamilyrel]{code:bosonic_q-ary_expansion}{Bosonic \(q\)-ary expansion} --- The bosonic \(q\)-ary expansion allows one to map between prime-dimensional qudit states and a Fock subspace of a single mode.
\item\relax
\flmRefsHyperref[eczindexfamilyrel]{code:chi2}{\(\chi^{(2)}\) code}\item\relax
\flmRefsHyperref[eczindexfamilyrel]{code:chuang-leung-yamamoto}{Chuang-Leung-Yamamoto (CLY) code} --- Chuang-Leung-Yamamoto codes are multi-mode Fock-state codes.
\item\relax
\flmRefsHyperref[eczindexfamilyrel]{code:constant_excitation_permutation_invariant}{Ouyang-Chao constant-excitation PI code}\item\relax
\flmRefsHyperref[eczindexfamilyrel]{code:icosahedral_fock}{Icosahedral Fock-state code}\item\relax
\flmRefsHyperref[eczindexfamilyrel]{code:very-small-logical-qubit}{Very small logical qubit (VSLQ) code}\item\relax
\flmRefsHyperref[eczindexfamilyrel]{code:matrix_qm}{Matrix-model code} --- Matrix-model logical states lie in a low-energy Fock-state subspace.
\item\relax
\flmRefsHyperref[eczindexfamilyrel]{code:paircat}{Pair-cat code}\item\relax
\flmRefsHyperref[eczindexfamilyrel]{code:bosonic_rotation}{Bosonic rotation code} --- Single-mode Fock-state codes are typically rotationally invariant.
\end{eczvaluelist}
\codefieldsection{Cousins}
\begin{eczvaluelist}
\item\relax
\flmRefsHyperref[eczindexfamilyrel]{code:bits_into_bits}{Binary code} --- Fock-state code distance is a natural extension of Hamming distance between binary strings.
\item\relax
\flmRefsHyperref[eczindexfamilyrel]{code:fock_state_ook}{Fock-state OOK c-q modulation format} --- Fock-state OOK transmits classical information using Fock states, while Fock-state bosonic codes store quantum information in subspaces built from Fock states.
\item\relax
\flmRefsHyperref[eczindexfamilyrel]{code:tiger}{Tiger code} --- Tiger codes encoding logical qudits are Fock-state codes.
\item\relax
\flmRefsHyperref[eczindexfamilyrel]{code:permutation_invariant}{Permutation-invariant (PI) code} --- Modular-qudit PI codes can be converted to constant-excitation Fock-state codes via the \flmRefsHyperref{ref499}{simplex mapping} \NoCaseChange{\protect\cite[{Prop. V.2}]{cite500}}. Any transversal gates are mapped to Gaussian gates on the Fock-state codes \NoCaseChange{\protect\cite{cite500}}.
\item\relax
\flmRefsHyperref[eczindexfamilyrel]{code:t_group}{Twisted \(1\)-group code} --- Twisted \(1\)-group codes can be converted to constant-excitation Fock-state codes via the \flmRefsHyperref{ref499}{simplex mapping} \NoCaseChange{\protect\cite[{Prop. V.2}]{cite500}}. Any transversal gates are mapped to Gaussian gates on the Fock-state codes \NoCaseChange{\protect\cite{cite500}}.
\end{eczvaluelist}
\eczhbkcontributors{ \eczhuVVA }
\endeczcode

\eczcode{gkp-cluster-state}{GKP CV-cluster-state code}{~\NoCaseChange{\protect\cite{cite508}}}
\codefieldsection{Alternative Names}
\begin{eczvaluelist}
\item\relax Hybrid cluster-state code
\end{eczvaluelist}
\eczhIndexCodeAliasName{gkp-cluster-state}{Hybrid cluster-state code}
\codefieldsection{Description}
A cluster-state code that utilizes a generalized analog cluster state with some of its physical modes initialized in GKP (resource) states.
Alternatively, it can be thought of as a multimode GKP code whose encoding consists of initializing \(k\) modes in momentum states (or, in the normalizable case, squeezed vacua), \(n-k\) modes in (normalizable) GKP states, and applying a Gaussian circuit consisting of two-body gates \(e^{i V_{jk} \hat{x}_j \hat{x}_k }\) for some angles \(V_{jk}\).
The code provides a way to perform fault-tolerant MBQC, with the required number \(n-k\) of GKP-encoded physical modes determined by the particular protocol \NoCaseChange{\protect\cite{cite508,cite509,cite415,cite510}}.

\codefieldsection{Encoding}
\begin{eczvaluelist}
\item\relax Initializing \(k\) modes in momentum states (or, in the normalizable case, squeezed vacua), \(n-k\) modes in (normalizable) GKP states, and applying a Gaussian circuit consisting of two-body gates \(e^{i V_{jk} \hat{x}_j \hat{x}_k }\) for some angles \(V_{jk}\).
\end{eczvaluelist}
\codefieldsection{Gates}
\begin{eczvaluelist}
\item\relax Logical Clifford gates are performed on the cluster state via a combination of linear-optical gates and homodyne measurements on subsets of vertices \NoCaseChange{\protect\cite{cite4682,cite4683}}. Magic-state distillation is required for universal computation.
\item\relax Single-mode logical Clifford gates can be performed using Gaussian operations and measurements on a 1D GKP cluster state, while two-mode logical Clifford gates require a 2D cluster state. Magic-state distillation using photon-counting can be used for a non-Clifford logical \(\pi/8\) gate.
\item\relax Gate teleportation and error correction can be performed without active squeezing \NoCaseChange{\protect\cite{cite4882}}.
\end{eczvaluelist}
\codefieldsection{Decoding}
\begin{eczvaluelist}
\item\relax GKP error correction can be naturally combined with CV MBQC protocols since the performance of both is quantified by a squeezing parameter \NoCaseChange{\protect\cite{cite508}}.
\end{eczvaluelist}
\codefieldsection{Threshold}
\begin{eczvaluelist}
\item\relax A lower bound on the squeezing required to obtain a particular error rate can be formulated in terms of the displacement noise strength. This in turn determines how much squeezing is required in order to be below \flmRefsHyperref{ref515}{threshold} for a particular concatenated code. For a qubit-level fault-tolerance threshold of \(10^{-6}\), the required squeezing is above 20.5 dB \NoCaseChange{\protect\cite{cite508}}. Anti-squeezing does not affect this threshold estimate \NoCaseChange{\protect\cite{cite4883}}.
\item\relax For topologically protected MBQC on 3D cluster states built from finitely squeezed GKP qubits, analog QEC together with postselected measurements during cluster-state construction reduces the required squeezing to \(9.8\) dB \NoCaseChange{\protect\cite{cite415}}.
\end{eczvaluelist}
\codefieldsection{Realizations}
\begin{eczvaluelist}
\item\relax A distance-two repetition code using a GKP CV cluster state consisting of squeezed states and GKP states has been implemented in a photonic device by Xanadu \NoCaseChange{\protect\cite{cite4691}}.
\end{eczvaluelist}
\codefieldsection{Parents}
\begin{eczvaluelist}
\item\relax
\flmRefsHyperref[eczindexfamilyrel]{code:gkp-stabilizer}{Oscillator-into-oscillator GKP code} --- A GKP CV-cluster-state code can be created by initializing \(k\) modes in momentum states (or, in the normalizable case, squeezed vacua), \(n-k\) modes in (normalizable) GKP states, and applying a Gaussian circuit consisting of two-body \(e^{i V_{jk} \hat{x}_j \hat{x}_k }\) for some angles \(V_{jk}\).
\item\relax
\flmRefsHyperref[eczindexfamilyrel]{code:qudits_into_oscillators}{Qudit-into-oscillator code}\end{eczvaluelist}
\codefieldsection{Cousins}
\begin{eczvaluelist}
\item\relax
\flmRefsHyperref[eczindexfamilyrel]{code:cv_cluster_state}{Analog cluster-state code} --- GKP CV-cluster-state codes reduce to analog-cluster-state codes when all physical modes are initialized in momentum states.
\item\relax
\flmRefsHyperref[eczindexfamilyrel]{code:cluster_state}{Cluster-state code} --- GKP CV-cluster-state codes reduce to cluster-state codes concatenated with single-mode GKP codes \NoCaseChange{\protect\cite{cite415}} when all physical modes are initialized in GKP states.
\item\relax
\flmRefsHyperref[eczindexfamilyrel]{code:gkp_concatenated}{Concatenated GKP code} --- GKP CV-cluster-state codes reduce to cluster-state codes concatenated with single-mode GKP codes \NoCaseChange{\protect\cite{cite415}} when all physical modes are initialized in GKP states. Analog QEC with GKP codes concatenated with surface-code and 3D-cluster-state fault-tolerant schemes can reach a threshold displacement standard deviation of about \(0.607\) for ideal syndrome measurements and reduce the squeezing required for topologically protected MBQC to \(9.8\) dB \NoCaseChange{\protect\cite{cite415}}.
\item\relax
\flmRefsHyperref[eczindexfamilyrel]{code:distance_balanced}{Distance-balanced code} --- \flmRefsHyperref{ref491}{Weight reduction} has been studied in the context of GKP CV-cluster-state codes \NoCaseChange{\protect\cite{cite4438}}.
\end{eczvaluelist}
\eczhbkcontributors{ \eczhuVVA }
\endeczcode

\eczcode{gkp_surface_concatenated}{GKP-surface code}{~\NoCaseChange{\protect\cite{cite415,cite416}}}
\codefieldsection{Description}
A concatenated code whose outer code is a GKP code and whose inner code is a surface code, including toric surface-code variants \NoCaseChange{\protect\cite{cite415,cite416}}, rotated surface codes \NoCaseChange{\protect\cite{cite417,cite418,cite419,cite420}}, and XZZX surface codes \NoCaseChange{\protect\cite{cite421}}.

\codefieldsection{Rate}
The error threshold under ML decoding of GKP-rotated-surface codes comes close to \(\sigma\approx 0.6065\), at which the best-known lower bound \NoCaseChange{\protect\cite{cite4765}} on the capacity vanishes \NoCaseChange{\protect\cite{cite3323}}.
\codefieldsection{Decoding}
\begin{eczvaluelist}
\item\relax Minimum-energy and random-plaquette-gauge-model decoders for the toric-GKP code \NoCaseChange{\protect\cite[{Secs. IV-V}]{cite416}}.
\item\relax MWPM closest point decoder \NoCaseChange{\protect\cite{cite420}}.
\end{eczvaluelist}
\codefieldsection{Code Capacity Threshold}
\begin{eczvaluelist}
\item\relax \(0.55\) (\(0.54\)) threshold displacement standard deviation for GKP-toric (GKP-surface) codes without using GKP analog information \NoCaseChange{\protect\cite{cite415}\protect\cite[{Sec. IV.B}]{cite416}}. Using the continuous GKP syndrome information raises the GKP-toric threshold to \(\sigma_0\approx 0.6\), corresponding to a qubit error rate of about \(14\%\) \NoCaseChange{\protect\cite[{Sec. IV.B}]{cite416}}.
\item\relax Analog QEC on GKP-surface codes with ideal syndrome measurements yields a threshold displacement standard deviation of about \(0.607\), close to the hashing bound for the Gaussian quantum channel \NoCaseChange{\protect\cite{cite415}}.
\item\relax \(0.67\) threshold displacement standard deviation for GKP-XZZX-surface code \NoCaseChange{\protect\cite{cite421}}.
\item\relax \(0.602\) threshold displacement standard deviation for GKP-surface codes with analog side information using MWPM closest point decoder \NoCaseChange{\protect\cite{cite420}}.
\end{eczvaluelist}
\codefieldsection{Threshold}
\begin{eczvaluelist}
\item\relax The ML decoding problem for the toric-GKP code maps to a 3D compact QED model in the presence of a quenched random gauge field \NoCaseChange{\protect\cite[{Secs. V.B-V.C}]{cite416}}. A decoder based on this mapping yields a threshold displacement standard deviation of \(\sigma_0\approx 0.243\) when toric-code measurements, data errors, and GKP ancilla errors are all noisy \NoCaseChange{\protect\cite[{Sec. V.D.3}]{cite416}}, but this noise model did not properly take into account error propagation \NoCaseChange{\protect\cite{cite417}}.
\item\relax \(11.2\)dB of squeezing under displacement noise using MWPM decoding for GKP-rotated-surface codes \NoCaseChange{\protect\cite{cite417,cite419}}. The error threshold under ML decoding of GKP-rotated-surface codes comes close to \(\sigma\approx 0.6065\), at which the best-known lower bound \NoCaseChange{\protect\cite{cite4765}} on the capacity vanishes \NoCaseChange{\protect\cite{cite3323}}.
\end{eczvaluelist}
\codefieldsection{Parents}
\begin{eczvaluelist}
\item\relax
\flmRefsHyperref[eczindexfamilyrel]{code:gkp_concatenated}{Concatenated GKP code}\item\relax
\flmRefsHyperref[eczindexfamilyrel]{code:2d_stabilizer}{2D lattice stabilizer code}\end{eczvaluelist}
\codefieldsection{Cousins}
\begin{eczvaluelist}
\item\relax
\flmRefsHyperref[eczindexfamilyrel]{code:toric}{Toric code} --- GKP codes have been concatenated with toric codes \NoCaseChange{\protect\cite{cite416}}.
\item\relax
\flmRefsHyperref[eczindexfamilyrel]{code:rotated_surface}{Rotated surface code} --- GKP codes have been concatenated with rotated surface codes \NoCaseChange{\protect\cite{cite417,cite418,cite419,cite420}}.
\item\relax
\flmRefsHyperref[eczindexfamilyrel]{code:xzzx}{XZZX surface code} --- GKP codes have been concatenated with XZZX surface codes \NoCaseChange{\protect\cite{cite421}}.
\item\relax
\flmRefsHyperref[eczindexfamilyrel]{code:analog_surface}{Analog surface code} --- \flmRefsHyperref{ref410}{Condensing} pure fluxes and charges in the analog surface code yields toric-GKP codes \NoCaseChange{\protect\cite{cite411}}.
\end{eczvaluelist}
\eczhbkcontributors{ \eczhuVVA }
\endeczcode

\eczcode{multimodegkp}{Gottesman-Kitaev-Preskill (GKP) code}{~\NoCaseChange{\protect\cite{cite513,cite4764}}}
\codefieldsection{Description}
Quantum lattice code for a non-degenerate lattice, thereby admitting a finite-dimensional logical subspace.
Codes on \(n\) modes can be constructed from lattices with \(2n\)-dimensional full-rank Gram matrices \(A\).
Any GKP code can be generated from a Gram matrix in standard form via a Gaussian unitary transformation \NoCaseChange{\protect\cite[{Corr. 1}]{cite511}}.

The centralizer for the stabilizer group within the displacement operators case can be identified with the symplectic dual lattice \({\mathcal{L}}^{\perp}\) (i.e. all points in \(\mathbb{R}^{2n}\) that have integer symplectic inner product with all points in \({\mathcal{L}}\) ), such that logical operations are identified with the dual quotients \({\mathcal{L}}^{\perp}/{\mathcal{L}}\). The size of this dual quotient is the determinant of the Gram matrix, yielding the logical dimension \(d=\sqrt{\| \det{A}\|}\) \NoCaseChange{\protect\cite{cite513}}.
Stabilizer generator matrices equivalent under symplectic transformations are classified by distinct Hermite normal forms \NoCaseChange{\protect\cite{cite4884}}.

The space of all single-mode GKP codes is the moduli space of elliptic curves, i.e., the three sphere with a trefoil knot removed \NoCaseChange{\protect\cite{cite4885}}.

\codefieldsection{Protection}
The level of protection against displacement errors is quantified by the Euclidean code distance \(\Delta=\min_{x\in {\mathcal{L}}^{\perp}\setminus {\mathcal{L}}} \|x\|_2\) \NoCaseChange{\protect\cite{cite4884}}. There are upper bounds on this distance \NoCaseChange{\protect\cite{cite4884,cite4886}}.
\codefieldsection{Rate}
Transmission schemes with multimode GKP codes achieve a lower bound on displacement noise and a lower bound on the thermal-noise Gaussian channel capacities \NoCaseChange{\protect\cite{cite4764,cite4887,cite4888,cite2608}}. 
Particular random lattice families of multimode GKP codes achieve the hashing bound of the displacement noise channel \NoCaseChange{\protect\cite{cite4764}}.
Particular families of GKP codes achieve the capacity of \flmRefsHyperref{ref498}{AD} and amplification channels for some loss rates \NoCaseChange{\protect\cite{cite4858}}.

\codefieldsection{Encoding}
\begin{eczvaluelist}
\item\relax GKP codes with fixed \(n\) and prime-dimensional logical Hilbert space are symplectically related to a disjoint product of single-mode GKP codes on \(n\) modes, such that encoding via Gaussian unitaries is possible.
\item\relax Dissipative stabilization of finite-energy GKP states using stabilizers conjugated by \textit{cooling} (\NoCaseChange{\protect\cite{cite508}}, Appx. B) or \textit{damping} operator, i.e., a damped exponential of the total occupation number \NoCaseChange{\protect\cite{cite4889,cite482}}.
\item\relax Logical Bell state can be created from two canonical GKP states by applying a beamsplitter \NoCaseChange{\protect\cite{cite4882}}.
\end{eczvaluelist}
\codefieldsection{Gates}
\begin{eczvaluelist}
\item\relax Gaussian operations and homodyne measurements on non-magic GKP states are classically simulable \NoCaseChange{\protect\cite{cite4890}}, and there is a sufficient condition for an additional element to achieve universal quantum computation \NoCaseChange{\protect\cite{cite4891}}. There is an algorithm for GKP circuit simulation whose runtime scales with the amount of negativity of the Zak-Gross Wigner function \NoCaseChange{\protect\cite{cite4892}}.
\item\relax There is a relation between magic (i.e., how far away a state is from being a stabilizer state) and non-Gaussianity for GKP codewords \NoCaseChange{\protect\cite{cite4893,cite4890}}. In particular, implementing a non-Clifford logical gate requires a higher degree of non-Gaussianity than that expressed by ideal non-normalizable GKP states \NoCaseChange{\protect\cite{cite4890}}.
\item\relax By applying GKP error correction to Gaussian input states, computational universality can be achieved without additional non-Gaussian elements  \NoCaseChange{\protect\cite{cite4894}}. This procedure can be alternatively described as performing heterodyne detection on one half of a GKP encoded Bell pair. The cubic phase gate is not a suitable gate \NoCaseChange{\protect\cite{cite4895}}.
\item\relax Logical shadow tomography protocol \NoCaseChange{\protect\cite{cite4896}}.
\item\relax Some gates have a constant logical gate error even in the limit of infinite squeezing \NoCaseChange{\protect\cite{cite4897}}.
\end{eczvaluelist}
\codefieldsection{Decoding}
\begin{eczvaluelist}
\item\relax Syndrome extraction is performed by measuring stabilizers and correcting. Issues arising from the use of finite-energy approximate states can be mitigated \NoCaseChange{\protect\cite{cite4898}}.
\item\relax The MLD decoder for Gaussian displacement errors is realized by evaluating a lattice theta function, and in general the decision can be approximated by either solving (approximating) the closest vector problem (CVP) \NoCaseChange{\protect\cite{cite4899}} (a.k.a. closest lattice point problem) or by using other effective iterative schemes when, e.g., the lattice represents a concatenated GKP code \NoCaseChange{\protect\cite{cite416,cite417,cite4884,cite1556}}. While the decoder time scales exponentially with number of modes \(n\) generically, the time can be polynomial in \(n\) for certain codes \NoCaseChange{\protect\cite{cite420}}.
\item\relax Babai's nearest plane algorithm \NoCaseChange{\protect\cite{cite4900}} can be used for bounded-distance decoding \NoCaseChange{\protect\cite{cite420}}.
\item\relax Combining \flmRefsHyperref{ref498}{AD} noise with amplification yields displacement noise, the noise that GKP codes are designed to correct \NoCaseChange{\protect\cite{cite4901,cite2608}}.
\item\relax ML decoder for correcting shift errors in GKP two-qubit gates \NoCaseChange{\protect\cite{cite419}}.
\end{eczvaluelist}
\codefieldsection{Fault Tolerance}
\begin{eczvaluelist}
\item\relax Logical Clifford operations are given by Gaussian unitaries, which map bounded-size errors to bounded-size errors \NoCaseChange{\protect\cite{cite513}}. For single-mode GKP codes, these operations correspond to non-trivial loops in the space of all single-mode GKP codes (the moduli space of elliptic curves, i.e., the three sphere with a trefoil knot removed) \NoCaseChange{\protect\cite{cite4885}}. Such gates provide another example of monodromy under the particular notion of parallel transport introduced in Ref. \NoCaseChange{\protect\cite{cite809}}.
\item\relax The 4D square-lattice GKP code admits the isthmus property, which allows certain ancilla errors to be detectable \NoCaseChange{\protect\cite{cite482}}.
\end{eczvaluelist}
\codefieldsection{Notes}
\begin{eczvaluelist}
\item\relax Reviews on GKP codes presented in Refs. \NoCaseChange{\protect\cite{cite4819,cite4902,cite4903,cite511,cite4904}}.
\end{eczvaluelist}
\codefieldsection{Parent}
\begin{eczvaluelist}
\item\relax
\flmRefsHyperref[eczindexfamilyrel]{code:quantum_lattice}{Quantum lattice code} --- GKP codes are \(n\)-mode quantum lattice codes with \(2n\) stabilizers, i.e., constructed using a non-degenerate lattice.
\end{eczvaluelist}
\codefieldsection{Children}
\begin{eczvaluelist}
\item\relax
\flmRefsHyperref[eczindexfamilyrel]{code:chern_simons_gkp}{\(U(1)_{2n} \times U(1)_{-2m}\) Chern-Simons GKP code}\item\relax
\flmRefsHyperref[eczindexfamilyrel]{code:gkp}{Square-lattice GKP code}\item\relax
\flmRefsHyperref[eczindexfamilyrel]{code:gkp_concatenated}{Concatenated GKP code}\item\relax
\flmRefsHyperref[eczindexfamilyrel]{code:hexagonal_gkp}{Hexagonal GKP code}\item\relax
\flmRefsHyperref[eczindexfamilyrel]{code:ntru_gkp}{NTRU-GKP code}\end{eczvaluelist}
\codefieldsection{Cousins}
\begin{eczvaluelist}
\item\relax
\flmRefsHyperref[eczindexfamilyrel]{code:t-designs}{\(t\)-design} --- GKP states on \(n\) modes and their displaced versions for all possible lattices form a rigged 2-design for all \(n\) \NoCaseChange{\protect\cite{cite933}}.
\item\relax
\flmRefsHyperref[eczindexfamilyrel]{code:holographic}{Holographic code} --- GKP codespaces exist in the CFT dual of a particular holographic framework \NoCaseChange{\protect\cite{cite2854,cite2855}}.
\item\relax
\flmRefsHyperref[eczindexfamilyrel]{code:4d_stabilizer}{4D lattice stabilizer code} --- The 4D square-lattice GKP code admits the isthmus property, which allows certain ancilla errors to be detectable \NoCaseChange{\protect\cite{cite482}}.
\item\relax
\flmRefsHyperref[eczindexfamilyrel]{code:qam}{Quadrature-amplitude modulation (QAM) format} --- Finite-energy GKP codes are quantum counterparts of lattice-based QAM codes in that both use a subset of points on a lattice.
\item\relax
\flmRefsHyperref[eczindexfamilyrel]{code:numopt}{Numerically optimized bosonic code} --- Numerically optimizing GKP code lattices yields codes for three, seven, and nine modes with larger distances and fidelities than known GKP codes \NoCaseChange{\protect\cite{cite420}}. Neural networks can be used to optimize approximate GKP states \NoCaseChange{\protect\cite{cite4905}}.
\item\relax
\flmRefsHyperref[eczindexfamilyrel]{code:qutrit_pauli_gkp_subcode}{Qutrit-Pauli tessellation code} --- The qutrit-Pauli tessellation code is a subcode of a two-mode GKP code with GKP-like stabilizers \NoCaseChange{\protect\cite{cite2811}}.
\end{eczvaluelist}
\eczhbkcontributors{ Jonathan Conrad, \eczhuVVA }
\endeczcode

\eczcode{hnss}{Hayden-Nezami-Salton-Sanders bosonic code}{~\NoCaseChange{\protect\cite{cite2172}}}
\codefieldsection{Description}
An \(\llbracket n,1\rrbracket _{\mathbb{R}}\) analog CSS code defined using homological structures associated with an \(n-1\) simplex. Relevant to the study of spacetime replication of quantum information \NoCaseChange{\protect\cite{cite512}}.

Stabilizer generators are defined by two orthogonal subspaces of \(C_1\) in the chain complex. \(C_X = \partial_2 C_2\) and \(C_P = \partial_1^T Q\) for some \(Q \subset C_0\). The standard approach would use \(Q = C_0\), which would mean the logical dimension would be the dimension of the 1st cohomology group \(H^1\). However, \(H^1\) is trivial for the \(n-1\) simplex, so one chooses \(Q \neq C_0\) such that exactly one stabilizer is removed, yielding a stabilizer code instead of a single stabilized state.

\codefieldsection{Protection}
Protects against certain types of erasure errors (depending on the specific dimension). Certain constructions also protect arbitrarily sized errors on multiple-photon states.
\codefieldsection{Encoding}
\begin{eczvaluelist}
\item\relax Encoding depends on the specific dimension, but can generally be done using generalized conditional-rotation and Fourier-transform gates.
\end{eczvaluelist}
\codefieldsection{Decoding}
\begin{eczvaluelist}
\item\relax Decoding requires a different circuit for each possible erasure error, with no general circuit decoding any possible erasure error. Every circuit relies on a generalized conditional rotation, which Ref. \NoCaseChange{\protect\cite{cite2172}} calls the \textit{QND Gate} and which is defined as \(QND_c | x , y \rangle = |x + c y, y \rangle\).
\end{eczvaluelist}
\codefieldsection{Notes}
\begin{eczvaluelist}
\item\relax Proposed experimental optical procedure for realizing the simplest non-trivial code with 5 modes \NoCaseChange{\protect\cite{cite2172}}.
\end{eczvaluelist}
\codefieldsection{Parents}
\begin{eczvaluelist}
\item\relax
\flmRefsHyperref[eczindexfamilyrel]{code:analog_stabilizer}{Analog stabilizer code}\item\relax
\flmRefsHyperref[eczindexfamilyrel]{code:oscillator_css}{Bosonic CSS code}\end{eczvaluelist}
\codefieldsection{Cousins}
\begin{eczvaluelist}
\item\relax
\flmRefsHyperref[eczindexfamilyrel]{code:generalized_homological_product_css}{Generalized homological-product CSS code} --- Hayden-Nezami-Salton-Sanders codes utilize chain complexes in code construction, but the complexes have trivial homology.
\item\relax
\flmRefsHyperref[eczindexfamilyrel]{code:niset_andersen_cerf}{Niset-Andersen-Cerf code} --- The Niset-Andersen-Cerf code can be viewed as a scheme to replicate quantum information in multiple regions \NoCaseChange{\protect\cite{cite2172}}.
\item\relax
\flmRefsHyperref[eczindexfamilyrel]{code:spacetime}{Spacetime code (STC)} --- Hayden-Nezami-Salton-Sanders codes have been considered in the context of spacetime replication of quantum data \NoCaseChange{\protect\cite{cite512,cite2172}}, while STCs are designed to replicate classical data.
\end{eczvaluelist}
\eczhbkcontributors{ Siddharth Taneja, \eczhuVVA }
\endeczcode

\eczcode{hessian_qsc}{Hessian QSC}{~\NoCaseChange{\protect\cite{cite382}}}
\codefieldsection{Description}
Quantum spherical code encoding a logical qubit, with each codeword an equal superposition of vertices of a Hessian complex polyhedron.

For the unit sphere, the codewords are
\flmMathEnvironment{align}{}{
  |\overline{0}\rangle &= \frac{1}{\sqrt{27}}\left( \sum_{\mu,\nu=0}^{2} |0,\omega^{\mu},-\omega^{\nu}\rangle + |-\omega^{\nu},0,\omega^{\mu}\rangle + |\omega^{\mu},-\omega^{\nu},0\rangle   \right) \\
  |\overline{1}\rangle &= \frac{1}{\sqrt{27}}\left( \sum_{\mu,\nu=0}^{2} |0,-\omega^{\mu},\omega^{\nu}\rangle + |\omega^{\nu},0,-\omega^{\mu}\rangle + |-\omega^{\mu},\omega^{\nu},0\rangle   \right)~,
}
where \(\omega = e^{\frac{2\pi i}{3}}\).
\begin{flmFloat}{figure}{NumCap}\includegraphics[width=324bp,max width=\linewidth]{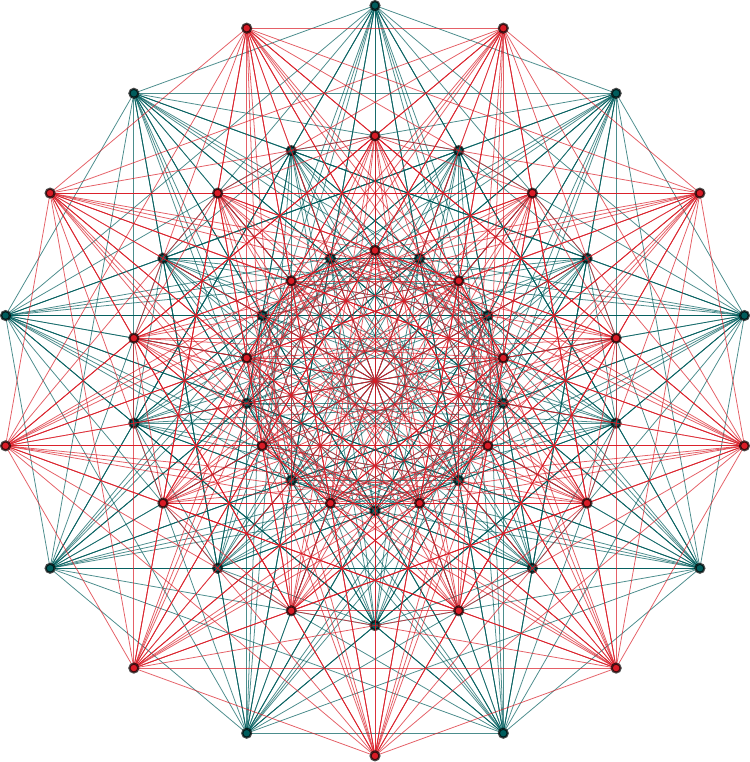}\caption{Projection of the \textit{double Hessian} code constellation with each copy of the \textit{Hessian} logical constellation marked in a different colour.}\label{ref4906}\end{flmFloat}

\codefieldsection{Protection}
The Hessian QSC is a \(\langle 4, 5, 9 \rangle\) code, i.e. it detects 8 \flmRefsHyperref{ref498}{photon losses} and protects against 3. The code also detects up to 4 ladder errors (losses or gains). The code resolution \( d_E = 1.0\).
\codefieldsection{Parent}
\begin{eczvaluelist}
\item\relax
\flmRefsHyperref[eczindexfamilyrel]{code:qsc}{Quantum spherical code (QSC)} --- The Hessian QSC is an example of a QSC with logical constellation built from the Hessian complex polyhedron.
\end{eczvaluelist}
\codefieldsection{Cousin}
\begin{eczvaluelist}
\item\relax
\flmRefsHyperref[eczindexfamilyrel]{code:hessian_polyhedron}{Hessian polyhedron code} --- Each codeword of the Hessian QSC is a quantum superposition of vertices of a Hessian complex polyhedron.
\end{eczvaluelist}
\eczhbkcontributors{ Shubham P. Jain, \eczhuVVA }
\endeczcode

\eczcode{hexagonal_gkp}{Hexagonal GKP code}{~\NoCaseChange{\protect\cite[{Sec. VI}]{cite513}}}
\codefieldsection{Description}
Single-mode GKP qudit-into-oscillator code based on the triangular lattice. Offers the best error correction against displacement noise in a single mode due to the optimal packing of the underlying lattice \NoCaseChange{\protect\cite[{Sec. VI}]{cite513}}.

\codefieldsection{Realizations}
\begin{eczvaluelist}
\item\relax Microwave cavity coupled to superconducting circuits: reduced form of GKP error correction, where displacement error syndromes are measured to one bit of precision using an ancillary transmon \NoCaseChange{\protect\cite{cite4907}}.
\end{eczvaluelist}
\codefieldsection{Notes}
\begin{eczvaluelist}
\item\relax Hexagonal GKP codes were obtained after iterative numerical optimization of encoding and recovery against \flmRefsHyperref{ref498}{photon loss}, starting with Haar-random states \NoCaseChange{\protect\cite{cite2608}}.
\end{eczvaluelist}
\codefieldsection{Parents}
\begin{eczvaluelist}
\item\relax
\flmRefsHyperref[eczindexfamilyrel]{code:multimodegkp}{Gottesman-Kitaev-Preskill (GKP) code}\item\relax
\flmRefsHyperref[eczindexfamilyrel]{code:single-mode}{Single-mode bosonic code}\end{eczvaluelist}
\codefieldsection{Cousins}
\begin{eczvaluelist}
\item\relax
\flmRefsHyperref[eczindexfamilyrel]{code:hexagonal}{\(A_2\) triangular lattice} --- The hexagonal GKP code is based on the triangular lattice.
\item\relax
\flmRefsHyperref[eczindexfamilyrel]{code:gkp-stabilizer}{Oscillator-into-oscillator GKP code} --- Hexagonal GKP codes may be optimal for oscillator-into-oscillator GKP codes utilizing one ancilla mode \NoCaseChange{\protect\cite{cite4668}}.
\end{eczvaluelist}
\eczhbkcontributors{ \eczhuVVA }
\endeczcode

\eczcode{homological_number-phase}{Homological number-phase code}{~\NoCaseChange{\protect\cite{cite2699}}}
\codefieldsection{Description}
A \flmRefsHyperref{code:homological_rotor}{homological \(n\)-rotor code} mapped into the Fock-state space of \(n\) oscillators by identifying non-negative rotor angular-momentum states with oscillator Fock states.
The resulting oscillator code can encode logical rotors or qudits due to the presence of torsion in the chain complex defining the original rotor code.
These codes are tailored to settings in which photon loss is present but random rotations, i.e., dephasing, are the dominant noise mechanism \NoCaseChange{\protect\cite{cite2699}}.

Due to the absence of negative Fock states, a given homological rotor code first has to be rotated such that it has non-trivial support in the all-positive momentum orthant.
This can be done by flipping the signs of the angular momenta of some of the rotors \NoCaseChange{\protect\cite[{Prop. 1}]{cite2699}}.
Ideal codewords are not normalizable, and approximate versions have to be constructed.

Since homological rotor codes use an extension of the \flmRefsHyperref{ref683}{qubit CSS-to-homology correspondence} to rotors, the mapping into oscillators makes such homological encodings possible for oscillators.

\codefieldsection{Protection}
The homology group of the logical operators has a torsion component because the chain complexes are defined over the ring of integers, which yields codes with finite logical dimension.
Products of chain complexes can also yield rotor codes.

The distances of the original homological rotor code are preserved, although the resulting number-phase code is approximately error-correcting due to the non-orthogonality of Pegg-Barnett phase states \NoCaseChange{\protect\cite{cite501}}, which act as the angular position states in the number-phase interpretation of the oscillator.

\codefieldsection{Parent}
\begin{eczvaluelist}
\item\relax
\flmRefsHyperref[eczindexfamilyrel]{code:oscillators}{Bosonic code} --- Homological number-phase codes are bosonic codes encoding logical qudits and/or logical rotors.
\end{eczvaluelist}
\codefieldsection{Cousins}
\begin{eczvaluelist}
\item\relax
\flmRefsHyperref[eczindexfamilyrel]{code:homological_rotor}{Homological rotor code} --- Homological number-phase codes can be thought of as homological rotor codes but whose underlying rotors consist of the number and phase degrees of freedom of physical modes.
\item\relax
\flmRefsHyperref[eczindexfamilyrel]{code:oscillator_stabilizer}{Bosonic stabilizer code} --- Homological number-phase codewords span the joint right eigenspace of powers of the non-unitary Susskind–Glogower phase operators and unitary bosonic rotation operators.
\item\relax
\flmRefsHyperref[eczindexfamilyrel]{code:gkp-stabilizer}{Oscillator-into-oscillator GKP code} --- Homological number-phase codes are finite-dimensional cousins of number-phase-rotor GKP-stabilizer codes: both use number-phase resource states and Clifford-semigroup encoders to protect oscillator information against photon loss and dephasing \NoCaseChange{\protect\cite[{Secs. 6-7}]{cite2699}}.
\item\relax
\flmRefsHyperref[eczindexfamilyrel]{code:number_phase}{Number-phase code} --- Homological number-phase codes are multi-mode generalizations of number-phase codes, obtained by projecting suitably parity-flipped homological rotor codes onto the non-negative angular-momentum orthant \NoCaseChange{\protect\cite[{Prop. 1}]{cite2699}}.
\item\relax
\flmRefsHyperref[eczindexfamilyrel]{code:generalized_homological_product_css}{Generalized homological-product CSS code} --- Homological number-phase codes are non-stabilizer codes constructed from chain complexes over the integers. The homology group of the logical operators has a torsion component because the chain complexes are defined over the ring of integers, which yields codes with finite logical dimension.
\item\relax
\flmRefsHyperref[eczindexfamilyrel]{code:rotor_4_2_2}{Four-rotor code} --- After suitable rotor-parity flips and projection onto the non-negative angular-momentum orthant, the four-rotor current-mirror code yields a homological number-phase code \NoCaseChange{\protect\cite[{Ex. 4}]{cite2699}}.
\item\relax
\flmRefsHyperref[eczindexfamilyrel]{code:tiger}{Tiger code} --- Tiger codes of infinite Fock-state support can be thought of as appropriately regularized homological number-phase codes \NoCaseChange{\protect\cite{cite4667}}.
\end{eczvaluelist}
\eczhbkcontributors{ \eczhuVVA }
\endeczcode

\eczcode{icosahedral_fock}{Icosahedral Fock-state code}{~\NoCaseChange{\protect\cite{cite500}}}
\codefieldsection{Description}
A constant-excitation Fock-state code designed to realize the \(2I\) group of gates using Gaussian rotations.
It is obtained from the corresponding icosahedral spin code via the \flmRefsHyperref{ref499}{simplex mapping} between spin and constant-excitation Fock spaces \NoCaseChange{\protect\cite{cite500}}.

The code has unnormalized logical states
\flmMathEnvironment{align}{}{
  \begin{split}
    |0_{L}\rangle&\propto\sqrt{3}|07\rangle+\sqrt{7}|52\rangle\\
    |1_{L}\rangle&\propto\sqrt{7}|25\rangle-\sqrt{3}|70\rangle\,.
  \end{split}
}

\codefieldsection{Parents}
\begin{eczvaluelist}
\item\relax
\flmRefsHyperref[eczindexfamilyrel]{code:fock_state}{Fock-state bosonic code}\item\relax
\flmRefsHyperref[eczindexfamilyrel]{code:constant_excitation}{Constant-excitation (CE) code}\item\relax
\flmRefsHyperref[eczindexfamilyrel]{code:group_representation}{Group-representation code} --- Icosahedral Fock-state codes are group-representation codes with the \(G = 2I\) subgroup of Gaussian rotations \NoCaseChange{\protect\cite{cite646}}.
\end{eczvaluelist}
\codefieldsection{Cousins}
\begin{eczvaluelist}
\item\relax
\flmRefsHyperref[eczindexfamilyrel]{code:icosahedral_spin}{Icosahedral spin code} --- The icosahedral spin code maps to the icosahedral Fock-state code via the \flmRefsHyperref{ref499}{simplex mapping} \NoCaseChange{\protect\cite{cite500}}.
\item\relax
\flmRefsHyperref[eczindexfamilyrel]{code:icosahedral_permutation_invariant}{\(\llparenthesis 7,2,3\rrparenthesis \) Pollatsek-Ruskai code} --- The \(\llparenthesis 7,2,3\rrparenthesis \) Pollatsek-Ruskai code maps to the icosahedral Fock-state code via the \flmRefsHyperref{ref499}{simplex mapping} \NoCaseChange{\protect\cite{cite500}}.
\end{eczvaluelist}
\eczhbkcontributors{ \eczhuVVA }
\endeczcode

\eczcode{homological_cv}{Integer-homology bosonic CSS code}{~\NoCaseChange{\protect\cite{cite411}}}
\codefieldsection{Description}
A bosonic stabilizer code whose physical modes have been restricted, via a single GKP stabilizer, from the space of functions on the real line to the space of periodic functions.
This restriction effectively realizes a rotor on each physical mode, allowing one to construct homological rotor codes out of displacement stabilizer groups.
The stabilizer group is continuous, but contains discrete components in the form of the single-mode GKP stabilizers.
The homology group of the logical operators has a torsion component because the chain complexes are defined over the ring of integers, which yields codes with finite logical dimension.

\codefieldsection{Parents}
\begin{eczvaluelist}
\item\relax
\flmRefsHyperref[eczindexfamilyrel]{code:oscillator_css}{Bosonic CSS code} --- Integer-homology bosonic CSS codes are constructed from chain complexes over the integers and realize homological rotor codes out of continuous displacement stabilizer groups. The stabilizer group is continuous, but contains discrete components in the form of the single-mode GKP stabilizers.
\item\relax
\flmRefsHyperref[eczindexfamilyrel]{code:generalized_homological_product_css}{Generalized homological-product CSS code} --- Integer-homology bosonic CSS codes are constructed from chain complexes over the integers and realize homological rotor codes out of continuous displacement stabilizer groups. The homology group of the logical operators has a torsion component because the chain complexes are defined over the ring of integers, which yields codes with finite logical dimension.
\end{eczvaluelist}
\codefieldsection{Child}
\begin{eczvaluelist}
\item\relax
\flmRefsHyperref[eczindexfamilyrel]{code:compactified_r}{Compactified \(\mathbb{R}\) gauge theory code} --- The compactified \(\mathbb{R}\) gauge theory code realizes \(U(1)\) gauge theory on bosonic modes.
\end{eczvaluelist}
\codefieldsection{Cousin}
\begin{eczvaluelist}
\item\relax
\flmRefsHyperref[eczindexfamilyrel]{code:homological_rotor}{Homological rotor code} --- Integer-homology bosonic CSS codes are constructed from chain complexes over the integers and realize homological rotor codes out of continuous displacement stabilizer groups \NoCaseChange{\protect\cite{cite411}}.
\end{eczvaluelist}
\eczhbkcontributors{ \eczhuVVA }
\endeczcode

\eczcode{matrix_qm}{Matrix-model code}{~\NoCaseChange{\protect\cite{cite4908,cite2851}}}
\codefieldsection{Description}
Multimode Fock-state bosonic approximate code derived from a matrix model, i.e., a bosonic theory with a large non-Abelian gauge group.
The model's degrees of freedom are matrix-valued bosons \(a\), each consisting of \(N^2\) harmonic oscillator modes and subject to an \(SU(N)\) gauge symmetry.

A simple matrix-model code \NoCaseChange{\protect\cite{cite2851}} consists of two spatially separated bosons with codewords
\flmMathEnvironment{align}{}{
    |\mathcal{I}\rangle :=\prod_{(m,n)\in \mathcal{I} } \frac{\text{Tr}(a_1^{\dagger m}a_2^{\dagger n})}{N^{\frac{m+n}{2}}}|0\rangle_{12}~,
}
where \(\cal I\) is some set of integer two-tuples, and \(n,m\geq 0\).

Gauge symmetry is assumed to be enforced in the above model.
In other variants \NoCaseChange{\protect\cite{cite2851}}, gauge symmetry is enforced energetically, requiring an energy penalty to scale as \(\log(N)\) in order to obtain a polynomial memory lifetime below a critical temperature.

\codefieldsection{Protection}
For the spatially separated boson code, logical errors stemming from gauge-invariant physical errors are suppressed polynomially with the number of modes \(N\), as shown by the approximate error-correction conditions.
For sufficiently low temperature, the memory time scales as \(N^2\) when the model is subject to a thermal bath \NoCaseChange{\protect\cite{cite2851}}.

\codefieldsection{Parents}
\begin{eczvaluelist}
\item\relax
\flmRefsHyperref[eczindexfamilyrel]{code:fock_state}{Fock-state bosonic code} --- Matrix-model logical states lie in a low-energy Fock-state subspace.
\item\relax
\flmRefsHyperref[eczindexfamilyrel]{code:hamiltonian}{Hamiltonian-based code} --- Matrix-model codewords for simple codes are eigenstates of a matrix-model Hamiltonian.
\item\relax
\flmRefsHyperref[eczindexfamilyrel]{code:holographic}{Holographic code} --- Matrix-model codes are motivated by the AdS/CFT correspondence because it is manifest in continuous non-Abelian gauge theories with large gauge groups \NoCaseChange{\protect\cite{cite2851}}.
\end{eczvaluelist}
\codefieldsection{Cousin}
\begin{eczvaluelist}
\item\relax
\flmRefsHyperref[eczindexfamilyrel]{code:self_correct}{Self-correcting quantum code} --- Matrix-model codes are similar to self-correcting memories in the sense that memory time becomes infinite in the thermodynamic limit, but with corrections being polynomial in \(N\).
\end{eczvaluelist}
\eczhbkcontributors{ \eczhuVVA }
\endeczcode

\eczcode{ntru_gkp}{NTRU-GKP code}{~\NoCaseChange{\protect\cite{cite4909}}}
\codefieldsection{Description}
Multi-mode GKP code whose underlying lattice is utilized in variations of the NTRU cryptosystem \NoCaseChange{\protect\cite{cite288}}.
Randomized constructions yield constant-rate GKP code families whose largest decodable displacement length scales as \(O(\sqrt{n})\) with high probability.

The integer-valued \(q\)-symplectic Gram matrix for an \(n\)-mode \(k\)-qubit good NTRU-GKP code is
\flmMathEnvironment{align}{}{
  A = \sqrt{\frac{2}{q}}\begin{pmatrix}I & Q\\
  0 & qI
  \end{pmatrix}~,
}
where \(Q\) is a circulant matrix constructed from coefficients of a cyclic polynomial used in the NTRU cryptosystem, and \(I\) is the \(n\)-dimensional identity matrix \NoCaseChange{\protect\cite[{Prop. 2}]{cite4909}}.

\codefieldsection{Rate}
Randomized constructions yield constant-rate GKP code families whose largest decodable displacement length scales as \(O(\sqrt{n})\) with high probability.
\codefieldsection{Decoding}
\begin{eczvaluelist}
\item\relax Babai's nearest plane algorithm \NoCaseChange{\protect\cite{cite4900}} can be used for bounded-distance decoding.
\item\relax An NTRU-based decoder against stochastic displacement noise is efficient because the decoding problem is equivalent to decrypting the NTRU cryptosystem with knowledge of the encoder.
\end{eczvaluelist}
\codefieldsection{Code Capacity Threshold}
\begin{eczvaluelist}
\item\relax A lower bound on the threshold for displacement noise can be formulated in terms of code parameters \NoCaseChange{\protect\cite[{Appx. B}]{cite4909}}.
\end{eczvaluelist}
\codefieldsection{Realizations}
\begin{eczvaluelist}
\item\relax Public-key NTRU-based quantum communication protocol \NoCaseChange{\protect\cite{cite4909}}.
\end{eczvaluelist}
\codefieldsection{Parents}
\begin{eczvaluelist}
\item\relax
\flmRefsHyperref[eczindexfamilyrel]{code:multimodegkp}{Gottesman-Kitaev-Preskill (GKP) code}\item\relax
\flmRefsHyperref[eczindexfamilyrel]{code:qudits_into_oscillators}{Qudit-into-oscillator code}\end{eczvaluelist}
\codefieldsection{Cousin}
\begin{eczvaluelist}
\item\relax
\flmRefsHyperref[eczindexfamilyrel]{code:quantum_random}{Random quantum code} --- Several NTRU lattices come from randomized constructions, yielding constant-rate GKP code families whose largest decodable displacement length scales as \(O(\sqrt{n})\) with high probability.
\end{eczvaluelist}
\eczhbkcontributors{ \eczhuVVA }
\endeczcode

\eczcode{number_phase}{Number-phase code}{~\NoCaseChange{\protect\cite{cite4722}}}
\codefieldsection{Alternative Names}
\begin{eczvaluelist}
\item\relax Single-mode translationally invariant Fock-state code
\end{eczvaluelist}
\eczhIndexCodeAliasName{number_phase}{Single-mode translationally invariant Fock-state code}
\codefieldsection{Description}
Bosonic rotation code consisting of superpositions of Pegg-Barnett phase states \NoCaseChange{\protect\cite{cite501}}.

Pegg-Barnett phase states are expressed in terms of Fock states as
\flmMathEnvironment{align}{}{
|\phi\rangle \equiv \frac{1}{\sqrt{2\pi}}\sum_{n=0}^{\infty} \mathrm{e}^{\mathrm{i} n \phi} \ket{n}.
}
Since phase states and thus the ideal codewords are not normalizable, approximate versions need to be constructed. The codes' key feature is that, in the ideal case, phase measurement has zero uncertainty, making it a good candidate for a syndrome measurement.

Logical states of an order-\(N\) number-phase qubit encoding are \(|\overline{0}\rangle= \sum_{m=0}^{2N-1} |\phi = m\pi/N\rangle\) and \(|\overline{1}\rangle = \sum_{m=0}^{2N-1} (-1)^m |\phi=m\pi/N\rangle\). By performing the summation over \(m\), one finds that \(|\overline{0}\rangle\) is supported on Fock states \(|2kN\rangle\), while \(|\overline{1}\rangle\) is supported on states \(|(2k+1)N\rangle\), for \(k \geq 0\).

\codefieldsection{Protection}
Number-phase codes detect up to \(N\) \flmRefsHyperref{ref498}{photon loss} or gain errors, and approximately correct rotations up to \(\theta = \pi/N\).
However, the code is only approximately error-correcting due to the non-orthogonality of Pegg-Barnett phase states \NoCaseChange{\protect\cite{cite501}}, which act as the angular position states in the number-phase interpretation of the oscillator.

\codefieldsection{Decoding}
\begin{eczvaluelist}
\item\relax Modular phase measurement done in the logical \(X\), or dual, basis has zero uncertainty in the case of ideal number-phase codes. This is equivalent to a quantum measurement of the spectrum of the Susskind-Glogower phase operator \NoCaseChange{\protect\cite{cite4722}}. Approximate number-phase codes are characterized by vanishing phase uncertainty \NoCaseChange{\protect\cite{cite4722}}. Such measurements can be utilized for Knill error correction (a.k.a. telecorrection \NoCaseChange{\protect\cite{cite3185}}), which is based on teleportation \NoCaseChange{\protect\cite{cite448,cite4369}}. This type of error correction avoids the complicated correction procedures typical in Fock-state codes, but requires a supply of clean codewords \NoCaseChange{\protect\cite{cite4722}}. Performance of this method was analyzed in Ref. \NoCaseChange{\protect\cite{cite4910}}, and it was extended in Ref. \NoCaseChange{\protect\cite{cite4911}}.
\item\relax Number measurement can be done by extracting modular number information using a CROT gate \(\mathrm{e}^{(2\pi \mathrm{i} / NM) \hat n \otimes \hat n}\) and performing phase measurements \NoCaseChange{\protect\cite{cite4912,cite4913}} on an ancillary mode \NoCaseChange{\protect\cite{cite4722}}.
\end{eczvaluelist}
\codefieldsection{Fault Tolerance}
\begin{eczvaluelist}
\item\relax Fault-tolerant computation schemes with number-phase codes have been proposed based on concatenation with Bacon-Shor subsystem codes \NoCaseChange{\protect\cite{cite4722}}.
\end{eczvaluelist}
\codefieldsection{Realizations}
\begin{eczvaluelist}
\item\relax Motional degree of freedom of a trapped ion: state initialization \NoCaseChange{\protect\cite{cite4914}}.
\end{eczvaluelist}
\codefieldsection{Parent}
\begin{eczvaluelist}
\item\relax
\flmRefsHyperref[eczindexfamilyrel]{code:bosonic_rotation}{Bosonic rotation code} --- Number-phase codes are bosonic rotation codes whose primitive state is a Pegg-Barnett phase state \NoCaseChange{\protect\cite{cite501}}.
\end{eczvaluelist}
\codefieldsection{Cousins}
\begin{eczvaluelist}
\item\relax
\flmRefsHyperref[eczindexfamilyrel]{code:rotor_gkp}{Rotor GKP code} --- Number-phase codes are obtained by projecting planar-rotor GKP codes onto the non-negative angular-momentum subspace and identifying that subspace with oscillator Fock space \NoCaseChange{\protect\cite[{Ex. 3}]{cite2699}}.
\item\relax
\flmRefsHyperref[eczindexfamilyrel]{code:oscillator_stabilizer}{Bosonic stabilizer code} --- Number-phase codewords span the joint right eigenspace of the \(N\)th power of the Susskind-Glogower phase operator and the bosonic rotation operator \NoCaseChange{\protect\cite{cite4722}}. These operators no longer form a group since the phase operator is not unitary.
\item\relax
\flmRefsHyperref[eczindexfamilyrel]{code:gkp}{Square-lattice GKP code} --- Square-lattice GKP codes utilize translational symmetry in phase space, while number-phase codes utilize rotational symmetry. The two are related via a mapping \NoCaseChange{\protect\cite{cite4915}}.
\item\relax
\flmRefsHyperref[eczindexfamilyrel]{code:gkp-stabilizer}{Oscillator-into-oscillator GKP code} --- Number-phase codes can serve as resource states for number-phase-rotor GKP-stabilizer codes, a polar analogue of oscillator-into-oscillator GKP codes that protects against photon loss and dephasing \NoCaseChange{\protect\cite[{Sec. 7}]{cite2699}}.
\item\relax
\flmRefsHyperref[eczindexfamilyrel]{code:t-designs}{\(t\)-design} --- Pegg-Barnett phase states undergoing Kerr evolution, together with Fock states, form a rigged 2-design for a single mode \NoCaseChange{\protect\cite{cite886}}.
\item\relax
\flmRefsHyperref[eczindexfamilyrel]{code:binomial}{Binomial code} --- In the limit as \(N,S \to \infty\), phase measurement in the binomial code has vanishing variance, just like in a number-phase code \NoCaseChange{\protect\cite{cite4722}}.
\item\relax
\flmRefsHyperref[eczindexfamilyrel]{code:squeezed_vacuum}{Squeezed-vacuum code} --- Squeezed-vacuum codes are approximate \flmRefsHyperref{code:number_phase}{number-phase codes} whose logical Fock-space support lies in the congruence classes \(n \equiv 0 \pmod{2m}\) and \(n \equiv m \pmod{2m}\), approaching ideal \flmRefsHyperref{code:number_phase}{number-phase codes} as squeezing strength \(r \to \infty\)~\NoCaseChange{\protect\cite{cite4916}}.
\item\relax
\flmRefsHyperref[eczindexfamilyrel]{code:cat}{Cat code} --- In the limit as \(N,S \to \infty\), phase measurement in the cat code has vanishing variance, just like in a number-phase code \NoCaseChange{\protect\cite{cite4722}}. Conversely, a cat code can be thought of as an appropriately regularized number-phase code.
\item\relax
\flmRefsHyperref[eczindexfamilyrel]{code:homological_number-phase}{Homological number-phase code} --- Homological number-phase codes are multi-mode generalizations of number-phase codes, obtained by projecting suitably parity-flipped homological rotor codes onto the non-negative angular-momentum orthant \NoCaseChange{\protect\cite[{Prop. 1}]{cite2699}}.
\end{eczvaluelist}
\eczhbkcontributors{ Joseph T. Iosue, \eczhuVVA }
\endeczcode

\eczcode{numopt}{Numerically optimized bosonic code}{~\NoCaseChange{\protect\cite{cite4707,cite496}}}
\codefieldsection{Description}
Bosonic Fock-state code obtained from a numerical minimization procedure, e.g., from enforcing error-correction criteria against some number of losses while minimizing average occupation number. Useful single-mode codes can be determined using basic numerical optimization \NoCaseChange{\protect\cite{cite4707,cite496}}, semidefinite-program recovery/encoding optimization \NoCaseChange{\protect\cite{cite2608,cite4917}}, or reinforcement learning \NoCaseChange{\protect\cite{cite4404,cite4405}}.

The smallest numerically optimized Fock-state code protecting against a single loss error is the \(\sqrt(17)\) code \NoCaseChange{\protect\cite{cite4707}},
\flmMathEnvironment{align}{}{
\begin{split}
|\overline{0}\rangle&=\frac{1}{\sqrt{6}}\left(\sqrt{7-\sqrt{17}}|0\rangle+\sqrt{\sqrt{17}-1}|3\rangle\right)\\
|\overline{1}\rangle&=\frac{1}{\sqrt{6}}\left(\sqrt{9-\sqrt{17}}|1\rangle-\sqrt{\sqrt{17}-3}|4\rangle\right)~,
\end{split}
}
correcting a single loss error. The average occupation number of the codewords is \(\approx 1.6\), which is \(0.4\) photons lower than that of the smallest binomial code with the same level of protection.

\codefieldsection{Protection}
Numerically optimized bosonic codes can be designed to protect against a finite number of loss events while minimizing resources such as the average occupation number. Many such codes are approximate QECCs.
\codefieldsection{Parent}
\begin{eczvaluelist}
\item\relax
\flmRefsHyperref[eczindexfamilyrel]{code:oscillators}{Bosonic code}\end{eczvaluelist}
\codefieldsection{Child}
\begin{eczvaluelist}
\item\relax
\flmRefsHyperref[eczindexfamilyrel]{code:ampdamp_numopt}{Numerically optimized four-qubit AD code} --- Numerically optimized four-qubit AD codes can be obtained from a biconvex optimization of the entanglement fidelity \NoCaseChange{\protect\cite{cite3995}}.
\end{eczvaluelist}
\codefieldsection{Cousins}
\begin{eczvaluelist}
\item\relax
\flmRefsHyperref[eczindexfamilyrel]{code:multimodegkp}{Gottesman-Kitaev-Preskill (GKP) code} --- Numerically optimizing GKP code lattices yields codes for three, seven, and nine modes with larger distances and fidelities than known GKP codes \NoCaseChange{\protect\cite{cite420}}. Neural networks can be used to optimize approximate GKP states \NoCaseChange{\protect\cite{cite4905}}.
\item\relax
\flmRefsHyperref[eczindexfamilyrel]{code:approximate_qecc}{Approximate quantum error-correcting code (AQECC)} --- Numerically optimized codes arising from optimization routines are often approximate QECCs.
\item\relax
\flmRefsHyperref[eczindexfamilyrel]{code:reinforcement_learning}{Reinforcement-learning quantum code} --- Numerically optimized bosonic codes can be obtained via reinforcement learning \NoCaseChange{\protect\cite{cite4404,cite4405}}.
\end{eczvaluelist}
\eczhbkcontributors{ \eczhuVVA }
\endeczcode

\eczcode{one_hot_quantum}{One-hot quantum code}{~\NoCaseChange{\protect\cite{cite4918}}}
\codefieldsection{Alternative Names}
\begin{eczvaluelist}
\item\relax Single-excitation subspace code
\item\relax Direct mapping
\item\relax Multi-rail code
\end{eczvaluelist}
\eczhIndexCodeAliasName{one_hot_quantum}{Single-excitation subspace code}
\eczhIndexCodeAliasName{one_hot_quantum}{Direct mapping}
\eczhIndexCodeAliasName{one_hot_quantum}{Multi-rail code}
\codefieldsection{Description}
Encoding of a \(q\)-dimensional qudit into the single-excitation subspace of \(q\) modes. The \(j\)th logical state is the multi-mode Fock state with one photon in mode \(j\) and zero photons in the other modes.
This code is useful for encoding and performing operations on qudits in multiple modes \NoCaseChange{\protect\cite{cite502,cite503,cite504,cite505,cite506}}.

Another name for this code \NoCaseChange{\protect\cite{cite4919}}, not used here, is a unary code. This term is reserved for a mapping between the natural numbers \(N\) and binary strings with the first \(N\) coordinates being 1 and the rest 0.

\codefieldsection{Protection}
This is an error-detecting code against one \flmRefsHyperref{ref498}{photon loss} event.

\codefieldsection{Gates}
\begin{eczvaluelist}
\item\relax Non-deterministic gates using linear optics and photon-number resolving detectors \NoCaseChange{\protect\cite{cite4918}}.
\item\relax The group \(SU(q)\) can be realized via Gaussian rotations \NoCaseChange{\protect\cite{cite2810}}.
\end{eczvaluelist}
\codefieldsection{Parents}
\begin{eczvaluelist}
\item\relax
\flmRefsHyperref[eczindexfamilyrel]{code:chuang-leung-yamamoto}{Chuang-Leung-Yamamoto (CLY) code}\item\relax
\flmRefsHyperref[eczindexfamilyrel]{code:group_representation}{Group-representation code} --- One-hot quantum codes are group-representation codes with the \(G = SU(q)\) subgroup of Gaussian rotations \NoCaseChange{\protect\cite{cite2810}}.
\end{eczvaluelist}
\codefieldsection{Child}
\begin{eczvaluelist}
\item\relax
\flmRefsHyperref[eczindexfamilyrel]{code:dual_rail}{Dual-rail quantum code}\end{eczvaluelist}
\codefieldsection{Cousin}
\begin{eczvaluelist}
\item\relax
\flmRefsHyperref[eczindexfamilyrel]{code:one_hot}{One-hot code} --- The one-hot quantum code is the quantum version of the one-hot code.
\end{eczvaluelist}
\eczhbkcontributors{ \eczhuVVA }
\endeczcode

\eczcode{oscillators_into_oscillators}{Oscillator-into-oscillator code}{~\NoCaseChange{\protect\cite{cite4664,cite4661}}}
\codefieldsection{Description}
Encodes \(k\) bosonic modes into \(n\) bosonic modes.
\codefieldsection{Parent}
\begin{eczvaluelist}
\item\relax
\flmRefsHyperref[eczindexfamilyrel]{code:oscillators}{Bosonic code} --- Oscillator-into-oscillator codes are bosonic codes with an infinite-dimensional logical subspace.
\end{eczvaluelist}
\codefieldsection{Children}
\begin{eczvaluelist}
\item\relax
\flmRefsHyperref[eczindexfamilyrel]{code:analog_stabilizer}{Analog stabilizer code}\item\relax
\flmRefsHyperref[eczindexfamilyrel]{code:gkp-stabilizer}{Oscillator-into-oscillator GKP code}\end{eczvaluelist}
\codefieldsection{Cousin}
\begin{eczvaluelist}
\item\relax
\flmRefsHyperref[eczindexfamilyrel]{code:tiger}{Tiger code} --- In general tiger codes, encodings of logical qubits, qudits, modes, and rotors are all possible.
\end{eczvaluelist}
\eczhbkcontributors{ \eczhuVVA }
\endeczcode

\eczcode{gkp-stabilizer}{Oscillator-into-oscillator GKP code}{~\NoCaseChange{\protect\cite{cite4778}}}
\codefieldsection{Alternative Names}
\begin{eczvaluelist}
\item\relax GKP-stabilizer code
\end{eczvaluelist}
\eczhIndexCodeAliasName{gkp-stabilizer}{GKP-stabilizer code}
\codefieldsection{Description}
Multimode GKP code with an infinite-dimensional logical space. Can be obtained by considering an \(n\)-mode GKP code with a finite-dimensional logical space, removing stabilizers such that the logical space becomes infinite dimensional, and applying a Gaussian circuit.

Simple oscillator-into-oscillator GKP codes include GKP-repetition codes and GKP two-mode-squeezing (TMS) codes \NoCaseChange{\protect\cite{cite4778}}.
Arbitrary oscillator-into-oscillator GKP codes can be reduced to a standard form consisting of a direct sum of GKP TMS codes, up to symplectic transformations \NoCaseChange{\protect\cite{cite4668}}.
The optimal code design problem of determining the squeezing parameters can be efficiently solved \NoCaseChange{\protect\cite{cite4668}}.

\codefieldsection{Protection}
Oscillator-into-oscillator GKP codes to protect one or more modes against displacement noise using GKP resource states.

\codefieldsection{Encoding}
\begin{eczvaluelist}
\item\relax Gaussian circuit applied to \(k\) modes storing logical information and \(n-k\) modes initialized in a fixed GKP state.
\end{eczvaluelist}
\codefieldsection{Decoding}
\begin{eczvaluelist}
\item\relax Syndromes can be read off using ancilla modes, yielding partial information about noise in the logical modes that can then be used in an efficient ML decoding procedure \NoCaseChange{\protect\cite{cite4863}}.
\end{eczvaluelist}
\codefieldsection{Threshold}
\begin{eczvaluelist}
\item\relax Thresholds against displacement noise cannot be obtained without ideal (i.e., non-normalizable) codewords \NoCaseChange{\protect\cite{cite4920}}.
\end{eczvaluelist}
\codefieldsection{Notes}
\begin{eczvaluelist}
\item\relax Introduction to and examples of oscillator-into-oscillator GKP codes \NoCaseChange{\protect\cite{cite4727}}.
\end{eczvaluelist}
\codefieldsection{Parents}
\begin{eczvaluelist}
\item\relax
\flmRefsHyperref[eczindexfamilyrel]{code:quantum_lattice}{Quantum lattice code} --- Oscillator-into-oscillator GKP codes are \(n\)-mode quantum lattice codes with less than \(2n\) stabilizers, i.e., constructed using a degenerate lattice (see Appx. A of Ref. \NoCaseChange{\protect\cite{cite4884}}).
\item\relax
\flmRefsHyperref[eczindexfamilyrel]{code:oscillators_into_oscillators}{Oscillator-into-oscillator code}\end{eczvaluelist}
\codefieldsection{Child}
\begin{eczvaluelist}
\item\relax
\flmRefsHyperref[eczindexfamilyrel]{code:gkp-cluster-state}{GKP CV-cluster-state code} --- A GKP CV-cluster-state code can be created by initializing \(k\) modes in momentum states (or, in the normalizable case, squeezed vacua), \(n-k\) modes in (normalizable) GKP states, and applying a Gaussian circuit consisting of two-body \(e^{i V_{jk} \hat{x}_j \hat{x}_k }\) for some angles \(V_{jk}\).
\end{eczvaluelist}
\codefieldsection{Cousins}
\begin{eczvaluelist}
\item\relax
\flmRefsHyperref[eczindexfamilyrel]{code:gkp_concatenated}{Concatenated GKP code} --- Oscillator-into-oscillator GKP codes concatenated with qubit-into-oscillator GKP codes can outperform more conventional concatenations of qubit-into-oscillator GKP codes with qubit stabilizer codes \NoCaseChange{\protect\cite{cite4863}}.
\item\relax
\flmRefsHyperref[eczindexfamilyrel]{code:dfour_gkp}{\(D_4\) hyper-diamond GKP code} --- \(D_4\) hyper-diamond GKP codes may be optimal for oscillator-into-oscillator GKP codes utilizing two ancilla modes \NoCaseChange{\protect\cite{cite4668}}.
\item\relax
\flmRefsHyperref[eczindexfamilyrel]{code:hexagonal_gkp}{Hexagonal GKP code} --- Hexagonal GKP codes may be optimal for oscillator-into-oscillator GKP codes utilizing one ancilla mode \NoCaseChange{\protect\cite{cite4668}}.
\item\relax
\flmRefsHyperref[eczindexfamilyrel]{code:number_phase}{Number-phase code} --- Number-phase codes can serve as resource states for number-phase-rotor GKP-stabilizer codes, a polar analogue of oscillator-into-oscillator GKP codes that protects against photon loss and dephasing \NoCaseChange{\protect\cite[{Sec. 7}]{cite2699}}.
\item\relax
\flmRefsHyperref[eczindexfamilyrel]{code:analog_stabilizer}{Analog stabilizer code} --- Analog stabilizer codes protect logical modes against arbitrarily large displacements on a few modes, while oscillator-into-oscillator GKP codes protect an infinite-dimensional logical space against sufficiently small displacements in any number of modes. Encoding in analog-stabilizer (oscillator-into-oscillator GKP) codes can be done by a Gaussian operation acting on a tensor product of an arbitrary state in the first mode and position states (GKP states) on the remaining modes. For Gaussian displacement noise, linear oscillator encodings only squeeze the logical noise between conjugate quadratures \NoCaseChange{\protect\cite[{Sec. VI}]{cite416}}, and protection of logical modes against small displacements cannot be done using only Gaussian resources \NoCaseChange{\protect\cite{cite4702,cite4701}}, so oscillator-into-oscillator GKP codes can be thought of as analog stabilizer encodings utilizing non-Gaussian GKP resource states.
\item\relax
\flmRefsHyperref[eczindexfamilyrel]{code:homological_number-phase}{Homological number-phase code} --- Homological number-phase codes are finite-dimensional cousins of number-phase-rotor GKP-stabilizer codes: both use number-phase resource states and Clifford-semigroup encoders to protect oscillator information against photon loss and dephasing \NoCaseChange{\protect\cite[{Secs. 6-7}]{cite2699}}.
\end{eczvaluelist}
\eczhbkcontributors{ Armin Gerami, \eczhuVVA }
\endeczcode

\eczcode{constant_excitation_permutation_invariant}{Ouyang-Chao constant-excitation PI code}{~\NoCaseChange{\protect\cite{cite2946}}}
\codefieldsection{Description}
A constant-excitation PI Fock-state code whose construction is based on integer partitions.

\codefieldsection{Parents}
\begin{eczvaluelist}
\item\relax
\flmRefsHyperref[eczindexfamilyrel]{code:fock_state}{Fock-state bosonic code}\item\relax
\flmRefsHyperref[eczindexfamilyrel]{code:permutation_invariant}{Permutation-invariant (PI) code}\item\relax
\flmRefsHyperref[eczindexfamilyrel]{code:constant_excitation}{Constant-excitation (CE) code}\end{eczvaluelist}
\codefieldsection{Child}
\begin{eczvaluelist}
\item\relax
\flmRefsHyperref[eczindexfamilyrel]{code:wasilewski-banaszek}{Wasilewski-Banaszek code} --- The Wasilewski-Banaszek code is a simple example of an Ouyang-Chao PI code \NoCaseChange{\protect\cite{cite2946}}.
\end{eczvaluelist}
\eczhbkcontributors{ \eczhuVVA }
\endeczcode

\eczcode{paircat}{Pair-cat code}{~\NoCaseChange{\protect\cite{cite4921}}}
\codefieldsection{Description}
Two- or higher-mode extension of cat codes whose codewords are right eigenstates of powers of products of the modes' lowering operators. Many gadgets for cat codes have two-mode pair-cat analogues, with the advantage being that such gates can be done in parallel with a dissipative error-correction process.

Two-mode codewords are supported by Fock states with occupation number \(\hat{n}_2-\hat{n}_1\) fixed to some integer \(\Delta\). 
In the \textit{two-component} case, \(|\overline{0}_{\gamma,\Delta}\rangle \sim (|\gamma_\Delta\rangle + (-1)^\Delta |i\gamma_\Delta\rangle)/\sqrt{2}\) and \(|\overline{1}_{\gamma,\Delta}\rangle \sim (|\gamma_\Delta\rangle - (-1)^\Delta |i\gamma_\Delta\rangle)/\sqrt{2}\) \NoCaseChange{\protect\cite{cite4922,cite4921}}, where
\flmMathEnvironment{align}{}{
|\gamma_\Delta \rangle \propto \sum_{n=0}^\infty \frac{\gamma^{2n+\Delta}}{\sqrt{n! (n+\Delta)!}} |n,n+\Delta\rangle
}
is the corresponding pair-coherent state \NoCaseChange{\protect\cite{cite4923,cite4924,cite4925,cite4926}} with amplitude \(\gamma > 0\), up to normalization.
The asymptotic expression of the codewords is valid in the limit of large energy, \(|\gamma|^2\to\infty\).

\codefieldsection{Protection}
The occupation-number differences form the syndromes, as opposed to the photon-number parity for the single-mode cat code. Any loss event or combination of losses that changes the relative occupation-number differences between modes is detectable.

\codefieldsection{Gates}
\begin{eczvaluelist}
\item\relax Hamiltonian \(X\), \(XX\), \(Z\) gates, holonomic \(Z\) gate, control-phase gate.
\item\relax Bias-preserving gates \NoCaseChange{\protect\cite{cite4927}}.
\end{eczvaluelist}
\codefieldsection{Decoding}
\begin{eczvaluelist}
\item\relax Lindbladian-based dissipative encoding and autonomous QEC utilizing two-mode two-photon absorption \NoCaseChange{\protect\cite{cite4925}}. Encoding passively protects against cavity dephasing, suppressing dephasing noise exponentially with \(\gamma^2\).
\end{eczvaluelist}
\codefieldsection{Realizations}
\begin{eczvaluelist}
\item\relax Microwave cavities coupled to superconducting circuits by the Wang group \NoCaseChange{\protect\cite{cite4928}}.
\end{eczvaluelist}
\codefieldsection{Parents}
\begin{eczvaluelist}
\item\relax
\flmRefsHyperref[eczindexfamilyrel]{code:tiger}{Tiger code} --- The pair-cat code is a tiger code with \(G = (2,2)\) and \(H = (1,-1)\) \NoCaseChange{\protect\cite{cite4667}}.
\item\relax
\flmRefsHyperref[eczindexfamilyrel]{code:fock_state}{Fock-state bosonic code}\end{eczvaluelist}
\codefieldsection{Cousins}
\begin{eczvaluelist}
\item\relax
\flmRefsHyperref[eczindexfamilyrel]{code:cat}{Cat code} --- Cat (pair-cat) codewords are superpositions of coherent (pair-coherent) states. Many cat-code protocols have analogues for the two-mode pair-cat codes.
\item\relax
\flmRefsHyperref[eczindexfamilyrel]{code:hamiltonian}{Hamiltonian-based code} --- Two-component pair-cat codewords form ground-state subspace of a multimode Kerr Hamiltonian.
\end{eczvaluelist}
\eczhbkcontributors{ Yijia Xu, \eczhuVVA }
\endeczcode

\eczcode{pauli_qsc}{Pauli tessellation QSC}{~\NoCaseChange{\protect\cite{cite2811}}}
\codefieldsection{Description}
Two-mode non-uniform QSC whose projection is onto a copy of an irreducible representation of the \flmRefsHyperref{ref663}{single-qubit Pauli group}, realized geometrically by the \(\{2,2,4\}\) tessellation of the sphere \NoCaseChange{\protect\cite{cite2811}}.
For the canonical choice in the paper, each logical codeword is a \(\pm 1\) superposition of four vertices of a cube, i.e., of one tetrahedron in the cube decomposition.

\codefieldsection{Protection}
For the \(\theta_0=\arccos(1/\sqrt{3})\), \(\phi_0=\pi/4\) configuration of Ref. \NoCaseChange{\protect\cite{cite2811}}, the code corrects arbitrary spherical rotation errors of angle smaller than \(\frac{1}{2}\arccos(1/3)\).
The lowest uncorrectable spherical-harmonic momentum-error pair is \(Y_1^{0\dagger}Y_2^{\pm 2}\), so all momentum errors with \(\ell \leq 1\) are correctable.

\codefieldsection{Gates}
\begin{eczvaluelist}
\item\relax The \flmRefsHyperref{ref663}{single-qubit Pauli group} is realized by geometric rotations of the sphere, namely \(\pi\) rotations for \(X\) and \(Z\) and a \(\pi/2\) rotation for \(XZ\) \NoCaseChange{\protect\cite{cite2811}}.
\end{eczvaluelist}
\codefieldsection{Parents}
\begin{eczvaluelist}
\item\relax
\flmRefsHyperref[eczindexfamilyrel]{code:qsc}{Quantum spherical code (QSC)} --- The Pauli tessellation QSC has non-uniform \(\pm 1\) coefficients.
\item\relax
\flmRefsHyperref[eczindexfamilyrel]{code:group_representation}{Group-representation code} --- The Pauli tessellation QSC is a group-representation code with \(G\) being the single-qubit Pauli group.
\end{eczvaluelist}
\codefieldsection{Cousins}
\begin{eczvaluelist}
\item\relax
\flmRefsHyperref[eczindexfamilyrel]{code:simplex_spherical}{Simplex spherical code} --- Each codeword of the Pauli tessellation QSC is a quantum superposition of vertices of a tetrahedron with \(\pm 1\) coefficients.
\item\relax
\flmRefsHyperref[eczindexfamilyrel]{code:tesselation}{Hyperbolic tessellation code} --- The tessellation-code framework spans spherical, Euclidean, and hyperbolic geometries; the Pauli tessellation QSC is the spherical member \NoCaseChange{\protect\cite{cite2811}}.
\item\relax
\flmRefsHyperref[eczindexfamilyrel]{code:fourier_bosonic}{Bosonic quantum Fourier code} --- The bosonic quantum Fourier code and the Pauli group-representation QSC are both group-representation codes with \(G\) being the single-qubit Pauli group.
\end{eczvaluelist}
\eczhbkcontributors{ \eczhuVVA }
\endeczcode

\eczcode{penrose}{Penrose tiling code}{~\NoCaseChange{\protect\cite{cite4929}}}
\codefieldsection{Description}
Encodes quantum information into superpositions of rotated and translated versions of different Penrose tilings of \(\mathbb{R}^n\).

Letting \(|T\rangle\) be a Penrose tiling, the codeword corresponding to this tiling is a superposition of all tilings in the tiling's orbit under Euclidean transformations,
\flmMathEnvironment{align}{}{
  |\overline{T}\rangle=\int \textnormal{d}g|gT\rangle~,
}
where \(g\) is a Euclidean transformation.

\codefieldsection{Protection}
Properties of Penrose tilings such as local indistinguishability and local recoverability ensure that Penrose tiling codes can correct erasures of any finite region of space.

\codefieldsection{Notes}
\begin{eczvaluelist}
\item\relax Popular summary of Penrose tiling codes in \flmHref{https://www.quantamagazine.org/never-repeating-tiles-can-safeguard-quantum-information-20240223}{Quanta Magazine}.
\end{eczvaluelist}
\codefieldsection{Parent}
\begin{eczvaluelist}
\item\relax
\flmRefsHyperref[eczindexfamilyrel]{code:oscillators}{Bosonic code} --- Penrose tiling codes encode information into Penrose tilings, which are non-periodic tilings of \(\mathbb{R}^n\).
\end{eczvaluelist}
\eczhbkcontributors{ \eczhuVVA }
\endeczcode

\eczcode{quantum_lattice}{Quantum lattice code}{}
\codefieldsection{Description}
Bosonic stabilizer code on \(n\) bosonic modes whose stabilizer group is an infinite countable group of oscillator displacement operators which implement lattice translations in phase space.

Displacement operators on \(n\) modes can be written as
\flmMathEnvironment{align}{}{
D(\xi) = \exp \left\{-i \sqrt{2\pi} {\xi}^\mathrm{T} J \hat{q} \right\} , \quad \xi \in \mathbb{R}^{2n}~,
}
where \(\hat{q}\) is a \(2n\)-dimensional vector of position and momentum operators of the modes, the symplectic form
\flmMathEnvironment{align}{}{
J = \begin{pmatrix} 0 & 1 \\ -1 & 0 \end{pmatrix} \otimes I_n = \begin{pmatrix}
0 &  I_n \\
-I_n  & 0 \end{pmatrix}~,
}
and \(I_n\) is the identity matrix. A group generated by a set of independent displacement operators is given by a lattice \({\mathcal{L}}\)
\flmMathEnvironment{align}{}{
\langle D(\xi_1) ,\dots,  D(\xi_{m})  \rangle = \{ e^{ i \phi_M (\xi) } D(\xi) ~\vert~ \xi \in {\mathcal{L}} \}
}
and becomes a valid stabilizer group when every symplectic inner product between lattice vectors yields an integer. In other words, the corresponding lattice is symplectically integral, corresponding to an integer-valued symplectic Gram matrix \(A\),
\flmMathEnvironment{align}{}{
A_{ij}={\xi}^T_i J \xi_j \in \mathbb{Z}~.
}
The \(m=2n\) case yields multimode GKP codes encoding a finite-dimensional logical subspace, while removing some displacements yields oscillator-into-oscillator GKP codes encoding an infinite-dimensional logical subspace. Codes defined on a hyper-rectangular lattice are \textit{CSS GKP} codes, and more general lattices, obtained by Gaussian transformations, yield non-CSS codes.

\codefieldsection{Notes}
\begin{eczvaluelist}
\item\relax Quantum lattice states are featured in the proof of hardness of LWE \NoCaseChange{\protect\cite[{pg. 12}]{cite320}} and the hidden subgroup problem over the reals \NoCaseChange{\protect\cite{cite4930}}.
\item\relax Single-mode quantum lattice states on a square lattice, otherwise known as square-lattice GKP states, are relevant to signal processing and condensed-matter physics; see the corresponding \flmRefsHyperref{code:gkp}{code entry} for details.
\end{eczvaluelist}
\codefieldsection{Parents}
\begin{eczvaluelist}
\item\relax
\flmRefsHyperref[eczindexfamilyrel]{code:oscillator_stabilizer}{Bosonic stabilizer code} --- Quantum lattice codes are bosonic stabilizer codes with a countably infinite stabilizer group, corresponding to modular constraints on positions and momenta.
\item\relax
\flmRefsHyperref[eczindexfamilyrel]{code:coherent_constellation}{Coherent-state constellation code} --- Quantum lattice codewords can be written as superpositions of coherent states lying on a lattice in phase space \NoCaseChange{\protect\cite{cite513,cite496}}.
\item\relax
\flmRefsHyperref[eczindexfamilyrel]{code:lca_stabilizer}{Locally compact Abelian (LCA) stabilizer code} --- LCA stabilizer codes defined on only oscillators reduce to quantum lattice codes.
\end{eczvaluelist}
\codefieldsection{Children}
\begin{eczvaluelist}
\item\relax
\flmRefsHyperref[eczindexfamilyrel]{code:gkp-stabilizer}{Oscillator-into-oscillator GKP code} --- Oscillator-into-oscillator GKP codes are \(n\)-mode quantum lattice codes with less than \(2n\) stabilizers, i.e., constructed using a degenerate lattice (see Appx. A of Ref. \NoCaseChange{\protect\cite{cite4884}}).
\item\relax
\flmRefsHyperref[eczindexfamilyrel]{code:multimodegkp}{Gottesman-Kitaev-Preskill (GKP) code} --- GKP codes are \(n\)-mode quantum lattice codes with \(2n\) stabilizers, i.e., constructed using a non-degenerate lattice.
\end{eczvaluelist}
\codefieldsection{Cousins}
\begin{eczvaluelist}
\item\relax
\flmRefsHyperref[eczindexfamilyrel]{code:points_into_lattices}{Lattice} --- Quantum lattice codes can be thought of as quantum analogues of lattices because they store information in quantum superpositions of points on a lattice in quantum phase space.
\item\relax
\flmRefsHyperref[eczindexfamilyrel]{code:css}{Calderbank-Shor-Steane (CSS) stabilizer code} --- Quantum lattice codes defined on rectangular lattices are CSS codes. There is no known relation to chain complexes for such codes. More general lattices, obtained from rectangular lattices by Gaussian transformations, yield non-CSS codes.
\end{eczvaluelist}
\eczhbkcontributors{ Jonathan Conrad, \eczhuVVA }
\endeczcode

\eczcode{qsc}{Quantum spherical code (QSC)}{~\NoCaseChange{\protect\cite{cite382}}}
\codefieldsection{Description}
Code whose codewords are superpositions of points on an \(n\)-dimensional real or complex sphere.
Such codes can in principle be defined on any configuration space housing a sphere, but the focus of this entry is on QSCs constructed out of coherent-state constellations.

More technically, a QSC is a collection \(\{\mathcal{C}_k\}_{k=1}^K\) of \textit{logical constellations}, each of which yields a codeword by taking a quantum superposition of all points \(\mathbf{x}\in \mathcal{C}_k\).
Taken together, the logical constellations yield the \textit{code constellation}, \(\mathcal{C}=\bigcup_{k=1}^{K}\mathcal{C}_{k}\).

Codewords of coherent-state QSCs of uniform superposition are defined as
\flmMathEnvironment{align}{}{
  |\mathcal{C}_{k}\rangle\sim\frac{1}{\sqrt{|{\mathcal{C}}_{k}|}}\sum_{\boldsymbol{\alpha}\in\mathcal{C}_{k}}|\sqrt{\bar{N}}\boldsymbol{\alpha}\rangle~,
}
where \( |\boldsymbol{\alpha} \rangle = |\alpha_1,\alpha_2,...\alpha_n \rangle \) is an \(n\)-mode coherent state.
This asymptotic expression is valid in the limit of large energy \(\bar{N}\to\infty\).

Coherent-state QSCs on \(n\) modes are denoted by
\(\llparenthesis n,K,d_E,\langle t_{\downarrow},d_{\updownarrow},d_{\downarrow}\rangle\rrparenthesis \),
where \(K\) is codespace dimension, \(d_E\) is the \textit{squared minimum distance}, i.e., the smallest Euclidean distance between pairs of distinct points across all codewords, and \( t_{\downarrow},d_{\updownarrow},d_{\downarrow} \) are the number of \textit{correctable} losses (plus 1), the degree distance, and the number of \textit{detectable} losses (plus 1), respectively.

\codefieldsection{Protection}
The \textit{resolution} \(d_E\) of the code is defined as
  \flmMathEnvironment{align}{}{
    d_E = \min_{\boldsymbol{\alpha},\boldsymbol{\beta}\in\mathcal{C}} \Vert\boldsymbol{\alpha}-\boldsymbol{\beta}\Vert^2~.
  }
The code protects against passive Gaussian transformations, which manifest as rotations on the sphere, \( |\boldsymbol{\alpha}\rangle \rightarrow |\mathbf{R}\boldsymbol{\alpha}\rangle \) for all \(\mathbf{R}\).
Detectable transformations correspond to rotations for which
  \flmMathEnvironment{align}{}{
    \Vert \mathbf{R}\boldsymbol{\alpha} - \boldsymbol{\alpha}\Vert^2 < d_E~,
  }
in the large \(\bar{N}\) limit.

The code also protects against general ladder errors, which are defined as
\flmMathEnvironment{align}{}{
  \mathbf{a}^{\dagger\mathbf{p}}\mathbf{a}^{\mathbf{q}}=\prod_{j=1}^{n}a_{j}^{\dagger p_{j}}a_{j}^{q_{j}}~.
}
Any \flmRefsHyperref{ref498}{AD} ladder error \(\mathbf{a}^{\mathbf{q}}\) with \(|\mathbf{q}|<d_{\downarrow}\) is detectable.
Any ladder error \(\mathbf{a}^{\dagger\mathbf{p}}\mathbf{a}^{\mathbf{q}}\) with \(|\mathbf{p}|,|\mathbf{q}|<t_{\downarrow}\) is detectable, implying that up to \(t_{\downarrow}-1\) losses are correctable.
Any ladder error with degree \(|\mathbf{p}+\mathbf{q}|<d_{\updownarrow}\) is detectable.

\codefieldsection{Decoding}
\begin{eczvaluelist}
\item\relax Lindbladian scheme stabilizing all points in the constellation and protecting from the \flmRefsHyperref{ref498}{AD} operator \(E_{0}^{\otimes n}\) \NoCaseChange{\protect\cite{cite382}}.
\end{eczvaluelist}
\codefieldsection{Parents}
\begin{eczvaluelist}
\item\relax
\flmRefsHyperref[eczindexfamilyrel]{code:coherent_constellation}{Coherent-state constellation code} --- Coherent-state QSCs are coherent-state constellation codes constrained to lie on a sphere.
\item\relax
\flmRefsHyperref[eczindexfamilyrel]{code:ampdamp}{Amplitude-damping (AD) code} --- QSC codewords are superpositions of coherent states with the same energy, but coherent states are not eigenstates of the energy Hamiltonian. The \flmRefsHyperref{ref498}{AD} Kraus operator \(E_{0}^{\otimes n}\) acts identically on each coherent state by shrinking the radius of the QSC's sphere.
\end{eczvaluelist}
\codefieldsection{Children}
\begin{eczvaluelist}
\item\relax
\flmRefsHyperref[eczindexfamilyrel]{code:2t_qutrit}{2T-qutrit code} --- The \(2T\)-qutrit is a QSC on the two-dimensional complex sphere whose code constellation is the \(4\{3\}4\) complex polytope.
\item\relax
\flmRefsHyperref[eczindexfamilyrel]{code:cat_concatenated}{Concatenated cat code}\item\relax
\flmRefsHyperref[eczindexfamilyrel]{code:clifford_qsc}{Clifford group-representation QSC} --- The Clifford group-representation QSC has non-uniform coefficients.
\item\relax
\flmRefsHyperref[eczindexfamilyrel]{code:fourier_bosonic}{Bosonic quantum Fourier code} --- The bosonic quantum Fourier code has non-uniform \(\pm 1\) coefficients.
\item\relax
\flmRefsHyperref[eczindexfamilyrel]{code:hessian_qsc}{Hessian QSC} --- The Hessian QSC is an example of a QSC with logical constellation built from the Hessian complex polyhedron.
\item\relax
\flmRefsHyperref[eczindexfamilyrel]{code:pauli_qsc}{Pauli tessellation QSC} --- The Pauli tessellation QSC has non-uniform \(\pm 1\) coefficients.
\item\relax
\flmRefsHyperref[eczindexfamilyrel]{code:quantum_sidelnikov}{Clifford subgroup-orbit QSC}\end{eczvaluelist}
\codefieldsection{Cousins}
\begin{eczvaluelist}
\item\relax
\flmRefsHyperref[eczindexfamilyrel]{code:group_representation}{Group-representation code} --- QSCs should be able to be formulated as group-representation codes whose group is that formed by the permutation representation of the code polytope symmetry group, but this representation may be reducible.
\item\relax
\flmRefsHyperref[eczindexfamilyrel]{code:points_into_spheres}{Constant-energy spherical code} --- QSCs are quantum counterparts of spherical and constant-energy codes because they store information in quantum superpositions of points on a sphere in quantum phase space.
\item\relax
\flmRefsHyperref[eczindexfamilyrel]{code:spherical}{Spherical code} --- QSCs are quantum counterparts of spherical and constant-energy codes because they store information in quantum superpositions of points on a sphere in quantum phase space.
\item\relax
\flmRefsHyperref[eczindexfamilyrel]{code:single_spin}{Single-spin code} --- Single-spin codes whose codewords are expressed in terms of discrete sets of spin-coherent states may also be interpreted as QSCs.
\item\relax
\flmRefsHyperref[eczindexfamilyrel]{code:polytope}{Polytope code} --- QSCs can be constructed by using vertices of polytopes for logical constellations. The logical constellations form the vertices of the code constellation, a polytope compound.
\item\relax
\flmRefsHyperref[eczindexfamilyrel]{code:tiger}{Tiger code} --- Tiger (quantum spherical) codewords consist of continuous and compact (discrete and finite) coherent-state constellations. Both codes protect against losses and gains of occupation numbers along with rotation noise stemming from modal dephasing. Protection against the latter type of noise is characterized by the minimum Euclidean distance between coherent states in different logical constellations.
\item\relax
\flmRefsHyperref[eczindexfamilyrel]{code:asymmetric_qecc}{Asymmetric quantum code (AQC)} --- QSC code parameters against loss/gain errors and Gaussian rotations can be tuned.
\end{eczvaluelist}
\eczhbkcontributors{ Shubham P. Jain, \eczhuVVA }
\endeczcode

\eczcode{qudits_into_oscillators}{Qudit-into-oscillator code}{}
\codefieldsection{Description}
Encodes \(K\)-dimensional Hilbert space into \(n\) bosonic modes.
\codefieldsection{Decoding}
\begin{eczvaluelist}
\item\relax Given an encoding of a finite-dimensional code, a decoder that yields the optimal entanglement fidelity can be obtained by solving a semi-definite program \NoCaseChange{\protect\cite{cite2568,cite2547}} (see also Ref. \NoCaseChange{\protect\cite{cite2569}}). This approximate QEC technique can be adapted to bosonic codes as long as they are restricted to a finite-dimensional subspace of the oscillator Hilbert space \NoCaseChange{\protect\cite{cite496}}.
\end{eczvaluelist}
\codefieldsection{Parent}
\begin{eczvaluelist}
\item\relax
\flmRefsHyperref[eczindexfamilyrel]{code:oscillators}{Bosonic code} --- Qudit-into-oscillator codes are bosonic codes with a finite-dimensional logical subspace.
\end{eczvaluelist}
\codefieldsection{Children}
\begin{eczvaluelist}
\item\relax
\flmRefsHyperref[eczindexfamilyrel]{code:fock_state}{Fock-state bosonic code}\item\relax
\flmRefsHyperref[eczindexfamilyrel]{code:single-mode}{Single-mode bosonic code}\item\relax
\flmRefsHyperref[eczindexfamilyrel]{code:dfour_gkp}{\(D_4\) hyper-diamond GKP code}\item\relax
\flmRefsHyperref[eczindexfamilyrel]{code:gkp-cluster-state}{GKP CV-cluster-state code}\item\relax
\flmRefsHyperref[eczindexfamilyrel]{code:ntru_gkp}{NTRU-GKP code}\end{eczvaluelist}
\codefieldsection{Cousin}
\begin{eczvaluelist}
\item\relax
\flmRefsHyperref[eczindexfamilyrel]{code:approximate_qecc}{Approximate quantum error-correcting code (AQECC)} --- Approximate QEC techniques of finding the entanglement fidelity can be adapted to bosonic codes with a finite-dimensional codespace \NoCaseChange{\protect\cite{cite496}}.
\end{eczvaluelist}
\eczhbkcontributors{ \eczhuVVA }
\endeczcode

\eczcode{qutrit_pauli_gkp_subcode}{Qutrit-Pauli tessellation code}{~\NoCaseChange{\protect\cite{cite2811}}}
\codefieldsection{Description}
Euclidean-plane tessellation code whose projection is onto a copy of an irreducible representation of the \flmRefsHyperref{ref2198}{single-qutrit Pauli group}, realized by the \(\{3,3,3\}\) tessellation \NoCaseChange{\protect\cite{cite2811}}.
The code is a subcode of a two-mode GKP code and has GKP-like stabilizers. Logical \(X\), \(Z\), and \((ZX)^{-1}\) are implemented by \(2\pi/3\) rotations around tessellation vertices, while only a GKP-like logical \(Z\) is available via real-space displacement.

\codefieldsection{Protection}
The code corrects translation errors of Euclidean norm less than \(\sqrt{3}/2\).
It also corrects momentum errors in any direction with \(|\vec{k}| < 2\pi/9\) \NoCaseChange{\protect\cite{cite2811}}.

\codefieldsection{Parents}
\begin{eczvaluelist}
\item\relax
\flmRefsHyperref[eczindexfamilyrel]{code:oscillators}{Bosonic code}\item\relax
\flmRefsHyperref[eczindexfamilyrel]{code:group_representation}{Group-representation code} --- The qutrit-Pauli tessellation code is a group-representation code with \(G\) being the \flmRefsHyperref{ref2198}{single-qutrit Pauli group}.
\end{eczvaluelist}
\codefieldsection{Cousins}
\begin{eczvaluelist}
\item\relax
\flmRefsHyperref[eczindexfamilyrel]{code:multimodegkp}{Gottesman-Kitaev-Preskill (GKP) code} --- The qutrit-Pauli tessellation code is a subcode of a two-mode GKP code with GKP-like stabilizers \NoCaseChange{\protect\cite{cite2811}}.
\item\relax
\flmRefsHyperref[eczindexfamilyrel]{code:tesselation}{Hyperbolic tessellation code} --- The qutrit-Pauli tessellation code is the Euclidean \(\{3,3,3\}\) member of the same curvature-dependent tessellation-code framework \NoCaseChange{\protect\cite{cite2811}}.
\end{eczvaluelist}
\eczhbkcontributors{ \eczhuVVA }
\endeczcode

\eczcode{rg_cat}{Renormalization group (RG) cat code}{~\NoCaseChange{\protect\cite{cite2545,cite4931,cite4932}}}
\codefieldsection{Description}
Code whose codespace is spanned by \(q\) field-theoretic coherent states which are flowing under the renormalization group (RG) flow of massive free fields. The code approximately protects against displacements that represent local (i.e., short-distance, ultraviolet, or UV) operators. Intuitively, this is because RG cat codewords represent non-local (i.e., long-distance) degrees of freedom, which should only be excitable by acting on a macroscopically large number of short-distance degrees of freedom.

\codefieldsection{Protection}
Approximately protects against displacements that represent ultraviolet coherent operators, i.e., short-distance degrees of freedom of the field theory.

\codefieldsection{Parents}
\begin{eczvaluelist}
\item\relax
\flmRefsHyperref[eczindexfamilyrel]{code:coherent_constellation}{Coherent-state constellation code}\item\relax
\flmRefsHyperref[eczindexfamilyrel]{code:holographic}{Holographic code} --- The RG cat code encoder has coarse-graining features reminiscent of holography \NoCaseChange{\protect\cite{cite2545}}.
\item\relax
\flmRefsHyperref[eczindexfamilyrel]{code:approximate_qecc}{Approximate quantum error-correcting code (AQECC)} --- RG cat codes approximately protect against displacements that represent ultraviolet coherent operators.
\end{eczvaluelist}
\eczhbkcontributors{ \eczhuVVA }
\endeczcode

\eczcode{single-mode}{Single-mode bosonic code}{}
\codefieldsection{Description}
Encodes a \(K\)-dimensional Hilbert space into a single bosonic mode. A trivial single-mode code encoding a qubit into the first two Fock states \(\{|0\rangle,|1\rangle\}\) is called the \textit{single-rail} encoding \NoCaseChange{\protect\cite{cite649,cite650}}.
\codefieldsection{Parents}
\begin{eczvaluelist}
\item\relax
\flmRefsHyperref[eczindexfamilyrel]{code:qudits_into_oscillators}{Qudit-into-oscillator code}\item\relax
\flmRefsHyperref[eczindexfamilyrel]{code:single_subsystem}{Monolithic quantum code}\end{eczvaluelist}
\codefieldsection{Children}
\begin{eczvaluelist}
\item\relax
\flmRefsHyperref[eczindexfamilyrel]{code:bosonic_q-ary_expansion}{Bosonic \(q\)-ary expansion}\item\relax
\flmRefsHyperref[eczindexfamilyrel]{code:chebyshev}{Chebyshev code}\item\relax
\flmRefsHyperref[eczindexfamilyrel]{code:bosonic_rotation}{Bosonic rotation code}\item\relax
\flmRefsHyperref[eczindexfamilyrel]{code:squeezed_cat}{Squeezed cat code}\item\relax
\flmRefsHyperref[eczindexfamilyrel]{code:gkp}{Square-lattice GKP code}\item\relax
\flmRefsHyperref[eczindexfamilyrel]{code:hexagonal_gkp}{Hexagonal GKP code}\item\relax
\flmRefsHyperref[eczindexfamilyrel]{code:squeezed_fock_state}{Squeezed Fock-state code}\end{eczvaluelist}
\codefieldsection{Cousin}
\begin{eczvaluelist}
\item\relax
\flmRefsHyperref[eczindexfamilyrel]{code:dual_rail}{Dual-rail quantum code} --- Using the concatenation convention of the Zoo, concatenating the inner dual-rail code with an outer single-mode bosonic code yields several gates that are independent of the outer code \NoCaseChange{\protect\cite{cite4856}}.
\end{eczvaluelist}
\eczhbkcontributors{ \eczhuVVA }
\endeczcode

\eczcode{gkp}{Square-lattice GKP code}{~\NoCaseChange{\protect\cite{cite513}}}
\codefieldsection{Description}
Single-mode GKP qudit-into-oscillator CSS code based on the rectangular lattice.
Its stabilizer generators are oscillator displacement operators \(\hat{S}_q(2\alpha)=e^{-2i\alpha \hat{p}}\) and \(\hat{S}_p(2\beta)=e^{2i\beta \hat{x}}\).
To ensure \(\hat{S}_q(2\alpha)\) and \(\hat{S}_p(2\beta)\) generate a stabilizer group that is Abelian, there is a constraint that \(\alpha\beta=2q\pi\) where \(q\) is an integer denoting the logical dimension.

Codewords can be expressed as equal weight superpositions of coherent states on a rectangular lattice in phase space with spatial period \(2\sqrt{\pi}\).
The exact GKP state is non-normalizable, so approximate constructions have to be considered.

The \(q=1\) trivial encoding is spanned by the \textit{canonical GKP state} (a.k.a. \textit{grid state} or qunaught state \NoCaseChange{\protect\cite{cite4882}}),
\flmMathEnvironment{align}{}{
  |GKP\rangle=\sum_{\ell\in\mathbb{Z}}|x=\ell\sqrt{2\pi}\rangle~,
}
where \(|x\rangle\) are single-mode position states.

Single-mode GKP states have been introduced in quantum foundations research defining modular conjugate variables \NoCaseChange{\protect\cite{cite4933}} and in coherent-state theory associated with the Heisenberg-Weyl group \NoCaseChange{\protect\cite{cite4934,cite4935,cite4936}\protect\cite[{Sec. 1.5 and 3.2}]{cite4937}}.
The Dirac-delta orthonormal and complete \NoCaseChange{\protect\cite{cite4938}} basis formed by the GKP canonical states and their error states is known as the \textit{Zak basis}, discovered independently by Gelfand \NoCaseChange{\protect\cite{cite4939}} and Zak \NoCaseChange{\protect\cite{cite4940}}.
It is also called the Gelfand mapping, Weil-Brezin transform, and \(kq\) representation in condensed-matter physics and signal processing \NoCaseChange{\protect\cite{cite4941}\protect\cite[{Ch. 1}]{cite4942}\protect\cite[{Eq. (1.112)}]{cite4943}} (see Refs. \NoCaseChange{\protect\cite{cite4944,cite4945}} for more history).
Expansion of a function on \(\mathbb{R}\) in terms of this basis is called the \textit{Zak transform}. 
The Segal-Bargmann representations of GKP states are the theta functions of the lowest Landau level on a torus \NoCaseChange{\protect\cite[{Sec. V}]{cite3073}\protect\cite[{Prop. 6.3}]{cite3074}} (see also Refs. \NoCaseChange{\protect\cite{cite3075,cite3076}}).

\codefieldsection{Protection}
For stabilizers \(\hat{S}_q(2\alpha),\hat{S}_p(2\beta)\), the code can correct displacement errors up to \(\alpha/2\) in the \(q\)-direction and \(\beta/2\) in the \(p\)-direction. Approximately protects against \flmRefsHyperref{ref498}{photon loss} errors \NoCaseChange{\protect\cite{cite2607,cite496}}, outperforming most other codes designed to explicitly protect against loss \NoCaseChange{\protect\cite{cite496}}. An analytical expression can be derived for the effective logical channel after loss \NoCaseChange{\protect\cite{cite4946}}. Very sensitive to dephasing errors \NoCaseChange{\protect\cite{cite4902}}. A biased-noise GKP error correcting code can be prepared by choosing \(\alpha\neq \beta\). Expectation values of observables versus energy may be fit to a power law \NoCaseChange{\protect\cite{cite4947}}.
\codefieldsection{Encoding}
\begin{eczvaluelist}
\item\relax Dissipative stabilization of finite-energy square-lattice GKP states using stabilizers conjugated by a \textit{cooling} (\NoCaseChange{\protect\cite{cite508}}, Appx. B) or \textit{damping} operator, i.e., a damped exponential of the total occupation number \NoCaseChange{\protect\cite{cite4889,cite4948}}. Preparation of approximate square-lattice GKP states has been studied both theoretically and experimentally \NoCaseChange{\protect\cite{cite2607,cite4949,cite4907,cite4950}}. Various damped versions of GKP states are equivalent \NoCaseChange{\protect\cite{cite4951,cite4952}}, and there exists a Fock-state expansion \NoCaseChange{\protect\cite[{Appx. A}]{cite4819}}.
\item\relax Two Josephson junctions coupled by a gyrator \NoCaseChange{\protect\cite{cite3073}}.
\item\relax Periodic driving (a.k.a. Floquet engineering) \NoCaseChange{\protect\cite{cite4953}}.
\item\relax Approximate GKP states can be prepared using Gaussian operations and photon detectors \NoCaseChange{\protect\cite{cite4847}}.
\item\relax An optimal-size circuit using ancillary qubits can be used to prepare an approximate GKP state \NoCaseChange{\protect\cite{cite4954}}. The size of the circuit is linear in the logarithm of the approximation parameters of the GKP codes.
\item\relax Numerically optimized preparation from the vacuum Fock state using a universal bosonic gate set \NoCaseChange{\protect\cite{cite4955}}.
\item\relax Dissipative stabilization using a high-impedance LRC circuit and a Josephson junction \NoCaseChange{\protect\cite{cite4956}}.
\end{eczvaluelist}
\codefieldsection{Gates}
\begin{eczvaluelist}
\item\relax Clifford gates can be realized by performing linear-optical operations, symplectic transformations and displacements, all of which are Gaussian operations. Pauli gates can be performed using displacement operators. Clifford gates are fault tolerant in the sense that they map bounded-size errors to bounded-size errors \NoCaseChange{\protect\cite{cite513}}.
\item\relax By applying square-lattice GKP error correction to Gaussian input states, universality can be achieved without non-Gaussian elements \NoCaseChange{\protect\cite{cite4894}}.
\item\relax Square-root of the Hadamard gate performed via a Kerr interaction \NoCaseChange{\protect\cite{cite4957}}.
\item\relax \(\sqrt{T}\) gate using a quadratic potential \NoCaseChange{\protect\cite{cite4958}}.
\end{eczvaluelist}
\codefieldsection{Decoding}
\begin{eczvaluelist}
\item\relax Syndrome measurement can be done by applying a controlled displacement controlled by an ancilla qubit. The syndrome information can be obtained by measuring the ancilla qubit after the controlled-displacement operation; see \NoCaseChange{\protect\cite[{Sec. 2D}]{cite4902}}.
\item\relax Decoder \NoCaseChange{\protect\cite{cite4882}} based on Knill error correction (a.k.a. telecorrection \NoCaseChange{\protect\cite{cite3185}}), which is based on teleportation \NoCaseChange{\protect\cite{cite448,cite4369}}.
\item\relax Pauli \(X\), \(Y\), and \(Z\) measurements can be performed by measuring \(-\hat{p}\), \(\hat{x}-\hat{p}\), and \(\hat{x}\), respectively. If the measurement outcome is close to an even multiple of \(\sqrt{\pi}\), then the outcome is +1. If the measurement outcome is close to an odd multiple of \(\sqrt{\pi}\), then the outcome is -1; see \NoCaseChange{\protect\cite[{Sec. 2D}]{cite4902}}.
\item\relax Reinforcement learning decoder that uses only one ancilla qubit \NoCaseChange{\protect\cite{cite4718}}. It has been extended to utilize previously measured syndrome information \NoCaseChange{\protect\cite{cite4959}}.
\item\relax Knill and Steane error correction have been analytically compared \NoCaseChange{\protect\cite{cite4960}}.
\end{eczvaluelist}
\codefieldsection{Fault Tolerance}
\begin{eczvaluelist}
\item\relax Clifford gates can be realized by performing linear-optical operations, symplectic transformations and displacements, all of which are Gaussian operations. Pauli gates can be performed using displacement operators. Clifford gates are fault tolerant in the sense that they map bounded-size errors to bounded-size errors \NoCaseChange{\protect\cite{cite513}}.
\item\relax Error correction scheme is fault-tolerant to displacement noise as long as all input states have displacement errors less than \(\sqrt{\pi}/6\) \NoCaseChange{\protect\cite{cite4961}}.
\end{eczvaluelist}
\codefieldsection{Realizations}
\begin{eczvaluelist}
\item\relax Motional degree of freedom of a trapped ion: square-lattice GKP encoding realized with the help of post-selection by Home group \NoCaseChange{\protect\cite{cite4962,cite4963}}, followed by realization of reduced form of GKP error correction, where displacement error syndromes are measured to one bit of precision using an ion electronic state \NoCaseChange{\protect\cite{cite4948}}. State preparation also realized by Tan group \NoCaseChange{\protect\cite{cite4721}}. Universal gate set, including a two-qubit entangling gate, realized by Tan group \NoCaseChange{\protect\cite{cite4964}}. State initialization and application to measuring displacements \NoCaseChange{\protect\cite{cite4914}}.
\item\relax Microwave cavity coupled to superconducting circuits: reduced form of square-lattice GKP error correction, where displacement error syndromes are measured to one bit of precision using an ancillary transmon \NoCaseChange{\protect\cite{cite4907}}. Subsequent paper by Devoret group \NoCaseChange{\protect\cite{cite4718}} uses reinforcement learning for error-correction cycle design and is the first to go beyond break-even error-correction, with the lifetime of a logical qubit exceeding the cavity lifetime by about a factor of two (see also \NoCaseChange{\protect\cite{cite4717}}). See Ref. \NoCaseChange{\protect\cite{cite4719}} for another experiment. A feed-forward-free, i.e., fully autonomous protocol has also been implemented by Nord Quantique \NoCaseChange{\protect\cite{cite4965}}. Qudit encodings with \(q=3,4\) have been realized, with logical error rates also beyond break even \NoCaseChange{\protect\cite{cite4966}}.
\item\relax Optical systems: GKP states and homodyne measurements have been realized in propagating telecom light by the Furusawa group \NoCaseChange{\protect\cite{cite4967}} and on-chip by Xanadu Quantum Technologies \NoCaseChange{\protect\cite{cite4968}}.
\item\relax Single-qubit \(Z\)-gate has been demonstrated \NoCaseChange{\protect\cite{cite4969}} in the single-photon subspace of an infinite-mode space \NoCaseChange{\protect\cite{cite4970}}, in which time and frequency become bosonic conjugate variables of a single effective bosonic mode. In this context, GKP position-state wavefunctions are called Dirac combs or frequency combs.
\end{eczvaluelist}
\codefieldsection{Notes}
\begin{eczvaluelist}
\item\relax GKP syndrome extraction can be used for QKD with squeezed states \NoCaseChange{\protect\cite{cite4901}}.
\end{eczvaluelist}
\codefieldsection{Parents}
\begin{eczvaluelist}
\item\relax
\flmRefsHyperref[eczindexfamilyrel]{code:multimodegkp}{Gottesman-Kitaev-Preskill (GKP) code}\item\relax
\flmRefsHyperref[eczindexfamilyrel]{code:oscillator_css}{Bosonic CSS code}\item\relax
\flmRefsHyperref[eczindexfamilyrel]{code:single-mode}{Single-mode bosonic code}\end{eczvaluelist}
\codefieldsection{Cousins}
\begin{eczvaluelist}
\item\relax
\flmRefsHyperref[eczindexfamilyrel]{code:approximate_qecc}{Approximate quantum error-correcting code (AQECC)} --- Square-lattice GKP codes approximately protect against \flmRefsHyperref{ref498}{photon loss} \NoCaseChange{\protect\cite{cite2607,cite496,cite2608}}.
\item\relax
\flmRefsHyperref[eczindexfamilyrel]{code:rotor}{Rotor code} --- Because square-lattice GKP error states are parameterized by two modular (i.e., periodic) variables of position and momentum, measuring one of the GKP stabilizers constrains the oscillator Hilbert space into that of a rotor.
\item\relax
\flmRefsHyperref[eczindexfamilyrel]{code:hypercubic}{\(\mathbb{Z}^n\) hypercubic lattice} --- GKP codewords, when written in terms of coherent states, form a square lattice in phase space.
\item\relax
\flmRefsHyperref[eczindexfamilyrel]{code:fusion}{Fusion-based quantum computing (FBQC) code} --- GKP states can be used to perform computation in a fusion-based encoding \NoCaseChange{\protect\cite{cite3688}}.
\item\relax
\flmRefsHyperref[eczindexfamilyrel]{code:spt}{Symmetry-protected topological (SPT) code} --- The Segal-Bargmann representations of GKP states are the theta functions of the lowest Landau level on a torus \NoCaseChange{\protect\cite[{Sec. V}]{cite3073}\protect\cite[{Prop. 6.3}]{cite3074}} (see also Refs. \NoCaseChange{\protect\cite{cite3075,cite3076,cite3077}}).
\item\relax
\flmRefsHyperref[eczindexfamilyrel]{code:current_mirror}{Kitaev current-mirror qubit code} --- Current-mirror code phase gates utilize ancillary oscillators in square-lattice GKP states \NoCaseChange{\protect\cite{cite4971,cite4972}}.
\item\relax
\flmRefsHyperref[eczindexfamilyrel]{code:zero_pi}{Zero-pi qubit code} --- Zero-pi code phase gates utilize ancillary oscillators in square-lattice GKP states \NoCaseChange{\protect\cite{cite4971,cite4972}}.
\item\relax
\flmRefsHyperref[eczindexfamilyrel]{code:rotor_gkp}{Rotor GKP code} --- GKP (rotor GKP) codes protect against shifts in linear (angular) degrees of freedom.
\item\relax
\flmRefsHyperref[eczindexfamilyrel]{code:number_phase}{Number-phase code} --- Square-lattice GKP codes utilize translational symmetry in phase space, while number-phase codes utilize rotational symmetry. The two are related via a mapping \NoCaseChange{\protect\cite{cite4915}}.
\item\relax
\flmRefsHyperref[eczindexfamilyrel]{code:asymmetric_qecc}{Asymmetric quantum code (AQC)} --- GKP code parameters against position and momentum displacements can be tuned by the choice of lattice (e.g., square vs rectangular).
\item\relax
\flmRefsHyperref[eczindexfamilyrel]{code:qudit_gkp}{Modular-qudit GKP code} --- The square-lattice GKP code can be obtained from the modular-qudit code by taking the physical qudit dimension to be infinite \NoCaseChange{\protect\cite[{Sec. II}]{cite513}}.
\item\relax
\flmRefsHyperref[eczindexfamilyrel]{code:spin_gkp}{Spin GKP code} --- Spin-GKP code constructions utilize the Holstein-Primakoff mapping \NoCaseChange{\protect\cite{cite651,cite652,cite653}} to convert various expressions for square-lattice GKP states into codes for spin systems.
\end{eczvaluelist}
\eczhbkcontributors{ Yijia Xu, \eczhuVVA }
\endeczcode

\eczcode{squeezed_cat}{Squeezed cat code}{~\NoCaseChange{\protect\cite{cite4973,cite4974,cite4975}}}
\codefieldsection{Description}
Two-component cat code whose two coherent states have been squeezed in a direction perpendicular to the segment formed by the two coherent state values \(\pm\alpha\).

\codefieldsection{Protection}
Squeezing of coherent states allows for approximate protection against a single \flmRefsHyperref{ref498}{photon loss}.

\codefieldsection{Encoding}
\begin{eczvaluelist}
\item\relax Lindbladian-based dissipative encoding and autonomous QEC \NoCaseChange{\protect\cite{cite4973,cite4974,cite4975,cite4976}}.
\end{eczvaluelist}
\codefieldsection{Realizations}
\begin{eczvaluelist}
\item\relax Quantum optics: the Laurat group \NoCaseChange{\protect\cite{cite4977}}.
\item\relax Approximately squeezed cat states ("compressed cats") have been realized in a superconducting circuit device by the Gao group \NoCaseChange{\protect\cite{cite4978}} (see also \NoCaseChange{\protect\cite{cite4977}}). Dissipative stabilization has been demonstrated by Alice and Bob \NoCaseChange{\protect\cite{cite4979}}.
\end{eczvaluelist}
\codefieldsection{Parents}
\begin{eczvaluelist}
\item\relax
\flmRefsHyperref[eczindexfamilyrel]{code:single-mode}{Single-mode bosonic code}\item\relax
\flmRefsHyperref[eczindexfamilyrel]{code:ampdamp}{Amplitude-damping (AD) code} --- Squeezing of coherent states allows for approximate protection against a single \flmRefsHyperref{ref498}{photon loss}.
\end{eczvaluelist}
\codefieldsection{Child}
\begin{eczvaluelist}
\item\relax
\flmRefsHyperref[eczindexfamilyrel]{code:two-legged-cat}{Two-component cat code} --- The squeezed cat code reduces to the two-component cat code when there is no squeezing.
\end{eczvaluelist}
\codefieldsection{Cousins}
\begin{eczvaluelist}
\item\relax
\flmRefsHyperref[eczindexfamilyrel]{code:squeezed_fock_state}{Squeezed Fock-state code} --- Squeezed Fock-state codes and squeezed cat codes both utilize squeezing to approximately protect against loss errors.
\item\relax
\flmRefsHyperref[eczindexfamilyrel]{code:squeezed_coherent_bpsk}{Squeezed-coherent BPSK c-q modulation format} --- Squeezed-coherent BPSK c-q modulation transmits classical information using displaced-squeezed states, while squeezed cat codes store quantum information in superpositions of squeezed coherent states.
\end{eczvaluelist}
\eczhbkcontributors{ \eczhuVVA }
\endeczcode

\eczcode{squeezed_fock_state}{Squeezed Fock-state code}{~\NoCaseChange{\protect\cite{cite4980,cite4981,cite654,cite4982}}}
\codefieldsection{Description}
Approximate bosonic code that encodes a qubit into a superposition of one or a few squeezed Fock states, some of which are the result of a photon-number resolving measurement \NoCaseChange{\protect\cite{cite654}}.

The simplest code encodes a qubit into the same Fock state, but one which is squeezed in opposite directions \NoCaseChange{\protect\cite{cite4980}}.
Taking the Fock state \(|1\rangle\), the codewords are
\flmMathEnvironment{align}{}{
\begin{split}
|\overline{0}\rangle&=S(r)|1\rangle \\
|\overline{1}\rangle&=S(-r)|1\rangle~,
\end{split}
}
where \(S(\pm r)\) is the squeezing operator with squeezing parameter \(\pm r\).

\codefieldsection{Protection}
The code approximately protects against loss and dephasing errors, becoming exact in the \(r\to\infty\) limit.
\codefieldsection{Parents}
\begin{eczvaluelist}
\item\relax
\flmRefsHyperref[eczindexfamilyrel]{code:single-mode}{Single-mode bosonic code}\item\relax
\flmRefsHyperref[eczindexfamilyrel]{code:ampdamp}{Amplitude-damping (AD) code} --- The squeezed Fock-state code approximately protects against loss and dephasing errors, becoming exact in the \(r\to\infty\) limit.
\end{eczvaluelist}
\codefieldsection{Child}
\begin{eczvaluelist}
\item\relax
\flmRefsHyperref[eczindexfamilyrel]{code:squeezed_vacuum}{Squeezed-vacuum code} --- Squeezed-vacuum codewords are a special case of \flmRefsHyperref{code:squeezed_fock_state}{squeezed Fock states} with Fock state \(\ket{n=0}\).
\end{eczvaluelist}
\codefieldsection{Cousin}
\begin{eczvaluelist}
\item\relax
\flmRefsHyperref[eczindexfamilyrel]{code:squeezed_cat}{Squeezed cat code} --- Squeezed Fock-state codes and squeezed cat codes both utilize squeezing to approximately protect against loss errors.
\end{eczvaluelist}
\eczhbkcontributors{ \eczhuVVA }
\endeczcode

\eczcode{squeezed_vacuum}{Squeezed-vacuum code}{~\NoCaseChange{\protect\cite{cite4983,cite4916}}}
\codefieldsection{Alternative Names}
\begin{eczvaluelist}
\item\relax Multi-legged squeezed code
\end{eczvaluelist}
\eczhIndexCodeAliasName{squeezed_vacuum}{Multi-legged squeezed code}
\codefieldsection{Description}
A squeezed Fock-state code constructed from a coherent superposition of \(m\) squeezed vacuum states, each squeezed along equiangular axes in phase space. 

For an even integer \(m > 0\) (the number of ``legs'') and squeezing strength \(r\), the two logical codewords are defined as
\flmMathEnvironment{align}{}{
  |0_L\rangle & \propto \sum_{j=0}^{m-1} S\left(r, \frac{\pi j}{m}\right) \ket{0}, \\
  |1_L\rangle & \propto \sum_{j=0}^{m-1} (-1)^{j} S\left(r, \frac{\pi j}{m}\right) \ket{0},
}
where \(S(r,\theta) \equiv S(r e^{i\phi(\theta)})\) is the squeezing operator with \(\phi(\theta) = 2\theta + \pi \pmod{2\pi}\), defined as
\flmMathEnvironment{equation}{}{
S(\zeta) = \exp\!\left[\frac{1}{2}\left(\zeta^{*} a^{2}-\zeta\, a^{\dagger 2}\right)\right],
}
and where \(\zeta = r e^{i\phi}\) and \(r\) is the squeezing strength. 
This operator elongates the vacuum state along direction \(\theta\) in phase space.

\codefieldsection{Protection}
In Fock space, the logical states occupy photon numbers in distinct congruence classes \(n \equiv 0 \pmod{2m}\) and \(n \equiv m \pmod{2m}\), yielding photon-number distributions that are interleaved by \(\Delta n = m\). 
The code distance against single-photon loss is \(d = m\), where \(m\) is the number of legs (squeezed vacuum states in superposition). 
The code provides protection against both photon loss and dephasing noise, with a fundamental trade-off: increasing \(m\) improves loss tolerance at the cost of higher dephasing sensitivity.

\codefieldsection{Encoding}
\begin{eczvaluelist}
\item\relax Probabilistic preparation for \(m=2\): 
The simplest circuit uses a Hadamard-Controlled-Squeezing-Hadamard (\(H\)-\(CS\)-\(H\)) sequence on a single ancilla qubit coupled to a bosonic mode initially in vacuum. 
The controlled squeezing gate is defined as
\flmMathEnvironment{equation}{}{
CS(r;\theta_0,\theta_1) = \ket{0}\!\bra{0}\otimes S(r,\theta_0) + \ket{1}\!\bra{1}\otimes S(r,\theta_1).
}
After applying \(H\)-\(CS(r;0,\pi/2)\)-\(H\) and measuring the ancilla in the \(Z\) basis, the bosonic mode collapses to either \(\ket{0_L}\) or \(\ket{1_L}\) with probabilities
\flmMathEnvironment{equation}{}{
\text{prob}(L \mid r) = \frac{1}{2} + \frac{(-1)^L}{2 \cosh r \sqrt{\tanh^2 r + 1}},
}
where \(L \in \{0, 1\}\). Post-selection on the measurement outcome yields the desired logical state.

\item\relax Deterministic preparation for general \(m\):
For \(m > 2\), the codes can be prepared using sequences of conditional rotations \(CR(\theta)\) that rotate the bosonic mode in phase space by angle \(\theta\) conditioned on the qubit state, combined with logical-\(X\) gates that flip between the computational basis states. 
The circuit involves creating superpositions with single-qubit gates and applying conditional operations in a recursive manner. 
With full single-qubit control and controlled-squeezing operations, these circuits provide universal control over the joint qubit-oscillator system.

\item\relax Recursive construction:
Higher-\(m\) codes can be generated by feeding back the output of lower-\(m\) code preparation circuits, applying additional conditional rotations, and selecting appropriate measurement outcomes. This allows systematic construction of the entire code family.

\end{eczvaluelist}
\codefieldsection{Gates}
\begin{eczvaluelist}
\item\relax Universal computation requires multiple bosonic modes and entangling gates between them, such as a controllable cross-Kerr interaction \(a^\dagger a \otimes a^\dagger a\) that enables generation of the entangling \(CZ\) gate.
\item\relax Logical operations within a single mode can be performed using Gaussian operations (displacement, rotation, squeezing) combined with conditional control from an ancilla qubit.
\end{eczvaluelist}
\codefieldsection{Decoding}
\begin{eczvaluelist}
\item\relax The interleaved photon-number structure, with support on distinct classes modulo \(2m\), enables photon-number-resolving measurements to identify single-photon loss events, which can then be corrected.
\end{eczvaluelist}
\codefieldsection{Parents}
\begin{eczvaluelist}
\item\relax
\flmRefsHyperref[eczindexfamilyrel]{code:bosonic_rotation}{Bosonic rotation code} --- Squeezed-vacuum codes are qubit \flmRefsHyperref{code:bosonic_rotation}{rotation-symmetric bosonic codes} with \(m\)-fold rotational symmetry in phase space, constructed from \(m\) primitive squeezed vacuum states arranged at evenly-spaced angles.
\item\relax
\flmRefsHyperref[eczindexfamilyrel]{code:squeezed_fock_state}{Squeezed Fock-state code} --- Squeezed-vacuum codewords are a special case of \flmRefsHyperref{code:squeezed_fock_state}{squeezed Fock states} with Fock state \(\ket{n=0}\).
\end{eczvaluelist}
\codefieldsection{Cousin}
\begin{eczvaluelist}
\item\relax
\flmRefsHyperref[eczindexfamilyrel]{code:number_phase}{Number-phase code} --- Squeezed-vacuum codes are approximate \flmRefsHyperref{code:number_phase}{number-phase codes} whose logical Fock-space support lies in the congruence classes \(n \equiv 0 \pmod{2m}\) and \(n \equiv m \pmod{2m}\), approaching ideal \flmRefsHyperref{code:number_phase}{number-phase codes} as squeezing strength \(r \to \infty\)~\NoCaseChange{\protect\cite{cite4916}}.
\end{eczvaluelist}
\eczhbkcontributors{ Nir Gutman, \eczhuVVA }
\endeczcode

\eczcode{tiger}{Tiger code}{~\NoCaseChange{\protect\cite{cite4667}}}
\codefieldsection{Description}
A CSS-like multi-mode bosonic non-stabilizer code that generalizes the pair-cat code and whose syndromes are linear combinations of occupation-number operators.

A tiger code is defined for a pair of integer matrices, \(G\) and \(H\), satisfying a homological constraint \(GH^{\text{T}} = 0\). 
Stabilizer-like operators of the code are either linear combinations of occupation-number operators defined by rows of \(H\), or products of annihilation/creation operators whose powers are defined by rows of \(G\).

The structure of the logical space is determined from the homology of the integer chain complex defined by \(G\) and \(H\). The homology group of the logical operators has a torsion component because the chain complexes are defined over the ring of integers, which yields codes with finite logical dimension.
In general tiger codes, encodings of logical qubits, qudits, modes, and rotors are all possible.

Codewords are coherent states projected into the subspace defined by the \(H\)-induced constraint, and their corresponding normalizations are Gelfand-Kapranov-Zelevinsky hypergeometric functions \NoCaseChange{\protect\cite{cite4984,cite4985,cite4986}}.
Codewords can be finitely or infinitely supported in Fock space, depending on the \(H\)-induced constraint.
When written in terms of coherent states, codewords are orbits of a set of fiducial coherent states (determined by the aforementioned homology calculation) under a group of tensor-product rotations generated by \(H\).
Therefore, codewords consist of continuous but compact coherent-state constellations. 

Using multi-index notation, a projected coherent state can be written in two ways,
\flmMathEnvironment{align}{}{
  |\boldsymbol{\alpha}\rangle_{\boldsymbol{\Delta}}^{H}&\propto\int \textnormal{d}\boldsymbol{\phi}e^{i\boldsymbol{\phi}(H\hat{\mathbf{n}}-\boldsymbol{\Delta})}|\boldsymbol{\alpha}\rangle\\&\propto\sum_{H\mathbf{n}=\boldsymbol{\Delta}}\frac{\boldsymbol{\alpha}^{\mathbf{n}}}{\sqrt{\mathbf{n}!}}|\mathbf{n}\rangle~,
}
where \(\boldsymbol{\alpha}\) is a complex vector, \(\boldsymbol{\Delta}\) is an integer vector, and \(\boldsymbol{\phi}\) is a vector of phases iterating over the elements of the group generated by \(H\).
Tiger codewords are of the above form, and their phase-space values \(\boldsymbol{\alpha}\) lie on a torus embedded in the complex sphere of fixed-energy coherent states, satisfying \(|\alpha_j|^2 = 1\).

\codefieldsection{Protection}
Tiger codes protect against losses and gains of occupation numbers along with rotation noise stemming from modal dephasing.
For infinite-support tiger codes, protection against the latter type of noise is characterized by the minimum Euclidean distance \(d_Z\) between coherent states in different (continuous) logical constellations \NoCaseChange{\protect\cite{cite4667}}.
Dephasing protection in finite-support tiger codes is conjectured to be governed by the Euclidean distance, but some examples realize exact dephasing detection to finite order or suppression of dephasing errors with increasing \(\Delta\) \NoCaseChange{\protect\cite{cite4667}}.

\codefieldsection{Parent}
\begin{eczvaluelist}
\item\relax
\flmRefsHyperref[eczindexfamilyrel]{code:oscillators}{Bosonic code} --- Tiger codewords are superpositions of coherent states with the same energy, but coherent states are not eigenstates of the energy Hamiltonian. The \flmRefsHyperref{ref498}{AD} Kraus operator \(E_{0}^{\otimes n}\) acts identically on each coherent state by shrinking the radius of the QSC's sphere.
\end{eczvaluelist}
\codefieldsection{Children}
\begin{eczvaluelist}
\item\relax
\flmRefsHyperref[eczindexfamilyrel]{code:paircat}{Pair-cat code} --- The pair-cat code is a tiger code with \(G = (2,2)\) and \(H = (1,-1)\) \NoCaseChange{\protect\cite{cite4667}}.
\item\relax
\flmRefsHyperref[eczindexfamilyrel]{code:coherent_state_repetition}{Coherent-state repetition code} --- For odd \(n\), the coherent-state repetition code is a tiger code whose matrix \(G\) is the cyclic repetition generator matrix over the integers and whose matrix \(H\) is zero \NoCaseChange{\protect\cite{cite4667}}. For even \(n\), or after removing the last row to impose open boundaries, the construction yields a logical-rotor variant instead of a logical qubit.
\item\relax
\flmRefsHyperref[eczindexfamilyrel]{code:tiger_surface}{Tiger surface code} --- The tiger surface code is constructed from a hypergraph product of two repetition codes over the integers.
\end{eczvaluelist}
\codefieldsection{Cousins}
\begin{eczvaluelist}
\item\relax
\flmRefsHyperref[eczindexfamilyrel]{code:fock_state}{Fock-state bosonic code} --- Tiger codes encoding logical qudits are Fock-state codes.
\item\relax
\flmRefsHyperref[eczindexfamilyrel]{code:coherent_constellation}{Coherent-state constellation code} --- Tiger codewords consist of continuous and compact coherent-state constellations \NoCaseChange{\protect\cite{cite4667}}.
\item\relax
\flmRefsHyperref[eczindexfamilyrel]{code:qsc}{Quantum spherical code (QSC)} --- Tiger (quantum spherical) codewords consist of continuous and compact (discrete and finite) coherent-state constellations. Both codes protect against losses and gains of occupation numbers along with rotation noise stemming from modal dephasing. Protection against the latter type of noise is characterized by the minimum Euclidean distance between coherent states in different logical constellations.
\item\relax
\flmRefsHyperref[eczindexfamilyrel]{code:generalized_homological_product_css}{Generalized homological-product CSS code} --- Tiger codes are CSS-like multi-mode bosonic non-stabilizer codes constructed from chain complexes over the integers \NoCaseChange{\protect\cite{cite4667}}. The homology group of the logical operators has a torsion component because the chain complexes are defined over the ring of integers, which yields codes with finite logical dimension.
\item\relax
\flmRefsHyperref[eczindexfamilyrel]{code:homological_number-phase}{Homological number-phase code} --- Tiger codes of infinite Fock-state support can be thought of as appropriately regularized homological number-phase codes \NoCaseChange{\protect\cite{cite4667}}.
\item\relax
\flmRefsHyperref[eczindexfamilyrel]{code:oscillators_into_oscillators}{Oscillator-into-oscillator code} --- In general tiger codes, encodings of logical qubits, qudits, modes, and rotors are all possible.
\item\relax
\flmRefsHyperref[eczindexfamilyrel]{code:chi2}{\(\chi^{(2)}\) code} --- A three-mode tiger code with \(G=(2,2,-2)\), \(H=\left(\begin{smallmatrix}0&1&1\\1&0&1\end{smallmatrix}\right)\), and equal syndrome parameters has the same Fock-state support as one of the \(\chi^{(2)}\) codes \NoCaseChange{\protect\cite{cite4667}}.
\item\relax
\flmRefsHyperref[eczindexfamilyrel]{code:two-mode_binomial}{Two-mode binomial code} --- The two-mode binomial code for \(S=0\) is a tiger code with \(G = (2,-2)\) and \(H = (1,1)\) \NoCaseChange{\protect\cite{cite4667}}. It generalizes to the multinomial code, an \(n\)-mode tiger code encoding a qu\(n\)it in generalized \(SU(n)\) coherent states \NoCaseChange{\protect\cite{cite4667,cite4987}}.
\end{eczvaluelist}
\eczhbkcontributors{ \eczhuVVA }
\endeczcode

\eczcode{tiger_surface}{Tiger surface code}{~\NoCaseChange{\protect\cite{cite4667}}}
\codefieldsection{Description}
A tiger-code family constructed from a hypergraph product of two repetition codes over the integers, rather than from concatenating a cat code with a qubit surface code.
The code is conjectured to realize phases of \(U(1)\) gauge theory. 

An \(r \times (2m-1)\) lattice encodes a logical qubit into \(2rm-r\) bosonic modes with \(d_X=m\) and \(d_Z \geq 4rm\sin^2\!\left(\frac{\pi}{2m}\right)\) \NoCaseChange{\protect\cite{cite4667}}.
The \(m=2\) case is the liger (long-tiger) surface code, which has \(d_X=2\) and \(d_Z=4r\).

\codefieldsection{Protection}
The code corrects at least \(\lfloor (m-1)/2 \rfloor\) losses on arbitrary modes, and it detects pure-loss error patterns of total weight up to \(2m-2\) \NoCaseChange{\protect\cite{cite4667}}.
Its Euclidean distance obeys
\flmMathEnvironment{align}{}{
  4rm\sin^2\!\left(\frac{\pi}{2m}\right)\leq d_Z \leq 4rm\sin^2\!\left(\frac{\pi}{2m}\right)+4\sum_{j=1}^{m-1}\sin^2\!\left(\frac{(m-j)\pi}{m}\right)
}
so choosing \(r=\Omega(m^2)\) makes both \(d_X\) and \(d_Z\) grow with system size. For suitable nonzero syndrome choices, the liger surface code has exact orthogonality in both logical \(X\)- and \(Z\)-bases at arbitrary energy \NoCaseChange{\protect\cite{cite4667}}.

\codefieldsection{Parent}
\begin{eczvaluelist}
\item\relax
\flmRefsHyperref[eczindexfamilyrel]{code:tiger}{Tiger code} --- The tiger surface code is constructed from a hypergraph product of two repetition codes over the integers.
\end{eczvaluelist}
\codefieldsection{Cousins}
\begin{eczvaluelist}
\item\relax
\flmRefsHyperref[eczindexfamilyrel]{code:compactified_r}{Compactified \(\mathbb{R}\) gauge theory code} --- Both the compactified \(\mathbb{R}\) gauge theory and tiger surface code are constructed from a hypergraph product of two repetition codes over the integers.
\item\relax
\flmRefsHyperref[eczindexfamilyrel]{code:topological_abelian}{Abelian topological code} --- The tiger surface code is conjectured to realize phases of \(U(1)\) gauge theory.
\item\relax
\flmRefsHyperref[eczindexfamilyrel]{code:qudit_surface}{Modular-qudit surface code} --- The tiger surface code can be thought of as a realization of the \(q\to\infty\) \(U(1)\) rotor limit \NoCaseChange{\protect\cite{cite2531}} of the qudit surface code as a tiger code.
\item\relax
\flmRefsHyperref[eczindexfamilyrel]{code:hypergraph_product}{Hypergraph product (HGP) code} --- The tiger surface code is constructed from a hypergraph product of two repetition codes over the integers.
\item\relax
\flmRefsHyperref[eczindexfamilyrel]{code:repetition}{Repetition code} --- The tiger surface code is constructed from a hypergraph product of two repetition codes over the integers.
\end{eczvaluelist}
\eczhbkcontributors{ \eczhuVVA }
\endeczcode

\eczcode{two-legged-cat}{Two-component cat code}{~\NoCaseChange{\protect\cite{cite4988}}}
\codefieldsection{Description}
Code whose codespace is spanned by two coherent states \(\left|\pm\alpha\right\rangle\) for nonzero complex \(\alpha\).

An orthonormal basis for the codespace consists of the bosonic \textit{cat states} \NoCaseChange{\protect\cite{cite4840}}
\flmMathEnvironment{align}{}{
  |\overline{\pm}\rangle=\frac{\left|\alpha\right\rangle \pm\left|-\alpha\right\rangle }{\sqrt{2\left(1\pm e^{-2|\alpha|^{2}}\right)}}
}
for any complex \(\alpha\).

A closely related approximate cat code is called \textit{T4C code} \NoCaseChange{\protect\cite{cite4989}}.

\codefieldsection{Protection}
Two-component cat codes for large \(\alpha\) provide protection against modal dephasing, i.e., diffusion of the angular degree of freedom of the mode. A single \flmRefsHyperref{ref498}{photon loss} event maps the even and odd cat states approximately into each other and therefore acts as a logical bit flip rather than being corrected by the code. There exist modifications based on sign alternation \NoCaseChange{\protect\cite{cite4990}}, squeezing (yielding squeezed cat codes) \NoCaseChange{\protect\cite{cite4973,cite4974,cite4975}}, detuning \NoCaseChange{\protect\cite{cite4991}}, and addition of higher-order nonlinearities \NoCaseChange{\protect\cite{cite4992}} that can add such protection.
\codefieldsection{Encoding}
\begin{eczvaluelist}
\item\relax Lindbladian-based dissipative encoding and autonomous QEC \NoCaseChange{\protect\cite{cite4846}} utilizing two-photon absorption \NoCaseChange{\protect\cite{cite4842,cite4843,cite4844,cite4845,cite2794}}. Encoding passively protects against cavity dephasing, suppressing dephasing noise exponentially with \(|\alpha|^2\) \NoCaseChange{\protect\cite{cite4846}}. See Refs. \NoCaseChange{\protect\cite{cite4072,cite4993}} for analyses using displaced Fock states \NoCaseChange{\protect\cite{cite4994,cite4995}}. The Keldysh formalism yields non-perturbative bit-flip rates under various types of noise \NoCaseChange{\protect\cite{cite4996}}.
\item\relax Hamiltonian-based 'Kerr-cat' encoding utilizing the Kerr-effect Hamiltonian \NoCaseChange{\protect\cite{cite2840}} (see also Ref. \NoCaseChange{\protect\cite{cite4997}}).
\item\relax Colored dissipation \NoCaseChange{\protect\cite{cite4998}}.
\item\relax Combined dissipative and Hamiltonian-based encoding utilizing two-photon exchange with an ancillary qubit \NoCaseChange{\protect\cite{cite4999}}.
\item\relax Critical encoding at nonzero detuning \NoCaseChange{\protect\cite{cite5000}}.
\end{eczvaluelist}
\codefieldsection{Gates}
\begin{eczvaluelist}
\item\relax Universal gates in the quantum optical setting can be performed using teleportation, Bell measurements, displacements, and rotations \NoCaseChange{\protect\cite{cite4118}}. An earlier protocol requires a nonlinear interaction and uses state teleportation \NoCaseChange{\protect\cite{cite4117}}.
\item\relax Universal gates in the microwave setting can be performed using displacement operators and a rotation based on the Kerr nonlinearity \NoCaseChange{\protect\cite{cite4846}}. Kerr nonlinearity converts coherent states into Yurke-Stoler states \NoCaseChange{\protect\cite{cite5001}}.
\item\relax Bias-preserving \(X\), CNOT, and Toffoli gates \NoCaseChange{\protect\cite{cite2616,cite2646}}. A bias-preserving SWAP gate has also been proposed \NoCaseChange{\protect\cite{cite2647}}.
\item\relax Cat-transmon entangling gate using an ancillary qubit \NoCaseChange{\protect\cite{cite5002}}.
\end{eczvaluelist}
\codefieldsection{Decoding}
\begin{eczvaluelist}
\item\relax All-optical decoder \NoCaseChange{\protect\cite{cite5003}} based on Knill error correction (a.k.a. telecorrection \NoCaseChange{\protect\cite{cite3185}}), which is based on teleportation \NoCaseChange{\protect\cite{cite448,cite4369}}.
\end{eczvaluelist}
\codefieldsection{Fault Tolerance}
\begin{eczvaluelist}
\item\relax Fault-tolerant error-correction procedure using small amplitude coherent states \NoCaseChange{\protect\cite{cite3231}}.
\item\relax Bias-preserving \(X\), CNOT, and Toffoli gates \NoCaseChange{\protect\cite{cite2616,cite2646}}. A bias-preserving SWAP gate has also been proposed \NoCaseChange{\protect\cite{cite2647}}.
\end{eczvaluelist}
\codefieldsection{Realizations}
\begin{eczvaluelist}
\item\relax Lindbladian-based dissipative \NoCaseChange{\protect\cite{cite5004,cite5005}} and Hamiltonian-based 'Kerr-cat' \NoCaseChange{\protect\cite{cite5006}} encodings have been achieved in superconducting circuit devices by the Devoret group; Ref. \NoCaseChange{\protect\cite{cite5005}} also demonstrated a displacement-based gate. The Lindbladian-based scheme has further achieved a suppression of bit-flip errors that is exponential in the average photon number up to a bit-flip time of 1ms \NoCaseChange{\protect\cite{cite5007}}. A bit-flip time of up to 10s has been achieved for the two-component cat code in the classical-bit regime \NoCaseChange{\protect\cite{cite5008,cite5009,cite5010}}. A holonomic gate has been repurposed as a logical measurement \NoCaseChange{\protect\cite{cite4848}}. The 'Kerr-cat' encoding and a \(\pi/2\) gate have been realized with the help of a band-block filter, yielding a bit-flip lifetime of 1 ms in the 10-photon regime \NoCaseChange{\protect\cite{cite5011}} (see also Ref. \NoCaseChange{\protect\cite{cite5012}}). Lindblad-based encoding achieved in a 2D cavity by AWS \NoCaseChange{\protect\cite{cite5013}}.
\item\relax T4C code realized in a superconducting circuit device by the Wang group \NoCaseChange{\protect\cite{cite4989}}.
\end{eczvaluelist}
\codefieldsection{Notes}
\begin{eczvaluelist}
\item\relax Pedagogical introduction to cat codes in the context of microwave cavities can be found in Refs. \NoCaseChange{\protect\cite{cite5014,cite5015}}, and in the context of optical systems in books \NoCaseChange{\protect\cite{cite5016,cite5017,cite5018}}.
\item\relax Ground states of the fluxonium superconducting qubit resemble two-component cat codewords \NoCaseChange{\protect\cite{cite5019}}.
\end{eczvaluelist}
\codefieldsection{Parents}
\begin{eczvaluelist}
\item\relax
\flmRefsHyperref[eczindexfamilyrel]{code:cat}{Cat code} --- The cat code reduces to its two-component version for \(S=0\).
\item\relax
\flmRefsHyperref[eczindexfamilyrel]{code:coherent_state_repetition}{Coherent-state repetition code} --- The coherent-state repetition code for \(n=1\) reduces to the two-component cat code.
\item\relax
\flmRefsHyperref[eczindexfamilyrel]{code:squeezed_cat}{Squeezed cat code} --- The squeezed cat code reduces to the two-component cat code when there is no squeezing.
\end{eczvaluelist}
\codefieldsection{Cousins}
\begin{eczvaluelist}
\item\relax
\flmRefsHyperref[eczindexfamilyrel]{code:hamiltonian}{Hamiltonian-based code} --- The two-component cat code forms the ground-state subspace of a Kerr Hamiltonian \NoCaseChange{\protect\cite{cite2840}}.
\item\relax
\flmRefsHyperref[eczindexfamilyrel]{code:quantum_repetition}{Quantum repetition code} --- Two-component cat and quantum repetition codes can be thought of as classical codes because they protect against only one type of noise. Two-component cat codes (quantum repetition) codes suppress cavity dephasing (bit-flip) noise exponentially with \(|\alpha|^2\) (\(n\)). The stability offered by cat codes has been linked to several favorable properties of phases of matter associated with the repetition-code Hamiltonian \NoCaseChange{\protect\cite{cite4114,cite4115}}.
\item\relax
\flmRefsHyperref[eczindexfamilyrel]{code:coherent_state_c-q}{Coherent-state c-q modulation format} --- Two-component cat codes can be thought of as coherent-state c-q codes because they protect against only one type of noise and thus only reliably store classical information.
\item\relax
\flmRefsHyperref[eczindexfamilyrel]{code:asymmetric_qecc}{Asymmetric quantum code (AQC)} --- Cat qubits provide an asymmetric-noise platform admitting bias-preserving \(X\), CNOT, and Toffoli gates \NoCaseChange{\protect\cite{cite2616,cite2646}}. A bias-preserving SWAP gate has also been proposed \NoCaseChange{\protect\cite{cite2647}}.
\item\relax
\flmRefsHyperref[eczindexfamilyrel]{code:bpsk}{Binary PSK (BPSK) modulation format} --- BPSK (two-component cat) codes are used to transmit classical (quantum) information using (superpositions of) antipodal coherent states over classical (quantum) channels.
\item\relax
\flmRefsHyperref[eczindexfamilyrel]{code:quantum_bpsk}{BPSK c-q modulation format} --- BPSK c-q (two-component cat) codes are used to transmit classical (quantum) information using (superpositions of) antipodal coherent states over quantum channels.
\item\relax
\flmRefsHyperref[eczindexfamilyrel]{code:qubit_stabilizer}{Qubit stabilizer code} --- Ancilla modes can be used for syndrome extraction instead of ancilla qubits \NoCaseChange{\protect\cite{cite4348}}, and using two-component cat codes \NoCaseChange{\protect\cite{cite4349}} yields fault-tolerant syndrome extraction circuits.
\item\relax
\flmRefsHyperref[eczindexfamilyrel]{code:spin_cat}{Spin cat code} --- The spin-cat code construction utilizes the Holstein-Primakoff mapping \NoCaseChange{\protect\cite{cite651,cite652,cite653}} to convert cat codes into codes for spin systems.
\end{eczvaluelist}
\eczhbkcontributors{ Hyunseok Jeong, \eczhuVVA }
\endeczcode

\eczcode{two-mode_binomial}{Two-mode binomial code}{~\NoCaseChange{\protect\cite{cite5020,cite2600}}}
\codefieldsection{Description}
Two-mode constant-energy CLY code whose coefficients are square-roots of binomial coefficients.

The simplest two-mode \(S=1\) code is an analogue of the "0-2-4" single-mode binomial code \NoCaseChange{\protect\cite{cite5020}}, with codewords
\flmMathEnvironment{align}{}{
\begin{split}
  |\overline{0}\rangle&=\frac{1}{\sqrt{2}}\left(|40\rangle+|04\rangle\right)\\
  |\overline{1}\rangle&=|22\rangle~.
\end{split}
}

An alternative basis for general codewords is
\flmMathEnvironment{align}{}{
  |\overline{\mu}\rangle=\frac{1}{2^{J}}\sum_{m=0}^{2J}\left(-1\right)^{\mu m}\sqrt{{2J \choose m}}\left|2J-(S+1)m,(S+1)m\right\rangle~,
}
with spacing \(S\) and dephasing error parameter \(N\) such that \(J = \frac{1}{2}(N+1)(S+1)\) \NoCaseChange{\protect\cite{cite4707}}.
The \(S=0\) version can be obtained by applying a \(50:50\) beamsplitter to the highest-weight Fock states \(|2J,0\rangle\) and \(|0,2J\rangle\) \NoCaseChange{\protect\cite{cite4049}}; in this case, codewords are two-mode binomial coherent states \NoCaseChange{\protect\cite{cite5021,cite5022,cite5023}}.

\codefieldsection{Protection}
The code exactly detects dephasing errors \(\hat{\mathbf{a}}^{\dagger \mathbf{p}}\hat{\mathbf{a}}^{\mathbf{p}}\) whenever \(p_1+p_2<\Delta\) \NoCaseChange{\protect\cite{cite4667}}.
It also detects \(p_1\) photon loss errors \(\hat{\mathbf{a}}^{\mathbf{p}}\) and \(p_2\) photon gain errors \(\hat{\mathbf{a}}^{\dagger \mathbf{p}}\) whenever \(p_1+p_2\leq S\) \NoCaseChange{\protect\cite{cite4667}}.

\codefieldsection{Gates}
\begin{eczvaluelist}
\item\relax A beamsplitter is a logical operation for the \(S=0\) two-mode binomial code \NoCaseChange{\protect\cite{cite4667}}.
\end{eczvaluelist}
\codefieldsection{Parent}
\begin{eczvaluelist}
\item\relax
\flmRefsHyperref[eczindexfamilyrel]{code:chuang-leung-yamamoto}{Chuang-Leung-Yamamoto (CLY) code}\end{eczvaluelist}
\codefieldsection{Child}
\begin{eczvaluelist}
\item\relax
\flmRefsHyperref[eczindexfamilyrel]{code:dual_rail}{Dual-rail quantum code} --- The two-mode binomial code for \(S=N=0\) reduces to the dual-rail code.
\end{eczvaluelist}
\codefieldsection{Cousins}
\begin{eczvaluelist}
\item\relax
\flmRefsHyperref[eczindexfamilyrel]{code:tiger}{Tiger code} --- The two-mode binomial code for \(S=0\) is a tiger code with \(G = (2,-2)\) and \(H = (1,1)\) \NoCaseChange{\protect\cite{cite4667}}. It generalizes to the multinomial code, an \(n\)-mode tiger code encoding a qu\(n\)it in generalized \(SU(n)\) coherent states \NoCaseChange{\protect\cite{cite4667,cite4987}}.
\item\relax
\flmRefsHyperref[eczindexfamilyrel]{code:binomial}{Binomial code} --- Two-mode binomial codes are two-mode analogues of binomial codes.
\item\relax
\flmRefsHyperref[eczindexfamilyrel]{code:chi2}{\(\chi^{(2)}\) code} --- Two-mode binomial codes \NoCaseChange{\protect\cite[{Eqs. (90-91)}]{cite4666}} are closely related to three-mode \(\chi^2\) binomial codes \NoCaseChange{\protect\cite[{Eqs. (61-62)}]{cite4666}}.
\item\relax
\flmRefsHyperref[eczindexfamilyrel]{code:oscillators_concatenated}{Concatenated bosonic code} --- Two-mode binomial codes can be concatenated with repetition codes to yield bosonic analogues of QPCs \NoCaseChange{\protect\cite{cite4049}}.
\item\relax
\flmRefsHyperref[eczindexfamilyrel]{code:quantum_parity}{Quantum parity code (QPC)} --- Two-mode binomial codes can be concatenated with repetition codes to yield bosonic analogues of QPCs \NoCaseChange{\protect\cite{cite4049}}.
\item\relax
\flmRefsHyperref[eczindexfamilyrel]{code:group_representation}{Group-representation code} --- An application of the group-representation encoding construction \NoCaseChange{\protect\cite[{Lemma 1}]{cite2810}} yields a family of two-mode codes that closely resemble the two-mode binomial codes \NoCaseChange{\protect\cite{cite2816}}.
\item\relax
\flmRefsHyperref[eczindexfamilyrel]{code:2t_qutrit}{2T-qutrit code} --- The \(2T\)-qutrit code reduces to the two-mode "0-2-4" binomial code as \(\alpha\to 0\).
\end{eczvaluelist}
\eczhbkcontributors{ \eczhuVVA }
\endeczcode

\eczcode{very-small-logical-qubit}{Very small logical qubit (VSLQ) code}{~\NoCaseChange{\protect\cite{cite5024,cite5025}}}
\codefieldsection{Description}
A code consisting of two logical codewords \(|\pm\rangle \propto (|0\rangle\pm|2\rangle)(|0\rangle\pm|2\rangle)\), where the total Hilbert space is the tensor product of two transmon qudits (whose ground states \(|0\rangle\) and second excited states \(|2\rangle\) are used in the codewords).
Since the code is intended to protect against losses, the qutrits can equivalently be thought of as oscillator Fock-state subspaces.

In the original proposal for autonomous stabilization \NoCaseChange{\protect\cite{cite5024}}, the single logical qubit is given by the two lowest energy states of a time-dependent Hamiltonian acting on two transmon qutrits and two lossy oscillators.

\codefieldsection{Protection}
Protects against a single \flmRefsHyperref{ref498}{photon loss}.
\codefieldsection{Encoding}
\begin{eczvaluelist}
\item\relax Engineering a circuit made of two transmons and two oscillators coupled through three driven superconducting quantum interference devices (SQUIDs) results in passive stabilization of the logical states.
\end{eczvaluelist}
\codefieldsection{Gates}
\begin{eczvaluelist}
\item\relax Single logical qubit operations implemented by resonant physical qubit driving and phase shifting the SQUID drives.
\item\relax A CZ gate between two logical qubits implemented by coupling devices through another driven SQUID and applying a pulse to the coupling squid simultaneously with a single qubit operation on one of the logical qubits.
\end{eczvaluelist}
\codefieldsection{Decoding}
\begin{eczvaluelist}
\item\relax Logical qubit can be measured with physical qubit measurements along \(X\). Can be implemented by engineering a coupling of one of the qubits to a readout cavity via the interaction \(\sigma_x (a+a^\dagger)\) \NoCaseChange{\protect\cite{cite5026}}. This results in an \(X\)-dependent shift of the readout cavity resonance which can be measured.
\item\relax Star-code autonomous correction scheme \NoCaseChange{\protect\cite{cite5025}}.
\end{eczvaluelist}
\codefieldsection{Realizations}
\begin{eczvaluelist}
\item\relax Star-code autonomous correction scheme realized using superconducting circuits \NoCaseChange{\protect\cite{cite5025}}.
\end{eczvaluelist}
\codefieldsection{Parents}
\begin{eczvaluelist}
\item\relax
\flmRefsHyperref[eczindexfamilyrel]{code:fock_state}{Fock-state bosonic code}\item\relax
\flmRefsHyperref[eczindexfamilyrel]{code:constant_excitation}{Constant-excitation (CE) code}\item\relax
\flmRefsHyperref[eczindexfamilyrel]{code:permutation_invariant}{Permutation-invariant (PI) code}\end{eczvaluelist}
\codefieldsection{Cousins}
\begin{eczvaluelist}
\item\relax
\flmRefsHyperref[eczindexfamilyrel]{code:hybrid_qudit_oscillator}{Mixed oscillator code} --- VSLQ decoder utilizes two ancillary oscillators.
\item\relax
\flmRefsHyperref[eczindexfamilyrel]{code:quantum_repetition}{Quantum repetition code} --- Parts of the VSLQ codewords resemble the two-qubit phase-flip repetition code, though the code cannot correct phase errors. Unlike the phase-flip code, the VSLQ code can correct for single \flmRefsHyperref{ref498}{photon loss} because it uses the second excited state in the construction, which remains distinct from the vacuum even after \flmRefsHyperref{ref498}{photon loss}.
\end{eczvaluelist}
\eczhbkcontributors{ Jonathan Kunjummen, \eczhuVVA }
\endeczcode

\eczcode{wasilewski-banaszek}{Wasilewski-Banaszek code}{~\NoCaseChange{\protect\cite{cite5027}}}
\codefieldsection{Description}
Three-oscillator constant-excitation Fock-state code encoding a single logical qubit.

A basis of codewords is
\flmMathEnvironment{align}{}{
\begin{split}
|\overline{0}\rangle &= \frac{1}{\sqrt{3}}(|003\rangle+|030\rangle+|300\rangle)\\
|\overline{1}\rangle &= |111\rangle
\end{split}.
}

\codefieldsection{Protection}
Protects against single \flmRefsHyperref{ref498}{photon loss} in any one mode.
\codefieldsection{Encoding}
\begin{eczvaluelist}
\item\relax A qubit in the dual-rail code can be transferred to this code via a linear optical network using four ancillary modes, each with one photon input. Successful encoding is conditioned on measuring the state \(|110\rangle\) on the last three modes.
\end{eczvaluelist}
\codefieldsection{Gates}
\begin{eczvaluelist}
\item\relax Single-qubit gates implemented using linear optical networks, sometimes with the addition of auxiliary modes with vacuum input and (conditional) output.
\end{eczvaluelist}
\codefieldsection{Decoding}
\begin{eczvaluelist}
\item\relax Destructive measurement with photon number measurements on each mode.
\end{eczvaluelist}
\codefieldsection{Parent}
\begin{eczvaluelist}
\item\relax
\flmRefsHyperref[eczindexfamilyrel]{code:constant_excitation_permutation_invariant}{Ouyang-Chao constant-excitation PI code} --- The Wasilewski-Banaszek code is a simple example of an Ouyang-Chao PI code \NoCaseChange{\protect\cite{cite2946}}.
\end{eczvaluelist}
\codefieldsection{Cousin}
\begin{eczvaluelist}
\item\relax
\flmRefsHyperref[eczindexfamilyrel]{code:three_qutrit_permutation_invariant}{\(\llparenthesis 3,2,2\rrparenthesis _3\) Three-qutrit single-deletion code} --- The three-qutrit single-deletion code maps to the Wasilewski-Banaszek code via the \flmRefsHyperref{ref499}{simplex mapping} \NoCaseChange{\protect\cite{cite500}}.
\end{eczvaluelist}
\eczhbkcontributors{ Jonathan Kunjummen, \eczhuVVA }
\endeczcode

\onecolumngrid
\clearpage

\section{Spin Kingdom}

\begin{eczEpigraph}
\begin{quote}
\flmQuoteSetup{quote}%
Imagine it's a ball that's rotating, except it's not a ball and it's not rotating.
\end{quote}
\end{eczEpigraph}

\twocolumngrid

\eczcode{su3_sigma360}{\(\llparenthesis 5,3,2\rrparenthesis _3\) qutrit code}{~\NoCaseChange{\protect\cite{cite2199}}}
\eczhIndexCodeAliasName{su3_sigma360}{qutrit code}
\codefieldsection{Description}
Smallest qutrit block code realizing the \(\Sigma(360\phi)=3.A_6\) subgroup of \(SU(3)\) transversally.
The next smallest code is \(\llparenthesis 7,3,2\rrparenthesis _3\).

\codefieldsection{Transversal and Permutation-Based Gates}
\begin{eczvaluelist}
\item\relax \(\Sigma(360\phi)=3.A_6\) group gates can be realized transversally.
\end{eczvaluelist}
\codefieldsection{Parent}
\begin{eczvaluelist}
\item\relax
\flmRefsHyperref[eczindexfamilyrel]{code:t_group}{Twisted \(1\)-group code} --- The \(\llparenthesis 5,3,2\rrparenthesis _3\) qutrit code admits a transversal representation of the twisted \(1\)-group \(\Sigma(360\phi)=3.A_6\) \NoCaseChange{\protect\cite{cite2199}}.
\end{eczvaluelist}
\codefieldsection{Cousin}
\begin{eczvaluelist}
\item\relax
\flmRefsHyperref[eczindexfamilyrel]{code:su3_spin}{\(SU(3)\) spin code} --- The \(\llparenthesis 5,3,2\rrparenthesis _3\) qutrit code can be interpreted as a \(SU(3)\) single-spin code via the \flmRefsHyperref{ref499}{simplex mapping} \NoCaseChange{\protect\cite[{Prop. VI.2}]{cite500}}.
\end{eczvaluelist}
\eczhbkcontributors{ \eczhuVVA }
\endeczcode

\eczcode{su3_spin}{\(SU(3)\) spin code}{~\NoCaseChange{\protect\cite{cite2815}}}
\eczhIndexCodeAliasName{su3_spin}{spin code}
\codefieldsection{Description}
An extension of Clifford single-spin codes to the group \(SU(3)\), whose codespace is a projection onto a particular irrep of a subgroup of \(SU(3)\) of an underlying spin that houses some particular irrep of \(SU(3)\).

\codefieldsection{Parents}
\begin{eczvaluelist}
\item\relax
\flmRefsHyperref[eczindexfamilyrel]{code:single_spin}{Single-spin code}\item\relax
\flmRefsHyperref[eczindexfamilyrel]{code:group_representation}{Group-representation code} --- \(SU(3)\) spin codes are group-representation codes with \(G\) being a subgroup of \(SU(3)\) \NoCaseChange{\protect\cite{cite2815}}.
\end{eczvaluelist}
\codefieldsection{Cousin}
\begin{eczvaluelist}
\item\relax
\flmRefsHyperref[eczindexfamilyrel]{code:su3_sigma360}{\(\llparenthesis 5,3,2\rrparenthesis _3\) qutrit code} --- The \(\llparenthesis 5,3,2\rrparenthesis _3\) qutrit code can be interpreted as a \(SU(3)\) single-spin code via the \flmRefsHyperref{ref499}{simplex mapping} \NoCaseChange{\protect\cite[{Prop. VI.2}]{cite500}}.
\end{eczvaluelist}
\eczhbkcontributors{ \eczhuVVA }
\endeczcode

\eczcode{su3_tverberg_spin}{\(SU(3)\) Tverberg spin code}{~\NoCaseChange{\protect\cite{cite5028}\protect\cite[{Ex. 6.4}]{cite5029}}}
\eczhIndexCodeAliasName{su3_tverberg_spin}{Tverberg spin code}
\codefieldsection{Description}
\(SU(3)\) single-spin code family obtained from the two-step Tverberg construction \NoCaseChange{\protect\cite{cite648}} in the totally symmetric \(N\)-particle irrep \(\mathcal{H}=\mathrm{Sym}^N(\mathbb{C}^3)\) of \(\mathfrak{su}(3)\), which has dimension \(\binom{N+2}{2}\).
Its weight basis is indexed by the \flmRefsHyperref{ref655}{discrete simplex} \(\Delta_{3,N}\), whose centered form is the triangular \(A_2\) lattice.

A distance-two intermediate subspace can be chosen as
\flmMathEnvironment{align}{}{
  \mathcal{B}=\mathrm{span}\{|a_1a_2a_3\rangle: a_1-a_2\equiv0\pmod 3\}~,
}
giving \(\dim\mathcal{B}\approx\binom{N+2}{2}/3\) \NoCaseChange{\protect\cite{cite5028}\protect\cite[{Ex. 6.4}]{cite5029}}.
A symmetric partition of this sublattice into pairs in the central hexagon and triples near the corners yields an error-detecting code for single Lie-algebra errors of dimension \(\frac{4}{27}\binom{N+2}{2}+O(N)\) \NoCaseChange{\protect\cite[{Sec. 6.2.1}]{cite5030}}.

\codefieldsection{Protection}
Detects errors in the \(\mathfrak{su}(3)\) Lie-algebra error set.
More generally, for the Lie-type graph metric \(V_t=\mathrm{span}(\mathfrak{su}(3)\oplus \mathbb{C}I)^t\), a distance-\(d\) construction can be obtained by first choosing a graph-distance-\(d\) subset of the discrete simplex \(\Delta_{3,N}\) \NoCaseChange{\protect\cite[{Secs. 5.1,6.2}]{cite5030}}.

\codefieldsection{Rate}
For general distance \(d\), the two-step construction uses an asymptotically optimal distance-\(d\) sublattice of the centered \(\Delta_{3,N}\) discrete simplex for the intermediate space.
If \(d=2t\), then \(\dim\mathcal{B}=\dim\mathcal{H}/(3t^2)+O(N)\); if \(d=2t+1\), then \(\dim\mathcal{B}=\dim\mathcal{H}/(3t^2+3t+1)+O(N)\) \NoCaseChange{\protect\cite[{Sec. 6.2.2}]{cite5030}}.

\codefieldsection{Parent}
\begin{eczvaluelist}
\item\relax
\flmRefsHyperref[eczindexfamilyrel]{code:single_spin}{Single-spin code}\end{eczvaluelist}
\eczhbkcontributors{ \eczhuVVA }
\endeczcode

\eczcode{su4_tverberg_spin}{\(SU(4)\) Tverberg spin code}{~\NoCaseChange{\protect\cite[{Sec. 6.3}]{cite5030}}}
\eczhIndexCodeAliasName{su4_tverberg_spin}{Tverberg spin code}
\codefieldsection{Description}
Single-spin code family in the totally symmetric \(N\)-particle irrep \(\mathcal{H}=\mathrm{Sym}^N(\mathbb{C}^4)\) of \(\mathfrak{su}(4)\), whose weight diagram is the \flmRefsHyperref{ref655}{discrete simplex} \(\Delta_{4,N}\), equivalently its centered tetrahedral realization.
The construction uses the two-step Tverberg-theorem method \NoCaseChange{\protect\cite{cite648}}: first choose an intermediate subspace from a distance-two subset of \(\Delta_{4,N}\), then combine basis states whose convex hulls contain the origin.

For \(N\) divisible by eight, an optimal distance-two intermediate lattice is obtained as the kernel of
\flmMathEnvironment{align}{}{
  a_1L_1+a_2L_2+a_3L_3+a_4L_4 \mapsto a_1+2a_2+3a_3 \pmod 4~.
}
The second step pairs opposite points in the central octahedral region and can form additional triples and quadruples near the tetrahedron's corners; for \(N=8\), this partition is optimal within the two-step framework \NoCaseChange{\protect\cite[{Sec. 6.3 and Ex. 6.10}]{cite5030}}.

More explicitly, write the normalized monomial basis of \(\mathrm{Sym}^N(\mathbb{C}^4)\) as \(|a_1a_2a_3a_4\rangle\), where \((a_1,a_2,a_3,a_4)\in\Delta_{4,N}\), and let \(\Lambda_B^N\) be the kernel above.
For each block \(Y\subset\Lambda_B^N\) whose convex hull contains the origin, choose barycentric weights \(\{\beta_{\mathbf{a}}\}_{\mathbf{a}\in Y}\) satisfying \(\beta_{\mathbf{a}}\geq0\), \(\sum_{\mathbf{a}\in Y}\beta_{\mathbf{a}}=1\), and \(\sum_{\mathbf{a}\in Y}\beta_{\mathbf{a}}\mathbf{a}=0\) in centered \(\Delta_{4,N}\) coordinates.
The corresponding codeword is
\flmMathEnvironment{align}{}{
  |\psi_Y\rangle=\sum_{\mathbf{a}\in Y}\sqrt{\beta_{\mathbf{a}}}\,|\mathbf{a}\rangle~.
}
Thus opposite pairs give \((|p\rangle+|-p\rangle)/\sqrt{2}\); triples of the form \(\{p,p',-(p+p')\}\) give \((|p\rangle+|p'\rangle+|-(p+p')\rangle)/\sqrt{3}\); and zero-sum quadruples give equal four-term superpositions.

\codefieldsection{Protection}
Detects single Lie-algebra errors from the \(\mathfrak{su}(4)\) error set, equivalently errors in \(V_1=\mathfrak{su}(4)\oplus \mathbb{C}I\) in the Lie-type graph metric.
The first step suppresses off-diagonal root-space errors by choosing weight vectors separated by graph distance at least two in the centered \(\Delta_{4,N}\) discrete simplex, leaving only a commuting diagonal error algebra for the convex-geometric second step \NoCaseChange{\protect\cite[{Secs. 5.1,5.2,6.3}]{cite5030}}.

\codefieldsection{Rate}
The distance-two intermediate subset has asymptotic density \(1/4\) in \(\Delta_{4,N}\).
The generic Tverberg step gives an asymptotic code dimension at least \(\dim\mathcal{H}/16\), while the extra triangle constructions improve finite-size instances but do not change the leading asymptotic rate in the construction described in Ref. \NoCaseChange{\protect\cite[{Sec. 6.3}]{cite5030}}.

\codefieldsection{Parent}
\begin{eczvaluelist}
\item\relax
\flmRefsHyperref[eczindexfamilyrel]{code:single_spin}{Single-spin code}\end{eczvaluelist}
\eczhbkcontributors{ \eczhuVVA }
\endeczcode

\eczcode{ae}{Æ code}{~\NoCaseChange{\protect\cite{cite5031}}}
\codefieldsection{Description}
Code defined in a single angular-momentum subspace that is embedded in a larger direct-sum space of different angular momenta, which can arise from combinations of spin, electronic, or rotational, or nuclear angular momenta of an atom or molecule.
A code is obtained by solving an over-constrained system of equations, and many solutions can be mapped into existing codes defined on other state spaces.

A simple example of an Æ code is the error-detecting code with codewords
\flmMathEnvironment{align}{}{
\begin{split}
  |\overline{0}\rangle&=\frac{1}{\sqrt{2}}\left(|_{-m}^{J}\rangle+|_{m}^{J}\rangle\right)\\|\overline{1}\rangle&=|_{0}^{J}\rangle~,
\end{split}
}
constructed out of states of total angular momentum \(J\) and its projection \(m\) for any \(J,m\geq 2\).
This code detects a single change in \(m\) or \(J\).

\codefieldsection{Protection}
Protects against noise native to atomic and molecular platforms, such as spontaneous emission, stray electromagnetic fields, and Raman scattering.
Noise operators arising from these processes, when restricted to the angular momentum degrees of freedom, change either the total angular momentum or its projection and correspond to matrices whose elements are particular combinations of Clebsch-Gordan coefficients.

\codefieldsection{Realizations}
\begin{eczvaluelist}
\item\relax Trapped ions: smallest antisymmetric code protecting against dephasing has been realized by the Du group \NoCaseChange{\protect\cite{cite5032}}.
\end{eczvaluelist}
\codefieldsection{Parent}
\begin{eczvaluelist}
\item\relax
\flmRefsHyperref[eczindexfamilyrel]{code:spins_into_spins}{Spin code} --- Æ codes protect against changes in both the total angular momentum \(J\) and its projection \(m\), with the former type necessarily causing the information to leak out of the space of a single spin.
\end{eczvaluelist}
\codefieldsection{Cousins}
\begin{eczvaluelist}
\item\relax
\flmRefsHyperref[eczindexfamilyrel]{code:single_spin}{Single-spin code} --- Since Æ codes are defined in a subspace of fixed total angular momentum and protect against errors linear in the angular-momentum generators, they can also be thought of as single-spin codes.
\item\relax
\flmRefsHyperref[eczindexfamilyrel]{code:diatomic_molecular}{Diatomic molecular code} --- Diatomic molecular codes are supported on states with various total angular momenta, while Æ codes are supported on only one subspace of fixed total momentum. The latter codes are more practical and applicable to other spin spaces.
\item\relax
\flmRefsHyperref[eczindexfamilyrel]{code:binomial}{Binomial code} --- Many well-performing Æ codes can be mapped into shifted versions of binomial codes via the Holstein-Primakoff mapping.
\item\relax
\flmRefsHyperref[eczindexfamilyrel]{code:gnu_permutation_invariant}{GNU PI code} --- Many well-performing Æ codes can be mapped into GNU codes via the \flmRefsHyperref{ref526}{Dicke state mapping}.
\end{eczvaluelist}
\eczhbkcontributors{ \eczhuVVA }
\endeczcode

\eczcode{j_gross}{Clifford-group spin code}{~\NoCaseChange{\protect\cite{cite646,cite2814}}}
\codefieldsection{Description}
A single-spin code designed to realize a discrete group of gates using \(SU(2)\) rotations.
Codewords are subspaces of a spin's Hilbert space that house irreducible representations (irreps) of a discrete subgroup of \(SU(2)\).

The first realization \NoCaseChange{\protect\cite{cite646}} used the \flmRefsHyperref{ref409}{single-qubit Clifford group} (effectively, the binary octahedral, or \(2O\) subgroup of \(SU(2)\)).
Code construction is done by restricting the \(SU(2)\) irrep to \(2O\), and determining the carrier spaces of any nontrivial irreps of \(2O\). Since irreps of \(2O\) do not appear in integer spins, half-integer spins are used.

A simple example of a codespace is a projection onto an instance of a particular irrep of \(2O\), referred to as either \( \varrho_4 \) or \( \varrho_5 \).
In the case of only one instance of the desired irrep present in the spin, the projection is created as follows:
\flmMathEnvironment{align}{}{
  P_\varrho = \frac{\text{dim} \varrho}{|2O|} \sum_{g \in 2O} \chi_\varrho (g)^* D(g)~,
}
where \(D(g)\) is the \(SU(2)\) Wigner matrix corresponding to group element \(g\), and the character \(\chi_\varrho (g) = \text{tr}(\varrho(g) )\) is the trace of the matrix of the desired irrep evaluated at a group element.
In cases where multiple copies of the irrep are present, one can try to optimize the distance of the code inside the multiplicity space.

Logical Pauli matrices \(\overline{\sigma}_w\) are defined using the above projection and the angular momentum operators:
\flmMathEnvironment{align}{}{
  \overline{\sigma}_w = i P_\varrho e^{-i \pi J_w} P_\varrho~.
}
Finally, \(|\overline{0} \rangle\) is defined as the \(+1\) eigenvalue of \(\overline{\sigma}_z\) and \(|\overline{1} \rangle = \overline{\sigma}_x |\overline{0} \rangle \).

\codefieldsection{Encoding}
\begin{eczvaluelist}
\item\relax Encoder for a modified code that protects against electric and magnetic field fluctuations \NoCaseChange{\protect\cite{cite5033}}.
\end{eczvaluelist}
\codefieldsection{Gates}
\begin{eczvaluelist}
\item\relax Universal computation results from being able to prepare a single logical state, perform one measurement, and the following logical gates: the phase gate (\( \overline{S} \)), the Hadamard gate (\(\overline{H}\)), the conditional phase gate (\(\overline{CZ}\)), and the square root of the phase gate (\(\overline{T}\)). Single-qubit Cliffords can be generated using \(\overline{S}\) and \(\overline{H}\), the extension to multiple-qubit Cliffords is done using \(\overline{CZ}\), and \(\overline{T}\) is to transform to non-Clifford states. Together these gates can be used to create all logical unitaries, while preparation and measurement complete universal quantum computation.
\end{eczvaluelist}
\codefieldsection{Parents}
\begin{eczvaluelist}
\item\relax
\flmRefsHyperref[eczindexfamilyrel]{code:single_spin}{Single-spin code}\item\relax
\flmRefsHyperref[eczindexfamilyrel]{code:group_representation}{Group-representation code} --- Clifford-group spin codes are group-representation codes with \(G\) being a subgroup of \(SU(2)\) \NoCaseChange{\protect\cite{cite646,cite2814}}.
\end{eczvaluelist}
\codefieldsection{Child}
\begin{eczvaluelist}
\item\relax
\flmRefsHyperref[eczindexfamilyrel]{code:icosahedral_spin}{Icosahedral spin code} --- The icosahedral spin code is the \(2I\) example of a Clifford-group spin code \NoCaseChange{\protect\cite{cite646}}.
\end{eczvaluelist}
\codefieldsection{Cousins}
\begin{eczvaluelist}
\item\relax
\flmRefsHyperref[eczindexfamilyrel]{code:qubit_permutation_invariant}{PI qubit code} --- Clifford codes for spins housing representations of \(SU(2)\) yield PI qubit codes with non-trivial distance when the single spin-\(n/2\) is treated as the permutationally invariant subspace of \(n\) qubits via the \flmRefsHyperref{ref526}{Dicke-state mapping}. The subgroup of gates of a Clifford-group spin code is implemented transversally via this mapping \NoCaseChange{\protect\cite{cite2814}}.
\item\relax
\flmRefsHyperref[eczindexfamilyrel]{code:binary_dihedral_permutation_invariant}{Binary dihedral PI code} --- Binary dihedral PI codes can be interpreted as Clifford single-spin codes via the \flmRefsHyperref{ref526}{Dicke-state mapping}.
\end{eczvaluelist}
\eczhbkcontributors{ Thomas Wrona, \eczhuVVA }
\endeczcode

\eczcode{icosahedral_spin}{Icosahedral spin code}{~\NoCaseChange{\protect\cite{cite646}}}
\codefieldsection{Description}
A spin-\(7/2\) single-spin code designed to realize the binary icosahedral group \(2I\) using \(SU(2)\) rotations \NoCaseChange{\protect\cite{cite646}}.
The codespace is a two-dimensional irrep subspace obtained by restricting the spin-\(7/2\) representation of \(SU(2)\) to the subgroup \(2I\).
Under the \flmRefsHyperref{ref526}{Dicke-state mapping}, this code is equivalent to the \(\llparenthesis 7,2,3\rrparenthesis \) Pollatsek-Ruskai permutation-invariant code \NoCaseChange{\protect\cite{cite646,cite647}}.

The code has unnormalized logical states
\flmMathEnvironment{align}{}{
  \begin{split}
    |0_{L}\rangle&\propto\sqrt{3}|_{7/2}^{7/2}\rangle+\sqrt{7}|_{-3/2}^{7/2}\rangle\\
    |1_{L}\rangle&\propto\sqrt{7}|_{3/2}^{7/2}\rangle-\sqrt{3}|_{-7/2}^{7/2}\rangle\,.
  \end{split}
}

\codefieldsection{Parent}
\begin{eczvaluelist}
\item\relax
\flmRefsHyperref[eczindexfamilyrel]{code:j_gross}{Clifford-group spin code} --- The icosahedral spin code is the \(2I\) example of a Clifford-group spin code \NoCaseChange{\protect\cite{cite646}}.
\end{eczvaluelist}
\codefieldsection{Cousins}
\begin{eczvaluelist}
\item\relax
\flmRefsHyperref[eczindexfamilyrel]{code:icosahedral_fock}{Icosahedral Fock-state code} --- The icosahedral spin code maps to the icosahedral Fock-state code via the \flmRefsHyperref{ref499}{simplex mapping} \NoCaseChange{\protect\cite{cite500}}.
\item\relax
\flmRefsHyperref[eczindexfamilyrel]{code:icosahedral_permutation_invariant}{\(\llparenthesis 7,2,3\rrparenthesis \) Pollatsek-Ruskai code} --- The \(\llparenthesis 7,2,3\rrparenthesis \) Pollatsek-Ruskai code maps to the icosahedral spin code via the \flmRefsHyperref{ref526}{Dicke state mapping} \NoCaseChange{\protect\cite{cite647}}.
\end{eczvaluelist}
\eczhbkcontributors{ \eczhuVVA }
\endeczcode

\eczcode{landau_level}{Landau-level spin code}{~\NoCaseChange{\protect\cite{cite5034}}}
\codefieldsection{Description}
Approximate quantum code that encodes a qudit in the finite-dimensional Hilbert space of a single spin, i.e., a spherical Landau level.
Codewords are approximately orthogonal spin coherent states whose orientations are spaced maximally far apart along a great circle (equator) of the sphere.
The larger the spin, the better the performance.

\codefieldsection{Protection}
Protects against equatorial rotational errors acting on the overall spin.

\codefieldsection{Parents}
\begin{eczvaluelist}
\item\relax
\flmRefsHyperref[eczindexfamilyrel]{code:single_spin}{Single-spin code} --- The Landau-level spin code lies in a particular irrep present in the induced representation \(\text{Ind}_{U(1)}^{SU(2)} \lambda\), where \(\lambda\in \mathbb{Z}\) labels irreps of \(U(1)\) and quantifies the monopole strength \NoCaseChange{\protect\cite{cite5035}}.
\item\relax
\flmRefsHyperref[eczindexfamilyrel]{code:approximate_qecc}{Approximate quantum error-correcting code (AQECC)} --- The Landau-level spin code approximately protects against rotational errors.
\end{eczvaluelist}
\codefieldsection{Cousins}
\begin{eczvaluelist}
\item\relax
\flmRefsHyperref[eczindexfamilyrel]{code:homogeneous_space_quantum}{Homogeneous-space quantum code} --- The Landau-level spin code lies in a particular irrep present in the induced representation \(\text{Ind}_{U(1)}^{SU(2)} \lambda\), where \(\lambda\in \mathbb{Z}\) labels irreps of \(U(1)\) and quantifies the monopole strength \NoCaseChange{\protect\cite{cite5035}}.
\item\relax
\flmRefsHyperref[eczindexfamilyrel]{code:spin_gkp}{Spin GKP code} --- The Landau-level (spin-GKP) code are both GKP-like encodings expressed as superpositions of (squeezed) spin coherent states.
\end{eczvaluelist}
\eczhbkcontributors{ Yale Fan, \eczhuVVA }
\endeczcode

\eczcode{mps}{Magnon code}{~\NoCaseChange{\protect\cite{cite595}}}
\codefieldsection{Description}
An \(n\)-spin approximate code whose codespace of \(k=\Omega(\log n)\) qubits is efficiently described in terms of particular matrix product states or Bethe ansatz tensor networks.
Magnon codewords are low-energy excited states of the frustration-free Heisenberg-XXX model Hamiltonian \NoCaseChange{\protect\cite{cite595}}.

\codefieldsection{Protection}
Distance \(d=\Omega(n^{1-\nu})\) for any \(\nu\in(0,1)\).
\codefieldsection{Parents}
\begin{eczvaluelist}
\item\relax
\flmRefsHyperref[eczindexfamilyrel]{code:spins_into_spins}{Spin code} --- Magnon codewords are low-energy excited states of the frustration-free Heisenberg-XXX model Hamiltonian \NoCaseChange{\protect\cite{cite595}}.
\item\relax
\flmRefsHyperref[eczindexfamilyrel]{code:frustration_free}{Frustration-free Hamiltonian code} --- Magnon codewords are low-energy excited states of the frustration-free Heisenberg-XXX model Hamiltonian \NoCaseChange{\protect\cite{cite595}}.
\item\relax
\flmRefsHyperref[eczindexfamilyrel]{code:approximate_qecc}{Approximate quantum error-correcting code (AQECC)} --- Magnon codes approximately protect against erasures in the thermodynamic limit.
\end{eczvaluelist}
\codefieldsection{Cousins}
\begin{eczvaluelist}
\item\relax
\flmRefsHyperref[eczindexfamilyrel]{code:eth}{Eigenstate thermalization hypothesis (ETH) code} --- Magnon codes have been shown to protect against non-geometrically local noise, while ETH codes protect only against erasures on geometrically local patches.
\item\relax
\flmRefsHyperref[eczindexfamilyrel]{code:spt}{Symmetry-protected topological (SPT) code} --- Magnon codewords \NoCaseChange{\protect\cite{cite595}} are associated with 1D SPT orders \NoCaseChange{\protect\cite{cite3097,cite3098,cite3099,cite3100}}.
\end{eczvaluelist}
\eczhbkcontributors{ \eczhuVVA }
\endeczcode

\eczcode{okada}{Okada spin code}{~\NoCaseChange{\protect\cite{cite564}}}
\codefieldsection{Description}
Non-diagonal \(SU(2)\) single-spin code in the spin-\(J = 3m\) irrep for integer \(m \geq 1\), encoding a logical \((2m+1)\)-dimensional space.
The construction uses a \textit{non-diagonal} subspace (one for which the projected error space \(P_{\mathcal{B}}\mathcal{E}P_{\mathcal{B}}\) is block-diagonal rather than diagonal) to exceed the dimension bound achievable by the Tverberg theorem construction \NoCaseChange{\protect\cite{cite648}}.

The code is defined in terms of a subspace \(\mathcal{B} = \mathrm{span}\{|k, n-k\rangle : k \equiv 0 \text{ or } 1 \pmod{3}\}\) of the spin-\(n/2\) Hilbert space \(\mathcal{H}_n\) (with \(n = 2J = 6m\)), where \(|k, n-k\rangle\) denotes the state with \(k\) particles in the first mode and \(n-k\) in the second \NoCaseChange{\protect\cite[{Ex. 7.1}]{cite5029}}.
The codespace has dimension \(\dim \mathcal{C} = 2m+1\), which is approximately \((n+1)/3\).

The \(m = 1\) (\(J = 3\)) instance encodes a logical qutrit and admits the unnormalized codewords
\flmMathEnvironment{align}{}{
\begin{split}
  |\overline{0}\rangle&=|_{0}^{3}\rangle\\|\overline{1}\rangle&\propto\sqrt{2}|_{-2}^{3}\rangle-|_{4}^{3}\rangle\\|\overline{2}\rangle&\propto|_{-4}^{3}\rangle+\sqrt{2}|_{2}^{3}\rangle~.
\end{split}
}

\codefieldsection{Protection}
Detects distance-1 errors from the \(\mathfrak{su}(2)\) Lie algebra, i.e., any linear combination of the angular momentum operators \(\{E, F, H\}\) \NoCaseChange{\protect\cite{cite564}}.
The codespace dimension \(2m+1 \approx (n+1)/3\) improves on the Tverberg theorem construction \NoCaseChange{\protect\cite{cite648}}, which gives \(\lceil (n+1)/4 \rceil\) \NoCaseChange{\protect\cite{cite5028}}.

\codefieldsection{Parent}
\begin{eczvaluelist}
\item\relax
\flmRefsHyperref[eczindexfamilyrel]{code:single_spin}{Single-spin code}\end{eczvaluelist}
\eczhbkcontributors{ \eczhuVVA }
\endeczcode

\eczcode{single_spin}{Single-spin code}{~\NoCaseChange{\protect\cite{cite5028}}}
\codefieldsection{Description}
An encoding into a monolithic (i.e. non-tensor-product) Hilbert space that houses an irreducible representation of \(SU(2)\) or, more generally, another Lie group.
In some cases, this space can be thought of as the permutation invariant subspace of a particular tensor-product space.

The analogue of oscillator coherent states for single spins are the spin coherent states \NoCaseChange{\protect\cite{cite5021}}. 

\codefieldsection{Protection}
For the \(SU(2)\) case, a continuous-time single-spin noise channel akin to the depolarizing channel is the Landau-Streater channel \NoCaseChange{\protect\cite{cite5036}}.
A particular error basis of interest consists of the spherical tensors \NoCaseChange{\protect\cite{cite2814}}.

The \(SU(2)\) Lie Algebra can also be used as a noise model; it connects states whose angular momentum projections differ by at most an integer \NoCaseChange{\protect\cite{cite646}}.
More generally, the group's Lie algebra induces a metric on the carrying vector space, and its operators can be chosen as a noise basis \NoCaseChange{\protect\cite{cite5028}}. Code existence is guaranteed by the Tverberg theorem \NoCaseChange{\protect\cite{cite5028}}.
There are \flmRefsHyperref{ref672}{quantum MacWilliams identities} for such metric spaces \NoCaseChange{\protect\cite{cite5028}}.

\codefieldsection{Rate}
For every \(K,t \geq 2\), there are explicitly constructible \(K\)-dimensional single-spin codes for \(SU(q=N)\) with total spin \(N=(K-1)t(t+1)\) and distance \(t+1\); there also exist families with logical dimension \(K = o(2^N)\) and distance of \flmRefsHyperref{ref65}{order} \(o(N/\log N)\) \NoCaseChange{\protect\cite{cite500}}.

For the spin-\(n/2\) irrep of \(\mathfrak{su}(2)\), the Tverberg theorem construction \NoCaseChange{\protect\cite{cite648}} yields a distance-1 error-detecting code of dimension \(\lceil (n+1)/4 \rceil\) against the \(\mathfrak{su}(2)\) Lie algebra error set \(\{E, F, H\}\) \NoCaseChange{\protect\cite{cite5028}\protect\cite[{Ex. 6.3}]{cite5029}}.

\codefieldsection{Parents}
\begin{eczvaluelist}
\item\relax
\flmRefsHyperref[eczindexfamilyrel]{code:spins_into_spins}{Spin code}\item\relax
\flmRefsHyperref[eczindexfamilyrel]{code:single_subsystem}{Monolithic quantum code}\end{eczvaluelist}
\codefieldsection{Children}
\begin{eczvaluelist}
\item\relax
\flmRefsHyperref[eczindexfamilyrel]{code:j_gross}{Clifford-group spin code}\item\relax
\flmRefsHyperref[eczindexfamilyrel]{code:landau_level}{Landau-level spin code} --- The Landau-level spin code lies in a particular irrep present in the induced representation \(\text{Ind}_{U(1)}^{SU(2)} \lambda\), where \(\lambda\in \mathbb{Z}\) labels irreps of \(U(1)\) and quantifies the monopole strength \NoCaseChange{\protect\cite{cite5035}}.
\item\relax
\flmRefsHyperref[eczindexfamilyrel]{code:okada}{Okada spin code}\item\relax
\flmRefsHyperref[eczindexfamilyrel]{code:spin_cat}{Spin cat code}\item\relax
\flmRefsHyperref[eczindexfamilyrel]{code:spin_gkp}{Spin GKP code}\item\relax
\flmRefsHyperref[eczindexfamilyrel]{code:su3_spin}{\(SU(3)\) spin code}\item\relax
\flmRefsHyperref[eczindexfamilyrel]{code:su3_tverberg_spin}{\(SU(3)\) Tverberg spin code}\item\relax
\flmRefsHyperref[eczindexfamilyrel]{code:su4_tverberg_spin}{\(SU(4)\) Tverberg spin code}\end{eczvaluelist}
\codefieldsection{Cousins}
\begin{eczvaluelist}
\item\relax
\flmRefsHyperref[eczindexfamilyrel]{code:qubit_permutation_invariant}{PI qubit code} --- Single-spin codes are subspaces of a single large \(SU(2)\) spin, which can be either standalone or correspond to the PI subspace of a set of spins via the \flmRefsHyperref{ref526}{Dicke state mapping}.
\item\relax
\flmRefsHyperref[eczindexfamilyrel]{code:oscillators}{Bosonic code} --- Bosonic states are typically represented with the assumption that a common phase reference exists, and the superselection rule compliant (SSRC) framework yields expressions without this assumption \NoCaseChange{\protect\cite{cite4825,cite4826,cite4827,cite4828,cite4829,cite4830,cite4831}}. In this framework, single-mode states can be treated as two-mode states in a fixed subspace of total occupation number \(N\) in the limit \(N \to \infty\). Passive Gaussian operations acting on the fixed-photon subspace of two modes realize \(U(2)\) transformations in the Jordan-Schwinger boson mapping \NoCaseChange{\protect\cite{cite4832,cite4833,cite4834,cite653}}.
\item\relax
\flmRefsHyperref[eczindexfamilyrel]{code:qsc}{Quantum spherical code (QSC)} --- Single-spin codes whose codewords are expressed in terms of discrete sets of spin-coherent states may also be interpreted as QSCs.
\item\relax
\flmRefsHyperref[eczindexfamilyrel]{code:permutation_invariant}{Permutation-invariant (PI) code} --- Modular-qudit PI codes can be converted to spin codes defined on the completely symmetric irrep of \(SU(q)\) via the \flmRefsHyperref{ref499}{simplex mapping} \NoCaseChange{\protect\cite[{Prop. VI.2}]{cite500}}. Any transversal gates are mapped to \(SU(q)\) gates on the spin codes \NoCaseChange{\protect\cite{cite500}}.
\item\relax
\flmRefsHyperref[eczindexfamilyrel]{code:ae}{Æ code} --- Since Æ codes are defined in a subspace of fixed total angular momentum and protect against errors linear in the angular-momentum generators, they can also be thought of as single-spin codes.
\end{eczvaluelist}
\eczhbkcontributors{ \eczhuVVA }
\endeczcode

\eczcode{spin_cat}{Spin cat code}{~\NoCaseChange{\protect\cite{cite5037,cite5038}}}
\codefieldsection{Description}
An analogue of the two-component cat code for a large spin, which is often realized in the PI subspace of atomic ensembles.

The encoding was designed by using the Holstein-Primakoff mapping \NoCaseChange{\protect\cite{cite651,cite652,cite653}} to pull back the phase-space structure of a bosonic system to the compact phase space of a quantum spin.

The codewords can be approximated by two spin-coherent states.
The version where the two spin-coherent states are antipodal has been considered in Ref. \NoCaseChange{\protect\cite{cite5038}}.

An extended version of the spin cat code, the dark spin-cat code, encodes in two spins, both thought of as hyperfine manifolds \NoCaseChange{\protect\cite{cite5039}}.

\codefieldsection{Gates}
\begin{eczvaluelist}
\item\relax CNOT gate preserving the rank of spherical-tensor noise operators \NoCaseChange{\protect\cite{cite5038}}.
\end{eczvaluelist}
\codefieldsection{Decoding}
\begin{eczvaluelist}
\item\relax Measurement-free error correction protocol \NoCaseChange{\protect\cite{cite5038}}.
\end{eczvaluelist}
\codefieldsection{Realizations}
\begin{eczvaluelist}
\item\relax Trapped ions: autonomous error-correction scheme reduces errors by a factor up to 2.2, as demonstrated by the Chiaverini group \NoCaseChange{\protect\cite{cite5040}}.
\item\relax Silicon spin qubits: cat-state initialization \NoCaseChange{\protect\cite{cite5041}}.
\item\relax Synthetic spin system inside a microwave cavity: universal control using linear and nonlinear pulses \NoCaseChange{\protect\cite{cite5042}}.
\item\relax Neutral atoms: cat-state initialization, universal single-qubit gates, and benchmarking of gate fidelity, coherence, and biased-noise properties \NoCaseChange{\protect\cite{cite5043}}.
\end{eczvaluelist}
\codefieldsection{Parent}
\begin{eczvaluelist}
\item\relax
\flmRefsHyperref[eczindexfamilyrel]{code:single_spin}{Single-spin code}\end{eczvaluelist}
\codefieldsection{Cousins}
\begin{eczvaluelist}
\item\relax
\flmRefsHyperref[eczindexfamilyrel]{code:two-legged-cat}{Two-component cat code} --- The spin-cat code construction utilizes the Holstein-Primakoff mapping \NoCaseChange{\protect\cite{cite651,cite652,cite653}} to convert cat codes into codes for spin systems.
\item\relax
\flmRefsHyperref[eczindexfamilyrel]{code:spins_into_spins}{Spin code} --- An extended version of the spin cat code, the dark spin-cat code, encodes in two spins, both thought of as hyperfine manifolds \NoCaseChange{\protect\cite{cite5039}}.
\end{eczvaluelist}
\eczhbkcontributors{ \eczhuVVA }
\endeczcode

\eczcode{spins_into_spins}{Spin code}{}

\codefieldsection{Kingdom root code}
for the \flmRefsHyperref{kingdom:spins_into_spins}{Spin Kingdom}.
\codefieldsection{Description}
Encodes a \(K\)-dimensional Hilbert space into a tensor-product or direct sum of factors, with each factor spanned by states of a quantum mechanical spin or, more generally, an irreducible representation of a compact Lie group.

In the simplest case of a single-spin \(SU(2)\) system, the canonical states \(|^J_m\rangle\) of a single \(2J+1\)-dimensional factor are labeled by total angular momentum \(J\) (either integer or half-integer) and its \(z\)-axis projection \(m\).
There can be multiple factors of the same size, as in the case of atomic or molecular state spaces, and the number of factors can be infinite.
In contrast to other qudit codes, spin codes are closely associated with the angular momentum Lie algebra and/or \(SU(2)\), \(SO(3)\), or more general Lie groups.

\codefieldsection{Protection}
Codes can be designed to protect against rotations by small angles, which effectively means they protect against low-order products of powers of the Lie algebra generators.
In the molecular (\(SU(2)\)) setting, there is a larger basis of error operators causing changes in the total angular momentum \(J\) and its projection \(m\) and modeling processes such as spontaneous emission, stray electromagnetic fields, and Raman scattering \NoCaseChange{\protect\cite{cite5031}}.

\codefieldsection{Parent}
\begin{eczvaluelist}
\item\relax
\flmRefsHyperref[eczindexfamilyrel]{code:qecc_finite}{Finite-dimensional quantum error-correcting code}\end{eczvaluelist}
\codefieldsection{Children}
\begin{eczvaluelist}
\item\relax
\flmRefsHyperref[eczindexfamilyrel]{code:qubits_into_qubits}{Qubit code} --- Spin codes with spin \(\ell=1/2\) correspond to qubit codes since the single-qubit Pauli matrices generate the Lie algebra of \(SU(2)\).
\item\relax
\flmRefsHyperref[eczindexfamilyrel]{code:ae}{Æ code} --- Æ codes protect against changes in both the total angular momentum \(J\) and its projection \(m\), with the former type necessarily causing the information to leak out of the space of a single spin.
\item\relax
\flmRefsHyperref[eczindexfamilyrel]{code:mps}{Magnon code} --- Magnon codewords are low-energy excited states of the frustration-free Heisenberg-XXX model Hamiltonian \NoCaseChange{\protect\cite{cite595}}.
\item\relax
\flmRefsHyperref[eczindexfamilyrel]{code:t_group}{Twisted \(1\)-group code}\item\relax
\flmRefsHyperref[eczindexfamilyrel]{code:vbs}{Valence-bond-solid (VBS) code} --- VBS codewords are eigenstates of the frustration-free VBS Hamiltonian \NoCaseChange{\protect\cite{cite2808,cite790}}.
\item\relax
\flmRefsHyperref[eczindexfamilyrel]{code:single_spin}{Single-spin code}\end{eczvaluelist}
\codefieldsection{Cousins}
\begin{eczvaluelist}
\item\relax
\flmRefsHyperref[eczindexfamilyrel]{code:eth}{Eigenstate thermalization hypothesis (ETH) code} --- Relevant many-body systems housing ETH codes include 1D non-interacting spin chains or frustration-free systems such as Motzkin chains and Heisenberg models.
\item\relax
\flmRefsHyperref[eczindexfamilyrel]{code:movassagh_ouyang}{Movassagh-Ouyang Hamiltonian code} --- Justesen codes can be used to build a family of \(n\)-qubit Movassagh-Ouyang Hamiltonian spin codes encoding one logical qubit with linear distance. These codes form the ground-state subspace of a frustration-free geometrically local Hamiltonian \NoCaseChange{\protect\cite{cite1407}}.
\item\relax
\flmRefsHyperref[eczindexfamilyrel]{code:spin_cat}{Spin cat code} --- An extended version of the spin cat code, the dark spin-cat code, encodes in two spins, both thought of as hyperfine manifolds \NoCaseChange{\protect\cite{cite5039}}.
\end{eczvaluelist}
\eczhbkcontributors{ Thomas Wrona, \eczhuVVA }
\endeczcode

\eczcode{spin_gkp}{Spin GKP code}{~\NoCaseChange{\protect\cite{cite5044}}}
\codefieldsection{Description}
An analogue of the single-mode GKP code designed for atomic ensembles. It was designed using the Holstein-Primakoff mapping \NoCaseChange{\protect\cite{cite651,cite652,cite653}} to pull back the phase-space structure of a bosonic system to the compact phase space of a quantum spin. A different construction emerges depending on which particular expression for GKP codewords is pulled back.

\codefieldsection{Protection}
Protects against errors native to spin systems like random rotations and stochastic relaxation.
\codefieldsection{Encoding}
\begin{eczvaluelist}
\item\relax Linear combination of unitaries method \NoCaseChange{\protect\cite{cite5045,cite5046,cite5047}}, which may be applicable to more general codewords.
\end{eczvaluelist}
\codefieldsection{Gates}
\begin{eczvaluelist}
\item\relax Approximate Clifford-group generators are composed of Hamiltonians at most quadratic in angular momentum operators of two spin systems. Assuming that these generators can be implemented with high fidelity, a magic state can be prepared from an atomic ensemble analog of the vacuum state.
\end{eczvaluelist}
\codefieldsection{Parent}
\begin{eczvaluelist}
\item\relax
\flmRefsHyperref[eczindexfamilyrel]{code:single_spin}{Single-spin code}\end{eczvaluelist}
\codefieldsection{Cousins}
\begin{eczvaluelist}
\item\relax
\flmRefsHyperref[eczindexfamilyrel]{code:gkp}{Square-lattice GKP code} --- Spin-GKP code constructions utilize the Holstein-Primakoff mapping \NoCaseChange{\protect\cite{cite651,cite652,cite653}} to convert various expressions for square-lattice GKP states into codes for spin systems.
\item\relax
\flmRefsHyperref[eczindexfamilyrel]{code:landau_level}{Landau-level spin code} --- The Landau-level (spin-GKP) code are both GKP-like encodings expressed as superpositions of (squeezed) spin coherent states.
\end{eczvaluelist}
\eczhbkcontributors{ Sivaprasad Omanakuttan, \eczhuVVA }
\endeczcode

\eczcode{t_group}{Twisted \(1\)-group code}{~\NoCaseChange{\protect\cite{cite2199,cite789}}}
\codefieldsection{Description}
Block group-representation code realizing particular irreps of particular groups such that a distance of two is automatically guaranteed.
Groups which admit irreps with this property are called \textit{twisted (unitary) \(1\)-groups} and include the binary icosahedral group \(2I\), the \(\Sigma(360\phi)\) subgroup of \(SU(3)\), the family \(\{PSp(2b, 3), b \geq 1\}\), and the alternating groups \(A_{5,6}\).
Groups whose irreps are images of the appropriate irreps of twisted \(1\)-groups also yield such properties, e.g., the binary tetrahedral group \(2T\) or qutrit Pauli group \(\Sigma(72\phi)\).

A \(\llparenthesis 3,2,2\rrparenthesis _3\) code can implement the qutrit Pauli group \(\Sigma(72\phi)\) transversally, a \(\llparenthesis 6,3,2\rrparenthesis \) code can implement \(A_5\) transversally, a \(\llparenthesis 6,2,2\rrparenthesis _3\) implements \(2T\) transversally, and a \(\llparenthesis 6,5,2\rrparenthesis _3\) code implements \(A_6\) transversally.

\codefieldsection{Transversal and Permutation-Based Gates}
\begin{eczvaluelist}
\item\relax All gates in the underlying twisted \(1\)-group. See \NoCaseChange{\protect\cite[{Table II}]{cite789}} for other notable groups including the sporadic groups.
\end{eczvaluelist}
\codefieldsection{Parents}
\begin{eczvaluelist}
\item\relax
\flmRefsHyperref[eczindexfamilyrel]{code:spins_into_spins}{Spin code}\item\relax
\flmRefsHyperref[eczindexfamilyrel]{code:permutation_invariant}{Permutation-invariant (PI) code}\item\relax
\flmRefsHyperref[eczindexfamilyrel]{code:group_representation}{Group-representation code} --- Twisted \(1\)-group codes are group-representation codes with \(G\) being a twisted \(1\)-group.
\item\relax
\flmRefsHyperref[eczindexfamilyrel]{code:small_distance_quantum}{Small-distance block quantum code} --- All twisted \(1\)-group codes have a distance \(d \geq 2\).
\end{eczvaluelist}
\codefieldsection{Children}
\begin{eczvaluelist}
\item\relax
\flmRefsHyperref[eczindexfamilyrel]{code:icosahedral_permutation_invariant}{\(\llparenthesis 7,2,3\rrparenthesis \) Pollatsek-Ruskai code} --- The \(\llparenthesis 7,2,3\rrparenthesis \) Pollatsek-Ruskai code admits a transversal representation of the twisted \(1\)-group \(2I\) \NoCaseChange{\protect\cite{cite2199}}.
\item\relax
\flmRefsHyperref[eczindexfamilyrel]{code:su3_sigma360}{\(\llparenthesis 5,3,2\rrparenthesis _3\) qutrit code} --- The \(\llparenthesis 5,3,2\rrparenthesis _3\) qutrit code admits a transversal representation of the twisted \(1\)-group \(\Sigma(360\phi)=3.A_6\) \NoCaseChange{\protect\cite{cite2199}}.
\end{eczvaluelist}
\codefieldsection{Cousins}
\begin{eczvaluelist}
\item\relax
\flmRefsHyperref[eczindexfamilyrel]{code:unitary_design}{Unitary \(t\)-design} --- Twisted unitary \(t\)-groups \NoCaseChange{\protect\cite{cite2199}} generalize the idea of unitary \(t\)-groups \NoCaseChange{\protect\cite{cite939,cite2193,cite2200}}, which are subgroups of the unitary group that form unitary \(t\)-designs.
\item\relax
\flmRefsHyperref[eczindexfamilyrel]{code:fock_state}{Fock-state bosonic code} --- Twisted \(1\)-group codes can be converted to constant-excitation Fock-state codes via the \flmRefsHyperref{ref499}{simplex mapping} \NoCaseChange{\protect\cite[{Prop. V.2}]{cite500}}. Any transversal gates are mapped to Gaussian gates on the Fock-state codes \NoCaseChange{\protect\cite{cite500}}.
\end{eczvaluelist}
\eczhbkcontributors{ \eczhuVVA }
\endeczcode

\eczcode{vbs}{Valence-bond-solid (VBS) code}{~\NoCaseChange{\protect\cite{cite627,cite2808,cite790}}}
\codefieldsection{Description}
A member of an approximate \(q\)-dimensional spin-code family whose codespace is described in terms of \(SU(q)\) valence-bond-solid (VBS) \NoCaseChange{\protect\cite{cite626}} matrix product states with various boundary conditions.
The codes become exact when either \(n\) or \(q\) go to infinity.
The original work on these codes studied the \(q=2\) case \NoCaseChange{\protect\cite{cite627}}.

\codefieldsection{Protection}
VBS codes approximately protect against erasures, with the approximation becoming exact in the thermodynamic limit \NoCaseChange{\protect\cite{cite2808,cite790}}.

\codefieldsection{Transversal and Permutation-Based Gates}
\begin{eczvaluelist}
\item\relax Two classes of (approximate) VBS codes have \(SU(q)\) transversal gates \NoCaseChange{\protect\cite[{Tab. III}]{cite790}}.
\end{eczvaluelist}
\codefieldsection{Parents}
\begin{eczvaluelist}
\item\relax
\flmRefsHyperref[eczindexfamilyrel]{code:spins_into_spins}{Spin code} --- VBS codewords are eigenstates of the frustration-free VBS Hamiltonian \NoCaseChange{\protect\cite{cite2808,cite790}}.
\item\relax
\flmRefsHyperref[eczindexfamilyrel]{code:frustration_free}{Frustration-free Hamiltonian code} --- VBS codewords are eigenstates of the frustration-free VBS Hamiltonian \NoCaseChange{\protect\cite{cite2808,cite790}}.
\item\relax
\flmRefsHyperref[eczindexfamilyrel]{code:approximate_qecc}{Approximate quantum error-correcting code (AQECC)} --- VBS codes approximately protect against erasures in the thermodynamic limit.
\end{eczvaluelist}
\codefieldsection{Cousins}
\begin{eczvaluelist}
\item\relax
\flmRefsHyperref[eczindexfamilyrel]{code:covariant}{Covariant block quantum code} --- Two classes of (approximate) VBS codes have \(SU(q)\) transversal gates, i.e., are \(SU(q)\)-covariant \NoCaseChange{\protect\cite[{Tab. III}]{cite790}}.
\item\relax
\flmRefsHyperref[eczindexfamilyrel]{code:spt}{Symmetry-protected topological (SPT) code} --- VBS codewords \NoCaseChange{\protect\cite{cite2808}} are associated with 1D SPT orders \NoCaseChange{\protect\cite{cite3097,cite3098,cite3099,cite3100}}.
\end{eczvaluelist}
\eczhbkcontributors{ \eczhuVVA }
\endeczcode

\onecolumngrid
\clearpage

\section{Group quantum Kingdom}

\begin{eczEpigraph}
\begin{quote}
\flmQuoteSetup{quote}%
The practical consequences appeared to be negligible, but everyone felt that to be in the mainstream one had to learn about it. Yet there were no good texts from which one could learn group theory. It was a frustrating experience, worthy of the name of a pest.
\flmQuoteAttributed{John C. Slater on the “Gruppenpest”}
\end{quote}
\end{eczEpigraph}

\twocolumngrid

\eczcode{group_10_1_4}{\(\llbracket 10,1,4\rrbracket _{G}\) tenfold code}{~\NoCaseChange{\protect\cite[{Prop. V.1}]{cite866}}}
\eczhIndexCodeAliasName{group_10_1_4}{tenfold code}
\codefieldsection{Description}
A \(\llbracket 10,1,4\rrbracket _{G}\) group code for finite Abelian \(G\).
The code is defined using a graph that is closely related to the \(\llbracket 5,1,3\rrbracket \) code.

\codefieldsection{Parents}
\begin{eczvaluelist}
\item\relax
\flmRefsHyperref[eczindexfamilyrel]{code:graph_quantum}{Graph quantum code} --- The \(\llbracket 10,1,4\rrbracket _{G}\) group code is defined using a graph \NoCaseChange{\protect\cite[{Prop. V.1}]{cite866}}.
\item\relax
\flmRefsHyperref[eczindexfamilyrel]{code:small_distance_quantum}{Small-distance block quantum code}\end{eczvaluelist}
\codefieldsection{Cousins}
\begin{eczvaluelist}
\item\relax
\flmRefsHyperref[eczindexfamilyrel]{code:qudit_5_1_3}{\(\llbracket 5,1,3\rrbracket _{\mathbb{Z}_q}\) modular-qudit code} --- The \(\llbracket 10,1,4\rrbracket _{G}\) Abelian group code for \(G=\mathbb{Z}_q\) is defined using a graph that is closely related to the \(\llbracket 5,1,3\rrbracket _{\mathbb{Z}_q}\) modular-qudit code \NoCaseChange{\protect\cite{cite866}}.
\item\relax
\flmRefsHyperref[eczindexfamilyrel]{code:galois_5_1_3}{\(\llbracket 5,1,3\rrbracket _q\) Galois-qudit code} --- The \(\llbracket 10,1,4\rrbracket _{G}\) Abelian group code for \(G=\mathbb{F}_q\) is defined using a graph that is closely related to the \(\llbracket 5,1,3\rrbracket _{q}\) Galois-qudit code \NoCaseChange{\protect\cite{cite866}}.
\item\relax
\flmRefsHyperref[eczindexfamilyrel]{code:stab_5_1_3}{\(\llbracket 5,1,3\rrbracket \) Five-qubit perfect code} --- The \(\llbracket 10,1,4\rrbracket _{G}\) Abelian group code for \(G=\mathbb{Z}_2\) is defined using a graph that is closely related to the \(\llbracket 5,1,3\rrbracket \) five-qubit code \NoCaseChange{\protect\cite{cite866}}. The former code can be obtained by converting the latter into a code that is oblivious to collective \(Z\)-type rotations \NoCaseChange{\protect\cite[{Exam. 6}]{cite808}}.
\end{eczvaluelist}
\eczhbkcontributors{ \eczhuVVA }
\endeczcode

\eczcode{rotor_3_1_2}{\(\llbracket 3,1,2\rrbracket _{\mathbb{Z}}\) Three-rotor code}{~\NoCaseChange{\protect\cite{cite2514}}}
\eczhIndexCodeAliasName{rotor_3_1_2}{Three-rotor code}
\codefieldsection{Description}
\(\llbracket 3,1,2\rrbracket _{\mathbb Z}\) rotor code that is an extension of the \(\llbracket 3,1,2\rrbracket _3\) qutrit CSS code to the integer alphabet, i.e., the angular momentum states of a rotor.

The code is \(U(1)\)-covariant and its ideal codewords,
\flmMathEnvironment{align}{}{
  |\overline{x}\rangle = \sum_{y\in\mathbb{Z}} \left| -3y,y-x,2(y+x) \right\rangle~,
}
where \(x\in\mathbb{Z}\), are not normalizable.

\codefieldsection{Protection}
Normalized codewords approximately protect against erasure while maintaining covariance \NoCaseChange{\protect\cite{cite2720}}.

\codefieldsection{Parents}
\begin{eczvaluelist}
\item\relax
\flmRefsHyperref[eczindexfamilyrel]{code:homological_rotor}{Homological rotor code} --- Taking \(H_X=\begin{pmatrix}-3 & 1 & 2\end{pmatrix}\) and \(H_Z=\begin{pmatrix}4&6&3\end{pmatrix}\) yields the three-rotor code.
\item\relax
\flmRefsHyperref[eczindexfamilyrel]{code:ame}{Perfect-tensor code} --- Three-rotor codewords are CV AME states \NoCaseChange{\protect\cite{cite507}}.
\item\relax
\flmRefsHyperref[eczindexfamilyrel]{code:small_distance_quantum}{Small-distance block quantum code}\end{eczvaluelist}
\codefieldsection{Cousins}
\begin{eczvaluelist}
\item\relax
\flmRefsHyperref[eczindexfamilyrel]{code:covariant}{Covariant block quantum code} --- The three-rotor code is \(U(1)\)-covariant.
\item\relax
\flmRefsHyperref[eczindexfamilyrel]{code:stab_3_1_2}{\(\llbracket 3,1,2\rrbracket _3\) Three-qutrit code} --- The three-rotor code is a rotor analogue of the three-qutrit code.
\end{eczvaluelist}
\eczhbkcontributors{ \eczhuVVA }
\endeczcode

\eczcode{group_4_2_2}{\(\llbracket 4,2,2\rrbracket _{G}\) four group-qudit code}{~\NoCaseChange{\protect\cite{cite423}\protect\cite[{Sec. VIII}]{cite2720}}}
\eczhIndexCodeAliasName{group_4_2_2}{four group-qudit code}
\codefieldsection{Description}
\(\llbracket 4,2,2\rrbracket _{G}\) group quantum code that is an extension of the four-qubit code to group-valued qudits.

For elements \(g_1, g_2\) of any finite group \(G\), a set of codewords is
\flmMathEnvironment{align}{}{
  |\overline{g_{1},g_{2}}\rangle=\frac{1}{\sqrt{|G|}}\sum_{g\in G}|g,gg_{1},gg_{2},gg_{1}g_{2}\rangle~.
}

See Ref. \NoCaseChange{\protect\cite{cite2581}} for a \(\llbracket 4,1,2\rrbracket _{\mathbb{Z}_q}\) subcode.

\codefieldsection{Parents}
\begin{eczvaluelist}
\item\relax
\flmRefsHyperref[eczindexfamilyrel]{code:quantum_double}{Quantum-double code} --- The four group-qudit code is the smallest quantum double code.
\item\relax
\flmRefsHyperref[eczindexfamilyrel]{code:covariant}{Covariant block quantum code} --- The four group-qudit code is \((G\times G)\)-covariant, with transversal logical left and right multiplication gates \NoCaseChange{\protect\cite[{Sec. VIII.B}]{cite2720}}.
\item\relax
\flmRefsHyperref[eczindexfamilyrel]{code:small_distance_quantum}{Small-distance block quantum code}\end{eczvaluelist}
\codefieldsection{Children}
\begin{eczvaluelist}
\item\relax
\flmRefsHyperref[eczindexfamilyrel]{code:rotor_4_2_2}{Four-rotor code} --- The four group-qudit code reduces to the four-rotor code for \(G= \mathbb{Z}\).
\item\relax
\flmRefsHyperref[eczindexfamilyrel]{code:stab_4_2_2}{\(\llbracket 4,2,2\rrbracket \) Four-qubit code} --- The four group-qudit code reduces to the four-qubit code for \(G=\mathbb{Z}_2\).
\end{eczvaluelist}
\codefieldsection{Cousins}
\begin{eczvaluelist}
\item\relax
\flmRefsHyperref[eczindexfamilyrel]{code:iceberg}{\(\llbracket 2m,2m-2,2\rrbracket \) error-detecting code} --- The four group-qudit code can be extended to the \(\llbracket 2m,2m-2,2\rrbracket _{G}\) group-qudit code \NoCaseChange{\protect\cite[{Sec. VIII}]{cite2720}}. The latter reduces to the \(\llbracket 2m,2m-2,2\rrbracket \) error-detecting code for \(G=\mathbb{Z}_2\).
\item\relax
\flmRefsHyperref[eczindexfamilyrel]{code:group_quantum_parity}{Group-based QPC} --- The \(|\overline{g_1=1,g_2}\rangle\) \(\llbracket 4,1,2\rrbracket _{G}\) subcode is the smallest group-based QPC, i.e., a concatenation of a bit-flip with a phase-flip group-based repetition code for that group.
\item\relax
\flmRefsHyperref[eczindexfamilyrel]{code:quantum_concatenated}{Concatenated quantum code} --- The \(|\overline{g_1=1,g_2}\rangle\) \(\llbracket 4,1,2\rrbracket _{G}\) subcode is the smallest group-based QPC, i.e., a concatenation of a bit-flip with a phase-flip group-based repetition code for that group.
\item\relax
\flmRefsHyperref[eczindexfamilyrel]{code:holographic_hyperinvariant}{Hyperinvariant tensor-network (HTN) code} --- The explicit 4-ququart encoding tensor \(A'\) used in the HTN code is a \(\llbracket 4,1,2\rrbracket _{\mathbb{Z}_4}\) subcode of the \(\llbracket 4,2,2\rrbracket _{\mathbb{Z}_4}\) four group-qudit code \NoCaseChange{\protect\cite[{Sec. IID}]{cite3788}}.
\end{eczvaluelist}
\eczhbkcontributors{ \eczhuVVA }
\endeczcode

\eczcode{rotor_5_1_3}{\(\llbracket 5,1,3\rrbracket _{\mathbb{Z}}\) Five-rotor code}{~\NoCaseChange{\protect\cite{cite2720}}}
\eczhIndexCodeAliasName{rotor_5_1_3}{Five-rotor code}
\codefieldsection{Description}
Extension of the five-qubit stabilizer code to the integer alphabet, i.e., the angular momentum states of a rotor. The code is \(U(1)\)-covariant and ideal codewords are not normalizable.

\codefieldsection{Protection}
Normalized codewords approximately protect against erasure while maintaining covariance \NoCaseChange{\protect\cite{cite2720}}.
For smooth cutoff width \(w\) and logical charge range \(2h+1\), the worst-case entanglement infidelity scales as \(O(h/w)\) for a known single erasure and for any known two-subsystem erasure \NoCaseChange{\protect\cite[{Eqs. (45)-(46)}]{cite2720}}.

\codefieldsection{Parents}
\begin{eczvaluelist}
\item\relax
\flmRefsHyperref[eczindexfamilyrel]{code:rotor_stabilizer}{Rotor stabilizer code}\item\relax
\flmRefsHyperref[eczindexfamilyrel]{code:ame}{Perfect-tensor code} --- Five-rotor codewords are CV AME \NoCaseChange{\protect\cite{cite507}}.
\item\relax
\flmRefsHyperref[eczindexfamilyrel]{code:quantum_cyclic}{Cyclic quantum code}\item\relax
\flmRefsHyperref[eczindexfamilyrel]{code:small_distance_quantum}{Small-distance block quantum code}\end{eczvaluelist}
\codefieldsection{Cousins}
\begin{eczvaluelist}
\item\relax
\flmRefsHyperref[eczindexfamilyrel]{code:covariant}{Covariant block quantum code} --- The five-rotor code is \(U(1)\)-covariant.
\item\relax
\flmRefsHyperref[eczindexfamilyrel]{code:qudit_5_1_3}{\(\llbracket 5,1,3\rrbracket _{\mathbb{Z}_q}\) modular-qudit code} --- The five-rotor code is a rotor analogue of the five-qudit code.
\end{eczvaluelist}
\eczhbkcontributors{ \eczhuVVA }
\endeczcode

\eczcode{1d_stabilizer}{1D lattice stabilizer code}{}
\codefieldsection{Description}
Lattice stabilizer code in one Euclidean dimension, using either the ordinary block notion of locality or the fermionic/Majorana notion of locality.

Any modular-qudit code can be converted to several copies of the 1D repetition code along with some trivial codes via a local constant-depth Clifford circuit \NoCaseChange{\protect\cite{cite3963}}.
There is no 1D bosonic topological order at nonzero temperature \NoCaseChange{\protect\cite{cite3126,cite3127,cite5048}}.

\codefieldsection{Fault Tolerance}
\begin{eczvaluelist}
\item\relax A fault-tolerant quantum computation scheme on a 1D nearest-neighbor qubit line can be built from a modified tower of interleaved quantum Hamming codes. It achieves coding rate above \(5\%\), constant space overhead, quasi-polylogarithmic time overhead, and a threshold \NoCaseChange{\protect\cite{cite3217}}.
\end{eczvaluelist}
\codefieldsection{Parent}
\begin{eczvaluelist}
\item\relax
\flmRefsHyperref[eczindexfamilyrel]{code:translationally_invariant_stabilizer}{Lattice stabilizer code}\end{eczvaluelist}
\codefieldsection{Children}
\begin{eczvaluelist}
\item\relax
\flmRefsHyperref[eczindexfamilyrel]{code:current_mirror}{Kitaev current-mirror qubit code}\item\relax
\flmRefsHyperref[eczindexfamilyrel]{code:analog_repetition}{Analog repetition code}\item\relax
\flmRefsHyperref[eczindexfamilyrel]{code:mbq}{Majorana box qubit} --- The Majorana box qubit is a 1D qubit stabilizer code with respect to the Majorana operator basis.
\item\relax
\flmRefsHyperref[eczindexfamilyrel]{code:quantum_repetition}{Quantum repetition code} --- The codespace of the quantum repetition code is the ground-state space of a frustration-free 1D classical Ising model with nearest-neighbor interactions.
\item\relax
\flmRefsHyperref[eczindexfamilyrel]{code:tfim}{Transverse-field Ising model (TFIM) code}\item\relax
\flmRefsHyperref[eczindexfamilyrel]{code:quantum_convolutional}{Quantum convolutional code} --- Quantum convolutional codes are lattice stabilizer codes on a semi-infinite or infinite lattice in one dimension \NoCaseChange{\protect\cite{cite4026}}. Some notions may be extendable to non-stabilizer codes \NoCaseChange{\protect\cite[{Sec. 4}]{cite4016}}.
Any prime-qudit code can be converted using a constant-depth \flmRefsHyperref{ref409}{Clifford circuit} to several copies of the 1D repetition code along with some trivial codes \NoCaseChange{\protect\cite{cite3963}}.

\end{eczvaluelist}
\codefieldsection{Cousins}
\begin{eczvaluelist}
\item\relax
\flmRefsHyperref[eczindexfamilyrel]{code:quantum_cyclic}{Cyclic quantum code} --- A 1D lattice stabilizer code with periodic boundary conditions is a quantum cyclic code.
\item\relax
\flmRefsHyperref[eczindexfamilyrel]{code:cv_cluster_state}{Analog cluster-state code} --- Analog cluster states defined on a 1D array of modes are called quantum wires \NoCaseChange{\protect\cite{cite4685,cite4686}}, not to be confused with the Kitaev quantum wire, a fermion code. Analog cluster states defined on a 1D ladder are sometimes called dual-rail, not to be confused with the dual-rail code.
\item\relax
\flmRefsHyperref[eczindexfamilyrel]{code:approximate_log_depth}{Log-depth geometrically local Clifford-circuit code} --- Log-depth \flmRefsHyperref{ref409}{Clifford circuits} on a 1D geometry yield approximate codes whose encoding rate achieves the hashing bound for Pauli noise and the channel capacity for erasure errors \NoCaseChange{\protect\cite{cite3972,cite3973}}.
\item\relax
\flmRefsHyperref[eczindexfamilyrel]{code:local_haar_random}{Local Haar-random circuit qubit code} --- In a 1D geometry, the local Haar-random circuit qubit code approximately detects any error with support on a segment of length \(\leq n/4\), with deviations exponentially suppressed in \(n\). There is a phase transition in error correction power vs error rate \(p\), with a critical depth of \flmRefsHyperref{ref65}{order} \(O(1/p)\) \NoCaseChange{\protect\cite{cite3970}}.
\item\relax
\flmRefsHyperref[eczindexfamilyrel]{code:qudit_cluster_state}{Modular-qudit cluster-state code} --- Qudit cluster states defined on 1D lattices are representatives of various 1D SPT phases \NoCaseChange{\protect\cite{cite3096}}. 

\end{eczvaluelist}
\eczhbkcontributors{ \eczhuVVA }
\endeczcode

\eczcode{2d_stabilizer}{2D lattice stabilizer code}{}
\codefieldsection{Description}
Lattice stabilizer code in two Euclidean dimensions, using either the ordinary block notion of locality or the fermionic/Majorana notion of locality.

Any translation-invariant 2D prime-qudit lattice stabilizer code can be converted to several copies of the prime-qudit 2D surface code along with some trivial codes \NoCaseChange{\protect\cite{cite4513}}.
Any 2D topological order requires weight-four Hamiltonian terms, i.e., it cannot be stabilized via weight-two or weight-three terms on 2D lattices of qubits or qutrits \NoCaseChange{\protect\cite{cite2684,cite2685,cite2686}}.

Translation-invariant 2D prime-qudit lattice stabilizer codes are equivalent to several copies of the prime-qudit surface code and a trivial code via a local constant-depth Clifford circuit \NoCaseChange{\protect\cite{cite4513}}.
There are algorithms which determine the fusion and braiding rules \NoCaseChange{\protect\cite{cite5049}} as well as boundaries and twist defects \NoCaseChange{\protect\cite{cite443}} of a 2D translationally invariant modular-qudit stabilizer code for any qudit dimension.

\codefieldsection{Encoding}
\begin{eczvaluelist}
\item\relax The geometric entanglement measure of a 2D stabilizer codeword with sufficiently high distance \(d\) scales as \flmRefsHyperref{ref65}{order} \(\Omega(d^2)\) \NoCaseChange{\protect\cite{cite3804}}.
\end{eczvaluelist}
\codefieldsection{Decoding}
\begin{eczvaluelist}
\item\relax Renormalization group (RG) decoder \NoCaseChange{\protect\cite{cite5050}}.
\item\relax Tensor-network based decoder for 2D codes subject to correlated noise \NoCaseChange{\protect\cite{cite4377}}.
\item\relax Standard stabilizer-based error correction can be performed even in the presence of perturbations to the codespace \NoCaseChange{\protect\cite{cite5051,cite5052,cite5053}}; see also Refs. \NoCaseChange{\protect\cite{cite3020,cite3009,cite5054}}.
\item\relax Real-time geometrically local decoder based on introducing an ancillary buffer and confining spacetime interactions between anyons  \NoCaseChange{\protect\cite{cite3030}}.
\end{eczvaluelist}
\codefieldsection{Code Capacity Threshold}
\begin{eczvaluelist}
\item\relax Noise thresholds can be formulated as anyon \flmRefsHyperref{ref410}{condensation} transitions in a topological field theory \NoCaseChange{\protect\cite{cite5055}}, generalizing the mapping of the effect of noise on a code state to a statistical mechanical model \NoCaseChange{\protect\cite{cite480,cite3441,cite4295,cite4377}}. Namely, the noise threshold for a noise channel \(\cal{E}\) acting on a 2D stabilizer state \(|\psi\rangle\) can be obtained from the properties of the resulting (mixed) state \(\mathcal{E}(|\psi\rangle\langle\psi|)\) \NoCaseChange{\protect\cite{cite5056,cite5055,cite5057,cite5058,cite4235}}.
\end{eczvaluelist}
\codefieldsection{Parent}
\begin{eczvaluelist}
\item\relax
\flmRefsHyperref[eczindexfamilyrel]{code:translationally_invariant_stabilizer}{Lattice stabilizer code}\end{eczvaluelist}
\codefieldsection{Children}
\begin{eczvaluelist}
\item\relax
\flmRefsHyperref[eczindexfamilyrel]{code:compactified_r}{Compactified \(\mathbb{R}\) gauge theory code}\item\relax
\flmRefsHyperref[eczindexfamilyrel]{code:analog_surface}{Analog surface code}\item\relax
\flmRefsHyperref[eczindexfamilyrel]{code:chern_simons_gkp}{\(U(1)_{2n} \times U(1)_{-2m}\) Chern-Simons GKP code}\item\relax
\flmRefsHyperref[eczindexfamilyrel]{code:gkp_surface_concatenated}{GKP-surface code}\item\relax
\flmRefsHyperref[eczindexfamilyrel]{code:majorana_color}{Majorana color code} --- The Majorana color code is a 2D qubit stabilizer code with respect to the Majorana operator basis.
\item\relax
\flmRefsHyperref[eczindexfamilyrel]{code:majorana_surface}{Majorana surface code} --- The Majorana surface code is a 2D qubit stabilizer code with respect to the Majorana operator basis.
\item\relax
\flmRefsHyperref[eczindexfamilyrel]{code:2d_bosonization}{2D bosonization code} --- The 2D bosonization code encodes fermionic modes into a 2D qubit stabilizer code.
\item\relax
\flmRefsHyperref[eczindexfamilyrel]{code:square_lattice_cluster}{Square-lattice cluster-state code}\item\relax
\flmRefsHyperref[eczindexfamilyrel]{code:qcga}{Bivariate bicycle (BB) code} --- Bivariate bicycle codes are defined on 2D lattices with periodic boundary conditions, and versions with open boundary conditions have been investigated \NoCaseChange{\protect\cite{cite443,cite3501}}. Bivariate bicycle codes are on par with the surface code in terms of threshold, but admit a much higher ancilla-added encoding rate at the expense of having non-geometrically local weight-six check operators. BB codes have been investigated in terms of their anyons and topological order \NoCaseChange{\protect\cite{cite2533}}.
\item\relax
\flmRefsHyperref[eczindexfamilyrel]{code:twist_defect_color}{Twist-defect color code}\item\relax
\flmRefsHyperref[eczindexfamilyrel]{code:twist_defect_surface}{Twist-defect surface code}\item\relax
\flmRefsHyperref[eczindexfamilyrel]{code:tqd_abelian_stabilizer}{Abelian TQD stabilizer code} --- For every finite Abelian group \(G=\prod_i \mathbb{Z}_{N_i}\) and every product of Type-I and Type-II cocycles, there is a 2D modular-qudit Pauli stabilizer Hamiltonian on composite-dimensional qudits realizing the corresponding Abelian TQD \NoCaseChange{\protect\cite{cite405}}.
\item\relax
\flmRefsHyperref[eczindexfamilyrel]{code:galois_color}{Galois-qudit color code}\item\relax
\flmRefsHyperref[eczindexfamilyrel]{code:galois_topological}{Galois-qudit surface code}\end{eczvaluelist}
\codefieldsection{Cousins}
\begin{eczvaluelist}
\item\relax
\flmRefsHyperref[eczindexfamilyrel]{code:surface}{Kitaev surface code} --- Translation-invariant 2D qubit lattice stabilizer codes are equivalent to several copies of the Kitaev surface code via a local constant-depth qudit Clifford circuit \NoCaseChange{\protect\cite{cite603,cite3962,cite3963}}.
\item\relax
\flmRefsHyperref[eczindexfamilyrel]{code:quantum_double_abelian}{Abelian quantum-double stabilizer code} --- Translation-invariant 2D prime-qudit lattice stabilizer codes are equivalent to several copies of the prime-qudit surface code and a trivial code via a local constant-depth qudit Clifford circuit \NoCaseChange{\protect\cite{cite4513}}.
\item\relax
\flmRefsHyperref[eczindexfamilyrel]{code:holographic}{Holographic code} --- 2D lattice stabilizer codes admit a bulk-boundary correspondence similar to that of holographic codes, namely, the boundary Hilbert space of the former cannot be realized via local degrees of freedom \NoCaseChange{\protect\cite{cite2852}}.
\item\relax
\flmRefsHyperref[eczindexfamilyrel]{code:qldpc}{Qubit QLDPC code} --- Chain complexes describing qubit QLDPC codes can be converted to 2D lattice stabilizer codes \NoCaseChange{\protect\cite{cite489}}.
\item\relax
\flmRefsHyperref[eczindexfamilyrel]{code:self_correct}{Self-correcting quantum code} --- 2D stabilizer codes \NoCaseChange{\protect\cite{cite3000}} and encodings of frustration-free code Hamiltonians \NoCaseChange{\protect\cite{cite3016}} admit only constant-energy excitations, and so do not have an energy barrier.
\end{eczvaluelist}
\eczhbkcontributors{ \eczhuVVA }
\endeczcode

\eczcode{3d_stabilizer}{3D lattice stabilizer code}{}
\codefieldsection{Description}
Lattice stabilizer code in three Euclidean dimensions, using either the ordinary block notion of locality or the fermionic/Majorana notion of locality.

For translation-invariant qubit topological stabilizer models in 3D, bulk commutation data can be used to coarsely sort phases into topological quantum field theory (TQFT), foliated type-I, fractal type-I, and type-II sectors \NoCaseChange{\protect\cite{cite456}}.
For the TQFT sector, the paper conjectures equivalence under a locality-preserving unitary to copies of the 3D surface code and/or the 3D fermionic surface code, together with trivial ancillas \NoCaseChange{\protect\cite{cite456}}.

\codefieldsection{Code Capacity Threshold}
\begin{eczvaluelist}
\item\relax Applying Clifford deformations to various 3D stabilizer codes, including the 3D surface code, 3D color code, X-cube model code, and Sierpinski prism model code, yields a \(50\%\) code capacity threshold under infinitely biased Pauli noise \NoCaseChange{\protect\cite{cite2626}}.
\end{eczvaluelist}
\codefieldsection{Parent}
\begin{eczvaluelist}
\item\relax
\flmRefsHyperref[eczindexfamilyrel]{code:translationally_invariant_stabilizer}{Lattice stabilizer code}\end{eczvaluelist}
\codefieldsection{Children}
\begin{eczvaluelist}
\item\relax
\flmRefsHyperref[eczindexfamilyrel]{code:3d_bosonization}{3D bosonization code} --- The 3D bosonization code encodes fermionic modes into a 3D qubit stabilizer code.
\item\relax
\flmRefsHyperref[eczindexfamilyrel]{code:rbh}{Raussendorf-Bravyi-Harrington (RBH) cluster-state code}\item\relax
\flmRefsHyperref[eczindexfamilyrel]{code:3d_color}{3D color code}\item\relax
\flmRefsHyperref[eczindexfamilyrel]{code:3d_fermionic_surface}{3D fermionic surface code}\item\relax
\flmRefsHyperref[eczindexfamilyrel]{code:three_fermion}{Three-fermion (3F) Walker-Wang model code}\item\relax
\flmRefsHyperref[eczindexfamilyrel]{code:fracton}{Fracton stabilizer code}\item\relax
\flmRefsHyperref[eczindexfamilyrel]{code:3d_semion}{Chiral semion Walker-Wang model code}\item\relax
\flmRefsHyperref[eczindexfamilyrel]{code:qudit_3d_surface}{Modular-qudit 3D surface code}\end{eczvaluelist}
\codefieldsection{Cousins}
\begin{eczvaluelist}
\item\relax
\flmRefsHyperref[eczindexfamilyrel]{code:clifford-deformed_surface}{Clifford-deformed surface code (CDSC)} --- Applying Clifford deformations to various 3D stabilizer codes, including the 3D surface code, 3D color code, X-cube model code, and Sierpinski prism model code, yields a \(50\%\) code capacity threshold under infinitely biased Pauli noise \NoCaseChange{\protect\cite{cite2626}}.
\item\relax
\flmRefsHyperref[eczindexfamilyrel]{code:asymmetric_qecc}{Asymmetric quantum code (AQC)} --- Applying Clifford deformations to various 3D stabilizer codes, including the 3D surface code, 3D color code, X-cube model code, and Sierpinski prism model code, yields a \(50\%\) code capacity threshold under infinitely biased Pauli noise \NoCaseChange{\protect\cite{cite2626}}.
\item\relax
\flmRefsHyperref[eczindexfamilyrel]{code:topological_abelian}{Abelian topological code} --- Translation-invariant qubit 3D TQFT stabilizer models are conjectured to be equivalent, under a locality-preserving unitary, to multiple copies of the 3D surface code and/or the 3D fermionic surface code together with trivial ancillas \NoCaseChange{\protect\cite{cite456}}.
\item\relax
\flmRefsHyperref[eczindexfamilyrel]{code:self_correct}{Self-correcting quantum code} --- 3D translationally-invariant qubit stabilizer code families with constant \(k\) support logical string operators and thus cannot be self-correcting \NoCaseChange{\protect\cite{cite3031}}. For non-constant \(k\), such families can support at most a logarithmic energy barrier \NoCaseChange{\protect\cite{cite3032}}.
\end{eczvaluelist}
\eczhbkcontributors{ \eczhuVVA }
\endeczcode

\eczcode{4d_stabilizer}{4D lattice stabilizer code}{}
\codefieldsection{Description}
Lattice stabilizer code in four Euclidean dimensions, using either the ordinary block notion of locality or the fermionic/Majorana notion of locality.

\codefieldsection{Parent}
\begin{eczvaluelist}
\item\relax
\flmRefsHyperref[eczindexfamilyrel]{code:translationally_invariant_stabilizer}{Lattice stabilizer code}\end{eczvaluelist}
\codefieldsection{Children}
\begin{eczvaluelist}
\item\relax
\flmRefsHyperref[eczindexfamilyrel]{code:dfour_gkp}{\(D_4\) hyper-diamond GKP code}\item\relax
\flmRefsHyperref[eczindexfamilyrel]{code:stab_16_6_4}{\(\llbracket 16,6,4\rrbracket \) Tesseract color code} --- The tesseract color code is a 4D color code defined on a tesseract \NoCaseChange{\protect\cite{cite3212,cite101,cite862,cite2362}}.
\item\relax
\flmRefsHyperref[eczindexfamilyrel]{code:4d_13_surface}{\((1,3)\) 4D toric code}\item\relax
\flmRefsHyperref[eczindexfamilyrel]{code:4d_surface}{\((2,2)\) Loop toric code}\end{eczvaluelist}
\codefieldsection{Cousin}
\begin{eczvaluelist}
\item\relax
\flmRefsHyperref[eczindexfamilyrel]{code:multimodegkp}{Gottesman-Kitaev-Preskill (GKP) code} --- The 4D square-lattice GKP code admits the isthmus property, which allows certain ancilla errors to be detectable \NoCaseChange{\protect\cite{cite482}}.
\end{eczvaluelist}
\eczhbkcontributors{ \eczhuVVA }
\endeczcode

\eczcode{tqd_abelian}{Abelian TQD code}{~\NoCaseChange{\protect\cite{cite573,cite571}}}
\codefieldsection{Description}
TQD code whose codewords realize a 2D Abelian twisted-quantum-double topological order.
For Abelian TQDs, the corresponding anyon theory is defined by an Abelian group and a group cocycle built from Type-I, Type-II, or Type-III 3-cocycles \NoCaseChange{\protect\cite{cite571,cite405,cite572}}.
Abelian TQDs with Type-I and -II cocycles account for all 2D Abelian topological orders that admit gapped boundaries \NoCaseChange{\protect\cite{cite573}}.
Abelian TQDs with Type-III cocycles may admit non-Abelian topological orders. 

Type-I and -II Abelian TQD codes can be realized as modular-qudit lattice stabilizer codes on composite-dimensional qudits by starting from a stack of Abelian quantum double models (it suffices to take \(\prod_i \mathbb{Z}_{N_i^2}\) toric codes for \(G=\prod_i \mathbb{Z}_{N_i}\)) and \flmRefsHyperref{ref410}{condensing} certain bosonic anyons \NoCaseChange{\protect\cite{cite405}}.
Many Abelian TQD code Hamiltonians were originally formulated as commuting-projector models \NoCaseChange{\protect\cite{cite2694,cite2693}}.

\codefieldsection{Encoding}
\begin{eczvaluelist}
\item\relax Fault-tolerant state-preparation circuits for all non-chiral abelian topological phases \NoCaseChange{\protect\cite{cite572}}.
\end{eczvaluelist}
\codefieldsection{Fault Tolerance}
\begin{eczvaluelist}
\item\relax Fault-tolerant state-preparation circuits for all non-chiral abelian topological phases \NoCaseChange{\protect\cite{cite572}}.
\end{eczvaluelist}
\codefieldsection{Parent}
\begin{eczvaluelist}
\item\relax
\flmRefsHyperref[eczindexfamilyrel]{code:tqd}{Twisted quantum double (TQD) code} --- The anyon theory corresponding to Abelian TQD codes is defined by an Abelian group and a Type-I, Type-II, or Type-III 3-cocycle.
Abelian TQDs with Type-I and -II cocycles account for all 2D Abelian topological orders that admit gapped boundaries \NoCaseChange{\protect\cite{cite573}}.

\end{eczvaluelist}
\codefieldsection{Children}
\begin{eczvaluelist}
\item\relax
\flmRefsHyperref[eczindexfamilyrel]{code:brickwork}{Brickwork \(XS\) stabilizer code} --- The ground-state subspace of the brickwork \(XS\) stabilizer code realizes the topological order of the Type-III \(G=\mathbb{Z}^3_2\) TQD model \NoCaseChange{\protect\cite{cite575,cite576}}, which is the same topological order as the \(G=D_4\) quantum double \NoCaseChange{\protect\cite{cite577}}.
\item\relax
\flmRefsHyperref[eczindexfamilyrel]{code:double_semion_string_net}{Double-semion string-net code} --- When treated as ground states of the code Hamiltonian, the double-semion string-net code states realize 2D double-semion topological order, a topological phase of matter that exists as the deconfined phase of the 2D twisted \(\mathbb{Z}_2\) gauge theory \NoCaseChange{\protect\cite{cite584}}.
\item\relax
\flmRefsHyperref[eczindexfamilyrel]{code:tqd_abelian_stabilizer}{Abelian TQD stabilizer code} --- Every Abelian TQD code with Type-I and -II cocycles can be realized as a modular-qudit Pauli stabilizer code by starting from a stack of Abelian quantum double models (it suffices to take \(\prod_i \mathbb{Z}_{N_i^2}\) toric codes) and \flmRefsHyperref{ref410}{condensing} certain bosonic anyons \NoCaseChange{\protect\cite{cite405}}.
\end{eczvaluelist}
\codefieldsection{Cousins}
\begin{eczvaluelist}
\item\relax
\flmRefsHyperref[eczindexfamilyrel]{code:commuting_projector}{Commuting-projector Hamiltonian code} --- Many Abelian TQD code Hamiltonians were originally formulated as commuting-projector models \NoCaseChange{\protect\cite{cite2693}}.
\item\relax
\flmRefsHyperref[eczindexfamilyrel]{code:topological_abelian}{Abelian topological code} --- Abelian TQDs with Type-I and -II cocycles account for all 2D Abelian topological orders that admit gapped boundaries \NoCaseChange{\protect\cite{cite573}}.
Conversely, every Abelian anyon theory is a subtheory of some TQD \NoCaseChange{\protect\cite[{Sec. 6.2}]{cite414}}. 
Any Abelian anyon theory \(A\) can be realized at one of the surfaces of a 3D Walker-Wang model whose underlying theory is an Abelian TQD containing \(A\) as a subtheory \NoCaseChange{\protect\cite{cite471,cite472}\protect\cite[{Appx. H}]{cite414}}.

\item\relax
\flmRefsHyperref[eczindexfamilyrel]{code:quantum_double}{Quantum-double code} --- A Type-III \(\mathbb{Z}_2^3\) Abelian TQD realizes the same topological order as the \(G=D_4\) quantum double model \NoCaseChange{\protect\cite{cite577,cite575}}. There is a sufficient condition for when a Type-III TQD can be realized as a quantum double model \NoCaseChange{\protect\cite{cite5059}}.
\item\relax
\flmRefsHyperref[eczindexfamilyrel]{code:quantum_double_dihedral}{Dihedral \(G=D_m\) quantum-double code} --- A Type-III \(\mathbb{Z}_2^3\) Abelian TQD realizes the same topological order as the \(G=D_4\) quantum double model \NoCaseChange{\protect\cite{cite577,cite575}}.
\item\relax
\flmRefsHyperref[eczindexfamilyrel]{code:xs_stabilizer}{XS stabilizer code} --- Abelian TQD models for the groups \(\mathbb{Z}_2^k\) can be realized as XS stabilizer codes \NoCaseChange{\protect\cite{cite3625}}. Upon gauging some symmetries \NoCaseChange{\protect\cite{cite462,cite463,cite233,cite464,cite465,cite466,cite467,cite468,cite469,cite470}}, a Type-III \(\mathbb{Z}_2^3\) TQD realizes the same topological order as the \(G=D_4\) quantum double model \NoCaseChange{\protect\cite{cite577,cite575}}.
\end{eczvaluelist}
\eczhbkcontributors{ \eczhuVVA }
\endeczcode

\eczcode{css}{Calderbank-Shor-Steane (CSS) stabilizer code}{}
\codefieldsection{Description}
A stabilizer code admitting a set of stabilizer generators that are either \(Z\)-type or \(X\)-type operators.
The two sets of stabilizer generators can often be related to parts of a chain complex over the appropriate ring or field.

CSS codes can also be viewed as an instance of a two-step convex-geometric construction: one first chooses an intermediate subspace stabilized by one type of generator, and then imposes the other type of generator so that the remaining diagonal error slopes vanish via an application of the Tverberg theorem \NoCaseChange{\protect\cite[{Sec. 4.3}]{cite5030}}.

\codefieldsection{Parents}
\begin{eczvaluelist}
\item\relax
\flmRefsHyperref[eczindexfamilyrel]{code:stabilizer}{Stabilizer code}\item\relax
\flmRefsHyperref[eczindexfamilyrel]{code:group_gkp}{Group GKP code} --- CSS codes are Abelian group GKP codes, i.e., group GKP codes constructed out of Pauli-type operators.
\end{eczvaluelist}
\codefieldsection{Children}
\begin{eczvaluelist}
\item\relax
\flmRefsHyperref[eczindexfamilyrel]{code:homological_rotor}{Homological rotor code} --- Homological rotor codes are rotor CSS codes constructed from chain complexes over the integers in an extension of the \flmRefsHyperref{ref683}{qubit CSS-to-homology correspondence} to rotors. The homology group of the logical operators has a torsion component because the chain complexes are defined over the ring of integers, which yields codes with finite logical dimension. Products of chain complexes can also yield rotor codes.
\item\relax
\flmRefsHyperref[eczindexfamilyrel]{code:rotor_gkp}{Rotor GKP code} --- Rotor GKP code stabilizers are purely position and purely momentum rotor Pauli-type operators, making these codes CSS.
\item\relax
\flmRefsHyperref[eczindexfamilyrel]{code:generalized_homological_product_css}{Generalized homological-product CSS code}\item\relax
\flmRefsHyperref[eczindexfamilyrel]{code:oscillator_css}{Bosonic CSS code}\item\relax
\flmRefsHyperref[eczindexfamilyrel]{code:qudit_css}{Modular-qudit CSS code}\item\relax
\flmRefsHyperref[eczindexfamilyrel]{code:galois_css}{Galois-qudit CSS code}\end{eczvaluelist}
\codefieldsection{Cousins}
\begin{eczvaluelist}
\item\relax
\flmRefsHyperref[eczindexfamilyrel]{code:rotor_stabilizer}{Rotor stabilizer code} --- A rotor stabilizer code admitting a set of generators such that each generator consists of either angular position or angular momentum operators is a CSS code.
\item\relax
\flmRefsHyperref[eczindexfamilyrel]{code:subsystem_css}{Subsystem CSS code} --- Subsystem CSS codes reduce to CSS stabilizer codes when there are no gauge degrees of freedom.
\item\relax
\flmRefsHyperref[eczindexfamilyrel]{code:quantum_lattice}{Quantum lattice code} --- Quantum lattice codes defined on rectangular lattices are CSS codes. There is no known relation to chain complexes for such codes. More general lattices, obtained from rectangular lattices by Gaussian transformations, yield non-CSS codes.
\item\relax
\flmRefsHyperref[eczindexfamilyrel]{code:asymmetric_qecc}{Asymmetric quantum code (AQC)} --- In the context of comparing weight as well as of determining distances for noise models biased toward \(X\)- or \(Z\)-type errors, an extended notation for asymmetric CSS block quantum codes is \(\llbracket n,k,(d_X,d_Z),w\rrbracket \) or \(\llbracket n,k,d_X/d_Z,w\rrbracket \).
\end{eczvaluelist}
\eczhbkcontributors{ \eczhuVVA }
\endeczcode

\eczcode{quantum_double_dihedral}{Dihedral \(G=D_m\) quantum-double code}{~\NoCaseChange{\protect\cite{cite423}}}
\codefieldsection{Description}
Quantum-double code whose codewords realize topological order associated with the dihedral group \(D_m\) of order \(2m\).
For \(m \geq 3\), these codes are non-Abelian, with the simplest case given by \(D_3=S_3\), the permutation group on three objects.
On an oriented lattice, each edge hosts a \(2m\)-dimensional group qudit, and the codespace is the ground-state subspace of the corresponding quantum double Hamiltonian \NoCaseChange{\protect\cite{cite423}}.

\codefieldsection{Transversal and Permutation-Based Gates}
\begin{eczvaluelist}
\item\relax On a triangular patch encoding one logical qubit, a transversal \(T^{1/N}=\mathrm{diag}(1,e^{i\pi/(4N)})\) gate can be implemented when \(G=D_{4N}\) \NoCaseChange{\protect\cite{cite729,cite730}}.
\end{eczvaluelist}
\codefieldsection{Gates}
\begin{eczvaluelist}
\item\relax Universal topological quantum computation is possible for certain groups such as \(G=D_3=S_3\) \NoCaseChange{\protect\cite{cite5060,cite5061}}.
\item\relax For \(G=S_3\), Ref. \NoCaseChange{\protect\cite{cite5062}} gives measurement-assisted universal gate sets; circuit-level fault tolerance can be improved using an anyon interferometer \NoCaseChange{\protect\cite{cite5063}}.
\end{eczvaluelist}
\codefieldsection{Fault Tolerance}
\begin{eczvaluelist}
\item\relax Universal topological quantum computation is possible for certain groups such as \(G=D_3=S_3\) \NoCaseChange{\protect\cite{cite5060,cite5061}}.
\item\relax For \(G=S_3\), Ref. \NoCaseChange{\protect\cite{cite5062}} gives measurement-assisted universal gate sets; circuit-level fault tolerance can be improved using an anyon interferometer \NoCaseChange{\protect\cite{cite5063}}.
\end{eczvaluelist}
\codefieldsection{Code Capacity Threshold}
\begin{eczvaluelist}
\item\relax Behavior under \(X\)-type noise (namely, diffusion of certain anyons) for the \(G=D_4\) case is related to the phase diagram of a disordered net model \NoCaseChange{\protect\cite{cite5064}}.
\end{eczvaluelist}
\codefieldsection{Realizations}
\begin{eczvaluelist}
\item\relax Trapped ions: ground state of the model for \(G = D_3\), encoding of logical qutrits in the anyon fusion space, and a universal anyon-based gate set realized on 54 qubits by Quantinuum \NoCaseChange{\protect\cite{cite5065}}.
\end{eczvaluelist}
\codefieldsection{Notes}
\begin{eczvaluelist}
\item\relax See \NoCaseChange{\protect\cite{cite5067}\protect\cite[{Sec. 5.4}]{cite5066}} for introductions to this code.
\item\relax The \(\Phi,\Lambda\) \flmHref{https://web.archive.org/web/20161223121819/http://citizensciencegames.com/games/decodoku/}{Decodoku game} is based on the quantum double model for the group \(D_3=S_3\) of permutations on three letters.
\end{eczvaluelist}
\codefieldsection{Parent}
\begin{eczvaluelist}
\item\relax
\flmRefsHyperref[eczindexfamilyrel]{code:quantum_double}{Quantum-double code}\end{eczvaluelist}
\codefieldsection{Cousins}
\begin{eczvaluelist}
\item\relax
\flmRefsHyperref[eczindexfamilyrel]{code:tqd_abelian}{Abelian TQD code} --- A Type-III \(\mathbb{Z}_2^3\) Abelian TQD realizes the same topological order as the \(G=D_4\) quantum double model \NoCaseChange{\protect\cite{cite577,cite575}}.
\item\relax
\flmRefsHyperref[eczindexfamilyrel]{code:spt}{Symmetry-protected topological (SPT) code} --- The \(D_4\) quantum double model can be obtained by gauging \NoCaseChange{\protect\cite{cite462,cite463,cite233,cite464,cite465,cite466,cite467,cite468,cite469,cite470}} symmetries of a Type III \(\mathbb{Z}_2^3\) SPT \NoCaseChange{\protect\cite{cite575,cite3071}}.
\item\relax
\flmRefsHyperref[eczindexfamilyrel]{code:cubic_theory}{Cubic theory code} --- For \(D=3\) with \(l=m=n=1\), the cubic theory is equivalent to the \(G=D_4\) quantum double, i.e. the non-Abelian Type-III \(\mathbb{Z}_2^3\) twisted quantum double \NoCaseChange{\protect\cite{cite576}}.
\item\relax
\flmRefsHyperref[eczindexfamilyrel]{code:hexagonal_cz}{Hexagonal \(CZ\) code} --- The ground-state subspace of the hexagonal \(CZ\) code realizes the topological order of the Type-III \(G=\mathbb{Z}^3_2\) Abelian TQD model \NoCaseChange{\protect\cite{cite575,cite576}}, which is the same topological order as the \(G=D_4\) quantum double \NoCaseChange{\protect\cite{cite577}}. There is a constant-depth circuit implementing a transversal logical \(T\) gate via an emergent automorphism symmetry of the underlying \(\mathbb{D}_4\) topological order \NoCaseChange{\protect\cite{cite725}}.
\item\relax
\flmRefsHyperref[eczindexfamilyrel]{code:brickwork}{Brickwork \(XS\) stabilizer code} --- The ground-state subspace of the brickwork \(XS\) stabilizer code realizes the topological order of the non-Abelian Type-III \(G=\mathbb{Z}^3_2\) TQD model \NoCaseChange{\protect\cite{cite575,cite576}}, which is the same topological order as the ordinary \(G=D_4\) quantum double \NoCaseChange{\protect\cite{cite577}}.
\item\relax
\flmRefsHyperref[eczindexfamilyrel]{code:qudit_surface}{Modular-qudit surface code} --- The \(D_3\) quantum double model can be obtained by gauging \NoCaseChange{\protect\cite{cite462,cite463,cite233,cite464,cite465,cite466,cite467,cite468,cite469,cite470}} the charge conjugation symmetry of the \(\mathbb{Z}_3\) surface code \NoCaseChange{\protect\cite{cite3071}}. A magic-state preparation routine for the \(\mathbb{Z}_4\) surface code traverses through the \(D_4\) quantum double model \NoCaseChange{\protect\cite{cite4584}}.
\end{eczvaluelist}
\eczhbkcontributors{ \eczhuVVA }
\endeczcode

\eczcode{dijkgraaf_witten}{Dijkgraaf-Witten gauge theory code}{~\NoCaseChange{\protect\cite{cite584,cite585,cite586}}}
\codefieldsection{Alternative Names}
\begin{eczvaluelist}
\item\relax Cohomological gauge theory code
\item\relax Twisted \(G\) gauge theory code
\end{eczvaluelist}
\eczhIndexCodeAliasName{dijkgraaf_witten}{Cohomological gauge theory code}
\eczhIndexCodeAliasName{dijkgraaf_witten}{Twisted \(G\) gauge theory code}
\codefieldsection{Description}
A code whose codewords realize \(D\)-dimensional lattice Dijkgraaf-Witten gauge theory \NoCaseChange{\protect\cite{cite584,cite585}} for a finite group \(G\) and a \(D+1\)-cocycle \(\omega\) in the cohomology class \(H^{D+1}( G, U(1) )\).
When the cocycle is non-trivial, the gauge theory is called a \textit{twisted gauge theory}.
There exist lattice-model formulations in arbitrary spatial dimension \NoCaseChange{\protect\cite{cite586}}.
Boundaries and excitations have been studied for arbitrary dimension \NoCaseChange{\protect\cite{cite587}}.

\codefieldsection{Parents}
\begin{eczvaluelist}
\item\relax
\flmRefsHyperref[eczindexfamilyrel]{code:group_quantum}{Group-based quantum code}\item\relax
\flmRefsHyperref[eczindexfamilyrel]{code:yetter_gauge_theory}{Two-gauge theory code} --- Replacing the two-group data in a two-gauge theory with a group and a cocycle reproduces the phase of the Dijkgraaf-Witten gauge theory, with the two theories equivalent in 2D \NoCaseChange{\protect\cite{cite616}}. Generalizations of Ocneanu's tube algebras \NoCaseChange{\protect\cite{cite5068,cite5069}} can be used to characterize excitations in both theories \NoCaseChange{\protect\cite[{Sec. 4.2}]{cite5070}}. A Dijkgraaf-Witten Lagrangian can also be re-expressed as a two-group gauge theory Lagrangian by relating the electric and magnetic gauge fields via the equations of motion \NoCaseChange{\protect\cite{cite616,cite5071}}.
\end{eczvaluelist}
\codefieldsection{Children}
\begin{eczvaluelist}
\item\relax
\flmRefsHyperref[eczindexfamilyrel]{code:tqd}{Twisted quantum double (TQD) code} --- Restricting Dijkgraaf-Witten gauge theory to a 2D manifold reproduces the phase of the TQD model \NoCaseChange{\protect\cite{cite605}}.
The Drinfeld center of the category \(\text{Vec}^{\omega}(G)\) is used to describe bulk excitations of 3D Dijkgraaf-Witten models, and this center is equivalent to the twisted quantum double \(D^{\omega}(G)\) \NoCaseChange{\protect\cite[{pg. 41}]{cite587}}.
TQD codewords are gauge-invariant boundary states of a 3D Dijkgraaf-Witten theory \NoCaseChange{\protect\cite[{Sec. IX}]{cite571}}.

\item\relax
\flmRefsHyperref[eczindexfamilyrel]{code:tqt}{Twisted quantum triple (TQT) code} --- Restricting Dijkgraaf-Witten gauge theory to a 3D manifold reproduces the phase of the TQT model.

\item\relax
\flmRefsHyperref[eczindexfamilyrel]{code:4d_13_surface}{\((1,3)\) 4D toric code} --- An untwisted Dijkgraaf-Witten theory in 4D for the group \(G=\mathbb{Z}_2\) is a \((1,3)\) 4D toric code.
\end{eczvaluelist}
\eczhbkcontributors{ \eczhuVVA }
\endeczcode

\eczcode{rotor_4_2_2}{Four-rotor code}{~\NoCaseChange{\protect\cite[{Sec. VIII}]{cite2720}}}
\codefieldsection{Description}
\(\llbracket 4,2,2\rrbracket _{\mathbb Z}\) CSS rotor code that is an extension of the four-qubit code to the integer alphabet, i.e., the angular momentum states of a rotor.

The code is \(U(1)\)-covariant and its ideal logical-rotor codewords,
\flmMathEnvironment{align}{}{
  |\overline{a,b}\rangle = \sum_{j,k,l\in\mathbb{Z}} \delta_{a,j+k}\delta_{b,l} \left| j,k,j+l,k+l \right\rangle~,
}
where \(a,b\in\mathbb{Z}\), are not normalizable.

\codefieldsection{Parents}
\begin{eczvaluelist}
\item\relax
\flmRefsHyperref[eczindexfamilyrel]{code:homological_rotor}{Homological rotor code}\item\relax
\flmRefsHyperref[eczindexfamilyrel]{code:group_4_2_2}{\(\llbracket 4,2,2\rrbracket _{G}\) four group-qudit code} --- The four group-qudit code reduces to the four-rotor code for \(G= \mathbb{Z}\).
\end{eczvaluelist}
\codefieldsection{Cousin}
\begin{eczvaluelist}
\item\relax
\flmRefsHyperref[eczindexfamilyrel]{code:homological_number-phase}{Homological number-phase code} --- After suitable rotor-parity flips and projection onto the non-negative angular-momentum orthant, the four-rotor current-mirror code yields a homological number-phase code \NoCaseChange{\protect\cite[{Ex. 4}]{cite2699}}.
\end{eczvaluelist}
\eczhbkcontributors{ \eczhuVVA }
\endeczcode

\eczcode{generalized_color}{Generalized 2D color code}{~\NoCaseChange{\protect\cite{cite5072}}}
\codefieldsection{Description}
Member of a family of non-Abelian 2D topological codes, defined by a finite group \( G \), that serves as a generalization of the 2D color code (for which \(G=\mathbb{Z}_2 \times \mathbb{Z}_2\)).
Hamiltonian terms are built from \flmRefsHyperref{ref20}{group-based right- and left-multiplication \(X\)-type operators together with \(Z\)-type operators}.

\codefieldsection{Parents}
\begin{eczvaluelist}
\item\relax
\flmRefsHyperref[eczindexfamilyrel]{code:group_quantum}{Group-based quantum code}\item\relax
\flmRefsHyperref[eczindexfamilyrel]{code:topological}{Topological code} --- A generalized color code for group \(G\) on the 4.8.8 lattice is equivalent to a \(G\) quantum double model and another \(G/[G,G]\) quantum double model defined using the Abelianization of \(G\).
\end{eczvaluelist}
\codefieldsection{Child}
\begin{eczvaluelist}
\item\relax
\flmRefsHyperref[eczindexfamilyrel]{code:2d_color}{2D color code} --- The generalized color code for \(G=\mathbb{Z}_2\) reduces to the 2D color code.
\end{eczvaluelist}
\codefieldsection{Cousins}
\begin{eczvaluelist}
\item\relax
\flmRefsHyperref[eczindexfamilyrel]{code:quantum_double}{Quantum-double code} --- A generalized color code for group \(G\) on the 4.8.8 lattice is equivalent to a \(G\) quantum double model and another \(G/[G,G]\) quantum double model defined using the Abelianization of \(G\).
\item\relax
\flmRefsHyperref[eczindexfamilyrel]{code:qudit_color}{Modular-qudit lattice color code} --- The generalized color code for \(G=\mathbb{Z}_q\) reduces to the 2D modular-qudit color code.
\end{eczvaluelist}
\eczhbkcontributors{ \eczhuVVA }
\endeczcode

\eczcode{generalized_homological_product}{Generalized homological-product code}{}
\codefieldsection{Description}
Stabilizer code whose properties are determined from an underlying chain complex, which often consists of some type of product of other chain complexes.
The \flmRefsCref{ref683} yields an interpretation of codes in terms of chain complexes, thus allowing for the use of various products from homology in constructing codes.

The codes participating in the product can be quantum, classical, or mixed.
Homology can be used to design codes for qubits, modular qudits, Galois qudits, as well as rotors; most codes are CSS codes.
However, products can be of more than two underlying codes, in which case the output code need not be CSS (e.g., for \flmRefsHyperref{code:xyz_product}{XYZ product codes}).

The simplest product is a tensor product, with more general products imposing equivalence or symmetry relations on the outputs of the tensor product.
A product of two codes can be interpreted as a fiber bundle, with one element of the product being the base and the other being the fiber.

\codefieldsection{Parent}
\begin{eczvaluelist}
\item\relax
\flmRefsHyperref[eczindexfamilyrel]{code:general_qldpc}{QLDPC code} --- Homological products are a primary tool for generating QLDPC codes with favorable parameters. Typically, whenever the input codes are LDPC or QLDPC, the resulting code will be QLDPC with non geometrically local stabilizer generators.
\end{eczvaluelist}
\codefieldsection{Children}
\begin{eczvaluelist}
\item\relax
\flmRefsHyperref[eczindexfamilyrel]{code:generalized_homological_product_css}{Generalized homological-product CSS code} --- The notion of homological products arises from interpreting CSS codes in terms of chain complexes over manifolds, but some products no longer yield CSS codes.
\item\relax
\flmRefsHyperref[eczindexfamilyrel]{code:xyz_product}{XYZ product code} --- The XYZ product code is a non-CSS three-fold variant of the hypergraph product built from three classical linear binary codes \NoCaseChange{\protect\cite{cite645}}.
\end{eczvaluelist}
\codefieldsection{Cousin}
\begin{eczvaluelist}
\item\relax
\flmRefsHyperref[eczindexfamilyrel]{code:homological_classical}{Cycle code} --- Cycle codes have been known in classical coding theory, and have been rediscovered in the quantum context; see Ref. \NoCaseChange{\protect\cite{cite1304}} for a brief exposition.
\end{eczvaluelist}
\eczhbkcontributors{ Nikolas Breuckmann, \eczhuVVA }
\endeczcode

\eczcode{generalized_homological_product_css}{Generalized homological-product CSS code}{}
\codefieldsection{Description}
CSS code whose properties are determined from an underlying chain complex, which often consists of some type of product of other chain complexes.

\codefieldsection{Parents}
\begin{eczvaluelist}
\item\relax
\flmRefsHyperref[eczindexfamilyrel]{code:css}{Calderbank-Shor-Steane (CSS) stabilizer code}\item\relax
\flmRefsHyperref[eczindexfamilyrel]{code:generalized_homological_product}{Generalized homological-product code} --- The notion of homological products arises from interpreting CSS codes in terms of chain complexes over manifolds, but some products no longer yield CSS codes.
\end{eczvaluelist}
\codefieldsection{Children}
\begin{eczvaluelist}
\item\relax
\flmRefsHyperref[eczindexfamilyrel]{code:homological_cv}{Integer-homology bosonic CSS code} --- Integer-homology bosonic CSS codes are constructed from chain complexes over the integers and realize homological rotor codes out of continuous displacement stabilizer groups. The homology group of the logical operators has a torsion component because the chain complexes are defined over the ring of integers, which yields codes with finite logical dimension.
\item\relax
\flmRefsHyperref[eczindexfamilyrel]{code:qubit_generalized_homological_product_css}{Generalized homological-product qubit CSS code}\item\relax
\flmRefsHyperref[eczindexfamilyrel]{code:qudit_3d_surface}{Modular-qudit 3D surface code}\item\relax
\flmRefsHyperref[eczindexfamilyrel]{code:qudit_color}{Modular-qudit lattice color code}\item\relax
\flmRefsHyperref[eczindexfamilyrel]{code:qudit_surface}{Modular-qudit surface code}\item\relax
\flmRefsHyperref[eczindexfamilyrel]{code:balanced_product}{Balanced product (BP) code} --- Balanced product codes result from a tensor product of two classical-code chain complexes, followed by a factoring out of certain symmetries.
\item\relax
\flmRefsHyperref[eczindexfamilyrel]{code:distance_balanced}{Distance-balanced code}\item\relax
\flmRefsHyperref[eczindexfamilyrel]{code:galois_color}{Galois-qudit color code}\item\relax
\flmRefsHyperref[eczindexfamilyrel]{code:galois_topological}{Galois-qudit surface code}\end{eczvaluelist}
\codefieldsection{Cousins}
\begin{eczvaluelist}
\item\relax
\flmRefsHyperref[eczindexfamilyrel]{code:tanner}{Tanner code} --- Tanner codes can be generalized to \textit{sheaf codes}, whose local codes satisfy a certain hierarchy. This allows for a way to formulate and understand many generalized homological-product CSS codes \NoCaseChange{\protect\cite{cite1101}} and LTCs \NoCaseChange{\protect\cite{cite1102}}.
\item\relax
\flmRefsHyperref[eczindexfamilyrel]{code:hnss}{Hayden-Nezami-Salton-Sanders bosonic code} --- Hayden-Nezami-Salton-Sanders codes utilize chain complexes in code construction, but the complexes have trivial homology.
\item\relax
\flmRefsHyperref[eczindexfamilyrel]{code:tiger}{Tiger code} --- Tiger codes are CSS-like multi-mode bosonic non-stabilizer codes constructed from chain complexes over the integers \NoCaseChange{\protect\cite{cite4667}}. The homology group of the logical operators has a torsion component because the chain complexes are defined over the ring of integers, which yields codes with finite logical dimension.
\item\relax
\flmRefsHyperref[eczindexfamilyrel]{code:homological_number-phase}{Homological number-phase code} --- Homological number-phase codes are non-stabilizer codes constructed from chain complexes over the integers. The homology group of the logical operators has a torsion component because the chain complexes are defined over the ring of integers, which yields codes with finite logical dimension.
\end{eczvaluelist}
\eczhbkcontributors{ \eczhuVVA }
\endeczcode

\eczcode{good_qldpc}{Good QLDPC code}{}
\codefieldsection{Description}
Also called \textit{asymptotically good QLDPC codes}. A family of QLDPC codes \(\llbracket n_i,k_i,d_i\rrbracket \) whose asymptotic rate \(\lim_{i\to\infty} k_i/n_i\) and asymptotic distance \(\lim_{i\to\infty} d_i/n_i\) are both positive.

Known constructions of good QLDPC codes can be understood as closely related balanced-product-type constructions on left-right Cayley complexes \NoCaseChange{\protect\cite{cite1501}}.
Three prominent constructions assign qubits and check operators to vertices, edges, and faces of the left-right Cayley complex in different ways:
  \begin{flmFloat}{table}{NumCap}\flmCellsBeginCenter
\long\def\flmTempTypesetThisTable#1{%
\begin{tblr}{#1,
  hspan=minimal,
  cell{1}{1}={}{c, font={\flmCellsHeaderFont}},
  cell{1}{2}={}{c, font={\flmCellsHeaderFont}},
  cell{1}{3}={}{c, font={\flmCellsHeaderFont}},
  cell{1}{4}={}{c, font={\flmCellsHeaderFont}},
  cell{2}{1}={}{c},
  cell{2}{2}={}{c},
  cell{2}{3}={}{c},
  cell{2}{4}={}{c},
  cell{3}{1}={}{c},
  cell{3}{2}={}{c},
  cell{3}{4}={}{c},
  cell{4}{1}={}{c},
  cell{4}{2}={}{c},
  cell{4}{3}={}{c},
  cell{4}{4}={}{c},
  hline{2}={1}{.4pt,solid},
  hline{2}={2}{.4pt,solid},
  hline{2}={3}{.4pt,solid},
  hline{2}={4}{.4pt,solid}}%
\toprule
Code & vertices & edges & faces\\

    \flmRefsHyperref{code:expander_lifted_product}{expander lifted-product} & qubits & \(X,Z\) checks & qubits
        \\

    \flmRefsHyperref{code:quantum_tanner}{quantum Tanner} & \(X,Z\) checks && qubits
        \\

    \flmRefsHyperref{code:dhlv}{Dinur-Hsieh-Lin-Vidick} & \(X\) checks & qubits & \(Z\) checks
    \\
\bottomrule
\end{tblr}%
}%
\def\flmTmpMaxW{\dimexpr 0.96\linewidth\relax}%
\setbox0=\hbox{\flmTempTypesetThisTable{colspec={cccc}}}%
\ifdim\wd0<\flmTmpMaxW\relax
  \leavevmode\box0 
\else
  \flmTempTypesetThisTable{width=\flmTmpMaxW,colspec={X[-1]X[-1]X[-1]X[-1]}}
\fi
\flmCellsEndCenter \caption{Assignment of qubits and checks for three asymptotically good QLDPC codes.}\label{ref5073}\end{flmFloat}
  The left-right Cayley complex can itself be understood as a balanced product of two Cayley graphs, and this viewpoint organizes its relation to balanced-product and quantum Tanner constructions \NoCaseChange{\protect\cite{cite1501}}.
  See \NoCaseChange{\protect\cite[{Fig. 12}]{cite1501}} for more relationships between the constructions.

\codefieldsection{Rate}
The codes' rate and distance are both separated from zero as block length goes to infinity. 
Rains shadow enumerators can be used to show that the distance of an asymptotically good QLDPC code should be bounded as \(d\leq n/3\) \NoCaseChange{\protect\cite{cite2663}}; see Ref. \NoCaseChange{\protect\cite{cite3360}}.
AEL distance amplification \NoCaseChange{\protect\cite{cite493,cite494}} can be used to construct constant-alphabet QLDPC CSS codes of any target rate \(R\) and relative distance \((1-R-\gamma)/2\) that are decodable in linear time up to half that distance \NoCaseChange{\protect\cite[{Corr. 5.3}]{cite495}}.

\codefieldsection{Parent}
\begin{eczvaluelist}
\item\relax
\flmRefsHyperref[eczindexfamilyrel]{code:general_qldpc}{QLDPC code}\end{eczvaluelist}
\codefieldsection{Cousins}
\begin{eczvaluelist}
\item\relax
\flmRefsHyperref[eczindexfamilyrel]{code:translationally_invariant_stabilizer}{Lattice stabilizer code} --- Chain complexes describing some QLDPC codes \NoCaseChange{\protect\cite{cite484,cite485}}, and, more generally, CSS codes \NoCaseChange{\protect\cite{cite486}} can be 'lifted' into higher-dimensional manifolds admitting some notion of geometric locality. Applying this procedure to good QLDPC codes yields \(\llbracket n,n^{1-2/D},n^{1-1/D}\rrbracket \) lattice stabilizer codes in \(D\) spatial dimensions that saturate the \flmRefsHyperref{ref487}{BPT bound} \NoCaseChange{\protect\cite{cite488,cite485,cite489}}.
\item\relax
\flmRefsHyperref[eczindexfamilyrel]{code:translationally_invariant_subsystem}{Lattice subsystem code} --- An \(\llbracket n,k,d\rrbracket \) qubit stabilizer code can be converted into an \flmRefsHyperref{ref65}{order} \(\llbracket O(\ell \delta n),k,\Omega(d/w)\rrbracket \) subsystem qubit stabilizer code with weight-three gauge operators via the wire-code mapping \NoCaseChange{\protect\cite{cite490}}, which uses \flmRefsHyperref{ref491}{weight reduction}. 
Here, \(w\) and \(\delta\) are the weight and degree of the input code's Tanner graph, while \(\ell\) is the length of the longest edge of a particular embedding of that graph.
Applying this procedure to good QLDPC codes and using an embedding into \(D\)-dimensional Euclidean space yields lattice subsystem codes whose logical-qubit number and distance both scale as \(\Theta(n^{1-1/D})\) as functions of block length \(n\), saturating the \flmRefsHyperref{ref492}{subsystem BT bound} \NoCaseChange{\protect\cite{cite490}}.

\item\relax
\flmRefsHyperref[eczindexfamilyrel]{code:quantum_mds}{Quantum maximum-distance-separable (MDS) code} --- AEL distance amplification \NoCaseChange{\protect\cite{cite493,cite494}} can be used to construct asymptotically good QLDPC codes that approach the quantum Singleton bound \NoCaseChange{\protect\cite[{Corr. 5.3}]{cite495}}.
\item\relax
\flmRefsHyperref[eczindexfamilyrel]{code:quantum_singleton}{Singleton-bound approaching AQECC} --- AEL distance amplification \NoCaseChange{\protect\cite{cite493,cite494}} can be used to construct constant-alphabet QLDPC CSS codes of any target rate \(R\) and relative distance \((1-R-\gamma)/2\) that are decodable in linear time up to half that distance \NoCaseChange{\protect\cite[{Corr. 5.3}]{cite495}}. The AEL distance-amplification framework also yields constant-alphabet approximate quantum codes that decode nearly up to the quantum Singleton bound \NoCaseChange{\protect\cite{cite495}}.
\item\relax
\flmRefsHyperref[eczindexfamilyrel]{code:layer}{Layer code} --- Layer codes achieve the 3D \flmRefsHyperref{ref487}{BPT bound}, with parameters  \(\llbracket n,\Theta(n^{1/3}),\Theta(n^{1/3})\rrbracket \), when asymptotically good QLDPC codes are used in the construction.
\item\relax
\flmRefsHyperref[eczindexfamilyrel]{code:lossless_expander}{Lossless expander balanced-product code} --- Taking a balanced product of two-sided expanders \NoCaseChange{\protect\cite{cite187}} yields an asymptotically good QLDPC code family \NoCaseChange{\protect\cite{cite188}}.
\item\relax
\flmRefsHyperref[eczindexfamilyrel]{code:dhlv}{Dinur-Hsieh-Lin-Vidick (DHLV) code} --- DHLV code construction yields asymptotically good QLDPC codes.
\item\relax
\flmRefsHyperref[eczindexfamilyrel]{code:quantum_tanner}{Quantum Tanner code} --- Quantum Tanner code construction yields asymptotically good QLDPC codes.
\item\relax
\flmRefsHyperref[eczindexfamilyrel]{code:expander_lifted_product}{Expander LP code} --- Lifted products of certain classical Tanner codes are the first asymptotically good QLDPC codes.
\end{eczvaluelist}
\eczhbkcontributors{ \eczhuVVA }
\endeczcode

\eczcode{graph_quantum}{Graph quantum code}{~\NoCaseChange{\protect\cite{cite866}}}
\codefieldsection{Description}
A stabilizer code on tensor products of \(G\)-valued qudits for Abelian \(G\) whose encoding isometry is defined using a graph \NoCaseChange{\protect\cite[{Eqs. (4-5)}]{cite866}}.
An analytical form of the codewords exists in terms of the adjacency matrix of the graph and bicharacters of the Abelian group \NoCaseChange{\protect\cite{cite866}}; see \NoCaseChange{\protect\cite[{Eq. (1)}]{cite867}}.
A graph quantum code for \(G=\mathbb{Z}_2\) contains a cluster state as one of its codewords and reduces to a cluster state when its logical dimension is one \NoCaseChange{\protect\cite{cite868}}.

\codefieldsection{Protection}
The \flmTerm{term}{ref1043}{}{Knill-Laflamme conditions} have a graph-based analogue \NoCaseChange{\protect\cite{cite866}}; see Ref. \NoCaseChange{\protect\cite[{Sec. III}]{cite3266}}.

\codefieldsection{Parents}
\begin{eczvaluelist}
\item\relax
\flmRefsHyperref[eczindexfamilyrel]{code:stabilizer}{Stabilizer code} --- Graph quantum codes are a subset of stabilizer codes over \(G\)-valued qudits for Abelian \(G\) \NoCaseChange{\protect\cite{cite3561}}. Any stabilizer code over Abelian \(G\) is locally equivalent to a graph quantum code \NoCaseChange{\protect\cite{cite3561}} (see also \NoCaseChange{\protect\cite{cite3536,cite867}}).
\item\relax
\flmRefsHyperref[eczindexfamilyrel]{code:group_cluster_state}{Group-based cluster-state code} --- Group-based cluster-state codes reduce to graph codes for Abelian \(G\).
\end{eczvaluelist}
\codefieldsection{Children}
\begin{eczvaluelist}
\item\relax
\flmRefsHyperref[eczindexfamilyrel]{code:rotor_cluster}{Rotor cluster-state code} --- Graph quantum codes for \(G=\mathbb{Z}\) reduce to rotor cluster-state codes.
\item\relax
\flmRefsHyperref[eczindexfamilyrel]{code:group_10_1_4}{\(\llbracket 10,1,4\rrbracket _{G}\) tenfold code} --- The \(\llbracket 10,1,4\rrbracket _{G}\) group code is defined using a graph \NoCaseChange{\protect\cite[{Prop. V.1}]{cite866}}.
\item\relax
\flmRefsHyperref[eczindexfamilyrel]{code:cv_cluster_state}{Analog cluster-state code} --- Graph quantum codes for \(G=\mathbb{R}\) reduce to analog cluster-state codes.
\item\relax
\flmRefsHyperref[eczindexfamilyrel]{code:qudit_cluster_state}{Modular-qudit cluster-state code} --- Graph quantum codes for \(G=\mathbb{Z}_q\) reduce to modular-qudit cluster-state codes.
\end{eczvaluelist}
\codefieldsection{Cousins}
\begin{eczvaluelist}
\item\relax
\flmRefsHyperref[eczindexfamilyrel]{code:galois_stabilizer}{Galois-qudit stabilizer code} --- Graph quantum codes for \(G=\mathbb{F}_q\) are a subset of Galois-qudit stabilizer codes \NoCaseChange{\protect\cite{cite3561}}. Any Galois-qudit stabilizer code is equivalent to a graph quantum code for \(G=\mathbb{F}_q\) via a single-Galois-qudit Clifford circuit \NoCaseChange{\protect\cite{cite3561}} (see also \NoCaseChange{\protect\cite{cite3536,cite867}}).
\item\relax
\flmRefsHyperref[eczindexfamilyrel]{code:qudit_5_1_3}{\(\llbracket 5,1,3\rrbracket _{\mathbb{Z}_q}\) modular-qudit code} --- The \(\llbracket 5,1,3\rrbracket _{\mathbb{Z}_q}\) code admits a graph-quantum-code realization for \(G=\mathbb{Z}_q\) \NoCaseChange{\protect\cite{cite866}}.
\item\relax
\flmRefsHyperref[eczindexfamilyrel]{code:galois_cws}{Galois-qudit CWS code} --- A type of Galois-qudit cluster-state code can be built from a Galois-qudit cluster state by applying the conjectural Galois-qudit CWS construction using a linear \(q\)-ary code, in which codewords are obtained by applying Galois-qudit \(Z\)-type operators defined by the code to the Galois-qudit cluster state.
\item\relax
\flmRefsHyperref[eczindexfamilyrel]{code:galois_5_1_3}{\(\llbracket 5,1,3\rrbracket _q\) Galois-qudit code} --- The \(\llbracket 5,1,3\rrbracket _q\) code admits a graph-quantum-code realization for the group \(G=\mathbb{F}_q\) \NoCaseChange{\protect\cite{cite866}}.
\item\relax
\flmRefsHyperref[eczindexfamilyrel]{code:galois_6_2_3}{\(\llbracket 6,2,3\rrbracket _{q}\) code} --- The \(\llbracket 6,2,3\rrbracket _{q}\) code family contains examples of graph quantum codes \NoCaseChange{\protect\cite{cite830}}.
\item\relax
\flmRefsHyperref[eczindexfamilyrel]{code:galois_7_3_3}{\(\llbracket 7,3,3\rrbracket _{q}\) code} --- The \(\llbracket 7,3,3\rrbracket _{q}\) code family contains examples of graph quantum codes \NoCaseChange{\protect\cite{cite830}}.
\end{eczvaluelist}
\eczhbkcontributors{ \eczhuVVA }
\endeczcode

\eczcode{group_gkp}{Group GKP code}{~\NoCaseChange{\protect\cite{cite735}}}
\codefieldsection{Alternative Names}
\begin{eczvaluelist}
\item\relax Non-Abelian CSS code
\item\relax Group coset code
\end{eczvaluelist}
\eczhIndexCodeAliasName{group_gkp}{Non-Abelian CSS code}
\eczhIndexCodeAliasName{group_gkp}{Group coset code}
\codefieldsection{Description}
Group-based quantum code whose construction is based on nested subgroups \(H\subset K
\subset G\). 
The group GKP code was originally formulated as an extension of the \flmRefsHyperref{code:gkp}{GKP code} construction to other group-valued spaces.
In other words, the only requirement to construct group GKP codes is that the configuration space \(G\) is a group under some operation.

Group GKP codes are based on the following decomposition of the underlying group,
\flmMathEnvironment{align}{}{
  G \cong \frac{G}{K} \times \frac{K}{H} \times \widehat{H}~,
}  
where \(\widehat{H}\) is the Fourier space of \(H\) obtained from the Peter-Weyl theorem.
For generalized GKP codes, detectable position shifts are labeled by elements of the coset space \(G/K\), logical position shifts are labeled by elements of \(K/H\) (i.e., cosets of \(H\) in \(K\)), and undetectable logical phase errors are labeled by representations of \(G\) that represent \(H\) trivially but \(K\) nontrivially \NoCaseChange{\protect\cite{cite735}}.

The codes' logical subspace is spanned by basis states that are equal
superpositions of elements of \(K/H\), and can be finite- or
infinite-dimensional depending on the chosen groups.
Codes are stabilized by \(X\)-type \flmRefsHyperref{ref20}{group-based right-multiplication error operators} representing \(H\), all \(Z\)-type operators that are constant on \(K\), and any left-multiplication error operators that are constant on the codespace.
The \(Z\)-type operators are non-unitary matrix product operators for non-Abelian groups.

The construction encompasses all Pauli-type CSS codes since those are Abelian group GKP codes; relevant \(G\) and \(H\) are tabulated in \flmRefsCref{ref5074}.
An \(n\)-qudit Galois CSS code corresponds to the \(\mathbb{F}_q^{k_1} \subseteq \mathbb{F}_q^{k_2} \subset \mathbb{F}_q^{n}\) group construction, where \(k=k_2-k_1\), and where the group operation is addition; this construction should be extendable to additive \(q\)-ary codes since those are also groups under addition.  
An \(\llbracket n,k,d\rrbracket _{\mathbb{R}}\) analog CSS code corresponds to the \(\mathbb{R}^{ k_1} \subseteq \mathbb{R}^{ k_2} \subset \mathbb{R}^{n}\) group construction, where \(k=k_2-k_1\).
A single-mode qubit GKP CSS code corresponds to the \(2\mathbb{Z}\subset\mathbb{Z}\subset\mathbb{R}\) group construction, and multimode GKP CSS codes can be similarly described.
Oscillator-into-oscillator GKP CSS codes for \(n\) modes correspond to subgroups \(\mathbb{Z}^m\) for \(m<n\).
Rotor GKP codes correspond to the \(\mathbb{Z}_{k_1} \subseteq \mathbb{Z}_{k_2} \subset U(1)\) group construction, where \(k=k_2/k_1\).
\begin{flmFloat}{table}{NumCap}\flmCellsBeginCenter
\long\def\flmTempTypesetThisTable#1{%
\begin{tblr}{#1,
  hspan=minimal,
  cell{1}{1}={}{c, font={\flmCellsHeaderFont}},
  cell{1}{2}={}{c, font={\flmCellsHeaderFont}},
  cell{1}{3}={}{c, font={\flmCellsHeaderFont}},
  cell{1}{4}={}{l, font={\flmCellsHeaderFont}},
  cell{2}{1}={}{c},
  cell{2}{2}={}{c},
  cell{2}{3}={}{c},
  cell{2}{4}={}{l},
  cell{3}{1}={}{c},
  cell{3}{2}={}{c},
  cell{3}{3}={}{c},
  cell{3}{4}={}{l},
  cell{4}{1}={}{c},
  cell{4}{2}={}{c},
  cell{4}{3}={}{c},
  cell{4}{4}={}{l},
  cell{5}{1}={}{c},
  cell{5}{2}={}{c},
  cell{5}{3}={}{c},
  cell{5}{4}={}{l},
  cell{6}{1}={}{c},
  cell{6}{2}={}{c},
  cell{6}{3}={}{c},
  cell{6}{4}={}{l},
  cell{7}{1}={}{c},
  cell{7}{2}={}{c},
  cell{7}{3}={}{c},
  cell{7}{4}={}{l},
  cell{8}{1}={}{c},
  cell{8}{2}={}{c},
  cell{8}{3}={}{c},
  cell{8}{4}={}{l},
  cell{9}{1}={}{c},
  cell{9}{2}={}{c},
  cell{9}{3}={}{c},
  cell{9}{4}={}{l},
  cell{10}{1}={}{c},
  cell{10}{2}={}{c},
  cell{10}{3}={}{c},
  cell{10}{4}={}{l},
  cell{11}{1}={}{c},
  cell{11}{2}={}{c},
  cell{11}{3}={}{c},
  cell{11}{4}={}{l},
  hline{2}={1}{.4pt,solid},
  hline{2}={2}{.4pt,solid},
  hline{2}={3}{.4pt,solid},
  hline{2}={4}{.4pt,solid}}%
\toprule
Space & \(G\) & \(K\) & Related code\\

  \(n\) qubits & \(\mathbb{Z}_2^n\) & \(\mathbb{Z}_2^m\)
      & qubit CSS
      \\

  \(n\) modular qudits & \(\mathbb{Z}_q^n\) & \(\mathbb{Z}_q^m\)
      & modular-qudit CSS
      \\

  \(n\) Galois qudits & \(\mathbb{F}_q^n\) & \(\mathbb{F}_q^m\)
      & Galois-qudit CSS
      \\

  \(n\) modes & \( \mathbb{R}^n \) & \( \mathbb{R}^m \)
      & analog CSS
      \\

  \(n\) modes & \( \mathbb{R}^n \) & \( \mathbb{Z}^n \)
      & multimode GKP
      \\

  \(n\) modes & \( \mathbb{R}^n \) & \( \mathbb{Z}^m \)
      & oscillator-into-oscillator GKP
      \\

  \(n\) rotors & \( \mathbb{Z}^n \) & \( \mathbb{Z}^m \)
      & homological rotor
      \\

  rotor & \(U(1)\) & \(\mathbb{Z}_n\)
      & rotor GKP
      \\

  rigid body & \(SO(3)\) & \(K\)
      & molecular
      \\

  1 mode, 1 qudit & \( \mathbb{R} \times \mathbb{Z}_q \) & \( \mathbb{Z} \)
      & simple LCA
  \\
\bottomrule
\end{tblr}%
}%
\def\flmTmpMaxW{\dimexpr 0.96\linewidth\relax}%
\setbox0=\hbox{\flmTempTypesetThisTable{colspec={cccc}}}%
\ifdim\wd0<\flmTmpMaxW\relax
  \leavevmode\box0 
\else
  \flmTempTypesetThisTable{width=\flmTmpMaxW,colspec={X[-1]X[-1]X[-1]X[-1]}}
\fi
\flmCellsEndCenter \caption{
  Special cases of group GKP codes.
  }\label{ref5074}\end{flmFloat}

\codefieldsection{Protection}
Protects against generalized bit-flip errors \(g\in G\) that are inside the fundamental domain of \(G/K\). Protection against phase-flip errors determined by branching rules of irreps of \(G\) into those of \(K\), and further into those of \(H\).
\codefieldsection{Transversal and Permutation-Based Gates}
\begin{eczvaluelist}
\item\relax Group-GKP codes corresponding to the \(G^{k_1} \subseteq G^{k_2} \subset G^{n}\) group construction admit \(X\)-type logical group-multiplication gates, and are thus covariant with respect to the induced \(G^{k_2}\)-action \NoCaseChange{\protect\cite{cite735}}.
\end{eczvaluelist}
\codefieldsection{Realizations}
\begin{eczvaluelist}
\item\relax Cryptographic applications stemming from the monogamy of entanglement of group GKP codes and their error words \NoCaseChange{\protect\cite{cite4705}}.
\end{eczvaluelist}
\codefieldsection{Parent}
\begin{eczvaluelist}
\item\relax
\flmRefsHyperref[eczindexfamilyrel]{code:group_quantum}{Group-based quantum code}\end{eczvaluelist}
\codefieldsection{Children}
\begin{eczvaluelist}
\item\relax
\flmRefsHyperref[eczindexfamilyrel]{code:molecular}{Molecular code}\item\relax
\flmRefsHyperref[eczindexfamilyrel]{code:css}{Calderbank-Shor-Steane (CSS) stabilizer code} --- CSS codes are Abelian group GKP codes, i.e., group GKP codes constructed out of Pauli-type operators.
\end{eczvaluelist}
\codefieldsection{Cousins}
\begin{eczvaluelist}
\item\relax
\flmRefsHyperref[eczindexfamilyrel]{code:covariant}{Covariant block quantum code} --- Group-GKP codes corresponding to the \(G^{k_1} \subseteq G^{k_2} \subset G^{n}\) group construction admit \(X\)-type logical group-multiplication gates, and are thus covariant with respect to the induced \(G^{k_2}\)-action \NoCaseChange{\protect\cite{cite735}}.
\item\relax
\flmRefsHyperref[eczindexfamilyrel]{code:group_linear}{Linear code over \(G\)} --- Group GKP codes are quantum analogues of linear codes over groups.
\item\relax
\flmRefsHyperref[eczindexfamilyrel]{code:lca_stabilizer}{Locally compact Abelian (LCA) stabilizer code} --- Simple single-mode single-qudit LCA codes are Abelian group-GKP codes with \(Kc \mathbb{Z} \subset \mathbb{Z} \subset \mathbb{R} \times \mathbb{Z}_c\), where the logical dimension \(K\) is coprime to the physical qudit dimension \(c\) \NoCaseChange{\protect\cite{cite1428}}.
\end{eczvaluelist}
\eczhbkcontributors{ Alexander Cowtan, \eczhuPhF, \eczhuVVA }
\endeczcode

\eczcode{group_cluster_state}{Group-based cluster-state code}{~\NoCaseChange{\protect\cite{cite5075}}}
\codefieldsection{Description}
A code based on a group-based cluster state for a group \(G\) \NoCaseChange{\protect\cite{cite5075}}.
Such cluster states can be defined using a graph and conditional group multiplication operations.
A group-based cluster state for \(G=\mathbb{F}_q\) for prime-power \(q\) is called a \textit{Galois-qudit cluster state}, while the state for \(G=\mathbb{Z}_q\) for positive \(q\) is called a modular-qudit cluster state.

\codefieldsection{Gates}
\begin{eczvaluelist}
\item\relax 1D group-based cluster states for certain non-Abelian groups \NoCaseChange{\protect\cite{cite5076}} are resources for universal MBQC.
\end{eczvaluelist}
\codefieldsection{Parent}
\begin{eczvaluelist}
\item\relax
\flmRefsHyperref[eczindexfamilyrel]{code:group_quantum}{Group-based quantum code} --- Group-based cluster states are stabilized by \flmRefsHyperref{ref20}{group-based right- and left-multiplication error} operators \NoCaseChange{\protect\cite{cite5075,cite5076}}.
\end{eczvaluelist}
\codefieldsection{Child}
\begin{eczvaluelist}
\item\relax
\flmRefsHyperref[eczindexfamilyrel]{code:graph_quantum}{Graph quantum code} --- Group-based cluster-state codes reduce to graph codes for Abelian \(G\).
\end{eczvaluelist}
\codefieldsection{Cousin}
\begin{eczvaluelist}
\item\relax
\flmRefsHyperref[eczindexfamilyrel]{code:hopf_cluster_state}{Hopf-algebra cluster-state code} --- Hopf-algebra cluster-state codes reduce to group-based cluster-state codes for finite groups when the Hopf algebra reduces to a finite group.
\end{eczvaluelist}
\eczhbkcontributors{ \eczhuVVA }
\endeczcode

\eczcode{group_quantum_parity}{Group-based QPC}{~\NoCaseChange{\protect\cite{cite2720}}}
\codefieldsection{Description}
An \(\llbracket m r,1,\min(m,r)\rrbracket _G\) generalization of the QPC.

Logical codewords for each group element \(g\) are
\flmMathEnvironment{align}{}{
  |\overline{g}\rangle=\left({\textstyle \frac{1}{\sqrt{|G|^{m-1}}}}\sum_{h_{1},h_{2},\cdots,h_{m}\in G}\delta^{G}_{g,h_{1}h_{2}\cdots h_{m}}|h_{1},h_{2},\cdots,h_{m}\rangle\right)^{\otimes r}~.
}
where \(\delta^{G}_{g,h}\) is the \flmRefsHyperref{ref20}{group Kronecker-delta function}.
For non-compact groups, the sum becomes an integral, and ideal codewords are no longer normalizable.

\codefieldsection{Parents}
\begin{eczvaluelist}
\item\relax
\flmRefsHyperref[eczindexfamilyrel]{code:group_quantum}{Group-based quantum code}\item\relax
\flmRefsHyperref[eczindexfamilyrel]{code:quantum_concatenated}{Concatenated quantum code} --- A group-based QPC is a concatenation of a phase-flip group-based repetition code with a bit-flip group-based repetition code.
\end{eczvaluelist}
\codefieldsection{Children}
\begin{eczvaluelist}
\item\relax
\flmRefsHyperref[eczindexfamilyrel]{code:group_quantum_repetition}{Group-based quantum repetition code} --- A \(\llbracket m_1 m_2,1,\min(m_1,m_2)\rrbracket _G\) group-based QPC reduces to a group-based quantum repetition code when \(m_1\) or \(m_2\) is one.
\item\relax
\flmRefsHyperref[eczindexfamilyrel]{code:lloyd_slotine}{\(\llbracket 9,1,3\rrbracket _{\mathbb{R}}\) Lloyd-Slotine code} --- The \(\llbracket 9,1,3\rrbracket _{G}\) group-based QPC reduces to the \(\llbracket 9,1,3\rrbracket _{\mathbb{R}}\) Lloyd-Slotine code for \(G=\mathbb{R}\).
\item\relax
\flmRefsHyperref[eczindexfamilyrel]{code:quantum_parity}{Quantum parity code (QPC)} --- A \(\llbracket m_1 m_2,1,\min(m_1,m_2)\rrbracket _G\) group-based QPC reduces to a QPC for \(G=\mathbb{Z}_2\).
\item\relax
\flmRefsHyperref[eczindexfamilyrel]{code:stab_9_1_3}{\(\llbracket 9,1,3\rrbracket _{\mathbb{Z}_q}\) modular-qudit code} --- The \(\llbracket 9,1,3\rrbracket _{G}\) group-based QPC reduces to the \(\llbracket 9,1,3\rrbracket _{\mathbb{Z}_q}\) modular-qudit code for \(G=\mathbb{Z}_q\).
\end{eczvaluelist}
\codefieldsection{Cousin}
\begin{eczvaluelist}
\item\relax
\flmRefsHyperref[eczindexfamilyrel]{code:group_4_2_2}{\(\llbracket 4,2,2\rrbracket _{G}\) four group-qudit code} --- The \(|\overline{g_1=1,g_2}\rangle\) \(\llbracket 4,1,2\rrbracket _{G}\) subcode is the smallest group-based QPC, i.e., a concatenation of a bit-flip with a phase-flip group-based repetition code for that group.
\end{eczvaluelist}
\eczhbkcontributors{ \eczhuVVA }
\endeczcode

\eczcode{group_quantum}{Group-based quantum code}{}

\codefieldsection{Kingdom root code}
for the \flmRefsHyperref{kingdom:group_quantum}{Group quantum Kingdom}.
\codefieldsection{Description}
Encodes a \textit{logical} Hilbert space, finite- or infinite-dimensional, into a \textit{physical} Hilbert space of \(L^2\)-normalizable functions on a second-countable unimodular group \(G\), i.e., a \(G\)\textit{-valued qudit} or \(G\)-qudit.
In other words, a group-valued qudit is a vector space whose canonical basis states \(|g\rangle\) are labeled by elements \(g\) of a group \(G\).
For \(K\)-dimensional logical subspace and for block codes defined on groups \(G^{n}\), can be denoted as \(\llparenthesis n,K\rrparenthesis _G\).
When the logical subspace is the Hilbert space of \(L^2\)-normalizable functions on \(G^{ k}\), can be denoted as \(\llbracket n,k\rrbracket _G\).
Ideal codewords may not be normalizable, depending on whether \(G\) is continuous and/or noncompact, so approximate versions have to be constructed in practice.

A notion of Gaussian states and Hudson's theorem have been developed for arbitrary locally compact Abelian \(G\) \NoCaseChange{\protect\cite{cite5077}}.
A Wigner function formalism has also been developed \NoCaseChange{\protect\cite{cite3662}}.

\codefieldsection{Protection}
\subsection{Group-based error basis}
A convenient error set is the group-based analogue of the Pauli string set for \flmRefsHyperref{code:qubits_into_qubits}{qubit} codes.
For a single group-valued qudit, this set consists of products of \(X\)-type operators labeled by group elements \(g\), and \(Z\)-type operators labeled by matrix elements of \(G\)-irreps \(\lambda\) \NoCaseChange{\protect\cite{cite5075,cite735,cite598}}.
The outline below is for finite groups, but can be extended to compact unimodular groups or to oscillators and rotors by substituting the sum over the group with a group integral.

\begin{defterm}{Group-based error basis}\label{ref5078}\label{ref20}
There are two types of \(X\)-type operators, corresponding to left and right group multiplication.
These act on computational basis states \(|h\rangle\) as
\flmMathEnvironment{align}{}{
  \overrightarrow{X}_{g}|h\rangle&=|gh\rangle\\
  \overleftarrow{X}_{g}|h\rangle&=|hg^{-1}\rangle
}
for any group elements \(h,g\).
The \(Z\)-type operators can be thought of as matrix-product operators (MPOs) \NoCaseChange{\protect\cite{cite5076}} whose virtual dimension is the dimension \(d_{\lambda}\) of their corresponding irrep.
They are diagonal in the group-valued basis, yielding the \(d_{\lambda}\)-dimensional irrep matrix \(Z_{\lambda}(g)\) evaluated at the given group element,
\flmMathEnvironment{align}{}{
  \hat{Z}_{\lambda}\otimes|g\rangle=Z_{\lambda}(g)\otimes|g\rangle~.
}
Each matrix element of this irrep matrix is a generally non-unitary operator on the group-valued qudit.
For 1D irreps, the matrix reduces to a single unitary \(Z\)-type operator, and the direct-product symbol is no longer needed.
For special cases of Abelian \(G\) being \(\mathbb{Z}_2\), \(\mathbb{Z}_q\), \(\mathbb{F}_q\), \(U(1)\cong \mathbb{Z}\), or \(\mathbb{R}\), the group-based error basis reduces to the familiar \flmRefsHyperref{ref663}{qubit Pauli}, \flmRefsHyperref{ref2198}{qudit Pauli}, \flmRefsHyperref{ref4618}{Galois-qudit Pauli}, \flmRefsHyperref{ref5079}{rotor generalized Pauli}, or \flmRefsHyperref{ref4745}{oscillator displacement} error basis, respectively.
\end{defterm}

Products of either left- or right-multiplication \(X\)-type operators with all \(Z\)-type operators form a basis for linear operators on the group-valued qudit space that is complete and orthonormal under the Hilbert-Schmidt inner product \NoCaseChange{\protect\cite[{Eq. (123)}]{cite735}}.
In particular,
\flmMathEnvironment{align}{}{
  \text{tr}(\overrightarrow{X}_{g}^{\dagger}\overrightarrow{X}_{h})=\delta_{g,h}^{G}~,
}
where the group Kronecker delta function \(\delta^{G}_{g,h}=1\) if \(g=h\) and zero otherwise.

\codefieldsection{Gates}
\begin{eczvaluelist}
\item\relax Various gates for a single \(G\)-valued qudit include the Fourier gate (which is a re-expression of the group basis in terms of \(G\)-irreps), the \(g \to g^{-1}\) inversion gate, conditional multiplication gates, and conditional multiplication gates in the irrep basis \NoCaseChange{\protect\cite{cite5075,cite735,cite5080,cite5081,cite5082}}.
\item\relax The \flmTerm{term}{ref694}{}{Clifford hierarchy} can be extended to arbitrary Abelian \(G\) \NoCaseChange{\protect\cite{cite3219,cite673,cite3066}}.
\end{eczvaluelist}
\codefieldsection{Notes}
\begin{eczvaluelist}
\item\relax See Refs. \NoCaseChange{\protect\cite{cite5083,cite2531}} for introductions to Hilbert spaces for Abelian groups.
\item\relax Group-based \(Z\)-type operators correspond to group-valued fields in the continuum limit \NoCaseChange{\protect\cite{cite598}} and Wilson link operators in lattice gauge theory \NoCaseChange{\protect\cite{cite468}}.
\end{eczvaluelist}
\codefieldsection{Parents}
\begin{eczvaluelist}
\item\relax
\flmRefsHyperref[eczindexfamilyrel]{code:homogeneous_space_quantum}{Homogeneous-space quantum code} --- Homogeneous spaces \(G/H\) for trivial \(H\) reduce to group spaces. A group-\(G\) space can also be thought of as a multiplicity-free homogeneous space \((G\times G) / G\) \NoCaseChange{\protect\cite[{pg. 60}]{cite2474}}.
\item\relax
\flmRefsHyperref[eczindexfamilyrel]{code:category_quantum}{Category-based quantum code} --- Finite-group-based quantum codes, whose basis states are parameterized by a finite group, correspond to category-based codes for the fusion category \(Vec G\). Extensions of such categories to Lie groups can also be done \NoCaseChange{\protect\cite{cite5084,cite5085,cite5086,cite5087}} (see also \NoCaseChange{\protect\cite{cite5088}}).
\end{eczvaluelist}
\codefieldsection{Children}
\begin{eczvaluelist}
\item\relax
\flmRefsHyperref[eczindexfamilyrel]{code:hybrid_qudit_oscillator}{Mixed oscillator code} --- Group quantum codes whose physical spaces are constructed using some number of modular qudits, some number of rotors, and a nonzero number of oscillators, which together constitute a general locally compact Abelian (LCA) group, are mixed oscillator codes.
\item\relax
\flmRefsHyperref[eczindexfamilyrel]{code:generalized_color}{Generalized 2D color code}\item\relax
\flmRefsHyperref[eczindexfamilyrel]{code:dijkgraaf_witten}{Dijkgraaf-Witten gauge theory code}\item\relax
\flmRefsHyperref[eczindexfamilyrel]{code:group_cluster_state}{Group-based cluster-state code} --- Group-based cluster states are stabilized by \flmRefsHyperref{ref20}{group-based right- and left-multiplication error} operators \NoCaseChange{\protect\cite{cite5075,cite5076}}.
\item\relax
\flmRefsHyperref[eczindexfamilyrel]{code:group_gkp}{Group GKP code}\item\relax
\flmRefsHyperref[eczindexfamilyrel]{code:group_quantum_parity}{Group-based QPC}\item\relax
\flmRefsHyperref[eczindexfamilyrel]{code:rotor}{Rotor code} --- Group quantum codes whose physical spaces are constructed using either the group of the integers \(\mathbb{Z}\) or the circle group \(U(1)\) are rotor codes.
\item\relax
\flmRefsHyperref[eczindexfamilyrel]{code:stabilizer}{Stabilizer code} --- Stabilizer codes are constructed out of Pauli strings, modular-qudit Pauli strings, Galois-qudit Pauli strings, oscillator displacement operators, or rotor generalized Pauli strings. All of these are examples of the Weyl-Heisenberg group on Manin's quantum plane, which is defined on a configuration space that is generally a free Abelian group \NoCaseChange{\protect\cite{cite4537,cite4538,cite4539,cite4540}}.
\item\relax
\flmRefsHyperref[eczindexfamilyrel]{code:qudits_into_qudits}{Modular-qudit code} --- Group quantum codes whose physical spaces are constructed using modular-integer groups \(\mathbb{Z}_q\) are modular-qudit codes.
\item\relax
\flmRefsHyperref[eczindexfamilyrel]{code:galois_into_galois}{Galois-qudit code} --- A Galois qudit for \(q=p^m\) can be decomposed into a Kronecker product of \(m\) modular qudits \NoCaseChange{\protect\cite{cite696}}; see \NoCaseChange{\protect\cite[{Sec. 5.3}]{cite697}}.
Interpreted this way, Galois-qudit codes are group quantum codes whose physical spaces are constructed using Galois fields \(\mathbb{F}_q\) as groups. More general versions of such qudits can be valued in a Galois ring \NoCaseChange{\protect\cite{cite4621}}, over which there also exists a Fourier transform \NoCaseChange{\protect\cite{cite4622}}.

\end{eczvaluelist}
\codefieldsection{Cousins}
\begin{eczvaluelist}
\item\relax
\flmRefsHyperref[eczindexfamilyrel]{code:group_classical}{Group-alphabet code} --- Group-based quantum codes are quantum counterparts of group-alphabet codes.
\item\relax
\flmRefsHyperref[eczindexfamilyrel]{code:subsystem_group_quantum}{Subsystem group-based quantum code} --- Subsystem group-based quantum codes reduce to (subspace) group-based quantum codes when there is no gauge subsystem.
\end{eczvaluelist}
\eczhbkcontributors{ David Aasen, \eczhuVVA }
\endeczcode

\eczcode{group_quantum_repetition}{Group-based quantum repetition code}{~\NoCaseChange{\protect\cite{cite2720}}}
\codefieldsection{Description}
An \(\llbracket n,1\rrbracket _G\) generalization of the quantum repetition code.

The code encodes one group-valued qudit into \(n\).
There are two variants, a bit- and a phase-flip code, whose codewords for any \(g\in G\) and for \(n=3\) are
\flmMathEnvironment{align}{}{
  |\overline{g}_{\text{bit}}\rangle&\rightarrow|g,g,g\rangle\\
  |\overline{g}_{\text{phase}}\rangle&\rightarrow\frac{1}{|G|}\sum_{h_{1},h_{2},h_{3}\in G}\delta^{G}_{g,h_{1}h_{2}h_{3}}|h_{1},h_{2},h_{3}\rangle~,
}
where \(\delta^{G}_{g,h}\) is the \flmRefsHyperref{ref20}{group Kronecker-delta function}.
For non-compact groups, the sum becomes an integral, and ideal codewords are no longer normalizable.

\codefieldsection{Parents}
\begin{eczvaluelist}
\item\relax
\flmRefsHyperref[eczindexfamilyrel]{code:group_quantum_parity}{Group-based QPC} --- A \(\llbracket m_1 m_2,1,\min(m_1,m_2)\rrbracket _G\) group-based QPC reduces to a group-based quantum repetition code when \(m_1\) or \(m_2\) is one.
\item\relax
\flmRefsHyperref[eczindexfamilyrel]{code:quantum_cyclic}{Cyclic quantum code}\end{eczvaluelist}
\codefieldsection{Children}
\begin{eczvaluelist}
\item\relax
\flmRefsHyperref[eczindexfamilyrel]{code:analog_repetition}{Analog repetition code} --- Group-based quantum repetition codes reduce to analog repetition codes for \(G = \mathbb{R}\).
\item\relax
\flmRefsHyperref[eczindexfamilyrel]{code:quantum_repetition}{Quantum repetition code} --- Group-based quantum repetition codes reduce to quantum repetition codes for \(G = \mathbb{Z}_2\).
\end{eczvaluelist}
\codefieldsection{Cousin}
\begin{eczvaluelist}
\item\relax
\flmRefsHyperref[eczindexfamilyrel]{code:stab_9_1_3}{\(\llbracket 9,1,3\rrbracket _{\mathbb{Z}_q}\) modular-qudit code} --- The \(\llbracket 9,1,3\rrbracket _{\mathbb{Z}_q}\) modular-qudit code is a concatenation of a bit-flip with a phase-flip group repetition code for \(G=\mathbb{Z}_q\).
\end{eczvaluelist}
\eczhbkcontributors{ \eczhuVVA }
\endeczcode

\eczcode{homological_rotor}{Homological rotor code}{~\NoCaseChange{\protect\cite{cite397}}}
\codefieldsection{Description}
A CSS rotor code stabilized by a group of rotor \(X\)-type and \(Z\)-type generalized Pauli operators.
Codes are formulated using an extension of the \flmRefsHyperref{ref683}{qubit CSS-to-homology correspondence} to rotors.
The homology group of the logical operators has a torsion component because the chain complexes are defined over the ring of integers, which yields codes with finite logical dimension, i.e., encoding logical qudits instead of only logical rotors.

A homological rotor code encoding \(k\) logical rotors and a \(q\)-dimensional logical qudit is denoted as \(\llbracket n,(k,q)\rrbracket _{\mathbb{Z}}\) or \(\llbracket n,(k,q),(d_X,\delta_Z)\rrbracket _{\mathbb{Z}}\), where \(d_X\) and \(\delta_Z\) are the code's \(X\) and \(Z\) distances, respectively.
The subscript \(\mathbb{Z}\) refers to the label used for the rotor's angular momentum, but shifts in the dual angular position degree of freedom are also used to construct stabilizers (the alternative subscript \(U(1)\) is used in some cases).

The stabilizer group is defined using two integer matrices \(H_X\in\mathbb{Z}^{r_X\times n}\) and \(H_Z\in\mathbb{Z}^{r_Z\times n}\) which are such that
\flmMathEnvironment{align}{}{
      H_XH_Z^T = 0.\label{ref5089}
}
The stabilizer is then defined as
\flmMathEnvironment{align}{}{
  \mathsf{S}=\left\langle e^{-i\boldsymbol{s}H_{X}\cdot\hat{\boldsymbol{L}}}e^{i\boldsymbol{\varphi}H_{Z}\cdot\hat{\boldsymbol{\phi}}}\middle\vert\forall\boldsymbol{s}\in\mathbb{Z}^{r_{X}},\forall\boldsymbol{\varphi}\in U(1)^{r_{Z}}\right\rangle .\label{ref5090}
}
The condition \eqref{ref5089} ensures that \(\mathsf{S}\) has a common +1 eigenspace.

As with \flmRefsHyperref{code:css}{CSS} codes, there is a natural connection to a length-3 integer chain complex,
\flmMathEnvironment{align}{}{
  \mathcal{A}:~\mathbb{Z}^{r_X} \xrightarrow{H_X} \mathbb{Z}^n \xrightarrow{H_Z^T} \mathbb{Z}^{r_Z}~,
}
whose middle homology group describes the logical \(X\) operators of the code.
The logical \(Z\) operators are defined by the middle cohomology group where the cohomology is taken with phase coefficients, \(\mathbb{T} = \mathbb{R}/2\pi\mathbb{Z}\),
\flmMathEnvironment{align}{}{
  \mathcal{A}^*:~\mathbb{T}^{r_X} \xleftarrow{H_X^T} \mathbb{T}^n \xleftarrow{H_Z} \mathbb{T}^{r_Z}.
}

The logical subspace can contain logical rotors as well as logical qudits.
The former correspond to the so-called free part of the homology group while the latter correspond to the torsion part,
\flmMathEnvironment{align}{}{
  H_1(\mathcal{A},\mathbb{Z}) = \mathbb{Z}^{k^\prime}\oplus\mathbb{Z}_{d_1}\oplus\cdots\oplus\mathbb{Z}_{d_{k^{\prime\prime}}}.
}
Stabilizer generator matrices equivalent under CSS rotor Clifford group transformations are classified by distinct Smith normal forms \NoCaseChange{\protect\cite{cite397,cite2699}}.

\codefieldsection{Protection}
The \(X\) distance \(d_X\) of a code is the (slightly generalized notion of) weight of the smallest logical operator constructed out of angular position shifts.
The \(Z\) distance \(\delta_Z\) depends on whether or not the code encodes logical rotors, but a similar notion exists in the case of only a logical qudit encoding.
One can extend the idea of disjointness \NoCaseChange{\protect\cite{cite775}} to rotors to obtain distance bounds \NoCaseChange{\protect\cite{cite397}}.

\codefieldsection{Transversal and Permutation-Based Gates}
\begin{eczvaluelist}
\item\relax All generalized Pauli gates are realized transversally.
\end{eczvaluelist}
\codefieldsection{Gates}
\begin{eczvaluelist}
\item\relax Some logical gates come from the rotor Clifford group \NoCaseChange{\protect\cite{cite2699}}.
\end{eczvaluelist}
\codefieldsection{Notes}
\begin{eczvaluelist}
\item\relax A \flmHref{https://github.com/cianibegood/quantum-rotor-codes}{Sage notebook} of small examples.
\end{eczvaluelist}
\codefieldsection{Parents}
\begin{eczvaluelist}
\item\relax
\flmRefsHyperref[eczindexfamilyrel]{code:rotor_stabilizer}{Rotor stabilizer code} --- Homological rotor codes are rotor CSS codes constructed from chain complexes over the integers in an extension of the \flmRefsHyperref{ref683}{qubit CSS-to-homology correspondence} to rotors.
\item\relax
\flmRefsHyperref[eczindexfamilyrel]{code:css}{Calderbank-Shor-Steane (CSS) stabilizer code} --- Homological rotor codes are rotor CSS codes constructed from chain complexes over the integers in an extension of the \flmRefsHyperref{ref683}{qubit CSS-to-homology correspondence} to rotors. The homology group of the logical operators has a torsion component because the chain complexes are defined over the ring of integers, which yields codes with finite logical dimension. Products of chain complexes can also yield rotor codes.
\end{eczvaluelist}
\codefieldsection{Children}
\begin{eczvaluelist}
\item\relax
\flmRefsHyperref[eczindexfamilyrel]{code:current_mirror}{Kitaev current-mirror qubit code}\item\relax
\flmRefsHyperref[eczindexfamilyrel]{code:rotor_3_1_2}{\(\llbracket 3,1,2\rrbracket _{\mathbb{Z}}\) Three-rotor code} --- Taking \(H_X=\begin{pmatrix}-3 & 1 & 2\end{pmatrix}\) and \(H_Z=\begin{pmatrix}4&6&3\end{pmatrix}\) yields the three-rotor code.
\item\relax
\flmRefsHyperref[eczindexfamilyrel]{code:rotor_4_2_2}{Four-rotor code}\item\relax
\flmRefsHyperref[eczindexfamilyrel]{code:zero_pi}{Zero-pi qubit code}\end{eczvaluelist}
\codefieldsection{Cousins}
\begin{eczvaluelist}
\item\relax
\flmRefsHyperref[eczindexfamilyrel]{code:homological_cv}{Integer-homology bosonic CSS code} --- Integer-homology bosonic CSS codes are constructed from chain complexes over the integers and realize homological rotor codes out of continuous displacement stabilizer groups \NoCaseChange{\protect\cite{cite411}}.
\item\relax
\flmRefsHyperref[eczindexfamilyrel]{code:homological_number-phase}{Homological number-phase code} --- Homological number-phase codes can be thought of as homological rotor codes but whose underlying rotors consist of the number and phase degrees of freedom of physical modes.
\item\relax
\flmRefsHyperref[eczindexfamilyrel]{code:qudit_cubic}{Qudit cubic code} --- The qudit cubic code can be generalized to rotors \NoCaseChange{\protect\cite{cite4590,cite2531}}.
\end{eczvaluelist}
\eczhbkcontributors{ Christophe Vuillot, \eczhuVVA }
\endeczcode

\eczcode{hybrid_cat}{Hybrid cat code}{~\NoCaseChange{\protect\cite{cite5091,cite5092,cite5093,cite4401}}}
\codefieldsection{Description}
A mixed oscillator code admitting codewords that are tensor products of a single-qubit (e.g., photon polarization) state with either a cat state or a coherent state.

Codewords of the coherent-state version \NoCaseChange{\protect\cite{cite5091}} are \(|\alpha\rangle|+\rangle\) and \(|-\alpha\rangle|-\rangle\), i.e., hyper-entangled states of the occupation-number and polarization degrees of freedom of a photon.
Codewords of the cat-state version \NoCaseChange{\protect\cite{cite5094,cite4401}} are proportional to \((\left|\alpha\right\rangle +\left|-\alpha\right\rangle )|+\rangle\) and \((\left|i\alpha\right\rangle -\left|-i\alpha\right\rangle )|-\rangle\).

\codefieldsection{Fault Tolerance}
\begin{eczvaluelist}
\item\relax Photonic architecture based on concatenation with RBH codes \NoCaseChange{\protect\cite{cite4401}}.
\end{eczvaluelist}
\codefieldsection{Notes}
\begin{eczvaluelist}
\item\relax See reviews \NoCaseChange{\protect\cite{cite5095,cite5096}} for introductions to mixed oscillator platforms.
\end{eczvaluelist}
\codefieldsection{Parent}
\begin{eczvaluelist}
\item\relax
\flmRefsHyperref[eczindexfamilyrel]{code:hybrid_qudit_oscillator}{Mixed oscillator code}\end{eczvaluelist}
\codefieldsection{Cousins}
\begin{eczvaluelist}
\item\relax
\flmRefsHyperref[eczindexfamilyrel]{code:cat}{Cat code} --- Hybrid cat codewords consist of a bosonic mode in either coherent or cat states.
\item\relax
\flmRefsHyperref[eczindexfamilyrel]{code:rbh}{Raussendorf-Bravyi-Harrington (RBH) cluster-state code} --- Hybrid cat codes can be concatenated with RBH codes \NoCaseChange{\protect\cite{cite4401}}.
\item\relax
\flmRefsHyperref[eczindexfamilyrel]{code:oscillators_concatenated}{Concatenated bosonic code} --- Hybrid cat codes can be concatenated with RBH codes \NoCaseChange{\protect\cite{cite4401}}.
\end{eczvaluelist}
\eczhbkcontributors{ \eczhuVVA }
\endeczcode

\eczcode{current_mirror}{Kitaev current-mirror qubit code}{~\NoCaseChange{\protect\cite{cite4971,cite5097,cite397}}}
\codefieldsection{Description}
Member of the family of \(\llbracket 2n,(0,2),(2,n)\rrbracket _{\mathbb{Z}}\) homological rotor codes storing a logical qubit on a thin Möbius strip.
The ideal code can be obtained from a Josephson-junction \NoCaseChange{\protect\cite{cite396}} system \NoCaseChange{\protect\cite{cite397}}.

Logical codewords can be expressed in the basis of angular momentum states as
\flmMathEnvironment{align}{}{
\begin{split}
  |\overline{0}\rangle&=\sum_{\overset{\ell_{1},\dots,\ell_{n}\in\mathbb{Z}}{\sum_{k=1}^{n}\ell_{k}=\mathrm{even}}}\left|\ell_{1},\dots,\ell_{n},-\ell_{1},\dots,-\ell_{n}\right\rangle \\|\overline{1}\rangle&=\sum_{\overset{\ell_{1},\dots,\ell_{n}\in\mathbb{Z}}{\sum_{k=1}^{n}\ell_{k}=\mathrm{odd}}}\left|\ell_{1},\dots,\ell_{n},-\ell_{1},\dots,-\ell_{n}\right\rangle~.
\end{split}
}

\codefieldsection{Protection}
Protection in the context of superconducting circuits investigated in Ref. \NoCaseChange{\protect\cite{cite5098}}.

\codefieldsection{Gates}
\begin{eczvaluelist}
\item\relax One- and two-qubit phase gates utilizing ancillary oscillators in GKP states \NoCaseChange{\protect\cite{cite4971}}.
\end{eczvaluelist}
\codefieldsection{Parents}
\begin{eczvaluelist}
\item\relax
\flmRefsHyperref[eczindexfamilyrel]{code:homological_rotor}{Homological rotor code}\item\relax
\flmRefsHyperref[eczindexfamilyrel]{code:1d_stabilizer}{1D lattice stabilizer code}\item\relax
\flmRefsHyperref[eczindexfamilyrel]{code:small_distance_quantum}{Small-distance block quantum code}\end{eczvaluelist}
\codefieldsection{Cousin}
\begin{eczvaluelist}
\item\relax
\flmRefsHyperref[eczindexfamilyrel]{code:gkp}{Square-lattice GKP code} --- Current-mirror code phase gates utilize ancillary oscillators in square-lattice GKP states \NoCaseChange{\protect\cite{cite4971,cite4972}}.
\end{eczvaluelist}
\eczhbkcontributors{ \eczhuVVA }
\endeczcode

\eczcode{translationally_invariant_stabilizer}{Lattice stabilizer code}{~\NoCaseChange{\protect\cite{cite3032,cite3456,cite5099,cite411}}}
\codefieldsection{Alternative Names}
\begin{eczvaluelist}
\item\relax Topological stabilizer code
\end{eczvaluelist}
\eczhIndexCodeAliasName{translationally_invariant_stabilizer}{Topological stabilizer code}
\codefieldsection{Description}
A geometrically local stabilizer code with sites organized on a lattice modeled by the additive group \(\mathbb{Z}^D\) for spatial dimension \(D\), using either the ordinary block notion of locality or the fermionic/Majorana notion of locality.
On an infinite lattice, its stabilizer group is generated by few-site Pauli-type operators and their translations, in which case the code is called \textit{translationally invariant stabilizer code}.
Boundary conditions have to be imposed on the lattice in order to obtain finite-dimensional versions.
Lattice defects and boundaries between different codes can also be introduced.

\subsection{Modular- and Galois-qudit lattice stabilizer codes}

Translationally-invariant prime-qudit (\(q=p\)) stabilizer codes with \(m\) qudits per unit cell have been classified in dimensions \(D\in\{1,2\}\) in the thermodynamic limit, up to equivalence under local constant-depth Clifford circuits.
Any 1D (2D) code can be converted to several copies of the 1D repetition code (prime-qudit 2D surface code) along with some trivial codes \NoCaseChange{\protect\cite{cite3963,cite4513}}.
See 3D lattice stabilizer code entry for the 3D classification.

\begin{defterm}{Pauli-to-polynomial mapping}\label{ref5100}\label{ref4121}
A single modular- or Galois-qudit Pauli operator can be specified by the lattice coordinate of the site and the symplectic vector
representation of the Pauli operator within the site.
In an extension of the symplectic representation, each lattice coordinate can be represented by a Laurent monomial of \(D\) formal variables. For example, when \(D=2\) and \(m=1\), the product of an \(X\) acting on the qubit at lattice coordinate \((-1,2)\) and a \(Z\) acting on the qubit at \((1,0)\) can be represented by the vector \( (x^{-1} y^2 | x) \). The multiplicative group of finitely supported Pauli operators modulo phase factors on the lattice of dimension \(D\) with \(m\) prime-dimensional qudits per site is isomorphic to the additive group of Laurent polynomial column vectors of length \(2m\) in \(D\) formal variables (see \NoCaseChange{\protect\cite{cite3963}\protect\cite[{Sec. IV}]{cite3068}}).

For periodic boundary conditions, this mapping can be thought of as a quantum extension of the \flmRefsHyperref{ref67}{cyclic-to-polynomial correspondence}.
For open boundary conditions, this mapping extends the mapping used in quantum convolutional codes to multiple spatial dimensions.
\end{defterm}

\subsection{Bosonic lattice stabilizer codes}

Bosonic lattice stabilizer codes can contain discrete or continuous subgroups and can admit logical qudit and/or oscillator logical subspaces.
Such codes can realize topological phases of matter that are expected not to be realizable with qudit stabilizer codes \NoCaseChange{\protect\cite{cite411}}. 

\codefieldsection{Protection}
A quantity called the recoverable information, defined for lattice stabilizer codes, is a measure of the non-locality of the encoding that is complementary to the topological entanglement entropy \NoCaseChange{\protect\cite{cite5101}}.

\codefieldsection{Rate}
\begin{defterm}{BPT bound}\label{ref3768}\label{ref487}
Qubit stabilizer code parameters on \(D\)-dimensional Euclidean lattices are limited by the \textit{Bravyi-Poulin-Terhal (BPT) bound} \NoCaseChange{\protect\cite{cite1419}} (see also \NoCaseChange{\protect\cite{cite3000,cite3328,cite2567}}), which states that \(d = O(n^{1-1/D})\) (the original \textit{Bravyi-Terhal (BT) bound} \NoCaseChange{\protect\cite{cite3000}}) and that \(k d^{2/(D-1)} = O(n)\) (using \flmRefsHyperref{ref65}{asymptotic notation}).
Codes on a \(D\)-dimensional homogeneous Riemannian manifold with diameter \(L\) satisfy \(k = O(L^{D-2})\) \NoCaseChange{\protect\cite{cite3141}}.
Some non-locality is necessary to circumvent these bounds \NoCaseChange{\protect\cite{cite5102}}.
The BPT bound does not apply to stabilizer circuits: a modified tower of interleaved quantum Hamming codes yields a 1D nearest-neighbor scheme with coding rate above \(5\%\), constant space overhead, quasi-polylogarithmic time overhead, and a threshold \NoCaseChange{\protect\cite{cite3217}}.
Codes for \(D < 2\) have constant distance \NoCaseChange{\protect\cite{cite5103}}.
\end{defterm}

\codefieldsection{Transversal and Permutation-Based Gates}
\begin{eczvaluelist}
\item\relax Fold-transversal logical gates can be obtained from crystalline symmetries \NoCaseChange{\protect\cite{cite749}}.
\end{eczvaluelist}
\codefieldsection{Gates}
\begin{eczvaluelist}
\item\relax \begin{defterm}{Bravyi-Koenig bound}\label{ref5104}\label{ref3630} Logical gates implemented via constant-depth quantum circuits on a \(D\)-dimensional lattice stabilizer code whose distance increases at least logarithmically with \(n\) lie in the \(D\)th level of the \flmTerm{term}{ref694}{}{Clifford hierarchy} \NoCaseChange{\protect\cite{cite5105}}. A refinement can be made that expresses the bound in terms of higher-group symmetries of the topological phases underlying the codes \NoCaseChange{\protect\cite[{Sec. 5.4.2}]{cite3459}}. The theorem has been made robust \NoCaseChange{\protect\cite{cite2691}}. Conversely, the distance of a code on an \(L^{D}\) lattice is upper bounded by \flmRefsHyperref{ref65}{order} \(O(L^{D+1-\nu})\) if the code implements an \(\nu\)th-level \flmTerm{term}{ref694}{}{Clifford hierarchy} gate \NoCaseChange{\protect\cite{cite3018}}. The code capacity threshold of such a code family is upper bounded by \(1/\nu\) \NoCaseChange{\protect\cite{cite3018}}. \end{defterm}
\end{eczvaluelist}
\codefieldsection{Decoding}
\begin{eczvaluelist}
\item\relax A \textit{local automaton decoder} (a.k.a. measurement-free local error correction) applies local rules to each small region of sites in a lattice geometry. Such decoders do not require any potentially non-local classical post-processing of error syndromes.
\item\relax Clustering decoder \NoCaseChange{\protect\cite{cite3890,cite3036}}.
\item\relax Quantum neural-network (QNN) decoder \NoCaseChange{\protect\cite{cite5052}}.
\item\relax Almost linear-time decoder \NoCaseChange{\protect\cite{cite5106}}.
\item\relax Offline message-passing decoder \NoCaseChange{\protect\cite{cite5107}}.
\end{eczvaluelist}
\codefieldsection{Code Capacity Threshold}
\begin{eczvaluelist}
\item\relax A threshold exists for the offline message-passing decoder \NoCaseChange{\protect\cite{cite5107}}.
\end{eczvaluelist}
\codefieldsection{Parent}
\begin{eczvaluelist}
\item\relax
\flmRefsHyperref[eczindexfamilyrel]{code:general_qldpc}{QLDPC code} --- Lattice stabilizer codes are QLDPC codes that are defined on Euclidean lattices.
\end{eczvaluelist}
\codefieldsection{Children}
\begin{eczvaluelist}
\item\relax
\flmRefsHyperref[eczindexfamilyrel]{code:1d_stabilizer}{1D lattice stabilizer code}\item\relax
\flmRefsHyperref[eczindexfamilyrel]{code:2d_stabilizer}{2D lattice stabilizer code}\item\relax
\flmRefsHyperref[eczindexfamilyrel]{code:3d_stabilizer}{3D lattice stabilizer code}\item\relax
\flmRefsHyperref[eczindexfamilyrel]{code:4d_stabilizer}{4D lattice stabilizer code}\item\relax
\flmRefsHyperref[eczindexfamilyrel]{code:crystalline_dynamic_gen}{Crystalline-circuit qubit code}\item\relax
\flmRefsHyperref[eczindexfamilyrel]{code:bosonization}{Bosonization code} --- The \(D\)-dimensional bosonization code encodes fermionic modes into a \(D\)-dimensional qubit stabilizer code.
\item\relax
\flmRefsHyperref[eczindexfamilyrel]{code:sc_qldpc}{Quantum spatially coupled (SC-QLDPC) code} --- Stabilizer generator matrices of SC-QLDPC codes on infinite-length chains or grids define a class of lattice stabilizer codes.
\item\relax
\flmRefsHyperref[eczindexfamilyrel]{code:higher_dimensional_toric}{\(D\)-dimensional twisted toric code}\item\relax
\flmRefsHyperref[eczindexfamilyrel]{code:qudit_color}{Modular-qudit lattice color code} --- Modular-qudit lattice color codes are defined analogous to qubit color codes on suitable lattices of any spatial dimension, but a directionality is required in order to make the modular-qudit stabilizers commute \NoCaseChange{\protect\cite[{Sec. III}]{cite673}}.
\item\relax
\flmRefsHyperref[eczindexfamilyrel]{code:generalized_bicycle}{Generalized bicycle (GB) code} --- Incommensurate GB codes of row weight \(w\) are equivalent to CSS codes local in \(D \leq w-1\) Euclidean dimensions, or in \(D \leq w-2\) dimensions when \(\ell\) is prime \NoCaseChange{\protect\cite[{Statement 13}]{cite3183}\protect\cite[{Sec. II.A}]{cite4639}}.
\end{eczvaluelist}
\codefieldsection{Cousins}
\begin{eczvaluelist}
\item\relax
\flmRefsHyperref[eczindexfamilyrel]{code:quasi_cyclic_qldpc}{Quasi-cyclic QLDPC (QC-QLDPC) code} --- Lattice stabilizer codes are QLDPC codes that are invariant under translations by a lattice unit cell in the bulk.
\item\relax
\flmRefsHyperref[eczindexfamilyrel]{code:spt}{Symmetry-protected topological (SPT) code} --- Lattice CSS codes in \(D\) dimensions can be converted into SPT Hamiltonians in one less dimension \NoCaseChange{\protect\cite{cite466}}.
\item\relax
\flmRefsHyperref[eczindexfamilyrel]{code:qldpc}{Qubit QLDPC code} --- Chain complexes describing some QLDPC codes \NoCaseChange{\protect\cite{cite484,cite485}}, and, more generally, CSS codes \NoCaseChange{\protect\cite{cite486}} can be 'lifted' into higher-dimensional manifolds admitting some notion of geometric locality. In addition, chain complexes describing QLDPC codes can be converted to 2D lattice stabilizer codes \NoCaseChange{\protect\cite{cite489}}.
\item\relax
\flmRefsHyperref[eczindexfamilyrel]{code:holographic}{Holographic code} --- Lattice stabilizer codes admit a bulk-boundary correspondence similar to that of holographic codes \NoCaseChange{\protect\cite{cite2853}}.
\item\relax
\flmRefsHyperref[eczindexfamilyrel]{code:topological}{Topological code} --- Topological phases are not realizable using lattice stabilizer codes iff they have long-range magic \NoCaseChange{\protect\cite{cite2691}}.
\item\relax
\flmRefsHyperref[eczindexfamilyrel]{code:plaquette_ising}{Plaquette Ising code}\item\relax
\flmRefsHyperref[eczindexfamilyrel]{code:good_qldpc}{Good QLDPC code} --- Chain complexes describing some QLDPC codes \NoCaseChange{\protect\cite{cite484,cite485}}, and, more generally, CSS codes \NoCaseChange{\protect\cite{cite486}} can be 'lifted' into higher-dimensional manifolds admitting some notion of geometric locality. Applying this procedure to good QLDPC codes yields \(\llbracket n,n^{1-2/D},n^{1-1/D}\rrbracket \) lattice stabilizer codes in \(D\) spatial dimensions that saturate the \flmRefsHyperref{ref487}{BPT bound} \NoCaseChange{\protect\cite{cite488,cite485,cite489}}.
\item\relax
\flmRefsHyperref[eczindexfamilyrel]{code:translationally_invariant_subsystem}{Lattice subsystem code} --- Lattice subsystem codes reduce to lattice stabilizer codes when there are no gauge qudits. The former (latter) is required to admit few-site gauge-group (stabilizer-group) generators on a lattice with boundary conditions.
\item\relax
\flmRefsHyperref[eczindexfamilyrel]{code:self_correct}{Self-correcting quantum code} --- Translationally invariant CSS codes are not self-correcting at high temperature \NoCaseChange{\protect\cite{cite3019}}.
\item\relax
\flmRefsHyperref[eczindexfamilyrel]{code:cluster_state}{Cluster-state code} --- Cluster states defined on various lattices are representatives of SPT phases, and states realizing these phases can be resources for MBQC. 
In 1D, cluster states are examples of SPT phases with global symmetries \NoCaseChange{\protect\cite{cite3078,cite3079,cite3080,cite3081,cite3072}} and enable MBQC on a single qubit \NoCaseChange{\protect\cite{cite428,cite429}}. 
The square-lattice cluster state, which is the prototypical resource for universal MBQC \NoCaseChange{\protect\cite{cite428,cite429}}, and other 2D cluster states \NoCaseChange{\protect\cite{cite3082,cite3083,cite3084}} have SPT order protected by subsystem symmetries \NoCaseChange{\protect\cite{cite3085,cite3086,cite3082}}.
States like AKLT states and SPT fixed-point states can be efficiently converted into cluster states using local measurements and subsequently used as resources for MBQC \NoCaseChange{\protect\cite{cite3087,cite3079,cite3088,cite3089,cite3090,cite3091}}.
In 3D, cluster states belong to SPT phases protected by higher-form symmetries \NoCaseChange{\protect\cite{cite3092}} and enable universal fault-tolerant MBQC \NoCaseChange{\protect\cite{cite3093}}.
A cluster-like state, or a state that is in the same SPT phase as a cluster state, can be prepared in finite time \NoCaseChange{\protect\cite{cite3094}}. Cluster states can be created on various lattices \NoCaseChange{\protect\cite{cite3095}}.

\item\relax
\flmRefsHyperref[eczindexfamilyrel]{code:qubit_css}{Qubit CSS code} --- The \flmRefsHyperref{ref683}{mapping of qubit CSS codes to chain complexes} allows the application of structures from topology to error correction. Chain complexes describing some QLDPC codes \NoCaseChange{\protect\cite{cite484,cite485}}, and, more generally, CSS codes \NoCaseChange{\protect\cite{cite486}} can be "lifted" into higher-dimensional manifolds admitting some notion of geometric locality. Qubit CSS codes admit several dualities \NoCaseChange{\protect\cite{cite2534,cite469}}. In particular, a CSS code and two dual classical codes can be organized by the same 2-complex, and gauging \NoCaseChange{\protect\cite{cite462,cite463,cite233,cite464,cite465,cite466,cite467,cite468,cite469,cite470}} either classical code yields the same CSS code up to Hadamard \NoCaseChange{\protect\cite{cite469}}.
\item\relax
\flmRefsHyperref[eczindexfamilyrel]{code:higher_dimensional_surface}{Homological code} --- Lattice surface codes in \(D\) spatial dimensions can be partially classified by the dimension of their stabilizer generators (and corresponding excitations).
There are \((p,q)\) \textit{surface codes} for \(p+q=D\) realized by \(Z\)-type stabilizer generators of dimension \(p\) and \(X\)-type stabilizer generators of dimension \(q\).
The two corresponding types of excitations are of dimension \(p-1\) and \(q-1\), respectively.

\item\relax
\flmRefsHyperref[eczindexfamilyrel]{code:qudit_da}{Modular-qudit dynamical code} --- Dynamical codes are typically defined on 2D and 3D lattices, but they are not conventional stabilizer codes in that they use \flmRefsHyperref{ref410}{code switching} for error correction and gates.
\end{eczvaluelist}
\eczhbkcontributors{ Tony Lau, \eczhuVVA }
\endeczcode

\eczcode{translationally_invariant_subsystem}{Lattice subsystem code}{~\NoCaseChange{\protect\cite{cite604}}}
\codefieldsection{Alternative Names}
\begin{eczvaluelist}
\item\relax Topological subsystem code
\end{eczvaluelist}
\eczhIndexCodeAliasName{translationally_invariant_subsystem}{Topological subsystem code}
\codefieldsection{Description}
A geometrically local qubit, modular-qudit, or Galois-qudit subsystem stabilizer code with qudits organized on a lattice modeled by the additive group \(\mathbb{Z}^D\) for spatial dimension \(D\), using either the ordinary block notion of locality or the fermionic/Majorana notion of locality.
On an infinite lattice, its gauge group is generated by few-site Pauli operators and their translations, in which case the code is called \textit{translationally invariant subsystem code}.
The stabilizer group may contain generators of unbounded weight, distinguishing these codes from stabilizer codes with bounded-weight generators for which some logical qubits were re-assigned to be gauge qubits.

Boundary conditions have to be imposed on the lattice in order to obtain finite-dimensional versions, in which case the stabilizer group may no longer be generated by few-site Pauli operators.
Lattice defects and boundaries between different codes can also be introduced.
Lattice subsystem stabilizer code Hamiltonians described by an Abelian anyon theory do not always realize the corresponding anyonic topological order in their ground-state subspace and may exhibit a rich phase diagram.

\codefieldsection{Rate}
\begin{defterm}{Subsystem BT bound}\label{ref5108}\label{ref492}
Subsystem qubit stabilizer code parameters on \(D\)-dimensional Euclidean lattices are limited by the \textit{subsystem Bravyi-Terhal (BT) bound} \NoCaseChange{\protect\cite{cite3000}}, which states that \(d = O(n^{1-1/D})\) and that \(k d^{1/(D-1)} = O(n)\) (using \flmRefsHyperref{ref65}{asymptotic notation}). 
The second equation is different from the (subspace) \flmRefsHyperref{ref487}{BPT bound}.
In particular, \(D=2\)-dimensional subsystem codes satisfy \(kd = O(n)\) \NoCaseChange{\protect\cite{cite3328}}.
More generally, there is a tradeoff theorem \NoCaseChange{\protect\cite{cite5109}} stating that, for any logical operator, there is an equivalent logical operator with weight \(\tilde{d}\) such that \(\tilde{d}d^{1/(D-1)}=O(L^{D})\).
\end{defterm}

\codefieldsection{Gates}
\begin{eczvaluelist}
\item\relax \begin{defterm}{Subsystem PYBK bound}\label{ref5110}\label{ref5111} The \flmRefsHyperref{ref3630}{Bravyi-Koenig bound} can be extended to subsystem codes by Pastawski and Yoshida. Namely, logical gates implemented via constant-depth quantum circuits on a \(D\)-dimensional lattice subsystem code whose distance increases at least logarithmically with \(n\) lie in the \(D\)th level of the \flmTerm{term}{ref694}{}{Clifford hierarchy} \NoCaseChange{\protect\cite{cite3018}}. \end{defterm}
\end{eczvaluelist}
\codefieldsection{Parent}
\begin{eczvaluelist}
\item\relax
\flmRefsHyperref[eczindexfamilyrel]{code:sparse_subsystem}{QLDPC subsystem code} --- Lattice subsystem codes are QLDPC subsystem codes that are defined on Euclidean lattices.
\end{eczvaluelist}
\codefieldsection{Children}
\begin{eczvaluelist}
\item\relax
\flmRefsHyperref[eczindexfamilyrel]{code:compass_model}{Compass code}\item\relax
\flmRefsHyperref[eczindexfamilyrel]{code:trapezoid}{Trapezoid subsystem code}\item\relax
\flmRefsHyperref[eczindexfamilyrel]{code:five_squares}{Generalized five-squares code}\item\relax
\flmRefsHyperref[eczindexfamilyrel]{code:2d_subsystem_color}{2D subsystem color code}\item\relax
\flmRefsHyperref[eczindexfamilyrel]{code:3d_subsystem_color}{3D subsystem color code}\item\relax
\flmRefsHyperref[eczindexfamilyrel]{code:3d_kitaev_honeycomb}{3D Kitaev honeycomb code}\item\relax
\flmRefsHyperref[eczindexfamilyrel]{code:3d_bacon_shor}{3D Bacon-Shor code}\item\relax
\flmRefsHyperref[eczindexfamilyrel]{code:css_plaquette}{CSS-Plaquette code}\item\relax
\flmRefsHyperref[eczindexfamilyrel]{code:subsystem_three_fermion}{Three-fermion (3F) subsystem code}\item\relax
\flmRefsHyperref[eczindexfamilyrel]{code:3d_subsystem_surface}{3D subsystem surface code}\item\relax
\flmRefsHyperref[eczindexfamilyrel]{code:subsystem_rotated_surface}{Subsystem rotated surface code}\item\relax
\flmRefsHyperref[eczindexfamilyrel]{code:subsystem_surface}{Subsystem surface code}\item\relax
\flmRefsHyperref[eczindexfamilyrel]{code:qudit_znone}{\(\mathbb{Z}_q^{(1)}\) subsystem code}\item\relax
\flmRefsHyperref[eczindexfamilyrel]{code:semion}{Chiral semion subsystem code}\item\relax
\flmRefsHyperref[eczindexfamilyrel]{code:zthree_znine}{\(\mathbb{Z}_3\times\mathbb{Z}_9\)-fusion subsystem code}\end{eczvaluelist}
\codefieldsection{Cousins}
\begin{eczvaluelist}
\item\relax
\flmRefsHyperref[eczindexfamilyrel]{code:translationally_invariant_stabilizer}{Lattice stabilizer code} --- Lattice subsystem codes reduce to lattice stabilizer codes when there are no gauge qudits. The former (latter) is required to admit few-site gauge-group (stabilizer-group) generators on a lattice with boundary conditions.
\item\relax
\flmRefsHyperref[eczindexfamilyrel]{code:topological_abelian}{Abelian topological code} --- All 2D Abelian bosonic topological orders can be realized as modular-qudit lattice subsystem codes by starting with a topological stabilizer code whose anyon theory contains the target Abelian anyon theory and then \flmRefsHyperref{ref666}{gauging out} complementary anyon types \NoCaseChange{\protect\cite{cite414}}.
One convenient choice for the parent stabilizer code is an Abelian TQD containing the target theory as a subtheory \NoCaseChange{\protect\cite[{Sec. 6.2}]{cite414}}.
The stabilizer generators of the new subsystem code may no longer be geometrically local.
Lattice subsystem stabilizer code Hamiltonians described by an Abelian anyon theory do not always realize the corresponding anyonic topological order in their ground-state subspace and may exhibit a rich phase diagram.
Non-Abelian topological orders are purported not to be realizable with Pauli stabilizer codes \NoCaseChange{\protect\cite{cite2530}}.

\item\relax
\flmRefsHyperref[eczindexfamilyrel]{code:good_qldpc}{Good QLDPC code} --- An \(\llbracket n,k,d\rrbracket \) qubit stabilizer code can be converted into an \flmRefsHyperref{ref65}{order} \(\llbracket O(\ell \delta n),k,\Omega(d/w)\rrbracket \) subsystem qubit stabilizer code with weight-three gauge operators via the wire-code mapping \NoCaseChange{\protect\cite{cite490}}, which uses \flmRefsHyperref{ref491}{weight reduction}. 
Here, \(w\) and \(\delta\) are the weight and degree of the input code's Tanner graph, while \(\ell\) is the length of the longest edge of a particular embedding of that graph.
Applying this procedure to good QLDPC codes and using an embedding into \(D\)-dimensional Euclidean space yields lattice subsystem codes whose logical-qubit number and distance both scale as \(\Theta(n^{1-1/D})\) as functions of block length \(n\), saturating the \flmRefsHyperref{ref492}{subsystem BT bound} \NoCaseChange{\protect\cite{cite490}}.

\item\relax
\flmRefsHyperref[eczindexfamilyrel]{code:majorana_subsystem}{Majorana subsystem stabilizer code} --- Translationally invariant subsystem codes have been formulated in terms of Majorana operators \NoCaseChange{\protect\cite{cite3482}}.
\item\relax
\flmRefsHyperref[eczindexfamilyrel]{code:qudit_subsystem_color}{Modular-qudit subsystem color code} --- Modular-qudit subsystem lattice color codes are defined analogously to qubit subsystem lattice color codes on suitable lattices of any spatial dimension, but a directionality is required in order to make the modular-qudit stabilizers commute \NoCaseChange{\protect\cite[{Sec. VII}]{cite673}}.
\end{eczvaluelist}
\eczhbkcontributors{ \eczhuVVA }
\endeczcode

\eczcode{lca_stabilizer}{Locally compact Abelian (LCA) stabilizer code}{~\NoCaseChange{\protect\cite{cite1428}}}
\codefieldsection{Alternative Names}
\begin{eczvaluelist}
\item\relax Mixed GKP code
\end{eczvaluelist}
\eczhIndexCodeAliasName{lca_stabilizer}{Mixed GKP code}
\codefieldsection{Description}
A mixed oscillator stabilizer code whose codewords are quantum lattice states defined on any number of qudits and a nonzero number of oscillators.
Its stabilizers are countably infinite subgroups of the qudit Pauli and oscillator displacement groups.
Codewords are entangled across the qudit-oscillator bipartition.

The simplest LCA state is a Bell state of a single physical qubit and a GKP-encoded qubit,
\flmMathEnvironment{align}{}{
  |\text{LCA}\rangle&=\sum_{\ell\in\mathbb{Z}}|x=\ell\sqrt{\pi}\rangle\left|\ell\text{ mod }2\right\rangle \\&=\sum_{s\in\mathbb{Z}}{|{x=(2s)\sqrt{\pi}}\rangle}\left|0\right\rangle +{|{x=(2s+1)\sqrt{\pi}}\rangle}\left|1\right\rangle ~.
}
LCA stabilizers with such codewords are called simple.

\textit{Simple} single-mode single-qudit LCA codewords can be viewed as \(Kc\)-dimensional GKP codewords whose logical subsystem decomposition \(\mathbb{Z}_{Kc}\cong\mathbb{Z}_{K}\times\mathbb{Z}_{c}\) entangles the \(\mathbb{Z}_{c}\) factor with the physical qudit \NoCaseChange{\protect\cite{cite1428}}.

\codefieldsection{Protection}
Simple LCA stabilizers can protect against either a larger set of displacements than GKP codes or a smaller set along with all single-qudit Pauli errors \NoCaseChange{\protect\cite{cite1428}}.

The syndrome space of a simple LCA stabilizer can be characterized entirely by a unit cell of displacements. For a single \(c\)-dimensional qudit and single-oscillator code, the area of this cell is \(2\pi c\). Displacement values that keep the codewords inside this unit cell can be measured simultaneously, meaning that one can simultaneously measure an arbitrary range of values of two non-commuting displacements given sufficient qudit dimension \NoCaseChange{\protect\cite{cite1428}}. 

\codefieldsection{Rate}
Single-mode single-qudit LCA codes have logical dimension \(K=c\theta+d\) for integers \(\theta\geq 0\) and \(d\in\mathbb{Z}_{c}^{\times}\), the multiplicative group of integers modulo \(c\) \NoCaseChange{\protect\cite{cite1428}}.

\codefieldsection{Encoding}
\begin{eczvaluelist}
\item\relax Codewords of a simple single-qudit single-oscillator code can be initialized by applying a conditional oscillator-qudit displacement to a GKP state and a qudit \(|+\rangle\) state \NoCaseChange{\protect\cite{cite1428}}. Alternatively, one can prepare a two-qudit Bell state and encode one subsystem into a GKP qudit code \NoCaseChange{\protect\cite{cite1428}}.
\item\relax General LCA stabilizers can be created by using a general embedding of their stabilizer algebra, a non-commutative torus \NoCaseChange{\protect\cite{cite5112,cite5113,cite5114,cite5115}}, into LCA groups \NoCaseChange{\protect\cite{cite1428}}. Logical operators can be obtained via Morita equivalence \NoCaseChange{\protect\cite{cite1428}}.
\end{eczvaluelist}
\codefieldsection{Gates}
\begin{eczvaluelist}
\item\relax Several LCA code families admit logical Clifford gates via Gaussian transformations on the oscillators together with Clifford gates on the qudits; for single-mode \((c,d)\)-LCA codes, this includes a logical Hadamard implemented by an oscillator Fourier transform and a corresponding qudit Clifford operation \NoCaseChange{\protect\cite{cite1428}}.
\item\relax Adding a conditional oscillator-qudit displacement makes the gate set universal \NoCaseChange{\protect\cite{cite5116}}.
\end{eczvaluelist}
\codefieldsection{Decoding}
\begin{eczvaluelist}
\item\relax Simple LCA codes admit a decoder for pure displacement noise, a decoder for all single-qudit Pauli errors together with a smaller displacement range, and for physical qubits a balanced decoder that trades oscillator against qudit error tolerance by moving syndrome-region boundaries \NoCaseChange{\protect\cite{cite1428}}.
\item\relax Under Petz (transpose) recovery against photon loss and qubit amplitude damping, \(c=2\) LCA codes can match or outperform comparable GKP codes at low energy for sufficiently large logical dimension \NoCaseChange{\protect\cite{cite1428}}.
\end{eczvaluelist}
\codefieldsection{Parents}
\begin{eczvaluelist}
\item\relax
\flmRefsHyperref[eczindexfamilyrel]{code:hybrid_qudit_oscillator}{Mixed oscillator code}\item\relax
\flmRefsHyperref[eczindexfamilyrel]{code:stabilizer}{Stabilizer code}\end{eczvaluelist}
\codefieldsection{Child}
\begin{eczvaluelist}
\item\relax
\flmRefsHyperref[eczindexfamilyrel]{code:quantum_lattice}{Quantum lattice code} --- LCA stabilizer codes defined on only oscillators reduce to quantum lattice codes.
\end{eczvaluelist}
\codefieldsection{Cousins}
\begin{eczvaluelist}
\item\relax
\flmRefsHyperref[eczindexfamilyrel]{code:binary_linear}{Linear binary code} --- Linear binary codes can be used to construct LCA stabilizer codes \NoCaseChange{\protect\cite{cite1428}}.
\item\relax
\flmRefsHyperref[eczindexfamilyrel]{code:eeight}{\(E_8\) Gosset lattice} --- Integer symplectic matrices like the symplectic \(E_8\) generator matrix can be used to construct LCA stabilizer codes \NoCaseChange{\protect\cite{cite1428}}.
\item\relax
\flmRefsHyperref[eczindexfamilyrel]{code:group_gkp}{Group GKP code} --- Simple single-mode single-qudit LCA codes are Abelian group-GKP codes with \(Kc \mathbb{Z} \subset \mathbb{Z} \subset \mathbb{R} \times \mathbb{Z}_c\), where the logical dimension \(K\) is coprime to the physical qudit dimension \(c\) \NoCaseChange{\protect\cite{cite1428}}.
\end{eczvaluelist}
\eczhbkcontributors{ \eczhuVVA }
\endeczcode

\eczcode{hybrid_qudit_oscillator}{Mixed oscillator code}{}
\codefieldsection{Alternative Names}
\begin{eczvaluelist}
\item\relax LCA code
\end{eczvaluelist}
\eczhIndexCodeAliasName{hybrid_qudit_oscillator}{LCA code}
\codefieldsection{Description}
Encodes a logical Hilbert space into some number of modular qudits, some number of rotors, and a nonzero number of oscillators, i.e., the Hilbert space of \(L^2\)-normalizable functions on a locally compact Abelian (LCA) group.
In photonic systems, photonic states of multiple degrees of freedom of a photon (e.g., frequency, amplitude, and polarization) are called \textit{hyper-entangled states} \NoCaseChange{\protect\cite{cite5117}}.

\codefieldsection{Gates}
\begin{eczvaluelist}
\item\relax Symplectic transformations (i.e., transformations that preserve the qudit Pauli and oscillator displacement group structure) are tensor products of qudit Clifford and oscillator Gaussian operations, and there are no entangling symplectic operations \NoCaseChange{\protect\cite{cite5118,cite5119,cite1428}}.
\item\relax Adding a conditional oscillator-qudit displacement makes the symplectic gate set universal \NoCaseChange{\protect\cite{cite5116}}.
\end{eczvaluelist}
\codefieldsection{Notes}
\begin{eczvaluelist}
\item\relax See reviews \NoCaseChange{\protect\cite{cite5095,cite3557,cite5096}} for introductions to mixed oscillator platforms.
\end{eczvaluelist}
\codefieldsection{Parent}
\begin{eczvaluelist}
\item\relax
\flmRefsHyperref[eczindexfamilyrel]{code:group_quantum}{Group-based quantum code} --- Group quantum codes whose physical spaces are constructed using some number of modular qudits, some number of rotors, and a nonzero number of oscillators, which together constitute a general locally compact Abelian (LCA) group, are mixed oscillator codes.
\end{eczvaluelist}
\codefieldsection{Children}
\begin{eczvaluelist}
\item\relax
\flmRefsHyperref[eczindexfamilyrel]{code:hybrid_cat}{Hybrid cat code}\item\relax
\flmRefsHyperref[eczindexfamilyrel]{code:lca_stabilizer}{Locally compact Abelian (LCA) stabilizer code}\item\relax
\flmRefsHyperref[eczindexfamilyrel]{code:oscillators}{Bosonic code} --- Mixed oscillator codes defined only on oscillators reduce to oscillator codes.
\end{eczvaluelist}
\codefieldsection{Cousins}
\begin{eczvaluelist}
\item\relax
\flmRefsHyperref[eczindexfamilyrel]{code:mixed}{Mixed code} --- Mixed oscillator codes are examples of quantum analogues of mixed codes.
\item\relax
\flmRefsHyperref[eczindexfamilyrel]{code:very-small-logical-qubit}{Very small logical qubit (VSLQ) code} --- VSLQ decoder utilizes two ancillary oscillators.
\item\relax
\flmRefsHyperref[eczindexfamilyrel]{code:eastab}{EA qubit stabilizer code} --- Encoders and decoders of a minimal EA qubit stabilizer code should be realizable using hyper-entangled states \NoCaseChange{\protect\cite{cite3645}}.
\end{eczvaluelist}
\eczhbkcontributors{ \eczhuVVA }
\endeczcode

\eczcode{molecular}{Molecular code}{~\NoCaseChange{\protect\cite{cite735}}}
\codefieldsection{Description}
Approximate quantum code that encodes a finite-dimensional logical space into the Hilbert space of \(L^2\)-normalizable functions on \(SO(3)\), i.e., rotational states of an asymmetric rigid body such as a polyatomic molecule.

Construction is based on nested finite subgroups \(H\subset K \subset SO(3)\).
The \(|K|/|H|\)-dimensional logical subspace is spanned by basis states that are uniform superpositions of elements of cosets of \(H\) in \(K\).
Examples discussed in the original work include cyclic, dihedral, tetrahedral-octahedral, and tetrahedral-icosahedral subgroup embeddings.

\codefieldsection{Protection}
Protects against generalized bit-flip errors \(g\in SO(3)\) that are inside the fundamental domain of \(SO(3)/K\).
In the cyclic \(Z_N\subset Z_{dN}\) family, the code corrects sufficiently small rigid-body rotations about any axis and angular-momentum kicks with \(\delta\ell<N/2\).
Protection against phase-flip and more general momentum-kick errors is determined by the branching rules of irreps of \(SO(3)\) into those of \(K\), and further into those of \(H\).

\codefieldsection{Notes}
\begin{eczvaluelist}
\item\relax Physical space characterizes orientations of a rigid body in 3D, which correspond to rotational states of an asymmetric molecule. See APS Physics Synopsis \NoCaseChange{\protect\cite{cite5120}} and \flmHref{https://www.youtube.com/watch?v=gjBbMMZ3L1k}{Physical Review Journal club} discussing molecular applications.
\item\relax Each ideal molecular code has a parent Hamiltonian whose ground space is the codespace, and normalizable approximate codewords can be obtained by damping in total angular momentum.
\end{eczvaluelist}
\codefieldsection{Parents}
\begin{eczvaluelist}
\item\relax
\flmRefsHyperref[eczindexfamilyrel]{code:group_gkp}{Group GKP code}\item\relax
\flmRefsHyperref[eczindexfamilyrel]{code:single_subsystem}{Monolithic quantum code}\end{eczvaluelist}
\codefieldsection{Cousins}
\begin{eczvaluelist}
\item\relax
\flmRefsHyperref[eczindexfamilyrel]{code:diatomic_molecular}{Diatomic molecular code} --- Molecular codes live on \(SO(3)\) for asymmetric rigid bodies, whereas diatomic molecular codes live on the homogeneous space \(S^2=SO(3)/SO(2)\) for linear rotors.
\item\relax
\flmRefsHyperref[eczindexfamilyrel]{code:fiber}{Fiber code} --- Molecular codes encode quantum information into superpositions of multiple orientations of an asymmetric molecule \NoCaseChange{\protect\cite{cite735}}, while fiber codes encode into the fiber associated with a single orientation of certain symmetric molecules \NoCaseChange{\protect\cite{cite5121}}.
\end{eczvaluelist}
\eczhbkcontributors{ \eczhuVVA }
\endeczcode

\eczcode{general_qldpc}{QLDPC code}{}
\codefieldsection{Alternative Names}
\begin{eczvaluelist}
\item\relax Sparse stabilizer code
\end{eczvaluelist}
\eczhIndexCodeAliasName{general_qldpc}{Sparse stabilizer code}
\codefieldsection{Description}
Member of a family of stabilizer codes for which the number of sites participating in each stabilizer generator and the number of stabilizer generators that each site participates in are both bounded by a constant as \(n\to\infty\).
Sometimes, the two parameters are explicitly stated: each site of an \((l,w)\)\textit{-regular QLDPC code} is acted on by \(\leq l\) generators of weight \(\leq w\).

Notable QLDPC codes are summarized in \flmRefsCref{ref5122}, demonstrating the steady improvement in code parameters that culminated in the first asymptotically good QLDPC codes.
  \begin{flmFloat}{table}{NumCap}\flmCellsBeginCenter
\long\def\flmTempTypesetThisTable#1{%
\begin{tblr}{#1,
  hspan=minimal,
  cell{1}{1}={}{c, font={\flmCellsHeaderFont}},
  cell{1}{2}={}{c, font={\flmCellsHeaderFont}},
  cell{1}{3}={}{c, font={\flmCellsHeaderFont}},
  cell{2}{1}={}{c},
  cell{2}{2}={}{c},
  cell{2}{3}={}{c},
  cell{3}{1}={}{c},
  cell{3}{2}={}{c},
  cell{3}{3}={}{c},
  cell{4}{1}={}{c},
  cell{4}{2}={}{c},
  cell{4}{3}={}{c},
  cell{5}{1}={}{c},
  cell{5}{2}={}{c},
  cell{5}{3}={}{c},
  cell{6}{1}={}{c},
  cell{6}{2}={}{c},
  cell{6}{3}={}{c},
  cell{7}{1}={}{c},
  cell{7}{2}={}{c},
  cell{7}{3}={}{c},
  cell{8}{1}={}{c},
  cell{8}{2}={}{c},
  cell{8}{3}={}{c},
  cell{9}{1}={}{c},
  cell{9}{2}={}{c},
  cell{9}{3}={}{c},
  cell{10}{1}={}{c},
  cell{10}{2}={}{c},
  cell{10}{3}={}{c},
  cell{11}{1}={}{c},
  cell{11}{2}={}{c},
  cell{11}{3}={}{c},
  cell{12}{1}={}{c},
  cell{12}{2}={}{c},
  cell{12}{3}={}{c},
  cell{13}{1}={}{c},
  cell{13}{2}={}{c},
  cell{13}{3}={}{c},
  cell{14}{1}={}{c},
  cell{14}{2}={}{c},
  cell{14}{3}={}{c},
  hline{2}={1}{.4pt,solid},
  hline{2}={2}{.4pt,solid},
  hline{2}={3}{.4pt,solid}}%
\toprule
\(k\) & \(d\) & Code\\

    \(2\) & \(\sqrt{n}\) & \flmRefsHyperref{code:surface}{Kitaev toric}
        \\

    \(\Theta(n)\) & \(\Theta(\log n)\) & \flmRefsHyperref{code:two_dimensional_hyperbolic_surface}{2D hyperbolic surface}
        \\

    \(\Theta(n)\) & \(\Omega (n^{1/10})\) & \flmRefsHyperref{code:four_dimensional_hyperbolic}{Guth-Lubotzky}
        \\

    \(2\) & \(\Omega (\sqrt{n\sqrt{\log n}})\) & \flmRefsHyperref{code:freedman_meyer_luo}{Freedman-Meyer-Luo}
        \\

    \(\Theta(n)\) & \(\Theta(\sqrt{n})\) & \flmRefsHyperref{code:hypergraph_product}{hypergraph product}
        \\

    \(\Theta (\sqrt{n}/\log n)\) & \(\Omega (\sqrt{n} \log n)\) & \flmRefsHyperref{code:ramanujan_tensor_product}{high-dimensional expander (HDX)}
        \\

    \(\Theta (\sqrt{n})\) & \(\Omega (\sqrt{n} \log^c n)\) & \flmRefsHyperref{code:iterated_ramanujan}{tensor-product HDX}
        \\

    \(\Theta (n^{3/5}/\text{polylog}(n) )\) & \(\Omega (n^{3/5}/\text{polylog}(n) )\) & \flmRefsHyperref{code:fiber_bundle}{fiber-bundle}
        \\

    \(\Theta (\log n)\) & \(\Omega (n/\log n)\) & \flmRefsHyperref{code:lifted_product}{lifted-product (LP)}
        \\

    \(\Theta (n^{4/5})\) & \(\Omega (n^{3/5})\) & \flmRefsHyperref{code:balanced_product}{balanced product (BP)}
        \\

    \(\Theta(n)\) & \(\Theta(n)\) & \flmRefsHyperref{code:expander_lifted_product}{expander LP}
        \\

    \(\Theta(n)\) & \(\Theta(n)\) & \flmRefsHyperref{code:quantum_tanner}{quantum Tanner}
        \\

    \(\Theta(n)\) & \(\Theta(n)\) & \flmRefsHyperref{code:dhlv}{Dinur-Hsieh-Lin-Vidick}
    \\
\bottomrule
\end{tblr}%
}%
\def\flmTmpMaxW{\dimexpr 0.96\linewidth\relax}%
\setbox0=\hbox{\flmTempTypesetThisTable{colspec={ccc}}}%
\ifdim\wd0<\flmTmpMaxW\relax
  \leavevmode\box0 
\else
  \flmTempTypesetThisTable{width=\flmTmpMaxW,colspec={X[-1]X[-1]X[-1]}}
\fi
\flmCellsEndCenter \caption{Notable QLDPC codes and their \flmRefsHyperref{ref65}{asymptotic scaling} (see also Ref. \NoCaseChange{\protect\cite{cite3442}}); \(c\) is a positive integer.}\label{ref5122}\end{flmFloat}

A \textit{geometrically local stabilizer code} is a QLDPC code where the sites involved in any syndrome value are contained in a fixed volume that does not scale with \(n\).
As opposed to general stabilizer codes, syndrome extraction of the constant-weight check operators of a QLDPC code can be done using a constant-depth circuit.

Strictly speaking, the term \textit{parity check} describes only bitwise qubit error syndromes. Nevertheless, qudit and bosonic stabilizer codes satisfying the above criteria are also called QLDPC codes.
This entry includes general code constructions which are generally intended to yield QLDPC codes, but may include code families that have non-QLDPC members.

\codefieldsection{Protection}
Detects errors on \(d-1\) sites, corrects errors on \(\left\lfloor (d-1)/2 \right\rfloor\) sites.
Code distance may not be a reliable marker of code performance.

\codefieldsection{Decoding}
\begin{eczvaluelist}
\item\relax Non-binary decoding algorithm for CSS-type QLDPC codes \NoCaseChange{\protect\cite{cite3634}}.
\item\relax GD-CSS Decoder for Galois-qudit CSS QLDPC codes \NoCaseChange{\protect\cite{cite5123,cite5124}}
\end{eczvaluelist}
\codefieldsection{Notes}
\begin{eczvaluelist}
\item\relax Infleqtion QLDPC software library for estimating distance and creating various qubit and Galois-qudit QLDPC CSS codes \NoCaseChange{\protect\cite{cite5125}}
\item\relax LDPC Python software library for decoding LDPC and QLDPC codes \NoCaseChange{\protect\cite{cite1481,cite1482}}.
\item\relax Reviews of QLDPC codes provided in Refs. \NoCaseChange{\protect\cite{cite3634,cite3442,cite5126}}.
\end{eczvaluelist}
\codefieldsection{Parent}
\begin{eczvaluelist}
\item\relax
\flmRefsHyperref[eczindexfamilyrel]{code:qlwc}{Quantum low-weight check (QLWC) code} --- QLDPC codes are QLWC codes for which the number of stabilizer generators that each site participates in is bounded by a constant as \(n\to\infty\).
\end{eczvaluelist}
\codefieldsection{Children}
\begin{eczvaluelist}
\item\relax
\flmRefsHyperref[eczindexfamilyrel]{code:translationally_invariant_stabilizer}{Lattice stabilizer code} --- Lattice stabilizer codes are QLDPC codes that are defined on Euclidean lattices.
\item\relax
\flmRefsHyperref[eczindexfamilyrel]{code:generalized_homological_product}{Generalized homological-product code} --- Homological products are a primary tool for generating QLDPC codes with favorable parameters. Typically, whenever the input codes are LDPC or QLDPC, the resulting code will be QLDPC with non geometrically local stabilizer generators.
\item\relax
\flmRefsHyperref[eczindexfamilyrel]{code:good_qldpc}{Good QLDPC code}\item\relax
\flmRefsHyperref[eczindexfamilyrel]{code:quasi_cyclic_qldpc}{Quasi-cyclic QLDPC (QC-QLDPC) code}\item\relax
\flmRefsHyperref[eczindexfamilyrel]{code:qldpc}{Qubit QLDPC code}\end{eczvaluelist}
\codefieldsection{Cousins}
\begin{eczvaluelist}
\item\relax
\flmRefsHyperref[eczindexfamilyrel]{code:quantum_locally_recoverable}{Quantum locally recoverable code (QLRC)} --- Finite-dimensional block QLDPC stabilizer codes are QLRCs whose locality \(r \leq w\), where \(w\) is the maximum stabilizer-generator weight \NoCaseChange{\protect\cite{cite812}}.
\item\relax
\flmRefsHyperref[eczindexfamilyrel]{code:q-ary_ldpc}{\(q\)-ary LDPC code} --- Galois-qudit QLDPC codes are quantum analogues of \(q\)-ary LDPC codes.
\item\relax
\flmRefsHyperref[eczindexfamilyrel]{code:topological}{Topological code} --- Topological codes are not generally defined using Pauli strings or their qudit and bosonic generalizations. However, for appropriate tessellations, the codespace is the ground-state subspace of a geometrically local Hamiltonian. In this sense, topological codes are QLDPC codes. Geometrically local commuting-projector code Hamiltonians on Euclidean manifolds are stable with respect to small perturbations when they satisfy the \flmRefsHyperref{ref2675}{TQO conditions}, meaning that a notion of a phase can be defined \NoCaseChange{\protect\cite{cite2676,cite2677,cite2678,cite2679}}. This notion can be extended to semi-hyperbolic manifolds \NoCaseChange{\protect\cite{cite2680}} and non-geometrically local QLDPC codes exhibiting check soundness \NoCaseChange{\protect\cite{cite2681}} (see also \NoCaseChange{\protect\cite{cite2682}}).
\item\relax
\flmRefsHyperref[eczindexfamilyrel]{code:dynamic_gen}{Dynamically generated QECC} --- QLDPC codes can arise from a dynamical process \NoCaseChange{\protect\cite{cite2734}}.
\item\relax
\flmRefsHyperref[eczindexfamilyrel]{code:hamiltonian}{Hamiltonian-based code} --- QLDPC code Hamiltonians can be simulated, with the help of perturbation theory, by two-dimensional Hamiltonians with non-commuting terms whose interactions scale with \(n\) \NoCaseChange{\protect\cite{cite2839}}.
\item\relax
\flmRefsHyperref[eczindexfamilyrel]{code:sparse_subsystem}{QLDPC subsystem code} --- QLDPC subsystem codes reduce to QLDPC codes when there are no gauge degrees of freedom.
\item\relax
\flmRefsHyperref[eczindexfamilyrel]{code:qltc}{Quantum locally testable code (QLTC)} --- Stabilizer LTCs are QLDPC. More general QLTCs are not defined using Pauli strings, but the codespace is the ground-state subspace of a local Hamiltonian. In this sense, QLTCs are QLDPC codes.
\item\relax
\flmRefsHyperref[eczindexfamilyrel]{code:self_correct}{Self-correcting quantum code} --- Linear confinement of QLDPC (LDPC) codes implies (classical) self-correction \NoCaseChange{\protect\cite{cite849}}.
\item\relax
\flmRefsHyperref[eczindexfamilyrel]{code:galois_bch}{Galois-qudit BCH code} --- Some Galois-qudit BCH codes are QLDPC \NoCaseChange{\protect\cite{cite4616}\protect\cite[{Ch. 16}]{cite872}}.
\item\relax
\flmRefsHyperref[eczindexfamilyrel]{code:two_block_quantum}{Two-block CSS code} --- When matrices \(A\) and \(B\) have row and column weights bounded by \(W\), a two-block CSS code is a quantum LDPC code with stabilizer generators bounded by \(2W\).
\item\relax
\flmRefsHyperref[eczindexfamilyrel]{code:2bga}{Two-block group-algebra (2BGA) codes} --- Given \flmRefsHyperref{ref205}{group algebra} elements \(a,b\in \mathbb{F}_q[G]\) with weights \(W_a\) and \(W_b\) (i.e., number of nonzero terms in the expansion), the 2BGA code LP\((a,b)\) has stabilizer
generators of uniform weight \(W_a+W_b\).

\item\relax
\flmRefsHyperref[eczindexfamilyrel]{code:generalized_bicycle}{Generalized bicycle (GB) code} --- Stabilizer generators of the code GB\((a,b)\) have weights given by the sum of weights of polynomials \(a(x)\) and \(b(x)\).
The GB code ansatz is convenient for designing QLDPC codes and several extensions exist \NoCaseChange{\protect\cite{cite4640}}.

\item\relax
\flmRefsHyperref[eczindexfamilyrel]{code:distance_balanced}{Distance-balanced code} --- Lattice surgery techniques for QLDPC codes \NoCaseChange{\protect\cite{cite3499,cite848}} utilize \flmRefsHyperref{ref491}{weight reduction}. Single-ancilla syndrome extraction circuits that, for the most part, preserve the \flmRefsHyperref{ref3496}{effective distance} of weight-reduced qLDPC codes \NoCaseChange{\protect\cite{cite3776}}.
\end{eczvaluelist}
\eczhbkcontributors{ \eczhuVVA }
\endeczcode

\eczcode{sparse_subsystem}{QLDPC subsystem code}{}
\codefieldsection{Alternative Names}
\begin{eczvaluelist}
\item\relax Sparse subsystem code
\end{eczvaluelist}
\eczhIndexCodeAliasName{sparse_subsystem}{Sparse subsystem code}
\codefieldsection{Description}
Member of a family of subsystem stabilizer codes for which the number of sites participating in each gauge generator and the number of gauge generators that each site participates in are both bounded by a constant as \(n\to\infty\).
The stabilizer group may contain generators of unbounded weight, distinguishing these codes from stabilizer codes with bounded-weight generators for which some logical qubits were re-assigned to be gauge qubits.

\codefieldsection{Rate}
There exists a family of QLDPC subsystem codes with \(d = n^{1-\epsilon}\), where \(\epsilon = O(1/\sqrt{\log n})\) \NoCaseChange{\protect\cite{cite668}}.
Spatially local subsystem codes also exist in \(D\geq 2\) dimensions with \(d = n^{1-\epsilon-1/D}\), where \(\epsilon = O(1/\sqrt{\log n})\), nearly saturating the \flmRefsHyperref{ref492}{subsystem BT bound} \NoCaseChange{\protect\cite{cite668}}.

\codefieldsection{Parent}
\begin{eczvaluelist}
\item\relax
\flmRefsHyperref[eczindexfamilyrel]{code:subsystem_stabilizer}{Subsystem stabilizer code}\end{eczvaluelist}
\codefieldsection{Children}
\begin{eczvaluelist}
\item\relax
\flmRefsHyperref[eczindexfamilyrel]{code:translationally_invariant_subsystem}{Lattice subsystem code} --- Lattice subsystem codes are QLDPC subsystem codes that are defined on Euclidean lattices.
\item\relax
\flmRefsHyperref[eczindexfamilyrel]{code:subsystem_spacetime_circuit}{Subsystem spacetime circuit code}\item\relax
\flmRefsHyperref[eczindexfamilyrel]{code:subsystem_higher_dimensional_surface}{Subsystem homological code}\item\relax
\flmRefsHyperref[eczindexfamilyrel]{code:qudit_subsystem_color}{Modular-qudit subsystem color code}\end{eczvaluelist}
\codefieldsection{Cousins}
\begin{eczvaluelist}
\item\relax
\flmRefsHyperref[eczindexfamilyrel]{code:general_qldpc}{QLDPC code} --- QLDPC subsystem codes reduce to QLDPC codes when there are no gauge degrees of freedom.
\item\relax
\flmRefsHyperref[eczindexfamilyrel]{code:qldpc}{Qubit QLDPC code} --- Any qubit QLDPC code with stabilizer-generator weights \(w_i\) can be mapped constructively to a sparse subsystem qubit code with the same number of logical qubits and distance, using \(n=O(\sum_i w_i)\) physical qubits and constant-weight gauge generators \NoCaseChange{\protect\cite{cite668}}.
\end{eczvaluelist}
\eczhbkcontributors{ Xiaozhen Fu, \eczhuVVA }
\endeczcode

\eczcode{qlwc}{Quantum low-weight check (QLWC) code}{~\NoCaseChange{\protect\cite{cite1633}}}
\codefieldsection{Description}
Member of a family of \(\llbracket n,k,d\rrbracket \) stabilizer codes for which the number of sites participating in each stabilizer generator is bounded by a constant as \(n\to\infty\).

\codefieldsection{Parent}
\begin{eczvaluelist}
\item\relax
\flmRefsHyperref[eczindexfamilyrel]{code:stabilizer}{Stabilizer code}\end{eczvaluelist}
\codefieldsection{Child}
\begin{eczvaluelist}
\item\relax
\flmRefsHyperref[eczindexfamilyrel]{code:general_qldpc}{QLDPC code} --- QLDPC codes are QLWC codes for which the number of stabilizer generators that each site participates in is bounded by a constant as \(n\to\infty\).
\end{eczvaluelist}
\codefieldsection{Cousins}
\begin{eczvaluelist}
\item\relax
\flmRefsHyperref[eczindexfamilyrel]{code:unary}{Unary code} --- A family of approximate non-stabilizer qubit QLWC codes with linear distance and rate has been constructed \NoCaseChange{\protect\cite{cite1633}} using unary codes that arise from the Feynman-Kitaev clock construction \NoCaseChange{\protect\cite{cite1634}}.
\item\relax
\flmRefsHyperref[eczindexfamilyrel]{code:approximate_qecc}{Approximate quantum error-correcting code (AQECC)} --- A family of approximate non-stabilizer qubit QLWC codes with linear distance and rate has been constructed \NoCaseChange{\protect\cite{cite1633}} using unary codes that arise from the Feynman-Kitaev clock construction \NoCaseChange{\protect\cite{cite1634}}.
\item\relax
\flmRefsHyperref[eczindexfamilyrel]{code:circuit_to_hamiltonian}{Circuit-to-Hamiltonian approximate code} --- The circuit-to-Hamiltonian code construction yields approximate codes whose distance and logical-qubit number are both of \flmRefsHyperref{ref65}{order} \(\Omega(n/\log^5 n)\) \NoCaseChange{\protect\cite[{Thm. 3.1}]{cite581}}.
These codes are approximate non-stabilizer QLWC codes since the Hamiltonian consists of non-commuting 9-local non-Pauli projectors, with each qubit acted on by \flmRefsHyperref{ref65}{order} \(O( \text{polylog}(n) )\) projectors.
\end{eczvaluelist}
\eczhbkcontributors{ \eczhuVVA }
\endeczcode

\eczcode{quantum_double}{Quantum-double code}{~\NoCaseChange{\protect\cite{cite423}}}
\codefieldsection{Alternative Names}
\begin{eczvaluelist}
\item\relax Non-Abelian surface code
\end{eczvaluelist}
\eczhIndexCodeAliasName{quantum_double}{Non-Abelian surface code}
\codefieldsection{Description}
Group-based code whose codewords realize 2D modular gapped topological order defined by a finite group \(G\).
The code's generators are few-body operators associated to the stars and plaquettes, respectively, of a tessellation of a two-dimensional surface (with a qudit of dimension \( |G| \) located at each edge of the tessellation).
The original Hamiltonian can be re-expressed via \flmRefsHyperref{ref20}{group-based right- and left-multiplication \(X\)-type as well as \(Z\)-type error} operators \NoCaseChange{\protect\cite[{Sec. 3.3}]{cite598}}.

The physical Hilbert space has dimension \( |G|^E  \), where \( E \) is the number of  edges in the tessellation. The dimension of the code space is the number of orbits of the conjugation action of \( G \) on \( \text{Hom}(\pi_1(\Sigma),G) \), the set of group homomorphisms from the fundamental group of the surface \( \Sigma \) into the finite group \( G \) \NoCaseChange{\protect\cite{cite5127,cite5128}} (see also Ref. \NoCaseChange{\protect\cite{cite5129}}). When \( G \) is Abelian, the formula for the dimension simplifies to \( |G|^{2g} \), where \( g \) is the genus of the surface \( \Sigma \).

The codespace is the ground-state subspace of the quantum double model Hamiltonian, while local excitations are characterized by anyons.
Different types of anyons are labeled by irreducible representations of the group's quantum double algebra, \(D(G)\) (a.k.a. Drinfeld center) \NoCaseChange{\protect\cite{cite5130,cite5131}}.
Ribbon operators create particle-antiparticle pairs at their endpoints, and braiding followed by fusion of the resulting anyons furnishes the fault-tolerant computational primitive of the original model \NoCaseChange{\protect\cite{cite423}}.
Not all isomorphic non-Abelian groups give rise to different quantum doubles \NoCaseChange{\protect\cite{cite5132}}.

For non-Abelian groups, alternative constructions are possible, encoding information in the fusion space of the low-energy anyonic quasiparticle excitations of the model \NoCaseChange{\protect\cite{cite5133,cite5060,cite5066}}.
The fusion space of such non-Abelian anyons has dimension greater than one, allowing for topological quantum computation of logical information stored in the fusion outcomes.

Gapped boundaries of the models are classified by a subgroup \(K \subseteq G\) and a two-cocycle \NoCaseChange{\protect\cite{cite5134,cite5130,cite5135,cite5136}}.

\codefieldsection{Protection}
Error-correcting properties established in Ref. \NoCaseChange{\protect\cite{cite5127}}.
The code distance is the number of edges in the shortest non-contractible cycle in the tessellation or dual tessellation \NoCaseChange{\protect\cite{cite480}}.
These models realize local topological order (LTO) \NoCaseChange{\protect\cite{cite5137}}.

\codefieldsection{Encoding}
\begin{eczvaluelist}
\item\relax A depth-\(L^2\) circuit that grows the code out of a small patch on an \(L\times L\) square lattice using CMULT gates (i.e., "local moves") \NoCaseChange{\protect\cite{cite3818,cite5138}}.
\item\relax For an \(L\times L\) lattice, deterministic state preparation can be done with a geometrically local unitary \(O(L)\)-depth circuit \NoCaseChange{\protect\cite{cite5061,cite5138}} or an \(O(\log{L})\)-depth unitary circuit with non-local two-qubit gates \NoCaseChange{\protect\cite{cite3818,cite3823}}.
\item\relax For any group \(G\) of nilpotency class two, states can be initialized with a single round of adaptive measurements \NoCaseChange{\protect\cite{cite5139}}.
\item\relax For any solvable group \(G\), ground-state preparation and anyon-pair creation can be done with an adaptive constant-depth circuit with geometrically local gates and measurements throughout \NoCaseChange{\protect\cite{cite5140,cite5141}} (see Ref. \NoCaseChange{\protect\cite{cite3235}} for specific dihedral groups). Anyon-pair creation requires an adaptive circuit for any non-Abelian \(G\) \NoCaseChange{\protect\cite{cite5141}}.
\item\relax For non-solvable groups, states may not be preparable with an adaptive constant-depth circuit with geometrically local gates and measurements throughout \NoCaseChange{\protect\cite{cite5139}}.
\end{eczvaluelist}
\codefieldsection{Gates}
\begin{eczvaluelist}
\item\relax Universal topological quantum computation possible for certain groups \NoCaseChange{\protect\cite{cite5060,cite5061}}.
\end{eczvaluelist}
\codefieldsection{Decoding}
\begin{eczvaluelist}
\item\relax For any solvable group \(G\), topological charge measurements can be done with an adaptive constant-depth circuit with geometrically local gates and measurements throughout \NoCaseChange{\protect\cite{cite5141}}.
\end{eczvaluelist}
\codefieldsection{Code Capacity Threshold}
\begin{eczvaluelist}
\item\relax Behavior under particular \(X\)-type noise (namely, diffusion of an anyon that squares to the trivial anyon) is related to the phase diagram of a disordered \(D_4\) rotor model \NoCaseChange{\protect\cite{cite5142,cite5064}}.
\end{eczvaluelist}
\codefieldsection{Notes}
\begin{eczvaluelist}
\item\relax See Ref. \NoCaseChange{\protect\cite{cite5143}} for a review of gauge theory, which admits quantum-double topological phases. See Ref. \NoCaseChange{\protect\cite{cite5144}} for another review.
\end{eczvaluelist}
\codefieldsection{Parents}
\begin{eczvaluelist}
\item\relax
\flmRefsHyperref[eczindexfamilyrel]{code:tqd}{Twisted quantum double (TQD) code} --- The anyon theory corresponding to a quantum-double code is a TQD with trivial cocycle.
\item\relax
\flmRefsHyperref[eczindexfamilyrel]{code:hopf_quantum_double}{Hopf-algebra quantum-double code} --- Hopf-algebra quantum-double codes reduce to quantum-double codes when the Hopf algebra is a \flmRefsHyperref{ref205}{group algebra}. Quantum-double codes for non-Abelian groups \(G\) are dual to Hopf-algebra quantum-double codes for Hopf algebras based on \(\text{Rep}(G)\) under the Tannaka-Krein duality \NoCaseChange{\protect\cite{cite5145}\protect\cite[{Fig. 1}]{cite5146}}.
\end{eczvaluelist}
\codefieldsection{Children}
\begin{eczvaluelist}
\item\relax
\flmRefsHyperref[eczindexfamilyrel]{code:quantum_double_dihedral}{Dihedral \(G=D_m\) quantum-double code}\item\relax
\flmRefsHyperref[eczindexfamilyrel]{code:group_4_2_2}{\(\llbracket 4,2,2\rrbracket _{G}\) four group-qudit code} --- The four group-qudit code is the smallest quantum double code.
\item\relax
\flmRefsHyperref[eczindexfamilyrel]{code:quantum_double_abelian}{Abelian quantum-double stabilizer code} --- The anyon theory corresponding to (Abelian) quantum double codes is defined by an (Abelian) group.
\item\relax
\flmRefsHyperref[eczindexfamilyrel]{code:galois_color}{Galois-qudit color code} --- A Galois qudit for \(q=p^m\) can be decomposed into a Kronecker product of \(m\) modular qudits \NoCaseChange{\protect\cite{cite696,cite398,cite698,cite699,cite700}\protect\cite[{Sec. 5.3}]{cite697}}. Galois-qudit color codes yield Abelian quantum-double codes with Abelian-group topological order via this decomposition.
\item\relax
\flmRefsHyperref[eczindexfamilyrel]{code:galois_topological}{Galois-qudit surface code} --- A Galois qudit for \(q=p^m\) can be decomposed into a Kronecker product of \(m\) modular qudits \NoCaseChange{\protect\cite{cite696,cite398,cite698,cite699,cite700}\protect\cite[{Sec. 5.3}]{cite697}}. Galois-qudit surface codes yield Abelian quantum-double codes with \(\mathbb{F}_{p^m}\cong \mathbb{Z}_p^m\) topological order via this decomposition.
\end{eczvaluelist}
\codefieldsection{Cousins}
\begin{eczvaluelist}
\item\relax
\flmRefsHyperref[eczindexfamilyrel]{code:hamiltonian}{Hamiltonian-based code} --- Quantum double code Hamiltonians can be simulated, with the help of perturbation theory and the \(\llbracket 4,1,1,2\rrbracket \) subsystem code, by two-dimensional two-body Hamiltonians with non-commuting terms \NoCaseChange{\protect\cite{cite2838}}.
\item\relax
\flmRefsHyperref[eczindexfamilyrel]{code:subsystem_group_quantum}{Subsystem group-based quantum code} --- Subsystem versions of quantum-double codes have been formulated \NoCaseChange{\protect\cite{cite5147}}.
\item\relax
\flmRefsHyperref[eczindexfamilyrel]{code:bacon_shor_4}{\(\llbracket 4,1,1,2\rrbracket \) Four-qubit subsystem code} --- Quantum double code Hamiltonians can be simulated, with the help of perturbation theory and the four-qubit subsystem code, by two-dimensional two-body Hamiltonians with non-commuting terms \NoCaseChange{\protect\cite{cite2838}}.
\item\relax
\flmRefsHyperref[eczindexfamilyrel]{code:spt}{Symmetry-protected topological (SPT) code} --- The \(Q\) quantum double model can be obtained by gauging \NoCaseChange{\protect\cite{cite462,cite463,cite233,cite464,cite465,cite466,cite467,cite468,cite469,cite470}} symmetries of a Type I and Type III \(\mathbb{Z}_2^3\) SPT \NoCaseChange{\protect\cite{cite3070,cite3071}}.
\item\relax
\flmRefsHyperref[eczindexfamilyrel]{code:tqd_abelian}{Abelian TQD code} --- A Type-III \(\mathbb{Z}_2^3\) Abelian TQD realizes the same topological order as the \(G=D_4\) quantum double model \NoCaseChange{\protect\cite{cite577,cite575}}. There is a sufficient condition for when a Type-III TQD can be realized as a quantum double model \NoCaseChange{\protect\cite{cite5059}}.
\item\relax
\flmRefsHyperref[eczindexfamilyrel]{code:generalized_color}{Generalized 2D color code} --- A generalized color code for group \(G\) on the 4.8.8 lattice is equivalent to a \(G\) quantum double model and another \(G/[G,G]\) quantum double model defined using the Abelianization of \(G\).
\item\relax
\flmRefsHyperref[eczindexfamilyrel]{code:quantum_triple}{Quantum-triple code} --- The quantum triple model can be thought of as a 3D version of the quantum double model.
\item\relax
\flmRefsHyperref[eczindexfamilyrel]{code:self_correct}{Self-correcting quantum code} --- An \(n\)-dependent energy barrier to creating all logical errors is likely necessary for a thermally stable memory, having been shown as such for a large class of 2D topological phases \NoCaseChange{\protect\cite{cite3012,cite3013}} including Abelian \NoCaseChange{\protect\cite{cite3014}} and non-Abelian \NoCaseChange{\protect\cite{cite3015}} quantum doubles.
\item\relax
\flmRefsHyperref[eczindexfamilyrel]{code:surface}{Kitaev surface code} --- On closed surfaces, a quantum-double model with \(G=\mathbb{Z}_2\) reduces to the surface code; on a torus, this is the toric code. Quantum doubles with open boundary conditions also reduce to surface codes on open surfaces \NoCaseChange{\protect\cite{cite3957,cite3958,cite3959,cite3960,cite730}}. Non-stabilizer surface-code states can be prepared by augmenting the surface code with a quantum double model \NoCaseChange{\protect\cite{cite3842,cite3843,cite3961}}.
\end{eczvaluelist}
\eczhbkcontributors{ Ian Teixeira, \eczhuVVA }
\endeczcode

\eczcode{quantum_triple}{Quantum-triple code}{~\NoCaseChange{\protect\cite{cite459}}}
\codefieldsection{Alternative Names}
\begin{eczvaluelist}
\item\relax 3D quantum-double code
\end{eczvaluelist}
\eczhIndexCodeAliasName{quantum_triple}{3D quantum-double code}
\codefieldsection{Description}
Group-based code whose codewords realize 3D topological order defined by a finite group \(G\).

Excitations are characterized by irreducible representations of a generalization of the quantum double (algebra), often called the quantum triple \NoCaseChange{\protect\cite{cite459,cite5148}}.

\codefieldsection{Parent}
\begin{eczvaluelist}
\item\relax
\flmRefsHyperref[eczindexfamilyrel]{code:tqt}{Twisted quantum triple (TQT) code} --- The anyon theory corresponding to a quantum-triple code is a TQT with trivial cocycle.
\end{eczvaluelist}
\codefieldsection{Child}
\begin{eczvaluelist}
\item\relax
\flmRefsHyperref[eczindexfamilyrel]{code:qudit_3d_surface}{Modular-qudit 3D surface code} --- A quantum triple model for the group \(G=\mathbb{Z}_q\) is a modular-qudit 3D surface code.
\end{eczvaluelist}
\codefieldsection{Cousin}
\begin{eczvaluelist}
\item\relax
\flmRefsHyperref[eczindexfamilyrel]{code:quantum_double}{Quantum-double code} --- The quantum triple model can be thought of as a 3D version of the quantum double model.
\end{eczvaluelist}
\eczhbkcontributors{ \eczhuVVA }
\endeczcode

\eczcode{quasi_cyclic_qldpc}{Quasi-cyclic QLDPC (QC-QLDPC) code}{~\NoCaseChange{\protect\cite{cite821,cite822}}}
\codefieldsection{Description}
A QLDPC code such that cyclic shifts of the subsystems by a fixed \(\ell\geq 1\) leave the codespace invariant.
Stabilizer generator matrices of such codes can be put into block form, where each nonzero block is a circulant matrix \NoCaseChange{\protect\cite{cite821,cite822}}. 

\codefieldsection{Parents}
\begin{eczvaluelist}
\item\relax
\flmRefsHyperref[eczindexfamilyrel]{code:general_qldpc}{QLDPC code}\item\relax
\flmRefsHyperref[eczindexfamilyrel]{code:quantum_quasi_cyclic}{Quasi-cyclic quantum code}\end{eczvaluelist}
\codefieldsection{Cousins}
\begin{eczvaluelist}
\item\relax
\flmRefsHyperref[eczindexfamilyrel]{code:qc_ldpc}{Quasi-cyclic LDPC (QC-LDPC) code} --- QC-QLDPC codes are quantum counterparts of QC-LDPC codes. QC-LDPC codes can be used to make qubit QLDPC codes using various non-CSS constructions \NoCaseChange{\protect\cite{cite1553}}. There exist explicit constructions of both whose parity-check (stabilizer generator) matrices have column weight 2 and girth 12 \NoCaseChange{\protect\cite{cite1554}}.
\item\relax
\flmRefsHyperref[eczindexfamilyrel]{code:translationally_invariant_stabilizer}{Lattice stabilizer code} --- Lattice stabilizer codes are QLDPC codes that are invariant under translations by a lattice unit cell in the bulk.
\item\relax
\flmRefsHyperref[eczindexfamilyrel]{code:ea_qc_qldpc}{EA QC-QLDPC code} --- EA QC-QLDPC codes are entanglement-assisted versions of QC-QLDPC codes.
\item\relax
\flmRefsHyperref[eczindexfamilyrel]{code:galois_bch}{Galois-qudit BCH code} --- Some Galois-qudit BCH codes are QC-QLDPC \NoCaseChange{\protect\cite[{Ch. 16}]{cite872}}.
\end{eczvaluelist}
\eczhbkcontributors{ \eczhuVVA }
\endeczcode

\eczcode{random_stabilizer}{Random stabilizer code}{~\NoCaseChange{\protect\cite{cite3369,cite3196,cite736}}}
\codefieldsection{Alternative Names}
\begin{eczvaluelist}
\item\relax Random Clifford-circuit code
\end{eczvaluelist}
\eczhIndexCodeAliasName{random_stabilizer}{Random Clifford-circuit code}
\codefieldsection{Description}
An \(n\)-qubit, modular-qudit, or Galois-qudit stabilizer code whose construction is non-deterministic.
Since stabilizer encoders are Clifford circuits, such codes can be thought of as arising from random Clifford circuits.

\codefieldsection{Rate}
Random qubit stabilizer codes asymptotically saturate the non-degenerate quantum Hamming bound, and hence achieve the  \flmRefsHyperref{ref1729}{quantum GV bound}, because a typical random stabilizer has negligible degeneracy  \NoCaseChange{\protect\cite[{Sec. 7.6}]{cite736}}; see notes \NoCaseChange{\protect\cite{cite2764}}. In fact, sampling random CSS codes is sufficient  \NoCaseChange{\protect\cite{cite3196}}; see also the original random-coding argument \NoCaseChange{\protect\cite{cite3369}}.
\codefieldsection{Parent}
\begin{eczvaluelist}
\item\relax
\flmRefsHyperref[eczindexfamilyrel]{code:stabilizer}{Stabilizer code}\end{eczvaluelist}
\codefieldsection{Children}
\begin{eczvaluelist}
\item\relax
\flmRefsHyperref[eczindexfamilyrel]{code:approximate_log_depth}{Log-depth geometrically local Clifford-circuit code} --- Log-depth \flmRefsHyperref{ref409}{Clifford circuits} on a 1D geometry yield approximate codes whose encoding rate achieves the hashing bound for Pauli noise and the channel capacity for erasure errors \NoCaseChange{\protect\cite{cite3972,cite3973}}.
\item\relax
\flmRefsHyperref[eczindexfamilyrel]{code:crystalline_dynamic_gen}{Crystalline-circuit qubit code}\item\relax
\flmRefsHyperref[eczindexfamilyrel]{code:nonlocal_lowdepth}{Brown-Fawzi Clifford-circuit code}\end{eczvaluelist}
\codefieldsection{Cousins}
\begin{eczvaluelist}
\item\relax
\flmRefsHyperref[eczindexfamilyrel]{code:random_circuit}{Random-circuit code} --- Random stabilizer codes can be constructed by sampling random Clifford circuits.
\item\relax
\flmRefsHyperref[eczindexfamilyrel]{code:monitored_random_circuits}{Monitored random-circuit code} --- An important sub-family of monitored random-circuit codes is the Clifford monitored random-circuit family, where unitaries are sampled from the \flmRefsHyperref{ref409}{Clifford group} \NoCaseChange{\protect\cite{cite2888}}.
\item\relax
\flmRefsHyperref[eczindexfamilyrel]{code:holographic_tensor}{Holographic tensor-network code} --- Random holographic tensor-network codes reproduce many aspects of holography \NoCaseChange{\protect\cite{cite2856,cite2857,cite2865}}.
\item\relax
\flmRefsHyperref[eczindexfamilyrel]{code:generalized_quantum_divisible}{Generalized quantum divisible code} --- Random CSS codes \NoCaseChange{\protect\cite{cite3196}} can be used to construct families of \(\llbracket O(d^{\nu−1}), \Omega(d), d\rrbracket \) level-\(\nu\) generalized quantum divisible codes \NoCaseChange{\protect\cite[{Sec. VI.A}]{cite734}}.
\item\relax
\flmRefsHyperref[eczindexfamilyrel]{code:fiber_bundle}{Fiber-bundle code} --- Taking a random LDPC code as the base and a cyclic repetition code as the fiber yields, after distance balancing, a QLDPC code with distance of \flmRefsHyperref{ref65}{order} \(\Omega( n^{3/5}\text{polylog}(n) )\) and rate of \flmRefsHyperref{ref65}{order} \(\Omega( n^{-2/5}\text{polylog}(n) )\).
\item\relax
\flmRefsHyperref[eczindexfamilyrel]{code:homological_product}{Homological product code} --- Random homological codes are asymptotically good with high probability \NoCaseChange{\protect\cite[{Thm. 1}]{cite2562}}.
\item\relax
\flmRefsHyperref[eczindexfamilyrel]{code:qldpc}{Qubit QLDPC code} --- Random qubit QLDPC codes found by solving certain constraint satisfaction problems (CSPs) practically achieve the capacity of the erasure channel \NoCaseChange{\protect\cite{cite4254}}.
\item\relax
\flmRefsHyperref[eczindexfamilyrel]{code:qubit_css}{Qubit CSS code} --- Random CSS codes asymptotically achieve linear distance with high probability, achieving the \flmRefsHyperref{ref1729}{quantum GV bound} \NoCaseChange{\protect\cite{cite3196}}.
\item\relax
\flmRefsHyperref[eczindexfamilyrel]{code:clifford-deformed_surface}{Clifford-deformed surface code (CDSC)} --- Many useful CDSCs are constructed using random \flmRefsHyperref{ref409}{Clifford circuits}.
\item\relax
\flmRefsHyperref[eczindexfamilyrel]{code:compass_model}{Compass code} --- Compass code families are constructed by randomly assigning stabilizers to plaquettes of a square lattice.
\item\relax
\flmRefsHyperref[eczindexfamilyrel]{code:qudit_da}{Modular-qudit dynamical code} --- Dynamical codes admit instantaneous stabilizer groups, and dynamical code state initialization, logical gates, and error correction are done by a sequence of different (usually weight-two) stabilizer measurements.
\item\relax
\flmRefsHyperref[eczindexfamilyrel]{code:galois_grs}{Galois-qudit GRS code} --- Concatenations of Galois-qudit GRS codes and random stabilizer codes can achieve the \flmRefsHyperref{ref1729}{quantum GV bound} \NoCaseChange{\protect\cite{cite2704}}.
\end{eczvaluelist}
\eczhbkcontributors{ \eczhuVVA }
\endeczcode

\eczcode{rotor_cluster}{Rotor cluster-state code}{~\NoCaseChange{\protect\cite{cite507}}}
\codefieldsection{Description}
Rotor analogue of the qubit and analog cluster-state codes.
The exact rotor cluster state is non-normalizable, so approximate constructions have to be considered.
Defined from a real-valued weighted adjacency matrix of a graph \NoCaseChange{\protect\cite{cite507}}.

\codefieldsection{Parents}
\begin{eczvaluelist}
\item\relax
\flmRefsHyperref[eczindexfamilyrel]{code:rotor_stabilizer}{Rotor stabilizer code} --- Rotor cluster-state codes are particular rotor stabilizer codes.
\item\relax
\flmRefsHyperref[eczindexfamilyrel]{code:graph_quantum}{Graph quantum code} --- Graph quantum codes for \(G=\mathbb{Z}\) reduce to rotor cluster-state codes.
\end{eczvaluelist}
\codefieldsection{Cousin}
\begin{eczvaluelist}
\item\relax
\flmRefsHyperref[eczindexfamilyrel]{code:ame}{Perfect-tensor code} --- Rotor AME cluster states exist for any number of modes \NoCaseChange{\protect\cite{cite507}}.
\end{eczvaluelist}
\eczhbkcontributors{ \eczhuVVA }
\endeczcode

\eczcode{rotor}{Rotor code}{}
\codefieldsection{Alternative Names}
\begin{eczvaluelist}
\item\relax Angle-number code
\end{eczvaluelist}
\eczhIndexCodeAliasName{rotor}{Angle-number code}
\codefieldsection{Description}
Encodes a \textit{logical} Hilbert space, finite- or infinite-dimensional, into a \textit{physical} Hilbert space of \(L^2\)-normalizable functions on either the integers \(\mathbb Z\) or the circle group \(U(1)\).
This space is colloquially referred to as a (planar) rotor.
Ideal codewords may not be normalizable because the space is infinite-dimensional, so approximate versions have to be constructed in practice.

\codefieldsection{Protection}

\subsection{Rotor generalized Pauli error basis}
A rotor analogue of the Pauli string basis for \flmRefsHyperref{code:qubits_into_qubits}{qubit} codes consists of rotor \textit{generalized Pauli operators}.

\begin{defterm}{Rotor generalized Pauli strings}\label{ref5149}\label{ref5079}
For a single rotor, its elements are products of exponentials of the rotor's angular position (\(\hat\phi\)) and angular momentum (\(\hat L\)) operators, acting on the rotor's angular position states \(|\phi\rangle\) for \(\phi\in U(1)\) as
\flmMathEnvironment{align}{}{
  e^{-i\varphi\hat{L}}\left|\phi\right\rangle =\left|\phi+\varphi\right\rangle \,\,\text{ and }\,\,e^{i\ell\hat{\phi}}\left|\phi\right\rangle =e^{i\ell\phi}\left|\phi\right\rangle ~,
}
where \(\varphi\in U(1)\) and \(\ell\in\mathbb{Z}\).
For multiple rotors, error set elements are tensor products of elements of the single-rotor error set, characterized by vectors of angle and integer coefficients multiplying vectors of angular momentum \(\hat{\boldsymbol{L}}\) and angular position \(\hat{\boldsymbol{\phi}}\) operators.
These satisfy the usual Weyl-type commutation relations but do not violate the Stone-von Neumann theorem because \(\ell\) is restricted to be an integer (cf. \NoCaseChange{\protect\cite[{Exam. 14.5}]{cite5150}}).
\end{defterm}

\codefieldsection{Gates}
\begin{eczvaluelist}
\item\relax The normalizer of the \flmRefsHyperref{ref5079}{rotor Pauli group} is the \(n\)-\textit{rotor Clifford group} \NoCaseChange{\protect\cite{cite5151,cite735}}. The rotor Clifford group permutes \flmRefsHyperref{ref5079}{rotor Pauli} operators amongst themselves, and, up to any phases, is equivalent to \(U(1)^{n(n+1)/2} \rtimes GL(n,\mathbb{Z})\) \NoCaseChange{\protect\cite{cite2699}}.
\end{eczvaluelist}
\codefieldsection{Notes}
\begin{eczvaluelist}
\item\relax See Refs. \NoCaseChange{\protect\cite{cite5083,cite2531}\protect\cite[{Sec. IV}]{cite735}} for introductions to rotor Hilbert spaces.
\end{eczvaluelist}
\codefieldsection{Parent}
\begin{eczvaluelist}
\item\relax
\flmRefsHyperref[eczindexfamilyrel]{code:group_quantum}{Group-based quantum code} --- Group quantum codes whose physical spaces are constructed using either the group of the integers \(\mathbb{Z}\) or the circle group \(U(1)\) are rotor codes.
\end{eczvaluelist}
\codefieldsection{Child}
\begin{eczvaluelist}
\item\relax
\flmRefsHyperref[eczindexfamilyrel]{code:rotor_stabilizer}{Rotor stabilizer code}\end{eczvaluelist}
\codefieldsection{Cousins}
\begin{eczvaluelist}
\item\relax
\flmRefsHyperref[eczindexfamilyrel]{code:integers_into_integers}{Integer-based code} --- Rotor codes are quantum counterparts of integer codes.
\item\relax
\flmRefsHyperref[eczindexfamilyrel]{code:gkp}{Square-lattice GKP code} --- Because square-lattice GKP error states are parameterized by two modular (i.e., periodic) variables of position and momentum, measuring one of the GKP stabilizers constrains the oscillator Hilbert space into that of a rotor.
\end{eczvaluelist}
\eczhbkcontributors{ Daniel Burgarth, Austin Yubo He, \eczhuVVA }
\endeczcode

\eczcode{rotor_gkp}{Rotor GKP code}{~\NoCaseChange{\protect\cite{cite513,cite5152,cite735}}}
\codefieldsection{Description}
GKP code protecting against small angular position and momentum shifts of a planar rotor.
\codefieldsection{Protection}
In the standard \(Z_N\subset Z_{dN}\) construction, the \(d\)-dimensional codespace corrects angular shifts \(|\delta\phi|<\pi/(dN)\) and angular-momentum shifts \(|\delta\ell|<N/2\).

\codefieldsection{Parents}
\begin{eczvaluelist}
\item\relax
\flmRefsHyperref[eczindexfamilyrel]{code:rotor_stabilizer}{Rotor stabilizer code}\item\relax
\flmRefsHyperref[eczindexfamilyrel]{code:css}{Calderbank-Shor-Steane (CSS) stabilizer code} --- Rotor GKP code stabilizers are purely position and purely momentum rotor Pauli-type operators, making these codes CSS.
\item\relax
\flmRefsHyperref[eczindexfamilyrel]{code:single_subsystem}{Monolithic quantum code}\end{eczvaluelist}
\codefieldsection{Cousins}
\begin{eczvaluelist}
\item\relax
\flmRefsHyperref[eczindexfamilyrel]{code:gkp}{Square-lattice GKP code} --- GKP (rotor GKP) codes protect against shifts in linear (angular) degrees of freedom.
\item\relax
\flmRefsHyperref[eczindexfamilyrel]{code:quantum_concatenated}{Concatenated quantum code} --- The rotor GKP code can be thought of as a concatenation of a homological rotor code and a modular-qudit GKP code \NoCaseChange{\protect\cite[{Fig. 3}]{cite2699}}.
\item\relax
\flmRefsHyperref[eczindexfamilyrel]{code:number_phase}{Number-phase code} --- Number-phase codes are obtained by projecting planar-rotor GKP codes onto the non-negative angular-momentum subspace and identifying that subspace with oscillator Fock space \NoCaseChange{\protect\cite[{Ex. 3}]{cite2699}}.
\item\relax
\flmRefsHyperref[eczindexfamilyrel]{code:qudit_gkp}{Modular-qudit GKP code} --- The rotor GKP code can be thought of as a concatenation of a homological rotor code and a modular-qudit GKP code \NoCaseChange{\protect\cite[{Fig. 3}]{cite2699}}.
\end{eczvaluelist}
\eczhbkcontributors{ \eczhuVVA }
\endeczcode

\eczcode{rotor_stabilizer}{Rotor stabilizer code}{~\NoCaseChange{\protect\cite{cite5151}}}
\codefieldsection{Description}
Rotor code whose codespace is defined as the common \(+1\) eigenspace of a group of mutually commuting rotor generalized Pauli operators.
The stabilizer group can be either discrete or continuous, corresponding to modular or linear constraints on angular positions and momenta.
Both cases can yield finite or infinite logical dimension.
Exact codewords are non-normalizable, so approximate constructions have to be considered.

\codefieldsection{Parents}
\begin{eczvaluelist}
\item\relax
\flmRefsHyperref[eczindexfamilyrel]{code:rotor}{Rotor code}\item\relax
\flmRefsHyperref[eczindexfamilyrel]{code:stabilizer}{Stabilizer code}\end{eczvaluelist}
\codefieldsection{Children}
\begin{eczvaluelist}
\item\relax
\flmRefsHyperref[eczindexfamilyrel]{code:homological_rotor}{Homological rotor code} --- Homological rotor codes are rotor CSS codes constructed from chain complexes over the integers in an extension of the \flmRefsHyperref{ref683}{qubit CSS-to-homology correspondence} to rotors.
\item\relax
\flmRefsHyperref[eczindexfamilyrel]{code:rotor_5_1_3}{\(\llbracket 5,1,3\rrbracket _{\mathbb{Z}}\) Five-rotor code}\item\relax
\flmRefsHyperref[eczindexfamilyrel]{code:rotor_cluster}{Rotor cluster-state code} --- Rotor cluster-state codes are particular rotor stabilizer codes.
\item\relax
\flmRefsHyperref[eczindexfamilyrel]{code:rotor_gkp}{Rotor GKP code}\end{eczvaluelist}
\codefieldsection{Cousins}
\begin{eczvaluelist}
\item\relax
\flmRefsHyperref[eczindexfamilyrel]{code:css}{Calderbank-Shor-Steane (CSS) stabilizer code} --- A rotor stabilizer code admitting a set of generators such that each generator consists of either angular position or angular momentum operators is a CSS code.
\item\relax
\flmRefsHyperref[eczindexfamilyrel]{code:qudit_stabilizer}{Modular-qudit stabilizer code} --- By combining the paper's bounded-phase-space and integer-local-dimension constructions, prime-qudit stabilizer codes can be algebraically imported into rotor-code settings \NoCaseChange{\protect\cite[{Sec. 3.2}]{cite4580}}.
\item\relax
\flmRefsHyperref[eczindexfamilyrel]{code:galois_stabilizer}{Galois-qudit stabilizer code} --- Galois-qudit stabilizer codes can be imported into integral-domain settings, and rotor codes and their parameters can be obtained through a synthesis of these cases \NoCaseChange{\protect\cite[{Thm. 13; Sec. 3.2}]{cite4580}}.
\end{eczvaluelist}
\eczhbkcontributors{ \eczhuVVA }
\endeczcode

\eczcode{stabilizer}{Stabilizer code}{}
\codefieldsection{Description}
A code whose logical subspace is the joint eigenspace (usually with eigenvalue \(+1\)) of a set of commuting unitary Pauli-type operators forming the code's stabilizer group.
They can be block codes defined on tensor-product spaces of qubits or qudits, or non-block codes defined on single sufficiently large Hilbert spaces such as bosonic modes or other Abelian group spaces.

The coding theory motivation for stabilizer codes came from linear binary codes, whose codewords form a closed subspace in the space of binary strings.
Stabilizer codes extend this property, in various ways, to quantum error correction.
Stabilizer codes can be defined succinctly using the stabilizer group generators and without explicitly writing out a basis of codewords.

Stabilizer codes were originally defined for qubits, where the relevant commuting operators are tensor products of Pauli matrices.
The Pauli stabilizer structure is useful in providing standardized encoding, gates, decoding, and performance bounds.
Elements of this structure remain in qudit extensions, in particular for prime-dimensional modular qudits and Galois qudits.
Infinite-dimensional Pauli-type bases yield the bosonic stabilizer and rotor stabilizer codes.

One can switch between stabilizer codes by appending another stabilizer group and taking the center of the resulting larger group.
\begin{defterm}{Stabilizer code switching, code deformation, update rule, or code rewiring}\label{ref5153}\label{ref410}
Stabilizer code switching is a map between stabilizer codes that is done using a stabilizer group \(\mathsf{F}\),
\flmMathEnvironment{align}{}{
\mathsf{S}\to\mathsf{N}_{\left\langle \mathsf{S},\mathsf{F}\right\rangle }\left(\mathsf{F}\right)~,
}
where \(\mathsf{N}\) denotes taking the normalizer of a group (e.g., see \NoCaseChange{\protect\cite{cite2579,cite3211}} for proofs).
Code switching may not preserve the logical information and instead implement logical measurements; conditions on \(\mathsf{S}\) and \(\mathsf{F}\) such that qubit stabilizer \flmRefsHyperref{ref410}{code switching} preserves logical information are derived in \NoCaseChange{\protect\cite[{Prop. II.1}]{cite5154}}.
The stabilizer rewiring algorithm (SRA) allows for code switching between a pair of compatible stabilizer codes \NoCaseChange{\protect\cite{cite5155}} (see also Ref. \NoCaseChange{\protect\cite{cite475,cite5156}}), and ancillary qubits may be used to maintain minimum distance of any intermediate codes \NoCaseChange{\protect\cite{cite5157}}.
Clifford operations and Pauli measurements can be expressed as sequences of \flmRefsHyperref{ref410}{code switching} \NoCaseChange{\protect\cite{cite4420}}.
In the context of stabilizer codes realizing Abelian topological phases, \flmRefsHyperref{ref410}{code switching} implements \textit{anyon condensation} of anyons represented by operators in the group \(\mathsf{F}\).
Code switching can be done using only transversal gates for qubit stabilizer codes \NoCaseChange{\protect\cite{cite4374}}.
\end{defterm}

Extensions of the stabilizer formalism, such as \flmRefsHyperref{code:xs_stabilizer}{XS} and \flmRefsHyperref{code:xp_stabilizer}{XP} stabilizer codes, relax the mutual commutation property.
Other extensions, such as \flmRefsHyperref{code:cws}{CWS} and \flmRefsHyperref{code:non_stabilizer}{union stabilizer} codes, enlarge the codespace by re-assigning error words as codewords.

\codefieldsection{Protection}
The group of all Pauli-type operators typically serves as the set of noise operators for stabilizer codes.

\codefieldsection{Gates}
\begin{eczvaluelist}
\item\relax A Gottesman-Knill-type theorem exists for qubits, modular qudits, Galois qudits, and rotors \NoCaseChange{\protect\cite{cite5158,cite5159}}, as well as oscillators \NoCaseChange{\protect\cite{cite4785,cite4786,cite4787}}. Stabilizer codes can be described by quadratic functions over Abelian groups \NoCaseChange{\protect\cite{cite5160}}.
\end{eczvaluelist}
\codefieldsection{Decoding}
\begin{eczvaluelist}
\item\relax The structure of stabilizer codes allows for straightforward syndrome-based decoding because the stabilizer generators serve as the code's check operators, and their eigenvalues serve as the error syndromes. The error correction process involves measuring the stabilizer generators and applying correcting Pauli-type operators based on the measurement outcomes.
\end{eczvaluelist}
\codefieldsection{Notes}
\begin{eczvaluelist}
\item\relax See \NoCaseChange{\protect\cite{cite2733}} for a pedagogical introduction to stabilizer codes.
\end{eczvaluelist}
\codefieldsection{Parents}
\begin{eczvaluelist}
\item\relax
\flmRefsHyperref[eczindexfamilyrel]{code:group_quantum}{Group-based quantum code} --- Stabilizer codes are constructed out of Pauli strings, modular-qudit Pauli strings, Galois-qudit Pauli strings, oscillator displacement operators, or rotor generalized Pauli strings. All of these are examples of the Weyl-Heisenberg group on Manin's quantum plane, which is defined on a configuration space that is generally a free Abelian group \NoCaseChange{\protect\cite{cite4537,cite4538,cite4539,cite4540}}.
\item\relax
\flmRefsHyperref[eczindexfamilyrel]{code:commuting_projector}{Commuting-projector Hamiltonian code} --- Codespace is the ground-state space of the \textit{code Hamiltonian}, which consists of an equal linear combination of stabilizer generators and which can be made into a frustration-free commuting-projector Hamiltonian.
\item\relax
\flmRefsHyperref[eczindexfamilyrel]{code:frustration_free}{Frustration-free Hamiltonian code} --- Codespace is the ground-state space of the \textit{code Hamiltonian}, which consists of an equal linear combination of stabilizer generators and which can be made into a frustration-free commuting-projector Hamiltonian.
\item\relax
\flmRefsHyperref[eczindexfamilyrel]{code:knill}{Knill code} --- Stabilizer codes are Knill codes whose \flmRefsHyperref{ref2812}{nice error basis} is either the Pauli strings, modular-qudit Pauli strings, Galois-qudit Pauli strings, oscillator displacement operators, or rotor generalized Pauli strings.
\end{eczvaluelist}
\codefieldsection{Children}
\begin{eczvaluelist}
\item\relax
\flmRefsHyperref[eczindexfamilyrel]{code:lca_stabilizer}{Locally compact Abelian (LCA) stabilizer code}\item\relax
\flmRefsHyperref[eczindexfamilyrel]{code:rotor_stabilizer}{Rotor stabilizer code}\item\relax
\flmRefsHyperref[eczindexfamilyrel]{code:css}{Calderbank-Shor-Steane (CSS) stabilizer code}\item\relax
\flmRefsHyperref[eczindexfamilyrel]{code:graph_quantum}{Graph quantum code} --- Graph quantum codes are a subset of stabilizer codes over \(G\)-valued qudits for Abelian \(G\) \NoCaseChange{\protect\cite{cite3561}}. Any stabilizer code over Abelian \(G\) is locally equivalent to a graph quantum code \NoCaseChange{\protect\cite{cite3561}} (see also \NoCaseChange{\protect\cite{cite3536,cite867}}).
\item\relax
\flmRefsHyperref[eczindexfamilyrel]{code:qlwc}{Quantum low-weight check (QLWC) code}\item\relax
\flmRefsHyperref[eczindexfamilyrel]{code:random_stabilizer}{Random stabilizer code}\item\relax
\flmRefsHyperref[eczindexfamilyrel]{code:oscillator_stabilizer}{Bosonic stabilizer code}\item\relax
\flmRefsHyperref[eczindexfamilyrel]{code:qudit_stabilizer}{Modular-qudit stabilizer code}\item\relax
\flmRefsHyperref[eczindexfamilyrel]{code:galois_stabilizer}{Galois-qudit stabilizer code}\end{eczvaluelist}
\codefieldsection{Cousins}
\begin{eczvaluelist}
\item\relax
\flmRefsHyperref[eczindexfamilyrel]{code:topological_abelian}{Abelian topological code} --- There is a general correspondence between stabilizer codes and gauge theory, with the stabilizer group playing the role of the gauge group \NoCaseChange{\protect\cite{cite1365}}.
\item\relax
\flmRefsHyperref[eczindexfamilyrel]{code:subsystem_stabilizer}{Subsystem stabilizer code} --- Subsystem stabilizer codes reduce to stabilizer codes when there are no gauge degrees of freedom.
\item\relax
\flmRefsHyperref[eczindexfamilyrel]{code:majorana_stab}{Majorana stabilizer code} --- Majorana stabilizer codes are useful for Majorana-based architectures, where the degrees of freedom are electrons, and the notion of locality is different than all other code kingdoms.
\end{eczvaluelist}
\eczhbkcontributors{ \eczhuVVA }
\endeczcode

\eczcode{subsystem_css}{Subsystem CSS code}{}
\codefieldsection{Description}
A subsystem stabilizer code admitting a set of gauge-group generators that are either \(Z\)-type or \(X\)-type operators.
This ensures that the associated stabilizer group is also CSS.

\codefieldsection{Parent}
\begin{eczvaluelist}
\item\relax
\flmRefsHyperref[eczindexfamilyrel]{code:subsystem_stabilizer}{Subsystem stabilizer code}\end{eczvaluelist}
\codefieldsection{Children}
\begin{eczvaluelist}
\item\relax
\flmRefsHyperref[eczindexfamilyrel]{code:qudit_subsystem_css}{Subsystem modular-qudit CSS code}\item\relax
\flmRefsHyperref[eczindexfamilyrel]{code:galois_subsystem_css}{Subsystem Galois-qudit CSS code}\end{eczvaluelist}
\codefieldsection{Cousin}
\begin{eczvaluelist}
\item\relax
\flmRefsHyperref[eczindexfamilyrel]{code:css}{Calderbank-Shor-Steane (CSS) stabilizer code} --- Subsystem CSS codes reduce to CSS stabilizer codes when there are no gauge degrees of freedom.
\end{eczvaluelist}
\eczhbkcontributors{ \eczhuVVA }
\endeczcode

\eczcode{subsystem_group_quantum}{Subsystem group-based quantum code}{}

\codefieldsection{Kingdom root code}
for the \flmRefsHyperref{kingdom:group_quantum}{Group quantum Kingdom}.
\codefieldsection{Description}
Group-based quantum code whose codespace admits a tensor-product decomposition into logical and gauge factors.

\codefieldsection{Parent}
\begin{eczvaluelist}
\item\relax
\flmRefsHyperref[eczindexfamilyrel]{code:oecc}{Subsystem QECC}\end{eczvaluelist}
\codefieldsection{Children}
\begin{eczvaluelist}
\item\relax
\flmRefsHyperref[eczindexfamilyrel]{code:subsystem_stabilizer}{Subsystem stabilizer code}\item\relax
\flmRefsHyperref[eczindexfamilyrel]{code:subsystem_qudits_into_qudits}{Subsystem modular-qudit code} --- Subsystem group quantum codes whose physical spaces are constructed using modular-integer groups \(\mathbb{Z}_q\) are subsystem modular-qudit codes.
\item\relax
\flmRefsHyperref[eczindexfamilyrel]{code:subsystem_galois_into_galois}{Subsystem Galois-qudit code} --- A Galois qudit for \(q=p^m\) can be decomposed into a Kronecker product of \(m\) modular qudits \NoCaseChange{\protect\cite{cite696}}; see \NoCaseChange{\protect\cite[{Sec. 5.3}]{cite697}}.
Interpreted this way, subsystem Galois-qudit codes are subsystem group quantum codes whose physical spaces are constructed using Galois fields \(\mathbb{F}_q\) as groups. More general versions of such qudits can be valued in a Galois ring \NoCaseChange{\protect\cite{cite4621}}, over which there also exists a Fourier transform \NoCaseChange{\protect\cite{cite4622}}.

\end{eczvaluelist}
\codefieldsection{Cousins}
\begin{eczvaluelist}
\item\relax
\flmRefsHyperref[eczindexfamilyrel]{code:group_quantum}{Group-based quantum code} --- Subsystem group-based quantum codes reduce to (subspace) group-based quantum codes when there is no gauge subsystem.
\item\relax
\flmRefsHyperref[eczindexfamilyrel]{code:quantum_double}{Quantum-double code} --- Subsystem versions of quantum-double codes have been formulated \NoCaseChange{\protect\cite{cite5147}}.
\end{eczvaluelist}
\eczhbkcontributors{ \eczhuVVA }
\endeczcode

\eczcode{subsystem_stabilizer}{Subsystem stabilizer code}{}
\codefieldsection{Description}
A subsystem code that is derived from a stabilizer code by assigning some factors of the stabilizer code's logical tensor-product structure to gauge degrees of freedom.

\codefieldsection{Parent}
\begin{eczvaluelist}
\item\relax
\flmRefsHyperref[eczindexfamilyrel]{code:subsystem_group_quantum}{Subsystem group-based quantum code}\end{eczvaluelist}
\codefieldsection{Children}
\begin{eczvaluelist}
\item\relax
\flmRefsHyperref[eczindexfamilyrel]{code:sparse_subsystem}{QLDPC subsystem code}\item\relax
\flmRefsHyperref[eczindexfamilyrel]{code:subsystem_css}{Subsystem CSS code}\item\relax
\flmRefsHyperref[eczindexfamilyrel]{code:qudit_subsystem_stabilizer}{Subsystem modular-qudit stabilizer code}\item\relax
\flmRefsHyperref[eczindexfamilyrel]{code:galois_subsystem_stabilizer}{Subsystem Galois-qudit stabilizer code}\end{eczvaluelist}
\codefieldsection{Cousin}
\begin{eczvaluelist}
\item\relax
\flmRefsHyperref[eczindexfamilyrel]{code:stabilizer}{Stabilizer code} --- Subsystem stabilizer codes reduce to stabilizer codes when there are no gauge degrees of freedom.
\end{eczvaluelist}
\eczhbkcontributors{ \eczhuVVA }
\endeczcode

\eczcode{tqd}{Twisted quantum double (TQD) code}{~\NoCaseChange{\protect\cite{cite5161,cite571,cite406}}}
\codefieldsection{Alternative Names}
\begin{eczvaluelist}
\item\relax 2D Dijkgraaf-Witten gauge theory code
\end{eczvaluelist}
\eczhIndexCodeAliasName{tqd}{2D Dijkgraaf-Witten gauge theory code}
\codefieldsection{Description}
Code whose codewords realize a 2D topological order rendered by a Chern-Simons topological field theory.
The corresponding anyon theory is defined by a finite group \(G\) and a 3-cocycle \(\omega\in H^3( G, U(1) )\) \NoCaseChange{\protect\cite{cite605,cite571,cite405}}.
Canonical TQD models \NoCaseChange{\protect\cite{cite571}} are defined on group-valued qudits.

Logical dimension is determined by the genus of the underlying surface (for closed surfaces), types of boundaries (for open surfaces), and any twist defects present.
Excitations are described by the twisted quantum double (a.k.a. twisted Drinfeld double) \(D^{\omega}(G)\).
Gapped boundaries of the models are classified by a subgroup \(K \subseteq G\) and a particular two-cochain \NoCaseChange{\protect\cite{cite5135}}.

\codefieldsection{Protection}
These models realize local topological order (LTO) \NoCaseChange{\protect\cite{cite5162}}.

\codefieldsection{Encoding}
\begin{eczvaluelist}
\item\relax For any solvable group \(G\), ground-state preparation can be done with an adaptive constant-depth circuit with geometrically local gates and measurements throughout \NoCaseChange{\protect\cite{cite5139}}.
\end{eczvaluelist}
\codefieldsection{Parents}
\begin{eczvaluelist}
\item\relax
\flmRefsHyperref[eczindexfamilyrel]{code:dijkgraaf_witten}{Dijkgraaf-Witten gauge theory code} --- Restricting Dijkgraaf-Witten gauge theory to a 2D manifold reproduces the phase of the TQD model \NoCaseChange{\protect\cite{cite605}}.
The Drinfeld center of the category \(\text{Vec}^{\omega}(G)\) is used to describe bulk excitations of 3D Dijkgraaf-Witten models, and this center is equivalent to the twisted quantum double \(D^{\omega}(G)\) \NoCaseChange{\protect\cite[{pg. 41}]{cite587}}.
TQD codewords are gauge-invariant boundary states of a 3D Dijkgraaf-Witten theory \NoCaseChange{\protect\cite[{Sec. IX}]{cite571}}.

\item\relax
\flmRefsHyperref[eczindexfamilyrel]{code:string_net}{String-net code} --- String-net models realize TQDs for categories \(\text{Vec}^{\omega}G\), where \(G\) is a finite group and \(\omega\) is a 3-cocycle on \(G\). There is a duality between a large class of string-net models and certain TQD models \NoCaseChange{\protect\cite{cite571}}.
\end{eczvaluelist}
\codefieldsection{Children}
\begin{eczvaluelist}
\item\relax
\flmRefsHyperref[eczindexfamilyrel]{code:quantum_double}{Quantum-double code} --- The anyon theory corresponding to a quantum-double code is a TQD with trivial cocycle.
\item\relax
\flmRefsHyperref[eczindexfamilyrel]{code:tqd_abelian}{Abelian TQD code} --- The anyon theory corresponding to Abelian TQD codes is defined by an Abelian group and a Type-I, Type-II, or Type-III 3-cocycle.
Abelian TQDs with Type-I and -II cocycles account for all 2D Abelian topological orders that admit gapped boundaries \NoCaseChange{\protect\cite{cite573}}.

\item\relax
\flmRefsHyperref[eczindexfamilyrel]{code:hexagonal_cz}{Hexagonal \(CZ\) code} --- The ground-state subspace of the hexagonal \(CZ\) code realizes the topological order of the Type-III \(G=\mathbb{Z}^3_2\) Abelian TQD model \NoCaseChange{\protect\cite{cite575,cite576}}, which is the same topological order as the \(G=D_4\) quantum double \NoCaseChange{\protect\cite{cite577}}. There is a constant-depth circuit implementing a transversal logical \(T\) gate via an emergent automorphism symmetry of the underlying \(\mathbb{D}_4\) topological order \NoCaseChange{\protect\cite{cite725}}.
\end{eczvaluelist}
\codefieldsection{Cousins}
\begin{eczvaluelist}
\item\relax
\flmRefsHyperref[eczindexfamilyrel]{code:spt}{Symmetry-protected topological (SPT) code} --- A TQD code Hamiltonian can be obtained by gauging \NoCaseChange{\protect\cite{cite462,cite463,cite233,cite464,cite465,cite466,cite467,cite468,cite469,cite470}} the symmetry of a particular 2D SPT model. The same group and cocycle data classifies both 2D SPTs and TQDs \NoCaseChange{\protect\cite{cite462,cite3072}}.
\item\relax
\flmRefsHyperref[eczindexfamilyrel]{code:tqt}{Twisted quantum triple (TQT) code} --- The TQT model can be thought of as a 3D version of the TQD model.
\end{eczvaluelist}
\eczhbkcontributors{ \eczhuVVA }
\endeczcode

\eczcode{tqt}{Twisted quantum triple (TQT) code}{~\NoCaseChange{\protect\cite{cite586,cite606,cite5163,cite5164}}}
\codefieldsection{Alternative Names}
\begin{eczvaluelist}
\item\relax 3D Dijkgraaf-Witten gauge theory code
\item\relax 3D twisted quantum-double code
\end{eczvaluelist}
\eczhIndexCodeAliasName{tqt}{3D Dijkgraaf-Witten gauge theory code}
\eczhIndexCodeAliasName{tqt}{3D twisted quantum-double code}
\codefieldsection{Description}
Group-based code realizing a 3D topological order rendered by a Dijkgraaf-Witten gauge theory.
The corresponding anyon theory is defined by a finite group \(G\) and a Type-IV four-cocycle \(\omega\).
Canonical TQT models \NoCaseChange{\protect\cite{cite586,cite606}} and other formulations whose ground states are in the same phase are all defined on group-valued qudits.

Boundaries and excitations have been studied in Refs. \NoCaseChange{\protect\cite{cite5165,cite5166,cite5167}}.
Gapped boundaries are classified by a subgroup \(K \subseteq G\) and a particular three-cochain \NoCaseChange{\protect\cite{cite5165}}.
Generalizations of Ocneanu's tube algebras \NoCaseChange{\protect\cite{cite5068,cite5069}} can be used to characterize excitations, which are described by the tube algebra of the category \(\text{Vec}^{\omega}(G)\) \NoCaseChange{\protect\cite{cite587,cite5168}}.

\codefieldsection{Parent}
\begin{eczvaluelist}
\item\relax
\flmRefsHyperref[eczindexfamilyrel]{code:dijkgraaf_witten}{Dijkgraaf-Witten gauge theory code} --- Restricting Dijkgraaf-Witten gauge theory to a 3D manifold reproduces the phase of the TQT model.

\end{eczvaluelist}
\codefieldsection{Children}
\begin{eczvaluelist}
\item\relax
\flmRefsHyperref[eczindexfamilyrel]{code:quantum_triple}{Quantum-triple code} --- The anyon theory corresponding to a quantum-triple code is a TQT with trivial cocycle.
\item\relax
\flmRefsHyperref[eczindexfamilyrel]{code:3d_semion}{Chiral semion Walker-Wang model code} --- When treated as ground states of the code Hamiltonian, the code states realize 3D double-semion topological order, a topological phase of matter that exists as the deconfined phase of the 3D twisted \(\mathbb{Z}_2\) gauge theory \NoCaseChange{\protect\cite{cite584}}.
\end{eczvaluelist}
\codefieldsection{Cousin}
\begin{eczvaluelist}
\item\relax
\flmRefsHyperref[eczindexfamilyrel]{code:tqd}{Twisted quantum double (TQD) code} --- The TQT model can be thought of as a 3D version of the TQD model.
\end{eczvaluelist}
\eczhbkcontributors{ \eczhuVVA }
\endeczcode

\eczcode{zero_pi}{Zero-pi qubit code}{~\NoCaseChange{\protect\cite{cite4972,cite5169,cite397}}}
\codefieldsection{Description}
A \(\llbracket 2,(0,2),(2,1)\rrbracket _{\mathbb{Z}}\) homological rotor code on the smallest tiling of the projective plane \(\mathbb{R}P^2\).
The ideal code can be obtained from a four-rotor Josephson-junction \NoCaseChange{\protect\cite{cite396}} system after a choice of grounding \NoCaseChange{\protect\cite{cite397}}.

Logical codewords can be expressed in the basis of angular momentum states as
\flmMathEnvironment{align}{}{
\begin{split}
  |\overline{0}\rangle&=\sum_{\ell\in\mathbb{Z}}\left|2\ell,-2\ell\right\rangle \\|\overline{1}\rangle&=\sum_{\ell\in\mathbb{Z}}\left|2\ell+1,-2\ell-1\right\rangle~.
\end{split}
}
An alternative codeword basis in terms of angular position states is
\flmMathEnvironment{align}{}{
\begin{split}
  |\overline{+}\rangle&=\intop_{U(1)}\textnormal{d}\phi\left|\phi,\phi\right\rangle \\|\overline{-}\rangle&=\intop_{U(1)}\textnormal{d}\phi\left|\phi,\phi+\pi\right\rangle~.
\end{split}
}

\codefieldsection{Protection}
Protection in the context of superconducting circuits was investigated in Ref. \NoCaseChange{\protect\cite{cite5170}}.

\codefieldsection{Gates}
\begin{eczvaluelist}
\item\relax One- and two-qubit phase gates utilizing ancillary oscillators in GKP states \NoCaseChange{\protect\cite{cite4971,cite4972}}.
\item\relax Protected phase gate \NoCaseChange{\protect\cite{cite5171}}.
\end{eczvaluelist}
\codefieldsection{Fault Tolerance}
\begin{eczvaluelist}
\item\relax One- and two-qubit phase gate errors can be suppressed \NoCaseChange{\protect\cite{cite4972}}.
\end{eczvaluelist}
\codefieldsection{Realizations}
\begin{eczvaluelist}
\item\relax A related superconducting circuit has been realized by the Houck group \NoCaseChange{\protect\cite{cite5172}}.
\end{eczvaluelist}
\codefieldsection{Notes}
\begin{eczvaluelist}
\item\relax The zero-pi qubit is based on earlier blueprints for protected subspaces using superconducting circuits \NoCaseChange{\protect\cite{cite5173,cite5174}}.
\end{eczvaluelist}
\codefieldsection{Parents}
\begin{eczvaluelist}
\item\relax
\flmRefsHyperref[eczindexfamilyrel]{code:homological_rotor}{Homological rotor code}\item\relax
\flmRefsHyperref[eczindexfamilyrel]{code:small_distance_quantum}{Small-distance block quantum code}\end{eczvaluelist}
\codefieldsection{Cousin}
\begin{eczvaluelist}
\item\relax
\flmRefsHyperref[eczindexfamilyrel]{code:gkp}{Square-lattice GKP code} --- Zero-pi code phase gates utilize ancillary oscillators in square-lattice GKP states \NoCaseChange{\protect\cite{cite4971,cite4972}}.
\end{eczvaluelist}
\eczhbkcontributors{ \eczhuVVA }
\endeczcode

\onecolumngrid
\clearpage

\section{Homogeneous-space quantum Kingdom}

\begin{eczEpigraph}
\begin{quote}
\flmQuoteSetup{quote}%
Always remember that you are absolutely unique. Just like everyone else.
\flmQuoteAttributed{Margaret Mead}
\end{quote}
\end{eczEpigraph}

\twocolumngrid

\eczcode{diatomic_molecular}{Diatomic molecular code}{~\NoCaseChange{\protect\cite[{Sec. VI}]{cite735}}}
\codefieldsection{Description}
Approximate quantum code that encodes a qudit in the infinite-dimensional Hilbert space of a rigid body with \(SO(2)\) symmetry, e.g., a heteronuclear diatomic molecule.
The physical space is \(L^2(S^2)\), equivalently the orientation space \(SO(3)/SO(2)\) of a linear rotor, consisting of a direct sum of all non-negative integer angular momenta.
Ideal codewords may not be normalizable because the space is infinite-dimensional, so approximate versions have to be constructed in practice.

Construction is based on nested subgroups \(H\subset K \subset SO(3)\), where \(H,K\) are finite.
Codewords consist of orbits of particular position states under \(H\), while some elements of \(K\) can cycle between codewords.

\codefieldsection{Protection}
Protects against sufficiently small rotations about any axis and small kicks in angular momentum.
In the simplest cyclic family, angular-momentum kicks with \(\ell<N/2\) are correctable.
But unlike molecular codes on \(SO(3)\), these codes cannot in general protect against arbitrary products of such rotations and kicks because the underlying state space \(S^2\) is not a group and rotations on \(S^2\) have fixed points.

\codefieldsection{Parent}
\begin{eczvaluelist}
\item\relax
\flmRefsHyperref[eczindexfamilyrel]{code:homogeneous_space_quantum}{Homogeneous-space quantum code} --- Diatomic molecular codes are defined on the space of orientations of a heteronuclear diatomic molecule, equivalently the space of normalizable functions on the two-sphere homogeneous space \(SO(3)/SO(2)=S^2\).
\end{eczvaluelist}
\codefieldsection{Cousins}
\begin{eczvaluelist}
\item\relax
\flmRefsHyperref[eczindexfamilyrel]{code:molecular}{Molecular code} --- Molecular codes live on \(SO(3)\) for asymmetric rigid bodies, whereas diatomic molecular codes live on the homogeneous space \(S^2=SO(3)/SO(2)\) for linear rotors.
\item\relax
\flmRefsHyperref[eczindexfamilyrel]{code:ae}{Æ code} --- Diatomic molecular codes are supported on states with various total angular momenta, while Æ codes are supported on only one subspace of fixed total momentum. The latter codes are more practical and applicable to other spin spaces.
\end{eczvaluelist}
\eczhbkcontributors{ \eczhuVVA }
\endeczcode

\eczcode{fiber}{Fiber code}{~\NoCaseChange{\protect\cite{cite5121}}}
\codefieldsection{Description}
Approximate quantum code that encodes a qudit in an induced representation \(\text{Ind}_G^{SO(3)} \Gamma\), induced by the irrep \(\Gamma\) of a subgroup \(G\subset SO(3)\). Such spaces correspond to orientation state spaces of certain symmetric molecules \NoCaseChange{\protect\cite{cite5121}}, with the base space \(SO(3)/G\) labeling the orientations and the fiber carrying the \(\Gamma\)-irrep encoding the logical information.

\codefieldsection{Protection}
A basis of noise operators is developed for general induced representations in Ref. \NoCaseChange{\protect\cite{cite5121}}.

\codefieldsection{Parent}
\begin{eczvaluelist}
\item\relax
\flmRefsHyperref[eczindexfamilyrel]{code:homogeneous_space_quantum}{Homogeneous-space quantum code} --- Fiber codes are defined on an induced representation \(\text{Ind}_G^{SO(3)} \Gamma\), induced by the irrep \(\Gamma\) of a subgroup \(G\subset SO(3)\).
\end{eczvaluelist}
\codefieldsection{Cousin}
\begin{eczvaluelist}
\item\relax
\flmRefsHyperref[eczindexfamilyrel]{code:molecular}{Molecular code} --- Molecular codes encode quantum information into superpositions of multiple orientations of an asymmetric molecule \NoCaseChange{\protect\cite{cite735}}, while fiber codes encode into the fiber associated with a single orientation of certain symmetric molecules \NoCaseChange{\protect\cite{cite5121}}.
\end{eczvaluelist}
\eczhbkcontributors{ \eczhuVVA }
\endeczcode

\eczcode{homogeneous_space_quantum}{Homogeneous-space quantum code}{}
\codefieldsection{Alternative Names}
\begin{eczvaluelist}
\item\relax Coset-space quantum code
\item\relax \(G/H\) quantum code
\end{eczvaluelist}
\eczhIndexCodeAliasName{homogeneous_space_quantum}{Coset-space quantum code}
\eczhIndexCodeAliasName{homogeneous_space_quantum}{\(G/H\) quantum code}

\codefieldsection{Kingdom root code}
for the \flmRefsHyperref{kingdom:homogeneous_space_quantum}{Homogeneous-space quantum Kingdom}.
\codefieldsection{Description}
Encodes a \textit{logical} Hilbert space, finite- or infinite-dimensional, into a \textit{physical} Hilbert space of \(L^2\)-normalizable functions on a homogeneous space \(G/H\) or, more generally, induced representations whose base space is \(G/H\) \NoCaseChange{\protect\cite{cite21,cite22,cite23,cite24,cite25}}. Here, \(G\) is a second-countable unimodular group, and \(H\) is a closed subgroup of \(G\).

\codefieldsection{Protection}
\flmRefsHyperref{ref672}{Quantum weight enumerators}, linear programming bounds, and Rains shadow enumerators have been extended to quantum codes defined on multiplicity-free two-point homogeneous spaces \NoCaseChange{\protect\cite{cite564}}.
In this multiplicity-free setting (the analogue of a Gelfand-pair decomposition), irreducible-representation labels entirely parameterize the Fourier basis useful for these bounds.

\codefieldsection{Parent}
\begin{eczvaluelist}
\item\relax
\flmRefsHyperref[eczindexfamilyrel]{code:qecc}{Quantum error-correcting code (QECC)}\end{eczvaluelist}
\codefieldsection{Children}
\begin{eczvaluelist}
\item\relax
\flmRefsHyperref[eczindexfamilyrel]{code:group_quantum}{Group-based quantum code} --- Homogeneous spaces \(G/H\) for trivial \(H\) reduce to group spaces. A group-\(G\) space can also be thought of as a multiplicity-free homogeneous space \((G\times G) / G\) \NoCaseChange{\protect\cite[{pg. 60}]{cite2474}}.
\item\relax
\flmRefsHyperref[eczindexfamilyrel]{code:diatomic_molecular}{Diatomic molecular code} --- Diatomic molecular codes are defined on the space of orientations of a heteronuclear diatomic molecule, equivalently the space of normalizable functions on the two-sphere homogeneous space \(SO(3)/SO(2)=S^2\).
\item\relax
\flmRefsHyperref[eczindexfamilyrel]{code:fiber}{Fiber code} --- Fiber codes are defined on an induced representation \(\text{Ind}_G^{SO(3)} \Gamma\), induced by the irrep \(\Gamma\) of a subgroup \(G\subset SO(3)\).
\item\relax
\flmRefsHyperref[eczindexfamilyrel]{code:tesselation}{Hyperbolic tessellation code} --- Hyperbolic tessellation codes are defined on the space of functions on the hyperbolic plane, the symmetric space \(G/H\) for \(G = SO(2,1)\) the proper Lorentz group and \(H = O(2)\).
\end{eczvaluelist}
\codefieldsection{Cousins}
\begin{eczvaluelist}
\item\relax
\flmRefsHyperref[eczindexfamilyrel]{code:homogeneous_space_classical}{Homogeneous-space code} --- Homogeneous-space quantum codes are quantum counterparts of homogeneous-space codes.
\item\relax
\flmRefsHyperref[eczindexfamilyrel]{code:landau_level}{Landau-level spin code} --- The Landau-level spin code lies in a particular irrep present in the induced representation \(\text{Ind}_{U(1)}^{SU(2)} \lambda\), where \(\lambda\in \mathbb{Z}\) labels irreps of \(U(1)\) and quantifies the monopole strength \NoCaseChange{\protect\cite{cite5035}}.
\end{eczvaluelist}
\eczhbkcontributors{ \eczhuVVA }
\endeczcode

\eczcode{tesselation}{Hyperbolic tessellation code}{~\NoCaseChange{\protect\cite{cite2811}}}
\codefieldsection{Description}
Homogeneous-space code whose codewords are superpositions of positional delta functions on the hyperbolic plane.
The positions are chosen according to regular triangle tessellations, and the code projector picks out an irreducible representation of the corresponding proper triangle group.

Examples in Ref. \NoCaseChange{\protect\cite{cite2811}} are the \(\{5,5,5\}\) tessellation realizing the Pauli group of a five-dimensional qudit, the \(\{6,4,8\}\) tessellation realizing the \flmRefsHyperref{ref409}{single-qubit Clifford group}, and the \(\{4,3,5\}\) tessellation realizing the binary icosahedral group \(2I\).

\codefieldsection{Protection}
Protects against sufficiently small position and momentum shifts on the hyperbolic plane.
In the three explicit examples of Ref. \NoCaseChange{\protect\cite{cite2811}}, position-translation errors are correctable up to resolutions \(d_x \approx 1.6169\), \(0.6605\), and \(0.5011\), respectively.
Momentum errors are represented by hyperbolic Laplacian eigenfunctions; the \(\{5,5,5\}\) example corrects all modes with angular index \(n<5\), and more generally the compact quotient \(\mathbb{H}^2/\Gamma\) provides a nonzero Laplacian gap that acts as a momentum-error distance scale.

\codefieldsection{Gates}
\begin{eczvaluelist}
\item\relax Logical operations are realized by rotations around selected vertices of the hyperbolic tessellation. In the \(\{6,4,8\}\) example, \(S\) and \(U\) are implemented by \(\pi/4\) and \(\pi/3\) rotations, while the \(\{4,3,5\}\) example realizes binary-icosahedral non-Clifford gates by \(2\pi/3\) and \(2\pi/5\) rotations \NoCaseChange{\protect\cite{cite2811}}.
\end{eczvaluelist}
\codefieldsection{Parents}
\begin{eczvaluelist}
\item\relax
\flmRefsHyperref[eczindexfamilyrel]{code:homogeneous_space_quantum}{Homogeneous-space quantum code} --- Hyperbolic tessellation codes are defined on the space of functions on the hyperbolic plane, the symmetric space \(G/H\) for \(G = SO(2,1)\) the proper Lorentz group and \(H = O(2)\).
\item\relax
\flmRefsHyperref[eczindexfamilyrel]{code:group_representation}{Group-representation code} --- Hyperbolic tessellation-code projections are onto a copy of an irreducible representation of the proper triangle group associated with the tessellation, and the resulting logical gates are implemented geometrically by hyperbolic rotations \NoCaseChange{\protect\cite{cite2811}}.
\end{eczvaluelist}
\codefieldsection{Cousins}
\begin{eczvaluelist}
\item\relax
\flmRefsHyperref[eczindexfamilyrel]{code:hyperbolic}{Hyperbolic sphere packing} --- Hyperbolic tessellation codes are quantum counterparts of hyperbolic sphere packings because they store information in quantum superpositions of points on the hyperbolic plane.
\item\relax
\flmRefsHyperref[eczindexfamilyrel]{code:pauli_qsc}{Pauli tessellation QSC} --- The tessellation-code framework spans spherical, Euclidean, and hyperbolic geometries; the Pauli tessellation QSC is the spherical member \NoCaseChange{\protect\cite{cite2811}}.
\item\relax
\flmRefsHyperref[eczindexfamilyrel]{code:qutrit_pauli_gkp_subcode}{Qutrit-Pauli tessellation code} --- The qutrit-Pauli tessellation code is the Euclidean \(\{3,3,3\}\) member of the same curvature-dependent tessellation-code framework \NoCaseChange{\protect\cite{cite2811}}.
\item\relax
\flmRefsHyperref[eczindexfamilyrel]{code:gkp_concatenated}{Concatenated GKP code} --- The qubit-Pauli tessellation GKP code \NoCaseChange{\protect\cite{cite2811}} is the Euclidean \(\{2,4,4\}\) member of the curvature-dependent tessellation-code framework. It is a two-mode code in which each Cartesian direction is a single-mode qubit GKP code, making the full code a 2-to-1 concatenated qubit encoding. The logical \flmRefsHyperref{ref663}{single-qubit Pauli group} is implemented geometrically by one \(\pi\) rotation and two \(\pi/2\) rotations on the Euclidean tessellation \NoCaseChange{\protect\cite{cite2811}}.
\end{eczvaluelist}
\eczhbkcontributors{ Yixu Wang, \eczhuVVA }
\endeczcode

\onecolumngrid
\clearpage

\section{Category Kingdom}

\begin{eczEpigraph}
\begin{quote}
\flmQuoteSetup{quote}%
First quantization is a mystery, but second quantization is a functor.
\flmQuoteAttributed{Edward Nelson}
\end{quote}
\end{eczEpigraph}

\twocolumngrid

\eczcode{enriched_walker_wang}{\(G\)-enriched Walker-Wang model code}{~\NoCaseChange{\protect\cite{cite5175}}}
\codefieldsection{Alternative Names}
\begin{eczvaluelist}
\item\relax Williamson-Wang model code
\end{eczvaluelist}
\eczhIndexCodeAliasName{enriched_walker_wang}{Williamson-Wang model code}
\codefieldsection{Description}
A 3D topological code defined by a unitary \(G\)-crossed braided fusion category \( \mathcal{C} \) \NoCaseChange{\protect\cite{cite632,cite633}}, where \(G\) is a finite group.
The model realizes TQFTs that include two-gauge theories, those behind Walker-Wang models, as well as the Kashaev TQFT \NoCaseChange{\protect\cite{cite634,cite635}}.
It has been generalized to include domain walls \NoCaseChange{\protect\cite{cite636}}.

\codefieldsection{Parents}
\begin{eczvaluelist}
\item\relax
\flmRefsHyperref[eczindexfamilyrel]{code:category_quantum}{Category-based quantum code}\item\relax
\flmRefsHyperref[eczindexfamilyrel]{code:commuting_projector}{Commuting-projector Hamiltonian code} --- \(G\)-enriched Walker-Wang model codewords form ground-state subspaces of frustration-free commuting projector Hamiltonians.
\item\relax
\flmRefsHyperref[eczindexfamilyrel]{code:frustration_free}{Frustration-free Hamiltonian code} --- \(G\)-enriched Walker-Wang model codewords form ground-state subspaces of frustration-free commuting projector Hamiltonians.
\item\relax
\flmRefsHyperref[eczindexfamilyrel]{code:topological}{Topological code} --- \(G\)-enriched Walker-Wang models realize 3D topological phases based on unitary \(G\)-crossed braided fusion categories.
\end{eczvaluelist}
\codefieldsection{Child}
\begin{eczvaluelist}
\item\relax
\flmRefsHyperref[eczindexfamilyrel]{code:walker_wang}{Walker-Wang model code} --- \(G\)-enriched Walker-Wang models reduce to Walker-Wang models for trivial \(G\) \NoCaseChange{\protect\cite{cite5175}}.
\end{eczvaluelist}
\codefieldsection{Cousin}
\begin{eczvaluelist}
\item\relax
\flmRefsHyperref[eczindexfamilyrel]{code:yetter_gauge_theory}{Two-gauge theory code} --- \(G\)-enriched Walker-Wang models realize 3D two-gauge theories \NoCaseChange{\protect\cite{cite5175}}.
\end{eczvaluelist}
\eczhbkcontributors{ \eczhuVVA }
\endeczcode

\eczcode{cage_net}{Cage-net code}{~\NoCaseChange{\protect\cite{cite567,cite569}}}
\codefieldsection{Alternative Names}
\begin{eczvaluelist}
\item\relax String-membrane-net code
\end{eczvaluelist}
\eczhIndexCodeAliasName{cage_net}{String-membrane-net code}
\codefieldsection{Description}
A commuting-projector code family obtained by coupling layers of two-dimensional topological orders and condensing extended one-dimensional flux strings \NoCaseChange{\protect\cite{cite567}}.
A modern lattice realization starts from isotropic stacks of \(G\)-graded string-net models \NoCaseChange{\protect\cite{cite568}}.
The family includes stabilizer examples such as the \(\mathbb{Z}_2\) string-membrane-net realization of the X-cube model \NoCaseChange{\protect\cite{cite569}} as well as non-stabilizer examples with non-Abelian restricted-mobility excitations \NoCaseChange{\protect\cite{cite567}}.
String-membrane-net and cage-net constructions can realize the same fracton phases; for isotropic stacks, this equivalence can be understood via generalized local unitaries \NoCaseChange{\protect\cite{cite568}}.
The cage-net construction can be used to realize various fracton phases, stabilizer and otherwise.

\codefieldsection{Parents}
\begin{eczvaluelist}
\item\relax
\flmRefsHyperref[eczindexfamilyrel]{code:category_quantum}{Category-based quantum code}\item\relax
\flmRefsHyperref[eczindexfamilyrel]{code:commuting_projector}{Commuting-projector Hamiltonian code} --- Cage-net codewords form ground-state subspaces of frustration-free commuting projector Hamiltonians.
\item\relax
\flmRefsHyperref[eczindexfamilyrel]{code:frustration_free}{Frustration-free Hamiltonian code} --- Cage-net codewords form ground-state subspaces of frustration-free commuting projector Hamiltonians.
\end{eczvaluelist}
\codefieldsection{Child}
\begin{eczvaluelist}
\item\relax
\flmRefsHyperref[eczindexfamilyrel]{code:qudit_xcube}{Qudit X-cube model code} --- A field-theoretic description of the qudit X-cube model can be obtained by coupling layers of 2D \(\mathbb{Z}_q\) gauge theory \NoCaseChange{\protect\cite{cite568}}. For three orthogonal foliations with \(\mathbb{Z}_q\) layers, the string-membrane-net model is equivalent to the \(\mathbb{Z}_q\) X-cube model \NoCaseChange{\protect\cite{cite569}}. String-membrane-net models are phase-equivalent to cage-net models under generalized local unitaries \NoCaseChange{\protect\cite{cite568}}.
\end{eczvaluelist}
\codefieldsection{Cousins}
\begin{eczvaluelist}
\item\relax
\flmRefsHyperref[eczindexfamilyrel]{code:fracton}{Fracton stabilizer code} --- The cage-net construction can be used to realize various fracton phases, stabilizer and otherwise.
\item\relax
\flmRefsHyperref[eczindexfamilyrel]{code:string_net}{String-net code} --- Cage-net codes are obtained by coupling layers of \(G\)-graded string-net models \NoCaseChange{\protect\cite{cite568}}.
\end{eczvaluelist}
\eczhbkcontributors{ \eczhuVVA }
\endeczcode

\eczcode{category_quantum}{Category-based quantum code}{}

\codefieldsection{Kingdom root code}
for the \flmRefsHyperref{kingdom:category_quantum}{Category Kingdom}.
\codefieldsection{Description}
Encodes a finite-dimensional \textit{logical} Hilbert space into a \textit{physical} Hilbert space associated with a category.
The categories of interest are typically fusion categories, which subsume all finite groups and many state spaces associated with topological codes. 
Codes on modular fusion categories are often associated with a particular topological quantum field theory (TQFT), as the data of such theories is described by such categories.

\codefieldsection{Parent}
\begin{eczvaluelist}
\item\relax
\flmRefsHyperref[eczindexfamilyrel]{code:qecc}{Quantum error-correcting code (QECC)}\end{eczvaluelist}
\codefieldsection{Children}
\begin{eczvaluelist}
\item\relax
\flmRefsHyperref[eczindexfamilyrel]{code:cage_net}{Cage-net code}\item\relax
\flmRefsHyperref[eczindexfamilyrel]{code:yetter_gauge_theory}{Two-gauge theory code}\item\relax
\flmRefsHyperref[eczindexfamilyrel]{code:hopf_cluster_state}{Hopf-algebra cluster-state code}\item\relax
\flmRefsHyperref[eczindexfamilyrel]{code:enriched_string_net}{Multi-fusion string-net code}\item\relax
\flmRefsHyperref[eczindexfamilyrel]{code:enriched_walker_wang}{\(G\)-enriched Walker-Wang model code}\item\relax
\flmRefsHyperref[eczindexfamilyrel]{code:group_quantum}{Group-based quantum code} --- Finite-group-based quantum codes, whose basis states are parameterized by a finite group, correspond to category-based codes for the fusion category \(Vec G\). Extensions of such categories to Lie groups can also be done \NoCaseChange{\protect\cite{cite5084,cite5085,cite5086,cite5087}} (see also \NoCaseChange{\protect\cite{cite5088}}).
\end{eczvaluelist}
\codefieldsection{Cousin}
\begin{eczvaluelist}
\item\relax
\flmRefsHyperref[eczindexfamilyrel]{code:block_perfect}{Planar-perfect-tensor code} --- Several modular fusion categories can be used to define \flmRefsHyperref{code:block_perfect}{planar-perfect tensor}s \NoCaseChange{\protect\cite{cite2951}}.
\end{eczvaluelist}
\eczhbkcontributors{ \eczhuVVA }
\endeczcode

\eczcode{fibonacci}{Fibonacci string-net code}{~\NoCaseChange{\protect\cite{cite3624,cite599}}}
\codefieldsection{Description}
Quantum error correcting code associated with the Levin-Wen string-net model with the Fibonacci input category, admitting two types of encodings.

The first type of encoding is into the ground-state subspace of the Levin-Wen model Hamiltonian, defined on a cell decomposition (dual to a triangulation) of a surface with a qubit on each link. The code space is the simultaneous \(+1\) eigenspace of a set of vertex operators and plaquette operators, which are defined by the fusion rules and the numerical data of the Fibonacci category, respectively.
In the corresponding doubled Fibonacci theory, the anyon types are \(1 \times 1\), \(1 \times \tau\), \(\tau \times 1\), and \(\tau \times \tau\) \NoCaseChange{\protect\cite{cite599}}.

The second type of encoding is into the degenerate fusion space of a number of anyonic quasiparticle excitations of the Levin-Wen model. In the computational scheme of Ref. \NoCaseChange{\protect\cite{cite599}}, logical information is encoded into pairs of \(\tau \times \tau\) anyons that fuse to \(1 \times 1\). This can equivalently be constructed by braiding holes in a spherical geometry \NoCaseChange{\protect\cite[{Sec. 5}]{cite599}}.

\codefieldsection{Protection}
When defined on a \(L \times L\) tailed honeycomb tiling on a torus, the code distance for ground-state encoding is \(L\).
\codefieldsection{Encoding}
\begin{eczvaluelist}
\item\relax Code states may not be preparable with an adaptive constant-depth circuit with geometrically local gates and measurements throughout \NoCaseChange{\protect\cite{cite5139}}.
\item\relax Ground-state initialization on small lattice \NoCaseChange{\protect\cite{cite5176}}.
\item\relax Fusion-space initialization can be reduced to preparing pairs of \(\tau \times \tau\) anyons by local cut-and-glue procedures and small projective measurements \NoCaseChange{\protect\cite{cite599}}.
\end{eczvaluelist}
\codefieldsection{Transversal and Permutation-Based Gates}
\begin{eczvaluelist}
\item\relax A universal transversal gate set could be implemented in a folded version of this code using the techniques introduced in Ref \NoCaseChange{\protect\cite{cite732}}.
\end{eczvaluelist}
\codefieldsection{Gates}
\begin{eczvaluelist}
\item\relax Universal gate set for the ground-state encoding is implemented through topological operations corresponding to elements of the mapping class group, which is generated by Dehn-twists along non-contractible cycles. These Dehn-twists can be implemented using constant-depth circuits when allowing long-range permutations of qubits \NoCaseChange{\protect\cite{cite5177,cite3839}}. The mapping class group of a disk with \(m\) punctures is the braid group of \(m\) objects.
\item\relax Universal gate set for the fusion-space encoding is implemented through braiding of the computational anyons \NoCaseChange{\protect\cite{cite5178,cite599}}. In the lattice realization of Ref. \NoCaseChange{\protect\cite{cite599}}, these braids and the associated Dehn twists are enacted by local \(F\)-moves, and readout reduces to measuring anyon labels on a fixed edge after local retriangulations. Circuit-based gates can be converted into braid patterns via quantum compiling algorithms \NoCaseChange{\protect\cite{cite5179,cite3983}}.
\end{eczvaluelist}
\codefieldsection{Decoding}
\begin{eczvaluelist}
\item\relax Clustering decoder (provides best known threshold for this code) \NoCaseChange{\protect\cite{cite5180,cite5181,cite5182}}.
\item\relax Fusion-aware iterative minimum-weight perfect matching decoder. Note that ordinary MWPM decoders do not produce a threshold with this code \NoCaseChange{\protect\cite{cite5182}}.
\item\relax Cellular automaton decoder \NoCaseChange{\protect\cite{cite5183}}.
\end{eczvaluelist}
\codefieldsection{Code Capacity Threshold}
\begin{eczvaluelist}
\item\relax \(4.7\%\) for depolarizing noise, \(7.3\%\) for dephasing noise, and \(3.8\%\) for bit-flip noise with clustering decoder, assuming perfect measurements and gates \NoCaseChange{\protect\cite{cite5182}}. See also Ref. \NoCaseChange{\protect\cite{cite5180}}.
\item\relax \(3.0\%\) for depolarizing noise, \(6.0\%\) for dephasing noise, and \(2.5\%\) for bit-flip noise with fusion-aware iterative MWPM decoder, assuming perfect measurements and gates \NoCaseChange{\protect\cite{cite5182}}.
\end{eczvaluelist}
\codefieldsection{Threshold}
\begin{eczvaluelist}
\item\relax Between \(10^{-2}\%\) and \(5\cdot 10^{-2}\%\) for pair-creation and measurement noise \NoCaseChange{\protect\cite{cite5183}}.
\end{eczvaluelist}
\codefieldsection{Realizations}
\begin{eczvaluelist}
\item\relax NMR: Implementation of braiding-based Hadamard gate on two qubits \NoCaseChange{\protect\cite{cite5184}}.
\item\relax Superconducting qubits: state preparation, fusion, and braiding \NoCaseChange{\protect\cite{cite5185,cite5186}}. The latter work utilized DSNP and sampled the string-net wavefunction to estimate the underlying chromatic polynomial.
\end{eczvaluelist}
\codefieldsection{Parent}
\begin{eczvaluelist}
\item\relax
\flmRefsHyperref[eczindexfamilyrel]{code:string_net}{String-net code} --- The string-net model code for the category \(\text{Fib}\) is the Fibonacci string-net code.
\end{eczvaluelist}
\eczhbkcontributors{ Alexis Schotte, \eczhuVVA }
\endeczcode

\eczcode{groupoid_surface}{Groupoid toric code}{~\NoCaseChange{\protect\cite{cite5187}}}
\codefieldsection{Description}
Extension of the Kitaev surface code from Abelian groups to groupoids, i.e., multi-fusion categories in which every morphism is an isomorphism \NoCaseChange{\protect\cite{cite570}}.
Some models admit fracton-like features such as extensive ground-state degeneracy and excitations with restricted mobility.
The robustness of these features has not yet been established.

\codefieldsection{Parent}
\begin{eczvaluelist}
\item\relax
\flmRefsHyperref[eczindexfamilyrel]{code:enriched_string_net}{Multi-fusion string-net code} --- Groupoid toric-code categories are unitary multi-fusion categories based on matrix algebras \NoCaseChange{\protect\cite{cite5188}}, so groupoid toric codes can equivalently be formulated as multi-fusion string-net codes.
\end{eczvaluelist}
\codefieldsection{Cousin}
\begin{eczvaluelist}
\item\relax
\flmRefsHyperref[eczindexfamilyrel]{code:fracton}{Fracton stabilizer code} --- Some groupoid toric code models admit fracton-like features such as extensive ground-state degeneracy and excitations with restricted mobility.
\end{eczvaluelist}
\eczhbkcontributors{ Meng Cheng (程蒙), \eczhuVVA }
\endeczcode

\eczcode{hopf_cluster_state}{Hopf-algebra cluster-state code}{~\NoCaseChange{\protect\cite{cite5189,cite5190}}}
\codefieldsection{Description}
Code based on a cluster state defined on qudits valued in a Hopf algebra.
This code has also been extended to weak Hopf algebras \NoCaseChange{\protect\cite{cite5190}}.

\codefieldsection{Parent}
\begin{eczvaluelist}
\item\relax
\flmRefsHyperref[eczindexfamilyrel]{code:category_quantum}{Category-based quantum code}\end{eczvaluelist}
\codefieldsection{Child}
\begin{eczvaluelist}
\item\relax
\flmRefsHyperref[eczindexfamilyrel]{code:qudit_cluster_state}{Modular-qudit cluster-state code} --- Hopf-algebra cluster-state codes reduce to modular-qudit cluster-state codes when the Hopf algebra reduces to the group \(\mathbb{Z}_q\).
\end{eczvaluelist}
\codefieldsection{Cousins}
\begin{eczvaluelist}
\item\relax
\flmRefsHyperref[eczindexfamilyrel]{code:hopf_quantum_double}{Hopf-algebra quantum-double code} --- Both Hopf-algebra quantum-double and Hopf-algebra cluster-state codes are defined on qudits valued in a Hopf algebra. A set of Pauli-type operators can be defined for Hopf algebras \NoCaseChange{\protect\cite{cite5189,cite5191}}.
\item\relax
\flmRefsHyperref[eczindexfamilyrel]{code:group_cluster_state}{Group-based cluster-state code} --- Hopf-algebra cluster-state codes reduce to group-based cluster-state codes for finite groups when the Hopf algebra reduces to a finite group.
\end{eczvaluelist}
\eczhbkcontributors{ \eczhuVVA }
\endeczcode

\eczcode{hopf_quantum_double}{Hopf-algebra quantum-double code}{~\NoCaseChange{\protect\cite{cite590,cite591}}}
\codefieldsection{Description}
Code whose codewords realize 2D gapped topological order defined on qudits valued in a Hopf algebra \(H\).
The code Hamiltonian is a generalization \NoCaseChange{\protect\cite{cite590,cite591}} of the quantum double model from group algebras to Hopf algebras, as anticipated by Kitaev \NoCaseChange{\protect\cite{cite423}}.
Boundaries of these models have been examined \NoCaseChange{\protect\cite{cite592,cite593}}.

\codefieldsection{Parent}
\begin{eczvaluelist}
\item\relax
\flmRefsHyperref[eczindexfamilyrel]{code:string_net}{String-net code} --- String-net model ground states reduce to Hopf-algebra quantum-double ground states for categories \(\text{Rep}(H)\), where \(H\) is a Hopf algebra \NoCaseChange{\protect\cite{cite591}}.
\end{eczvaluelist}
\codefieldsection{Child}
\begin{eczvaluelist}
\item\relax
\flmRefsHyperref[eczindexfamilyrel]{code:quantum_double}{Quantum-double code} --- Hopf-algebra quantum-double codes reduce to quantum-double codes when the Hopf algebra is a \flmRefsHyperref{ref205}{group algebra}. Quantum-double codes for non-Abelian groups \(G\) are dual to Hopf-algebra quantum-double codes for Hopf algebras based on \(\text{Rep}(G)\) under the Tannaka-Krein duality \NoCaseChange{\protect\cite{cite5145}\protect\cite[{Fig. 1}]{cite5146}}.
\end{eczvaluelist}
\codefieldsection{Cousins}
\begin{eczvaluelist}
\item\relax
\flmRefsHyperref[eczindexfamilyrel]{code:enriched_string_net}{Multi-fusion string-net code} --- Extending the Hopf algebra quantum-double construction to a weak Hopf algebra construction yields an alternative formulation \NoCaseChange{\protect\cite{cite5192}\protect\cite[{Fig. 1}]{cite5146}} for realizing multi-fusion string-net topological orders because of the relationship between representations of weak Hopf algebras and multi-fusion categories \NoCaseChange{\protect\cite{cite5193}}. Tensor network constructions can be done for either formulation \NoCaseChange{\protect\cite{cite5194,cite5195}}.
\item\relax
\flmRefsHyperref[eczindexfamilyrel]{code:qudit_surface}{Modular-qudit surface code} --- The modular-qudit surface code can be generalized to a Hopf-algebra quantum-double code whose ground states remain the same but whose excitations are based on quasitriangular semisimple Hopf algebras of \(\mathbb{Z}_q\) \NoCaseChange{\protect\cite{cite4587}}.
\item\relax
\flmRefsHyperref[eczindexfamilyrel]{code:hopf_cluster_state}{Hopf-algebra cluster-state code} --- Both Hopf-algebra quantum-double and Hopf-algebra cluster-state codes are defined on qudits valued in a Hopf algebra. A set of Pauli-type operators can be defined for Hopf algebras \NoCaseChange{\protect\cite{cite5189,cite5191}}.
\end{eczvaluelist}
\eczhbkcontributors{ Laurens Lootens, \eczhuVVA }
\endeczcode

\eczcode{enriched_string_net}{Multi-fusion string-net code}{~\NoCaseChange{\protect\cite{cite5188}}}
\codefieldsection{Description}
Family of codes resulting from the string-net construction but whose input is a unitary multi-fusion category (as opposed to a unitary fusion category).

\codefieldsection{Parents}
\begin{eczvaluelist}
\item\relax
\flmRefsHyperref[eczindexfamilyrel]{code:category_quantum}{Category-based quantum code}\item\relax
\flmRefsHyperref[eczindexfamilyrel]{code:topological}{Topological code} --- Enriched string-net codes realize 2D topological phases based on unitary multi-fusion categories.
\item\relax
\flmRefsHyperref[eczindexfamilyrel]{code:commuting_projector}{Commuting-projector Hamiltonian code} --- Multi-fusion string-net codes form eigenspaces of frustration-free commuting projector Hamiltonians.
\item\relax
\flmRefsHyperref[eczindexfamilyrel]{code:frustration_free}{Frustration-free Hamiltonian code} --- Multi-fusion string-net codes form eigenspaces of frustration-free commuting projector Hamiltonians.
\end{eczvaluelist}
\codefieldsection{Children}
\begin{eczvaluelist}
\item\relax
\flmRefsHyperref[eczindexfamilyrel]{code:groupoid_surface}{Groupoid toric code} --- Groupoid toric-code categories are unitary multi-fusion categories based on matrix algebras \NoCaseChange{\protect\cite{cite5188}}, so groupoid toric codes can equivalently be formulated as multi-fusion string-net codes.
\item\relax
\flmRefsHyperref[eczindexfamilyrel]{code:string_net}{String-net code}\end{eczvaluelist}
\codefieldsection{Cousin}
\begin{eczvaluelist}
\item\relax
\flmRefsHyperref[eczindexfamilyrel]{code:hopf_quantum_double}{Hopf-algebra quantum-double code} --- Extending the Hopf algebra quantum-double construction to a weak Hopf algebra construction yields an alternative formulation \NoCaseChange{\protect\cite{cite5192}\protect\cite[{Fig. 1}]{cite5146}} for realizing multi-fusion string-net topological orders because of the relationship between representations of weak Hopf algebras and multi-fusion categories \NoCaseChange{\protect\cite{cite5193}}. Tensor network constructions can be done for either formulation \NoCaseChange{\protect\cite{cite5194,cite5195}}.
\end{eczvaluelist}
\eczhbkcontributors{ \eczhuVVA }
\endeczcode

\eczcode{string_net}{String-net code}{~\NoCaseChange{\protect\cite{cite3624,cite5196,cite599,cite5197}}}
\codefieldsection{Alternative Names}
\begin{eczvaluelist}
\item\relax Levin-Wen model code
\item\relax Turaev-Viro code
\end{eczvaluelist}
\eczhIndexCodeAliasName{string_net}{Levin-Wen model code}
\eczhIndexCodeAliasName{string_net}{Turaev-Viro code}
\codefieldsection{Description}
A non-stabilizer commuting-projector code whose codewords realize a 2D topological order rendered by a Turaev-Viro topological field theory.
The corresponding anyon theory is defined by a (multiplicity-free) unitary fusion category \( \mathcal{C} \).
The code is defined on a cell decomposition dual to a triangulation of a two-dimensional surface, with a qudit of dimension \( |\mathcal{C}| \) located at each edge of the decomposition.
For modular input categories, the emergent anyon theory is the doubled category \( \mathcal{C}\otimes\mathcal{C}^{*} \) \NoCaseChange{\protect\cite{cite599}}.
These models realize local topological order (LTO) \NoCaseChange{\protect\cite{cite600}}.

The codespace is the ground-state subspace of the Levin-Wen model commuting-projector Hamiltonian \NoCaseChange{\protect\cite{cite3624}}, a many-body Hamiltonian realizing the 3-manifold Turaev-Viro invariant \NoCaseChange{\protect\cite{cite599,cite5198}}.
For the \(\mathbb{Z}_2\) input category on a genus-one handlebody, this construction yields the toric code \NoCaseChange{\protect\cite{cite599}}.
This Hamiltonian is defined to act on a constrained Hilbert space defined by vertex constraints.
Alternative constructions are possible, encoding information in the fusion space of the low-energy anyonic quasiparticle excitations of the model \NoCaseChange{\protect\cite{cite5178,cite599}}.
The fusion space can have dimension greater than one, allowing for topological quantum computation of logical information stored in the fusion outcomes.
Domain walls of string nets can be formulated using bimodule categories \NoCaseChange{\protect\cite{cite5199}}.
Morita-equivalent string-net models are connected by constant-depth local unitary circuits built from invertible bimodule categories; on a hexagonal lattice, the circuit has depth three and also induces a map between tube-algebra excitations upon adding ancillas \NoCaseChange{\protect\cite{cite3158}}.
Anyon condensation of general string net models is studied in Ref. \NoCaseChange{\protect\cite{cite5200}}.

The initial formulation of string-net models \NoCaseChange{\protect\cite{cite3624}} considered unitary fusion categories with an extra tetrahedral symmetry, but this was realized not to be necessary \NoCaseChange{\protect\cite{cite5196,cite599}}.
Explicit Hamiltonians for the more general categories have been studied \NoCaseChange{\protect\cite{cite5197}}.

\codefieldsection{Protection}
Error-correcting properties established in Ref. \NoCaseChange{\protect\cite{cite5201}}.
\codefieldsection{Encoding}
\begin{eczvaluelist}
\item\relax For an \(L\times L\) lattice, deterministic state preparation can be done with a geometrically local unitary \(O(L)\)-depth circuit \NoCaseChange{\protect\cite{cite3822}} or an \(O(\log{L})\)-depth unitary circuit with non-local two-qubit gates \NoCaseChange{\protect\cite{cite3818,cite3823}}.
\item\relax Scalable dynamic string-net preparation (DSNP) \NoCaseChange{\protect\cite{cite5186}}.
\item\relax String nets with solvable anyons and gappable boundaries \NoCaseChange{\protect\cite[{Def. 1.2}]{cite5202}} can be prepared via an adaptive finite-depth local unitary circuit \NoCaseChange{\protect\cite{cite5203}}.
\item\relax Suitable initial states can be prepared by locally cutting and gluing regions of the triangulation, reducing the task to preparing small fusion-space states \NoCaseChange{\protect\cite{cite599}}.
\end{eczvaluelist}
\codefieldsection{Gates}
\begin{eczvaluelist}
\item\relax Gates can be implemented through topological operations corresponding to elements of the mapping class group, which is generated by Dehn-twists along non-contractible cycles for triangulations of toroidal \NoCaseChange{\protect\cite{cite5177,cite3839}} and hyperbolic \NoCaseChange{\protect\cite{cite5204}} manifolds. Whether or not a gate set is universal depends on the choice of input category; in some cases such as the Ising category, gates can be complemented by topological charge measurements to obtain a universal gate set.
\item\relax Alternatively, one could encode the logical quantum information into the degenerate fusion space of a number of computational anyons. In this case, a universal logical gate set can be implemented through the braiding of the computational anyons \NoCaseChange{\protect\cite{cite5178,cite5205,cite599}}, e.g., for the case of the \flmRefsHyperref{code:fibonacci}{Fibonacci} input category.
\item\relax In the Levin-Wen realization, braid moves and Dehn twists can be implemented by sequences of local \(F\)-moves; for general multiplicity-free modular categories these are fixed five-qudit gates, and the same local moves reduce topological-charge measurements to measurements of anyon labels on a fixed edge \NoCaseChange{\protect\cite{cite599}}.
\end{eczvaluelist}
\codefieldsection{Decoding}
\begin{eczvaluelist}
\item\relax Syndrome measurement circuits analyzed in Ref. \NoCaseChange{\protect\cite{cite5206}}.
\item\relax Clustering decoder \NoCaseChange{\protect\cite{cite5181}}.
\end{eczvaluelist}
\codefieldsection{Parent}
\begin{eczvaluelist}
\item\relax
\flmRefsHyperref[eczindexfamilyrel]{code:enriched_string_net}{Multi-fusion string-net code}\end{eczvaluelist}
\codefieldsection{Children}
\begin{eczvaluelist}
\item\relax
\flmRefsHyperref[eczindexfamilyrel]{code:hopf_quantum_double}{Hopf-algebra quantum-double code} --- String-net model ground states reduce to Hopf-algebra quantum-double ground states for categories \(\text{Rep}(H)\), where \(H\) is a Hopf algebra \NoCaseChange{\protect\cite{cite591}}.
\item\relax
\flmRefsHyperref[eczindexfamilyrel]{code:fibonacci}{Fibonacci string-net code} --- The string-net model code for the category \(\text{Fib}\) is the Fibonacci string-net code.
\item\relax
\flmRefsHyperref[eczindexfamilyrel]{code:tqd}{Twisted quantum double (TQD) code} --- String-net models realize TQDs for categories \(\text{Vec}^{\omega}G\), where \(G\) is a finite group and \(\omega\) is a 3-cocycle on \(G\). There is a duality between a large class of string-net models and certain TQD models \NoCaseChange{\protect\cite{cite571}}.
\end{eczvaluelist}
\codefieldsection{Cousins}
\begin{eczvaluelist}
\item\relax
\flmRefsHyperref[eczindexfamilyrel]{code:topological}{Topological code} --- String-net codes realize 2D topological phases based on unitary fusion categories. Any 2D many-body state satisfying the entanglement bootstrap axioms can be mapped into the ground-state subspace of a string-net model via a constant-depth unitary circuit \NoCaseChange{\protect\cite{cite3157}}. Different string-net models with Morita-equivalent input fusion categories and the same topological order are connected by constant-depth unitary circuits and therefore lie in the same phase \NoCaseChange{\protect\cite{cite3158,cite3159}}.
\item\relax
\flmRefsHyperref[eczindexfamilyrel]{code:cage_net}{Cage-net code} --- Cage-net codes are obtained by coupling layers of \(G\)-graded string-net models \NoCaseChange{\protect\cite{cite568}}.
\item\relax
\flmRefsHyperref[eczindexfamilyrel]{code:walker_wang}{Walker-Wang model code} --- The Walker-Wang model is a generalization of the 3D version of the Levin-Wen model \NoCaseChange{\protect\cite[{Sec. 5}]{cite3624}}, which realizes gauge theories coupled to bosons and fermions.
\item\relax
\flmRefsHyperref[eczindexfamilyrel]{code:double_semion_string_net}{Double-semion string-net code} --- The string-net model code for the category \(\text{Vec}^{\omega}\mathbb{Z}_2\) for a nontrivial cocycle is the double semion string-net code.
\item\relax
\flmRefsHyperref[eczindexfamilyrel]{code:xcube}{X-cube model code} --- A non-stabilizer commuting-projector code constructed by stacking layers of the double-semion string-net model, called the semionic X-cube model \NoCaseChange{\protect\cite{cite534}}, is equivalent to the X-cube model \NoCaseChange{\protect\cite{cite4492}} (see also Refs. \NoCaseChange{\protect\cite{cite3980,cite4497}}).
\item\relax
\flmRefsHyperref[eczindexfamilyrel]{code:toric}{Toric code} --- The toric code is the Turaev-Viro/Levin-Wen string-net code for the \(\mathbb{Z}_2\) input category; equivalently, the construction of Ref. \NoCaseChange{\protect\cite{cite599}} on a genus-one handlebody yields the toric code.
\end{eczvaluelist}
\eczhbkcontributors{ Alexis Schotte, David Aasen, \eczhuVVA }
\endeczcode

\eczcode{yetter_gauge_theory}{Two-gauge theory code}{~\NoCaseChange{\protect\cite{cite616}}}
\codefieldsection{Alternative Names}
\begin{eczvaluelist}
\item\relax Higher gauge theory code
\end{eczvaluelist}
\eczhIndexCodeAliasName{yetter_gauge_theory}{Higher gauge theory code}
\codefieldsection{Description}
A code whose codewords realize lattice two-gauge theory \NoCaseChange{\protect\cite{cite607,cite608,cite609,cite610,cite611,cite612,cite613,cite614,cite615}} for a finite \textit{two-group} (a.k.a. a \textit{crossed module}) in arbitrary spatial dimension.
There exist several lattice-model formulations in arbitrary spatial dimension \NoCaseChange{\protect\cite{cite616,cite617}} as well as explicitly in 3D \NoCaseChange{\protect\cite{cite618,cite619,cite620,cite621}} and 4D \NoCaseChange{\protect\cite{cite621}}, with the 3D case realizing the Yetter model \NoCaseChange{\protect\cite{cite622,cite623,cite624,cite625}}.

A two-gauge theory generalizes ordinary gauge theory by replacing the gauge group with a two-group (a.k.a. finite crossed module).
Lattice formulations place gauge fields not only on edges of a lattice (as they do in ordinary gauge theory), but also on higher-dimensional structures such as faces.

Ground-state degeneracy is a topological invariant for 3D manifolds \NoCaseChange{\protect\cite{cite617}}; more precisely, it equals the number of homotopy classes of maps from the spatial manifold to the classifying space of the underlying finite two-group.
Excitations of the 3D models are studied in Refs. \NoCaseChange{\protect\cite{cite5070,cite5207,cite5208}}.
Generalizations of Ocneanu's tube algebras \NoCaseChange{\protect\cite{cite5068,cite5069}} can be used to characterize excitations \NoCaseChange{\protect\cite{cite5070,cite5168}}.

\codefieldsection{Parents}
\begin{eczvaluelist}
\item\relax
\flmRefsHyperref[eczindexfamilyrel]{code:category_quantum}{Category-based quantum code}\item\relax
\flmRefsHyperref[eczindexfamilyrel]{code:commuting_projector}{Commuting-projector Hamiltonian code} --- Two-gauge theory codewords span ground-state subspaces of frustration-free commuting-projector Hamiltonians.
\item\relax
\flmRefsHyperref[eczindexfamilyrel]{code:frustration_free}{Frustration-free Hamiltonian code} --- Two-gauge theory codewords span ground-state subspaces of frustration-free commuting-projector Hamiltonians.
\item\relax
\flmRefsHyperref[eczindexfamilyrel]{code:topological}{Topological code} --- Two-gauge theory codes realize lattice two-gauge theory for a finite two-group.
\end{eczvaluelist}
\codefieldsection{Children}
\begin{eczvaluelist}
\item\relax
\flmRefsHyperref[eczindexfamilyrel]{code:dijkgraaf_witten}{Dijkgraaf-Witten gauge theory code} --- Replacing the two-group data in a two-gauge theory with a group and a cocycle reproduces the phase of the Dijkgraaf-Witten gauge theory, with the two theories equivalent in 2D \NoCaseChange{\protect\cite{cite616}}. Generalizations of Ocneanu's tube algebras \NoCaseChange{\protect\cite{cite5068,cite5069}} can be used to characterize excitations in both theories \NoCaseChange{\protect\cite[{Sec. 4.2}]{cite5070}}. A Dijkgraaf-Witten Lagrangian can also be re-expressed as a two-group gauge theory Lagrangian by relating the electric and magnetic gauge fields via the equations of motion \NoCaseChange{\protect\cite{cite616,cite5071}}.
\item\relax
\flmRefsHyperref[eczindexfamilyrel]{code:cubic_theory}{Cubic theory code} --- Cubic theory codes realize higher-form \(\mathbb{Z}_2^3\) gauge theories with non-Abelian excitations in arbitrary dimensions.
\item\relax
\flmRefsHyperref[eczindexfamilyrel]{code:invertible}{Chen-Hsin invertible-order code} --- Chen-Hsin invertible-order codes realize beyond-group-cohomology invertible topological phases of order two and four in arbitrary dimensions. These phases are described by invertible two-gauge theories \NoCaseChange{\protect\cite[{pg. 11}]{cite578}}.
\end{eczvaluelist}
\codefieldsection{Cousins}
\begin{eczvaluelist}
\item\relax
\flmRefsHyperref[eczindexfamilyrel]{code:enriched_walker_wang}{\(G\)-enriched Walker-Wang model code} --- \(G\)-enriched Walker-Wang models realize 3D two-gauge theories \NoCaseChange{\protect\cite{cite5175}}.
\item\relax
\flmRefsHyperref[eczindexfamilyrel]{code:walker_wang}{Walker-Wang model code} --- Two-gauge theory codes for particular two-groups are dual to certain Walker-Wang models based on Abelian groups \NoCaseChange{\protect\cite[{Sec. V}]{cite618}\protect\cite[{Sec. 7}]{cite620}}.
\item\relax
\flmRefsHyperref[eczindexfamilyrel]{code:clifford_hierarchy}{Clifford-hierarchy stabilizer code} --- Clifford-hierarchy codes in \(D\) spatial dimensions include \((D+1)\)-dimensional Dijkgraaf-Witten gauge theories with non-Abelian topological order \NoCaseChange{\protect\cite{cite725}}.
A \(D\)-dimensional code can be constructed from a twisted \(\mathbb{Z}_2^{D+1}\) gauge theory with Dijkgraaf-Witten twist \((-1)^{\int a_1 \cup a_2 \cup \cdots \cup a_{D+1}}\), where the stabilizers include gates at the \(D\)th level of the Clifford hierarchy in addition to Pauli \(X\) operators.

\end{eczvaluelist}
\eczhbkcontributors{ \eczhuVVA }
\endeczcode

\eczcode{walker_wang}{Walker-Wang model code}{~\NoCaseChange{\protect\cite{cite628}}}
\codefieldsection{Description}
A non-stabilizer commuting-projector 3D topological code defined by a unitary braided fusion category \( \mathcal{C} \) (also known as a unitary premodular category).
The code is defined on a cubic lattice that is resolved to be trivalent, with a qudit of dimension \( |\mathcal{C}| \) located at each edge.
The codespace is the ground-state subspace of the Walker-Wang model Hamiltonian \NoCaseChange{\protect\cite{cite628}} and realizes the Crane-Yetter model \NoCaseChange{\protect\cite{cite629,cite630,cite631}}.
A single-state version of the code provides a resource state for MBQC \NoCaseChange{\protect\cite{cite478}}.

\codefieldsection{Protection}
Codespace dimensions (i.e., ground-state degeneracy) has been calculated for various boundary conditions \NoCaseChange{\protect\cite{cite473}}.

\codefieldsection{Encoding}
\begin{eczvaluelist}
\item\relax For modular chiral anyon theories, a unitary encoder is conjectured to not be implementable in constant depth because it is believed to be an example of a \textit{quantum cellular automaton} (QCA) (i.e., causal or locality-preserving automorphism) that cannot be locally implemented \NoCaseChange{\protect\cite{cite3068,cite3069}}. States of modular gapped theories can be initialized in constant depth \NoCaseChange{\protect\cite{cite5209}}.
\end{eczvaluelist}
\codefieldsection{Parent}
\begin{eczvaluelist}
\item\relax
\flmRefsHyperref[eczindexfamilyrel]{code:enriched_walker_wang}{\(G\)-enriched Walker-Wang model code} --- \(G\)-enriched Walker-Wang models reduce to Walker-Wang models for trivial \(G\) \NoCaseChange{\protect\cite{cite5175}}.
\end{eczvaluelist}
\codefieldsection{Children}
\begin{eczvaluelist}
\item\relax
\flmRefsHyperref[eczindexfamilyrel]{code:rbh}{Raussendorf-Bravyi-Harrington (RBH) cluster-state code} --- The Walker-Wang model code reduces to the RBH cluster-state code when the input category \(\mathcal{C}\) is that of the surface code \NoCaseChange{\protect\cite[{Sec. V.A}]{cite478}}.
\item\relax
\flmRefsHyperref[eczindexfamilyrel]{code:3d_fermionic_surface}{3D fermionic surface code} --- The 3D fermionic surface code is a Walker-Wang model code with premodular input category \(\mathcal{C} = \text{sVec}\) consisting of a trivial anyon and a fermion.
\item\relax
\flmRefsHyperref[eczindexfamilyrel]{code:three_fermion}{Three-fermion (3F) Walker-Wang model code} --- The Walker-Wang model code reduces to the 3F model code when the input category \(\mathcal{C}=3F\) \NoCaseChange{\protect\cite{cite478}}. When treated as ground states of the code Hamiltonian, 3F Walker-Wang model code states realize a 3D time-reversal SPT order \NoCaseChange{\protect\cite{cite477}}, while the gapped boundary supports the 3F anyon theory.
\item\relax
\flmRefsHyperref[eczindexfamilyrel]{code:3d_semion}{Chiral semion Walker-Wang model code} --- The Walker-Wang model code reduces to the chiral semion model code when the input category is \(\mathcal{C}=\mathbb{Z}_{2}^{(1/2)}\), or alternatively \(\mathcal{C}=\mathbb{Z}_{4}^{(1)}\) after condensing a \(\mathbb{Z}_{2}\)-transparent boson.
\end{eczvaluelist}
\codefieldsection{Cousins}
\begin{eczvaluelist}
\item\relax
\flmRefsHyperref[eczindexfamilyrel]{code:string_net}{String-net code} --- The Walker-Wang model is a generalization of the 3D version of the Levin-Wen model \NoCaseChange{\protect\cite[{Sec. 5}]{cite3624}}, which realizes gauge theories coupled to bosons and fermions.
\item\relax
\flmRefsHyperref[eczindexfamilyrel]{code:yetter_gauge_theory}{Two-gauge theory code} --- Two-gauge theory codes for particular two-groups are dual to certain Walker-Wang models based on Abelian groups \NoCaseChange{\protect\cite[{Sec. V}]{cite618}\protect\cite[{Sec. 7}]{cite620}}.
\item\relax
\flmRefsHyperref[eczindexfamilyrel]{code:topological_abelian}{Abelian topological code} --- Any Abelian anyon theory \(A\) can be realized at one of the surfaces of a 3D Walker-Wang model whose underlying theory is an Abelian TQD containing \(A\) as a subtheory \NoCaseChange{\protect\cite{cite471,cite472}\protect\cite[{Appx. H}]{cite414}}.
\end{eczvaluelist}
\eczhbkcontributors{ \eczhuVVA }
\endeczcode

\part{Codes in the Classical-quantum Domain}
\onecolumngrid

\begin{eczEpigraph}
\begin{quote}
\flmQuoteSetup{quote}%
The ultimate detectability of optical signals and the accuracy with which their parameters can be estimated cannot be ascertained by the methods of detection theory that apply at radio frequencies; the fundamental concepts of the theory must be revised.
\flmQuoteAttributed{Carl W. Helstrom, 1976}
\end{quote}
\end{eczEpigraph}

\section{Property codes}

\twocolumngrid

\eczcode{classical_into_quantum}{Classical-quantum (c-q) code}{}
\codefieldsection{Description}
Code designed specifically for transmission of classical information through non-classical channels, e.g., quantum channels, hybrid classical-quantum channels, or channels with classical inputs and quantum outputs. 
Such codes include maps from a classical alphabet into a quantum Hilbert space.

\codefieldsection{Rate}
The Holevo channel capacity,
\flmMathEnvironment{align}{}{
  C=\lim_{n\to\infty}\frac{1}{n}\chi\left({\cal N}^{\otimes n}\right)~,
}
where \(\chi\) is the Holevo information, is the highest rate of classical information transmission through a quantum channel with arbitrarily small error rate \NoCaseChange{\protect\cite{cite5210,cite5211,cite5212}}.

For an ensemble \(\{p_i,\rho_i\}\) of input states the single-letter Holevo quantity is
\flmMathEnvironment{align}{}{
  \chi(\{p_i,\rho_i\},\mathcal N)
  &= S\Bigl(\mathcal N\Bigl(\sum_i p_i\rho_i\Bigr)\Bigr)
   - \sum_i p_i S\bigl(\mathcal N(\rho_i)\bigr)
}
and maximization over ensembles defines \(\chi(\mathcal N)\).  The capacity
above is the regularized version of this because \(\chi\) can be superadditive;
Hastings' counterexample shows strict superadditivity for certain random channels \NoCaseChange{\protect\cite{cite5213}}.

Corrections to the Holevo capacity and tradeoff between decoding error, code rate and code length are determined in quantum generalizations of small \NoCaseChange{\protect\cite{cite5214}}, moderate \NoCaseChange{\protect\cite{cite1023,cite5215}}, and large \NoCaseChange{\protect\cite{cite5216}} deviation analysis.
Bounds exist on the one-shot capacity, i.e., the achievability of classical codes given only one use of the quantum channel.
The ideal decoding error is suppressed exponentially with the number of subsystems \(n\) (for c-q block codes), and the achievable exponent has been studied in Refs. \NoCaseChange{\protect\cite{cite5217,cite5218,cite5219,cite5220,cite5221,cite5222,cite5223,cite5224,cite5225,cite5226}}; see \NoCaseChange{\protect\cite[{Table 2}]{cite5224}} for a summary.
Achievable error exponents for communication are related to those for privacy amplification \NoCaseChange{\protect\cite{cite5227}}.
In the high-rate case, a lower \NoCaseChange{\protect\cite{cite5228}} and upper \NoCaseChange{\protect\cite{cite5229}} bound on the error exponent for general channels matches a conjecture by Holevo \NoCaseChange{\protect\cite{cite5218}}.
A one-shot bound for random codes \NoCaseChange{\protect\cite{cite5230}} resolves a conjecture by Burnashev and Holevo \NoCaseChange{\protect\cite{cite5217}}. 

Unambiguous state discrimination (USD) can be used to achieve Holevo capacity on a general pure-state c-q channel \NoCaseChange{\protect\cite{cite5231}}.

\codefieldsection{Decoding}
\begin{eczvaluelist}
\item\relax Unambiguous state discrimination (USD) \NoCaseChange{\protect\cite{cite5231}}.
\end{eczvaluelist}
\codefieldsection{Parent}
\begin{eczvaluelist}
\item\relax
\flmRefsHyperref[eczindexfamilyrel]{code:oaecc}{Operator-algebra QECC (OAQECC)} --- An OAQECC that retains its block structure for storing classical information but stores no quantum information and has no gauge degrees of freedom (e.g., gauge qubits) is a c-q code.
\end{eczvaluelist}
\codefieldsection{Children}
\begin{eczvaluelist}
\item\relax
\flmRefsHyperref[eczindexfamilyrel]{code:ecc}{Error-correcting code (ECC)} --- Any ECC can be embedded into a quantum Hilbert space, and thus passed through a quantum channel, by associating elements of the alphabet with basis vectors in a Hilbert space over the complex numbers. In other words, classical codewords are elements of an alphabet, while quantum codewords are functions on the alphabet. Classical codes can be unified with quantum codes using various algebraic frameworks \NoCaseChange{\protect\cite{cite1039,cite1040}}.
\item\relax
\flmRefsHyperref[eczindexfamilyrel]{code:concatenated_c-q}{Concatenated c-q code}\item\relax
\flmRefsHyperref[eczindexfamilyrel]{code:bosonic_classical_into_quantum}{Bosonic c-q code}\item\relax
\flmRefsHyperref[eczindexfamilyrel]{code:qubit_classical_into_quantum}{Qubit c-q code}\end{eczvaluelist}
\codefieldsection{Cousin}
\begin{eczvaluelist}
\item\relax
\flmRefsHyperref[eczindexfamilyrel]{code:hybridqecc}{Hybrid QECC} --- A hybrid QECC storing no quantum information reduces to a c-q code.
\end{eczvaluelist}
\eczhbkcontributors{ \eczhuVVA }
\endeczcode

\eczcode{concatenated_c-q}{Concatenated c-q code}{}
\codefieldsection{Description}
A c-q code constructed out of two classical or quantum codes for the purposes of transmission of classical information over quantum channels.
\codefieldsection{Rate}
Concatenated codes can achieve Holevo capacity \NoCaseChange{\protect\cite{cite5232}}.
\codefieldsection{Parent}
\begin{eczvaluelist}
\item\relax
\flmRefsHyperref[eczindexfamilyrel]{code:classical_into_quantum}{Classical-quantum (c-q) code}\end{eczvaluelist}
\codefieldsection{Child}
\begin{eczvaluelist}
\item\relax
\flmRefsHyperref[eczindexfamilyrel]{code:quantum_hadamard_bpsk}{Hadamard BPSK c-q modulation format} --- The Hadamard BPSK c-q code can be thought of as a concatenation of the Hadamard binary linear code with BPSK for the purposes of transmission of classical information over quantum channels.
\end{eczvaluelist}
\codefieldsection{Cousins}
\begin{eczvaluelist}
\item\relax
\flmRefsHyperref[eczindexfamilyrel]{code:generalized_concatenated}{Generalized concatenated code (GCC)} --- Concatenated c-q codes are c-q analogues of generalized concatenated codes.
\item\relax
\flmRefsHyperref[eczindexfamilyrel]{code:quantum_concatenated}{Concatenated quantum code} --- Concatenated c-q codes are hybrid classical-into-quantum analogues of concatenated quantum codes.
\end{eczvaluelist}
\eczhbkcontributors{ \eczhuVVA }
\endeczcode

\eczcode{ea_mixed_alphabet_reed_solomon}{EA mixed-alphabet Reed-Solomon c-q code}{~\NoCaseChange{\protect\cite{cite1922}}}
\codefieldsection{Alternative Names}
\begin{eczvaluelist}
\item\relax Mixed-alphabet Reed-Solomon EACC code
\item\relax Mixed-alphabet RS entanglement-assisted classical code
\end{eczvaluelist}
\eczhIndexCodeAliasName{ea_mixed_alphabet_reed_solomon}{Mixed-alphabet Reed-Solomon EACC code}
\eczhIndexCodeAliasName{ea_mixed_alphabet_reed_solomon}{Mixed-alphabet RS entanglement-assisted classical code}
\codefieldsection{Description}
Entanglement-assisted c-q code obtained from a mixed-alphabet Reed-Solomon construction over \(\mathbb{F}_q\) and \(\mathbb{F}_{q^2}\).
A codeword of an \([n,k,d;c]_q\) code consists of \(n-c\) symbols transmitted directly over \(q\)-dimensional quantum systems and \(c\) symbols transmitted through super-dense coding using \(c\) pre-shared maximally entangled qudit pairs.

More explicitly, the code evaluates all polynomials \(f\in\mathbb{F}_q[x]\) of degree at most \(k-1\) at \(n-c\) distinct points \(\alpha_i\in\mathbb{F}_q\) and \(c\) representatives \(\gamma_j\in\mathbb{F}_{q^2}\setminus\mathbb{F}_q\), choosing at most one element from each conjugate pair \(\{\gamma,\gamma^q\}\).
This yields codewords in \(\mathbb{F}_q^{n-c}\times\mathbb{F}_{q^2}^{c}\), where each \(\mathbb{F}_{q^2}\) symbol is identified with two \(q\)-ary symbols for dense coding.

\codefieldsection{Protection}
Let \(n_1=n-c\) and \(n_2=c\).
If \(n_1\geq k-1\), then the minimum distance is \(d=n-k+1\), saturating the classical Singleton bound.
If \(n_1<k-1\), then
\flmMathEnvironment{align}{}{
  d=\left\lceil\frac{n-k+1+n_2}{2}\right\rceil~,
}
which can exceed the classical Singleton bound because a known erasure of an \(\mathbb{F}_{q^2}\) position removes a two-symbol block \NoCaseChange{\protect\cite{cite1922}}.

\codefieldsection{Rate}
The construction can have \(k>n\) when \(c>0\), since each dense-coded position carries two \(q\)-ary symbols.
Its length is bounded by \(n\leq q+(q^2-q)/2=(q^2+q)/2\), with a possible one-symbol extension using the point at infinity \NoCaseChange{\protect\cite{cite1922}}.
In the range \(n\leq q+(q^2-q)/2\) and \(n-q\leq c\), the distance formula above meets the block-erasure bound of Ref. \NoCaseChange{\protect\cite{cite1922}}.

\codefieldsection{Parent}
\begin{eczvaluelist}
\item\relax
\flmRefsHyperref[eczindexfamilyrel]{code:ea_classical_into_quantum}{Entanglement-assisted (EA) c-q code}\end{eczvaluelist}
\codefieldsection{Cousins}
\begin{eczvaluelist}
\item\relax
\flmRefsHyperref[eczindexfamilyrel]{code:reed_solomon}{Reed-Solomon (RS) code} --- EA mixed-alphabet RS c-q codes use Reed-Solomon polynomial evaluation, but evaluate over both \(\mathbb{F}_q\) and selected representatives from \(\mathbb{F}_{q^2}\setminus\mathbb{F}_q\) to support direct and dense-coded channel uses \NoCaseChange{\protect\cite{cite1922}}.
\item\relax
\flmRefsHyperref[eczindexfamilyrel]{code:mds}{Maximum distance separable (MDS) code} --- EA mixed-alphabet RS c-q codes can saturate a block-erasure bound and, in some parameter ranges, exceed the classical Singleton bound for ordinary \(q\)-ary codes \NoCaseChange{\protect\cite{cite1922}}.
\end{eczvaluelist}
\eczhbkcontributors{ \eczhuVVA }
\endeczcode

\eczcode{ea_classical_into_quantum}{Entanglement-assisted (EA) c-q code}{}
\codefieldsection{Alternative Names}
\begin{eczvaluelist}
\item\relax Entanglement-assisted classical communication (EACC) code
\item\relax Entanglement-assisted classical code
\end{eczvaluelist}
\eczhIndexCodeAliasName{ea_classical_into_quantum}{Entanglement-assisted classical communication (EACC) code}
\eczhIndexCodeAliasName{ea_classical_into_quantum}{Entanglement-assisted classical code}
\codefieldsection{Description}
Classical-quantum code whose encoding and decoding utilize pre-shared entanglement between sender and receiver.
The sender encodes classical information into quantum systems sent through a quantum channel, while the receiver decodes using the channel outputs together with retained halves of pre-shared entangled states.

\codefieldsection{Protection}
A finite-block EACC code is often denoted by \([n,k,d;c]_q\), where \(n\) is the number of \(q\)-dimensional channel uses, \(q^k\) is the number of classical messages, \(d\) is the minimum distance, and \(c\) is the number of pre-shared maximally entangled qudit pairs.
Such a code corrects \(d-1\) erasures or \(\left\lfloor(d-1)/2\right\rfloor\) errors in the setting of Ref. \NoCaseChange{\protect\cite{cite1922}}.

\codefieldsection{Rate}
The entanglement-assisted classical capacity \(C^{\rm ea}(T)\) is the highest asymptotic rate for reliable classical communication through a quantum channel \(T\) when arbitrary pre-shared entanglement is available \NoCaseChange{\protect\cite{cite5233,cite5234}}.
For lossy bosonic channels with high thermal noise and low transmitted photon number, pre-shared entanglement can yield a capacity ratio scaling as \(\log(1/N_S)\) relative to the unassisted Holevo capacity \NoCaseChange{\protect\cite{cite5235,cite5236}}.
If the encoding and decoding circuits themselves are noisy, the fault-tolerant EA capacity approaches the usual EA capacity as the gate error tends to zero \NoCaseChange{\protect\cite{cite2747}}.

\codefieldsection{Encoding}
\begin{eczvaluelist}
\item\relax Super-dense coding maps two \(q\)-ary classical symbols to one transmitted qudit when one maximally entangled qudit pair is available \NoCaseChange{\protect\cite{cite5237}}.
\end{eczvaluelist}
\codefieldsection{Parent}
\begin{eczvaluelist}
\item\relax
\flmRefsHyperref[eczindexfamilyrel]{code:eaoaecc}{Entanglement-assisted operator-algebra QECC (EAOA QECC)} --- An EAOA QECC that has no gauge structure (e.g., gauge qubits), that has a block structure that corresponds to a classical code, that stores no quantum information, and that utilizes pre-shared entanglement is an EA c-q code.
\end{eczvaluelist}
\codefieldsection{Child}
\begin{eczvaluelist}
\item\relax
\flmRefsHyperref[eczindexfamilyrel]{code:ea_mixed_alphabet_reed_solomon}{EA mixed-alphabet Reed-Solomon c-q code}\end{eczvaluelist}
\codefieldsection{Cousins}
\begin{eczvaluelist}
\item\relax
\flmRefsHyperref[eczindexfamilyrel]{code:bosonic_classical_into_quantum}{Bosonic c-q code} --- Bosonic EA c-q schemes use pre-shared continuous-variable entanglement to assist bosonic c-q communication, including structured transceivers for lossy thermal-noise channels \NoCaseChange{\protect\cite{cite5235,cite5236}}.
\item\relax
\flmRefsHyperref[eczindexfamilyrel]{code:eacq}{Entanglement-assisted (EA) hybrid QECC} --- EA c-q codes transmit only classical information with entanglement assistance, while EA hybrid QECCs transmit both classical and quantum information with entanglement assistance.
\item\relax
\flmRefsHyperref[eczindexfamilyrel]{code:eaqecc}{Entanglement-assisted (EA) QECC} --- EA c-q codes transmit classical information with entanglement assistance, while EAQECCs transmit quantum information with entanglement assistance.
\end{eczvaluelist}
\eczhbkcontributors{ \eczhuVVA }
\endeczcode

\onecolumngrid
\clearpage

\section{Binary c-q Kingdom}

\begin{eczEpigraph}
\begin{quote}
\flmQuoteSetup{quote}%
“It from bit” symbolizes the idea that every item of the physical world has at bottom — a very deep bottom, in most instances — an immaterial source and explanation; that which we call reality arises in the last analysis from the posing of yes-or-no questions and the registering of equipment-evoked responses; in short, that all things physical are information-theoretic in origin and that this is a participatory universe.
\flmQuoteAttributed{John A. Wheeler}
\end{quote}
\end{eczEpigraph}

\twocolumngrid

\eczcode{lhz}{Lechner-Hauke-Zoller (LHZ) code}{~\NoCaseChange{\protect\cite{cite5238,cite5239}}}
\codefieldsection{Alternative Names}
\begin{eczvaluelist}
\item\relax Lechner-Hauke-Zoller (LHZ) parity code
\end{eczvaluelist}
\eczhIndexCodeAliasName{lhz}{Lechner-Hauke-Zoller (LHZ) parity code}
\codefieldsection{Description}
LDPC c-q code designed to convert the long-range interactions of a quantum annealer into local constraints.
The code maps the pairwise couplings of a fully connected classical Ising model into local fields together with local parity constraints on a lattice of physical qubits.
An extension maps more general models onto the same lattice \NoCaseChange{\protect\cite{cite5240}}.

\codefieldsection{Encoding}
\begin{eczvaluelist}
\item\relax Arbitrary quantum states \NoCaseChange{\protect\cite{cite5241}}.
\end{eczvaluelist}
\codefieldsection{Gates}
\begin{eczvaluelist}
\item\relax Universal gate set \NoCaseChange{\protect\cite{cite5242}}.
\end{eczvaluelist}
\codefieldsection{Decoding}
\begin{eczvaluelist}
\item\relax BP decoder \NoCaseChange{\protect\cite{cite5239}}.
\end{eczvaluelist}
\codefieldsection{Parent}
\begin{eczvaluelist}
\item\relax
\flmRefsHyperref[eczindexfamilyrel]{code:qubit_classical_into_quantum}{Qubit c-q code} --- The LHZ code is an LDPC c-q code designed to convert the long-range interactions of a quantum annealer into local constraints.
\end{eczvaluelist}
\codefieldsection{Cousins}
\begin{eczvaluelist}
\item\relax
\flmRefsHyperref[eczindexfamilyrel]{code:ldpc}{Low-density parity-check (LDPC) code} --- The LHZ code is an LDPC c-q code designed to convert the long-range interactions of a quantum annealer into local constraints.
\item\relax
\flmRefsHyperref[eczindexfamilyrel]{code:cat_concatenated}{Concatenated cat code} --- LHZ parity-codes have been concatenated with cat codes \NoCaseChange{\protect\cite{cite4857}}.
\end{eczvaluelist}
\eczhbkcontributors{ \eczhuVVA }
\endeczcode

\eczcode{polar_for_quantum}{Polar c-q code}{~\NoCaseChange{\protect\cite{cite5243,cite5244}}}
\codefieldsection{Description}
Polar code adapted to transmit classical information over channels with classical inputs and quantum outputs.

\codefieldsection{Rate}
Codes achieve the symmetric Holevo information for sending classical information over channels with classical inputs and quantum outputs \NoCaseChange{\protect\cite{cite5243}}.
\codefieldsection{Decoding}
\begin{eczvaluelist}
\item\relax Quantum-limited successive-cancellation (SC) joint-detection receiver \NoCaseChange{\protect\cite{cite5243}}.
\end{eczvaluelist}
\codefieldsection{Parent}
\begin{eczvaluelist}
\item\relax
\flmRefsHyperref[eczindexfamilyrel]{code:qubit_classical_into_quantum}{Qubit c-q code}\end{eczvaluelist}
\codefieldsection{Cousins}
\begin{eczvaluelist}
\item\relax
\flmRefsHyperref[eczindexfamilyrel]{code:polar}{Polar code} --- Quantum-classical polar codes generalize polar codes for transmission through channels with quantum output.
\item\relax
\flmRefsHyperref[eczindexfamilyrel]{code:bpsk}{Binary PSK (BPSK) modulation format} --- BPSK concatenated with classical-quantum polar codes approaches the Holevo capacity of the \flmRefsHyperref{ref498}{pure-loss} optical channel in the low-photon-number regime \NoCaseChange{\protect\cite{cite2343}}.
\end{eczvaluelist}
\eczhbkcontributors{ \eczhuVVA }
\endeczcode

\eczcode{qubit_classical_into_quantum}{Qubit c-q code}{}

\codefieldsection{Kingdom root code}
for the \flmRefsHyperref{kingdom:qubit_classical_into_quantum}{Binary c-q Kingdom}.
\codefieldsection{Description}
A qubit code designed for transmission of classical information in the form of bits through non-classical channels.

\codefieldsection{Protection}
Performance of a linear binary code over a channel is dual to the performance of its dual over a particular dual channel \NoCaseChange{\protect\cite{cite5245}}.

\codefieldsection{Decoding}
\begin{eczvaluelist}
\item\relax Belief propagation with quantum messages (BPQM) decoder for c-q channel communication \NoCaseChange{\protect\cite{cite5246}}.
The original BPQM proposal used quantum messages to decode pure-state c-q channels, had c-q polar codes in mind, and yielded explicit capacity-achieving decoders for non-Pauli channels \NoCaseChange{\protect\cite{cite5246}}.
BPQM was analyzed for a BPSK-modulated \flmRefsHyperref{ref498}{pure-loss} channel and shown to be optimal for small tree codes \NoCaseChange{\protect\cite{cite5247}}.
A later quantum message-passing formulation proved optimality of BPQM for any binary linear code with a tree Tanner graph, and proposed an extension to factor graphs with cycles using approximate cloning \NoCaseChange{\protect\cite{cite5248}}.
BPQM has also been extended to symmetric classical-quantum channels using paired measurements \NoCaseChange{\protect\cite{cite5249}}.

\end{eczvaluelist}
\codefieldsection{Realizations}
\begin{eczvaluelist}
\item\relax Quantum enhancement was demonstrated using a polarization-based non-error-correcting c-q encoding \NoCaseChange{\protect\cite{cite5250}}.
\end{eczvaluelist}
\codefieldsection{Parent}
\begin{eczvaluelist}
\item\relax
\flmRefsHyperref[eczindexfamilyrel]{code:classical_into_quantum}{Classical-quantum (c-q) code}\end{eczvaluelist}
\codefieldsection{Children}
\begin{eczvaluelist}
\item\relax
\flmRefsHyperref[eczindexfamilyrel]{code:lhz}{Lechner-Hauke-Zoller (LHZ) code} --- The LHZ code is an LDPC c-q code designed to convert the long-range interactions of a quantum annealer into local constraints.
\item\relax
\flmRefsHyperref[eczindexfamilyrel]{code:polar_for_quantum}{Polar c-q code}\end{eczvaluelist}
\codefieldsection{Cousins}
\begin{eczvaluelist}
\item\relax
\flmRefsHyperref[eczindexfamilyrel]{code:bits_into_bits}{Binary code} --- Any binary code can be embedded into a qubit Hilbert space, and thus passed through a qubit channel, by associating length-\(n\) bitstrings with basis vectors in a Hilbert space over \(\mathbb{Z}_2^n\). For example, a bit of information can be embedded into a two-dimensional vector space by associating the two bit values with two basis vectors for the space.
\item\relax
\flmRefsHyperref[eczindexfamilyrel]{code:hybrid_qubits_into_qubits}{Hybrid qubit code} --- A hybrid qubit code storing no quantum information reduces to a qubit c-q code.
\item\relax
\flmRefsHyperref[eczindexfamilyrel]{code:qubits_into_qubits}{Qubit code} --- Qubit c-q codes are qubit codes designed to transmit classical information.
\end{eczvaluelist}
\eczhbkcontributors{ Saikat Guha, \eczhuVVA }
\endeczcode

\onecolumngrid
\clearpage

\section{Analog c-q Kingdom}

\begin{eczEpigraph}
\begin{quote}
\flmQuoteSetup{quote}%
I like to tell my students: “There is no such thing as classical mechanics!” What I mean by this is that all physical systems are quantum mechanical in nature; the only question is if the classical theory is a sufficiently accurate approximation to reality so that it can be used instead of quantum theory. To use or not to use quantum theory? — That is the question.
\flmQuoteAttributed{Jonathan P. Dowling}
\end{quote}
\end{eczEpigraph}

\twocolumngrid

\eczcode{bosonic_classical_into_quantum}{Bosonic c-q code}{}
\codefieldsection{Alternative Names}
\begin{eczvaluelist}
\item\relax Bosonic c-q modulation format
\item\relax Bosonic c-q modulation scheme
\item\relax Bosonic c-q modulation code
\item\relax Bosonic c-q signaling format
\end{eczvaluelist}
\eczhIndexCodeAliasName{bosonic_classical_into_quantum}{Bosonic c-q modulation format}
\eczhIndexCodeAliasName{bosonic_classical_into_quantum}{Bosonic c-q modulation scheme}
\eczhIndexCodeAliasName{bosonic_classical_into_quantum}{Bosonic c-q modulation code}
\eczhIndexCodeAliasName{bosonic_classical_into_quantum}{Bosonic c-q signaling format}

\codefieldsection{Kingdom root code}
for the \flmRefsHyperref{kingdom:bosonic_classical_into_quantum}{Analog c-q Kingdom}.
\codefieldsection{Description}
Bosonic code designed for transmission of classical information through non-classical channels.
Encodes classical symbols into bosonic quantum states for transmission over a quantum channel and decoding with a quantum-enhanced \textit{receiver}.
This entry includes bosonic c-q modulation formats and is distinct from a \flmRefsHyperref{code:modulation}{classical modulation scheme}, which maps classical symbols into classical electromagnetic signals for transmission over classical channels.
A bosonic c-q modulation format instead treats the transmitted signals as quantum states and allows the receiver to use quantum measurements.

\codefieldsection{Rate}
The Holevo capacity has been calculated for various bosonic quantum channels \NoCaseChange{\protect\cite{cite5251,cite5252,cite5253}} such as the \flmRefsHyperref{ref498}{pure-loss bosonic channel} \NoCaseChange{\protect\cite{cite5254}} or quantum AWGN \NoCaseChange{\protect\cite{cite5255}}. The energy-constrained capacity of the noiseless bosonic c-q channel is finite due to quantum effects \NoCaseChange{\protect\cite{cite5256,cite5257}}, while the Shannon capacity can be infinite. Gordon was the first to calculate such capacities (in a published work) for a specific case \NoCaseChange{\protect\cite{cite5258,cite5259,cite5260}}, and a related discussion is given by Forney \NoCaseChange{\protect\cite{cite5261}}. The most information-efficient format of a transmitted message is indistinguishable from black-body radiation \NoCaseChange{\protect\cite{cite5262}}.
\codefieldsection{Parent}
\begin{eczvaluelist}
\item\relax
\flmRefsHyperref[eczindexfamilyrel]{code:classical_into_quantum}{Classical-quantum (c-q) code}\end{eczvaluelist}
\codefieldsection{Children}
\begin{eczvaluelist}
\item\relax
\flmRefsHyperref[eczindexfamilyrel]{code:coherent_state_c-q}{Coherent-state c-q modulation format}\item\relax
\flmRefsHyperref[eczindexfamilyrel]{code:fock_state_ook}{Fock-state OOK c-q modulation format}\item\relax
\flmRefsHyperref[eczindexfamilyrel]{code:niset_andersen_cerf}{Niset-Andersen-Cerf code}\item\relax
\flmRefsHyperref[eczindexfamilyrel]{code:squeezed_coherent_bpsk}{Squeezed-coherent BPSK c-q modulation format}\end{eczvaluelist}
\codefieldsection{Cousins}
\begin{eczvaluelist}
\item\relax
\flmRefsHyperref[eczindexfamilyrel]{code:oscillators}{Bosonic code} --- Bosonic c-q codes are bosonic codes designed to transmit classical information.
\item\relax
\flmRefsHyperref[eczindexfamilyrel]{code:modulation}{Modulation scheme} --- Classical modulation schemes transmit classical signals over classical channels, while bosonic c-q modulation formats transmit quantum states over quantum channels and can use quantum-enhanced receivers.
\item\relax
\flmRefsHyperref[eczindexfamilyrel]{code:analog}{Analog code} --- Any analog code can be embedded into a bosonic Hilbert space, and thus passed through a bosonic channel, by associating the reals with the configuration space of position states of bosonic modes.
\item\relax
\flmRefsHyperref[eczindexfamilyrel]{code:ea_classical_into_quantum}{Entanglement-assisted (EA) c-q code} --- Bosonic EA c-q schemes use pre-shared continuous-variable entanglement to assist bosonic c-q communication, including structured transceivers for lossy thermal-noise channels \NoCaseChange{\protect\cite{cite5235,cite5236}}.
\end{eczvaluelist}
\eczhbkcontributors{ Jasminder Sidhu, \eczhuVVA }
\endeczcode

\eczcode{quantum_bpsk}{BPSK c-q modulation format}{}
\codefieldsection{Alternative Names}
\begin{eczvaluelist}
\item\relax BPSK c-q modulation code
\item\relax BPSK c-q modulation scheme
\item\relax BPSK c-q signaling format
\end{eczvaluelist}
\eczhIndexCodeAliasName{quantum_bpsk}{BPSK c-q modulation code}
\eczhIndexCodeAliasName{quantum_bpsk}{BPSK c-q modulation scheme}
\eczhIndexCodeAliasName{quantum_bpsk}{BPSK c-q signaling format}
\codefieldsection{Description}
Coherent-state c-q binary code encoding into two coherent states \(|\pm\alpha\rangle\) for complex \(\alpha\). A shifted version, with codewords \(\{|0\rangle,|\alpha\rangle\}\), is called binary amplitude modulation (BAM), The three-state subcode \(\{|\alpha,\alpha\rangle,|-\alpha,\alpha\rangle,|\alpha,-\alpha\rangle\}\) of two-mode BPSK is called the \textit{single-degeneracy code} \NoCaseChange{\protect\cite{cite874}}.
\codefieldsection{Rate}
The single-degeneracy code yields an improved PIE by \(2.8\%\) over BPSK \NoCaseChange{\protect\cite{cite5263}} (see \NoCaseChange{\protect\cite{cite874}}).
\codefieldsection{Decoding}
\begin{eczvaluelist}
\item\relax Linear-optical quantum receiver \NoCaseChange{\protect\cite{cite5264}}.
\item\relax Kennedy receiver \NoCaseChange{\protect\cite{cite5265,cite5266}}.
\item\relax Photon-number resolving detector \NoCaseChange{\protect\cite{cite5267}}.
\item\relax Non-Gaussian near-optimal receiver \NoCaseChange{\protect\cite{cite5266}}.
\item\relax Multi-stage quantum receiver \NoCaseChange{\protect\cite{cite5268}}.
\item\relax Quantum receiver attaining the Helstrom bound in the low-photon regime \NoCaseChange{\protect\cite{cite5269}}.
\item\relax Green machine receiver \NoCaseChange{\protect\cite{cite5270}}.
\end{eczvaluelist}
\codefieldsection{Realizations}
\begin{eczvaluelist}
\item\relax Linear-optical quantum receiver \NoCaseChange{\protect\cite{cite5264}}.
\item\relax Homodyne receiver \NoCaseChange{\protect\cite{cite5271}}.
\item\relax Kennedy receiver \NoCaseChange{\protect\cite{cite5271,cite5272}}.
\item\relax Photon-number resolving detector \NoCaseChange{\protect\cite{cite5267}}.
\item\relax Communication over dephasing \NoCaseChange{\protect\cite{cite5273}}, time-varying phase-noise \NoCaseChange{\protect\cite{cite5274}}, and thermal-noise \NoCaseChange{\protect\cite{cite5275}} channels.
\item\relax Adaptive decoder using displacements and photon detection \NoCaseChange{\protect\cite{cite5276}}.
\item\relax BPQM detector on a BPSK-modulated tree code \NoCaseChange{\protect\cite{cite5277}}.
\item\relax Superadditivity of the green machine receiver has been demonstrated in the photon starved regime \NoCaseChange{\protect\cite{cite5270}}.
\end{eczvaluelist}
\codefieldsection{Parents}
\begin{eczvaluelist}
\item\relax
\flmRefsHyperref[eczindexfamilyrel]{code:quantum_psk}{PSK c-q modulation format}\item\relax
\flmRefsHyperref[eczindexfamilyrel]{code:quantum_hadamard_bpsk}{Hadamard BPSK c-q modulation format}\item\relax
\flmRefsHyperref[eczindexfamilyrel]{code:squeezed_coherent_bpsk}{Squeezed-coherent BPSK c-q modulation format} --- Squeezed-coherent BPSK c-q modulation reduces to coherent-state BPSK c-q modulation when the squeezing parameter is zero.
\end{eczvaluelist}
\codefieldsection{Cousins}
\begin{eczvaluelist}
\item\relax
\flmRefsHyperref[eczindexfamilyrel]{code:bpsk}{Binary PSK (BPSK) modulation format} --- BPSK (BPSK c-q) codes are used to transmit classical information using antipodal coherent states over classical (quantum) channels.
\item\relax
\flmRefsHyperref[eczindexfamilyrel]{code:two-legged-cat}{Two-component cat code} --- BPSK c-q (two-component cat) codes are used to transmit classical (quantum) information using (superpositions of) antipodal coherent states over quantum channels.
\item\relax
\flmRefsHyperref[eczindexfamilyrel]{code:quantum_ook}{On-off keyed (OOK) c-q modulation format} --- OOK c-q codewords are related to BPSK c-q codewords by a displacement in phase space.
\end{eczvaluelist}
\eczhbkcontributors{ \eczhuVVA }
\endeczcode

\eczcode{quantum_fsk}{Coherent FSK (CFSK) c-q modulation format}{~\NoCaseChange{\protect\cite{cite5278,cite5279}}}
\codefieldsection{Alternative Names}
\begin{eczvaluelist}
\item\relax Coherent FSK (CFSK) c-q modulation code
\item\relax Coherent FSK (CFSK) c-q modulation scheme
\item\relax Coherent FSK (CFSK) c-q signaling format
\end{eczvaluelist}
\eczhIndexCodeAliasName{quantum_fsk}{Coherent FSK (CFSK) c-q modulation code}
\eczhIndexCodeAliasName{quantum_fsk}{Coherent FSK (CFSK) c-q modulation scheme}
\eczhIndexCodeAliasName{quantum_fsk}{Coherent FSK (CFSK) c-q signaling format}
\codefieldsection{Description}
\flmRefsHyperref{code:coherent_state_c-q}{Coherent-state c-q code} encoding into coherent states that are frequency-shifted with certain initial relative phase.

Codewords are coherent states \(|\alpha_m\rangle\), where \flmMathEnvironment{align}{}{ \alpha_m
= \alpha e^{i(\omega_0+[m-1]\Delta\omega)t+i(m-1)\Delta\theta} } for common
frequency \(\omega_0\), frequency shift \(\Delta\omega < 2\pi/T\), total time \(T\),
and phase shift \(\Delta\theta\).

\codefieldsection{Protection}
The square-root measurement is not optimal for CFSK c-q codes, unlike for PSK c-q codes \NoCaseChange{\protect\cite{cite5280}}.

\codefieldsection{Decoding}
\begin{eczvaluelist}
\item\relax Bondurant receiver \NoCaseChange{\protect\cite{cite5281}}.
\item\relax Cyclic receiver \NoCaseChange{\protect\cite{cite5282}}.
\item\relax Time-resolving receiver \NoCaseChange{\protect\cite{cite5278,cite5279,cite5283}}.
\item\relax Bayesian inference \NoCaseChange{\protect\cite{cite5278}}.
\end{eczvaluelist}
\codefieldsection{Realizations}
\begin{eczvaluelist}
\item\relax Time-resolving quantum receiver \NoCaseChange{\protect\cite{cite5283}}.
\item\relax Bondurant receiver \NoCaseChange{\protect\cite{cite5282}}.
\item\relax Bayesian inference \NoCaseChange{\protect\cite{cite5279}}.
\end{eczvaluelist}
\codefieldsection{Parent}
\begin{eczvaluelist}
\item\relax
\flmRefsHyperref[eczindexfamilyrel]{code:coherent_state_c-q}{Coherent-state c-q modulation format}\end{eczvaluelist}
\codefieldsection{Child}
\begin{eczvaluelist}
\item\relax
\flmRefsHyperref[eczindexfamilyrel]{code:quantum_psk}{PSK c-q modulation format} --- The CFSK c-q code reduces to the \(q\)-ary PSK c-q code when \(\Delta\omega = 0\) and \(\Delta\theta = 2\pi/q\).
\end{eczvaluelist}
\codefieldsection{Cousin}
\begin{eczvaluelist}
\item\relax
\flmRefsHyperref[eczindexfamilyrel]{code:fsk}{Frequency-shift keying (FSK) modulation format} --- Coherent FSK c-q codes are classical-quantum analogues of FSK codes.
\end{eczvaluelist}
\eczhbkcontributors{ Ivan A. Burenkov, \eczhuVVA }
\endeczcode

\eczcode{coherent_state_c-q}{Coherent-state c-q modulation format}{}
\codefieldsection{Alternative Names}
\begin{eczvaluelist}
\item\relax Coherent-state c-q modulation code
\item\relax Coherent-state c-q modulation scheme
\item\relax Coherent-state c-q signaling format
\end{eczvaluelist}
\eczhIndexCodeAliasName{coherent_state_c-q}{Coherent-state c-q modulation code}
\eczhIndexCodeAliasName{coherent_state_c-q}{Coherent-state c-q modulation scheme}
\eczhIndexCodeAliasName{coherent_state_c-q}{Coherent-state c-q signaling format}
\codefieldsection{Description}
Bosonic c-q code whose codewords form a constellation of coherent states.
Encodes classical symbols into coherent states for transmission over a quantum channel and decoding with a quantum-enhanced \textit{receiver}.

The code consists of \(K\) coherent states on \(n\) modes, where the \(j\)th state, or codeword, is uniquely defined through the amplitude vector \(\boldsymbol{\alpha}^j=(\alpha_1^j, \alpha_2^j, \cdots, \alpha_n^j)\).
The \textit{codebook},
\flmMathEnvironment{align}{}{
C=\left(\begin{array}{c}
\boldsymbol{\alpha}^{1}\\
\vdots\\
\boldsymbol{\alpha}^{K}
\end{array}\right)=\left(\begin{array}{cccc}
\alpha_{1}^{1} & \alpha_{2}^{1} & \dots & \alpha_{n}^{1}\\
\vdots & \vdots & \ddots & \vdots\\
\alpha_{1}^{K} & \alpha_{2}^{K} & \dots & \alpha_{n}^{K}
\end{array}\right)~,
}
collects each codeword into the matrix \(C\) that characterizes the system of states to discriminate.

From the properties of \(C\), we can assess whether it is possible to discriminate the codebook unambiguously.
For a finite constellation, unambiguous state discrimination is possible only if the coherent states in the codebook are linearly independent.

\codefieldsection{Rate}
Random Gaussian-distributed coherent-state c-q codes achieve the capacity of the \flmRefsHyperref{ref498}{pure-loss bosonic channel} \NoCaseChange{\protect\cite{cite5254}}.
\codefieldsection{Decoding}
\begin{eczvaluelist}
\item\relax Optimal receiver performance in ambiguous state discrimination is determined using the \textit{Yuen-Kennedy-Lax (YKL) conditions} \NoCaseChange{\protect\cite{cite5284}}. See review \NoCaseChange{\protect\cite{cite5285}} for details on receivers used for coherent-state c-q codes.
\item\relax Joint-detection receiver that can attain channel capacity \NoCaseChange{\protect\cite{cite874}}.
\item\relax Various near-optimal receiver designs that can handle arbitrary constellations of coherent states with possible degeneracies \NoCaseChange{\protect\cite{cite5286}}.
\item\relax The \textit{square-root measurement} (a.k.a. pretty good measurement) \NoCaseChange{\protect\cite{cite5287,cite5288,cite5289}} is optimal for geometrically uniform \NoCaseChange{\protect\cite{cite5290,cite5291,cite5280,cite5292}}, direct sums of geometrically uniform \NoCaseChange{\protect\cite{cite5293}}, and compound geometrically uniform \NoCaseChange{\protect\cite{cite5294}} constellations.
\end{eczvaluelist}
\codefieldsection{Realizations}
\begin{eczvaluelist}
\item\relax Continuous-variable quantum key distribution (CV-QKD) \NoCaseChange{\protect\cite{cite5295,cite5296,cite5297}}.
\end{eczvaluelist}
\codefieldsection{Notes}
\begin{eczvaluelist}
\item\relax See book \NoCaseChange{\protect\cite{cite3659}}.
\end{eczvaluelist}
\codefieldsection{Parent}
\begin{eczvaluelist}
\item\relax
\flmRefsHyperref[eczindexfamilyrel]{code:bosonic_classical_into_quantum}{Bosonic c-q code}\end{eczvaluelist}
\codefieldsection{Children}
\begin{eczvaluelist}
\item\relax
\flmRefsHyperref[eczindexfamilyrel]{code:quantum_hadamard_bpsk}{Hadamard BPSK c-q modulation format}\item\relax
\flmRefsHyperref[eczindexfamilyrel]{code:quantum_fsk}{Coherent FSK (CFSK) c-q modulation format}\item\relax
\flmRefsHyperref[eczindexfamilyrel]{code:quantum_ook}{On-off keyed (OOK) c-q modulation format}\item\relax
\flmRefsHyperref[eczindexfamilyrel]{code:quantum_ppm}{Pulse-position (PPM) c-q modulation format}\end{eczvaluelist}
\codefieldsection{Cousins}
\begin{eczvaluelist}
\item\relax
\flmRefsHyperref[eczindexfamilyrel]{code:coherent_constellation}{Coherent-state constellation code} --- Coherent-state c-q codes encode classical alphabets into constellations of coherent states, while coherent-state constellation codes encode quantum information into superpositions of coherent states.
\item\relax
\flmRefsHyperref[eczindexfamilyrel]{code:niset_andersen_cerf}{Niset-Andersen-Cerf code} --- The Niset-Andersen-Cerf code uses coherent states but functions as an erasure-protection code rather than as a modulation format.
\item\relax
\flmRefsHyperref[eczindexfamilyrel]{code:two-legged-cat}{Two-component cat code} --- Two-component cat codes can be thought of as coherent-state c-q codes because they protect against only one type of noise and thus only reliably store classical information.
\end{eczvaluelist}
\eczhbkcontributors{ \eczhuVVA }
\endeczcode

\eczcode{fock_state_ook}{Fock-state OOK c-q modulation format}{~\NoCaseChange{\protect\cite{cite5298}}}
\codefieldsection{Alternative Names}
\begin{eczvaluelist}
\item\relax Number-state OOK c-q modulation format
\item\relax Fock-state OOK c-q modulation code
\item\relax Fock-state OOK c-q modulation scheme
\item\relax Fock-state OOK c-q signaling format
\item\relax Single-photon OOK c-q modulation format
\item\relax Single-rail c-q code
\end{eczvaluelist}
\eczhIndexCodeAliasName{fock_state_ook}{Number-state OOK c-q modulation format}
\eczhIndexCodeAliasName{fock_state_ook}{Fock-state OOK c-q modulation code}
\eczhIndexCodeAliasName{fock_state_ook}{Fock-state OOK c-q modulation scheme}
\eczhIndexCodeAliasName{fock_state_ook}{Fock-state OOK c-q signaling format}
\eczhIndexCodeAliasName{fock_state_ook}{Single-photon OOK c-q modulation format}
\eczhIndexCodeAliasName{fock_state_ook}{Single-rail c-q code}
\codefieldsection{Description}
Bosonic c-q on-off keying (OOK) modulation format whose binary alphabet consists of the vacuum state \(|0\rangle\) and the single-photon Fock state \(|1\rangle\) of one mode.
More generally, the nonzero OOK symbol can be a number state or a mixture of adjacent number states.
Fock-state OOK with photon-number detection was analyzed as a nonclassical alternative to coherent-state OOK for photon-efficient communication \NoCaseChange{\protect\cite{cite5298}}.

\codefieldsection{Decoding}
\begin{eczvaluelist}
\item\relax Photon-number detection distinguishes the erasure/loss-degraded output alphabet by detecting whether photons are present in the received mode \NoCaseChange{\protect\cite{cite5298,cite5299}}.
\item\relax Number-state modulation with photon detection, including the numerical observation that the optimal prior distribution can be multimodal, was studied in Ref. \NoCaseChange{\protect\cite[{Ch. 5}]{cite5300}}. Later work on incoherent OOK and PPM quantified improvements from sub-Poissonian nonclassical light and used Fock states as the nonclassical low-photon-number alphabet \NoCaseChange{\protect\cite{cite5299}}.
\end{eczvaluelist}
\codefieldsection{Parent}
\begin{eczvaluelist}
\item\relax
\flmRefsHyperref[eczindexfamilyrel]{code:bosonic_classical_into_quantum}{Bosonic c-q code}\end{eczvaluelist}
\codefieldsection{Cousins}
\begin{eczvaluelist}
\item\relax
\flmRefsHyperref[eczindexfamilyrel]{code:quantum_ook}{On-off keyed (OOK) c-q modulation format} --- Fock-state OOK and coherent-state OOK both use a vacuum off symbol, but their on symbols are respectively a single-photon number state and a coherent state.
\item\relax
\flmRefsHyperref[eczindexfamilyrel]{code:fock_state}{Fock-state bosonic code} --- Fock-state OOK transmits classical information using Fock states, while Fock-state bosonic codes store quantum information in subspaces built from Fock states.
\end{eczvaluelist}
\eczhbkcontributors{ Saikat Guha, \eczhuVVA }
\endeczcode

\eczcode{quantum_hadamard_bpsk}{Hadamard BPSK c-q modulation format}{~\NoCaseChange{\protect\cite{cite874}}}
\codefieldsection{Alternative Names}
\begin{eczvaluelist}
\item\relax Hadamard BPSK c-q modulation code
\item\relax Hadamard BPSK c-q modulation scheme
\item\relax Hadamard BPSK c-q signaling format
\end{eczvaluelist}
\eczhIndexCodeAliasName{quantum_hadamard_bpsk}{Hadamard BPSK c-q modulation code}
\eczhIndexCodeAliasName{quantum_hadamard_bpsk}{Hadamard BPSK c-q modulation scheme}
\eczhIndexCodeAliasName{quantum_hadamard_bpsk}{Hadamard BPSK c-q signaling format}
\codefieldsection{Description}
Multimode coherent-state c-q code that is a concatenation of a Hadamard code with a BPSK c-q code.
Its codewords are \(n\)-mode coherent states whose components \(\pm\alpha\) are arranged according to rows of a Hadamard matrix.

\codefieldsection{Rate}
Using a joint-detection receiver, the code exhibits superadditive capacity relative to symbol-by-symbol detection and approaches the Holevo limit in the low-photon regime \NoCaseChange{\protect\cite{cite874}}.
\codefieldsection{Parents}
\begin{eczvaluelist}
\item\relax
\flmRefsHyperref[eczindexfamilyrel]{code:coherent_state_c-q}{Coherent-state c-q modulation format}\item\relax
\flmRefsHyperref[eczindexfamilyrel]{code:concatenated_c-q}{Concatenated c-q code} --- The Hadamard BPSK c-q code can be thought of as a concatenation of the Hadamard binary linear code with BPSK for the purposes of transmission of classical information over quantum channels.
\end{eczvaluelist}
\codefieldsection{Child}
\begin{eczvaluelist}
\item\relax
\flmRefsHyperref[eczindexfamilyrel]{code:quantum_bpsk}{BPSK c-q modulation format}\end{eczvaluelist}
\codefieldsection{Cousins}
\begin{eczvaluelist}
\item\relax
\flmRefsHyperref[eczindexfamilyrel]{code:hadamard}{\([2^m,m,2^{m-1}]\) Hadamard code} --- The Hadamard BPSK c-q code can be thought of as a concatenation of the Hadamard binary linear code with BPSK for the purposes of transmission of classical information over quantum channels.
\item\relax
\flmRefsHyperref[eczindexfamilyrel]{code:bpsk}{Binary PSK (BPSK) modulation format} --- The Hadamard BPSK c-q code can be thought of as a concatenation of the Hadamard binary linear code with BPSK for the purposes of transmission of classical information over quantum channels.
\end{eczvaluelist}
\eczhbkcontributors{ \eczhuVVA }
\endeczcode

\eczcode{niset_andersen_cerf}{Niset-Andersen-Cerf code}{~\NoCaseChange{\protect\cite{cite4700}}}
\codefieldsection{Description}
Coherent-state c-q code encoding two-mode coherent states \(\{|\alpha\rangle, |\beta\rangle\}\) into four modes such that the complex values \((\alpha,\beta)\) are recoverable after a single-mode erasure. There are two variations of the storage procedure: a deterministic protocol that offers recovery against a single-mode erasure, and a probabilistic one that can protect against multiple errors with post-selection. This code effectively protects classical information stored in \((\alpha,\beta)\) using quantum operations.
\codefieldsection{Protection}
The deterministic protocol protects against a single erasure error on a known mode. This recovers one state perfectly and the other state with fidelity \(F = \frac{1}{1 + e^{-2 r}}\) for an initial EPR pair squeezed with variance \(e^{-2r}\). The probabilistic protocol utilizes post-selection to protect against multiple erasures with state-dependent fidelity.
\codefieldsection{Encoding}
\begin{eczvaluelist}
\item\relax After an EPR pair preparation, use 2 continuous CNOT and 2 continuous inverse CNOT gates to entangle a bosonic EPR pair with initial states \(|\alpha \rangle\) and \(|\beta \rangle\).
\item\relax Alternate optical encoder using a two-mode squeezed vacuum state and two balanced beam splitters to mix the input coherent states with the EPR pair.
\end{eczvaluelist}
\codefieldsection{Decoding}
\begin{eczvaluelist}
\item\relax Optical decoder using three beam splitters, electronic gain detectors, and two phase-insensitive amplifiers as described in Ref. \NoCaseChange{\protect\cite{cite4700}}.
\end{eczvaluelist}
\codefieldsection{Realizations}
\begin{eczvaluelist}
\item\relax Realized in Ref. \NoCaseChange{\protect\cite{cite5301}} in an optical system with 3 beam splitters. The fidelity peaked around \(0.6\) for the deterministic approach, and around \(0.77\) for the probabilistic approach (with a 25\% chance of error).
\end{eczvaluelist}
\codefieldsection{Parent}
\begin{eczvaluelist}
\item\relax
\flmRefsHyperref[eczindexfamilyrel]{code:bosonic_classical_into_quantum}{Bosonic c-q code}\end{eczvaluelist}
\codefieldsection{Cousins}
\begin{eczvaluelist}
\item\relax
\flmRefsHyperref[eczindexfamilyrel]{code:coherent_state_c-q}{Coherent-state c-q modulation format} --- The Niset-Andersen-Cerf code uses coherent states but functions as an erasure-protection code rather than as a modulation format.
\item\relax
\flmRefsHyperref[eczindexfamilyrel]{code:analog_repetition}{Analog repetition code} --- EPR pairs are used in an encoding of the Niset-Andersen-Cerf code \NoCaseChange{\protect\cite{cite4700}}.
\item\relax
\flmRefsHyperref[eczindexfamilyrel]{code:hnss}{Hayden-Nezami-Salton-Sanders bosonic code} --- The Niset-Andersen-Cerf code can be viewed as a scheme to replicate quantum information in multiple regions \NoCaseChange{\protect\cite{cite2172}}.
\end{eczvaluelist}
\eczhbkcontributors{ Siddharth Taneja, \eczhuVVA }
\endeczcode

\eczcode{quantum_ook}{On-off keyed (OOK) c-q modulation format}{~\NoCaseChange{\protect\cite{cite5302}}}
\codefieldsection{Alternative Names}
\begin{eczvaluelist}
\item\relax On-off keyed (OOK) c-q modulation code
\item\relax On-off keyed (OOK) c-q modulation scheme
\item\relax On-off keyed (OOK) c-q signaling format
\end{eczvaluelist}
\eczhIndexCodeAliasName{quantum_ook}{On-off keyed (OOK) c-q modulation code}
\eczhIndexCodeAliasName{quantum_ook}{On-off keyed (OOK) c-q modulation scheme}
\eczhIndexCodeAliasName{quantum_ook}{On-off keyed (OOK) c-q signaling format}
\codefieldsection{Description}
Coherent-state c-q binary code whose encoding is either in the vacuum \(|0\rangle\) or in a nonzero coherent state \(|\alpha\rangle\).
\codefieldsection{Protection}
For equal priors on the binary alphabet \(\{|0\rangle,|\alpha\rangle\}\), the minimum error probability is given by the Helstrom expression
\flmMathEnvironment{align}{}{
  P_{\mathrm{err}}^{\star}=\frac{1}{2}\left(1-\sqrt{1-e^{-|\alpha|^2}}\right)~,
}
since \(|\langle 0|\alpha\rangle|^2=e^{-|\alpha|^2}\).
The mean photon number of the non-vacuum symbol is \(|\alpha|^2\).

\codefieldsection{Decoding}
\begin{eczvaluelist}
\item\relax Dolinar receiver \NoCaseChange{\protect\cite{cite5303}}, which attains the Helstrom limit in the ideal model and has been demonstrated in proof-of-principle experiments \NoCaseChange{\protect\cite{cite5302}}.
\item\relax Superconducting transition edge sensor (TES) photon-number resolving detector \NoCaseChange{\protect\cite{cite5304}}.
\end{eczvaluelist}
\codefieldsection{Realizations}
\begin{eczvaluelist}
\item\relax Proof-of-principle experiments using Dolinar \NoCaseChange{\protect\cite{cite5302}} and TES receivers \NoCaseChange{\protect\cite{cite5304}}.
\end{eczvaluelist}
\codefieldsection{Parent}
\begin{eczvaluelist}
\item\relax
\flmRefsHyperref[eczindexfamilyrel]{code:coherent_state_c-q}{Coherent-state c-q modulation format}\end{eczvaluelist}
\codefieldsection{Cousins}
\begin{eczvaluelist}
\item\relax
\flmRefsHyperref[eczindexfamilyrel]{code:quantum_bpsk}{BPSK c-q modulation format} --- OOK c-q codewords are related to BPSK c-q codewords by a displacement in phase space.
\item\relax
\flmRefsHyperref[eczindexfamilyrel]{code:fock_state_ook}{Fock-state OOK c-q modulation format} --- Fock-state OOK and coherent-state OOK both use a vacuum off symbol, but their on symbols are respectively a single-photon number state and a coherent state.
\end{eczvaluelist}
\eczhbkcontributors{ \eczhuVVA }
\endeczcode

\eczcode{quantum_psk}{PSK c-q modulation format}{~\NoCaseChange{\protect\cite{cite5305}}}
\codefieldsection{Alternative Names}
\begin{eczvaluelist}
\item\relax PSK c-q modulation code
\item\relax PSK c-q modulation scheme
\item\relax PSK c-q signaling format
\end{eczvaluelist}
\eczhIndexCodeAliasName{quantum_psk}{PSK c-q modulation code}
\eczhIndexCodeAliasName{quantum_psk}{PSK c-q modulation scheme}
\eczhIndexCodeAliasName{quantum_psk}{PSK c-q signaling format}
\codefieldsection{Description}
Coherent-state c-q \(q\)-ary code whose \(j\)th codeword corresponds to a coherent state whose phase is the \(j\)th multiple of \(2\pi/q\). These states are also called geometrically uniform states (GUS) \NoCaseChange{\protect\cite{cite875}}.
\codefieldsection{Protection}
The error probability for \(q=4\) under an optimal quantum detector is worked out in \NoCaseChange{\protect\cite[{Sec. IV.3}]{cite4912}}; see also \NoCaseChange{\protect\cite{cite875,cite5306,cite5307}}.

\codefieldsection{Decoding}
\begin{eczvaluelist}
\item\relax Multi-stage quantum receivers \NoCaseChange{\protect\cite{cite5308,cite5309,cite5310,cite5311,cite5312,cite5313}}.
\item\relax Bayesian inference \NoCaseChange{\protect\cite{cite5278}}.
\end{eczvaluelist}
\codefieldsection{Realizations}
\begin{eczvaluelist}
\item\relax Unambiguous state discrimination using displacement-based receiver for 4-PSK \NoCaseChange{\protect\cite{cite5314}}.
\item\relax Multi-stage quantum receivers \NoCaseChange{\protect\cite{cite5315,cite5316,cite5305,cite5317}}.
\item\relax Bayesian inference \NoCaseChange{\protect\cite{cite5318}}.
\item\relax Time resolving quantum receiver operating in the telecom C band \NoCaseChange{\protect\cite{cite5319}}.
\item\relax Displacements and photon detection \NoCaseChange{\protect\cite{cite5320}}.
\item\relax Adaptive decoder using linear-optical elements and photon detection \NoCaseChange{\protect\cite{cite5276}}.
\end{eczvaluelist}
\codefieldsection{Parent}
\begin{eczvaluelist}
\item\relax
\flmRefsHyperref[eczindexfamilyrel]{code:quantum_fsk}{Coherent FSK (CFSK) c-q modulation format} --- The CFSK c-q code reduces to the \(q\)-ary PSK c-q code when \(\Delta\omega = 0\) and \(\Delta\theta = 2\pi/q\).
\end{eczvaluelist}
\codefieldsection{Child}
\begin{eczvaluelist}
\item\relax
\flmRefsHyperref[eczindexfamilyrel]{code:quantum_bpsk}{BPSK c-q modulation format}\end{eczvaluelist}
\codefieldsection{Cousins}
\begin{eczvaluelist}
\item\relax
\flmRefsHyperref[eczindexfamilyrel]{code:psk}{Phase-shift keying (PSK) modulation format} --- PSK (PSK c-q) codes are used to transmit classical information using single-mode coherent states distributed on a circle over classical (quantum) channels.
\item\relax
\flmRefsHyperref[eczindexfamilyrel]{code:cat}{Cat code} --- PSK c-q (cat) codes are used to transmit classical (quantum) information using (superpositions of) single-mode coherent states distributed on a circle over quantum channels.
\item\relax
\flmRefsHyperref[eczindexfamilyrel]{code:polygon}{Polygon code} --- The PSK coherent-state constellation forms the vertices of a \(q\)-gon.
\end{eczvaluelist}
\eczhbkcontributors{ \eczhuVVA }
\endeczcode

\eczcode{quantum_ppm}{Pulse-position (PPM) c-q modulation format}{~\NoCaseChange{\protect\cite{cite5321}}}
\codefieldsection{Alternative Names}
\begin{eczvaluelist}
\item\relax Pulse-position (PPM) c-q modulation code
\item\relax Pulse-position (PPM) c-q modulation scheme
\item\relax Pulse-position (PPM) c-q signaling format
\end{eczvaluelist}
\eczhIndexCodeAliasName{quantum_ppm}{Pulse-position (PPM) c-q modulation code}
\eczhIndexCodeAliasName{quantum_ppm}{Pulse-position (PPM) c-q modulation scheme}
\eczhIndexCodeAliasName{quantum_ppm}{Pulse-position (PPM) c-q signaling format}
\codefieldsection{Description}
A \(q\)-PPM c-q code is a coherent-state c-q code whose \(j\)th codeword corresponds to a tensor-product state of zero-amplitude coherent states at all modes except mode \(j\).
For example, a 3-PPM encoding corresponds to the three-mode states \(|\alpha\rangle|0\rangle|0\rangle\), \(|0\rangle|\alpha\rangle|0\rangle\), and \(|0\rangle|0\rangle|\alpha\rangle\) for some complex \(\alpha\).
The dual of a PPM code is obtained by the exchange \(0\leftrightarrow\alpha\).

\codefieldsection{Protection}
The error probability under an optimal quantum detector is worked out in \NoCaseChange{\protect\cite[{Sec. IV.2}]{cite4912}}.

\codefieldsection{Decoding}
\begin{eczvaluelist}
\item\relax Conditional pulse nulling (CPN) receiver \NoCaseChange{\protect\cite{cite5322}}.
\end{eczvaluelist}
\codefieldsection{Realizations}
\begin{eczvaluelist}
\item\relax Conditional pulse nulling (CPN) receiver \NoCaseChange{\protect\cite{cite5321}}.
\end{eczvaluelist}
\codefieldsection{Parent}
\begin{eczvaluelist}
\item\relax
\flmRefsHyperref[eczindexfamilyrel]{code:coherent_state_c-q}{Coherent-state c-q modulation format}\end{eczvaluelist}
\codefieldsection{Cousins}
\begin{eczvaluelist}
\item\relax
\flmRefsHyperref[eczindexfamilyrel]{code:ppm}{Pulse-position modulation (PPM) format} --- PPM c-q codes are quantum analogues of PPM codes.
\item\relax
\flmRefsHyperref[eczindexfamilyrel]{code:biorthogonal_spherical}{Biorthogonal spherical code} --- PPM c-q codewords are c-q spherical codes whose constellation consists of the standard basis vectors. Adjoining negatives yields the corresponding biorthogonal c-q spherical code.
\item\relax
\flmRefsHyperref[eczindexfamilyrel]{code:one_hot}{One-hot code} --- The PPM c-q code is a continuous analogue of the one-hot code designed for transmission through quantum channels.
\end{eczvaluelist}
\eczhbkcontributors{ \eczhuVVA }
\endeczcode

\eczcode{squeezed_coherent_bpsk}{Squeezed-coherent BPSK c-q modulation format}{~\NoCaseChange{\protect\cite{cite5323}}}
\codefieldsection{Alternative Names}
\begin{eczvaluelist}
\item\relax Displaced-squeezed BPSK c-q modulation format
\item\relax Squeezed-state BPSK c-q modulation code
\item\relax Squeezed-state BPSK c-q modulation scheme
\item\relax Squeezed-state BPSK c-q signaling format
\item\relax Two-photon coherent-state BPSK c-q modulation format
\end{eczvaluelist}
\eczhIndexCodeAliasName{squeezed_coherent_bpsk}{Displaced-squeezed BPSK c-q modulation format}
\eczhIndexCodeAliasName{squeezed_coherent_bpsk}{Squeezed-state BPSK c-q modulation code}
\eczhIndexCodeAliasName{squeezed_coherent_bpsk}{Squeezed-state BPSK c-q modulation scheme}
\eczhIndexCodeAliasName{squeezed_coherent_bpsk}{Squeezed-state BPSK c-q signaling format}
\eczhIndexCodeAliasName{squeezed_coherent_bpsk}{Two-photon coherent-state BPSK c-q modulation format}
\codefieldsection{Description}
Bosonic c-q binary modulation format whose codewords are antipodal displaced-squeezed states, i.e., states of the form \(D(\pm\alpha)S(\zeta)|0\rangle\) for a common squeezing parameter \(\zeta\).
The format was originally formulated using the term \textit{two-photon coherent states} (TCS), an early name for squeezed states.

\codefieldsection{Protection}
Optimizing the displacement and squeezing for fixed mean photon number \(N\) changes the coherent-state BPSK overlap exponent from \(4N\) to \(4N(N+1)\) \NoCaseChange{\protect\cite[{Eq. (3.41)}]{cite5323}}.
Thus, in the ideal binary pure-state discrimination expression, the coherent-state BPSK factor \(\exp[-4N]\) is replaced by \(\exp[-4N(N+1)]\).

\codefieldsection{Rate}
The c-q capacity of squeezed-state modulation over a \flmRefsHyperref{ref498}{pure-loss bosonic channel} with coherent homodyne or heterodyne receivers, including closed-form optima for homodyne detection, was studied in Ref. \NoCaseChange{\protect\cite[{Ch. 5}]{cite5300}}.
\codefieldsection{Decoding}
\begin{eczvaluelist}
\item\relax Homodyne detection realizes the field-quadrature measurement advantage of antipodal TCS signals in the ideal model \NoCaseChange{\protect\cite{cite5323}}.
\item\relax Homodyne and heterodyne receivers for squeezed-state modulation are treated in Ref. \NoCaseChange{\protect\cite[{Ch. 5}]{cite5300}}.
\end{eczvaluelist}
\codefieldsection{Parent}
\begin{eczvaluelist}
\item\relax
\flmRefsHyperref[eczindexfamilyrel]{code:bosonic_classical_into_quantum}{Bosonic c-q code}\end{eczvaluelist}
\codefieldsection{Child}
\begin{eczvaluelist}
\item\relax
\flmRefsHyperref[eczindexfamilyrel]{code:quantum_bpsk}{BPSK c-q modulation format} --- Squeezed-coherent BPSK c-q modulation reduces to coherent-state BPSK c-q modulation when the squeezing parameter is zero.
\end{eczvaluelist}
\codefieldsection{Cousin}
\begin{eczvaluelist}
\item\relax
\flmRefsHyperref[eczindexfamilyrel]{code:squeezed_cat}{Squeezed cat code} --- Squeezed-coherent BPSK c-q modulation transmits classical information using displaced-squeezed states, while squeezed cat codes store quantum information in superpositions of squeezed coherent states.
\end{eczvaluelist}
\eczhbkcontributors{ Saikat Guha, \eczhuVVA }
\endeczcode

\part*{Team and Contributors}
\onecolumngrid

\section{Core team}

\begin{eczPersonList}

\eczPrintPerson{Victor V. Albert}{}{VictorVAlbert}{valbert4}{hyddAfkAAAAJ}{https://sites.google.com/site/victorvalbert}{zookeeper}{core}

\eczPrintPerson{Philippe Faist}{}{PhilippeFaist}{phfaist}{rdeNGqMAAAAJ}{https://phfaist.com}{architect}{core}

\end{eczPersonList}

\section{Veterinarians}

\begin{eczPersonList}

\eczPrintPerson{Alexander Barg}{}{AlexanderBarg}{AlexanderBarg}{nvU3IvEAAAAJ}{https://user.eng.umd.edu/~abarg/}{}{veterinarians}

\eczPrintPerson{Daniel Gottesman}{}{DanielGottesman}{}{373HnhYAAAAJ}{https://quics.umd.edu/people/daniel-gottesman}{}{veterinarians}

\eczPrintPerson{Leonid Pryadko}{}{LeonidPryadko}{LeonidPryadko}{S8iAOo4AAAAJ}{https://faculty.ucr.edu/~leonid/}{}{veterinarians}

\end{eczPersonList}

\section{Code contributors}

\noindent Sorted chronologically, by first recorded contribution.

\twocolumngrid

\begin{eczPersonList}

\eczPrintPerson{Michael Vasmer}{}{MichaelVasmer}{MikeVasmer}{YWXgW2wAAAAJ}{}{}{code_contributors}

\eczPrintPerson{Khalil Guy}{}{khalilguy}{}{KTjDjOEAAAAJ}{}{}{code_contributors}

\eczPrintPerson{Sam Gunn}{}{SamGunn}{}{KOUwHVYAAAAJ}{}{}{code_contributors}

\eczPrintPerson{Sarah Meng Li}{}{SarahMengLi}{SarahMLi}{PvIx1RcAAAAJ}{}{}{code_contributors}

\eczPrintPerson{Hassan Shapourian}{}{HassanShapourian}{}{Jv7tPW0AAAAJ}{}{}{code_contributors}

\eczPrintPerson{Eugene Tang}{}{EugeneTang}{}{q9Pxs4oAAAAJ}{}{}{code_contributors}

\eczPrintPerson{Hyunseok Jeong}{}{HyunseokJeong}{}{__73pJ0AAAAJ}{}{}{code_contributors}

\eczPrintPerson{Sivaprasad Omanakuttan}{}{SivaprasadOmanakuttan}{}{4_fuQdUAAAAJ}{}{}{code_contributors}

\eczPrintPerson{Hengyun (Harry) Zhou}{}{HarryZhou}{}{XLHpQy8AAAAJ}{}{}{code_contributors}

\eczPrintPerson{Ivan A. Burenkov}{}{IvanBurenkov}{}{KTjDjOEAAAAJ}{}{}{code_contributors}

\eczPrintPerson{Meng Cheng (程蒙)}{}{MengCheng}{}{3lyNWR0AAAAJ}{}{}{code_contributors}

\eczPrintPerson{Vikram Elijah Amin}{}{VikramAmin}{}{}{}{}{code_contributors}

\eczPrintPerson{Shouzhen (Bailey) Gu}{}{BaileyGu}{}{h3W0RfUAAAAJ}{}{}{code_contributors}

\eczPrintPerson{Bao Bach}{}{Baobach}{bachbao}{M1GL0SEAAAAJ}{}{}{code_contributors}

\eczPrintPerson{Shubham P. Jain}{}{ShubhamPJain}{shubhamj810}{wahii2QAAAAJ}{}{}{code_contributors}

\eczPrintPerson{Yaron Jarach}{}{YaronJarach}{}{}{}{}{code_contributors}

\eczPrintPerson{Jasminder Sidhu}{}{JasSidhu}{}{jFHb8uQAAAAJ}{}{}{code_contributors}

\eczPrintPerson{Nathanan Tantivasadakarn}{}{nathanan}{}{QUj3lAIAAAAJ}{}{}{code_contributors}

\eczPrintPerson{Christophe Vuillot}{}{ChristopheVuillot}{ChristopheVuillot}{G6IqzFIAAAAJ}{}{}{code_contributors}

\eczPrintPerson{Ke Liu (刘科 子竞)}{}{KeLiu}{}{9Xn3T6EAAAAJ}{}{}{code_contributors}

\eczPrintPerson{Renyu Wang}{}{RenyuWang}{}{}{}{}{code_contributors}

\eczPrintPerson{Yale Fan}{}{YaleFan}{}{}{}{}{code_contributors}

\eczPrintPerson{Alexander Cowtan}{}{AlexCowtan}{alexcowtan}{lBTgt3cAAAAJ}{}{}{code_contributors}

\eczPrintPerson{Milan Tenn}{}{M10T}{M10T}{}{}{}{code_contributors}

\eczPrintPerson{Shankar N. Balasubramanian}{}{ShankarBalasubramanian}{}{38YTpfkAAAAJ}{}{}{code_contributors}

\eczPrintPerson{Gage Erwin}{}{GageErwin}{Gerwinlab}{}{}{}{code_contributors}

\eczPrintPerson{Weilei Zeng}{}{WeileiZeng}{WeileiZeng}{WbmLp6QAAAAJ}{}{}{code_contributors}

\eczPrintPerson{Ali Fahimniya}{}{AliFahimniya}{fahimniya}{MwxRkAsAAAAJ}{}{}{code_contributors}

\eczPrintPerson{Daniel Zhang}{}{DanielZhang}{}{UcjzDqQAAAAJ}{}{}{code_contributors}

\eczPrintPerson{Toby S. Cubitt}{}{TobyCubitt}{}{3b-ny5YAAAAJ}{}{}{code_contributors}

\eczPrintPerson{Madhura Pankaja}{}{MadhuraPankaja}{MadhuraPankaja}{}{}{}{code_contributors}

\eczPrintPerson{Aniket Maiti}{}{AniketMaiti}{}{mzOY1a0AAAAJ}{}{}{code_contributors}

\eczPrintPerson{Esha Swaroop}{}{EshaSwaroop}{eshaspark}{j_IVnz0AAAAJ}{}{}{code_contributors}

\eczPrintPerson{Eric Huang}{}{EricHuang}{ehua3765}{HPSt6CsAAAAJ}{}{}{code_contributors}

\eczPrintPerson{Raley Roberts}{}{RaleyRoberts}{raleyr}{}{}{}{code_contributors}

\eczPrintPerson{Yinchen Liu}{}{YinchenLiu}{imbacalvin}{6JJ1BYAAAAJ}{}{}{code_contributors}

\eczPrintPerson{Connor Clayton}{}{ConnorClayton}{cbclayton}{INT-vAYAAAAJ}{}{}{code_contributors}

\eczPrintPerson{Fengxing Zhu}{}{FengxingZhu}{fengxing-zhu}{}{}{}{code_contributors}

\eczPrintPerson{En-Jui Kuo}{}{EnJuiKuo}{ericntunctu}{zLoWNsoAAAAJ}{}{}{code_contributors}

\eczPrintPerson{Adway Patra}{}{AdwayPatra}{adwayp007}{gu9cVgYAAAAJ}{}{}{code_contributors}

\eczPrintPerson{Laurens Lootens}{}{LaurensLootens}{}{AriTElQAAAAJ}{}{}{code_contributors}

\eczPrintPerson{Guanyu Zhu}{}{GuanyuZhu}{}{49-A-XQAAAAJ}{}{}{code_contributors}

\eczPrintPerson{Matthew Steinberg}{}{MatthewSteinberg}{}{lGZwMRAAAAAJ}{}{}{code_contributors}

\eczPrintPerson{Greg Kuperberg}{}{GregKuperberg}{}{OrKdXCgAAAAJ}{}{}{code_contributors}

\eczPrintPerson{Simon Burton}{}{SimonBurton}{punkdit}{}{}{}{code_contributors}

\eczPrintPerson{Markus Grassl}{}{MarkusGrassl}{}{3ufeWtMAAAAJ}{}{}{code_contributors}

\eczPrintPerson{Julio Carlos Magdalena De La Fuente}{}{JulioCarlosMagdalenaDeLaFuente}{}{4NhHYSkAAAAJ}{}{}{code_contributors}

\eczPrintPerson{Earl T. Campbell}{}{EarlCampbell}{}{HVKsi_sAAAAJ}{}{}{code_contributors}

\eczPrintPerson{Narayanan Rengaswamy}{}{NarayananRengaswamy}{nrenga}{qkAERWAAAAAJ}{}{}{code_contributors}

\eczPrintPerson{Adam Wills}{}{AdamWills}{}{hyatxQ4AAAAJ}{}{}{code_contributors}

\eczPrintPerson{Shi Jie Samuel Tan}{}{ShiJieSamuelTan}{}{57rSajgAAAAJ}{}{}{code_contributors}

\eczPrintPerson{Prabha Mandayam}{}{PrabhaMandayam}{}{kHDUY0QAAAAJ}{}{}{code_contributors}

\eczPrintPerson{Aditya Jain}{}{AdityaJain}{}{29yVv_cAAAAJ}{}{}{code_contributors}

\eczPrintPerson{Sourav Dutta}{}{SouravDutta}{}{STv-HycAAAAJ}{}{}{code_contributors}

\eczPrintPerson{David T. Stephen}{}{DavidTStephen}{}{KBuD0aMAAAAJ}{}{}{code_contributors}

\eczPrintPerson{Andrey Boris Khesin}{}{AndreyBorisKhesin}{AndreyBorisKhesin}{PwP7294AAAAJ}{}{}{code_contributors}

\eczPrintPerson{Alexander Schuckert}{}{AlexanderSchuckert}{}{l9EmNDYAAAAJ}{}{}{code_contributors}

\eczPrintPerson{Mark Webster}{}{MarkWebster}{}{k_Kbf_kAAAAJ}{}{}{code_contributors}

\eczPrintPerson{Zongyuan Wang}{}{ZongyuanWang}{}{KQgFLxEAAAAJ}{}{}{code_contributors}

\eczPrintPerson{Saikat Guha}{}{SaikatGuha}{}{E4JS8pQAAAAJ}{}{}{code_contributors}

\eczPrintPerson{Benjamin J. Brown}{}{BenjaminBrown}{}{25TjBooAAAAJ}{}{}{code_contributors}

\eczPrintPerson{Andrew J. Landahl}{}{AndrewJLandahl}{}{}{}{}{code_contributors}

\eczPrintPerson{Jim Harrington}{}{JimHarrington}{}{jrLPKn8AAAAJ}{}{}{code_contributors}

\eczPrintPerson{Yixu Wang}{}{YixuWang}{}{0dNog3kAAAAJ}{}{}{code_contributors}

\eczPrintPerson{Rui Okada}{}{RuiOkada}{}{52liFT8AAAAJ}{}{}{code_contributors}

\eczPrintPerson{Sergiy Borodachov}{}{SergiyBorodachov}{}{}{}{}{code_contributors}

\eczPrintPerson{Karol Życzkowski}{}{KarolZyczkowski}{}{Cfk8KuMAAAAJ}{}{}{code_contributors}

\eczPrintPerson{David Gross}{}{DavidGross}{}{MQuwOP4AAAAJ}{}{}{code_contributors}

\eczPrintPerson{Daniel Burgarth}{}{DanielBurgarth}{}{Rn762kMAAAAJ}{}{}{code_contributors}

\eczPrintPerson{Arda Aydin}{}{ArdaAydin}{}{B1IwuFcAAAAJ}{}{}{code_contributors}

\eczPrintPerson{Michael Liaofan Liu}{}{MichaelLiu}{}{oLt4IHUAAAAJ}{}{}{code_contributors}

\eczPrintPerson{Joschka Roffe}{}{JoschkaRoffe}{quantumgizmos}{vGDLxeoAAAAJ}{}{}{code_contributors}

\eczPrintPerson{Ian Teixeira}{}{IanTeixeira}{iangteixeira}{}{}{}{code_contributors}

\eczPrintPerson{Chris Fechisin}{}{ChrisFechisin}{fechisin}{uQFhtHMAAAAJ}{}{}{code_contributors}

\eczPrintPerson{Nolan Coble}{}{NolanCoble}{nolanjcoble}{}{https://nolanjcoble.com}{}{code_contributors}

\eczPrintPerson{Pavel Panteleev}{}{PavelPanteleev}{}{adjeaUkAAAAJ}{}{}{code_contributors}

\eczPrintPerson{Michael Gullans}{}{MichaelGullans}{mgullans}{YP7rWbkAAAAJ}{}{}{code_contributors}

\eczPrintPerson{Finnegan Voichick}{}{FinneganVoichick}{fvoichick}{}{}{}{code_contributors}

\eczPrintPerson{Elizabeth R. Bennewitz}{}{ElizabethRBennewitz}{ebennewitz}{}{}{}{code_contributors}

\eczPrintPerson{Manasi Shingane}{}{ManasiShingane}{}{}{}{}{code_contributors}

\eczPrintPerson{Jonathan Kunjummen}{}{JonathanKunjummen}{jkunjummen}{S4JlzYsAAAAJ}{}{}{code_contributors}

\eczPrintPerson{Qingfeng (Kee) Wang}{}{QingfengKeeWang}{kee-wang}{}{}{}{code_contributors}

\eczPrintPerson{Xiaozhen Fu}{}{XiaozhenFu}{merrysalad}{odxJdfYAAAAJ}{}{}{code_contributors}

\eczPrintPerson{Srilekha Gandhari}{}{SrilekhaGandhari}{gndhari}{xg7GZVIAAAAJ}{}{}{code_contributors}

\eczPrintPerson{Marianna Podzorova}{}{MariannaPodzorova}{podzy-pow}{}{}{}{code_contributors}

\eczPrintPerson{Nikolas Breuckmann}{}{NikolasBreuckmann}{nikobreu}{en1CbxsAAAAJ}{}{}{code_contributors}

\eczPrintPerson{Seyed Sajjad Nezhadi}{}{SeyedSajjadNezhadi}{sajjad-nezhadi}{}{https://sajjad-nezhadi.github.io}{}{code_contributors}

\eczPrintPerson{Yijia Xu}{}{YijiaXu}{xuyij}{}{https://xuyij.github.io/homepage/}{}{code_contributors}

\eczPrintPerson{Eric Kubischta}{}{EricKubischta}{eric-kub}{}{}{}{code_contributors}

\eczPrintPerson{Jon Nelson}{}{JonNelson}{jnelson1}{}{}{}{code_contributors}

\eczPrintPerson{Xinyuan Zheng}{}{XinyuanZheng}{xzheng16}{}{}{}{code_contributors}

\eczPrintPerson{Benjamin Quiring}{}{BenjaminQuiring}{bquiring}{bq5XpzsAAAAJ}{}{}{code_contributors}

\eczPrintPerson{Dhruv Devulapalli}{}{DhruvDevulapalli}{dhruvdevu}{KnD_CEYAAAAJ}{}{}{code_contributors}

\eczPrintPerson{Joseph T. Iosue}{}{JosephTIosue}{jtiosue}{}{https://jtiosue.github.io/}{}{code_contributors}

\eczPrintPerson{Siddharth Taneja}{}{SiddharthTaneja}{staneja1}{}{}{}{code_contributors}

\eczPrintPerson{Joel Rajakumar}{}{JoelRajakumar}{joeldraj4}{wyC53DgAAAAJ}{}{}{code_contributors}

\eczPrintPerson{Arpit Dua}{}{ArpitDua}{dua-arpit}{w4ASA38AAAAJ}{}{}{code_contributors}

\eczPrintPerson{Alexis Schotte}{}{AlexisSchotte}{AlexisSchotte}{}{}{}{code_contributors}

\eczPrintPerson{David Aasen}{}{DavidAasen}{davidaasen}{vNuuzc8AAAAJ}{}{}{code_contributors}

\eczPrintPerson{Lane G. Gunderman}{}{LaneGGunderman}{}{L8OpvIIAAAAJ}{}{}{code_contributors}

\eczPrintPerson{Noah Berthusen}{}{NoahBerthusen}{noahberthusen}{QjupysQAAAAJ}{}{}{code_contributors}

\eczPrintPerson{Antonio D. Córcoles}{}{AntonioDCorcoles}{}{_nYGfnIAAAAJ}{}{}{code_contributors}

\eczPrintPerson{Shashank Sule}{}{ShashankSule}{ShashankSule}{}{}{}{code_contributors}

\eczPrintPerson{Aleksander Kubica}{}{AleksanderKubica}{}{D8VDo7YAAAAJ}{}{}{code_contributors}

\eczPrintPerson{Basudha Srivastava}{}{BasudhaSrivastava}{}{W61wyBUAAAAJ}{}{}{code_contributors}

\eczPrintPerson{Christopher Chamberland}{}{ChristopherChamberland}{}{Rb9SG3wAAAAJ}{}{}{code_contributors}

\eczPrintPerson{Xiao Xiao}{}{XiaoXiao}{xiaoxiao9689}{}{}{}{code_contributors}

\eczPrintPerson{Yi-Ting (Rick) Tu}{}{YiTingRickTu}{ricktu288}{Dt8I_IYAAAAJ}{}{}{code_contributors}

\eczPrintPerson{Mustafa Doger}{}{MustafaDoger}{mustafadgr}{JmaQvEIAAAAJ}{}{}{code_contributors}

\eczPrintPerson{Tony Lau}{}{TonyLau}{Tonexus}{}{}{}{code_contributors}

\eczPrintPerson{Jack Davis}{}{JackDavis}{JackGDavis}{}{}{}{code_contributors}

\eczPrintPerson{Richard Barney}{}{RichardBarney}{rbarney4}{}{}{}{code_contributors}

\eczPrintPerson{Jiaxin Huang}{}{JiaxinHuang}{jackkyyh}{}{}{}{code_contributors}

\eczPrintPerson{Thomas Wrona}{}{ThomasWrona}{thomaswrona}{}{}{}{code_contributors}

\eczPrintPerson{Muhammad Junaid Aftab}{}{MuhammadJunaidAftab}{MJunaidAftab}{}{}{}{code_contributors}

\eczPrintPerson{Armin Gerami}{}{ArminGerami}{armingerami}{XKrFIXMAAAAJ}{}{}{code_contributors}

\eczPrintPerson{Jingzhen Hu}{}{JingzhenHu}{}{TjPWHhcAAAAJ}{}{}{code_contributors}

\eczPrintPerson{Shuubham Ojha}{}{ShuubhamOjha}{shuubham4}{}{}{}{code_contributors}

\eczPrintPerson{Micah Shaw}{}{MicahShaw}{MicahShaw}{}{}{}{code_contributors}

\eczPrintPerson{Mazin Karjikar}{}{MazinKarjikar}{MazinKarjikar}{}{}{}{code_contributors}

\eczPrintPerson{Kenneth R. Brown}{}{KennethRBrown}{}{EFVX84sAAAAJ}{}{}{code_contributors}

\eczPrintPerson{Alexander Grimm}{}{AlexanderGrimm}{}{_9aGEQ4AAAAJ}{}{}{code_contributors}

\eczPrintPerson{Jonathan Conrad}{}{JonathanConrad}{}{ETjTNJ8AAAAJ}{}{}{code_contributors}

\eczPrintPerson{João Ribeiro}{}{JoaoRibeiro}{}{_v-WdQUAAAAJ}{}{}{code_contributors}

\eczPrintPerson{Anqi Gong}{}{AnqiGong}{gongaa}{gfvhQH0AAAAJ}{}{}{code_contributors}

\eczPrintPerson{Sisi Zhou}{}{SisiZhou}{}{i-m2j4oAAAAJ}{}{}{code_contributors}

\eczPrintPerson{Felix Huber}{}{FelixHuber}{}{1O58f3gAAAAJ}{}{}{code_contributors}

\eczPrintPerson{Liang Jiang}{}{LiangJiang}{}{-4vWibwAAAAJ}{}{}{code_contributors}

\eczPrintPerson{Steven T. Flammia}{}{StevenTFlammia}{}{-VnX0xYAAAAJ}{}{}{code_contributors}

\eczPrintPerson{Grace M. Sommers}{}{GraceSommers}{}{ChUE7jwAAAAJ}{}{}{code_contributors}

\eczPrintPerson{Balint Pato}{}{balopat}{balopat}{mnNvWjkAAAAJ}{https://refactorium.com/}{}{code_contributors}

\eczPrintPerson{Marcus P da Silva}{}{MarcusPS}{marcusps}{qsKgI6AAAAAJ}{}{}{code_contributors}

\eczPrintPerson{Christopher A. Pattison}{}{ChrisPattison}{chrispattison}{4neYf8oAAAAJ}{}{}{code_contributors}

\eczPrintPerson{Cella Kove}{}{CellaKove}{viola1963}{}{}{}{code_contributors}

\eczPrintPerson{Feroz Ahmed Mian}{}{FerozAhmedMian}{Fe-r-oz}{xdGGmiUAAAAJ}{}{}{code_contributors}

\eczPrintPerson{Victory Omole}{}{vtomole}{vtomole}{CmosrmsAAAAJ}{https://vtomole.com/}{}{code_contributors}

\eczPrintPerson{Austin Yubo He}{}{AustinHe}{Austin-He}{MOYd3tAAAAAJ}{}{}{code_contributors}

\eczPrintPerson{Remmy Zen}{}{remmyzen}{remmyzen}{YgCYIrIAAAAJ}{https://remmyzen.github.io/}{}{code_contributors}

\eczPrintPerson{Marc Serra-Peralta}{}{MarcSerraPeralta}{MarcSerraPeralta}{cfFbFYUAAAAJ}{}{}{code_contributors}

\eczPrintPerson{Nir Gutman}{}{NirGutman}{NGBigField}{azmlG_EAAAAJ}{}{}{code_contributors}

\end{eczPersonList}

\part*{Index}

\eczhIndexPrologue{%
  \emph{Page numbers in bold italic underline
  indicate the definition of the code.
  A page number in upright boldface refers to a participation
  in the definition of a notable code list.  
  Page numbers in italic indicate
  reference in a code family relationship.}%
}
\printindex[codes]

\eczWorkAroundSectionSpacingBefore{5ex} 

\eczhIndexPrologue{%
  \emph{Page numbers in bold italic underline indicate definition of the concept.}%
}
\printindex[concepts]

\bibliography{EcZooHandbookEcZooBibRefs.bib}

\end{document}